# The AFLOW Library of Crystallographic Prototypes: Part 3

David Hicks[a,b], Michael J. Mehl[a,b], Marco Esters[a,b], Corey Oses[a,b],
Ohad Levy[a,b,c], Gus L. W. Hart[d], Cormac Toher[a,b], Stefano Curtarolo[a,e]

[a]*Center for Autonomous Materials Design, Duke University, Durham, North Carolina 27708, USA*
[b]*Department of Mechanical Engineering and Materials Science, Duke University, Durham, North Carolina 27708, USA*
[c]*Department of Physics, NRCN, P.O. Box 9001, Beer-Sheva 84190, Israel*
[d]*Department of Physics and Astronomy, Brigham Young University, Provo, Utah 84602, USA*
[e]*Materials Science, Electrical Engineering, and Physics, Duke University, Durham, North Carolina 27708, USA*

---

**Abstract**

The *AFLOW Library of Crystallographic Prototypes* has been extended to include a total of 1,100 common crystal structural prototypes (510 new ones with Part 3), comprising all of the inorganic crystal structures defined in the seven-volume *Strukturbericht* series published in Germany from 1937 through 1943. We cover a history of the *Strukturbericht* designation system, the evolution of the system over time, and the first comprehensive index of inorganic *Strukturbericht* designations ever published.

---

*Email addresses:* `michael.mehl@duke.edu` (Michael J. Mehl), `stefano@duke.edu` (Stefano Curtarolo)

## Table of Contents

[‖] $\gamma$-$Y_2Si_2O_7$ and $AgMnO_4$ ($H0_9$) have similar AFLOW prototype labels (*i.e.*, same symmetry and set of Wyckoff positions with different stoichiometry labels due to alphabetic ordering of atomic species). They are generated by the same symmetry operations with different sets of parameters.

[□] $\beta$-$B_2H_6$ and $B_2H_6$ ($P2_1/c$) have the same AFLOW prototype label. They are generated by the same symmetry operations with different sets of parameters.

[‡] Ammonium Persulfate [$(NH_4)_2S_2O_8$, $K4_1$] and Monasite ($LaPO_4$) have the same AFLOW prototype label. They are generated by the same symmetry operations with different sets of parameters.

[°] $KNO_2$ III and Manganite ($\gamma$-MnO(OH), $E0_6$) have the same AFLOW prototype label. They are generated by the same symmetry operations with different sets of parameters.

[×] $\alpha$-ICl and LiAs have the same AFLOW prototype label. They are generated by the same symmetry operations with different sets of parameters.

[*] Pararealgar (AsS) and Realgar (AsS, $B_l$) have the same AFLOW prototype label. They are generated by the same symmetry operations with different sets of parameters.

** $Ag_2O_3$ and $Zr_2Al_3$ have similar AFLOW prototype labels (*i.e.*, same symmetry and set of Wyckoff positions with different stoichiometry labels due to alphabetic ordering of atomic species). They are generated by the same symmetry operations with different sets of parameters.

⊗ $\zeta$-$Fe_2N$ and $\alpha$-$PbO_2$ have the same AFLOW prototype label. They are generated by the same symmetry operations with different sets of parameters.

[†] $\alpha$-Potassium Nitrate ($KNO_3$) I, $NH_4NO_3$ III ($G0_{10}$), and Aragonite ($CaCO_3$, $G0_2$) have the same AFLOW prototype label. They are generated by the same symmetry operations with different sets of parameters.

[††] Pseudobrookite ($Fe_2TiO_5$, $E4_1$) and $Ta_2NiS_5$ have similar AFLOW prototype labels (*i.e.*, same symmetry and set of Wyckoff positions with different stoichiometry labels due to alphabetic ordering of atomic species). They are generated by the same symmetry operations with different sets of parameters.

[¶] $TlCo_2S_2$ and $ThCr_2Si_2$ have the same AFLOW prototype label. They are generated by the same symmetry operations with different sets of parameters.

[§] $Li_2CN_2$ and $H5_9$ [Autunite, $Ca(UO_2)_2(PO_4)_2{\cdot}10\frac{1}{2}H_2O$] (*obsolete*) have the same AFLOW prototype label. They are generated by the same symmetry operations with different sets of parameters.

[∂] $KBrO_3$ ($G0_7$) and $\gamma$-Potassium Nitrate ($KNO_3$) have the same AFLOW prototype label. They are generated by the same symmetry operations with different sets of parameters.

[‡‡] $I1_3$ ($SrCl_2{\cdot}(H_2O)_6$) (*obsolete*) and Rosiaite ($PbSb_2O_6$) have similar AFLOW prototype labels (*i.e.*, same symmetry and set of Wyckoff positions with different stoichiometry labels due to alphabetic ordering of atomic species). They are generated by the same symmetry operations with different sets of parameters.

[§§] $Ce_2O_2S$ and Brucite [$Mg(OH)_2$] have similar AFLOW prototype labels (*i.e.*, same symmetry and set of Wyckoff positions with different stoichiometry labels due to alphabetic ordering of atomic species). They are generated by the same symmetry operations with different sets of parameters.

[××] $K_2Pt(SCN)_6$ ($H6_3$) and $K_2GeF_6$ ($J1_{13}$) have similar AFLOW prototype labels (*i.e.*, same symmetry and set of Wyckoff positions with different stoichiometry labels due to alphabetic ordering of atomic species). They are generated by the same symmetry operations with different sets of parameters.

°°Bararite (Trigonal $(NH_4)_2SiF_6$, $J1_6$) and $K_2GeF_6$ ($J1_{13}$) have similar AFLOW prototype labels (*i.e.*, same symmetry and set of Wyckoff positions with different stoichiometry labels due to alphabetic ordering of atomic species). They are generated by the same symmetry operations with different sets of parameters.

∂∂$MnBi_2Te_4$ and $CaCu_4P_2$ have similar AFLOW prototype labels (*i.e.*, same symmetry and set of Wyckoff positions with different stoichiometry labels due to alphabetic ordering of atomic species). They are generated by the same symmetry operations with different sets of parameters.

⊗⊗$LiZn_2$ ($C_k$) and $Fe_2N$ (approximate, $L'3_0$) have similar AFLOW prototype labels (*i.e.*, same symmetry and set of Wyckoff positions with different stoichiometry labels due to alphabetic ordering of atomic species). They are generated by the same symmetry operations with different sets of parameters.

□□$Zn(BrO_3)_2{\cdot}6H_2O$ ($J1_{10}$) and $H6_4$ [$Ni(NO_3)_2(NH_3)_6$] (*obsolete*) have similar AFLOW prototype labels (*i.e.*, same symmetry and set of Wyckoff positions with different stoichiometry labels due to alphabetic ordering of atomic species). They are generated by the same symmetry operations with different sets of parameters.

**Index**

## 1. Introduction

Crystal structure classification began near the dawn of X-ray Crystallography with early works by Wyckoff [1], Ewald's 1927 *Handbuch der Physik* article [2], and the Landolt-Börnstein 1927 supplement [3]. In the next few decades, these catalogs proliferated, including the *Strukturbericht* (Structure Reports) series [4], Pearson's *Handbook of Lattice Spacings and Structures of Metals and Alloys* [5, 6], Smithells's *Metals Reference Book* [7, 8], and Wyckoff's updated *Structure of Crystals* [9]. In later years, *Pearson's Handbook* [10] and *Smithells Metals Reference Book* [11] acquired new editors and even today produce periodic editions. Subsequent works, such as the *Gmelin Handbook* [12], also appear regularly. As online access became the norm at the beginning of this century, electronic resources appeared, including the *American Mineralogist Crystal Structure Database* [13], which compiles the structures of minerals from a variety of sources, and the SpringerMaterials website [14], an extension of *Pearson's Handbook* that also incorporates the Landolt-Börnstein series.

Computational material-property databases have also become available: for example AFLOW (Automatic FLOW for Materials Discovery) [15, 16, 17], Novel Materials Discovery (NoMaD) [18], the Materials Project [19], and the Open Quantum Materials Database (OQMD) [20].

While these print and electronic databases contain a vast number of crystal structures, many are not optimized for use in materials computations, *e.g.*, by providing structural information in a format which can be read by one of the common density functional computer codes. In addition, while most of these resources make their data available in standard formats, such as space group/Wyckoff positions or a Crystallographic Information File (CIF) [21], these forms still pose a challenge for novice researchers. Therefore, it is desirable to have a database that addresses both of these concerns.

The AFLOW Library of Crystallographic Prototypes (hereafter AFLOW Prototype Encyclopedia or Prototype Encyclopedia for brevity) [22, 23] has been designed with these objectives in mind. Integrated with the AFLOW computational framework, it can provide structural input for a wide variety of materials computational programs, including VASP [24], QUANTUM ESPRESSO [25], FHI-AIMS [26], ABINIT [27], and ELK [28] as well as in CIF format. Each structure's web page provides all the necessary crystallographic information, including space group, lattice vectors, and Wyckoff positions. In addition, the formulas for the primitive vectors and basis vectors of the structure are shown. This is particularly useful for students and novice researchers, as it shows the relationship between the standard crystallographic coordinates needed for computational work and the Cartesian positions of the atom in the primitive and/or conventional unit cell. If the desired chemical com-position for the structure is not the prototype, the chemistry and primitive cell dimensions can be changed by the user before generating the structural input. Finally, the structure may be viewed as a primitive cell, conventional cell, or supercell at a variety of angles and magnifications via JSmol [29].

This article is about the ongoing development of the AFLOW Prototype Encyclopedia. In Part 1 [22], we introduced the Prototype Encyclopedia, which began as an updated version of the now-defunct Naval Research Laboratory "*Structure of Crystals*" web site. This was a collection of crystal structures that we found useful as a starting point for computational studies of materials. One feature of the site was that it included a number of well-known *Strukturbericht* designations, from the German effort to classify common crystal structures [4]. This article also provided a brief introduction of the crystallography needed to understand crystal structures, as well as a description of the entries in the AFLOW Prototype Encyclopedia for each structure. In all, 288 structures were included in the initial Prototype Encyclopedia.

Part 2 [23] continued the discussion of crystallography by reviewing the enantiomorphic space groups, which are pairs of space groups that are mirror images of each other. As part of our teaching mission, we also described the Wigner-Seitz version of the primitive unit cell of a crystal, and discussed the 17 two-dimensional "plane groups," the analog of the 230 three-dimensional space groups. Another 302 structures were added to the database to bring the total to 590, with at least one structure shown in each of the 230 space groups.

In this article, Part 3, we add 510 new structures to the encyclopedia. These include all of the inorganic *Strukturbericht* designations which had not previously been cataloged by us. The *Strukturbericht* classification scheme was one of the earliest attempts to group elements[1] and compounds into common crystal structures. However, despite its historical importance there is no available history of its development, and until now there has been no comprehensive index of all the *Strukturbericht* designations that were defined in the series or added by others.

Recently, M. J. Mehl published a brief history of the development of the *Strukturbericht* designations [31]. In Part 3, we improve and extend that work, beginning with the introduction of these labels by Ewald [2] through the publication of the seven *Strukturbericht* volumes and the post-World War II extensions to the scheme made by Smithells [8], Pearson [5], and others. The *Strukturbericht* classifications changed over the run of the series, and several structures were published under multiple designations. Here, we track all of those changes.

The article is organized as follows: in Sections 2 and 3, we present common prototype classification schemes, followed by a short history of the development and presentation of *Strukturbericht* designations, respectively. As the *Strukturbericht* designations changed over time, Section 4 discusses the history of the major categories and subcategories, as well as our editorial decisions as to which label should be used for a given prototype. Section 5 presents what we believe to be the first comprehensive list of *Strukturbericht* designations for elements and inorganic compounds, including all references to labels which we could find in the literature, with a link to the corresponding online pages in the AFLOW Prototype Encyclopedia. Section 6 describes AFLOW software modules that will be used to identify and classify new prototypes for future iterations of this work. Additionally, Section 7 introduces an online form for users to suggest new prototypes to enrich the Encyclopedia. Section 8 provides information about AFLOW Schools: hands-on workshops for students and reseachers to learn about structural analysis methods — especially those discussed in Parts 1, 2, and 3 of this work — as well as other topics related to computational materials science. Section 9 summarizes our conclusions. This is followed by a list of all 510 new prototypes we have included in the Prototype Encyclopedia since the publication of Part 2.

## 2. Crystallographic Classification Schemes

Just before the First World War, Max von Laue suggested that diffraction of X-rays through crystals would produce a pattern which could be used to determine the crystal structure [32]. Shortly thereafter, this technique was also applied by Bragg and Bragg [33] and Hull [34] to analyze the structures of diamond and graphite, respectively. Over the next decade, the structures of hundreds of elements and compounds were determined [2, 3, 35, 36].

It was soon obvious that nature is in the habit of repeating itself. For example, the metallic elements mostly take close-packed face-centered-cubic (fcc) or hexagonal (hcp) structures, or the nearly close-packed body-centered cubic (bcc) structure. (The major exceptions being manganese, which takes on two unique atomic arrangements, and the actinides). Of the remaining elements, the early group 14 all take on the diamond structure. We only see other patterns when we go to the far right-hand side of the periodic table.

Similar trends occur in compounds. Many $AB$ type binaries have ground states in the sodium chloride, cesium chloride or zincblende structures, $A_2B$ compounds can frequently have the positions found in fluorite ($CaF_2$), and metallic compounds of the form $A_2BC$ often appear in the Heusler structure. This continues into more complicated chemistries. Data-mining sources such as AFLOW-ICSD show that a large number of compounds exist in a relatively small number of crystal structures [37].

[1] Of interest is the work of Davey [30], who found that tungsten was too opaque for standard X-ray diffraction techniques, and so diluted his sample with ten parts wheat flour in order to get a diffraction pattern.

This duplication of structures leads to the assumption that most elements and compounds exist in a relatively small number of crystallographic forms, and that compounds having similar structures are somehow related to one another. If true, then it should be possible to classify crystal structures according to their periodicity and arrangement of atoms in the unit cell. We now show four common classification schemes.

The simplest crystal structure classification is by their space group and the occupied Wyckoff positions within that space group [38]. This is certainly efficient, and we have used a variant of it to index crystal structures [22, 23]. Unfortunately, this description is neither particularly informative nor unique. The ground states of gallium, iodine, and phosphorous are all in space group *Cmca* (#64), and all the atoms occupy a single (8*f*) Wyckoff position [39], but their structures are all different: the iodine atoms form dimers, while gallium and phosphorous have very different conventional unit cells.

A second way of defining classes of crystal structures is to group all similar elements or compounds into one class, and refer to it by one "prototype". Ionic compounds such as NaCl, MgO, BaS, and many others all take the same structure: fcc space group $Fm\overline{3}m$ (#225), with one atom on the (4*a*) Wyckoff site and the other on the (4*b*) site. This is universally known as the "sodium chloride" or "rock salt" structure [35]. This approach works well if there is an agreement on a single prototype for each structure, but can lead to confusion if different researchers use different prototypes: for example, $\epsilon$-$FeN_2$ and $\beta$-$VN_2$ are both claimed as prototypes for the same structure [10, 40], while the mineral rosiaite ($PbSb_2O_6$), with similar lattice parameters and atomic coordinates [41], is considered the prototype of the ternary form.

A third method was introduced by Pearson [6]: specify the lattice type (cubic, hexagonal, tetragonal, etc.), followed by the number of atoms in the conventional unit cell. These *Pearson Symbols* are a compact version of the first method, and suffer from the same problems.

A fourth method is derived from the second: assign a label to each structure class and refer to it by that label. These came to be called *Strukturbericht* designations, for reasons discussed below. This scheme can prove unwieldy, as there are a potentially infinite number of possible structures, making it impossible for any labeling scheme to keep up, especially if there is no systematic method for choosing the labels. It does have the advantage that there is a unique label for every structure class, independent of chemical composition.

In practice, one identifies the structure using several of these methods. Of the four, the *Strukturbericht* scheme is the most obscure, and so we now turn our attention to it.

## 3. A Brief History of *Strukturbericht* Symbols [2]

This *Strukturbericht* scheme was introduced by Ewald in his review of X-ray crystallography for the 1927 edition of the *Handbuch der Physik* [2]. The 1927 supplement to the 1923 edition of Landolt-Börnstein [3] also contains a shortened version of Ewald's list, presumably taken from the *Handbuch*, but we have not been able to determine which one was actually published first.

The structural cataloging method devised by Ewald was rather simple:

- Purely elemental structures were given designations beginning with the letter *A*, followed by a number. Thus, the face-centered cubic lattice, prototype copper, was given the designation *A*1, body-centered cubic (tungsten) *A*2, hexagonal close-packed (magnesium) *A*3, diamond *A*4, and so on. Each new structure was labeled with the next available number, making the scheme rather arbitrary, as there may be no relation between structure $An$ and $An + 1$.

- Binary compounds with composition *AB* were given designations starting with *B*: *B*1 (NaCl), *B*2 (CsCl), *B*3 (zincblende ZnS), *B*4 (wurtzite ZnS), *etc.* Thus, we can easily distinguish between the two forms of zinc sulfide.

- Binary compounds with composition $AB_2$ had designations starting with *C*, *e.g.*, fluorite ($CaF_2$) is *C*1.

- Binary compounds $A_mB_n$ start with *D*. The original *D*1 structure was corundum ($Al_2O_3$).

- Ternary and higher compounds that do not contain radicals such as $SiO_4$ or OH have labels beginning with *E*. The only example was *E*1 ($KMgF_3$).

- The letter *F* was used for compounds with linear radicals, such as *F*1 [K(CN)].

- "Flat" (two dimensional) radicals start with *G*: *G*1, calcite ($CaCO_3$).

- "Complex" (three dimensional) radicals were assigned to *H*: *H*1, anhydrite ($CaSO_4$).

The *Strukturbericht* scheme was refined and extended by various literature reviews. The earliest crystallographic critical review that we have found was published by Wyckoff as part of his seminal 1924 work, *The Structure of Crystals* [1], followed by the 1931 second edition [35], the 1934 supplement edition [42], and culminating in the six-volume second edition of *The Structure of Crystals* from 1963 through 1971 [9].

[2] A somewhat different version of this section appeared in Ref. [31].

In 1931, Ewald and C. Hermann [36] published an exhaustive review of the structural literature from 1913-1928, "*Strukturbericht* (Structure Reports) 1913-1928" (we will refer to this as SB-I) as a supplement to the journal *Zeitschrift für Kristallographie*. In their introduction, they say that the purpose was (translated from German)

> ... to provide a compilation and critical review of *all* [our emphasis] previously published articles on X-ray structure determination. The structures are arranged according to their chemical formulas, at the same time a summary by "types" provides an overview over the known structures and their representatives. Unlike other published structural compositions, the present one stands out by discussing individual works, not just the results of the structures of the crystal types. It is expected that this will improve the accuracy of the structural determinations, the exact nature of the material used, the corrections to be made to older reports due to new experiences, and that other similar points can be assessed better.

The given dates show that this was intended to cover the field of structural studies from the beginning. This volume was primarily a critical review of the current X-ray crystallographic literature, but it also grouped crystal structures into types using the classification scheme defined in the *Handbuch*, leading to the structural prototypes becoming known as *Strukturbericht* designations.

The second (SB-II) and third (SB-III) volumes of *Strukturbericht* were published in 1937, encompassing the years 1928-1932 and 1933-1935 respectively. Each was designed to cover all published crystal structures, or at least all structures which the editors could obtain, for the indicated timeframe. Later volumes (SB-IV through SB-VII) span one year of research each, with the seventh and final volume, including 1939, published in 1943 during the Second World War. At the end of the war, the International Union of Crystallography (IUCr) continued this project, publishing it as *Structure Reports*, in English, through 1990. The first IUCr volume contained research published in 1945-1946 and was designated as volume 10 [43]. Volumes 8 and 9, covering 1940-1941 [44] and 1942-1944, respectively, were published later, providing continuity with the original *Strukturbericht*.

Volumes SB-I-VII modified and extended Ewald's classification scheme, but after the war, it was realized that the number of *Strukturbericht* symbols would keep growing uncontrollably and so the practice of giving new structures symbols was discontinued by the IUCr. A few new symbols were introduced by C. J. Smithells in his 1955 *Metals Reference Book* [8], and by W. B. Pearson in his 1958 *Handbook of Lattice Spacings and Structures of Metals and Alloys* [5]. Several other symbols have been added by other authors, noted in Section 5.

Even though the generation of new symbols ceased, the ones found in the *Strukturbericht* volumes, Smithells, and Pearson are still in use. For example, compounds with the $A15$ structure ($Cr_3Si$ or $\beta$-W) became known as the first "high temperature" (20 K) superconductors [45]; and there are frequent references to "$B1$" (NaCl) or "$B2$" (CsCl) structures. Despite this, there is no complete index of all the *Strukturbericht* symbols: an index was published as part of the table of contents for SB-II, but this was not continued into the later volumes. In 1976, Trotter and Bree [46] published a complete index of all the compounds mentioned in all seven volumes, but it was by formula and by author, with no reference to the *Strukturbericht* designations. Smithells, Pearson, and other works such as Villars [10], and Parthé *et al.* [12] have published partial lists, but these are weighted toward the intermetallics and mostly ignore silicates (*Strukturbericht S*) and other non-metallic categories.

While we do not intend to write a complete history of *Strukturbericht* designations, it is worthwhile to give a brief description of the changes in the system that occurred throughout its run. The remainder of this section summarizes the changes to the system made in each volume. Following that, Section 4 delves more deeply into the changes made in each *Strukturbericht* category. Finally, we provide the first complete list of inorganic *Strukturbericht* designations in Section 5. As part of this effort we added all of the the inorganic *Strukturbericht* designations as entries in the AFLOW Prototype Encyclopedia, leaving out the very few where the definition was so vague that we could not determine a structure.

### 3.1. *Strukturbericht* Band (Volume) I 1913-1928

SB-I [36] classified structures using a modified version of the scheme presented in Ewald's *Handbuch* article. This included the introduction of three new *Strukturbericht* categories:

- Type *L*, "alloys" (*Legierungen*): These are ordered structures based on the fcc and bcc lattices, such as $Cu_3Au$ (*Strukturbericht* $L1_2$), which can be considered as an fcc lattice with an ordered substitution of gold atoms for copper, and the Heusler prototype $AlCu_2Mn$ ($L2_1$), a decoration of the bcc lattice.
- Type *M*, "solid solutions" (*Mischkristalle*). Presumably this was to be used for solid solutions, but no *M*-type structures were ever defined.
- Type *O*, "organic:" These are categories of organic crystals cataloged in a manner similar to the inorganic categories *A*-*L*. Type *O* structures were defined in every volume of *Strukturbericht*, but the labels never became common and we will not discuss them in this article.

In addition to the new categories, SB-I subdivided Ewald's original categories. For example, while type *D* still represented binary compounds of the form $A_mB_n$, the type was divided into ranges, with $D1 - D10$ reserved for compounds $AB_3$, $D11 - D20$ for $AB_n$ with $n > 3$, and finally, $D50 - D100$ for the more general compounds. Thus, the original $D1$, $Al_2O_3$, now became $D51$, while $D1$ was used for ammonia, $NH_3$. Similar divisions were made in the $F$, $G$, $H$ and $L$ categories.

The volume itself was divided by *Strukturbericht* category. Each section began with a description of the structures associated with each *Strukturbericht* designation and ended with a review of the literature for these and similar structures. These reviews would be present in all of the *Strukturbericht* volumes.

### 3.2. *Strukturbericht* Band II, 1928-1932

SB-II [47] was published in 1937 and covered the years 1928 through 1932. By the, Ewald, protesting the firing of all Jewish professors at the Technische Hochschule Stuttgart, resigned his post and emigrated to the United Kingdom [48], so the volume was edited by Hermann, Otto Lohrmann, and Hans Philipp.

The new publication included a table of contents with the only index of *Strukturbericht* symbols ever published in the original volumes. Scheme changes developed in the *Handbuch* and SB-I were made. The most notable was in the format of the labels for classes $D$ and above, due to the large numbers of new structures being found. This caused a problem in the $D$ category, where designations $D1$ through $D10$ were reserved for structures with stoichiometry $AB_3$. By 1932, thirteen unique structures of this type were known, exceeding the category limit. To remedy this, the editors changed the SB-I $D1 - D100$ letter-number format to letter-number-subscript, where now the letter-number combination reveals the stoichiometry of the crystal and the subscript the specific structure. Thus, ammonia ($NH_3$) formerly $D1$, became $D0_1$; skutterudite, formerly $D2$, became $D0_2$. (Eventually the $D0$ range would extend to $D0_{24}$, $Ni_3Ti$). This practice was applied to all of the categories beyond $D$.

SB-II also introduced the first fourteen entries in the rather general $E$ category, including $E0_1$, PbFCl. Other $E$ entries included chalcopyrite ($CuFeS_2$, $F6_1$), cubic perovskite ($CaTiO_3$, $G0_5$), and ilmenite ($FeTiO_3$, $G0_4$); were reclassified to $E1_1$, $E2_1$, and $E2_2$, respectively, presumably because they were insufficiently radical.

Other new categories were added:

- Type $I$, compounds with $BC_6$ radicals. These were mostly renamed from type $H$, where in SB-I $H50$ and above had been reserved for large radicals. Now, *e.g.*, $K_2PtCl_6$, originally $H6_1$, was reclassified as $I1_1$, though the SB-II table of contents listed it under both designations. Compounds with radicals of the form $BC_m, m < 6$, remained in the $H$ category, although many $SiO_5$ silicates were co-listed in the new $S$ category.
- Type $K$, "complex radicals," including $K0_1$ ($K_2S_2O_5$), $K1_1$ ($K_2S_2O_6$), and $K1_2$ ($Cs_2S_2O_6$). At least that is how they were listed in the text. In the SB-II table of contents, they were listed as $K1_1$, $K2_1$, and $K2_2$, respectively. (We use the names in the text rather than the names in the table of contents below.)
- Type $S$, the silicates, which were now moved to a new category. The silicates defined in SB-I had mostly been given $H$ designations, but now were also listed under $S$, at least in the index of SB-II. Thus, kyanite ($Al_2SiO_5$, originally labeled $H5_1$ or $H51$, acquired the new label $S0_1$. In all, thirty-two compounds were identified, all but six of these new to SB-II.

### 3.3. *Strukturbericht* Band III, 1933-1935

SB-III was the last multi-year volume of the series, published in 1937 [49]. Edited by C. Gottfried and F. Schlossberger, it covered papers from 1933 through 1935.

This volume did not include any new *Strukturbericht* categories, but, perhaps to avoid confusion between the letters in all Latin alphabets, category $I$ was renamed as category $J$. Since some of the previously defined $I$ compounds were never mentioned in this or later volumes, we will attempt to clarify some confusion in Sections 4 and 5.

### 3.4. The War Years: *Strukturbericht* Band IV-VII

Each remaining volumes covered a single year and did not introduce any new categories. Remarkably, most were published while World War II was in progress, despite the fact that many German scientists, following Ewald, had fled the country, or like Hermann, had been imprisoned by the Nazi regime [50]. In spite of the War, these later volumes were distributed throughout the world. In 1943 in the United States, Edwards Brothers (Ann Arbor) published all seven volumes "in the Public Interest by Authority of the Alien Property Custodian," an office created by Presidential order to manage enemy alien property [51].

- SB-IV (covering 1936) [52], edited by Gottfried, was published in 1938.
- SB-V (1937) [53], again edited by Gottfried, was published in 1940.
- SB-VI (1938) [54], edited by K. Herrmann (not to be confused with C. Hermann), was published in 1941.
- SB-VII (1939) [55] — the last volume — was again edited by Herrmann and published in 1943, while the Eastern Front with the Soviet Union was collapsing, the Allies were invading Italy, and preparations for D-Day were underway. SB-VII begins by noting that scandium is face-centered cubic and so is an $A1$ structure, and ends with an entry for cronstedtite, $Fe^{II}_4Fe^{III}_2[(OH)_8|Fe^{III}_2Si_2O_{10}]$, assigning it the designation $S5_7$. This is the final inorganic *Strukturbericht* designation given in the original series.

### 3.5. Post-War: Continuation and Extension of *Strukturbericht*

The classification of compounds into *Strukturbericht* designations discontinued in the IUCr's post-war *Structure Reports*, but two authors created new designations as part of their own structural compilations. In 1949, Colin J. Smithells published his first *Metals Reference Book* [7], which included a partial index of *Strukturbericht* designations and the crystal structure associated with each. This list concentrated on metallic systems, and so left out some elements and compounds. This was not just a list of previously defined types, however. Smithells listed two new compounds: $ThSi_2$, and $TeCu_2$, provisionally giving them the *Strukturbericht* designation $C_x$. In the second edition of *Metals Reference* [8], Smithells extended this, adding many new *Strukturbericht* designations. While still following the letter-number-subscript scheme started in *Strukturbericht* II, the subscripts for the new entries were alphabetic rather than numeric. Thus, $ThSi_2$ became *Strukturbericht* type $C_c$ and $TeCu_2$ became $C_g$. Reflecting the advances in physics in the war and post-war era, the list included the unique structures of the actinides: $A_a$ (Pa), $A_b$ ($\beta$-U), $A_c$ ($\alpha$-Np), $A_d$ ($\beta$-Np), etc. Altogether Smithells added seventy-five new structures to the list.

In 1958, Pearson published the first edition of his influential *Handbook of Lattice Spacings and Structures of Metals and Alloys* [5]. It included most of Smithells' *Strukturbericht* index and added ten new structures, using the letter-number-alphabetic subscript format, with subscripts chosen to avoid overlap with Smithells. Pearson soon realized that further extensions would become unwieldy. In the next edition of his Handbook [6], he dropped the *Strukturbericht* index and listed structures by their Pearson symbols, which are not at all arbitrary and have the advantage of making the crystal type and number of atoms readily apparent, at the cost of having numerous structures of wildly different types using the same symbol. He still, however, give the *Strukturbericht* designation for systems he was describing, a practice that continues in the modern editions of *Pearson's Handbook.*

### 3.6. Modern indices of compounds

Despite the demise of the *Strukturbericht* labeling system, it is still useful to have lists which classify crystal structures. Several authors and groups have done this over the years:

- Trotter and Bree [46] published an index to the original seven volumes of *Strukturbericht*, including both organic and inorganic structures. Regrettably, it did not include an index by *Strukturbericht* designation.

- Modern versions of *Smithells Metals Reference Book* [11] and *Pearson's Handbook* [10] include indices of metallic structures by *Strukturbericht* designation, Pearson symbol, and space group.

- The *Gmelin Handbook of Inorganic and Organometallic Chemistry* [12] also includes a list of *Strukturbericht* designations, as well as the Pearson symbol, space group, and occupied Wyckoff positions for each structure. Unlike the Pearson and Smithells Handbooks, this list includes many non-metallic systems. The list also notes when two *Strukturbericht* designations actually refer to the same structure. This happened several times in the original works, undoubtedly due to the frequent changes of editors and the lack of computers to keep accurate databases.

- The modern versions of Smithells, Pearson and Gmelin also include other structures not given a *Strukturbericht* designation.

Electronic databases are also available:

- *The American Mineralogist Crystal Structure Database* [13] gives structural information for a large number of minerals (including mineral names), but does not include the *Strukturbericht* designation. It does list the source publication, and when possible links to its PDF version. The site includes all structures published in *American Mineralogist*, *The Canadian Mineralogist*, *European Journal of Mineralogy*, and *Physics and Chemistry of Minerals*, as well as other sources.

- *Springer Materials* [14] maintains a constantly updated list of crystal structures. Edited by P. Villars, this is essentially an electronic form of *Pearson's Handbook*, but it includes non-metallic systems as well, unfortunately neglecting *Strukturbericht* designations.

None of these lists and databases give a comprehensive list of *Strukturbericht* symbols. In 1995, R. Benjamin Young, a summer intern at the U.S. Naval Research Laboratory in Washington D.C., and M. J. Mehl created an online database of *Strukturbericht* and related structures, in what was called the *Crystal Lattice Structures* web site, currently offline. In 2016, Mehl joined with the Curtarolo Research Group at Duke University to create the AFLOW Prototype Encyclopedia as an updated version of of the original web-site.

The original database contained 288 structures, including many, but not nearly all, of the inorganic *Strukturbericht* designations. It was later expanded [23] to 590 structures, and with this publication increases the count to well over 1,000, including all of the inorganic *Strukturbericht* designations from the original volumes, Smithells, Pearson, and other sources (see Section 5). We will continue to add new structures (Section 6), and invite readers to suggest structures (Section 7).

AFLOW Prototype Encyclopedia lists structures by *Strukturbericht* (if any), Pearson symbol, space group and prototype material. Each entry includes:

- The *Strukturbericht* designation of the structure, if one has been given,
- the Pearson symbol
- the space group,
- occupied Wyckoff positions,
- primitive vectors for the lattice,
- basis vectors for each atom in the primitive unit cell,
- a rotatable view of the primitive or conventional unit cell, which can be expanded to show multiple unit cells,
- a standardized Crystallographic Information File (CIF), and
- a structure file which can be used as input for a variety of electronic structure codes.

## 4. *Strukturbericht* categories and subcategories.

The original *Strukturbericht* scheme developed by Ewald was modified and extended over the years, and until now there has been no comprehensive list of *Strukturbericht* categories or designations. This section and the next correct this omission. Here we describe all the primary ($A$, $B$, $C$, ...) *Strukturbericht* categories, while the next section will list every *Strukturbericht* designation.

Primary categories from $D$ through $S$ are broken into smaller secondary categories, *e.g.*, $D0$, $E1$, $F2$, which we will describe as well. As the various primary and secondary categories changed over time, we offer a brief description of the changes in each category introduced over the course of the *Strukturbericht* publication run.

***A*:** Elements – this is essentially unchanged from Ewald's original *Handbuch* article.

***B*:** Binary compounds with stoichiometry $AB$ – again unchanged from the original list.

***C*:** Binary compounds with stoichiometry $AB_2$ – also unchanged.

***D*:** Binary compounds with stoichiometry $A_mB_n$, beyond $AB$ and $AB_2$.

In the *Handbuch*, this was just a list of compounds, in no particular order. SB-I changed this by introducing ranges corresponding to a given stoichiometry, so that compounds with the same $m$ and $n$ were in the same secondary category. Originally, these categories held at most ten compounds, *e.g.*, $D1 - D10$ for $AB_3$. In SB-II this was changed to the letter-numeral-subscript format described in the previous section:

***D*0:** (originally $D1 - D10$) Binary compounds with stoichiometry $AB_3$.

***D*1:** ($D11-D20$) Binary compounds with stoichiometry $AB_4$. In SB-I, it was defined for all $AB_n$ structure with $n > 3$, but all the examples given were $AB_4$.

***D*2:** This should have been the range $D21 - D30$, but this was not listed in SB-I. Starting with SB-II, it was used for binary compounds of stoichiometry $AB_n$, $n > 4$.

***D*3:** Binary compounds with stoichiometry $(AB)_2$.

***D*4:** Binary compounds with stoichiometry $(AB_n)_2$, $n > 1$.

Obviously, these last two categories can also be indexed in $B$, $C$ or $D0 - D2$ by reducing the stoichiometries. The stoichiometry for $D3$ ($A_2B_2$) reduces to that of $B$ ($AB$). Similarly, the stoichiometry for $D4$ can be reduced ($A_2B_{2n}$ to $AB_n$), which is equivalent to $C$ when $n = 2$, $D0$ when $n = 3$, $D1$ when $n = 4$, and $D2$ when $n > 4$. Therefore, only one compound was defined in $D3$ and $D4$.

The remaining secondary categories in $D$ were listed together in SB-I as $D50 - D100$, and only implicitly divided into other categories in later volumes. This is our best interpretation of the method:

***D*5:** Binary compounds with stoichiometry $A_2B_3$.

***D*6:** Binary compounds with stoichiometry $A_2B_4$ – this is violated in SB-I by the first entry $D61$, but it seems to be latter editors' intent to define it in this way.

***D*7:** Binary compounds with stoichiometry $A_3B_4$.

***D*8:** Binary compounds with stoichiometry $A_mB_n$ that do not fit into previous categories.

***D*10:** This was only defined post-war by Pearson, who used it for compounds with stoichiometry $A_3B_7$.

***E*:** Ternary and more complex compounds.

In the original *Handbuch* article, the $E$ label was to be used for general multi-atom compounds that contained radicals. No $E$ structures were introduced in SB-I, but subsequent volumes defined several secondary types, all involving ternary compounds:

***E*0:** Ternary compounds with stoichiometry $ABC$.

***E*1:** Ternary compounds with stoichiometry $ABC_2$.

***E*2:** Ternary compounds with stoichiometry $ABC_n$, $n > 2$.

***E*3:** Ternary compounds with stoichiometry $AB_2C_n$, $1 < n < 5$.

Pearson [5] gives $CdAl_2S_4$ the designation $E3$, without any subscript. It should be defined as $E3_a$, but as he did not put any other structures into this subcategory, we will leave it as is.

- ***E*4:** Ternary compounds with stoichiometry $AB_2C_5$.
- ***E*5:** Ternary compounds with stoichiometry $AB_2C_n$, $5 < n < 8$.
- ***E*6:** Ternary compounds with stoichiometry $AB_2C_8$.
- ***E*7:** Never defined.
- ***E*8:** Ternary compounds with stoichiometry $A_2B_2C_n$.
- ***E*9:** Other ternary and quaternary compounds.

***F*:** Compounds with two- and three- atom radicals.

Originally, compounds of the form $A_m(BC)_n$, where "*BC*" form a radical, were to be given designations in the range $F1-F50$, with higher numbers for $A_m(BC_2)_n$ and $A_m(BCD)_n$ compounds. With the publication of SB-II, this became:

- ***F*0:** Compounds with stoichiometry $A(BC)$.
- ***F*1:** Compounds with stoichiometry $A(BC)_2$.
- ***F*2:** Compounds with stoichiometry $A(BC)_m, m > 2$.
- ***F*3:** Never defined.
- ***F*4:** Compounds with stoichiometry $A_2(BC)_m$.
- ***F*5:** Compounds with stoichiometry $A_m(BC_2)$ or $A_m(BCD)$.
- ***F*6:** Additional compounds with stoichiometry $A_m(BC_2)$ or $A_m(BCD)$.

$F6$ was only used in SB-I for $F61$ (now $F6_1$), $CuFeS_2$ chalcopyrite.

Here, as well as in categories *G*-*J*, the letter "*A*" may represent more than one atom.

***G*:** Compounds with radicals of the form $(BC_3)$:

- ***G*0:** Compounds with stoichiometry $A(BC_3)$.
- ***G*1:** Compounds with stoichiometry $AD(BC_3)$.
- ***G*2:** Compounds with stoichiometry $A(BC_3)_2$.
- ***G*3:** Compounds with higher order $(BC_3)$ radicals.
- ***G*4:** Never defined.
- ***G*5:** Compounds with $BO_3$ and $PO_3$ radicals.

***H*:** Compounds with radicals of the form $(BC_n, n > 3)$.

This was originally a catch-all category, but in SB-II, compounds which would have been labeled as $H6$ and up were moved to a new class, *I*, and most silicates ($SiO_n$) were moved to *S*. This means that some structures have multiple *Strukturbericht* listings, so we will describe the complete *H* class here. The SB-II table of contents listed some compounds in both the *H* and *I* categories:

- ***H*0:** Compounds with stoichiometry $A(BC_4)$.
- ***H*1:** Compounds with stoichiometry $A_2(BC_4)$ or $AD(BC_4)$.
- ***H*2:** More complex compounds with only one $(BC_4)$ radical.
- ***H*3:** Compounds with multiple $(BC_4)$ radicals.
- ***H*4:** Hydrated compounds with $(BC_4)$ radicals.
- ***H*5:** Compounds with $(BC_5)$ radicals.
- ***H*6:** Compounds with stoichiometry $A_2(BC_6)$.

  This became $I1$ in SB-II and $J1$ in SB-III. $H6_2$ and $H6_4$ were only referred to as $I1_2$ and $I1_4$ in the SB-II table of contents, so we use the $H6$ designation in the list below. The examples given for $H6_3$ and $I1_3$ are rather different, so we use $H6_3$ for the original prototype $K_2(SCN)_6$ and $I1_3$ for $SrCl_2{\cdot}(H_2O)_6$ as given in SB-II.
- ***H*7:** Compounds with stoichiometry $A_3(BC_6)$.

  This became $I2$ in SB-II and $J2$ in SB-III.

***I*:** Compounds with radicals of the form $A_m(BC_6)$.

This was introduced in SB-II, essentially transferring $H6$ to $I1$, $H7$ to $I2$, etc.:

- ***I*1:** Compounds with stoichiometry $A_2(BC_6)$.
- ***I*2:** Compounds with stoichiometry $A_3(BC_6)$.
- ***I*3:** Never defined.
- ***I*4:** Never defined.
- ***I*5:** Never defined.
- ***I*6:** Compounds with stoichiometry $A_6(BC_6)$.

  The only structure listed here is $Ni(H_2O)_5SnCl_6$ ($I6_1$), which should have been subcategory $I5$ if the scheme laid out for $I1$ and $I2$ was used. It may have been the editors' intent to use this for hydrated systems. The $I6_1$ structure was never listed as $J6_1$, so we leave it in the *I* category in the AFLOW Prototype Encyclopedia.

***J*:** Compounds with radicals of the form $A_m(BC_6)$.

As noted above, this is the same category as *I* in SB-II, renamed in SB-III. Some structures in the *I* category were never renamed, and so we have left them with the *I* designation in Section 5:

- ***J*1:** Compounds with stoichiometry $A_2(BC_6)$.
- ***J*2:** Compounds with stoichiometry $A_3(BC_6)$.
- ***J*3:** Compounds with stoichiometry $A_3(BC_6) \cdot m\ H_2O$.

  The only structure with this designation is $K_3TlC_6{\cdot}\ 2H_2O$, from which we infer the editors' intent.

***K*:** Compounds with more complicated radicals.

The intended range of these labels is open to question, as many of the entries have only one or two compounds:

- ***K*0:** Compounds with radicals of the form $B_2C_n$, $n < 6$.
- ***K*1:** Compounds with radicals of the form $B_2C_n$, $n < 6$.

  There is some confusion in this category. SB-II defines $K1_1$ $KSO_3$ and $K1_2$ $CaSO_3$ in the text but uses $K2_1$ and $K2_2$ for the same structures in the table of contents. We use the labels in the actual text.
- ***K*2:** Never defined, except in the index of SB-II.
- ***K*3:** Compounds with radicals of the form $B_mC_5$.
- ***K*4:** Compounds with radicals of the form $B_2C_n$, $n > 5$.
- ***K*5:** Compounds with radicals of the form $B_3C_n$.
- ***K*6:** Compounds with radicals of the form $B_mC_7$.
- ***K*7:** Compounds with radicals of the form $B_mC_n$, $m < n$ , $n > 7$.

***L*:** Alloys.

These are ordered intermetallic systems. To determine, the subcategory start with an elemental lattice, say fcc ($A1$), construct a supercell of that lattice, and populate that supercell with various atomic types, so that on average the structure looks like the original fcc lattice, even though it is an ordered crystal. This category could in principle be used with any of the lattices found in part A, *e.g.*, zincblende $B3$, a derivative of the diamond structure, and could have been listed as $L4_1$. In practice, though, only a few such structures were defined:

- ***L*1:** Supercells of the face-centered cubic (fcc) structure, $A1$.
- ***L*2:** Supercells of the body-centered cubic (bcc) structure, $A2$.
- ***L*3:** Supercells of the hexagonal-close-packed (hcp) structure, $A3$.
- ***L*6:** Supercells of the body-centered tetragonal (bct) structure, $A6$.

There are also labels such as $L'1_0$, $Fe_4N$. These indicate interstitial additions to the supercell. In this case, the atoms in $Fe_3N$ are on fcc sites, and the extra iron atom is in the octahedral interstitial between the iron atoms.

***S*:** Silicates.

These compounds were broken off from the $H$ category in SB-II, and some of the compounds have both $H$ and $S$ labels:

- ***S*0:** Silicates with isolated $SiO_4$ tetrahedra and other anions.
- ***S*1:** Silicates with isolated $(SiO_4)_n$ tetrahedra, $n > 1$.
- ***S*2:** Silicates with connected groups of $SiO_4$ tetrahedra.

  In these compounds, the oxygen ions are shared between two tetrahedra.
- ***S*3:** Silicates with one-dimensional Si-O chains.
- ***S*4:** Silicates with two-dimensional Si-O networks.
- ***S*5:** Silicates with three-dimensional Si-O networks.
- ***S*6:** Other silicates.

## 5. Index of *Strukturbericht* symbols in the Prototype Encyclopedia

Here we list all of the “inorganic” (not beginning with $O$) *Strukturbericht* designations we have found in the literature, with a link to the corresponding web page in the Prototype Encyclopedia. Our primary source is the seven volumes of *Strukturbericht*, but we include designations from other sources as well.

The list has five columns:

**SB:** This is the *Strukturbericht* designation appearing in the literature. When a structure has multiple designations, we choose one of these, usually the latest, to be the primary designation. If the current designation is the primary one, it will appear in bold and italics, otherwise it will be in italics (unbolded).

In several cases, the literature did not give sufficient information to allow us to reconstruct the structure. In those cases the designation is in italics and there is no primary designation nor any page in the Prototype Encyclopedia.

**Prototype:** This is the chemical formula of the prototype for this *Strukturbericht* designation. When available, the common name for this structure is also listed here. In most cases we use the prototype from the source in the next column. When we use another prototype, this column will have two entries: the first is the compound we use as the prototype, and the second is the compound defined as the prototype in the literature.

In some cases, later experiments have shown that a structure defined in the literature is incorrect. We will still have an Prototype Encyclopedia entry for these structures, but they will be marked “obsolete.” Here we indicate an obsolete structure by listing the prototype in italics. If data for the currently accepted structure is available, there will be a link to it from the obsolete structure's Prototype Encyclopedia page.

**Source:** This is where we found the definition of the current *Strukturbericht* designation. The entry, or entries, consist of a label defining the source followed by the page number, where available. We use the following abbreviations:

- **SB-I – SB-VII:** The original *Strukturbericht* volumes [36, 47, 49, 52, 53, 54, 55].
- **Smi:** *Smithells Metals Reference, $2^{nd}$ Edition* (1955) [8]. This is the source of forty-five designations not found in the original *Strukturbericht* series.
- **Per:** *Pearson's Handbook of Lattice Spacings and Structures of Metals and Alloys* (1958) [5] adds ten more designations, and repeats some of Smithells's list. Pearson's list is repeated in later editions of the Handbook, but often not in an easily searchable form.
- **Per91:** *Pearson's Handbook* (1991 edition) [10].
- **Gmelin:** *Gmelin Handbook* (1993) [12].
- **Hart:** G. L. W. Hart and coworkers (2009, 2013) [56, 57].

This reference here is the source which defines the *Strukturbericht designation*. The source we used to obtain the actual crystal structure is given in the prototype's Encyclopedia entry.

**Other SB:** We list other *Strukturbericht* designations for the current structure, as well as their sources. If the designation in the SB column is not the primary one, the primary will be highlighted in bold here. These duplications occurred because of shifting definition of the main *Strukturbericht* categories in the first three volumes, but there were also several editorial errors, including missing *Strukturbericht* designations ($C20$) and multiple definitions of the $S3_3$ and $S3_4$ structures.

**AFLOW:** The AFLOW label for this structure. These are only the primary designations. In the online version of this article, clicking on the structure label will open the corresponding web page in the AFLOW Prototype Encyclopedia.

| *SB* | Prototype | Source | Other SB | AFLOW |
|---|---|---|---|---|
| ***A*1** | *fcc* Cu | SB-I 13 | | A_cF4_225_a |
| ***A*2** | *bcc* W | SB-I 15 | | A_cI2_229_a |
| ***A*3** | *hcp* Mg | SB-I 16 | | A_hP2_194_c |
| ***A*3′** | $\alpha$-La | Per91 | | A_hP4_194_ac |
| ***A*4** | diamond | SB-I 19 | | A_cF8_227_a |
| ***A*5** | $\beta$-Sn | SB-I 21 | | A_tI4_141_a |
| ***A*6** | In | SB-I 23 | | A_tI2_139_a.In |
| ***A*7** | $\alpha$-As | SB-I 25 | | A_hR2_166_c.alpha-As |
| ***A*8** | $\gamma$-Se | SB-I 27 | | A_hP3_152_a |
| ***A*9** | hexagonal graphite | SB-I 28 | | A_hP4_194_bc |
| ***A*10** | $\alpha$-Hg | SB-I 31 | | A_hR1_166_a.alpha-Hg |
| ***A*11** | $\alpha$-Ga | SB-I 738 | | A_oC8_64_f.alpha-Ga |
| ***A*12** | $\alpha$-Mn | SB-II 2 | | A_cI58_217_ac2g |
| ***A*13** | $\beta$-Mn | SB-II 3 | | A_cP20_213_cd |
| ***A*14** | $I_2$ | SB-II 5 | | A_oC8_64_f.I |
| ***A*15** | $\beta$-W / $Cr_3Si$ | SB-II 6 | | A3B_cP8_223_c_a |
| ***A*16** | $\alpha$-S | SB-III 4 | | A_oF128_70_4h |
| ***A*17** | P | SB-III 6 | | A_oC8_64_f.P |
| ***A*18** | $Cl_2$ | SB-IV 1 | | A_tP16_138_j |
| ***A*19** | *Po* | SB-IV 4 | | A_mC12_5_3c |
| ***A*20** | $\alpha$-U | SB-VI 1 | | A_oC4_63_c |
| $\boldsymbol{A_a}$ | $\alpha$-Pr | Smi 214 | | A_tI2_139_a.alpha-Pa |
| $\boldsymbol{A_b}$ | $\beta$-U | Smi 214 | | A_tP30_136_bf2ij |
| $\boldsymbol{A_c}$ | $\alpha$-Np | Smi 214 | | A_oP8_62_2c |
| $\boldsymbol{A_d}$ | $\beta$-Np | Smi 214 | | A_tP4_129_ac |
| $A_e$ | CuTi (alloy) | Smi 214 | ***A*20** SB-VI 1 | |
| $\boldsymbol{A_f}$ | simple hexagonal | Smi 214 | | A_hP1_191_a |
| $\boldsymbol{A_g}$ | T-50 B | Smi 214 | | A_tP50_134_b2m2n |
| $\boldsymbol{A_h}$ | $\alpha$-Po | Smi 215 | | A_cP1_221_a |
| $\boldsymbol{A_i}$ | $\beta$-Po | Smi 215 | | A_hR1_166_a.beta-Po |
| $\boldsymbol{A_k}$ | Se | Smi 215 | | A_mP64_14_16e |
| $\boldsymbol{A_l}$ | Se | Smi 215 | | A_mP32_14_8e |
| ***B*1** | NaCl (rock salt) | SB-I 72 | | AB_cF8_225_a_b |
| ***B*2** | CsCl | SB-I 74 | | AB_cP2_221_b_a |
| ***B*3** | ZnS (zincblende) | SB-I 76 | | AB_cF8_216_c_a |
| ***B*4** | ZnS (wurtzite) | SB-I 78 | | AB_hP4_186_b_b |
| ***B*5** | SiC (moissanite-4H) | SB-I 80 | | AB_hP8_186_ab_ab |
| ***B*6** | SiC (moissanite-6H) | SB-I 82 | | AB_hP12_186_a2b_a2b |
| ***B*7** | SiC (moissanite-15R) | SB-I 83 | | AB_hR10_160_5a_5a |
| $\boldsymbol{B8_1}$ | NiAs | Smi67 184 | *B*8 SB-I 84 | AB_hP4_194_c_a |
| $\boldsymbol{B8_2}$ | $Ni_2In$ | Smi67 184 | | AB2_hP6_194_c_ad |
| ***B*9** | HgS (cinnabar) | SB-I 86 | | AB_hP6_154_a_b |
| ***B*10** | PbO (litharge) | SB-I 89 | | AB_tP4_129_a_c |
| ***B*11** | $\gamma$-CuTi | SB-I 94 | | AB_tP4_129_c_c |
| ***B*12** | *BN* | SB-I 95 | | AB_hP4_186_b_a |
| ***B*13** | NiS (millerite) | SB-I 740 | | AB_hR6_160_b_b |
| ***B*14** | FeAs (westerveldite) | SB-II 7 | | AB_oP8_62_c_c.FeAs |
| *B*15 | FeB | SB-II 7 | ***B*27** SB-III 12 | |
| ***B*16** | GeS | SB-II 8 | | AB_oP8_62_c_c.GeS |
| ***B*17** | PtS (cooperite) | SB-II 9 | | AB_tP4_131_c_e |
| ***B*18** | CuS (covellite) | SB-II 9 | | AB_tP12_194_df_ce |
| ***B*19** | $\beta'$-AuCd | SB-II 11 | | AB_oP4_51_e_f |
| ***B*20** | FeSi | SB-II 13 | | AB_cP8_198_a_a.FeSi |

| *SB* | Prototype | Source | Other SB | AFLOW |
|---|---|---|---|---|
| ***B21*** | $\alpha$-CO | SB-II 13 | | AB_cP8_198_a_a.alpha-CO |
| ***B22*** | KSH | SB-III 8 | | AB_hR2_166_a_b.KSH |
| ***B23*** | $\alpha$-AgI | SB-III 8 | | A21B_cI44_229_bdh_a |
| ***B24*** | TlF | SB-III 9 | | AB_oF8_69_a_b |
| ***B25*** | $NH_4Br$ | SB-III 10 | | AB4C_tP12_129_c_i_a |
| ***B26*** | CuO (tenorite) | SB-III 11 | | AB_mC8_15_c_e |
| ***B27*** | FeB | SB-III 12 | *B*15 SB-II 7 | AB_oP8_62_c_c.FeB |
| *B*28 | FeSi | SB-III 13 | ***B20*** SB-II 13 | |
| ***B29*** | SnS | SB-III 14 | | AB_oP8_62_c_c.SnS |
| ***B30*** | *MgZn* | SB-III 16 | | AB_oI48_44_6d_ab2cde |
| ***B31*** | MnP | SB-III 17 | | AB_oP8_62_c_c.MnP |
| ***B32*** | NaTl | SB-III 19 | | AB_cF16_227_a_b |
| ***B33*** | TlI | SB-IV 6 | | AB_oC8_63_c_c |
| ***B34*** | PdS | SB-V 1 | | AB_tP16_84_cej_k |
| ***B35*** | CoSn | SB-VI 4 | | AB_hP6_191_f_ad |
| ***B36*** | $LiOH{\cdot}H_2O$ | SB-VII 4 | | A3BC2_mC24_12_ij_h_gi |
| ***B37*** | TlSe | SB-VII 6 | | AB_tI16_140_ab_h |
| $\boldsymbol{B_a}$ | CoU | Smi 217 | | AB_cI16_199_a_a |
| $\boldsymbol{B_b}$ | $\zeta$-AgZn | Smi 217 | | A2B_hP9_147_g_ad |
| $B_c$ | CaSi | Smi 218 | ***B33*** SB-IV 6, $B_f$ Smi 218 | |
| $\boldsymbol{B_d}$ | $\eta$-NiSi | Smi 218 | | AB_oP8_62_c_c.NiSi |
| $\boldsymbol{B_e}$ | CdSb | Smi 218 | | AB_oP16_61_c_c |
| $B_f$ | CrB | Smi 218 | ***B33*** SB-IV 6, $B_c$ Smi 218 | |
| $\boldsymbol{B_g}$ | MoB | Smi 218 | | AB_tI16_141_e_e |
| $\boldsymbol{B_h}$ | WC | Smi 218 | | AB_hP2_187_d_a |
| $\boldsymbol{B_i}$ | TiAs/$\gamma$'-MoC | Smi 218 per 95 | | AB_hP8_194_ad_f |
| $\boldsymbol{B_k}$ | BN | Smi 218 | | AB_hP4_194_c_d |
| $\boldsymbol{B_l}$ | AsS (realgar) | Smi 219 | | AB_mP32_14_4e_4e |
| $B_m$ | TiB | Smi 219 | ***B27*** SB-III 12 | |
| ***C1*** | $CaF_2$ (fluorite) | SB-I 148 | | AB2_cF12_225_a_c |
| $\boldsymbol{C1_b}$ | MgAgAs (half-Heusler) | Per 96 | | ABC_cF12_216_b_c_a |
| ***C2*** | $FeS_2$ (pyrite) | SB-I 150 | | AB2_cP12_205_a_c |
| ***C3*** | $Cu_2O$ (cuprite) | SB-I 153 | | A2B_cP6_224_b_a |
| ***C4*** | $TiO_2$ (rutile) | SB-I 155 | | A2B_tP6_136_f_a |
| ***C5*** | $TiO_2$ (anatase) | SB-I 158 | | A2B_tI12_141_e_a |
| ***C6*** | $CI_2$ (hexagonal $\omega$-phase) | SB-I 161 | | AB2_hP3_164_a_d |
| ***C7*** | $MoS_2$ (molybdenite) | SB-I 164 | | AB2_hP6_194_c_f |
| ***C8*** | $SiO_2$ ($\beta$-quartz) | SB-I 166 | | A2B_hP9_180_j_c |
| ***C9*** | $SiO_2$ ($\beta$-cristobalite) | SB-I 169 | | A2B_cF24_227_c_a |
| ***C10*** | $SiO_2$ ($\beta$-tridymite) | SB-I 171 | | A2B_hP12_194_cg_f |
| *C*11 | $CaC_2$ | SB-I 174 | $\boldsymbol{C11_b}$ Smi 219 | |
| $\boldsymbol{C11_a}$ | $CaC_2$ | Smi 219 | *C*11 SB-I 174, 740 | A2B_tI6_139_e_a |
| $\boldsymbol{C11_b}$ | $MoSi_2$ | Smi 219 | *C*11 SB-I 740 | AB2_tI6_139_a_e |
| ***C12*** | $CaSi_2$ | SB-I 175 | | AB2_hR6_166_c_2c |
| ***C13*** | $HgI_2$ | SB-I 177 | | AB2_tP6_137_a_d |
| ***C14*** | $MgZn_2$ (hexagonal laves) | SB-I 180 | | AB2_hP12_194_f_ah |
| ***C15*** | $Cu_2Mg$ (cubic laves) | SB-I 490 | | A2B_cF24_227_d_a |
| $\boldsymbol{C15_b}$ | $AuBe_5$ | Per 99 | | AB5_cF24_216_a_ce |
| ***C16*** | $CuAl_2$ (khatyrkite) | SB-I 491 | | A2B_tI12_140_h_a |
| ***C17*** | *Fe*$_2$*B* | SB-I 493 | | AB2_tI12_121_ab_i |
| ***C18*** | $FeS_2$ (marcasite) | SB-I 495 | | AB2_oP6_58_a_g.FeS2 |

| *SB* | Prototype | Source | Other SB | AFLOW |
|---|---|---|---|---|
| ***C19*** | $\alpha$-Sm | SB-I 742 | | A_hR3_166_ac |
| ***C21*** | $TiO_2$ (brookite) | SB-II 14 | | A2B_oP24_61_2c_c |
| ***C22*** | *$Fe_2P$* | SB-II 15 | | A2B_hP9_150_ef_bd |
| ***C23*** | $PbCl_2$ (cotunnite) | SB-II 16 | | A2B_oP12_62_2c_c.PbCl2 |
| ***C24*** | $HgBr_2$ | SB-II 18 | | A2B_oC12_36_2a_a |
| ***C25*** | $HgCl_2$ | SB-II 19 | | A2B_oP12_62_2c_c.HgCl2 |
| ***C26*** | $NO_2$ | SB-II 21 | $C26_b$ SB-II 21 | AB2_cI36_204_d_g |
| $\boldsymbol{C26_a}$ | *$NO_2$* | SB-II 21 | | AB2_cI36_199_b_c |
| $C26_b$ | $NO_2$ | SB-II 21 | ***C26*** SB-II 21 | |
| ***C27*** | $CdI_2$ | SB-III 22 | | AB2_hP6_186_b_ab |
| ***C28*** | *$HgCl_2$* | SB-III 23 | ***C25*** SB-II 19 | |
| ***C29*** | $SrH_2$ | SB-III 24 | | A2B_oP12_62_2c_c.SrH2 |
| ***C30*** | $SiO_2$ ($\alpha$-cristobalite) | SB-III 25 | | A2B_tP12_92_b_a |
| ***C31*** | $\epsilon$-$Zn(OH)_2$ (wülfingite) | SB-III 26 | | A2B2C_oP20_19_2a_2a_a |
| ***C32*** | $AlB_2$ (hexagonal $\omega$) | SB-III 28 | | AB2_hP3_191_a_d |
| ***C33*** | $Bi_2Te_3$/$Bi_2Te_2S$ | Sb-III 28 | | A2B3_hR5_166_c_ac |
| ***C34*** | $AuTe_2$ (calaverite) | SB-III 30 | | AB2_mC6_12_a_i |
| ***C35*** | $CaCl_2$ (hydrophilite) | SB-III 30 | | AB2_oP6_58_a_g.CaCl2 |
| ***C36*** | $MgNi_2$ (hexagonal laves) | SB-III 31 | | AB2_hP24_194_ef_fgh |
| ***C37*** | $Co_2Si$ | SB-III 32 | | A2B_oP12_62_2c_c.Co2Si |
| ***C38*** | $Cu_2Sb$ | SB-III. 33 | | A2B_tP6_129_ac_c |
| *C39* | $ZrW_2$ | SB-III 34 | ***C15*** Sb-I 660 | |
| ***C40*** | $CrSi_2$ | SB-III 35 | | AB2_hP9_180_d_j |
| *C41* | $Fe_2W$ | SB-III 36 | ***C14*** SB-I 180 | |
| ***C42*** | $SiS_2$ | SB-III 37 | | A2B_oI12_72_j_a |
| ***C43*** | $ZrO_2$ (baddeleyite) | SB-IV 9 | | A2B_mP12_14_2e_e |
| ***C44*** | $GeS_2$ | SB-IV 11 | | AB2_oF72_43_ab_3b |
| ***C45*** | $CuCl_2 \cdot 2H_2O$ (eriochalcite) | SB-IV 13 | | A2BC4D2_oP18_53_h_a_i_e |
| ***C46*** | $AuTe_2$ (krennerite) | SB-IV 15 | | AB2_oP24_28_acd_2c3d |
| ***C47*** | $SeO_2$ (downeyite) | SB-V 4 | | A2B_tP24_135_gh_h |
| *C48* | $Cr_2Al$ | SB-V 5 | $\boldsymbol{C11_b}$ SB-I 147, 740; Smi 219 | |
| ***C49*** | $ZrSi_2$ | SB-V 5 | | A2B_oC12_63_2c_c |
| ***C50*** | $\alpha$-$PdCl_2$ | SB-VI 5 | | A2B_oP6_58_g_a |
| *C51* | $CoSn_2$ | SB-VI 7 | ***C16*** SB-I 491 | |
| ***C52*** | $\beta$-$TeO_2$ (tellurite) | SB-VII 8 | | A2B_oP24_61_2c_c.TeO2 |
| ***C53*** | *$Br_2Sr$* | SB-VII 10 | | A2B_oP12_62_2c_c.SrBr2 |
| ***C54*** | $TiSi_2$ | SB-VII 12 | | A2B_oF24_70_e_a |
| $\boldsymbol{C_a}$ | $Mg_2Ni$ | Smi 223 | | A2B_hP18_180_fi_bd |
| $\boldsymbol{C_b}$ | $Mg_2Cu$ | Smi 223 | | AB2_oF48_70_g_fg |
| $\boldsymbol{C_c}$ | $\alpha$-$ThSi_2$ | Smi 223 | | A2B_tI12_141_e_a.ThSi2 |
| $\boldsymbol{C_e}$ | $PdSn_2$/$CoGe_2$ | Smi 223 | | AB2_oC24_41_2a_2b |
| $C_f$ | $RbSn_2$/$PdSn_2$ | Smi 233 | insufficient structural information | |
| $\boldsymbol{C_g}$ | $ThC_2$ | Smi 223 | | A2B_mC12_15_f_e |
| $\boldsymbol{C_h}$ | $Cu_2Te$ | Smi 223 | | A2B_hP6_191_h_e |
| $\boldsymbol{C_k}$ | $LiZn_2$ | Smi 223 | | AB_hP4_194_a_c |
| $\boldsymbol{D0_1}$ | $NH_3$ (ammonia) | SB-I 230 | *D1* | A3B_cP16_198_b_a |
| $\boldsymbol{D0_2}$ | $CoAs_3$ (skutterudite) | SB-I 232 | *D2* | A3B_cI32_204_g_c |
| $\boldsymbol{D0_3}$ | $BiF_3$ | SB-II 22 | | AB3_cF16_225_a_bc |
| $\boldsymbol{D0_4}$ | $CrCl_3$ | SB-II 23 | | A3B_hP24_151_3c_2a |
| $\boldsymbol{D0_5}$ | $BiI_3$ | SB-I 25 | | AB3_hR8_148_c_f |
| $\boldsymbol{D0_6}$ | $LaF_3$ | SB-II 27 | | A3B_hP24_193_ack_g |

| *SB* | Prototype | Source | Other SB | AFLOW |
|---|---|---|---|---|
| **$D0_7$** | $CrO_3$ | SB-II 29 | | AB3_oC16_20_a_bc |
| **$D0_8$** | $MoO_3$ | SB-II 30 | | AB3_oP16_62_c_3c |
| **$D0_9$** | $\alpha$-$ReO_3$ | SB-II 31 | | A3B_cP4_221_d_a |
| **$D0_{10}$** | $WO_3$ | SB-II 32 | | A3B_oP16_57_a2d_d |
| **$D0_{11}$** | $Fe_3C$ (cementite) | SB-II 33 | | AB3_oP16_62_c_cd |
| **$D0_{12}$** | $FeF_3$ | SB-II 34 | | A3B_hR8_167_e_b |
| **$D0_{13}$** | $AlCl_3$ | SB-II 36 | | AB3_hP4_164_b_ad |
| **$D0_{14}$** | $AlFe_3$ | SB-III 40 | | AB3_hR8_155_c_de |
| **$D0_{15}$** | $AlCl_3$ | SB-III 41 | | AB3_mC16_5_c_3c |
| **$D0_{16}$** | $NH_4I_3$ | SB-III 43 | | A3B_oP16_62_3c_c |
| **$D0_{17}$** | $BaS_3$ | SB-IV 18 | | AB3_tP8_113_a_ce |
| **$D0_{18}$** | $Na_3As$ | SB-V 6 | | AB3_hP8_194_c_bf |
| **$D0_{19}$** | $Na_3Sn$ | SB-V 7 | | A3B_hP8_194_h_c |
| **$D0_{20}$** | $\epsilon$-$NiAl_3$ | SB-V 8 | | A3B_oP16_62_cd_c |
| **$D0_{21}$** | $Cu_3P$ | SB-VI 7 | | A3B_hP24_165_bdg_f |
| **$D0_{22}$** | $TiAl_3$ | SB-VII 13 | | A3B_tI8_139_bd_a |
| **$D0_{23}$** | $ZrAl_3$ | SB-VII 14 | | A3B_tI16_139_cde_e |
| **$D0_{24}$** | $Ni_3Ti$ | SB-VII 14 | | A3B_hP16_194_gh_ac |
| **$D0_a$** | $\beta$-$TiCu_3$ | Smi 225 | | A3B_oP8_59_bf_a |
| $D0_b$ | $\zeta$-$Ag_3Ga$ | Smi 225 | **$B_b$** Smi. p 217 | |
| **$D0_c$** | $SiU_3$ | Smi 225 | $D0'_c$ Gmelin 367 | AB3_tI16_140_b_ah |
| **$D0_d$** | $Mn_3As$ | Smi 225 | | AB3_oC16_63_c_3c |
| **$D0_e$** | $Ni_3P$ | Per 105 | | A3B_tI32_82_3g_g |
| **$D1_1$** | $SnI_4$ | SB-I 234 | *D*11 | A4B_cP40_205_cd_c |
| **$D1_2$** | $SiF_4$ | SB-II 37 | | A4B_cI10_217_c_a |
| **$D1_3$** | $Al_4Ba$ | SB-VII 15 | | A4B_tI10_139_de_a |
| **$D1_a$** | $Ni_4Mo$ | Smi 225 | | AB4_tI10_87_a_h |
| **$D1_b$** | $Al_4U$ | Smi 226 | | A4B_oI20_74_beh_e |
| **$D1_c$** | $PtSn_4$ | Smi 226 | | AB4_oC20_41_a_2b |
| **$D1_d$** | $PbPt_4$ | Smi 226 | | A4B_tP10_125_m_a |
| **$D1_e$** | $ThB_f/UB_4$ | Smi 226 | | A4B_tP20_127_ehj_g |
| **$D1_f$** | $Mn_2B$ | Smi 226 | | AB2_oF48_70_f_fg |
| **$D1_g$** | $B_{13}C_2$ | Smi 226 | | A13B2_hR15_166_b2h_c |
| **$D2_1$** | $CaB_6$ | SB-II 37 | | A6B_cP7_221_f_a |
| **$D2_2$** | $MgZn_5$ | SB III 47 | | AB5_mC48_12_2i_ac5i2j |
| **$D2_3$** | $NaZn_{13}$ | SB VI 8 | | AB13_cF112_226_a_bi |
| **$D2_a$** | $Be_{12}Ti$ | Smi 227 | | A12B_hP13_191_cdei_a |
| **$D2_b$** | $ThMn_{12}$ | Smi 227 | | A12B_tI26_139_fij_a |
| **$D2_c$** | $MnU_6$ | Smi 227 | | AB6_tI28_140_a_hk |
| **$D2_d$** | $CaCu_5$ | Smi 227 | | AB5_hP6_191_a_cg |
| **$D2_e$** | $BaHg_{11}$ | Smi 227 | | AB11_cP36_221_c_agij |
| **$D2_f$** | $UB_{12}$ | Smi 227 | | A12B_cF52_225_i_a |
| **$D2_g$** | $Fe_8N$ | Smi 227 | | A8B_tI18_139_deh_a |
| **$D2_h$** | $MnAl_6$ | Per 109 | | A6B_oC28_63_efg_c |
| **$D3_1$** | $HgCl$ | SB I 237 | *D*31 | AB_tI8_139_e_e |
| $D4_1$ | $BH_3$ | SB I 239 | *D*41, ***A*10** SB-I 31 | |
| **$D5_1$** | $Al_2O_3$ (corundum) | SB-I 240 | *D*51 | A2B3_hR10_167_c_e |
| **$D5_2$** | $La_2O_3$ | SB-I 242 | *D*52 | A2B3_hP5_164_d_ad |
| **$D5_3$** | $Mn_2O_3$ (bixbyite) | SB-II 38 | | AB3C6_cI80_206_b_d_e |
| $D5_4$ | $Sb_2O_3$ (senarmonite) | Gmelin 368 | **$D6_1$** SB-I 245 | |
| **$D5_5$** | $Mg_3P_2$ | SB-II 40 | | A3B2_cP10_224_d_b |
| **$D5_6$** | $Al_2O_3$ ($\beta$-alumina) | SB-II 41 | | A2B3_hP60_194_3fk_cdef2k |
| **$D5_7$** | $Al_2O_3/Fe_2O_3$ ($\gamma$-corundum) | SB-II 43 | | A2B3_cP60_212_bcd_ace |

| *SB* | Prototype | Source | Other SB | AFLOW |
|---|---|---|---|---|
| ***D*5$_8$** | $Sb_2S_3$ (stibnite) | SB-III 49 | | A3B2_oP20_62_3c_2c |
| ***D*5$_9$** | $Zn_3P_2$ | SB-III 51 | | A2B3_tP40_137_cdf_3g |
| ***D*5$_{10}$** | $Cr_3C_2$ (tongbaite) | SB-III 53 | | A2B3_oP20_62_2c_3c |
| ***D*5$_{11}$** | $Sb_2O_3$ (valentinite) | SB-IV 20 | | A3B2_oP20_56_ce_e |
| ***D*5$_{12}$** | $\beta$-$Bi_2O_3$ | SB-V 9 | | A2B3_tP20_117_i_adgh |
| ***D*5$_{13}$** | $Ni_2Al_3$ | SB-V 10 | | A3B2_hP5_164_ad_d |
| ***D*5$_a$** | $Si_2U_3$ | Smi 228 | | A2B3_tP10_127_g_ah |
| ***D*5$_b$** | $Pt_2Sn_3$ | Smi 229 | | A2B3_hP10_194_f_bf |
| ***D*5$_c$** | $Pu_2C_3$ | Smi 229 | | A3B2_cI40_220_d_c |
| ***D*5$_e$** | $Ni_3S_2$ | Smi 229 | | A3B2_hR5_155_e_c |
| ***D*5$_f$** | $As_2S_3$ (orpiment) | Smi 229 | | A2B3_mP20_14_2e_3e |
| ***D*6$_1$** | $Sm_2O_3$ (senarmontite) | SB-I 245 | | A3B2_cF80_227_f_e |
| ***D*6$_2$** | *$Sb_2O_4$* | SB-III 54 | | A2B_cF96_227_abf_cd |
| ***D*7$_1$** | $Al_4C_3$ | SB-III 56 | | A4B3_hR7_166_2c_ac |
| ***D*7$_2$** | $Co_3O_4$ (spinel) | SB-VI 9 | | A3B4_cF56_227_ad_e |
| ***D*7$_3$** | $Th_3P_4$ | SB-VII 15 | | A4B3_cI28_220_c_a |
| ***D*7$_a$** | $Nb_3Sn_4$ | Smi 229 | | A3B4_mC14_12_ai_2i |
| ***D*7$_b$** | $Ta_3B_4$ | Smi 230 | | A4B3_oI14_71_gh_cg |
| ***D*8$_1$** | $Fe_3Zn_{10}$ ($\gamma$-brass) | SB-I 497 | | A3B10_cI52_229_e_fh |
| ***D*8$_2$** | $Cu_5Zn_8$ ($\gamma$-brass) | SB-I 497 | | A5B8_cI52_217_ce_cg |
| ***D*8$_3$** | $Cu_9Al_4$ ($\gamma$-brass) | SB-I 498 | | A4B9_cP52_215_ei_3efgi |
| ***D*8$_4$** | $Cr_{23}C_6$ | SB-III 59 | | A6B23_cF116_225_e_acfh |
| ***D*8$_5$** | $Fe_7W_6$ ($\mu$-phase) | SB-III 61 | | A7B6_hR13_166_ah_3c |
| ***D*8$_6$** | $Cu_{15}Si_4$ | SB-III 62 | | A15B4_cI76_220_ae_c |
| ***D*8$_7$** | *$V_2O_5$ (shcherbinaite)* | SB-IV 22 | | A5B2_oP14_31_a2b_b |
| ***D*8$_8$** | $Mn_5Si_3$ (mavlyanovite) | SB-IV 24 | | A5B3_hP16_193_dg_g |
| ***D*8$_9$** | $Co_9S_8$ | SB-IV 26 | | A9B8_cF68_225_af_ce |
| ***D*8$_{10}$** | $Cr_5Al_8$ | SB-V 10 | | A8B5_hR26_160_a3bc_a3b |
| ***D*8$_{11}$** | $Co_2Al_5$ | SB-VI 11 | | A5B2_hP28_194_ahk_ch |
| ***D*8$_a$** | $Th_6Mn_{23}$ | Smi 231 | | A23B6_cF116_225_bd2f_e |
| ***D*8$_b$** | $\sigma$-CrFe ($\sigma$-phase) | Smi 231 | | sigma_tP30_136_bf2ij |
| ***D*8$_c$** | $Mg_2Zn_{11}$ or $Mg_2Cu_6Al_5$ | Smi 232 | | A2B11_cP39_200_f_aghij |
| ***D*8$_d$** | $Co_2Al_9$ | Smi 232 | | A9B2_mP22_14_a4e_e |
| ***D*8$_e$** | $Mg_{32}(Al,Zn)_{49}$ (Bergmann) | Smi 232 | | AB32C48_cI162_204_a_2efg_2gh |
| ***D*8$_f$** | $Ir_3Ge_7$ | Smi 233 | | A7B3_cI40_229_df_e |
| ***D*8$_g$** | $Mg_5Ga_2$ | Smi 233 | | A2B5_oI28_72_j_bfj |
| ***D*8$_h$** | $W_2B_5$ | Smi 233 | | A5B2_hP14_194_abdf_f |
| ***D*8$_i$** | $Mo_2B_5$ | Smi 233 | | A5B2_hR7_166_a2c_c |
| ***D*8$_k$** | $Th_7S_{12}$ | Smi 233 | | A3B2_hP20_176_2h_ah |
| ***D*8$_l$** | $Cr_5B_3$ | Per 117 | | A3B5_tI32_140_ah_cl |
| ***D*8$_m$** | $W_5Si_3$ | Per 117 | | A3B5_tI32_140_ah_bk |
| ***D*10$_1$** | $C_3Cr_7$ | Per 117 | | A3B7_oP40_62_cd_3c2d |
| ***D*10$_2$** | $Fe_3Th_7$ | Per 117 | | A3B7_hP20_186_c_b2c |
| ***E*0$_1$** | PbFCl (matlockite) | SB-II 45 | | ABC_tP6_129_c_a_c |
| ***E*0$_2$** | AlO(OH) (diaspore) | SB-II 46 | | ABC2_oP16_62_c_c_2c |
| ***E*0$_3$** | Cd(OH)Al | SB-III 65 | | ABCD_hP8_186_b_b_a_a |
| ***E*0$_4$** | $\gamma$-FeO(OH) | SB-III 66 | | AB2C2_oC20_63_c_f_2c |
| ***E*0$_5$** | FeOCl | SB-III 67 | | ABC_oP6_59_a_b_a |
| ***E*0$_6$** | Mn(OH)O (manganite) | SB-IV 28 | | ABC2_mP16_14_e_e_2e |
| ***E*0$_7$** | FeAsS (arsenopyrite) | SB-IV 30 | | ABC_mP12_14_e_e_e |
| ***E*1$_1$** | $CuFeS_2$ (chalcopyrite) | SB-II 48 | | ABC2_tI16_122_a_b_d |
| ***E*1$_2$** | $Zn(NH_3)_2Cl_2$ | SB-IV 31 | | A2B6C2D_oI44_74_h_ij_i_e |
| ***E*1$_3$** | $Mg(NH_3)_2Cl_2$ or $Cd(NH_3)_2Cl_2$ | SB-IV 33 | | A2B8CD2_oC26_65_h_r_a_i |

| *SB* | Prototype | Source | Other SB | AFLOW |
|---|---|---|---|---|
| $E1_4$ | $PNCl_2$ | SB-VII 17 | | A2BC_tP32_86_2g_g_g |
| $E1_a$ | $MgCuAl_2$ | Smi 233 | | A2BC_oC16_63_f_c_c |
| $E1_b$ | $AuAgTe_4$ (sylvanite) | Smi 234 | | ABC4_mP12_13_e_a_2g |
| $E2_1$ | $CaTiO_3$ (perovskite) | SB-II 49 | $G5/G0_5$ SB-I 300 | AB3C_cP5_221_a_c_b |
| $E2_2$ | $FeTiO_3$ (ilmenite) | Gmelin 369 | $G4/G0_4$ SB-I 300 | AB3C_hR10_148_c_f_c |
| $E2_3$ | $\alpha$-$LiIO_3$ | SB-II 49 | | ABC3_hP10_182_c_b_g |
| $E2_4$ | $(NH_4)CdCl_3$ | SB-VI 13 | | AB3C_oP20_62_c_3c_c |
| $E2_5$ | $(NH_4)HgCl_3$ | SB-VI 15 | | A3BC_tP5_123_cg_a_d |
| $E2_6$ | $KMg(H_2O)_6(Cl,Br)_3$ (bromocarnallite) | SB-VII 19 | | A3B6CD_tP44_85_bcg_3g_ac_e |
| $E3$ | $CdAl_2S_4$ | Per 118 | | A2BC4_tI14_82_bc_a_g |
| $E3_1$ | $\beta$-$Ag_2HgI_4$ | SB-II 50 | | A2BC4_tP7_111_f_a_n |
| $E3_2$ | $CaB_2O_4$ | SB-II 54 | | A2BC4_oP28_60_d_c_2d |
| $E3_3$ | $FeSb_2S_4$ (berthierite) | SB-IV 35 | | AB4C2_oP28_62_c_4c_2c |
| $E3_4$ | $K_2HgCl_4{\cdot}H_2O$ | SB-VI 16 | | A4BCD2_oP32_55_ghi_f_e_gh |
| $E3_5$ | $K_2SnCl_4{\cdot}H_2O$ | SB-VII 22 | | A4BC2D_oP32_62_2cd_b_2c_a |
| $E4_1$ | $Fe_2TiO_5$ (pseudobrookite) | SB-II 53 | | A2B5C_oC32_63_f_c2f_c |
| $E5_1$ | $FeNb_2O_6$ (columbite or niobite) | SB-II 55 | | AB2C6_oP36_60_c_d_3d |
| $E6_1$ | $Sr(OH)_2(H_2O)_8$ | SB-II 56 | | A8B2C_tP11_123_r_f_a |
| $E6_2$ | $SrO_2(H_2O)_8$ | SB-II 57 | | A8B2C_tP11_123_r_h_a |
| $E8_1$ | $Eu_2Ir_2O_7$ (cubic pyrochlore) | SB-II 58 | | A2B2C7_cF88_227_c_d_af |
| $E9_1$ | $Ca_2Al_2O_6$ | SB-II 60 | | A2B3C6_cP33_221_cd_ag_fh |
| $E9_2$ | $NaBe_4SbO_7$ (swedenborgite) | SB-III 69 | | A4BC7D_hP26_186_ac_b_a2c_b |
| $E9_3$ | $Fe_3W_3C$ | SB-III 71 | | AB3C3_cF112_227_c_de_f |
| $E9_4$ | $Al_5C_3N$ | SB-III 73 | | A5B3C_hP18_186_2a3b_2ab_b |
| $E9_a$ | $FeCu_2Al_7$ | Smi 234 | | A7B2C_tP40_128_egi_h_e |
| $E9_b$ | $\pi$-$FeMg_3Al_8Si_6$ | Smi 235 | | A8BC3D6_hP18_189_bfh_a_g_i |
| $E9_c$ | $Al_9Mn_3Si$ | Smi 234 | | A9B3C_hP26_194_hk_h_a |
| $E9_d$ | $AlLi_3N_2$ | Smi 235 | | AB3C2_cI96_206_c_e_ad |
| $E9_e$ | $CuFe_2S_3$ (cubanite) | Smi 235 | | AB2C3_oP24_62_c_d_cd |
| $F0_1$ | NiSSb (ullmannite) | SB-I 269 | | ABC_cP12_198_a_a_a |
| $F0_2$ | COS | SB-II 62 | | ABC_hR3_160_a_a_a |
| $F1_1$ | $Hg(CN)_2$ | SB-I 271 | $F11$ | A2BC2_tI40_122_e_d_e |
| $F2_1$ | $K_4[Mo(CN)_8]{\cdot}2H_2O$ | SB-VII 25 | | A8B4C4DE8F2_oP108_62_4c2d_2d_2cd_c_4c2d_d |
| $F4_1$ | $Fe_2(CO)_8$ | SB-VII 28 | | A9B2C9_hP40_176_hi_f_hi |
| $F5_1$ | $NaCrS_2$ (caswellsilverite)/$NaHF4_2$ | SB-I 271 | $F51$ | ABC2_hR4_166_a_b_c |
| $F5_2$ | $KHF_2$ | SB-I 276 | $F52$ | A2BC_tI16_140_h_d_a |
| $F5_3$ | $CuFeS_2$ (chalcopyrite) | SB-II XXII | $E1_1$ SB-II 48 | |
| $F5_4$ | $NH_4ClO_2$ | SB-II 64 | | ABC2_tP8_100_b_a_c |
| $F5_5$ | $NaNO_2$ | SB-II 65 | | ABC2_oI8_44_a_a_c |
| $F5_6$ | $CuSbS_2$ (chalcostibite) | SB-III 75 | | AB2C_oP16_62_c_2c_c |
| $F5_7$ | $NH_4H_2PO_2$ | SB-III 77 | | A2BC2D_oC24_67_m_a_n_g |
| $F5_8$ | $NH_4HF_2$ | SB-III 79 | | A2BC_oP16_53_eh_ab_g |
| $F5_9$ | KCNS | SB-III 80 | | ABCD_oP16_57_d_c_d_d |
| $F5_{10}$ | $KAg(CN)_2$ | SB-III 82 | | AB2CD2_hP36_163_h_i_bf_i |
| $F5_{11}$ | $KNO_2$ | SB-IV 38 | | ABC2_mC8_8_a_a_b |
| $F5_{12}$ | $AgNO_2$ | SB-IV 41 | | ABC2_oI8_44_a_a_d |
| $F5_{13}$ | $KBO_2$ | SB-V 12 | | ABC2_hR24_167_e_e_2e |
| $F5_a$ | $FeKS_2$ | Smi 235 | | ABC2_mC16_15_e_e_f |
| $F6_1$ | *$CuFeS_2$ (chalcopyrite)* | SB-I 279 | | ABC2_tP4_115_a_c_g |
| $G0_1$ | $CaCO_3$ (calcite) | SB-I 292 | $G1$ | ABC3_hR10_167_a_b_e.CaCO3 |
| $G0_2$ | $CaCO_3$ (aragonite) | SB-I 295 | $G2$ | ABC3_oP20_62_c_c_cd |

| *SB* | Prototype | Source | Other SB | AFLOW |
|---|---|---|---|---|
| **$G0_3$** | $NaClO_3$ | SB-I 297 | $G3$ | ABC3_cP20_198_a_a_b |
| $G0_4$ | $FeTiO_3$ | SB-I 300 | $G4$ **$E2_2$** Gmelin 369 | |
| $G0_5$ | $CaTiO_3$ (perovskite) | SB-I 300 | $G5$ **$E2_1$** SB-II 49 | |
| **$G0_6$** | $KClO_3$ | SB-II 66 | | ABC3_mP10_11_e_e_ef |
| **$G0_7$** | $KBrO_3$ | SB-II 68 | | ABC3_hR5_160_a_a_b.KBrO3 |
| **$G0_8$** | $NH_4NO_3$ I | SB-II 69 | | AB_cP2_221_a_b.NH4.NO3 |
| **$G0_9$** | $NH_4NO_3$ II | SB-II 69 | | ABC3_tP10_100_b_a_bc |
| **$G0_{10}$** | $NH_4NO_3$ III | SB-II 70 | | ABC3_oP20_62_c_c_cd.N.NH4.O |
| **$G0_{11}$** | $NH_4NO_3$ IV | SB-II 71 | | A4B2C3_oP18_59_ef_ab_af |
| **$G0_{12}$** | $NaHCO_3$ | SB-III 86 | | ABCD3_mP24_14_e_e_e_3e |
| **$G1_1$** | $MgCa(CO_3)_2$ (dolomite) | SB-I 303 | $G11$ | A2BCD6_hR10_148_c_a_b_f |
| **$G2_1$** | $Pb(NO_3)_2$ | SB-II 73 | $G21$ | A2B6C_cP36_205_c_d_a |
| **$G2_2$** | $Nd(BrO_3)_3{\cdot}9H_2O$ | SB-VII 29 | | A3B9CD9_hP44_186_c_3c_b_cd |
| **$G3_1$** | $Be_3Al_2(SiO_3)_6$ (beryl) | SB-I 305 | $G31$ **$S3_1$** SB-II XXIV | |
| **$G3_2$** | $Na_2(SO_3)$ | SB-II 74 | | A2B3C_hP12_147_abd_g_d |
| **$G5_1$** | $H_3BO_3$ (boric acid) | SB-III 87 | | AB3C3_aP28_2_2i_6i_6i |
| **$G5_2$** | $Al(PO_3)_3$ | SB-V 15 | | AB9C3_cI208_220_c_3e_e |
| **$G7_1$** | $CeF(CO_3)$ (bastnäsite) | SB-II 75 | | ABCD3_hP36_190_h_g_af_hi |
| **$G7_2$** | $Be_2(BO_3)(OH)$ (hambergite) | SB-II 78 | | AB2CD4_oP64_61_c_2c_c_4c |
| **$G7_3$** | *$Na_3MgCl(CO_3)_2$ (northupite)* | SB-II 80 | | A2BCD3E6_cF208_227_e_c_d_f_g |
| **$G7_4$** | $Cu_3(CO_3)_2(OH)_2$ (azurite) | SB-II 81 | | A2B3C2D8_mP30_14_e_ce_e_4e |
| **$G7_5$** | *$PbCO_3{\cdot}PbCl_2$ (phosgenite)* | SB-III 89 | | AB2C3D2_tP16_90_c_f_ce_e |
| **$H0_1$** | $CaSO_4$ (anhydrite) | SB-I 340 | $H1$ | AB4C_oC24_63_c_fg_c |
| **$H0_2$** | $BaSO_4$ (barite) | SB-I 313 | $H2$ | AB4C_oP24_62_c_2cd_c |
| $H0_3$ | $ZrSiO_4$ (zircon) | SB-I 345 | $H3$ **$S1_1$** SB-II XXIV | |
| **$H0_4$** | $CaWO_4$ (scheelite) | SB-I 347 | | AB4C_tI24_88_b_f_a |
| **$H0_5$** | $KClO_4$ (high-temperature phase) | SB-II 84 | | ABC4_cF24_216_b_a_e |
| **$H0_6$** | $MgWO_4$ (magnesium wolframite) | SB-II 85 | | AB4C_mP12_13_f_2g_e |
| **$H0_7$** | $BPO_4$ | SB-III 92 | | AB4C_tI12_82_c_g_a |
| **$H0_8$** | $TlAlF_4$ | SB-V 17 | | AB4C_tP6_123_d_eh_a |
| **$H0_9$** | $AgMnO_4$ | SB-VI 19 | | ABC4_mP24_14_e_e_4e |
| **$H0_{10}$** | $KICl_4{\cdot}H_2O$ | SB-VI 20 | | A4BCD_mP28_14_4e_e_e_e |
| **$H1_1$** | $Al_2MgO_4$ (spinel) | SB-I 350 | $H11$ | A2BC4_cF56_227_d_a_e |
| **$H1_2$** | $Mg_2SiO_4$ (fosterite) | SB-I 352 | $H12$ **$S1_2$** | |
| **$H1_3$** | $Be_2SiO_4$ (phnakite) | SB-I 356 | $H13$ **$S1_3$** | |
| **$H1_4$** | $KLiSO_4$ | SB-I 357 | $H14$ | ABC4D_hP14_173_a_b_bc_b |
| **$H1_5$** | $K_2PtCl_4$ | SB-I 358 | $H15$ | A4B2C_tP7_123_j_e_a |
| **$H1_6$** | $K_2SO_4$ | SB-II 86 | | A2B4C_oP28_62_2c_2cd_c |
| **$H1_7$** | $Na_2SO_4$ (thenardite) | SB-II 88 | | A2B4C_oF56_70_g_h_a |
| **$H1_8$** | $Na_2CrO_4$ | SB-IV 45 | | AB2C4_oC28_63_c_bc_fg |
| **$H2_1$** | $Ag_3PO_4$ | SB-I 360 | $H21$ | A3B4C_cP16_218_c_e_a |
| **$H2_2$** | $KH_2PO_4$ | SB-I 362 | $H22$ | A4BC4D_tI40_122_e_b_e_a |
| **$H2_4$** | $Cu_3VS_4$ (sulvanite) | SB-III 94 | | A3B4C_cP8_215_d_e_a |
| **$H2_5$** | $Cu_3AsS_4$ (enargite) | SB-III 96 | | AB3C4_oP16_31_a_ab_2ab |
| **$H2_6$** | $Cu_2FeSnS_4$ | SB-III 96 | | A2BC4D_tI16_121_d_a_i_b |
| **$H2_7$** | $Zn_2(OH)AsO_4$ (adamite) | SB-V 17 | | ABC5D2_oP36_58_g_g_3gh_eg |
| **$H2_8$** | $BaAl_2O_4$ | SB-V 19 | | A2BC6_hP18_182_f_b_gh |
| **$H3_1$** | $Al_2Ca_3(SiO_4)_3$ (garnet) | SB-I 363 | $H31$ **$S1_4$** SB-III 150 | |
| **$H3_2$** | $AlK(SO_4)_2$ (steklite) | SB-II 90 | | ABC8D2_hP12_150_b_a_dg_d |
| **$H4_1$** | $K_2CuCl_4{\cdot}2H_2O$ | SB-I 366 | $H41$ | A4BC4D2E2_tP26_136_fg_a_j_d_e |
| $H4_2$ | $KAl(SO_4)_2{\cdot}12H_2O$ ($\alpha$-alum) | SB-I 369 | $H42$ **$H4_{13}$** SB-III 108 | |
| **$H4_3$** | $BeSO_4{\cdot}4H_2O$ | SB-II 91 | | AB8C8D_tI72_120_c_2i_2i_b |

| *SB* | Prototype | Source | Other SB | AFLOW |
|---|---|---|---|---|
| $\boldsymbol{H4_4}$ | $Cu(NH_4)_2(SO_4)_2\cdot H_2O$ | SB-II 93 | | AB20C2D14E2_mP78_14_a_10e_e_7e_e |
| $\boldsymbol{H4_5}$ | $\alpha$-$NiSo_4\cdot 6H_2O$ (retgersite) | SB-II 95 | | A12BC10D_tP96_92_6b_a_5b_a |
| $\boldsymbol{H4_6}$ | $CaSO_4\cdot 2H_2O$ (gypsum) | SB-II 97/IV 47 | | AB4C6D_mC48_15_e_2f_3f_e |
| $\boldsymbol{H4_7}$ | $CaSO_4(H_2O)_{0.5}$ (bassanite) | SB-III 97 | | A2B2C9D2_mC90_5_ab2c_3c_b13c_3c |
| $\boldsymbol{H4_8}$ | $Li_2SO_4\cdot H_2O$ | SB-III 98 | | A2B2C5D_mP20_4_2a_2a_5a_a |
| $\boldsymbol{H4_9}$ | $Pd(NH_3)_4Cl_2\cdot H_2O$ | SB-III 100 | | A2BC4D_tP16_127_h_d_i_a |
| $\boldsymbol{H4_{10}}$ | $CuSO_4\cdot 5H_2O$ (chalcanthite) | SB-III 102 | | AB10C9D_aP42_2_ae_10i_9i_i |
| $\boldsymbol{H4_{11}}$ | $Mg(ClO_4)_2\cdot 6H_2O$ | SB-III 103 | | A2B6CD8_oP34_31_2a_2a2b_a_4a2b |
| $\boldsymbol{H4_{12}}$ | $NiSO_4\cdot 7H_2O$ (morenosite) | SB-III 105 | | A14BC11D_oP108_19_14a_a_11a_a |
| $\boldsymbol{H4_{13}}$ | $KAl(SO_4)_2\cdot 12H_2O$ ($\alpha$-alum) | SB-III 108 | | AB24CD28E2_cP224_205_a_4d_b_2c4d_c |
| $\boldsymbol{H4_{14}}$ | $Al(NH_3CH_3)_2(SO_4)_2\cdot 12H_2O$ ($\beta$-alum) | SB-III 111 | | AB2C36D2E20F2_cP252_205_a_c_6d_c_c3d_c |
| $\boldsymbol{H4_{15}}$ | $AlNa(SO_4)_2\cdot 12H_2O$ ($\gamma$-alum) | SB-III 112 | | AB24CD20E2_cP192_205_a_4d_b_c3d_c |
| $\boldsymbol{H4_{16}}$ | $H_3PW_{12}O_{40}\cdot 5H_2O$ | SB-III 113 | | A5B40CD12_cP116_224_cd_e3k_a_k |
| $\boldsymbol{H4_{17}}$ | $Ag_2SO_4\cdot 4NH_3$ | SB-III 115 | | A2B12C4D4E_tP46_114_d_3e_e_e_a |
| $\boldsymbol{H4_{18}}$ | $LiClO_4\cdot 3H_2O$ | SB-III 117 | | AB6CD7_hP30_186_b_d_a_b2c |
| $\boldsymbol{H4_{19}}$ | $KAuBr_4\cdot H_2O$ | SB-IV 50 | | AB4C2D_mP32_14_e_4e_2e_e |
| $\boldsymbol{H4_{20}}$ | $(CdSO_4)_3\cdot 8H_2O$ | SB-IV 52 | | A3B16C20D3_mC168_15_ef_8f_10f_ef |
| $\boldsymbol{H4_{21}}$ | $H_3PW_{12}O_{40}\cdot 29H_2O$ | SB-IV 56 | | A29B40CD12_cF656_227_ae2fg_e3g_b_g |
| $\boldsymbol{H4_{22}}$ | $BaNi(CN)_4\cdot 4H_2O$ | SB-VI 23 | | AB4C4D4E_mC56_15_e_2f_2f_2f_a |
| $\boldsymbol{H4_{23}}$ | $K_2Mn(SO_4)_2\cdot 4H_2O$ (leonite) | SB-VII 32 | | A8B2CD15E2_mC112_12_2i3j_j_ad_g4i5j_2i |
| $H5_1$ | $Al_2SiO_5$ (kyanite) | SB-II 109 | $\boldsymbol{S0_1}$ SB-II XXIII | |
| $H5_2$ | $Al_2SiO_5$ (sillimanite) | SB-II 112 | $\boldsymbol{S0_3}$ SB-II XXIII | |
| $H5_3$ | $Al_2SiO_5$ (andalusite) | SB-II 110 | $\boldsymbol{S0_2}$ SB-II XXIII | |
| $H5_4$ | $Fe(OH)_2Al_4SiO_{10}$ | SB-II 113 | $\boldsymbol{S0_4}$ SB-II XXIV | |
| $H5_5$ | $Al_2SiO_4F_2$ (topaz) | SB-II 117 | $\boldsymbol{S0_5}$ SB-II XXIV | |
| $\boldsymbol{H5_6}$ | $Na_6Mg_2SO_4(CO_3)_4$ (tychite) | SB-II 98 | | A4B2C6D16E_cF232_227_e_d_f_eg_a |
| $\boldsymbol{H5_7}$ | $Ca_5F(PO_4)_3$ (fluorapatite) | SB-II 99 | | A5BC12D3_hP42_176_fh_a_2hi_h |
| $\boldsymbol{H5_8}$ | $Na_6ClF(SO_4)_2$ (sulphohalite) | SB-III 118 | | ABC6D8E2_cF72_225_b_a_e_f_c |
| $\boldsymbol{H5_9}$ | $Ca(UO_2)_2(PO_4)_2\cdot \frac{1}{2}H_2O$ (autunite) | SB-VI 24 | | AB2C2_tI10_139_a_d_e |
| $\boldsymbol{H5_{10}}$ | $Ca(UO_2)_2(PO_4)_2\cdot 6H_2O$ (meta-autunite I) | SB-VI 25 | | AB4C6DE_tP26_129_c_j_2ci_a_c |
| $H6_1$ | $K_2PtCl_6$ | SB-I 429 | $H61$ $I1_1$ SB-II XXIII<br>$\boldsymbol{J1_1}$ SB-III 121 | |
| $\boldsymbol{H6_2}$ | $K_2Sn(OH)_6$ | SB-I 431 | $I1_2$ SB-II XXIII | A6B2C6D_hR15_148_f_c_f_a |
| $\boldsymbol{H6_3}$ | $K_2Pt(SCN)_6$ | SB-I 433 | $I1_3$ SB-II 102 | A2BC6_hP9_164_d_a_i |
| $\boldsymbol{H6_4}$ | $Ni(NO_3)_2(NH_3)_6$ | SB-I 435 | $H64$ $I1_4$ SB-II XXIII | A2B6CD6_cP60_205_c_d_a_d |
| $H7_1$ | $(NH_4)_3AlF_6$ | SB-I 437 | $H71$ $I2_1$ SB-II XXIII<br>$\boldsymbol{J2_1}$ SB-III 127 | |
| $I1_1$ | $K_2PtCl_6$ | SB-II XXIII | $H61/H6_1$ SB-I 429<br>$\boldsymbol{J1_1}$ SB-III 121 | |
| $I1_2$ | $K_2Sn(OH)_6$ | SB-II XXIII | $H62/H6_2$ SB-I 431<br>$\boldsymbol{I1_2}$ SB-II XXIII | |
| $\boldsymbol{I1_3}$ | $SrCl_2\cdot(H_2O)_6$ | SB-II 102 | $H63$ $H6_3$ SB-I 433 | A2B6C_hP9_162_d_k_a |
| $I1_4$ | $Ni(NO_3)_2(NH_3)_6$ | SB-II XXIII | $H64$ $\boldsymbol{H6_4}$ SB-I 435 | |
| $I2_1$ | $(NH_4)_3AlF_6$ | SB-II XXIII | $H71/H7_1$ SB-I 437<br>SB-II XXIII $\boldsymbol{J2_1}$<br>SB-III 127 | |
| $\boldsymbol{I6_1}$ | $Ni(H_2O)_6SnCl_6$ | SB-II 102 | | A6B6CD_hR14_148_f_f_b_a |
| $\boldsymbol{J1_1}$ | $K_2PtCl_6$ | SB-III 121 | $H61/H6_1$ SB-I 429<br>$I1_1$ SB-II XXIII | A6B2C_cF36_225_e_c_a |
| $\boldsymbol{J1_5}$ | $K_2OsO_2Cl_4$ | SB-III 122 | | A4B2C2D_tI18_139_h_d_e_a |
| $\boldsymbol{J1_6}$ | $(NH_4)_2SiF_6$ (bararite) | SB-III 123 | | A6B2C_hP9_164_i_d_a |
| $\boldsymbol{J1_7}$ | $MgCl_2\cdot 6H_2O$ (bischofite) | SB-III 124 | | A2B12CD6_mC42_12_i_2i2j_a_ij |

| *SB* | Prototype | Source | Other SB | AFLOW |
|---|---|---|---|---|
| $J1_8$ | $RhCl_2(NH_3)_5Cl$ | SB-III 125 | | A3B15C5D_oP96_62_cd_3c6d_3cd_c |
| $J1_9$ | $Ag[Co(NH_3)_2(NO_2)_4]$ | SB-IV 60 | | ABC4D2E8_tP32_126_a_b_h_e_k |
| $J1_{10}$ | $Zn(BrO_3)_2 \cdot 6H_2O$ | SB-IV 63 | | A2B6C6D_cP60_205_c_d_d_a |
| $J1_{11}$ | $NaSb(OH)_6$ | SB-VI 26 | | AB6C_tP32_86_d_3g_c |
| $J1_{12}$ | $NaSbF_4(OH)_2$ | SB-VI 27 | | A6BC_hP16_163_i_b_c |
| $J1_{13}$ | $K_2GeF_6$ | SB-VII 34 | | A6BC2_hP9_164_i_a_d |
| $J2_1$ | $(NH_4)_3AlF_6$ | SB-III 34 | $H71/H7_1$ SB-I 437<br>$I2_1$ SB-II XXIII | AB30C16D3_cF200_225_a_ej_2f_bc |
| $J2_2$ | $CrCl_3(H_2O)_6$ | SB-III 127 | | A3BC6_hR20_167_e_b_f |
| $J2_3$ | $Ca_3Al_2(OH)_{12}$ | SB-III 129 | | A2B3C12D12_cI232_230_a_c_h_h |
| $J2_4$ | $K_3Co(NO_2)_6$ | SB-IV 65 | | AB3C6D12_cF88_202_a_bc_e_h |
| $J2_5$ | $Cu_3[Fe(CN)_6]_2 \cdot xH_2O$ | SB-VI 28 | | A6B9CD2E6_cF96_225_e_bf_a_c_e |
| $J2_6$ | $Na_3AlF_6$ (cryolite) | SB-VI 29 | | AB6C3_mP20_14_a_3e_de |
| $J3_1$ | $K_3TlCl_6{\cdot}2H_2O$ | SB-III 131 | | A6B2C3D_tI168_139_egikl2m_ejn_bh2n_acf |
| $K0_1$ | $K_2S_2O_5$ | SB-II 104 | $K1_1$ SB-II XXIII | A2B5C2_mP18_11_2e_e2f_2e |
| $K1_1$ | $KSO_3$ | SB-II 106 | $K2_1$ SB-II XXIII | AB3C_hP30_150_ef_3g_c2d |
| $K1_2$ | $CaSO_3$ | SB-II 107 | $K2_2$ SB-II XXIII | AB3C_hP20_190_ac_i_f |
| $K3_1$ | $Cs_3CoCl_5$ | SB-III 134 | | A5BC3_tI36_140_cl_b_ah |
| $K3_3$ | $Tl_2AlF_5$ | SB-V 20 | | AB5C2_oC32_20_b_a2bc_c |
| $K3_4$ | $NH_4Pb_2Br_5$ | SB-V 22 | | A5BC2_tI32_140_bl_a_h |
| $K3_5$ | $KB_5O_8 \cdot 4H_2O$ (santite) | SB-V 23 | | A5B8CD12_oC104_41_a2b_4b_a_6b |
| $K4_1$ | $(NH_4)SO_4$ | SB-III 136 | | AB4C_mP24_14_e_4e_e.K41 |
| $K5_1$ | $K_2S_3O_6$ | SB-III 138 | | A2B6C3_oP44_62_2c_2c2d_3c |
| $K6_1$ | $ZrP_2O_7$ | SB-III 140 | | A7B2C_cP40_205_bd_c_a |
| $K6_2$ | $K_2NbF_7$ | SB-VII 36 | | A7B2C_mP40_14_7e_2e_e |
| $K7_1$ | $K_3W_2Cl_9$ | SB-III 142 | | A9B3C2_hP28_176_hi_af_f |
| $K7_2$ | $Cs_3Tl_2Cl_9$ | SB-III 145 | | A9B3C2_hR28_167_ef_e_c |
| $K7_3$ | $Cs_3As_2Cl_9$ | SB-III 145 | | A2B9C3_hP14_150_d_eg_ad |
| $K7_4$ | $12CaO \cdot 7Al_2O_3$ (mayenite) | SB-IV 68 | | A7B12C19_cI152_220_bc_2d_ace |
| $K7_5$ | $Na_5Al_3F_{14}$ (chiolite) | SB-VI 30 | | A3B14C5_tP44_128_ac_ehi_bg |
| $K7_6$ | $AuCsCl_3$ | SB-VI 33 | | AB3C_tI20_139_ab_eh_d |
| $L1_0$ | CuAu | SB-I 484 | $L10$ | AB_tP2_123_a_d |
| $L1_1$ | CuPt (rhombohedral) | SB-I 485 | $L11$ | AB_hR2_166_a_b |
| $L1_2$ | $Cu_3Au$ (bogdanovite) | SB-I 496 | $L12$ | AB3_cP4_221_a_c |
| $L1_3(I)$ | CuPt (cubic) | SB-I 486 | $L13$ | AB_cF32_227_c_d. |
| $L1_3(II)$ | $CdPt_3$ | Hart | | AB3_oC8_65_a_bf |
| $L1_a$ | $CuPt_3$ | Smi 236 | | AB7_cF32_225_b_ad |
| $L'1_0$ | $\gamma$-$Fe_4N$ | SB-I 487 | | A4B_cP5_221_bc_a |
| $L'1_2$ | $AlFe_3C$ | Smi 237 | $E2_1$ SB-II 49 | |
| $L2_0$ | PdCu | Sb-I 487 | **B2** SB-I 72 | |
| $L2_1$ | $AlCu_2Mn$ (heusler) | SB-I 488 | | AB2C_cF16_225_a_c_b |
| $L2_2$ | $Sb_2Tl_7$ | SB-I 489 | | A2B7_cI54_229_e_afh |
| $L2_a$ | $\delta$TiCu | Smi 237 | | AB_tP2_123_a_d.CuTi |
| $L'2_0$ | $C_xFe$ ("martensite type") | SB-I 489 | | AB_tI4_139_b_a |
| $L'3_0$ | $Fe_2N$ | Smi 237 Per 121 | | AB_hP4_194_c_a.Fe2N |
| $L'3_2$ | $\beta$-$V_2N$ | Gmelin 371 | | AB2_hP9_162_ad_k |
| $L6_0$ | $CuTi_3$ | Smi 237 | | AB3_tP4_123_a_ce |
| $L'6_0$ | $M_4N_x$ | Smi 237 | insufficient structural information | |
| $S0_1$ | $Al_2SiO_5$ (kyanite) | SB-II 109 | $H51$ $H5_1$ SB-I 427 | A2B5C_aP32_2_4i_10i_2i |
| $S0_2$ | $Al_2SiO_5$ (andalusite) | SB-II 110 | $H5_3$ SB-II XXIII | A2B5C_oP32_58_eg_3gh_g |
| $S0_3$ | $Al_2SiO_5$ (sillimanite) | SB-II 112 | $H5_2$ SB-II XXIII | A2B5C_oP32_62_bc_3cd_c |
| $S0_4$ | *$Fe(OH)_2Al_4SiO_{10}$ (staurolite)* | SB-II 113 | $H5_4$ SB-II XXIV | A4BC12D2_oC76_63_eg_c_f3gh_g |

| *SB* | Prototype | Source | Other SB | AFLOW |
|---|---|---|---|---|
| ***S*0$_5$** | $Al_2SiO_4F_2$ (topaz) | SB-II 116 | *H*5$_5$ SB-II XXIV | A2B2C4D_oP36_62_d_d_2cd_c |
| ***S*0$_6$** | $CaTiSiO_5$ (titanite) | SB-II 117 | | AB5CD_mC32_15_e_e2f_e_b |
| ***S*0$_7$** | $Mg(F,OH)_2{\cdot}Mg_2SiO_6$ (norbergite) | SB-II 117 | | A2B3C4D_oP40_62_d_cd_2cd_c |
| ***S*0$_8$** | $Al_{13}(OH,F)_{18}Si_5O_{20}Cl$ (zunyite) | SB-III 147 | | A13BC18D20E5_cF228_216_dh_b_fh_2eh_ce |
| ***S*1$_1$** | $ZrSiO_4$ (zircon) | SB-II XXIV | *H*3 *H*0$_3$ SB-I 345 | A4BC_tI24_141_h_b_a |
| ***S*1$_2$** | $Mg_2SiO_4$ (fosterite/olivine) | SB-II XXIV | *H*12 *H*1$_2$ SB-I 352 | A2B4C_oP28_62_ac_2cd_c |
| ***S*1$_3$** | $Be_2SiO_4$ (phenakite) | SB-II XXIV | *H*13 *H*1$_3$ SB-I 356 | A2B4C_hR42_148_2f_4f_f |
| ***S*1$_4$** | $Co_3Al_2(SiO_4)_3$ (garnet) | SB-II XXIV | *H*31 *H*3$_1$ SB-I 363 | A2B3C12D3_cI160_230_a_c_h_d |
| ***S*1$_5$** | $Bi_4(SiO_4)_3$ (eulytine) | SB-II 122 | | A4B12C3_cI76_220_c_e_a |
| ***S*2$_1$** | $(Sc,Y)_2Si_2O_7$ (thorveitite) | SB-II 124 | | A7B2C2_mC22_12_aij_h_i |
| ***S*2$_2$** | $Zn_4Si_2O_7(OH)_2{\cdot}H_2O$ (hemimorphite) | SB-II 125 | | A2B5CD2_oI40_44_2c_abcde_d_e |
| ***S*2$_3$** | $Ca_{10}Al_4(Mg,Fe)_2Si_9O_{34}(OH)_4$ | SB-II 126 | | A4B10C2D34E4F9_tP252_126_k_ce2k_f_h8k_k_d2k |
| ***S*3$_1$** | $Be_3Al_2(SiO_3)_6$ (beryl) | SB-II XXIV | *G*31 *G*3$_1$ SB-I 305 | A2B3C18D6_hP58_192_c_f_lm_l |
| ***S*3$_2$** | $BaSi_4O_9$ | SB-II 128 | | AB9C4_hP28_188_e_kl_ak |
| ***S*3$_3$(*I*)** | $Na_6Ca_2Al_6Si_6O_{24}(CO_3)_2$ (crancrinite) | SB-III 150 | | A3BCD3E15F3_hP52_173_c_b_b_c_5c_c |
| ***S*3$_3$(*II*)** | $CaSiO_3$ (parawollastonite) | SB-IV 71 | | AB3C_mP60_14_3e_9e_3e |
| ***S*3$_4$(*I*)** | $Ca_{1.4}Sr_{0.3}Al_{3.8}Si_{8.3}O_{24}{\cdot}13H_2O$ (chabazite) | SB-III 151 | | A5B21C24D12_hR62_166_a2c_ehi_fg2h_i |
| ***S*3$_4$(*II*)** | *$Na_2Zr(SiO_2)_3{\cdot}H_2O$ (catapleiite)* | SB-V 24 | | A3B2C9D3E_hP36_194_g_f_hk_h_a |
| ***S*4$_1$** | $CaMg(SiO_3)_2$ (diopside) | SB-II 130 | | ABC6D2_mC40_15_e_e_3f_f.S4_1 |
| ***S*4$_2$** | $Ca_2Mg_5Si_8O_{22}(OH)_2$ (tremolite) | SB-II 131 | | A2B2C5D24E8_mC82_12_h_i_agh_2i5j_2j |
| ***S*4$_3$** | $MgSiO_3$ (enstatite) | SB-II 134 | | AB3C_oP80_61_2c_6c_2c |
| ***S*4$_4$** | $H_2Mg_5Fe_2Si_8O_{24}$ (anthophyllite) | SB-II 137 | | A2B5C22D2E8_oP156_62_d_c2d_2c10d_2c_4d |
| ***S*4$_5$** | $H_4Mg_3Si_2O_9$ (chrysotile) | SB-II 139 | | AB6C11D6E4_mC112_12_e_gi2j_i5j_2i2j_2j |
| ***S*4$_6$** | $Be_4Si_2O_7(OH)_2$ (bertrandite) | SB-II 141 | | A4B7C2D2_oC60_36_2b_a3b_2a_b |
| ***S*4$_7$** | $BeHNaO_8Si_3$ (epididymite) | SB-III 153 | | ABCD8E3_oP112_62_d_2c_d_4c6d_3d |
| ***S*5$_1$** | $KH_2Al_3Si_3O_{12}$ (muscovite) | SB-II 143 | | A2BC10D2E4_mC76_15_f_e_5f_f_2f |
| ***S*5$_2$** | $KCa_4Si_8O_{20}F{\cdot}8H_2O$ (apophyllite) | SB-II 145 | | A4BC16DE28F8_tP116_128_h_a_2i_b_g3i_i |
| ***S*5$_3$** | $Ca_2MgSi_2O_7$ (akermanite) | SB-II 146 | | A2BC7D2_tP24_113_e_a_cef_e |
| ***S*5$_4$** | $Al_2Si_2O_5(OH)_4$ (nacrite) | SB-III 155/VII 37 | | A2B4C9D2_mC68_9_2a_4a_9a_2a |
| ***S*5$_5$** | $Al_2Mg_5Si_3O_{10}(OH)_8$ | SB-III 156 | | A5B10C8D4_mC108_15_a2ef_5f_4f_2f |
| ***S*5$_6$** | $AlSi_2O_5(OH)$ (pyrophyllite) | SB-III 159 | | AB5CD2_mC72_15_f_5f_f_2f |
| ***S*5$_7$** | $Fe(Fe,Si)[(OH)_2O]O_3OH$ (cronstedtite) | SB-VII 40 | | AB3C2D_hR7_160_a_b_2a_a |
| ***S*6$_1$** | $NaAlSi_2O_6{\cdot}H_2O$ (analcime) | SB-II 148 | | A2B2C3D12E4_tI184_142_f_f_be_3g_g |
| ***S*6$_2$** | $Na_4(AlSiO_4)_3Cl$ (sodalite) | SB-II 150 | | A3BC4D12E3_cP46_218_d_a_e_i_c |
| ***S*6$_3$** | $CaB_2Si_2O_8$ (danburite) | SB-II 153 | | A2BC8D2_oP52_62_d_c_2c3d_d |
| ***S*6$_4$** | $Na_4Cl(AlSi_3)_3O_{24}$ (scapolite) | SB-II 155 | | AB4C24D12_tI82_87_a_h_2h2i_hi |
| ***S*6$_5$** | $NaAlSiO_4$ ($\alpha$-carnegieite) | SB-II 158 | | ABC4D_cP28_198_a_a_ab_a |
| ***S*6$_6$** | $Na_2CaSiO_4$ | SB-II 159 | | AB2C4D_cP32_198_a_2a_ab_a |
| ***S*6$_7$** | $KAlSi_3O_8$ (sanidine) | SB-III 161 | | AB8C4_mC52_12_i_gi3j_2j |
| ***S*6$_8$** | $NaAlSi_3O_8$ (albite) | SB-III 164 | | ABC8D3_aP26_2_i_i_8i_3i |
| ***S*6$_9$** | $(Na_{0.5}Ca_{0.3}K_{0.2})_8(Al_6Si_6O_{24})(SO_4)_{1.5}$ (hauyne) | SB-III 166 | | A3B4C4D4E16F4G3_cP76_218_c_e_e_e_ei_e_d |
| ***S*6$_{10}$** | $Na_2Al_2Si_3O_{10}{\cdot}2H_2O$ (natrolite) | SB-III 168 | | A2B4C2D12E3_oF184_43_b_2b_b_6b_ab |

## 6. Future work

New structure-types are constantly being investigated for inclusion into the Prototype Encyclopedia. AFLOW's symmetry (AFLOW-SYM) [58] and crystal prototype finder (AFLOW-XtalFinder) [37] software modules are actively being employed to identify and classify prototypes from literature (*e.g.*, ICSD [59]). Unique prototypes — after comparing against those in Parts 1, 2, and 3 — will be added to future installments of the Prototype Encyclopedia. The symmetries of the prototypes (*i.e.*, Pearson symbol, space group, and Wyckoff positions) will be calculated autonomously and self-consistently via AFLOW-SYM [58]. AFLOW-XtalFinder will map structures into the degrees of freedom consistent with the International Tables for Crystallography, determine the AFLOW prototype label and parameters, and create geometry files for structures in this representation with a symbolic prototype generator [37].

## 7. Users contributions

Online functionality has been developed to analyze the prototype of a user's input structure (http://aflow.org/prototype-encyclopedia/xtal-finder.html). The structure is automatically compared to all existing prototype entries via AFLOW-XtalFinder [37]. If a matching entry is found, the relevant prototype information is returned (*e.g.*, prototype label and parameters, *Strukturbericht* designation, symmetry descriptions, and weblink to the entry). If no matching entry is found — signaling a new prototype — users can report the structure by providing the following information in an online form: **i.** geometry file (in any standardized format), **ii.** reference(s) that reported the structure, **iii.** supplemental reference(s) used to find the structure (*e.g.*, catalogs or online databases), **iv.** any notable comments, and **v.** uploader's information (to credit the contributor on that structure's entry page, if desired). Unique prototype suggestions will be collected for future enrichment of the encyclopedia.

## 8. Outreach and education

AFLOW Schools (http://aflow.org/aflow-school/) provide learning modules to teach the theory of and provide hands-on examples for analyzing the structure of materials; including crystallographic symmetry, prototype classification, and structural similarity. Modules are also available for studying thermodynamics (polar and non-polar materials), accessing data in AFLOW.org (APIs), calculating thermomechanical properties, modeling disordered compounds, and predicting the properties of materials with machine learning.

## 9. Conclusion

This article presents the newest additions to the *AFLOW Library of Crystallographic Prototypes*, which now contains 1,100 entries. This includes all of the inorganic crystal structures that were given *Strukturbericht* designations in the original volumes or by references found in the literature. We also provide a brief history of the development of the *Strukturbericht* designations and the changes to the system over time.

## Acknowledgments

This work has been supported by DOD-ONR (N00014-15-1-2863, N00014-16-1-2326, N00014-17-1-2876). David Hicks acknowledges support from the Department of Defense through the National Defense Science and Engineering Graduate (NDSEG) Fellowship Program. Michael Mehl would like to thank Lauren Mehl, Stephanie Mehl, and Gus Hart for the donation of several of the original *Strukturbericht* volumes as well as the Trotter and Bree cumulative index. The authors thank Eric Gossett, Demet Usanmaz, Andriy Smolyanyuk, Rico Friedrich, Frisco Rose, Yoav Lederer, Robert Hanson, Matthias Scheffler, Christian Carbogno, Paul Saxe, and Xiomara Campilongo for valuable discussions.

# $NaC_5H_{11}O_8S$ Structure: A5B11CD8E_aP26_1_5a_11a_a_8a_a

http://aflow.org/prototype-encyclopedia/A5B11CD8E_aP26_1_5a_11a_a_8a_a

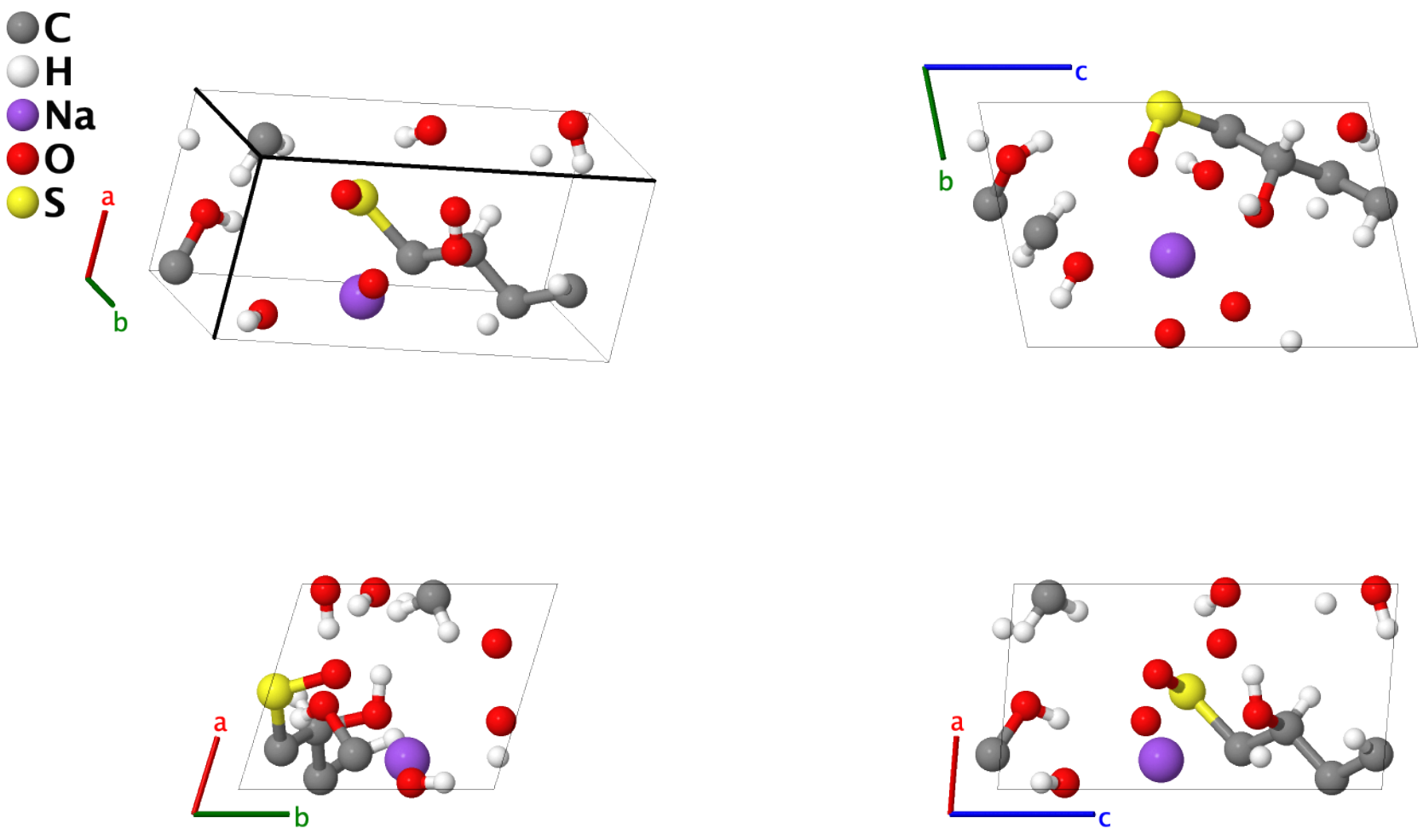

| | | |
|---|---|---|
| **Prototype** | : | $C_5H_{11}NaO_8S$ |
| **AFLOW prototype label** | : | A5B11CD8E_aP26_1_5a_11a_a_8a_a |
| ***Strukturbericht* designation** | : | None |
| **Pearson symbol** | : | aP26 |
| **Space group number** | : | 1 |
| **Space group symbol** | : | $P1$ |
| **AFLOW prototype command** | : | `aflow --proto=A5B11CD8E_aP26_1_5a_11a_a_8a_a --params=`$a, b/a, c/a, \alpha, \beta, \gamma, x_1, y_1, z_1, x_2, y_2, z_2, x_3, y_3, z_3, x_4, y_4, z_4, x_5, y_5, z_5, x_6, y_6, z_6, x_7, y_7, z_7, x_8, y_8, z_8, x_9, y_9, z_9, x_{10}, y_{10}, z_{10}, x_{11}, y_{11}, z_{11}, x_{12}, y_{12}, z_{12}, x_{13}, y_{13}, z_{13}, x_{14}, y_{14}, z_{14}, x_{15}, y_{15}, z_{15}, x_{16}, y_{16}, z_{16}, x_{17}, y_{17}, z_{17}, x_{18}, y_{18}, z_{18}, x_{19}, y_{19}, z_{19}, x_{20}, y_{20}, z_{20}, x_{21}, y_{21}, z_{21}, x_{22}, y_{22}, z_{22}, x_{23}, y_{23}, z_{23}, x_{24}, y_{24}, z_{24}, x_{25}, y_{25}, z_{25}, x_{26}, y_{26}, z_{26}$ |

**Triclinic primitive vectors:**

$$\mathbf{a}_1 = a\,\hat{\mathbf{x}}$$

$$\mathbf{a}_2 = b\cos\gamma\,\hat{\mathbf{x}} + b\sin\gamma\,\hat{\mathbf{y}}$$

$$\mathbf{a}_3 = c_x\,\hat{\mathbf{x}} + c_y\,\hat{\mathbf{y}} + c_z\,\hat{\mathbf{z}}$$

$$c_x = c\,\cos\beta$$

$$c_y = c\,(\cos\alpha - \cos\beta\cos\gamma)/\sin\gamma$$

$$c_z = \sqrt{c^2 - c_x^2 - c_y^2}$$

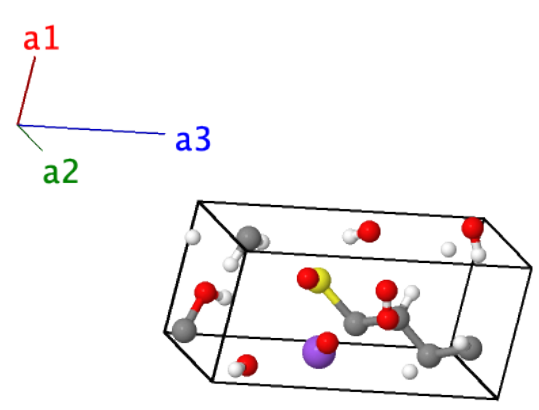

**Basis vectors:**

| | | Lattice Coordinates | | Cartesian Coordinates | Wyckoff Position | Atom Type |
|---|---|---|---|---|---|---|
| $\mathbf{B}_1$ | = | $x_1\,\mathbf{a}_1 + y_1\,\mathbf{a}_2 + z_1\,\mathbf{a}_3$ | = | $(x_1 a + y_1 b\cos\gamma + z_1 c_x)\,\hat{\mathbf{x}} + (y_1 b\sin\gamma + z_1 c_y)\,\hat{\mathbf{y}} + z_1 c_z\,\hat{\mathbf{z}}$ | (1$a$) | C I |
| $\mathbf{B}_{2}$ | = | $x_{2}\,\mathbf{a}_{1} + y_{2}\,\mathbf{a}_{2} + z_{2}\,\mathbf{a}_{3}$ | = | $\left(x_{2}a + y_{2}b\cos\gamma + z_{2}c_{x}\right)\hat{\mathbf{x}} + \left(y_{2}b\sin\gamma + z_{2}c_{y}\right)\hat{\mathbf{y}} + z_{2}c_{z}\hat{\mathbf{z}}$ | $(1a)$ | C II |
| $\mathbf{B}_{3}$ | = | $x_{3}\,\mathbf{a}_{1} + y_{3}\,\mathbf{a}_{2} + z_{3}\,\mathbf{a}_{3}$ | = | $\left(x_{3}a + y_{3}b\cos\gamma + z_{3}c_{x}\right)\hat{\mathbf{x}} + \left(y_{3}b\sin\gamma + z_{3}c_{y}\right)\hat{\mathbf{y}} + z_{3}c_{z}\hat{\mathbf{z}}$ | $(1a)$ | C III |
| $\mathbf{B}_{4}$ | = | $x_{4}\,\mathbf{a}_{1} + y_{4}\,\mathbf{a}_{2} + z_{4}\,\mathbf{a}_{3}$ | = | $\left(x_{4}a + y_{4}b\cos\gamma + z_{4}c_{x}\right)\hat{\mathbf{x}} + \left(y_{4}b\sin\gamma + z_{4}c_{y}\right)\hat{\mathbf{y}} + z_{4}c_{z}\hat{\mathbf{z}}$ | $(1a)$ | C IV |
| $\mathbf{B}_{5}$ | = | $x_{5}\,\mathbf{a}_{1} + y_{5}\,\mathbf{a}_{2} + z_{5}\,\mathbf{a}_{3}$ | = | $\left(x_{5}a + y_{5}b\cos\gamma + z_{5}c_{x}\right)\hat{\mathbf{x}} + \left(y_{5}b\sin\gamma + z_{5}c_{y}\right)\hat{\mathbf{y}} + z_{5}c_{z}\hat{\mathbf{z}}$ | $(1a)$ | C V |
| $\mathbf{B}_{6}$ | = | $x_{6}\,\mathbf{a}_{1} + y_{6}\,\mathbf{a}_{2} + z_{6}\,\mathbf{a}_{3}$ | = | $\left(x_{6}a + y_{6}b\cos\gamma + z_{6}c_{x}\right)\hat{\mathbf{x}} + \left(y_{6}b\sin\gamma + z_{6}c_{y}\right)\hat{\mathbf{y}} + z_{6}c_{z}\hat{\mathbf{z}}$ | $(1a)$ | H I |
| $\mathbf{B}_{7}$ | = | $x_{7}\,\mathbf{a}_{1} + y_{7}\,\mathbf{a}_{2} + z_{7}\,\mathbf{a}_{3}$ | = | $\left(x_{7}a + y_{7}b\cos\gamma + z_{7}c_{x}\right)\hat{\mathbf{x}} + \left(y_{7}b\sin\gamma + z_{7}c_{y}\right)\hat{\mathbf{y}} + z_{7}c_{z}\hat{\mathbf{z}}$ | $(1a)$ | H II |
| $\mathbf{B}_{8}$ | = | $x_{8}\,\mathbf{a}_{1} + y_{8}\,\mathbf{a}_{2} + z_{8}\,\mathbf{a}_{3}$ | = | $\left(x_{8}a + y_{8}b\cos\gamma + z_{8}c_{x}\right)\hat{\mathbf{x}} + \left(y_{8}b\sin\gamma + z_{8}c_{y}\right)\hat{\mathbf{y}} + z_{8}c_{z}\hat{\mathbf{z}}$ | $(1a)$ | H III |
| $\mathbf{B}_{9}$ | = | $x_{9}\,\mathbf{a}_{1} + y_{9}\,\mathbf{a}_{2} + z_{9}\,\mathbf{a}_{3}$ | = | $\left(x_{9}a + y_{9}b\cos\gamma + z_{9}c_{x}\right)\hat{\mathbf{x}} + \left(y_{9}b\sin\gamma + z_{9}c_{y}\right)\hat{\mathbf{y}} + z_{9}c_{z}\hat{\mathbf{z}}$ | $(1a)$ | H IV |
| $\mathbf{B}_{10}$ | = | $x_{10}\,\mathbf{a}_{1} + y_{10}\,\mathbf{a}_{2} + z_{10}\,\mathbf{a}_{3}$ | = | $\left(x_{10}a + y_{10}b\cos\gamma + z_{10}c_{x}\right)\hat{\mathbf{x}} + \left(y_{10}b\sin\gamma + z_{10}c_{y}\right)\hat{\mathbf{y}} + z_{10}c_{z}\hat{\mathbf{z}}$ | $(1a)$ | H V |
| $\mathbf{B}_{11}$ | = | $x_{11}\,\mathbf{a}_{1} + y_{11}\,\mathbf{a}_{2} + z_{11}\,\mathbf{a}_{3}$ | = | $\left(x_{11}a + y_{11}b\cos\gamma + z_{11}c_{x}\right)\hat{\mathbf{x}} + \left(y_{11}b\sin\gamma + z_{11}c_{y}\right)\hat{\mathbf{y}} + z_{11}c_{z}\hat{\mathbf{z}}$ | $(1a)$ | H VI |
| $\mathbf{B}_{12}$ | = | $x_{12}\,\mathbf{a}_{1} + y_{12}\,\mathbf{a}_{2} + z_{12}\,\mathbf{a}_{3}$ | = | $\left(x_{12}a + y_{12}b\cos\gamma + z_{12}c_{x}\right)\hat{\mathbf{x}} + \left(y_{12}b\sin\gamma + z_{12}c_{y}\right)\hat{\mathbf{y}} + z_{12}c_{z}\hat{\mathbf{z}}$ | $(1a)$ | H VII |
| $\mathbf{B}_{13}$ | = | $x_{13}\,\mathbf{a}_{1} + y_{13}\,\mathbf{a}_{2} + z_{13}\,\mathbf{a}_{3}$ | = | $\left(x_{13}a + y_{13}b\cos\gamma + z_{13}c_{x}\right)\hat{\mathbf{x}} + \left(y_{13}b\sin\gamma + z_{13}c_{y}\right)\hat{\mathbf{y}} + z_{13}c_{z}\hat{\mathbf{z}}$ | $(1a)$ | H VIII |
| $\mathbf{B}_{14}$ | = | $x_{14}\,\mathbf{a}_{1} + y_{14}\,\mathbf{a}_{2} + z_{14}\,\mathbf{a}_{3}$ | = | $\left(x_{14}a + y_{14}b\cos\gamma + z_{14}c_{x}\right)\hat{\mathbf{x}} + \left(y_{14}b\sin\gamma + z_{14}c_{y}\right)\hat{\mathbf{y}} + z_{14}c_{z}\hat{\mathbf{z}}$ | $(1a)$ | H IX |
| $\mathbf{B}_{15}$ | = | $x_{15}\,\mathbf{a}_{1} + y_{15}\,\mathbf{a}_{2} + z_{15}\,\mathbf{a}_{3}$ | = | $\left(x_{15}a + y_{15}b\cos\gamma + z_{15}c_{x}\right)\hat{\mathbf{x}} + \left(y_{15}b\sin\gamma + z_{15}c_{y}\right)\hat{\mathbf{y}} + z_{15}c_{z}\hat{\mathbf{z}}$ | $(1a)$ | H X |
| $\mathbf{B}_{16}$ | = | $x_{16}\,\mathbf{a}_{1} + y_{16}\,\mathbf{a}_{2} + z_{16}\,\mathbf{a}_{3}$ | = | $\left(x_{16}a + y_{16}b\cos\gamma + z_{16}c_{x}\right)\hat{\mathbf{x}} + \left(y_{16}b\sin\gamma + z_{16}c_{y}\right)\hat{\mathbf{y}} + z_{16}c_{z}\hat{\mathbf{z}}$ | $(1a)$ | H XI |
| $\mathbf{B}_{17}$ | = | $x_{17}\,\mathbf{a}_{1} + y_{17}\,\mathbf{a}_{2} + z_{17}\,\mathbf{a}_{3}$ | = | $\left(x_{17}a + y_{17}b\cos\gamma + z_{17}c_{x}\right)\hat{\mathbf{x}} + \left(y_{17}b\sin\gamma + z_{17}c_{y}\right)\hat{\mathbf{y}} + z_{17}c_{z}\hat{\mathbf{z}}$ | $(1a)$ | Na |
| $\mathbf{B}_{18}$ | = | $x_{18}\,\mathbf{a}_{1} + y_{18}\,\mathbf{a}_{2} + z_{18}\,\mathbf{a}_{3}$ | = | $\left(x_{18}a + y_{18}b\cos\gamma + z_{18}c_{x}\right)\hat{\mathbf{x}} + \left(y_{18}b\sin\gamma + z_{18}c_{y}\right)\hat{\mathbf{y}} + z_{18}c_{z}\hat{\mathbf{z}}$ | $(1a)$ | O I |
| $\mathbf{B}_{19}$ | = | $x_{19}\,\mathbf{a}_{1} + y_{19}\,\mathbf{a}_{2} + z_{19}\,\mathbf{a}_{3}$ | = | $\left(x_{19}a + y_{19}b\cos\gamma + z_{19}c_{x}\right)\hat{\mathbf{x}} + \left(y_{19}b\sin\gamma + z_{19}c_{y}\right)\hat{\mathbf{y}} + z_{19}c_{z}\hat{\mathbf{z}}$ | $(1a)$ | O II |
| $\mathbf{B}_{20}$ | = | $x_{20}\,\mathbf{a}_{1} + y_{20}\,\mathbf{a}_{2} + z_{20}\,\mathbf{a}_{3}$ | = | $\left(x_{20}a + y_{20}b\cos\gamma + z_{20}c_{x}\right)\hat{\mathbf{x}} + \left(y_{20}b\sin\gamma + z_{20}c_{y}\right)\hat{\mathbf{y}} + z_{20}c_{z}\hat{\mathbf{z}}$ | $(1a)$ | O III |
| $\mathbf{B}_{21}$ | = | $x_{21}\,\mathbf{a}_{1} + y_{21}\,\mathbf{a}_{2} + z_{21}\,\mathbf{a}_{3}$ | = | $\left(x_{21}a + y_{21}b\cos\gamma + z_{21}c_{x}\right)\hat{\mathbf{x}} + \left(y_{21}b\sin\gamma + z_{21}c_{y}\right)\hat{\mathbf{y}} + z_{21}c_{z}\hat{\mathbf{z}}$ | $(1a)$ | O IV |
| $\mathbf{B}_{22}$ | = | $x_{22}\,\mathbf{a}_{1} + y_{22}\,\mathbf{a}_{2} + z_{22}\,\mathbf{a}_{3}$ | = | $\left(x_{22}a + y_{22}b\cos\gamma + z_{22}c_{x}\right)\hat{\mathbf{x}} + \left(y_{22}b\sin\gamma + z_{22}c_{y}\right)\hat{\mathbf{y}} + z_{22}c_{z}\hat{\mathbf{z}}$ | $(1a)$ | O V |
| $\mathbf{B}_{23}$ | = | $x_{23}\,\mathbf{a}_{1} + y_{23}\,\mathbf{a}_{2} + z_{23}\,\mathbf{a}_{3}$ | = | $\left(x_{23}a + y_{23}b\cos\gamma + z_{23}c_{x}\right)\hat{\mathbf{x}} + \left(y_{23}b\sin\gamma + z_{23}c_{y}\right)\hat{\mathbf{y}} + z_{23}c_{z}\hat{\mathbf{z}}$ | $(1a)$ | O VI |
| $\mathbf{B}_{24}$ | = | $x_{24}\,\mathbf{a}_1 + y_{24}\,\mathbf{a}_2 + z_{24}\,\mathbf{a}_3$ | = | $(x_{24}a + y_{24}b\cos\gamma + z_{24}c_x)\,\hat{\mathbf{x}} + \left(y_{24}b\sin\gamma + z_{24}c_y\right)\,\hat{\mathbf{y}} + z_{24}c_z\,\hat{\mathbf{z}}$ | (1$a$) | O VII |
| $\mathbf{B}_{25}$ | = | $x_{25}\,\mathbf{a}_1 + y_{25}\,\mathbf{a}_2 + z_{25}\,\mathbf{a}_3$ | = | $(x_{25}a + y_{25}b\cos\gamma + z_{25}c_x)\,\hat{\mathbf{x}} + \left(y_{25}b\sin\gamma + z_{25}c_y\right)\,\hat{\mathbf{y}} + z_{25}c_z\,\hat{\mathbf{z}}$ | (1$a$) | O VIII |
| $\mathbf{B}_{26}$ | = | $x_{26}\,\mathbf{a}_1 + y_{26}\,\mathbf{a}_2 + z_{26}\,\mathbf{a}_3$ | = | $(x_{26}a + y_{26}b\cos\gamma + z_{26}c_x)\,\hat{\mathbf{x}} + \left(y_{26}b\sin\gamma + z_{26}c_y\right)\,\hat{\mathbf{y}} + z_{26}c_z\,\hat{\mathbf{z}}$ | (1$a$) | S |

---

---

**Geometry files:**

- CIF: pp. 1500
- POSCAR: pp. 1500

# $Ni(NO_3)_2(H_2O)_6$ Structure: A12B2CD12_aP54_2_12i_2i_i_12i

http://aflow.org/prototype-encyclopedia/A12B2CD12_aP54_2_12i_2i_i_12i

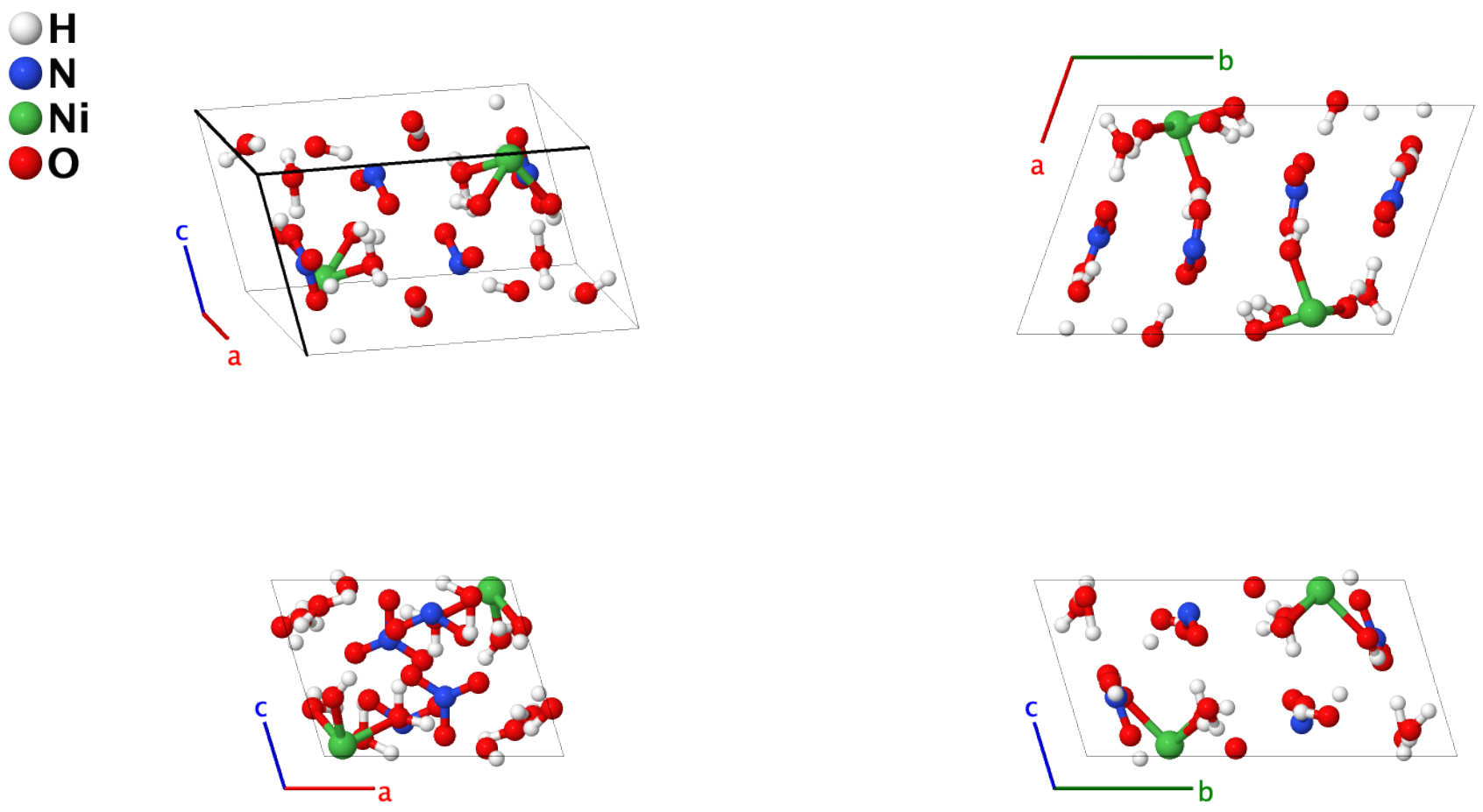

| | | |
|---|---|---|
| **Prototype** | : | $H_{12}N_2NiO_{12}$ |
| **AFLOW prototype label** | : | A12B2CD12_aP54_2_12i_2i_i_12i |
| ***Strukturbericht* designation** | : | None |
| **Pearson symbol** | : | aP54 |
| **Space group number** | : | 2 |
| **Space group symbol** | : | $P\bar{1}$ |
| **AFLOW prototype command** | : | `aflow --proto=A12B2CD12_aP54_2_12i_2i_i_12i --params=`$a, b/a, c/a, \alpha, \beta, \gamma, x_1, y_1, z_1, x_2, y_2, z_2, x_3, y_3, z_3, x_4, y_4, z_4, x_5, y_5, z_5, x_6, y_6, z_6, x_7, y_7, z_7, x_8, y_8, z_8, x_9, y_9, z_9, x_{10}, y_{10}, z_{10}, x_{11}, y_{11}, z_{11}, x_{12}, y_{12}, z_{12}, x_{13}, y_{13}, z_{13}, x_{14}, y_{14}, z_{14}, x_{15}, y_{15}, z_{15}, x_{16}, y_{16}, z_{16}, x_{17}, y_{17}, z_{17}, x_{18}, y_{18}, z_{18}, x_{19}, y_{19}, z_{19}, x_{20}, y_{20}, z_{20}, x_{21}, y_{21}, z_{21}, x_{22}, y_{22}, z_{22}, x_{23}, y_{23}, z_{23}, x_{24}, y_{24}, z_{24}, x_{25}, y_{25}, z_{25}, x_{26}, y_{26}, z_{26}, x_{27}, y_{27}, z_{27}$ |

- This is the accepted structure for $Ni(NO_3)_2(H_2O)_6$, replacing the $H6_4$ structure. The positions of the hydrogen atoms are approximate.

**Triclinic primitive vectors:**

$$\mathbf{a}_1 = a\,\hat{\mathbf{x}}$$

$$\mathbf{a}_2 = b\cos\gamma\,\hat{\mathbf{x}} + b\sin\gamma\,\hat{\mathbf{y}}$$

$$\mathbf{a}_3 = c_x\,\hat{\mathbf{x}} + c_y\,\hat{\mathbf{y}} + c_z\,\hat{\mathbf{z}}$$

$$c_x = c\,\cos\beta$$

$$c_y = c\,(\cos\alpha - \cos\beta\cos\gamma)/\sin\gamma$$

$$c_z = \sqrt{c^2 - c_x^2 - c_y^2}$$

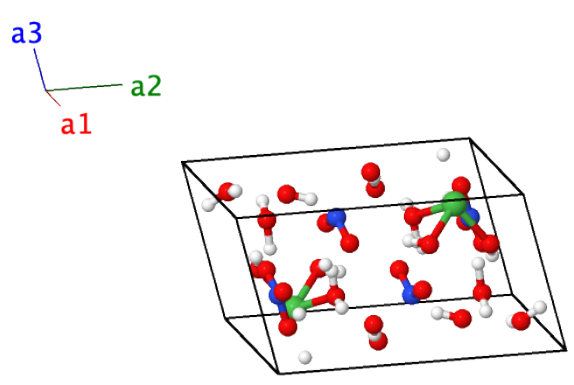

**Basis vectors:**

| | | Lattice Coordinates | | Cartesian Coordinates | Wyckoff Position | Atom Type |
|---|---|---|---|---|---|---|
| $\mathbf{B}_{1}$ | = | $x_{1}\,\mathbf{a}_{1}+y_{1}\,\mathbf{a}_{2}+z_{1}\,\mathbf{a}_{3}$ | = | $\left(x_{1}a+y_{1}b\cos\gamma+z_{1}c_{x}\right)\,\mathbf{\hat{x}}+\left(y_{1}b\sin\gamma+z_{1}c_{y}\right)\,\mathbf{\hat{y}}+z_{1}c_{z}\,\mathbf{\hat{z}}$ | $(2i)$ | H I |
| $\mathbf{B}_{2}$ | = | $-x_{1}\,\mathbf{a}_{1}-y_{1}\,\mathbf{a}_{2}-z_{1}\,\mathbf{a}_{3}$ | = | $\left(-x_{1}a-y_{1}b\cos\gamma-z_{1}c_{x}\right)\,\mathbf{\hat{x}}+\left(-y_{1}b\sin\gamma-z_{1}c_{y}\right)\,\mathbf{\hat{y}}-z_{1}c_{z}\,\mathbf{\hat{z}}$ | $(2i)$ | H I |
| $\mathbf{B}_{3}$ | = | $x_{2}\,\mathbf{a}_{1}+y_{2}\,\mathbf{a}_{2}+z_{2}\,\mathbf{a}_{3}$ | = | $\left(x_{2}a+y_{2}b\cos\gamma+z_{2}c_{x}\right)\,\mathbf{\hat{x}}+\left(y_{2}b\sin\gamma+z_{2}c_{y}\right)\,\mathbf{\hat{y}}+z_{2}c_{z}\,\mathbf{\hat{z}}$ | $(2i)$ | H II |
| $\mathbf{B}_{4}$ | = | $-x_{2}\,\mathbf{a}_{1}-y_{2}\,\mathbf{a}_{2}-z_{2}\,\mathbf{a}_{3}$ | = | $\left(-x_{2}a-y_{2}b\cos\gamma-z_{2}c_{x}\right)\,\mathbf{\hat{x}}+\left(-y_{2}b\sin\gamma-z_{2}c_{y}\right)\,\mathbf{\hat{y}}-z_{2}c_{z}\,\mathbf{\hat{z}}$ | $(2i)$ | H II |
| $\mathbf{B}_{5}$ | = | $x_{3}\,\mathbf{a}_{1}+y_{3}\,\mathbf{a}_{2}+z_{3}\,\mathbf{a}_{3}$ | = | $\left(x_{3}a+y_{3}b\cos\gamma+z_{3}c_{x}\right)\,\mathbf{\hat{x}}+\left(y_{3}b\sin\gamma+z_{3}c_{y}\right)\,\mathbf{\hat{y}}+z_{3}c_{z}\,\mathbf{\hat{z}}$ | $(2i)$ | H III |
| $\mathbf{B}_{6}$ | = | $-x_{3}\,\mathbf{a}_{1}-y_{3}\,\mathbf{a}_{2}-z_{3}\,\mathbf{a}_{3}$ | = | $\left(-x_{3}a-y_{3}b\cos\gamma-z_{3}c_{x}\right)\,\mathbf{\hat{x}}+\left(-y_{3}b\sin\gamma-z_{3}c_{y}\right)\,\mathbf{\hat{y}}-z_{3}c_{z}\,\mathbf{\hat{z}}$ | $(2i)$ | H III |
| $\mathbf{B}_{7}$ | = | $x_{4}\,\mathbf{a}_{1}+y_{4}\,\mathbf{a}_{2}+z_{4}\,\mathbf{a}_{3}$ | = | $\left(x_{4}a+y_{4}b\cos\gamma+z_{4}c_{x}\right)\,\mathbf{\hat{x}}+\left(y_{4}b\sin\gamma+z_{4}c_{y}\right)\,\mathbf{\hat{y}}+z_{4}c_{z}\,\mathbf{\hat{z}}$ | $(2i)$ | H IV |
| $\mathbf{B}_{8}$ | = | $-x_{4}\,\mathbf{a}_{1}-y_{4}\,\mathbf{a}_{2}-z_{4}\,\mathbf{a}_{3}$ | = | $\left(-x_{4}a-y_{4}b\cos\gamma-z_{4}c_{x}\right)\,\mathbf{\hat{x}}+\left(-y_{4}b\sin\gamma-z_{4}c_{y}\right)\,\mathbf{\hat{y}}-z_{4}c_{z}\,\mathbf{\hat{z}}$ | $(2i)$ | H IV |
| $\mathbf{B}_{9}$ | = | $x_{5}\,\mathbf{a}_{1}+y_{5}\,\mathbf{a}_{2}+z_{5}\,\mathbf{a}_{3}$ | = | $\left(x_{5}a+y_{5}b\cos\gamma+z_{5}c_{x}\right)\,\mathbf{\hat{x}}+\left(y_{5}b\sin\gamma+z_{5}c_{y}\right)\,\mathbf{\hat{y}}+z_{5}c_{z}\,\mathbf{\hat{z}}$ | $(2i)$ | H V |
| $\mathbf{B}_{10}$ | = | $-x_{5}\,\mathbf{a}_{1}-y_{5}\,\mathbf{a}_{2}-z_{5}\,\mathbf{a}_{3}$ | = | $\left(-x_{5}a-y_{5}b\cos\gamma-z_{5}c_{x}\right)\,\mathbf{\hat{x}}+\left(-y_{5}b\sin\gamma-z_{5}c_{y}\right)\,\mathbf{\hat{y}}-z_{5}c_{z}\,\mathbf{\hat{z}}$ | $(2i)$ | H V |
| $\mathbf{B}_{11}$ | = | $x_{6}\,\mathbf{a}_{1}+y_{6}\,\mathbf{a}_{2}+z_{6}\,\mathbf{a}_{3}$ | = | $\left(x_{6}a+y_{6}b\cos\gamma+z_{6}c_{x}\right)\,\mathbf{\hat{x}}+\left(y_{6}b\sin\gamma+z_{6}c_{y}\right)\,\mathbf{\hat{y}}+z_{6}c_{z}\,\mathbf{\hat{z}}$ | $(2i)$ | H VI |
| $\mathbf{B}_{12}$ | = | $-x_{6}\,\mathbf{a}_{1}-y_{6}\,\mathbf{a}_{2}-z_{6}\,\mathbf{a}_{3}$ | = | $\left(-x_{6}a-y_{6}b\cos\gamma-z_{6}c_{x}\right)\,\mathbf{\hat{x}}+\left(-y_{6}b\sin\gamma-z_{6}c_{y}\right)\,\mathbf{\hat{y}}-z_{6}c_{z}\,\mathbf{\hat{z}}$ | $(2i)$ | H VI |
| $\mathbf{B}_{13}$ | = | $x_{7}\,\mathbf{a}_{1}+y_{7}\,\mathbf{a}_{2}+z_{7}\,\mathbf{a}_{3}$ | = | $\left(x_{7}a+y_{7}b\cos\gamma+z_{7}c_{x}\right)\,\mathbf{\hat{x}}+\left(y_{7}b\sin\gamma+z_{7}c_{y}\right)\,\mathbf{\hat{y}}+z_{7}c_{z}\,\mathbf{\hat{z}}$ | $(2i)$ | H VII |
| $\mathbf{B}_{14}$ | = | $-x_{7}\,\mathbf{a}_{1}-y_{7}\,\mathbf{a}_{2}-z_{7}\,\mathbf{a}_{3}$ | = | $\left(-x_{7}a-y_{7}b\cos\gamma-z_{7}c_{x}\right)\,\mathbf{\hat{x}}+\left(-y_{7}b\sin\gamma-z_{7}c_{y}\right)\,\mathbf{\hat{y}}-z_{7}c_{z}\,\mathbf{\hat{z}}$ | $(2i)$ | H VII |
| $\mathbf{B}_{15}$ | = | $x_{8}\,\mathbf{a}_{1}+y_{8}\,\mathbf{a}_{2}+z_{8}\,\mathbf{a}_{3}$ | = | $\left(x_{8}a+y_{8}b\cos\gamma+z_{8}c_{x}\right)\,\mathbf{\hat{x}}+\left(y_{8}b\sin\gamma+z_{8}c_{y}\right)\,\mathbf{\hat{y}}+z_{8}c_{z}\,\mathbf{\hat{z}}$ | $(2i)$ | H VIII |
| $\mathbf{B}_{16}$ | = | $-x_{8}\,\mathbf{a}_{1}-y_{8}\,\mathbf{a}_{2}-z_{8}\,\mathbf{a}_{3}$ | = | $\left(-x_{8}a-y_{8}b\cos\gamma-z_{8}c_{x}\right)\,\mathbf{\hat{x}}+\left(-y_{8}b\sin\gamma-z_{8}c_{y}\right)\,\mathbf{\hat{y}}-z_{8}c_{z}\,\mathbf{\hat{z}}$ | $(2i)$ | H VIII |
| $\mathbf{B}_{17}$ | = | $x_{9}\,\mathbf{a}_{1}+y_{9}\,\mathbf{a}_{2}+z_{9}\,\mathbf{a}_{3}$ | = | $\left(x_{9}a+y_{9}b\cos\gamma+z_{9}c_{x}\right)\,\mathbf{\hat{x}}+\left(y_{9}b\sin\gamma+z_{9}c_{y}\right)\,\mathbf{\hat{y}}+z_{9}c_{z}\,\mathbf{\hat{z}}$ | $(2i)$ | H IX |
| $\mathbf{B}_{18}$ | = | $-x_{9}\,\mathbf{a}_{1}-y_{9}\,\mathbf{a}_{2}-z_{9}\,\mathbf{a}_{3}$ | = | $\left(-x_{9}a-y_{9}b\cos\gamma-z_{9}c_{x}\right)\,\mathbf{\hat{x}}+\left(-y_{9}b\sin\gamma-z_{9}c_{y}\right)\,\mathbf{\hat{y}}-z_{9}c_{z}\,\mathbf{\hat{z}}$ | $(2i)$ | H IX |
| $\mathbf{B}_{19}$ | = | $x_{10}\,\mathbf{a}_{1}+y_{10}\,\mathbf{a}_{2}+z_{10}\,\mathbf{a}_{3}$ | = | $\left(x_{10}a+y_{10}b\cos\gamma+z_{10}c_{x}\right)\,\mathbf{\hat{x}}+\left(y_{10}b\sin\gamma+z_{10}c_{y}\right)\,\mathbf{\hat{y}}+z_{10}c_{z}\,\mathbf{\hat{z}}$ | $(2i)$ | H X |
| $\mathbf{B}_{20}$ | = | $-x_{10}\,\mathbf{a}_{1}-y_{10}\,\mathbf{a}_{2}-z_{10}\,\mathbf{a}_{3}$ | = | $\left(-x_{10}a-y_{10}b\cos\gamma-z_{10}c_{x}\right)\,\mathbf{\hat{x}}+\left(-y_{10}b\sin\gamma-z_{10}c_{y}\right)\,\mathbf{\hat{y}}-z_{10}c_{z}\,\mathbf{\hat{z}}$ | $(2i)$ | H X |
| $\mathbf{B}_{21}$ | = | $x_{11}\,\mathbf{a}_{1}+y_{11}\,\mathbf{a}_{2}+z_{11}\,\mathbf{a}_{3}$ | = | $\left(x_{11}a+y_{11}b\cos\gamma+z_{11}c_{x}\right)\,\mathbf{\hat{x}}+\left(y_{11}b\sin\gamma+z_{11}c_{y}\right)\,\mathbf{\hat{y}}+z_{11}c_{z}\,\mathbf{\hat{z}}$ | $(2i)$ | H XI |
| $\mathbf{B}_{22}$ | = | $-x_{11}\,\mathbf{a}_{1}-y_{11}\,\mathbf{a}_{2}-z_{11}\,\mathbf{a}_{3}$ | = | $\left(-x_{11}a-y_{11}b\cos\gamma-z_{11}c_{x}\right)\,\mathbf{\hat{x}}+\left(-y_{11}b\sin\gamma-z_{11}c_{y}\right)\,\mathbf{\hat{y}}-z_{11}c_{z}\,\mathbf{\hat{z}}$ | $(2i)$ | H XI |
| $\mathbf{B}_{23}$ | = | $x_{12}\,\mathbf{a}_1 + y_{12}\,\mathbf{a}_2 + z_{12}\,\mathbf{a}_3$ | = | $\left(x_{12}a + y_{12}b\cos\gamma + z_{12}c_x\right)\,\hat{\mathbf{x}} + \left(y_{12}b\sin\gamma + z_{12}c_y\right)\,\hat{\mathbf{y}} + z_{12}c_z\,\hat{\mathbf{z}}$ | (2*i*) | H XII |
| $\mathbf{B}_{24}$ | = | $-x_{12}\,\mathbf{a}_1 - y_{12}\,\mathbf{a}_2 - z_{12}\,\mathbf{a}_3$ | = | $\left(-x_{12}a - y_{12}b\cos\gamma - z_{12}c_x\right)\,\hat{\mathbf{x}} + \left(-y_{12}b\sin\gamma - z_{12}c_y\right)\,\hat{\mathbf{y}} - z_{12}c_z\,\hat{\mathbf{z}}$ | (2*i*) | H XII |
| $\mathbf{B}_{25}$ | = | $x_{13}\,\mathbf{a}_1 + y_{13}\,\mathbf{a}_2 + z_{13}\,\mathbf{a}_3$ | = | $\left(x_{13}a + y_{13}b\cos\gamma + z_{13}c_x\right)\,\hat{\mathbf{x}} + \left(y_{13}b\sin\gamma + z_{13}c_y\right)\,\hat{\mathbf{y}} + z_{13}c_z\,\hat{\mathbf{z}}$ | (2*i*) | N I |
| $\mathbf{B}_{26}$ | = | $-x_{13}\,\mathbf{a}_1 - y_{13}\,\mathbf{a}_2 - z_{13}\,\mathbf{a}_3$ | = | $\left(-x_{13}a - y_{13}b\cos\gamma - z_{13}c_x\right)\,\hat{\mathbf{x}} + \left(-y_{13}b\sin\gamma - z_{13}c_y\right)\,\hat{\mathbf{y}} - z_{13}c_z\,\hat{\mathbf{z}}$ | (2*i*) | N I |
| $\mathbf{B}_{27}$ | = | $x_{14}\,\mathbf{a}_1 + y_{14}\,\mathbf{a}_2 + z_{14}\,\mathbf{a}_3$ | = | $\left(x_{14}a + y_{14}b\cos\gamma + z_{14}c_x\right)\,\hat{\mathbf{x}} + \left(y_{14}b\sin\gamma + z_{14}c_y\right)\,\hat{\mathbf{y}} + z_{14}c_z\,\hat{\mathbf{z}}$ | (2*i*) | N II |
| $\mathbf{B}_{28}$ | = | $-x_{14}\,\mathbf{a}_1 - y_{14}\,\mathbf{a}_2 - z_{14}\,\mathbf{a}_3$ | = | $\left(-x_{14}a - y_{14}b\cos\gamma - z_{14}c_x\right)\,\hat{\mathbf{x}} + \left(-y_{14}b\sin\gamma - z_{14}c_y\right)\,\hat{\mathbf{y}} - z_{14}c_z\,\hat{\mathbf{z}}$ | (2*i*) | N II |
| $\mathbf{B}_{29}$ | = | $x_{15}\,\mathbf{a}_1 + y_{15}\,\mathbf{a}_2 + z_{15}\,\mathbf{a}_3$ | = | $\left(x_{15}a + y_{15}b\cos\gamma + z_{15}c_x\right)\,\hat{\mathbf{x}} + \left(y_{15}b\sin\gamma + z_{15}c_y\right)\,\hat{\mathbf{y}} + z_{15}c_z\,\hat{\mathbf{z}}$ | (2*i*) | Ni |
| $\mathbf{B}_{30}$ | = | $-x_{15}\,\mathbf{a}_1 - y_{15}\,\mathbf{a}_2 - z_{15}\,\mathbf{a}_3$ | = | $\left(-x_{15}a - y_{15}b\cos\gamma - z_{15}c_x\right)\,\hat{\mathbf{x}} + \left(-y_{15}b\sin\gamma - z_{15}c_y\right)\,\hat{\mathbf{y}} - z_{15}c_z\,\hat{\mathbf{z}}$ | (2*i*) | Ni |
| $\mathbf{B}_{31}$ | = | $x_{16}\,\mathbf{a}_1 + y_{16}\,\mathbf{a}_2 + z_{16}\,\mathbf{a}_3$ | = | $\left(x_{16}a + y_{16}b\cos\gamma + z_{16}c_x\right)\,\hat{\mathbf{x}} + \left(y_{16}b\sin\gamma + z_{16}c_y\right)\,\hat{\mathbf{y}} + z_{16}c_z\,\hat{\mathbf{z}}$ | (2*i*) | O I |
| $\mathbf{B}_{32}$ | = | $-x_{16}\,\mathbf{a}_1 - y_{16}\,\mathbf{a}_2 - z_{16}\,\mathbf{a}_3$ | = | $\left(-x_{16}a - y_{16}b\cos\gamma - z_{16}c_x\right)\,\hat{\mathbf{x}} + \left(-y_{16}b\sin\gamma - z_{16}c_y\right)\,\hat{\mathbf{y}} - z_{16}c_z\,\hat{\mathbf{z}}$ | (2*i*) | O I |
| $\mathbf{B}_{33}$ | = | $x_{17}\,\mathbf{a}_1 + y_{17}\,\mathbf{a}_2 + z_{17}\,\mathbf{a}_3$ | = | $\left(x_{17}a + y_{17}b\cos\gamma + z_{17}c_x\right)\,\hat{\mathbf{x}} + \left(y_{17}b\sin\gamma + z_{17}c_y\right)\,\hat{\mathbf{y}} + z_{17}c_z\,\hat{\mathbf{z}}$ | (2*i*) | O II |
| $\mathbf{B}_{34}$ | = | $-x_{17}\,\mathbf{a}_1 - y_{17}\,\mathbf{a}_2 - z_{17}\,\mathbf{a}_3$ | = | $\left(-x_{17}a - y_{17}b\cos\gamma - z_{17}c_x\right)\,\hat{\mathbf{x}} + \left(-y_{17}b\sin\gamma - z_{17}c_y\right)\,\hat{\mathbf{y}} - z_{17}c_z\,\hat{\mathbf{z}}$ | (2*i*) | O II |
| $\mathbf{B}_{35}$ | = | $x_{18}\,\mathbf{a}_1 + y_{18}\,\mathbf{a}_2 + z_{18}\,\mathbf{a}_3$ | = | $\left(x_{18}a + y_{18}b\cos\gamma + z_{18}c_x\right)\,\hat{\mathbf{x}} + \left(y_{18}b\sin\gamma + z_{18}c_y\right)\,\hat{\mathbf{y}} + z_{18}c_z\,\hat{\mathbf{z}}$ | (2*i*) | O III |
| $\mathbf{B}_{36}$ | = | $-x_{18}\,\mathbf{a}_1 - y_{18}\,\mathbf{a}_2 - z_{18}\,\mathbf{a}_3$ | = | $\left(-x_{18}a - y_{18}b\cos\gamma - z_{18}c_x\right)\,\hat{\mathbf{x}} + \left(-y_{18}b\sin\gamma - z_{18}c_y\right)\,\hat{\mathbf{y}} - z_{18}c_z\,\hat{\mathbf{z}}$ | (2*i*) | O III |
| $\mathbf{B}_{37}$ | = | $x_{19}\,\mathbf{a}_1 + y_{19}\,\mathbf{a}_2 + z_{19}\,\mathbf{a}_3$ | = | $\left(x_{19}a + y_{19}b\cos\gamma + z_{19}c_x\right)\,\hat{\mathbf{x}} + \left(y_{19}b\sin\gamma + z_{19}c_y\right)\,\hat{\mathbf{y}} + z_{19}c_z\,\hat{\mathbf{z}}$ | (2*i*) | O IV |
| $\mathbf{B}_{38}$ | = | $-x_{19}\,\mathbf{a}_1 - y_{19}\,\mathbf{a}_2 - z_{19}\,\mathbf{a}_3$ | = | $\left(-x_{19}a - y_{19}b\cos\gamma - z_{19}c_x\right)\,\hat{\mathbf{x}} + \left(-y_{19}b\sin\gamma - z_{19}c_y\right)\,\hat{\mathbf{y}} - z_{19}c_z\,\hat{\mathbf{z}}$ | (2*i*) | O IV |
| $\mathbf{B}_{39}$ | = | $x_{20}\,\mathbf{a}_1 + y_{20}\,\mathbf{a}_2 + z_{20}\,\mathbf{a}_3$ | = | $\left(x_{20}a + y_{20}b\cos\gamma + z_{20}c_x\right)\,\hat{\mathbf{x}} + \left(y_{20}b\sin\gamma + z_{20}c_y\right)\,\hat{\mathbf{y}} + z_{20}c_z\,\hat{\mathbf{z}}$ | (2*i*) | O V |
| $\mathbf{B}_{40}$ | = | $-x_{20}\,\mathbf{a}_1 - y_{20}\,\mathbf{a}_2 - z_{20}\,\mathbf{a}_3$ | = | $\left(-x_{20}a - y_{20}b\cos\gamma - z_{20}c_x\right)\,\hat{\mathbf{x}} + \left(-y_{20}b\sin\gamma - z_{20}c_y\right)\,\hat{\mathbf{y}} - z_{20}c_z\,\hat{\mathbf{z}}$ | (2*i*) | O V |
| $\mathbf{B}_{41}$ | = | $x_{21}\,\mathbf{a}_1 + y_{21}\,\mathbf{a}_2 + z_{21}\,\mathbf{a}_3$ | = | $\left(x_{21}a + y_{21}b\cos\gamma + z_{21}c_x\right)\,\hat{\mathbf{x}} + \left(y_{21}b\sin\gamma + z_{21}c_y\right)\,\hat{\mathbf{y}} + z_{21}c_z\,\hat{\mathbf{z}}$ | (2*i*) | O VI |
| $\mathbf{B}_{42}$ | = | $-x_{21}\,\mathbf{a}_1 - y_{21}\,\mathbf{a}_2 - z_{21}\,\mathbf{a}_3$ | = | $\left(-x_{21}a - y_{21}b\cos\gamma - z_{21}c_x\right)\,\hat{\mathbf{x}} + \left(-y_{21}b\sin\gamma - z_{21}c_y\right)\,\hat{\mathbf{y}} - z_{21}c_z\,\hat{\mathbf{z}}$ | (2*i*) | O VI |
| $\mathbf{B}_{43}$ | = | $x_{22}\,\mathbf{a}_1 + y_{22}\,\mathbf{a}_2 + z_{22}\,\mathbf{a}_3$ | = | $\left(x_{22}a + y_{22}b\cos\gamma + z_{22}c_x\right)\,\hat{\mathbf{x}} + \left(y_{22}b\sin\gamma + z_{22}c_y\right)\,\hat{\mathbf{y}} + z_{22}c_z\,\hat{\mathbf{z}}$ | (2*i*) | O VII |
| $\mathbf{B}_{44}$ | = | $-x_{22}\,\mathbf{a}_1 - y_{22}\,\mathbf{a}_2 - z_{22}\,\mathbf{a}_3$ | = | $\left(-x_{22}a - y_{22}b\cos\gamma - z_{22}c_x\right)\,\hat{\mathbf{x}} + \left(-y_{22}b\sin\gamma - z_{22}c_y\right)\,\hat{\mathbf{y}} - z_{22}c_z\,\hat{\mathbf{z}}$ | (2*i*) | O VII |
| $\mathbf{B}_{45}$ | = | $x_{23}\,\mathbf{a}_1 + y_{23}\,\mathbf{a}_2 + z_{23}\,\mathbf{a}_3$ | = | $(x_{23}a + y_{23}b\cos\gamma + z_{23}c_x)\,\hat{\mathbf{x}} + \left(y_{23}b\sin\gamma + z_{23}c_y\right)\,\hat{\mathbf{y}} + z_{23}c_z\,\hat{\mathbf{z}}$ | (2$i$) | O VIII |
| $\mathbf{B}_{46}$ | = | $-x_{23}\,\mathbf{a}_1 - y_{23}\,\mathbf{a}_2 - z_{23}\,\mathbf{a}_3$ | = | $(-x_{23}a - y_{23}b\cos\gamma - z_{23}c_x)\,\hat{\mathbf{x}} + \left(-y_{23}b\sin\gamma - z_{23}c_y\right)\,\hat{\mathbf{y}} - z_{23}c_z\,\hat{\mathbf{z}}$ | (2$i$) | O VIII |
| $\mathbf{B}_{47}$ | = | $x_{24}\,\mathbf{a}_1 + y_{24}\,\mathbf{a}_2 + z_{24}\,\mathbf{a}_3$ | = | $(x_{24}a + y_{24}b\cos\gamma + z_{24}c_x)\,\hat{\mathbf{x}} + \left(y_{24}b\sin\gamma + z_{24}c_y\right)\,\hat{\mathbf{y}} + z_{24}c_z\,\hat{\mathbf{z}}$ | (2$i$) | O IX |
| $\mathbf{B}_{48}$ | = | $-x_{24}\,\mathbf{a}_1 - y_{24}\,\mathbf{a}_2 - z_{24}\,\mathbf{a}_3$ | = | $(-x_{24}a - y_{24}b\cos\gamma - z_{24}c_x)\,\hat{\mathbf{x}} + \left(-y_{24}b\sin\gamma - z_{24}c_y\right)\,\hat{\mathbf{y}} - z_{24}c_z\,\hat{\mathbf{z}}$ | (2$i$) | O IX |
| $\mathbf{B}_{49}$ | = | $x_{25}\,\mathbf{a}_1 + y_{25}\,\mathbf{a}_2 + z_{25}\,\mathbf{a}_3$ | = | $(x_{25}a + y_{25}b\cos\gamma + z_{25}c_x)\,\hat{\mathbf{x}} + \left(y_{25}b\sin\gamma + z_{25}c_y\right)\,\hat{\mathbf{y}} + z_{25}c_z\,\hat{\mathbf{z}}$ | (2$i$) | O X |
| $\mathbf{B}_{50}$ | = | $-x_{25}\,\mathbf{a}_1 - y_{25}\,\mathbf{a}_2 - z_{25}\,\mathbf{a}_3$ | = | $(-x_{25}a - y_{25}b\cos\gamma - z_{25}c_x)\,\hat{\mathbf{x}} + \left(-y_{25}b\sin\gamma - z_{25}c_y\right)\,\hat{\mathbf{y}} - z_{25}c_z\,\hat{\mathbf{z}}$ | (2$i$) | O X |
| $\mathbf{B}_{51}$ | = | $x_{26}\,\mathbf{a}_1 + y_{26}\,\mathbf{a}_2 + z_{26}\,\mathbf{a}_3$ | = | $(x_{26}a + y_{26}b\cos\gamma + z_{26}c_x)\,\hat{\mathbf{x}} + \left(y_{26}b\sin\gamma + z_{26}c_y\right)\,\hat{\mathbf{y}} + z_{26}c_z\,\hat{\mathbf{z}}$ | (2$i$) | O XI |
| $\mathbf{B}_{52}$ | = | $-x_{26}\,\mathbf{a}_1 - y_{26}\,\mathbf{a}_2 - z_{26}\,\mathbf{a}_3$ | = | $(-x_{26}a - y_{26}b\cos\gamma - z_{26}c_x)\,\hat{\mathbf{x}} + \left(-y_{26}b\sin\gamma - z_{26}c_y\right)\,\hat{\mathbf{y}} - z_{26}c_z\,\hat{\mathbf{z}}$ | (2$i$) | O XI |
| $\mathbf{B}_{53}$ | = | $x_{27}\,\mathbf{a}_1 + y_{27}\,\mathbf{a}_2 + z_{27}\,\mathbf{a}_3$ | = | $(x_{27}a + y_{27}b\cos\gamma + z_{27}c_x)\,\hat{\mathbf{x}} + \left(y_{27}b\sin\gamma + z_{27}c_y\right)\,\hat{\mathbf{y}} + z_{27}c_z\,\hat{\mathbf{z}}$ | (2$i$) | O XII |
| $\mathbf{B}_{54}$ | = | $-x_{27}\,\mathbf{a}_1 - y_{27}\,\mathbf{a}_2 - z_{27}\,\mathbf{a}_3$ | = | $(-x_{27}a - y_{27}b\cos\gamma - z_{27}c_x)\,\hat{\mathbf{x}} + \left(-y_{27}b\sin\gamma - z_{27}c_y\right)\,\hat{\mathbf{y}} - z_{27}c_z\,\hat{\mathbf{z}}$ | (2$i$) | O XII |

**Geometry files:**

- CIF: pp. 1500
- POSCAR: pp. 1501

# $Co_2B_2O_5$ Structure: A2B2C5_aP18_2_2i_2i_5i

http://aflow.org/prototype-encyclopedia/A2B2C5_aP18_2_2i_2i_5i

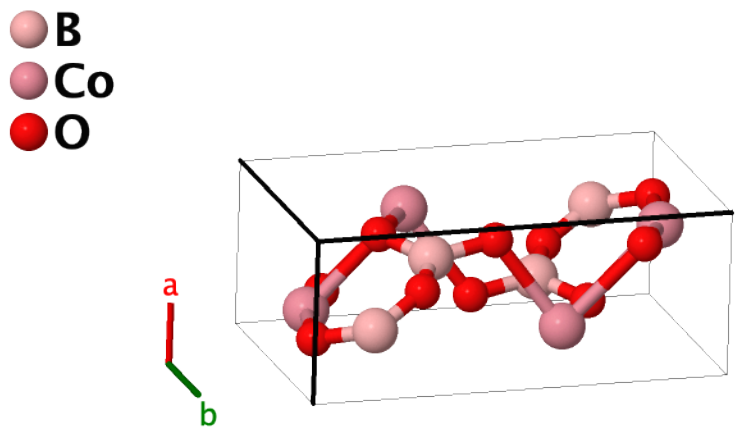

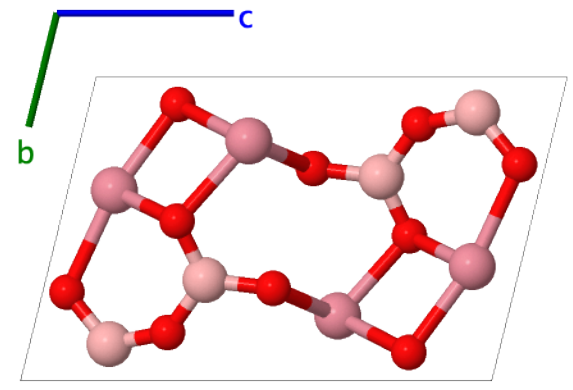

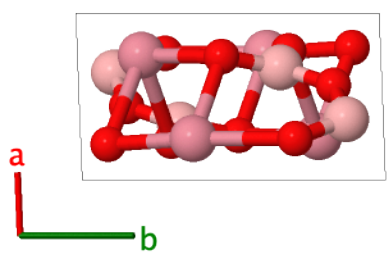

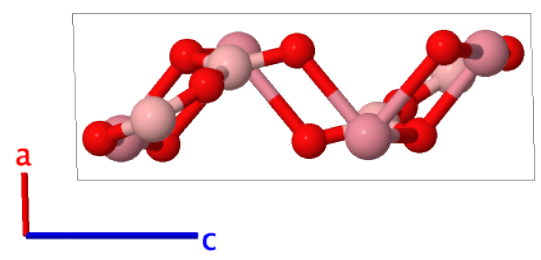

| | | |
|---|---|---|
| **Prototype** | : | $B_2Co_2O_5$ |
| **AFLOW prototype label** | : | A2B2C5_aP18_2_2i_2i_5i |
| ***Strukturbericht* designation** | : | None |
| **Pearson symbol** | : | aP18 |
| **Space group number** | : | 2 |
| **Space group symbol** | : | $P\bar{1}$ |
| **AFLOW prototype command** | : | `aflow --proto=A2B2C5_aP18_2_2i_2i_5i` `--params=`$a, b/a, c/a, \alpha, \beta, \gamma, x_1, y_1, z_1, x_2, y_2, z_2, x_3, y_3, z_3, x_4, y_4, z_4, x_5, y_5, z_5, x_6, y_6, z_6, x_7, y_7, z_7, x_8, y_8, z_8, x_9, y_9, z_9$ |

**Triclinic primitive vectors:**

$$\mathbf{a}_1 = a\,\hat{\mathbf{x}}$$

$$\mathbf{a}_2 = b\cos\gamma\,\hat{\mathbf{x}} + b\sin\gamma\,\hat{\mathbf{y}}$$

$$\mathbf{a}_3 = c_x\,\hat{\mathbf{x}} + c_y\,\hat{\mathbf{y}} + c_z\,\hat{\mathbf{z}}$$

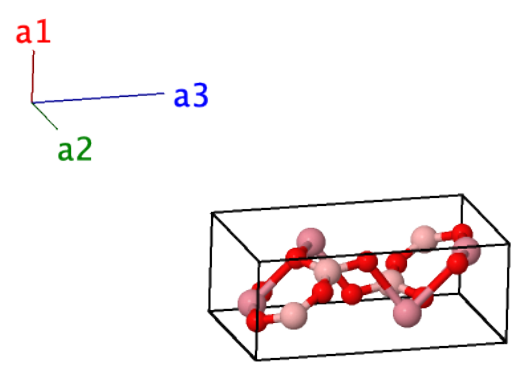

$$c_x = c\,\cos\beta$$

$$c_y = c\,(\cos\alpha - \cos\beta\cos\gamma)/\sin\gamma$$

$$c_z = \sqrt{c^2 - c_x^2 - c_y^2}$$

**Basis vectors:**

| | | Lattice Coordinates | | Cartesian Coordinates | Wyckoff Position | Atom Type |
|---|---|---|---|---|---|---|
| $\mathbf{B}_1$ | = | $x_1\,\mathbf{a}_1 + y_1\,\mathbf{a}_2 + z_1\,\mathbf{a}_3$ | = | $(x_1 a + y_1 b\cos\gamma + z_1 c_x)\,\hat{\mathbf{x}} + \left(y_1 b\sin\gamma + z_1 c_y\right)\,\hat{\mathbf{y}} + z_1 c_z\,\hat{\mathbf{z}}$ | (2i) | B I |
| $\mathbf{B}_2$ | = | $-x_1\,\mathbf{a}_1 - y_1\,\mathbf{a}_2 - z_1\,\mathbf{a}_3$ | = | $(-x_1 a - y_1 b\cos\gamma - z_1 c_x)\,\hat{\mathbf{x}} + \left(-y_1 b\sin\gamma - z_1 c_y\right)\,\hat{\mathbf{y}} - z_1 c_z\,\hat{\mathbf{z}}$ | (2i) | B I |
| $\mathbf{B}_3$ | = | $x_2\,\mathbf{a}_1 + y_2\,\mathbf{a}_2 + z_2\,\mathbf{a}_3$ | = | $(x_2 a + y_2 b\cos\gamma + z_2 c_x)\,\hat{\mathbf{x}} + \left(y_2 b\sin\gamma + z_2 c_y\right)\,\hat{\mathbf{y}} + z_2 c_z\,\hat{\mathbf{z}}$ | (2i) | B II |
| $\mathbf{B}_4$ | = | $-x_2\,\mathbf{a}_1 - y_2\,\mathbf{a}_2 - z_2\,\mathbf{a}_3$ | = | $(-x_2 a - y_2 b\cos\gamma - z_2 c_x)\,\hat{\mathbf{x}} + \left(-y_2 b\sin\gamma - z_2 c_y\right)\,\hat{\mathbf{y}} - z_2 c_z\,\hat{\mathbf{z}}$ | (2*i*) | B II |
| $\mathbf{B}_5$ | = | $x_3\,\mathbf{a}_1 + y_3\,\mathbf{a}_2 + z_3\,\mathbf{a}_3$ | = | $(x_3 a + y_3 b\cos\gamma + z_3 c_x)\,\hat{\mathbf{x}} + \left(y_3 b\sin\gamma + z_3 c_y\right)\,\hat{\mathbf{y}} + z_3 c_z\,\hat{\mathbf{z}}$ | (2*i*) | Co I |
| $\mathbf{B}_6$ | = | $-x_3\,\mathbf{a}_1 - y_3\,\mathbf{a}_2 - z_3\,\mathbf{a}_3$ | = | $(-x_3 a - y_3 b\cos\gamma - z_3 c_x)\,\hat{\mathbf{x}} + \left(-y_3 b\sin\gamma - z_3 c_y\right)\,\hat{\mathbf{y}} - z_3 c_z\,\hat{\mathbf{z}}$ | (2*i*) | Co I |
| $\mathbf{B}_7$ | = | $x_4\,\mathbf{a}_1 + y_4\,\mathbf{a}_2 + z_4\,\mathbf{a}_3$ | = | $(x_4 a + y_4 b\cos\gamma + z_4 c_x)\,\hat{\mathbf{x}} + \left(y_4 b\sin\gamma + z_4 c_y\right)\,\hat{\mathbf{y}} + z_4 c_z\,\hat{\mathbf{z}}$ | (2*i*) | Co II |
| $\mathbf{B}_8$ | = | $-x_4\,\mathbf{a}_1 - y_4\,\mathbf{a}_2 - z_4\,\mathbf{a}_3$ | = | $(-x_4 a - y_4 b\cos\gamma - z_4 c_x)\,\hat{\mathbf{x}} + \left(-y_4 b\sin\gamma - z_4 c_y\right)\,\hat{\mathbf{y}} - z_4 c_z\,\hat{\mathbf{z}}$ | (2*i*) | Co II |
| $\mathbf{B}_9$ | = | $x_5\,\mathbf{a}_1 + y_5\,\mathbf{a}_2 + z_5\,\mathbf{a}_3$ | = | $(x_5 a + y_5 b\cos\gamma + z_5 c_x)\,\hat{\mathbf{x}} + \left(y_5 b\sin\gamma + z_5 c_y\right)\,\hat{\mathbf{y}} + z_5 c_z\,\hat{\mathbf{z}}$ | (2*i*) | O I |
| $\mathbf{B}_{10}$ | = | $-x_5\,\mathbf{a}_1 - y_5\,\mathbf{a}_2 - z_5\,\mathbf{a}_3$ | = | $(-x_5 a - y_5 b\cos\gamma - z_5 c_x)\,\hat{\mathbf{x}} + \left(-y_5 b\sin\gamma - z_5 c_y\right)\,\hat{\mathbf{y}} - z_5 c_z\,\hat{\mathbf{z}}$ | (2*i*) | O I |
| $\mathbf{B}_{11}$ | = | $x_6\,\mathbf{a}_1 + y_6\,\mathbf{a}_2 + z_6\,\mathbf{a}_3$ | = | $(x_6 a + y_6 b\cos\gamma + z_6 c_x)\,\hat{\mathbf{x}} + \left(y_6 b\sin\gamma + z_6 c_y\right)\,\hat{\mathbf{y}} + z_6 c_z\,\hat{\mathbf{z}}$ | (2*i*) | O II |
| $\mathbf{B}_{12}$ | = | $-x_6\,\mathbf{a}_1 - y_6\,\mathbf{a}_2 - z_6\,\mathbf{a}_3$ | = | $(-x_6 a - y_6 b\cos\gamma - z_6 c_x)\,\hat{\mathbf{x}} + \left(-y_6 b\sin\gamma - z_6 c_y\right)\,\hat{\mathbf{y}} - z_6 c_z\,\hat{\mathbf{z}}$ | (2*i*) | O II |
| $\mathbf{B}_{13}$ | = | $x_7\,\mathbf{a}_1 + y_7\,\mathbf{a}_2 + z_7\,\mathbf{a}_3$ | = | $(x_7 a + y_7 b\cos\gamma + z_7 c_x)\,\hat{\mathbf{x}} + \left(y_7 b\sin\gamma + z_7 c_y\right)\,\hat{\mathbf{y}} + z_7 c_z\,\hat{\mathbf{z}}$ | (2*i*) | O III |
| $\mathbf{B}_{14}$ | = | $-x_7\,\mathbf{a}_1 - y_7\,\mathbf{a}_2 - z_7\,\mathbf{a}_3$ | = | $(-x_7 a - y_7 b\cos\gamma - z_7 c_x)\,\hat{\mathbf{x}} + \left(-y_7 b\sin\gamma - z_7 c_y\right)\,\hat{\mathbf{y}} - z_7 c_z\,\hat{\mathbf{z}}$ | (2*i*) | O III |
| $\mathbf{B}_{15}$ | = | $x_8\,\mathbf{a}_1 + y_8\,\mathbf{a}_2 + z_8\,\mathbf{a}_3$ | = | $(x_8 a + y_8 b\cos\gamma + z_8 c_x)\,\hat{\mathbf{x}} + \left(y_8 b\sin\gamma + z_8 c_y\right)\,\hat{\mathbf{y}} + z_8 c_z\,\hat{\mathbf{z}}$ | (2*i*) | O IV |
| $\mathbf{B}_{16}$ | = | $-x_8\,\mathbf{a}_1 - y_8\,\mathbf{a}_2 - z_8\,\mathbf{a}_3$ | = | $(-x_8 a - y_8 b\cos\gamma - z_8 c_x)\,\hat{\mathbf{x}} + \left(-y_8 b\sin\gamma - z_8 c_y\right)\,\hat{\mathbf{y}} - z_8 c_z\,\hat{\mathbf{z}}$ | (2*i*) | O IV |
| $\mathbf{B}_{17}$ | = | $x_9\,\mathbf{a}_1 + y_9\,\mathbf{a}_2 + z_9\,\mathbf{a}_3$ | = | $(x_9 a + y_9 b\cos\gamma + z_9 c_x)\,\hat{\mathbf{x}} + \left(y_9 b\sin\gamma + z_9 c_y\right)\,\hat{\mathbf{y}} + z_9 c_z\,\hat{\mathbf{z}}$ | (2*i*) | O V |
| $\mathbf{B}_{18}$ | = | $-x_9\,\mathbf{a}_1 - y_9\,\mathbf{a}_2 - z_9\,\mathbf{a}_3$ | = | $(-x_9 a - y_9 b\cos\gamma - z_9 c_x)\,\hat{\mathbf{x}} + \left(-y_9 b\sin\gamma - z_9 c_y\right)\,\hat{\mathbf{y}} - z_9 c_z\,\hat{\mathbf{z}}$ | (2*i*) | O V |

---

---

**Geometry files:**

- CIF: pp. 1501
- POSCAR: pp. 1502

# Kyanite ($Al_2SiO_5$, $S0_1$) Structure: A2B5C_aP32_2_4i_10i_2i

http://aflow.org/prototype-encyclopedia/A2B5C_aP32_2_4i_10i_2i

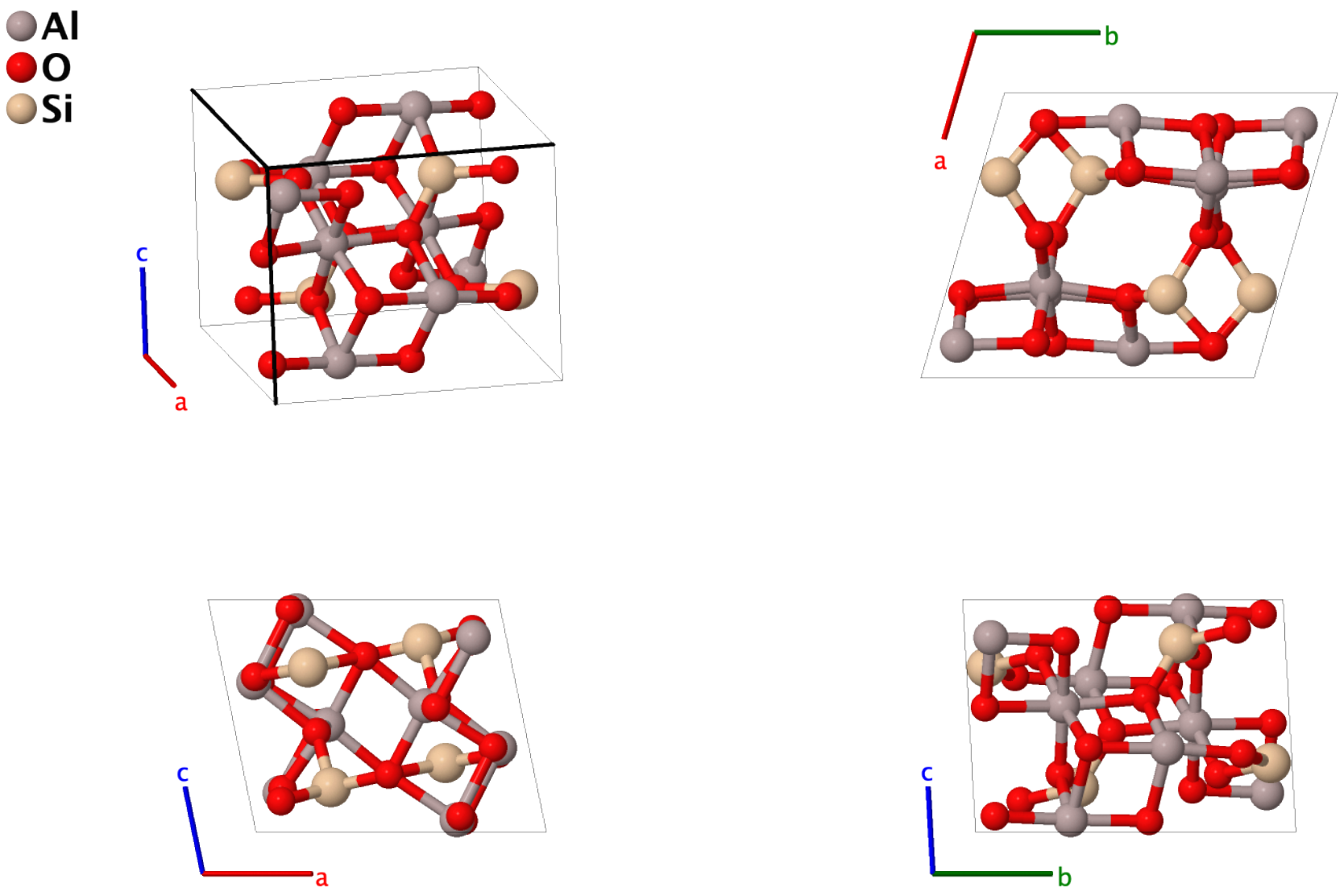

| | | |
|---|---|---|
| **Prototype** | : | $Al_2O_5Si$ |
| **AFLOW prototype label** | : | A2B5C_aP32_2_4i_10i_2i |
| ***Strukturbericht* designation** | : | $S0_1$ |
| **Pearson symbol** | : | aP32 |
| **Space group number** | : | 2 |
| **Space group symbol** | : | $P\bar{1}$ |
| **AFLOW prototype command** | : | `aflow --proto=A2B5C_aP32_2_4i_10i_2i` `--params=`$a, b/a, c/a, \alpha, \beta, \gamma, x_1, y_1, z_1, x_2, y_2, z_2, x_3, y_3, z_3, x_4, y_4, z_4, x_5, y_5, z_5, x_6, y_6, z_6, x_7, y_7, z_7, x_8, y_8, z_8, x_9, y_9, z_9, x_{10}, y_{10}, z_{10}, x_{11}, y_{11}, z_{11}, x_{12}, y_{12}, z_{12}, x_{13}, y_{13}, z_{13}, x_{14}, y_{14}, z_{14}, x_{15}, y_{15}, z_{15}, x_{16}, y_{16}, z_{16}$ |

- Three crystal polymorphs of $Al_2SiO_5$ have been characterized: kyanite ($S0_1$), space group $P\bar{1}$ #2, andalusite ($S0_2$), space group *Pnnm* #58, and sillimanite ($S0_3$), space group *Pnma* #62. All are characterized by chains of edge-sharing $SiO_6$ tetrahedra and Al octahedra.
- (Ewald, 1931) originally gave Kyanite the $H5_1$ *Strukturbericht* symbol, but this was changed to $S0_1$ in (Hermann, 1937).
- We use the ambient pressure data of (Yang, 1997).

**Triclinic primitive vectors:**

$$
\begin{aligned}
\mathbf{a}_1 &= a\,\hat{\mathbf{x}} \\
\mathbf{a}_2 &= b\cos\gamma\,\hat{\mathbf{x}} + b\sin\gamma\,\hat{\mathbf{y}} \\
\mathbf{a}_3 &= c_x\,\hat{\mathbf{x}} + c_y\,\hat{\mathbf{y}} + c_z\,\hat{\mathbf{z}} \\
\\
c_x &= c\,\cos\beta \\
c_y &= c\,(\cos\alpha - \cos\beta\cos\gamma)/\sin\gamma \\
c_z &= \sqrt{c^2 - c_x^2 - c_y^2}
\end{aligned}
$$

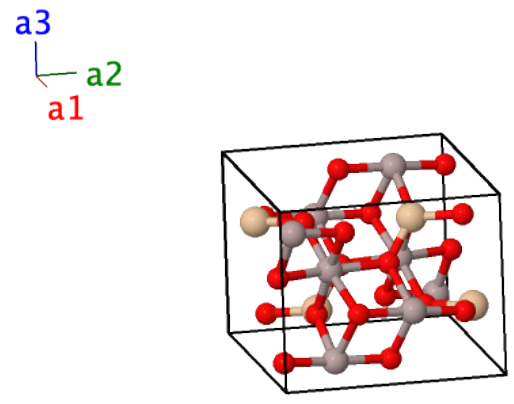

**Basis vectors:**

| | | Lattice Coordinates | | Cartesian Coordinates | Wyckoff Position | Atom Type |
|---|---|---|---|---|---|---|
| $\mathbf{B}_1$ | = | $x_1\,\mathbf{a}_1 + y_1\,\mathbf{a}_2 + z_1\,\mathbf{a}_3$ | = | $(x_1 a + y_1 b\cos\gamma + z_1 c_x)\,\hat{\mathbf{x}} + \left(y_1 b\sin\gamma + z_1 c_y\right)\hat{\mathbf{y}} + z_1 c_z\,\hat{\mathbf{z}}$ | (2$i$) | Al I |
| $\mathbf{B}_2$ | = | $-x_1\,\mathbf{a}_1 - y_1\,\mathbf{a}_2 - z_1\,\mathbf{a}_3$ | = | $(-x_1 a - y_1 b\cos\gamma - z_1 c_x)\,\hat{\mathbf{x}} + \left(-y_1 b\sin\gamma - z_1 c_y\right)\hat{\mathbf{y}} - z_1 c_z\,\hat{\mathbf{z}}$ | (2$i$) | Al I |
| $\mathbf{B}_3$ | = | $x_2\,\mathbf{a}_1 + y_2\,\mathbf{a}_2 + z_2\,\mathbf{a}_3$ | = | $(x_2 a + y_2 b\cos\gamma + z_2 c_x)\,\hat{\mathbf{x}} + \left(y_2 b\sin\gamma + z_2 c_y\right)\hat{\mathbf{y}} + z_2 c_z\,\hat{\mathbf{z}}$ | (2$i$) | Al II |
| $\mathbf{B}_4$ | = | $-x_2\,\mathbf{a}_1 - y_2\,\mathbf{a}_2 - z_2\,\mathbf{a}_3$ | = | $(-x_2 a - y_2 b\cos\gamma - z_2 c_x)\,\hat{\mathbf{x}} + \left(-y_2 b\sin\gamma - z_2 c_y\right)\hat{\mathbf{y}} - z_2 c_z\,\hat{\mathbf{z}}$ | (2$i$) | Al II |
| $\mathbf{B}_5$ | = | $x_3\,\mathbf{a}_1 + y_3\,\mathbf{a}_2 + z_3\,\mathbf{a}_3$ | = | $(x_3 a + y_3 b\cos\gamma + z_3 c_x)\,\hat{\mathbf{x}} + \left(y_3 b\sin\gamma + z_3 c_y\right)\hat{\mathbf{y}} + z_3 c_z\,\hat{\mathbf{z}}$ | (2$i$) | Al III |
| $\mathbf{B}_6$ | = | $-x_3\,\mathbf{a}_1 - y_3\,\mathbf{a}_2 - z_3\,\mathbf{a}_3$ | = | $(-x_3 a - y_3 b\cos\gamma - z_3 c_x)\,\hat{\mathbf{x}} + \left(-y_3 b\sin\gamma - z_3 c_y\right)\hat{\mathbf{y}} - z_3 c_z\,\hat{\mathbf{z}}$ | (2$i$) | Al III |
| $\mathbf{B}_7$ | = | $x_4\,\mathbf{a}_1 + y_4\,\mathbf{a}_2 + z_4\,\mathbf{a}_3$ | = | $(x_4 a + y_4 b\cos\gamma + z_4 c_x)\,\hat{\mathbf{x}} + \left(y_4 b\sin\gamma + z_4 c_y\right)\hat{\mathbf{y}} + z_4 c_z\,\hat{\mathbf{z}}$ | (2$i$) | Al IV |
| $\mathbf{B}_8$ | = | $-x_4\,\mathbf{a}_1 - y_4\,\mathbf{a}_2 - z_4\,\mathbf{a}_3$ | = | $(-x_4 a - y_4 b\cos\gamma - z_4 c_x)\,\hat{\mathbf{x}} + \left(-y_4 b\sin\gamma - z_4 c_y\right)\hat{\mathbf{y}} - z_4 c_z\,\hat{\mathbf{z}}$ | (2$i$) | Al IV |
| $\mathbf{B}_9$ | = | $x_5\,\mathbf{a}_1 + y_5\,\mathbf{a}_2 + z_5\,\mathbf{a}_3$ | = | $(x_5 a + y_5 b\cos\gamma + z_5 c_x)\,\hat{\mathbf{x}} + \left(y_5 b\sin\gamma + z_5 c_y\right)\hat{\mathbf{y}} + z_5 c_z\,\hat{\mathbf{z}}$ | (2$i$) | O I |
| $\mathbf{B}_{10}$ | = | $-x_5\,\mathbf{a}_1 - y_5\,\mathbf{a}_2 - z_5\,\mathbf{a}_3$ | = | $(-x_5 a - y_5 b\cos\gamma - z_5 c_x)\,\hat{\mathbf{x}} + \left(-y_5 b\sin\gamma - z_5 c_y\right)\hat{\mathbf{y}} - z_5 c_z\,\hat{\mathbf{z}}$ | (2$i$) | O I |
| $\mathbf{B}_{11}$ | = | $x_6\,\mathbf{a}_1 + y_6\,\mathbf{a}_2 + z_6\,\mathbf{a}_3$ | = | $(x_6 a + y_6 b\cos\gamma + z_6 c_x)\,\hat{\mathbf{x}} + \left(y_6 b\sin\gamma + z_6 c_y\right)\hat{\mathbf{y}} + z_6 c_z\,\hat{\mathbf{z}}$ | (2$i$) | O II |
| $\mathbf{B}_{12}$ | = | $-x_6\,\mathbf{a}_1 - y_6\,\mathbf{a}_2 - z_6\,\mathbf{a}_3$ | = | $(-x_6 a - y_6 b\cos\gamma - z_6 c_x)\,\hat{\mathbf{x}} + \left(-y_6 b\sin\gamma - z_6 c_y\right)\hat{\mathbf{y}} - z_6 c_z\,\hat{\mathbf{z}}$ | (2$i$) | O II |
| $\mathbf{B}_{13}$ | = | $x_7\,\mathbf{a}_1 + y_7\,\mathbf{a}_2 + z_7\,\mathbf{a}_3$ | = | $(x_7 a + y_7 b\cos\gamma + z_7 c_x)\,\hat{\mathbf{x}} + \left(y_7 b\sin\gamma + z_7 c_y\right)\hat{\mathbf{y}} + z_7 c_z\,\hat{\mathbf{z}}$ | (2$i$) | O III |
| $\mathbf{B}_{14}$ | = | $-x_7\,\mathbf{a}_1 - y_7\,\mathbf{a}_2 - z_7\,\mathbf{a}_3$ | = | $(-x_7 a - y_7 b\cos\gamma - z_7 c_x)\,\hat{\mathbf{x}} + \left(-y_7 b\sin\gamma - z_7 c_y\right)\hat{\mathbf{y}} - z_7 c_z\,\hat{\mathbf{z}}$ | (2$i$) | O III |
| $\mathbf{B}_{15}$ | = | $x_8\,\mathbf{a}_1 + y_8\,\mathbf{a}_2 + z_8\,\mathbf{a}_3$ | = | $(x_8 a + y_8 b\cos\gamma + z_8 c_x)\,\hat{\mathbf{x}} + \left(y_8 b\sin\gamma + z_8 c_y\right)\hat{\mathbf{y}} + z_8 c_z\,\hat{\mathbf{z}}$ | (2$i$) | O IV |
| $\mathbf{B}_{16}$ | = | $-x_8\,\mathbf{a}_1 - y_8\,\mathbf{a}_2 - z_8\,\mathbf{a}_3$ | = | $(-x_8 a - y_8 b\cos\gamma - z_8 c_x)\,\hat{\mathbf{x}} + \left(-y_8 b\sin\gamma - z_8 c_y\right)\hat{\mathbf{y}} - z_8 c_z\,\hat{\mathbf{z}}$ | (2$i$) | O IV |
| $\mathbf{B}_{17}$ | = | $x_9\,\mathbf{a}_1 + y_9\,\mathbf{a}_2 + z_9\,\mathbf{a}_3$ | = | $(x_9 a + y_9 b\cos\gamma + z_9 c_x)\,\hat{\mathbf{x}} + \left(y_9 b\sin\gamma + z_9 c_y\right)\,\hat{\mathbf{y}} + z_9 c_z\,\hat{\mathbf{z}}$ | (2i) | O V |
| $\mathbf{B}_{18}$ | = | $-x_9\,\mathbf{a}_1 - y_9\,\mathbf{a}_2 - z_9\,\mathbf{a}_3$ | = | $(-x_9 a - y_9 b\cos\gamma - z_9 c_x)\,\hat{\mathbf{x}} + \left(-y_9 b\sin\gamma - z_9 c_y\right)\,\hat{\mathbf{y}} - z_9 c_z\,\hat{\mathbf{z}}$ | (2i) | O V |
| $\mathbf{B}_{19}$ | = | $x_{10}\,\mathbf{a}_1 + y_{10}\,\mathbf{a}_2 + z_{10}\,\mathbf{a}_3$ | = | $(x_{10} a + y_{10} b\cos\gamma + z_{10} c_x)\,\hat{\mathbf{x}} + \left(y_{10} b\sin\gamma + z_{10} c_y\right)\,\hat{\mathbf{y}} + z_{10} c_z\,\hat{\mathbf{z}}$ | (2i) | O VI |
| $\mathbf{B}_{20}$ | = | $-x_{10}\,\mathbf{a}_1 - y_{10}\,\mathbf{a}_2 - z_{10}\,\mathbf{a}_3$ | = | $(-x_{10} a - y_{10} b\cos\gamma - z_{10} c_x)\,\hat{\mathbf{x}} + \left(-y_{10} b\sin\gamma - z_{10} c_y\right)\,\hat{\mathbf{y}} - z_{10} c_z\,\hat{\mathbf{z}}$ | (2i) | O VI |
| $\mathbf{B}_{21}$ | = | $x_{11}\,\mathbf{a}_1 + y_{11}\,\mathbf{a}_2 + z_{11}\,\mathbf{a}_3$ | = | $(x_{11} a + y_{11} b\cos\gamma + z_{11} c_x)\,\hat{\mathbf{x}} + \left(y_{11} b\sin\gamma + z_{11} c_y\right)\,\hat{\mathbf{y}} + z_{11} c_z\,\hat{\mathbf{z}}$ | (2i) | O VII |
| $\mathbf{B}_{22}$ | = | $-x_{11}\,\mathbf{a}_1 - y_{11}\,\mathbf{a}_2 - z_{11}\,\mathbf{a}_3$ | = | $(-x_{11} a - y_{11} b\cos\gamma - z_{11} c_x)\,\hat{\mathbf{x}} + \left(-y_{11} b\sin\gamma - z_{11} c_y\right)\,\hat{\mathbf{y}} - z_{11} c_z\,\hat{\mathbf{z}}$ | (2i) | O VII |
| $\mathbf{B}_{23}$ | = | $x_{12}\,\mathbf{a}_1 + y_{12}\,\mathbf{a}_2 + z_{12}\,\mathbf{a}_3$ | = | $(x_{12} a + y_{12} b\cos\gamma + z_{12} c_x)\,\hat{\mathbf{x}} + \left(y_{12} b\sin\gamma + z_{12} c_y\right)\,\hat{\mathbf{y}} + z_{12} c_z\,\hat{\mathbf{z}}$ | (2i) | O VIII |
| $\mathbf{B}_{24}$ | = | $-x_{12}\,\mathbf{a}_1 - y_{12}\,\mathbf{a}_2 - z_{12}\,\mathbf{a}_3$ | = | $(-x_{12} a - y_{12} b\cos\gamma - z_{12} c_x)\,\hat{\mathbf{x}} + \left(-y_{12} b\sin\gamma - z_{12} c_y\right)\,\hat{\mathbf{y}} - z_{12} c_z\,\hat{\mathbf{z}}$ | (2i) | O VIII |
| $\mathbf{B}_{25}$ | = | $x_{13}\,\mathbf{a}_1 + y_{13}\,\mathbf{a}_2 + z_{13}\,\mathbf{a}_3$ | = | $(x_{13} a + y_{13} b\cos\gamma + z_{13} c_x)\,\hat{\mathbf{x}} + \left(y_{13} b\sin\gamma + z_{13} c_y\right)\,\hat{\mathbf{y}} + z_{13} c_z\,\hat{\mathbf{z}}$ | (2i) | O IX |
| $\mathbf{B}_{26}$ | = | $-x_{13}\,\mathbf{a}_1 - y_{13}\,\mathbf{a}_2 - z_{13}\,\mathbf{a}_3$ | = | $(-x_{13} a - y_{13} b\cos\gamma - z_{13} c_x)\,\hat{\mathbf{x}} + \left(-y_{13} b\sin\gamma - z_{13} c_y\right)\,\hat{\mathbf{y}} - z_{13} c_z\,\hat{\mathbf{z}}$ | (2i) | O IX |
| $\mathbf{B}_{27}$ | = | $x_{14}\,\mathbf{a}_1 + y_{14}\,\mathbf{a}_2 + z_{14}\,\mathbf{a}_3$ | = | $(x_{14} a + y_{14} b\cos\gamma + z_{14} c_x)\,\hat{\mathbf{x}} + \left(y_{14} b\sin\gamma + z_{14} c_y\right)\,\hat{\mathbf{y}} + z_{14} c_z\,\hat{\mathbf{z}}$ | (2i) | O X |
| $\mathbf{B}_{28}$ | = | $-x_{14}\,\mathbf{a}_1 - y_{14}\,\mathbf{a}_2 - z_{14}\,\mathbf{a}_3$ | = | $(-x_{14} a - y_{14} b\cos\gamma - z_{14} c_x)\,\hat{\mathbf{x}} + \left(-y_{14} b\sin\gamma - z_{14} c_y\right)\,\hat{\mathbf{y}} - z_{14} c_z\,\hat{\mathbf{z}}$ | (2i) | O X |
| $\mathbf{B}_{29}$ | = | $x_{15}\,\mathbf{a}_1 + y_{15}\,\mathbf{a}_2 + z_{15}\,\mathbf{a}_3$ | = | $(x_{15} a + y_{15} b\cos\gamma + z_{15} c_x)\,\hat{\mathbf{x}} + \left(y_{15} b\sin\gamma + z_{15} c_y\right)\,\hat{\mathbf{y}} + z_{15} c_z\,\hat{\mathbf{z}}$ | (2i) | Si I |
| $\mathbf{B}_{30}$ | = | $-x_{15}\,\mathbf{a}_1 - y_{15}\,\mathbf{a}_2 - z_{15}\,\mathbf{a}_3$ | = | $(-x_{15} a - y_{15} b\cos\gamma - z_{15} c_x)\,\hat{\mathbf{x}} + \left(-y_{15} b\sin\gamma - z_{15} c_y\right)\,\hat{\mathbf{y}} - z_{15} c_z\,\hat{\mathbf{z}}$ | (2i) | Si I |
| $\mathbf{B}_{31}$ | = | $x_{16}\,\mathbf{a}_1 + y_{16}\,\mathbf{a}_2 + z_{16}\,\mathbf{a}_3$ | = | $(x_{16} a + y_{16} b\cos\gamma + z_{16} c_x)\,\hat{\mathbf{x}} + \left(y_{16} b\sin\gamma + z_{16} c_y\right)\,\hat{\mathbf{y}} + z_{16} c_z\,\hat{\mathbf{z}}$ | (2i) | Si II |
| $\mathbf{B}_{32}$ | = | $-x_{16}\,\mathbf{a}_1 - y_{16}\,\mathbf{a}_2 - z_{16}\,\mathbf{a}_3$ | = | $(-x_{16} a - y_{16} b\cos\gamma - z_{16} c_x)\,\hat{\mathbf{x}} + \left(-y_{16} b\sin\gamma - z_{16} c_y\right)\,\hat{\mathbf{y}} - z_{16} c_z\,\hat{\mathbf{z}}$ | (2i) | Si II |

**Geometry files:**

- CIF: pp. 1502
- POSCAR: pp. 1502

# $\alpha$-$Ho_2Si_2O_7$ Structure: A2B7C2_aP44_2_4i_14i_4i

http://aflow.org/prototype-encyclopedia/A2B7C2_aP44_2_4i_14i_4i

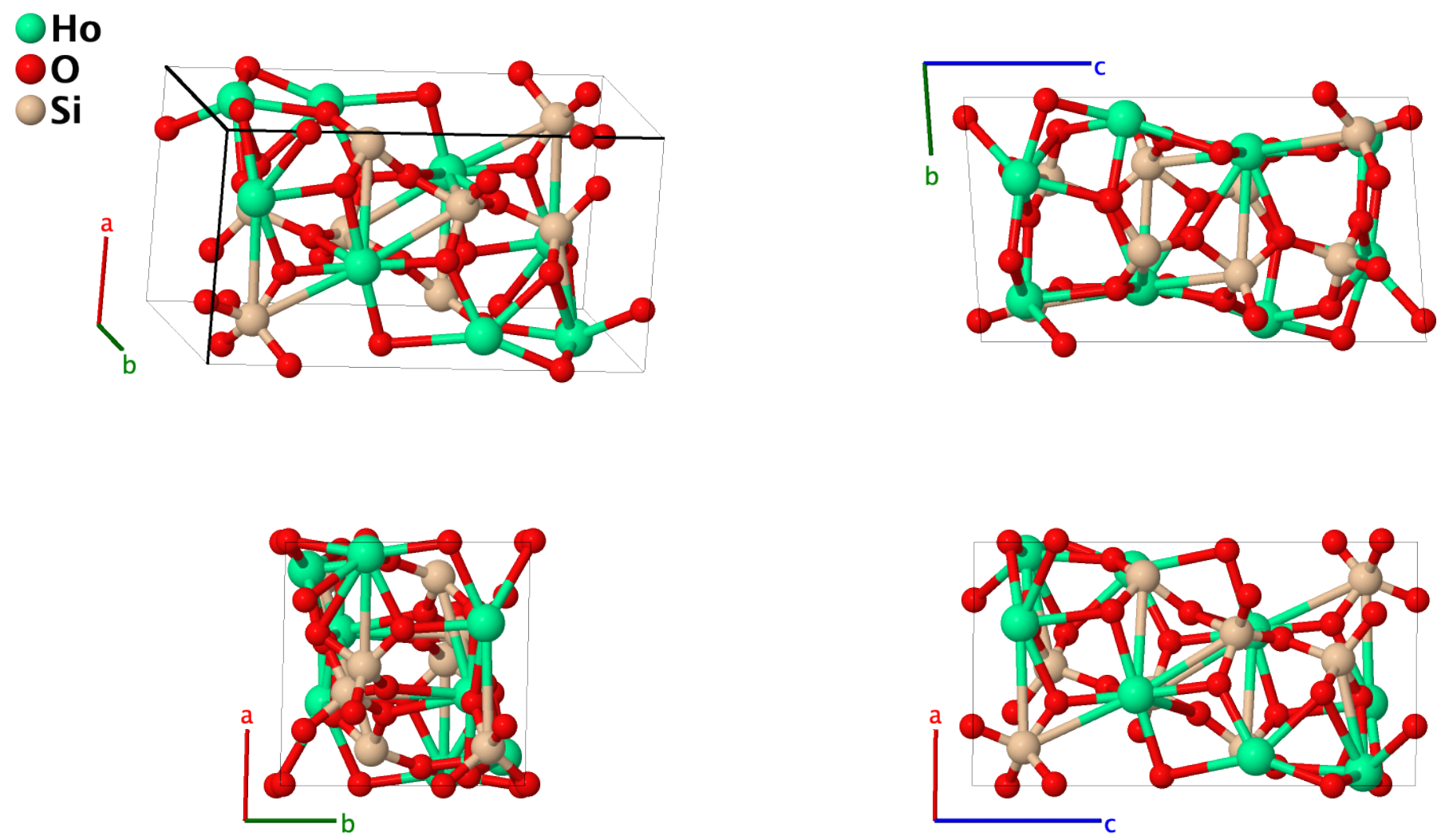

**Prototype** : $Ho_2O_7Si_2$

**AFLOW prototype label** : A2B7C2_aP44_2_4i_14i_4i

***Strukturbericht* designation** : None

**Pearson symbol** : aP44

**Space group number** : 2

**Space group symbol** : $P\bar{1}$

**AFLOW prototype command** : `aflow --proto=A2B7C2_aP44_2_4i_14i_4i`
`--params=`$a, b/a, c/a, \alpha, \beta, \gamma, x_1, y_1, z_1, x_2, y_2, z_2, x_3, y_3, z_3, x_4, y_4, z_4, x_5, y_5, z_5, x_6, y_6, z_6, x_7, y_7, z_7, x_8, y_8, z_8, x_9, y_9, z_9, x_{10}, y_{10}, z_{10}, x_{11}, y_{11}, z_{11}, x_{12}, y_{12}, z_{12}, x_{13}, y_{13}, z_{13}, x_{14}, y_{14}, z_{14}, x_{15}, y_{15}, z_{15}, x_{16}, y_{16}, z_{16}, x_{17}, y_{17}, z_{17}, x_{18}, y_{18}, z_{18}, x_{19}, y_{19}, z_{19}, x_{20}, y_{20}, z_{20}, x_{21}, y_{21}, z_{21}, x_{22}, y_{22}, z_{22}$

---

**Triclinic primitive vectors:**

$$\mathbf{a}_1 = a\,\hat{\mathbf{x}}$$
$$\mathbf{a}_2 = b\cos\gamma\,\hat{\mathbf{x}} + b\sin\gamma\,\hat{\mathbf{y}}$$
$$\mathbf{a}_3 = c_x\,\hat{\mathbf{x}} + c_y\,\hat{\mathbf{y}} + c_z\,\hat{\mathbf{z}}$$

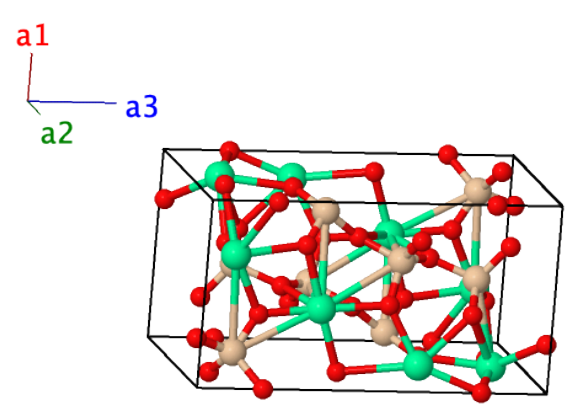

$$c_x = c\,\cos\beta$$
$$c_y = c\,(\cos\alpha - \cos\beta\cos\gamma)/\sin\gamma$$
$$c_z = \sqrt{c^2 - c_x^2 - c_y^2}$$

---

**Basis vectors:**

| | | Lattice Coordinates | | Cartesian Coordinates | Wyckoff Position | Atom Type |
|---|---|---|---|---|---|---|
| $\mathbf{B}_1$ | = | $x_1\,\mathbf{a}_1 + y_1\,\mathbf{a}_2 + z_1\,\mathbf{a}_3$ | = | $(x_1 a + y_1 b\cos\gamma + z_1 c_x)\,\hat{\mathbf{x}} + \left(y_1 b\sin\gamma + z_1 c_y\right)\,\hat{\mathbf{y}} + z_1 c_z\,\hat{\mathbf{z}}$ | (2*i*) | Ho I |
| $\mathbf{B}_2$ | = | $-x_1\,\mathbf{a}_1 - y_1\,\mathbf{a}_2 - z_1\,\mathbf{a}_3$ | = | $(-x_1 a - y_1 b\cos\gamma - z_1 c_x)\,\hat{\mathbf{x}} + \left(-y_1 b\sin\gamma - z_1 c_y\right)\,\hat{\mathbf{y}} - z_1 c_z\,\hat{\mathbf{z}}$ | (2*i*) | Ho I |
| $\mathbf{B}_3$ | = | $x_2\,\mathbf{a}_1 + y_2\,\mathbf{a}_2 + z_2\,\mathbf{a}_3$ | = | $(x_2a + y_2b\cos\gamma + z_2c_x)\,\hat{\mathbf{x}} + \left(y_2b\sin\gamma + z_2c_y\right)\hat{\mathbf{y}} + z_2c_z\,\hat{\mathbf{z}}$ | (2$i$) | Ho II |
| $\mathbf{B}_4$ | = | $-x_2\,\mathbf{a}_1 - y_2\,\mathbf{a}_2 - z_2\,\mathbf{a}_3$ | = | $(-x_2a - y_2b\cos\gamma - z_2c_x)\,\hat{\mathbf{x}} + \left(-y_2b\sin\gamma - z_2c_y\right)\hat{\mathbf{y}} - z_2c_z\,\hat{\mathbf{z}}$ | (2$i$) | Ho II |
| $\mathbf{B}_5$ | = | $x_3\,\mathbf{a}_1 + y_3\,\mathbf{a}_2 + z_3\,\mathbf{a}_3$ | = | $(x_3a + y_3b\cos\gamma + z_3c_x)\,\hat{\mathbf{x}} + \left(y_3b\sin\gamma + z_3c_y\right)\hat{\mathbf{y}} + z_3c_z\,\hat{\mathbf{z}}$ | (2$i$) | Ho III |
| $\mathbf{B}_6$ | = | $-x_3\,\mathbf{a}_1 - y_3\,\mathbf{a}_2 - z_3\,\mathbf{a}_3$ | = | $(-x_3a - y_3b\cos\gamma - z_3c_x)\,\hat{\mathbf{x}} + \left(-y_3b\sin\gamma - z_3c_y\right)\hat{\mathbf{y}} - z_3c_z\,\hat{\mathbf{z}}$ | (2$i$) | Ho III |
| $\mathbf{B}_7$ | = | $x_4\,\mathbf{a}_1 + y_4\,\mathbf{a}_2 + z_4\,\mathbf{a}_3$ | = | $(x_4a + y_4b\cos\gamma + z_4c_x)\,\hat{\mathbf{x}} + \left(y_4b\sin\gamma + z_4c_y\right)\hat{\mathbf{y}} + z_4c_z\,\hat{\mathbf{z}}$ | (2$i$) | Ho IV |
| $\mathbf{B}_8$ | = | $-x_4\,\mathbf{a}_1 - y_4\,\mathbf{a}_2 - z_4\,\mathbf{a}_3$ | = | $(-x_4a - y_4b\cos\gamma - z_4c_x)\,\hat{\mathbf{x}} + \left(-y_4b\sin\gamma - z_4c_y\right)\hat{\mathbf{y}} - z_4c_z\,\hat{\mathbf{z}}$ | (2$i$) | Ho IV |
| $\mathbf{B}_9$ | = | $x_5\,\mathbf{a}_1 + y_5\,\mathbf{a}_2 + z_5\,\mathbf{a}_3$ | = | $(x_5a + y_5b\cos\gamma + z_5c_x)\,\hat{\mathbf{x}} + \left(y_5b\sin\gamma + z_5c_y\right)\hat{\mathbf{y}} + z_5c_z\,\hat{\mathbf{z}}$ | (2$i$) | O I |
| $\mathbf{B}_{10}$ | = | $-x_5\,\mathbf{a}_1 - y_5\,\mathbf{a}_2 - z_5\,\mathbf{a}_3$ | = | $(-x_5a - y_5b\cos\gamma - z_5c_x)\,\hat{\mathbf{x}} + \left(-y_5b\sin\gamma - z_5c_y\right)\hat{\mathbf{y}} - z_5c_z\,\hat{\mathbf{z}}$ | (2$i$) | O I |
| $\mathbf{B}_{11}$ | = | $x_6\,\mathbf{a}_1 + y_6\,\mathbf{a}_2 + z_6\,\mathbf{a}_3$ | = | $(x_6a + y_6b\cos\gamma + z_6c_x)\,\hat{\mathbf{x}} + \left(y_6b\sin\gamma + z_6c_y\right)\hat{\mathbf{y}} + z_6c_z\,\hat{\mathbf{z}}$ | (2$i$) | O II |
| $\mathbf{B}_{12}$ | = | $-x_6\,\mathbf{a}_1 - y_6\,\mathbf{a}_2 - z_6\,\mathbf{a}_3$ | = | $(-x_6a - y_6b\cos\gamma - z_6c_x)\,\hat{\mathbf{x}} + \left(-y_6b\sin\gamma - z_6c_y\right)\hat{\mathbf{y}} - z_6c_z\,\hat{\mathbf{z}}$ | (2$i$) | O II |
| $\mathbf{B}_{13}$ | = | $x_7\,\mathbf{a}_1 + y_7\,\mathbf{a}_2 + z_7\,\mathbf{a}_3$ | = | $(x_7a + y_7b\cos\gamma + z_7c_x)\,\hat{\mathbf{x}} + \left(y_7b\sin\gamma + z_7c_y\right)\hat{\mathbf{y}} + z_7c_z\,\hat{\mathbf{z}}$ | (2$i$) | O III |
| $\mathbf{B}_{14}$ | = | $-x_7\,\mathbf{a}_1 - y_7\,\mathbf{a}_2 - z_7\,\mathbf{a}_3$ | = | $(-x_7a - y_7b\cos\gamma - z_7c_x)\,\hat{\mathbf{x}} + \left(-y_7b\sin\gamma - z_7c_y\right)\hat{\mathbf{y}} - z_7c_z\,\hat{\mathbf{z}}$ | (2$i$) | O III |
| $\mathbf{B}_{15}$ | = | $x_8\,\mathbf{a}_1 + y_8\,\mathbf{a}_2 + z_8\,\mathbf{a}_3$ | = | $(x_8a + y_8b\cos\gamma + z_8c_x)\,\hat{\mathbf{x}} + \left(y_8b\sin\gamma + z_8c_y\right)\hat{\mathbf{y}} + z_8c_z\,\hat{\mathbf{z}}$ | (2$i$) | O IV |
| $\mathbf{B}_{16}$ | = | $-x_8\,\mathbf{a}_1 - y_8\,\mathbf{a}_2 - z_8\,\mathbf{a}_3$ | = | $(-x_8a - y_8b\cos\gamma - z_8c_x)\,\hat{\mathbf{x}} + \left(-y_8b\sin\gamma - z_8c_y\right)\hat{\mathbf{y}} - z_8c_z\,\hat{\mathbf{z}}$ | (2$i$) | O IV |
| $\mathbf{B}_{17}$ | = | $x_9\,\mathbf{a}_1 + y_9\,\mathbf{a}_2 + z_9\,\mathbf{a}_3$ | = | $(x_9a + y_9b\cos\gamma + z_9c_x)\,\hat{\mathbf{x}} + \left(y_9b\sin\gamma + z_9c_y\right)\hat{\mathbf{y}} + z_9c_z\,\hat{\mathbf{z}}$ | (2$i$) | O V |
| $\mathbf{B}_{18}$ | = | $-x_9\,\mathbf{a}_1 - y_9\,\mathbf{a}_2 - z_9\,\mathbf{a}_3$ | = | $(-x_9a - y_9b\cos\gamma - z_9c_x)\,\hat{\mathbf{x}} + \left(-y_9b\sin\gamma - z_9c_y\right)\hat{\mathbf{y}} - z_9c_z\,\hat{\mathbf{z}}$ | (2$i$) | O V |
| $\mathbf{B}_{19}$ | = | $x_{10}\,\mathbf{a}_1 + y_{10}\,\mathbf{a}_2 + z_{10}\,\mathbf{a}_3$ | = | $(x_{10}a + y_{10}b\cos\gamma + z_{10}c_x)\,\hat{\mathbf{x}} + \left(y_{10}b\sin\gamma + z_{10}c_y\right)\hat{\mathbf{y}} + z_{10}c_z\,\hat{\mathbf{z}}$ | (2$i$) | O VI |
| $\mathbf{B}_{20}$ | = | $-x_{10}\,\mathbf{a}_1 - y_{10}\,\mathbf{a}_2 - z_{10}\,\mathbf{a}_3$ | = | $(-x_{10}a - y_{10}b\cos\gamma - z_{10}c_x)\,\hat{\mathbf{x}} + \left(-y_{10}b\sin\gamma - z_{10}c_y\right)\hat{\mathbf{y}} - z_{10}c_z\,\hat{\mathbf{z}}$ | (2$i$) | O VI |
| $\mathbf{B}_{21}$ | = | $x_{11}\,\mathbf{a}_1 + y_{11}\,\mathbf{a}_2 + z_{11}\,\mathbf{a}_3$ | = | $(x_{11}a + y_{11}b\cos\gamma + z_{11}c_x)\,\hat{\mathbf{x}} + \left(y_{11}b\sin\gamma + z_{11}c_y\right)\hat{\mathbf{y}} + z_{11}c_z\,\hat{\mathbf{z}}$ | (2$i$) | O VII |
| $\mathbf{B}_{22}$ | = | $-x_{11}\,\mathbf{a}_1 - y_{11}\,\mathbf{a}_2 - z_{11}\,\mathbf{a}_3$ | = | $(-x_{11}a - y_{11}b\cos\gamma - z_{11}c_x)\,\hat{\mathbf{x}} + \left(-y_{11}b\sin\gamma - z_{11}c_y\right)\hat{\mathbf{y}} - z_{11}c_z\,\hat{\mathbf{z}}$ | (2$i$) | O VII |
| $\mathbf{B}_{23}$ | = | $x_{12}\,\mathbf{a}_1 + y_{12}\,\mathbf{a}_2 + z_{12}\,\mathbf{a}_3$ | = | $(x_{12}a + y_{12}b\cos\gamma + z_{12}c_x)\,\hat{\mathbf{x}} + \left(y_{12}b\sin\gamma + z_{12}c_y\right)\hat{\mathbf{y}} + z_{12}c_z\,\hat{\mathbf{z}}$ | (2$i$) | O VIII |
| $\mathbf{B}_{24}$ | = | $-x_{12}\,\mathbf{a}_1 - y_{12}\,\mathbf{a}_2 - z_{12}\,\mathbf{a}_3$ | = | $(-x_{12}a - y_{12}b\cos\gamma - z_{12}c_x)\,\hat{\mathbf{x}} + \left(-y_{12}b\sin\gamma - z_{12}c_y\right)\hat{\mathbf{y}} - z_{12}c_z\,\hat{\mathbf{z}}$ | (2$i$) | O VIII |
| $\mathbf{B}_{25}$ | = | $x_{13}\,\mathbf{a}_1 + y_{13}\,\mathbf{a}_2 + z_{13}\,\mathbf{a}_3$ | = | $(x_{13}a + y_{13}b\cos\gamma + z_{13}c_x)\,\hat{\mathbf{x}} + \left(y_{13}b\sin\gamma + z_{13}c_y\right)\hat{\mathbf{y}} + z_{13}c_z\,\hat{\mathbf{z}}$ | (2$i$) | O IX |
| $\mathbf{B}_{26}$ | = | $-x_{13}\,\mathbf{a}_1 - y_{13}\,\mathbf{a}_2 - z_{13}\,\mathbf{a}_3$ | = | $(-x_{13}a - y_{13}b\cos\gamma - z_{13}c_x)\,\hat{\mathbf{x}} + \left(-y_{13}b\sin\gamma - z_{13}c_y\right)\hat{\mathbf{y}} - z_{13}c_z\,\hat{\mathbf{z}}$ | (2$i$) | O IX |
| $\mathbf{B}_{27}$ | = | $x_{14}\,\mathbf{a}_1 + y_{14}\,\mathbf{a}_2 + z_{14}\,\mathbf{a}_3$ | = | $(x_{14}a + y_{14}b\cos\gamma + z_{14}c_x)\,\hat{\mathbf{x}} + \left(y_{14}b\sin\gamma + z_{14}c_y\right)\hat{\mathbf{y}} + z_{14}c_z\,\hat{\mathbf{z}}$ | (2$i$) | O X |
| $\mathbf{B}_{28}$ | = | $-x_{14}\,\mathbf{a}_1 - y_{14}\,\mathbf{a}_2 - z_{14}\,\mathbf{a}_3$ | = | $(-x_{14}a - y_{14}b\cos\gamma - z_{14}c_x)\,\hat{\mathbf{x}} + \left(-y_{14}b\sin\gamma - z_{14}c_y\right)\hat{\mathbf{y}} - z_{14}c_z\,\hat{\mathbf{z}}$ | (2$i$) | O X |
| $\mathbf{B}_{29}$ | = | $x_{15}\,\mathbf{a}_1 + y_{15}\,\mathbf{a}_2 + z_{15}\,\mathbf{a}_3$ | = | $(x_{15}a + y_{15}b\cos\gamma + z_{15}c_x)\,\hat{\mathbf{x}} + \left(y_{15}b\sin\gamma + z_{15}c_y\right)\hat{\mathbf{y}} + z_{15}c_z\,\hat{\mathbf{z}}$ | (2$i$) | O XI |
| $\mathbf{B}_{30}$ | = | $-x_{15}\,\mathbf{a}_1 - y_{15}\,\mathbf{a}_2 - z_{15}\,\mathbf{a}_3$ | = | $(-x_{15}a - y_{15}b\cos\gamma - z_{15}c_x)\,\hat{\mathbf{x}} + \left(-y_{15}b\sin\gamma - z_{15}c_y\right)\hat{\mathbf{y}} - z_{15}c_z\,\hat{\mathbf{z}}$ | (2$i$) | O XI |
| $\mathbf{B}_{31}$ | = | $x_{16}\,\mathbf{a}_1 + y_{16}\,\mathbf{a}_2 + z_{16}\,\mathbf{a}_3$ | = | $(x_{16}a + y_{16}b\cos\gamma + z_{16}c_x)\,\hat{\mathbf{x}} + \left(y_{16}b\sin\gamma + z_{16}c_y\right)\hat{\mathbf{y}} + z_{16}c_z\,\hat{\mathbf{z}}$ | (2$i$) | O XII |
| $\mathbf{B}_{32}$ | = | $-x_{16}\,\mathbf{a}_1 - y_{16}\,\mathbf{a}_2 - z_{16}\,\mathbf{a}_3$ | = | $(-x_{16}a - y_{16}b\cos\gamma - z_{16}c_x)\,\hat{\mathbf{x}} + \left(-y_{16}b\sin\gamma - z_{16}c_y\right)\hat{\mathbf{y}} - z_{16}c_z\,\hat{\mathbf{z}}$ | (2$i$) | O XII |
| $\mathbf{B}_{33}$ | = | $x_{17}\,\mathbf{a}_1 + y_{17}\,\mathbf{a}_2 + z_{17}\,\mathbf{a}_3$ | = | $(x_{17}a + y_{17}b\cos\gamma + z_{17}c_x)\,\hat{\mathbf{x}} + \left(y_{17}b\sin\gamma + z_{17}c_y\right)\hat{\mathbf{y}} + z_{17}c_z\,\hat{\mathbf{z}}$ | (2$i$) | O XIII |
| $\mathbf{B}_{34}$ | = | $-x_{17}\,\mathbf{a}_1 - y_{17}\,\mathbf{a}_2 - z_{17}\,\mathbf{a}_3$ | = | $(-x_{17}a - y_{17}b\cos\gamma - z_{17}c_x)\,\hat{\mathbf{x}} + \left(-y_{17}b\sin\gamma - z_{17}c_y\right)\hat{\mathbf{y}} - z_{17}c_z\,\hat{\mathbf{z}}$ | (2$i$) | O XIII |
| $\mathbf{B}_{35}$ | = | $x_{18}\,\mathbf{a}_1 + y_{18}\,\mathbf{a}_2 + z_{18}\,\mathbf{a}_3$ | = | $(x_{18}a + y_{18}b\cos\gamma + z_{18}c_x)\,\hat{\mathbf{x}} + \left(y_{18}b\sin\gamma + z_{18}c_y\right)\hat{\mathbf{y}} + z_{18}c_z\,\hat{\mathbf{z}}$ | (2$i$) | O XIV |
| $\mathbf{B}_{36}$ | = | $-x_{18}\,\mathbf{a}_1 - y_{18}\,\mathbf{a}_2 - z_{18}\,\mathbf{a}_3$ | = | $(-x_{18}a - y_{18}b\cos\gamma - z_{18}c_x)\,\hat{\mathbf{x}} + \left(-y_{18}b\sin\gamma - z_{18}c_y\right)\hat{\mathbf{y}} - z_{18}c_z\,\hat{\mathbf{z}}$ | (2$i$) | O XIV |
| $\mathbf{B}_{37}$ | = | $x_{19}\,\mathbf{a}_1 + y_{19}\,\mathbf{a}_2 + z_{19}\,\mathbf{a}_3$ | = | $(x_{19}a + y_{19}b\cos\gamma + z_{19}c_x)\,\hat{\mathbf{x}} + \left(y_{19}b\sin\gamma + z_{19}c_y\right)\hat{\mathbf{y}} + z_{19}c_z\,\hat{\mathbf{z}}$ | (2$i$) | Si I |
| $\mathbf{B}_{38}$ | = | $-x_{19}\,\mathbf{a}_1 - y_{19}\,\mathbf{a}_2 - z_{19}\,\mathbf{a}_3$ | = | $(-x_{19}a - y_{19}b\cos\gamma - z_{19}c_x)\,\hat{\mathbf{x}} + \left(-y_{19}b\sin\gamma - z_{19}c_y\right)\hat{\mathbf{y}} - z_{19}c_z\,\hat{\mathbf{z}}$ | (2$i$) | Si I |
| $\mathbf{B}_{39}$ | = | $x_{20}\,\mathbf{a}_1 + y_{20}\,\mathbf{a}_2 + z_{20}\,\mathbf{a}_3$ | = | $(x_{20}a + y_{20}b\cos\gamma + z_{20}c_x)\,\hat{\mathbf{x}} + \left(y_{20}b\sin\gamma + z_{20}c_y\right)\hat{\mathbf{y}} + z_{20}c_z\,\hat{\mathbf{z}}$ | (2$i$) | Si II |
| $\mathbf{B}_{40}$ | = | $-x_{20}\,\mathbf{a}_1 - y_{20}\,\mathbf{a}_2 - z_{20}\,\mathbf{a}_3$ | = | $(-x_{20}a - y_{20}b\cos\gamma - z_{20}c_x)\,\hat{\mathbf{x}} + \left(-y_{20}b\sin\gamma - z_{20}c_y\right)\hat{\mathbf{y}} - z_{20}c_z\,\hat{\mathbf{z}}$ | (2$i$) | Si II |
| $\mathbf{B}_{41}$ | = | $x_{21}\,\mathbf{a}_1 + y_{21}\,\mathbf{a}_2 + z_{21}\,\mathbf{a}_3$ | = | $(x_{21}a + y_{21}b\cos\gamma + z_{21}c_x)\,\hat{\mathbf{x}} + \left(y_{21}b\sin\gamma + z_{21}c_y\right)\hat{\mathbf{y}} + z_{21}c_z\,\hat{\mathbf{z}}$ | (2$i$) | Si III |
| $\mathbf{B}_{42}$ | = | $-x_{21}\,\mathbf{a}_1 - y_{21}\,\mathbf{a}_2 - z_{21}\,\mathbf{a}_3$ | = | $(-x_{21}a - y_{21}b\cos\gamma - z_{21}c_x)\,\hat{\mathbf{x}} + \left(-y_{21}b\sin\gamma - z_{21}c_y\right)\hat{\mathbf{y}} - z_{21}c_z\,\hat{\mathbf{z}}$ | (2$i$) | Si III |
| $\mathbf{B}_{43}$ | = | $x_{22}\,\mathbf{a}_1 + y_{22}\,\mathbf{a}_2 + z_{22}\,\mathbf{a}_3$ | = | $(x_{22}a + y_{22}b\cos\gamma + z_{22}c_x)\,\hat{\mathbf{x}} + \left(y_{22}b\sin\gamma + z_{22}c_y\right)\hat{\mathbf{y}} + z_{22}c_z\,\hat{\mathbf{z}}$ | (2$i$) | Si IV |
| $\mathbf{B}_{44}$ | = | $-x_{22}\,\mathbf{a}_1 - y_{22}\,\mathbf{a}_2 - z_{22}\,\mathbf{a}_3$ | = | $(-x_{22}a - y_{22}b\cos\gamma - z_{22}c_x)\,\hat{\mathbf{x}} + \left(-y_{22}b\sin\gamma - z_{22}c_y\right)\hat{\mathbf{y}} - z_{22}c_z\,\hat{\mathbf{z}}$ | (2$i$) | Si IV |

**Found in:**

---

**Geometry files:**

- CIF: pp. 1502
- POSCAR: pp. 1503

# $Co_3(SeO_3)_3{\cdot}H_2O$ Structure: A3B2C10D3_aP36_2_ah2i_2i_10i_3i

http://aflow.org/prototype-encyclopedia/A3B2C10D3_aP36_2_ah2i_2i_10i_3i

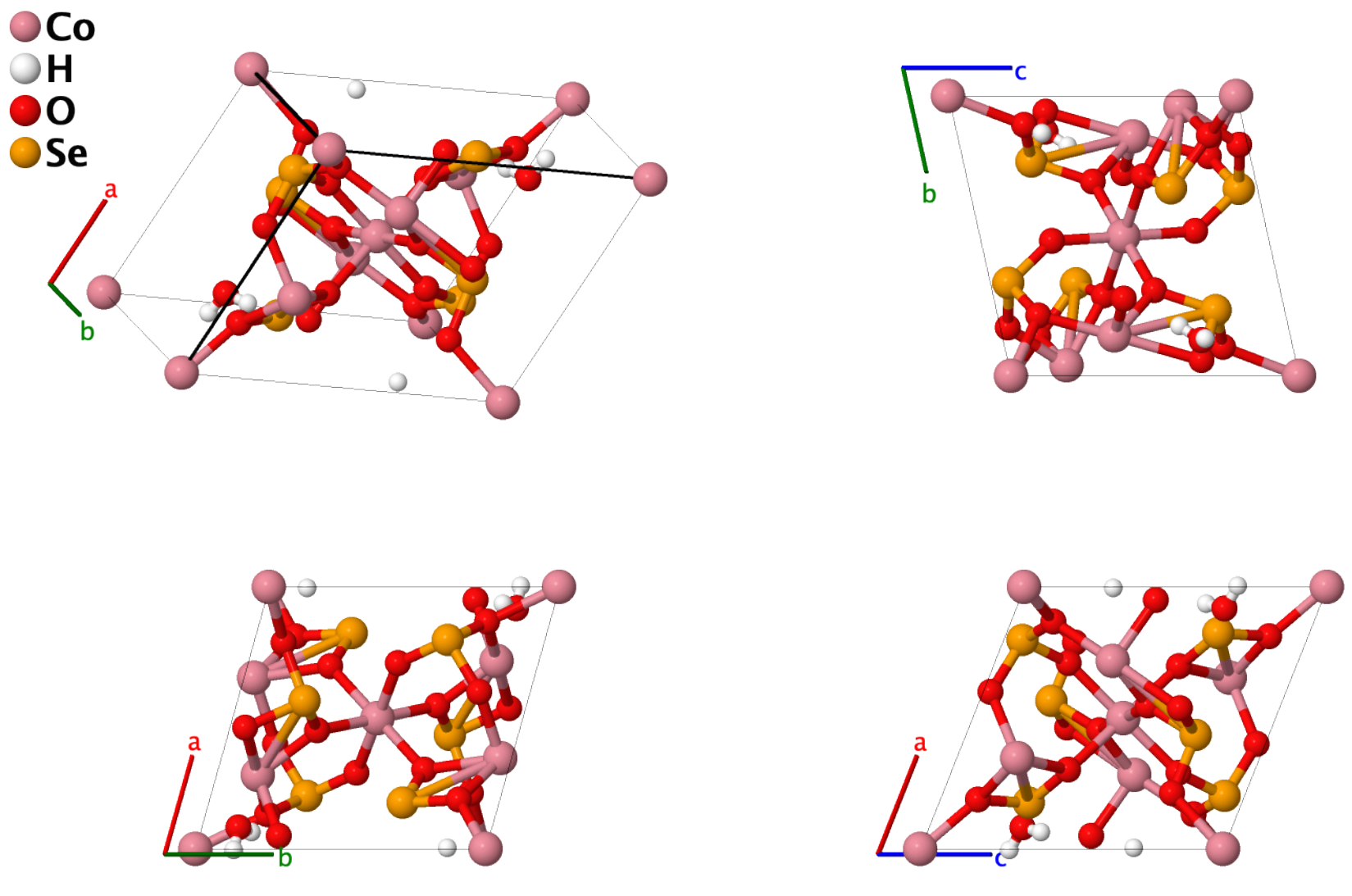

| | | |
|---|---|---|
| **Prototype** | : | $Co_3H_2O_{10}Se_3$ |
| **AFLOW prototype label** | : | A3B2C10D3_aP36_2_ah2i_2i_10i_3i |
| ***Strukturbericht* designation** | : | None |
| **Pearson symbol** | : | aP36 |
| **Space group number** | : | 2 |
| **Space group symbol** | : | $P\bar{1}$ |
| **AFLOW prototype command** | : | `aflow --proto=A3B2C10D3_aP36_2_ah2i_2i_10i_3i --params=`$a, b/a, c/a, \alpha, \beta, \gamma, x_3, y_3, z_3, x_4, y_4, z_4, x_5, y_5, z_5, x_6, y_6, z_6, x_7, y_7, z_7, x_8, y_8, z_8, x_9, y_9, z_9, x_{10}, y_{10}, z_{10}, x_{11}, y_{11}, z_{11}, x_{12}, y_{12}, z_{12}, x_{13}, y_{13}, z_{13}, x_{14}, y_{14}, z_{14}, x_{15}, y_{15}, z_{15}, x_{16}, y_{16}, z_{16}, x_{17}, y_{17}, z_{17}, x_{18}, y_{18}, z_{18}, x_{19}, y_{19}, z_{19}$ |

### Other compounds with this structure

- $Mn_3(SeO_3)_3{\cdot}H_2O$ and $Ni_3(SeO_3)_3{\cdot}H_2O$

### Triclinic primitive vectors:

$$\mathbf{a}_1 = a\,\hat{\mathbf{x}}$$

$$\mathbf{a}_2 = b\cos\gamma\,\hat{\mathbf{x}} + b\sin\gamma\,\hat{\mathbf{y}}$$

$$\mathbf{a}_3 = c_x\,\hat{\mathbf{x}} + c_y\,\hat{\mathbf{y}} + c_z\,\hat{\mathbf{z}}$$

$$c_x = c\,\cos\beta$$

$$c_y = c\,(\cos\alpha - \cos\beta\cos\gamma)/\sin\gamma$$

$$c_z = \sqrt{c^2 - c_x^2 - c_y^2}$$

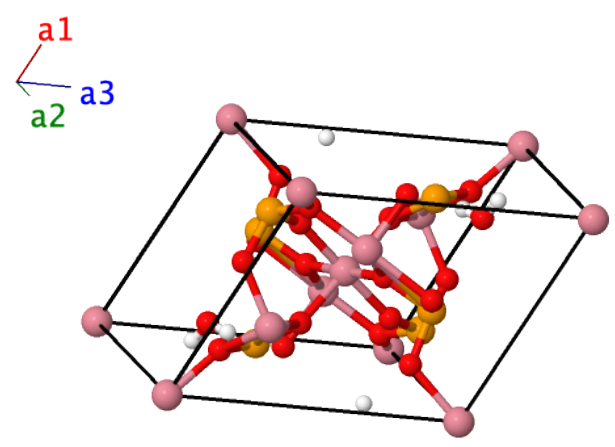

### Basis vectors:

| | | Lattice Coordinates | | Cartesian Coordinates | Wyckoff Position | Atom Type |
|---|---|---|---|---|---|---|
| $\mathbf{B}_{1}$ | = | $0\,\mathbf{a}_{1}+0\,\mathbf{a}_{2}+0\,\mathbf{a}_{3}$ | = | $0\,\hat{\mathbf{x}}+0\,\hat{\mathbf{y}}+0\,\hat{\mathbf{z}}$ | (1$a$) | Co I |
| $\mathbf{B}_{2}$ | = | $\frac{1}{2}\,\mathbf{a}_{1}+\frac{1}{2}\,\mathbf{a}_{2}+\frac{1}{2}\,\mathbf{a}_{3}$ | = | $\frac{1}{2}\left(a+b\cos\gamma+c_{x}\right)\,\hat{\mathbf{x}}+\frac{1}{2}\left(b\sin\gamma+c_{y}\right)\,\hat{\mathbf{y}}+\frac{1}{2}c_{z}\,\hat{\mathbf{z}}$ | (1$h$) | Co II |
| $\mathbf{B}_{3}$ | = | $x_{3}\,\mathbf{a}_{1}+y_{3}\,\mathbf{a}_{2}+z_{3}\,\mathbf{a}_{3}$ | = | $\left(x_{3}a+y_{3}b\cos\gamma+z_{3}c_{x}\right)\,\hat{\mathbf{x}}+\left(y_{3}b\sin\gamma+z_{3}c_{y}\right)\,\hat{\mathbf{y}}+z_{3}c_{z}\,\hat{\mathbf{z}}$ | (2$i$) | Co III |
| $\mathbf{B}_{4}$ | = | $-x_{3}\,\mathbf{a}_{1}-y_{3}\,\mathbf{a}_{2}-z_{3}\,\mathbf{a}_{3}$ | = | $\left(-x_{3}a-y_{3}b\cos\gamma-z_{3}c_{x}\right)\,\hat{\mathbf{x}}+\left(-y_{3}b\sin\gamma-z_{3}c_{y}\right)\,\hat{\mathbf{y}}-z_{3}c_{z}\,\hat{\mathbf{z}}$ | (2$i$) | Co III |
| $\mathbf{B}_{5}$ | = | $x_{4}\,\mathbf{a}_{1}+y_{4}\,\mathbf{a}_{2}+z_{4}\,\mathbf{a}_{3}$ | = | $\left(x_{4}a+y_{4}b\cos\gamma+z_{4}c_{x}\right)\,\hat{\mathbf{x}}+\left(y_{4}b\sin\gamma+z_{4}c_{y}\right)\,\hat{\mathbf{y}}+z_{4}c_{z}\,\hat{\mathbf{z}}$ | (2$i$) | Co IV |
| $\mathbf{B}_{6}$ | = | $-x_{4}\,\mathbf{a}_{1}-y_{4}\,\mathbf{a}_{2}-z_{4}\,\mathbf{a}_{3}$ | = | $\left(-x_{4}a-y_{4}b\cos\gamma-z_{4}c_{x}\right)\,\hat{\mathbf{x}}+\left(-y_{4}b\sin\gamma-z_{4}c_{y}\right)\,\hat{\mathbf{y}}-z_{4}c_{z}\,\hat{\mathbf{z}}$ | (2$i$) | Co IV |
| $\mathbf{B}_{7}$ | = | $x_{5}\,\mathbf{a}_{1}+y_{5}\,\mathbf{a}_{2}+z_{5}\,\mathbf{a}_{3}$ | = | $\left(x_{5}a+y_{5}b\cos\gamma+z_{5}c_{x}\right)\,\hat{\mathbf{x}}+\left(y_{5}b\sin\gamma+z_{5}c_{y}\right)\,\hat{\mathbf{y}}+z_{5}c_{z}\,\hat{\mathbf{z}}$ | (2$i$) | H I |
| $\mathbf{B}_{8}$ | = | $-x_{5}\,\mathbf{a}_{1}-y_{5}\,\mathbf{a}_{2}-z_{5}\,\mathbf{a}_{3}$ | = | $\left(-x_{5}a-y_{5}b\cos\gamma-z_{5}c_{x}\right)\,\hat{\mathbf{x}}+\left(-y_{5}b\sin\gamma-z_{5}c_{y}\right)\,\hat{\mathbf{y}}-z_{5}c_{z}\,\hat{\mathbf{z}}$ | (2$i$) | H I |
| $\mathbf{B}_{9}$ | = | $x_{6}\,\mathbf{a}_{1}+y_{6}\,\mathbf{a}_{2}+z_{6}\,\mathbf{a}_{3}$ | = | $\left(x_{6}a+y_{6}b\cos\gamma+z_{6}c_{x}\right)\,\hat{\mathbf{x}}+\left(y_{6}b\sin\gamma+z_{6}c_{y}\right)\,\hat{\mathbf{y}}+z_{6}c_{z}\,\hat{\mathbf{z}}$ | (2$i$) | H II |
| $\mathbf{B}_{10}$ | = | $-x_{6}\,\mathbf{a}_{1}-y_{6}\,\mathbf{a}_{2}-z_{6}\,\mathbf{a}_{3}$ | = | $\left(-x_{6}a-y_{6}b\cos\gamma-z_{6}c_{x}\right)\,\hat{\mathbf{x}}+\left(-y_{6}b\sin\gamma-z_{6}c_{y}\right)\,\hat{\mathbf{y}}-z_{6}c_{z}\,\hat{\mathbf{z}}$ | (2$i$) | H II |
| $\mathbf{B}_{11}$ | = | $x_{7}\,\mathbf{a}_{1}+y_{7}\,\mathbf{a}_{2}+z_{7}\,\mathbf{a}_{3}$ | = | $\left(x_{7}a+y_{7}b\cos\gamma+z_{7}c_{x}\right)\,\hat{\mathbf{x}}+\left(y_{7}b\sin\gamma+z_{7}c_{y}\right)\,\hat{\mathbf{y}}+z_{7}c_{z}\,\hat{\mathbf{z}}$ | (2$i$) | O I |
| $\mathbf{B}_{12}$ | = | $-x_{7}\,\mathbf{a}_{1}-y_{7}\,\mathbf{a}_{2}-z_{7}\,\mathbf{a}_{3}$ | = | $\left(-x_{7}a-y_{7}b\cos\gamma-z_{7}c_{x}\right)\,\hat{\mathbf{x}}+\left(-y_{7}b\sin\gamma-z_{7}c_{y}\right)\,\hat{\mathbf{y}}-z_{7}c_{z}\,\hat{\mathbf{z}}$ | (2$i$) | O I |
| $\mathbf{B}_{13}$ | = | $x_{8}\,\mathbf{a}_{1}+y_{8}\,\mathbf{a}_{2}+z_{8}\,\mathbf{a}_{3}$ | = | $\left(x_{8}a+y_{8}b\cos\gamma+z_{8}c_{x}\right)\,\hat{\mathbf{x}}+\left(y_{8}b\sin\gamma+z_{8}c_{y}\right)\,\hat{\mathbf{y}}+z_{8}c_{z}\,\hat{\mathbf{z}}$ | (2$i$) | O II |
| $\mathbf{B}_{14}$ | = | $-x_{8}\,\mathbf{a}_{1}-y_{8}\,\mathbf{a}_{2}-z_{8}\,\mathbf{a}_{3}$ | = | $\left(-x_{8}a-y_{8}b\cos\gamma-z_{8}c_{x}\right)\,\hat{\mathbf{x}}+\left(-y_{8}b\sin\gamma-z_{8}c_{y}\right)\,\hat{\mathbf{y}}-z_{8}c_{z}\,\hat{\mathbf{z}}$ | (2$i$) | O II |
| $\mathbf{B}_{15}$ | = | $x_{9}\,\mathbf{a}_{1}+y_{9}\,\mathbf{a}_{2}+z_{9}\,\mathbf{a}_{3}$ | = | $\left(x_{9}a+y_{9}b\cos\gamma+z_{9}c_{x}\right)\,\hat{\mathbf{x}}+\left(y_{9}b\sin\gamma+z_{9}c_{y}\right)\,\hat{\mathbf{y}}+z_{9}c_{z}\,\hat{\mathbf{z}}$ | (2$i$) | O III |
| $\mathbf{B}_{16}$ | = | $-x_{9}\,\mathbf{a}_{1}-y_{9}\,\mathbf{a}_{2}-z_{9}\,\mathbf{a}_{3}$ | = | $\left(-x_{9}a-y_{9}b\cos\gamma-z_{9}c_{x}\right)\,\hat{\mathbf{x}}+\left(-y_{9}b\sin\gamma-z_{9}c_{y}\right)\,\hat{\mathbf{y}}-z_{9}c_{z}\,\hat{\mathbf{z}}$ | (2$i$) | O III |
| $\mathbf{B}_{17}$ | = | $x_{10}\,\mathbf{a}_{1}+y_{10}\,\mathbf{a}_{2}+z_{10}\,\mathbf{a}_{3}$ | = | $\left(x_{10}a+y_{10}b\cos\gamma+z_{10}c_{x}\right)\,\hat{\mathbf{x}}+\left(y_{10}b\sin\gamma+z_{10}c_{y}\right)\,\hat{\mathbf{y}}+z_{10}c_{z}\,\hat{\mathbf{z}}$ | (2$i$) | O IV |
| $\mathbf{B}_{18}$ | = | $-x_{10}\,\mathbf{a}_{1}-y_{10}\,\mathbf{a}_{2}-z_{10}\,\mathbf{a}_{3}$ | = | $\left(-x_{10}a-y_{10}b\cos\gamma-z_{10}c_{x}\right)\,\hat{\mathbf{x}}+\left(-y_{10}b\sin\gamma-z_{10}c_{y}\right)\,\hat{\mathbf{y}}-z_{10}c_{z}\,\hat{\mathbf{z}}$ | (2$i$) | O IV |
| $\mathbf{B}_{19}$ | = | $x_{11}\,\mathbf{a}_{1}+y_{11}\,\mathbf{a}_{2}+z_{11}\,\mathbf{a}_{3}$ | = | $\left(x_{11}a+y_{11}b\cos\gamma+z_{11}c_{x}\right)\,\hat{\mathbf{x}}+\left(y_{11}b\sin\gamma+z_{11}c_{y}\right)\,\hat{\mathbf{y}}+z_{11}c_{z}\,\hat{\mathbf{z}}$ | (2$i$) | O V |
| $\mathbf{B}_{20}$ | = | $-x_{11}\,\mathbf{a}_{1}-y_{11}\,\mathbf{a}_{2}-z_{11}\,\mathbf{a}_{3}$ | = | $\left(-x_{11}a-y_{11}b\cos\gamma-z_{11}c_{x}\right)\,\hat{\mathbf{x}}+\left(-y_{11}b\sin\gamma-z_{11}c_{y}\right)\,\hat{\mathbf{y}}-z_{11}c_{z}\,\hat{\mathbf{z}}$ | (2$i$) | O V |
| $\mathbf{B}_{21}$ | = | $x_{12}\,\mathbf{a}_{1}+y_{12}\,\mathbf{a}_{2}+z_{12}\,\mathbf{a}_{3}$ | = | $\left(x_{12}a+y_{12}b\cos\gamma+z_{12}c_{x}\right)\,\hat{\mathbf{x}}+\left(y_{12}b\sin\gamma+z_{12}c_{y}\right)\,\hat{\mathbf{y}}+z_{12}c_{z}\,\hat{\mathbf{z}}$ | (2$i$) | O VI |
| $\mathbf{B}_{22}$ | = | $-x_{12}\,\mathbf{a}_{1}-y_{12}\,\mathbf{a}_{2}-z_{12}\,\mathbf{a}_{3}$ | = | $\left(-x_{12}a-y_{12}b\cos\gamma-z_{12}c_{x}\right)\,\hat{\mathbf{x}}+\left(-y_{12}b\sin\gamma-z_{12}c_{y}\right)\,\hat{\mathbf{y}}-z_{12}c_{z}\,\hat{\mathbf{z}}$ | (2$i$) | O VI |
| $\mathbf{B}_{23}$ | = | $x_{13}\,\mathbf{a}_1 + y_{13}\,\mathbf{a}_2 + z_{13}\,\mathbf{a}_3$ | = | $(x_{13}a + y_{13}b\cos\gamma + z_{13}c_x)\,\hat{\mathbf{x}} + \left(y_{13}b\sin\gamma + z_{13}c_y\right)\,\hat{\mathbf{y}} + z_{13}c_z\,\hat{\mathbf{z}}$ | (2i) | O VII |
| $\mathbf{B}_{24}$ | = | $-x_{13}\,\mathbf{a}_1 - y_{13}\,\mathbf{a}_2 - z_{13}\,\mathbf{a}_3$ | = | $(-x_{13}a - y_{13}b\cos\gamma - z_{13}c_x)\,\hat{\mathbf{x}} + \left(-y_{13}b\sin\gamma - z_{13}c_y\right)\,\hat{\mathbf{y}} - z_{13}c_z\,\hat{\mathbf{z}}$ | (2i) | O VII |
| $\mathbf{B}_{25}$ | = | $x_{14}\,\mathbf{a}_1 + y_{14}\,\mathbf{a}_2 + z_{14}\,\mathbf{a}_3$ | = | $(x_{14}a + y_{14}b\cos\gamma + z_{14}c_x)\,\hat{\mathbf{x}} + \left(y_{14}b\sin\gamma + z_{14}c_y\right)\,\hat{\mathbf{y}} + z_{14}c_z\,\hat{\mathbf{z}}$ | (2i) | O VIII |
| $\mathbf{B}_{26}$ | = | $-x_{14}\,\mathbf{a}_1 - y_{14}\,\mathbf{a}_2 - z_{14}\,\mathbf{a}_3$ | = | $(-x_{14}a - y_{14}b\cos\gamma - z_{14}c_x)\,\hat{\mathbf{x}} + \left(-y_{14}b\sin\gamma - z_{14}c_y\right)\,\hat{\mathbf{y}} - z_{14}c_z\,\hat{\mathbf{z}}$ | (2i) | O VIII |
| $\mathbf{B}_{27}$ | = | $x_{15}\,\mathbf{a}_1 + y_{15}\,\mathbf{a}_2 + z_{15}\,\mathbf{a}_3$ | = | $(x_{15}a + y_{15}b\cos\gamma + z_{15}c_x)\,\hat{\mathbf{x}} + \left(y_{15}b\sin\gamma + z_{15}c_y\right)\,\hat{\mathbf{y}} + z_{15}c_z\,\hat{\mathbf{z}}$ | (2i) | O IX |
| $\mathbf{B}_{28}$ | = | $-x_{15}\,\mathbf{a}_1 - y_{15}\,\mathbf{a}_2 - z_{15}\,\mathbf{a}_3$ | = | $(-x_{15}a - y_{15}b\cos\gamma - z_{15}c_x)\,\hat{\mathbf{x}} + \left(-y_{15}b\sin\gamma - z_{15}c_y\right)\,\hat{\mathbf{y}} - z_{15}c_z\,\hat{\mathbf{z}}$ | (2i) | O IX |
| $\mathbf{B}_{29}$ | = | $x_{16}\,\mathbf{a}_1 + y_{16}\,\mathbf{a}_2 + z_{16}\,\mathbf{a}_3$ | = | $(x_{16}a + y_{16}b\cos\gamma + z_{16}c_x)\,\hat{\mathbf{x}} + \left(y_{16}b\sin\gamma + z_{16}c_y\right)\,\hat{\mathbf{y}} + z_{16}c_z\,\hat{\mathbf{z}}$ | (2i) | O X |
| $\mathbf{B}_{30}$ | = | $-x_{16}\,\mathbf{a}_1 - y_{16}\,\mathbf{a}_2 - z_{16}\,\mathbf{a}_3$ | = | $(-x_{16}a - y_{16}b\cos\gamma - z_{16}c_x)\,\hat{\mathbf{x}} + \left(-y_{16}b\sin\gamma - z_{16}c_y\right)\,\hat{\mathbf{y}} - z_{16}c_z\,\hat{\mathbf{z}}$ | (2i) | O X |
| $\mathbf{B}_{31}$ | = | $x_{17}\,\mathbf{a}_1 + y_{17}\,\mathbf{a}_2 + z_{17}\,\mathbf{a}_3$ | = | $(x_{17}a + y_{17}b\cos\gamma + z_{17}c_x)\,\hat{\mathbf{x}} + \left(y_{17}b\sin\gamma + z_{17}c_y\right)\,\hat{\mathbf{y}} + z_{17}c_z\,\hat{\mathbf{z}}$ | (2i) | Se I |
| $\mathbf{B}_{32}$ | = | $-x_{17}\,\mathbf{a}_1 - y_{17}\,\mathbf{a}_2 - z_{17}\,\mathbf{a}_3$ | = | $(-x_{17}a - y_{17}b\cos\gamma - z_{17}c_x)\,\hat{\mathbf{x}} + \left(-y_{17}b\sin\gamma - z_{17}c_y\right)\,\hat{\mathbf{y}} - z_{17}c_z\,\hat{\mathbf{z}}$ | (2i) | Se I |
| $\mathbf{B}_{33}$ | = | $x_{18}\,\mathbf{a}_1 + y_{18}\,\mathbf{a}_2 + z_{18}\,\mathbf{a}_3$ | = | $(x_{18}a + y_{18}b\cos\gamma + z_{18}c_x)\,\hat{\mathbf{x}} + \left(y_{18}b\sin\gamma + z_{18}c_y\right)\,\hat{\mathbf{y}} + z_{18}c_z\,\hat{\mathbf{z}}$ | (2i) | Se II |
| $\mathbf{B}_{34}$ | = | $-x_{18}\,\mathbf{a}_1 - y_{18}\,\mathbf{a}_2 - z_{18}\,\mathbf{a}_3$ | = | $(-x_{18}a - y_{18}b\cos\gamma - z_{18}c_x)\,\hat{\mathbf{x}} + \left(-y_{18}b\sin\gamma - z_{18}c_y\right)\,\hat{\mathbf{y}} - z_{18}c_z\,\hat{\mathbf{z}}$ | (2i) | Se II |
| $\mathbf{B}_{35}$ | = | $x_{19}\,\mathbf{a}_1 + y_{19}\,\mathbf{a}_2 + z_{19}\,\mathbf{a}_3$ | = | $(x_{19}a + y_{19}b\cos\gamma + z_{19}c_x)\,\hat{\mathbf{x}} + \left(y_{19}b\sin\gamma + z_{19}c_y\right)\,\hat{\mathbf{y}} + z_{19}c_z\,\hat{\mathbf{z}}$ | (2i) | Se III |
| $\mathbf{B}_{36}$ | = | $-x_{19}\,\mathbf{a}_1 - y_{19}\,\mathbf{a}_2 - z_{19}\,\mathbf{a}_3$ | = | $(-x_{19}a - y_{19}b\cos\gamma - z_{19}c_x)\,\hat{\mathbf{x}} + \left(-y_{19}b\sin\gamma - z_{19}c_y\right)\,\hat{\mathbf{y}} - z_{19}c_z\,\hat{\mathbf{z}}$ | (2i) | Se III |

---

**Found in:**

---

**Geometry files:**

- CIF: pp. 1503
- POSCAR: pp. 1504

# $\delta$-$WO_3$ Structure: A3B_aP32_2_12i_4i

http://aflow.org/prototype-encyclopedia/A3B_aP32_2_12i_4i

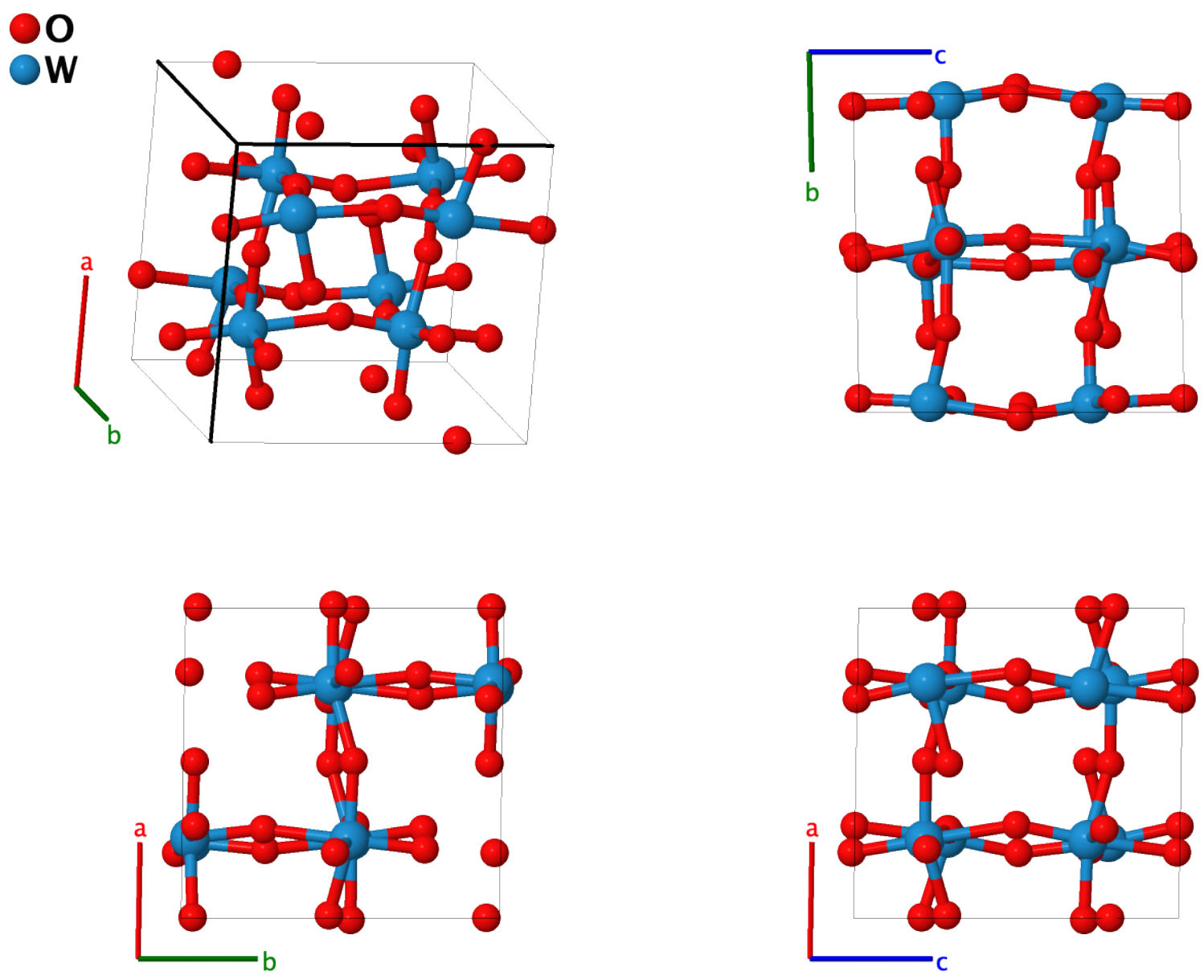

| | | |
|---|---|---|
| **Prototype** | : | $O_3W$ |
| **AFLOW prototype label** | : | A3B_aP32_2_12i_4i |
| ***Strukturbericht* designation** | : | None |
| **Pearson symbol** | : | aP32 |
| **Space group number** | : | 2 |
| **Space group symbol** | : | $P\bar{1}$ |
| **AFLOW prototype command** | : | `aflow --proto=A3B_aP32_2_12i_4i` `--params=`$a, b/a, c/a, \alpha, \beta, \gamma, x_1, y_1, z_1, x_2, y_2, z_2, x_3, y_3, z_3, x_4, y_4, z_4, x_5, y_5, z_5, x_6, y_6, z_6, x_7, y_7, z_7, x_8, y_8, z_8, x_9, y_9, z_9, x_{10}, y_{10}, z_{10}, x_{11}, y_{11}, z_{11}, x_{12}, y_{12}, z_{12}, x_{13}, y_{13}, z_{13}, x_{14}, y_{14}, z_{14}, x_{15}, y_{15}, z_{15}, x_{16}, y_{16}, z_{16}$ |

- All stable phases of $WO_3$ are distortions of the cubic $\alpha$-$ReO_3$ ($D0_9$) phase. (Woodward, 1997 and Vogt, 1999) The known stable phases and their approximate temperature ranges are:
  - $\alpha$-$WO_3$ (1010-1170K) (Vogt, 1999)
  - $\beta$-$WO_3$ (600-1170K) (Vogt, 1999)
  - $\gamma$-$WO_3$ (290-600K) (Vogt, 1999)
  - $\delta$-$WO_3$ (230-290K) (Diehl, 1978), this structure
  - $\epsilon$-$WO_3$ (below 23K) (Woodward, 1997)
- In addition, several other structures have been proposed and/or found:
  - The original $D0_{10}$ structure (Bräkken, 1931), (Hermann, 1937) superseded by $\delta$-$WO_3$
  - Original $\beta$-$WO_3$ (Salje, 1977)
  - Hexagonal $WO_3$ (Gerand, 1979) (metastable)

**Triclinic primitive vectors:**

$$
\begin{aligned}
\mathbf{a}_1 &= a\,\hat{\mathbf{x}} \\
\mathbf{a}_2 &= b\cos\gamma\,\hat{\mathbf{x}} + b\sin\gamma\,\hat{\mathbf{y}} \\
\mathbf{a}_3 &= c_x\,\hat{\mathbf{x}} + c_y\,\hat{\mathbf{y}} + c_z\,\hat{\mathbf{z}} \\
\\
c_x &= c\,\cos\beta \\
c_y &= c\,(\cos\alpha - \cos\beta\cos\gamma)/\sin\gamma \\
c_z &= \sqrt{c^2 - c_x^2 - c_y^2}
\end{aligned}
$$

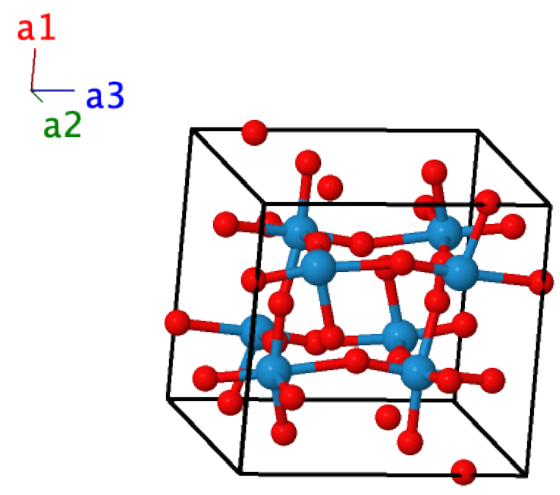

**Basis vectors:**

| | | Lattice Coordinates | | Cartesian Coordinates | Wyckoff Position | Atom Type |
|---|---|---|---|---|---|---|
| $\mathbf{B}_1$ | = | $x_1\,\mathbf{a}_1 + y_1\,\mathbf{a}_2 + z_1\,\mathbf{a}_3$ | = | $(x_1 a + y_1 b\cos\gamma + z_1 c_x)\,\hat{\mathbf{x}} + \left(y_1 b\sin\gamma + z_1 c_y\right)\hat{\mathbf{y}} + z_1 c_z\,\hat{\mathbf{z}}$ | (2*i*) | O I |
| $\mathbf{B}_2$ | = | $-x_1\,\mathbf{a}_1 - y_1\,\mathbf{a}_2 - z_1\,\mathbf{a}_3$ | = | $(-x_1 a - y_1 b\cos\gamma - z_1 c_x)\,\hat{\mathbf{x}} + \left(-y_1 b\sin\gamma - z_1 c_y\right)\hat{\mathbf{y}} - z_1 c_z\,\hat{\mathbf{z}}$ | (2*i*) | O I |
| $\mathbf{B}_3$ | = | $x_2\,\mathbf{a}_1 + y_2\,\mathbf{a}_2 + z_2\,\mathbf{a}_3$ | = | $(x_2 a + y_2 b\cos\gamma + z_2 c_x)\,\hat{\mathbf{x}} + \left(y_2 b\sin\gamma + z_2 c_y\right)\hat{\mathbf{y}} + z_2 c_z\,\hat{\mathbf{z}}$ | (2*i*) | O II |
| $\mathbf{B}_4$ | = | $-x_2\,\mathbf{a}_1 - y_2\,\mathbf{a}_2 - z_2\,\mathbf{a}_3$ | = | $(-x_2 a - y_2 b\cos\gamma - z_2 c_x)\,\hat{\mathbf{x}} + \left(-y_2 b\sin\gamma - z_2 c_y\right)\hat{\mathbf{y}} - z_2 c_z\,\hat{\mathbf{z}}$ | (2*i*) | O II |
| $\mathbf{B}_5$ | = | $x_3\,\mathbf{a}_1 + y_3\,\mathbf{a}_2 + z_3\,\mathbf{a}_3$ | = | $(x_3 a + y_3 b\cos\gamma + z_3 c_x)\,\hat{\mathbf{x}} + \left(y_3 b\sin\gamma + z_3 c_y\right)\hat{\mathbf{y}} + z_3 c_z\,\hat{\mathbf{z}}$ | (2*i*) | O III |
| $\mathbf{B}_6$ | = | $-x_3\,\mathbf{a}_1 - y_3\,\mathbf{a}_2 - z_3\,\mathbf{a}_3$ | = | $(-x_3 a - y_3 b\cos\gamma - z_3 c_x)\,\hat{\mathbf{x}} + \left(-y_3 b\sin\gamma - z_3 c_y\right)\hat{\mathbf{y}} - z_3 c_z\,\hat{\mathbf{z}}$ | (2*i*) | O III |
| $\mathbf{B}_7$ | = | $x_4\,\mathbf{a}_1 + y_4\,\mathbf{a}_2 + z_4\,\mathbf{a}_3$ | = | $(x_4 a + y_4 b\cos\gamma + z_4 c_x)\,\hat{\mathbf{x}} + \left(y_4 b\sin\gamma + z_4 c_y\right)\hat{\mathbf{y}} + z_4 c_z\,\hat{\mathbf{z}}$ | (2*i*) | O IV |
| $\mathbf{B}_8$ | = | $-x_4\,\mathbf{a}_1 - y_4\,\mathbf{a}_2 - z_4\,\mathbf{a}_3$ | = | $(-x_4 a - y_4 b\cos\gamma - z_4 c_x)\,\hat{\mathbf{x}} + \left(-y_4 b\sin\gamma - z_4 c_y\right)\hat{\mathbf{y}} - z_4 c_z\,\hat{\mathbf{z}}$ | (2*i*) | O IV |
| $\mathbf{B}_9$ | = | $x_5\,\mathbf{a}_1 + y_5\,\mathbf{a}_2 + z_5\,\mathbf{a}_3$ | = | $(x_5 a + y_5 b\cos\gamma + z_5 c_x)\,\hat{\mathbf{x}} + \left(y_5 b\sin\gamma + z_5 c_y\right)\hat{\mathbf{y}} + z_5 c_z\,\hat{\mathbf{z}}$ | (2*i*) | O V |
| $\mathbf{B}_{10}$ | = | $-x_5\,\mathbf{a}_1 - y_5\,\mathbf{a}_2 - z_5\,\mathbf{a}_3$ | = | $(-x_5 a - y_5 b\cos\gamma - z_5 c_x)\,\hat{\mathbf{x}} + \left(-y_5 b\sin\gamma - z_5 c_y\right)\hat{\mathbf{y}} - z_5 c_z\,\hat{\mathbf{z}}$ | (2*i*) | O V |
| $\mathbf{B}_{11}$ | = | $x_6\,\mathbf{a}_1 + y_6\,\mathbf{a}_2 + z_6\,\mathbf{a}_3$ | = | $(x_6 a + y_6 b\cos\gamma + z_6 c_x)\,\hat{\mathbf{x}} + \left(y_6 b\sin\gamma + z_6 c_y\right)\hat{\mathbf{y}} + z_6 c_z\,\hat{\mathbf{z}}$ | (2*i*) | O VI |
| $\mathbf{B}_{12}$ | = | $-x_6\,\mathbf{a}_1 - y_6\,\mathbf{a}_2 - z_6\,\mathbf{a}_3$ | = | $(-x_6 a - y_6 b\cos\gamma - z_6 c_x)\,\hat{\mathbf{x}} + \left(-y_6 b\sin\gamma - z_6 c_y\right)\hat{\mathbf{y}} - z_6 c_z\,\hat{\mathbf{z}}$ | (2*i*) | O VI |
| $\mathbf{B}_{13}$ | = | $x_7\,\mathbf{a}_1 + y_7\,\mathbf{a}_2 + z_7\,\mathbf{a}_3$ | = | $(x_7 a + y_7 b\cos\gamma + z_7 c_x)\,\hat{\mathbf{x}} + \left(y_7 b\sin\gamma + z_7 c_y\right)\hat{\mathbf{y}} + z_7 c_z\,\hat{\mathbf{z}}$ | (2*i*) | O VII |
| $\mathbf{B}_{14}$ | = | $-x_7\,\mathbf{a}_1 - y_7\,\mathbf{a}_2 - z_7\,\mathbf{a}_3$ | = | $(-x_7 a - y_7 b\cos\gamma - z_7 c_x)\,\hat{\mathbf{x}} + \left(-y_7 b\sin\gamma - z_7 c_y\right)\hat{\mathbf{y}} - z_7 c_z\,\hat{\mathbf{z}}$ | (2*i*) | O VII |
| $\mathbf{B}_{15}$ | = | $x_8\,\mathbf{a}_1 + y_8\,\mathbf{a}_2 + z_8\,\mathbf{a}_3$ | = | $(x_8 a + y_8 b\cos\gamma + z_8 c_x)\,\hat{\mathbf{x}} + \left(y_8 b\sin\gamma + z_8 c_y\right)\hat{\mathbf{y}} + z_8 c_z\,\hat{\mathbf{z}}$ | (2*i*) | O VIII |
| $\mathbf{B}_{16}$ | = | $-x_8\,\mathbf{a}_1 - y_8\,\mathbf{a}_2 - z_8\,\mathbf{a}_3$ | = | $(-x_8 a - y_8 b\cos\gamma - z_8 c_x)\,\hat{\mathbf{x}} + \left(-y_8 b\sin\gamma - z_8 c_y\right)\hat{\mathbf{y}} - z_8 c_z\,\hat{\mathbf{z}}$ | (2*i*) | O VIII |
| $\mathbf{B}_{17}$ | = | $x_9\,\mathbf{a}_1 + y_9\,\mathbf{a}_2 + z_9\,\mathbf{a}_3$ | = | $(x_9 a + y_9 b\cos\gamma + z_9 c_x)\,\mathbf{\hat{x}} + \left(y_9 b\sin\gamma + z_9 c_y\right)\,\mathbf{\hat{y}} + z_9 c_z\,\mathbf{\hat{z}}$ | (2$i$) | O IX |
| $\mathbf{B}_{18}$ | = | $-x_9\,\mathbf{a}_1 - y_9\,\mathbf{a}_2 - z_9\,\mathbf{a}_3$ | = | $(-x_9 a - y_9 b\cos\gamma - z_9 c_x)\,\mathbf{\hat{x}} + \left(-y_9 b\sin\gamma - z_9 c_y\right)\,\mathbf{\hat{y}} - z_9 c_z\,\mathbf{\hat{z}}$ | (2$i$) | O IX |
| $\mathbf{B}_{19}$ | = | $x_{10}\,\mathbf{a}_1 + y_{10}\,\mathbf{a}_2 + z_{10}\,\mathbf{a}_3$ | = | $(x_{10} a + y_{10} b\cos\gamma + z_{10} c_x)\,\mathbf{\hat{x}} + \left(y_{10} b\sin\gamma + z_{10} c_y\right)\,\mathbf{\hat{y}} + z_{10} c_z\,\mathbf{\hat{z}}$ | (2$i$) | O X |
| $\mathbf{B}_{20}$ | = | $-x_{10}\,\mathbf{a}_1 - y_{10}\,\mathbf{a}_2 - z_{10}\,\mathbf{a}_3$ | = | $(-x_{10} a - y_{10} b\cos\gamma - z_{10} c_x)\,\mathbf{\hat{x}} + \left(-y_{10} b\sin\gamma - z_{10} c_y\right)\,\mathbf{\hat{y}} - z_{10} c_z\,\mathbf{\hat{z}}$ | (2$i$) | O X |
| $\mathbf{B}_{21}$ | = | $x_{11}\,\mathbf{a}_1 + y_{11}\,\mathbf{a}_2 + z_{11}\,\mathbf{a}_3$ | = | $(x_{11} a + y_{11} b\cos\gamma + z_{11} c_x)\,\mathbf{\hat{x}} + \left(y_{11} b\sin\gamma + z_{11} c_y\right)\,\mathbf{\hat{y}} + z_{11} c_z\,\mathbf{\hat{z}}$ | (2$i$) | O XI |
| $\mathbf{B}_{22}$ | = | $-x_{11}\,\mathbf{a}_1 - y_{11}\,\mathbf{a}_2 - z_{11}\,\mathbf{a}_3$ | = | $(-x_{11} a - y_{11} b\cos\gamma - z_{11} c_x)\,\mathbf{\hat{x}} + \left(-y_{11} b\sin\gamma - z_{11} c_y\right)\,\mathbf{\hat{y}} - z_{11} c_z\,\mathbf{\hat{z}}$ | (2$i$) | O XI |
| $\mathbf{B}_{23}$ | = | $x_{12}\,\mathbf{a}_1 + y_{12}\,\mathbf{a}_2 + z_{12}\,\mathbf{a}_3$ | = | $(x_{12} a + y_{12} b\cos\gamma + z_{12} c_x)\,\mathbf{\hat{x}} + \left(y_{12} b\sin\gamma + z_{12} c_y\right)\,\mathbf{\hat{y}} + z_{12} c_z\,\mathbf{\hat{z}}$ | (2$i$) | O XII |
| $\mathbf{B}_{24}$ | = | $-x_{12}\,\mathbf{a}_1 - y_{12}\,\mathbf{a}_2 - z_{12}\,\mathbf{a}_3$ | = | $(-x_{12} a - y_{12} b\cos\gamma - z_{12} c_x)\,\mathbf{\hat{x}} + \left(-y_{12} b\sin\gamma - z_{12} c_y\right)\,\mathbf{\hat{y}} - z_{12} c_z\,\mathbf{\hat{z}}$ | (2$i$) | O XII |
| $\mathbf{B}_{25}$ | = | $x_{13}\,\mathbf{a}_1 + y_{13}\,\mathbf{a}_2 + z_{13}\,\mathbf{a}_3$ | = | $(x_{13} a + y_{13} b\cos\gamma + z_{13} c_x)\,\mathbf{\hat{x}} + \left(y_{13} b\sin\gamma + z_{13} c_y\right)\,\mathbf{\hat{y}} + z_{13} c_z\,\mathbf{\hat{z}}$ | (2$i$) | W I |
| $\mathbf{B}_{26}$ | = | $-x_{13}\,\mathbf{a}_1 - y_{13}\,\mathbf{a}_2 - z_{13}\,\mathbf{a}_3$ | = | $(-x_{13} a - y_{13} b\cos\gamma - z_{13} c_x)\,\mathbf{\hat{x}} + \left(-y_{13} b\sin\gamma - z_{13} c_y\right)\,\mathbf{\hat{y}} - z_{13} c_z\,\mathbf{\hat{z}}$ | (2$i$) | W I |
| $\mathbf{B}_{27}$ | = | $x_{14}\,\mathbf{a}_1 + y_{14}\,\mathbf{a}_2 + z_{14}\,\mathbf{a}_3$ | = | $(x_{14} a + y_{14} b\cos\gamma + z_{14} c_x)\,\mathbf{\hat{x}} + \left(y_{14} b\sin\gamma + z_{14} c_y\right)\,\mathbf{\hat{y}} + z_{14} c_z\,\mathbf{\hat{z}}$ | (2$i$) | W II |
| $\mathbf{B}_{28}$ | = | $-x_{14}\,\mathbf{a}_1 - y_{14}\,\mathbf{a}_2 - z_{14}\,\mathbf{a}_3$ | = | $(-x_{14} a - y_{14} b\cos\gamma - z_{14} c_x)\,\mathbf{\hat{x}} + \left(-y_{14} b\sin\gamma - z_{14} c_y\right)\,\mathbf{\hat{y}} - z_{14} c_z\,\mathbf{\hat{z}}$ | (2$i$) | W II |
| $\mathbf{B}_{29}$ | = | $x_{15}\,\mathbf{a}_1 + y_{15}\,\mathbf{a}_2 + z_{15}\,\mathbf{a}_3$ | = | $(x_{15} a + y_{15} b\cos\gamma + z_{15} c_x)\,\mathbf{\hat{x}} + \left(y_{15} b\sin\gamma + z_{15} c_y\right)\,\mathbf{\hat{y}} + z_{15} c_z\,\mathbf{\hat{z}}$ | (2$i$) | W III |
| $\mathbf{B}_{30}$ | = | $-x_{15}\,\mathbf{a}_1 - y_{15}\,\mathbf{a}_2 - z_{15}\,\mathbf{a}_3$ | = | $(-x_{15} a - y_{15} b\cos\gamma - z_{15} c_x)\,\mathbf{\hat{x}} + \left(-y_{15} b\sin\gamma - z_{15} c_y\right)\,\mathbf{\hat{y}} - z_{15} c_z\,\mathbf{\hat{z}}$ | (2$i$) | W III |
| $\mathbf{B}_{31}$ | = | $x_{16}\,\mathbf{a}_1 + y_{16}\,\mathbf{a}_2 + z_{16}\,\mathbf{a}_3$ | = | $(x_{16} a + y_{16} b\cos\gamma + z_{16} c_x)\,\mathbf{\hat{x}} + \left(y_{16} b\sin\gamma + z_{16} c_y\right)\,\mathbf{\hat{y}} + z_{16} c_z\,\mathbf{\hat{z}}$ | (2$i$) | W IV |
| $\mathbf{B}_{32}$ | = | $-x_{16}\,\mathbf{a}_1 - y_{16}\,\mathbf{a}_2 - z_{16}\,\mathbf{a}_3$ | = | $(-x_{16} a - y_{16} b\cos\gamma - z_{16} c_x)\,\mathbf{\hat{x}} + \left(-y_{16} b\sin\gamma - z_{16} c_y\right)\,\mathbf{\hat{y}} - z_{16} c_z\,\mathbf{\hat{z}}$ | (2$i$) | W IV |

---

**Geometry files:**
- CIF: pp. 1504
- POSCAR: pp. 1504

# Chalcanthite ($CuSO_4 \cdot 5H_2O$, $H4_{10}$) Structure: AB10C9D_aP42_2_ae_10i_9i_i

http://aflow.org/prototype-encyclopedia/AB10C9D_aP42_2_ae_10i_9i_i

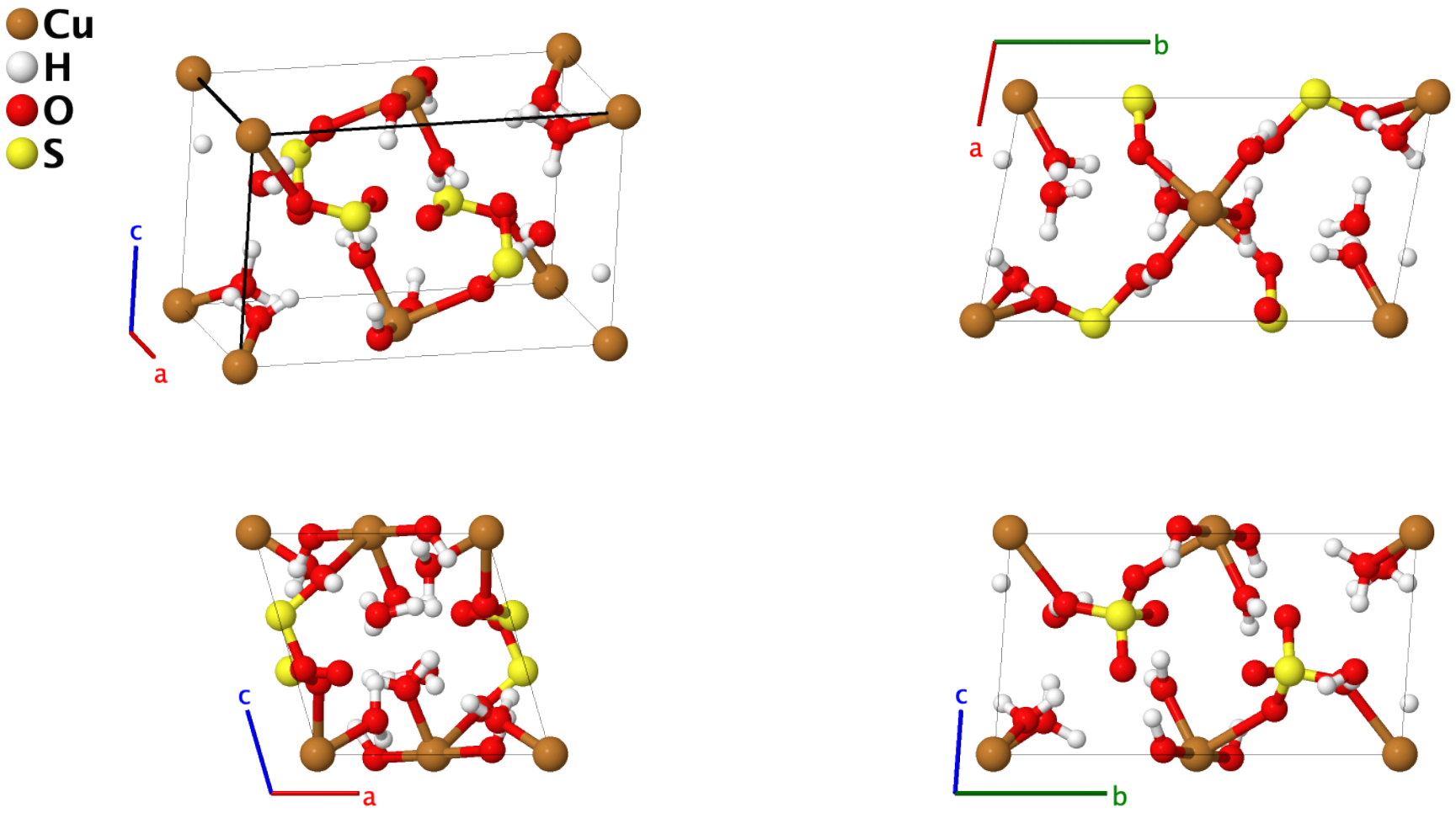

| | | |
|---|---|---|
| **Prototype** | : | $CuH_{10}O_9S$ |
| **AFLOW prototype label** | : | AB10C9D_aP42_2_ae_10i_9i_i |
| ***Strukturbericht* designation** | : | $H4_{10}$ |
| **Pearson symbol** | : | aP42 |
| **Space group number** | : | 2 |
| **Space group symbol** | : | $P\bar{1}$ |
| **AFLOW prototype command** | : | `aflow --proto=AB10C9D_aP42_2_ae_10i_9i_i --params=`$a, b/a, c/a, \alpha, \beta, \gamma, x_3, y_3, z_3, x_4, y_4, z_4, x_5, y_5, z_5, x_6, y_6, z_6, x_7, y_7, z_7, x_8, y_8, z_8, x_9, y_9, z_9, x_{10}, y_{10}, z_{10}, x_{11}, y_{11}, z_{11}, x_{12}, y_{12}, z_{12}, x_{13}, y_{13}, z_{13}, x_{14}, y_{14}, z_{14}, x_{15}, y_{15}, z_{15}, x_{16}, y_{16}, z_{16}, x_{17}, y_{17}, z_{17}, x_{18}, y_{18}, z_{18}, x_{19}, y_{19}, z_{19}, x_{20}, y_{20}, z_{20}, x_{21}, y_{21}, z_{21}, x_{22}, y_{22}, z_{22}$ |

- We reference (Bacon, 1975) in order to include the positions of the hydrogen atoms.

**Triclinic primitive vectors:**

$$\mathbf{a}_1 = a\,\hat{\mathbf{x}}$$

$$\mathbf{a}_2 = b\cos\gamma\,\hat{\mathbf{x}} + b\sin\gamma\,\hat{\mathbf{y}}$$

$$\mathbf{a}_3 = c_x\hat{\mathbf{x}} + c_y\,\hat{\mathbf{y}} + c_z\,\hat{\mathbf{z}}$$

$$c_x = c\,\cos\beta$$

$$c_y = c\,(\cos\alpha - \cos\beta\cos\gamma)/\,\sin\gamma$$

$$c_z = \sqrt{c^2 - c_x^2 - c_y^2}$$

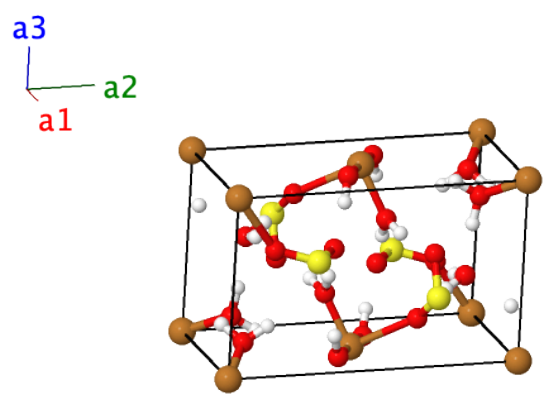

**Basis vectors:**

| | | Lattice Coordinates | | Cartesian Coordinates | Wyckoff Position | Atom Type |
|---|---|---|---|---|---|---|
| $\mathbf{B}_1$ | = | $0\,\mathbf{a}_1 + 0\,\mathbf{a}_2 + 0\,\mathbf{a}_3$ | = | $0\,\hat{\mathbf{x}} + 0\,\hat{\mathbf{y}} + 0\,\hat{\mathbf{z}}$ | (1$a$) | Cu I |
| $\mathbf{B}_2$ | = | $\frac{1}{2}\,\mathbf{a}_1 + \frac{1}{2}\,\mathbf{a}_2$ | = | $\frac{1}{2}\,(a + b\cos\gamma)\,\mathbf{\hat{x}} + \frac{1}{2}b\sin\gamma\,\mathbf{\hat{y}}$ | (1$e$) | Cu II |
| $\mathbf{B}_3$ | = | $x_3\,\mathbf{a}_1 + y_3\,\mathbf{a}_2 + z_3\,\mathbf{a}_3$ | = | $(x_3a + y_3b\cos\gamma + z_3c_x)\,\mathbf{\hat{x}} + \left(y_3b\sin\gamma + z_3c_y\right)\,\mathbf{\hat{y}} + z_3c_z\,\mathbf{\hat{z}}$ | (2$i$) | H I |
| $\mathbf{B}_4$ | = | $-x_3\,\mathbf{a}_1 - y_3\,\mathbf{a}_2 - z_3\,\mathbf{a}_3$ | = | $(-x_3a - y_3b\cos\gamma - z_3c_x)\,\mathbf{\hat{x}} + \left(-y_3b\sin\gamma - z_3c_y\right)\,\mathbf{\hat{y}} - z_3c_z\,\mathbf{\hat{z}}$ | (2$i$) | H I |
| $\mathbf{B}_5$ | = | $x_4\,\mathbf{a}_1 + y_4\,\mathbf{a}_2 + z_4\,\mathbf{a}_3$ | = | $(x_4a + y_4b\cos\gamma + z_4c_x)\,\mathbf{\hat{x}} + \left(y_4b\sin\gamma + z_4c_y\right)\,\mathbf{\hat{y}} + z_4c_z\,\mathbf{\hat{z}}$ | (2$i$) | H II |
| $\mathbf{B}_6$ | = | $-x_4\,\mathbf{a}_1 - y_4\,\mathbf{a}_2 - z_4\,\mathbf{a}_3$ | = | $(-x_4a - y_4b\cos\gamma - z_4c_x)\,\mathbf{\hat{x}} + \left(-y_4b\sin\gamma - z_4c_y\right)\,\mathbf{\hat{y}} - z_4c_z\,\mathbf{\hat{z}}$ | (2$i$) | H II |
| $\mathbf{B}_7$ | = | $x_5\,\mathbf{a}_1 + y_5\,\mathbf{a}_2 + z_5\,\mathbf{a}_3$ | = | $(x_5a + y_5b\cos\gamma + z_5c_x)\,\mathbf{\hat{x}} + \left(y_5b\sin\gamma + z_5c_y\right)\,\mathbf{\hat{y}} + z_5c_z\,\mathbf{\hat{z}}$ | (2$i$) | H III |
| $\mathbf{B}_8$ | = | $-x_5\,\mathbf{a}_1 - y_5\,\mathbf{a}_2 - z_5\,\mathbf{a}_3$ | = | $(-x_5a - y_5b\cos\gamma - z_5c_x)\,\mathbf{\hat{x}} + \left(-y_5b\sin\gamma - z_5c_y\right)\,\mathbf{\hat{y}} - z_5c_z\,\mathbf{\hat{z}}$ | (2$i$) | H III |
| $\mathbf{B}_9$ | = | $x_6\,\mathbf{a}_1 + y_6\,\mathbf{a}_2 + z_6\,\mathbf{a}_3$ | = | $(x_6a + y_6b\cos\gamma + z_6c_x)\,\mathbf{\hat{x}} + \left(y_6b\sin\gamma + z_6c_y\right)\,\mathbf{\hat{y}} + z_6c_z\,\mathbf{\hat{z}}$ | (2$i$) | H IV |
| $\mathbf{B}_{10}$ | = | $-x_6\,\mathbf{a}_1 - y_6\,\mathbf{a}_2 - z_6\,\mathbf{a}_3$ | = | $(-x_6a - y_6b\cos\gamma - z_6c_x)\,\mathbf{\hat{x}} + \left(-y_6b\sin\gamma - z_6c_y\right)\,\mathbf{\hat{y}} - z_6c_z\,\mathbf{\hat{z}}$ | (2$i$) | H IV |
| $\mathbf{B}_{11}$ | = | $x_7\,\mathbf{a}_1 + y_7\,\mathbf{a}_2 + z_7\,\mathbf{a}_3$ | = | $(x_7a + y_7b\cos\gamma + z_7c_x)\,\mathbf{\hat{x}} + \left(y_7b\sin\gamma + z_7c_y\right)\,\mathbf{\hat{y}} + z_7c_z\,\mathbf{\hat{z}}$ | (2$i$) | H V |
| $\mathbf{B}_{12}$ | = | $-x_7\,\mathbf{a}_1 - y_7\,\mathbf{a}_2 - z_7\,\mathbf{a}_3$ | = | $(-x_7a - y_7b\cos\gamma - z_7c_x)\,\mathbf{\hat{x}} + \left(-y_7b\sin\gamma - z_7c_y\right)\,\mathbf{\hat{y}} - z_7c_z\,\mathbf{\hat{z}}$ | (2$i$) | H V |
| $\mathbf{B}_{13}$ | = | $x_8\,\mathbf{a}_1 + y_8\,\mathbf{a}_2 + z_8\,\mathbf{a}_3$ | = | $(x_8a + y_8b\cos\gamma + z_8c_x)\,\mathbf{\hat{x}} + \left(y_8b\sin\gamma + z_8c_y\right)\,\mathbf{\hat{y}} + z_8c_z\,\mathbf{\hat{z}}$ | (2$i$) | H VI |
| $\mathbf{B}_{14}$ | = | $-x_8\,\mathbf{a}_1 - y_8\,\mathbf{a}_2 - z_8\,\mathbf{a}_3$ | = | $(-x_8a - y_8b\cos\gamma - z_8c_x)\,\mathbf{\hat{x}} + \left(-y_8b\sin\gamma - z_8c_y\right)\,\mathbf{\hat{y}} - z_8c_z\,\mathbf{\hat{z}}$ | (2$i$) | H VI |
| $\mathbf{B}_{15}$ | = | $x_9\,\mathbf{a}_1 + y_9\,\mathbf{a}_2 + z_9\,\mathbf{a}_3$ | = | $(x_9a + y_9b\cos\gamma + z_9c_x)\,\mathbf{\hat{x}} + \left(y_9b\sin\gamma + z_9c_y\right)\,\mathbf{\hat{y}} + z_9c_z\,\mathbf{\hat{z}}$ | (2$i$) | H VII |
| $\mathbf{B}_{16}$ | = | $-x_9\,\mathbf{a}_1 - y_9\,\mathbf{a}_2 - z_9\,\mathbf{a}_3$ | = | $(-x_9a - y_9b\cos\gamma - z_9c_x)\,\mathbf{\hat{x}} + \left(-y_9b\sin\gamma - z_9c_y\right)\,\mathbf{\hat{y}} - z_9c_z\,\mathbf{\hat{z}}$ | (2$i$) | H VII |
| $\mathbf{B}_{17}$ | = | $x_{10}\,\mathbf{a}_1 + y_{10}\,\mathbf{a}_2 + z_{10}\,\mathbf{a}_3$ | = | $(x_{10}a + y_{10}b\cos\gamma + z_{10}c_x)\,\mathbf{\hat{x}} + \left(y_{10}b\sin\gamma + z_{10}c_y\right)\,\mathbf{\hat{y}} + z_{10}c_z\,\mathbf{\hat{z}}$ | (2$i$) | H VIII |
| $\mathbf{B}_{18}$ | = | $-x_{10}\,\mathbf{a}_1 - y_{10}\,\mathbf{a}_2 - z_{10}\,\mathbf{a}_3$ | = | $(-x_{10}a - y_{10}b\cos\gamma - z_{10}c_x)\,\mathbf{\hat{x}} + \left(-y_{10}b\sin\gamma - z_{10}c_y\right)\,\mathbf{\hat{y}} - z_{10}c_z\,\mathbf{\hat{z}}$ | (2$i$) | H VIII |
| $\mathbf{B}_{19}$ | = | $x_{11}\,\mathbf{a}_1 + y_{11}\,\mathbf{a}_2 + z_{11}\,\mathbf{a}_3$ | = | $(x_{11}a + y_{11}b\cos\gamma + z_{11}c_x)\,\mathbf{\hat{x}} + \left(y_{11}b\sin\gamma + z_{11}c_y\right)\,\mathbf{\hat{y}} + z_{11}c_z\,\mathbf{\hat{z}}$ | (2$i$) | H IX |
| $\mathbf{B}_{20}$ | = | $-x_{11}\,\mathbf{a}_1 - y_{11}\,\mathbf{a}_2 - z_{11}\,\mathbf{a}_3$ | = | $(-x_{11}a - y_{11}b\cos\gamma - z_{11}c_x)\,\mathbf{\hat{x}} + \left(-y_{11}b\sin\gamma - z_{11}c_y\right)\,\mathbf{\hat{y}} - z_{11}c_z\,\mathbf{\hat{z}}$ | (2$i$) | H IX |
| $\mathbf{B}_{21}$ | = | $x_{12}\,\mathbf{a}_1 + y_{12}\,\mathbf{a}_2 + z_{12}\,\mathbf{a}_3$ | = | $(x_{12}a + y_{12}b\cos\gamma + z_{12}c_x)\,\mathbf{\hat{x}} + \left(y_{12}b\sin\gamma + z_{12}c_y\right)\,\mathbf{\hat{y}} + z_{12}c_z\,\mathbf{\hat{z}}$ | (2$i$) | H X |
| $\mathbf{B}_{22}$ | = | $-x_{12}\,\mathbf{a}_1 - y_{12}\,\mathbf{a}_2 - z_{12}\,\mathbf{a}_3$ | = | $(-x_{12}a - y_{12}b\cos\gamma - z_{12}c_x)\,\mathbf{\hat{x}} + \left(-y_{12}b\sin\gamma - z_{12}c_y\right)\,\mathbf{\hat{y}} - z_{12}c_z\,\mathbf{\hat{z}}$ | (2$i$) | H X |
| $\mathbf{B}_{23}$ | = | $x_{13}\,\mathbf{a}_1 + y_{13}\,\mathbf{a}_2 + z_{13}\,\mathbf{a}_3$ | = | $(x_{13}a + y_{13}b\cos\gamma + z_{13}c_x)\,\mathbf{\hat{x}} + \left(y_{13}b\sin\gamma + z_{13}c_y\right)\,\mathbf{\hat{y}} + z_{13}c_z\,\mathbf{\hat{z}}$ | (2$i$) | O I |
| $\mathbf{B}_{24}$ | = | $-x_{13}\,\mathbf{a}_1 - y_{13}\,\mathbf{a}_2 - z_{13}\,\mathbf{a}_3$ | = | $(-x_{13}a - y_{13}b\cos\gamma - z_{13}c_x)\,\mathbf{\hat{x}} + \left(-y_{13}b\sin\gamma - z_{13}c_y\right)\,\mathbf{\hat{y}} - z_{13}c_z\,\mathbf{\hat{z}}$ | (2$i$) | O I |
| $\mathbf{B}_{25}$ | = | $x_{14}\,\mathbf{a}_1 + y_{14}\,\mathbf{a}_2 + z_{14}\,\mathbf{a}_3$ | = | $(x_{14}a + y_{14}b\cos\gamma + z_{14}c_x)\,\hat{\mathbf{x}} + \left(y_{14}b\sin\gamma + z_{14}c_y\right)\hat{\mathbf{y}} + z_{14}c_z\,\hat{\mathbf{z}}$ | (2$i$) | O II |
| $\mathbf{B}_{26}$ | = | $-x_{14}\,\mathbf{a}_1 - y_{14}\,\mathbf{a}_2 - z_{14}\,\mathbf{a}_3$ | = | $(-x_{14}a - y_{14}b\cos\gamma - z_{14}c_x)\,\hat{\mathbf{x}} + \left(-y_{14}b\sin\gamma - z_{14}c_y\right)\hat{\mathbf{y}} - z_{14}c_z\,\hat{\mathbf{z}}$ | (2$i$) | O II |
| $\mathbf{B}_{27}$ | = | $x_{15}\,\mathbf{a}_1 + y_{15}\,\mathbf{a}_2 + z_{15}\,\mathbf{a}_3$ | = | $(x_{15}a + y_{15}b\cos\gamma + z_{15}c_x)\,\hat{\mathbf{x}} + \left(y_{15}b\sin\gamma + z_{15}c_y\right)\hat{\mathbf{y}} + z_{15}c_z\,\hat{\mathbf{z}}$ | (2$i$) | O III |
| $\mathbf{B}_{28}$ | = | $-x_{15}\,\mathbf{a}_1 - y_{15}\,\mathbf{a}_2 - z_{15}\,\mathbf{a}_3$ | = | $(-x_{15}a - y_{15}b\cos\gamma - z_{15}c_x)\,\hat{\mathbf{x}} + \left(-y_{15}b\sin\gamma - z_{15}c_y\right)\hat{\mathbf{y}} - z_{15}c_z\,\hat{\mathbf{z}}$ | (2$i$) | O III |
| $\mathbf{B}_{29}$ | = | $x_{16}\,\mathbf{a}_1 + y_{16}\,\mathbf{a}_2 + z_{16}\,\mathbf{a}_3$ | = | $(x_{16}a + y_{16}b\cos\gamma + z_{16}c_x)\,\hat{\mathbf{x}} + \left(y_{16}b\sin\gamma + z_{16}c_y\right)\hat{\mathbf{y}} + z_{16}c_z\,\hat{\mathbf{z}}$ | (2$i$) | O IV |
| $\mathbf{B}_{30}$ | = | $-x_{16}\,\mathbf{a}_1 - y_{16}\,\mathbf{a}_2 - z_{16}\,\mathbf{a}_3$ | = | $(-x_{16}a - y_{16}b\cos\gamma - z_{16}c_x)\,\hat{\mathbf{x}} + \left(-y_{16}b\sin\gamma - z_{16}c_y\right)\hat{\mathbf{y}} - z_{16}c_z\,\hat{\mathbf{z}}$ | (2$i$) | O IV |
| $\mathbf{B}_{31}$ | = | $x_{17}\,\mathbf{a}_1 + y_{17}\,\mathbf{a}_2 + z_{17}\,\mathbf{a}_3$ | = | $(x_{17}a + y_{17}b\cos\gamma + z_{17}c_x)\,\hat{\mathbf{x}} + \left(y_{17}b\sin\gamma + z_{17}c_y\right)\hat{\mathbf{y}} + z_{17}c_z\,\hat{\mathbf{z}}$ | (2$i$) | O V |
| $\mathbf{B}_{32}$ | = | $-x_{17}\,\mathbf{a}_1 - y_{17}\,\mathbf{a}_2 - z_{17}\,\mathbf{a}_3$ | = | $(-x_{17}a - y_{17}b\cos\gamma - z_{17}c_x)\,\hat{\mathbf{x}} + \left(-y_{17}b\sin\gamma - z_{17}c_y\right)\hat{\mathbf{y}} - z_{17}c_z\,\hat{\mathbf{z}}$ | (2$i$) | O V |
| $\mathbf{B}_{33}$ | = | $x_{18}\,\mathbf{a}_1 + y_{18}\,\mathbf{a}_2 + z_{18}\,\mathbf{a}_3$ | = | $(x_{18}a + y_{18}b\cos\gamma + z_{18}c_x)\,\hat{\mathbf{x}} + \left(y_{18}b\sin\gamma + z_{18}c_y\right)\hat{\mathbf{y}} + z_{18}c_z\,\hat{\mathbf{z}}$ | (2$i$) | O VI |
| $\mathbf{B}_{34}$ | = | $-x_{18}\,\mathbf{a}_1 - y_{18}\,\mathbf{a}_2 - z_{18}\,\mathbf{a}_3$ | = | $(-x_{18}a - y_{18}b\cos\gamma - z_{18}c_x)\,\hat{\mathbf{x}} + \left(-y_{18}b\sin\gamma - z_{18}c_y\right)\hat{\mathbf{y}} - z_{18}c_z\,\hat{\mathbf{z}}$ | (2$i$) | O VI |
| $\mathbf{B}_{35}$ | = | $x_{19}\,\mathbf{a}_1 + y_{19}\,\mathbf{a}_2 + z_{19}\,\mathbf{a}_3$ | = | $(x_{19}a + y_{19}b\cos\gamma + z_{19}c_x)\,\hat{\mathbf{x}} + \left(y_{19}b\sin\gamma + z_{19}c_y\right)\hat{\mathbf{y}} + z_{19}c_z\,\hat{\mathbf{z}}$ | (2$i$) | O VII |
| $\mathbf{B}_{36}$ | = | $-x_{19}\,\mathbf{a}_1 - y_{19}\,\mathbf{a}_2 - z_{19}\,\mathbf{a}_3$ | = | $(-x_{19}a - y_{19}b\cos\gamma - z_{19}c_x)\,\hat{\mathbf{x}} + \left(-y_{19}b\sin\gamma - z_{19}c_y\right)\hat{\mathbf{y}} - z_{19}c_z\,\hat{\mathbf{z}}$ | (2$i$) | O VII |
| $\mathbf{B}_{37}$ | = | $x_{20}\,\mathbf{a}_1 + y_{20}\,\mathbf{a}_2 + z_{20}\,\mathbf{a}_3$ | = | $(x_{20}a + y_{20}b\cos\gamma + z_{20}c_x)\,\hat{\mathbf{x}} + \left(y_{20}b\sin\gamma + z_{20}c_y\right)\hat{\mathbf{y}} + z_{20}c_z\,\hat{\mathbf{z}}$ | (2$i$) | O VIII |
| $\mathbf{B}_{38}$ | = | $-x_{20}\,\mathbf{a}_1 - y_{20}\,\mathbf{a}_2 - z_{20}\,\mathbf{a}_3$ | = | $(-x_{20}a - y_{20}b\cos\gamma - z_{20}c_x)\,\hat{\mathbf{x}} + \left(-y_{20}b\sin\gamma - z_{20}c_y\right)\hat{\mathbf{y}} - z_{20}c_z\,\hat{\mathbf{z}}$ | (2$i$) | O VIII |
| $\mathbf{B}_{39}$ | = | $x_{21}\,\mathbf{a}_1 + y_{21}\,\mathbf{a}_2 + z_{21}\,\mathbf{a}_3$ | = | $(x_{21}a + y_{21}b\cos\gamma + z_{21}c_x)\,\hat{\mathbf{x}} + \left(y_{21}b\sin\gamma + z_{21}c_y\right)\hat{\mathbf{y}} + z_{21}c_z\,\hat{\mathbf{z}}$ | (2$i$) | O IX |
| $\mathbf{B}_{40}$ | = | $-x_{21}\,\mathbf{a}_1 - y_{21}\,\mathbf{a}_2 - z_{21}\,\mathbf{a}_3$ | = | $(-x_{21}a - y_{21}b\cos\gamma - z_{21}c_x)\,\hat{\mathbf{x}} + \left(-y_{21}b\sin\gamma - z_{21}c_y\right)\hat{\mathbf{y}} - z_{21}c_z\,\hat{\mathbf{z}}$ | (2$i$) | O IX |
| $\mathbf{B}_{41}$ | = | $x_{22}\,\mathbf{a}_1 + y_{22}\,\mathbf{a}_2 + z_{22}\,\mathbf{a}_3$ | = | $(x_{22}a + y_{22}b\cos\gamma + z_{22}c_x)\,\hat{\mathbf{x}} + \left(y_{22}b\sin\gamma + z_{22}c_y\right)\hat{\mathbf{y}} + z_{22}c_z\,\hat{\mathbf{z}}$ | (2$i$) | S |
| $\mathbf{B}_{42}$ | = | $-x_{22}\,\mathbf{a}_1 - y_{22}\,\mathbf{a}_2 - z_{22}\,\mathbf{a}_3$ | = | $(-x_{22}a - y_{22}b\cos\gamma - z_{22}c_x)\,\hat{\mathbf{x}} + \left(-y_{22}b\sin\gamma - z_{22}c_y\right)\hat{\mathbf{y}} - z_{22}c_z\,\hat{\mathbf{z}}$ | (2$i$) | S |

**Geometry files:**

- CIF: pp. 1505
- POSCAR: pp. 1505

# Boric Acid ($H_3BO_3$, $G5_1$) Structure: AB3C3_aP28_2_2i_6i_6i

http://aflow.org/prototype-encyclopedia/AB3C3_aP28_2_2i_6i_6i

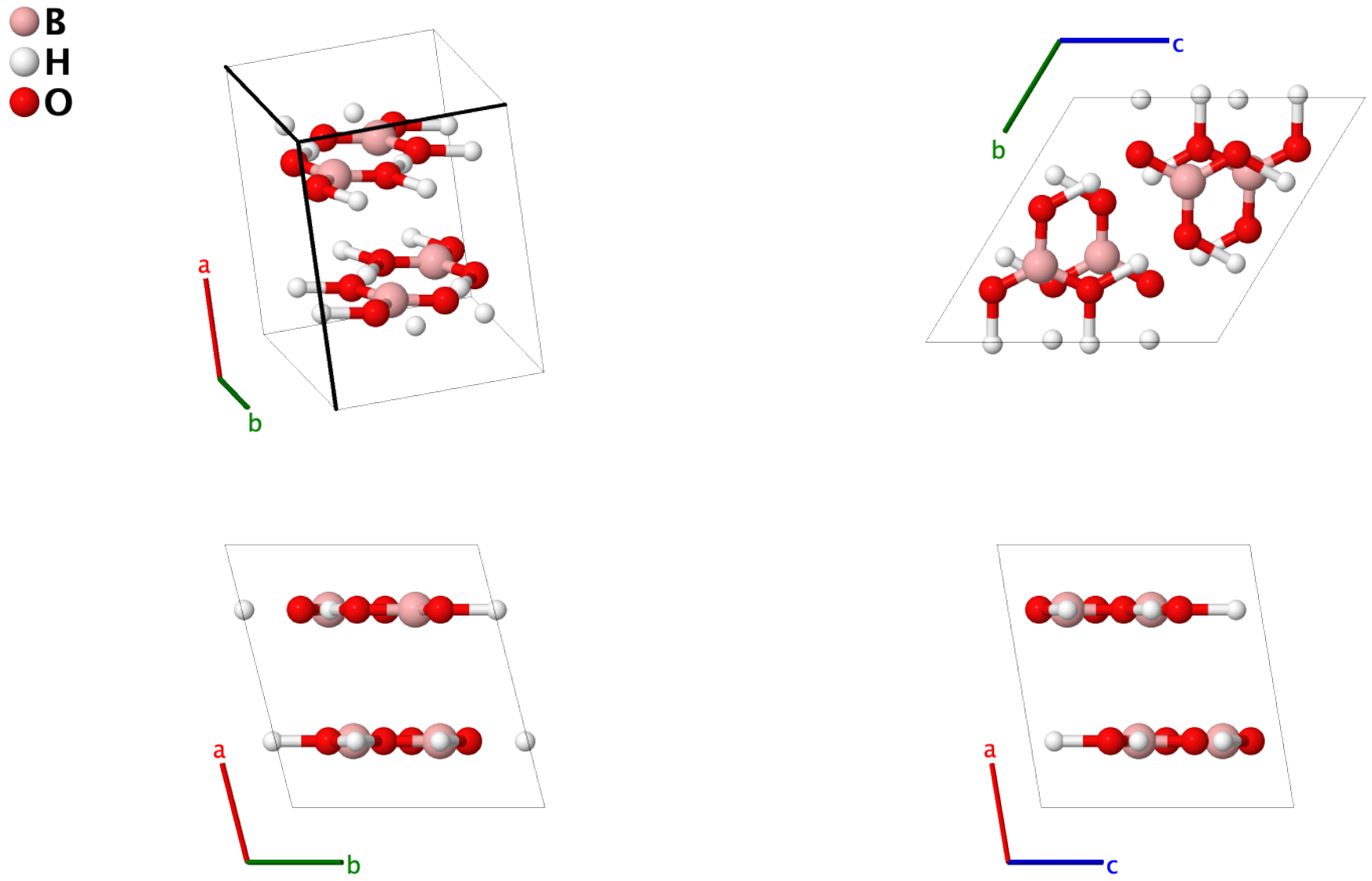

| | | |
|---|---|---|
| **Prototype** | : | $BH_3O_3$ |
| **AFLOW prototype label** | : | AB3C3_aP28_2_2i_6i_6i |
| ***Strukturbericht* designation** | : | $G5_1$ |
| **Pearson symbol** | : | aP28 |
| **Space group number** | : | 2 |
| **Space group symbol** | : | $P\bar{1}$ |
| **AFLOW prototype command** | : | `aflow --proto=AB3C3_aP28_2_2i_6i_6i` `--params=`$a, b/a, c/a, \alpha, \beta, \gamma, x_1, y_1, z_1, x_2, y_2, z_2, x_3, y_3, z_3, x_4, y_4, z_4, x_5, y_5, z_5, x_6, y_6, z_6, x_7, y_7, z_7, x_8, y_8, z_8, x_9, y_9, z_9, x_{10}, y_{10}, z_{10}, x_{11}, y_{11}, z_{11}, x_{12}, y_{12}, z_{12}, x_{13}, y_{13}, z_{13}, x_{14}, y_{14}, z_{14}$ |

**Triclinic primitive vectors:**

$$\mathbf{a}_1 = a\,\hat{\mathbf{x}}$$
$$\mathbf{a}_2 = b\cos\gamma\,\hat{\mathbf{x}} + b\sin\gamma\,\hat{\mathbf{y}}$$
$$\mathbf{a}_3 = c_x\,\hat{\mathbf{x}} + c_y\,\hat{\mathbf{y}} + c_z\,\hat{\mathbf{z}}$$

$$c_x = c\,\cos\beta$$
$$c_y = c\,(\cos\alpha - \cos\beta\cos\gamma)/\sin\gamma$$
$$c_z = \sqrt{c^2 - c_x^2 - c_y^2}$$

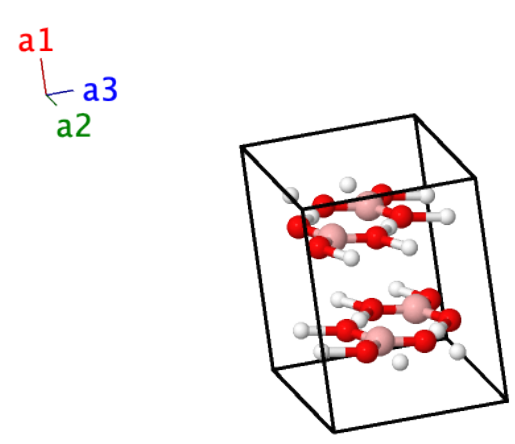

**Basis vectors:**

| | | Lattice Coordinates | | Cartesian Coordinates | Wyckoff Position | Atom Type |
|---|---|---|---|---|---|---|
| $\mathbf{B}_1$ | = | $x_1\,\mathbf{a}_1 + y_1\,\mathbf{a}_2 + z_1\,\mathbf{a}_3$ | = | $(x_1 a + y_1 b\cos\gamma + z_1 c_x)\,\hat{\mathbf{x}} + \left(y_1 b\sin\gamma + z_1 c_y\right)\,\hat{\mathbf{y}} + z_1 c_z\,\hat{\mathbf{z}}$ | $(2i)$ | B I |
| $\mathbf{B}_{2}$ | = | $-x_{1}\,\mathbf{a}_{1} - y_{1}\,\mathbf{a}_{2} - z_{1}\,\mathbf{a}_{3}$ | = | $\left(-x_{1}a - y_{1}b\cos\gamma - z_{1}c_{x}\right)\,\mathbf{\hat{x}} + \left(-y_{1}b\sin\gamma - z_{1}c_{y}\right)\,\mathbf{\hat{y}} - z_{1}c_{z}\,\mathbf{\hat{z}}$ | (2*i*) | B I |
| $\mathbf{B}_{3}$ | = | $x_{2}\,\mathbf{a}_{1} + y_{2}\,\mathbf{a}_{2} + z_{2}\,\mathbf{a}_{3}$ | = | $\left(x_{2}a + y_{2}b\cos\gamma + z_{2}c_{x}\right)\,\mathbf{\hat{x}} + \left(y_{2}b\sin\gamma + z_{2}c_{y}\right)\,\mathbf{\hat{y}} + z_{2}c_{z}\,\mathbf{\hat{z}}$ | (2*i*) | B II |
| $\mathbf{B}_{4}$ | = | $-x_{2}\,\mathbf{a}_{1} - y_{2}\,\mathbf{a}_{2} - z_{2}\,\mathbf{a}_{3}$ | = | $\left(-x_{2}a - y_{2}b\cos\gamma - z_{2}c_{x}\right)\,\mathbf{\hat{x}} + \left(-y_{2}b\sin\gamma - z_{2}c_{y}\right)\,\mathbf{\hat{y}} - z_{2}c_{z}\,\mathbf{\hat{z}}$ | (2*i*) | B II |
| $\mathbf{B}_{5}$ | = | $x_{3}\,\mathbf{a}_{1} + y_{3}\,\mathbf{a}_{2} + z_{3}\,\mathbf{a}_{3}$ | = | $\left(x_{3}a + y_{3}b\cos\gamma + z_{3}c_{x}\right)\,\mathbf{\hat{x}} + \left(y_{3}b\sin\gamma + z_{3}c_{y}\right)\,\mathbf{\hat{y}} + z_{3}c_{z}\,\mathbf{\hat{z}}$ | (2*i*) | H I |
| $\mathbf{B}_{6}$ | = | $-x_{3}\,\mathbf{a}_{1} - y_{3}\,\mathbf{a}_{2} - z_{3}\,\mathbf{a}_{3}$ | = | $\left(-x_{3}a - y_{3}b\cos\gamma - z_{3}c_{x}\right)\,\mathbf{\hat{x}} + \left(-y_{3}b\sin\gamma - z_{3}c_{y}\right)\,\mathbf{\hat{y}} - z_{3}c_{z}\,\mathbf{\hat{z}}$ | (2*i*) | H I |
| $\mathbf{B}_{7}$ | = | $x_{4}\,\mathbf{a}_{1} + y_{4}\,\mathbf{a}_{2} + z_{4}\,\mathbf{a}_{3}$ | = | $\left(x_{4}a + y_{4}b\cos\gamma + z_{4}c_{x}\right)\,\mathbf{\hat{x}} + \left(y_{4}b\sin\gamma + z_{4}c_{y}\right)\,\mathbf{\hat{y}} + z_{4}c_{z}\,\mathbf{\hat{z}}$ | (2*i*) | H II |
| $\mathbf{B}_{8}$ | = | $-x_{4}\,\mathbf{a}_{1} - y_{4}\,\mathbf{a}_{2} - z_{4}\,\mathbf{a}_{3}$ | = | $\left(-x_{4}a - y_{4}b\cos\gamma - z_{4}c_{x}\right)\,\mathbf{\hat{x}} + \left(-y_{4}b\sin\gamma - z_{4}c_{y}\right)\,\mathbf{\hat{y}} - z_{4}c_{z}\,\mathbf{\hat{z}}$ | (2*i*) | H II |
| $\mathbf{B}_{9}$ | = | $x_{5}\,\mathbf{a}_{1} + y_{5}\,\mathbf{a}_{2} + z_{5}\,\mathbf{a}_{3}$ | = | $\left(x_{5}a + y_{5}b\cos\gamma + z_{5}c_{x}\right)\,\mathbf{\hat{x}} + \left(y_{5}b\sin\gamma + z_{5}c_{y}\right)\,\mathbf{\hat{y}} + z_{5}c_{z}\,\mathbf{\hat{z}}$ | (2*i*) | H III |
| $\mathbf{B}_{10}$ | = | $-x_{5}\,\mathbf{a}_{1} - y_{5}\,\mathbf{a}_{2} - z_{5}\,\mathbf{a}_{3}$ | = | $\left(-x_{5}a - y_{5}b\cos\gamma - z_{5}c_{x}\right)\,\mathbf{\hat{x}} + \left(-y_{5}b\sin\gamma - z_{5}c_{y}\right)\,\mathbf{\hat{y}} - z_{5}c_{z}\,\mathbf{\hat{z}}$ | (2*i*) | H III |
| $\mathbf{B}_{11}$ | = | $x_{6}\,\mathbf{a}_{1} + y_{6}\,\mathbf{a}_{2} + z_{6}\,\mathbf{a}_{3}$ | = | $\left(x_{6}a + y_{6}b\cos\gamma + z_{6}c_{x}\right)\,\mathbf{\hat{x}} + \left(y_{6}b\sin\gamma + z_{6}c_{y}\right)\,\mathbf{\hat{y}} + z_{6}c_{z}\,\mathbf{\hat{z}}$ | (2*i*) | H IV |
| $\mathbf{B}_{12}$ | = | $-x_{6}\,\mathbf{a}_{1} - y_{6}\,\mathbf{a}_{2} - z_{6}\,\mathbf{a}_{3}$ | = | $\left(-x_{6}a - y_{6}b\cos\gamma - z_{6}c_{x}\right)\,\mathbf{\hat{x}} + \left(-y_{6}b\sin\gamma - z_{6}c_{y}\right)\,\mathbf{\hat{y}} - z_{6}c_{z}\,\mathbf{\hat{z}}$ | (2*i*) | H IV |
| $\mathbf{B}_{13}$ | = | $x_{7}\,\mathbf{a}_{1} + y_{7}\,\mathbf{a}_{2} + z_{7}\,\mathbf{a}_{3}$ | = | $\left(x_{7}a + y_{7}b\cos\gamma + z_{7}c_{x}\right)\,\mathbf{\hat{x}} + \left(y_{7}b\sin\gamma + z_{7}c_{y}\right)\,\mathbf{\hat{y}} + z_{7}c_{z}\,\mathbf{\hat{z}}$ | (2*i*) | H V |
| $\mathbf{B}_{14}$ | = | $-x_{7}\,\mathbf{a}_{1} - y_{7}\,\mathbf{a}_{2} - z_{7}\,\mathbf{a}_{3}$ | = | $\left(-x_{7}a - y_{7}b\cos\gamma - z_{7}c_{x}\right)\,\mathbf{\hat{x}} + \left(-y_{7}b\sin\gamma - z_{7}c_{y}\right)\,\mathbf{\hat{y}} - z_{7}c_{z}\,\mathbf{\hat{z}}$ | (2*i*) | H V |
| $\mathbf{B}_{15}$ | = | $x_{8}\,\mathbf{a}_{1} + y_{8}\,\mathbf{a}_{2} + z_{8}\,\mathbf{a}_{3}$ | = | $\left(x_{8}a + y_{8}b\cos\gamma + z_{8}c_{x}\right)\,\mathbf{\hat{x}} + \left(y_{8}b\sin\gamma + z_{8}c_{y}\right)\,\mathbf{\hat{y}} + z_{8}c_{z}\,\mathbf{\hat{z}}$ | (2*i*) | H VI |
| $\mathbf{B}_{16}$ | = | $-x_{8}\,\mathbf{a}_{1} - y_{8}\,\mathbf{a}_{2} - z_{8}\,\mathbf{a}_{3}$ | = | $\left(-x_{8}a - y_{8}b\cos\gamma - z_{8}c_{x}\right)\,\mathbf{\hat{x}} + \left(-y_{8}b\sin\gamma - z_{8}c_{y}\right)\,\mathbf{\hat{y}} - z_{8}c_{z}\,\mathbf{\hat{z}}$ | (2*i*) | H VI |
| $\mathbf{B}_{17}$ | = | $x_{9}\,\mathbf{a}_{1} + y_{9}\,\mathbf{a}_{2} + z_{9}\,\mathbf{a}_{3}$ | = | $\left(x_{9}a + y_{9}b\cos\gamma + z_{9}c_{x}\right)\,\mathbf{\hat{x}} + \left(y_{9}b\sin\gamma + z_{9}c_{y}\right)\,\mathbf{\hat{y}} + z_{9}c_{z}\,\mathbf{\hat{z}}$ | (2*i*) | O I |
| $\mathbf{B}_{18}$ | = | $-x_{9}\,\mathbf{a}_{1} - y_{9}\,\mathbf{a}_{2} - z_{9}\,\mathbf{a}_{3}$ | = | $\left(-x_{9}a - y_{9}b\cos\gamma - z_{9}c_{x}\right)\,\mathbf{\hat{x}} + \left(-y_{9}b\sin\gamma - z_{9}c_{y}\right)\,\mathbf{\hat{y}} - z_{9}c_{z}\,\mathbf{\hat{z}}$ | (2*i*) | O I |
| $\mathbf{B}_{19}$ | = | $x_{10}\,\mathbf{a}_{1} + y_{10}\,\mathbf{a}_{2} + z_{10}\,\mathbf{a}_{3}$ | = | $\left(x_{10}a + y_{10}b\cos\gamma + z_{10}c_{x}\right)\,\mathbf{\hat{x}} + \left(y_{10}b\sin\gamma + z_{10}c_{y}\right)\,\mathbf{\hat{y}} + z_{10}c_{z}\,\mathbf{\hat{z}}$ | (2*i*) | O II |
| $\mathbf{B}_{20}$ | = | $-x_{10}\,\mathbf{a}_{1} - y_{10}\,\mathbf{a}_{2} - z_{10}\,\mathbf{a}_{3}$ | = | $\left(-x_{10}a - y_{10}b\cos\gamma - z_{10}c_{x}\right)\,\mathbf{\hat{x}} + \left(-y_{10}b\sin\gamma - z_{10}c_{y}\right)\,\mathbf{\hat{y}} - z_{10}c_{z}\,\mathbf{\hat{z}}$ | (2*i*) | O II |
| $\mathbf{B}_{21}$ | = | $x_{11}\,\mathbf{a}_{1} + y_{11}\,\mathbf{a}_{2} + z_{11}\,\mathbf{a}_{3}$ | = | $\left(x_{11}a + y_{11}b\cos\gamma + z_{11}c_{x}\right)\,\mathbf{\hat{x}} + \left(y_{11}b\sin\gamma + z_{11}c_{y}\right)\,\mathbf{\hat{y}} + z_{11}c_{z}\,\mathbf{\hat{z}}$ | (2*i*) | O III |
| $\mathbf{B}_{22}$ | = | $-x_{11}\,\mathbf{a}_{1} - y_{11}\,\mathbf{a}_{2} - z_{11}\,\mathbf{a}_{3}$ | = | $\left(-x_{11}a - y_{11}b\cos\gamma - z_{11}c_{x}\right)\,\mathbf{\hat{x}} + \left(-y_{11}b\sin\gamma - z_{11}c_{y}\right)\,\mathbf{\hat{y}} - z_{11}c_{z}\,\mathbf{\hat{z}}$ | (2*i*) | O III |
| $\mathbf{B}_{23}$ | = | $x_{12}\,\mathbf{a}_{1} + y_{12}\,\mathbf{a}_{2} + z_{12}\,\mathbf{a}_{3}$ | = | $\left(x_{12}a + y_{12}b\cos\gamma + z_{12}c_{x}\right)\,\mathbf{\hat{x}} + \left(y_{12}b\sin\gamma + z_{12}c_{y}\right)\,\mathbf{\hat{y}} + z_{12}c_{z}\,\mathbf{\hat{z}}$ | (2*i*) | O IV |
| $\mathbf{B}_{24}$ | = | $-x_{12}\,\mathbf{a}_1 - y_{12}\,\mathbf{a}_2 - z_{12}\,\mathbf{a}_3$ | = | $(-x_{12}a - y_{12}b\cos\gamma - z_{12}c_x)\,\mathbf{\hat{x}} + \left(-y_{12}b\sin\gamma - z_{12}c_y\right)\,\mathbf{\hat{y}} - z_{12}c_z\,\mathbf{\hat{z}}$ | (2$i$) | O IV |
| $\mathbf{B}_{25}$ | = | $x_{13}\,\mathbf{a}_1 + y_{13}\,\mathbf{a}_2 + z_{13}\,\mathbf{a}_3$ | = | $(x_{13}a + y_{13}b\cos\gamma + z_{13}c_x)\,\mathbf{\hat{x}} + \left(y_{13}b\sin\gamma + z_{13}c_y\right)\,\mathbf{\hat{y}} + z_{13}c_z\,\mathbf{\hat{z}}$ | (2$i$) | O V |
| $\mathbf{B}_{26}$ | = | $-x_{13}\,\mathbf{a}_1 - y_{13}\,\mathbf{a}_2 - z_{13}\,\mathbf{a}_3$ | = | $(-x_{13}a - y_{13}b\cos\gamma - z_{13}c_x)\,\mathbf{\hat{x}} + \left(-y_{13}b\sin\gamma - z_{13}c_y\right)\,\mathbf{\hat{y}} - z_{13}c_z\,\mathbf{\hat{z}}$ | (2$i$) | O V |
| $\mathbf{B}_{27}$ | = | $x_{14}\,\mathbf{a}_1 + y_{14}\,\mathbf{a}_2 + z_{14}\,\mathbf{a}_3$ | = | $(x_{14}a + y_{14}b\cos\gamma + z_{14}c_x)\,\mathbf{\hat{x}} + \left(y_{14}b\sin\gamma + z_{14}c_y\right)\,\mathbf{\hat{y}} + z_{14}c_z\,\mathbf{\hat{z}}$ | (2$i$) | O VI |
| $\mathbf{B}_{28}$ | = | $-x_{14}\,\mathbf{a}_1 - y_{14}\,\mathbf{a}_2 - z_{14}\,\mathbf{a}_3$ | = | $(-x_{14}a - y_{14}b\cos\gamma - z_{14}c_x)\,\mathbf{\hat{x}} + \left(-y_{14}b\sin\gamma - z_{14}c_y\right)\,\mathbf{\hat{y}} - z_{14}c_z\,\mathbf{\hat{z}}$ | (2$i$) | O VI |

**Geometry files:**

- CIF: pp. 1505
- POSCAR: pp. 1506

# Wollastonite ($CaSiO_3$) Structure: AB3C_aP30_2_3i_9i_3i

http://aflow.org/prototype-encyclopedia/AB3C_aP30_2_3i_9i_3i

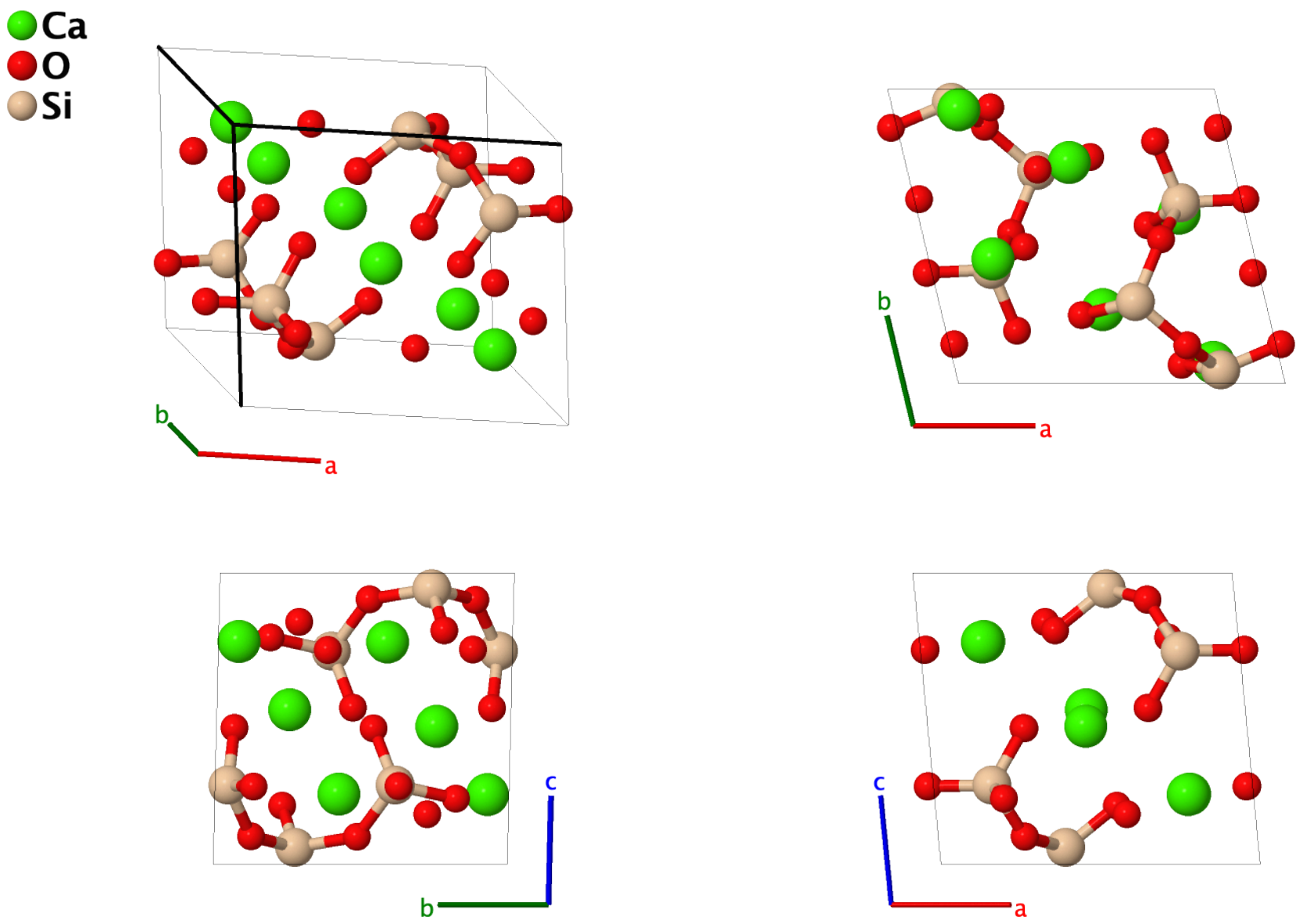

| | | |
|---|---|---|
| **Prototype** | : | $CaO_3Si$ |
| **AFLOW prototype label** | : | AB3C_aP30_2_3i_9i_3i |
| ***Strukturbericht* designation** | : | None |
| **Pearson symbol** | : | aP30 |
| **Space group number** | : | 2 |
| **Space group symbol** | : | $P\bar{1}$ |
| **AFLOW prototype command** | : | `aflow --proto=AB3C_aP30_2_3i_9i_3i` `--params=`$a, b/a, c/a, \alpha, \beta, \gamma, x_1, y_1, z_1, x_2, y_2, z_2, x_3, y_3, z_3, x_4, y_4, z_4, x_5, y_5, z_5, x_6, y_6, z_6, x_7, y_7, z_7, x_8, y_8, z_8, x_9, y_9, z_9, x_{10}, y_{10}, z_{10}, x_{11}, y_{11}, z_{11}, x_{12}, y_{12}, z_{12}, x_{13}, y_{13}, z_{13}, x_{14}, y_{14}, z_{14}, x_{15}, y_{15}, z_{15}$ |

### Other compounds with this structure

- $Ca_2NaHSi_2O_9$ (pectolite)

---

- (Barnick, 1936) found a monoclinic structure for wollastonite, but that structure is now referred to as parawollastonite, with the current triclinic structure receiving the original name.

### Triclinic primitive vectors:

$$\mathbf{a}_1 = a\,\hat{\mathbf{x}}$$

$$\mathbf{a}_2 = b\cos\gamma\,\hat{\mathbf{x}} + b\sin\gamma\,\hat{\mathbf{y}}$$

$$\mathbf{a}_3 = c_x\,\hat{\mathbf{x}} + c_y\,\hat{\mathbf{y}} + c_z\,\hat{\mathbf{z}}$$

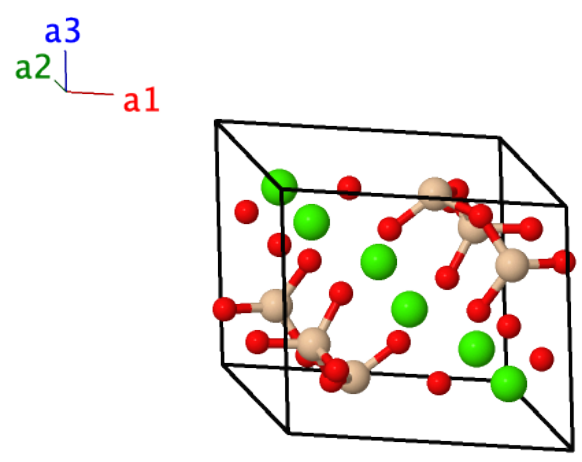

$$c_x = c\,\cos\beta$$

$$c_y = c\,(\cos\alpha - \cos\beta\cos\gamma)/\sin\gamma$$

$$c_z = \sqrt{c^2 - c_x^2 - c_y^2}$$

**Basis vectors:**

| | | Lattice Coordinates | | Cartesian Coordinates | Wyckoff Position | Atom Type |
|---|---|---|---|---|---|---|
| $\mathbf{B}_{1}$ | = | $x_{1}\,\mathbf{a}_{1} + y_{1}\,\mathbf{a}_{2} + z_{1}\,\mathbf{a}_{3}$ | = | $\left(x_{1}a + y_{1}b\cos\gamma + z_{1}c_{x}\right)\,\mathbf{\hat{x}} + \left(y_{1}b\sin\gamma + z_{1}c_{y}\right)\,\mathbf{\hat{y}} + z_{1}c_{z}\,\mathbf{\hat{z}}$ | (2$i$) | Ca I |
| $\mathbf{B}_{2}$ | = | $-x_{1}\,\mathbf{a}_{1} - y_{1}\,\mathbf{a}_{2} - z_{1}\,\mathbf{a}_{3}$ | = | $\left(-x_{1}a - y_{1}b\cos\gamma - z_{1}c_{x}\right)\,\mathbf{\hat{x}} + \left(-y_{1}b\sin\gamma - z_{1}c_{y}\right)\,\mathbf{\hat{y}} - z_{1}c_{z}\,\mathbf{\hat{z}}$ | (2$i$) | Ca I |
| $\mathbf{B}_{3}$ | = | $x_{2}\,\mathbf{a}_{1} + y_{2}\,\mathbf{a}_{2} + z_{2}\,\mathbf{a}_{3}$ | = | $\left(x_{2}a + y_{2}b\cos\gamma + z_{2}c_{x}\right)\,\mathbf{\hat{x}} + \left(y_{2}b\sin\gamma + z_{2}c_{y}\right)\,\mathbf{\hat{y}} + z_{2}c_{z}\,\mathbf{\hat{z}}$ | (2$i$) | Ca II |
| $\mathbf{B}_{4}$ | = | $-x_{2}\,\mathbf{a}_{1} - y_{2}\,\mathbf{a}_{2} - z_{2}\,\mathbf{a}_{3}$ | = | $\left(-x_{2}a - y_{2}b\cos\gamma - z_{2}c_{x}\right)\,\mathbf{\hat{x}} + \left(-y_{2}b\sin\gamma - z_{2}c_{y}\right)\,\mathbf{\hat{y}} - z_{2}c_{z}\,\mathbf{\hat{z}}$ | (2$i$) | Ca II |
| $\mathbf{B}_{5}$ | = | $x_{3}\,\mathbf{a}_{1} + y_{3}\,\mathbf{a}_{2} + z_{3}\,\mathbf{a}_{3}$ | = | $\left(x_{3}a + y_{3}b\cos\gamma + z_{3}c_{x}\right)\,\mathbf{\hat{x}} + \left(y_{3}b\sin\gamma + z_{3}c_{y}\right)\,\mathbf{\hat{y}} + z_{3}c_{z}\,\mathbf{\hat{z}}$ | (2$i$) | Ca III |
| $\mathbf{B}_{6}$ | = | $-x_{3}\,\mathbf{a}_{1} - y_{3}\,\mathbf{a}_{2} - z_{3}\,\mathbf{a}_{3}$ | = | $\left(-x_{3}a - y_{3}b\cos\gamma - z_{3}c_{x}\right)\,\mathbf{\hat{x}} + \left(-y_{3}b\sin\gamma - z_{3}c_{y}\right)\,\mathbf{\hat{y}} - z_{3}c_{z}\,\mathbf{\hat{z}}$ | (2$i$) | Ca III |
| $\mathbf{B}_{7}$ | = | $x_{4}\,\mathbf{a}_{1} + y_{4}\,\mathbf{a}_{2} + z_{4}\,\mathbf{a}_{3}$ | = | $\left(x_{4}a + y_{4}b\cos\gamma + z_{4}c_{x}\right)\,\mathbf{\hat{x}} + \left(y_{4}b\sin\gamma + z_{4}c_{y}\right)\,\mathbf{\hat{y}} + z_{4}c_{z}\,\mathbf{\hat{z}}$ | (2$i$) | O I |
| $\mathbf{B}_{8}$ | = | $-x_{4}\,\mathbf{a}_{1} - y_{4}\,\mathbf{a}_{2} - z_{4}\,\mathbf{a}_{3}$ | = | $\left(-x_{4}a - y_{4}b\cos\gamma - z_{4}c_{x}\right)\,\mathbf{\hat{x}} + \left(-y_{4}b\sin\gamma - z_{4}c_{y}\right)\,\mathbf{\hat{y}} - z_{4}c_{z}\,\mathbf{\hat{z}}$ | (2$i$) | O I |
| $\mathbf{B}_{9}$ | = | $x_{5}\,\mathbf{a}_{1} + y_{5}\,\mathbf{a}_{2} + z_{5}\,\mathbf{a}_{3}$ | = | $\left(x_{5}a + y_{5}b\cos\gamma + z_{5}c_{x}\right)\,\mathbf{\hat{x}} + \left(y_{5}b\sin\gamma + z_{5}c_{y}\right)\,\mathbf{\hat{y}} + z_{5}c_{z}\,\mathbf{\hat{z}}$ | (2$i$) | O II |
| $\mathbf{B}_{10}$ | = | $-x_{5}\,\mathbf{a}_{1} - y_{5}\,\mathbf{a}_{2} - z_{5}\,\mathbf{a}_{3}$ | = | $\left(-x_{5}a - y_{5}b\cos\gamma - z_{5}c_{x}\right)\,\mathbf{\hat{x}} + \left(-y_{5}b\sin\gamma - z_{5}c_{y}\right)\,\mathbf{\hat{y}} - z_{5}c_{z}\,\mathbf{\hat{z}}$ | (2$i$) | O II |
| $\mathbf{B}_{11}$ | = | $x_{6}\,\mathbf{a}_{1} + y_{6}\,\mathbf{a}_{2} + z_{6}\,\mathbf{a}_{3}$ | = | $\left(x_{6}a + y_{6}b\cos\gamma + z_{6}c_{x}\right)\,\mathbf{\hat{x}} + \left(y_{6}b\sin\gamma + z_{6}c_{y}\right)\,\mathbf{\hat{y}} + z_{6}c_{z}\,\mathbf{\hat{z}}$ | (2$i$) | O III |
| $\mathbf{B}_{12}$ | = | $-x_{6}\,\mathbf{a}_{1} - y_{6}\,\mathbf{a}_{2} - z_{6}\,\mathbf{a}_{3}$ | = | $\left(-x_{6}a - y_{6}b\cos\gamma - z_{6}c_{x}\right)\,\mathbf{\hat{x}} + \left(-y_{6}b\sin\gamma - z_{6}c_{y}\right)\,\mathbf{\hat{y}} - z_{6}c_{z}\,\mathbf{\hat{z}}$ | (2$i$) | O III |
| $\mathbf{B}_{13}$ | = | $x_{7}\,\mathbf{a}_{1} + y_{7}\,\mathbf{a}_{2} + z_{7}\,\mathbf{a}_{3}$ | = | $\left(x_{7}a + y_{7}b\cos\gamma + z_{7}c_{x}\right)\,\mathbf{\hat{x}} + \left(y_{7}b\sin\gamma + z_{7}c_{y}\right)\,\mathbf{\hat{y}} + z_{7}c_{z}\,\mathbf{\hat{z}}$ | (2$i$) | O IV |
| $\mathbf{B}_{14}$ | = | $-x_{7}\,\mathbf{a}_{1} - y_{7}\,\mathbf{a}_{2} - z_{7}\,\mathbf{a}_{3}$ | = | $\left(-x_{7}a - y_{7}b\cos\gamma - z_{7}c_{x}\right)\,\mathbf{\hat{x}} + \left(-y_{7}b\sin\gamma - z_{7}c_{y}\right)\,\mathbf{\hat{y}} - z_{7}c_{z}\,\mathbf{\hat{z}}$ | (2$i$) | O IV |
| $\mathbf{B}_{15}$ | = | $x_{8}\,\mathbf{a}_{1} + y_{8}\,\mathbf{a}_{2} + z_{8}\,\mathbf{a}_{3}$ | = | $\left(x_{8}a + y_{8}b\cos\gamma + z_{8}c_{x}\right)\,\mathbf{\hat{x}} + \left(y_{8}b\sin\gamma + z_{8}c_{y}\right)\,\mathbf{\hat{y}} + z_{8}c_{z}\,\mathbf{\hat{z}}$ | (2$i$) | O V |
| $\mathbf{B}_{16}$ | = | $-x_{8}\,\mathbf{a}_{1} - y_{8}\,\mathbf{a}_{2} - z_{8}\,\mathbf{a}_{3}$ | = | $\left(-x_{8}a - y_{8}b\cos\gamma - z_{8}c_{x}\right)\,\mathbf{\hat{x}} + \left(-y_{8}b\sin\gamma - z_{8}c_{y}\right)\,\mathbf{\hat{y}} - z_{8}c_{z}\,\mathbf{\hat{z}}$ | (2$i$) | O V |
| $\mathbf{B}_{17}$ | = | $x_{9}\,\mathbf{a}_{1} + y_{9}\,\mathbf{a}_{2} + z_{9}\,\mathbf{a}_{3}$ | = | $\left(x_{9}a + y_{9}b\cos\gamma + z_{9}c_{x}\right)\,\mathbf{\hat{x}} + \left(y_{9}b\sin\gamma + z_{9}c_{y}\right)\,\mathbf{\hat{y}} + z_{9}c_{z}\,\mathbf{\hat{z}}$ | (2$i$) | O VI |
| $\mathbf{B}_{18}$ | = | $-x_{9}\,\mathbf{a}_{1} - y_{9}\,\mathbf{a}_{2} - z_{9}\,\mathbf{a}_{3}$ | = | $\left(-x_{9}a - y_{9}b\cos\gamma - z_{9}c_{x}\right)\,\mathbf{\hat{x}} + \left(-y_{9}b\sin\gamma - z_{9}c_{y}\right)\,\mathbf{\hat{y}} - z_{9}c_{z}\,\mathbf{\hat{z}}$ | (2$i$) | O VI |
| $\mathbf{B}_{19}$ | = | $x_{10}\,\mathbf{a}_{1} + y_{10}\,\mathbf{a}_{2} + z_{10}\,\mathbf{a}_{3}$ | = | $\left(x_{10}a + y_{10}b\cos\gamma + z_{10}c_{x}\right)\,\mathbf{\hat{x}} + \left(y_{10}b\sin\gamma + z_{10}c_{y}\right)\,\mathbf{\hat{y}} + z_{10}c_{z}\,\mathbf{\hat{z}}$ | (2$i$) | O VII |
| $\mathbf{B}_{20}$ | = | $-x_{10}\,\mathbf{a}_{1} - y_{10}\,\mathbf{a}_{2} - z_{10}\,\mathbf{a}_{3}$ | = | $\left(-x_{10}a - y_{10}b\cos\gamma - z_{10}c_{x}\right)\,\mathbf{\hat{x}} + \left(-y_{10}b\sin\gamma - z_{10}c_{y}\right)\,\mathbf{\hat{y}} - z_{10}c_{z}\,\mathbf{\hat{z}}$ | (2$i$) | O VII |
| $\mathbf{B}_{21}$ | = | $x_{11}\,\mathbf{a}_{1} + y_{11}\,\mathbf{a}_{2} + z_{11}\,\mathbf{a}_{3}$ | = | $\left(x_{11}a + y_{11}b\cos\gamma + z_{11}c_{x}\right)\,\mathbf{\hat{x}} + \left(y_{11}b\sin\gamma + z_{11}c_{y}\right)\,\mathbf{\hat{y}} + z_{11}c_{z}\,\mathbf{\hat{z}}$ | (2$i$) | O VIII |
| $\mathbf{B}_{22}$ | = | $-x_{11}\,\mathbf{a}_1 - y_{11}\,\mathbf{a}_2 - z_{11}\,\mathbf{a}_3$ | = | $(-x_{11}a - y_{11}b\cos\gamma - z_{11}c_x)\,\mathbf{\hat{x}} + \left(-y_{11}b\sin\gamma - z_{11}c_y\right)\,\mathbf{\hat{y}} - z_{11}c_z\,\mathbf{\hat{z}}$ | (2*i*) | O VIII |
| $\mathbf{B}_{23}$ | = | $x_{12}\,\mathbf{a}_1 + y_{12}\,\mathbf{a}_2 + z_{12}\,\mathbf{a}_3$ | = | $(x_{12}a + y_{12}b\cos\gamma + z_{12}c_x)\,\mathbf{\hat{x}} + \left(y_{12}b\sin\gamma + z_{12}c_y\right)\,\mathbf{\hat{y}} + z_{12}c_z\,\mathbf{\hat{z}}$ | (2*i*) | O IX |
| $\mathbf{B}_{24}$ | = | $-x_{12}\,\mathbf{a}_1 - y_{12}\,\mathbf{a}_2 - z_{12}\,\mathbf{a}_3$ | = | $(-x_{12}a - y_{12}b\cos\gamma - z_{12}c_x)\,\mathbf{\hat{x}} + \left(-y_{12}b\sin\gamma - z_{12}c_y\right)\,\mathbf{\hat{y}} - z_{12}c_z\,\mathbf{\hat{z}}$ | (2*i*) | O IX |
| $\mathbf{B}_{25}$ | = | $x_{13}\,\mathbf{a}_1 + y_{13}\,\mathbf{a}_2 + z_{13}\,\mathbf{a}_3$ | = | $(x_{13}a + y_{13}b\cos\gamma + z_{13}c_x)\,\mathbf{\hat{x}} + \left(y_{13}b\sin\gamma + z_{13}c_y\right)\,\mathbf{\hat{y}} + z_{13}c_z\,\mathbf{\hat{z}}$ | (2*i*) | Si I |
| $\mathbf{B}_{26}$ | = | $-x_{13}\,\mathbf{a}_1 - y_{13}\,\mathbf{a}_2 - z_{13}\,\mathbf{a}_3$ | = | $(-x_{13}a - y_{13}b\cos\gamma - z_{13}c_x)\,\mathbf{\hat{x}} + \left(-y_{13}b\sin\gamma - z_{13}c_y\right)\,\mathbf{\hat{y}} - z_{13}c_z\,\mathbf{\hat{z}}$ | (2*i*) | Si I |
| $\mathbf{B}_{27}$ | = | $x_{14}\,\mathbf{a}_1 + y_{14}\,\mathbf{a}_2 + z_{14}\,\mathbf{a}_3$ | = | $(x_{14}a + y_{14}b\cos\gamma + z_{14}c_x)\,\mathbf{\hat{x}} + \left(y_{14}b\sin\gamma + z_{14}c_y\right)\,\mathbf{\hat{y}} + z_{14}c_z\,\mathbf{\hat{z}}$ | (2*i*) | Si II |
| $\mathbf{B}_{28}$ | = | $-x_{14}\,\mathbf{a}_1 - y_{14}\,\mathbf{a}_2 - z_{14}\,\mathbf{a}_3$ | = | $(-x_{14}a - y_{14}b\cos\gamma - z_{14}c_x)\,\mathbf{\hat{x}} + \left(-y_{14}b\sin\gamma - z_{14}c_y\right)\,\mathbf{\hat{y}} - z_{14}c_z\,\mathbf{\hat{z}}$ | (2*i*) | Si II |
| $\mathbf{B}_{29}$ | = | $x_{15}\,\mathbf{a}_1 + y_{15}\,\mathbf{a}_2 + z_{15}\,\mathbf{a}_3$ | = | $(x_{15}a + y_{15}b\cos\gamma + z_{15}c_x)\,\mathbf{\hat{x}} + \left(y_{15}b\sin\gamma + z_{15}c_y\right)\,\mathbf{\hat{y}} + z_{15}c_z\,\mathbf{\hat{z}}$ | (2*i*) | Si III |
| $\mathbf{B}_{30}$ | = | $-x_{15}\,\mathbf{a}_1 - y_{15}\,\mathbf{a}_2 - z_{15}\,\mathbf{a}_3$ | = | $(-x_{15}a - y_{15}b\cos\gamma - z_{15}c_x)\,\mathbf{\hat{x}} + \left(-y_{15}b\sin\gamma - z_{15}c_y\right)\,\mathbf{\hat{y}} - z_{15}c_z\,\mathbf{\hat{z}}$ | (2*i*) | Si III |

---

---

**Geometry files:**

- CIF: pp. 1506
- POSCAR: pp. 1506

# Albite ($NaAlSi_3O_8$, $S6_8$) Structure: ABC8D3_aP26_2_i_i_8i_3i

http://aflow.org/prototype-encyclopedia/ABC8D3_aP26_2_i_i_8i_3i

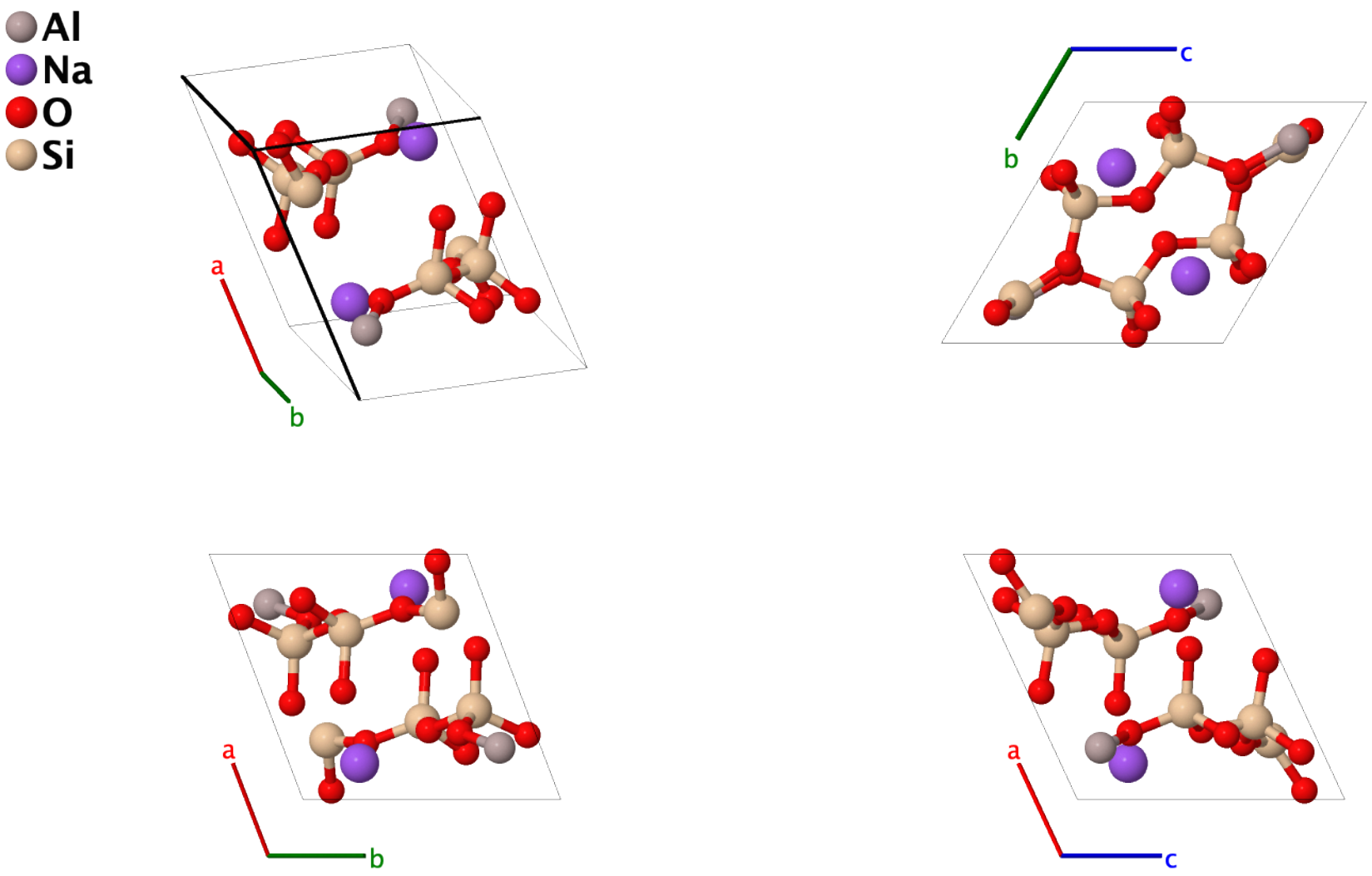

| | | |
|---|---|---|
| **Prototype** | : | $AlNaO_8Si_3$ |
| **AFLOW prototype label** | : | ABC8D3_aP26_2_i_i_8i_3i |
| ***Strukturbericht* designation** | : | $S6_8$ |
| **Pearson symbol** | : | aP26 |
| **Space group number** | : | 2 |
| **Space group symbol** | : | $P\bar{1}$ |
| **AFLOW prototype command** | : | `aflow --proto=ABC8D3_aP26_2_i_i_8i_3i` `--params=`$a, b/a, c/a, \alpha, \beta, \gamma, x_1, y_1, z_1, x_2, y_2, z_2, x_3, y_3, z_3, x_4, y_4, z_4, x_5, y_5, z_5, x_6, y_6, z_6, x_7, y_7, z_7, x_8, y_8, z_8, x_9, y_9, z_9, x_{10}, y_{10}, z_{10}, x_{11}, y_{11}, z_{11}, x_{12}, y_{12}, z_{12}, x_{13}, y_{13}, z_{13}$ |

- We used the 13 K data from (Smith, 1986), however they present their data in space group $C\bar{1}$, which doubles the primitive unit cell compared to the standard space group $P\bar{1}$ #2. We used FINDSYM to convert from the presented cell to the conventional cell. This involved a rotation of the cell, *e.g.*, the original $c$ axis is the $a$ axis in our standard primitive cell.

- Technically this is "low" albite. In high albite the silicon and aluminum atoms are mixed over all four sites, as in sandine ($S6_7$). Indeed, under some conditions albite crystals are seen in the sandine structure (Winter, 1979). See the albite entry in (Downs, 2003) for other experimental work.

**Triclinic primitive vectors:**

$$
\begin{aligned}
\mathbf{a}_1 &= a\,\hat{\mathbf{x}} \\
\mathbf{a}_2 &= b\cos\gamma\,\hat{\mathbf{x}} + b\sin\gamma\,\hat{\mathbf{y}} \\
\mathbf{a}_3 &= c_x\,\hat{\mathbf{x}} + c_y\,\hat{\mathbf{y}} + c_z\,\hat{\mathbf{z}} \\
\\
c_x &= c\,\cos\beta \\
c_y &= c\,(\cos\alpha - \cos\beta\cos\gamma)/\sin\gamma \\
c_z &= \sqrt{c^2 - c_x^2 - c_y^2}
\end{aligned}
$$

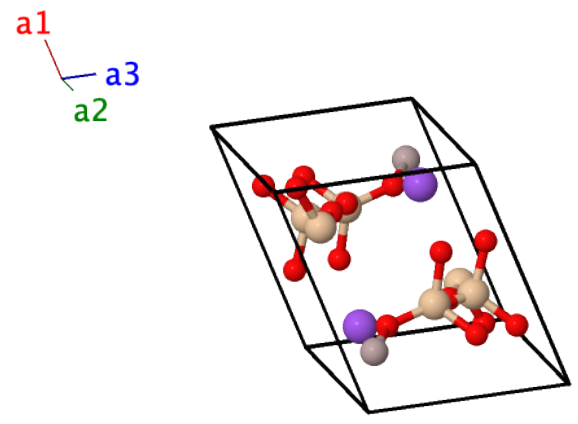

**Basis vectors:**

| | | Lattice Coordinates | | Cartesian Coordinates | Wyckoff Position | Atom Type |
|---|---|---|---|---|---|---|
| $\mathbf{B}_1$ | = | $x_1\,\mathbf{a}_1 + y_1\,\mathbf{a}_2 + z_1\,\mathbf{a}_3$ | = | $(x_1 a + y_1 b\cos\gamma + z_1 c_x)\,\hat{\mathbf{x}} + \left(y_1 b\sin\gamma + z_1 c_y\right)\hat{\mathbf{y}} + z_1 c_z\,\hat{\mathbf{z}}$ | (2$i$) | Al |
| $\mathbf{B}_2$ | = | $-x_1\,\mathbf{a}_1 - y_1\,\mathbf{a}_2 - z_1\,\mathbf{a}_3$ | = | $(-x_1 a - y_1 b\cos\gamma - z_1 c_x)\,\hat{\mathbf{x}} + \left(-y_1 b\sin\gamma - z_1 c_y\right)\hat{\mathbf{y}} - z_1 c_z\,\hat{\mathbf{z}}$ | (2$i$) | Al |
| $\mathbf{B}_3$ | = | $x_2\,\mathbf{a}_1 + y_2\,\mathbf{a}_2 + z_2\,\mathbf{a}_3$ | = | $(x_2 a + y_2 b\cos\gamma + z_2 c_x)\,\hat{\mathbf{x}} + \left(y_2 b\sin\gamma + z_2 c_y\right)\hat{\mathbf{y}} + z_2 c_z\,\hat{\mathbf{z}}$ | (2$i$) | Na |
| $\mathbf{B}_4$ | = | $-x_2\,\mathbf{a}_1 - y_2\,\mathbf{a}_2 - z_2\,\mathbf{a}_3$ | = | $(-x_2 a - y_2 b\cos\gamma - z_2 c_x)\,\hat{\mathbf{x}} + \left(-y_2 b\sin\gamma - z_2 c_y\right)\hat{\mathbf{y}} - z_2 c_z\,\hat{\mathbf{z}}$ | (2$i$) | Na |
| $\mathbf{B}_5$ | = | $x_3\,\mathbf{a}_1 + y_3\,\mathbf{a}_2 + z_3\,\mathbf{a}_3$ | = | $(x_3 a + y_3 b\cos\gamma + z_3 c_x)\,\hat{\mathbf{x}} + \left(y_3 b\sin\gamma + z_3 c_y\right)\hat{\mathbf{y}} + z_3 c_z\,\hat{\mathbf{z}}$ | (2$i$) | O I |
| $\mathbf{B}_6$ | = | $-x_3\,\mathbf{a}_1 - y_3\,\mathbf{a}_2 - z_3\,\mathbf{a}_3$ | = | $(-x_3 a - y_3 b\cos\gamma - z_3 c_x)\,\hat{\mathbf{x}} + \left(-y_3 b\sin\gamma - z_3 c_y\right)\hat{\mathbf{y}} - z_3 c_z\,\hat{\mathbf{z}}$ | (2$i$) | O I |
| $\mathbf{B}_7$ | = | $x_4\,\mathbf{a}_1 + y_4\,\mathbf{a}_2 + z_4\,\mathbf{a}_3$ | = | $(x_4 a + y_4 b\cos\gamma + z_4 c_x)\,\hat{\mathbf{x}} + \left(y_4 b\sin\gamma + z_4 c_y\right)\hat{\mathbf{y}} + z_4 c_z\,\hat{\mathbf{z}}$ | (2$i$) | O II |
| $\mathbf{B}_8$ | = | $-x_4\,\mathbf{a}_1 - y_4\,\mathbf{a}_2 - z_4\,\mathbf{a}_3$ | = | $(-x_4 a - y_4 b\cos\gamma - z_4 c_x)\,\hat{\mathbf{x}} + \left(-y_4 b\sin\gamma - z_4 c_y\right)\hat{\mathbf{y}} - z_4 c_z\,\hat{\mathbf{z}}$ | (2$i$) | O II |
| $\mathbf{B}_9$ | = | $x_5\,\mathbf{a}_1 + y_5\,\mathbf{a}_2 + z_5\,\mathbf{a}_3$ | = | $(x_5 a + y_5 b\cos\gamma + z_5 c_x)\,\hat{\mathbf{x}} + \left(y_5 b\sin\gamma + z_5 c_y\right)\hat{\mathbf{y}} + z_5 c_z\,\hat{\mathbf{z}}$ | (2$i$) | O III |
| $\mathbf{B}_{10}$ | = | $-x_5\,\mathbf{a}_1 - y_5\,\mathbf{a}_2 - z_5\,\mathbf{a}_3$ | = | $(-x_5 a - y_5 b\cos\gamma - z_5 c_x)\,\hat{\mathbf{x}} + \left(-y_5 b\sin\gamma - z_5 c_y\right)\hat{\mathbf{y}} - z_5 c_z\,\hat{\mathbf{z}}$ | (2$i$) | O III |
| $\mathbf{B}_{11}$ | = | $x_6\,\mathbf{a}_1 + y_6\,\mathbf{a}_2 + z_6\,\mathbf{a}_3$ | = | $(x_6 a + y_6 b\cos\gamma + z_6 c_x)\,\hat{\mathbf{x}} + \left(y_6 b\sin\gamma + z_6 c_y\right)\hat{\mathbf{y}} + z_6 c_z\,\hat{\mathbf{z}}$ | (2$i$) | O IV |
| $\mathbf{B}_{12}$ | = | $-x_6\,\mathbf{a}_1 - y_6\,\mathbf{a}_2 - z_6\,\mathbf{a}_3$ | = | $(-x_6 a - y_6 b\cos\gamma - z_6 c_x)\,\hat{\mathbf{x}} + \left(-y_6 b\sin\gamma - z_6 c_y\right)\hat{\mathbf{y}} - z_6 c_z\,\hat{\mathbf{z}}$ | (2$i$) | O IV |
| $\mathbf{B}_{13}$ | = | $x_7\,\mathbf{a}_1 + y_7\,\mathbf{a}_2 + z_7\,\mathbf{a}_3$ | = | $(x_7 a + y_7 b\cos\gamma + z_7 c_x)\,\hat{\mathbf{x}} + \left(y_7 b\sin\gamma + z_7 c_y\right)\hat{\mathbf{y}} + z_7 c_z\,\hat{\mathbf{z}}$ | (2$i$) | O V |
| $\mathbf{B}_{14}$ | = | $-x_7\,\mathbf{a}_1 - y_7\,\mathbf{a}_2 - z_7\,\mathbf{a}_3$ | = | $(-x_7 a - y_7 b\cos\gamma - z_7 c_x)\,\hat{\mathbf{x}} + \left(-y_7 b\sin\gamma - z_7 c_y\right)\hat{\mathbf{y}} - z_7 c_z\,\hat{\mathbf{z}}$ | (2$i$) | O V |
| $\mathbf{B}_{15}$ | = | $x_8\,\mathbf{a}_1 + y_8\,\mathbf{a}_2 + z_8\,\mathbf{a}_3$ | = | $(x_8 a + y_8 b\cos\gamma + z_8 c_x)\,\hat{\mathbf{x}} + \left(y_8 b\sin\gamma + z_8 c_y\right)\hat{\mathbf{y}} + z_8 c_z\,\hat{\mathbf{z}}$ | (2$i$) | O VI |
| $\mathbf{B}_{16}$ | = | $-x_8\,\mathbf{a}_1 - y_8\,\mathbf{a}_2 - z_8\,\mathbf{a}_3$ | = | $(-x_8 a - y_8 b\cos\gamma - z_8 c_x)\,\hat{\mathbf{x}} + \left(-y_8 b\sin\gamma - z_8 c_y\right)\hat{\mathbf{y}} - z_8 c_z\,\hat{\mathbf{z}}$ | (2$i$) | O VI |
| $\mathbf{B}_{17}$ | = | $x_9\,\mathbf{a}_1 + y_9\,\mathbf{a}_2 + z_9\,\mathbf{a}_3$ | = | $(x_9 a + y_9 b\cos\gamma + z_9 c_x)\,\hat{\mathbf{x}} + \left(y_9 b\sin\gamma + z_9 c_y\right)\,\hat{\mathbf{y}} + z_9 c_z\,\hat{\mathbf{z}}$ | (2*i*) | O VII |
| $\mathbf{B}_{18}$ | = | $-x_9\,\mathbf{a}_1 - y_9\,\mathbf{a}_2 - z_9\,\mathbf{a}_3$ | = | $(-x_9 a - y_9 b\cos\gamma - z_9 c_x)\,\hat{\mathbf{x}} + \left(-y_9 b\sin\gamma - z_9 c_y\right)\,\hat{\mathbf{y}} - z_9 c_z\,\hat{\mathbf{z}}$ | (2*i*) | O VII |
| $\mathbf{B}_{19}$ | = | $x_{10}\,\mathbf{a}_1 + y_{10}\,\mathbf{a}_2 + z_{10}\,\mathbf{a}_3$ | = | $(x_{10} a + y_{10} b\cos\gamma + z_{10} c_x)\,\hat{\mathbf{x}} + \left(y_{10} b\sin\gamma + z_{10} c_y\right)\,\hat{\mathbf{y}} + z_{10} c_z\,\hat{\mathbf{z}}$ | (2*i*) | O VIII |
| $\mathbf{B}_{20}$ | = | $-x_{10}\,\mathbf{a}_1 - y_{10}\,\mathbf{a}_2 - z_{10}\,\mathbf{a}_3$ | = | $(-x_{10} a - y_{10} b\cos\gamma - z_{10} c_x)\,\hat{\mathbf{x}} + \left(-y_{10} b\sin\gamma - z_{10} c_y\right)\,\hat{\mathbf{y}} - z_{10} c_z\,\hat{\mathbf{z}}$ | (2*i*) | O VIII |
| $\mathbf{B}_{21}$ | = | $x_{11}\,\mathbf{a}_1 + y_{11}\,\mathbf{a}_2 + z_{11}\,\mathbf{a}_3$ | = | $(x_{11} a + y_{11} b\cos\gamma + z_{11} c_x)\,\hat{\mathbf{x}} + \left(y_{11} b\sin\gamma + z_{11} c_y\right)\,\hat{\mathbf{y}} + z_{11} c_z\,\hat{\mathbf{z}}$ | (2*i*) | Si I |
| $\mathbf{B}_{22}$ | = | $-x_{11}\,\mathbf{a}_1 - y_{11}\,\mathbf{a}_2 - z_{11}\,\mathbf{a}_3$ | = | $(-x_{11} a - y_{11} b\cos\gamma - z_{11} c_x)\,\hat{\mathbf{x}} + \left(-y_{11} b\sin\gamma - z_{11} c_y\right)\,\hat{\mathbf{y}} - z_{11} c_z\,\hat{\mathbf{z}}$ | (2*i*) | Si I |
| $\mathbf{B}_{23}$ | = | $x_{12}\,\mathbf{a}_1 + y_{12}\,\mathbf{a}_2 + z_{12}\,\mathbf{a}_3$ | = | $(x_{12} a + y_{12} b\cos\gamma + z_{12} c_x)\,\hat{\mathbf{x}} + \left(y_{12} b\sin\gamma + z_{12} c_y\right)\,\hat{\mathbf{y}} + z_{12} c_z\,\hat{\mathbf{z}}$ | (2*i*) | Si II |
| $\mathbf{B}_{24}$ | = | $-x_{12}\,\mathbf{a}_1 - y_{12}\,\mathbf{a}_2 - z_{12}\,\mathbf{a}_3$ | = | $(-x_{12} a - y_{12} b\cos\gamma - z_{12} c_x)\,\hat{\mathbf{x}} + \left(-y_{12} b\sin\gamma - z_{12} c_y\right)\,\hat{\mathbf{y}} - z_{12} c_z\,\hat{\mathbf{z}}$ | (2*i*) | Si II |
| $\mathbf{B}_{25}$ | = | $x_{13}\,\mathbf{a}_1 + y_{13}\,\mathbf{a}_2 + z_{13}\,\mathbf{a}_3$ | = | $(x_{13} a + y_{13} b\cos\gamma + z_{13} c_x)\,\hat{\mathbf{x}} + \left(y_{13} b\sin\gamma + z_{13} c_y\right)\,\hat{\mathbf{y}} + z_{13} c_z\,\hat{\mathbf{z}}$ | (2*i*) | Si III |
| $\mathbf{B}_{26}$ | = | $-x_{13}\,\mathbf{a}_1 - y_{13}\,\mathbf{a}_2 - z_{13}\,\mathbf{a}_3$ | = | $(-x_{13} a - y_{13} b\cos\gamma - z_{13} c_x)\,\hat{\mathbf{x}} + \left(-y_{13} b\sin\gamma - z_{13} c_y\right)\,\hat{\mathbf{y}} - z_{13} c_z\,\hat{\mathbf{z}}$ | (2*i*) | Si III |

**Geometry files:**

- CIF: pp. 1507
- POSCAR: pp. 1507

# TaTi (BCC SQS-16) Structure: AB_aP16_2_4i_4i

http://aflow.org/prototype-encyclopedia/AB_aP16_2_4i_4i

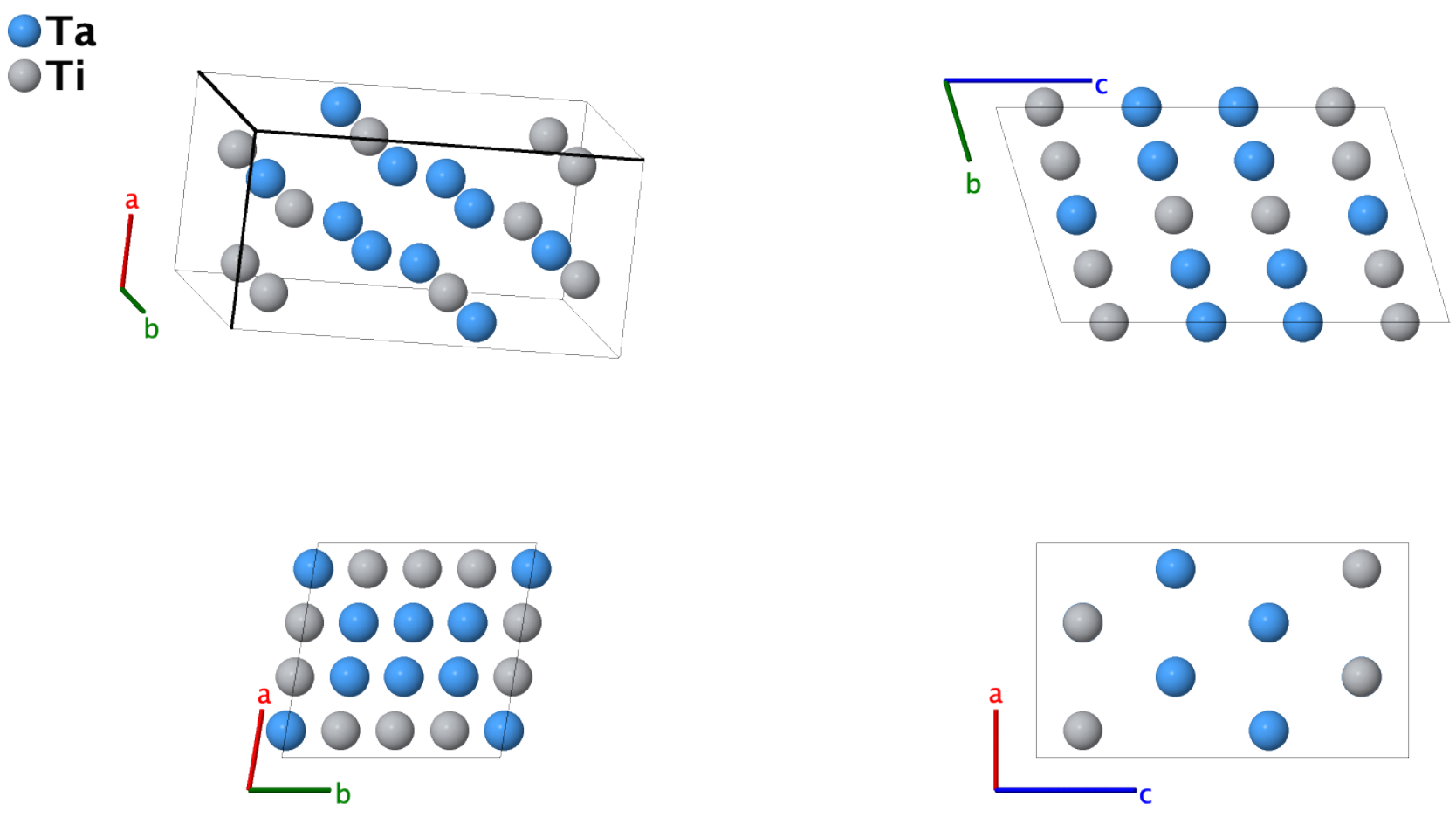

| | | |
|---|---|---|
| **Prototype** | : | TaTi |
| **AFLOW prototype label** | : | AB_aP16_2_4i_4i |
| ***Strukturbericht* designation** | : | None |
| **Pearson symbol** | : | aP16 |
| **Space group number** | : | 2 |
| **Space group symbol** | : | $P\bar{1}$ |
| **AFLOW prototype command** | : | `aflow --proto=AB_aP16_2_4i_4i`<br>`--params=`$a, b/a, c/a, \alpha, \beta, \gamma, x_1, y_1, z_1, x_2, y_2, z_2, x_3, y_3, z_3, x_4, y_4, z_4, x_5, y_5, z_5, x_6, y_6, z_6, x_7, y_7, z_7, x_8, y_8, z_8$ |

- This is a special quasirandom structure with 16 atoms per unit cell (SQS-16) for a bcc binary substitutional alloy $A_xB_{1-x}$ (Jiang, 2004). This prototype is the equicompositional structure ($x = 0.5$). The $x = 0.25$ and $x = 0.75$ structures are given by AB3_mC32_8_4a_12a.

**Triclinic primitive vectors:**

$$\mathbf{a}_1 = a\,\hat{\mathbf{x}}$$
$$\mathbf{a}_2 = b\cos\gamma\,\hat{\mathbf{x}} + b\sin\gamma\,\hat{\mathbf{y}}$$
$$\mathbf{a}_3 = c_x\hat{\mathbf{x}} + c_y\,\hat{\mathbf{y}} + c_z\,\hat{\mathbf{z}}$$

$$c_x = c\,\cos\beta$$
$$c_y = c\,(\cos\alpha - \cos\beta\cos\gamma)/\sin\gamma$$
$$c_z = \sqrt{c^2 - c_x^2 - c_y^2}$$

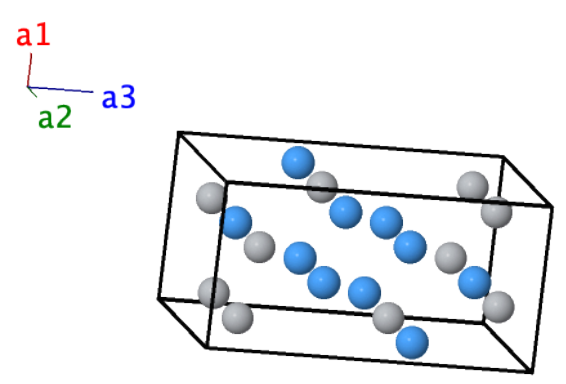

**Basis vectors:**

| | | Lattice Coordinates | | Cartesian Coordinates | Wyckoff Position | Atom Type |
|---|---|---|---|---|---|---|
| $\mathbf{B}_1$ | = | $x_1\,\mathbf{a}_1 + y_1\,\mathbf{a}_2 + z_1\,\mathbf{a}_3$ | = | $(x_1 a + y_1 b\cos\gamma + z_1 c_x)\,\hat{\mathbf{x}} + \left(y_1 b\sin\gamma + z_1 c_y\right)\,\hat{\mathbf{y}} + z_1 c_z\,\hat{\mathbf{z}}$ | (2$i$) | Ta I |
| $\mathbf{B}_2$ | = | $-x_1\,\mathbf{a}_1 - y_1\,\mathbf{a}_2 - z_1\,\mathbf{a}_3$ | = | $(-x_1 a - y_1 b\cos\gamma - z_1 c_x)\,\hat{\mathbf{x}} + \left(-y_1 b\sin\gamma - z_1 c_y\right)\,\hat{\mathbf{y}} - z_1 c_z\,\hat{\mathbf{z}}$ | (2$i$) | Ta I |
| $\mathbf{B}_3$ | = | $x_2\,\mathbf{a}_1 + y_2\,\mathbf{a}_2 + z_2\,\mathbf{a}_3$ | = | $(x_2 a + y_2 b\cos\gamma + z_2 c_x)\,\hat{\mathbf{x}} + \left(y_2 b\sin\gamma + z_2 c_y\right)\hat{\mathbf{y}} + z_2 c_z\,\hat{\mathbf{z}}$ | (2i) | Ta II |
| $\mathbf{B}_4$ | = | $-x_2\,\mathbf{a}_1 - y_2\,\mathbf{a}_2 - z_2\,\mathbf{a}_3$ | = | $(-x_2 a - y_2 b\cos\gamma - z_2 c_x)\,\hat{\mathbf{x}} + \left(-y_2 b\sin\gamma - z_2 c_y\right)\hat{\mathbf{y}} - z_2 c_z\,\hat{\mathbf{z}}$ | (2i) | Ta II |
| $\mathbf{B}_5$ | = | $x_3\,\mathbf{a}_1 + y_3\,\mathbf{a}_2 + z_3\,\mathbf{a}_3$ | = | $(x_3 a + y_3 b\cos\gamma + z_3 c_x)\,\hat{\mathbf{x}} + \left(y_3 b\sin\gamma + z_3 c_y\right)\hat{\mathbf{y}} + z_3 c_z\,\hat{\mathbf{z}}$ | (2i) | Ta III |
| $\mathbf{B}_6$ | = | $-x_3\,\mathbf{a}_1 - y_3\,\mathbf{a}_2 - z_3\,\mathbf{a}_3$ | = | $(-x_3 a - y_3 b\cos\gamma - z_3 c_x)\,\hat{\mathbf{x}} + \left(-y_3 b\sin\gamma - z_3 c_y\right)\hat{\mathbf{y}} - z_3 c_z\,\hat{\mathbf{z}}$ | (2i) | Ta III |
| $\mathbf{B}_7$ | = | $x_4\,\mathbf{a}_1 + y_4\,\mathbf{a}_2 + z_4\,\mathbf{a}_3$ | = | $(x_4 a + y_4 b\cos\gamma + z_4 c_x)\,\hat{\mathbf{x}} + \left(y_4 b\sin\gamma + z_4 c_y\right)\hat{\mathbf{y}} + z_4 c_z\,\hat{\mathbf{z}}$ | (2i) | Ta IV |
| $\mathbf{B}_8$ | = | $-x_4\,\mathbf{a}_1 - y_4\,\mathbf{a}_2 - z_4\,\mathbf{a}_3$ | = | $(-x_4 a - y_4 b\cos\gamma - z_4 c_x)\,\hat{\mathbf{x}} + \left(-y_4 b\sin\gamma - z_4 c_y\right)\hat{\mathbf{y}} - z_4 c_z\,\hat{\mathbf{z}}$ | (2i) | Ta IV |
| $\mathbf{B}_9$ | = | $x_5\,\mathbf{a}_1 + y_5\,\mathbf{a}_2 + z_5\,\mathbf{a}_3$ | = | $(x_5 a + y_5 b\cos\gamma + z_5 c_x)\,\hat{\mathbf{x}} + \left(y_5 b\sin\gamma + z_5 c_y\right)\hat{\mathbf{y}} + z_5 c_z\,\hat{\mathbf{z}}$ | (2i) | Ti I |
| $\mathbf{B}_{10}$ | = | $-x_5\,\mathbf{a}_1 - y_5\,\mathbf{a}_2 - z_5\,\mathbf{a}_3$ | = | $(-x_5 a - y_5 b\cos\gamma - z_5 c_x)\,\hat{\mathbf{x}} + \left(-y_5 b\sin\gamma - z_5 c_y\right)\hat{\mathbf{y}} - z_5 c_z\,\hat{\mathbf{z}}$ | (2i) | Ti I |
| $\mathbf{B}_{11}$ | = | $x_6\,\mathbf{a}_1 + y_6\,\mathbf{a}_2 + z_6\,\mathbf{a}_3$ | = | $(x_6 a + y_6 b\cos\gamma + z_6 c_x)\,\hat{\mathbf{x}} + \left(y_6 b\sin\gamma + z_6 c_y\right)\hat{\mathbf{y}} + z_6 c_z\,\hat{\mathbf{z}}$ | (2i) | Ti II |
| $\mathbf{B}_{12}$ | = | $-x_6\,\mathbf{a}_1 - y_6\,\mathbf{a}_2 - z_6\,\mathbf{a}_3$ | = | $(-x_6 a - y_6 b\cos\gamma - z_6 c_x)\,\hat{\mathbf{x}} + \left(-y_6 b\sin\gamma - z_6 c_y\right)\hat{\mathbf{y}} - z_6 c_z\,\hat{\mathbf{z}}$ | (2i) | Ti II |
| $\mathbf{B}_{13}$ | = | $x_7\,\mathbf{a}_1 + y_7\,\mathbf{a}_2 + z_7\,\mathbf{a}_3$ | = | $(x_7 a + y_7 b\cos\gamma + z_7 c_x)\,\hat{\mathbf{x}} + \left(y_7 b\sin\gamma + z_7 c_y\right)\hat{\mathbf{y}} + z_7 c_z\,\hat{\mathbf{z}}$ | (2i) | Ti III |
| $\mathbf{B}_{14}$ | = | $-x_7\,\mathbf{a}_1 - y_7\,\mathbf{a}_2 - z_7\,\mathbf{a}_3$ | = | $(-x_7 a - y_7 b\cos\gamma - z_7 c_x)\,\hat{\mathbf{x}} + \left(-y_7 b\sin\gamma - z_7 c_y\right)\hat{\mathbf{y}} - z_7 c_z\,\hat{\mathbf{z}}$ | (2i) | Ti III |
| $\mathbf{B}_{15}$ | = | $x_8\,\mathbf{a}_1 + y_8\,\mathbf{a}_2 + z_8\,\mathbf{a}_3$ | = | $(x_8 a + y_8 b\cos\gamma + z_8 c_x)\,\hat{\mathbf{x}} + \left(y_8 b\sin\gamma + z_8 c_y\right)\hat{\mathbf{y}} + z_8 c_z\,\hat{\mathbf{z}}$ | (2i) | Ti IV |
| $\mathbf{B}_{16}$ | = | $-x_8\,\mathbf{a}_1 - y_8\,\mathbf{a}_2 - z_8\,\mathbf{a}_3$ | = | $(-x_8 a - y_8 b\cos\gamma - z_8 c_x)\,\hat{\mathbf{x}} + \left(-y_8 b\sin\gamma - z_8 c_y\right)\hat{\mathbf{y}} - z_8 c_z\,\hat{\mathbf{z}}$ | (2i) | Ti IV |

**Geometry files:**

- CIF: pp. 1507
- POSCAR: pp. 1508

# $W_2O_3(PO_4)_2$ Structure: A11B2C2_mP60_4_22a_4a_4a

http://aflow.org/prototype-encyclopedia/A11B2C2_mP60_4_22a_4a_4a

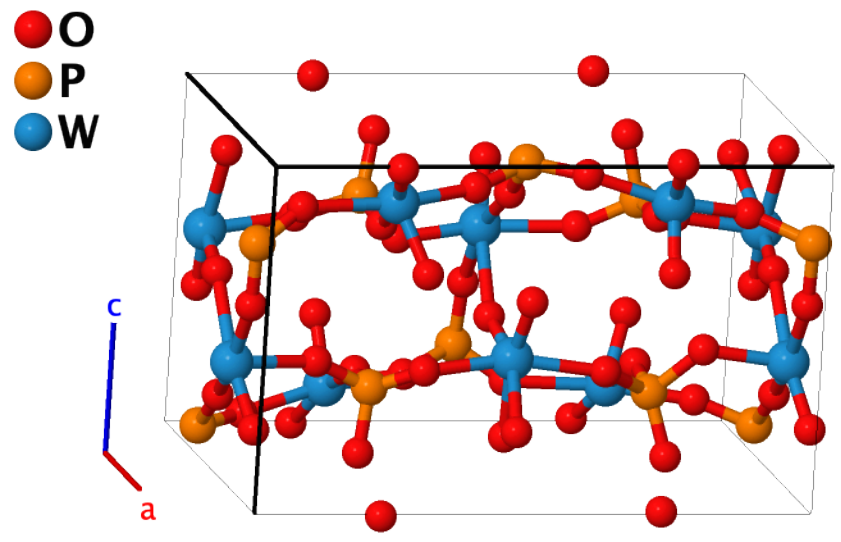

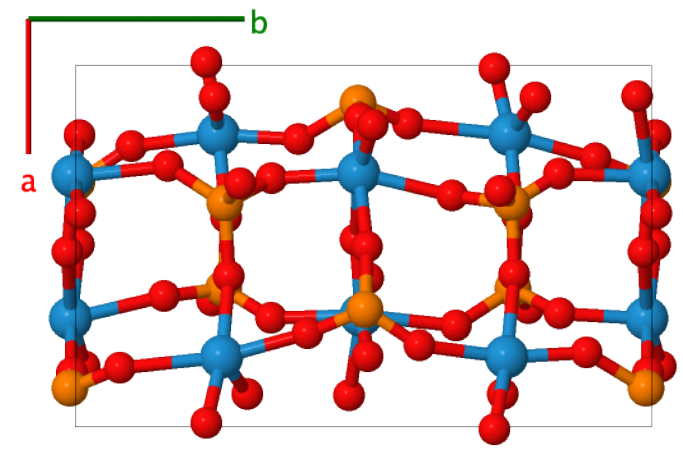

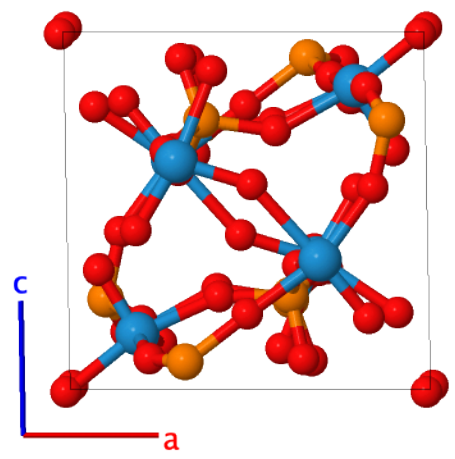

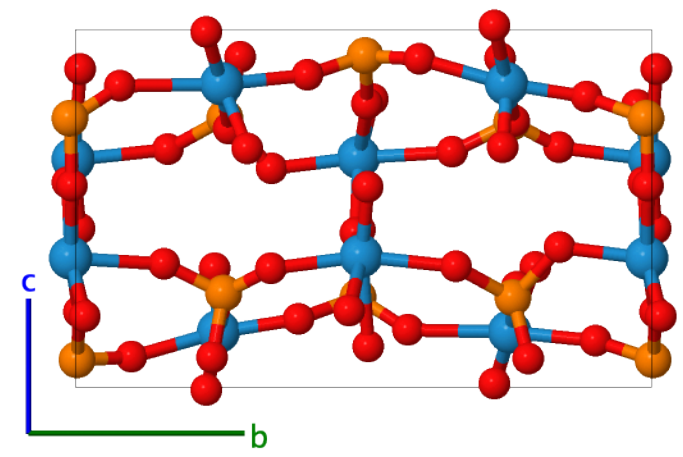

| | | |
|---|---|---|
| **Prototype** | : | $O_{11}P_2W_2$ |
| **AFLOW prototype label** | : | A11B2C2_mP60_4_22a_4a_4a |
| ***Strukturbericht* designation** | : | None |
| **Pearson symbol** | : | mP60 |
| **Space group number** | : | 4 |
| **Space group symbol** | : | $P2_1$ |
| **AFLOW prototype command** | : | `aflow --proto=A11B2C2_mP60_4_22a_4a_4a` `--params=`$a, b/a, c/a, \beta, x_1, y_1, z_1, x_2, y_2, z_2, x_3, y_3, z_3, x_4, y_4, z_4, x_5, y_5, z_5, x_6, y_6, z_6, x_7, y_7, z_7, x_8, y_8, z_8, x_9, y_9, z_9, x_{10}, y_{10}, z_{10}, x_{11}, y_{11}, z_{11}, x_{12}, y_{12}, z_{12}, x_{13}, y_{13}, z_{13}, x_{14}, y_{14}, z_{14}, x_{15}, y_{15}, z_{15}, x_{16}, y_{16}, z_{16}, x_{17}, y_{17}, z_{17}, x_{18}, y_{18}, z_{18}, x_{19}, y_{19}, z_{19}, x_{20}, y_{20}, z_{20}, x_{21}, y_{21}, z_{21}, x_{22}, y_{22}, z_{22}, x_{23}, y_{23}, z_{23}, x_{24}, y_{24}, z_{24}, x_{25}, y_{25}, z_{25}, x_{26}, y_{26}, z_{26}, x_{27}, y_{27}, z_{27}, x_{28}, y_{28}, z_{28}, x_{29}, y_{29}, z_{29}, x_{30}, y_{30}, z_{30}$ |

**Simple Monoclinic primitive vectors:**

$$\mathbf{a}_1 = a\,\hat{\mathbf{x}}$$
$$\mathbf{a}_2 = b\,\hat{\mathbf{y}}$$
$$\mathbf{a}_3 = c\cos\beta\,\hat{\mathbf{x}} + c\sin\beta\,\hat{\mathbf{z}}$$

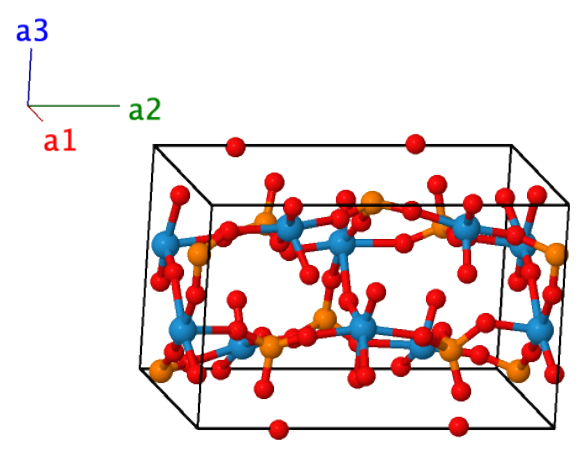

**Basis vectors:**

| | | Lattice Coordinates | | Cartesian Coordinates | Wyckoff Position | Atom Type |
|---|---|---|---|---|---|---|
| $\mathbf{B}_1$ | = | $x_1\,\mathbf{a}_1 + y_1\,\mathbf{a}_2 + z_1\,\mathbf{a}_3$ | = | $(x_1 a + z_1 c\cos\beta)\,\hat{\mathbf{x}} + y_1 b\,\hat{\mathbf{y}} + z_1 c\sin\beta\,\hat{\mathbf{z}}$ | (2a) | O I |
| $\mathbf{B}_2$ | = | $-x_1\,\mathbf{a}_1 + \left(\frac{1}{2} + y_1\right)\mathbf{a}_2 - z_1\,\mathbf{a}_3$ | = | $(-x_1 a - z_1 c\cos\beta)\,\hat{\mathbf{x}} + \left(\frac{1}{2} + y_1\right) b\,\hat{\mathbf{y}} - z_1 c\sin\beta\,\hat{\mathbf{z}}$ | (2a) | O I |
| $\mathbf{B}_3$ | = | $x_2\,\mathbf{a}_1 + y_2\,\mathbf{a}_2 + z_2\,\mathbf{a}_3$ | = | $(x_2 a + z_2 c\cos\beta)\,\hat{\mathbf{x}} + y_2 b\,\hat{\mathbf{y}} + z_2 c\sin\beta\,\hat{\mathbf{z}}$ | (2a) | O II |
| $\mathbf{B}_{4}$ | = | $-x_{2}\,\mathbf{a}_{1}+\left(\frac{1}{2}+y_{2}\right)\,\mathbf{a}_{2}-z_{2}\,\mathbf{a}_{3}$ | = | $\left(-x_{2}a-z_{2}c\cos\beta\right)\,\mathbf{\hat{x}}+\left(\frac{1}{2}+y_{2}\right)b\,\mathbf{\hat{y}}-z_{2}c\sin\beta\,\mathbf{\hat{z}}$ | (2a) | O II |
| $\mathbf{B}_{5}$ | = | $x_{3}\,\mathbf{a}_{1}+y_{3}\,\mathbf{a}_{2}+z_{3}\,\mathbf{a}_{3}$ | = | $\left(x_{3}a+z_{3}c\cos\beta\right)\,\mathbf{\hat{x}}+y_{3}b\,\mathbf{\hat{y}}+z_{3}c\sin\beta\,\mathbf{\hat{z}}$ | (2a) | O III |
| $\mathbf{B}_{6}$ | = | $-x_{3}\,\mathbf{a}_{1}+\left(\frac{1}{2}+y_{3}\right)\,\mathbf{a}_{2}-z_{3}\,\mathbf{a}_{3}$ | = | $\left(-x_{3}a-z_{3}c\cos\beta\right)\,\mathbf{\hat{x}}+\left(\frac{1}{2}+y_{3}\right)b\,\mathbf{\hat{y}}-z_{3}c\sin\beta\,\mathbf{\hat{z}}$ | (2a) | O III |
| $\mathbf{B}_{7}$ | = | $x_{4}\,\mathbf{a}_{1}+y_{4}\,\mathbf{a}_{2}+z_{4}\,\mathbf{a}_{3}$ | = | $\left(x_{4}a+z_{4}c\cos\beta\right)\,\mathbf{\hat{x}}+y_{4}b\,\mathbf{\hat{y}}+z_{4}c\sin\beta\,\mathbf{\hat{z}}$ | (2a) | O IV |
| $\mathbf{B}_{8}$ | = | $-x_{4}\,\mathbf{a}_{1}+\left(\frac{1}{2}+y_{4}\right)\,\mathbf{a}_{2}-z_{4}\,\mathbf{a}_{3}$ | = | $\left(-x_{4}a-z_{4}c\cos\beta\right)\,\mathbf{\hat{x}}+\left(\frac{1}{2}+y_{4}\right)b\,\mathbf{\hat{y}}-z_{4}c\sin\beta\,\mathbf{\hat{z}}$ | (2a) | O IV |
| $\mathbf{B}_{9}$ | = | $x_{5}\,\mathbf{a}_{1}+y_{5}\,\mathbf{a}_{2}+z_{5}\,\mathbf{a}_{3}$ | = | $\left(x_{5}a+z_{5}c\cos\beta\right)\,\mathbf{\hat{x}}+y_{5}b\,\mathbf{\hat{y}}+z_{5}c\sin\beta\,\mathbf{\hat{z}}$ | (2a) | O V |
| $\mathbf{B}_{10}$ | = | $-x_{5}\,\mathbf{a}_{1}+\left(\frac{1}{2}+y_{5}\right)\,\mathbf{a}_{2}-z_{5}\,\mathbf{a}_{3}$ | = | $\left(-x_{5}a-z_{5}c\cos\beta\right)\,\mathbf{\hat{x}}+\left(\frac{1}{2}+y_{5}\right)b\,\mathbf{\hat{y}}-z_{5}c\sin\beta\,\mathbf{\hat{z}}$ | (2a) | O V |
| $\mathbf{B}_{11}$ | = | $x_{6}\,\mathbf{a}_{1}+y_{6}\,\mathbf{a}_{2}+z_{6}\,\mathbf{a}_{3}$ | = | $\left(x_{6}a+z_{6}c\cos\beta\right)\,\mathbf{\hat{x}}+y_{6}b\,\mathbf{\hat{y}}+z_{6}c\sin\beta\,\mathbf{\hat{z}}$ | (2a) | O VI |
| $\mathbf{B}_{12}$ | = | $-x_{6}\,\mathbf{a}_{1}+\left(\frac{1}{2}+y_{6}\right)\,\mathbf{a}_{2}-z_{6}\,\mathbf{a}_{3}$ | = | $\left(-x_{6}a-z_{6}c\cos\beta\right)\,\mathbf{\hat{x}}+\left(\frac{1}{2}+y_{6}\right)b\,\mathbf{\hat{y}}-z_{6}c\sin\beta\,\mathbf{\hat{z}}$ | (2a) | O VI |
| $\mathbf{B}_{13}$ | = | $x_{7}\,\mathbf{a}_{1}+y_{7}\,\mathbf{a}_{2}+z_{7}\,\mathbf{a}_{3}$ | = | $\left(x_{7}a+z_{7}c\cos\beta\right)\,\mathbf{\hat{x}}+y_{7}b\,\mathbf{\hat{y}}+z_{7}c\sin\beta\,\mathbf{\hat{z}}$ | (2a) | O VII |
| $\mathbf{B}_{14}$ | = | $-x_{7}\,\mathbf{a}_{1}+\left(\frac{1}{2}+y_{7}\right)\,\mathbf{a}_{2}-z_{7}\,\mathbf{a}_{3}$ | = | $\left(-x_{7}a-z_{7}c\cos\beta\right)\,\mathbf{\hat{x}}+\left(\frac{1}{2}+y_{7}\right)b\,\mathbf{\hat{y}}-z_{7}c\sin\beta\,\mathbf{\hat{z}}$ | (2a) | O VII |
| $\mathbf{B}_{15}$ | = | $x_{8}\,\mathbf{a}_{1}+y_{8}\,\mathbf{a}_{2}+z_{8}\,\mathbf{a}_{3}$ | = | $\left(x_{8}a+z_{8}c\cos\beta\right)\,\mathbf{\hat{x}}+y_{8}b\,\mathbf{\hat{y}}+z_{8}c\sin\beta\,\mathbf{\hat{z}}$ | (2a) | O VIII |
| $\mathbf{B}_{16}$ | = | $-x_{8}\,\mathbf{a}_{1}+\left(\frac{1}{2}+y_{8}\right)\,\mathbf{a}_{2}-z_{8}\,\mathbf{a}_{3}$ | = | $\left(-x_{8}a-z_{8}c\cos\beta\right)\,\mathbf{\hat{x}}+\left(\frac{1}{2}+y_{8}\right)b\,\mathbf{\hat{y}}-z_{8}c\sin\beta\,\mathbf{\hat{z}}$ | (2a) | O VIII |
| $\mathbf{B}_{17}$ | = | $x_{9}\,\mathbf{a}_{1}+y_{9}\,\mathbf{a}_{2}+z_{9}\,\mathbf{a}_{3}$ | = | $\left(x_{9}a+z_{9}c\cos\beta\right)\,\mathbf{\hat{x}}+y_{9}b\,\mathbf{\hat{y}}+z_{9}c\sin\beta\,\mathbf{\hat{z}}$ | (2a) | O IX |
| $\mathbf{B}_{18}$ | = | $-x_{9}\,\mathbf{a}_{1}+\left(\frac{1}{2}+y_{9}\right)\,\mathbf{a}_{2}-z_{9}\,\mathbf{a}_{3}$ | = | $\left(-x_{9}a-z_{9}c\cos\beta\right)\,\mathbf{\hat{x}}+\left(\frac{1}{2}+y_{9}\right)b\,\mathbf{\hat{y}}-z_{9}c\sin\beta\,\mathbf{\hat{z}}$ | (2a) | O IX |
| $\mathbf{B}_{19}$ | = | $x_{10}\,\mathbf{a}_{1}+y_{10}\,\mathbf{a}_{2}+z_{10}\,\mathbf{a}_{3}$ | = | $\left(x_{10}a+z_{10}c\cos\beta\right)\,\mathbf{\hat{x}}+y_{10}b\,\mathbf{\hat{y}}+z_{10}c\sin\beta\,\mathbf{\hat{z}}$ | (2a) | O X |
| $\mathbf{B}_{20}$ | = | $-x_{10}\,\mathbf{a}_{1}+\left(\frac{1}{2}+y_{10}\right)\,\mathbf{a}_{2}-z_{10}\,\mathbf{a}_{3}$ | = | $\left(-x_{10}a-z_{10}c\cos\beta\right)\,\mathbf{\hat{x}}+\left(\frac{1}{2}+y_{10}\right)b\,\mathbf{\hat{y}}-z_{10}c\sin\beta\,\mathbf{\hat{z}}$ | (2a) | O X |
| $\mathbf{B}_{21}$ | = | $x_{11}\,\mathbf{a}_{1}+y_{11}\,\mathbf{a}_{2}+z_{11}\,\mathbf{a}_{3}$ | = | $\left(x_{11}a+z_{11}c\cos\beta\right)\,\mathbf{\hat{x}}+y_{11}b\,\mathbf{\hat{y}}+z_{11}c\sin\beta\,\mathbf{\hat{z}}$ | (2a) | O XI |
| $\mathbf{B}_{22}$ | = | $-x_{11}\,\mathbf{a}_{1}+\left(\frac{1}{2}+y_{11}\right)\,\mathbf{a}_{2}-z_{11}\,\mathbf{a}_{3}$ | = | $\left(-x_{11}a-z_{11}c\cos\beta\right)\,\mathbf{\hat{x}}+\left(\frac{1}{2}+y_{11}\right)b\,\mathbf{\hat{y}}-z_{11}c\sin\beta\,\mathbf{\hat{z}}$ | (2a) | O XI |
| $\mathbf{B}_{23}$ | = | $x_{12}\,\mathbf{a}_{1}+y_{12}\,\mathbf{a}_{2}+z_{12}\,\mathbf{a}_{3}$ | = | $\left(x_{12}a+z_{12}c\cos\beta\right)\,\mathbf{\hat{x}}+y_{12}b\,\mathbf{\hat{y}}+z_{12}c\sin\beta\,\mathbf{\hat{z}}$ | (2a) | O XII |
| $\mathbf{B}_{24}$ | = | $-x_{12}\,\mathbf{a}_{1}+\left(\frac{1}{2}+y_{12}\right)\,\mathbf{a}_{2}-z_{12}\,\mathbf{a}_{3}$ | = | $\left(-x_{12}a-z_{12}c\cos\beta\right)\,\mathbf{\hat{x}}+\left(\frac{1}{2}+y_{12}\right)b\,\mathbf{\hat{y}}-z_{12}c\sin\beta\,\mathbf{\hat{z}}$ | (2a) | O XII |
| $\mathbf{B}_{25}$ | = | $x_{13}\,\mathbf{a}_{1}+y_{13}\,\mathbf{a}_{2}+z_{13}\,\mathbf{a}_{3}$ | = | $\left(x_{13}a+z_{13}c\cos\beta\right)\,\mathbf{\hat{x}}+y_{13}b\,\mathbf{\hat{y}}+z_{13}c\sin\beta\,\mathbf{\hat{z}}$ | (2a) | O XIII |
| $\mathbf{B}_{26}$ | = | $-x_{13}\,\mathbf{a}_{1}+\left(\frac{1}{2}+y_{13}\right)\,\mathbf{a}_{2}-z_{13}\,\mathbf{a}_{3}$ | = | $\left(-x_{13}a-z_{13}c\cos\beta\right)\,\mathbf{\hat{x}}+\left(\frac{1}{2}+y_{13}\right)b\,\mathbf{\hat{y}}-z_{13}c\sin\beta\,\mathbf{\hat{z}}$ | (2a) | O XIII |
| $\mathbf{B}_{27}$ | = | $x_{14}\,\mathbf{a}_{1}+y_{14}\,\mathbf{a}_{2}+z_{14}\,\mathbf{a}_{3}$ | = | $\left(x_{14}a+z_{14}c\cos\beta\right)\,\mathbf{\hat{x}}+y_{14}b\,\mathbf{\hat{y}}+z_{14}c\sin\beta\,\mathbf{\hat{z}}$ | (2a) | O XIV |
| $\mathbf{B}_{28}$ | = | $-x_{14}\,\mathbf{a}_{1}+\left(\frac{1}{2}+y_{14}\right)\,\mathbf{a}_{2}-z_{14}\,\mathbf{a}_{3}$ | = | $\left(-x_{14}a-z_{14}c\cos\beta\right)\,\mathbf{\hat{x}}+\left(\frac{1}{2}+y_{14}\right)b\,\mathbf{\hat{y}}-z_{14}c\sin\beta\,\mathbf{\hat{z}}$ | (2a) | O XIV |
| $\mathbf{B}_{29}$ | = | $x_{15}\,\mathbf{a}_{1}+y_{15}\,\mathbf{a}_{2}+z_{15}\,\mathbf{a}_{3}$ | = | $\left(x_{15}a+z_{15}c\cos\beta\right)\,\mathbf{\hat{x}}+y_{15}b\,\mathbf{\hat{y}}+z_{15}c\sin\beta\,\mathbf{\hat{z}}$ | (2a) | O XV |

$$\mathbf{B}_{30} = -x_{15}\,\mathbf{a}_1 + \left(\frac{1}{2} + y_{15}\right)\mathbf{a}_2 - z_{15}\,\mathbf{a}_3 = (-x_{15}a - z_{15}c\cos\beta)\,\hat{\mathbf{x}} + \left(\frac{1}{2} + y_{15}\right)b\,\hat{\mathbf{y}} - z_{15}c\sin\beta\,\hat{\mathbf{z}} \quad (2a) \quad \text{O XV}$$

$$\mathbf{B}_{31} = x_{16}\,\mathbf{a}_1 + y_{16}\,\mathbf{a}_2 + z_{16}\,\mathbf{a}_3 = (x_{16}a + z_{16}c\cos\beta)\,\hat{\mathbf{x}} + y_{16}b\,\hat{\mathbf{y}} + z_{16}c\sin\beta\,\hat{\mathbf{z}} \quad (2a) \quad \text{O XVI}$$

$$\mathbf{B}_{32} = -x_{16}\,\mathbf{a}_1 + \left(\frac{1}{2} + y_{16}\right)\mathbf{a}_2 - z_{16}\,\mathbf{a}_3 = (-x_{16}a - z_{16}c\cos\beta)\,\hat{\mathbf{x}} + \left(\frac{1}{2} + y_{16}\right)b\,\hat{\mathbf{y}} - z_{16}c\sin\beta\,\hat{\mathbf{z}} \quad (2a) \quad \text{O XVI}$$

$$\mathbf{B}_{33} = x_{17}\,\mathbf{a}_1 + y_{17}\,\mathbf{a}_2 + z_{17}\,\mathbf{a}_3 = (x_{17}a + z_{17}c\cos\beta)\,\hat{\mathbf{x}} + y_{17}b\,\hat{\mathbf{y}} + z_{17}c\sin\beta\,\hat{\mathbf{z}} \quad (2a) \quad \text{O XVII}$$

$$\mathbf{B}_{34} = -x_{17}\,\mathbf{a}_1 + \left(\frac{1}{2} + y_{17}\right)\mathbf{a}_2 - z_{17}\,\mathbf{a}_3 = (-x_{17}a - z_{17}c\cos\beta)\,\hat{\mathbf{x}} + \left(\frac{1}{2} + y_{17}\right)b\,\hat{\mathbf{y}} - z_{17}c\sin\beta\,\hat{\mathbf{z}} \quad (2a) \quad \text{O XVII}$$

$$\mathbf{B}_{35} = x_{18}\,\mathbf{a}_1 + y_{18}\,\mathbf{a}_2 + z_{18}\,\mathbf{a}_3 = (x_{18}a + z_{18}c\cos\beta)\,\hat{\mathbf{x}} + y_{18}b\,\hat{\mathbf{y}} + z_{18}c\sin\beta\,\hat{\mathbf{z}} \quad (2a) \quad \text{O XVIII}$$

$$\mathbf{B}_{36} = -x_{18}\,\mathbf{a}_1 + \left(\frac{1}{2} + y_{18}\right)\mathbf{a}_2 - z_{18}\,\mathbf{a}_3 = (-x_{18}a - z_{18}c\cos\beta)\,\hat{\mathbf{x}} + \left(\frac{1}{2} + y_{18}\right)b\,\hat{\mathbf{y}} - z_{18}c\sin\beta\,\hat{\mathbf{z}} \quad (2a) \quad \text{O XVIII}$$

$$\mathbf{B}_{37} = x_{19}\,\mathbf{a}_1 + y_{19}\,\mathbf{a}_2 + z_{19}\,\mathbf{a}_3 = (x_{19}a + z_{19}c\cos\beta)\,\hat{\mathbf{x}} + y_{19}b\,\hat{\mathbf{y}} + z_{19}c\sin\beta\,\hat{\mathbf{z}} \quad (2a) \quad \text{O XIX}$$

$$\mathbf{B}_{38} = -x_{19}\,\mathbf{a}_1 + \left(\frac{1}{2} + y_{19}\right)\mathbf{a}_2 - z_{19}\,\mathbf{a}_3 = (-x_{19}a - z_{19}c\cos\beta)\,\hat{\mathbf{x}} + \left(\frac{1}{2} + y_{19}\right)b\,\hat{\mathbf{y}} - z_{19}c\sin\beta\,\hat{\mathbf{z}} \quad (2a) \quad \text{O XIX}$$

$$\mathbf{B}_{39} = x_{20}\,\mathbf{a}_1 + y_{20}\,\mathbf{a}_2 + z_{20}\,\mathbf{a}_3 = (x_{20}a + z_{20}c\cos\beta)\,\hat{\mathbf{x}} + y_{20}b\,\hat{\mathbf{y}} + z_{20}c\sin\beta\,\hat{\mathbf{z}} \quad (2a) \quad \text{O XX}$$

$$\mathbf{B}_{40} = -x_{20}\,\mathbf{a}_1 + \left(\frac{1}{2} + y_{20}\right)\mathbf{a}_2 - z_{20}\,\mathbf{a}_3 = (-x_{20}a - z_{20}c\cos\beta)\,\hat{\mathbf{x}} + \left(\frac{1}{2} + y_{20}\right)b\,\hat{\mathbf{y}} - z_{20}c\sin\beta\,\hat{\mathbf{z}} \quad (2a) \quad \text{O XX}$$

$$\mathbf{B}_{41} = x_{21}\,\mathbf{a}_1 + y_{21}\,\mathbf{a}_2 + z_{21}\,\mathbf{a}_3 = (x_{21}a + z_{21}c\cos\beta)\,\hat{\mathbf{x}} + y_{21}b\,\hat{\mathbf{y}} + z_{21}c\sin\beta\,\hat{\mathbf{z}} \quad (2a) \quad \text{O XXI}$$

$$\mathbf{B}_{42} = -x_{21}\,\mathbf{a}_1 + \left(\frac{1}{2} + y_{21}\right)\mathbf{a}_2 - z_{21}\,\mathbf{a}_3 = (-x_{21}a - z_{21}c\cos\beta)\,\hat{\mathbf{x}} + \left(\frac{1}{2} + y_{21}\right)b\,\hat{\mathbf{y}} - z_{21}c\sin\beta\,\hat{\mathbf{z}} \quad (2a) \quad \text{O XXI}$$

$$\mathbf{B}_{43} = x_{22}\,\mathbf{a}_1 + y_{22}\,\mathbf{a}_2 + z_{22}\,\mathbf{a}_3 = (x_{22}a + z_{22}c\cos\beta)\,\hat{\mathbf{x}} + y_{22}b\,\hat{\mathbf{y}} + z_{22}c\sin\beta\,\hat{\mathbf{z}} \quad (2a) \quad \text{O XXII}$$

$$\mathbf{B}_{44} = -x_{22}\,\mathbf{a}_1 + \left(\frac{1}{2} + y_{22}\right)\mathbf{a}_2 - z_{22}\,\mathbf{a}_3 = (-x_{22}a - z_{22}c\cos\beta)\,\hat{\mathbf{x}} + \left(\frac{1}{2} + y_{22}\right)b\,\hat{\mathbf{y}} - z_{22}c\sin\beta\,\hat{\mathbf{z}} \quad (2a) \quad \text{O XXII}$$

$$\mathbf{B}_{45} = x_{23}\,\mathbf{a}_1 + y_{23}\,\mathbf{a}_2 + z_{23}\,\mathbf{a}_3 = (x_{23}a + z_{23}c\cos\beta)\,\hat{\mathbf{x}} + y_{23}b\,\hat{\mathbf{y}} + z_{23}c\sin\beta\,\hat{\mathbf{z}} \quad (2a) \quad \text{P I}$$

$$\mathbf{B}_{46} = -x_{23}\,\mathbf{a}_1 + \left(\frac{1}{2} + y_{23}\right)\mathbf{a}_2 - z_{23}\,\mathbf{a}_3 = (-x_{23}a - z_{23}c\cos\beta)\,\hat{\mathbf{x}} + \left(\frac{1}{2} + y_{23}\right)b\,\hat{\mathbf{y}} - z_{23}c\sin\beta\,\hat{\mathbf{z}} \quad (2a) \quad \text{P I}$$

$$\mathbf{B}_{47} = x_{24}\,\mathbf{a}_1 + y_{24}\,\mathbf{a}_2 + z_{24}\,\mathbf{a}_3 = (x_{24}a + z_{24}c\cos\beta)\,\hat{\mathbf{x}} + y_{24}b\,\hat{\mathbf{y}} + z_{24}c\sin\beta\,\hat{\mathbf{z}} \quad (2a) \quad \text{P II}$$

$$\mathbf{B}_{48} = -x_{24}\,\mathbf{a}_1 + \left(\frac{1}{2} + y_{24}\right)\mathbf{a}_2 - z_{24}\,\mathbf{a}_3 = (-x_{24}a - z_{24}c\cos\beta)\,\hat{\mathbf{x}} + \left(\frac{1}{2} + y_{24}\right)b\,\hat{\mathbf{y}} - z_{24}c\sin\beta\,\hat{\mathbf{z}} \quad (2a) \quad \text{P II}$$

$$\mathbf{B}_{49} = x_{25}\,\mathbf{a}_1 + y_{25}\,\mathbf{a}_2 + z_{25}\,\mathbf{a}_3 = (x_{25}a + z_{25}c\cos\beta)\,\hat{\mathbf{x}} + y_{25}b\,\hat{\mathbf{y}} + z_{25}c\sin\beta\,\hat{\mathbf{z}} \quad (2a) \quad \text{P III}$$

$$\mathbf{B}_{50} = -x_{25}\,\mathbf{a}_1 + \left(\frac{1}{2} + y_{25}\right)\mathbf{a}_2 - z_{25}\,\mathbf{a}_3 = (-x_{25}a - z_{25}c\cos\beta)\,\hat{\mathbf{x}} + \left(\frac{1}{2} + y_{25}\right)b\,\hat{\mathbf{y}} - z_{25}c\sin\beta\,\hat{\mathbf{z}} \quad (2a) \quad \text{P III}$$

$$\mathbf{B}_{51} = x_{26}\,\mathbf{a}_1 + y_{26}\,\mathbf{a}_2 + z_{26}\,\mathbf{a}_3 = (x_{26}a + z_{26}c\cos\beta)\,\hat{\mathbf{x}} + y_{26}b\,\hat{\mathbf{y}} + z_{26}c\sin\beta\,\hat{\mathbf{z}} \quad (2a) \quad \text{P IV}$$

$$\mathbf{B}_{52} = -x_{26}\,\mathbf{a}_1 + \left(\frac{1}{2} + y_{26}\right)\mathbf{a}_2 - z_{26}\,\mathbf{a}_3 = (-x_{26}a - z_{26}c\cos\beta)\,\hat{\mathbf{x}} + \left(\frac{1}{2} + y_{26}\right)b\,\hat{\mathbf{y}} - z_{26}c\sin\beta\,\hat{\mathbf{z}} \quad (2a) \quad \text{P IV}$$

| | | | | | | |
|---|---|---|---|---|---|---|
| $\mathbf{B}_{53}$ | = | $x_{27}\,\mathbf{a}_1 + y_{27}\,\mathbf{a}_2 + z_{27}\,\mathbf{a}_3$ | = | $(x_{27}a + z_{27}c\cos\beta)\,\hat{\mathbf{x}} + y_{27}b\,\hat{\mathbf{y}} + z_{27}c\sin\beta\,\hat{\mathbf{z}}$ | (2a) | W I |
| $\mathbf{B}_{54}$ | = | $-x_{27}\,\mathbf{a}_1 + \left(\frac{1}{2} + y_{27}\right)\,\mathbf{a}_2 - z_{27}\,\mathbf{a}_3$ | = | $(-x_{27}a - z_{27}c\cos\beta)\,\hat{\mathbf{x}} + \left(\frac{1}{2} + y_{27}\right)b\,\hat{\mathbf{y}} - z_{27}c\sin\beta\,\hat{\mathbf{z}}$ | (2a) | W I |
| $\mathbf{B}_{55}$ | = | $x_{28}\,\mathbf{a}_1 + y_{28}\,\mathbf{a}_2 + z_{28}\,\mathbf{a}_3$ | = | $(x_{28}a + z_{28}c\cos\beta)\,\hat{\mathbf{x}} + y_{28}b\,\hat{\mathbf{y}} + z_{28}c\sin\beta\,\hat{\mathbf{z}}$ | (2a) | W II |
| $\mathbf{B}_{56}$ | = | $-x_{28}\,\mathbf{a}_1 + \left(\frac{1}{2} + y_{28}\right)\,\mathbf{a}_2 - z_{28}\,\mathbf{a}_3$ | = | $(-x_{28}a - z_{28}c\cos\beta)\,\hat{\mathbf{x}} + \left(\frac{1}{2} + y_{28}\right)b\,\hat{\mathbf{y}} - z_{28}c\sin\beta\,\hat{\mathbf{z}}$ | (2a) | W II |
| $\mathbf{B}_{57}$ | = | $x_{29}\,\mathbf{a}_1 + y_{29}\,\mathbf{a}_2 + z_{29}\,\mathbf{a}_3$ | = | $(x_{29}a + z_{29}c\cos\beta)\,\hat{\mathbf{x}} + y_{29}b\,\hat{\mathbf{y}} + z_{29}c\sin\beta\,\hat{\mathbf{z}}$ | (2a) | W III |
| $\mathbf{B}_{58}$ | = | $-x_{29}\,\mathbf{a}_1 + \left(\frac{1}{2} + y_{29}\right)\,\mathbf{a}_2 - z_{29}\,\mathbf{a}_3$ | = | $(-x_{29}a - z_{29}c\cos\beta)\,\hat{\mathbf{x}} + \left(\frac{1}{2} + y_{29}\right)b\,\hat{\mathbf{y}} - z_{29}c\sin\beta\,\hat{\mathbf{z}}$ | (2a) | W III |
| $\mathbf{B}_{59}$ | = | $x_{30}\,\mathbf{a}_1 + y_{30}\,\mathbf{a}_2 + z_{30}\,\mathbf{a}_3$ | = | $(x_{30}a + z_{30}c\cos\beta)\,\hat{\mathbf{x}} + y_{30}b\,\hat{\mathbf{y}} + z_{30}c\sin\beta\,\hat{\mathbf{z}}$ | (2a) | W IV |
| $\mathbf{B}_{60}$ | = | $-x_{30}\,\mathbf{a}_1 + \left(\frac{1}{2} + y_{30}\right)\,\mathbf{a}_2 - z_{30}\,\mathbf{a}_3$ | = | $(-x_{30}a - z_{30}c\cos\beta)\,\hat{\mathbf{x}} + \left(\frac{1}{2} + y_{30}\right)b\,\hat{\mathbf{y}} - z_{30}c\sin\beta\,\hat{\mathbf{z}}$ | (2a) | W IV |

**Geometry files:**

- CIF: pp. 1508
- POSCAR: pp. 1508

# $Li_2SO_4 \cdot H_2O$ ($H4_8$) Structure: A2B2C5D_mP20_4_2a_2a_5a_a

http://aflow.org/prototype-encyclopedia/A2B2C5D_mP20_4_2a_2a_5a_a

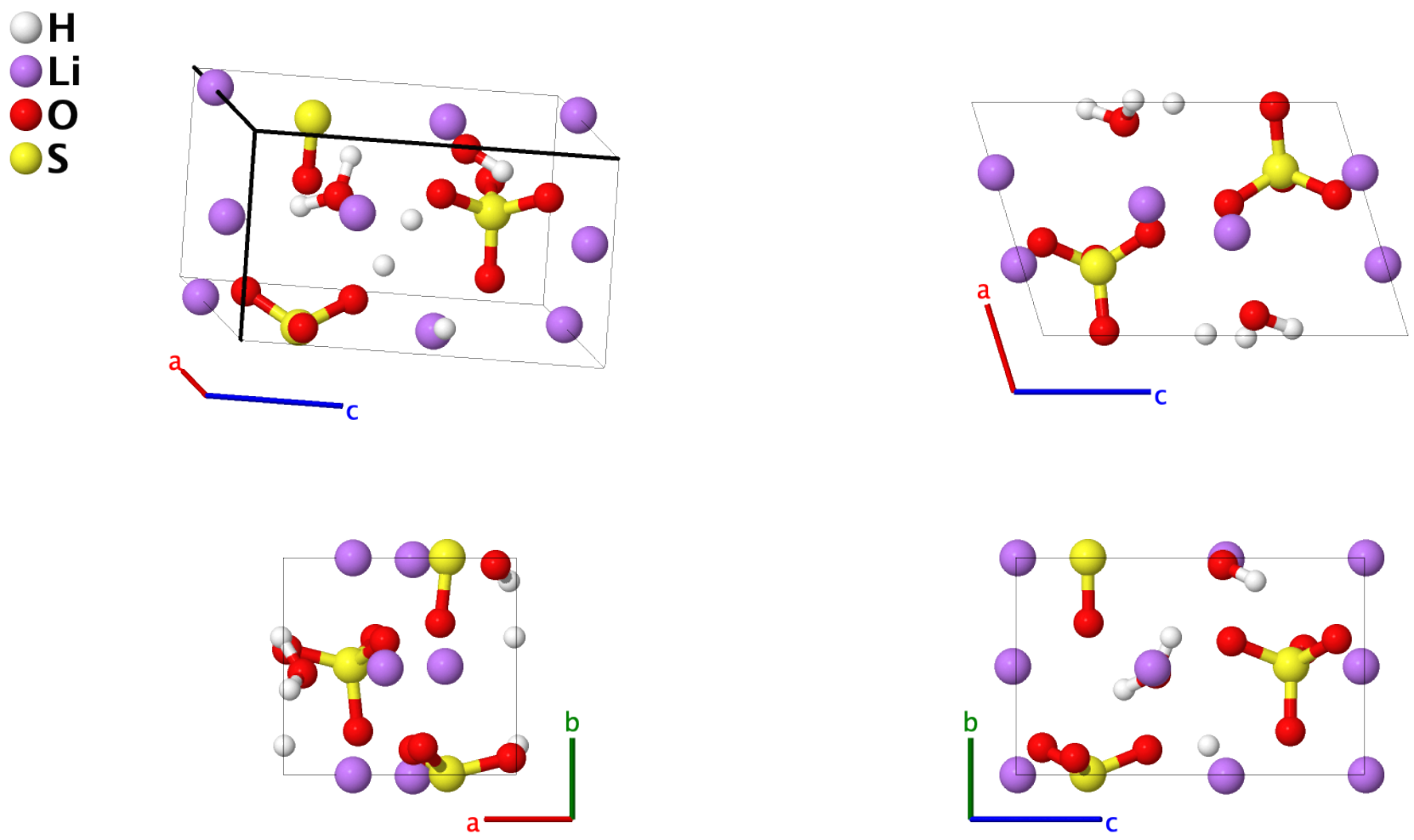

| | | |
|---|---|---|
| **Prototype** | : | $H_2Li_2O_5S$ |
| **AFLOW prototype label** | : | A2B2C5D_mP20_4_2a_2a_5a_a |
| ***Strukturbericht* designation** | : | $H4_8$ |
| **Pearson symbol** | : | mP20 |
| **Space group number** | : | 4 |
| **Space group symbol** | : | $P2_1$ |
| **AFLOW prototype command** | : | `aflow --proto=A2B2C5D_mP20_4_2a_2a_5a_a --params=`$a, b/a, c/a, \beta, x_1, y_1, z_1, x_2, y_2, z_2, x_3, y_3, z_3, x_4, y_4, z_4, x_5, y_5, z_5, x_6, y_6, z_6, x_7, y_7, z_7, x_8, y_8, z_8, x_9, y_9, z_9, x_{10}, y_{10}, z_{10}$ |

- We use the data from (Lundgren, 1984) at 20 K, including the positions of the hydrogen atoms not found in the original $H4_8$ structure in (Gottfried, 1937).
- Space group $P2_1$ #4 allows the $y$ coordinates to have an arbitrary origin. We follow (Lundgren, 1984) and set the $y$ coordinate of the sulfur atom, $y_{10}$, to zero.

**Simple Monoclinic primitive vectors:**

$$\mathbf{a}_1 = a\,\hat{\mathbf{x}}$$
$$\mathbf{a}_2 = b\,\hat{\mathbf{y}}$$
$$\mathbf{a}_3 = c\cos\beta\,\hat{\mathbf{x}} + c\sin\beta\,\hat{\mathbf{z}}$$

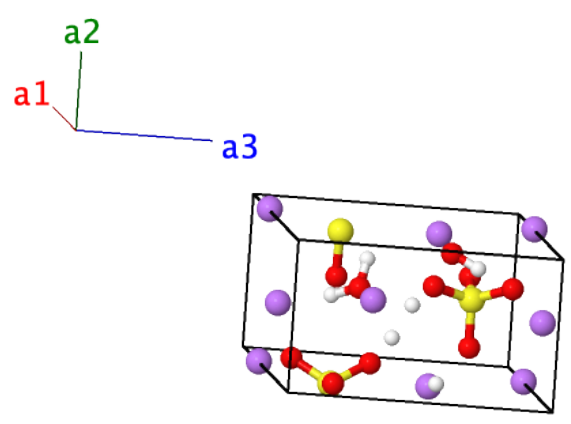

**Basis vectors:**

| | | Lattice Coordinates | | Cartesian Coordinates | Wyckoff Position | Atom Type |
|---|---|---|---|---|---|---|
| $\mathbf{B}_1$ | = | $x_1\,\mathbf{a}_1 + y_1\,\mathbf{a}_2 + z_1\,\mathbf{a}_3$ | = | $(x_1 a + z_1 c\cos\beta)\,\hat{\mathbf{x}} + y_1 b\,\hat{\mathbf{y}} + z_1 c\sin\beta\,\hat{\mathbf{z}}$ | (2a) | H I |
| $\mathbf{B}_2$ | = | $-x_1\,\mathbf{a}_1 + \left(\frac{1}{2}+y_1\right)\mathbf{a}_2 - z_1\,\mathbf{a}_3$ | = | $(-x_1a - z_1c\cos\beta)\,\hat{\mathbf{x}} + \left(\frac{1}{2}+y_1\right)b\,\hat{\mathbf{y}} - z_1c\sin\beta\,\hat{\mathbf{z}}$ | (2a) | H I |
| $\mathbf{B}_3$ | = | $x_2\,\mathbf{a}_1 + y_2\,\mathbf{a}_2 + z_2\,\mathbf{a}_3$ | = | $(x_2a + z_2c\cos\beta)\,\hat{\mathbf{x}} + y_2b\,\hat{\mathbf{y}} + z_2c\sin\beta\,\hat{\mathbf{z}}$ | (2a) | H II |
| $\mathbf{B}_4$ | = | $-x_2\,\mathbf{a}_1 + \left(\frac{1}{2}+y_2\right)\mathbf{a}_2 - z_2\,\mathbf{a}_3$ | = | $(-x_2a - z_2c\cos\beta)\,\hat{\mathbf{x}} + \left(\frac{1}{2}+y_2\right)b\,\hat{\mathbf{y}} - z_2c\sin\beta\,\hat{\mathbf{z}}$ | (2a) | H II |
| $\mathbf{B}_5$ | = | $x_3\,\mathbf{a}_1 + y_3\,\mathbf{a}_2 + z_3\,\mathbf{a}_3$ | = | $(x_3a + z_3c\cos\beta)\,\hat{\mathbf{x}} + y_3b\,\hat{\mathbf{y}} + z_3c\sin\beta\,\hat{\mathbf{z}}$ | (2a) | Li I |
| $\mathbf{B}_6$ | = | $-x_3\,\mathbf{a}_1 + \left(\frac{1}{2}+y_3\right)\mathbf{a}_2 - z_3\,\mathbf{a}_3$ | = | $(-x_3a - z_3c\cos\beta)\,\hat{\mathbf{x}} + \left(\frac{1}{2}+y_3\right)b\,\hat{\mathbf{y}} - z_3c\sin\beta\,\hat{\mathbf{z}}$ | (2a) | Li I |
| $\mathbf{B}_7$ | = | $x_4\,\mathbf{a}_1 + y_4\,\mathbf{a}_2 + z_4\,\mathbf{a}_3$ | = | $(x_4a + z_4c\cos\beta)\,\hat{\mathbf{x}} + y_4b\,\hat{\mathbf{y}} + z_4c\sin\beta\,\hat{\mathbf{z}}$ | (2a) | Li II |
| $\mathbf{B}_8$ | = | $-x_4\,\mathbf{a}_1 + \left(\frac{1}{2}+y_4\right)\mathbf{a}_2 - z_4\,\mathbf{a}_3$ | = | $(-x_4a - z_4c\cos\beta)\,\hat{\mathbf{x}} + \left(\frac{1}{2}+y_4\right)b\,\hat{\mathbf{y}} - z_4c\sin\beta\,\hat{\mathbf{z}}$ | (2a) | Li II |
| $\mathbf{B}_9$ | = | $x_5\,\mathbf{a}_1 + y_5\,\mathbf{a}_2 + z_5\,\mathbf{a}_3$ | = | $(x_5a + z_5c\cos\beta)\,\hat{\mathbf{x}} + y_5b\,\hat{\mathbf{y}} + z_5c\sin\beta\,\hat{\mathbf{z}}$ | (2a) | O I |
| $\mathbf{B}_{10}$ | = | $-x_5\,\mathbf{a}_1 + \left(\frac{1}{2}+y_5\right)\mathbf{a}_2 - z_5\,\mathbf{a}_3$ | = | $(-x_5a - z_5c\cos\beta)\,\hat{\mathbf{x}} + \left(\frac{1}{2}+y_5\right)b\,\hat{\mathbf{y}} - z_5c\sin\beta\,\hat{\mathbf{z}}$ | (2a) | O I |
| $\mathbf{B}_{11}$ | = | $x_6\,\mathbf{a}_1 + y_6\,\mathbf{a}_2 + z_6\,\mathbf{a}_3$ | = | $(x_6a + z_6c\cos\beta)\,\hat{\mathbf{x}} + y_6b\,\hat{\mathbf{y}} + z_6c\sin\beta\,\hat{\mathbf{z}}$ | (2a) | O II |
| $\mathbf{B}_{12}$ | = | $-x_6\,\mathbf{a}_1 + \left(\frac{1}{2}+y_6\right)\mathbf{a}_2 - z_6\,\mathbf{a}_3$ | = | $(-x_6a - z_6c\cos\beta)\,\hat{\mathbf{x}} + \left(\frac{1}{2}+y_6\right)b\,\hat{\mathbf{y}} - z_6c\sin\beta\,\hat{\mathbf{z}}$ | (2a) | O II |
| $\mathbf{B}_{13}$ | = | $x_7\,\mathbf{a}_1 + y_7\,\mathbf{a}_2 + z_7\,\mathbf{a}_3$ | = | $(x_7a + z_7c\cos\beta)\,\hat{\mathbf{x}} + y_7b\,\hat{\mathbf{y}} + z_7c\sin\beta\,\hat{\mathbf{z}}$ | (2a) | O III |
| $\mathbf{B}_{14}$ | = | $-x_7\,\mathbf{a}_1 + \left(\frac{1}{2}+y_7\right)\mathbf{a}_2 - z_7\,\mathbf{a}_3$ | = | $(-x_7a - z_7c\cos\beta)\,\hat{\mathbf{x}} + \left(\frac{1}{2}+y_7\right)b\,\hat{\mathbf{y}} - z_7c\sin\beta\,\hat{\mathbf{z}}$ | (2a) | O III |
| $\mathbf{B}_{15}$ | = | $x_8\,\mathbf{a}_1 + y_8\,\mathbf{a}_2 + z_8\,\mathbf{a}_3$ | = | $(x_8a + z_8c\cos\beta)\,\hat{\mathbf{x}} + y_8b\,\hat{\mathbf{y}} + z_8c\sin\beta\,\hat{\mathbf{z}}$ | (2a) | O IV |
| $\mathbf{B}_{16}$ | = | $-x_8\,\mathbf{a}_1 + \left(\frac{1}{2}+y_8\right)\mathbf{a}_2 - z_8\,\mathbf{a}_3$ | = | $(-x_8a - z_8c\cos\beta)\,\hat{\mathbf{x}} + \left(\frac{1}{2}+y_8\right)b\,\hat{\mathbf{y}} - z_8c\sin\beta\,\hat{\mathbf{z}}$ | (2a) | O IV |
| $\mathbf{B}_{17}$ | = | $x_9\,\mathbf{a}_1 + y_9\,\mathbf{a}_2 + z_9\,\mathbf{a}_3$ | = | $(x_9a + z_9c\cos\beta)\,\hat{\mathbf{x}} + y_9b\,\hat{\mathbf{y}} + z_9c\sin\beta\,\hat{\mathbf{z}}$ | (2a) | O V |
| $\mathbf{B}_{18}$ | = | $-x_9\,\mathbf{a}_1 + \left(\frac{1}{2}+y_9\right)\mathbf{a}_2 - z_9\,\mathbf{a}_3$ | = | $(-x_9a - z_9c\cos\beta)\,\hat{\mathbf{x}} + \left(\frac{1}{2}+y_9\right)b\,\hat{\mathbf{y}} - z_9c\sin\beta\,\hat{\mathbf{z}}$ | (2a) | O V |
| $\mathbf{B}_{19}$ | = | $x_{10}\,\mathbf{a}_1 + y_{10}\,\mathbf{a}_2 + z_{10}\,\mathbf{a}_3$ | = | $(x_{10}a + z_{10}c\cos\beta)\,\hat{\mathbf{x}} + y_{10}b\,\hat{\mathbf{y}} + z_{10}c\sin\beta\,\hat{\mathbf{z}}$ | (2a) | S |
| $\mathbf{B}_{20}$ | = | $-x_{10}\,\mathbf{a}_1 + \left(\frac{1}{2}+y_{10}\right)\mathbf{a}_2 - z_{10}\,\mathbf{a}_3$ | = | $(-x_{10}a - z_{10}c\cos\beta)\,\hat{\mathbf{x}} + \left(\frac{1}{2}+y_{10}\right)b\,\hat{\mathbf{y}} - z_{10}c\sin\beta\,\hat{\mathbf{z}}$ | (2a) | S |

---

---

**Geometry files:**

- CIF: pp. 1509
- POSCAR: pp. 1509

# $Ca_3UO_6$ Structure: A3B6C_mP20_4_3a_6a_a

http://aflow.org/prototype-encyclopedia/A3B6C_mP20_4_3a_6a_a

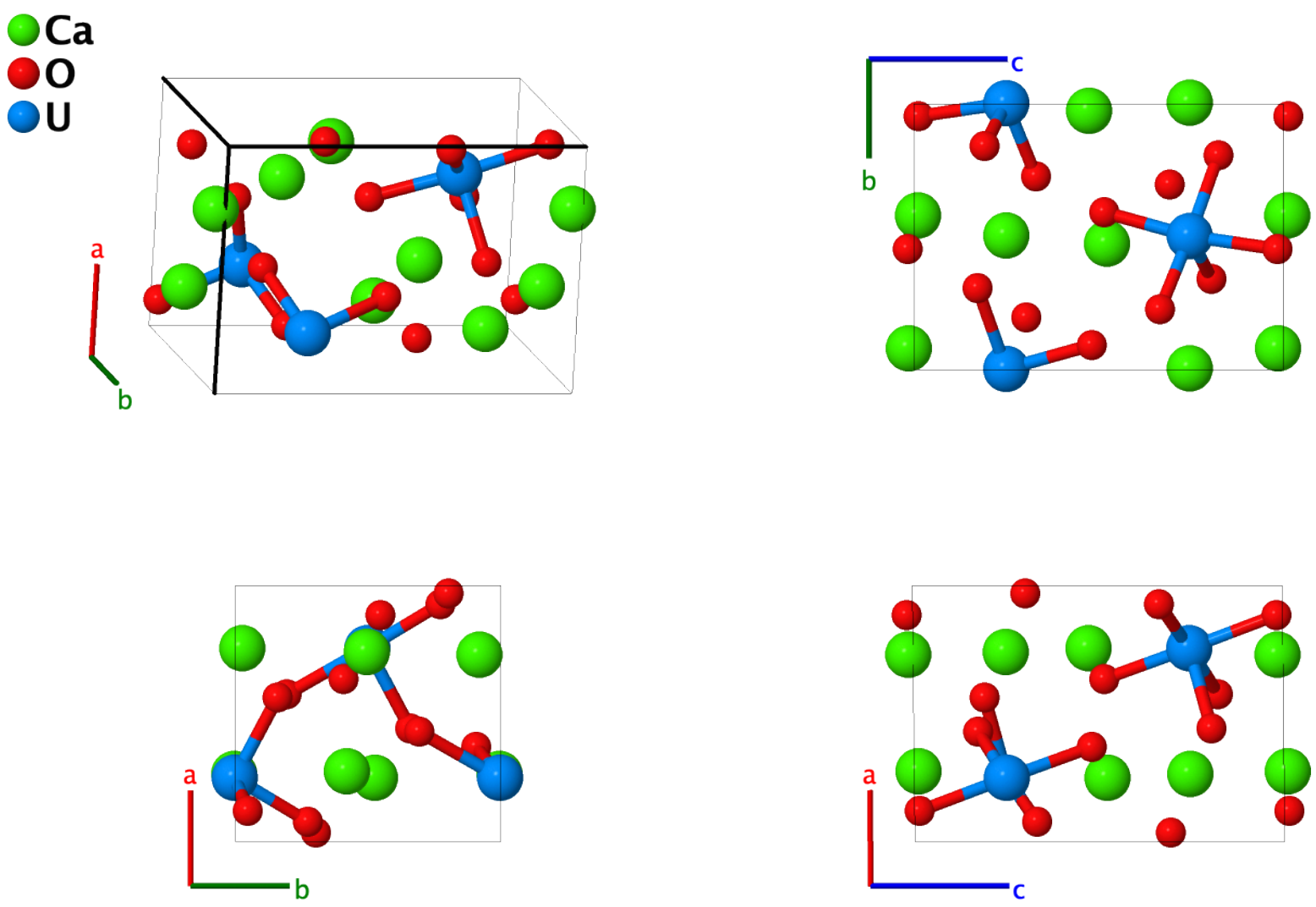

| | | |
|---|---|---|
| **Prototype** | : | $Ca_3O_6U$ |
| **AFLOW prototype label** | : | A3B6C_mP20_4_3a_6a_a |
| ***Strukturbericht* designation** | : | None |
| **Pearson symbol** | : | mP20 |
| **Space group number** | : | 4 |
| **Space group symbol** | : | $P2_1$ |
| **AFLOW prototype command** | : | `aflow --proto=A3B6C_mP20_4_3a_6a_a` `--params=`$a, b/a, c/a, \beta, x_1, y_1, z_1, x_2, y_2, z_2, x_3, y_3, z_3, x_4, y_4, z_4, x_5, y_5, z_5, x_6, y_6, z_6, x_7, y_7, z_7, x_8, y_8, z_8, x_9, y_9, z_9, x_{10}, y_{10}, z_{10}$ |

### Other compounds with this structure

- $Sr_3UO_6$

- Space group $P2_1$ #4 allows the $y$ coordinates to have an arbitrary origin. Here, this freedom is used to set $y_{10} = 0$ for the uranium position.

### Simple Monoclinic primitive vectors:

$$\mathbf{a}_1 = a\,\hat{\mathbf{x}}$$
$$\mathbf{a}_2 = b\,\hat{\mathbf{y}}$$
$$\mathbf{a}_3 = c\cos\beta\,\hat{\mathbf{x}} + c\sin\beta\,\hat{\mathbf{z}}$$

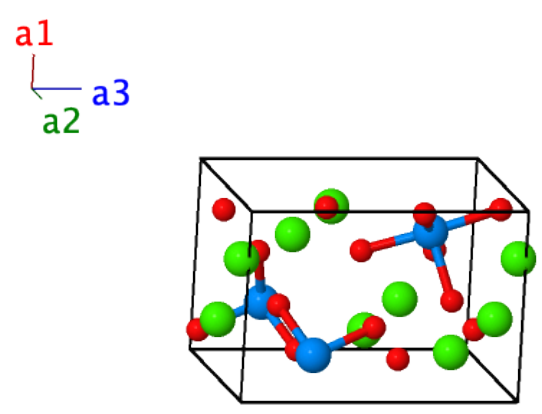

### Basis vectors:

| | | Lattice Coordinates | | Cartesian Coordinates | Wyckoff Position | Atom Type |
|---|---|---|---|---|---|---|
| $\mathbf{B}_1$ | = | $x_1\,\mathbf{a}_1 + y_1\,\mathbf{a}_2 + z_1\,\mathbf{a}_3$ | = | $(x_1 a + z_1 c\cos\beta)\,\hat{\mathbf{x}} + y_1 b\,\hat{\mathbf{y}} + z_1 c\sin\beta\,\hat{\mathbf{z}}$ | (2a) | Ca I |
| $\mathbf{B}_2$ | = | $-x_1\,\mathbf{a}_1 + \left(\frac{1}{2}+y_1\right)\mathbf{a}_2 - z_1\,\mathbf{a}_3$ | = | $(-x_1 a - z_1 c\cos\beta)\,\hat{\mathbf{x}} + \left(\frac{1}{2}+y_1\right)b\,\hat{\mathbf{y}} - z_1 c\sin\beta\,\hat{\mathbf{z}}$ | (2a) | Ca I |
| $\mathbf{B}_3$ | = | $x_2\,\mathbf{a}_1 + y_2\,\mathbf{a}_2 + z_2\,\mathbf{a}_3$ | = | $(x_2 a + z_2 c\cos\beta)\,\hat{\mathbf{x}} + y_2 b\,\hat{\mathbf{y}} + z_2 c\sin\beta\,\hat{\mathbf{z}}$ | (2a) | Ca II |
| $\mathbf{B}_4$ | = | $-x_2\,\mathbf{a}_1 + \left(\frac{1}{2}+y_2\right)\mathbf{a}_2 - z_2\,\mathbf{a}_3$ | = | $(-x_2 a - z_2 c\cos\beta)\,\hat{\mathbf{x}} + \left(\frac{1}{2}+y_2\right)b\,\hat{\mathbf{y}} - z_2 c\sin\beta\,\hat{\mathbf{z}}$ | (2a) | Ca II |
| $\mathbf{B}_5$ | = | $x_3\,\mathbf{a}_1 + y_3\,\mathbf{a}_2 + z_3\,\mathbf{a}_3$ | = | $(x_3 a + z_3 c\cos\beta)\,\hat{\mathbf{x}} + y_3 b\,\hat{\mathbf{y}} + z_3 c\sin\beta\,\hat{\mathbf{z}}$ | (2a) | Ca III |
| $\mathbf{B}_6$ | = | $-x_3\,\mathbf{a}_1 + \left(\frac{1}{2}+y_3\right)\mathbf{a}_2 - z_3\,\mathbf{a}_3$ | = | $(-x_3 a - z_3 c\cos\beta)\,\hat{\mathbf{x}} + \left(\frac{1}{2}+y_3\right)b\,\hat{\mathbf{y}} - z_3 c\sin\beta\,\hat{\mathbf{z}}$ | (2a) | Ca III |
| $\mathbf{B}_7$ | = | $x_4\,\mathbf{a}_1 + y_4\,\mathbf{a}_2 + z_4\,\mathbf{a}_3$ | = | $(x_4 a + z_4 c\cos\beta)\,\hat{\mathbf{x}} + y_4 b\,\hat{\mathbf{y}} + z_4 c\sin\beta\,\hat{\mathbf{z}}$ | (2a) | O I |
| $\mathbf{B}_8$ | = | $-x_4\,\mathbf{a}_1 + \left(\frac{1}{2}+y_4\right)\mathbf{a}_2 - z_4\,\mathbf{a}_3$ | = | $(-x_4 a - z_4 c\cos\beta)\,\hat{\mathbf{x}} + \left(\frac{1}{2}+y_4\right)b\,\hat{\mathbf{y}} - z_4 c\sin\beta\,\hat{\mathbf{z}}$ | (2a) | O I |
| $\mathbf{B}_9$ | = | $x_5\,\mathbf{a}_1 + y_5\,\mathbf{a}_2 + z_5\,\mathbf{a}_3$ | = | $(x_5 a + z_5 c\cos\beta)\,\hat{\mathbf{x}} + y_5 b\,\hat{\mathbf{y}} + z_5 c\sin\beta\,\hat{\mathbf{z}}$ | (2a) | O II |
| $\mathbf{B}_{10}$ | = | $-x_5\,\mathbf{a}_1 + \left(\frac{1}{2}+y_5\right)\mathbf{a}_2 - z_5\,\mathbf{a}_3$ | = | $(-x_5 a - z_5 c\cos\beta)\,\hat{\mathbf{x}} + \left(\frac{1}{2}+y_5\right)b\,\hat{\mathbf{y}} - z_5 c\sin\beta\,\hat{\mathbf{z}}$ | (2a) | O II |
| $\mathbf{B}_{11}$ | = | $x_6\,\mathbf{a}_1 + y_6\,\mathbf{a}_2 + z_6\,\mathbf{a}_3$ | = | $(x_6 a + z_6 c\cos\beta)\,\hat{\mathbf{x}} + y_6 b\,\hat{\mathbf{y}} + z_6 c\sin\beta\,\hat{\mathbf{z}}$ | (2a) | O III |
| $\mathbf{B}_{12}$ | = | $-x_6\,\mathbf{a}_1 + \left(\frac{1}{2}+y_6\right)\mathbf{a}_2 - z_6\,\mathbf{a}_3$ | = | $(-x_6 a - z_6 c\cos\beta)\,\hat{\mathbf{x}} + \left(\frac{1}{2}+y_6\right)b\,\hat{\mathbf{y}} - z_6 c\sin\beta\,\hat{\mathbf{z}}$ | (2a) | O III |
| $\mathbf{B}_{13}$ | = | $x_7\,\mathbf{a}_1 + y_7\,\mathbf{a}_2 + z_7\,\mathbf{a}_3$ | = | $(x_7 a + z_7 c\cos\beta)\,\hat{\mathbf{x}} + y_7 b\,\hat{\mathbf{y}} + z_7 c\sin\beta\,\hat{\mathbf{z}}$ | (2a) | O IV |
| $\mathbf{B}_{14}$ | = | $-x_7\,\mathbf{a}_1 + \left(\frac{1}{2}+y_7\right)\mathbf{a}_2 - z_7\,\mathbf{a}_3$ | = | $(-x_7 a - z_7 c\cos\beta)\,\hat{\mathbf{x}} + \left(\frac{1}{2}+y_7\right)b\,\hat{\mathbf{y}} - z_7 c\sin\beta\,\hat{\mathbf{z}}$ | (2a) | O IV |
| $\mathbf{B}_{15}$ | = | $x_8\,\mathbf{a}_1 + y_8\,\mathbf{a}_2 + z_8\,\mathbf{a}_3$ | = | $(x_8 a + z_8 c\cos\beta)\,\hat{\mathbf{x}} + y_8 b\,\hat{\mathbf{y}} + z_8 c\sin\beta\,\hat{\mathbf{z}}$ | (2a) | O V |
| $\mathbf{B}_{16}$ | = | $-x_8\,\mathbf{a}_1 + \left(\frac{1}{2}+y_8\right)\mathbf{a}_2 - z_8\,\mathbf{a}_3$ | = | $(-x_8 a - z_8 c\cos\beta)\,\hat{\mathbf{x}} + \left(\frac{1}{2}+y_8\right)b\,\hat{\mathbf{y}} - z_8 c\sin\beta\,\hat{\mathbf{z}}$ | (2a) | O V |
| $\mathbf{B}_{17}$ | = | $x_9\,\mathbf{a}_1 + y_9\,\mathbf{a}_2 + z_9\,\mathbf{a}_3$ | = | $(x_9 a + z_9 c\cos\beta)\,\hat{\mathbf{x}} + y_9 b\,\hat{\mathbf{y}} + z_9 c\sin\beta\,\hat{\mathbf{z}}$ | (2a) | O VI |
| $\mathbf{B}_{18}$ | = | $-x_9\,\mathbf{a}_1 + \left(\frac{1}{2}+y_9\right)\mathbf{a}_2 - z_9\,\mathbf{a}_3$ | = | $(-x_9 a - z_9 c\cos\beta)\,\hat{\mathbf{x}} + \left(\frac{1}{2}+y_9\right)b\,\hat{\mathbf{y}} - z_9 c\sin\beta\,\hat{\mathbf{z}}$ | (2a) | O VI |
| $\mathbf{B}_{19}$ | = | $x_{10}\,\mathbf{a}_1 + y_{10}\,\mathbf{a}_2 + z_{10}\,\mathbf{a}_3$ | = | $(x_{10} a + z_{10} c\cos\beta)\,\hat{\mathbf{x}} + y_{10} b\,\hat{\mathbf{y}} + z_{10} c\sin\beta\,\hat{\mathbf{z}}$ | (2a) | U |
| $\mathbf{B}_{20}$ | = | $-x_{10}\,\mathbf{a}_1 + \left(\frac{1}{2}+y_{10}\right)\mathbf{a}_2 - z_{10}\,\mathbf{a}_3$ | = | $(-x_{10} a - z_{10} c\cos\beta)\,\hat{\mathbf{x}} + \left(\frac{1}{2}+y_{10}\right)b\,\hat{\mathbf{y}} - z_{10} c\sin\beta\,\hat{\mathbf{z}}$ | (2a) | U |

**Geometry files:**

- CIF: pp. 1509
- POSCAR: pp. 1510

# Bassanite [$CaSO_4(H_2O)_{0.5}$, $H4_7$] Structure: A2B2C9D2_mC90_5_ab2c_3c_b13c_3c

http://aflow.org/prototype-encyclopedia/A2B2C9D2_mC90_5_ab2c_3c_b13c_3c

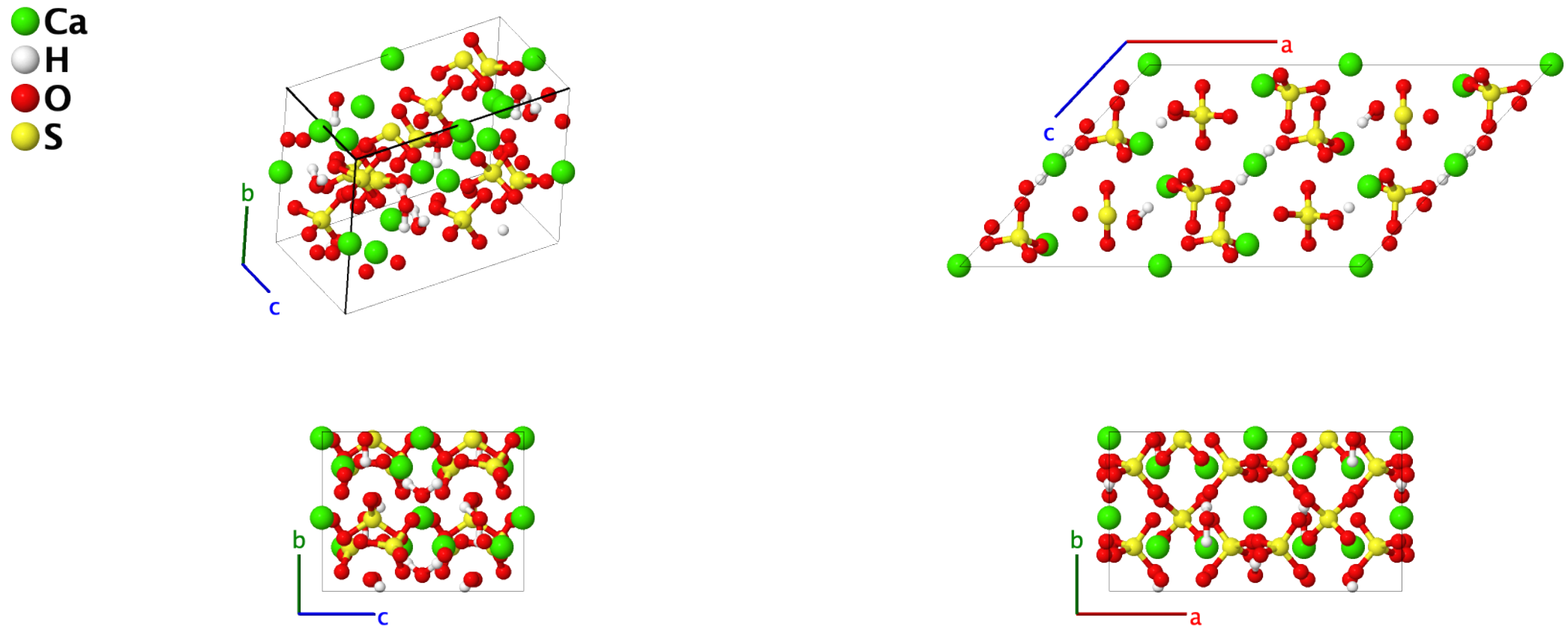

| | | |
|---|---|---|
| **Prototype** | : | $Ca_2H_2O_9S_2$ |
| **AFLOW prototype label** | : | A2B2C9D2_mC90_5_ab2c_3c_b13c_3c |
| ***Strukturbericht* designation** | : | $H4_7$ |
| **Pearson symbol** | : | mC90 |
| **Space group number** | : | 5 |
| **Space group symbol** | : | $C2$ |
| **AFLOW prototype command** | : | `aflow --proto=A2B2C9D2_mC90_5_ab2c_3c_b13c_3c` `--params=`$a, b/a, c/a, \beta, y_1, y_2, y_3, x_4, y_4, z_4, x_5, y_5, z_5, x_6, y_6, z_6, x_7, y_7, z_7, x_8, y_8, z_8, x_9, y_9, z_9, x_{10}, y_{10}, z_{10}, x_{11}, y_{11}, z_{11}, x_{12}, y_{12}, z_{12}, x_{13}, y_{13}, z_{13}, x_{14}, y_{14}, z_{14}, x_{15}, y_{15}, z_{15}, x_{16}, y_{16}, z_{16}, x_{17}, y_{17}, z_{17}, x_{18}, y_{18}, z_{18}, x_{19}, y_{19}, z_{19}, x_{20}, y_{20}, z_{20}, x_{21}, y_{21}, z_{21}, x_{22}, y_{22}, z_{22}, x_{23}, y_{23}, z_{23}, x_{24}, y_{24}, z_{24}$ |

- (Gottfried, 1937) gave this the *Strukturbericht* designation $H4_7$. They listed the system as monoclinic, with space group $C2$ #5, but noted that it was pseudo-hexagonal and gave the coordinates for the all of the atoms except the water molecules in terms of the trigonal space group $P3_121$ #152. (Abriel, 1993) found a complete determination of the structure, in space group $I2$ #5, which we have converted to the standard $C2$ setting. The exact structure of this system seems to depend on the actual amount of water and the preparation mechanism (Singh, 2007).
- The $P3_121$ structure can be obtained from these positions by removing all of the water molecules (the hydrogen atoms plus O-I and O-XIV), and allowing for a small uncertainty in the positions of the remaining atoms.

**Base-centered Monoclinic primitive vectors:**

$$\mathbf{a}_1 = \frac{1}{2}\,a\,\hat{\mathbf{x}} - \frac{1}{2}\,b\,\hat{\mathbf{y}}$$
$$\mathbf{a}_2 = \frac{1}{2}\,a\,\hat{\mathbf{x}} + \frac{1}{2}\,b\,\hat{\mathbf{y}}$$
$$\mathbf{a}_3 = c\cos\beta\,\hat{\mathbf{x}} + c\sin\beta\,\hat{\mathbf{z}}$$

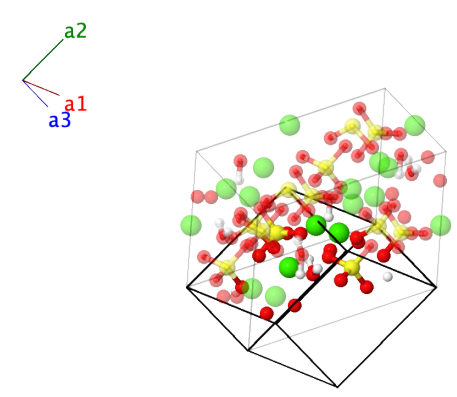

**Basis vectors:**

| | | Lattice Coordinates | | Cartesian Coordinates | Wyckoff Position | Atom Type |
|---|---|---|---|---|---|---|
| $\mathbf{B}_1$ | = | $-y_1\,\mathbf{a}_1 + y_1\,\mathbf{a}_2$ | = | $y_1\,b\,\hat{\mathbf{y}}$ | (2a) | Ca I |
| $\mathbf{B}_2$ | = | $-y_2\,\mathbf{a}_1 + y_2\,\mathbf{a}_2 + \frac{1}{2}\,\mathbf{a}_3$ | = | $\frac{1}{2}c\cos\beta\,\hat{\mathbf{x}} + y_2 b\,\hat{\mathbf{y}} + \frac{1}{2}c\sin\beta\,\hat{\mathbf{z}}$ | $(2b)$ | Ca II |
| $\mathbf{B}_3$ | = | $-y_3\,\mathbf{a}_1 + y_3\,\mathbf{a}_2 + \frac{1}{2}\,\mathbf{a}_3$ | = | $\frac{1}{2}c\cos\beta\,\hat{\mathbf{x}} + y_3 b\,\hat{\mathbf{y}} + \frac{1}{2}c\sin\beta\,\hat{\mathbf{z}}$ | $(2b)$ | O I |
| $\mathbf{B}_4$ | = | $(x_4 - y_4)\,\mathbf{a}_1 + (x_4 + y_4)\,\mathbf{a}_2 + z_4\,\mathbf{a}_3$ | = | $(x_4 a + z_4 c\cos\beta)\,\hat{\mathbf{x}} + y_4 b\,\hat{\mathbf{y}} + z_4 c\sin\beta\,\hat{\mathbf{z}}$ | $(4c)$ | Ca III |
| $\mathbf{B}_5$ | = | $(-x_4 - y_4)\,\mathbf{a}_1 + (-x_4 + y_4)\,\mathbf{a}_2 - z_4\,\mathbf{a}_3$ | = | $(-x_4 a - z_4 c\cos\beta)\,\hat{\mathbf{x}} + y_4 b\,\hat{\mathbf{y}} - z_4 c\sin\beta\,\hat{\mathbf{z}}$ | $(4c)$ | Ca III |
| $\mathbf{B}_6$ | = | $(x_5 - y_5)\,\mathbf{a}_1 + (x_5 + y_5)\,\mathbf{a}_2 + z_5\,\mathbf{a}_3$ | = | $(x_5 a + z_5 c\cos\beta)\,\hat{\mathbf{x}} + y_5 b\,\hat{\mathbf{y}} + z_5 c\sin\beta\,\hat{\mathbf{z}}$ | $(4c)$ | Ca IV |
| $\mathbf{B}_7$ | = | $(-x_5 - y_5)\,\mathbf{a}_1 + (-x_5 + y_5)\,\mathbf{a}_2 - z_5\,\mathbf{a}_3$ | = | $(-x_5 a - z_5 c\cos\beta)\,\hat{\mathbf{x}} + y_5 b\,\hat{\mathbf{y}} - z_5 c\sin\beta\,\hat{\mathbf{z}}$ | $(4c)$ | Ca IV |
| $\mathbf{B}_8$ | = | $(x_6 - y_6)\,\mathbf{a}_1 + (x_6 + y_6)\,\mathbf{a}_2 + z_6\,\mathbf{a}_3$ | = | $(x_6 a + z_6 c\cos\beta)\,\hat{\mathbf{x}} + y_6 b\,\hat{\mathbf{y}} + z_6 c\sin\beta\,\hat{\mathbf{z}}$ | $(4c)$ | H I |
| $\mathbf{B}_9$ | = | $(-x_6 - y_6)\,\mathbf{a}_1 + (-x_6 + y_6)\,\mathbf{a}_2 - z_6\,\mathbf{a}_3$ | = | $(-x_6 a - z_6 c\cos\beta)\,\hat{\mathbf{x}} + y_6 b\,\hat{\mathbf{y}} - z_6 c\sin\beta\,\hat{\mathbf{z}}$ | $(4c)$ | H I |
| $\mathbf{B}_{10}$ | = | $(x_7 - y_7)\,\mathbf{a}_1 + (x_7 + y_7)\,\mathbf{a}_2 + z_7\,\mathbf{a}_3$ | = | $(x_7 a + z_7 c\cos\beta)\,\hat{\mathbf{x}} + y_7 b\,\hat{\mathbf{y}} + z_7 c\sin\beta\,\hat{\mathbf{z}}$ | $(4c)$ | H II |
| $\mathbf{B}_{11}$ | = | $(-x_7 - y_7)\,\mathbf{a}_1 + (-x_7 + y_7)\,\mathbf{a}_2 - z_7\,\mathbf{a}_3$ | = | $(-x_7 a - z_7 c\cos\beta)\,\hat{\mathbf{x}} + y_7 b\,\hat{\mathbf{y}} - z_7 c\sin\beta\,\hat{\mathbf{z}}$ | $(4c)$ | H II |
| $\mathbf{B}_{12}$ | = | $(x_8 - y_8)\,\mathbf{a}_1 + (x_8 + y_8)\,\mathbf{a}_2 + z_8\,\mathbf{a}_3$ | = | $(x_8 a + z_8 c\cos\beta)\,\hat{\mathbf{x}} + y_8 b\,\hat{\mathbf{y}} + z_8 c\sin\beta\,\hat{\mathbf{z}}$ | $(4c)$ | H III |
| $\mathbf{B}_{13}$ | = | $(-x_8 - y_8)\,\mathbf{a}_1 + (-x_8 + y_8)\,\mathbf{a}_2 - z_8\,\mathbf{a}_3$ | = | $(-x_8 a - z_8 c\cos\beta)\,\hat{\mathbf{x}} + y_8 b\,\hat{\mathbf{y}} - z_8 c\sin\beta\,\hat{\mathbf{z}}$ | $(4c)$ | H III |
| $\mathbf{B}_{14}$ | = | $(x_9 - y_9)\,\mathbf{a}_1 + (x_9 + y_9)\,\mathbf{a}_2 + z_9\,\mathbf{a}_3$ | = | $(x_9 a + z_9 c\cos\beta)\,\hat{\mathbf{x}} + y_9 b\,\hat{\mathbf{y}} + z_9 c\sin\beta\,\hat{\mathbf{z}}$ | $(4c)$ | O II |
| $\mathbf{B}_{15}$ | = | $(-x_9 - y_9)\,\mathbf{a}_1 + (-x_9 + y_9)\,\mathbf{a}_2 - z_9\,\mathbf{a}_3$ | = | $(-x_9 a - z_9 c\cos\beta)\,\hat{\mathbf{x}} + y_9 b\,\hat{\mathbf{y}} - z_9 c\sin\beta\,\hat{\mathbf{z}}$ | $(4c)$ | O II |
| $\mathbf{B}_{16}$ | = | $(x_{10} - y_{10})\,\mathbf{a}_1 + (x_{10} + y_{10})\,\mathbf{a}_2 + z_{10}\,\mathbf{a}_3$ | = | $(x_{10} a + z_{10} c\cos\beta)\,\hat{\mathbf{x}} + y_{10} b\,\hat{\mathbf{y}} + z_{10} c\sin\beta\,\hat{\mathbf{z}}$ | $(4c)$ | O III |
| $\mathbf{B}_{17}$ | = | $(-x_{10} - y_{10})\,\mathbf{a}_1 + (-x_{10} + y_{10})\,\mathbf{a}_2 - z_{10}\,\mathbf{a}_3$ | = | $(-x_{10} a - z_{10} c\cos\beta)\,\hat{\mathbf{x}} + y_{10} b\,\hat{\mathbf{y}} - z_{10} c\sin\beta\,\hat{\mathbf{z}}$ | $(4c)$ | O III |
| $\mathbf{B}_{18}$ | = | $(x_{11} - y_{11})\,\mathbf{a}_1 + (x_{11} + y_{11})\,\mathbf{a}_2 + z_{11}\,\mathbf{a}_3$ | = | $(x_{11} a + z_{11} c\cos\beta)\,\hat{\mathbf{x}} + y_{11} b\,\hat{\mathbf{y}} + z_{11} c\sin\beta\,\hat{\mathbf{z}}$ | $(4c)$ | O IV |
| $\mathbf{B}_{19}$ | = | $(-x_{11} - y_{11})\,\mathbf{a}_1 + (-x_{11} + y_{11})\,\mathbf{a}_2 - z_{11}\,\mathbf{a}_3$ | = | $(-x_{11} a - z_{11} c\cos\beta)\,\hat{\mathbf{x}} + y_{11} b\,\hat{\mathbf{y}} - z_{11} c\sin\beta\,\hat{\mathbf{z}}$ | $(4c)$ | O IV |
| $\mathbf{B}_{20}$ | = | $(x_{12} - y_{12})\,\mathbf{a}_1 + (x_{12} + y_{12})\,\mathbf{a}_2 + z_{12}\,\mathbf{a}_3$ | = | $(x_{12} a + z_{12} c\cos\beta)\,\hat{\mathbf{x}} + y_{12} b\,\hat{\mathbf{y}} + z_{12} c\sin\beta\,\hat{\mathbf{z}}$ | $(4c)$ | O V |
| $\mathbf{B}_{21}$ | = | $(-x_{12} - y_{12})\,\mathbf{a}_1 + (-x_{12} + y_{12})\,\mathbf{a}_2 - z_{12}\,\mathbf{a}_3$ | = | $(-x_{12} a - z_{12} c\cos\beta)\,\hat{\mathbf{x}} + y_{12} b\,\hat{\mathbf{y}} - z_{12} c\sin\beta\,\hat{\mathbf{z}}$ | $(4c)$ | O V |
| $\mathbf{B}_{22}$ | = | $(x_{13} - y_{13})\,\mathbf{a}_1 + (x_{13} + y_{13})\,\mathbf{a}_2 + z_{13}\,\mathbf{a}_3$ | = | $(x_{13} a + z_{13} c\cos\beta)\,\hat{\mathbf{x}} + y_{13} b\,\hat{\mathbf{y}} + z_{13} c\sin\beta\,\hat{\mathbf{z}}$ | $(4c)$ | O VI |
| $\mathbf{B}_{23}$ | = | $(-x_{13} - y_{13})\,\mathbf{a}_1 + (-x_{13} + y_{13})\,\mathbf{a}_2 - z_{13}\,\mathbf{a}_3$ | = | $(-x_{13} a - z_{13} c\cos\beta)\,\hat{\mathbf{x}} + y_{13} b\,\hat{\mathbf{y}} - z_{13} c\sin\beta\,\hat{\mathbf{z}}$ | $(4c)$ | O VI |
| $\mathbf{B}_{24}$ | = | $(x_{14} - y_{14})\,\mathbf{a}_1 + (x_{14} + y_{14})\,\mathbf{a}_2 + z_{14}\,\mathbf{a}_3$ | = | $(x_{14} a + z_{14} c\cos\beta)\,\hat{\mathbf{x}} + y_{14} b\,\hat{\mathbf{y}} + z_{14} c\sin\beta\,\hat{\mathbf{z}}$ | $(4c)$ | O VII |
| $\mathbf{B}_{25}$ | = | $(-x_{14} - y_{14})\,\mathbf{a}_1 + (-x_{14} + y_{14})\,\mathbf{a}_2 - z_{14}\,\mathbf{a}_3$ | = | $(-x_{14} a - z_{14} c\cos\beta)\,\hat{\mathbf{x}} + y_{14} b\,\hat{\mathbf{y}} - z_{14} c\sin\beta\,\hat{\mathbf{z}}$ | $(4c)$ | O VII |
| $\mathbf{B}_{26}$ | = | $(x_{15} - y_{15})\,\mathbf{a}_1 + (x_{15} + y_{15})\,\mathbf{a}_2 + z_{15}\,\mathbf{a}_3$ | = | $(x_{15} a + z_{15} c\cos\beta)\,\hat{\mathbf{x}} + y_{15} b\,\hat{\mathbf{y}} + z_{15} c\sin\beta\,\hat{\mathbf{z}}$ | $(4c)$ | O VIII |
| $\mathbf{B}_{27}$ | = | $(-x_{15}-y_{15})\,\mathbf{a}_1 + (-x_{15}+y_{15})\,\mathbf{a}_2 - z_{15}\,\mathbf{a}_3$ | = | $(-x_{15}a - z_{15}c\cos\beta)\,\hat{\mathbf{x}} + y_{15}b\,\hat{\mathbf{y}} - z_{15}c\sin\beta\,\hat{\mathbf{z}}$ | (4c) | O VIII |
| $\mathbf{B}_{28}$ | = | $(x_{16}-y_{16})\,\mathbf{a}_1 + (x_{16}+y_{16})\,\mathbf{a}_2 + z_{16}\,\mathbf{a}_3$ | = | $(x_{16}a + z_{16}c\cos\beta)\,\hat{\mathbf{x}} + y_{16}b\,\hat{\mathbf{y}} + z_{16}c\sin\beta\,\hat{\mathbf{z}}$ | (4c) | O IX |
| $\mathbf{B}_{29}$ | = | $(-x_{16}-y_{16})\,\mathbf{a}_1 + (-x_{16}+y_{16})\,\mathbf{a}_2 - z_{16}\,\mathbf{a}_3$ | = | $(-x_{16}a - z_{16}c\cos\beta)\,\hat{\mathbf{x}} + y_{16}b\,\hat{\mathbf{y}} - z_{16}c\sin\beta\,\hat{\mathbf{z}}$ | (4c) | O IX |
| $\mathbf{B}_{30}$ | = | $(x_{17}-y_{17})\,\mathbf{a}_1 + (x_{17}+y_{17})\,\mathbf{a}_2 + z_{17}\,\mathbf{a}_3$ | = | $(x_{17}a + z_{17}c\cos\beta)\,\hat{\mathbf{x}} + y_{17}b\,\hat{\mathbf{y}} + z_{17}c\sin\beta\,\hat{\mathbf{z}}$ | (4c) | O X |
| $\mathbf{B}_{31}$ | = | $(-x_{17}-y_{17})\,\mathbf{a}_1 + (-x_{17}+y_{17})\,\mathbf{a}_2 - z_{17}\,\mathbf{a}_3$ | = | $(-x_{17}a - z_{17}c\cos\beta)\,\hat{\mathbf{x}} + y_{17}b\,\hat{\mathbf{y}} - z_{17}c\sin\beta\,\hat{\mathbf{z}}$ | (4c) | O X |
| $\mathbf{B}_{32}$ | = | $(x_{18}-y_{18})\,\mathbf{a}_1 + (x_{18}+y_{18})\,\mathbf{a}_2 + z_{18}\,\mathbf{a}_3$ | = | $(x_{18}a + z_{18}c\cos\beta)\,\hat{\mathbf{x}} + y_{18}b\,\hat{\mathbf{y}} + z_{18}c\sin\beta\,\hat{\mathbf{z}}$ | (4c) | O XI |
| $\mathbf{B}_{33}$ | = | $(-x_{18}-y_{18})\,\mathbf{a}_1 + (-x_{18}+y_{18})\,\mathbf{a}_2 - z_{18}\,\mathbf{a}_3$ | = | $(-x_{18}a - z_{18}c\cos\beta)\,\hat{\mathbf{x}} + y_{18}b\,\hat{\mathbf{y}} - z_{18}c\sin\beta\,\hat{\mathbf{z}}$ | (4c) | O XI |
| $\mathbf{B}_{34}$ | = | $(x_{19}-y_{19})\,\mathbf{a}_1 + (x_{19}+y_{19})\,\mathbf{a}_2 + z_{19}\,\mathbf{a}_3$ | = | $(x_{19}a + z_{19}c\cos\beta)\,\hat{\mathbf{x}} + y_{19}b\,\hat{\mathbf{y}} + z_{19}c\sin\beta\,\hat{\mathbf{z}}$ | (4c) | O XII |
| $\mathbf{B}_{35}$ | = | $(-x_{19}-y_{19})\,\mathbf{a}_1 + (-x_{19}+y_{19})\,\mathbf{a}_2 - z_{19}\,\mathbf{a}_3$ | = | $(-x_{19}a - z_{19}c\cos\beta)\,\hat{\mathbf{x}} + y_{19}b\,\hat{\mathbf{y}} - z_{19}c\sin\beta\,\hat{\mathbf{z}}$ | (4c) | O XII |
| $\mathbf{B}_{36}$ | = | $(x_{20}-y_{20})\,\mathbf{a}_1 + (x_{20}+y_{20})\,\mathbf{a}_2 + z_{20}\,\mathbf{a}_3$ | = | $(x_{20}a + z_{20}c\cos\beta)\,\hat{\mathbf{x}} + y_{20}b\,\hat{\mathbf{y}} + z_{20}c\sin\beta\,\hat{\mathbf{z}}$ | (4c) | O XIII |
| $\mathbf{B}_{37}$ | = | $(-x_{20}-y_{20})\,\mathbf{a}_1 + (-x_{20}+y_{20})\,\mathbf{a}_2 - z_{20}\,\mathbf{a}_3$ | = | $(-x_{20}a - z_{20}c\cos\beta)\,\hat{\mathbf{x}} + y_{20}b\,\hat{\mathbf{y}} - z_{20}c\sin\beta\,\hat{\mathbf{z}}$ | (4c) | O XIII |
| $\mathbf{B}_{38}$ | = | $(x_{21}-y_{21})\,\mathbf{a}_1 + (x_{21}+y_{21})\,\mathbf{a}_2 + z_{21}\,\mathbf{a}_3$ | = | $(x_{21}a + z_{21}c\cos\beta)\,\hat{\mathbf{x}} + y_{21}b\,\hat{\mathbf{y}} + z_{21}c\sin\beta\,\hat{\mathbf{z}}$ | (4c) | O XIV |
| $\mathbf{B}_{39}$ | = | $(-x_{21}-y_{21})\,\mathbf{a}_1 + (-x_{21}+y_{21})\,\mathbf{a}_2 - z_{21}\,\mathbf{a}_3$ | = | $(-x_{21}a - z_{21}c\cos\beta)\,\hat{\mathbf{x}} + y_{21}b\,\hat{\mathbf{y}} - z_{21}c\sin\beta\,\hat{\mathbf{z}}$ | (4c) | O XIV |
| $\mathbf{B}_{40}$ | = | $(x_{22}-y_{22})\,\mathbf{a}_1 + (x_{22}+y_{22})\,\mathbf{a}_2 + z_{22}\,\mathbf{a}_3$ | = | $(x_{22}a + z_{22}c\cos\beta)\,\hat{\mathbf{x}} + y_{22}b\,\hat{\mathbf{y}} + z_{22}c\sin\beta\,\hat{\mathbf{z}}$ | (4c) | S I |
| $\mathbf{B}_{41}$ | = | $(-x_{22}-y_{22})\,\mathbf{a}_1 + (-x_{22}+y_{22})\,\mathbf{a}_2 - z_{22}\,\mathbf{a}_3$ | = | $(-x_{22}a - z_{22}c\cos\beta)\,\hat{\mathbf{x}} + y_{22}b\,\hat{\mathbf{y}} - z_{22}c\sin\beta\,\hat{\mathbf{z}}$ | (4c) | S I |
| $\mathbf{B}_{42}$ | = | $(x_{23}-y_{23})\,\mathbf{a}_1 + (x_{23}+y_{23})\,\mathbf{a}_2 + z_{23}\,\mathbf{a}_3$ | = | $(x_{23}a + z_{23}c\cos\beta)\,\hat{\mathbf{x}} + y_{23}b\,\hat{\mathbf{y}} + z_{23}c\sin\beta\,\hat{\mathbf{z}}$ | (4c) | S II |
| $\mathbf{B}_{43}$ | = | $(-x_{23}-y_{23})\,\mathbf{a}_1 + (-x_{23}+y_{23})\,\mathbf{a}_2 - z_{23}\,\mathbf{a}_3$ | = | $(-x_{23}a - z_{23}c\cos\beta)\,\hat{\mathbf{x}} + y_{23}b\,\hat{\mathbf{y}} - z_{23}c\sin\beta\,\hat{\mathbf{z}}$ | (4c) | S II |
| $\mathbf{B}_{44}$ | = | $(x_{24}-y_{24})\,\mathbf{a}_1 + (x_{24}+y_{24})\,\mathbf{a}_2 + z_{24}\,\mathbf{a}_3$ | = | $(x_{24}a + z_{24}c\cos\beta)\,\hat{\mathbf{x}} + y_{24}b\,\hat{\mathbf{y}} + z_{24}c\sin\beta\,\hat{\mathbf{z}}$ | (4c) | S III |
| $\mathbf{B}_{45}$ | = | $(-x_{24}-y_{24})\,\mathbf{a}_1 + (-x_{24}+y_{24})\,\mathbf{a}_2 - z_{24}\,\mathbf{a}_3$ | = | $(-x_{24}a - z_{24}c\cos\beta)\,\hat{\mathbf{x}} + y_{24}b\,\hat{\mathbf{y}} - z_{24}c\sin\beta\,\hat{\mathbf{z}}$ | (4c) | S III |

**Geometry files:**

- CIF: pp. 1510
- POSCAR: pp. 1510

# $NbAs_2$ Structure: A2B_mC12_5_2c_c

http://aflow.org/prototype-encyclopedia/A2B_mC12_5_2c_c

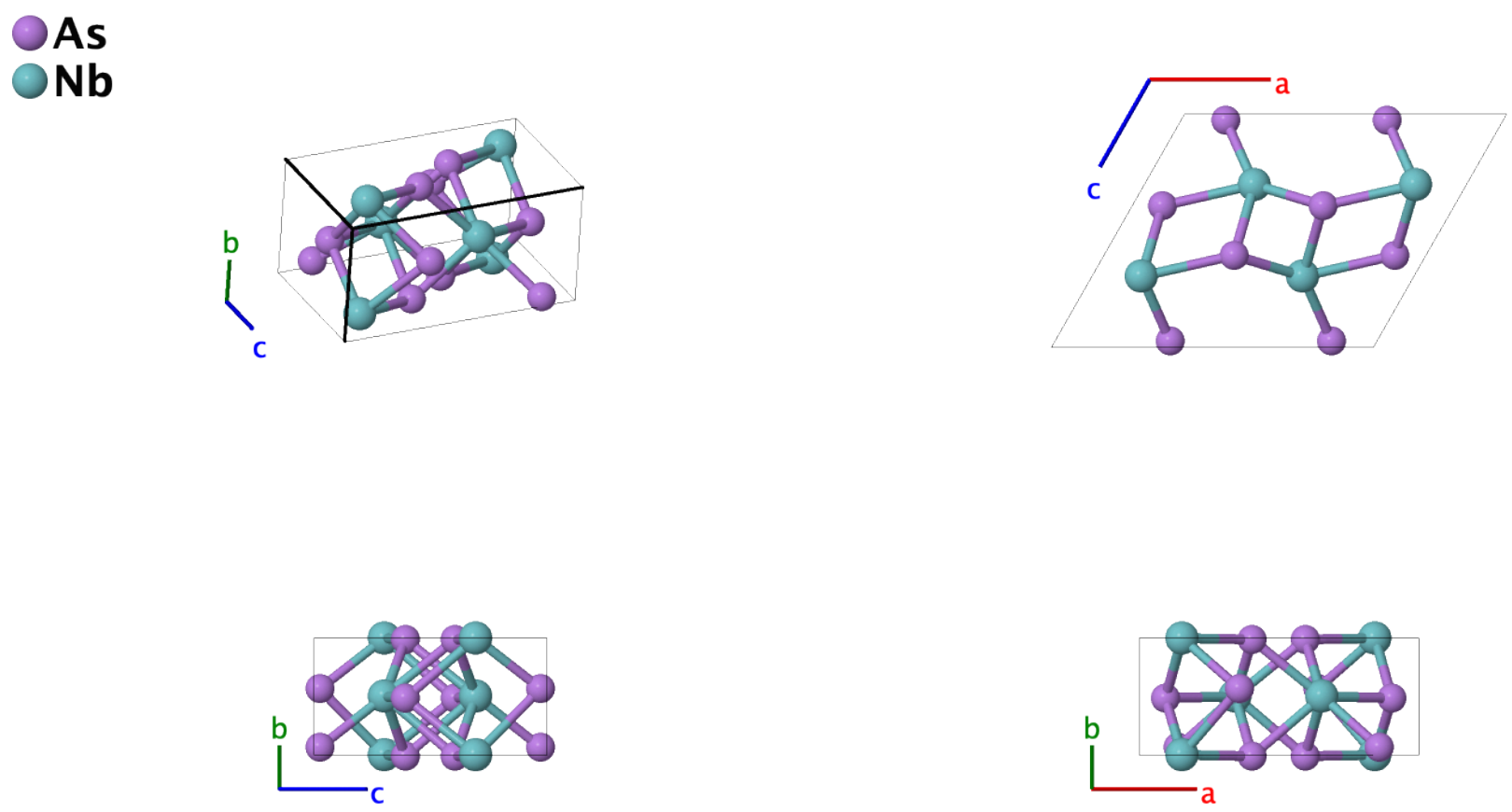

| | | |
|---|---|---|
| **Prototype** | : | $As_2Nb$ |
| **AFLOW prototype label** | : | A2B_mC12_5_2c_c |
| ***Strukturbericht* designation** | : | None |
| **Pearson symbol** | : | mC12 |
| **Space group number** | : | 5 |
| **Space group symbol** | : | $C2$ |
| **AFLOW prototype command** | : | `aflow --proto=A2B_mC12_5_2c_c`<br>`--params=`$a, b/a, c/a, \beta, x_1, y_1, z_1, x_2, y_2, z_2, x_3, y_3, z_3$ |

### Other compounds with this structure

- $MoAs_2$, $TaAs_2$, $NbSb_2$, and $TaSb_2$

---

- The zero of the $y$-axis can be chosen arbitrarily in space group $C2$ #5. Here we chose it so that $y_3 = 1/2$ for the niobium atom.

### Base-centered Monoclinic primitive vectors:

$$\mathbf{a}_1 = \frac{1}{2}\,a\,\hat{\mathbf{x}} - \frac{1}{2}\,b\,\hat{\mathbf{y}}$$
$$\mathbf{a}_2 = \frac{1}{2}\,a\,\hat{\mathbf{x}} + \frac{1}{2}\,b\,\hat{\mathbf{y}}$$
$$\mathbf{a}_3 = c\cos\beta\,\hat{\mathbf{x}} + c\sin\beta\,\hat{\mathbf{z}}$$

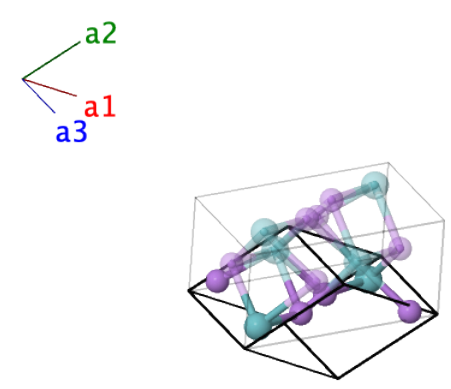

### Basis vectors:

| | | Lattice Coordinates | | Cartesian Coordinates | Wyckoff Position | Atom Type |
|---|---|---|---|---|---|---|
| $\mathbf{B}_1$ | = | $(x_1 - y_1)\,\mathbf{a}_1 + (x_1 + y_1)\,\mathbf{a}_2 + z_1\,\mathbf{a}_3$ | = | $(x_1 a + z_1 c\cos\beta)\,\hat{\mathbf{x}} + y_1 b\,\hat{\mathbf{y}} + z_1 c\sin\beta\,\hat{\mathbf{z}}$ | (4$c$) | As I |
| $\mathbf{B}_2$ | = | $(-x_1 - y_1)\,\mathbf{a}_1 + (-x_1 + y_1)\,\mathbf{a}_2 - z_1\,\mathbf{a}_3$ | = | $(-x_1 a - z_1 c\cos\beta)\,\hat{\mathbf{x}} + y_1 b\,\hat{\mathbf{y}} - z_1 c\sin\beta\,\hat{\mathbf{z}}$ | (4$c$) | As I |
| $\mathbf{B}_3$ | = | $(x_2 - y_2)\,\mathbf{a}_1 + (x_2 + y_2)\,\mathbf{a}_2 + z_2\,\mathbf{a}_3$ | = | $(x_2 a + z_2 c\cos\beta)\,\hat{\mathbf{x}} + y_2 b\,\hat{\mathbf{y}} + z_2 c\sin\beta\,\hat{\mathbf{z}}$ | (4$c$) | As II |
| $\mathbf{B}_4$ | = | $(-x_2 - y_2)\,\mathbf{a}_1 + (-x_2 + y_2)\,\mathbf{a}_2 - z_2\,\mathbf{a}_3$ | = | $(-x_2 a - z_2 c\cos\beta)\,\hat{\mathbf{x}} + y_2 b\,\hat{\mathbf{y}} - z_2 c\sin\beta\,\hat{\mathbf{z}}$ | (4$c$) | As II |
| $\mathbf{B}_5$ | = | $(x_3 - y_3)\,\mathbf{a}_1 + (x_3 + y_3)\,\mathbf{a}_2 + z_3\,\mathbf{a}_3$ | = | $(x_3 a + z_3 c\cos\beta)\,\hat{\mathbf{x}} + y_3 b\,\hat{\mathbf{y}} + z_3 c\sin\beta\,\hat{\mathbf{z}}$ | (4$c$) | Nb |
| $\mathbf{B}_6$ | = | $(-x_3 - y_3)\,\mathbf{a}_1 + (-x_3 + y_3)\,\mathbf{a}_2 - z_3\,\mathbf{a}_3$ | = | $(-x_3 a - z_3 c\cos\beta)\,\hat{\mathbf{x}} + y_3 b\,\hat{\mathbf{y}} - z_3 c\sin\beta\,\hat{\mathbf{z}}$ | (4$c$) | Nb |

**Geometry files:**

- CIF: pp. 1511
- POSCAR: pp. 1511

# $D0_{15}$ ($AlCl_3$) (*obsolete*) Structure: AB3_mC16_5_c_3c

http://aflow.org/prototype-encyclopedia/AB3_mC16_5_c_3c

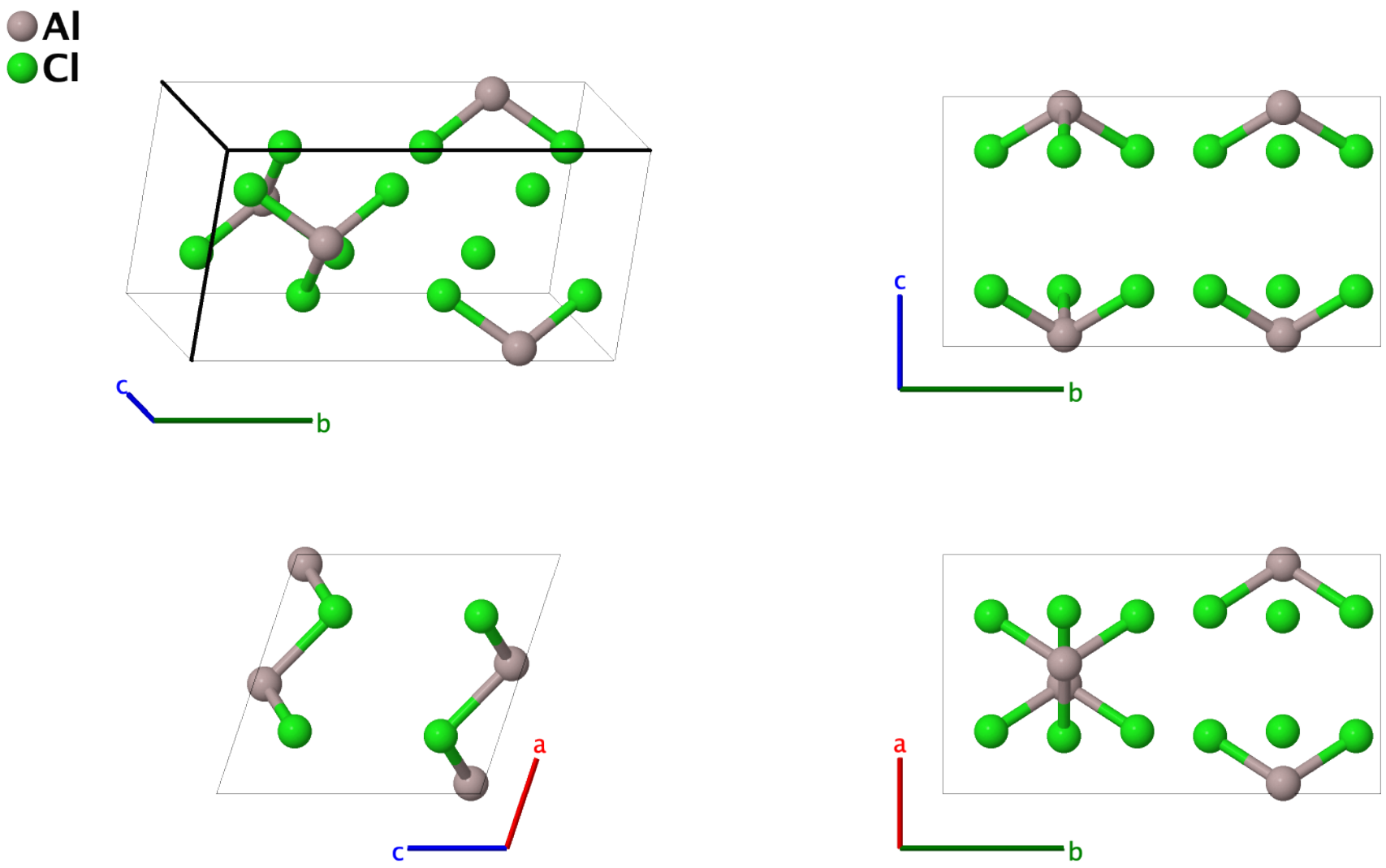

| | | |
|---|---|---|
| **Prototype** | : | $AlCl_3$ |
| **AFLOW prototype label** | : | AB3_mC16_5_c_3c |
| ***Strukturbericht* designation** | : | $D0_{15}$ |
| **Pearson symbol** | : | mC16 |
| **Space group number** | : | 5 |
| **Space group symbol** | : | $C2$ |
| **AFLOW prototype command** | : | `aflow --proto=AB3_mC16_5_c_3c`<br>`--params=`$a, b/a, c/a, \beta, x_1, y_1, z_1, x_2, y_2, z_2, x_3, y_3, z_3, x_4, y_4, z_4$ |

- This structure was suggested by (Ketelaar, 1935) and given the *Strukturbericht* designation $D0_{15}$ by (Gottfried, 1395) with no reference to the $D0_{13}$ structure of (Laschkarew, 1930) found in (Hermann, 1937).
- Ketelaar gave a description of the unit cell a hexagonal setting, but the atomic positions resolve into the body-centered monoclinic cell shown here. This structure, however, has a lower symmetry than the currently accepted structure, which is body-centered monoclinic, space group $C2/m$ #12. We had previously designated that structure $D0_{15}$, but the current structure was given that designation by (Hermann, 1937).

**Base-centered Monoclinic primitive vectors:**

$$\mathbf{a}_1 = \tfrac{1}{2}\,a\,\hat{\mathbf{x}} - \tfrac{1}{2}\,b\,\hat{\mathbf{y}}$$
$$\mathbf{a}_2 = \tfrac{1}{2}\,a\,\hat{\mathbf{x}} + \tfrac{1}{2}\,b\,\hat{\mathbf{y}}$$
$$\mathbf{a}_3 = c\cos\beta\,\hat{\mathbf{x}} + c\sin\beta\,\hat{\mathbf{z}}$$

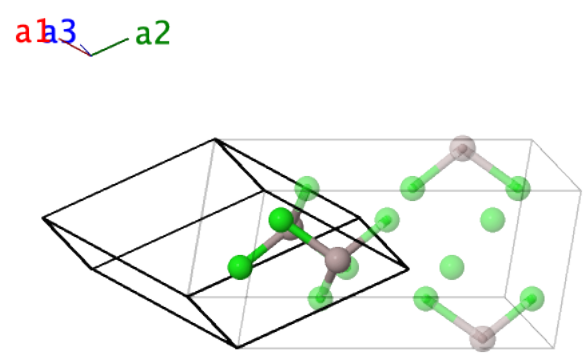

**Basis vectors:**

| | | Lattice Coordinates | | Cartesian Coordinates | Wyckoff Position | Atom Type |
|---|---|---|---|---|---|---|
| $\mathbf{B}_1$ | = | $(x_1 - y_1)\,\mathbf{a}_1 + (x_1 + y_1)\,\mathbf{a}_2 + z_1\,\mathbf{a}_3$ | = | $(x_1 a + z_1 c\cos\beta)\,\hat{\mathbf{x}} + y_1 b\,\hat{\mathbf{y}} + z_1 c\sin\beta\,\hat{\mathbf{z}}$ | (4$c$) | Al |
| $\mathbf{B}_2$ | = | $(-x_1 - y_1)\,\mathbf{a}_1 + (-x_1 + y_1)\,\mathbf{a}_2 - z_1\,\mathbf{a}_3$ | = | $(-x_1 a - z_1 c\cos\beta)\,\hat{\mathbf{x}} + y_1 b\,\hat{\mathbf{y}} - z_1 c\sin\beta\,\hat{\mathbf{z}}$ | (4$c$) | Al |
| $\mathbf{B}_3$ | = | $(x_2 - y_2)\,\mathbf{a}_1 + (x_2 + y_2)\,\mathbf{a}_2 + z_2\,\mathbf{a}_3$ | = | $(x_2 a + z_2 c\cos\beta)\,\hat{\mathbf{x}} + y_2 b\,\hat{\mathbf{y}} + z_2 c\sin\beta\,\hat{\mathbf{z}}$ | (4$c$) | Cl I |
| $\mathbf{B}_4$ | = | $(-x_2 - y_2)\,\mathbf{a}_1 + (-x_2 + y_2)\,\mathbf{a}_2 - z_2\,\mathbf{a}_3$ | = | $(-x_2 a - z_2 c\cos\beta)\,\hat{\mathbf{x}} + y_2 b\,\hat{\mathbf{y}} - z_2 c\sin\beta\,\hat{\mathbf{z}}$ | (4$c$) | Cl I |
| $\mathbf{B}_5$ | = | $(x_3 - y_3)\,\mathbf{a}_1 + (x_3 + y_3)\,\mathbf{a}_2 + z_3\,\mathbf{a}_3$ | = | $(x_3 a + z_3 c\cos\beta)\,\hat{\mathbf{x}} + y_3 b\,\hat{\mathbf{y}} + z_3 c\sin\beta\,\hat{\mathbf{z}}$ | (4$c$) | Cl II |
| $\mathbf{B}_6$ | = | $(-x_3 - y_3)\,\mathbf{a}_1 + (-x_3 + y_3)\,\mathbf{a}_2 - z_3\,\mathbf{a}_3$ | = | $(-x_3 a - z_3 c\cos\beta)\,\hat{\mathbf{x}} + y_3 b\,\hat{\mathbf{y}} - z_3 c\sin\beta\,\hat{\mathbf{z}}$ | (4$c$) | Cl II |
| $\mathbf{B}_7$ | = | $(x_4 - y_4)\,\mathbf{a}_1 + (x_4 + y_4)\,\mathbf{a}_2 + z_4\,\mathbf{a}_3$ | = | $(x_4 a + z_4 c\cos\beta)\,\hat{\mathbf{x}} + y_4 b\,\hat{\mathbf{y}} + z_4 c\sin\beta\,\hat{\mathbf{z}}$ | (4$c$) | Cl III |
| $\mathbf{B}_8$ | = | $(-x_4 - y_4)\,\mathbf{a}_1 + (-x_4 + y_4)\,\mathbf{a}_2 - z_4\,\mathbf{a}_3$ | = | $(-x_4 a - z_4 c\cos\beta)\,\hat{\mathbf{x}} + y_4 b\,\hat{\mathbf{y}} - z_4 c\sin\beta\,\hat{\mathbf{z}}$ | (4$c$) | Cl III |

**Geometry files:**

- CIF: pp. 1511
- POSCAR: pp. 1511

# $Rb_2CaCu_6(PO_4)_4O_2$ Structure: AB6C18D4E2_mC62_5_a_2b2c_9c_2c_c

http://aflow.org/prototype-encyclopedia/AB6C18D4E2_mC62_5_a_2b2c_9c_2c_c

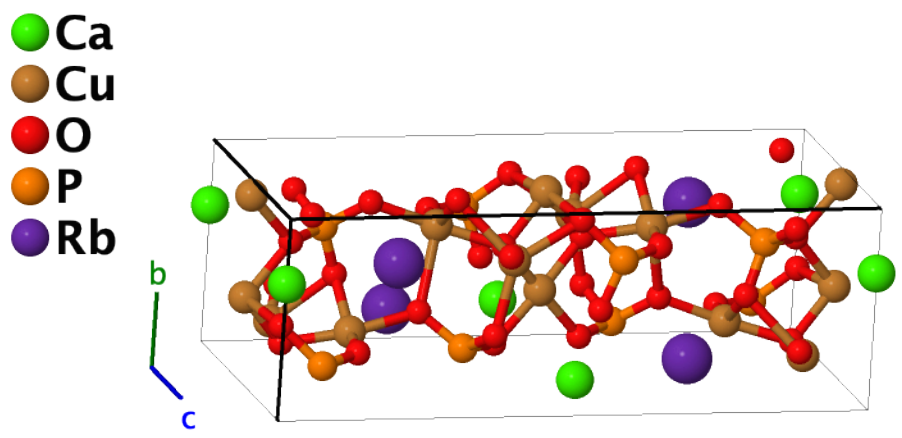

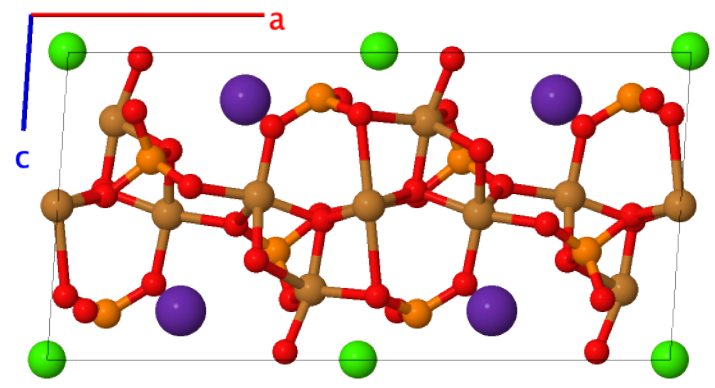

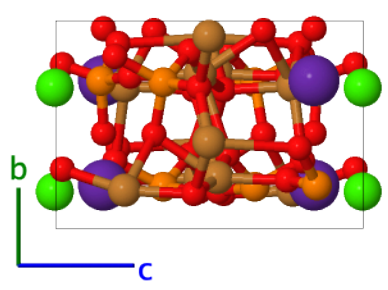

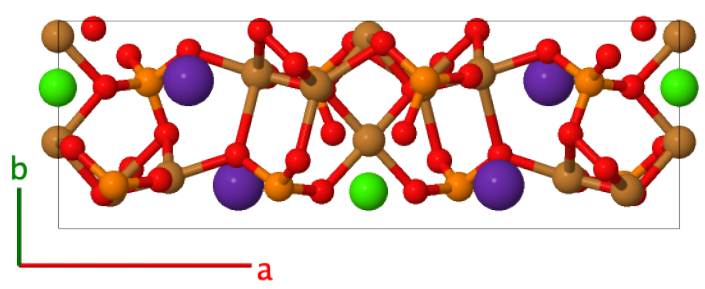

| | | |
|---|---|---|
| **Prototype** | : | $CaCu_6O_{18}P_4Rb_2$ |
| **AFLOW prototype label** | : | AB6C18D4E2_mC62_5_a_2b2c_9c_2c_c |
| ***Strukturbericht* designation** | : | None |
| **Pearson symbol** | : | mC62 |
| **Space group number** | : | 5 |
| **Space group symbol** | : | $C2$ |
| **AFLOW prototype command** | : | `aflow --proto=AB6C18D4E2_mC62_5_a_2b2c_9c_2c_c --params=`$a, b/a, c/a, \beta, y_1, y_2, y_3, x_4, y_4, z_4, x_5, y_5, z_5, x_6, y_6, z_6, x_7, y_7, z_7, x_8, y_8, z_8, x_9, y_9, z_9, x_{10}, y_{10}, z_{10}, x_{11}, y_{11}, z_{11}, x_{12}, y_{12}, z_{12}, x_{13}, y_{13}, z_{13}, x_{14}, y_{14}, z_{14}, x_{15}, y_{15}, z_{15}, x_{16}, y_{16}, z_{16}, x_{17}, y_{17}, z_{17}$ |

### Base-centered Monoclinic primitive vectors:

$$\mathbf{a}_1 = \frac{1}{2}\,a\,\hat{\mathbf{x}} - \frac{1}{2}\,b\,\hat{\mathbf{y}}$$
$$\mathbf{a}_2 = \frac{1}{2}\,a\,\hat{\mathbf{x}} + \frac{1}{2}\,b\,\hat{\mathbf{y}}$$
$$\mathbf{a}_3 = c\cos\beta\,\hat{\mathbf{x}} + c\sin\beta\,\hat{\mathbf{z}}$$

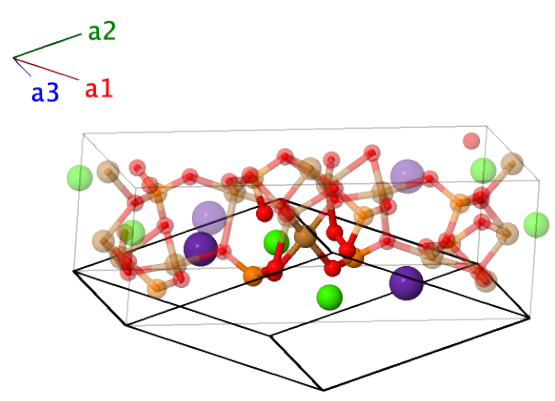

### Basis vectors:

| | | Lattice Coordinates | | Cartesian Coordinates | Wyckoff Position | Atom Type |
|---|---|---|---|---|---|---|
| $\mathbf{B}_1$ | = | $-y_1\,\mathbf{a}_1 + y_1\,\mathbf{a}_2$ | = | $y_1\,b\,\hat{\mathbf{y}}$ | (2$a$) | Ca |
| $\mathbf{B}_2$ | = | $-y_2\,\mathbf{a}_1 + y_2\,\mathbf{a}_2 + \frac{1}{2}\,\mathbf{a}_3$ | = | $\frac{1}{2}c\cos\beta\,\hat{\mathbf{x}} + y_2 b\,\hat{\mathbf{y}} + \frac{1}{2}c\sin\beta\,\hat{\mathbf{z}}$ | (2$b$) | Cu I |
| $\mathbf{B}_3$ | = | $-y_3\,\mathbf{a}_1 + y_3\,\mathbf{a}_2 + \frac{1}{2}\,\mathbf{a}_3$ | = | $\frac{1}{2}c\cos\beta\,\hat{\mathbf{x}} + y_3 b\,\hat{\mathbf{y}} + \frac{1}{2}c\sin\beta\,\hat{\mathbf{z}}$ | (2$b$) | Cu II |
| $\mathbf{B}_4$ | = | $(x_4 - y_4)\,\mathbf{a}_1 + (x_4 + y_4)\,\mathbf{a}_2 + z_4\,\mathbf{a}_3$ | = | $(x_4 a + z_4 c\cos\beta)\,\hat{\mathbf{x}} + y_4 b\,\hat{\mathbf{y}} + z_4 c\sin\beta\,\hat{\mathbf{z}}$ | (4$c$) | Cu III |
| $\mathbf{B}_5$ | = | $(-x_4 - y_4)\,\mathbf{a}_1 + (-x_4 + y_4)\,\mathbf{a}_2 - z_4\,\mathbf{a}_3$ | = | $(-x_4 a - z_4 c\cos\beta)\,\hat{\mathbf{x}} + y_4 b\,\hat{\mathbf{y}} - z_4 c\sin\beta\,\hat{\mathbf{z}}$ | (4$c$) | Cu III |
| $\mathbf{B}_6$ | = | $(x_5 - y_5)\,\mathbf{a}_1 + (x_5 + y_5)\,\mathbf{a}_2 + z_5\,\mathbf{a}_3$ | = | $(x_5 a + z_5 c\cos\beta)\,\hat{\mathbf{x}} + y_5 b\,\hat{\mathbf{y}} + z_5 c\sin\beta\,\hat{\mathbf{z}}$ | (4$c$) | Cu IV |
| $\mathbf{B}_7$ | = | $(-x_5 - y_5)\,\mathbf{a}_1 + (-x_5 + y_5)\,\mathbf{a}_2 - z_5\,\mathbf{a}_3$ | = | $(-x_5 a - z_5 c\cos\beta)\,\hat{\mathbf{x}} + y_5 b\,\hat{\mathbf{y}} - z_5 c\sin\beta\,\hat{\mathbf{z}}$ | (4$c$) | Cu IV |
| $\mathbf{B}_8$ | = | $(x_6 - y_6)\,\mathbf{a}_1 + (x_6 + y_6)\,\mathbf{a}_2 + z_6\,\mathbf{a}_3$ | = | $(x_6 a + z_6 c\cos\beta)\,\hat{\mathbf{x}} + y_6 b\,\hat{\mathbf{y}} + z_6 c\sin\beta\,\hat{\mathbf{z}}$ | (4$c$) | O I |
| $\mathbf{B}_9$ | = | $(-x_6 - y_6)\,\mathbf{a}_1 + (-x_6 + y_6)\,\mathbf{a}_2 - z_6\,\mathbf{a}_3$ | = | $(-x_6 a - z_6 c\cos\beta)\,\hat{\mathbf{x}} + y_6 b\,\hat{\mathbf{y}} - z_6 c\sin\beta\,\hat{\mathbf{z}}$ | (4$c$) | O I |
| $\mathbf{B}_{10}$ | = | $(x_7 - y_7)\,\mathbf{a}_1 + (x_7 + y_7)\,\mathbf{a}_2 + z_7\,\mathbf{a}_3$ | = | $(x_7 a + z_7 c\cos\beta)\,\hat{\mathbf{x}} + y_7 b\,\hat{\mathbf{y}} + z_7 c\sin\beta\,\hat{\mathbf{z}}$ | (4$c$) | O II |
| $\mathbf{B}_{11}$ | = | $(-x_7 - y_7)\,\mathbf{a}_1 + (-x_7 + y_7)\,\mathbf{a}_2 - z_7\,\mathbf{a}_3$ | = | $(-x_7 a - z_7 c\cos\beta)\,\hat{\mathbf{x}} + y_7 b\,\hat{\mathbf{y}} - z_7 c\sin\beta\,\hat{\mathbf{z}}$ | (4$c$) | O II |
| $\mathbf{B}_{12}$ | = | $(x_8 - y_8)\,\mathbf{a}_1 + (x_8 + y_8)\,\mathbf{a}_2 + z_8\,\mathbf{a}_3$ | = | $(x_8 a + z_8 c\cos\beta)\,\hat{\mathbf{x}} + y_8 b\,\hat{\mathbf{y}} + z_8 c\sin\beta\,\hat{\mathbf{z}}$ | (4$c$) | O III |
| $\mathbf{B}_{13}$ | = | $(-x_8 - y_8)\,\mathbf{a}_1 + (-x_8 + y_8)\,\mathbf{a}_2 - z_8\,\mathbf{a}_3$ | = | $(-x_8 a - z_8 c\cos\beta)\,\hat{\mathbf{x}} + y_8 b\,\hat{\mathbf{y}} - z_8 c\sin\beta\,\hat{\mathbf{z}}$ | (4$c$) | O III |
| $\mathbf{B}_{14}$ | = | $(x_9 - y_9)\,\mathbf{a}_1 + (x_9 + y_9)\,\mathbf{a}_2 + z_9\,\mathbf{a}_3$ | = | $(x_9 a + z_9 c\cos\beta)\,\hat{\mathbf{x}} + y_9 b\,\hat{\mathbf{y}} + z_9 c\sin\beta\,\hat{\mathbf{z}}$ | (4$c$) | O IV |
| $\mathbf{B}_{15}$ | = | $(-x_9 - y_9)\,\mathbf{a}_1 + (-x_9 + y_9)\,\mathbf{a}_2 - z_9\,\mathbf{a}_3$ | = | $(-x_9 a - z_9 c\cos\beta)\,\hat{\mathbf{x}} + y_9 b\,\hat{\mathbf{y}} - z_9 c\sin\beta\,\hat{\mathbf{z}}$ | (4$c$) | O IV |
| $\mathbf{B}_{16}$ | = | $(x_{10} - y_{10})\,\mathbf{a}_1 + (x_{10} + y_{10})\,\mathbf{a}_2 + z_{10}\,\mathbf{a}_3$ | = | $(x_{10} a + z_{10} c\cos\beta)\,\hat{\mathbf{x}} + y_{10} b\,\hat{\mathbf{y}} + z_{10} c\sin\beta\,\hat{\mathbf{z}}$ | (4$c$) | O V |
| $\mathbf{B}_{17}$ | = | $(-x_{10} - y_{10})\,\mathbf{a}_1 + (-x_{10} + y_{10})\,\mathbf{a}_2 - z_{10}\,\mathbf{a}_3$ | = | $(-x_{10} a - z_{10} c\cos\beta)\,\hat{\mathbf{x}} + y_{10} b\,\hat{\mathbf{y}} - z_{10} c\sin\beta\,\hat{\mathbf{z}}$ | (4$c$) | O V |
| $\mathbf{B}_{18}$ | = | $(x_{11} - y_{11})\,\mathbf{a}_1 + (x_{11} + y_{11})\,\mathbf{a}_2 + z_{11}\,\mathbf{a}_3$ | = | $(x_{11} a + z_{11} c\cos\beta)\,\hat{\mathbf{x}} + y_{11} b\,\hat{\mathbf{y}} + z_{11} c\sin\beta\,\hat{\mathbf{z}}$ | (4$c$) | O VI |
| $\mathbf{B}_{19}$ | = | $(-x_{11} - y_{11})\,\mathbf{a}_1 + (-x_{11} + y_{11})\,\mathbf{a}_2 - z_{11}\,\mathbf{a}_3$ | = | $(-x_{11} a - z_{11} c\cos\beta)\,\hat{\mathbf{x}} + y_{11} b\,\hat{\mathbf{y}} - z_{11} c\sin\beta\,\hat{\mathbf{z}}$ | (4$c$) | O VI |
| $\mathbf{B}_{20}$ | = | $(x_{12} - y_{12})\,\mathbf{a}_1 + (x_{12} + y_{12})\,\mathbf{a}_2 + z_{12}\,\mathbf{a}_3$ | = | $(x_{12} a + z_{12} c\cos\beta)\,\hat{\mathbf{x}} + y_{12} b\,\hat{\mathbf{y}} + z_{12} c\sin\beta\,\hat{\mathbf{z}}$ | (4$c$) | O VII |
| $\mathbf{B}_{21}$ | = | $(-x_{12} - y_{12})\,\mathbf{a}_1 + (-x_{12} + y_{12})\,\mathbf{a}_2 - z_{12}\,\mathbf{a}_3$ | = | $(-x_{12} a - z_{12} c\cos\beta)\,\hat{\mathbf{x}} + y_{12} b\,\hat{\mathbf{y}} - z_{12} c\sin\beta\,\hat{\mathbf{z}}$ | (4$c$) | O VII |
| $\mathbf{B}_{22}$ | = | $(x_{13} - y_{13})\,\mathbf{a}_1 + (x_{13} + y_{13})\,\mathbf{a}_2 + z_{13}\,\mathbf{a}_3$ | = | $(x_{13} a + z_{13} c\cos\beta)\,\hat{\mathbf{x}} + y_{13} b\,\hat{\mathbf{y}} + z_{13} c\sin\beta\,\hat{\mathbf{z}}$ | (4$c$) | O VIII |
| $\mathbf{B}_{23}$ | = | $(-x_{13} - y_{13})\,\mathbf{a}_1 + (-x_{13} + y_{13})\,\mathbf{a}_2 - z_{13}\,\mathbf{a}_3$ | = | $(-x_{13} a - z_{13} c\cos\beta)\,\hat{\mathbf{x}} + y_{13} b\,\hat{\mathbf{y}} - z_{13} c\sin\beta\,\hat{\mathbf{z}}$ | (4$c$) | O VIII |
| $\mathbf{B}_{24}$ | = | $(x_{14} - y_{14})\,\mathbf{a}_1 + (x_{14} + y_{14})\,\mathbf{a}_2 + z_{14}\,\mathbf{a}_3$ | = | $(x_{14} a + z_{14} c\cos\beta)\,\hat{\mathbf{x}} + y_{14} b\,\hat{\mathbf{y}} + z_{14} c\sin\beta\,\hat{\mathbf{z}}$ | (4$c$) | O IX |
| $\mathbf{B}_{25}$ | = | $(-x_{14} - y_{14})\,\mathbf{a}_1 + (-x_{14} + y_{14})\,\mathbf{a}_2 - z_{14}\,\mathbf{a}_3$ | = | $(-x_{14} a - z_{14} c\cos\beta)\,\hat{\mathbf{x}} + y_{14} b\,\hat{\mathbf{y}} - z_{14} c\sin\beta\,\hat{\mathbf{z}}$ | (4$c$) | O IX |
| $\mathbf{B}_{26}$ | = | $(x_{15} - y_{15})\,\mathbf{a}_1 + (x_{15} + y_{15})\,\mathbf{a}_2 + z_{15}\,\mathbf{a}_3$ | = | $(x_{15} a + z_{15} c\cos\beta)\,\hat{\mathbf{x}} + y_{15} b\,\hat{\mathbf{y}} + z_{15} c\sin\beta\,\hat{\mathbf{z}}$ | (4$c$) | P I |
| $\mathbf{B}_{27}$ | = | $(-x_{15} - y_{15})\,\mathbf{a}_1 + (-x_{15} + y_{15})\,\mathbf{a}_2 - z_{15}\,\mathbf{a}_3$ | = | $(-x_{15} a - z_{15} c\cos\beta)\,\hat{\mathbf{x}} + y_{15} b\,\hat{\mathbf{y}} - z_{15} c\sin\beta\,\hat{\mathbf{z}}$ | (4$c$) | P I |
| $\mathbf{B}_{28}$ | = | $(x_{16} - y_{16})\,\mathbf{a}_1 + (x_{16} + y_{16})\,\mathbf{a}_2 + z_{16}\,\mathbf{a}_3$ | = | $(x_{16} a + z_{16} c\cos\beta)\,\hat{\mathbf{x}} + y_{16} b\,\hat{\mathbf{y}} + z_{16} c\sin\beta\,\hat{\mathbf{z}}$ | (4$c$) | P II |
| $\mathbf{B}_{29}$ | = | $(-x_{16} - y_{16})\,\mathbf{a}_1 + (-x_{16} + y_{16})\,\mathbf{a}_2 - z_{16}\,\mathbf{a}_3$ | = | $(-x_{16}a - z_{16}c\cos\beta)\,\hat{\mathbf{x}} + y_{16}b\,\hat{\mathbf{y}} - z_{16}c\sin\beta\,\hat{\mathbf{z}}$ | (4$c$) | P II |
| $\mathbf{B}_{30}$ | = | $(x_{17} - y_{17})\,\mathbf{a}_1 + (x_{17} + y_{17})\,\mathbf{a}_2 + z_{17}\,\mathbf{a}_3$ | = | $(x_{17}a + z_{17}c\cos\beta)\,\hat{\mathbf{x}} + y_{17}b\,\hat{\mathbf{y}} + z_{17}c\sin\beta\,\hat{\mathbf{z}}$ | (4$c$) | Rb |
| $\mathbf{B}_{31}$ | = | $(-x_{17} - y_{17})\,\mathbf{a}_1 + (-x_{17} + y_{17})\,\mathbf{a}_2 - z_{17}\,\mathbf{a}_3$ | = | $(-x_{17}a - z_{17}c\cos\beta)\,\hat{\mathbf{x}} + y_{17}b\,\hat{\mathbf{y}} - z_{17}c\sin\beta\,\hat{\mathbf{z}}$ | (4$c$) | Rb |

**Geometry files:**

- CIF: pp. 1512
- POSCAR: pp. 1512

# $C2$ (Ba,Ca)$CO_3$ Structure: ABC3_mC10_5_b_a_ac

http://aflow.org/prototype-encyclopedia/ABC3_mC10_5_b_a_ac

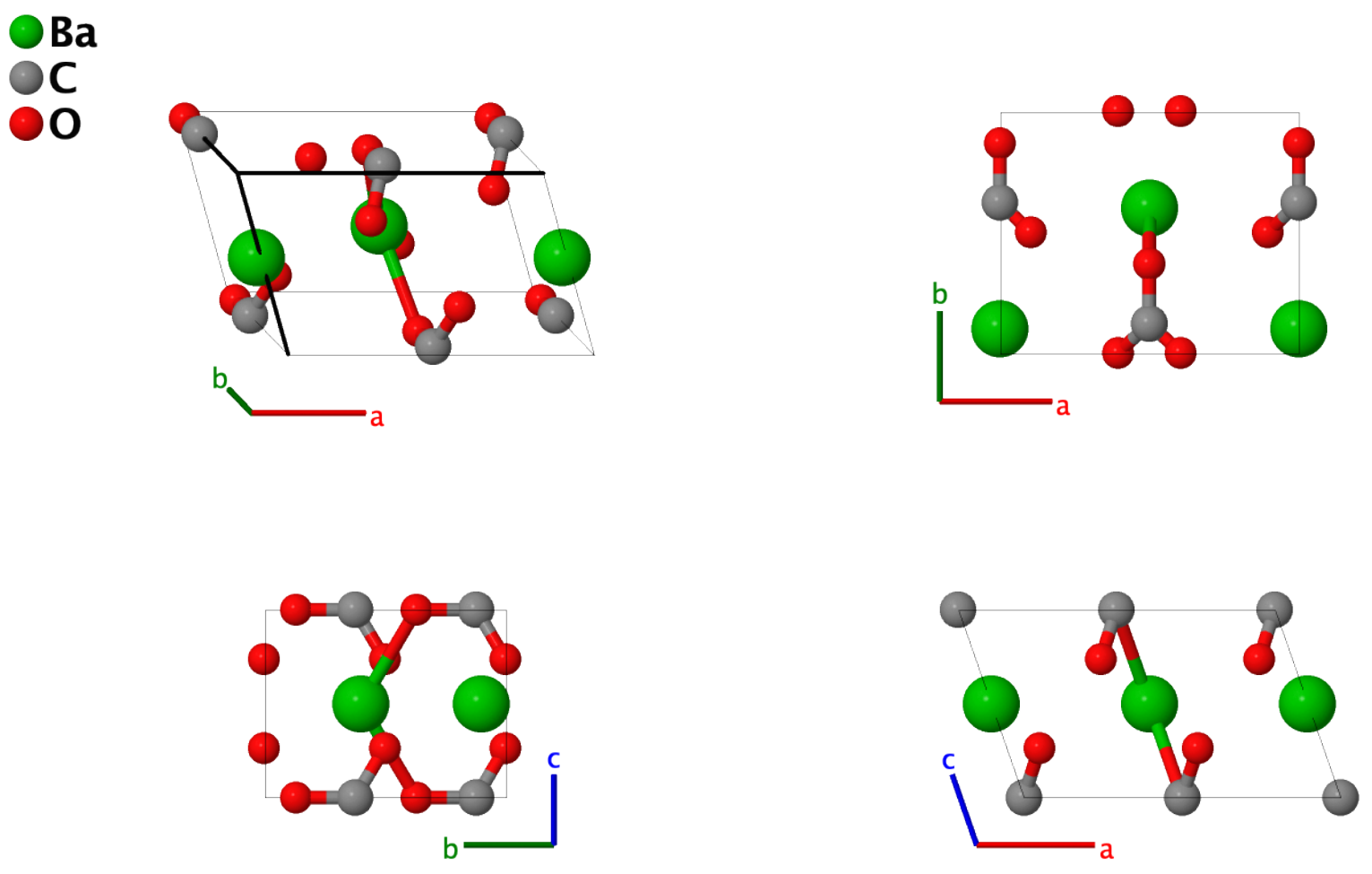

| | | |
|---|---|---|
| **Prototype** | : | (Ba,Ca)$CO_3$ |
| **AFLOW prototype label** | : | ABC3_mC10_5_b_a_ac |
| ***Strukturbericht* designation** | : | None |
| **Pearson symbol** | : | mC10 |
| **Space group number** | : | 5 |
| **Space group symbol** | : | $C2$ |
| **AFLOW prototype command** | : | `aflow --proto=ABC3_mC10_5_b_a_ac` <br> `--params=`$a, b/a, c/a, \beta, y_1, y_2, y_3, x_4, y_4, z_4$ |

- (Ba,Ca)$(CO_3)_2$ comes in a variety of crystal structures (Spahr, 2019):
  - monoclinic barytocalcite, space group $P2_1/m$ #11 (the current structure).
  - trigonal paralstonite, space group $P321$ #150,
  - Triclinic alstonite, space group $P1$ #1 or $P\bar{1}$ #2 (Sartori, 1975), and
  - a new monoclinic structure, space group $C2$ #5, synthesized by (Spahr, 2019), and lacking the centrosymmetric character of barytocalcite.
- The site we have labeled Ba is actually a mixture of barium and calcium atoms.

**Base-centered Monoclinic primitive vectors:**

$$\mathbf{a}_1 = \frac{1}{2}\,a\,\mathbf{\hat{x}} - \frac{1}{2}\,b\,\mathbf{\hat{y}}$$
$$\mathbf{a}_2 = \frac{1}{2}\,a\,\mathbf{\hat{x}} + \frac{1}{2}\,b\,\mathbf{\hat{y}}$$
$$\mathbf{a}_3 = c\cos\beta\,\mathbf{\hat{x}} + c\sin\beta\,\mathbf{\hat{z}}$$

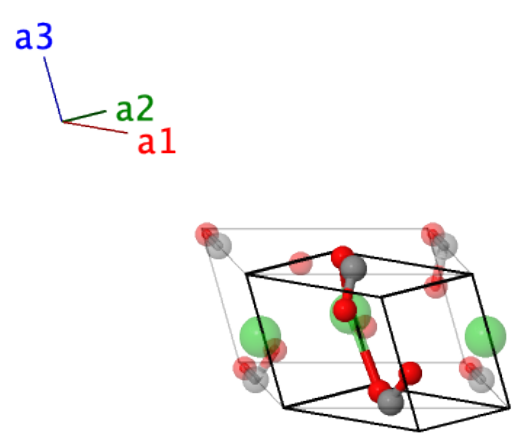

**Basis vectors:**

| | | Lattice Coordinates | | Cartesian Coordinates | Wyckoff Position | Atom Type |
|---|---|---|---|---|---|---|
| $\mathbf{B}_1$ | = | $-y_1\,\mathbf{a}_1 + y_1\,\mathbf{a}_2$ | = | $y_1 b\,\hat{\mathbf{y}}$ | (2$a$) | C |
| $\mathbf{B}_2$ | = | $-y_2\,\mathbf{a}_1 + y_2\,\mathbf{a}_2$ | = | $y_2 b\,\hat{\mathbf{y}}$ | (2$a$) | O I |
| $\mathbf{B}_3$ | = | $-y_3\,\mathbf{a}_1 + y_3\,\mathbf{a}_2 + \frac{1}{2}\,\mathbf{a}_3$ | = | $\frac{1}{2}c\cos\beta\,\hat{\mathbf{x}} + y_3 b\,\hat{\mathbf{y}} + \frac{1}{2}c\sin\beta\,\hat{\mathbf{z}}$ | (2$b$) | Ba |
| $\mathbf{B}_4$ | = | $(x_4 - y_4)\,\mathbf{a}_1 + (x_4 + y_4)\,\mathbf{a}_2 + z_4\,\mathbf{a}_3$ | = | $(x_4 a + z_4 c\cos\beta)\,\hat{\mathbf{x}} + y_4 b\,\hat{\mathbf{y}} + z_4 c\sin\beta\,\hat{\mathbf{z}}$ | (4$c$) | O II |
| $\mathbf{B}_5$ | = | $(-x_4 - y_4)\,\mathbf{a}_1 + (-x_4 + y_4)\,\mathbf{a}_2 - z_4\,\mathbf{a}_3$ | = | $(-x_4 a - z_4 c\cos\beta)\,\hat{\mathbf{x}} + y_4 b\,\hat{\mathbf{y}} - z_4 c\sin\beta\,\hat{\mathbf{z}}$ | (4$c$) | O II |

---

---

**Geometry files:**

- CIF: pp. 1512
- POSCAR: pp. 1513

# $Ta_5Ti_{11}$ (BCC SQS-16) Structure: A5B11_mP16_6_2abc_2a3b3c

http://aflow.org/prototype-encyclopedia/A5B11_mP16_6_2abc_2a3b3c

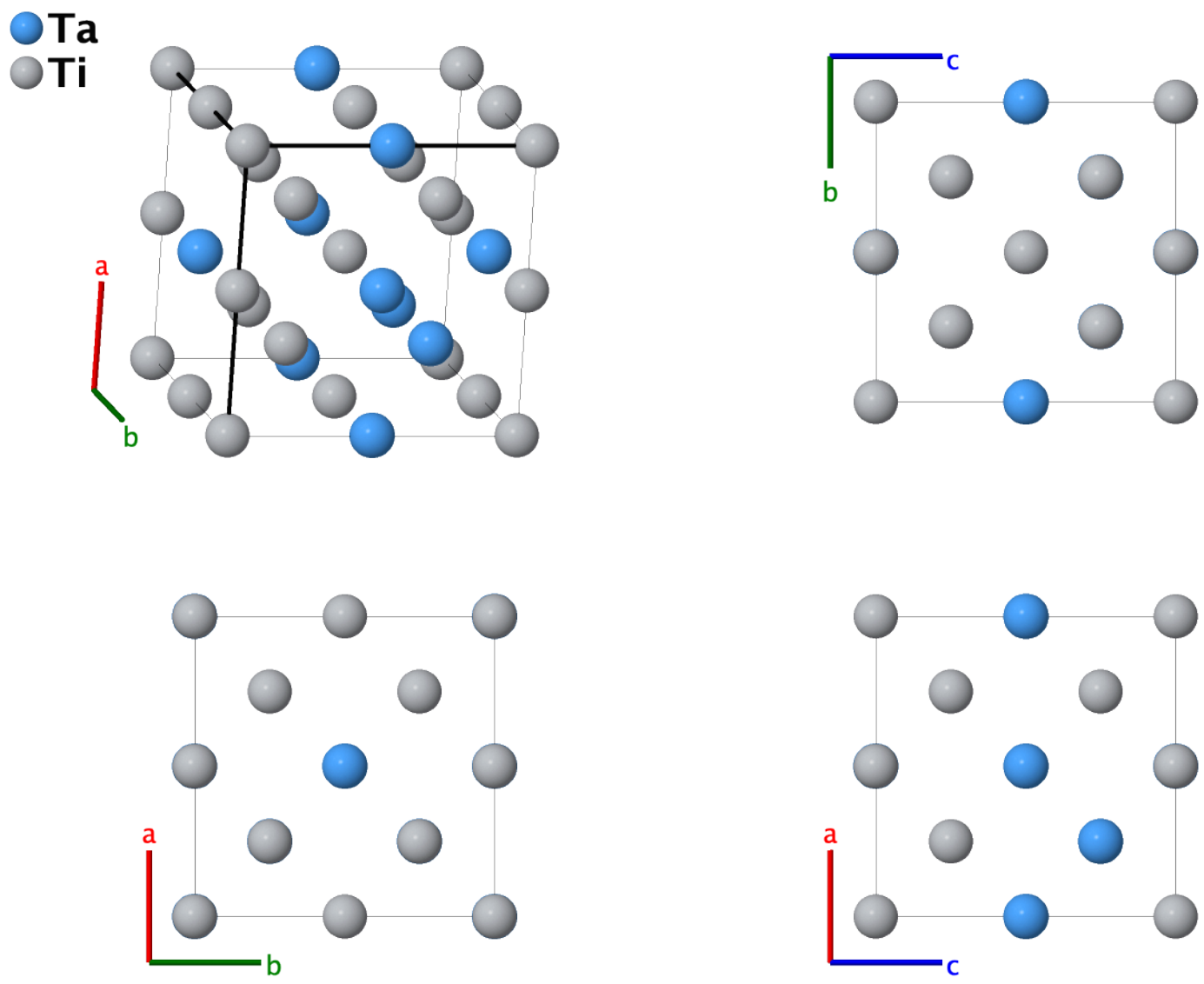

| | | |
|---|---|---|
| **Prototype** | : | $Ta_5Ti_{11}$ |
| **AFLOW prototype label** | : | A5B11_mP16_6_2abc_2a3b3c |
| ***Strukturbericht* designation** | : | None |
| **Pearson symbol** | : | mP16 |
| **Space group number** | : | 6 |
| **Space group symbol** | : | $Pm$ |
| **AFLOW prototype command** | : | `aflow --proto=A5B11_mP16_6_2abc_2a3b3c` `--params=`$a, b/a, c/a, \beta, x_1, z_1, x_2, z_2, x_3, z_3, x_4, z_4, x_5, z_5, x_6, z_6, x_7, z_7, x_8, z_8, x_9, y_9, z_9, x_{10}, y_{10}, z_{10}, x_{11}, y_{11}, z_{11}, x_{12}, y_{12}, z_{12}$ |

- This is a special quasirandom structure with 16 atoms per unit cell (SQS-16) for the $\beta$-phase (high-temperature austenite) bcc substitutional Ti-Ta alloy (Chakraborty, 2016). This protoype contains 31.25% Ta. Prototypes are listed for other Ta-Ti concentrations: 12.5% Ta (AB7_hR16_166_c_c2h), 18.75% Ta (A3B13_oC32_38_ac_a2bcdef), 25% Ta (AB3_mC32_8_4a_4a4b), and 37.5% Ta (A3B5_oC32_38_abce_abcdf).

**Simple Monoclinic primitive vectors:**

$$
\begin{aligned}
\mathbf{a}_1 &= a\,\hat{\mathbf{x}} \\
\mathbf{a}_2 &= b\,\hat{\mathbf{y}} \\
\mathbf{a}_3 &= c\cos\beta\,\hat{\mathbf{x}} + c\sin\beta\,\hat{\mathbf{z}}
\end{aligned}
$$

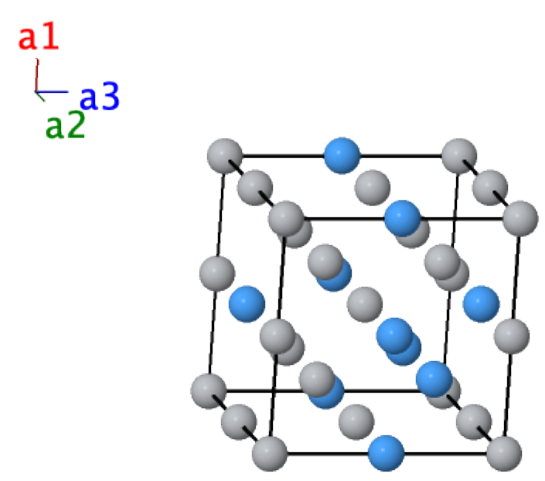

**Basis vectors:**

| | | Lattice Coordinates | | Cartesian Coordinates | Wyckoff Position | Atom Type |
|---|---|---|---|---|---|---|
| $\mathbf{B}_1$ | = | $x_1\,\mathbf{a}_1 + z_1\,\mathbf{a}_3$ | = | $(x_1 a + z_1 c\cos\beta)\,\mathbf{\hat{x}} + z_1 c\sin\beta\,\mathbf{\hat{z}}$ | $(1a)$ | Ta I |
| $\mathbf{B}_2$ | = | $x_2\,\mathbf{a}_1 + z_2\,\mathbf{a}_3$ | = | $(x_2 a + z_2 c\cos\beta)\,\mathbf{\hat{x}} + z_2 c\sin\beta\,\mathbf{\hat{z}}$ | $(1a)$ | Ta II |
| $\mathbf{B}_3$ | = | $x_3\,\mathbf{a}_1 + z_3\,\mathbf{a}_3$ | = | $(x_3 a + z_3 c\cos\beta)\,\mathbf{\hat{x}} + z_3 c\sin\beta\,\mathbf{\hat{z}}$ | $(1a)$ | Ti I |
| $\mathbf{B}_4$ | = | $x_4\,\mathbf{a}_1 + z_4\,\mathbf{a}_3$ | = | $(x_4 a + z_4 c\cos\beta)\,\mathbf{\hat{x}} + z_4 c\sin\beta\,\mathbf{\hat{z}}$ | $(1a)$ | Ti II |
| $\mathbf{B}_5$ | = | $x_5\,\mathbf{a}_1 + \frac{1}{2}\,\mathbf{a}_2 + z_5\,\mathbf{a}_3$ | = | $(x_5 a + z_5 c\cos\beta)\,\mathbf{\hat{x}} + \frac{1}{2}b\,\mathbf{\hat{y}} + z_5 c\sin\beta\,\mathbf{\hat{z}}$ | $(1b)$ | Ta III |
| $\mathbf{B}_6$ | = | $x_6\,\mathbf{a}_1 + \frac{1}{2}\,\mathbf{a}_2 + z_6\,\mathbf{a}_3$ | = | $(x_6 a + z_6 c\cos\beta)\,\mathbf{\hat{x}} + \frac{1}{2}b\,\mathbf{\hat{y}} + z_6 c\sin\beta\,\mathbf{\hat{z}}$ | $(1b)$ | Ti III |
| $\mathbf{B}_7$ | = | $x_7\,\mathbf{a}_1 + \frac{1}{2}\,\mathbf{a}_2 + z_7\,\mathbf{a}_3$ | = | $(x_7 a + z_7 c\cos\beta)\,\mathbf{\hat{x}} + \frac{1}{2}b\,\mathbf{\hat{y}} + z_7 c\sin\beta\,\mathbf{\hat{z}}$ | $(1b)$ | Ti IV |
| $\mathbf{B}_8$ | = | $x_8\,\mathbf{a}_1 + \frac{1}{2}\,\mathbf{a}_2 + z_8\,\mathbf{a}_3$ | = | $(x_8 a + z_8 c\cos\beta)\,\mathbf{\hat{x}} + \frac{1}{2}b\,\mathbf{\hat{y}} + z_8 c\sin\beta\,\mathbf{\hat{z}}$ | $(1b)$ | Ti V |
| $\mathbf{B}_9$ | = | $x_9\,\mathbf{a}_1 + y_9\,\mathbf{a}_2 + z_9\,\mathbf{a}_3$ | = | $(x_9 a + z_9 c\cos\beta)\,\mathbf{\hat{x}} + y_9 b\,\mathbf{\hat{y}} + z_9 c\sin\beta\,\mathbf{\hat{z}}$ | $(2c)$ | Ta IV |
| $\mathbf{B}_{10}$ | = | $x_9\,\mathbf{a}_1 - y_9\,\mathbf{a}_2 + z_9\,\mathbf{a}_3$ | = | $(x_9 a + z_9 c\cos\beta)\,\mathbf{\hat{x}} - y_9 b\,\mathbf{\hat{y}} + z_9 c\sin\beta\,\mathbf{\hat{z}}$ | $(2c)$ | Ta IV |
| $\mathbf{B}_{11}$ | = | $x_{10}\,\mathbf{a}_1 + y_{10}\,\mathbf{a}_2 + z_{10}\,\mathbf{a}_3$ | = | $(x_{10} a + z_{10} c\cos\beta)\,\mathbf{\hat{x}} + y_{10} b\,\mathbf{\hat{y}} + z_{10} c\sin\beta\,\mathbf{\hat{z}}$ | $(2c)$ | Ti VI |
| $\mathbf{B}_{12}$ | = | $x_{10}\,\mathbf{a}_1 - y_{10}\,\mathbf{a}_2 + z_{10}\,\mathbf{a}_3$ | = | $(x_{10} a + z_{10} c\cos\beta)\,\mathbf{\hat{x}} - y_{10} b\,\mathbf{\hat{y}} + z_{10} c\sin\beta\,\mathbf{\hat{z}}$ | $(2c)$ | Ti VI |
| $\mathbf{B}_{13}$ | = | $x_{11}\,\mathbf{a}_1 + y_{11}\,\mathbf{a}_2 + z_{11}\,\mathbf{a}_3$ | = | $(x_{11} a + z_{11} c\cos\beta)\,\mathbf{\hat{x}} + y_{11} b\,\mathbf{\hat{y}} + z_{11} c\sin\beta\,\mathbf{\hat{z}}$ | $(2c)$ | Ti VII |
| $\mathbf{B}_{14}$ | = | $x_{11}\,\mathbf{a}_1 - y_{11}\,\mathbf{a}_2 + z_{11}\,\mathbf{a}_3$ | = | $(x_{11} a + z_{11} c\cos\beta)\,\mathbf{\hat{x}} - y_{11} b\,\mathbf{\hat{y}} + z_{11} c\sin\beta\,\mathbf{\hat{z}}$ | $(2c)$ | Ti VII |
| $\mathbf{B}_{15}$ | = | $x_{12}\,\mathbf{a}_1 + y_{12}\,\mathbf{a}_2 + z_{12}\,\mathbf{a}_3$ | = | $(x_{12} a + z_{12} c\cos\beta)\,\mathbf{\hat{x}} + y_{12} b\,\mathbf{\hat{y}} + z_{12} c\sin\beta\,\mathbf{\hat{z}}$ | $(2c)$ | Ti VIII |
| $\mathbf{B}_{16}$ | = | $x_{12}\,\mathbf{a}_1 - y_{12}\,\mathbf{a}_2 + z_{12}\,\mathbf{a}_3$ | = | $(x_{12} a + z_{12} c\cos\beta)\,\mathbf{\hat{x}} - y_{12} b\,\mathbf{\hat{y}} + z_{12} c\sin\beta\,\mathbf{\hat{z}}$ | $(2c)$ | Ti VIII |

**Geometry files:**

- CIF: pp. 1513
- POSCAR: pp. 1513

# $Na_2Ca_6Si_4O_{15}$ Structure: A6B2C15D4_mP54_7_6a_2a_15a_4a

http://aflow.org/prototype-encyclopedia/A6B2C15D4_mP54_7_6a_2a_15a_4a

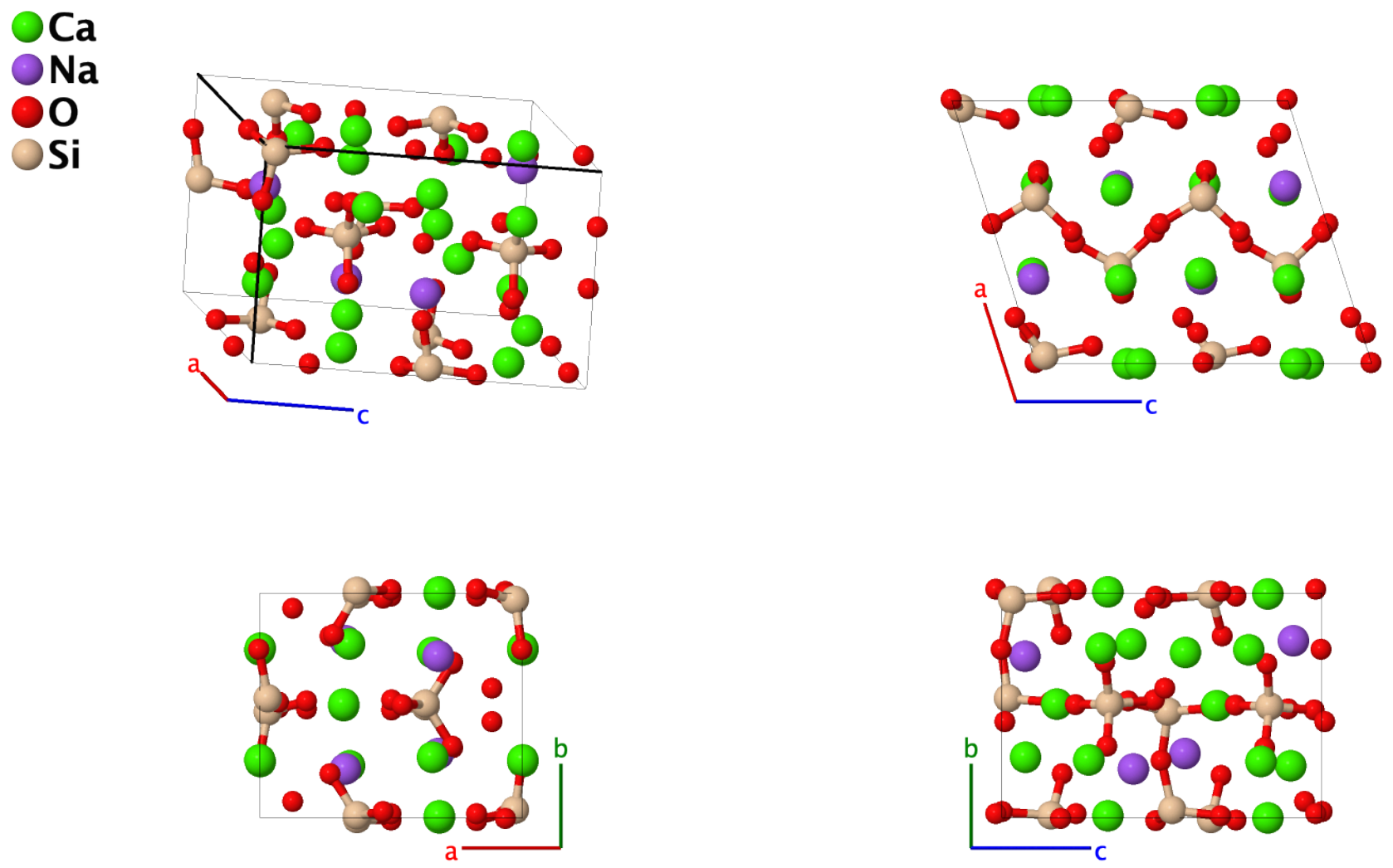

| | | |
|---|---|---|
| **Prototype** | : | $Ca_6Na_2O_{15}Si_4$ |
| **AFLOW prototype label** | : | A6B2C15D4_mP54_7_6a_2a_15a_4a |
| ***Strukturbericht* designation** | : | None |
| **Pearson symbol** | : | mP54 |
| **Space group number** | : | 7 |
| **Space group symbol** | : | $Pc$ |
| **AFLOW prototype command** | : | `aflow --proto=A6B2C15D4_mP54_7_6a_2a_15a_4a` `--params=`$a, b/a, c/a, \beta, x_1, y_1, z_1, x_2, y_2, z_2, x_3, y_3, z_3, x_4, y_4, z_4, x_5, y_5, z_5, x_6, y_6, z_6, x_7, y_7, z_7, x_8, y_8, z_8, x_9, y_9, z_9, x_{10}, y_{10}, z_{10}, x_{11}, y_{11}, z_{11}, x_{12}, y_{12}, z_{12}, x_{13}, y_{13}, z_{13}, x_{14}, y_{14}, z_{14}, x_{15}, y_{15}, z_{15}, x_{16}, y_{16}, z_{16}, x_{17}, y_{17}, z_{17}, x_{18}, y_{18}, z_{18}, x_{19}, y_{19}, z_{19}, x_{20}, y_{20}, z_{20}, x_{21}, y_{21}, z_{21}, x_{22}, y_{22}, z_{22}, x_{23}, y_{23}, z_{23}, x_{24}, y_{24}, z_{24}, x_{25}, y_{25}, z_{25}, x_{26}, y_{26}, z_{26}, x_{27}, y_{27}, z_{27}$ |

- Although this structure was formed at 1300 °C, the data was taken after cooling to room temperature, 295 K.
- The true composition of the Ca-I site is $Ca_{0.84}Na_{0.16}$, and the Na-I site has composition $Na_{0.84}Ca_{0.16}$.
- The Si-III and Si-IV atoms share the O-$X$ atom between them, accounting for the "missing" oxygen atom.

**Simple Monoclinic primitive vectors:**

$$\mathbf{a}_1 = a\,\hat{\mathbf{x}}$$
$$\mathbf{a}_2 = b\,\hat{\mathbf{y}}$$
$$\mathbf{a}_3 = c\cos\beta\,\hat{\mathbf{x}} + c\sin\beta\,\hat{\mathbf{z}}$$

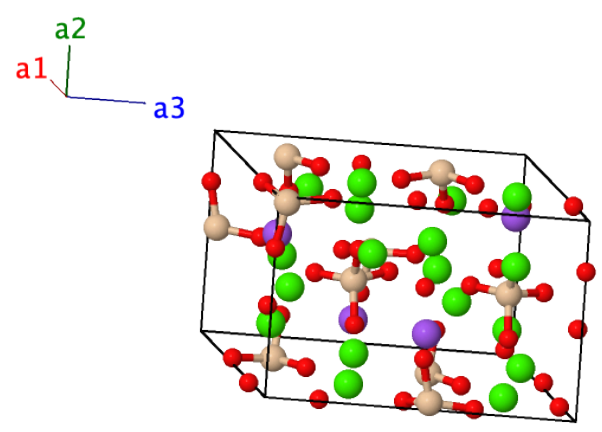

**Basis vectors:**

| | | Lattice Coordinates | | Cartesian Coordinates | Wyckoff Position | Atom Type |
|---|---|---|---|---|---|---|
| $\mathbf{B}_1$ | = | $x_1\,\mathbf{a}_1 + y_1\,\mathbf{a}_2 + z_1\,\mathbf{a}_3$ | = | $(x_1a + z_1c\cos\beta)\,\hat{\mathbf{x}} + y_1b\,\hat{\mathbf{y}} + z_1c\sin\beta\,\hat{\mathbf{z}}$ | (2a) | Ca I |
| $\mathbf{B}_2$ | = | $x_1\,\mathbf{a}_1 - y_1\,\mathbf{a}_2 + \left(\frac{1}{2} + z_1\right)\mathbf{a}_3$ | = | $\left(\frac{1}{2}c\cos\beta + x_1a + z_1c\cos\beta\right)\hat{\mathbf{x}} - y_1b\,\hat{\mathbf{y}} + \left(\frac{1}{2} + z_1\right)c\sin\beta\,\hat{\mathbf{z}}$ | (2a) | Ca I |
| $\mathbf{B}_3$ | = | $x_2\,\mathbf{a}_1 + y_2\,\mathbf{a}_2 + z_2\,\mathbf{a}_3$ | = | $(x_2a + z_2c\cos\beta)\,\hat{\mathbf{x}} + y_2b\,\hat{\mathbf{y}} + z_2c\sin\beta\,\hat{\mathbf{z}}$ | (2a) | Ca II |
| $\mathbf{B}_4$ | = | $x_2\,\mathbf{a}_1 - y_2\,\mathbf{a}_2 + \left(\frac{1}{2} + z_2\right)\mathbf{a}_3$ | = | $\left(\frac{1}{2}c\cos\beta + x_2a + z_2c\cos\beta\right)\hat{\mathbf{x}} - y_2b\,\hat{\mathbf{y}} + \left(\frac{1}{2} + z_2\right)c\sin\beta\,\hat{\mathbf{z}}$ | (2a) | Ca II |
| $\mathbf{B}_5$ | = | $x_3\,\mathbf{a}_1 + y_3\,\mathbf{a}_2 + z_3\,\mathbf{a}_3$ | = | $(x_3a + z_3c\cos\beta)\,\hat{\mathbf{x}} + y_3b\,\hat{\mathbf{y}} + z_3c\sin\beta\,\hat{\mathbf{z}}$ | (2a) | Ca III |
| $\mathbf{B}_6$ | = | $x_3\,\mathbf{a}_1 - y_3\,\mathbf{a}_2 + \left(\frac{1}{2} + z_3\right)\mathbf{a}_3$ | = | $\left(\frac{1}{2}c\cos\beta + x_3a + z_3c\cos\beta\right)\hat{\mathbf{x}} - y_3b\,\hat{\mathbf{y}} + \left(\frac{1}{2} + z_3\right)c\sin\beta\,\hat{\mathbf{z}}$ | (2a) | Ca III |
| $\mathbf{B}_7$ | = | $x_4\,\mathbf{a}_1 + y_4\,\mathbf{a}_2 + z_4\,\mathbf{a}_3$ | = | $(x_4a + z_4c\cos\beta)\,\hat{\mathbf{x}} + y_4b\,\hat{\mathbf{y}} + z_4c\sin\beta\,\hat{\mathbf{z}}$ | (2a) | Ca IV |
| $\mathbf{B}_8$ | = | $x_4\,\mathbf{a}_1 - y_4\,\mathbf{a}_2 + \left(\frac{1}{2} + z_4\right)\mathbf{a}_3$ | = | $\left(\frac{1}{2}c\cos\beta + x_4a + z_4c\cos\beta\right)\hat{\mathbf{x}} - y_4b\,\hat{\mathbf{y}} + \left(\frac{1}{2} + z_4\right)c\sin\beta\,\hat{\mathbf{z}}$ | (2a) | Ca IV |
| $\mathbf{B}_9$ | = | $x_5\,\mathbf{a}_1 + y_5\,\mathbf{a}_2 + z_5\,\mathbf{a}_3$ | = | $(x_5a + z_5c\cos\beta)\,\hat{\mathbf{x}} + y_5b\,\hat{\mathbf{y}} + z_5c\sin\beta\,\hat{\mathbf{z}}$ | (2a) | Ca V |
| $\mathbf{B}_{10}$ | = | $x_5\,\mathbf{a}_1 - y_5\,\mathbf{a}_2 + \left(\frac{1}{2} + z_5\right)\mathbf{a}_3$ | = | $\left(\frac{1}{2}c\cos\beta + x_5a + z_5c\cos\beta\right)\hat{\mathbf{x}} - y_5b\,\hat{\mathbf{y}} + \left(\frac{1}{2} + z_5\right)c\sin\beta\,\hat{\mathbf{z}}$ | (2a) | Ca V |
| $\mathbf{B}_{11}$ | = | $x_6\,\mathbf{a}_1 + y_6\,\mathbf{a}_2 + z_6\,\mathbf{a}_3$ | = | $(x_6a + z_6c\cos\beta)\,\hat{\mathbf{x}} + y_6b\,\hat{\mathbf{y}} + z_6c\sin\beta\,\hat{\mathbf{z}}$ | (2a) | Ca VI |
| $\mathbf{B}_{12}$ | = | $x_6\,\mathbf{a}_1 - y_6\,\mathbf{a}_2 + \left(\frac{1}{2} + z_6\right)\mathbf{a}_3$ | = | $\left(\frac{1}{2}c\cos\beta + x_6a + z_6c\cos\beta\right)\hat{\mathbf{x}} - y_6b\,\hat{\mathbf{y}} + \left(\frac{1}{2} + z_6\right)c\sin\beta\,\hat{\mathbf{z}}$ | (2a) | Ca VI |
| $\mathbf{B}_{13}$ | = | $x_7\,\mathbf{a}_1 + y_7\,\mathbf{a}_2 + z_7\,\mathbf{a}_3$ | = | $(x_7a + z_7c\cos\beta)\,\hat{\mathbf{x}} + y_7b\,\hat{\mathbf{y}} + z_7c\sin\beta\,\hat{\mathbf{z}}$ | (2a) | Na I |
| $\mathbf{B}_{14}$ | = | $x_7\,\mathbf{a}_1 - y_7\,\mathbf{a}_2 + \left(\frac{1}{2} + z_7\right)\mathbf{a}_3$ | = | $\left(\frac{1}{2}c\cos\beta + x_7a + z_7c\cos\beta\right)\hat{\mathbf{x}} - y_7b\,\hat{\mathbf{y}} + \left(\frac{1}{2} + z_7\right)c\sin\beta\,\hat{\mathbf{z}}$ | (2a) | Na I |
| $\mathbf{B}_{15}$ | = | $x_8\,\mathbf{a}_1 + y_8\,\mathbf{a}_2 + z_8\,\mathbf{a}_3$ | = | $(x_8a + z_8c\cos\beta)\,\hat{\mathbf{x}} + y_8b\,\hat{\mathbf{y}} + z_8c\sin\beta\,\hat{\mathbf{z}}$ | (2a) | Na II |
| $\mathbf{B}_{16}$ | = | $x_8\,\mathbf{a}_1 - y_8\,\mathbf{a}_2 + \left(\frac{1}{2} + z_8\right)\mathbf{a}_3$ | = | $\left(\frac{1}{2}c\cos\beta + x_8a + z_8c\cos\beta\right)\hat{\mathbf{x}} - y_8b\,\hat{\mathbf{y}} + \left(\frac{1}{2} + z_8\right)c\sin\beta\,\hat{\mathbf{z}}$ | (2a) | Na II |
| $\mathbf{B}_{17}$ | = | $x_9\,\mathbf{a}_1 + y_9\,\mathbf{a}_2 + z_9\,\mathbf{a}_3$ | = | $(x_9a + z_9c\cos\beta)\,\hat{\mathbf{x}} + y_9b\,\hat{\mathbf{y}} + z_9c\sin\beta\,\hat{\mathbf{z}}$ | (2a) | O I |
| $\mathbf{B}_{18}$ | = | $x_9\,\mathbf{a}_1 - y_9\,\mathbf{a}_2 + \left(\frac{1}{2} + z_9\right)\mathbf{a}_3$ | = | $\left(\frac{1}{2}c\cos\beta + x_9a + z_9c\cos\beta\right)\hat{\mathbf{x}} - y_9b\,\hat{\mathbf{y}} + \left(\frac{1}{2} + z_9\right)c\sin\beta\,\hat{\mathbf{z}}$ | (2a) | O I |
| $\mathbf{B}_{19}$ | = | $x_{10}\,\mathbf{a}_1 + y_{10}\,\mathbf{a}_2 + z_{10}\,\mathbf{a}_3$ | = | $(x_{10}a + z_{10}c\cos\beta)\,\hat{\mathbf{x}} + y_{10}b\,\hat{\mathbf{y}} + z_{10}c\sin\beta\,\hat{\mathbf{z}}$ | (2a) | O II |
| $\mathbf{B}_{20}$ | = | $x_{10}\,\mathbf{a}_1 - y_{10}\,\mathbf{a}_2 + \left(\frac{1}{2} + z_{10}\right)\mathbf{a}_3$ | = | $\left(\frac{1}{2}c\cos\beta + x_{10}a + z_{10}c\cos\beta\right)\hat{\mathbf{x}} - y_{10}b\,\hat{\mathbf{y}} + \left(\frac{1}{2} + z_{10}\right)c\sin\beta\,\hat{\mathbf{z}}$ | (2a) | O II |
| $\mathbf{B}_{21}$ | = | $x_{11}\,\mathbf{a}_1 + y_{11}\,\mathbf{a}_2 + z_{11}\,\mathbf{a}_3$ | = | $(x_{11}a + z_{11}c\cos\beta)\,\hat{\mathbf{x}} + y_{11}b\,\hat{\mathbf{y}} + z_{11}c\sin\beta\,\hat{\mathbf{z}}$ | (2a) | O III |
| $\mathbf{B}_{22}$ | = | $x_{11}\,\mathbf{a}_1 - y_{11}\,\mathbf{a}_2 + \left(\frac{1}{2} + z_{11}\right)\mathbf{a}_3$ | = | $\left(\frac{1}{2}c\cos\beta + x_{11}a + z_{11}c\cos\beta\right)\hat{\mathbf{x}} - y_{11}b\,\hat{\mathbf{y}} + \left(\frac{1}{2} + z_{11}\right)c\sin\beta\,\hat{\mathbf{z}}$ | (2a) | O III |
| $\mathbf{B}_{23}$ | = | $x_{12}\,\mathbf{a}_1 + y_{12}\,\mathbf{a}_2 + z_{12}\,\mathbf{a}_3$ | = | $(x_{12}a + z_{12}c\cos\beta)\,\hat{\mathbf{x}} + y_{12}b\,\hat{\mathbf{y}} + z_{12}c\sin\beta\,\hat{\mathbf{z}}$ | (2a) | O IV |
| $\mathbf{B}_{24}$ | = | $x_{12}\,\mathbf{a}_1 - y_{12}\,\mathbf{a}_2 + \left(\frac{1}{2} + z_{12}\right)\mathbf{a}_3$ | = | $\left(\frac{1}{2}c\cos\beta + x_{12}a + z_{12}c\cos\beta\right)\hat{\mathbf{x}} - y_{12}b\,\hat{\mathbf{y}} + \left(\frac{1}{2} + z_{12}\right)c\sin\beta\,\hat{\mathbf{z}}$ | (2a) | O IV |
| $\mathbf{B}_{25}$ | = | $x_{13}\,\mathbf{a}_1 + y_{13}\,\mathbf{a}_2 + z_{13}\,\mathbf{a}_3$ | = | $(x_{13}a + z_{13}c\cos\beta)\,\hat{\mathbf{x}} + y_{13}b\,\hat{\mathbf{y}} + z_{13}c\sin\beta\,\hat{\mathbf{z}}$ | (2a) | O V |
| $\mathbf{B}_{26}$ | = | $x_{13}\,\mathbf{a}_1 - y_{13}\,\mathbf{a}_2 + \left(\frac{1}{2} + z_{13}\right)\mathbf{a}_3$ | = | $\left(\frac{1}{2}c\cos\beta + x_{13}a + z_{13}c\cos\beta\right)\hat{\mathbf{x}} - y_{13}b\,\hat{\mathbf{y}} + \left(\frac{1}{2} + z_{13}\right)c\sin\beta\,\hat{\mathbf{z}}$ | (2a) | O V |
| $\mathbf{B}_{27}$ | = | $x_{14}\,\mathbf{a}_1 + y_{14}\,\mathbf{a}_2 + z_{14}\,\mathbf{a}_3$ | = | $(x_{14}a + z_{14}c\cos\beta)\,\hat{\mathbf{x}} + y_{14}b\,\hat{\mathbf{y}} + z_{14}c\sin\beta\,\hat{\mathbf{z}}$ | (2a) | O VI |
| $\mathbf{B}_{28}$ | = | $x_{14}\,\mathbf{a}_1 - y_{14}\,\mathbf{a}_2 + \left(\frac{1}{2} + z_{14}\right)\mathbf{a}_3$ | = | $\left(\frac{1}{2}c\cos\beta + x_{14}a + z_{14}c\cos\beta\right)\hat{\mathbf{x}} - y_{14}b\,\hat{\mathbf{y}} + \left(\frac{1}{2} + z_{14}\right)c\sin\beta\,\hat{\mathbf{z}}$ | (2a) | O VI |
| $\mathbf{B}_{29}$ | = | $x_{15}\,\mathbf{a}_1 + y_{15}\,\mathbf{a}_2 + z_{15}\,\mathbf{a}_3$ | = | $(x_{15}a + z_{15}c\cos\beta)\,\hat{\mathbf{x}} + y_{15}b\,\hat{\mathbf{y}} + z_{15}c\sin\beta\,\hat{\mathbf{z}}$ | (2a) | O VII |
| $\mathbf{B}_{30}$ | = | $x_{15}\,\mathbf{a}_1 - y_{15}\,\mathbf{a}_2 + \left(\frac{1}{2} + z_{15}\right)\mathbf{a}_3$ | = | $\left(\frac{1}{2}c\cos\beta + x_{15}a + z_{15}c\cos\beta\right)\hat{\mathbf{x}} - y_{15}b\,\hat{\mathbf{y}} + \left(\frac{1}{2} + z_{15}\right)c\sin\beta\,\hat{\mathbf{z}}$ | (2a) | O VII |
| $\mathbf{B}_{31}$ | = | $x_{16}\,\mathbf{a}_1 + y_{16}\,\mathbf{a}_2 + z_{16}\,\mathbf{a}_3$ | = | $(x_{16}a + z_{16}c\cos\beta)\,\hat{\mathbf{x}} + y_{16}b\,\hat{\mathbf{y}} + z_{16}c\sin\beta\,\hat{\mathbf{z}}$ | (2a) | O VIII |
| $\mathbf{B}_{32}$ | = | $x_{16}\,\mathbf{a}_1 - y_{16}\,\mathbf{a}_2 + \left(\frac{1}{2} + z_{16}\right)\mathbf{a}_3$ | = | $\left(\frac{1}{2}c\cos\beta + x_{16}a + z_{16}c\cos\beta\right)\hat{\mathbf{x}} - y_{16}b\,\hat{\mathbf{y}} + \left(\frac{1}{2} + z_{16}\right)c\sin\beta\,\hat{\mathbf{z}}$ | (2a) | O VIII |
| $\mathbf{B}_{33}$ | = | $x_{17}\,\mathbf{a}_1 + y_{17}\,\mathbf{a}_2 + z_{17}\,\mathbf{a}_3$ | = | $(x_{17}a + z_{17}c\cos\beta)\,\hat{\mathbf{x}} + y_{17}b\,\hat{\mathbf{y}} + z_{17}c\sin\beta\,\hat{\mathbf{z}}$ | (2a) | O IX |
| $\mathbf{B}_{34}$ | = | $x_{17}\,\mathbf{a}_1 - y_{17}\,\mathbf{a}_2 + \left(\frac{1}{2} + z_{17}\right)\mathbf{a}_3$ | = | $\left(\frac{1}{2}c\cos\beta + x_{17}a + z_{17}c\cos\beta\right)\hat{\mathbf{x}} - y_{17}b\,\hat{\mathbf{y}} + \left(\frac{1}{2} + z_{17}\right)c\sin\beta\,\hat{\mathbf{z}}$ | (2a) | O IX |
| $\mathbf{B}_{35}$ | = | $x_{18}\,\mathbf{a}_1 + y_{18}\,\mathbf{a}_2 + z_{18}\,\mathbf{a}_3$ | = | $(x_{18}a + z_{18}c\cos\beta)\,\hat{\mathbf{x}} + y_{18}b\,\hat{\mathbf{y}} + z_{18}c\sin\beta\,\hat{\mathbf{z}}$ | (2a) | O X |
| $\mathbf{B}_{36}$ | = | $x_{18}\,\mathbf{a}_1 - y_{18}\,\mathbf{a}_2 + \left(\frac{1}{2} + z_{18}\right)\mathbf{a}_3$ | = | $\left(\frac{1}{2}c\cos\beta + x_{18}a + z_{18}c\cos\beta\right)\hat{\mathbf{x}} - y_{18}b\,\hat{\mathbf{y}} + \left(\frac{1}{2} + z_{18}\right)c\sin\beta\,\hat{\mathbf{z}}$ | (2a) | O X |
| $\mathbf{B}_{37}$ | = | $x_{19}\,\mathbf{a}_1 + y_{19}\,\mathbf{a}_2 + z_{19}\,\mathbf{a}_3$ | = | $(x_{19}a + z_{19}c\cos\beta)\,\hat{\mathbf{x}} + y_{19}b\,\hat{\mathbf{y}} + z_{19}c\sin\beta\,\hat{\mathbf{z}}$ | (2a) | O XI |
| $\mathbf{B}_{38}$ | = | $x_{19}\,\mathbf{a}_1 - y_{19}\,\mathbf{a}_2 + \left(\frac{1}{2} + z_{19}\right)\mathbf{a}_3$ | = | $\left(\frac{1}{2}c\cos\beta + x_{19}a + z_{19}c\cos\beta\right)\hat{\mathbf{x}} - y_{19}b\,\hat{\mathbf{y}} + \left(\frac{1}{2} + z_{19}\right)c\sin\beta\,\hat{\mathbf{z}}$ | (2a) | O XI |
| $\mathbf{B}_{39}$ | = | $x_{20}\,\mathbf{a}_1 + y_{20}\,\mathbf{a}_2 + z_{20}\,\mathbf{a}_3$ | = | $(x_{20}a + z_{20}c\cos\beta)\,\hat{\mathbf{x}} + y_{20}b\,\hat{\mathbf{y}} + z_{20}c\sin\beta\,\hat{\mathbf{z}}$ | (2a) | O XII |
| $\mathbf{B}_{40}$ | = | $x_{20}\,\mathbf{a}_1 - y_{20}\,\mathbf{a}_2 + \left(\frac{1}{2} + z_{20}\right)\mathbf{a}_3$ | = | $\left(\frac{1}{2}c\cos\beta + x_{20}a + z_{20}c\cos\beta\right)\hat{\mathbf{x}} - y_{20}b\,\hat{\mathbf{y}} + \left(\frac{1}{2} + z_{20}\right)c\sin\beta\,\hat{\mathbf{z}}$ | (2a) | O XII |
| $\mathbf{B}_{41}$ | = | $x_{21}\,\mathbf{a}_1 + y_{21}\,\mathbf{a}_2 + z_{21}\,\mathbf{a}_3$ | = | $(x_{21}a + z_{21}c\cos\beta)\,\hat{\mathbf{x}} + y_{21}b\,\hat{\mathbf{y}} + z_{21}c\sin\beta\,\hat{\mathbf{z}}$ | (2a) | O XIII |
| $\mathbf{B}_{42}$ | = | $x_{21}\,\mathbf{a}_1 - y_{21}\,\mathbf{a}_2 + \left(\frac{1}{2} + z_{21}\right)\mathbf{a}_3$ | = | $\left(\frac{1}{2}c\cos\beta + x_{21}a + z_{21}c\cos\beta\right)\hat{\mathbf{x}} - y_{21}b\,\hat{\mathbf{y}} + \left(\frac{1}{2} + z_{21}\right)c\sin\beta\,\hat{\mathbf{z}}$ | (2a) | O XIII |
| $\mathbf{B}_{43}$ | = | $x_{22}\,\mathbf{a}_1 + y_{22}\,\mathbf{a}_2 + z_{22}\,\mathbf{a}_3$ | = | $(x_{22}a + z_{22}c\cos\beta)\,\hat{\mathbf{x}} + y_{22}b\,\hat{\mathbf{y}} + z_{22}c\sin\beta\,\hat{\mathbf{z}}$ | (2a) | O XIV |
| $\mathbf{B}_{44}$ | = | $x_{22}\,\mathbf{a}_1 - y_{22}\,\mathbf{a}_2 + \left(\frac{1}{2} + z_{22}\right)\mathbf{a}_3$ | = | $\left(\frac{1}{2}c\cos\beta + x_{22}a + z_{22}c\cos\beta\right)\hat{\mathbf{x}} - y_{22}b\,\hat{\mathbf{y}} + \left(\frac{1}{2} + z_{22}\right)c\sin\beta\,\hat{\mathbf{z}}$ | (2a) | O XIV |
| $\mathbf{B}_{45}$ | = | $x_{23}\,\mathbf{a}_1 + y_{23}\,\mathbf{a}_2 + z_{23}\,\mathbf{a}_3$ | = | $(x_{23}a + z_{23}c\cos\beta)\,\hat{\mathbf{x}} + y_{23}b\,\hat{\mathbf{y}} + z_{23}c\sin\beta\,\hat{\mathbf{z}}$ | (2a) | O XV |
| $\mathbf{B}_{46}$ | = | $x_{23}\,\mathbf{a}_1 - y_{23}\,\mathbf{a}_2 + \left(\frac{1}{2} + z_{23}\right)\mathbf{a}_3$ | = | $\left(\frac{1}{2}c\cos\beta + x_{23}a + z_{23}c\cos\beta\right)\hat{\mathbf{x}} - y_{23}b\,\hat{\mathbf{y}} + \left(\frac{1}{2} + z_{23}\right)c\sin\beta\,\hat{\mathbf{z}}$ | (2a) | O XV |

$$\mathbf{B}_{47} = x_{24}\,\mathbf{a}_1 + y_{24}\,\mathbf{a}_2 + z_{24}\,\mathbf{a}_3 = \left(x_{24}a + z_{24}c\cos\beta\right)\,\hat{\mathbf{x}} + y_{24}b\,\hat{\mathbf{y}} + z_{24}c\sin\beta\,\hat{\mathbf{z}} \quad (2a) \quad \text{Si I}$$

$$\mathbf{B}_{48} = x_{24}\,\mathbf{a}_1 - y_{24}\,\mathbf{a}_2 + \left(\frac{1}{2} + z_{24}\right)\,\mathbf{a}_3 = \left(\frac{1}{2}c\cos\beta + x_{24}a + z_{24}c\cos\beta\right)\,\hat{\mathbf{x}} - y_{24}b\,\hat{\mathbf{y}} + \left(\frac{1}{2} + z_{24}\right)c\sin\beta\,\hat{\mathbf{z}} \quad (2a) \quad \text{Si I}$$

$$\mathbf{B}_{49} = x_{25}\,\mathbf{a}_1 + y_{25}\,\mathbf{a}_2 + z_{25}\,\mathbf{a}_3 = \left(x_{25}a + z_{25}c\cos\beta\right)\,\hat{\mathbf{x}} + y_{25}b\,\hat{\mathbf{y}} + z_{25}c\sin\beta\,\hat{\mathbf{z}} \quad (2a) \quad \text{Si II}$$

$$\mathbf{B}_{50} = x_{25}\,\mathbf{a}_1 - y_{25}\,\mathbf{a}_2 + \left(\frac{1}{2} + z_{25}\right)\,\mathbf{a}_3 = \left(\frac{1}{2}c\cos\beta + x_{25}a + z_{25}c\cos\beta\right)\,\hat{\mathbf{x}} - y_{25}b\,\hat{\mathbf{y}} + \left(\frac{1}{2} + z_{25}\right)c\sin\beta\,\hat{\mathbf{z}} \quad (2a) \quad \text{Si II}$$

$$\mathbf{B}_{51} = x_{26}\,\mathbf{a}_1 + y_{26}\,\mathbf{a}_2 + z_{26}\,\mathbf{a}_3 = \left(x_{26}a + z_{26}c\cos\beta\right)\,\hat{\mathbf{x}} + y_{26}b\,\hat{\mathbf{y}} + z_{26}c\sin\beta\,\hat{\mathbf{z}} \quad (2a) \quad \text{Si III}$$

$$\mathbf{B}_{52} = x_{26}\,\mathbf{a}_1 - y_{26}\,\mathbf{a}_2 + \left(\frac{1}{2} + z_{26}\right)\,\mathbf{a}_3 = \left(\frac{1}{2}c\cos\beta + x_{26}a + z_{26}c\cos\beta\right)\,\hat{\mathbf{x}} - y_{26}b\,\hat{\mathbf{y}} + \left(\frac{1}{2} + z_{26}\right)c\sin\beta\,\hat{\mathbf{z}} \quad (2a) \quad \text{Si III}$$

$$\mathbf{B}_{53} = x_{27}\,\mathbf{a}_1 + y_{27}\,\mathbf{a}_2 + z_{27}\,\mathbf{a}_3 = \left(x_{27}a + z_{27}c\cos\beta\right)\,\hat{\mathbf{x}} + y_{27}b\,\hat{\mathbf{y}} + z_{27}c\sin\beta\,\hat{\mathbf{z}} \quad (2a) \quad \text{Si IV}$$

$$\mathbf{B}_{54} = x_{27}\,\mathbf{a}_1 - y_{27}\,\mathbf{a}_2 + \left(\frac{1}{2} + z_{27}\right)\,\mathbf{a}_3 = \left(\frac{1}{2}c\cos\beta + x_{27}a + z_{27}c\cos\beta\right)\,\hat{\mathbf{x}} - y_{27}b\,\hat{\mathbf{y}} + \left(\frac{1}{2} + z_{27}\right)c\sin\beta\,\hat{\mathbf{z}} \quad (2a) \quad \text{Si IV}$$

**Geometry files:**

- CIF: pp. 1513
- POSCAR: pp. 1514

# Low-Temperature $Mo_8O_{23}$ Structure: A8B23_mP124_7_16a_46a

http://aflow.org/prototype-encyclopedia/A8B23_mP124_7_16a_46a

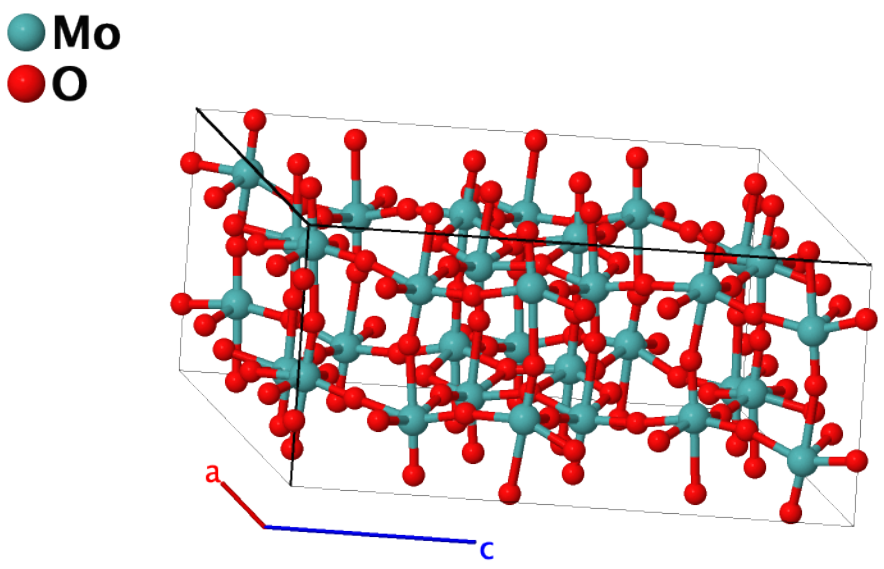

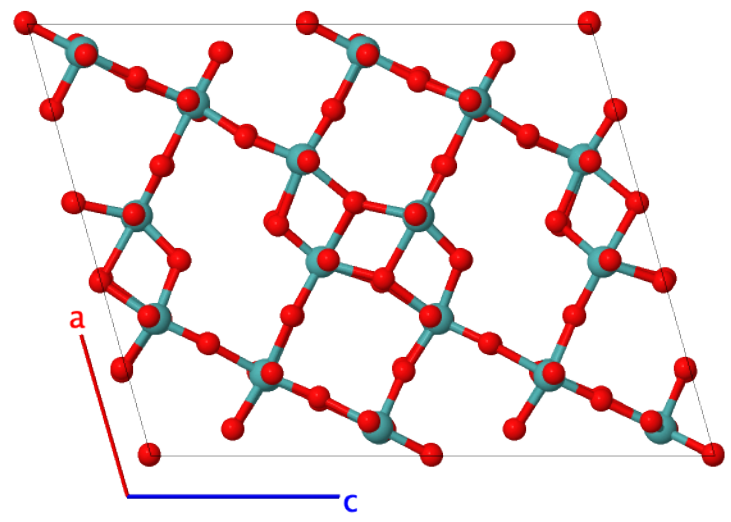

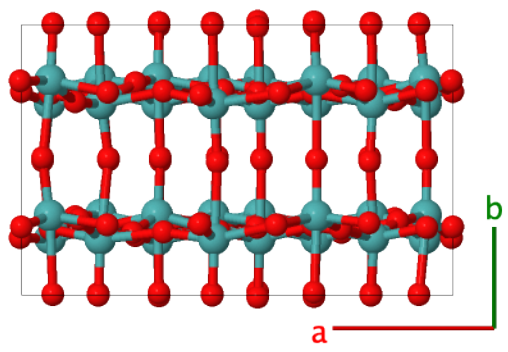

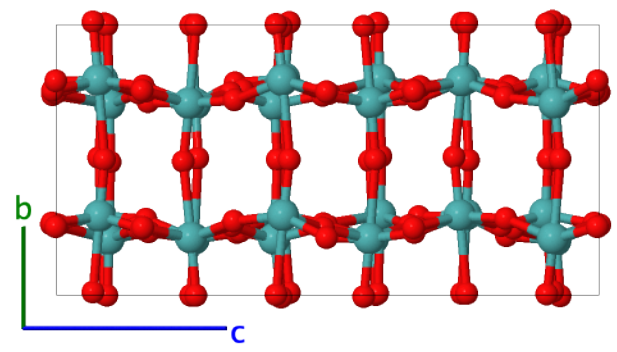

| | | |
|---|---|---|
| **Prototype** | : | $Mo_8O_{23}$ |
| **AFLOW prototype label** | : | A8B23_mP124_7_16a_46a |
| ***Strukturbericht* designation** | : | None |
| **Pearson symbol** | : | mP124 |
| **Space group number** | : | 7 |
| **Space group symbol** | : | *Pc* |
| **AFLOW prototype command** | : | `aflow --proto=A8B23_mP124_7_16a_46a` `--params=`$a, b/a, c/a, \beta, x_1, y_1, z_1, x_2, y_2, z_2, x_3, y_3, z_3, x_4, y_4, z_4, x_5, y_5, z_5, x_6, y_6, z_6, x_7, y_7, z_7, x_8, y_8, z_8, x_9, y_9, z_9, x_{10}, y_{10}, z_{10}, x_{11}, y_{11}, z_{11}, x_{12}, y_{12}, z_{12}, x_{13}, y_{13}, z_{13}, x_{14}, y_{14}, z_{14}, x_{15}, y_{15}, z_{15}, x_{16}, y_{16}, z_{16}, x_{17}, y_{17}, z_{17}, x_{18}, y_{18}, z_{18}, x_{19}, y_{19}, z_{19}, x_{20}, y_{20}, z_{20}, x_{21}, y_{21}, z_{21}, x_{22}, y_{22}, z_{22}, x_{23}, y_{23}, z_{23}, x_{24}, y_{24}, z_{24}, x_{25}, y_{25}, z_{25}, x_{26}, y_{26}, z_{26}, x_{27}, y_{27}, z_{27}, x_{28}, y_{28}, z_{28}, x_{29}, y_{29}, z_{29}, x_{30}, y_{30}, z_{30}, x_{31}, y_{31}, z_{31}, x_{32}, y_{32}, z_{32}, x_{33}, y_{33}, z_{33}, x_{34}, y_{34}, z_{34}, x_{35}, y_{35}, z_{35}, x_{36}, y_{36}, z_{36}, x_{37}, y_{37}, z_{37}, x_{38}, y_{38}, z_{38}, x_{39}, y_{39}, z_{39}, x_{40}, y_{40}, z_{40}, x_{41}, y_{41}, z_{41}, x_{42}, y_{42}, z_{42}, x_{43}, y_{43}, z_{43}, x_{44}, y_{44}, z_{44}, x_{45}, y_{45}, z_{45}, x_{46}, y_{46}, z_{46}, x_{47}, y_{47}, z_{47}, x_{48}, y_{48}, z_{48}, x_{49}, y_{49}, z_{49}, x_{50}, y_{50}, z_{50}, x_{51}, y_{51}, z_{51}, x_{52}, y_{52}, z_{52}, x_{53}, y_{53}, z_{53}, x_{54}, y_{54}, z_{54}, x_{55}, y_{55}, z_{55}, x_{56}, y_{56}, z_{56}, x_{57}, y_{57}, z_{57}, x_{58}, y_{58}, z_{58}, x_{59}, y_{59}, z_{59}, x_{60}, y_{60}, z_{60}, x_{61}, y_{61}, z_{61}, x_{62}, y_{62}, z_{62}$ |

- This data was taken at 100 K. Above 285 K the structure exhibits an incommensurate charge density wave, which is approximated by the high temperature $Mo_8O_{23}$ structure.

**Simple Monoclinic primitive vectors:**

$$\mathbf{a}_1 = a\,\hat{\mathbf{x}}$$
$$\mathbf{a}_2 = b\,\hat{\mathbf{y}}$$
$$\mathbf{a}_3 = c\cos\beta\,\hat{\mathbf{x}} + c\sin\beta\,\hat{\mathbf{z}}$$

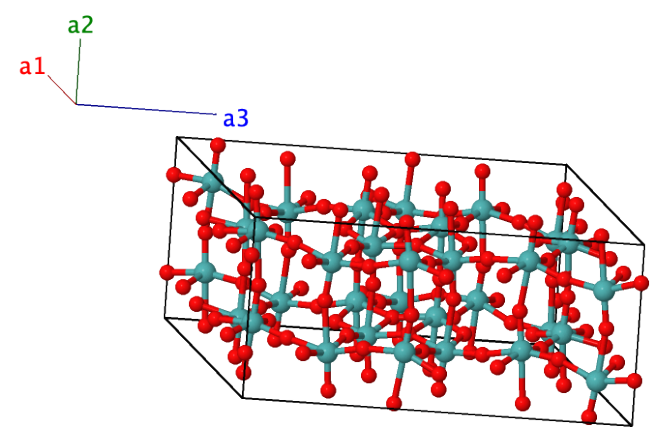

**Basis vectors:**

| | | Lattice Coordinates | | Cartesian Coordinates | Wyckoff Position | Atom Type |
|---|---|---|---|---|---|---|
| $\mathbf{B}_1$ | = | $x_1\,\mathbf{a}_1 + y_1\,\mathbf{a}_2 + z_1\,\mathbf{a}_3$ | = | $(x_1 a + z_1 c\cos\beta)\,\hat{\mathbf{x}} + y_1 b\,\hat{\mathbf{y}} + z_1 c\sin\beta\,\hat{\mathbf{z}}$ | (2a) | Mo I |
| $\mathbf{B}_2$ | = | $x_1\,\mathbf{a}_1 - y_1\,\mathbf{a}_2 + \left(\frac{1}{2} + z_1\right)\mathbf{a}_3$ | = | $\left(\frac{1}{2}c\cos\beta + x_1 a + z_1 c\cos\beta\right)\hat{\mathbf{x}} - y_1 b\,\hat{\mathbf{y}} + \left(\frac{1}{2} + z_1\right)c\sin\beta\,\hat{\mathbf{z}}$ | (2a) | Mo I |
| $\mathbf{B}_3$ | = | $x_2\,\mathbf{a}_1 + y_2\,\mathbf{a}_2 + z_2\,\mathbf{a}_3$ | = | $(x_2 a + z_2 c\cos\beta)\,\hat{\mathbf{x}} + y_2 b\,\hat{\mathbf{y}} + z_2 c\sin\beta\,\hat{\mathbf{z}}$ | (2a) | Mo II |
| $\mathbf{B}_4$ | = | $x_2\,\mathbf{a}_1 - y_2\,\mathbf{a}_2 + \left(\frac{1}{2} + z_2\right)\mathbf{a}_3$ | = | $\left(\frac{1}{2}c\cos\beta + x_2 a + z_2 c\cos\beta\right)\hat{\mathbf{x}} - y_2 b\,\hat{\mathbf{y}} + \left(\frac{1}{2} + z_2\right)c\sin\beta\,\hat{\mathbf{z}}$ | (2a) | Mo II |
| $\mathbf{B}_5$ | = | $x_3\,\mathbf{a}_1 + y_3\,\mathbf{a}_2 + z_3\,\mathbf{a}_3$ | = | $(x_3 a + z_3 c\cos\beta)\,\hat{\mathbf{x}} + y_3 b\,\hat{\mathbf{y}} + z_3 c\sin\beta\,\hat{\mathbf{z}}$ | (2a) | Mo III |
| $\mathbf{B}_6$ | = | $x_3\,\mathbf{a}_1 - y_3\,\mathbf{a}_2 + \left(\frac{1}{2} + z_3\right)\mathbf{a}_3$ | = | $\left(\frac{1}{2}c\cos\beta + x_3 a + z_3 c\cos\beta\right)\hat{\mathbf{x}} - y_3 b\,\hat{\mathbf{y}} + \left(\frac{1}{2} + z_3\right)c\sin\beta\,\hat{\mathbf{z}}$ | (2a) | Mo III |
| $\mathbf{B}_7$ | = | $x_4\,\mathbf{a}_1 + y_4\,\mathbf{a}_2 + z_4\,\mathbf{a}_3$ | = | $(x_4 a + z_4 c\cos\beta)\,\hat{\mathbf{x}} + y_4 b\,\hat{\mathbf{y}} + z_4 c\sin\beta\,\hat{\mathbf{z}}$ | (2a) | Mo IV |
| $\mathbf{B}_8$ | = | $x_4\,\mathbf{a}_1 - y_4\,\mathbf{a}_2 + \left(\frac{1}{2} + z_4\right)\mathbf{a}_3$ | = | $\left(\frac{1}{2}c\cos\beta + x_4 a + z_4 c\cos\beta\right)\hat{\mathbf{x}} - y_4 b\,\hat{\mathbf{y}} + \left(\frac{1}{2} + z_4\right)c\sin\beta\,\hat{\mathbf{z}}$ | (2a) | Mo IV |
| $\mathbf{B}_9$ | = | $x_5\,\mathbf{a}_1 + y_5\,\mathbf{a}_2 + z_5\,\mathbf{a}_3$ | = | $(x_5 a + z_5 c\cos\beta)\,\hat{\mathbf{x}} + y_5 b\,\hat{\mathbf{y}} + z_5 c\sin\beta\,\hat{\mathbf{z}}$ | (2a) | Mo V |
| $\mathbf{B}_{10}$ | = | $x_5\,\mathbf{a}_1 - y_5\,\mathbf{a}_2 + \left(\frac{1}{2} + z_5\right)\mathbf{a}_3$ | = | $\left(\frac{1}{2}c\cos\beta + x_5 a + z_5 c\cos\beta\right)\hat{\mathbf{x}} - y_5 b\,\hat{\mathbf{y}} + \left(\frac{1}{2} + z_5\right)c\sin\beta\,\hat{\mathbf{z}}$ | (2a) | Mo V |
| $\mathbf{B}_{11}$ | = | $x_6\,\mathbf{a}_1 + y_6\,\mathbf{a}_2 + z_6\,\mathbf{a}_3$ | = | $(x_6 a + z_6 c\cos\beta)\,\hat{\mathbf{x}} + y_6 b\,\hat{\mathbf{y}} + z_6 c\sin\beta\,\hat{\mathbf{z}}$ | (2a) | Mo VI |
| $\mathbf{B}_{12}$ | = | $x_6\,\mathbf{a}_1 - y_6\,\mathbf{a}_2 + \left(\frac{1}{2} + z_6\right)\mathbf{a}_3$ | = | $\left(\frac{1}{2}c\cos\beta + x_6 a + z_6 c\cos\beta\right)\hat{\mathbf{x}} - y_6 b\,\hat{\mathbf{y}} + \left(\frac{1}{2} + z_6\right)c\sin\beta\,\hat{\mathbf{z}}$ | (2a) | Mo VI |
| $\mathbf{B}_{13}$ | = | $x_7\,\mathbf{a}_1 + y_7\,\mathbf{a}_2 + z_7\,\mathbf{a}_3$ | = | $(x_7 a + z_7 c\cos\beta)\,\hat{\mathbf{x}} + y_7 b\,\hat{\mathbf{y}} + z_7 c\sin\beta\,\hat{\mathbf{z}}$ | (2a) | Mo VII |
| $\mathbf{B}_{14}$ | = | $x_7\,\mathbf{a}_1 - y_7\,\mathbf{a}_2 + \left(\frac{1}{2} + z_7\right)\mathbf{a}_3$ | = | $\left(\frac{1}{2}c\cos\beta + x_7 a + z_7 c\cos\beta\right)\hat{\mathbf{x}} - y_7 b\,\hat{\mathbf{y}} + \left(\frac{1}{2} + z_7\right)c\sin\beta\,\hat{\mathbf{z}}$ | (2a) | Mo VII |
| $\mathbf{B}_{15}$ | = | $x_8\,\mathbf{a}_1 + y_8\,\mathbf{a}_2 + z_8\,\mathbf{a}_3$ | = | $(x_8 a + z_8 c\cos\beta)\,\hat{\mathbf{x}} + y_8 b\,\hat{\mathbf{y}} + z_8 c\sin\beta\,\hat{\mathbf{z}}$ | (2a) | Mo VIII |
| $\mathbf{B}_{16}$ | = | $x_8\,\mathbf{a}_1 - y_8\,\mathbf{a}_2 + \left(\frac{1}{2} + z_8\right)\mathbf{a}_3$ | = | $\left(\frac{1}{2}c\cos\beta + x_8 a + z_8 c\cos\beta\right)\hat{\mathbf{x}} - y_8 b\,\hat{\mathbf{y}} + \left(\frac{1}{2} + z_8\right)c\sin\beta\,\hat{\mathbf{z}}$ | (2a) | Mo VIII |
| $\mathbf{B}_{17}$ | = | $x_9\,\mathbf{a}_1 + y_9\,\mathbf{a}_2 + z_9\,\mathbf{a}_3$ | = | $(x_9 a + z_9 c\cos\beta)\,\hat{\mathbf{x}} + y_9 b\,\hat{\mathbf{y}} + z_9 c\sin\beta\,\hat{\mathbf{z}}$ | (2a) | Mo IX |
| $\mathbf{B}_{18}$ | = | $x_9\,\mathbf{a}_1 - y_9\,\mathbf{a}_2 + \left(\frac{1}{2} + z_9\right)\mathbf{a}_3$ | = | $\left(\frac{1}{2}c\cos\beta + x_9 a + z_9 c\cos\beta\right)\hat{\mathbf{x}} - y_9 b\,\hat{\mathbf{y}} + \left(\frac{1}{2} + z_9\right)c\sin\beta\,\hat{\mathbf{z}}$ | (2a) | Mo IX |
| $\mathbf{B}_{19}$ | = | $x_{10}\,\mathbf{a}_1 + y_{10}\,\mathbf{a}_2 + z_{10}\,\mathbf{a}_3$ | = | $(x_{10} a + z_{10} c\cos\beta)\,\hat{\mathbf{x}} + y_{10} b\,\hat{\mathbf{y}} + z_{10} c\sin\beta\,\hat{\mathbf{z}}$ | (2a) | Mo X |
| $\mathbf{B}_{20}$ | = | $x_{10}\,\mathbf{a}_1 - y_{10}\,\mathbf{a}_2 + \left(\frac{1}{2} + z_{10}\right)\mathbf{a}_3$ | = | $\left(\frac{1}{2}c\cos\beta + x_{10} a + z_{10} c\cos\beta\right)\hat{\mathbf{x}} - y_{10} b\,\hat{\mathbf{y}} + \left(\frac{1}{2} + z_{10}\right)c\sin\beta\,\hat{\mathbf{z}}$ | (2a) | Mo X |
| $\mathbf{B}_{21}$ | = | $x_{11}\,\mathbf{a}_{1}+y_{11}\,\mathbf{a}_{2}+z_{11}\,\mathbf{a}_{3}$ | = | $\left(x_{11}a+z_{11}c\cos\beta\right)\,\mathbf{\hat{x}}+y_{11}b\,\mathbf{\hat{y}}+z_{11}c\sin\beta\,\mathbf{\hat{z}}$ | $(2a)$ | Mo XI |
| $\mathbf{B}_{22}$ | = | $x_{11}\,\mathbf{a}_{1}-y_{11}\,\mathbf{a}_{2}+\left(\frac{1}{2}+z_{11}\right)\,\mathbf{a}_{3}$ | = | $\left(\frac{1}{2}c\cos\beta+x_{11}a+z_{11}c\cos\beta\right)\,\mathbf{\hat{x}}-y_{11}b\,\mathbf{\hat{y}}+\left(\frac{1}{2}+z_{11}\right)c\sin\beta\,\mathbf{\hat{z}}$ | $(2a)$ | Mo XI |
| $\mathbf{B}_{23}$ | = | $x_{12}\,\mathbf{a}_{1}+y_{12}\,\mathbf{a}_{2}+z_{12}\,\mathbf{a}_{3}$ | = | $\left(x_{12}a+z_{12}c\cos\beta\right)\,\mathbf{\hat{x}}+y_{12}b\,\mathbf{\hat{y}}+z_{12}c\sin\beta\,\mathbf{\hat{z}}$ | $(2a)$ | Mo XII |
| $\mathbf{B}_{24}$ | = | $x_{12}\,\mathbf{a}_{1}-y_{12}\,\mathbf{a}_{2}+\left(\frac{1}{2}+z_{12}\right)\,\mathbf{a}_{3}$ | = | $\left(\frac{1}{2}c\cos\beta+x_{12}a+z_{12}c\cos\beta\right)\,\mathbf{\hat{x}}-y_{12}b\,\mathbf{\hat{y}}+\left(\frac{1}{2}+z_{12}\right)c\sin\beta\,\mathbf{\hat{z}}$ | $(2a)$ | Mo XII |
| $\mathbf{B}_{25}$ | = | $x_{13}\,\mathbf{a}_{1}+y_{13}\,\mathbf{a}_{2}+z_{13}\,\mathbf{a}_{3}$ | = | $\left(x_{13}a+z_{13}c\cos\beta\right)\,\mathbf{\hat{x}}+y_{13}b\,\mathbf{\hat{y}}+z_{13}c\sin\beta\,\mathbf{\hat{z}}$ | $(2a)$ | Mo XIII |
| $\mathbf{B}_{26}$ | = | $x_{13}\,\mathbf{a}_{1}-y_{13}\,\mathbf{a}_{2}+\left(\frac{1}{2}+z_{13}\right)\,\mathbf{a}_{3}$ | = | $\left(\frac{1}{2}c\cos\beta+x_{13}a+z_{13}c\cos\beta\right)\,\mathbf{\hat{x}}-y_{13}b\,\mathbf{\hat{y}}+\left(\frac{1}{2}+z_{13}\right)c\sin\beta\,\mathbf{\hat{z}}$ | $(2a)$ | Mo XIII |
| $\mathbf{B}_{27}$ | = | $x_{14}\,\mathbf{a}_{1}+y_{14}\,\mathbf{a}_{2}+z_{14}\,\mathbf{a}_{3}$ | = | $\left(x_{14}a+z_{14}c\cos\beta\right)\,\mathbf{\hat{x}}+y_{14}b\,\mathbf{\hat{y}}+z_{14}c\sin\beta\,\mathbf{\hat{z}}$ | $(2a)$ | Mo XIV |
| $\mathbf{B}_{28}$ | = | $x_{14}\,\mathbf{a}_{1}-y_{14}\,\mathbf{a}_{2}+\left(\frac{1}{2}+z_{14}\right)\,\mathbf{a}_{3}$ | = | $\left(\frac{1}{2}c\cos\beta+x_{14}a+z_{14}c\cos\beta\right)\,\mathbf{\hat{x}}-y_{14}b\,\mathbf{\hat{y}}+\left(\frac{1}{2}+z_{14}\right)c\sin\beta\,\mathbf{\hat{z}}$ | $(2a)$ | Mo XIV |
| $\mathbf{B}_{29}$ | = | $x_{15}\,\mathbf{a}_{1}+y_{15}\,\mathbf{a}_{2}+z_{15}\,\mathbf{a}_{3}$ | = | $\left(x_{15}a+z_{15}c\cos\beta\right)\,\mathbf{\hat{x}}+y_{15}b\,\mathbf{\hat{y}}+z_{15}c\sin\beta\,\mathbf{\hat{z}}$ | $(2a)$ | Mo XV |
| $\mathbf{B}_{30}$ | = | $x_{15}\,\mathbf{a}_{1}-y_{15}\,\mathbf{a}_{2}+\left(\frac{1}{2}+z_{15}\right)\,\mathbf{a}_{3}$ | = | $\left(\frac{1}{2}c\cos\beta+x_{15}a+z_{15}c\cos\beta\right)\,\mathbf{\hat{x}}-y_{15}b\,\mathbf{\hat{y}}+\left(\frac{1}{2}+z_{15}\right)c\sin\beta\,\mathbf{\hat{z}}$ | $(2a)$ | Mo XV |
| $\mathbf{B}_{31}$ | = | $x_{16}\,\mathbf{a}_{1}+y_{16}\,\mathbf{a}_{2}+z_{16}\,\mathbf{a}_{3}$ | = | $\left(x_{16}a+z_{16}c\cos\beta\right)\,\mathbf{\hat{x}}+y_{16}b\,\mathbf{\hat{y}}+z_{16}c\sin\beta\,\mathbf{\hat{z}}$ | $(2a)$ | Mo XVI |
| $\mathbf{B}_{32}$ | = | $x_{16}\,\mathbf{a}_{1}-y_{16}\,\mathbf{a}_{2}+\left(\frac{1}{2}+z_{16}\right)\,\mathbf{a}_{3}$ | = | $\left(\frac{1}{2}c\cos\beta+x_{16}a+z_{16}c\cos\beta\right)\,\mathbf{\hat{x}}-y_{16}b\,\mathbf{\hat{y}}+\left(\frac{1}{2}+z_{16}\right)c\sin\beta\,\mathbf{\hat{z}}$ | $(2a)$ | Mo XVI |
| $\mathbf{B}_{33}$ | = | $x_{17}\,\mathbf{a}_{1}+y_{17}\,\mathbf{a}_{2}+z_{17}\,\mathbf{a}_{3}$ | = | $\left(x_{17}a+z_{17}c\cos\beta\right)\,\mathbf{\hat{x}}+y_{17}b\,\mathbf{\hat{y}}+z_{17}c\sin\beta\,\mathbf{\hat{z}}$ | $(2a)$ | O I |
| $\mathbf{B}_{34}$ | = | $x_{17}\,\mathbf{a}_{1}-y_{17}\,\mathbf{a}_{2}+\left(\frac{1}{2}+z_{17}\right)\,\mathbf{a}_{3}$ | = | $\left(\frac{1}{2}c\cos\beta+x_{17}a+z_{17}c\cos\beta\right)\,\mathbf{\hat{x}}-y_{17}b\,\mathbf{\hat{y}}+\left(\frac{1}{2}+z_{17}\right)c\sin\beta\,\mathbf{\hat{z}}$ | $(2a)$ | O I |
| $\mathbf{B}_{35}$ | = | $x_{18}\,\mathbf{a}_{1}+y_{18}\,\mathbf{a}_{2}+z_{18}\,\mathbf{a}_{3}$ | = | $\left(x_{18}a+z_{18}c\cos\beta\right)\,\mathbf{\hat{x}}+y_{18}b\,\mathbf{\hat{y}}+z_{18}c\sin\beta\,\mathbf{\hat{z}}$ | $(2a)$ | O II |
| $\mathbf{B}_{36}$ | = | $x_{18}\,\mathbf{a}_{1}-y_{18}\,\mathbf{a}_{2}+\left(\frac{1}{2}+z_{18}\right)\,\mathbf{a}_{3}$ | = | $\left(\frac{1}{2}c\cos\beta+x_{18}a+z_{18}c\cos\beta\right)\,\mathbf{\hat{x}}-y_{18}b\,\mathbf{\hat{y}}+\left(\frac{1}{2}+z_{18}\right)c\sin\beta\,\mathbf{\hat{z}}$ | $(2a)$ | O II |
| $\mathbf{B}_{37}$ | = | $x_{19}\,\mathbf{a}_{1}+y_{19}\,\mathbf{a}_{2}+z_{19}\,\mathbf{a}_{3}$ | = | $\left(x_{19}a+z_{19}c\cos\beta\right)\,\mathbf{\hat{x}}+y_{19}b\,\mathbf{\hat{y}}+z_{19}c\sin\beta\,\mathbf{\hat{z}}$ | $(2a)$ | O III |
| $\mathbf{B}_{38}$ | = | $x_{19}\,\mathbf{a}_{1}-y_{19}\,\mathbf{a}_{2}+\left(\frac{1}{2}+z_{19}\right)\,\mathbf{a}_{3}$ | = | $\left(\frac{1}{2}c\cos\beta+x_{19}a+z_{19}c\cos\beta\right)\,\mathbf{\hat{x}}-y_{19}b\,\mathbf{\hat{y}}+\left(\frac{1}{2}+z_{19}\right)c\sin\beta\,\mathbf{\hat{z}}$ | $(2a)$ | O III |
| $\mathbf{B}_{39}$ | = | $x_{20}\,\mathbf{a}_{1}+y_{20}\,\mathbf{a}_{2}+z_{20}\,\mathbf{a}_{3}$ | = | $\left(x_{20}a+z_{20}c\cos\beta\right)\,\mathbf{\hat{x}}+y_{20}b\,\mathbf{\hat{y}}+z_{20}c\sin\beta\,\mathbf{\hat{z}}$ | $(2a)$ | O IV |
| $\mathbf{B}_{40}$ | = | $x_{20}\,\mathbf{a}_{1}-y_{20}\,\mathbf{a}_{2}+\left(\frac{1}{2}+z_{20}\right)\,\mathbf{a}_{3}$ | = | $\left(\frac{1}{2}c\cos\beta+x_{20}a+z_{20}c\cos\beta\right)\,\mathbf{\hat{x}}-y_{20}b\,\mathbf{\hat{y}}+\left(\frac{1}{2}+z_{20}\right)c\sin\beta\,\mathbf{\hat{z}}$ | $(2a)$ | O IV |
| $\mathbf{B}_{41}$ | = | $x_{21}\,\mathbf{a}_{1}+y_{21}\,\mathbf{a}_{2}+z_{21}\,\mathbf{a}_{3}$ | = | $\left(x_{21}a+z_{21}c\cos\beta\right)\,\mathbf{\hat{x}}+y_{21}b\,\mathbf{\hat{y}}+z_{21}c\sin\beta\,\mathbf{\hat{z}}$ | $(2a)$ | O V |
| $\mathbf{B}_{42}$ | = | $x_{21}\,\mathbf{a}_{1}-y_{21}\,\mathbf{a}_{2}+\left(\frac{1}{2}+z_{21}\right)\,\mathbf{a}_{3}$ | = | $\left(\frac{1}{2}c\cos\beta+x_{21}a+z_{21}c\cos\beta\right)\,\mathbf{\hat{x}}-y_{21}b\,\mathbf{\hat{y}}+\left(\frac{1}{2}+z_{21}\right)c\sin\beta\,\mathbf{\hat{z}}$ | $(2a)$ | O V |

$$\mathbf{B}_{43} = x_{22}\,\mathbf{a}_1 + y_{22}\,\mathbf{a}_2 + z_{22}\,\mathbf{a}_3 = (x_{22}a + z_{22}c\cos\beta)\,\hat{\mathbf{x}} + y_{22}b\,\hat{\mathbf{y}} + z_{22}c\sin\beta\,\hat{\mathbf{z}} \quad (2a) \quad \text{O VI}$$

$$\mathbf{B}_{44} = x_{22}\,\mathbf{a}_1 - y_{22}\,\mathbf{a}_2 + \left(\tfrac{1}{2} + z_{22}\right)\mathbf{a}_3 = \left(\tfrac{1}{2}c\cos\beta + x_{22}a + z_{22}c\cos\beta\right)\hat{\mathbf{x}} - y_{22}b\,\hat{\mathbf{y}} + \left(\tfrac{1}{2} + z_{22}\right)c\sin\beta\,\hat{\mathbf{z}} \quad (2a) \quad \text{O VI}$$

$$\mathbf{B}_{45} = x_{23}\,\mathbf{a}_1 + y_{23}\,\mathbf{a}_2 + z_{23}\,\mathbf{a}_3 = (x_{23}a + z_{23}c\cos\beta)\,\hat{\mathbf{x}} + y_{23}b\,\hat{\mathbf{y}} + z_{23}c\sin\beta\,\hat{\mathbf{z}} \quad (2a) \quad \text{O VII}$$

$$\mathbf{B}_{46} = x_{23}\,\mathbf{a}_1 - y_{23}\,\mathbf{a}_2 + \left(\tfrac{1}{2} + z_{23}\right)\mathbf{a}_3 = \left(\tfrac{1}{2}c\cos\beta + x_{23}a + z_{23}c\cos\beta\right)\hat{\mathbf{x}} - y_{23}b\,\hat{\mathbf{y}} + \left(\tfrac{1}{2} + z_{23}\right)c\sin\beta\,\hat{\mathbf{z}} \quad (2a) \quad \text{O VII}$$

$$\mathbf{B}_{47} = x_{24}\,\mathbf{a}_1 + y_{24}\,\mathbf{a}_2 + z_{24}\,\mathbf{a}_3 = (x_{24}a + z_{24}c\cos\beta)\,\hat{\mathbf{x}} + y_{24}b\,\hat{\mathbf{y}} + z_{24}c\sin\beta\,\hat{\mathbf{z}} \quad (2a) \quad \text{O VIII}$$

$$\mathbf{B}_{48} = x_{24}\,\mathbf{a}_1 - y_{24}\,\mathbf{a}_2 + \left(\tfrac{1}{2} + z_{24}\right)\mathbf{a}_3 = \left(\tfrac{1}{2}c\cos\beta + x_{24}a + z_{24}c\cos\beta\right)\hat{\mathbf{x}} - y_{24}b\,\hat{\mathbf{y}} + \left(\tfrac{1}{2} + z_{24}\right)c\sin\beta\,\hat{\mathbf{z}} \quad (2a) \quad \text{O VIII}$$

$$\mathbf{B}_{49} = x_{25}\,\mathbf{a}_1 + y_{25}\,\mathbf{a}_2 + z_{25}\,\mathbf{a}_3 = (x_{25}a + z_{25}c\cos\beta)\,\hat{\mathbf{x}} + y_{25}b\,\hat{\mathbf{y}} + z_{25}c\sin\beta\,\hat{\mathbf{z}} \quad (2a) \quad \text{O IX}$$

$$\mathbf{B}_{50} = x_{25}\,\mathbf{a}_1 - y_{25}\,\mathbf{a}_2 + \left(\tfrac{1}{2} + z_{25}\right)\mathbf{a}_3 = \left(\tfrac{1}{2}c\cos\beta + x_{25}a + z_{25}c\cos\beta\right)\hat{\mathbf{x}} - y_{25}b\,\hat{\mathbf{y}} + \left(\tfrac{1}{2} + z_{25}\right)c\sin\beta\,\hat{\mathbf{z}} \quad (2a) \quad \text{O IX}$$

$$\mathbf{B}_{51} = x_{26}\,\mathbf{a}_1 + y_{26}\,\mathbf{a}_2 + z_{26}\,\mathbf{a}_3 = (x_{26}a + z_{26}c\cos\beta)\,\hat{\mathbf{x}} + y_{26}b\,\hat{\mathbf{y}} + z_{26}c\sin\beta\,\hat{\mathbf{z}} \quad (2a) \quad \text{O X}$$

$$\mathbf{B}_{52} = x_{26}\,\mathbf{a}_1 - y_{26}\,\mathbf{a}_2 + \left(\tfrac{1}{2} + z_{26}\right)\mathbf{a}_3 = \left(\tfrac{1}{2}c\cos\beta + x_{26}a + z_{26}c\cos\beta\right)\hat{\mathbf{x}} - y_{26}b\,\hat{\mathbf{y}} + \left(\tfrac{1}{2} + z_{26}\right)c\sin\beta\,\hat{\mathbf{z}} \quad (2a) \quad \text{O X}$$

$$\mathbf{B}_{53} = x_{27}\,\mathbf{a}_1 + y_{27}\,\mathbf{a}_2 + z_{27}\,\mathbf{a}_3 = (x_{27}a + z_{27}c\cos\beta)\,\hat{\mathbf{x}} + y_{27}b\,\hat{\mathbf{y}} + z_{27}c\sin\beta\,\hat{\mathbf{z}} \quad (2a) \quad \text{O XI}$$

$$\mathbf{B}_{54} = x_{27}\,\mathbf{a}_1 - y_{27}\,\mathbf{a}_2 + \left(\tfrac{1}{2} + z_{27}\right)\mathbf{a}_3 = \left(\tfrac{1}{2}c\cos\beta + x_{27}a + z_{27}c\cos\beta\right)\hat{\mathbf{x}} - y_{27}b\,\hat{\mathbf{y}} + \left(\tfrac{1}{2} + z_{27}\right)c\sin\beta\,\hat{\mathbf{z}} \quad (2a) \quad \text{O XI}$$

$$\mathbf{B}_{55} = x_{28}\,\mathbf{a}_1 + y_{28}\,\mathbf{a}_2 + z_{28}\,\mathbf{a}_3 = (x_{28}a + z_{28}c\cos\beta)\,\hat{\mathbf{x}} + y_{28}b\,\hat{\mathbf{y}} + z_{28}c\sin\beta\,\hat{\mathbf{z}} \quad (2a) \quad \text{O XII}$$

$$\mathbf{B}_{56} = x_{28}\,\mathbf{a}_1 - y_{28}\,\mathbf{a}_2 + \left(\tfrac{1}{2} + z_{28}\right)\mathbf{a}_3 = \left(\tfrac{1}{2}c\cos\beta + x_{28}a + z_{28}c\cos\beta\right)\hat{\mathbf{x}} - y_{28}b\,\hat{\mathbf{y}} + \left(\tfrac{1}{2} + z_{28}\right)c\sin\beta\,\hat{\mathbf{z}} \quad (2a) \quad \text{O XII}$$

$$\mathbf{B}_{57} = x_{29}\,\mathbf{a}_1 + y_{29}\,\mathbf{a}_2 + z_{29}\,\mathbf{a}_3 = (x_{29}a + z_{29}c\cos\beta)\,\hat{\mathbf{x}} + y_{29}b\,\hat{\mathbf{y}} + z_{29}c\sin\beta\,\hat{\mathbf{z}} \quad (2a) \quad \text{O XIII}$$

$$\mathbf{B}_{58} = x_{29}\,\mathbf{a}_1 - y_{29}\,\mathbf{a}_2 + \left(\tfrac{1}{2} + z_{29}\right)\mathbf{a}_3 = \left(\tfrac{1}{2}c\cos\beta + x_{29}a + z_{29}c\cos\beta\right)\hat{\mathbf{x}} - y_{29}b\,\hat{\mathbf{y}} + \left(\tfrac{1}{2} + z_{29}\right)c\sin\beta\,\hat{\mathbf{z}} \quad (2a) \quad \text{O XIII}$$

$$\mathbf{B}_{59} = x_{30}\,\mathbf{a}_1 + y_{30}\,\mathbf{a}_2 + z_{30}\,\mathbf{a}_3 = (x_{30}a + z_{30}c\cos\beta)\,\hat{\mathbf{x}} + y_{30}b\,\hat{\mathbf{y}} + z_{30}c\sin\beta\,\hat{\mathbf{z}} \quad (2a) \quad \text{O XIV}$$

$$\mathbf{B}_{60} = x_{30}\,\mathbf{a}_1 - y_{30}\,\mathbf{a}_2 + \left(\tfrac{1}{2} + z_{30}\right)\mathbf{a}_3 = \left(\tfrac{1}{2}c\cos\beta + x_{30}a + z_{30}c\cos\beta\right)\hat{\mathbf{x}} - y_{30}b\,\hat{\mathbf{y}} + \left(\tfrac{1}{2} + z_{30}\right)c\sin\beta\,\hat{\mathbf{z}} \quad (2a) \quad \text{O XIV}$$

$$\mathbf{B}_{61} = x_{31}\,\mathbf{a}_1 + y_{31}\,\mathbf{a}_2 + z_{31}\,\mathbf{a}_3 = (x_{31}a + z_{31}c\cos\beta)\,\hat{\mathbf{x}} + y_{31}b\,\hat{\mathbf{y}} + z_{31}c\sin\beta\,\hat{\mathbf{z}} \quad (2a) \quad \text{O XV}$$

$$\mathbf{B}_{62} = x_{31}\,\mathbf{a}_1 - y_{31}\,\mathbf{a}_2 + \left(\tfrac{1}{2} + z_{31}\right)\mathbf{a}_3 = \left(\tfrac{1}{2}c\cos\beta + x_{31}a + z_{31}c\cos\beta\right)\hat{\mathbf{x}} - y_{31}b\,\hat{\mathbf{y}} + \left(\tfrac{1}{2} + z_{31}\right)c\sin\beta\,\hat{\mathbf{z}} \quad (2a) \quad \text{O XV}$$

$$\mathbf{B}_{63} = x_{32}\,\mathbf{a}_1 + y_{32}\,\mathbf{a}_2 + z_{32}\,\mathbf{a}_3 = (x_{32}a + z_{32}c\cos\beta)\,\hat{\mathbf{x}} + y_{32}b\,\hat{\mathbf{y}} + z_{32}c\sin\beta\,\hat{\mathbf{z}} \quad (2a) \quad \text{O XVI}$$

$$\mathbf{B}_{64} = x_{32}\,\mathbf{a}_1 - y_{32}\,\mathbf{a}_2 + \left(\tfrac{1}{2} + z_{32}\right)\mathbf{a}_3 = \left(\tfrac{1}{2}c\cos\beta + x_{32}a + z_{32}c\cos\beta\right)\hat{\mathbf{x}} - y_{32}b\,\hat{\mathbf{y}} + \left(\tfrac{1}{2} + z_{32}\right)c\sin\beta\,\hat{\mathbf{z}} \quad (2a) \quad \text{O XVI}$$

| | | | | | | |
|---|---|---|---|---|---|---|
| $\mathbf{B}_{65}$ | = | $x_{33}\,\mathbf{a}_1 + y_{33}\,\mathbf{a}_2 + z_{33}\,\mathbf{a}_3$ | = | $(x_{33}a + z_{33}c\cos\beta)\,\hat{\mathbf{x}} + y_{33}b\,\hat{\mathbf{y}} + z_{33}c\sin\beta\,\hat{\mathbf{z}}$ | (2$a$) | O XVII |
| $\mathbf{B}_{66}$ | = | $x_{33}\,\mathbf{a}_1 - y_{33}\,\mathbf{a}_2 + \left(\frac{1}{2} + z_{33}\right)\mathbf{a}_3$ | = | $\left(\frac{1}{2}c\cos\beta + x_{33}a + z_{33}c\cos\beta\right)\hat{\mathbf{x}} - y_{33}b\,\hat{\mathbf{y}} + \left(\frac{1}{2} + z_{33}\right)c\sin\beta\,\hat{\mathbf{z}}$ | (2$a$) | O XVII |
| $\mathbf{B}_{67}$ | = | $x_{34}\,\mathbf{a}_1 + y_{34}\,\mathbf{a}_2 + z_{34}\,\mathbf{a}_3$ | = | $(x_{34}a + z_{34}c\cos\beta)\,\hat{\mathbf{x}} + y_{34}b\,\hat{\mathbf{y}} + z_{34}c\sin\beta\,\hat{\mathbf{z}}$ | (2$a$) | O XVIII |
| $\mathbf{B}_{68}$ | = | $x_{34}\,\mathbf{a}_1 - y_{34}\,\mathbf{a}_2 + \left(\frac{1}{2} + z_{34}\right)\mathbf{a}_3$ | = | $\left(\frac{1}{2}c\cos\beta + x_{34}a + z_{34}c\cos\beta\right)\hat{\mathbf{x}} - y_{34}b\,\hat{\mathbf{y}} + \left(\frac{1}{2} + z_{34}\right)c\sin\beta\,\hat{\mathbf{z}}$ | (2$a$) | O XVIII |
| $\mathbf{B}_{69}$ | = | $x_{35}\,\mathbf{a}_1 + y_{35}\,\mathbf{a}_2 + z_{35}\,\mathbf{a}_3$ | = | $(x_{35}a + z_{35}c\cos\beta)\,\hat{\mathbf{x}} + y_{35}b\,\hat{\mathbf{y}} + z_{35}c\sin\beta\,\hat{\mathbf{z}}$ | (2$a$) | O XIX |
| $\mathbf{B}_{70}$ | = | $x_{35}\,\mathbf{a}_1 - y_{35}\,\mathbf{a}_2 + \left(\frac{1}{2} + z_{35}\right)\mathbf{a}_3$ | = | $\left(\frac{1}{2}c\cos\beta + x_{35}a + z_{35}c\cos\beta\right)\hat{\mathbf{x}} - y_{35}b\,\hat{\mathbf{y}} + \left(\frac{1}{2} + z_{35}\right)c\sin\beta\,\hat{\mathbf{z}}$ | (2$a$) | O XIX |
| $\mathbf{B}_{71}$ | = | $x_{36}\,\mathbf{a}_1 + y_{36}\,\mathbf{a}_2 + z_{36}\,\mathbf{a}_3$ | = | $(x_{36}a + z_{36}c\cos\beta)\,\hat{\mathbf{x}} + y_{36}b\,\hat{\mathbf{y}} + z_{36}c\sin\beta\,\hat{\mathbf{z}}$ | (2$a$) | O XX |
| $\mathbf{B}_{72}$ | = | $x_{36}\,\mathbf{a}_1 - y_{36}\,\mathbf{a}_2 + \left(\frac{1}{2} + z_{36}\right)\mathbf{a}_3$ | = | $\left(\frac{1}{2}c\cos\beta + x_{36}a + z_{36}c\cos\beta\right)\hat{\mathbf{x}} - y_{36}b\,\hat{\mathbf{y}} + \left(\frac{1}{2} + z_{36}\right)c\sin\beta\,\hat{\mathbf{z}}$ | (2$a$) | O XX |
| $\mathbf{B}_{73}$ | = | $x_{37}\,\mathbf{a}_1 + y_{37}\,\mathbf{a}_2 + z_{37}\,\mathbf{a}_3$ | = | $(x_{37}a + z_{37}c\cos\beta)\,\hat{\mathbf{x}} + y_{37}b\,\hat{\mathbf{y}} + z_{37}c\sin\beta\,\hat{\mathbf{z}}$ | (2$a$) | O XXI |
| $\mathbf{B}_{74}$ | = | $x_{37}\,\mathbf{a}_1 - y_{37}\,\mathbf{a}_2 + \left(\frac{1}{2} + z_{37}\right)\mathbf{a}_3$ | = | $\left(\frac{1}{2}c\cos\beta + x_{37}a + z_{37}c\cos\beta\right)\hat{\mathbf{x}} - y_{37}b\,\hat{\mathbf{y}} + \left(\frac{1}{2} + z_{37}\right)c\sin\beta\,\hat{\mathbf{z}}$ | (2$a$) | O XXI |
| $\mathbf{B}_{75}$ | = | $x_{38}\,\mathbf{a}_1 + y_{38}\,\mathbf{a}_2 + z_{38}\,\mathbf{a}_3$ | = | $(x_{38}a + z_{38}c\cos\beta)\,\hat{\mathbf{x}} + y_{38}b\,\hat{\mathbf{y}} + z_{38}c\sin\beta\,\hat{\mathbf{z}}$ | (2$a$) | O XXII |
| $\mathbf{B}_{76}$ | = | $x_{38}\,\mathbf{a}_1 - y_{38}\,\mathbf{a}_2 + \left(\frac{1}{2} + z_{38}\right)\mathbf{a}_3$ | = | $\left(\frac{1}{2}c\cos\beta + x_{38}a + z_{38}c\cos\beta\right)\hat{\mathbf{x}} - y_{38}b\,\hat{\mathbf{y}} + \left(\frac{1}{2} + z_{38}\right)c\sin\beta\,\hat{\mathbf{z}}$ | (2$a$) | O XXII |
| $\mathbf{B}_{77}$ | = | $x_{39}\,\mathbf{a}_1 + y_{39}\,\mathbf{a}_2 + z_{39}\,\mathbf{a}_3$ | = | $(x_{39}a + z_{39}c\cos\beta)\,\hat{\mathbf{x}} + y_{39}b\,\hat{\mathbf{y}} + z_{39}c\sin\beta\,\hat{\mathbf{z}}$ | (2$a$) | O XXIII |
| $\mathbf{B}_{78}$ | = | $x_{39}\,\mathbf{a}_1 - y_{39}\,\mathbf{a}_2 + \left(\frac{1}{2} + z_{39}\right)\mathbf{a}_3$ | = | $\left(\frac{1}{2}c\cos\beta + x_{39}a + z_{39}c\cos\beta\right)\hat{\mathbf{x}} - y_{39}b\,\hat{\mathbf{y}} + \left(\frac{1}{2} + z_{39}\right)c\sin\beta\,\hat{\mathbf{z}}$ | (2$a$) | O XXIII |
| $\mathbf{B}_{79}$ | = | $x_{40}\,\mathbf{a}_1 + y_{40}\,\mathbf{a}_2 + z_{40}\,\mathbf{a}_3$ | = | $(x_{40}a + z_{40}c\cos\beta)\,\hat{\mathbf{x}} + y_{40}b\,\hat{\mathbf{y}} + z_{40}c\sin\beta\,\hat{\mathbf{z}}$ | (2$a$) | O XXIV |
| $\mathbf{B}_{80}$ | = | $x_{40}\,\mathbf{a}_1 - y_{40}\,\mathbf{a}_2 + \left(\frac{1}{2} + z_{40}\right)\mathbf{a}_3$ | = | $\left(\frac{1}{2}c\cos\beta + x_{40}a + z_{40}c\cos\beta\right)\hat{\mathbf{x}} - y_{40}b\,\hat{\mathbf{y}} + \left(\frac{1}{2} + z_{40}\right)c\sin\beta\,\hat{\mathbf{z}}$ | (2$a$) | O XXIV |
| $\mathbf{B}_{81}$ | = | $x_{41}\,\mathbf{a}_1 + y_{41}\,\mathbf{a}_2 + z_{41}\,\mathbf{a}_3$ | = | $(x_{41}a + z_{41}c\cos\beta)\,\hat{\mathbf{x}} + y_{41}b\,\hat{\mathbf{y}} + z_{41}c\sin\beta\,\hat{\mathbf{z}}$ | (2$a$) | O XXV |
| $\mathbf{B}_{82}$ | = | $x_{41}\,\mathbf{a}_1 - y_{41}\,\mathbf{a}_2 + \left(\frac{1}{2} + z_{41}\right)\mathbf{a}_3$ | = | $\left(\frac{1}{2}c\cos\beta + x_{41}a + z_{41}c\cos\beta\right)\hat{\mathbf{x}} - y_{41}b\,\hat{\mathbf{y}} + \left(\frac{1}{2} + z_{41}\right)c\sin\beta\,\hat{\mathbf{z}}$ | (2$a$) | O XXV |
| $\mathbf{B}_{83}$ | = | $x_{42}\,\mathbf{a}_1 + y_{42}\,\mathbf{a}_2 + z_{42}\,\mathbf{a}_3$ | = | $(x_{42}a + z_{42}c\cos\beta)\,\hat{\mathbf{x}} + y_{42}b\,\hat{\mathbf{y}} + z_{42}c\sin\beta\,\hat{\mathbf{z}}$ | (2$a$) | O XXVI |
| $\mathbf{B}_{84}$ | = | $x_{42}\,\mathbf{a}_1 - y_{42}\,\mathbf{a}_2 + \left(\frac{1}{2} + z_{42}\right)\mathbf{a}_3$ | = | $\left(\frac{1}{2}c\cos\beta + x_{42}a + z_{42}c\cos\beta\right)\hat{\mathbf{x}} - y_{42}b\,\hat{\mathbf{y}} + \left(\frac{1}{2} + z_{42}\right)c\sin\beta\,\hat{\mathbf{z}}$ | (2$a$) | O XXVI |
| $\mathbf{B}_{85}$ | = | $x_{43}\,\mathbf{a}_1 + y_{43}\,\mathbf{a}_2 + z_{43}\,\mathbf{a}_3$ | = | $(x_{43}a + z_{43}c\cos\beta)\,\hat{\mathbf{x}} + y_{43}b\,\hat{\mathbf{y}} + z_{43}c\sin\beta\,\hat{\mathbf{z}}$ | (2$a$) | O XXVII |
| $\mathbf{B}_{86}$ | = | $x_{43}\,\mathbf{a}_1 - y_{43}\,\mathbf{a}_2 + \left(\frac{1}{2} + z_{43}\right)\mathbf{a}_3$ | = | $\left(\frac{1}{2}c\cos\beta + x_{43}a + z_{43}c\cos\beta\right)\hat{\mathbf{x}} - y_{43}b\,\hat{\mathbf{y}} + \left(\frac{1}{2} + z_{43}\right)c\sin\beta\,\hat{\mathbf{z}}$ | (2$a$) | O XXVII |
| $\mathbf{B}_{87}$ | = | $x_{44}\,\mathbf{a}_1 + y_{44}\,\mathbf{a}_2 + z_{44}\,\mathbf{a}_3$ | = | $(x_{44}a + z_{44}c\cos\beta)\,\hat{\mathbf{x}} + y_{44}b\,\hat{\mathbf{y}} + z_{44}c\sin\beta\,\hat{\mathbf{z}}$ | (2$a$) | O XXVIII |
| $\mathbf{B}_{88}$ | = | $x_{44}\,\mathbf{a}_1 - y_{44}\,\mathbf{a}_2 + \left(\frac{1}{2} + z_{44}\right)\mathbf{a}_3$ | = | $\left(\frac{1}{2}c\cos\beta + x_{44}a + z_{44}c\cos\beta\right)\hat{\mathbf{x}} - y_{44}b\,\hat{\mathbf{y}} + \left(\frac{1}{2} + z_{44}\right)c\sin\beta\,\hat{\mathbf{z}}$ | (2$a$) | O XXVIII |
| $\mathbf{B}_{89}$ | = | $x_{45}\,\mathbf{a}_1 + y_{45}\,\mathbf{a}_2 + z_{45}\,\mathbf{a}_3$ | = | $(x_{45}a + z_{45}c\cos\beta)\,\hat{\mathbf{x}} + y_{45}b\,\hat{\mathbf{y}} + z_{45}c\sin\beta\,\hat{\mathbf{z}}$ | (2$a$) | O XXIX |
| $\mathbf{B}_{90}$ | = | $x_{45}\,\mathbf{a}_1 - y_{45}\,\mathbf{a}_2 + \left(\frac{1}{2} + z_{45}\right)\mathbf{a}_3$ | = | $\left(\frac{1}{2}c\cos\beta + x_{45}a + z_{45}c\cos\beta\right)\hat{\mathbf{x}} - y_{45}b\,\hat{\mathbf{y}} + \left(\frac{1}{2} + z_{45}\right)c\sin\beta\,\hat{\mathbf{z}}$ | (2$a$) | O XXIX |
| $\mathbf{B}_{91}$ | = | $x_{46}\,\mathbf{a}_1 + y_{46}\,\mathbf{a}_2 + z_{46}\,\mathbf{a}_3$ | = | $(x_{46}a + z_{46}c\cos\beta)\,\hat{\mathbf{x}} + y_{46}b\,\hat{\mathbf{y}} + z_{46}c\sin\beta\,\hat{\mathbf{z}}$ | (2$a$) | O XXX |
| $\mathbf{B}_{92}$ | = | $x_{46}\,\mathbf{a}_1 - y_{46}\,\mathbf{a}_2 + \left(\frac{1}{2} + z_{46}\right)\mathbf{a}_3$ | = | $\left(\frac{1}{2}c\cos\beta + x_{46}a + z_{46}c\cos\beta\right)\hat{\mathbf{x}} - y_{46}b\,\hat{\mathbf{y}} + \left(\frac{1}{2} + z_{46}\right)c\sin\beta\,\hat{\mathbf{z}}$ | (2$a$) | O XXX |
| $\mathbf{B}_{93}$ | = | $x_{47}\,\mathbf{a}_1 + y_{47}\,\mathbf{a}_2 + z_{47}\,\mathbf{a}_3$ | = | $(x_{47}a + z_{47}c\cos\beta)\,\hat{\mathbf{x}} + y_{47}b\,\hat{\mathbf{y}} + z_{47}c\sin\beta\,\hat{\mathbf{z}}$ | (2$a$) | O XXXI |
| $\mathbf{B}_{94}$ | = | $x_{47}\,\mathbf{a}_1 - y_{47}\,\mathbf{a}_2 + \left(\frac{1}{2} + z_{47}\right)\mathbf{a}_3$ | = | $\left(\frac{1}{2}c\cos\beta + x_{47}a + z_{47}c\cos\beta\right)\hat{\mathbf{x}} - y_{47}b\,\hat{\mathbf{y}} + \left(\frac{1}{2} + z_{47}\right)c\sin\beta\,\hat{\mathbf{z}}$ | (2$a$) | O XXXI |
| $\mathbf{B}_{95}$ | = | $x_{48}\,\mathbf{a}_1 + y_{48}\,\mathbf{a}_2 + z_{48}\,\mathbf{a}_3$ | = | $(x_{48}a + z_{48}c\cos\beta)\,\hat{\mathbf{x}} + y_{48}b\,\hat{\mathbf{y}} + z_{48}c\sin\beta\,\hat{\mathbf{z}}$ | (2$a$) | O XXXII |
| $\mathbf{B}_{96}$ | = | $x_{48}\,\mathbf{a}_1 - y_{48}\,\mathbf{a}_2 + \left(\frac{1}{2} + z_{48}\right)\mathbf{a}_3$ | = | $\left(\frac{1}{2}c\cos\beta + x_{48}a + z_{48}c\cos\beta\right)\hat{\mathbf{x}} - y_{48}b\,\hat{\mathbf{y}} + \left(\frac{1}{2} + z_{48}\right)c\sin\beta\,\hat{\mathbf{z}}$ | (2$a$) | O XXXII |
| $\mathbf{B}_{97}$ | = | $x_{49}\,\mathbf{a}_1 + y_{49}\,\mathbf{a}_2 + z_{49}\,\mathbf{a}_3$ | = | $(x_{49}a + z_{49}c\cos\beta)\,\hat{\mathbf{x}} + y_{49}b\,\hat{\mathbf{y}} + z_{49}c\sin\beta\,\hat{\mathbf{z}}$ | (2$a$) | O XXXIII |
| $\mathbf{B}_{98}$ | = | $x_{49}\,\mathbf{a}_1 - y_{49}\,\mathbf{a}_2 + \left(\frac{1}{2} + z_{49}\right)\mathbf{a}_3$ | = | $\left(\frac{1}{2}c\cos\beta + x_{49}a + z_{49}c\cos\beta\right)\hat{\mathbf{x}} - y_{49}b\,\hat{\mathbf{y}} + \left(\frac{1}{2} + z_{49}\right)c\sin\beta\,\hat{\mathbf{z}}$ | (2$a$) | O XXXIII |
| $\mathbf{B}_{99}$ | = | $x_{50}\,\mathbf{a}_1 + y_{50}\,\mathbf{a}_2 + z_{50}\,\mathbf{a}_3$ | = | $(x_{50}a + z_{50}c\cos\beta)\,\hat{\mathbf{x}} + y_{50}b\,\hat{\mathbf{y}} + z_{50}c\sin\beta\,\hat{\mathbf{z}}$ | (2$a$) | O XXXIV |
| $\mathbf{B}_{100}$ | = | $x_{50}\,\mathbf{a}_1 - y_{50}\,\mathbf{a}_2 + \left(\frac{1}{2} + z_{50}\right)\mathbf{a}_3$ | = | $\left(\frac{1}{2}c\cos\beta + x_{50}a + z_{50}c\cos\beta\right)\hat{\mathbf{x}} - y_{50}b\,\hat{\mathbf{y}} + \left(\frac{1}{2} + z_{50}\right)c\sin\beta\,\hat{\mathbf{z}}$ | (2$a$) | O XXXIV |
| $\mathbf{B}_{101}$ | = | $x_{51}\,\mathbf{a}_1 + y_{51}\,\mathbf{a}_2 + z_{51}\,\mathbf{a}_3$ | = | $(x_{51}a + z_{51}c\cos\beta)\,\hat{\mathbf{x}} + y_{51}b\,\hat{\mathbf{y}} + z_{51}c\sin\beta\,\hat{\mathbf{z}}$ | (2$a$) | O XXXV |
| $\mathbf{B}_{102}$ | = | $x_{51}\,\mathbf{a}_1 - y_{51}\,\mathbf{a}_2 + \left(\frac{1}{2} + z_{51}\right)\mathbf{a}_3$ | = | $\left(\frac{1}{2}c\cos\beta + x_{51}a + z_{51}c\cos\beta\right)\hat{\mathbf{x}} - y_{51}b\,\hat{\mathbf{y}} + \left(\frac{1}{2} + z_{51}\right)c\sin\beta\,\hat{\mathbf{z}}$ | (2$a$) | O XXXV |
| $\mathbf{B}_{103}$ | = | $x_{52}\,\mathbf{a}_1 + y_{52}\,\mathbf{a}_2 + z_{52}\,\mathbf{a}_3$ | = | $(x_{52}a + z_{52}c\cos\beta)\,\hat{\mathbf{x}} + y_{52}b\,\hat{\mathbf{y}} + z_{52}c\sin\beta\,\hat{\mathbf{z}}$ | (2$a$) | O XXXVI |
| $\mathbf{B}_{104}$ | = | $x_{52}\,\mathbf{a}_1 - y_{52}\,\mathbf{a}_2 + \left(\frac{1}{2} + z_{52}\right)\mathbf{a}_3$ | = | $\left(\frac{1}{2}c\cos\beta + x_{52}a + z_{52}c\cos\beta\right)\hat{\mathbf{x}} - y_{52}b\,\hat{\mathbf{y}} + \left(\frac{1}{2} + z_{52}\right)c\sin\beta\,\hat{\mathbf{z}}$ | (2$a$) | O XXXVI |
| $\mathbf{B}_{105}$ | = | $x_{53}\,\mathbf{a}_1 + y_{53}\,\mathbf{a}_2 + z_{53}\,\mathbf{a}_3$ | = | $(x_{53}a + z_{53}c\cos\beta)\,\hat{\mathbf{x}} + y_{53}b\,\hat{\mathbf{y}} + z_{53}c\sin\beta\,\hat{\mathbf{z}}$ | (2$a$) | O XXXVII |
| $\mathbf{B}_{106}$ | = | $x_{53}\,\mathbf{a}_1 - y_{53}\,\mathbf{a}_2 + \left(\frac{1}{2} + z_{53}\right)\mathbf{a}_3$ | = | $\left(\frac{1}{2}c\cos\beta + x_{53}a + z_{53}c\cos\beta\right)\hat{\mathbf{x}} - y_{53}b\,\hat{\mathbf{y}} + \left(\frac{1}{2} + z_{53}\right)c\sin\beta\,\hat{\mathbf{z}}$ | (2$a$) | O XXXVII |
| $\mathbf{B}_{107}$ | = | $x_{54}\,\mathbf{a}_1 + y_{54}\,\mathbf{a}_2 + z_{54}\,\mathbf{a}_3$ | = | $(x_{54}a + z_{54}c\cos\beta)\,\hat{\mathbf{x}} + y_{54}b\,\hat{\mathbf{y}} + z_{54}c\sin\beta\,\hat{\mathbf{z}}$ | (2$a$) | O XXXVIII |
| $\mathbf{B}_{108}$ | = | $x_{54}\,\mathbf{a}_1 - y_{54}\,\mathbf{a}_2 + \left(\frac{1}{2} + z_{54}\right)\mathbf{a}_3$ | = | $\left(\frac{1}{2}c\cos\beta + x_{54}a + z_{54}c\cos\beta\right)\hat{\mathbf{x}} - y_{54}b\,\hat{\mathbf{y}} + \left(\frac{1}{2} + z_{54}\right)c\sin\beta\,\hat{\mathbf{z}}$ | (2$a$) | O XXXVIII |
| $\mathbf{B}_{109}$ | = | $x_{55}\,\mathbf{a}_1 + y_{55}\,\mathbf{a}_2 + z_{55}\,\mathbf{a}_3$ | = | $(x_{55}a + z_{55}c\cos\beta)\,\hat{\mathbf{x}} + y_{55}b\,\hat{\mathbf{y}} + z_{55}c\sin\beta\,\hat{\mathbf{z}}$ | (2a) | O XXXIX |
| $\mathbf{B}_{110}$ | = | $x_{55}\,\mathbf{a}_1 - y_{55}\,\mathbf{a}_2 + \left(\frac{1}{2} + z_{55}\right)\mathbf{a}_3$ | = | $\left(\frac{1}{2}c\cos\beta + x_{55}a + z_{55}c\cos\beta\right)\hat{\mathbf{x}} - y_{55}b\,\hat{\mathbf{y}} + \left(\frac{1}{2} + z_{55}\right)c\sin\beta\,\hat{\mathbf{z}}$ | (2a) | O XXXIX |
| $\mathbf{B}_{111}$ | = | $x_{56}\,\mathbf{a}_1 + y_{56}\,\mathbf{a}_2 + z_{56}\,\mathbf{a}_3$ | = | $(x_{56}a + z_{56}c\cos\beta)\,\hat{\mathbf{x}} + y_{56}b\,\hat{\mathbf{y}} + z_{56}c\sin\beta\,\hat{\mathbf{z}}$ | (2a) | O XL |
| $\mathbf{B}_{112}$ | = | $x_{56}\,\mathbf{a}_1 - y_{56}\,\mathbf{a}_2 + \left(\frac{1}{2} + z_{56}\right)\mathbf{a}_3$ | = | $\left(\frac{1}{2}c\cos\beta + x_{56}a + z_{56}c\cos\beta\right)\hat{\mathbf{x}} - y_{56}b\,\hat{\mathbf{y}} + \left(\frac{1}{2} + z_{56}\right)c\sin\beta\,\hat{\mathbf{z}}$ | (2a) | O XL |
| $\mathbf{B}_{113}$ | = | $x_{57}\,\mathbf{a}_1 + y_{57}\,\mathbf{a}_2 + z_{57}\,\mathbf{a}_3$ | = | $(x_{57}a + z_{57}c\cos\beta)\,\hat{\mathbf{x}} + y_{57}b\,\hat{\mathbf{y}} + z_{57}c\sin\beta\,\hat{\mathbf{z}}$ | (2a) | O XLI |
| $\mathbf{B}_{114}$ | = | $x_{57}\,\mathbf{a}_1 - y_{57}\,\mathbf{a}_2 + \left(\frac{1}{2} + z_{57}\right)\mathbf{a}_3$ | = | $\left(\frac{1}{2}c\cos\beta + x_{57}a + z_{57}c\cos\beta\right)\hat{\mathbf{x}} - y_{57}b\,\hat{\mathbf{y}} + \left(\frac{1}{2} + z_{57}\right)c\sin\beta\,\hat{\mathbf{z}}$ | (2a) | O XLI |
| $\mathbf{B}_{115}$ | = | $x_{58}\,\mathbf{a}_1 + y_{58}\,\mathbf{a}_2 + z_{58}\,\mathbf{a}_3$ | = | $(x_{58}a + z_{58}c\cos\beta)\,\hat{\mathbf{x}} + y_{58}b\,\hat{\mathbf{y}} + z_{58}c\sin\beta\,\hat{\mathbf{z}}$ | (2a) | O XLII |
| $\mathbf{B}_{116}$ | = | $x_{58}\,\mathbf{a}_1 - y_{58}\,\mathbf{a}_2 + \left(\frac{1}{2} + z_{58}\right)\mathbf{a}_3$ | = | $\left(\frac{1}{2}c\cos\beta + x_{58}a + z_{58}c\cos\beta\right)\hat{\mathbf{x}} - y_{58}b\,\hat{\mathbf{y}} + \left(\frac{1}{2} + z_{58}\right)c\sin\beta\,\hat{\mathbf{z}}$ | (2a) | O XLII |
| $\mathbf{B}_{117}$ | = | $x_{59}\,\mathbf{a}_1 + y_{59}\,\mathbf{a}_2 + z_{59}\,\mathbf{a}_3$ | = | $(x_{59}a + z_{59}c\cos\beta)\,\hat{\mathbf{x}} + y_{59}b\,\hat{\mathbf{y}} + z_{59}c\sin\beta\,\hat{\mathbf{z}}$ | (2a) | O XLIII |
| $\mathbf{B}_{118}$ | = | $x_{59}\,\mathbf{a}_1 - y_{59}\,\mathbf{a}_2 + \left(\frac{1}{2} + z_{59}\right)\mathbf{a}_3$ | = | $\left(\frac{1}{2}c\cos\beta + x_{59}a + z_{59}c\cos\beta\right)\hat{\mathbf{x}} - y_{59}b\,\hat{\mathbf{y}} + \left(\frac{1}{2} + z_{59}\right)c\sin\beta\,\hat{\mathbf{z}}$ | (2a) | O XLIII |
| $\mathbf{B}_{119}$ | = | $x_{60}\,\mathbf{a}_1 + y_{60}\,\mathbf{a}_2 + z_{60}\,\mathbf{a}_3$ | = | $(x_{60}a + z_{60}c\cos\beta)\,\hat{\mathbf{x}} + y_{60}b\,\hat{\mathbf{y}} + z_{60}c\sin\beta\,\hat{\mathbf{z}}$ | (2a) | O XLIV |
| $\mathbf{B}_{120}$ | = | $x_{60}\,\mathbf{a}_1 - y_{60}\,\mathbf{a}_2 + \left(\frac{1}{2} + z_{60}\right)\mathbf{a}_3$ | = | $\left(\frac{1}{2}c\cos\beta + x_{60}a + z_{60}c\cos\beta\right)\hat{\mathbf{x}} - y_{60}b\,\hat{\mathbf{y}} + \left(\frac{1}{2} + z_{60}\right)c\sin\beta\,\hat{\mathbf{z}}$ | (2a) | O XLIV |
| $\mathbf{B}_{121}$ | = | $x_{61}\,\mathbf{a}_1 + y_{61}\,\mathbf{a}_2 + z_{61}\,\mathbf{a}_3$ | = | $(x_{61}a + z_{61}c\cos\beta)\,\hat{\mathbf{x}} + y_{61}b\,\hat{\mathbf{y}} + z_{61}c\sin\beta\,\hat{\mathbf{z}}$ | (2a) | O XLV |
| $\mathbf{B}_{122}$ | = | $x_{61}\,\mathbf{a}_1 - y_{61}\,\mathbf{a}_2 + \left(\frac{1}{2} + z_{61}\right)\mathbf{a}_3$ | = | $\left(\frac{1}{2}c\cos\beta + x_{61}a + z_{61}c\cos\beta\right)\hat{\mathbf{x}} - y_{61}b\,\hat{\mathbf{y}} + \left(\frac{1}{2} + z_{61}\right)c\sin\beta\,\hat{\mathbf{z}}$ | (2a) | O XLV |
| $\mathbf{B}_{123}$ | = | $x_{62}\,\mathbf{a}_1 + y_{62}\,\mathbf{a}_2 + z_{62}\,\mathbf{a}_3$ | = | $(x_{62}a + z_{62}c\cos\beta)\,\hat{\mathbf{x}} + y_{62}b\,\hat{\mathbf{y}} + z_{62}c\sin\beta\,\hat{\mathbf{z}}$ | (2a) | O XLVI |
| $\mathbf{B}_{124}$ | = | $x_{62}\,\mathbf{a}_1 - y_{62}\,\mathbf{a}_2 + \left(\frac{1}{2} + z_{62}\right)\mathbf{a}_3$ | = | $\left(\frac{1}{2}c\cos\beta + x_{62}a + z_{62}c\cos\beta\right)\hat{\mathbf{x}} - y_{62}b\,\hat{\mathbf{y}} + \left(\frac{1}{2} + z_{62}\right)c\sin\beta\,\hat{\mathbf{z}}$ | (2a) | O XLVI |

**Geometry files:**

- CIF: pp. 1514
- POSCAR: pp. 1515

# Calaverite ($AuTe_2$) Structure: AB2_mP12_7_2a_4a

http://aflow.org/prototype-encyclopedia/AB2_mP12_7_2a_4a

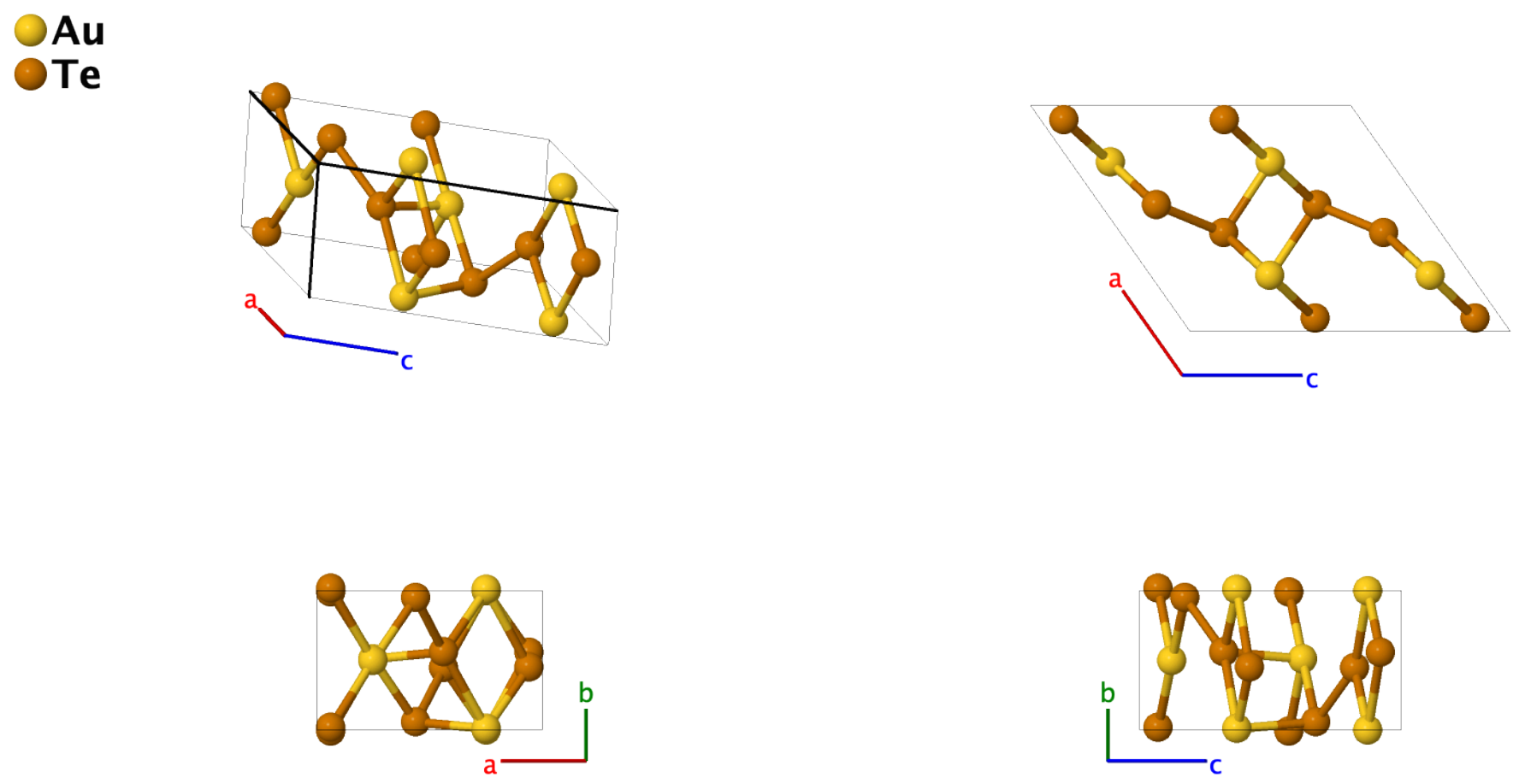

| | | |
|---|---|---|
| **Prototype** | : | $AuTe_2$ |
| **AFLOW prototype label** | : | AB2_mP12_7_2a_4a |
| ***Strukturbericht* designation** | : | None |
| **Pearson symbol** | : | mP12 |
| **Space group number** | : | 7 |
| **Space group symbol** | : | $Pc$ |
| **AFLOW prototype command** | : | `aflow --proto=AB2_mP12_7_2a_4a --params=`$a, b/a, c/a, \beta, x_1, y_1, z_1, x_2, y_2, z_2, x_3, y_3, z_3, x_4, y_4, z_4, x_5, y_5, z_5, x_6, y_6, z_6$ |

**Simple Monoclinic primitive vectors:**

$$\mathbf{a}_1 = a\,\hat{\mathbf{x}}$$
$$\mathbf{a}_2 = b\,\hat{\mathbf{y}}$$
$$\mathbf{a}_3 = c\cos\beta\,\hat{\mathbf{x}} + c\sin\beta\,\hat{\mathbf{z}}$$

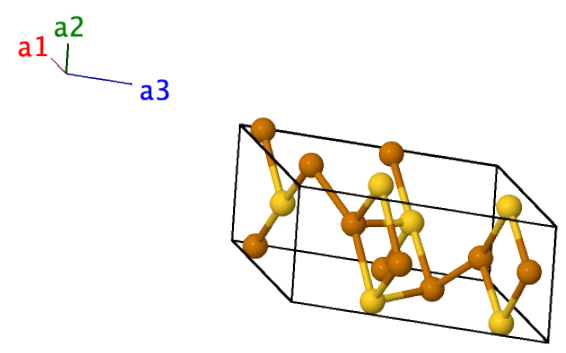

**Basis vectors:**

| | | Lattice Coordinates | | Cartesian Coordinates | Wyckoff Position | Atom Type |
|---|---|---|---|---|---|---|
| $\mathbf{B}_1$ | = | $x_1\,\mathbf{a}_1 + y_1\,\mathbf{a}_2 + z_1\,\mathbf{a}_3$ | = | $(x_1 a + z_1 c\cos\beta)\,\hat{\mathbf{x}} + y_1 b\,\hat{\mathbf{y}} + z_1 c\sin\beta\,\hat{\mathbf{z}}$ | (2$a$) | Au I |
| $\mathbf{B}_2$ | = | $x_1\,\mathbf{a}_1 - y_1\,\mathbf{a}_2 + \left(\frac{1}{2} + z_1\right)\mathbf{a}_3$ | = | $\left(\frac{1}{2}c\cos\beta + x_1 a + z_1 c\cos\beta\right)\hat{\mathbf{x}} - y_1 b\,\hat{\mathbf{y}} + \left(\frac{1}{2} + z_1\right)c\sin\beta\,\hat{\mathbf{z}}$ | (2$a$) | Au I |
| $\mathbf{B}_3$ | = | $x_2\,\mathbf{a}_1 + y_2\,\mathbf{a}_2 + z_2\,\mathbf{a}_3$ | = | $(x_2 a + z_2 c\cos\beta)\,\hat{\mathbf{x}} + y_2 b\,\hat{\mathbf{y}} + z_2 c\sin\beta\,\hat{\mathbf{z}}$ | (2$a$) | Au II |
| $\mathbf{B}_4$ | = | $x_2\,\mathbf{a}_1 - y_2\,\mathbf{a}_2 + \left(\frac{1}{2} + z_2\right)\mathbf{a}_3$ | = | $\left(\frac{1}{2}c\cos\beta + x_2 a + z_2 c\cos\beta\right)\hat{\mathbf{x}} - y_2 b\,\hat{\mathbf{y}} + \left(\frac{1}{2} + z_2\right)c\sin\beta\,\hat{\mathbf{z}}$ | (2$a$) | Au II |
| $\mathbf{B}_5$ | = | $x_3\,\mathbf{a}_1 + y_3\,\mathbf{a}_2 + z_3\,\mathbf{a}_3$ | = | $(x_3 a + z_3 c\cos\beta)\,\hat{\mathbf{x}} + y_3 b\,\hat{\mathbf{y}} + z_3 c\sin\beta\,\hat{\mathbf{z}}$ | (2$a$) | Te I |
| $\mathbf{B}_6$ | = | $x_3\,\mathbf{a}_1 - y_3\,\mathbf{a}_2 + \left(\frac{1}{2} + z_3\right)\,\mathbf{a}_3$ | = | $\left(\frac{1}{2}c\cos\beta + x_3a + z_3c\cos\beta\right)\,\hat{\mathbf{x}} - y_3b\,\hat{\mathbf{y}} + \left(\frac{1}{2} + z_3\right)c\sin\beta\,\hat{\mathbf{z}}$ | (2a) | Te I |
| $\mathbf{B}_7$ | = | $x_4\,\mathbf{a}_1 + y_4\,\mathbf{a}_2 + z_4\,\mathbf{a}_3$ | = | $(x_4a + z_4c\cos\beta)\,\hat{\mathbf{x}} + y_4b\,\hat{\mathbf{y}} + z_4c\sin\beta\,\hat{\mathbf{z}}$ | (2a) | Te II |
| $\mathbf{B}_8$ | = | $x_4\,\mathbf{a}_1 - y_4\,\mathbf{a}_2 + \left(\frac{1}{2} + z_4\right)\,\mathbf{a}_3$ | = | $\left(\frac{1}{2}c\cos\beta + x_4a + z_4c\cos\beta\right)\,\hat{\mathbf{x}} - y_4b\,\hat{\mathbf{y}} + \left(\frac{1}{2} + z_4\right)c\sin\beta\,\hat{\mathbf{z}}$ | (2a) | Te II |
| $\mathbf{B}_9$ | = | $x_5\,\mathbf{a}_1 + y_5\,\mathbf{a}_2 + z_5\,\mathbf{a}_3$ | = | $(x_5a + z_5c\cos\beta)\,\hat{\mathbf{x}} + y_5b\,\hat{\mathbf{y}} + z_5c\sin\beta\,\hat{\mathbf{z}}$ | (2a) | Te III |
| $\mathbf{B}_{10}$ | = | $x_5\,\mathbf{a}_1 - y_5\,\mathbf{a}_2 + \left(\frac{1}{2} + z_5\right)\,\mathbf{a}_3$ | = | $\left(\frac{1}{2}c\cos\beta + x_5a + z_5c\cos\beta\right)\,\hat{\mathbf{x}} - y_5b\,\hat{\mathbf{y}} + \left(\frac{1}{2} + z_5\right)c\sin\beta\,\hat{\mathbf{z}}$ | (2a) | Te III |
| $\mathbf{B}_{11}$ | = | $x_6\,\mathbf{a}_1 + y_6\,\mathbf{a}_2 + z_6\,\mathbf{a}_3$ | = | $(x_6a + z_6c\cos\beta)\,\hat{\mathbf{x}} + y_6b\,\hat{\mathbf{y}} + z_6c\sin\beta\,\hat{\mathbf{z}}$ | (2a) | Te IV |
| $\mathbf{B}_{12}$ | = | $x_6\,\mathbf{a}_1 - y_6\,\mathbf{a}_2 + \left(\frac{1}{2} + z_6\right)\,\mathbf{a}_3$ | = | $\left(\frac{1}{2}c\cos\beta + x_6a + z_6c\cos\beta\right)\,\hat{\mathbf{x}} - y_6b\,\hat{\mathbf{y}} + \left(\frac{1}{2} + z_6\right)c\sin\beta\,\hat{\mathbf{z}}$ | (2a) | Te IV |

---

---

**Geometry files:**

- CIF: pp. 1516
- POSCAR: pp. 1516

# Monoclinic $Co_4Al_{13}$ Structure: A13B4_mC102_8_17a11b_8a2b

http://aflow.org/prototype-encyclopedia/A13B4_mC102_8_17a11b_8a2b

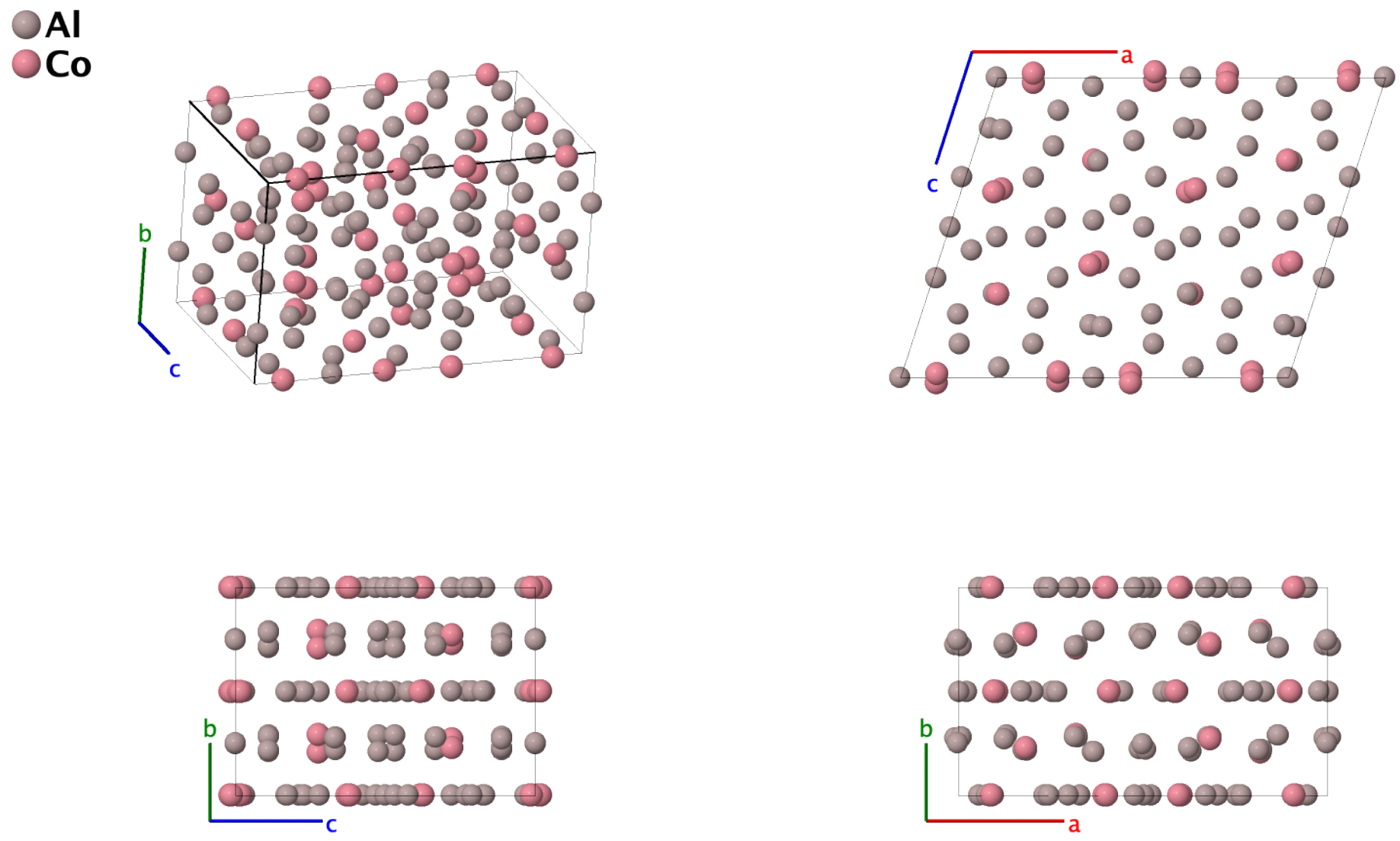

| | | |
|---|---|---|
| **Prototype** | : | $Al_{13}Co_4$ |
| **AFLOW prototype label** | : | A13B4_mC102_8_17a11b_8a2b |
| ***Strukturbericht* designation** | : | None |
| **Pearson symbol** | : | mC102 |
| **Space group number** | : | 8 |
| **Space group symbol** | : | *Cm* |
| **AFLOW prototype command** | : | `aflow --proto=A13B4_mC102_8_17a11b_8a2b` `--params=`$a, b/a, c/a, \beta, x_1, z_1, x_2, z_2, x_3, z_3, x_4, z_4, x_5, z_5, x_6, z_6, x_7, z_7, x_8, z_8, x_9, z_9, x_{10}, z_{10}, x_{11}, z_{11}, x_{12}, z_{12}, x_{13}, z_{13}, x_{14}, z_{14}, x_{15}, z_{15}, x_{16}, z_{16}, x_{17}, z_{17}, x_{18}, z_{18}, x_{19}, z_{19}, x_{20}, z_{20}, x_{21}, z_{21}, x_{22}, z_{22}, x_{23}, z_{23}, x_{24}, z_{24}, x_{25}, z_{25}, x_{26}, y_{26}, z_{26}, x_{27}, y_{27}, z_{27}, x_{28}, y_{28}, z_{28}, x_{29}, y_{29}, z_{29}, x_{30}, y_{30}, z_{30}, x_{31}, y_{31}, z_{31}, x_{32}, y_{32}, z_{32}, x_{33}, y_{33}, z_{33}, x_{34}, y_{34}, z_{34}, x_{35}, y_{35}, z_{35}, x_{36}, y_{36}, z_{36}, x_{37}, y_{37}, z_{37}, x_{38}, y_{38}, z_{38}$ |

- Following (Hudd, 1962), the Al-IV and Al-XIII sites are occupied 30% of the time, while the occupation of Al-VI, Al-IX, Al-XIV, and Al-XVII is 70%. This gives a nominial occupationof $Al_{91}Co_{30}$, though the authors state the actual composition is $Al_{68.3}Co_{24.4}$.
- Space group *Cm* #8 allows an arbitrary choice for the origin of the $z$-axis. We follow (Hudd, 1962) and set the $z_{26} = 0$.
- AFLOW-SYM yields space group *Cm* #8 consistently for small and large tolerances. However, if we allow a rather large uncertainty of 0.3 Å in the atomic positions with FINDSYM, the symmetry is set as $C2/m$ #12. That crystal has the $Al_{13}Fe_4$ prototype.

**Base-centered Monoclinic primitive vectors:**

$$\mathbf{a}_1 = \tfrac{1}{2}\,a\,\hat{\mathbf{x}} - \tfrac{1}{2}\,b\,\hat{\mathbf{y}}$$
$$\mathbf{a}_2 = \tfrac{1}{2}\,a\,\hat{\mathbf{x}} + \tfrac{1}{2}\,b\,\hat{\mathbf{y}}$$
$$\mathbf{a}_3 = c\cos\beta\,\hat{\mathbf{x}} + c\sin\beta\,\hat{\mathbf{z}}$$

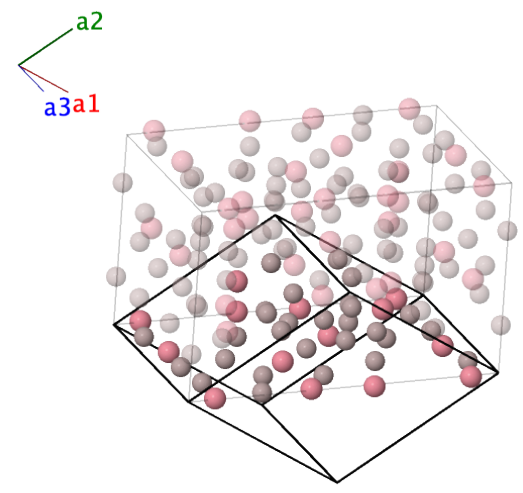

**Basis vectors:**

| | | Lattice Coordinates | | Cartesian Coordinates | Wyckoff Position | Atom Type |
|---|---|---|---|---|---|---|
| $\mathbf{B}_1$ | = | $x_1\,\mathbf{a}_1 + x_1\,\mathbf{a}_2 + z_1\,\mathbf{a}_3$ | = | $(x_1 a + z_1 c\cos\beta)\,\hat{\mathbf{x}} + z_1 c\sin\beta\,\hat{\mathbf{z}}$ | (2$a$) | Al I |
| $\mathbf{B}_2$ | = | $x_2\,\mathbf{a}_1 + x_2\,\mathbf{a}_2 + z_2\,\mathbf{a}_3$ | = | $(x_2 a + z_2 c\cos\beta)\,\hat{\mathbf{x}} + z_2 c\sin\beta\,\hat{\mathbf{z}}$ | (2$a$) | Al II |
| $\mathbf{B}_3$ | = | $x_3\,\mathbf{a}_1 + x_3\,\mathbf{a}_2 + z_3\,\mathbf{a}_3$ | = | $(x_3 a + z_3 c\cos\beta)\,\hat{\mathbf{x}} + z_3 c\sin\beta\,\hat{\mathbf{z}}$ | (2$a$) | Al III |
| $\mathbf{B}_4$ | = | $x_4\,\mathbf{a}_1 + x_4\,\mathbf{a}_2 + z_4\,\mathbf{a}_3$ | = | $(x_4 a + z_4 c\cos\beta)\,\hat{\mathbf{x}} + z_4 c\sin\beta\,\hat{\mathbf{z}}$ | (2$a$) | Al IV |
| $\mathbf{B}_5$ | = | $x_5\,\mathbf{a}_1 + x_5\,\mathbf{a}_2 + z_5\,\mathbf{a}_3$ | = | $(x_5 a + z_5 c\cos\beta)\,\hat{\mathbf{x}} + z_5 c\sin\beta\,\hat{\mathbf{z}}$ | (2$a$) | Al V |
| $\mathbf{B}_6$ | = | $x_6\,\mathbf{a}_1 + x_6\,\mathbf{a}_2 + z_6\,\mathbf{a}_3$ | = | $(x_6 a + z_6 c\cos\beta)\,\hat{\mathbf{x}} + z_6 c\sin\beta\,\hat{\mathbf{z}}$ | (2$a$) | Al VI |
| $\mathbf{B}_7$ | = | $x_7\,\mathbf{a}_1 + x_7\,\mathbf{a}_2 + z_7\,\mathbf{a}_3$ | = | $(x_7 a + z_7 c\cos\beta)\,\hat{\mathbf{x}} + z_7 c\sin\beta\,\hat{\mathbf{z}}$ | (2$a$) | Al VII |
| $\mathbf{B}_8$ | = | $x_8\,\mathbf{a}_1 + x_8\,\mathbf{a}_2 + z_8\,\mathbf{a}_3$ | = | $(x_8 a + z_8 c\cos\beta)\,\hat{\mathbf{x}} + z_8 c\sin\beta\,\hat{\mathbf{z}}$ | (2$a$) | Al VIII |
| $\mathbf{B}_9$ | = | $x_9\,\mathbf{a}_1 + x_9\,\mathbf{a}_2 + z_9\,\mathbf{a}_3$ | = | $(x_9 a + z_9 c\cos\beta)\,\hat{\mathbf{x}} + z_9 c\sin\beta\,\hat{\mathbf{z}}$ | (2$a$) | Al IX |
| $\mathbf{B}_{10}$ | = | $x_{10}\,\mathbf{a}_1 + x_{10}\,\mathbf{a}_2 + z_{10}\,\mathbf{a}_3$ | = | $(x_{10} a + z_{10} c\cos\beta)\,\hat{\mathbf{x}} + z_{10} c\sin\beta\,\hat{\mathbf{z}}$ | (2$a$) | Al X |
| $\mathbf{B}_{11}$ | = | $x_{11}\,\mathbf{a}_1 + x_{11}\,\mathbf{a}_2 + z_{11}\,\mathbf{a}_3$ | = | $(x_{11} a + z_{11} c\cos\beta)\,\hat{\mathbf{x}} + z_{11} c\sin\beta\,\hat{\mathbf{z}}$ | (2$a$) | Al XI |
| $\mathbf{B}_{12}$ | = | $x_{12}\,\mathbf{a}_1 + x_{12}\,\mathbf{a}_2 + z_{12}\,\mathbf{a}_3$ | = | $(x_{12} a + z_{12} c\cos\beta)\,\hat{\mathbf{x}} + z_{12} c\sin\beta\,\hat{\mathbf{z}}$ | (2$a$) | Al XII |
| $\mathbf{B}_{13}$ | = | $x_{13}\,\mathbf{a}_1 + x_{13}\,\mathbf{a}_2 + z_{13}\,\mathbf{a}_3$ | = | $(x_{13} a + z_{13} c\cos\beta)\,\hat{\mathbf{x}} + z_{13} c\sin\beta\,\hat{\mathbf{z}}$ | (2$a$) | Al XIII |
| $\mathbf{B}_{14}$ | = | $x_{14}\,\mathbf{a}_1 + x_{14}\,\mathbf{a}_2 + z_{14}\,\mathbf{a}_3$ | = | $(x_{14} a + z_{14} c\cos\beta)\,\hat{\mathbf{x}} + z_{14} c\sin\beta\,\hat{\mathbf{z}}$ | (2$a$) | Al XIV |
| $\mathbf{B}_{15}$ | = | $x_{15}\,\mathbf{a}_1 + x_{15}\,\mathbf{a}_2 + z_{15}\,\mathbf{a}_3$ | = | $(x_{15} a + z_{15} c\cos\beta)\,\hat{\mathbf{x}} + z_{15} c\sin\beta\,\hat{\mathbf{z}}$ | (2$a$) | Al XV |
| $\mathbf{B}_{16}$ | = | $x_{16}\,\mathbf{a}_1 + x_{16}\,\mathbf{a}_2 + z_{16}\,\mathbf{a}_3$ | = | $(x_{16} a + z_{16} c\cos\beta)\,\hat{\mathbf{x}} + z_{16} c\sin\beta\,\hat{\mathbf{z}}$ | (2$a$) | Al XVI |
| $\mathbf{B}_{17}$ | = | $x_{17}\,\mathbf{a}_1 + x_{17}\,\mathbf{a}_2 + z_{17}\,\mathbf{a}_3$ | = | $(x_{17} a + z_{17} c\cos\beta)\,\hat{\mathbf{x}} + z_{17} c\sin\beta\,\hat{\mathbf{z}}$ | (2$a$) | Al XVII |
| $\mathbf{B}_{18}$ | = | $x_{18}\,\mathbf{a}_1 + x_{18}\,\mathbf{a}_2 + z_{18}\,\mathbf{a}_3$ | = | $(x_{18} a + z_{18} c\cos\beta)\,\hat{\mathbf{x}} + z_{18} c\sin\beta\,\hat{\mathbf{z}}$ | (2$a$) | Co I |
| $\mathbf{B}_{19}$ | = | $x_{19}\,\mathbf{a}_1 + x_{19}\,\mathbf{a}_2 + z_{19}\,\mathbf{a}_3$ | = | $(x_{19} a + z_{19} c\cos\beta)\,\hat{\mathbf{x}} + z_{19} c\sin\beta\,\hat{\mathbf{z}}$ | (2$a$) | Co II |
| $\mathbf{B}_{20}$ | = | $x_{20}\,\mathbf{a}_1 + x_{20}\,\mathbf{a}_2 + z_{20}\,\mathbf{a}_3$ | = | $(x_{20} a + z_{20} c\cos\beta)\,\hat{\mathbf{x}} + z_{20} c\sin\beta\,\hat{\mathbf{z}}$ | (2$a$) | Co III |
| $\mathbf{B}_{21}$ | = | $x_{21}\,\mathbf{a}_1 + x_{21}\,\mathbf{a}_2 + z_{21}\,\mathbf{a}_3$ | = | $(x_{21} a + z_{21} c\cos\beta)\,\hat{\mathbf{x}} + z_{21} c\sin\beta\,\hat{\mathbf{z}}$ | (2$a$) | Co IV |
| $\mathbf{B}_{22}$ | = | $x_{22}\,\mathbf{a}_1 + x_{22}\,\mathbf{a}_2 + z_{22}\,\mathbf{a}_3$ | = | $(x_{22} a + z_{22} c\cos\beta)\,\hat{\mathbf{x}} + z_{22} c\sin\beta\,\hat{\mathbf{z}}$ | (2$a$) | Co V |
| $\mathbf{B}_{23}$ | = | $x_{23}\,\mathbf{a}_1 + x_{23}\,\mathbf{a}_2 + z_{23}\,\mathbf{a}_3$ | = | $(x_{23} a + z_{23} c\cos\beta)\,\hat{\mathbf{x}} + z_{23} c\sin\beta\,\hat{\mathbf{z}}$ | (2$a$) | Co VI |
| $\mathbf{B}_{24}$ | = | $x_{24}\,\mathbf{a}_1 + x_{24}\,\mathbf{a}_2 + z_{24}\,\mathbf{a}_3$ | = | $(x_{24} a + z_{24} c\cos\beta)\,\hat{\mathbf{x}} + z_{24} c\sin\beta\,\hat{\mathbf{z}}$ | (2$a$) | Co VII |
| $\mathbf{B}_{25}$ | = | $x_{25}\,\mathbf{a}_1 + x_{25}\,\mathbf{a}_2 + z_{25}\,\mathbf{a}_3$ | = | $(x_{25} a + z_{25} c\cos\beta)\,\hat{\mathbf{x}} + z_{25} c\sin\beta\,\hat{\mathbf{z}}$ | (2$a$) | Co VIII |
| $\mathbf{B}_{26}$ | = | $(x_{26} - y_{26})\,\mathbf{a}_1 + (x_{26} + y_{26})\,\mathbf{a}_2 + z_{26}\,\mathbf{a}_3$ | = | $(x_{26} a + z_{26} c\cos\beta)\,\hat{\mathbf{x}} + y_{26} b\,\hat{\mathbf{y}} + z_{26} c\sin\beta\,\hat{\mathbf{z}}$ | (4$b$) | Al XVIII |
| $\mathbf{B}_{27}$ | = | $(x_{26} + y_{26})\,\mathbf{a}_1 + (x_{26} - y_{26})\,\mathbf{a}_2 + z_{26}\,\mathbf{a}_3$ | = | $(x_{26} a + z_{26} c\cos\beta)\,\hat{\mathbf{x}} - y_{26} b\,\hat{\mathbf{y}} + z_{26} c\sin\beta\,\hat{\mathbf{z}}$ | (4$b$) | Al XVIII |
| $\mathbf{B}_{28}$ | = | $(x_{27} - y_{27})\,\mathbf{a}_1 + (x_{27} + y_{27})\,\mathbf{a}_2 + z_{27}\,\mathbf{a}_3$ | = | $(x_{27}a + z_{27}c\cos\beta)\,\hat{\mathbf{x}} + y_{27}b\,\hat{\mathbf{y}} + z_{27}c\sin\beta\,\hat{\mathbf{z}}$ | (4*b*) | Al XIX |
| $\mathbf{B}_{29}$ | = | $(x_{27} + y_{27})\,\mathbf{a}_1 + (x_{27} - y_{27})\,\mathbf{a}_2 + z_{27}\,\mathbf{a}_3$ | = | $(x_{27}a + z_{27}c\cos\beta)\,\hat{\mathbf{x}} - y_{27}b\,\hat{\mathbf{y}} + z_{27}c\sin\beta\,\hat{\mathbf{z}}$ | (4*b*) | Al XIX |
| $\mathbf{B}_{30}$ | = | $(x_{28} - y_{28})\,\mathbf{a}_1 + (x_{28} + y_{28})\,\mathbf{a}_2 + z_{28}\,\mathbf{a}_3$ | = | $(x_{28}a + z_{28}c\cos\beta)\,\hat{\mathbf{x}} + y_{28}b\,\hat{\mathbf{y}} + z_{28}c\sin\beta\,\hat{\mathbf{z}}$ | (4*b*) | Al XX |
| $\mathbf{B}_{31}$ | = | $(x_{28} + y_{28})\,\mathbf{a}_1 + (x_{28} - y_{28})\,\mathbf{a}_2 + z_{28}\,\mathbf{a}_3$ | = | $(x_{28}a + z_{28}c\cos\beta)\,\hat{\mathbf{x}} - y_{28}b\,\hat{\mathbf{y}} + z_{28}c\sin\beta\,\hat{\mathbf{z}}$ | (4*b*) | Al XX |
| $\mathbf{B}_{32}$ | = | $(x_{29} - y_{29})\,\mathbf{a}_1 + (x_{29} + y_{29})\,\mathbf{a}_2 + z_{29}\,\mathbf{a}_3$ | = | $(x_{29}a + z_{29}c\cos\beta)\,\hat{\mathbf{x}} + y_{29}b\,\hat{\mathbf{y}} + z_{29}c\sin\beta\,\hat{\mathbf{z}}$ | (4*b*) | Al XXI |
| $\mathbf{B}_{33}$ | = | $(x_{29} + y_{29})\,\mathbf{a}_1 + (x_{29} - y_{29})\,\mathbf{a}_2 + z_{29}\,\mathbf{a}_3$ | = | $(x_{29}a + z_{29}c\cos\beta)\,\hat{\mathbf{x}} - y_{29}b\,\hat{\mathbf{y}} + z_{29}c\sin\beta\,\hat{\mathbf{z}}$ | (4*b*) | Al XXI |
| $\mathbf{B}_{34}$ | = | $(x_{30} - y_{30})\,\mathbf{a}_1 + (x_{30} + y_{30})\,\mathbf{a}_2 + z_{30}\,\mathbf{a}_3$ | = | $(x_{30}a + z_{30}c\cos\beta)\,\hat{\mathbf{x}} + y_{30}b\,\hat{\mathbf{y}} + z_{30}c\sin\beta\,\hat{\mathbf{z}}$ | (4*b*) | Al XXII |
| $\mathbf{B}_{35}$ | = | $(x_{30} + y_{30})\,\mathbf{a}_1 + (x_{30} - y_{30})\,\mathbf{a}_2 + z_{30}\,\mathbf{a}_3$ | = | $(x_{30}a + z_{30}c\cos\beta)\,\hat{\mathbf{x}} - y_{30}b\,\hat{\mathbf{y}} + z_{30}c\sin\beta\,\hat{\mathbf{z}}$ | (4*b*) | Al XXII |
| $\mathbf{B}_{36}$ | = | $(x_{31} - y_{31})\,\mathbf{a}_1 + (x_{31} + y_{31})\,\mathbf{a}_2 + z_{31}\,\mathbf{a}_3$ | = | $(x_{31}a + z_{31}c\cos\beta)\,\hat{\mathbf{x}} + y_{31}b\,\hat{\mathbf{y}} + z_{31}c\sin\beta\,\hat{\mathbf{z}}$ | (4*b*) | Al XXIII |
| $\mathbf{B}_{37}$ | = | $(x_{31} + y_{31})\,\mathbf{a}_1 + (x_{31} - y_{31})\,\mathbf{a}_2 + z_{31}\,\mathbf{a}_3$ | = | $(x_{31}a + z_{31}c\cos\beta)\,\hat{\mathbf{x}} - y_{31}b\,\hat{\mathbf{y}} + z_{31}c\sin\beta\,\hat{\mathbf{z}}$ | (4*b*) | Al XXIII |
| $\mathbf{B}_{38}$ | = | $(x_{32} - y_{32})\,\mathbf{a}_1 + (x_{32} + y_{32})\,\mathbf{a}_2 + z_{32}\,\mathbf{a}_3$ | = | $(x_{32}a + z_{32}c\cos\beta)\,\hat{\mathbf{x}} + y_{32}b\,\hat{\mathbf{y}} + z_{32}c\sin\beta\,\hat{\mathbf{z}}$ | (4*b*) | Al XXIV |
| $\mathbf{B}_{39}$ | = | $(x_{32} + y_{32})\,\mathbf{a}_1 + (x_{32} - y_{32})\,\mathbf{a}_2 + z_{32}\,\mathbf{a}_3$ | = | $(x_{32}a + z_{32}c\cos\beta)\,\hat{\mathbf{x}} - y_{32}b\,\hat{\mathbf{y}} + z_{32}c\sin\beta\,\hat{\mathbf{z}}$ | (4*b*) | Al XXIV |
| $\mathbf{B}_{40}$ | = | $(x_{33} - y_{33})\,\mathbf{a}_1 + (x_{33} + y_{33})\,\mathbf{a}_2 + z_{33}\,\mathbf{a}_3$ | = | $(x_{33}a + z_{33}c\cos\beta)\,\hat{\mathbf{x}} + y_{33}b\,\hat{\mathbf{y}} + z_{33}c\sin\beta\,\hat{\mathbf{z}}$ | (4*b*) | Al XXV |
| $\mathbf{B}_{41}$ | = | $(x_{33} + y_{33})\,\mathbf{a}_1 + (x_{33} - y_{33})\,\mathbf{a}_2 + z_{33}\,\mathbf{a}_3$ | = | $(x_{33}a + z_{33}c\cos\beta)\,\hat{\mathbf{x}} - y_{33}b\,\hat{\mathbf{y}} + z_{33}c\sin\beta\,\hat{\mathbf{z}}$ | (4*b*) | Al XXV |
| $\mathbf{B}_{42}$ | = | $(x_{34} - y_{34})\,\mathbf{a}_1 + (x_{34} + y_{34})\,\mathbf{a}_2 + z_{34}\,\mathbf{a}_3$ | = | $(x_{34}a + z_{34}c\cos\beta)\,\hat{\mathbf{x}} + y_{34}b\,\hat{\mathbf{y}} + z_{34}c\sin\beta\,\hat{\mathbf{z}}$ | (4*b*) | Al XXVI |
| $\mathbf{B}_{43}$ | = | $(x_{34} + y_{34})\,\mathbf{a}_1 + (x_{34} - y_{34})\,\mathbf{a}_2 + z_{34}\,\mathbf{a}_3$ | = | $(x_{34}a + z_{34}c\cos\beta)\,\hat{\mathbf{x}} - y_{34}b\,\hat{\mathbf{y}} + z_{34}c\sin\beta\,\hat{\mathbf{z}}$ | (4*b*) | Al XXVI |
| $\mathbf{B}_{44}$ | = | $(x_{35} - y_{35})\,\mathbf{a}_1 + (x_{35} + y_{35})\,\mathbf{a}_2 + z_{35}\,\mathbf{a}_3$ | = | $(x_{35}a + z_{35}c\cos\beta)\,\hat{\mathbf{x}} + y_{35}b\,\hat{\mathbf{y}} + z_{35}c\sin\beta\,\hat{\mathbf{z}}$ | (4*b*) | Al XXVII |
| $\mathbf{B}_{45}$ | = | $(x_{35} + y_{35})\,\mathbf{a}_1 + (x_{35} - y_{35})\,\mathbf{a}_2 + z_{35}\,\mathbf{a}_3$ | = | $(x_{35}a + z_{35}c\cos\beta)\,\hat{\mathbf{x}} - y_{35}b\,\hat{\mathbf{y}} + z_{35}c\sin\beta\,\hat{\mathbf{z}}$ | (4*b*) | Al XXVII |
| $\mathbf{B}_{46}$ | = | $(x_{36} - y_{36})\,\mathbf{a}_1 + (x_{36} + y_{36})\,\mathbf{a}_2 + z_{36}\,\mathbf{a}_3$ | = | $(x_{36}a + z_{36}c\cos\beta)\,\hat{\mathbf{x}} + y_{36}b\,\hat{\mathbf{y}} + z_{36}c\sin\beta\,\hat{\mathbf{z}}$ | (4*b*) | Al XXVIII |
| $\mathbf{B}_{47}$ | = | $(x_{36} + y_{36})\,\mathbf{a}_1 + (x_{36} - y_{36})\,\mathbf{a}_2 + z_{36}\,\mathbf{a}_3$ | = | $(x_{36}a + z_{36}c\cos\beta)\,\hat{\mathbf{x}} - y_{36}b\,\hat{\mathbf{y}} + z_{36}c\sin\beta\,\hat{\mathbf{z}}$ | (4*b*) | Al XXVIII |
| $\mathbf{B}_{48}$ | = | $(x_{37} - y_{37})\,\mathbf{a}_1 + (x_{37} + y_{37})\,\mathbf{a}_2 + z_{37}\,\mathbf{a}_3$ | = | $(x_{37}a + z_{37}c\cos\beta)\,\hat{\mathbf{x}} + y_{37}b\,\hat{\mathbf{y}} + z_{37}c\sin\beta\,\hat{\mathbf{z}}$ | (4*b*) | Co IX |
| $\mathbf{B}_{49}$ | = | $(x_{37} + y_{37})\,\mathbf{a}_1 + (x_{37} - y_{37})\,\mathbf{a}_2 + z_{37}\,\mathbf{a}_3$ | = | $(x_{37}a + z_{37}c\cos\beta)\,\hat{\mathbf{x}} - y_{37}b\,\hat{\mathbf{y}} + z_{37}c\sin\beta\,\hat{\mathbf{z}}$ | (4*b*) | Co IX |
| $\mathbf{B}_{50}$ | = | $(x_{38} - y_{38})\,\mathbf{a}_1 + (x_{38} + y_{38})\,\mathbf{a}_2 + z_{38}\,\mathbf{a}_3$ | = | $(x_{38}a + z_{38}c\cos\beta)\,\hat{\mathbf{x}} + y_{38}b\,\hat{\mathbf{y}} + z_{38}c\sin\beta\,\hat{\mathbf{z}}$ | (4*b*) | Co X |
| $\mathbf{B}_{51}$ | = | $(x_{38} + y_{38})\,\mathbf{a}_1 + (x_{38} - y_{38})\,\mathbf{a}_2 + z_{38}\,\mathbf{a}_3$ | = | $(x_{38}a + z_{38}c\cos\beta)\,\hat{\mathbf{x}} - y_{38}b\,\hat{\mathbf{y}} + z_{38}c\sin\beta\,\hat{\mathbf{z}}$ | (4*b*) | Co X |

---

**Geometry files:**
- CIF: pp. 1516
- POSCAR: pp. 1517

# $TaTi_3$ (BCC SQS-16) Structure: AB3_mC32_8_4a_12a

http://aflow.org/prototype-encyclopedia/AB3_mC32_8_4a_12a

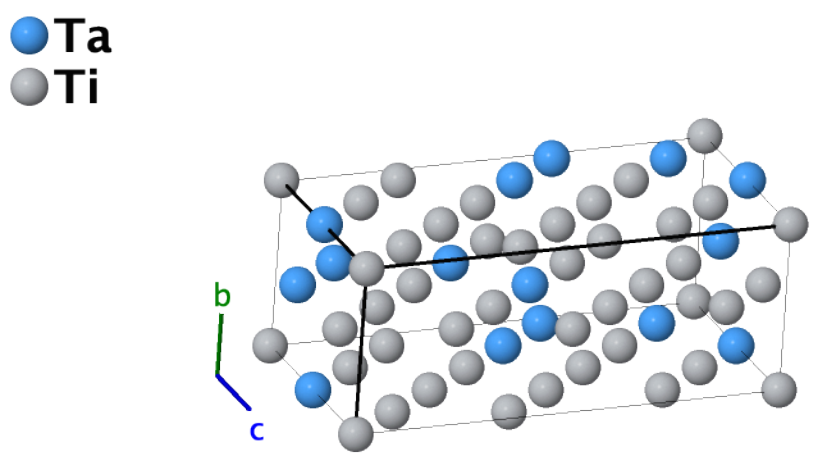

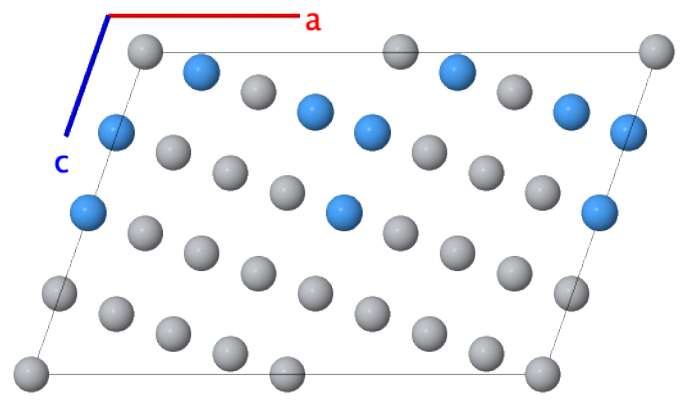

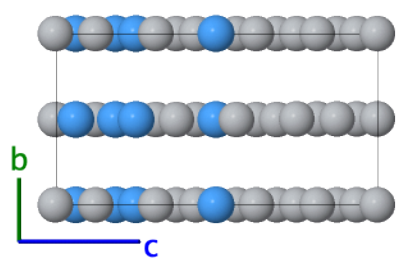

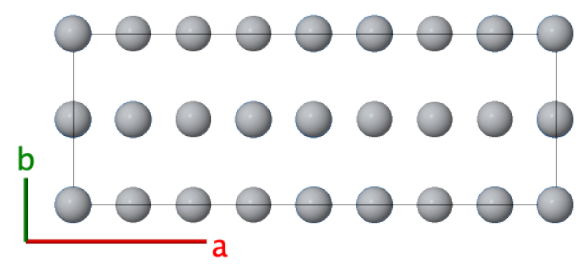

| | | |
|---|---|---|
| **Prototype** | : | $TaTi_3$ |
| **AFLOW prototype label** | : | AB3_mC32_8_4a_12a |
| ***Strukturbericht* designation** | : | None |
| **Pearson symbol** | : | mC32 |
| **Space group number** | : | 8 |
| **Space group symbol** | : | $Cm$ |
| **AFLOW prototype command** | : | `aflow --proto=AB3_mC32_8_4a_12a` `--params=`$a, b/a, c/a, \beta, x_1, z_1, x_2, z_2, x_3, z_3, x_4, z_4, x_5, z_5, x_6, z_6, x_7, z_7, x_8, z_8, x_9, z_9, x_{10}, z_{10}, x_{11}, z_{11}, x_{12}, z_{12}, x_{13}, z_{13}, x_{14}, z_{14}, x_{15}, z_{15}, x_{16}, z_{16}$ |

- This is a special quasirandom structure with 16 atoms per unit cell (SQS-16) for a bcc binary substitutional alloy $A_xB_{1-x}$ (Jiang, 2004). This prototype represents the $x = 0.25$ and $x = 0.75$ structures (change concentration by swapping elements). The $x = 0.5$ structure is given by AB_aP16_2_4i_4i.

**Base-centered Monoclinic primitive vectors:**

$$
\begin{aligned}
\mathbf{a}_1 &= \tfrac{1}{2}\,a\,\hat{\mathbf{x}} - \tfrac{1}{2}\,b\,\hat{\mathbf{y}} \\
\mathbf{a}_2 &= \tfrac{1}{2}\,a\,\hat{\mathbf{x}} + \tfrac{1}{2}\,b\,\hat{\mathbf{y}} \\
\mathbf{a}_3 &= c\cos\beta\,\hat{\mathbf{x}} + c\sin\beta\,\hat{\mathbf{z}}
\end{aligned}
$$

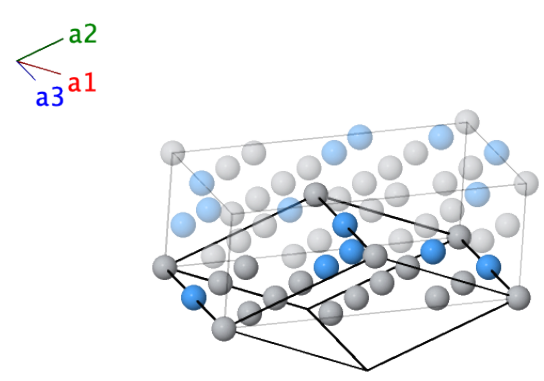

**Basis vectors:**

| | | Lattice Coordinates | | Cartesian Coordinates | Wyckoff Position | Atom Type |
|---|---|---|---|---|---|---|
| $\mathbf{B}_1$ | = | $x_1\,\mathbf{a}_1 + x_1\,\mathbf{a}_2 + z_1\,\mathbf{a}_3$ | = | $(x_1 a + z_1 c\cos\beta)\,\hat{\mathbf{x}} + z_1 c\sin\beta\,\hat{\mathbf{z}}$ | $(2a)$ | Ta I |
| $\mathbf{B}_2$ | = | $x_2\,\mathbf{a}_1 + x_2\,\mathbf{a}_2 + z_2\,\mathbf{a}_3$ | = | $(x_2 a + z_2 c\cos\beta)\,\hat{\mathbf{x}} + z_2 c\sin\beta\,\hat{\mathbf{z}}$ | $(2a)$ | Ta II |
| $\mathbf{B}_3$ | = | $x_3\,\mathbf{a}_1 + x_3\,\mathbf{a}_2 + z_3\,\mathbf{a}_3$ | = | $(x_3 a + z_3 c\cos\beta)\,\hat{\mathbf{x}} + z_3 c\sin\beta\,\hat{\mathbf{z}}$ | $(2a)$ | Ta III |
| $\mathbf{B}_4$ | = | $x_4\,\mathbf{a}_1 + x_4\,\mathbf{a}_2 + z_4\,\mathbf{a}_3$ | = | $(x_4 a + z_4 c\cos\beta)\,\hat{\mathbf{x}} + z_4 c\sin\beta\,\hat{\mathbf{z}}$ | (2a) | Ta IV |
| $\mathbf{B}_5$ | = | $x_5\,\mathbf{a}_1 + x_5\,\mathbf{a}_2 + z_5\,\mathbf{a}_3$ | = | $(x_5 a + z_5 c\cos\beta)\,\hat{\mathbf{x}} + z_5 c\sin\beta\,\hat{\mathbf{z}}$ | (2a) | Ti I |
| $\mathbf{B}_6$ | = | $x_6\,\mathbf{a}_1 + x_6\,\mathbf{a}_2 + z_6\,\mathbf{a}_3$ | = | $(x_6 a + z_6 c\cos\beta)\,\hat{\mathbf{x}} + z_6 c\sin\beta\,\hat{\mathbf{z}}$ | (2a) | Ti II |
| $\mathbf{B}_7$ | = | $x_7\,\mathbf{a}_1 + x_7\,\mathbf{a}_2 + z_7\,\mathbf{a}_3$ | = | $(x_7 a + z_7 c\cos\beta)\,\hat{\mathbf{x}} + z_7 c\sin\beta\,\hat{\mathbf{z}}$ | (2a) | Ti III |
| $\mathbf{B}_8$ | = | $x_8\,\mathbf{a}_1 + x_8\,\mathbf{a}_2 + z_8\,\mathbf{a}_3$ | = | $(x_8 a + z_8 c\cos\beta)\,\hat{\mathbf{x}} + z_8 c\sin\beta\,\hat{\mathbf{z}}$ | (2a) | Ti IV |
| $\mathbf{B}_9$ | = | $x_9\,\mathbf{a}_1 + x_9\,\mathbf{a}_2 + z_9\,\mathbf{a}_3$ | = | $(x_9 a + z_9 c\cos\beta)\,\hat{\mathbf{x}} + z_9 c\sin\beta\,\hat{\mathbf{z}}$ | (2a) | Ti V |
| $\mathbf{B}_{10}$ | = | $x_{10}\,\mathbf{a}_1 + x_{10}\,\mathbf{a}_2 + z_{10}\,\mathbf{a}_3$ | = | $(x_{10} a + z_{10} c\cos\beta)\,\hat{\mathbf{x}} + z_{10} c\sin\beta\,\hat{\mathbf{z}}$ | (2a) | Ti VI |
| $\mathbf{B}_{11}$ | = | $x_{11}\,\mathbf{a}_1 + x_{11}\,\mathbf{a}_2 + z_{11}\,\mathbf{a}_3$ | = | $(x_{11} a + z_{11} c\cos\beta)\,\hat{\mathbf{x}} + z_{11} c\sin\beta\,\hat{\mathbf{z}}$ | (2a) | Ti VII |
| $\mathbf{B}_{12}$ | = | $x_{12}\,\mathbf{a}_1 + x_{12}\,\mathbf{a}_2 + z_{12}\,\mathbf{a}_3$ | = | $(x_{12} a + z_{12} c\cos\beta)\,\hat{\mathbf{x}} + z_{12} c\sin\beta\,\hat{\mathbf{z}}$ | (2a) | Ti VIII |
| $\mathbf{B}_{13}$ | = | $x_{13}\,\mathbf{a}_1 + x_{13}\,\mathbf{a}_2 + z_{13}\,\mathbf{a}_3$ | = | $(x_{13} a + z_{13} c\cos\beta)\,\hat{\mathbf{x}} + z_{13} c\sin\beta\,\hat{\mathbf{z}}$ | (2a) | Ti IX |
| $\mathbf{B}_{14}$ | = | $x_{14}\,\mathbf{a}_1 + x_{14}\,\mathbf{a}_2 + z_{14}\,\mathbf{a}_3$ | = | $(x_{14} a + z_{14} c\cos\beta)\,\hat{\mathbf{x}} + z_{14} c\sin\beta\,\hat{\mathbf{z}}$ | (2a) | Ti X |
| $\mathbf{B}_{15}$ | = | $x_{15}\,\mathbf{a}_1 + x_{15}\,\mathbf{a}_2 + z_{15}\,\mathbf{a}_3$ | = | $(x_{15} a + z_{15} c\cos\beta)\,\hat{\mathbf{x}} + z_{15} c\sin\beta\,\hat{\mathbf{z}}$ | (2a) | Ti XI |
| $\mathbf{B}_{16}$ | = | $x_{16}\,\mathbf{a}_1 + x_{16}\,\mathbf{a}_2 + z_{16}\,\mathbf{a}_3$ | = | $(x_{16} a + z_{16} c\cos\beta)\,\hat{\mathbf{x}} + z_{16} c\sin\beta\,\hat{\mathbf{z}}$ | (2a) | Ti XII |

**Geometry files:**

- CIF: pp. 1517
- POSCAR: pp. 1518

# $TaTi_3$ (BCC SQS-16) Structure: AB3_mC32_8_4a_4a4b

http://aflow.org/prototype-encyclopedia/AB3_mC32_8_4a_4a4b

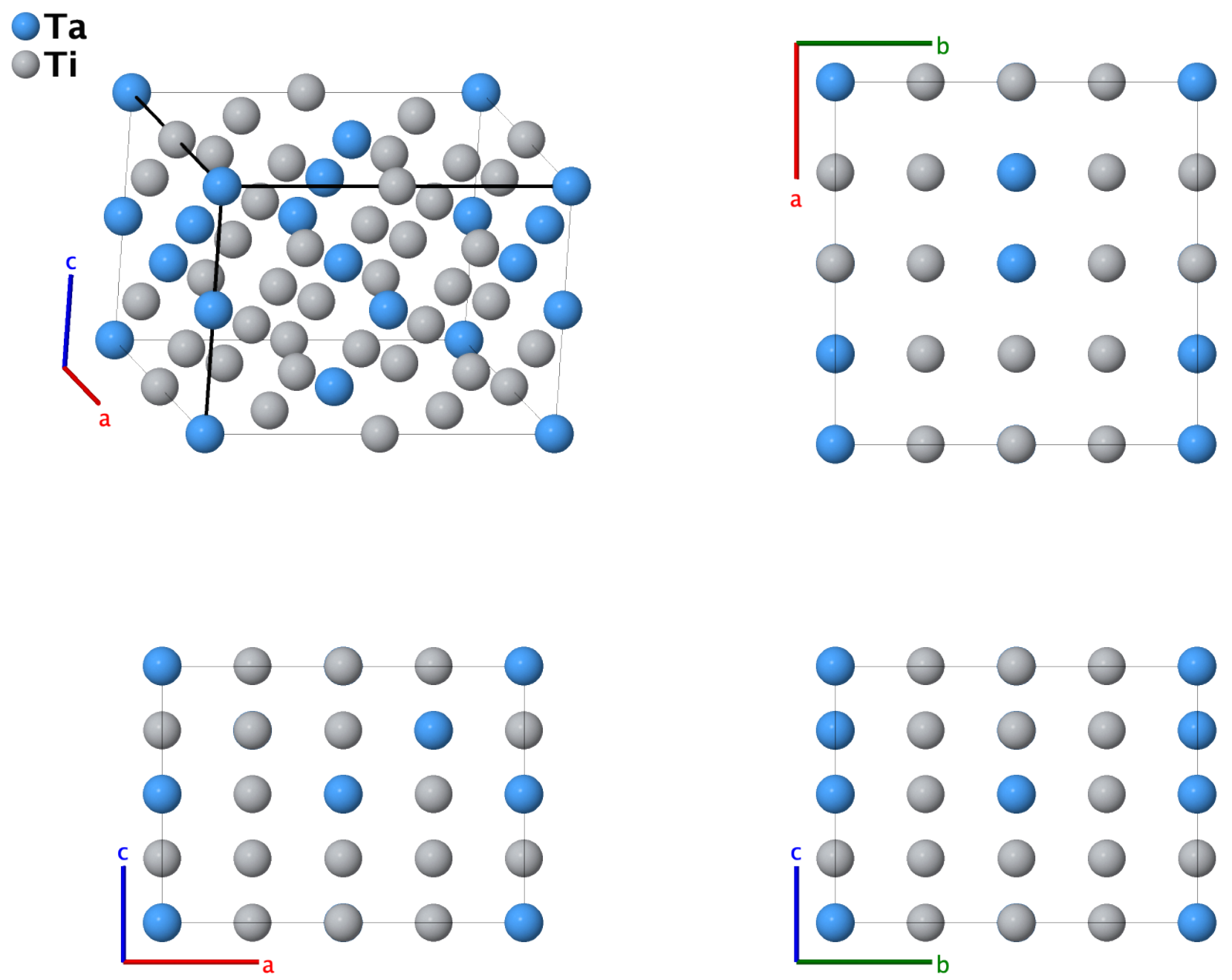

| | | |
|---|---|---|
| **Prototype** | : | $TaTi_3$ |
| **AFLOW prototype label** | : | AB3_mC32_8_4a_4a4b |
| ***Strukturbericht* designation** | : | None |
| **Pearson symbol** | : | mC32 |
| **Space group number** | : | 8 |
| **Space group symbol** | : | *Cm* |
| **AFLOW prototype command** | : | `aflow --proto=AB3_mC32_8_4a_4a4b` `--params=`$a, b/a, c/a, \beta, x_1, z_1, x_2, z_2, x_3, z_3, x_4, z_4, x_5, z_5, x_6, z_6, x_7, z_7, x_8, z_8, x_9, y_9, z_9, x_{10}, y_{10}, z_{10}, x_{11}, y_{11}, z_{11}, x_{12}, y_{12}, z_{12}$ |

- This is a special quasirandom structure with 16 atoms per unit cell (SQS-16) for the $\beta$-phase (high-temperature austenite) bcc substitutional Ti-Ta alloy (Chakraborty, 2016). This protoype contains 25% Ta. Prototypes are listed for other Ta-Ti concentrations: 12.5% Ta (AB7_hR16_166_c_c2h), 18.75% Ta (A3B13_oC32_38_ac_a2bcdef), 31.25% Ta (A5B11_mP16_6_2abc_2a3b3c), and 37.5% Ta (A3B5_oC32_38_abce_abcdf).

**Base-centered Monoclinic primitive vectors:**

$$\mathbf{a}_1 = \frac{1}{2}\,a\,\hat{\mathbf{x}} - \frac{1}{2}\,b\,\hat{\mathbf{y}}$$

$$\mathbf{a}_2 = \frac{1}{2}\,a\,\hat{\mathbf{x}} + \frac{1}{2}\,b\,\hat{\mathbf{y}}$$

$$\mathbf{a}_3 = c\cos\beta\,\hat{\mathbf{x}} + c\sin\beta\,\hat{\mathbf{z}}$$

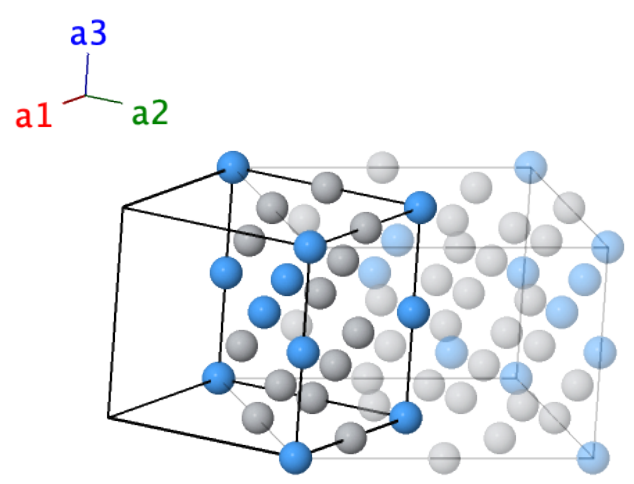

**Basis vectors:**

| | | Lattice Coordinates | | Cartesian Coordinates | Wyckoff Position | Atom Type |
|---|---|---|---|---|---|---|
| $\mathbf{B}_1$ | = | $x_1\,\mathbf{a}_1 + x_1\,\mathbf{a}_2 + z_1\,\mathbf{a}_3$ | = | $(x_1 a + z_1 c\cos\beta)\,\hat{\mathbf{x}} + z_1 c\sin\beta\,\hat{\mathbf{z}}$ | (2a) | Ta I |
| $\mathbf{B}_2$ | = | $x_2\,\mathbf{a}_1 + x_2\,\mathbf{a}_2 + z_2\,\mathbf{a}_3$ | = | $(x_2 a + z_2 c\cos\beta)\,\hat{\mathbf{x}} + z_2 c\sin\beta\,\hat{\mathbf{z}}$ | (2a) | Ta II |
| $\mathbf{B}_3$ | = | $x_3\,\mathbf{a}_1 + x_3\,\mathbf{a}_2 + z_3\,\mathbf{a}_3$ | = | $(x_3 a + z_3 c\cos\beta)\,\hat{\mathbf{x}} + z_3 c\sin\beta\,\hat{\mathbf{z}}$ | (2a) | Ta III |
| $\mathbf{B}_4$ | = | $x_4\,\mathbf{a}_1 + x_4\,\mathbf{a}_2 + z_4\,\mathbf{a}_3$ | = | $(x_4 a + z_4 c\cos\beta)\,\hat{\mathbf{x}} + z_4 c\sin\beta\,\hat{\mathbf{z}}$ | (2a) | Ta IV |
| $\mathbf{B}_5$ | = | $x_5\,\mathbf{a}_1 + x_5\,\mathbf{a}_2 + z_5\,\mathbf{a}_3$ | = | $(x_5 a + z_5 c\cos\beta)\,\hat{\mathbf{x}} + z_5 c\sin\beta\,\hat{\mathbf{z}}$ | (2a) | Ti I |
| $\mathbf{B}_6$ | = | $x_6\,\mathbf{a}_1 + x_6\,\mathbf{a}_2 + z_6\,\mathbf{a}_3$ | = | $(x_6 a + z_6 c\cos\beta)\,\hat{\mathbf{x}} + z_6 c\sin\beta\,\hat{\mathbf{z}}$ | (2a) | Ti II |
| $\mathbf{B}_7$ | = | $x_7\,\mathbf{a}_1 + x_7\,\mathbf{a}_2 + z_7\,\mathbf{a}_3$ | = | $(x_7 a + z_7 c\cos\beta)\,\hat{\mathbf{x}} + z_7 c\sin\beta\,\hat{\mathbf{z}}$ | (2a) | Ti III |
| $\mathbf{B}_8$ | = | $x_8\,\mathbf{a}_1 + x_8\,\mathbf{a}_2 + z_8\,\mathbf{a}_3$ | = | $(x_8 a + z_8 c\cos\beta)\,\hat{\mathbf{x}} + z_8 c\sin\beta\,\hat{\mathbf{z}}$ | (2a) | Ti IV |
| $\mathbf{B}_9$ | = | $(x_9 - y_9)\,\mathbf{a}_1 + (x_9 + y_9)\,\mathbf{a}_2 + z_9\,\mathbf{a}_3$ | = | $(x_9 a + z_9 c\cos\beta)\,\hat{\mathbf{x}} + y_9 b\,\hat{\mathbf{y}} + z_9 c\sin\beta\,\hat{\mathbf{z}}$ | (4b) | Ti V |
| $\mathbf{B}_{10}$ | = | $(x_9 + y_9)\,\mathbf{a}_1 + (x_9 - y_9)\,\mathbf{a}_2 + z_9\,\mathbf{a}_3$ | = | $(x_9 a + z_9 c\cos\beta)\,\hat{\mathbf{x}} - y_9 b\,\hat{\mathbf{y}} + z_9 c\sin\beta\,\hat{\mathbf{z}}$ | (4b) | Ti V |
| $\mathbf{B}_{11}$ | = | $(x_{10} - y_{10})\,\mathbf{a}_1 + (x_{10} + y_{10})\,\mathbf{a}_2 + z_{10}\,\mathbf{a}_3$ | = | $(x_{10} a + z_{10} c\cos\beta)\,\hat{\mathbf{x}} + y_{10} b\,\hat{\mathbf{y}} + z_{10} c\sin\beta\,\hat{\mathbf{z}}$ | (4b) | Ti VI |
| $\mathbf{B}_{12}$ | = | $(x_{10} + y_{10})\,\mathbf{a}_1 + (x_{10} - y_{10})\,\mathbf{a}_2 + z_{10}\,\mathbf{a}_3$ | = | $(x_{10} a + z_{10} c\cos\beta)\,\hat{\mathbf{x}} - y_{10} b\,\hat{\mathbf{y}} + z_{10} c\sin\beta\,\hat{\mathbf{z}}$ | (4b) | Ti VI |
| $\mathbf{B}_{13}$ | = | $(x_{11} - y_{11})\,\mathbf{a}_1 + (x_{11} + y_{11})\,\mathbf{a}_2 + z_{11}\,\mathbf{a}_3$ | = | $(x_{11} a + z_{11} c\cos\beta)\,\hat{\mathbf{x}} + y_{11} b\,\hat{\mathbf{y}} + z_{11} c\sin\beta\,\hat{\mathbf{z}}$ | (4b) | Ti VII |
| $\mathbf{B}_{14}$ | = | $(x_{11} + y_{11})\,\mathbf{a}_1 + (x_{11} - y_{11})\,\mathbf{a}_2 + z_{11}\,\mathbf{a}_3$ | = | $(x_{11} a + z_{11} c\cos\beta)\,\hat{\mathbf{x}} - y_{11} b\,\hat{\mathbf{y}} + z_{11} c\sin\beta\,\hat{\mathbf{z}}$ | (4b) | Ti VII |
| $\mathbf{B}_{15}$ | = | $(x_{12} - y_{12})\,\mathbf{a}_1 + (x_{12} + y_{12})\,\mathbf{a}_2 + z_{12}\,\mathbf{a}_3$ | = | $(x_{12} a + z_{12} c\cos\beta)\,\hat{\mathbf{x}} + y_{12} b\,\hat{\mathbf{y}} + z_{12} c\sin\beta\,\hat{\mathbf{z}}$ | (4b) | Ti VIII |
| $\mathbf{B}_{16}$ | = | $(x_{12} + y_{12})\,\mathbf{a}_1 + (x_{12} - y_{12})\,\mathbf{a}_2 + z_{12}\,\mathbf{a}_3$ | = | $(x_{12} a + z_{12} c\cos\beta)\,\hat{\mathbf{x}} - y_{12} b\,\hat{\mathbf{y}} + z_{12} c\sin\beta\,\hat{\mathbf{z}}$ | (4b) | Ti VIII |

**Geometry files:**

- CIF: pp. 1518
- POSCAR: pp. 1518

# $F5_{11}$ ($KNO_2$) (*obsolete*) Structure: ABC2_mC8_8_a_a_b

http://aflow.org/prototype-encyclopedia/ABC2_mC8_8_a_a_b

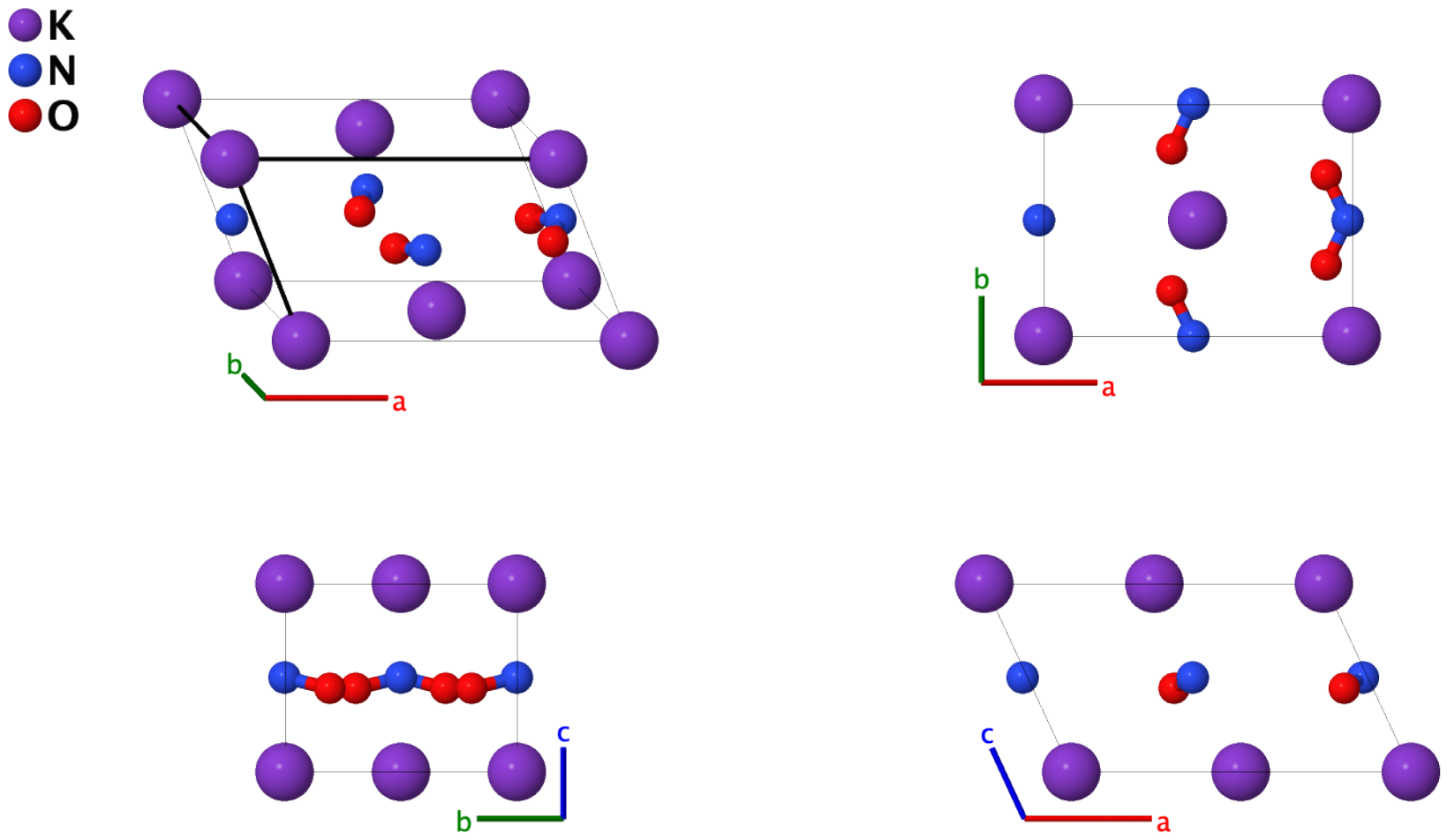

| | | |
|---|---|---|
| **Prototype** | : | $KNO_2$ |
| **AFLOW prototype label** | : | ABC2_mC8_8_a_a_b |
| ***Strukturbericht* designation** | : | $F5_{11}$ |
| **Pearson symbol** | : | mC8 |
| **Space group number** | : | 8 |
| **Space group symbol** | : | $Cm$ |
| **AFLOW prototype command** | : | `aflow --proto=ABC2_mC8_8_a_a_b` `--params=`$a, b/a, c/a, \beta, x_1, z_1, x_2, z_2, x_3, y_3, z_3$ |

- "The room-temperature structure of $KNO_2$ was first considered to have monoclinic symmetry, ..., but recent studies have established the structure to be rhombohedral ..." (Rao, 1975). The $F5_{11}$ structure is thus neither the ground state structure of $KNO_2$ nor the room-temperature structure, which is somewhat disordered with space group $R\bar{3}m$. We present this structure as part of the historical record.
- (Ziegler, 1936) gave this structure in the $Am$ setting of space group #8. We used FINDSYM to transform it to the standard $Cm$ setting, which involved a considerable change in the orientation and length of the primitive lattice vectors.

**Base-centered Monoclinic primitive vectors:**

$$\mathbf{a}_1 = \frac{1}{2}\,a\,\hat{\mathbf{x}} - \frac{1}{2}\,b\,\hat{\mathbf{y}}$$
$$\mathbf{a}_2 = \frac{1}{2}\,a\,\hat{\mathbf{x}} + \frac{1}{2}\,b\,\hat{\mathbf{y}}$$
$$\mathbf{a}_3 = c\cos\beta\,\hat{\mathbf{x}} + c\sin\beta\,\hat{\mathbf{z}}$$

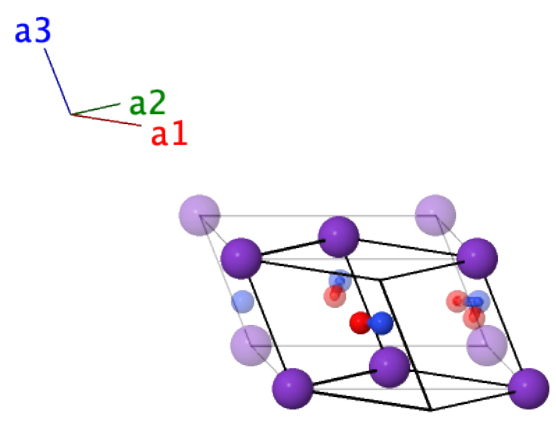

**Basis vectors:**

| | | Lattice Coordinates | | Cartesian Coordinates | Wyckoff Position | Atom Type |
|---|---|---|---|---|---|---|
| $\mathbf{B}_1$ | = | $x_1\,\mathbf{a}_1 + x_1\,\mathbf{a}_2 + z_1\,\mathbf{a}_3$ | = | $(x_1 a + z_1 c\cos\beta)\,\hat{\mathbf{x}} + z_1 c\sin\beta\,\hat{\mathbf{z}}$ | (2$a$) | K |
| $\mathbf{B}_2$ | = | $x_2\,\mathbf{a}_1 + x_2\,\mathbf{a}_2 + z_2\,\mathbf{a}_3$ | = | $(x_2 a + z_2 c\cos\beta)\,\hat{\mathbf{x}} + z_2 c\sin\beta\,\hat{\mathbf{z}}$ | (2$a$) | N |
| $\mathbf{B}_3$ | = | $(x_3 - y_3)\,\mathbf{a}_1 + (x_3 + y_3)\,\mathbf{a}_2 + z_3\,\mathbf{a}_3$ | = | $(x_3 a + z_3 c\cos\beta)\,\hat{\mathbf{x}} + y_3 b\,\hat{\mathbf{y}} + z_3 c\sin\beta\,\hat{\mathbf{z}}$ | (4$b$) | O |
| $\mathbf{B}_4$ | = | $(x_3 + y_3)\,\mathbf{a}_1 + (x_3 - y_3)\,\mathbf{a}_2 + z_3\,\mathbf{a}_3$ | = | $(x_3 a + z_3 c\cos\beta)\,\hat{\mathbf{x}} - y_3 b\,\hat{\mathbf{y}} + z_3 c\sin\beta\,\hat{\mathbf{z}}$ | (4$b$) | O |

**Geometry files:**

- CIF: pp. 1518
- POSCAR: pp. 1519

# Nacrite [$Al_2Si_2O_5(OH)_4$, $S5_4$] Structure: A2B4C9D2_mC68_9_2a_4a_9a_2a

http://aflow.org/prototype-encyclopedia/A2B4C9D2_mC68_9_2a_4a_9a_2a

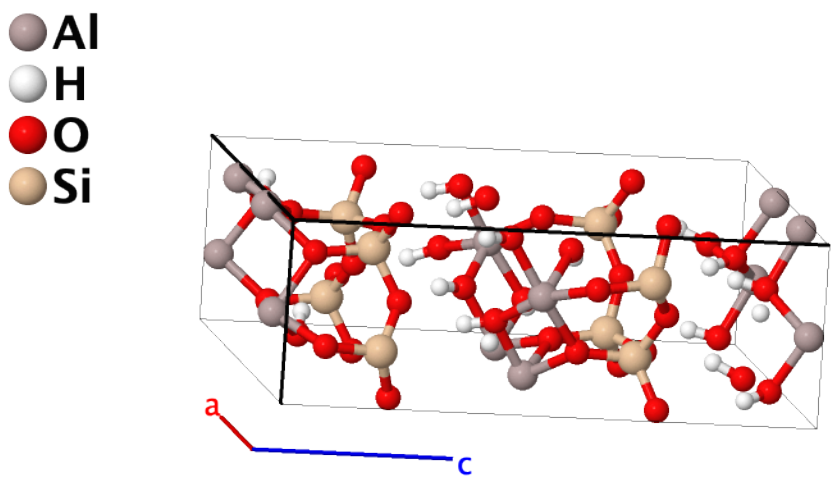

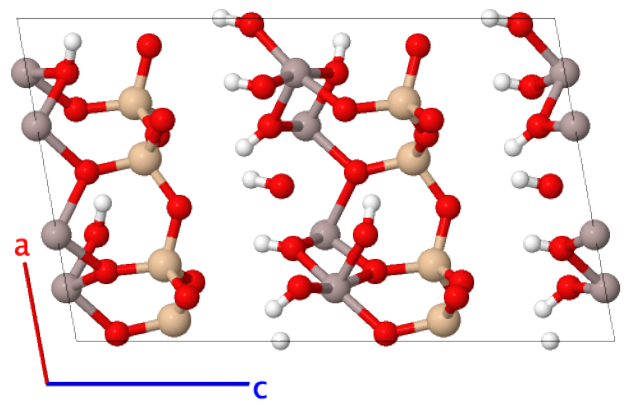

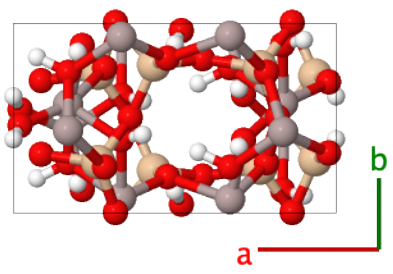

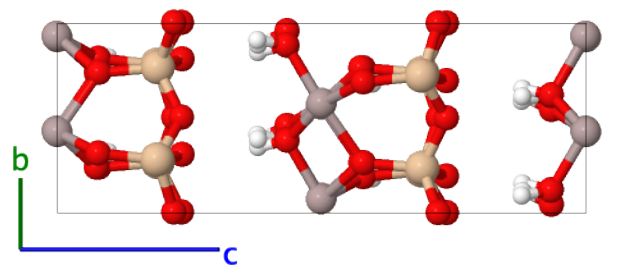

| | | |
|---|---|---|
| **Prototype** | : | $Al_2H_4O_9Si_2$ |
| **AFLOW prototype label** | : | A2B4C9D2_mC68_9_2a_4a_9a_2a |
| ***Strukturbericht* designation** | : | $S5_4$ |
| **Pearson symbol** | : | mC68 |
| **Space group number** | : | 9 |
| **Space group symbol** | : | $Cc$ |
| **AFLOW prototype command** | : | `aflow --proto=A2B4C9D2_mC68_9_2a_4a_9a_2a --params=`$a, b/a, c/a, \beta, x_1, y_1, z_1, x_2, y_2, z_2, x_3, y_3, z_3, x_4, y_4, z_4, x_5, y_5, z_5, x_6, y_6, z_6, x_7, y_7, z_7, x_8, y_8, z_8, x_9, y_9, z_9, x_{10}, y_{10}, z_{10}, x_{11}, y_{11}, z_{11}, x_{12}, y_{12}, z_{12}, x_{13}, y_{13}, z_{13}, x_{14}, y_{14}, z_{14}, x_{15}, y_{15}, z_{15}, x_{16}, y_{16}, z_{16}, x_{17}, y_{17}, z_{17}$ |

- (Herrmann, 1943) gave the nacrite structure determined by (Hendricks, 1939) the *Strukturbericht* symbol $S5_4$. This structure was somewhat uncertain, as Hendricks knew that the structure had space group $Cc$ but could only determine the structure in terms of a pseudo-rhombohedral unit cell. This led to a tripling of the unit cell compared to the work of (Zhukhlistov, 2008), who was also able to determine the positions of the hydrogen atoms. We use the modern structure, noting that we can essentially recover the original structure by tripling the unit cell along the $c$-axis.

**Base-centered Monoclinic primitive vectors:**

$$\mathbf{a}_1 = \frac{1}{2}\,a\,\hat{\mathbf{x}} - \frac{1}{2}\,b\,\hat{\mathbf{y}}$$
$$\mathbf{a}_2 = \frac{1}{2}\,a\,\hat{\mathbf{x}} + \frac{1}{2}\,b\,\hat{\mathbf{y}}$$
$$\mathbf{a}_3 = c\cos\beta\,\hat{\mathbf{x}} + c\sin\beta\,\hat{\mathbf{z}}$$

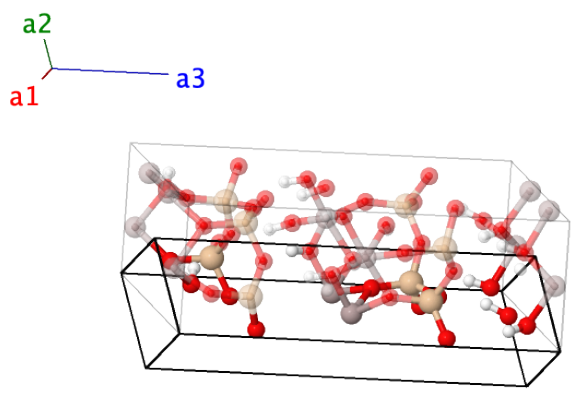

**Basis vectors:**

| | | Lattice Coordinates | | Cartesian Coordinates | Wyckoff Position | Atom Type |
|---|---|---|---|---|---|---|
| $\mathbf{B}_1$ | = | $(x_1 - y_1)\,\mathbf{a}_1 + (x_1 + y_1)\,\mathbf{a}_2 + z_1\,\mathbf{a}_3$ | = | $(x_1 a + z_1 c\cos\beta)\,\hat{\mathbf{x}} + y_1 b\,\hat{\mathbf{y}} + z_1 c\sin\beta\,\hat{\mathbf{z}}$ | (4$a$) | Al I |
| $\mathbf{B}_2$ | = | $(x_1 + y_1)\,\mathbf{a}_1 + (x_1 - y_1)\,\mathbf{a}_2 + \left(\frac{1}{2} + z_1\right)\mathbf{a}_3$ | = | $\left(\frac{1}{2}c\cos\beta + x_1 a + z_1 c\cos\beta\right)\hat{\mathbf{x}} - y_1 b\,\hat{\mathbf{y}} + \left(\frac{1}{2} + z_1\right)c\sin\beta\,\hat{\mathbf{z}}$ | (4$a$) | Al I |
| $\mathbf{B}_3$ | = | $(x_2 - y_2)\,\mathbf{a}_1 + (x_2 + y_2)\,\mathbf{a}_2 + z_2\,\mathbf{a}_3$ | = | $(x_2 a + z_2 c\cos\beta)\,\hat{\mathbf{x}} + y_2 b\,\hat{\mathbf{y}} + z_2 c\sin\beta\,\hat{\mathbf{z}}$ | (4$a$) | Al II |
| $\mathbf{B}_4$ | = | $(x_2 + y_2)\,\mathbf{a}_1 + (x_2 - y_2)\,\mathbf{a}_2 + \left(\frac{1}{2} + z_2\right)\mathbf{a}_3$ | = | $\left(\frac{1}{2}c\cos\beta + x_2 a + z_2 c\cos\beta\right)\hat{\mathbf{x}} - y_2 b\,\hat{\mathbf{y}} + \left(\frac{1}{2} + z_2\right)c\sin\beta\,\hat{\mathbf{z}}$ | (4$a$) | Al II |
| $\mathbf{B}_5$ | = | $(x_3 - y_3)\,\mathbf{a}_1 + (x_3 + y_3)\,\mathbf{a}_2 + z_3\,\mathbf{a}_3$ | = | $(x_3 a + z_3 c\cos\beta)\,\hat{\mathbf{x}} + y_3 b\,\hat{\mathbf{y}} + z_3 c\sin\beta\,\hat{\mathbf{z}}$ | (4$a$) | H I |
| $\mathbf{B}_6$ | = | $(x_3 + y_3)\,\mathbf{a}_1 + (x_3 - y_3)\,\mathbf{a}_2 + \left(\frac{1}{2} + z_3\right)\mathbf{a}_3$ | = | $\left(\frac{1}{2}c\cos\beta + x_3 a + z_3 c\cos\beta\right)\hat{\mathbf{x}} - y_3 b\,\hat{\mathbf{y}} + \left(\frac{1}{2} + z_3\right)c\sin\beta\,\hat{\mathbf{z}}$ | (4$a$) | H I |
| $\mathbf{B}_7$ | = | $(x_4 - y_4)\,\mathbf{a}_1 + (x_4 + y_4)\,\mathbf{a}_2 + z_4\,\mathbf{a}_3$ | = | $(x_4 a + z_4 c\cos\beta)\,\hat{\mathbf{x}} + y_4 b\,\hat{\mathbf{y}} + z_4 c\sin\beta\,\hat{\mathbf{z}}$ | (4$a$) | H II |
| $\mathbf{B}_8$ | = | $(x_4 + y_4)\,\mathbf{a}_1 + (x_4 - y_4)\,\mathbf{a}_2 + \left(\frac{1}{2} + z_4\right)\mathbf{a}_3$ | = | $\left(\frac{1}{2}c\cos\beta + x_4 a + z_4 c\cos\beta\right)\hat{\mathbf{x}} - y_4 b\,\hat{\mathbf{y}} + \left(\frac{1}{2} + z_4\right)c\sin\beta\,\hat{\mathbf{z}}$ | (4$a$) | H II |
| $\mathbf{B}_9$ | = | $(x_5 - y_5)\,\mathbf{a}_1 + (x_5 + y_5)\,\mathbf{a}_2 + z_5\,\mathbf{a}_3$ | = | $(x_5 a + z_5 c\cos\beta)\,\hat{\mathbf{x}} + y_5 b\,\hat{\mathbf{y}} + z_5 c\sin\beta\,\hat{\mathbf{z}}$ | (4$a$) | H III |
| $\mathbf{B}_{10}$ | = | $(x_5 + y_5)\,\mathbf{a}_1 + (x_5 - y_5)\,\mathbf{a}_2 + \left(\frac{1}{2} + z_5\right)\mathbf{a}_3$ | = | $\left(\frac{1}{2}c\cos\beta + x_5 a + z_5 c\cos\beta\right)\hat{\mathbf{x}} - y_5 b\,\hat{\mathbf{y}} + \left(\frac{1}{2} + z_5\right)c\sin\beta\,\hat{\mathbf{z}}$ | (4$a$) | H III |
| $\mathbf{B}_{11}$ | = | $(x_6 - y_6)\,\mathbf{a}_1 + (x_6 + y_6)\,\mathbf{a}_2 + z_6\,\mathbf{a}_3$ | = | $(x_6 a + z_6 c\cos\beta)\,\hat{\mathbf{x}} + y_6 b\,\hat{\mathbf{y}} + z_6 c\sin\beta\,\hat{\mathbf{z}}$ | (4$a$) | H IV |
| $\mathbf{B}_{12}$ | = | $(x_6 + y_6)\,\mathbf{a}_1 + (x_6 - y_6)\,\mathbf{a}_2 + \left(\frac{1}{2} + z_6\right)\mathbf{a}_3$ | = | $\left(\frac{1}{2}c\cos\beta + x_6 a + z_6 c\cos\beta\right)\hat{\mathbf{x}} - y_6 b\,\hat{\mathbf{y}} + \left(\frac{1}{2} + z_6\right)c\sin\beta\,\hat{\mathbf{z}}$ | (4$a$) | H IV |
| $\mathbf{B}_{13}$ | = | $(x_7 - y_7)\,\mathbf{a}_1 + (x_7 + y_7)\,\mathbf{a}_2 + z_7\,\mathbf{a}_3$ | = | $(x_7 a + z_7 c\cos\beta)\,\hat{\mathbf{x}} + y_7 b\,\hat{\mathbf{y}} + z_7 c\sin\beta\,\hat{\mathbf{z}}$ | (4$a$) | O I |
| $\mathbf{B}_{14}$ | = | $(x_7 + y_7)\,\mathbf{a}_1 + (x_7 - y_7)\,\mathbf{a}_2 + \left(\frac{1}{2} + z_7\right)\mathbf{a}_3$ | = | $\left(\frac{1}{2}c\cos\beta + x_7 a + z_7 c\cos\beta\right)\hat{\mathbf{x}} - y_7 b\,\hat{\mathbf{y}} + \left(\frac{1}{2} + z_7\right)c\sin\beta\,\hat{\mathbf{z}}$ | (4$a$) | O I |
| $\mathbf{B}_{15}$ | = | $(x_8 - y_8)\,\mathbf{a}_1 + (x_8 + y_8)\,\mathbf{a}_2 + z_8\,\mathbf{a}_3$ | = | $(x_8 a + z_8 c\cos\beta)\,\hat{\mathbf{x}} + y_8 b\,\hat{\mathbf{y}} + z_8 c\sin\beta\,\hat{\mathbf{z}}$ | (4$a$) | O II |
| $\mathbf{B}_{16}$ | = | $(x_8 + y_8)\,\mathbf{a}_1 + (x_8 - y_8)\,\mathbf{a}_2 + \left(\frac{1}{2} + z_8\right)\mathbf{a}_3$ | = | $\left(\frac{1}{2}c\cos\beta + x_8 a + z_8 c\cos\beta\right)\hat{\mathbf{x}} - y_8 b\,\hat{\mathbf{y}} + \left(\frac{1}{2} + z_8\right)c\sin\beta\,\hat{\mathbf{z}}$ | (4$a$) | O II |
| $\mathbf{B}_{17}$ | = | $(x_9 - y_9)\,\mathbf{a}_1 + (x_9 + y_9)\,\mathbf{a}_2 + z_9\,\mathbf{a}_3$ | = | $(x_9 a + z_9 c\cos\beta)\,\hat{\mathbf{x}} + y_9 b\,\hat{\mathbf{y}} + z_9 c\sin\beta\,\hat{\mathbf{z}}$ | (4$a$) | O III |
| $\mathbf{B}_{18}$ | = | $(x_9 + y_9)\,\mathbf{a}_1 + (x_9 - y_9)\,\mathbf{a}_2 + \left(\frac{1}{2} + z_9\right)\mathbf{a}_3$ | = | $\left(\frac{1}{2}c\cos\beta + x_9 a + z_9 c\cos\beta\right)\hat{\mathbf{x}} - y_9 b\,\hat{\mathbf{y}} + \left(\frac{1}{2} + z_9\right)c\sin\beta\,\hat{\mathbf{z}}$ | (4$a$) | O III |
| $\mathbf{B}_{19}$ | = | $(x_{10} - y_{10})\,\mathbf{a}_1 + (x_{10} + y_{10})\,\mathbf{a}_2 + z_{10}\,\mathbf{a}_3$ | = | $(x_{10} a + z_{10} c\cos\beta)\,\hat{\mathbf{x}} + y_{10} b\,\hat{\mathbf{y}} + z_{10} c\sin\beta\,\hat{\mathbf{z}}$ | (4$a$) | O IV |
| $\mathbf{B}_{20}$ | = | $(x_{10} + y_{10})\,\mathbf{a}_1 + (x_{10} - y_{10})\,\mathbf{a}_2 + \left(\frac{1}{2} + z_{10}\right)\mathbf{a}_3$ | = | $\left(\frac{1}{2}c\cos\beta + x_{10} a + z_{10} c\cos\beta\right)\hat{\mathbf{x}} - y_{10} b\,\hat{\mathbf{y}} + \left(\frac{1}{2} + z_{10}\right)c\sin\beta\,\hat{\mathbf{z}}$ | (4$a$) | O IV |
| $\mathbf{B}_{21}$ | = | $(x_{11} - y_{11})\,\mathbf{a}_1 + (x_{11} + y_{11})\,\mathbf{a}_2 + z_{11}\,\mathbf{a}_3$ | = | $(x_{11} a + z_{11} c\cos\beta)\,\hat{\mathbf{x}} + y_{11} b\,\hat{\mathbf{y}} + z_{11} c\sin\beta\,\hat{\mathbf{z}}$ | (4$a$) | O V |
| $\mathbf{B}_{22}$ | = | $(x_{11}+y_{11})\,\mathbf{a}_1+(x_{11}-y_{11})\,\mathbf{a}_2+\left(\frac{1}{2}+z_{11}\right)\,\mathbf{a}_3$ | = | $\left(\frac{1}{2}c\cos\beta+x_{11}a+z_{11}c\cos\beta\right)\,\hat{\mathbf{x}}-y_{11}b\,\hat{\mathbf{y}}+\left(\frac{1}{2}+z_{11}\right)c\sin\beta\,\hat{\mathbf{z}}$ | (4$a$) | O V |
| $\mathbf{B}_{23}$ | = | $(x_{12}-y_{12})\,\mathbf{a}_1+(x_{12}+y_{12})\,\mathbf{a}_2+z_{12}\,\mathbf{a}_3$ | = | $(x_{12}a+z_{12}c\cos\beta)\,\hat{\mathbf{x}}+y_{12}b\,\hat{\mathbf{y}}+z_{12}c\sin\beta\,\hat{\mathbf{z}}$ | (4$a$) | O VI |
| $\mathbf{B}_{24}$ | = | $(x_{12}+y_{12})\,\mathbf{a}_1+(x_{12}-y_{12})\,\mathbf{a}_2+\left(\frac{1}{2}+z_{12}\right)\,\mathbf{a}_3$ | = | $\left(\frac{1}{2}c\cos\beta+x_{12}a+z_{12}c\cos\beta\right)\,\hat{\mathbf{x}}-y_{12}b\,\hat{\mathbf{y}}+\left(\frac{1}{2}+z_{12}\right)c\sin\beta\,\hat{\mathbf{z}}$ | (4$a$) | O VI |
| $\mathbf{B}_{25}$ | = | $(x_{13}-y_{13})\,\mathbf{a}_1+(x_{13}+y_{13})\,\mathbf{a}_2+z_{13}\,\mathbf{a}_3$ | = | $(x_{13}a+z_{13}c\cos\beta)\,\hat{\mathbf{x}}+y_{13}b\,\hat{\mathbf{y}}+z_{13}c\sin\beta\,\hat{\mathbf{z}}$ | (4$a$) | O VII |
| $\mathbf{B}_{26}$ | = | $(x_{13}+y_{13})\,\mathbf{a}_1+(x_{13}-y_{13})\,\mathbf{a}_2+\left(\frac{1}{2}+z_{13}\right)\,\mathbf{a}_3$ | = | $\left(\frac{1}{2}c\cos\beta+x_{13}a+z_{13}c\cos\beta\right)\,\hat{\mathbf{x}}-y_{13}b\,\hat{\mathbf{y}}+\left(\frac{1}{2}+z_{13}\right)c\sin\beta\,\hat{\mathbf{z}}$ | (4$a$) | O VII |
| $\mathbf{B}_{27}$ | = | $(x_{14}-y_{14})\,\mathbf{a}_1+(x_{14}+y_{14})\,\mathbf{a}_2+z_{14}\,\mathbf{a}_3$ | = | $(x_{14}a+z_{14}c\cos\beta)\,\hat{\mathbf{x}}+y_{14}b\,\hat{\mathbf{y}}+z_{14}c\sin\beta\,\hat{\mathbf{z}}$ | (4$a$) | O VIII |
| $\mathbf{B}_{28}$ | = | $(x_{14}+y_{14})\,\mathbf{a}_1+(x_{14}-y_{14})\,\mathbf{a}_2+\left(\frac{1}{2}+z_{14}\right)\,\mathbf{a}_3$ | = | $\left(\frac{1}{2}c\cos\beta+x_{14}a+z_{14}c\cos\beta\right)\,\hat{\mathbf{x}}-y_{14}b\,\hat{\mathbf{y}}+\left(\frac{1}{2}+z_{14}\right)c\sin\beta\,\hat{\mathbf{z}}$ | (4$a$) | O VIII |
| $\mathbf{B}_{29}$ | = | $(x_{15}-y_{15})\,\mathbf{a}_1+(x_{15}+y_{15})\,\mathbf{a}_2+z_{15}\,\mathbf{a}_3$ | = | $(x_{15}a+z_{15}c\cos\beta)\,\hat{\mathbf{x}}+y_{15}b\,\hat{\mathbf{y}}+z_{15}c\sin\beta\,\hat{\mathbf{z}}$ | (4$a$) | O IX |
| $\mathbf{B}_{30}$ | = | $(x_{15}+y_{15})\,\mathbf{a}_1+(x_{15}-y_{15})\,\mathbf{a}_2+\left(\frac{1}{2}+z_{15}\right)\,\mathbf{a}_3$ | = | $\left(\frac{1}{2}c\cos\beta+x_{15}a+z_{15}c\cos\beta\right)\,\hat{\mathbf{x}}-y_{15}b\,\hat{\mathbf{y}}+\left(\frac{1}{2}+z_{15}\right)c\sin\beta\,\hat{\mathbf{z}}$ | (4$a$) | O IX |
| $\mathbf{B}_{31}$ | = | $(x_{16}-y_{16})\,\mathbf{a}_1+(x_{16}+y_{16})\,\mathbf{a}_2+z_{16}\,\mathbf{a}_3$ | = | $(x_{16}a+z_{16}c\cos\beta)\,\hat{\mathbf{x}}+y_{16}b\,\hat{\mathbf{y}}+z_{16}c\sin\beta\,\hat{\mathbf{z}}$ | (4$a$) | Si I |
| $\mathbf{B}_{32}$ | = | $(x_{16}+y_{16})\,\mathbf{a}_1+(x_{16}-y_{16})\,\mathbf{a}_2+\left(\frac{1}{2}+z_{16}\right)\,\mathbf{a}_3$ | = | $\left(\frac{1}{2}c\cos\beta+x_{16}a+z_{16}c\cos\beta\right)\,\hat{\mathbf{x}}-y_{16}b\,\hat{\mathbf{y}}+\left(\frac{1}{2}+z_{16}\right)c\sin\beta\,\hat{\mathbf{z}}$ | (4$a$) | Si I |
| $\mathbf{B}_{33}$ | = | $(x_{17}-y_{17})\,\mathbf{a}_1+(x_{17}+y_{17})\,\mathbf{a}_2+z_{17}\,\mathbf{a}_3$ | = | $(x_{17}a+z_{17}c\cos\beta)\,\hat{\mathbf{x}}+y_{17}b\,\hat{\mathbf{y}}+z_{17}c\sin\beta\,\hat{\mathbf{z}}$ | (4$a$) | Si II |
| $\mathbf{B}_{34}$ | = | $(x_{17}+y_{17})\,\mathbf{a}_1+(x_{17}-y_{17})\,\mathbf{a}_2+\left(\frac{1}{2}+z_{17}\right)\,\mathbf{a}_3$ | = | $\left(\frac{1}{2}c\cos\beta+x_{17}a+z_{17}c\cos\beta\right)\,\hat{\mathbf{x}}-y_{17}b\,\hat{\mathbf{y}}+\left(\frac{1}{2}+z_{17}\right)c\sin\beta\,\hat{\mathbf{z}}$ | (4$a$) | Si II |

**Geometry files:**

- CIF: pp. 1519
- POSCAR: pp. 1519

# Chrysotile ($Mg_3Si_2O_5(OH)_4$) Structure: A3B5C4D2_mC56_9_3a_5a_4a_2a

http://aflow.org/prototype-encyclopedia/A3B5C4D2_mC56_9_3a_5a_4a_2a

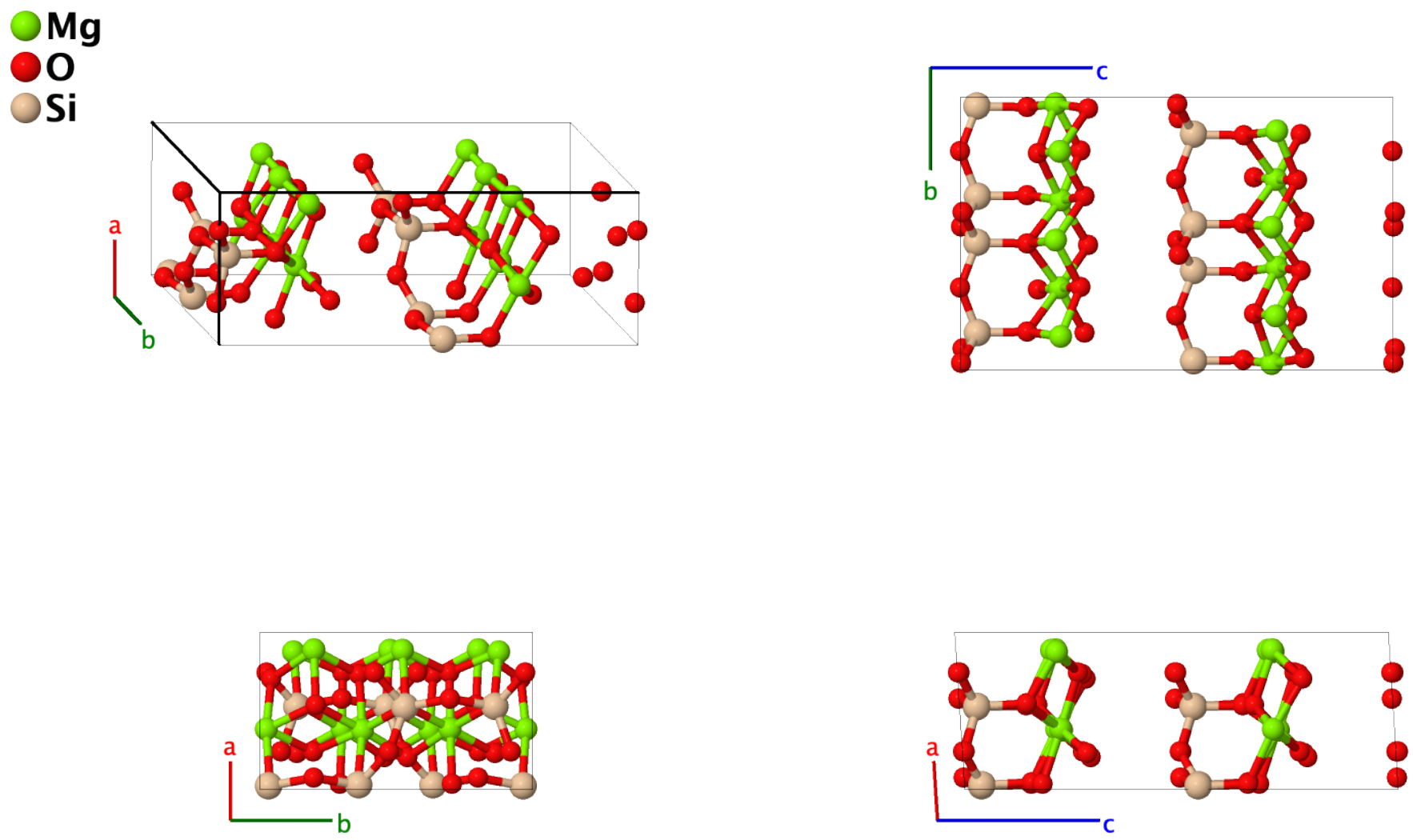

| | | |
|---|---|---|
| **Prototype** | : | $Mg_3O_5(OH)_4Si_2$ |
| **AFLOW prototype label** | : | A3B5C4D2_mC56_9_3a_5a_4a_2a |
| ***Strukturbericht* designation** | : | None |
| **Pearson symbol** | : | mC56 |
| **Space group number** | : | 9 |
| **Space group symbol** | : | $Cc$ |
| **AFLOW prototype command** | : | `aflow --proto=A3B5C4D2_mC56_9_3a_5a_4a_2a` `--params=`$a, b/a, c/a, \beta, x_1, y_1, z_1, x_2, y_2, z_2, x_3, y_3, z_3, x_4, y_4, z_4, x_5, y_5, z_5, x_6, y_6, z_6, x_7, y_7, z_7, x_8, y_8, z_8, x_9, y_9, z_9, x_{10}, y_{10}, z_{10}, x_{11}, y_{11}, z_{11}, x_{12}, y_{12}, z_{12}, x_{13}, y_{13}, z_{13}, x_{14}, y_{14}, z_{14}$ |

- Chrysotile is more commonly known as "white asbestos." Chrysotile sheets typically curl into tubular fibers and the crystal structure is difficult to determine. (Yada, 1967) has a partial list of the experiments performed to determine this structure. This differs from the early work of (Warren, 1930), $S4_5$, which halves the size of the unit cell and removes the inversion site. For this prototype, the OH molecules are centered on four $a$ Wyckoff positions.

**Base-centered Monoclinic primitive vectors:**

$$\mathbf{a}_1 = \frac{1}{2}\, a\, \hat{\mathbf{x}} - \frac{1}{2}\, b\, \hat{\mathbf{y}}$$
$$\mathbf{a}_2 = \frac{1}{2}\, a\, \hat{\mathbf{x}} + \frac{1}{2}\, b\, \hat{\mathbf{y}}$$
$$\mathbf{a}_3 = c \cos\beta\, \hat{\mathbf{x}} + c \sin\beta\, \hat{\mathbf{z}}$$

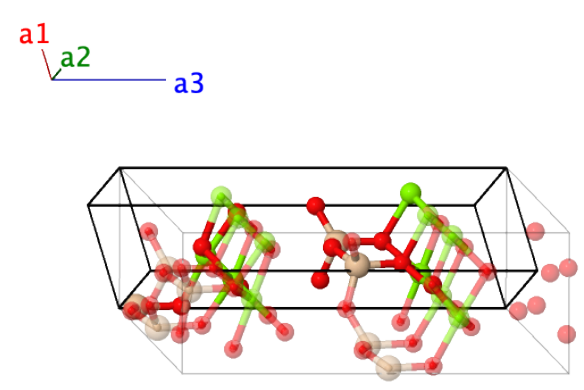

**Basis vectors:**

| | | Lattice Coordinates | | Cartesian Coordinates | Wyckoff Position | Atom Type |
|---|---|---|---|---|---|---|
| $\mathbf{B}_{1}$ | = | $(x_1 - y_1)\,\mathbf{a}_1 + (x_1 + y_1)\,\mathbf{a}_2 + z_1\,\mathbf{a}_3$ | = | $(x_1 a + z_1 c\cos\beta)\,\hat{\mathbf{x}} + y_1 b\,\hat{\mathbf{y}} + z_1 c\sin\beta\,\hat{\mathbf{z}}$ | (4a) | Mg I |
| $\mathbf{B}_{2}$ | = | $(x_1 + y_1)\,\mathbf{a}_1 + (x_1 - y_1)\,\mathbf{a}_2 + \left(\frac{1}{2} + z_1\right)\mathbf{a}_3$ | = | $\left(\frac{1}{2}c\cos\beta + x_1 a + z_1 c\cos\beta\right)\hat{\mathbf{x}} - y_1 b\,\hat{\mathbf{y}} + \left(\frac{1}{2} + z_1\right)c\sin\beta\,\hat{\mathbf{z}}$ | (4a) | Mg I |
| $\mathbf{B}_{3}$ | = | $(x_2 - y_2)\,\mathbf{a}_1 + (x_2 + y_2)\,\mathbf{a}_2 + z_2\,\mathbf{a}_3$ | = | $(x_2 a + z_2 c\cos\beta)\,\hat{\mathbf{x}} + y_2 b\,\hat{\mathbf{y}} + z_2 c\sin\beta\,\hat{\mathbf{z}}$ | (4a) | Mg II |
| $\mathbf{B}_{4}$ | = | $(x_2 + y_2)\,\mathbf{a}_1 + (x_2 - y_2)\,\mathbf{a}_2 + \left(\frac{1}{2} + z_2\right)\mathbf{a}_3$ | = | $\left(\frac{1}{2}c\cos\beta + x_2 a + z_2 c\cos\beta\right)\hat{\mathbf{x}} - y_2 b\,\hat{\mathbf{y}} + \left(\frac{1}{2} + z_2\right)c\sin\beta\,\hat{\mathbf{z}}$ | (4a) | Mg II |
| $\mathbf{B}_{5}$ | = | $(x_3 - y_3)\,\mathbf{a}_1 + (x_3 + y_3)\,\mathbf{a}_2 + z_3\,\mathbf{a}_3$ | = | $(x_3 a + z_3 c\cos\beta)\,\hat{\mathbf{x}} + y_3 b\,\hat{\mathbf{y}} + z_3 c\sin\beta\,\hat{\mathbf{z}}$ | (4a) | Mg III |
| $\mathbf{B}_{6}$ | = | $(x_3 + y_3)\,\mathbf{a}_1 + (x_3 - y_3)\,\mathbf{a}_2 + \left(\frac{1}{2} + z_3\right)\mathbf{a}_3$ | = | $\left(\frac{1}{2}c\cos\beta + x_3 a + z_3 c\cos\beta\right)\hat{\mathbf{x}} - y_3 b\,\hat{\mathbf{y}} + \left(\frac{1}{2} + z_3\right)c\sin\beta\,\hat{\mathbf{z}}$ | (4a) | Mg III |
| $\mathbf{B}_{7}$ | = | $(x_4 - y_4)\,\mathbf{a}_1 + (x_4 + y_4)\,\mathbf{a}_2 + z_4\,\mathbf{a}_3$ | = | $(x_4 a + z_4 c\cos\beta)\,\hat{\mathbf{x}} + y_4 b\,\hat{\mathbf{y}} + z_4 c\sin\beta\,\hat{\mathbf{z}}$ | (4a) | O I |
| $\mathbf{B}_{8}$ | = | $(x_4 + y_4)\,\mathbf{a}_1 + (x_4 - y_4)\,\mathbf{a}_2 + \left(\frac{1}{2} + z_4\right)\mathbf{a}_3$ | = | $\left(\frac{1}{2}c\cos\beta + x_4 a + z_4 c\cos\beta\right)\hat{\mathbf{x}} - y_4 b\,\hat{\mathbf{y}} + \left(\frac{1}{2} + z_4\right)c\sin\beta\,\hat{\mathbf{z}}$ | (4a) | O I |
| $\mathbf{B}_{9}$ | = | $(x_5 - y_5)\,\mathbf{a}_1 + (x_5 + y_5)\,\mathbf{a}_2 + z_5\,\mathbf{a}_3$ | = | $(x_5 a + z_5 c\cos\beta)\,\hat{\mathbf{x}} + y_5 b\,\hat{\mathbf{y}} + z_5 c\sin\beta\,\hat{\mathbf{z}}$ | (4a) | O II |
| $\mathbf{B}_{10}$ | = | $(x_5 + y_5)\,\mathbf{a}_1 + (x_5 - y_5)\,\mathbf{a}_2 + \left(\frac{1}{2} + z_5\right)\mathbf{a}_3$ | = | $\left(\frac{1}{2}c\cos\beta + x_5 a + z_5 c\cos\beta\right)\hat{\mathbf{x}} - y_5 b\,\hat{\mathbf{y}} + \left(\frac{1}{2} + z_5\right)c\sin\beta\,\hat{\mathbf{z}}$ | (4a) | O II |
| $\mathbf{B}_{11}$ | = | $(x_6 - y_6)\,\mathbf{a}_1 + (x_6 + y_6)\,\mathbf{a}_2 + z_6\,\mathbf{a}_3$ | = | $(x_6 a + z_6 c\cos\beta)\,\hat{\mathbf{x}} + y_6 b\,\hat{\mathbf{y}} + z_6 c\sin\beta\,\hat{\mathbf{z}}$ | (4a) | O III |
| $\mathbf{B}_{12}$ | = | $(x_6 + y_6)\,\mathbf{a}_1 + (x_6 - y_6)\,\mathbf{a}_2 + \left(\frac{1}{2} + z_6\right)\mathbf{a}_3$ | = | $\left(\frac{1}{2}c\cos\beta + x_6 a + z_6 c\cos\beta\right)\hat{\mathbf{x}} - y_6 b\,\hat{\mathbf{y}} + \left(\frac{1}{2} + z_6\right)c\sin\beta\,\hat{\mathbf{z}}$ | (4a) | O III |
| $\mathbf{B}_{13}$ | = | $(x_7 - y_7)\,\mathbf{a}_1 + (x_7 + y_7)\,\mathbf{a}_2 + z_7\,\mathbf{a}_3$ | = | $(x_7 a + z_7 c\cos\beta)\,\hat{\mathbf{x}} + y_7 b\,\hat{\mathbf{y}} + z_7 c\sin\beta\,\hat{\mathbf{z}}$ | (4a) | O IV |
| $\mathbf{B}_{14}$ | = | $(x_7 + y_7)\,\mathbf{a}_1 + (x_7 - y_7)\,\mathbf{a}_2 + \left(\frac{1}{2} + z_7\right)\mathbf{a}_3$ | = | $\left(\frac{1}{2}c\cos\beta + x_7 a + z_7 c\cos\beta\right)\hat{\mathbf{x}} - y_7 b\,\hat{\mathbf{y}} + \left(\frac{1}{2} + z_7\right)c\sin\beta\,\hat{\mathbf{z}}$ | (4a) | O IV |
| $\mathbf{B}_{15}$ | = | $(x_8 - y_8)\,\mathbf{a}_1 + (x_8 + y_8)\,\mathbf{a}_2 + z_8\,\mathbf{a}_3$ | = | $(x_8 a + z_8 c\cos\beta)\,\hat{\mathbf{x}} + y_8 b\,\hat{\mathbf{y}} + z_8 c\sin\beta\,\hat{\mathbf{z}}$ | (4a) | O V |
| $\mathbf{B}_{16}$ | = | $(x_8 + y_8)\,\mathbf{a}_1 + (x_8 - y_8)\,\mathbf{a}_2 + \left(\frac{1}{2} + z_8\right)\mathbf{a}_3$ | = | $\left(\frac{1}{2}c\cos\beta + x_8 a + z_8 c\cos\beta\right)\hat{\mathbf{x}} - y_8 b\,\hat{\mathbf{y}} + \left(\frac{1}{2} + z_8\right)c\sin\beta\,\hat{\mathbf{z}}$ | (4a) | O V |
| $\mathbf{B}_{17}$ | = | $(x_9 - y_9)\,\mathbf{a}_1 + (x_9 + y_9)\,\mathbf{a}_2 + z_9\,\mathbf{a}_3$ | = | $(x_9 a + z_9 c\cos\beta)\,\hat{\mathbf{x}} + y_9 b\,\hat{\mathbf{y}} + z_9 c\sin\beta\,\hat{\mathbf{z}}$ | (4a) | OH I |
| $\mathbf{B}_{18}$ | = | $(x_9 + y_9)\,\mathbf{a}_1 + (x_9 - y_9)\,\mathbf{a}_2 + \left(\frac{1}{2} + z_9\right)\mathbf{a}_3$ | = | $\left(\frac{1}{2}c\cos\beta + x_9 a + z_9 c\cos\beta\right)\hat{\mathbf{x}} - y_9 b\,\hat{\mathbf{y}} + \left(\frac{1}{2} + z_9\right)c\sin\beta\,\hat{\mathbf{z}}$ | (4a) | OH I |
| $\mathbf{B}_{19}$ | = | $(x_{10} - y_{10})\,\mathbf{a}_1 + (x_{10} + y_{10})\,\mathbf{a}_2 + z_{10}\,\mathbf{a}_3$ | = | $(x_{10} a + z_{10} c\cos\beta)\,\hat{\mathbf{x}} + y_{10} b\,\hat{\mathbf{y}} + z_{10} c\sin\beta\,\hat{\mathbf{z}}$ | (4a) | OH II |
| $\mathbf{B}_{20}$ | = | $(x_{10} + y_{10})\,\mathbf{a}_1 + (x_{10} - y_{10})\,\mathbf{a}_2 + \left(\frac{1}{2} + z_{10}\right)\mathbf{a}_3$ | = | $\left(\frac{1}{2}c\cos\beta + x_{10} a + z_{10} c\cos\beta\right)\hat{\mathbf{x}} - y_{10} b\,\hat{\mathbf{y}} + \left(\frac{1}{2} + z_{10}\right)c\sin\beta\,\hat{\mathbf{z}}$ | (4a) | OH II |
| $\mathbf{B}_{21}$ | = | $(x_{11} - y_{11})\,\mathbf{a}_1 + (x_{11} + y_{11})\,\mathbf{a}_2 + z_{11}\,\mathbf{a}_3$ | = | $(x_{11} a + z_{11} c\cos\beta)\,\hat{\mathbf{x}} + y_{11} b\,\hat{\mathbf{y}} + z_{11} c\sin\beta\,\hat{\mathbf{z}}$ | (4a) | OH III |

| | | | | | |
|---|---|---|---|---|---|
| $\mathbf{B}_{22}$ | = | $(x_{11}+y_{11})\,\mathbf{a}_1+(x_{11}-y_{11})\,\mathbf{a}_2+\left(\frac{1}{2}+z_{11}\right)\mathbf{a}_3$ | = | $\left(\frac{1}{2}c\cos\beta+x_{11}a+z_{11}c\cos\beta\right)\hat{\mathbf{x}}-y_{11}b\,\hat{\mathbf{y}}+\left(\frac{1}{2}+z_{11}\right)c\sin\beta\,\hat{\mathbf{z}}$ | (4$a$) | OH III |
| $\mathbf{B}_{23}$ | = | $(x_{12}-y_{12})\,\mathbf{a}_1+(x_{12}+y_{12})\,\mathbf{a}_2+z_{12}\,\mathbf{a}_3$ | = | $(x_{12}a+z_{12}c\cos\beta)\,\hat{\mathbf{x}}+y_{12}b\,\hat{\mathbf{y}}+z_{12}c\sin\beta\,\hat{\mathbf{z}}$ | (4$a$) | OH IV |
| $\mathbf{B}_{24}$ | = | $(x_{12}+y_{12})\,\mathbf{a}_1+(x_{12}-y_{12})\,\mathbf{a}_2+\left(\frac{1}{2}+z_{12}\right)\mathbf{a}_3$ | = | $\left(\frac{1}{2}c\cos\beta+x_{12}a+z_{12}c\cos\beta\right)\hat{\mathbf{x}}-y_{12}b\,\hat{\mathbf{y}}+\left(\frac{1}{2}+z_{12}\right)c\sin\beta\,\hat{\mathbf{z}}$ | (4$a$) | OH IV |
| $\mathbf{B}_{25}$ | = | $(x_{13}-y_{13})\,\mathbf{a}_1+(x_{13}+y_{13})\,\mathbf{a}_2+z_{13}\,\mathbf{a}_3$ | = | $(x_{13}a+z_{13}c\cos\beta)\,\hat{\mathbf{x}}+y_{13}b\,\hat{\mathbf{y}}+z_{13}c\sin\beta\,\hat{\mathbf{z}}$ | (4$a$) | Si I |
| $\mathbf{B}_{26}$ | = | $(x_{13}+y_{13})\,\mathbf{a}_1+(x_{13}-y_{13})\,\mathbf{a}_2+\left(\frac{1}{2}+z_{13}\right)\mathbf{a}_3$ | = | $\left(\frac{1}{2}c\cos\beta+x_{13}a+z_{13}c\cos\beta\right)\hat{\mathbf{x}}-y_{13}b\,\hat{\mathbf{y}}+\left(\frac{1}{2}+z_{13}\right)c\sin\beta\,\hat{\mathbf{z}}$ | (4$a$) | Si I |
| $\mathbf{B}_{27}$ | = | $(x_{14}-y_{14})\,\mathbf{a}_1+(x_{14}+y_{14})\,\mathbf{a}_2+z_{14}\,\mathbf{a}_3$ | = | $(x_{14}a+z_{14}c\cos\beta)\,\hat{\mathbf{x}}+y_{14}b\,\hat{\mathbf{y}}+z_{14}c\sin\beta\,\hat{\mathbf{z}}$ | (4$a$) | Si II |
| $\mathbf{B}_{28}$ | = | $(x_{14}+y_{14})\,\mathbf{a}_1+(x_{14}-y_{14})\,\mathbf{a}_2+\left(\frac{1}{2}+z_{14}\right)\mathbf{a}_3$ | = | $\left(\frac{1}{2}c\cos\beta+x_{14}a+z_{14}c\cos\beta\right)\hat{\mathbf{x}}-y_{14}b\,\hat{\mathbf{y}}+\left(\frac{1}{2}+z_{14}\right)c\sin\beta\,\hat{\mathbf{z}}$ | (4$a$) | Si II |

**Geometry files:**

- CIF: pp. 1519
- POSCAR: pp. 1520

# $Cs_6W_{11}O_{36}$ Structure: A6B36C11_mC212_9_6a_36a_11a

http://aflow.org/prototype-encyclopedia/A6B36C11_mC212_9_6a_36a_11a

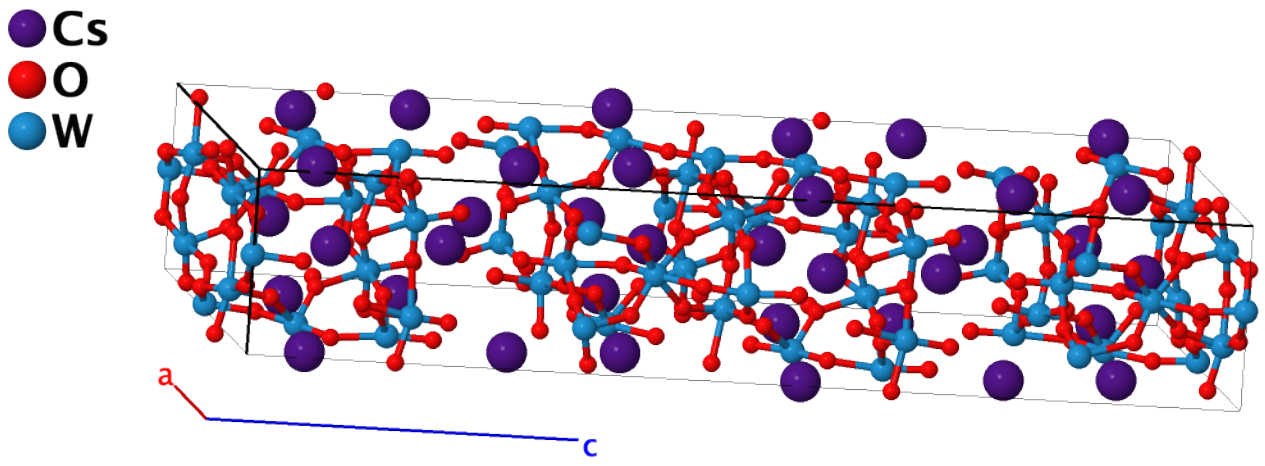

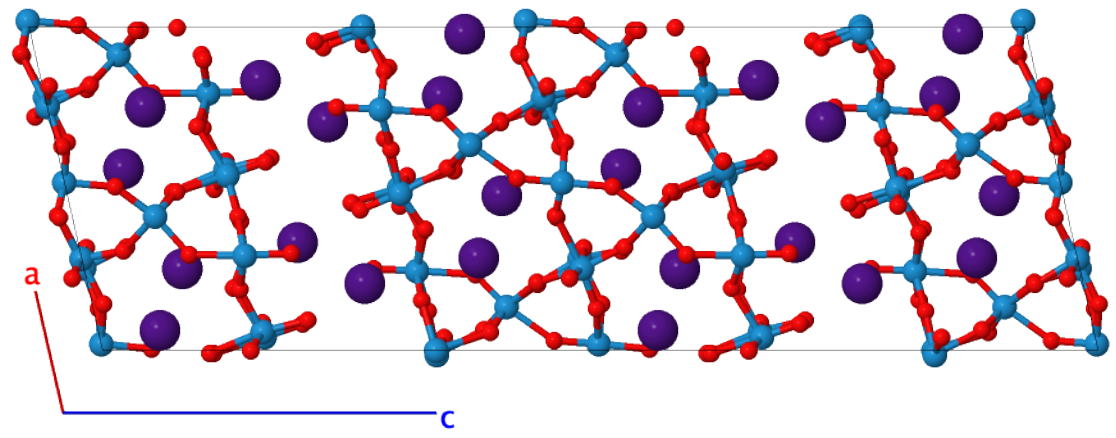

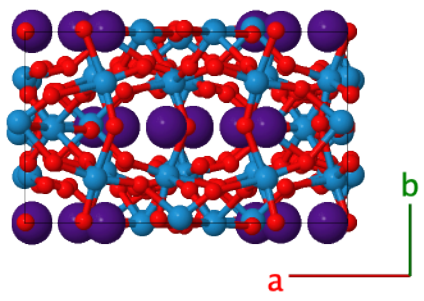

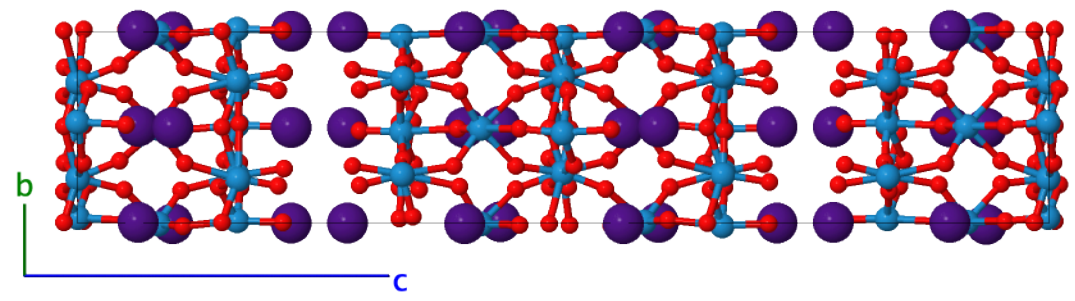

| | | |
|---|---|---|
| **Prototype** | : | $Cs_6O_{36}W_{11}$ |
| **AFLOW prototype label** | : | A6B36C11_mC212_9_6a_36a_11a |
| ***Strukturbericht* designation** | : | None |
| **Pearson symbol** | : | mC212 |
| **Space group number** | : | 9 |
| **Space group symbol** | : | $Cc$ |
| **AFLOW prototype command** | : | `aflow --proto=A6B36C11_mC212_9_6a_36a_11a` `--params=`$a, b/a, c/a, \beta, x_1, y_1, z_1, x_2, y_2, z_2, x_3, y_3, z_3, x_4, y_4, z_4, x_5, y_5, z_5, x_6, y_6, z_6, x_7, y_7, z_7, x_8, y_8, z_8, x_9, y_9, z_9, x_{10}, y_{10}, z_{10}, x_{11}, y_{11}, z_{11}, x_{12}, y_{12}, z_{12}, x_{13}, y_{13}, z_{13}, x_{14}, y_{14}, z_{14}, x_{15}, y_{15}, z_{15}, x_{16}, y_{16}, z_{16}, x_{17}, y_{17}, z_{17}, x_{18}, y_{18}, z_{18}, x_{19}, y_{19}, z_{19}, x_{20}, y_{20}, z_{20}, x_{21}, y_{21}, z_{21}, x_{22}, y_{22}, z_{22}, x_{23}, y_{23}, z_{23}, x_{24}, y_{24}, z_{24}, x_{25}, y_{25}, z_{25}, x_{26}, y_{26}, z_{26}, x_{27}, y_{27}, z_{27}, x_{28}, y_{28}, z_{28}, x_{29}, y_{29}, z_{29}, x_{30}, y_{30}, z_{30}, x_{31}, y_{31}, z_{31}, x_{32}, y_{32}, z_{32}, x_{33}, y_{33}, z_{33}, x_{34}, y_{34}, z_{34}, x_{35}, y_{35}, z_{35}, x_{36}, y_{36}, z_{36}, x_{37}, y_{37}, z_{37}, x_{38}, y_{38}, z_{38}, x_{39}, y_{39}, z_{39}, x_{40}, y_{40}, z_{40}, x_{41}, y_{41}, z_{41}, x_{42}, y_{42}, z_{42}, x_{43}, y_{43}, z_{43}, x_{44}, y_{44}, z_{44}, x_{45}, y_{45}, z_{45}, x_{46}, y_{46}, z_{46}, x_{47}, y_{47}, z_{47}, x_{48}, y_{48}, z_{48}, x_{49}, y_{49}, z_{49}, x_{50}, y_{50}, z_{50}, x_{51}, y_{51}, z_{51}, x_{52}, y_{52}, z_{52}, x_{53}, y_{53}, z_{53}$ |

**Base-centered Monoclinic primitive vectors:**

$$\mathbf{a}_1 = \frac{1}{2}\,a\,\hat{\mathbf{x}} - \frac{1}{2}\,b\,\hat{\mathbf{y}}$$
$$\mathbf{a}_2 = \frac{1}{2}\,a\,\hat{\mathbf{x}} + \frac{1}{2}\,b\,\hat{\mathbf{y}}$$
$$\mathbf{a}_3 = c\cos\beta\,\hat{\mathbf{x}} + c\sin\beta\,\hat{\mathbf{z}}$$

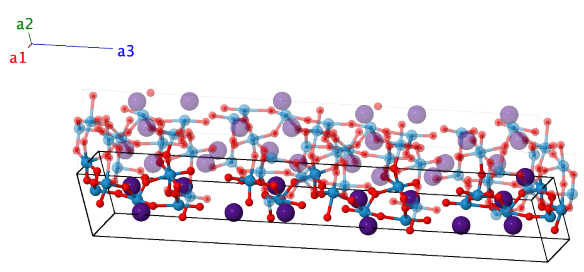

**Basis vectors:**

| | | Lattice Coordinates | | Cartesian Coordinates | Wyckoff Position | Atom Type |
|---|---|---|---|---|---|---|
| $\mathbf{B}_1$ | = | $(x_1 - y_1)\,\mathbf{a}_1 + (x_1 + y_1)\,\mathbf{a}_2 + z_1\,\mathbf{a}_3$ | = | $(x_1 a + z_1 c\cos\beta)\,\hat{\mathbf{x}} + y_1 b\,\hat{\mathbf{y}} + z_1 c\sin\beta\,\hat{\mathbf{z}}$ | (4$a$) | Cs I |
| $\mathbf{B}_2$ | = | $(x_1 + y_1)\,\mathbf{a}_1 + (x_1 - y_1)\,\mathbf{a}_2 + \left(\frac{1}{2} + z_1\right)\,\mathbf{a}_3$ | = | $\left(\frac{1}{2}c\cos\beta + x_1 a + z_1 c\cos\beta\right)\,\hat{\mathbf{x}} - y_1 b\,\hat{\mathbf{y}} + \left(\frac{1}{2} + z_1\right)c\sin\beta\,\hat{\mathbf{z}}$ | (4$a$) | Cs I |
| $\mathbf{B}_3$ | = | $(x_2 - y_2)\,\mathbf{a}_1 + (x_2 + y_2)\,\mathbf{a}_2 + z_2\,\mathbf{a}_3$ | = | $(x_2 a + z_2 c\cos\beta)\,\hat{\mathbf{x}} + y_2 b\,\hat{\mathbf{y}} + z_2 c\sin\beta\,\hat{\mathbf{z}}$ | (4$a$) | Cs II |
| $\mathbf{B}_{4}$ | = | $(x_{2}+y_{2})\,\mathbf{a}_{1}+(x_{2}-y_{2})\,\mathbf{a}_{2}+\left(\frac{1}{2}+z_{2}\right)\mathbf{a}_{3}$ | = | $\left(\frac{1}{2}c\cos\beta+x_{2}a+z_{2}c\cos\beta\right)\hat{\mathbf{x}}-y_{2}b\,\hat{\mathbf{y}}+\left(\frac{1}{2}+z_{2}\right)c\sin\beta\,\hat{\mathbf{z}}$ | (4$a$) | Cs II |
| $\mathbf{B}_{5}$ | = | $(x_{3}-y_{3})\,\mathbf{a}_{1}+(x_{3}+y_{3})\,\mathbf{a}_{2}+z_{3}\,\mathbf{a}_{3}$ | = | $(x_{3}a+z_{3}c\cos\beta)\,\hat{\mathbf{x}}+y_{3}b\,\hat{\mathbf{y}}+z_{3}c\sin\beta\,\hat{\mathbf{z}}$ | (4$a$) | Cs III |
| $\mathbf{B}_{6}$ | = | $(x_{3}+y_{3})\,\mathbf{a}_{1}+(x_{3}-y_{3})\,\mathbf{a}_{2}+\left(\frac{1}{2}+z_{3}\right)\mathbf{a}_{3}$ | = | $\left(\frac{1}{2}c\cos\beta+x_{3}a+z_{3}c\cos\beta\right)\hat{\mathbf{x}}-y_{3}b\,\hat{\mathbf{y}}+\left(\frac{1}{2}+z_{3}\right)c\sin\beta\,\hat{\mathbf{z}}$ | (4$a$) | Cs III |
| $\mathbf{B}_{7}$ | = | $(x_{4}-y_{4})\,\mathbf{a}_{1}+(x_{4}+y_{4})\,\mathbf{a}_{2}+z_{4}\,\mathbf{a}_{3}$ | = | $(x_{4}a+z_{4}c\cos\beta)\,\hat{\mathbf{x}}+y_{4}b\,\hat{\mathbf{y}}+z_{4}c\sin\beta\,\hat{\mathbf{z}}$ | (4$a$) | Cs IV |
| $\mathbf{B}_{8}$ | = | $(x_{4}+y_{4})\,\mathbf{a}_{1}+(x_{4}-y_{4})\,\mathbf{a}_{2}+\left(\frac{1}{2}+z_{4}\right)\mathbf{a}_{3}$ | = | $\left(\frac{1}{2}c\cos\beta+x_{4}a+z_{4}c\cos\beta\right)\hat{\mathbf{x}}-y_{4}b\,\hat{\mathbf{y}}+\left(\frac{1}{2}+z_{4}\right)c\sin\beta\,\hat{\mathbf{z}}$ | (4$a$) | Cs IV |
| $\mathbf{B}_{9}$ | = | $(x_{5}-y_{5})\,\mathbf{a}_{1}+(x_{5}+y_{5})\,\mathbf{a}_{2}+z_{5}\,\mathbf{a}_{3}$ | = | $(x_{5}a+z_{5}c\cos\beta)\,\hat{\mathbf{x}}+y_{5}b\,\hat{\mathbf{y}}+z_{5}c\sin\beta\,\hat{\mathbf{z}}$ | (4$a$) | Cs V |
| $\mathbf{B}_{10}$ | = | $(x_{5}+y_{5})\,\mathbf{a}_{1}+(x_{5}-y_{5})\,\mathbf{a}_{2}+\left(\frac{1}{2}+z_{5}\right)\mathbf{a}_{3}$ | = | $\left(\frac{1}{2}c\cos\beta+x_{5}a+z_{5}c\cos\beta\right)\hat{\mathbf{x}}-y_{5}b\,\hat{\mathbf{y}}+\left(\frac{1}{2}+z_{5}\right)c\sin\beta\,\hat{\mathbf{z}}$ | (4$a$) | Cs V |
| $\mathbf{B}_{11}$ | = | $(x_{6}-y_{6})\,\mathbf{a}_{1}+(x_{6}+y_{6})\,\mathbf{a}_{2}+z_{6}\,\mathbf{a}_{3}$ | = | $(x_{6}a+z_{6}c\cos\beta)\,\hat{\mathbf{x}}+y_{6}b\,\hat{\mathbf{y}}+z_{6}c\sin\beta\,\hat{\mathbf{z}}$ | (4$a$) | Cs VI |
| $\mathbf{B}_{12}$ | = | $(x_{6}+y_{6})\,\mathbf{a}_{1}+(x_{6}-y_{6})\,\mathbf{a}_{2}+\left(\frac{1}{2}+z_{6}\right)\mathbf{a}_{3}$ | = | $\left(\frac{1}{2}c\cos\beta+x_{6}a+z_{6}c\cos\beta\right)\hat{\mathbf{x}}-y_{6}b\,\hat{\mathbf{y}}+\left(\frac{1}{2}+z_{6}\right)c\sin\beta\,\hat{\mathbf{z}}$ | (4$a$) | Cs VI |
| $\mathbf{B}_{13}$ | = | $(x_{7}-y_{7})\,\mathbf{a}_{1}+(x_{7}+y_{7})\,\mathbf{a}_{2}+z_{7}\,\mathbf{a}_{3}$ | = | $(x_{7}a+z_{7}c\cos\beta)\,\hat{\mathbf{x}}+y_{7}b\,\hat{\mathbf{y}}+z_{7}c\sin\beta\,\hat{\mathbf{z}}$ | (4$a$) | O I |
| $\mathbf{B}_{14}$ | = | $(x_{7}+y_{7})\,\mathbf{a}_{1}+(x_{7}-y_{7})\,\mathbf{a}_{2}+\left(\frac{1}{2}+z_{7}\right)\mathbf{a}_{3}$ | = | $\left(\frac{1}{2}c\cos\beta+x_{7}a+z_{7}c\cos\beta\right)\hat{\mathbf{x}}-y_{7}b\,\hat{\mathbf{y}}+\left(\frac{1}{2}+z_{7}\right)c\sin\beta\,\hat{\mathbf{z}}$ | (4$a$) | O I |
| $\mathbf{B}_{15}$ | = | $(x_{8}-y_{8})\,\mathbf{a}_{1}+(x_{8}+y_{8})\,\mathbf{a}_{2}+z_{8}\,\mathbf{a}_{3}$ | = | $(x_{8}a+z_{8}c\cos\beta)\,\hat{\mathbf{x}}+y_{8}b\,\hat{\mathbf{y}}+z_{8}c\sin\beta\,\hat{\mathbf{z}}$ | (4$a$) | O II |
| $\mathbf{B}_{16}$ | = | $(x_{8}+y_{8})\,\mathbf{a}_{1}+(x_{8}-y_{8})\,\mathbf{a}_{2}+\left(\frac{1}{2}+z_{8}\right)\mathbf{a}_{3}$ | = | $\left(\frac{1}{2}c\cos\beta+x_{8}a+z_{8}c\cos\beta\right)\hat{\mathbf{x}}-y_{8}b\,\hat{\mathbf{y}}+\left(\frac{1}{2}+z_{8}\right)c\sin\beta\,\hat{\mathbf{z}}$ | (4$a$) | O II |
| $\mathbf{B}_{17}$ | = | $(x_{9}-y_{9})\,\mathbf{a}_{1}+(x_{9}+y_{9})\,\mathbf{a}_{2}+z_{9}\,\mathbf{a}_{3}$ | = | $(x_{9}a+z_{9}c\cos\beta)\,\hat{\mathbf{x}}+y_{9}b\,\hat{\mathbf{y}}+z_{9}c\sin\beta\,\hat{\mathbf{z}}$ | (4$a$) | O III |
| $\mathbf{B}_{18}$ | = | $(x_{9}+y_{9})\,\mathbf{a}_{1}+(x_{9}-y_{9})\,\mathbf{a}_{2}+\left(\frac{1}{2}+z_{9}\right)\mathbf{a}_{3}$ | = | $\left(\frac{1}{2}c\cos\beta+x_{9}a+z_{9}c\cos\beta\right)\hat{\mathbf{x}}-y_{9}b\,\hat{\mathbf{y}}+\left(\frac{1}{2}+z_{9}\right)c\sin\beta\,\hat{\mathbf{z}}$ | (4$a$) | O III |
| $\mathbf{B}_{19}$ | = | $(x_{10}-y_{10})\,\mathbf{a}_{1}+(x_{10}+y_{10})\,\mathbf{a}_{2}+z_{10}\,\mathbf{a}_{3}$ | = | $(x_{10}a+z_{10}c\cos\beta)\,\hat{\mathbf{x}}+y_{10}b\,\hat{\mathbf{y}}+z_{10}c\sin\beta\,\hat{\mathbf{z}}$ | (4$a$) | O IV |
| $\mathbf{B}_{20}$ | = | $(x_{10}+y_{10})\,\mathbf{a}_{1}+(x_{10}-y_{10})\,\mathbf{a}_{2}+\left(\frac{1}{2}+z_{10}\right)\mathbf{a}_{3}$ | = | $\left(\frac{1}{2}c\cos\beta+x_{10}a+z_{10}c\cos\beta\right)\hat{\mathbf{x}}-y_{10}b\,\hat{\mathbf{y}}+\left(\frac{1}{2}+z_{10}\right)c\sin\beta\,\hat{\mathbf{z}}$ | (4$a$) | O IV |
| $\mathbf{B}_{21}$ | = | $(x_{11}-y_{11})\,\mathbf{a}_{1}+(x_{11}+y_{11})\,\mathbf{a}_{2}+z_{11}\,\mathbf{a}_{3}$ | = | $(x_{11}a+z_{11}c\cos\beta)\,\hat{\mathbf{x}}+y_{11}b\,\hat{\mathbf{y}}+z_{11}c\sin\beta\,\hat{\mathbf{z}}$ | (4$a$) | O V |
| $\mathbf{B}_{22}$ | = | $(x_{11}+y_{11})\,\mathbf{a}_{1}+(x_{11}-y_{11})\,\mathbf{a}_{2}+\left(\frac{1}{2}+z_{11}\right)\mathbf{a}_{3}$ | = | $\left(\frac{1}{2}c\cos\beta+x_{11}a+z_{11}c\cos\beta\right)\hat{\mathbf{x}}-y_{11}b\,\hat{\mathbf{y}}+\left(\frac{1}{2}+z_{11}\right)c\sin\beta\,\hat{\mathbf{z}}$ | (4$a$) | O V |
| $\mathbf{B}_{23}$ | = | $(x_{12}-y_{12})\,\mathbf{a}_{1}+(x_{12}+y_{12})\,\mathbf{a}_{2}+z_{12}\,\mathbf{a}_{3}$ | = | $(x_{12}a+z_{12}c\cos\beta)\,\hat{\mathbf{x}}+y_{12}b\,\hat{\mathbf{y}}+z_{12}c\sin\beta\,\hat{\mathbf{z}}$ | (4$a$) | O VI |
| $\mathbf{B}_{24}$ | = | $(x_{12}+y_{12})\,\mathbf{a}_{1}+(x_{12}-y_{12})\,\mathbf{a}_{2}+\left(\frac{1}{2}+z_{12}\right)\mathbf{a}_{3}$ | = | $\left(\frac{1}{2}c\cos\beta+x_{12}a+z_{12}c\cos\beta\right)\hat{\mathbf{x}}-y_{12}b\,\hat{\mathbf{y}}+\left(\frac{1}{2}+z_{12}\right)c\sin\beta\,\hat{\mathbf{z}}$ | (4$a$) | O VI |
| $\mathbf{B}_{25}$ | = | $(x_{13}-y_{13})\,\mathbf{a}_{1}+(x_{13}+y_{13})\,\mathbf{a}_{2}+z_{13}\,\mathbf{a}_{3}$ | = | $(x_{13}a+z_{13}c\cos\beta)\,\hat{\mathbf{x}}+y_{13}b\,\hat{\mathbf{y}}+z_{13}c\sin\beta\,\hat{\mathbf{z}}$ | (4$a$) | O VII |
| $\mathbf{B}_{26}$ | = | $(x_{13} + y_{13})\,\mathbf{a}_1 + (x_{13} - y_{13})\,\mathbf{a}_2 + \left(\frac{1}{2} + z_{13}\right)\mathbf{a}_3$ | = | $\left(\frac{1}{2}c\cos\beta + x_{13}a + z_{13}c\cos\beta\right)\hat{\mathbf{x}} - y_{13}b\,\hat{\mathbf{y}} + \left(\frac{1}{2} + z_{13}\right)c\sin\beta\,\hat{\mathbf{z}}$ | (4$a$) | O VII |
| $\mathbf{B}_{27}$ | = | $(x_{14} - y_{14})\,\mathbf{a}_1 + (x_{14} + y_{14})\,\mathbf{a}_2 + z_{14}\,\mathbf{a}_3$ | = | $(x_{14}a + z_{14}c\cos\beta)\,\hat{\mathbf{x}} + y_{14}b\,\hat{\mathbf{y}} + z_{14}c\sin\beta\,\hat{\mathbf{z}}$ | (4$a$) | O VIII |
| $\mathbf{B}_{28}$ | = | $(x_{14} + y_{14})\,\mathbf{a}_1 + (x_{14} - y_{14})\,\mathbf{a}_2 + \left(\frac{1}{2} + z_{14}\right)\mathbf{a}_3$ | = | $\left(\frac{1}{2}c\cos\beta + x_{14}a + z_{14}c\cos\beta\right)\hat{\mathbf{x}} - y_{14}b\,\hat{\mathbf{y}} + \left(\frac{1}{2} + z_{14}\right)c\sin\beta\,\hat{\mathbf{z}}$ | (4$a$) | O VIII |
| $\mathbf{B}_{29}$ | = | $(x_{15} - y_{15})\,\mathbf{a}_1 + (x_{15} + y_{15})\,\mathbf{a}_2 + z_{15}\,\mathbf{a}_3$ | = | $(x_{15}a + z_{15}c\cos\beta)\,\hat{\mathbf{x}} + y_{15}b\,\hat{\mathbf{y}} + z_{15}c\sin\beta\,\hat{\mathbf{z}}$ | (4$a$) | O IX |
| $\mathbf{B}_{30}$ | = | $(x_{15} + y_{15})\,\mathbf{a}_1 + (x_{15} - y_{15})\,\mathbf{a}_2 + \left(\frac{1}{2} + z_{15}\right)\mathbf{a}_3$ | = | $\left(\frac{1}{2}c\cos\beta + x_{15}a + z_{15}c\cos\beta\right)\hat{\mathbf{x}} - y_{15}b\,\hat{\mathbf{y}} + \left(\frac{1}{2} + z_{15}\right)c\sin\beta\,\hat{\mathbf{z}}$ | (4$a$) | O IX |
| $\mathbf{B}_{31}$ | = | $(x_{16} - y_{16})\,\mathbf{a}_1 + (x_{16} + y_{16})\,\mathbf{a}_2 + z_{16}\,\mathbf{a}_3$ | = | $(x_{16}a + z_{16}c\cos\beta)\,\hat{\mathbf{x}} + y_{16}b\,\hat{\mathbf{y}} + z_{16}c\sin\beta\,\hat{\mathbf{z}}$ | (4$a$) | O X |
| $\mathbf{B}_{32}$ | = | $(x_{16} + y_{16})\,\mathbf{a}_1 + (x_{16} - y_{16})\,\mathbf{a}_2 + \left(\frac{1}{2} + z_{16}\right)\mathbf{a}_3$ | = | $\left(\frac{1}{2}c\cos\beta + x_{16}a + z_{16}c\cos\beta\right)\hat{\mathbf{x}} - y_{16}b\,\hat{\mathbf{y}} + \left(\frac{1}{2} + z_{16}\right)c\sin\beta\,\hat{\mathbf{z}}$ | (4$a$) | O X |
| $\mathbf{B}_{33}$ | = | $(x_{17} - y_{17})\,\mathbf{a}_1 + (x_{17} + y_{17})\,\mathbf{a}_2 + z_{17}\,\mathbf{a}_3$ | = | $(x_{17}a + z_{17}c\cos\beta)\,\hat{\mathbf{x}} + y_{17}b\,\hat{\mathbf{y}} + z_{17}c\sin\beta\,\hat{\mathbf{z}}$ | (4$a$) | O XI |
| $\mathbf{B}_{34}$ | = | $(x_{17} + y_{17})\,\mathbf{a}_1 + (x_{17} - y_{17})\,\mathbf{a}_2 + \left(\frac{1}{2} + z_{17}\right)\mathbf{a}_3$ | = | $\left(\frac{1}{2}c\cos\beta + x_{17}a + z_{17}c\cos\beta\right)\hat{\mathbf{x}} - y_{17}b\,\hat{\mathbf{y}} + \left(\frac{1}{2} + z_{17}\right)c\sin\beta\,\hat{\mathbf{z}}$ | (4$a$) | O XI |
| $\mathbf{B}_{35}$ | = | $(x_{18} - y_{18})\,\mathbf{a}_1 + (x_{18} + y_{18})\,\mathbf{a}_2 + z_{18}\,\mathbf{a}_3$ | = | $(x_{18}a + z_{18}c\cos\beta)\,\hat{\mathbf{x}} + y_{18}b\,\hat{\mathbf{y}} + z_{18}c\sin\beta\,\hat{\mathbf{z}}$ | (4$a$) | O XII |
| $\mathbf{B}_{36}$ | = | $(x_{18} + y_{18})\,\mathbf{a}_1 + (x_{18} - y_{18})\,\mathbf{a}_2 + \left(\frac{1}{2} + z_{18}\right)\mathbf{a}_3$ | = | $\left(\frac{1}{2}c\cos\beta + x_{18}a + z_{18}c\cos\beta\right)\hat{\mathbf{x}} - y_{18}b\,\hat{\mathbf{y}} + \left(\frac{1}{2} + z_{18}\right)c\sin\beta\,\hat{\mathbf{z}}$ | (4$a$) | O XII |
| $\mathbf{B}_{37}$ | = | $(x_{19} - y_{19})\,\mathbf{a}_1 + (x_{19} + y_{19})\,\mathbf{a}_2 + z_{19}\,\mathbf{a}_3$ | = | $(x_{19}a + z_{19}c\cos\beta)\,\hat{\mathbf{x}} + y_{19}b\,\hat{\mathbf{y}} + z_{19}c\sin\beta\,\hat{\mathbf{z}}$ | (4$a$) | O XIII |
| $\mathbf{B}_{38}$ | = | $(x_{19} + y_{19})\,\mathbf{a}_1 + (x_{19} - y_{19})\,\mathbf{a}_2 + \left(\frac{1}{2} + z_{19}\right)\mathbf{a}_3$ | = | $\left(\frac{1}{2}c\cos\beta + x_{19}a + z_{19}c\cos\beta\right)\hat{\mathbf{x}} - y_{19}b\,\hat{\mathbf{y}} + \left(\frac{1}{2} + z_{19}\right)c\sin\beta\,\hat{\mathbf{z}}$ | (4$a$) | O XIII |
| $\mathbf{B}_{39}$ | = | $(x_{20} - y_{20})\,\mathbf{a}_1 + (x_{20} + y_{20})\,\mathbf{a}_2 + z_{20}\,\mathbf{a}_3$ | = | $(x_{20}a + z_{20}c\cos\beta)\,\hat{\mathbf{x}} + y_{20}b\,\hat{\mathbf{y}} + z_{20}c\sin\beta\,\hat{\mathbf{z}}$ | (4$a$) | O XIV |
| $\mathbf{B}_{40}$ | = | $(x_{20} + y_{20})\,\mathbf{a}_1 + (x_{20} - y_{20})\,\mathbf{a}_2 + \left(\frac{1}{2} + z_{20}\right)\mathbf{a}_3$ | = | $\left(\frac{1}{2}c\cos\beta + x_{20}a + z_{20}c\cos\beta\right)\hat{\mathbf{x}} - y_{20}b\,\hat{\mathbf{y}} + \left(\frac{1}{2} + z_{20}\right)c\sin\beta\,\hat{\mathbf{z}}$ | (4$a$) | O XIV |
| $\mathbf{B}_{41}$ | = | $(x_{21} - y_{21})\,\mathbf{a}_1 + (x_{21} + y_{21})\,\mathbf{a}_2 + z_{21}\,\mathbf{a}_3$ | = | $(x_{21}a + z_{21}c\cos\beta)\,\hat{\mathbf{x}} + y_{21}b\,\hat{\mathbf{y}} + z_{21}c\sin\beta\,\hat{\mathbf{z}}$ | (4$a$) | O XV |
| $\mathbf{B}_{42}$ | = | $(x_{21} + y_{21})\,\mathbf{a}_1 + (x_{21} - y_{21})\,\mathbf{a}_2 + \left(\frac{1}{2} + z_{21}\right)\mathbf{a}_3$ | = | $\left(\frac{1}{2}c\cos\beta + x_{21}a + z_{21}c\cos\beta\right)\hat{\mathbf{x}} - y_{21}b\,\hat{\mathbf{y}} + \left(\frac{1}{2} + z_{21}\right)c\sin\beta\,\hat{\mathbf{z}}$ | (4$a$) | O XV |
| $\mathbf{B}_{43}$ | = | $(x_{22} - y_{22})\,\mathbf{a}_1 + (x_{22} + y_{22})\,\mathbf{a}_2 + z_{22}\,\mathbf{a}_3$ | = | $(x_{22}a + z_{22}c\cos\beta)\,\hat{\mathbf{x}} + y_{22}b\,\hat{\mathbf{y}} + z_{22}c\sin\beta\,\hat{\mathbf{z}}$ | (4$a$) | O XVI |
| $\mathbf{B}_{44}$ | = | $(x_{22} + y_{22})\,\mathbf{a}_1 + (x_{22} - y_{22})\,\mathbf{a}_2 + \left(\frac{1}{2} + z_{22}\right)\mathbf{a}_3$ | = | $\left(\frac{1}{2}c\cos\beta + x_{22}a + z_{22}c\cos\beta\right)\hat{\mathbf{x}} - y_{22}b\,\hat{\mathbf{y}} + \left(\frac{1}{2} + z_{22}\right)c\sin\beta\,\hat{\mathbf{z}}$ | (4$a$) | O XVI |
| $\mathbf{B}_{45}$ | = | $(x_{23} - y_{23})\,\mathbf{a}_1 + (x_{23} + y_{23})\,\mathbf{a}_2 + z_{23}\,\mathbf{a}_3$ | = | $(x_{23}a + z_{23}c\cos\beta)\,\hat{\mathbf{x}} + y_{23}b\,\hat{\mathbf{y}} + z_{23}c\sin\beta\,\hat{\mathbf{z}}$ | (4$a$) | O XVII |
| $\mathbf{B}_{46}$ | = | $(x_{23} + y_{23})\,\mathbf{a}_1 + (x_{23} - y_{23})\,\mathbf{a}_2 + \left(\frac{1}{2} + z_{23}\right)\mathbf{a}_3$ | = | $\left(\frac{1}{2}c\cos\beta + x_{23}a + z_{23}c\cos\beta\right)\hat{\mathbf{x}} - y_{23}b\,\hat{\mathbf{y}} + \left(\frac{1}{2} + z_{23}\right)c\sin\beta\,\hat{\mathbf{z}}$ | (4$a$) | O XVII |
| $\mathbf{B}_{47}$ | = | $(x_{24} - y_{24})\,\mathbf{a}_1 + (x_{24} + y_{24})\,\mathbf{a}_2 + z_{24}\,\mathbf{a}_3$ | = | $(x_{24}a + z_{24}c\cos\beta)\,\hat{\mathbf{x}} + y_{24}b\,\hat{\mathbf{y}} + z_{24}c\sin\beta\,\hat{\mathbf{z}}$ | (4$a$) | O XVIII |
| $\mathbf{B}_{48}$ | = | $(x_{24}+y_{24})\,\mathbf{a}_1+(x_{24}-y_{24})\,\mathbf{a}_2+\left(\frac{1}{2}+z_{24}\right)\mathbf{a}_3$ | = | $\left(\frac{1}{2}c\cos\beta+x_{24}a+z_{24}c\cos\beta\right)\hat{\mathbf{x}}-y_{24}b\,\hat{\mathbf{y}}+\left(\frac{1}{2}+z_{24}\right)c\sin\beta\,\hat{\mathbf{z}}$ | (4$a$) | O XVIII |
| $\mathbf{B}_{49}$ | = | $(x_{25}-y_{25})\,\mathbf{a}_1+(x_{25}+y_{25})\,\mathbf{a}_2+z_{25}\,\mathbf{a}_3$ | = | $(x_{25}a+z_{25}c\cos\beta)\,\hat{\mathbf{x}}+y_{25}b\,\hat{\mathbf{y}}+z_{25}c\sin\beta\,\hat{\mathbf{z}}$ | (4$a$) | O XIX |
| $\mathbf{B}_{50}$ | = | $(x_{25}+y_{25})\,\mathbf{a}_1+(x_{25}-y_{25})\,\mathbf{a}_2+\left(\frac{1}{2}+z_{25}\right)\mathbf{a}_3$ | = | $\left(\frac{1}{2}c\cos\beta+x_{25}a+z_{25}c\cos\beta\right)\hat{\mathbf{x}}-y_{25}b\,\hat{\mathbf{y}}+\left(\frac{1}{2}+z_{25}\right)c\sin\beta\,\hat{\mathbf{z}}$ | (4$a$) | O XIX |
| $\mathbf{B}_{51}$ | = | $(x_{26}-y_{26})\,\mathbf{a}_1+(x_{26}+y_{26})\,\mathbf{a}_2+z_{26}\,\mathbf{a}_3$ | = | $(x_{26}a+z_{26}c\cos\beta)\,\hat{\mathbf{x}}+y_{26}b\,\hat{\mathbf{y}}+z_{26}c\sin\beta\,\hat{\mathbf{z}}$ | (4$a$) | O XX |
| $\mathbf{B}_{52}$ | = | $(x_{26}+y_{26})\,\mathbf{a}_1+(x_{26}-y_{26})\,\mathbf{a}_2+\left(\frac{1}{2}+z_{26}\right)\mathbf{a}_3$ | = | $\left(\frac{1}{2}c\cos\beta+x_{26}a+z_{26}c\cos\beta\right)\hat{\mathbf{x}}-y_{26}b\,\hat{\mathbf{y}}+\left(\frac{1}{2}+z_{26}\right)c\sin\beta\,\hat{\mathbf{z}}$ | (4$a$) | O XX |
| $\mathbf{B}_{53}$ | = | $(x_{27}-y_{27})\,\mathbf{a}_1+(x_{27}+y_{27})\,\mathbf{a}_2+z_{27}\,\mathbf{a}_3$ | = | $(x_{27}a+z_{27}c\cos\beta)\,\hat{\mathbf{x}}+y_{27}b\,\hat{\mathbf{y}}+z_{27}c\sin\beta\,\hat{\mathbf{z}}$ | (4$a$) | O XXI |
| $\mathbf{B}_{54}$ | = | $(x_{27}+y_{27})\,\mathbf{a}_1+(x_{27}-y_{27})\,\mathbf{a}_2+\left(\frac{1}{2}+z_{27}\right)\mathbf{a}_3$ | = | $\left(\frac{1}{2}c\cos\beta+x_{27}a+z_{27}c\cos\beta\right)\hat{\mathbf{x}}-y_{27}b\,\hat{\mathbf{y}}+\left(\frac{1}{2}+z_{27}\right)c\sin\beta\,\hat{\mathbf{z}}$ | (4$a$) | O XXI |
| $\mathbf{B}_{55}$ | = | $(x_{28}-y_{28})\,\mathbf{a}_1+(x_{28}+y_{28})\,\mathbf{a}_2+z_{28}\,\mathbf{a}_3$ | = | $(x_{28}a+z_{28}c\cos\beta)\,\hat{\mathbf{x}}+y_{28}b\,\hat{\mathbf{y}}+z_{28}c\sin\beta\,\hat{\mathbf{z}}$ | (4$a$) | O XXII |
| $\mathbf{B}_{56}$ | = | $(x_{28}+y_{28})\,\mathbf{a}_1+(x_{28}-y_{28})\,\mathbf{a}_2+\left(\frac{1}{2}+z_{28}\right)\mathbf{a}_3$ | = | $\left(\frac{1}{2}c\cos\beta+x_{28}a+z_{28}c\cos\beta\right)\hat{\mathbf{x}}-y_{28}b\,\hat{\mathbf{y}}+\left(\frac{1}{2}+z_{28}\right)c\sin\beta\,\hat{\mathbf{z}}$ | (4$a$) | O XXII |
| $\mathbf{B}_{57}$ | = | $(x_{29}-y_{29})\,\mathbf{a}_1+(x_{29}+y_{29})\,\mathbf{a}_2+z_{29}\,\mathbf{a}_3$ | = | $(x_{29}a+z_{29}c\cos\beta)\,\hat{\mathbf{x}}+y_{29}b\,\hat{\mathbf{y}}+z_{29}c\sin\beta\,\hat{\mathbf{z}}$ | (4$a$) | O XXIII |
| $\mathbf{B}_{58}$ | = | $(x_{29}+y_{29})\,\mathbf{a}_1+(x_{29}-y_{29})\,\mathbf{a}_2+\left(\frac{1}{2}+z_{29}\right)\mathbf{a}_3$ | = | $\left(\frac{1}{2}c\cos\beta+x_{29}a+z_{29}c\cos\beta\right)\hat{\mathbf{x}}-y_{29}b\,\hat{\mathbf{y}}+\left(\frac{1}{2}+z_{29}\right)c\sin\beta\,\hat{\mathbf{z}}$ | (4$a$) | O XXIII |
| $\mathbf{B}_{59}$ | = | $(x_{30}-y_{30})\,\mathbf{a}_1+(x_{30}+y_{30})\,\mathbf{a}_2+z_{30}\,\mathbf{a}_3$ | = | $(x_{30}a+z_{30}c\cos\beta)\,\hat{\mathbf{x}}+y_{30}b\,\hat{\mathbf{y}}+z_{30}c\sin\beta\,\hat{\mathbf{z}}$ | (4$a$) | O XXIV |
| $\mathbf{B}_{60}$ | = | $(x_{30}+y_{30})\,\mathbf{a}_1+(x_{30}-y_{30})\,\mathbf{a}_2+\left(\frac{1}{2}+z_{30}\right)\mathbf{a}_3$ | = | $\left(\frac{1}{2}c\cos\beta+x_{30}a+z_{30}c\cos\beta\right)\hat{\mathbf{x}}-y_{30}b\,\hat{\mathbf{y}}+\left(\frac{1}{2}+z_{30}\right)c\sin\beta\,\hat{\mathbf{z}}$ | (4$a$) | O XXIV |
| $\mathbf{B}_{61}$ | = | $(x_{31}-y_{31})\,\mathbf{a}_1+(x_{31}+y_{31})\,\mathbf{a}_2+z_{31}\,\mathbf{a}_3$ | = | $(x_{31}a+z_{31}c\cos\beta)\,\hat{\mathbf{x}}+y_{31}b\,\hat{\mathbf{y}}+z_{31}c\sin\beta\,\hat{\mathbf{z}}$ | (4$a$) | O XXV |
| $\mathbf{B}_{62}$ | = | $(x_{31}+y_{31})\,\mathbf{a}_1+(x_{31}-y_{31})\,\mathbf{a}_2+\left(\frac{1}{2}+z_{31}\right)\mathbf{a}_3$ | = | $\left(\frac{1}{2}c\cos\beta+x_{31}a+z_{31}c\cos\beta\right)\hat{\mathbf{x}}-y_{31}b\,\hat{\mathbf{y}}+\left(\frac{1}{2}+z_{31}\right)c\sin\beta\,\hat{\mathbf{z}}$ | (4$a$) | O XXV |
| $\mathbf{B}_{63}$ | = | $(x_{32}-y_{32})\,\mathbf{a}_1+(x_{32}+y_{32})\,\mathbf{a}_2+z_{32}\,\mathbf{a}_3$ | = | $(x_{32}a+z_{32}c\cos\beta)\,\hat{\mathbf{x}}+y_{32}b\,\hat{\mathbf{y}}+z_{32}c\sin\beta\,\hat{\mathbf{z}}$ | (4$a$) | O XXVI |
| $\mathbf{B}_{64}$ | = | $(x_{32}+y_{32})\,\mathbf{a}_1+(x_{32}-y_{32})\,\mathbf{a}_2+\left(\frac{1}{2}+z_{32}\right)\mathbf{a}_3$ | = | $\left(\frac{1}{2}c\cos\beta+x_{32}a+z_{32}c\cos\beta\right)\hat{\mathbf{x}}-y_{32}b\,\hat{\mathbf{y}}+\left(\frac{1}{2}+z_{32}\right)c\sin\beta\,\hat{\mathbf{z}}$ | (4$a$) | O XXVI |
| $\mathbf{B}_{65}$ | = | $(x_{33}-y_{33})\,\mathbf{a}_1+(x_{33}+y_{33})\,\mathbf{a}_2+z_{33}\,\mathbf{a}_3$ | = | $(x_{33}a+z_{33}c\cos\beta)\,\hat{\mathbf{x}}+y_{33}b\,\hat{\mathbf{y}}+z_{33}c\sin\beta\,\hat{\mathbf{z}}$ | (4$a$) | O XXVII |
| $\mathbf{B}_{66}$ | = | $(x_{33}+y_{33})\,\mathbf{a}_1+(x_{33}-y_{33})\,\mathbf{a}_2+\left(\frac{1}{2}+z_{33}\right)\mathbf{a}_3$ | = | $\left(\frac{1}{2}c\cos\beta+x_{33}a+z_{33}c\cos\beta\right)\hat{\mathbf{x}}-y_{33}b\,\hat{\mathbf{y}}+\left(\frac{1}{2}+z_{33}\right)c\sin\beta\,\hat{\mathbf{z}}$ | (4$a$) | O XXVII |
| $\mathbf{B}_{67}$ | = | $(x_{34}-y_{34})\,\mathbf{a}_1+(x_{34}+y_{34})\,\mathbf{a}_2+z_{34}\,\mathbf{a}_3$ | = | $(x_{34}a+z_{34}c\cos\beta)\,\hat{\mathbf{x}}+y_{34}b\,\hat{\mathbf{y}}+z_{34}c\sin\beta\,\hat{\mathbf{z}}$ | (4$a$) | O XXVIII |
| $\mathbf{B}_{68}$ | = | $(x_{34}+y_{34})\,\mathbf{a}_1+(x_{34}-y_{34})\,\mathbf{a}_2+\left(\frac{1}{2}+z_{34}\right)\mathbf{a}_3$ | = | $\left(\frac{1}{2}c\cos\beta+x_{34}a+z_{34}c\cos\beta\right)\hat{\mathbf{x}}-y_{34}b\,\hat{\mathbf{y}}+\left(\frac{1}{2}+z_{34}\right)c\sin\beta\,\hat{\mathbf{z}}$ | (4$a$) | O XXVIII |
| $\mathbf{B}_{69}$ | = | $(x_{35}-y_{35})\,\mathbf{a}_1+(x_{35}+y_{35})\,\mathbf{a}_2+z_{35}\,\mathbf{a}_3$ | = | $(x_{35}a+z_{35}c\cos\beta)\,\hat{\mathbf{x}}+y_{35}b\,\hat{\mathbf{y}}+z_{35}c\sin\beta\,\hat{\mathbf{z}}$ | (4$a$) | O XXIX |
| $\mathbf{B}_{70}$ | = | $\left(x_{35}+y_{35}\right)\,\mathbf{a}_{1}+\left(x_{35}-y_{35}\right)\,\mathbf{a}_{2}+\left(\frac{1}{2}+z_{35}\right)\,\mathbf{a}_{3}$ | = | $\left(\frac{1}{2}c\cos\beta+x_{35}a+z_{35}c\cos\beta\right)\,\mathbf{\hat{x}}-y_{35}b\,\mathbf{\hat{y}}+\left(\frac{1}{2}+z_{35}\right)c\sin\beta\,\mathbf{\hat{z}}$ | (4$a$) | O XXIX |
| $\mathbf{B}_{71}$ | = | $\left(x_{36}-y_{36}\right)\,\mathbf{a}_{1}+\left(x_{36}+y_{36}\right)\,\mathbf{a}_{2}+z_{36}\,\mathbf{a}_{3}$ | = | $\left(x_{36}a+z_{36}c\cos\beta\right)\,\mathbf{\hat{x}}+y_{36}b\,\mathbf{\hat{y}}+z_{36}c\sin\beta\,\mathbf{\hat{z}}$ | (4$a$) | O XXX |
| $\mathbf{B}_{72}$ | = | $\left(x_{36}+y_{36}\right)\,\mathbf{a}_{1}+\left(x_{36}-y_{36}\right)\,\mathbf{a}_{2}+\left(\frac{1}{2}+z_{36}\right)\,\mathbf{a}_{3}$ | = | $\left(\frac{1}{2}c\cos\beta+x_{36}a+z_{36}c\cos\beta\right)\,\mathbf{\hat{x}}-y_{36}b\,\mathbf{\hat{y}}+\left(\frac{1}{2}+z_{36}\right)c\sin\beta\,\mathbf{\hat{z}}$ | (4$a$) | O XXX |
| $\mathbf{B}_{73}$ | = | $\left(x_{37}-y_{37}\right)\,\mathbf{a}_{1}+\left(x_{37}+y_{37}\right)\,\mathbf{a}_{2}+z_{37}\,\mathbf{a}_{3}$ | = | $\left(x_{37}a+z_{37}c\cos\beta\right)\,\mathbf{\hat{x}}+y_{37}b\,\mathbf{\hat{y}}+z_{37}c\sin\beta\,\mathbf{\hat{z}}$ | (4$a$) | O XXXI |
| $\mathbf{B}_{74}$ | = | $\left(x_{37}+y_{37}\right)\,\mathbf{a}_{1}+\left(x_{37}-y_{37}\right)\,\mathbf{a}_{2}+\left(\frac{1}{2}+z_{37}\right)\,\mathbf{a}_{3}$ | = | $\left(\frac{1}{2}c\cos\beta+x_{37}a+z_{37}c\cos\beta\right)\,\mathbf{\hat{x}}-y_{37}b\,\mathbf{\hat{y}}+\left(\frac{1}{2}+z_{37}\right)c\sin\beta\,\mathbf{\hat{z}}$ | (4$a$) | O XXXI |
| $\mathbf{B}_{75}$ | = | $\left(x_{38}-y_{38}\right)\,\mathbf{a}_{1}+\left(x_{38}+y_{38}\right)\,\mathbf{a}_{2}+z_{38}\,\mathbf{a}_{3}$ | = | $\left(x_{38}a+z_{38}c\cos\beta\right)\,\mathbf{\hat{x}}+y_{38}b\,\mathbf{\hat{y}}+z_{38}c\sin\beta\,\mathbf{\hat{z}}$ | (4$a$) | O XXXII |
| $\mathbf{B}_{76}$ | = | $\left(x_{38}+y_{38}\right)\,\mathbf{a}_{1}+\left(x_{38}-y_{38}\right)\,\mathbf{a}_{2}+\left(\frac{1}{2}+z_{38}\right)\,\mathbf{a}_{3}$ | = | $\left(\frac{1}{2}c\cos\beta+x_{38}a+z_{38}c\cos\beta\right)\,\mathbf{\hat{x}}-y_{38}b\,\mathbf{\hat{y}}+\left(\frac{1}{2}+z_{38}\right)c\sin\beta\,\mathbf{\hat{z}}$ | (4$a$) | O XXXII |
| $\mathbf{B}_{77}$ | = | $\left(x_{39}-y_{39}\right)\,\mathbf{a}_{1}+\left(x_{39}+y_{39}\right)\,\mathbf{a}_{2}+z_{39}\,\mathbf{a}_{3}$ | = | $\left(x_{39}a+z_{39}c\cos\beta\right)\,\mathbf{\hat{x}}+y_{39}b\,\mathbf{\hat{y}}+z_{39}c\sin\beta\,\mathbf{\hat{z}}$ | (4$a$) | O XXXIII |
| $\mathbf{B}_{78}$ | = | $\left(x_{39}+y_{39}\right)\,\mathbf{a}_{1}+\left(x_{39}-y_{39}\right)\,\mathbf{a}_{2}+\left(\frac{1}{2}+z_{39}\right)\,\mathbf{a}_{3}$ | = | $\left(\frac{1}{2}c\cos\beta+x_{39}a+z_{39}c\cos\beta\right)\,\mathbf{\hat{x}}-y_{39}b\,\mathbf{\hat{y}}+\left(\frac{1}{2}+z_{39}\right)c\sin\beta\,\mathbf{\hat{z}}$ | (4$a$) | O XXXIII |
| $\mathbf{B}_{79}$ | = | $\left(x_{40}-y_{40}\right)\,\mathbf{a}_{1}+\left(x_{40}+y_{40}\right)\,\mathbf{a}_{2}+z_{40}\,\mathbf{a}_{3}$ | = | $\left(x_{40}a+z_{40}c\cos\beta\right)\,\mathbf{\hat{x}}+y_{40}b\,\mathbf{\hat{y}}+z_{40}c\sin\beta\,\mathbf{\hat{z}}$ | (4$a$) | O XXXIV |
| $\mathbf{B}_{80}$ | = | $\left(x_{40}+y_{40}\right)\,\mathbf{a}_{1}+\left(x_{40}-y_{40}\right)\,\mathbf{a}_{2}+\left(\frac{1}{2}+z_{40}\right)\,\mathbf{a}_{3}$ | = | $\left(\frac{1}{2}c\cos\beta+x_{40}a+z_{40}c\cos\beta\right)\,\mathbf{\hat{x}}-y_{40}b\,\mathbf{\hat{y}}+\left(\frac{1}{2}+z_{40}\right)c\sin\beta\,\mathbf{\hat{z}}$ | (4$a$) | O XXXIV |
| $\mathbf{B}_{81}$ | = | $\left(x_{41}-y_{41}\right)\,\mathbf{a}_{1}+\left(x_{41}+y_{41}\right)\,\mathbf{a}_{2}+z_{41}\,\mathbf{a}_{3}$ | = | $\left(x_{41}a+z_{41}c\cos\beta\right)\,\mathbf{\hat{x}}+y_{41}b\,\mathbf{\hat{y}}+z_{41}c\sin\beta\,\mathbf{\hat{z}}$ | (4$a$) | O XXXV |
| $\mathbf{B}_{82}$ | = | $\left(x_{41}+y_{41}\right)\,\mathbf{a}_{1}+\left(x_{41}-y_{41}\right)\,\mathbf{a}_{2}+\left(\frac{1}{2}+z_{41}\right)\,\mathbf{a}_{3}$ | = | $\left(\frac{1}{2}c\cos\beta+x_{41}a+z_{41}c\cos\beta\right)\,\mathbf{\hat{x}}-y_{41}b\,\mathbf{\hat{y}}+\left(\frac{1}{2}+z_{41}\right)c\sin\beta\,\mathbf{\hat{z}}$ | (4$a$) | O XXXV |
| $\mathbf{B}_{83}$ | = | $\left(x_{42}-y_{42}\right)\,\mathbf{a}_{1}+\left(x_{42}+y_{42}\right)\,\mathbf{a}_{2}+z_{42}\,\mathbf{a}_{3}$ | = | $\left(x_{42}a+z_{42}c\cos\beta\right)\,\mathbf{\hat{x}}+y_{42}b\,\mathbf{\hat{y}}+z_{42}c\sin\beta\,\mathbf{\hat{z}}$ | (4$a$) | O XXXVI |
| $\mathbf{B}_{84}$ | = | $\left(x_{42}+y_{42}\right)\,\mathbf{a}_{1}+\left(x_{42}-y_{42}\right)\,\mathbf{a}_{2}+\left(\frac{1}{2}+z_{42}\right)\,\mathbf{a}_{3}$ | = | $\left(\frac{1}{2}c\cos\beta+x_{42}a+z_{42}c\cos\beta\right)\,\mathbf{\hat{x}}-y_{42}b\,\mathbf{\hat{y}}+\left(\frac{1}{2}+z_{42}\right)c\sin\beta\,\mathbf{\hat{z}}$ | (4$a$) | O XXXVI |
| $\mathbf{B}_{85}$ | = | $\left(x_{43}-y_{43}\right)\,\mathbf{a}_{1}+\left(x_{43}+y_{43}\right)\,\mathbf{a}_{2}+z_{43}\,\mathbf{a}_{3}$ | = | $\left(x_{43}a+z_{43}c\cos\beta\right)\,\mathbf{\hat{x}}+y_{43}b\,\mathbf{\hat{y}}+z_{43}c\sin\beta\,\mathbf{\hat{z}}$ | (4$a$) | W I |
| $\mathbf{B}_{86}$ | = | $\left(x_{43}+y_{43}\right)\,\mathbf{a}_{1}+\left(x_{43}-y_{43}\right)\,\mathbf{a}_{2}+\left(\frac{1}{2}+z_{43}\right)\,\mathbf{a}_{3}$ | = | $\left(\frac{1}{2}c\cos\beta+x_{43}a+z_{43}c\cos\beta\right)\,\mathbf{\hat{x}}-y_{43}b\,\mathbf{\hat{y}}+\left(\frac{1}{2}+z_{43}\right)c\sin\beta\,\mathbf{\hat{z}}$ | (4$a$) | W I |
| $\mathbf{B}_{87}$ | = | $\left(x_{44}-y_{44}\right)\,\mathbf{a}_{1}+\left(x_{44}+y_{44}\right)\,\mathbf{a}_{2}+z_{44}\,\mathbf{a}_{3}$ | = | $\left(x_{44}a+z_{44}c\cos\beta\right)\,\mathbf{\hat{x}}+y_{44}b\,\mathbf{\hat{y}}+z_{44}c\sin\beta\,\mathbf{\hat{z}}$ | (4$a$) | W II |
| $\mathbf{B}_{88}$ | = | $\left(x_{44}+y_{44}\right)\,\mathbf{a}_{1}+\left(x_{44}-y_{44}\right)\,\mathbf{a}_{2}+\left(\frac{1}{2}+z_{44}\right)\,\mathbf{a}_{3}$ | = | $\left(\frac{1}{2}c\cos\beta+x_{44}a+z_{44}c\cos\beta\right)\,\mathbf{\hat{x}}-y_{44}b\,\mathbf{\hat{y}}+\left(\frac{1}{2}+z_{44}\right)c\sin\beta\,\mathbf{\hat{z}}$ | (4$a$) | W II |
| $\mathbf{B}_{89}$ | = | $\left(x_{45}-y_{45}\right)\,\mathbf{a}_{1}+\left(x_{45}+y_{45}\right)\,\mathbf{a}_{2}+z_{45}\,\mathbf{a}_{3}$ | = | $\left(x_{45}a+z_{45}c\cos\beta\right)\,\mathbf{\hat{x}}+y_{45}b\,\mathbf{\hat{y}}+z_{45}c\sin\beta\,\mathbf{\hat{z}}$ | (4$a$) | W III |
| $\mathbf{B}_{90}$ | = | $\left(x_{45}+y_{45}\right)\,\mathbf{a}_{1}+\left(x_{45}-y_{45}\right)\,\mathbf{a}_{2}+\left(\frac{1}{2}+z_{45}\right)\,\mathbf{a}_{3}$ | = | $\left(\frac{1}{2}c\cos\beta+x_{45}a+z_{45}c\cos\beta\right)\,\mathbf{\hat{x}}-y_{45}b\,\mathbf{\hat{y}}+\left(\frac{1}{2}+z_{45}\right)c\sin\beta\,\mathbf{\hat{z}}$ | (4$a$) | W III |
| $\mathbf{B}_{91}$ | = | $\left(x_{46}-y_{46}\right)\,\mathbf{a}_{1}+\left(x_{46}+y_{46}\right)\,\mathbf{a}_{2}+z_{46}\,\mathbf{a}_{3}$ | = | $\left(x_{46}a+z_{46}c\cos\beta\right)\,\mathbf{\hat{x}}+y_{46}b\,\mathbf{\hat{y}}+z_{46}c\sin\beta\,\mathbf{\hat{z}}$ | (4$a$) | W IV |
| $\mathbf{B}_{92}$ | = | $(x_{46}+y_{46})\,\mathbf{a}_1+(x_{46}-y_{46})\,\mathbf{a}_2+\left(\frac{1}{2}+z_{46}\right)\mathbf{a}_3$ | = | $\left(\frac{1}{2}c\cos\beta+x_{46}a+z_{46}c\cos\beta\right)\hat{\mathbf{x}}-y_{46}b\,\hat{\mathbf{y}}+\left(\frac{1}{2}+z_{46}\right)c\sin\beta\,\hat{\mathbf{z}}$ | (4a) | W IV |
| $\mathbf{B}_{93}$ | = | $(x_{47}-y_{47})\,\mathbf{a}_1+(x_{47}+y_{47})\,\mathbf{a}_2+z_{47}\,\mathbf{a}_3$ | = | $(x_{47}a+z_{47}c\cos\beta)\,\hat{\mathbf{x}}+y_{47}b\,\hat{\mathbf{y}}+z_{47}c\sin\beta\,\hat{\mathbf{z}}$ | (4a) | W V |
| $\mathbf{B}_{94}$ | = | $(x_{47}+y_{47})\,\mathbf{a}_1+(x_{47}-y_{47})\,\mathbf{a}_2+\left(\frac{1}{2}+z_{47}\right)\mathbf{a}_3$ | = | $\left(\frac{1}{2}c\cos\beta+x_{47}a+z_{47}c\cos\beta\right)\hat{\mathbf{x}}-y_{47}b\,\hat{\mathbf{y}}+\left(\frac{1}{2}+z_{47}\right)c\sin\beta\,\hat{\mathbf{z}}$ | (4a) | W V |
| $\mathbf{B}_{95}$ | = | $(x_{48}-y_{48})\,\mathbf{a}_1+(x_{48}+y_{48})\,\mathbf{a}_2+z_{48}\,\mathbf{a}_3$ | = | $(x_{48}a+z_{48}c\cos\beta)\,\hat{\mathbf{x}}+y_{48}b\,\hat{\mathbf{y}}+z_{48}c\sin\beta\,\hat{\mathbf{z}}$ | (4a) | W VI |
| $\mathbf{B}_{96}$ | = | $(x_{48}+y_{48})\,\mathbf{a}_1+(x_{48}-y_{48})\,\mathbf{a}_2+\left(\frac{1}{2}+z_{48}\right)\mathbf{a}_3$ | = | $\left(\frac{1}{2}c\cos\beta+x_{48}a+z_{48}c\cos\beta\right)\hat{\mathbf{x}}-y_{48}b\,\hat{\mathbf{y}}+\left(\frac{1}{2}+z_{48}\right)c\sin\beta\,\hat{\mathbf{z}}$ | (4a) | W VI |
| $\mathbf{B}_{97}$ | = | $(x_{49}-y_{49})\,\mathbf{a}_1+(x_{49}+y_{49})\,\mathbf{a}_2+z_{49}\,\mathbf{a}_3$ | = | $(x_{49}a+z_{49}c\cos\beta)\,\hat{\mathbf{x}}+y_{49}b\,\hat{\mathbf{y}}+z_{49}c\sin\beta\,\hat{\mathbf{z}}$ | (4a) | W VII |
| $\mathbf{B}_{98}$ | = | $(x_{49}+y_{49})\,\mathbf{a}_1+(x_{49}-y_{49})\,\mathbf{a}_2+\left(\frac{1}{2}+z_{49}\right)\mathbf{a}_3$ | = | $\left(\frac{1}{2}c\cos\beta+x_{49}a+z_{49}c\cos\beta\right)\hat{\mathbf{x}}-y_{49}b\,\hat{\mathbf{y}}+\left(\frac{1}{2}+z_{49}\right)c\sin\beta\,\hat{\mathbf{z}}$ | (4a) | W VII |
| $\mathbf{B}_{99}$ | = | $(x_{50}-y_{50})\,\mathbf{a}_1+(x_{50}+y_{50})\,\mathbf{a}_2+z_{50}\,\mathbf{a}_3$ | = | $(x_{50}a+z_{50}c\cos\beta)\,\hat{\mathbf{x}}+y_{50}b\,\hat{\mathbf{y}}+z_{50}c\sin\beta\,\hat{\mathbf{z}}$ | (4a) | W VIII |
| $\mathbf{B}_{100}$ | = | $(x_{50}+y_{50})\,\mathbf{a}_1+(x_{50}-y_{50})\,\mathbf{a}_2+\left(\frac{1}{2}+z_{50}\right)\mathbf{a}_3$ | = | $\left(\frac{1}{2}c\cos\beta+x_{50}a+z_{50}c\cos\beta\right)\hat{\mathbf{x}}-y_{50}b\,\hat{\mathbf{y}}+\left(\frac{1}{2}+z_{50}\right)c\sin\beta\,\hat{\mathbf{z}}$ | (4a) | W VIII |
| $\mathbf{B}_{101}$ | = | $(x_{51}-y_{51})\,\mathbf{a}_1+(x_{51}+y_{51})\,\mathbf{a}_2+z_{51}\,\mathbf{a}_3$ | = | $(x_{51}a+z_{51}c\cos\beta)\,\hat{\mathbf{x}}+y_{51}b\,\hat{\mathbf{y}}+z_{51}c\sin\beta\,\hat{\mathbf{z}}$ | (4a) | W IX |
| $\mathbf{B}_{102}$ | = | $(x_{51}+y_{51})\,\mathbf{a}_1+(x_{51}-y_{51})\,\mathbf{a}_2+\left(\frac{1}{2}+z_{51}\right)\mathbf{a}_3$ | = | $\left(\frac{1}{2}c\cos\beta+x_{51}a+z_{51}c\cos\beta\right)\hat{\mathbf{x}}-y_{51}b\,\hat{\mathbf{y}}+\left(\frac{1}{2}+z_{51}\right)c\sin\beta\,\hat{\mathbf{z}}$ | (4a) | W IX |
| $\mathbf{B}_{103}$ | = | $(x_{52}-y_{52})\,\mathbf{a}_1+(x_{52}+y_{52})\,\mathbf{a}_2+z_{52}\,\mathbf{a}_3$ | = | $(x_{52}a+z_{52}c\cos\beta)\,\hat{\mathbf{x}}+y_{52}b\,\hat{\mathbf{y}}+z_{52}c\sin\beta\,\hat{\mathbf{z}}$ | (4a) | W X |
| $\mathbf{B}_{104}$ | = | $(x_{52}+y_{52})\,\mathbf{a}_1+(x_{52}-y_{52})\,\mathbf{a}_2+\left(\frac{1}{2}+z_{52}\right)\mathbf{a}_3$ | = | $\left(\frac{1}{2}c\cos\beta+x_{52}a+z_{52}c\cos\beta\right)\hat{\mathbf{x}}-y_{52}b\,\hat{\mathbf{y}}+\left(\frac{1}{2}+z_{52}\right)c\sin\beta\,\hat{\mathbf{z}}$ | (4a) | W X |
| $\mathbf{B}_{105}$ | = | $(x_{53}-y_{53})\,\mathbf{a}_1+(x_{53}+y_{53})\,\mathbf{a}_2+z_{53}\,\mathbf{a}_3$ | = | $(x_{53}a+z_{53}c\cos\beta)\,\hat{\mathbf{x}}+y_{53}b\,\hat{\mathbf{y}}+z_{53}c\sin\beta\,\hat{\mathbf{z}}$ | (4a) | W XI |
| $\mathbf{B}_{106}$ | = | $(x_{53}+y_{53})\,\mathbf{a}_1+(x_{53}-y_{53})\,\mathbf{a}_2+\left(\frac{1}{2}+z_{53}\right)\mathbf{a}_3$ | = | $\left(\frac{1}{2}c\cos\beta+x_{53}a+z_{53}c\cos\beta\right)\hat{\mathbf{x}}-y_{53}b\,\hat{\mathbf{y}}+\left(\frac{1}{2}+z_{53}\right)c\sin\beta\,\hat{\mathbf{z}}$ | (4a) | W XI |

**Geometry files:**

- CIF: pp. 1520
- POSCAR: pp. 1521

# $K_2S_2O_5$ ($K0_1$) Structure: A2B5C2_mP18_11_2e_e2f_2e

http://aflow.org/prototype-encyclopedia/A2B5C2_mP18_11_2e_e2f_2e

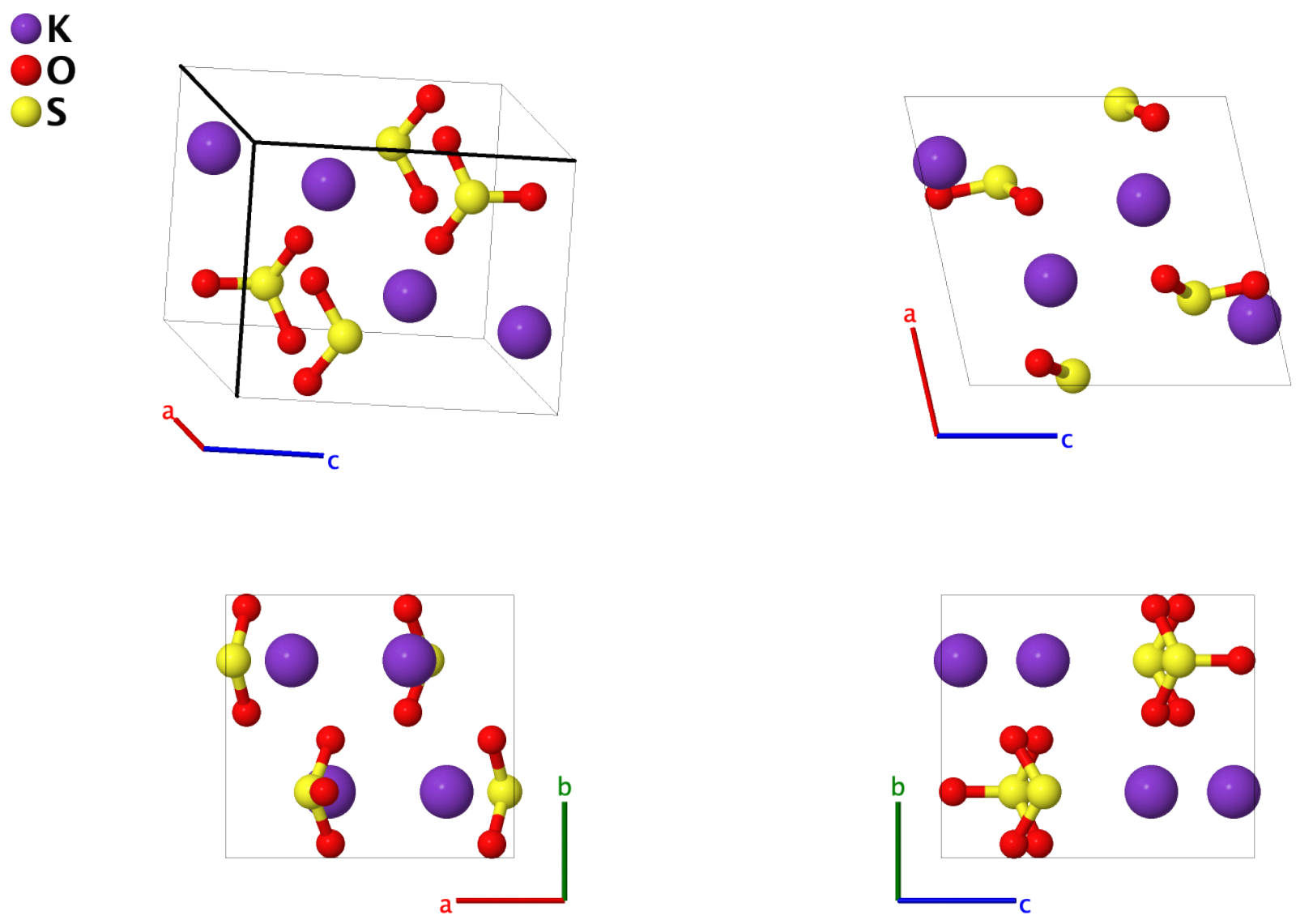

| | | |
|---|---|---|
| **Prototype** | : | $K_2O_5S_2$ |
| **AFLOW prototype label** | : | A2B5C2_mP18_11_2e_e2f_2e |
| ***Strukturbericht* designation** | : | $K0_1$ |
| **Pearson symbol** | : | mP18 |
| **Space group number** | : | 11 |
| **Space group symbol** | : | $P2_1/m$ |
| **AFLOW prototype command** | : | `aflow --proto=A2B5C2_mP18_11_2e_e2f_2e` `--params=`$a, b/a, c/a, \beta, x_1, z_1, x_2, z_2, x_3, z_3, x_4, z_4, x_5, z_5, x_6, y_6, z_6, x_7, y_7, z_7$ |

### Other compounds with this structure

- $K_2S_2P_5$ and $(NH_4)_2S_2P_5$

### Simple Monoclinic primitive vectors:

$$\mathbf{a}_1 = a\,\hat{\mathbf{x}}$$
$$\mathbf{a}_2 = b\,\hat{\mathbf{y}}$$
$$\mathbf{a}_3 = c\cos\beta\,\hat{\mathbf{x}} + c\sin\beta\,\hat{\mathbf{z}}$$

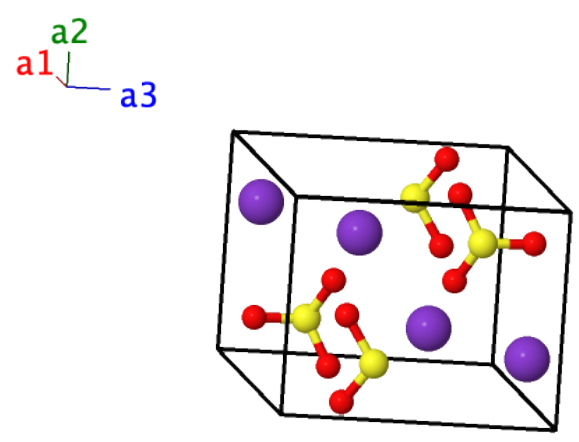

### Basis vectors:

| | | Lattice Coordinates | | Cartesian Coordinates | Wyckoff Position | Atom Type |
|---|---|---|---|---|---|---|
| $\mathbf{B}_1$ | = | $x_1\,\mathbf{a}_1 + \frac{1}{4}\,\mathbf{a}_2 + z_1\,\mathbf{a}_3$ | = | $(x_1 a + z_1 c\cos\beta)\,\hat{\mathbf{x}} + \frac{1}{4} b\,\hat{\mathbf{y}} + z_1 c\sin\beta\,\hat{\mathbf{z}}$ | (2e) | K I |
| $\mathbf{B}_2$ | = | $-x_1\,\mathbf{a}_1 + \frac{3}{4}\,\mathbf{a}_2 - z_1\,\mathbf{a}_3$ | = | $(-x_1 a - z_1 c\cos\beta)\,\hat{\mathbf{x}} + \frac{3}{4} b\,\hat{\mathbf{y}} - z_1 c\sin\beta\,\hat{\mathbf{z}}$ | (2e) | K I |
| $\mathbf{B}_3$ | = | $x_2\,\mathbf{a}_1 + \frac{1}{4}\,\mathbf{a}_2 + z_2\,\mathbf{a}_3$ | = | $(x_2 a + z_2 c\cos\beta)\,\hat{\mathbf{x}} + \frac{1}{4} b\,\hat{\mathbf{y}} + z_2 c\sin\beta\,\hat{\mathbf{z}}$ | (2e) | K II |
| $\mathbf{B}_{4}$ | = | $-x_{2}\,\mathbf{a}_{1}+\frac{3}{4}\,\mathbf{a}_{2}-z_{2}\,\mathbf{a}_{3}$ | = | $\left(-x_{2}a-z_{2}c\cos\beta\right)\,\mathbf{\hat{x}}+\frac{3}{4}b\,\mathbf{\hat{y}}-z_{2}c\sin\beta\,\mathbf{\hat{z}}$ | (2e) | K II |
| $\mathbf{B}_{5}$ | = | $x_{3}\,\mathbf{a}_{1}+\frac{1}{4}\,\mathbf{a}_{2}+z_{3}\,\mathbf{a}_{3}$ | = | $\left(x_{3}a+z_{3}c\cos\beta\right)\,\mathbf{\hat{x}}+\frac{1}{4}b\,\mathbf{\hat{y}}+z_{3}c\sin\beta\,\mathbf{\hat{z}}$ | (2e) | O I |
| $\mathbf{B}_{6}$ | = | $-x_{3}\,\mathbf{a}_{1}+\frac{3}{4}\,\mathbf{a}_{2}-z_{3}\,\mathbf{a}_{3}$ | = | $\left(-x_{3}a-z_{3}c\cos\beta\right)\,\mathbf{\hat{x}}+\frac{3}{4}b\,\mathbf{\hat{y}}-z_{3}c\sin\beta\,\mathbf{\hat{z}}$ | (2e) | O I |
| $\mathbf{B}_{7}$ | = | $x_{4}\,\mathbf{a}_{1}+\frac{1}{4}\,\mathbf{a}_{2}+z_{4}\,\mathbf{a}_{3}$ | = | $\left(x_{4}a+z_{4}c\cos\beta\right)\,\mathbf{\hat{x}}+\frac{1}{4}b\,\mathbf{\hat{y}}+z_{4}c\sin\beta\,\mathbf{\hat{z}}$ | (2e) | S I |
| $\mathbf{B}_{8}$ | = | $-x_{4}\,\mathbf{a}_{1}+\frac{3}{4}\,\mathbf{a}_{2}-z_{4}\,\mathbf{a}_{3}$ | = | $\left(-x_{4}a-z_{4}c\cos\beta\right)\,\mathbf{\hat{x}}+\frac{3}{4}b\,\mathbf{\hat{y}}-z_{4}c\sin\beta\,\mathbf{\hat{z}}$ | (2e) | S I |
| $\mathbf{B}_{9}$ | = | $x_{5}\,\mathbf{a}_{1}+\frac{1}{4}\,\mathbf{a}_{2}+z_{5}\,\mathbf{a}_{3}$ | = | $\left(x_{5}a+z_{5}c\cos\beta\right)\,\mathbf{\hat{x}}+\frac{1}{4}b\,\mathbf{\hat{y}}+z_{5}c\sin\beta\,\mathbf{\hat{z}}$ | (2e) | S II |
| $\mathbf{B}_{10}$ | = | $-x_{5}\,\mathbf{a}_{1}+\frac{3}{4}\,\mathbf{a}_{2}-z_{5}\,\mathbf{a}_{3}$ | = | $\left(-x_{5}a-z_{5}c\cos\beta\right)\,\mathbf{\hat{x}}+\frac{3}{4}b\,\mathbf{\hat{y}}-z_{5}c\sin\beta\,\mathbf{\hat{z}}$ | (2e) | S II |
| $\mathbf{B}_{11}$ | = | $x_{6}\,\mathbf{a}_{1}+y_{6}\,\mathbf{a}_{2}+z_{6}\,\mathbf{a}_{3}$ | = | $\left(x_{6}a+z_{6}c\cos\beta\right)\,\mathbf{\hat{x}}+y_{6}b\,\mathbf{\hat{y}}+z_{6}c\sin\beta\,\mathbf{\hat{z}}$ | (4f) | O II |
| $\mathbf{B}_{12}$ | = | $-x_{6}\,\mathbf{a}_{1}+\left(\frac{1}{2}+y_{6}\right)\,\mathbf{a}_{2}-z_{6}\,\mathbf{a}_{3}$ | = | $\left(-x_{6}a-z_{6}c\cos\beta\right)\,\mathbf{\hat{x}}+\left(\frac{1}{2}+y_{6}\right)b\,\mathbf{\hat{y}}-z_{6}c\sin\beta\,\mathbf{\hat{z}}$ | (4f) | O II |
| $\mathbf{B}_{13}$ | = | $-x_{6}\,\mathbf{a}_{1}-y_{6}\,\mathbf{a}_{2}-z_{6}\,\mathbf{a}_{3}$ | = | $\left(-x_{6}a-z_{6}c\cos\beta\right)\,\mathbf{\hat{x}}-y_{6}b\,\mathbf{\hat{y}}-z_{6}c\sin\beta\,\mathbf{\hat{z}}$ | (4f) | O II |
| $\mathbf{B}_{14}$ | = | $x_{6}\,\mathbf{a}_{1}+\left(\frac{1}{2}-y_{6}\right)\,\mathbf{a}_{2}+z_{6}\,\mathbf{a}_{3}$ | = | $\left(x_{6}a+z_{6}c\cos\beta\right)\,\mathbf{\hat{x}}+\left(\frac{1}{2}-y_{6}\right)b\,\mathbf{\hat{y}}+z_{6}c\sin\beta\,\mathbf{\hat{z}}$ | (4f) | O II |
| $\mathbf{B}_{15}$ | = | $x_{7}\,\mathbf{a}_{1}+y_{7}\,\mathbf{a}_{2}+z_{7}\,\mathbf{a}_{3}$ | = | $\left(x_{7}a+z_{7}c\cos\beta\right)\,\mathbf{\hat{x}}+y_{7}b\,\mathbf{\hat{y}}+z_{7}c\sin\beta\,\mathbf{\hat{z}}$ | (4f) | O III |
| $\mathbf{B}_{16}$ | = | $-x_{7}\,\mathbf{a}_{1}+\left(\frac{1}{2}+y_{7}\right)\,\mathbf{a}_{2}-z_{7}\,\mathbf{a}_{3}$ | = | $\left(-x_{7}a-z_{7}c\cos\beta\right)\,\mathbf{\hat{x}}+\left(\frac{1}{2}+y_{7}\right)b\,\mathbf{\hat{y}}-z_{7}c\sin\beta\,\mathbf{\hat{z}}$ | (4f) | O III |
| $\mathbf{B}_{17}$ | = | $-x_{7}\,\mathbf{a}_{1}-y_{7}\,\mathbf{a}_{2}-z_{7}\,\mathbf{a}_{3}$ | = | $\left(-x_{7}a-z_{7}c\cos\beta\right)\,\mathbf{\hat{x}}-y_{7}b\,\mathbf{\hat{y}}-z_{7}c\sin\beta\,\mathbf{\hat{z}}$ | (4f) | O III |
| $\mathbf{B}_{18}$ | = | $x_{7}\,\mathbf{a}_{1}+\left(\frac{1}{2}-y_{7}\right)\,\mathbf{a}_{2}+z_{7}\,\mathbf{a}_{3}$ | = | $\left(x_{7}a+z_{7}c\cos\beta\right)\,\mathbf{\hat{x}}+\left(\frac{1}{2}-y_{7}\right)b\,\mathbf{\hat{y}}+z_{7}c\sin\beta\,\mathbf{\hat{z}}$ | (4f) | O III |

**Geometry files:**

- CIF: pp. 1522
- POSCAR: pp. 1522

# $ZrSe_3$ Structure: A3B_mP8_11_3e_e

http://aflow.org/prototype-encyclopedia/A3B_mP8_11_3e_e

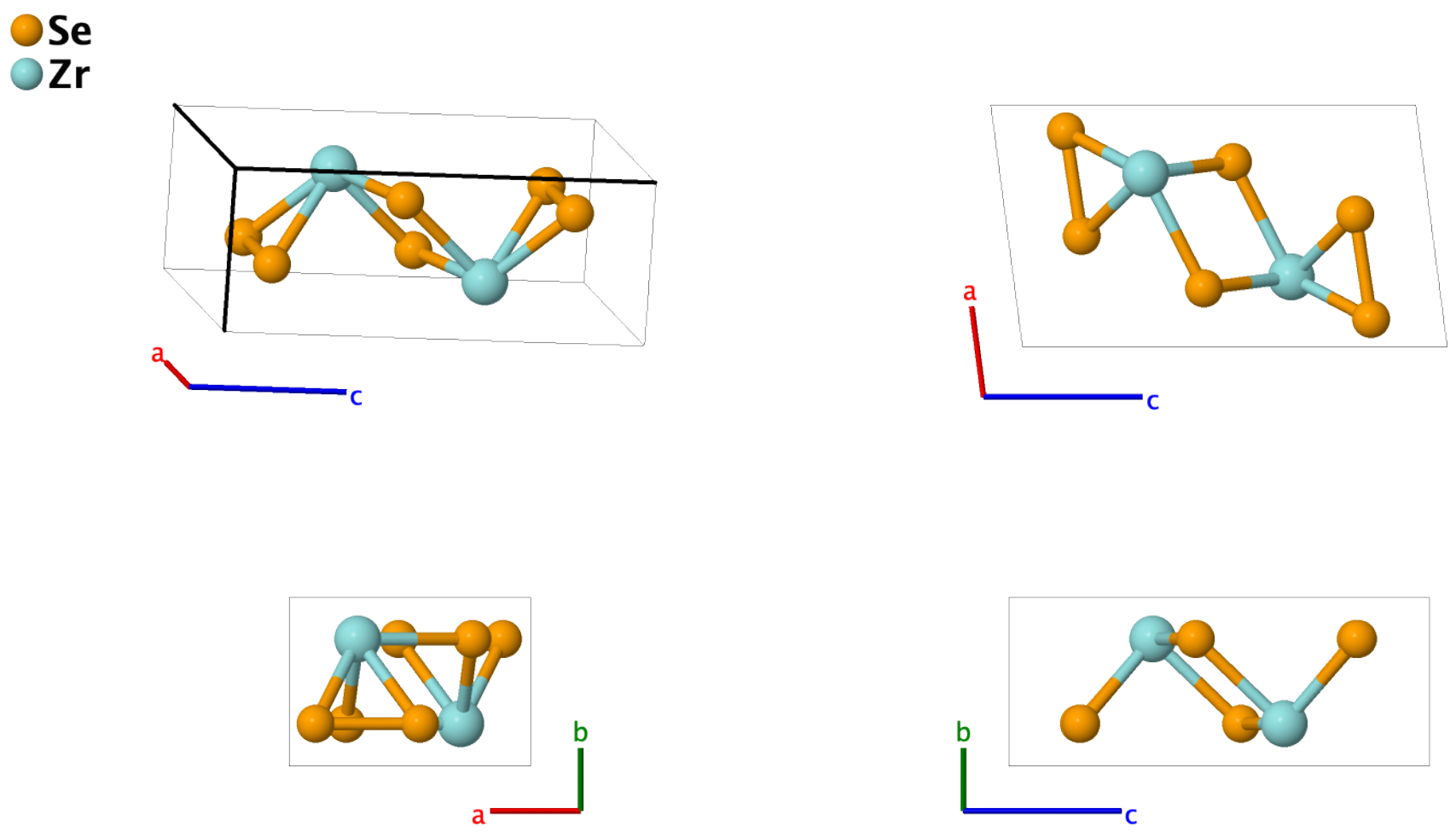

| | | |
|---|---|---|
| **Prototype** | : | $Se_3Zr$ |
| **AFLOW prototype label** | : | A3B_mP8_11_3e_e |
| ***Strukturbericht* designation** | : | None |
| **Pearson symbol** | : | mP8 |
| **Space group number** | : | 11 |
| **Space group symbol** | : | $P2_1/m$ |
| **AFLOW prototype command** | : | `aflow --proto=A3B_mP8_11_3e_e` `--params=`$a, b/a, c/a, \beta, x_1, z_1, x_2, z_2, x_3, z_3, x_4, z_4$ |

### Other compounds with this structure

- $HfS_3$, $HfSe_3$, $HfTe_3$, $TaSe_3$, $TiS_3$, $ZrS_3$, and $ZrTe_3$

### Simple Monoclinic primitive vectors:

$$\mathbf{a}_1 = a\,\hat{\mathbf{x}}$$
$$\mathbf{a}_2 = b\,\hat{\mathbf{y}}$$
$$\mathbf{a}_3 = c\cos\beta\,\hat{\mathbf{x}} + c\sin\beta\,\hat{\mathbf{z}}$$

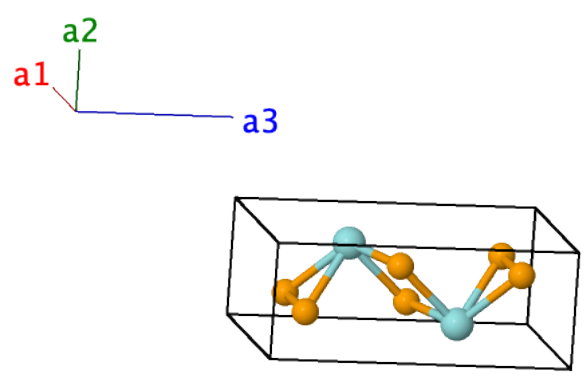

### Basis vectors:

| | | Lattice Coordinates | | Cartesian Coordinates | Wyckoff Position | Atom Type |
|---|---|---|---|---|---|---|
| $\mathbf{B}_1$ | = | $x_1\,\mathbf{a}_1 + \frac{1}{4}\,\mathbf{a}_2 + z_1\,\mathbf{a}_3$ | = | $(x_1 a + z_1 c\cos\beta)\,\hat{\mathbf{x}} + \frac{1}{4}b\,\hat{\mathbf{y}} + z_1 c\sin\beta\,\hat{\mathbf{z}}$ | (2e) | Se I |
| $\mathbf{B}_2$ | = | $-x_1\,\mathbf{a}_1 + \frac{3}{4}\,\mathbf{a}_2 - z_1\,\mathbf{a}_3$ | = | $(-x_1 a - z_1 c\cos\beta)\,\hat{\mathbf{x}} + \frac{3}{4}b\,\hat{\mathbf{y}} - z_1 c\sin\beta\,\hat{\mathbf{z}}$ | (2e) | Se I |
| $\mathbf{B}_3$ | = | $x_2\,\mathbf{a}_1 + \frac{1}{4}\,\mathbf{a}_2 + z_2\,\mathbf{a}_3$ | = | $(x_2 a + z_2 c\cos\beta)\,\hat{\mathbf{x}} + \frac{1}{4}b\,\hat{\mathbf{y}} + z_2 c\sin\beta\,\hat{\mathbf{z}}$ | (2e) | Se II |
| $\mathbf{B}_4$ | = | $-x_2\,\mathbf{a}_1 + \frac{3}{4}\,\mathbf{a}_2 - z_2\,\mathbf{a}_3$ | = | $(-x_2 a - z_2 c\cos\beta)\,\hat{\mathbf{x}} + \frac{3}{4}b\,\hat{\mathbf{y}} - z_2 c\sin\beta\,\hat{\mathbf{z}}$ | (2e) | Se II |
| $\mathbf{B}_5$ | = | $x_3\,\mathbf{a}_1 + \frac{1}{4}\,\mathbf{a}_2 + z_3\,\mathbf{a}_3$ | = | $(x_3 a + z_3 c\cos\beta)\,\hat{\mathbf{x}} + \frac{1}{4}b\,\hat{\mathbf{y}} + z_3 c\sin\beta\,\hat{\mathbf{z}}$ | (2e) | Se III |
| $\mathbf{B}_6$ | = | $-x_3\,\mathbf{a}_1 + \frac{3}{4}\,\mathbf{a}_2 - z_3\,\mathbf{a}_3$ | = | $(-x_3 a - z_3 c\cos\beta)\,\hat{\mathbf{x}} + \frac{3}{4} b\,\hat{\mathbf{y}} - z_3 c\sin\beta\,\hat{\mathbf{z}}$ | (2$e$) | Se III |
| $\mathbf{B}_7$ | = | $x_4\,\mathbf{a}_1 + \frac{1}{4}\,\mathbf{a}_2 + z_4\,\mathbf{a}_3$ | = | $(x_4 a + z_4 c\cos\beta)\,\hat{\mathbf{x}} + \frac{1}{4} b\,\hat{\mathbf{y}} + z_4 c\sin\beta\,\hat{\mathbf{z}}$ | (2$e$) | Zr |
| $\mathbf{B}_8$ | = | $-x_4\,\mathbf{a}_1 + \frac{3}{4}\,\mathbf{a}_2 - z_4\,\mathbf{a}_3$ | = | $(-x_4 a - z_4 c\cos\beta)\,\hat{\mathbf{x}} + \frac{3}{4} b\,\hat{\mathbf{y}} - z_4 c\sin\beta\,\hat{\mathbf{z}}$ | (2$e$) | Zr |

---

---

**Geometry files:**

- CIF: pp. 1522
- POSCAR: pp. 1522

# $y$-$Y_2Si_2O_7$ Structure: A7B2C2_mP22_11_3e2f_2e_ab

http://aflow.org/prototype-encyclopedia/A7B2C2_mP22_11_3e2f_2e_ab

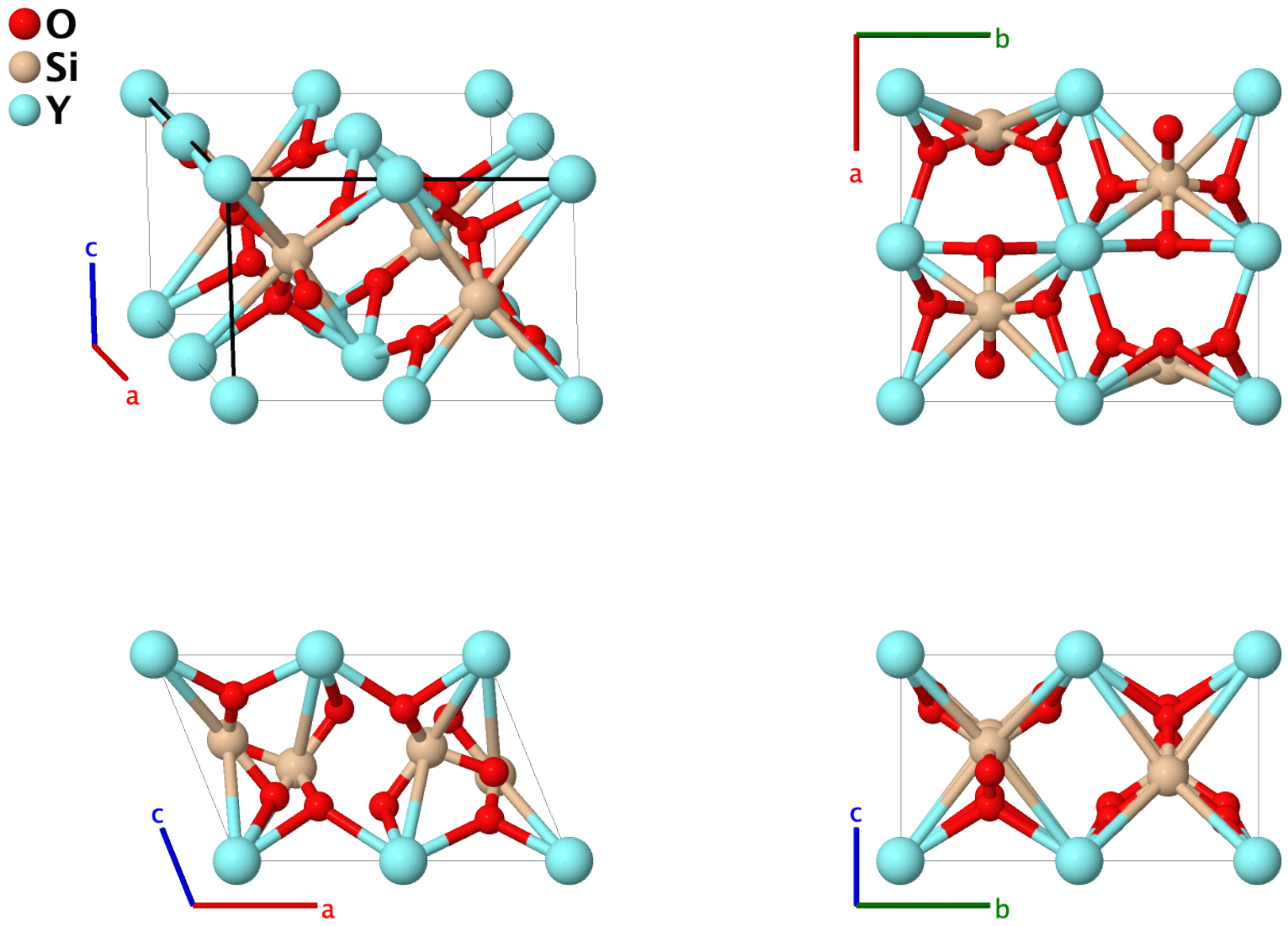

| | | |
|---|---|---|
| **Prototype** | : | $O_7Si_2Y_2$ |
| **AFLOW prototype label** | : | A7B2C2_mP22_11_3e2f_2e_ab |
| ***Strukturbericht* designation** | : | None |
| **Pearson symbol** | : | mP22 |
| **Space group number** | : | 11 |
| **Space group symbol** | : | $P2_1/m$ |
| **AFLOW prototype command** | : | `aflow --proto=A7B2C2_mP22_11_3e2f_2e_ab`<br>`--params=`$a, b/a, c/a, \beta, x_3, z_3, x_4, z_4, x_5, z_5, x_6, z_6, x_7, z_7, x_8, y_8, z_8, x_9, y_9, z_9$ |

- (Becerro, 2003) found that the NMR spectrum of $y$-$Y_2Si_2O_7$ showed only one peak for yttrium. If that is the case, then this crystal structure is incorrect.

**Simple Monoclinic primitive vectors:**

$$\mathbf{a}_1 = a\,\hat{\mathbf{x}}$$
$$\mathbf{a}_2 = b\,\hat{\mathbf{y}}$$
$$\mathbf{a}_3 = c\cos\beta\,\hat{\mathbf{x}} + c\sin\beta\,\hat{\mathbf{z}}$$

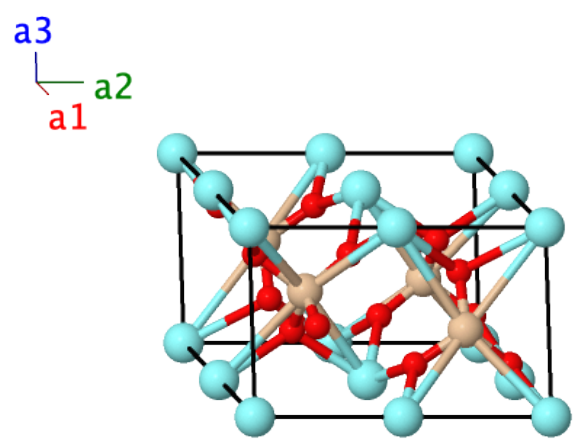

**Basis vectors:**

| | | Lattice Coordinates | | Cartesian Coordinates | Wyckoff Position | Atom Type |
|---|---|---|---|---|---|---|
| $\mathbf{B}_1$ | = | $0\,\mathbf{a}_1 + 0\,\mathbf{a}_2 + 0\,\mathbf{a}_3$ | = | $0\,\hat{\mathbf{x}} + 0\,\hat{\mathbf{y}} + 0\,\hat{\mathbf{z}}$ | (2$a$) | Y I |
| $\mathbf{B}_2$ | = | $\frac{1}{2}\,\mathbf{a}_2$ | = | $\frac{1}{2}b\,\hat{\mathbf{y}}$ | (2$a$) | Y I |
| $\mathbf{B}_3$ | = | $\frac{1}{2}\,\mathbf{a}_1$ | = | $\frac{1}{2}a\,\hat{\mathbf{x}}$ | (2$b$) | Y II |
| $\mathbf{B}_4$ | = | $\frac{1}{2}\,\mathbf{a}_1 + \frac{1}{2}\,\mathbf{a}_2$ | = | $\frac{1}{2}a\,\hat{\mathbf{x}} + \frac{1}{2}b\,\hat{\mathbf{y}}$ | (2b) | Y II |
| $\mathbf{B}_5$ | = | $x_3\,\mathbf{a}_1 + \frac{1}{4}\,\mathbf{a}_2 + z_3\,\mathbf{a}_3$ | = | $(x_3a + z_3c\cos\beta)\,\hat{\mathbf{x}} + \frac{1}{4}b\,\hat{\mathbf{y}} + z_3c\sin\beta\,\hat{\mathbf{z}}$ | (2e) | O I |
| $\mathbf{B}_6$ | = | $-x_3\,\mathbf{a}_1 + \frac{3}{4}\,\mathbf{a}_2 - z_3\,\mathbf{a}_3$ | = | $(-x_3a - z_3c\cos\beta)\,\hat{\mathbf{x}} + \frac{3}{4}b\,\hat{\mathbf{y}} - z_3c\sin\beta\,\hat{\mathbf{z}}$ | (2e) | O I |
| $\mathbf{B}_7$ | = | $x_4\,\mathbf{a}_1 + \frac{1}{4}\,\mathbf{a}_2 + z_4\,\mathbf{a}_3$ | = | $(x_4a + z_4c\cos\beta)\,\hat{\mathbf{x}} + \frac{1}{4}b\,\hat{\mathbf{y}} + z_4c\sin\beta\,\hat{\mathbf{z}}$ | (2e) | O II |
| $\mathbf{B}_8$ | = | $-x_4\,\mathbf{a}_1 + \frac{3}{4}\,\mathbf{a}_2 - z_4\,\mathbf{a}_3$ | = | $(-x_4a - z_4c\cos\beta)\,\hat{\mathbf{x}} + \frac{3}{4}b\,\hat{\mathbf{y}} - z_4c\sin\beta\,\hat{\mathbf{z}}$ | (2e) | O II |
| $\mathbf{B}_9$ | = | $x_5\,\mathbf{a}_1 + \frac{1}{4}\,\mathbf{a}_2 + z_5\,\mathbf{a}_3$ | = | $(x_5a + z_5c\cos\beta)\,\hat{\mathbf{x}} + \frac{1}{4}b\,\hat{\mathbf{y}} + z_5c\sin\beta\,\hat{\mathbf{z}}$ | (2e) | O III |
| $\mathbf{B}_{10}$ | = | $-x_5\,\mathbf{a}_1 + \frac{3}{4}\,\mathbf{a}_2 - z_5\,\mathbf{a}_3$ | = | $(-x_5a - z_5c\cos\beta)\,\hat{\mathbf{x}} + \frac{3}{4}b\,\hat{\mathbf{y}} - z_5c\sin\beta\,\hat{\mathbf{z}}$ | (2e) | O III |
| $\mathbf{B}_{11}$ | = | $x_6\,\mathbf{a}_1 + \frac{1}{4}\,\mathbf{a}_2 + z_6\,\mathbf{a}_3$ | = | $(x_6a + z_6c\cos\beta)\,\hat{\mathbf{x}} + \frac{1}{4}b\,\hat{\mathbf{y}} + z_6c\sin\beta\,\hat{\mathbf{z}}$ | (2e) | Si I |
| $\mathbf{B}_{12}$ | = | $-x_6\,\mathbf{a}_1 + \frac{3}{4}\,\mathbf{a}_2 - z_6\,\mathbf{a}_3$ | = | $(-x_6a - z_6c\cos\beta)\,\hat{\mathbf{x}} + \frac{3}{4}b\,\hat{\mathbf{y}} - z_6c\sin\beta\,\hat{\mathbf{z}}$ | (2e) | Si I |
| $\mathbf{B}_{13}$ | = | $x_7\,\mathbf{a}_1 + \frac{1}{4}\,\mathbf{a}_2 + z_7\,\mathbf{a}_3$ | = | $(x_7a + z_7c\cos\beta)\,\hat{\mathbf{x}} + \frac{1}{4}b\,\hat{\mathbf{y}} + z_7c\sin\beta\,\hat{\mathbf{z}}$ | (2e) | Si II |
| $\mathbf{B}_{14}$ | = | $-x_7\,\mathbf{a}_1 + \frac{3}{4}\,\mathbf{a}_2 - z_7\,\mathbf{a}_3$ | = | $(-x_7a - z_7c\cos\beta)\,\hat{\mathbf{x}} + \frac{3}{4}b\,\hat{\mathbf{y}} - z_7c\sin\beta\,\hat{\mathbf{z}}$ | (2e) | Si II |
| $\mathbf{B}_{15}$ | = | $x_8\,\mathbf{a}_1 + y_8\,\mathbf{a}_2 + z_8\,\mathbf{a}_3$ | = | $(x_8a + z_8c\cos\beta)\,\hat{\mathbf{x}} + y_8b\,\hat{\mathbf{y}} + z_8c\sin\beta\,\hat{\mathbf{z}}$ | (4f) | O IV |
| $\mathbf{B}_{16}$ | = | $-x_8\,\mathbf{a}_1 + \left(\frac{1}{2} + y_8\right)\,\mathbf{a}_2 - z_8\,\mathbf{a}_3$ | = | $(-x_8a - z_8c\cos\beta)\,\hat{\mathbf{x}} + \left(\frac{1}{2} + y_8\right)b\,\hat{\mathbf{y}} - z_8c\sin\beta\,\hat{\mathbf{z}}$ | (4f) | O IV |
| $\mathbf{B}_{17}$ | = | $-x_8\,\mathbf{a}_1 - y_8\,\mathbf{a}_2 - z_8\,\mathbf{a}_3$ | = | $(-x_8a - z_8c\cos\beta)\,\hat{\mathbf{x}} - y_8b\,\hat{\mathbf{y}} - z_8c\sin\beta\,\hat{\mathbf{z}}$ | (4f) | O IV |
| $\mathbf{B}_{18}$ | = | $x_8\,\mathbf{a}_1 + \left(\frac{1}{2} - y_8\right)\,\mathbf{a}_2 + z_8\,\mathbf{a}_3$ | = | $(x_8a + z_8c\cos\beta)\,\hat{\mathbf{x}} + \left(\frac{1}{2} - y_8\right)b\,\hat{\mathbf{y}} + z_8c\sin\beta\,\hat{\mathbf{z}}$ | (4f) | O IV |
| $\mathbf{B}_{19}$ | = | $x_9\,\mathbf{a}_1 + y_9\,\mathbf{a}_2 + z_9\,\mathbf{a}_3$ | = | $(x_9a + z_9c\cos\beta)\,\hat{\mathbf{x}} + y_9b\,\hat{\mathbf{y}} + z_9c\sin\beta\,\hat{\mathbf{z}}$ | (4f) | O V |
| $\mathbf{B}_{20}$ | = | $-x_9\,\mathbf{a}_1 + \left(\frac{1}{2} + y_9\right)\,\mathbf{a}_2 - z_9\,\mathbf{a}_3$ | = | $(-x_9a - z_9c\cos\beta)\,\hat{\mathbf{x}} + \left(\frac{1}{2} + y_9\right)b\,\hat{\mathbf{y}} - z_9c\sin\beta\,\hat{\mathbf{z}}$ | (4f) | O V |
| $\mathbf{B}_{21}$ | = | $-x_9\,\mathbf{a}_1 - y_9\,\mathbf{a}_2 - z_9\,\mathbf{a}_3$ | = | $(-x_9a - z_9c\cos\beta)\,\hat{\mathbf{x}} - y_9b\,\hat{\mathbf{y}} - z_9c\sin\beta\,\hat{\mathbf{z}}$ | (4f) | O V |
| $\mathbf{B}_{22}$ | = | $x_9\,\mathbf{a}_1 + \left(\frac{1}{2} - y_9\right)\,\mathbf{a}_2 + z_9\,\mathbf{a}_3$ | = | $(x_9a + z_9c\cos\beta)\,\hat{\mathbf{x}} + \left(\frac{1}{2} - y_9\right)b\,\hat{\mathbf{y}} + z_9c\sin\beta\,\hat{\mathbf{z}}$ | (4f) | O V |

**Found in:**

**Geometry files:**

- CIF: pp. 1522
- POSCAR: pp. 1523

# Barytocalcite ($BaCa(CO_3)_2$) Structure: AB2CD6_mP20_11_e_2e_e_2e2f

http://aflow.org/prototype-encyclopedia/AB2CD6_mP20_11_e_2e_e_2e2f

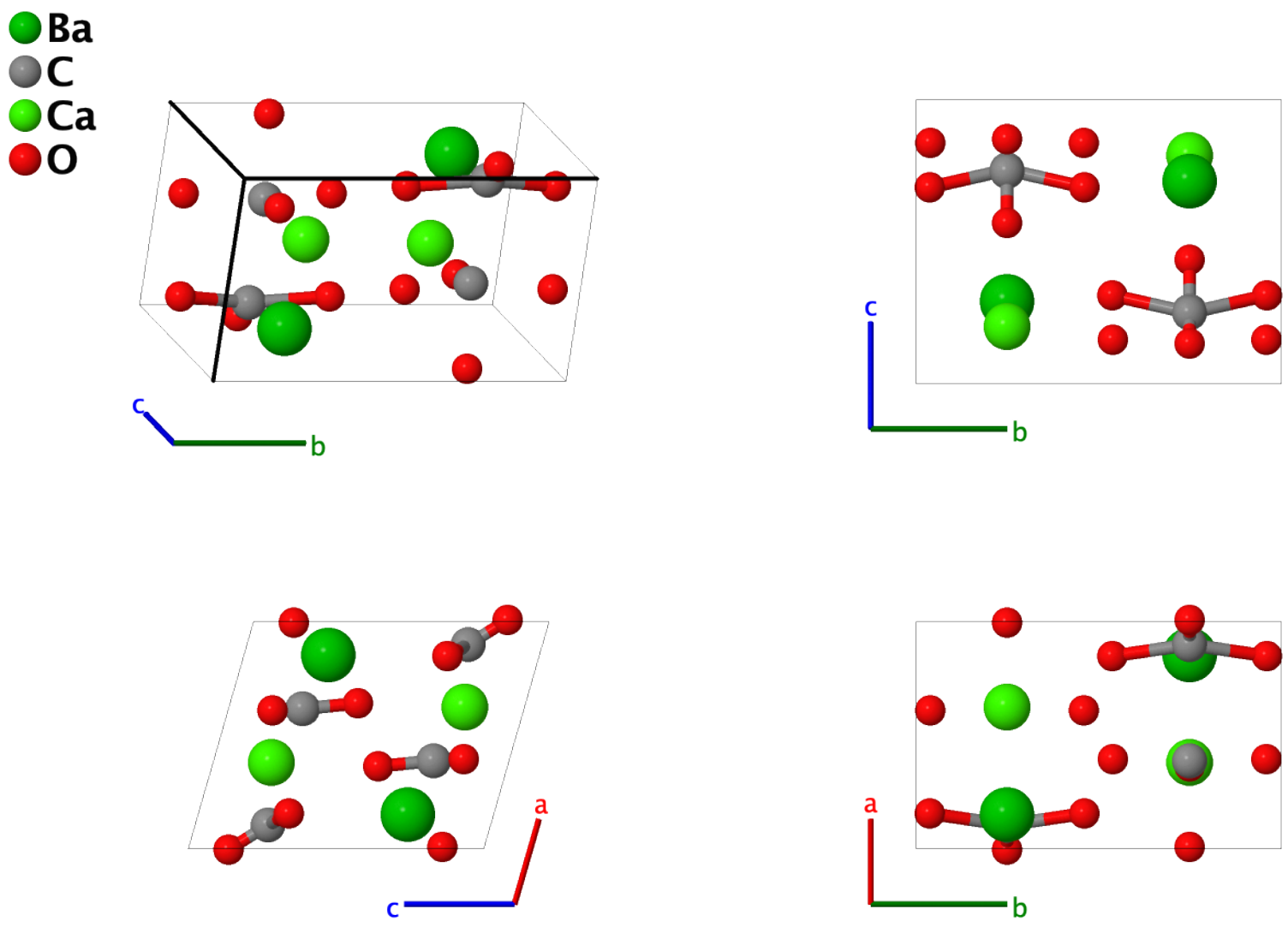

| | | |
|---|---|---|
| **Prototype** | : | $BaC_2CaO_6$ |
| **AFLOW prototype label** | : | AB2CD6_mP20_11_e_2e_e_2e2f |
| ***Strukturbericht* designation** | : | None |
| **Pearson symbol** | : | mP20 |
| **Space group number** | : | 11 |
| **Space group symbol** | : | $P2_1/m$ |
| **AFLOW prototype command** | : | `aflow --proto=AB2CD6_mP20_11_e_2e_e_2e2f`<br>`--params=`$a, b/a, c/a, \beta, x_1, z_1, x_2, z_2, x_3, z_3, x_4, z_4, x_5, z_5, x_6, z_6, x_7, y_7, z_7, x_8, y_8, z_8$ |

- $BaCa(CO_3)_2$ comes in a variety of crystal structures (Spahr, 2019):
  - monoclinic barytocalcite, space group $P2_1/m$ #11 (the current structure)
  - trigonal paralstonite, space group $P321$ #150,
  - Triclinic alstonite, space group $P1$ #1 or $P\overline{1}$ #2 (Sartori, 1975), and
  - a new monoclinic structure, space group $C2$ #5, synthesized by (Spahr, 2019), and lacking the centrosymmetric character of barytocalcite.

**Simple Monoclinic primitive vectors:**

$$\mathbf{a}_1 = a\,\hat{\mathbf{x}}$$
$$\mathbf{a}_2 = b\,\hat{\mathbf{y}}$$
$$\mathbf{a}_3 = c\cos\beta\,\hat{\mathbf{x}} + c\sin\beta\,\hat{\mathbf{z}}$$

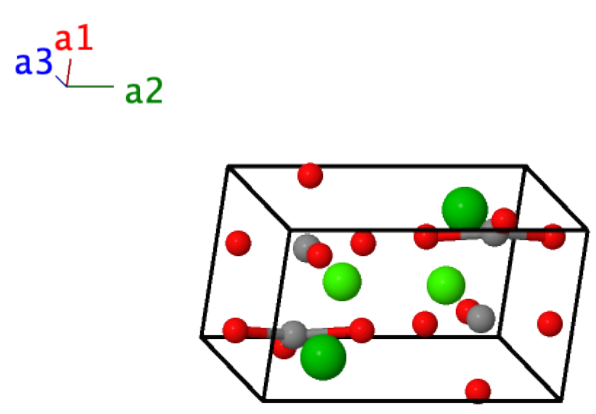

**Basis vectors:**

| | | Lattice Coordinates | | Cartesian Coordinates | Wyckoff Position | Atom Type |
|---|---|---|---|---|---|---|
| $\mathbf{B}_1$ | = | $x_1\,\mathbf{a}_1 + \frac{1}{4}\,\mathbf{a}_2 + z_1\,\mathbf{a}_3$ | = | $(x_1 a + z_1 c\cos\beta)\,\hat{\mathbf{x}} + \frac{1}{4} b\,\hat{\mathbf{y}} + z_1 c\sin\beta\,\hat{\mathbf{z}}$ | (2e) | Ba |
| $\mathbf{B}_2$ | = | $-x_1\,\mathbf{a}_1 + \frac{3}{4}\,\mathbf{a}_2 - z_1\,\mathbf{a}_3$ | = | $(-x_1 a - z_1 c\cos\beta)\,\hat{\mathbf{x}} + \frac{3}{4} b\,\hat{\mathbf{y}} - z_1 c\sin\beta\,\hat{\mathbf{z}}$ | (2e) | Ba |
| $\mathbf{B}_3$ | = | $x_2\,\mathbf{a}_1 + \frac{1}{4}\,\mathbf{a}_2 + z_2\,\mathbf{a}_3$ | = | $(x_2 a + z_2 c\cos\beta)\,\hat{\mathbf{x}} + \frac{1}{4} b\,\hat{\mathbf{y}} + z_2 c\sin\beta\,\hat{\mathbf{z}}$ | (2e) | C I |
| $\mathbf{B}_4$ | = | $-x_2\,\mathbf{a}_1 + \frac{3}{4}\,\mathbf{a}_2 - z_2\,\mathbf{a}_3$ | = | $(-x_2 a - z_2 c\cos\beta)\,\hat{\mathbf{x}} + \frac{3}{4} b\,\hat{\mathbf{y}} - z_2 c\sin\beta\,\hat{\mathbf{z}}$ | (2e) | C I |
| $\mathbf{B}_5$ | = | $x_3\,\mathbf{a}_1 + \frac{1}{4}\,\mathbf{a}_2 + z_3\,\mathbf{a}_3$ | = | $(x_3 a + z_3 c\cos\beta)\,\hat{\mathbf{x}} + \frac{1}{4} b\,\hat{\mathbf{y}} + z_3 c\sin\beta\,\hat{\mathbf{z}}$ | (2e) | C II |
| $\mathbf{B}_6$ | = | $-x_3\,\mathbf{a}_1 + \frac{3}{4}\,\mathbf{a}_2 - z_3\,\mathbf{a}_3$ | = | $(-x_3 a - z_3 c\cos\beta)\,\hat{\mathbf{x}} + \frac{3}{4} b\,\hat{\mathbf{y}} - z_3 c\sin\beta\,\hat{\mathbf{z}}$ | (2e) | C II |
| $\mathbf{B}_7$ | = | $x_4\,\mathbf{a}_1 + \frac{1}{4}\,\mathbf{a}_2 + z_4\,\mathbf{a}_3$ | = | $(x_4 a + z_4 c\cos\beta)\,\hat{\mathbf{x}} + \frac{1}{4} b\,\hat{\mathbf{y}} + z_4 c\sin\beta\,\hat{\mathbf{z}}$ | (2e) | Ca |
| $\mathbf{B}_8$ | = | $-x_4\,\mathbf{a}_1 + \frac{3}{4}\,\mathbf{a}_2 - z_4\,\mathbf{a}_3$ | = | $(-x_4 a - z_4 c\cos\beta)\,\hat{\mathbf{x}} + \frac{3}{4} b\,\hat{\mathbf{y}} - z_4 c\sin\beta\,\hat{\mathbf{z}}$ | (2e) | Ca |
| $\mathbf{B}_9$ | = | $x_5\,\mathbf{a}_1 + \frac{1}{4}\,\mathbf{a}_2 + z_5\,\mathbf{a}_3$ | = | $(x_5 a + z_5 c\cos\beta)\,\hat{\mathbf{x}} + \frac{1}{4} b\,\hat{\mathbf{y}} + z_5 c\sin\beta\,\hat{\mathbf{z}}$ | (2e) | O I |
| $\mathbf{B}_{10}$ | = | $-x_5\,\mathbf{a}_1 + \frac{3}{4}\,\mathbf{a}_2 - z_5\,\mathbf{a}_3$ | = | $(-x_5 a - z_5 c\cos\beta)\,\hat{\mathbf{x}} + \frac{3}{4} b\,\hat{\mathbf{y}} - z_5 c\sin\beta\,\hat{\mathbf{z}}$ | (2e) | O I |
| $\mathbf{B}_{11}$ | = | $x_6\,\mathbf{a}_1 + \frac{1}{4}\,\mathbf{a}_2 + z_6\,\mathbf{a}_3$ | = | $(x_6 a + z_6 c\cos\beta)\,\hat{\mathbf{x}} + \frac{1}{4} b\,\hat{\mathbf{y}} + z_6 c\sin\beta\,\hat{\mathbf{z}}$ | (2e) | O II |
| $\mathbf{B}_{12}$ | = | $-x_6\,\mathbf{a}_1 + \frac{3}{4}\,\mathbf{a}_2 - z_6\,\mathbf{a}_3$ | = | $(-x_6 a - z_6 c\cos\beta)\,\hat{\mathbf{x}} + \frac{3}{4} b\,\hat{\mathbf{y}} - z_6 c\sin\beta\,\hat{\mathbf{z}}$ | (2e) | O II |
| $\mathbf{B}_{13}$ | = | $x_7\,\mathbf{a}_1 + y_7\,\mathbf{a}_2 + z_7\,\mathbf{a}_3$ | = | $(x_7 a + z_7 c\cos\beta)\,\hat{\mathbf{x}} + y_7 b\,\hat{\mathbf{y}} + z_7 c\sin\beta\,\hat{\mathbf{z}}$ | (4f) | O III |
| $\mathbf{B}_{14}$ | = | $-x_7\,\mathbf{a}_1 + \left(\frac{1}{2} + y_7\right)\,\mathbf{a}_2 - z_7\,\mathbf{a}_3$ | = | $(-x_7 a - z_7 c\cos\beta)\,\hat{\mathbf{x}} + \left(\frac{1}{2} + y_7\right) b\,\hat{\mathbf{y}} - z_7 c\sin\beta\,\hat{\mathbf{z}}$ | (4f) | O III |
| $\mathbf{B}_{15}$ | = | $-x_7\,\mathbf{a}_1 - y_7\,\mathbf{a}_2 - z_7\,\mathbf{a}_3$ | = | $(-x_7 a - z_7 c\cos\beta)\,\hat{\mathbf{x}} - y_7 b\,\hat{\mathbf{y}} - z_7 c\sin\beta\,\hat{\mathbf{z}}$ | (4f) | O III |
| $\mathbf{B}_{16}$ | = | $x_7\,\mathbf{a}_1 + \left(\frac{1}{2} - y_7\right)\,\mathbf{a}_2 + z_7\,\mathbf{a}_3$ | = | $(x_7 a + z_7 c\cos\beta)\,\hat{\mathbf{x}} + \left(\frac{1}{2} - y_7\right) b\,\hat{\mathbf{y}} + z_7 c\sin\beta\,\hat{\mathbf{z}}$ | (4f) | O III |
| $\mathbf{B}_{17}$ | = | $x_8\,\mathbf{a}_1 + y_8\,\mathbf{a}_2 + z_8\,\mathbf{a}_3$ | = | $(x_8 a + z_8 c\cos\beta)\,\hat{\mathbf{x}} + y_8 b\,\hat{\mathbf{y}} + z_8 c\sin\beta\,\hat{\mathbf{z}}$ | (4f) | O IV |
| $\mathbf{B}_{18}$ | = | $-x_8\,\mathbf{a}_1 + \left(\frac{1}{2} + y_8\right)\,\mathbf{a}_2 - z_8\,\mathbf{a}_3$ | = | $(-x_8 a - z_8 c\cos\beta)\,\hat{\mathbf{x}} + \left(\frac{1}{2} + y_8\right) b\,\hat{\mathbf{y}} - z_8 c\sin\beta\,\hat{\mathbf{z}}$ | (4f) | O IV |
| $\mathbf{B}_{19}$ | = | $-x_8\,\mathbf{a}_1 - y_8\,\mathbf{a}_2 - z_8\,\mathbf{a}_3$ | = | $(-x_8 a - z_8 c\cos\beta)\,\hat{\mathbf{x}} - y_8 b\,\hat{\mathbf{y}} - z_8 c\sin\beta\,\hat{\mathbf{z}}$ | (4f) | O IV |
| $\mathbf{B}_{20}$ | = | $x_8\,\mathbf{a}_1 + \left(\frac{1}{2} - y_8\right)\,\mathbf{a}_2 + z_8\,\mathbf{a}_3$ | = | $(x_8 a + z_8 c\cos\beta)\,\hat{\mathbf{x}} + \left(\frac{1}{2} - y_8\right) b\,\hat{\mathbf{y}} + z_8 c\sin\beta\,\hat{\mathbf{z}}$ | (4f) | O IV |

**Geometry files:**

- CIF: pp. 1523
- POSCAR: pp. 1523

# O(OH)Y Structure: ABC_mP6_11_e_e_e

http://aflow.org/prototype-encyclopedia/ABC_mP6_11_e_e_e

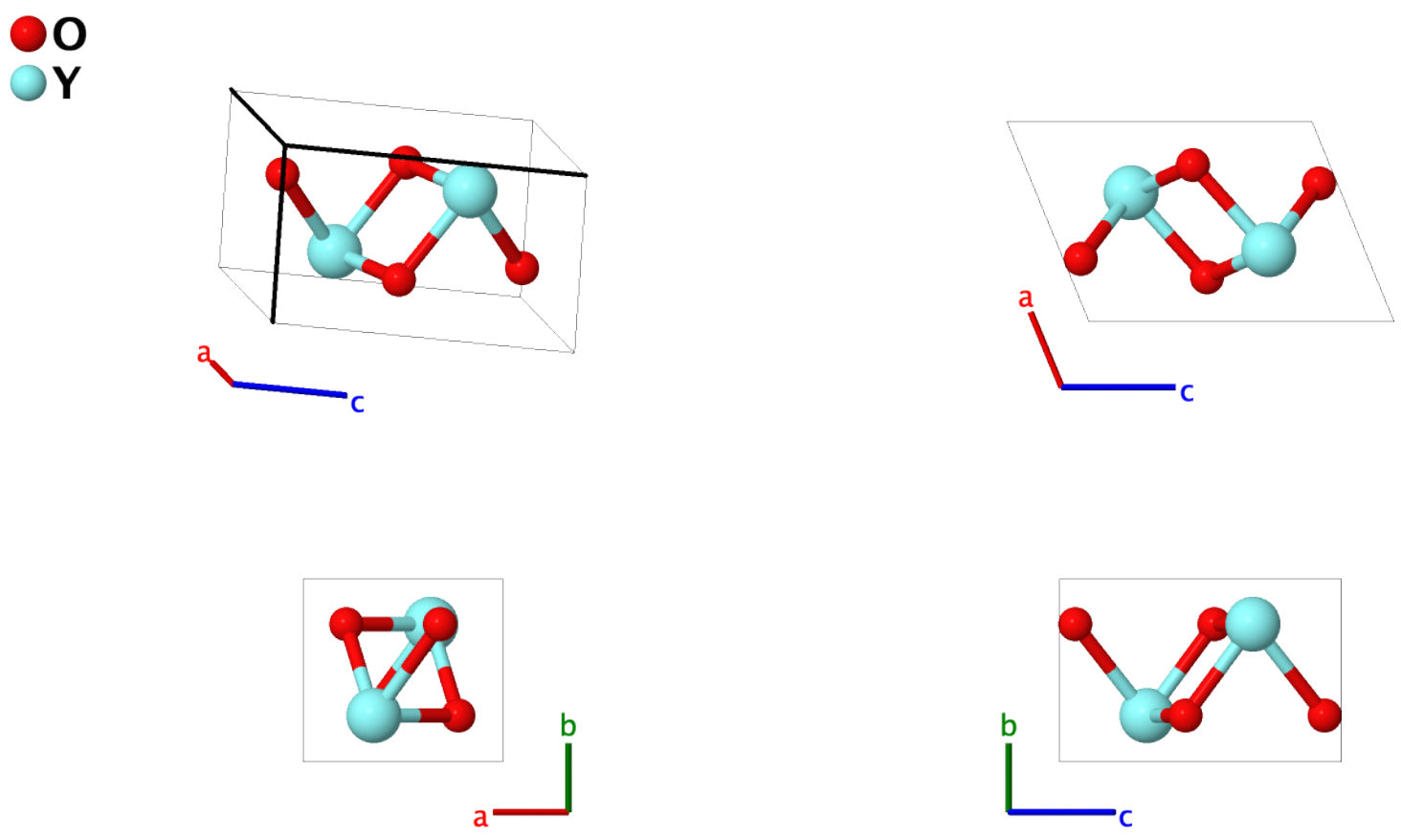

| | | |
|---|---|---|
| **Prototype** | : | $HO_2Y$ |
| **AFLOW prototype label** | : | ABC_mP6_11_e_e_e |
| ***Strukturbericht* designation** | : | None |
| **Pearson symbol** | : | mP6 |
| **Space group number** | : | 11 |
| **Space group symbol** | : | $P2_1/m$ |
| **AFLOW prototype command** | : | `aflow --proto=ABC_mP6_11_e_e_e`<br>`--params=`$a, b/a, c/a, \beta, x_1, z_1, x_2, z_2, x_3, z_3$ |

### Other compounds with this structure

- HoO(OH), ErO(OH), and YbO(OH)

---

- For this prototype, the OH molecules are centered on the (2*e*) Wyckoff position.

### Simple Monoclinic primitive vectors:

$$\mathbf{a}_1 = a\,\mathbf{\hat{x}}$$
$$\mathbf{a}_2 = b\,\mathbf{\hat{y}}$$
$$\mathbf{a}_3 = c\cos\beta\,\mathbf{\hat{x}} + c\sin\beta\,\mathbf{\hat{z}}$$

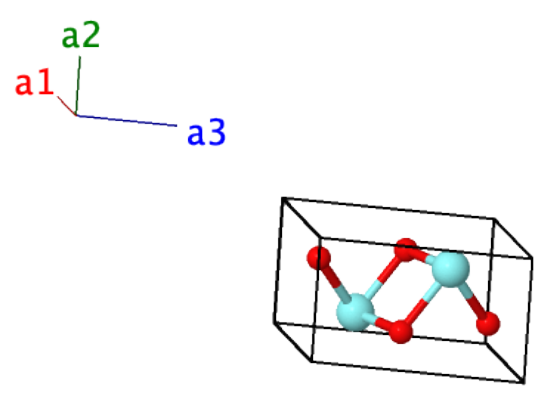

### Basis vectors:

| | | Lattice Coordinates | | Cartesian Coordinates | Wyckoff Position | Atom Type |
|---|---|---|---|---|---|---|
| $\mathbf{B}_1$ | = | $x_1\,\mathbf{a}_1 + \frac{1}{4}\,\mathbf{a}_2 + z_1\,\mathbf{a}_3$ | = | $(x_1 a + z_1 c\cos\beta)\,\mathbf{\hat{x}} + \frac{1}{4}b\,\mathbf{\hat{y}} + z_1 c\sin\beta\,\mathbf{\hat{z}}$ | (2*e*) | O |
| $\mathbf{B}_2$ | = | $-x_1\,\mathbf{a}_1 + \frac{3}{4}\,\mathbf{a}_2 - z_1\,\mathbf{a}_3$ | = | $(-x_1 a - z_1 c\cos\beta)\,\mathbf{\hat{x}} + \frac{3}{4}b\,\mathbf{\hat{y}} - z_1 c\sin\beta\,\mathbf{\hat{z}}$ | (2*e*) | O |
| $\mathbf{B}_3$ | = | $x_2\,\mathbf{a}_1 + \frac{1}{4}\,\mathbf{a}_2 + z_2\,\mathbf{a}_3$ | = | $(x_2 a + z_2 c\cos\beta)\,\mathbf{\hat{x}} + \frac{1}{4}b\,\mathbf{\hat{y}} + z_2 c\sin\beta\,\mathbf{\hat{z}}$ | (2*e*) | OH |
| $\mathbf{B}_4$ | = | $-x_2\,\mathbf{a}_1 + \frac{3}{4}\,\mathbf{a}_2 - z_2\,\mathbf{a}_3$ | = | $(-x_2 a - z_2 c\cos\beta)\,\mathbf{\hat{x}} + \frac{3}{4}b\,\mathbf{\hat{y}} - z_2 c\sin\beta\,\mathbf{\hat{z}}$ | (2*e*) | OH |
| $\mathbf{B}_5$ | = | $x_3\,\mathbf{a}_1 + \frac{1}{4}\,\mathbf{a}_2 + z_3\,\mathbf{a}_3$ | = | $(x_3a + z_3c\cos\beta)\,\hat{\mathbf{x}} + \frac{1}{4}b\,\hat{\mathbf{y}} + z_3c\sin\beta\,\hat{\mathbf{z}}$ | (2e) | Y |
| $\mathbf{B}_6$ | = | $-x_3\,\mathbf{a}_1 + \frac{3}{4}\,\mathbf{a}_2 - z_3\,\mathbf{a}_3$ | = | $(-x_3a - z_3c\cos\beta)\,\hat{\mathbf{x}} + \frac{3}{4}b\,\hat{\mathbf{y}} - z_3c\sin\beta\,\hat{\mathbf{z}}$ | (2e) | Y |

**Geometry files:**

- CIF: pp. 1524
- POSCAR: pp. 1524

# $Al_{13}Fe_4$ Structure: A13B4_mC102_12_dg8i5j_4ij

http://aflow.org/prototype-encyclopedia/A13B4_mC102_12_dg8i5j_4ij

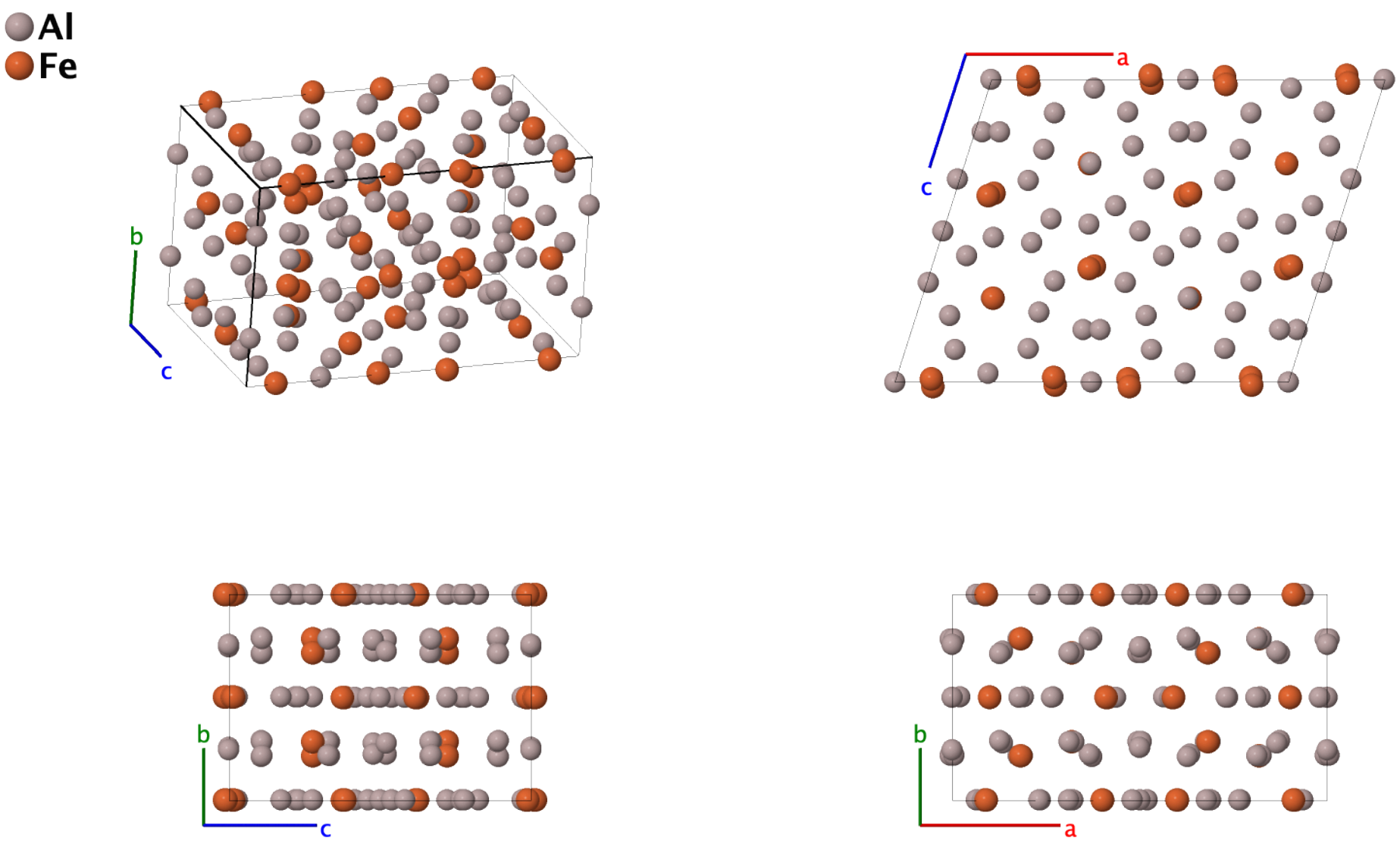

| | | |
|---|---|---|
| **Prototype** | : | $Al_{13}Fe_4$ |
| **AFLOW prototype label** | : | A13B4_mC102_12_dg8i5j_4ij |
| ***Strukturbericht* designation** | : | None |
| **Pearson symbol** | : | mC102 |
| **Space group number** | : | 12 |
| **Space group symbol** | : | $C2/m$ |
| **AFLOW prototype command** | : | `aflow --proto=A13B4_mC102_12_dg8i5j_4ij` `--params=`$a, b/a, c/a, \beta, y_2, x_3, z_3, x_4, z_4, x_5, z_5, x_6, z_6, x_7, z_7, x_8, z_8, x_9, z_9, x_{10}, z_{10}, x_{11}, z_{11}, x_{12}, z_{12}, x_{13}, z_{13}, x_{14}, z_{14}, x_{15}, y_{15}, z_{15}, x_{16}, y_{16}, z_{16}, x_{17}, y_{17}, z_{17}, x_{18}, y_{18}, z_{18}, x_{19}, y_{19}, z_{19}, x_{20}, y_{20}, z_{20}$ |

- The Al-IV site is only occupied 70% of the time. This makes the stoichiometry $Al_{12.8}Fe_4$, which (Black I and II, 1955) rounds to $Al_{13}Fe_4$.

**Base-centered Monoclinic primitive vectors:**

$$\mathbf{a}_1 = \frac{1}{2}\,a\,\hat{\mathbf{x}} - \frac{1}{2}\,b\,\hat{\mathbf{y}}$$
$$\mathbf{a}_2 = \frac{1}{2}\,a\,\hat{\mathbf{x}} + \frac{1}{2}\,b\,\hat{\mathbf{y}}$$
$$\mathbf{a}_3 = c\cos\beta\,\hat{\mathbf{x}} + c\sin\beta\,\hat{\mathbf{z}}$$

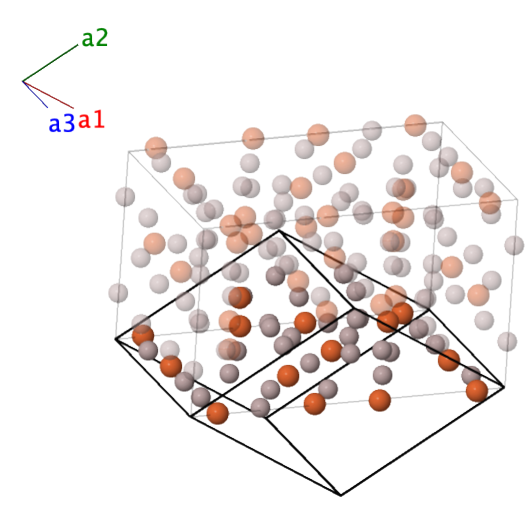

**Basis vectors:**

| | | Lattice Coordinates | | Cartesian Coordinates | Wyckoff Position | Atom Type |
|---|---|---|---|---|---|---|
| $\mathbf{B}_1$ | = | $\frac{1}{2}\,\mathbf{a}_1 + \frac{1}{2}\,\mathbf{a}_2 + \frac{1}{2}\,\mathbf{a}_3$ | = | $\frac{1}{2}\,(a + c\cos\beta)\,\hat{\mathbf{x}} + \frac{1}{2}c\sin\beta\,\hat{\mathbf{z}}$ | (2$d$) | Al I |
| $\mathbf{B}_2$ | = | $-y_2\,\mathbf{a}_1 + y_2\,\mathbf{a}_2$ | = | $y_2 b\,\hat{\mathbf{y}}$ | (4$g$) | Al II |
| $\mathbf{B}_3$ | = | $y_2\,\mathbf{a}_1 - y_2\,\mathbf{a}_2$ | = | $-y_2 b\,\hat{\mathbf{y}}$ | (4$g$) | Al II |
| $\mathbf{B}_4$ | = | $x_3\,\mathbf{a}_1 + x_3\,\mathbf{a}_2 + z_3\,\mathbf{a}_3$ | = | $(x_3 a + z_3 c\cos\beta)\,\hat{\mathbf{x}} + z_3 c\sin\beta\,\hat{\mathbf{z}}$ | (4$i$) | Al III |
| $\mathbf{B}_5$ | = | $-x_3\,\mathbf{a}_1 - x_3\,\mathbf{a}_2 - z_3\,\mathbf{a}_3$ | = | $(-x_3 a - z_3 c\cos\beta)\,\hat{\mathbf{x}} - z_3 c\sin\beta\,\hat{\mathbf{z}}$ | (4$i$) | Al III |
| $\mathbf{B}_6$ | = | $x_4\,\mathbf{a}_1 + x_4\,\mathbf{a}_2 + z_4\,\mathbf{a}_3$ | = | $(x_4 a + z_4 c\cos\beta)\,\hat{\mathbf{x}} + z_4 c\sin\beta\,\hat{\mathbf{z}}$ | (4$i$) | Al IV |
| $\mathbf{B}_7$ | = | $-x_4\,\mathbf{a}_1 - x_4\,\mathbf{a}_2 - z_4\,\mathbf{a}_3$ | = | $(-x_4 a - z_4 c\cos\beta)\,\hat{\mathbf{x}} - z_4 c\sin\beta\,\hat{\mathbf{z}}$ | (4$i$) | Al IV |
| $\mathbf{B}_8$ | = | $x_5\,\mathbf{a}_1 + x_5\,\mathbf{a}_2 + z_5\,\mathbf{a}_3$ | = | $(x_5 a + z_5 c\cos\beta)\,\hat{\mathbf{x}} + z_5 c\sin\beta\,\hat{\mathbf{z}}$ | (4$i$) | Al V |
| $\mathbf{B}_9$ | = | $-x_5\,\mathbf{a}_1 - x_5\,\mathbf{a}_2 - z_5\,\mathbf{a}_3$ | = | $(-x_5 a - z_5 c\cos\beta)\,\hat{\mathbf{x}} - z_5 c\sin\beta\,\hat{\mathbf{z}}$ | (4$i$) | Al V |
| $\mathbf{B}_{10}$ | = | $x_6\,\mathbf{a}_1 + x_6\,\mathbf{a}_2 + z_6\,\mathbf{a}_3$ | = | $(x_6 a + z_6 c\cos\beta)\,\hat{\mathbf{x}} + z_6 c\sin\beta\,\hat{\mathbf{z}}$ | (4$i$) | Al VI |
| $\mathbf{B}_{11}$ | = | $-x_6\,\mathbf{a}_1 - x_6\,\mathbf{a}_2 - z_6\,\mathbf{a}_3$ | = | $(-x_6 a - z_6 c\cos\beta)\,\hat{\mathbf{x}} - z_6 c\sin\beta\,\hat{\mathbf{z}}$ | (4$i$) | Al VI |
| $\mathbf{B}_{12}$ | = | $x_7\,\mathbf{a}_1 + x_7\,\mathbf{a}_2 + z_7\,\mathbf{a}_3$ | = | $(x_7 a + z_7 c\cos\beta)\,\hat{\mathbf{x}} + z_7 c\sin\beta\,\hat{\mathbf{z}}$ | (4$i$) | Al VII |
| $\mathbf{B}_{13}$ | = | $-x_7\,\mathbf{a}_1 - x_7\,\mathbf{a}_2 - z_7\,\mathbf{a}_3$ | = | $(-x_7 a - z_7 c\cos\beta)\,\hat{\mathbf{x}} - z_7 c\sin\beta\,\hat{\mathbf{z}}$ | (4$i$) | Al VII |
| $\mathbf{B}_{14}$ | = | $x_8\,\mathbf{a}_1 + x_8\,\mathbf{a}_2 + z_8\,\mathbf{a}_3$ | = | $(x_8 a + z_8 c\cos\beta)\,\hat{\mathbf{x}} + z_8 c\sin\beta\,\hat{\mathbf{z}}$ | (4$i$) | Al VIII |
| $\mathbf{B}_{15}$ | = | $-x_8\,\mathbf{a}_1 - x_8\,\mathbf{a}_2 - z_8\,\mathbf{a}_3$ | = | $(-x_8 a - z_8 c\cos\beta)\,\hat{\mathbf{x}} - z_8 c\sin\beta\,\hat{\mathbf{z}}$ | (4$i$) | Al VIII |
| $\mathbf{B}_{16}$ | = | $x_9\,\mathbf{a}_1 + x_9\,\mathbf{a}_2 + z_9\,\mathbf{a}_3$ | = | $(x_9 a + z_9 c\cos\beta)\,\hat{\mathbf{x}} + z_9 c\sin\beta\,\hat{\mathbf{z}}$ | (4$i$) | Al IX |
| $\mathbf{B}_{17}$ | = | $-x_9\,\mathbf{a}_1 - x_9\,\mathbf{a}_2 - z_9\,\mathbf{a}_3$ | = | $(-x_9 a - z_9 c\cos\beta)\,\hat{\mathbf{x}} - z_9 c\sin\beta\,\hat{\mathbf{z}}$ | (4$i$) | Al IX |
| $\mathbf{B}_{18}$ | = | $x_{10}\,\mathbf{a}_1 + x_{10}\,\mathbf{a}_2 + z_{10}\,\mathbf{a}_3$ | = | $(x_{10} a + z_{10} c\cos\beta)\,\hat{\mathbf{x}} + z_{10} c\sin\beta\,\hat{\mathbf{z}}$ | (4$i$) | Al X |
| $\mathbf{B}_{19}$ | = | $-x_{10}\,\mathbf{a}_1 - x_{10}\,\mathbf{a}_2 - z_{10}\,\mathbf{a}_3$ | = | $(-x_{10} a - z_{10} c\cos\beta)\,\hat{\mathbf{x}} - z_{10} c\sin\beta\,\hat{\mathbf{z}}$ | (4$i$) | Al X |
| $\mathbf{B}_{20}$ | = | $x_{11}\,\mathbf{a}_1 + x_{11}\,\mathbf{a}_2 + z_{11}\,\mathbf{a}_3$ | = | $(x_{11} a + z_{11} c\cos\beta)\,\hat{\mathbf{x}} + z_{11} c\sin\beta\,\hat{\mathbf{z}}$ | (4$i$) | Fe I |
| $\mathbf{B}_{21}$ | = | $-x_{11}\,\mathbf{a}_1 - x_{11}\,\mathbf{a}_2 - z_{11}\,\mathbf{a}_3$ | = | $(-x_{11} a - z_{11} c\cos\beta)\,\hat{\mathbf{x}} - z_{11} c\sin\beta\,\hat{\mathbf{z}}$ | (4$i$) | Fe I |
| $\mathbf{B}_{22}$ | = | $x_{12}\,\mathbf{a}_1 + x_{12}\,\mathbf{a}_2 + z_{12}\,\mathbf{a}_3$ | = | $(x_{12} a + z_{12} c\cos\beta)\,\hat{\mathbf{x}} + z_{12} c\sin\beta\,\hat{\mathbf{z}}$ | (4$i$) | Fe II |
| $\mathbf{B}_{23}$ | = | $-x_{12}\,\mathbf{a}_1 - x_{12}\,\mathbf{a}_2 - z_{12}\,\mathbf{a}_3$ | = | $(-x_{12} a - z_{12} c\cos\beta)\,\hat{\mathbf{x}} - z_{12} c\sin\beta\,\hat{\mathbf{z}}$ | (4$i$) | Fe II |
| $\mathbf{B}_{24}$ | = | $x_{13}\,\mathbf{a}_1 + x_{13}\,\mathbf{a}_2 + z_{13}\,\mathbf{a}_3$ | = | $(x_{13} a + z_{13} c\cos\beta)\,\hat{\mathbf{x}} + z_{13} c\sin\beta\,\hat{\mathbf{z}}$ | (4$i$) | Fe III |
| $\mathbf{B}_{25}$ | = | $-x_{13}\,\mathbf{a}_1 - x_{13}\,\mathbf{a}_2 - z_{13}\,\mathbf{a}_3$ | = | $(-x_{13} a - z_{13} c\cos\beta)\,\hat{\mathbf{x}} - z_{13} c\sin\beta\,\hat{\mathbf{z}}$ | (4$i$) | Fe III |
| $\mathbf{B}_{26}$ | = | $x_{14}\,\mathbf{a}_1 + x_{14}\,\mathbf{a}_2 + z_{14}\,\mathbf{a}_3$ | = | $(x_{14} a + z_{14} c\cos\beta)\,\hat{\mathbf{x}} + z_{14} c\sin\beta\,\hat{\mathbf{z}}$ | (4$i$) | Fe IV |
| $\mathbf{B}_{27}$ | = | $-x_{14}\,\mathbf{a}_1 - x_{14}\,\mathbf{a}_2 - z_{14}\,\mathbf{a}_3$ | = | $(-x_{14} a - z_{14} c\cos\beta)\,\hat{\mathbf{x}} - z_{14} c\sin\beta\,\hat{\mathbf{z}}$ | (4$i$) | Fe IV |
| $\mathbf{B}_{28}$ | = | $(x_{15} - y_{15})\,\mathbf{a}_1 + (x_{15} + y_{15})\,\mathbf{a}_2 + z_{15}\,\mathbf{a}_3$ | = | $(x_{15} a + z_{15} c\cos\beta)\,\hat{\mathbf{x}} + y_{15} b\,\hat{\mathbf{y}} + z_{15} c\sin\beta\,\hat{\mathbf{z}}$ | (8$j$) | Al XI |
| $\mathbf{B}_{29}$ | = | $(-x_{15} - y_{15})\,\mathbf{a}_1 + (-x_{15} + y_{15})\,\mathbf{a}_2 - z_{15}\,\mathbf{a}_3$ | = | $(-x_{15} a - z_{15} c\cos\beta)\,\hat{\mathbf{x}} + y_{15} b\,\hat{\mathbf{y}} - z_{15} c\sin\beta\,\hat{\mathbf{z}}$ | (8$j$) | Al XI |
| $\mathbf{B}_{30}$ | = | $(-x_{15} + y_{15})\,\mathbf{a}_1 + (-x_{15} - y_{15})\,\mathbf{a}_2 - z_{15}\,\mathbf{a}_3$ | = | $(-x_{15} a - z_{15} c\cos\beta)\,\hat{\mathbf{x}} - y_{15} b\,\hat{\mathbf{y}} - z_{15} c\sin\beta\,\hat{\mathbf{z}}$ | (8$j$) | Al XI |
| $\mathbf{B}_{31}$ | = | $(x_{15} + y_{15})\,\mathbf{a}_1 + (x_{15} - y_{15})\,\mathbf{a}_2 + z_{15}\,\mathbf{a}_3$ | = | $(x_{15} a + z_{15} c\cos\beta)\,\hat{\mathbf{x}} - y_{15} b\,\hat{\mathbf{y}} + z_{15} c\sin\beta\,\hat{\mathbf{z}}$ | (8$j$) | Al XI |
| $\mathbf{B}_{32}$ | = | $(x_{16} - y_{16})\,\mathbf{a}_1 + (x_{16} + y_{16})\,\mathbf{a}_2 + z_{16}\,\mathbf{a}_3$ | = | $(x_{16} a + z_{16} c\cos\beta)\,\hat{\mathbf{x}} + y_{16} b\,\hat{\mathbf{y}} + z_{16} c\sin\beta\,\hat{\mathbf{z}}$ | (8$j$) | Al XII |
| $\mathbf{B}_{33}$ | = | $(-x_{16} - y_{16})\,\mathbf{a}_1 + (-x_{16} + y_{16})\,\mathbf{a}_2 - z_{16}\,\mathbf{a}_3$ | = | $(-x_{16} a - z_{16} c\cos\beta)\,\hat{\mathbf{x}} + y_{16} b\,\hat{\mathbf{y}} - z_{16} c\sin\beta\,\hat{\mathbf{z}}$ | (8$j$) | Al XII |
| $\mathbf{B}_{34}$ | = | $(-x_{16} + y_{16})\,\mathbf{a}_1 + (-x_{16} - y_{16})\,\mathbf{a}_2 - z_{16}\,\mathbf{a}_3$ | = | $(-x_{16} a - z_{16} c\cos\beta)\,\hat{\mathbf{x}} - y_{16} b\,\hat{\mathbf{y}} - z_{16} c\sin\beta\,\hat{\mathbf{z}}$ | (8$j$) | Al XII |
| $\mathbf{B}_{35}$ | = | $(x_{16}+y_{16})\,\mathbf{a}_1+(x_{16}-y_{16})\,\mathbf{a}_2+z_{16}\,\mathbf{a}_3$ | = | $(x_{16}a+z_{16}c\cos\beta)\,\hat{\mathbf{x}}-y_{16}b\,\hat{\mathbf{y}}+z_{16}c\sin\beta\,\hat{\mathbf{z}}$ | (8$j$) | Al XII |
| $\mathbf{B}_{36}$ | = | $(x_{17}-y_{17})\,\mathbf{a}_1+(x_{17}+y_{17})\,\mathbf{a}_2+z_{17}\,\mathbf{a}_3$ | = | $(x_{17}a+z_{17}c\cos\beta)\,\hat{\mathbf{x}}+y_{17}b\,\hat{\mathbf{y}}+z_{17}c\sin\beta\,\hat{\mathbf{z}}$ | (8$j$) | Al XIII |
| $\mathbf{B}_{37}$ | = | $(-x_{17}-y_{17})\,\mathbf{a}_1+(-x_{17}+y_{17})\,\mathbf{a}_2-z_{17}\,\mathbf{a}_3$ | = | $(-x_{17}a-z_{17}c\cos\beta)\,\hat{\mathbf{x}}+y_{17}b\,\hat{\mathbf{y}}-z_{17}c\sin\beta\,\hat{\mathbf{z}}$ | (8$j$) | Al XIII |
| $\mathbf{B}_{38}$ | = | $(-x_{17}+y_{17})\,\mathbf{a}_1+(-x_{17}-y_{17})\,\mathbf{a}_2-z_{17}\,\mathbf{a}_3$ | = | $(-x_{17}a-z_{17}c\cos\beta)\,\hat{\mathbf{x}}-y_{17}b\,\hat{\mathbf{y}}-z_{17}c\sin\beta\,\hat{\mathbf{z}}$ | (8$j$) | Al XIII |
| $\mathbf{B}_{39}$ | = | $(x_{17}+y_{17})\,\mathbf{a}_1+(x_{17}-y_{17})\,\mathbf{a}_2+z_{17}\,\mathbf{a}_3$ | = | $(x_{17}a+z_{17}c\cos\beta)\,\hat{\mathbf{x}}-y_{17}b\,\hat{\mathbf{y}}+z_{17}c\sin\beta\,\hat{\mathbf{z}}$ | (8$j$) | Al XIII |
| $\mathbf{B}_{40}$ | = | $(x_{18}-y_{18})\,\mathbf{a}_1+(x_{18}+y_{18})\,\mathbf{a}_2+z_{18}\,\mathbf{a}_3$ | = | $(x_{18}a+z_{18}c\cos\beta)\,\hat{\mathbf{x}}+y_{18}b\,\hat{\mathbf{y}}+z_{18}c\sin\beta\,\hat{\mathbf{z}}$ | (8$j$) | Al XIV |
| $\mathbf{B}_{41}$ | = | $(-x_{18}-y_{18})\,\mathbf{a}_1+(-x_{18}+y_{18})\,\mathbf{a}_2-z_{18}\,\mathbf{a}_3$ | = | $(-x_{18}a-z_{18}c\cos\beta)\,\hat{\mathbf{x}}+y_{18}b\,\hat{\mathbf{y}}-z_{18}c\sin\beta\,\hat{\mathbf{z}}$ | (8$j$) | Al XIV |
| $\mathbf{B}_{42}$ | = | $(-x_{18}+y_{18})\,\mathbf{a}_1+(-x_{18}-y_{18})\,\mathbf{a}_2-z_{18}\,\mathbf{a}_3$ | = | $(-x_{18}a-z_{18}c\cos\beta)\,\hat{\mathbf{x}}-y_{18}b\,\hat{\mathbf{y}}-z_{18}c\sin\beta\,\hat{\mathbf{z}}$ | (8$j$) | Al XIV |
| $\mathbf{B}_{43}$ | = | $(x_{18}+y_{18})\,\mathbf{a}_1+(x_{18}-y_{18})\,\mathbf{a}_2+z_{18}\,\mathbf{a}_3$ | = | $(x_{18}a+z_{18}c\cos\beta)\,\hat{\mathbf{x}}-y_{18}b\,\hat{\mathbf{y}}+z_{18}c\sin\beta\,\hat{\mathbf{z}}$ | (8$j$) | Al XIV |
| $\mathbf{B}_{44}$ | = | $(x_{19}-y_{19})\,\mathbf{a}_1+(x_{19}+y_{19})\,\mathbf{a}_2+z_{19}\,\mathbf{a}_3$ | = | $(x_{19}a+z_{19}c\cos\beta)\,\hat{\mathbf{x}}+y_{19}b\,\hat{\mathbf{y}}+z_{19}c\sin\beta\,\hat{\mathbf{z}}$ | (8$j$) | Al XV |
| $\mathbf{B}_{45}$ | = | $(-x_{19}-y_{19})\,\mathbf{a}_1+(-x_{19}+y_{19})\,\mathbf{a}_2-z_{19}\,\mathbf{a}_3$ | = | $(-x_{19}a-z_{19}c\cos\beta)\,\hat{\mathbf{x}}+y_{19}b\,\hat{\mathbf{y}}-z_{19}c\sin\beta\,\hat{\mathbf{z}}$ | (8$j$) | Al XV |
| $\mathbf{B}_{46}$ | = | $(-x_{19}+y_{19})\,\mathbf{a}_1+(-x_{19}-y_{19})\,\mathbf{a}_2-z_{19}\,\mathbf{a}_3$ | = | $(-x_{19}a-z_{19}c\cos\beta)\,\hat{\mathbf{x}}-y_{19}b\,\hat{\mathbf{y}}-z_{19}c\sin\beta\,\hat{\mathbf{z}}$ | (8$j$) | Al XV |
| $\mathbf{B}_{47}$ | = | $(x_{19}+y_{19})\,\mathbf{a}_1+(x_{19}-y_{19})\,\mathbf{a}_2+z_{19}\,\mathbf{a}_3$ | = | $(x_{19}a+z_{19}c\cos\beta)\,\hat{\mathbf{x}}-y_{19}b\,\hat{\mathbf{y}}+z_{19}c\sin\beta\,\hat{\mathbf{z}}$ | (8$j$) | Al XV |
| $\mathbf{B}_{48}$ | = | $(x_{20}-y_{20})\,\mathbf{a}_1+(x_{20}+y_{20})\,\mathbf{a}_2+z_{20}\,\mathbf{a}_3$ | = | $(x_{20}a+z_{20}c\cos\beta)\,\hat{\mathbf{x}}+y_{20}b\,\hat{\mathbf{y}}+z_{20}c\sin\beta\,\hat{\mathbf{z}}$ | (8$j$) | Fe V |
| $\mathbf{B}_{49}$ | = | $(-x_{20}-y_{20})\,\mathbf{a}_1+(-x_{20}+y_{20})\,\mathbf{a}_2-z_{20}\,\mathbf{a}_3$ | = | $(-x_{20}a-z_{20}c\cos\beta)\,\hat{\mathbf{x}}+y_{20}b\,\hat{\mathbf{y}}-z_{20}c\sin\beta\,\hat{\mathbf{z}}$ | (8$j$) | Fe V |
| $\mathbf{B}_{50}$ | = | $(-x_{20}+y_{20})\,\mathbf{a}_1+(-x_{20}-y_{20})\,\mathbf{a}_2-z_{20}\,\mathbf{a}_3$ | = | $(-x_{20}a-z_{20}c\cos\beta)\,\hat{\mathbf{x}}-y_{20}b\,\hat{\mathbf{y}}-z_{20}c\sin\beta\,\hat{\mathbf{z}}$ | (8$j$) | Fe V |
| $\mathbf{B}_{51}$ | = | $(x_{20}+y_{20})\,\mathbf{a}_1+(x_{20}-y_{20})\,\mathbf{a}_2+z_{20}\,\mathbf{a}_3$ | = | $(x_{20}a+z_{20}c\cos\beta)\,\hat{\mathbf{x}}-y_{20}b\,\hat{\mathbf{y}}+z_{20}c\sin\beta\,\hat{\mathbf{z}}$ | (8$j$) | Fe V |

**Geometry files:**

- CIF: pp. 1524
- POSCAR: pp. 1524

# $Os_4Al_{13}$ Structure: A13B4_mC34_12_b6i_2i

http://aflow.org/prototype-encyclopedia/A13B4_mC34_12_b6i_2i

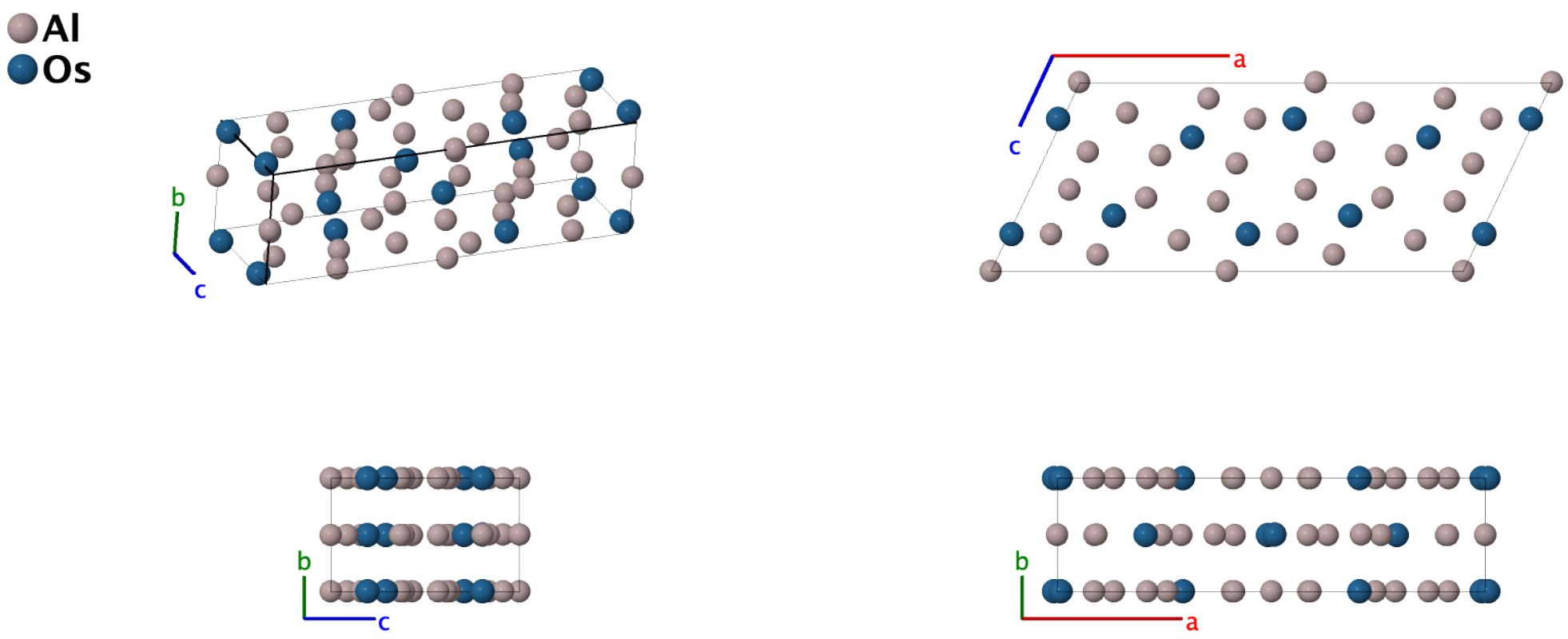

| | | |
|---|---|---|
| **Prototype** | : | $Al_{13}Os_4$ |
| **AFLOW prototype label** | : | A13B4_mC34_12_b6i_2i |
| ***Strukturbericht* designation** | : | None |
| **Pearson symbol** | : | mC34 |
| **Space group number** | : | 12 |
| **Space group symbol** | : | $C2/m$ |
| **AFLOW prototype command** | : | `aflow --proto=A13B4_mC34_12_b6i_2i` `--params=`$a, b/a, c/a, \beta, x_2, z_2, x_3, z_3, x_4, z_4, x_5, z_5, x_6, z_6, x_7, z_7, x_8, z_8, x_9, z_9$ |

**Base-centered Monoclinic primitive vectors:**

$$\mathbf{a}_1 = \frac{1}{2}\,a\,\hat{\mathbf{x}} - \frac{1}{2}\,b\,\hat{\mathbf{y}}$$
$$\mathbf{a}_2 = \frac{1}{2}\,a\,\hat{\mathbf{x}} + \frac{1}{2}\,b\,\hat{\mathbf{y}}$$
$$\mathbf{a}_3 = c\cos\beta\,\hat{\mathbf{x}} + c\sin\beta\,\hat{\mathbf{z}}$$

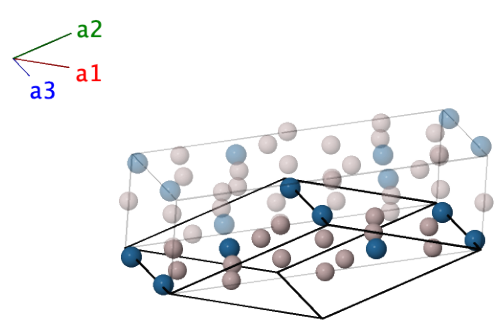

**Basis vectors:**

| | | Lattice Coordinates | | Cartesian Coordinates | Wyckoff Position | Atom Type |
|---|---|---|---|---|---|---|
| $\mathbf{B}_1$ | = | $\frac{1}{2}\,\mathbf{a}_1 + \frac{1}{2}\,\mathbf{a}_2$ | = | $\frac{1}{2}a\,\hat{\mathbf{x}}$ | (2b) | Al I |
| $\mathbf{B}_2$ | = | $x_2\,\mathbf{a}_1 + x_2\,\mathbf{a}_2 + z_2\,\mathbf{a}_3$ | = | $(x_2a + z_2c\cos\beta)\,\hat{\mathbf{x}} + z_2c\sin\beta\,\hat{\mathbf{z}}$ | (4i) | Al II |
| $\mathbf{B}_3$ | = | $-x_2\,\mathbf{a}_1 - x_2\,\mathbf{a}_2 - z_2\,\mathbf{a}_3$ | = | $(-x_2a - z_2c\cos\beta)\,\hat{\mathbf{x}} - z_2c\sin\beta\,\hat{\mathbf{z}}$ | (4i) | Al II |
| $\mathbf{B}_4$ | = | $x_3\,\mathbf{a}_1 + x_3\,\mathbf{a}_2 + z_3\,\mathbf{a}_3$ | = | $(x_3a + z_3c\cos\beta)\,\hat{\mathbf{x}} + z_3c\sin\beta\,\hat{\mathbf{z}}$ | (4i) | Al III |
| $\mathbf{B}_5$ | = | $-x_3\,\mathbf{a}_1 - x_3\,\mathbf{a}_2 - z_3\,\mathbf{a}_3$ | = | $(-x_3a - z_3c\cos\beta)\,\hat{\mathbf{x}} - z_3c\sin\beta\,\hat{\mathbf{z}}$ | (4i) | Al III |
| $\mathbf{B}_6$ | = | $x_4\,\mathbf{a}_1 + x_4\,\mathbf{a}_2 + z_4\,\mathbf{a}_3$ | = | $(x_4a + z_4c\cos\beta)\,\hat{\mathbf{x}} + z_4c\sin\beta\,\hat{\mathbf{z}}$ | (4i) | Al IV |
| $\mathbf{B}_7$ | = | $-x_4\,\mathbf{a}_1 - x_4\,\mathbf{a}_2 - z_4\,\mathbf{a}_3$ | = | $(-x_4a - z_4c\cos\beta)\,\hat{\mathbf{x}} - z_4c\sin\beta\,\hat{\mathbf{z}}$ | (4i) | Al IV |
| $\mathbf{B}_8$ | = | $x_5\,\mathbf{a}_1 + x_5\,\mathbf{a}_2 + z_5\,\mathbf{a}_3$ | = | $(x_5a + z_5c\cos\beta)\,\hat{\mathbf{x}} + z_5c\sin\beta\,\hat{\mathbf{z}}$ | (4i) | Al V |
| $\mathbf{B}_9$ | = | $-x_5\,\mathbf{a}_1 - x_5\,\mathbf{a}_2 - z_5\,\mathbf{a}_3$ | = | $(-x_5a - z_5c\cos\beta)\,\hat{\mathbf{x}} - z_5c\sin\beta\,\hat{\mathbf{z}}$ | (4i) | Al V |
| $\mathbf{B}_{10}$ | = | $x_6\,\mathbf{a}_1 + x_6\,\mathbf{a}_2 + z_6\,\mathbf{a}_3$ | = | $(x_6a + z_6c\cos\beta)\,\hat{\mathbf{x}} + z_6c\sin\beta\,\hat{\mathbf{z}}$ | (4i) | Al VI |
| $\mathbf{B}_{11}$ | = | $-x_6\,\mathbf{a}_1 - x_6\,\mathbf{a}_2 - z_6\,\mathbf{a}_3$ | = | $(-x_6 a - z_6 c\cos\beta)\,\hat{\mathbf{x}} - z_6 c\sin\beta\,\hat{\mathbf{z}}$ | (4$i$) | Al VI |
| $\mathbf{B}_{12}$ | = | $x_7\,\mathbf{a}_1 + x_7\,\mathbf{a}_2 + z_7\,\mathbf{a}_3$ | = | $(x_7 a + z_7 c\cos\beta)\,\hat{\mathbf{x}} + z_7 c\sin\beta\,\hat{\mathbf{z}}$ | (4$i$) | Al VII |
| $\mathbf{B}_{13}$ | = | $-x_7\,\mathbf{a}_1 - x_7\,\mathbf{a}_2 - z_7\,\mathbf{a}_3$ | = | $(-x_7 a - z_7 c\cos\beta)\,\hat{\mathbf{x}} - z_7 c\sin\beta\,\hat{\mathbf{z}}$ | (4$i$) | Al VII |
| $\mathbf{B}_{14}$ | = | $x_8\,\mathbf{a}_1 + x_8\,\mathbf{a}_2 + z_8\,\mathbf{a}_3$ | = | $(x_8 a + z_8 c\cos\beta)\,\hat{\mathbf{x}} + z_8 c\sin\beta\,\hat{\mathbf{z}}$ | (4$i$) | Os I |
| $\mathbf{B}_{15}$ | = | $-x_8\,\mathbf{a}_1 - x_8\,\mathbf{a}_2 - z_8\,\mathbf{a}_3$ | = | $(-x_8 a - z_8 c\cos\beta)\,\hat{\mathbf{x}} - z_8 c\sin\beta\,\hat{\mathbf{z}}$ | (4$i$) | Os I |
| $\mathbf{B}_{16}$ | = | $x_9\,\mathbf{a}_1 + x_9\,\mathbf{a}_2 + z_9\,\mathbf{a}_3$ | = | $(x_9 a + z_9 c\cos\beta)\,\hat{\mathbf{x}} + z_9 c\sin\beta\,\hat{\mathbf{z}}$ | (4$i$) | Os II |
| $\mathbf{B}_{17}$ | = | $-x_9\,\mathbf{a}_1 - x_9\,\mathbf{a}_2 - z_9\,\mathbf{a}_3$ | = | $(-x_9 a - z_9 c\cos\beta)\,\hat{\mathbf{x}} - z_9 c\sin\beta\,\hat{\mathbf{z}}$ | (4$i$) | Os II |

**Geometry files:**

- CIF: pp. 1525
- POSCAR: pp. 1525

# Bischofite ($MgCl_2 \cdot 6H_2O$, $J1_7$) Structure: A2B12CD6_mC42_12_i_2i2j_a_ij

http://aflow.org/prototype-encyclopedia/A2B12CD6_mC42_12_i_2i2j_a_ij

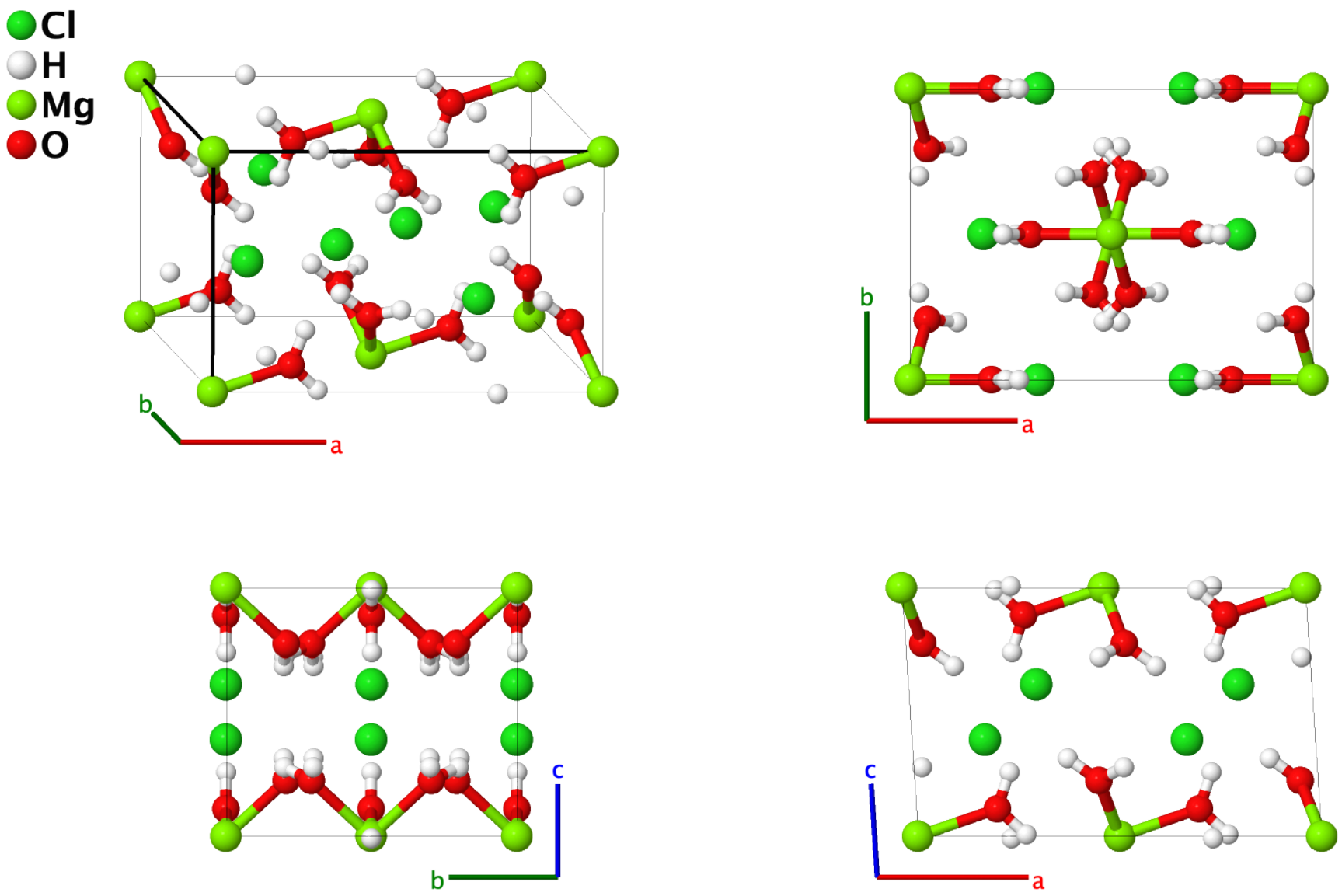

|  |  |  |
|---|---|---|
| **Prototype** | : | $Cl_2H_{12}MgO_6$ |
| **AFLOW prototype label** | : | A2B12CD6_mC42_12_i_2i2j_a_ij |
| ***Strukturbericht* designation** | : | $J1_7$ |
| **Pearson symbol** | : | mC42 |
| **Space group number** | : | 12 |
| **Space group symbol** | : | $C2/m$ |
| **AFLOW prototype command** | : | `aflow --proto=A2B12CD6_mC42_12_i_2i2j_a_ij` `--params=`$a, b/a, c/a, \beta, x_2, z_2, x_3, z_3, x_4, z_4, x_5, z_5, x_6, y_6, z_6, x_7, y_7, z_7, x_8, y_8, z_8$ |

**Other compounds with this structure**

- $MgBr_2 \cdot 6H_2O$

- This structure is nearly identical to the one presented in (Gottfried, 1937) as $J1_7$, but now includes the positions of the hydrogen atoms.

**Base-centered Monoclinic primitive vectors:**

$$\mathbf{a}_1 = \frac{1}{2}\,a\,\hat{\mathbf{x}} - \frac{1}{2}\,b\,\hat{\mathbf{y}}$$
$$\mathbf{a}_2 = \frac{1}{2}\,a\,\hat{\mathbf{x}} + \frac{1}{2}\,b\,\hat{\mathbf{y}}$$
$$\mathbf{a}_3 = c\cos\beta\,\hat{\mathbf{x}} + c\sin\beta\,\hat{\mathbf{z}}$$

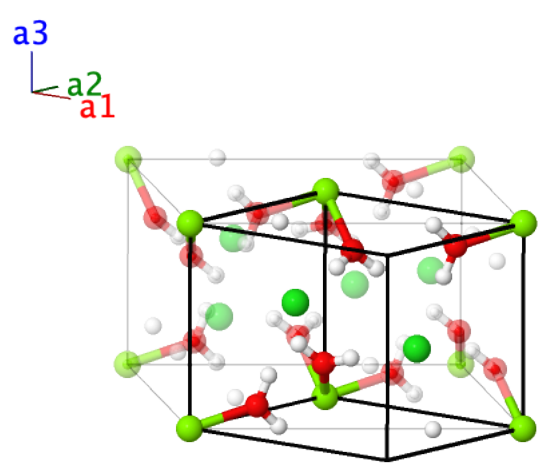

**Basis vectors:**

| | | Lattice Coordinates | | Cartesian Coordinates | Wyckoff Position | Atom Type |
|---|---|---|---|---|---|---|
| $\mathbf{B}_1$ | = | $0\,\mathbf{a}_1 + 0\,\mathbf{a}_2 + 0\,\mathbf{a}_3$ | = | $0\,\hat{\mathbf{x}} + 0\,\hat{\mathbf{y}} + 0\,\hat{\mathbf{z}}$ | (2$a$) | Mg |
| $\mathbf{B}_2$ | = | $x_2\,\mathbf{a}_1 + x_2\,\mathbf{a}_2 + z_2\,\mathbf{a}_3$ | = | $(x_2 a + z_2 c\cos\beta)\,\hat{\mathbf{x}} + z_2 c\sin\beta\,\hat{\mathbf{z}}$ | (4$i$) | Cl |
| $\mathbf{B}_3$ | = | $-x_2\,\mathbf{a}_1 - x_2\,\mathbf{a}_2 - z_2\,\mathbf{a}_3$ | = | $(-x_2 a - z_2 c\cos\beta)\,\hat{\mathbf{x}} - z_2 c\sin\beta\,\hat{\mathbf{z}}$ | (4$i$) | Cl |
| $\mathbf{B}_4$ | = | $x_3\,\mathbf{a}_1 + x_3\,\mathbf{a}_2 + z_3\,\mathbf{a}_3$ | = | $(x_3 a + z_3 c\cos\beta)\,\hat{\mathbf{x}} + z_3 c\sin\beta\,\hat{\mathbf{z}}$ | (4$i$) | H I |
| $\mathbf{B}_5$ | = | $-x_3\,\mathbf{a}_1 - x_3\,\mathbf{a}_2 - z_3\,\mathbf{a}_3$ | = | $(-x_3 a - z_3 c\cos\beta)\,\hat{\mathbf{x}} - z_3 c\sin\beta\,\hat{\mathbf{z}}$ | (4$i$) | H I |
| $\mathbf{B}_6$ | = | $x_4\,\mathbf{a}_1 + x_4\,\mathbf{a}_2 + z_4\,\mathbf{a}_3$ | = | $(x_4 a + z_4 c\cos\beta)\,\hat{\mathbf{x}} + z_4 c\sin\beta\,\hat{\mathbf{z}}$ | (4$i$) | H II |
| $\mathbf{B}_7$ | = | $-x_4\,\mathbf{a}_1 - x_4\,\mathbf{a}_2 - z_4\,\mathbf{a}_3$ | = | $(-x_4 a - z_4 c\cos\beta)\,\hat{\mathbf{x}} - z_4 c\sin\beta\,\hat{\mathbf{z}}$ | (4$i$) | H II |
| $\mathbf{B}_8$ | = | $x_5\,\mathbf{a}_1 + x_5\,\mathbf{a}_2 + z_5\,\mathbf{a}_3$ | = | $(x_5 a + z_5 c\cos\beta)\,\hat{\mathbf{x}} + z_5 c\sin\beta\,\hat{\mathbf{z}}$ | (4$i$) | O I |
| $\mathbf{B}_9$ | = | $-x_5\,\mathbf{a}_1 - x_5\,\mathbf{a}_2 - z_5\,\mathbf{a}_3$ | = | $(-x_5 a - z_5 c\cos\beta)\,\hat{\mathbf{x}} - z_5 c\sin\beta\,\hat{\mathbf{z}}$ | (4$i$) | O I |
| $\mathbf{B}_{10}$ | = | $(x_6 - y_6)\,\mathbf{a}_1 + (x_6 + y_6)\,\mathbf{a}_2 + z_6\,\mathbf{a}_3$ | = | $(x_6 a + z_6 c\cos\beta)\,\hat{\mathbf{x}} + y_6 b\,\hat{\mathbf{y}} + z_6 c\sin\beta\,\hat{\mathbf{z}}$ | (8$j$) | H III |
| $\mathbf{B}_{11}$ | = | $(-x_6 - y_6)\,\mathbf{a}_1 + (-x_6 + y_6)\,\mathbf{a}_2 - z_6\,\mathbf{a}_3$ | = | $(-x_6 a - z_6 c\cos\beta)\,\hat{\mathbf{x}} + y_6 b\,\hat{\mathbf{y}} - z_6 c\sin\beta\,\hat{\mathbf{z}}$ | (8$j$) | H III |
| $\mathbf{B}_{12}$ | = | $(-x_6 + y_6)\,\mathbf{a}_1 + (-x_6 - y_6)\,\mathbf{a}_2 - z_6\,\mathbf{a}_3$ | = | $(-x_6 a - z_6 c\cos\beta)\,\hat{\mathbf{x}} - y_6 b\,\hat{\mathbf{y}} - z_6 c\sin\beta\,\hat{\mathbf{z}}$ | (8$j$) | H III |
| $\mathbf{B}_{13}$ | = | $(x_6 + y_6)\,\mathbf{a}_1 + (x_6 - y_6)\,\mathbf{a}_2 + z_6\,\mathbf{a}_3$ | = | $(x_6 a + z_6 c\cos\beta)\,\hat{\mathbf{x}} - y_6 b\,\hat{\mathbf{y}} + z_6 c\sin\beta\,\hat{\mathbf{z}}$ | (8$j$) | H III |
| $\mathbf{B}_{14}$ | = | $(x_7 - y_7)\,\mathbf{a}_1 + (x_7 + y_7)\,\mathbf{a}_2 + z_7\,\mathbf{a}_3$ | = | $(x_7 a + z_7 c\cos\beta)\,\hat{\mathbf{x}} + y_7 b\,\hat{\mathbf{y}} + z_7 c\sin\beta\,\hat{\mathbf{z}}$ | (8$j$) | H IV |
| $\mathbf{B}_{15}$ | = | $(-x_7 - y_7)\,\mathbf{a}_1 + (-x_7 + y_7)\,\mathbf{a}_2 - z_7\,\mathbf{a}_3$ | = | $(-x_7 a - z_7 c\cos\beta)\,\hat{\mathbf{x}} + y_7 b\,\hat{\mathbf{y}} - z_7 c\sin\beta\,\hat{\mathbf{z}}$ | (8$j$) | H IV |
| $\mathbf{B}_{16}$ | = | $(-x_7 + y_7)\,\mathbf{a}_1 + (-x_7 - y_7)\,\mathbf{a}_2 - z_7\,\mathbf{a}_3$ | = | $(-x_7 a - z_7 c\cos\beta)\,\hat{\mathbf{x}} - y_7 b\,\hat{\mathbf{y}} - z_7 c\sin\beta\,\hat{\mathbf{z}}$ | (8$j$) | H IV |
| $\mathbf{B}_{17}$ | = | $(x_7 + y_7)\,\mathbf{a}_1 + (x_7 - y_7)\,\mathbf{a}_2 + z_7\,\mathbf{a}_3$ | = | $(x_7 a + z_7 c\cos\beta)\,\hat{\mathbf{x}} - y_7 b\,\hat{\mathbf{y}} + z_7 c\sin\beta\,\hat{\mathbf{z}}$ | (8$j$) | H IV |
| $\mathbf{B}_{18}$ | = | $(x_8 - y_8)\,\mathbf{a}_1 + (x_8 + y_8)\,\mathbf{a}_2 + z_8\,\mathbf{a}_3$ | = | $(x_8 a + z_8 c\cos\beta)\,\hat{\mathbf{x}} + y_8 b\,\hat{\mathbf{y}} + z_8 c\sin\beta\,\hat{\mathbf{z}}$ | (8$j$) | O II |
| $\mathbf{B}_{19}$ | = | $(-x_8 - y_8)\,\mathbf{a}_1 + (-x_8 + y_8)\,\mathbf{a}_2 - z_8\,\mathbf{a}_3$ | = | $(-x_8 a - z_8 c\cos\beta)\,\hat{\mathbf{x}} + y_8 b\,\hat{\mathbf{y}} - z_8 c\sin\beta\,\hat{\mathbf{z}}$ | (8$j$) | O II |
| $\mathbf{B}_{20}$ | = | $(-x_8 + y_8)\,\mathbf{a}_1 + (-x_8 - y_8)\,\mathbf{a}_2 - z_8\,\mathbf{a}_3$ | = | $(-x_8 a - z_8 c\cos\beta)\,\hat{\mathbf{x}} - y_8 b\,\hat{\mathbf{y}} - z_8 c\sin\beta\,\hat{\mathbf{z}}$ | (8$j$) | O II |
| $\mathbf{B}_{21}$ | = | $(x_8 + y_8)\,\mathbf{a}_1 + (x_8 - y_8)\,\mathbf{a}_2 + z_8\,\mathbf{a}_3$ | = | $(x_8 a + z_8 c\cos\beta)\,\hat{\mathbf{x}} - y_8 b\,\hat{\mathbf{y}} + z_8 c\sin\beta\,\hat{\mathbf{z}}$ | (8$j$) | O II |

**Geometry files:**

- CIF: pp. 1525
- POSCAR: pp. 1526

# Tremolite ($Ca_2Mg_5Si_8O_{22}(OH)_2$, $S4_2$) Structure: A2B2C5D24E8_mC82_12_h_i_agh_2i5j_2j

http://aflow.org/prototype-encyclopedia/A2B2C5D24E8_mC82_12_h_i_agh_2i5j_2j

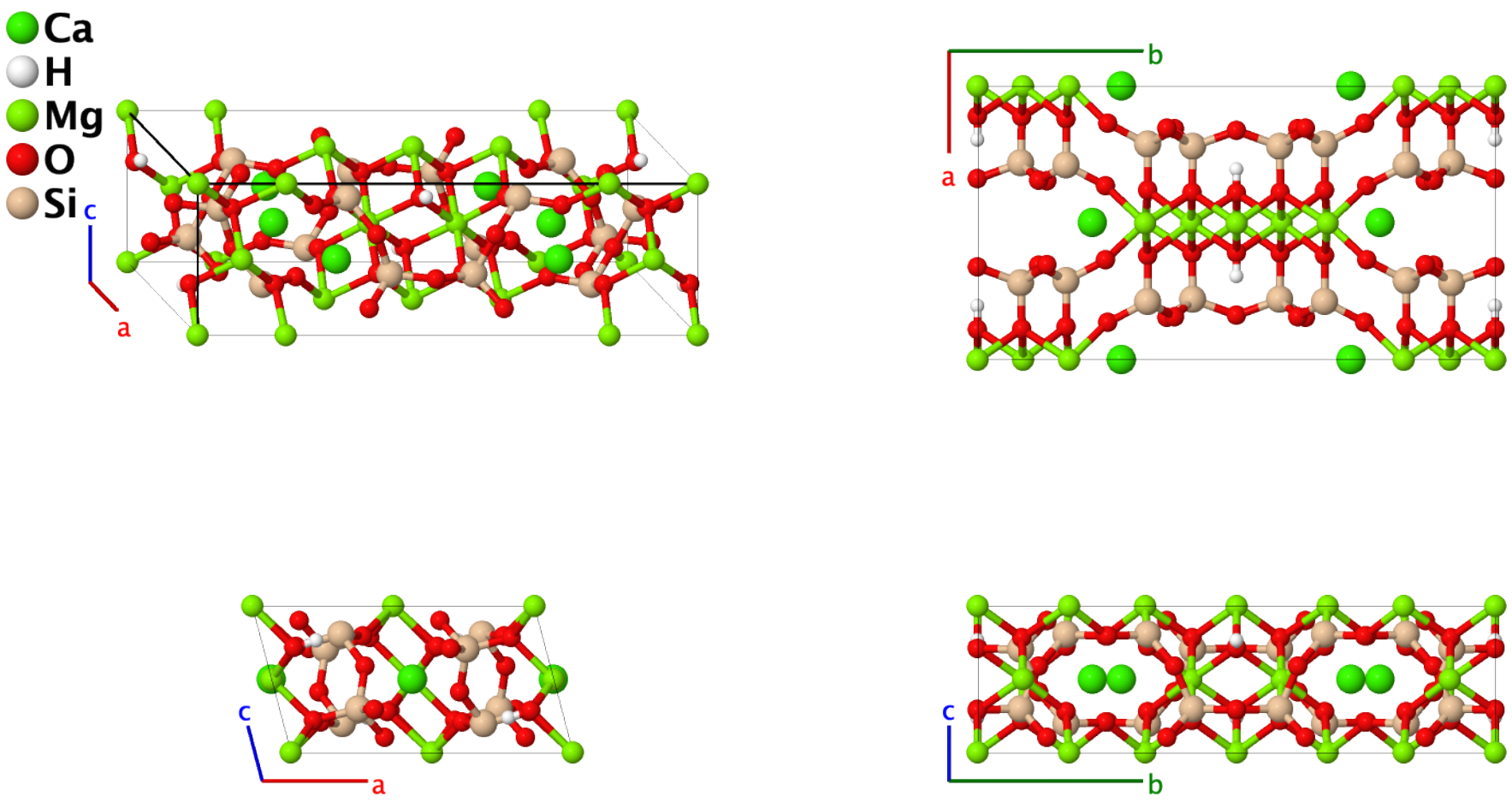

| | | |
|---|---|---|
| **Prototype** | : | $Ca_2H_2Mg_5O_{24}Si_8$ |
| **AFLOW prototype label** | : | A2B2C5D24E8_mC82_12_h_i_agh_2i5j_2j |
| ***Strukturbericht* designation** | : | $S4_2$ |
| **Pearson symbol** | : | mC82 |
| **Space group number** | : | 12 |
| **Space group symbol** | : | $C2/m$ |
| **AFLOW prototype command** | : | `aflow --proto=A2B2C5D24E8_mC82_12_h_i_agh_2i5j_2j` `--params=`$a, b/a, c/a, \beta, y_2, y_3, y_4, x_5, z_5, x_6, z_6, x_7, z_7, x_8, y_8, z_8, x_9, y_9, z_9, x_{10}, y_{10}, z_{10}, x_{11}, y_{11}, z_{11}, x_{12}, y_{12}, z_{12}, x_{13}, y_{13}, z_{13}, x_{14}, y_{14}, z_{14}$ |

### Other compounds with this structure

- $Ca_2H_2(Mg_{5-x},Fe_x)O_{24}Si_8$ and $Ca_2F_2Mg_5O_{22}Si_8$

### Base-centered Monoclinic primitive vectors:

$$\mathbf{a}_1 = \frac{1}{2}a\,\hat{\mathbf{x}} - \frac{1}{2}b\,\hat{\mathbf{y}}$$
$$\mathbf{a}_2 = \frac{1}{2}a\,\hat{\mathbf{x}} + \frac{1}{2}b\,\hat{\mathbf{y}}$$
$$\mathbf{a}_3 = c\cos\beta\,\hat{\mathbf{x}} + c\sin\beta\,\hat{\mathbf{z}}$$

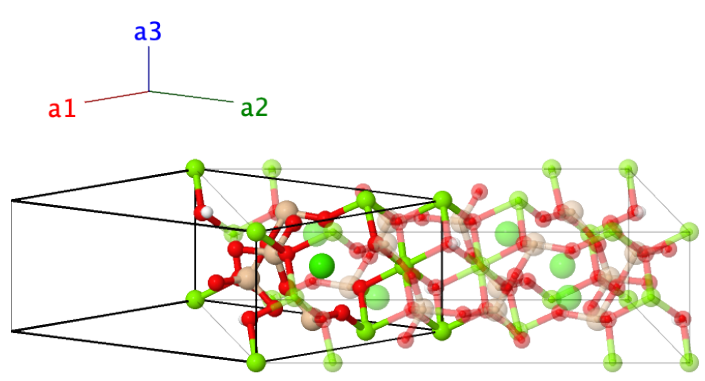

### Basis vectors:

| | | Lattice Coordinates | | Cartesian Coordinates | Wyckoff Position | Atom Type |
|---|---|---|---|---|---|---|
| $\mathbf{B}_1$ | = | $0\,\mathbf{a}_1 + 0\,\mathbf{a}_2 + 0\,\mathbf{a}_3$ | = | $0\,\hat{\mathbf{x}} + 0\,\hat{\mathbf{y}} + 0\,\hat{\mathbf{z}}$ | (2$a$) | Mg I |
| $\mathbf{B}_2$ | = | $-y_2\,\mathbf{a}_1 + y_2\,\mathbf{a}_2$ | = | $y_2 b\,\hat{\mathbf{y}}$ | (4$g$) | Mg II |
| $\mathbf{B}_3$ | = | $y_2\,\mathbf{a}_1 - y_2\,\mathbf{a}_2$ | = | $-y_2 b\,\hat{\mathbf{y}}$ | (4$g$) | Mg II |
| $\mathbf{B}_4$ | = | $-y_3\,\mathbf{a}_1 + y_3\,\mathbf{a}_2 + \frac{1}{2}\,\mathbf{a}_3$ | = | $\frac{1}{2}c\cos\beta\,\hat{\mathbf{x}} + y_3 b\,\hat{\mathbf{y}} + \frac{1}{2}c\sin\beta\,\hat{\mathbf{z}}$ | (4$h$) | Ca |
| $\mathbf{B}_5$ | = | $y_3\,\mathbf{a}_1 - y_3\,\mathbf{a}_2 + \frac{1}{2}\,\mathbf{a}_3$ | = | $\frac{1}{2}c\cos\beta\,\hat{\mathbf{x}} - y_3 b\,\hat{\mathbf{y}} + \frac{1}{2}c\sin\beta\,\hat{\mathbf{z}}$ | (4$h$) | Ca |
| $\mathbf{B}_6$ | = | $-y_4\,\mathbf{a}_1 + y_4\,\mathbf{a}_2 + \frac{1}{2}\,\mathbf{a}_3$ | = | $\frac{1}{2}c\cos\beta\,\hat{\mathbf{x}} + y_4 b\,\hat{\mathbf{y}} + \frac{1}{2}c\sin\beta\,\hat{\mathbf{z}}$ | (4$h$) | Mg III |
| $\mathbf{B}_7$ | = | $y_4\,\mathbf{a}_1 - y_4\,\mathbf{a}_2 + \frac{1}{2}\,\mathbf{a}_3$ | = | $\frac{1}{2}c\cos\beta\,\hat{\mathbf{x}} - y_4 b\,\hat{\mathbf{y}} + \frac{1}{2}c\sin\beta\,\hat{\mathbf{z}}$ | (4$h$) | Mg III |
| $\mathbf{B}_8$ | = | $x_5\,\mathbf{a}_1 + x_5\,\mathbf{a}_2 + z_5\,\mathbf{a}_3$ | = | $(x_5 a + z_5 c\cos\beta)\,\hat{\mathbf{x}} + z_5 c\sin\beta\,\hat{\mathbf{z}}$ | (4$i$) | H |
| $\mathbf{B}_9$ | = | $-x_5\,\mathbf{a}_1 - x_5\,\mathbf{a}_2 - z_5\,\mathbf{a}_3$ | = | $(-x_5 a - z_5 c\cos\beta)\,\hat{\mathbf{x}} - z_5 c\sin\beta\,\hat{\mathbf{z}}$ | (4$i$) | H |
| $\mathbf{B}_{10}$ | = | $x_6\,\mathbf{a}_1 + x_6\,\mathbf{a}_2 + z_6\,\mathbf{a}_3$ | = | $(x_6 a + z_6 c\cos\beta)\,\hat{\mathbf{x}} + z_6 c\sin\beta\,\hat{\mathbf{z}}$ | (4$i$) | O I |
| $\mathbf{B}_{11}$ | = | $-x_6\,\mathbf{a}_1 - x_6\,\mathbf{a}_2 - z_6\,\mathbf{a}_3$ | = | $(-x_6 a - z_6 c\cos\beta)\,\hat{\mathbf{x}} - z_6 c\sin\beta\,\hat{\mathbf{z}}$ | (4$i$) | O I |
| $\mathbf{B}_{12}$ | = | $x_7\,\mathbf{a}_1 + x_7\,\mathbf{a}_2 + z_7\,\mathbf{a}_3$ | = | $(x_7 a + z_7 c\cos\beta)\,\hat{\mathbf{x}} + z_7 c\sin\beta\,\hat{\mathbf{z}}$ | (4$i$) | O II |
| $\mathbf{B}_{13}$ | = | $-x_7\,\mathbf{a}_1 - x_7\,\mathbf{a}_2 - z_7\,\mathbf{a}_3$ | = | $(-x_7 a - z_7 c\cos\beta)\,\hat{\mathbf{x}} - z_7 c\sin\beta\,\hat{\mathbf{z}}$ | (4$i$) | O II |
| $\mathbf{B}_{14}$ | = | $(x_8 - y_8)\,\mathbf{a}_1 + (x_8 + y_8)\,\mathbf{a}_2 + z_8\,\mathbf{a}_3$ | = | $(x_8 a + z_8 c\cos\beta)\,\hat{\mathbf{x}} + y_8 b\,\hat{\mathbf{y}} + z_8 c\sin\beta\,\hat{\mathbf{z}}$ | (8$j$) | O III |
| $\mathbf{B}_{15}$ | = | $(-x_8 - y_8)\,\mathbf{a}_1 + (-x_8 + y_8)\,\mathbf{a}_2 - z_8\,\mathbf{a}_3$ | = | $(-x_8 a - z_8 c\cos\beta)\,\hat{\mathbf{x}} + y_8 b\,\hat{\mathbf{y}} - z_8 c\sin\beta\,\hat{\mathbf{z}}$ | (8$j$) | O III |
| $\mathbf{B}_{16}$ | = | $(-x_8 + y_8)\,\mathbf{a}_1 + (-x_8 - y_8)\,\mathbf{a}_2 - z_8\,\mathbf{a}_3$ | = | $(-x_8 a - z_8 c\cos\beta)\,\hat{\mathbf{x}} - y_8 b\,\hat{\mathbf{y}} - z_8 c\sin\beta\,\hat{\mathbf{z}}$ | (8$j$) | O III |
| $\mathbf{B}_{17}$ | = | $(x_8 + y_8)\,\mathbf{a}_1 + (x_8 - y_8)\,\mathbf{a}_2 + z_8\,\mathbf{a}_3$ | = | $(x_8 a + z_8 c\cos\beta)\,\hat{\mathbf{x}} - y_8 b\,\hat{\mathbf{y}} + z_8 c\sin\beta\,\hat{\mathbf{z}}$ | (8$j$) | O III |
| $\mathbf{B}_{18}$ | = | $(x_9 - y_9)\,\mathbf{a}_1 + (x_9 + y_9)\,\mathbf{a}_2 + z_9\,\mathbf{a}_3$ | = | $(x_9 a + z_9 c\cos\beta)\,\hat{\mathbf{x}} + y_9 b\,\hat{\mathbf{y}} + z_9 c\sin\beta\,\hat{\mathbf{z}}$ | (8$j$) | O IV |
| $\mathbf{B}_{19}$ | = | $(-x_9 - y_9)\,\mathbf{a}_1 + (-x_9 + y_9)\,\mathbf{a}_2 - z_9\,\mathbf{a}_3$ | = | $(-x_9 a - z_9 c\cos\beta)\,\hat{\mathbf{x}} + y_9 b\,\hat{\mathbf{y}} - z_9 c\sin\beta\,\hat{\mathbf{z}}$ | (8$j$) | O IV |
| $\mathbf{B}_{20}$ | = | $(-x_9 + y_9)\,\mathbf{a}_1 + (-x_9 - y_9)\,\mathbf{a}_2 - z_9\,\mathbf{a}_3$ | = | $(-x_9 a - z_9 c\cos\beta)\,\hat{\mathbf{x}} - y_9 b\,\hat{\mathbf{y}} - z_9 c\sin\beta\,\hat{\mathbf{z}}$ | (8$j$) | O IV |
| $\mathbf{B}_{21}$ | = | $(x_9 + y_9)\,\mathbf{a}_1 + (x_9 - y_9)\,\mathbf{a}_2 + z_9\,\mathbf{a}_3$ | = | $(x_9 a + z_9 c\cos\beta)\,\hat{\mathbf{x}} - y_9 b\,\hat{\mathbf{y}} + z_9 c\sin\beta\,\hat{\mathbf{z}}$ | (8$j$) | O IV |
| $\mathbf{B}_{22}$ | = | $(x_{10} - y_{10})\,\mathbf{a}_1 + (x_{10} + y_{10})\,\mathbf{a}_2 + z_{10}\,\mathbf{a}_3$ | = | $(x_{10} a + z_{10} c\cos\beta)\,\hat{\mathbf{x}} + y_{10} b\,\hat{\mathbf{y}} + z_{10} c\sin\beta\,\hat{\mathbf{z}}$ | (8$j$) | O V |
| $\mathbf{B}_{23}$ | = | $(-x_{10} - y_{10})\,\mathbf{a}_1 + (-x_{10} + y_{10})\,\mathbf{a}_2 - z_{10}\,\mathbf{a}_3$ | = | $(-x_{10} a - z_{10} c\cos\beta)\,\hat{\mathbf{x}} + y_{10} b\,\hat{\mathbf{y}} - z_{10} c\sin\beta\,\hat{\mathbf{z}}$ | (8$j$) | O V |
| $\mathbf{B}_{24}$ | = | $(-x_{10} + y_{10})\,\mathbf{a}_1 + (-x_{10} - y_{10})\,\mathbf{a}_2 - z_{10}\,\mathbf{a}_3$ | = | $(-x_{10} a - z_{10} c\cos\beta)\,\hat{\mathbf{x}} - y_{10} b\,\hat{\mathbf{y}} - z_{10} c\sin\beta\,\hat{\mathbf{z}}$ | (8$j$) | O V |
| $\mathbf{B}_{25}$ | = | $(x_{10} + y_{10})\,\mathbf{a}_1 + (x_{10} - y_{10})\,\mathbf{a}_2 + z_{10}\,\mathbf{a}_3$ | = | $(x_{10} a + z_{10} c\cos\beta)\,\hat{\mathbf{x}} - y_{10} b\,\hat{\mathbf{y}} + z_{10} c\sin\beta\,\hat{\mathbf{z}}$ | (8$j$) | O V |
| $\mathbf{B}_{26}$ | = | $(x_{11} - y_{11})\,\mathbf{a}_1 + (x_{11} + y_{11})\,\mathbf{a}_2 + z_{11}\,\mathbf{a}_3$ | = | $(x_{11} a + z_{11} c\cos\beta)\,\hat{\mathbf{x}} + y_{11} b\,\hat{\mathbf{y}} + z_{11} c\sin\beta\,\hat{\mathbf{z}}$ | (8$j$) | O VI |
| $\mathbf{B}_{27}$ | = | $(-x_{11} - y_{11})\,\mathbf{a}_1 + (-x_{11} + y_{11})\,\mathbf{a}_2 - z_{11}\,\mathbf{a}_3$ | = | $(-x_{11} a - z_{11} c\cos\beta)\,\hat{\mathbf{x}} + y_{11} b\,\hat{\mathbf{y}} - z_{11} c\sin\beta\,\hat{\mathbf{z}}$ | (8$j$) | O VI |
| $\mathbf{B}_{28}$ | = | $(-x_{11} + y_{11})\,\mathbf{a}_1 + (-x_{11} - y_{11})\,\mathbf{a}_2 - z_{11}\,\mathbf{a}_3$ | = | $(-x_{11} a - z_{11} c\cos\beta)\,\hat{\mathbf{x}} - y_{11} b\,\hat{\mathbf{y}} - z_{11} c\sin\beta\,\hat{\mathbf{z}}$ | (8$j$) | O VI |
| $\mathbf{B}_{29}$ | = | $(x_{11} + y_{11})\,\mathbf{a}_1 + (x_{11} - y_{11})\,\mathbf{a}_2 + z_{11}\,\mathbf{a}_3$ | = | $(x_{11} a + z_{11} c\cos\beta)\,\hat{\mathbf{x}} - y_{11} b\,\hat{\mathbf{y}} + z_{11} c\sin\beta\,\hat{\mathbf{z}}$ | (8$j$) | O VI |
| $\mathbf{B}_{30}$ | = | $(x_{12} - y_{12})\,\mathbf{a}_1 + (x_{12} + y_{12})\,\mathbf{a}_2 + z_{12}\,\mathbf{a}_3$ | = | $(x_{12} a + z_{12} c\cos\beta)\,\hat{\mathbf{x}} + y_{12} b\,\hat{\mathbf{y}} + z_{12} c\sin\beta\,\hat{\mathbf{z}}$ | (8$j$) | O VII |
| $\mathbf{B}_{31}$ | = | $(-x_{12} - y_{12})\,\mathbf{a}_1 + (-x_{12} + y_{12})\,\mathbf{a}_2 - z_{12}\,\mathbf{a}_3$ | = | $(-x_{12} a - z_{12} c\cos\beta)\,\hat{\mathbf{x}} + y_{12} b\,\hat{\mathbf{y}} - z_{12} c\sin\beta\,\hat{\mathbf{z}}$ | (8$j$) | O VII |
| $\mathbf{B}_{32}$ | = | $\left(-x_{12}+y_{12}\right)\,\mathbf{a}_{1}+\left(-x_{12}-y_{12}\right)\,\mathbf{a}_{2}-z_{12}\,\mathbf{a}_{3}$ | = | $\left(-x_{12}a-z_{12}c\cos\beta\right)\,\mathbf{\hat{x}}-y_{12}b\,\mathbf{\hat{y}}-z_{12}c\sin\beta\,\mathbf{\hat{z}}$ | (8*j*) | O VII |
| $\mathbf{B}_{33}$ | = | $\left(x_{12}+y_{12}\right)\,\mathbf{a}_{1}+\left(x_{12}-y_{12}\right)\,\mathbf{a}_{2}+z_{12}\,\mathbf{a}_{3}$ | = | $\left(x_{12}a+z_{12}c\cos\beta\right)\,\mathbf{\hat{x}}-y_{12}b\,\mathbf{\hat{y}}+z_{12}c\sin\beta\,\mathbf{\hat{z}}$ | (8*j*) | O VII |
| $\mathbf{B}_{34}$ | = | $\left(x_{13}-y_{13}\right)\,\mathbf{a}_{1}+\left(x_{13}+y_{13}\right)\,\mathbf{a}_{2}+z_{13}\,\mathbf{a}_{3}$ | = | $\left(x_{13}a+z_{13}c\cos\beta\right)\,\mathbf{\hat{x}}+y_{13}b\,\mathbf{\hat{y}}+z_{13}c\sin\beta\,\mathbf{\hat{z}}$ | (8*j*) | Si I |
| $\mathbf{B}_{35}$ | = | $\left(-x_{13}-y_{13}\right)\,\mathbf{a}_{1}+\left(-x_{13}+y_{13}\right)\,\mathbf{a}_{2}-z_{13}\,\mathbf{a}_{3}$ | = | $\left(-x_{13}a-z_{13}c\cos\beta\right)\,\mathbf{\hat{x}}+y_{13}b\,\mathbf{\hat{y}}-z_{13}c\sin\beta\,\mathbf{\hat{z}}$ | (8*j*) | Si I |
| $\mathbf{B}_{36}$ | = | $\left(-x_{13}+y_{13}\right)\,\mathbf{a}_{1}+\left(-x_{13}-y_{13}\right)\,\mathbf{a}_{2}-z_{13}\,\mathbf{a}_{3}$ | = | $\left(-x_{13}a-z_{13}c\cos\beta\right)\,\mathbf{\hat{x}}-y_{13}b\,\mathbf{\hat{y}}-z_{13}c\sin\beta\,\mathbf{\hat{z}}$ | (8*j*) | Si I |
| $\mathbf{B}_{37}$ | = | $\left(x_{13}+y_{13}\right)\,\mathbf{a}_{1}+\left(x_{13}-y_{13}\right)\,\mathbf{a}_{2}+z_{13}\,\mathbf{a}_{3}$ | = | $\left(x_{13}a+z_{13}c\cos\beta\right)\,\mathbf{\hat{x}}-y_{13}b\,\mathbf{\hat{y}}+z_{13}c\sin\beta\,\mathbf{\hat{z}}$ | (8*j*) | Si I |
| $\mathbf{B}_{38}$ | = | $\left(x_{14}-y_{14}\right)\,\mathbf{a}_{1}+\left(x_{14}+y_{14}\right)\,\mathbf{a}_{2}+z_{14}\,\mathbf{a}_{3}$ | = | $\left(x_{14}a+z_{14}c\cos\beta\right)\,\mathbf{\hat{x}}+y_{14}b\,\mathbf{\hat{y}}+z_{14}c\sin\beta\,\mathbf{\hat{z}}$ | (8*j*) | Si II |
| $\mathbf{B}_{39}$ | = | $\left(-x_{14}-y_{14}\right)\,\mathbf{a}_{1}+\left(-x_{14}+y_{14}\right)\,\mathbf{a}_{2}-z_{14}\,\mathbf{a}_{3}$ | = | $\left(-x_{14}a-z_{14}c\cos\beta\right)\,\mathbf{\hat{x}}+y_{14}b\,\mathbf{\hat{y}}-z_{14}c\sin\beta\,\mathbf{\hat{z}}$ | (8*j*) | Si II |
| $\mathbf{B}_{40}$ | = | $\left(-x_{14}+y_{14}\right)\,\mathbf{a}_{1}+\left(-x_{14}-y_{14}\right)\,\mathbf{a}_{2}-z_{14}\,\mathbf{a}_{3}$ | = | $\left(-x_{14}a-z_{14}c\cos\beta\right)\,\mathbf{\hat{x}}-y_{14}b\,\mathbf{\hat{y}}-z_{14}c\sin\beta\,\mathbf{\hat{z}}$ | (8*j*) | Si II |
| $\mathbf{B}_{41}$ | = | $\left(x_{14}+y_{14}\right)\,\mathbf{a}_{1}+\left(x_{14}-y_{14}\right)\,\mathbf{a}_{2}+z_{14}\,\mathbf{a}_{3}$ | = | $\left(x_{14}a+z_{14}c\cos\beta\right)\,\mathbf{\hat{x}}-y_{14}b\,\mathbf{\hat{y}}+z_{14}c\sin\beta\,\mathbf{\hat{z}}$ | (8*j*) | Si II |

**Found in:**

**Geometry files:**

- CIF: pp. 1526
- POSCAR: pp. 1526

# $\beta$-$Ga_2O_3$ Structure: A2B3_mC20_12_2i_3i

http://aflow.org/prototype-encyclopedia/A2B3_mC20_12_2i_3i

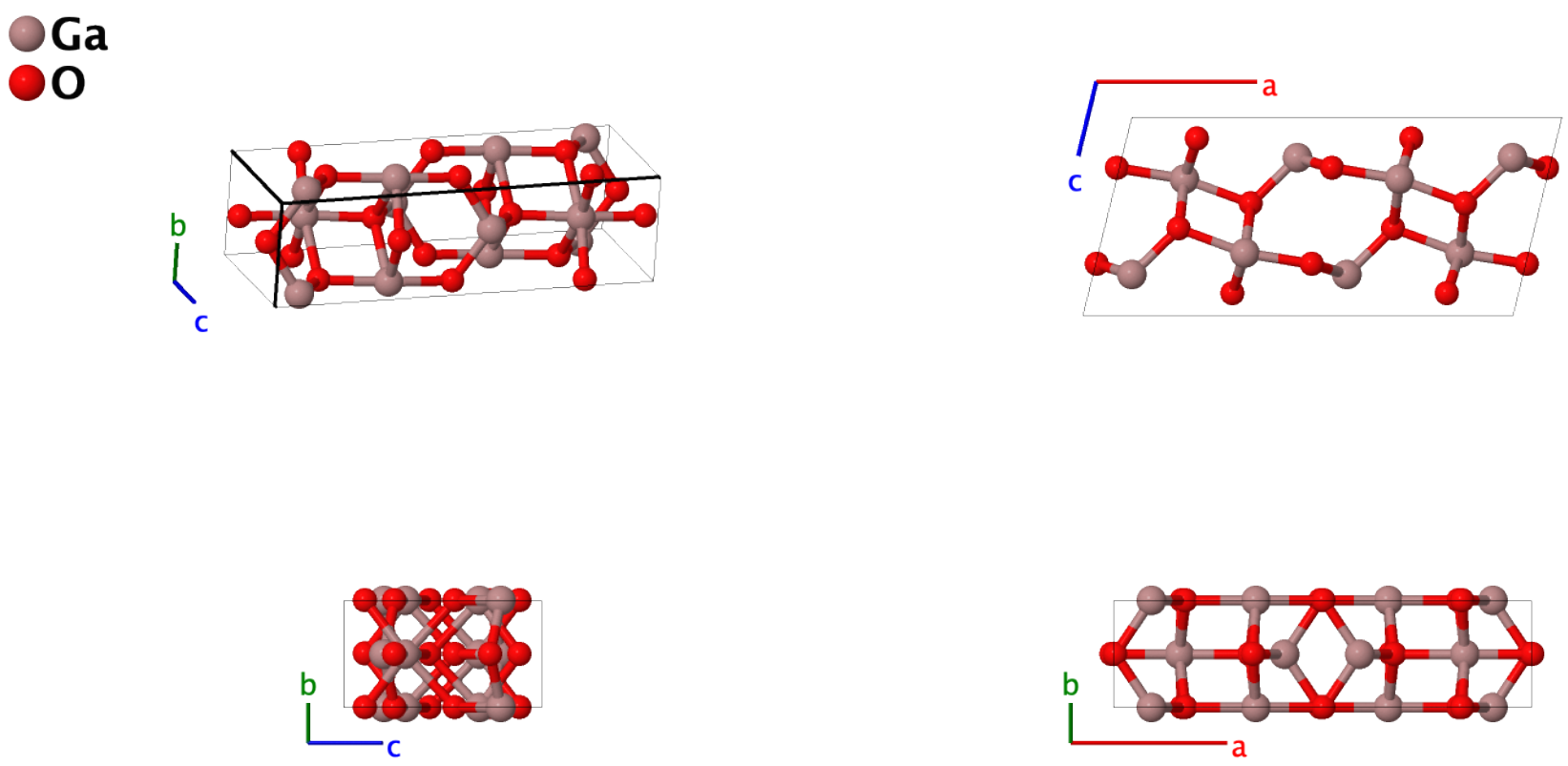

| | | |
|---|---|---|
| **Prototype** | : | $Ga_2O_3$ |
| **AFLOW prototype label** | : | A2B3_mC20_12_2i_3i |
| ***Strukturbericht* designation** | : | None |
| **Pearson symbol** | : | mC20 |
| **Space group number** | : | 12 |
| **Space group symbol** | : | $C2/m$ |
| **AFLOW prototype command** | : | `aflow --proto=A2B3_mC20_12_2i_3i` <br> `--params=`$a, b/a, c/a, \beta, x_1, z_1, x_2, z_2, x_3, z_3, x_4, z_4, x_5, z_5$ |

- $Ga_2O3$ takes on a variety of structures:
  - $\alpha$-$Ga_2O_3$, which has the corundum ($D5_1$) structure,
  - $\beta$-$Ga_2O_3$, this structure,
  - $\gamma$-$Ga_2O_3$, and
  - $\epsilon$-$Ga_2O_3$, a structure with many vacancies which can be approximated by the $\kappa$-alumina structure.

**Base-centered Monoclinic primitive vectors:**

$$\mathbf{a}_1 = \frac{1}{2}\,a\,\hat{\mathbf{x}} - \frac{1}{2}\,b\,\hat{\mathbf{y}}$$
$$\mathbf{a}_2 = \frac{1}{2}\,a\,\hat{\mathbf{x}} + \frac{1}{2}\,b\,\hat{\mathbf{y}}$$
$$\mathbf{a}_3 = c\cos\beta\,\hat{\mathbf{x}} + c\sin\beta\,\hat{\mathbf{z}}$$

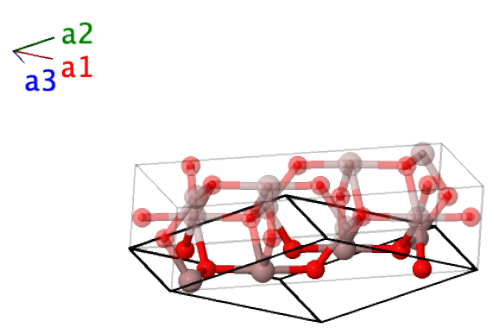

**Basis vectors:**

| | | Lattice Coordinates | | Cartesian Coordinates | Wyckoff Position | Atom Type |
|---|---|---|---|---|---|---|
| $\mathbf{B}_1$ | = | $x_1\,\mathbf{a}_1 + x_1\,\mathbf{a}_2 + z_1\,\mathbf{a}_3$ | = | $(x_1 a + z_1 c\cos\beta)\,\hat{\mathbf{x}} + z_1 c\sin\beta\,\hat{\mathbf{z}}$ | (4$i$) | Ga I |
| $\mathbf{B}_2$ | = | $-x_1\,\mathbf{a}_1 - x_1\,\mathbf{a}_2 - z_1\,\mathbf{a}_3$ | = | $(-x_1 a - z_1 c\cos\beta)\,\hat{\mathbf{x}} - z_1 c\sin\beta\,\hat{\mathbf{z}}$ | (4$i$) | Ga I |
| $\mathbf{B}_3$ | = | $x_2\,\mathbf{a}_1 + x_2\,\mathbf{a}_2 + z_2\,\mathbf{a}_3$ | = | $(x_2 a + z_2 c\cos\beta)\,\hat{\mathbf{x}} + z_2 c\sin\beta\,\hat{\mathbf{z}}$ | (4$i$) | Ga II |
| $\mathbf{B}_4$ | = | $-x_2\,\mathbf{a}_1 - x_2\,\mathbf{a}_2 - z_2\,\mathbf{a}_3$ | = | $(-x_2 a - z_2 c\cos\beta)\,\hat{\mathbf{x}} - z_2 c\sin\beta\,\hat{\mathbf{z}}$ | (4$i$) | Ga II |
| $\mathbf{B}_5$ | = | $x_3\,\mathbf{a}_1 + x_3\,\mathbf{a}_2 + z_3\,\mathbf{a}_3$ | = | $(x_3 a + z_3 c\cos\beta)\,\hat{\mathbf{x}} + z_3 c\sin\beta\,\hat{\mathbf{z}}$ | (4$i$) | O I |
| $\mathbf{B}_6$ | = | $-x_3\,\mathbf{a}_1 - x_3\,\mathbf{a}_2 - z_3\,\mathbf{a}_3$ | = | $(-x_3 a - z_3 c\cos\beta)\,\hat{\mathbf{x}} - z_3 c\sin\beta\,\hat{\mathbf{z}}$ | (4$i$) | O I |
| $\mathbf{B}_7$ | = | $x_4\,\mathbf{a}_1 + x_4\,\mathbf{a}_2 + z_4\,\mathbf{a}_3$ | = | $(x_4 a + z_4 c\cos\beta)\,\hat{\mathbf{x}} + z_4 c\sin\beta\,\hat{\mathbf{z}}$ | (4$i$) | O II |
| $\mathbf{B}_8$ | = | $-x_4\,\mathbf{a}_1 - x_4\,\mathbf{a}_2 - z_4\,\mathbf{a}_3$ | = | $(-x_4 a - z_4 c\cos\beta)\,\hat{\mathbf{x}} - z_4 c\sin\beta\,\hat{\mathbf{z}}$ | (4$i$) | O II |
| $\mathbf{B}_9$ | = | $x_5\,\mathbf{a}_1 + x_5\,\mathbf{a}_2 + z_5\,\mathbf{a}_3$ | = | $(x_5 a + z_5 c\cos\beta)\,\hat{\mathbf{x}} + z_5 c\sin\beta\,\hat{\mathbf{z}}$ | (4$i$) | O III |
| $\mathbf{B}_{10}$ | = | $-x_5\,\mathbf{a}_1 - x_5\,\mathbf{a}_2 - z_5\,\mathbf{a}_3$ | = | $(-x_5 a - z_5 c\cos\beta)\,\hat{\mathbf{x}} - z_5 c\sin\beta\,\hat{\mathbf{z}}$ | (4$i$) | O III |

**Geometry files:**

- CIF: pp. 1527
- POSCAR: pp. 1527

# $K_2Ti_2O_5$ Structure: A2B5C2_mC18_12_i_a2i_i

http://aflow.org/prototype-encyclopedia/A2B5C2_mC18_12_i_a2i_i

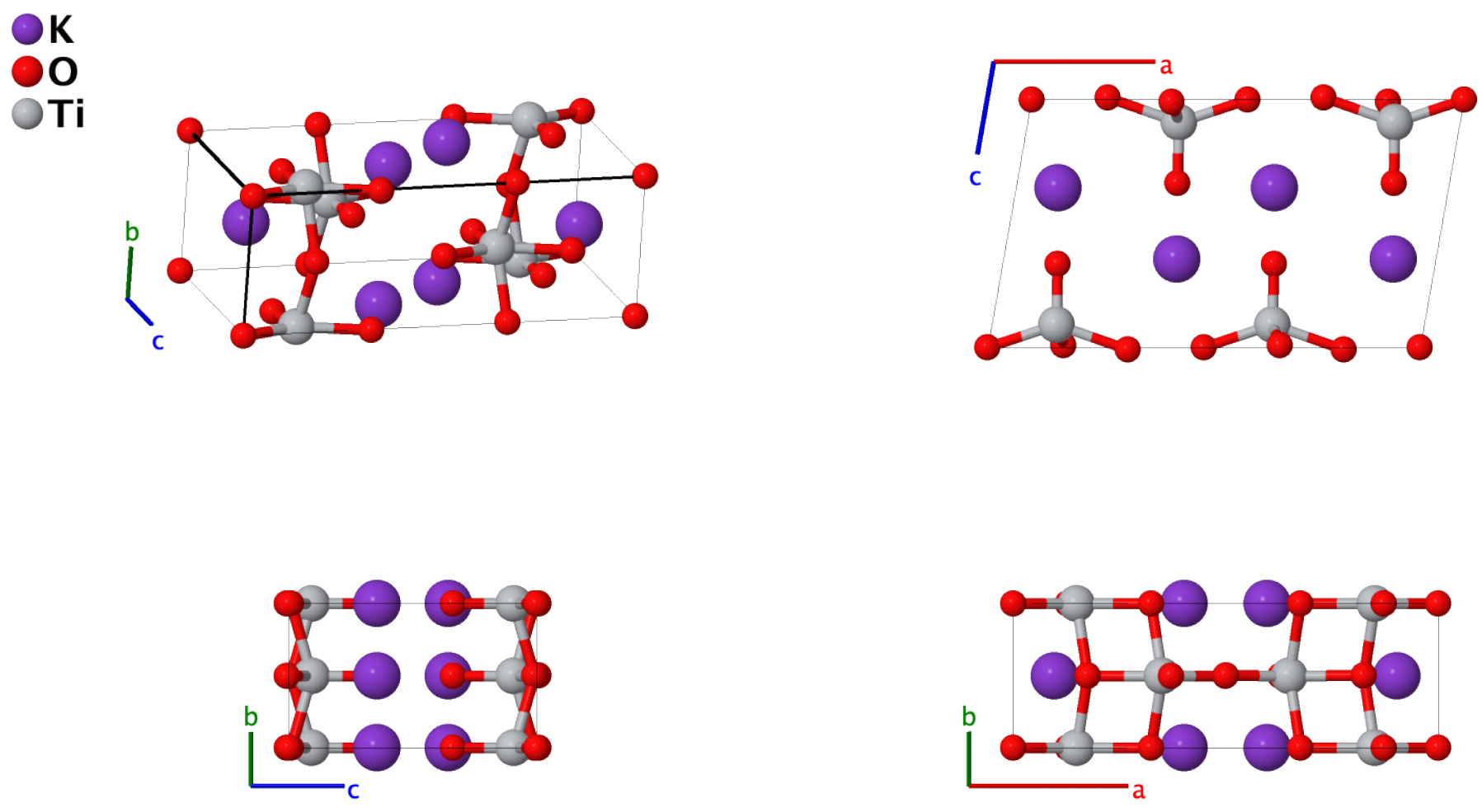

| | | |
|---|---|---|
| **Prototype** | : | $K_2O_5Ti_2$ |
| **AFLOW prototype label** | : | A2B5C2_mC18_12_i_a2i_i |
| ***Strukturbericht* designation** | : | None |
| **Pearson symbol** | : | mC18 |
| **Space group number** | : | 12 |
| **Space group symbol** | : | $C2/m$ |
| **AFLOW prototype command** | : | `aflow --proto=A2B5C2_mC18_12_i_a2i_i` `--params=`$a, b/a, c/a, \beta, x_2, z_2, x_3, z_3, x_4, z_4, x_5, z_5$ |

**Base-centered Monoclinic primitive vectors:**

$$\mathbf{a}_1 = \frac{1}{2}\,a\,\hat{\mathbf{x}} - \frac{1}{2}\,b\,\hat{\mathbf{y}}$$
$$\mathbf{a}_2 = \frac{1}{2}\,a\,\hat{\mathbf{x}} + \frac{1}{2}\,b\,\hat{\mathbf{y}}$$
$$\mathbf{a}_3 = c\cos\beta\,\hat{\mathbf{x}} + c\sin\beta\,\hat{\mathbf{z}}$$

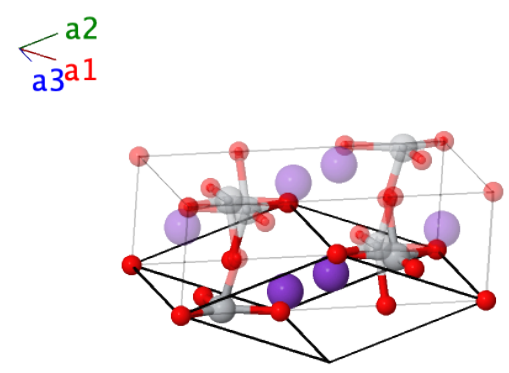

**Basis vectors:**

| | | Lattice Coordinates | | Cartesian Coordinates | Wyckoff Position | Atom Type |
|---|---|---|---|---|---|---|
| $\mathbf{B}_1$ | = | $0\,\mathbf{a}_1 + 0\,\mathbf{a}_2 + 0\,\mathbf{a}_3$ | = | $0\,\hat{\mathbf{x}} + 0\,\hat{\mathbf{y}} + 0\,\hat{\mathbf{z}}$ | (2a) | O I |
| $\mathbf{B}_2$ | = | $x_2\,\mathbf{a}_1 + x_2\,\mathbf{a}_2 + z_2\,\mathbf{a}_3$ | = | $(x_2 a + z_2 c\cos\beta)\,\hat{\mathbf{x}} + z_2 c\sin\beta\,\hat{\mathbf{z}}$ | (4i) | K |
| $\mathbf{B}_3$ | = | $-x_2\,\mathbf{a}_1 - x_2\,\mathbf{a}_2 - z_2\,\mathbf{a}_3$ | = | $(-x_2 a - z_2 c\cos\beta)\,\hat{\mathbf{x}} - z_2 c\sin\beta\,\hat{\mathbf{z}}$ | (4i) | K |
| $\mathbf{B}_4$ | = | $x_3\,\mathbf{a}_1 + x_3\,\mathbf{a}_2 + z_3\,\mathbf{a}_3$ | = | $(x_3 a + z_3 c\cos\beta)\,\hat{\mathbf{x}} + z_3 c\sin\beta\,\hat{\mathbf{z}}$ | (4i) | O II |
| $\mathbf{B}_5$ | = | $-x_3\,\mathbf{a}_1 - x_3\,\mathbf{a}_2 - z_3\,\mathbf{a}_3$ | = | $(-x_3 a - z_3 c\cos\beta)\,\hat{\mathbf{x}} - z_3 c\sin\beta\,\hat{\mathbf{z}}$ | (4i) | O II |
| $\mathbf{B}_6$ | = | $x_4\,\mathbf{a}_1 + x_4\,\mathbf{a}_2 + z_4\,\mathbf{a}_3$ | = | $(x_4 a + z_4 c\cos\beta)\,\hat{\mathbf{x}} + z_4 c\sin\beta\,\hat{\mathbf{z}}$ | (4i) | O III |
| $\mathbf{B}_7$ | = | $-x_4\,\mathbf{a}_1 - x_4\,\mathbf{a}_2 - z_4\,\mathbf{a}_3$ | = | $(-x_4 a - z_4 c\cos\beta)\,\hat{\mathbf{x}} - z_4 c\sin\beta\,\hat{\mathbf{z}}$ | (4i) | O III |
| $\mathbf{B}_8$ | = | $x_5\,\mathbf{a}_1 + x_5\,\mathbf{a}_2 + z_5\,\mathbf{a}_3$ | = | $(x_5 a + z_5 c\cos\beta)\,\hat{\mathbf{x}} + z_5 c\sin\beta\,\hat{\mathbf{z}}$ | (4$i$) | Ti |
| $\mathbf{B}_9$ | = | $-x_5\,\mathbf{a}_1 - x_5\,\mathbf{a}_2 - z_5\,\mathbf{a}_3$ | = | $(-x_5 a - z_5 c\cos\beta)\,\hat{\mathbf{x}} - z_5 c\sin\beta\,\hat{\mathbf{z}}$ | (4$i$) | Ti |

**Geometry files:**

- CIF: pp. 1527
- POSCAR: pp. 1527

# $CaC_2$-III Structure: A2B_mC12_12_2i_i

http://aflow.org/prototype-encyclopedia/A2B_mC12_12_2i_i

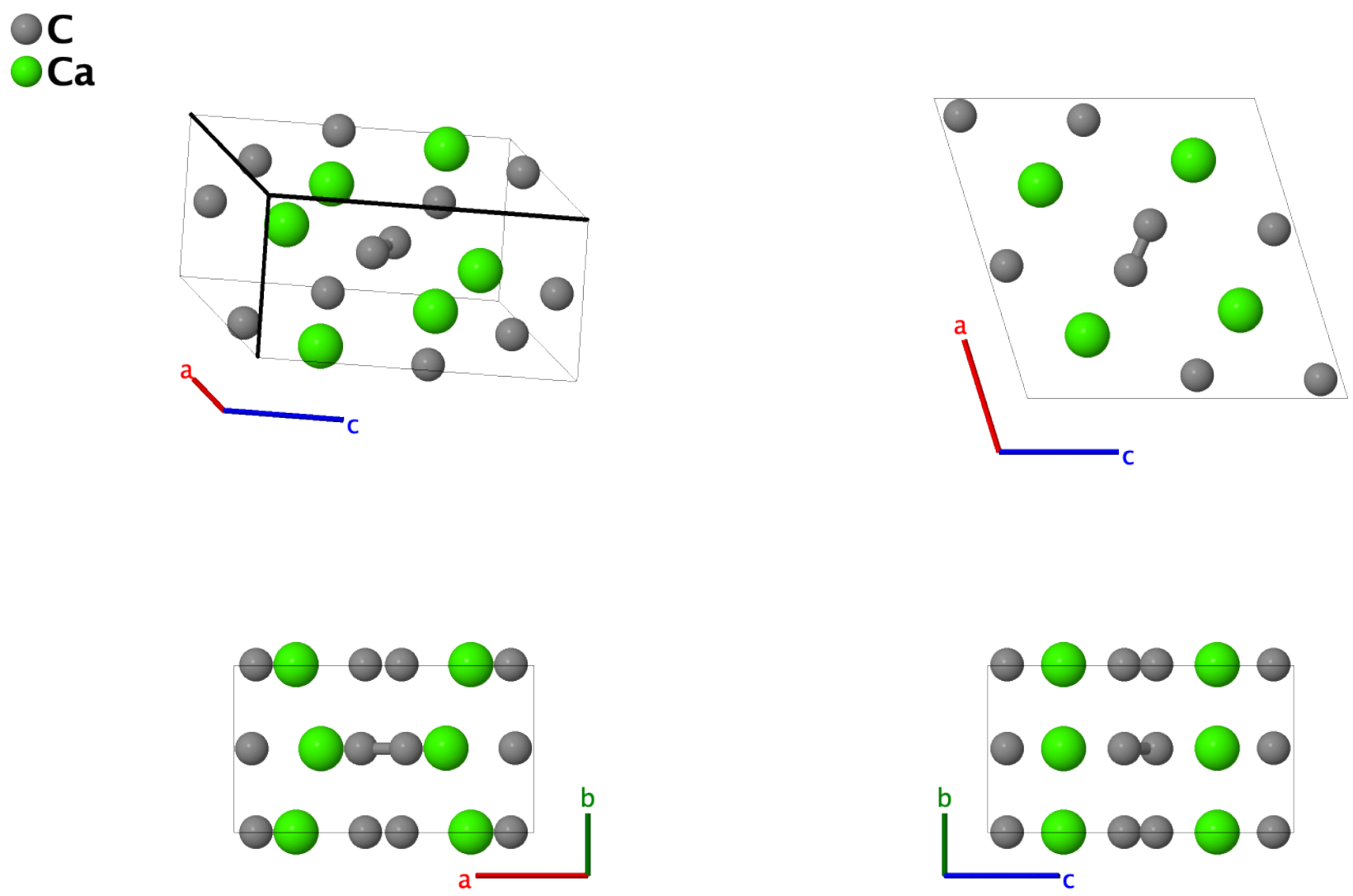

| | | |
|---|---|---|
| **Prototype** | : | $C_2Ca$ |
| **AFLOW prototype label** | : | A2B_mC12_12_2i_i |
| ***Strukturbericht* designation** | : | None |
| **Pearson symbol** | : | mC12 |
| **Space group number** | : | 12 |
| **Space group symbol** | : | $C2/m$ |
| **AFLOW prototype command** | : | `aflow --proto=A2B_mC12_12_2i_i`<br>`--params=`$a, b/a, c/a, \beta, x_1, z_1, x_2, z_2, x_3, z_3$ |

- This is a meta-stable room temperature structure.
- The stable room temperature configuration has the $C11_a$ structure.
- The low-temperature $CaC_2$ crystallizes in the $ThC_2$ ($C_g$) structure.

**Base-centered Monoclinic primitive vectors:**

$$\mathbf{a}_1 = \frac{1}{2}\,a\,\hat{\mathbf{x}} - \frac{1}{2}\,b\,\hat{\mathbf{y}}$$
$$\mathbf{a}_2 = \frac{1}{2}\,a\,\hat{\mathbf{x}} + \frac{1}{2}\,b\,\hat{\mathbf{y}}$$
$$\mathbf{a}_3 = c\cos\beta\,\hat{\mathbf{x}} + c\sin\beta\,\hat{\mathbf{z}}$$

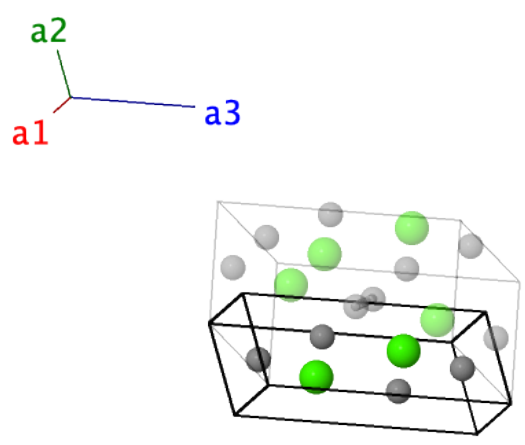

**Basis vectors:**

| | | Lattice Coordinates | | Cartesian Coordinates | Wyckoff Position | Atom Type |
|---|---|---|---|---|---|---|
| $\mathbf{B}_1$ | = | $x_1\,\mathbf{a}_1 + x_1\,\mathbf{a}_2 + z_1\,\mathbf{a}_3$ | = | $(x_1 a + z_1 c\cos\beta)\,\hat{\mathbf{x}} + z_1 c\sin\beta\,\hat{\mathbf{z}}$ | (4$i$) | C I |
| $\mathbf{B}_2$ | = | $-x_1\,\mathbf{a}_1 - x_1\,\mathbf{a}_2 - z_1\,\mathbf{a}_3$ | = | $(-x_1 a - z_1 c \cos\beta)\,\hat{\mathbf{x}} - z_1 c \sin\beta\,\hat{\mathbf{z}}$ | (4i) | C I |
| $\mathbf{B}_3$ | = | $x_2\,\mathbf{a}_1 + x_2\,\mathbf{a}_2 + z_2\,\mathbf{a}_3$ | = | $(x_2 a + z_2 c \cos\beta)\,\hat{\mathbf{x}} + z_2 c \sin\beta\,\hat{\mathbf{z}}$ | (4i) | C II |
| $\mathbf{B}_4$ | = | $-x_2\,\mathbf{a}_1 - x_2\,\mathbf{a}_2 - z_2\,\mathbf{a}_3$ | = | $(-x_2 a - z_2 c \cos\beta)\,\hat{\mathbf{x}} - z_2 c \sin\beta\,\hat{\mathbf{z}}$ | (4i) | C II |
| $\mathbf{B}_5$ | = | $x_3\,\mathbf{a}_1 + x_3\,\mathbf{a}_2 + z_3\,\mathbf{a}_3$ | = | $(x_3 a + z_3 c \cos\beta)\,\hat{\mathbf{x}} + z_3 c \sin\beta\,\hat{\mathbf{z}}$ | (4i) | Ca |
| $\mathbf{B}_6$ | = | $-x_3\,\mathbf{a}_1 - x_3\,\mathbf{a}_2 - z_3\,\mathbf{a}_3$ | = | $(-x_3 a - z_3 c \cos\beta)\,\hat{\mathbf{x}} - z_3 c \sin\beta\,\hat{\mathbf{z}}$ | (4i) | Ca |

---

---

**Geometry files:**

- CIF: pp. 1527
- POSCAR: pp. 1528

# Tolbachite ($CuCl_2$) Structure: A2B_mC6_12_i_a

http://aflow.org/prototype-encyclopedia/A2B_mC6_12_i_a

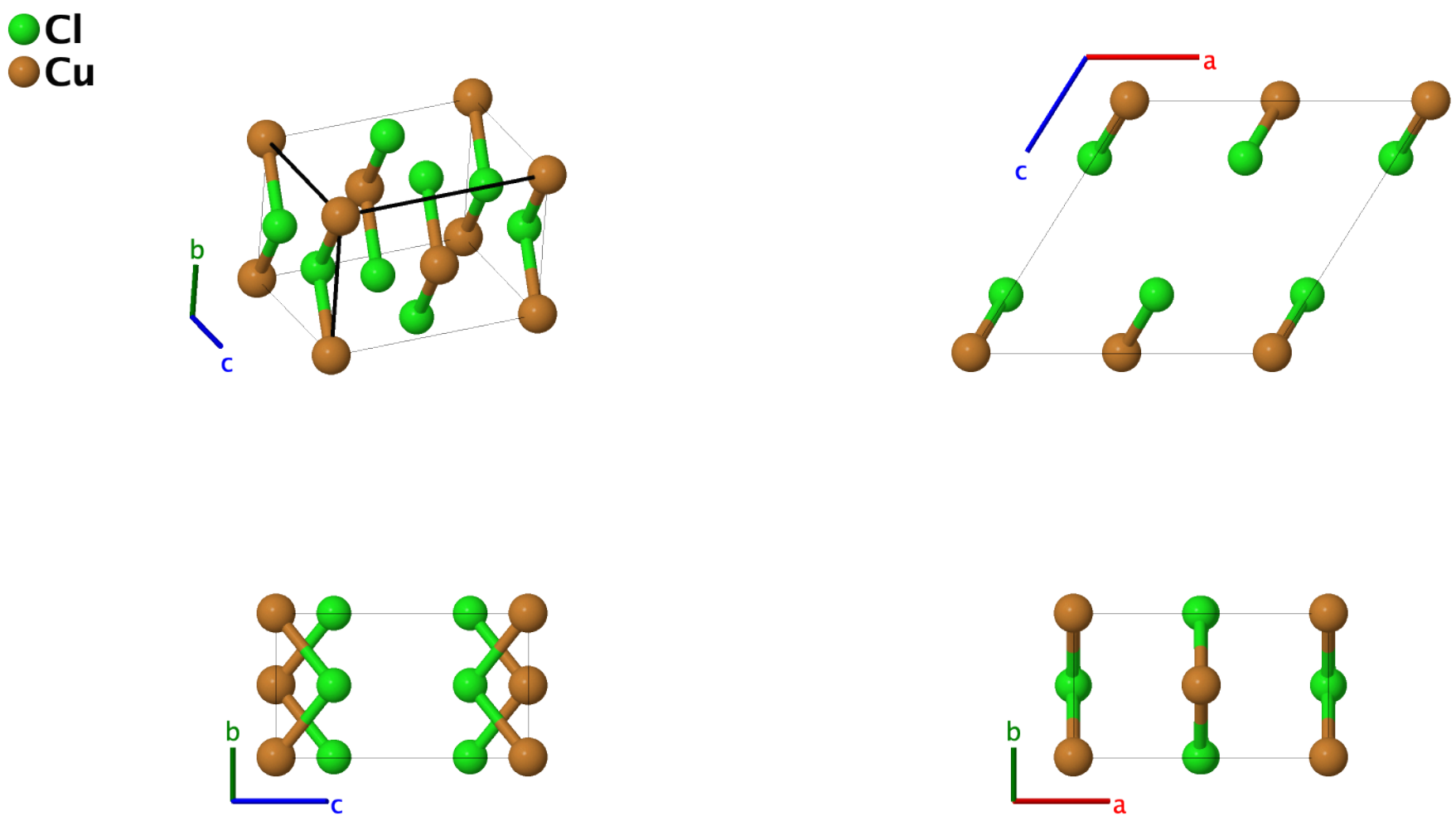

| | | |
|---|---|---|
| **Prototype** | : | $Cl_2Cu$ |
| **AFLOW prototype label** | : | A2B_mC6_12_i_a |
| ***Strukturbericht* designation** | : | None |
| **Pearson symbol** | : | mC6 |
| **Space group number** | : | 12 |
| **Space group symbol** | : | $C2/m$ |
| **AFLOW prototype command** | : | `aflow --proto=A2B_mC6_12_i_a`<br>`--params=`$a, b/a, c/a, \beta, x_2, z_2$ |

### Other compounds with this structure

- $CuBr_2$

### Base-centered Monoclinic primitive vectors:

$$
\begin{aligned}
\mathbf{a}_1 &= \tfrac{1}{2}\, a\,\hat{\mathbf{x}} - \tfrac{1}{2}\, b\,\hat{\mathbf{y}} \\
\mathbf{a}_2 &= \tfrac{1}{2}\, a\,\hat{\mathbf{x}} + \tfrac{1}{2}\, b\,\hat{\mathbf{y}} \\
\mathbf{a}_3 &= c\cos\beta\,\hat{\mathbf{x}} + c\sin\beta\,\hat{\mathbf{z}}
\end{aligned}
$$

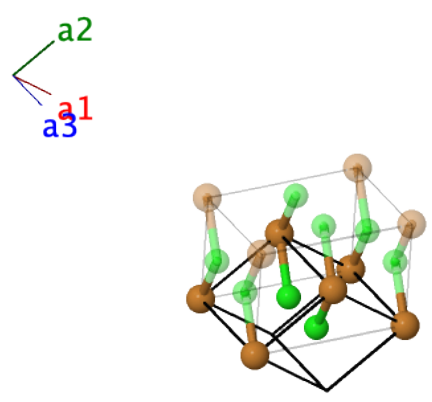

### Basis vectors:

| | | Lattice Coordinates | | Cartesian Coordinates | Wyckoff Position | Atom Type |
|---|---|---|---|---|---|---|
| $\mathbf{B}_1$ | = | $0\,\mathbf{a}_1 + 0\,\mathbf{a}_2 + 0\,\mathbf{a}_3$ | = | $0\,\hat{\mathbf{x}} + 0\,\hat{\mathbf{y}} + 0\,\hat{\mathbf{z}}$ | (2$a$) | Cu |
| $\mathbf{B}_2$ | = | $x_2\,\mathbf{a}_1 + x_2\,\mathbf{a}_2 + z_2\,\mathbf{a}_3$ | = | $(x_2 a + z_2 c\cos\beta)\,\hat{\mathbf{x}} + z_2 c\sin\beta\,\hat{\mathbf{z}}$ | (4$i$) | Cl |
| $\mathbf{B}_3$ | = | $-x_2\,\mathbf{a}_1 - x_2\,\mathbf{a}_2 - z_2\,\mathbf{a}_3$ | = | $(-x_2 a - z_2 c\cos\beta)\,\hat{\mathbf{x}} - z_2 c\sin\beta\,\hat{\mathbf{z}}$ | (4$i$) | Cl |

**Found in:**

---

**Geometry files:**

- CIF: pp. 1528
- POSCAR: pp. 1528

# $\delta$-$Ni_3Sn_4$ ($D7_a$) Structure: A3B4_mC14_12_ai_2i

http://aflow.org/prototype-encyclopedia/A3B4_mC14_12_ai_2i

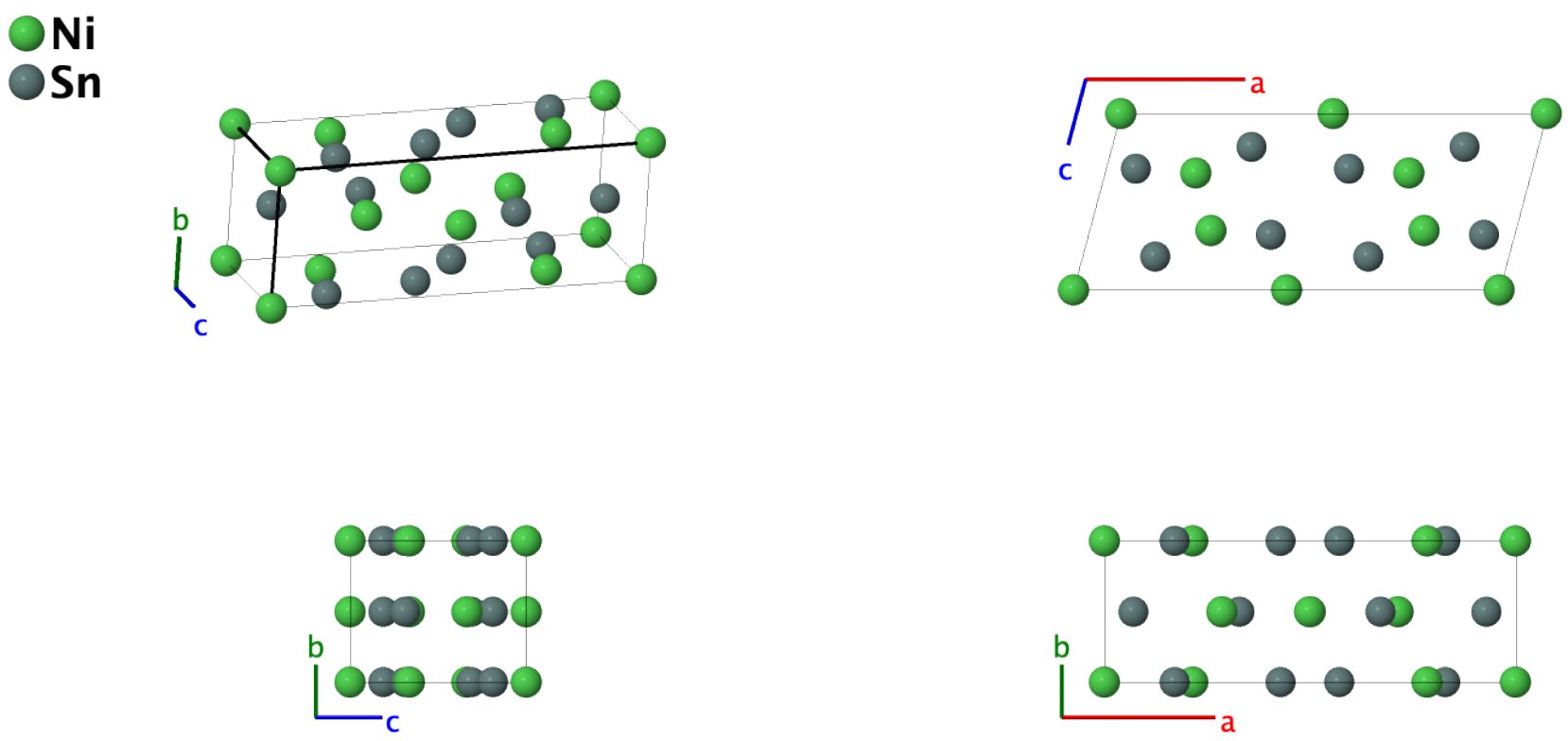

| | | |
|---|---|---|
| **Prototype** | : | $Ni_3Sn_4$ |
| **AFLOW prototype label** | : | A3B4_mC14_12_ai_2i |
| ***Strukturbericht* designation** | : | $D7_a$ |
| **Pearson symbol** | : | mC14 |
| **Space group number** | : | 12 |
| **Space group symbol** | : | $C2/m$ |
| **AFLOW prototype command** | : | `aflow --proto=A3B4_mC14_12_ai_2i` `--params=`$a, b/a, c/a, \beta, x_2, z_2, x_3, z_3, x_4, z_4$ |

**Other compounds with this structure**

- $Cr_3S_4$, $Cr_3Se_4$, $Ni_3Sn_4$, $Fe_3Se_4$, $Ni_3Se_4$, $Ti_3Se_4$, and $V_3Se_4$

**Base-centered Monoclinic primitive vectors:**

$$\mathbf{a}_1 = \frac{1}{2}\,a\,\hat{\mathbf{x}} - \frac{1}{2}\,b\,\hat{\mathbf{y}}$$
$$\mathbf{a}_2 = \frac{1}{2}\,a\,\hat{\mathbf{x}} + \frac{1}{2}\,b\,\hat{\mathbf{y}}$$
$$\mathbf{a}_3 = c\cos\beta\,\hat{\mathbf{x}} + c\sin\beta\,\hat{\mathbf{z}}$$

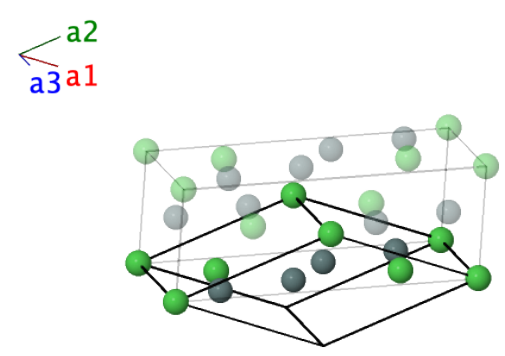

**Basis vectors:**

| | | Lattice Coordinates | | Cartesian Coordinates | Wyckoff Position | Atom Type |
|---|---|---|---|---|---|---|
| $\mathbf{B}_1$ | = | $0\,\mathbf{a}_1 + 0\,\mathbf{a}_2 + 0\,\mathbf{a}_3$ | = | $0\,\hat{\mathbf{x}} + 0\,\hat{\mathbf{y}} + 0\,\hat{\mathbf{z}}$ | (2$a$) | Ni I |
| $\mathbf{B}_2$ | = | $x_2\,\mathbf{a}_1 + x_2\,\mathbf{a}_2 + z_2\,\mathbf{a}_3$ | = | $(x_2a + z_2c\cos\beta)\,\hat{\mathbf{x}} + z_2c\sin\beta\,\hat{\mathbf{z}}$ | (4$i$) | Ni II |
| $\mathbf{B}_3$ | = | $-x_2\,\mathbf{a}_1 - x_2\,\mathbf{a}_2 - z_2\,\mathbf{a}_3$ | = | $(-x_2a - z_2c\cos\beta)\,\hat{\mathbf{x}} - z_2c\sin\beta\,\hat{\mathbf{z}}$ | (4$i$) | Ni II |
| $\mathbf{B}_4$ | = | $x_3\,\mathbf{a}_1 + x_3\,\mathbf{a}_2 + z_3\,\mathbf{a}_3$ | = | $(x_3a + z_3c\cos\beta)\,\hat{\mathbf{x}} + z_3c\sin\beta\,\hat{\mathbf{z}}$ | (4$i$) | Sn I |
| $\mathbf{B}_5$ | = | $-x_3\,\mathbf{a}_1 - x_3\,\mathbf{a}_2 - z_3\,\mathbf{a}_3$ | = | $(-x_3a - z_3c\cos\beta)\,\hat{\mathbf{x}} - z_3c\sin\beta\,\hat{\mathbf{z}}$ | (4$i$) | Sn I |
| $\mathbf{B}_6$ | = | $x_4\,\mathbf{a}_1 + x_4\,\mathbf{a}_2 + z_4\,\mathbf{a}_3$ | = | $(x_4a + z_4c\cos\beta)\,\hat{\mathbf{x}} + z_4c\sin\beta\,\hat{\mathbf{z}}$ | (4$i$) | Sn II |
| $\mathbf{B}_7$ | = | $-x_4\,\mathbf{a}_1 - x_4\,\mathbf{a}_2 - z_4\,\mathbf{a}_3$ | = | $(-x_4a - z_4c\cos\beta)\,\hat{\mathbf{x}} - z_4c\sin\beta\,\hat{\mathbf{z}}$ | (4$i$) | Sn II |

**Geometry files:**

- CIF: pp. 1528
- POSCAR: pp. 1529

# LiOH·$H_2O$ ($B36$) Structure: A3BC2_mC24_12_ij_h_gi

http://aflow.org/prototype-encyclopedia/A3BC2_mC24_12_ij_h_gi

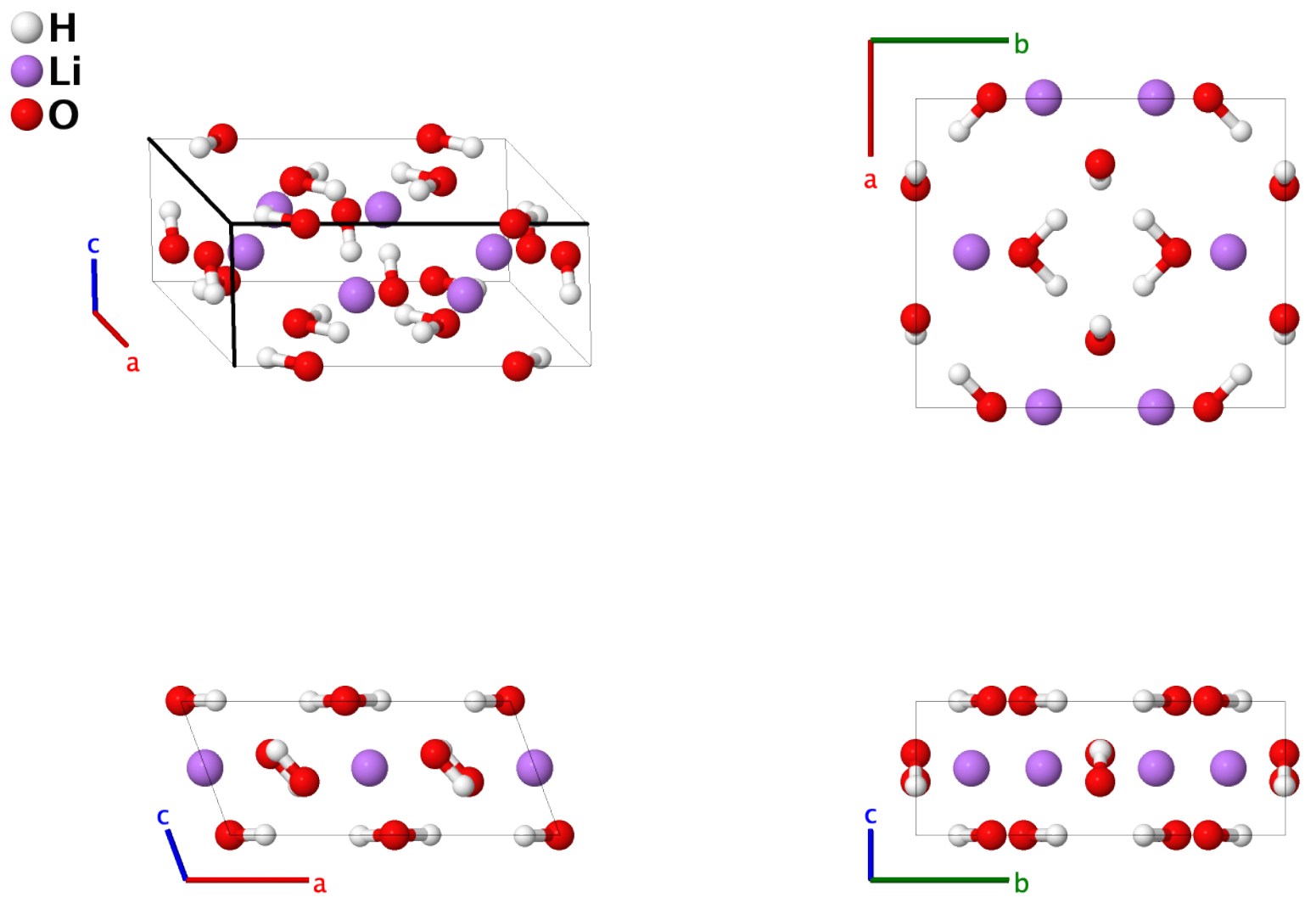

| | | |
|---|---|---|
| **Prototype** | : | $H_3LiO_2$ |
| **AFLOW prototype label** | : | A3BC2_mC24_12_ij_h_gi |
| ***Strukturbericht* designation** | : | $B36$ |
| **Pearson symbol** | : | mC24 |
| **Space group number** | : | 12 |
| **Space group symbol** | : | $C2/m$ |
| **AFLOW prototype command** | : | `aflow --proto=A3BC2_mC24_12_ij_h_gi` <br> `--params=`$a, b/a, c/a, \beta, y_1, y_2, x_3, z_3, x_4, z_4, x_5, y_5, z_5$ |

- The structure given in *Strukturbericht* (Herrmann, 1943) does not give the hydrogen positions, instead giving positions for Li, OH and $H_2O$.

**Base-centered Monoclinic primitive vectors:**

$$\mathbf{a}_1 = \frac{1}{2}\,a\,\hat{\mathbf{x}} - \frac{1}{2}\,b\,\hat{\mathbf{y}}$$
$$\mathbf{a}_2 = \frac{1}{2}\,a\,\hat{\mathbf{x}} + \frac{1}{2}\,b\,\hat{\mathbf{y}}$$
$$\mathbf{a}_3 = c\cos\beta\,\hat{\mathbf{x}} + c\sin\beta\,\hat{\mathbf{z}}$$

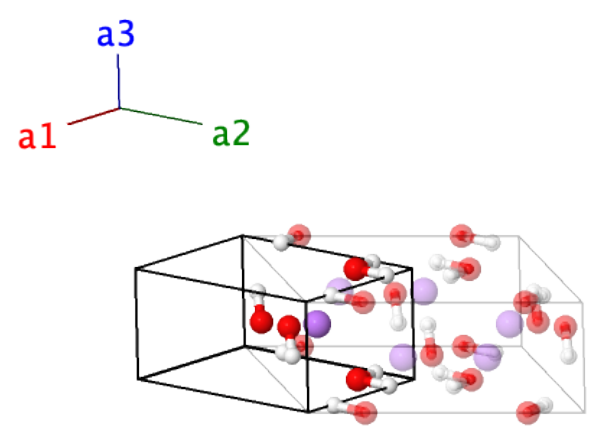

**Basis vectors:**

| | | Lattice Coordinates | | Cartesian Coordinates | Wyckoff Position | Atom Type |
|---|---|---|---|---|---|---|
| $\mathbf{B}_1$ | = | $-y_1\,\mathbf{a}_1 + y_1\,\mathbf{a}_2$ | = | $y_1\,b\,\hat{\mathbf{y}}$ | (4$g$) | O I |
| $\mathbf{B}_2$ | = | $y_1\,\mathbf{a}_1 - y_1\,\mathbf{a}_2$ | = | $-y_1\,b\,\hat{\mathbf{y}}$ | (4$g$) | O I |
| $\mathbf{B}_3$ | = | $-y_2\,\mathbf{a}_1 + y_2\,\mathbf{a}_2 + \frac{1}{2}\,\mathbf{a}_3$ | = | $\frac{1}{2}c\cos\beta\,\hat{\mathbf{x}} + y_2 b\,\hat{\mathbf{y}} + \frac{1}{2}c\sin\beta\,\hat{\mathbf{z}}$ | (4$h$) | Li |
| $\mathbf{B}_4$ | = | $y_2\,\mathbf{a}_1 - y_2\,\mathbf{a}_2 + \frac{1}{2}\,\mathbf{a}_3$ | = | $\frac{1}{2}c\cos\beta\,\hat{\mathbf{x}} - y_2 b\,\hat{\mathbf{y}} + \frac{1}{2}c\sin\beta\,\hat{\mathbf{z}}$ | (4$h$) | Li |
| $\mathbf{B}_5$ | = | $x_3\,\mathbf{a}_1 + x_3\,\mathbf{a}_2 + z_3\,\mathbf{a}_3$ | = | $(x_3 a + z_3 c\cos\beta)\,\hat{\mathbf{x}} + z_3 c\sin\beta\,\hat{\mathbf{z}}$ | (4$i$) | H I |
| $\mathbf{B}_6$ | = | $-x_3\,\mathbf{a}_1 - x_3\,\mathbf{a}_2 - z_3\,\mathbf{a}_3$ | = | $(-x_3 a - z_3 c\cos\beta)\,\hat{\mathbf{x}} - z_3 c\sin\beta\,\hat{\mathbf{z}}$ | (4$i$) | H I |
| $\mathbf{B}_7$ | = | $x_4\,\mathbf{a}_1 + x_4\,\mathbf{a}_2 + z_4\,\mathbf{a}_3$ | = | $(x_4 a + z_4 c\cos\beta)\,\hat{\mathbf{x}} + z_4 c\sin\beta\,\hat{\mathbf{z}}$ | (4$i$) | O II |
| $\mathbf{B}_8$ | = | $-x_4\,\mathbf{a}_1 - x_4\,\mathbf{a}_2 - z_4\,\mathbf{a}_3$ | = | $(-x_4 a - z_4 c\cos\beta)\,\hat{\mathbf{x}} - z_4 c\sin\beta\,\hat{\mathbf{z}}$ | (4$i$) | O II |
| $\mathbf{B}_9$ | = | $(x_5 - y_5)\,\mathbf{a}_1 + (x_5 + y_5)\,\mathbf{a}_2 + z_5\,\mathbf{a}_3$ | = | $(x_5 a + z_5 c\cos\beta)\,\hat{\mathbf{x}} + y_5 b\,\hat{\mathbf{y}} + z_5 c\sin\beta\,\hat{\mathbf{z}}$ | (8$j$) | H II |
| $\mathbf{B}_{10}$ | = | $(-x_5 - y_5)\,\mathbf{a}_1 + (-x_5 + y_5)\,\mathbf{a}_2 - z_5\,\mathbf{a}_3$ | = | $(-x_5 a - z_5 c\cos\beta)\,\hat{\mathbf{x}} + y_5 b\,\hat{\mathbf{y}} - z_5 c\sin\beta\,\hat{\mathbf{z}}$ | (8$j$) | H II |
| $\mathbf{B}_{11}$ | = | $(-x_5 + y_5)\,\mathbf{a}_1 + (-x_5 - y_5)\,\mathbf{a}_2 - z_5\,\mathbf{a}_3$ | = | $(-x_5 a - z_5 c\cos\beta)\,\hat{\mathbf{x}} - y_5 b\,\hat{\mathbf{y}} - z_5 c\sin\beta\,\hat{\mathbf{z}}$ | (8$j$) | H II |
| $\mathbf{B}_{12}$ | = | $(x_5 + y_5)\,\mathbf{a}_1 + (x_5 - y_5)\,\mathbf{a}_2 + z_5\,\mathbf{a}_3$ | = | $(x_5 a + z_5 c\cos\beta)\,\hat{\mathbf{x}} - y_5 b\,\hat{\mathbf{y}} + z_5 c\sin\beta\,\hat{\mathbf{z}}$ | (8$j$) | H II |

**Geometry files:**

- CIF: pp. 1529
- POSCAR: pp. 1529

# Staurolite ($Al_5Fe_2O_{10}(OH)_2Si_2$) Structure: A5B2C10D2E2_mC84_12_acghj_bdi_5j_2i_j

http://aflow.org/prototype-encyclopedia/A5B2C10D2E2_mC84_12_acghj_bdi_5j_2i_j

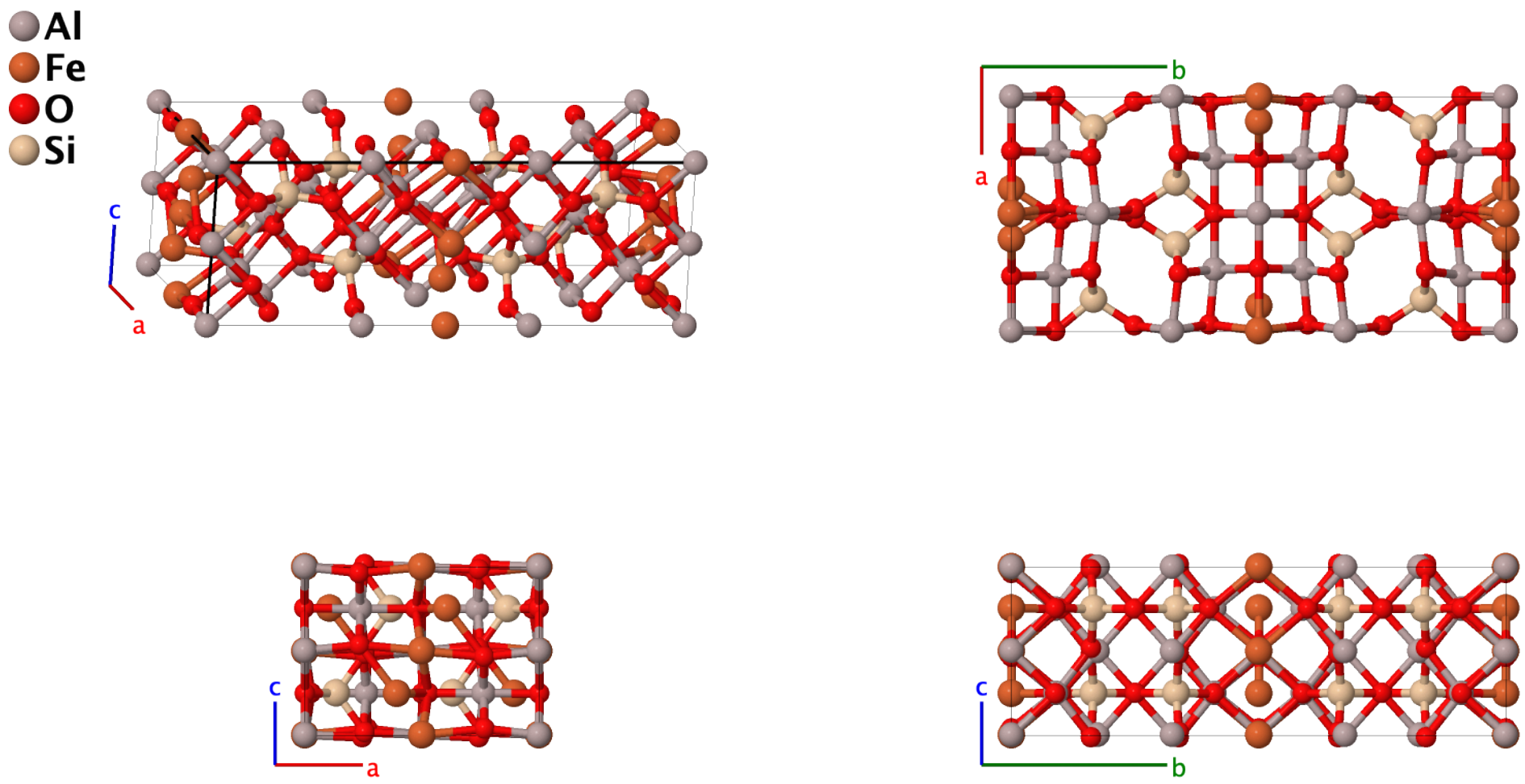

| | | |
|---|---|---|
| **Prototype** | : | $Al_5Fe_2O_{10}(OH)_2Si_2$ |
| **AFLOW prototype label** | : | A5B2C10D2E2_mC84_12_acghj_bdi_5j_2i_j |
| ***Strukturbericht* designation** | : | None |
| **Pearson symbol** | : | mC84 |
| **Space group number** | : | 12 |
| **Space group symbol** | : | $C2/m$ |
| **AFLOW prototype command** | : | `aflow --proto=A5B2C10D2E2_mC84_12_acghj_bdi_5j_2i_j --params=`$a, b/a, c/a, \beta, y_5, y_6, x_7, z_7, x_8, z_8, x_9, z_9, x_{10}, y_{10}, z_{10}, x_{11}, y_{11}, z_{11}, x_{12}, y_{12}, z_{12}, x_{13}, y_{13}, z_{13}, x_{14}, y_{14}, z_{14}, x_{15}, y_{15}, z_{15}, x_{16}, y_{16}, z_{16}$ |

- The orthorhombic structure of staurolite determined by (Náray-Szabó, 1929) was given the *Strukturbericht* designation $S0_4$ by (Hermann, 1937). (Smith, 1968) showed that the structure is actually monoclinic with $\beta \approx 90°$, rather than orthorhombic. This paper also corrected the chemical composition of the mineral.
- The hydrogen positions are undetermined, but they part of a "complex distribution of OH ions," and are "probably" associated with the atoms on the (4*i*) sites (Smith, 1968). We therefore label the (4*i*) sites as OH.
- The metallic sites are actually somewhat disordered. (Smith, 1968) gives the composition of the various sites as
  - Al (2*a*) $Al_{0.67}$ $Fe_{0.33}$
  - Fe (2*b*) $Fe_{0.68}$ $Mn_{0.32}$
  - Al (2*c*) $Al_{0.67}$ $Fe_{0.33}$
  - Fe (2*d*) $Fe_{0.68}$ $Mn_{0.32}$
  - Al (4*g*) $Al_{0.95}$ $Mg_{0.05}$
  - Al (4*h*) $Al_{0.95}$ $Mg_{0.05}$
  - Fe (4*i*) $Fe_{0.64}$ $Al_{0.32}$ $Ti_{0.04}$
  - Si (8*j*) $Si_{0.936}$ $Al_{0.064}$
- (Donnay, 1983) presents a history of the difficulties in determining the staurolite structure.

**Base-centered Monoclinic primitive vectors:**

$$\mathbf{a}_1 = \tfrac{1}{2}\,a\,\hat{\mathbf{x}} - \tfrac{1}{2}\,b\,\hat{\mathbf{y}}$$

$$\mathbf{a}_2 = \tfrac{1}{2}\,a\,\hat{\mathbf{x}} + \tfrac{1}{2}\,b\,\hat{\mathbf{y}}$$

$$\mathbf{a}_3 = c\cos\beta\,\hat{\mathbf{x}} + c\sin\beta\,\hat{\mathbf{z}}$$

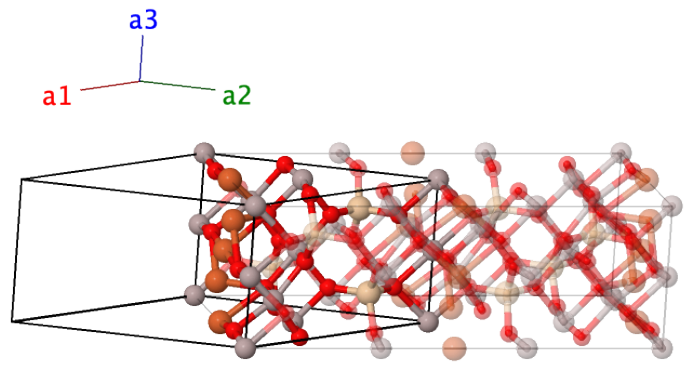

**Basis vectors:**

| | | Lattice Coordinates | | Cartesian Coordinates | Wyckoff Position | Atom Type |
|---|---|---|---|---|---|---|
| $\mathbf{B}_1$ | = | $0\,\mathbf{a}_1 + 0\,\mathbf{a}_2 + 0\,\mathbf{a}_3$ | = | $0\,\hat{\mathbf{x}} + 0\,\hat{\mathbf{y}} + 0\,\hat{\mathbf{z}}$ | (2$a$) | Al I |
| $\mathbf{B}_2$ | = | $\frac{1}{2}\,\mathbf{a}_1 + \frac{1}{2}\,\mathbf{a}_2$ | = | $\frac{1}{2}a\,\hat{\mathbf{x}}$ | (2$b$) | Fe I |
| $\mathbf{B}_3$ | = | $\frac{1}{2}\,\mathbf{a}_3$ | = | $\frac{1}{2}c\cos\beta\,\hat{\mathbf{x}} + \frac{1}{2}c\sin\beta\,\hat{\mathbf{z}}$ | (2$c$) | Al II |
| $\mathbf{B}_4$ | = | $\frac{1}{2}\,\mathbf{a}_1 + \frac{1}{2}\,\mathbf{a}_2 + \frac{1}{2}\,\mathbf{a}_3$ | = | $\frac{1}{2}\left(a + c\cos\beta\right)\,\hat{\mathbf{x}} + \frac{1}{2}c\sin\beta\,\hat{\mathbf{z}}$ | (2$d$) | Fe II |
| $\mathbf{B}_5$ | = | $-y_5\,\mathbf{a}_1 + y_5\,\mathbf{a}_2$ | = | $y_5 b\,\hat{\mathbf{y}}$ | (4$g$) | Al III |
| $\mathbf{B}_6$ | = | $y_5\,\mathbf{a}_1 - y_5\,\mathbf{a}_2$ | = | $-y_5 b\,\hat{\mathbf{y}}$ | (4$g$) | Al III |
| $\mathbf{B}_7$ | = | $-y_6\,\mathbf{a}_1 + y_6\,\mathbf{a}_2 + \frac{1}{2}\,\mathbf{a}_3$ | = | $\frac{1}{2}c\cos\beta\,\hat{\mathbf{x}} + y_6 b\,\hat{\mathbf{y}} + \frac{1}{2}c\sin\beta\,\hat{\mathbf{z}}$ | (4$h$) | Al IV |
| $\mathbf{B}_8$ | = | $y_6\,\mathbf{a}_1 - y_6\,\mathbf{a}_2 + \frac{1}{2}\,\mathbf{a}_3$ | = | $\frac{1}{2}c\cos\beta\,\hat{\mathbf{x}} - y_6 b\,\hat{\mathbf{y}} + \frac{1}{2}c\sin\beta\,\hat{\mathbf{z}}$ | (4$h$) | Al IV |
| $\mathbf{B}_9$ | = | $x_7\,\mathbf{a}_1 + x_7\,\mathbf{a}_2 + z_7\,\mathbf{a}_3$ | = | $\left(x_7 a + z_7 c\cos\beta\right)\,\hat{\mathbf{x}} + z_7 c\sin\beta\,\hat{\mathbf{z}}$ | (4$i$) | Fe III |
| $\mathbf{B}_{10}$ | = | $-x_7\,\mathbf{a}_1 - x_7\,\mathbf{a}_2 - z_7\,\mathbf{a}_3$ | = | $\left(-x_7 a - z_7 c\cos\beta\right)\,\hat{\mathbf{x}} - z_7 c\sin\beta\,\hat{\mathbf{z}}$ | (4$i$) | Fe III |
| $\mathbf{B}_{11}$ | = | $x_8\,\mathbf{a}_1 + x_8\,\mathbf{a}_2 + z_8\,\mathbf{a}_3$ | = | $\left(x_8 a + z_8 c\cos\beta\right)\,\hat{\mathbf{x}} + z_8 c\sin\beta\,\hat{\mathbf{z}}$ | (4$i$) | OH I |
| $\mathbf{B}_{12}$ | = | $-x_8\,\mathbf{a}_1 - x_8\,\mathbf{a}_2 - z_8\,\mathbf{a}_3$ | = | $\left(-x_8 a - z_8 c\cos\beta\right)\,\hat{\mathbf{x}} - z_8 c\sin\beta\,\hat{\mathbf{z}}$ | (4$i$) | OH I |
| $\mathbf{B}_{13}$ | = | $x_9\,\mathbf{a}_1 + x_9\,\mathbf{a}_2 + z_9\,\mathbf{a}_3$ | = | $\left(x_9 a + z_9 c\cos\beta\right)\,\hat{\mathbf{x}} + z_9 c\sin\beta\,\hat{\mathbf{z}}$ | (4$i$) | OH II |
| $\mathbf{B}_{14}$ | = | $-x_9\,\mathbf{a}_1 - x_9\,\mathbf{a}_2 - z_9\,\mathbf{a}_3$ | = | $\left(-x_9 a - z_9 c\cos\beta\right)\,\hat{\mathbf{x}} - z_9 c\sin\beta\,\hat{\mathbf{z}}$ | (4$i$) | OH II |
| $\mathbf{B}_{15}$ | = | $\left(x_{10} - y_{10}\right)\,\mathbf{a}_1 + \left(x_{10} + y_{10}\right)\,\mathbf{a}_2 + z_{10}\,\mathbf{a}_3$ | = | $\left(x_{10} a + z_{10} c\cos\beta\right)\,\hat{\mathbf{x}} + y_{10} b\,\hat{\mathbf{y}} + z_{10} c\sin\beta\,\hat{\mathbf{z}}$ | (8$j$) | Al V |
| $\mathbf{B}_{16}$ | = | $\left(-x_{10} - y_{10}\right)\,\mathbf{a}_1 + \left(-x_{10} + y_{10}\right)\,\mathbf{a}_2 - z_{10}\,\mathbf{a}_3$ | = | $\left(-x_{10} a - z_{10} c\cos\beta\right)\,\hat{\mathbf{x}} + y_{10} b\,\hat{\mathbf{y}} - z_{10} c\sin\beta\,\hat{\mathbf{z}}$ | (8$j$) | Al V |
| $\mathbf{B}_{17}$ | = | $\left(-x_{10} + y_{10}\right)\,\mathbf{a}_1 + \left(-x_{10} - y_{10}\right)\,\mathbf{a}_2 - z_{10}\,\mathbf{a}_3$ | = | $\left(-x_{10} a - z_{10} c\cos\beta\right)\,\hat{\mathbf{x}} - y_{10} b\,\hat{\mathbf{y}} - z_{10} c\sin\beta\,\hat{\mathbf{z}}$ | (8$j$) | Al V |
| $\mathbf{B}_{18}$ | = | $\left(x_{10} + y_{10}\right)\,\mathbf{a}_1 + \left(x_{10} - y_{10}\right)\,\mathbf{a}_2 + z_{10}\,\mathbf{a}_3$ | = | $\left(x_{10} a + z_{10} c\cos\beta\right)\,\hat{\mathbf{x}} - y_{10} b\,\hat{\mathbf{y}} + z_{10} c\sin\beta\,\hat{\mathbf{z}}$ | (8$j$) | Al V |
| $\mathbf{B}_{19}$ | = | $\left(x_{11} - y_{11}\right)\,\mathbf{a}_1 + \left(x_{11} + y_{11}\right)\,\mathbf{a}_2 + z_{11}\,\mathbf{a}_3$ | = | $\left(x_{11} a + z_{11} c\cos\beta\right)\,\hat{\mathbf{x}} + y_{11} b\,\hat{\mathbf{y}} + z_{11} c\sin\beta\,\hat{\mathbf{z}}$ | (8$j$) | O I |
| $\mathbf{B}_{20}$ | = | $\left(-x_{11} - y_{11}\right)\,\mathbf{a}_1 + \left(-x_{11} + y_{11}\right)\,\mathbf{a}_2 - z_{11}\,\mathbf{a}_3$ | = | $\left(-x_{11} a - z_{11} c\cos\beta\right)\,\hat{\mathbf{x}} + y_{11} b\,\hat{\mathbf{y}} - z_{11} c\sin\beta\,\hat{\mathbf{z}}$ | (8$j$) | O I |
| $\mathbf{B}_{21}$ | = | $\left(-x_{11} + y_{11}\right)\,\mathbf{a}_1 + \left(-x_{11} - y_{11}\right)\,\mathbf{a}_2 - z_{11}\,\mathbf{a}_3$ | = | $\left(-x_{11} a - z_{11} c\cos\beta\right)\,\hat{\mathbf{x}} - y_{11} b\,\hat{\mathbf{y}} - z_{11} c\sin\beta\,\hat{\mathbf{z}}$ | (8$j$) | O I |
| $\mathbf{B}_{22}$ | = | $\left(x_{11} + y_{11}\right)\,\mathbf{a}_1 + \left(x_{11} - y_{11}\right)\,\mathbf{a}_2 + z_{11}\,\mathbf{a}_3$ | = | $\left(x_{11} a + z_{11} c\cos\beta\right)\,\hat{\mathbf{x}} - y_{11} b\,\hat{\mathbf{y}} + z_{11} c\sin\beta\,\hat{\mathbf{z}}$ | (8$j$) | O I |
| $\mathbf{B}_{23}$ | = | $\left(x_{12} - y_{12}\right)\,\mathbf{a}_1 + \left(x_{12} + y_{12}\right)\,\mathbf{a}_2 + z_{12}\,\mathbf{a}_3$ | = | $\left(x_{12} a + z_{12} c\cos\beta\right)\,\hat{\mathbf{x}} + y_{12} b\,\hat{\mathbf{y}} + z_{12} c\sin\beta\,\hat{\mathbf{z}}$ | (8$j$) | O II |
| $\mathbf{B}_{24}$ | = | $\left(-x_{12} - y_{12}\right)\,\mathbf{a}_1 + \left(-x_{12} + y_{12}\right)\,\mathbf{a}_2 - z_{12}\,\mathbf{a}_3$ | = | $\left(-x_{12} a - z_{12} c\cos\beta\right)\,\hat{\mathbf{x}} + y_{12} b\,\hat{\mathbf{y}} - z_{12} c\sin\beta\,\hat{\mathbf{z}}$ | (8$j$) | O II |
| $\mathbf{B}_{25}$ | = | $(-x_{12}+y_{12})\,\mathbf{a}_1 + (-x_{12}-y_{12})\,\mathbf{a}_2 - z_{12}\,\mathbf{a}_3$ | = | $(-x_{12}a - z_{12}c\cos\beta)\,\hat{\mathbf{x}} - y_{12}b\,\hat{\mathbf{y}} - z_{12}c\sin\beta\,\hat{\mathbf{z}}$ | $(8j)$ | O II |
| $\mathbf{B}_{26}$ | = | $(x_{12}+y_{12})\,\mathbf{a}_1 + (x_{12}-y_{12})\,\mathbf{a}_2 + z_{12}\,\mathbf{a}_3$ | = | $(x_{12}a + z_{12}c\cos\beta)\,\hat{\mathbf{x}} - y_{12}b\,\hat{\mathbf{y}} + z_{12}c\sin\beta\,\hat{\mathbf{z}}$ | $(8j)$ | O II |
| $\mathbf{B}_{27}$ | = | $(x_{13}-y_{13})\,\mathbf{a}_1 + (x_{13}+y_{13})\,\mathbf{a}_2 + z_{13}\,\mathbf{a}_3$ | = | $(x_{13}a + z_{13}c\cos\beta)\,\hat{\mathbf{x}} + y_{13}b\,\hat{\mathbf{y}} + z_{13}c\sin\beta\,\hat{\mathbf{z}}$ | $(8j)$ | O III |
| $\mathbf{B}_{28}$ | = | $(-x_{13}-y_{13})\,\mathbf{a}_1 + (-x_{13}+y_{13})\,\mathbf{a}_2 - z_{13}\,\mathbf{a}_3$ | = | $(-x_{13}a - z_{13}c\cos\beta)\,\hat{\mathbf{x}} + y_{13}b\,\hat{\mathbf{y}} - z_{13}c\sin\beta\,\hat{\mathbf{z}}$ | $(8j)$ | O III |
| $\mathbf{B}_{29}$ | = | $(-x_{13}+y_{13})\,\mathbf{a}_1 + (-x_{13}-y_{13})\,\mathbf{a}_2 - z_{13}\,\mathbf{a}_3$ | = | $(-x_{13}a - z_{13}c\cos\beta)\,\hat{\mathbf{x}} - y_{13}b\,\hat{\mathbf{y}} - z_{13}c\sin\beta\,\hat{\mathbf{z}}$ | $(8j)$ | O III |
| $\mathbf{B}_{30}$ | = | $(x_{13}+y_{13})\,\mathbf{a}_1 + (x_{13}-y_{13})\,\mathbf{a}_2 + z_{13}\,\mathbf{a}_3$ | = | $(x_{13}a + z_{13}c\cos\beta)\,\hat{\mathbf{x}} - y_{13}b\,\hat{\mathbf{y}} + z_{13}c\sin\beta\,\hat{\mathbf{z}}$ | $(8j)$ | O III |
| $\mathbf{B}_{31}$ | = | $(x_{14}-y_{14})\,\mathbf{a}_1 + (x_{14}+y_{14})\,\mathbf{a}_2 + z_{14}\,\mathbf{a}_3$ | = | $(x_{14}a + z_{14}c\cos\beta)\,\hat{\mathbf{x}} + y_{14}b\,\hat{\mathbf{y}} + z_{14}c\sin\beta\,\hat{\mathbf{z}}$ | $(8j)$ | O IV |
| $\mathbf{B}_{32}$ | = | $(-x_{14}-y_{14})\,\mathbf{a}_1 + (-x_{14}+y_{14})\,\mathbf{a}_2 - z_{14}\,\mathbf{a}_3$ | = | $(-x_{14}a - z_{14}c\cos\beta)\,\hat{\mathbf{x}} + y_{14}b\,\hat{\mathbf{y}} - z_{14}c\sin\beta\,\hat{\mathbf{z}}$ | $(8j)$ | O IV |
| $\mathbf{B}_{33}$ | = | $(-x_{14}+y_{14})\,\mathbf{a}_1 + (-x_{14}-y_{14})\,\mathbf{a}_2 - z_{14}\,\mathbf{a}_3$ | = | $(-x_{14}a - z_{14}c\cos\beta)\,\hat{\mathbf{x}} - y_{14}b\,\hat{\mathbf{y}} - z_{14}c\sin\beta\,\hat{\mathbf{z}}$ | $(8j)$ | O IV |
| $\mathbf{B}_{34}$ | = | $(x_{14}+y_{14})\,\mathbf{a}_1 + (x_{14}-y_{14})\,\mathbf{a}_2 + z_{14}\,\mathbf{a}_3$ | = | $(x_{14}a + z_{14}c\cos\beta)\,\hat{\mathbf{x}} - y_{14}b\,\hat{\mathbf{y}} + z_{14}c\sin\beta\,\hat{\mathbf{z}}$ | $(8j)$ | O IV |
| $\mathbf{B}_{35}$ | = | $(x_{15}-y_{15})\,\mathbf{a}_1 + (x_{15}+y_{15})\,\mathbf{a}_2 + z_{15}\,\mathbf{a}_3$ | = | $(x_{15}a + z_{15}c\cos\beta)\,\hat{\mathbf{x}} + y_{15}b\,\hat{\mathbf{y}} + z_{15}c\sin\beta\,\hat{\mathbf{z}}$ | $(8j)$ | O V |
| $\mathbf{B}_{36}$ | = | $(-x_{15}-y_{15})\,\mathbf{a}_1 + (-x_{15}+y_{15})\,\mathbf{a}_2 - z_{15}\,\mathbf{a}_3$ | = | $(-x_{15}a - z_{15}c\cos\beta)\,\hat{\mathbf{x}} + y_{15}b\,\hat{\mathbf{y}} - z_{15}c\sin\beta\,\hat{\mathbf{z}}$ | $(8j)$ | O V |
| $\mathbf{B}_{37}$ | = | $(-x_{15}+y_{15})\,\mathbf{a}_1 + (-x_{15}-y_{15})\,\mathbf{a}_2 - z_{15}\,\mathbf{a}_3$ | = | $(-x_{15}a - z_{15}c\cos\beta)\,\hat{\mathbf{x}} - y_{15}b\,\hat{\mathbf{y}} - z_{15}c\sin\beta\,\hat{\mathbf{z}}$ | $(8j)$ | O V |
| $\mathbf{B}_{38}$ | = | $(x_{15}+y_{15})\,\mathbf{a}_1 + (x_{15}-y_{15})\,\mathbf{a}_2 + z_{15}\,\mathbf{a}_3$ | = | $(x_{15}a + z_{15}c\cos\beta)\,\hat{\mathbf{x}} - y_{15}b\,\hat{\mathbf{y}} + z_{15}c\sin\beta\,\hat{\mathbf{z}}$ | $(8j)$ | O V |
| $\mathbf{B}_{39}$ | = | $(x_{16}-y_{16})\,\mathbf{a}_1 + (x_{16}+y_{16})\,\mathbf{a}_2 + z_{16}\,\mathbf{a}_3$ | = | $(x_{16}a + z_{16}c\cos\beta)\,\hat{\mathbf{x}} + y_{16}b\,\hat{\mathbf{y}} + z_{16}c\sin\beta\,\hat{\mathbf{z}}$ | $(8j)$ | Si |
| $\mathbf{B}_{40}$ | = | $(-x_{16}-y_{16})\,\mathbf{a}_1 + (-x_{16}+y_{16})\,\mathbf{a}_2 - z_{16}\,\mathbf{a}_3$ | = | $(-x_{16}a - z_{16}c\cos\beta)\,\hat{\mathbf{x}} + y_{16}b\,\hat{\mathbf{y}} - z_{16}c\sin\beta\,\hat{\mathbf{z}}$ | $(8j)$ | Si |
| $\mathbf{B}_{41}$ | = | $(-x_{16}+y_{16})\,\mathbf{a}_1 + (-x_{16}-y_{16})\,\mathbf{a}_2 - z_{16}\,\mathbf{a}_3$ | = | $(-x_{16}a - z_{16}c\cos\beta)\,\hat{\mathbf{x}} - y_{16}b\,\hat{\mathbf{y}} - z_{16}c\sin\beta\,\hat{\mathbf{z}}$ | $(8j)$ | Si |
| $\mathbf{B}_{42}$ | = | $(x_{16}+y_{16})\,\mathbf{a}_1 + (x_{16}-y_{16})\,\mathbf{a}_2 + z_{16}\,\mathbf{a}_3$ | = | $(x_{16}a + z_{16}c\cos\beta)\,\hat{\mathbf{x}} - y_{16}b\,\hat{\mathbf{y}} + z_{16}c\sin\beta\,\hat{\mathbf{z}}$ | $(8j)$ | Si |

**Geometry files:**

- CIF: pp. 1529
- POSCAR: pp. 1530

# Manganese-leonite [$K_2Mn(SO_4)_2{\cdot}4H_2O$, $H4_{23}$] Structure: A8B2CD15E2_mC112_12_2i3j_j_ad_g4i5j_2i

http://aflow.org/prototype-encyclopedia/A8B2CD15E2_mC112_12_2i3j_j_ad_g4i5j_2i

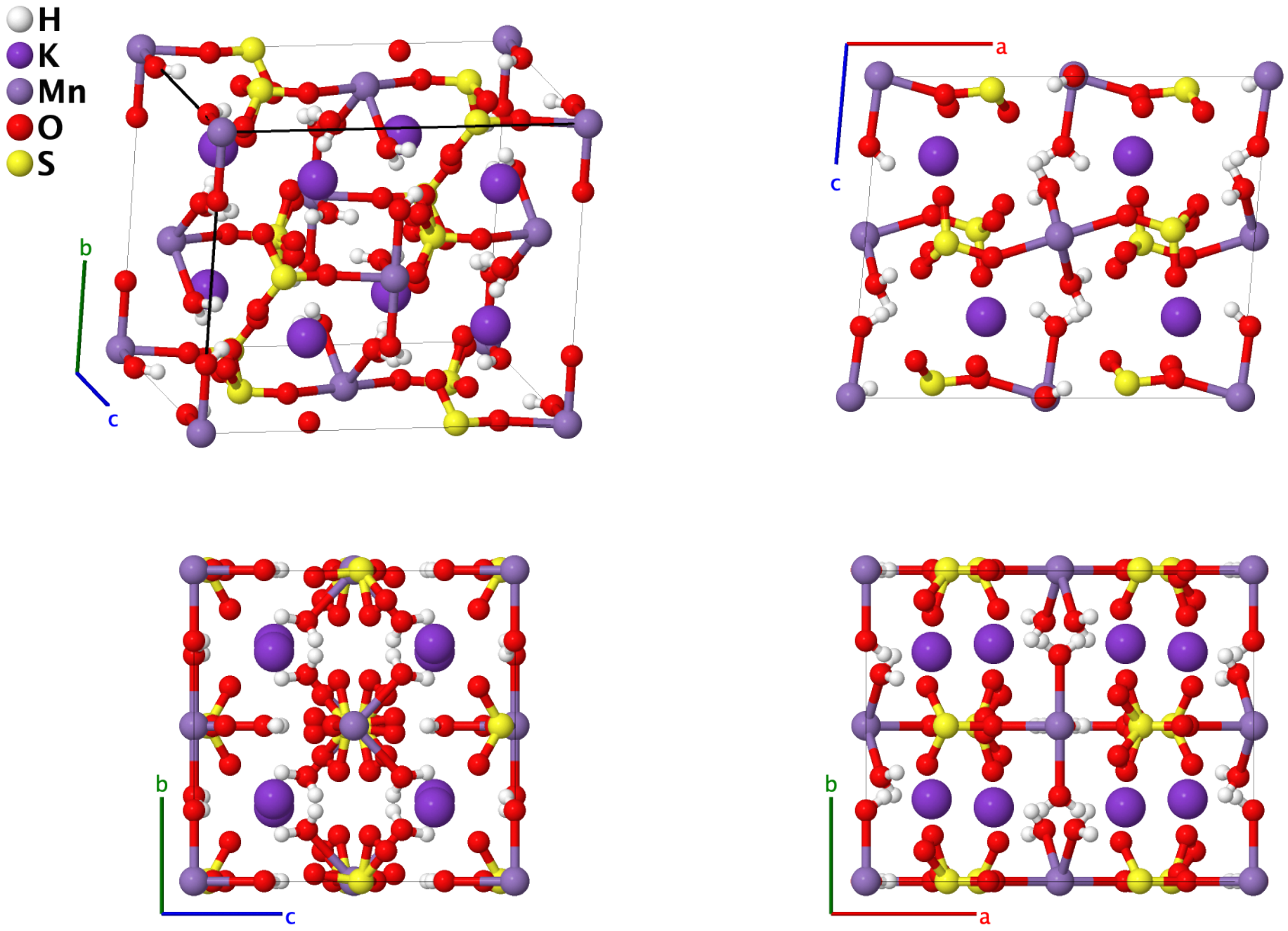

| | | |
|---|---|---|
| **Prototype** | : | $H_8K_2MnO_{12}S_2$ |
| **AFLOW prototype label** | : | A8B2CD15E2_mC112_12_2i3j_j_ad_g4i5j_2i |
| ***Strukturbericht* designation** | : | $H4_{23}$ |
| **Pearson symbol** | : | mC112 |
| **Space group number** | : | 12 |
| **Space group symbol** | : | $C2/m$ |
| **AFLOW prototype command** | : | `aflow --proto=A8B2CD15E2_mC112_12_2i3j_j_ad_g4i5j_2i` `--params=`$a, b/a, c/a, \beta, y_3, x_4, z_4, x_5, z_5, x_6, z_6, x_7, z_7, x_8, z_8, x_9, z_9, x_{10}, z_{10}, x_{11}, z_{11}, x_{12}, y_{12}, z_{12}, x_{13}, y_{13}, z_{13}, x_{14}, y_{14}, z_{14}, x_{15}, y_{15}, z_{15}, x_{16}, y_{16}, z_{16}, x_{17}, y_{17}, z_{17}, x_{18}, y_{18}, z_{18}, x_{19}, y_{19}, z_{19}, x_{20}, y_{20}, z_{20}$ |

### Other compounds with this structure

- $(NH_2)_2Co(SO_4)_2{\cdot}4H_2O$, $(NH_2)_2Cu(SO_4)_2{\cdot}4H_2O$, $(NH_2)_2Fe(SO_4)_2{\cdot}4H_2O$, $(NH_2)_2Mg(SO_4)_2{\cdot}4H_2O$, $(NH_2)_2Mn(SO_4)_2{\cdot}4H_2O$, $Cd_2Co(SO_4)_2{\cdot}4H_2O$, $Cd_2Cu(SO_4)_2{\cdot}4H_2O$, $Cd_2Fe(SO_4)_2{\cdot}4H_2O$, $Cd_2Mg(SO_4)_2{\cdot}4H_2O$, $Cd_2Mn(SO_4)_2{\cdot}4H_2O$, $Cs_2Co(SO_4)_2{\cdot}4H_2O$, $Cs_2Cu(SO_4)_2{\cdot}4H_2O$, $Cs_2Fe(SO_4)_2{\cdot}4H_2O$, $Cs_2Mg(SO_4)_2{\cdot}4H_2O$, $Cs_2Mn(SO_4)_2{\cdot}4H_2O$, $K_2Co(SO_4)_2{\cdot}4H_2O$, $K_2Cu(SO_4)_2{\cdot}4H_2O$, $K_2Fe(SO_4)_2{\cdot}4H_2O$ (mereiterite), $K_2Mg(SO_4)_2{\cdot}4H_2O$ (leonite), $Rb_2Co(SO_4)_2{\cdot}4H_2O$, $Rb_2Cu(SO_4)_2{\cdot}4H_2O$, $Rb_2Fe(SO_4)_2{\cdot}4H_2O$, $Rb_2Mg(SO_4)_2{\cdot}4H_2O$, $Rb_2Mn(SO_4)_2{\cdot}4H_2O$, $Tl_2Co(SO_4)_2{\cdot}4H_2O$, $Tl_2Cu(SO_4)_2{\cdot}4H_2O$, $Tl_2Fe(SO_4)_2{\cdot}4H_2O$, $Tl_2Mg(SO_4)_2{\cdot}4H_2O$, and $Tl_2Mn(SO_4)_2{\cdot}4H_2O$

---

- Properly, the prototype of leonite is $K_2Mg(SO_4)_2{\cdot}4H_2O$, but (Herrmann, 1943) gives manganese-leonite *Strukturbericht* symbol $H4_{23}$, so we will use Mn-leonite as the prototype.
- The room temperature leonite structure, shown here, is in space group $C2/m$ #12. It is characterized by the disorder of the second sulfate group, centered on atom S-II (Hertweck, 2001). The oxygen sites around S-II, labeled O-VII, O-VIII, and O-IX in our notation, are only occupied 50% of the time. (Anspach, 1939), who did the original determination of this structure, was not able to see the disorder, and so gives an ordered structure for both sulfates.

- We use the structure determined by (Hertweck, 2001) at room temperature, 293 K. At lower temperatures, the leonites undergo phase transitions. The exact transition temperature depends on the compound.
- At somewhat lower than room temperature (205 K for Mn-leonite) the sulfate group orders, doubling the size of the unit cell and changing the space group from $C2/m$ to $C2/c$ #15. At still lower temperatures (168 K for Mn-leonite) further ordering takes place and the structure transforms into the $P2_1/c$ #14 space group.

---

**Base-centered Monoclinic primitive vectors:**

$$\mathbf{a}_1 = \frac{1}{2}\,a\,\hat{\mathbf{x}} - \frac{1}{2}\,b\,\hat{\mathbf{y}}$$
$$\mathbf{a}_2 = \frac{1}{2}\,a\,\hat{\mathbf{x}} + \frac{1}{2}\,b\,\hat{\mathbf{y}}$$
$$\mathbf{a}_3 = c\cos\beta\,\hat{\mathbf{x}} + c\sin\beta\,\hat{\mathbf{z}}$$

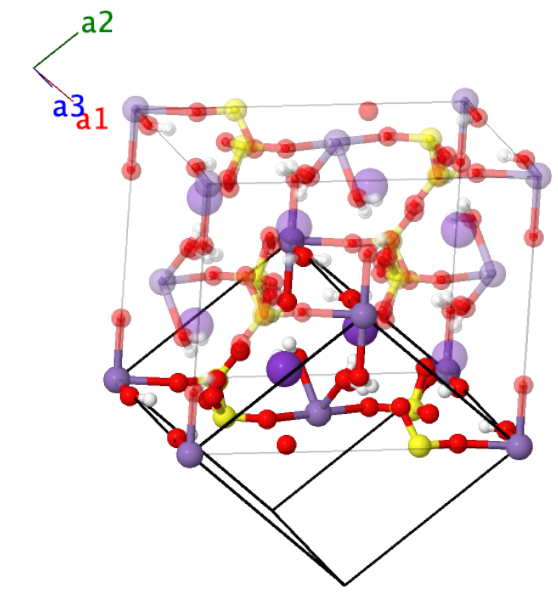

---

**Basis vectors:**

| | | Lattice Coordinates | | Cartesian Coordinates | Wyckoff Position | Atom Type |
|---|---|---|---|---|---|---|
| $\mathbf{B}_1$ | = | $0\,\mathbf{a}_1 + 0\,\mathbf{a}_2 + 0\,\mathbf{a}_3$ | = | $0\,\hat{\mathbf{x}} + 0\,\hat{\mathbf{y}} + 0\,\hat{\mathbf{z}}$ | (2$a$) | Mn I |
| $\mathbf{B}_2$ | = | $\frac{1}{2}\,\mathbf{a}_1 + \frac{1}{2}\,\mathbf{a}_2 + \frac{1}{2}\,\mathbf{a}_3$ | = | $\frac{1}{2}\left(a + c\cos\beta\right)\,\hat{\mathbf{x}} + \frac{1}{2}c\sin\beta\,\hat{\mathbf{z}}$ | (2$d$) | Mn II |
| $\mathbf{B}_3$ | = | $-y_3\,\mathbf{a}_1 + y_3\,\mathbf{a}_2$ | = | $y_3 b\,\hat{\mathbf{y}}$ | (4$g$) | O I |
| $\mathbf{B}_4$ | = | $y_3\,\mathbf{a}_1 - y_3\,\mathbf{a}_2$ | = | $-y_3 b\,\hat{\mathbf{y}}$ | (4$g$) | O I |
| $\mathbf{B}_5$ | = | $x_4\,\mathbf{a}_1 + x_4\,\mathbf{a}_2 + z_4\,\mathbf{a}_3$ | = | $\left(x_4 a + z_4 c\cos\beta\right)\,\hat{\mathbf{x}} + z_4 c\sin\beta\,\hat{\mathbf{z}}$ | (4$i$) | H I |
| $\mathbf{B}_6$ | = | $-x_4\,\mathbf{a}_1 - x_4\,\mathbf{a}_2 - z_4\,\mathbf{a}_3$ | = | $\left(-x_4 a - z_4 c\cos\beta\right)\,\hat{\mathbf{x}} - z_4 c\sin\beta\,\hat{\mathbf{z}}$ | (4$i$) | H I |
| $\mathbf{B}_7$ | = | $x_5\,\mathbf{a}_1 + x_5\,\mathbf{a}_2 + z_5\,\mathbf{a}_3$ | = | $\left(x_5 a + z_5 c\cos\beta\right)\,\hat{\mathbf{x}} + z_5 c\sin\beta\,\hat{\mathbf{z}}$ | (4$i$) | H II |
| $\mathbf{B}_8$ | = | $-x_5\,\mathbf{a}_1 - x_5\,\mathbf{a}_2 - z_5\,\mathbf{a}_3$ | = | $\left(-x_5 a - z_5 c\cos\beta\right)\,\hat{\mathbf{x}} - z_5 c\sin\beta\,\hat{\mathbf{z}}$ | (4$i$) | H II |
| $\mathbf{B}_9$ | = | $x_6\,\mathbf{a}_1 + x_6\,\mathbf{a}_2 + z_6\,\mathbf{a}_3$ | = | $\left(x_6 a + z_6 c\cos\beta\right)\,\hat{\mathbf{x}} + z_6 c\sin\beta\,\hat{\mathbf{z}}$ | (4$i$) | O II |
| $\mathbf{B}_{10}$ | = | $-x_6\,\mathbf{a}_1 - x_6\,\mathbf{a}_2 - z_6\,\mathbf{a}_3$ | = | $\left(-x_6 a - z_6 c\cos\beta\right)\,\hat{\mathbf{x}} - z_6 c\sin\beta\,\hat{\mathbf{z}}$ | (4$i$) | O II |
| $\mathbf{B}_{11}$ | = | $x_7\,\mathbf{a}_1 + x_7\,\mathbf{a}_2 + z_7\,\mathbf{a}_3$ | = | $\left(x_7 a + z_7 c\cos\beta\right)\,\hat{\mathbf{x}} + z_7 c\sin\beta\,\hat{\mathbf{z}}$ | (4$i$) | O III |
| $\mathbf{B}_{12}$ | = | $-x_7\,\mathbf{a}_1 - x_7\,\mathbf{a}_2 - z_7\,\mathbf{a}_3$ | = | $\left(-x_7 a - z_7 c\cos\beta\right)\,\hat{\mathbf{x}} - z_7 c\sin\beta\,\hat{\mathbf{z}}$ | (4$i$) | O III |
| $\mathbf{B}_{13}$ | = | $x_8\,\mathbf{a}_1 + x_8\,\mathbf{a}_2 + z_8\,\mathbf{a}_3$ | = | $\left(x_8 a + z_8 c\cos\beta\right)\,\hat{\mathbf{x}} + z_8 c\sin\beta\,\hat{\mathbf{z}}$ | (4$i$) | O IV |
| $\mathbf{B}_{14}$ | = | $-x_8\,\mathbf{a}_1 - x_8\,\mathbf{a}_2 - z_8\,\mathbf{a}_3$ | = | $\left(-x_8 a - z_8 c\cos\beta\right)\,\hat{\mathbf{x}} - z_8 c\sin\beta\,\hat{\mathbf{z}}$ | (4$i$) | O IV |
| $\mathbf{B}_{15}$ | = | $x_9\,\mathbf{a}_1 + x_9\,\mathbf{a}_2 + z_9\,\mathbf{a}_3$ | = | $\left(x_9 a + z_9 c\cos\beta\right)\,\hat{\mathbf{x}} + z_9 c\sin\beta\,\hat{\mathbf{z}}$ | (4$i$) | O V |
| $\mathbf{B}_{16}$ | = | $-x_9\,\mathbf{a}_1 - x_9\,\mathbf{a}_2 - z_9\,\mathbf{a}_3$ | = | $\left(-x_9 a - z_9 c\cos\beta\right)\,\hat{\mathbf{x}} - z_9 c\sin\beta\,\hat{\mathbf{z}}$ | (4$i$) | O V |
| $\mathbf{B}_{17}$ | = | $x_{10}\,\mathbf{a}_1 + x_{10}\,\mathbf{a}_2 + z_{10}\,\mathbf{a}_3$ | = | $\left(x_{10} a + z_{10} c\cos\beta\right)\,\hat{\mathbf{x}} + z_{10} c\sin\beta\,\hat{\mathbf{z}}$ | (4$i$) | S I |
| $\mathbf{B}_{18}$ | = | $-x_{10}\,\mathbf{a}_1 - x_{10}\,\mathbf{a}_2 - z_{10}\,\mathbf{a}_3$ | = | $\left(-x_{10} a - z_{10} c\cos\beta\right)\,\hat{\mathbf{x}} - z_{10} c\sin\beta\,\hat{\mathbf{z}}$ | (4$i$) | S I |
| $\mathbf{B}_{19}$ | = | $x_{11}\,\mathbf{a}_1 + x_{11}\,\mathbf{a}_2 + z_{11}\,\mathbf{a}_3$ | = | $\left(x_{11} a + z_{11} c\cos\beta\right)\,\hat{\mathbf{x}} + z_{11} c\sin\beta\,\hat{\mathbf{z}}$ | (4$i$) | S II |
| $\mathbf{B}_{20}$ | = | $-x_{11}\,\mathbf{a}_1 - x_{11}\,\mathbf{a}_2 - z_{11}\,\mathbf{a}_3$ | = | $\left(-x_{11} a - z_{11} c\cos\beta\right)\,\hat{\mathbf{x}} - z_{11} c\sin\beta\,\hat{\mathbf{z}}$ | (4$i$) | S II |
| $\mathbf{B}_{21}$ | = | $\left(x_{12} - y_{12}\right)\,\mathbf{a}_1 + \left(x_{12} + y_{12}\right)\,\mathbf{a}_2 + z_{12}\,\mathbf{a}_3$ | = | $\left(x_{12} a + z_{12} c\cos\beta\right)\,\hat{\mathbf{x}} + y_{12} b\,\hat{\mathbf{y}} + z_{12} c\sin\beta\,\hat{\mathbf{z}}$ | (8$j$) | H III |
| $\mathbf{B}_{22}$ | = | $\left(-x_{12} - y_{12}\right)\,\mathbf{a}_1 + \left(-x_{12} + y_{12}\right)\,\mathbf{a}_2 - z_{12}\,\mathbf{a}_3$ | = | $\left(-x_{12} a - z_{12} c\cos\beta\right)\,\hat{\mathbf{x}} + y_{12} b\,\hat{\mathbf{y}} - z_{12} c\sin\beta\,\hat{\mathbf{z}}$ | (8$j$) | H III |

$$\mathbf{B}_{23} = (-x_{12}+y_{12})\,\mathbf{a}_1 + (-x_{12}-y_{12})\,\mathbf{a}_2 - z_{12}\,\mathbf{a}_3 = (-x_{12}a - z_{12}c\cos\beta)\,\hat{\mathbf{x}} - y_{12}b\,\hat{\mathbf{y}} - z_{12}c\sin\beta\,\hat{\mathbf{z}} \quad (8j) \quad \text{H III}$$

$$\mathbf{B}_{24} = (x_{12}+y_{12})\,\mathbf{a}_1 + (x_{12}-y_{12})\,\mathbf{a}_2 + z_{12}\,\mathbf{a}_3 = (x_{12}a + z_{12}c\cos\beta)\,\hat{\mathbf{x}} - y_{12}b\,\hat{\mathbf{y}} + z_{12}c\sin\beta\,\hat{\mathbf{z}} \quad (8j) \quad \text{H III}$$

$$\mathbf{B}_{25} = (x_{13}-y_{13})\,\mathbf{a}_1 + (x_{13}+y_{13})\,\mathbf{a}_2 + z_{13}\,\mathbf{a}_3 = (x_{13}a + z_{13}c\cos\beta)\,\hat{\mathbf{x}} + y_{13}b\,\hat{\mathbf{y}} + z_{13}c\sin\beta\,\hat{\mathbf{z}} \quad (8j) \quad \text{H IV}$$

$$\mathbf{B}_{26} = (-x_{13}-y_{13})\,\mathbf{a}_1 + (-x_{13}+y_{13})\,\mathbf{a}_2 - z_{13}\,\mathbf{a}_3 = (-x_{13}a - z_{13}c\cos\beta)\,\hat{\mathbf{x}} + y_{13}b\,\hat{\mathbf{y}} - z_{13}c\sin\beta\,\hat{\mathbf{z}} \quad (8j) \quad \text{H IV}$$

$$\mathbf{B}_{27} = (-x_{13}+y_{13})\,\mathbf{a}_1 + (-x_{13}-y_{13})\,\mathbf{a}_2 - z_{13}\,\mathbf{a}_3 = (-x_{13}a - z_{13}c\cos\beta)\,\hat{\mathbf{x}} - y_{13}b\,\hat{\mathbf{y}} - z_{13}c\sin\beta\,\hat{\mathbf{z}} \quad (8j) \quad \text{H IV}$$

$$\mathbf{B}_{28} = (x_{13}+y_{13})\,\mathbf{a}_1 + (x_{13}-y_{13})\,\mathbf{a}_2 + z_{13}\,\mathbf{a}_3 = (x_{13}a + z_{13}c\cos\beta)\,\hat{\mathbf{x}} - y_{13}b\,\hat{\mathbf{y}} + z_{13}c\sin\beta\,\hat{\mathbf{z}} \quad (8j) \quad \text{H IV}$$

$$\mathbf{B}_{29} = (x_{14}-y_{14})\,\mathbf{a}_1 + (x_{14}+y_{14})\,\mathbf{a}_2 + z_{14}\,\mathbf{a}_3 = (x_{14}a + z_{14}c\cos\beta)\,\hat{\mathbf{x}} + y_{14}b\,\hat{\mathbf{y}} + z_{14}c\sin\beta\,\hat{\mathbf{z}} \quad (8j) \quad \text{H V}$$

$$\mathbf{B}_{30} = (-x_{14}-y_{14})\,\mathbf{a}_1 + (-x_{14}+y_{14})\,\mathbf{a}_2 - z_{14}\,\mathbf{a}_3 = (-x_{14}a - z_{14}c\cos\beta)\,\hat{\mathbf{x}} + y_{14}b\,\hat{\mathbf{y}} - z_{14}c\sin\beta\,\hat{\mathbf{z}} \quad (8j) \quad \text{H V}$$

$$\mathbf{B}_{31} = (-x_{14}+y_{14})\,\mathbf{a}_1 + (-x_{14}-y_{14})\,\mathbf{a}_2 - z_{14}\,\mathbf{a}_3 = (-x_{14}a - z_{14}c\cos\beta)\,\hat{\mathbf{x}} - y_{14}b\,\hat{\mathbf{y}} - z_{14}c\sin\beta\,\hat{\mathbf{z}} \quad (8j) \quad \text{H V}$$

$$\mathbf{B}_{32} = (x_{14}+y_{14})\,\mathbf{a}_1 + (x_{14}-y_{14})\,\mathbf{a}_2 + z_{14}\,\mathbf{a}_3 = (x_{14}a + z_{14}c\cos\beta)\,\hat{\mathbf{x}} - y_{14}b\,\hat{\mathbf{y}} + z_{14}c\sin\beta\,\hat{\mathbf{z}} \quad (8j) \quad \text{H V}$$

$$\mathbf{B}_{33} = (x_{15}-y_{15})\,\mathbf{a}_1 + (x_{15}+y_{15})\,\mathbf{a}_2 + z_{15}\,\mathbf{a}_3 = (x_{15}a + z_{15}c\cos\beta)\,\hat{\mathbf{x}} + y_{15}b\,\hat{\mathbf{y}} + z_{15}c\sin\beta\,\hat{\mathbf{z}} \quad (8j) \quad \text{K}$$

$$\mathbf{B}_{34} = (-x_{15}-y_{15})\,\mathbf{a}_1 + (-x_{15}+y_{15})\,\mathbf{a}_2 - z_{15}\,\mathbf{a}_3 = (-x_{15}a - z_{15}c\cos\beta)\,\hat{\mathbf{x}} + y_{15}b\,\hat{\mathbf{y}} - z_{15}c\sin\beta\,\hat{\mathbf{z}} \quad (8j) \quad \text{K}$$

$$\mathbf{B}_{35} = (-x_{15}+y_{15})\,\mathbf{a}_1 + (-x_{15}-y_{15})\,\mathbf{a}_2 - z_{15}\,\mathbf{a}_3 = (-x_{15}a - z_{15}c\cos\beta)\,\hat{\mathbf{x}} - y_{15}b\,\hat{\mathbf{y}} - z_{15}c\sin\beta\,\hat{\mathbf{z}} \quad (8j) \quad \text{K}$$

$$\mathbf{B}_{36} = (x_{15}+y_{15})\,\mathbf{a}_1 + (x_{15}-y_{15})\,\mathbf{a}_2 + z_{15}\,\mathbf{a}_3 = (x_{15}a + z_{15}c\cos\beta)\,\hat{\mathbf{x}} - y_{15}b\,\hat{\mathbf{y}} + z_{15}c\sin\beta\,\hat{\mathbf{z}} \quad (8j) \quad \text{K}$$

$$\mathbf{B}_{37} = (x_{16}-y_{16})\,\mathbf{a}_1 + (x_{16}+y_{16})\,\mathbf{a}_2 + z_{16}\,\mathbf{a}_3 = (x_{16}a + z_{16}c\cos\beta)\,\hat{\mathbf{x}} + y_{16}b\,\hat{\mathbf{y}} + z_{16}c\sin\beta\,\hat{\mathbf{z}} \quad (8j) \quad \text{O VI}$$

$$\mathbf{B}_{38} = (-x_{16}-y_{16})\,\mathbf{a}_1 + (-x_{16}+y_{16})\,\mathbf{a}_2 - z_{16}\,\mathbf{a}_3 = (-x_{16}a - z_{16}c\cos\beta)\,\hat{\mathbf{x}} + y_{16}b\,\hat{\mathbf{y}} - z_{16}c\sin\beta\,\hat{\mathbf{z}} \quad (8j) \quad \text{O VI}$$

$$\mathbf{B}_{39} = (-x_{16}+y_{16})\,\mathbf{a}_1 + (-x_{16}-y_{16})\,\mathbf{a}_2 - z_{16}\,\mathbf{a}_3 = (-x_{16}a - z_{16}c\cos\beta)\,\hat{\mathbf{x}} - y_{16}b\,\hat{\mathbf{y}} - z_{16}c\sin\beta\,\hat{\mathbf{z}} \quad (8j) \quad \text{O VI}$$

$$\mathbf{B}_{40} = (x_{16}+y_{16})\,\mathbf{a}_1 + (x_{16}-y_{16})\,\mathbf{a}_2 + z_{16}\,\mathbf{a}_3 = (x_{16}a + z_{16}c\cos\beta)\,\hat{\mathbf{x}} - y_{16}b\,\hat{\mathbf{y}} + z_{16}c\sin\beta\,\hat{\mathbf{z}} \quad (8j) \quad \text{O VI}$$

$$\mathbf{B}_{41} = (x_{17}-y_{17})\,\mathbf{a}_1 + (x_{17}+y_{17})\,\mathbf{a}_2 + z_{17}\,\mathbf{a}_3 = (x_{17}a + z_{17}c\cos\beta)\,\hat{\mathbf{x}} + y_{17}b\,\hat{\mathbf{y}} + z_{17}c\sin\beta\,\hat{\mathbf{z}} \quad (8j) \quad \text{O VII}$$

$$\mathbf{B}_{42} = (-x_{17}-y_{17})\,\mathbf{a}_1 + (-x_{17}+y_{17})\,\mathbf{a}_2 - z_{17}\,\mathbf{a}_3 = (-x_{17}a - z_{17}c\cos\beta)\,\hat{\mathbf{x}} + y_{17}b\,\hat{\mathbf{y}} - z_{17}c\sin\beta\,\hat{\mathbf{z}} \quad (8j) \quad \text{O VII}$$

$$\mathbf{B}_{43} = (-x_{17}+y_{17})\,\mathbf{a}_1 + (-x_{17}-y_{17})\,\mathbf{a}_2 - z_{17}\,\mathbf{a}_3 = (-x_{17}a - z_{17}c\cos\beta)\,\hat{\mathbf{x}} - y_{17}b\,\hat{\mathbf{y}} - z_{17}c\sin\beta\,\hat{\mathbf{z}} \quad (8j) \quad \text{O VII}$$

$$\mathbf{B}_{44} = (x_{17}+y_{17})\,\mathbf{a}_1 + (x_{17}-y_{17})\,\mathbf{a}_2 + z_{17}\,\mathbf{a}_3 = (x_{17}a + z_{17}c\cos\beta)\,\hat{\mathbf{x}} - y_{17}b\,\hat{\mathbf{y}} + z_{17}c\sin\beta\,\hat{\mathbf{z}} \quad (8j) \quad \text{O VII}$$

$$\mathbf{B}_{45} = (x_{18}-y_{18})\,\mathbf{a}_1 + (x_{18}+y_{18})\,\mathbf{a}_2 + z_{18}\,\mathbf{a}_3 = (x_{18}a + z_{18}c\cos\beta)\,\hat{\mathbf{x}} + y_{18}b\,\hat{\mathbf{y}} + z_{18}c\sin\beta\,\hat{\mathbf{z}} \quad (8j) \quad \text{O VIII}$$

$$\mathbf{B}_{46} = (-x_{18}-y_{18})\,\mathbf{a}_1 + (-x_{18}+y_{18})\,\mathbf{a}_2 - z_{18}\,\mathbf{a}_3 = (-x_{18}a - z_{18}c\cos\beta)\,\hat{\mathbf{x}} + y_{18}b\,\hat{\mathbf{y}} - z_{18}c\sin\beta\,\hat{\mathbf{z}} \quad (8j) \quad \text{O VIII}$$

| | | | | | | |
|---|---|---|---|---|---|---|
| $\mathbf{B}_{47}$ | = | $(-x_{18}+y_{18})\,\mathbf{a}_1 + (-x_{18}-y_{18})\,\mathbf{a}_2 - z_{18}\,\mathbf{a}_3$ | = | $(-x_{18}a - z_{18}c\cos\beta)\,\hat{\mathbf{x}} - y_{18}b\,\hat{\mathbf{y}} - z_{18}c\sin\beta\,\hat{\mathbf{z}}$ | (8*j*) | O VIII |
| $\mathbf{B}_{48}$ | = | $(x_{18}+y_{18})\,\mathbf{a}_1 + (x_{18}-y_{18})\,\mathbf{a}_2 + z_{18}\,\mathbf{a}_3$ | = | $(x_{18}a + z_{18}c\cos\beta)\,\hat{\mathbf{x}} - y_{18}b\,\hat{\mathbf{y}} + z_{18}c\sin\beta\,\hat{\mathbf{z}}$ | (8*j*) | O VIII |
| $\mathbf{B}_{49}$ | = | $(x_{19}-y_{19})\,\mathbf{a}_1 + (x_{19}+y_{19})\,\mathbf{a}_2 + z_{19}\,\mathbf{a}_3$ | = | $(x_{19}a + z_{19}c\cos\beta)\,\hat{\mathbf{x}} + y_{19}b\,\hat{\mathbf{y}} + z_{19}c\sin\beta\,\hat{\mathbf{z}}$ | (8*j*) | O IX |
| $\mathbf{B}_{50}$ | = | $(-x_{19}-y_{19})\,\mathbf{a}_1 + (-x_{19}+y_{19})\,\mathbf{a}_2 - z_{19}\,\mathbf{a}_3$ | = | $(-x_{19}a - z_{19}c\cos\beta)\,\hat{\mathbf{x}} + y_{19}b\,\hat{\mathbf{y}} - z_{19}c\sin\beta\,\hat{\mathbf{z}}$ | (8*j*) | O IX |
| $\mathbf{B}_{51}$ | = | $(-x_{19}+y_{19})\,\mathbf{a}_1 + (-x_{19}-y_{19})\,\mathbf{a}_2 - z_{19}\,\mathbf{a}_3$ | = | $(-x_{19}a - z_{19}c\cos\beta)\,\hat{\mathbf{x}} - y_{19}b\,\hat{\mathbf{y}} - z_{19}c\sin\beta\,\hat{\mathbf{z}}$ | (8*j*) | O IX |
| $\mathbf{B}_{52}$ | = | $(x_{19}+y_{19})\,\mathbf{a}_1 + (x_{19}-y_{19})\,\mathbf{a}_2 + z_{19}\,\mathbf{a}_3$ | = | $(x_{19}a + z_{19}c\cos\beta)\,\hat{\mathbf{x}} - y_{19}b\,\hat{\mathbf{y}} + z_{19}c\sin\beta\,\hat{\mathbf{z}}$ | (8*j*) | O IX |
| $\mathbf{B}_{53}$ | = | $(x_{20}-y_{20})\,\mathbf{a}_1 + (x_{20}+y_{20})\,\mathbf{a}_2 + z_{20}\,\mathbf{a}_3$ | = | $(x_{20}a + z_{20}c\cos\beta)\,\hat{\mathbf{x}} + y_{20}b\,\hat{\mathbf{y}} + z_{20}c\sin\beta\,\hat{\mathbf{z}}$ | (8*j*) | O X |
| $\mathbf{B}_{54}$ | = | $(-x_{20}-y_{20})\,\mathbf{a}_1 + (-x_{20}+y_{20})\,\mathbf{a}_2 - z_{20}\,\mathbf{a}_3$ | = | $(-x_{20}a - z_{20}c\cos\beta)\,\hat{\mathbf{x}} + y_{20}b\,\hat{\mathbf{y}} - z_{20}c\sin\beta\,\hat{\mathbf{z}}$ | (8*j*) | O X |
| $\mathbf{B}_{55}$ | = | $(-x_{20}+y_{20})\,\mathbf{a}_1 + (-x_{20}-y_{20})\,\mathbf{a}_2 - z_{20}\,\mathbf{a}_3$ | = | $(-x_{20}a - z_{20}c\cos\beta)\,\hat{\mathbf{x}} - y_{20}b\,\hat{\mathbf{y}} - z_{20}c\sin\beta\,\hat{\mathbf{z}}$ | (8*j*) | O X |
| $\mathbf{B}_{56}$ | = | $(x_{20}+y_{20})\,\mathbf{a}_1 + (x_{20}-y_{20})\,\mathbf{a}_2 + z_{20}\,\mathbf{a}_3$ | = | $(x_{20}a + z_{20}c\cos\beta)\,\hat{\mathbf{x}} - y_{20}b\,\hat{\mathbf{y}} + z_{20}c\sin\beta\,\hat{\mathbf{z}}$ | (8*j*) | O X |

**Geometry files:**

- CIF: pp. 1530
- POSCAR: pp. 1530

# Monoclinic $FeTlSe_2$ Structure: AB2C_mC16_12_g_2i_i

http://aflow.org/prototype-encyclopedia/AB2C_mC16_12_g_2i_i

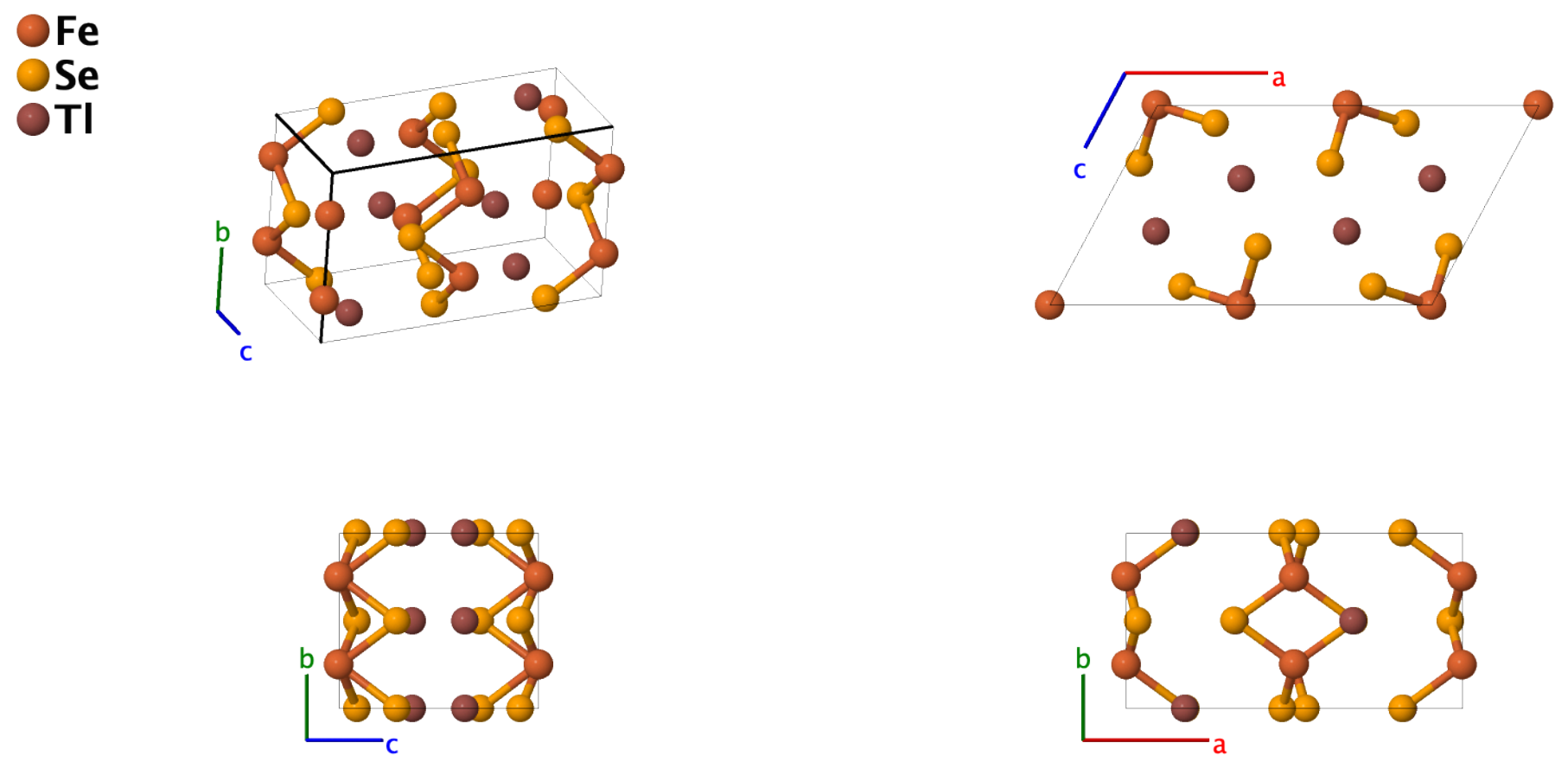

| | | |
|---|---|---|
| **Prototype** | : | $FeSe_2Tl$ |
| **AFLOW prototype label** | : | AB2C_mC16_12_g_2i_i |
| ***Strukturbericht* designation** | : | None |
| **Pearson symbol** | : | mC16 |
| **Space group number** | : | 12 |
| **Space group symbol** | : | $C2/m$ |
| **AFLOW prototype command** | : | `aflow --proto=AB2C_mC16_12_g_2i_i` `--params=`$a, b/a, c/a, \beta, y_1, x_2, z_2, x_3, z_3, x_4, z_4$ |

### Other compounds with this structure

- $FeTlS_2$

- This is the low temperature structure of $FeTlSe_2/FeTlS_2$. At high temperatures, it transforms to the tetragonal $FeTlS_2$ structure.

### Base-centered Monoclinic primitive vectors:

$$\mathbf{a}_1 = \frac{1}{2}\,a\,\hat{\mathbf{x}} - \frac{1}{2}\,b\,\hat{\mathbf{y}}$$
$$\mathbf{a}_2 = \frac{1}{2}\,a\,\hat{\mathbf{x}} + \frac{1}{2}\,b\,\hat{\mathbf{y}}$$
$$\mathbf{a}_3 = c\cos\beta\,\hat{\mathbf{x}} + c\sin\beta\,\hat{\mathbf{z}}$$

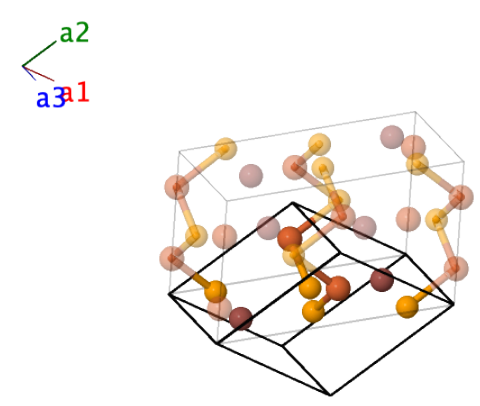

### Basis vectors:

| | | Lattice Coordinates | | Cartesian Coordinates | Wyckoff Position | Atom Type |
|---|---|---|---|---|---|---|
| $\mathbf{B}_1$ | = | $-y_1\,\mathbf{a}_1 + y_1\,\mathbf{a}_2$ | = | $y_1 b\,\hat{\mathbf{y}}$ | (4g) | Fe |
| $\mathbf{B}_2$ | = | $y_1\,\mathbf{a}_1 - y_1\,\mathbf{a}_2$ | = | $-y_1 b\,\hat{\mathbf{y}}$ | (4g) | Fe |
| $\mathbf{B}_3$ | = | $x_2\,\mathbf{a}_1 + x_2\,\mathbf{a}_2 + z_2\,\mathbf{a}_3$ | = | $(x_2 a + z_2 c\cos\beta)\,\hat{\mathbf{x}} + z_2 c\sin\beta\,\hat{\mathbf{z}}$ | (4i) | Se I |
| $\mathbf{B}_4$ | = | $-x_2\,\mathbf{a}_1 - x_2\,\mathbf{a}_2 - z_2\,\mathbf{a}_3$ | = | $(-x_2 a - z_2 c\cos\beta)\,\hat{\mathbf{x}} - z_2 c\sin\beta\,\hat{\mathbf{z}}$ | (4i) | Se I |
| $\mathbf{B}_5$ | = | $x_3\,\mathbf{a}_1 + x_3\,\mathbf{a}_2 + z_3\,\mathbf{a}_3$ | = | $(x_3 a + z_3 c\cos\beta)\,\hat{\mathbf{x}} + z_3 c\sin\beta\,\hat{\mathbf{z}}$ | (4$i$) | Se II |
| $\mathbf{B}_6$ | = | $-x_3\,\mathbf{a}_1 - x_3\,\mathbf{a}_2 - z_3\,\mathbf{a}_3$ | = | $(-x_3 a - z_3 c\cos\beta)\,\hat{\mathbf{x}} - z_3 c\sin\beta\,\hat{\mathbf{z}}$ | (4$i$) | Se II |
| $\mathbf{B}_7$ | = | $x_4\,\mathbf{a}_1 + x_4\,\mathbf{a}_2 + z_4\,\mathbf{a}_3$ | = | $(x_4 a + z_4 c\cos\beta)\,\hat{\mathbf{x}} + z_4 c\sin\beta\,\hat{\mathbf{z}}$ | (4$i$) | Tl |
| $\mathbf{B}_8$ | = | $-x_4\,\mathbf{a}_1 - x_4\,\mathbf{a}_2 - z_4\,\mathbf{a}_3$ | = | $(-x_4 a - z_4 c\cos\beta)\,\hat{\mathbf{x}} - z_4 c\sin\beta\,\hat{\mathbf{z}}$ | (4$i$) | Tl |

**Geometry files:**

- CIF: pp. 1531
- POSCAR: pp. 1531

# $NbTe_2$ Structure: AB2_mC18_12_ai_3i

http://aflow.org/prototype-encyclopedia/AB2_mC18_12_ai_3i

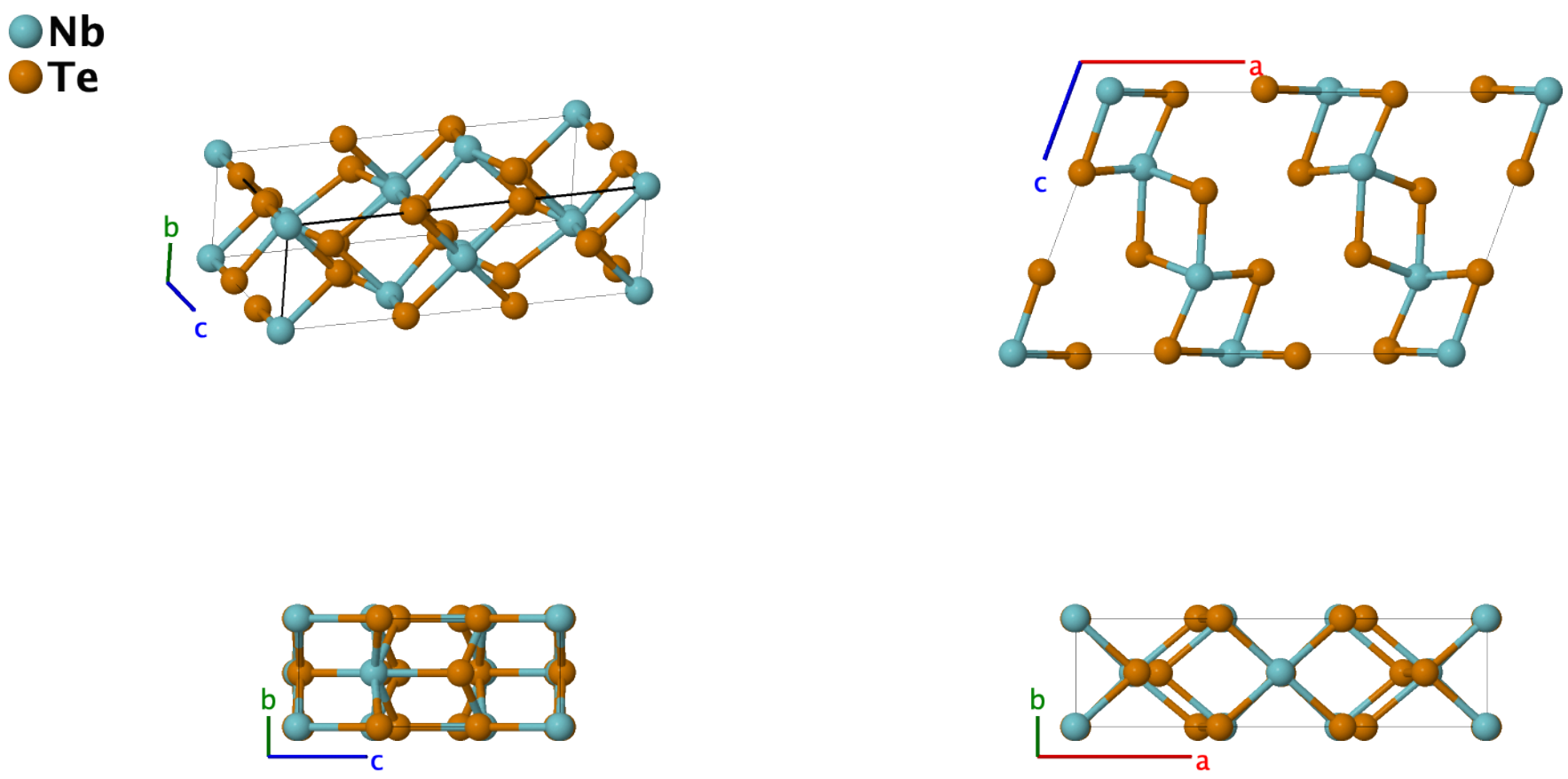

| | | |
|---|---|---|
| **Prototype** | : | $NbTe_2$ |
| **AFLOW prototype label** | : | AB2_mC18_12_ai_3i |
| ***Strukturbericht* designation** | : | None |
| **Pearson symbol** | : | mC18 |
| **Space group number** | : | 12 |
| **Space group symbol** | : | $C2/m$ |
| **AFLOW prototype command** | : | `aflow --proto=AB2_mC18_12_ai_3i` <br> `--params=`$a, b/a, c/a, \beta, x_2, z_2, x_3, z_3, x_4, z_4, x_5, z_5$ |

### Other compounds with this structure

- $TaTe_2$

- FINDSYM found a somewhat more compact representation of the unit cell (smaller $a$ and $\beta$) than that given in (Brown, 1966), so the lattice constants and Wyckoff parameters differ from those given in the reference.

### Base-centered Monoclinic primitive vectors:

$$\mathbf{a}_1 = \frac{1}{2}\,a\,\hat{\mathbf{x}} - \frac{1}{2}\,b\,\hat{\mathbf{y}}$$
$$\mathbf{a}_2 = \frac{1}{2}\,a\,\hat{\mathbf{x}} + \frac{1}{2}\,b\,\hat{\mathbf{y}}$$
$$\mathbf{a}_3 = c\cos\beta\,\hat{\mathbf{x}} + c\sin\beta\,\hat{\mathbf{z}}$$

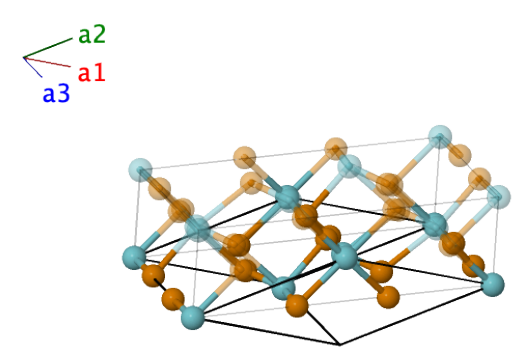

### Basis vectors:

| | | Lattice Coordinates | | Cartesian Coordinates | Wyckoff Position | Atom Type |
|---|---|---|---|---|---|---|
| $\mathbf{B}_1$ | = | $0\,\mathbf{a}_1 + 0\,\mathbf{a}_2 + 0\,\mathbf{a}_3$ | = | $0\,\hat{\mathbf{x}} + 0\,\hat{\mathbf{y}} + 0\,\hat{\mathbf{z}}$ | (2$a$) | Nb I |
| $\mathbf{B}_2$ | = | $x_2\,\mathbf{a}_1 + x_2\,\mathbf{a}_2 + z_2\,\mathbf{a}_3$ | = | $(x_2 a + z_2 c\cos\beta)\,\hat{\mathbf{x}} + z_2 c\sin\beta\,\hat{\mathbf{z}}$ | (4$i$) | Nb II |
| $\mathbf{B}_3$ | = | $-x_2\,\mathbf{a}_1 - x_2\,\mathbf{a}_2 - z_2\,\mathbf{a}_3$ | = | $(-x_2 a - z_2 c\cos\beta)\,\hat{\mathbf{x}} - z_2 c\sin\beta\,\hat{\mathbf{z}}$ | (4$i$) | Nb II |
| $\mathbf{B}_4$ | = | $x_3\,\mathbf{a}_1 + x_3\,\mathbf{a}_2 + z_3\,\mathbf{a}_3$ | = | $(x_3 a + z_3 c\cos\beta)\,\hat{\mathbf{x}} + z_3 c\sin\beta\,\hat{\mathbf{z}}$ | $(4i)$ | Te I |
| $\mathbf{B}_5$ | = | $-x_3\,\mathbf{a}_1 - x_3\,\mathbf{a}_2 - z_3\,\mathbf{a}_3$ | = | $(-x_3 a - z_3 c\cos\beta)\,\hat{\mathbf{x}} - z_3 c\sin\beta\,\hat{\mathbf{z}}$ | $(4i)$ | Te I |
| $\mathbf{B}_6$ | = | $x_4\,\mathbf{a}_1 + x_4\,\mathbf{a}_2 + z_4\,\mathbf{a}_3$ | = | $(x_4 a + z_4 c\cos\beta)\,\hat{\mathbf{x}} + z_4 c\sin\beta\,\hat{\mathbf{z}}$ | $(4i)$ | Te II |
| $\mathbf{B}_7$ | = | $-x_4\,\mathbf{a}_1 - x_4\,\mathbf{a}_2 - z_4\,\mathbf{a}_3$ | = | $(-x_4 a - z_4 c\cos\beta)\,\hat{\mathbf{x}} - z_4 c\sin\beta\,\hat{\mathbf{z}}$ | $(4i)$ | Te II |
| $\mathbf{B}_8$ | = | $x_5\,\mathbf{a}_1 + x_5\,\mathbf{a}_2 + z_5\,\mathbf{a}_3$ | = | $(x_5 a + z_5 c\cos\beta)\,\hat{\mathbf{x}} + z_5 c\sin\beta\,\hat{\mathbf{z}}$ | $(4i)$ | Te III |
| $\mathbf{B}_9$ | = | $-x_5\,\mathbf{a}_1 - x_5\,\mathbf{a}_2 - z_5\,\mathbf{a}_3$ | = | $(-x_5 a - z_5 c\cos\beta)\,\hat{\mathbf{x}} - z_5 c\sin\beta\,\hat{\mathbf{z}}$ | $(4i)$ | Te III |

**Geometry files:**

- CIF: pp. 1531
- POSCAR: pp. 1532

# $D2_2$ ($MgZn_5$?) (*Problematic*) Structure: AB5_mC48_12_2i_ac5i2j

http://aflow.org/prototype-encyclopedia/AB5_mC48_12_2i_ac5i2j

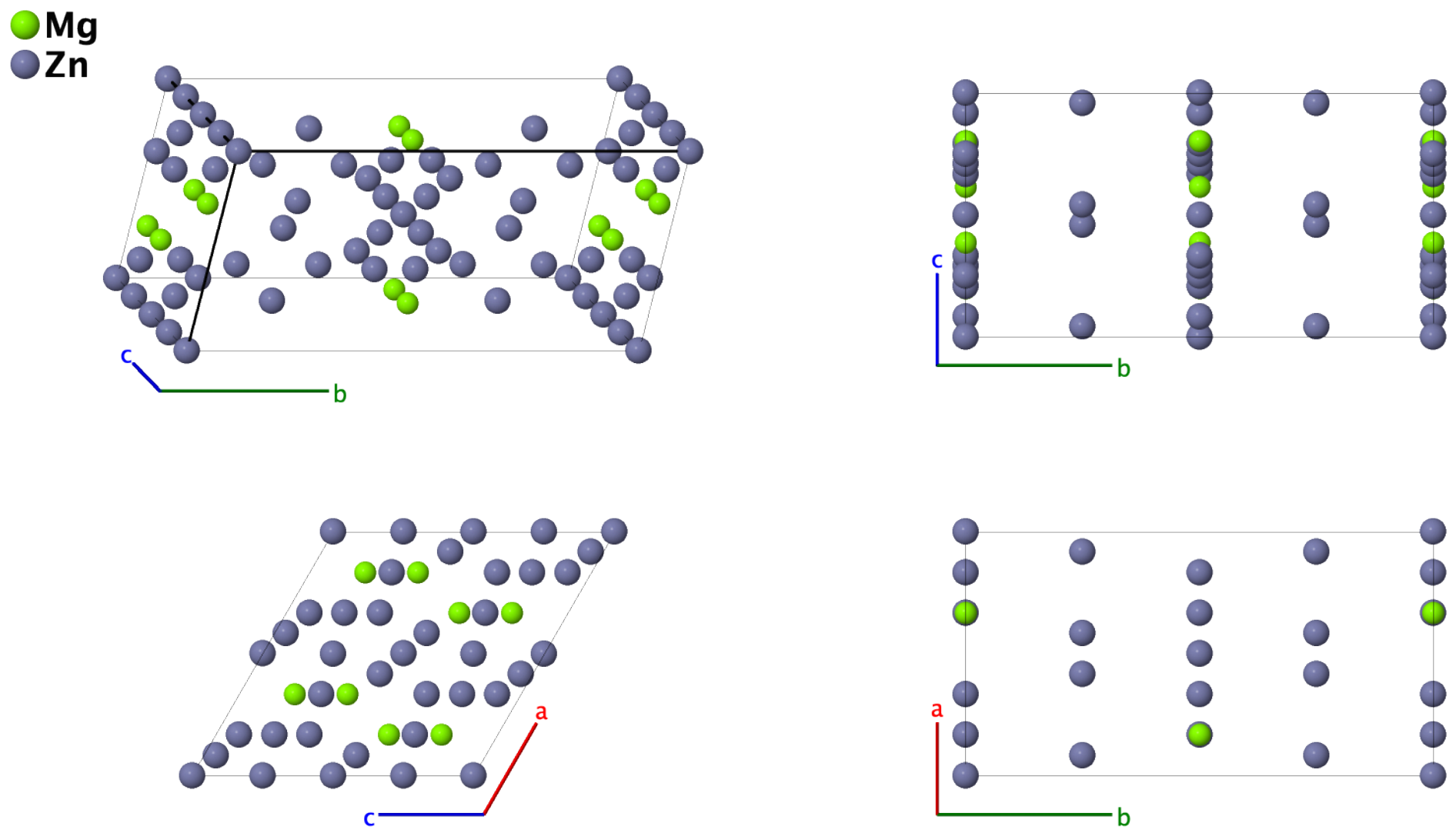

| | | |
|---|---|---|
| **Prototype** | : | $MgZn_5$ |
| **AFLOW prototype label** | : | AB5_mC48_12_2i_ac5i2j |
| ***Strukturbericht* designation** | : | $D2_2$ |
| **Pearson symbol** | : | mC48 |
| **Space group number** | : | 12 |
| **Space group symbol** | : | $C2/m$ |
| **AFLOW prototype command** | : | `aflow --proto=AB5_mC48_12_2i_ac5i2j` `--params=`$a, b/a, c/a, \beta, x_3, z_3, x_4, z_4, x_5, z_5, x_6, z_6, x_7, z_7, x_8, z_8, x_9, z_9, x_{10}, y_{10}, z_{10}, x_{11}, y_{11}, z_{11}$ |

- This structure has problems similar to the *B*30 MgZn structure, and for the same reasons. (Hermann, 1937) assigned this the *Strukturbericht* designation $D2_2$, based on the paper of (Tarschish, 1933), who derived it from the hexagonal Laves structure $MgZn_2$ (C14), eventually resulting in a 48 atom cell with composition $MgZn_5$. As with MgZn, he assumed that the space group remained $P6_3/mmc$ #194.
- (McKeehan, 1935) again pointed out that this is impossible. (Hermann, 1937) referenced both papers, giving the space group as $P6_3/mmc$ but listing the atomic coordinates enumerated by McKeehan.
- The McKeehan structure has space group $C2/m$ #12, with 48 atoms in the conventional cell, and 24 atoms in the primitive cell. As with *B*30, this agrees with the structure (Parthé, 1993) designate as $D2_2$.
- It is not clear that any $MgZn_5$ compound actually exists. It does not appear in the assessed Mg-Zn binary phase diagram (Massalski, 1990). It *may* actually be the $Mg_2Zn_{11}$ $D8_c$ structure, but we have found no literature supporting this claim.

## Base-centered Monoclinic primitive vectors:

$$\mathbf{a}_1 = \frac{1}{2}\,a\,\hat{\mathbf{x}} - \frac{1}{2}\,b\,\hat{\mathbf{y}}$$
$$\mathbf{a}_2 = \frac{1}{2}\,a\,\hat{\mathbf{x}} + \frac{1}{2}\,b\,\hat{\mathbf{y}}$$
$$\mathbf{a}_3 = c\cos\beta\,\hat{\mathbf{x}} + c\sin\beta\,\hat{\mathbf{z}}$$

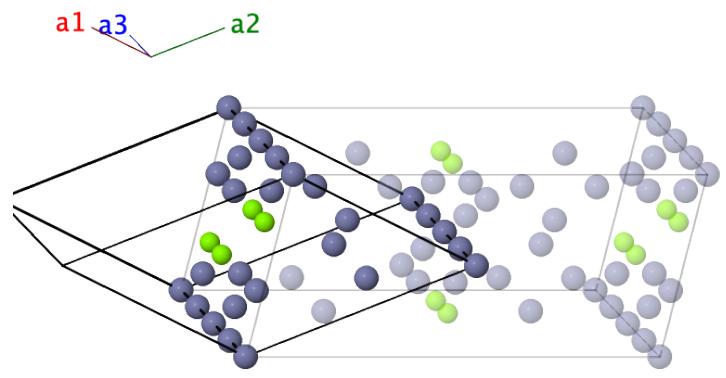

## Basis vectors:

| | | Lattice Coordinates | | Cartesian Coordinates | Wyckoff Position | Atom Type |
|---|---|---|---|---|---|---|
| $\mathbf{B}_1$ | = | $0\,\mathbf{a}_1 + 0\,\mathbf{a}_2 + 0\,\mathbf{a}_3$ | = | $0\,\hat{\mathbf{x}} + 0\,\hat{\mathbf{y}} + 0\,\hat{\mathbf{z}}$ | (2a) | Zn I |
| $\mathbf{B}_2$ | = | $\frac{1}{2}\,\mathbf{a}_3$ | = | $\frac{1}{2}c\cos\beta\,\hat{\mathbf{x}} + \frac{1}{2}c\sin\beta\,\hat{\mathbf{z}}$ | (2c) | Zn II |
| $\mathbf{B}_3$ | = | $x_3\,\mathbf{a}_1 + x_3\,\mathbf{a}_2 + z_3\,\mathbf{a}_3$ | = | $(x_3a + z_3c\cos\beta)\,\hat{\mathbf{x}} + z_3c\sin\beta\,\hat{\mathbf{z}}$ | (4i) | Mg I |
| $\mathbf{B}_4$ | = | $-x_3\,\mathbf{a}_1 - x_3\,\mathbf{a}_2 - z_3\,\mathbf{a}_3$ | = | $(-x_3a - z_3c\cos\beta)\,\hat{\mathbf{x}} - z_3c\sin\beta\,\hat{\mathbf{z}}$ | (4i) | Mg I |
| $\mathbf{B}_5$ | = | $x_4\,\mathbf{a}_1 + x_4\,\mathbf{a}_2 + z_4\,\mathbf{a}_3$ | = | $(x_4a + z_4c\cos\beta)\,\hat{\mathbf{x}} + z_4c\sin\beta\,\hat{\mathbf{z}}$ | (4i) | Mg II |
| $\mathbf{B}_6$ | = | $-x_4\,\mathbf{a}_1 - x_4\,\mathbf{a}_2 - z_4\,\mathbf{a}_3$ | = | $(-x_4a - z_4c\cos\beta)\,\hat{\mathbf{x}} - z_4c\sin\beta\,\hat{\mathbf{z}}$ | (4i) | Mg II |
| $\mathbf{B}_7$ | = | $x_5\,\mathbf{a}_1 + x_5\,\mathbf{a}_2 + z_5\,\mathbf{a}_3$ | = | $(x_5a + z_5c\cos\beta)\,\hat{\mathbf{x}} + z_5c\sin\beta\,\hat{\mathbf{z}}$ | (4i) | Zn III |
| $\mathbf{B}_8$ | = | $-x_5\,\mathbf{a}_1 - x_5\,\mathbf{a}_2 - z_5\,\mathbf{a}_3$ | = | $(-x_5a - z_5c\cos\beta)\,\hat{\mathbf{x}} - z_5c\sin\beta\,\hat{\mathbf{z}}$ | (4i) | Zn III |
| $\mathbf{B}_9$ | = | $x_6\,\mathbf{a}_1 + x_6\,\mathbf{a}_2 + z_6\,\mathbf{a}_3$ | = | $(x_6a + z_6c\cos\beta)\,\hat{\mathbf{x}} + z_6c\sin\beta\,\hat{\mathbf{z}}$ | (4i) | Zn IV |
| $\mathbf{B}_{10}$ | = | $-x_6\,\mathbf{a}_1 - x_6\,\mathbf{a}_2 - z_6\,\mathbf{a}_3$ | = | $(-x_6a - z_6c\cos\beta)\,\hat{\mathbf{x}} - z_6c\sin\beta\,\hat{\mathbf{z}}$ | (4i) | Zn IV |
| $\mathbf{B}_{11}$ | = | $x_7\,\mathbf{a}_1 + x_7\,\mathbf{a}_2 + z_7\,\mathbf{a}_3$ | = | $(x_7a + z_7c\cos\beta)\,\hat{\mathbf{x}} + z_7c\sin\beta\,\hat{\mathbf{z}}$ | (4i) | Zn V |
| $\mathbf{B}_{12}$ | = | $-x_7\,\mathbf{a}_1 - x_7\,\mathbf{a}_2 - z_7\,\mathbf{a}_3$ | = | $(-x_7a - z_7c\cos\beta)\,\hat{\mathbf{x}} - z_7c\sin\beta\,\hat{\mathbf{z}}$ | (4i) | Zn V |
| $\mathbf{B}_{13}$ | = | $x_8\,\mathbf{a}_1 + x_8\,\mathbf{a}_2 + z_8\,\mathbf{a}_3$ | = | $(x_8a + z_8c\cos\beta)\,\hat{\mathbf{x}} + z_8c\sin\beta\,\hat{\mathbf{z}}$ | (4i) | Zn VI |
| $\mathbf{B}_{14}$ | = | $-x_8\,\mathbf{a}_1 - x_8\,\mathbf{a}_2 - z_8\,\mathbf{a}_3$ | = | $(-x_8a - z_8c\cos\beta)\,\hat{\mathbf{x}} - z_8c\sin\beta\,\hat{\mathbf{z}}$ | (4i) | Zn VI |
| $\mathbf{B}_{15}$ | = | $x_9\,\mathbf{a}_1 + x_9\,\mathbf{a}_2 + z_9\,\mathbf{a}_3$ | = | $(x_9a + z_9c\cos\beta)\,\hat{\mathbf{x}} + z_9c\sin\beta\,\hat{\mathbf{z}}$ | (4i) | Zn VII |
| $\mathbf{B}_{16}$ | = | $-x_9\,\mathbf{a}_1 - x_9\,\mathbf{a}_2 - z_9\,\mathbf{a}_3$ | = | $(-x_9a - z_9c\cos\beta)\,\hat{\mathbf{x}} - z_9c\sin\beta\,\hat{\mathbf{z}}$ | (4i) | Zn VII |
| $\mathbf{B}_{17}$ | = | $(x_{10} - y_{10})\,\mathbf{a}_1 + (x_{10} + y_{10})\,\mathbf{a}_2 + z_{10}\,\mathbf{a}_3$ | = | $(x_{10}a + z_{10}c\cos\beta)\,\hat{\mathbf{x}} + y_{10}b\,\hat{\mathbf{y}} + z_{10}c\sin\beta\,\hat{\mathbf{z}}$ | (8j) | Zn VIII |
| $\mathbf{B}_{18}$ | = | $(-x_{10} - y_{10})\,\mathbf{a}_1 + (-x_{10} + y_{10})\,\mathbf{a}_2 - z_{10}\,\mathbf{a}_3$ | = | $(-x_{10}a - z_{10}c\cos\beta)\,\hat{\mathbf{x}} + y_{10}b\,\hat{\mathbf{y}} - z_{10}c\sin\beta\,\hat{\mathbf{z}}$ | (8j) | Zn VIII |
| $\mathbf{B}_{19}$ | = | $(-x_{10} + y_{10})\,\mathbf{a}_1 + (-x_{10} - y_{10})\,\mathbf{a}_2 - z_{10}\,\mathbf{a}_3$ | = | $(-x_{10}a - z_{10}c\cos\beta)\,\hat{\mathbf{x}} - y_{10}b\,\hat{\mathbf{y}} - z_{10}c\sin\beta\,\hat{\mathbf{z}}$ | (8j) | Zn VIII |
| $\mathbf{B}_{20}$ | = | $(x_{10} + y_{10})\,\mathbf{a}_1 + (x_{10} - y_{10})\,\mathbf{a}_2 + z_{10}\,\mathbf{a}_3$ | = | $(x_{10}a + z_{10}c\cos\beta)\,\hat{\mathbf{x}} - y_{10}b\,\hat{\mathbf{y}} + z_{10}c\sin\beta\,\hat{\mathbf{z}}$ | (8j) | Zn VIII |
| $\mathbf{B}_{21}$ | = | $(x_{11} - y_{11})\,\mathbf{a}_1 + (x_{11} + y_{11})\,\mathbf{a}_2 + z_{11}\,\mathbf{a}_3$ | = | $(x_{11}a + z_{11}c\cos\beta)\,\hat{\mathbf{x}} + y_{11}b\,\hat{\mathbf{y}} + z_{11}c\sin\beta\,\hat{\mathbf{z}}$ | (8j) | Zn IX |
| $\mathbf{B}_{22}$ | = | $(-x_{11} - y_{11})\,\mathbf{a}_1 + (-x_{11} + y_{11})\,\mathbf{a}_2 - z_{11}\,\mathbf{a}_3$ | = | $(-x_{11}a - z_{11}c\cos\beta)\,\hat{\mathbf{x}} + y_{11}b\,\hat{\mathbf{y}} - z_{11}c\sin\beta\,\hat{\mathbf{z}}$ | (8j) | Zn IX |
| $\mathbf{B}_{23}$ | = | $(-x_{11} + y_{11})\,\mathbf{a}_1 + (-x_{11} - y_{11})\,\mathbf{a}_2 - z_{11}\,\mathbf{a}_3$ | = | $(-x_{11}a - z_{11}c\cos\beta)\,\hat{\mathbf{x}} - y_{11}b\,\hat{\mathbf{y}} - z_{11}c\sin\beta\,\hat{\mathbf{z}}$ | (8j) | Zn IX |
| $\mathbf{B}_{24}$ | = | $(x_{11} + y_{11})\,\mathbf{a}_1 + (x_{11} - y_{11})\,\mathbf{a}_2 + z_{11}\,\mathbf{a}_3$ | = | $(x_{11}a + z_{11}c\cos\beta)\,\hat{\mathbf{x}} - y_{11}b\,\hat{\mathbf{y}} + z_{11}c\sin\beta\,\hat{\mathbf{z}}$ | (8j) | Zn IX |

**Geometry files:**
- CIF: pp. 1532
- POSCAR: pp. 1532

# Chrysotile ($H_4Mg_3Si_2O_9$, $S4_5$) Structure: AB6C11D6E4_mC112_12_e_gi2j_i5j_2i2j_2j

http://aflow.org/prototype-encyclopedia/AB6C11D6E4_mC112_12_e_gi2j_i5j_2i2j_2j

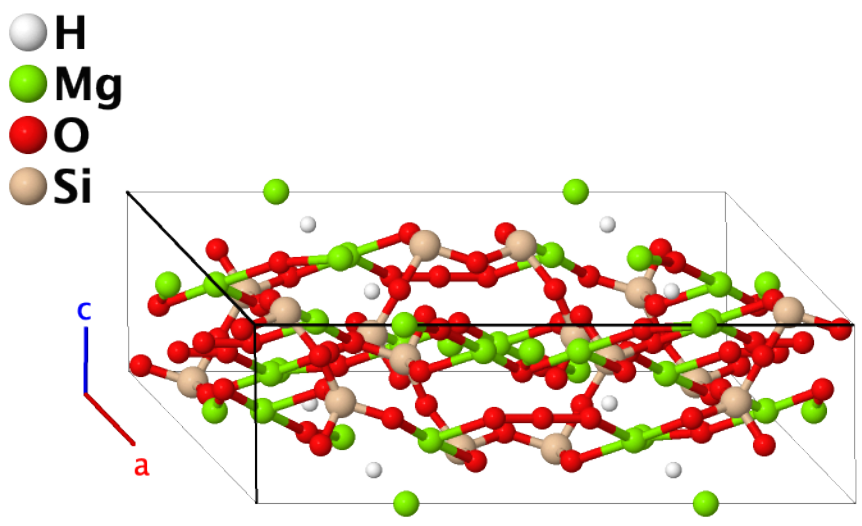

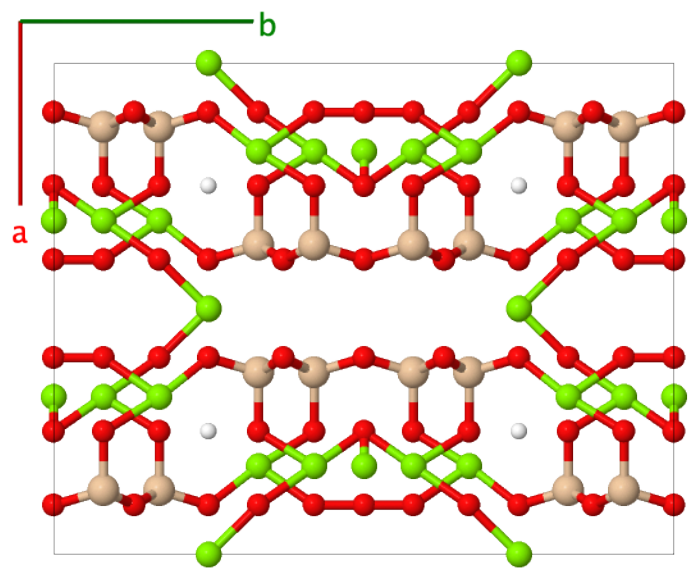

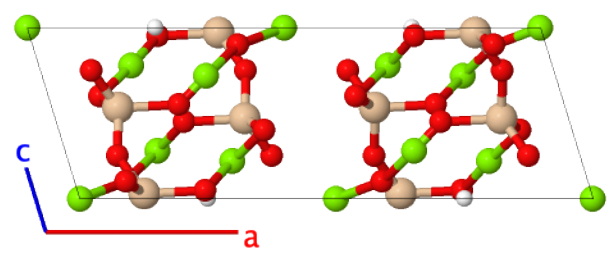

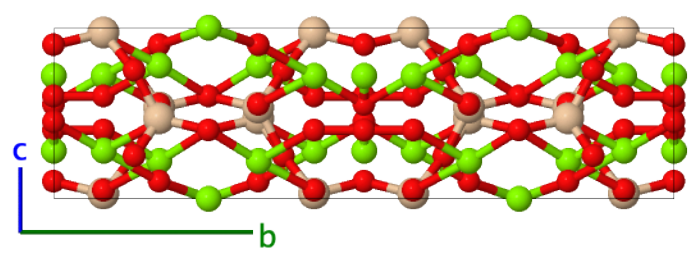

| | | |
|---|---|---|
| **Prototype** | : | $(H_2O)(OH)_6O_{11}Mg_6Si_4$ |
| **AFLOW prototype label** | : | AB6C11D6E4_mC112_12_e_gi2j_i5j_2i2j_2j |
| ***Strukturbericht* designation** | : | $S4_5$ |
| **Pearson symbol** | : | mC112 |
| **Space group number** | : | 12 |
| **Space group symbol** | : | $C2/m$ |
| **AFLOW prototype command** | : | `aflow --proto=AB6C11D6E4_mC112_12_e_gi2j_i5j_2i2j_2j --params=`$a, b/a, c/a, \beta, y_2, x_3, z_3, x_4, z_4, x_5, z_5, x_6, z_6, x_7, y_7, z_7, x_8, y_8, z_8, x_9, y_9, z_9, x_{10}, y_{10}, z_{10}, x_{11}, y_{11}, z_{11}, x_{12}, y_{12}, z_{12}, x_{13}, y_{13}, z_{13}, x_{14}, y_{14}, z_{14}, x_{15}, y_{15}, z_{15}, x_{16}, y_{16}, z_{16}, x_{17}, y_{17}, z_{17}$ |

- Chrysotile is more commonly known as "white asbestos." Chrysotile sheets typically curl into tubular fibers and the crystal structure is difficult to determine. (Yada, 1967) has a partial list of the experiments performed to determine this structure. The current structure was given the *Strukturbericht* designation $S4_5$ by (Hermann, 1937). It doubles the unit cell of our other chrysotile structure, by (Falini, 2004) and includes an inversion site.
- The structure was originally given in the $I2/m$ setting of space group #12. We used FINDSYM to place it in the standard $C2/m$ setting.

## Base-centered Monoclinic primitive vectors:

$$\mathbf{a}_1 = \frac{1}{2}\,a\,\hat{\mathbf{x}} - \frac{1}{2}\,b\,\hat{\mathbf{y}}$$
$$\mathbf{a}_2 = \frac{1}{2}\,a\,\hat{\mathbf{x}} + \frac{1}{2}\,b\,\hat{\mathbf{y}}$$
$$\mathbf{a}_3 = c\cos\beta\,\hat{\mathbf{x}} + c\sin\beta\,\hat{\mathbf{z}}$$

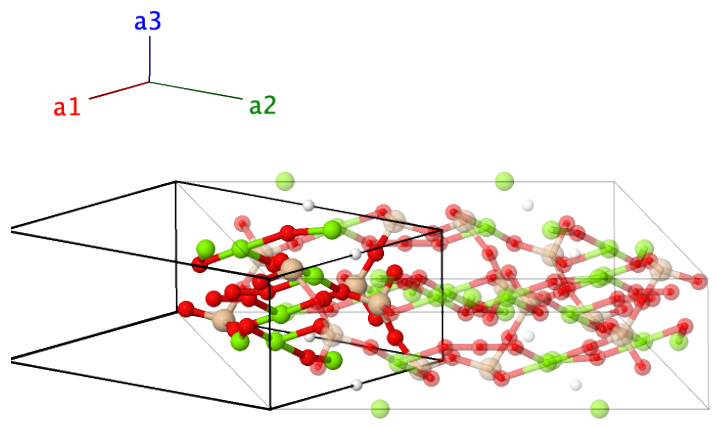

## Basis vectors:

| | | Lattice Coordinates | | Cartesian Coordinates | Wyckoff Position | Atom Type |
|---|---|---|---|---|---|---|
| $\mathbf{B}_1$ | = | $\frac{1}{2}\,\mathbf{a}_2$ | = | $\frac{1}{4}a\,\hat{\mathbf{x}} + \frac{1}{4}b\,\hat{\mathbf{y}}$ | (4$e$) | $H_2O$ |
| $\mathbf{B}_2$ | = | $\frac{1}{2}\,\mathbf{a}_1$ | = | $\frac{1}{4}a\,\hat{\mathbf{x}} - \frac{1}{4}b\,\hat{\mathbf{y}}$ | (4$e$) | $H_2O$ |
| $\mathbf{B}_3$ | = | $-y_2\,\mathbf{a}_1 + y_2\,\mathbf{a}_2$ | = | $y_2 b\,\hat{\mathbf{y}}$ | (4$g$) | Mg I |
| $\mathbf{B}_4$ | = | $y_2\,\mathbf{a}_1 - y_2\,\mathbf{a}_2$ | = | $-y_2 b\,\hat{\mathbf{y}}$ | (4$g$) | Mg I |
| $\mathbf{B}_5$ | = | $x_3\,\mathbf{a}_1 + x_3\,\mathbf{a}_2 + z_3\,\mathbf{a}_3$ | = | $(x_3 a + z_3 c\cos\beta)\,\hat{\mathbf{x}} + z_3 c\sin\beta\,\hat{\mathbf{z}}$ | (4$i$) | Mg II |
| $\mathbf{B}_6$ | = | $-x_3\,\mathbf{a}_1 - x_3\,\mathbf{a}_2 - z_3\,\mathbf{a}_3$ | = | $(-x_3 a - z_3 c\cos\beta)\,\hat{\mathbf{x}} - z_3 c\sin\beta\,\hat{\mathbf{z}}$ | (4$i$) | Mg II |
| $\mathbf{B}_7$ | = | $x_4\,\mathbf{a}_1 + x_4\,\mathbf{a}_2 + z_4\,\mathbf{a}_3$ | = | $(x_4 a + z_4 c\cos\beta)\,\hat{\mathbf{x}} + z_4 c\sin\beta\,\hat{\mathbf{z}}$ | (4$i$) | O I |
| $\mathbf{B}_8$ | = | $-x_4\,\mathbf{a}_1 - x_4\,\mathbf{a}_2 - z_4\,\mathbf{a}_3$ | = | $(-x_4 a - z_4 c\cos\beta)\,\hat{\mathbf{x}} - z_4 c\sin\beta\,\hat{\mathbf{z}}$ | (4$i$) | O I |
| $\mathbf{B}_9$ | = | $x_5\,\mathbf{a}_1 + x_5\,\mathbf{a}_2 + z_5\,\mathbf{a}_3$ | = | $(x_5 a + z_5 c\cos\beta)\,\hat{\mathbf{x}} + z_5 c\sin\beta\,\hat{\mathbf{z}}$ | (4$i$) | OH I |
| $\mathbf{B}_{10}$ | = | $-x_5\,\mathbf{a}_1 - x_5\,\mathbf{a}_2 - z_5\,\mathbf{a}_3$ | = | $(-x_5 a - z_5 c\cos\beta)\,\hat{\mathbf{x}} - z_5 c\sin\beta\,\hat{\mathbf{z}}$ | (4$i$) | OH I |
| $\mathbf{B}_{11}$ | = | $x_6\,\mathbf{a}_1 + x_6\,\mathbf{a}_2 + z_6\,\mathbf{a}_3$ | = | $(x_6 a + z_6 c\cos\beta)\,\hat{\mathbf{x}} + z_6 c\sin\beta\,\hat{\mathbf{z}}$ | (4$i$) | OH II |
| $\mathbf{B}_{12}$ | = | $-x_6\,\mathbf{a}_1 - x_6\,\mathbf{a}_2 - z_6\,\mathbf{a}_3$ | = | $(-x_6 a - z_6 c\cos\beta)\,\hat{\mathbf{x}} - z_6 c\sin\beta\,\hat{\mathbf{z}}$ | (4$i$) | OH II |
| $\mathbf{B}_{13}$ | = | $(x_7 - y_7)\,\mathbf{a}_1 + (x_7 + y_7)\,\mathbf{a}_2 + z_7\,\mathbf{a}_3$ | = | $(x_7 a + z_7 c\cos\beta)\,\hat{\mathbf{x}} + y_7 b\,\hat{\mathbf{y}} + z_7 c\sin\beta\,\hat{\mathbf{z}}$ | (8$j$) | Mg III |
| $\mathbf{B}_{14}$ | = | $(-x_7 - y_7)\,\mathbf{a}_1 + (-x_7 + y_7)\,\mathbf{a}_2 - z_7\,\mathbf{a}_3$ | = | $(-x_7 a - z_7 c\cos\beta)\,\hat{\mathbf{x}} + y_7 b\,\hat{\mathbf{y}} - z_7 c\sin\beta\,\hat{\mathbf{z}}$ | (8$j$) | Mg III |
| $\mathbf{B}_{15}$ | = | $(-x_7 + y_7)\,\mathbf{a}_1 + (-x_7 - y_7)\,\mathbf{a}_2 - z_7\,\mathbf{a}_3$ | = | $(-x_7 a - z_7 c\cos\beta)\,\hat{\mathbf{x}} - y_7 b\,\hat{\mathbf{y}} - z_7 c\sin\beta\,\hat{\mathbf{z}}$ | (8$j$) | Mg III |
| $\mathbf{B}_{16}$ | = | $(x_7 + y_7)\,\mathbf{a}_1 + (x_7 - y_7)\,\mathbf{a}_2 + z_7\,\mathbf{a}_3$ | = | $(x_7 a + z_7 c\cos\beta)\,\hat{\mathbf{x}} - y_7 b\,\hat{\mathbf{y}} + z_7 c\sin\beta\,\hat{\mathbf{z}}$ | (8$j$) | Mg III |
| $\mathbf{B}_{17}$ | = | $(x_8 - y_8)\,\mathbf{a}_1 + (x_8 + y_8)\,\mathbf{a}_2 + z_8\,\mathbf{a}_3$ | = | $(x_8 a + z_8 c\cos\beta)\,\hat{\mathbf{x}} + y_8 b\,\hat{\mathbf{y}} + z_8 c\sin\beta\,\hat{\mathbf{z}}$ | (8$j$) | Mg IV |
| $\mathbf{B}_{18}$ | = | $(-x_8 - y_8)\,\mathbf{a}_1 + (-x_8 + y_8)\,\mathbf{a}_2 - z_8\,\mathbf{a}_3$ | = | $(-x_8 a - z_8 c\cos\beta)\,\hat{\mathbf{x}} + y_8 b\,\hat{\mathbf{y}} - z_8 c\sin\beta\,\hat{\mathbf{z}}$ | (8$j$) | Mg IV |
| $\mathbf{B}_{19}$ | = | $(-x_8 + y_8)\,\mathbf{a}_1 + (-x_8 - y_8)\,\mathbf{a}_2 - z_8\,\mathbf{a}_3$ | = | $(-x_8 a - z_8 c\cos\beta)\,\hat{\mathbf{x}} - y_8 b\,\hat{\mathbf{y}} - z_8 c\sin\beta\,\hat{\mathbf{z}}$ | (8$j$) | Mg IV |
| $\mathbf{B}_{20}$ | = | $(x_8 + y_8)\,\mathbf{a}_1 + (x_8 - y_8)\,\mathbf{a}_2 + z_8\,\mathbf{a}_3$ | = | $(x_8 a + z_8 c\cos\beta)\,\hat{\mathbf{x}} - y_8 b\,\hat{\mathbf{y}} + z_8 c\sin\beta\,\hat{\mathbf{z}}$ | (8$j$) | Mg IV |
| $\mathbf{B}_{21}$ | = | $(x_9 - y_9)\,\mathbf{a}_1 + (x_9 + y_9)\,\mathbf{a}_2 + z_9\,\mathbf{a}_3$ | = | $(x_9 a + z_9 c\cos\beta)\,\hat{\mathbf{x}} + y_9 b\,\hat{\mathbf{y}} + z_9 c\sin\beta\,\hat{\mathbf{z}}$ | (8$j$) | O II |
| $\mathbf{B}_{22}$ | = | $(-x_9 - y_9)\,\mathbf{a}_1 + (-x_9 + y_9)\,\mathbf{a}_2 - z_9\,\mathbf{a}_3$ | = | $(-x_9 a - z_9 c\cos\beta)\,\hat{\mathbf{x}} + y_9 b\,\hat{\mathbf{y}} - z_9 c\sin\beta\,\hat{\mathbf{z}}$ | (8$j$) | O II |
| $\mathbf{B}_{23}$ | = | $\left(-x_{9}+y_{9}\right)\,\mathbf{a}_{1}+\left(-x_{9}-y_{9}\right)\,\mathbf{a}_{2}-z_{9}\,\mathbf{a}_{3}$ | = | $\left(-x_{9}a-z_{9}c\cos\beta\right)\,\mathbf{\hat{x}}-y_{9}b\,\mathbf{\hat{y}}-z_{9}c\sin\beta\,\mathbf{\hat{z}}$ | $(8j)$ | O II |
| $\mathbf{B}_{24}$ | = | $\left(x_{9}+y_{9}\right)\,\mathbf{a}_{1}+\left(x_{9}-y_{9}\right)\,\mathbf{a}_{2}+z_{9}\,\mathbf{a}_{3}$ | = | $\left(x_{9}a+z_{9}c\cos\beta\right)\,\mathbf{\hat{x}}-y_{9}b\,\mathbf{\hat{y}}+z_{9}c\sin\beta\,\mathbf{\hat{z}}$ | $(8j)$ | O II |
| $\mathbf{B}_{25}$ | = | $\left(x_{10}-y_{10}\right)\,\mathbf{a}_{1}+\left(x_{10}+y_{10}\right)\,\mathbf{a}_{2}+z_{10}\,\mathbf{a}_{3}$ | = | $\left(x_{10}a+z_{10}c\cos\beta\right)\,\mathbf{\hat{x}}+y_{10}b\,\mathbf{\hat{y}}+z_{10}c\sin\beta\,\mathbf{\hat{z}}$ | $(8j)$ | O III |
| $\mathbf{B}_{26}$ | = | $\left(-x_{10}-y_{10}\right)\,\mathbf{a}_{1}+\left(-x_{10}+y_{10}\right)\,\mathbf{a}_{2}-z_{10}\,\mathbf{a}_{3}$ | = | $\left(-x_{10}a-z_{10}c\cos\beta\right)\,\mathbf{\hat{x}}+y_{10}b\,\mathbf{\hat{y}}-z_{10}c\sin\beta\,\mathbf{\hat{z}}$ | $(8j)$ | O III |
| $\mathbf{B}_{27}$ | = | $\left(-x_{10}+y_{10}\right)\,\mathbf{a}_{1}+\left(-x_{10}-y_{10}\right)\,\mathbf{a}_{2}-z_{10}\,\mathbf{a}_{3}$ | = | $\left(-x_{10}a-z_{10}c\cos\beta\right)\,\mathbf{\hat{x}}-y_{10}b\,\mathbf{\hat{y}}-z_{10}c\sin\beta\,\mathbf{\hat{z}}$ | $(8j)$ | O III |
| $\mathbf{B}_{28}$ | = | $\left(x_{10}+y_{10}\right)\,\mathbf{a}_{1}+\left(x_{10}-y_{10}\right)\,\mathbf{a}_{2}+z_{10}\,\mathbf{a}_{3}$ | = | $\left(x_{10}a+z_{10}c\cos\beta\right)\,\mathbf{\hat{x}}-y_{10}b\,\mathbf{\hat{y}}+z_{10}c\sin\beta\,\mathbf{\hat{z}}$ | $(8j)$ | O III |
| $\mathbf{B}_{29}$ | = | $\left(x_{11}-y_{11}\right)\,\mathbf{a}_{1}+\left(x_{11}+y_{11}\right)\,\mathbf{a}_{2}+z_{11}\,\mathbf{a}_{3}$ | = | $\left(x_{11}a+z_{11}c\cos\beta\right)\,\mathbf{\hat{x}}+y_{11}b\,\mathbf{\hat{y}}+z_{11}c\sin\beta\,\mathbf{\hat{z}}$ | $(8j)$ | O IV |
| $\mathbf{B}_{30}$ | = | $\left(-x_{11}-y_{11}\right)\,\mathbf{a}_{1}+\left(-x_{11}+y_{11}\right)\,\mathbf{a}_{2}-z_{11}\,\mathbf{a}_{3}$ | = | $\left(-x_{11}a-z_{11}c\cos\beta\right)\,\mathbf{\hat{x}}+y_{11}b\,\mathbf{\hat{y}}-z_{11}c\sin\beta\,\mathbf{\hat{z}}$ | $(8j)$ | O IV |
| $\mathbf{B}_{31}$ | = | $\left(-x_{11}+y_{11}\right)\,\mathbf{a}_{1}+\left(-x_{11}-y_{11}\right)\,\mathbf{a}_{2}-z_{11}\,\mathbf{a}_{3}$ | = | $\left(-x_{11}a-z_{11}c\cos\beta\right)\,\mathbf{\hat{x}}-y_{11}b\,\mathbf{\hat{y}}-z_{11}c\sin\beta\,\mathbf{\hat{z}}$ | $(8j)$ | O IV |
| $\mathbf{B}_{32}$ | = | $\left(x_{11}+y_{11}\right)\,\mathbf{a}_{1}+\left(x_{11}-y_{11}\right)\,\mathbf{a}_{2}+z_{11}\,\mathbf{a}_{3}$ | = | $\left(x_{11}a+z_{11}c\cos\beta\right)\,\mathbf{\hat{x}}-y_{11}b\,\mathbf{\hat{y}}+z_{11}c\sin\beta\,\mathbf{\hat{z}}$ | $(8j)$ | O IV |
| $\mathbf{B}_{33}$ | = | $\left(x_{12}-y_{12}\right)\,\mathbf{a}_{1}+\left(x_{12}+y_{12}\right)\,\mathbf{a}_{2}+z_{12}\,\mathbf{a}_{3}$ | = | $\left(x_{12}a+z_{12}c\cos\beta\right)\,\mathbf{\hat{x}}+y_{12}b\,\mathbf{\hat{y}}+z_{12}c\sin\beta\,\mathbf{\hat{z}}$ | $(8j)$ | O V |
| $\mathbf{B}_{34}$ | = | $\left(-x_{12}-y_{12}\right)\,\mathbf{a}_{1}+\left(-x_{12}+y_{12}\right)\,\mathbf{a}_{2}-z_{12}\,\mathbf{a}_{3}$ | = | $\left(-x_{12}a-z_{12}c\cos\beta\right)\,\mathbf{\hat{x}}+y_{12}b\,\mathbf{\hat{y}}-z_{12}c\sin\beta\,\mathbf{\hat{z}}$ | $(8j)$ | O V |
| $\mathbf{B}_{35}$ | = | $\left(-x_{12}+y_{12}\right)\,\mathbf{a}_{1}+\left(-x_{12}-y_{12}\right)\,\mathbf{a}_{2}-z_{12}\,\mathbf{a}_{3}$ | = | $\left(-x_{12}a-z_{12}c\cos\beta\right)\,\mathbf{\hat{x}}-y_{12}b\,\mathbf{\hat{y}}-z_{12}c\sin\beta\,\mathbf{\hat{z}}$ | $(8j)$ | O V |
| $\mathbf{B}_{36}$ | = | $\left(x_{12}+y_{12}\right)\,\mathbf{a}_{1}+\left(x_{12}-y_{12}\right)\,\mathbf{a}_{2}+z_{12}\,\mathbf{a}_{3}$ | = | $\left(x_{12}a+z_{12}c\cos\beta\right)\,\mathbf{\hat{x}}-y_{12}b\,\mathbf{\hat{y}}+z_{12}c\sin\beta\,\mathbf{\hat{z}}$ | $(8j)$ | O V |
| $\mathbf{B}_{37}$ | = | $\left(x_{13}-y_{13}\right)\,\mathbf{a}_{1}+\left(x_{13}+y_{13}\right)\,\mathbf{a}_{2}+z_{13}\,\mathbf{a}_{3}$ | = | $\left(x_{13}a+z_{13}c\cos\beta\right)\,\mathbf{\hat{x}}+y_{13}b\,\mathbf{\hat{y}}+z_{13}c\sin\beta\,\mathbf{\hat{z}}$ | $(8j)$ | O VI |
| $\mathbf{B}_{38}$ | = | $\left(-x_{13}-y_{13}\right)\,\mathbf{a}_{1}+\left(-x_{13}+y_{13}\right)\,\mathbf{a}_{2}-z_{13}\,\mathbf{a}_{3}$ | = | $\left(-x_{13}a-z_{13}c\cos\beta\right)\,\mathbf{\hat{x}}+y_{13}b\,\mathbf{\hat{y}}-z_{13}c\sin\beta\,\mathbf{\hat{z}}$ | $(8j)$ | O VI |
| $\mathbf{B}_{39}$ | = | $\left(-x_{13}+y_{13}\right)\,\mathbf{a}_{1}+\left(-x_{13}-y_{13}\right)\,\mathbf{a}_{2}-z_{13}\,\mathbf{a}_{3}$ | = | $\left(-x_{13}a-z_{13}c\cos\beta\right)\,\mathbf{\hat{x}}-y_{13}b\,\mathbf{\hat{y}}-z_{13}c\sin\beta\,\mathbf{\hat{z}}$ | $(8j)$ | O VI |
| $\mathbf{B}_{40}$ | = | $\left(x_{13}+y_{13}\right)\,\mathbf{a}_{1}+\left(x_{13}-y_{13}\right)\,\mathbf{a}_{2}+z_{13}\,\mathbf{a}_{3}$ | = | $\left(x_{13}a+z_{13}c\cos\beta\right)\,\mathbf{\hat{x}}-y_{13}b\,\mathbf{\hat{y}}+z_{13}c\sin\beta\,\mathbf{\hat{z}}$ | $(8j)$ | O VI |
| $\mathbf{B}_{41}$ | = | $\left(x_{14}-y_{14}\right)\,\mathbf{a}_{1}+\left(x_{14}+y_{14}\right)\,\mathbf{a}_{2}+z_{14}\,\mathbf{a}_{3}$ | = | $\left(x_{14}a+z_{14}c\cos\beta\right)\,\mathbf{\hat{x}}+y_{14}b\,\mathbf{\hat{y}}+z_{14}c\sin\beta\,\mathbf{\hat{z}}$ | $(8j)$ | OH III |
| $\mathbf{B}_{42}$ | = | $\left(-x_{14}-y_{14}\right)\,\mathbf{a}_{1}+\left(-x_{14}+y_{14}\right)\,\mathbf{a}_{2}-z_{14}\,\mathbf{a}_{3}$ | = | $\left(-x_{14}a-z_{14}c\cos\beta\right)\,\mathbf{\hat{x}}+y_{14}b\,\mathbf{\hat{y}}-z_{14}c\sin\beta\,\mathbf{\hat{z}}$ | $(8j)$ | OH III |
| $\mathbf{B}_{43}$ | = | $\left(-x_{14}+y_{14}\right)\,\mathbf{a}_{1}+\left(-x_{14}-y_{14}\right)\,\mathbf{a}_{2}-z_{14}\,\mathbf{a}_{3}$ | = | $\left(-x_{14}a-z_{14}c\cos\beta\right)\,\mathbf{\hat{x}}-y_{14}b\,\mathbf{\hat{y}}-z_{14}c\sin\beta\,\mathbf{\hat{z}}$ | $(8j)$ | OH III |
| $\mathbf{B}_{44}$ | = | $\left(x_{14}+y_{14}\right)\,\mathbf{a}_{1}+\left(x_{14}-y_{14}\right)\,\mathbf{a}_{2}+z_{14}\,\mathbf{a}_{3}$ | = | $\left(x_{14}a+z_{14}c\cos\beta\right)\,\mathbf{\hat{x}}-y_{14}b\,\mathbf{\hat{y}}+z_{14}c\sin\beta\,\mathbf{\hat{z}}$ | $(8j)$ | OH III |
| $\mathbf{B}_{45}$ | = | $\left(x_{15}-y_{15}\right)\,\mathbf{a}_{1}+\left(x_{15}+y_{15}\right)\,\mathbf{a}_{2}+z_{15}\,\mathbf{a}_{3}$ | = | $\left(x_{15}a+z_{15}c\cos\beta\right)\,\mathbf{\hat{x}}+y_{15}b\,\mathbf{\hat{y}}+z_{15}c\sin\beta\,\mathbf{\hat{z}}$ | $(8j)$ | OH IV |
| $\mathbf{B}_{46}$ | = | $\left(-x_{15}-y_{15}\right)\,\mathbf{a}_{1}+\left(-x_{15}+y_{15}\right)\,\mathbf{a}_{2}-z_{15}\,\mathbf{a}_{3}$ | = | $\left(-x_{15}a-z_{15}c\cos\beta\right)\,\mathbf{\hat{x}}+y_{15}b\,\mathbf{\hat{y}}-z_{15}c\sin\beta\,\mathbf{\hat{z}}$ | $(8j)$ | OH IV |
| $\mathbf{B}_{47}$ | = | $\left(-x_{15}+y_{15}\right)\,\mathbf{a}_{1}+\left(-x_{15}-y_{15}\right)\,\mathbf{a}_{2}-z_{15}\,\mathbf{a}_{3}$ | = | $\left(-x_{15}a-z_{15}c\cos\beta\right)\,\mathbf{\hat{x}}-y_{15}b\,\mathbf{\hat{y}}-z_{15}c\sin\beta\,\mathbf{\hat{z}}$ | (8$j$) | OH IV |
| $\mathbf{B}_{48}$ | = | $\left(x_{15}+y_{15}\right)\,\mathbf{a}_{1}+\left(x_{15}-y_{15}\right)\,\mathbf{a}_{2}+z_{15}\,\mathbf{a}_{3}$ | = | $\left(x_{15}a+z_{15}c\cos\beta\right)\,\mathbf{\hat{x}}-y_{15}b\,\mathbf{\hat{y}}+z_{15}c\sin\beta\,\mathbf{\hat{z}}$ | (8$j$) | OH IV |
| $\mathbf{B}_{49}$ | = | $\left(x_{16}-y_{16}\right)\,\mathbf{a}_{1}+\left(x_{16}+y_{16}\right)\,\mathbf{a}_{2}+z_{16}\,\mathbf{a}_{3}$ | = | $\left(x_{16}a+z_{16}c\cos\beta\right)\,\mathbf{\hat{x}}+y_{16}b\,\mathbf{\hat{y}}+z_{16}c\sin\beta\,\mathbf{\hat{z}}$ | (8$j$) | Si I |
| $\mathbf{B}_{50}$ | = | $\left(-x_{16}-y_{16}\right)\,\mathbf{a}_{1}+\left(-x_{16}+y_{16}\right)\,\mathbf{a}_{2}-z_{16}\,\mathbf{a}_{3}$ | = | $\left(-x_{16}a-z_{16}c\cos\beta\right)\,\mathbf{\hat{x}}+y_{16}b\,\mathbf{\hat{y}}-z_{16}c\sin\beta\,\mathbf{\hat{z}}$ | (8$j$) | Si I |
| $\mathbf{B}_{51}$ | = | $\left(-x_{16}+y_{16}\right)\,\mathbf{a}_{1}+\left(-x_{16}-y_{16}\right)\,\mathbf{a}_{2}-z_{16}\,\mathbf{a}_{3}$ | = | $\left(-x_{16}a-z_{16}c\cos\beta\right)\,\mathbf{\hat{x}}-y_{16}b\,\mathbf{\hat{y}}-z_{16}c\sin\beta\,\mathbf{\hat{z}}$ | (8$j$) | Si I |
| $\mathbf{B}_{52}$ | = | $\left(x_{16}+y_{16}\right)\,\mathbf{a}_{1}+\left(x_{16}-y_{16}\right)\,\mathbf{a}_{2}+z_{16}\,\mathbf{a}_{3}$ | = | $\left(x_{16}a+z_{16}c\cos\beta\right)\,\mathbf{\hat{x}}-y_{16}b\,\mathbf{\hat{y}}+z_{16}c\sin\beta\,\mathbf{\hat{z}}$ | (8$j$) | Si I |
| $\mathbf{B}_{53}$ | = | $\left(x_{17}-y_{17}\right)\,\mathbf{a}_{1}+\left(x_{17}+y_{17}\right)\,\mathbf{a}_{2}+z_{17}\,\mathbf{a}_{3}$ | = | $\left(x_{17}a+z_{17}c\cos\beta\right)\,\mathbf{\hat{x}}+y_{17}b\,\mathbf{\hat{y}}+z_{17}c\sin\beta\,\mathbf{\hat{z}}$ | (8$j$) | Si II |
| $\mathbf{B}_{54}$ | = | $\left(-x_{17}-y_{17}\right)\,\mathbf{a}_{1}+\left(-x_{17}+y_{17}\right)\,\mathbf{a}_{2}-z_{17}\,\mathbf{a}_{3}$ | = | $\left(-x_{17}a-z_{17}c\cos\beta\right)\,\mathbf{\hat{x}}+y_{17}b\,\mathbf{\hat{y}}-z_{17}c\sin\beta\,\mathbf{\hat{z}}$ | (8$j$) | Si II |
| $\mathbf{B}_{55}$ | = | $\left(-x_{17}+y_{17}\right)\,\mathbf{a}_{1}+\left(-x_{17}-y_{17}\right)\,\mathbf{a}_{2}-z_{17}\,\mathbf{a}_{3}$ | = | $\left(-x_{17}a-z_{17}c\cos\beta\right)\,\mathbf{\hat{x}}-y_{17}b\,\mathbf{\hat{y}}-z_{17}c\sin\beta\,\mathbf{\hat{z}}$ | (8$j$) | Si II |
| $\mathbf{B}_{56}$ | = | $\left(x_{17}+y_{17}\right)\,\mathbf{a}_{1}+\left(x_{17}-y_{17}\right)\,\mathbf{a}_{2}+z_{17}\,\mathbf{a}_{3}$ | = | $\left(x_{17}a+z_{17}c\cos\beta\right)\,\mathbf{\hat{x}}-y_{17}b\,\mathbf{\hat{y}}+z_{17}c\sin\beta\,\mathbf{\hat{z}}$ | (8$j$) | Si II |

**Found in:**

**Geometry files:**

- CIF: pp. 1532
- POSCAR: pp. 1533

# $Sr_2NiTeO_6$ Structure: AB6C2D_mC40_12_ad_gh4i_j_bc

http://aflow.org/prototype-encyclopedia/AB6C2D_mC40_12_ad_gh4i_j_bc

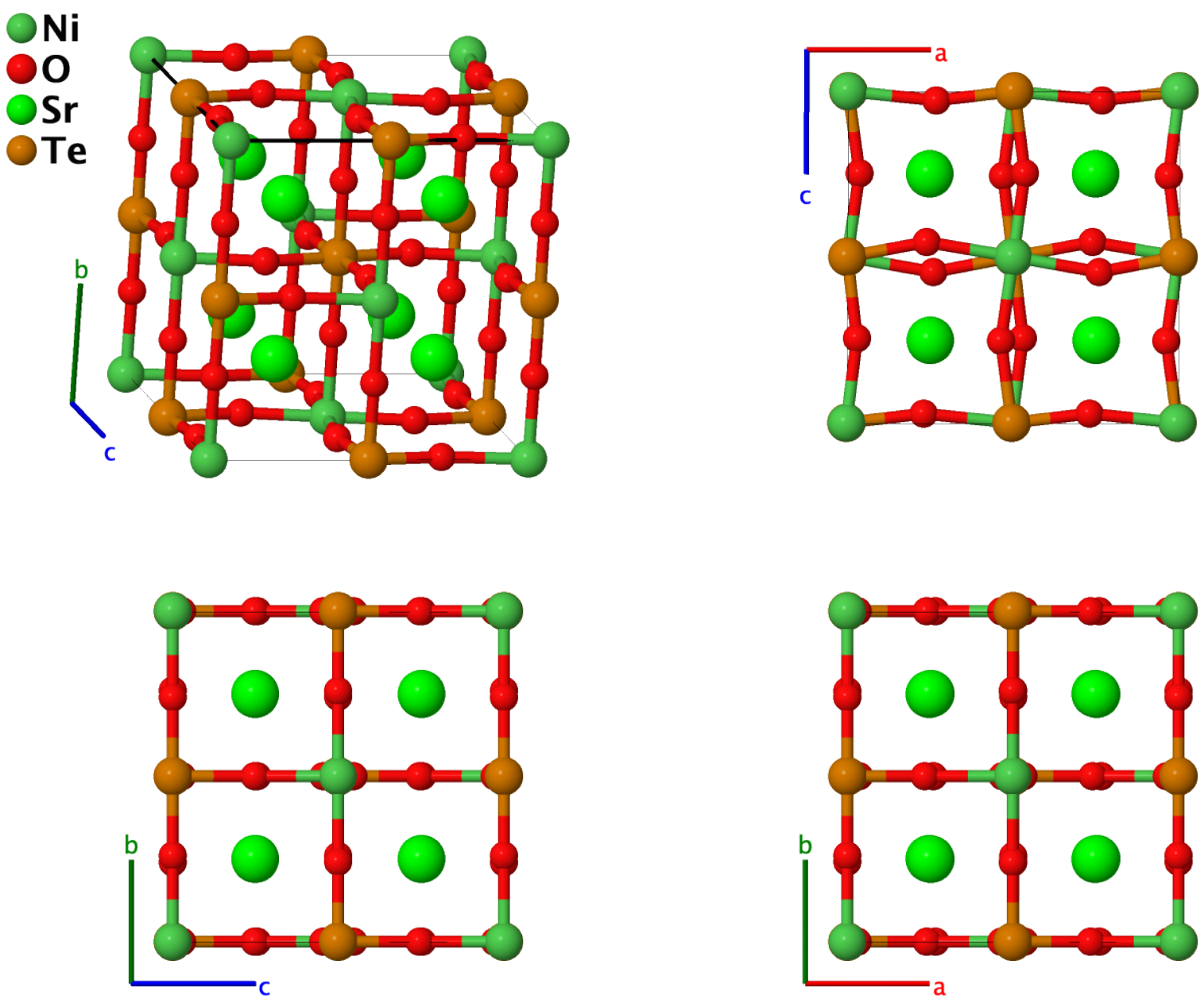

| | | |
|---|---|---|
| **Prototype** | : | $NiO_6Sr_2Te$ |
| **AFLOW prototype label** | : | AB6C2D_mC40_12_ad_gh4i_j_bc |
| ***Strukturbericht* designation** | : | None |
| **Pearson symbol** | : | mC40 |
| **Space group number** | : | 12 |
| **Space group symbol** | : | $C2/m$ |
| **AFLOW prototype command** | : | `aflow --proto=AB6C2D_mC40_12_ad_gh4i_j_bc`<br>`--params=`$a, b/a, c/a, \beta, y_5, y_6, x_7, z_7, x_8, z_8, x_9, z_9, x_{10}, z_{10}, x_{11}, y_{11}, z_{11}$ |

**Other compounds with this structure**

- $Sr_2NiTeO_6$ and $Cs_2RbDy_6$

- At high temperatures $Sr_2NiTeO_6$ transforms into the cubic perovskite $E2_1$ structure.

**Base-centered Monoclinic primitive vectors:**

$$\mathbf{a}_1 = \frac{1}{2}\,a\,\hat{\mathbf{x}} - \frac{1}{2}\,b\,\hat{\mathbf{y}}$$
$$\mathbf{a}_2 = \frac{1}{2}\,a\,\hat{\mathbf{x}} + \frac{1}{2}\,b\,\hat{\mathbf{y}}$$
$$\mathbf{a}_3 = c\cos\beta\,\hat{\mathbf{x}} + c\sin\beta\,\hat{\mathbf{z}}$$

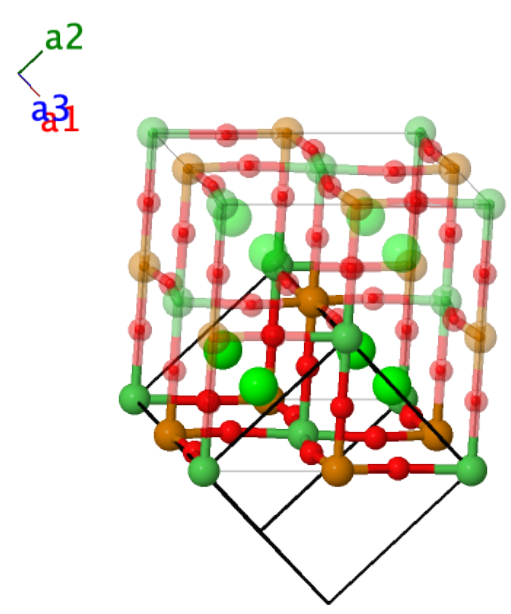

**Basis vectors:**

| | | Lattice Coordinates | | Cartesian Coordinates | Wyckoff Position | Atom Type |
|---|---|---|---|---|---|---|
| $\mathbf{B}_1$ | = | $0\,\mathbf{a}_1 + 0\,\mathbf{a}_2 + 0\,\mathbf{a}_3$ | = | $0\,\hat{\mathbf{x}} + 0\,\hat{\mathbf{y}} + 0\,\hat{\mathbf{z}}$ | $(2a)$ | Ni I |
| $\mathbf{B}_2$ | = | $\frac{1}{2}\,\mathbf{a}_1 + \frac{1}{2}\,\mathbf{a}_2$ | = | $\frac{1}{2}a\,\hat{\mathbf{x}}$ | $(2b)$ | Te I |
| $\mathbf{B}_3$ | = | $\frac{1}{2}\,\mathbf{a}_3$ | = | $\frac{1}{2}c\cos\beta\,\hat{\mathbf{x}} + \frac{1}{2}c\sin\beta\,\hat{\mathbf{z}}$ | $(2c)$ | Te II |
| $\mathbf{B}_4$ | = | $\frac{1}{2}\,\mathbf{a}_1 + \frac{1}{2}\,\mathbf{a}_2 + \frac{1}{2}\,\mathbf{a}_3$ | = | $\frac{1}{2}\,(a + c\cos\beta)\,\hat{\mathbf{x}} + \frac{1}{2}c\sin\beta\,\hat{\mathbf{z}}$ | $(2d)$ | Ni II |
| $\mathbf{B}_5$ | = | $-y_5\,\mathbf{a}_1 + y_5\,\mathbf{a}_2$ | = | $y_5b\,\hat{\mathbf{y}}$ | $(4g)$ | O I |
| $\mathbf{B}_6$ | = | $y_5\,\mathbf{a}_1 - y_5\,\mathbf{a}_2$ | = | $-y_5b\,\hat{\mathbf{y}}$ | $(4g)$ | O I |
| $\mathbf{B}_7$ | = | $-y_6\,\mathbf{a}_1 + y_6\,\mathbf{a}_2 + \frac{1}{2}\,\mathbf{a}_3$ | = | $\frac{1}{2}c\cos\beta\,\hat{\mathbf{x}} + y_6b\,\hat{\mathbf{y}} + \frac{1}{2}c\sin\beta\,\hat{\mathbf{z}}$ | $(4h)$ | O II |
| $\mathbf{B}_8$ | = | $y_6\,\mathbf{a}_1 - y_6\,\mathbf{a}_2 + \frac{1}{2}\,\mathbf{a}_3$ | = | $\frac{1}{2}c\cos\beta\,\hat{\mathbf{x}} - y_6b\,\hat{\mathbf{y}} + \frac{1}{2}c\sin\beta\,\hat{\mathbf{z}}$ | $(4h)$ | O II |
| $\mathbf{B}_9$ | = | $x_7\,\mathbf{a}_1 + x_7\,\mathbf{a}_2 + z_7\,\mathbf{a}_3$ | = | $(x_7a + z_7c\cos\beta)\,\hat{\mathbf{x}} + z_7c\sin\beta\,\hat{\mathbf{z}}$ | $(4i)$ | O III |
| $\mathbf{B}_{10}$ | = | $-x_7\,\mathbf{a}_1 - x_7\,\mathbf{a}_2 - z_7\,\mathbf{a}_3$ | = | $(-x_7a - z_7c\cos\beta)\,\hat{\mathbf{x}} - z_7c\sin\beta\,\hat{\mathbf{z}}$ | $(4i)$ | O III |
| $\mathbf{B}_{11}$ | = | $x_8\,\mathbf{a}_1 + x_8\,\mathbf{a}_2 + z_8\,\mathbf{a}_3$ | = | $(x_8a + z_8c\cos\beta)\,\hat{\mathbf{x}} + z_8c\sin\beta\,\hat{\mathbf{z}}$ | $(4i)$ | O IV |
| $\mathbf{B}_{12}$ | = | $-x_8\,\mathbf{a}_1 - x_8\,\mathbf{a}_2 - z_8\,\mathbf{a}_3$ | = | $(-x_8a - z_8c\cos\beta)\,\hat{\mathbf{x}} - z_8c\sin\beta\,\hat{\mathbf{z}}$ | $(4i)$ | O IV |
| $\mathbf{B}_{13}$ | = | $x_9\,\mathbf{a}_1 + x_9\,\mathbf{a}_2 + z_9\,\mathbf{a}_3$ | = | $(x_9a + z_9c\cos\beta)\,\hat{\mathbf{x}} + z_9c\sin\beta\,\hat{\mathbf{z}}$ | $(4i)$ | O V |
| $\mathbf{B}_{14}$ | = | $-x_9\,\mathbf{a}_1 - x_9\,\mathbf{a}_2 - z_9\,\mathbf{a}_3$ | = | $(-x_9a - z_9c\cos\beta)\,\hat{\mathbf{x}} - z_9c\sin\beta\,\hat{\mathbf{z}}$ | $(4i)$ | O V |
| $\mathbf{B}_{15}$ | = | $x_{10}\,\mathbf{a}_1 + x_{10}\,\mathbf{a}_2 + z_{10}\,\mathbf{a}_3$ | = | $(x_{10}a + z_{10}c\cos\beta)\,\hat{\mathbf{x}} + z_{10}c\sin\beta\,\hat{\mathbf{z}}$ | $(4i)$ | O VI |
| $\mathbf{B}_{16}$ | = | $-x_{10}\,\mathbf{a}_1 - x_{10}\,\mathbf{a}_2 - z_{10}\,\mathbf{a}_3$ | = | $(-x_{10}a - z_{10}c\cos\beta)\,\hat{\mathbf{x}} - z_{10}c\sin\beta\,\hat{\mathbf{z}}$ | $(4i)$ | O VI |
| $\mathbf{B}_{17}$ | = | $(x_{11} - y_{11})\,\mathbf{a}_1 + (x_{11} + y_{11})\,\mathbf{a}_2 + z_{11}\,\mathbf{a}_3$ | = | $(x_{11}a + z_{11}c\cos\beta)\,\hat{\mathbf{x}} + y_{11}b\,\hat{\mathbf{y}} + z_{11}c\sin\beta\,\hat{\mathbf{z}}$ | $(8j)$ | Sr |
| $\mathbf{B}_{18}$ | = | $(-x_{11} - y_{11})\,\mathbf{a}_1 + (-x_{11} + y_{11})\,\mathbf{a}_2 - z_{11}\,\mathbf{a}_3$ | = | $(-x_{11}a - z_{11}c\cos\beta)\,\hat{\mathbf{x}} + y_{11}b\,\hat{\mathbf{y}} - z_{11}c\sin\beta\,\hat{\mathbf{z}}$ | $(8j)$ | Sr |
| $\mathbf{B}_{19}$ | = | $(-x_{11} + y_{11})\,\mathbf{a}_1 + (-x_{11} - y_{11})\,\mathbf{a}_2 - z_{11}\,\mathbf{a}_3$ | = | $(-x_{11}a - z_{11}c\cos\beta)\,\hat{\mathbf{x}} - y_{11}b\,\hat{\mathbf{y}} - z_{11}c\sin\beta\,\hat{\mathbf{z}}$ | $(8j)$ | Sr |
| $\mathbf{B}_{20}$ | = | $(x_{11} + y_{11})\,\mathbf{a}_1 + (x_{11} - y_{11})\,\mathbf{a}_2 + z_{11}\,\mathbf{a}_3$ | = | $(x_{11}a + z_{11}c\cos\beta)\,\hat{\mathbf{x}} - y_{11}b\,\hat{\mathbf{y}} + z_{11}c\sin\beta\,\hat{\mathbf{z}}$ | $(8j)$ | Sr |

**Geometry files:**

- CIF: pp. 1533
- POSCAR: pp. 1533

# $Ta_2PdSe_6$ Structure: AB6C2_mC18_12_a_3i_i

http://aflow.org/prototype-encyclopedia/AB6C2_mC18_12_a_3i_i

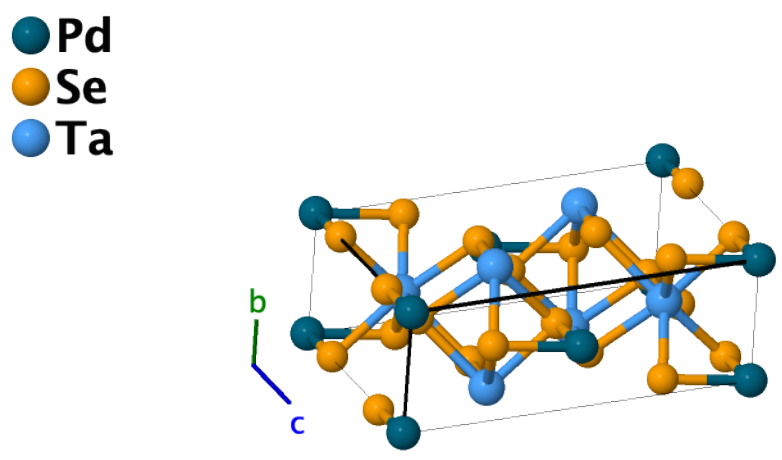

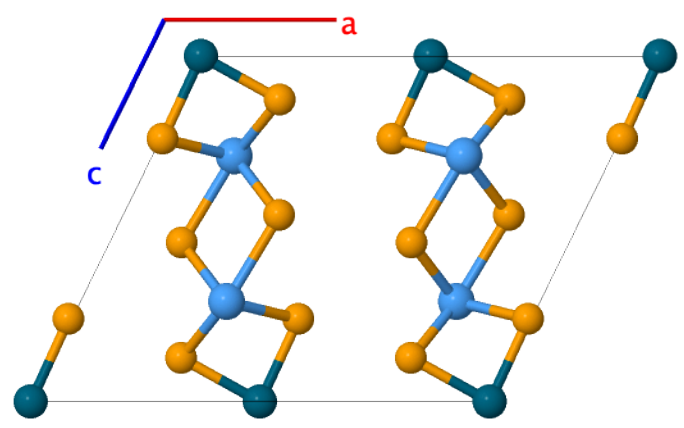

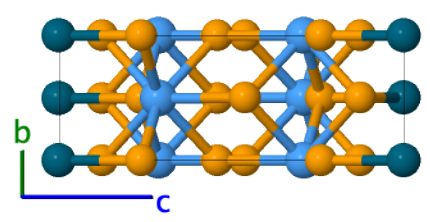

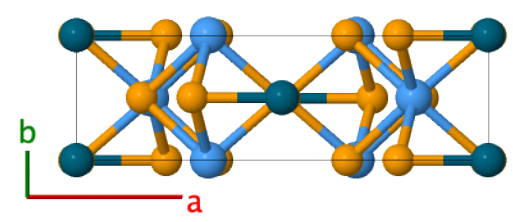

| | | |
|---|---|---|
| **Prototype** | : | $PdSe_6Ta_2$ |
| **AFLOW prototype label** | : | AB6C2_mC18_12_a_3i_i |
| ***Strukturbericht* designation** | : | None |
| **Pearson symbol** | : | mC18 |
| **Space group number** | : | 12 |
| **Space group symbol** | : | $C2/m$ |
| **AFLOW prototype command** | : | `aflow --proto=AB6C2_mC18_12_a_3i_i`<br>`--params=`$a, b/a, c/a, \beta, x_2, z_2, x_3, z_3, x_4, z_4, x_5, z_5$ |

**Other compounds with this structure**

- $Nb_2PdS_6$, $Nb_2PdSe_6$, and $Ta_2PdS_6$

- (Keszler, 1985) gave the structure in the $I2/m$ setting of space group #12. We used FINDSYM to change this to the standard $C2/m$ setting. Because of this change, our primitive vectors are linear combinations of the original ones, and the lattice has been rotated.

**Base-centered Monoclinic primitive vectors:**

$$\mathbf{a}_1 = \frac{1}{2}\,a\,\hat{\mathbf{x}} - \frac{1}{2}\,b\,\hat{\mathbf{y}}$$
$$\mathbf{a}_2 = \frac{1}{2}\,a\,\hat{\mathbf{x}} + \frac{1}{2}\,b\,\hat{\mathbf{y}}$$
$$\mathbf{a}_3 = c\cos\beta\,\hat{\mathbf{x}} + c\sin\beta\,\hat{\mathbf{z}}$$

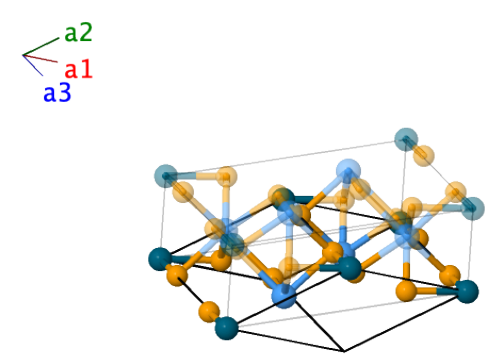

**Basis vectors:**

| | | Lattice Coordinates | | Cartesian Coordinates | Wyckoff Position | Atom Type |
|---|---|---|---|---|---|---|
| $\mathbf{B}_1$ | = | $0\,\mathbf{a}_1 + 0\,\mathbf{a}_2 + 0\,\mathbf{a}_3$ | = | $0\,\hat{\mathbf{x}} + 0\,\hat{\mathbf{y}} + 0\,\hat{\mathbf{z}}$ | (2$a$) | Pd |
| $\mathbf{B}_2$ | = | $x_2\,\mathbf{a}_1 + x_2\,\mathbf{a}_2 + z_2\,\mathbf{a}_3$ | = | $(x_2 a + z_2 c\cos\beta)\,\hat{\mathbf{x}} + z_2 c\sin\beta\,\hat{\mathbf{z}}$ | (4$i$) | Se I |
| $\mathbf{B}_3$ | = | $-x_2\,\mathbf{a}_1 - x_2\,\mathbf{a}_2 - z_2\,\mathbf{a}_3$ | = | $(-x_2 a - z_2 c\cos\beta)\,\hat{\mathbf{x}} - z_2 c\sin\beta\,\hat{\mathbf{z}}$ | (4$i$) | Se I |
| $\mathbf{B}_4$ | = | $x_3\,\mathbf{a}_1 + x_3\,\mathbf{a}_2 + z_3\,\mathbf{a}_3$ | = | $(x_3 a + z_3 c\cos\beta)\,\hat{\mathbf{x}} + z_3 c\sin\beta\,\hat{\mathbf{z}}$ | (4$i$) | Se II |
| $\mathbf{B}_5$ | = | $-x_3\,\mathbf{a}_1 - x_3\,\mathbf{a}_2 - z_3\,\mathbf{a}_3$ | = | $(-x_3 a - z_3 c\cos\beta)\,\hat{\mathbf{x}} - z_3 c\sin\beta\,\hat{\mathbf{z}}$ | (4$i$) | Se II |
| $\mathbf{B}_6$ | = | $x_4\,\mathbf{a}_1 + x_4\,\mathbf{a}_2 + z_4\,\mathbf{a}_3$ | = | $(x_4 a + z_4 c\cos\beta)\,\hat{\mathbf{x}} + z_4 c\sin\beta\,\hat{\mathbf{z}}$ | (4$i$) | Se III |
| $\mathbf{B}_7$ | = | $-x_4\,\mathbf{a}_1 - x_4\,\mathbf{a}_2 - z_4\,\mathbf{a}_3$ | = | $(-x_4 a - z_4 c\cos\beta)\,\hat{\mathbf{x}} - z_4 c\sin\beta\,\hat{\mathbf{z}}$ | (4$i$) | Se III |
| $\mathbf{B}_8$ | = | $x_5\,\mathbf{a}_1 + x_5\,\mathbf{a}_2 + z_5\,\mathbf{a}_3$ | = | $(x_5 a + z_5 c\cos\beta)\,\hat{\mathbf{x}} + z_5 c\sin\beta\,\hat{\mathbf{z}}$ | (4$i$) | Ta |
| $\mathbf{B}_9$ | = | $-x_5\,\mathbf{a}_1 - x_5\,\mathbf{a}_2 - z_5\,\mathbf{a}_3$ | = | $(-x_5 a - z_5 c\cos\beta)\,\hat{\mathbf{x}} - z_5 c\sin\beta\,\hat{\mathbf{z}}$ | (4$i$) | Ta |

**Geometry files:**

- CIF: pp. 1534
- POSCAR: pp. 1534

# Sanidine ($KAlSi_3O_8$, $S6_7$) Structure: AB8C4_mC52_12_i_gi3j_2j

http://aflow.org/prototype-encyclopedia/AB8C4_mC52_12_i_gi3j_2j

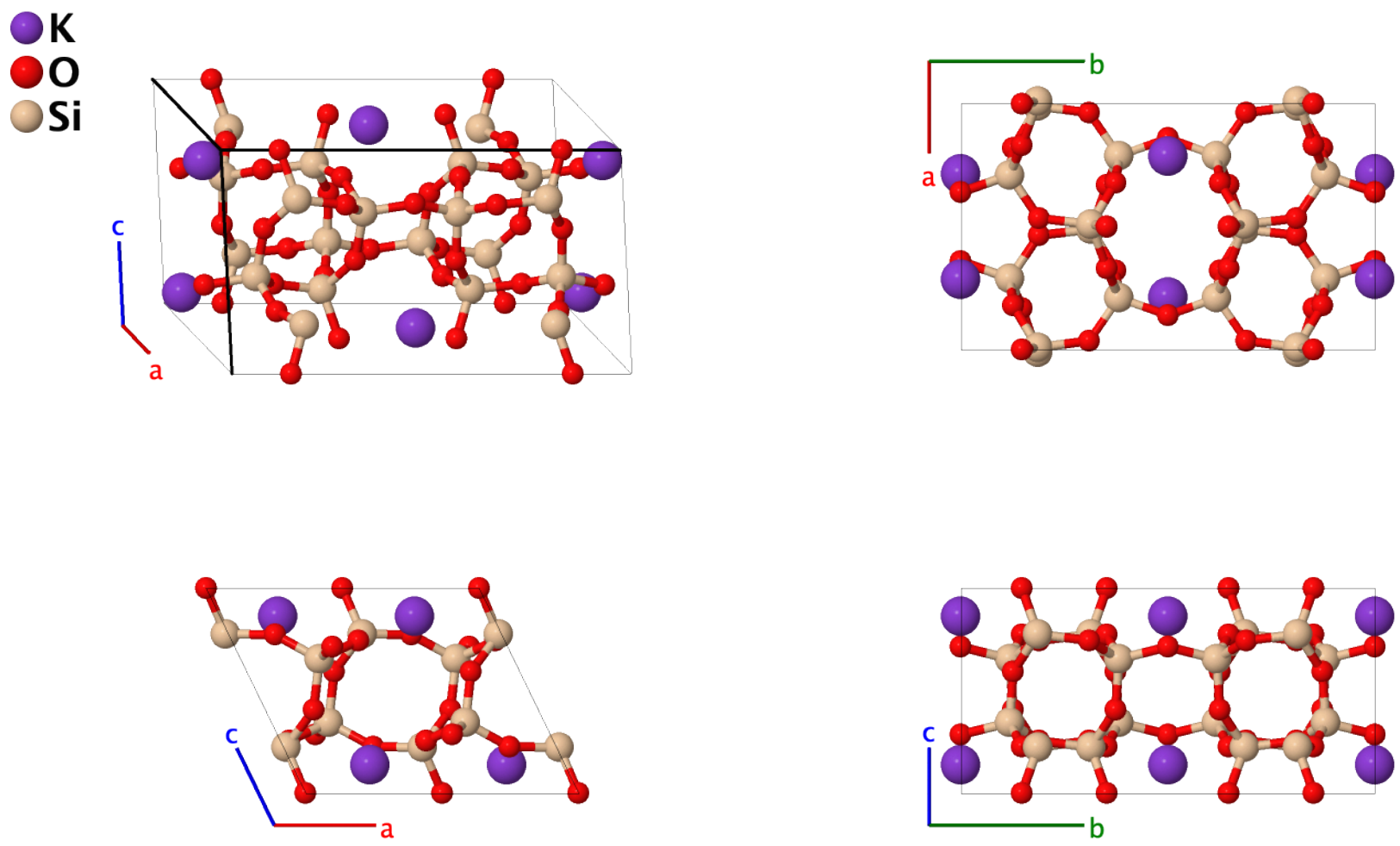

| | | |
|---|---|---|
| **Prototype** | : | $AlKO_8Si_3$ |
| **AFLOW prototype label** | : | AB8C4_mC52_12_i_gi3j_2j |
| ***Strukturbericht* designation** | : | $S6_7$ |
| **Pearson symbol** | : | mC52 |
| **Space group number** | : | 12 |
| **Space group symbol** | : | $C2/m$ |
| **AFLOW prototype command** | : | `aflow --proto=AB8C4_mC52_12_i_gi3j_2j` `--params=`$a, b/a, c/a, \beta, y_1, x_2, z_2, x_3, z_3, x_4, y_4, z_4, x_5, y_5, z_5, x_6, y_6, z_6, x_7, y_7, z_7, x_8, y_8, z_8$ |

**Other compounds with this structure**

- $NaAlSi_3O_8$ and $(BaKNa)AlSi_3O_8$ (orthoclase)

- Begining with the $S6_7$ defining paper of (Taylor, 1933), all researchers agree that the sites Si-I and Si-II are actually a random mixture of silicon and aluminum with overall stoichiometry $AlSi_3$. Most results found in (Downs, 2003) assume that this composition holds independently for both sites, but (Scambos, 1987) states that the Si-I site is $Al_{1.064}Si_{2.936}$ while the Si-II site is $Al_{0.936}Si_{3.064}$.
- (Scambos, 1987) list the coordinate $y_8 = 0.18813$ for the Si-II site in Table 2. This position does not result in the expected $SiO_4$ tetrahedra, and the interatomic distances and angles found do not agree with their Table 5. (Downs, 2003) correct this to $y_8 = 0.11813$, and this value results in good tetrahedra and reproduces the distances and angles reported by (Scambos, 1987). We use this later value on this page.

**Base-centered Monoclinic primitive vectors:**

$$
\begin{aligned}
\mathbf{a}_1 &= \tfrac{1}{2}\,a\,\hat{\mathbf{x}} - \tfrac{1}{2}\,b\,\hat{\mathbf{y}} \\
\mathbf{a}_2 &= \tfrac{1}{2}\,a\,\hat{\mathbf{x}} + \tfrac{1}{2}\,b\,\hat{\mathbf{y}} \\
\mathbf{a}_3 &= c\cos\beta\,\hat{\mathbf{x}} + c\sin\beta\,\hat{\mathbf{z}}
\end{aligned}
$$

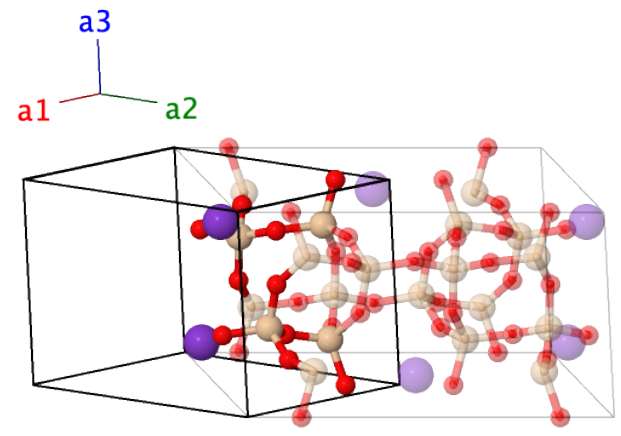

**Basis vectors:**

| | | Lattice Coordinates | | Cartesian Coordinates | Wyckoff Position | Atom Type |
|---|---|---|---|---|---|---|
| $\mathbf{B}_1$ | = | $-y_1\,\mathbf{a}_1 + y_1\,\mathbf{a}_2$ | = | $y_1 b\,\hat{\mathbf{y}}$ | (4$g$) | O I |
| $\mathbf{B}_2$ | = | $y_1\,\mathbf{a}_1 - y_1\,\mathbf{a}_2$ | = | $-y_1 b\,\hat{\mathbf{y}}$ | (4$g$) | O I |
| $\mathbf{B}_3$ | = | $x_2\,\mathbf{a}_1 + x_2\,\mathbf{a}_2 + z_2\,\mathbf{a}_3$ | = | $(x_2 a + z_2 c\cos\beta)\,\hat{\mathbf{x}} + z_2 c\sin\beta\,\hat{\mathbf{z}}$ | (4$i$) | K |
| $\mathbf{B}_4$ | = | $-x_2\,\mathbf{a}_1 - x_2\,\mathbf{a}_2 - z_2\,\mathbf{a}_3$ | = | $(-x_2 a - z_2 c\cos\beta)\,\hat{\mathbf{x}} - z_2 c\sin\beta\,\hat{\mathbf{z}}$ | (4$i$) | K |
| $\mathbf{B}_5$ | = | $x_3\,\mathbf{a}_1 + x_3\,\mathbf{a}_2 + z_3\,\mathbf{a}_3$ | = | $(x_3 a + z_3 c\cos\beta)\,\hat{\mathbf{x}} + z_3 c\sin\beta\,\hat{\mathbf{z}}$ | (4$i$) | O II |
| $\mathbf{B}_6$ | = | $-x_3\,\mathbf{a}_1 - x_3\,\mathbf{a}_2 - z_3\,\mathbf{a}_3$ | = | $(-x_3 a - z_3 c\cos\beta)\,\hat{\mathbf{x}} - z_3 c\sin\beta\,\hat{\mathbf{z}}$ | (4$i$) | O II |
| $\mathbf{B}_7$ | = | $(x_4 - y_4)\,\mathbf{a}_1 + (x_4 + y_4)\,\mathbf{a}_2 + z_4\,\mathbf{a}_3$ | = | $(x_4 a + z_4 c\cos\beta)\,\hat{\mathbf{x}} + y_4 b\,\hat{\mathbf{y}} + z_4 c\sin\beta\,\hat{\mathbf{z}}$ | (8$j$) | O III |
| $\mathbf{B}_8$ | = | $(-x_4 - y_4)\,\mathbf{a}_1 + (-x_4 + y_4)\,\mathbf{a}_2 - z_4\,\mathbf{a}_3$ | = | $(-x_4 a - z_4 c\cos\beta)\,\hat{\mathbf{x}} + y_4 b\,\hat{\mathbf{y}} - z_4 c\sin\beta\,\hat{\mathbf{z}}$ | (8$j$) | O III |
| $\mathbf{B}_9$ | = | $(-x_4 + y_4)\,\mathbf{a}_1 + (-x_4 - y_4)\,\mathbf{a}_2 - z_4\,\mathbf{a}_3$ | = | $(-x_4 a - z_4 c\cos\beta)\,\hat{\mathbf{x}} - y_4 b\,\hat{\mathbf{y}} - z_4 c\sin\beta\,\hat{\mathbf{z}}$ | (8$j$) | O III |
| $\mathbf{B}_{10}$ | = | $(x_4 + y_4)\,\mathbf{a}_1 + (x_4 - y_4)\,\mathbf{a}_2 + z_4\,\mathbf{a}_3$ | = | $(x_4 a + z_4 c\cos\beta)\,\hat{\mathbf{x}} - y_4 b\,\hat{\mathbf{y}} + z_4 c\sin\beta\,\hat{\mathbf{z}}$ | (8$j$) | O III |
| $\mathbf{B}_{11}$ | = | $(x_5 - y_5)\,\mathbf{a}_1 + (x_5 + y_5)\,\mathbf{a}_2 + z_5\,\mathbf{a}_3$ | = | $(x_5 a + z_5 c\cos\beta)\,\hat{\mathbf{x}} + y_5 b\,\hat{\mathbf{y}} + z_5 c\sin\beta\,\hat{\mathbf{z}}$ | (8$j$) | O IV |
| $\mathbf{B}_{12}$ | = | $(-x_5 - y_5)\,\mathbf{a}_1 + (-x_5 + y_5)\,\mathbf{a}_2 - z_5\,\mathbf{a}_3$ | = | $(-x_5 a - z_5 c\cos\beta)\,\hat{\mathbf{x}} + y_5 b\,\hat{\mathbf{y}} - z_5 c\sin\beta\,\hat{\mathbf{z}}$ | (8$j$) | O IV |
| $\mathbf{B}_{13}$ | = | $(-x_5 + y_5)\,\mathbf{a}_1 + (-x_5 - y_5)\,\mathbf{a}_2 - z_5\,\mathbf{a}_3$ | = | $(-x_5 a - z_5 c\cos\beta)\,\hat{\mathbf{x}} - y_5 b\,\hat{\mathbf{y}} - z_5 c\sin\beta\,\hat{\mathbf{z}}$ | (8$j$) | O IV |
| $\mathbf{B}_{14}$ | = | $(x_5 + y_5)\,\mathbf{a}_1 + (x_5 - y_5)\,\mathbf{a}_2 + z_5\,\mathbf{a}_3$ | = | $(x_5 a + z_5 c\cos\beta)\,\hat{\mathbf{x}} - y_5 b\,\hat{\mathbf{y}} + z_5 c\sin\beta\,\hat{\mathbf{z}}$ | (8$j$) | O IV |
| $\mathbf{B}_{15}$ | = | $(x_6 - y_6)\,\mathbf{a}_1 + (x_6 + y_6)\,\mathbf{a}_2 + z_6\,\mathbf{a}_3$ | = | $(x_6 a + z_6 c\cos\beta)\,\hat{\mathbf{x}} + y_6 b\,\hat{\mathbf{y}} + z_6 c\sin\beta\,\hat{\mathbf{z}}$ | (8$j$) | O V |
| $\mathbf{B}_{16}$ | = | $(-x_6 - y_6)\,\mathbf{a}_1 + (-x_6 + y_6)\,\mathbf{a}_2 - z_6\,\mathbf{a}_3$ | = | $(-x_6 a - z_6 c\cos\beta)\,\hat{\mathbf{x}} + y_6 b\,\hat{\mathbf{y}} - z_6 c\sin\beta\,\hat{\mathbf{z}}$ | (8$j$) | O V |
| $\mathbf{B}_{17}$ | = | $(-x_6 + y_6)\,\mathbf{a}_1 + (-x_6 - y_6)\,\mathbf{a}_2 - z_6\,\mathbf{a}_3$ | = | $(-x_6 a - z_6 c\cos\beta)\,\hat{\mathbf{x}} - y_6 b\,\hat{\mathbf{y}} - z_6 c\sin\beta\,\hat{\mathbf{z}}$ | (8$j$) | O V |
| $\mathbf{B}_{18}$ | = | $(x_6 + y_6)\,\mathbf{a}_1 + (x_6 - y_6)\,\mathbf{a}_2 + z_6\,\mathbf{a}_3$ | = | $(x_6 a + z_6 c\cos\beta)\,\hat{\mathbf{x}} - y_6 b\,\hat{\mathbf{y}} + z_6 c\sin\beta\,\hat{\mathbf{z}}$ | (8$j$) | O V |
| $\mathbf{B}_{19}$ | = | $(x_7 - y_7)\,\mathbf{a}_1 + (x_7 + y_7)\,\mathbf{a}_2 + z_7\,\mathbf{a}_3$ | = | $(x_7 a + z_7 c\cos\beta)\,\hat{\mathbf{x}} + y_7 b\,\hat{\mathbf{y}} + z_7 c\sin\beta\,\hat{\mathbf{z}}$ | (8$j$) | Si I |
| $\mathbf{B}_{20}$ | = | $(-x_7 - y_7)\,\mathbf{a}_1 + (-x_7 + y_7)\,\mathbf{a}_2 - z_7\,\mathbf{a}_3$ | = | $(-x_7 a - z_7 c\cos\beta)\,\hat{\mathbf{x}} + y_7 b\,\hat{\mathbf{y}} - z_7 c\sin\beta\,\hat{\mathbf{z}}$ | (8$j$) | Si I |
| $\mathbf{B}_{21}$ | = | $(-x_7 + y_7)\,\mathbf{a}_1 + (-x_7 - y_7)\,\mathbf{a}_2 - z_7\,\mathbf{a}_3$ | = | $(-x_7 a - z_7 c\cos\beta)\,\hat{\mathbf{x}} - y_7 b\,\hat{\mathbf{y}} - z_7 c\sin\beta\,\hat{\mathbf{z}}$ | (8$j$) | Si I |
| $\mathbf{B}_{22}$ | = | $(x_7 + y_7)\,\mathbf{a}_1 + (x_7 - y_7)\,\mathbf{a}_2 + z_7\,\mathbf{a}_3$ | = | $(x_7 a + z_7 c\cos\beta)\,\hat{\mathbf{x}} - y_7 b\,\hat{\mathbf{y}} + z_7 c\sin\beta\,\hat{\mathbf{z}}$ | (8*j*) | Si I |
| $\mathbf{B}_{23}$ | = | $(x_8 - y_8)\,\mathbf{a}_1 + (x_8 + y_8)\,\mathbf{a}_2 + z_8\,\mathbf{a}_3$ | = | $(x_8 a + z_8 c\cos\beta)\,\hat{\mathbf{x}} + y_8 b\,\hat{\mathbf{y}} + z_8 c\sin\beta\,\hat{\mathbf{z}}$ | (8*j*) | Si II |
| $\mathbf{B}_{24}$ | = | $(-x_8 - y_8)\,\mathbf{a}_1 + (-x_8 + y_8)\,\mathbf{a}_2 - z_8\,\mathbf{a}_3$ | = | $(-x_8 a - z_8 c\cos\beta)\,\hat{\mathbf{x}} + y_8 b\,\hat{\mathbf{y}} - z_8 c\sin\beta\,\hat{\mathbf{z}}$ | (8*j*) | Si II |
| $\mathbf{B}_{25}$ | = | $(-x_8 + y_8)\,\mathbf{a}_1 + (-x_8 - y_8)\,\mathbf{a}_2 - z_8\,\mathbf{a}_3$ | = | $(-x_8 a - z_8 c\cos\beta)\,\hat{\mathbf{x}} - y_8 b\,\hat{\mathbf{y}} - z_8 c\sin\beta\,\hat{\mathbf{z}}$ | (8*j*) | Si II |
| $\mathbf{B}_{26}$ | = | $(x_8 + y_8)\,\mathbf{a}_1 + (x_8 - y_8)\,\mathbf{a}_2 + z_8\,\mathbf{a}_3$ | = | $(x_8 a + z_8 c\cos\beta)\,\hat{\mathbf{x}} - y_8 b\,\hat{\mathbf{y}} + z_8 c\sin\beta\,\hat{\mathbf{z}}$ | (8*j*) | Si II |

**Geometry files:**

- CIF: pp. 1534
- POSCAR: pp. 1534

# $MnPS_3$ Structure: ABC3_mC20_12_g_i_ij

http://aflow.org/prototype-encyclopedia/ABC3_mC20_12_g_i_ij

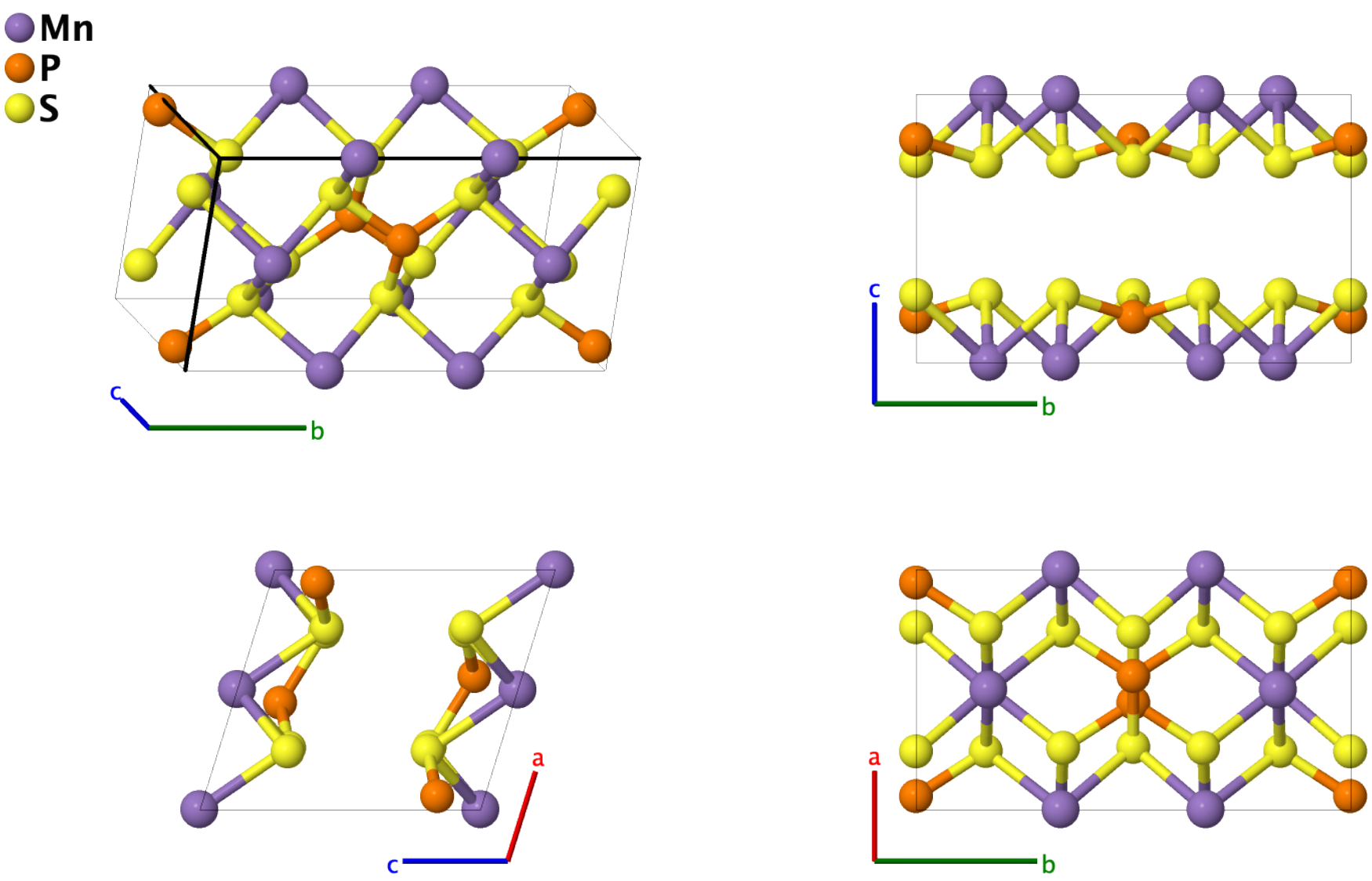

|  |  |  |
|---|---|---|
| **Prototype** | : | $MnPS_3$ |
| **AFLOW prototype label** | : | ABC3_mC20_12_g_i_ij |
| ***Strukturbericht* designation** | : | None |
| **Pearson symbol** | : | mC20 |
| **Space group number** | : | 12 |
| **Space group symbol** | : | $C2/m$ |
| **AFLOW prototype command** | : | `aflow --proto=ABC3_mC20_12_g_i_ij`<br>`--params=`$a, b/a, c/a, \beta, y_1, x_2, z_2, x_3, z_3, x_4, y_4, z_4$ |

**Other compounds with this structure**

- $CdPS_3$, $CdPSe_3$, $CoPS_3$, $CoPSe_3$, $FePS_3$, $FePSe_3$, $MnPSe_3$, $NiPS_3$, and $NiPSe_3$

**Base-centered Monoclinic primitive vectors:**

$$\mathbf{a}_1 = \frac{1}{2}\,a\,\hat{\mathbf{x}} - \frac{1}{2}\,b\,\hat{\mathbf{y}}$$
$$\mathbf{a}_2 = \frac{1}{2}\,a\,\hat{\mathbf{x}} + \frac{1}{2}\,b\,\hat{\mathbf{y}}$$
$$\mathbf{a}_3 = c\cos\beta\,\hat{\mathbf{x}} + c\sin\beta\,\hat{\mathbf{z}}$$

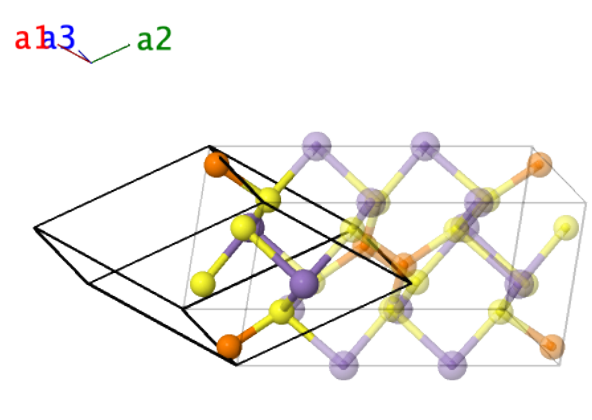

**Basis vectors:**

|  |  | Lattice Coordinates |  | Cartesian Coordinates | Wyckoff Position | Atom Type |
|---|---|---|---|---|---|---|
| $\mathbf{B}_1$ | = | $-y_1\,\mathbf{a}_1 + y_1\,\mathbf{a}_2$ | = | $y_1\,b\,\hat{\mathbf{y}}$ | (4$g$) | Mn |
| $\mathbf{B}_2$ | = | $y_1\,\mathbf{a}_1 - y_1\,\mathbf{a}_2$ | = | $-y_1\,b\,\hat{\mathbf{y}}$ | (4$g$) | Mn |
| $\mathbf{B}_3$ | = | $x_2\,\mathbf{a}_1 + x_2\,\mathbf{a}_2 + z_2\,\mathbf{a}_3$ | = | $(x_2 a + z_2 c\cos\beta)\,\hat{\mathbf{x}} + z_2 c\sin\beta\,\hat{\mathbf{z}}$ | (4$i$) | P |
| $\mathbf{B}_4$ | = | $-x_2\,\mathbf{a}_1 - x_2\,\mathbf{a}_2 - z_2\,\mathbf{a}_3$ | = | $(-x_2 a - z_2 c\cos\beta)\,\hat{\mathbf{x}} - z_2 c\sin\beta\,\hat{\mathbf{z}}$ | (4$i$) | P |
| $\mathbf{B}_5$ | = | $x_3\,\mathbf{a}_1 + x_3\,\mathbf{a}_2 + z_3\,\mathbf{a}_3$ | = | $(x_3 a + z_3 c\cos\beta)\,\hat{\mathbf{x}} + z_3 c\sin\beta\,\hat{\mathbf{z}}$ | (4$i$) | S I |
| $\mathbf{B}_6$ | = | $-x_3\,\mathbf{a}_1 - x_3\,\mathbf{a}_2 - z_3\,\mathbf{a}_3$ | = | $(-x_3 a - z_3 c\cos\beta)\,\hat{\mathbf{x}} - z_3 c\sin\beta\,\hat{\mathbf{z}}$ | (4$i$) | S I |
| $\mathbf{B}_7$ | = | $(x_4 - y_4)\,\mathbf{a}_1 + (x_4 + y_4)\,\mathbf{a}_2 + z_4\,\mathbf{a}_3$ | = | $(x_4 a + z_4 c\cos\beta)\,\hat{\mathbf{x}} + y_4 b\,\hat{\mathbf{y}} + z_4 c\sin\beta\,\hat{\mathbf{z}}$ | (8$j$) | S II |
| $\mathbf{B}_8$ | = | $(-x_4 - y_4)\,\mathbf{a}_1 + (-x_4 + y_4)\,\mathbf{a}_2 - z_4\,\mathbf{a}_3$ | = | $(-x_4 a - z_4 c\cos\beta)\,\hat{\mathbf{x}} + y_4 b\,\hat{\mathbf{y}} - z_4 c\sin\beta\,\hat{\mathbf{z}}$ | (8$j$) | S II |
| $\mathbf{B}_9$ | = | $(-x_4 + y_4)\,\mathbf{a}_1 + (-x_4 - y_4)\,\mathbf{a}_2 - z_4\,\mathbf{a}_3$ | = | $(-x_4 a - z_4 c\cos\beta)\,\hat{\mathbf{x}} - y_4 b\,\hat{\mathbf{y}} - z_4 c\sin\beta\,\hat{\mathbf{z}}$ | (8$j$) | S II |
| $\mathbf{B}_{10}$ | = | $(x_4 + y_4)\,\mathbf{a}_1 + (x_4 - y_4)\,\mathbf{a}_2 + z_4\,\mathbf{a}_3$ | = | $(x_4 a + z_4 c\cos\beta)\,\hat{\mathbf{x}} - y_4 b\,\hat{\mathbf{y}} + z_4 c\sin\beta\,\hat{\mathbf{z}}$ | (8$j$) | S II |

**Found in:**

**Geometry files:**

- CIF: pp. 1535
- POSCAR: pp. 1535

# $AlNbO_4$ Structure: ABC4_mC24_12_i_i_4i

http://aflow.org/prototype-encyclopedia/ABC4_mC24_12_i_i_4i

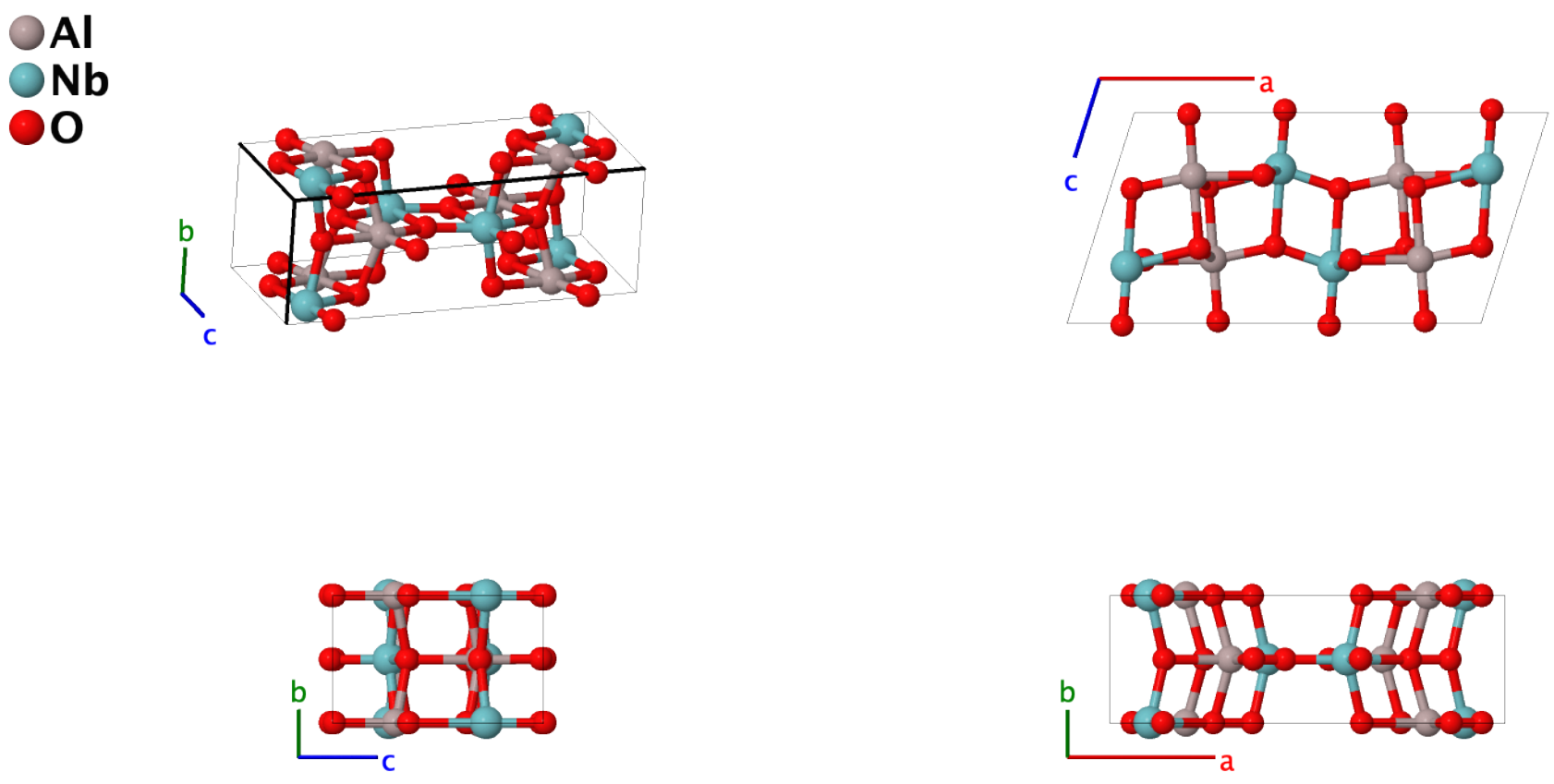

| | | |
|---|---|---|
| **Prototype** | : | $AlNbO_4$ |
| **AFLOW prototype label** | : | ABC4_mC24_12_i_i_4i |
| ***Strukturbericht* designation** | : | None |
| **Pearson symbol** | : | mC24 |
| **Space group number** | : | 12 |
| **Space group symbol** | : | $C2/m$ |
| **AFLOW prototype command** | : | `aflow --proto=ABC4_mC24_12_i_i_4i`<br>`--params=`$a, b/a, c/a, \beta, x_1, z_1, x_2, z_2, x_3, z_3, x_4, z_4, x_5, z_5, x_6, z_6$ |

- (Ardit, 2012) looked at samples with obvious disordering on the sites we label Al and Nb. For their sample N00, the occupation of our Al site was 80% aluminum and 20% niobium, with the reverse concentrations on our Nb site.
- (Pederson, 1962) analyzed a sample of $AlNbO_4$ assuming that these sites were fully ordered, however the atomic positions listed there have some errors, so we used the data found by Ardit *et al.*
- The atomic positions for this structure are not found in the main text. They are filed as item #AM-12-035 in the American Mineralogist repository.

**Base-centered Monoclinic primitive vectors:**

$$\mathbf{a}_1 = \frac{1}{2}\,a\,\hat{\mathbf{x}} - \frac{1}{2}\,b\,\hat{\mathbf{y}}$$
$$\mathbf{a}_2 = \frac{1}{2}\,a\,\hat{\mathbf{x}} + \frac{1}{2}\,b\,\hat{\mathbf{y}}$$
$$\mathbf{a}_3 = c\cos\beta\,\hat{\mathbf{x}} + c\sin\beta\,\hat{\mathbf{z}}$$

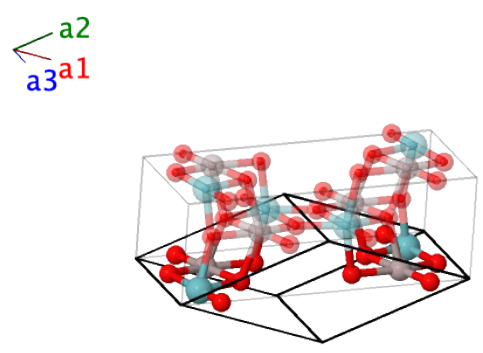

**Basis vectors:**

| | | Lattice Coordinates | | Cartesian Coordinates | Wyckoff Position | Atom Type |
|---|---|---|---|---|---|---|
| $\mathbf{B}_1$ | = | $x_1\,\mathbf{a}_1 + x_1\,\mathbf{a}_2 + z_1\,\mathbf{a}_3$ | = | $(x_1 a + z_1 c\cos\beta)\,\hat{\mathbf{x}} + z_1 c\sin\beta\,\hat{\mathbf{z}}$ | (4$i$) | Al |
| $\mathbf{B}_2$ | = | $-x_1\,\mathbf{a}_1 - x_1\,\mathbf{a}_2 - z_1\,\mathbf{a}_3$ | = | $(-x_1 a - z_1 c\cos\beta)\,\hat{\mathbf{x}} - z_1 c\sin\beta\,\hat{\mathbf{z}}$ | (4$i$) | Al |
| $\mathbf{B}_3$ | = | $x_2\,\mathbf{a}_1 + x_2\,\mathbf{a}_2 + z_2\,\mathbf{a}_3$ | = | $(x_2 a + z_2 c\cos\beta)\,\hat{\mathbf{x}} + z_2 c\sin\beta\,\hat{\mathbf{z}}$ | (4$i$) | Nb |
| $\mathbf{B}_4$ | = | $-x_2\,\mathbf{a}_1 - x_2\,\mathbf{a}_2 - z_2\,\mathbf{a}_3$ | = | $(-x_2 a - z_2 c\cos\beta)\,\hat{\mathbf{x}} - z_2 c\sin\beta\,\hat{\mathbf{z}}$ | $(4i)$ | Nb |
| $\mathbf{B}_5$ | = | $x_3\,\mathbf{a}_1 + x_3\,\mathbf{a}_2 + z_3\,\mathbf{a}_3$ | = | $(x_3 a + z_3 c\cos\beta)\,\hat{\mathbf{x}} + z_3 c\sin\beta\,\hat{\mathbf{z}}$ | $(4i)$ | O I |
| $\mathbf{B}_6$ | = | $-x_3\,\mathbf{a}_1 - x_3\,\mathbf{a}_2 - z_3\,\mathbf{a}_3$ | = | $(-x_3 a - z_3 c\cos\beta)\,\hat{\mathbf{x}} - z_3 c\sin\beta\,\hat{\mathbf{z}}$ | $(4i)$ | O I |
| $\mathbf{B}_7$ | = | $x_4\,\mathbf{a}_1 + x_4\,\mathbf{a}_2 + z_4\,\mathbf{a}_3$ | = | $(x_4 a + z_4 c\cos\beta)\,\hat{\mathbf{x}} + z_4 c\sin\beta\,\hat{\mathbf{z}}$ | $(4i)$ | O II |
| $\mathbf{B}_8$ | = | $-x_4\,\mathbf{a}_1 - x_4\,\mathbf{a}_2 - z_4\,\mathbf{a}_3$ | = | $(-x_4 a - z_4 c\cos\beta)\,\hat{\mathbf{x}} - z_4 c\sin\beta\,\hat{\mathbf{z}}$ | $(4i)$ | O II |
| $\mathbf{B}_9$ | = | $x_5\,\mathbf{a}_1 + x_5\,\mathbf{a}_2 + z_5\,\mathbf{a}_3$ | = | $(x_5 a + z_5 c\cos\beta)\,\hat{\mathbf{x}} + z_5 c\sin\beta\,\hat{\mathbf{z}}$ | $(4i)$ | O III |
| $\mathbf{B}_{10}$ | = | $-x_5\,\mathbf{a}_1 - x_5\,\mathbf{a}_2 - z_5\,\mathbf{a}_3$ | = | $(-x_5 a - z_5 c\cos\beta)\,\hat{\mathbf{x}} - z_5 c\sin\beta\,\hat{\mathbf{z}}$ | $(4i)$ | O III |
| $\mathbf{B}_{11}$ | = | $x_6\,\mathbf{a}_1 + x_6\,\mathbf{a}_2 + z_6\,\mathbf{a}_3$ | = | $(x_6 a + z_6 c\cos\beta)\,\hat{\mathbf{x}} + z_6 c\sin\beta\,\hat{\mathbf{z}}$ | $(4i)$ | O IV |
| $\mathbf{B}_{12}$ | = | $-x_6\,\mathbf{a}_1 - x_6\,\mathbf{a}_2 - z_6\,\mathbf{a}_3$ | = | $(-x_6 a - z_6 c\cos\beta)\,\hat{\mathbf{x}} - z_6 c\sin\beta\,\hat{\mathbf{z}}$ | $(4i)$ | O IV |

---

**Found in:**

---

**Geometry files:**

- CIF: pp. 1535
- POSCAR: pp. 1536

# SiAs Structure: AB_mC24_12_3i_3i

http://aflow.org/prototype-encyclopedia/AB_mC24_12_3i_3i

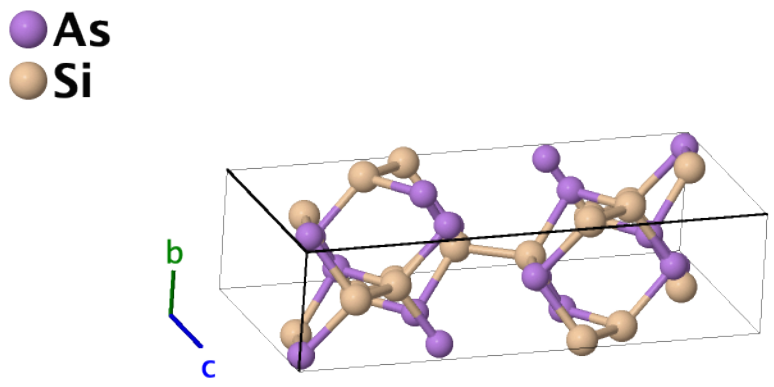

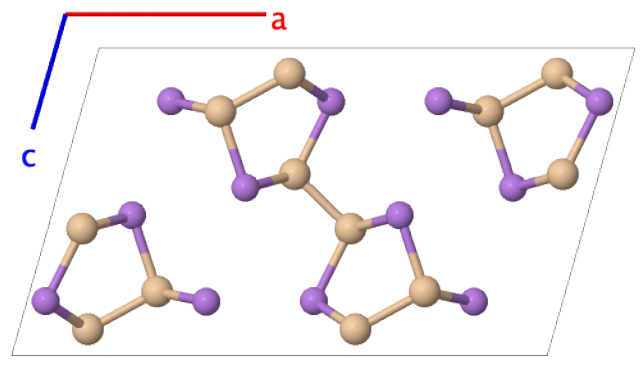

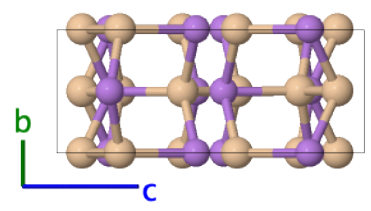

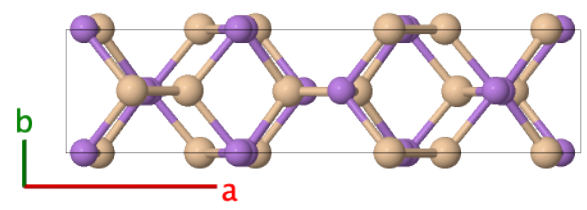

| | | |
|---|---|---|
| **Prototype** | : | AsSi |
| **AFLOW prototype label** | : | AB_mC24_12_3i_3i |
| ***Strukturbericht* designation** | : | None |
| **Pearson symbol** | : | mC24 |
| **Space group number** | : | 12 |
| **Space group symbol** | : | $C2/m$ |
| **AFLOW prototype command** | : | `aflow --proto=AB_mC24_12_3i_3i` <br> `--params=`$a, b/a, c/a, \beta, x_1, z_1, x_2, z_2, x_3, z_3, x_4, z_4, x_5, z_5, x_6, z_6$ |

### Other compounds with this structure

- GaTe and GeAs

- The structures of GaTe and GeAs were apparently determined by (Bryden, 1965) before (Wadsten, 1965) found the structure of SiAs, but they were never published (Pearson, 1964; Wadsten, 1965; Mead, 1982). As SiAs was the first published determination of this structure, we use it as the prototype.

### Base-centered Monoclinic primitive vectors:

$$\mathbf{a}_1 = \frac{1}{2}\,a\,\hat{\mathbf{x}} - \frac{1}{2}\,b\,\hat{\mathbf{y}}$$
$$\mathbf{a}_2 = \frac{1}{2}\,a\,\hat{\mathbf{x}} + \frac{1}{2}\,b\,\hat{\mathbf{y}}$$
$$\mathbf{a}_3 = c\cos\beta\,\hat{\mathbf{x}} + c\sin\beta\,\hat{\mathbf{z}}$$

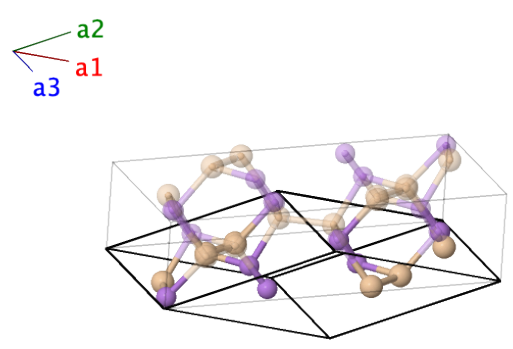

### Basis vectors:

| | | Lattice Coordinates | | Cartesian Coordinates | Wyckoff Position | Atom Type |
|---|---|---|---|---|---|---|
| $\mathbf{B}_1$ | = | $x_1\,\mathbf{a}_1 + x_1\,\mathbf{a}_2 + z_1\,\mathbf{a}_3$ | = | $(x_1 a + z_1 c\cos\beta)\,\hat{\mathbf{x}} + z_1 c\sin\beta\,\hat{\mathbf{z}}$ | (4$i$) | As I |
| $\mathbf{B}_2$ | = | $-x_1\,\mathbf{a}_1 - x_1\,\mathbf{a}_2 - z_1\,\mathbf{a}_3$ | = | $(-x_1 a - z_1 c\cos\beta)\,\hat{\mathbf{x}} - z_1 c\sin\beta\,\hat{\mathbf{z}}$ | (4$i$) | As I |
| $\mathbf{B}_3$ | = | $x_2\,\mathbf{a}_1 + x_2\,\mathbf{a}_2 + z_2\,\mathbf{a}_3$ | = | $(x_2 a + z_2 c\cos\beta)\,\hat{\mathbf{x}} + z_2 c\sin\beta\,\hat{\mathbf{z}}$ | (4$i$) | As II |
| $\mathbf{B}_4$ | = | $-x_2\,\mathbf{a}_1 - x_2\,\mathbf{a}_2 - z_2\,\mathbf{a}_3$ | = | $(-x_2 a - z_2 c\cos\beta)\,\hat{\mathbf{x}} - z_2 c\sin\beta\,\hat{\mathbf{z}}$ | (4$i$) | As II |
| $\mathbf{B}_5$ | = | $x_3\,\mathbf{a}_1 + x_3\,\mathbf{a}_2 + z_3\,\mathbf{a}_3$ | = | $(x_3 a + z_3 c\cos\beta)\,\hat{\mathbf{x}} + z_3 c\sin\beta\,\hat{\mathbf{z}}$ | (4$i$) | As III |
| $\mathbf{B}_6$ | = | $-x_3\,\mathbf{a}_1 - x_3\,\mathbf{a}_2 - z_3\,\mathbf{a}_3$ | = | $(-x_3 a - z_3 c\cos\beta)\,\hat{\mathbf{x}} - z_3 c\sin\beta\,\hat{\mathbf{z}}$ | (4$i$) | As III |
| $\mathbf{B}_7$ | = | $x_4\,\mathbf{a}_1 + x_4\,\mathbf{a}_2 + z_4\,\mathbf{a}_3$ | = | $(x_4 a + z_4 c\cos\beta)\,\hat{\mathbf{x}} + z_4 c\sin\beta\,\hat{\mathbf{z}}$ | (4$i$) | Si I |
| $\mathbf{B}_8$ | = | $-x_4\,\mathbf{a}_1 - x_4\,\mathbf{a}_2 - z_4\,\mathbf{a}_3$ | = | $(-x_4 a - z_4 c\cos\beta)\,\hat{\mathbf{x}} - z_4 c\sin\beta\,\hat{\mathbf{z}}$ | (4$i$) | Si I |
| $\mathbf{B}_9$ | = | $x_5\,\mathbf{a}_1 + x_5\,\mathbf{a}_2 + z_5\,\mathbf{a}_3$ | = | $(x_5 a + z_5 c\cos\beta)\,\hat{\mathbf{x}} + z_5 c\sin\beta\,\hat{\mathbf{z}}$ | (4$i$) | Si II |
| $\mathbf{B}_{10}$ | = | $-x_5\,\mathbf{a}_1 - x_5\,\mathbf{a}_2 - z_5\,\mathbf{a}_3$ | = | $(-x_5 a - z_5 c\cos\beta)\,\hat{\mathbf{x}} - z_5 c\sin\beta\,\hat{\mathbf{z}}$ | (4$i$) | Si II |
| $\mathbf{B}_{11}$ | = | $x_6\,\mathbf{a}_1 + x_6\,\mathbf{a}_2 + z_6\,\mathbf{a}_3$ | = | $(x_6 a + z_6 c\cos\beta)\,\hat{\mathbf{x}} + z_6 c\sin\beta\,\hat{\mathbf{z}}$ | (4$i$) | Si III |
| $\mathbf{B}_{12}$ | = | $-x_6\,\mathbf{a}_1 - x_6\,\mathbf{a}_2 - z_6\,\mathbf{a}_3$ | = | $(-x_6 a - z_6 c\cos\beta)\,\hat{\mathbf{x}} - z_6 c\sin\beta\,\hat{\mathbf{z}}$ | (4$i$) | Si III |

**Geometry files:**

- CIF: pp. 1536
- POSCAR: pp. 1536

# High-Temperature $Mo_8O_{23}$ Structure: A8B23_mP62_13_4g_c11g

http://aflow.org/prototype-encyclopedia/A8B23_mP62_13_4g_c11g

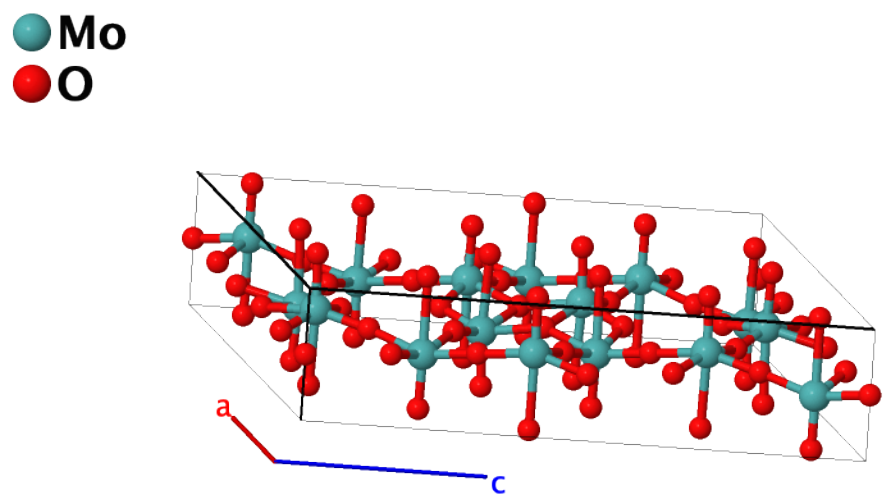

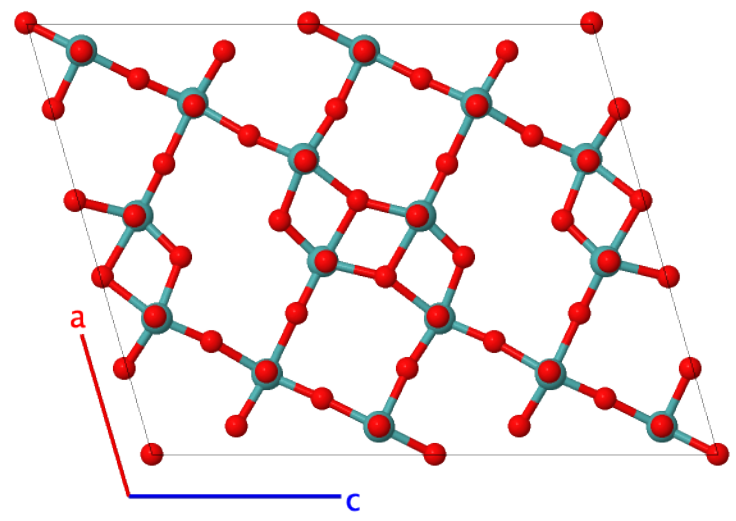

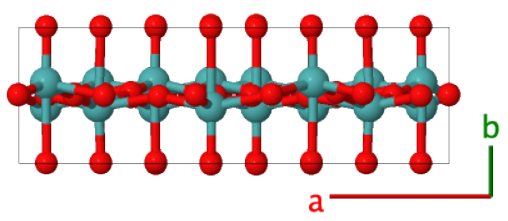

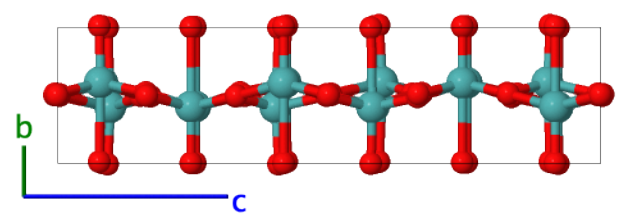

| | | |
|---|---|---|
| **Prototype** | : | $Mo_8O_{23}$ |
| **AFLOW prototype label** | : | A8B23_mP62_13_4g_c11g |
| ***Strukturbericht* designation** | : | None |
| **Pearson symbol** | : | mP62 |
| **Space group number** | : | 13 |
| **Space group symbol** | : | $P2/c$ |
| **AFLOW prototype command** | : | `aflow --proto=A8B23_mP62_13_4g_c11g` `--params=`$a, b/a, c/a, \beta, x_2, y_2, z_2, x_3, y_3, z_3, x_4, y_4, z_4, x_5, y_5, z_5, x_6, y_6, z_6, x_7, y_7, z_7, x_8, y_8, z_8, x_9, y_9, z_9, x_{10}, y_{10}, z_{10}, x_{11}, y_{11}, z_{11}, x_{12}, y_{12}, z_{12}, x_{13}, y_{13}, z_{13}, x_{14}, y_{14}, z_{14}, x_{15}, y_{15}, z_{15}, x_{16}, y_{16}, z_{16}$ |

- Above 285 K, the structure exhibits an incommensurate charge density wave. This data was taken at 370 K, so the structure given here is only approximate. Below 285 K the CDW is locked in, leading to the low-temperature $Mo_8O_{23}$ structure.

**Simple Monoclinic primitive vectors:**

$$\mathbf{a}_1 = a\,\hat{\mathbf{x}}$$
$$\mathbf{a}_2 = b\,\hat{\mathbf{y}}$$
$$\mathbf{a}_3 = c\cos\beta\,\hat{\mathbf{x}} + c\sin\beta\,\hat{\mathbf{z}}$$

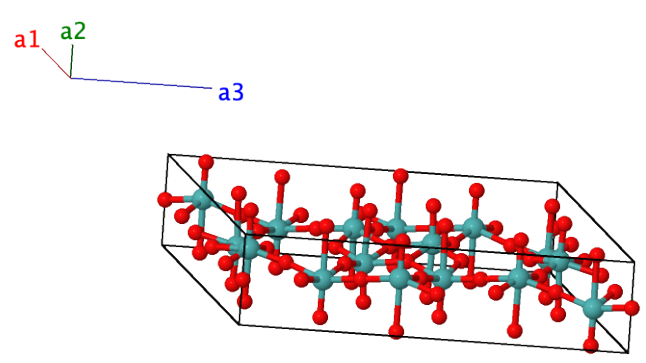

**Basis vectors:**

| | | Lattice Coordinates | | Cartesian Coordinates | Wyckoff Position | Atom Type |
|---|---|---|---|---|---|---|
| $\mathbf{B}_1$ | = | $\frac{1}{2}\,\mathbf{a}_2$ | = | $\frac{1}{2}b\,\hat{\mathbf{y}}$ | (2c) | O I |
| $\mathbf{B}_2$ | = | $\frac{1}{2}\,\mathbf{a}_2+\frac{1}{2}\,\mathbf{a}_3$ | = | $\frac{1}{2}c\cos\beta\,\hat{\mathbf{x}}+\frac{1}{2}b\,\hat{\mathbf{y}}+\frac{1}{2}c\sin\beta\,\hat{\mathbf{z}}$ | (2c) | O I |
| $\mathbf{B}_3$ | = | $x_2\,\mathbf{a}_1+y_2\,\mathbf{a}_2+z_2\,\mathbf{a}_3$ | = | $(x_2a+z_2c\cos\beta)\,\hat{\mathbf{x}}+y_2b\,\hat{\mathbf{y}}+z_2c\sin\beta\,\hat{\mathbf{z}}$ | (4g) | Mo I |
| $\mathbf{B}_4$ | = | $-x_2\,\mathbf{a}_1+y_2\,\mathbf{a}_2+\left(\frac{1}{2}-z_2\right)\,\mathbf{a}_3$ | = | $\left(\frac{1}{2}c\cos\beta-x_2a-z_2c\cos\beta\right)\,\hat{\mathbf{x}}+y_2b\,\hat{\mathbf{y}}+\left(\frac{1}{2}-z_2\right)c\sin\beta\,\hat{\mathbf{z}}$ | (4g) | Mo I |
| $\mathbf{B}_5$ | = | $-x_2\,\mathbf{a}_1-y_2\,\mathbf{a}_2-z_2\,\mathbf{a}_3$ | = | $(-x_2a-z_2c\cos\beta)\,\hat{\mathbf{x}}-y_2b\,\hat{\mathbf{y}}-z_2c\sin\beta\,\hat{\mathbf{z}}$ | (4g) | Mo I |
| $\mathbf{B}_6$ | = | $x_2\,\mathbf{a}_1-y_2\,\mathbf{a}_2+\left(\frac{1}{2}+z_2\right)\,\mathbf{a}_3$ | = | $\left(\frac{1}{2}c\cos\beta+x_2a+z_2c\cos\beta\right)\,\hat{\mathbf{x}}-y_2b\,\hat{\mathbf{y}}+\left(\frac{1}{2}+z_2\right)c\sin\beta\,\hat{\mathbf{z}}$ | (4g) | Mo I |
| $\mathbf{B}_7$ | = | $x_3\,\mathbf{a}_1+y_3\,\mathbf{a}_2+z_3\,\mathbf{a}_3$ | = | $(x_3a+z_3c\cos\beta)\,\hat{\mathbf{x}}+y_3b\,\hat{\mathbf{y}}+z_3c\sin\beta\,\hat{\mathbf{z}}$ | (4g) | Mo II |
| $\mathbf{B}_8$ | = | $-x_3\,\mathbf{a}_1+y_3\,\mathbf{a}_2+\left(\frac{1}{2}-z_3\right)\,\mathbf{a}_3$ | = | $\left(\frac{1}{2}c\cos\beta-x_3a-z_3c\cos\beta\right)\,\hat{\mathbf{x}}+y_3b\,\hat{\mathbf{y}}+\left(\frac{1}{2}-z_3\right)c\sin\beta\,\hat{\mathbf{z}}$ | (4g) | Mo II |
| $\mathbf{B}_9$ | = | $-x_3\,\mathbf{a}_1-y_3\,\mathbf{a}_2-z_3\,\mathbf{a}_3$ | = | $(-x_3a-z_3c\cos\beta)\,\hat{\mathbf{x}}-y_3b\,\hat{\mathbf{y}}-z_3c\sin\beta\,\hat{\mathbf{z}}$ | (4g) | Mo II |
| $\mathbf{B}_{10}$ | = | $x_3\,\mathbf{a}_1-y_3\,\mathbf{a}_2+\left(\frac{1}{2}+z_3\right)\,\mathbf{a}_3$ | = | $\left(\frac{1}{2}c\cos\beta+x_3a+z_3c\cos\beta\right)\,\hat{\mathbf{x}}-y_3b\,\hat{\mathbf{y}}+\left(\frac{1}{2}+z_3\right)c\sin\beta\,\hat{\mathbf{z}}$ | (4g) | Mo II |
| $\mathbf{B}_{11}$ | = | $x_4\,\mathbf{a}_1+y_4\,\mathbf{a}_2+z_4\,\mathbf{a}_3$ | = | $(x_4a+z_4c\cos\beta)\,\hat{\mathbf{x}}+y_4b\,\hat{\mathbf{y}}+z_4c\sin\beta\,\hat{\mathbf{z}}$ | (4g) | Mo III |
| $\mathbf{B}_{12}$ | = | $-x_4\,\mathbf{a}_1+y_4\,\mathbf{a}_2+\left(\frac{1}{2}-z_4\right)\,\mathbf{a}_3$ | = | $\left(\frac{1}{2}c\cos\beta-x_4a-z_4c\cos\beta\right)\,\hat{\mathbf{x}}+y_4b\,\hat{\mathbf{y}}+\left(\frac{1}{2}-z_4\right)c\sin\beta\,\hat{\mathbf{z}}$ | (4g) | Mo III |
| $\mathbf{B}_{13}$ | = | $-x_4\,\mathbf{a}_1-y_4\,\mathbf{a}_2-z_4\,\mathbf{a}_3$ | = | $(-x_4a-z_4c\cos\beta)\,\hat{\mathbf{x}}-y_4b\,\hat{\mathbf{y}}-z_4c\sin\beta\,\hat{\mathbf{z}}$ | (4g) | Mo III |
| $\mathbf{B}_{14}$ | = | $x_4\,\mathbf{a}_1-y_4\,\mathbf{a}_2+\left(\frac{1}{2}+z_4\right)\,\mathbf{a}_3$ | = | $\left(\frac{1}{2}c\cos\beta+x_4a+z_4c\cos\beta\right)\,\hat{\mathbf{x}}-y_4b\,\hat{\mathbf{y}}+\left(\frac{1}{2}+z_4\right)c\sin\beta\,\hat{\mathbf{z}}$ | (4g) | Mo III |
| $\mathbf{B}_{15}$ | = | $x_5\,\mathbf{a}_1+y_5\,\mathbf{a}_2+z_5\,\mathbf{a}_3$ | = | $(x_5a+z_5c\cos\beta)\,\hat{\mathbf{x}}+y_5b\,\hat{\mathbf{y}}+z_5c\sin\beta\,\hat{\mathbf{z}}$ | (4g) | Mo IV |
| $\mathbf{B}_{16}$ | = | $-x_5\,\mathbf{a}_1+y_5\,\mathbf{a}_2+\left(\frac{1}{2}-z_5\right)\,\mathbf{a}_3$ | = | $\left(\frac{1}{2}c\cos\beta-x_5a-z_5c\cos\beta\right)\,\hat{\mathbf{x}}+y_5b\,\hat{\mathbf{y}}+\left(\frac{1}{2}-z_5\right)c\sin\beta\,\hat{\mathbf{z}}$ | (4g) | Mo IV |
| $\mathbf{B}_{17}$ | = | $-x_5\,\mathbf{a}_1-y_5\,\mathbf{a}_2-z_5\,\mathbf{a}_3$ | = | $(-x_5a-z_5c\cos\beta)\,\hat{\mathbf{x}}-y_5b\,\hat{\mathbf{y}}-z_5c\sin\beta\,\hat{\mathbf{z}}$ | (4g) | Mo IV |
| $\mathbf{B}_{18}$ | = | $x_5\,\mathbf{a}_1-y_5\,\mathbf{a}_2+\left(\frac{1}{2}+z_5\right)\,\mathbf{a}_3$ | = | $\left(\frac{1}{2}c\cos\beta+x_5a+z_5c\cos\beta\right)\,\hat{\mathbf{x}}-y_5b\,\hat{\mathbf{y}}+\left(\frac{1}{2}+z_5\right)c\sin\beta\,\hat{\mathbf{z}}$ | (4g) | Mo IV |
| $\mathbf{B}_{19}$ | = | $x_6\,\mathbf{a}_1+y_6\,\mathbf{a}_2+z_6\,\mathbf{a}_3$ | = | $(x_6a+z_6c\cos\beta)\,\hat{\mathbf{x}}+y_6b\,\hat{\mathbf{y}}+z_6c\sin\beta\,\hat{\mathbf{z}}$ | (4g) | O II |
| $\mathbf{B}_{20}$ | = | $-x_6\,\mathbf{a}_1+y_6\,\mathbf{a}_2+\left(\frac{1}{2}-z_6\right)\,\mathbf{a}_3$ | = | $\left(\frac{1}{2}c\cos\beta-x_6a-z_6c\cos\beta\right)\,\hat{\mathbf{x}}+y_6b\,\hat{\mathbf{y}}+\left(\frac{1}{2}-z_6\right)c\sin\beta\,\hat{\mathbf{z}}$ | (4g) | O II |
| $\mathbf{B}_{21}$ | = | $-x_6\,\mathbf{a}_1-y_6\,\mathbf{a}_2-z_6\,\mathbf{a}_3$ | = | $(-x_6a-z_6c\cos\beta)\,\hat{\mathbf{x}}-y_6b\,\hat{\mathbf{y}}-z_6c\sin\beta\,\hat{\mathbf{z}}$ | (4g) | O II |
| $\mathbf{B}_{22}$ | = | $x_6\,\mathbf{a}_1-y_6\,\mathbf{a}_2+\left(\frac{1}{2}+z_6\right)\,\mathbf{a}_3$ | = | $\left(\frac{1}{2}c\cos\beta+x_6a+z_6c\cos\beta\right)\,\hat{\mathbf{x}}-y_6b\,\hat{\mathbf{y}}+\left(\frac{1}{2}+z_6\right)c\sin\beta\,\hat{\mathbf{z}}$ | (4g) | O II |
| $\mathbf{B}_{23}$ | = | $x_7\,\mathbf{a}_1+y_7\,\mathbf{a}_2+z_7\,\mathbf{a}_3$ | = | $(x_7a+z_7c\cos\beta)\,\hat{\mathbf{x}}+y_7b\,\hat{\mathbf{y}}+z_7c\sin\beta\,\hat{\mathbf{z}}$ | (4g) | O III |
| $\mathbf{B}_{24}$ | = | $-x_7\,\mathbf{a}_1+y_7\,\mathbf{a}_2+\left(\frac{1}{2}-z_7\right)\,\mathbf{a}_3$ | = | $\left(\frac{1}{2}c\cos\beta-x_7a-z_7c\cos\beta\right)\,\hat{\mathbf{x}}+y_7b\,\hat{\mathbf{y}}+\left(\frac{1}{2}-z_7\right)c\sin\beta\,\hat{\mathbf{z}}$ | (4g) | O III |
| $\mathbf{B}_{25}$ | = | $-x_7\,\mathbf{a}_1-y_7\,\mathbf{a}_2-z_7\,\mathbf{a}_3$ | = | $(-x_7a-z_7c\cos\beta)\,\hat{\mathbf{x}}-y_7b\,\hat{\mathbf{y}}-z_7c\sin\beta\,\hat{\mathbf{z}}$ | (4g) | O III |
| $\mathbf{B}_{26}$ | = | $x_7\,\mathbf{a}_1-y_7\,\mathbf{a}_2+\left(\frac{1}{2}+z_7\right)\,\mathbf{a}_3$ | = | $\left(\frac{1}{2}c\cos\beta+x_7a+z_7c\cos\beta\right)\,\hat{\mathbf{x}}-y_7b\,\hat{\mathbf{y}}+\left(\frac{1}{2}+z_7\right)c\sin\beta\,\hat{\mathbf{z}}$ | (4g) | O III |

$$\mathbf{B}_{27} = x_8\,\mathbf{a}_1 + y_8\,\mathbf{a}_2 + z_8\,\mathbf{a}_3 = (x_8 a + z_8 c\cos\beta)\,\hat{\mathbf{x}} + y_8 b\,\hat{\mathbf{y}} + z_8 c\sin\beta\,\hat{\mathbf{z}} \quad (4g) \quad \text{O IV}$$

$$\mathbf{B}_{28} = -x_8\,\mathbf{a}_1 + y_8\,\mathbf{a}_2 + \left(\tfrac{1}{2} - z_8\right)\mathbf{a}_3 = \left(\tfrac{1}{2}c\cos\beta - x_8 a - z_8 c\cos\beta\right)\hat{\mathbf{x}} + y_8 b\,\hat{\mathbf{y}} + \left(\tfrac{1}{2} - z_8\right)c\sin\beta\,\hat{\mathbf{z}} \quad (4g) \quad \text{O IV}$$

$$\mathbf{B}_{29} = -x_8\,\mathbf{a}_1 - y_8\,\mathbf{a}_2 - z_8\,\mathbf{a}_3 = (-x_8 a - z_8 c\cos\beta)\,\hat{\mathbf{x}} - y_8 b\,\hat{\mathbf{y}} - z_8 c\sin\beta\,\hat{\mathbf{z}} \quad (4g) \quad \text{O IV}$$

$$\mathbf{B}_{30} = x_8\,\mathbf{a}_1 - y_8\,\mathbf{a}_2 + \left(\tfrac{1}{2} + z_8\right)\mathbf{a}_3 = \left(\tfrac{1}{2}c\cos\beta + x_8 a + z_8 c\cos\beta\right)\hat{\mathbf{x}} - y_8 b\,\hat{\mathbf{y}} + \left(\tfrac{1}{2} + z_8\right)c\sin\beta\,\hat{\mathbf{z}} \quad (4g) \quad \text{O IV}$$

$$\mathbf{B}_{31} = x_9\,\mathbf{a}_1 + y_9\,\mathbf{a}_2 + z_9\,\mathbf{a}_3 = (x_9 a + z_9 c\cos\beta)\,\hat{\mathbf{x}} + y_9 b\,\hat{\mathbf{y}} + z_9 c\sin\beta\,\hat{\mathbf{z}} \quad (4g) \quad \text{O V}$$

$$\mathbf{B}_{32} = -x_9\,\mathbf{a}_1 + y_9\,\mathbf{a}_2 + \left(\tfrac{1}{2} - z_9\right)\mathbf{a}_3 = \left(\tfrac{1}{2}c\cos\beta - x_9 a - z_9 c\cos\beta\right)\hat{\mathbf{x}} + y_9 b\,\hat{\mathbf{y}} + \left(\tfrac{1}{2} - z_9\right)c\sin\beta\,\hat{\mathbf{z}} \quad (4g) \quad \text{O V}$$

$$\mathbf{B}_{33} = -x_9\,\mathbf{a}_1 - y_9\,\mathbf{a}_2 - z_9\,\mathbf{a}_3 = (-x_9 a - z_9 c\cos\beta)\,\hat{\mathbf{x}} - y_9 b\,\hat{\mathbf{y}} - z_9 c\sin\beta\,\hat{\mathbf{z}} \quad (4g) \quad \text{O V}$$

$$\mathbf{B}_{34} = x_9\,\mathbf{a}_1 - y_9\,\mathbf{a}_2 + \left(\tfrac{1}{2} + z_9\right)\mathbf{a}_3 = \left(\tfrac{1}{2}c\cos\beta + x_9 a + z_9 c\cos\beta\right)\hat{\mathbf{x}} - y_9 b\,\hat{\mathbf{y}} + \left(\tfrac{1}{2} + z_9\right)c\sin\beta\,\hat{\mathbf{z}} \quad (4g) \quad \text{O V}$$

$$\mathbf{B}_{35} = x_{10}\,\mathbf{a}_1 + y_{10}\,\mathbf{a}_2 + z_{10}\,\mathbf{a}_3 = (x_{10} a + z_{10} c\cos\beta)\,\hat{\mathbf{x}} + y_{10} b\,\hat{\mathbf{y}} + z_{10} c\sin\beta\,\hat{\mathbf{z}} \quad (4g) \quad \text{O VI}$$

$$\mathbf{B}_{36} = -x_{10}\,\mathbf{a}_1 + y_{10}\,\mathbf{a}_2 + \left(\tfrac{1}{2} - z_{10}\right)\mathbf{a}_3 = \left(\tfrac{1}{2}c\cos\beta - x_{10} a - z_{10} c\cos\beta\right)\hat{\mathbf{x}} + y_{10} b\,\hat{\mathbf{y}} + \left(\tfrac{1}{2} - z_{10}\right)c\sin\beta\,\hat{\mathbf{z}} \quad (4g) \quad \text{O VI}$$

$$\mathbf{B}_{37} = -x_{10}\,\mathbf{a}_1 - y_{10}\,\mathbf{a}_2 - z_{10}\,\mathbf{a}_3 = (-x_{10} a - z_{10} c\cos\beta)\,\hat{\mathbf{x}} - y_{10} b\,\hat{\mathbf{y}} - z_{10} c\sin\beta\,\hat{\mathbf{z}} \quad (4g) \quad \text{O VI}$$

$$\mathbf{B}_{38} = x_{10}\,\mathbf{a}_1 - y_{10}\,\mathbf{a}_2 + \left(\tfrac{1}{2} + z_{10}\right)\mathbf{a}_3 = \left(\tfrac{1}{2}c\cos\beta + x_{10} a + z_{10} c\cos\beta\right)\hat{\mathbf{x}} - y_{10} b\,\hat{\mathbf{y}} + \left(\tfrac{1}{2} + z_{10}\right)c\sin\beta\,\hat{\mathbf{z}} \quad (4g) \quad \text{O VI}$$

$$\mathbf{B}_{39} = x_{11}\,\mathbf{a}_1 + y_{11}\,\mathbf{a}_2 + z_{11}\,\mathbf{a}_3 = (x_{11} a + z_{11} c\cos\beta)\,\hat{\mathbf{x}} + y_{11} b\,\hat{\mathbf{y}} + z_{11} c\sin\beta\,\hat{\mathbf{z}} \quad (4g) \quad \text{O VII}$$

$$\mathbf{B}_{40} = -x_{11}\,\mathbf{a}_1 + y_{11}\,\mathbf{a}_2 + \left(\tfrac{1}{2} - z_{11}\right)\mathbf{a}_3 = \left(\tfrac{1}{2}c\cos\beta - x_{11} a - z_{11} c\cos\beta\right)\hat{\mathbf{x}} + y_{11} b\,\hat{\mathbf{y}} + \left(\tfrac{1}{2} - z_{11}\right)c\sin\beta\,\hat{\mathbf{z}} \quad (4g) \quad \text{O VII}$$

$$\mathbf{B}_{41} = -x_{11}\,\mathbf{a}_1 - y_{11}\,\mathbf{a}_2 - z_{11}\,\mathbf{a}_3 = (-x_{11} a - z_{11} c\cos\beta)\,\hat{\mathbf{x}} - y_{11} b\,\hat{\mathbf{y}} - z_{11} c\sin\beta\,\hat{\mathbf{z}} \quad (4g) \quad \text{O VII}$$

$$\mathbf{B}_{42} = x_{11}\,\mathbf{a}_1 - y_{11}\,\mathbf{a}_2 + \left(\tfrac{1}{2} + z_{11}\right)\mathbf{a}_3 = \left(\tfrac{1}{2}c\cos\beta + x_{11} a + z_{11} c\cos\beta\right)\hat{\mathbf{x}} - y_{11} b\,\hat{\mathbf{y}} + \left(\tfrac{1}{2} + z_{11}\right)c\sin\beta\,\hat{\mathbf{z}} \quad (4g) \quad \text{O VII}$$

$$\mathbf{B}_{43} = x_{12}\,\mathbf{a}_1 + y_{12}\,\mathbf{a}_2 + z_{12}\,\mathbf{a}_3 = (x_{12} a + z_{12} c\cos\beta)\,\hat{\mathbf{x}} + y_{12} b\,\hat{\mathbf{y}} + z_{12} c\sin\beta\,\hat{\mathbf{z}} \quad (4g) \quad \text{O VIII}$$

$$\mathbf{B}_{44} = -x_{12}\,\mathbf{a}_1 + y_{12}\,\mathbf{a}_2 + \left(\tfrac{1}{2} - z_{12}\right)\mathbf{a}_3 = \left(\tfrac{1}{2}c\cos\beta - x_{12} a - z_{12} c\cos\beta\right)\hat{\mathbf{x}} + y_{12} b\,\hat{\mathbf{y}} + \left(\tfrac{1}{2} - z_{12}\right)c\sin\beta\,\hat{\mathbf{z}} \quad (4g) \quad \text{O VIII}$$

$$\mathbf{B}_{45} = -x_{12}\,\mathbf{a}_1 - y_{12}\,\mathbf{a}_2 - z_{12}\,\mathbf{a}_3 = (-x_{12} a - z_{12} c\cos\beta)\,\hat{\mathbf{x}} - y_{12} b\,\hat{\mathbf{y}} - z_{12} c\sin\beta\,\hat{\mathbf{z}} \quad (4g) \quad \text{O VIII}$$

$$\mathbf{B}_{46} = x_{12}\,\mathbf{a}_1 - y_{12}\,\mathbf{a}_2 + \left(\tfrac{1}{2} + z_{12}\right)\mathbf{a}_3 = \left(\tfrac{1}{2}c\cos\beta + x_{12} a + z_{12} c\cos\beta\right)\hat{\mathbf{x}} - y_{12} b\,\hat{\mathbf{y}} + \left(\tfrac{1}{2} + z_{12}\right)c\sin\beta\,\hat{\mathbf{z}} \quad (4g) \quad \text{O VIII}$$

$$\mathbf{B}_{47} = x_{13}\,\mathbf{a}_1 + y_{13}\,\mathbf{a}_2 + z_{13}\,\mathbf{a}_3 = (x_{13} a + z_{13} c\cos\beta)\,\hat{\mathbf{x}} + y_{13} b\,\hat{\mathbf{y}} + z_{13} c\sin\beta\,\hat{\mathbf{z}} \quad (4g) \quad \text{O IX}$$

$$\mathbf{B}_{48} = -x_{13}\,\mathbf{a}_1 + y_{13}\,\mathbf{a}_2 + \left(\tfrac{1}{2} - z_{13}\right)\mathbf{a}_3 = \left(\tfrac{1}{2}c\cos\beta - x_{13} a - z_{13} c\cos\beta\right)\hat{\mathbf{x}} + y_{13} b\,\hat{\mathbf{y}} + \left(\tfrac{1}{2} - z_{13}\right)c\sin\beta\,\hat{\mathbf{z}} \quad (4g) \quad \text{O IX}$$

$$\mathbf{B}_{49} = -x_{13}\,\mathbf{a}_1 - y_{13}\,\mathbf{a}_2 - z_{13}\,\mathbf{a}_3 = (-x_{13} a - z_{13} c\cos\beta)\,\hat{\mathbf{x}} - y_{13} b\,\hat{\mathbf{y}} - z_{13} c\sin\beta\,\hat{\mathbf{z}} \quad (4g) \quad \text{O IX}$$

| | | | | | | |
|---|---|---|---|---|---|---|
| $\mathbf{B}_{50}$ | = | $x_{13}\,\mathbf{a}_1 - y_{13}\,\mathbf{a}_2 + \left(\frac{1}{2} + z_{13}\right)\,\mathbf{a}_3$ | = | $\left(\frac{1}{2}c\cos\beta + x_{13}a + z_{13}c\cos\beta\right)\,\hat{\mathbf{x}} - y_{13}b\,\hat{\mathbf{y}} + \left(\frac{1}{2} + z_{13}\right)c\sin\beta\,\hat{\mathbf{z}}$ | (4g) | O IX |
| $\mathbf{B}_{51}$ | = | $x_{14}\,\mathbf{a}_1 + y_{14}\,\mathbf{a}_2 + z_{14}\,\mathbf{a}_3$ | = | $(x_{14}a + z_{14}c\cos\beta)\,\hat{\mathbf{x}} + y_{14}b\,\hat{\mathbf{y}} + z_{14}c\sin\beta\,\hat{\mathbf{z}}$ | (4g) | O X |
| $\mathbf{B}_{52}$ | = | $-x_{14}\,\mathbf{a}_1 + y_{14}\,\mathbf{a}_2 + \left(\frac{1}{2} - z_{14}\right)\,\mathbf{a}_3$ | = | $\left(\frac{1}{2}c\cos\beta - x_{14}a - z_{14}c\cos\beta\right)\,\hat{\mathbf{x}} + y_{14}b\,\hat{\mathbf{y}} + \left(\frac{1}{2} - z_{14}\right)c\sin\beta\,\hat{\mathbf{z}}$ | (4g) | O X |
| $\mathbf{B}_{53}$ | = | $-x_{14}\,\mathbf{a}_1 - y_{14}\,\mathbf{a}_2 - z_{14}\,\mathbf{a}_3$ | = | $(-x_{14}a - z_{14}c\cos\beta)\,\hat{\mathbf{x}} - y_{14}b\,\hat{\mathbf{y}} - z_{14}c\sin\beta\,\hat{\mathbf{z}}$ | (4g) | O X |
| $\mathbf{B}_{54}$ | = | $x_{14}\,\mathbf{a}_1 - y_{14}\,\mathbf{a}_2 + \left(\frac{1}{2} + z_{14}\right)\,\mathbf{a}_3$ | = | $\left(\frac{1}{2}c\cos\beta + x_{14}a + z_{14}c\cos\beta\right)\,\hat{\mathbf{x}} - y_{14}b\,\hat{\mathbf{y}} + \left(\frac{1}{2} + z_{14}\right)c\sin\beta\,\hat{\mathbf{z}}$ | (4g) | O X |
| $\mathbf{B}_{55}$ | = | $x_{15}\,\mathbf{a}_1 + y_{15}\,\mathbf{a}_2 + z_{15}\,\mathbf{a}_3$ | = | $(x_{15}a + z_{15}c\cos\beta)\,\hat{\mathbf{x}} + y_{15}b\,\hat{\mathbf{y}} + z_{15}c\sin\beta\,\hat{\mathbf{z}}$ | (4g) | O XI |
| $\mathbf{B}_{56}$ | = | $-x_{15}\,\mathbf{a}_1 + y_{15}\,\mathbf{a}_2 + \left(\frac{1}{2} - z_{15}\right)\,\mathbf{a}_3$ | = | $\left(\frac{1}{2}c\cos\beta - x_{15}a - z_{15}c\cos\beta\right)\,\hat{\mathbf{x}} + y_{15}b\,\hat{\mathbf{y}} + \left(\frac{1}{2} - z_{15}\right)c\sin\beta\,\hat{\mathbf{z}}$ | (4g) | O XI |
| $\mathbf{B}_{57}$ | = | $-x_{15}\,\mathbf{a}_1 - y_{15}\,\mathbf{a}_2 - z_{15}\,\mathbf{a}_3$ | = | $(-x_{15}a - z_{15}c\cos\beta)\,\hat{\mathbf{x}} - y_{15}b\,\hat{\mathbf{y}} - z_{15}c\sin\beta\,\hat{\mathbf{z}}$ | (4g) | O XI |
| $\mathbf{B}_{58}$ | = | $x_{15}\,\mathbf{a}_1 - y_{15}\,\mathbf{a}_2 + \left(\frac{1}{2} + z_{15}\right)\,\mathbf{a}_3$ | = | $\left(\frac{1}{2}c\cos\beta + x_{15}a + z_{15}c\cos\beta\right)\,\hat{\mathbf{x}} - y_{15}b\,\hat{\mathbf{y}} + \left(\frac{1}{2} + z_{15}\right)c\sin\beta\,\hat{\mathbf{z}}$ | (4g) | O XI |
| $\mathbf{B}_{59}$ | = | $x_{16}\,\mathbf{a}_1 + y_{16}\,\mathbf{a}_2 + z_{16}\,\mathbf{a}_3$ | = | $(x_{16}a + z_{16}c\cos\beta)\,\hat{\mathbf{x}} + y_{16}b\,\hat{\mathbf{y}} + z_{16}c\sin\beta\,\hat{\mathbf{z}}$ | (4g) | O XII |
| $\mathbf{B}_{60}$ | = | $-x_{16}\,\mathbf{a}_1 + y_{16}\,\mathbf{a}_2 + \left(\frac{1}{2} - z_{16}\right)\,\mathbf{a}_3$ | = | $\left(\frac{1}{2}c\cos\beta - x_{16}a - z_{16}c\cos\beta\right)\,\hat{\mathbf{x}} + y_{16}b\,\hat{\mathbf{y}} + \left(\frac{1}{2} - z_{16}\right)c\sin\beta\,\hat{\mathbf{z}}$ | (4g) | O XII |
| $\mathbf{B}_{61}$ | = | $-x_{16}\,\mathbf{a}_1 - y_{16}\,\mathbf{a}_2 - z_{16}\,\mathbf{a}_3$ | = | $(-x_{16}a - z_{16}c\cos\beta)\,\hat{\mathbf{x}} - y_{16}b\,\hat{\mathbf{y}} - z_{16}c\sin\beta\,\hat{\mathbf{z}}$ | (4g) | O XII |
| $\mathbf{B}_{62}$ | = | $x_{16}\,\mathbf{a}_1 - y_{16}\,\mathbf{a}_2 + \left(\frac{1}{2} + z_{16}\right)\,\mathbf{a}_3$ | = | $\left(\frac{1}{2}c\cos\beta + x_{16}a + z_{16}c\cos\beta\right)\,\hat{\mathbf{x}} - y_{16}b\,\hat{\mathbf{y}} + \left(\frac{1}{2} + z_{16}\right)c\sin\beta\,\hat{\mathbf{z}}$ | (4g) | O XII |

---

---

**Geometry files:**

- CIF: pp. 1536
- POSCAR: pp. 1537

# Huanzalaite ($MgWO_4$, $H0_6$) Structure: AB4C_mP12_13_f_2g_e

http://aflow.org/prototype-encyclopedia/AB4C_mP12_13_f_2g_e

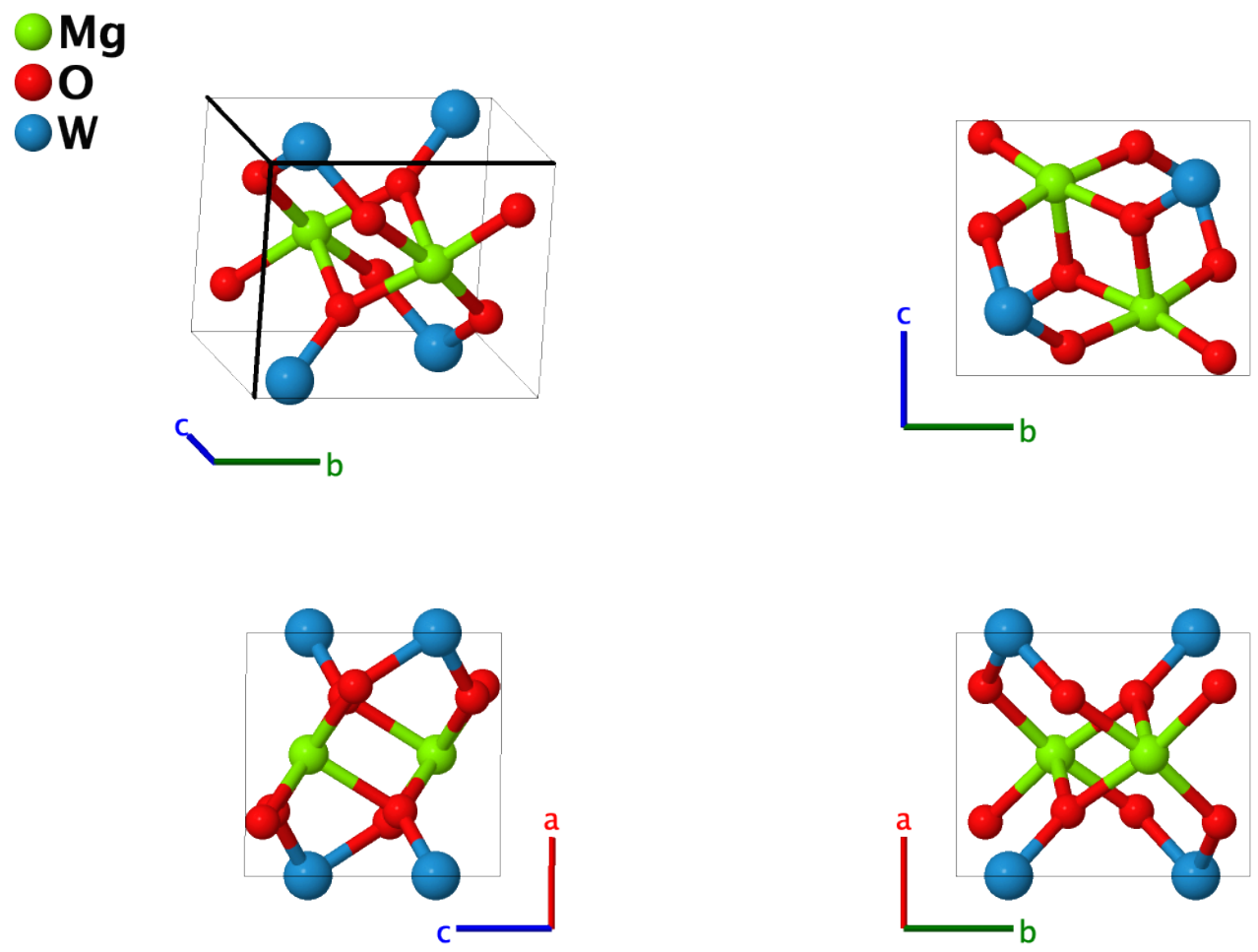

| | | |
|---|---|---|
| **Prototype** | : | $MgO_4W$ |
| **AFLOW prototype label** | : | AB4C_mP12_13_f_2g_e |
| ***Strukturbericht* designation** | : | $H0_6$ |
| **Pearson symbol** | : | mP12 |
| **Space group number** | : | 13 |
| **Space group symbol** | : | $P2/c$ |
| **AFLOW prototype command** | : | `aflow --proto=AB4C_mP12_13_f_2g_e` <br> `--params=`$a, b/a, c/a, \beta, y_1, y_2, x_3, y_3, z_3, x_4, y_4, z_4$ |

### Other compounds with this structure

- $CoWO_4$, $FeWO_4$ (ferberite), $MnWO_4$ (hübernite), $NiWO_4$, and $ZnWO_4$

---

- Most authors refer to this as the wolframite structure, but that name properly belongs to the solid solution (Mn,Fe)$WO_4$. (Herman, 1937) chose this "magnesium wolframite" compound to represent the $H0_6$ structures.

### Simple Monoclinic primitive vectors:

$$\mathbf{a}_1 = a\,\hat{\mathbf{x}}$$
$$\mathbf{a}_2 = b\,\hat{\mathbf{y}}$$
$$\mathbf{a}_3 = c\cos\beta\,\hat{\mathbf{x}} + c\sin\beta\,\hat{\mathbf{z}}$$

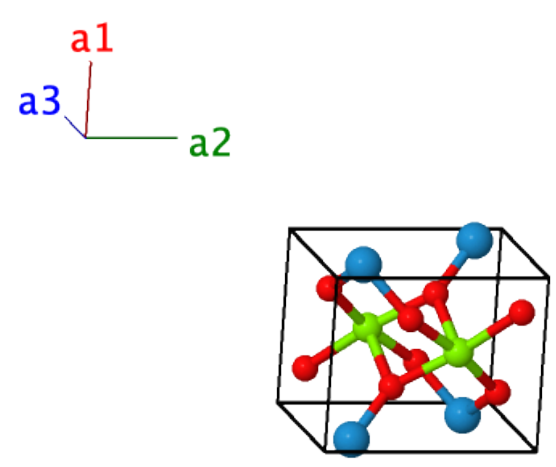

### Basis vectors:

| | | Lattice Coordinates | | Cartesian Coordinates | Wyckoff Position | Atom Type |
|---|---|---|---|---|---|---|
| $\mathbf{B}_1$ | = | $y_1\,\mathbf{a}_2 + \frac{1}{4}\,\mathbf{a}_3$ | = | $\frac{1}{4}c\cos\beta\,\hat{\mathbf{x}} + y_1 b\,\hat{\mathbf{y}} + \frac{1}{4}c\sin\beta\,\hat{\mathbf{z}}$ | (2$e$) | W |
| $\mathbf{B}_2$ | = | $-y_1\,\mathbf{a}_2 + \frac{3}{4}\,\mathbf{a}_3$ | = | $\frac{3}{4}c\cos\beta\,\hat{\mathbf{x}} - y_1 b\,\hat{\mathbf{y}} + \frac{3}{4}c\sin\beta\,\hat{\mathbf{z}}$ | (2e) | W |
| $\mathbf{B}_3$ | = | $\frac{1}{2}\,\mathbf{a}_1 + y_2\,\mathbf{a}_2 + \frac{1}{4}\,\mathbf{a}_3$ | = | $\left(\frac{1}{2}a + \frac{1}{4}c\cos\beta\right)\,\hat{\mathbf{x}} + y_2 b\,\hat{\mathbf{y}} + \frac{1}{4}c\sin\beta\,\hat{\mathbf{z}}$ | (2f) | Mg |
| $\mathbf{B}_4$ | = | $\frac{1}{2}\,\mathbf{a}_1 - y_2\,\mathbf{a}_2 + \frac{3}{4}\,\mathbf{a}_3$ | = | $\left(\frac{1}{2}a + \frac{3}{4}c\cos\beta\right)\,\hat{\mathbf{x}} - y_2 b\,\hat{\mathbf{y}} + \frac{3}{4}c\sin\beta\,\hat{\mathbf{z}}$ | (2f) | Mg |
| $\mathbf{B}_5$ | = | $x_3\,\mathbf{a}_1 + y_3\,\mathbf{a}_2 + z_3\,\mathbf{a}_3$ | = | $(x_3 a + z_3 c\cos\beta)\,\hat{\mathbf{x}} + y_3 b\,\hat{\mathbf{y}} + z_3 c\sin\beta\,\hat{\mathbf{z}}$ | (4g) | O I |
| $\mathbf{B}_6$ | = | $-x_3\,\mathbf{a}_1 + y_3\,\mathbf{a}_2 + \left(\frac{1}{2} - z_3\right)\,\mathbf{a}_3$ | = | $\left(\frac{1}{2}c\cos\beta - x_3 a - z_3 c\cos\beta\right)\,\hat{\mathbf{x}} + y_3 b\,\hat{\mathbf{y}} + \left(\frac{1}{2} - z_3\right)c\sin\beta\,\hat{\mathbf{z}}$ | (4g) | O I |
| $\mathbf{B}_7$ | = | $-x_3\,\mathbf{a}_1 - y_3\,\mathbf{a}_2 - z_3\,\mathbf{a}_3$ | = | $(-x_3 a - z_3 c\cos\beta)\,\hat{\mathbf{x}} - y_3 b\,\hat{\mathbf{y}} - z_3 c\sin\beta\,\hat{\mathbf{z}}$ | (4g) | O I |
| $\mathbf{B}_8$ | = | $x_3\,\mathbf{a}_1 - y_3\,\mathbf{a}_2 + \left(\frac{1}{2} + z_3\right)\,\mathbf{a}_3$ | = | $\left(\frac{1}{2}c\cos\beta + x_3 a + z_3 c\cos\beta\right)\,\hat{\mathbf{x}} - y_3 b\,\hat{\mathbf{y}} + \left(\frac{1}{2} + z_3\right)c\sin\beta\,\hat{\mathbf{z}}$ | (4g) | O I |
| $\mathbf{B}_9$ | = | $x_4\,\mathbf{a}_1 + y_4\,\mathbf{a}_2 + z_4\,\mathbf{a}_3$ | = | $(x_4 a + z_4 c\cos\beta)\,\hat{\mathbf{x}} + y_4 b\,\hat{\mathbf{y}} + z_4 c\sin\beta\,\hat{\mathbf{z}}$ | (4g) | O II |
| $\mathbf{B}_{10}$ | = | $-x_4\,\mathbf{a}_1 + y_4\,\mathbf{a}_2 + \left(\frac{1}{2} - z_4\right)\,\mathbf{a}_3$ | = | $\left(\frac{1}{2}c\cos\beta - x_4 a - z_4 c\cos\beta\right)\,\hat{\mathbf{x}} + y_4 b\,\hat{\mathbf{y}} + \left(\frac{1}{2} - z_4\right)c\sin\beta\,\hat{\mathbf{z}}$ | (4g) | O II |
| $\mathbf{B}_{11}$ | = | $-x_4\,\mathbf{a}_1 - y_4\,\mathbf{a}_2 - z_4\,\mathbf{a}_3$ | = | $(-x_4 a - z_4 c\cos\beta)\,\hat{\mathbf{x}} - y_4 b\,\hat{\mathbf{y}} - z_4 c\sin\beta\,\hat{\mathbf{z}}$ | (4g) | O II |
| $\mathbf{B}_{12}$ | = | $x_4\,\mathbf{a}_1 - y_4\,\mathbf{a}_2 + \left(\frac{1}{2} + z_4\right)\,\mathbf{a}_3$ | = | $\left(\frac{1}{2}c\cos\beta + x_4 a + z_4 c\cos\beta\right)\,\hat{\mathbf{x}} - y_4 b\,\hat{\mathbf{y}} + \left(\frac{1}{2} + z_4\right)c\sin\beta\,\hat{\mathbf{z}}$ | (4g) | O II |

**Geometry files:**

- CIF: pp. 1537
- POSCAR: pp. 1537

# $Cs_{11}O_3$ Structure: A11B3_mP56_14_11e_3e

http://aflow.org/prototype-encyclopedia/A11B3_mP56_14_11e_3e

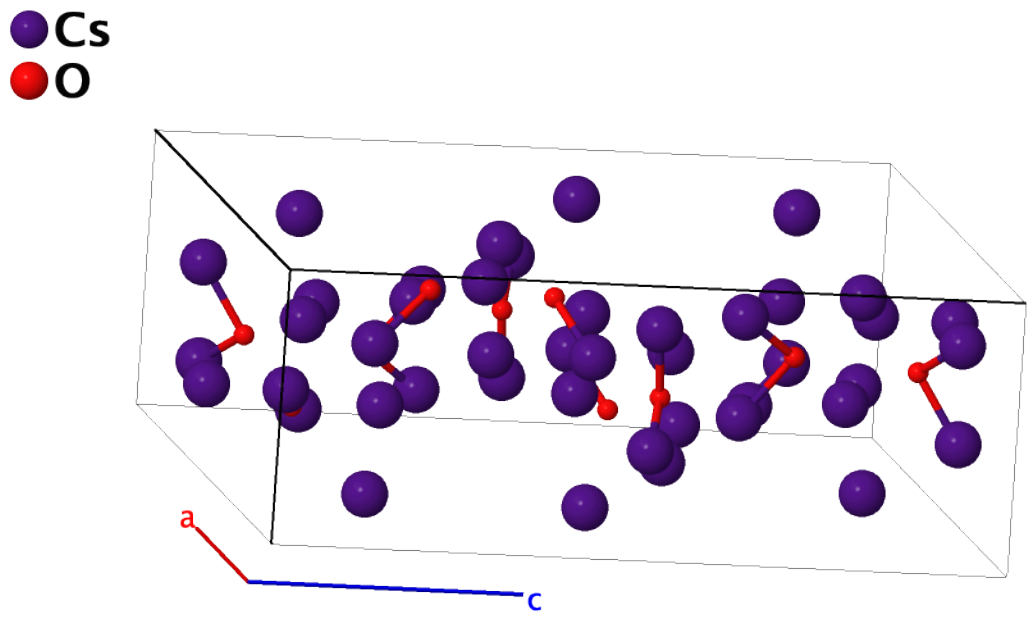

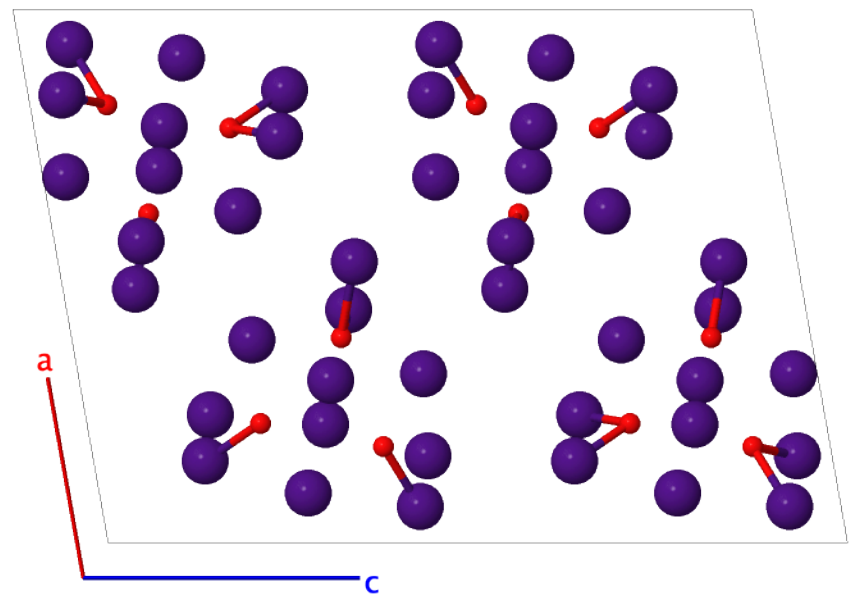

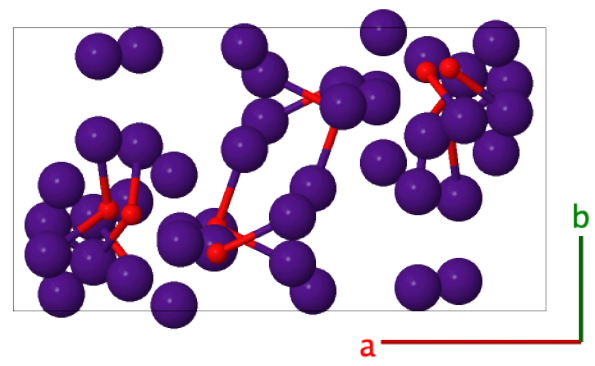

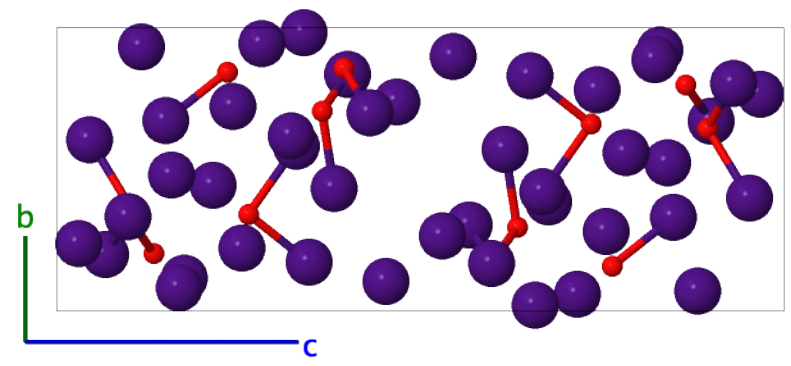

| | | |
|---|---|---|
| **Prototype** | : | $Cs_{11}O_3$ |
| **AFLOW prototype label** | : | A11B3_mP56_14_11e_3e |
| ***Strukturbericht* designation** | : | None |
| **Pearson symbol** | : | mP56 |
| **Space group number** | : | 14 |
| **Space group symbol** | : | $P2_1/c$ |
| **AFLOW prototype command** | : | `aflow --proto=A11B3_mP56_14_11e_3e` `--params=`$a,b/a,c/a,\beta,x_1,y_1,z_1,x_2,y_2,z_2,x_3,y_3,z_3,x_4,y_4,z_4,x_5,y_5,z_5,x_6,y_6,z_6,x_7,y_7,z_7,x_8,y_8,z_8,x_9,y_9,z_9,x_{10},y_{10},z_{10},x_{11},y_{11},z_{11},x_{12},y_{12},z_{12},x_{13},y_{13},z_{13},x_{14},y_{14},z_{14}$ |

### Simple Monoclinic primitive vectors:

$$\mathbf{a}_1 = a\,\hat{\mathbf{x}}$$
$$\mathbf{a}_2 = b\,\hat{\mathbf{y}}$$
$$\mathbf{a}_3 = c\cos\beta\,\hat{\mathbf{x}} + c\sin\beta\,\hat{\mathbf{z}}$$

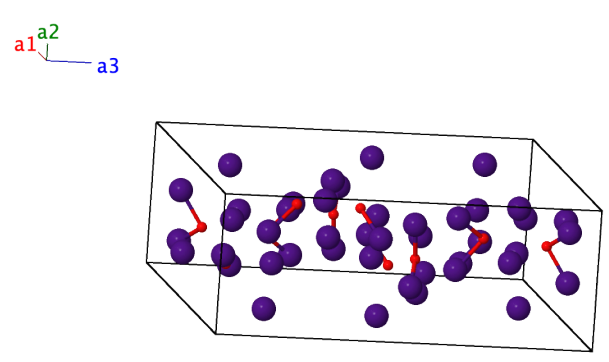

### Basis vectors:

| | | Lattice Coordinates | | Cartesian Coordinates | Wyckoff Position | Atom Type |
|---|---|---|---|---|---|---|
| $\mathbf{B}_1$ | = | $x_1\,\mathbf{a}_1 + y_1\,\mathbf{a}_2 + z_1\,\mathbf{a}_3$ | = | $(x_1 a + z_1 c\cos\beta)\,\hat{\mathbf{x}} + y_1 b\,\hat{\mathbf{y}} + z_1 c\sin\beta\,\hat{\mathbf{z}}$ | (4e) | Cs I |
| $\mathbf{B}_{2}$ | = | $-x_{1}\,\mathbf{a}_{1}+\left(\frac{1}{2}+y_{1}\right)\,\mathbf{a}_{2}+\left(\frac{1}{2}-z_{1}\right)\,\mathbf{a}_{3}$ | = | $\left(\frac{1}{2}c\cos\beta-x_{1}a-z_{1}c\cos\beta\right)\,\hat{\mathbf{x}}+\left(\frac{1}{2}+y_{1}\right)b\,\hat{\mathbf{y}}+\left(\frac{1}{2}-z_{1}\right)c\sin\beta\,\hat{\mathbf{z}}$ | (4$e$) | Cs I |
| $\mathbf{B}_{3}$ | = | $-x_{1}\,\mathbf{a}_{1}-y_{1}\,\mathbf{a}_{2}-z_{1}\,\mathbf{a}_{3}$ | = | $\left(-x_{1}a-z_{1}c\cos\beta\right)\,\hat{\mathbf{x}}-y_{1}b\,\hat{\mathbf{y}}-z_{1}c\sin\beta\,\hat{\mathbf{z}}$ | (4$e$) | Cs I |
| $\mathbf{B}_{4}$ | = | $x_{1}\,\mathbf{a}_{1}+\left(\frac{1}{2}-y_{1}\right)\,\mathbf{a}_{2}+\left(\frac{1}{2}+z_{1}\right)\,\mathbf{a}_{3}$ | = | $\left(\frac{1}{2}c\cos\beta+x_{1}a+z_{1}c\cos\beta\right)\,\hat{\mathbf{x}}+\left(\frac{1}{2}-y_{1}\right)b\,\hat{\mathbf{y}}+\left(\frac{1}{2}+z_{1}\right)c\sin\beta\,\hat{\mathbf{z}}$ | (4$e$) | Cs I |
| $\mathbf{B}_{5}$ | = | $x_{2}\,\mathbf{a}_{1}+y_{2}\,\mathbf{a}_{2}+z_{2}\,\mathbf{a}_{3}$ | = | $\left(x_{2}a+z_{2}c\cos\beta\right)\,\hat{\mathbf{x}}+y_{2}b\,\hat{\mathbf{y}}+z_{2}c\sin\beta\,\hat{\mathbf{z}}$ | (4$e$) | Cs II |
| $\mathbf{B}_{6}$ | = | $-x_{2}\,\mathbf{a}_{1}+\left(\frac{1}{2}+y_{2}\right)\,\mathbf{a}_{2}+\left(\frac{1}{2}-z_{2}\right)\,\mathbf{a}_{3}$ | = | $\left(\frac{1}{2}c\cos\beta-x_{2}a-z_{2}c\cos\beta\right)\,\hat{\mathbf{x}}+\left(\frac{1}{2}+y_{2}\right)b\,\hat{\mathbf{y}}+\left(\frac{1}{2}-z_{2}\right)c\sin\beta\,\hat{\mathbf{z}}$ | (4$e$) | Cs II |
| $\mathbf{B}_{7}$ | = | $-x_{2}\,\mathbf{a}_{1}-y_{2}\,\mathbf{a}_{2}-z_{2}\,\mathbf{a}_{3}$ | = | $\left(-x_{2}a-z_{2}c\cos\beta\right)\,\hat{\mathbf{x}}-y_{2}b\,\hat{\mathbf{y}}-z_{2}c\sin\beta\,\hat{\mathbf{z}}$ | (4$e$) | Cs II |
| $\mathbf{B}_{8}$ | = | $x_{2}\,\mathbf{a}_{1}+\left(\frac{1}{2}-y_{2}\right)\,\mathbf{a}_{2}+\left(\frac{1}{2}+z_{2}\right)\,\mathbf{a}_{3}$ | = | $\left(\frac{1}{2}c\cos\beta+x_{2}a+z_{2}c\cos\beta\right)\,\hat{\mathbf{x}}+\left(\frac{1}{2}-y_{2}\right)b\,\hat{\mathbf{y}}+\left(\frac{1}{2}+z_{2}\right)c\sin\beta\,\hat{\mathbf{z}}$ | (4$e$) | Cs II |
| $\mathbf{B}_{9}$ | = | $x_{3}\,\mathbf{a}_{1}+y_{3}\,\mathbf{a}_{2}+z_{3}\,\mathbf{a}_{3}$ | = | $\left(x_{3}a+z_{3}c\cos\beta\right)\,\hat{\mathbf{x}}+y_{3}b\,\hat{\mathbf{y}}+z_{3}c\sin\beta\,\hat{\mathbf{z}}$ | (4$e$) | Cs III |
| $\mathbf{B}_{10}$ | = | $-x_{3}\,\mathbf{a}_{1}+\left(\frac{1}{2}+y_{3}\right)\,\mathbf{a}_{2}+\left(\frac{1}{2}-z_{3}\right)\,\mathbf{a}_{3}$ | = | $\left(\frac{1}{2}c\cos\beta-x_{3}a-z_{3}c\cos\beta\right)\,\hat{\mathbf{x}}+\left(\frac{1}{2}+y_{3}\right)b\,\hat{\mathbf{y}}+\left(\frac{1}{2}-z_{3}\right)c\sin\beta\,\hat{\mathbf{z}}$ | (4$e$) | Cs III |
| $\mathbf{B}_{11}$ | = | $-x_{3}\,\mathbf{a}_{1}-y_{3}\,\mathbf{a}_{2}-z_{3}\,\mathbf{a}_{3}$ | = | $\left(-x_{3}a-z_{3}c\cos\beta\right)\,\hat{\mathbf{x}}-y_{3}b\,\hat{\mathbf{y}}-z_{3}c\sin\beta\,\hat{\mathbf{z}}$ | (4$e$) | Cs III |
| $\mathbf{B}_{12}$ | = | $x_{3}\,\mathbf{a}_{1}+\left(\frac{1}{2}-y_{3}\right)\,\mathbf{a}_{2}+\left(\frac{1}{2}+z_{3}\right)\,\mathbf{a}_{3}$ | = | $\left(\frac{1}{2}c\cos\beta+x_{3}a+z_{3}c\cos\beta\right)\,\hat{\mathbf{x}}+\left(\frac{1}{2}-y_{3}\right)b\,\hat{\mathbf{y}}+\left(\frac{1}{2}+z_{3}\right)c\sin\beta\,\hat{\mathbf{z}}$ | (4$e$) | Cs III |
| $\mathbf{B}_{13}$ | = | $x_{4}\,\mathbf{a}_{1}+y_{4}\,\mathbf{a}_{2}+z_{4}\,\mathbf{a}_{3}$ | = | $\left(x_{4}a+z_{4}c\cos\beta\right)\,\hat{\mathbf{x}}+y_{4}b\,\hat{\mathbf{y}}+z_{4}c\sin\beta\,\hat{\mathbf{z}}$ | (4$e$) | Cs IV |
| $\mathbf{B}_{14}$ | = | $-x_{4}\,\mathbf{a}_{1}+\left(\frac{1}{2}+y_{4}\right)\,\mathbf{a}_{2}+\left(\frac{1}{2}-z_{4}\right)\,\mathbf{a}_{3}$ | = | $\left(\frac{1}{2}c\cos\beta-x_{4}a-z_{4}c\cos\beta\right)\,\hat{\mathbf{x}}+\left(\frac{1}{2}+y_{4}\right)b\,\hat{\mathbf{y}}+\left(\frac{1}{2}-z_{4}\right)c\sin\beta\,\hat{\mathbf{z}}$ | (4$e$) | Cs IV |
| $\mathbf{B}_{15}$ | = | $-x_{4}\,\mathbf{a}_{1}-y_{4}\,\mathbf{a}_{2}-z_{4}\,\mathbf{a}_{3}$ | = | $\left(-x_{4}a-z_{4}c\cos\beta\right)\,\hat{\mathbf{x}}-y_{4}b\,\hat{\mathbf{y}}-z_{4}c\sin\beta\,\hat{\mathbf{z}}$ | (4$e$) | Cs IV |
| $\mathbf{B}_{16}$ | = | $x_{4}\,\mathbf{a}_{1}+\left(\frac{1}{2}-y_{4}\right)\,\mathbf{a}_{2}+\left(\frac{1}{2}+z_{4}\right)\,\mathbf{a}_{3}$ | = | $\left(\frac{1}{2}c\cos\beta+x_{4}a+z_{4}c\cos\beta\right)\,\hat{\mathbf{x}}+\left(\frac{1}{2}-y_{4}\right)b\,\hat{\mathbf{y}}+\left(\frac{1}{2}+z_{4}\right)c\sin\beta\,\hat{\mathbf{z}}$ | (4$e$) | Cs IV |
| $\mathbf{B}_{17}$ | = | $x_{5}\,\mathbf{a}_{1}+y_{5}\,\mathbf{a}_{2}+z_{5}\,\mathbf{a}_{3}$ | = | $\left(x_{5}a+z_{5}c\cos\beta\right)\,\hat{\mathbf{x}}+y_{5}b\,\hat{\mathbf{y}}+z_{5}c\sin\beta\,\hat{\mathbf{z}}$ | (4$e$) | Cs V |
| $\mathbf{B}_{18}$ | = | $-x_{5}\,\mathbf{a}_{1}+\left(\frac{1}{2}+y_{5}\right)\,\mathbf{a}_{2}+\left(\frac{1}{2}-z_{5}\right)\,\mathbf{a}_{3}$ | = | $\left(\frac{1}{2}c\cos\beta-x_{5}a-z_{5}c\cos\beta\right)\,\hat{\mathbf{x}}+\left(\frac{1}{2}+y_{5}\right)b\,\hat{\mathbf{y}}+\left(\frac{1}{2}-z_{5}\right)c\sin\beta\,\hat{\mathbf{z}}$ | (4$e$) | Cs V |
| $\mathbf{B}_{19}$ | = | $-x_{5}\,\mathbf{a}_{1}-y_{5}\,\mathbf{a}_{2}-z_{5}\,\mathbf{a}_{3}$ | = | $\left(-x_{5}a-z_{5}c\cos\beta\right)\,\hat{\mathbf{x}}-y_{5}b\,\hat{\mathbf{y}}-z_{5}c\sin\beta\,\hat{\mathbf{z}}$ | (4$e$) | Cs V |
| $\mathbf{B}_{20}$ | = | $x_{5}\,\mathbf{a}_{1}+\left(\frac{1}{2}-y_{5}\right)\,\mathbf{a}_{2}+\left(\frac{1}{2}+z_{5}\right)\,\mathbf{a}_{3}$ | = | $\left(\frac{1}{2}c\cos\beta+x_{5}a+z_{5}c\cos\beta\right)\,\hat{\mathbf{x}}+\left(\frac{1}{2}-y_{5}\right)b\,\hat{\mathbf{y}}+\left(\frac{1}{2}+z_{5}\right)c\sin\beta\,\hat{\mathbf{z}}$ | (4$e$) | Cs V |
| $\mathbf{B}_{21}$ | = | $x_{6}\,\mathbf{a}_{1}+y_{6}\,\mathbf{a}_{2}+z_{6}\,\mathbf{a}_{3}$ | = | $\left(x_{6}a+z_{6}c\cos\beta\right)\,\hat{\mathbf{x}}+y_{6}b\,\hat{\mathbf{y}}+z_{6}c\sin\beta\,\hat{\mathbf{z}}$ | (4$e$) | Cs VI |
| $\mathbf{B}_{22}$ | = | $-x_{6}\,\mathbf{a}_{1}+\left(\frac{1}{2}+y_{6}\right)\,\mathbf{a}_{2}+\left(\frac{1}{2}-z_{6}\right)\,\mathbf{a}_{3}$ | = | $\left(\frac{1}{2}c\cos\beta-x_{6}a-z_{6}c\cos\beta\right)\,\hat{\mathbf{x}}+\left(\frac{1}{2}+y_{6}\right)b\,\hat{\mathbf{y}}+\left(\frac{1}{2}-z_{6}\right)c\sin\beta\,\hat{\mathbf{z}}$ | (4$e$) | Cs VI |
| $\mathbf{B}_{23}$ | = | $-x_{6}\,\mathbf{a}_{1}-y_{6}\,\mathbf{a}_{2}-z_{6}\,\mathbf{a}_{3}$ | = | $\left(-x_{6}a-z_{6}c\cos\beta\right)\,\hat{\mathbf{x}}-y_{6}b\,\hat{\mathbf{y}}-z_{6}c\sin\beta\,\hat{\mathbf{z}}$ | (4$e$) | Cs VI |
| $\mathbf{B}_{24}$ | = | $x_6\,\mathbf{a}_1 + \left(\frac{1}{2} - y_6\right)\mathbf{a}_2 + \left(\frac{1}{2} + z_6\right)\mathbf{a}_3$ | = | $\left(\frac{1}{2}c\cos\beta + x_6 a + z_6 c\cos\beta\right)\hat{\mathbf{x}} + \left(\frac{1}{2} - y_6\right)b\,\hat{\mathbf{y}} + \left(\frac{1}{2} + z_6\right)c\sin\beta\,\hat{\mathbf{z}}$ | (4$e$) | Cs VI |
| $\mathbf{B}_{25}$ | = | $x_7\,\mathbf{a}_1 + y_7\,\mathbf{a}_2 + z_7\,\mathbf{a}_3$ | = | $(x_7 a + z_7 c\cos\beta)\,\hat{\mathbf{x}} + y_7 b\,\hat{\mathbf{y}} + z_7 c\sin\beta\,\hat{\mathbf{z}}$ | (4$e$) | Cs VII |
| $\mathbf{B}_{26}$ | = | $-x_7\,\mathbf{a}_1 + \left(\frac{1}{2} + y_7\right)\mathbf{a}_2 + \left(\frac{1}{2} - z_7\right)\mathbf{a}_3$ | = | $\left(\frac{1}{2}c\cos\beta - x_7 a - z_7 c\cos\beta\right)\hat{\mathbf{x}} + \left(\frac{1}{2} + y_7\right)b\,\hat{\mathbf{y}} + \left(\frac{1}{2} - z_7\right)c\sin\beta\,\hat{\mathbf{z}}$ | (4$e$) | Cs VII |
| $\mathbf{B}_{27}$ | = | $-x_7\,\mathbf{a}_1 - y_7\,\mathbf{a}_2 - z_7\,\mathbf{a}_3$ | = | $(-x_7 a - z_7 c\cos\beta)\,\hat{\mathbf{x}} - y_7 b\,\hat{\mathbf{y}} - z_7 c\sin\beta\,\hat{\mathbf{z}}$ | (4$e$) | Cs VII |
| $\mathbf{B}_{28}$ | = | $x_7\,\mathbf{a}_1 + \left(\frac{1}{2} - y_7\right)\mathbf{a}_2 + \left(\frac{1}{2} + z_7\right)\mathbf{a}_3$ | = | $\left(\frac{1}{2}c\cos\beta + x_7 a + z_7 c\cos\beta\right)\hat{\mathbf{x}} + \left(\frac{1}{2} - y_7\right)b\,\hat{\mathbf{y}} + \left(\frac{1}{2} + z_7\right)c\sin\beta\,\hat{\mathbf{z}}$ | (4$e$) | Cs VII |
| $\mathbf{B}_{29}$ | = | $x_8\,\mathbf{a}_1 + y_8\,\mathbf{a}_2 + z_8\,\mathbf{a}_3$ | = | $(x_8 a + z_8 c\cos\beta)\,\hat{\mathbf{x}} + y_8 b\,\hat{\mathbf{y}} + z_8 c\sin\beta\,\hat{\mathbf{z}}$ | (4$e$) | Cs VIII |
| $\mathbf{B}_{30}$ | = | $-x_8\,\mathbf{a}_1 + \left(\frac{1}{2} + y_8\right)\mathbf{a}_2 + \left(\frac{1}{2} - z_8\right)\mathbf{a}_3$ | = | $\left(\frac{1}{2}c\cos\beta - x_8 a - z_8 c\cos\beta\right)\hat{\mathbf{x}} + \left(\frac{1}{2} + y_8\right)b\,\hat{\mathbf{y}} + \left(\frac{1}{2} - z_8\right)c\sin\beta\,\hat{\mathbf{z}}$ | (4$e$) | Cs VIII |
| $\mathbf{B}_{31}$ | = | $-x_8\,\mathbf{a}_1 - y_8\,\mathbf{a}_2 - z_8\,\mathbf{a}_3$ | = | $(-x_8 a - z_8 c\cos\beta)\,\hat{\mathbf{x}} - y_8 b\,\hat{\mathbf{y}} - z_8 c\sin\beta\,\hat{\mathbf{z}}$ | (4$e$) | Cs VIII |
| $\mathbf{B}_{32}$ | = | $x_8\,\mathbf{a}_1 + \left(\frac{1}{2} - y_8\right)\mathbf{a}_2 + \left(\frac{1}{2} + z_8\right)\mathbf{a}_3$ | = | $\left(\frac{1}{2}c\cos\beta + x_8 a + z_8 c\cos\beta\right)\hat{\mathbf{x}} + \left(\frac{1}{2} - y_8\right)b\,\hat{\mathbf{y}} + \left(\frac{1}{2} + z_8\right)c\sin\beta\,\hat{\mathbf{z}}$ | (4$e$) | Cs VIII |
| $\mathbf{B}_{33}$ | = | $x_9\,\mathbf{a}_1 + y_9\,\mathbf{a}_2 + z_9\,\mathbf{a}_3$ | = | $(x_9 a + z_9 c\cos\beta)\,\hat{\mathbf{x}} + y_9 b\,\hat{\mathbf{y}} + z_9 c\sin\beta\,\hat{\mathbf{z}}$ | (4$e$) | Cs IX |
| $\mathbf{B}_{34}$ | = | $-x_9\,\mathbf{a}_1 + \left(\frac{1}{2} + y_9\right)\mathbf{a}_2 + \left(\frac{1}{2} - z_9\right)\mathbf{a}_3$ | = | $\left(\frac{1}{2}c\cos\beta - x_9 a - z_9 c\cos\beta\right)\hat{\mathbf{x}} + \left(\frac{1}{2} + y_9\right)b\,\hat{\mathbf{y}} + \left(\frac{1}{2} - z_9\right)c\sin\beta\,\hat{\mathbf{z}}$ | (4$e$) | Cs IX |
| $\mathbf{B}_{35}$ | = | $-x_9\,\mathbf{a}_1 - y_9\,\mathbf{a}_2 - z_9\,\mathbf{a}_3$ | = | $(-x_9 a - z_9 c\cos\beta)\,\hat{\mathbf{x}} - y_9 b\,\hat{\mathbf{y}} - z_9 c\sin\beta\,\hat{\mathbf{z}}$ | (4$e$) | Cs IX |
| $\mathbf{B}_{36}$ | = | $x_9\,\mathbf{a}_1 + \left(\frac{1}{2} - y_9\right)\mathbf{a}_2 + \left(\frac{1}{2} + z_9\right)\mathbf{a}_3$ | = | $\left(\frac{1}{2}c\cos\beta + x_9 a + z_9 c\cos\beta\right)\hat{\mathbf{x}} + \left(\frac{1}{2} - y_9\right)b\,\hat{\mathbf{y}} + \left(\frac{1}{2} + z_9\right)c\sin\beta\,\hat{\mathbf{z}}$ | (4$e$) | Cs IX |
| $\mathbf{B}_{37}$ | = | $x_{10}\,\mathbf{a}_1 + y_{10}\,\mathbf{a}_2 + z_{10}\,\mathbf{a}_3$ | = | $(x_{10} a + z_{10} c\cos\beta)\,\hat{\mathbf{x}} + y_{10} b\,\hat{\mathbf{y}} + z_{10} c\sin\beta\,\hat{\mathbf{z}}$ | (4$e$) | Cs X |
| $\mathbf{B}_{38}$ | = | $-x_{10}\,\mathbf{a}_1 + \left(\frac{1}{2} + y_{10}\right)\mathbf{a}_2 + \left(\frac{1}{2} - z_{10}\right)\mathbf{a}_3$ | = | $\left(\frac{1}{2}c\cos\beta - x_{10} a - z_{10} c\cos\beta\right)\hat{\mathbf{x}} + \left(\frac{1}{2} + y_{10}\right)b\,\hat{\mathbf{y}} + \left(\frac{1}{2} - z_{10}\right)c\sin\beta\,\hat{\mathbf{z}}$ | (4$e$) | Cs X |
| $\mathbf{B}_{39}$ | = | $-x_{10}\,\mathbf{a}_1 - y_{10}\,\mathbf{a}_2 - z_{10}\,\mathbf{a}_3$ | = | $(-x_{10} a - z_{10} c\cos\beta)\,\hat{\mathbf{x}} - y_{10} b\,\hat{\mathbf{y}} - z_{10} c\sin\beta\,\hat{\mathbf{z}}$ | (4$e$) | Cs X |
| $\mathbf{B}_{40}$ | = | $x_{10}\,\mathbf{a}_1 + \left(\frac{1}{2} - y_{10}\right)\mathbf{a}_2 + \left(\frac{1}{2} + z_{10}\right)\mathbf{a}_3$ | = | $\left(\frac{1}{2}c\cos\beta + x_{10} a + z_{10} c\cos\beta\right)\hat{\mathbf{x}} + \left(\frac{1}{2} - y_{10}\right)b\,\hat{\mathbf{y}} + \left(\frac{1}{2} + z_{10}\right)c\sin\beta\,\hat{\mathbf{z}}$ | (4$e$) | Cs X |
| $\mathbf{B}_{41}$ | = | $x_{11}\,\mathbf{a}_1 + y_{11}\,\mathbf{a}_2 + z_{11}\,\mathbf{a}_3$ | = | $(x_{11} a + z_{11} c\cos\beta)\,\hat{\mathbf{x}} + y_{11} b\,\hat{\mathbf{y}} + z_{11} c\sin\beta\,\hat{\mathbf{z}}$ | (4$e$) | Cs XI |
| $\mathbf{B}_{42}$ | = | $-x_{11}\,\mathbf{a}_1 + \left(\frac{1}{2} + y_{11}\right)\mathbf{a}_2 + \left(\frac{1}{2} - z_{11}\right)\mathbf{a}_3$ | = | $\left(\frac{1}{2}c\cos\beta - x_{11} a - z_{11} c\cos\beta\right)\hat{\mathbf{x}} + \left(\frac{1}{2} + y_{11}\right)b\,\hat{\mathbf{y}} + \left(\frac{1}{2} - z_{11}\right)c\sin\beta\,\hat{\mathbf{z}}$ | (4$e$) | Cs XI |
| $\mathbf{B}_{43}$ | = | $-x_{11}\,\mathbf{a}_1 - y_{11}\,\mathbf{a}_2 - z_{11}\,\mathbf{a}_3$ | = | $(-x_{11} a - z_{11} c\cos\beta)\,\hat{\mathbf{x}} - y_{11} b\,\hat{\mathbf{y}} - z_{11} c\sin\beta\,\hat{\mathbf{z}}$ | (4$e$) | Cs XI |
| $\mathbf{B}_{44}$ | = | $x_{11}\,\mathbf{a}_1 + \left(\frac{1}{2} - y_{11}\right)\mathbf{a}_2 + \left(\frac{1}{2} + z_{11}\right)\mathbf{a}_3$ | = | $\left(\frac{1}{2}c\cos\beta + x_{11} a + z_{11} c\cos\beta\right)\hat{\mathbf{x}} + \left(\frac{1}{2} - y_{11}\right)b\,\hat{\mathbf{y}} + \left(\frac{1}{2} + z_{11}\right)c\sin\beta\,\hat{\mathbf{z}}$ | (4$e$) | Cs XI |
| $\mathbf{B}_{45}$ | = | $x_{12}\,\mathbf{a}_1 + y_{12}\,\mathbf{a}_2 + z_{12}\,\mathbf{a}_3$ | = | $(x_{12} a + z_{12} c\cos\beta)\,\hat{\mathbf{x}} + y_{12} b\,\hat{\mathbf{y}} + z_{12} c\sin\beta\,\hat{\mathbf{z}}$ | (4$e$) | O I |
| $\mathbf{B}_{46}$ | = | $-x_{12}\,\mathbf{a}_1 + \left(\frac{1}{2} + y_{12}\right)\,\mathbf{a}_2 + \left(\frac{1}{2} - z_{12}\right)\,\mathbf{a}_3$ | = | $\left(\frac{1}{2}c\cos\beta - x_{12}a - z_{12}c\cos\beta\right)\,\hat{\mathbf{x}} + \left(\frac{1}{2} + y_{12}\right)b\,\hat{\mathbf{y}} + \left(\frac{1}{2} - z_{12}\right)c\sin\beta\,\hat{\mathbf{z}}$ | (4$e$) | O I |
| $\mathbf{B}_{47}$ | = | $-x_{12}\,\mathbf{a}_1 - y_{12}\,\mathbf{a}_2 - z_{12}\,\mathbf{a}_3$ | = | $(-x_{12}a - z_{12}c\cos\beta)\,\hat{\mathbf{x}} - y_{12}b\,\hat{\mathbf{y}} - z_{12}c\sin\beta\,\hat{\mathbf{z}}$ | (4$e$) | O I |
| $\mathbf{B}_{48}$ | = | $x_{12}\,\mathbf{a}_1 + \left(\frac{1}{2} - y_{12}\right)\,\mathbf{a}_2 + \left(\frac{1}{2} + z_{12}\right)\,\mathbf{a}_3$ | = | $\left(\frac{1}{2}c\cos\beta + x_{12}a + z_{12}c\cos\beta\right)\,\hat{\mathbf{x}} + \left(\frac{1}{2} - y_{12}\right)b\,\hat{\mathbf{y}} + \left(\frac{1}{2} + z_{12}\right)c\sin\beta\,\hat{\mathbf{z}}$ | (4$e$) | O I |
| $\mathbf{B}_{49}$ | = | $x_{13}\,\mathbf{a}_1 + y_{13}\,\mathbf{a}_2 + z_{13}\,\mathbf{a}_3$ | = | $(x_{13}a + z_{13}c\cos\beta)\,\hat{\mathbf{x}} + y_{13}b\,\hat{\mathbf{y}} + z_{13}c\sin\beta\,\hat{\mathbf{z}}$ | (4$e$) | O II |
| $\mathbf{B}_{50}$ | = | $-x_{13}\,\mathbf{a}_1 + \left(\frac{1}{2} + y_{13}\right)\,\mathbf{a}_2 + \left(\frac{1}{2} - z_{13}\right)\,\mathbf{a}_3$ | = | $\left(\frac{1}{2}c\cos\beta - x_{13}a - z_{13}c\cos\beta\right)\,\hat{\mathbf{x}} + \left(\frac{1}{2} + y_{13}\right)b\,\hat{\mathbf{y}} + \left(\frac{1}{2} - z_{13}\right)c\sin\beta\,\hat{\mathbf{z}}$ | (4$e$) | O II |
| $\mathbf{B}_{51}$ | = | $-x_{13}\,\mathbf{a}_1 - y_{13}\,\mathbf{a}_2 - z_{13}\,\mathbf{a}_3$ | = | $(-x_{13}a - z_{13}c\cos\beta)\,\hat{\mathbf{x}} - y_{13}b\,\hat{\mathbf{y}} - z_{13}c\sin\beta\,\hat{\mathbf{z}}$ | (4$e$) | O II |
| $\mathbf{B}_{52}$ | = | $x_{13}\,\mathbf{a}_1 + \left(\frac{1}{2} - y_{13}\right)\,\mathbf{a}_2 + \left(\frac{1}{2} + z_{13}\right)\,\mathbf{a}_3$ | = | $\left(\frac{1}{2}c\cos\beta + x_{13}a + z_{13}c\cos\beta\right)\,\hat{\mathbf{x}} + \left(\frac{1}{2} - y_{13}\right)b\,\hat{\mathbf{y}} + \left(\frac{1}{2} + z_{13}\right)c\sin\beta\,\hat{\mathbf{z}}$ | (4$e$) | O II |
| $\mathbf{B}_{53}$ | = | $x_{14}\,\mathbf{a}_1 + y_{14}\,\mathbf{a}_2 + z_{14}\,\mathbf{a}_3$ | = | $(x_{14}a + z_{14}c\cos\beta)\,\hat{\mathbf{x}} + y_{14}b\,\hat{\mathbf{y}} + z_{14}c\sin\beta\,\hat{\mathbf{z}}$ | (4$e$) | O III |
| $\mathbf{B}_{54}$ | = | $-x_{14}\,\mathbf{a}_1 + \left(\frac{1}{2} + y_{14}\right)\,\mathbf{a}_2 + \left(\frac{1}{2} - z_{14}\right)\,\mathbf{a}_3$ | = | $\left(\frac{1}{2}c\cos\beta - x_{14}a - z_{14}c\cos\beta\right)\,\hat{\mathbf{x}} + \left(\frac{1}{2} + y_{14}\right)b\,\hat{\mathbf{y}} + \left(\frac{1}{2} - z_{14}\right)c\sin\beta\,\hat{\mathbf{z}}$ | (4$e$) | O III |
| $\mathbf{B}_{55}$ | = | $-x_{14}\,\mathbf{a}_1 - y_{14}\,\mathbf{a}_2 - z_{14}\,\mathbf{a}_3$ | = | $(-x_{14}a - z_{14}c\cos\beta)\,\hat{\mathbf{x}} - y_{14}b\,\hat{\mathbf{y}} - z_{14}c\sin\beta\,\hat{\mathbf{z}}$ | (4$e$) | O III |
| $\mathbf{B}_{56}$ | = | $x_{14}\,\mathbf{a}_1 + \left(\frac{1}{2} - y_{14}\right)\,\mathbf{a}_2 + \left(\frac{1}{2} + z_{14}\right)\,\mathbf{a}_3$ | = | $\left(\frac{1}{2}c\cos\beta + x_{14}a + z_{14}c\cos\beta\right)\,\hat{\mathbf{x}} + \left(\frac{1}{2} - y_{14}\right)b\,\hat{\mathbf{y}} + \left(\frac{1}{2} + z_{14}\right)c\sin\beta\,\hat{\mathbf{z}}$ | (4$e$) | O III |

---

---

**Geometry files:**

- CIF: pp. 1537
- POSCAR: pp. 1538

# Azurite [$Cu_3(CO_3)_2(OH)_2$, $G7_4$] Structure: A2B3C2D8_mP30_14_e_ce_e_4e

http://aflow.org/prototype-encyclopedia/A2B3C2D8_mP30_14_e_ce_e_4e

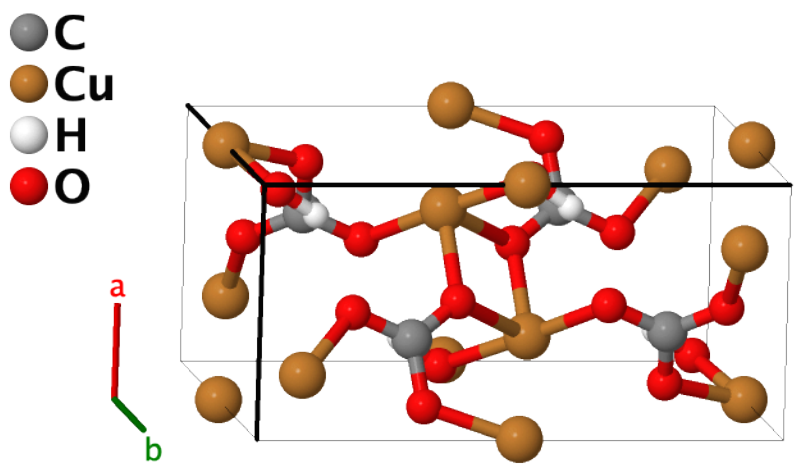

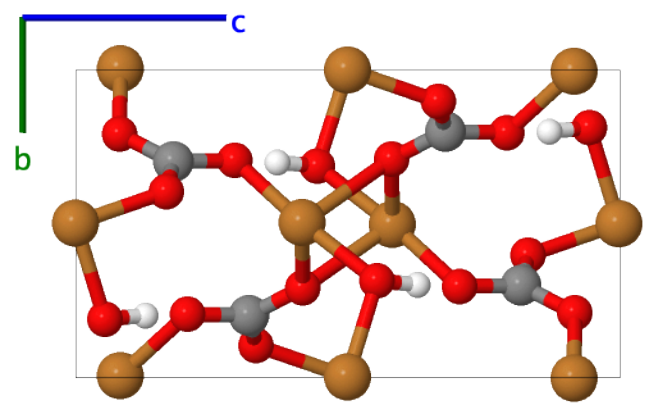

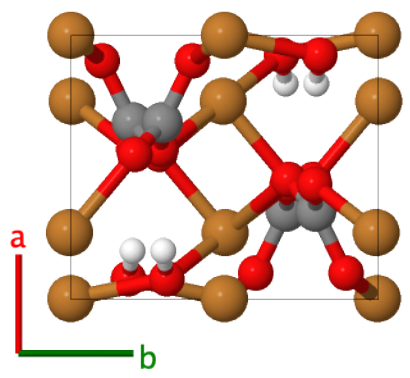

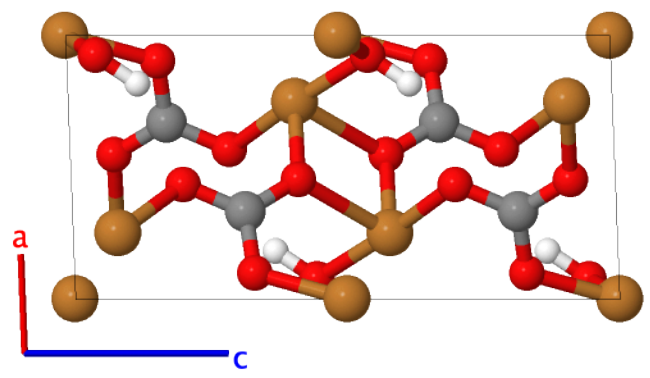

| | | |
|---|---|---|
| **Prototype** | : | $C_2Cu_3H_2O_8$ |
| **AFLOW prototype label** | : | A2B3C2D8_mP30_14_e_ce_e_4e |
| ***Strukturbericht* designation** | : | $G7_4$ |
| **Pearson symbol** | : | mP30 |
| **Space group number** | : | 14 |
| **Space group symbol** | : | $P2_1/c$ |
| **AFLOW prototype command** | : | `aflow --proto=A2B3C2D8_mP30_14_e_ce_e_4e` `--params=`$a, b/a, c/a, \beta, x_2, y_2, z_2, x_3, y_3, z_3, x_4, y_4, z_4, x_5, y_5, z_5, x_6, y_6, z_6, x_7, y_7, z_7, x_8, y_8, z_8$ |

- (Rule, 2011) actually present two possible structures for azurite: the current structure, with space group $P2_1/c$ #14, which is identical to the structure designated $G7_4$ by (Hermann, 1937), except that the OH radical is now separated into the O(IV) and H atom sites; and a structure with the same unit cell dimensions but without an inversion site, leading to space group $P2_1$ #4. The actual difference between the two structures is extremely small, and we choose to use the traditional structure.
- Structural information was taken from powder diffraction data at 1.28 K.

**Simple Monoclinic primitive vectors:**

$$\mathbf{a}_1 = a\,\hat{\mathbf{x}}$$
$$\mathbf{a}_2 = b\,\hat{\mathbf{y}}$$
$$\mathbf{a}_3 = c\cos\beta\,\hat{\mathbf{x}} + c\sin\beta\,\hat{\mathbf{z}}$$

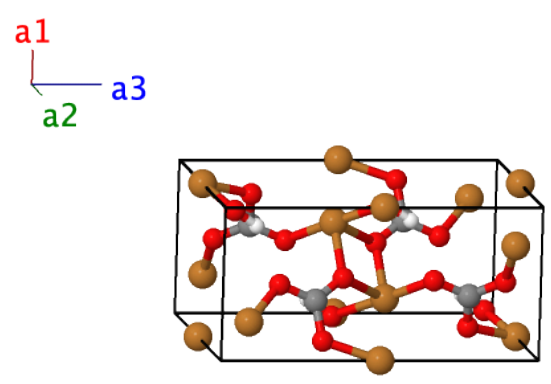

**Basis vectors:**

| | | Lattice Coordinates | | Cartesian Coordinates | Wyckoff Position | Atom Type |
|---|---|---|---|---|---|---|
| $\mathbf{B}_{1}$ | = | $\frac{1}{2}\,\mathbf{a}_{3}$ | = | $\frac{1}{2}c\cos\beta\,\hat{\mathbf{x}}+\frac{1}{2}c\sin\beta\,\hat{\mathbf{z}}$ | (2$c$) | Cu I |
| $\mathbf{B}_{2}$ | = | $\frac{1}{2}\,\mathbf{a}_{2}$ | = | $\frac{1}{2}b\,\hat{\mathbf{y}}$ | (2$c$) | Cu I |
| $\mathbf{B}_{3}$ | = | $x_{2}\,\mathbf{a}_{1}+y_{2}\,\mathbf{a}_{2}+z_{2}\,\mathbf{a}_{3}$ | = | $(x_{2}a+z_{2}c\cos\beta)\,\hat{\mathbf{x}}+y_{2}b\,\hat{\mathbf{y}}+z_{2}c\sin\beta\,\hat{\mathbf{z}}$ | (4$e$) | C |
| $\mathbf{B}_{4}$ | = | $-x_{2}\,\mathbf{a}_{1}+\left(\frac{1}{2}+y_{2}\right)\mathbf{a}_{2}+\left(\frac{1}{2}-z_{2}\right)\mathbf{a}_{3}$ | = | $\left(\frac{1}{2}c\cos\beta-x_{2}a-z_{2}c\cos\beta\right)\hat{\mathbf{x}}+\left(\frac{1}{2}+y_{2}\right)b\,\hat{\mathbf{y}}+\left(\frac{1}{2}-z_{2}\right)c\sin\beta\,\hat{\mathbf{z}}$ | (4$e$) | C |
| $\mathbf{B}_{5}$ | = | $-x_{2}\,\mathbf{a}_{1}-y_{2}\,\mathbf{a}_{2}-z_{2}\,\mathbf{a}_{3}$ | = | $(-x_{2}a-z_{2}c\cos\beta)\,\hat{\mathbf{x}}-y_{2}b\,\hat{\mathbf{y}}-z_{2}c\sin\beta\,\hat{\mathbf{z}}$ | (4$e$) | C |
| $\mathbf{B}_{6}$ | = | $x_{2}\,\mathbf{a}_{1}+\left(\frac{1}{2}-y_{2}\right)\mathbf{a}_{2}+\left(\frac{1}{2}+z_{2}\right)\mathbf{a}_{3}$ | = | $\left(\frac{1}{2}c\cos\beta+x_{2}a+z_{2}c\cos\beta\right)\hat{\mathbf{x}}+\left(\frac{1}{2}-y_{2}\right)b\,\hat{\mathbf{y}}+\left(\frac{1}{2}+z_{2}\right)c\sin\beta\,\hat{\mathbf{z}}$ | (4$e$) | C |
| $\mathbf{B}_{7}$ | = | $x_{3}\,\mathbf{a}_{1}+y_{3}\,\mathbf{a}_{2}+z_{3}\,\mathbf{a}_{3}$ | = | $(x_{3}a+z_{3}c\cos\beta)\,\hat{\mathbf{x}}+y_{3}b\,\hat{\mathbf{y}}+z_{3}c\sin\beta\,\hat{\mathbf{z}}$ | (4$e$) | Cu II |
| $\mathbf{B}_{8}$ | = | $-x_{3}\,\mathbf{a}_{1}+\left(\frac{1}{2}+y_{3}\right)\mathbf{a}_{2}+\left(\frac{1}{2}-z_{3}\right)\mathbf{a}_{3}$ | = | $\left(\frac{1}{2}c\cos\beta-x_{3}a-z_{3}c\cos\beta\right)\hat{\mathbf{x}}+\left(\frac{1}{2}+y_{3}\right)b\,\hat{\mathbf{y}}+\left(\frac{1}{2}-z_{3}\right)c\sin\beta\,\hat{\mathbf{z}}$ | (4$e$) | Cu II |
| $\mathbf{B}_{9}$ | = | $-x_{3}\,\mathbf{a}_{1}-y_{3}\,\mathbf{a}_{2}-z_{3}\,\mathbf{a}_{3}$ | = | $(-x_{3}a-z_{3}c\cos\beta)\,\hat{\mathbf{x}}-y_{3}b\,\hat{\mathbf{y}}-z_{3}c\sin\beta\,\hat{\mathbf{z}}$ | (4$e$) | Cu II |
| $\mathbf{B}_{10}$ | = | $x_{3}\,\mathbf{a}_{1}+\left(\frac{1}{2}-y_{3}\right)\mathbf{a}_{2}+\left(\frac{1}{2}+z_{3}\right)\mathbf{a}_{3}$ | = | $\left(\frac{1}{2}c\cos\beta+x_{3}a+z_{3}c\cos\beta\right)\hat{\mathbf{x}}+\left(\frac{1}{2}-y_{3}\right)b\,\hat{\mathbf{y}}+\left(\frac{1}{2}+z_{3}\right)c\sin\beta\,\hat{\mathbf{z}}$ | (4$e$) | Cu II |
| $\mathbf{B}_{11}$ | = | $x_{4}\,\mathbf{a}_{1}+y_{4}\,\mathbf{a}_{2}+z_{4}\,\mathbf{a}_{3}$ | = | $(x_{4}a+z_{4}c\cos\beta)\,\hat{\mathbf{x}}+y_{4}b\,\hat{\mathbf{y}}+z_{4}c\sin\beta\,\hat{\mathbf{z}}$ | (4$e$) | H |
| $\mathbf{B}_{12}$ | = | $-x_{4}\,\mathbf{a}_{1}+\left(\frac{1}{2}+y_{4}\right)\mathbf{a}_{2}+\left(\frac{1}{2}-z_{4}\right)\mathbf{a}_{3}$ | = | $\left(\frac{1}{2}c\cos\beta-x_{4}a-z_{4}c\cos\beta\right)\hat{\mathbf{x}}+\left(\frac{1}{2}+y_{4}\right)b\,\hat{\mathbf{y}}+\left(\frac{1}{2}-z_{4}\right)c\sin\beta\,\hat{\mathbf{z}}$ | (4$e$) | H |
| $\mathbf{B}_{13}$ | = | $-x_{4}\,\mathbf{a}_{1}-y_{4}\,\mathbf{a}_{2}-z_{4}\,\mathbf{a}_{3}$ | = | $(-x_{4}a-z_{4}c\cos\beta)\,\hat{\mathbf{x}}-y_{4}b\,\hat{\mathbf{y}}-z_{4}c\sin\beta\,\hat{\mathbf{z}}$ | (4$e$) | H |
| $\mathbf{B}_{14}$ | = | $x_{4}\,\mathbf{a}_{1}+\left(\frac{1}{2}-y_{4}\right)\mathbf{a}_{2}+\left(\frac{1}{2}+z_{4}\right)\mathbf{a}_{3}$ | = | $\left(\frac{1}{2}c\cos\beta+x_{4}a+z_{4}c\cos\beta\right)\hat{\mathbf{x}}+\left(\frac{1}{2}-y_{4}\right)b\,\hat{\mathbf{y}}+\left(\frac{1}{2}+z_{4}\right)c\sin\beta\,\hat{\mathbf{z}}$ | (4$e$) | H |
| $\mathbf{B}_{15}$ | = | $x_{5}\,\mathbf{a}_{1}+y_{5}\,\mathbf{a}_{2}+z_{5}\,\mathbf{a}_{3}$ | = | $(x_{5}a+z_{5}c\cos\beta)\,\hat{\mathbf{x}}+y_{5}b\,\hat{\mathbf{y}}+z_{5}c\sin\beta\,\hat{\mathbf{z}}$ | (4$e$) | O I |
| $\mathbf{B}_{16}$ | = | $-x_{5}\,\mathbf{a}_{1}+\left(\frac{1}{2}+y_{5}\right)\mathbf{a}_{2}+\left(\frac{1}{2}-z_{5}\right)\mathbf{a}_{3}$ | = | $\left(\frac{1}{2}c\cos\beta-x_{5}a-z_{5}c\cos\beta\right)\hat{\mathbf{x}}+\left(\frac{1}{2}+y_{5}\right)b\,\hat{\mathbf{y}}+\left(\frac{1}{2}-z_{5}\right)c\sin\beta\,\hat{\mathbf{z}}$ | (4$e$) | O I |
| $\mathbf{B}_{17}$ | = | $-x_{5}\,\mathbf{a}_{1}-y_{5}\,\mathbf{a}_{2}-z_{5}\,\mathbf{a}_{3}$ | = | $(-x_{5}a-z_{5}c\cos\beta)\,\hat{\mathbf{x}}-y_{5}b\,\hat{\mathbf{y}}-z_{5}c\sin\beta\,\hat{\mathbf{z}}$ | (4$e$) | O I |
| $\mathbf{B}_{18}$ | = | $x_{5}\,\mathbf{a}_{1}+\left(\frac{1}{2}-y_{5}\right)\mathbf{a}_{2}+\left(\frac{1}{2}+z_{5}\right)\mathbf{a}_{3}$ | = | $\left(\frac{1}{2}c\cos\beta+x_{5}a+z_{5}c\cos\beta\right)\hat{\mathbf{x}}+\left(\frac{1}{2}-y_{5}\right)b\,\hat{\mathbf{y}}+\left(\frac{1}{2}+z_{5}\right)c\sin\beta\,\hat{\mathbf{z}}$ | (4$e$) | O I |
| $\mathbf{B}_{19}$ | = | $x_{6}\,\mathbf{a}_{1}+y_{6}\,\mathbf{a}_{2}+z_{6}\,\mathbf{a}_{3}$ | = | $(x_{6}a+z_{6}c\cos\beta)\,\hat{\mathbf{x}}+y_{6}b\,\hat{\mathbf{y}}+z_{6}c\sin\beta\,\hat{\mathbf{z}}$ | (4$e$) | O II |
| $\mathbf{B}_{20}$ | = | $-x_{6}\,\mathbf{a}_{1}+\left(\frac{1}{2}+y_{6}\right)\mathbf{a}_{2}+\left(\frac{1}{2}-z_{6}\right)\mathbf{a}_{3}$ | = | $\left(\frac{1}{2}c\cos\beta-x_{6}a-z_{6}c\cos\beta\right)\hat{\mathbf{x}}+\left(\frac{1}{2}+y_{6}\right)b\,\hat{\mathbf{y}}+\left(\frac{1}{2}-z_{6}\right)c\sin\beta\,\hat{\mathbf{z}}$ | (4$e$) | O II |
| $\mathbf{B}_{21}$ | = | $-x_{6}\,\mathbf{a}_{1}-y_{6}\,\mathbf{a}_{2}-z_{6}\,\mathbf{a}_{3}$ | = | $(-x_{6}a-z_{6}c\cos\beta)\,\hat{\mathbf{x}}-y_{6}b\,\hat{\mathbf{y}}-z_{6}c\sin\beta\,\hat{\mathbf{z}}$ | (4$e$) | O II |
| $\mathbf{B}_{22}$ | = | $x_{6}\,\mathbf{a}_{1}+\left(\frac{1}{2}-y_{6}\right)\mathbf{a}_{2}+\left(\frac{1}{2}+z_{6}\right)\mathbf{a}_{3}$ | = | $\left(\frac{1}{2}c\cos\beta+x_{6}a+z_{6}c\cos\beta\right)\hat{\mathbf{x}}+\left(\frac{1}{2}-y_{6}\right)b\,\hat{\mathbf{y}}+\left(\frac{1}{2}+z_{6}\right)c\sin\beta\,\hat{\mathbf{z}}$ | (4$e$) | O II |
| $\mathbf{B}_{23}$ | = | $x_7\,\mathbf{a}_1 + y_7\,\mathbf{a}_2 + z_7\,\mathbf{a}_3$ | = | $(x_7 a + z_7 c\cos\beta)\,\hat{\mathbf{x}} + y_7 b\,\hat{\mathbf{y}} + z_7 c\sin\beta\,\hat{\mathbf{z}}$ | (4e) | O III |
| $\mathbf{B}_{24}$ | = | $-x_7\,\mathbf{a}_1 + \left(\frac{1}{2} + y_7\right)\mathbf{a}_2 + \left(\frac{1}{2} - z_7\right)\mathbf{a}_3$ | = | $\left(\frac{1}{2}c\cos\beta - x_7 a - z_7 c\cos\beta\right)\hat{\mathbf{x}} + \left(\frac{1}{2} + y_7\right)b\,\hat{\mathbf{y}} + \left(\frac{1}{2} - z_7\right)c\sin\beta\,\hat{\mathbf{z}}$ | (4e) | O III |
| $\mathbf{B}_{25}$ | = | $-x_7\,\mathbf{a}_1 - y_7\,\mathbf{a}_2 - z_7\,\mathbf{a}_3$ | = | $(-x_7 a - z_7 c\cos\beta)\,\hat{\mathbf{x}} - y_7 b\,\hat{\mathbf{y}} - z_7 c\sin\beta\,\hat{\mathbf{z}}$ | (4e) | O III |
| $\mathbf{B}_{26}$ | = | $x_7\,\mathbf{a}_1 + \left(\frac{1}{2} - y_7\right)\mathbf{a}_2 + \left(\frac{1}{2} + z_7\right)\mathbf{a}_3$ | = | $\left(\frac{1}{2}c\cos\beta + x_7 a + z_7 c\cos\beta\right)\hat{\mathbf{x}} + \left(\frac{1}{2} - y_7\right)b\,\hat{\mathbf{y}} + \left(\frac{1}{2} + z_7\right)c\sin\beta\,\hat{\mathbf{z}}$ | (4e) | O III |
| $\mathbf{B}_{27}$ | = | $x_8\,\mathbf{a}_1 + y_8\,\mathbf{a}_2 + z_8\,\mathbf{a}_3$ | = | $(x_8 a + z_8 c\cos\beta)\,\hat{\mathbf{x}} + y_8 b\,\hat{\mathbf{y}} + z_8 c\sin\beta\,\hat{\mathbf{z}}$ | (4e) | O IV |
| $\mathbf{B}_{28}$ | = | $-x_8\,\mathbf{a}_1 + \left(\frac{1}{2} + y_8\right)\mathbf{a}_2 + \left(\frac{1}{2} - z_8\right)\mathbf{a}_3$ | = | $\left(\frac{1}{2}c\cos\beta - x_8 a - z_8 c\cos\beta\right)\hat{\mathbf{x}} + \left(\frac{1}{2} + y_8\right)b\,\hat{\mathbf{y}} + \left(\frac{1}{2} - z_8\right)c\sin\beta\,\hat{\mathbf{z}}$ | (4e) | O IV |
| $\mathbf{B}_{29}$ | = | $-x_8\,\mathbf{a}_1 - y_8\,\mathbf{a}_2 - z_8\,\mathbf{a}_3$ | = | $(-x_8 a - z_8 c\cos\beta)\,\hat{\mathbf{x}} - y_8 b\,\hat{\mathbf{y}} - z_8 c\sin\beta\,\hat{\mathbf{z}}$ | (4e) | O IV |
| $\mathbf{B}_{30}$ | = | $x_8\,\mathbf{a}_1 + \left(\frac{1}{2} - y_8\right)\mathbf{a}_2 + \left(\frac{1}{2} + z_8\right)\mathbf{a}_3$ | = | $\left(\frac{1}{2}c\cos\beta + x_8 a + z_8 c\cos\beta\right)\hat{\mathbf{x}} + \left(\frac{1}{2} - y_8\right)b\,\hat{\mathbf{y}} + \left(\frac{1}{2} + z_8\right)c\sin\beta\,\hat{\mathbf{z}}$ | (4e) | O IV |

---

---

**Geometry files:**

- CIF: pp. 1538
- POSCAR: pp. 1539

# $HgCl_2 \cdot 2HgO$ Structure: A2B3C2_mP14_14_e_ae_e

http://aflow.org/prototype-encyclopedia/A2B3C2_mP14_14_e_ae_e

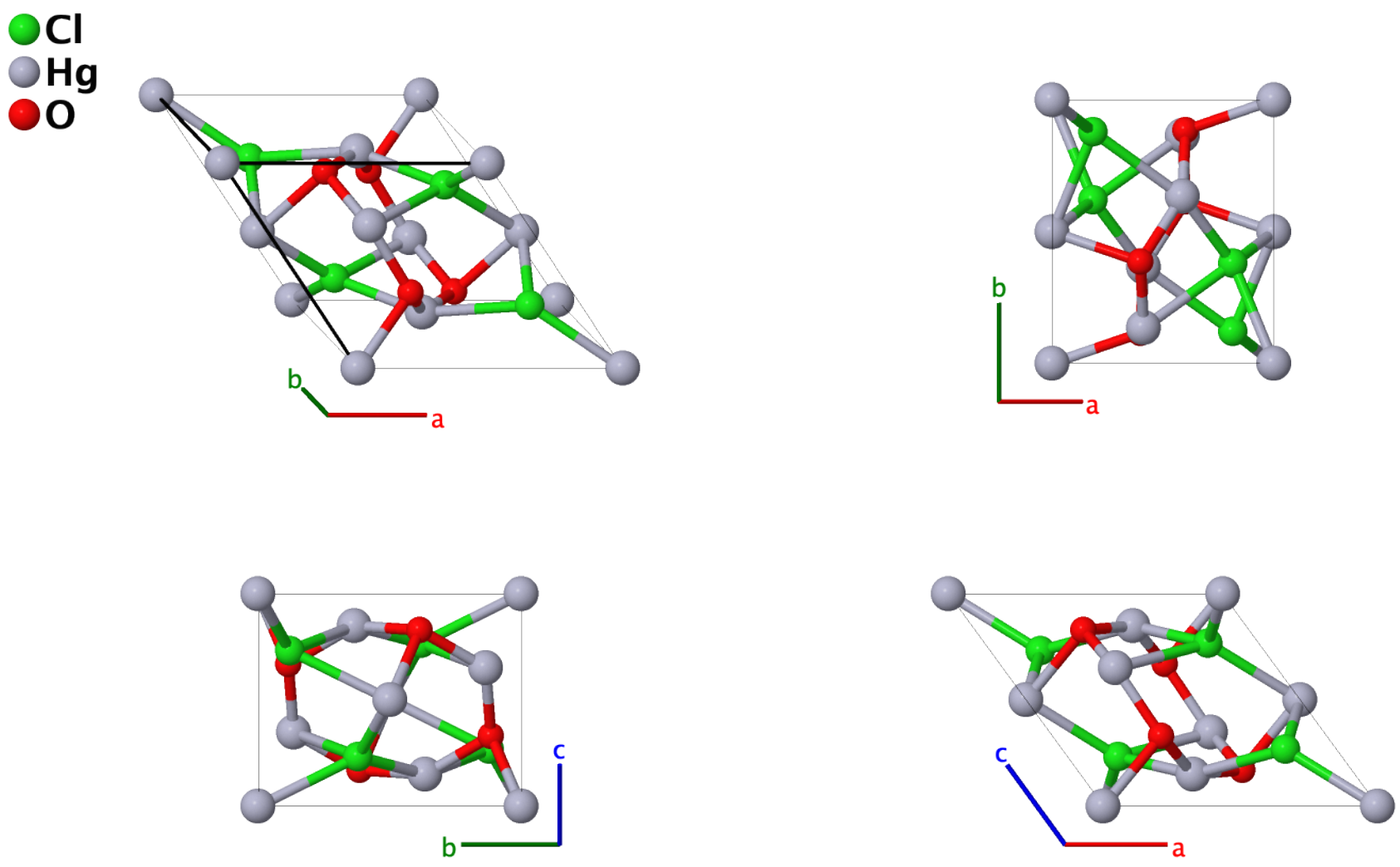

| | | |
|---|---|---|
| **Prototype** | : | $Cl_2Hg_3O_2$ |
| **AFLOW prototype label** | : | A2B3C2_mP14_14_e_ae_e |
| ***Strukturbericht* designation** | : | None |
| **Pearson symbol** | : | mP14 |
| **Space group number** | : | 14 |
| **Space group symbol** | : | $P2_1/c$ |
| **AFLOW prototype command** | : | `aflow --proto=A2B3C2_mP14_14_e_ae_e` <br> `--params=`$a, b/a, c/a, \beta, x_2, y_2, z_2, x_3, y_3, z_3, x_4, y_4, z_4$ |

**Simple Monoclinic primitive vectors:**

$$\mathbf{a}_1 = a\,\hat{\mathbf{x}}$$
$$\mathbf{a}_2 = b\,\hat{\mathbf{y}}$$
$$\mathbf{a}_3 = c\cos\beta\,\hat{\mathbf{x}} + c\sin\beta\,\hat{\mathbf{z}}$$

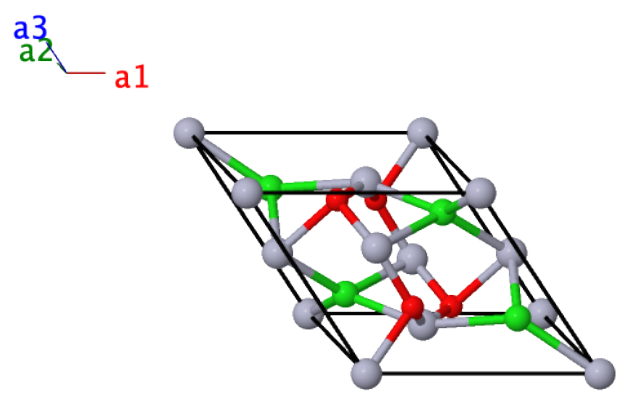

**Basis vectors:**

| | | Lattice Coordinates | | Cartesian Coordinates | Wyckoff Position | Atom Type |
|---|---|---|---|---|---|---|
| $\mathbf{B}_1$ | = | $0\,\mathbf{a}_1 + 0\,\mathbf{a}_2 + 0\,\mathbf{a}_3$ | = | $0\,\hat{\mathbf{x}} + 0\,\hat{\mathbf{y}} + 0\,\hat{\mathbf{z}}$ | (2a) | Hg I |
| $\mathbf{B}_2$ | = | $\frac{1}{2}\,\mathbf{a}_2 + \frac{1}{2}\,\mathbf{a}_3$ | = | $\frac{1}{2}c\cos\beta\,\hat{\mathbf{x}} + \frac{1}{2}b\,\hat{\mathbf{y}} + \frac{1}{2}c\sin\beta\,\hat{\mathbf{z}}$ | (2a) | Hg I |
| $\mathbf{B}_3$ | = | $x_2\,\mathbf{a}_1 + y_2\,\mathbf{a}_2 + z_2\,\mathbf{a}_3$ | = | $(x_2a + z_2c\cos\beta)\,\hat{\mathbf{x}} + y_2b\,\hat{\mathbf{y}} + z_2c\sin\beta\,\hat{\mathbf{z}}$ | (4e) | Cl |
| $\mathbf{B}_4$ | = | $-x_2\,\mathbf{a}_1 + \left(\frac{1}{2} + y_2\right)\,\mathbf{a}_2 + \left(\frac{1}{2} - z_2\right)\,\mathbf{a}_3$ | = | $\left(\frac{1}{2}c\cos\beta - x_2a - z_2c\cos\beta\right)\,\hat{\mathbf{x}} + \left(\frac{1}{2} + y_2\right)b\,\hat{\mathbf{y}} + \left(\frac{1}{2} - z_2\right)c\sin\beta\,\hat{\mathbf{z}}$ | (4e) | Cl |
| $\mathbf{B}_5$ | = | $-x_2\,\mathbf{a}_1 - y_2\,\mathbf{a}_2 - z_2\,\mathbf{a}_3$ | = | $(-x_2a - z_2c\cos\beta)\,\hat{\mathbf{x}} - y_2b\,\hat{\mathbf{y}} - z_2c\sin\beta\,\hat{\mathbf{z}}$ | (4e) | Cl |
| $\mathbf{B}_{6}$ | = | $x_2\,\mathbf{a}_1+\left(\frac{1}{2}-y_2\right)\mathbf{a}_2+\left(\frac{1}{2}+z_2\right)\mathbf{a}_3$ | = | $\left(\frac{1}{2}c\cos\beta+x_2a+z_2c\cos\beta\right)\hat{\mathbf{x}}+\left(\frac{1}{2}-y_2\right)b\,\hat{\mathbf{y}}+\left(\frac{1}{2}+z_2\right)c\sin\beta\,\hat{\mathbf{z}}$ | (4e) | Cl |
| $\mathbf{B}_{7}$ | = | $x_3\,\mathbf{a}_1+y_3\,\mathbf{a}_2+z_3\,\mathbf{a}_3$ | = | $(x_3a+z_3c\cos\beta)\,\hat{\mathbf{x}}+y_3b\,\hat{\mathbf{y}}+z_3c\sin\beta\,\hat{\mathbf{z}}$ | (4e) | Hg II |
| $\mathbf{B}_{8}$ | = | $-x_3\,\mathbf{a}_1+\left(\frac{1}{2}+y_3\right)\mathbf{a}_2+\left(\frac{1}{2}-z_3\right)\mathbf{a}_3$ | = | $\left(\frac{1}{2}c\cos\beta-x_3a-z_3c\cos\beta\right)\hat{\mathbf{x}}+\left(\frac{1}{2}+y_3\right)b\,\hat{\mathbf{y}}+\left(\frac{1}{2}-z_3\right)c\sin\beta\,\hat{\mathbf{z}}$ | (4e) | Hg II |
| $\mathbf{B}_{9}$ | = | $-x_3\,\mathbf{a}_1-y_3\,\mathbf{a}_2-z_3\,\mathbf{a}_3$ | = | $(-x_3a-z_3c\cos\beta)\,\hat{\mathbf{x}}-y_3b\,\hat{\mathbf{y}}-z_3c\sin\beta\,\hat{\mathbf{z}}$ | (4e) | Hg II |
| $\mathbf{B}_{10}$ | = | $x_3\,\mathbf{a}_1+\left(\frac{1}{2}-y_3\right)\mathbf{a}_2+\left(\frac{1}{2}+z_3\right)\mathbf{a}_3$ | = | $\left(\frac{1}{2}c\cos\beta+x_3a+z_3c\cos\beta\right)\hat{\mathbf{x}}+\left(\frac{1}{2}-y_3\right)b\,\hat{\mathbf{y}}+\left(\frac{1}{2}+z_3\right)c\sin\beta\,\hat{\mathbf{z}}$ | (4e) | Hg II |
| $\mathbf{B}_{11}$ | = | $x_4\,\mathbf{a}_1+y_4\,\mathbf{a}_2+z_4\,\mathbf{a}_3$ | = | $(x_4a+z_4c\cos\beta)\,\hat{\mathbf{x}}+y_4b\,\hat{\mathbf{y}}+z_4c\sin\beta\,\hat{\mathbf{z}}$ | (4e) | O |
| $\mathbf{B}_{12}$ | = | $-x_4\,\mathbf{a}_1+\left(\frac{1}{2}+y_4\right)\mathbf{a}_2+\left(\frac{1}{2}-z_4\right)\mathbf{a}_3$ | = | $\left(\frac{1}{2}c\cos\beta-x_4a-z_4c\cos\beta\right)\hat{\mathbf{x}}+\left(\frac{1}{2}+y_4\right)b\,\hat{\mathbf{y}}+\left(\frac{1}{2}-z_4\right)c\sin\beta\,\hat{\mathbf{z}}$ | (4e) | O |
| $\mathbf{B}_{13}$ | = | $-x_4\,\mathbf{a}_1-y_4\,\mathbf{a}_2-z_4\,\mathbf{a}_3$ | = | $(-x_4a-z_4c\cos\beta)\,\hat{\mathbf{x}}-y_4b\,\hat{\mathbf{y}}-z_4c\sin\beta\,\hat{\mathbf{z}}$ | (4e) | O |
| $\mathbf{B}_{14}$ | = | $x_4\,\mathbf{a}_1+\left(\frac{1}{2}-y_4\right)\mathbf{a}_2+\left(\frac{1}{2}+z_4\right)\mathbf{a}_3$ | = | $\left(\frac{1}{2}c\cos\beta+x_4a+z_4c\cos\beta\right)\hat{\mathbf{x}}+\left(\frac{1}{2}-y_4\right)b\,\hat{\mathbf{y}}+\left(\frac{1}{2}+z_4\right)c\sin\beta\,\hat{\mathbf{z}}$ | (4e) | O |

**Geometry files:**

- CIF: pp. 1539
- POSCAR: pp. 1539

# Orpiment ($As_2S_3$, $D5_f$) Structure: A2B3_mP20_14_2e_3e

http://aflow.org/prototype-encyclopedia/A2B3_mP20_14_2e_3e

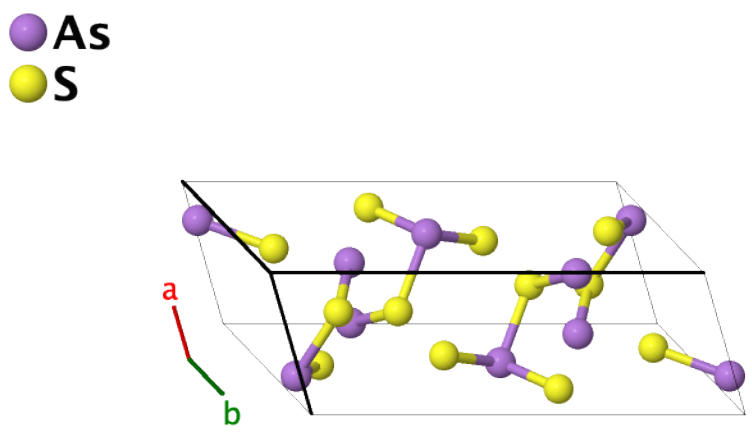

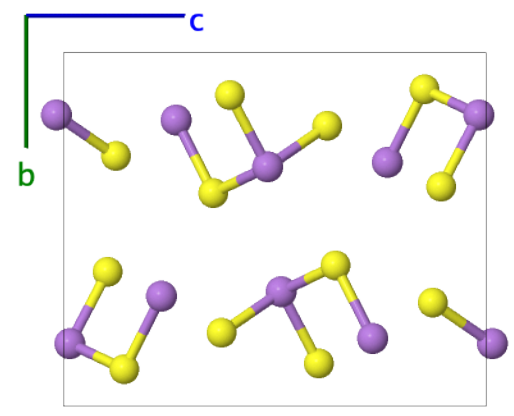

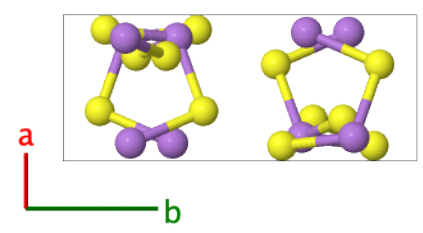

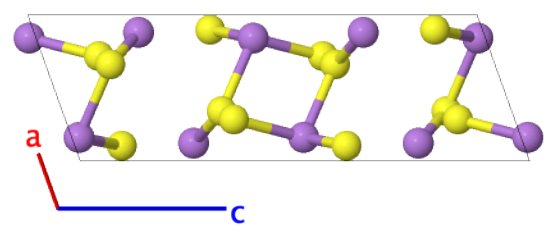

| | | |
|---|---|---|
| **Prototype** | : | $As_2S_3$ |
| **AFLOW prototype label** | : | A2B3_mP20_14_2e_3e |
| ***Strukturbericht* designation** | : | $D5_f$ |
| **Pearson symbol** | : | mP20 |
| **Space group number** | : | 14 |
| **Space group symbol** | : | $P2_1/c$ |
| **AFLOW prototype command** | : | `aflow --proto=A2B3_mP20_14_2e_3e` <br> `--params=`$a, b/a, c/a, \beta, x_1, y_1, z_1, x_2, y_2, z_2, x_3, y_3, z_3, x_4, y_4, z_4, x_5, y_5, z_5$ |

### Other compounds with this structure

- $As_2Se_3$ and $As_2Te_3$

- (Morimoto, 1954) gives the structure in setting $P2_1/n$ of space group #14. We used FINDSYM to translate the structure to the standard $P2_1/c$ setting. This included a redefinition of the primitive vectors of the lattice.

### Simple Monoclinic primitive vectors:

$$\mathbf{a}_1 = a\,\hat{\mathbf{x}}$$
$$\mathbf{a}_2 = b\,\hat{\mathbf{y}}$$
$$\mathbf{a}_3 = c\cos\beta\,\hat{\mathbf{x}} + c\sin\beta\,\hat{\mathbf{z}}$$

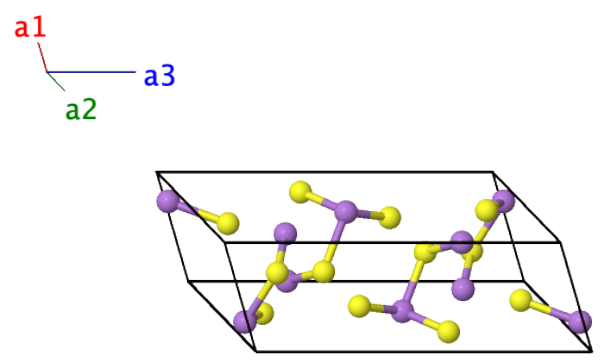

### Basis vectors:

| | | Lattice Coordinates | | Cartesian Coordinates | Wyckoff Position | Atom Type |
|---|---|---|---|---|---|---|
| $\mathbf{B}_1$ | = | $x_1\,\mathbf{a}_1 + y_1\,\mathbf{a}_2 + z_1\,\mathbf{a}_3$ | = | $(x_1 a + z_1 c\cos\beta)\,\hat{\mathbf{x}} + y_1 b\,\hat{\mathbf{y}} + z_1 c\sin\beta\,\hat{\mathbf{z}}$ | (4e) | As I |
| $\mathbf{B}_2$ | = | $-x_1\,\mathbf{a}_1 + \left(\frac{1}{2} + y_1\right)\mathbf{a}_2 + \left(\frac{1}{2} - z_1\right)\mathbf{a}_3$ | = | $\left(\frac{1}{2}c\cos\beta - x_1 a - z_1 c\cos\beta\right)\hat{\mathbf{x}} + \left(\frac{1}{2} + y_1\right)b\,\hat{\mathbf{y}} + \left(\frac{1}{2} - z_1\right)c\sin\beta\,\hat{\mathbf{z}}$ | (4e) | As I |
| $\mathbf{B}_3$ | = | $-x_1\,\mathbf{a}_1 - y_1\,\mathbf{a}_2 - z_1\,\mathbf{a}_3$ | = | $(-x_1 a - z_1 c\cos\beta)\,\hat{\mathbf{x}} - y_1 b\,\hat{\mathbf{y}} - z_1 c\sin\beta\,\hat{\mathbf{z}}$ | (4e) | As I |
| $\mathbf{B}_4$ | = | $x_1\,\mathbf{a}_1 + \left(\frac{1}{2} - y_1\right)\mathbf{a}_2 + \left(\frac{1}{2} + z_1\right)\mathbf{a}_3$ | = | $\left(\frac{1}{2}c\cos\beta + x_1 a + z_1 c\cos\beta\right)\hat{\mathbf{x}} + \left(\frac{1}{2} - y_1\right)b\,\hat{\mathbf{y}} + \left(\frac{1}{2} + z_1\right)c\sin\beta\,\hat{\mathbf{z}}$ | (4e) | As I |
| $\mathbf{B}_5$ | = | $x_2\,\mathbf{a}_1 + y_2\,\mathbf{a}_2 + z_2\,\mathbf{a}_3$ | = | $(x_2 a + z_2 c\cos\beta)\,\hat{\mathbf{x}} + y_2 b\,\hat{\mathbf{y}} + z_2 c\sin\beta\,\hat{\mathbf{z}}$ | (4e) | As II |
| $\mathbf{B}_6$ | = | $-x_2\,\mathbf{a}_1 + \left(\frac{1}{2} + y_2\right)\mathbf{a}_2 + \left(\frac{1}{2} - z_2\right)\mathbf{a}_3$ | = | $\left(\frac{1}{2}c\cos\beta - x_2 a - z_2 c\cos\beta\right)\hat{\mathbf{x}} + \left(\frac{1}{2} + y_2\right)b\,\hat{\mathbf{y}} + \left(\frac{1}{2} - z_2\right)c\sin\beta\,\hat{\mathbf{z}}$ | (4e) | As II |
| $\mathbf{B}_7$ | = | $-x_2\,\mathbf{a}_1 - y_2\,\mathbf{a}_2 - z_2\,\mathbf{a}_3$ | = | $(-x_2 a - z_2 c\cos\beta)\,\hat{\mathbf{x}} - y_2 b\,\hat{\mathbf{y}} - z_2 c\sin\beta\,\hat{\mathbf{z}}$ | (4e) | As II |
| $\mathbf{B}_8$ | = | $x_2\,\mathbf{a}_1 + \left(\frac{1}{2} - y_2\right)\mathbf{a}_2 + \left(\frac{1}{2} + z_2\right)\mathbf{a}_3$ | = | $\left(\frac{1}{2}c\cos\beta + x_2 a + z_2 c\cos\beta\right)\hat{\mathbf{x}} + \left(\frac{1}{2} - y_2\right)b\,\hat{\mathbf{y}} + \left(\frac{1}{2} + z_2\right)c\sin\beta\,\hat{\mathbf{z}}$ | (4e) | As II |
| $\mathbf{B}_9$ | = | $x_3\,\mathbf{a}_1 + y_3\,\mathbf{a}_2 + z_3\,\mathbf{a}_3$ | = | $(x_3 a + z_3 c\cos\beta)\,\hat{\mathbf{x}} + y_3 b\,\hat{\mathbf{y}} + z_3 c\sin\beta\,\hat{\mathbf{z}}$ | (4e) | S I |
| $\mathbf{B}_{10}$ | = | $-x_3\,\mathbf{a}_1 + \left(\frac{1}{2} + y_3\right)\mathbf{a}_2 + \left(\frac{1}{2} - z_3\right)\mathbf{a}_3$ | = | $\left(\frac{1}{2}c\cos\beta - x_3 a - z_3 c\cos\beta\right)\hat{\mathbf{x}} + \left(\frac{1}{2} + y_3\right)b\,\hat{\mathbf{y}} + \left(\frac{1}{2} - z_3\right)c\sin\beta\,\hat{\mathbf{z}}$ | (4e) | S I |
| $\mathbf{B}_{11}$ | = | $-x_3\,\mathbf{a}_1 - y_3\,\mathbf{a}_2 - z_3\,\mathbf{a}_3$ | = | $(-x_3 a - z_3 c\cos\beta)\,\hat{\mathbf{x}} - y_3 b\,\hat{\mathbf{y}} - z_3 c\sin\beta\,\hat{\mathbf{z}}$ | (4e) | S I |
| $\mathbf{B}_{12}$ | = | $x_3\,\mathbf{a}_1 + \left(\frac{1}{2} - y_3\right)\mathbf{a}_2 + \left(\frac{1}{2} + z_3\right)\mathbf{a}_3$ | = | $\left(\frac{1}{2}c\cos\beta + x_3 a + z_3 c\cos\beta\right)\hat{\mathbf{x}} + \left(\frac{1}{2} - y_3\right)b\,\hat{\mathbf{y}} + \left(\frac{1}{2} + z_3\right)c\sin\beta\,\hat{\mathbf{z}}$ | (4e) | S I |
| $\mathbf{B}_{13}$ | = | $x_4\,\mathbf{a}_1 + y_4\,\mathbf{a}_2 + z_4\,\mathbf{a}_3$ | = | $(x_4 a + z_4 c\cos\beta)\,\hat{\mathbf{x}} + y_4 b\,\hat{\mathbf{y}} + z_4 c\sin\beta\,\hat{\mathbf{z}}$ | (4e) | S II |
| $\mathbf{B}_{14}$ | = | $-x_4\,\mathbf{a}_1 + \left(\frac{1}{2} + y_4\right)\mathbf{a}_2 + \left(\frac{1}{2} - z_4\right)\mathbf{a}_3$ | = | $\left(\frac{1}{2}c\cos\beta - x_4 a - z_4 c\cos\beta\right)\hat{\mathbf{x}} + \left(\frac{1}{2} + y_4\right)b\,\hat{\mathbf{y}} + \left(\frac{1}{2} - z_4\right)c\sin\beta\,\hat{\mathbf{z}}$ | (4e) | S II |
| $\mathbf{B}_{15}$ | = | $-x_4\,\mathbf{a}_1 - y_4\,\mathbf{a}_2 - z_4\,\mathbf{a}_3$ | = | $(-x_4 a - z_4 c\cos\beta)\,\hat{\mathbf{x}} - y_4 b\,\hat{\mathbf{y}} - z_4 c\sin\beta\,\hat{\mathbf{z}}$ | (4e) | S II |
| $\mathbf{B}_{16}$ | = | $x_4\,\mathbf{a}_1 + \left(\frac{1}{2} - y_4\right)\mathbf{a}_2 + \left(\frac{1}{2} + z_4\right)\mathbf{a}_3$ | = | $\left(\frac{1}{2}c\cos\beta + x_4 a + z_4 c\cos\beta\right)\hat{\mathbf{x}} + \left(\frac{1}{2} - y_4\right)b\,\hat{\mathbf{y}} + \left(\frac{1}{2} + z_4\right)c\sin\beta\,\hat{\mathbf{z}}$ | (4e) | S II |
| $\mathbf{B}_{17}$ | = | $x_5\,\mathbf{a}_1 + y_5\,\mathbf{a}_2 + z_5\,\mathbf{a}_3$ | = | $(x_5 a + z_5 c\cos\beta)\,\hat{\mathbf{x}} + y_5 b\,\hat{\mathbf{y}} + z_5 c\sin\beta\,\hat{\mathbf{z}}$ | (4e) | S III |
| $\mathbf{B}_{18}$ | = | $-x_5\,\mathbf{a}_1 + \left(\frac{1}{2} + y_5\right)\mathbf{a}_2 + \left(\frac{1}{2} - z_5\right)\mathbf{a}_3$ | = | $\left(\frac{1}{2}c\cos\beta - x_5 a - z_5 c\cos\beta\right)\hat{\mathbf{x}} + \left(\frac{1}{2} + y_5\right)b\,\hat{\mathbf{y}} + \left(\frac{1}{2} - z_5\right)c\sin\beta\,\hat{\mathbf{z}}$ | (4e) | S III |
| $\mathbf{B}_{19}$ | = | $-x_5\,\mathbf{a}_1 - y_5\,\mathbf{a}_2 - z_5\,\mathbf{a}_3$ | = | $(-x_5 a - z_5 c\cos\beta)\,\hat{\mathbf{x}} - y_5 b\,\hat{\mathbf{y}} - z_5 c\sin\beta\,\hat{\mathbf{z}}$ | (4e) | S III |
| $\mathbf{B}_{20}$ | = | $x_5\,\mathbf{a}_1 + \left(\frac{1}{2} - y_5\right)\mathbf{a}_2 + \left(\frac{1}{2} + z_5\right)\mathbf{a}_3$ | = | $\left(\frac{1}{2}c\cos\beta + x_5 a + z_5 c\cos\beta\right)\hat{\mathbf{x}} + \left(\frac{1}{2} - y_5\right)b\,\hat{\mathbf{y}} + \left(\frac{1}{2} + z_5\right)c\sin\beta\,\hat{\mathbf{z}}$ | (4e) | S III |

**Geometry files:**
- CIF: pp. 1539
- POSCAR: pp. 1540

# Monoclinic $Cu_2OSeO_3$ Structure: A2B4C_mP28_14_abe_4e_e

http://aflow.org/prototype-encyclopedia/A2B4C_mP28_14_abe_4e_e

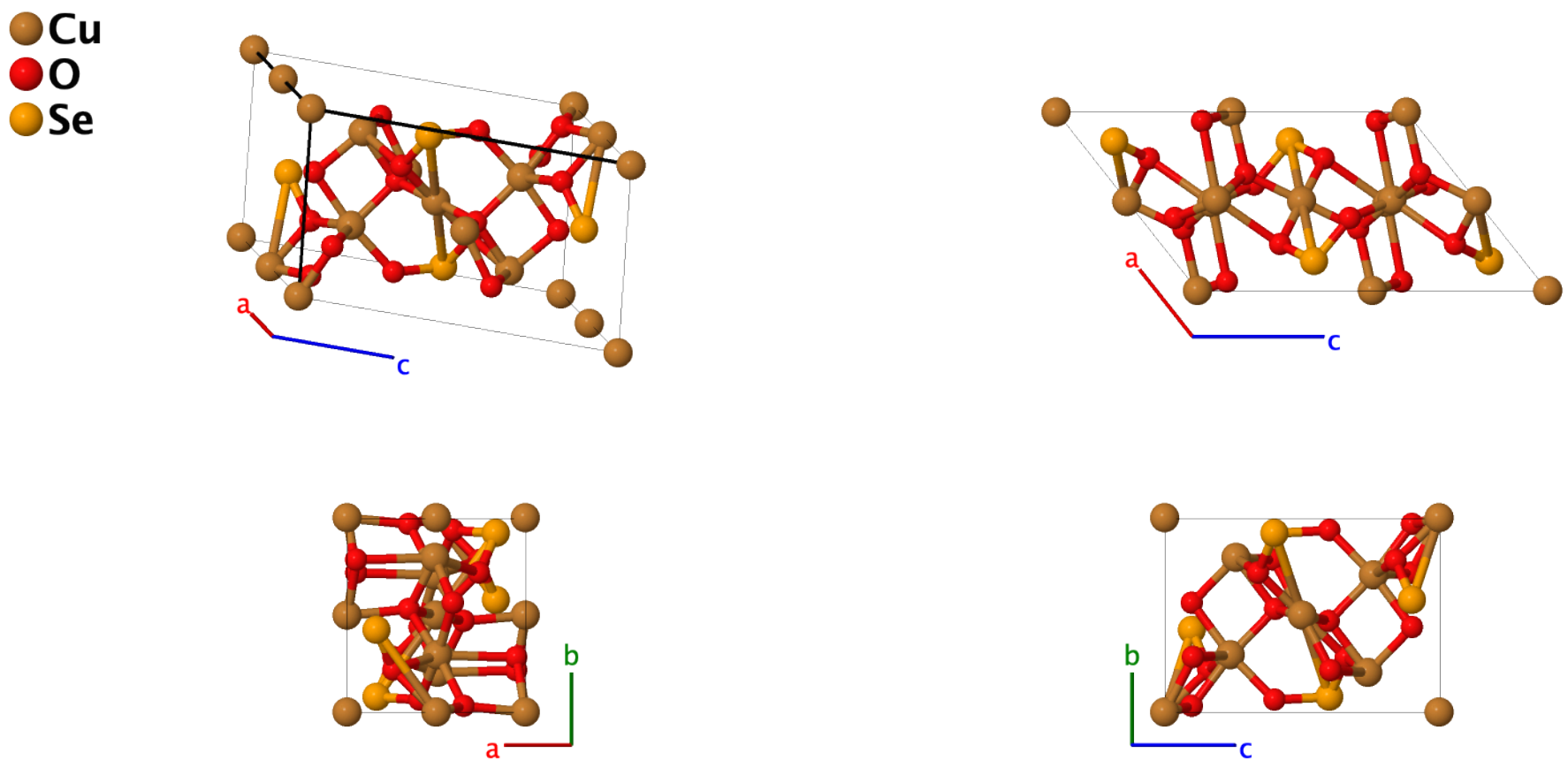

| | | |
|---|---|---|
| **Prototype** | : | $Cu_2O_4Se$ |
| **AFLOW prototype label** | : | A2B4C_mP28_14_abe_4e_e |
| ***Strukturbericht* designation** | : | None |
| **Pearson symbol** | : | mP28 |
| **Space group number** | : | 14 |
| **Space group symbol** | : | $P2_1/c$ |
| **AFLOW prototype command** | : | `aflow --proto=A2B4C_mP28_14_abe_4e_e --params=`$a, b/a, c/a, \beta, x_3, y_3, z_3, x_4, y_4, z_4, x_5, y_5, z_5, x_6, y_6, z_6, x_7, y_7, z_7, x_8, y_8, z_8$ |

- This is the monoclinic phase of $Cu_2OSeO_3$. There is also a cubic phase.

**Simple Monoclinic primitive vectors:**

$$\mathbf{a}_1 = a\,\hat{\mathbf{x}}$$
$$\mathbf{a}_2 = b\,\hat{\mathbf{y}}$$
$$\mathbf{a}_3 = c\cos\beta\,\hat{\mathbf{x}} + c\sin\beta\,\hat{\mathbf{z}}$$

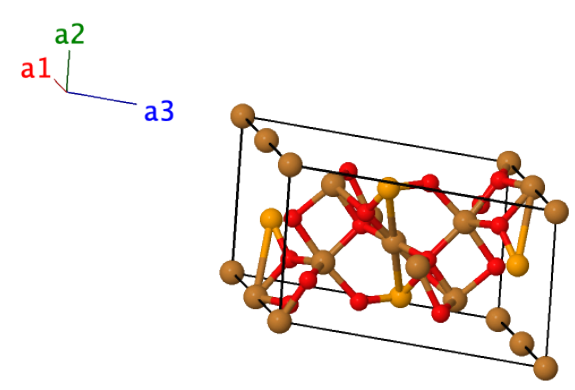

**Basis vectors:**

| | | Lattice Coordinates | | Cartesian Coordinates | Wyckoff Position | Atom Type |
|---|---|---|---|---|---|---|
| $\mathbf{B}_1$ | = | $0\,\mathbf{a}_1 + 0\,\mathbf{a}_2 + 0\,\mathbf{a}_3$ | = | $0\,\hat{\mathbf{x}} + 0\,\hat{\mathbf{y}} + 0\,\hat{\mathbf{z}}$ | (2a) | Cu I |
| $\mathbf{B}_2$ | = | $\frac{1}{2}\,\mathbf{a}_2 + \frac{1}{2}\,\mathbf{a}_3$ | = | $\frac{1}{2}c\cos\beta\,\hat{\mathbf{x}} + \frac{1}{2}b\,\hat{\mathbf{y}} + \frac{1}{2}c\sin\beta\,\hat{\mathbf{z}}$ | (2a) | Cu I |
| $\mathbf{B}_3$ | = | $\frac{1}{2}\,\mathbf{a}_1$ | = | $\frac{1}{2}a\,\hat{\mathbf{x}}$ | (2b) | Cu II |
| $\mathbf{B}_4$ | = | $\frac{1}{2}\,\mathbf{a}_1 + \frac{1}{2}\,\mathbf{a}_2 + \frac{1}{2}\,\mathbf{a}_3$ | = | $\frac{1}{2}(a + c\cos\beta)\,\hat{\mathbf{x}} + \frac{1}{2}b\,\hat{\mathbf{y}} + \frac{1}{2}c\sin\beta\,\hat{\mathbf{z}}$ | (2b) | Cu II |
| $\mathbf{B}_5$ | = | $x_3\,\mathbf{a}_1 + y_3\,\mathbf{a}_2 + z_3\,\mathbf{a}_3$ | = | $(x_3 a + z_3 c\cos\beta)\,\hat{\mathbf{x}} + y_3 b\,\hat{\mathbf{y}} + z_3 c\sin\beta\,\hat{\mathbf{z}}$ | (4e) | Cu III |

$$\mathbf{B}_{6} = -x_{3}\,\mathbf{a}_{1} + \left(\frac{1}{2} + y_{3}\right)\,\mathbf{a}_{2} + \left(\frac{1}{2} - z_{3}\right)\,\mathbf{a}_{3} = \left(\frac{1}{2}c\cos\beta - x_{3}a - z_{3}c\cos\beta\right)\,\hat{\mathbf{x}} + \left(\frac{1}{2} + y_{3}\right)b\,\hat{\mathbf{y}} + \left(\frac{1}{2} - z_{3}\right)c\sin\beta\,\hat{\mathbf{z}} \quad (4e) \quad \text{Cu III}$$

$$\mathbf{B}_{7} = -x_{3}\,\mathbf{a}_{1} - y_{3}\,\mathbf{a}_{2} - z_{3}\,\mathbf{a}_{3} = (-x_{3}a - z_{3}c\cos\beta)\,\hat{\mathbf{x}} - y_{3}b\,\hat{\mathbf{y}} - z_{3}c\sin\beta\,\hat{\mathbf{z}} \quad (4e) \quad \text{Cu III}$$

$$\mathbf{B}_{8} = x_{3}\,\mathbf{a}_{1} + \left(\frac{1}{2} - y_{3}\right)\,\mathbf{a}_{2} + \left(\frac{1}{2} + z_{3}\right)\,\mathbf{a}_{3} = \left(\frac{1}{2}c\cos\beta + x_{3}a + z_{3}c\cos\beta\right)\,\hat{\mathbf{x}} + \left(\frac{1}{2} - y_{3}\right)b\,\hat{\mathbf{y}} + \left(\frac{1}{2} + z_{3}\right)c\sin\beta\,\hat{\mathbf{z}} \quad (4e) \quad \text{Cu III}$$

$$\mathbf{B}_{9} = x_{4}\,\mathbf{a}_{1} + y_{4}\,\mathbf{a}_{2} + z_{4}\,\mathbf{a}_{3} = (x_{4}a + z_{4}c\cos\beta)\,\hat{\mathbf{x}} + y_{4}b\,\hat{\mathbf{y}} + z_{4}c\sin\beta\,\hat{\mathbf{z}} \quad (4e) \quad \text{O I}$$

$$\mathbf{B}_{10} = -x_{4}\,\mathbf{a}_{1} + \left(\frac{1}{2} + y_{4}\right)\,\mathbf{a}_{2} + \left(\frac{1}{2} - z_{4}\right)\,\mathbf{a}_{3} = \left(\frac{1}{2}c\cos\beta - x_{4}a - z_{4}c\cos\beta\right)\,\hat{\mathbf{x}} + \left(\frac{1}{2} + y_{4}\right)b\,\hat{\mathbf{y}} + \left(\frac{1}{2} - z_{4}\right)c\sin\beta\,\hat{\mathbf{z}} \quad (4e) \quad \text{O I}$$

$$\mathbf{B}_{11} = -x_{4}\,\mathbf{a}_{1} - y_{4}\,\mathbf{a}_{2} - z_{4}\,\mathbf{a}_{3} = (-x_{4}a - z_{4}c\cos\beta)\,\hat{\mathbf{x}} - y_{4}b\,\hat{\mathbf{y}} - z_{4}c\sin\beta\,\hat{\mathbf{z}} \quad (4e) \quad \text{O I}$$

$$\mathbf{B}_{12} = x_{4}\,\mathbf{a}_{1} + \left(\frac{1}{2} - y_{4}\right)\,\mathbf{a}_{2} + \left(\frac{1}{2} + z_{4}\right)\,\mathbf{a}_{3} = \left(\frac{1}{2}c\cos\beta + x_{4}a + z_{4}c\cos\beta\right)\,\hat{\mathbf{x}} + \left(\frac{1}{2} - y_{4}\right)b\,\hat{\mathbf{y}} + \left(\frac{1}{2} + z_{4}\right)c\sin\beta\,\hat{\mathbf{z}} \quad (4e) \quad \text{O I}$$

$$\mathbf{B}_{13} = x_{5}\,\mathbf{a}_{1} + y_{5}\,\mathbf{a}_{2} + z_{5}\,\mathbf{a}_{3} = (x_{5}a + z_{5}c\cos\beta)\,\hat{\mathbf{x}} + y_{5}b\,\hat{\mathbf{y}} + z_{5}c\sin\beta\,\hat{\mathbf{z}} \quad (4e) \quad \text{O II}$$

$$\mathbf{B}_{14} = -x_{5}\,\mathbf{a}_{1} + \left(\frac{1}{2} + y_{5}\right)\,\mathbf{a}_{2} + \left(\frac{1}{2} - z_{5}\right)\,\mathbf{a}_{3} = \left(\frac{1}{2}c\cos\beta - x_{5}a - z_{5}c\cos\beta\right)\,\hat{\mathbf{x}} + \left(\frac{1}{2} + y_{5}\right)b\,\hat{\mathbf{y}} + \left(\frac{1}{2} - z_{5}\right)c\sin\beta\,\hat{\mathbf{z}} \quad (4e) \quad \text{O II}$$

$$\mathbf{B}_{15} = -x_{5}\,\mathbf{a}_{1} - y_{5}\,\mathbf{a}_{2} - z_{5}\,\mathbf{a}_{3} = (-x_{5}a - z_{5}c\cos\beta)\,\hat{\mathbf{x}} - y_{5}b\,\hat{\mathbf{y}} - z_{5}c\sin\beta\,\hat{\mathbf{z}} \quad (4e) \quad \text{O II}$$

$$\mathbf{B}_{16} = x_{5}\,\mathbf{a}_{1} + \left(\frac{1}{2} - y_{5}\right)\,\mathbf{a}_{2} + \left(\frac{1}{2} + z_{5}\right)\,\mathbf{a}_{3} = \left(\frac{1}{2}c\cos\beta + x_{5}a + z_{5}c\cos\beta\right)\,\hat{\mathbf{x}} + \left(\frac{1}{2} - y_{5}\right)b\,\hat{\mathbf{y}} + \left(\frac{1}{2} + z_{5}\right)c\sin\beta\,\hat{\mathbf{z}} \quad (4e) \quad \text{O II}$$

$$\mathbf{B}_{17} = x_{6}\,\mathbf{a}_{1} + y_{6}\,\mathbf{a}_{2} + z_{6}\,\mathbf{a}_{3} = (x_{6}a + z_{6}c\cos\beta)\,\hat{\mathbf{x}} + y_{6}b\,\hat{\mathbf{y}} + z_{6}c\sin\beta\,\hat{\mathbf{z}} \quad (4e) \quad \text{O III}$$

$$\mathbf{B}_{18} = -x_{6}\,\mathbf{a}_{1} + \left(\frac{1}{2} + y_{6}\right)\,\mathbf{a}_{2} + \left(\frac{1}{2} - z_{6}\right)\,\mathbf{a}_{3} = \left(\frac{1}{2}c\cos\beta - x_{6}a - z_{6}c\cos\beta\right)\,\hat{\mathbf{x}} + \left(\frac{1}{2} + y_{6}\right)b\,\hat{\mathbf{y}} + \left(\frac{1}{2} - z_{6}\right)c\sin\beta\,\hat{\mathbf{z}} \quad (4e) \quad \text{O III}$$

$$\mathbf{B}_{19} = -x_{6}\,\mathbf{a}_{1} - y_{6}\,\mathbf{a}_{2} - z_{6}\,\mathbf{a}_{3} = (-x_{6}a - z_{6}c\cos\beta)\,\hat{\mathbf{x}} - y_{6}b\,\hat{\mathbf{y}} - z_{6}c\sin\beta\,\hat{\mathbf{z}} \quad (4e) \quad \text{O III}$$

$$\mathbf{B}_{20} = x_{6}\,\mathbf{a}_{1} + \left(\frac{1}{2} - y_{6}\right)\,\mathbf{a}_{2} + \left(\frac{1}{2} + z_{6}\right)\,\mathbf{a}_{3} = \left(\frac{1}{2}c\cos\beta + x_{6}a + z_{6}c\cos\beta\right)\,\hat{\mathbf{x}} + \left(\frac{1}{2} - y_{6}\right)b\,\hat{\mathbf{y}} + \left(\frac{1}{2} + z_{6}\right)c\sin\beta\,\hat{\mathbf{z}} \quad (4e) \quad \text{O III}$$

$$\mathbf{B}_{21} = x_{7}\,\mathbf{a}_{1} + y_{7}\,\mathbf{a}_{2} + z_{7}\,\mathbf{a}_{3} = (x_{7}a + z_{7}c\cos\beta)\,\hat{\mathbf{x}} + y_{7}b\,\hat{\mathbf{y}} + z_{7}c\sin\beta\,\hat{\mathbf{z}} \quad (4e) \quad \text{O IV}$$

$$\mathbf{B}_{22} = -x_{7}\,\mathbf{a}_{1} + \left(\frac{1}{2} + y_{7}\right)\,\mathbf{a}_{2} + \left(\frac{1}{2} - z_{7}\right)\,\mathbf{a}_{3} = \left(\frac{1}{2}c\cos\beta - x_{7}a - z_{7}c\cos\beta\right)\,\hat{\mathbf{x}} + \left(\frac{1}{2} + y_{7}\right)b\,\hat{\mathbf{y}} + \left(\frac{1}{2} - z_{7}\right)c\sin\beta\,\hat{\mathbf{z}} \quad (4e) \quad \text{O IV}$$

$$\mathbf{B}_{23} = -x_{7}\,\mathbf{a}_{1} - y_{7}\,\mathbf{a}_{2} - z_{7}\,\mathbf{a}_{3} = (-x_{7}a - z_{7}c\cos\beta)\,\hat{\mathbf{x}} - y_{7}b\,\hat{\mathbf{y}} - z_{7}c\sin\beta\,\hat{\mathbf{z}} \quad (4e) \quad \text{O IV}$$

$$\mathbf{B}_{24} = x_{7}\,\mathbf{a}_{1} + \left(\frac{1}{2} - y_{7}\right)\,\mathbf{a}_{2} + \left(\frac{1}{2} + z_{7}\right)\,\mathbf{a}_{3} = \left(\frac{1}{2}c\cos\beta + x_{7}a + z_{7}c\cos\beta\right)\,\hat{\mathbf{x}} + \left(\frac{1}{2} - y_{7}\right)b\,\hat{\mathbf{y}} + \left(\frac{1}{2} + z_{7}\right)c\sin\beta\,\hat{\mathbf{z}} \quad (4e) \quad \text{O IV}$$

$$\mathbf{B}_{25} = x_{8}\,\mathbf{a}_{1} + y_{8}\,\mathbf{a}_{2} + z_{8}\,\mathbf{a}_{3} = (x_{8}a + z_{8}c\cos\beta)\,\hat{\mathbf{x}} + y_{8}b\,\hat{\mathbf{y}} + z_{8}c\sin\beta\,\hat{\mathbf{z}} \quad (4e) \quad \text{Se}$$

$$\mathbf{B}_{26} = -x_{8}\,\mathbf{a}_{1} + \left(\frac{1}{2} + y_{8}\right)\,\mathbf{a}_{2} + \left(\frac{1}{2} - z_{8}\right)\,\mathbf{a}_{3} = \left(\frac{1}{2}c\cos\beta - x_{8}a - z_{8}c\cos\beta\right)\,\hat{\mathbf{x}} + \left(\frac{1}{2} + y_{8}\right)b\,\hat{\mathbf{y}} + \left(\frac{1}{2} - z_{8}\right)c\sin\beta\,\hat{\mathbf{z}} \quad (4e) \quad \text{Se}$$

$$\mathbf{B}_{27} = -x_{8}\,\mathbf{a}_{1} - y_{8}\,\mathbf{a}_{2} - z_{8}\,\mathbf{a}_{3} = (-x_{8}a - z_{8}c\cos\beta)\,\hat{\mathbf{x}} - y_{8}b\,\hat{\mathbf{y}} - z_{8}c\sin\beta\,\hat{\mathbf{z}} \quad (4e) \quad \text{Se}$$

$$\mathbf{B}_{28} = x_8\,\mathbf{a}_1 + \left(\frac{1}{2} - y_8\right)\mathbf{a}_2 + \left(\frac{1}{2} + z_8\right)\mathbf{a}_3 = \left(\frac{1}{2}c\cos\beta + x_8 a + z_8 c\cos\beta\right)\hat{\mathbf{x}} + \left(\frac{1}{2} - y_8\right)b\,\hat{\mathbf{y}} + \left(\frac{1}{2} + z_8\right)c\sin\beta\,\hat{\mathbf{z}} \qquad (4e) \qquad \text{Se}$$

---

**Found in:**

---

**Geometry files:**

- CIF: pp. 1540
- POSCAR: pp. 1540

# $Sb_4O_5Cl_2$ Structure: A2B5C4_mP22_14_e_c2e_2e

http://aflow.org/prototype-encyclopedia/A2B5C4_mP22_14_e_c2e_2e

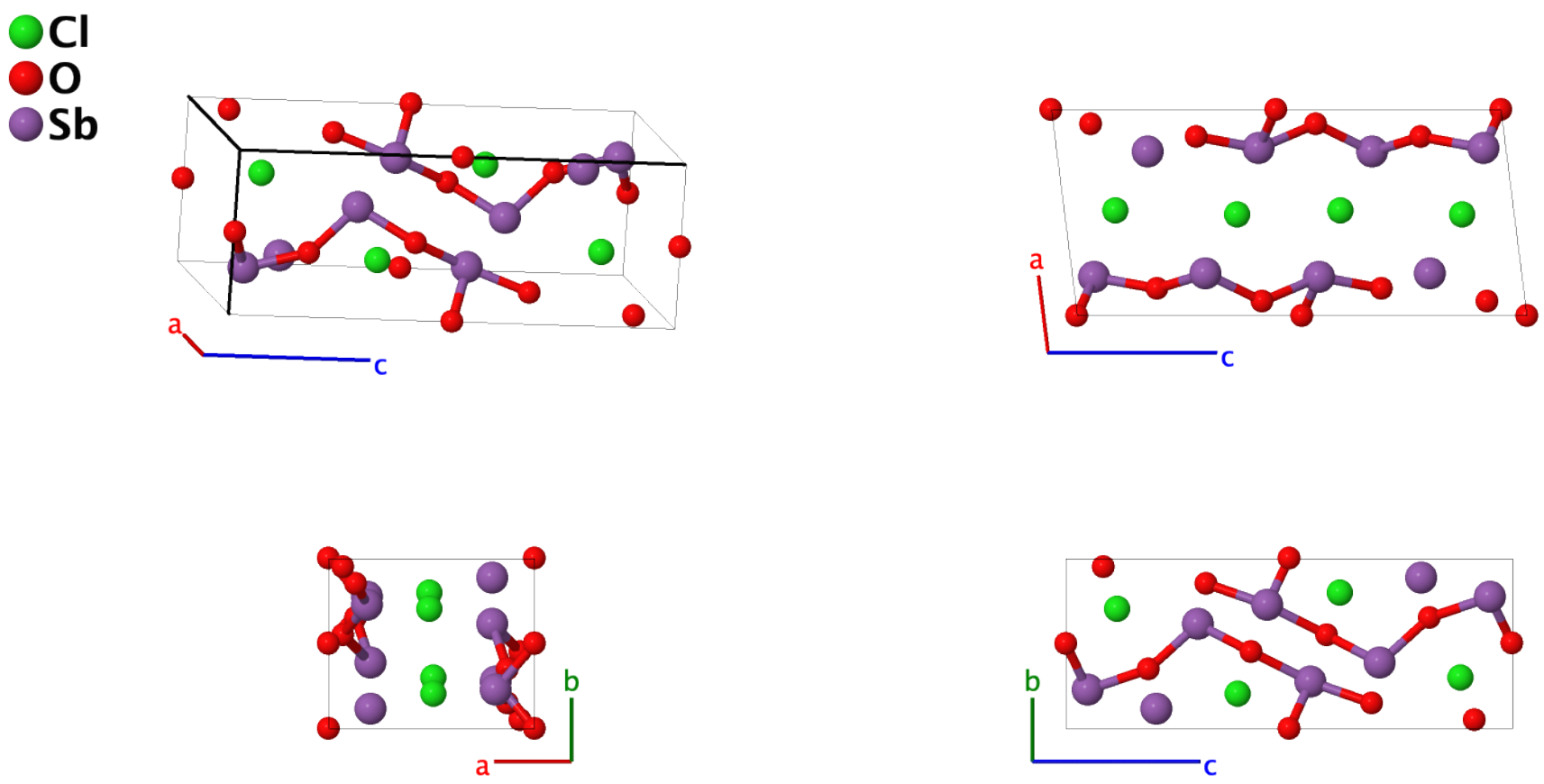

| | | |
|---|---|---|
| **Prototype** | : | $Cl_2O_5Sb_4$ |
| **AFLOW prototype label** | : | A2B5C4_mP22_14_e_c2e_2e |
| ***Strukturbericht* designation** | : | None |
| **Pearson symbol** | : | mP22 |
| **Space group number** | : | 14 |
| **Space group symbol** | : | $P2_1/c$ |
| **AFLOW prototype command** | : | `aflow --proto=A2B5C4_mP22_14_e_c2e_2e` `--params=`$a, b/a, c/a, \beta, x_2, y_2, z_2, x_3, y_3, z_3, x_4, y_4, z_4, x_5, y_5, z_5, x_6, y_6, z_6$ |

### Other compounds with this structure

- $Sb_4O_5Br_2$

### Simple Monoclinic primitive vectors:

$$\mathbf{a}_1 = a\,\hat{\mathbf{x}}$$
$$\mathbf{a}_2 = b\,\hat{\mathbf{y}}$$
$$\mathbf{a}_3 = c\cos\beta\,\hat{\mathbf{x}} + c\sin\beta\,\hat{\mathbf{z}}$$

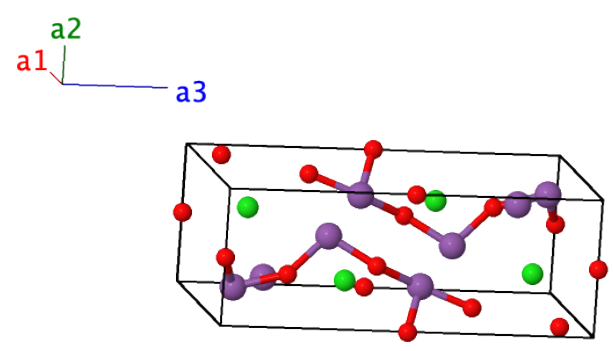

### Basis vectors:

| | | Lattice Coordinates | | Cartesian Coordinates | Wyckoff Position | Atom Type |
|---|---|---|---|---|---|---|
| $\mathbf{B}_1$ | = | $\frac{1}{2}\,\mathbf{a}_3$ | = | $\frac{1}{2}c\cos\beta\,\hat{\mathbf{x}} + \frac{1}{2}c\sin\beta\,\hat{\mathbf{z}}$ | (2c) | O I |
| $\mathbf{B}_2$ | = | $\frac{1}{2}\,\mathbf{a}_2$ | = | $\frac{1}{2}b\,\hat{\mathbf{y}}$ | (2c) | O I |
| $\mathbf{B}_3$ | = | $x_2\,\mathbf{a}_1 + y_2\,\mathbf{a}_2 + z_2\,\mathbf{a}_3$ | = | $(x_2a + z_2c\cos\beta)\,\hat{\mathbf{x}} + y_2b\,\hat{\mathbf{y}} + z_2c\sin\beta\,\hat{\mathbf{z}}$ | (4e) | Cl |
| $\mathbf{B}_4$ | = | $-x_2\,\mathbf{a}_1 + \left(\frac{1}{2} + y_2\right)\mathbf{a}_2 + \left(\frac{1}{2} - z_2\right)\mathbf{a}_3$ | = | $\left(\frac{1}{2}c\cos\beta - x_2a - z_2c\cos\beta\right)\hat{\mathbf{x}} + \left(\frac{1}{2} + y_2\right)b\,\hat{\mathbf{y}} + \left(\frac{1}{2} - z_2\right)c\sin\beta\,\hat{\mathbf{z}}$ | (4e) | Cl |
| $\mathbf{B}_5$ | = | $-x_2\,\mathbf{a}_1 - y_2\,\mathbf{a}_2 - z_2\,\mathbf{a}_3$ | = | $(-x_2a - z_2c\cos\beta)\,\hat{\mathbf{x}} - y_2b\,\hat{\mathbf{y}} - z_2c\sin\beta\,\hat{\mathbf{z}}$ | (4e) | Cl |
| $\mathbf{B}_6$ | = | $x_2\,\mathbf{a}_1 + \left(\frac{1}{2} - y_2\right)\,\mathbf{a}_2 + \left(\frac{1}{2} + z_2\right)\,\mathbf{a}_3$ | = | $\left(\frac{1}{2}c\cos\beta + x_2 a + z_2 c\cos\beta\right)\,\hat{\mathbf{x}} + \left(\frac{1}{2} - y_2\right) b\,\hat{\mathbf{y}} + \left(\frac{1}{2} + z_2\right) c\sin\beta\,\hat{\mathbf{z}}$ | (4e) | Cl |
| $\mathbf{B}_7$ | = | $x_3\,\mathbf{a}_1 + y_3\,\mathbf{a}_2 + z_3\,\mathbf{a}_3$ | = | $(x_3 a + z_3 c\cos\beta)\,\hat{\mathbf{x}} + y_3 b\,\hat{\mathbf{y}} + z_3 c\sin\beta\,\hat{\mathbf{z}}$ | (4e) | O II |
| $\mathbf{B}_8$ | = | $-x_3\,\mathbf{a}_1 + \left(\frac{1}{2} + y_3\right)\,\mathbf{a}_2 + \left(\frac{1}{2} - z_3\right)\,\mathbf{a}_3$ | = | $\left(\frac{1}{2}c\cos\beta - x_3 a - z_3 c\cos\beta\right)\,\hat{\mathbf{x}} + \left(\frac{1}{2} + y_3\right) b\,\hat{\mathbf{y}} + \left(\frac{1}{2} - z_3\right) c\sin\beta\,\hat{\mathbf{z}}$ | (4e) | O II |
| $\mathbf{B}_9$ | = | $-x_3\,\mathbf{a}_1 - y_3\,\mathbf{a}_2 - z_3\,\mathbf{a}_3$ | = | $(-x_3 a - z_3 c\cos\beta)\,\hat{\mathbf{x}} - y_3 b\,\hat{\mathbf{y}} - z_3 c\sin\beta\,\hat{\mathbf{z}}$ | (4e) | O II |
| $\mathbf{B}_{10}$ | = | $x_3\,\mathbf{a}_1 + \left(\frac{1}{2} - y_3\right)\,\mathbf{a}_2 + \left(\frac{1}{2} + z_3\right)\,\mathbf{a}_3$ | = | $\left(\frac{1}{2}c\cos\beta + x_3 a + z_3 c\cos\beta\right)\,\hat{\mathbf{x}} + \left(\frac{1}{2} - y_3\right) b\,\hat{\mathbf{y}} + \left(\frac{1}{2} + z_3\right) c\sin\beta\,\hat{\mathbf{z}}$ | (4e) | O II |
| $\mathbf{B}_{11}$ | = | $x_4\,\mathbf{a}_1 + y_4\,\mathbf{a}_2 + z_4\,\mathbf{a}_3$ | = | $(x_4 a + z_4 c\cos\beta)\,\hat{\mathbf{x}} + y_4 b\,\hat{\mathbf{y}} + z_4 c\sin\beta\,\hat{\mathbf{z}}$ | (4e) | O III |
| $\mathbf{B}_{12}$ | = | $-x_4\,\mathbf{a}_1 + \left(\frac{1}{2} + y_4\right)\,\mathbf{a}_2 + \left(\frac{1}{2} - z_4\right)\,\mathbf{a}_3$ | = | $\left(\frac{1}{2}c\cos\beta - x_4 a - z_4 c\cos\beta\right)\,\hat{\mathbf{x}} + \left(\frac{1}{2} + y_4\right) b\,\hat{\mathbf{y}} + \left(\frac{1}{2} - z_4\right) c\sin\beta\,\hat{\mathbf{z}}$ | (4e) | O III |
| $\mathbf{B}_{13}$ | = | $-x_4\,\mathbf{a}_1 - y_4\,\mathbf{a}_2 - z_4\,\mathbf{a}_3$ | = | $(-x_4 a - z_4 c\cos\beta)\,\hat{\mathbf{x}} - y_4 b\,\hat{\mathbf{y}} - z_4 c\sin\beta\,\hat{\mathbf{z}}$ | (4e) | O III |
| $\mathbf{B}_{14}$ | = | $x_4\,\mathbf{a}_1 + \left(\frac{1}{2} - y_4\right)\,\mathbf{a}_2 + \left(\frac{1}{2} + z_4\right)\,\mathbf{a}_3$ | = | $\left(\frac{1}{2}c\cos\beta + x_4 a + z_4 c\cos\beta\right)\,\hat{\mathbf{x}} + \left(\frac{1}{2} - y_4\right) b\,\hat{\mathbf{y}} + \left(\frac{1}{2} + z_4\right) c\sin\beta\,\hat{\mathbf{z}}$ | (4e) | O III |
| $\mathbf{B}_{15}$ | = | $x_5\,\mathbf{a}_1 + y_5\,\mathbf{a}_2 + z_5\,\mathbf{a}_3$ | = | $(x_5 a + z_5 c\cos\beta)\,\hat{\mathbf{x}} + y_5 b\,\hat{\mathbf{y}} + z_5 c\sin\beta\,\hat{\mathbf{z}}$ | (4e) | Sb I |
| $\mathbf{B}_{16}$ | = | $-x_5\,\mathbf{a}_1 + \left(\frac{1}{2} + y_5\right)\,\mathbf{a}_2 + \left(\frac{1}{2} - z_5\right)\,\mathbf{a}_3$ | = | $\left(\frac{1}{2}c\cos\beta - x_5 a - z_5 c\cos\beta\right)\,\hat{\mathbf{x}} + \left(\frac{1}{2} + y_5\right) b\,\hat{\mathbf{y}} + \left(\frac{1}{2} - z_5\right) c\sin\beta\,\hat{\mathbf{z}}$ | (4e) | Sb I |
| $\mathbf{B}_{17}$ | = | $-x_5\,\mathbf{a}_1 - y_5\,\mathbf{a}_2 - z_5\,\mathbf{a}_3$ | = | $(-x_5 a - z_5 c\cos\beta)\,\hat{\mathbf{x}} - y_5 b\,\hat{\mathbf{y}} - z_5 c\sin\beta\,\hat{\mathbf{z}}$ | (4e) | Sb I |
| $\mathbf{B}_{18}$ | = | $x_5\,\mathbf{a}_1 + \left(\frac{1}{2} - y_5\right)\,\mathbf{a}_2 + \left(\frac{1}{2} + z_5\right)\,\mathbf{a}_3$ | = | $\left(\frac{1}{2}c\cos\beta + x_5 a + z_5 c\cos\beta\right)\,\hat{\mathbf{x}} + \left(\frac{1}{2} - y_5\right) b\,\hat{\mathbf{y}} + \left(\frac{1}{2} + z_5\right) c\sin\beta\,\hat{\mathbf{z}}$ | (4e) | Sb I |
| $\mathbf{B}_{19}$ | = | $x_6\,\mathbf{a}_1 + y_6\,\mathbf{a}_2 + z_6\,\mathbf{a}_3$ | = | $(x_6 a + z_6 c\cos\beta)\,\hat{\mathbf{x}} + y_6 b\,\hat{\mathbf{y}} + z_6 c\sin\beta\,\hat{\mathbf{z}}$ | (4e) | Sb II |
| $\mathbf{B}_{20}$ | = | $-x_6\,\mathbf{a}_1 + \left(\frac{1}{2} + y_6\right)\,\mathbf{a}_2 + \left(\frac{1}{2} - z_6\right)\,\mathbf{a}_3$ | = | $\left(\frac{1}{2}c\cos\beta - x_6 a - z_6 c\cos\beta\right)\,\hat{\mathbf{x}} + \left(\frac{1}{2} + y_6\right) b\,\hat{\mathbf{y}} + \left(\frac{1}{2} - z_6\right) c\sin\beta\,\hat{\mathbf{z}}$ | (4e) | Sb II |
| $\mathbf{B}_{21}$ | = | $-x_6\,\mathbf{a}_1 - y_6\,\mathbf{a}_2 - z_6\,\mathbf{a}_3$ | = | $(-x_6 a - z_6 c\cos\beta)\,\hat{\mathbf{x}} - y_6 b\,\hat{\mathbf{y}} - z_6 c\sin\beta\,\hat{\mathbf{z}}$ | (4e) | Sb II |
| $\mathbf{B}_{22}$ | = | $x_6\,\mathbf{a}_1 + \left(\frac{1}{2} - y_6\right)\,\mathbf{a}_2 + \left(\frac{1}{2} + z_6\right)\,\mathbf{a}_3$ | = | $\left(\frac{1}{2}c\cos\beta + x_6 a + z_6 c\cos\beta\right)\,\hat{\mathbf{x}} + \left(\frac{1}{2} - y_6\right) b\,\hat{\mathbf{y}} + \left(\frac{1}{2} + z_6\right) c\sin\beta\,\hat{\mathbf{z}}$ | (4e) | Sb II |

**Geometry files:**

- CIF: pp. 1540
- POSCAR: pp. 1541

# $Ca_2UO_5$ Structure: A2B5C_mP32_14_2e_5e_ab

http://aflow.org/prototype-encyclopedia/A2B5C_mP32_14_2e_5e_ab

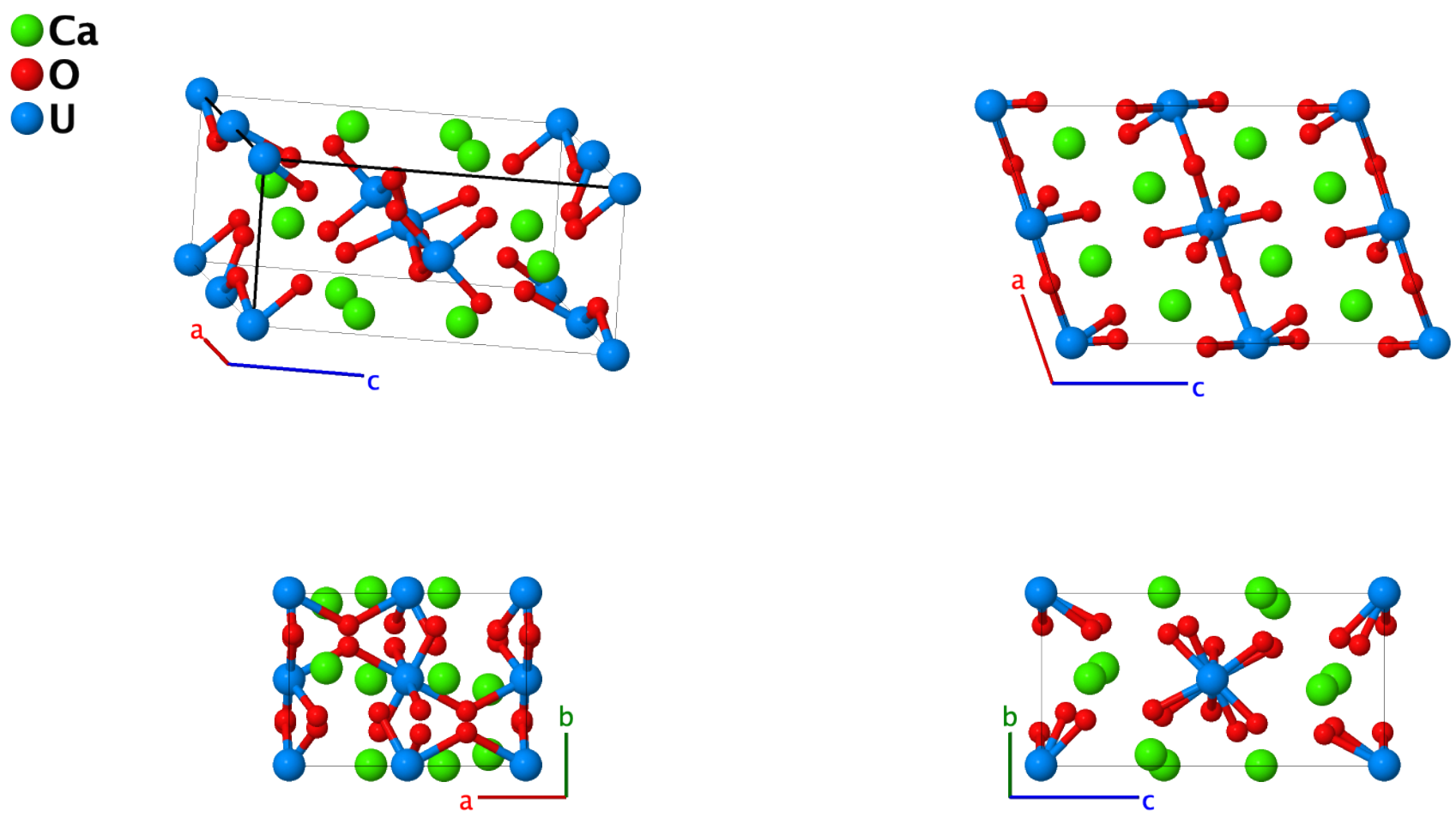

| | | |
|---|---|---|
| **Prototype** | : | $Ca_2O_5U$ |
| **AFLOW prototype label** | : | A2B5C_mP32_14_2e_5e_ab |
| ***Strukturbericht* designation** | : | None |
| **Pearson symbol** | : | mP32 |
| **Space group number** | : | 14 |
| **Space group symbol** | : | $P2_1/c$ |
| **AFLOW prototype command** | : | `aflow --proto=A2B5C_mP32_14_2e_5e_ab --params=`$a, b/a, c/a, \beta, x_3, y_3, z_3, x_4, y_4, z_4, x_5, y_5, z_5, x_6, y_6, z_6, x_7, y_7, z_7, x_8, y_8, z_8, x_9, y_9, z_9$ |

### Other compounds with this structure

- $Sr_2UO_5$

### Simple Monoclinic primitive vectors:

$$\mathbf{a}_1 = a\,\hat{\mathbf{x}}$$
$$\mathbf{a}_2 = b\,\hat{\mathbf{y}}$$
$$\mathbf{a}_3 = c\cos\beta\,\hat{\mathbf{x}} + c\sin\beta\,\hat{\mathbf{z}}$$

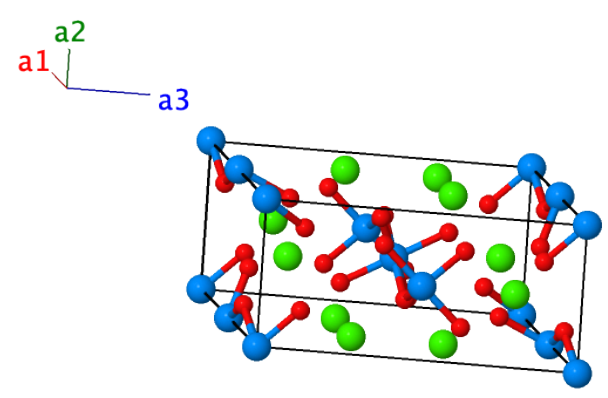

### Basis vectors:

| | | Lattice Coordinates | | Cartesian Coordinates | Wyckoff Position | Atom Type |
|---|---|---|---|---|---|---|
| $\mathbf{B}_1$ | = | $0\,\mathbf{a}_1 + 0\,\mathbf{a}_2 + 0\,\mathbf{a}_3$ | = | $0\,\hat{\mathbf{x}} + 0\,\hat{\mathbf{y}} + 0\,\hat{\mathbf{z}}$ | $(2a)$ | U I |
| $\mathbf{B}_2$ | = | $\frac{1}{2}\,\mathbf{a}_2 + \frac{1}{2}\,\mathbf{a}_3$ | = | $\frac{1}{2}c\cos\beta\,\hat{\mathbf{x}} + \frac{1}{2}b\,\hat{\mathbf{y}} + \frac{1}{2}c\sin\beta\,\hat{\mathbf{z}}$ | $(2a)$ | U I |
| $\mathbf{B}_3$ | = | $\frac{1}{2}\,\mathbf{a}_1$ | = | $\frac{1}{2}a\,\hat{\mathbf{x}}$ | $(2b)$ | U II |
| $\mathbf{B}_4$ | = | $\frac{1}{2}\,\mathbf{a}_1 + \frac{1}{2}\,\mathbf{a}_2 + \frac{1}{2}\,\mathbf{a}_3$ | = | $\frac{1}{2}(a + c\cos\beta)\,\hat{\mathbf{x}} + \frac{1}{2}b\,\hat{\mathbf{y}} + \frac{1}{2}c\sin\beta\,\hat{\mathbf{z}}$ | $(2b)$ | U II |
| $\mathbf{B}_{5}$ | = | $x_{3} \, \mathbf{a}_{1} + y_{3} \, \mathbf{a}_{2} + z_{3} \, \mathbf{a}_{3}$ | = | $\left(x_{3} a + z_{3} c \cos\beta\right) \, \mathbf{\hat{x}} + y_{3} b \, \mathbf{\hat{y}} + z_{3} c \sin\beta \, \mathbf{\hat{z}}$ | (4e) | Ca I |
| $\mathbf{B}_{6}$ | = | $-x_{3} \, \mathbf{a}_{1} + \left(\frac{1}{2} + y_{3}\right) \, \mathbf{a}_{2} + \left(\frac{1}{2} - z_{3}\right) \, \mathbf{a}_{3}$ | = | $\left(\frac{1}{2} c \cos\beta - x_{3} a - z_{3} c \cos\beta\right) \, \mathbf{\hat{x}} + \left(\frac{1}{2} + y_{3}\right) b \, \mathbf{\hat{y}} + \left(\frac{1}{2} - z_{3}\right) c \sin\beta \, \mathbf{\hat{z}}$ | (4e) | Ca I |
| $\mathbf{B}_{7}$ | = | $-x_{3} \, \mathbf{a}_{1} - y_{3} \, \mathbf{a}_{2} - z_{3} \, \mathbf{a}_{3}$ | = | $\left(-x_{3} a - z_{3} c \cos\beta\right) \, \mathbf{\hat{x}} - y_{3} b \, \mathbf{\hat{y}} - z_{3} c \sin\beta \, \mathbf{\hat{z}}$ | (4e) | Ca I |
| $\mathbf{B}_{8}$ | = | $x_{3} \, \mathbf{a}_{1} + \left(\frac{1}{2} - y_{3}\right) \, \mathbf{a}_{2} + \left(\frac{1}{2} + z_{3}\right) \, \mathbf{a}_{3}$ | = | $\left(\frac{1}{2} c \cos\beta + x_{3} a + z_{3} c \cos\beta\right) \, \mathbf{\hat{x}} + \left(\frac{1}{2} - y_{3}\right) b \, \mathbf{\hat{y}} + \left(\frac{1}{2} + z_{3}\right) c \sin\beta \, \mathbf{\hat{z}}$ | (4e) | Ca I |
| $\mathbf{B}_{9}$ | = | $x_{4} \, \mathbf{a}_{1} + y_{4} \, \mathbf{a}_{2} + z_{4} \, \mathbf{a}_{3}$ | = | $\left(x_{4} a + z_{4} c \cos\beta\right) \, \mathbf{\hat{x}} + y_{4} b \, \mathbf{\hat{y}} + z_{4} c \sin\beta \, \mathbf{\hat{z}}$ | (4e) | Ca II |
| $\mathbf{B}_{10}$ | = | $-x_{4} \, \mathbf{a}_{1} + \left(\frac{1}{2} + y_{4}\right) \, \mathbf{a}_{2} + \left(\frac{1}{2} - z_{4}\right) \, \mathbf{a}_{3}$ | = | $\left(\frac{1}{2} c \cos\beta - x_{4} a - z_{4} c \cos\beta\right) \, \mathbf{\hat{x}} + \left(\frac{1}{2} + y_{4}\right) b \, \mathbf{\hat{y}} + \left(\frac{1}{2} - z_{4}\right) c \sin\beta \, \mathbf{\hat{z}}$ | (4e) | Ca II |
| $\mathbf{B}_{11}$ | = | $-x_{4} \, \mathbf{a}_{1} - y_{4} \, \mathbf{a}_{2} - z_{4} \, \mathbf{a}_{3}$ | = | $\left(-x_{4} a - z_{4} c \cos\beta\right) \, \mathbf{\hat{x}} - y_{4} b \, \mathbf{\hat{y}} - z_{4} c \sin\beta \, \mathbf{\hat{z}}$ | (4e) | Ca II |
| $\mathbf{B}_{12}$ | = | $x_{4} \, \mathbf{a}_{1} + \left(\frac{1}{2} - y_{4}\right) \, \mathbf{a}_{2} + \left(\frac{1}{2} + z_{4}\right) \, \mathbf{a}_{3}$ | = | $\left(\frac{1}{2} c \cos\beta + x_{4} a + z_{4} c \cos\beta\right) \, \mathbf{\hat{x}} + \left(\frac{1}{2} - y_{4}\right) b \, \mathbf{\hat{y}} + \left(\frac{1}{2} + z_{4}\right) c \sin\beta \, \mathbf{\hat{z}}$ | (4e) | Ca II |
| $\mathbf{B}_{13}$ | = | $x_{5} \, \mathbf{a}_{1} + y_{5} \, \mathbf{a}_{2} + z_{5} \, \mathbf{a}_{3}$ | = | $\left(x_{5} a + z_{5} c \cos\beta\right) \, \mathbf{\hat{x}} + y_{5} b \, \mathbf{\hat{y}} + z_{5} c \sin\beta \, \mathbf{\hat{z}}$ | (4e) | O I |
| $\mathbf{B}_{14}$ | = | $-x_{5} \, \mathbf{a}_{1} + \left(\frac{1}{2} + y_{5}\right) \, \mathbf{a}_{2} + \left(\frac{1}{2} - z_{5}\right) \, \mathbf{a}_{3}$ | = | $\left(\frac{1}{2} c \cos\beta - x_{5} a - z_{5} c \cos\beta\right) \, \mathbf{\hat{x}} + \left(\frac{1}{2} + y_{5}\right) b \, \mathbf{\hat{y}} + \left(\frac{1}{2} - z_{5}\right) c \sin\beta \, \mathbf{\hat{z}}$ | (4e) | O I |
| $\mathbf{B}_{15}$ | = | $-x_{5} \, \mathbf{a}_{1} - y_{5} \, \mathbf{a}_{2} - z_{5} \, \mathbf{a}_{3}$ | = | $\left(-x_{5} a - z_{5} c \cos\beta\right) \, \mathbf{\hat{x}} - y_{5} b \, \mathbf{\hat{y}} - z_{5} c \sin\beta \, \mathbf{\hat{z}}$ | (4e) | O I |
| $\mathbf{B}_{16}$ | = | $x_{5} \, \mathbf{a}_{1} + \left(\frac{1}{2} - y_{5}\right) \, \mathbf{a}_{2} + \left(\frac{1}{2} + z_{5}\right) \, \mathbf{a}_{3}$ | = | $\left(\frac{1}{2} c \cos\beta + x_{5} a + z_{5} c \cos\beta\right) \, \mathbf{\hat{x}} + \left(\frac{1}{2} - y_{5}\right) b \, \mathbf{\hat{y}} + \left(\frac{1}{2} + z_{5}\right) c \sin\beta \, \mathbf{\hat{z}}$ | (4e) | O I |
| $\mathbf{B}_{17}$ | = | $x_{6} \, \mathbf{a}_{1} + y_{6} \, \mathbf{a}_{2} + z_{6} \, \mathbf{a}_{3}$ | = | $\left(x_{6} a + z_{6} c \cos\beta\right) \, \mathbf{\hat{x}} + y_{6} b \, \mathbf{\hat{y}} + z_{6} c \sin\beta \, \mathbf{\hat{z}}$ | (4e) | O II |
| $\mathbf{B}_{18}$ | = | $-x_{6} \, \mathbf{a}_{1} + \left(\frac{1}{2} + y_{6}\right) \, \mathbf{a}_{2} + \left(\frac{1}{2} - z_{6}\right) \, \mathbf{a}_{3}$ | = | $\left(\frac{1}{2} c \cos\beta - x_{6} a - z_{6} c \cos\beta\right) \, \mathbf{\hat{x}} + \left(\frac{1}{2} + y_{6}\right) b \, \mathbf{\hat{y}} + \left(\frac{1}{2} - z_{6}\right) c \sin\beta \, \mathbf{\hat{z}}$ | (4e) | O II |
| $\mathbf{B}_{19}$ | = | $-x_{6} \, \mathbf{a}_{1} - y_{6} \, \mathbf{a}_{2} - z_{6} \, \mathbf{a}_{3}$ | = | $\left(-x_{6} a - z_{6} c \cos\beta\right) \, \mathbf{\hat{x}} - y_{6} b \, \mathbf{\hat{y}} - z_{6} c \sin\beta \, \mathbf{\hat{z}}$ | (4e) | O II |
| $\mathbf{B}_{20}$ | = | $x_{6} \, \mathbf{a}_{1} + \left(\frac{1}{2} - y_{6}\right) \, \mathbf{a}_{2} + \left(\frac{1}{2} + z_{6}\right) \, \mathbf{a}_{3}$ | = | $\left(\frac{1}{2} c \cos\beta + x_{6} a + z_{6} c \cos\beta\right) \, \mathbf{\hat{x}} + \left(\frac{1}{2} - y_{6}\right) b \, \mathbf{\hat{y}} + \left(\frac{1}{2} + z_{6}\right) c \sin\beta \, \mathbf{\hat{z}}$ | (4e) | O II |
| $\mathbf{B}_{21}$ | = | $x_{7} \, \mathbf{a}_{1} + y_{7} \, \mathbf{a}_{2} + z_{7} \, \mathbf{a}_{3}$ | = | $\left(x_{7} a + z_{7} c \cos\beta\right) \, \mathbf{\hat{x}} + y_{7} b \, \mathbf{\hat{y}} + z_{7} c \sin\beta \, \mathbf{\hat{z}}$ | (4e) | O III |
| $\mathbf{B}_{22}$ | = | $-x_{7} \, \mathbf{a}_{1} + \left(\frac{1}{2} + y_{7}\right) \, \mathbf{a}_{2} + \left(\frac{1}{2} - z_{7}\right) \, \mathbf{a}_{3}$ | = | $\left(\frac{1}{2} c \cos\beta - x_{7} a - z_{7} c \cos\beta\right) \, \mathbf{\hat{x}} + \left(\frac{1}{2} + y_{7}\right) b \, \mathbf{\hat{y}} + \left(\frac{1}{2} - z_{7}\right) c \sin\beta \, \mathbf{\hat{z}}$ | (4e) | O III |
| $\mathbf{B}_{23}$ | = | $-x_{7} \, \mathbf{a}_{1} - y_{7} \, \mathbf{a}_{2} - z_{7} \, \mathbf{a}_{3}$ | = | $\left(-x_{7} a - z_{7} c \cos\beta\right) \, \mathbf{\hat{x}} - y_{7} b \, \mathbf{\hat{y}} - z_{7} c \sin\beta \, \mathbf{\hat{z}}$ | (4e) | O III |
| $\mathbf{B}_{24}$ | = | $x_{7} \, \mathbf{a}_{1} + \left(\frac{1}{2} - y_{7}\right) \, \mathbf{a}_{2} + \left(\frac{1}{2} + z_{7}\right) \, \mathbf{a}_{3}$ | = | $\left(\frac{1}{2} c \cos\beta + x_{7} a + z_{7} c \cos\beta\right) \, \mathbf{\hat{x}} + \left(\frac{1}{2} - y_{7}\right) b \, \mathbf{\hat{y}} + \left(\frac{1}{2} + z_{7}\right) c \sin\beta \, \mathbf{\hat{z}}$ | (4e) | O III |
| $\mathbf{B}_{25}$ | = | $x_{8} \, \mathbf{a}_{1} + y_{8} \, \mathbf{a}_{2} + z_{8} \, \mathbf{a}_{3}$ | = | $\left(x_{8} a + z_{8} c \cos\beta\right) \, \mathbf{\hat{x}} + y_{8} b \, \mathbf{\hat{y}} + z_{8} c \sin\beta \, \mathbf{\hat{z}}$ | (4e) | O IV |
| $\mathbf{B}_{26}$ | = | $-x_{8} \, \mathbf{a}_{1} + \left(\frac{1}{2} + y_{8}\right) \, \mathbf{a}_{2} + \left(\frac{1}{2} - z_{8}\right) \, \mathbf{a}_{3}$ | = | $\left(\frac{1}{2} c \cos\beta - x_{8} a - z_{8} c \cos\beta\right) \, \mathbf{\hat{x}} + \left(\frac{1}{2} + y_{8}\right) b \, \mathbf{\hat{y}} + \left(\frac{1}{2} - z_{8}\right) c \sin\beta \, \mathbf{\hat{z}}$ | (4e) | O IV |
| $\mathbf{B}_{27}$ | = | $-x_8\,\mathbf{a}_1 - y_8\,\mathbf{a}_2 - z_8\,\mathbf{a}_3$ | = | $(-x_8 a - z_8 c\cos\beta)\,\hat{\mathbf{x}} - y_8 b\,\hat{\mathbf{y}} - z_8 c\sin\beta\,\hat{\mathbf{z}}$ | (4$e$) | O IV |
| $\mathbf{B}_{28}$ | = | $x_8\,\mathbf{a}_1 + \left(\frac{1}{2} - y_8\right)\mathbf{a}_2 + \left(\frac{1}{2} + z_8\right)\mathbf{a}_3$ | = | $\left(\frac{1}{2}c\cos\beta + x_8 a + z_8 c\cos\beta\right)\hat{\mathbf{x}} + \left(\frac{1}{2} - y_8\right)b\,\hat{\mathbf{y}} + \left(\frac{1}{2} + z_8\right)c\sin\beta\,\hat{\mathbf{z}}$ | (4$e$) | O IV |
| $\mathbf{B}_{29}$ | = | $x_9\,\mathbf{a}_1 + y_9\,\mathbf{a}_2 + z_9\,\mathbf{a}_3$ | = | $(x_9 a + z_9 c\cos\beta)\,\hat{\mathbf{x}} + y_9 b\,\hat{\mathbf{y}} + z_9 c\sin\beta\,\hat{\mathbf{z}}$ | (4$e$) | O V |
| $\mathbf{B}_{30}$ | = | $-x_9\,\mathbf{a}_1 + \left(\frac{1}{2} + y_9\right)\mathbf{a}_2 + \left(\frac{1}{2} - z_9\right)\mathbf{a}_3$ | = | $\left(\frac{1}{2}c\cos\beta - x_9 a - z_9 c\cos\beta\right)\hat{\mathbf{x}} + \left(\frac{1}{2} + y_9\right)b\,\hat{\mathbf{y}} + \left(\frac{1}{2} - z_9\right)c\sin\beta\,\hat{\mathbf{z}}$ | (4$e$) | O V |
| $\mathbf{B}_{31}$ | = | $-x_9\,\mathbf{a}_1 - y_9\,\mathbf{a}_2 - z_9\,\mathbf{a}_3$ | = | $(-x_9 a - z_9 c\cos\beta)\,\hat{\mathbf{x}} - y_9 b\,\hat{\mathbf{y}} - z_9 c\sin\beta\,\hat{\mathbf{z}}$ | (4$e$) | O V |
| $\mathbf{B}_{32}$ | = | $x_9\,\mathbf{a}_1 + \left(\frac{1}{2} - y_9\right)\mathbf{a}_2 + \left(\frac{1}{2} + z_9\right)\mathbf{a}_3$ | = | $\left(\frac{1}{2}c\cos\beta + x_9 a + z_9 c\cos\beta\right)\hat{\mathbf{x}} + \left(\frac{1}{2} - y_9\right)b\,\hat{\mathbf{y}} + \left(\frac{1}{2} + z_9\right)c\sin\beta\,\hat{\mathbf{z}}$ | (4$e$) | O V |

**Geometry files:**

- CIF: pp. 1541
- POSCAR: pp. 1541

# $Gd_2SiO_5$ ($RE_2SiO_5$ X1) Structure: A2B5C_mP32_14_2e_5e_e

http://aflow.org/prototype-encyclopedia/A2B5C_mP32_14_2e_5e_e

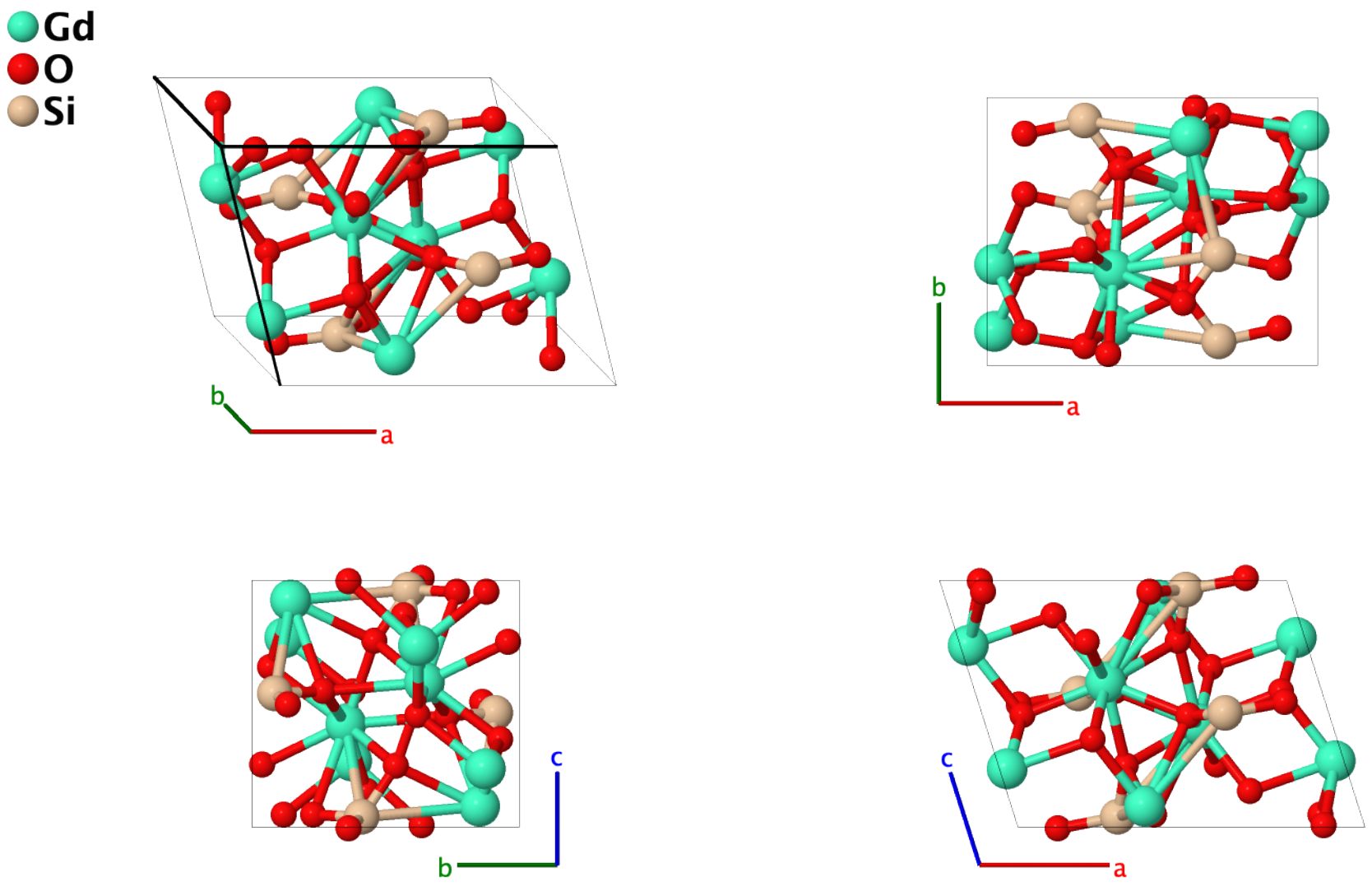

| | | |
|---|---|---|
| **Prototype** | : | $Gd_2O_5Si$ |
| **AFLOW prototype label** | : | A2B5C_mP32_14_2e_5e_e |
| ***Strukturbericht* designation** | : | None |
| **Pearson symbol** | : | mP32 |
| **Space group number** | : | 14 |
| **Space group symbol** | : | $P2_1/c$ |
| **AFLOW prototype command** | : | `aflow --proto=A2B5C_mP32_14_2e_5e_e --params=`$a, b/a, c/a, \beta, x_1, y_1, z_1, x_2, y_2, z_2, x_3, y_3, z_3, x_4, y_4, z_4, x_5, y_5, z_5, x_6, y_6, z_6, x_7, y_7, z_7, x_8, y_8, z_8$ |

### Other compounds with this structure

- $Dy_2SiO_5$, $La_2SiO_5$, $Tb_2SiO_5$, and $Y_2SiO_5$

---

- Compounds of the form $RESiO_5$ ($RE$ = Rare Earth and related elements) crystallize in one of two forms (Wang, 2001 and Tian, 2016): X1, shown here, for rare earths between lanthanum and gadolinium, and X2, space group $C2/c$ #15 for the later rare earths, with the dysprosium, yttrium and ytterbium compounds existing in both structures.

### Simple Monoclinic primitive vectors:

$$\mathbf{a}_1 = a\,\hat{\mathbf{x}}$$
$$\mathbf{a}_2 = b\,\hat{\mathbf{y}}$$
$$\mathbf{a}_3 = c\cos\beta\,\hat{\mathbf{x}} + c\sin\beta\,\hat{\mathbf{z}}$$

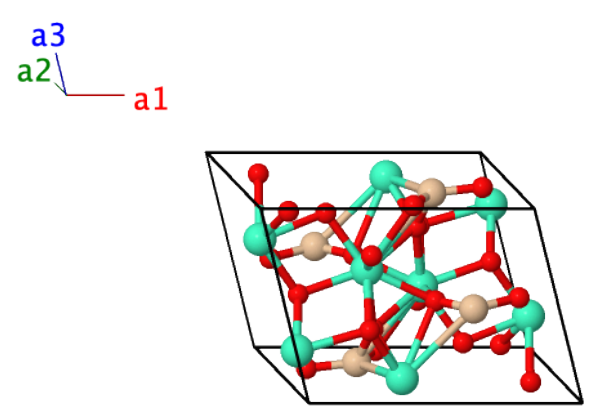

### Basis vectors:

| | | Lattice Coordinates | | Cartesian Coordinates | Wyckoff Position | Atom Type |
|---|---|---|---|---|---|---|
| $\mathbf{B}_{1}$ | = | $x_{1}\,\mathbf{a}_{1}+y_{1}\,\mathbf{a}_{2}+z_{1}\,\mathbf{a}_{3}$ | = | $\left(x_{1}a+z_{1}c\cos\beta\right)\,\mathbf{\hat{x}}+y_{1}b\,\mathbf{\hat{y}}+z_{1}c\sin\beta\,\mathbf{\hat{z}}$ | (4$e$) | Gd I |
| $\mathbf{B}_{2}$ | = | $-x_{1}\,\mathbf{a}_{1}+\left(\frac{1}{2}+y_{1}\right)\,\mathbf{a}_{2}+\left(\frac{1}{2}-z_{1}\right)\,\mathbf{a}_{3}$ | = | $\left(\frac{1}{2}c\cos\beta-x_{1}a-z_{1}c\cos\beta\right)\,\mathbf{\hat{x}}+\left(\frac{1}{2}+y_{1}\right)b\,\mathbf{\hat{y}}+\left(\frac{1}{2}-z_{1}\right)c\sin\beta\,\mathbf{\hat{z}}$ | (4$e$) | Gd I |
| $\mathbf{B}_{3}$ | = | $-x_{1}\,\mathbf{a}_{1}-y_{1}\,\mathbf{a}_{2}-z_{1}\,\mathbf{a}_{3}$ | = | $\left(-x_{1}a-z_{1}c\cos\beta\right)\,\mathbf{\hat{x}}-y_{1}b\,\mathbf{\hat{y}}-z_{1}c\sin\beta\,\mathbf{\hat{z}}$ | (4$e$) | Gd I |
| $\mathbf{B}_{4}$ | = | $x_{1}\,\mathbf{a}_{1}+\left(\frac{1}{2}-y_{1}\right)\,\mathbf{a}_{2}+\left(\frac{1}{2}+z_{1}\right)\,\mathbf{a}_{3}$ | = | $\left(\frac{1}{2}c\cos\beta+x_{1}a+z_{1}c\cos\beta\right)\,\mathbf{\hat{x}}+\left(\frac{1}{2}-y_{1}\right)b\,\mathbf{\hat{y}}+\left(\frac{1}{2}+z_{1}\right)c\sin\beta\,\mathbf{\hat{z}}$ | (4$e$) | Gd I |
| $\mathbf{B}_{5}$ | = | $x_{2}\,\mathbf{a}_{1}+y_{2}\,\mathbf{a}_{2}+z_{2}\,\mathbf{a}_{3}$ | = | $\left(x_{2}a+z_{2}c\cos\beta\right)\,\mathbf{\hat{x}}+y_{2}b\,\mathbf{\hat{y}}+z_{2}c\sin\beta\,\mathbf{\hat{z}}$ | (4$e$) | Gd II |
| $\mathbf{B}_{6}$ | = | $-x_{2}\,\mathbf{a}_{1}+\left(\frac{1}{2}+y_{2}\right)\,\mathbf{a}_{2}+\left(\frac{1}{2}-z_{2}\right)\,\mathbf{a}_{3}$ | = | $\left(\frac{1}{2}c\cos\beta-x_{2}a-z_{2}c\cos\beta\right)\,\mathbf{\hat{x}}+\left(\frac{1}{2}+y_{2}\right)b\,\mathbf{\hat{y}}+\left(\frac{1}{2}-z_{2}\right)c\sin\beta\,\mathbf{\hat{z}}$ | (4$e$) | Gd II |
| $\mathbf{B}_{7}$ | = | $-x_{2}\,\mathbf{a}_{1}-y_{2}\,\mathbf{a}_{2}-z_{2}\,\mathbf{a}_{3}$ | = | $\left(-x_{2}a-z_{2}c\cos\beta\right)\,\mathbf{\hat{x}}-y_{2}b\,\mathbf{\hat{y}}-z_{2}c\sin\beta\,\mathbf{\hat{z}}$ | (4$e$) | Gd II |
| $\mathbf{B}_{8}$ | = | $x_{2}\,\mathbf{a}_{1}+\left(\frac{1}{2}-y_{2}\right)\,\mathbf{a}_{2}+\left(\frac{1}{2}+z_{2}\right)\,\mathbf{a}_{3}$ | = | $\left(\frac{1}{2}c\cos\beta+x_{2}a+z_{2}c\cos\beta\right)\,\mathbf{\hat{x}}+\left(\frac{1}{2}-y_{2}\right)b\,\mathbf{\hat{y}}+\left(\frac{1}{2}+z_{2}\right)c\sin\beta\,\mathbf{\hat{z}}$ | (4$e$) | Gd II |
| $\mathbf{B}_{9}$ | = | $x_{3}\,\mathbf{a}_{1}+y_{3}\,\mathbf{a}_{2}+z_{3}\,\mathbf{a}_{3}$ | = | $\left(x_{3}a+z_{3}c\cos\beta\right)\,\mathbf{\hat{x}}+y_{3}b\,\mathbf{\hat{y}}+z_{3}c\sin\beta\,\mathbf{\hat{z}}$ | (4$e$) | O I |
| $\mathbf{B}_{10}$ | = | $-x_{3}\,\mathbf{a}_{1}+\left(\frac{1}{2}+y_{3}\right)\,\mathbf{a}_{2}+\left(\frac{1}{2}-z_{3}\right)\,\mathbf{a}_{3}$ | = | $\left(\frac{1}{2}c\cos\beta-x_{3}a-z_{3}c\cos\beta\right)\,\mathbf{\hat{x}}+\left(\frac{1}{2}+y_{3}\right)b\,\mathbf{\hat{y}}+\left(\frac{1}{2}-z_{3}\right)c\sin\beta\,\mathbf{\hat{z}}$ | (4$e$) | O I |
| $\mathbf{B}_{11}$ | = | $-x_{3}\,\mathbf{a}_{1}-y_{3}\,\mathbf{a}_{2}-z_{3}\,\mathbf{a}_{3}$ | = | $\left(-x_{3}a-z_{3}c\cos\beta\right)\,\mathbf{\hat{x}}-y_{3}b\,\mathbf{\hat{y}}-z_{3}c\sin\beta\,\mathbf{\hat{z}}$ | (4$e$) | O I |
| $\mathbf{B}_{12}$ | = | $x_{3}\,\mathbf{a}_{1}+\left(\frac{1}{2}-y_{3}\right)\,\mathbf{a}_{2}+\left(\frac{1}{2}+z_{3}\right)\,\mathbf{a}_{3}$ | = | $\left(\frac{1}{2}c\cos\beta+x_{3}a+z_{3}c\cos\beta\right)\,\mathbf{\hat{x}}+\left(\frac{1}{2}-y_{3}\right)b\,\mathbf{\hat{y}}+\left(\frac{1}{2}+z_{3}\right)c\sin\beta\,\mathbf{\hat{z}}$ | (4$e$) | O I |
| $\mathbf{B}_{13}$ | = | $x_{4}\,\mathbf{a}_{1}+y_{4}\,\mathbf{a}_{2}+z_{4}\,\mathbf{a}_{3}$ | = | $\left(x_{4}a+z_{4}c\cos\beta\right)\,\mathbf{\hat{x}}+y_{4}b\,\mathbf{\hat{y}}+z_{4}c\sin\beta\,\mathbf{\hat{z}}$ | (4$e$) | O II |
| $\mathbf{B}_{14}$ | = | $-x_{4}\,\mathbf{a}_{1}+\left(\frac{1}{2}+y_{4}\right)\,\mathbf{a}_{2}+\left(\frac{1}{2}-z_{4}\right)\,\mathbf{a}_{3}$ | = | $\left(\frac{1}{2}c\cos\beta-x_{4}a-z_{4}c\cos\beta\right)\,\mathbf{\hat{x}}+\left(\frac{1}{2}+y_{4}\right)b\,\mathbf{\hat{y}}+\left(\frac{1}{2}-z_{4}\right)c\sin\beta\,\mathbf{\hat{z}}$ | (4$e$) | O II |
| $\mathbf{B}_{15}$ | = | $-x_{4}\,\mathbf{a}_{1}-y_{4}\,\mathbf{a}_{2}-z_{4}\,\mathbf{a}_{3}$ | = | $\left(-x_{4}a-z_{4}c\cos\beta\right)\,\mathbf{\hat{x}}-y_{4}b\,\mathbf{\hat{y}}-z_{4}c\sin\beta\,\mathbf{\hat{z}}$ | (4$e$) | O II |
| $\mathbf{B}_{16}$ | = | $x_{4}\,\mathbf{a}_{1}+\left(\frac{1}{2}-y_{4}\right)\,\mathbf{a}_{2}+\left(\frac{1}{2}+z_{4}\right)\,\mathbf{a}_{3}$ | = | $\left(\frac{1}{2}c\cos\beta+x_{4}a+z_{4}c\cos\beta\right)\,\mathbf{\hat{x}}+\left(\frac{1}{2}-y_{4}\right)b\,\mathbf{\hat{y}}+\left(\frac{1}{2}+z_{4}\right)c\sin\beta\,\mathbf{\hat{z}}$ | (4$e$) | O II |
| $\mathbf{B}_{17}$ | = | $x_{5}\,\mathbf{a}_{1}+y_{5}\,\mathbf{a}_{2}+z_{5}\,\mathbf{a}_{3}$ | = | $\left(x_{5}a+z_{5}c\cos\beta\right)\,\mathbf{\hat{x}}+y_{5}b\,\mathbf{\hat{y}}+z_{5}c\sin\beta\,\mathbf{\hat{z}}$ | (4$e$) | O III |
| $\mathbf{B}_{18}$ | = | $-x_{5}\,\mathbf{a}_{1}+\left(\frac{1}{2}+y_{5}\right)\,\mathbf{a}_{2}+\left(\frac{1}{2}-z_{5}\right)\,\mathbf{a}_{3}$ | = | $\left(\frac{1}{2}c\cos\beta-x_{5}a-z_{5}c\cos\beta\right)\,\mathbf{\hat{x}}+\left(\frac{1}{2}+y_{5}\right)b\,\mathbf{\hat{y}}+\left(\frac{1}{2}-z_{5}\right)c\sin\beta\,\mathbf{\hat{z}}$ | (4$e$) | O III |
| $\mathbf{B}_{19}$ | = | $-x_{5}\,\mathbf{a}_{1}-y_{5}\,\mathbf{a}_{2}-z_{5}\,\mathbf{a}_{3}$ | = | $\left(-x_{5}a-z_{5}c\cos\beta\right)\,\mathbf{\hat{x}}-y_{5}b\,\mathbf{\hat{y}}-z_{5}c\sin\beta\,\mathbf{\hat{z}}$ | (4$e$) | O III |
| $\mathbf{B}_{20}$ | = | $x_{5}\,\mathbf{a}_{1}+\left(\frac{1}{2}-y_{5}\right)\,\mathbf{a}_{2}+\left(\frac{1}{2}+z_{5}\right)\,\mathbf{a}_{3}$ | = | $\left(\frac{1}{2}c\cos\beta+x_{5}a+z_{5}c\cos\beta\right)\,\mathbf{\hat{x}}+\left(\frac{1}{2}-y_{5}\right)b\,\mathbf{\hat{y}}+\left(\frac{1}{2}+z_{5}\right)c\sin\beta\,\mathbf{\hat{z}}$ | (4$e$) | O III |
| $\mathbf{B}_{21}$ | = | $x_{6}\,\mathbf{a}_{1}+y_{6}\,\mathbf{a}_{2}+z_{6}\,\mathbf{a}_{3}$ | = | $\left(x_{6}a+z_{6}c\cos\beta\right)\,\mathbf{\hat{x}}+y_{6}b\,\mathbf{\hat{y}}+z_{6}c\sin\beta\,\mathbf{\hat{z}}$ | (4$e$) | O IV |
| $\mathbf{B}_{22}$ | = | $-x_6\,\mathbf{a}_1 + \left(\frac{1}{2} + y_6\right)\mathbf{a}_2 + \left(\frac{1}{2} - z_6\right)\mathbf{a}_3$ | = | $\left(\frac{1}{2}c\cos\beta - x_6 a - z_6 c\cos\beta\right)\hat{\mathbf{x}} + \left(\frac{1}{2} + y_6\right)b\,\hat{\mathbf{y}} + \left(\frac{1}{2} - z_6\right)c\sin\beta\,\hat{\mathbf{z}}$ | (4e) | O IV |
| $\mathbf{B}_{23}$ | = | $-x_6\,\mathbf{a}_1 - y_6\,\mathbf{a}_2 - z_6\,\mathbf{a}_3$ | = | $(-x_6 a - z_6 c\cos\beta)\,\hat{\mathbf{x}} - y_6 b\,\hat{\mathbf{y}} - z_6 c\sin\beta\,\hat{\mathbf{z}}$ | (4e) | O IV |
| $\mathbf{B}_{24}$ | = | $x_6\,\mathbf{a}_1 + \left(\frac{1}{2} - y_6\right)\mathbf{a}_2 + \left(\frac{1}{2} + z_6\right)\mathbf{a}_3$ | = | $\left(\frac{1}{2}c\cos\beta + x_6 a + z_6 c\cos\beta\right)\hat{\mathbf{x}} + \left(\frac{1}{2} - y_6\right)b\,\hat{\mathbf{y}} + \left(\frac{1}{2} + z_6\right)c\sin\beta\,\hat{\mathbf{z}}$ | (4e) | O IV |
| $\mathbf{B}_{25}$ | = | $x_7\,\mathbf{a}_1 + y_7\,\mathbf{a}_2 + z_7\,\mathbf{a}_3$ | = | $(x_7 a + z_7 c\cos\beta)\,\hat{\mathbf{x}} + y_7 b\,\hat{\mathbf{y}} + z_7 c\sin\beta\,\hat{\mathbf{z}}$ | (4e) | O V |
| $\mathbf{B}_{26}$ | = | $-x_7\,\mathbf{a}_1 + \left(\frac{1}{2} + y_7\right)\mathbf{a}_2 + \left(\frac{1}{2} - z_7\right)\mathbf{a}_3$ | = | $\left(\frac{1}{2}c\cos\beta - x_7 a - z_7 c\cos\beta\right)\hat{\mathbf{x}} + \left(\frac{1}{2} + y_7\right)b\,\hat{\mathbf{y}} + \left(\frac{1}{2} - z_7\right)c\sin\beta\,\hat{\mathbf{z}}$ | (4e) | O V |
| $\mathbf{B}_{27}$ | = | $-x_7\,\mathbf{a}_1 - y_7\,\mathbf{a}_2 - z_7\,\mathbf{a}_3$ | = | $(-x_7 a - z_7 c\cos\beta)\,\hat{\mathbf{x}} - y_7 b\,\hat{\mathbf{y}} - z_7 c\sin\beta\,\hat{\mathbf{z}}$ | (4e) | O V |
| $\mathbf{B}_{28}$ | = | $x_7\,\mathbf{a}_1 + \left(\frac{1}{2} - y_7\right)\mathbf{a}_2 + \left(\frac{1}{2} + z_7\right)\mathbf{a}_3$ | = | $\left(\frac{1}{2}c\cos\beta + x_7 a + z_7 c\cos\beta\right)\hat{\mathbf{x}} + \left(\frac{1}{2} - y_7\right)b\,\hat{\mathbf{y}} + \left(\frac{1}{2} + z_7\right)c\sin\beta\,\hat{\mathbf{z}}$ | (4e) | O V |
| $\mathbf{B}_{29}$ | = | $x_8\,\mathbf{a}_1 + y_8\,\mathbf{a}_2 + z_8\,\mathbf{a}_3$ | = | $(x_8 a + z_8 c\cos\beta)\,\hat{\mathbf{x}} + y_8 b\,\hat{\mathbf{y}} + z_8 c\sin\beta\,\hat{\mathbf{z}}$ | (4e) | Si |
| $\mathbf{B}_{30}$ | = | $-x_8\,\mathbf{a}_1 + \left(\frac{1}{2} + y_8\right)\mathbf{a}_2 + \left(\frac{1}{2} - z_8\right)\mathbf{a}_3$ | = | $\left(\frac{1}{2}c\cos\beta - x_8 a - z_8 c\cos\beta\right)\hat{\mathbf{x}} + \left(\frac{1}{2} + y_8\right)b\,\hat{\mathbf{y}} + \left(\frac{1}{2} - z_8\right)c\sin\beta\,\hat{\mathbf{z}}$ | (4e) | Si |
| $\mathbf{B}_{31}$ | = | $-x_8\,\mathbf{a}_1 - y_8\,\mathbf{a}_2 - z_8\,\mathbf{a}_3$ | = | $(-x_8 a - z_8 c\cos\beta)\,\hat{\mathbf{x}} - y_8 b\,\hat{\mathbf{y}} - z_8 c\sin\beta\,\hat{\mathbf{z}}$ | (4e) | Si |
| $\mathbf{B}_{32}$ | = | $x_8\,\mathbf{a}_1 + \left(\frac{1}{2} - y_8\right)\mathbf{a}_2 + \left(\frac{1}{2} + z_8\right)\mathbf{a}_3$ | = | $\left(\frac{1}{2}c\cos\beta + x_8 a + z_8 c\cos\beta\right)\hat{\mathbf{x}} + \left(\frac{1}{2} - y_8\right)b\,\hat{\mathbf{y}} + \left(\frac{1}{2} + z_8\right)c\sin\beta\,\hat{\mathbf{z}}$ | (4e) | Si |

**Geometry files:**

- CIF: pp. 1541
- POSCAR: pp. 1542

# Sanguite ($KCuCl_3$) Structure: A3BC_mP20_14_3e_e_e

http://aflow.org/prototype-encyclopedia/A3BC_mP20_14_3e_e_e

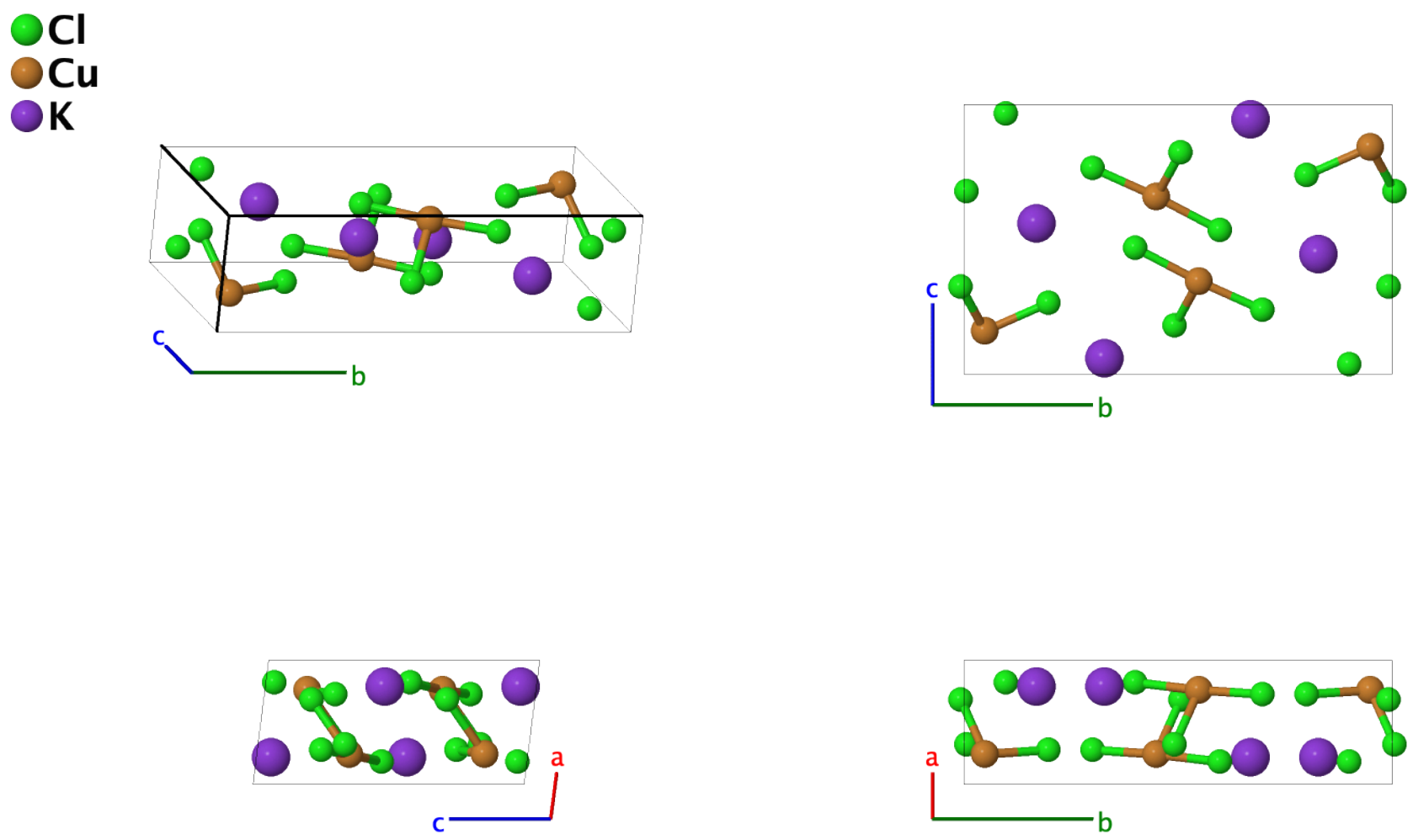

| | | |
|---|---|---|
| **Prototype** | : | $Cl_3CuK$ |
| **AFLOW prototype label** | : | A3BC_mP20_14_3e_e_e |
| ***Strukturbericht* designation** | : | None |
| **Pearson symbol** | : | mP20 |
| **Space group number** | : | 14 |
| **Space group symbol** | : | $P2_1/c$ |
| **AFLOW prototype command** | : | `aflow --proto=A3BC_mP20_14_3e_e_e`<br>`--params=`$a, b/a, c/a, \beta, x_1, y_1, z_1, x_2, y_2, z_2, x_3, y_3, z_3, x_4, y_4, z_4, x_5, y_5, z_5$ |

### Other compounds with this structure

- $NH_4CuCl_3$

### Simple Monoclinic primitive vectors:

$$\mathbf{a}_1 = a\,\hat{\mathbf{x}}$$
$$\mathbf{a}_2 = b\,\hat{\mathbf{y}}$$
$$\mathbf{a}_3 = c\cos\beta\,\hat{\mathbf{x}} + c\sin\beta\,\hat{\mathbf{z}}$$

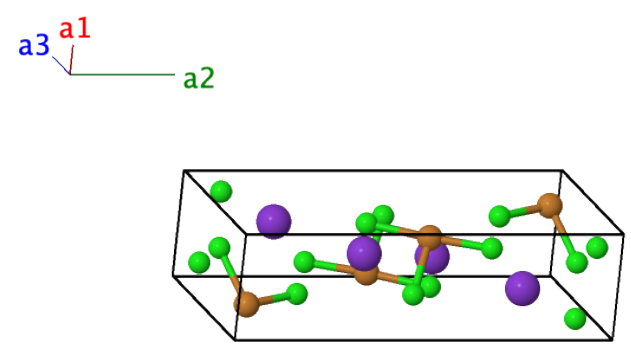

### Basis vectors:

| | | Lattice Coordinates | | Cartesian Coordinates | Wyckoff Position | Atom Type |
|---|---|---|---|---|---|---|
| $\mathbf{B}_1$ | = | $x_1\,\mathbf{a}_1 + y_1\,\mathbf{a}_2 + z_1\,\mathbf{a}_3$ | = | $(x_1 a + z_1 c\cos\beta)\,\hat{\mathbf{x}} + y_1 b\,\hat{\mathbf{y}} + z_1 c\sin\beta\,\hat{\mathbf{z}}$ | (4$e$) | Cl I |
| $\mathbf{B}_2$ | = | $-x_1\,\mathbf{a}_1 + \left(\frac{1}{2}+y_1\right)\mathbf{a}_2 + \left(\frac{1}{2}-z_1\right)\mathbf{a}_3$ | = | $\left(\frac{1}{2}c\cos\beta - x_1 a - z_1 c\cos\beta\right)\hat{\mathbf{x}} + \left(\frac{1}{2}+y_1\right)b\,\hat{\mathbf{y}} + \left(\frac{1}{2}-z_1\right)c\sin\beta\,\hat{\mathbf{z}}$ | (4$e$) | Cl I |
| $\mathbf{B}_3$ | = | $-x_1\,\mathbf{a}_1 - y_1\,\mathbf{a}_2 - z_1\,\mathbf{a}_3$ | = | $(-x_1 a - z_1 c\cos\beta)\,\hat{\mathbf{x}} - y_1 b\,\hat{\mathbf{y}} - z_1 c\sin\beta\,\hat{\mathbf{z}}$ | (4$e$) | Cl I |
| $\mathbf{B}_{4}$ | = | $x_1\,\mathbf{a}_1 + \left(\frac{1}{2} - y_1\right)\,\mathbf{a}_2 + \left(\frac{1}{2} + z_1\right)\,\mathbf{a}_3$ | = | $\left(\frac{1}{2}c\cos\beta + x_1 a + z_1 c\cos\beta\right)\,\hat{\mathbf{x}} + \left(\frac{1}{2} - y_1\right) b\,\hat{\mathbf{y}} + \left(\frac{1}{2} + z_1\right) c\sin\beta\,\hat{\mathbf{z}}$ | (4$e$) | Cl I |
| $\mathbf{B}_{5}$ | = | $x_2\,\mathbf{a}_1 + y_2\,\mathbf{a}_2 + z_2\,\mathbf{a}_3$ | = | $(x_2 a + z_2 c\cos\beta)\,\hat{\mathbf{x}} + y_2 b\,\hat{\mathbf{y}} + z_2 c\sin\beta\,\hat{\mathbf{z}}$ | (4$e$) | Cl II |
| $\mathbf{B}_{6}$ | = | $-x_2\,\mathbf{a}_1 + \left(\frac{1}{2} + y_2\right)\,\mathbf{a}_2 + \left(\frac{1}{2} - z_2\right)\,\mathbf{a}_3$ | = | $\left(\frac{1}{2}c\cos\beta - x_2 a - z_2 c\cos\beta\right)\,\hat{\mathbf{x}} + \left(\frac{1}{2} + y_2\right) b\,\hat{\mathbf{y}} + \left(\frac{1}{2} - z_2\right) c\sin\beta\,\hat{\mathbf{z}}$ | (4$e$) | Cl II |
| $\mathbf{B}_{7}$ | = | $-x_2\,\mathbf{a}_1 - y_2\,\mathbf{a}_2 - z_2\,\mathbf{a}_3$ | = | $(-x_2 a - z_2 c\cos\beta)\,\hat{\mathbf{x}} - y_2 b\,\hat{\mathbf{y}} - z_2 c\sin\beta\,\hat{\mathbf{z}}$ | (4$e$) | Cl II |
| $\mathbf{B}_{8}$ | = | $x_2\,\mathbf{a}_1 + \left(\frac{1}{2} - y_2\right)\,\mathbf{a}_2 + \left(\frac{1}{2} + z_2\right)\,\mathbf{a}_3$ | = | $\left(\frac{1}{2}c\cos\beta + x_2 a + z_2 c\cos\beta\right)\,\hat{\mathbf{x}} + \left(\frac{1}{2} - y_2\right) b\,\hat{\mathbf{y}} + \left(\frac{1}{2} + z_2\right) c\sin\beta\,\hat{\mathbf{z}}$ | (4$e$) | Cl II |
| $\mathbf{B}_{9}$ | = | $x_3\,\mathbf{a}_1 + y_3\,\mathbf{a}_2 + z_3\,\mathbf{a}_3$ | = | $(x_3 a + z_3 c\cos\beta)\,\hat{\mathbf{x}} + y_3 b\,\hat{\mathbf{y}} + z_3 c\sin\beta\,\hat{\mathbf{z}}$ | (4$e$) | Cl III |
| $\mathbf{B}_{10}$ | = | $-x_3\,\mathbf{a}_1 + \left(\frac{1}{2} + y_3\right)\,\mathbf{a}_2 + \left(\frac{1}{2} - z_3\right)\,\mathbf{a}_3$ | = | $\left(\frac{1}{2}c\cos\beta - x_3 a - z_3 c\cos\beta\right)\,\hat{\mathbf{x}} + \left(\frac{1}{2} + y_3\right) b\,\hat{\mathbf{y}} + \left(\frac{1}{2} - z_3\right) c\sin\beta\,\hat{\mathbf{z}}$ | (4$e$) | Cl III |
| $\mathbf{B}_{11}$ | = | $-x_3\,\mathbf{a}_1 - y_3\,\mathbf{a}_2 - z_3\,\mathbf{a}_3$ | = | $(-x_3 a - z_3 c\cos\beta)\,\hat{\mathbf{x}} - y_3 b\,\hat{\mathbf{y}} - z_3 c\sin\beta\,\hat{\mathbf{z}}$ | (4$e$) | Cl III |
| $\mathbf{B}_{12}$ | = | $x_3\,\mathbf{a}_1 + \left(\frac{1}{2} - y_3\right)\,\mathbf{a}_2 + \left(\frac{1}{2} + z_3\right)\,\mathbf{a}_3$ | = | $\left(\frac{1}{2}c\cos\beta + x_3 a + z_3 c\cos\beta\right)\,\hat{\mathbf{x}} + \left(\frac{1}{2} - y_3\right) b\,\hat{\mathbf{y}} + \left(\frac{1}{2} + z_3\right) c\sin\beta\,\hat{\mathbf{z}}$ | (4$e$) | Cl III |
| $\mathbf{B}_{13}$ | = | $x_4\,\mathbf{a}_1 + y_4\,\mathbf{a}_2 + z_4\,\mathbf{a}_3$ | = | $(x_4 a + z_4 c\cos\beta)\,\hat{\mathbf{x}} + y_4 b\,\hat{\mathbf{y}} + z_4 c\sin\beta\,\hat{\mathbf{z}}$ | (4$e$) | Cu |
| $\mathbf{B}_{14}$ | = | $-x_4\,\mathbf{a}_1 + \left(\frac{1}{2} + y_4\right)\,\mathbf{a}_2 + \left(\frac{1}{2} - z_4\right)\,\mathbf{a}_3$ | = | $\left(\frac{1}{2}c\cos\beta - x_4 a - z_4 c\cos\beta\right)\,\hat{\mathbf{x}} + \left(\frac{1}{2} + y_4\right) b\,\hat{\mathbf{y}} + \left(\frac{1}{2} - z_4\right) c\sin\beta\,\hat{\mathbf{z}}$ | (4$e$) | Cu |
| $\mathbf{B}_{15}$ | = | $-x_4\,\mathbf{a}_1 - y_4\,\mathbf{a}_2 - z_4\,\mathbf{a}_3$ | = | $(-x_4 a - z_4 c\cos\beta)\,\hat{\mathbf{x}} - y_4 b\,\hat{\mathbf{y}} - z_4 c\sin\beta\,\hat{\mathbf{z}}$ | (4$e$) | Cu |
| $\mathbf{B}_{16}$ | = | $x_4\,\mathbf{a}_1 + \left(\frac{1}{2} - y_4\right)\,\mathbf{a}_2 + \left(\frac{1}{2} + z_4\right)\,\mathbf{a}_3$ | = | $\left(\frac{1}{2}c\cos\beta + x_4 a + z_4 c\cos\beta\right)\,\hat{\mathbf{x}} + \left(\frac{1}{2} - y_4\right) b\,\hat{\mathbf{y}} + \left(\frac{1}{2} + z_4\right) c\sin\beta\,\hat{\mathbf{z}}$ | (4$e$) | Cu |
| $\mathbf{B}_{17}$ | = | $x_5\,\mathbf{a}_1 + y_5\,\mathbf{a}_2 + z_5\,\mathbf{a}_3$ | = | $(x_5 a + z_5 c\cos\beta)\,\hat{\mathbf{x}} + y_5 b\,\hat{\mathbf{y}} + z_5 c\sin\beta\,\hat{\mathbf{z}}$ | (4$e$) | K |
| $\mathbf{B}_{18}$ | = | $-x_5\,\mathbf{a}_1 + \left(\frac{1}{2} + y_5\right)\,\mathbf{a}_2 + \left(\frac{1}{2} - z_5\right)\,\mathbf{a}_3$ | = | $\left(\frac{1}{2}c\cos\beta - x_5 a - z_5 c\cos\beta\right)\,\hat{\mathbf{x}} + \left(\frac{1}{2} + y_5\right) b\,\hat{\mathbf{y}} + \left(\frac{1}{2} - z_5\right) c\sin\beta\,\hat{\mathbf{z}}$ | (4$e$) | K |
| $\mathbf{B}_{19}$ | = | $-x_5\,\mathbf{a}_1 - y_5\,\mathbf{a}_2 - z_5\,\mathbf{a}_3$ | = | $(-x_5 a - z_5 c\cos\beta)\,\hat{\mathbf{x}} - y_5 b\,\hat{\mathbf{y}} - z_5 c\sin\beta\,\hat{\mathbf{z}}$ | (4$e$) | K |
| $\mathbf{B}_{20}$ | = | $x_5\,\mathbf{a}_1 + \left(\frac{1}{2} - y_5\right)\,\mathbf{a}_2 + \left(\frac{1}{2} + z_5\right)\,\mathbf{a}_3$ | = | $\left(\frac{1}{2}c\cos\beta + x_5 a + z_5 c\cos\beta\right)\,\hat{\mathbf{x}} + \left(\frac{1}{2} - y_5\right) b\,\hat{\mathbf{y}} + \left(\frac{1}{2} + z_5\right) c\sin\beta\,\hat{\mathbf{z}}$ | (4$e$) | K |

**Geometry files:**

- CIF: pp. 1542
- POSCAR: pp. 1542

# $\gamma$-$WO_3$ Structure: A3B_mP32_14_6e_2e

http://aflow.org/prototype-encyclopedia/A3B_mP32_14_6e_2e

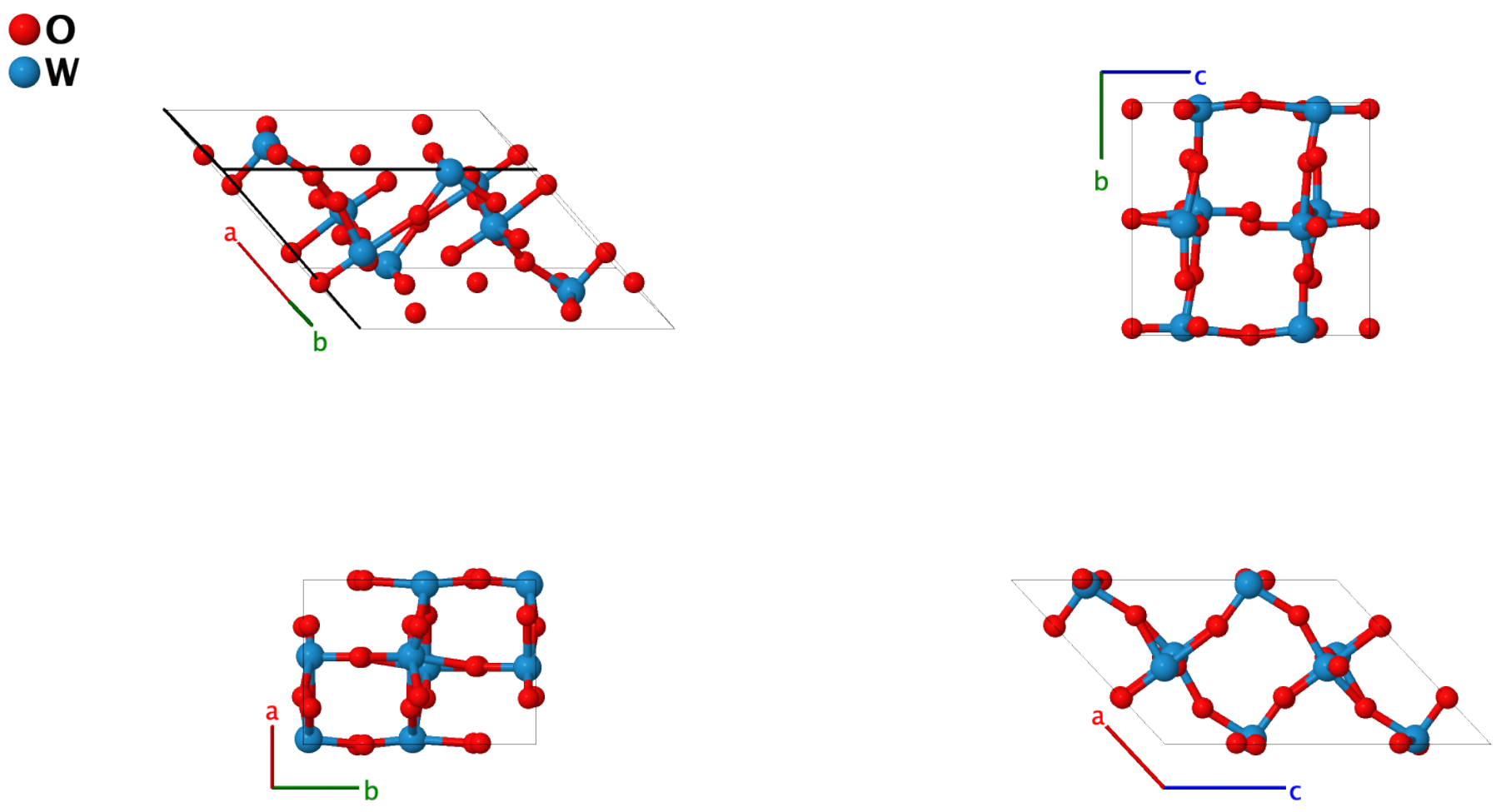

| | | |
|---|---|---|
| **Prototype** | : | $O_3W$ |
| **AFLOW prototype label** | : | A3B_mP32_14_6e_2e |
| ***Strukturbericht* designation** | : | None |
| **Pearson symbol** | : | mP32 |
| **Space group number** | : | 14 |
| **Space group symbol** | : | $P2_1/c$ |
| **AFLOW prototype command** | : | `aflow --proto=A3B_mP32_14_6e_2e` `--params=`$a, b/a, c/a, \beta, x_1, y_1, z_1, x_2, y_2, z_2, x_3, y_3, z_3, x_4, y_4, z_4, x_5, y_5, z_5, x_6, y_6, z_6, x_7, y_7, z_7, x_8, y_8, z_8$ |

- All stable phases of $WO_3$ are distortions of the cubic $\alpha$-$ReO_3$ ($D0_9$) phase. (Woodward, 1997 and Vogt, 1999) The known stable phases and their approximate temperature ranges are:
  - $\alpha$-$WO_3$ (1010-1170 K) (Vogt, 1999)
  - $\beta$-$WO_3$ (600-1170 K) (Vogt, 1999)
  - $\gamma$-$WO_3$ (290-600 K) (Vogt, 1999), this structure
  - $\delta$-$WO_3$ (230-290 K) (Diehl, 1978)
  - $\epsilon$-$WO_3$ (below 23 K) (Woodward, 1997)
- In addition, several other structures have been proposed and/or found:
  - The original $D0_{10}$ structure (Bräkken, 1931), (Hermann, 1937) superseded by $\delta$-$WO_3$
  - Original $\beta$-$WO_3$ (Salje, 1977)
  - Hexagonal $WO_3$ (Gerand, 1979) (metastable)

**Simple Monoclinic primitive vectors:**

$$
\begin{aligned}
\mathbf{a}_1 &= a\,\hat{\mathbf{x}} \\
\mathbf{a}_2 &= b\,\hat{\mathbf{y}} \\
\mathbf{a}_3 &= c\cos\beta\,\hat{\mathbf{x}} + c\sin\beta\,\hat{\mathbf{z}}
\end{aligned}
$$

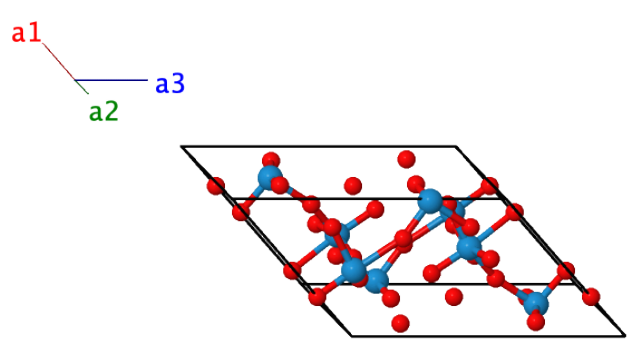

**Basis vectors:**

| | | Lattice Coordinates | | Cartesian Coordinates | Wyckoff Position | Atom Type |
|---|---|---|---|---|---|---|
| $\mathbf{B}_1$ | = | $x_1\,\mathbf{a}_1 + y_1\,\mathbf{a}_2 + z_1\,\mathbf{a}_3$ | = | $(x_1 a + z_1 c\cos\beta)\,\hat{\mathbf{x}} + y_1 b\,\hat{\mathbf{y}} + z_1 c\sin\beta\,\hat{\mathbf{z}}$ | (4$e$) | O I |
| $\mathbf{B}_2$ | = | $-x_1\,\mathbf{a}_1 + \left(\frac{1}{2} + y_1\right)\mathbf{a}_2 + \left(\frac{1}{2} - z_1\right)\mathbf{a}_3$ | = | $\left(\frac{1}{2}c\cos\beta - x_1 a - z_1 c\cos\beta\right)\hat{\mathbf{x}} + \left(\frac{1}{2} + y_1\right)b\,\hat{\mathbf{y}} + \left(\frac{1}{2} - z_1\right)c\sin\beta\,\hat{\mathbf{z}}$ | (4$e$) | O I |
| $\mathbf{B}_3$ | = | $-x_1\,\mathbf{a}_1 - y_1\,\mathbf{a}_2 - z_1\,\mathbf{a}_3$ | = | $(-x_1 a - z_1 c\cos\beta)\,\hat{\mathbf{x}} - y_1 b\,\hat{\mathbf{y}} - z_1 c\sin\beta\,\hat{\mathbf{z}}$ | (4$e$) | O I |
| $\mathbf{B}_4$ | = | $x_1\,\mathbf{a}_1 + \left(\frac{1}{2} - y_1\right)\mathbf{a}_2 + \left(\frac{1}{2} + z_1\right)\mathbf{a}_3$ | = | $\left(\frac{1}{2}c\cos\beta + x_1 a + z_1 c\cos\beta\right)\hat{\mathbf{x}} + \left(\frac{1}{2} - y_1\right)b\,\hat{\mathbf{y}} + \left(\frac{1}{2} + z_1\right)c\sin\beta\,\hat{\mathbf{z}}$ | (4$e$) | O I |
| $\mathbf{B}_5$ | = | $x_2\,\mathbf{a}_1 + y_2\,\mathbf{a}_2 + z_2\,\mathbf{a}_3$ | = | $(x_2 a + z_2 c\cos\beta)\,\hat{\mathbf{x}} + y_2 b\,\hat{\mathbf{y}} + z_2 c\sin\beta\,\hat{\mathbf{z}}$ | (4$e$) | O II |
| $\mathbf{B}_6$ | = | $-x_2\,\mathbf{a}_1 + \left(\frac{1}{2} + y_2\right)\mathbf{a}_2 + \left(\frac{1}{2} - z_2\right)\mathbf{a}_3$ | = | $\left(\frac{1}{2}c\cos\beta - x_2 a - z_2 c\cos\beta\right)\hat{\mathbf{x}} + \left(\frac{1}{2} + y_2\right)b\,\hat{\mathbf{y}} + \left(\frac{1}{2} - z_2\right)c\sin\beta\,\hat{\mathbf{z}}$ | (4$e$) | O II |
| $\mathbf{B}_7$ | = | $-x_2\,\mathbf{a}_1 - y_2\,\mathbf{a}_2 - z_2\,\mathbf{a}_3$ | = | $(-x_2 a - z_2 c\cos\beta)\,\hat{\mathbf{x}} - y_2 b\,\hat{\mathbf{y}} - z_2 c\sin\beta\,\hat{\mathbf{z}}$ | (4$e$) | O II |
| $\mathbf{B}_8$ | = | $x_2\,\mathbf{a}_1 + \left(\frac{1}{2} - y_2\right)\mathbf{a}_2 + \left(\frac{1}{2} + z_2\right)\mathbf{a}_3$ | = | $\left(\frac{1}{2}c\cos\beta + x_2 a + z_2 c\cos\beta\right)\hat{\mathbf{x}} + \left(\frac{1}{2} - y_2\right)b\,\hat{\mathbf{y}} + \left(\frac{1}{2} + z_2\right)c\sin\beta\,\hat{\mathbf{z}}$ | (4$e$) | O II |
| $\mathbf{B}_9$ | = | $x_3\,\mathbf{a}_1 + y_3\,\mathbf{a}_2 + z_3\,\mathbf{a}_3$ | = | $(x_3 a + z_3 c\cos\beta)\,\hat{\mathbf{x}} + y_3 b\,\hat{\mathbf{y}} + z_3 c\sin\beta\,\hat{\mathbf{z}}$ | (4$e$) | O III |
| $\mathbf{B}_{10}$ | = | $-x_3\,\mathbf{a}_1 + \left(\frac{1}{2} + y_3\right)\mathbf{a}_2 + \left(\frac{1}{2} - z_3\right)\mathbf{a}_3$ | = | $\left(\frac{1}{2}c\cos\beta - x_3 a - z_3 c\cos\beta\right)\hat{\mathbf{x}} + \left(\frac{1}{2} + y_3\right)b\,\hat{\mathbf{y}} + \left(\frac{1}{2} - z_3\right)c\sin\beta\,\hat{\mathbf{z}}$ | (4$e$) | O III |
| $\mathbf{B}_{11}$ | = | $-x_3\,\mathbf{a}_1 - y_3\,\mathbf{a}_2 - z_3\,\mathbf{a}_3$ | = | $(-x_3 a - z_3 c\cos\beta)\,\hat{\mathbf{x}} - y_3 b\,\hat{\mathbf{y}} - z_3 c\sin\beta\,\hat{\mathbf{z}}$ | (4$e$) | O III |
| $\mathbf{B}_{12}$ | = | $x_3\,\mathbf{a}_1 + \left(\frac{1}{2} - y_3\right)\mathbf{a}_2 + \left(\frac{1}{2} + z_3\right)\mathbf{a}_3$ | = | $\left(\frac{1}{2}c\cos\beta + x_3 a + z_3 c\cos\beta\right)\hat{\mathbf{x}} + \left(\frac{1}{2} - y_3\right)b\,\hat{\mathbf{y}} + \left(\frac{1}{2} + z_3\right)c\sin\beta\,\hat{\mathbf{z}}$ | (4$e$) | O III |
| $\mathbf{B}_{13}$ | = | $x_4\,\mathbf{a}_1 + y_4\,\mathbf{a}_2 + z_4\,\mathbf{a}_3$ | = | $(x_4 a + z_4 c\cos\beta)\,\hat{\mathbf{x}} + y_4 b\,\hat{\mathbf{y}} + z_4 c\sin\beta\,\hat{\mathbf{z}}$ | (4$e$) | O IV |
| $\mathbf{B}_{14}$ | = | $-x_4\,\mathbf{a}_1 + \left(\frac{1}{2} + y_4\right)\mathbf{a}_2 + \left(\frac{1}{2} - z_4\right)\mathbf{a}_3$ | = | $\left(\frac{1}{2}c\cos\beta - x_4 a - z_4 c\cos\beta\right)\hat{\mathbf{x}} + \left(\frac{1}{2} + y_4\right)b\,\hat{\mathbf{y}} + \left(\frac{1}{2} - z_4\right)c\sin\beta\,\hat{\mathbf{z}}$ | (4$e$) | O IV |
| $\mathbf{B}_{15}$ | = | $-x_4\,\mathbf{a}_1 - y_4\,\mathbf{a}_2 - z_4\,\mathbf{a}_3$ | = | $(-x_4 a - z_4 c\cos\beta)\,\hat{\mathbf{x}} - y_4 b\,\hat{\mathbf{y}} - z_4 c\sin\beta\,\hat{\mathbf{z}}$ | (4$e$) | O IV |
| $\mathbf{B}_{16}$ | = | $x_4\,\mathbf{a}_1 + \left(\frac{1}{2} - y_4\right)\mathbf{a}_2 + \left(\frac{1}{2} + z_4\right)\mathbf{a}_3$ | = | $\left(\frac{1}{2}c\cos\beta + x_4 a + z_4 c\cos\beta\right)\hat{\mathbf{x}} + \left(\frac{1}{2} - y_4\right)b\,\hat{\mathbf{y}} + \left(\frac{1}{2} + z_4\right)c\sin\beta\,\hat{\mathbf{z}}$ | (4$e$) | O IV |
| $\mathbf{B}_{17}$ | = | $x_5\,\mathbf{a}_1 + y_5\,\mathbf{a}_2 + z_5\,\mathbf{a}_3$ | = | $(x_5 a + z_5 c\cos\beta)\,\hat{\mathbf{x}} + y_5 b\,\hat{\mathbf{y}} + z_5 c\sin\beta\,\hat{\mathbf{z}}$ | (4$e$) | O V |
| $\mathbf{B}_{18}$ | = | $-x_5\,\mathbf{a}_1 + \left(\frac{1}{2} + y_5\right)\mathbf{a}_2 + \left(\frac{1}{2} - z_5\right)\mathbf{a}_3$ | = | $\left(\frac{1}{2}c\cos\beta - x_5 a - z_5 c\cos\beta\right)\hat{\mathbf{x}} + \left(\frac{1}{2} + y_5\right)b\,\hat{\mathbf{y}} + \left(\frac{1}{2} - z_5\right)c\sin\beta\,\hat{\mathbf{z}}$ | (4$e$) | O V |
| $\mathbf{B}_{19}$ | = | $-x_5\,\mathbf{a}_1 - y_5\,\mathbf{a}_2 - z_5\,\mathbf{a}_3$ | = | $(-x_5 a - z_5 c\cos\beta)\,\hat{\mathbf{x}} - y_5 b\,\hat{\mathbf{y}} - z_5 c\sin\beta\,\hat{\mathbf{z}}$ | (4$e$) | O V |
| $\mathbf{B}_{20}$ | = | $x_5\,\mathbf{a}_1 + \left(\frac{1}{2} - y_5\right)\mathbf{a}_2 + \left(\frac{1}{2} + z_5\right)\mathbf{a}_3$ | = | $\left(\frac{1}{2}c\cos\beta + x_5 a + z_5 c\cos\beta\right)\hat{\mathbf{x}} + \left(\frac{1}{2} - y_5\right)b\,\hat{\mathbf{y}} + \left(\frac{1}{2} + z_5\right)c\sin\beta\,\hat{\mathbf{z}}$ | (4$e$) | O V |
| $\mathbf{B}_{21}$ | = | $x_6\,\mathbf{a}_1 + y_6\,\mathbf{a}_2 + z_6\,\mathbf{a}_3$ | = | $(x_6 a + z_6 c\cos\beta)\,\hat{\mathbf{x}} + y_6 b\,\hat{\mathbf{y}} + z_6 c\sin\beta\,\hat{\mathbf{z}}$ | (4$e$) | O VI |
| $\mathbf{B}_{22}$ | = | $-x_6\,\mathbf{a}_1 + \left(\frac{1}{2} + y_6\right)\,\mathbf{a}_2 + \left(\frac{1}{2} - z_6\right)\,\mathbf{a}_3$ | = | $\left(\frac{1}{2}c\cos\beta - x_6 a - z_6 c\cos\beta\right)\,\hat{\mathbf{x}} + \left(\frac{1}{2} + y_6\right)b\,\hat{\mathbf{y}} + \left(\frac{1}{2} - z_6\right)c\sin\beta\,\hat{\mathbf{z}}$ | (4e) | O VI |
| $\mathbf{B}_{23}$ | = | $-x_6\,\mathbf{a}_1 - y_6\,\mathbf{a}_2 - z_6\,\mathbf{a}_3$ | = | $(-x_6 a - z_6 c\cos\beta)\,\hat{\mathbf{x}} - y_6 b\,\hat{\mathbf{y}} - z_6 c\sin\beta\,\hat{\mathbf{z}}$ | (4e) | O VI |
| $\mathbf{B}_{24}$ | = | $x_6\,\mathbf{a}_1 + \left(\frac{1}{2} - y_6\right)\,\mathbf{a}_2 + \left(\frac{1}{2} + z_6\right)\,\mathbf{a}_3$ | = | $\left(\frac{1}{2}c\cos\beta + x_6 a + z_6 c\cos\beta\right)\,\hat{\mathbf{x}} + \left(\frac{1}{2} - y_6\right)b\,\hat{\mathbf{y}} + \left(\frac{1}{2} + z_6\right)c\sin\beta\,\hat{\mathbf{z}}$ | (4e) | O VI |
| $\mathbf{B}_{25}$ | = | $x_7\,\mathbf{a}_1 + y_7\,\mathbf{a}_2 + z_7\,\mathbf{a}_3$ | = | $(x_7 a + z_7 c\cos\beta)\,\hat{\mathbf{x}} + y_7 b\,\hat{\mathbf{y}} + z_7 c\sin\beta\,\hat{\mathbf{z}}$ | (4e) | W I |
| $\mathbf{B}_{26}$ | = | $-x_7\,\mathbf{a}_1 + \left(\frac{1}{2} + y_7\right)\,\mathbf{a}_2 + \left(\frac{1}{2} - z_7\right)\,\mathbf{a}_3$ | = | $\left(\frac{1}{2}c\cos\beta - x_7 a - z_7 c\cos\beta\right)\,\hat{\mathbf{x}} + \left(\frac{1}{2} + y_7\right)b\,\hat{\mathbf{y}} + \left(\frac{1}{2} - z_7\right)c\sin\beta\,\hat{\mathbf{z}}$ | (4e) | W I |
| $\mathbf{B}_{27}$ | = | $-x_7\,\mathbf{a}_1 - y_7\,\mathbf{a}_2 - z_7\,\mathbf{a}_3$ | = | $(-x_7 a - z_7 c\cos\beta)\,\hat{\mathbf{x}} - y_7 b\,\hat{\mathbf{y}} - z_7 c\sin\beta\,\hat{\mathbf{z}}$ | (4e) | W I |
| $\mathbf{B}_{28}$ | = | $x_7\,\mathbf{a}_1 + \left(\frac{1}{2} - y_7\right)\,\mathbf{a}_2 + \left(\frac{1}{2} + z_7\right)\,\mathbf{a}_3$ | = | $\left(\frac{1}{2}c\cos\beta + x_7 a + z_7 c\cos\beta\right)\,\hat{\mathbf{x}} + \left(\frac{1}{2} - y_7\right)b\,\hat{\mathbf{y}} + \left(\frac{1}{2} + z_7\right)c\sin\beta\,\hat{\mathbf{z}}$ | (4e) | W I |
| $\mathbf{B}_{29}$ | = | $x_8\,\mathbf{a}_1 + y_8\,\mathbf{a}_2 + z_8\,\mathbf{a}_3$ | = | $(x_8 a + z_8 c\cos\beta)\,\hat{\mathbf{x}} + y_8 b\,\hat{\mathbf{y}} + z_8 c\sin\beta\,\hat{\mathbf{z}}$ | (4e) | W II |
| $\mathbf{B}_{30}$ | = | $-x_8\,\mathbf{a}_1 + \left(\frac{1}{2} + y_8\right)\,\mathbf{a}_2 + \left(\frac{1}{2} - z_8\right)\,\mathbf{a}_3$ | = | $\left(\frac{1}{2}c\cos\beta - x_8 a - z_8 c\cos\beta\right)\,\hat{\mathbf{x}} + \left(\frac{1}{2} + y_8\right)b\,\hat{\mathbf{y}} + \left(\frac{1}{2} - z_8\right)c\sin\beta\,\hat{\mathbf{z}}$ | (4e) | W II |
| $\mathbf{B}_{31}$ | = | $-x_8\,\mathbf{a}_1 - y_8\,\mathbf{a}_2 - z_8\,\mathbf{a}_3$ | = | $(-x_8 a - z_8 c\cos\beta)\,\hat{\mathbf{x}} - y_8 b\,\hat{\mathbf{y}} - z_8 c\sin\beta\,\hat{\mathbf{z}}$ | (4e) | W II |
| $\mathbf{B}_{32}$ | = | $x_8\,\mathbf{a}_1 + \left(\frac{1}{2} - y_8\right)\,\mathbf{a}_2 + \left(\frac{1}{2} + z_8\right)\,\mathbf{a}_3$ | = | $\left(\frac{1}{2}c\cos\beta + x_8 a + z_8 c\cos\beta\right)\,\hat{\mathbf{x}} + \left(\frac{1}{2} - y_8\right)b\,\hat{\mathbf{y}} + \left(\frac{1}{2} + z_8\right)c\sin\beta\,\hat{\mathbf{z}}$ | (4e) | W II |

**Geometry files:**

- CIF: pp. 1543
- POSCAR: pp. 1543

# $K_2Ni(CN)_4$ Structure: A4B2C4D_mP22_14_2e_e_2e_a

http://aflow.org/prototype-encyclopedia/A4B2C4D_mP22_14_2e_e_2e_a

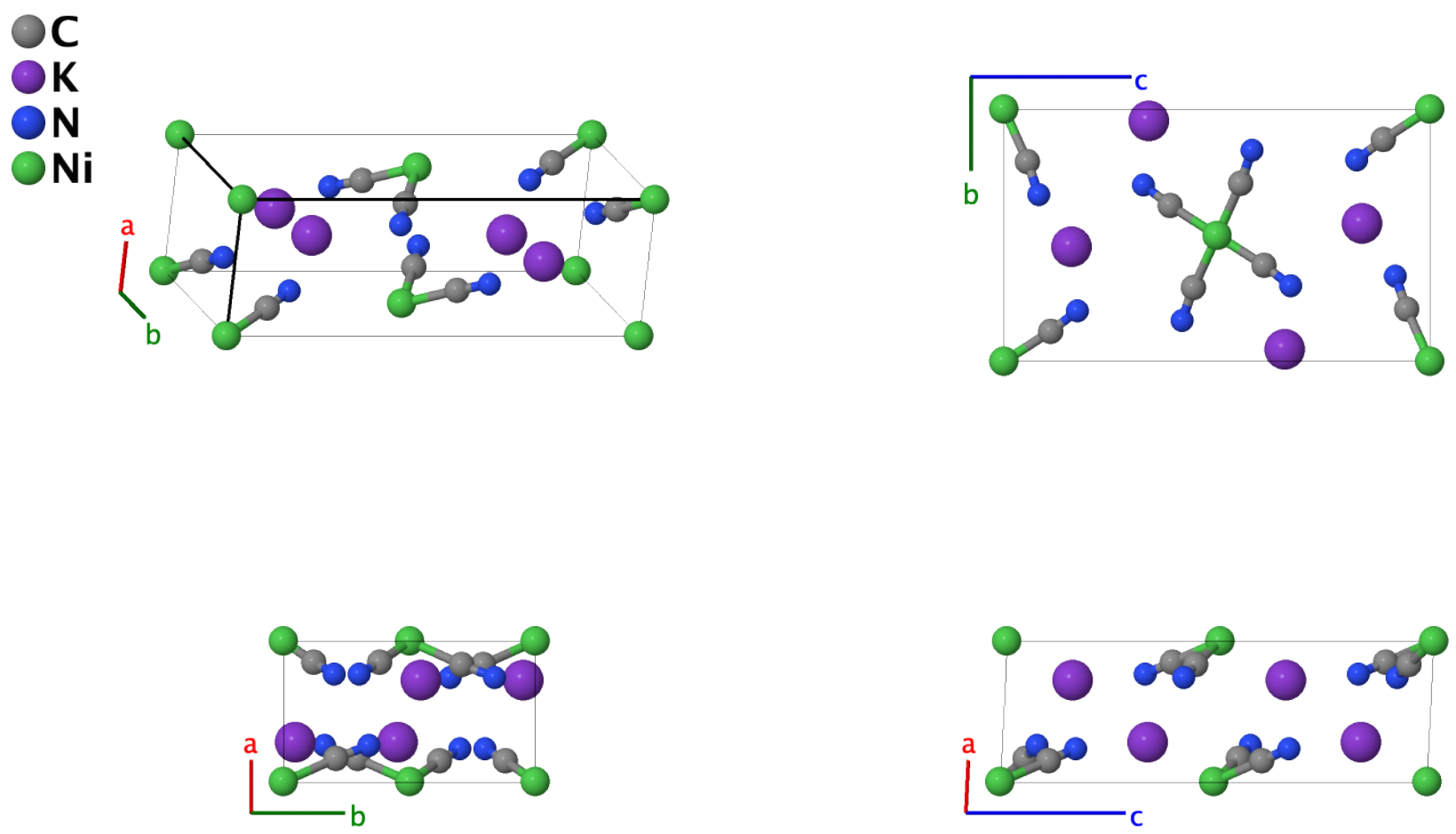

| | | |
|---|---|---|
| **Prototype** | : | $C_4K_2N_4Ni$ |
| **AFLOW prototype label** | : | A4B2C4D_mP22_14_2e_e_2e_a |
| ***Strukturbericht* designation** | : | None |
| **Pearson symbol** | : | mP22 |
| **Space group number** | : | 14 |
| **Space group symbol** | : | $P2_1/c$ |
| **AFLOW prototype command** | : | `aflow --proto=A4B2C4D_mP22_14_2e_e_2e_a`<br>`--params=`$a, b/a, c/a, \beta, x_2, y_2, z_2, x_3, y_3, z_3, x_4, y_4, z_4, x_5, y_5, z_5, x_6, y_6, z_6$ |

**Simple Monoclinic primitive vectors:**

$$\mathbf{a}_1 = a\,\hat{\mathbf{x}}$$
$$\mathbf{a}_2 = b\,\hat{\mathbf{y}}$$
$$\mathbf{a}_3 = c\cos\beta\,\hat{\mathbf{x}} + c\sin\beta\,\hat{\mathbf{z}}$$

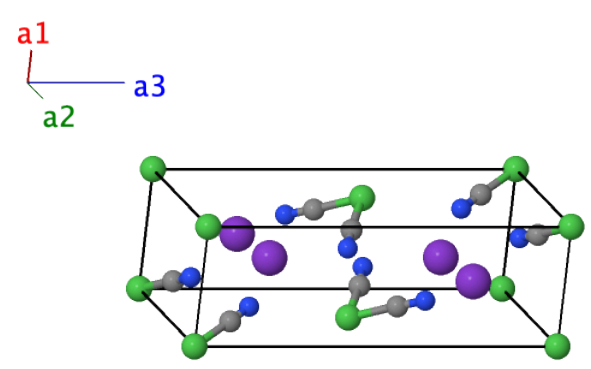

**Basis vectors:**

| | | Lattice Coordinates | | Cartesian Coordinates | Wyckoff Position | Atom Type |
|---|---|---|---|---|---|---|
| $\mathbf{B}_1$ | = | $0\,\mathbf{a}_1 + 0\,\mathbf{a}_2 + 0\,\mathbf{a}_3$ | = | $0\,\hat{\mathbf{x}} + 0\,\hat{\mathbf{y}} + 0\,\hat{\mathbf{z}}$ | (2$a$) | Ni |
| $\mathbf{B}_2$ | = | $\frac{1}{2}\,\mathbf{a}_2 + \frac{1}{2}\,\mathbf{a}_3$ | = | $\frac{1}{2}c\cos\beta\,\hat{\mathbf{x}} + \frac{1}{2}b\,\hat{\mathbf{y}} + \frac{1}{2}c\sin\beta\,\hat{\mathbf{z}}$ | (2$a$) | Ni |
| $\mathbf{B}_3$ | = | $x_2\,\mathbf{a}_1 + y_2\,\mathbf{a}_2 + z_2\,\mathbf{a}_3$ | = | $(x_2a + z_2c\cos\beta)\,\hat{\mathbf{x}} + y_2b\,\hat{\mathbf{y}} + z_2c\sin\beta\,\hat{\mathbf{z}}$ | (4$e$) | C I |
| $\mathbf{B}_4$ | = | $-x_2\,\mathbf{a}_1 + \left(\frac{1}{2} + y_2\right)\mathbf{a}_2 + \left(\frac{1}{2} - z_2\right)\mathbf{a}_3$ | = | $\left(\frac{1}{2}c\cos\beta - x_2a - z_2c\cos\beta\right)\hat{\mathbf{x}} + \left(\frac{1}{2} + y_2\right)b\,\hat{\mathbf{y}} + \left(\frac{1}{2} - z_2\right)c\sin\beta\,\hat{\mathbf{z}}$ | (4$e$) | C I |
| $\mathbf{B}_5$ | = | $-x_2\,\mathbf{a}_1 - y_2\,\mathbf{a}_2 - z_2\,\mathbf{a}_3$ | = | $(-x_2a - z_2c\cos\beta)\,\hat{\mathbf{x}} - y_2b\,\hat{\mathbf{y}} - z_2c\sin\beta\,\hat{\mathbf{z}}$ | (4$e$) | C I |
| $\mathbf{B}_{6}$ | = | $x_{2}\,\mathbf{a}_{1}+\left(\frac{1}{2}-y_{2}\right)\,\mathbf{a}_{2}+\left(\frac{1}{2}+z_{2}\right)\,\mathbf{a}_{3}$ | = | $\left(\frac{1}{2}c\cos\beta+x_{2}a+z_{2}c\cos\beta\right)\,\hat{\mathbf{x}}+\left(\frac{1}{2}-y_{2}\right)b\,\hat{\mathbf{y}}+\left(\frac{1}{2}+z_{2}\right)c\sin\beta\,\hat{\mathbf{z}}$ | (4$e$) | C I |
| $\mathbf{B}_{7}$ | = | $x_{3}\,\mathbf{a}_{1}+y_{3}\,\mathbf{a}_{2}+z_{3}\,\mathbf{a}_{3}$ | = | $\left(x_{3}a+z_{3}c\cos\beta\right)\,\hat{\mathbf{x}}+y_{3}b\,\hat{\mathbf{y}}+z_{3}c\sin\beta\,\hat{\mathbf{z}}$ | (4$e$) | C II |
| $\mathbf{B}_{8}$ | = | $-x_{3}\,\mathbf{a}_{1}+\left(\frac{1}{2}+y_{3}\right)\,\mathbf{a}_{2}+\left(\frac{1}{2}-z_{3}\right)\,\mathbf{a}_{3}$ | = | $\left(\frac{1}{2}c\cos\beta-x_{3}a-z_{3}c\cos\beta\right)\,\hat{\mathbf{x}}+\left(\frac{1}{2}+y_{3}\right)b\,\hat{\mathbf{y}}+\left(\frac{1}{2}-z_{3}\right)c\sin\beta\,\hat{\mathbf{z}}$ | (4$e$) | C II |
| $\mathbf{B}_{9}$ | = | $-x_{3}\,\mathbf{a}_{1}-y_{3}\,\mathbf{a}_{2}-z_{3}\,\mathbf{a}_{3}$ | = | $\left(-x_{3}a-z_{3}c\cos\beta\right)\,\hat{\mathbf{x}}-y_{3}b\,\hat{\mathbf{y}}-z_{3}c\sin\beta\,\hat{\mathbf{z}}$ | (4$e$) | C II |
| $\mathbf{B}_{10}$ | = | $x_{3}\,\mathbf{a}_{1}+\left(\frac{1}{2}-y_{3}\right)\,\mathbf{a}_{2}+\left(\frac{1}{2}+z_{3}\right)\,\mathbf{a}_{3}$ | = | $\left(\frac{1}{2}c\cos\beta+x_{3}a+z_{3}c\cos\beta\right)\,\hat{\mathbf{x}}+\left(\frac{1}{2}-y_{3}\right)b\,\hat{\mathbf{y}}+\left(\frac{1}{2}+z_{3}\right)c\sin\beta\,\hat{\mathbf{z}}$ | (4$e$) | C II |
| $\mathbf{B}_{11}$ | = | $x_{4}\,\mathbf{a}_{1}+y_{4}\,\mathbf{a}_{2}+z_{4}\,\mathbf{a}_{3}$ | = | $\left(x_{4}a+z_{4}c\cos\beta\right)\,\hat{\mathbf{x}}+y_{4}b\,\hat{\mathbf{y}}+z_{4}c\sin\beta\,\hat{\mathbf{z}}$ | (4$e$) | K |
| $\mathbf{B}_{12}$ | = | $-x_{4}\,\mathbf{a}_{1}+\left(\frac{1}{2}+y_{4}\right)\,\mathbf{a}_{2}+\left(\frac{1}{2}-z_{4}\right)\,\mathbf{a}_{3}$ | = | $\left(\frac{1}{2}c\cos\beta-x_{4}a-z_{4}c\cos\beta\right)\,\hat{\mathbf{x}}+\left(\frac{1}{2}+y_{4}\right)b\,\hat{\mathbf{y}}+\left(\frac{1}{2}-z_{4}\right)c\sin\beta\,\hat{\mathbf{z}}$ | (4$e$) | K |
| $\mathbf{B}_{13}$ | = | $-x_{4}\,\mathbf{a}_{1}-y_{4}\,\mathbf{a}_{2}-z_{4}\,\mathbf{a}_{3}$ | = | $\left(-x_{4}a-z_{4}c\cos\beta\right)\,\hat{\mathbf{x}}-y_{4}b\,\hat{\mathbf{y}}-z_{4}c\sin\beta\,\hat{\mathbf{z}}$ | (4$e$) | K |
| $\mathbf{B}_{14}$ | = | $x_{4}\,\mathbf{a}_{1}+\left(\frac{1}{2}-y_{4}\right)\,\mathbf{a}_{2}+\left(\frac{1}{2}+z_{4}\right)\,\mathbf{a}_{3}$ | = | $\left(\frac{1}{2}c\cos\beta+x_{4}a+z_{4}c\cos\beta\right)\,\hat{\mathbf{x}}+\left(\frac{1}{2}-y_{4}\right)b\,\hat{\mathbf{y}}+\left(\frac{1}{2}+z_{4}\right)c\sin\beta\,\hat{\mathbf{z}}$ | (4$e$) | K |
| $\mathbf{B}_{15}$ | = | $x_{5}\,\mathbf{a}_{1}+y_{5}\,\mathbf{a}_{2}+z_{5}\,\mathbf{a}_{3}$ | = | $\left(x_{5}a+z_{5}c\cos\beta\right)\,\hat{\mathbf{x}}+y_{5}b\,\hat{\mathbf{y}}+z_{5}c\sin\beta\,\hat{\mathbf{z}}$ | (4$e$) | N I |
| $\mathbf{B}_{16}$ | = | $-x_{5}\,\mathbf{a}_{1}+\left(\frac{1}{2}+y_{5}\right)\,\mathbf{a}_{2}+\left(\frac{1}{2}-z_{5}\right)\,\mathbf{a}_{3}$ | = | $\left(\frac{1}{2}c\cos\beta-x_{5}a-z_{5}c\cos\beta\right)\,\hat{\mathbf{x}}+\left(\frac{1}{2}+y_{5}\right)b\,\hat{\mathbf{y}}+\left(\frac{1}{2}-z_{5}\right)c\sin\beta\,\hat{\mathbf{z}}$ | (4$e$) | N I |
| $\mathbf{B}_{17}$ | = | $-x_{5}\,\mathbf{a}_{1}-y_{5}\,\mathbf{a}_{2}-z_{5}\,\mathbf{a}_{3}$ | = | $\left(-x_{5}a-z_{5}c\cos\beta\right)\,\hat{\mathbf{x}}-y_{5}b\,\hat{\mathbf{y}}-z_{5}c\sin\beta\,\hat{\mathbf{z}}$ | (4$e$) | N I |
| $\mathbf{B}_{18}$ | = | $x_{5}\,\mathbf{a}_{1}+\left(\frac{1}{2}-y_{5}\right)\,\mathbf{a}_{2}+\left(\frac{1}{2}+z_{5}\right)\,\mathbf{a}_{3}$ | = | $\left(\frac{1}{2}c\cos\beta+x_{5}a+z_{5}c\cos\beta\right)\,\hat{\mathbf{x}}+\left(\frac{1}{2}-y_{5}\right)b\,\hat{\mathbf{y}}+\left(\frac{1}{2}+z_{5}\right)c\sin\beta\,\hat{\mathbf{z}}$ | (4$e$) | N I |
| $\mathbf{B}_{19}$ | = | $x_{6}\,\mathbf{a}_{1}+y_{6}\,\mathbf{a}_{2}+z_{6}\,\mathbf{a}_{3}$ | = | $\left(x_{6}a+z_{6}c\cos\beta\right)\,\hat{\mathbf{x}}+y_{6}b\,\hat{\mathbf{y}}+z_{6}c\sin\beta\,\hat{\mathbf{z}}$ | (4$e$) | N II |
| $\mathbf{B}_{20}$ | = | $-x_{6}\,\mathbf{a}_{1}+\left(\frac{1}{2}+y_{6}\right)\,\mathbf{a}_{2}+\left(\frac{1}{2}-z_{6}\right)\,\mathbf{a}_{3}$ | = | $\left(\frac{1}{2}c\cos\beta-x_{6}a-z_{6}c\cos\beta\right)\,\hat{\mathbf{x}}+\left(\frac{1}{2}+y_{6}\right)b\,\hat{\mathbf{y}}+\left(\frac{1}{2}-z_{6}\right)c\sin\beta\,\hat{\mathbf{z}}$ | (4$e$) | N II |
| $\mathbf{B}_{21}$ | = | $-x_{6}\,\mathbf{a}_{1}-y_{6}\,\mathbf{a}_{2}-z_{6}\,\mathbf{a}_{3}$ | = | $\left(-x_{6}a-z_{6}c\cos\beta\right)\,\hat{\mathbf{x}}-y_{6}b\,\hat{\mathbf{y}}-z_{6}c\sin\beta\,\hat{\mathbf{z}}$ | (4$e$) | N II |
| $\mathbf{B}_{22}$ | = | $x_{6}\,\mathbf{a}_{1}+\left(\frac{1}{2}-y_{6}\right)\,\mathbf{a}_{2}+\left(\frac{1}{2}+z_{6}\right)\,\mathbf{a}_{3}$ | = | $\left(\frac{1}{2}c\cos\beta+x_{6}a+z_{6}c\cos\beta\right)\,\hat{\mathbf{x}}+\left(\frac{1}{2}-y_{6}\right)b\,\hat{\mathbf{y}}+\left(\frac{1}{2}+z_{6}\right)c\sin\beta\,\hat{\mathbf{z}}$ | (4$e$) | N II |

**Geometry files:**

- CIF: pp. 1543
- POSCAR: pp. 1544

# $KICl_4 \cdot H_2O$ ($H0_{10}$) Structure: A4BCD_mP28_14_4e_e_e_e

http://aflow.org/prototype-encyclopedia/A4BCD_mP28_14_4e_e_e_e

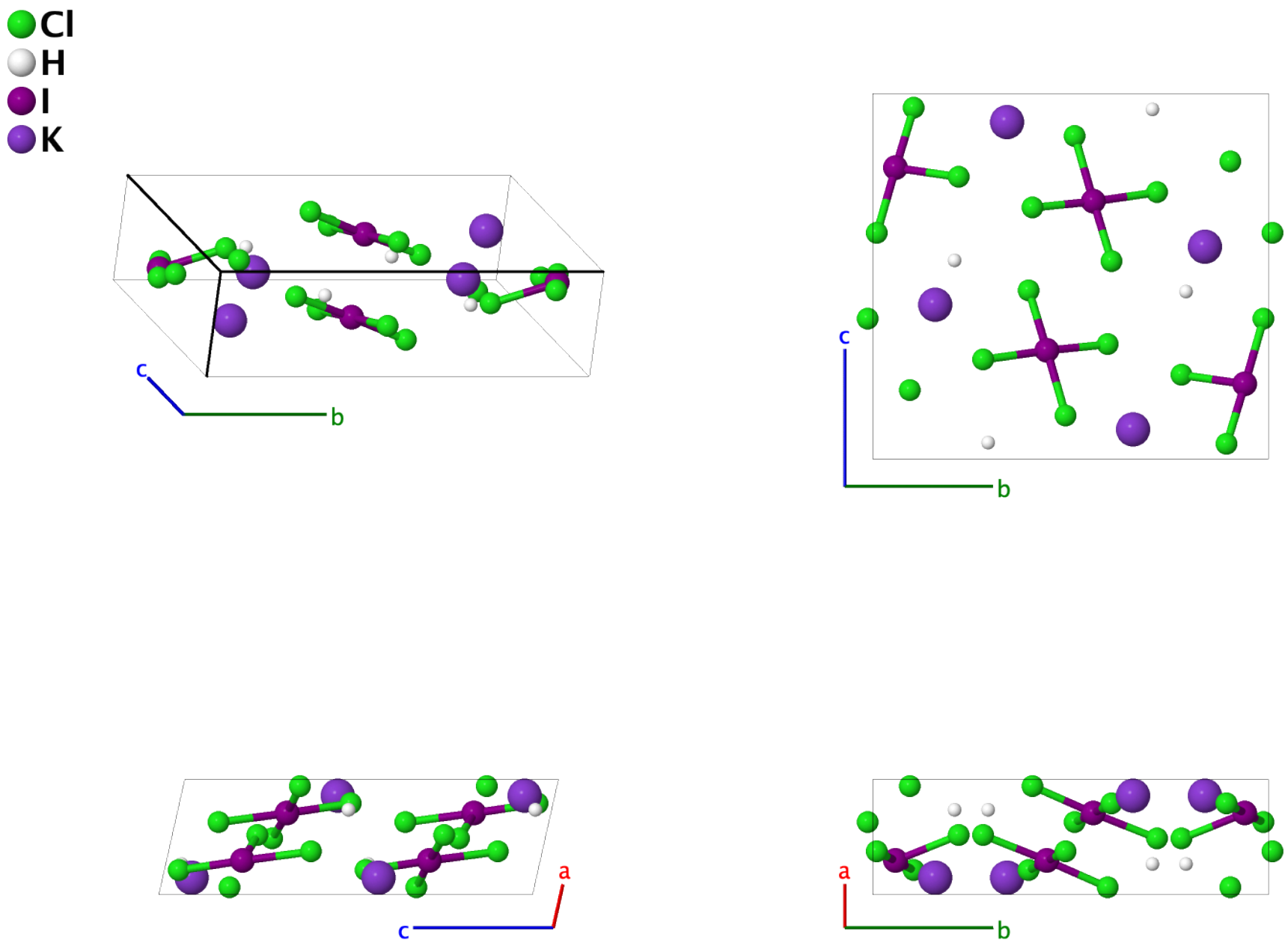

| | | |
|---|---|---|
| **Prototype** | : | $Cl_4IK \cdot H_2O$ |
| **AFLOW prototype label** | : | A4BCD_mP28_14_4e_e_e_e |
| ***Strukturbericht* designation** | : | $H0_{10}$ |
| **Pearson symbol** | : | mP28 |
| **Space group number** | : | 14 |
| **Space group symbol** | : | $P2_1/c$ |
| **AFLOW prototype command** | : | `aflow --proto=A4BCD_mP28_14_4e_e_e_e` `--params=`$a, b/a, c/a, \beta, x_1, y_1, z_1, x_2, y_2, z_2, x_3, y_3, z_3, x_4, y_4, z_4, x_5, y_5, z_5, x_6, y_6, z_6, x_7, y_7, z_7$ |

- The structure of $KICl_4$ was originally determined by (Mooney, 1938) and assigned *Strukturbericht* designation $H0_{10}$ by (Herrmann, 1941). "During Mooney's structure determination of KIC14 · $H_2O$ it was not realized that the crystals contain water of crystallization." (Elema, 1963) The correct structure is essentially the same as Mooney's with water molecules added into the vacancies, and we use $KICl_4 \cdot H_2O$ as our prototype for $H0_{10}$. To return to Mooney's structure, remove the water molecules, but of course this will not be a stable structure.

- The Wyckoff positions given by both (Mooney, 1938) and (Elema, 1963) are in the $P2_1/n$ setting of space group #14. We have used FINDSYM to translate this to the standard $P2_1/c$ setting.

**Simple Monoclinic primitive vectors:**

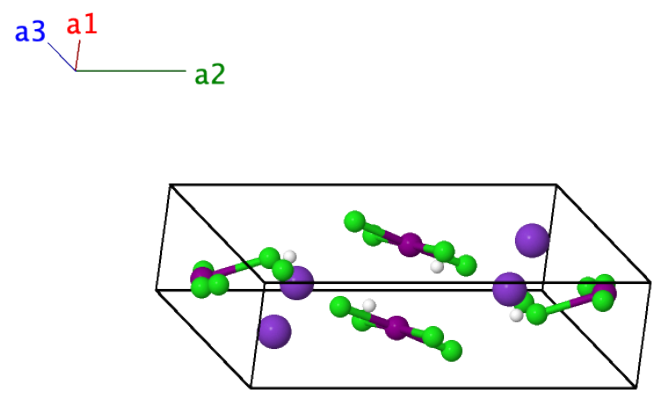

$$\mathbf{a}_1 = a\,\hat{\mathbf{x}}$$
$$\mathbf{a}_2 = b\,\hat{\mathbf{y}}$$
$$\mathbf{a}_3 = c\cos\beta\,\hat{\mathbf{x}} + c\sin\beta\,\hat{\mathbf{z}}$$

**Basis vectors:**

| | | Lattice Coordinates | | Cartesian Coordinates | Wyckoff Position | Atom Type |
|---|---|---|---|---|---|---|
| $\mathbf{B}_1$ | = | $x_1\,\mathbf{a}_1 + y_1\,\mathbf{a}_2 + z_1\,\mathbf{a}_3$ | = | $(x_1 a + z_1 c\cos\beta)\,\hat{\mathbf{x}} + y_1 b\,\hat{\mathbf{y}} + z_1 c\sin\beta\,\hat{\mathbf{z}}$ | (4e) | Cl I |
| $\mathbf{B}_2$ | = | $-x_1\,\mathbf{a}_1 + \left(\frac{1}{2} + y_1\right)\mathbf{a}_2 + \left(\frac{1}{2} - z_1\right)\mathbf{a}_3$ | = | $\left(\frac{1}{2}c\cos\beta - x_1 a - z_1 c\cos\beta\right)\hat{\mathbf{x}} + \left(\frac{1}{2} + y_1\right)b\,\hat{\mathbf{y}} + \left(\frac{1}{2} - z_1\right)c\sin\beta\,\hat{\mathbf{z}}$ | (4e) | Cl I |
| $\mathbf{B}_3$ | = | $-x_1\,\mathbf{a}_1 - y_1\,\mathbf{a}_2 - z_1\,\mathbf{a}_3$ | = | $(-x_1 a - z_1 c\cos\beta)\,\hat{\mathbf{x}} - y_1 b\,\hat{\mathbf{y}} - z_1 c\sin\beta\,\hat{\mathbf{z}}$ | (4e) | Cl I |
| $\mathbf{B}_4$ | = | $x_1\,\mathbf{a}_1 + \left(\frac{1}{2} - y_1\right)\mathbf{a}_2 + \left(\frac{1}{2} + z_1\right)\mathbf{a}_3$ | = | $\left(\frac{1}{2}c\cos\beta + x_1 a + z_1 c\cos\beta\right)\hat{\mathbf{x}} + \left(\frac{1}{2} - y_1\right)b\,\hat{\mathbf{y}} + \left(\frac{1}{2} + z_1\right)c\sin\beta\,\hat{\mathbf{z}}$ | (4e) | Cl I |
| $\mathbf{B}_5$ | = | $x_2\,\mathbf{a}_1 + y_2\,\mathbf{a}_2 + z_2\,\mathbf{a}_3$ | = | $(x_2 a + z_2 c\cos\beta)\,\hat{\mathbf{x}} + y_2 b\,\hat{\mathbf{y}} + z_2 c\sin\beta\,\hat{\mathbf{z}}$ | (4e) | Cl II |
| $\mathbf{B}_6$ | = | $-x_2\,\mathbf{a}_1 + \left(\frac{1}{2} + y_2\right)\mathbf{a}_2 + \left(\frac{1}{2} - z_2\right)\mathbf{a}_3$ | = | $\left(\frac{1}{2}c\cos\beta - x_2 a - z_2 c\cos\beta\right)\hat{\mathbf{x}} + \left(\frac{1}{2} + y_2\right)b\,\hat{\mathbf{y}} + \left(\frac{1}{2} - z_2\right)c\sin\beta\,\hat{\mathbf{z}}$ | (4e) | Cl II |
| $\mathbf{B}_7$ | = | $-x_2\,\mathbf{a}_1 - y_2\,\mathbf{a}_2 - z_2\,\mathbf{a}_3$ | = | $(-x_2 a - z_2 c\cos\beta)\,\hat{\mathbf{x}} - y_2 b\,\hat{\mathbf{y}} - z_2 c\sin\beta\,\hat{\mathbf{z}}$ | (4e) | Cl II |
| $\mathbf{B}_8$ | = | $x_2\,\mathbf{a}_1 + \left(\frac{1}{2} - y_2\right)\mathbf{a}_2 + \left(\frac{1}{2} + z_2\right)\mathbf{a}_3$ | = | $\left(\frac{1}{2}c\cos\beta + x_2 a + z_2 c\cos\beta\right)\hat{\mathbf{x}} + \left(\frac{1}{2} - y_2\right)b\,\hat{\mathbf{y}} + \left(\frac{1}{2} + z_2\right)c\sin\beta\,\hat{\mathbf{z}}$ | (4e) | Cl II |
| $\mathbf{B}_9$ | = | $x_3\,\mathbf{a}_1 + y_3\,\mathbf{a}_2 + z_3\,\mathbf{a}_3$ | = | $(x_3 a + z_3 c\cos\beta)\,\hat{\mathbf{x}} + y_3 b\,\hat{\mathbf{y}} + z_3 c\sin\beta\,\hat{\mathbf{z}}$ | (4e) | Cl III |
| $\mathbf{B}_{10}$ | = | $-x_3\,\mathbf{a}_1 + \left(\frac{1}{2} + y_3\right)\mathbf{a}_2 + \left(\frac{1}{2} - z_3\right)\mathbf{a}_3$ | = | $\left(\frac{1}{2}c\cos\beta - x_3 a - z_3 c\cos\beta\right)\hat{\mathbf{x}} + \left(\frac{1}{2} + y_3\right)b\,\hat{\mathbf{y}} + \left(\frac{1}{2} - z_3\right)c\sin\beta\,\hat{\mathbf{z}}$ | (4e) | Cl III |
| $\mathbf{B}_{11}$ | = | $-x_3\,\mathbf{a}_1 - y_3\,\mathbf{a}_2 - z_3\,\mathbf{a}_3$ | = | $(-x_3 a - z_3 c\cos\beta)\,\hat{\mathbf{x}} - y_3 b\,\hat{\mathbf{y}} - z_3 c\sin\beta\,\hat{\mathbf{z}}$ | (4e) | Cl III |
| $\mathbf{B}_{12}$ | = | $x_3\,\mathbf{a}_1 + \left(\frac{1}{2} - y_3\right)\mathbf{a}_2 + \left(\frac{1}{2} + z_3\right)\mathbf{a}_3$ | = | $\left(\frac{1}{2}c\cos\beta + x_3 a + z_3 c\cos\beta\right)\hat{\mathbf{x}} + \left(\frac{1}{2} - y_3\right)b\,\hat{\mathbf{y}} + \left(\frac{1}{2} + z_3\right)c\sin\beta\,\hat{\mathbf{z}}$ | (4e) | Cl III |
| $\mathbf{B}_{13}$ | = | $x_4\,\mathbf{a}_1 + y_4\,\mathbf{a}_2 + z_4\,\mathbf{a}_3$ | = | $(x_4 a + z_4 c\cos\beta)\,\hat{\mathbf{x}} + y_4 b\,\hat{\mathbf{y}} + z_4 c\sin\beta\,\hat{\mathbf{z}}$ | (4e) | Cl IV |
| $\mathbf{B}_{14}$ | = | $-x_4\,\mathbf{a}_1 + \left(\frac{1}{2} + y_4\right)\mathbf{a}_2 + \left(\frac{1}{2} - z_4\right)\mathbf{a}_3$ | = | $\left(\frac{1}{2}c\cos\beta - x_4 a - z_4 c\cos\beta\right)\hat{\mathbf{x}} + \left(\frac{1}{2} + y_4\right)b\,\hat{\mathbf{y}} + \left(\frac{1}{2} - z_4\right)c\sin\beta\,\hat{\mathbf{z}}$ | (4e) | Cl IV |
| $\mathbf{B}_{15}$ | = | $-x_4\,\mathbf{a}_1 - y_4\,\mathbf{a}_2 - z_4\,\mathbf{a}_3$ | = | $(-x_4 a - z_4 c\cos\beta)\,\hat{\mathbf{x}} - y_4 b\,\hat{\mathbf{y}} - z_4 c\sin\beta\,\hat{\mathbf{z}}$ | (4e) | Cl IV |
| $\mathbf{B}_{16}$ | = | $x_4\,\mathbf{a}_1 + \left(\frac{1}{2} - y_4\right)\mathbf{a}_2 + \left(\frac{1}{2} + z_4\right)\mathbf{a}_3$ | = | $\left(\frac{1}{2}c\cos\beta + x_4 a + z_4 c\cos\beta\right)\hat{\mathbf{x}} + \left(\frac{1}{2} - y_4\right)b\,\hat{\mathbf{y}} + \left(\frac{1}{2} + z_4\right)c\sin\beta\,\hat{\mathbf{z}}$ | (4e) | Cl IV |
| $\mathbf{B}_{17}$ | = | $x_5\,\mathbf{a}_1 + y_5\,\mathbf{a}_2 + z_5\,\mathbf{a}_3$ | = | $(x_5 a + z_5 c\cos\beta)\,\hat{\mathbf{x}} + y_5 b\,\hat{\mathbf{y}} + z_5 c\sin\beta\,\hat{\mathbf{z}}$ | (4e) | $H_2O$ |
| $\mathbf{B}_{18}$ | = | $-x_5\,\mathbf{a}_1 + \left(\frac{1}{2} + y_5\right)\mathbf{a}_2 + \left(\frac{1}{2} - z_5\right)\mathbf{a}_3$ | = | $\left(\frac{1}{2}c\cos\beta - x_5 a - z_5 c\cos\beta\right)\hat{\mathbf{x}} + \left(\frac{1}{2} + y_5\right) b\,\hat{\mathbf{y}} + \left(\frac{1}{2} - z_5\right) c\sin\beta\,\hat{\mathbf{z}}$ | (4e) | $H_2O$ |
| $\mathbf{B}_{19}$ | = | $-x_5\,\mathbf{a}_1 - y_5\,\mathbf{a}_2 - z_5\,\mathbf{a}_3$ | = | $(-x_5 a - z_5 c\cos\beta)\,\hat{\mathbf{x}} - y_5 b\,\hat{\mathbf{y}} - z_5 c\sin\beta\,\hat{\mathbf{z}}$ | (4e) | $H_2O$ |
| $\mathbf{B}_{20}$ | = | $x_5\,\mathbf{a}_1 + \left(\frac{1}{2} - y_5\right)\mathbf{a}_2 + \left(\frac{1}{2} + z_5\right)\mathbf{a}_3$ | = | $\left(\frac{1}{2}c\cos\beta + x_5 a + z_5 c\cos\beta\right)\hat{\mathbf{x}} + \left(\frac{1}{2} - y_5\right) b\,\hat{\mathbf{y}} + \left(\frac{1}{2} + z_5\right) c\sin\beta\,\hat{\mathbf{z}}$ | (4e) | $H_2O$ |
| $\mathbf{B}_{21}$ | = | $x_6\,\mathbf{a}_1 + y_6\,\mathbf{a}_2 + z_6\,\mathbf{a}_3$ | = | $(x_6 a + z_6 c\cos\beta)\,\hat{\mathbf{x}} + y_6 b\,\hat{\mathbf{y}} + z_6 c\sin\beta\,\hat{\mathbf{z}}$ | (4e) | I |
| $\mathbf{B}_{22}$ | = | $-x_6\,\mathbf{a}_1 + \left(\frac{1}{2} + y_6\right)\mathbf{a}_2 + \left(\frac{1}{2} - z_6\right)\mathbf{a}_3$ | = | $\left(\frac{1}{2}c\cos\beta - x_6 a - z_6 c\cos\beta\right)\hat{\mathbf{x}} + \left(\frac{1}{2} + y_6\right) b\,\hat{\mathbf{y}} + \left(\frac{1}{2} - z_6\right) c\sin\beta\,\hat{\mathbf{z}}$ | (4e) | I |
| $\mathbf{B}_{23}$ | = | $-x_6\,\mathbf{a}_1 - y_6\,\mathbf{a}_2 - z_6\,\mathbf{a}_3$ | = | $(-x_6 a - z_6 c\cos\beta)\,\hat{\mathbf{x}} - y_6 b\,\hat{\mathbf{y}} - z_6 c\sin\beta\,\hat{\mathbf{z}}$ | (4e) | I |
| $\mathbf{B}_{24}$ | = | $x_6\,\mathbf{a}_1 + \left(\frac{1}{2} - y_6\right)\mathbf{a}_2 + \left(\frac{1}{2} + z_6\right)\mathbf{a}_3$ | = | $\left(\frac{1}{2}c\cos\beta + x_6 a + z_6 c\cos\beta\right)\hat{\mathbf{x}} + \left(\frac{1}{2} - y_6\right) b\,\hat{\mathbf{y}} + \left(\frac{1}{2} + z_6\right) c\sin\beta\,\hat{\mathbf{z}}$ | (4e) | I |
| $\mathbf{B}_{25}$ | = | $x_7\,\mathbf{a}_1 + y_7\,\mathbf{a}_2 + z_7\,\mathbf{a}_3$ | = | $(x_7 a + z_7 c\cos\beta)\,\hat{\mathbf{x}} + y_7 b\,\hat{\mathbf{y}} + z_7 c\sin\beta\,\hat{\mathbf{z}}$ | (4e) | K |
| $\mathbf{B}_{26}$ | = | $-x_7\,\mathbf{a}_1 + \left(\frac{1}{2} + y_7\right)\mathbf{a}_2 + \left(\frac{1}{2} - z_7\right)\mathbf{a}_3$ | = | $\left(\frac{1}{2}c\cos\beta - x_7 a - z_7 c\cos\beta\right)\hat{\mathbf{x}} + \left(\frac{1}{2} + y_7\right) b\,\hat{\mathbf{y}} + \left(\frac{1}{2} - z_7\right) c\sin\beta\,\hat{\mathbf{z}}$ | (4e) | K |
| $\mathbf{B}_{27}$ | = | $-x_7\,\mathbf{a}_1 - y_7\,\mathbf{a}_2 - z_7\,\mathbf{a}_3$ | = | $(-x_7 a - z_7 c\cos\beta)\,\hat{\mathbf{x}} - y_7 b\,\hat{\mathbf{y}} - z_7 c\sin\beta\,\hat{\mathbf{z}}$ | (4e) | K |
| $\mathbf{B}_{28}$ | = | $x_7\,\mathbf{a}_1 + \left(\frac{1}{2} - y_7\right)\mathbf{a}_2 + \left(\frac{1}{2} + z_7\right)\mathbf{a}_3$ | = | $\left(\frac{1}{2}c\cos\beta + x_7 a + z_7 c\cos\beta\right)\hat{\mathbf{x}} + \left(\frac{1}{2} - y_7\right) b\,\hat{\mathbf{y}} + \left(\frac{1}{2} + z_7\right) c\sin\beta\,\hat{\mathbf{z}}$ | (4e) | K |

**Geometry files:**

- CIF: pp. 1544
- POSCAR: pp. 1544

# $\gamma$-$Y_2Si_2O_7$ Structure: A4BC_mP24_14_4e_e_e

http://aflow.org/prototype-encyclopedia/A4BC_mP24_14_4e_e_e

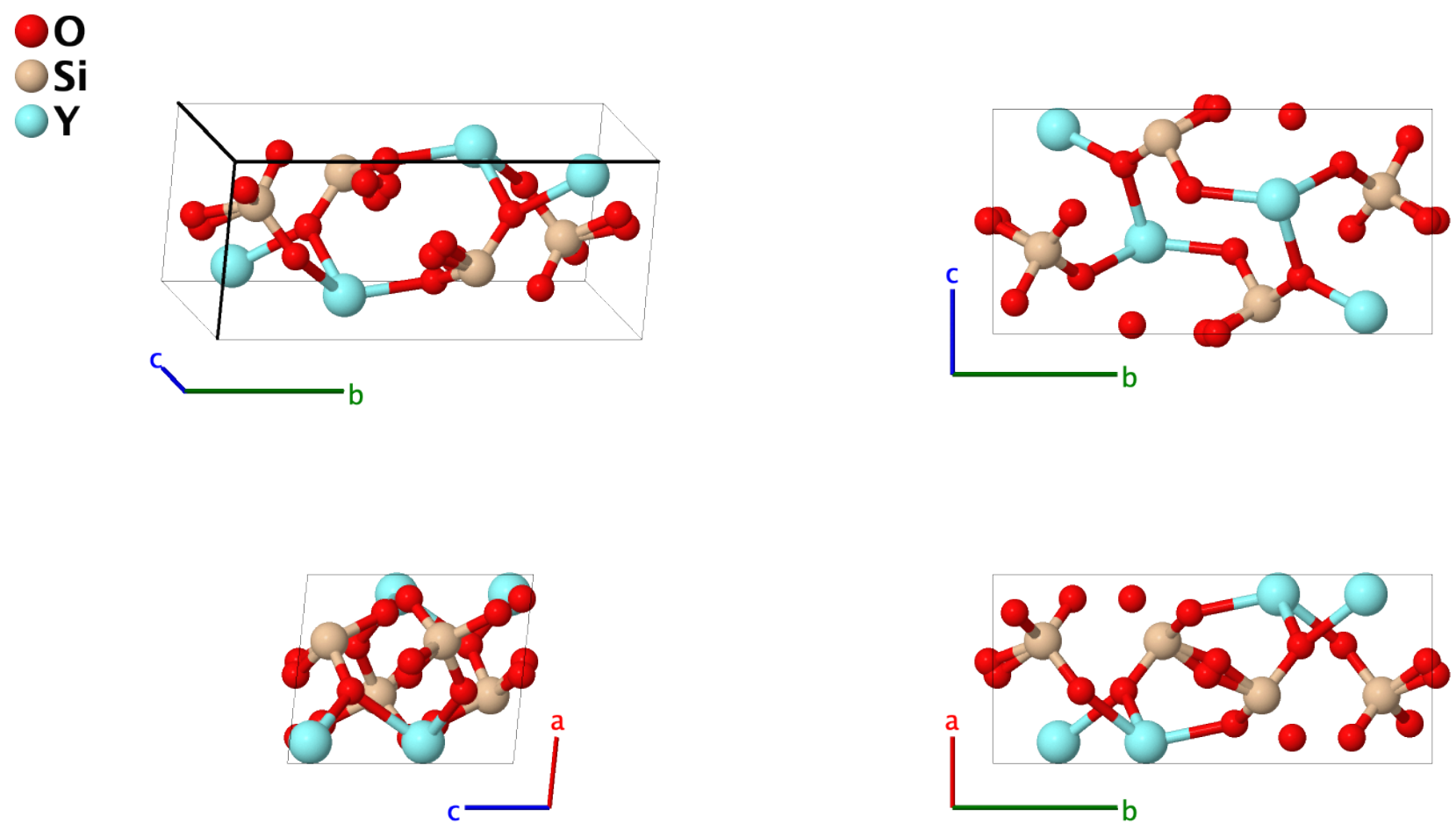

| | | |
|---|---|---|
| **Prototype** | : | $O_7Si_2Y_2$ |
| **AFLOW prototype label** | : | A4BC_mP24_14_4e_e_e |
| ***Strukturbericht* designation** | : | None |
| **Pearson symbol** | : | mP24 |
| **Space group number** | : | 14 |
| **Space group symbol** | : | $P2_1/c$ |
| **AFLOW prototype command** | : | `aflow --proto=A4BC_mP24_14_4e_e_e` `--params=`$a, b/a, c/a, \beta, x_1, y_1, z_1, x_2, y_2, z_2, x_3, y_3, z_3, x_4, y_4, z_4, x_5, y_5, z_5, x_6, y_6, z_6$ |

### Other compounds with this structure

- $\gamma$-$Er_2Si_2O_7$ and $\gamma$-$Ho_2Si_2O_7$

---

- The O-III site is only 50% occupied. The Jmol image shows the two possible sites as a pair of very closely spaced oxygen atoms.
- (Christensen, 1997) refer to this as $D$-$Y_2Si_2O_7$, but we follow the classificatio of the $RE_2Si_2O_7$ structures in (Becerro, 2004) and refer to this as $\gamma$-$Y_2Si_2O_7$.

### Simple Monoclinic primitive vectors:

$$\mathbf{a}_1 = a\,\hat{\mathbf{x}}$$
$$\mathbf{a}_2 = b\,\hat{\mathbf{y}}$$
$$\mathbf{a}_3 = c\cos\beta\,\hat{\mathbf{x}} + c\sin\beta\,\hat{\mathbf{z}}$$

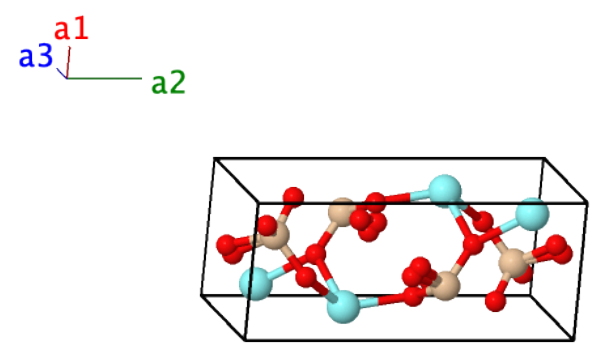

### Basis vectors:

Lattice Coordinates | Cartesian Coordinates | Wyckoff Position | Atom Type

| | | | | | | |
|---|---|---|---|---|---|---|
| $\mathbf{B}_{1}$ | = | $x_{1}\,\mathbf{a}_{1}+y_{1}\,\mathbf{a}_{2}+z_{1}\,\mathbf{a}_{3}$ | = | $(x_{1}a+z_{1}c\cos\beta)\,\mathbf{\hat{x}}+y_{1}b\,\mathbf{\hat{y}}+z_{1}c\sin\beta\,\mathbf{\hat{z}}$ | (4$e$) | O I |
| $\mathbf{B}_{2}$ | = | $-x_{1}\,\mathbf{a}_{1}+\left(\frac{1}{2}+y_{1}\right)\,\mathbf{a}_{2}+\left(\frac{1}{2}-z_{1}\right)\,\mathbf{a}_{3}$ | = | $\left(\frac{1}{2}c\cos\beta-x_{1}a-z_{1}c\cos\beta\right)\,\mathbf{\hat{x}}+\left(\frac{1}{2}+y_{1}\right)b\,\mathbf{\hat{y}}+\left(\frac{1}{2}-z_{1}\right)c\sin\beta\,\mathbf{\hat{z}}$ | (4$e$) | O I |
| $\mathbf{B}_{3}$ | = | $-x_{1}\,\mathbf{a}_{1}-y_{1}\,\mathbf{a}_{2}-z_{1}\,\mathbf{a}_{3}$ | = | $(-x_{1}a-z_{1}c\cos\beta)\,\mathbf{\hat{x}}-y_{1}b\,\mathbf{\hat{y}}-z_{1}c\sin\beta\,\mathbf{\hat{z}}$ | (4$e$) | O I |
| $\mathbf{B}_{4}$ | = | $x_{1}\,\mathbf{a}_{1}+\left(\frac{1}{2}-y_{1}\right)\,\mathbf{a}_{2}+\left(\frac{1}{2}+z_{1}\right)\,\mathbf{a}_{3}$ | = | $\left(\frac{1}{2}c\cos\beta+x_{1}a+z_{1}c\cos\beta\right)\,\mathbf{\hat{x}}+\left(\frac{1}{2}-y_{1}\right)b\,\mathbf{\hat{y}}+\left(\frac{1}{2}+z_{1}\right)c\sin\beta\,\mathbf{\hat{z}}$ | (4$e$) | O I |
| $\mathbf{B}_{5}$ | = | $x_{2}\,\mathbf{a}_{1}+y_{2}\,\mathbf{a}_{2}+z_{2}\,\mathbf{a}_{3}$ | = | $(x_{2}a+z_{2}c\cos\beta)\,\mathbf{\hat{x}}+y_{2}b\,\mathbf{\hat{y}}+z_{2}c\sin\beta\,\mathbf{\hat{z}}$ | (4$e$) | O II |
| $\mathbf{B}_{6}$ | = | $-x_{2}\,\mathbf{a}_{1}+\left(\frac{1}{2}+y_{2}\right)\,\mathbf{a}_{2}+\left(\frac{1}{2}-z_{2}\right)\,\mathbf{a}_{3}$ | = | $\left(\frac{1}{2}c\cos\beta-x_{2}a-z_{2}c\cos\beta\right)\,\mathbf{\hat{x}}+\left(\frac{1}{2}+y_{2}\right)b\,\mathbf{\hat{y}}+\left(\frac{1}{2}-z_{2}\right)c\sin\beta\,\mathbf{\hat{z}}$ | (4$e$) | O II |
| $\mathbf{B}_{7}$ | = | $-x_{2}\,\mathbf{a}_{1}-y_{2}\,\mathbf{a}_{2}-z_{2}\,\mathbf{a}_{3}$ | = | $(-x_{2}a-z_{2}c\cos\beta)\,\mathbf{\hat{x}}-y_{2}b\,\mathbf{\hat{y}}-z_{2}c\sin\beta\,\mathbf{\hat{z}}$ | (4$e$) | O II |
| $\mathbf{B}_{8}$ | = | $x_{2}\,\mathbf{a}_{1}+\left(\frac{1}{2}-y_{2}\right)\,\mathbf{a}_{2}+\left(\frac{1}{2}+z_{2}\right)\,\mathbf{a}_{3}$ | = | $\left(\frac{1}{2}c\cos\beta+x_{2}a+z_{2}c\cos\beta\right)\,\mathbf{\hat{x}}+\left(\frac{1}{2}-y_{2}\right)b\,\mathbf{\hat{y}}+\left(\frac{1}{2}+z_{2}\right)c\sin\beta\,\mathbf{\hat{z}}$ | (4$e$) | O II |
| $\mathbf{B}_{9}$ | = | $x_{3}\,\mathbf{a}_{1}+y_{3}\,\mathbf{a}_{2}+z_{3}\,\mathbf{a}_{3}$ | = | $(x_{3}a+z_{3}c\cos\beta)\,\mathbf{\hat{x}}+y_{3}b\,\mathbf{\hat{y}}+z_{3}c\sin\beta\,\mathbf{\hat{z}}$ | (4$e$) | O III |
| $\mathbf{B}_{10}$ | = | $-x_{3}\,\mathbf{a}_{1}+\left(\frac{1}{2}+y_{3}\right)\,\mathbf{a}_{2}+\left(\frac{1}{2}-z_{3}\right)\,\mathbf{a}_{3}$ | = | $\left(\frac{1}{2}c\cos\beta-x_{3}a-z_{3}c\cos\beta\right)\,\mathbf{\hat{x}}+\left(\frac{1}{2}+y_{3}\right)b\,\mathbf{\hat{y}}+\left(\frac{1}{2}-z_{3}\right)c\sin\beta\,\mathbf{\hat{z}}$ | (4$e$) | O III |
| $\mathbf{B}_{11}$ | = | $-x_{3}\,\mathbf{a}_{1}-y_{3}\,\mathbf{a}_{2}-z_{3}\,\mathbf{a}_{3}$ | = | $(-x_{3}a-z_{3}c\cos\beta)\,\mathbf{\hat{x}}-y_{3}b\,\mathbf{\hat{y}}-z_{3}c\sin\beta\,\mathbf{\hat{z}}$ | (4$e$) | O III |
| $\mathbf{B}_{12}$ | = | $x_{3}\,\mathbf{a}_{1}+\left(\frac{1}{2}-y_{3}\right)\,\mathbf{a}_{2}+\left(\frac{1}{2}+z_{3}\right)\,\mathbf{a}_{3}$ | = | $\left(\frac{1}{2}c\cos\beta+x_{3}a+z_{3}c\cos\beta\right)\,\mathbf{\hat{x}}+\left(\frac{1}{2}-y_{3}\right)b\,\mathbf{\hat{y}}+\left(\frac{1}{2}+z_{3}\right)c\sin\beta\,\mathbf{\hat{z}}$ | (4$e$) | O III |
| $\mathbf{B}_{13}$ | = | $x_{4}\,\mathbf{a}_{1}+y_{4}\,\mathbf{a}_{2}+z_{4}\,\mathbf{a}_{3}$ | = | $(x_{4}a+z_{4}c\cos\beta)\,\mathbf{\hat{x}}+y_{4}b\,\mathbf{\hat{y}}+z_{4}c\sin\beta\,\mathbf{\hat{z}}$ | (4$e$) | O IV |
| $\mathbf{B}_{14}$ | = | $-x_{4}\,\mathbf{a}_{1}+\left(\frac{1}{2}+y_{4}\right)\,\mathbf{a}_{2}+\left(\frac{1}{2}-z_{4}\right)\,\mathbf{a}_{3}$ | = | $\left(\frac{1}{2}c\cos\beta-x_{4}a-z_{4}c\cos\beta\right)\,\mathbf{\hat{x}}+\left(\frac{1}{2}+y_{4}\right)b\,\mathbf{\hat{y}}+\left(\frac{1}{2}-z_{4}\right)c\sin\beta\,\mathbf{\hat{z}}$ | (4$e$) | O IV |
| $\mathbf{B}_{15}$ | = | $-x_{4}\,\mathbf{a}_{1}-y_{4}\,\mathbf{a}_{2}-z_{4}\,\mathbf{a}_{3}$ | = | $(-x_{4}a-z_{4}c\cos\beta)\,\mathbf{\hat{x}}-y_{4}b\,\mathbf{\hat{y}}-z_{4}c\sin\beta\,\mathbf{\hat{z}}$ | (4$e$) | O IV |
| $\mathbf{B}_{16}$ | = | $x_{4}\,\mathbf{a}_{1}+\left(\frac{1}{2}-y_{4}\right)\,\mathbf{a}_{2}+\left(\frac{1}{2}+z_{4}\right)\,\mathbf{a}_{3}$ | = | $\left(\frac{1}{2}c\cos\beta+x_{4}a+z_{4}c\cos\beta\right)\,\mathbf{\hat{x}}+\left(\frac{1}{2}-y_{4}\right)b\,\mathbf{\hat{y}}+\left(\frac{1}{2}+z_{4}\right)c\sin\beta\,\mathbf{\hat{z}}$ | (4$e$) | O IV |
| $\mathbf{B}_{17}$ | = | $x_{5}\,\mathbf{a}_{1}+y_{5}\,\mathbf{a}_{2}+z_{5}\,\mathbf{a}_{3}$ | = | $(x_{5}a+z_{5}c\cos\beta)\,\mathbf{\hat{x}}+y_{5}b\,\mathbf{\hat{y}}+z_{5}c\sin\beta\,\mathbf{\hat{z}}$ | (4$e$) | Si |
| $\mathbf{B}_{18}$ | = | $-x_{5}\,\mathbf{a}_{1}+\left(\frac{1}{2}+y_{5}\right)\,\mathbf{a}_{2}+\left(\frac{1}{2}-z_{5}\right)\,\mathbf{a}_{3}$ | = | $\left(\frac{1}{2}c\cos\beta-x_{5}a-z_{5}c\cos\beta\right)\,\mathbf{\hat{x}}+\left(\frac{1}{2}+y_{5}\right)b\,\mathbf{\hat{y}}+\left(\frac{1}{2}-z_{5}\right)c\sin\beta\,\mathbf{\hat{z}}$ | (4$e$) | Si |
| $\mathbf{B}_{19}$ | = | $-x_{5}\,\mathbf{a}_{1}-y_{5}\,\mathbf{a}_{2}-z_{5}\,\mathbf{a}_{3}$ | = | $(-x_{5}a-z_{5}c\cos\beta)\,\mathbf{\hat{x}}-y_{5}b\,\mathbf{\hat{y}}-z_{5}c\sin\beta\,\mathbf{\hat{z}}$ | (4$e$) | Si |
| $\mathbf{B}_{20}$ | = | $x_{5}\,\mathbf{a}_{1}+\left(\frac{1}{2}-y_{5}\right)\,\mathbf{a}_{2}+\left(\frac{1}{2}+z_{5}\right)\,\mathbf{a}_{3}$ | = | $\left(\frac{1}{2}c\cos\beta+x_{5}a+z_{5}c\cos\beta\right)\,\mathbf{\hat{x}}+\left(\frac{1}{2}-y_{5}\right)b\,\mathbf{\hat{y}}+\left(\frac{1}{2}+z_{5}\right)c\sin\beta\,\mathbf{\hat{z}}$ | (4$e$) | Si |
| $\mathbf{B}_{21}$ | = | $x_{6}\,\mathbf{a}_{1}+y_{6}\,\mathbf{a}_{2}+z_{6}\,\mathbf{a}_{3}$ | = | $(x_{6}a+z_{6}c\cos\beta)\,\mathbf{\hat{x}}+y_{6}b\,\mathbf{\hat{y}}+z_{6}c\sin\beta\,\mathbf{\hat{z}}$ | (4$e$) | Y |
| $\mathbf{B}_{22}$ | = | $-x_{6}\,\mathbf{a}_{1}+\left(\frac{1}{2}+y_{6}\right)\,\mathbf{a}_{2}+\left(\frac{1}{2}-z_{6}\right)\,\mathbf{a}_{3}$ | = | $\left(\frac{1}{2}c\cos\beta-x_{6}a-z_{6}c\cos\beta\right)\,\mathbf{\hat{x}}+\left(\frac{1}{2}+y_{6}\right)b\,\mathbf{\hat{y}}+\left(\frac{1}{2}-z_{6}\right)c\sin\beta\,\mathbf{\hat{z}}$ | (4$e$) | Y |
| $\mathbf{B}_{23}$ | = | $-x_6\,\mathbf{a}_1 - y_6\,\mathbf{a}_2 - z_6\,\mathbf{a}_3$ | = | $(-x_6 a - z_6 c\cos\beta)\,\hat{\mathbf{x}} - y_6 b\,\hat{\mathbf{y}} - z_6 c\sin\beta\,\hat{\mathbf{z}}$ | (4$e$) | Y |
| $\mathbf{B}_{24}$ | = | $x_6\,\mathbf{a}_1 + \left(\frac{1}{2} - y_6\right)\mathbf{a}_2 + \left(\frac{1}{2} + z_6\right)\mathbf{a}_3$ | = | $\left(\frac{1}{2}c\cos\beta + x_6 a + z_6 c\cos\beta\right)\hat{\mathbf{x}} + \left(\frac{1}{2} - y_6\right)b\,\hat{\mathbf{y}} + \left(\frac{1}{2} + z_6\right)c\sin\beta\,\hat{\mathbf{z}}$ | (4$e$) | Y |

**Geometry files:**

- CIF: pp. 1544
- POSCAR: pp. 1545

# $K_2Pt(SCN)_6 \cdot 2H_2O$ Structure: A6B4C2D6E2FG6_mP54_14_3e_2e_e_3e_e_a_3e

http://aflow.org/prototype-encyclopedia/A6B4C2D6E2FG6_mP54_14_3e_2e_e_3e_e_a_3e

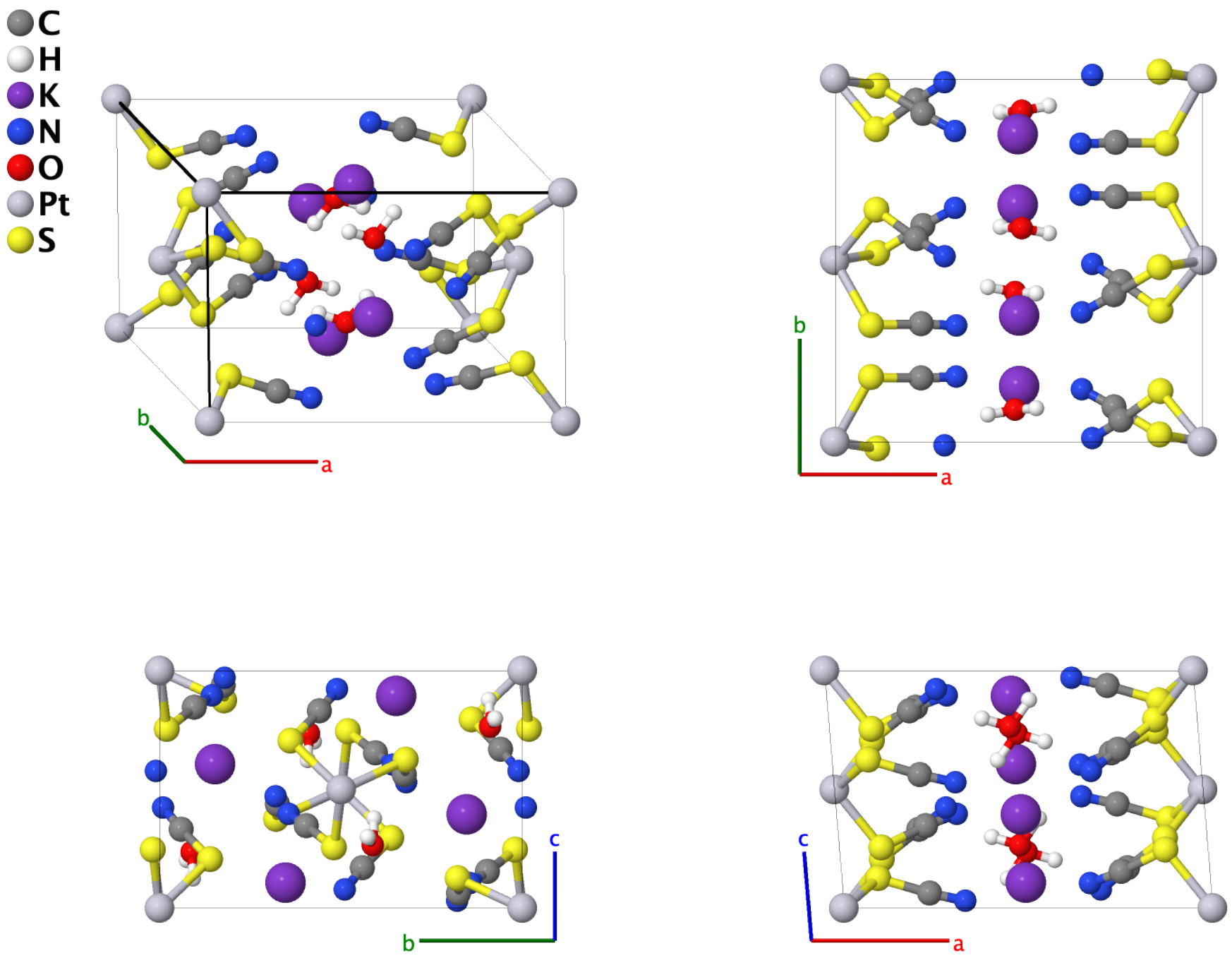

| | | |
|---|---|---|
| **Prototype** | : | $C_6H_4K_2N_6O_2PtS_6$ |
| **AFLOW prototype label** | : | A6B4C2D6E2FG6_mP54_14_3e_2e_e_3e_e_a_3e |
| ***Strukturbericht* designation** | : | None |
| **Pearson symbol** | : | mP54 |
| **Space group number** | : | 14 |
| **Space group symbol** | : | $P2_1/c$ |
| **AFLOW prototype command** | : | `aflow --proto=A6B4C2D6E2FG6_mP54_14_3e_2e_e_3e_e_a_3e --params=`$a, b/a, c/a, \beta, x_2, y_2, z_2, x_3, y_3, z_3, x_4, y_4, z_4, x_5, y_5, z_5, x_6, y_6, z_6, x_7, y_7, z_7, x_8, y_8, z_8, x_9, y_9, z_9, x_{10}, y_{10}, z_{10}, x_{11}, y_{11}, z_{11}, x_{12}, y_{12}, z_{12}, x_{13}, y_{13}, z_{13}, x_{14}, y_{14}, z_{14}$ |

### Other compounds with this structure

- $Rb_2Pt(SCN)_6 \cdot 2H_2O$ and $NH_{42}Pt(SCN)_6 \cdot 2H_2O$

- This is the hydrated form. For the anhydrous structure, see the $H6_3$ structure page.

**Simple Monoclinic primitive vectors:**

$$\mathbf{a}_1 = a\,\hat{\mathbf{x}}$$
$$\mathbf{a}_2 = b\,\hat{\mathbf{y}}$$
$$\mathbf{a}_3 = c\cos\beta\,\hat{\mathbf{x}} + c\sin\beta\,\hat{\mathbf{z}}$$

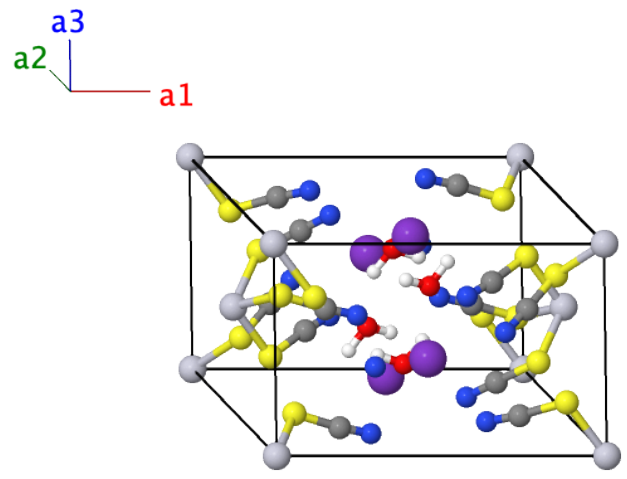

**Basis vectors:**

| | | Lattice Coordinates | | Cartesian Coordinates | Wyckoff Position | Atom Type |
|---|---|---|---|---|---|---|
| $\mathbf{B}_1$ | = | $0\,\mathbf{a}_1 + 0\,\mathbf{a}_2 + 0\,\mathbf{a}_3$ | = | $0\,\hat{\mathbf{x}} + 0\,\hat{\mathbf{y}} + 0\,\hat{\mathbf{z}}$ | (2a) | Pt |
| $\mathbf{B}_2$ | = | $\frac{1}{2}\,\mathbf{a}_2 + \frac{1}{2}\,\mathbf{a}_3$ | = | $\frac{1}{2}c\cos\beta\,\hat{\mathbf{x}} + \frac{1}{2}b\,\hat{\mathbf{y}} + \frac{1}{2}c\sin\beta\,\hat{\mathbf{z}}$ | (2a) | Pt |
| $\mathbf{B}_3$ | = | $x_2\,\mathbf{a}_1 + y_2\,\mathbf{a}_2 + z_2\,\mathbf{a}_3$ | = | $(x_2a + z_2c\cos\beta)\,\hat{\mathbf{x}} + y_2b\,\hat{\mathbf{y}} + z_2c\sin\beta\,\hat{\mathbf{z}}$ | (4e) | C I |
| $\mathbf{B}_4$ | = | $-x_2\,\mathbf{a}_1 + \left(\frac{1}{2} + y_2\right)\mathbf{a}_2 + \left(\frac{1}{2} - z_2\right)\mathbf{a}_3$ | = | $\left(\frac{1}{2}c\cos\beta - x_2a - z_2c\cos\beta\right)\hat{\mathbf{x}} + \left(\frac{1}{2} + y_2\right)b\,\hat{\mathbf{y}} + \left(\frac{1}{2} - z_2\right)c\sin\beta\,\hat{\mathbf{z}}$ | (4e) | C I |
| $\mathbf{B}_5$ | = | $-x_2\,\mathbf{a}_1 - y_2\,\mathbf{a}_2 - z_2\,\mathbf{a}_3$ | = | $(-x_2a - z_2c\cos\beta)\,\hat{\mathbf{x}} - y_2b\,\hat{\mathbf{y}} - z_2c\sin\beta\,\hat{\mathbf{z}}$ | (4e) | C I |
| $\mathbf{B}_6$ | = | $x_2\,\mathbf{a}_1 + \left(\frac{1}{2} - y_2\right)\mathbf{a}_2 + \left(\frac{1}{2} + z_2\right)\mathbf{a}_3$ | = | $\left(\frac{1}{2}c\cos\beta + x_2a + z_2c\cos\beta\right)\hat{\mathbf{x}} + \left(\frac{1}{2} - y_2\right)b\,\hat{\mathbf{y}} + \left(\frac{1}{2} + z_2\right)c\sin\beta\,\hat{\mathbf{z}}$ | (4e) | C I |
| $\mathbf{B}_7$ | = | $x_3\,\mathbf{a}_1 + y_3\,\mathbf{a}_2 + z_3\,\mathbf{a}_3$ | = | $(x_3a + z_3c\cos\beta)\,\hat{\mathbf{x}} + y_3b\,\hat{\mathbf{y}} + z_3c\sin\beta\,\hat{\mathbf{z}}$ | (4e) | C II |
| $\mathbf{B}_8$ | = | $-x_3\,\mathbf{a}_1 + \left(\frac{1}{2} + y_3\right)\mathbf{a}_2 + \left(\frac{1}{2} - z_3\right)\mathbf{a}_3$ | = | $\left(\frac{1}{2}c\cos\beta - x_3a - z_3c\cos\beta\right)\hat{\mathbf{x}} + \left(\frac{1}{2} + y_3\right)b\,\hat{\mathbf{y}} + \left(\frac{1}{2} - z_3\right)c\sin\beta\,\hat{\mathbf{z}}$ | (4e) | C II |
| $\mathbf{B}_9$ | = | $-x_3\,\mathbf{a}_1 - y_3\,\mathbf{a}_2 - z_3\,\mathbf{a}_3$ | = | $(-x_3a - z_3c\cos\beta)\,\hat{\mathbf{x}} - y_3b\,\hat{\mathbf{y}} - z_3c\sin\beta\,\hat{\mathbf{z}}$ | (4e) | C II |
| $\mathbf{B}_{10}$ | = | $x_3\,\mathbf{a}_1 + \left(\frac{1}{2} - y_3\right)\mathbf{a}_2 + \left(\frac{1}{2} + z_3\right)\mathbf{a}_3$ | = | $\left(\frac{1}{2}c\cos\beta + x_3a + z_3c\cos\beta\right)\hat{\mathbf{x}} + \left(\frac{1}{2} - y_3\right)b\,\hat{\mathbf{y}} + \left(\frac{1}{2} + z_3\right)c\sin\beta\,\hat{\mathbf{z}}$ | (4e) | C II |
| $\mathbf{B}_{11}$ | = | $x_4\,\mathbf{a}_1 + y_4\,\mathbf{a}_2 + z_4\,\mathbf{a}_3$ | = | $(x_4a + z_4c\cos\beta)\,\hat{\mathbf{x}} + y_4b\,\hat{\mathbf{y}} + z_4c\sin\beta\,\hat{\mathbf{z}}$ | (4e) | C III |
| $\mathbf{B}_{12}$ | = | $-x_4\,\mathbf{a}_1 + \left(\frac{1}{2} + y_4\right)\mathbf{a}_2 + \left(\frac{1}{2} - z_4\right)\mathbf{a}_3$ | = | $\left(\frac{1}{2}c\cos\beta - x_4a - z_4c\cos\beta\right)\hat{\mathbf{x}} + \left(\frac{1}{2} + y_4\right)b\,\hat{\mathbf{y}} + \left(\frac{1}{2} - z_4\right)c\sin\beta\,\hat{\mathbf{z}}$ | (4e) | C III |
| $\mathbf{B}_{13}$ | = | $-x_4\,\mathbf{a}_1 - y_4\,\mathbf{a}_2 - z_4\,\mathbf{a}_3$ | = | $(-x_4a - z_4c\cos\beta)\,\hat{\mathbf{x}} - y_4b\,\hat{\mathbf{y}} - z_4c\sin\beta\,\hat{\mathbf{z}}$ | (4e) | C III |
| $\mathbf{B}_{14}$ | = | $x_4\,\mathbf{a}_1 + \left(\frac{1}{2} - y_4\right)\mathbf{a}_2 + \left(\frac{1}{2} + z_4\right)\mathbf{a}_3$ | = | $\left(\frac{1}{2}c\cos\beta + x_4a + z_4c\cos\beta\right)\hat{\mathbf{x}} + \left(\frac{1}{2} - y_4\right)b\,\hat{\mathbf{y}} + \left(\frac{1}{2} + z_4\right)c\sin\beta\,\hat{\mathbf{z}}$ | (4e) | C III |
| $\mathbf{B}_{15}$ | = | $x_5\,\mathbf{a}_1 + y_5\,\mathbf{a}_2 + z_5\,\mathbf{a}_3$ | = | $(x_5a + z_5c\cos\beta)\,\hat{\mathbf{x}} + y_5b\,\hat{\mathbf{y}} + z_5c\sin\beta\,\hat{\mathbf{z}}$ | (4e) | H I |
| $\mathbf{B}_{16}$ | = | $-x_5\,\mathbf{a}_1 + \left(\frac{1}{2} + y_5\right)\mathbf{a}_2 + \left(\frac{1}{2} - z_5\right)\mathbf{a}_3$ | = | $\left(\frac{1}{2}c\cos\beta - x_5a - z_5c\cos\beta\right)\hat{\mathbf{x}} + \left(\frac{1}{2} + y_5\right)b\,\hat{\mathbf{y}} + \left(\frac{1}{2} - z_5\right)c\sin\beta\,\hat{\mathbf{z}}$ | (4e) | H I |
| $\mathbf{B}_{17}$ | = | $-x_5\,\mathbf{a}_1 - y_5\,\mathbf{a}_2 - z_5\,\mathbf{a}_3$ | = | $(-x_5a - z_5c\cos\beta)\,\hat{\mathbf{x}} - y_5b\,\hat{\mathbf{y}} - z_5c\sin\beta\,\hat{\mathbf{z}}$ | (4e) | H I |
| $\mathbf{B}_{18}$ | = | $x_5\,\mathbf{a}_1 + \left(\frac{1}{2} - y_5\right)\,\mathbf{a}_2 + \left(\frac{1}{2} + z_5\right)\,\mathbf{a}_3$ | = | $\left(\frac{1}{2}c\cos\beta + x_5a + z_5c\cos\beta\right)\,\mathbf{\hat{x}} + \left(\frac{1}{2} - y_5\right)b\,\mathbf{\hat{y}} + \left(\frac{1}{2} + z_5\right)c\sin\beta\,\mathbf{\hat{z}}$ | (4$e$) | H I |
| $\mathbf{B}_{19}$ | = | $x_6\,\mathbf{a}_1 + y_6\,\mathbf{a}_2 + z_6\,\mathbf{a}_3$ | = | $(x_6a + z_6c\cos\beta)\,\mathbf{\hat{x}} + y_6b\,\mathbf{\hat{y}} + z_6c\sin\beta\,\mathbf{\hat{z}}$ | (4$e$) | H II |
| $\mathbf{B}_{20}$ | = | $-x_6\,\mathbf{a}_1 + \left(\frac{1}{2} + y_6\right)\,\mathbf{a}_2 + \left(\frac{1}{2} - z_6\right)\,\mathbf{a}_3$ | = | $\left(\frac{1}{2}c\cos\beta - x_6a - z_6c\cos\beta\right)\,\mathbf{\hat{x}} + \left(\frac{1}{2} + y_6\right)b\,\mathbf{\hat{y}} + \left(\frac{1}{2} - z_6\right)c\sin\beta\,\mathbf{\hat{z}}$ | (4$e$) | H II |
| $\mathbf{B}_{21}$ | = | $-x_6\,\mathbf{a}_1 - y_6\,\mathbf{a}_2 - z_6\,\mathbf{a}_3$ | = | $(-x_6a - z_6c\cos\beta)\,\mathbf{\hat{x}} - y_6b\,\mathbf{\hat{y}} - z_6c\sin\beta\,\mathbf{\hat{z}}$ | (4$e$) | H II |
| $\mathbf{B}_{22}$ | = | $x_6\,\mathbf{a}_1 + \left(\frac{1}{2} - y_6\right)\,\mathbf{a}_2 + \left(\frac{1}{2} + z_6\right)\,\mathbf{a}_3$ | = | $\left(\frac{1}{2}c\cos\beta + x_6a + z_6c\cos\beta\right)\,\mathbf{\hat{x}} + \left(\frac{1}{2} - y_6\right)b\,\mathbf{\hat{y}} + \left(\frac{1}{2} + z_6\right)c\sin\beta\,\mathbf{\hat{z}}$ | (4$e$) | H II |
| $\mathbf{B}_{23}$ | = | $x_7\,\mathbf{a}_1 + y_7\,\mathbf{a}_2 + z_7\,\mathbf{a}_3$ | = | $(x_7a + z_7c\cos\beta)\,\mathbf{\hat{x}} + y_7b\,\mathbf{\hat{y}} + z_7c\sin\beta\,\mathbf{\hat{z}}$ | (4$e$) | K |
| $\mathbf{B}_{24}$ | = | $-x_7\,\mathbf{a}_1 + \left(\frac{1}{2} + y_7\right)\,\mathbf{a}_2 + \left(\frac{1}{2} - z_7\right)\,\mathbf{a}_3$ | = | $\left(\frac{1}{2}c\cos\beta - x_7a - z_7c\cos\beta\right)\,\mathbf{\hat{x}} + \left(\frac{1}{2} + y_7\right)b\,\mathbf{\hat{y}} + \left(\frac{1}{2} - z_7\right)c\sin\beta\,\mathbf{\hat{z}}$ | (4$e$) | K |
| $\mathbf{B}_{25}$ | = | $-x_7\,\mathbf{a}_1 - y_7\,\mathbf{a}_2 - z_7\,\mathbf{a}_3$ | = | $(-x_7a - z_7c\cos\beta)\,\mathbf{\hat{x}} - y_7b\,\mathbf{\hat{y}} - z_7c\sin\beta\,\mathbf{\hat{z}}$ | (4$e$) | K |
| $\mathbf{B}_{26}$ | = | $x_7\,\mathbf{a}_1 + \left(\frac{1}{2} - y_7\right)\,\mathbf{a}_2 + \left(\frac{1}{2} + z_7\right)\,\mathbf{a}_3$ | = | $\left(\frac{1}{2}c\cos\beta + x_7a + z_7c\cos\beta\right)\,\mathbf{\hat{x}} + \left(\frac{1}{2} - y_7\right)b\,\mathbf{\hat{y}} + \left(\frac{1}{2} + z_7\right)c\sin\beta\,\mathbf{\hat{z}}$ | (4$e$) | K |
| $\mathbf{B}_{27}$ | = | $x_8\,\mathbf{a}_1 + y_8\,\mathbf{a}_2 + z_8\,\mathbf{a}_3$ | = | $(x_8a + z_8c\cos\beta)\,\mathbf{\hat{x}} + y_8b\,\mathbf{\hat{y}} + z_8c\sin\beta\,\mathbf{\hat{z}}$ | (4$e$) | N I |
| $\mathbf{B}_{28}$ | = | $-x_8\,\mathbf{a}_1 + \left(\frac{1}{2} + y_8\right)\,\mathbf{a}_2 + \left(\frac{1}{2} - z_8\right)\,\mathbf{a}_3$ | = | $\left(\frac{1}{2}c\cos\beta - x_8a - z_8c\cos\beta\right)\,\mathbf{\hat{x}} + \left(\frac{1}{2} + y_8\right)b\,\mathbf{\hat{y}} + \left(\frac{1}{2} - z_8\right)c\sin\beta\,\mathbf{\hat{z}}$ | (4$e$) | N I |
| $\mathbf{B}_{29}$ | = | $-x_8\,\mathbf{a}_1 - y_8\,\mathbf{a}_2 - z_8\,\mathbf{a}_3$ | = | $(-x_8a - z_8c\cos\beta)\,\mathbf{\hat{x}} - y_8b\,\mathbf{\hat{y}} - z_8c\sin\beta\,\mathbf{\hat{z}}$ | (4$e$) | N I |
| $\mathbf{B}_{30}$ | = | $x_8\,\mathbf{a}_1 + \left(\frac{1}{2} - y_8\right)\,\mathbf{a}_2 + \left(\frac{1}{2} + z_8\right)\,\mathbf{a}_3$ | = | $\left(\frac{1}{2}c\cos\beta + x_8a + z_8c\cos\beta\right)\,\mathbf{\hat{x}} + \left(\frac{1}{2} - y_8\right)b\,\mathbf{\hat{y}} + \left(\frac{1}{2} + z_8\right)c\sin\beta\,\mathbf{\hat{z}}$ | (4$e$) | N I |
| $\mathbf{B}_{31}$ | = | $x_9\,\mathbf{a}_1 + y_9\,\mathbf{a}_2 + z_9\,\mathbf{a}_3$ | = | $(x_9a + z_9c\cos\beta)\,\mathbf{\hat{x}} + y_9b\,\mathbf{\hat{y}} + z_9c\sin\beta\,\mathbf{\hat{z}}$ | (4$e$) | N II |
| $\mathbf{B}_{32}$ | = | $-x_9\,\mathbf{a}_1 + \left(\frac{1}{2} + y_9\right)\,\mathbf{a}_2 + \left(\frac{1}{2} - z_9\right)\,\mathbf{a}_3$ | = | $\left(\frac{1}{2}c\cos\beta - x_9a - z_9c\cos\beta\right)\,\mathbf{\hat{x}} + \left(\frac{1}{2} + y_9\right)b\,\mathbf{\hat{y}} + \left(\frac{1}{2} - z_9\right)c\sin\beta\,\mathbf{\hat{z}}$ | (4$e$) | N II |
| $\mathbf{B}_{33}$ | = | $-x_9\,\mathbf{a}_1 - y_9\,\mathbf{a}_2 - z_9\,\mathbf{a}_3$ | = | $(-x_9a - z_9c\cos\beta)\,\mathbf{\hat{x}} - y_9b\,\mathbf{\hat{y}} - z_9c\sin\beta\,\mathbf{\hat{z}}$ | (4$e$) | N II |
| $\mathbf{B}_{34}$ | = | $x_9\,\mathbf{a}_1 + \left(\frac{1}{2} - y_9\right)\,\mathbf{a}_2 + \left(\frac{1}{2} + z_9\right)\,\mathbf{a}_3$ | = | $\left(\frac{1}{2}c\cos\beta + x_9a + z_9c\cos\beta\right)\,\mathbf{\hat{x}} + \left(\frac{1}{2} - y_9\right)b\,\mathbf{\hat{y}} + \left(\frac{1}{2} + z_9\right)c\sin\beta\,\mathbf{\hat{z}}$ | (4$e$) | N II |
| $\mathbf{B}_{35}$ | = | $x_{10}\,\mathbf{a}_1 + y_{10}\,\mathbf{a}_2 + z_{10}\,\mathbf{a}_3$ | = | $(x_{10}a + z_{10}c\cos\beta)\,\mathbf{\hat{x}} + y_{10}b\,\mathbf{\hat{y}} + z_{10}c\sin\beta\,\mathbf{\hat{z}}$ | (4$e$) | N III |
| $\mathbf{B}_{36}$ | = | $-x_{10}\,\mathbf{a}_1 + \left(\frac{1}{2} + y_{10}\right)\,\mathbf{a}_2 + \left(\frac{1}{2} - z_{10}\right)\,\mathbf{a}_3$ | = | $\left(\frac{1}{2}c\cos\beta - x_{10}a - z_{10}c\cos\beta\right)\,\mathbf{\hat{x}} + \left(\frac{1}{2} + y_{10}\right)b\,\mathbf{\hat{y}} + \left(\frac{1}{2} - z_{10}\right)c\sin\beta\,\mathbf{\hat{z}}$ | (4$e$) | N III |
| $\mathbf{B}_{37}$ | = | $-x_{10}\,\mathbf{a}_1 - y_{10}\,\mathbf{a}_2 - z_{10}\,\mathbf{a}_3$ | = | $(-x_{10}a - z_{10}c\cos\beta)\,\mathbf{\hat{x}} - y_{10}b\,\mathbf{\hat{y}} - z_{10}c\sin\beta\,\mathbf{\hat{z}}$ | (4$e$) | N III |
| $\mathbf{B}_{38}$ | = | $x_{10}\,\mathbf{a}_1 + \left(\frac{1}{2} - y_{10}\right)\,\mathbf{a}_2 + \left(\frac{1}{2} + z_{10}\right)\,\mathbf{a}_3$ | = | $\left(\frac{1}{2}c\cos\beta + x_{10}a + z_{10}c\cos\beta\right)\,\mathbf{\hat{x}} + \left(\frac{1}{2} - y_{10}\right)b\,\mathbf{\hat{y}} + \left(\frac{1}{2} + z_{10}\right)c\sin\beta\,\mathbf{\hat{z}}$ | (4$e$) | N III |
| $\mathbf{B}_{39}$ | = | $x_{11}\,\mathbf{a}_1 + y_{11}\,\mathbf{a}_2 + z_{11}\,\mathbf{a}_3$ | = | $(x_{11}a + z_{11}c\cos\beta)\,\mathbf{\hat{x}} + y_{11}b\,\mathbf{\hat{y}} + z_{11}c\sin\beta\,\mathbf{\hat{z}}$ | (4$e$) | O |
| $\mathbf{B}_{40}$ | = | $-x_{11}\,\mathbf{a}_1 + \left(\frac{1}{2}+y_{11}\right)\,\mathbf{a}_2 + \left(\frac{1}{2}-z_{11}\right)\,\mathbf{a}_3$ | = | $\left(\frac{1}{2}c\cos\beta - x_{11}a - z_{11}c\cos\beta\right)\,\mathbf{\hat{x}} + \left(\frac{1}{2}+y_{11}\right)b\,\mathbf{\hat{y}} + \left(\frac{1}{2}-z_{11}\right)c\sin\beta\,\mathbf{\hat{z}}$ | (4e) | O |
| $\mathbf{B}_{41}$ | = | $-x_{11}\,\mathbf{a}_1 - y_{11}\,\mathbf{a}_2 - z_{11}\,\mathbf{a}_3$ | = | $(-x_{11}a - z_{11}c\cos\beta)\,\mathbf{\hat{x}} - y_{11}b\,\mathbf{\hat{y}} - z_{11}c\sin\beta\,\mathbf{\hat{z}}$ | (4e) | O |
| $\mathbf{B}_{42}$ | = | $x_{11}\,\mathbf{a}_1 + \left(\frac{1}{2}-y_{11}\right)\,\mathbf{a}_2 + \left(\frac{1}{2}+z_{11}\right)\,\mathbf{a}_3$ | = | $\left(\frac{1}{2}c\cos\beta + x_{11}a + z_{11}c\cos\beta\right)\,\mathbf{\hat{x}} + \left(\frac{1}{2}-y_{11}\right)b\,\mathbf{\hat{y}} + \left(\frac{1}{2}+z_{11}\right)c\sin\beta\,\mathbf{\hat{z}}$ | (4e) | O |
| $\mathbf{B}_{43}$ | = | $x_{12}\,\mathbf{a}_1 + y_{12}\,\mathbf{a}_2 + z_{12}\,\mathbf{a}_3$ | = | $(x_{12}a + z_{12}c\cos\beta)\,\mathbf{\hat{x}} + y_{12}b\,\mathbf{\hat{y}} + z_{12}c\sin\beta\,\mathbf{\hat{z}}$ | (4e) | S I |
| $\mathbf{B}_{44}$ | = | $-x_{12}\,\mathbf{a}_1 + \left(\frac{1}{2}+y_{12}\right)\,\mathbf{a}_2 + \left(\frac{1}{2}-z_{12}\right)\,\mathbf{a}_3$ | = | $\left(\frac{1}{2}c\cos\beta - x_{12}a - z_{12}c\cos\beta\right)\,\mathbf{\hat{x}} + \left(\frac{1}{2}+y_{12}\right)b\,\mathbf{\hat{y}} + \left(\frac{1}{2}-z_{12}\right)c\sin\beta\,\mathbf{\hat{z}}$ | (4e) | S I |
| $\mathbf{B}_{45}$ | = | $-x_{12}\,\mathbf{a}_1 - y_{12}\,\mathbf{a}_2 - z_{12}\,\mathbf{a}_3$ | = | $(-x_{12}a - z_{12}c\cos\beta)\,\mathbf{\hat{x}} - y_{12}b\,\mathbf{\hat{y}} - z_{12}c\sin\beta\,\mathbf{\hat{z}}$ | (4e) | S I |
| $\mathbf{B}_{46}$ | = | $x_{12}\,\mathbf{a}_1 + \left(\frac{1}{2}-y_{12}\right)\,\mathbf{a}_2 + \left(\frac{1}{2}+z_{12}\right)\,\mathbf{a}_3$ | = | $\left(\frac{1}{2}c\cos\beta + x_{12}a + z_{12}c\cos\beta\right)\,\mathbf{\hat{x}} + \left(\frac{1}{2}-y_{12}\right)b\,\mathbf{\hat{y}} + \left(\frac{1}{2}+z_{12}\right)c\sin\beta\,\mathbf{\hat{z}}$ | (4e) | S I |
| $\mathbf{B}_{47}$ | = | $x_{13}\,\mathbf{a}_1 + y_{13}\,\mathbf{a}_2 + z_{13}\,\mathbf{a}_3$ | = | $(x_{13}a + z_{13}c\cos\beta)\,\mathbf{\hat{x}} + y_{13}b\,\mathbf{\hat{y}} + z_{13}c\sin\beta\,\mathbf{\hat{z}}$ | (4e) | S II |
| $\mathbf{B}_{48}$ | = | $-x_{13}\,\mathbf{a}_1 + \left(\frac{1}{2}+y_{13}\right)\,\mathbf{a}_2 + \left(\frac{1}{2}-z_{13}\right)\,\mathbf{a}_3$ | = | $\left(\frac{1}{2}c\cos\beta - x_{13}a - z_{13}c\cos\beta\right)\,\mathbf{\hat{x}} + \left(\frac{1}{2}+y_{13}\right)b\,\mathbf{\hat{y}} + \left(\frac{1}{2}-z_{13}\right)c\sin\beta\,\mathbf{\hat{z}}$ | (4e) | S II |
| $\mathbf{B}_{49}$ | = | $-x_{13}\,\mathbf{a}_1 - y_{13}\,\mathbf{a}_2 - z_{13}\,\mathbf{a}_3$ | = | $(-x_{13}a - z_{13}c\cos\beta)\,\mathbf{\hat{x}} - y_{13}b\,\mathbf{\hat{y}} - z_{13}c\sin\beta\,\mathbf{\hat{z}}$ | (4e) | S II |
| $\mathbf{B}_{50}$ | = | $x_{13}\,\mathbf{a}_1 + \left(\frac{1}{2}-y_{13}\right)\,\mathbf{a}_2 + \left(\frac{1}{2}+z_{13}\right)\,\mathbf{a}_3$ | = | $\left(\frac{1}{2}c\cos\beta + x_{13}a + z_{13}c\cos\beta\right)\,\mathbf{\hat{x}} + \left(\frac{1}{2}-y_{13}\right)b\,\mathbf{\hat{y}} + \left(\frac{1}{2}+z_{13}\right)c\sin\beta\,\mathbf{\hat{z}}$ | (4e) | S II |
| $\mathbf{B}_{51}$ | = | $x_{14}\,\mathbf{a}_1 + y_{14}\,\mathbf{a}_2 + z_{14}\,\mathbf{a}_3$ | = | $(x_{14}a + z_{14}c\cos\beta)\,\mathbf{\hat{x}} + y_{14}b\,\mathbf{\hat{y}} + z_{14}c\sin\beta\,\mathbf{\hat{z}}$ | (4e) | S III |
| $\mathbf{B}_{52}$ | = | $-x_{14}\,\mathbf{a}_1 + \left(\frac{1}{2}+y_{14}\right)\,\mathbf{a}_2 + \left(\frac{1}{2}-z_{14}\right)\,\mathbf{a}_3$ | = | $\left(\frac{1}{2}c\cos\beta - x_{14}a - z_{14}c\cos\beta\right)\,\mathbf{\hat{x}} + \left(\frac{1}{2}+y_{14}\right)b\,\mathbf{\hat{y}} + \left(\frac{1}{2}-z_{14}\right)c\sin\beta\,\mathbf{\hat{z}}$ | (4e) | S III |
| $\mathbf{B}_{53}$ | = | $-x_{14}\,\mathbf{a}_1 - y_{14}\,\mathbf{a}_2 - z_{14}\,\mathbf{a}_3$ | = | $(-x_{14}a - z_{14}c\cos\beta)\,\mathbf{\hat{x}} - y_{14}b\,\mathbf{\hat{y}} - z_{14}c\sin\beta\,\mathbf{\hat{z}}$ | (4e) | S III |
| $\mathbf{B}_{54}$ | = | $x_{14}\,\mathbf{a}_1 + \left(\frac{1}{2}-y_{14}\right)\,\mathbf{a}_2 + \left(\frac{1}{2}+z_{14}\right)\,\mathbf{a}_3$ | = | $\left(\frac{1}{2}c\cos\beta + x_{14}a + z_{14}c\cos\beta\right)\,\mathbf{\hat{x}} + \left(\frac{1}{2}-y_{14}\right)b\,\mathbf{\hat{y}} + \left(\frac{1}{2}+z_{14}\right)c\sin\beta\,\mathbf{\hat{z}}$ | (4e) | S III |

**Geometry files:**

- CIF: pp. 1545
- POSCAR: pp. 1545

# $K_2NbF_7$ ($K6_2$) Structure: A7B2C_mP40_14_7e_2e_e

http://aflow.org/prototype-encyclopedia/A7B2C_mP40_14_7e_2e_e

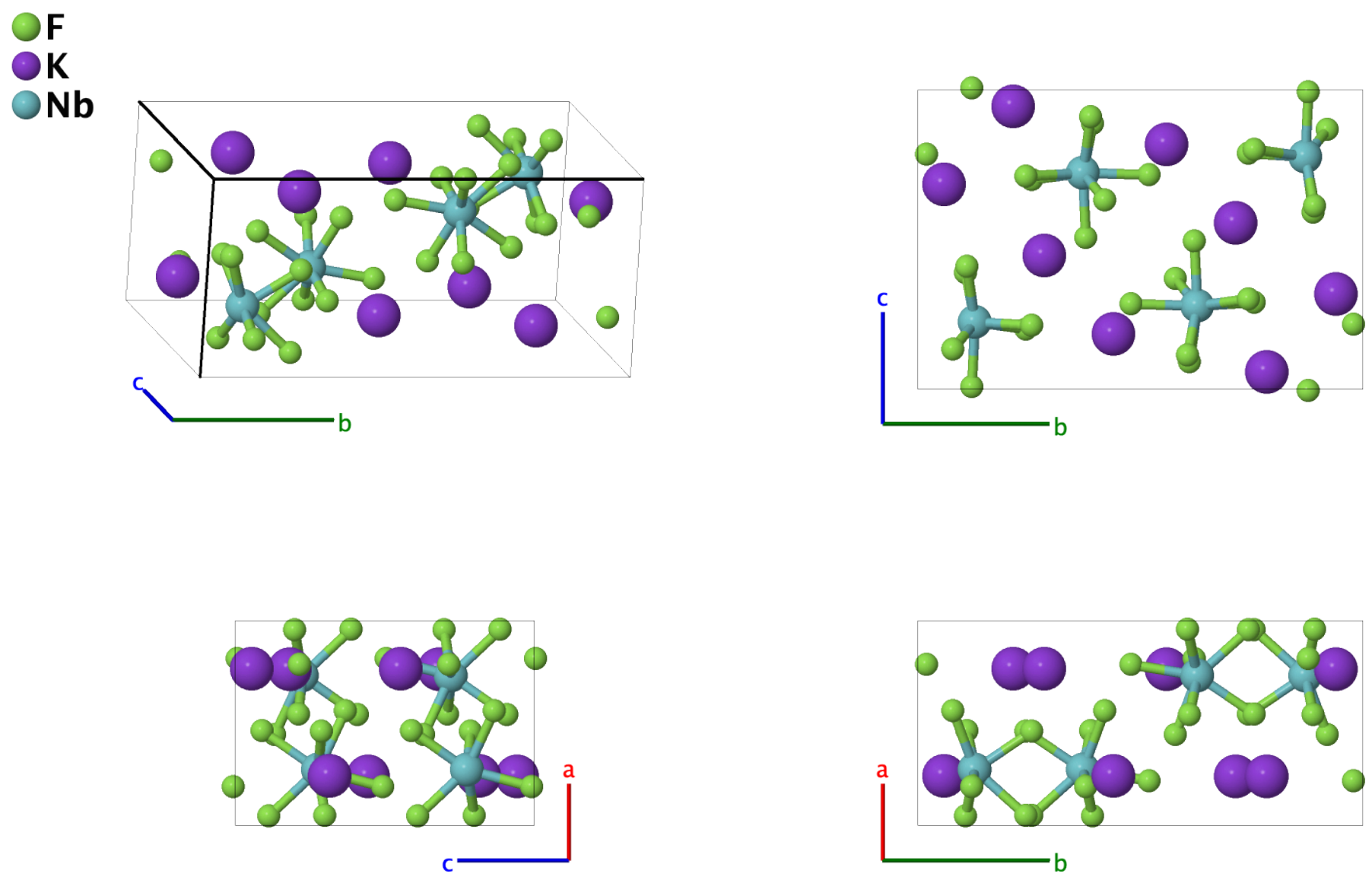

| | | |
|---|---|---|
| **Prototype** | : | $F_7K_2Nb$ |
| **AFLOW prototype label** | : | A7B2C_mP40_14_7e_2e_e |
| ***Strukturbericht* designation** | : | $K6_2$ |
| **Pearson symbol** | : | mP40 |
| **Space group number** | : | 14 |
| **Space group symbol** | : | $P2_1/c$ |
| **AFLOW prototype command** | : | `aflow --proto=A7B2C_mP40_14_7e_2e_e` `--params=`$a, b/a, c/a, \beta, x_1, y_1, z_1, x_2, y_2, z_2, x_3, y_3, z_3, x_4, y_4, z_4, x_5, y_5, z_5, x_6, y_6, z_6, x_7, y_7, z_7, x_8, y_8, z_8, x_9, y_9, z_9, x_{10}, y_{10}, z_{10}$ |

### Other compounds with this structure

- $K_2TaF_7$

### Simple Monoclinic primitive vectors:

$$\mathbf{a}_1 = a\,\hat{\mathbf{x}}$$
$$\mathbf{a}_2 = b\,\hat{\mathbf{y}}$$
$$\mathbf{a}_3 = c\cos\beta\,\hat{\mathbf{x}} + c\sin\beta\,\hat{\mathbf{z}}$$

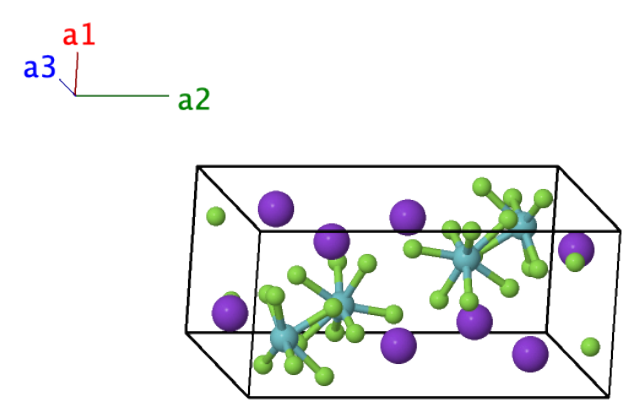

### Basis vectors:

| | | Lattice Coordinates | | Cartesian Coordinates | Wyckoff Position | Atom Type |
|---|---|---|---|---|---|---|
| $\mathbf{B}_1$ | = | $x_1\,\mathbf{a}_1 + y_1\,\mathbf{a}_2 + z_1\,\mathbf{a}_3$ | = | $(x_1 a + z_1 c\cos\beta)\,\hat{\mathbf{x}} + y_1 b\,\hat{\mathbf{y}} + z_1 c\sin\beta\,\hat{\mathbf{z}}$ | (4$e$) | F I |
| $\mathbf{B}_{2}$ | = | $-x_{1}\,\mathbf{a}_{1}+\left(\frac{1}{2}+y_{1}\right)\,\mathbf{a}_{2}+\left(\frac{1}{2}-z_{1}\right)\,\mathbf{a}_{3}$ | = | $\left(\frac{1}{2}c\cos\beta-x_{1}a-z_{1}c\cos\beta\right)\,\mathbf{\hat{x}}+\left(\frac{1}{2}+y_{1}\right)b\,\mathbf{\hat{y}}+\left(\frac{1}{2}-z_{1}\right)c\sin\beta\,\mathbf{\hat{z}}$ | $(4e)$ | F I |
| $\mathbf{B}_{3}$ | = | $-x_{1}\,\mathbf{a}_{1}-y_{1}\,\mathbf{a}_{2}-z_{1}\,\mathbf{a}_{3}$ | = | $(-x_{1}a-z_{1}c\cos\beta)\,\mathbf{\hat{x}}-y_{1}b\,\mathbf{\hat{y}}-z_{1}c\sin\beta\,\mathbf{\hat{z}}$ | $(4e)$ | F I |
| $\mathbf{B}_{4}$ | = | $x_{1}\,\mathbf{a}_{1}+\left(\frac{1}{2}-y_{1}\right)\,\mathbf{a}_{2}+\left(\frac{1}{2}+z_{1}\right)\,\mathbf{a}_{3}$ | = | $\left(\frac{1}{2}c\cos\beta+x_{1}a+z_{1}c\cos\beta\right)\,\mathbf{\hat{x}}+\left(\frac{1}{2}-y_{1}\right)b\,\mathbf{\hat{y}}+\left(\frac{1}{2}+z_{1}\right)c\sin\beta\,\mathbf{\hat{z}}$ | $(4e)$ | F I |
| $\mathbf{B}_{5}$ | = | $x_{2}\,\mathbf{a}_{1}+y_{2}\,\mathbf{a}_{2}+z_{2}\,\mathbf{a}_{3}$ | = | $(x_{2}a+z_{2}c\cos\beta)\,\mathbf{\hat{x}}+y_{2}b\,\mathbf{\hat{y}}+z_{2}c\sin\beta\,\mathbf{\hat{z}}$ | $(4e)$ | F II |
| $\mathbf{B}_{6}$ | = | $-x_{2}\,\mathbf{a}_{1}+\left(\frac{1}{2}+y_{2}\right)\,\mathbf{a}_{2}+\left(\frac{1}{2}-z_{2}\right)\,\mathbf{a}_{3}$ | = | $\left(\frac{1}{2}c\cos\beta-x_{2}a-z_{2}c\cos\beta\right)\,\mathbf{\hat{x}}+\left(\frac{1}{2}+y_{2}\right)b\,\mathbf{\hat{y}}+\left(\frac{1}{2}-z_{2}\right)c\sin\beta\,\mathbf{\hat{z}}$ | $(4e)$ | F II |
| $\mathbf{B}_{7}$ | = | $-x_{2}\,\mathbf{a}_{1}-y_{2}\,\mathbf{a}_{2}-z_{2}\,\mathbf{a}_{3}$ | = | $(-x_{2}a-z_{2}c\cos\beta)\,\mathbf{\hat{x}}-y_{2}b\,\mathbf{\hat{y}}-z_{2}c\sin\beta\,\mathbf{\hat{z}}$ | $(4e)$ | F II |
| $\mathbf{B}_{8}$ | = | $x_{2}\,\mathbf{a}_{1}+\left(\frac{1}{2}-y_{2}\right)\,\mathbf{a}_{2}+\left(\frac{1}{2}+z_{2}\right)\,\mathbf{a}_{3}$ | = | $\left(\frac{1}{2}c\cos\beta+x_{2}a+z_{2}c\cos\beta\right)\,\mathbf{\hat{x}}+\left(\frac{1}{2}-y_{2}\right)b\,\mathbf{\hat{y}}+\left(\frac{1}{2}+z_{2}\right)c\sin\beta\,\mathbf{\hat{z}}$ | $(4e)$ | F II |
| $\mathbf{B}_{9}$ | = | $x_{3}\,\mathbf{a}_{1}+y_{3}\,\mathbf{a}_{2}+z_{3}\,\mathbf{a}_{3}$ | = | $(x_{3}a+z_{3}c\cos\beta)\,\mathbf{\hat{x}}+y_{3}b\,\mathbf{\hat{y}}+z_{3}c\sin\beta\,\mathbf{\hat{z}}$ | $(4e)$ | F III |
| $\mathbf{B}_{10}$ | = | $-x_{3}\,\mathbf{a}_{1}+\left(\frac{1}{2}+y_{3}\right)\,\mathbf{a}_{2}+\left(\frac{1}{2}-z_{3}\right)\,\mathbf{a}_{3}$ | = | $\left(\frac{1}{2}c\cos\beta-x_{3}a-z_{3}c\cos\beta\right)\,\mathbf{\hat{x}}+\left(\frac{1}{2}+y_{3}\right)b\,\mathbf{\hat{y}}+\left(\frac{1}{2}-z_{3}\right)c\sin\beta\,\mathbf{\hat{z}}$ | $(4e)$ | F III |
| $\mathbf{B}_{11}$ | = | $-x_{3}\,\mathbf{a}_{1}-y_{3}\,\mathbf{a}_{2}-z_{3}\,\mathbf{a}_{3}$ | = | $(-x_{3}a-z_{3}c\cos\beta)\,\mathbf{\hat{x}}-y_{3}b\,\mathbf{\hat{y}}-z_{3}c\sin\beta\,\mathbf{\hat{z}}$ | $(4e)$ | F III |
| $\mathbf{B}_{12}$ | = | $x_{3}\,\mathbf{a}_{1}+\left(\frac{1}{2}-y_{3}\right)\,\mathbf{a}_{2}+\left(\frac{1}{2}+z_{3}\right)\,\mathbf{a}_{3}$ | = | $\left(\frac{1}{2}c\cos\beta+x_{3}a+z_{3}c\cos\beta\right)\,\mathbf{\hat{x}}+\left(\frac{1}{2}-y_{3}\right)b\,\mathbf{\hat{y}}+\left(\frac{1}{2}+z_{3}\right)c\sin\beta\,\mathbf{\hat{z}}$ | $(4e)$ | F III |
| $\mathbf{B}_{13}$ | = | $x_{4}\,\mathbf{a}_{1}+y_{4}\,\mathbf{a}_{2}+z_{4}\,\mathbf{a}_{3}$ | = | $(x_{4}a+z_{4}c\cos\beta)\,\mathbf{\hat{x}}+y_{4}b\,\mathbf{\hat{y}}+z_{4}c\sin\beta\,\mathbf{\hat{z}}$ | $(4e)$ | F IV |
| $\mathbf{B}_{14}$ | = | $-x_{4}\,\mathbf{a}_{1}+\left(\frac{1}{2}+y_{4}\right)\,\mathbf{a}_{2}+\left(\frac{1}{2}-z_{4}\right)\,\mathbf{a}_{3}$ | = | $\left(\frac{1}{2}c\cos\beta-x_{4}a-z_{4}c\cos\beta\right)\,\mathbf{\hat{x}}+\left(\frac{1}{2}+y_{4}\right)b\,\mathbf{\hat{y}}+\left(\frac{1}{2}-z_{4}\right)c\sin\beta\,\mathbf{\hat{z}}$ | $(4e)$ | F IV |
| $\mathbf{B}_{15}$ | = | $-x_{4}\,\mathbf{a}_{1}-y_{4}\,\mathbf{a}_{2}-z_{4}\,\mathbf{a}_{3}$ | = | $(-x_{4}a-z_{4}c\cos\beta)\,\mathbf{\hat{x}}-y_{4}b\,\mathbf{\hat{y}}-z_{4}c\sin\beta\,\mathbf{\hat{z}}$ | $(4e)$ | F IV |
| $\mathbf{B}_{16}$ | = | $x_{4}\,\mathbf{a}_{1}+\left(\frac{1}{2}-y_{4}\right)\,\mathbf{a}_{2}+\left(\frac{1}{2}+z_{4}\right)\,\mathbf{a}_{3}$ | = | $\left(\frac{1}{2}c\cos\beta+x_{4}a+z_{4}c\cos\beta\right)\,\mathbf{\hat{x}}+\left(\frac{1}{2}-y_{4}\right)b\,\mathbf{\hat{y}}+\left(\frac{1}{2}+z_{4}\right)c\sin\beta\,\mathbf{\hat{z}}$ | $(4e)$ | F IV |
| $\mathbf{B}_{17}$ | = | $x_{5}\,\mathbf{a}_{1}+y_{5}\,\mathbf{a}_{2}+z_{5}\,\mathbf{a}_{3}$ | = | $(x_{5}a+z_{5}c\cos\beta)\,\mathbf{\hat{x}}+y_{5}b\,\mathbf{\hat{y}}+z_{5}c\sin\beta\,\mathbf{\hat{z}}$ | $(4e)$ | F V |
| $\mathbf{B}_{18}$ | = | $-x_{5}\,\mathbf{a}_{1}+\left(\frac{1}{2}+y_{5}\right)\,\mathbf{a}_{2}+\left(\frac{1}{2}-z_{5}\right)\,\mathbf{a}_{3}$ | = | $\left(\frac{1}{2}c\cos\beta-x_{5}a-z_{5}c\cos\beta\right)\,\mathbf{\hat{x}}+\left(\frac{1}{2}+y_{5}\right)b\,\mathbf{\hat{y}}+\left(\frac{1}{2}-z_{5}\right)c\sin\beta\,\mathbf{\hat{z}}$ | $(4e)$ | F V |
| $\mathbf{B}_{19}$ | = | $-x_{5}\,\mathbf{a}_{1}-y_{5}\,\mathbf{a}_{2}-z_{5}\,\mathbf{a}_{3}$ | = | $(-x_{5}a-z_{5}c\cos\beta)\,\mathbf{\hat{x}}-y_{5}b\,\mathbf{\hat{y}}-z_{5}c\sin\beta\,\mathbf{\hat{z}}$ | $(4e)$ | F V |
| $\mathbf{B}_{20}$ | = | $x_{5}\,\mathbf{a}_{1}+\left(\frac{1}{2}-y_{5}\right)\,\mathbf{a}_{2}+\left(\frac{1}{2}+z_{5}\right)\,\mathbf{a}_{3}$ | = | $\left(\frac{1}{2}c\cos\beta+x_{5}a+z_{5}c\cos\beta\right)\,\mathbf{\hat{x}}+\left(\frac{1}{2}-y_{5}\right)b\,\mathbf{\hat{y}}+\left(\frac{1}{2}+z_{5}\right)c\sin\beta\,\mathbf{\hat{z}}$ | $(4e)$ | F V |
| $\mathbf{B}_{21}$ | = | $x_{6}\,\mathbf{a}_{1}+y_{6}\,\mathbf{a}_{2}+z_{6}\,\mathbf{a}_{3}$ | = | $(x_{6}a+z_{6}c\cos\beta)\,\mathbf{\hat{x}}+y_{6}b\,\mathbf{\hat{y}}+z_{6}c\sin\beta\,\mathbf{\hat{z}}$ | $(4e)$ | F VI |
| $\mathbf{B}_{22}$ | = | $-x_{6}\,\mathbf{a}_{1}+\left(\frac{1}{2}+y_{6}\right)\,\mathbf{a}_{2}+\left(\frac{1}{2}-z_{6}\right)\,\mathbf{a}_{3}$ | = | $\left(\frac{1}{2}c\cos\beta-x_{6}a-z_{6}c\cos\beta\right)\,\mathbf{\hat{x}}+\left(\frac{1}{2}+y_{6}\right)b\,\mathbf{\hat{y}}+\left(\frac{1}{2}-z_{6}\right)c\sin\beta\,\mathbf{\hat{z}}$ | $(4e)$ | F VI |
| $\mathbf{B}_{23}$ | = | $-x_{6}\,\mathbf{a}_{1}-y_{6}\,\mathbf{a}_{2}-z_{6}\,\mathbf{a}_{3}$ | = | $(-x_{6}a-z_{6}c\cos\beta)\,\mathbf{\hat{x}}-y_{6}b\,\mathbf{\hat{y}}-z_{6}c\sin\beta\,\mathbf{\hat{z}}$ | $(4e)$ | F VI |
| $\mathbf{B}_{24}$ | = | $x_6\,\mathbf{a}_1 + \left(\frac{1}{2} - y_6\right)\mathbf{a}_2 + \left(\frac{1}{2} + z_6\right)\mathbf{a}_3$ | = | $\left(\frac{1}{2}c\cos\beta + x_6 a + z_6 c\cos\beta\right)\hat{\mathbf{x}} + \left(\frac{1}{2} - y_6\right)b\,\hat{\mathbf{y}} + \left(\frac{1}{2} + z_6\right)c\sin\beta\,\hat{\mathbf{z}}$ | (4$e$) | F VI |
| $\mathbf{B}_{25}$ | = | $x_7\,\mathbf{a}_1 + y_7\,\mathbf{a}_2 + z_7\,\mathbf{a}_3$ | = | $(x_7 a + z_7 c\cos\beta)\,\hat{\mathbf{x}} + y_7 b\,\hat{\mathbf{y}} + z_7 c\sin\beta\,\hat{\mathbf{z}}$ | (4$e$) | F VII |
| $\mathbf{B}_{26}$ | = | $-x_7\,\mathbf{a}_1 + \left(\frac{1}{2} + y_7\right)\mathbf{a}_2 + \left(\frac{1}{2} - z_7\right)\mathbf{a}_3$ | = | $\left(\frac{1}{2}c\cos\beta - x_7 a - z_7 c\cos\beta\right)\hat{\mathbf{x}} + \left(\frac{1}{2} + y_7\right)b\,\hat{\mathbf{y}} + \left(\frac{1}{2} - z_7\right)c\sin\beta\,\hat{\mathbf{z}}$ | (4$e$) | F VII |
| $\mathbf{B}_{27}$ | = | $-x_7\,\mathbf{a}_1 - y_7\,\mathbf{a}_2 - z_7\,\mathbf{a}_3$ | = | $(-x_7 a - z_7 c\cos\beta)\,\hat{\mathbf{x}} - y_7 b\,\hat{\mathbf{y}} - z_7 c\sin\beta\,\hat{\mathbf{z}}$ | (4$e$) | F VII |
| $\mathbf{B}_{28}$ | = | $x_7\,\mathbf{a}_1 + \left(\frac{1}{2} - y_7\right)\mathbf{a}_2 + \left(\frac{1}{2} + z_7\right)\mathbf{a}_3$ | = | $\left(\frac{1}{2}c\cos\beta + x_7 a + z_7 c\cos\beta\right)\hat{\mathbf{x}} + \left(\frac{1}{2} - y_7\right)b\,\hat{\mathbf{y}} + \left(\frac{1}{2} + z_7\right)c\sin\beta\,\hat{\mathbf{z}}$ | (4$e$) | F VII |
| $\mathbf{B}_{29}$ | = | $x_8\,\mathbf{a}_1 + y_8\,\mathbf{a}_2 + z_8\,\mathbf{a}_3$ | = | $(x_8 a + z_8 c\cos\beta)\,\hat{\mathbf{x}} + y_8 b\,\hat{\mathbf{y}} + z_8 c\sin\beta\,\hat{\mathbf{z}}$ | (4$e$) | K I |
| $\mathbf{B}_{30}$ | = | $-x_8\,\mathbf{a}_1 + \left(\frac{1}{2} + y_8\right)\mathbf{a}_2 + \left(\frac{1}{2} - z_8\right)\mathbf{a}_3$ | = | $\left(\frac{1}{2}c\cos\beta - x_8 a - z_8 c\cos\beta\right)\hat{\mathbf{x}} + \left(\frac{1}{2} + y_8\right)b\,\hat{\mathbf{y}} + \left(\frac{1}{2} - z_8\right)c\sin\beta\,\hat{\mathbf{z}}$ | (4$e$) | K I |
| $\mathbf{B}_{31}$ | = | $-x_8\,\mathbf{a}_1 - y_8\,\mathbf{a}_2 - z_8\,\mathbf{a}_3$ | = | $(-x_8 a - z_8 c\cos\beta)\,\hat{\mathbf{x}} - y_8 b\,\hat{\mathbf{y}} - z_8 c\sin\beta\,\hat{\mathbf{z}}$ | (4$e$) | K I |
| $\mathbf{B}_{32}$ | = | $x_8\,\mathbf{a}_1 + \left(\frac{1}{2} - y_8\right)\mathbf{a}_2 + \left(\frac{1}{2} + z_8\right)\mathbf{a}_3$ | = | $\left(\frac{1}{2}c\cos\beta + x_8 a + z_8 c\cos\beta\right)\hat{\mathbf{x}} + \left(\frac{1}{2} - y_8\right)b\,\hat{\mathbf{y}} + \left(\frac{1}{2} + z_8\right)c\sin\beta\,\hat{\mathbf{z}}$ | (4$e$) | K I |
| $\mathbf{B}_{33}$ | = | $x_9\,\mathbf{a}_1 + y_9\,\mathbf{a}_2 + z_9\,\mathbf{a}_3$ | = | $(x_9 a + z_9 c\cos\beta)\,\hat{\mathbf{x}} + y_9 b\,\hat{\mathbf{y}} + z_9 c\sin\beta\,\hat{\mathbf{z}}$ | (4$e$) | K II |
| $\mathbf{B}_{34}$ | = | $-x_9\,\mathbf{a}_1 + \left(\frac{1}{2} + y_9\right)\mathbf{a}_2 + \left(\frac{1}{2} - z_9\right)\mathbf{a}_3$ | = | $\left(\frac{1}{2}c\cos\beta - x_9 a - z_9 c\cos\beta\right)\hat{\mathbf{x}} + \left(\frac{1}{2} + y_9\right)b\,\hat{\mathbf{y}} + \left(\frac{1}{2} - z_9\right)c\sin\beta\,\hat{\mathbf{z}}$ | (4$e$) | K II |
| $\mathbf{B}_{35}$ | = | $-x_9\,\mathbf{a}_1 - y_9\,\mathbf{a}_2 - z_9\,\mathbf{a}_3$ | = | $(-x_9 a - z_9 c\cos\beta)\,\hat{\mathbf{x}} - y_9 b\,\hat{\mathbf{y}} - z_9 c\sin\beta\,\hat{\mathbf{z}}$ | (4$e$) | K II |
| $\mathbf{B}_{36}$ | = | $x_9\,\mathbf{a}_1 + \left(\frac{1}{2} - y_9\right)\mathbf{a}_2 + \left(\frac{1}{2} + z_9\right)\mathbf{a}_3$ | = | $\left(\frac{1}{2}c\cos\beta + x_9 a + z_9 c\cos\beta\right)\hat{\mathbf{x}} + \left(\frac{1}{2} - y_9\right)b\,\hat{\mathbf{y}} + \left(\frac{1}{2} + z_9\right)c\sin\beta\,\hat{\mathbf{z}}$ | (4$e$) | K II |
| $\mathbf{B}_{37}$ | = | $x_{10}\,\mathbf{a}_1 + y_{10}\,\mathbf{a}_2 + z_{10}\,\mathbf{a}_3$ | = | $(x_{10} a + z_{10} c\cos\beta)\,\hat{\mathbf{x}} + y_{10} b\,\hat{\mathbf{y}} + z_{10} c\sin\beta\,\hat{\mathbf{z}}$ | (4$e$) | Nb |
| $\mathbf{B}_{38}$ | = | $-x_{10}\,\mathbf{a}_1 + \left(\frac{1}{2} + y_{10}\right)\mathbf{a}_2 + \left(\frac{1}{2} - z_{10}\right)\mathbf{a}_3$ | = | $\left(\frac{1}{2}c\cos\beta - x_{10} a - z_{10} c\cos\beta\right)\hat{\mathbf{x}} + \left(\frac{1}{2} + y_{10}\right)b\,\hat{\mathbf{y}} + \left(\frac{1}{2} - z_{10}\right)c\sin\beta\,\hat{\mathbf{z}}$ | (4$e$) | Nb |
| $\mathbf{B}_{39}$ | = | $-x_{10}\,\mathbf{a}_1 - y_{10}\,\mathbf{a}_2 - z_{10}\,\mathbf{a}_3$ | = | $(-x_{10} a - z_{10} c\cos\beta)\,\hat{\mathbf{x}} - y_{10} b\,\hat{\mathbf{y}} - z_{10} c\sin\beta\,\hat{\mathbf{z}}$ | (4$e$) | Nb |
| $\mathbf{B}_{40}$ | = | $x_{10}\,\mathbf{a}_1 + \left(\frac{1}{2} - y_{10}\right)\mathbf{a}_2 + \left(\frac{1}{2} + z_{10}\right)\mathbf{a}_3$ | = | $\left(\frac{1}{2}c\cos\beta + x_{10} a + z_{10} c\cos\beta\right)\hat{\mathbf{x}} + \left(\frac{1}{2} - y_{10}\right)b\,\hat{\mathbf{y}} + \left(\frac{1}{2} + z_{10}\right)c\sin\beta\,\hat{\mathbf{z}}$ | (4$e$) | Nb |

**Geometry files:**
- CIF: pp. 1546
- POSCAR: pp. 1546

# Manganese-leonite 110 K [$K_2Mn(SO_4)_2{\cdot}4H_2O$] Structure: A8B2CD12E2_mP100_14_8e_2e_ad_12e_2e

http://aflow.org/prototype-encyclopedia/A8B2CD12E2_mP100_14_8e_2e_ad_12e_2e

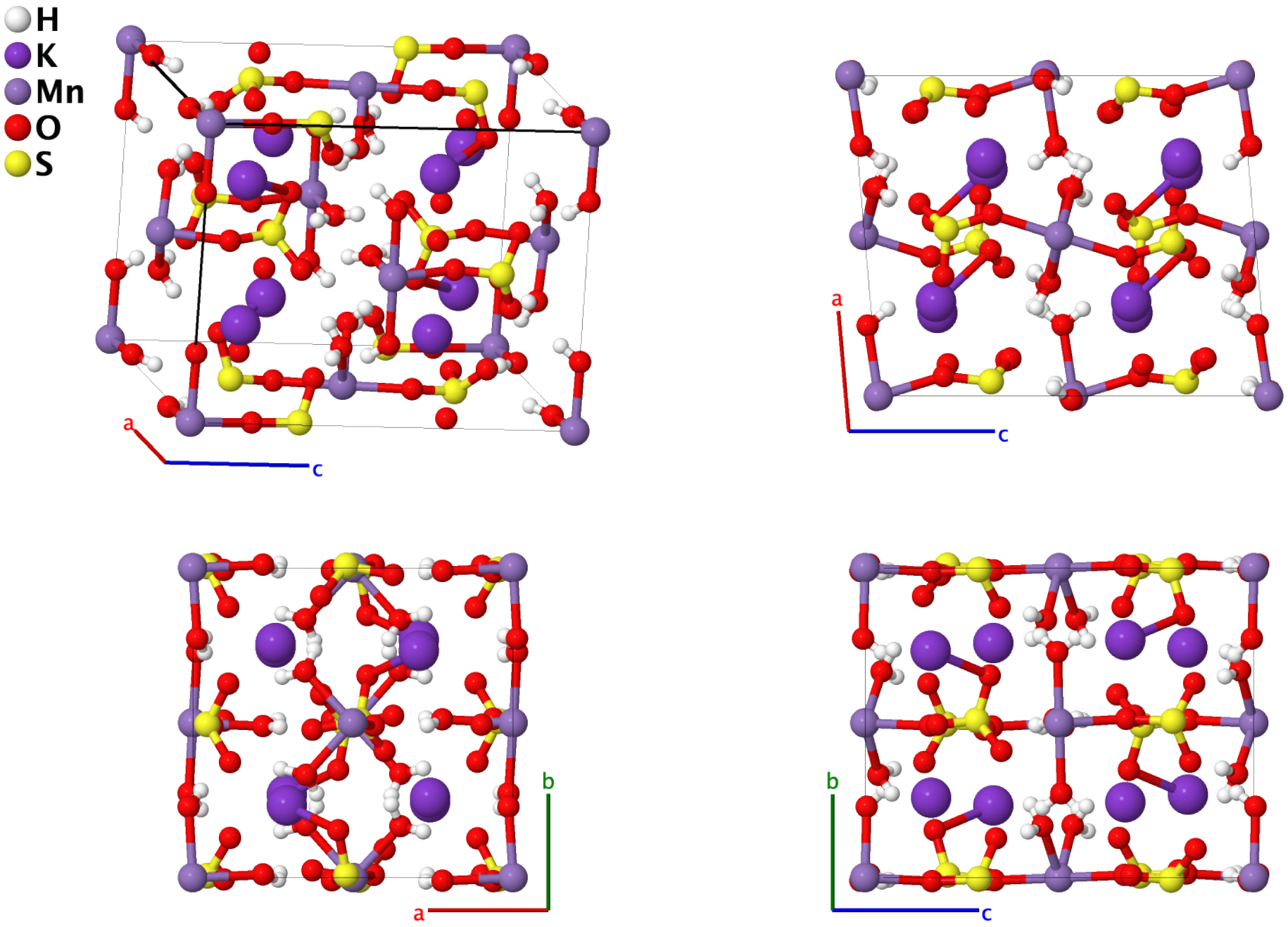

| | | |
|---|---|---|
| **Prototype** | : | $H_8K_2MnO_{12}S_2$ |
| **AFLOW prototype label** | : | A8B2CD12E2_mP100_14_8e_2e_ad_12e_2e |
| ***Strukturbericht* designation** | : | None |
| **Pearson symbol** | : | mP100 |
| **Space group number** | : | 14 |
| **Space group symbol** | : | $P2_1/c$ |
| **AFLOW prototype command** | : | `aflow --proto=A8B2CD12E2_mP100_14_8e_2e_ad_12e_2e` `--params=`$a, b/a, c/a, \beta, x_3, y_3, z_3, x_4, y_4, z_4, x_5, y_5, z_5, x_6, y_6, z_6, x_7, y_7, z_7, x_8, y_8, z_8, x_9, y_9, z_9, x_{10}, y_{10}, z_{10}, x_{11}, y_{11}, z_{11}, x_{12}, y_{12}, z_{12}, x_{13}, y_{13}, z_{13}, x_{14}, y_{14}, z_{14}, x_{15}, y_{15}, z_{15}, x_{16}, y_{16}, z_{16}, x_{17}, y_{17}, z_{17}, x_{18}, y_{18}, z_{18}, x_{19}, y_{19}, z_{19}, x_{20}, y_{20}, z_{20}, x_{21}, y_{21}, z_{21}, x_{22}, y_{22}, z_{22}, x_{23}, y_{23}, z_{23}, x_{24}, y_{24}, z_{24}, x_{25}, y_{25}, z_{25}, x_{26}, y_{26}, z_{26}$ |

### Other compounds with this structure

- $K_2Mg(SO_4)_2{\cdot}4H_2O$ (leonite)

- This is the intermediate-low temperature structure of leonite. For Mn-leonite, discussed here, it is stable below 168 K. The structure shown here was determined at 110 K.
- The room temperature structure is *Strukturbericht* $H4_{23}$, space group $C2/m$ #13. The current structure doubles the size of the unit cell and orders all of the $SO_4$ radicals.
- The intermediate temperature structure (168-205 K) has space group $P2_1/c$ #14.
- (Hertweck, 2001) give crystallographic information of this phase in the $P2_1/a$ setting of space group #14. We used FINDSYM to transform this to the $P2_1/c$ setting, which resulted in a switching of the $a$- and $c$-axes.

**Simple Monoclinic primitive vectors:**

$$\mathbf{a}_1 = a\,\hat{\mathbf{x}}$$
$$\mathbf{a}_2 = b\,\hat{\mathbf{y}}$$
$$\mathbf{a}_3 = c\cos\beta\,\hat{\mathbf{x}} + c\sin\beta\,\hat{\mathbf{z}}$$

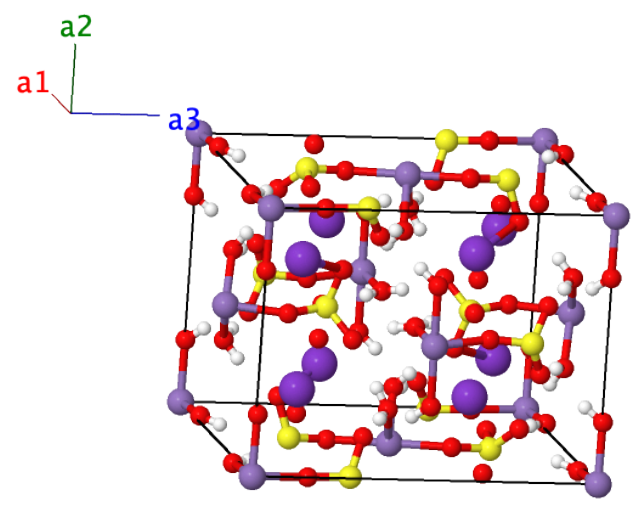

**Basis vectors:**

| | | Lattice Coordinates | | Cartesian Coordinates | Wyckoff Position | Atom Type |
|---|---|---|---|---|---|---|
| $\mathbf{B}_1$ | = | $0\,\mathbf{a}_1 + 0\,\mathbf{a}_2 + 0\,\mathbf{a}_3$ | = | $0\,\hat{\mathbf{x}} + 0\,\hat{\mathbf{y}} + 0\,\hat{\mathbf{z}}$ | (2a) | Mn I |
| $\mathbf{B}_2$ | = | $\frac{1}{2}\,\mathbf{a}_2 + \frac{1}{2}\,\mathbf{a}_3$ | = | $\frac{1}{2}c\cos\beta\,\hat{\mathbf{x}} + \frac{1}{2}b\,\hat{\mathbf{y}} + \frac{1}{2}c\sin\beta\,\hat{\mathbf{z}}$ | (2a) | Mn I |
| $\mathbf{B}_3$ | = | $\frac{1}{2}\,\mathbf{a}_1 + \frac{1}{2}\,\mathbf{a}_3$ | = | $\frac{1}{2}(a + c\cos\beta)\,\hat{\mathbf{x}} + \frac{1}{2}c\sin\beta\,\hat{\mathbf{z}}$ | (2d) | Mn II |
| $\mathbf{B}_4$ | = | $\frac{1}{2}\,\mathbf{a}_1 + \frac{1}{2}\,\mathbf{a}_2$ | = | $\frac{1}{2}a\,\hat{\mathbf{x}} + \frac{1}{2}b\,\hat{\mathbf{y}}$ | (2d) | Mn II |
| $\mathbf{B}_5$ | = | $x_3\,\mathbf{a}_1 + y_3\,\mathbf{a}_2 + z_3\,\mathbf{a}_3$ | = | $(x_3a + z_3c\cos\beta)\,\hat{\mathbf{x}} + y_3b\,\hat{\mathbf{y}} + z_3c\sin\beta\,\hat{\mathbf{z}}$ | (4e) | H I |
| $\mathbf{B}_6$ | = | $-x_3\,\mathbf{a}_1 + \left(\frac{1}{2} + y_3\right)\mathbf{a}_2 + \left(\frac{1}{2} - z_3\right)\mathbf{a}_3$ | = | $\left(\frac{1}{2}c\cos\beta - x_3a - z_3c\cos\beta\right)\hat{\mathbf{x}} + \left(\frac{1}{2} + y_3\right)b\,\hat{\mathbf{y}} + \left(\frac{1}{2} - z_3\right)c\sin\beta\,\hat{\mathbf{z}}$ | (4e) | H I |
| $\mathbf{B}_7$ | = | $-x_3\,\mathbf{a}_1 - y_3\,\mathbf{a}_2 - z_3\,\mathbf{a}_3$ | = | $(-x_3a - z_3c\cos\beta)\,\hat{\mathbf{x}} - y_3b\,\hat{\mathbf{y}} - z_3c\sin\beta\,\hat{\mathbf{z}}$ | (4e) | H I |
| $\mathbf{B}_8$ | = | $x_3\,\mathbf{a}_1 + \left(\frac{1}{2} - y_3\right)\mathbf{a}_2 + \left(\frac{1}{2} + z_3\right)\mathbf{a}_3$ | = | $\left(\frac{1}{2}c\cos\beta + x_3a + z_3c\cos\beta\right)\hat{\mathbf{x}} + \left(\frac{1}{2} - y_3\right)b\,\hat{\mathbf{y}} + \left(\frac{1}{2} + z_3\right)c\sin\beta\,\hat{\mathbf{z}}$ | (4e) | H I |
| $\mathbf{B}_9$ | = | $x_4\,\mathbf{a}_1 + y_4\,\mathbf{a}_2 + z_4\,\mathbf{a}_3$ | = | $(x_4a + z_4c\cos\beta)\,\hat{\mathbf{x}} + y_4b\,\hat{\mathbf{y}} + z_4c\sin\beta\,\hat{\mathbf{z}}$ | (4e) | H II |
| $\mathbf{B}_{10}$ | = | $-x_4\,\mathbf{a}_1 + \left(\frac{1}{2} + y_4\right)\mathbf{a}_2 + \left(\frac{1}{2} - z_4\right)\mathbf{a}_3$ | = | $\left(\frac{1}{2}c\cos\beta - x_4a - z_4c\cos\beta\right)\hat{\mathbf{x}} + \left(\frac{1}{2} + y_4\right)b\,\hat{\mathbf{y}} + \left(\frac{1}{2} - z_4\right)c\sin\beta\,\hat{\mathbf{z}}$ | (4e) | H II |
| $\mathbf{B}_{11}$ | = | $-x_4\,\mathbf{a}_1 - y_4\,\mathbf{a}_2 - z_4\,\mathbf{a}_3$ | = | $(-x_4a - z_4c\cos\beta)\,\hat{\mathbf{x}} - y_4b\,\hat{\mathbf{y}} - z_4c\sin\beta\,\hat{\mathbf{z}}$ | (4e) | H II |
| $\mathbf{B}_{12}$ | = | $x_4\,\mathbf{a}_1 + \left(\frac{1}{2} - y_4\right)\mathbf{a}_2 + \left(\frac{1}{2} + z_4\right)\mathbf{a}_3$ | = | $\left(\frac{1}{2}c\cos\beta + x_4a + z_4c\cos\beta\right)\hat{\mathbf{x}} + \left(\frac{1}{2} - y_4\right)b\,\hat{\mathbf{y}} + \left(\frac{1}{2} + z_4\right)c\sin\beta\,\hat{\mathbf{z}}$ | (4e) | H II |
| $\mathbf{B}_{13}$ | = | $x_5\,\mathbf{a}_1 + y_5\,\mathbf{a}_2 + z_5\,\mathbf{a}_3$ | = | $(x_5a + z_5c\cos\beta)\,\hat{\mathbf{x}} + y_5b\,\hat{\mathbf{y}} + z_5c\sin\beta\,\hat{\mathbf{z}}$ | (4e) | H III |
| $\mathbf{B}_{14}$ | = | $-x_5\,\mathbf{a}_1 + \left(\frac{1}{2} + y_5\right)\mathbf{a}_2 + \left(\frac{1}{2} - z_5\right)\mathbf{a}_3$ | = | $\left(\frac{1}{2}c\cos\beta - x_5a - z_5c\cos\beta\right)\hat{\mathbf{x}} + \left(\frac{1}{2} + y_5\right)b\,\hat{\mathbf{y}} + \left(\frac{1}{2} - z_5\right)c\sin\beta\,\hat{\mathbf{z}}$ | (4e) | H III |
| $\mathbf{B}_{15}$ | = | $-x_5\,\mathbf{a}_1 - y_5\,\mathbf{a}_2 - z_5\,\mathbf{a}_3$ | = | $(-x_5a - z_5c\cos\beta)\,\hat{\mathbf{x}} - y_5b\,\hat{\mathbf{y}} - z_5c\sin\beta\,\hat{\mathbf{z}}$ | (4e) | H III |
| $\mathbf{B}_{16}$ | = | $x_5\,\mathbf{a}_1 + \left(\frac{1}{2} - y_5\right)\mathbf{a}_2 + \left(\frac{1}{2} + z_5\right)\mathbf{a}_3$ | = | $\left(\frac{1}{2}c\cos\beta + x_5a + z_5c\cos\beta\right)\hat{\mathbf{x}} + \left(\frac{1}{2} - y_5\right)b\,\hat{\mathbf{y}} + \left(\frac{1}{2} + z_5\right)c\sin\beta\,\hat{\mathbf{z}}$ | (4e) | H III |
| $\mathbf{B}_{17}$ | = | $x_6\,\mathbf{a}_1 + y_6\,\mathbf{a}_2 + z_6\,\mathbf{a}_3$ | = | $(x_6a + z_6c\cos\beta)\,\hat{\mathbf{x}} + y_6b\,\hat{\mathbf{y}} + z_6c\sin\beta\,\hat{\mathbf{z}}$ | (4e) | H IV |
| $\mathbf{B}_{18}$ | = | $-x_6\,\mathbf{a}_1 + \left(\frac{1}{2} + y_6\right)\mathbf{a}_2 + \left(\frac{1}{2} - z_6\right)\mathbf{a}_3$ | = | $\left(\frac{1}{2}c\cos\beta - x_6a - z_6c\cos\beta\right)\hat{\mathbf{x}} + \left(\frac{1}{2} + y_6\right)b\,\hat{\mathbf{y}} + \left(\frac{1}{2} - z_6\right)c\sin\beta\,\hat{\mathbf{z}}$ | (4e) | H IV |
| $\mathbf{B}_{19}$ | = | $-x_6\,\mathbf{a}_1 - y_6\,\mathbf{a}_2 - z_6\,\mathbf{a}_3$ | = | $(-x_6 a - z_6 c\cos\beta)\,\hat{\mathbf{x}} - y_6 b\,\hat{\mathbf{y}} - z_6 c\sin\beta\,\hat{\mathbf{z}}$ | (4$e$) | H IV |
| $\mathbf{B}_{20}$ | = | $x_6\,\mathbf{a}_1 + \left(\frac{1}{2} - y_6\right)\mathbf{a}_2 + \left(\frac{1}{2} + z_6\right)\mathbf{a}_3$ | = | $\left(\frac{1}{2}c\cos\beta + x_6 a + z_6 c\cos\beta\right)\hat{\mathbf{x}} + \left(\frac{1}{2} - y_6\right)b\,\hat{\mathbf{y}} + \left(\frac{1}{2} + z_6\right)c\sin\beta\,\hat{\mathbf{z}}$ | (4$e$) | H IV |
| $\mathbf{B}_{21}$ | = | $x_7\,\mathbf{a}_1 + y_7\,\mathbf{a}_2 + z_7\,\mathbf{a}_3$ | = | $(x_7 a + z_7 c\cos\beta)\,\hat{\mathbf{x}} + y_7 b\,\hat{\mathbf{y}} + z_7 c\sin\beta\,\hat{\mathbf{z}}$ | (4$e$) | H V |
| $\mathbf{B}_{22}$ | = | $-x_7\,\mathbf{a}_1 + \left(\frac{1}{2} + y_7\right)\mathbf{a}_2 + \left(\frac{1}{2} - z_7\right)\mathbf{a}_3$ | = | $\left(\frac{1}{2}c\cos\beta - x_7 a - z_7 c\cos\beta\right)\hat{\mathbf{x}} + \left(\frac{1}{2} + y_7\right)b\,\hat{\mathbf{y}} + \left(\frac{1}{2} - z_7\right)c\sin\beta\,\hat{\mathbf{z}}$ | (4$e$) | H V |
| $\mathbf{B}_{23}$ | = | $-x_7\,\mathbf{a}_1 - y_7\,\mathbf{a}_2 - z_7\,\mathbf{a}_3$ | = | $(-x_7 a - z_7 c\cos\beta)\,\hat{\mathbf{x}} - y_7 b\,\hat{\mathbf{y}} - z_7 c\sin\beta\,\hat{\mathbf{z}}$ | (4$e$) | H V |
| $\mathbf{B}_{24}$ | = | $x_7\,\mathbf{a}_1 + \left(\frac{1}{2} - y_7\right)\mathbf{a}_2 + \left(\frac{1}{2} + z_7\right)\mathbf{a}_3$ | = | $\left(\frac{1}{2}c\cos\beta + x_7 a + z_7 c\cos\beta\right)\hat{\mathbf{x}} + \left(\frac{1}{2} - y_7\right)b\,\hat{\mathbf{y}} + \left(\frac{1}{2} + z_7\right)c\sin\beta\,\hat{\mathbf{z}}$ | (4$e$) | H V |
| $\mathbf{B}_{25}$ | = | $x_8\,\mathbf{a}_1 + y_8\,\mathbf{a}_2 + z_8\,\mathbf{a}_3$ | = | $(x_8 a + z_8 c\cos\beta)\,\hat{\mathbf{x}} + y_8 b\,\hat{\mathbf{y}} + z_8 c\sin\beta\,\hat{\mathbf{z}}$ | (4$e$) | H VI |
| $\mathbf{B}_{26}$ | = | $-x_8\,\mathbf{a}_1 + \left(\frac{1}{2} + y_8\right)\mathbf{a}_2 + \left(\frac{1}{2} - z_8\right)\mathbf{a}_3$ | = | $\left(\frac{1}{2}c\cos\beta - x_8 a - z_8 c\cos\beta\right)\hat{\mathbf{x}} + \left(\frac{1}{2} + y_8\right)b\,\hat{\mathbf{y}} + \left(\frac{1}{2} - z_8\right)c\sin\beta\,\hat{\mathbf{z}}$ | (4$e$) | H VI |
| $\mathbf{B}_{27}$ | = | $-x_8\,\mathbf{a}_1 - y_8\,\mathbf{a}_2 - z_8\,\mathbf{a}_3$ | = | $(-x_8 a - z_8 c\cos\beta)\,\hat{\mathbf{x}} - y_8 b\,\hat{\mathbf{y}} - z_8 c\sin\beta\,\hat{\mathbf{z}}$ | (4$e$) | H VI |
| $\mathbf{B}_{28}$ | = | $x_8\,\mathbf{a}_1 + \left(\frac{1}{2} - y_8\right)\mathbf{a}_2 + \left(\frac{1}{2} + z_8\right)\mathbf{a}_3$ | = | $\left(\frac{1}{2}c\cos\beta + x_8 a + z_8 c\cos\beta\right)\hat{\mathbf{x}} + \left(\frac{1}{2} - y_8\right)b\,\hat{\mathbf{y}} + \left(\frac{1}{2} + z_8\right)c\sin\beta\,\hat{\mathbf{z}}$ | (4$e$) | H VI |
| $\mathbf{B}_{29}$ | = | $x_9\,\mathbf{a}_1 + y_9\,\mathbf{a}_2 + z_9\,\mathbf{a}_3$ | = | $(x_9 a + z_9 c\cos\beta)\,\hat{\mathbf{x}} + y_9 b\,\hat{\mathbf{y}} + z_9 c\sin\beta\,\hat{\mathbf{z}}$ | (4$e$) | H VII |
| $\mathbf{B}_{30}$ | = | $-x_9\,\mathbf{a}_1 + \left(\frac{1}{2} + y_9\right)\mathbf{a}_2 + \left(\frac{1}{2} - z_9\right)\mathbf{a}_3$ | = | $\left(\frac{1}{2}c\cos\beta - x_9 a - z_9 c\cos\beta\right)\hat{\mathbf{x}} + \left(\frac{1}{2} + y_9\right)b\,\hat{\mathbf{y}} + \left(\frac{1}{2} - z_9\right)c\sin\beta\,\hat{\mathbf{z}}$ | (4$e$) | H VII |
| $\mathbf{B}_{31}$ | = | $-x_9\,\mathbf{a}_1 - y_9\,\mathbf{a}_2 - z_9\,\mathbf{a}_3$ | = | $(-x_9 a - z_9 c\cos\beta)\,\hat{\mathbf{x}} - y_9 b\,\hat{\mathbf{y}} - z_9 c\sin\beta\,\hat{\mathbf{z}}$ | (4$e$) | H VII |
| $\mathbf{B}_{32}$ | = | $x_9\,\mathbf{a}_1 + \left(\frac{1}{2} - y_9\right)\mathbf{a}_2 + \left(\frac{1}{2} + z_9\right)\mathbf{a}_3$ | = | $\left(\frac{1}{2}c\cos\beta + x_9 a + z_9 c\cos\beta\right)\hat{\mathbf{x}} + \left(\frac{1}{2} - y_9\right)b\,\hat{\mathbf{y}} + \left(\frac{1}{2} + z_9\right)c\sin\beta\,\hat{\mathbf{z}}$ | (4$e$) | H VII |
| $\mathbf{B}_{33}$ | = | $x_{10}\,\mathbf{a}_1 + y_{10}\,\mathbf{a}_2 + z_{10}\,\mathbf{a}_3$ | = | $(x_{10} a + z_{10} c\cos\beta)\,\hat{\mathbf{x}} + y_{10} b\,\hat{\mathbf{y}} + z_{10} c\sin\beta\,\hat{\mathbf{z}}$ | (4$e$) | H VIII |
| $\mathbf{B}_{34}$ | = | $-x_{10}\,\mathbf{a}_1 + \left(\frac{1}{2} + y_{10}\right)\mathbf{a}_2 + \left(\frac{1}{2} - z_{10}\right)\mathbf{a}_3$ | = | $\left(\frac{1}{2}c\cos\beta - x_{10} a - z_{10} c\cos\beta\right)\hat{\mathbf{x}} + \left(\frac{1}{2} + y_{10}\right)b\,\hat{\mathbf{y}} + \left(\frac{1}{2} - z_{10}\right)c\sin\beta\,\hat{\mathbf{z}}$ | (4$e$) | H VIII |
| $\mathbf{B}_{35}$ | = | $-x_{10}\,\mathbf{a}_1 - y_{10}\,\mathbf{a}_2 - z_{10}\,\mathbf{a}_3$ | = | $(-x_{10} a - z_{10} c\cos\beta)\,\hat{\mathbf{x}} - y_{10} b\,\hat{\mathbf{y}} - z_{10} c\sin\beta\,\hat{\mathbf{z}}$ | (4$e$) | H VIII |
| $\mathbf{B}_{36}$ | = | $x_{10}\,\mathbf{a}_1 + \left(\frac{1}{2} - y_{10}\right)\mathbf{a}_2 + \left(\frac{1}{2} + z_{10}\right)\mathbf{a}_3$ | = | $\left(\frac{1}{2}c\cos\beta + x_{10} a + z_{10} c\cos\beta\right)\hat{\mathbf{x}} + \left(\frac{1}{2} - y_{10}\right)b\,\hat{\mathbf{y}} + \left(\frac{1}{2} + z_{10}\right)c\sin\beta\,\hat{\mathbf{z}}$ | (4$e$) | H VIII |
| $\mathbf{B}_{37}$ | = | $x_{11}\,\mathbf{a}_1 + y_{11}\,\mathbf{a}_2 + z_{11}\,\mathbf{a}_3$ | = | $(x_{11} a + z_{11} c\cos\beta)\,\hat{\mathbf{x}} + y_{11} b\,\hat{\mathbf{y}} + z_{11} c\sin\beta\,\hat{\mathbf{z}}$ | (4$e$) | K I |
| $\mathbf{B}_{38}$ | = | $-x_{11}\,\mathbf{a}_1 + \left(\frac{1}{2} + y_{11}\right)\mathbf{a}_2 + \left(\frac{1}{2} - z_{11}\right)\mathbf{a}_3$ | = | $\left(\frac{1}{2}c\cos\beta - x_{11} a - z_{11} c\cos\beta\right)\hat{\mathbf{x}} + \left(\frac{1}{2} + y_{11}\right)b\,\hat{\mathbf{y}} + \left(\frac{1}{2} - z_{11}\right)c\sin\beta\,\hat{\mathbf{z}}$ | (4$e$) | K I |
| $\mathbf{B}_{39}$ | = | $-x_{11}\,\mathbf{a}_1 - y_{11}\,\mathbf{a}_2 - z_{11}\,\mathbf{a}_3$ | = | $(-x_{11} a - z_{11} c\cos\beta)\,\hat{\mathbf{x}} - y_{11} b\,\hat{\mathbf{y}} - z_{11} c\sin\beta\,\hat{\mathbf{z}}$ | (4$e$) | K I |
| $\mathbf{B}_{40}$ | = | $x_{11}\,\mathbf{a}_1 + \left(\frac{1}{2} - y_{11}\right)\mathbf{a}_2 + \left(\frac{1}{2} + z_{11}\right)\mathbf{a}_3$ | = | $\left(\frac{1}{2}c\cos\beta + x_{11} a + z_{11} c\cos\beta\right)\hat{\mathbf{x}} + \left(\frac{1}{2} - y_{11}\right)b\,\hat{\mathbf{y}} + \left(\frac{1}{2} + z_{11}\right)c\sin\beta\,\hat{\mathbf{z}}$ | (4$e$) | K I |
| $\mathbf{B}_{41}$ | = | $x_{12}\,\mathbf{a}_1 + y_{12}\,\mathbf{a}_2 + z_{12}\,\mathbf{a}_3$ | = | $(x_{12}a + z_{12}c\cos\beta)\,\hat{\mathbf{x}} + y_{12}b\,\hat{\mathbf{y}} + z_{12}c\sin\beta\,\hat{\mathbf{z}}$ | (4$e$) | K II |
| $\mathbf{B}_{42}$ | = | $-x_{12}\,\mathbf{a}_1 + \left(\frac{1}{2} + y_{12}\right)\mathbf{a}_2 + \left(\frac{1}{2} - z_{12}\right)\mathbf{a}_3$ | = | $\left(\frac{1}{2}c\cos\beta - x_{12}a - z_{12}c\cos\beta\right)\hat{\mathbf{x}} + \left(\frac{1}{2} + y_{12}\right)b\,\hat{\mathbf{y}} + \left(\frac{1}{2} - z_{12}\right)c\sin\beta\,\hat{\mathbf{z}}$ | (4$e$) | K II |
| $\mathbf{B}_{43}$ | = | $-x_{12}\,\mathbf{a}_1 - y_{12}\,\mathbf{a}_2 - z_{12}\,\mathbf{a}_3$ | = | $(-x_{12}a - z_{12}c\cos\beta)\,\hat{\mathbf{x}} - y_{12}b\,\hat{\mathbf{y}} - z_{12}c\sin\beta\,\hat{\mathbf{z}}$ | (4$e$) | K II |
| $\mathbf{B}_{44}$ | = | $x_{12}\,\mathbf{a}_1 + \left(\frac{1}{2} - y_{12}\right)\mathbf{a}_2 + \left(\frac{1}{2} + z_{12}\right)\mathbf{a}_3$ | = | $\left(\frac{1}{2}c\cos\beta + x_{12}a + z_{12}c\cos\beta\right)\hat{\mathbf{x}} + \left(\frac{1}{2} - y_{12}\right)b\,\hat{\mathbf{y}} + \left(\frac{1}{2} + z_{12}\right)c\sin\beta\,\hat{\mathbf{z}}$ | (4$e$) | K II |
| $\mathbf{B}_{45}$ | = | $x_{13}\,\mathbf{a}_1 + y_{13}\,\mathbf{a}_2 + z_{13}\,\mathbf{a}_3$ | = | $(x_{13}a + z_{13}c\cos\beta)\,\hat{\mathbf{x}} + y_{13}b\,\hat{\mathbf{y}} + z_{13}c\sin\beta\,\hat{\mathbf{z}}$ | (4$e$) | O I |
| $\mathbf{B}_{46}$ | = | $-x_{13}\,\mathbf{a}_1 + \left(\frac{1}{2} + y_{13}\right)\mathbf{a}_2 + \left(\frac{1}{2} - z_{13}\right)\mathbf{a}_3$ | = | $\left(\frac{1}{2}c\cos\beta - x_{13}a - z_{13}c\cos\beta\right)\hat{\mathbf{x}} + \left(\frac{1}{2} + y_{13}\right)b\,\hat{\mathbf{y}} + \left(\frac{1}{2} - z_{13}\right)c\sin\beta\,\hat{\mathbf{z}}$ | (4$e$) | O I |
| $\mathbf{B}_{47}$ | = | $-x_{13}\,\mathbf{a}_1 - y_{13}\,\mathbf{a}_2 - z_{13}\,\mathbf{a}_3$ | = | $(-x_{13}a - z_{13}c\cos\beta)\,\hat{\mathbf{x}} - y_{13}b\,\hat{\mathbf{y}} - z_{13}c\sin\beta\,\hat{\mathbf{z}}$ | (4$e$) | O I |
| $\mathbf{B}_{48}$ | = | $x_{13}\,\mathbf{a}_1 + \left(\frac{1}{2} - y_{13}\right)\mathbf{a}_2 + \left(\frac{1}{2} + z_{13}\right)\mathbf{a}_3$ | = | $\left(\frac{1}{2}c\cos\beta + x_{13}a + z_{13}c\cos\beta\right)\hat{\mathbf{x}} + \left(\frac{1}{2} - y_{13}\right)b\,\hat{\mathbf{y}} + \left(\frac{1}{2} + z_{13}\right)c\sin\beta\,\hat{\mathbf{z}}$ | (4$e$) | O I |
| $\mathbf{B}_{49}$ | = | $x_{14}\,\mathbf{a}_1 + y_{14}\,\mathbf{a}_2 + z_{14}\,\mathbf{a}_3$ | = | $(x_{14}a + z_{14}c\cos\beta)\,\hat{\mathbf{x}} + y_{14}b\,\hat{\mathbf{y}} + z_{14}c\sin\beta\,\hat{\mathbf{z}}$ | (4$e$) | O II |
| $\mathbf{B}_{50}$ | = | $-x_{14}\,\mathbf{a}_1 + \left(\frac{1}{2} + y_{14}\right)\mathbf{a}_2 + \left(\frac{1}{2} - z_{14}\right)\mathbf{a}_3$ | = | $\left(\frac{1}{2}c\cos\beta - x_{14}a - z_{14}c\cos\beta\right)\hat{\mathbf{x}} + \left(\frac{1}{2} + y_{14}\right)b\,\hat{\mathbf{y}} + \left(\frac{1}{2} - z_{14}\right)c\sin\beta\,\hat{\mathbf{z}}$ | (4$e$) | O II |
| $\mathbf{B}_{51}$ | = | $-x_{14}\,\mathbf{a}_1 - y_{14}\,\mathbf{a}_2 - z_{14}\,\mathbf{a}_3$ | = | $(-x_{14}a - z_{14}c\cos\beta)\,\hat{\mathbf{x}} - y_{14}b\,\hat{\mathbf{y}} - z_{14}c\sin\beta\,\hat{\mathbf{z}}$ | (4$e$) | O II |
| $\mathbf{B}_{52}$ | = | $x_{14}\,\mathbf{a}_1 + \left(\frac{1}{2} - y_{14}\right)\mathbf{a}_2 + \left(\frac{1}{2} + z_{14}\right)\mathbf{a}_3$ | = | $\left(\frac{1}{2}c\cos\beta + x_{14}a + z_{14}c\cos\beta\right)\hat{\mathbf{x}} + \left(\frac{1}{2} - y_{14}\right)b\,\hat{\mathbf{y}} + \left(\frac{1}{2} + z_{14}\right)c\sin\beta\,\hat{\mathbf{z}}$ | (4$e$) | O II |
| $\mathbf{B}_{53}$ | = | $x_{15}\,\mathbf{a}_1 + y_{15}\,\mathbf{a}_2 + z_{15}\,\mathbf{a}_3$ | = | $(x_{15}a + z_{15}c\cos\beta)\,\hat{\mathbf{x}} + y_{15}b\,\hat{\mathbf{y}} + z_{15}c\sin\beta\,\hat{\mathbf{z}}$ | (4$e$) | O III |
| $\mathbf{B}_{54}$ | = | $-x_{15}\,\mathbf{a}_1 + \left(\frac{1}{2} + y_{15}\right)\mathbf{a}_2 + \left(\frac{1}{2} - z_{15}\right)\mathbf{a}_3$ | = | $\left(\frac{1}{2}c\cos\beta - x_{15}a - z_{15}c\cos\beta\right)\hat{\mathbf{x}} + \left(\frac{1}{2} + y_{15}\right)b\,\hat{\mathbf{y}} + \left(\frac{1}{2} - z_{15}\right)c\sin\beta\,\hat{\mathbf{z}}$ | (4$e$) | O III |
| $\mathbf{B}_{55}$ | = | $-x_{15}\,\mathbf{a}_1 - y_{15}\,\mathbf{a}_2 - z_{15}\,\mathbf{a}_3$ | = | $(-x_{15}a - z_{15}c\cos\beta)\,\hat{\mathbf{x}} - y_{15}b\,\hat{\mathbf{y}} - z_{15}c\sin\beta\,\hat{\mathbf{z}}$ | (4$e$) | O III |
| $\mathbf{B}_{56}$ | = | $x_{15}\,\mathbf{a}_1 + \left(\frac{1}{2} - y_{15}\right)\mathbf{a}_2 + \left(\frac{1}{2} + z_{15}\right)\mathbf{a}_3$ | = | $\left(\frac{1}{2}c\cos\beta + x_{15}a + z_{15}c\cos\beta\right)\hat{\mathbf{x}} + \left(\frac{1}{2} - y_{15}\right)b\,\hat{\mathbf{y}} + \left(\frac{1}{2} + z_{15}\right)c\sin\beta\,\hat{\mathbf{z}}$ | (4$e$) | O III |
| $\mathbf{B}_{57}$ | = | $x_{16}\,\mathbf{a}_1 + y_{16}\,\mathbf{a}_2 + z_{16}\,\mathbf{a}_3$ | = | $(x_{16}a + z_{16}c\cos\beta)\,\hat{\mathbf{x}} + y_{16}b\,\hat{\mathbf{y}} + z_{16}c\sin\beta\,\hat{\mathbf{z}}$ | (4$e$) | O IV |
| $\mathbf{B}_{58}$ | = | $-x_{16}\,\mathbf{a}_1 + \left(\frac{1}{2} + y_{16}\right)\mathbf{a}_2 + \left(\frac{1}{2} - z_{16}\right)\mathbf{a}_3$ | = | $\left(\frac{1}{2}c\cos\beta - x_{16}a - z_{16}c\cos\beta\right)\hat{\mathbf{x}} + \left(\frac{1}{2} + y_{16}\right)b\,\hat{\mathbf{y}} + \left(\frac{1}{2} - z_{16}\right)c\sin\beta\,\hat{\mathbf{z}}$ | (4$e$) | O IV |
| $\mathbf{B}_{59}$ | = | $-x_{16}\,\mathbf{a}_1 - y_{16}\,\mathbf{a}_2 - z_{16}\,\mathbf{a}_3$ | = | $(-x_{16}a - z_{16}c\cos\beta)\,\hat{\mathbf{x}} - y_{16}b\,\hat{\mathbf{y}} - z_{16}c\sin\beta\,\hat{\mathbf{z}}$ | (4$e$) | O IV |
| $\mathbf{B}_{60}$ | = | $x_{16}\,\mathbf{a}_1 + \left(\frac{1}{2} - y_{16}\right)\mathbf{a}_2 + \left(\frac{1}{2} + z_{16}\right)\mathbf{a}_3$ | = | $\left(\frac{1}{2}c\cos\beta + x_{16}a + z_{16}c\cos\beta\right)\hat{\mathbf{x}} + \left(\frac{1}{2} - y_{16}\right)b\,\hat{\mathbf{y}} + \left(\frac{1}{2} + z_{16}\right)c\sin\beta\,\hat{\mathbf{z}}$ | (4$e$) | O IV |
| $\mathbf{B}_{61}$ | = | $x_{17}\,\mathbf{a}_1 + y_{17}\,\mathbf{a}_2 + z_{17}\,\mathbf{a}_3$ | = | $(x_{17}a + z_{17}c\cos\beta)\,\hat{\mathbf{x}} + y_{17}b\,\hat{\mathbf{y}} + z_{17}c\sin\beta\,\hat{\mathbf{z}}$ | (4$e$) | O V |
| $\mathbf{B}_{62}$ | = | $-x_{17}\,\mathbf{a}_1 + \left(\frac{1}{2} + y_{17}\right)\mathbf{a}_2 + \left(\frac{1}{2} - z_{17}\right)\mathbf{a}_3$ | = | $\left(\frac{1}{2}c\cos\beta - x_{17}a - z_{17}c\cos\beta\right)\hat{\mathbf{x}} + \left(\frac{1}{2} + y_{17}\right)b\,\hat{\mathbf{y}} + \left(\frac{1}{2} - z_{17}\right)c\sin\beta\,\hat{\mathbf{z}}$ | (4$e$) | O V |
| $\mathbf{B}_{63}$ | = | $-x_{17}\,\mathbf{a}_1 - y_{17}\,\mathbf{a}_2 - z_{17}\,\mathbf{a}_3$ | = | $(-x_{17}a - z_{17}c\cos\beta)\,\hat{\mathbf{x}} - y_{17}b\,\hat{\mathbf{y}} - z_{17}c\sin\beta\,\hat{\mathbf{z}}$ | (4$e$) | O V |
| $\mathbf{B}_{64}$ | = | $x_{17}\,\mathbf{a}_1 + \left(\frac{1}{2} - y_{17}\right)\,\mathbf{a}_2 + \left(\frac{1}{2} + z_{17}\right)\,\mathbf{a}_3$ | = | $\left(\frac{1}{2}c\cos\beta + x_{17}a + z_{17}c\cos\beta\right)\,\hat{\mathbf{x}} + \left(\frac{1}{2} - y_{17}\right)b\,\hat{\mathbf{y}} + \left(\frac{1}{2} + z_{17}\right)c\sin\beta\,\hat{\mathbf{z}}$ | (4$e$) | O V |
| $\mathbf{B}_{65}$ | = | $x_{18}\,\mathbf{a}_1 + y_{18}\,\mathbf{a}_2 + z_{18}\,\mathbf{a}_3$ | = | $(x_{18}a + z_{18}c\cos\beta)\,\hat{\mathbf{x}} + y_{18}b\,\hat{\mathbf{y}} + z_{18}c\sin\beta\,\hat{\mathbf{z}}$ | (4$e$) | O VI |
| $\mathbf{B}_{66}$ | = | $-x_{18}\,\mathbf{a}_1 + \left(\frac{1}{2} + y_{18}\right)\,\mathbf{a}_2 + \left(\frac{1}{2} - z_{18}\right)\,\mathbf{a}_3$ | = | $\left(\frac{1}{2}c\cos\beta - x_{18}a - z_{18}c\cos\beta\right)\,\hat{\mathbf{x}} + \left(\frac{1}{2} + y_{18}\right)b\,\hat{\mathbf{y}} + \left(\frac{1}{2} - z_{18}\right)c\sin\beta\,\hat{\mathbf{z}}$ | (4$e$) | O VI |
| $\mathbf{B}_{67}$ | = | $-x_{18}\,\mathbf{a}_1 - y_{18}\,\mathbf{a}_2 - z_{18}\,\mathbf{a}_3$ | = | $(-x_{18}a - z_{18}c\cos\beta)\,\hat{\mathbf{x}} - y_{18}b\,\hat{\mathbf{y}} - z_{18}c\sin\beta\,\hat{\mathbf{z}}$ | (4$e$) | O VI |
| $\mathbf{B}_{68}$ | = | $x_{18}\,\mathbf{a}_1 + \left(\frac{1}{2} - y_{18}\right)\,\mathbf{a}_2 + \left(\frac{1}{2} + z_{18}\right)\,\mathbf{a}_3$ | = | $\left(\frac{1}{2}c\cos\beta + x_{18}a + z_{18}c\cos\beta\right)\,\hat{\mathbf{x}} + \left(\frac{1}{2} - y_{18}\right)b\,\hat{\mathbf{y}} + \left(\frac{1}{2} + z_{18}\right)c\sin\beta\,\hat{\mathbf{z}}$ | (4$e$) | O VI |
| $\mathbf{B}_{69}$ | = | $x_{19}\,\mathbf{a}_1 + y_{19}\,\mathbf{a}_2 + z_{19}\,\mathbf{a}_3$ | = | $(x_{19}a + z_{19}c\cos\beta)\,\hat{\mathbf{x}} + y_{19}b\,\hat{\mathbf{y}} + z_{19}c\sin\beta\,\hat{\mathbf{z}}$ | (4$e$) | O VII |
| $\mathbf{B}_{70}$ | = | $-x_{19}\,\mathbf{a}_1 + \left(\frac{1}{2} + y_{19}\right)\,\mathbf{a}_2 + \left(\frac{1}{2} - z_{19}\right)\,\mathbf{a}_3$ | = | $\left(\frac{1}{2}c\cos\beta - x_{19}a - z_{19}c\cos\beta\right)\,\hat{\mathbf{x}} + \left(\frac{1}{2} + y_{19}\right)b\,\hat{\mathbf{y}} + \left(\frac{1}{2} - z_{19}\right)c\sin\beta\,\hat{\mathbf{z}}$ | (4$e$) | O VII |
| $\mathbf{B}_{71}$ | = | $-x_{19}\,\mathbf{a}_1 - y_{19}\,\mathbf{a}_2 - z_{19}\,\mathbf{a}_3$ | = | $(-x_{19}a - z_{19}c\cos\beta)\,\hat{\mathbf{x}} - y_{19}b\,\hat{\mathbf{y}} - z_{19}c\sin\beta\,\hat{\mathbf{z}}$ | (4$e$) | O VII |
| $\mathbf{B}_{72}$ | = | $x_{19}\,\mathbf{a}_1 + \left(\frac{1}{2} - y_{19}\right)\,\mathbf{a}_2 + \left(\frac{1}{2} + z_{19}\right)\,\mathbf{a}_3$ | = | $\left(\frac{1}{2}c\cos\beta + x_{19}a + z_{19}c\cos\beta\right)\,\hat{\mathbf{x}} + \left(\frac{1}{2} - y_{19}\right)b\,\hat{\mathbf{y}} + \left(\frac{1}{2} + z_{19}\right)c\sin\beta\,\hat{\mathbf{z}}$ | (4$e$) | O VII |
| $\mathbf{B}_{73}$ | = | $x_{20}\,\mathbf{a}_1 + y_{20}\,\mathbf{a}_2 + z_{20}\,\mathbf{a}_3$ | = | $(x_{20}a + z_{20}c\cos\beta)\,\hat{\mathbf{x}} + y_{20}b\,\hat{\mathbf{y}} + z_{20}c\sin\beta\,\hat{\mathbf{z}}$ | (4$e$) | O VIII |
| $\mathbf{B}_{74}$ | = | $-x_{20}\,\mathbf{a}_1 + \left(\frac{1}{2} + y_{20}\right)\,\mathbf{a}_2 + \left(\frac{1}{2} - z_{20}\right)\,\mathbf{a}_3$ | = | $\left(\frac{1}{2}c\cos\beta - x_{20}a - z_{20}c\cos\beta\right)\,\hat{\mathbf{x}} + \left(\frac{1}{2} + y_{20}\right)b\,\hat{\mathbf{y}} + \left(\frac{1}{2} - z_{20}\right)c\sin\beta\,\hat{\mathbf{z}}$ | (4$e$) | O VIII |
| $\mathbf{B}_{75}$ | = | $-x_{20}\,\mathbf{a}_1 - y_{20}\,\mathbf{a}_2 - z_{20}\,\mathbf{a}_3$ | = | $(-x_{20}a - z_{20}c\cos\beta)\,\hat{\mathbf{x}} - y_{20}b\,\hat{\mathbf{y}} - z_{20}c\sin\beta\,\hat{\mathbf{z}}$ | (4$e$) | O VIII |
| $\mathbf{B}_{76}$ | = | $x_{20}\,\mathbf{a}_1 + \left(\frac{1}{2} - y_{20}\right)\,\mathbf{a}_2 + \left(\frac{1}{2} + z_{20}\right)\,\mathbf{a}_3$ | = | $\left(\frac{1}{2}c\cos\beta + x_{20}a + z_{20}c\cos\beta\right)\,\hat{\mathbf{x}} + \left(\frac{1}{2} - y_{20}\right)b\,\hat{\mathbf{y}} + \left(\frac{1}{2} + z_{20}\right)c\sin\beta\,\hat{\mathbf{z}}$ | (4$e$) | O VIII |
| $\mathbf{B}_{77}$ | = | $x_{21}\,\mathbf{a}_1 + y_{21}\,\mathbf{a}_2 + z_{21}\,\mathbf{a}_3$ | = | $(x_{21}a + z_{21}c\cos\beta)\,\hat{\mathbf{x}} + y_{21}b\,\hat{\mathbf{y}} + z_{21}c\sin\beta\,\hat{\mathbf{z}}$ | (4$e$) | O IX |
| $\mathbf{B}_{78}$ | = | $-x_{21}\,\mathbf{a}_1 + \left(\frac{1}{2} + y_{21}\right)\,\mathbf{a}_2 + \left(\frac{1}{2} - z_{21}\right)\,\mathbf{a}_3$ | = | $\left(\frac{1}{2}c\cos\beta - x_{21}a - z_{21}c\cos\beta\right)\,\hat{\mathbf{x}} + \left(\frac{1}{2} + y_{21}\right)b\,\hat{\mathbf{y}} + \left(\frac{1}{2} - z_{21}\right)c\sin\beta\,\hat{\mathbf{z}}$ | (4$e$) | O IX |
| $\mathbf{B}_{79}$ | = | $-x_{21}\,\mathbf{a}_1 - y_{21}\,\mathbf{a}_2 - z_{21}\,\mathbf{a}_3$ | = | $(-x_{21}a - z_{21}c\cos\beta)\,\hat{\mathbf{x}} - y_{21}b\,\hat{\mathbf{y}} - z_{21}c\sin\beta\,\hat{\mathbf{z}}$ | (4$e$) | O IX |
| $\mathbf{B}_{80}$ | = | $x_{21}\,\mathbf{a}_1 + \left(\frac{1}{2} - y_{21}\right)\,\mathbf{a}_2 + \left(\frac{1}{2} + z_{21}\right)\,\mathbf{a}_3$ | = | $\left(\frac{1}{2}c\cos\beta + x_{21}a + z_{21}c\cos\beta\right)\,\hat{\mathbf{x}} + \left(\frac{1}{2} - y_{21}\right)b\,\hat{\mathbf{y}} + \left(\frac{1}{2} + z_{21}\right)c\sin\beta\,\hat{\mathbf{z}}$ | (4$e$) | O IX |
| $\mathbf{B}_{81}$ | = | $x_{22}\,\mathbf{a}_1 + y_{22}\,\mathbf{a}_2 + z_{22}\,\mathbf{a}_3$ | = | $(x_{22}a + z_{22}c\cos\beta)\,\hat{\mathbf{x}} + y_{22}b\,\hat{\mathbf{y}} + z_{22}c\sin\beta\,\hat{\mathbf{z}}$ | (4$e$) | O X |
| $\mathbf{B}_{82}$ | = | $-x_{22}\,\mathbf{a}_1 + \left(\frac{1}{2} + y_{22}\right)\,\mathbf{a}_2 + \left(\frac{1}{2} - z_{22}\right)\,\mathbf{a}_3$ | = | $\left(\frac{1}{2}c\cos\beta - x_{22}a - z_{22}c\cos\beta\right)\,\hat{\mathbf{x}} + \left(\frac{1}{2} + y_{22}\right)b\,\hat{\mathbf{y}} + \left(\frac{1}{2} - z_{22}\right)c\sin\beta\,\hat{\mathbf{z}}$ | (4$e$) | O X |
| $\mathbf{B}_{83}$ | = | $-x_{22}\,\mathbf{a}_1 - y_{22}\,\mathbf{a}_2 - z_{22}\,\mathbf{a}_3$ | = | $(-x_{22}a - z_{22}c\cos\beta)\,\hat{\mathbf{x}} - y_{22}b\,\hat{\mathbf{y}} - z_{22}c\sin\beta\,\hat{\mathbf{z}}$ | (4$e$) | O X |
| $\mathbf{B}_{84}$ | = | $x_{22}\,\mathbf{a}_1 + \left(\frac{1}{2} - y_{22}\right)\,\mathbf{a}_2 + \left(\frac{1}{2} + z_{22}\right)\,\mathbf{a}_3$ | = | $\left(\frac{1}{2}c\cos\beta + x_{22}a + z_{22}c\cos\beta\right)\,\hat{\mathbf{x}} + \left(\frac{1}{2} - y_{22}\right)b\,\hat{\mathbf{y}} + \left(\frac{1}{2} + z_{22}\right)c\sin\beta\,\hat{\mathbf{z}}$ | (4$e$) | O X |
| $\mathbf{B}_{85}$ | = | $x_{23}\,\mathbf{a}_1 + y_{23}\,\mathbf{a}_2 + z_{23}\,\mathbf{a}_3$ | = | $(x_{23}a + z_{23}c\cos\beta)\,\hat{\mathbf{x}} + y_{23}b\,\hat{\mathbf{y}} + z_{23}c\sin\beta\,\hat{\mathbf{z}}$ | (4e) | O XI |
| $\mathbf{B}_{86}$ | = | $-x_{23}\,\mathbf{a}_1 + \left(\frac{1}{2} + y_{23}\right)\mathbf{a}_2 + \left(\frac{1}{2} - z_{23}\right)\mathbf{a}_3$ | = | $\left(\frac{1}{2}c\cos\beta - x_{23}a - z_{23}c\cos\beta\right)\hat{\mathbf{x}} + \left(\frac{1}{2} + y_{23}\right)b\,\hat{\mathbf{y}} + \left(\frac{1}{2} - z_{23}\right)c\sin\beta\,\hat{\mathbf{z}}$ | (4e) | O XI |
| $\mathbf{B}_{87}$ | = | $-x_{23}\,\mathbf{a}_1 - y_{23}\,\mathbf{a}_2 - z_{23}\,\mathbf{a}_3$ | = | $(-x_{23}a - z_{23}c\cos\beta)\,\hat{\mathbf{x}} - y_{23}b\,\hat{\mathbf{y}} - z_{23}c\sin\beta\,\hat{\mathbf{z}}$ | (4e) | O XI |
| $\mathbf{B}_{88}$ | = | $x_{23}\,\mathbf{a}_1 + \left(\frac{1}{2} - y_{23}\right)\mathbf{a}_2 + \left(\frac{1}{2} + z_{23}\right)\mathbf{a}_3$ | = | $\left(\frac{1}{2}c\cos\beta + x_{23}a + z_{23}c\cos\beta\right)\hat{\mathbf{x}} + \left(\frac{1}{2} - y_{23}\right)b\,\hat{\mathbf{y}} + \left(\frac{1}{2} + z_{23}\right)c\sin\beta\,\hat{\mathbf{z}}$ | (4e) | O XI |
| $\mathbf{B}_{89}$ | = | $x_{24}\,\mathbf{a}_1 + y_{24}\,\mathbf{a}_2 + z_{24}\,\mathbf{a}_3$ | = | $(x_{24}a + z_{24}c\cos\beta)\,\hat{\mathbf{x}} + y_{24}b\,\hat{\mathbf{y}} + z_{24}c\sin\beta\,\hat{\mathbf{z}}$ | (4e) | O XII |
| $\mathbf{B}_{90}$ | = | $-x_{24}\,\mathbf{a}_1 + \left(\frac{1}{2} + y_{24}\right)\mathbf{a}_2 + \left(\frac{1}{2} - z_{24}\right)\mathbf{a}_3$ | = | $\left(\frac{1}{2}c\cos\beta - x_{24}a - z_{24}c\cos\beta\right)\hat{\mathbf{x}} + \left(\frac{1}{2} + y_{24}\right)b\,\hat{\mathbf{y}} + \left(\frac{1}{2} - z_{24}\right)c\sin\beta\,\hat{\mathbf{z}}$ | (4e) | O XII |
| $\mathbf{B}_{91}$ | = | $-x_{24}\,\mathbf{a}_1 - y_{24}\,\mathbf{a}_2 - z_{24}\,\mathbf{a}_3$ | = | $(-x_{24}a - z_{24}c\cos\beta)\,\hat{\mathbf{x}} - y_{24}b\,\hat{\mathbf{y}} - z_{24}c\sin\beta\,\hat{\mathbf{z}}$ | (4e) | O XII |
| $\mathbf{B}_{92}$ | = | $x_{24}\,\mathbf{a}_1 + \left(\frac{1}{2} - y_{24}\right)\mathbf{a}_2 + \left(\frac{1}{2} + z_{24}\right)\mathbf{a}_3$ | = | $\left(\frac{1}{2}c\cos\beta + x_{24}a + z_{24}c\cos\beta\right)\hat{\mathbf{x}} + \left(\frac{1}{2} - y_{24}\right)b\,\hat{\mathbf{y}} + \left(\frac{1}{2} + z_{24}\right)c\sin\beta\,\hat{\mathbf{z}}$ | (4e) | O XII |
| $\mathbf{B}_{93}$ | = | $x_{25}\,\mathbf{a}_1 + y_{25}\,\mathbf{a}_2 + z_{25}\,\mathbf{a}_3$ | = | $(x_{25}a + z_{25}c\cos\beta)\,\hat{\mathbf{x}} + y_{25}b\,\hat{\mathbf{y}} + z_{25}c\sin\beta\,\hat{\mathbf{z}}$ | (4e) | S I |
| $\mathbf{B}_{94}$ | = | $-x_{25}\,\mathbf{a}_1 + \left(\frac{1}{2} + y_{25}\right)\mathbf{a}_2 + \left(\frac{1}{2} - z_{25}\right)\mathbf{a}_3$ | = | $\left(\frac{1}{2}c\cos\beta - x_{25}a - z_{25}c\cos\beta\right)\hat{\mathbf{x}} + \left(\frac{1}{2} + y_{25}\right)b\,\hat{\mathbf{y}} + \left(\frac{1}{2} - z_{25}\right)c\sin\beta\,\hat{\mathbf{z}}$ | (4e) | S I |
| $\mathbf{B}_{95}$ | = | $-x_{25}\,\mathbf{a}_1 - y_{25}\,\mathbf{a}_2 - z_{25}\,\mathbf{a}_3$ | = | $(-x_{25}a - z_{25}c\cos\beta)\,\hat{\mathbf{x}} - y_{25}b\,\hat{\mathbf{y}} - z_{25}c\sin\beta\,\hat{\mathbf{z}}$ | (4e) | S I |
| $\mathbf{B}_{96}$ | = | $x_{25}\,\mathbf{a}_1 + \left(\frac{1}{2} - y_{25}\right)\mathbf{a}_2 + \left(\frac{1}{2} + z_{25}\right)\mathbf{a}_3$ | = | $\left(\frac{1}{2}c\cos\beta + x_{25}a + z_{25}c\cos\beta\right)\hat{\mathbf{x}} + \left(\frac{1}{2} - y_{25}\right)b\,\hat{\mathbf{y}} + \left(\frac{1}{2} + z_{25}\right)c\sin\beta\,\hat{\mathbf{z}}$ | (4e) | S I |
| $\mathbf{B}_{97}$ | = | $x_{26}\,\mathbf{a}_1 + y_{26}\,\mathbf{a}_2 + z_{26}\,\mathbf{a}_3$ | = | $(x_{26}a + z_{26}c\cos\beta)\,\hat{\mathbf{x}} + y_{26}b\,\hat{\mathbf{y}} + z_{26}c\sin\beta\,\hat{\mathbf{z}}$ | (4e) | S II |
| $\mathbf{B}_{98}$ | = | $-x_{26}\,\mathbf{a}_1 + \left(\frac{1}{2} + y_{26}\right)\mathbf{a}_2 + \left(\frac{1}{2} - z_{26}\right)\mathbf{a}_3$ | = | $\left(\frac{1}{2}c\cos\beta - x_{26}a - z_{26}c\cos\beta\right)\hat{\mathbf{x}} + \left(\frac{1}{2} + y_{26}\right)b\,\hat{\mathbf{y}} + \left(\frac{1}{2} - z_{26}\right)c\sin\beta\,\hat{\mathbf{z}}$ | (4e) | S II |
| $\mathbf{B}_{99}$ | = | $-x_{26}\,\mathbf{a}_1 - y_{26}\,\mathbf{a}_2 - z_{26}\,\mathbf{a}_3$ | = | $(-x_{26}a - z_{26}c\cos\beta)\,\hat{\mathbf{x}} - y_{26}b\,\hat{\mathbf{y}} - z_{26}c\sin\beta\,\hat{\mathbf{z}}$ | (4e) | S II |
| $\mathbf{B}_{100}$ | = | $x_{26}\,\mathbf{a}_1 + \left(\frac{1}{2} - y_{26}\right)\mathbf{a}_2 + \left(\frac{1}{2} + z_{26}\right)\mathbf{a}_3$ | = | $\left(\frac{1}{2}c\cos\beta + x_{26}a + z_{26}c\cos\beta\right)\hat{\mathbf{x}} + \left(\frac{1}{2} - y_{26}\right)b\,\hat{\mathbf{y}} + \left(\frac{1}{2} + z_{26}\right)c\sin\beta\,\hat{\mathbf{z}}$ | (4e) | S II |

**Geometry files:**

- CIF: pp. 1546
- POSCAR: pp. 1547

# $Co_2Al_9$ ($D8_d$) Structure: A9B2_mP22_14_a4e_e

http://aflow.org/prototype-encyclopedia/A9B2_mP22_14_a4e_e

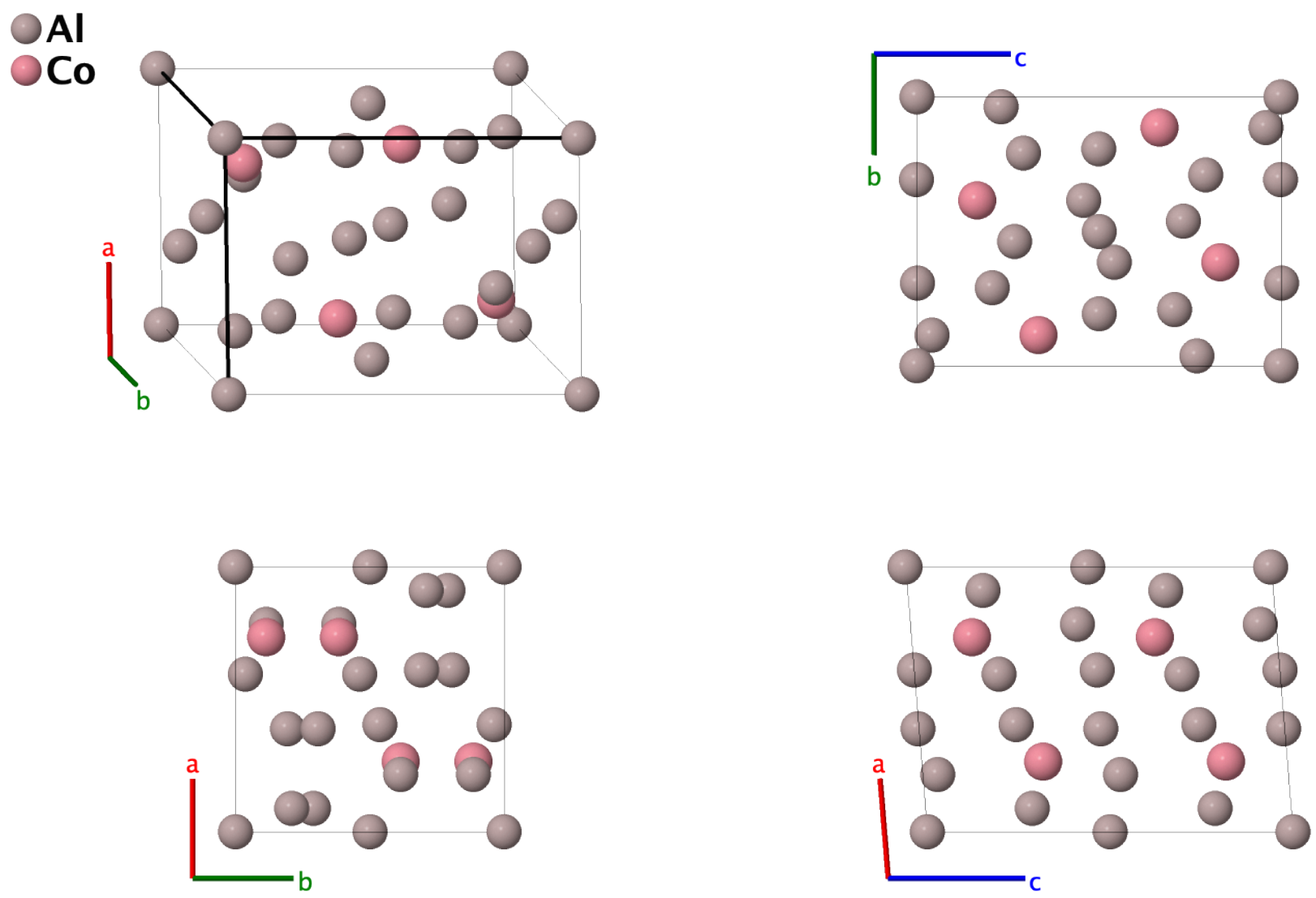

| | | |
|---|---|---|
| **Prototype** | : | $Co_2Al_9$ |
| **AFLOW prototype label** | : | A9B2_mP22_14_a4e_e |
| ***Strukturbericht* designation** | : | $D8_d$ |
| **Pearson symbol** | : | mP22 |
| **Space group number** | : | 14 |
| **Space group symbol** | : | $P2_1/c$ |
| **AFLOW prototype command** | : | `aflow --proto=A9B2_mP22_14_a4e_e` <br> `--params=`$a, b/a, c/a, \beta, x_2, y_2, z_2, x_3, y_3, z_3, x_4, y_4, z_4, x_5, y_5, z_5, x_6, y_6, z_6$ |

**Other compounds with this structure**

- $Ir_2Al_9$ and $Rh_2Al_9$

- (Douglas, 1950) gives the Wyckoff positions in terms of the $P2_1/a$ setting of space group #14. We have rotated the axes to change this to our standard $P2_1/c$ orientation.

**Simple Monoclinic primitive vectors:**

$$\mathbf{a}_1 = a\,\hat{\mathbf{x}}$$
$$\mathbf{a}_2 = b\,\hat{\mathbf{y}}$$
$$\mathbf{a}_3 = c\cos\beta\,\hat{\mathbf{x}} + c\sin\beta\,\hat{\mathbf{z}}$$

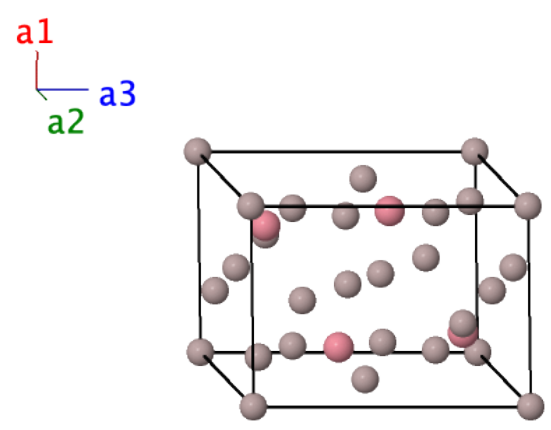

**Basis vectors:**

| | | Lattice Coordinates | | Cartesian Coordinates | Wyckoff Position | Atom Type |
|---|---|---|---|---|---|---|
| $\mathbf{B}_1$ | = | $0\,\mathbf{a}_1 + 0\,\mathbf{a}_2 + 0\,\mathbf{a}_3$ | = | $0\,\hat{\mathbf{x}} + 0\,\hat{\mathbf{y}} + 0\,\hat{\mathbf{z}}$ | $(2a)$ | Al I |
| $\mathbf{B}_{2}$ | = | $\frac{1}{2}\,\mathbf{a}_{2} + \frac{1}{2}\,\mathbf{a}_{3}$ | = | $\frac{1}{2}c\cos\beta\,\hat{\mathbf{x}} + \frac{1}{2}b\,\hat{\mathbf{y}} + \frac{1}{2}c\sin\beta\,\hat{\mathbf{z}}$ | (2*a*) | Al I |
| $\mathbf{B}_{3}$ | = | $x_{2}\,\mathbf{a}_{1} + y_{2}\,\mathbf{a}_{2} + z_{2}\,\mathbf{a}_{3}$ | = | $(x_{2}a + z_{2}c\cos\beta)\,\hat{\mathbf{x}} + y_{2}b\,\hat{\mathbf{y}} + z_{2}c\sin\beta\,\hat{\mathbf{z}}$ | (4*e*) | Al II |
| $\mathbf{B}_{4}$ | = | $-x_{2}\,\mathbf{a}_{1} + \left(\frac{1}{2} + y_{2}\right)\,\mathbf{a}_{2} + \left(\frac{1}{2} - z_{2}\right)\,\mathbf{a}_{3}$ | = | $\left(\frac{1}{2}c\cos\beta - x_{2}a - z_{2}c\cos\beta\right)\,\hat{\mathbf{x}} + \left(\frac{1}{2} + y_{2}\right)b\,\hat{\mathbf{y}} + \left(\frac{1}{2} - z_{2}\right)c\sin\beta\,\hat{\mathbf{z}}$ | (4*e*) | Al II |
| $\mathbf{B}_{5}$ | = | $-x_{2}\,\mathbf{a}_{1} - y_{2}\,\mathbf{a}_{2} - z_{2}\,\mathbf{a}_{3}$ | = | $(-x_{2}a - z_{2}c\cos\beta)\,\hat{\mathbf{x}} - y_{2}b\,\hat{\mathbf{y}} - z_{2}c\sin\beta\,\hat{\mathbf{z}}$ | (4*e*) | Al II |
| $\mathbf{B}_{6}$ | = | $x_{2}\,\mathbf{a}_{1} + \left(\frac{1}{2} - y_{2}\right)\,\mathbf{a}_{2} + \left(\frac{1}{2} + z_{2}\right)\,\mathbf{a}_{3}$ | = | $\left(\frac{1}{2}c\cos\beta + x_{2}a + z_{2}c\cos\beta\right)\,\hat{\mathbf{x}} + \left(\frac{1}{2} - y_{2}\right)b\,\hat{\mathbf{y}} + \left(\frac{1}{2} + z_{2}\right)c\sin\beta\,\hat{\mathbf{z}}$ | (4*e*) | Al II |
| $\mathbf{B}_{7}$ | = | $x_{3}\,\mathbf{a}_{1} + y_{3}\,\mathbf{a}_{2} + z_{3}\,\mathbf{a}_{3}$ | = | $(x_{3}a + z_{3}c\cos\beta)\,\hat{\mathbf{x}} + y_{3}b\,\hat{\mathbf{y}} + z_{3}c\sin\beta\,\hat{\mathbf{z}}$ | (4*e*) | Al III |
| $\mathbf{B}_{8}$ | = | $-x_{3}\,\mathbf{a}_{1} + \left(\frac{1}{2} + y_{3}\right)\,\mathbf{a}_{2} + \left(\frac{1}{2} - z_{3}\right)\,\mathbf{a}_{3}$ | = | $\left(\frac{1}{2}c\cos\beta - x_{3}a - z_{3}c\cos\beta\right)\,\hat{\mathbf{x}} + \left(\frac{1}{2} + y_{3}\right)b\,\hat{\mathbf{y}} + \left(\frac{1}{2} - z_{3}\right)c\sin\beta\,\hat{\mathbf{z}}$ | (4*e*) | Al III |
| $\mathbf{B}_{9}$ | = | $-x_{3}\,\mathbf{a}_{1} - y_{3}\,\mathbf{a}_{2} - z_{3}\,\mathbf{a}_{3}$ | = | $(-x_{3}a - z_{3}c\cos\beta)\,\hat{\mathbf{x}} - y_{3}b\,\hat{\mathbf{y}} - z_{3}c\sin\beta\,\hat{\mathbf{z}}$ | (4*e*) | Al III |
| $\mathbf{B}_{10}$ | = | $x_{3}\,\mathbf{a}_{1} + \left(\frac{1}{2} - y_{3}\right)\,\mathbf{a}_{2} + \left(\frac{1}{2} + z_{3}\right)\,\mathbf{a}_{3}$ | = | $\left(\frac{1}{2}c\cos\beta + x_{3}a + z_{3}c\cos\beta\right)\,\hat{\mathbf{x}} + \left(\frac{1}{2} - y_{3}\right)b\,\hat{\mathbf{y}} + \left(\frac{1}{2} + z_{3}\right)c\sin\beta\,\hat{\mathbf{z}}$ | (4*e*) | Al III |
| $\mathbf{B}_{11}$ | = | $x_{4}\,\mathbf{a}_{1} + y_{4}\,\mathbf{a}_{2} + z_{4}\,\mathbf{a}_{3}$ | = | $(x_{4}a + z_{4}c\cos\beta)\,\hat{\mathbf{x}} + y_{4}b\,\hat{\mathbf{y}} + z_{4}c\sin\beta\,\hat{\mathbf{z}}$ | (4*e*) | Al IV |
| $\mathbf{B}_{12}$ | = | $-x_{4}\,\mathbf{a}_{1} + \left(\frac{1}{2} + y_{4}\right)\,\mathbf{a}_{2} + \left(\frac{1}{2} - z_{4}\right)\,\mathbf{a}_{3}$ | = | $\left(\frac{1}{2}c\cos\beta - x_{4}a - z_{4}c\cos\beta\right)\,\hat{\mathbf{x}} + \left(\frac{1}{2} + y_{4}\right)b\,\hat{\mathbf{y}} + \left(\frac{1}{2} - z_{4}\right)c\sin\beta\,\hat{\mathbf{z}}$ | (4*e*) | Al IV |
| $\mathbf{B}_{13}$ | = | $-x_{4}\,\mathbf{a}_{1} - y_{4}\,\mathbf{a}_{2} - z_{4}\,\mathbf{a}_{3}$ | = | $(-x_{4}a - z_{4}c\cos\beta)\,\hat{\mathbf{x}} - y_{4}b\,\hat{\mathbf{y}} - z_{4}c\sin\beta\,\hat{\mathbf{z}}$ | (4*e*) | Al IV |
| $\mathbf{B}_{14}$ | = | $x_{4}\,\mathbf{a}_{1} + \left(\frac{1}{2} - y_{4}\right)\,\mathbf{a}_{2} + \left(\frac{1}{2} + z_{4}\right)\,\mathbf{a}_{3}$ | = | $\left(\frac{1}{2}c\cos\beta + x_{4}a + z_{4}c\cos\beta\right)\,\hat{\mathbf{x}} + \left(\frac{1}{2} - y_{4}\right)b\,\hat{\mathbf{y}} + \left(\frac{1}{2} + z_{4}\right)c\sin\beta\,\hat{\mathbf{z}}$ | (4*e*) | Al IV |
| $\mathbf{B}_{15}$ | = | $x_{5}\,\mathbf{a}_{1} + y_{5}\,\mathbf{a}_{2} + z_{5}\,\mathbf{a}_{3}$ | = | $(x_{5}a + z_{5}c\cos\beta)\,\hat{\mathbf{x}} + y_{5}b\,\hat{\mathbf{y}} + z_{5}c\sin\beta\,\hat{\mathbf{z}}$ | (4*e*) | Al V |
| $\mathbf{B}_{16}$ | = | $-x_{5}\,\mathbf{a}_{1} + \left(\frac{1}{2} + y_{5}\right)\,\mathbf{a}_{2} + \left(\frac{1}{2} - z_{5}\right)\,\mathbf{a}_{3}$ | = | $\left(\frac{1}{2}c\cos\beta - x_{5}a - z_{5}c\cos\beta\right)\,\hat{\mathbf{x}} + \left(\frac{1}{2} + y_{5}\right)b\,\hat{\mathbf{y}} + \left(\frac{1}{2} - z_{5}\right)c\sin\beta\,\hat{\mathbf{z}}$ | (4*e*) | Al V |
| $\mathbf{B}_{17}$ | = | $-x_{5}\,\mathbf{a}_{1} - y_{5}\,\mathbf{a}_{2} - z_{5}\,\mathbf{a}_{3}$ | = | $(-x_{5}a - z_{5}c\cos\beta)\,\hat{\mathbf{x}} - y_{5}b\,\hat{\mathbf{y}} - z_{5}c\sin\beta\,\hat{\mathbf{z}}$ | (4*e*) | Al V |
| $\mathbf{B}_{18}$ | = | $x_{5}\,\mathbf{a}_{1} + \left(\frac{1}{2} - y_{5}\right)\,\mathbf{a}_{2} + \left(\frac{1}{2} + z_{5}\right)\,\mathbf{a}_{3}$ | = | $\left(\frac{1}{2}c\cos\beta + x_{5}a + z_{5}c\cos\beta\right)\,\hat{\mathbf{x}} + \left(\frac{1}{2} - y_{5}\right)b\,\hat{\mathbf{y}} + \left(\frac{1}{2} + z_{5}\right)c\sin\beta\,\hat{\mathbf{z}}$ | (4*e*) | Al V |
| $\mathbf{B}_{19}$ | = | $x_{6}\,\mathbf{a}_{1} + y_{6}\,\mathbf{a}_{2} + z_{6}\,\mathbf{a}_{3}$ | = | $(x_{6}a + z_{6}c\cos\beta)\,\hat{\mathbf{x}} + y_{6}b\,\hat{\mathbf{y}} + z_{6}c\sin\beta\,\hat{\mathbf{z}}$ | (4*e*) | Co |
| $\mathbf{B}_{20}$ | = | $-x_{6}\,\mathbf{a}_{1} + \left(\frac{1}{2} + y_{6}\right)\,\mathbf{a}_{2} + \left(\frac{1}{2} - z_{6}\right)\,\mathbf{a}_{3}$ | = | $\left(\frac{1}{2}c\cos\beta - x_{6}a - z_{6}c\cos\beta\right)\,\hat{\mathbf{x}} + \left(\frac{1}{2} + y_{6}\right)b\,\hat{\mathbf{y}} + \left(\frac{1}{2} - z_{6}\right)c\sin\beta\,\hat{\mathbf{z}}$ | (4*e*) | Co |
| $\mathbf{B}_{21}$ | = | $-x_{6}\,\mathbf{a}_{1} - y_{6}\,\mathbf{a}_{2} - z_{6}\,\mathbf{a}_{3}$ | = | $(-x_{6}a - z_{6}c\cos\beta)\,\hat{\mathbf{x}} - y_{6}b\,\hat{\mathbf{y}} - z_{6}c\sin\beta\,\hat{\mathbf{z}}$ | (4*e*) | Co |
| $\mathbf{B}_{22}$ | = | $x_{6}\,\mathbf{a}_{1} + \left(\frac{1}{2} - y_{6}\right)\,\mathbf{a}_{2} + \left(\frac{1}{2} + z_{6}\right)\,\mathbf{a}_{3}$ | = | $\left(\frac{1}{2}c\cos\beta + x_{6}a + z_{6}c\cos\beta\right)\,\hat{\mathbf{x}} + \left(\frac{1}{2} - y_{6}\right)b\,\hat{\mathbf{y}} + \left(\frac{1}{2} + z_{6}\right)c\sin\beta\,\hat{\mathbf{z}}$ | (4*e*) | Co |

**Geometry files:**
- CIF: pp. 1547
- POSCAR: pp. 1548

# Tutton salt [$Cu(NH_4)_2(SO_4)_2{\cdot}6H_2O$, $H4_4$] Structure: AB20C2D14E2_mP78_14_a_10e_e_7e_e

http://aflow.org/prototype-encyclopedia/AB20C2D14E2_mP78_14_a_10e_e_7e_e

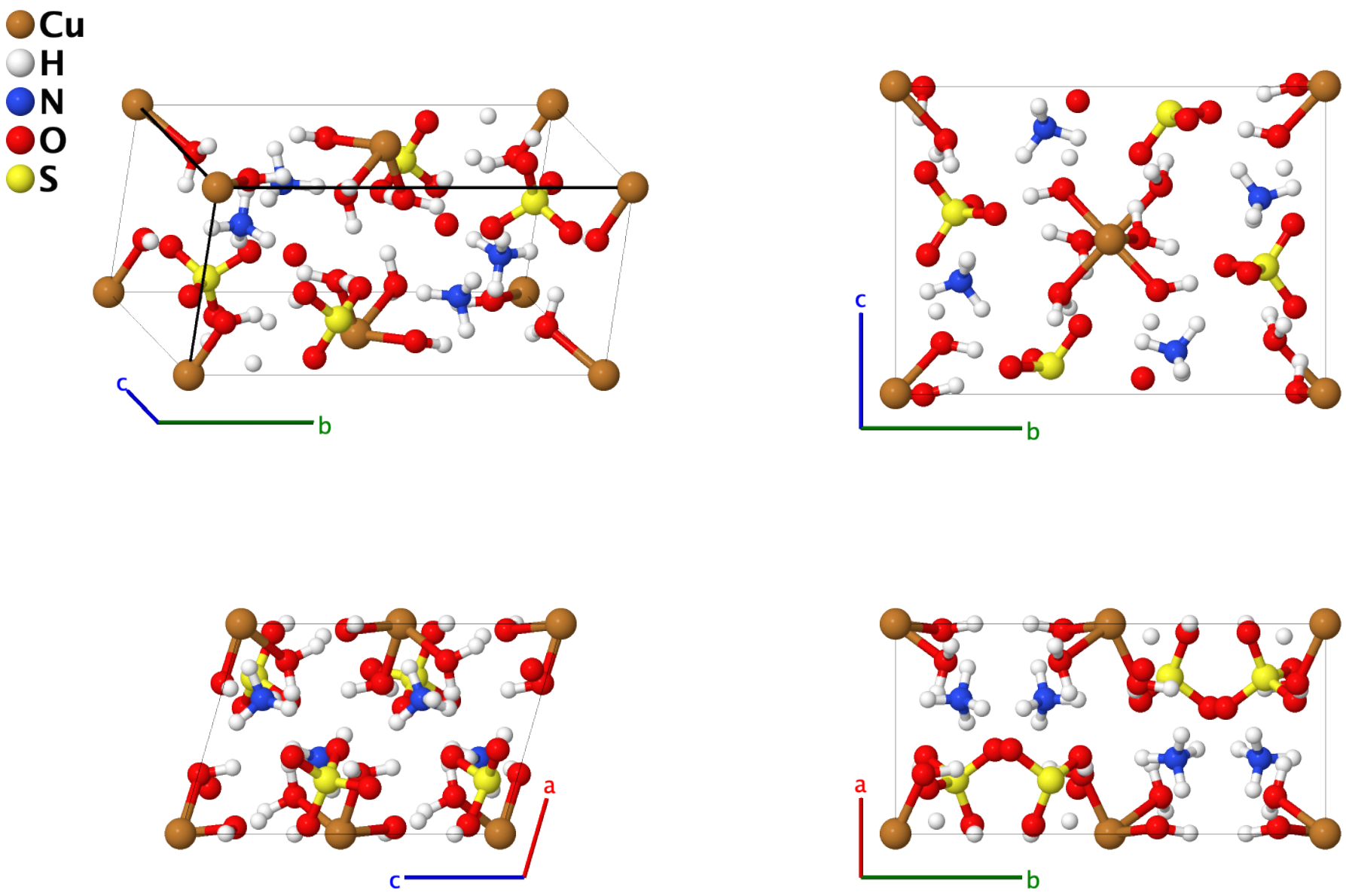

| | | |
|---|---|---|
| **Prototype** | : | $CuH_{20}N_2O_{14}S_2$ |
| **AFLOW prototype label** | : | AB20C2D14E2_mP78_14_a_10e_e_7e_e |
| ***Strukturbericht* designation** | : | $H4_4$ |
| **Pearson symbol** | : | mP78 |
| **Space group number** | : | 14 |
| **Space group symbol** | : | $P2_1/c$ |
| **AFLOW prototype command** | : | `aflow --proto=AB20C2D14E2_mP78_14_a_10e_e_7e_e --params=`$a, b/a, c/a, \beta, x_2, y_2, z_2, x_3, y_3, z_3, x_4, y_4, z_4, x_5, y_5, z_5, x_6, y_6, z_6, x_7, y_7, z_7, x_8, y_8, z_8, x_9, y_9, z_9, x_{10}, y_{10}, z_{10}, x_{11}, y_{11}, z_{11}, x_{12}, y_{12}, z_{12}, x_{13}, y_{13}, z_{13}, x_{14}, y_{14}, z_{14}, x_{15}, y_{15}, z_{15}, x_{16}, y_{16}, z_{16}, x_{17}, y_{17}, z_{17}, x_{18}, y_{18}, z_{18}, x_{19}, y_{19}, z_{19}, x_{20}, y_{20}, z_{20}$ |

### Other compounds with this structure

- $Cd(NH_4)_2(SO_4)_2{\cdot}6H_2O$, $Co(NH_4)_2(SO_4)_2{\cdot}6H_2O$, $Fe(NH_4)_2(SO_4)_2{\cdot}6H_2O$, $Mg(NH_4)_2(SO_4)_2{\cdot}6H_2O$, $Mn(NH_4)_2(SO_4)_2{\cdot}6H_2O$, $Ni(NH_4)_2(SO_4)_2{\cdot}6H_2O$, $V(NH_4)_2(SO_4)_2{\cdot}6H_2O$, $Zn(NH_4)_2(SO_4)_2{\cdot}6H_2O$, $Mg(NH_4)_2(SeO_4)_2{\cdot}6H_2O$, $Co(KSO_4)_2{\cdot}6H_2O$, $Cu(KSeO_4)_2{\cdot}6H_2O$, $Cu(KSO_4)_2{\cdot}6H_2O$, $Mg(KSO_4)_2{\cdot}6H_2O$, $Mg(TlSO_4)_2{\cdot}6H_2O$, and $Ni(KSO_4)_2{\cdot}6H_2O$

---

- Tutton Salts have the form $AB_2(SO_4)_2{\cdot}6H_2O$, were $A$ is a divalent ion and $B$ is a monovalent ion. (Hermann, 1937) formally lists $Mg(NH_4)_2(SO_4)_2{\cdot}6H_2O$ as the prototype, but notes that one may have "other monovalent metals in place of $NH_4$ and divalent [metals] in place of Mg." We choose $Cu(NH_4)_2(SO_4)_2{\cdot}6H_2O$ as the prototype because (Brown, 1969) were able to locate all of the hydrogen ions in agreement with expectation, *i.e.*, $H_2O$ molecules have properly bent H-O bonds, and the hydrogen atoms in the ammonium ion form a tetrahedron around the nitrogen atom.
- The atomic positions were originally given in the $P2_1/a$ setting of space group #14. We used FINDSYM to convert this to the standard $P2_1/c$ setting. This resulted in the $a$ and $c$ axes being swapped.

**Simple Monoclinic primitive vectors:**

$$\mathbf{a}_1 = a\,\hat{\mathbf{x}}$$
$$\mathbf{a}_2 = b\,\hat{\mathbf{y}}$$
$$\mathbf{a}_3 = c\cos\beta\,\hat{\mathbf{x}} + c\sin\beta\,\hat{\mathbf{z}}$$

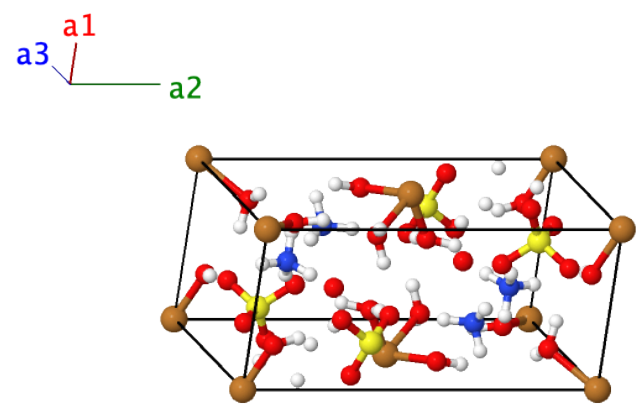

**Basis vectors:**

| | | Lattice Coordinates | | Cartesian Coordinates | Wyckoff Position | Atom Type |
|---|---|---|---|---|---|---|
| $\mathbf{B}_1$ | = | $0\,\mathbf{a}_1 + 0\,\mathbf{a}_2 + 0\,\mathbf{a}_3$ | = | $0\,\hat{\mathbf{x}} + 0\,\hat{\mathbf{y}} + 0\,\hat{\mathbf{z}}$ | (2$a$) | Cu |
| $\mathbf{B}_2$ | = | $\frac{1}{2}\,\mathbf{a}_2 + \frac{1}{2}\,\mathbf{a}_3$ | = | $\frac{1}{2}c\cos\beta\,\hat{\mathbf{x}} + \frac{1}{2}b\,\hat{\mathbf{y}} + \frac{1}{2}c\sin\beta\,\hat{\mathbf{z}}$ | (2$a$) | Cu |
| $\mathbf{B}_3$ | = | $x_2\,\mathbf{a}_1 + y_2\,\mathbf{a}_2 + z_2\,\mathbf{a}_3$ | = | $(x_2a + z_2c\cos\beta)\,\hat{\mathbf{x}} + y_2b\,\hat{\mathbf{y}} + z_2c\sin\beta\,\hat{\mathbf{z}}$ | (4$e$) | H I |
| $\mathbf{B}_4$ | = | $-x_2\,\mathbf{a}_1 + \left(\frac{1}{2} + y_2\right)\mathbf{a}_2 + \left(\frac{1}{2} - z_2\right)\mathbf{a}_3$ | = | $\left(\frac{1}{2}c\cos\beta - x_2a - z_2c\cos\beta\right)\hat{\mathbf{x}} + \left(\frac{1}{2} + y_2\right)b\,\hat{\mathbf{y}} + \left(\frac{1}{2} - z_2\right)c\sin\beta\,\hat{\mathbf{z}}$ | (4$e$) | H I |
| $\mathbf{B}_5$ | = | $-x_2\,\mathbf{a}_1 - y_2\,\mathbf{a}_2 - z_2\,\mathbf{a}_3$ | = | $(-x_2a - z_2c\cos\beta)\,\hat{\mathbf{x}} - y_2b\,\hat{\mathbf{y}} - z_2c\sin\beta\,\hat{\mathbf{z}}$ | (4$e$) | H I |
| $\mathbf{B}_6$ | = | $x_2\,\mathbf{a}_1 + \left(\frac{1}{2} - y_2\right)\mathbf{a}_2 + \left(\frac{1}{2} + z_2\right)\mathbf{a}_3$ | = | $\left(\frac{1}{2}c\cos\beta + x_2a + z_2c\cos\beta\right)\hat{\mathbf{x}} + \left(\frac{1}{2} - y_2\right)b\,\hat{\mathbf{y}} + \left(\frac{1}{2} + z_2\right)c\sin\beta\,\hat{\mathbf{z}}$ | (4$e$) | H I |
| $\mathbf{B}_7$ | = | $x_3\,\mathbf{a}_1 + y_3\,\mathbf{a}_2 + z_3\,\mathbf{a}_3$ | = | $(x_3a + z_3c\cos\beta)\,\hat{\mathbf{x}} + y_3b\,\hat{\mathbf{y}} + z_3c\sin\beta\,\hat{\mathbf{z}}$ | (4$e$) | H II |
| $\mathbf{B}_8$ | = | $-x_3\,\mathbf{a}_1 + \left(\frac{1}{2} + y_3\right)\mathbf{a}_2 + \left(\frac{1}{2} - z_3\right)\mathbf{a}_3$ | = | $\left(\frac{1}{2}c\cos\beta - x_3a - z_3c\cos\beta\right)\hat{\mathbf{x}} + \left(\frac{1}{2} + y_3\right)b\,\hat{\mathbf{y}} + \left(\frac{1}{2} - z_3\right)c\sin\beta\,\hat{\mathbf{z}}$ | (4$e$) | H II |
| $\mathbf{B}_9$ | = | $-x_3\,\mathbf{a}_1 - y_3\,\mathbf{a}_2 - z_3\,\mathbf{a}_3$ | = | $(-x_3a - z_3c\cos\beta)\,\hat{\mathbf{x}} - y_3b\,\hat{\mathbf{y}} - z_3c\sin\beta\,\hat{\mathbf{z}}$ | (4$e$) | H II |
| $\mathbf{B}_{10}$ | = | $x_3\,\mathbf{a}_1 + \left(\frac{1}{2} - y_3\right)\mathbf{a}_2 + \left(\frac{1}{2} + z_3\right)\mathbf{a}_3$ | = | $\left(\frac{1}{2}c\cos\beta + x_3a + z_3c\cos\beta\right)\hat{\mathbf{x}} + \left(\frac{1}{2} - y_3\right)b\,\hat{\mathbf{y}} + \left(\frac{1}{2} + z_3\right)c\sin\beta\,\hat{\mathbf{z}}$ | (4$e$) | H II |
| $\mathbf{B}_{11}$ | = | $x_4\,\mathbf{a}_1 + y_4\,\mathbf{a}_2 + z_4\,\mathbf{a}_3$ | = | $(x_4a + z_4c\cos\beta)\,\hat{\mathbf{x}} + y_4b\,\hat{\mathbf{y}} + z_4c\sin\beta\,\hat{\mathbf{z}}$ | (4$e$) | H III |
| $\mathbf{B}_{12}$ | = | $-x_4\,\mathbf{a}_1 + \left(\frac{1}{2} + y_4\right)\mathbf{a}_2 + \left(\frac{1}{2} - z_4\right)\mathbf{a}_3$ | = | $\left(\frac{1}{2}c\cos\beta - x_4a - z_4c\cos\beta\right)\hat{\mathbf{x}} + \left(\frac{1}{2} + y_4\right)b\,\hat{\mathbf{y}} + \left(\frac{1}{2} - z_4\right)c\sin\beta\,\hat{\mathbf{z}}$ | (4$e$) | H III |
| $\mathbf{B}_{13}$ | = | $-x_4\,\mathbf{a}_1 - y_4\,\mathbf{a}_2 - z_4\,\mathbf{a}_3$ | = | $(-x_4a - z_4c\cos\beta)\,\hat{\mathbf{x}} - y_4b\,\hat{\mathbf{y}} - z_4c\sin\beta\,\hat{\mathbf{z}}$ | (4$e$) | H III |
| $\mathbf{B}_{14}$ | = | $x_4\,\mathbf{a}_1 + \left(\frac{1}{2} - y_4\right)\mathbf{a}_2 + \left(\frac{1}{2} + z_4\right)\mathbf{a}_3$ | = | $\left(\frac{1}{2}c\cos\beta + x_4a + z_4c\cos\beta\right)\hat{\mathbf{x}} + \left(\frac{1}{2} - y_4\right)b\,\hat{\mathbf{y}} + \left(\frac{1}{2} + z_4\right)c\sin\beta\,\hat{\mathbf{z}}$ | (4$e$) | H III |
| $\mathbf{B}_{15}$ | = | $x_5\,\mathbf{a}_1 + y_5\,\mathbf{a}_2 + z_5\,\mathbf{a}_3$ | = | $(x_5a + z_5c\cos\beta)\,\hat{\mathbf{x}} + y_5b\,\hat{\mathbf{y}} + z_5c\sin\beta\,\hat{\mathbf{z}}$ | (4$e$) | H IV |
| $\mathbf{B}_{16}$ | = | $-x_5\,\mathbf{a}_1 + \left(\frac{1}{2} + y_5\right)\mathbf{a}_2 + \left(\frac{1}{2} - z_5\right)\mathbf{a}_3$ | = | $\left(\frac{1}{2}c\cos\beta - x_5a - z_5c\cos\beta\right)\hat{\mathbf{x}} + \left(\frac{1}{2} + y_5\right)b\,\hat{\mathbf{y}} + \left(\frac{1}{2} - z_5\right)c\sin\beta\,\hat{\mathbf{z}}$ | (4$e$) | H IV |
| $\mathbf{B}_{17}$ | = | $-x_5\,\mathbf{a}_1 - y_5\,\mathbf{a}_2 - z_5\,\mathbf{a}_3$ | = | $(-x_5a - z_5c\cos\beta)\,\hat{\mathbf{x}} - y_5b\,\hat{\mathbf{y}} - z_5c\sin\beta\,\hat{\mathbf{z}}$ | (4$e$) | H IV |
| $\mathbf{B}_{18}$ | = | $x_5\,\mathbf{a}_1 + \left(\frac{1}{2} - y_5\right)\mathbf{a}_2 + \left(\frac{1}{2} + z_5\right)\mathbf{a}_3$ | = | $\left(\frac{1}{2}c\cos\beta + x_5a + z_5c\cos\beta\right)\hat{\mathbf{x}} + \left(\frac{1}{2} - y_5\right)b\,\hat{\mathbf{y}} + \left(\frac{1}{2} + z_5\right)c\sin\beta\,\hat{\mathbf{z}}$ | (4$e$) | H IV |
| $\mathbf{B}_{19}$ | = | $x_6\,\mathbf{a}_1 + y_6\,\mathbf{a}_2 + z_6\,\mathbf{a}_3$ | = | $(x_6 a + z_6 c\cos\beta)\,\hat{\mathbf{x}} + y_6 b\,\hat{\mathbf{y}} + z_6 c\sin\beta\,\hat{\mathbf{z}}$ | (4*e*) | H V |
| $\mathbf{B}_{20}$ | = | $-x_6\,\mathbf{a}_1 + \left(\frac{1}{2} + y_6\right)\mathbf{a}_2 + \left(\frac{1}{2} - z_6\right)\mathbf{a}_3$ | = | $\left(\frac{1}{2}c\cos\beta - x_6 a - z_6 c\cos\beta\right)\hat{\mathbf{x}} + \left(\frac{1}{2} + y_6\right)b\,\hat{\mathbf{y}} + \left(\frac{1}{2} - z_6\right)c\sin\beta\,\hat{\mathbf{z}}$ | (4*e*) | H V |
| $\mathbf{B}_{21}$ | = | $-x_6\,\mathbf{a}_1 - y_6\,\mathbf{a}_2 - z_6\,\mathbf{a}_3$ | = | $(-x_6 a - z_6 c\cos\beta)\,\hat{\mathbf{x}} - y_6 b\,\hat{\mathbf{y}} - z_6 c\sin\beta\,\hat{\mathbf{z}}$ | (4*e*) | H V |
| $\mathbf{B}_{22}$ | = | $x_6\,\mathbf{a}_1 + \left(\frac{1}{2} - y_6\right)\mathbf{a}_2 + \left(\frac{1}{2} + z_6\right)\mathbf{a}_3$ | = | $\left(\frac{1}{2}c\cos\beta + x_6 a + z_6 c\cos\beta\right)\hat{\mathbf{x}} + \left(\frac{1}{2} - y_6\right)b\,\hat{\mathbf{y}} + \left(\frac{1}{2} + z_6\right)c\sin\beta\,\hat{\mathbf{z}}$ | (4*e*) | H V |
| $\mathbf{B}_{23}$ | = | $x_7\,\mathbf{a}_1 + y_7\,\mathbf{a}_2 + z_7\,\mathbf{a}_3$ | = | $(x_7 a + z_7 c\cos\beta)\,\hat{\mathbf{x}} + y_7 b\,\hat{\mathbf{y}} + z_7 c\sin\beta\,\hat{\mathbf{z}}$ | (4*e*) | H VI |
| $\mathbf{B}_{24}$ | = | $-x_7\,\mathbf{a}_1 + \left(\frac{1}{2} + y_7\right)\mathbf{a}_2 + \left(\frac{1}{2} - z_7\right)\mathbf{a}_3$ | = | $\left(\frac{1}{2}c\cos\beta - x_7 a - z_7 c\cos\beta\right)\hat{\mathbf{x}} + \left(\frac{1}{2} + y_7\right)b\,\hat{\mathbf{y}} + \left(\frac{1}{2} - z_7\right)c\sin\beta\,\hat{\mathbf{z}}$ | (4*e*) | H VI |
| $\mathbf{B}_{25}$ | = | $-x_7\,\mathbf{a}_1 - y_7\,\mathbf{a}_2 - z_7\,\mathbf{a}_3$ | = | $(-x_7 a - z_7 c\cos\beta)\,\hat{\mathbf{x}} - y_7 b\,\hat{\mathbf{y}} - z_7 c\sin\beta\,\hat{\mathbf{z}}$ | (4*e*) | H VI |
| $\mathbf{B}_{26}$ | = | $x_7\,\mathbf{a}_1 + \left(\frac{1}{2} - y_7\right)\mathbf{a}_2 + \left(\frac{1}{2} + z_7\right)\mathbf{a}_3$ | = | $\left(\frac{1}{2}c\cos\beta + x_7 a + z_7 c\cos\beta\right)\hat{\mathbf{x}} + \left(\frac{1}{2} - y_7\right)b\,\hat{\mathbf{y}} + \left(\frac{1}{2} + z_7\right)c\sin\beta\,\hat{\mathbf{z}}$ | (4*e*) | H VI |
| $\mathbf{B}_{27}$ | = | $x_8\,\mathbf{a}_1 + y_8\,\mathbf{a}_2 + z_8\,\mathbf{a}_3$ | = | $(x_8 a + z_8 c\cos\beta)\,\hat{\mathbf{x}} + y_8 b\,\hat{\mathbf{y}} + z_8 c\sin\beta\,\hat{\mathbf{z}}$ | (4*e*) | H VII |
| $\mathbf{B}_{28}$ | = | $-x_8\,\mathbf{a}_1 + \left(\frac{1}{2} + y_8\right)\mathbf{a}_2 + \left(\frac{1}{2} - z_8\right)\mathbf{a}_3$ | = | $\left(\frac{1}{2}c\cos\beta - x_8 a - z_8 c\cos\beta\right)\hat{\mathbf{x}} + \left(\frac{1}{2} + y_8\right)b\,\hat{\mathbf{y}} + \left(\frac{1}{2} - z_8\right)c\sin\beta\,\hat{\mathbf{z}}$ | (4*e*) | H VII |
| $\mathbf{B}_{29}$ | = | $-x_8\,\mathbf{a}_1 - y_8\,\mathbf{a}_2 - z_8\,\mathbf{a}_3$ | = | $(-x_8 a - z_8 c\cos\beta)\,\hat{\mathbf{x}} - y_8 b\,\hat{\mathbf{y}} - z_8 c\sin\beta\,\hat{\mathbf{z}}$ | (4*e*) | H VII |
| $\mathbf{B}_{30}$ | = | $x_8\,\mathbf{a}_1 + \left(\frac{1}{2} - y_8\right)\mathbf{a}_2 + \left(\frac{1}{2} + z_8\right)\mathbf{a}_3$ | = | $\left(\frac{1}{2}c\cos\beta + x_8 a + z_8 c\cos\beta\right)\hat{\mathbf{x}} + \left(\frac{1}{2} - y_8\right)b\,\hat{\mathbf{y}} + \left(\frac{1}{2} + z_8\right)c\sin\beta\,\hat{\mathbf{z}}$ | (4*e*) | H VII |
| $\mathbf{B}_{31}$ | = | $x_9\,\mathbf{a}_1 + y_9\,\mathbf{a}_2 + z_9\,\mathbf{a}_3$ | = | $(x_9 a + z_9 c\cos\beta)\,\hat{\mathbf{x}} + y_9 b\,\hat{\mathbf{y}} + z_9 c\sin\beta\,\hat{\mathbf{z}}$ | (4*e*) | H VIII |
| $\mathbf{B}_{32}$ | = | $-x_9\,\mathbf{a}_1 + \left(\frac{1}{2} + y_9\right)\mathbf{a}_2 + \left(\frac{1}{2} - z_9\right)\mathbf{a}_3$ | = | $\left(\frac{1}{2}c\cos\beta - x_9 a - z_9 c\cos\beta\right)\hat{\mathbf{x}} + \left(\frac{1}{2} + y_9\right)b\,\hat{\mathbf{y}} + \left(\frac{1}{2} - z_9\right)c\sin\beta\,\hat{\mathbf{z}}$ | (4*e*) | H VIII |
| $\mathbf{B}_{33}$ | = | $-x_9\,\mathbf{a}_1 - y_9\,\mathbf{a}_2 - z_9\,\mathbf{a}_3$ | = | $(-x_9 a - z_9 c\cos\beta)\,\hat{\mathbf{x}} - y_9 b\,\hat{\mathbf{y}} - z_9 c\sin\beta\,\hat{\mathbf{z}}$ | (4*e*) | H VIII |
| $\mathbf{B}_{34}$ | = | $x_9\,\mathbf{a}_1 + \left(\frac{1}{2} - y_9\right)\mathbf{a}_2 + \left(\frac{1}{2} + z_9\right)\mathbf{a}_3$ | = | $\left(\frac{1}{2}c\cos\beta + x_9 a + z_9 c\cos\beta\right)\hat{\mathbf{x}} + \left(\frac{1}{2} - y_9\right)b\,\hat{\mathbf{y}} + \left(\frac{1}{2} + z_9\right)c\sin\beta\,\hat{\mathbf{z}}$ | (4*e*) | H VIII |
| $\mathbf{B}_{35}$ | = | $x_{10}\,\mathbf{a}_1 + y_{10}\,\mathbf{a}_2 + z_{10}\,\mathbf{a}_3$ | = | $(x_{10} a + z_{10} c\cos\beta)\,\hat{\mathbf{x}} + y_{10} b\,\hat{\mathbf{y}} + z_{10} c\sin\beta\,\hat{\mathbf{z}}$ | (4*e*) | H IX |
| $\mathbf{B}_{36}$ | = | $-x_{10}\,\mathbf{a}_1 + \left(\frac{1}{2} + y_{10}\right)\mathbf{a}_2 + \left(\frac{1}{2} - z_{10}\right)\mathbf{a}_3$ | = | $\left(\frac{1}{2}c\cos\beta - x_{10} a - z_{10} c\cos\beta\right)\hat{\mathbf{x}} + \left(\frac{1}{2} + y_{10}\right)b\,\hat{\mathbf{y}} + \left(\frac{1}{2} - z_{10}\right)c\sin\beta\,\hat{\mathbf{z}}$ | (4*e*) | H IX |
| $\mathbf{B}_{37}$ | = | $-x_{10}\,\mathbf{a}_1 - y_{10}\,\mathbf{a}_2 - z_{10}\,\mathbf{a}_3$ | = | $(-x_{10} a - z_{10} c\cos\beta)\,\hat{\mathbf{x}} - y_{10} b\,\hat{\mathbf{y}} - z_{10} c\sin\beta\,\hat{\mathbf{z}}$ | (4*e*) | H IX |
| $\mathbf{B}_{38}$ | = | $x_{10}\,\mathbf{a}_1 + \left(\frac{1}{2} - y_{10}\right)\mathbf{a}_2 + \left(\frac{1}{2} + z_{10}\right)\mathbf{a}_3$ | = | $\left(\frac{1}{2}c\cos\beta + x_{10} a + z_{10} c\cos\beta\right)\hat{\mathbf{x}} + \left(\frac{1}{2} - y_{10}\right)b\,\hat{\mathbf{y}} + \left(\frac{1}{2} + z_{10}\right)c\sin\beta\,\hat{\mathbf{z}}$ | (4*e*) | H IX |
| $\mathbf{B}_{39}$ | = | $x_{11}\,\mathbf{a}_1 + y_{11}\,\mathbf{a}_2 + z_{11}\,\mathbf{a}_3$ | = | $(x_{11} a + z_{11} c\cos\beta)\,\hat{\mathbf{x}} + y_{11} b\,\hat{\mathbf{y}} + z_{11} c\sin\beta\,\hat{\mathbf{z}}$ | (4*e*) | H X |
| $\mathbf{B}_{40}$ | = | $-x_{11}\,\mathbf{a}_1 + \left(\frac{1}{2} + y_{11}\right)\mathbf{a}_2 + \left(\frac{1}{2} - z_{11}\right)\mathbf{a}_3$ | = | $\left(\frac{1}{2}c\cos\beta - x_{11} a - z_{11} c\cos\beta\right)\hat{\mathbf{x}} + \left(\frac{1}{2} + y_{11}\right)b\,\hat{\mathbf{y}} + \left(\frac{1}{2} - z_{11}\right)c\sin\beta\,\hat{\mathbf{z}}$ | (4*e*) | H X |
| $\mathbf{B}_{41}$ | = | $-x_{11}\,\mathbf{a}_1 - y_{11}\,\mathbf{a}_2 - z_{11}\,\mathbf{a}_3$ | = | $(-x_{11}a - z_{11}c\cos\beta)\,\hat{\mathbf{x}} - y_{11}b\,\hat{\mathbf{y}} - z_{11}c\sin\beta\,\hat{\mathbf{z}}$ | (4$e$) | H X |
| $\mathbf{B}_{42}$ | = | $x_{11}\,\mathbf{a}_1 + \left(\frac{1}{2} - y_{11}\right)\mathbf{a}_2 + \left(\frac{1}{2} + z_{11}\right)\mathbf{a}_3$ | = | $\left(\frac{1}{2}c\cos\beta + x_{11}a + z_{11}c\cos\beta\right)\hat{\mathbf{x}} + \left(\frac{1}{2} - y_{11}\right)b\,\hat{\mathbf{y}} + \left(\frac{1}{2} + z_{11}\right)c\sin\beta\,\hat{\mathbf{z}}$ | (4$e$) | H X |
| $\mathbf{B}_{43}$ | = | $x_{12}\,\mathbf{a}_1 + y_{12}\,\mathbf{a}_2 + z_{12}\,\mathbf{a}_3$ | = | $(x_{12}a + z_{12}c\cos\beta)\,\hat{\mathbf{x}} + y_{12}b\,\hat{\mathbf{y}} + z_{12}c\sin\beta\,\hat{\mathbf{z}}$ | (4$e$) | N |
| $\mathbf{B}_{44}$ | = | $-x_{12}\,\mathbf{a}_1 + \left(\frac{1}{2} + y_{12}\right)\mathbf{a}_2 + \left(\frac{1}{2} - z_{12}\right)\mathbf{a}_3$ | = | $\left(\frac{1}{2}c\cos\beta - x_{12}a - z_{12}c\cos\beta\right)\hat{\mathbf{x}} + \left(\frac{1}{2} + y_{12}\right)b\,\hat{\mathbf{y}} + \left(\frac{1}{2} - z_{12}\right)c\sin\beta\,\hat{\mathbf{z}}$ | (4$e$) | N |
| $\mathbf{B}_{45}$ | = | $-x_{12}\,\mathbf{a}_1 - y_{12}\,\mathbf{a}_2 - z_{12}\,\mathbf{a}_3$ | = | $(-x_{12}a - z_{12}c\cos\beta)\,\hat{\mathbf{x}} - y_{12}b\,\hat{\mathbf{y}} - z_{12}c\sin\beta\,\hat{\mathbf{z}}$ | (4$e$) | N |
| $\mathbf{B}_{46}$ | = | $x_{12}\,\mathbf{a}_1 + \left(\frac{1}{2} - y_{12}\right)\mathbf{a}_2 + \left(\frac{1}{2} + z_{12}\right)\mathbf{a}_3$ | = | $\left(\frac{1}{2}c\cos\beta + x_{12}a + z_{12}c\cos\beta\right)\hat{\mathbf{x}} + \left(\frac{1}{2} - y_{12}\right)b\,\hat{\mathbf{y}} + \left(\frac{1}{2} + z_{12}\right)c\sin\beta\,\hat{\mathbf{z}}$ | (4$e$) | N |
| $\mathbf{B}_{47}$ | = | $x_{13}\,\mathbf{a}_1 + y_{13}\,\mathbf{a}_2 + z_{13}\,\mathbf{a}_3$ | = | $(x_{13}a + z_{13}c\cos\beta)\,\hat{\mathbf{x}} + y_{13}b\,\hat{\mathbf{y}} + z_{13}c\sin\beta\,\hat{\mathbf{z}}$ | (4$e$) | O I |
| $\mathbf{B}_{48}$ | = | $-x_{13}\,\mathbf{a}_1 + \left(\frac{1}{2} + y_{13}\right)\mathbf{a}_2 + \left(\frac{1}{2} - z_{13}\right)\mathbf{a}_3$ | = | $\left(\frac{1}{2}c\cos\beta - x_{13}a - z_{13}c\cos\beta\right)\hat{\mathbf{x}} + \left(\frac{1}{2} + y_{13}\right)b\,\hat{\mathbf{y}} + \left(\frac{1}{2} - z_{13}\right)c\sin\beta\,\hat{\mathbf{z}}$ | (4$e$) | O I |
| $\mathbf{B}_{49}$ | = | $-x_{13}\,\mathbf{a}_1 - y_{13}\,\mathbf{a}_2 - z_{13}\,\mathbf{a}_3$ | = | $(-x_{13}a - z_{13}c\cos\beta)\,\hat{\mathbf{x}} - y_{13}b\,\hat{\mathbf{y}} - z_{13}c\sin\beta\,\hat{\mathbf{z}}$ | (4$e$) | O I |
| $\mathbf{B}_{50}$ | = | $x_{13}\,\mathbf{a}_1 + \left(\frac{1}{2} - y_{13}\right)\mathbf{a}_2 + \left(\frac{1}{2} + z_{13}\right)\mathbf{a}_3$ | = | $\left(\frac{1}{2}c\cos\beta + x_{13}a + z_{13}c\cos\beta\right)\hat{\mathbf{x}} + \left(\frac{1}{2} - y_{13}\right)b\,\hat{\mathbf{y}} + \left(\frac{1}{2} + z_{13}\right)c\sin\beta\,\hat{\mathbf{z}}$ | (4$e$) | O I |
| $\mathbf{B}_{51}$ | = | $x_{14}\,\mathbf{a}_1 + y_{14}\,\mathbf{a}_2 + z_{14}\,\mathbf{a}_3$ | = | $(x_{14}a + z_{14}c\cos\beta)\,\hat{\mathbf{x}} + y_{14}b\,\hat{\mathbf{y}} + z_{14}c\sin\beta\,\hat{\mathbf{z}}$ | (4$e$) | O II |
| $\mathbf{B}_{52}$ | = | $-x_{14}\,\mathbf{a}_1 + \left(\frac{1}{2} + y_{14}\right)\mathbf{a}_2 + \left(\frac{1}{2} - z_{14}\right)\mathbf{a}_3$ | = | $\left(\frac{1}{2}c\cos\beta - x_{14}a - z_{14}c\cos\beta\right)\hat{\mathbf{x}} + \left(\frac{1}{2} + y_{14}\right)b\,\hat{\mathbf{y}} + \left(\frac{1}{2} - z_{14}\right)c\sin\beta\,\hat{\mathbf{z}}$ | (4$e$) | O II |
| $\mathbf{B}_{53}$ | = | $-x_{14}\,\mathbf{a}_1 - y_{14}\,\mathbf{a}_2 - z_{14}\,\mathbf{a}_3$ | = | $(-x_{14}a - z_{14}c\cos\beta)\,\hat{\mathbf{x}} - y_{14}b\,\hat{\mathbf{y}} - z_{14}c\sin\beta\,\hat{\mathbf{z}}$ | (4$e$) | O II |
| $\mathbf{B}_{54}$ | = | $x_{14}\,\mathbf{a}_1 + \left(\frac{1}{2} - y_{14}\right)\mathbf{a}_2 + \left(\frac{1}{2} + z_{14}\right)\mathbf{a}_3$ | = | $\left(\frac{1}{2}c\cos\beta + x_{14}a + z_{14}c\cos\beta\right)\hat{\mathbf{x}} + \left(\frac{1}{2} - y_{14}\right)b\,\hat{\mathbf{y}} + \left(\frac{1}{2} + z_{14}\right)c\sin\beta\,\hat{\mathbf{z}}$ | (4$e$) | O II |
| $\mathbf{B}_{55}$ | = | $x_{15}\,\mathbf{a}_1 + y_{15}\,\mathbf{a}_2 + z_{15}\,\mathbf{a}_3$ | = | $(x_{15}a + z_{15}c\cos\beta)\,\hat{\mathbf{x}} + y_{15}b\,\hat{\mathbf{y}} + z_{15}c\sin\beta\,\hat{\mathbf{z}}$ | (4$e$) | O III |
| $\mathbf{B}_{56}$ | = | $-x_{15}\,\mathbf{a}_1 + \left(\frac{1}{2} + y_{15}\right)\mathbf{a}_2 + \left(\frac{1}{2} - z_{15}\right)\mathbf{a}_3$ | = | $\left(\frac{1}{2}c\cos\beta - x_{15}a - z_{15}c\cos\beta\right)\hat{\mathbf{x}} + \left(\frac{1}{2} + y_{15}\right)b\,\hat{\mathbf{y}} + \left(\frac{1}{2} - z_{15}\right)c\sin\beta\,\hat{\mathbf{z}}$ | (4$e$) | O III |
| $\mathbf{B}_{57}$ | = | $-x_{15}\,\mathbf{a}_1 - y_{15}\,\mathbf{a}_2 - z_{15}\,\mathbf{a}_3$ | = | $(-x_{15}a - z_{15}c\cos\beta)\,\hat{\mathbf{x}} - y_{15}b\,\hat{\mathbf{y}} - z_{15}c\sin\beta\,\hat{\mathbf{z}}$ | (4$e$) | O III |
| $\mathbf{B}_{58}$ | = | $x_{15}\,\mathbf{a}_1 + \left(\frac{1}{2} - y_{15}\right)\mathbf{a}_2 + \left(\frac{1}{2} + z_{15}\right)\mathbf{a}_3$ | = | $\left(\frac{1}{2}c\cos\beta + x_{15}a + z_{15}c\cos\beta\right)\hat{\mathbf{x}} + \left(\frac{1}{2} - y_{15}\right)b\,\hat{\mathbf{y}} + \left(\frac{1}{2} + z_{15}\right)c\sin\beta\,\hat{\mathbf{z}}$ | (4$e$) | O III |
| $\mathbf{B}_{59}$ | = | $x_{16}\,\mathbf{a}_1 + y_{16}\,\mathbf{a}_2 + z_{16}\,\mathbf{a}_3$ | = | $(x_{16}a + z_{16}c\cos\beta)\,\hat{\mathbf{x}} + y_{16}b\,\hat{\mathbf{y}} + z_{16}c\sin\beta\,\hat{\mathbf{z}}$ | (4$e$) | O IV |
| $\mathbf{B}_{60}$ | = | $-x_{16}\,\mathbf{a}_1 + \left(\frac{1}{2} + y_{16}\right)\mathbf{a}_2 + \left(\frac{1}{2} - z_{16}\right)\mathbf{a}_3$ | = | $\left(\frac{1}{2}c\cos\beta - x_{16}a - z_{16}c\cos\beta\right)\hat{\mathbf{x}} + \left(\frac{1}{2} + y_{16}\right)b\,\hat{\mathbf{y}} + \left(\frac{1}{2} - z_{16}\right)c\sin\beta\,\hat{\mathbf{z}}$ | (4$e$) | O IV |
| $\mathbf{B}_{61}$ | = | $-x_{16}\,\mathbf{a}_1 - y_{16}\,\mathbf{a}_2 - z_{16}\,\mathbf{a}_3$ | = | $(-x_{16}a - z_{16}c\cos\beta)\,\hat{\mathbf{x}} - y_{16}b\,\hat{\mathbf{y}} - z_{16}c\sin\beta\,\hat{\mathbf{z}}$ | (4$e$) | O IV |
| $\mathbf{B}_{62}$ | = | $x_{16}\,\mathbf{a}_1 + \left(\frac{1}{2} - y_{16}\right)\mathbf{a}_2 + \left(\frac{1}{2} + z_{16}\right)\mathbf{a}_3$ | = | $\left(\frac{1}{2}c\cos\beta + x_{16}a + z_{16}c\cos\beta\right)\hat{\mathbf{x}} + \left(\frac{1}{2} - y_{16}\right)b\,\hat{\mathbf{y}} + \left(\frac{1}{2} + z_{16}\right)c\sin\beta\,\hat{\mathbf{z}}$ | (4$e$) | O IV |
| $\mathbf{B}_{63}$ | = | $x_{17}\,\mathbf{a}_{1} + y_{17}\,\mathbf{a}_{2} + z_{17}\,\mathbf{a}_{3}$ | = | $\left(x_{17}a + z_{17}c\cos\beta\right)\,\mathbf{\hat{x}} + y_{17}b\,\mathbf{\hat{y}} + z_{17}c\sin\beta\,\mathbf{\hat{z}}$ | (4$e$) | O V |
| $\mathbf{B}_{64}$ | = | $-x_{17}\,\mathbf{a}_{1} + \left(\frac{1}{2} + y_{17}\right)\,\mathbf{a}_{2} + \left(\frac{1}{2} - z_{17}\right)\,\mathbf{a}_{3}$ | = | $\left(\frac{1}{2}c\cos\beta - x_{17}a - z_{17}c\cos\beta\right)\,\mathbf{\hat{x}} + \left(\frac{1}{2} + y_{17}\right)b\,\mathbf{\hat{y}} + \left(\frac{1}{2} - z_{17}\right)c\sin\beta\,\mathbf{\hat{z}}$ | (4$e$) | O V |
| $\mathbf{B}_{65}$ | = | $-x_{17}\,\mathbf{a}_{1} - y_{17}\,\mathbf{a}_{2} - z_{17}\,\mathbf{a}_{3}$ | = | $\left(-x_{17}a - z_{17}c\cos\beta\right)\,\mathbf{\hat{x}} - y_{17}b\,\mathbf{\hat{y}} - z_{17}c\sin\beta\,\mathbf{\hat{z}}$ | (4$e$) | O V |
| $\mathbf{B}_{66}$ | = | $x_{17}\,\mathbf{a}_{1} + \left(\frac{1}{2} - y_{17}\right)\,\mathbf{a}_{2} + \left(\frac{1}{2} + z_{17}\right)\,\mathbf{a}_{3}$ | = | $\left(\frac{1}{2}c\cos\beta + x_{17}a + z_{17}c\cos\beta\right)\,\mathbf{\hat{x}} + \left(\frac{1}{2} - y_{17}\right)b\,\mathbf{\hat{y}} + \left(\frac{1}{2} + z_{17}\right)c\sin\beta\,\mathbf{\hat{z}}$ | (4$e$) | O V |
| $\mathbf{B}_{67}$ | = | $x_{18}\,\mathbf{a}_{1} + y_{18}\,\mathbf{a}_{2} + z_{18}\,\mathbf{a}_{3}$ | = | $\left(x_{18}a + z_{18}c\cos\beta\right)\,\mathbf{\hat{x}} + y_{18}b\,\mathbf{\hat{y}} + z_{18}c\sin\beta\,\mathbf{\hat{z}}$ | (4$e$) | O VI |
| $\mathbf{B}_{68}$ | = | $-x_{18}\,\mathbf{a}_{1} + \left(\frac{1}{2} + y_{18}\right)\,\mathbf{a}_{2} + \left(\frac{1}{2} - z_{18}\right)\,\mathbf{a}_{3}$ | = | $\left(\frac{1}{2}c\cos\beta - x_{18}a - z_{18}c\cos\beta\right)\,\mathbf{\hat{x}} + \left(\frac{1}{2} + y_{18}\right)b\,\mathbf{\hat{y}} + \left(\frac{1}{2} - z_{18}\right)c\sin\beta\,\mathbf{\hat{z}}$ | (4$e$) | O VI |
| $\mathbf{B}_{69}$ | = | $-x_{18}\,\mathbf{a}_{1} - y_{18}\,\mathbf{a}_{2} - z_{18}\,\mathbf{a}_{3}$ | = | $\left(-x_{18}a - z_{18}c\cos\beta\right)\,\mathbf{\hat{x}} - y_{18}b\,\mathbf{\hat{y}} - z_{18}c\sin\beta\,\mathbf{\hat{z}}$ | (4$e$) | O VI |
| $\mathbf{B}_{70}$ | = | $x_{18}\,\mathbf{a}_{1} + \left(\frac{1}{2} - y_{18}\right)\,\mathbf{a}_{2} + \left(\frac{1}{2} + z_{18}\right)\,\mathbf{a}_{3}$ | = | $\left(\frac{1}{2}c\cos\beta + x_{18}a + z_{18}c\cos\beta\right)\,\mathbf{\hat{x}} + \left(\frac{1}{2} - y_{18}\right)b\,\mathbf{\hat{y}} + \left(\frac{1}{2} + z_{18}\right)c\sin\beta\,\mathbf{\hat{z}}$ | (4$e$) | O VI |
| $\mathbf{B}_{71}$ | = | $x_{19}\,\mathbf{a}_{1} + y_{19}\,\mathbf{a}_{2} + z_{19}\,\mathbf{a}_{3}$ | = | $\left(x_{19}a + z_{19}c\cos\beta\right)\,\mathbf{\hat{x}} + y_{19}b\,\mathbf{\hat{y}} + z_{19}c\sin\beta\,\mathbf{\hat{z}}$ | (4$e$) | O VII |
| $\mathbf{B}_{72}$ | = | $-x_{19}\,\mathbf{a}_{1} + \left(\frac{1}{2} + y_{19}\right)\,\mathbf{a}_{2} + \left(\frac{1}{2} - z_{19}\right)\,\mathbf{a}_{3}$ | = | $\left(\frac{1}{2}c\cos\beta - x_{19}a - z_{19}c\cos\beta\right)\,\mathbf{\hat{x}} + \left(\frac{1}{2} + y_{19}\right)b\,\mathbf{\hat{y}} + \left(\frac{1}{2} - z_{19}\right)c\sin\beta\,\mathbf{\hat{z}}$ | (4$e$) | O VII |
| $\mathbf{B}_{73}$ | = | $-x_{19}\,\mathbf{a}_{1} - y_{19}\,\mathbf{a}_{2} - z_{19}\,\mathbf{a}_{3}$ | = | $\left(-x_{19}a - z_{19}c\cos\beta\right)\,\mathbf{\hat{x}} - y_{19}b\,\mathbf{\hat{y}} - z_{19}c\sin\beta\,\mathbf{\hat{z}}$ | (4$e$) | O VII |
| $\mathbf{B}_{74}$ | = | $x_{19}\,\mathbf{a}_{1} + \left(\frac{1}{2} - y_{19}\right)\,\mathbf{a}_{2} + \left(\frac{1}{2} + z_{19}\right)\,\mathbf{a}_{3}$ | = | $\left(\frac{1}{2}c\cos\beta + x_{19}a + z_{19}c\cos\beta\right)\,\mathbf{\hat{x}} + \left(\frac{1}{2} - y_{19}\right)b\,\mathbf{\hat{y}} + \left(\frac{1}{2} + z_{19}\right)c\sin\beta\,\mathbf{\hat{z}}$ | (4$e$) | O VII |
| $\mathbf{B}_{75}$ | = | $x_{20}\,\mathbf{a}_{1} + y_{20}\,\mathbf{a}_{2} + z_{20}\,\mathbf{a}_{3}$ | = | $\left(x_{20}a + z_{20}c\cos\beta\right)\,\mathbf{\hat{x}} + y_{20}b\,\mathbf{\hat{y}} + z_{20}c\sin\beta\,\mathbf{\hat{z}}$ | (4$e$) | S |
| $\mathbf{B}_{76}$ | = | $-x_{20}\,\mathbf{a}_{1} + \left(\frac{1}{2} + y_{20}\right)\,\mathbf{a}_{2} + \left(\frac{1}{2} - z_{20}\right)\,\mathbf{a}_{3}$ | = | $\left(\frac{1}{2}c\cos\beta - x_{20}a - z_{20}c\cos\beta\right)\,\mathbf{\hat{x}} + \left(\frac{1}{2} + y_{20}\right)b\,\mathbf{\hat{y}} + \left(\frac{1}{2} - z_{20}\right)c\sin\beta\,\mathbf{\hat{z}}$ | (4$e$) | S |
| $\mathbf{B}_{77}$ | = | $-x_{20}\,\mathbf{a}_{1} - y_{20}\,\mathbf{a}_{2} - z_{20}\,\mathbf{a}_{3}$ | = | $\left(-x_{20}a - z_{20}c\cos\beta\right)\,\mathbf{\hat{x}} - y_{20}b\,\mathbf{\hat{y}} - z_{20}c\sin\beta\,\mathbf{\hat{z}}$ | (4$e$) | S |
| $\mathbf{B}_{78}$ | = | $x_{20}\,\mathbf{a}_{1} + \left(\frac{1}{2} - y_{20}\right)\,\mathbf{a}_{2} + \left(\frac{1}{2} + z_{20}\right)\,\mathbf{a}_{3}$ | = | $\left(\frac{1}{2}c\cos\beta + x_{20}a + z_{20}c\cos\beta\right)\,\mathbf{\hat{x}} + \left(\frac{1}{2} - y_{20}\right)b\,\mathbf{\hat{y}} + \left(\frac{1}{2} + z_{20}\right)c\sin\beta\,\mathbf{\hat{z}}$ | (4$e$) | S |

**Geometry files:**

- CIF: pp. 1548
- POSCAR: pp. 1548

# Parawollastonite ($CaSiO_3$, $S3_3$ (II)) Structure: AB3C_mP60_14_3e_9e_3e

http://aflow.org/prototype-encyclopedia/AB3C_mP60_14_3e_9e_3e

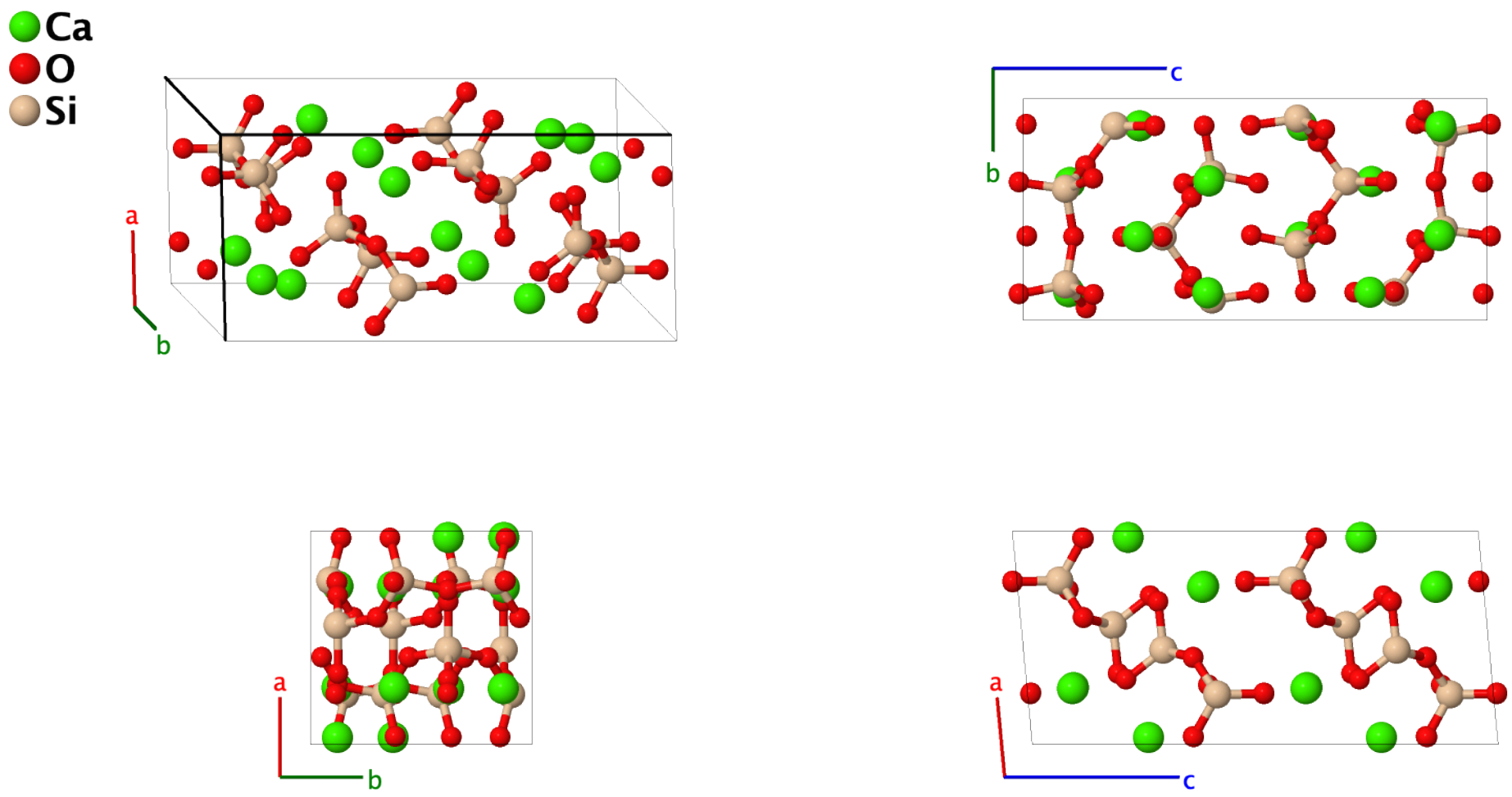

| | | |
|---|---|---|
| **Prototype** | : | $CaO_3Si$ |
| **AFLOW prototype label** | : | AB3C_mP60_14_3e_9e_3e |
| ***Strukturbericht* designation** | : | $S3_3$ (II) |
| **Pearson symbol** | : | mP60 |
| **Space group number** | : | 14 |
| **Space group symbol** | : | $P2_1/c$ |
| **AFLOW prototype command** | : | `aflow --proto=AB3C_mP60_14_3e_9e_3e`<br>`--params=`$a, b/a, c/a, \beta, x_1, y_1, z_1, x_2, y_2, z_2, x_3, y_3, z_3, x_4, y_4, z_4, x_5, y_5, z_5, x_6, y_6, z_6, x_7, y_7, z_7, x_8, y_8, z_8, x_9, y_9, z_9, x_{10}, y_{10}, z_{10}, x_{11}, y_{11}, z_{11}, x_{12}, y_{12}, z_{12}, x_{13}, y_{13}, z_{13}, x_{14}, y_{14}, z_{14}, x_{15}, y_{15}, z_{15}$ |

- (Trojer, 1968) refined the original structure of (Barnick, 1936). While Barnick referred to this structure as "wollastonite," in modern terminology it is called "parawollastonite," with the original name used for the triclinic $CaSiO_3$ form.
- (Gottfried, 1938) gave this structure the $S3_3$ designation, but (Gottfried, 1937) had already used this label for crancrinite, $Na_6Ca_2Al_6Si_6O_{24}(CO_3)_2$. We will refer to parawollastonite as $S3_3(II)$, and crancrinite as $S3_3(I)$.

**Simple Monoclinic primitive vectors:**

$$\mathbf{a}_1 = a\,\hat{\mathbf{x}}$$
$$\mathbf{a}_2 = b\,\hat{\mathbf{y}}$$
$$\mathbf{a}_3 = c\cos\beta\,\hat{\mathbf{x}} + c\sin\beta\,\hat{\mathbf{z}}$$

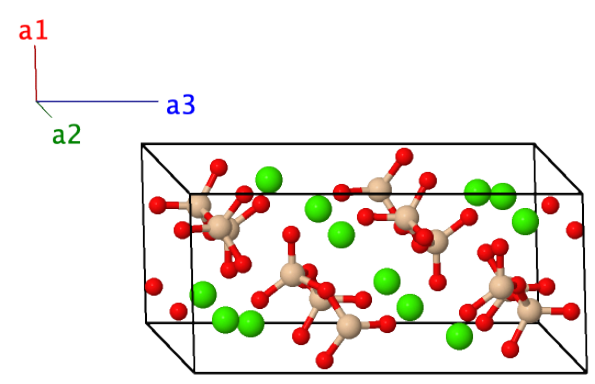

**Basis vectors:**

| | | Lattice Coordinates | | Cartesian Coordinates | Wyckoff Position | Atom Type |
|---|---|---|---|---|---|---|
| $\mathbf{B}_1$ | = | $x_1\,\mathbf{a}_1 + y_1\,\mathbf{a}_2 + z_1\,\mathbf{a}_3$ | = | $(x_1 a + z_1 c\cos\beta)\,\hat{\mathbf{x}} + y_1 b\,\hat{\mathbf{y}} + z_1 c\sin\beta\,\hat{\mathbf{z}}$ | (4e) | Ca I |
| $\mathbf{B}_2$ | = | $-x_1\,\mathbf{a}_1 + \left(\frac{1}{2} + y_1\right)\mathbf{a}_2 + \left(\frac{1}{2} - z_1\right)\mathbf{a}_3$ | = | $\left(\frac{1}{2}c\cos\beta - x_1 a - z_1 c\cos\beta\right)\hat{\mathbf{x}} + \left(\frac{1}{2} + y_1\right)b\,\hat{\mathbf{y}} + \left(\frac{1}{2} - z_1\right)c\sin\beta\,\hat{\mathbf{z}}$ | (4$e$) | Ca I |
| $\mathbf{B}_3$ | = | $-x_1\,\mathbf{a}_1 - y_1\,\mathbf{a}_2 - z_1\,\mathbf{a}_3$ | = | $(-x_1 a - z_1 c\cos\beta)\,\hat{\mathbf{x}} - y_1 b\,\hat{\mathbf{y}} - z_1 c\sin\beta\,\hat{\mathbf{z}}$ | (4$e$) | Ca I |
| $\mathbf{B}_4$ | = | $x_1\,\mathbf{a}_1 + \left(\frac{1}{2} - y_1\right)\mathbf{a}_2 + \left(\frac{1}{2} + z_1\right)\mathbf{a}_3$ | = | $\left(\frac{1}{2}c\cos\beta + x_1 a + z_1 c\cos\beta\right)\hat{\mathbf{x}} + \left(\frac{1}{2} - y_1\right)b\,\hat{\mathbf{y}} + \left(\frac{1}{2} + z_1\right)c\sin\beta\,\hat{\mathbf{z}}$ | (4$e$) | Ca I |
| $\mathbf{B}_5$ | = | $x_2\,\mathbf{a}_1 + y_2\,\mathbf{a}_2 + z_2\,\mathbf{a}_3$ | = | $(x_2 a + z_2 c\cos\beta)\,\hat{\mathbf{x}} + y_2 b\,\hat{\mathbf{y}} + z_2 c\sin\beta\,\hat{\mathbf{z}}$ | (4$e$) | Ca II |
| $\mathbf{B}_6$ | = | $-x_2\,\mathbf{a}_1 + \left(\frac{1}{2} + y_2\right)\mathbf{a}_2 + \left(\frac{1}{2} - z_2\right)\mathbf{a}_3$ | = | $\left(\frac{1}{2}c\cos\beta - x_2 a - z_2 c\cos\beta\right)\hat{\mathbf{x}} + \left(\frac{1}{2} + y_2\right)b\,\hat{\mathbf{y}} + \left(\frac{1}{2} - z_2\right)c\sin\beta\,\hat{\mathbf{z}}$ | (4$e$) | Ca II |
| $\mathbf{B}_7$ | = | $-x_2\,\mathbf{a}_1 - y_2\,\mathbf{a}_2 - z_2\,\mathbf{a}_3$ | = | $(-x_2 a - z_2 c\cos\beta)\,\hat{\mathbf{x}} - y_2 b\,\hat{\mathbf{y}} - z_2 c\sin\beta\,\hat{\mathbf{z}}$ | (4$e$) | Ca II |
| $\mathbf{B}_8$ | = | $x_2\,\mathbf{a}_1 + \left(\frac{1}{2} - y_2\right)\mathbf{a}_2 + \left(\frac{1}{2} + z_2\right)\mathbf{a}_3$ | = | $\left(\frac{1}{2}c\cos\beta + x_2 a + z_2 c\cos\beta\right)\hat{\mathbf{x}} + \left(\frac{1}{2} - y_2\right)b\,\hat{\mathbf{y}} + \left(\frac{1}{2} + z_2\right)c\sin\beta\,\hat{\mathbf{z}}$ | (4$e$) | Ca II |
| $\mathbf{B}_9$ | = | $x_3\,\mathbf{a}_1 + y_3\,\mathbf{a}_2 + z_3\,\mathbf{a}_3$ | = | $(x_3 a + z_3 c\cos\beta)\,\hat{\mathbf{x}} + y_3 b\,\hat{\mathbf{y}} + z_3 c\sin\beta\,\hat{\mathbf{z}}$ | (4$e$) | Ca III |
| $\mathbf{B}_{10}$ | = | $-x_3\,\mathbf{a}_1 + \left(\frac{1}{2} + y_3\right)\mathbf{a}_2 + \left(\frac{1}{2} - z_3\right)\mathbf{a}_3$ | = | $\left(\frac{1}{2}c\cos\beta - x_3 a - z_3 c\cos\beta\right)\hat{\mathbf{x}} + \left(\frac{1}{2} + y_3\right)b\,\hat{\mathbf{y}} + \left(\frac{1}{2} - z_3\right)c\sin\beta\,\hat{\mathbf{z}}$ | (4$e$) | Ca III |
| $\mathbf{B}_{11}$ | = | $-x_3\,\mathbf{a}_1 - y_3\,\mathbf{a}_2 - z_3\,\mathbf{a}_3$ | = | $(-x_3 a - z_3 c\cos\beta)\,\hat{\mathbf{x}} - y_3 b\,\hat{\mathbf{y}} - z_3 c\sin\beta\,\hat{\mathbf{z}}$ | (4$e$) | Ca III |
| $\mathbf{B}_{12}$ | = | $x_3\,\mathbf{a}_1 + \left(\frac{1}{2} - y_3\right)\mathbf{a}_2 + \left(\frac{1}{2} + z_3\right)\mathbf{a}_3$ | = | $\left(\frac{1}{2}c\cos\beta + x_3 a + z_3 c\cos\beta\right)\hat{\mathbf{x}} + \left(\frac{1}{2} - y_3\right)b\,\hat{\mathbf{y}} + \left(\frac{1}{2} + z_3\right)c\sin\beta\,\hat{\mathbf{z}}$ | (4$e$) | Ca III |
| $\mathbf{B}_{13}$ | = | $x_4\,\mathbf{a}_1 + y_4\,\mathbf{a}_2 + z_4\,\mathbf{a}_3$ | = | $(x_4 a + z_4 c\cos\beta)\,\hat{\mathbf{x}} + y_4 b\,\hat{\mathbf{y}} + z_4 c\sin\beta\,\hat{\mathbf{z}}$ | (4$e$) | O I |
| $\mathbf{B}_{14}$ | = | $-x_4\,\mathbf{a}_1 + \left(\frac{1}{2} + y_4\right)\mathbf{a}_2 + \left(\frac{1}{2} - z_4\right)\mathbf{a}_3$ | = | $\left(\frac{1}{2}c\cos\beta - x_4 a - z_4 c\cos\beta\right)\hat{\mathbf{x}} + \left(\frac{1}{2} + y_4\right)b\,\hat{\mathbf{y}} + \left(\frac{1}{2} - z_4\right)c\sin\beta\,\hat{\mathbf{z}}$ | (4$e$) | O I |
| $\mathbf{B}_{15}$ | = | $-x_4\,\mathbf{a}_1 - y_4\,\mathbf{a}_2 - z_4\,\mathbf{a}_3$ | = | $(-x_4 a - z_4 c\cos\beta)\,\hat{\mathbf{x}} - y_4 b\,\hat{\mathbf{y}} - z_4 c\sin\beta\,\hat{\mathbf{z}}$ | (4$e$) | O I |
| $\mathbf{B}_{16}$ | = | $x_4\,\mathbf{a}_1 + \left(\frac{1}{2} - y_4\right)\mathbf{a}_2 + \left(\frac{1}{2} + z_4\right)\mathbf{a}_3$ | = | $\left(\frac{1}{2}c\cos\beta + x_4 a + z_4 c\cos\beta\right)\hat{\mathbf{x}} + \left(\frac{1}{2} - y_4\right)b\,\hat{\mathbf{y}} + \left(\frac{1}{2} + z_4\right)c\sin\beta\,\hat{\mathbf{z}}$ | (4$e$) | O I |
| $\mathbf{B}_{17}$ | = | $x_5\,\mathbf{a}_1 + y_5\,\mathbf{a}_2 + z_5\,\mathbf{a}_3$ | = | $(x_5 a + z_5 c\cos\beta)\,\hat{\mathbf{x}} + y_5 b\,\hat{\mathbf{y}} + z_5 c\sin\beta\,\hat{\mathbf{z}}$ | (4$e$) | O II |
| $\mathbf{B}_{18}$ | = | $-x_5\,\mathbf{a}_1 + \left(\frac{1}{2} + y_5\right)\mathbf{a}_2 + \left(\frac{1}{2} - z_5\right)\mathbf{a}_3$ | = | $\left(\frac{1}{2}c\cos\beta - x_5 a - z_5 c\cos\beta\right)\hat{\mathbf{x}} + \left(\frac{1}{2} + y_5\right)b\,\hat{\mathbf{y}} + \left(\frac{1}{2} - z_5\right)c\sin\beta\,\hat{\mathbf{z}}$ | (4$e$) | O II |
| $\mathbf{B}_{19}$ | = | $-x_5\,\mathbf{a}_1 - y_5\,\mathbf{a}_2 - z_5\,\mathbf{a}_3$ | = | $(-x_5 a - z_5 c\cos\beta)\,\hat{\mathbf{x}} - y_5 b\,\hat{\mathbf{y}} - z_5 c\sin\beta\,\hat{\mathbf{z}}$ | (4$e$) | O II |
| $\mathbf{B}_{20}$ | = | $x_5\,\mathbf{a}_1 + \left(\frac{1}{2} - y_5\right)\mathbf{a}_2 + \left(\frac{1}{2} + z_5\right)\mathbf{a}_3$ | = | $\left(\frac{1}{2}c\cos\beta + x_5 a + z_5 c\cos\beta\right)\hat{\mathbf{x}} + \left(\frac{1}{2} - y_5\right)b\,\hat{\mathbf{y}} + \left(\frac{1}{2} + z_5\right)c\sin\beta\,\hat{\mathbf{z}}$ | (4$e$) | O II |
| $\mathbf{B}_{21}$ | = | $x_6\,\mathbf{a}_1 + y_6\,\mathbf{a}_2 + z_6\,\mathbf{a}_3$ | = | $(x_6 a + z_6 c\cos\beta)\,\hat{\mathbf{x}} + y_6 b\,\hat{\mathbf{y}} + z_6 c\sin\beta\,\hat{\mathbf{z}}$ | (4$e$) | O III |
| $\mathbf{B}_{22}$ | = | $-x_6\,\mathbf{a}_1 + \left(\frac{1}{2} + y_6\right)\mathbf{a}_2 + \left(\frac{1}{2} - z_6\right)\mathbf{a}_3$ | = | $\left(\frac{1}{2}c\cos\beta - x_6 a - z_6 c\cos\beta\right)\hat{\mathbf{x}} + \left(\frac{1}{2} + y_6\right)b\,\hat{\mathbf{y}} + \left(\frac{1}{2} - z_6\right)c\sin\beta\,\hat{\mathbf{z}}$ | (4$e$) | O III |
| $\mathbf{B}_{23}$ | = | $-x_6\,\mathbf{a}_1 - y_6\,\mathbf{a}_2 - z_6\,\mathbf{a}_3$ | = | $(-x_6 a - z_6 c\cos\beta)\,\hat{\mathbf{x}} - y_6 b\,\hat{\mathbf{y}} - z_6 c\sin\beta\,\hat{\mathbf{z}}$ | (4$e$) | O III |
| $\mathbf{B}_{24}$ | = | $x_6\,\mathbf{a}_1 + \left(\frac{1}{2} - y_6\right)\mathbf{a}_2 + \left(\frac{1}{2} + z_6\right)\mathbf{a}_3$ | = | $\left(\frac{1}{2}c\cos\beta + x_6 a + z_6 c\cos\beta\right)\hat{\mathbf{x}} + \left(\frac{1}{2} - y_6\right)b\,\hat{\mathbf{y}} + \left(\frac{1}{2} + z_6\right)c\sin\beta\,\hat{\mathbf{z}}$ | (4e) | O III |
| $\mathbf{B}_{25}$ | = | $x_7\,\mathbf{a}_1 + y_7\,\mathbf{a}_2 + z_7\,\mathbf{a}_3$ | = | $(x_7 a + z_7 c\cos\beta)\,\hat{\mathbf{x}} + y_7 b\,\hat{\mathbf{y}} + z_7 c\sin\beta\,\hat{\mathbf{z}}$ | (4e) | O IV |
| $\mathbf{B}_{26}$ | = | $-x_7\,\mathbf{a}_1 + \left(\frac{1}{2} + y_7\right)\mathbf{a}_2 + \left(\frac{1}{2} - z_7\right)\mathbf{a}_3$ | = | $\left(\frac{1}{2}c\cos\beta - x_7 a - z_7 c\cos\beta\right)\hat{\mathbf{x}} + \left(\frac{1}{2} + y_7\right)b\,\hat{\mathbf{y}} + \left(\frac{1}{2} - z_7\right)c\sin\beta\,\hat{\mathbf{z}}$ | (4e) | O IV |
| $\mathbf{B}_{27}$ | = | $-x_7\,\mathbf{a}_1 - y_7\,\mathbf{a}_2 - z_7\,\mathbf{a}_3$ | = | $(-x_7 a - z_7 c\cos\beta)\,\hat{\mathbf{x}} - y_7 b\,\hat{\mathbf{y}} - z_7 c\sin\beta\,\hat{\mathbf{z}}$ | (4e) | O IV |
| $\mathbf{B}_{28}$ | = | $x_7\,\mathbf{a}_1 + \left(\frac{1}{2} - y_7\right)\mathbf{a}_2 + \left(\frac{1}{2} + z_7\right)\mathbf{a}_3$ | = | $\left(\frac{1}{2}c\cos\beta + x_7 a + z_7 c\cos\beta\right)\hat{\mathbf{x}} + \left(\frac{1}{2} - y_7\right)b\,\hat{\mathbf{y}} + \left(\frac{1}{2} + z_7\right)c\sin\beta\,\hat{\mathbf{z}}$ | (4e) | O IV |
| $\mathbf{B}_{29}$ | = | $x_8\,\mathbf{a}_1 + y_8\,\mathbf{a}_2 + z_8\,\mathbf{a}_3$ | = | $(x_8 a + z_8 c\cos\beta)\,\hat{\mathbf{x}} + y_8 b\,\hat{\mathbf{y}} + z_8 c\sin\beta\,\hat{\mathbf{z}}$ | (4e) | O V |
| $\mathbf{B}_{30}$ | = | $-x_8\,\mathbf{a}_1 + \left(\frac{1}{2} + y_8\right)\mathbf{a}_2 + \left(\frac{1}{2} - z_8\right)\mathbf{a}_3$ | = | $\left(\frac{1}{2}c\cos\beta - x_8 a - z_8 c\cos\beta\right)\hat{\mathbf{x}} + \left(\frac{1}{2} + y_8\right)b\,\hat{\mathbf{y}} + \left(\frac{1}{2} - z_8\right)c\sin\beta\,\hat{\mathbf{z}}$ | (4e) | O V |
| $\mathbf{B}_{31}$ | = | $-x_8\,\mathbf{a}_1 - y_8\,\mathbf{a}_2 - z_8\,\mathbf{a}_3$ | = | $(-x_8 a - z_8 c\cos\beta)\,\hat{\mathbf{x}} - y_8 b\,\hat{\mathbf{y}} - z_8 c\sin\beta\,\hat{\mathbf{z}}$ | (4e) | O V |
| $\mathbf{B}_{32}$ | = | $x_8\,\mathbf{a}_1 + \left(\frac{1}{2} - y_8\right)\mathbf{a}_2 + \left(\frac{1}{2} + z_8\right)\mathbf{a}_3$ | = | $\left(\frac{1}{2}c\cos\beta + x_8 a + z_8 c\cos\beta\right)\hat{\mathbf{x}} + \left(\frac{1}{2} - y_8\right)b\,\hat{\mathbf{y}} + \left(\frac{1}{2} + z_8\right)c\sin\beta\,\hat{\mathbf{z}}$ | (4e) | O V |
| $\mathbf{B}_{33}$ | = | $x_9\,\mathbf{a}_1 + y_9\,\mathbf{a}_2 + z_9\,\mathbf{a}_3$ | = | $(x_9 a + z_9 c\cos\beta)\,\hat{\mathbf{x}} + y_9 b\,\hat{\mathbf{y}} + z_9 c\sin\beta\,\hat{\mathbf{z}}$ | (4e) | O VI |
| $\mathbf{B}_{34}$ | = | $-x_9\,\mathbf{a}_1 + \left(\frac{1}{2} + y_9\right)\mathbf{a}_2 + \left(\frac{1}{2} - z_9\right)\mathbf{a}_3$ | = | $\left(\frac{1}{2}c\cos\beta - x_9 a - z_9 c\cos\beta\right)\hat{\mathbf{x}} + \left(\frac{1}{2} + y_9\right)b\,\hat{\mathbf{y}} + \left(\frac{1}{2} - z_9\right)c\sin\beta\,\hat{\mathbf{z}}$ | (4e) | O VI |
| $\mathbf{B}_{35}$ | = | $-x_9\,\mathbf{a}_1 - y_9\,\mathbf{a}_2 - z_9\,\mathbf{a}_3$ | = | $(-x_9 a - z_9 c\cos\beta)\,\hat{\mathbf{x}} - y_9 b\,\hat{\mathbf{y}} - z_9 c\sin\beta\,\hat{\mathbf{z}}$ | (4e) | O VI |
| $\mathbf{B}_{36}$ | = | $x_9\,\mathbf{a}_1 + \left(\frac{1}{2} - y_9\right)\mathbf{a}_2 + \left(\frac{1}{2} + z_9\right)\mathbf{a}_3$ | = | $\left(\frac{1}{2}c\cos\beta + x_9 a + z_9 c\cos\beta\right)\hat{\mathbf{x}} + \left(\frac{1}{2} - y_9\right)b\,\hat{\mathbf{y}} + \left(\frac{1}{2} + z_9\right)c\sin\beta\,\hat{\mathbf{z}}$ | (4e) | O VI |
| $\mathbf{B}_{37}$ | = | $x_{10}\,\mathbf{a}_1 + y_{10}\,\mathbf{a}_2 + z_{10}\,\mathbf{a}_3$ | = | $(x_{10} a + z_{10} c\cos\beta)\,\hat{\mathbf{x}} + y_{10} b\,\hat{\mathbf{y}} + z_{10} c\sin\beta\,\hat{\mathbf{z}}$ | (4e) | O VII |
| $\mathbf{B}_{38}$ | = | $-x_{10}\,\mathbf{a}_1 + \left(\frac{1}{2} + y_{10}\right)\mathbf{a}_2 + \left(\frac{1}{2} - z_{10}\right)\mathbf{a}_3$ | = | $\left(\frac{1}{2}c\cos\beta - x_{10} a - z_{10} c\cos\beta\right)\hat{\mathbf{x}} + \left(\frac{1}{2} + y_{10}\right)b\,\hat{\mathbf{y}} + \left(\frac{1}{2} - z_{10}\right)c\sin\beta\,\hat{\mathbf{z}}$ | (4e) | O VII |
| $\mathbf{B}_{39}$ | = | $-x_{10}\,\mathbf{a}_1 - y_{10}\,\mathbf{a}_2 - z_{10}\,\mathbf{a}_3$ | = | $(-x_{10} a - z_{10} c\cos\beta)\,\hat{\mathbf{x}} - y_{10} b\,\hat{\mathbf{y}} - z_{10} c\sin\beta\,\hat{\mathbf{z}}$ | (4e) | O VII |
| $\mathbf{B}_{40}$ | = | $x_{10}\,\mathbf{a}_1 + \left(\frac{1}{2} - y_{10}\right)\mathbf{a}_2 + \left(\frac{1}{2} + z_{10}\right)\mathbf{a}_3$ | = | $\left(\frac{1}{2}c\cos\beta + x_{10} a + z_{10} c\cos\beta\right)\hat{\mathbf{x}} + \left(\frac{1}{2} - y_{10}\right)b\,\hat{\mathbf{y}} + \left(\frac{1}{2} + z_{10}\right)c\sin\beta\,\hat{\mathbf{z}}$ | (4e) | O VII |
| $\mathbf{B}_{41}$ | = | $x_{11}\,\mathbf{a}_1 + y_{11}\,\mathbf{a}_2 + z_{11}\,\mathbf{a}_3$ | = | $(x_{11} a + z_{11} c\cos\beta)\,\hat{\mathbf{x}} + y_{11} b\,\hat{\mathbf{y}} + z_{11} c\sin\beta\,\hat{\mathbf{z}}$ | (4e) | O VIII |
| $\mathbf{B}_{42}$ | = | $-x_{11}\,\mathbf{a}_1 + \left(\frac{1}{2} + y_{11}\right)\mathbf{a}_2 + \left(\frac{1}{2} - z_{11}\right)\mathbf{a}_3$ | = | $\left(\frac{1}{2}c\cos\beta - x_{11} a - z_{11} c\cos\beta\right)\hat{\mathbf{x}} + \left(\frac{1}{2} + y_{11}\right)b\,\hat{\mathbf{y}} + \left(\frac{1}{2} - z_{11}\right)c\sin\beta\,\hat{\mathbf{z}}$ | (4e) | O VIII |
| $\mathbf{B}_{43}$ | = | $-x_{11}\,\mathbf{a}_1 - y_{11}\,\mathbf{a}_2 - z_{11}\,\mathbf{a}_3$ | = | $(-x_{11} a - z_{11} c\cos\beta)\,\hat{\mathbf{x}} - y_{11} b\,\hat{\mathbf{y}} - z_{11} c\sin\beta\,\hat{\mathbf{z}}$ | (4e) | O VIII |
| $\mathbf{B}_{44}$ | = | $x_{11}\,\mathbf{a}_1 + \left(\frac{1}{2} - y_{11}\right)\mathbf{a}_2 + \left(\frac{1}{2} + z_{11}\right)\mathbf{a}_3$ | = | $\left(\frac{1}{2}c\cos\beta + x_{11} a + z_{11} c\cos\beta\right)\hat{\mathbf{x}} + \left(\frac{1}{2} - y_{11}\right)b\,\hat{\mathbf{y}} + \left(\frac{1}{2} + z_{11}\right)c\sin\beta\,\hat{\mathbf{z}}$ | (4e) | O VIII |
| $\mathbf{B}_{45}$ | = | $x_{12}\,\mathbf{a}_1 + y_{12}\,\mathbf{a}_2 + z_{12}\,\mathbf{a}_3$ | = | $(x_{12} a + z_{12} c\cos\beta)\,\hat{\mathbf{x}} + y_{12} b\,\hat{\mathbf{y}} + z_{12} c\sin\beta\,\hat{\mathbf{z}}$ | (4e) | O IX |
| $\mathbf{B}_{46}$ | = | $-x_{12}\,\mathbf{a}_1 + \left(\frac{1}{2} + y_{12}\right)\,\mathbf{a}_2 + \left(\frac{1}{2} - z_{12}\right)\,\mathbf{a}_3$ | = | $\left(\frac{1}{2}c\cos\beta - x_{12}a - z_{12}c\cos\beta\right)\,\hat{\mathbf{x}} + \left(\frac{1}{2} + y_{12}\right)b\,\hat{\mathbf{y}} + \left(\frac{1}{2} - z_{12}\right)c\sin\beta\,\hat{\mathbf{z}}$ | (4e) | O IX |
| $\mathbf{B}_{47}$ | = | $-x_{12}\,\mathbf{a}_1 - y_{12}\,\mathbf{a}_2 - z_{12}\,\mathbf{a}_3$ | = | $(-x_{12}a - z_{12}c\cos\beta)\,\hat{\mathbf{x}} - y_{12}b\,\hat{\mathbf{y}} - z_{12}c\sin\beta\,\hat{\mathbf{z}}$ | (4e) | O IX |
| $\mathbf{B}_{48}$ | = | $x_{12}\,\mathbf{a}_1 + \left(\frac{1}{2} - y_{12}\right)\,\mathbf{a}_2 + \left(\frac{1}{2} + z_{12}\right)\,\mathbf{a}_3$ | = | $\left(\frac{1}{2}c\cos\beta + x_{12}a + z_{12}c\cos\beta\right)\,\hat{\mathbf{x}} + \left(\frac{1}{2} - y_{12}\right)b\,\hat{\mathbf{y}} + \left(\frac{1}{2} + z_{12}\right)c\sin\beta\,\hat{\mathbf{z}}$ | (4e) | O IX |
| $\mathbf{B}_{49}$ | = | $x_{13}\,\mathbf{a}_1 + y_{13}\,\mathbf{a}_2 + z_{13}\,\mathbf{a}_3$ | = | $(x_{13}a + z_{13}c\cos\beta)\,\hat{\mathbf{x}} + y_{13}b\,\hat{\mathbf{y}} + z_{13}c\sin\beta\,\hat{\mathbf{z}}$ | (4e) | Si I |
| $\mathbf{B}_{50}$ | = | $-x_{13}\,\mathbf{a}_1 + \left(\frac{1}{2} + y_{13}\right)\,\mathbf{a}_2 + \left(\frac{1}{2} - z_{13}\right)\,\mathbf{a}_3$ | = | $\left(\frac{1}{2}c\cos\beta - x_{13}a - z_{13}c\cos\beta\right)\,\hat{\mathbf{x}} + \left(\frac{1}{2} + y_{13}\right)b\,\hat{\mathbf{y}} + \left(\frac{1}{2} - z_{13}\right)c\sin\beta\,\hat{\mathbf{z}}$ | (4e) | Si I |
| $\mathbf{B}_{51}$ | = | $-x_{13}\,\mathbf{a}_1 - y_{13}\,\mathbf{a}_2 - z_{13}\,\mathbf{a}_3$ | = | $(-x_{13}a - z_{13}c\cos\beta)\,\hat{\mathbf{x}} - y_{13}b\,\hat{\mathbf{y}} - z_{13}c\sin\beta\,\hat{\mathbf{z}}$ | (4e) | Si I |
| $\mathbf{B}_{52}$ | = | $x_{13}\,\mathbf{a}_1 + \left(\frac{1}{2} - y_{13}\right)\,\mathbf{a}_2 + \left(\frac{1}{2} + z_{13}\right)\,\mathbf{a}_3$ | = | $\left(\frac{1}{2}c\cos\beta + x_{13}a + z_{13}c\cos\beta\right)\,\hat{\mathbf{x}} + \left(\frac{1}{2} - y_{13}\right)b\,\hat{\mathbf{y}} + \left(\frac{1}{2} + z_{13}\right)c\sin\beta\,\hat{\mathbf{z}}$ | (4e) | Si I |
| $\mathbf{B}_{53}$ | = | $x_{14}\,\mathbf{a}_1 + y_{14}\,\mathbf{a}_2 + z_{14}\,\mathbf{a}_3$ | = | $(x_{14}a + z_{14}c\cos\beta)\,\hat{\mathbf{x}} + y_{14}b\,\hat{\mathbf{y}} + z_{14}c\sin\beta\,\hat{\mathbf{z}}$ | (4e) | Si II |
| $\mathbf{B}_{54}$ | = | $-x_{14}\,\mathbf{a}_1 + \left(\frac{1}{2} + y_{14}\right)\,\mathbf{a}_2 + \left(\frac{1}{2} - z_{14}\right)\,\mathbf{a}_3$ | = | $\left(\frac{1}{2}c\cos\beta - x_{14}a - z_{14}c\cos\beta\right)\,\hat{\mathbf{x}} + \left(\frac{1}{2} + y_{14}\right)b\,\hat{\mathbf{y}} + \left(\frac{1}{2} - z_{14}\right)c\sin\beta\,\hat{\mathbf{z}}$ | (4e) | Si II |
| $\mathbf{B}_{55}$ | = | $-x_{14}\,\mathbf{a}_1 - y_{14}\,\mathbf{a}_2 - z_{14}\,\mathbf{a}_3$ | = | $(-x_{14}a - z_{14}c\cos\beta)\,\hat{\mathbf{x}} - y_{14}b\,\hat{\mathbf{y}} - z_{14}c\sin\beta\,\hat{\mathbf{z}}$ | (4e) | Si II |
| $\mathbf{B}_{56}$ | = | $x_{14}\,\mathbf{a}_1 + \left(\frac{1}{2} - y_{14}\right)\,\mathbf{a}_2 + \left(\frac{1}{2} + z_{14}\right)\,\mathbf{a}_3$ | = | $\left(\frac{1}{2}c\cos\beta + x_{14}a + z_{14}c\cos\beta\right)\,\hat{\mathbf{x}} + \left(\frac{1}{2} - y_{14}\right)b\,\hat{\mathbf{y}} + \left(\frac{1}{2} + z_{14}\right)c\sin\beta\,\hat{\mathbf{z}}$ | (4e) | Si II |
| $\mathbf{B}_{57}$ | = | $x_{15}\,\mathbf{a}_1 + y_{15}\,\mathbf{a}_2 + z_{15}\,\mathbf{a}_3$ | = | $(x_{15}a + z_{15}c\cos\beta)\,\hat{\mathbf{x}} + y_{15}b\,\hat{\mathbf{y}} + z_{15}c\sin\beta\,\hat{\mathbf{z}}$ | (4e) | Si III |
| $\mathbf{B}_{58}$ | = | $-x_{15}\,\mathbf{a}_1 + \left(\frac{1}{2} + y_{15}\right)\,\mathbf{a}_2 + \left(\frac{1}{2} - z_{15}\right)\,\mathbf{a}_3$ | = | $\left(\frac{1}{2}c\cos\beta - x_{15}a - z_{15}c\cos\beta\right)\,\hat{\mathbf{x}} + \left(\frac{1}{2} + y_{15}\right)b\,\hat{\mathbf{y}} + \left(\frac{1}{2} - z_{15}\right)c\sin\beta\,\hat{\mathbf{z}}$ | (4e) | Si III |
| $\mathbf{B}_{59}$ | = | $-x_{15}\,\mathbf{a}_1 - y_{15}\,\mathbf{a}_2 - z_{15}\,\mathbf{a}_3$ | = | $(-x_{15}a - z_{15}c\cos\beta)\,\hat{\mathbf{x}} - y_{15}b\,\hat{\mathbf{y}} - z_{15}c\sin\beta\,\hat{\mathbf{z}}$ | (4e) | Si III |
| $\mathbf{B}_{60}$ | = | $x_{15}\,\mathbf{a}_1 + \left(\frac{1}{2} - y_{15}\right)\,\mathbf{a}_2 + \left(\frac{1}{2} + z_{15}\right)\,\mathbf{a}_3$ | = | $\left(\frac{1}{2}c\cos\beta + x_{15}a + z_{15}c\cos\beta\right)\,\hat{\mathbf{x}} + \left(\frac{1}{2} - y_{15}\right)b\,\hat{\mathbf{y}} + \left(\frac{1}{2} + z_{15}\right)c\sin\beta\,\hat{\mathbf{z}}$ | (4e) | Si III |

**Geometry files:**

- CIF: pp. 1549
- POSCAR: pp. 1549

# $\beta$-$B_2H_6$ Structure: AB3_mP16_14_e_3e

http://aflow.org/prototype-encyclopedia/AB3_mP16_14_e_3e.beta

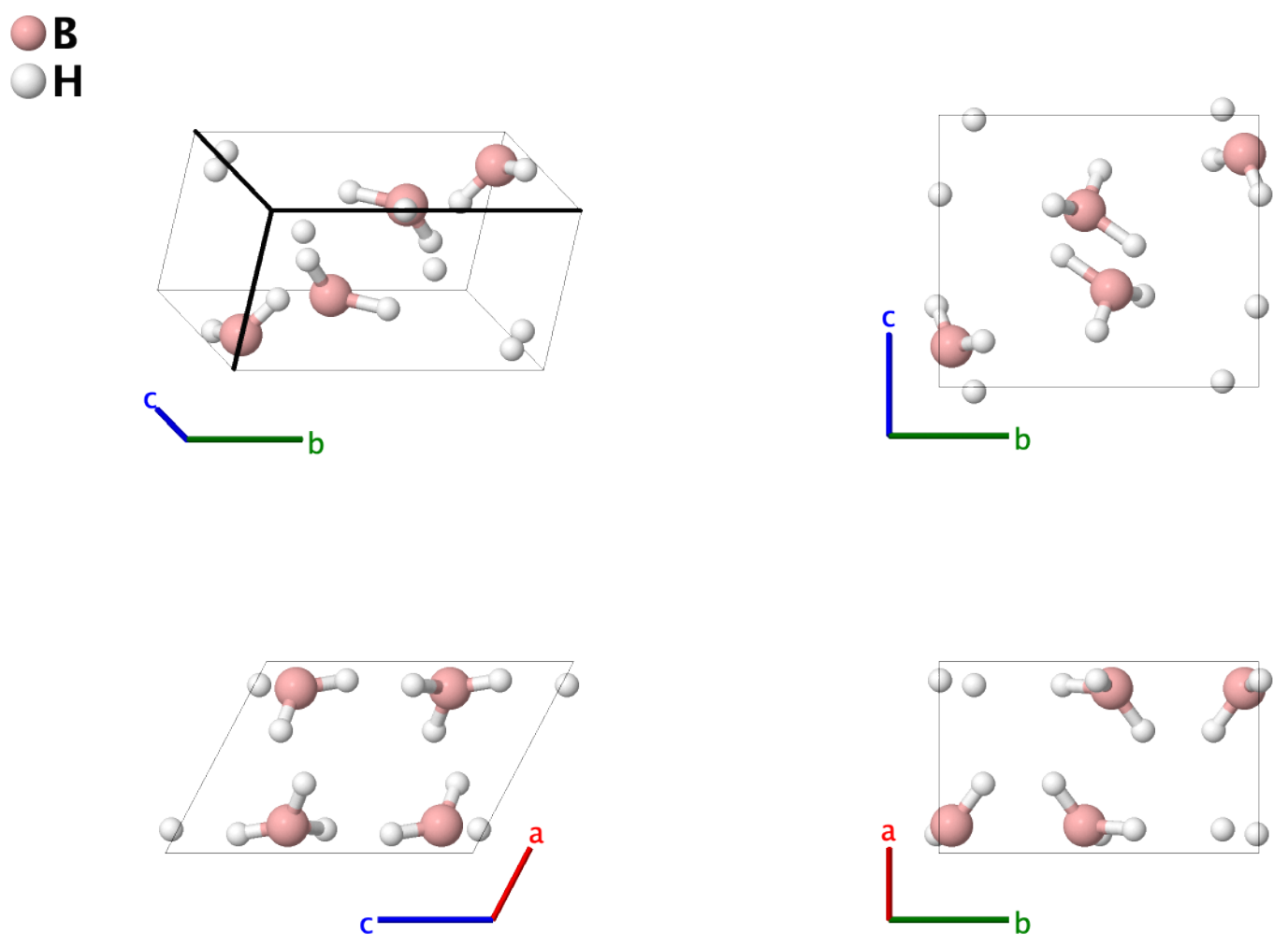

| | | |
|---|---|---|
| **Prototype** | : | $\beta$-$B_2H_6$ |
| **AFLOW prototype label** | : | AB3_mP16_14_e_3e |
| ***Strukturbericht* designation** | : | None |
| **Pearson symbol** | : | mP16 |
| **Space group number** | : | 14 |
| **Space group symbol** | : | $P2_1/c$ |
| **AFLOW prototype command** | : | `aflow --proto=AB3_mP16_14_e_3e`<br>`--params=`$a, b/a, c/a, \beta, x_1, y_1, z_1, x_2, y_2, z_2, x_3, y_3, z_3, x_4, y_4, z_4$ |

- (Mark, 1925) studied a variety of $B_2H_6$ structures, including the one given *Strukturbericht* designation $D4_1$ by (Ewald,1931). (Smith, 1965) refined this structure, including the hydrogen positions.
- This structure shares the same AFLOW designation, AB3_mP16_14_e_3e, as the $B_2H_6$ structure defined by (Yao, 2011). If we remove the hydrogen atoms from either of these structures we get the A10 ($\alpha$-Hg) structure, but that is more closely associated with the Yao structure than it is with this one.
- (Smith, 1965) presented the crystallographic information for this structure in the $P2_1/n$, unique axis $c$, setting of space group #14. We used FINDSYM to put the information into the standard $P2_1/c$, unique axis $b$, setting. This involved modifications of the primitive vectors beyond a simple rotation.

**Simple Monoclinic primitive vectors:**

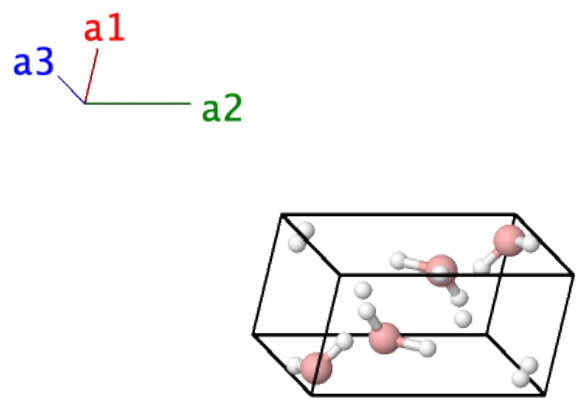

$$\mathbf{a}_1 = a\,\hat{\mathbf{x}}$$
$$\mathbf{a}_2 = b\,\hat{\mathbf{y}}$$
$$\mathbf{a}_3 = c\cos\beta\,\hat{\mathbf{x}} + c\sin\beta\,\hat{\mathbf{z}}$$

**Basis vectors:**

| | | Lattice Coordinates | | Cartesian Coordinates | Wyckoff Position | Atom Type |
|---|---|---|---|---|---|---|
| $\mathbf{B}_1$ | = | $x_1\,\mathbf{a}_1 + y_1\,\mathbf{a}_2 + z_1\,\mathbf{a}_3$ | = | $(x_1 a + z_1 c\cos\beta)\,\hat{\mathbf{x}} + y_1 b\,\hat{\mathbf{y}} + z_1 c\sin\beta\,\hat{\mathbf{z}}$ | (4$e$) | B |
| $\mathbf{B}_2$ | = | $-x_1\,\mathbf{a}_1 + \left(\frac{1}{2} + y_1\right)\mathbf{a}_2 + \left(\frac{1}{2} - z_1\right)\mathbf{a}_3$ | = | $\left(\frac{1}{2}c\cos\beta - x_1 a - z_1 c\cos\beta\right)\hat{\mathbf{x}} + \left(\frac{1}{2} + y_1\right)b\,\hat{\mathbf{y}} + \left(\frac{1}{2} - z_1\right)c\sin\beta\,\hat{\mathbf{z}}$ | (4$e$) | B |
| $\mathbf{B}_3$ | = | $-x_1\,\mathbf{a}_1 - y_1\,\mathbf{a}_2 - z_1\,\mathbf{a}_3$ | = | $(-x_1 a - z_1 c\cos\beta)\,\hat{\mathbf{x}} - y_1 b\,\hat{\mathbf{y}} - z_1 c\sin\beta\,\hat{\mathbf{z}}$ | (4$e$) | B |
| $\mathbf{B}_4$ | = | $x_1\,\mathbf{a}_1 + \left(\frac{1}{2} - y_1\right)\mathbf{a}_2 + \left(\frac{1}{2} + z_1\right)\mathbf{a}_3$ | = | $\left(\frac{1}{2}c\cos\beta + x_1 a + z_1 c\cos\beta\right)\hat{\mathbf{x}} + \left(\frac{1}{2} - y_1\right)b\,\hat{\mathbf{y}} + \left(\frac{1}{2} + z_1\right)c\sin\beta\,\hat{\mathbf{z}}$ | (4$e$) | B |
| $\mathbf{B}_5$ | = | $x_2\,\mathbf{a}_1 + y_2\,\mathbf{a}_2 + z_2\,\mathbf{a}_3$ | = | $(x_2 a + z_2 c\cos\beta)\,\hat{\mathbf{x}} + y_2 b\,\hat{\mathbf{y}} + z_2 c\sin\beta\,\hat{\mathbf{z}}$ | (4$e$) | H I |
| $\mathbf{B}_6$ | = | $-x_2\,\mathbf{a}_1 + \left(\frac{1}{2} + y_2\right)\mathbf{a}_2 + \left(\frac{1}{2} - z_2\right)\mathbf{a}_3$ | = | $\left(\frac{1}{2}c\cos\beta - x_2 a - z_2 c\cos\beta\right)\hat{\mathbf{x}} + \left(\frac{1}{2} + y_2\right)b\,\hat{\mathbf{y}} + \left(\frac{1}{2} - z_2\right)c\sin\beta\,\hat{\mathbf{z}}$ | (4$e$) | H I |
| $\mathbf{B}_7$ | = | $-x_2\,\mathbf{a}_1 - y_2\,\mathbf{a}_2 - z_2\,\mathbf{a}_3$ | = | $(-x_2 a - z_2 c\cos\beta)\,\hat{\mathbf{x}} - y_2 b\,\hat{\mathbf{y}} - z_2 c\sin\beta\,\hat{\mathbf{z}}$ | (4$e$) | H I |
| $\mathbf{B}_8$ | = | $x_2\,\mathbf{a}_1 + \left(\frac{1}{2} - y_2\right)\mathbf{a}_2 + \left(\frac{1}{2} + z_2\right)\mathbf{a}_3$ | = | $\left(\frac{1}{2}c\cos\beta + x_2 a + z_2 c\cos\beta\right)\hat{\mathbf{x}} + \left(\frac{1}{2} - y_2\right)b\,\hat{\mathbf{y}} + \left(\frac{1}{2} + z_2\right)c\sin\beta\,\hat{\mathbf{z}}$ | (4$e$) | H I |
| $\mathbf{B}_9$ | = | $x_3\,\mathbf{a}_1 + y_3\,\mathbf{a}_2 + z_3\,\mathbf{a}_3$ | = | $(x_3 a + z_3 c\cos\beta)\,\hat{\mathbf{x}} + y_3 b\,\hat{\mathbf{y}} + z_3 c\sin\beta\,\hat{\mathbf{z}}$ | (4$e$) | H II |
| $\mathbf{B}_{10}$ | = | $-x_3\,\mathbf{a}_1 + \left(\frac{1}{2} + y_3\right)\mathbf{a}_2 + \left(\frac{1}{2} - z_3\right)\mathbf{a}_3$ | = | $\left(\frac{1}{2}c\cos\beta - x_3 a - z_3 c\cos\beta\right)\hat{\mathbf{x}} + \left(\frac{1}{2} + y_3\right)b\,\hat{\mathbf{y}} + \left(\frac{1}{2} - z_3\right)c\sin\beta\,\hat{\mathbf{z}}$ | (4$e$) | H II |
| $\mathbf{B}_{11}$ | = | $-x_3\,\mathbf{a}_1 - y_3\,\mathbf{a}_2 - z_3\,\mathbf{a}_3$ | = | $(-x_3 a - z_3 c\cos\beta)\,\hat{\mathbf{x}} - y_3 b\,\hat{\mathbf{y}} - z_3 c\sin\beta\,\hat{\mathbf{z}}$ | (4$e$) | H II |
| $\mathbf{B}_{12}$ | = | $x_3\,\mathbf{a}_1 + \left(\frac{1}{2} - y_3\right)\mathbf{a}_2 + \left(\frac{1}{2} + z_3\right)\mathbf{a}_3$ | = | $\left(\frac{1}{2}c\cos\beta + x_3 a + z_3 c\cos\beta\right)\hat{\mathbf{x}} + \left(\frac{1}{2} - y_3\right)b\,\hat{\mathbf{y}} + \left(\frac{1}{2} + z_3\right)c\sin\beta\,\hat{\mathbf{z}}$ | (4$e$) | H II |
| $\mathbf{B}_{13}$ | = | $x_4\,\mathbf{a}_1 + y_4\,\mathbf{a}_2 + z_4\,\mathbf{a}_3$ | = | $(x_4 a + z_4 c\cos\beta)\,\hat{\mathbf{x}} + y_4 b\,\hat{\mathbf{y}} + z_4 c\sin\beta\,\hat{\mathbf{z}}$ | (4$e$) | H III |
| $\mathbf{B}_{14}$ | = | $-x_4\,\mathbf{a}_1 + \left(\frac{1}{2} + y_4\right)\mathbf{a}_2 + \left(\frac{1}{2} - z_4\right)\mathbf{a}_3$ | = | $\left(\frac{1}{2}c\cos\beta - x_4 a - z_4 c\cos\beta\right)\hat{\mathbf{x}} + \left(\frac{1}{2} + y_4\right)b\,\hat{\mathbf{y}} + \left(\frac{1}{2} - z_4\right)c\sin\beta\,\hat{\mathbf{z}}$ | (4$e$) | H III |
| $\mathbf{B}_{15}$ | = | $-x_4\,\mathbf{a}_1 - y_4\,\mathbf{a}_2 - z_4\,\mathbf{a}_3$ | = | $(-x_4 a - z_4 c\cos\beta)\,\hat{\mathbf{x}} - y_4 b\,\hat{\mathbf{y}} - z_4 c\sin\beta\,\hat{\mathbf{z}}$ | (4$e$) | H III |
| $\mathbf{B}_{16}$ | = | $x_4\,\mathbf{a}_1 + \left(\frac{1}{2} - y_4\right)\mathbf{a}_2 + \left(\frac{1}{2} + z_4\right)\mathbf{a}_3$ | = | $\left(\frac{1}{2}c\cos\beta + x_4 a + z_4 c\cos\beta\right)\hat{\mathbf{x}} + \left(\frac{1}{2} - y_4\right)b\,\hat{\mathbf{y}} + \left(\frac{1}{2} + z_4\right)c\sin\beta\,\hat{\mathbf{z}}$ | (4$e$) | H III |

**Geometry files:**

- CIF: pp. 1550
- POSCAR: pp. 1550

# $B_2H_6$ ($P2_1/c$) Structure: AB3_mP16_14_e_3e

http://aflow.org/prototype-encyclopedia/AB3_mP16_14_e_3e

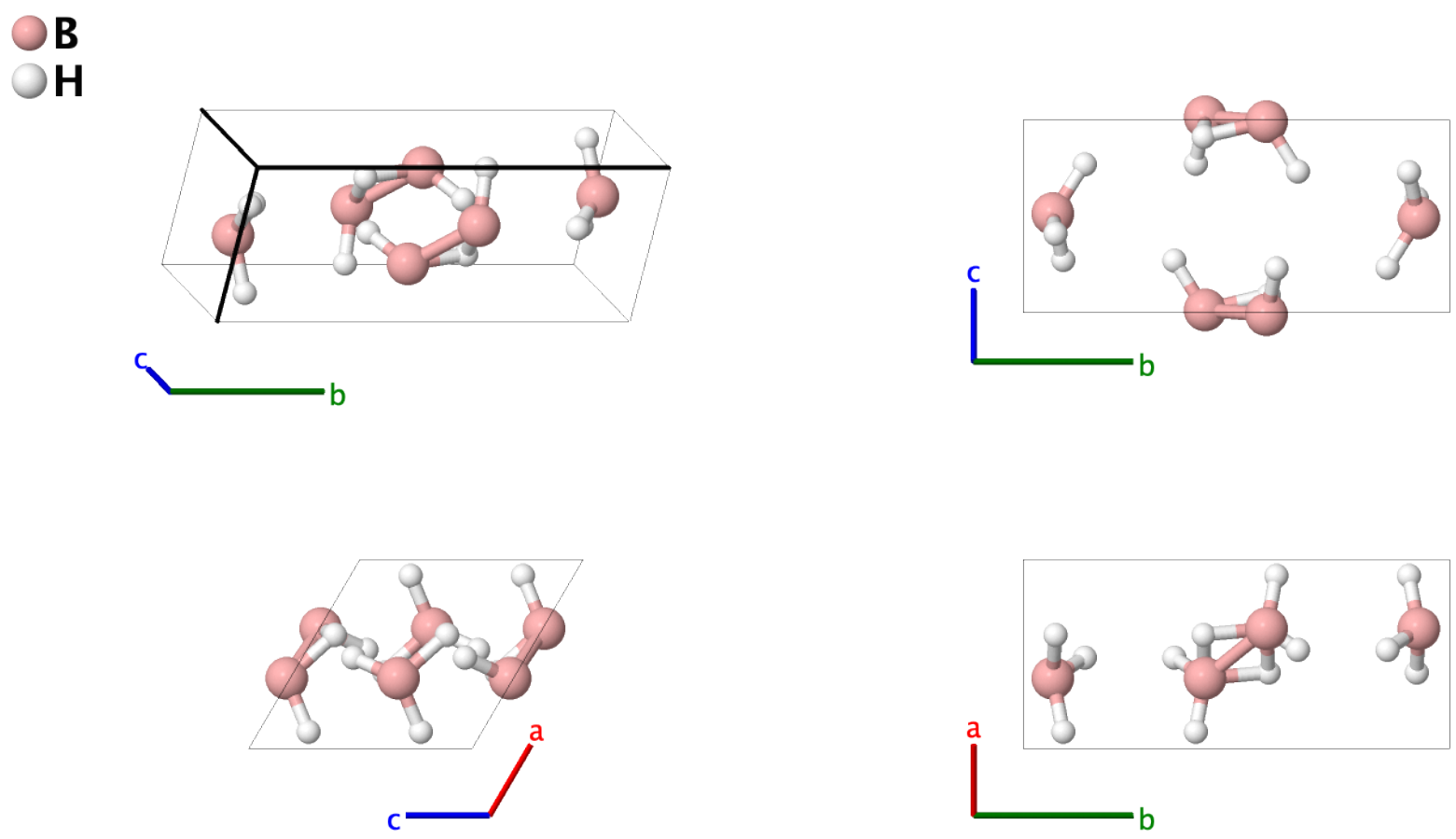

| | | |
|---|---|---|
| **Prototype** | : | $B_2H_6$ |
| **AFLOW prototype label** | : | AB3_mP16_14_e_3e |
| ***Strukturbericht* designation** | : | None |
| **Pearson symbol** | : | mP16 |
| **Space group number** | : | 14 |
| **Space group symbol** | : | $P2_1/c$ |
| **AFLOW prototype command** | : | `aflow --proto=AB3_mP16_14_e_3e`<br>`--params=`$a,b/a,c/a,\beta,x_1,y_1,z_1,x_2,y_2,z_2,x_3,y_3,z_3,x_4,y_4,z_4$ |

- (Mark, 1925) studied a variety of $B_2H_6$ structures, including the one given *Strukturbericht* designation $D4_1$ by (Ewald,1931). Very little information was given about this structure, other than Ewalds's statement "The position of the boron atoms is exactly the same as that of the Hg atoms in the A10 ($\alpha$-Hg) structure." (Yao, 2011) found the current structure computationally, and note that it is likely the structure found by (Mark, 1925).
- This structure shares the same AFLOW designation, AB3_mP16_14_e_3e, as the $\beta$-$B_2H_6$ structure described by (Smith, 1965).

**Simple Monoclinic primitive vectors:**

$$\mathbf{a}_1 = a\,\hat{\mathbf{x}}$$
$$\mathbf{a}_2 = b\,\hat{\mathbf{y}}$$
$$\mathbf{a}_3 = c\cos\beta\,\hat{\mathbf{x}} + c\sin\beta\,\hat{\mathbf{z}}$$

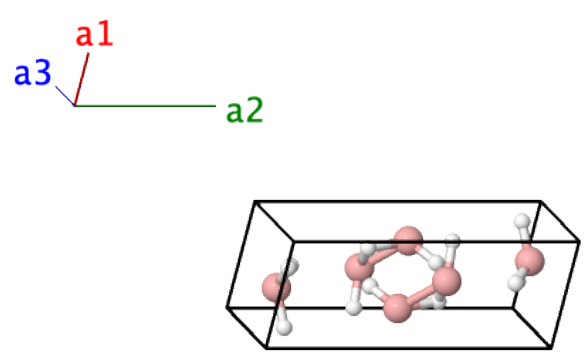

**Basis vectors:**

| | | Lattice Coordinates | | Cartesian Coordinates | Wyckoff Position | Atom Type |
|---|---|---|---|---|---|---|
| $\mathbf{B}_1$ | = | $x_1\,\mathbf{a}_1 + y_1\,\mathbf{a}_2 + z_1\,\mathbf{a}_3$ | = | $(x_1 a + z_1 c\cos\beta)\,\hat{\mathbf{x}} + y_1 b\,\hat{\mathbf{y}} + z_1 c\sin\beta\,\hat{\mathbf{z}}$ | (4$e$) | B |
| $\mathbf{B}_2$ | = | $-x_1\,\mathbf{a}_1 + \left(\frac{1}{2}+y_1\right)\mathbf{a}_2 + \left(\frac{1}{2}-z_1\right)\mathbf{a}_3$ | = | $\left(\frac{1}{2}c\cos\beta - x_1 a - z_1 c\cos\beta\right)\hat{\mathbf{x}} + \left(\frac{1}{2}+y_1\right)b\,\hat{\mathbf{y}} + \left(\frac{1}{2}-z_1\right)c\sin\beta\,\hat{\mathbf{z}}$ | (4$e$) | B |
| $\mathbf{B}_{3}$ | = | $-x_{1}\,\mathbf{a}_{1} - y_{1}\,\mathbf{a}_{2} - z_{1}\,\mathbf{a}_{3}$ | = | $(-x_{1}a - z_{1}c\cos\beta)\,\hat{\mathbf{x}} - y_{1}b\,\hat{\mathbf{y}} - z_{1}c\sin\beta\,\hat{\mathbf{z}}$ | (4$e$) | B |
| $\mathbf{B}_{4}$ | = | $x_{1}\,\mathbf{a}_{1} + \left(\frac{1}{2} - y_{1}\right)\mathbf{a}_{2} + \left(\frac{1}{2} + z_{1}\right)\mathbf{a}_{3}$ | = | $\left(\frac{1}{2}c\cos\beta + x_{1}a + z_{1}c\cos\beta\right)\hat{\mathbf{x}} + \left(\frac{1}{2} - y_{1}\right)b\,\hat{\mathbf{y}} + \left(\frac{1}{2} + z_{1}\right)c\sin\beta\,\hat{\mathbf{z}}$ | (4$e$) | B |
| $\mathbf{B}_{5}$ | = | $x_{2}\,\mathbf{a}_{1} + y_{2}\,\mathbf{a}_{2} + z_{2}\,\mathbf{a}_{3}$ | = | $(x_{2}a + z_{2}c\cos\beta)\,\hat{\mathbf{x}} + y_{2}b\,\hat{\mathbf{y}} + z_{2}c\sin\beta\,\hat{\mathbf{z}}$ | (4$e$) | H I |
| $\mathbf{B}_{6}$ | = | $-x_{2}\,\mathbf{a}_{1} + \left(\frac{1}{2} + y_{2}\right)\mathbf{a}_{2} + \left(\frac{1}{2} - z_{2}\right)\mathbf{a}_{3}$ | = | $\left(\frac{1}{2}c\cos\beta - x_{2}a - z_{2}c\cos\beta\right)\hat{\mathbf{x}} + \left(\frac{1}{2} + y_{2}\right)b\,\hat{\mathbf{y}} + \left(\frac{1}{2} - z_{2}\right)c\sin\beta\,\hat{\mathbf{z}}$ | (4$e$) | H I |
| $\mathbf{B}_{7}$ | = | $-x_{2}\,\mathbf{a}_{1} - y_{2}\,\mathbf{a}_{2} - z_{2}\,\mathbf{a}_{3}$ | = | $(-x_{2}a - z_{2}c\cos\beta)\,\hat{\mathbf{x}} - y_{2}b\,\hat{\mathbf{y}} - z_{2}c\sin\beta\,\hat{\mathbf{z}}$ | (4$e$) | H I |
| $\mathbf{B}_{8}$ | = | $x_{2}\,\mathbf{a}_{1} + \left(\frac{1}{2} - y_{2}\right)\mathbf{a}_{2} + \left(\frac{1}{2} + z_{2}\right)\mathbf{a}_{3}$ | = | $\left(\frac{1}{2}c\cos\beta + x_{2}a + z_{2}c\cos\beta\right)\hat{\mathbf{x}} + \left(\frac{1}{2} - y_{2}\right)b\,\hat{\mathbf{y}} + \left(\frac{1}{2} + z_{2}\right)c\sin\beta\,\hat{\mathbf{z}}$ | (4$e$) | H I |
| $\mathbf{B}_{9}$ | = | $x_{3}\,\mathbf{a}_{1} + y_{3}\,\mathbf{a}_{2} + z_{3}\,\mathbf{a}_{3}$ | = | $(x_{3}a + z_{3}c\cos\beta)\,\hat{\mathbf{x}} + y_{3}b\,\hat{\mathbf{y}} + z_{3}c\sin\beta\,\hat{\mathbf{z}}$ | (4$e$) | H II |
| $\mathbf{B}_{10}$ | = | $-x_{3}\,\mathbf{a}_{1} + \left(\frac{1}{2} + y_{3}\right)\mathbf{a}_{2} + \left(\frac{1}{2} - z_{3}\right)\mathbf{a}_{3}$ | = | $\left(\frac{1}{2}c\cos\beta - x_{3}a - z_{3}c\cos\beta\right)\hat{\mathbf{x}} + \left(\frac{1}{2} + y_{3}\right)b\,\hat{\mathbf{y}} + \left(\frac{1}{2} - z_{3}\right)c\sin\beta\,\hat{\mathbf{z}}$ | (4$e$) | H II |
| $\mathbf{B}_{11}$ | = | $-x_{3}\,\mathbf{a}_{1} - y_{3}\,\mathbf{a}_{2} - z_{3}\,\mathbf{a}_{3}$ | = | $(-x_{3}a - z_{3}c\cos\beta)\,\hat{\mathbf{x}} - y_{3}b\,\hat{\mathbf{y}} - z_{3}c\sin\beta\,\hat{\mathbf{z}}$ | (4$e$) | H II |
| $\mathbf{B}_{12}$ | = | $x_{3}\,\mathbf{a}_{1} + \left(\frac{1}{2} - y_{3}\right)\mathbf{a}_{2} + \left(\frac{1}{2} + z_{3}\right)\mathbf{a}_{3}$ | = | $\left(\frac{1}{2}c\cos\beta + x_{3}a + z_{3}c\cos\beta\right)\hat{\mathbf{x}} + \left(\frac{1}{2} - y_{3}\right)b\,\hat{\mathbf{y}} + \left(\frac{1}{2} + z_{3}\right)c\sin\beta\,\hat{\mathbf{z}}$ | (4$e$) | H II |
| $\mathbf{B}_{13}$ | = | $x_{4}\,\mathbf{a}_{1} + y_{4}\,\mathbf{a}_{2} + z_{4}\,\mathbf{a}_{3}$ | = | $(x_{4}a + z_{4}c\cos\beta)\,\hat{\mathbf{x}} + y_{4}b\,\hat{\mathbf{y}} + z_{4}c\sin\beta\,\hat{\mathbf{z}}$ | (4$e$) | H III |
| $\mathbf{B}_{14}$ | = | $-x_{4}\,\mathbf{a}_{1} + \left(\frac{1}{2} + y_{4}\right)\mathbf{a}_{2} + \left(\frac{1}{2} - z_{4}\right)\mathbf{a}_{3}$ | = | $\left(\frac{1}{2}c\cos\beta - x_{4}a - z_{4}c\cos\beta\right)\hat{\mathbf{x}} + \left(\frac{1}{2} + y_{4}\right)b\,\hat{\mathbf{y}} + \left(\frac{1}{2} - z_{4}\right)c\sin\beta\,\hat{\mathbf{z}}$ | (4$e$) | H III |
| $\mathbf{B}_{15}$ | = | $-x_{4}\,\mathbf{a}_{1} - y_{4}\,\mathbf{a}_{2} - z_{4}\,\mathbf{a}_{3}$ | = | $(-x_{4}a - z_{4}c\cos\beta)\,\hat{\mathbf{x}} - y_{4}b\,\hat{\mathbf{y}} - z_{4}c\sin\beta\,\hat{\mathbf{z}}$ | (4$e$) | H III |
| $\mathbf{B}_{16}$ | = | $x_{4}\,\mathbf{a}_{1} + \left(\frac{1}{2} - y_{4}\right)\mathbf{a}_{2} + \left(\frac{1}{2} + z_{4}\right)\mathbf{a}_{3}$ | = | $\left(\frac{1}{2}c\cos\beta + x_{4}a + z_{4}c\cos\beta\right)\hat{\mathbf{x}} + \left(\frac{1}{2} - y_{4}\right)b\,\hat{\mathbf{y}} + \left(\frac{1}{2} + z_{4}\right)c\sin\beta\,\hat{\mathbf{z}}$ | (4$e$) | H III |

---

---

**Geometry files:**

- CIF: pp. 1550
- POSCAR: pp. 1550

# $KAuBr_4 \cdot 2H_2O$ ($H4_{19}$) Structure: AB4C2D_mP32_14_e_4e_2e_e

http://aflow.org/prototype-encyclopedia/AB4C2D_mP32_14_e_4e_2e_e

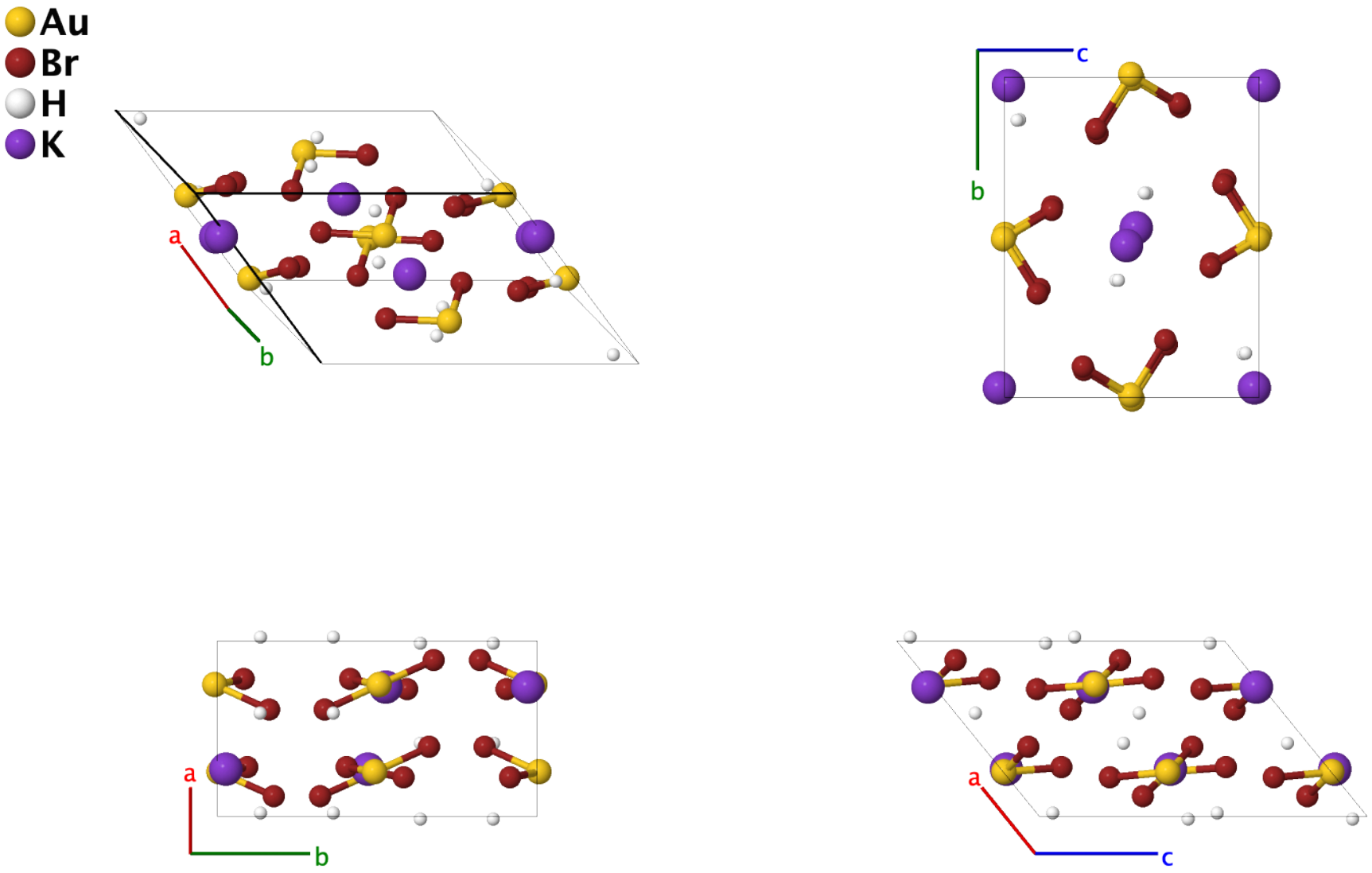

**Prototype** : $AuBr_4(H_2O)_2K$

**AFLOW prototype label** : AB4C2D_mP32_14_e_4e_2e_e

***Strukturbericht* designation** : $H4_{19}$

**Pearson symbol** : mP32

**Space group number** : 14

**Space group symbol** : $P2_1/c$

**AFLOW prototype command** : `aflow --proto=AB4C2D_mP32_14_e_4e_2e_e`
`--params=`$a, b/a, c/a, \beta, x_1, y_1, z_1, x_2, y_2, z_2, x_3, y_3, z_3, x_4, y_4, z_4, x_5, y_5, z_5, x_6, y_6, z_6, x_7, y_7, z_7, x_8, y_8, z_8$

- (Cox, 1936) determined lattice constants for $KAuBr_4 \cdot H_2O$, which were very close to those later found by (Omarni, 1986). However, Cox *et al.* found that the (100) planes (in our orientation) with odd indices showed very weak diffraction spots. Neglecting these spots gives a unit cell which has a lattice constant $a$ (in our orientation) which is half the size of that found by Omarni *et al*, and so the Pearson symbol mP16. Fitting this cell into space group $P2_1/c$ #14 then required that the gold atom be at the (2a) Wyckoff position and the potassium atom at (2b), while the later paper found that these atoms are slightly displaced from these points.

- The results of (Omarni, 1986) are very close to those of (Cox, 1936) and can indeed reduce to the Cox *et al.* structure by allowing some uncertainty in the atomic positions. Given this, and the fact that the former reference actually found the correct unit cell, we will use the more modern work as our prototype for *Strukturbericht* symbol $H4_{19}$.

- (Omarni, 1986) gave the unit cell and Wyckoff positions in terms of setting $P2_1/n$ of space group #14. Changing this to the standard $P2_1/c$ setting required a rotation of the lattice and a significant rewriting of the primitive vectors.

- The anhydrous form of this compound can be seen at $KAuBr_4$ (AB4C_mP24_14_ab_4e_e).

## Simple Monoclinic primitive vectors:

$$\mathbf{a}_1 = a\,\hat{\mathbf{x}}$$
$$\mathbf{a}_2 = b\,\hat{\mathbf{y}}$$
$$\mathbf{a}_3 = c\cos\beta\,\hat{\mathbf{x}} + c\sin\beta\,\hat{\mathbf{z}}$$

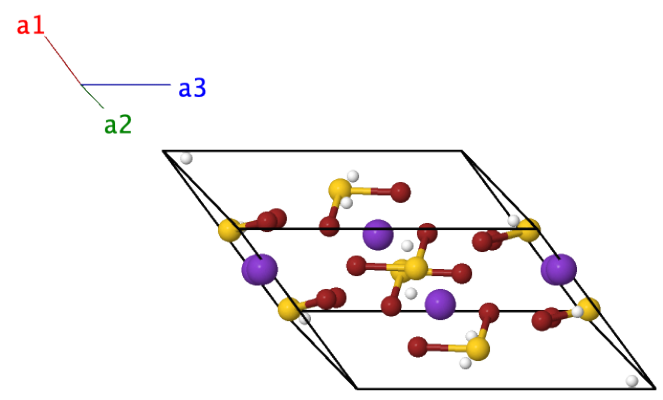

## Basis vectors:

| | | Lattice Coordinates | | Cartesian Coordinates | Wyckoff Position | Atom Type |
|---|---|---|---|---|---|---|
| $\mathbf{B}_1$ | = | $x_1\,\mathbf{a}_1 + y_1\,\mathbf{a}_2 + z_1\,\mathbf{a}_3$ | = | $(x_1 a + z_1 c\cos\beta)\,\hat{\mathbf{x}} + y_1 b\,\hat{\mathbf{y}} + z_1 c\sin\beta\,\hat{\mathbf{z}}$ | (4e) | Au |
| $\mathbf{B}_2$ | = | $-x_1\,\mathbf{a}_1 + \left(\frac{1}{2} + y_1\right)\mathbf{a}_2 + \left(\frac{1}{2} - z_1\right)\mathbf{a}_3$ | = | $\left(\frac{1}{2}c\cos\beta - x_1 a - z_1 c\cos\beta\right)\hat{\mathbf{x}} + \left(\frac{1}{2} + y_1\right)b\,\hat{\mathbf{y}} + \left(\frac{1}{2} - z_1\right)c\sin\beta\,\hat{\mathbf{z}}$ | (4e) | Au |
| $\mathbf{B}_3$ | = | $-x_1\,\mathbf{a}_1 - y_1\,\mathbf{a}_2 - z_1\,\mathbf{a}_3$ | = | $(-x_1 a - z_1 c\cos\beta)\,\hat{\mathbf{x}} - y_1 b\,\hat{\mathbf{y}} - z_1 c\sin\beta\,\hat{\mathbf{z}}$ | (4e) | Au |
| $\mathbf{B}_4$ | = | $x_1\,\mathbf{a}_1 + \left(\frac{1}{2} - y_1\right)\mathbf{a}_2 + \left(\frac{1}{2} + z_1\right)\mathbf{a}_3$ | = | $\left(\frac{1}{2}c\cos\beta + x_1 a + z_1 c\cos\beta\right)\hat{\mathbf{x}} + \left(\frac{1}{2} - y_1\right)b\,\hat{\mathbf{y}} + \left(\frac{1}{2} + z_1\right)c\sin\beta\,\hat{\mathbf{z}}$ | (4e) | Au |
| $\mathbf{B}_5$ | = | $x_2\,\mathbf{a}_1 + y_2\,\mathbf{a}_2 + z_2\,\mathbf{a}_3$ | = | $(x_2 a + z_2 c\cos\beta)\,\hat{\mathbf{x}} + y_2 b\,\hat{\mathbf{y}} + z_2 c\sin\beta\,\hat{\mathbf{z}}$ | (4e) | Br I |
| $\mathbf{B}_6$ | = | $-x_2\,\mathbf{a}_1 + \left(\frac{1}{2} + y_2\right)\mathbf{a}_2 + \left(\frac{1}{2} - z_2\right)\mathbf{a}_3$ | = | $\left(\frac{1}{2}c\cos\beta - x_2 a - z_2 c\cos\beta\right)\hat{\mathbf{x}} + \left(\frac{1}{2} + y_2\right)b\,\hat{\mathbf{y}} + \left(\frac{1}{2} - z_2\right)c\sin\beta\,\hat{\mathbf{z}}$ | (4e) | Br I |
| $\mathbf{B}_7$ | = | $-x_2\,\mathbf{a}_1 - y_2\,\mathbf{a}_2 - z_2\,\mathbf{a}_3$ | = | $(-x_2 a - z_2 c\cos\beta)\,\hat{\mathbf{x}} - y_2 b\,\hat{\mathbf{y}} - z_2 c\sin\beta\,\hat{\mathbf{z}}$ | (4e) | Br I |
| $\mathbf{B}_8$ | = | $x_2\,\mathbf{a}_1 + \left(\frac{1}{2} - y_2\right)\mathbf{a}_2 + \left(\frac{1}{2} + z_2\right)\mathbf{a}_3$ | = | $\left(\frac{1}{2}c\cos\beta + x_2 a + z_2 c\cos\beta\right)\hat{\mathbf{x}} + \left(\frac{1}{2} - y_2\right)b\,\hat{\mathbf{y}} + \left(\frac{1}{2} + z_2\right)c\sin\beta\,\hat{\mathbf{z}}$ | (4e) | Br I |
| $\mathbf{B}_9$ | = | $x_3\,\mathbf{a}_1 + y_3\,\mathbf{a}_2 + z_3\,\mathbf{a}_3$ | = | $(x_3 a + z_3 c\cos\beta)\,\hat{\mathbf{x}} + y_3 b\,\hat{\mathbf{y}} + z_3 c\sin\beta\,\hat{\mathbf{z}}$ | (4e) | Br II |
| $\mathbf{B}_{10}$ | = | $-x_3\,\mathbf{a}_1 + \left(\frac{1}{2} + y_3\right)\mathbf{a}_2 + \left(\frac{1}{2} - z_3\right)\mathbf{a}_3$ | = | $\left(\frac{1}{2}c\cos\beta - x_3 a - z_3 c\cos\beta\right)\hat{\mathbf{x}} + \left(\frac{1}{2} + y_3\right)b\,\hat{\mathbf{y}} + \left(\frac{1}{2} - z_3\right)c\sin\beta\,\hat{\mathbf{z}}$ | (4e) | Br II |
| $\mathbf{B}_{11}$ | = | $-x_3\,\mathbf{a}_1 - y_3\,\mathbf{a}_2 - z_3\,\mathbf{a}_3$ | = | $(-x_3 a - z_3 c\cos\beta)\,\hat{\mathbf{x}} - y_3 b\,\hat{\mathbf{y}} - z_3 c\sin\beta\,\hat{\mathbf{z}}$ | (4e) | Br II |
| $\mathbf{B}_{12}$ | = | $x_3\,\mathbf{a}_1 + \left(\frac{1}{2} - y_3\right)\mathbf{a}_2 + \left(\frac{1}{2} + z_3\right)\mathbf{a}_3$ | = | $\left(\frac{1}{2}c\cos\beta + x_3 a + z_3 c\cos\beta\right)\hat{\mathbf{x}} + \left(\frac{1}{2} - y_3\right)b\,\hat{\mathbf{y}} + \left(\frac{1}{2} + z_3\right)c\sin\beta\,\hat{\mathbf{z}}$ | (4e) | Br II |
| $\mathbf{B}_{13}$ | = | $x_4\,\mathbf{a}_1 + y_4\,\mathbf{a}_2 + z_4\,\mathbf{a}_3$ | = | $(x_4 a + z_4 c\cos\beta)\,\hat{\mathbf{x}} + y_4 b\,\hat{\mathbf{y}} + z_4 c\sin\beta\,\hat{\mathbf{z}}$ | (4e) | Br III |
| $\mathbf{B}_{14}$ | = | $-x_4\,\mathbf{a}_1 + \left(\frac{1}{2} + y_4\right)\mathbf{a}_2 + \left(\frac{1}{2} - z_4\right)\mathbf{a}_3$ | = | $\left(\frac{1}{2}c\cos\beta - x_4 a - z_4 c\cos\beta\right)\hat{\mathbf{x}} + \left(\frac{1}{2} + y_4\right)b\,\hat{\mathbf{y}} + \left(\frac{1}{2} - z_4\right)c\sin\beta\,\hat{\mathbf{z}}$ | (4e) | Br III |
| $\mathbf{B}_{15}$ | = | $-x_4\,\mathbf{a}_1 - y_4\,\mathbf{a}_2 - z_4\,\mathbf{a}_3$ | = | $(-x_4 a - z_4 c\cos\beta)\,\hat{\mathbf{x}} - y_4 b\,\hat{\mathbf{y}} - z_4 c\sin\beta\,\hat{\mathbf{z}}$ | (4e) | Br III |
| $\mathbf{B}_{16}$ | = | $x_4\,\mathbf{a}_1 + \left(\frac{1}{2} - y_4\right)\mathbf{a}_2 + \left(\frac{1}{2} + z_4\right)\mathbf{a}_3$ | = | $\left(\frac{1}{2}c\cos\beta + x_4 a + z_4 c\cos\beta\right)\hat{\mathbf{x}} + \left(\frac{1}{2} - y_4\right)b\,\hat{\mathbf{y}} + \left(\frac{1}{2} + z_4\right)c\sin\beta\,\hat{\mathbf{z}}$ | (4e) | Br III |
| $\mathbf{B}_{17}$ | = | $x_5\,\mathbf{a}_1 + y_5\,\mathbf{a}_2 + z_5\,\mathbf{a}_3$ | = | $(x_5 a + z_5 c\cos\beta)\,\hat{\mathbf{x}} + y_5 b\,\hat{\mathbf{y}} + z_5 c\sin\beta\,\hat{\mathbf{z}}$ | (4e) | Br IV |
| $\mathbf{B}_{18}$ | = | $-x_5\,\mathbf{a}_1 + \left(\frac{1}{2} + y_5\right)\,\mathbf{a}_2 + \left(\frac{1}{2} - z_5\right)\,\mathbf{a}_3$ | = | $\left(\frac{1}{2}c\cos\beta - x_5 a - z_5 c\cos\beta\right)\,\mathbf{\hat{x}} + \left(\frac{1}{2} + y_5\right)b\,\mathbf{\hat{y}} + \left(\frac{1}{2} - z_5\right)c\sin\beta\,\mathbf{\hat{z}}$ | (4$e$) | Br IV |
| $\mathbf{B}_{19}$ | = | $-x_5\,\mathbf{a}_1 - y_5\,\mathbf{a}_2 - z_5\,\mathbf{a}_3$ | = | $(-x_5 a - z_5 c\cos\beta)\,\mathbf{\hat{x}} - y_5 b\,\mathbf{\hat{y}} - z_5 c\sin\beta\,\mathbf{\hat{z}}$ | (4$e$) | Br IV |
| $\mathbf{B}_{20}$ | = | $x_5\,\mathbf{a}_1 + \left(\frac{1}{2} - y_5\right)\,\mathbf{a}_2 + \left(\frac{1}{2} + z_5\right)\,\mathbf{a}_3$ | = | $\left(\frac{1}{2}c\cos\beta + x_5 a + z_5 c\cos\beta\right)\,\mathbf{\hat{x}} + \left(\frac{1}{2} - y_5\right)b\,\mathbf{\hat{y}} + \left(\frac{1}{2} + z_5\right)c\sin\beta\,\mathbf{\hat{z}}$ | (4$e$) | Br IV |
| $\mathbf{B}_{21}$ | = | $x_6\,\mathbf{a}_1 + y_6\,\mathbf{a}_2 + z_6\,\mathbf{a}_3$ | = | $(x_6 a + z_6 c\cos\beta)\,\mathbf{\hat{x}} + y_6 b\,\mathbf{\hat{y}} + z_6 c\sin\beta\,\mathbf{\hat{z}}$ | (4$e$) | $H_2O$ I |
| $\mathbf{B}_{22}$ | = | $-x_6\,\mathbf{a}_1 + \left(\frac{1}{2} + y_6\right)\,\mathbf{a}_2 + \left(\frac{1}{2} - z_6\right)\,\mathbf{a}_3$ | = | $\left(\frac{1}{2}c\cos\beta - x_6 a - z_6 c\cos\beta\right)\,\mathbf{\hat{x}} + \left(\frac{1}{2} + y_6\right)b\,\mathbf{\hat{y}} + \left(\frac{1}{2} - z_6\right)c\sin\beta\,\mathbf{\hat{z}}$ | (4$e$) | $H_2O$ I |
| $\mathbf{B}_{23}$ | = | $-x_6\,\mathbf{a}_1 - y_6\,\mathbf{a}_2 - z_6\,\mathbf{a}_3$ | = | $(-x_6 a - z_6 c\cos\beta)\,\mathbf{\hat{x}} - y_6 b\,\mathbf{\hat{y}} - z_6 c\sin\beta\,\mathbf{\hat{z}}$ | (4$e$) | $H_2O$ I |
| $\mathbf{B}_{24}$ | = | $x_6\,\mathbf{a}_1 + \left(\frac{1}{2} - y_6\right)\,\mathbf{a}_2 + \left(\frac{1}{2} + z_6\right)\,\mathbf{a}_3$ | = | $\left(\frac{1}{2}c\cos\beta + x_6 a + z_6 c\cos\beta\right)\,\mathbf{\hat{x}} + \left(\frac{1}{2} - y_6\right)b\,\mathbf{\hat{y}} + \left(\frac{1}{2} + z_6\right)c\sin\beta\,\mathbf{\hat{z}}$ | (4$e$) | $H_2O$ I |
| $\mathbf{B}_{25}$ | = | $x_7\,\mathbf{a}_1 + y_7\,\mathbf{a}_2 + z_7\,\mathbf{a}_3$ | = | $(x_7 a + z_7 c\cos\beta)\,\mathbf{\hat{x}} + y_7 b\,\mathbf{\hat{y}} + z_7 c\sin\beta\,\mathbf{\hat{z}}$ | (4$e$) | $H_2O$ II |
| $\mathbf{B}_{26}$ | = | $-x_7\,\mathbf{a}_1 + \left(\frac{1}{2} + y_7\right)\,\mathbf{a}_2 + \left(\frac{1}{2} - z_7\right)\,\mathbf{a}_3$ | = | $\left(\frac{1}{2}c\cos\beta - x_7 a - z_7 c\cos\beta\right)\,\mathbf{\hat{x}} + \left(\frac{1}{2} + y_7\right)b\,\mathbf{\hat{y}} + \left(\frac{1}{2} - z_7\right)c\sin\beta\,\mathbf{\hat{z}}$ | (4$e$) | $H_2O$ II |
| $\mathbf{B}_{27}$ | = | $-x_7\,\mathbf{a}_1 - y_7\,\mathbf{a}_2 - z_7\,\mathbf{a}_3$ | = | $(-x_7 a - z_7 c\cos\beta)\,\mathbf{\hat{x}} - y_7 b\,\mathbf{\hat{y}} - z_7 c\sin\beta\,\mathbf{\hat{z}}$ | (4$e$) | $H_2O$ II |
| $\mathbf{B}_{28}$ | = | $x_7\,\mathbf{a}_1 + \left(\frac{1}{2} - y_7\right)\,\mathbf{a}_2 + \left(\frac{1}{2} + z_7\right)\,\mathbf{a}_3$ | = | $\left(\frac{1}{2}c\cos\beta + x_7 a + z_7 c\cos\beta\right)\,\mathbf{\hat{x}} + \left(\frac{1}{2} - y_7\right)b\,\mathbf{\hat{y}} + \left(\frac{1}{2} + z_7\right)c\sin\beta\,\mathbf{\hat{z}}$ | (4$e$) | $H_2O$ II |
| $\mathbf{B}_{29}$ | = | $x_8\,\mathbf{a}_1 + y_8\,\mathbf{a}_2 + z_8\,\mathbf{a}_3$ | = | $(x_8 a + z_8 c\cos\beta)\,\mathbf{\hat{x}} + y_8 b\,\mathbf{\hat{y}} + z_8 c\sin\beta\,\mathbf{\hat{z}}$ | (4$e$) | K |
| $\mathbf{B}_{30}$ | = | $-x_8\,\mathbf{a}_1 + \left(\frac{1}{2} + y_8\right)\,\mathbf{a}_2 + \left(\frac{1}{2} - z_8\right)\,\mathbf{a}_3$ | = | $\left(\frac{1}{2}c\cos\beta - x_8 a - z_8 c\cos\beta\right)\,\mathbf{\hat{x}} + \left(\frac{1}{2} + y_8\right)b\,\mathbf{\hat{y}} + \left(\frac{1}{2} - z_8\right)c\sin\beta\,\mathbf{\hat{z}}$ | (4$e$) | K |
| $\mathbf{B}_{31}$ | = | $-x_8\,\mathbf{a}_1 - y_8\,\mathbf{a}_2 - z_8\,\mathbf{a}_3$ | = | $(-x_8 a - z_8 c\cos\beta)\,\mathbf{\hat{x}} - y_8 b\,\mathbf{\hat{y}} - z_8 c\sin\beta\,\mathbf{\hat{z}}$ | (4$e$) | K |
| $\mathbf{B}_{32}$ | = | $x_8\,\mathbf{a}_1 + \left(\frac{1}{2} - y_8\right)\,\mathbf{a}_2 + \left(\frac{1}{2} + z_8\right)\,\mathbf{a}_3$ | = | $\left(\frac{1}{2}c\cos\beta + x_8 a + z_8 c\cos\beta\right)\,\mathbf{\hat{x}} + \left(\frac{1}{2} - y_8\right)b\,\mathbf{\hat{y}} + \left(\frac{1}{2} + z_8\right)c\sin\beta\,\mathbf{\hat{z}}$ | (4$e$) | K |

**Geometry files:**

- CIF: pp. 1551
- POSCAR: pp. 1551

# Anhydrous $KAuBr_4$ Structure: AB4C_mP24_14_ab_4e_e

http://aflow.org/prototype-encyclopedia/AB4C_mP24_14_ab_4e_e

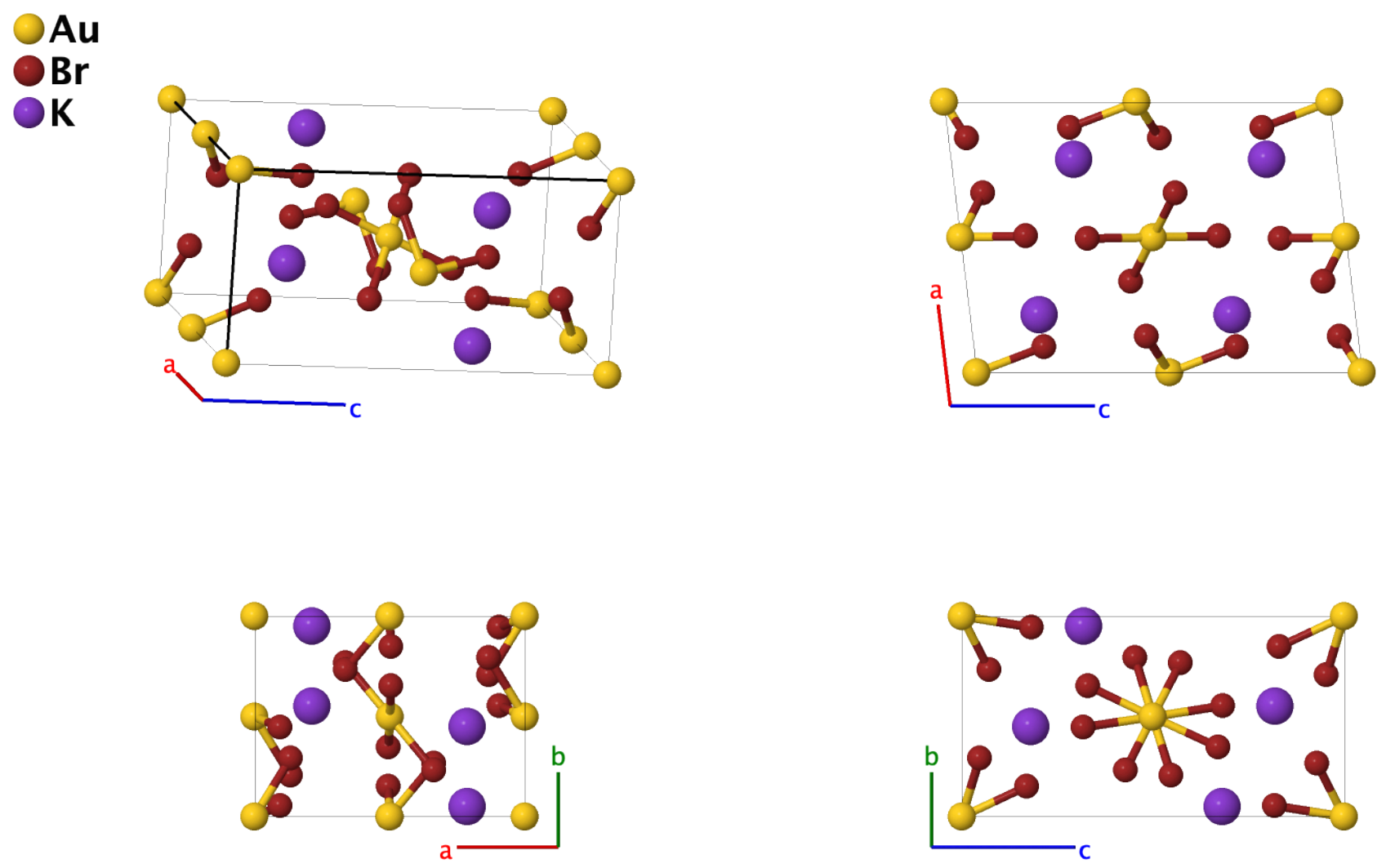

| | | |
|---|---|---|
| **Prototype** | : | $AuBr_4K$ |
| **AFLOW prototype label** | : | AB4C_mP24_14_ab_4e_e |
| ***Strukturbericht* designation** | : | None |
| **Pearson symbol** | : | mP24 |
| **Space group number** | : | 14 |
| **Space group symbol** | : | $P2_1/c$ |
| **AFLOW prototype command** | : | `aflow --proto=AB4C_mP24_14_ab_4e_e` `--params=`$a, b/a, c/a, \beta, x_3, y_3, z_3, x_4, y_4, z_4, x_5, y_5, z_5, x_6, y_6, z_6, x_7, y_7, z_7$ |

- This hydrated form of this compound is $KAuBr_4 \cdot 2H_2O$ ($H4_{19}$).

**Simple Monoclinic primitive vectors:**

$$\mathbf{a}_1 = a\,\hat{\mathbf{x}}$$
$$\mathbf{a}_2 = b\,\hat{\mathbf{y}}$$
$$\mathbf{a}_3 = c\cos\beta\,\hat{\mathbf{x}} + c\sin\beta\,\hat{\mathbf{z}}$$

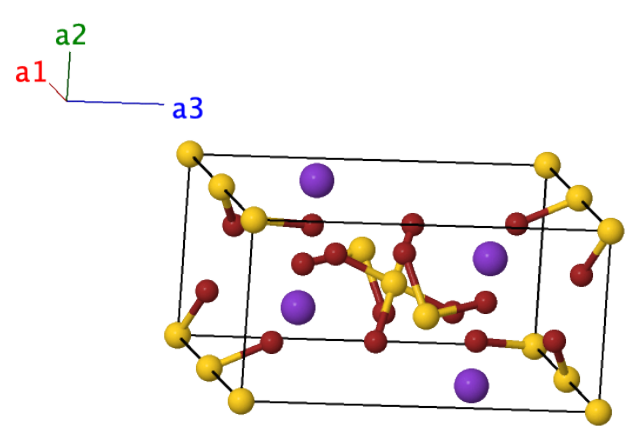

**Basis vectors:**

| | | Lattice Coordinates | | Cartesian Coordinates | Wyckoff Position | Atom Type |
|---|---|---|---|---|---|---|
| $\mathbf{B}_1$ | = | $0\,\mathbf{a}_1 + 0\,\mathbf{a}_2 + 0\,\mathbf{a}_3$ | = | $0\,\hat{\mathbf{x}} + 0\,\hat{\mathbf{y}} + 0\,\hat{\mathbf{z}}$ | (2$a$) | Au I |
| $\mathbf{B}_2$ | = | $\frac{1}{2}\,\mathbf{a}_2 + \frac{1}{2}\,\mathbf{a}_3$ | = | $\frac{1}{2}c\cos\beta\,\hat{\mathbf{x}} + \frac{1}{2}b\,\hat{\mathbf{y}} + \frac{1}{2}c\sin\beta\,\hat{\mathbf{z}}$ | (2$a$) | Au I |
| $\mathbf{B}_3$ | = | $\frac{1}{2}\,\mathbf{a}_1$ | = | $\frac{1}{2}a\,\hat{\mathbf{x}}$ | (2$b$) | Au II |
| $\mathbf{B}_4$ | = | $\frac{1}{2}\,\mathbf{a}_1 + \frac{1}{2}\,\mathbf{a}_2 + \frac{1}{2}\,\mathbf{a}_3$ | = | $\frac{1}{2}\left(a + c\cos\beta\right)\,\hat{\mathbf{x}} + \frac{1}{2}b\,\hat{\mathbf{y}} + \frac{1}{2}c\sin\beta\,\hat{\mathbf{z}}$ | (2$b$) | Au II |
| $\mathbf{B}_{5}$ | = | $x_{3}\,\mathbf{a}_{1} + y_{3}\,\mathbf{a}_{2} + z_{3}\,\mathbf{a}_{3}$ | = | $(x_{3}a + z_{3}c\cos\beta)\,\hat{\mathbf{x}} + y_{3}b\,\hat{\mathbf{y}} + z_{3}c\sin\beta\,\hat{\mathbf{z}}$ | (4$e$) | Br I |
| $\mathbf{B}_{6}$ | = | $-x_{3}\,\mathbf{a}_{1} + \left(\frac{1}{2} + y_{3}\right)\mathbf{a}_{2} + \left(\frac{1}{2} - z_{3}\right)\mathbf{a}_{3}$ | = | $\left(\frac{1}{2}c\cos\beta - x_{3}a - z_{3}c\cos\beta\right)\hat{\mathbf{x}} + \left(\frac{1}{2} + y_{3}\right)b\,\hat{\mathbf{y}} + \left(\frac{1}{2} - z_{3}\right)c\sin\beta\,\hat{\mathbf{z}}$ | (4$e$) | Br I |
| $\mathbf{B}_{7}$ | = | $-x_{3}\,\mathbf{a}_{1} - y_{3}\,\mathbf{a}_{2} - z_{3}\,\mathbf{a}_{3}$ | = | $(-x_{3}a - z_{3}c\cos\beta)\,\hat{\mathbf{x}} - y_{3}b\,\hat{\mathbf{y}} - z_{3}c\sin\beta\,\hat{\mathbf{z}}$ | (4$e$) | Br I |
| $\mathbf{B}_{8}$ | = | $x_{3}\,\mathbf{a}_{1} + \left(\frac{1}{2} - y_{3}\right)\mathbf{a}_{2} + \left(\frac{1}{2} + z_{3}\right)\mathbf{a}_{3}$ | = | $\left(\frac{1}{2}c\cos\beta + x_{3}a + z_{3}c\cos\beta\right)\hat{\mathbf{x}} + \left(\frac{1}{2} - y_{3}\right)b\,\hat{\mathbf{y}} + \left(\frac{1}{2} + z_{3}\right)c\sin\beta\,\hat{\mathbf{z}}$ | (4$e$) | Br I |
| $\mathbf{B}_{9}$ | = | $x_{4}\,\mathbf{a}_{1} + y_{4}\,\mathbf{a}_{2} + z_{4}\,\mathbf{a}_{3}$ | = | $(x_{4}a + z_{4}c\cos\beta)\,\hat{\mathbf{x}} + y_{4}b\,\hat{\mathbf{y}} + z_{4}c\sin\beta\,\hat{\mathbf{z}}$ | (4$e$) | Br II |
| $\mathbf{B}_{10}$ | = | $-x_{4}\,\mathbf{a}_{1} + \left(\frac{1}{2} + y_{4}\right)\mathbf{a}_{2} + \left(\frac{1}{2} - z_{4}\right)\mathbf{a}_{3}$ | = | $\left(\frac{1}{2}c\cos\beta - x_{4}a - z_{4}c\cos\beta\right)\hat{\mathbf{x}} + \left(\frac{1}{2} + y_{4}\right)b\,\hat{\mathbf{y}} + \left(\frac{1}{2} - z_{4}\right)c\sin\beta\,\hat{\mathbf{z}}$ | (4$e$) | Br II |
| $\mathbf{B}_{11}$ | = | $-x_{4}\,\mathbf{a}_{1} - y_{4}\,\mathbf{a}_{2} - z_{4}\,\mathbf{a}_{3}$ | = | $(-x_{4}a - z_{4}c\cos\beta)\,\hat{\mathbf{x}} - y_{4}b\,\hat{\mathbf{y}} - z_{4}c\sin\beta\,\hat{\mathbf{z}}$ | (4$e$) | Br II |
| $\mathbf{B}_{12}$ | = | $x_{4}\,\mathbf{a}_{1} + \left(\frac{1}{2} - y_{4}\right)\mathbf{a}_{2} + \left(\frac{1}{2} + z_{4}\right)\mathbf{a}_{3}$ | = | $\left(\frac{1}{2}c\cos\beta + x_{4}a + z_{4}c\cos\beta\right)\hat{\mathbf{x}} + \left(\frac{1}{2} - y_{4}\right)b\,\hat{\mathbf{y}} + \left(\frac{1}{2} + z_{4}\right)c\sin\beta\,\hat{\mathbf{z}}$ | (4$e$) | Br II |
| $\mathbf{B}_{13}$ | = | $x_{5}\,\mathbf{a}_{1} + y_{5}\,\mathbf{a}_{2} + z_{5}\,\mathbf{a}_{3}$ | = | $(x_{5}a + z_{5}c\cos\beta)\,\hat{\mathbf{x}} + y_{5}b\,\hat{\mathbf{y}} + z_{5}c\sin\beta\,\hat{\mathbf{z}}$ | (4$e$) | Br III |
| $\mathbf{B}_{14}$ | = | $-x_{5}\,\mathbf{a}_{1} + \left(\frac{1}{2} + y_{5}\right)\mathbf{a}_{2} + \left(\frac{1}{2} - z_{5}\right)\mathbf{a}_{3}$ | = | $\left(\frac{1}{2}c\cos\beta - x_{5}a - z_{5}c\cos\beta\right)\hat{\mathbf{x}} + \left(\frac{1}{2} + y_{5}\right)b\,\hat{\mathbf{y}} + \left(\frac{1}{2} - z_{5}\right)c\sin\beta\,\hat{\mathbf{z}}$ | (4$e$) | Br III |
| $\mathbf{B}_{15}$ | = | $-x_{5}\,\mathbf{a}_{1} - y_{5}\,\mathbf{a}_{2} - z_{5}\,\mathbf{a}_{3}$ | = | $(-x_{5}a - z_{5}c\cos\beta)\,\hat{\mathbf{x}} - y_{5}b\,\hat{\mathbf{y}} - z_{5}c\sin\beta\,\hat{\mathbf{z}}$ | (4$e$) | Br III |
| $\mathbf{B}_{16}$ | = | $x_{5}\,\mathbf{a}_{1} + \left(\frac{1}{2} - y_{5}\right)\mathbf{a}_{2} + \left(\frac{1}{2} + z_{5}\right)\mathbf{a}_{3}$ | = | $\left(\frac{1}{2}c\cos\beta + x_{5}a + z_{5}c\cos\beta\right)\hat{\mathbf{x}} + \left(\frac{1}{2} - y_{5}\right)b\,\hat{\mathbf{y}} + \left(\frac{1}{2} + z_{5}\right)c\sin\beta\,\hat{\mathbf{z}}$ | (4$e$) | Br III |
| $\mathbf{B}_{17}$ | = | $x_{6}\,\mathbf{a}_{1} + y_{6}\,\mathbf{a}_{2} + z_{6}\,\mathbf{a}_{3}$ | = | $(x_{6}a + z_{6}c\cos\beta)\,\hat{\mathbf{x}} + y_{6}b\,\hat{\mathbf{y}} + z_{6}c\sin\beta\,\hat{\mathbf{z}}$ | (4$e$) | Br IV |
| $\mathbf{B}_{18}$ | = | $-x_{6}\,\mathbf{a}_{1} + \left(\frac{1}{2} + y_{6}\right)\mathbf{a}_{2} + \left(\frac{1}{2} - z_{6}\right)\mathbf{a}_{3}$ | = | $\left(\frac{1}{2}c\cos\beta - x_{6}a - z_{6}c\cos\beta\right)\hat{\mathbf{x}} + \left(\frac{1}{2} + y_{6}\right)b\,\hat{\mathbf{y}} + \left(\frac{1}{2} - z_{6}\right)c\sin\beta\,\hat{\mathbf{z}}$ | (4$e$) | Br IV |
| $\mathbf{B}_{19}$ | = | $-x_{6}\,\mathbf{a}_{1} - y_{6}\,\mathbf{a}_{2} - z_{6}\,\mathbf{a}_{3}$ | = | $(-x_{6}a - z_{6}c\cos\beta)\,\hat{\mathbf{x}} - y_{6}b\,\hat{\mathbf{y}} - z_{6}c\sin\beta\,\hat{\mathbf{z}}$ | (4$e$) | Br IV |
| $\mathbf{B}_{20}$ | = | $x_{6}\,\mathbf{a}_{1} + \left(\frac{1}{2} - y_{6}\right)\mathbf{a}_{2} + \left(\frac{1}{2} + z_{6}\right)\mathbf{a}_{3}$ | = | $\left(\frac{1}{2}c\cos\beta + x_{6}a + z_{6}c\cos\beta\right)\hat{\mathbf{x}} + \left(\frac{1}{2} - y_{6}\right)b\,\hat{\mathbf{y}} + \left(\frac{1}{2} + z_{6}\right)c\sin\beta\,\hat{\mathbf{z}}$ | (4$e$) | Br IV |
| $\mathbf{B}_{21}$ | = | $x_{7}\,\mathbf{a}_{1} + y_{7}\,\mathbf{a}_{2} + z_{7}\,\mathbf{a}_{3}$ | = | $(x_{7}a + z_{7}c\cos\beta)\,\hat{\mathbf{x}} + y_{7}b\,\hat{\mathbf{y}} + z_{7}c\sin\beta\,\hat{\mathbf{z}}$ | (4$e$) | K |
| $\mathbf{B}_{22}$ | = | $-x_{7}\,\mathbf{a}_{1} + \left(\frac{1}{2} + y_{7}\right)\mathbf{a}_{2} + \left(\frac{1}{2} - z_{7}\right)\mathbf{a}_{3}$ | = | $\left(\frac{1}{2}c\cos\beta - x_{7}a - z_{7}c\cos\beta\right)\hat{\mathbf{x}} + \left(\frac{1}{2} + y_{7}\right)b\,\hat{\mathbf{y}} + \left(\frac{1}{2} - z_{7}\right)c\sin\beta\,\hat{\mathbf{z}}$ | (4$e$) | K |
| $\mathbf{B}_{23}$ | = | $-x_{7}\,\mathbf{a}_{1} - y_{7}\,\mathbf{a}_{2} - z_{7}\,\mathbf{a}_{3}$ | = | $(-x_{7}a - z_{7}c\cos\beta)\,\hat{\mathbf{x}} - y_{7}b\,\hat{\mathbf{y}} - z_{7}c\sin\beta\,\hat{\mathbf{z}}$ | (4$e$) | K |
| $\mathbf{B}_{24}$ | = | $x_{7}\,\mathbf{a}_{1} + \left(\frac{1}{2} - y_{7}\right)\mathbf{a}_{2} + \left(\frac{1}{2} + z_{7}\right)\mathbf{a}_{3}$ | = | $\left(\frac{1}{2}c\cos\beta + x_{7}a + z_{7}c\cos\beta\right)\hat{\mathbf{x}} + \left(\frac{1}{2} - y_{7}\right)b\,\hat{\mathbf{y}} + \left(\frac{1}{2} + z_{7}\right)c\sin\beta\,\hat{\mathbf{z}}$ | (4$e$) | K |

**Found in:**

---

**Geometry files:**
- CIF: pp. 1551
- POSCAR: pp. 1552

# Ammonium Persulfate [$(NH_4)_2S_2O_8$, $K4_1$] Structure: AB4C_mP24_14_e_4e_e

http://aflow.org/prototype-encyclopedia/AB4C_mP24_14_e_4e_e.K41

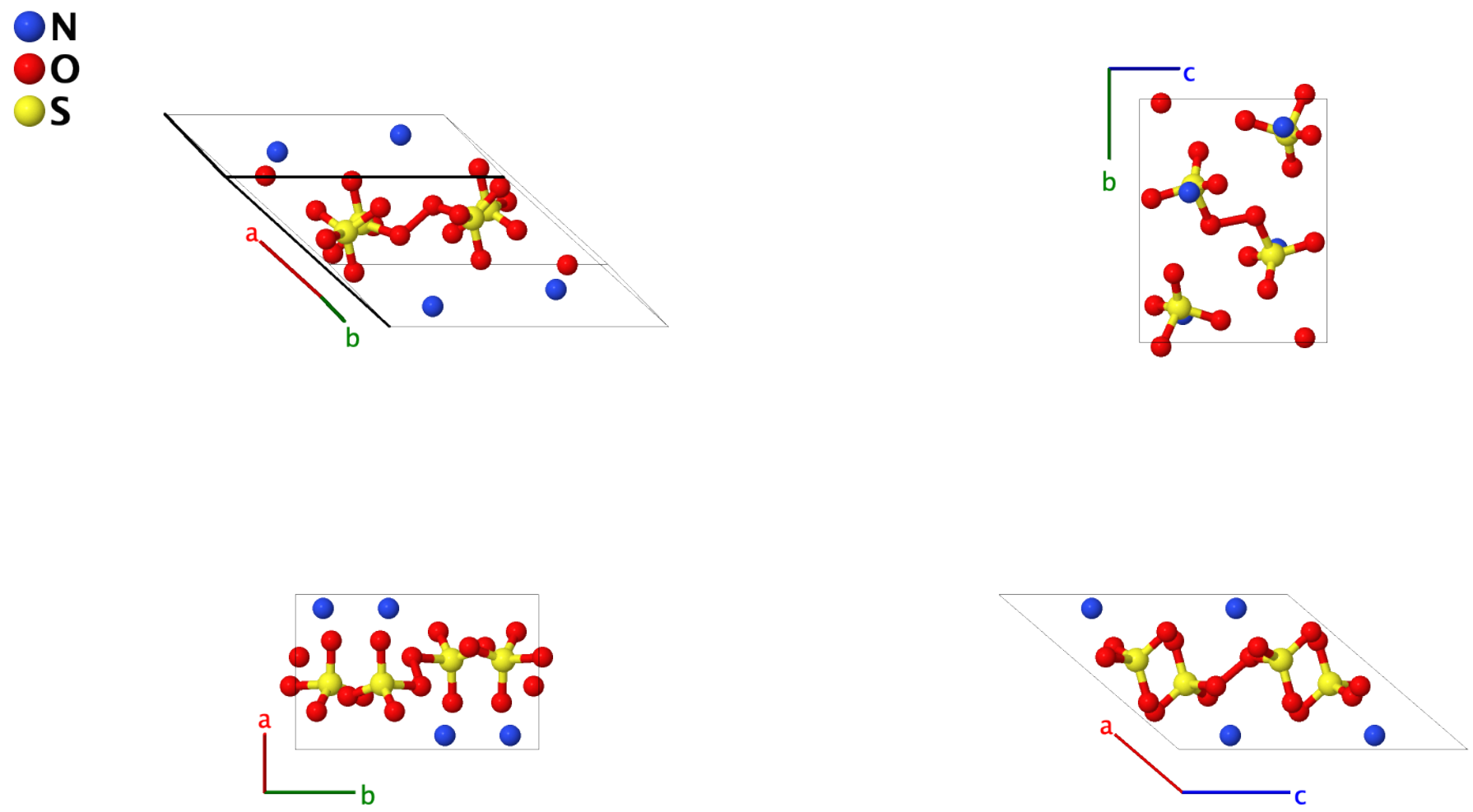

| | | |
|---|---|---|
| **Prototype** | : | $(NH_4)O_4S$ |
| **AFLOW prototype label** | : | AB4C_mP24_14_e_4e_e |
| ***Strukturbericht* designation** | : | $K4_1$ |
| **Pearson symbol** | : | mP24 |
| **Space group number** | : | 14 |
| **Space group symbol** | : | $P2_1/c$ |
| **AFLOW prototype command** | : | `aflow --proto=AB4C_mP24_14_e_4e_e` `--params=`$a, b/a, c/a, \beta, x_1, y_1, z_1, x_2, y_2, z_2, x_3, y_3, z_3, x_4, y_4, z_4, x_5, y_5, z_5, x_6, y_6, z_6$ |

**Other compounds with this structure**

- $CaSO_4$

- This structure has the same AFLOW label as monasite, $LaPO_4$. The structures are generated by the same symmetry operations with different sets of parameters (`--params`) specified in their corresponding CIF files.

**Simple Monoclinic primitive vectors:**

$$\mathbf{a}_1 = a\,\hat{\mathbf{x}}$$
$$\mathbf{a}_2 = b\,\hat{\mathbf{y}}$$
$$\mathbf{a}_3 = c\cos\beta\,\hat{\mathbf{x}} + c\sin\beta\,\hat{\mathbf{z}}$$

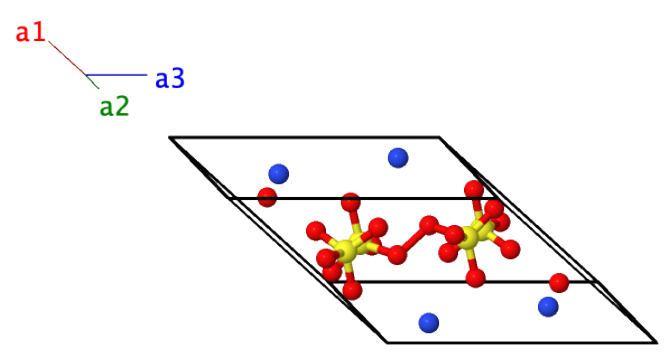

**Basis vectors:**

| Lattice Coordinates | Cartesian Coordinates | Wyckoff Position | Atom Type |
|---|---|---|---|

$$\mathbf{B}_1 = x_1\,\mathbf{a}_1 + y_1\,\mathbf{a}_2 + z_1\,\mathbf{a}_3 = (x_1 a + z_1 c\cos\beta)\,\hat{\mathbf{x}} + y_1 b\,\hat{\mathbf{y}} + z_1 c\sin\beta\,\hat{\mathbf{z}} \quad (4e) \quad \text{NH}_4$$

$$\mathbf{B}_2 = -x_1\,\mathbf{a}_1 + \left(\tfrac{1}{2} + y_1\right)\mathbf{a}_2 + \left(\tfrac{1}{2} - z_1\right)\mathbf{a}_3 = \left(\tfrac{1}{2}c\cos\beta - x_1 a - z_1 c\cos\beta\right)\hat{\mathbf{x}} + \left(\tfrac{1}{2} + y_1\right)b\,\hat{\mathbf{y}} + \left(\tfrac{1}{2} - z_1\right)c\sin\beta\,\hat{\mathbf{z}} \quad (4e) \quad \text{NH}_4$$

$$\mathbf{B}_3 = -x_1\,\mathbf{a}_1 - y_1\,\mathbf{a}_2 - z_1\,\mathbf{a}_3 = (-x_1 a - z_1 c\cos\beta)\,\hat{\mathbf{x}} - y_1 b\,\hat{\mathbf{y}} - z_1 c\sin\beta\,\hat{\mathbf{z}} \quad (4e) \quad \text{NH}_4$$

$$\mathbf{B}_4 = x_1\,\mathbf{a}_1 + \left(\tfrac{1}{2} - y_1\right)\mathbf{a}_2 + \left(\tfrac{1}{2} + z_1\right)\mathbf{a}_3 = \left(\tfrac{1}{2}c\cos\beta + x_1 a + z_1 c\cos\beta\right)\hat{\mathbf{x}} + \left(\tfrac{1}{2} - y_1\right)b\,\hat{\mathbf{y}} + \left(\tfrac{1}{2} + z_1\right)c\sin\beta\,\hat{\mathbf{z}} \quad (4e) \quad \text{NH}_4$$

$$\mathbf{B}_5 = x_2\,\mathbf{a}_1 + y_2\,\mathbf{a}_2 + z_2\,\mathbf{a}_3 = (x_2 a + z_2 c\cos\beta)\,\hat{\mathbf{x}} + y_2 b\,\hat{\mathbf{y}} + z_2 c\sin\beta\,\hat{\mathbf{z}} \quad (4e) \quad \text{O I}$$

$$\mathbf{B}_6 = -x_2\,\mathbf{a}_1 + \left(\tfrac{1}{2} + y_2\right)\mathbf{a}_2 + \left(\tfrac{1}{2} - z_2\right)\mathbf{a}_3 = \left(\tfrac{1}{2}c\cos\beta - x_2 a - z_2 c\cos\beta\right)\hat{\mathbf{x}} + \left(\tfrac{1}{2} + y_2\right)b\,\hat{\mathbf{y}} + \left(\tfrac{1}{2} - z_2\right)c\sin\beta\,\hat{\mathbf{z}} \quad (4e) \quad \text{O I}$$

$$\mathbf{B}_7 = -x_2\,\mathbf{a}_1 - y_2\,\mathbf{a}_2 - z_2\,\mathbf{a}_3 = (-x_2 a - z_2 c\cos\beta)\,\hat{\mathbf{x}} - y_2 b\,\hat{\mathbf{y}} - z_2 c\sin\beta\,\hat{\mathbf{z}} \quad (4e) \quad \text{O I}$$

$$\mathbf{B}_8 = x_2\,\mathbf{a}_1 + \left(\tfrac{1}{2} - y_2\right)\mathbf{a}_2 + \left(\tfrac{1}{2} + z_2\right)\mathbf{a}_3 = \left(\tfrac{1}{2}c\cos\beta + x_2 a + z_2 c\cos\beta\right)\hat{\mathbf{x}} + \left(\tfrac{1}{2} - y_2\right)b\,\hat{\mathbf{y}} + \left(\tfrac{1}{2} + z_2\right)c\sin\beta\,\hat{\mathbf{z}} \quad (4e) \quad \text{O I}$$

$$\mathbf{B}_9 = x_3\,\mathbf{a}_1 + y_3\,\mathbf{a}_2 + z_3\,\mathbf{a}_3 = (x_3 a + z_3 c\cos\beta)\,\hat{\mathbf{x}} + y_3 b\,\hat{\mathbf{y}} + z_3 c\sin\beta\,\hat{\mathbf{z}} \quad (4e) \quad \text{O II}$$

$$\mathbf{B}_{10} = -x_3\,\mathbf{a}_1 + \left(\tfrac{1}{2} + y_3\right)\mathbf{a}_2 + \left(\tfrac{1}{2} - z_3\right)\mathbf{a}_3 = \left(\tfrac{1}{2}c\cos\beta - x_3 a - z_3 c\cos\beta\right)\hat{\mathbf{x}} + \left(\tfrac{1}{2} + y_3\right)b\,\hat{\mathbf{y}} + \left(\tfrac{1}{2} - z_3\right)c\sin\beta\,\hat{\mathbf{z}} \quad (4e) \quad \text{O II}$$

$$\mathbf{B}_{11} = -x_3\,\mathbf{a}_1 - y_3\,\mathbf{a}_2 - z_3\,\mathbf{a}_3 = (-x_3 a - z_3 c\cos\beta)\,\hat{\mathbf{x}} - y_3 b\,\hat{\mathbf{y}} - z_3 c\sin\beta\,\hat{\mathbf{z}} \quad (4e) \quad \text{O II}$$

$$\mathbf{B}_{12} = x_3\,\mathbf{a}_1 + \left(\tfrac{1}{2} - y_3\right)\mathbf{a}_2 + \left(\tfrac{1}{2} + z_3\right)\mathbf{a}_3 = \left(\tfrac{1}{2}c\cos\beta + x_3 a + z_3 c\cos\beta\right)\hat{\mathbf{x}} + \left(\tfrac{1}{2} - y_3\right)b\,\hat{\mathbf{y}} + \left(\tfrac{1}{2} + z_3\right)c\sin\beta\,\hat{\mathbf{z}} \quad (4e) \quad \text{O II}$$

$$\mathbf{B}_{13} = x_4\,\mathbf{a}_1 + y_4\,\mathbf{a}_2 + z_4\,\mathbf{a}_3 = (x_4 a + z_4 c\cos\beta)\,\hat{\mathbf{x}} + y_4 b\,\hat{\mathbf{y}} + z_4 c\sin\beta\,\hat{\mathbf{z}} \quad (4e) \quad \text{O III}$$

$$\mathbf{B}_{14} = -x_4\,\mathbf{a}_1 + \left(\tfrac{1}{2} + y_4\right)\mathbf{a}_2 + \left(\tfrac{1}{2} - z_4\right)\mathbf{a}_3 = \left(\tfrac{1}{2}c\cos\beta - x_4 a - z_4 c\cos\beta\right)\hat{\mathbf{x}} + \left(\tfrac{1}{2} + y_4\right)b\,\hat{\mathbf{y}} + \left(\tfrac{1}{2} - z_4\right)c\sin\beta\,\hat{\mathbf{z}} \quad (4e) \quad \text{O III}$$

$$\mathbf{B}_{15} = -x_4\,\mathbf{a}_1 - y_4\,\mathbf{a}_2 - z_4\,\mathbf{a}_3 = (-x_4 a - z_4 c\cos\beta)\,\hat{\mathbf{x}} - y_4 b\,\hat{\mathbf{y}} - z_4 c\sin\beta\,\hat{\mathbf{z}} \quad (4e) \quad \text{O III}$$

$$\mathbf{B}_{16} = x_4\,\mathbf{a}_1 + \left(\tfrac{1}{2} - y_4\right)\mathbf{a}_2 + \left(\tfrac{1}{2} + z_4\right)\mathbf{a}_3 = \left(\tfrac{1}{2}c\cos\beta + x_4 a + z_4 c\cos\beta\right)\hat{\mathbf{x}} + \left(\tfrac{1}{2} - y_4\right)b\,\hat{\mathbf{y}} + \left(\tfrac{1}{2} + z_4\right)c\sin\beta\,\hat{\mathbf{z}} \quad (4e) \quad \text{O III}$$

$$\mathbf{B}_{17} = x_5\,\mathbf{a}_1 + y_5\,\mathbf{a}_2 + z_5\,\mathbf{a}_3 = (x_5 a + z_5 c\cos\beta)\,\hat{\mathbf{x}} + y_5 b\,\hat{\mathbf{y}} + z_5 c\sin\beta\,\hat{\mathbf{z}} \quad (4e) \quad \text{O IV}$$

$$\mathbf{B}_{18} = -x_5\,\mathbf{a}_1 + \left(\tfrac{1}{2} + y_5\right)\mathbf{a}_2 + \left(\tfrac{1}{2} - z_5\right)\mathbf{a}_3 = \left(\tfrac{1}{2}c\cos\beta - x_5 a - z_5 c\cos\beta\right)\hat{\mathbf{x}} + \left(\tfrac{1}{2} + y_5\right)b\,\hat{\mathbf{y}} + \left(\tfrac{1}{2} - z_5\right)c\sin\beta\,\hat{\mathbf{z}} \quad (4e) \quad \text{O IV}$$

$$\mathbf{B}_{19} = -x_5\,\mathbf{a}_1 - y_5\,\mathbf{a}_2 - z_5\,\mathbf{a}_3 = (-x_5 a - z_5 c\cos\beta)\,\hat{\mathbf{x}} - y_5 b\,\hat{\mathbf{y}} - z_5 c\sin\beta\,\hat{\mathbf{z}} \quad (4e) \quad \text{O IV}$$

$$\mathbf{B}_{20} = x_5\,\mathbf{a}_1 + \left(\tfrac{1}{2} - y_5\right)\mathbf{a}_2 + \left(\tfrac{1}{2} + z_5\right)\mathbf{a}_3 = \left(\tfrac{1}{2}c\cos\beta + x_5 a + z_5 c\cos\beta\right)\hat{\mathbf{x}} + \left(\tfrac{1}{2} - y_5\right)b\,\hat{\mathbf{y}} + \left(\tfrac{1}{2} + z_5\right)c\sin\beta\,\hat{\mathbf{z}} \quad (4e) \quad \text{O IV}$$

$$\mathbf{B}_{21} = x_6\,\mathbf{a}_1 + y_6\,\mathbf{a}_2 + z_6\,\mathbf{a}_3 = (x_6 a + z_6 c\cos\beta)\,\hat{\mathbf{x}} + y_6 b\,\hat{\mathbf{y}} + z_6 c\sin\beta\,\hat{\mathbf{z}} \quad (4e) \quad \text{S}$$

$$\mathbf{B}_{22} = -x_6\,\mathbf{a}_1 + \left(\tfrac{1}{2} + y_6\right)\mathbf{a}_2 + \left(\tfrac{1}{2} - z_6\right)\mathbf{a}_3 = \left(\tfrac{1}{2}c\cos\beta - x_6 a - z_6 c\cos\beta\right)\hat{\mathbf{x}} + \left(\tfrac{1}{2} + y_6\right)b\,\hat{\mathbf{y}} + \left(\tfrac{1}{2} - z_6\right)c\sin\beta\,\hat{\mathbf{z}} \quad (4e) \quad \text{S}$$

| | | | | | | |
|---|---|---|---|---|---|---|
| $\mathbf{B}_{23}$ | = | $-x_6\,\mathbf{a}_1 - y_6\,\mathbf{a}_2 - z_6\,\mathbf{a}_3$ | = | $(-x_6 a - z_6 c\cos\beta)\,\hat{\mathbf{x}} - y_6 b\,\hat{\mathbf{y}} - z_6 c\sin\beta\,\hat{\mathbf{z}}$ | (4$e$) | S |
| $\mathbf{B}_{24}$ | = | $x_6\,\mathbf{a}_1 + \left(\frac{1}{2} - y_6\right)\mathbf{a}_2 + \left(\frac{1}{2} + z_6\right)\mathbf{a}_3$ | = | $\left(\frac{1}{2}c\cos\beta + x_6 a + z_6 c\cos\beta\right)\hat{\mathbf{x}} + \left(\frac{1}{2} - y_6\right)b\,\hat{\mathbf{y}} + \left(\frac{1}{2} + z_6\right)c\sin\beta\,\hat{\mathbf{z}}$ | (4$e$) | S |

**Geometry files:**

- CIF: pp. 1552
- POSCAR: pp. 1552

# Monasite ($LaPO_4$) Structure: AB4C_mP24_14_e_4e_e

http://aflow.org/prototype-encyclopedia/AB4C_mP24_14_e_4e_e

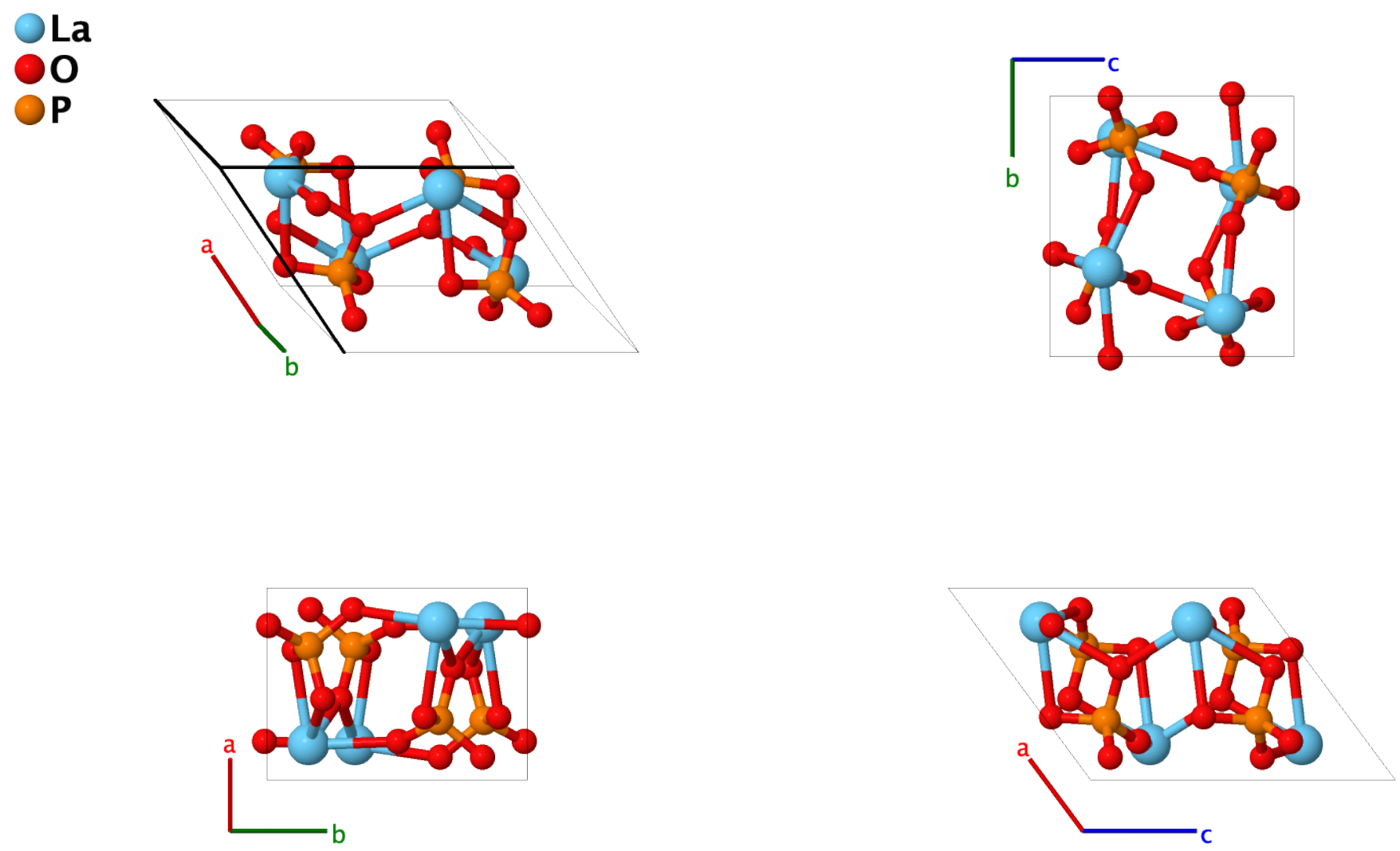

| | | |
|---|---|---|
| **Prototype** | : | $LaO_4P$ |
| **AFLOW prototype label** | : | AB4C_mP24_14_e_4e_e |
| ***Strukturbericht* designation** | : | None |
| **Pearson symbol** | : | mP24 |
| **Space group number** | : | 14 |
| **Space group symbol** | : | $P2_1/c$ |
| **AFLOW prototype command** | : | `aflow --proto=AB4C_mP24_14_e_4e_e` `--params=`$a, b/a, c/a, \beta, x_1, y_1, z_1, x_2, y_2, z_2, x_3, y_3, z_3, x_4, y_4, z_4, x_5, y_5, z_5, x_6, y_6, z_6$ |

**Other compounds with this structure**

- $CePO_4$ (monazite-Ce), $PrPO_4$ (monazite-Pr), $NdPO_4$ (monazite-Nd), $SmPO_4$ (monazite-Sm), $EuPO_4$ (monazite-Eu), $LaAsO_4$ (gasparite-La), and $CeAsO_4$ (gasparite-Ce)

- Monasites (*RE*-$PO_4$) and gasparites (*RE*-$AsO_4$) can have a mixture of rare earth elements in the *RE* slot of the formula. Technically these minerals are called monasite-(X) and gasparite-(X), where "X" is the predominant rare earth element in the sample. All of the structures are similar, but we must pick one as the prototype, so we take the first entry in (Ni, 1995) to define the class.

- (Ni, 1995) gives the structural data in the $P2_1/n$ setting of space group #14. We used FINDSYM to change this to our standard $P2_1/c$ setting. This involves a change of primitive vectors as well as a rotation of the crystal.

- This structure has the same AFLOW label as $NH_4SO_4$, $K4_1$. The structures are generated by the same symmetry operations with different sets of parameters (`--params`) specified in their corresponding CIF files.

**Simple Monoclinic primitive vectors:**

$$
\begin{aligned}
\mathbf{a}_1 &= a\,\hat{\mathbf{x}} \\
\mathbf{a}_2 &= b\,\hat{\mathbf{y}} \\
\mathbf{a}_3 &= c\cos\beta\,\hat{\mathbf{x}} + c\sin\beta\,\hat{\mathbf{z}}
\end{aligned}
$$

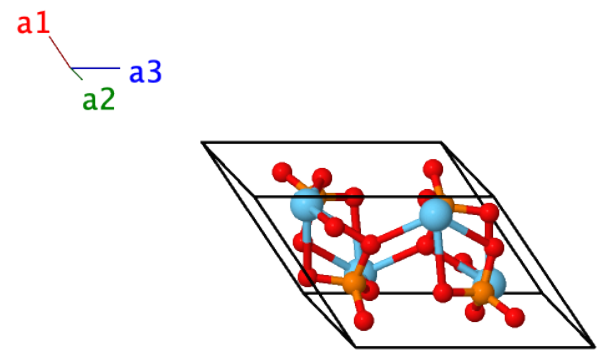

**Basis vectors:**

| | | Lattice Coordinates | | Cartesian Coordinates | Wyckoff Position | Atom Type |
|---|---|---|---|---|---|---|
| $\mathbf{B}_1$ | = | $x_1\,\mathbf{a}_1 + y_1\,\mathbf{a}_2 + z_1\,\mathbf{a}_3$ | = | $(x_1 a + z_1 c\cos\beta)\,\hat{\mathbf{x}} + y_1 b\,\hat{\mathbf{y}} + z_1 c\sin\beta\,\hat{\mathbf{z}}$ | (4$e$) | La |
| $\mathbf{B}_2$ | = | $-x_1\,\mathbf{a}_1 + \left(\frac{1}{2} + y_1\right)\mathbf{a}_2 + \left(\frac{1}{2} - z_1\right)\mathbf{a}_3$ | = | $\left(\frac{1}{2}c\cos\beta - x_1 a - z_1 c\cos\beta\right)\hat{\mathbf{x}} + \left(\frac{1}{2} + y_1\right)b\,\hat{\mathbf{y}} + \left(\frac{1}{2} - z_1\right)c\sin\beta\,\hat{\mathbf{z}}$ | (4$e$) | La |
| $\mathbf{B}_3$ | = | $-x_1\,\mathbf{a}_1 - y_1\,\mathbf{a}_2 - z_1\,\mathbf{a}_3$ | = | $(-x_1 a - z_1 c\cos\beta)\,\hat{\mathbf{x}} - y_1 b\,\hat{\mathbf{y}} - z_1 c\sin\beta\,\hat{\mathbf{z}}$ | (4$e$) | La |
| $\mathbf{B}_4$ | = | $x_1\,\mathbf{a}_1 + \left(\frac{1}{2} - y_1\right)\mathbf{a}_2 + \left(\frac{1}{2} + z_1\right)\mathbf{a}_3$ | = | $\left(\frac{1}{2}c\cos\beta + x_1 a + z_1 c\cos\beta\right)\hat{\mathbf{x}} + \left(\frac{1}{2} - y_1\right)b\,\hat{\mathbf{y}} + \left(\frac{1}{2} + z_1\right)c\sin\beta\,\hat{\mathbf{z}}$ | (4$e$) | La |
| $\mathbf{B}_5$ | = | $x_2\,\mathbf{a}_1 + y_2\,\mathbf{a}_2 + z_2\,\mathbf{a}_3$ | = | $(x_2 a + z_2 c\cos\beta)\,\hat{\mathbf{x}} + y_2 b\,\hat{\mathbf{y}} + z_2 c\sin\beta\,\hat{\mathbf{z}}$ | (4$e$) | O I |
| $\mathbf{B}_6$ | = | $-x_2\,\mathbf{a}_1 + \left(\frac{1}{2} + y_2\right)\mathbf{a}_2 + \left(\frac{1}{2} - z_2\right)\mathbf{a}_3$ | = | $\left(\frac{1}{2}c\cos\beta - x_2 a - z_2 c\cos\beta\right)\hat{\mathbf{x}} + \left(\frac{1}{2} + y_2\right)b\,\hat{\mathbf{y}} + \left(\frac{1}{2} - z_2\right)c\sin\beta\,\hat{\mathbf{z}}$ | (4$e$) | O I |
| $\mathbf{B}_7$ | = | $-x_2\,\mathbf{a}_1 - y_2\,\mathbf{a}_2 - z_2\,\mathbf{a}_3$ | = | $(-x_2 a - z_2 c\cos\beta)\,\hat{\mathbf{x}} - y_2 b\,\hat{\mathbf{y}} - z_2 c\sin\beta\,\hat{\mathbf{z}}$ | (4$e$) | O I |
| $\mathbf{B}_8$ | = | $x_2\,\mathbf{a}_1 + \left(\frac{1}{2} - y_2\right)\mathbf{a}_2 + \left(\frac{1}{2} + z_2\right)\mathbf{a}_3$ | = | $\left(\frac{1}{2}c\cos\beta + x_2 a + z_2 c\cos\beta\right)\hat{\mathbf{x}} + \left(\frac{1}{2} - y_2\right)b\,\hat{\mathbf{y}} + \left(\frac{1}{2} + z_2\right)c\sin\beta\,\hat{\mathbf{z}}$ | (4$e$) | O I |
| $\mathbf{B}_9$ | = | $x_3\,\mathbf{a}_1 + y_3\,\mathbf{a}_2 + z_3\,\mathbf{a}_3$ | = | $(x_3 a + z_3 c\cos\beta)\,\hat{\mathbf{x}} + y_3 b\,\hat{\mathbf{y}} + z_3 c\sin\beta\,\hat{\mathbf{z}}$ | (4$e$) | O II |
| $\mathbf{B}_{10}$ | = | $-x_3\,\mathbf{a}_1 + \left(\frac{1}{2} + y_3\right)\mathbf{a}_2 + \left(\frac{1}{2} - z_3\right)\mathbf{a}_3$ | = | $\left(\frac{1}{2}c\cos\beta - x_3 a - z_3 c\cos\beta\right)\hat{\mathbf{x}} + \left(\frac{1}{2} + y_3\right)b\,\hat{\mathbf{y}} + \left(\frac{1}{2} - z_3\right)c\sin\beta\,\hat{\mathbf{z}}$ | (4$e$) | O II |
| $\mathbf{B}_{11}$ | = | $-x_3\,\mathbf{a}_1 - y_3\,\mathbf{a}_2 - z_3\,\mathbf{a}_3$ | = | $(-x_3 a - z_3 c\cos\beta)\,\hat{\mathbf{x}} - y_3 b\,\hat{\mathbf{y}} - z_3 c\sin\beta\,\hat{\mathbf{z}}$ | (4$e$) | O II |
| $\mathbf{B}_{12}$ | = | $x_3\,\mathbf{a}_1 + \left(\frac{1}{2} - y_3\right)\mathbf{a}_2 + \left(\frac{1}{2} + z_3\right)\mathbf{a}_3$ | = | $\left(\frac{1}{2}c\cos\beta + x_3 a + z_3 c\cos\beta\right)\hat{\mathbf{x}} + \left(\frac{1}{2} - y_3\right)b\,\hat{\mathbf{y}} + \left(\frac{1}{2} + z_3\right)c\sin\beta\,\hat{\mathbf{z}}$ | (4$e$) | O II |
| $\mathbf{B}_{13}$ | = | $x_4\,\mathbf{a}_1 + y_4\,\mathbf{a}_2 + z_4\,\mathbf{a}_3$ | = | $(x_4 a + z_4 c\cos\beta)\,\hat{\mathbf{x}} + y_4 b\,\hat{\mathbf{y}} + z_4 c\sin\beta\,\hat{\mathbf{z}}$ | (4$e$) | O III |
| $\mathbf{B}_{14}$ | = | $-x_4\,\mathbf{a}_1 + \left(\frac{1}{2} + y_4\right)\mathbf{a}_2 + \left(\frac{1}{2} - z_4\right)\mathbf{a}_3$ | = | $\left(\frac{1}{2}c\cos\beta - x_4 a - z_4 c\cos\beta\right)\hat{\mathbf{x}} + \left(\frac{1}{2} + y_4\right)b\,\hat{\mathbf{y}} + \left(\frac{1}{2} - z_4\right)c\sin\beta\,\hat{\mathbf{z}}$ | (4$e$) | O III |
| $\mathbf{B}_{15}$ | = | $-x_4\,\mathbf{a}_1 - y_4\,\mathbf{a}_2 - z_4\,\mathbf{a}_3$ | = | $(-x_4 a - z_4 c\cos\beta)\,\hat{\mathbf{x}} - y_4 b\,\hat{\mathbf{y}} - z_4 c\sin\beta\,\hat{\mathbf{z}}$ | (4$e$) | O III |
| $\mathbf{B}_{16}$ | = | $x_4\,\mathbf{a}_1 + \left(\frac{1}{2} - y_4\right)\mathbf{a}_2 + \left(\frac{1}{2} + z_4\right)\mathbf{a}_3$ | = | $\left(\frac{1}{2}c\cos\beta + x_4 a + z_4 c\cos\beta\right)\hat{\mathbf{x}} + \left(\frac{1}{2} - y_4\right)b\,\hat{\mathbf{y}} + \left(\frac{1}{2} + z_4\right)c\sin\beta\,\hat{\mathbf{z}}$ | (4$e$) | O III |
| $\mathbf{B}_{17}$ | = | $x_5\,\mathbf{a}_1 + y_5\,\mathbf{a}_2 + z_5\,\mathbf{a}_3$ | = | $(x_5 a + z_5 c\cos\beta)\,\hat{\mathbf{x}} + y_5 b\,\hat{\mathbf{y}} + z_5 c\sin\beta\,\hat{\mathbf{z}}$ | (4$e$) | O IV |
| $\mathbf{B}_{18}$ | = | $-x_5\,\mathbf{a}_1 + \left(\frac{1}{2}+y_5\right)\mathbf{a}_2 + \left(\frac{1}{2}-z_5\right)\mathbf{a}_3$ | = | $\left(\frac{1}{2}c\cos\beta - x_5 a - z_5 c\cos\beta\right)\hat{\mathbf{x}} + \left(\frac{1}{2}+y_5\right)b\,\hat{\mathbf{y}} + \left(\frac{1}{2}-z_5\right)c\sin\beta\,\hat{\mathbf{z}}$ | (4$e$) | O IV |
| $\mathbf{B}_{19}$ | = | $-x_5\,\mathbf{a}_1 - y_5\,\mathbf{a}_2 - z_5\,\mathbf{a}_3$ | = | $(-x_5 a - z_5 c\cos\beta)\,\hat{\mathbf{x}} - y_5 b\,\hat{\mathbf{y}} - z_5 c\sin\beta\,\hat{\mathbf{z}}$ | (4$e$) | O IV |
| $\mathbf{B}_{20}$ | = | $x_5\,\mathbf{a}_1 + \left(\frac{1}{2}-y_5\right)\mathbf{a}_2 + \left(\frac{1}{2}+z_5\right)\mathbf{a}_3$ | = | $\left(\frac{1}{2}c\cos\beta + x_5 a + z_5 c\cos\beta\right)\hat{\mathbf{x}} + \left(\frac{1}{2}-y_5\right)b\,\hat{\mathbf{y}} + \left(\frac{1}{2}+z_5\right)c\sin\beta\,\hat{\mathbf{z}}$ | (4$e$) | O IV |
| $\mathbf{B}_{21}$ | = | $x_6\,\mathbf{a}_1 + y_6\,\mathbf{a}_2 + z_6\,\mathbf{a}_3$ | = | $(x_6 a + z_6 c\cos\beta)\,\hat{\mathbf{x}} + y_6 b\,\hat{\mathbf{y}} + z_6 c\sin\beta\,\hat{\mathbf{z}}$ | (4$e$) | P |
| $\mathbf{B}_{22}$ | = | $-x_6\,\mathbf{a}_1 + \left(\frac{1}{2}+y_6\right)\mathbf{a}_2 + \left(\frac{1}{2}-z_6\right)\mathbf{a}_3$ | = | $\left(\frac{1}{2}c\cos\beta - x_6 a - z_6 c\cos\beta\right)\hat{\mathbf{x}} + \left(\frac{1}{2}+y_6\right)b\,\hat{\mathbf{y}} + \left(\frac{1}{2}-z_6\right)c\sin\beta\,\hat{\mathbf{z}}$ | (4$e$) | P |
| $\mathbf{B}_{23}$ | = | $-x_6\,\mathbf{a}_1 - y_6\,\mathbf{a}_2 - z_6\,\mathbf{a}_3$ | = | $(-x_6 a - z_6 c\cos\beta)\,\hat{\mathbf{x}} - y_6 b\,\hat{\mathbf{y}} - z_6 c\sin\beta\,\hat{\mathbf{z}}$ | (4$e$) | P |
| $\mathbf{B}_{24}$ | = | $x_6\,\mathbf{a}_1 + \left(\frac{1}{2}-y_6\right)\mathbf{a}_2 + \left(\frac{1}{2}+z_6\right)\mathbf{a}_3$ | = | $\left(\frac{1}{2}c\cos\beta + x_6 a + z_6 c\cos\beta\right)\hat{\mathbf{x}} + \left(\frac{1}{2}-y_6\right)b\,\hat{\mathbf{y}} + \left(\frac{1}{2}+z_6\right)c\sin\beta\,\hat{\mathbf{z}}$ | (4$e$) | P |

---

**Found in:**

---

**Geometry files:**

- CIF: pp. 1552
- POSCAR: pp. 1553

# $Sr_2MnTeO_6$ Structure: AB6C2D_mP20_14_a_3e_e_d

http://aflow.org/prototype-encyclopedia/AB6C2D_mP20_14_a_3e_e_d

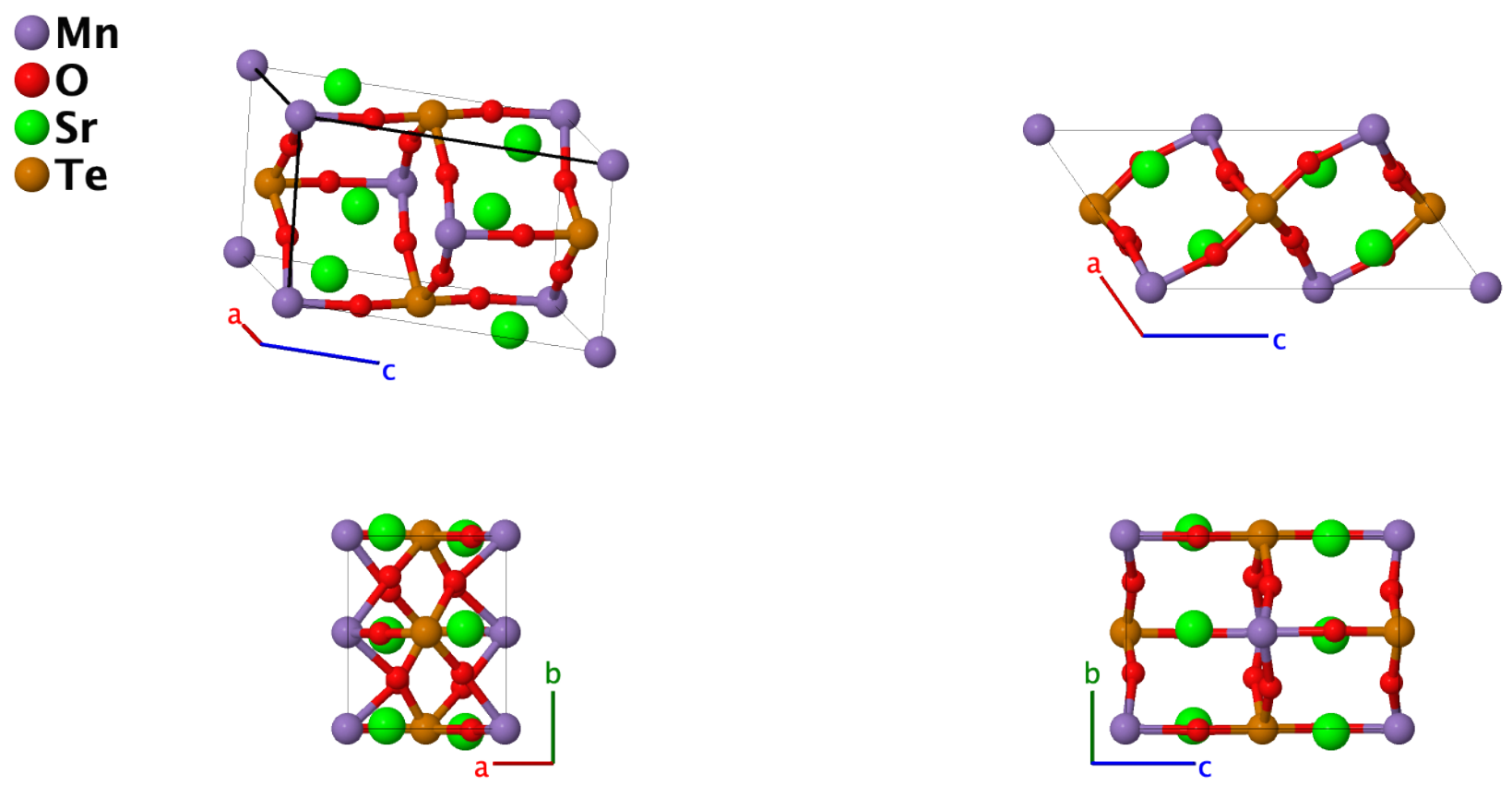

| | | |
|---|---|---|
| **Prototype** | : | $MnO_6Sr_2Te$ |
| **AFLOW prototype label** | : | AB6C2D_mP20_14_a_3e_e_d |
| ***Strukturbericht* designation** | : | None |
| **Pearson symbol** | : | mP20 |
| **Space group number** | : | 14 |
| **Space group symbol** | : | $P2_1/c$ |
| **AFLOW prototype command** | : | `aflow --proto=AB6C2D_mP20_14_a_3e_e_d`<br>`--params=`$a, b/a, c/a, \beta, x_3, y_3, z_3, x_4, y_4, z_4, x_5, y_5, z_5, x_6, y_6, z_6$ |

**Other compounds with this structure**

- $Cs_2RbDyF_6$

- This is the ground state structure of the double perovskite $Sr_2MnTeO_6$. Above 250 °C, it transforms into the $Sr_2NiTeO_6$ structure. Above 550 °C it transforms into the $Sr_2NiWO_6$ structure, and above 675 °C it transforms into the cubic perovskite $E2_1$ structure. (Ortega-San Martin, 2004).
- (Ortega-San Martin, 2004) give the structure in the $P2_1/n$ setting of space group #14. We used FINDSYM to transform this to the standard $P2_1/c$ setting, which uses a different set of basis vectors.

**Simple Monoclinic primitive vectors:**

$$\mathbf{a}_1 = a\,\hat{\mathbf{x}}$$
$$\mathbf{a}_2 = b\,\hat{\mathbf{y}}$$
$$\mathbf{a}_3 = c\cos\beta\,\hat{\mathbf{x}} + c\sin\beta\,\hat{\mathbf{z}}$$

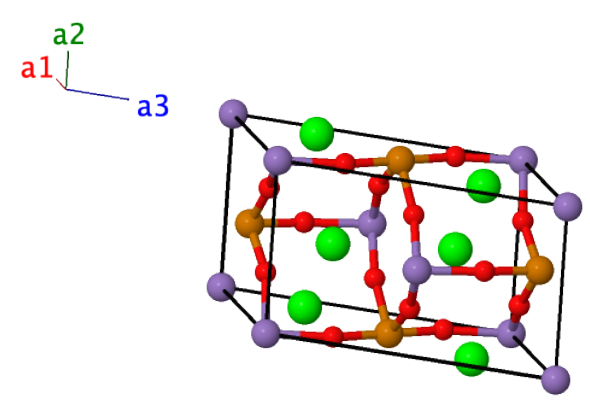

**Basis vectors:**

| | | Lattice Coordinates | | Cartesian Coordinates | Wyckoff Position | Atom Type |
|---|---|---|---|---|---|---|
| $\mathbf{B}_1$ | = | $0\,\mathbf{a}_1 + 0\,\mathbf{a}_2 + 0\,\mathbf{a}_3$ | = | $0\,\hat{\mathbf{x}} + 0\,\hat{\mathbf{y}} + 0\,\hat{\mathbf{z}}$ | (2a) | Mn |
| $\mathbf{B}_2$ | = | $\frac{1}{2}\,\mathbf{a}_2 + \frac{1}{2}\,\mathbf{a}_3$ | = | $\frac{1}{2}c\cos\beta\,\hat{\mathbf{x}} + \frac{1}{2}b\,\hat{\mathbf{y}} + \frac{1}{2}c\sin\beta\,\hat{\mathbf{z}}$ | (2a) | Mn |
| $\mathbf{B}_3$ | = | $\frac{1}{2}\,\mathbf{a}_1 + \frac{1}{2}\,\mathbf{a}_3$ | = | $\frac{1}{2}\,(a + c\cos\beta)\,\hat{\mathbf{x}} + \frac{1}{2}c\sin\beta\,\hat{\mathbf{z}}$ | (2$d$) | Te |
| $\mathbf{B}_4$ | = | $\frac{1}{2}\,\mathbf{a}_1 + \frac{1}{2}\,\mathbf{a}_2$ | = | $\frac{1}{2}a\,\hat{\mathbf{x}} + \frac{1}{2}b\,\hat{\mathbf{y}}$ | (2$d$) | Te |
| $\mathbf{B}_5$ | = | $x_3\,\mathbf{a}_1 + y_3\,\mathbf{a}_2 + z_3\,\mathbf{a}_3$ | = | $(x_3a + z_3c\cos\beta)\,\hat{\mathbf{x}} + y_3b\,\hat{\mathbf{y}} + z_3c\sin\beta\,\hat{\mathbf{z}}$ | (4$e$) | O I |
| $\mathbf{B}_6$ | = | $-x_3\,\mathbf{a}_1 + \left(\frac{1}{2} + y_3\right)\,\mathbf{a}_2 + \left(\frac{1}{2} - z_3\right)\,\mathbf{a}_3$ | = | $\left(\frac{1}{2}c\cos\beta - x_3a - z_3c\cos\beta\right)\,\hat{\mathbf{x}} + \left(\frac{1}{2} + y_3\right)b\,\hat{\mathbf{y}} + \left(\frac{1}{2} - z_3\right)c\sin\beta\,\hat{\mathbf{z}}$ | (4$e$) | O I |
| $\mathbf{B}_7$ | = | $-x_3\,\mathbf{a}_1 - y_3\,\mathbf{a}_2 - z_3\,\mathbf{a}_3$ | = | $(-x_3a - z_3c\cos\beta)\,\hat{\mathbf{x}} - y_3b\,\hat{\mathbf{y}} - z_3c\sin\beta\,\hat{\mathbf{z}}$ | (4$e$) | O I |
| $\mathbf{B}_8$ | = | $x_3\,\mathbf{a}_1 + \left(\frac{1}{2} - y_3\right)\,\mathbf{a}_2 + \left(\frac{1}{2} + z_3\right)\,\mathbf{a}_3$ | = | $\left(\frac{1}{2}c\cos\beta + x_3a + z_3c\cos\beta\right)\,\hat{\mathbf{x}} + \left(\frac{1}{2} - y_3\right)b\,\hat{\mathbf{y}} + \left(\frac{1}{2} + z_3\right)c\sin\beta\,\hat{\mathbf{z}}$ | (4$e$) | O I |
| $\mathbf{B}_9$ | = | $x_4\,\mathbf{a}_1 + y_4\,\mathbf{a}_2 + z_4\,\mathbf{a}_3$ | = | $(x_4a + z_4c\cos\beta)\,\hat{\mathbf{x}} + y_4b\,\hat{\mathbf{y}} + z_4c\sin\beta\,\hat{\mathbf{z}}$ | (4$e$) | O II |
| $\mathbf{B}_{10}$ | = | $-x_4\,\mathbf{a}_1 + \left(\frac{1}{2} + y_4\right)\,\mathbf{a}_2 + \left(\frac{1}{2} - z_4\right)\,\mathbf{a}_3$ | = | $\left(\frac{1}{2}c\cos\beta - x_4a - z_4c\cos\beta\right)\,\hat{\mathbf{x}} + \left(\frac{1}{2} + y_4\right)b\,\hat{\mathbf{y}} + \left(\frac{1}{2} - z_4\right)c\sin\beta\,\hat{\mathbf{z}}$ | (4$e$) | O II |
| $\mathbf{B}_{11}$ | = | $-x_4\,\mathbf{a}_1 - y_4\,\mathbf{a}_2 - z_4\,\mathbf{a}_3$ | = | $(-x_4a - z_4c\cos\beta)\,\hat{\mathbf{x}} - y_4b\,\hat{\mathbf{y}} - z_4c\sin\beta\,\hat{\mathbf{z}}$ | (4$e$) | O II |
| $\mathbf{B}_{12}$ | = | $x_4\,\mathbf{a}_1 + \left(\frac{1}{2} - y_4\right)\,\mathbf{a}_2 + \left(\frac{1}{2} + z_4\right)\,\mathbf{a}_3$ | = | $\left(\frac{1}{2}c\cos\beta + x_4a + z_4c\cos\beta\right)\,\hat{\mathbf{x}} + \left(\frac{1}{2} - y_4\right)b\,\hat{\mathbf{y}} + \left(\frac{1}{2} + z_4\right)c\sin\beta\,\hat{\mathbf{z}}$ | (4$e$) | O II |
| $\mathbf{B}_{13}$ | = | $x_5\,\mathbf{a}_1 + y_5\,\mathbf{a}_2 + z_5\,\mathbf{a}_3$ | = | $(x_5a + z_5c\cos\beta)\,\hat{\mathbf{x}} + y_5b\,\hat{\mathbf{y}} + z_5c\sin\beta\,\hat{\mathbf{z}}$ | (4$e$) | O III |
| $\mathbf{B}_{14}$ | = | $-x_5\,\mathbf{a}_1 + \left(\frac{1}{2} + y_5\right)\,\mathbf{a}_2 + \left(\frac{1}{2} - z_5\right)\,\mathbf{a}_3$ | = | $\left(\frac{1}{2}c\cos\beta - x_5a - z_5c\cos\beta\right)\,\hat{\mathbf{x}} + \left(\frac{1}{2} + y_5\right)b\,\hat{\mathbf{y}} + \left(\frac{1}{2} - z_5\right)c\sin\beta\,\hat{\mathbf{z}}$ | (4$e$) | O III |
| $\mathbf{B}_{15}$ | = | $-x_5\,\mathbf{a}_1 - y_5\,\mathbf{a}_2 - z_5\,\mathbf{a}_3$ | = | $(-x_5a - z_5c\cos\beta)\,\hat{\mathbf{x}} - y_5b\,\hat{\mathbf{y}} - z_5c\sin\beta\,\hat{\mathbf{z}}$ | (4$e$) | O III |
| $\mathbf{B}_{16}$ | = | $x_5\,\mathbf{a}_1 + \left(\frac{1}{2} - y_5\right)\,\mathbf{a}_2 + \left(\frac{1}{2} + z_5\right)\,\mathbf{a}_3$ | = | $\left(\frac{1}{2}c\cos\beta + x_5a + z_5c\cos\beta\right)\,\hat{\mathbf{x}} + \left(\frac{1}{2} - y_5\right)b\,\hat{\mathbf{y}} + \left(\frac{1}{2} + z_5\right)c\sin\beta\,\hat{\mathbf{z}}$ | (4$e$) | O III |
| $\mathbf{B}_{17}$ | = | $x_6\,\mathbf{a}_1 + y_6\,\mathbf{a}_2 + z_6\,\mathbf{a}_3$ | = | $(x_6a + z_6c\cos\beta)\,\hat{\mathbf{x}} + y_6b\,\hat{\mathbf{y}} + z_6c\sin\beta\,\hat{\mathbf{z}}$ | (4$e$) | Sr |
| $\mathbf{B}_{18}$ | = | $-x_6\,\mathbf{a}_1 + \left(\frac{1}{2} + y_6\right)\,\mathbf{a}_2 + \left(\frac{1}{2} - z_6\right)\,\mathbf{a}_3$ | = | $\left(\frac{1}{2}c\cos\beta - x_6a - z_6c\cos\beta\right)\,\hat{\mathbf{x}} + \left(\frac{1}{2} + y_6\right)b\,\hat{\mathbf{y}} + \left(\frac{1}{2} - z_6\right)c\sin\beta\,\hat{\mathbf{z}}$ | (4$e$) | Sr |
| $\mathbf{B}_{19}$ | = | $-x_6\,\mathbf{a}_1 - y_6\,\mathbf{a}_2 - z_6\,\mathbf{a}_3$ | = | $(-x_6a - z_6c\cos\beta)\,\hat{\mathbf{x}} - y_6b\,\hat{\mathbf{y}} - z_6c\sin\beta\,\hat{\mathbf{z}}$ | (4$e$) | Sr |
| $\mathbf{B}_{20}$ | = | $x_6\,\mathbf{a}_1 + \left(\frac{1}{2} - y_6\right)\,\mathbf{a}_2 + \left(\frac{1}{2} + z_6\right)\,\mathbf{a}_3$ | = | $\left(\frac{1}{2}c\cos\beta + x_6a + z_6c\cos\beta\right)\,\hat{\mathbf{x}} + \left(\frac{1}{2} - y_6\right)b\,\hat{\mathbf{y}} + \left(\frac{1}{2} + z_6\right)c\sin\beta\,\hat{\mathbf{z}}$ | (4$e$) | Sr |

**Geometry files:**

- CIF: pp. 1553
- POSCAR: pp. 1553

# Cryolite ($Na_3AlF_6$, $J2_6$) Structure: AB6C3_mP20_14_a_3e_de

http://aflow.org/prototype-encyclopedia/AB6C3_mP20_14_a_3e_de

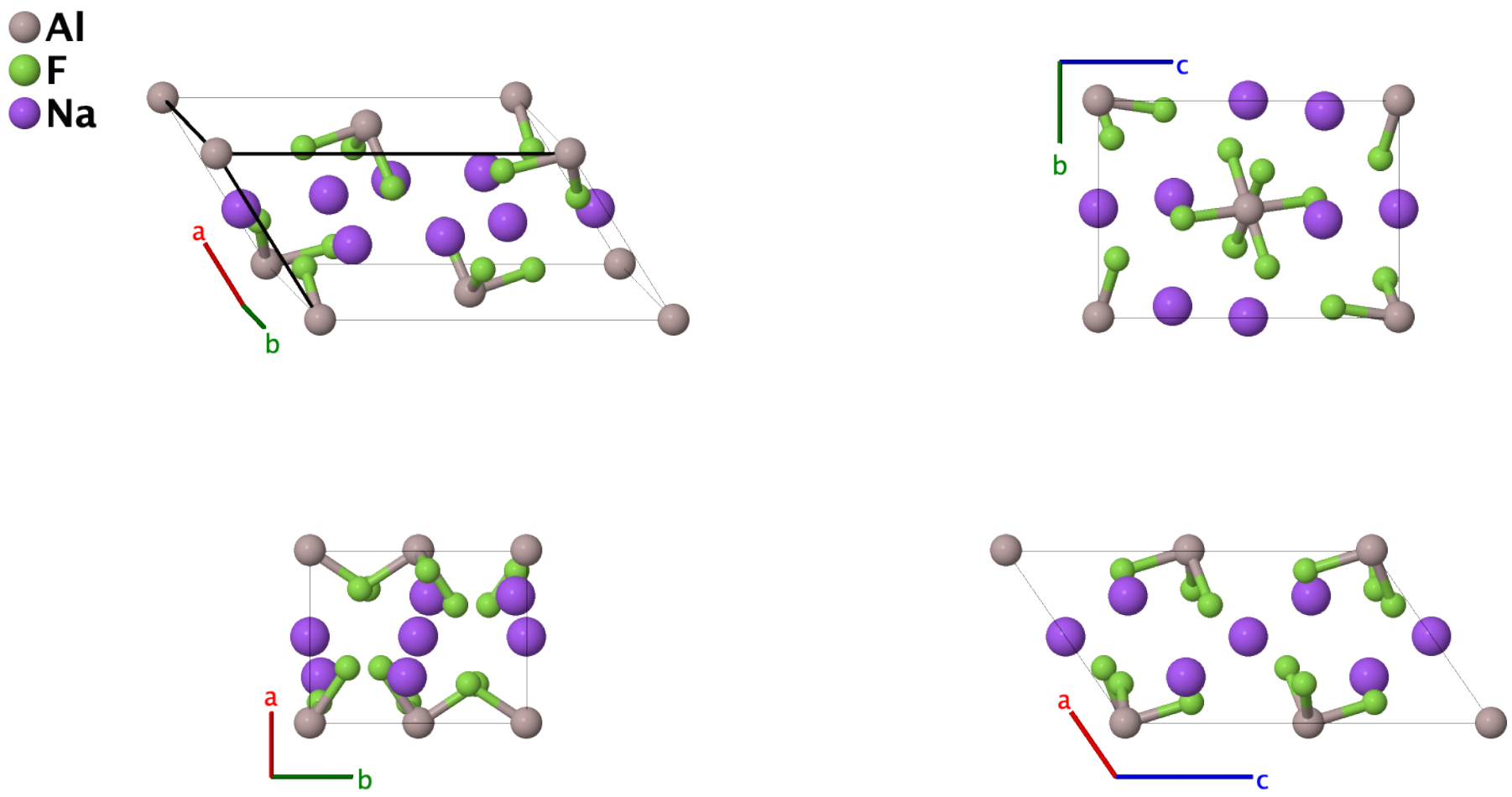

| | | |
|---|---|---|
| **Prototype** | : | $AlF_6Na_3$ |
| **AFLOW prototype label** | : | AB6C3_mP20_14_a_3e_de |
| ***Strukturbericht* designation** | : | $J2_6$ |
| **Pearson symbol** | : | mP20 |
| **Space group number** | : | 14 |
| **Space group symbol** | : | $P2_1/c$ |
| **AFLOW prototype command** | : | `aflow --proto=AB6C3_mP20_14_a_3e_de` `--params=`$a, b/a, c/a, \beta, x_3, y_3, z_3, x_4, y_4, z_4, x_5, y_5, z_5, x_6, y_6, z_6$ |

- Cryolite undergoes a phase transition to an orthorhombic *Immm* phase at 890 K. Here we use structural data taken at 295 K.
- (Yang, 1993) gives the Wyckoff positions for the orientation $P2_1/n$ of space group #14. We used FINDSYM to change this to the standard $P2_1/c$ orientation. This included a change in the primitive vectors and the shift of the sodium atom on the $(1b)$ site to the $(1d)$ site.

**Simple Monoclinic primitive vectors:**

$$\mathbf{a}_1 = a\,\hat{\mathbf{x}}$$
$$\mathbf{a}_2 = b\,\hat{\mathbf{y}}$$
$$\mathbf{a}_3 = c\cos\beta\,\hat{\mathbf{x}} + c\sin\beta\,\hat{\mathbf{z}}$$

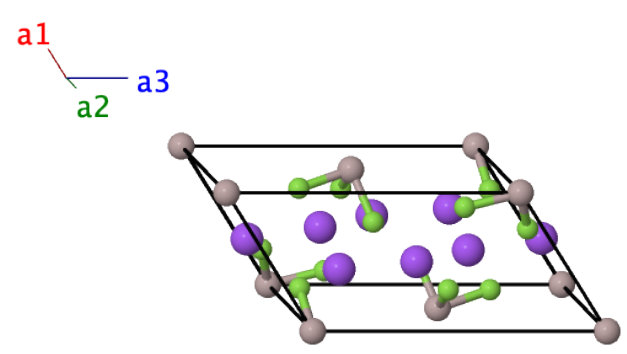

**Basis vectors:**

| | | Lattice Coordinates | | Cartesian Coordinates | Wyckoff Position | Atom Type |
|---|---|---|---|---|---|---|
| $\mathbf{B}_1$ | = | $0\,\mathbf{a}_1 + 0\,\mathbf{a}_2 + 0\,\mathbf{a}_3$ | = | $0\,\hat{\mathbf{x}} + 0\,\hat{\mathbf{y}} + 0\,\hat{\mathbf{z}}$ | $(2a)$ | Al |
| $\mathbf{B}_2$ | = | $\frac{1}{2}\,\mathbf{a}_2 + \frac{1}{2}\,\mathbf{a}_3$ | = | $\frac{1}{2}c\cos\beta\,\hat{\mathbf{x}} + \frac{1}{2}b\,\hat{\mathbf{y}} + \frac{1}{2}c\sin\beta\,\hat{\mathbf{z}}$ | $(2a)$ | Al |
| $\mathbf{B}_3$ | = | $\frac{1}{2}\,\mathbf{a}_1 + \frac{1}{2}\,\mathbf{a}_3$ | = | $\frac{1}{2}\,(a + c\cos\beta)\,\hat{\mathbf{x}} + \frac{1}{2}c\sin\beta\,\hat{\mathbf{z}}$ | (2$d$) | Na I |
| $\mathbf{B}_4$ | = | $\frac{1}{2}\,\mathbf{a}_1 + \frac{1}{2}\,\mathbf{a}_2$ | = | $\frac{1}{2}a\,\hat{\mathbf{x}} + \frac{1}{2}b\,\hat{\mathbf{y}}$ | (2$d$) | Na I |
| $\mathbf{B}_5$ | = | $x_3\,\mathbf{a}_1 + y_3\,\mathbf{a}_2 + z_3\,\mathbf{a}_3$ | = | $(x_3a + z_3c\cos\beta)\,\hat{\mathbf{x}} + y_3b\,\hat{\mathbf{y}} + z_3c\sin\beta\,\hat{\mathbf{z}}$ | (4$e$) | F I |
| $\mathbf{B}_6$ | = | $-x_3\,\mathbf{a}_1 + \left(\frac{1}{2} + y_3\right)\,\mathbf{a}_2 + \left(\frac{1}{2} - z_3\right)\,\mathbf{a}_3$ | = | $\left(\frac{1}{2}c\cos\beta - x_3a - z_3c\cos\beta\right)\,\hat{\mathbf{x}} + \left(\frac{1}{2} + y_3\right)b\,\hat{\mathbf{y}} + \left(\frac{1}{2} - z_3\right)c\sin\beta\,\hat{\mathbf{z}}$ | (4$e$) | F I |
| $\mathbf{B}_7$ | = | $-x_3\,\mathbf{a}_1 - y_3\,\mathbf{a}_2 - z_3\,\mathbf{a}_3$ | = | $(-x_3a - z_3c\cos\beta)\,\hat{\mathbf{x}} - y_3b\,\hat{\mathbf{y}} - z_3c\sin\beta\,\hat{\mathbf{z}}$ | (4$e$) | F I |
| $\mathbf{B}_8$ | = | $x_3\,\mathbf{a}_1 + \left(\frac{1}{2} - y_3\right)\,\mathbf{a}_2 + \left(\frac{1}{2} + z_3\right)\,\mathbf{a}_3$ | = | $\left(\frac{1}{2}c\cos\beta + x_3a + z_3c\cos\beta\right)\,\hat{\mathbf{x}} + \left(\frac{1}{2} - y_3\right)b\,\hat{\mathbf{y}} + \left(\frac{1}{2} + z_3\right)c\sin\beta\,\hat{\mathbf{z}}$ | (4$e$) | F I |
| $\mathbf{B}_9$ | = | $x_4\,\mathbf{a}_1 + y_4\,\mathbf{a}_2 + z_4\,\mathbf{a}_3$ | = | $(x_4a + z_4c\cos\beta)\,\hat{\mathbf{x}} + y_4b\,\hat{\mathbf{y}} + z_4c\sin\beta\,\hat{\mathbf{z}}$ | (4$e$) | F II |
| $\mathbf{B}_{10}$ | = | $-x_4\,\mathbf{a}_1 + \left(\frac{1}{2} + y_4\right)\,\mathbf{a}_2 + \left(\frac{1}{2} - z_4\right)\,\mathbf{a}_3$ | = | $\left(\frac{1}{2}c\cos\beta - x_4a - z_4c\cos\beta\right)\,\hat{\mathbf{x}} + \left(\frac{1}{2} + y_4\right)b\,\hat{\mathbf{y}} + \left(\frac{1}{2} - z_4\right)c\sin\beta\,\hat{\mathbf{z}}$ | (4$e$) | F II |
| $\mathbf{B}_{11}$ | = | $-x_4\,\mathbf{a}_1 - y_4\,\mathbf{a}_2 - z_4\,\mathbf{a}_3$ | = | $(-x_4a - z_4c\cos\beta)\,\hat{\mathbf{x}} - y_4b\,\hat{\mathbf{y}} - z_4c\sin\beta\,\hat{\mathbf{z}}$ | (4$e$) | F II |
| $\mathbf{B}_{12}$ | = | $x_4\,\mathbf{a}_1 + \left(\frac{1}{2} - y_4\right)\,\mathbf{a}_2 + \left(\frac{1}{2} + z_4\right)\,\mathbf{a}_3$ | = | $\left(\frac{1}{2}c\cos\beta + x_4a + z_4c\cos\beta\right)\,\hat{\mathbf{x}} + \left(\frac{1}{2} - y_4\right)b\,\hat{\mathbf{y}} + \left(\frac{1}{2} + z_4\right)c\sin\beta\,\hat{\mathbf{z}}$ | (4$e$) | F II |
| $\mathbf{B}_{13}$ | = | $x_5\,\mathbf{a}_1 + y_5\,\mathbf{a}_2 + z_5\,\mathbf{a}_3$ | = | $(x_5a + z_5c\cos\beta)\,\hat{\mathbf{x}} + y_5b\,\hat{\mathbf{y}} + z_5c\sin\beta\,\hat{\mathbf{z}}$ | (4$e$) | F III |
| $\mathbf{B}_{14}$ | = | $-x_5\,\mathbf{a}_1 + \left(\frac{1}{2} + y_5\right)\,\mathbf{a}_2 + \left(\frac{1}{2} - z_5\right)\,\mathbf{a}_3$ | = | $\left(\frac{1}{2}c\cos\beta - x_5a - z_5c\cos\beta\right)\,\hat{\mathbf{x}} + \left(\frac{1}{2} + y_5\right)b\,\hat{\mathbf{y}} + \left(\frac{1}{2} - z_5\right)c\sin\beta\,\hat{\mathbf{z}}$ | (4$e$) | F III |
| $\mathbf{B}_{15}$ | = | $-x_5\,\mathbf{a}_1 - y_5\,\mathbf{a}_2 - z_5\,\mathbf{a}_3$ | = | $(-x_5a - z_5c\cos\beta)\,\hat{\mathbf{x}} - y_5b\,\hat{\mathbf{y}} - z_5c\sin\beta\,\hat{\mathbf{z}}$ | (4$e$) | F III |
| $\mathbf{B}_{16}$ | = | $x_5\,\mathbf{a}_1 + \left(\frac{1}{2} - y_5\right)\,\mathbf{a}_2 + \left(\frac{1}{2} + z_5\right)\,\mathbf{a}_3$ | = | $\left(\frac{1}{2}c\cos\beta + x_5a + z_5c\cos\beta\right)\,\hat{\mathbf{x}} + \left(\frac{1}{2} - y_5\right)b\,\hat{\mathbf{y}} + \left(\frac{1}{2} + z_5\right)c\sin\beta\,\hat{\mathbf{z}}$ | (4$e$) | F III |
| $\mathbf{B}_{17}$ | = | $x_6\,\mathbf{a}_1 + y_6\,\mathbf{a}_2 + z_6\,\mathbf{a}_3$ | = | $(x_6a + z_6c\cos\beta)\,\hat{\mathbf{x}} + y_6b\,\hat{\mathbf{y}} + z_6c\sin\beta\,\hat{\mathbf{z}}$ | (4$e$) | Na II |
| $\mathbf{B}_{18}$ | = | $-x_6\,\mathbf{a}_1 + \left(\frac{1}{2} + y_6\right)\,\mathbf{a}_2 + \left(\frac{1}{2} - z_6\right)\,\mathbf{a}_3$ | = | $\left(\frac{1}{2}c\cos\beta - x_6a - z_6c\cos\beta\right)\,\hat{\mathbf{x}} + \left(\frac{1}{2} + y_6\right)b\,\hat{\mathbf{y}} + \left(\frac{1}{2} - z_6\right)c\sin\beta\,\hat{\mathbf{z}}$ | (4$e$) | Na II |
| $\mathbf{B}_{19}$ | = | $-x_6\,\mathbf{a}_1 - y_6\,\mathbf{a}_2 - z_6\,\mathbf{a}_3$ | = | $(-x_6a - z_6c\cos\beta)\,\hat{\mathbf{x}} - y_6b\,\hat{\mathbf{y}} - z_6c\sin\beta\,\hat{\mathbf{z}}$ | (4$e$) | Na II |
| $\mathbf{B}_{20}$ | = | $x_6\,\mathbf{a}_1 + \left(\frac{1}{2} - y_6\right)\,\mathbf{a}_2 + \left(\frac{1}{2} + z_6\right)\,\mathbf{a}_3$ | = | $\left(\frac{1}{2}c\cos\beta + x_6a + z_6c\cos\beta\right)\,\hat{\mathbf{x}} + \left(\frac{1}{2} - y_6\right)b\,\hat{\mathbf{y}} + \left(\frac{1}{2} + z_6\right)c\sin\beta\,\hat{\mathbf{z}}$ | (4$e$) | Na II |

**Found in:**

**Geometry files:**

- CIF: pp. 1553
- POSCAR: pp. 1554

# $KNO_2$ III Structure: ABC2_mP16_14_e_e_2e

http://aflow.org/prototype-encyclopedia/ABC2_mP16_14_e_e_2e.KNO2

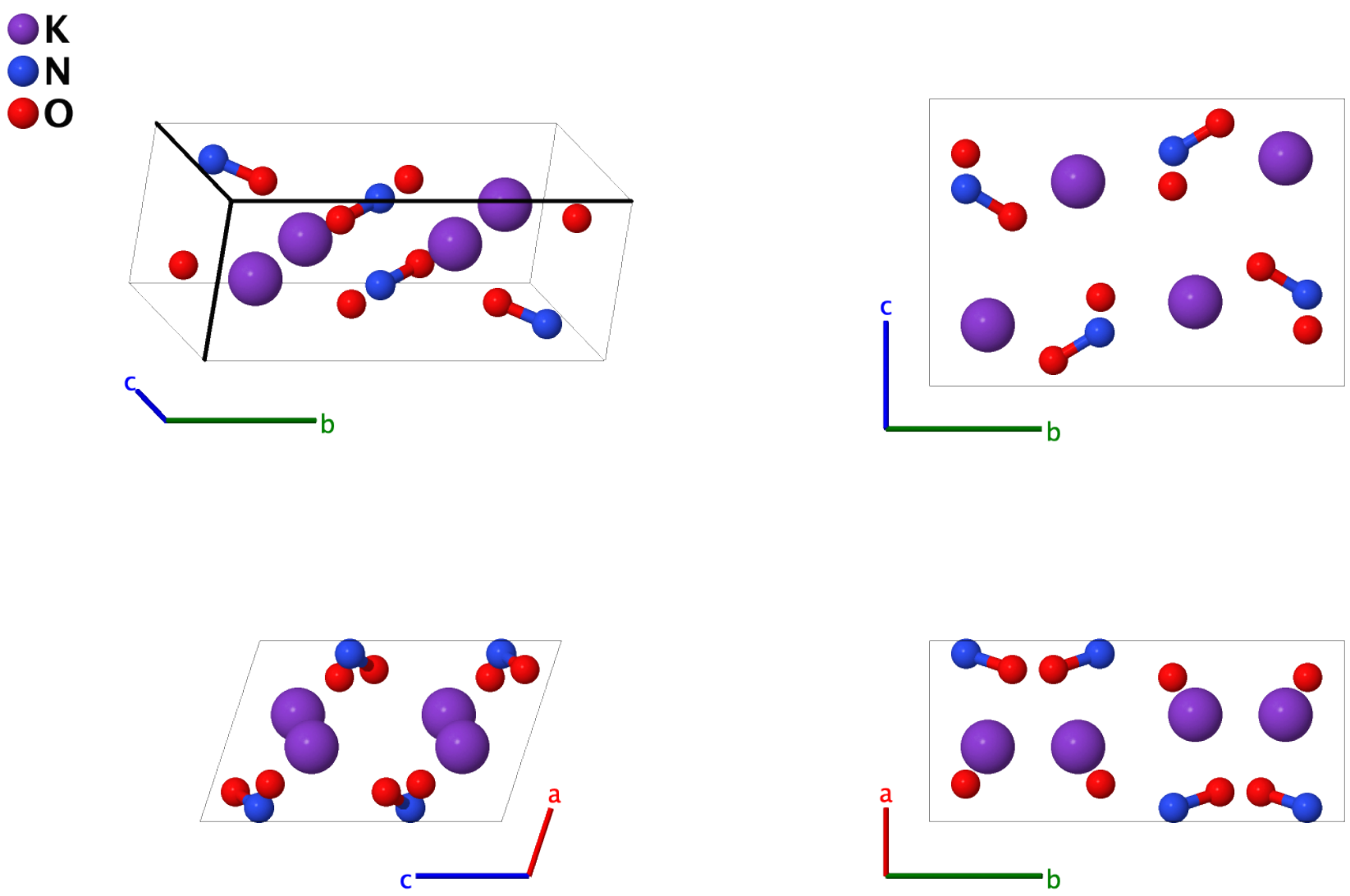

| | | |
|---|---|---|
| **Prototype** | : | $KNO_2$ |
| **AFLOW prototype label** | : | ABC2_mP16_14_e_e_2e |
| ***Strukturbericht* designation** | : | None |
| **Pearson symbol** | : | mP16 |
| **Space group number** | : | 14 |
| **Space group symbol** | : | $P2_1/c$ |
| **AFLOW prototype command** | : | `aflow --proto=ABC2_mP16_14_e_e_2e`<br>`--params=`$a, b/a, c/a, \beta, x_1, y_1, z_1, x_2, y_2, z_2, x_3, y_3, z_3, x_4, y_4, z_4$ |

- This is *not* the $KNO_2$ structure found by (Ziegler, 1936) and given the *Strukturbericht* designation $F5_{11}$. That structure is now considered obsolete.

**Simple Monoclinic primitive vectors:**

$$\mathbf{a}_1 = a\,\hat{\mathbf{x}}$$
$$\mathbf{a}_2 = b\,\hat{\mathbf{y}}$$
$$\mathbf{a}_3 = c\cos\beta\,\hat{\mathbf{x}} + c\sin\beta\,\hat{\mathbf{z}}$$

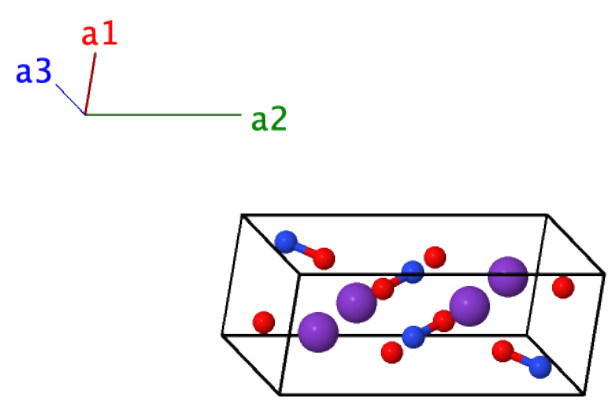

**Basis vectors:**

| | | Lattice Coordinates | | Cartesian Coordinates | Wyckoff Position | Atom Type |
|---|---|---|---|---|---|---|
| $\mathbf{B}_1$ | = | $x_1\,\mathbf{a}_1 + y_1\,\mathbf{a}_2 + z_1\,\mathbf{a}_3$ | = | $(x_1 a + z_1 c\cos\beta)\,\hat{\mathbf{x}} + y_1 b\,\hat{\mathbf{y}} + z_1 c\sin\beta\,\hat{\mathbf{z}}$ | (4*e*) | K |
| $\mathbf{B}_2$ | = | $-x_1\,\mathbf{a}_1 + \left(\frac{1}{2}+y_1\right)\mathbf{a}_2 + \left(\frac{1}{2}-z_1\right)\mathbf{a}_3$ | = | $\left(\frac{1}{2}c\cos\beta - x_1 a - z_1 c\cos\beta\right)\hat{\mathbf{x}} + \left(\frac{1}{2}+y_1\right)b\,\hat{\mathbf{y}} + \left(\frac{1}{2}-z_1\right)c\sin\beta\,\hat{\mathbf{z}}$ | (4*e*) | K |
| $\mathbf{B}_3$ | = | $-x_1\,\mathbf{a}_1 - y_1\,\mathbf{a}_2 - z_1\,\mathbf{a}_3$ | = | $(-x_1 a - z_1 c\cos\beta)\,\hat{\mathbf{x}} - y_1 b\,\hat{\mathbf{y}} - z_1 c\sin\beta\,\hat{\mathbf{z}}$ | (4e) | K |
| $\mathbf{B}_4$ | = | $x_1\,\mathbf{a}_1 + \left(\frac{1}{2} - y_1\right)\mathbf{a}_2 + \left(\frac{1}{2} + z_1\right)\mathbf{a}_3$ | = | $\left(\frac{1}{2}c\cos\beta + x_1 a + z_1 c\cos\beta\right)\hat{\mathbf{x}} + \left(\frac{1}{2} - y_1\right)b\,\hat{\mathbf{y}} + \left(\frac{1}{2} + z_1\right)c\sin\beta\,\hat{\mathbf{z}}$ | (4e) | K |
| $\mathbf{B}_5$ | = | $x_2\,\mathbf{a}_1 + y_2\,\mathbf{a}_2 + z_2\,\mathbf{a}_3$ | = | $(x_2 a + z_2 c\cos\beta)\,\hat{\mathbf{x}} + y_2 b\,\hat{\mathbf{y}} + z_2 c\sin\beta\,\hat{\mathbf{z}}$ | (4e) | N |
| $\mathbf{B}_6$ | = | $-x_2\,\mathbf{a}_1 + \left(\frac{1}{2} + y_2\right)\mathbf{a}_2 + \left(\frac{1}{2} - z_2\right)\mathbf{a}_3$ | = | $\left(\frac{1}{2}c\cos\beta - x_2 a - z_2 c\cos\beta\right)\hat{\mathbf{x}} + \left(\frac{1}{2} + y_2\right)b\,\hat{\mathbf{y}} + \left(\frac{1}{2} - z_2\right)c\sin\beta\,\hat{\mathbf{z}}$ | (4e) | N |
| $\mathbf{B}_7$ | = | $-x_2\,\mathbf{a}_1 - y_2\,\mathbf{a}_2 - z_2\,\mathbf{a}_3$ | = | $(-x_2 a - z_2 c\cos\beta)\,\hat{\mathbf{x}} - y_2 b\,\hat{\mathbf{y}} - z_2 c\sin\beta\,\hat{\mathbf{z}}$ | (4e) | N |
| $\mathbf{B}_8$ | = | $x_2\,\mathbf{a}_1 + \left(\frac{1}{2} - y_2\right)\mathbf{a}_2 + \left(\frac{1}{2} + z_2\right)\mathbf{a}_3$ | = | $\left(\frac{1}{2}c\cos\beta + x_2 a + z_2 c\cos\beta\right)\hat{\mathbf{x}} + \left(\frac{1}{2} - y_2\right)b\,\hat{\mathbf{y}} + \left(\frac{1}{2} + z_2\right)c\sin\beta\,\hat{\mathbf{z}}$ | (4e) | N |
| $\mathbf{B}_9$ | = | $x_3\,\mathbf{a}_1 + y_3\,\mathbf{a}_2 + z_3\,\mathbf{a}_3$ | = | $(x_3 a + z_3 c\cos\beta)\,\hat{\mathbf{x}} + y_3 b\,\hat{\mathbf{y}} + z_3 c\sin\beta\,\hat{\mathbf{z}}$ | (4e) | O I |
| $\mathbf{B}_{10}$ | = | $-x_3\,\mathbf{a}_1 + \left(\frac{1}{2} + y_3\right)\mathbf{a}_2 + \left(\frac{1}{2} - z_3\right)\mathbf{a}_3$ | = | $\left(\frac{1}{2}c\cos\beta - x_3 a - z_3 c\cos\beta\right)\hat{\mathbf{x}} + \left(\frac{1}{2} + y_3\right)b\,\hat{\mathbf{y}} + \left(\frac{1}{2} - z_3\right)c\sin\beta\,\hat{\mathbf{z}}$ | (4e) | O I |
| $\mathbf{B}_{11}$ | = | $-x_3\,\mathbf{a}_1 - y_3\,\mathbf{a}_2 - z_3\,\mathbf{a}_3$ | = | $(-x_3 a - z_3 c\cos\beta)\,\hat{\mathbf{x}} - y_3 b\,\hat{\mathbf{y}} - z_3 c\sin\beta\,\hat{\mathbf{z}}$ | (4e) | O I |
| $\mathbf{B}_{12}$ | = | $x_3\,\mathbf{a}_1 + \left(\frac{1}{2} - y_3\right)\mathbf{a}_2 + \left(\frac{1}{2} + z_3\right)\mathbf{a}_3$ | = | $\left(\frac{1}{2}c\cos\beta + x_3 a + z_3 c\cos\beta\right)\hat{\mathbf{x}} + \left(\frac{1}{2} - y_3\right)b\,\hat{\mathbf{y}} + \left(\frac{1}{2} + z_3\right)c\sin\beta\,\hat{\mathbf{z}}$ | (4e) | O I |
| $\mathbf{B}_{13}$ | = | $x_4\,\mathbf{a}_1 + y_4\,\mathbf{a}_2 + z_4\,\mathbf{a}_3$ | = | $(x_4 a + z_4 c\cos\beta)\,\hat{\mathbf{x}} + y_4 b\,\hat{\mathbf{y}} + z_4 c\sin\beta\,\hat{\mathbf{z}}$ | (4e) | O II |
| $\mathbf{B}_{14}$ | = | $-x_4\,\mathbf{a}_1 + \left(\frac{1}{2} + y_4\right)\mathbf{a}_2 + \left(\frac{1}{2} - z_4\right)\mathbf{a}_3$ | = | $\left(\frac{1}{2}c\cos\beta - x_4 a - z_4 c\cos\beta\right)\hat{\mathbf{x}} + \left(\frac{1}{2} + y_4\right)b\,\hat{\mathbf{y}} + \left(\frac{1}{2} - z_4\right)c\sin\beta\,\hat{\mathbf{z}}$ | (4e) | O II |
| $\mathbf{B}_{15}$ | = | $-x_4\,\mathbf{a}_1 - y_4\,\mathbf{a}_2 - z_4\,\mathbf{a}_3$ | = | $(-x_4 a - z_4 c\cos\beta)\,\hat{\mathbf{x}} - y_4 b\,\hat{\mathbf{y}} - z_4 c\sin\beta\,\hat{\mathbf{z}}$ | (4e) | O II |
| $\mathbf{B}_{16}$ | = | $x_4\,\mathbf{a}_1 + \left(\frac{1}{2} - y_4\right)\mathbf{a}_2 + \left(\frac{1}{2} + z_4\right)\mathbf{a}_3$ | = | $\left(\frac{1}{2}c\cos\beta + x_4 a + z_4 c\cos\beta\right)\hat{\mathbf{x}} + \left(\frac{1}{2} - y_4\right)b\,\hat{\mathbf{y}} + \left(\frac{1}{2} + z_4\right)c\sin\beta\,\hat{\mathbf{z}}$ | (4e) | O II |

**Geometry files:**

- CIF: pp. 1554
- POSCAR: pp. 1554

# Manganite ($\gamma$-MnO(OH), $E0_6$) Structure: ABC2_mP16_14_e_e_2e

http://aflow.org/prototype-encyclopedia/ABC2_mP16_14_e_e_2e

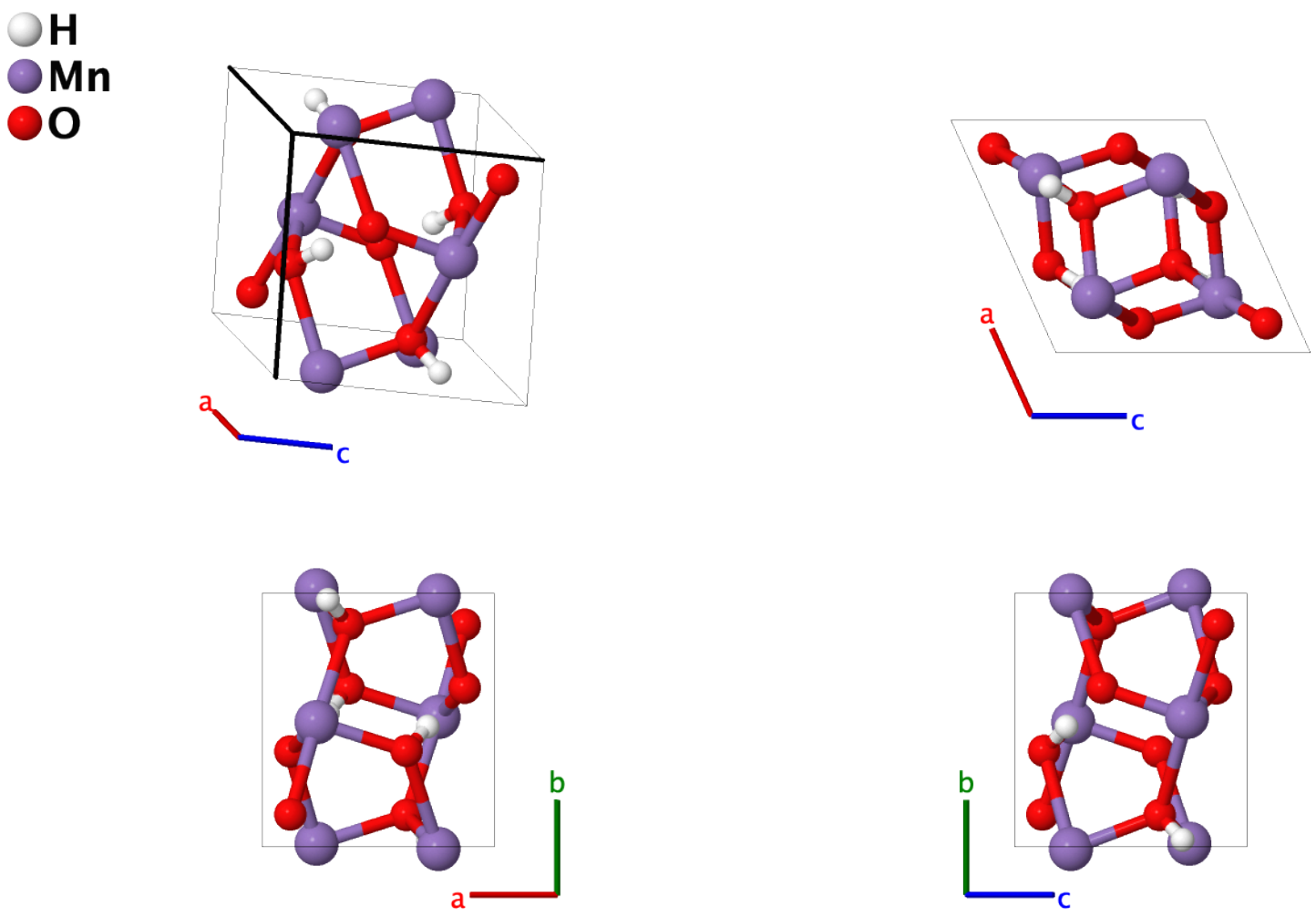

| | | |
|---|---|---|
| **Prototype** | : | $MnHO_2$ |
| **AFLOW prototype label** | : | ABC2_mP16_14_e_e_2e |
| ***Strukturbericht* designation** | : | $E0_6$ |
| **Pearson symbol** | : | mP16 |
| **Space group number** | : | 14 |
| **Space group symbol** | : | $P2_1/c$ |
| **AFLOW prototype command** | : | `aflow --proto=ABC2_mP16_14_e_e_2e` <br> `--params=`$a, b/a, c/a, \beta, x_1, y_1, z_1, x_2, y_2, z_2, x_3, y_3, z_3, x_4, y_4, z_4$ |

- (Buerger, 1936) originally determined the positions of the manganese and oxygen atoms using the $B2_1/d$ setting of space group #14. This structure was given the *Strukturbericht* designation $E0_6$ in (Gottfried, 1938). Much later, (Kohler, 1997) found the positions of the hydrogen atoms, and presented the data in the standard $P2_1/c$ setting of space group #14.

**Simple Monoclinic primitive vectors:**

$$\mathbf{a}_1 = a\,\hat{\mathbf{x}}$$
$$\mathbf{a}_2 = b\,\hat{\mathbf{y}}$$
$$\mathbf{a}_3 = c\cos\beta\,\hat{\mathbf{x}} + c\sin\beta\,\hat{\mathbf{z}}$$

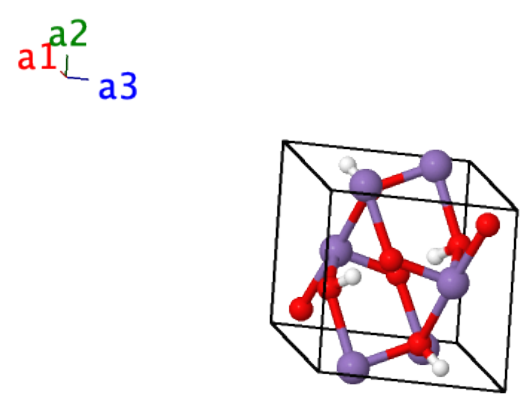

**Basis vectors:**

| | | Lattice Coordinates | | Cartesian Coordinates | Wyckoff Position | Atom Type |
|---|---|---|---|---|---|---|
| $\mathbf{B}_1$ | = | $x_1\,\mathbf{a}_1 + y_1\,\mathbf{a}_2 + z_1\,\mathbf{a}_3$ | = | $(x_1 a + z_1 c\cos\beta)\,\hat{\mathbf{x}} + y_1 b\,\hat{\mathbf{y}} + z_1 c\sin\beta\,\hat{\mathbf{z}}$ | (4$e$) | H |
| $\mathbf{B}_2$ | = | $-x_1\,\mathbf{a}_1 + \left(\frac{1}{2}+y_1\right)\,\mathbf{a}_2 + \left(\frac{1}{2}-z_1\right)\,\mathbf{a}_3$ | = | $\left(\frac{1}{2}c\cos\beta - x_1 a - z_1 c\cos\beta\right)\,\hat{\mathbf{x}} + \left(\frac{1}{2}+y_1\right)b\,\hat{\mathbf{y}} + \left(\frac{1}{2}-z_1\right)c\sin\beta\,\hat{\mathbf{z}}$ | (4$e$) | H |
| $\mathbf{B}_3$ | = | $-x_1\,\mathbf{a}_1 - y_1\,\mathbf{a}_2 - z_1\,\mathbf{a}_3$ | = | $(-x_1 a - z_1 c\cos\beta)\,\hat{\mathbf{x}} - y_1 b\,\hat{\mathbf{y}} - z_1 c\sin\beta\,\hat{\mathbf{z}}$ | (4$e$) | H |
| $\mathbf{B}_4$ | = | $x_1\,\mathbf{a}_1 + \left(\frac{1}{2}-y_1\right)\,\mathbf{a}_2 + \left(\frac{1}{2}+z_1\right)\,\mathbf{a}_3$ | = | $\left(\frac{1}{2}c\cos\beta + x_1 a + z_1 c\cos\beta\right)\,\hat{\mathbf{x}} + \left(\frac{1}{2}-y_1\right)b\,\hat{\mathbf{y}} + \left(\frac{1}{2}+z_1\right)c\sin\beta\,\hat{\mathbf{z}}$ | (4$e$) | H |
| $\mathbf{B}_5$ | = | $x_2\,\mathbf{a}_1 + y_2\,\mathbf{a}_2 + z_2\,\mathbf{a}_3$ | = | $(x_2 a + z_2 c\cos\beta)\,\hat{\mathbf{x}} + y_2 b\,\hat{\mathbf{y}} + z_2 c\sin\beta\,\hat{\mathbf{z}}$ | (4$e$) | Mn |
| $\mathbf{B}_6$ | = | $-x_2\,\mathbf{a}_1 + \left(\frac{1}{2}+y_2\right)\,\mathbf{a}_2 + \left(\frac{1}{2}-z_2\right)\,\mathbf{a}_3$ | = | $\left(\frac{1}{2}c\cos\beta - x_2 a - z_2 c\cos\beta\right)\,\hat{\mathbf{x}} + \left(\frac{1}{2}+y_2\right)b\,\hat{\mathbf{y}} + \left(\frac{1}{2}-z_2\right)c\sin\beta\,\hat{\mathbf{z}}$ | (4$e$) | Mn |
| $\mathbf{B}_7$ | = | $-x_2\,\mathbf{a}_1 - y_2\,\mathbf{a}_2 - z_2\,\mathbf{a}_3$ | = | $(-x_2 a - z_2 c\cos\beta)\,\hat{\mathbf{x}} - y_2 b\,\hat{\mathbf{y}} - z_2 c\sin\beta\,\hat{\mathbf{z}}$ | (4$e$) | Mn |
| $\mathbf{B}_8$ | = | $x_2\,\mathbf{a}_1 + \left(\frac{1}{2}-y_2\right)\,\mathbf{a}_2 + \left(\frac{1}{2}+z_2\right)\,\mathbf{a}_3$ | = | $\left(\frac{1}{2}c\cos\beta + x_2 a + z_2 c\cos\beta\right)\,\hat{\mathbf{x}} + \left(\frac{1}{2}-y_2\right)b\,\hat{\mathbf{y}} + \left(\frac{1}{2}+z_2\right)c\sin\beta\,\hat{\mathbf{z}}$ | (4$e$) | Mn |
| $\mathbf{B}_9$ | = | $x_3\,\mathbf{a}_1 + y_3\,\mathbf{a}_2 + z_3\,\mathbf{a}_3$ | = | $(x_3 a + z_3 c\cos\beta)\,\hat{\mathbf{x}} + y_3 b\,\hat{\mathbf{y}} + z_3 c\sin\beta\,\hat{\mathbf{z}}$ | (4$e$) | O I |
| $\mathbf{B}_{10}$ | = | $-x_3\,\mathbf{a}_1 + \left(\frac{1}{2}+y_3\right)\,\mathbf{a}_2 + \left(\frac{1}{2}-z_3\right)\,\mathbf{a}_3$ | = | $\left(\frac{1}{2}c\cos\beta - x_3 a - z_3 c\cos\beta\right)\,\hat{\mathbf{x}} + \left(\frac{1}{2}+y_3\right)b\,\hat{\mathbf{y}} + \left(\frac{1}{2}-z_3\right)c\sin\beta\,\hat{\mathbf{z}}$ | (4$e$) | O I |
| $\mathbf{B}_{11}$ | = | $-x_3\,\mathbf{a}_1 - y_3\,\mathbf{a}_2 - z_3\,\mathbf{a}_3$ | = | $(-x_3 a - z_3 c\cos\beta)\,\hat{\mathbf{x}} - y_3 b\,\hat{\mathbf{y}} - z_3 c\sin\beta\,\hat{\mathbf{z}}$ | (4$e$) | O I |
| $\mathbf{B}_{12}$ | = | $x_3\,\mathbf{a}_1 + \left(\frac{1}{2}-y_3\right)\,\mathbf{a}_2 + \left(\frac{1}{2}+z_3\right)\,\mathbf{a}_3$ | = | $\left(\frac{1}{2}c\cos\beta + x_3 a + z_3 c\cos\beta\right)\,\hat{\mathbf{x}} + \left(\frac{1}{2}-y_3\right)b\,\hat{\mathbf{y}} + \left(\frac{1}{2}+z_3\right)c\sin\beta\,\hat{\mathbf{z}}$ | (4$e$) | O I |
| $\mathbf{B}_{13}$ | = | $x_4\,\mathbf{a}_1 + y_4\,\mathbf{a}_2 + z_4\,\mathbf{a}_3$ | = | $(x_4 a + z_4 c\cos\beta)\,\hat{\mathbf{x}} + y_4 b\,\hat{\mathbf{y}} + z_4 c\sin\beta\,\hat{\mathbf{z}}$ | (4$e$) | O II |
| $\mathbf{B}_{14}$ | = | $-x_4\,\mathbf{a}_1 + \left(\frac{1}{2}+y_4\right)\,\mathbf{a}_2 + \left(\frac{1}{2}-z_4\right)\,\mathbf{a}_3$ | = | $\left(\frac{1}{2}c\cos\beta - x_4 a - z_4 c\cos\beta\right)\,\hat{\mathbf{x}} + \left(\frac{1}{2}+y_4\right)b\,\hat{\mathbf{y}} + \left(\frac{1}{2}-z_4\right)c\sin\beta\,\hat{\mathbf{z}}$ | (4$e$) | O II |
| $\mathbf{B}_{15}$ | = | $-x_4\,\mathbf{a}_1 - y_4\,\mathbf{a}_2 - z_4\,\mathbf{a}_3$ | = | $(-x_4 a - z_4 c\cos\beta)\,\hat{\mathbf{x}} - y_4 b\,\hat{\mathbf{y}} - z_4 c\sin\beta\,\hat{\mathbf{z}}$ | (4$e$) | O II |
| $\mathbf{B}_{16}$ | = | $x_4\,\mathbf{a}_1 + \left(\frac{1}{2}-y_4\right)\,\mathbf{a}_2 + \left(\frac{1}{2}+z_4\right)\,\mathbf{a}_3$ | = | $\left(\frac{1}{2}c\cos\beta + x_4 a + z_4 c\cos\beta\right)\,\hat{\mathbf{x}} + \left(\frac{1}{2}-y_4\right)b\,\hat{\mathbf{y}} + \left(\frac{1}{2}+z_4\right)c\sin\beta\,\hat{\mathbf{z}}$ | (4$e$) | O II |

---

---

**Geometry files:**

- CIF: pp. 1554
- POSCAR: pp. 1555

# $AgMnO_4$ ($H0_9$) Structure: ABC4_mP24_14_e_e_4e

http://aflow.org/prototype-encyclopedia/ABC4_mP24_14_e_e_4e

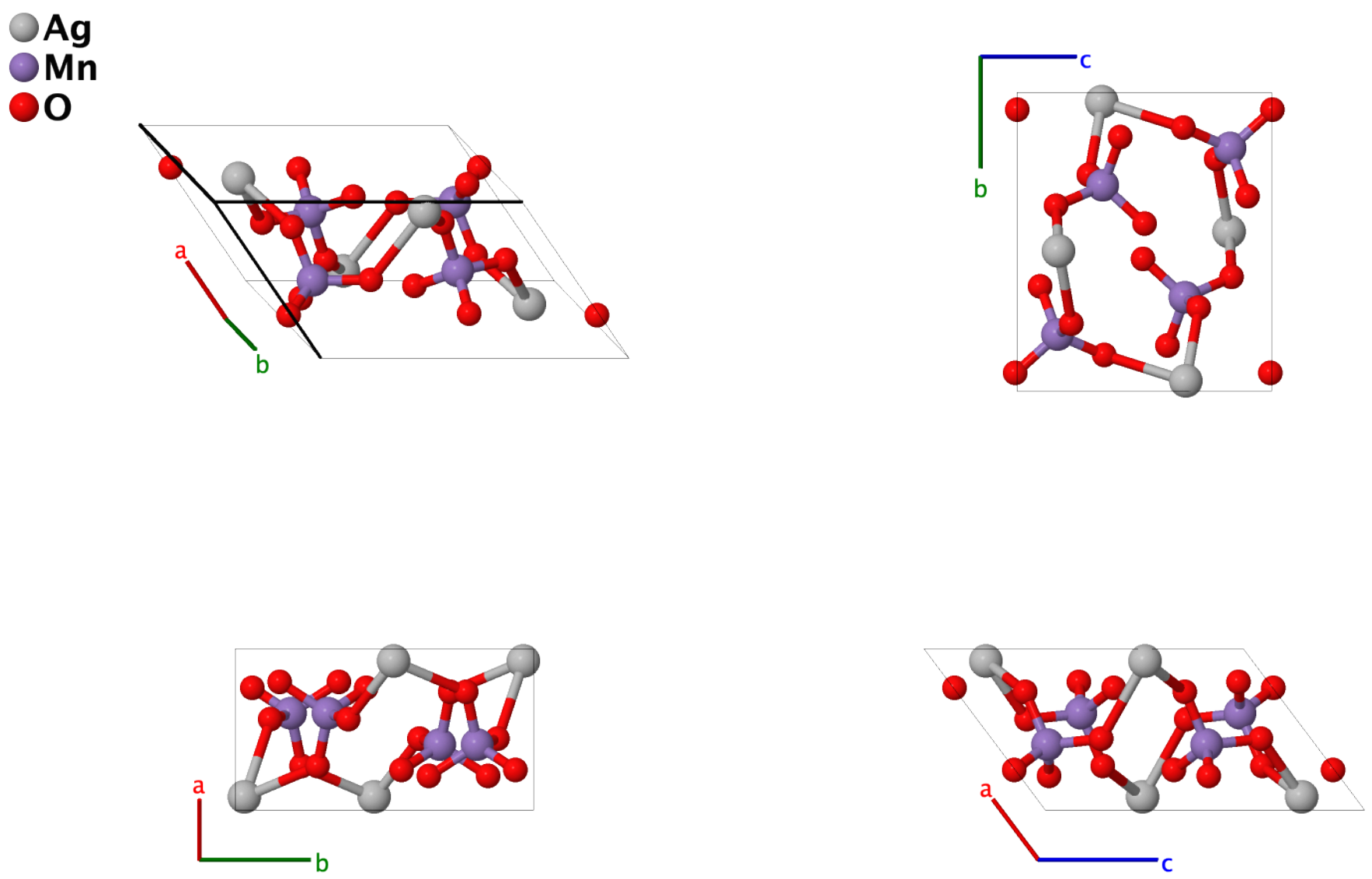

| | | |
|---|---|---|
| **Prototype** | : | $AgMnO_4$ |
| **AFLOW prototype label** | : | ABC4_mP24_14_e_e_4e |
| ***Strukturbericht* designation** | : | $H0_9$ |
| **Pearson symbol** | : | mP24 |
| **Space group number** | : | 14 |
| **Space group symbol** | : | $P2_1/c$ |
| **AFLOW prototype command** | : | `aflow --proto=ABC4_mP24_14_e_e_4e` `--params=`$a, b/a, c/a, \beta, x_1, y_1, z_1, x_2, y_2, z_2, x_3, y_3, z_3, x_4, y_4, z_4, x_5, y_5, z_5, x_6, y_6, z_6$ |

**Other compounds with this structure**

- $CaSO_4$, $SrCrO_4$, $InVO_4$, $CsSO_4$, and $(NH_4)SO_4$

- The structure of $AgMnO_4$ was originally determined by (Sasvári, 1938) and refined by (Boonstra, 1968). In particular, Boonstra found that the actual positions of the Ag and Mn atoms were switched compared to the structure found by Sasvári. As the lattice constants remain approximately the same and there was no change in the space group or the occupied Wyckoff positions, we use the Boonstra structure to define *Strukturbericht* designation $H0_9$.
- Boonstra gave the structure in setting $P2_1/n$ of space group #14. We used FINDSYM to translate the structure to the standard $P2_1/c$ setting. This included a redefinition of the primitive vectors of the lattice.

**Simple Monoclinic primitive vectors:**

$$
\begin{aligned}
\mathbf{a}_1 &= a\,\hat{\mathbf{x}} \\
\mathbf{a}_2 &= b\,\hat{\mathbf{y}} \\
\mathbf{a}_3 &= c\cos\beta\,\hat{\mathbf{x}} + c\sin\beta\,\hat{\mathbf{z}}
\end{aligned}
$$

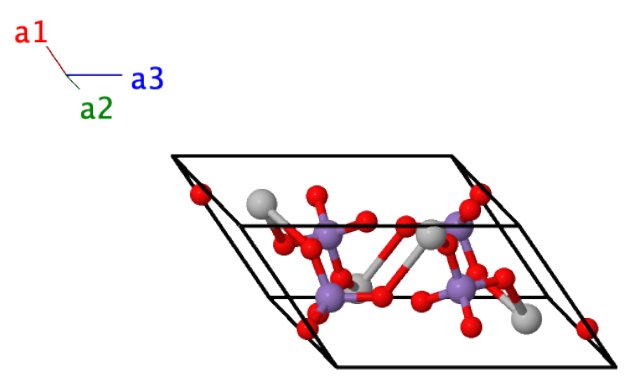

**Basis vectors:**

| | | Lattice Coordinates | | Cartesian Coordinates | Wyckoff Position | Atom Type |
|---|---|---|---|---|---|---|
| $\mathbf{B}_1$ | = | $x_1\,\mathbf{a}_1 + y_1\,\mathbf{a}_2 + z_1\,\mathbf{a}_3$ | = | $(x_1 a + z_1 c\cos\beta)\,\hat{\mathbf{x}} + y_1 b\,\hat{\mathbf{y}} + z_1 c\sin\beta\,\hat{\mathbf{z}}$ | (4$e$) | Ag |
| $\mathbf{B}_2$ | = | $-x_1\,\mathbf{a}_1 + \left(\frac{1}{2} + y_1\right)\mathbf{a}_2 + \left(\frac{1}{2} - z_1\right)\mathbf{a}_3$ | = | $\left(\frac{1}{2}c\cos\beta - x_1 a - z_1 c\cos\beta\right)\hat{\mathbf{x}} + \left(\frac{1}{2} + y_1\right)b\,\hat{\mathbf{y}} + \left(\frac{1}{2} - z_1\right)c\sin\beta\,\hat{\mathbf{z}}$ | (4$e$) | Ag |
| $\mathbf{B}_3$ | = | $-x_1\,\mathbf{a}_1 - y_1\,\mathbf{a}_2 - z_1\,\mathbf{a}_3$ | = | $(-x_1 a - z_1 c\cos\beta)\,\hat{\mathbf{x}} - y_1 b\,\hat{\mathbf{y}} - z_1 c\sin\beta\,\hat{\mathbf{z}}$ | (4$e$) | Ag |
| $\mathbf{B}_4$ | = | $x_1\,\mathbf{a}_1 + \left(\frac{1}{2} - y_1\right)\mathbf{a}_2 + \left(\frac{1}{2} + z_1\right)\mathbf{a}_3$ | = | $\left(\frac{1}{2}c\cos\beta + x_1 a + z_1 c\cos\beta\right)\hat{\mathbf{x}} + \left(\frac{1}{2} - y_1\right)b\,\hat{\mathbf{y}} + \left(\frac{1}{2} + z_1\right)c\sin\beta\,\hat{\mathbf{z}}$ | (4$e$) | Ag |
| $\mathbf{B}_5$ | = | $x_2\,\mathbf{a}_1 + y_2\,\mathbf{a}_2 + z_2\,\mathbf{a}_3$ | = | $(x_2 a + z_2 c\cos\beta)\,\hat{\mathbf{x}} + y_2 b\,\hat{\mathbf{y}} + z_2 c\sin\beta\,\hat{\mathbf{z}}$ | (4$e$) | Mn |
| $\mathbf{B}_6$ | = | $-x_2\,\mathbf{a}_1 + \left(\frac{1}{2} + y_2\right)\mathbf{a}_2 + \left(\frac{1}{2} - z_2\right)\mathbf{a}_3$ | = | $\left(\frac{1}{2}c\cos\beta - x_2 a - z_2 c\cos\beta\right)\hat{\mathbf{x}} + \left(\frac{1}{2} + y_2\right)b\,\hat{\mathbf{y}} + \left(\frac{1}{2} - z_2\right)c\sin\beta\,\hat{\mathbf{z}}$ | (4$e$) | Mn |
| $\mathbf{B}_7$ | = | $-x_2\,\mathbf{a}_1 - y_2\,\mathbf{a}_2 - z_2\,\mathbf{a}_3$ | = | $(-x_2 a - z_2 c\cos\beta)\,\hat{\mathbf{x}} - y_2 b\,\hat{\mathbf{y}} - z_2 c\sin\beta\,\hat{\mathbf{z}}$ | (4$e$) | Mn |
| $\mathbf{B}_8$ | = | $x_2\,\mathbf{a}_1 + \left(\frac{1}{2} - y_2\right)\mathbf{a}_2 + \left(\frac{1}{2} + z_2\right)\mathbf{a}_3$ | = | $\left(\frac{1}{2}c\cos\beta + x_2 a + z_2 c\cos\beta\right)\hat{\mathbf{x}} + \left(\frac{1}{2} - y_2\right)b\,\hat{\mathbf{y}} + \left(\frac{1}{2} + z_2\right)c\sin\beta\,\hat{\mathbf{z}}$ | (4$e$) | Mn |
| $\mathbf{B}_9$ | = | $x_3\,\mathbf{a}_1 + y_3\,\mathbf{a}_2 + z_3\,\mathbf{a}_3$ | = | $(x_3 a + z_3 c\cos\beta)\,\hat{\mathbf{x}} + y_3 b\,\hat{\mathbf{y}} + z_3 c\sin\beta\,\hat{\mathbf{z}}$ | (4$e$) | O I |
| $\mathbf{B}_{10}$ | = | $-x_3\,\mathbf{a}_1 + \left(\frac{1}{2} + y_3\right)\mathbf{a}_2 + \left(\frac{1}{2} - z_3\right)\mathbf{a}_3$ | = | $\left(\frac{1}{2}c\cos\beta - x_3 a - z_3 c\cos\beta\right)\hat{\mathbf{x}} + \left(\frac{1}{2} + y_3\right)b\,\hat{\mathbf{y}} + \left(\frac{1}{2} - z_3\right)c\sin\beta\,\hat{\mathbf{z}}$ | (4$e$) | O I |
| $\mathbf{B}_{11}$ | = | $-x_3\,\mathbf{a}_1 - y_3\,\mathbf{a}_2 - z_3\,\mathbf{a}_3$ | = | $(-x_3 a - z_3 c\cos\beta)\,\hat{\mathbf{x}} - y_3 b\,\hat{\mathbf{y}} - z_3 c\sin\beta\,\hat{\mathbf{z}}$ | (4$e$) | O I |
| $\mathbf{B}_{12}$ | = | $x_3\,\mathbf{a}_1 + \left(\frac{1}{2} - y_3\right)\mathbf{a}_2 + \left(\frac{1}{2} + z_3\right)\mathbf{a}_3$ | = | $\left(\frac{1}{2}c\cos\beta + x_3 a + z_3 c\cos\beta\right)\hat{\mathbf{x}} + \left(\frac{1}{2} - y_3\right)b\,\hat{\mathbf{y}} + \left(\frac{1}{2} + z_3\right)c\sin\beta\,\hat{\mathbf{z}}$ | (4$e$) | O I |
| $\mathbf{B}_{13}$ | = | $x_4\,\mathbf{a}_1 + y_4\,\mathbf{a}_2 + z_4\,\mathbf{a}_3$ | = | $(x_4 a + z_4 c\cos\beta)\,\hat{\mathbf{x}} + y_4 b\,\hat{\mathbf{y}} + z_4 c\sin\beta\,\hat{\mathbf{z}}$ | (4$e$) | O II |
| $\mathbf{B}_{14}$ | = | $-x_4\,\mathbf{a}_1 + \left(\frac{1}{2} + y_4\right)\mathbf{a}_2 + \left(\frac{1}{2} - z_4\right)\mathbf{a}_3$ | = | $\left(\frac{1}{2}c\cos\beta - x_4 a - z_4 c\cos\beta\right)\hat{\mathbf{x}} + \left(\frac{1}{2} + y_4\right)b\,\hat{\mathbf{y}} + \left(\frac{1}{2} - z_4\right)c\sin\beta\,\hat{\mathbf{z}}$ | (4$e$) | O II |
| $\mathbf{B}_{15}$ | = | $-x_4\,\mathbf{a}_1 - y_4\,\mathbf{a}_2 - z_4\,\mathbf{a}_3$ | = | $(-x_4 a - z_4 c\cos\beta)\,\hat{\mathbf{x}} - y_4 b\,\hat{\mathbf{y}} - z_4 c\sin\beta\,\hat{\mathbf{z}}$ | (4$e$) | O II |
| $\mathbf{B}_{16}$ | = | $x_4\,\mathbf{a}_1 + \left(\frac{1}{2} - y_4\right)\mathbf{a}_2 + \left(\frac{1}{2} + z_4\right)\mathbf{a}_3$ | = | $\left(\frac{1}{2}c\cos\beta + x_4 a + z_4 c\cos\beta\right)\hat{\mathbf{x}} + \left(\frac{1}{2} - y_4\right)b\,\hat{\mathbf{y}} + \left(\frac{1}{2} + z_4\right)c\sin\beta\,\hat{\mathbf{z}}$ | (4$e$) | O II |
| $\mathbf{B}_{17}$ | = | $x_5\,\mathbf{a}_1 + y_5\,\mathbf{a}_2 + z_5\,\mathbf{a}_3$ | = | $(x_5 a + z_5 c\cos\beta)\,\hat{\mathbf{x}} + y_5 b\,\hat{\mathbf{y}} + z_5 c\sin\beta\,\hat{\mathbf{z}}$ | (4$e$) | O III |
| $\mathbf{B}_{18}$ | = | $-x_5\,\mathbf{a}_1 + \left(\frac{1}{2} + y_5\right)\mathbf{a}_2 + \left(\frac{1}{2} - z_5\right)\mathbf{a}_3$ | = | $\left(\frac{1}{2}c\cos\beta - x_5 a - z_5 c\cos\beta\right)\hat{\mathbf{x}} + \left(\frac{1}{2} + y_5\right)b\,\hat{\mathbf{y}} + \left(\frac{1}{2} - z_5\right)c\sin\beta\,\hat{\mathbf{z}}$ | (4$e$) | O III |
| $\mathbf{B}_{19}$ | = | $-x_5\,\mathbf{a}_1 - y_5\,\mathbf{a}_2 - z_5\,\mathbf{a}_3$ | = | $(-x_5 a - z_5 c\cos\beta)\,\hat{\mathbf{x}} - y_5 b\,\hat{\mathbf{y}} - z_5 c\sin\beta\,\hat{\mathbf{z}}$ | (4$e$) | O III |
| $\mathbf{B}_{20}$ | = | $x_5\,\mathbf{a}_1 + \left(\frac{1}{2} - y_5\right)\mathbf{a}_2 + \left(\frac{1}{2} + z_5\right)\mathbf{a}_3$ | = | $\left(\frac{1}{2}c\cos\beta + x_5 a + z_5 c\cos\beta\right)\hat{\mathbf{x}} + \left(\frac{1}{2} - y_5\right)b\,\hat{\mathbf{y}} + \left(\frac{1}{2} + z_5\right)c\sin\beta\,\hat{\mathbf{z}}$ | (4$e$) | O III |
| $\mathbf{B}_{21}$ | = | $x_6\,\mathbf{a}_1 + y_6\,\mathbf{a}_2 + z_6\,\mathbf{a}_3$ | = | $(x_6 a + z_6 c\cos\beta)\,\hat{\mathbf{x}} + y_6 b\,\hat{\mathbf{y}} + z_6 c\sin\beta\,\hat{\mathbf{z}}$ | (4$e$) | O IV |
| $\mathbf{B}_{22}$ | = | $-x_6\,\mathbf{a}_1 + \left(\frac{1}{2} + y_6\right)\,\mathbf{a}_2 + \left(\frac{1}{2} - z_6\right)\,\mathbf{a}_3$ | = | $\left(\frac{1}{2}c\cos\beta - x_6 a - z_6 c\cos\beta\right)\,\hat{\mathbf{x}} + \left(\frac{1}{2} + y_6\right) b\,\hat{\mathbf{y}} + \left(\frac{1}{2} - z_6\right) c\sin\beta\,\hat{\mathbf{z}}$ | (4$e$) | O IV |
| $\mathbf{B}_{23}$ | = | $-x_6\,\mathbf{a}_1 - y_6\,\mathbf{a}_2 - z_6\,\mathbf{a}_3$ | = | $(-x_6 a - z_6 c\cos\beta)\,\hat{\mathbf{x}} - y_6 b\,\hat{\mathbf{y}} - z_6 c\sin\beta\,\hat{\mathbf{z}}$ | (4$e$) | O IV |
| $\mathbf{B}_{24}$ | = | $x_6\,\mathbf{a}_1 + \left(\frac{1}{2} - y_6\right)\,\mathbf{a}_2 + \left(\frac{1}{2} + z_6\right)\,\mathbf{a}_3$ | = | $\left(\frac{1}{2}c\cos\beta + x_6 a + z_6 c\cos\beta\right)\,\hat{\mathbf{x}} + \left(\frac{1}{2} - y_6\right) b\,\hat{\mathbf{y}} + \left(\frac{1}{2} + z_6\right) c\sin\beta\,\hat{\mathbf{z}}$ | (4$e$) | O IV |

**Geometry files:**

- CIF: pp. 1555
- POSCAR: pp. 1555

# Nahcolite ($NaHCO_3$, $G0_{12}$) Structure: ABCD3_mP24_14_e_e_e_3e

http://aflow.org/prototype-encyclopedia/ABCD3_mP24_14_e_e_e_3e

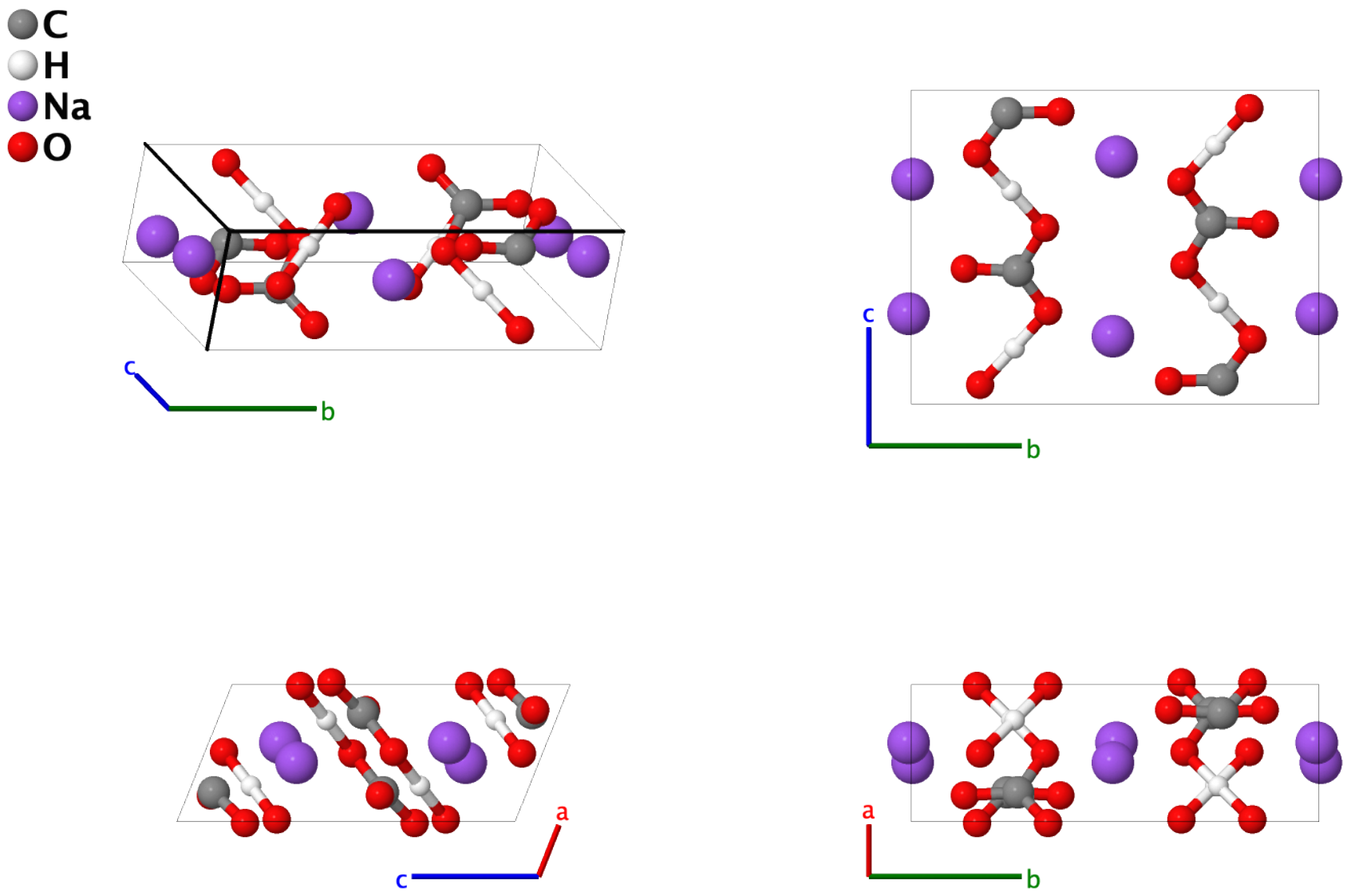

| | | |
|---|---|---|
| **Prototype** | : | $CHNaO_3$ |
| **AFLOW prototype label** | : | ABCD3_mP24_14_e_e_e_3e |
| ***Strukturbericht* designation** | : | $G0_{12}$ |
| **Pearson symbol** | : | mP24 |
| **Space group number** | : | 14 |
| **Space group symbol** | : | $P2_1/c$ |
| **AFLOW prototype command** | : | `aflow --proto=ABCD3_mP24_14_e_e_e_3e` `--params=`$a, b/a, c/a, \beta, x_1, y_1, z_1, x_2, y_2, z_2, x_3, y_3, z_3, x_4, y_4, z_4, x_5, y_5, z_5, x_6, y_6, z_6$ |

### Other compounds with this structure

- $KHCO_3$

---

- This is more commonly known as sodium bicarbonate or baking soda.
- (Sass, 1962) do not locate the hydrogen atoms.
- (Zachariasen, 1933) applied chemical and crystallographic reasoning and placed the hydrogen atoms halfway between each O II atom and the nearest O III atom. We use this structure, while realizing that the hydrogen positions are only approximate.

**Simple Monoclinic primitive vectors:**

$$\mathbf{a}_1 = a\,\hat{\mathbf{x}}$$
$$\mathbf{a}_2 = b\,\hat{\mathbf{y}}$$
$$\mathbf{a}_3 = c\cos\beta\,\hat{\mathbf{x}} + c\sin\beta\,\hat{\mathbf{z}}$$

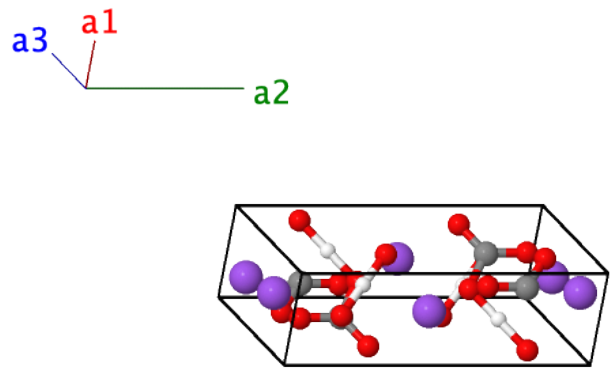

**Basis vectors:**

| | | Lattice Coordinates | | Cartesian Coordinates | Wyckoff Position | Atom Type |
|---|---|---|---|---|---|---|
| $\mathbf{B}_1$ | = | $x_1\,\mathbf{a}_1 + y_1\,\mathbf{a}_2 + z_1\,\mathbf{a}_3$ | = | $(x_1 a + z_1 c\cos\beta)\,\hat{\mathbf{x}} + y_1 b\,\hat{\mathbf{y}} + z_1 c\sin\beta\,\hat{\mathbf{z}}$ | (4e) | C |
| $\mathbf{B}_2$ | = | $-x_1\,\mathbf{a}_1 + \left(\frac{1}{2} + y_1\right)\mathbf{a}_2 + \left(\frac{1}{2} - z_1\right)\mathbf{a}_3$ | = | $\left(\frac{1}{2}c\cos\beta - x_1 a - z_1 c\cos\beta\right)\hat{\mathbf{x}} + \left(\frac{1}{2} + y_1\right)b\,\hat{\mathbf{y}} + \left(\frac{1}{2} - z_1\right)c\sin\beta\,\hat{\mathbf{z}}$ | (4e) | C |
| $\mathbf{B}_3$ | = | $-x_1\,\mathbf{a}_1 - y_1\,\mathbf{a}_2 - z_1\,\mathbf{a}_3$ | = | $(-x_1 a - z_1 c\cos\beta)\,\hat{\mathbf{x}} - y_1 b\,\hat{\mathbf{y}} - z_1 c\sin\beta\,\hat{\mathbf{z}}$ | (4e) | C |
| $\mathbf{B}_4$ | = | $x_1\,\mathbf{a}_1 + \left(\frac{1}{2} - y_1\right)\mathbf{a}_2 + \left(\frac{1}{2} + z_1\right)\mathbf{a}_3$ | = | $\left(\frac{1}{2}c\cos\beta + x_1 a + z_1 c\cos\beta\right)\hat{\mathbf{x}} + \left(\frac{1}{2} - y_1\right)b\,\hat{\mathbf{y}} + \left(\frac{1}{2} + z_1\right)c\sin\beta\,\hat{\mathbf{z}}$ | (4e) | C |
| $\mathbf{B}_5$ | = | $x_2\,\mathbf{a}_1 + y_2\,\mathbf{a}_2 + z_2\,\mathbf{a}_3$ | = | $(x_2 a + z_2 c\cos\beta)\,\hat{\mathbf{x}} + y_2 b\,\hat{\mathbf{y}} + z_2 c\sin\beta\,\hat{\mathbf{z}}$ | (4e) | H |
| $\mathbf{B}_6$ | = | $-x_2\,\mathbf{a}_1 + \left(\frac{1}{2} + y_2\right)\mathbf{a}_2 + \left(\frac{1}{2} - z_2\right)\mathbf{a}_3$ | = | $\left(\frac{1}{2}c\cos\beta - x_2 a - z_2 c\cos\beta\right)\hat{\mathbf{x}} + \left(\frac{1}{2} + y_2\right)b\,\hat{\mathbf{y}} + \left(\frac{1}{2} - z_2\right)c\sin\beta\,\hat{\mathbf{z}}$ | (4e) | H |
| $\mathbf{B}_7$ | = | $-x_2\,\mathbf{a}_1 - y_2\,\mathbf{a}_2 - z_2\,\mathbf{a}_3$ | = | $(-x_2 a - z_2 c\cos\beta)\,\hat{\mathbf{x}} - y_2 b\,\hat{\mathbf{y}} - z_2 c\sin\beta\,\hat{\mathbf{z}}$ | (4e) | H |
| $\mathbf{B}_8$ | = | $x_2\,\mathbf{a}_1 + \left(\frac{1}{2} - y_2\right)\mathbf{a}_2 + \left(\frac{1}{2} + z_2\right)\mathbf{a}_3$ | = | $\left(\frac{1}{2}c\cos\beta + x_2 a + z_2 c\cos\beta\right)\hat{\mathbf{x}} + \left(\frac{1}{2} - y_2\right)b\,\hat{\mathbf{y}} + \left(\frac{1}{2} + z_2\right)c\sin\beta\,\hat{\mathbf{z}}$ | (4e) | H |
| $\mathbf{B}_9$ | = | $x_3\,\mathbf{a}_1 + y_3\,\mathbf{a}_2 + z_3\,\mathbf{a}_3$ | = | $(x_3 a + z_3 c\cos\beta)\,\hat{\mathbf{x}} + y_3 b\,\hat{\mathbf{y}} + z_3 c\sin\beta\,\hat{\mathbf{z}}$ | (4e) | Na |
| $\mathbf{B}_{10}$ | = | $-x_3\,\mathbf{a}_1 + \left(\frac{1}{2} + y_3\right)\mathbf{a}_2 + \left(\frac{1}{2} - z_3\right)\mathbf{a}_3$ | = | $\left(\frac{1}{2}c\cos\beta - x_3 a - z_3 c\cos\beta\right)\hat{\mathbf{x}} + \left(\frac{1}{2} + y_3\right)b\,\hat{\mathbf{y}} + \left(\frac{1}{2} - z_3\right)c\sin\beta\,\hat{\mathbf{z}}$ | (4e) | Na |
| $\mathbf{B}_{11}$ | = | $-x_3\,\mathbf{a}_1 - y_3\,\mathbf{a}_2 - z_3\,\mathbf{a}_3$ | = | $(-x_3 a - z_3 c\cos\beta)\,\hat{\mathbf{x}} - y_3 b\,\hat{\mathbf{y}} - z_3 c\sin\beta\,\hat{\mathbf{z}}$ | (4e) | Na |
| $\mathbf{B}_{12}$ | = | $x_3\,\mathbf{a}_1 + \left(\frac{1}{2} - y_3\right)\mathbf{a}_2 + \left(\frac{1}{2} + z_3\right)\mathbf{a}_3$ | = | $\left(\frac{1}{2}c\cos\beta + x_3 a + z_3 c\cos\beta\right)\hat{\mathbf{x}} + \left(\frac{1}{2} - y_3\right)b\,\hat{\mathbf{y}} + \left(\frac{1}{2} + z_3\right)c\sin\beta\,\hat{\mathbf{z}}$ | (4e) | Na |
| $\mathbf{B}_{13}$ | = | $x_4\,\mathbf{a}_1 + y_4\,\mathbf{a}_2 + z_4\,\mathbf{a}_3$ | = | $(x_4 a + z_4 c\cos\beta)\,\hat{\mathbf{x}} + y_4 b\,\hat{\mathbf{y}} + z_4 c\sin\beta\,\hat{\mathbf{z}}$ | (4e) | O I |
| $\mathbf{B}_{14}$ | = | $-x_4\,\mathbf{a}_1 + \left(\frac{1}{2} + y_4\right)\mathbf{a}_2 + \left(\frac{1}{2} - z_4\right)\mathbf{a}_3$ | = | $\left(\frac{1}{2}c\cos\beta - x_4 a - z_4 c\cos\beta\right)\hat{\mathbf{x}} + \left(\frac{1}{2} + y_4\right)b\,\hat{\mathbf{y}} + \left(\frac{1}{2} - z_4\right)c\sin\beta\,\hat{\mathbf{z}}$ | (4e) | O I |
| $\mathbf{B}_{15}$ | = | $-x_4\,\mathbf{a}_1 - y_4\,\mathbf{a}_2 - z_4\,\mathbf{a}_3$ | = | $(-x_4 a - z_4 c\cos\beta)\,\hat{\mathbf{x}} - y_4 b\,\hat{\mathbf{y}} - z_4 c\sin\beta\,\hat{\mathbf{z}}$ | (4e) | O I |
| $\mathbf{B}_{16}$ | = | $x_4\,\mathbf{a}_1 + \left(\frac{1}{2} - y_4\right)\mathbf{a}_2 + \left(\frac{1}{2} + z_4\right)\mathbf{a}_3$ | = | $\left(\frac{1}{2}c\cos\beta + x_4 a + z_4 c\cos\beta\right)\hat{\mathbf{x}} + \left(\frac{1}{2} - y_4\right)b\,\hat{\mathbf{y}} + \left(\frac{1}{2} + z_4\right)c\sin\beta\,\hat{\mathbf{z}}$ | (4e) | O I |
| $\mathbf{B}_{17}$ | = | $x_5\,\mathbf{a}_1 + y_5\,\mathbf{a}_2 + z_5\,\mathbf{a}_3$ | = | $(x_5 a + z_5 c\cos\beta)\,\hat{\mathbf{x}} + y_5 b\,\hat{\mathbf{y}} + z_5 c\sin\beta\,\hat{\mathbf{z}}$ | (4e) | O II |
| $\mathbf{B}_{18}$ | = | $-x_5\,\mathbf{a}_1 + \left(\frac{1}{2} + y_5\right)\mathbf{a}_2 + \left(\frac{1}{2} - z_5\right)\mathbf{a}_3$ | = | $\left(\frac{1}{2}c\cos\beta - x_5 a - z_5 c\cos\beta\right)\hat{\mathbf{x}} + \left(\frac{1}{2} + y_5\right)b\,\hat{\mathbf{y}} + \left(\frac{1}{2} - z_5\right)c\sin\beta\,\hat{\mathbf{z}}$ | (4$e$) | O II |
| $\mathbf{B}_{19}$ | = | $-x_5\,\mathbf{a}_1 - y_5\,\mathbf{a}_2 - z_5\,\mathbf{a}_3$ | = | $(-x_5 a - z_5 c\cos\beta)\,\hat{\mathbf{x}} - y_5 b\,\hat{\mathbf{y}} - z_5 c\sin\beta\,\hat{\mathbf{z}}$ | (4$e$) | O II |
| $\mathbf{B}_{20}$ | = | $x_5\,\mathbf{a}_1 + \left(\frac{1}{2} - y_5\right)\mathbf{a}_2 + \left(\frac{1}{2} + z_5\right)\mathbf{a}_3$ | = | $\left(\frac{1}{2}c\cos\beta + x_5 a + z_5 c\cos\beta\right)\hat{\mathbf{x}} + \left(\frac{1}{2} - y_5\right)b\,\hat{\mathbf{y}} + \left(\frac{1}{2} + z_5\right)c\sin\beta\,\hat{\mathbf{z}}$ | (4$e$) | O II |
| $\mathbf{B}_{21}$ | = | $x_6\,\mathbf{a}_1 + y_6\,\mathbf{a}_2 + z_6\,\mathbf{a}_3$ | = | $(x_6 a + z_6 c\cos\beta)\,\hat{\mathbf{x}} + y_6 b\,\hat{\mathbf{y}} + z_6 c\sin\beta\,\hat{\mathbf{z}}$ | (4$e$) | O III |
| $\mathbf{B}_{22}$ | = | $-x_6\,\mathbf{a}_1 + \left(\frac{1}{2} + y_6\right)\mathbf{a}_2 + \left(\frac{1}{2} - z_6\right)\mathbf{a}_3$ | = | $\left(\frac{1}{2}c\cos\beta - x_6 a - z_6 c\cos\beta\right)\hat{\mathbf{x}} + \left(\frac{1}{2} + y_6\right)b\,\hat{\mathbf{y}} + \left(\frac{1}{2} - z_6\right)c\sin\beta\,\hat{\mathbf{z}}$ | (4$e$) | O III |
| $\mathbf{B}_{23}$ | = | $-x_6\,\mathbf{a}_1 - y_6\,\mathbf{a}_2 - z_6\,\mathbf{a}_3$ | = | $(-x_6 a - z_6 c\cos\beta)\,\hat{\mathbf{x}} - y_6 b\,\hat{\mathbf{y}} - z_6 c\sin\beta\,\hat{\mathbf{z}}$ | (4$e$) | O III |
| $\mathbf{B}_{24}$ | = | $x_6\,\mathbf{a}_1 + \left(\frac{1}{2} - y_6\right)\mathbf{a}_2 + \left(\frac{1}{2} + z_6\right)\mathbf{a}_3$ | = | $\left(\frac{1}{2}c\cos\beta + x_6 a + z_6 c\cos\beta\right)\hat{\mathbf{x}} + \left(\frac{1}{2} - y_6\right)b\,\hat{\mathbf{y}} + \left(\frac{1}{2} + z_6\right)c\sin\beta\,\hat{\mathbf{z}}$ | (4$e$) | O III |

**Found in:**

**Geometry files:**

- CIF: pp. 1555
- POSCAR: pp. 1556

# Cu(OH)Cl Structure: ABCD_mP16_14_e_e_e_e

http://aflow.org/prototype-encyclopedia/ABCD_mP16_14_e_e_e_e

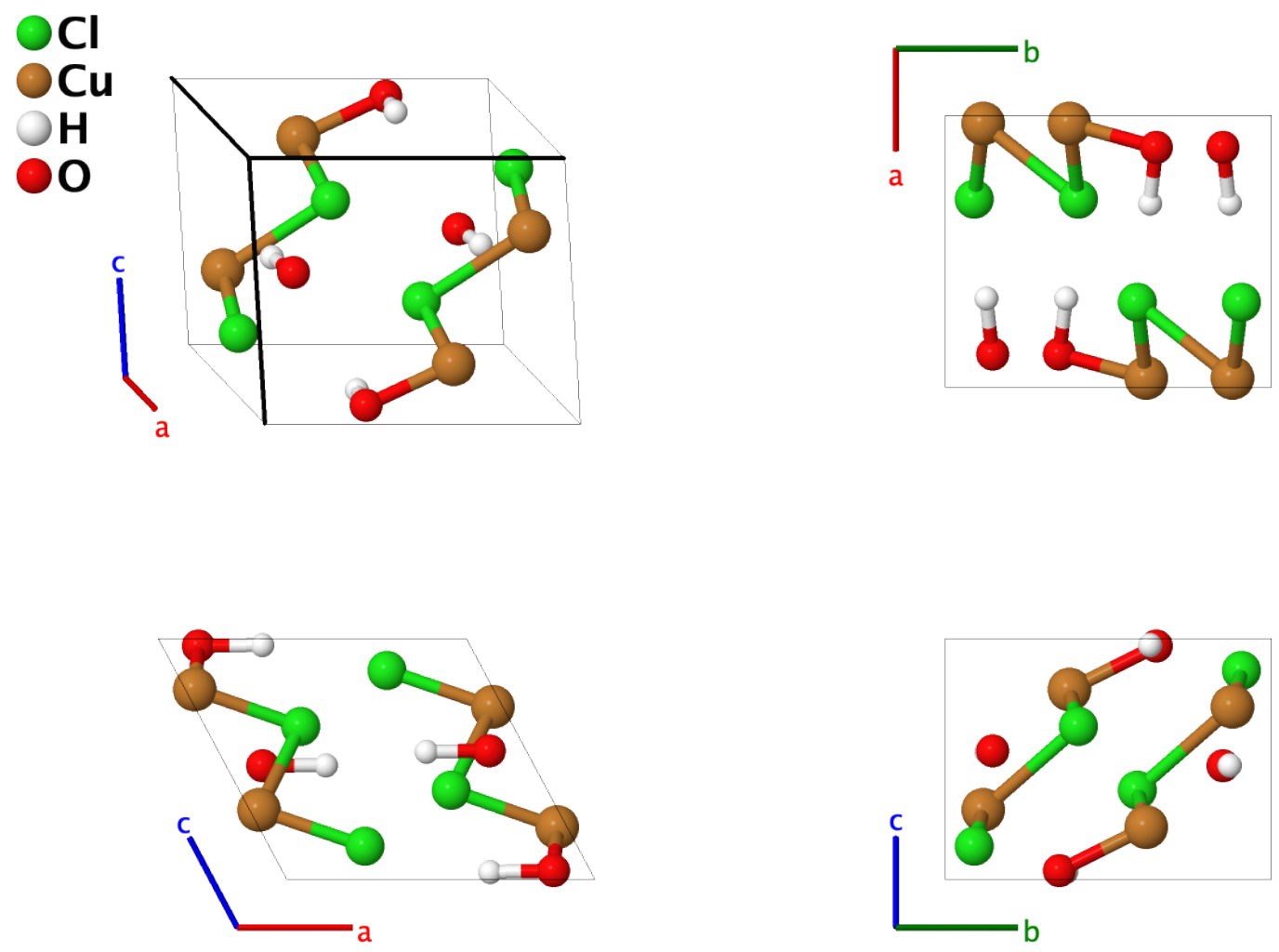

**Prototype** : ClCuHO

**AFLOW prototype label** : ABCD_mP16_14_e_e_e_e

***Strukturbericht* designation** : None

**Pearson symbol** : mP16

**Space group number** : 14

**Space group symbol** : $P2_1/c$

**AFLOW prototype command** : `aflow --proto=ABCD_mP16_14_e_e_e_e`
`--params=`$a, b/a, c/a, \beta, x_1, y_1, z_1, x_2, y_2, z_2, x_3, y_3, z_3, x_4, y_4, z_4$

---

**Simple Monoclinic primitive vectors:**

$$\mathbf{a}_1 = a\,\hat{\mathbf{x}}$$
$$\mathbf{a}_2 = b\,\hat{\mathbf{y}}$$
$$\mathbf{a}_3 = c\cos\beta\,\hat{\mathbf{x}} + c\sin\beta\,\hat{\mathbf{z}}$$

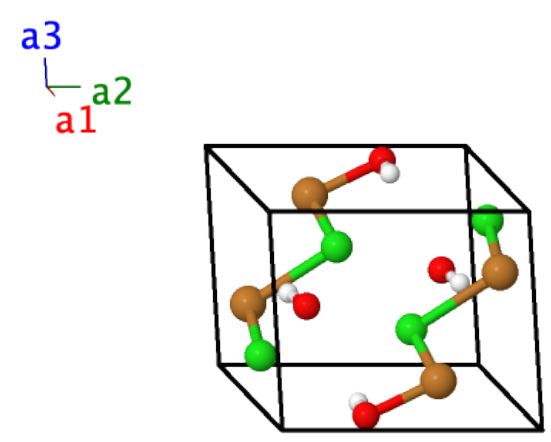

---

**Basis vectors:**

| | | Lattice Coordinates | | Cartesian Coordinates | Wyckoff Position | Atom Type |
|---|---|---|---|---|---|---|
| $\mathbf{B}_1$ | = | $x_1\,\mathbf{a}_1 + y_1\,\mathbf{a}_2 + z_1\,\mathbf{a}_3$ | = | $(x_1 a + z_1 c\cos\beta)\,\hat{\mathbf{x}} + y_1 b\,\hat{\mathbf{y}} + z_1 c\sin\beta\,\hat{\mathbf{z}}$ | (4*e*) | Cl |
| $\mathbf{B}_2$ | = | $-x_1\,\mathbf{a}_1 + \left(\frac{1}{2}+y_1\right)\mathbf{a}_2 + \left(\frac{1}{2}-z_1\right)\mathbf{a}_3$ | = | $\left(\frac{1}{2}c\cos\beta - x_1 a - z_1 c\cos\beta\right)\hat{\mathbf{x}} + \left(\frac{1}{2}+y_1\right)b\,\hat{\mathbf{y}} + \left(\frac{1}{2}-z_1\right)c\sin\beta\,\hat{\mathbf{z}}$ | (4*e*) | Cl |
| $\mathbf{B}_3$ | = | $-x_1\,\mathbf{a}_1 - y_1\,\mathbf{a}_2 - z_1\,\mathbf{a}_3$ | = | $(-x_1 a - z_1 c\cos\beta)\,\hat{\mathbf{x}} - y_1 b\,\hat{\mathbf{y}} - z_1 c\sin\beta\,\hat{\mathbf{z}}$ | (4*e*) | Cl |
| $\mathbf{B}_4$ | = | $x_1\,\mathbf{a}_1 + \left(\frac{1}{2}-y_1\right)\mathbf{a}_2 + \left(\frac{1}{2}+z_1\right)\mathbf{a}_3$ | = | $\left(\frac{1}{2}c\cos\beta + x_1 a + z_1 c\cos\beta\right)\hat{\mathbf{x}} + \left(\frac{1}{2}-y_1\right)b\,\hat{\mathbf{y}} + \left(\frac{1}{2}+z_1\right)c\sin\beta\,\hat{\mathbf{z}}$ | (4*e*) | Cl |
| $\mathbf{B}_5$ | = | $x_2\,\mathbf{a}_1 + y_2\,\mathbf{a}_2 + z_2\,\mathbf{a}_3$ | = | $(x_2 a + z_2 c\cos\beta)\,\hat{\mathbf{x}} + y_2 b\,\hat{\mathbf{y}} + z_2 c\sin\beta\,\hat{\mathbf{z}}$ | (4e) | Cu |
| $\mathbf{B}_6$ | = | $-x_2\,\mathbf{a}_1 + \left(\frac{1}{2} + y_2\right)\mathbf{a}_2 + \left(\frac{1}{2} - z_2\right)\mathbf{a}_3$ | = | $\left(\frac{1}{2}c\cos\beta - x_2 a - z_2 c\cos\beta\right)\hat{\mathbf{x}} + \left(\frac{1}{2} + y_2\right)b\,\hat{\mathbf{y}} + \left(\frac{1}{2} - z_2\right)c\sin\beta\,\hat{\mathbf{z}}$ | (4e) | Cu |
| $\mathbf{B}_7$ | = | $-x_2\,\mathbf{a}_1 - y_2\,\mathbf{a}_2 - z_2\,\mathbf{a}_3$ | = | $(-x_2 a - z_2 c\cos\beta)\,\hat{\mathbf{x}} - y_2 b\,\hat{\mathbf{y}} - z_2 c\sin\beta\,\hat{\mathbf{z}}$ | (4e) | Cu |
| $\mathbf{B}_8$ | = | $x_2\,\mathbf{a}_1 + \left(\frac{1}{2} - y_2\right)\mathbf{a}_2 + \left(\frac{1}{2} + z_2\right)\mathbf{a}_3$ | = | $\left(\frac{1}{2}c\cos\beta + x_2 a + z_2 c\cos\beta\right)\hat{\mathbf{x}} + \left(\frac{1}{2} - y_2\right)b\,\hat{\mathbf{y}} + \left(\frac{1}{2} + z_2\right)c\sin\beta\,\hat{\mathbf{z}}$ | (4e) | Cu |
| $\mathbf{B}_9$ | = | $x_3\,\mathbf{a}_1 + y_3\,\mathbf{a}_2 + z_3\,\mathbf{a}_3$ | = | $(x_3 a + z_3 c\cos\beta)\,\hat{\mathbf{x}} + y_3 b\,\hat{\mathbf{y}} + z_3 c\sin\beta\,\hat{\mathbf{z}}$ | (4e) | H |
| $\mathbf{B}_{10}$ | = | $-x_3\,\mathbf{a}_1 + \left(\frac{1}{2} + y_3\right)\mathbf{a}_2 + \left(\frac{1}{2} - z_3\right)\mathbf{a}_3$ | = | $\left(\frac{1}{2}c\cos\beta - x_3 a - z_3 c\cos\beta\right)\hat{\mathbf{x}} + \left(\frac{1}{2} + y_3\right)b\,\hat{\mathbf{y}} + \left(\frac{1}{2} - z_3\right)c\sin\beta\,\hat{\mathbf{z}}$ | (4e) | H |
| $\mathbf{B}_{11}$ | = | $-x_3\,\mathbf{a}_1 - y_3\,\mathbf{a}_2 - z_3\,\mathbf{a}_3$ | = | $(-x_3 a - z_3 c\cos\beta)\,\hat{\mathbf{x}} - y_3 b\,\hat{\mathbf{y}} - z_3 c\sin\beta\,\hat{\mathbf{z}}$ | (4e) | H |
| $\mathbf{B}_{12}$ | = | $x_3\,\mathbf{a}_1 + \left(\frac{1}{2} - y_3\right)\mathbf{a}_2 + \left(\frac{1}{2} + z_3\right)\mathbf{a}_3$ | = | $\left(\frac{1}{2}c\cos\beta + x_3 a + z_3 c\cos\beta\right)\hat{\mathbf{x}} + \left(\frac{1}{2} - y_3\right)b\,\hat{\mathbf{y}} + \left(\frac{1}{2} + z_3\right)c\sin\beta\,\hat{\mathbf{z}}$ | (4e) | H |
| $\mathbf{B}_{13}$ | = | $x_4\,\mathbf{a}_1 + y_4\,\mathbf{a}_2 + z_4\,\mathbf{a}_3$ | = | $(x_4 a + z_4 c\cos\beta)\,\hat{\mathbf{x}} + y_4 b\,\hat{\mathbf{y}} + z_4 c\sin\beta\,\hat{\mathbf{z}}$ | (4e) | O |
| $\mathbf{B}_{14}$ | = | $-x_4\,\mathbf{a}_1 + \left(\frac{1}{2} + y_4\right)\mathbf{a}_2 + \left(\frac{1}{2} - z_4\right)\mathbf{a}_3$ | = | $\left(\frac{1}{2}c\cos\beta - x_4 a - z_4 c\cos\beta\right)\hat{\mathbf{x}} + \left(\frac{1}{2} + y_4\right)b\,\hat{\mathbf{y}} + \left(\frac{1}{2} - z_4\right)c\sin\beta\,\hat{\mathbf{z}}$ | (4e) | O |
| $\mathbf{B}_{15}$ | = | $-x_4\,\mathbf{a}_1 - y_4\,\mathbf{a}_2 - z_4\,\mathbf{a}_3$ | = | $(-x_4 a - z_4 c\cos\beta)\,\hat{\mathbf{x}} - y_4 b\,\hat{\mathbf{y}} - z_4 c\sin\beta\,\hat{\mathbf{z}}$ | (4e) | O |
| $\mathbf{B}_{16}$ | = | $x_4\,\mathbf{a}_1 + \left(\frac{1}{2} - y_4\right)\mathbf{a}_2 + \left(\frac{1}{2} + z_4\right)\mathbf{a}_3$ | = | $\left(\frac{1}{2}c\cos\beta + x_4 a + z_4 c\cos\beta\right)\hat{\mathbf{x}} + \left(\frac{1}{2} - y_4\right)b\,\hat{\mathbf{y}} + \left(\frac{1}{2} + z_4\right)c\sin\beta\,\hat{\mathbf{z}}$ | (4e) | O |

**Geometry files:**

- CIF: pp. 1556
- POSCAR: pp. 1556

# Arsenopyrite (FeAsS, $E0_7$) Structure: ABC_mP12_14_e_e_e

http://aflow.org/prototype-encyclopedia/ABC_mP12_14_e_e_e

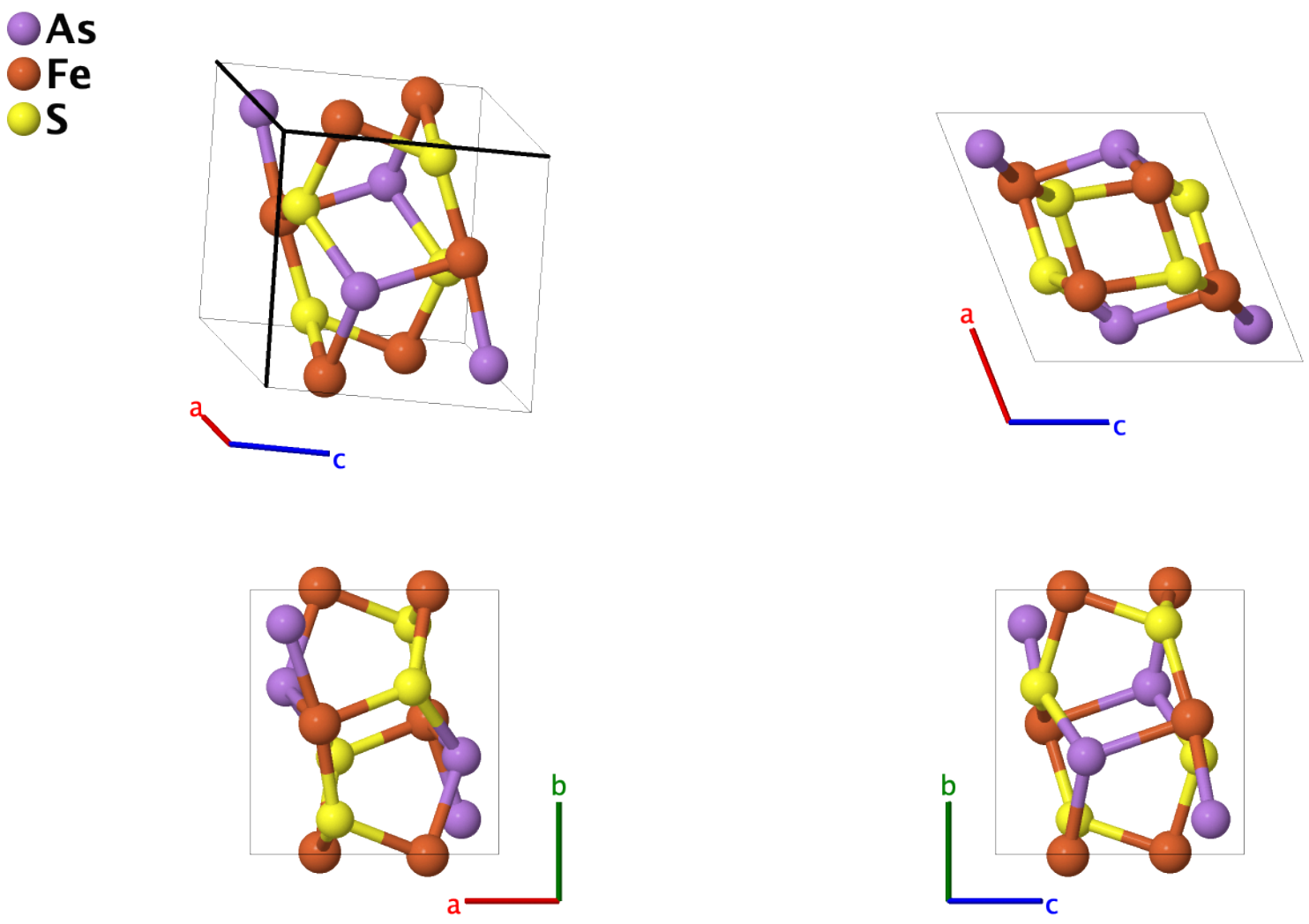

| | | |
|---|---|---|
| **Prototype** | : | AsFeS |
| **AFLOW prototype label** | : | ABC_mP12_14_e_e_e |
| ***Strukturbericht* designation** | : | $E0_7$ |
| **Pearson symbol** | : | mP12 |
| **Space group number** | : | 14 |
| **Space group symbol** | : | $P2_1/c$ |
| **AFLOW prototype command** | : | `aflow --proto=ABC_mP12_14_e_e_e`<br>`--params=`$a, b/a, c/a, \beta, x_1, y_1, z_1, x_2, y_2, z_2, x_3, y_3, z_3$ |

### Other compounds with this structure

- FeSbS (Gudmundite), $CoSb_2$, $IrN_2$, $MoO_2$ (Tugarinovite), $VO_2$, $IrAs_2$ (Iridarsenite), $CoAs_2$ (Clinosafflorite), and $HfO_2$

- This is very similar to $C43$ Baddeleyite, $ZrO_2$.

### Simple Monoclinic primitive vectors:

$$
\begin{aligned}
\mathbf{a}_1 &= a\,\hat{\mathbf{x}} \\
\mathbf{a}_2 &= b\,\hat{\mathbf{y}} \\
\mathbf{a}_3 &= c\cos\beta\,\hat{\mathbf{x}} + c\sin\beta\,\hat{\mathbf{z}}
\end{aligned}
$$

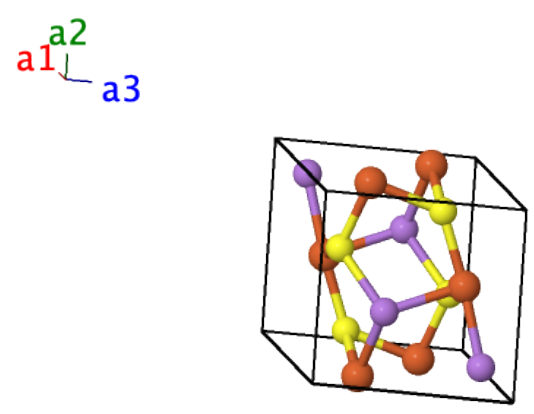

### Basis vectors:

| | | Lattice Coordinates | | Cartesian Coordinates | Wyckoff Position | Atom Type |
|---|---|---|---|---|---|---|
| $\mathbf{B}_1$ | = | $x_1\,\mathbf{a}_1 + y_1\,\mathbf{a}_2 + z_1\,\mathbf{a}_3$ | = | $(x_1 a + z_1 c\cos\beta)\,\hat{\mathbf{x}} + y_1 b\,\hat{\mathbf{y}} + z_1 c\sin\beta\,\hat{\mathbf{z}}$ | (4$e$) | As |
| $\mathbf{B}_{2}$ | = | $-x_{1}\,\mathbf{a}_{1}+\left(\frac{1}{2}+y_{1}\right)\,\mathbf{a}_{2}+\left(\frac{1}{2}-z_{1}\right)\,\mathbf{a}_{3}$ | = | $\left(\frac{1}{2}c\cos\beta-x_{1}a-z_{1}c\cos\beta\right)\,\mathbf{\hat{x}}+\left(\frac{1}{2}+y_{1}\right)b\,\mathbf{\hat{y}}+\left(\frac{1}{2}-z_{1}\right)c\sin\beta\,\mathbf{\hat{z}}$ | (4$e$) | As |
| $\mathbf{B}_{3}$ | = | $-x_{1}\,\mathbf{a}_{1}-y_{1}\,\mathbf{a}_{2}-z_{1}\,\mathbf{a}_{3}$ | = | $(-x_{1}a-z_{1}c\cos\beta)\,\mathbf{\hat{x}}-y_{1}b\,\mathbf{\hat{y}}-z_{1}c\sin\beta\,\mathbf{\hat{z}}$ | (4$e$) | As |
| $\mathbf{B}_{4}$ | = | $x_{1}\,\mathbf{a}_{1}+\left(\frac{1}{2}-y_{1}\right)\,\mathbf{a}_{2}+\left(\frac{1}{2}+z_{1}\right)\,\mathbf{a}_{3}$ | = | $\left(\frac{1}{2}c\cos\beta+x_{1}a+z_{1}c\cos\beta\right)\,\mathbf{\hat{x}}+\left(\frac{1}{2}-y_{1}\right)b\,\mathbf{\hat{y}}+\left(\frac{1}{2}+z_{1}\right)c\sin\beta\,\mathbf{\hat{z}}$ | (4$e$) | As |
| $\mathbf{B}_{5}$ | = | $x_{2}\,\mathbf{a}_{1}+y_{2}\,\mathbf{a}_{2}+z_{2}\,\mathbf{a}_{3}$ | = | $(x_{2}a+z_{2}c\cos\beta)\,\mathbf{\hat{x}}+y_{2}b\,\mathbf{\hat{y}}+z_{2}c\sin\beta\,\mathbf{\hat{z}}$ | (4$e$) | Fe |
| $\mathbf{B}_{6}$ | = | $-x_{2}\,\mathbf{a}_{1}+\left(\frac{1}{2}+y_{2}\right)\,\mathbf{a}_{2}+\left(\frac{1}{2}-z_{2}\right)\,\mathbf{a}_{3}$ | = | $\left(\frac{1}{2}c\cos\beta-x_{2}a-z_{2}c\cos\beta\right)\,\mathbf{\hat{x}}+\left(\frac{1}{2}+y_{2}\right)b\,\mathbf{\hat{y}}+\left(\frac{1}{2}-z_{2}\right)c\sin\beta\,\mathbf{\hat{z}}$ | (4$e$) | Fe |
| $\mathbf{B}_{7}$ | = | $-x_{2}\,\mathbf{a}_{1}-y_{2}\,\mathbf{a}_{2}-z_{2}\,\mathbf{a}_{3}$ | = | $(-x_{2}a-z_{2}c\cos\beta)\,\mathbf{\hat{x}}-y_{2}b\,\mathbf{\hat{y}}-z_{2}c\sin\beta\,\mathbf{\hat{z}}$ | (4$e$) | Fe |
| $\mathbf{B}_{8}$ | = | $x_{2}\,\mathbf{a}_{1}+\left(\frac{1}{2}-y_{2}\right)\,\mathbf{a}_{2}+\left(\frac{1}{2}+z_{2}\right)\,\mathbf{a}_{3}$ | = | $\left(\frac{1}{2}c\cos\beta+x_{2}a+z_{2}c\cos\beta\right)\,\mathbf{\hat{x}}+\left(\frac{1}{2}-y_{2}\right)b\,\mathbf{\hat{y}}+\left(\frac{1}{2}+z_{2}\right)c\sin\beta\,\mathbf{\hat{z}}$ | (4$e$) | Fe |
| $\mathbf{B}_{9}$ | = | $x_{3}\,\mathbf{a}_{1}+y_{3}\,\mathbf{a}_{2}+z_{3}\,\mathbf{a}_{3}$ | = | $(x_{3}a+z_{3}c\cos\beta)\,\mathbf{\hat{x}}+y_{3}b\,\mathbf{\hat{y}}+z_{3}c\sin\beta\,\mathbf{\hat{z}}$ | (4$e$) | S |
| $\mathbf{B}_{10}$ | = | $-x_{3}\,\mathbf{a}_{1}+\left(\frac{1}{2}+y_{3}\right)\,\mathbf{a}_{2}+\left(\frac{1}{2}-z_{3}\right)\,\mathbf{a}_{3}$ | = | $\left(\frac{1}{2}c\cos\beta-x_{3}a-z_{3}c\cos\beta\right)\,\mathbf{\hat{x}}+\left(\frac{1}{2}+y_{3}\right)b\,\mathbf{\hat{y}}+\left(\frac{1}{2}-z_{3}\right)c\sin\beta\,\mathbf{\hat{z}}$ | (4$e$) | S |
| $\mathbf{B}_{11}$ | = | $-x_{3}\,\mathbf{a}_{1}-y_{3}\,\mathbf{a}_{2}-z_{3}\,\mathbf{a}_{3}$ | = | $(-x_{3}a-z_{3}c\cos\beta)\,\mathbf{\hat{x}}-y_{3}b\,\mathbf{\hat{y}}-z_{3}c\sin\beta\,\mathbf{\hat{z}}$ | (4$e$) | S |
| $\mathbf{B}_{12}$ | = | $x_{3}\,\mathbf{a}_{1}+\left(\frac{1}{2}-y_{3}\right)\,\mathbf{a}_{2}+\left(\frac{1}{2}+z_{3}\right)\,\mathbf{a}_{3}$ | = | $\left(\frac{1}{2}c\cos\beta+x_{3}a+z_{3}c\cos\beta\right)\,\mathbf{\hat{x}}+\left(\frac{1}{2}-y_{3}\right)b\,\mathbf{\hat{y}}+\left(\frac{1}{2}+z_{3}\right)c\sin\beta\,\mathbf{\hat{z}}$ | (4$e$) | S |

---

---

**Geometry files:**

- CIF: pp. 1557
- POSCAR: pp. 1557

# $\alpha$-ICl Structure: AB_mP16_14_2e_2e

http://aflow.org/prototype-encyclopedia/AB_mP16_14_2e_2e.ICl

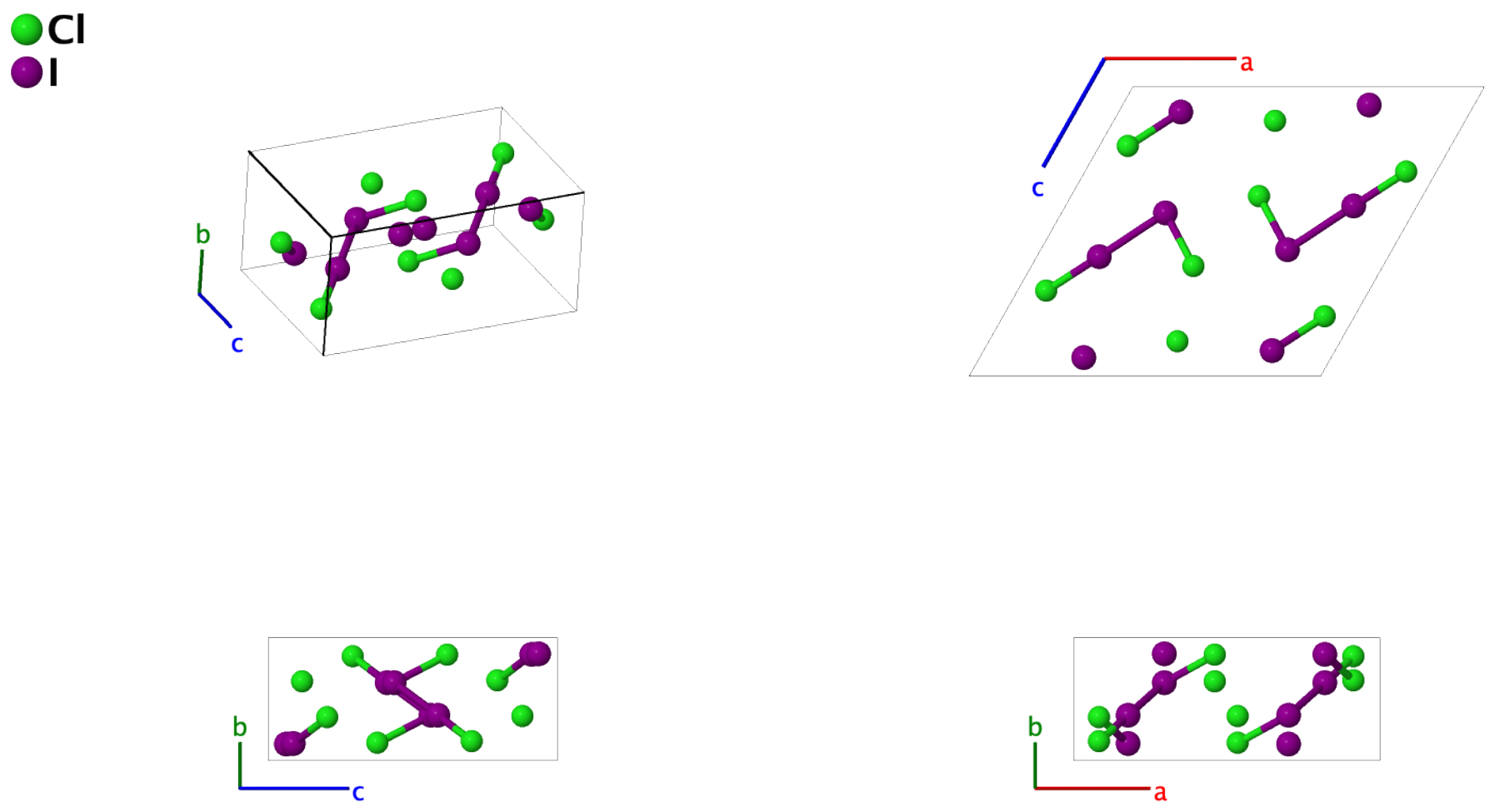

| | | |
|---|---|---|
| **Prototype** | : | ClI |
| **AFLOW prototype label** | : | AB_mP16_14_2e_2e |
| ***Strukturbericht* designation** | : | None |
| **Pearson symbol** | : | mP16 |
| **Space group number** | : | 14 |
| **Space group symbol** | : | $P2_1/c$ |
| **AFLOW prototype command** | : | `aflow --proto=AB_mP16_14_2e_2e`<br>`--params=`$a, b/a, c/a, \beta, x_1, y_1, z_1, x_2, y_2, z_2, x_3, y_3, z_3, x_4, y_4, z_4$ |

- LiAs (AB_mP16_14_2e_2e) and $\alpha$-ICl (AB_mP16_14_2e_2e) have the same AFLOW prototype label. They are generated by the same symmetry operations with different sets of parameters (`--params`) specified in their corresponding CIF files.

**Simple Monoclinic primitive vectors:**

$$\mathbf{a}_1 = a\,\hat{\mathbf{x}}$$
$$\mathbf{a}_2 = b\,\hat{\mathbf{y}}$$
$$\mathbf{a}_3 = c\cos\beta\,\hat{\mathbf{x}} + c\sin\beta\,\hat{\mathbf{z}}$$

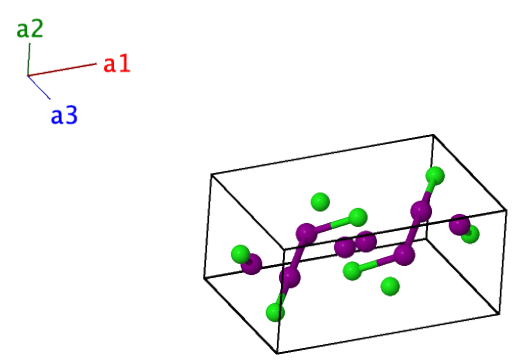

**Basis vectors:**

| | | Lattice Coordinates | | Cartesian Coordinates | Wyckoff Position | Atom Type |
|---|---|---|---|---|---|---|
| $\mathbf{B}_1$ | = | $x_1\,\mathbf{a}_1 + y_1\,\mathbf{a}_2 + z_1\,\mathbf{a}_3$ | = | $(x_1 a + z_1 c\cos\beta)\,\hat{\mathbf{x}} + y_1 b\,\hat{\mathbf{y}} + z_1 c\sin\beta\,\hat{\mathbf{z}}$ | (4*e*) | Cl I |
| $\mathbf{B}_2$ | = | $-x_1\,\mathbf{a}_1 + \left(\frac{1}{2} + y_1\right)\mathbf{a}_2 + \left(\frac{1}{2} - z_1\right)\mathbf{a}_3$ | = | $\left(\frac{1}{2}c\cos\beta - x_1 a - z_1 c\cos\beta\right)\hat{\mathbf{x}} + \left(\frac{1}{2} + y_1\right) b\,\hat{\mathbf{y}} + \left(\frac{1}{2} - z_1\right) c\sin\beta\,\hat{\mathbf{z}}$ | (4*e*) | Cl I |
| $\mathbf{B}_3$ | = | $-x_1\,\mathbf{a}_1 - y_1\,\mathbf{a}_2 - z_1\,\mathbf{a}_3$ | = | $(-x_1 a - z_1 c\cos\beta)\,\hat{\mathbf{x}} - y_1 b\,\hat{\mathbf{y}} - z_1 c\sin\beta\,\hat{\mathbf{z}}$ | (4*e*) | Cl I |
| $\mathbf{B}_{4}$ | = | $x_1\,\mathbf{a}_1 + \left(\frac{1}{2} - y_1\right)\,\mathbf{a}_2 + \left(\frac{1}{2} + z_1\right)\,\mathbf{a}_3$ | = | $\left(\frac{1}{2}c\cos\beta + x_1 a + z_1 c\cos\beta\right)\,\hat{\mathbf{x}} + \left(\frac{1}{2} - y_1\right)b\,\hat{\mathbf{y}} + \left(\frac{1}{2} + z_1\right)c\sin\beta\,\hat{\mathbf{z}}$ | (4e) | Cl I |
| $\mathbf{B}_{5}$ | = | $x_2\,\mathbf{a}_1 + y_2\,\mathbf{a}_2 + z_2\,\mathbf{a}_3$ | = | $(x_2 a + z_2 c\cos\beta)\,\hat{\mathbf{x}} + y_2 b\,\hat{\mathbf{y}} + z_2 c\sin\beta\,\hat{\mathbf{z}}$ | (4e) | Cl II |
| $\mathbf{B}_{6}$ | = | $-x_2\,\mathbf{a}_1 + \left(\frac{1}{2} + y_2\right)\,\mathbf{a}_2 + \left(\frac{1}{2} - z_2\right)\,\mathbf{a}_3$ | = | $\left(\frac{1}{2}c\cos\beta - x_2 a - z_2 c\cos\beta\right)\,\hat{\mathbf{x}} + \left(\frac{1}{2} + y_2\right)b\,\hat{\mathbf{y}} + \left(\frac{1}{2} - z_2\right)c\sin\beta\,\hat{\mathbf{z}}$ | (4e) | Cl II |
| $\mathbf{B}_{7}$ | = | $-x_2\,\mathbf{a}_1 - y_2\,\mathbf{a}_2 - z_2\,\mathbf{a}_3$ | = | $(-x_2 a - z_2 c\cos\beta)\,\hat{\mathbf{x}} - y_2 b\,\hat{\mathbf{y}} - z_2 c\sin\beta\,\hat{\mathbf{z}}$ | (4e) | Cl II |
| $\mathbf{B}_{8}$ | = | $x_2\,\mathbf{a}_1 + \left(\frac{1}{2} - y_2\right)\,\mathbf{a}_2 + \left(\frac{1}{2} + z_2\right)\,\mathbf{a}_3$ | = | $\left(\frac{1}{2}c\cos\beta + x_2 a + z_2 c\cos\beta\right)\,\hat{\mathbf{x}} + \left(\frac{1}{2} - y_2\right)b\,\hat{\mathbf{y}} + \left(\frac{1}{2} + z_2\right)c\sin\beta\,\hat{\mathbf{z}}$ | (4e) | Cl II |
| $\mathbf{B}_{9}$ | = | $x_3\,\mathbf{a}_1 + y_3\,\mathbf{a}_2 + z_3\,\mathbf{a}_3$ | = | $(x_3 a + z_3 c\cos\beta)\,\hat{\mathbf{x}} + y_3 b\,\hat{\mathbf{y}} + z_3 c\sin\beta\,\hat{\mathbf{z}}$ | (4e) | I I |
| $\mathbf{B}_{10}$ | = | $-x_3\,\mathbf{a}_1 + \left(\frac{1}{2} + y_3\right)\,\mathbf{a}_2 + \left(\frac{1}{2} - z_3\right)\,\mathbf{a}_3$ | = | $\left(\frac{1}{2}c\cos\beta - x_3 a - z_3 c\cos\beta\right)\,\hat{\mathbf{x}} + \left(\frac{1}{2} + y_3\right)b\,\hat{\mathbf{y}} + \left(\frac{1}{2} - z_3\right)c\sin\beta\,\hat{\mathbf{z}}$ | (4e) | I I |
| $\mathbf{B}_{11}$ | = | $-x_3\,\mathbf{a}_1 - y_3\,\mathbf{a}_2 - z_3\,\mathbf{a}_3$ | = | $(-x_3 a - z_3 c\cos\beta)\,\hat{\mathbf{x}} - y_3 b\,\hat{\mathbf{y}} - z_3 c\sin\beta\,\hat{\mathbf{z}}$ | (4e) | I I |
| $\mathbf{B}_{12}$ | = | $x_3\,\mathbf{a}_1 + \left(\frac{1}{2} - y_3\right)\,\mathbf{a}_2 + \left(\frac{1}{2} + z_3\right)\,\mathbf{a}_3$ | = | $\left(\frac{1}{2}c\cos\beta + x_3 a + z_3 c\cos\beta\right)\,\hat{\mathbf{x}} + \left(\frac{1}{2} - y_3\right)b\,\hat{\mathbf{y}} + \left(\frac{1}{2} + z_3\right)c\sin\beta\,\hat{\mathbf{z}}$ | (4e) | I I |
| $\mathbf{B}_{13}$ | = | $x_4\,\mathbf{a}_1 + y_4\,\mathbf{a}_2 + z_4\,\mathbf{a}_3$ | = | $(x_4 a + z_4 c\cos\beta)\,\hat{\mathbf{x}} + y_4 b\,\hat{\mathbf{y}} + z_4 c\sin\beta\,\hat{\mathbf{z}}$ | (4e) | I II |
| $\mathbf{B}_{14}$ | = | $-x_4\,\mathbf{a}_1 + \left(\frac{1}{2} + y_4\right)\,\mathbf{a}_2 + \left(\frac{1}{2} - z_4\right)\,\mathbf{a}_3$ | = | $\left(\frac{1}{2}c\cos\beta - x_4 a - z_4 c\cos\beta\right)\,\hat{\mathbf{x}} + \left(\frac{1}{2} + y_4\right)b\,\hat{\mathbf{y}} + \left(\frac{1}{2} - z_4\right)c\sin\beta\,\hat{\mathbf{z}}$ | (4e) | I II |
| $\mathbf{B}_{15}$ | = | $-x_4\,\mathbf{a}_1 - y_4\,\mathbf{a}_2 - z_4\,\mathbf{a}_3$ | = | $(-x_4 a - z_4 c\cos\beta)\,\hat{\mathbf{x}} - y_4 b\,\hat{\mathbf{y}} - z_4 c\sin\beta\,\hat{\mathbf{z}}$ | (4e) | I II |
| $\mathbf{B}_{16}$ | = | $x_4\,\mathbf{a}_1 + \left(\frac{1}{2} - y_4\right)\,\mathbf{a}_2 + \left(\frac{1}{2} + z_4\right)\,\mathbf{a}_3$ | = | $\left(\frac{1}{2}c\cos\beta + x_4 a + z_4 c\cos\beta\right)\,\hat{\mathbf{x}} + \left(\frac{1}{2} - y_4\right)b\,\hat{\mathbf{y}} + \left(\frac{1}{2} + z_4\right)c\sin\beta\,\hat{\mathbf{z}}$ | (4e) | I II |

**Geometry files:**

- CIF: pp. 1557
- POSCAR: pp. 1557

# LiAs Structure: AB_mP16_14_2e_2e

http://aflow.org/prototype-encyclopedia/AB_mP16_14_2e_2e

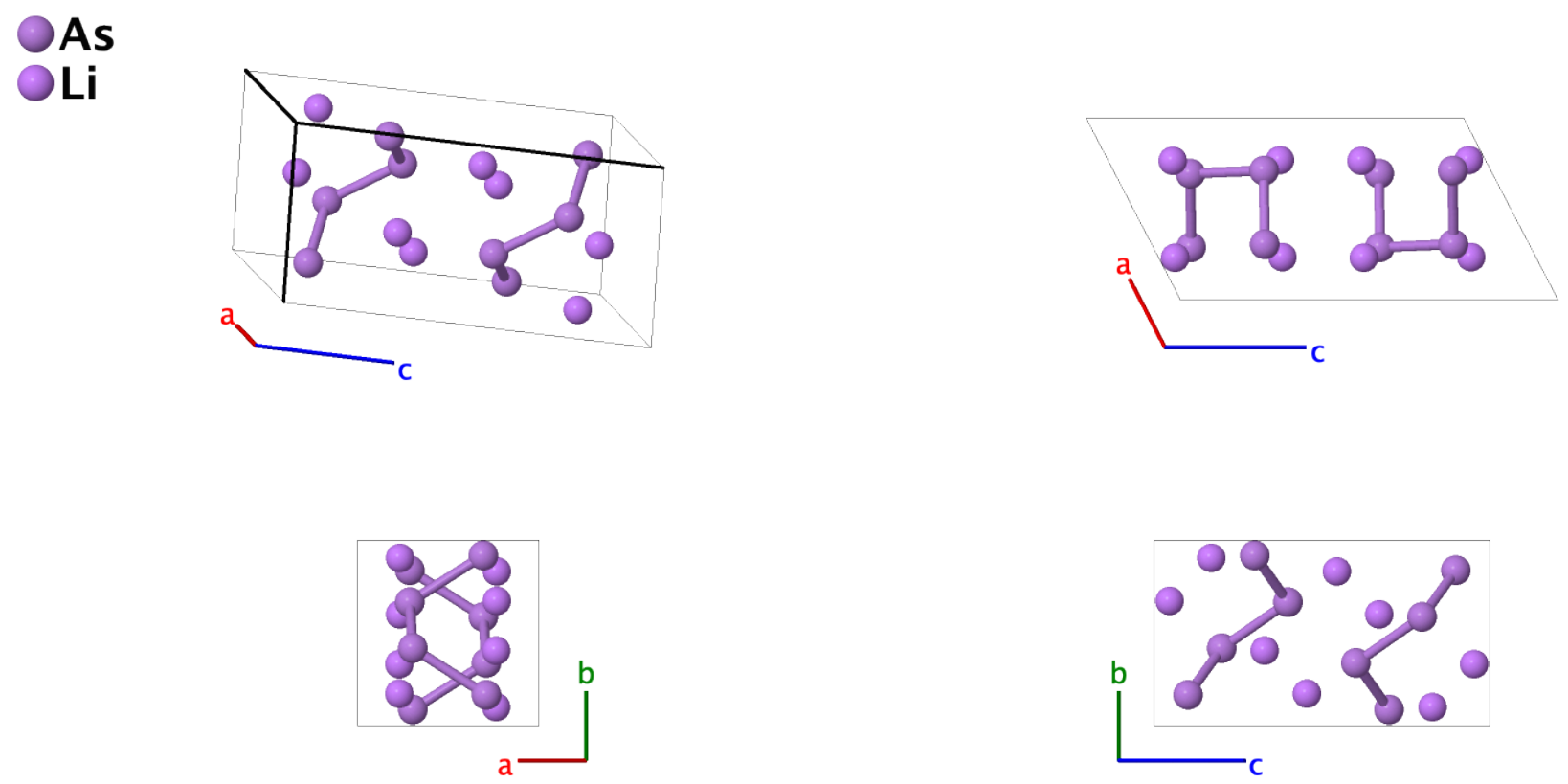

| | | |
|---|---|---|
| **Prototype** | : | AsLi |
| **AFLOW prototype label** | : | AB_mP16_14_2e_2e |
| ***Strukturbericht* designation** | : | None |
| **Pearson symbol** | : | mP16 |
| **Space group number** | : | 14 |
| **Space group symbol** | : | $P2_1/c$ |
| **AFLOW prototype command** | : | `aflow --proto=AB_mP16_14_2e_2e` `--params=`$a, b/a, c/a, \beta, x_1, y_1, z_1, x_2, y_2, z_2, x_3, y_3, z_3, x_4, y_4, z_4$ |

### Other compounds with this structure

- NaSb and KSb

- LiAs (AB_mP16_14_2e_2e) and $\alpha$-ICl (AB_mP16_14_2e_2e) have the same AFLOW prototype label. They are generated by the same symmetry operations with different sets of parameters (`--params`) specified in their corresponding CIF files.

### Simple Monoclinic primitive vectors:

$$\mathbf{a}_1 = a\,\hat{\mathbf{x}}$$
$$\mathbf{a}_2 = b\,\hat{\mathbf{y}}$$
$$\mathbf{a}_3 = c\cos\beta\,\hat{\mathbf{x}} + c\sin\beta\,\hat{\mathbf{z}}$$

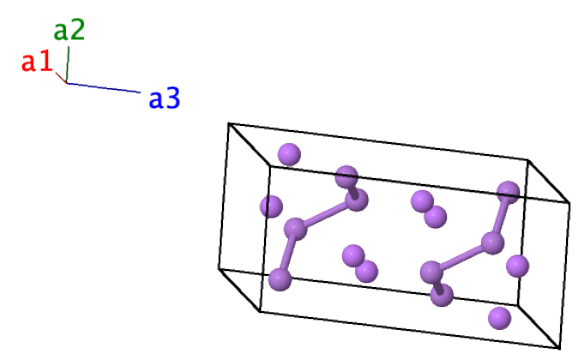

### Basis vectors:

| | | Lattice Coordinates | | Cartesian Coordinates | Wyckoff Position | Atom Type |
|---|---|---|---|---|---|---|
| $\mathbf{B}_1$ | = | $x_1\,\mathbf{a}_1 + y_1\,\mathbf{a}_2 + z_1\,\mathbf{a}_3$ | = | $(x_1 a + z_1 c\cos\beta)\,\hat{\mathbf{x}} + y_1 b\,\hat{\mathbf{y}} + z_1 c\sin\beta\,\hat{\mathbf{z}}$ | (4e) | As I |
| $\mathbf{B}_2$ | = | $-x_1\,\mathbf{a}_1 + \left(\frac{1}{2}+y_1\right)\mathbf{a}_2 + \left(\frac{1}{2}-z_1\right)\mathbf{a}_3$ | = | $\left(\frac{1}{2}c\cos\beta - x_1 a - z_1 c\cos\beta\right)\hat{\mathbf{x}} + \left(\frac{1}{2}+y_1\right)b\,\hat{\mathbf{y}} + \left(\frac{1}{2}-z_1\right)c\sin\beta\,\hat{\mathbf{z}}$ | (4e) | As I |
| $\mathbf{B}_3$ | = | $-x_1\,\mathbf{a}_1 - y_1\,\mathbf{a}_2 - z_1\,\mathbf{a}_3$ | = | $(-x_1 a - z_1 c\cos\beta)\,\hat{\mathbf{x}} - y_1 b\,\hat{\mathbf{y}} - z_1 c\sin\beta\,\hat{\mathbf{z}}$ | (4e) | As I |

$$\mathbf{B}_4 = x_1\,\mathbf{a}_1 + \left(\frac{1}{2} - y_1\right)\mathbf{a}_2 + \left(\frac{1}{2} + z_1\right)\mathbf{a}_3 = \left(\frac{1}{2}c\cos\beta + x_1 a + z_1 c\cos\beta\right)\hat{\mathbf{x}} + \left(\frac{1}{2} - y_1\right)b\,\hat{\mathbf{y}} + \left(\frac{1}{2} + z_1\right)c\sin\beta\,\hat{\mathbf{z}} \quad (4e) \quad \text{As I}$$

$$\mathbf{B}_5 = x_2\,\mathbf{a}_1 + y_2\,\mathbf{a}_2 + z_2\,\mathbf{a}_3 = (x_2 a + z_2 c\cos\beta)\,\hat{\mathbf{x}} + y_2 b\,\hat{\mathbf{y}} + z_2 c\sin\beta\,\hat{\mathbf{z}} \quad (4e) \quad \text{As II}$$

$$\mathbf{B}_6 = -x_2\,\mathbf{a}_1 + \left(\frac{1}{2} + y_2\right)\mathbf{a}_2 + \left(\frac{1}{2} - z_2\right)\mathbf{a}_3 = \left(\frac{1}{2}c\cos\beta - x_2 a - z_2 c\cos\beta\right)\hat{\mathbf{x}} + \left(\frac{1}{2} + y_2\right)b\,\hat{\mathbf{y}} + \left(\frac{1}{2} - z_2\right)c\sin\beta\,\hat{\mathbf{z}} \quad (4e) \quad \text{As II}$$

$$\mathbf{B}_7 = -x_2\,\mathbf{a}_1 - y_2\,\mathbf{a}_2 - z_2\,\mathbf{a}_3 = (-x_2 a - z_2 c\cos\beta)\,\hat{\mathbf{x}} - y_2 b\,\hat{\mathbf{y}} - z_2 c\sin\beta\,\hat{\mathbf{z}} \quad (4e) \quad \text{As II}$$

$$\mathbf{B}_8 = x_2\,\mathbf{a}_1 + \left(\frac{1}{2} - y_2\right)\mathbf{a}_2 + \left(\frac{1}{2} + z_2\right)\mathbf{a}_3 = \left(\frac{1}{2}c\cos\beta + x_2 a + z_2 c\cos\beta\right)\hat{\mathbf{x}} + \left(\frac{1}{2} - y_2\right)b\,\hat{\mathbf{y}} + \left(\frac{1}{2} + z_2\right)c\sin\beta\,\hat{\mathbf{z}} \quad (4e) \quad \text{As II}$$

$$\mathbf{B}_9 = x_3\,\mathbf{a}_1 + y_3\,\mathbf{a}_2 + z_3\,\mathbf{a}_3 = (x_3 a + z_3 c\cos\beta)\,\hat{\mathbf{x}} + y_3 b\,\hat{\mathbf{y}} + z_3 c\sin\beta\,\hat{\mathbf{z}} \quad (4e) \quad \text{Li I}$$

$$\mathbf{B}_{10} = -x_3\,\mathbf{a}_1 + \left(\frac{1}{2} + y_3\right)\mathbf{a}_2 + \left(\frac{1}{2} - z_3\right)\mathbf{a}_3 = \left(\frac{1}{2}c\cos\beta - x_3 a - z_3 c\cos\beta\right)\hat{\mathbf{x}} + \left(\frac{1}{2} + y_3\right)b\,\hat{\mathbf{y}} + \left(\frac{1}{2} - z_3\right)c\sin\beta\,\hat{\mathbf{z}} \quad (4e) \quad \text{Li I}$$

$$\mathbf{B}_{11} = -x_3\,\mathbf{a}_1 - y_3\,\mathbf{a}_2 - z_3\,\mathbf{a}_3 = (-x_3 a - z_3 c\cos\beta)\,\hat{\mathbf{x}} - y_3 b\,\hat{\mathbf{y}} - z_3 c\sin\beta\,\hat{\mathbf{z}} \quad (4e) \quad \text{Li I}$$

$$\mathbf{B}_{12} = x_3\,\mathbf{a}_1 + \left(\frac{1}{2} - y_3\right)\mathbf{a}_2 + \left(\frac{1}{2} + z_3\right)\mathbf{a}_3 = \left(\frac{1}{2}c\cos\beta + x_3 a + z_3 c\cos\beta\right)\hat{\mathbf{x}} + \left(\frac{1}{2} - y_3\right)b\,\hat{\mathbf{y}} + \left(\frac{1}{2} + z_3\right)c\sin\beta\,\hat{\mathbf{z}} \quad (4e) \quad \text{Li I}$$

$$\mathbf{B}_{13} = x_4\,\mathbf{a}_1 + y_4\,\mathbf{a}_2 + z_4\,\mathbf{a}_3 = (x_4 a + z_4 c\cos\beta)\,\hat{\mathbf{x}} + y_4 b\,\hat{\mathbf{y}} + z_4 c\sin\beta\,\hat{\mathbf{z}} \quad (4e) \quad \text{Li II}$$

$$\mathbf{B}_{14} = -x_4\,\mathbf{a}_1 + \left(\frac{1}{2} + y_4\right)\mathbf{a}_2 + \left(\frac{1}{2} - z_4\right)\mathbf{a}_3 = \left(\frac{1}{2}c\cos\beta - x_4 a - z_4 c\cos\beta\right)\hat{\mathbf{x}} + \left(\frac{1}{2} + y_4\right)b\,\hat{\mathbf{y}} + \left(\frac{1}{2} - z_4\right)c\sin\beta\,\hat{\mathbf{z}} \quad (4e) \quad \text{Li II}$$

$$\mathbf{B}_{15} = -x_4\,\mathbf{a}_1 - y_4\,\mathbf{a}_2 - z_4\,\mathbf{a}_3 = (-x_4 a - z_4 c\cos\beta)\,\hat{\mathbf{x}} - y_4 b\,\hat{\mathbf{y}} - z_4 c\sin\beta\,\hat{\mathbf{z}} \quad (4e) \quad \text{Li II}$$

$$\mathbf{B}_{16} = x_4\,\mathbf{a}_1 + \left(\frac{1}{2} - y_4\right)\mathbf{a}_2 + \left(\frac{1}{2} + z_4\right)\mathbf{a}_3 = \left(\frac{1}{2}c\cos\beta + x_4 a + z_4 c\cos\beta\right)\hat{\mathbf{x}} + \left(\frac{1}{2} - y_4\right)b\,\hat{\mathbf{y}} + \left(\frac{1}{2} + z_4\right)c\sin\beta\,\hat{\mathbf{z}} \quad (4e) \quad \text{Li II}$$

---

---

**Geometry files:**

- CIF: pp. 1557
- POSCAR: pp. 1558

# $\epsilon$-1,2,3,4,5,6-Hexachlorocyclohexane ($C_6Cl_6$) Structure: AB_mP24_14_3e_3e

http://aflow.org/prototype-encyclopedia/AB_mP24_14_3e_3e

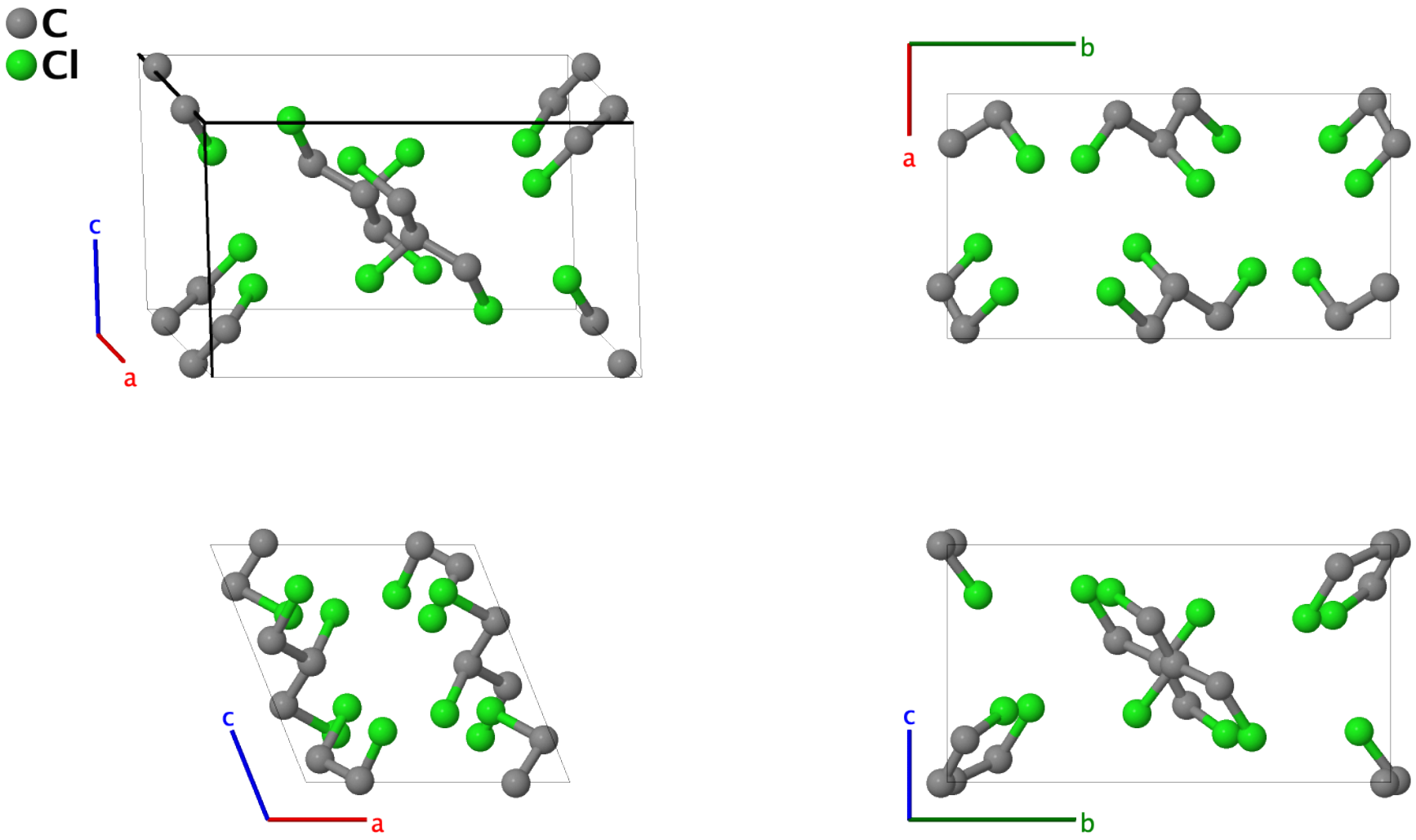

| | | |
|---|---|---|
| **Prototype** | : | $C_6Cl_6$ |
| **AFLOW prototype label** | : | AB_mP24_14_3e_3e |
| ***Strukturbericht* designation** | : | None |
| **Pearson symbol** | : | mP24 |
| **Space group number** | : | 14 |
| **Space group symbol** | : | $P2_1/c$ |
| **AFLOW prototype command** | : | `aflow --proto=AB_mP24_14_3e_3e` `--params=`$a, b/a, c/a, \beta, x_1, y_1, z_1, x_2, y_2, z_2, x_3, y_3, z_3, x_4, y_4, z_4, x_5, y_5, z_5, x_6, y_6, z_6$ |

- This unit cell contains two benzene molecules, with every hydrogen replaced by a chlorine atom, causing considerable distortion. To see this it will be necessary to use a supercell view of at least 212.

**Simple Monoclinic primitive vectors:**

$$\mathbf{a}_1 = a\,\hat{\mathbf{x}}$$
$$\mathbf{a}_2 = b\,\hat{\mathbf{y}}$$
$$\mathbf{a}_3 = c\cos\beta\,\hat{\mathbf{x}} + c\sin\beta\,\hat{\mathbf{z}}$$

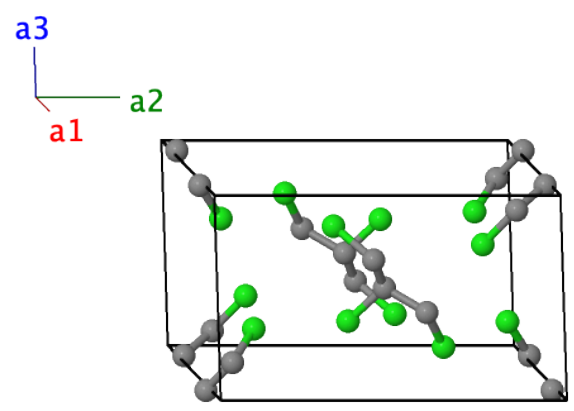

**Basis vectors:**

| | | Lattice Coordinates | | Cartesian Coordinates | Wyckoff Position | Atom Type |
|---|---|---|---|---|---|---|
| $\mathbf{B}_1$ | = | $x_1\,\mathbf{a}_1 + y_1\,\mathbf{a}_2 + z_1\,\mathbf{a}_3$ | = | $(x_1 a + z_1 c\cos\beta)\,\hat{\mathbf{x}} + y_1 b\,\hat{\mathbf{y}} + z_1 c\sin\beta\,\hat{\mathbf{z}}$ | (4e) | C I |
| $\mathbf{B}_2$ | = | $-x_1\,\mathbf{a}_1 + \left(\frac{1}{2} + y_1\right)\mathbf{a}_2 + \left(\frac{1}{2} - z_1\right)\mathbf{a}_3$ | = | $\left(\frac{1}{2}c\cos\beta - x_1 a - z_1 c\cos\beta\right)\hat{\mathbf{x}} + \left(\frac{1}{2} + y_1\right)b\,\hat{\mathbf{y}} + \left(\frac{1}{2} - z_1\right)c\sin\beta\,\hat{\mathbf{z}}$ | (4e) | C I |
| $\mathbf{B}_3$ | = | $-x_1\,\mathbf{a}_1 - y_1\,\mathbf{a}_2 - z_1\,\mathbf{a}_3$ | = | $(-x_1 a - z_1 c\cos\beta)\,\hat{\mathbf{x}} - y_1 b\,\hat{\mathbf{y}} - z_1 c\sin\beta\,\hat{\mathbf{z}}$ | (4$e$) | C I |
| $\mathbf{B}_4$ | = | $x_1\,\mathbf{a}_1 + \left(\frac{1}{2} - y_1\right)\mathbf{a}_2 + \left(\frac{1}{2} + z_1\right)\mathbf{a}_3$ | = | $\left(\frac{1}{2}c\cos\beta + x_1 a + z_1 c\cos\beta\right)\hat{\mathbf{x}} + \left(\frac{1}{2} - y_1\right)b\,\hat{\mathbf{y}} + \left(\frac{1}{2} + z_1\right)c\sin\beta\,\hat{\mathbf{z}}$ | (4$e$) | C I |
| $\mathbf{B}_5$ | = | $x_2\,\mathbf{a}_1 + y_2\,\mathbf{a}_2 + z_2\,\mathbf{a}_3$ | = | $(x_2 a + z_2 c\cos\beta)\,\hat{\mathbf{x}} + y_2 b\,\hat{\mathbf{y}} + z_2 c\sin\beta\,\hat{\mathbf{z}}$ | (4$e$) | C II |
| $\mathbf{B}_6$ | = | $-x_2\,\mathbf{a}_1 + \left(\frac{1}{2} + y_2\right)\mathbf{a}_2 + \left(\frac{1}{2} - z_2\right)\mathbf{a}_3$ | = | $\left(\frac{1}{2}c\cos\beta - x_2 a - z_2 c\cos\beta\right)\hat{\mathbf{x}} + \left(\frac{1}{2} + y_2\right)b\,\hat{\mathbf{y}} + \left(\frac{1}{2} - z_2\right)c\sin\beta\,\hat{\mathbf{z}}$ | (4$e$) | C II |
| $\mathbf{B}_7$ | = | $-x_2\,\mathbf{a}_1 - y_2\,\mathbf{a}_2 - z_2\,\mathbf{a}_3$ | = | $(-x_2 a - z_2 c\cos\beta)\,\hat{\mathbf{x}} - y_2 b\,\hat{\mathbf{y}} - z_2 c\sin\beta\,\hat{\mathbf{z}}$ | (4$e$) | C II |
| $\mathbf{B}_8$ | = | $x_2\,\mathbf{a}_1 + \left(\frac{1}{2} - y_2\right)\mathbf{a}_2 + \left(\frac{1}{2} + z_2\right)\mathbf{a}_3$ | = | $\left(\frac{1}{2}c\cos\beta + x_2 a + z_2 c\cos\beta\right)\hat{\mathbf{x}} + \left(\frac{1}{2} - y_2\right)b\,\hat{\mathbf{y}} + \left(\frac{1}{2} + z_2\right)c\sin\beta\,\hat{\mathbf{z}}$ | (4$e$) | C II |
| $\mathbf{B}_9$ | = | $x_3\,\mathbf{a}_1 + y_3\,\mathbf{a}_2 + z_3\,\mathbf{a}_3$ | = | $(x_3 a + z_3 c\cos\beta)\,\hat{\mathbf{x}} + y_3 b\,\hat{\mathbf{y}} + z_3 c\sin\beta\,\hat{\mathbf{z}}$ | (4$e$) | C III |
| $\mathbf{B}_{10}$ | = | $-x_3\,\mathbf{a}_1 + \left(\frac{1}{2} + y_3\right)\mathbf{a}_2 + \left(\frac{1}{2} - z_3\right)\mathbf{a}_3$ | = | $\left(\frac{1}{2}c\cos\beta - x_3 a - z_3 c\cos\beta\right)\hat{\mathbf{x}} + \left(\frac{1}{2} + y_3\right)b\,\hat{\mathbf{y}} + \left(\frac{1}{2} - z_3\right)c\sin\beta\,\hat{\mathbf{z}}$ | (4$e$) | C III |
| $\mathbf{B}_{11}$ | = | $-x_3\,\mathbf{a}_1 - y_3\,\mathbf{a}_2 - z_3\,\mathbf{a}_3$ | = | $(-x_3 a - z_3 c\cos\beta)\,\hat{\mathbf{x}} - y_3 b\,\hat{\mathbf{y}} - z_3 c\sin\beta\,\hat{\mathbf{z}}$ | (4$e$) | C III |
| $\mathbf{B}_{12}$ | = | $x_3\,\mathbf{a}_1 + \left(\frac{1}{2} - y_3\right)\mathbf{a}_2 + \left(\frac{1}{2} + z_3\right)\mathbf{a}_3$ | = | $\left(\frac{1}{2}c\cos\beta + x_3 a + z_3 c\cos\beta\right)\hat{\mathbf{x}} + \left(\frac{1}{2} - y_3\right)b\,\hat{\mathbf{y}} + \left(\frac{1}{2} + z_3\right)c\sin\beta\,\hat{\mathbf{z}}$ | (4$e$) | C III |
| $\mathbf{B}_{13}$ | = | $x_4\,\mathbf{a}_1 + y_4\,\mathbf{a}_2 + z_4\,\mathbf{a}_3$ | = | $(x_4 a + z_4 c\cos\beta)\,\hat{\mathbf{x}} + y_4 b\,\hat{\mathbf{y}} + z_4 c\sin\beta\,\hat{\mathbf{z}}$ | (4$e$) | Cl I |
| $\mathbf{B}_{14}$ | = | $-x_4\,\mathbf{a}_1 + \left(\frac{1}{2} + y_4\right)\mathbf{a}_2 + \left(\frac{1}{2} - z_4\right)\mathbf{a}_3$ | = | $\left(\frac{1}{2}c\cos\beta - x_4 a - z_4 c\cos\beta\right)\hat{\mathbf{x}} + \left(\frac{1}{2} + y_4\right)b\,\hat{\mathbf{y}} + \left(\frac{1}{2} - z_4\right)c\sin\beta\,\hat{\mathbf{z}}$ | (4$e$) | Cl I |
| $\mathbf{B}_{15}$ | = | $-x_4\,\mathbf{a}_1 - y_4\,\mathbf{a}_2 - z_4\,\mathbf{a}_3$ | = | $(-x_4 a - z_4 c\cos\beta)\,\hat{\mathbf{x}} - y_4 b\,\hat{\mathbf{y}} - z_4 c\sin\beta\,\hat{\mathbf{z}}$ | (4$e$) | Cl I |
| $\mathbf{B}_{16}$ | = | $x_4\,\mathbf{a}_1 + \left(\frac{1}{2} - y_4\right)\mathbf{a}_2 + \left(\frac{1}{2} + z_4\right)\mathbf{a}_3$ | = | $\left(\frac{1}{2}c\cos\beta + x_4 a + z_4 c\cos\beta\right)\hat{\mathbf{x}} + \left(\frac{1}{2} - y_4\right)b\,\hat{\mathbf{y}} + \left(\frac{1}{2} + z_4\right)c\sin\beta\,\hat{\mathbf{z}}$ | (4$e$) | Cl I |
| $\mathbf{B}_{17}$ | = | $x_5\,\mathbf{a}_1 + y_5\,\mathbf{a}_2 + z_5\,\mathbf{a}_3$ | = | $(x_5 a + z_5 c\cos\beta)\,\hat{\mathbf{x}} + y_5 b\,\hat{\mathbf{y}} + z_5 c\sin\beta\,\hat{\mathbf{z}}$ | (4$e$) | Cl II |
| $\mathbf{B}_{18}$ | = | $-x_5\,\mathbf{a}_1 + \left(\frac{1}{2} + y_5\right)\mathbf{a}_2 + \left(\frac{1}{2} - z_5\right)\mathbf{a}_3$ | = | $\left(\frac{1}{2}c\cos\beta - x_5 a - z_5 c\cos\beta\right)\hat{\mathbf{x}} + \left(\frac{1}{2} + y_5\right)b\,\hat{\mathbf{y}} + \left(\frac{1}{2} - z_5\right)c\sin\beta\,\hat{\mathbf{z}}$ | (4$e$) | Cl II |
| $\mathbf{B}_{19}$ | = | $-x_5\,\mathbf{a}_1 - y_5\,\mathbf{a}_2 - z_5\,\mathbf{a}_3$ | = | $(-x_5 a - z_5 c\cos\beta)\,\hat{\mathbf{x}} - y_5 b\,\hat{\mathbf{y}} - z_5 c\sin\beta\,\hat{\mathbf{z}}$ | (4$e$) | Cl II |
| $\mathbf{B}_{20}$ | = | $x_5\,\mathbf{a}_1 + \left(\frac{1}{2} - y_5\right)\mathbf{a}_2 + \left(\frac{1}{2} + z_5\right)\mathbf{a}_3$ | = | $\left(\frac{1}{2}c\cos\beta + x_5 a + z_5 c\cos\beta\right)\hat{\mathbf{x}} + \left(\frac{1}{2} - y_5\right)b\,\hat{\mathbf{y}} + \left(\frac{1}{2} + z_5\right)c\sin\beta\,\hat{\mathbf{z}}$ | (4$e$) | Cl II |
| $\mathbf{B}_{21}$ | = | $x_6\,\mathbf{a}_1 + y_6\,\mathbf{a}_2 + z_6\,\mathbf{a}_3$ | = | $(x_6 a + z_6 c\cos\beta)\,\hat{\mathbf{x}} + y_6 b\,\hat{\mathbf{y}} + z_6 c\sin\beta\,\hat{\mathbf{z}}$ | (4$e$) | Cl III |
| $\mathbf{B}_{22}$ | = | $-x_6\,\mathbf{a}_1 + \left(\frac{1}{2} + y_6\right)\mathbf{a}_2 + \left(\frac{1}{2} - z_6\right)\mathbf{a}_3$ | = | $\left(\frac{1}{2}c\cos\beta - x_6 a - z_6 c\cos\beta\right)\hat{\mathbf{x}} + \left(\frac{1}{2} + y_6\right)b\,\hat{\mathbf{y}} + \left(\frac{1}{2} - z_6\right)c\sin\beta\,\hat{\mathbf{z}}$ | (4$e$) | Cl III |
| $\mathbf{B}_{23}$ | = | $-x_6\,\mathbf{a}_1 - y_6\,\mathbf{a}_2 - z_6\,\mathbf{a}_3$ | = | $(-x_6 a - z_6 c\cos\beta)\,\hat{\mathbf{x}} - y_6 b\,\hat{\mathbf{y}} - z_6 c\sin\beta\,\hat{\mathbf{z}}$ | (4$e$) | Cl III |
| $\mathbf{B}_{24}$ | = | $x_6\,\mathbf{a}_1 + \left(\frac{1}{2} - y_6\right)\mathbf{a}_2 + \left(\frac{1}{2} + z_6\right)\mathbf{a}_3$ | = | $\left(\frac{1}{2}c\cos\beta + x_6 a + z_6 c\cos\beta\right)\hat{\mathbf{x}} + \left(\frac{1}{2} - y_6\right)b\,\hat{\mathbf{y}} + \left(\frac{1}{2} + z_6\right)c\sin\beta\,\hat{\mathbf{z}}$ | (4$e$) | Cl III |

**Geometry files:**

- CIF: pp. 1558
- POSCAR: pp. 1558

# Pararealgar (AsS) Structure: AB_mP32_14_4e_4e

http://aflow.org/prototype-encyclopedia/AB_mP32_14_4e_4e.pararealgar

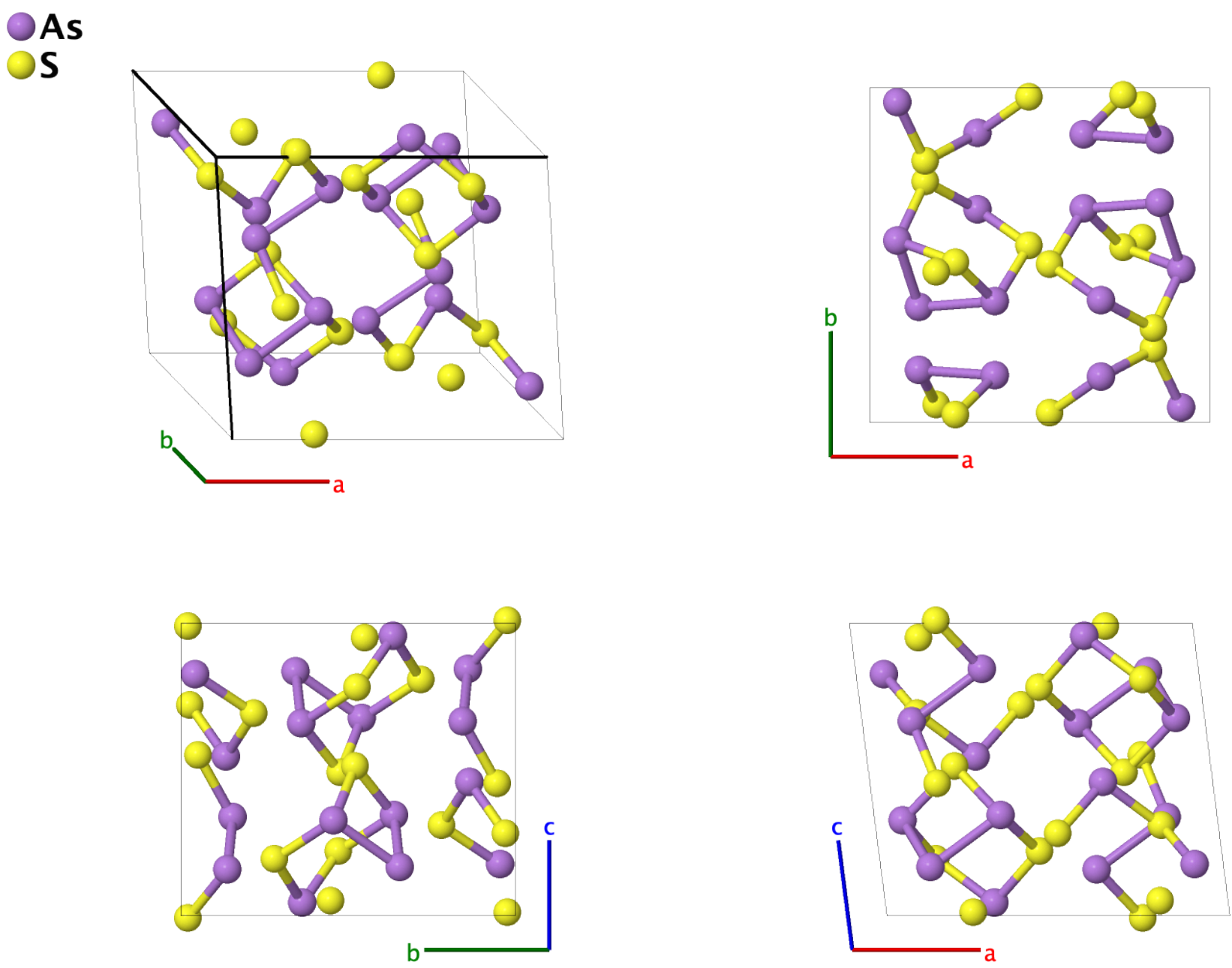

| | | |
|---|---|---|
| **Prototype** | : | AsS |
| **AFLOW prototype label** | : | AB_mP32_14_4e_4e |
| ***Strukturbericht* designation** | : | None |
| **Pearson symbol** | : | mP32 |
| **Space group number** | : | 14 |
| **Space group symbol** | : | $P2_1/c$ |
| **AFLOW prototype command** | : | `aflow --proto=AB_mP32_14_4e_4e` <br> `--params=`$a, b/a, c/a, \beta, x_1, y_1, z_1, x_2, y_2, z_2, x_3, y_3, z_3, x_4, y_4, z_4, x_5, y_5, z_5, x_6, y_6, z_6, x_7, y_7, z_7, x_8, y_8, z_8$ |

**Other compounds with this structure**

- AsSe

- Realgar ($B_l$) transforms to pararealgar on exposure to visible light with wavelengths in the range 500-670 nm (Douglass, 1992).

**Simple Monoclinic primitive vectors:**

$$\mathbf{a}_1 = a\,\hat{\mathbf{x}}$$
$$\mathbf{a}_2 = b\,\hat{\mathbf{y}}$$
$$\mathbf{a}_3 = c\cos\beta\,\hat{\mathbf{x}} + c\sin\beta\,\hat{\mathbf{z}}$$

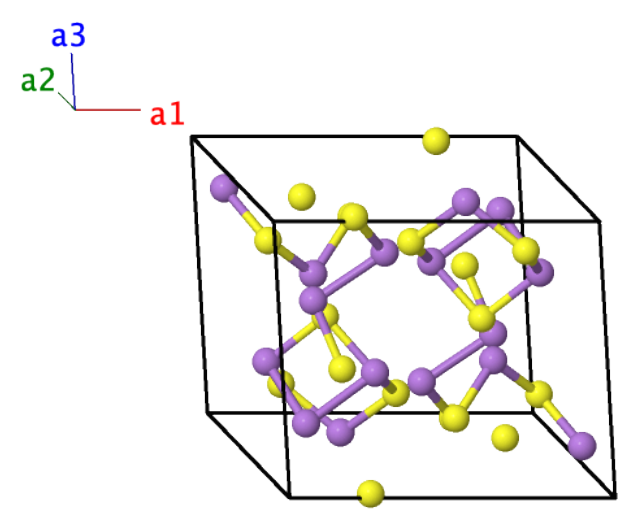

**Basis vectors:**

| | | Lattice Coordinates | | Cartesian Coordinates | Wyckoff Position | Atom Type |
|---|---|---|---|---|---|---|
| $\mathbf{B}_{1}$ | = | $x_{1}\,\mathbf{a}_{1}+y_{1}\,\mathbf{a}_{2}+z_{1}\,\mathbf{a}_{3}$ | = | $\left(x_{1}a+z_{1}c\cos\beta\right)\,\mathbf{\hat{x}}+y_{1}b\,\mathbf{\hat{y}}+z_{1}c\sin\beta\,\mathbf{\hat{z}}$ | (4$e$) | As I |
| $\mathbf{B}_{2}$ | = | $-x_{1}\,\mathbf{a}_{1}+\left(\frac{1}{2}+y_{1}\right)\,\mathbf{a}_{2}+\left(\frac{1}{2}-z_{1}\right)\,\mathbf{a}_{3}$ | = | $\left(\frac{1}{2}c\cos\beta-x_{1}a-z_{1}c\cos\beta\right)\,\mathbf{\hat{x}}+\left(\frac{1}{2}+y_{1}\right)b\,\mathbf{\hat{y}}+\left(\frac{1}{2}-z_{1}\right)c\sin\beta\,\mathbf{\hat{z}}$ | (4$e$) | As I |
| $\mathbf{B}_{3}$ | = | $-x_{1}\,\mathbf{a}_{1}-y_{1}\,\mathbf{a}_{2}-z_{1}\,\mathbf{a}_{3}$ | = | $\left(-x_{1}a-z_{1}c\cos\beta\right)\,\mathbf{\hat{x}}-y_{1}b\,\mathbf{\hat{y}}-z_{1}c\sin\beta\,\mathbf{\hat{z}}$ | (4$e$) | As I |
| $\mathbf{B}_{4}$ | = | $x_{1}\,\mathbf{a}_{1}+\left(\frac{1}{2}-y_{1}\right)\,\mathbf{a}_{2}+\left(\frac{1}{2}+z_{1}\right)\,\mathbf{a}_{3}$ | = | $\left(\frac{1}{2}c\cos\beta+x_{1}a+z_{1}c\cos\beta\right)\,\mathbf{\hat{x}}+\left(\frac{1}{2}-y_{1}\right)b\,\mathbf{\hat{y}}+\left(\frac{1}{2}+z_{1}\right)c\sin\beta\,\mathbf{\hat{z}}$ | (4$e$) | As I |
| $\mathbf{B}_{5}$ | = | $x_{2}\,\mathbf{a}_{1}+y_{2}\,\mathbf{a}_{2}+z_{2}\,\mathbf{a}_{3}$ | = | $\left(x_{2}a+z_{2}c\cos\beta\right)\,\mathbf{\hat{x}}+y_{2}b\,\mathbf{\hat{y}}+z_{2}c\sin\beta\,\mathbf{\hat{z}}$ | (4$e$) | As II |
| $\mathbf{B}_{6}$ | = | $-x_{2}\,\mathbf{a}_{1}+\left(\frac{1}{2}+y_{2}\right)\,\mathbf{a}_{2}+\left(\frac{1}{2}-z_{2}\right)\,\mathbf{a}_{3}$ | = | $\left(\frac{1}{2}c\cos\beta-x_{2}a-z_{2}c\cos\beta\right)\,\mathbf{\hat{x}}+\left(\frac{1}{2}+y_{2}\right)b\,\mathbf{\hat{y}}+\left(\frac{1}{2}-z_{2}\right)c\sin\beta\,\mathbf{\hat{z}}$ | (4$e$) | As II |
| $\mathbf{B}_{7}$ | = | $-x_{2}\,\mathbf{a}_{1}-y_{2}\,\mathbf{a}_{2}-z_{2}\,\mathbf{a}_{3}$ | = | $\left(-x_{2}a-z_{2}c\cos\beta\right)\,\mathbf{\hat{x}}-y_{2}b\,\mathbf{\hat{y}}-z_{2}c\sin\beta\,\mathbf{\hat{z}}$ | (4$e$) | As II |
| $\mathbf{B}_{8}$ | = | $x_{2}\,\mathbf{a}_{1}+\left(\frac{1}{2}-y_{2}\right)\,\mathbf{a}_{2}+\left(\frac{1}{2}+z_{2}\right)\,\mathbf{a}_{3}$ | = | $\left(\frac{1}{2}c\cos\beta+x_{2}a+z_{2}c\cos\beta\right)\,\mathbf{\hat{x}}+\left(\frac{1}{2}-y_{2}\right)b\,\mathbf{\hat{y}}+\left(\frac{1}{2}+z_{2}\right)c\sin\beta\,\mathbf{\hat{z}}$ | (4$e$) | As II |
| $\mathbf{B}_{9}$ | = | $x_{3}\,\mathbf{a}_{1}+y_{3}\,\mathbf{a}_{2}+z_{3}\,\mathbf{a}_{3}$ | = | $\left(x_{3}a+z_{3}c\cos\beta\right)\,\mathbf{\hat{x}}+y_{3}b\,\mathbf{\hat{y}}+z_{3}c\sin\beta\,\mathbf{\hat{z}}$ | (4$e$) | As III |
| $\mathbf{B}_{10}$ | = | $-x_{3}\,\mathbf{a}_{1}+\left(\frac{1}{2}+y_{3}\right)\,\mathbf{a}_{2}+\left(\frac{1}{2}-z_{3}\right)\,\mathbf{a}_{3}$ | = | $\left(\frac{1}{2}c\cos\beta-x_{3}a-z_{3}c\cos\beta\right)\,\mathbf{\hat{x}}+\left(\frac{1}{2}+y_{3}\right)b\,\mathbf{\hat{y}}+\left(\frac{1}{2}-z_{3}\right)c\sin\beta\,\mathbf{\hat{z}}$ | (4$e$) | As III |
| $\mathbf{B}_{11}$ | = | $-x_{3}\,\mathbf{a}_{1}-y_{3}\,\mathbf{a}_{2}-z_{3}\,\mathbf{a}_{3}$ | = | $\left(-x_{3}a-z_{3}c\cos\beta\right)\,\mathbf{\hat{x}}-y_{3}b\,\mathbf{\hat{y}}-z_{3}c\sin\beta\,\mathbf{\hat{z}}$ | (4$e$) | As III |
| $\mathbf{B}_{12}$ | = | $x_{3}\,\mathbf{a}_{1}+\left(\frac{1}{2}-y_{3}\right)\,\mathbf{a}_{2}+\left(\frac{1}{2}+z_{3}\right)\,\mathbf{a}_{3}$ | = | $\left(\frac{1}{2}c\cos\beta+x_{3}a+z_{3}c\cos\beta\right)\,\mathbf{\hat{x}}+\left(\frac{1}{2}-y_{3}\right)b\,\mathbf{\hat{y}}+\left(\frac{1}{2}+z_{3}\right)c\sin\beta\,\mathbf{\hat{z}}$ | (4$e$) | As III |
| $\mathbf{B}_{13}$ | = | $x_{4}\,\mathbf{a}_{1}+y_{4}\,\mathbf{a}_{2}+z_{4}\,\mathbf{a}_{3}$ | = | $\left(x_{4}a+z_{4}c\cos\beta\right)\,\mathbf{\hat{x}}+y_{4}b\,\mathbf{\hat{y}}+z_{4}c\sin\beta\,\mathbf{\hat{z}}$ | (4$e$) | As IV |
| $\mathbf{B}_{14}$ | = | $-x_{4}\,\mathbf{a}_{1}+\left(\frac{1}{2}+y_{4}\right)\,\mathbf{a}_{2}+\left(\frac{1}{2}-z_{4}\right)\,\mathbf{a}_{3}$ | = | $\left(\frac{1}{2}c\cos\beta-x_{4}a-z_{4}c\cos\beta\right)\,\mathbf{\hat{x}}+\left(\frac{1}{2}+y_{4}\right)b\,\mathbf{\hat{y}}+\left(\frac{1}{2}-z_{4}\right)c\sin\beta\,\mathbf{\hat{z}}$ | (4$e$) | As IV |
| $\mathbf{B}_{15}$ | = | $-x_{4}\,\mathbf{a}_{1}-y_{4}\,\mathbf{a}_{2}-z_{4}\,\mathbf{a}_{3}$ | = | $\left(-x_{4}a-z_{4}c\cos\beta\right)\,\mathbf{\hat{x}}-y_{4}b\,\mathbf{\hat{y}}-z_{4}c\sin\beta\,\mathbf{\hat{z}}$ | (4$e$) | As IV |
| $\mathbf{B}_{16}$ | = | $x_{4}\,\mathbf{a}_{1}+\left(\frac{1}{2}-y_{4}\right)\,\mathbf{a}_{2}+\left(\frac{1}{2}+z_{4}\right)\,\mathbf{a}_{3}$ | = | $\left(\frac{1}{2}c\cos\beta+x_{4}a+z_{4}c\cos\beta\right)\,\mathbf{\hat{x}}+\left(\frac{1}{2}-y_{4}\right)b\,\mathbf{\hat{y}}+\left(\frac{1}{2}+z_{4}\right)c\sin\beta\,\mathbf{\hat{z}}$ | (4$e$) | As IV |
| $\mathbf{B}_{17}$ | = | $x_{5}\,\mathbf{a}_{1}+y_{5}\,\mathbf{a}_{2}+z_{5}\,\mathbf{a}_{3}$ | = | $\left(x_{5}a+z_{5}c\cos\beta\right)\,\mathbf{\hat{x}}+y_{5}b\,\mathbf{\hat{y}}+z_{5}c\sin\beta\,\mathbf{\hat{z}}$ | (4$e$) | S I |
| $\mathbf{B}_{18}$ | = | $-x_{5}\,\mathbf{a}_{1}+\left(\frac{1}{2}+y_{5}\right)\,\mathbf{a}_{2}+\left(\frac{1}{2}-z_{5}\right)\,\mathbf{a}_{3}$ | = | $\left(\frac{1}{2}c\cos\beta-x_{5}a-z_{5}c\cos\beta\right)\,\mathbf{\hat{x}}+\left(\frac{1}{2}+y_{5}\right)b\,\mathbf{\hat{y}}+\left(\frac{1}{2}-z_{5}\right)c\sin\beta\,\mathbf{\hat{z}}$ | (4$e$) | S I |
| $\mathbf{B}_{19}$ | = | $-x_{5}\,\mathbf{a}_{1}-y_{5}\,\mathbf{a}_{2}-z_{5}\,\mathbf{a}_{3}$ | = | $\left(-x_{5}a-z_{5}c\cos\beta\right)\,\mathbf{\hat{x}}-y_{5}b\,\mathbf{\hat{y}}-z_{5}c\sin\beta\,\mathbf{\hat{z}}$ | (4$e$) | S I |
| $\mathbf{B}_{20}$ | = | $x_{5}\,\mathbf{a}_{1}+\left(\frac{1}{2}-y_{5}\right)\,\mathbf{a}_{2}+\left(\frac{1}{2}+z_{5}\right)\,\mathbf{a}_{3}$ | = | $\left(\frac{1}{2}c\cos\beta+x_{5}a+z_{5}c\cos\beta\right)\,\mathbf{\hat{x}}+\left(\frac{1}{2}-y_{5}\right)b\,\mathbf{\hat{y}}+\left(\frac{1}{2}+z_{5}\right)c\sin\beta\,\mathbf{\hat{z}}$ | (4$e$) | S I |
| $\mathbf{B}_{21}$ | = | $x_{6}\,\mathbf{a}_{1}+y_{6}\,\mathbf{a}_{2}+z_{6}\,\mathbf{a}_{3}$ | = | $\left(x_{6}a+z_{6}c\cos\beta\right)\,\mathbf{\hat{x}}+y_{6}b\,\mathbf{\hat{y}}+z_{6}c\sin\beta\,\mathbf{\hat{z}}$ | (4$e$) | S II |
| $\mathbf{B}_{22}$ | = | $-x_6\,\mathbf{a}_1 + \left(\frac{1}{2}+y_6\right)\mathbf{a}_2 + \left(\frac{1}{2}-z_6\right)\mathbf{a}_3$ | = | $\left(\frac{1}{2}c\cos\beta - x_6a - z_6c\cos\beta\right)\hat{\mathbf{x}} + \left(\frac{1}{2}+y_6\right)b\,\hat{\mathbf{y}} + \left(\frac{1}{2}-z_6\right)c\sin\beta\,\hat{\mathbf{z}}$ | (4$e$) | S II |
| $\mathbf{B}_{23}$ | = | $-x_6\,\mathbf{a}_1 - y_6\,\mathbf{a}_2 - z_6\,\mathbf{a}_3$ | = | $(-x_6a - z_6c\cos\beta)\,\hat{\mathbf{x}} - y_6b\,\hat{\mathbf{y}} - z_6c\sin\beta\,\hat{\mathbf{z}}$ | (4$e$) | S II |
| $\mathbf{B}_{24}$ | = | $x_6\,\mathbf{a}_1 + \left(\frac{1}{2}-y_6\right)\mathbf{a}_2 + \left(\frac{1}{2}+z_6\right)\mathbf{a}_3$ | = | $\left(\frac{1}{2}c\cos\beta + x_6a + z_6c\cos\beta\right)\hat{\mathbf{x}} + \left(\frac{1}{2}-y_6\right)b\,\hat{\mathbf{y}} + \left(\frac{1}{2}+z_6\right)c\sin\beta\,\hat{\mathbf{z}}$ | (4$e$) | S II |
| $\mathbf{B}_{25}$ | = | $x_7\,\mathbf{a}_1 + y_7\,\mathbf{a}_2 + z_7\,\mathbf{a}_3$ | = | $(x_7a + z_7c\cos\beta)\,\hat{\mathbf{x}} + y_7b\,\hat{\mathbf{y}} + z_7c\sin\beta\,\hat{\mathbf{z}}$ | (4$e$) | S III |
| $\mathbf{B}_{26}$ | = | $-x_7\,\mathbf{a}_1 + \left(\frac{1}{2}+y_7\right)\mathbf{a}_2 + \left(\frac{1}{2}-z_7\right)\mathbf{a}_3$ | = | $\left(\frac{1}{2}c\cos\beta - x_7a - z_7c\cos\beta\right)\hat{\mathbf{x}} + \left(\frac{1}{2}+y_7\right)b\,\hat{\mathbf{y}} + \left(\frac{1}{2}-z_7\right)c\sin\beta\,\hat{\mathbf{z}}$ | (4$e$) | S III |
| $\mathbf{B}_{27}$ | = | $-x_7\,\mathbf{a}_1 - y_7\,\mathbf{a}_2 - z_7\,\mathbf{a}_3$ | = | $(-x_7a - z_7c\cos\beta)\,\hat{\mathbf{x}} - y_7b\,\hat{\mathbf{y}} - z_7c\sin\beta\,\hat{\mathbf{z}}$ | (4$e$) | S III |
| $\mathbf{B}_{28}$ | = | $x_7\,\mathbf{a}_1 + \left(\frac{1}{2}-y_7\right)\mathbf{a}_2 + \left(\frac{1}{2}+z_7\right)\mathbf{a}_3$ | = | $\left(\frac{1}{2}c\cos\beta + x_7a + z_7c\cos\beta\right)\hat{\mathbf{x}} + \left(\frac{1}{2}-y_7\right)b\,\hat{\mathbf{y}} + \left(\frac{1}{2}+z_7\right)c\sin\beta\,\hat{\mathbf{z}}$ | (4$e$) | S III |
| $\mathbf{B}_{29}$ | = | $x_8\,\mathbf{a}_1 + y_8\,\mathbf{a}_2 + z_8\,\mathbf{a}_3$ | = | $(x_8a + z_8c\cos\beta)\,\hat{\mathbf{x}} + y_8b\,\hat{\mathbf{y}} + z_8c\sin\beta\,\hat{\mathbf{z}}$ | (4$e$) | S IV |
| $\mathbf{B}_{30}$ | = | $-x_8\,\mathbf{a}_1 + \left(\frac{1}{2}+y_8\right)\mathbf{a}_2 + \left(\frac{1}{2}-z_8\right)\mathbf{a}_3$ | = | $\left(\frac{1}{2}c\cos\beta - x_8a - z_8c\cos\beta\right)\hat{\mathbf{x}} + \left(\frac{1}{2}+y_8\right)b\,\hat{\mathbf{y}} + \left(\frac{1}{2}-z_8\right)c\sin\beta\,\hat{\mathbf{z}}$ | (4$e$) | S IV |
| $\mathbf{B}_{31}$ | = | $-x_8\,\mathbf{a}_1 - y_8\,\mathbf{a}_2 - z_8\,\mathbf{a}_3$ | = | $(-x_8a - z_8c\cos\beta)\,\hat{\mathbf{x}} - y_8b\,\hat{\mathbf{y}} - z_8c\sin\beta\,\hat{\mathbf{z}}$ | (4$e$) | S IV |
| $\mathbf{B}_{32}$ | = | $x_8\,\mathbf{a}_1 + \left(\frac{1}{2}-y_8\right)\mathbf{a}_2 + \left(\frac{1}{2}+z_8\right)\mathbf{a}_3$ | = | $\left(\frac{1}{2}c\cos\beta + x_8a + z_8c\cos\beta\right)\hat{\mathbf{x}} + \left(\frac{1}{2}-y_8\right)b\,\hat{\mathbf{y}} + \left(\frac{1}{2}+z_8\right)c\sin\beta\,\hat{\mathbf{z}}$ | (4$e$) | S IV |

**Geometry files:**

- CIF: pp. 1558
- POSCAR: pp. 1559

# Realgar (AsS, $B_l$) Structure: AB_mP32_14_4e_4e

http://aflow.org/prototype-encyclopedia/AB_mP32_14_4e_4e.realgar

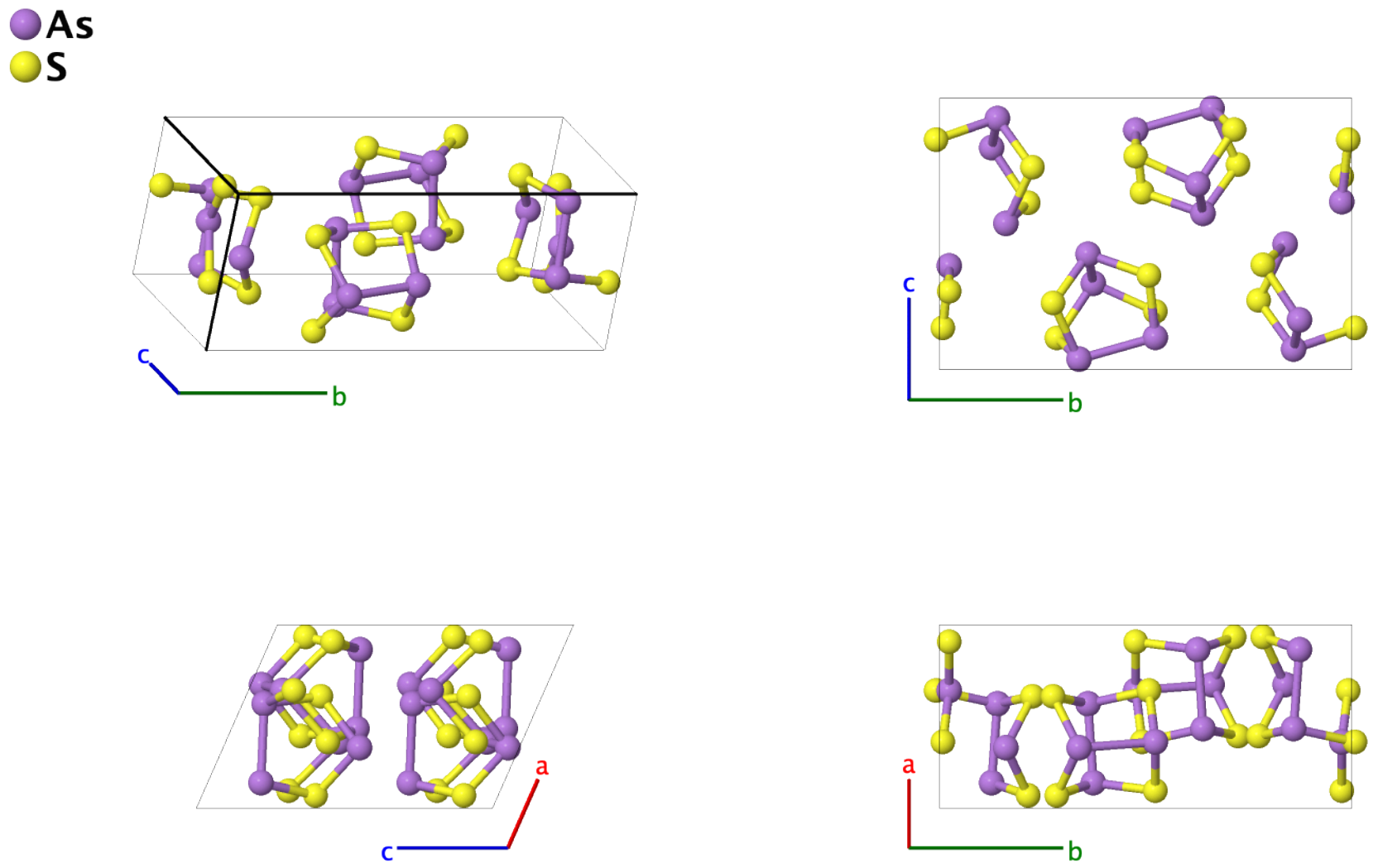

| | | |
|---|---|---|
| **Prototype** | : | AsS |
| **AFLOW prototype label** | : | AB_mP32_14_4e_4e |
| ***Strukturbericht* designation** | : | $B_l$ |
| **Pearson symbol** | : | mP32 |
| **Space group number** | : | 14 |
| **Space group symbol** | : | $P2_1/c$ |
| **AFLOW prototype command** | : | `aflow --proto=AB_mP32_14_4e_4e` <br> `--params=`$a, b/a, c/a, \beta, x_1, y_1, z_1, x_2, y_2, z_2, x_3, y_3, z_3, x_4, y_4, z_4, x_5, y_5, z_5, x_6, y_6, z_6, x_7, y_7, z_7, x_8, y_8, z_8$ |

### Other compounds with this structure

- AsSe

---

- Realgar transforms to pararealgar on exposure to visible light with wavelengths in the range 500-670 nm (Douglass, 1992).
- (Ito, 1952) give the crystal structure of realgar in the $P2_1/n$ setting of space group #14. We have used FINDSYM to transform this to the standard $P2_1/c$ setting. This involved both rotations and an origin shift.

### Simple Monoclinic primitive vectors:

$$\mathbf{a}_1 = a\,\hat{\mathbf{x}}$$
$$\mathbf{a}_2 = b\,\hat{\mathbf{y}}$$
$$\mathbf{a}_3 = c\cos\beta\,\hat{\mathbf{x}} + c\sin\beta\,\hat{\mathbf{z}}$$

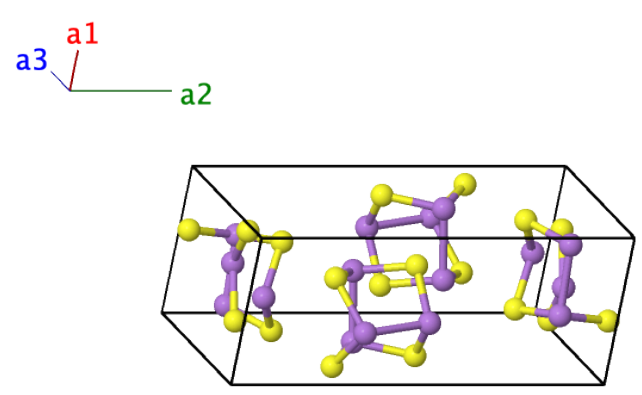

### Basis vectors:

| | | Lattice Coordinates | | Cartesian Coordinates | Wyckoff Position | Atom Type |
|---|---|---|---|---|---|---|
| $\mathbf{B}_{1}$ | = | $x_{1}\,\mathbf{a}_{1} + y_{1}\,\mathbf{a}_{2} + z_{1}\,\mathbf{a}_{3}$ | = | $\left(x_{1}a + z_{1}c\cos\beta\right)\,\mathbf{\hat{x}} + y_{1}b\,\mathbf{\hat{y}} + z_{1}c\sin\beta\,\mathbf{\hat{z}}$ | (4e) | As I |
| $\mathbf{B}_{2}$ | = | $-x_{1}\,\mathbf{a}_{1} + \left(\frac{1}{2} + y_{1}\right)\,\mathbf{a}_{2} + \left(\frac{1}{2} - z_{1}\right)\,\mathbf{a}_{3}$ | = | $\left(\frac{1}{2}c\cos\beta - x_{1}a - z_{1}c\cos\beta\right)\,\mathbf{\hat{x}} + \left(\frac{1}{2} + y_{1}\right)b\,\mathbf{\hat{y}} + \left(\frac{1}{2} - z_{1}\right)c\sin\beta\,\mathbf{\hat{z}}$ | (4e) | As I |
| $\mathbf{B}_{3}$ | = | $-x_{1}\,\mathbf{a}_{1} - y_{1}\,\mathbf{a}_{2} - z_{1}\,\mathbf{a}_{3}$ | = | $\left(-x_{1}a - z_{1}c\cos\beta\right)\,\mathbf{\hat{x}} - y_{1}b\,\mathbf{\hat{y}} - z_{1}c\sin\beta\,\mathbf{\hat{z}}$ | (4e) | As I |
| $\mathbf{B}_{4}$ | = | $x_{1}\,\mathbf{a}_{1} + \left(\frac{1}{2} - y_{1}\right)\,\mathbf{a}_{2} + \left(\frac{1}{2} + z_{1}\right)\,\mathbf{a}_{3}$ | = | $\left(\frac{1}{2}c\cos\beta + x_{1}a + z_{1}c\cos\beta\right)\,\mathbf{\hat{x}} + \left(\frac{1}{2} - y_{1}\right)b\,\mathbf{\hat{y}} + \left(\frac{1}{2} + z_{1}\right)c\sin\beta\,\mathbf{\hat{z}}$ | (4e) | As I |
| $\mathbf{B}_{5}$ | = | $x_{2}\,\mathbf{a}_{1} + y_{2}\,\mathbf{a}_{2} + z_{2}\,\mathbf{a}_{3}$ | = | $\left(x_{2}a + z_{2}c\cos\beta\right)\,\mathbf{\hat{x}} + y_{2}b\,\mathbf{\hat{y}} + z_{2}c\sin\beta\,\mathbf{\hat{z}}$ | (4e) | As II |
| $\mathbf{B}_{6}$ | = | $-x_{2}\,\mathbf{a}_{1} + \left(\frac{1}{2} + y_{2}\right)\,\mathbf{a}_{2} + \left(\frac{1}{2} - z_{2}\right)\,\mathbf{a}_{3}$ | = | $\left(\frac{1}{2}c\cos\beta - x_{2}a - z_{2}c\cos\beta\right)\,\mathbf{\hat{x}} + \left(\frac{1}{2} + y_{2}\right)b\,\mathbf{\hat{y}} + \left(\frac{1}{2} - z_{2}\right)c\sin\beta\,\mathbf{\hat{z}}$ | (4e) | As II |
| $\mathbf{B}_{7}$ | = | $-x_{2}\,\mathbf{a}_{1} - y_{2}\,\mathbf{a}_{2} - z_{2}\,\mathbf{a}_{3}$ | = | $\left(-x_{2}a - z_{2}c\cos\beta\right)\,\mathbf{\hat{x}} - y_{2}b\,\mathbf{\hat{y}} - z_{2}c\sin\beta\,\mathbf{\hat{z}}$ | (4e) | As II |
| $\mathbf{B}_{8}$ | = | $x_{2}\,\mathbf{a}_{1} + \left(\frac{1}{2} - y_{2}\right)\,\mathbf{a}_{2} + \left(\frac{1}{2} + z_{2}\right)\,\mathbf{a}_{3}$ | = | $\left(\frac{1}{2}c\cos\beta + x_{2}a + z_{2}c\cos\beta\right)\,\mathbf{\hat{x}} + \left(\frac{1}{2} - y_{2}\right)b\,\mathbf{\hat{y}} + \left(\frac{1}{2} + z_{2}\right)c\sin\beta\,\mathbf{\hat{z}}$ | (4e) | As II |
| $\mathbf{B}_{9}$ | = | $x_{3}\,\mathbf{a}_{1} + y_{3}\,\mathbf{a}_{2} + z_{3}\,\mathbf{a}_{3}$ | = | $\left(x_{3}a + z_{3}c\cos\beta\right)\,\mathbf{\hat{x}} + y_{3}b\,\mathbf{\hat{y}} + z_{3}c\sin\beta\,\mathbf{\hat{z}}$ | (4e) | As III |
| $\mathbf{B}_{10}$ | = | $-x_{3}\,\mathbf{a}_{1} + \left(\frac{1}{2} + y_{3}\right)\,\mathbf{a}_{2} + \left(\frac{1}{2} - z_{3}\right)\,\mathbf{a}_{3}$ | = | $\left(\frac{1}{2}c\cos\beta - x_{3}a - z_{3}c\cos\beta\right)\,\mathbf{\hat{x}} + \left(\frac{1}{2} + y_{3}\right)b\,\mathbf{\hat{y}} + \left(\frac{1}{2} - z_{3}\right)c\sin\beta\,\mathbf{\hat{z}}$ | (4e) | As III |
| $\mathbf{B}_{11}$ | = | $-x_{3}\,\mathbf{a}_{1} - y_{3}\,\mathbf{a}_{2} - z_{3}\,\mathbf{a}_{3}$ | = | $\left(-x_{3}a - z_{3}c\cos\beta\right)\,\mathbf{\hat{x}} - y_{3}b\,\mathbf{\hat{y}} - z_{3}c\sin\beta\,\mathbf{\hat{z}}$ | (4e) | As III |
| $\mathbf{B}_{12}$ | = | $x_{3}\,\mathbf{a}_{1} + \left(\frac{1}{2} - y_{3}\right)\,\mathbf{a}_{2} + \left(\frac{1}{2} + z_{3}\right)\,\mathbf{a}_{3}$ | = | $\left(\frac{1}{2}c\cos\beta + x_{3}a + z_{3}c\cos\beta\right)\,\mathbf{\hat{x}} + \left(\frac{1}{2} - y_{3}\right)b\,\mathbf{\hat{y}} + \left(\frac{1}{2} + z_{3}\right)c\sin\beta\,\mathbf{\hat{z}}$ | (4e) | As III |
| $\mathbf{B}_{13}$ | = | $x_{4}\,\mathbf{a}_{1} + y_{4}\,\mathbf{a}_{2} + z_{4}\,\mathbf{a}_{3}$ | = | $\left(x_{4}a + z_{4}c\cos\beta\right)\,\mathbf{\hat{x}} + y_{4}b\,\mathbf{\hat{y}} + z_{4}c\sin\beta\,\mathbf{\hat{z}}$ | (4e) | As IV |
| $\mathbf{B}_{14}$ | = | $-x_{4}\,\mathbf{a}_{1} + \left(\frac{1}{2} + y_{4}\right)\,\mathbf{a}_{2} + \left(\frac{1}{2} - z_{4}\right)\,\mathbf{a}_{3}$ | = | $\left(\frac{1}{2}c\cos\beta - x_{4}a - z_{4}c\cos\beta\right)\,\mathbf{\hat{x}} + \left(\frac{1}{2} + y_{4}\right)b\,\mathbf{\hat{y}} + \left(\frac{1}{2} - z_{4}\right)c\sin\beta\,\mathbf{\hat{z}}$ | (4e) | As IV |
| $\mathbf{B}_{15}$ | = | $-x_{4}\,\mathbf{a}_{1} - y_{4}\,\mathbf{a}_{2} - z_{4}\,\mathbf{a}_{3}$ | = | $\left(-x_{4}a - z_{4}c\cos\beta\right)\,\mathbf{\hat{x}} - y_{4}b\,\mathbf{\hat{y}} - z_{4}c\sin\beta\,\mathbf{\hat{z}}$ | (4e) | As IV |
| $\mathbf{B}_{16}$ | = | $x_{4}\,\mathbf{a}_{1} + \left(\frac{1}{2} - y_{4}\right)\,\mathbf{a}_{2} + \left(\frac{1}{2} + z_{4}\right)\,\mathbf{a}_{3}$ | = | $\left(\frac{1}{2}c\cos\beta + x_{4}a + z_{4}c\cos\beta\right)\,\mathbf{\hat{x}} + \left(\frac{1}{2} - y_{4}\right)b\,\mathbf{\hat{y}} + \left(\frac{1}{2} + z_{4}\right)c\sin\beta\,\mathbf{\hat{z}}$ | (4e) | As IV |
| $\mathbf{B}_{17}$ | = | $x_{5}\,\mathbf{a}_{1} + y_{5}\,\mathbf{a}_{2} + z_{5}\,\mathbf{a}_{3}$ | = | $\left(x_{5}a + z_{5}c\cos\beta\right)\,\mathbf{\hat{x}} + y_{5}b\,\mathbf{\hat{y}} + z_{5}c\sin\beta\,\mathbf{\hat{z}}$ | (4e) | S I |
| $\mathbf{B}_{18}$ | = | $-x_{5}\,\mathbf{a}_{1} + \left(\frac{1}{2} + y_{5}\right)\,\mathbf{a}_{2} + \left(\frac{1}{2} - z_{5}\right)\,\mathbf{a}_{3}$ | = | $\left(\frac{1}{2}c\cos\beta - x_{5}a - z_{5}c\cos\beta\right)\,\mathbf{\hat{x}} + \left(\frac{1}{2} + y_{5}\right)b\,\mathbf{\hat{y}} + \left(\frac{1}{2} - z_{5}\right)c\sin\beta\,\mathbf{\hat{z}}$ | (4e) | S I |
| $\mathbf{B}_{19}$ | = | $-x_{5}\,\mathbf{a}_{1} - y_{5}\,\mathbf{a}_{2} - z_{5}\,\mathbf{a}_{3}$ | = | $\left(-x_{5}a - z_{5}c\cos\beta\right)\,\mathbf{\hat{x}} - y_{5}b\,\mathbf{\hat{y}} - z_{5}c\sin\beta\,\mathbf{\hat{z}}$ | (4e) | S I |
| $\mathbf{B}_{20}$ | = | $x_{5}\,\mathbf{a}_{1} + \left(\frac{1}{2} - y_{5}\right)\,\mathbf{a}_{2} + \left(\frac{1}{2} + z_{5}\right)\,\mathbf{a}_{3}$ | = | $\left(\frac{1}{2}c\cos\beta + x_{5}a + z_{5}c\cos\beta\right)\,\mathbf{\hat{x}} + \left(\frac{1}{2} - y_{5}\right)b\,\mathbf{\hat{y}} + \left(\frac{1}{2} + z_{5}\right)c\sin\beta\,\mathbf{\hat{z}}$ | (4e) | S I |
| $\mathbf{B}_{21}$ | = | $x_{6}\,\mathbf{a}_{1} + y_{6}\,\mathbf{a}_{2} + z_{6}\,\mathbf{a}_{3}$ | = | $\left(x_{6}a + z_{6}c\cos\beta\right)\,\mathbf{\hat{x}} + y_{6}b\,\mathbf{\hat{y}} + z_{6}c\sin\beta\,\mathbf{\hat{z}}$ | (4e) | S II |
| $\mathbf{B}_{22}$ | = | $-x_6\,\mathbf{a}_1 + \left(\frac{1}{2} + y_6\right)\mathbf{a}_2 + \left(\frac{1}{2} - z_6\right)\mathbf{a}_3$ | = | $\left(\frac{1}{2}c\cos\beta - x_6 a - z_6 c\cos\beta\right)\hat{\mathbf{x}} + \left(\frac{1}{2} + y_6\right)b\,\hat{\mathbf{y}} + \left(\frac{1}{2} - z_6\right)c\sin\beta\,\hat{\mathbf{z}}$ | (4*e*) | S II |
| $\mathbf{B}_{23}$ | = | $-x_6\,\mathbf{a}_1 - y_6\,\mathbf{a}_2 - z_6\,\mathbf{a}_3$ | = | $(-x_6 a - z_6 c\cos\beta)\,\hat{\mathbf{x}} - y_6 b\,\hat{\mathbf{y}} - z_6 c\sin\beta\,\hat{\mathbf{z}}$ | (4*e*) | S II |
| $\mathbf{B}_{24}$ | = | $x_6\,\mathbf{a}_1 + \left(\frac{1}{2} - y_6\right)\mathbf{a}_2 + \left(\frac{1}{2} + z_6\right)\mathbf{a}_3$ | = | $\left(\frac{1}{2}c\cos\beta + x_6 a + z_6 c\cos\beta\right)\hat{\mathbf{x}} + \left(\frac{1}{2} - y_6\right)b\,\hat{\mathbf{y}} + \left(\frac{1}{2} + z_6\right)c\sin\beta\,\hat{\mathbf{z}}$ | (4*e*) | S II |
| $\mathbf{B}_{25}$ | = | $x_7\,\mathbf{a}_1 + y_7\,\mathbf{a}_2 + z_7\,\mathbf{a}_3$ | = | $(x_7 a + z_7 c\cos\beta)\,\hat{\mathbf{x}} + y_7 b\,\hat{\mathbf{y}} + z_7 c\sin\beta\,\hat{\mathbf{z}}$ | (4*e*) | S III |
| $\mathbf{B}_{26}$ | = | $-x_7\,\mathbf{a}_1 + \left(\frac{1}{2} + y_7\right)\mathbf{a}_2 + \left(\frac{1}{2} - z_7\right)\mathbf{a}_3$ | = | $\left(\frac{1}{2}c\cos\beta - x_7 a - z_7 c\cos\beta\right)\hat{\mathbf{x}} + \left(\frac{1}{2} + y_7\right)b\,\hat{\mathbf{y}} + \left(\frac{1}{2} - z_7\right)c\sin\beta\,\hat{\mathbf{z}}$ | (4*e*) | S III |
| $\mathbf{B}_{27}$ | = | $-x_7\,\mathbf{a}_1 - y_7\,\mathbf{a}_2 - z_7\,\mathbf{a}_3$ | = | $(-x_7 a - z_7 c\cos\beta)\,\hat{\mathbf{x}} - y_7 b\,\hat{\mathbf{y}} - z_7 c\sin\beta\,\hat{\mathbf{z}}$ | (4*e*) | S III |
| $\mathbf{B}_{28}$ | = | $x_7\,\mathbf{a}_1 + \left(\frac{1}{2} - y_7\right)\mathbf{a}_2 + \left(\frac{1}{2} + z_7\right)\mathbf{a}_3$ | = | $\left(\frac{1}{2}c\cos\beta + x_7 a + z_7 c\cos\beta\right)\hat{\mathbf{x}} + \left(\frac{1}{2} - y_7\right)b\,\hat{\mathbf{y}} + \left(\frac{1}{2} + z_7\right)c\sin\beta\,\hat{\mathbf{z}}$ | (4*e*) | S III |
| $\mathbf{B}_{29}$ | = | $x_8\,\mathbf{a}_1 + y_8\,\mathbf{a}_2 + z_8\,\mathbf{a}_3$ | = | $(x_8 a + z_8 c\cos\beta)\,\hat{\mathbf{x}} + y_8 b\,\hat{\mathbf{y}} + z_8 c\sin\beta\,\hat{\mathbf{z}}$ | (4*e*) | S IV |
| $\mathbf{B}_{30}$ | = | $-x_8\,\mathbf{a}_1 + \left(\frac{1}{2} + y_8\right)\mathbf{a}_2 + \left(\frac{1}{2} - z_8\right)\mathbf{a}_3$ | = | $\left(\frac{1}{2}c\cos\beta - x_8 a - z_8 c\cos\beta\right)\hat{\mathbf{x}} + \left(\frac{1}{2} + y_8\right)b\,\hat{\mathbf{y}} + \left(\frac{1}{2} - z_8\right)c\sin\beta\,\hat{\mathbf{z}}$ | (4*e*) | S IV |
| $\mathbf{B}_{31}$ | = | $-x_8\,\mathbf{a}_1 - y_8\,\mathbf{a}_2 - z_8\,\mathbf{a}_3$ | = | $(-x_8 a - z_8 c\cos\beta)\,\hat{\mathbf{x}} - y_8 b\,\hat{\mathbf{y}} - z_8 c\sin\beta\,\hat{\mathbf{z}}$ | (4*e*) | S IV |
| $\mathbf{B}_{32}$ | = | $x_8\,\mathbf{a}_1 + \left(\frac{1}{2} - y_8\right)\mathbf{a}_2 + \left(\frac{1}{2} + z_8\right)\mathbf{a}_3$ | = | $\left(\frac{1}{2}c\cos\beta + x_8 a + z_8 c\cos\beta\right)\hat{\mathbf{x}} + \left(\frac{1}{2} - y_8\right)b\,\hat{\mathbf{y}} + \left(\frac{1}{2} + z_8\right)c\sin\beta\,\hat{\mathbf{z}}$ | (4*e*) | S IV |

---

---

**Geometry files:**

- CIF: pp. 1559
- POSCAR: pp. 1559

# $Ag_2PbO_2$ Structure: A2B2C_mC20_15_ad_f_e

http://aflow.org/prototype-encyclopedia/A2B2C_mC20_15_ad_f_e

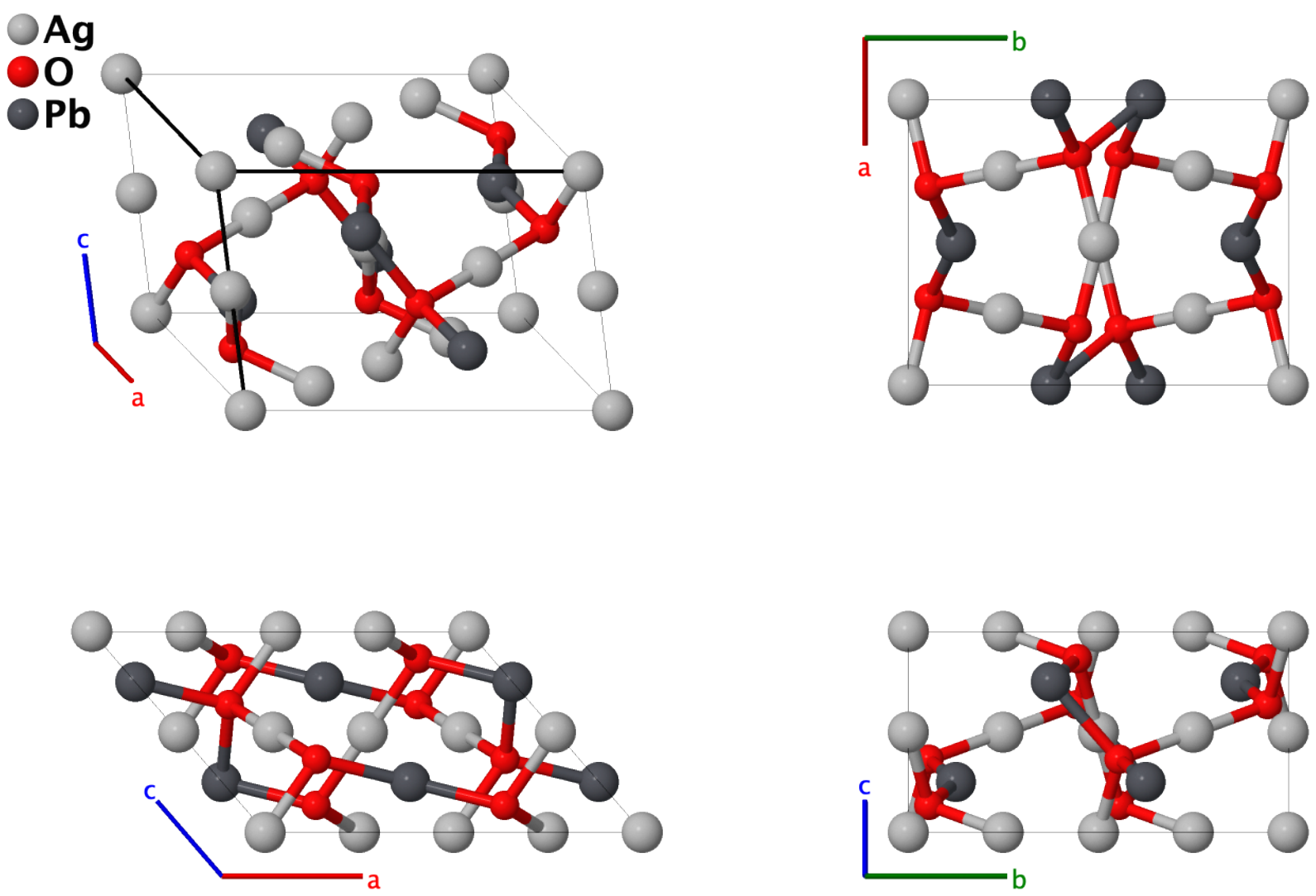

| | | |
|---|---|---|
| **Prototype** | : | $Ag_2O_2Pb$ |
| **AFLOW prototype label** | : | A2B2C_mC20_15_ad_f_e |
| ***Strukturbericht* designation** | : | None |
| **Pearson symbol** | : | mC20 |
| **Space group number** | : | 15 |
| **Space group symbol** | : | $C2/c$ |
| **AFLOW prototype command** | : | `aflow --proto=A2B2C_mC20_15_ad_f_e`<br>`--params=`$a, b/a, c/a, \beta, y_3, x_4, y_4, z_4$ |

- (Byström, 1950) gives the structure in the $I2/c$ setting of space group #15. We have used FINDSYM to change this to the standard $C2/c$ setting. This conversion involved a rotation of the axis, and placed the origin on what had been an Ag (4$d$) site, tranforming it to Ag (4$a$).

**Base-centered Monoclinic primitive vectors:**

$$\mathbf{a}_1 = \frac{1}{2}\,a\,\hat{\mathbf{x}} - \frac{1}{2}\,b\,\hat{\mathbf{y}}$$
$$\mathbf{a}_2 = \frac{1}{2}\,a\,\hat{\mathbf{x}} + \frac{1}{2}\,b\,\hat{\mathbf{y}}$$
$$\mathbf{a}_3 = c\cos\beta\,\hat{\mathbf{x}} + c\sin\beta\,\hat{\mathbf{z}}$$

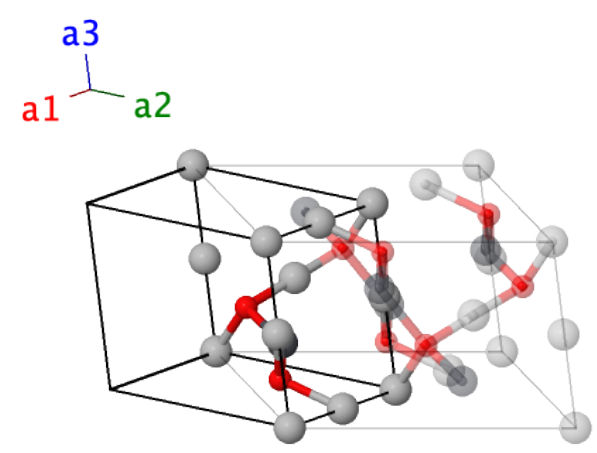

**Basis vectors:**

| | | Lattice Coordinates | | Cartesian Coordinates | Wyckoff Position | Atom Type |
|---|---|---|---|---|---|---|
| $\mathbf{B}_1$ | = | $0\,\mathbf{a}_1 + 0\,\mathbf{a}_2 + 0\,\mathbf{a}_3$ | = | $0\,\hat{\mathbf{x}} + 0\,\hat{\mathbf{y}} + 0\,\hat{\mathbf{z}}$ | (4$a$) | Ag I |
| $\mathbf{B}_2$ | = | $\frac{1}{2}\,\mathbf{a}_3$ | = | $\frac{1}{2}c\cos\beta\,\hat{\mathbf{x}} + \frac{1}{2}c\sin\beta\,\hat{\mathbf{z}}$ | (4$a$) | Ag I |
| $\mathbf{B}_3$ | = | $\frac{1}{2}\,\mathbf{a}_2 + \frac{1}{2}\,\mathbf{a}_3$ | = | $\left(\frac{1}{4}a + \frac{1}{2}c\cos\beta\right)\,\hat{\mathbf{x}} + \frac{1}{4}b\,\hat{\mathbf{y}} + \frac{1}{2}c\sin\beta\,\hat{\mathbf{z}}$ | (4$d$) | Ag II |
| $\mathbf{B}_4$ | = | $\frac{1}{2}\,\mathbf{a}_1$ | = | $\frac{1}{4}a\,\hat{\mathbf{x}} - \frac{1}{4}b\,\hat{\mathbf{y}}$ | (4$d$) | Ag II |
| $\mathbf{B}_5$ | = | $-y_3\,\mathbf{a}_1 + y_3\,\mathbf{a}_2 + \frac{1}{4}\,\mathbf{a}_3$ | = | $\frac{1}{4}c\cos\beta\,\hat{\mathbf{x}} + y_3 b\,\hat{\mathbf{y}} + \frac{1}{4}c\sin\beta\,\hat{\mathbf{z}}$ | (4$e$) | Pb |
| $\mathbf{B}_6$ | = | $y_3\,\mathbf{a}_1 - y_3\,\mathbf{a}_2 + \frac{3}{4}\,\mathbf{a}_3$ | = | $\frac{3}{4}c\cos\beta\,\hat{\mathbf{x}} - y_3 b\,\hat{\mathbf{y}} + \frac{3}{4}c\sin\beta\,\hat{\mathbf{z}}$ | (4$e$) | Pb |
| $\mathbf{B}_7$ | = | $(x_4 - y_4)\,\mathbf{a}_1 + (x_4 + y_4)\,\mathbf{a}_2 + z_4\,\mathbf{a}_3$ | = | $(x_4 a + z_4 c\cos\beta)\,\hat{\mathbf{x}} + y_4 b\,\hat{\mathbf{y}} + z_4 c\sin\beta\,\hat{\mathbf{z}}$ | (8$f$) | O |
| $\mathbf{B}_8$ | = | $(-x_4 - y_4)\,\mathbf{a}_1 + (-x_4 + y_4)\,\mathbf{a}_2 + \left(\frac{1}{2} - z_4\right)\,\mathbf{a}_3$ | = | $\left(\frac{1}{2}c\cos\beta - x_4 a - z_4 c\cos\beta\right)\,\hat{\mathbf{x}} + y_4 b\,\hat{\mathbf{y}} + \left(\frac{1}{2} - z_4\right)c\sin\beta\,\hat{\mathbf{z}}$ | (8$f$) | O |
| $\mathbf{B}_9$ | = | $(-x_4 + y_4)\,\mathbf{a}_1 + (-x_4 - y_4)\,\mathbf{a}_2 - z_4\,\mathbf{a}_3$ | = | $(-x_4 a - z_4 c\cos\beta)\,\hat{\mathbf{x}} - y_4 b\,\hat{\mathbf{y}} - z_4 c\sin\beta\,\hat{\mathbf{z}}$ | (8$f$) | O |
| $\mathbf{B}_{10}$ | = | $(x_4 + y_4)\,\mathbf{a}_1 + (x_4 - y_4)\,\mathbf{a}_2 + \left(\frac{1}{2} + z_4\right)\,\mathbf{a}_3$ | = | $\left(\frac{1}{2}c\cos\beta + x_4 a + z_4 c\cos\beta\right)\,\hat{\mathbf{x}} - y_4 b\,\hat{\mathbf{y}} + \left(\frac{1}{2} + z_4\right)c\sin\beta\,\hat{\mathbf{z}}$ | (8$f$) | O |

**Geometry files:**

- CIF: pp. 1560
- POSCAR: pp. 1560

# Catapleiite ($Na_2ZrSi_3O_9{\cdot}2H_2O$) Structure: A2B3C9D3E_mC144_15_2f_bcdef_9f_3f_ae

http://aflow.org/prototype-encyclopedia/A2B3C9D3E_mC144_15_2f_bcdef_9f_3f_ae

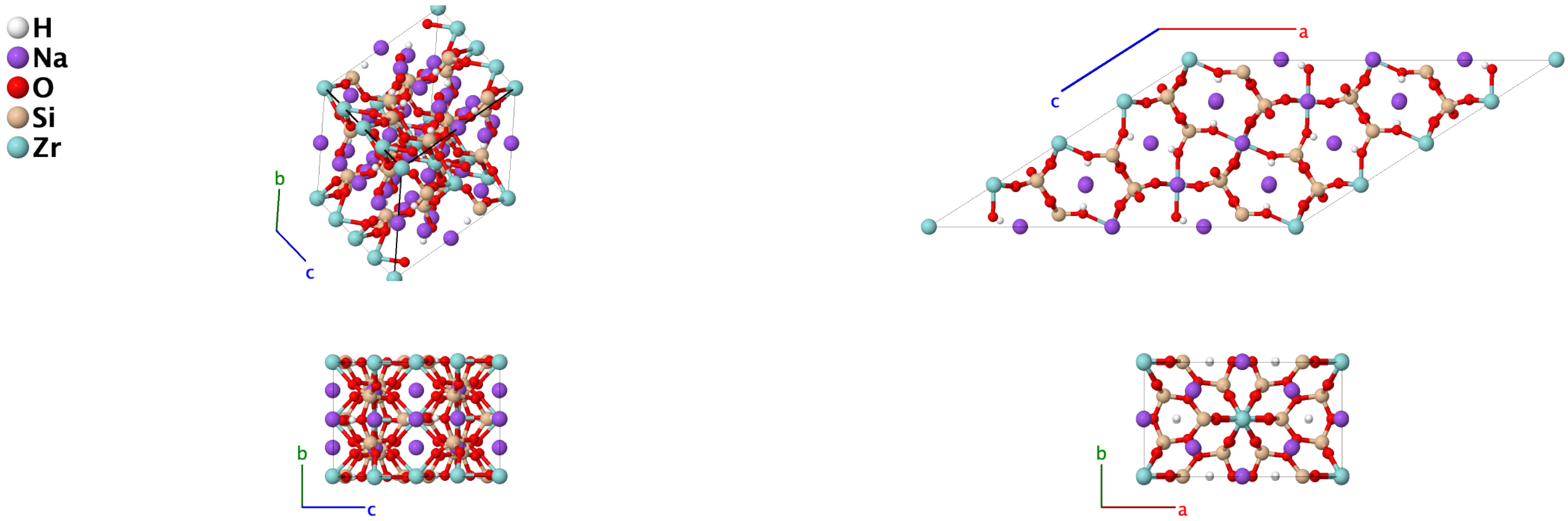

| | | |
|---|---|---|
| **Prototype** | : | $(H_2O)_2Na_2O_9Si_3Zr$ |
| **AFLOW prototype label** | : | A2B3C9D3E_mC144_15_2f_bcdef_9f_3f_ae |
| ***Strukturbericht* designation** | : | None |
| **Pearson symbol** | : | mC144 |
| **Space group number** | : | 15 |
| **Space group symbol** | : | $C2/c$ |
| **AFLOW prototype command** | : | `aflow --proto=A2B3C9D3E_mC144_15_2f_bcdef_9f_3f_ae` `--params=`$a, b/a, c/a, \beta, y_5, y_6, x_7, y_7, z_7, x_8, y_8, z_8, x_9, y_9, z_9, x_{10}, y_{10}, z_{10}, x_{11}, y_{11}, z_{11}, x_{12}, y_{12}, z_{12}, x_{13}, y_{13}, z_{13}, x_{14}, y_{14}, z_{14}, x_{15}, y_{15}, z_{15}, x_{16}, y_{16}, z_{16}, x_{17}, y_{17}, z_{17}, x_{18}, y_{18}, z_{18}, x_{19}, y_{19}, z_{19}, x_{20}, y_{20}, z_{20}, x_{21}, y_{21}, z_{21}$ |

- This is a refinement of the crystal structure of catapleiite. The original hexagonal structure was given the *Strukturbericht* designation $S3_4$ by (Gottfried, 1937).
- The sodium atom sites are only 66.7% occupied in this structure. However, the stoichiometry in the AFLOW label treats the Na sites as fully occupied (*i.e.*, a stoichiometric ratio of three as opposed to two).
- (Ilyushin, 1981) gave the lattice parameters and Wyckoff positions in terms of the $B2/b$ setting of space group #15. We used FINDSYM to transform this to the standard $C2/c$ setting. This required the $y-$ and $z-$axes to be swapped.

**Base-centered Monoclinic primitive vectors:**

$$\mathbf{a}_1 = \frac{1}{2}\,a\,\hat{\mathbf{x}} - \frac{1}{2}\,b\,\hat{\mathbf{y}}$$
$$\mathbf{a}_2 = \frac{1}{2}\,a\,\hat{\mathbf{x}} + \frac{1}{2}\,b\,\hat{\mathbf{y}}$$
$$\mathbf{a}_3 = c\cos\beta\,\hat{\mathbf{x}} + c\sin\beta\,\hat{\mathbf{z}}$$

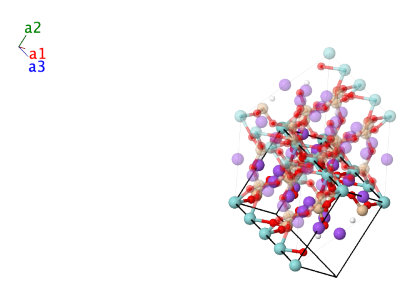

**Basis vectors:**

| | | Lattice Coordinates | | Cartesian Coordinates | Wyckoff Position | Atom Type |
|---|---|---|---|---|---|---|
| $\mathbf{B}_1$ | = | $0\,\mathbf{a}_1 + 0\,\mathbf{a}_2 + 0\,\mathbf{a}_3$ | = | $0\,\hat{\mathbf{x}} + 0\,\hat{\mathbf{y}} + 0\,\hat{\mathbf{z}}$ | (4$a$) | Zr I |
| $\mathbf{B}_2$ | = | $\frac{1}{2}\,\mathbf{a}_3$ | = | $\frac{1}{2}c\cos\beta\,\hat{\mathbf{x}} + \frac{1}{2}c\sin\beta\,\hat{\mathbf{z}}$ | (4$a$) | Zr I |
| $\mathbf{B}_3$ | = | $\frac{1}{2}\,\mathbf{a}_1 + \frac{1}{2}\,\mathbf{a}_2$ | = | $\frac{1}{2}a\,\hat{\mathbf{x}}$ | (4*b*) | Na I |
| $\mathbf{B}_4$ | = | $\frac{1}{2}\,\mathbf{a}_1 + \frac{1}{2}\,\mathbf{a}_2 + \frac{1}{2}\,\mathbf{a}_3$ | = | $\frac{1}{2}\left(a + c\cos\beta\right)\,\hat{\mathbf{x}} + \frac{1}{2}c\sin\beta\,\hat{\mathbf{z}}$ | (4*b*) | Na I |
| $\mathbf{B}_5$ | = | $\frac{1}{2}\,\mathbf{a}_2$ | = | $\frac{1}{4}a\,\hat{\mathbf{x}} + \frac{1}{4}b\,\hat{\mathbf{y}}$ | (4*c*) | Na II |
| $\mathbf{B}_6$ | = | $\frac{1}{2}\,\mathbf{a}_1 + \frac{1}{2}\,\mathbf{a}_3$ | = | $\left(\frac{1}{4}a + \frac{1}{2}c\cos\beta\right)\,\hat{\mathbf{x}} - \frac{1}{4}b\,\hat{\mathbf{y}} + \frac{1}{2}c\sin\beta\,\hat{\mathbf{z}}$ | (4*c*) | Na II |
| $\mathbf{B}_7$ | = | $\frac{1}{2}\,\mathbf{a}_2 + \frac{1}{2}\,\mathbf{a}_3$ | = | $\left(\frac{1}{4}a + \frac{1}{2}c\cos\beta\right)\,\hat{\mathbf{x}} + \frac{1}{4}b\,\hat{\mathbf{y}} + \frac{1}{2}c\sin\beta\,\hat{\mathbf{z}}$ | (4*d*) | Na III |
| $\mathbf{B}_8$ | = | $\frac{1}{2}\,\mathbf{a}_1$ | = | $\frac{1}{4}a\,\hat{\mathbf{x}} - \frac{1}{4}b\,\hat{\mathbf{y}}$ | (4*d*) | Na III |
| $\mathbf{B}_9$ | = | $-y_5\,\mathbf{a}_1 + y_5\,\mathbf{a}_2 + \frac{1}{4}\,\mathbf{a}_3$ | = | $\frac{1}{4}c\cos\beta\,\hat{\mathbf{x}} + y_5 b\,\hat{\mathbf{y}} + \frac{1}{4}c\sin\beta\,\hat{\mathbf{z}}$ | (4*e*) | Na IV |
| $\mathbf{B}_{10}$ | = | $y_5\,\mathbf{a}_1 - y_5\,\mathbf{a}_2 + \frac{3}{4}\,\mathbf{a}_3$ | = | $\frac{3}{4}c\cos\beta\,\hat{\mathbf{x}} - y_5 b\,\hat{\mathbf{y}} + \frac{3}{4}c\sin\beta\,\hat{\mathbf{z}}$ | (4*e*) | Na IV |
| $\mathbf{B}_{11}$ | = | $-y_6\,\mathbf{a}_1 + y_6\,\mathbf{a}_2 + \frac{1}{4}\,\mathbf{a}_3$ | = | $\frac{1}{4}c\cos\beta\,\hat{\mathbf{x}} + y_6 b\,\hat{\mathbf{y}} + \frac{1}{4}c\sin\beta\,\hat{\mathbf{z}}$ | (4*e*) | Zr II |
| $\mathbf{B}_{12}$ | = | $y_6\,\mathbf{a}_1 - y_6\,\mathbf{a}_2 + \frac{3}{4}\,\mathbf{a}_3$ | = | $\frac{3}{4}c\cos\beta\,\hat{\mathbf{x}} - y_6 b\,\hat{\mathbf{y}} + \frac{3}{4}c\sin\beta\,\hat{\mathbf{z}}$ | (4*e*) | Zr II |
| $\mathbf{B}_{13}$ | = | $\left(x_7 - y_7\right)\,\mathbf{a}_1 + \left(x_7 + y_7\right)\,\mathbf{a}_2 + z_7\,\mathbf{a}_3$ | = | $\left(x_7 a + z_7 c\cos\beta\right)\,\hat{\mathbf{x}} + y_7 b\,\hat{\mathbf{y}} + z_7 c\sin\beta\,\hat{\mathbf{z}}$ | (8*f*) | $H_2O$ I |
| $\mathbf{B}_{14}$ | = | $\left(-x_7 - y_7\right)\,\mathbf{a}_1 + \left(-x_7 + y_7\right)\,\mathbf{a}_2 + \left(\frac{1}{2} - z_7\right)\,\mathbf{a}_3$ | = | $\left(\frac{1}{2}c\cos\beta - x_7 a - z_7 c\cos\beta\right)\,\hat{\mathbf{x}} + y_7 b\,\hat{\mathbf{y}} + \left(\frac{1}{2} - z_7\right)c\sin\beta\,\hat{\mathbf{z}}$ | (8*f*) | $H_2O$ I |
| $\mathbf{B}_{15}$ | = | $\left(-x_7 + y_7\right)\,\mathbf{a}_1 + \left(-x_7 - y_7\right)\,\mathbf{a}_2 - z_7\,\mathbf{a}_3$ | = | $\left(-x_7 a - z_7 c\cos\beta\right)\,\hat{\mathbf{x}} - y_7 b\,\hat{\mathbf{y}} - z_7 c\sin\beta\,\hat{\mathbf{z}}$ | (8*f*) | $H_2O$ I |
| $\mathbf{B}_{16}$ | = | $\left(x_7 + y_7\right)\,\mathbf{a}_1 + \left(x_7 - y_7\right)\,\mathbf{a}_2 + \left(\frac{1}{2} + z_7\right)\,\mathbf{a}_3$ | = | $\left(\frac{1}{2}c\cos\beta + x_7 a + z_7 c\cos\beta\right)\,\hat{\mathbf{x}} - y_7 b\,\hat{\mathbf{y}} + \left(\frac{1}{2} + z_7\right)c\sin\beta\,\hat{\mathbf{z}}$ | (8*f*) | $H_2O$ I |
| $\mathbf{B}_{17}$ | = | $\left(x_8 - y_8\right)\,\mathbf{a}_1 + \left(x_8 + y_8\right)\,\mathbf{a}_2 + z_8\,\mathbf{a}_3$ | = | $\left(x_8 a + z_8 c\cos\beta\right)\,\hat{\mathbf{x}} + y_8 b\,\hat{\mathbf{y}} + z_8 c\sin\beta\,\hat{\mathbf{z}}$ | (8*f*) | $H_2O$ II |
| $\mathbf{B}_{18}$ | = | $\left(-x_8 - y_8\right)\,\mathbf{a}_1 + \left(-x_8 + y_8\right)\,\mathbf{a}_2 + \left(\frac{1}{2} - z_8\right)\,\mathbf{a}_3$ | = | $\left(\frac{1}{2}c\cos\beta - x_8 a - z_8 c\cos\beta\right)\,\hat{\mathbf{x}} + y_8 b\,\hat{\mathbf{y}} + \left(\frac{1}{2} - z_8\right)c\sin\beta\,\hat{\mathbf{z}}$ | (8*f*) | $H_2O$ II |
| $\mathbf{B}_{19}$ | = | $\left(-x_8 + y_8\right)\,\mathbf{a}_1 + \left(-x_8 - y_8\right)\,\mathbf{a}_2 - z_8\,\mathbf{a}_3$ | = | $\left(-x_8 a - z_8 c\cos\beta\right)\,\hat{\mathbf{x}} - y_8 b\,\hat{\mathbf{y}} - z_8 c\sin\beta\,\hat{\mathbf{z}}$ | (8*f*) | $H_2O$ II |
| $\mathbf{B}_{20}$ | = | $\left(x_8 + y_8\right)\,\mathbf{a}_1 + \left(x_8 - y_8\right)\,\mathbf{a}_2 + \left(\frac{1}{2} + z_8\right)\,\mathbf{a}_3$ | = | $\left(\frac{1}{2}c\cos\beta + x_8 a + z_8 c\cos\beta\right)\,\hat{\mathbf{x}} - y_8 b\,\hat{\mathbf{y}} + \left(\frac{1}{2} + z_8\right)c\sin\beta\,\hat{\mathbf{z}}$ | (8*f*) | $H_2O$ II |
| $\mathbf{B}_{21}$ | = | $\left(x_9 - y_9\right)\,\mathbf{a}_1 + \left(x_9 + y_9\right)\,\mathbf{a}_2 + z_9\,\mathbf{a}_3$ | = | $\left(x_9 a + z_9 c\cos\beta\right)\,\hat{\mathbf{x}} + y_9 b\,\hat{\mathbf{y}} + z_9 c\sin\beta\,\hat{\mathbf{z}}$ | (8*f*) | Na V |
| $\mathbf{B}_{22}$ | = | $\left(-x_9 - y_9\right)\,\mathbf{a}_1 + \left(-x_9 + y_9\right)\,\mathbf{a}_2 + \left(\frac{1}{2} - z_9\right)\,\mathbf{a}_3$ | = | $\left(\frac{1}{2}c\cos\beta - x_9 a - z_9 c\cos\beta\right)\,\hat{\mathbf{x}} + y_9 b\,\hat{\mathbf{y}} + \left(\frac{1}{2} - z_9\right)c\sin\beta\,\hat{\mathbf{z}}$ | (8*f*) | Na V |
| $\mathbf{B}_{23}$ | = | $\left(-x_9 + y_9\right)\,\mathbf{a}_1 + \left(-x_9 - y_9\right)\,\mathbf{a}_2 - z_9\,\mathbf{a}_3$ | = | $\left(-x_9 a - z_9 c\cos\beta\right)\,\hat{\mathbf{x}} - y_9 b\,\hat{\mathbf{y}} - z_9 c\sin\beta\,\hat{\mathbf{z}}$ | (8*f*) | Na V |
| $\mathbf{B}_{24}$ | = | $\left(x_9 + y_9\right)\,\mathbf{a}_1 + \left(x_9 - y_9\right)\,\mathbf{a}_2 + \left(\frac{1}{2} + z_9\right)\,\mathbf{a}_3$ | = | $\left(\frac{1}{2}c\cos\beta + x_9 a + z_9 c\cos\beta\right)\,\hat{\mathbf{x}} - y_9 b\,\hat{\mathbf{y}} + \left(\frac{1}{2} + z_9\right)c\sin\beta\,\hat{\mathbf{z}}$ | (8*f*) | Na V |
| $\mathbf{B}_{25}$ | = | $\left(x_{10} - y_{10}\right)\,\mathbf{a}_1 + \left(x_{10} + y_{10}\right)\,\mathbf{a}_2 + z_{10}\,\mathbf{a}_3$ | = | $\left(x_{10} a + z_{10} c\cos\beta\right)\,\hat{\mathbf{x}} + y_{10} b\,\hat{\mathbf{y}} + z_{10} c\sin\beta\,\hat{\mathbf{z}}$ | (8*f*) | O I |
| $\mathbf{B}_{26}$ | = | $\left(-x_{10} - y_{10}\right)\,\mathbf{a}_1 + \left(-x_{10} + y_{10}\right)\,\mathbf{a}_2 + \left(\frac{1}{2} - z_{10}\right)\,\mathbf{a}_3$ | = | $\left(\frac{1}{2}c\cos\beta - x_{10} a - z_{10} c\cos\beta\right)\,\hat{\mathbf{x}} + y_{10} b\,\hat{\mathbf{y}} + \left(\frac{1}{2} - z_{10}\right)c\sin\beta\,\hat{\mathbf{z}}$ | (8*f*) | O I |
| $\mathbf{B}_{27}$ | = | $\left(-x_{10} + y_{10}\right)\,\mathbf{a}_1 + \left(-x_{10} - y_{10}\right)\,\mathbf{a}_2 - z_{10}\,\mathbf{a}_3$ | = | $\left(-x_{10} a - z_{10} c\cos\beta\right)\,\hat{\mathbf{x}} - y_{10} b\,\hat{\mathbf{y}} - z_{10} c\sin\beta\,\hat{\mathbf{z}}$ | (8*f*) | O I |
| $\mathbf{B}_{28}$ | = | $\left(x_{10} + y_{10}\right)\,\mathbf{a}_1 + \left(x_{10} - y_{10}\right)\,\mathbf{a}_2 + \left(\frac{1}{2} + z_{10}\right)\,\mathbf{a}_3$ | = | $\left(\frac{1}{2}c\cos\beta + x_{10} a + z_{10} c\cos\beta\right)\,\hat{\mathbf{x}} - y_{10} b\,\hat{\mathbf{y}} + \left(\frac{1}{2} + z_{10}\right)c\sin\beta\,\hat{\mathbf{z}}$ | (8*f*) | O I |
| $\mathbf{B}_{29}$ | = | $\left(x_{11}-y_{11}\right)\,\mathbf{a}_1+\left(x_{11}+y_{11}\right)\,\mathbf{a}_2+z_{11}\,\mathbf{a}_3$ | = | $\left(x_{11}a+z_{11}c\cos\beta\right)\,\mathbf{\hat{x}}+y_{11}b\,\mathbf{\hat{y}}+z_{11}c\sin\beta\,\mathbf{\hat{z}}$ | (8$f$) | O II |
| $\mathbf{B}_{30}$ | = | $\left(-x_{11}-y_{11}\right)\,\mathbf{a}_1+\left(-x_{11}+y_{11}\right)\,\mathbf{a}_2+\left(\frac{1}{2}-z_{11}\right)\,\mathbf{a}_3$ | = | $\left(\frac{1}{2}c\cos\beta-x_{11}a-z_{11}c\cos\beta\right)\,\mathbf{\hat{x}}+y_{11}b\,\mathbf{\hat{y}}+\left(\frac{1}{2}-z_{11}\right)c\sin\beta\,\mathbf{\hat{z}}$ | (8$f$) | O II |
| $\mathbf{B}_{31}$ | = | $\left(-x_{11}+y_{11}\right)\,\mathbf{a}_1+\left(-x_{11}-y_{11}\right)\,\mathbf{a}_2-z_{11}\,\mathbf{a}_3$ | = | $\left(-x_{11}a-z_{11}c\cos\beta\right)\,\mathbf{\hat{x}}-y_{11}b\,\mathbf{\hat{y}}-z_{11}c\sin\beta\,\mathbf{\hat{z}}$ | (8$f$) | O II |
| $\mathbf{B}_{32}$ | = | $\left(x_{11}+y_{11}\right)\,\mathbf{a}_1+\left(x_{11}-y_{11}\right)\,\mathbf{a}_2+\left(\frac{1}{2}+z_{11}\right)\,\mathbf{a}_3$ | = | $\left(\frac{1}{2}c\cos\beta+x_{11}a+z_{11}c\cos\beta\right)\,\mathbf{\hat{x}}-y_{11}b\,\mathbf{\hat{y}}+\left(\frac{1}{2}+z_{11}\right)c\sin\beta\,\mathbf{\hat{z}}$ | (8$f$) | O II |
| $\mathbf{B}_{33}$ | = | $\left(x_{12}-y_{12}\right)\,\mathbf{a}_1+\left(x_{12}+y_{12}\right)\,\mathbf{a}_2+z_{12}\,\mathbf{a}_3$ | = | $\left(x_{12}a+z_{12}c\cos\beta\right)\,\mathbf{\hat{x}}+y_{12}b\,\mathbf{\hat{y}}+z_{12}c\sin\beta\,\mathbf{\hat{z}}$ | (8$f$) | O III |
| $\mathbf{B}_{34}$ | = | $\left(-x_{12}-y_{12}\right)\,\mathbf{a}_1+\left(-x_{12}+y_{12}\right)\,\mathbf{a}_2+\left(\frac{1}{2}-z_{12}\right)\,\mathbf{a}_3$ | = | $\left(\frac{1}{2}c\cos\beta-x_{12}a-z_{12}c\cos\beta\right)\,\mathbf{\hat{x}}+y_{12}b\,\mathbf{\hat{y}}+\left(\frac{1}{2}-z_{12}\right)c\sin\beta\,\mathbf{\hat{z}}$ | (8$f$) | O III |
| $\mathbf{B}_{35}$ | = | $\left(-x_{12}+y_{12}\right)\,\mathbf{a}_1+\left(-x_{12}-y_{12}\right)\,\mathbf{a}_2-z_{12}\,\mathbf{a}_3$ | = | $\left(-x_{12}a-z_{12}c\cos\beta\right)\,\mathbf{\hat{x}}-y_{12}b\,\mathbf{\hat{y}}-z_{12}c\sin\beta\,\mathbf{\hat{z}}$ | (8$f$) | O III |
| $\mathbf{B}_{36}$ | = | $\left(x_{12}+y_{12}\right)\,\mathbf{a}_1+\left(x_{12}-y_{12}\right)\,\mathbf{a}_2+\left(\frac{1}{2}+z_{12}\right)\,\mathbf{a}_3$ | = | $\left(\frac{1}{2}c\cos\beta+x_{12}a+z_{12}c\cos\beta\right)\,\mathbf{\hat{x}}-y_{12}b\,\mathbf{\hat{y}}+\left(\frac{1}{2}+z_{12}\right)c\sin\beta\,\mathbf{\hat{z}}$ | (8$f$) | O III |
| $\mathbf{B}_{37}$ | = | $\left(x_{13}-y_{13}\right)\,\mathbf{a}_1+\left(x_{13}+y_{13}\right)\,\mathbf{a}_2+z_{13}\,\mathbf{a}_3$ | = | $\left(x_{13}a+z_{13}c\cos\beta\right)\,\mathbf{\hat{x}}+y_{13}b\,\mathbf{\hat{y}}+z_{13}c\sin\beta\,\mathbf{\hat{z}}$ | (8$f$) | O IV |
| $\mathbf{B}_{38}$ | = | $\left(-x_{13}-y_{13}\right)\,\mathbf{a}_1+\left(-x_{13}+y_{13}\right)\,\mathbf{a}_2+\left(\frac{1}{2}-z_{13}\right)\,\mathbf{a}_3$ | = | $\left(\frac{1}{2}c\cos\beta-x_{13}a-z_{13}c\cos\beta\right)\,\mathbf{\hat{x}}+y_{13}b\,\mathbf{\hat{y}}+\left(\frac{1}{2}-z_{13}\right)c\sin\beta\,\mathbf{\hat{z}}$ | (8$f$) | O IV |
| $\mathbf{B}_{39}$ | = | $\left(-x_{13}+y_{13}\right)\,\mathbf{a}_1+\left(-x_{13}-y_{13}\right)\,\mathbf{a}_2-z_{13}\,\mathbf{a}_3$ | = | $\left(-x_{13}a-z_{13}c\cos\beta\right)\,\mathbf{\hat{x}}-y_{13}b\,\mathbf{\hat{y}}-z_{13}c\sin\beta\,\mathbf{\hat{z}}$ | (8$f$) | O IV |
| $\mathbf{B}_{40}$ | = | $\left(x_{13}+y_{13}\right)\,\mathbf{a}_1+\left(x_{13}-y_{13}\right)\,\mathbf{a}_2+\left(\frac{1}{2}+z_{13}\right)\,\mathbf{a}_3$ | = | $\left(\frac{1}{2}c\cos\beta+x_{13}a+z_{13}c\cos\beta\right)\,\mathbf{\hat{x}}-y_{13}b\,\mathbf{\hat{y}}+\left(\frac{1}{2}+z_{13}\right)c\sin\beta\,\mathbf{\hat{z}}$ | (8$f$) | O IV |
| $\mathbf{B}_{41}$ | = | $\left(x_{14}-y_{14}\right)\,\mathbf{a}_1+\left(x_{14}+y_{14}\right)\,\mathbf{a}_2+z_{14}\,\mathbf{a}_3$ | = | $\left(x_{14}a+z_{14}c\cos\beta\right)\,\mathbf{\hat{x}}+y_{14}b\,\mathbf{\hat{y}}+z_{14}c\sin\beta\,\mathbf{\hat{z}}$ | (8$f$) | O V |
| $\mathbf{B}_{42}$ | = | $\left(-x_{14}-y_{14}\right)\,\mathbf{a}_1+\left(-x_{14}+y_{14}\right)\,\mathbf{a}_2+\left(\frac{1}{2}-z_{14}\right)\,\mathbf{a}_3$ | = | $\left(\frac{1}{2}c\cos\beta-x_{14}a-z_{14}c\cos\beta\right)\,\mathbf{\hat{x}}+y_{14}b\,\mathbf{\hat{y}}+\left(\frac{1}{2}-z_{14}\right)c\sin\beta\,\mathbf{\hat{z}}$ | (8$f$) | O V |
| $\mathbf{B}_{43}$ | = | $\left(-x_{14}+y_{14}\right)\,\mathbf{a}_1+\left(-x_{14}-y_{14}\right)\,\mathbf{a}_2-z_{14}\,\mathbf{a}_3$ | = | $\left(-x_{14}a-z_{14}c\cos\beta\right)\,\mathbf{\hat{x}}-y_{14}b\,\mathbf{\hat{y}}-z_{14}c\sin\beta\,\mathbf{\hat{z}}$ | (8$f$) | O V |
| $\mathbf{B}_{44}$ | = | $\left(x_{14}+y_{14}\right)\,\mathbf{a}_1+\left(x_{14}-y_{14}\right)\,\mathbf{a}_2+\left(\frac{1}{2}+z_{14}\right)\,\mathbf{a}_3$ | = | $\left(\frac{1}{2}c\cos\beta+x_{14}a+z_{14}c\cos\beta\right)\,\mathbf{\hat{x}}-y_{14}b\,\mathbf{\hat{y}}+\left(\frac{1}{2}+z_{14}\right)c\sin\beta\,\mathbf{\hat{z}}$ | (8$f$) | O V |
| $\mathbf{B}_{45}$ | = | $\left(x_{15}-y_{15}\right)\,\mathbf{a}_1+\left(x_{15}+y_{15}\right)\,\mathbf{a}_2+z_{15}\,\mathbf{a}_3$ | = | $\left(x_{15}a+z_{15}c\cos\beta\right)\,\mathbf{\hat{x}}+y_{15}b\,\mathbf{\hat{y}}+z_{15}c\sin\beta\,\mathbf{\hat{z}}$ | (8$f$) | O VI |
| $\mathbf{B}_{46}$ | = | $\left(-x_{15}-y_{15}\right)\,\mathbf{a}_1+\left(-x_{15}+y_{15}\right)\,\mathbf{a}_2+\left(\frac{1}{2}-z_{15}\right)\,\mathbf{a}_3$ | = | $\left(\frac{1}{2}c\cos\beta-x_{15}a-z_{15}c\cos\beta\right)\,\mathbf{\hat{x}}+y_{15}b\,\mathbf{\hat{y}}+\left(\frac{1}{2}-z_{15}\right)c\sin\beta\,\mathbf{\hat{z}}$ | (8$f$) | O VI |
| $\mathbf{B}_{47}$ | = | $\left(-x_{15}+y_{15}\right)\,\mathbf{a}_1+\left(-x_{15}-y_{15}\right)\,\mathbf{a}_2-z_{15}\,\mathbf{a}_3$ | = | $\left(-x_{15}a-z_{15}c\cos\beta\right)\,\mathbf{\hat{x}}-y_{15}b\,\mathbf{\hat{y}}-z_{15}c\sin\beta\,\mathbf{\hat{z}}$ | (8$f$) | O VI |
| $\mathbf{B}_{48}$ | = | $\left(x_{15}+y_{15}\right)\,\mathbf{a}_1+\left(x_{15}-y_{15}\right)\,\mathbf{a}_2+\left(\frac{1}{2}+z_{15}\right)\,\mathbf{a}_3$ | = | $\left(\frac{1}{2}c\cos\beta+x_{15}a+z_{15}c\cos\beta\right)\,\mathbf{\hat{x}}-y_{15}b\,\mathbf{\hat{y}}+\left(\frac{1}{2}+z_{15}\right)c\sin\beta\,\mathbf{\hat{z}}$ | (8$f$) | O VI |
| $\mathbf{B}_{49}$ | = | $\left(x_{16}-y_{16}\right)\,\mathbf{a}_1+\left(x_{16}+y_{16}\right)\,\mathbf{a}_2+z_{16}\,\mathbf{a}_3$ | = | $\left(x_{16}a+z_{16}c\cos\beta\right)\,\mathbf{\hat{x}}+y_{16}b\,\mathbf{\hat{y}}+z_{16}c\sin\beta\,\mathbf{\hat{z}}$ | (8$f$) | O VII |
| $\mathbf{B}_{50}$ | = | $\left(-x_{16}-y_{16}\right)\,\mathbf{a}_1+\left(-x_{16}+y_{16}\right)\,\mathbf{a}_2+\left(\frac{1}{2}-z_{16}\right)\,\mathbf{a}_3$ | = | $\left(\frac{1}{2}c\cos\beta-x_{16}a-z_{16}c\cos\beta\right)\,\mathbf{\hat{x}}+y_{16}b\,\mathbf{\hat{y}}+\left(\frac{1}{2}-z_{16}\right)c\sin\beta\,\mathbf{\hat{z}}$ | (8$f$) | O VII |
| $\mathbf{B}_{51}$ | = | $\left(-x_{16}+y_{16}\right)\mathbf{a}_1 + \left(-x_{16}-y_{16}\right)\mathbf{a}_2 - z_{16}\,\mathbf{a}_3$ | = | $\left(-x_{16}a - z_{16}c\cos\beta\right)\hat{\mathbf{x}} - y_{16}b\,\hat{\mathbf{y}} - z_{16}c\sin\beta\,\hat{\mathbf{z}}$ | $(8f)$ | O VII |
| $\mathbf{B}_{52}$ | = | $\left(x_{16}+y_{16}\right)\mathbf{a}_1 + \left(x_{16}-y_{16}\right)\mathbf{a}_2 + \left(\frac{1}{2}+z_{16}\right)\mathbf{a}_3$ | = | $\left(\frac{1}{2}c\cos\beta + x_{16}a + z_{16}c\cos\beta\right)\hat{\mathbf{x}} - y_{16}b\,\hat{\mathbf{y}} + \left(\frac{1}{2}+z_{16}\right)c\sin\beta\,\hat{\mathbf{z}}$ | $(8f)$ | O VII |
| $\mathbf{B}_{53}$ | = | $\left(x_{17}-y_{17}\right)\mathbf{a}_1 + \left(x_{17}+y_{17}\right)\mathbf{a}_2 + z_{17}\,\mathbf{a}_3$ | = | $\left(x_{17}a + z_{17}c\cos\beta\right)\hat{\mathbf{x}} + y_{17}b\,\hat{\mathbf{y}} + z_{17}c\sin\beta\,\hat{\mathbf{z}}$ | $(8f)$ | O VIII |
| $\mathbf{B}_{54}$ | = | $\left(-x_{17}-y_{17}\right)\mathbf{a}_1 + \left(-x_{17}+y_{17}\right)\mathbf{a}_2 + \left(\frac{1}{2}-z_{17}\right)\mathbf{a}_3$ | = | $\left(\frac{1}{2}c\cos\beta - x_{17}a - z_{17}c\cos\beta\right)\hat{\mathbf{x}} + y_{17}b\,\hat{\mathbf{y}} + \left(\frac{1}{2}-z_{17}\right)c\sin\beta\,\hat{\mathbf{z}}$ | $(8f)$ | O VIII |
| $\mathbf{B}_{55}$ | = | $\left(-x_{17}+y_{17}\right)\mathbf{a}_1 + \left(-x_{17}-y_{17}\right)\mathbf{a}_2 - z_{17}\,\mathbf{a}_3$ | = | $\left(-x_{17}a - z_{17}c\cos\beta\right)\hat{\mathbf{x}} - y_{17}b\,\hat{\mathbf{y}} - z_{17}c\sin\beta\,\hat{\mathbf{z}}$ | $(8f)$ | O VIII |
| $\mathbf{B}_{56}$ | = | $\left(x_{17}+y_{17}\right)\mathbf{a}_1 + \left(x_{17}-y_{17}\right)\mathbf{a}_2 + \left(\frac{1}{2}+z_{17}\right)\mathbf{a}_3$ | = | $\left(\frac{1}{2}c\cos\beta + x_{17}a + z_{17}c\cos\beta\right)\hat{\mathbf{x}} - y_{17}b\,\hat{\mathbf{y}} + \left(\frac{1}{2}+z_{17}\right)c\sin\beta\,\hat{\mathbf{z}}$ | $(8f)$ | O VIII |
| $\mathbf{B}_{57}$ | = | $\left(x_{18}-y_{18}\right)\mathbf{a}_1 + \left(x_{18}+y_{18}\right)\mathbf{a}_2 + z_{18}\,\mathbf{a}_3$ | = | $\left(x_{18}a + z_{18}c\cos\beta\right)\hat{\mathbf{x}} + y_{18}b\,\hat{\mathbf{y}} + z_{18}c\sin\beta\,\hat{\mathbf{z}}$ | $(8f)$ | O IX |
| $\mathbf{B}_{58}$ | = | $\left(-x_{18}-y_{18}\right)\mathbf{a}_1 + \left(-x_{18}+y_{18}\right)\mathbf{a}_2 + \left(\frac{1}{2}-z_{18}\right)\mathbf{a}_3$ | = | $\left(\frac{1}{2}c\cos\beta - x_{18}a - z_{18}c\cos\beta\right)\hat{\mathbf{x}} + y_{18}b\,\hat{\mathbf{y}} + \left(\frac{1}{2}-z_{18}\right)c\sin\beta\,\hat{\mathbf{z}}$ | $(8f)$ | O IX |
| $\mathbf{B}_{59}$ | = | $\left(-x_{18}+y_{18}\right)\mathbf{a}_1 + \left(-x_{18}-y_{18}\right)\mathbf{a}_2 - z_{18}\,\mathbf{a}_3$ | = | $\left(-x_{18}a - z_{18}c\cos\beta\right)\hat{\mathbf{x}} - y_{18}b\,\hat{\mathbf{y}} - z_{18}c\sin\beta\,\hat{\mathbf{z}}$ | $(8f)$ | O IX |
| $\mathbf{B}_{60}$ | = | $\left(x_{18}+y_{18}\right)\mathbf{a}_1 + \left(x_{18}-y_{18}\right)\mathbf{a}_2 + \left(\frac{1}{2}+z_{18}\right)\mathbf{a}_3$ | = | $\left(\frac{1}{2}c\cos\beta + x_{18}a + z_{18}c\cos\beta\right)\hat{\mathbf{x}} - y_{18}b\,\hat{\mathbf{y}} + \left(\frac{1}{2}+z_{18}\right)c\sin\beta\,\hat{\mathbf{z}}$ | $(8f)$ | O IX |
| $\mathbf{B}_{61}$ | = | $\left(x_{19}-y_{19}\right)\mathbf{a}_1 + \left(x_{19}+y_{19}\right)\mathbf{a}_2 + z_{19}\,\mathbf{a}_3$ | = | $\left(x_{19}a + z_{19}c\cos\beta\right)\hat{\mathbf{x}} + y_{19}b\,\hat{\mathbf{y}} + z_{19}c\sin\beta\,\hat{\mathbf{z}}$ | $(8f)$ | Si I |
| $\mathbf{B}_{62}$ | = | $\left(-x_{19}-y_{19}\right)\mathbf{a}_1 + \left(-x_{19}+y_{19}\right)\mathbf{a}_2 + \left(\frac{1}{2}-z_{19}\right)\mathbf{a}_3$ | = | $\left(\frac{1}{2}c\cos\beta - x_{19}a - z_{19}c\cos\beta\right)\hat{\mathbf{x}} + y_{19}b\,\hat{\mathbf{y}} + \left(\frac{1}{2}-z_{19}\right)c\sin\beta\,\hat{\mathbf{z}}$ | $(8f)$ | Si I |
| $\mathbf{B}_{63}$ | = | $\left(-x_{19}+y_{19}\right)\mathbf{a}_1 + \left(-x_{19}-y_{19}\right)\mathbf{a}_2 - z_{19}\,\mathbf{a}_3$ | = | $\left(-x_{19}a - z_{19}c\cos\beta\right)\hat{\mathbf{x}} - y_{19}b\,\hat{\mathbf{y}} - z_{19}c\sin\beta\,\hat{\mathbf{z}}$ | $(8f)$ | Si I |
| $\mathbf{B}_{64}$ | = | $\left(x_{19}+y_{19}\right)\mathbf{a}_1 + \left(x_{19}-y_{19}\right)\mathbf{a}_2 + \left(\frac{1}{2}+z_{19}\right)\mathbf{a}_3$ | = | $\left(\frac{1}{2}c\cos\beta + x_{19}a + z_{19}c\cos\beta\right)\hat{\mathbf{x}} - y_{19}b\,\hat{\mathbf{y}} + \left(\frac{1}{2}+z_{19}\right)c\sin\beta\,\hat{\mathbf{z}}$ | $(8f)$ | Si I |
| $\mathbf{B}_{65}$ | = | $\left(x_{20}-y_{20}\right)\mathbf{a}_1 + \left(x_{20}+y_{20}\right)\mathbf{a}_2 + z_{20}\,\mathbf{a}_3$ | = | $\left(x_{20}a + z_{20}c\cos\beta\right)\hat{\mathbf{x}} + y_{20}b\,\hat{\mathbf{y}} + z_{20}c\sin\beta\,\hat{\mathbf{z}}$ | $(8f)$ | Si II |
| $\mathbf{B}_{66}$ | = | $\left(-x_{20}-y_{20}\right)\mathbf{a}_1 + \left(-x_{20}+y_{20}\right)\mathbf{a}_2 + \left(\frac{1}{2}-z_{20}\right)\mathbf{a}_3$ | = | $\left(\frac{1}{2}c\cos\beta - x_{20}a - z_{20}c\cos\beta\right)\hat{\mathbf{x}} + y_{20}b\,\hat{\mathbf{y}} + \left(\frac{1}{2}-z_{20}\right)c\sin\beta\,\hat{\mathbf{z}}$ | $(8f)$ | Si II |
| $\mathbf{B}_{67}$ | = | $\left(-x_{20}+y_{20}\right)\mathbf{a}_1 + \left(-x_{20}-y_{20}\right)\mathbf{a}_2 - z_{20}\,\mathbf{a}_3$ | = | $\left(-x_{20}a - z_{20}c\cos\beta\right)\hat{\mathbf{x}} - y_{20}b\,\hat{\mathbf{y}} - z_{20}c\sin\beta\,\hat{\mathbf{z}}$ | $(8f)$ | Si II |
| $\mathbf{B}_{68}$ | = | $\left(x_{20}+y_{20}\right)\mathbf{a}_1 + \left(x_{20}-y_{20}\right)\mathbf{a}_2 + \left(\frac{1}{2}+z_{20}\right)\mathbf{a}_3$ | = | $\left(\frac{1}{2}c\cos\beta + x_{20}a + z_{20}c\cos\beta\right)\hat{\mathbf{x}} - y_{20}b\,\hat{\mathbf{y}} + \left(\frac{1}{2}+z_{20}\right)c\sin\beta\,\hat{\mathbf{z}}$ | $(8f)$ | Si II |
| $\mathbf{B}_{69}$ | = | $\left(x_{21}-y_{21}\right)\mathbf{a}_1 + \left(x_{21}+y_{21}\right)\mathbf{a}_2 + z_{21}\,\mathbf{a}_3$ | = | $\left(x_{21}a + z_{21}c\cos\beta\right)\hat{\mathbf{x}} + y_{21}b\,\hat{\mathbf{y}} + z_{21}c\sin\beta\,\hat{\mathbf{z}}$ | $(8f)$ | Si III |
| $\mathbf{B}_{70}$ | = | $\left(-x_{21}-y_{21}\right)\mathbf{a}_1 + \left(-x_{21}+y_{21}\right)\mathbf{a}_2 + \left(\frac{1}{2}-z_{21}\right)\mathbf{a}_3$ | = | $\left(\frac{1}{2}c\cos\beta - x_{21}a - z_{21}c\cos\beta\right)\hat{\mathbf{x}} + y_{21}b\,\hat{\mathbf{y}} + \left(\frac{1}{2}-z_{21}\right)c\sin\beta\,\hat{\mathbf{z}}$ | $(8f)$ | Si III |
| $\mathbf{B}_{71}$ | = | $\left(-x_{21}+y_{21}\right)\mathbf{a}_1 + \left(-x_{21}-y_{21}\right)\mathbf{a}_2 - z_{21}\,\mathbf{a}_3$ | = | $\left(-x_{21}a - z_{21}c\cos\beta\right)\hat{\mathbf{x}} - y_{21}b\,\hat{\mathbf{y}} - z_{21}c\sin\beta\,\hat{\mathbf{z}}$ | $(8f)$ | Si III |
| $\mathbf{B}_{72}$ | = | $\left(x_{21}+y_{21}\right)\mathbf{a}_1 + \left(x_{21}-y_{21}\right)\mathbf{a}_2 + \left(\frac{1}{2}+z_{21}\right)\mathbf{a}_3$ | = | $\left(\frac{1}{2}c\cos\beta + x_{21}a + z_{21}c\cos\beta\right)\hat{\mathbf{x}} - y_{21}b\,\hat{\mathbf{y}} + \left(\frac{1}{2}+z_{21}\right)c\sin\beta\,\hat{\mathbf{z}}$ | $(8f)$ | Si III |

**Geometry files:**

- CIF: pp. 1560
- POSCAR: pp. 1561

# $Na_2PrO_3$ Structure: A2B3C_mC48_15_aef_3f_2e

http://aflow.org/prototype-encyclopedia/A2B3C_mC48_15_aef_3f_2e

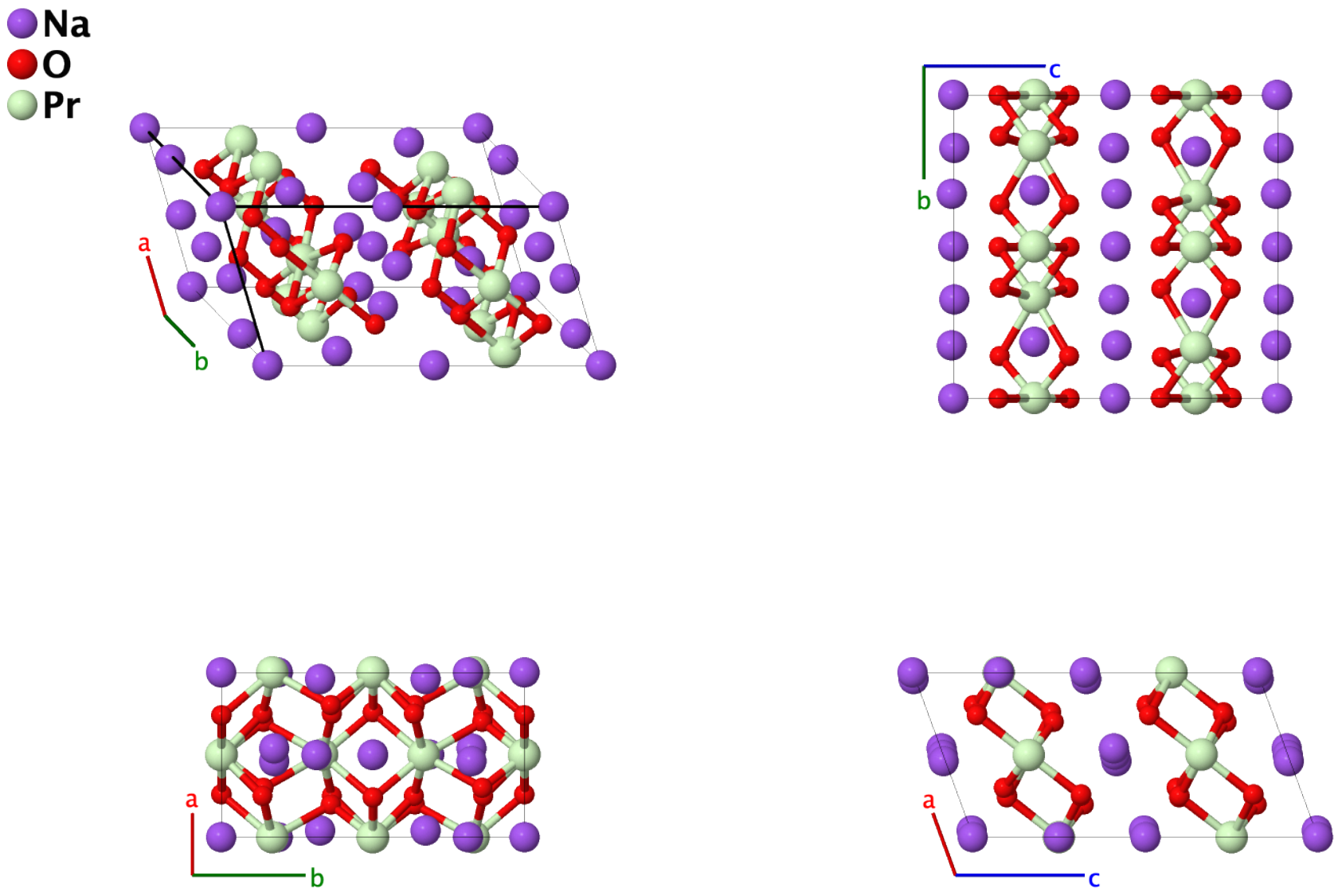

| | | |
|---|---|---|
| **Prototype** | : | $Na_2O_3Pr$ |
| **AFLOW prototype label** | : | A2B3C_mC48_15_aef_3f_2e |
| ***Strukturbericht* designation** | : | None |
| **Pearson symbol** | : | mC48 |
| **Space group number** | : | 15 |
| **Space group symbol** | : | $C2/c$ |
| **AFLOW prototype command** | : | `aflow --proto=A2B3C_mC48_15_aef_3f_2e` `--params=`$a, b/a, c/a, \beta, y_2, y_3, y_4, x_5, y_5, z_5, x_6, y_6, z_6, x_7, y_7, z_7, x_8, y_8, z_8$ |

**Other compounds with this structure**

- $Na_2CrO_3$ and $Na_2TbO_3$

- (Hinatsu, 2006) found that the site we have labeled Na-II is actually 2/3 sodium and 1/3 praseodymium, statistically distributed, while the site we label Pr-I is 1/3 sodium and 2/3 praseodymium.

**Base-centered Monoclinic primitive vectors:**

$$\mathbf{a}_1 = \frac{1}{2}\,a\,\hat{\mathbf{x}} - \frac{1}{2}\,b\,\hat{\mathbf{y}}$$
$$\mathbf{a}_2 = \frac{1}{2}\,a\,\hat{\mathbf{x}} + \frac{1}{2}\,b\,\hat{\mathbf{y}}$$
$$\mathbf{a}_3 = c\cos\beta\,\hat{\mathbf{x}} + c\sin\beta\,\hat{\mathbf{z}}$$

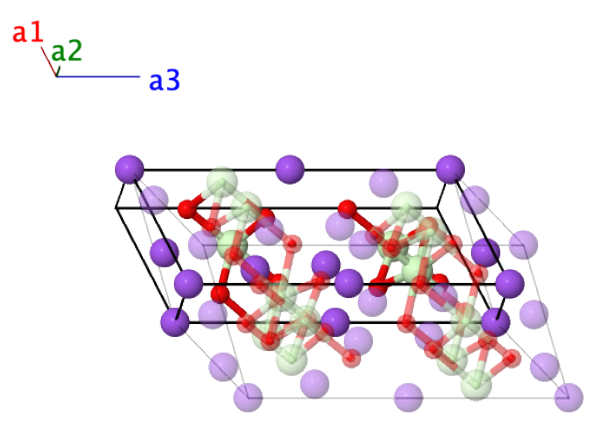

**Basis vectors:**

| | | Lattice Coordinates | | Cartesian Coordinates | Wyckoff Position | Atom Type |
|---|---|---|---|---|---|---|
| $\mathbf{B}_1$ | = | $0\,\mathbf{a}_1 + 0\,\mathbf{a}_2 + 0\,\mathbf{a}_3$ | = | $0\,\hat{\mathbf{x}} + 0\,\hat{\mathbf{y}} + 0\,\hat{\mathbf{z}}$ | (4$a$) | Na I |
| $\mathbf{B}_2$ | = | $\frac{1}{2}\,\mathbf{a}_3$ | = | $\frac{1}{2}c\cos\beta\,\hat{\mathbf{x}} + \frac{1}{2}c\sin\beta\,\hat{\mathbf{z}}$ | (4$a$) | Na I |
| $\mathbf{B}_3$ | = | $-y_2\,\mathbf{a}_1 + y_2\,\mathbf{a}_2 + \frac{1}{4}\,\mathbf{a}_3$ | = | $\frac{1}{4}c\cos\beta\,\hat{\mathbf{x}} + y_2 b\,\hat{\mathbf{y}} + \frac{1}{4}c\sin\beta\,\hat{\mathbf{z}}$ | (4$e$) | Na II |
| $\mathbf{B}_4$ | = | $y_2\,\mathbf{a}_1 - y_2\,\mathbf{a}_2 + \frac{3}{4}\,\mathbf{a}_3$ | = | $\frac{3}{4}c\cos\beta\,\hat{\mathbf{x}} - y_2 b\,\hat{\mathbf{y}} + \frac{3}{4}c\sin\beta\,\hat{\mathbf{z}}$ | (4$e$) | Na II |
| $\mathbf{B}_5$ | = | $-y_3\,\mathbf{a}_1 + y_3\,\mathbf{a}_2 + \frac{1}{4}\,\mathbf{a}_3$ | = | $\frac{1}{4}c\cos\beta\,\hat{\mathbf{x}} + y_3 b\,\hat{\mathbf{y}} + \frac{1}{4}c\sin\beta\,\hat{\mathbf{z}}$ | (4$e$) | Pr I |
| $\mathbf{B}_6$ | = | $y_3\,\mathbf{a}_1 - y_3\,\mathbf{a}_2 + \frac{3}{4}\,\mathbf{a}_3$ | = | $\frac{3}{4}c\cos\beta\,\hat{\mathbf{x}} - y_3 b\,\hat{\mathbf{y}} + \frac{3}{4}c\sin\beta\,\hat{\mathbf{z}}$ | (4$e$) | Pr I |
| $\mathbf{B}_7$ | = | $-y_4\,\mathbf{a}_1 + y_4\,\mathbf{a}_2 + \frac{1}{4}\,\mathbf{a}_3$ | = | $\frac{1}{4}c\cos\beta\,\hat{\mathbf{x}} + y_4 b\,\hat{\mathbf{y}} + \frac{1}{4}c\sin\beta\,\hat{\mathbf{z}}$ | (4$e$) | Pr II |
| $\mathbf{B}_8$ | = | $y_4\,\mathbf{a}_1 - y_4\,\mathbf{a}_2 + \frac{3}{4}\,\mathbf{a}_3$ | = | $\frac{3}{4}c\cos\beta\,\hat{\mathbf{x}} - y_4 b\,\hat{\mathbf{y}} + \frac{3}{4}c\sin\beta\,\hat{\mathbf{z}}$ | (4$e$) | Pr II |
| $\mathbf{B}_9$ | = | $(x_5 - y_5)\,\mathbf{a}_1 + (x_5 + y_5)\,\mathbf{a}_2 + z_5\,\mathbf{a}_3$ | = | $(x_5 a + z_5 c\cos\beta)\,\hat{\mathbf{x}} + y_5 b\,\hat{\mathbf{y}} + z_5 c\sin\beta\,\hat{\mathbf{z}}$ | (8$f$) | Na III |
| $\mathbf{B}_{10}$ | = | $(-x_5 - y_5)\,\mathbf{a}_1 + (-x_5 + y_5)\,\mathbf{a}_2 + \left(\frac{1}{2} - z_5\right)\mathbf{a}_3$ | = | $\left(\frac{1}{2}c\cos\beta - x_5 a - z_5 c\cos\beta\right)\hat{\mathbf{x}} + y_5 b\,\hat{\mathbf{y}} + \left(\frac{1}{2} - z_5\right)c\sin\beta\,\hat{\mathbf{z}}$ | (8$f$) | Na III |
| $\mathbf{B}_{11}$ | = | $(-x_5 + y_5)\,\mathbf{a}_1 + (-x_5 - y_5)\,\mathbf{a}_2 - z_5\,\mathbf{a}_3$ | = | $(-x_5 a - z_5 c\cos\beta)\,\hat{\mathbf{x}} - y_5 b\,\hat{\mathbf{y}} - z_5 c\sin\beta\,\hat{\mathbf{z}}$ | (8$f$) | Na III |
| $\mathbf{B}_{12}$ | = | $(x_5 + y_5)\,\mathbf{a}_1 + (x_5 - y_5)\,\mathbf{a}_2 + \left(\frac{1}{2} + z_5\right)\mathbf{a}_3$ | = | $\left(\frac{1}{2}c\cos\beta + x_5 a + z_5 c\cos\beta\right)\hat{\mathbf{x}} - y_5 b\,\hat{\mathbf{y}} + \left(\frac{1}{2} + z_5\right)c\sin\beta\,\hat{\mathbf{z}}$ | (8$f$) | Na III |
| $\mathbf{B}_{13}$ | = | $(x_6 - y_6)\,\mathbf{a}_1 + (x_6 + y_6)\,\mathbf{a}_2 + z_6\,\mathbf{a}_3$ | = | $(x_6 a + z_6 c\cos\beta)\,\hat{\mathbf{x}} + y_6 b\,\hat{\mathbf{y}} + z_6 c\sin\beta\,\hat{\mathbf{z}}$ | (8$f$) | O I |
| $\mathbf{B}_{14}$ | = | $(-x_6 - y_6)\,\mathbf{a}_1 + (-x_6 + y_6)\,\mathbf{a}_2 + \left(\frac{1}{2} - z_6\right)\mathbf{a}_3$ | = | $\left(\frac{1}{2}c\cos\beta - x_6 a - z_6 c\cos\beta\right)\hat{\mathbf{x}} + y_6 b\,\hat{\mathbf{y}} + \left(\frac{1}{2} - z_6\right)c\sin\beta\,\hat{\mathbf{z}}$ | (8$f$) | O I |
| $\mathbf{B}_{15}$ | = | $(-x_6 + y_6)\,\mathbf{a}_1 + (-x_6 - y_6)\,\mathbf{a}_2 - z_6\,\mathbf{a}_3$ | = | $(-x_6 a - z_6 c\cos\beta)\,\hat{\mathbf{x}} - y_6 b\,\hat{\mathbf{y}} - z_6 c\sin\beta\,\hat{\mathbf{z}}$ | (8$f$) | O I |
| $\mathbf{B}_{16}$ | = | $(x_6 + y_6)\,\mathbf{a}_1 + (x_6 - y_6)\,\mathbf{a}_2 + \left(\frac{1}{2} + z_6\right)\mathbf{a}_3$ | = | $\left(\frac{1}{2}c\cos\beta + x_6 a + z_6 c\cos\beta\right)\hat{\mathbf{x}} - y_6 b\,\hat{\mathbf{y}} + \left(\frac{1}{2} + z_6\right)c\sin\beta\,\hat{\mathbf{z}}$ | (8$f$) | O I |
| $\mathbf{B}_{17}$ | = | $(x_7 - y_7)\,\mathbf{a}_1 + (x_7 + y_7)\,\mathbf{a}_2 + z_7\,\mathbf{a}_3$ | = | $(x_7 a + z_7 c\cos\beta)\,\hat{\mathbf{x}} + y_7 b\,\hat{\mathbf{y}} + z_7 c\sin\beta\,\hat{\mathbf{z}}$ | (8$f$) | O II |
| $\mathbf{B}_{18}$ | = | $(-x_7 - y_7)\,\mathbf{a}_1 + (-x_7 + y_7)\,\mathbf{a}_2 + \left(\frac{1}{2} - z_7\right)\mathbf{a}_3$ | = | $\left(\frac{1}{2}c\cos\beta - x_7 a - z_7 c\cos\beta\right)\hat{\mathbf{x}} + y_7 b\,\hat{\mathbf{y}} + \left(\frac{1}{2} - z_7\right)c\sin\beta\,\hat{\mathbf{z}}$ | (8$f$) | O II |
| $\mathbf{B}_{19}$ | = | $(-x_7 + y_7)\,\mathbf{a}_1 + (-x_7 - y_7)\,\mathbf{a}_2 - z_7\,\mathbf{a}_3$ | = | $(-x_7 a - z_7 c\cos\beta)\,\hat{\mathbf{x}} - y_7 b\,\hat{\mathbf{y}} - z_7 c\sin\beta\,\hat{\mathbf{z}}$ | (8$f$) | O II |
| $\mathbf{B}_{20}$ | = | $(x_7 + y_7)\,\mathbf{a}_1 + (x_7 - y_7)\,\mathbf{a}_2 + \left(\frac{1}{2} + z_7\right)\mathbf{a}_3$ | = | $\left(\frac{1}{2}c\cos\beta + x_7 a + z_7 c\cos\beta\right)\hat{\mathbf{x}} - y_7 b\,\hat{\mathbf{y}} + \left(\frac{1}{2} + z_7\right)c\sin\beta\,\hat{\mathbf{z}}$ | (8$f$) | O II |
| $\mathbf{B}_{21}$ | = | $(x_8 - y_8)\,\mathbf{a}_1 + (x_8 + y_8)\,\mathbf{a}_2 + z_8\,\mathbf{a}_3$ | = | $(x_8 a + z_8 c\cos\beta)\,\hat{\mathbf{x}} + y_8 b\,\hat{\mathbf{y}} + z_8 c\sin\beta\,\hat{\mathbf{z}}$ | (8$f$) | O III |
| $\mathbf{B}_{22}$ | = | $(-x_8 - y_8)\,\mathbf{a}_1 + (-x_8 + y_8)\,\mathbf{a}_2 + \left(\frac{1}{2} - z_8\right)\mathbf{a}_3$ | = | $\left(\frac{1}{2}c\cos\beta - x_8 a - z_8 c\cos\beta\right)\hat{\mathbf{x}} + y_8 b\,\hat{\mathbf{y}} + \left(\frac{1}{2} - z_8\right)c\sin\beta\,\hat{\mathbf{z}}$ | (8$f$) | O III |
| $\mathbf{B}_{23}$ | = | $(-x_8 + y_8)\,\mathbf{a}_1 + (-x_8 - y_8)\,\mathbf{a}_2 - z_8\,\mathbf{a}_3$ | = | $(-x_8 a - z_8 c\cos\beta)\,\hat{\mathbf{x}} - y_8 b\,\hat{\mathbf{y}} - z_8 c\sin\beta\,\hat{\mathbf{z}}$ | (8$f$) | O III |
| $\mathbf{B}_{24}$ | = | $(x_8 + y_8)\,\mathbf{a}_1 + (x_8 - y_8)\,\mathbf{a}_2 + \left(\frac{1}{2} + z_8\right)\mathbf{a}_3$ | = | $\left(\frac{1}{2}c\cos\beta + x_8 a + z_8 c\cos\beta\right)\hat{\mathbf{x}} - y_8 b\,\hat{\mathbf{y}} + \left(\frac{1}{2} + z_8\right)c\sin\beta\,\hat{\mathbf{z}}$ | (8$f$) | O III |

---

**References:**

- Y. Hinatsu and Y. Doi, *Crystal structures and magnetic properties of alkali-metal lanthanide oxides $A_2LnO_3$ (A = Li, Na; Ln = Ce, Pr, Tb)*, J. Alloys Compd. **418**, 155–160 (2006), doi:10.1016/j.jallcom.2005.08.100.

---

**Geometry files:**
- CIF: pp. 1561
- POSCAR: pp. 1561

# Eudidymite ($BeHNaO_8Si_3$ Structure: A2B4C2D17E6_mC124_15_f_2f_f_e8f_3f

http://aflow.org/prototype-encyclopedia/A2B4C2D17E6_mC124_15_f_2f_f_e8f_3f

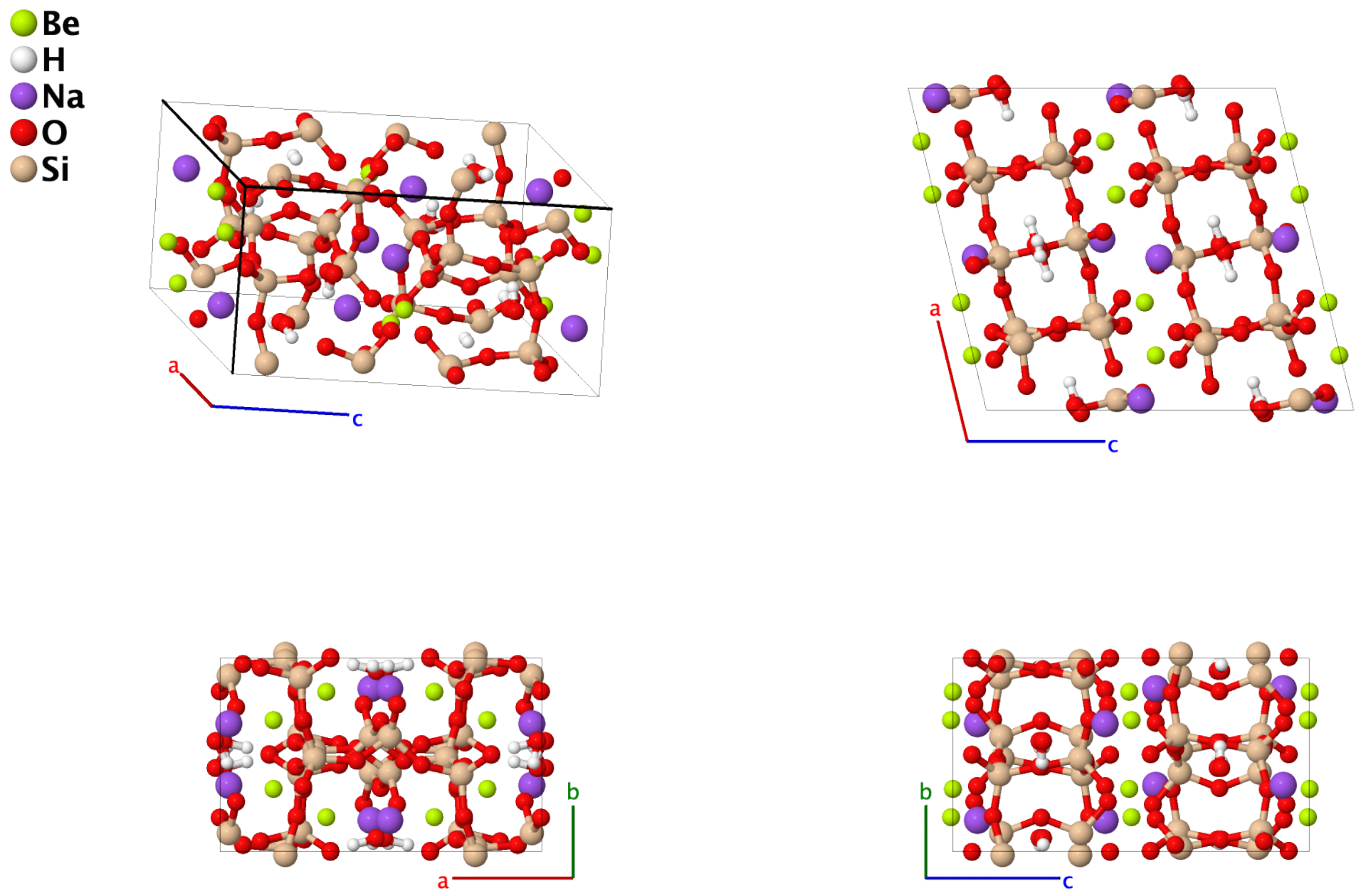

| | | |
|---|---|---|
| **Prototype** | : | $BeHNaO_8Si_3$ |
| **AFLOW prototype label** | : | A2B4C2D17E6_mC124_15_f_2f_f_e8f_3f |
| ***Strukturbericht* designation** | : | None |
| **Pearson symbol** | : | mC124 |
| **Space group number** | : | 15 |
| **Space group symbol** | : | $C2/c$ |
| **AFLOW prototype command** | : | `aflow --proto=A2B4C2D17E6_mC124_15_f_2f_f_e8f_3f` `--params=`$a, b/a, c/a, \beta, y_1, x_2, y_2, z_2, x_3, y_3, z_3, x_4, y_4, z_4, x_5, y_5, z_5, x_6, y_6, z_6, x_7, y_7, z_7, x_8, y_8, z_8, x_9, y_9, z_9, x_{10}, y_{10}, z_{10}, x_{11}, y_{11}, z_{11}, x_{12}, y_{12}, z_{12}, x_{13}, y_{13}, z_{13}, x_{14}, y_{14}, z_{14}, x_{15}, y_{15}, z_{15}, x_{16}, y_{16}, z_{16}$ |

- Eudidymite and its dimorph, epididymite ($S4_7$) are two forms of hydrated sodium beryllium silicate which are stable under ambient conditions. (Diego Gatta, 2008)
- The occupancy of the sites making up the water molecule (H-I, H-II and O-IX) is 50%. In the picture on this page the atoms forming the water molecules appear doubled because of this.

**Base-centered Monoclinic primitive vectors:**

$$\mathbf{a}_1 = \frac{1}{2}\,a\,\hat{\mathbf{x}} - \frac{1}{2}\,b\,\hat{\mathbf{y}}$$
$$\mathbf{a}_2 = \frac{1}{2}\,a\,\hat{\mathbf{x}} + \frac{1}{2}\,b\,\hat{\mathbf{y}}$$
$$\mathbf{a}_3 = c\cos\beta\,\hat{\mathbf{x}} + c\sin\beta\,\hat{\mathbf{z}}$$

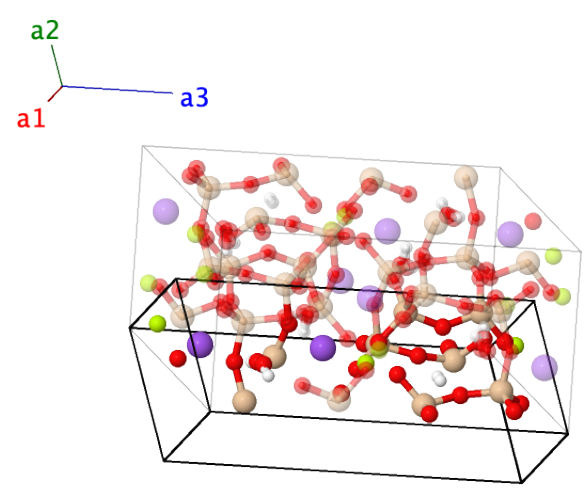

**Basis vectors:**

| | | Lattice Coordinates | | Cartesian Coordinates | Wyckoff Position | Atom Type |
|---|---|---|---|---|---|---|
| $\mathbf{B}_1$ | = | $-y_1\,\mathbf{a}_1 + y_1\,\mathbf{a}_2 + \frac{1}{4}\,\mathbf{a}_3$ | = | $\frac{1}{4}c\cos\beta\,\hat{\mathbf{x}} + y_1 b\,\hat{\mathbf{y}} + \frac{1}{4}c\sin\beta\,\hat{\mathbf{z}}$ | (4$e$) | O I |
| $\mathbf{B}_2$ | = | $y_1\,\mathbf{a}_1 - y_1\,\mathbf{a}_2 + \frac{3}{4}\,\mathbf{a}_3$ | = | $\frac{3}{4}c\cos\beta\,\hat{\mathbf{x}} - y_1 b\,\hat{\mathbf{y}} + \frac{3}{4}c\sin\beta\,\hat{\mathbf{z}}$ | (4$e$) | O I |
| $\mathbf{B}_3$ | = | $(x_2 - y_2)\,\mathbf{a}_1 + (x_2 + y_2)\,\mathbf{a}_2 + z_2\,\mathbf{a}_3$ | = | $(x_2 a + z_2 c\cos\beta)\,\hat{\mathbf{x}} + y_2 b\,\hat{\mathbf{y}} + z_2 c\sin\beta\,\hat{\mathbf{z}}$ | (8$f$) | Be |
| $\mathbf{B}_4$ | = | $(-x_2 - y_2)\,\mathbf{a}_1 + (-x_2 + y_2)\,\mathbf{a}_2 + \left(\frac{1}{2} - z_2\right)\mathbf{a}_3$ | = | $\left(\frac{1}{2}c\cos\beta - x_2 a - z_2 c\cos\beta\right)\hat{\mathbf{x}} + y_2 b\,\hat{\mathbf{y}} + \left(\frac{1}{2} - z_2\right)c\sin\beta\,\hat{\mathbf{z}}$ | (8$f$) | Be |
| $\mathbf{B}_5$ | = | $(-x_2 + y_2)\,\mathbf{a}_1 + (-x_2 - y_2)\,\mathbf{a}_2 - z_2\,\mathbf{a}_3$ | = | $(-x_2 a - z_2 c\cos\beta)\,\hat{\mathbf{x}} - y_2 b\,\hat{\mathbf{y}} - z_2 c\sin\beta\,\hat{\mathbf{z}}$ | (8$f$) | Be |
| $\mathbf{B}_6$ | = | $(x_2 + y_2)\,\mathbf{a}_1 + (x_2 - y_2)\,\mathbf{a}_2 + \left(\frac{1}{2} + z_2\right)\mathbf{a}_3$ | = | $\left(\frac{1}{2}c\cos\beta + x_2 a + z_2 c\cos\beta\right)\hat{\mathbf{x}} - y_2 b\,\hat{\mathbf{y}} + \left(\frac{1}{2} + z_2\right)c\sin\beta\,\hat{\mathbf{z}}$ | (8$f$) | Be |
| $\mathbf{B}_7$ | = | $(x_3 - y_3)\,\mathbf{a}_1 + (x_3 + y_3)\,\mathbf{a}_2 + z_3\,\mathbf{a}_3$ | = | $(x_3 a + z_3 c\cos\beta)\,\hat{\mathbf{x}} + y_3 b\,\hat{\mathbf{y}} + z_3 c\sin\beta\,\hat{\mathbf{z}}$ | (8$f$) | H I |
| $\mathbf{B}_8$ | = | $(-x_3 - y_3)\,\mathbf{a}_1 + (-x_3 + y_3)\,\mathbf{a}_2 + \left(\frac{1}{2} - z_3\right)\mathbf{a}_3$ | = | $\left(\frac{1}{2}c\cos\beta - x_3 a - z_3 c\cos\beta\right)\hat{\mathbf{x}} + y_3 b\,\hat{\mathbf{y}} + \left(\frac{1}{2} - z_3\right)c\sin\beta\,\hat{\mathbf{z}}$ | (8$f$) | H I |
| $\mathbf{B}_9$ | = | $(-x_3 + y_3)\,\mathbf{a}_1 + (-x_3 - y_3)\,\mathbf{a}_2 - z_3\,\mathbf{a}_3$ | = | $(-x_3 a - z_3 c\cos\beta)\,\hat{\mathbf{x}} - y_3 b\,\hat{\mathbf{y}} - z_3 c\sin\beta\,\hat{\mathbf{z}}$ | (8$f$) | H I |
| $\mathbf{B}_{10}$ | = | $(x_3 + y_3)\,\mathbf{a}_1 + (x_3 - y_3)\,\mathbf{a}_2 + \left(\frac{1}{2} + z_3\right)\mathbf{a}_3$ | = | $\left(\frac{1}{2}c\cos\beta + x_3 a + z_3 c\cos\beta\right)\hat{\mathbf{x}} - y_3 b\,\hat{\mathbf{y}} + \left(\frac{1}{2} + z_3\right)c\sin\beta\,\hat{\mathbf{z}}$ | (8$f$) | H I |
| $\mathbf{B}_{11}$ | = | $(x_4 - y_4)\,\mathbf{a}_1 + (x_4 + y_4)\,\mathbf{a}_2 + z_4\,\mathbf{a}_3$ | = | $(x_4 a + z_4 c\cos\beta)\,\hat{\mathbf{x}} + y_4 b\,\hat{\mathbf{y}} + z_4 c\sin\beta\,\hat{\mathbf{z}}$ | (8$f$) | H II |
| $\mathbf{B}_{12}$ | = | $(-x_4 - y_4)\,\mathbf{a}_1 + (-x_4 + y_4)\,\mathbf{a}_2 + \left(\frac{1}{2} - z_4\right)\mathbf{a}_3$ | = | $\left(\frac{1}{2}c\cos\beta - x_4 a - z_4 c\cos\beta\right)\hat{\mathbf{x}} + y_4 b\,\hat{\mathbf{y}} + \left(\frac{1}{2} - z_4\right)c\sin\beta\,\hat{\mathbf{z}}$ | (8$f$) | H II |
| $\mathbf{B}_{13}$ | = | $(-x_4 + y_4)\,\mathbf{a}_1 + (-x_4 - y_4)\,\mathbf{a}_2 - z_4\,\mathbf{a}_3$ | = | $(-x_4 a - z_4 c\cos\beta)\,\hat{\mathbf{x}} - y_4 b\,\hat{\mathbf{y}} - z_4 c\sin\beta\,\hat{\mathbf{z}}$ | (8$f$) | H II |
| $\mathbf{B}_{14}$ | = | $(x_4 + y_4)\,\mathbf{a}_1 + (x_4 - y_4)\,\mathbf{a}_2 + \left(\frac{1}{2} + z_4\right)\mathbf{a}_3$ | = | $\left(\frac{1}{2}c\cos\beta + x_4 a + z_4 c\cos\beta\right)\hat{\mathbf{x}} - y_4 b\,\hat{\mathbf{y}} + \left(\frac{1}{2} + z_4\right)c\sin\beta\,\hat{\mathbf{z}}$ | (8$f$) | H II |
| $\mathbf{B}_{15}$ | = | $(x_5 - y_5)\,\mathbf{a}_1 + (x_5 + y_5)\,\mathbf{a}_2 + z_5\,\mathbf{a}_3$ | = | $(x_5 a + z_5 c\cos\beta)\,\hat{\mathbf{x}} + y_5 b\,\hat{\mathbf{y}} + z_5 c\sin\beta\,\hat{\mathbf{z}}$ | (8$f$) | Na |
| $\mathbf{B}_{16}$ | = | $(-x_5 - y_5)\,\mathbf{a}_1 + (-x_5 + y_5)\,\mathbf{a}_2 + \left(\frac{1}{2} - z_5\right)\mathbf{a}_3$ | = | $\left(\frac{1}{2}c\cos\beta - x_5 a - z_5 c\cos\beta\right)\hat{\mathbf{x}} + y_5 b\,\hat{\mathbf{y}} + \left(\frac{1}{2} - z_5\right)c\sin\beta\,\hat{\mathbf{z}}$ | (8$f$) | Na |
| $\mathbf{B}_{17}$ | = | $(-x_5 + y_5)\,\mathbf{a}_1 + (-x_5 - y_5)\,\mathbf{a}_2 - z_5\,\mathbf{a}_3$ | = | $(-x_5 a - z_5 c\cos\beta)\,\hat{\mathbf{x}} - y_5 b\,\hat{\mathbf{y}} - z_5 c\sin\beta\,\hat{\mathbf{z}}$ | (8$f$) | Na |
| $\mathbf{B}_{18}$ | = | $(x_5 + y_5)\,\mathbf{a}_1 + (x_5 - y_5)\,\mathbf{a}_2 + \left(\frac{1}{2} + z_5\right)\mathbf{a}_3$ | = | $\left(\frac{1}{2}c\cos\beta + x_5 a + z_5 c\cos\beta\right)\hat{\mathbf{x}} - y_5 b\,\hat{\mathbf{y}} + \left(\frac{1}{2} + z_5\right)c\sin\beta\,\hat{\mathbf{z}}$ | (8$f$) | Na |
| $\mathbf{B}_{19}$ | = | $(x_6 - y_6)\,\mathbf{a}_1 + (x_6 + y_6)\,\mathbf{a}_2 + z_6\,\mathbf{a}_3$ | = | $(x_6 a + z_6 c\cos\beta)\,\hat{\mathbf{x}} + y_6 b\,\hat{\mathbf{y}} + z_6 c\sin\beta\,\hat{\mathbf{z}}$ | (8$f$) | O II |
| $\mathbf{B}_{20}$ | = | $(-x_6 - y_6)\,\mathbf{a}_1 + (-x_6 + y_6)\,\mathbf{a}_2 + \left(\frac{1}{2} - z_6\right)\mathbf{a}_3$ | = | $\left(\frac{1}{2}c\cos\beta - x_6 a - z_6 c\cos\beta\right)\hat{\mathbf{x}} + y_6 b\,\hat{\mathbf{y}} + \left(\frac{1}{2} - z_6\right)c\sin\beta\,\hat{\mathbf{z}}$ | (8$f$) | O II |
| $\mathbf{B}_{21}$ | = | $(-x_6 + y_6)\,\mathbf{a}_1 + (-x_6 - y_6)\,\mathbf{a}_2 - z_6\,\mathbf{a}_3$ | = | $(-x_6 a - z_6 c\cos\beta)\,\hat{\mathbf{x}} - y_6 b\,\hat{\mathbf{y}} - z_6 c\sin\beta\,\hat{\mathbf{z}}$ | (8$f$) | O II |
| $\mathbf{B}_{22}$ | = | $(x_6+y_6)\,\mathbf{a}_1+(x_6-y_6)\,\mathbf{a}_2+\left(\frac{1}{2}+z_6\right)\mathbf{a}_3$ | = | $\left(\frac{1}{2}c\cos\beta+x_6a+z_6c\cos\beta\right)\hat{\mathbf{x}}-y_6b\,\hat{\mathbf{y}}+\left(\frac{1}{2}+z_6\right)c\sin\beta\,\hat{\mathbf{z}}$ | $(8f)$ | O II |
| $\mathbf{B}_{23}$ | = | $(x_7-y_7)\,\mathbf{a}_1+(x_7+y_7)\,\mathbf{a}_2+z_7\,\mathbf{a}_3$ | = | $(x_7a+z_7c\cos\beta)\,\hat{\mathbf{x}}+y_7b\,\hat{\mathbf{y}}+z_7c\sin\beta\,\hat{\mathbf{z}}$ | $(8f)$ | O III |
| $\mathbf{B}_{24}$ | = | $(-x_7-y_7)\,\mathbf{a}_1+(-x_7+y_7)\,\mathbf{a}_2+\left(\frac{1}{2}-z_7\right)\mathbf{a}_3$ | = | $\left(\frac{1}{2}c\cos\beta-x_7a-z_7c\cos\beta\right)\hat{\mathbf{x}}+y_7b\,\hat{\mathbf{y}}+\left(\frac{1}{2}-z_7\right)c\sin\beta\,\hat{\mathbf{z}}$ | $(8f)$ | O III |
| $\mathbf{B}_{25}$ | = | $(-x_7+y_7)\,\mathbf{a}_1+(-x_7-y_7)\,\mathbf{a}_2-z_7\,\mathbf{a}_3$ | = | $(-x_7a-z_7c\cos\beta)\,\hat{\mathbf{x}}-y_7b\,\hat{\mathbf{y}}-z_7c\sin\beta\,\hat{\mathbf{z}}$ | $(8f)$ | O III |
| $\mathbf{B}_{26}$ | = | $(x_7+y_7)\,\mathbf{a}_1+(x_7-y_7)\,\mathbf{a}_2+\left(\frac{1}{2}+z_7\right)\mathbf{a}_3$ | = | $\left(\frac{1}{2}c\cos\beta+x_7a+z_7c\cos\beta\right)\hat{\mathbf{x}}-y_7b\,\hat{\mathbf{y}}+\left(\frac{1}{2}+z_7\right)c\sin\beta\,\hat{\mathbf{z}}$ | $(8f)$ | O III |
| $\mathbf{B}_{27}$ | = | $(x_8-y_8)\,\mathbf{a}_1+(x_8+y_8)\,\mathbf{a}_2+z_8\,\mathbf{a}_3$ | = | $(x_8a+z_8c\cos\beta)\,\hat{\mathbf{x}}+y_8b\,\hat{\mathbf{y}}+z_8c\sin\beta\,\hat{\mathbf{z}}$ | $(8f)$ | O IV |
| $\mathbf{B}_{28}$ | = | $(-x_8-y_8)\,\mathbf{a}_1+(-x_8+y_8)\,\mathbf{a}_2+\left(\frac{1}{2}-z_8\right)\mathbf{a}_3$ | = | $\left(\frac{1}{2}c\cos\beta-x_8a-z_8c\cos\beta\right)\hat{\mathbf{x}}+y_8b\,\hat{\mathbf{y}}+\left(\frac{1}{2}-z_8\right)c\sin\beta\,\hat{\mathbf{z}}$ | $(8f)$ | O IV |
| $\mathbf{B}_{29}$ | = | $(-x_8+y_8)\,\mathbf{a}_1+(-x_8-y_8)\,\mathbf{a}_2-z_8\,\mathbf{a}_3$ | = | $(-x_8a-z_8c\cos\beta)\,\hat{\mathbf{x}}-y_8b\,\hat{\mathbf{y}}-z_8c\sin\beta\,\hat{\mathbf{z}}$ | $(8f)$ | O IV |
| $\mathbf{B}_{30}$ | = | $(x_8+y_8)\,\mathbf{a}_1+(x_8-y_8)\,\mathbf{a}_2+\left(\frac{1}{2}+z_8\right)\mathbf{a}_3$ | = | $\left(\frac{1}{2}c\cos\beta+x_8a+z_8c\cos\beta\right)\hat{\mathbf{x}}-y_8b\,\hat{\mathbf{y}}+\left(\frac{1}{2}+z_8\right)c\sin\beta\,\hat{\mathbf{z}}$ | $(8f)$ | O IV |
| $\mathbf{B}_{31}$ | = | $(x_9-y_9)\,\mathbf{a}_1+(x_9+y_9)\,\mathbf{a}_2+z_9\,\mathbf{a}_3$ | = | $(x_9a+z_9c\cos\beta)\,\hat{\mathbf{x}}+y_9b\,\hat{\mathbf{y}}+z_9c\sin\beta\,\hat{\mathbf{z}}$ | $(8f)$ | O V |
| $\mathbf{B}_{32}$ | = | $(-x_9-y_9)\,\mathbf{a}_1+(-x_9+y_9)\,\mathbf{a}_2+\left(\frac{1}{2}-z_9\right)\mathbf{a}_3$ | = | $\left(\frac{1}{2}c\cos\beta-x_9a-z_9c\cos\beta\right)\hat{\mathbf{x}}+y_9b\,\hat{\mathbf{y}}+\left(\frac{1}{2}-z_9\right)c\sin\beta\,\hat{\mathbf{z}}$ | $(8f)$ | O V |
| $\mathbf{B}_{33}$ | = | $(-x_9+y_9)\,\mathbf{a}_1+(-x_9-y_9)\,\mathbf{a}_2-z_9\,\mathbf{a}_3$ | = | $(-x_9a-z_9c\cos\beta)\,\hat{\mathbf{x}}-y_9b\,\hat{\mathbf{y}}-z_9c\sin\beta\,\hat{\mathbf{z}}$ | $(8f)$ | O V |
| $\mathbf{B}_{34}$ | = | $(x_9+y_9)\,\mathbf{a}_1+(x_9-y_9)\,\mathbf{a}_2+\left(\frac{1}{2}+z_9\right)\mathbf{a}_3$ | = | $\left(\frac{1}{2}c\cos\beta+x_9a+z_9c\cos\beta\right)\hat{\mathbf{x}}-y_9b\,\hat{\mathbf{y}}+\left(\frac{1}{2}+z_9\right)c\sin\beta\,\hat{\mathbf{z}}$ | $(8f)$ | O V |
| $\mathbf{B}_{35}$ | = | $(x_{10}-y_{10})\,\mathbf{a}_1+(x_{10}+y_{10})\,\mathbf{a}_2+z_{10}\,\mathbf{a}_3$ | = | $(x_{10}a+z_{10}c\cos\beta)\,\hat{\mathbf{x}}+y_{10}b\,\hat{\mathbf{y}}+z_{10}c\sin\beta\,\hat{\mathbf{z}}$ | $(8f)$ | O VI |
| $\mathbf{B}_{36}$ | = | $(-x_{10}-y_{10})\,\mathbf{a}_1+(-x_{10}+y_{10})\,\mathbf{a}_2+\left(\frac{1}{2}-z_{10}\right)\mathbf{a}_3$ | = | $\left(\frac{1}{2}c\cos\beta-x_{10}a-z_{10}c\cos\beta\right)\hat{\mathbf{x}}+y_{10}b\,\hat{\mathbf{y}}+\left(\frac{1}{2}-z_{10}\right)c\sin\beta\,\hat{\mathbf{z}}$ | $(8f)$ | O VI |
| $\mathbf{B}_{37}$ | = | $(-x_{10}+y_{10})\,\mathbf{a}_1+(-x_{10}-y_{10})\,\mathbf{a}_2-z_{10}\,\mathbf{a}_3$ | = | $(-x_{10}a-z_{10}c\cos\beta)\,\hat{\mathbf{x}}-y_{10}b\,\hat{\mathbf{y}}-z_{10}c\sin\beta\,\hat{\mathbf{z}}$ | $(8f)$ | O VI |
| $\mathbf{B}_{38}$ | = | $(x_{10}+y_{10})\,\mathbf{a}_1+(x_{10}-y_{10})\,\mathbf{a}_2+\left(\frac{1}{2}+z_{10}\right)\mathbf{a}_3$ | = | $\left(\frac{1}{2}c\cos\beta+x_{10}a+z_{10}c\cos\beta\right)\hat{\mathbf{x}}-y_{10}b\,\hat{\mathbf{y}}+\left(\frac{1}{2}+z_{10}\right)c\sin\beta\,\hat{\mathbf{z}}$ | $(8f)$ | O VI |
| $\mathbf{B}_{39}$ | = | $(x_{11}-y_{11})\,\mathbf{a}_1+(x_{11}+y_{11})\,\mathbf{a}_2+z_{11}\,\mathbf{a}_3$ | = | $(x_{11}a+z_{11}c\cos\beta)\,\hat{\mathbf{x}}+y_{11}b\,\hat{\mathbf{y}}+z_{11}c\sin\beta\,\hat{\mathbf{z}}$ | $(8f)$ | O VII |
| $\mathbf{B}_{40}$ | = | $(-x_{11}-y_{11})\,\mathbf{a}_1+(-x_{11}+y_{11})\,\mathbf{a}_2+\left(\frac{1}{2}-z_{11}\right)\mathbf{a}_3$ | = | $\left(\frac{1}{2}c\cos\beta-x_{11}a-z_{11}c\cos\beta\right)\hat{\mathbf{x}}+y_{11}b\,\hat{\mathbf{y}}+\left(\frac{1}{2}-z_{11}\right)c\sin\beta\,\hat{\mathbf{z}}$ | $(8f)$ | O VII |
| $\mathbf{B}_{41}$ | = | $(-x_{11}+y_{11})\,\mathbf{a}_1+(-x_{11}-y_{11})\,\mathbf{a}_2-z_{11}\,\mathbf{a}_3$ | = | $(-x_{11}a-z_{11}c\cos\beta)\,\hat{\mathbf{x}}-y_{11}b\,\hat{\mathbf{y}}-z_{11}c\sin\beta\,\hat{\mathbf{z}}$ | $(8f)$ | O VII |
| $\mathbf{B}_{42}$ | = | $(x_{11}+y_{11})\,\mathbf{a}_1+(x_{11}-y_{11})\,\mathbf{a}_2+\left(\frac{1}{2}+z_{11}\right)\mathbf{a}_3$ | = | $\left(\frac{1}{2}c\cos\beta+x_{11}a+z_{11}c\cos\beta\right)\hat{\mathbf{x}}-y_{11}b\,\hat{\mathbf{y}}+\left(\frac{1}{2}+z_{11}\right)c\sin\beta\,\hat{\mathbf{z}}$ | $(8f)$ | O VII |
| $\mathbf{B}_{43}$ | = | $(x_{12}-y_{12})\,\mathbf{a}_1+(x_{12}+y_{12})\,\mathbf{a}_2+z_{12}\,\mathbf{a}_3$ | = | $(x_{12}a+z_{12}c\cos\beta)\,\hat{\mathbf{x}}+y_{12}b\,\hat{\mathbf{y}}+z_{12}c\sin\beta\,\hat{\mathbf{z}}$ | $(8f)$ | O VIII |
| $\mathbf{B}_{44}$ | = | $(-x_{12}-y_{12})\,\mathbf{a}_1 + (-x_{12}+y_{12})\,\mathbf{a}_2 + \left(\frac{1}{2}-z_{12}\right)\,\mathbf{a}_3$ | = | $\left(\frac{1}{2}c\cos\beta - x_{12}a - z_{12}c\cos\beta\right)\,\hat{\mathbf{x}} + y_{12}b\,\hat{\mathbf{y}} + \left(\frac{1}{2}-z_{12}\right)c\sin\beta\,\hat{\mathbf{z}}$ | (8$f$) | O VIII |
| $\mathbf{B}_{45}$ | = | $(-x_{12}+y_{12})\,\mathbf{a}_1 + (-x_{12}-y_{12})\,\mathbf{a}_2 - z_{12}\,\mathbf{a}_3$ | = | $(-x_{12}a - z_{12}c\cos\beta)\,\hat{\mathbf{x}} - y_{12}b\,\hat{\mathbf{y}} - z_{12}c\sin\beta\,\hat{\mathbf{z}}$ | (8$f$) | O VIII |
| $\mathbf{B}_{46}$ | = | $(x_{12}+y_{12})\,\mathbf{a}_1 + (x_{12}-y_{12})\,\mathbf{a}_2 + \left(\frac{1}{2}+z_{12}\right)\,\mathbf{a}_3$ | = | $\left(\frac{1}{2}c\cos\beta + x_{12}a + z_{12}c\cos\beta\right)\,\hat{\mathbf{x}} - y_{12}b\,\hat{\mathbf{y}} + \left(\frac{1}{2}+z_{12}\right)c\sin\beta\,\hat{\mathbf{z}}$ | (8$f$) | O VIII |
| $\mathbf{B}_{47}$ | = | $(x_{13}-y_{13})\,\mathbf{a}_1 + (x_{13}+y_{13})\,\mathbf{a}_2 + z_{13}\,\mathbf{a}_3$ | = | $(x_{13}a + z_{13}c\cos\beta)\,\hat{\mathbf{x}} + y_{13}b\,\hat{\mathbf{y}} + z_{13}c\sin\beta\,\hat{\mathbf{z}}$ | (8$f$) | O IX |
| $\mathbf{B}_{48}$ | = | $(-x_{13}-y_{13})\,\mathbf{a}_1 + (-x_{13}+y_{13})\,\mathbf{a}_2 + \left(\frac{1}{2}-z_{13}\right)\,\mathbf{a}_3$ | = | $\left(\frac{1}{2}c\cos\beta - x_{13}a - z_{13}c\cos\beta\right)\,\hat{\mathbf{x}} + y_{13}b\,\hat{\mathbf{y}} + \left(\frac{1}{2}-z_{13}\right)c\sin\beta\,\hat{\mathbf{z}}$ | (8$f$) | O IX |
| $\mathbf{B}_{49}$ | = | $(-x_{13}+y_{13})\,\mathbf{a}_1 + (-x_{13}-y_{13})\,\mathbf{a}_2 - z_{13}\,\mathbf{a}_3$ | = | $(-x_{13}a - z_{13}c\cos\beta)\,\hat{\mathbf{x}} - y_{13}b\,\hat{\mathbf{y}} - z_{13}c\sin\beta\,\hat{\mathbf{z}}$ | (8$f$) | O IX |
| $\mathbf{B}_{50}$ | = | $(x_{13}+y_{13})\,\mathbf{a}_1 + (x_{13}-y_{13})\,\mathbf{a}_2 + \left(\frac{1}{2}+z_{13}\right)\,\mathbf{a}_3$ | = | $\left(\frac{1}{2}c\cos\beta + x_{13}a + z_{13}c\cos\beta\right)\,\hat{\mathbf{x}} - y_{13}b\,\hat{\mathbf{y}} + \left(\frac{1}{2}+z_{13}\right)c\sin\beta\,\hat{\mathbf{z}}$ | (8$f$) | O IX |
| $\mathbf{B}_{51}$ | = | $(x_{14}-y_{14})\,\mathbf{a}_1 + (x_{14}+y_{14})\,\mathbf{a}_2 + z_{14}\,\mathbf{a}_3$ | = | $(x_{14}a + z_{14}c\cos\beta)\,\hat{\mathbf{x}} + y_{14}b\,\hat{\mathbf{y}} + z_{14}c\sin\beta\,\hat{\mathbf{z}}$ | (8$f$) | Si I |
| $\mathbf{B}_{52}$ | = | $(-x_{14}-y_{14})\,\mathbf{a}_1 + (-x_{14}+y_{14})\,\mathbf{a}_2 + \left(\frac{1}{2}-z_{14}\right)\,\mathbf{a}_3$ | = | $\left(\frac{1}{2}c\cos\beta - x_{14}a - z_{14}c\cos\beta\right)\,\hat{\mathbf{x}} + y_{14}b\,\hat{\mathbf{y}} + \left(\frac{1}{2}-z_{14}\right)c\sin\beta\,\hat{\mathbf{z}}$ | (8$f$) | Si I |
| $\mathbf{B}_{53}$ | = | $(-x_{14}+y_{14})\,\mathbf{a}_1 + (-x_{14}-y_{14})\,\mathbf{a}_2 - z_{14}\,\mathbf{a}_3$ | = | $(-x_{14}a - z_{14}c\cos\beta)\,\hat{\mathbf{x}} - y_{14}b\,\hat{\mathbf{y}} - z_{14}c\sin\beta\,\hat{\mathbf{z}}$ | (8$f$) | Si I |
| $\mathbf{B}_{54}$ | = | $(x_{14}+y_{14})\,\mathbf{a}_1 + (x_{14}-y_{14})\,\mathbf{a}_2 + \left(\frac{1}{2}+z_{14}\right)\,\mathbf{a}_3$ | = | $\left(\frac{1}{2}c\cos\beta + x_{14}a + z_{14}c\cos\beta\right)\,\hat{\mathbf{x}} - y_{14}b\,\hat{\mathbf{y}} + \left(\frac{1}{2}+z_{14}\right)c\sin\beta\,\hat{\mathbf{z}}$ | (8$f$) | Si I |
| $\mathbf{B}_{55}$ | = | $(x_{15}-y_{15})\,\mathbf{a}_1 + (x_{15}+y_{15})\,\mathbf{a}_2 + z_{15}\,\mathbf{a}_3$ | = | $(x_{15}a + z_{15}c\cos\beta)\,\hat{\mathbf{x}} + y_{15}b\,\hat{\mathbf{y}} + z_{15}c\sin\beta\,\hat{\mathbf{z}}$ | (8$f$) | Si II |
| $\mathbf{B}_{56}$ | = | $(-x_{15}-y_{15})\,\mathbf{a}_1 + (-x_{15}+y_{15})\,\mathbf{a}_2 + \left(\frac{1}{2}-z_{15}\right)\,\mathbf{a}_3$ | = | $\left(\frac{1}{2}c\cos\beta - x_{15}a - z_{15}c\cos\beta\right)\,\hat{\mathbf{x}} + y_{15}b\,\hat{\mathbf{y}} + \left(\frac{1}{2}-z_{15}\right)c\sin\beta\,\hat{\mathbf{z}}$ | (8$f$) | Si II |
| $\mathbf{B}_{57}$ | = | $(-x_{15}+y_{15})\,\mathbf{a}_1 + (-x_{15}-y_{15})\,\mathbf{a}_2 - z_{15}\,\mathbf{a}_3$ | = | $(-x_{15}a - z_{15}c\cos\beta)\,\hat{\mathbf{x}} - y_{15}b\,\hat{\mathbf{y}} - z_{15}c\sin\beta\,\hat{\mathbf{z}}$ | (8$f$) | Si II |
| $\mathbf{B}_{58}$ | = | $(x_{15}+y_{15})\,\mathbf{a}_1 + (x_{15}-y_{15})\,\mathbf{a}_2 + \left(\frac{1}{2}+z_{15}\right)\,\mathbf{a}_3$ | = | $\left(\frac{1}{2}c\cos\beta + x_{15}a + z_{15}c\cos\beta\right)\,\hat{\mathbf{x}} - y_{15}b\,\hat{\mathbf{y}} + \left(\frac{1}{2}+z_{15}\right)c\sin\beta\,\hat{\mathbf{z}}$ | (8$f$) | Si II |
| $\mathbf{B}_{59}$ | = | $(x_{16}-y_{16})\,\mathbf{a}_1 + (x_{16}+y_{16})\,\mathbf{a}_2 + z_{16}\,\mathbf{a}_3$ | = | $(x_{16}a + z_{16}c\cos\beta)\,\hat{\mathbf{x}} + y_{16}b\,\hat{\mathbf{y}} + z_{16}c\sin\beta\,\hat{\mathbf{z}}$ | (8$f$) | Si III |
| $\mathbf{B}_{60}$ | = | $(-x_{16}-y_{16})\,\mathbf{a}_1 + (-x_{16}+y_{16})\,\mathbf{a}_2 + \left(\frac{1}{2}-z_{16}\right)\,\mathbf{a}_3$ | = | $\left(\frac{1}{2}c\cos\beta - x_{16}a - z_{16}c\cos\beta\right)\,\hat{\mathbf{x}} + y_{16}b\,\hat{\mathbf{y}} + \left(\frac{1}{2}-z_{16}\right)c\sin\beta\,\hat{\mathbf{z}}$ | (8$f$) | Si III |
| $\mathbf{B}_{61}$ | = | $(-x_{16}+y_{16})\,\mathbf{a}_1 + (-x_{16}-y_{16})\,\mathbf{a}_2 - z_{16}\,\mathbf{a}_3$ | = | $(-x_{16}a - z_{16}c\cos\beta)\,\hat{\mathbf{x}} - y_{16}b\,\hat{\mathbf{y}} - z_{16}c\sin\beta\,\hat{\mathbf{z}}$ | (8$f$) | Si III |
| $\mathbf{B}_{62}$ | = | $(x_{16}+y_{16})\,\mathbf{a}_1 + (x_{16}-y_{16})\,\mathbf{a}_2 + \left(\frac{1}{2}+z_{16}\right)\,\mathbf{a}_3$ | = | $\left(\frac{1}{2}c\cos\beta + x_{16}a + z_{16}c\cos\beta\right)\,\hat{\mathbf{x}} - y_{16}b\,\hat{\mathbf{y}} + \left(\frac{1}{2}+z_{16}\right)c\sin\beta\,\hat{\mathbf{z}}$ | (8$f$) | Si III |

**Geometry files:**

- CIF: pp. 1562
- POSCAR: pp. 1562

# $\zeta$-$Nb_2O_5$ ($B$-$Nb_2O_5$) Structure: A2B5_mC28_15_f_e2f

http://aflow.org/prototype-encyclopedia/A2B5_mC28_15_f_e2f.Nb2O5

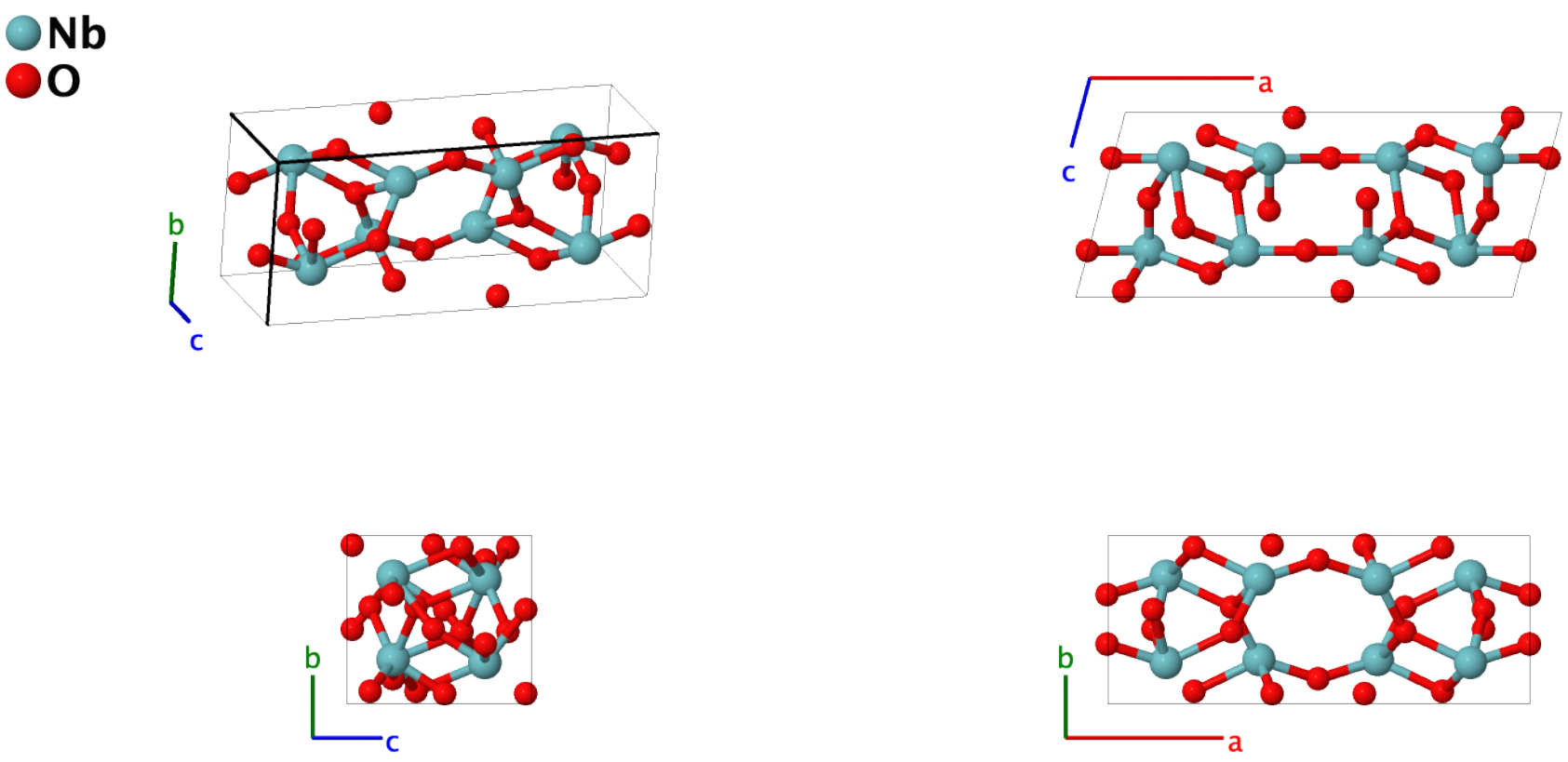

| | | |
|---|---|---|
| **Prototype** | : | $Nb_2O_5$ |
| **AFLOW prototype label** | : | A2B5_mC28_15_f_e2f |
| ***Strukturbericht* designation** | : | None |
| **Pearson symbol** | : | mC28 |
| **Space group number** | : | 15 |
| **Space group symbol** | : | $C2/c$ |
| **AFLOW prototype command** | : | `aflow --proto=A2B5_mC28_15_f_e2f`<br>`--params=`$a, b/a, c/a, \beta, y_1, x_2, y_2, z_2, x_3, y_3, z_3, x_4, y_4, z_4$ |

**Other compounds with this structure**

- $Sb_2O_5$

- This structure is referred to as both $\zeta$-$Nb_2O_5$ and B-$Nb_2O_5$.
- $Nb_2O_5$ and $B_2Pd_5$ share the same AFLOW prototype label, A2B5_mC28_15_f_e2f, but have substantially different environments around each atom. The structures are generated by the same symmetry operations with different sets of parameters (`--params`) specified in their corresponding CIF files.

**Base-centered Monoclinic primitive vectors:**

$$\mathbf{a}_1 = \frac{1}{2}\,a\,\mathbf{\hat{x}} - \frac{1}{2}\,b\,\mathbf{\hat{y}}$$
$$\mathbf{a}_2 = \frac{1}{2}\,a\,\mathbf{\hat{x}} + \frac{1}{2}\,b\,\mathbf{\hat{y}}$$
$$\mathbf{a}_3 = c\cos\beta\,\mathbf{\hat{x}} + c\sin\beta\,\mathbf{\hat{z}}$$

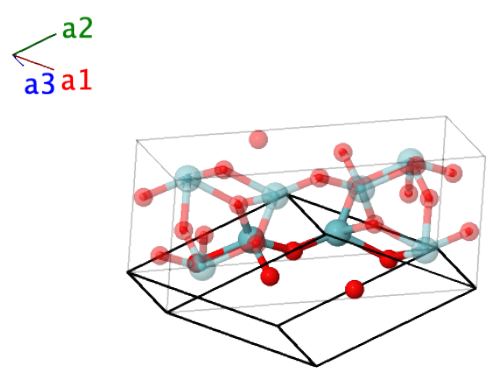

**Basis vectors:**

| | | Lattice Coordinates | | Cartesian Coordinates | Wyckoff Position | Atom Type |
|---|---|---|---|---|---|---|
| $\mathbf{B}_1$ | = | $-y_1\,\mathbf{a}_1 + y_1\,\mathbf{a}_2 + \frac{1}{4}\,\mathbf{a}_3$ | = | $\frac{1}{4}c\cos\beta\,\mathbf{\hat{x}} + y_1 b\,\mathbf{\hat{y}} + \frac{1}{4}c\sin\beta\,\mathbf{\hat{z}}$ | (4$e$) | O I |
| $\mathbf{B}_2$ | = | $y_1\,\mathbf{a}_1 - y_1\,\mathbf{a}_2 + \frac{3}{4}\,\mathbf{a}_3$ | = | $\frac{3}{4}c\cos\beta\,\mathbf{\hat{x}} - y_1 b\,\mathbf{\hat{y}} + \frac{3}{4}c\sin\beta\,\mathbf{\hat{z}}$ | (4$e$) | O I |
| $\mathbf{B}_3$ | = | $(x_2 - y_2)\,\mathbf{a}_1 + (x_2 + y_2)\,\mathbf{a}_2 + z_2\,\mathbf{a}_3$ | = | $(x_2 a + z_2 c\cos\beta)\,\mathbf{\hat{x}} + y_2 b\,\mathbf{\hat{y}} + z_2 c\sin\beta\,\mathbf{\hat{z}}$ | (8$f$) | Nb |
| $\mathbf{B}_{4}$ | = | $\left(-x_{2}-y_{2}\right)\,\mathbf{a}_{1}+\left(-x_{2}+y_{2}\right)\,\mathbf{a}_{2}+\left(\frac{1}{2}-z_{2}\right)\,\mathbf{a}_{3}$ | = | $\left(\frac{1}{2}c\cos\beta-x_{2}a-z_{2}c\cos\beta\right)\,\mathbf{\hat{x}}+y_{2}b\,\mathbf{\hat{y}}+\left(\frac{1}{2}-z_{2}\right)c\sin\beta\,\mathbf{\hat{z}}$ | $(8f)$ | Nb |
| $\mathbf{B}_{5}$ | = | $\left(-x_{2}+y_{2}\right)\,\mathbf{a}_{1}+\left(-x_{2}-y_{2}\right)\,\mathbf{a}_{2}-z_{2}\,\mathbf{a}_{3}$ | = | $\left(-x_{2}a-z_{2}c\cos\beta\right)\,\mathbf{\hat{x}}-y_{2}b\,\mathbf{\hat{y}}-z_{2}c\sin\beta\,\mathbf{\hat{z}}$ | $(8f)$ | Nb |
| $\mathbf{B}_{6}$ | = | $\left(x_{2}+y_{2}\right)\,\mathbf{a}_{1}+\left(x_{2}-y_{2}\right)\,\mathbf{a}_{2}+\left(\frac{1}{2}+z_{2}\right)\,\mathbf{a}_{3}$ | = | $\left(\frac{1}{2}c\cos\beta+x_{2}a+z_{2}c\cos\beta\right)\,\mathbf{\hat{x}}-y_{2}b\,\mathbf{\hat{y}}+\left(\frac{1}{2}+z_{2}\right)c\sin\beta\,\mathbf{\hat{z}}$ | $(8f)$ | Nb |
| $\mathbf{B}_{7}$ | = | $\left(x_{3}-y_{3}\right)\,\mathbf{a}_{1}+\left(x_{3}+y_{3}\right)\,\mathbf{a}_{2}+z_{3}\,\mathbf{a}_{3}$ | = | $\left(x_{3}a+z_{3}c\cos\beta\right)\,\mathbf{\hat{x}}+y_{3}b\,\mathbf{\hat{y}}+z_{3}c\sin\beta\,\mathbf{\hat{z}}$ | $(8f)$ | O II |
| $\mathbf{B}_{8}$ | = | $\left(-x_{3}-y_{3}\right)\,\mathbf{a}_{1}+\left(-x_{3}+y_{3}\right)\,\mathbf{a}_{2}+\left(\frac{1}{2}-z_{3}\right)\,\mathbf{a}_{3}$ | = | $\left(\frac{1}{2}c\cos\beta-x_{3}a-z_{3}c\cos\beta\right)\,\mathbf{\hat{x}}+y_{3}b\,\mathbf{\hat{y}}+\left(\frac{1}{2}-z_{3}\right)c\sin\beta\,\mathbf{\hat{z}}$ | $(8f)$ | O II |
| $\mathbf{B}_{9}$ | = | $\left(-x_{3}+y_{3}\right)\,\mathbf{a}_{1}+\left(-x_{3}-y_{3}\right)\,\mathbf{a}_{2}-z_{3}\,\mathbf{a}_{3}$ | = | $\left(-x_{3}a-z_{3}c\cos\beta\right)\,\mathbf{\hat{x}}-y_{3}b\,\mathbf{\hat{y}}-z_{3}c\sin\beta\,\mathbf{\hat{z}}$ | $(8f)$ | O II |
| $\mathbf{B}_{10}$ | = | $\left(x_{3}+y_{3}\right)\,\mathbf{a}_{1}+\left(x_{3}-y_{3}\right)\,\mathbf{a}_{2}+\left(\frac{1}{2}+z_{3}\right)\,\mathbf{a}_{3}$ | = | $\left(\frac{1}{2}c\cos\beta+x_{3}a+z_{3}c\cos\beta\right)\,\mathbf{\hat{x}}-y_{3}b\,\mathbf{\hat{y}}+\left(\frac{1}{2}+z_{3}\right)c\sin\beta\,\mathbf{\hat{z}}$ | $(8f)$ | O II |
| $\mathbf{B}_{11}$ | = | $\left(x_{4}-y_{4}\right)\,\mathbf{a}_{1}+\left(x_{4}+y_{4}\right)\,\mathbf{a}_{2}+z_{4}\,\mathbf{a}_{3}$ | = | $\left(x_{4}a+z_{4}c\cos\beta\right)\,\mathbf{\hat{x}}+y_{4}b\,\mathbf{\hat{y}}+z_{4}c\sin\beta\,\mathbf{\hat{z}}$ | $(8f)$ | O III |
| $\mathbf{B}_{12}$ | = | $\left(-x_{4}-y_{4}\right)\,\mathbf{a}_{1}+\left(-x_{4}+y_{4}\right)\,\mathbf{a}_{2}+\left(\frac{1}{2}-z_{4}\right)\,\mathbf{a}_{3}$ | = | $\left(\frac{1}{2}c\cos\beta-x_{4}a-z_{4}c\cos\beta\right)\,\mathbf{\hat{x}}+y_{4}b\,\mathbf{\hat{y}}+\left(\frac{1}{2}-z_{4}\right)c\sin\beta\,\mathbf{\hat{z}}$ | $(8f)$ | O III |
| $\mathbf{B}_{13}$ | = | $\left(-x_{4}+y_{4}\right)\,\mathbf{a}_{1}+\left(-x_{4}-y_{4}\right)\,\mathbf{a}_{2}-z_{4}\,\mathbf{a}_{3}$ | = | $\left(-x_{4}a-z_{4}c\cos\beta\right)\,\mathbf{\hat{x}}-y_{4}b\,\mathbf{\hat{y}}-z_{4}c\sin\beta\,\mathbf{\hat{z}}$ | $(8f)$ | O III |
| $\mathbf{B}_{14}$ | = | $\left(x_{4}+y_{4}\right)\,\mathbf{a}_{1}+\left(x_{4}-y_{4}\right)\,\mathbf{a}_{2}+\left(\frac{1}{2}+z_{4}\right)\,\mathbf{a}_{3}$ | = | $\left(\frac{1}{2}c\cos\beta+x_{4}a+z_{4}c\cos\beta\right)\,\mathbf{\hat{x}}-y_{4}b\,\mathbf{\hat{y}}+\left(\frac{1}{2}+z_{4}\right)c\sin\beta\,\mathbf{\hat{z}}$ | $(8f)$ | O III |

**Geometry files:**

- CIF: pp. 1562
- POSCAR: pp. 1563

# Muscovite ($KH_2Al_3Si_3O_{12}$, $S5_1$) Structure: A2BC10D2E4_mC76_15_f_e_5f_f_2f

http://aflow.org/prototype-encyclopedia/A2BC10D2E4_mC76_15_f_e_5f_f_2f

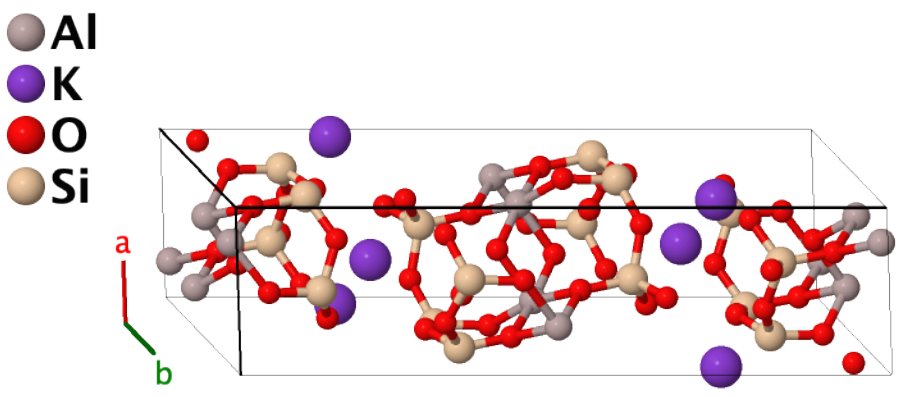

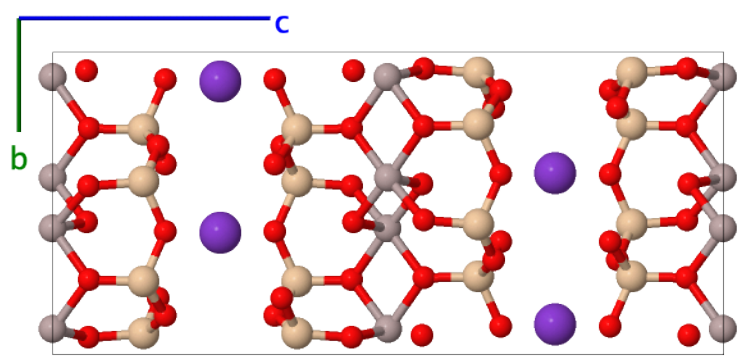

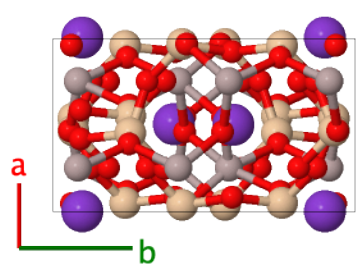

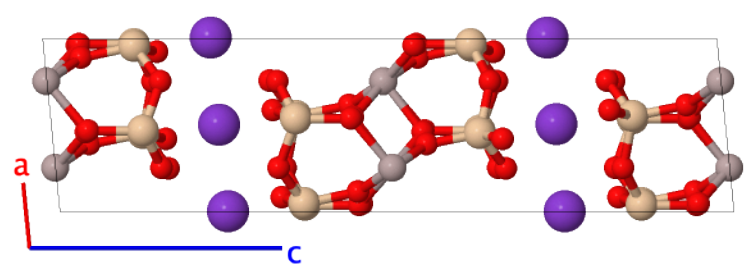

| | | |
|---|---|---|
| **Prototype** | : | $Al_3KO_{10}(OH)_2Si_3$ |
| **AFLOW prototype label** | : | A2BC10D2E4_mC76_15_f_e_5f_f_2f |
| ***Strukturbericht* designation** | : | $S5_1$ |
| **Pearson symbol** | : | mC76 |
| **Space group number** | : | 15 |
| **Space group symbol** | : | $C2/c$ |
| **AFLOW prototype command** | : | `aflow --proto=A2BC10D2E4_mC76_15_f_e_5f_f_2f --params=`$a, b/a, c/a, \beta, y_1, x_2, y_2, z_2, x_3, y_3, z_3, x_4, y_4, z_4, x_5, y_5, z_5, x_6, y_6, z_6, x_7, y_7, z_7, x_8, y_8, z_8, x_9, y_9, z_9, x_{10}, y_{10}, z_{10}$ |

- The sites we label Si-I and Si-II are actually approximately 75% silicon and 25% aluminum, but trace elements like manganese can appear on both sites. In addition, the potassium site (K) can be alloyed with small amounts of sodium, and the aluminum site (Al) with iron. These trace elements give muscovite a variety of colors. The sample studied by (Richardson, 1982) was pink.

**Base-centered Monoclinic primitive vectors:**

$$\mathbf{a}_1 = \frac{1}{2}a\,\hat{\mathbf{x}} - \frac{1}{2}b\,\hat{\mathbf{y}}$$
$$\mathbf{a}_2 = \frac{1}{2}a\,\hat{\mathbf{x}} + \frac{1}{2}b\,\hat{\mathbf{y}}$$
$$\mathbf{a}_3 = c\cos\beta\,\hat{\mathbf{x}} + c\sin\beta\,\hat{\mathbf{z}}$$

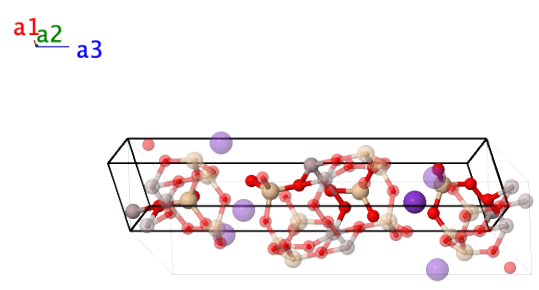

**Basis vectors:**

| | | Lattice Coordinates | | Cartesian Coordinates | Wyckoff Position | Atom Type |
|---|---|---|---|---|---|---|
| $\mathbf{B}_1$ | = | $-y_1\,\mathbf{a}_1 + y_1\,\mathbf{a}_2 + \frac{1}{4}\,\mathbf{a}_3$ | = | $\frac{1}{4}c\cos\beta\,\hat{\mathbf{x}} + y_1 b\,\hat{\mathbf{y}} + \frac{1}{4}c\sin\beta\,\hat{\mathbf{z}}$ | (4$e$) | K |
| $\mathbf{B}_2$ | = | $y_1\,\mathbf{a}_1 - y_1\,\mathbf{a}_2 + \frac{3}{4}\,\mathbf{a}_3$ | = | $\frac{3}{4}c\cos\beta\,\hat{\mathbf{x}} - y_1 b\,\hat{\mathbf{y}} + \frac{3}{4}c\sin\beta\,\hat{\mathbf{z}}$ | (4$e$) | K |
| $\mathbf{B}_3$ | = | $(x_2 - y_2)\,\mathbf{a}_1 + (x_2 + y_2)\,\mathbf{a}_2 + z_2\,\mathbf{a}_3$ | = | $(x_2 a + z_2 c\cos\beta)\,\hat{\mathbf{x}} + y_2 b\,\hat{\mathbf{y}} + z_2 c\sin\beta\,\hat{\mathbf{z}}$ | (8$f$) | Al |
| $\mathbf{B}_{4}$ | = | $(-x_{2} - y_{2})\,\mathbf{a}_{1} + (-x_{2} + y_{2})\,\mathbf{a}_{2} + \left(\frac{1}{2} - z_{2}\right)\mathbf{a}_{3}$ | = | $\left(\frac{1}{2}c\cos\beta - x_{2}a - z_{2}c\cos\beta\right)\hat{\mathbf{x}} + y_{2}b\,\hat{\mathbf{y}} + \left(\frac{1}{2} - z_{2}\right)c\sin\beta\,\hat{\mathbf{z}}$ | $(8f)$ | Al |
| $\mathbf{B}_{5}$ | = | $(-x_{2} + y_{2})\,\mathbf{a}_{1} + (-x_{2} - y_{2})\,\mathbf{a}_{2} - z_{2}\,\mathbf{a}_{3}$ | = | $(-x_{2}a - z_{2}c\cos\beta)\,\hat{\mathbf{x}} - y_{2}b\,\hat{\mathbf{y}} - z_{2}c\sin\beta\,\hat{\mathbf{z}}$ | $(8f)$ | Al |
| $\mathbf{B}_{6}$ | = | $(x_{2} + y_{2})\,\mathbf{a}_{1} + (x_{2} - y_{2})\,\mathbf{a}_{2} + \left(\frac{1}{2} + z_{2}\right)\mathbf{a}_{3}$ | = | $\left(\frac{1}{2}c\cos\beta + x_{2}a + z_{2}c\cos\beta\right)\hat{\mathbf{x}} - y_{2}b\,\hat{\mathbf{y}} + \left(\frac{1}{2} + z_{2}\right)c\sin\beta\,\hat{\mathbf{z}}$ | $(8f)$ | Al |
| $\mathbf{B}_{7}$ | = | $(x_{3} - y_{3})\,\mathbf{a}_{1} + (x_{3} + y_{3})\,\mathbf{a}_{2} + z_{3}\,\mathbf{a}_{3}$ | = | $(x_{3}a + z_{3}c\cos\beta)\,\hat{\mathbf{x}} + y_{3}b\,\hat{\mathbf{y}} + z_{3}c\sin\beta\,\hat{\mathbf{z}}$ | $(8f)$ | O I |
| $\mathbf{B}_{8}$ | = | $(-x_{3} - y_{3})\,\mathbf{a}_{1} + (-x_{3} + y_{3})\,\mathbf{a}_{2} + \left(\frac{1}{2} - z_{3}\right)\mathbf{a}_{3}$ | = | $\left(\frac{1}{2}c\cos\beta - x_{3}a - z_{3}c\cos\beta\right)\hat{\mathbf{x}} + y_{3}b\,\hat{\mathbf{y}} + \left(\frac{1}{2} - z_{3}\right)c\sin\beta\,\hat{\mathbf{z}}$ | $(8f)$ | O I |
| $\mathbf{B}_{9}$ | = | $(-x_{3} + y_{3})\,\mathbf{a}_{1} + (-x_{3} - y_{3})\,\mathbf{a}_{2} - z_{3}\,\mathbf{a}_{3}$ | = | $(-x_{3}a - z_{3}c\cos\beta)\,\hat{\mathbf{x}} - y_{3}b\,\hat{\mathbf{y}} - z_{3}c\sin\beta\,\hat{\mathbf{z}}$ | $(8f)$ | O I |
| $\mathbf{B}_{10}$ | = | $(x_{3} + y_{3})\,\mathbf{a}_{1} + (x_{3} - y_{3})\,\mathbf{a}_{2} + \left(\frac{1}{2} + z_{3}\right)\mathbf{a}_{3}$ | = | $\left(\frac{1}{2}c\cos\beta + x_{3}a + z_{3}c\cos\beta\right)\hat{\mathbf{x}} - y_{3}b\,\hat{\mathbf{y}} + \left(\frac{1}{2} + z_{3}\right)c\sin\beta\,\hat{\mathbf{z}}$ | $(8f)$ | O I |
| $\mathbf{B}_{11}$ | = | $(x_{4} - y_{4})\,\mathbf{a}_{1} + (x_{4} + y_{4})\,\mathbf{a}_{2} + z_{4}\,\mathbf{a}_{3}$ | = | $(x_{4}a + z_{4}c\cos\beta)\,\hat{\mathbf{x}} + y_{4}b\,\hat{\mathbf{y}} + z_{4}c\sin\beta\,\hat{\mathbf{z}}$ | $(8f)$ | O II |
| $\mathbf{B}_{12}$ | = | $(-x_{4} - y_{4})\,\mathbf{a}_{1} + (-x_{4} + y_{4})\,\mathbf{a}_{2} + \left(\frac{1}{2} - z_{4}\right)\mathbf{a}_{3}$ | = | $\left(\frac{1}{2}c\cos\beta - x_{4}a - z_{4}c\cos\beta\right)\hat{\mathbf{x}} + y_{4}b\,\hat{\mathbf{y}} + \left(\frac{1}{2} - z_{4}\right)c\sin\beta\,\hat{\mathbf{z}}$ | $(8f)$ | O II |
| $\mathbf{B}_{13}$ | = | $(-x_{4} + y_{4})\,\mathbf{a}_{1} + (-x_{4} - y_{4})\,\mathbf{a}_{2} - z_{4}\,\mathbf{a}_{3}$ | = | $(-x_{4}a - z_{4}c\cos\beta)\,\hat{\mathbf{x}} - y_{4}b\,\hat{\mathbf{y}} - z_{4}c\sin\beta\,\hat{\mathbf{z}}$ | $(8f)$ | O II |
| $\mathbf{B}_{14}$ | = | $(x_{4} + y_{4})\,\mathbf{a}_{1} + (x_{4} - y_{4})\,\mathbf{a}_{2} + \left(\frac{1}{2} + z_{4}\right)\mathbf{a}_{3}$ | = | $\left(\frac{1}{2}c\cos\beta + x_{4}a + z_{4}c\cos\beta\right)\hat{\mathbf{x}} - y_{4}b\,\hat{\mathbf{y}} + \left(\frac{1}{2} + z_{4}\right)c\sin\beta\,\hat{\mathbf{z}}$ | $(8f)$ | O II |
| $\mathbf{B}_{15}$ | = | $(x_{5} - y_{5})\,\mathbf{a}_{1} + (x_{5} + y_{5})\,\mathbf{a}_{2} + z_{5}\,\mathbf{a}_{3}$ | = | $(x_{5}a + z_{5}c\cos\beta)\,\hat{\mathbf{x}} + y_{5}b\,\hat{\mathbf{y}} + z_{5}c\sin\beta\,\hat{\mathbf{z}}$ | $(8f)$ | O III |
| $\mathbf{B}_{16}$ | = | $(-x_{5} - y_{5})\,\mathbf{a}_{1} + (-x_{5} + y_{5})\,\mathbf{a}_{2} + \left(\frac{1}{2} - z_{5}\right)\mathbf{a}_{3}$ | = | $\left(\frac{1}{2}c\cos\beta - x_{5}a - z_{5}c\cos\beta\right)\hat{\mathbf{x}} + y_{5}b\,\hat{\mathbf{y}} + \left(\frac{1}{2} - z_{5}\right)c\sin\beta\,\hat{\mathbf{z}}$ | $(8f)$ | O III |
| $\mathbf{B}_{17}$ | = | $(-x_{5} + y_{5})\,\mathbf{a}_{1} + (-x_{5} - y_{5})\,\mathbf{a}_{2} - z_{5}\,\mathbf{a}_{3}$ | = | $(-x_{5}a - z_{5}c\cos\beta)\,\hat{\mathbf{x}} - y_{5}b\,\hat{\mathbf{y}} - z_{5}c\sin\beta\,\hat{\mathbf{z}}$ | $(8f)$ | O III |
| $\mathbf{B}_{18}$ | = | $(x_{5} + y_{5})\,\mathbf{a}_{1} + (x_{5} - y_{5})\,\mathbf{a}_{2} + \left(\frac{1}{2} + z_{5}\right)\mathbf{a}_{3}$ | = | $\left(\frac{1}{2}c\cos\beta + x_{5}a + z_{5}c\cos\beta\right)\hat{\mathbf{x}} - y_{5}b\,\hat{\mathbf{y}} + \left(\frac{1}{2} + z_{5}\right)c\sin\beta\,\hat{\mathbf{z}}$ | $(8f)$ | O III |
| $\mathbf{B}_{19}$ | = | $(x_{6} - y_{6})\,\mathbf{a}_{1} + (x_{6} + y_{6})\,\mathbf{a}_{2} + z_{6}\,\mathbf{a}_{3}$ | = | $(x_{6}a + z_{6}c\cos\beta)\,\hat{\mathbf{x}} + y_{6}b\,\hat{\mathbf{y}} + z_{6}c\sin\beta\,\hat{\mathbf{z}}$ | $(8f)$ | O IV |
| $\mathbf{B}_{20}$ | = | $(-x_{6} - y_{6})\,\mathbf{a}_{1} + (-x_{6} + y_{6})\,\mathbf{a}_{2} + \left(\frac{1}{2} - z_{6}\right)\mathbf{a}_{3}$ | = | $\left(\frac{1}{2}c\cos\beta - x_{6}a - z_{6}c\cos\beta\right)\hat{\mathbf{x}} + y_{6}b\,\hat{\mathbf{y}} + \left(\frac{1}{2} - z_{6}\right)c\sin\beta\,\hat{\mathbf{z}}$ | $(8f)$ | O IV |
| $\mathbf{B}_{21}$ | = | $(-x_{6} + y_{6})\,\mathbf{a}_{1} + (-x_{6} - y_{6})\,\mathbf{a}_{2} - z_{6}\,\mathbf{a}_{3}$ | = | $(-x_{6}a - z_{6}c\cos\beta)\,\hat{\mathbf{x}} - y_{6}b\,\hat{\mathbf{y}} - z_{6}c\sin\beta\,\hat{\mathbf{z}}$ | $(8f)$ | O IV |
| $\mathbf{B}_{22}$ | = | $(x_{6} + y_{6})\,\mathbf{a}_{1} + (x_{6} - y_{6})\,\mathbf{a}_{2} + \left(\frac{1}{2} + z_{6}\right)\mathbf{a}_{3}$ | = | $\left(\frac{1}{2}c\cos\beta + x_{6}a + z_{6}c\cos\beta\right)\hat{\mathbf{x}} - y_{6}b\,\hat{\mathbf{y}} + \left(\frac{1}{2} + z_{6}\right)c\sin\beta\,\hat{\mathbf{z}}$ | $(8f)$ | O IV |
| $\mathbf{B}_{23}$ | = | $(x_{7} - y_{7})\,\mathbf{a}_{1} + (x_{7} + y_{7})\,\mathbf{a}_{2} + z_{7}\,\mathbf{a}_{3}$ | = | $(x_{7}a + z_{7}c\cos\beta)\,\hat{\mathbf{x}} + y_{7}b\,\hat{\mathbf{y}} + z_{7}c\sin\beta\,\hat{\mathbf{z}}$ | $(8f)$ | O V |
| $\mathbf{B}_{24}$ | = | $(-x_{7} - y_{7})\,\mathbf{a}_{1} + (-x_{7} + y_{7})\,\mathbf{a}_{2} + \left(\frac{1}{2} - z_{7}\right)\mathbf{a}_{3}$ | = | $\left(\frac{1}{2}c\cos\beta - x_{7}a - z_{7}c\cos\beta\right)\hat{\mathbf{x}} + y_{7}b\,\hat{\mathbf{y}} + \left(\frac{1}{2} - z_{7}\right)c\sin\beta\,\hat{\mathbf{z}}$ | $(8f)$ | O V |
| $\mathbf{B}_{25}$ | = | $(-x_{7} + y_{7})\,\mathbf{a}_{1} + (-x_{7} - y_{7})\,\mathbf{a}_{2} - z_{7}\,\mathbf{a}_{3}$ | = | $(-x_{7}a - z_{7}c\cos\beta)\,\hat{\mathbf{x}} - y_{7}b\,\hat{\mathbf{y}} - z_{7}c\sin\beta\,\hat{\mathbf{z}}$ | $(8f)$ | O V |
| $\mathbf{B}_{26}$ | = | $(x_7 + y_7)\,\mathbf{a}_1 + (x_7 - y_7)\,\mathbf{a}_2 + \left(\frac{1}{2} + z_7\right)\mathbf{a}_3$ | = | $\left(\frac{1}{2}c\cos\beta + x_7 a + z_7 c\cos\beta\right)\hat{\mathbf{x}} - y_7 b\,\hat{\mathbf{y}} + \left(\frac{1}{2} + z_7\right)c\sin\beta\,\hat{\mathbf{z}}$ | (8$f$) | O V |
| $\mathbf{B}_{27}$ | = | $(x_8 - y_8)\,\mathbf{a}_1 + (x_8 + y_8)\,\mathbf{a}_2 + z_8\,\mathbf{a}_3$ | = | $(x_8 a + z_8 c\cos\beta)\,\hat{\mathbf{x}} + y_8 b\,\hat{\mathbf{y}} + z_8 c\sin\beta\,\hat{\mathbf{z}}$ | (8$f$) | OH |
| $\mathbf{B}_{28}$ | = | $(-x_8 - y_8)\,\mathbf{a}_1 + (-x_8 + y_8)\,\mathbf{a}_2 + \left(\frac{1}{2} - z_8\right)\mathbf{a}_3$ | = | $\left(\frac{1}{2}c\cos\beta - x_8 a - z_8 c\cos\beta\right)\hat{\mathbf{x}} + y_8 b\,\hat{\mathbf{y}} + \left(\frac{1}{2} - z_8\right)c\sin\beta\,\hat{\mathbf{z}}$ | (8$f$) | OH |
| $\mathbf{B}_{29}$ | = | $(-x_8 + y_8)\,\mathbf{a}_1 + (-x_8 - y_8)\,\mathbf{a}_2 - z_8\,\mathbf{a}_3$ | = | $(-x_8 a - z_8 c\cos\beta)\,\hat{\mathbf{x}} - y_8 b\,\hat{\mathbf{y}} - z_8 c\sin\beta\,\hat{\mathbf{z}}$ | (8$f$) | OH |
| $\mathbf{B}_{30}$ | = | $(x_8 + y_8)\,\mathbf{a}_1 + (x_8 - y_8)\,\mathbf{a}_2 + \left(\frac{1}{2} + z_8\right)\mathbf{a}_3$ | = | $\left(\frac{1}{2}c\cos\beta + x_8 a + z_8 c\cos\beta\right)\hat{\mathbf{x}} - y_8 b\,\hat{\mathbf{y}} + \left(\frac{1}{2} + z_8\right)c\sin\beta\,\hat{\mathbf{z}}$ | (8$f$) | OH |
| $\mathbf{B}_{31}$ | = | $(x_9 - y_9)\,\mathbf{a}_1 + (x_9 + y_9)\,\mathbf{a}_2 + z_9\,\mathbf{a}_3$ | = | $(x_9 a + z_9 c\cos\beta)\,\hat{\mathbf{x}} + y_9 b\,\hat{\mathbf{y}} + z_9 c\sin\beta\,\hat{\mathbf{z}}$ | (8$f$) | Si I |
| $\mathbf{B}_{32}$ | = | $(-x_9 - y_9)\,\mathbf{a}_1 + (-x_9 + y_9)\,\mathbf{a}_2 + \left(\frac{1}{2} - z_9\right)\mathbf{a}_3$ | = | $\left(\frac{1}{2}c\cos\beta - x_9 a - z_9 c\cos\beta\right)\hat{\mathbf{x}} + y_9 b\,\hat{\mathbf{y}} + \left(\frac{1}{2} - z_9\right)c\sin\beta\,\hat{\mathbf{z}}$ | (8$f$) | Si I |
| $\mathbf{B}_{33}$ | = | $(-x_9 + y_9)\,\mathbf{a}_1 + (-x_9 - y_9)\,\mathbf{a}_2 - z_9\,\mathbf{a}_3$ | = | $(-x_9 a - z_9 c\cos\beta)\,\hat{\mathbf{x}} - y_9 b\,\hat{\mathbf{y}} - z_9 c\sin\beta\,\hat{\mathbf{z}}$ | (8$f$) | Si I |
| $\mathbf{B}_{34}$ | = | $(x_9 + y_9)\,\mathbf{a}_1 + (x_9 - y_9)\,\mathbf{a}_2 + \left(\frac{1}{2} + z_9\right)\mathbf{a}_3$ | = | $\left(\frac{1}{2}c\cos\beta + x_9 a + z_9 c\cos\beta\right)\hat{\mathbf{x}} - y_9 b\,\hat{\mathbf{y}} + \left(\frac{1}{2} + z_9\right)c\sin\beta\,\hat{\mathbf{z}}$ | (8$f$) | Si I |
| $\mathbf{B}_{35}$ | = | $(x_{10} - y_{10})\,\mathbf{a}_1 + (x_{10} + y_{10})\,\mathbf{a}_2 + z_{10}\,\mathbf{a}_3$ | = | $(x_{10} a + z_{10} c\cos\beta)\,\hat{\mathbf{x}} + y_{10} b\,\hat{\mathbf{y}} + z_{10} c\sin\beta\,\hat{\mathbf{z}}$ | (8$f$) | Si II |
| $\mathbf{B}_{36}$ | = | $(-x_{10} - y_{10})\,\mathbf{a}_1 + (-x_{10} + y_{10})\,\mathbf{a}_2 + \left(\frac{1}{2} - z_{10}\right)\mathbf{a}_3$ | = | $\left(\frac{1}{2}c\cos\beta - x_{10} a - z_{10} c\cos\beta\right)\hat{\mathbf{x}} + y_{10} b\,\hat{\mathbf{y}} + \left(\frac{1}{2} - z_{10}\right)c\sin\beta\,\hat{\mathbf{z}}$ | (8$f$) | Si II |
| $\mathbf{B}_{37}$ | = | $(-x_{10} + y_{10})\,\mathbf{a}_1 + (-x_{10} - y_{10})\,\mathbf{a}_2 - z_{10}\,\mathbf{a}_3$ | = | $(-x_{10} a - z_{10} c\cos\beta)\,\hat{\mathbf{x}} - y_{10} b\,\hat{\mathbf{y}} - z_{10} c\sin\beta\,\hat{\mathbf{z}}$ | (8$f$) | Si II |
| $\mathbf{B}_{38}$ | = | $(x_{10} + y_{10})\,\mathbf{a}_1 + (x_{10} - y_{10})\,\mathbf{a}_2 + \left(\frac{1}{2} + z_{10}\right)\mathbf{a}_3$ | = | $\left(\frac{1}{2}c\cos\beta + x_{10} a + z_{10} c\cos\beta\right)\hat{\mathbf{x}} - y_{10} b\,\hat{\mathbf{y}} + \left(\frac{1}{2} + z_{10}\right)c\sin\beta\,\hat{\mathbf{z}}$ | (8$f$) | Si II |

**Found in:**

**Geometry files:**

- CIF: pp. 1563
- POSCAR: pp. 1563

# $Rb_2C_2O_4{\cdot}H_2O$ Structure: A2BC4D2_mC36_15_f_e_2f_f

http://aflow.org/prototype-encyclopedia/A2BC4D2_mC36_15_f_e_2f_f

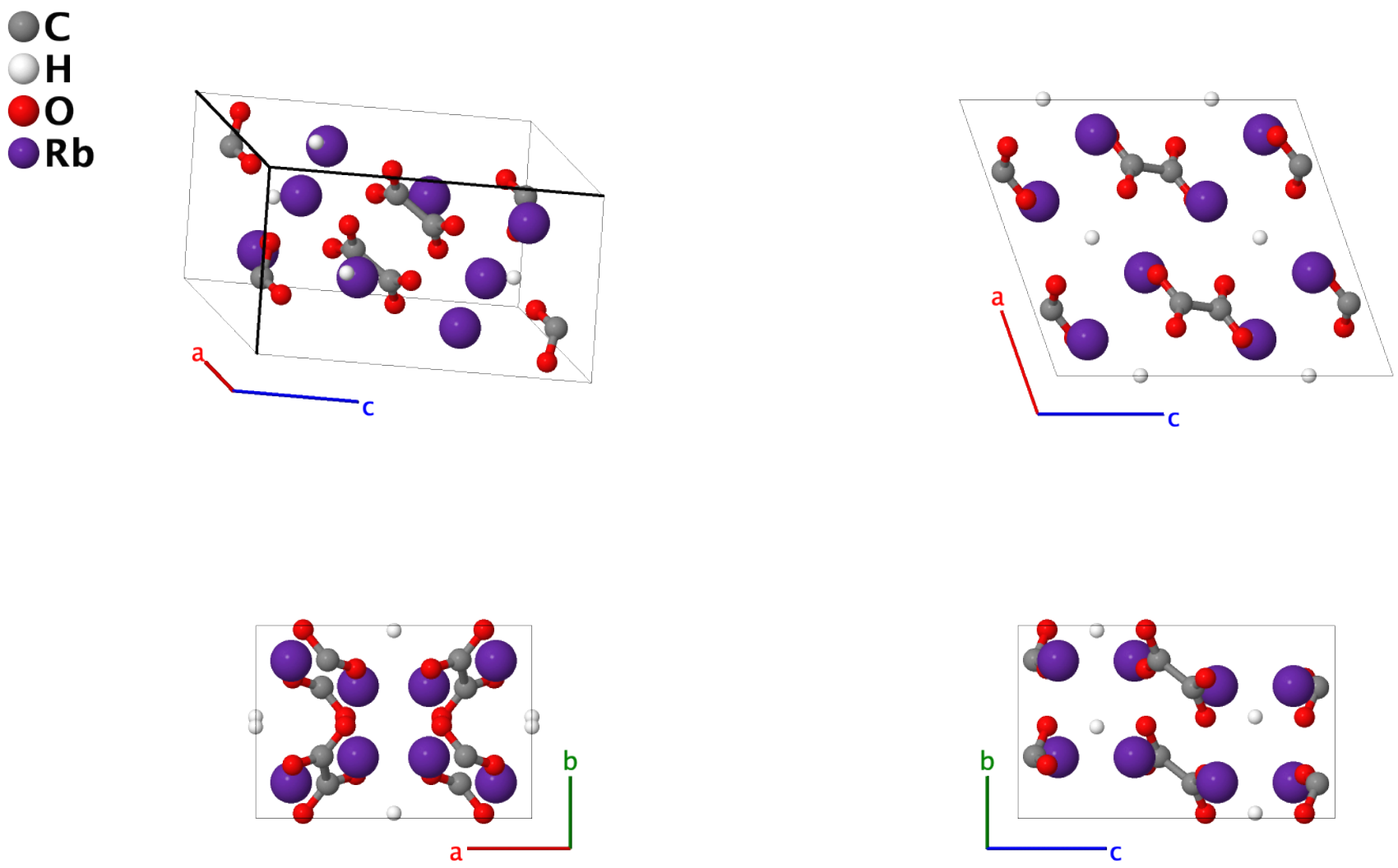

| | | |
|---|---|---|
| **Prototype** | : | $C_2(H_2O)O_4Rb_2$ |
| **AFLOW prototype label** | : | A2BC4D2_mC36_15_f_e_2f_f |
| ***Strukturbericht* designation** | : | None |
| **Pearson symbol** | : | mC36 |
| **Space group number** | : | 15 |
| **Space group symbol** | : | $C2/c$ |
| **AFLOW prototype command** | : | `aflow --proto=A2BC4D2_mC36_15_f_e_2f_f` `--params=`$a, b/a, c/a, \beta, y_1, x_2, y_2, z_2, x_3, y_3, z_3, x_4, y_4, z_4, x_5, y_5, z_5$ |

**Base-centered Monoclinic primitive vectors:**

$$\mathbf{a}_1 = \frac{1}{2}\,a\,\hat{\mathbf{x}} - \frac{1}{2}\,b\,\hat{\mathbf{y}}$$
$$\mathbf{a}_2 = \frac{1}{2}\,a\,\hat{\mathbf{x}} + \frac{1}{2}\,b\,\hat{\mathbf{y}}$$
$$\mathbf{a}_3 = c\cos\beta\,\hat{\mathbf{x}} + c\sin\beta\,\hat{\mathbf{z}}$$

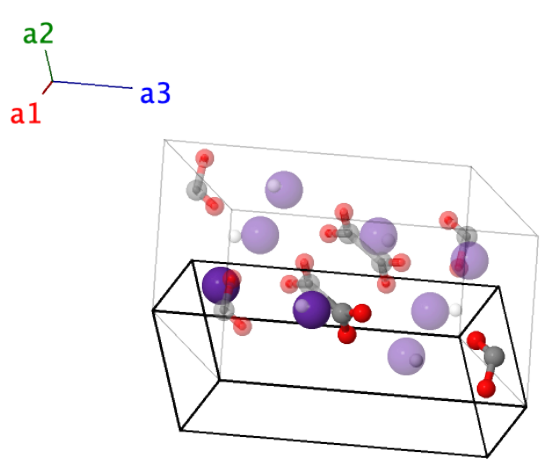

**Basis vectors:**

| | | Lattice Coordinates | | Cartesian Coordinates | Wyckoff Position | Atom Type |
|---|---|---|---|---|---|---|
| $\mathbf{B}_1$ | = | $-y_1\,\mathbf{a}_1 + y_1\,\mathbf{a}_2 + \frac{1}{4}\,\mathbf{a}_3$ | = | $\frac{1}{4}c\cos\beta\,\hat{\mathbf{x}} + y_1 b\,\hat{\mathbf{y}} + \frac{1}{4}c\sin\beta\,\hat{\mathbf{z}}$ | (4e) | $H_2O$ |
| $\mathbf{B}_2$ | = | $y_1\,\mathbf{a}_1 - y_1\,\mathbf{a}_2 + \frac{3}{4}\,\mathbf{a}_3$ | = | $\frac{3}{4}c\cos\beta\,\hat{\mathbf{x}} - y_1 b\,\hat{\mathbf{y}} + \frac{3}{4}c\sin\beta\,\hat{\mathbf{z}}$ | (4e) | $H_2O$ |
| $\mathbf{B}_3$ | = | $(x_2 - y_2)\,\mathbf{a}_1 + (x_2 + y_2)\,\mathbf{a}_2 + z_2\,\mathbf{a}_3$ | = | $(x_2 a + z_2 c\cos\beta)\,\hat{\mathbf{x}} + y_2 b\,\hat{\mathbf{y}} + z_2 c\sin\beta\,\hat{\mathbf{z}}$ | (8f) | C |
| $\mathbf{B}_4$ | = | $(-x_2 - y_2)\,\mathbf{a}_1 + (-x_2 + y_2)\,\mathbf{a}_2 + \left(\frac{1}{2} - z_2\right)\,\mathbf{a}_3$ | = | $\left(\frac{1}{2}c\cos\beta - x_2 a - z_2 c\cos\beta\right)\,\hat{\mathbf{x}} + y_2 b\,\hat{\mathbf{y}} + \left(\frac{1}{2} - z_2\right)c\sin\beta\,\hat{\mathbf{z}}$ | (8f) | C |
| $\mathbf{B}_{5}$ | = | $\left(-x_{2}+y_{2}\right)\,\mathbf{a}_{1}+\left(-x_{2}-y_{2}\right)\,\mathbf{a}_{2}-z_{2}\,\mathbf{a}_{3}$ | = | $\left(-x_{2}a-z_{2}c\cos\beta\right)\,\mathbf{\hat{x}}-y_{2}b\,\mathbf{\hat{y}}-z_{2}c\sin\beta\,\mathbf{\hat{z}}$ | (8$f$) | C |
| $\mathbf{B}_{6}$ | = | $\left(x_{2}+y_{2}\right)\,\mathbf{a}_{1}+\left(x_{2}-y_{2}\right)\,\mathbf{a}_{2}+\left(\frac{1}{2}+z_{2}\right)\,\mathbf{a}_{3}$ | = | $\left(\frac{1}{2}c\cos\beta+x_{2}a+z_{2}c\cos\beta\right)\,\mathbf{\hat{x}}-y_{2}b\,\mathbf{\hat{y}}+\left(\frac{1}{2}+z_{2}\right)c\sin\beta\,\mathbf{\hat{z}}$ | (8$f$) | C |
| $\mathbf{B}_{7}$ | = | $\left(x_{3}-y_{3}\right)\,\mathbf{a}_{1}+\left(x_{3}+y_{3}\right)\,\mathbf{a}_{2}+z_{3}\,\mathbf{a}_{3}$ | = | $\left(x_{3}a+z_{3}c\cos\beta\right)\,\mathbf{\hat{x}}+y_{3}b\,\mathbf{\hat{y}}+z_{3}c\sin\beta\,\mathbf{\hat{z}}$ | (8$f$) | O I |
| $\mathbf{B}_{8}$ | = | $\left(-x_{3}-y_{3}\right)\,\mathbf{a}_{1}+\left(-x_{3}+y_{3}\right)\,\mathbf{a}_{2}+\left(\frac{1}{2}-z_{3}\right)\,\mathbf{a}_{3}$ | = | $\left(\frac{1}{2}c\cos\beta-x_{3}a-z_{3}c\cos\beta\right)\,\mathbf{\hat{x}}+y_{3}b\,\mathbf{\hat{y}}+\left(\frac{1}{2}-z_{3}\right)c\sin\beta\,\mathbf{\hat{z}}$ | (8$f$) | O I |
| $\mathbf{B}_{9}$ | = | $\left(-x_{3}+y_{3}\right)\,\mathbf{a}_{1}+\left(-x_{3}-y_{3}\right)\,\mathbf{a}_{2}-z_{3}\,\mathbf{a}_{3}$ | = | $\left(-x_{3}a-z_{3}c\cos\beta\right)\,\mathbf{\hat{x}}-y_{3}b\,\mathbf{\hat{y}}-z_{3}c\sin\beta\,\mathbf{\hat{z}}$ | (8$f$) | O I |
| $\mathbf{B}_{10}$ | = | $\left(x_{3}+y_{3}\right)\,\mathbf{a}_{1}+\left(x_{3}-y_{3}\right)\,\mathbf{a}_{2}+\left(\frac{1}{2}+z_{3}\right)\,\mathbf{a}_{3}$ | = | $\left(\frac{1}{2}c\cos\beta+x_{3}a+z_{3}c\cos\beta\right)\,\mathbf{\hat{x}}-y_{3}b\,\mathbf{\hat{y}}+\left(\frac{1}{2}+z_{3}\right)c\sin\beta\,\mathbf{\hat{z}}$ | (8$f$) | O I |
| $\mathbf{B}_{11}$ | = | $\left(x_{4}-y_{4}\right)\,\mathbf{a}_{1}+\left(x_{4}+y_{4}\right)\,\mathbf{a}_{2}+z_{4}\,\mathbf{a}_{3}$ | = | $\left(x_{4}a+z_{4}c\cos\beta\right)\,\mathbf{\hat{x}}+y_{4}b\,\mathbf{\hat{y}}+z_{4}c\sin\beta\,\mathbf{\hat{z}}$ | (8$f$) | O II |
| $\mathbf{B}_{12}$ | = | $\left(-x_{4}-y_{4}\right)\,\mathbf{a}_{1}+\left(-x_{4}+y_{4}\right)\,\mathbf{a}_{2}+\left(\frac{1}{2}-z_{4}\right)\,\mathbf{a}_{3}$ | = | $\left(\frac{1}{2}c\cos\beta-x_{4}a-z_{4}c\cos\beta\right)\,\mathbf{\hat{x}}+y_{4}b\,\mathbf{\hat{y}}+\left(\frac{1}{2}-z_{4}\right)c\sin\beta\,\mathbf{\hat{z}}$ | (8$f$) | O II |
| $\mathbf{B}_{13}$ | = | $\left(-x_{4}+y_{4}\right)\,\mathbf{a}_{1}+\left(-x_{4}-y_{4}\right)\,\mathbf{a}_{2}-z_{4}\,\mathbf{a}_{3}$ | = | $\left(-x_{4}a-z_{4}c\cos\beta\right)\,\mathbf{\hat{x}}-y_{4}b\,\mathbf{\hat{y}}-z_{4}c\sin\beta\,\mathbf{\hat{z}}$ | (8$f$) | O II |
| $\mathbf{B}_{14}$ | = | $\left(x_{4}+y_{4}\right)\,\mathbf{a}_{1}+\left(x_{4}-y_{4}\right)\,\mathbf{a}_{2}+\left(\frac{1}{2}+z_{4}\right)\,\mathbf{a}_{3}$ | = | $\left(\frac{1}{2}c\cos\beta+x_{4}a+z_{4}c\cos\beta\right)\,\mathbf{\hat{x}}-y_{4}b\,\mathbf{\hat{y}}+\left(\frac{1}{2}+z_{4}\right)c\sin\beta\,\mathbf{\hat{z}}$ | (8$f$) | O II |
| $\mathbf{B}_{15}$ | = | $\left(x_{5}-y_{5}\right)\,\mathbf{a}_{1}+\left(x_{5}+y_{5}\right)\,\mathbf{a}_{2}+z_{5}\,\mathbf{a}_{3}$ | = | $\left(x_{5}a+z_{5}c\cos\beta\right)\,\mathbf{\hat{x}}+y_{5}b\,\mathbf{\hat{y}}+z_{5}c\sin\beta\,\mathbf{\hat{z}}$ | (8$f$) | Rb |
| $\mathbf{B}_{16}$ | = | $\left(-x_{5}-y_{5}\right)\,\mathbf{a}_{1}+\left(-x_{5}+y_{5}\right)\,\mathbf{a}_{2}+\left(\frac{1}{2}-z_{5}\right)\,\mathbf{a}_{3}$ | = | $\left(\frac{1}{2}c\cos\beta-x_{5}a-z_{5}c\cos\beta\right)\,\mathbf{\hat{x}}+y_{5}b\,\mathbf{\hat{y}}+\left(\frac{1}{2}-z_{5}\right)c\sin\beta\,\mathbf{\hat{z}}$ | (8$f$) | Rb |
| $\mathbf{B}_{17}$ | = | $\left(-x_{5}+y_{5}\right)\,\mathbf{a}_{1}+\left(-x_{5}-y_{5}\right)\,\mathbf{a}_{2}-z_{5}\,\mathbf{a}_{3}$ | = | $\left(-x_{5}a-z_{5}c\cos\beta\right)\,\mathbf{\hat{x}}-y_{5}b\,\mathbf{\hat{y}}-z_{5}c\sin\beta\,\mathbf{\hat{z}}$ | (8$f$) | Rb |
| $\mathbf{B}_{18}$ | = | $\left(x_{5}+y_{5}\right)\,\mathbf{a}_{1}+\left(x_{5}-y_{5}\right)\,\mathbf{a}_{2}+\left(\frac{1}{2}+z_{5}\right)\,\mathbf{a}_{3}$ | = | $\left(\frac{1}{2}c\cos\beta+x_{5}a+z_{5}c\cos\beta\right)\,\mathbf{\hat{x}}-y_{5}b\,\mathbf{\hat{y}}+\left(\frac{1}{2}+z_{5}\right)c\sin\beta\,\mathbf{\hat{z}}$ | (8$f$) | Rb |

---

---

**Geometry files:**

- CIF: pp. 1564
- POSCAR: pp. 1564

# Alluaudite [$NaMnFe_2(PO_4)_3$] Structure: A2BCD12E3_mC76_15_f_e_b_6f_ef

http://aflow.org/prototype-encyclopedia/A2BCD12E3_mC76_15_f_e_b_6f_ef

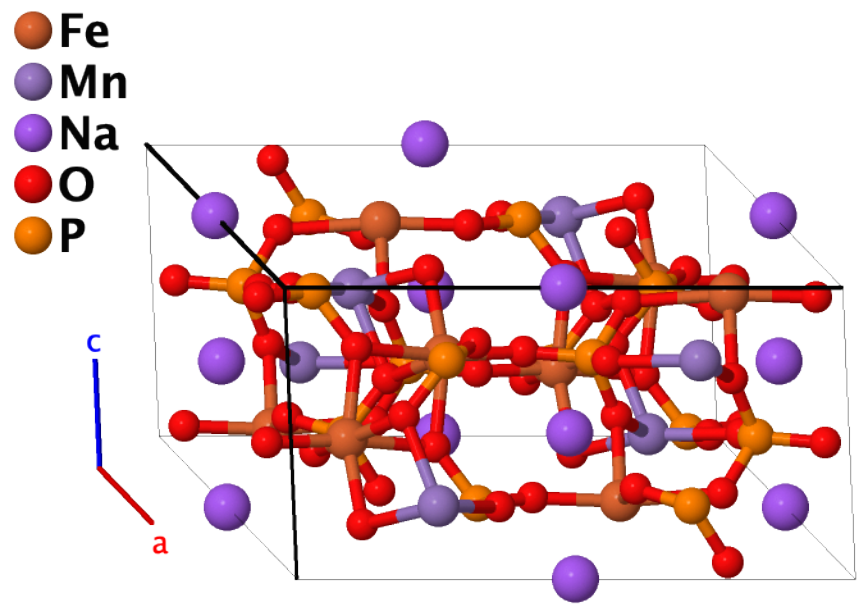

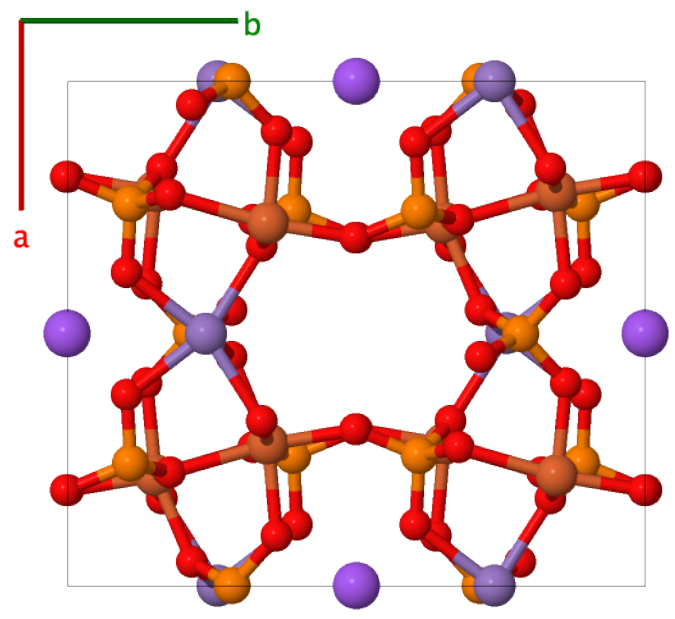

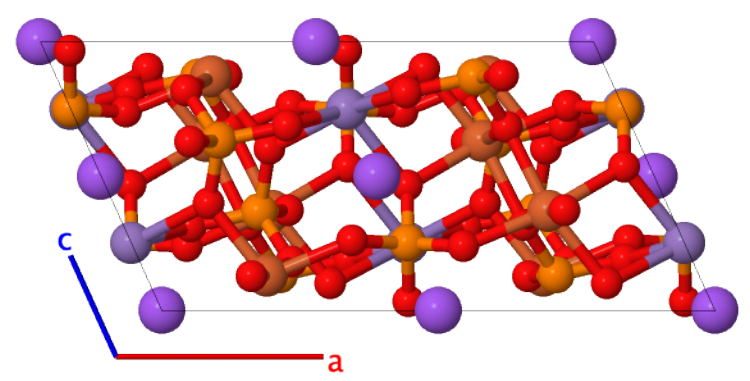

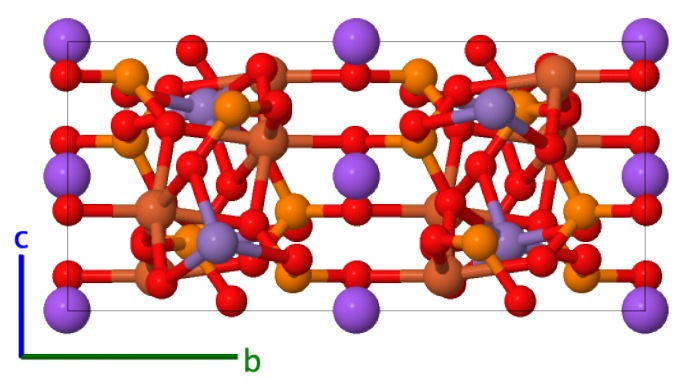

| | | |
|---|---|---|
| **Prototype** | : | $Fe_2MnNaO_{12}P_3$ |
| **AFLOW prototype label** | : | A2BCD12E3_mC76_15_f_e_b_6f_ef |
| ***Strukturbericht* designation** | : | None |
| **Pearson symbol** | : | mC76 |
| **Space group number** | : | 15 |
| **Space group symbol** | : | $C2/c$ |
| **AFLOW prototype command** | : | `aflow --proto=A2BCD12E3_mC76_15_f_e_b_6f_ef`<br>`--params=`$a, b/a, c/a, \beta, y_2, y_3, x_4, y_4, z_4, x_5, y_5, z_5, x_6, y_6, z_6, x_7, y_7, z_7, x_8, y_8, z_8, x_9, y_9, z_9, x_{10}, y_{10}, z_{10}, x_{11}, y_{11}, z_{11}$ |

### Other compounds with this structure

- $Ag_xNa_{1-x}Mn_3(PO_4)_3$, $Li_xNa_{1-x}CdIn_2(PO_4)_3$, $Li_xNa_{1-x}MnFe_2(PO_4)_3$, $Na_2$(Fe,Co)$Fe(VO_4)_3$, $Na_2Co_2Cr(PO_4)_3$, $Na_2Fe_2V(PO_4)_3$, $Na_2Zn_2Fe(VO_4)_3$, $Na_3Bi_2(AsO_4)_3$, $Na_4Co(MnO_4)_3$, and $Na_xMn_yAl_{5-x-y}(PO_4)_3$

- This is a rather idealized version of alluaudite. (Moore, 1971) gives the composition of the sodium site (Na) as $Na_{0.625}Mn_{0.175}Ca_{0.125}$, with vacancies on the remaining sites; the manganese (Mn) site is $Mn_{0.950}Mg_{0.025}Li_{0.025}$; and the iron (Fe) site as $Fe_{0.988}Mg_{0.012}$. (Hatert, 2005) puts some sodium atoms on another (4e) site, with $y \approx 0$; the exact placement and concentration depends upon the manganese content. We may expect similar variations in other compounds having this structure.

**Base-centered Monoclinic primitive vectors:**

$$\mathbf{a}_1 = \tfrac{1}{2}\,a\,\hat{\mathbf{x}} - \tfrac{1}{2}\,b\,\hat{\mathbf{y}}$$
$$\mathbf{a}_2 = \tfrac{1}{2}\,a\,\hat{\mathbf{x}} + \tfrac{1}{2}\,b\,\hat{\mathbf{y}}$$
$$\mathbf{a}_3 = c\cos\beta\,\hat{\mathbf{x}} + c\sin\beta\,\hat{\mathbf{z}}$$

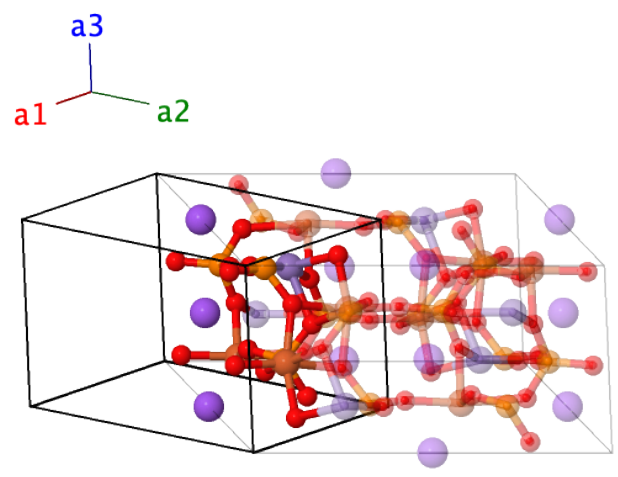

**Basis vectors:**

| | | Lattice Coordinates | | Cartesian Coordinates | Wyckoff Position | Atom Type |
|---|---|---|---|---|---|---|
| $\mathbf{B}_1$ | = | $\frac{1}{2}\,\mathbf{a}_1 + \frac{1}{2}\,\mathbf{a}_2$ | = | $\frac{1}{2}a\,\hat{\mathbf{x}}$ | (4$b$) | Na |
| $\mathbf{B}_2$ | = | $\frac{1}{2}\,\mathbf{a}_1 + \frac{1}{2}\,\mathbf{a}_2 + \frac{1}{2}\,\mathbf{a}_3$ | = | $\frac{1}{2}\,(a + c\cos\beta)\,\hat{\mathbf{x}} + \frac{1}{2}c\sin\beta\,\hat{\mathbf{z}}$ | (4$b$) | Na |
| $\mathbf{B}_3$ | = | $-y_2\,\mathbf{a}_1 + y_2\,\mathbf{a}_2 + \frac{1}{4}\,\mathbf{a}_3$ | = | $\frac{1}{4}c\cos\beta\,\hat{\mathbf{x}} + y_2 b\,\hat{\mathbf{y}} + \frac{1}{4}c\sin\beta\,\hat{\mathbf{z}}$ | (4$e$) | Mn |
| $\mathbf{B}_4$ | = | $y_2\,\mathbf{a}_1 - y_2\,\mathbf{a}_2 + \frac{3}{4}\,\mathbf{a}_3$ | = | $\frac{3}{4}c\cos\beta\,\hat{\mathbf{x}} - y_2 b\,\hat{\mathbf{y}} + \frac{3}{4}c\sin\beta\,\hat{\mathbf{z}}$ | (4$e$) | Mn |
| $\mathbf{B}_5$ | = | $-y_3\,\mathbf{a}_1 + y_3\,\mathbf{a}_2 + \frac{1}{4}\,\mathbf{a}_3$ | = | $\frac{1}{4}c\cos\beta\,\hat{\mathbf{x}} + y_3 b\,\hat{\mathbf{y}} + \frac{1}{4}c\sin\beta\,\hat{\mathbf{z}}$ | (4$e$) | P I |
| $\mathbf{B}_6$ | = | $y_3\,\mathbf{a}_1 - y_3\,\mathbf{a}_2 + \frac{3}{4}\,\mathbf{a}_3$ | = | $\frac{3}{4}c\cos\beta\,\hat{\mathbf{x}} - y_3 b\,\hat{\mathbf{y}} + \frac{3}{4}c\sin\beta\,\hat{\mathbf{z}}$ | (4$e$) | P I |
| $\mathbf{B}_7$ | = | $(x_4 - y_4)\,\mathbf{a}_1 + (x_4 + y_4)\,\mathbf{a}_2 + z_4\,\mathbf{a}_3$ | = | $(x_4 a + z_4 c\cos\beta)\,\hat{\mathbf{x}} + y_4 b\,\hat{\mathbf{y}} + z_4 c\sin\beta\,\hat{\mathbf{z}}$ | (8$f$) | Fe |
| $\mathbf{B}_8$ | = | $(-x_4 - y_4)\,\mathbf{a}_1 + (-x_4 + y_4)\,\mathbf{a}_2 + \left(\frac{1}{2} - z_4\right)\mathbf{a}_3$ | = | $\left(\frac{1}{2}c\cos\beta - x_4 a - z_4 c\cos\beta\right)\hat{\mathbf{x}} + y_4 b\,\hat{\mathbf{y}} + \left(\frac{1}{2} - z_4\right)c\sin\beta\,\hat{\mathbf{z}}$ | (8$f$) | Fe |
| $\mathbf{B}_9$ | = | $(-x_4 + y_4)\,\mathbf{a}_1 + (-x_4 - y_4)\,\mathbf{a}_2 - z_4\,\mathbf{a}_3$ | = | $(-x_4 a - z_4 c\cos\beta)\,\hat{\mathbf{x}} - y_4 b\,\hat{\mathbf{y}} - z_4 c\sin\beta\,\hat{\mathbf{z}}$ | (8$f$) | Fe |
| $\mathbf{B}_{10}$ | = | $(x_4 + y_4)\,\mathbf{a}_1 + (x_4 - y_4)\,\mathbf{a}_2 + \left(\frac{1}{2} + z_4\right)\mathbf{a}_3$ | = | $\left(\frac{1}{2}c\cos\beta + x_4 a + z_4 c\cos\beta\right)\hat{\mathbf{x}} - y_4 b\,\hat{\mathbf{y}} + \left(\frac{1}{2} + z_4\right)c\sin\beta\,\hat{\mathbf{z}}$ | (8$f$) | Fe |
| $\mathbf{B}_{11}$ | = | $(x_5 - y_5)\,\mathbf{a}_1 + (x_5 + y_5)\,\mathbf{a}_2 + z_5\,\mathbf{a}_3$ | = | $(x_5 a + z_5 c\cos\beta)\,\hat{\mathbf{x}} + y_5 b\,\hat{\mathbf{y}} + z_5 c\sin\beta\,\hat{\mathbf{z}}$ | (8$f$) | O I |
| $\mathbf{B}_{12}$ | = | $(-x_5 - y_5)\,\mathbf{a}_1 + (-x_5 + y_5)\,\mathbf{a}_2 + \left(\frac{1}{2} - z_5\right)\mathbf{a}_3$ | = | $\left(\frac{1}{2}c\cos\beta - x_5 a - z_5 c\cos\beta\right)\hat{\mathbf{x}} + y_5 b\,\hat{\mathbf{y}} + \left(\frac{1}{2} - z_5\right)c\sin\beta\,\hat{\mathbf{z}}$ | (8$f$) | O I |
| $\mathbf{B}_{13}$ | = | $(-x_5 + y_5)\,\mathbf{a}_1 + (-x_5 - y_5)\,\mathbf{a}_2 - z_5\,\mathbf{a}_3$ | = | $(-x_5 a - z_5 c\cos\beta)\,\hat{\mathbf{x}} - y_5 b\,\hat{\mathbf{y}} - z_5 c\sin\beta\,\hat{\mathbf{z}}$ | (8$f$) | O I |
| $\mathbf{B}_{14}$ | = | $(x_5 + y_5)\,\mathbf{a}_1 + (x_5 - y_5)\,\mathbf{a}_2 + \left(\frac{1}{2} + z_5\right)\mathbf{a}_3$ | = | $\left(\frac{1}{2}c\cos\beta + x_5 a + z_5 c\cos\beta\right)\hat{\mathbf{x}} - y_5 b\,\hat{\mathbf{y}} + \left(\frac{1}{2} + z_5\right)c\sin\beta\,\hat{\mathbf{z}}$ | (8$f$) | O I |
| $\mathbf{B}_{15}$ | = | $(x_6 - y_6)\,\mathbf{a}_1 + (x_6 + y_6)\,\mathbf{a}_2 + z_6\,\mathbf{a}_3$ | = | $(x_6 a + z_6 c\cos\beta)\,\hat{\mathbf{x}} + y_6 b\,\hat{\mathbf{y}} + z_6 c\sin\beta\,\hat{\mathbf{z}}$ | (8$f$) | O II |
| $\mathbf{B}_{16}$ | = | $(-x_6 - y_6)\,\mathbf{a}_1 + (-x_6 + y_6)\,\mathbf{a}_2 + \left(\frac{1}{2} - z_6\right)\mathbf{a}_3$ | = | $\left(\frac{1}{2}c\cos\beta - x_6 a - z_6 c\cos\beta\right)\hat{\mathbf{x}} + y_6 b\,\hat{\mathbf{y}} + \left(\frac{1}{2} - z_6\right)c\sin\beta\,\hat{\mathbf{z}}$ | (8$f$) | O II |
| $\mathbf{B}_{17}$ | = | $(-x_6 + y_6)\,\mathbf{a}_1 + (-x_6 - y_6)\,\mathbf{a}_2 - z_6\,\mathbf{a}_3$ | = | $(-x_6 a - z_6 c\cos\beta)\,\hat{\mathbf{x}} - y_6 b\,\hat{\mathbf{y}} - z_6 c\sin\beta\,\hat{\mathbf{z}}$ | (8$f$) | O II |
| $\mathbf{B}_{18}$ | = | $(x_6 + y_6)\,\mathbf{a}_1 + (x_6 - y_6)\,\mathbf{a}_2 + \left(\frac{1}{2} + z_6\right)\mathbf{a}_3$ | = | $\left(\frac{1}{2}c\cos\beta + x_6 a + z_6 c\cos\beta\right)\hat{\mathbf{x}} - y_6 b\,\hat{\mathbf{y}} + \left(\frac{1}{2} + z_6\right)c\sin\beta\,\hat{\mathbf{z}}$ | (8$f$) | O II |
| $\mathbf{B}_{19}$ | = | $(x_7 - y_7)\,\mathbf{a}_1 + (x_7 + y_7)\,\mathbf{a}_2 + z_7\,\mathbf{a}_3$ | = | $(x_7 a + z_7 c\cos\beta)\,\hat{\mathbf{x}} + y_7 b\,\hat{\mathbf{y}} + z_7 c\sin\beta\,\hat{\mathbf{z}}$ | (8$f$) | O III |
| $\mathbf{B}_{20}$ | = | $(-x_7 - y_7)\,\mathbf{a}_1 + (-x_7 + y_7)\,\mathbf{a}_2 + \left(\frac{1}{2} - z_7\right)\mathbf{a}_3$ | = | $\left(\frac{1}{2}c\cos\beta - x_7 a - z_7 c\cos\beta\right)\hat{\mathbf{x}} + y_7 b\,\hat{\mathbf{y}} + \left(\frac{1}{2} - z_7\right)c\sin\beta\,\hat{\mathbf{z}}$ | $(8f)$ | O III |
| $\mathbf{B}_{21}$ | = | $(-x_7 + y_7)\,\mathbf{a}_1 + (-x_7 - y_7)\,\mathbf{a}_2 - z_7\,\mathbf{a}_3$ | = | $(-x_7 a - z_7 c\cos\beta)\,\hat{\mathbf{x}} - y_7 b\,\hat{\mathbf{y}} - z_7 c\sin\beta\,\hat{\mathbf{z}}$ | $(8f)$ | O III |
| $\mathbf{B}_{22}$ | = | $(x_7 + y_7)\,\mathbf{a}_1 + (x_7 - y_7)\,\mathbf{a}_2 + \left(\frac{1}{2} + z_7\right)\mathbf{a}_3$ | = | $\left(\frac{1}{2}c\cos\beta + x_7 a + z_7 c\cos\beta\right)\hat{\mathbf{x}} - y_7 b\,\hat{\mathbf{y}} + \left(\frac{1}{2} + z_7\right)c\sin\beta\,\hat{\mathbf{z}}$ | $(8f)$ | O III |
| $\mathbf{B}_{23}$ | = | $(x_8 - y_8)\,\mathbf{a}_1 + (x_8 + y_8)\,\mathbf{a}_2 + z_8\,\mathbf{a}_3$ | = | $(x_8 a + z_8 c\cos\beta)\,\hat{\mathbf{x}} + y_8 b\,\hat{\mathbf{y}} + z_8 c\sin\beta\,\hat{\mathbf{z}}$ | $(8f)$ | O IV |
| $\mathbf{B}_{24}$ | = | $(-x_8 - y_8)\,\mathbf{a}_1 + (-x_8 + y_8)\,\mathbf{a}_2 + \left(\frac{1}{2} - z_8\right)\mathbf{a}_3$ | = | $\left(\frac{1}{2}c\cos\beta - x_8 a - z_8 c\cos\beta\right)\hat{\mathbf{x}} + y_8 b\,\hat{\mathbf{y}} + \left(\frac{1}{2} - z_8\right)c\sin\beta\,\hat{\mathbf{z}}$ | $(8f)$ | O IV |
| $\mathbf{B}_{25}$ | = | $(-x_8 + y_8)\,\mathbf{a}_1 + (-x_8 - y_8)\,\mathbf{a}_2 - z_8\,\mathbf{a}_3$ | = | $(-x_8 a - z_8 c\cos\beta)\,\hat{\mathbf{x}} - y_8 b\,\hat{\mathbf{y}} - z_8 c\sin\beta\,\hat{\mathbf{z}}$ | $(8f)$ | O IV |
| $\mathbf{B}_{26}$ | = | $(x_8 + y_8)\,\mathbf{a}_1 + (x_8 - y_8)\,\mathbf{a}_2 + \left(\frac{1}{2} + z_8\right)\mathbf{a}_3$ | = | $\left(\frac{1}{2}c\cos\beta + x_8 a + z_8 c\cos\beta\right)\hat{\mathbf{x}} - y_8 b\,\hat{\mathbf{y}} + \left(\frac{1}{2} + z_8\right)c\sin\beta\,\hat{\mathbf{z}}$ | $(8f)$ | O IV |
| $\mathbf{B}_{27}$ | = | $(x_9 - y_9)\,\mathbf{a}_1 + (x_9 + y_9)\,\mathbf{a}_2 + z_9\,\mathbf{a}_3$ | = | $(x_9 a + z_9 c\cos\beta)\,\hat{\mathbf{x}} + y_9 b\,\hat{\mathbf{y}} + z_9 c\sin\beta\,\hat{\mathbf{z}}$ | $(8f)$ | O V |
| $\mathbf{B}_{28}$ | = | $(-x_9 - y_9)\,\mathbf{a}_1 + (-x_9 + y_9)\,\mathbf{a}_2 + \left(\frac{1}{2} - z_9\right)\mathbf{a}_3$ | = | $\left(\frac{1}{2}c\cos\beta - x_9 a - z_9 c\cos\beta\right)\hat{\mathbf{x}} + y_9 b\,\hat{\mathbf{y}} + \left(\frac{1}{2} - z_9\right)c\sin\beta\,\hat{\mathbf{z}}$ | $(8f)$ | O V |
| $\mathbf{B}_{29}$ | = | $(-x_9 + y_9)\,\mathbf{a}_1 + (-x_9 - y_9)\,\mathbf{a}_2 - z_9\,\mathbf{a}_3$ | = | $(-x_9 a - z_9 c\cos\beta)\,\hat{\mathbf{x}} - y_9 b\,\hat{\mathbf{y}} - z_9 c\sin\beta\,\hat{\mathbf{z}}$ | $(8f)$ | O V |
| $\mathbf{B}_{30}$ | = | $(x_9 + y_9)\,\mathbf{a}_1 + (x_9 - y_9)\,\mathbf{a}_2 + \left(\frac{1}{2} + z_9\right)\mathbf{a}_3$ | = | $\left(\frac{1}{2}c\cos\beta + x_9 a + z_9 c\cos\beta\right)\hat{\mathbf{x}} - y_9 b\,\hat{\mathbf{y}} + \left(\frac{1}{2} + z_9\right)c\sin\beta\,\hat{\mathbf{z}}$ | $(8f)$ | O V |
| $\mathbf{B}_{31}$ | = | $(x_{10} - y_{10})\,\mathbf{a}_1 + (x_{10} + y_{10})\,\mathbf{a}_2 + z_{10}\,\mathbf{a}_3$ | = | $(x_{10} a + z_{10} c\cos\beta)\,\hat{\mathbf{x}} + y_{10} b\,\hat{\mathbf{y}} + z_{10} c\sin\beta\,\hat{\mathbf{z}}$ | $(8f)$ | O VI |
| $\mathbf{B}_{32}$ | = | $(-x_{10} - y_{10})\,\mathbf{a}_1 + (-x_{10} + y_{10})\,\mathbf{a}_2 + \left(\frac{1}{2} - z_{10}\right)\mathbf{a}_3$ | = | $\left(\frac{1}{2}c\cos\beta - x_{10} a - z_{10} c\cos\beta\right)\hat{\mathbf{x}} + y_{10} b\,\hat{\mathbf{y}} + \left(\frac{1}{2} - z_{10}\right)c\sin\beta\,\hat{\mathbf{z}}$ | $(8f)$ | O VI |
| $\mathbf{B}_{33}$ | = | $(-x_{10} + y_{10})\,\mathbf{a}_1 + (-x_{10} - y_{10})\,\mathbf{a}_2 - z_{10}\,\mathbf{a}_3$ | = | $(-x_{10} a - z_{10} c\cos\beta)\,\hat{\mathbf{x}} - y_{10} b\,\hat{\mathbf{y}} - z_{10} c\sin\beta\,\hat{\mathbf{z}}$ | $(8f)$ | O VI |
| $\mathbf{B}_{34}$ | = | $(x_{10} + y_{10})\,\mathbf{a}_1 + (x_{10} - y_{10})\,\mathbf{a}_2 + \left(\frac{1}{2} + z_{10}\right)\mathbf{a}_3$ | = | $\left(\frac{1}{2}c\cos\beta + x_{10} a + z_{10} c\cos\beta\right)\hat{\mathbf{x}} - y_{10} b\,\hat{\mathbf{y}} + \left(\frac{1}{2} + z_{10}\right)c\sin\beta\,\hat{\mathbf{z}}$ | $(8f)$ | O VI |
| $\mathbf{B}_{35}$ | = | $(x_{11} - y_{11})\,\mathbf{a}_1 + (x_{11} + y_{11})\,\mathbf{a}_2 + z_{11}\,\mathbf{a}_3$ | = | $(x_{11} a + z_{11} c\cos\beta)\,\hat{\mathbf{x}} + y_{11} b\,\hat{\mathbf{y}} + z_{11} c\sin\beta\,\hat{\mathbf{z}}$ | $(8f)$ | P II |
| $\mathbf{B}_{36}$ | = | $(-x_{11} - y_{11})\,\mathbf{a}_1 + (-x_{11} + y_{11})\,\mathbf{a}_2 + \left(\frac{1}{2} - z_{11}\right)\mathbf{a}_3$ | = | $\left(\frac{1}{2}c\cos\beta - x_{11} a - z_{11} c\cos\beta\right)\hat{\mathbf{x}} + y_{11} b\,\hat{\mathbf{y}} + \left(\frac{1}{2} - z_{11}\right)c\sin\beta\,\hat{\mathbf{z}}$ | $(8f)$ | P II |
| $\mathbf{B}_{37}$ | = | $(-x_{11} + y_{11})\,\mathbf{a}_1 + (-x_{11} - y_{11})\,\mathbf{a}_2 - z_{11}\,\mathbf{a}_3$ | = | $(-x_{11} a - z_{11} c\cos\beta)\,\hat{\mathbf{x}} - y_{11} b\,\hat{\mathbf{y}} - z_{11} c\sin\beta\,\hat{\mathbf{z}}$ | $(8f)$ | P II |
| $\mathbf{B}_{38}$ | = | $(x_{11} + y_{11})\,\mathbf{a}_1 + (x_{11} - y_{11})\,\mathbf{a}_2 + \left(\frac{1}{2} + z_{11}\right)\mathbf{a}_3$ | = | $\left(\frac{1}{2}c\cos\beta + x_{11} a + z_{11} c\cos\beta\right)\hat{\mathbf{x}} - y_{11} b\,\hat{\mathbf{y}} + \left(\frac{1}{2} + z_{11}\right)c\sin\beta\,\hat{\mathbf{z}}$ | $(8f)$ | P II |

**Geometry files:**
- CIF: pp. 1564
- POSCAR: pp. 1565

# $ThC_2$ ($C_g$) Structure: A2B_mC12_15_f_e

http://aflow.org/prototype-encyclopedia/A2B_mC12_15_f_e

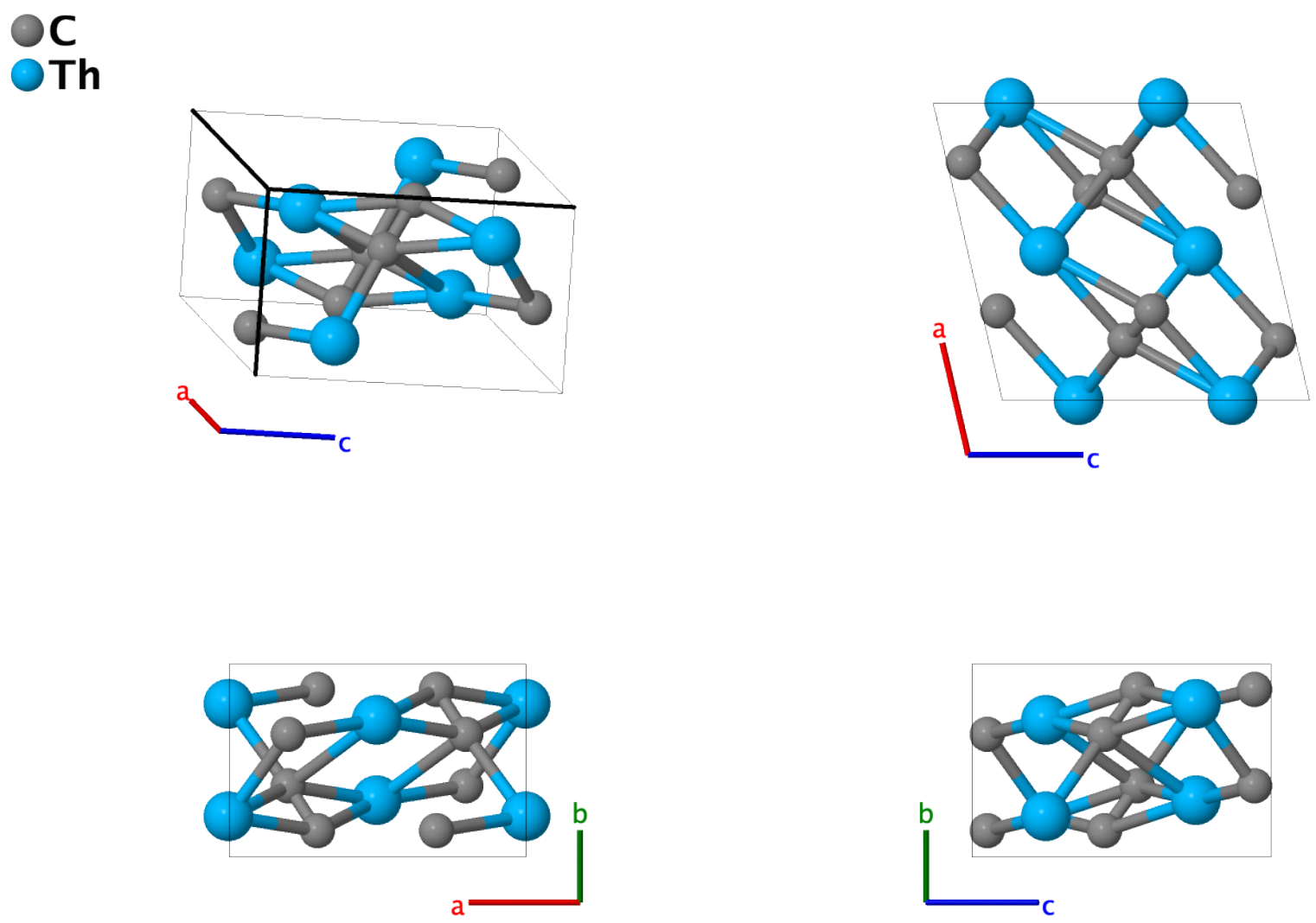

| | | |
|---|---|---|
| **Prototype** | : | $C_2Th$ |
| **AFLOW prototype label** | : | A2B_mC12_15_f_e |
| ***Strukturbericht* designation** | : | None |
| **Pearson symbol** | : | mC12 |
| **Space group number** | : | 15 |
| **Space group symbol** | : | $C2/c$ |
| **AFLOW prototype command** | : | `aflow --proto=A2B_mC12_15_f_e --params=`$a, b/a, c/a, \beta, y_1, x_2, y_2, z_2$ |

### Other compounds with this structure

- $YbS_2$ and $BaS_2$

---

- This is a metastable phase of $CaC_2$. The stable room-temperature phase is the $C11_a$ (A2B_tI6_139_e_a) structure.

### Base-centered Monoclinic primitive vectors:

$$\mathbf{a}_1 = \frac{1}{2}\,a\,\hat{\mathbf{x}} - \frac{1}{2}\,b\,\hat{\mathbf{y}}$$
$$\mathbf{a}_2 = \frac{1}{2}\,a\,\hat{\mathbf{x}} + \frac{1}{2}\,b\,\hat{\mathbf{y}}$$
$$\mathbf{a}_3 = c\cos\beta\,\hat{\mathbf{x}} + c\sin\beta\,\hat{\mathbf{z}}$$

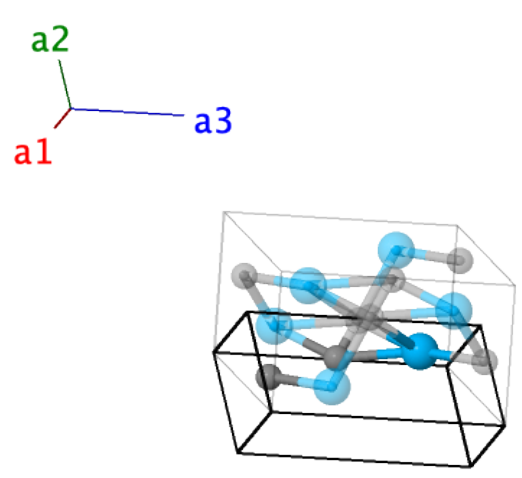

### Basis vectors:

| | | Lattice Coordinates | | Cartesian Coordinates | Wyckoff Position | Atom Type |
|---|---|---|---|---|---|---|
| $\mathbf{B}_1$ | = | $-y_1\,\mathbf{a}_1 + y_1\,\mathbf{a}_2 + \frac{1}{4}\,\mathbf{a}_3$ | = | $\frac{1}{4}c\cos\beta\,\hat{\mathbf{x}} + y_1 b\,\hat{\mathbf{y}} + \frac{1}{4}c\sin\beta\,\hat{\mathbf{z}}$ | (4$e$) | Th |
| $\mathbf{B}_2$ | = | $y_1\,\mathbf{a}_1 - y_1\,\mathbf{a}_2 + \frac{3}{4}\,\mathbf{a}_3$ | = | $\frac{3}{4}c\cos\beta\,\hat{\mathbf{x}} - y_1 b\,\hat{\mathbf{y}} + \frac{3}{4}c\sin\beta\,\hat{\mathbf{z}}$ | (4*e*) | Th |
| $\mathbf{B}_3$ | = | $(x_2 - y_2)\,\mathbf{a}_1 + (x_2 + y_2)\,\mathbf{a}_2 + z_2\,\mathbf{a}_3$ | = | $(x_2 a + z_2 c\cos\beta)\,\hat{\mathbf{x}} + y_2 b\,\hat{\mathbf{y}} + z_2 c\sin\beta\,\hat{\mathbf{z}}$ | (8*f*) | C |
| $\mathbf{B}_4$ | = | $(-x_2 - y_2)\,\mathbf{a}_1 + (-x_2 + y_2)\,\mathbf{a}_2 + \left(\frac{1}{2} - z_2\right)\,\mathbf{a}_3$ | = | $\left(\frac{1}{2}c\cos\beta - x_2 a - z_2 c\cos\beta\right)\,\hat{\mathbf{x}} + y_2 b\,\hat{\mathbf{y}} + \left(\frac{1}{2} - z_2\right)c\sin\beta\,\hat{\mathbf{z}}$ | (8*f*) | C |
| $\mathbf{B}_5$ | = | $(-x_2 + y_2)\,\mathbf{a}_1 + (-x_2 - y_2)\,\mathbf{a}_2 - z_2\,\mathbf{a}_3$ | = | $(-x_2 a - z_2 c\cos\beta)\,\hat{\mathbf{x}} - y_2 b\,\hat{\mathbf{y}} - z_2 c\sin\beta\,\hat{\mathbf{z}}$ | (8*f*) | C |
| $\mathbf{B}_6$ | = | $(x_2 + y_2)\,\mathbf{a}_1 + (x_2 - y_2)\,\mathbf{a}_2 + \left(\frac{1}{2} + z_2\right)\,\mathbf{a}_3$ | = | $\left(\frac{1}{2}c\cos\beta + x_2 a + z_2 c\cos\beta\right)\,\hat{\mathbf{x}} - y_2 b\,\hat{\mathbf{y}} + \left(\frac{1}{2} + z_2\right)c\sin\beta\,\hat{\mathbf{z}}$ | (8*f*) | C |

---

---

**Geometry files:**

- CIF: pp. 1565
- POSCAR: pp. 1565

# Clinocervantite ($\beta$-$Sb_2O_4$) Structure: A2B_mC24_15_2f_ce

http://aflow.org/prototype-encyclopedia/A2B_mC24_15_2f_ce

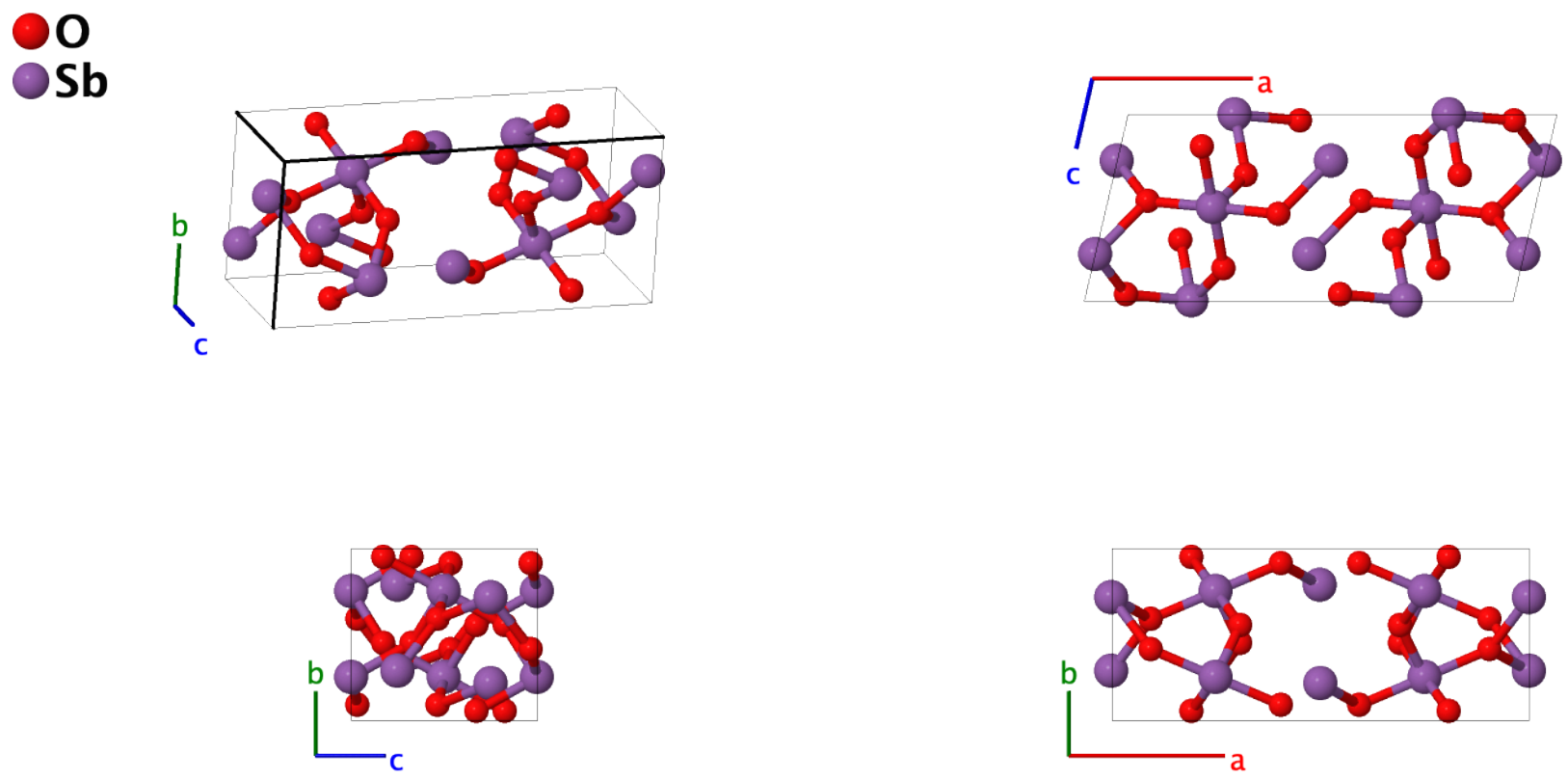

| | | |
|---|---|---|
| **Prototype** | : | $O_2Sb$ |
| **AFLOW prototype label** | : | A2B_mC24_15_2f_ce |
| ***Strukturbericht* designation** | : | None |
| **Pearson symbol** | : | mC24 |
| **Space group number** | : | 15 |
| **Space group symbol** | : | $C2/c$ |
| **AFLOW prototype command** | : | `aflow --proto=A2B_mC24_15_2f_ce` `--params=`$a, b/a, c/a, \beta, y_2, x_3, y_3, z_3, x_4, y_4, z_4$ |

- This is *not* the $D6_2$ structure of $SbO_2$, which was found to be erroneous. Clinocervantite is a naturally occurring monoclinic modification of cervantite ($\alpha$-$Sb_2O_4$)

**Base-centered Monoclinic primitive vectors:**

$$\mathbf{a}_1 = \frac{1}{2}\,a\,\hat{\mathbf{x}} - \frac{1}{2}\,b\,\hat{\mathbf{y}}$$
$$\mathbf{a}_2 = \frac{1}{2}\,a\,\hat{\mathbf{x}} + \frac{1}{2}\,b\,\hat{\mathbf{y}}$$
$$\mathbf{a}_3 = c\cos\beta\,\hat{\mathbf{x}} + c\sin\beta\,\hat{\mathbf{z}}$$

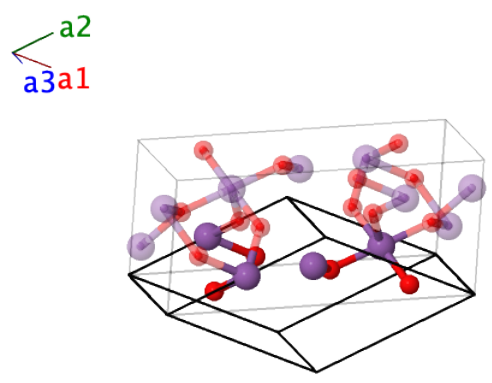

**Basis vectors:**

| | | Lattice Coordinates | | Cartesian Coordinates | Wyckoff Position | Atom Type |
|---|---|---|---|---|---|---|
| $\mathbf{B}_1$ | = | $\frac{1}{2}\,\mathbf{a}_2$ | = | $\frac{1}{4}a\,\hat{\mathbf{x}} + \frac{1}{4}b\,\hat{\mathbf{y}}$ | (4$c$) | Sb I |
| $\mathbf{B}_2$ | = | $\frac{1}{2}\,\mathbf{a}_1 + \frac{1}{2}\,\mathbf{a}_3$ | = | $\left(\frac{1}{4}a + \frac{1}{2}c\cos\beta\right)\hat{\mathbf{x}} - \frac{1}{4}b\,\hat{\mathbf{y}} + \frac{1}{2}c\sin\beta\,\hat{\mathbf{z}}$ | (4$c$) | Sb I |
| $\mathbf{B}_3$ | = | $-y_2\,\mathbf{a}_1 + y_2\,\mathbf{a}_2 + \frac{1}{4}\,\mathbf{a}_3$ | = | $\frac{1}{4}c\cos\beta\,\hat{\mathbf{x}} + y_2 b\,\hat{\mathbf{y}} + \frac{1}{4}c\sin\beta\,\hat{\mathbf{z}}$ | (4$e$) | Sb II |
| $\mathbf{B}_4$ | = | $y_2\,\mathbf{a}_1 - y_2\,\mathbf{a}_2 + \frac{3}{4}\,\mathbf{a}_3$ | = | $\frac{3}{4}c\cos\beta\,\hat{\mathbf{x}} - y_2 b\,\hat{\mathbf{y}} + \frac{3}{4}c\sin\beta\,\hat{\mathbf{z}}$ | (4$e$) | Sb II |
| $\mathbf{B}_5$ | = | $(x_3 - y_3)\,\mathbf{a}_1 + (x_3 + y_3)\,\mathbf{a}_2 + z_3\,\mathbf{a}_3$ | = | $(x_3 a + z_3 c\cos\beta)\,\hat{\mathbf{x}} + y_3 b\,\hat{\mathbf{y}} + z_3 c\sin\beta\,\hat{\mathbf{z}}$ | (8$f$) | O I |
| $\mathbf{B}_6$ | = | $(-x_3 - y_3)\,\mathbf{a}_1 + (-x_3 + y_3)\,\mathbf{a}_2 + \left(\frac{1}{2} - z_3\right)\mathbf{a}_3$ | = | $\left(\frac{1}{2}c\cos\beta - x_3 a - z_3 c\cos\beta\right)\hat{\mathbf{x}} + y_3 b\,\hat{\mathbf{y}} + \left(\frac{1}{2} - z_3\right)c\sin\beta\,\hat{\mathbf{z}}$ | $(8f)$ | O I |
| $\mathbf{B}_7$ | = | $(-x_3 + y_3)\,\mathbf{a}_1 + (-x_3 - y_3)\,\mathbf{a}_2 - z_3\,\mathbf{a}_3$ | = | $(-x_3 a - z_3 c\cos\beta)\,\hat{\mathbf{x}} - y_3 b\,\hat{\mathbf{y}} - z_3 c\sin\beta\,\hat{\mathbf{z}}$ | $(8f)$ | O I |
| $\mathbf{B}_8$ | = | $(x_3 + y_3)\,\mathbf{a}_1 + (x_3 - y_3)\,\mathbf{a}_2 + \left(\frac{1}{2} + z_3\right)\mathbf{a}_3$ | = | $\left(\frac{1}{2}c\cos\beta + x_3 a + z_3 c\cos\beta\right)\hat{\mathbf{x}} - y_3 b\,\hat{\mathbf{y}} + \left(\frac{1}{2} + z_3\right)c\sin\beta\,\hat{\mathbf{z}}$ | $(8f)$ | O I |
| $\mathbf{B}_9$ | = | $(x_4 - y_4)\,\mathbf{a}_1 + (x_4 + y_4)\,\mathbf{a}_2 + z_4\,\mathbf{a}_3$ | = | $(x_4 a + z_4 c\cos\beta)\,\hat{\mathbf{x}} + y_4 b\,\hat{\mathbf{y}} + z_4 c\sin\beta\,\hat{\mathbf{z}}$ | $(8f)$ | O II |
| $\mathbf{B}_{10}$ | = | $(-x_4 - y_4)\,\mathbf{a}_1 + (-x_4 + y_4)\,\mathbf{a}_2 + \left(\frac{1}{2} - z_4\right)\mathbf{a}_3$ | = | $\left(\frac{1}{2}c\cos\beta - x_4 a - z_4 c\cos\beta\right)\hat{\mathbf{x}} + y_4 b\,\hat{\mathbf{y}} + \left(\frac{1}{2} - z_4\right)c\sin\beta\,\hat{\mathbf{z}}$ | $(8f)$ | O II |
| $\mathbf{B}_{11}$ | = | $(-x_4 + y_4)\,\mathbf{a}_1 + (-x_4 - y_4)\,\mathbf{a}_2 - z_4\,\mathbf{a}_3$ | = | $(-x_4 a - z_4 c\cos\beta)\,\hat{\mathbf{x}} - y_4 b\,\hat{\mathbf{y}} - z_4 c\sin\beta\,\hat{\mathbf{z}}$ | $(8f)$ | O II |
| $\mathbf{B}_{12}$ | = | $(x_4 + y_4)\,\mathbf{a}_1 + (x_4 - y_4)\,\mathbf{a}_2 + \left(\frac{1}{2} + z_4\right)\mathbf{a}_3$ | = | $\left(\frac{1}{2}c\cos\beta + x_4 a + z_4 c\cos\beta\right)\hat{\mathbf{x}} - y_4 b\,\hat{\mathbf{y}} + \left(\frac{1}{2} + z_4\right)c\sin\beta\,\hat{\mathbf{z}}$ | $(8f)$ | O II |

**Found in:**

**Geometry files:**

- CIF: pp. 1565
- POSCAR: pp. 1566

# $(CdSO_4)_3 \cdot 8H_2O$ ($H4_{20}$) Structure: A3B16C20D3_mC168_15_ef_8f_10f_ef

http://aflow.org/prototype-encyclopedia/A3B16C20D3_mC168_15_ef_8f_10f_ef

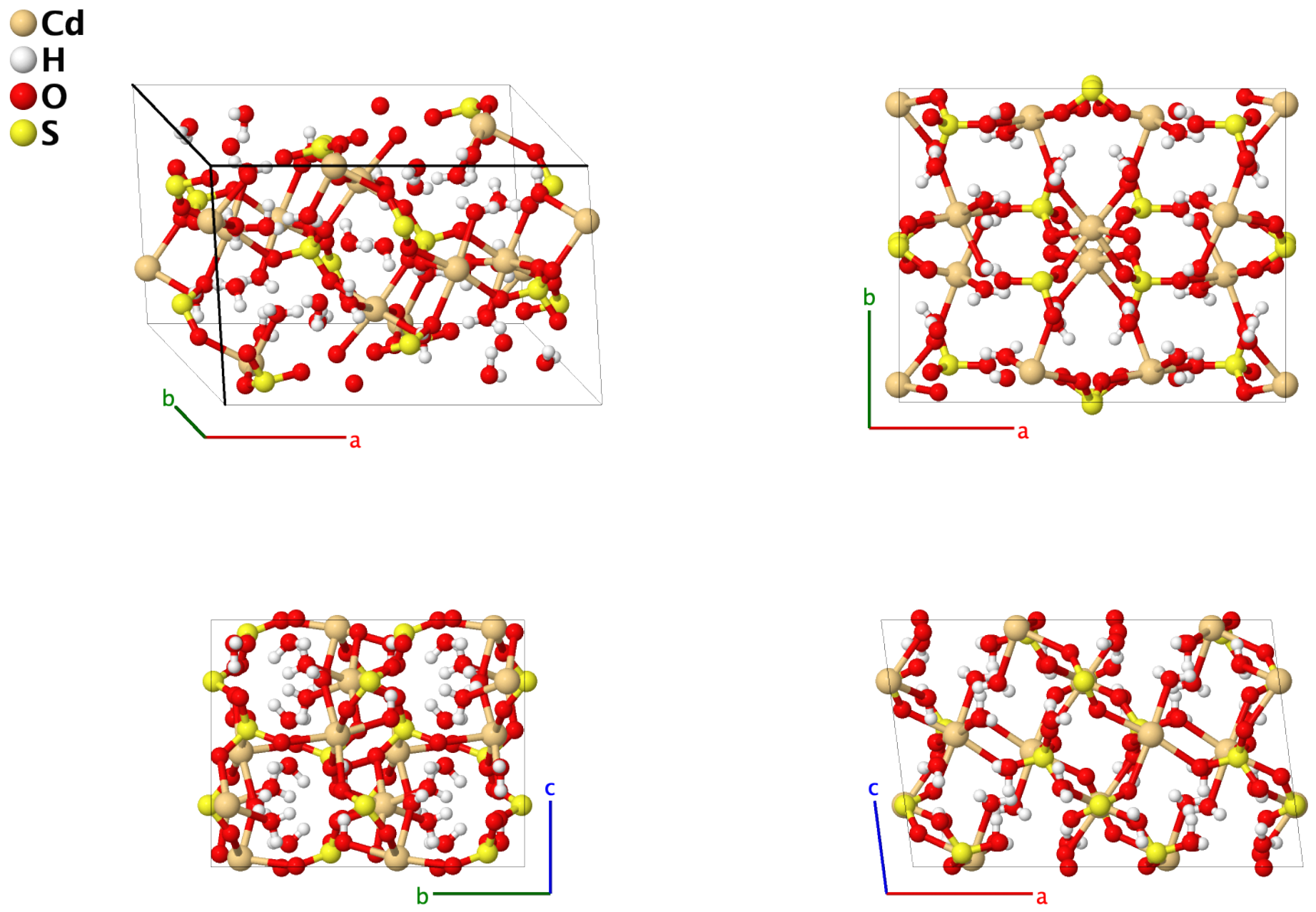

| | | |
|---|---|---|
| **Prototype** | : | $Cd_3H_{16}O_{20}S_3$ |
| **AFLOW prototype label** | : | A3B16C20D3_mC168_15_ef_8f_10f_ef |
| ***Strukturbericht* designation** | : | $H4_{20}$ |
| **Pearson symbol** | : | mC168 |
| **Space group number** | : | 15 |
| **Space group symbol** | : | $C2/c$ |
| **AFLOW prototype command** | : | `aflow --proto=A3B16C20D3_mC168_15_ef_8f_10f_ef` `--params=`$a, b/a, c/a, \beta, y_1, y_2, x_3, y_3, z_3, x_4, y_4, z_4, x_5, y_5, z_5, x_6, y_6, z_6, x_7, y_7, z_7, x_8, y_8, z_8, x_9, y_9, z_9, x_{10}, y_{10}, z_{10}, x_{11}, y_{11}, z_{11}, x_{12}, y_{12}, z_{12}, x_{13}, y_{13}, z_{13}, x_{14}, y_{14}, z_{14}, x_{15}, y_{15}, z_{15}, x_{16}, y_{16}, z_{16}, x_{17}, y_{17}, z_{17}, x_{18}, y_{18}, z_{18}, x_{19}, y_{19}, z_{19}, x_{20}, y_{20}, z_{20}, x_{21}, y_{21}, z_{21}, x_{22}, y_{22}, z_{22}$ |

- Some of the H-O distances appear to be very small. For example, the O-VIII – H-III distance is only 0.43Å. We have checked our inputs compared to (Caminiti, 1981) and found no error. Unfortunately they do not provide H-O distances for comparison.

**Base-centered Monoclinic primitive vectors:**

$$\mathbf{a}_1 = \frac{1}{2}\,a\,\hat{\mathbf{x}} - \frac{1}{2}\,b\,\hat{\mathbf{y}}$$
$$\mathbf{a}_2 = \frac{1}{2}\,a\,\hat{\mathbf{x}} + \frac{1}{2}\,b\,\hat{\mathbf{y}}$$
$$\mathbf{a}_3 = c\cos\beta\,\hat{\mathbf{x}} + c\sin\beta\,\hat{\mathbf{z}}$$

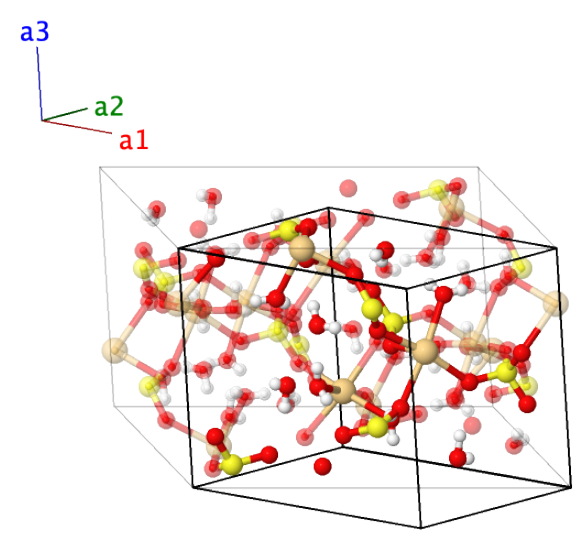

**Basis vectors:**

| | | Lattice Coordinates | | Cartesian Coordinates | Wyckoff Position | Atom Type |
|---|---|---|---|---|---|---|
| $\mathbf{B}_1$ | = | $-y_1\,\mathbf{a}_1 + y_1\,\mathbf{a}_2 + \frac{1}{4}\,\mathbf{a}_3$ | = | $\frac{1}{4}c\cos\beta\,\hat{\mathbf{x}} + y_1 b\,\hat{\mathbf{y}} + \frac{1}{4}c\sin\beta\,\hat{\mathbf{z}}$ | (4$e$) | Cd I |
| $\mathbf{B}_2$ | = | $y_1\,\mathbf{a}_1 - y_1\,\mathbf{a}_2 + \frac{3}{4}\,\mathbf{a}_3$ | = | $\frac{3}{4}c\cos\beta\,\hat{\mathbf{x}} - y_1 b\,\hat{\mathbf{y}} + \frac{3}{4}c\sin\beta\,\hat{\mathbf{z}}$ | (4$e$) | Cd I |
| $\mathbf{B}_3$ | = | $-y_2\,\mathbf{a}_1 + y_2\,\mathbf{a}_2 + \frac{1}{4}\,\mathbf{a}_3$ | = | $\frac{1}{4}c\cos\beta\,\hat{\mathbf{x}} + y_2 b\,\hat{\mathbf{y}} + \frac{1}{4}c\sin\beta\,\hat{\mathbf{z}}$ | (4$e$) | S I |
| $\mathbf{B}_4$ | = | $y_2\,\mathbf{a}_1 - y_2\,\mathbf{a}_2 + \frac{3}{4}\,\mathbf{a}_3$ | = | $\frac{3}{4}c\cos\beta\,\hat{\mathbf{x}} - y_2 b\,\hat{\mathbf{y}} + \frac{3}{4}c\sin\beta\,\hat{\mathbf{z}}$ | (4$e$) | S I |
| $\mathbf{B}_5$ | = | $(x_3 - y_3)\,\mathbf{a}_1 + (x_3 + y_3)\,\mathbf{a}_2 + z_3\,\mathbf{a}_3$ | = | $(x_3 a + z_3 c\cos\beta)\,\hat{\mathbf{x}} + y_3 b\,\hat{\mathbf{y}} + z_3 c\sin\beta\,\hat{\mathbf{z}}$ | (8$f$) | Cd II |
| $\mathbf{B}_6$ | = | $(-x_3 - y_3)\,\mathbf{a}_1 + (-x_3 + y_3)\,\mathbf{a}_2 + \left(\frac{1}{2} - z_3\right)\mathbf{a}_3$ | = | $\left(\frac{1}{2}c\cos\beta - x_3 a - z_3 c\cos\beta\right)\hat{\mathbf{x}} + y_3 b\,\hat{\mathbf{y}} + \left(\frac{1}{2} - z_3\right)c\sin\beta\,\hat{\mathbf{z}}$ | (8$f$) | Cd II |
| $\mathbf{B}_7$ | = | $(-x_3 + y_3)\,\mathbf{a}_1 + (-x_3 - y_3)\,\mathbf{a}_2 - z_3\,\mathbf{a}_3$ | = | $(-x_3 a - z_3 c\cos\beta)\,\hat{\mathbf{x}} - y_3 b\,\hat{\mathbf{y}} - z_3 c\sin\beta\,\hat{\mathbf{z}}$ | (8$f$) | Cd II |
| $\mathbf{B}_8$ | = | $(x_3 + y_3)\,\mathbf{a}_1 + (x_3 - y_3)\,\mathbf{a}_2 + \left(\frac{1}{2} + z_3\right)\mathbf{a}_3$ | = | $\left(\frac{1}{2}c\cos\beta + x_3 a + z_3 c\cos\beta\right)\hat{\mathbf{x}} - y_3 b\,\hat{\mathbf{y}} + \left(\frac{1}{2} + z_3\right)c\sin\beta\,\hat{\mathbf{z}}$ | (8$f$) | Cd II |
| $\mathbf{B}_9$ | = | $(x_4 - y_4)\,\mathbf{a}_1 + (x_4 + y_4)\,\mathbf{a}_2 + z_4\,\mathbf{a}_3$ | = | $(x_4 a + z_4 c\cos\beta)\,\hat{\mathbf{x}} + y_4 b\,\hat{\mathbf{y}} + z_4 c\sin\beta\,\hat{\mathbf{z}}$ | (8$f$) | H I |
| $\mathbf{B}_{10}$ | = | $(-x_4 - y_4)\,\mathbf{a}_1 + (-x_4 + y_4)\,\mathbf{a}_2 + \left(\frac{1}{2} - z_4\right)\mathbf{a}_3$ | = | $\left(\frac{1}{2}c\cos\beta - x_4 a - z_4 c\cos\beta\right)\hat{\mathbf{x}} + y_4 b\,\hat{\mathbf{y}} + \left(\frac{1}{2} - z_4\right)c\sin\beta\,\hat{\mathbf{z}}$ | (8$f$) | H I |
| $\mathbf{B}_{11}$ | = | $(-x_4 + y_4)\,\mathbf{a}_1 + (-x_4 - y_4)\,\mathbf{a}_2 - z_4\,\mathbf{a}_3$ | = | $(-x_4 a - z_4 c\cos\beta)\,\hat{\mathbf{x}} - y_4 b\,\hat{\mathbf{y}} - z_4 c\sin\beta\,\hat{\mathbf{z}}$ | (8$f$) | H I |
| $\mathbf{B}_{12}$ | = | $(x_4 + y_4)\,\mathbf{a}_1 + (x_4 - y_4)\,\mathbf{a}_2 + \left(\frac{1}{2} + z_4\right)\mathbf{a}_3$ | = | $\left(\frac{1}{2}c\cos\beta + x_4 a + z_4 c\cos\beta\right)\hat{\mathbf{x}} - y_4 b\,\hat{\mathbf{y}} + \left(\frac{1}{2} + z_4\right)c\sin\beta\,\hat{\mathbf{z}}$ | (8$f$) | H I |
| $\mathbf{B}_{13}$ | = | $(x_5 - y_5)\,\mathbf{a}_1 + (x_5 + y_5)\,\mathbf{a}_2 + z_5\,\mathbf{a}_3$ | = | $(x_5 a + z_5 c\cos\beta)\,\hat{\mathbf{x}} + y_5 b\,\hat{\mathbf{y}} + z_5 c\sin\beta\,\hat{\mathbf{z}}$ | (8$f$) | H II |
| $\mathbf{B}_{14}$ | = | $(-x_5 - y_5)\,\mathbf{a}_1 + (-x_5 + y_5)\,\mathbf{a}_2 + \left(\frac{1}{2} - z_5\right)\mathbf{a}_3$ | = | $\left(\frac{1}{2}c\cos\beta - x_5 a - z_5 c\cos\beta\right)\hat{\mathbf{x}} + y_5 b\,\hat{\mathbf{y}} + \left(\frac{1}{2} - z_5\right)c\sin\beta\,\hat{\mathbf{z}}$ | (8$f$) | H II |
| $\mathbf{B}_{15}$ | = | $(-x_5 + y_5)\,\mathbf{a}_1 + (-x_5 - y_5)\,\mathbf{a}_2 - z_5\,\mathbf{a}_3$ | = | $(-x_5 a - z_5 c\cos\beta)\,\hat{\mathbf{x}} - y_5 b\,\hat{\mathbf{y}} - z_5 c\sin\beta\,\hat{\mathbf{z}}$ | (8$f$) | H II |
| $\mathbf{B}_{16}$ | = | $(x_5 + y_5)\,\mathbf{a}_1 + (x_5 - y_5)\,\mathbf{a}_2 + \left(\frac{1}{2} + z_5\right)\mathbf{a}_3$ | = | $\left(\frac{1}{2}c\cos\beta + x_5 a + z_5 c\cos\beta\right)\hat{\mathbf{x}} - y_5 b\,\hat{\mathbf{y}} + \left(\frac{1}{2} + z_5\right)c\sin\beta\,\hat{\mathbf{z}}$ | (8$f$) | H II |
| $\mathbf{B}_{17}$ | = | $(x_6 - y_6)\,\mathbf{a}_1 + (x_6 + y_6)\,\mathbf{a}_2 + z_6\,\mathbf{a}_3$ | = | $(x_6 a + z_6 c\cos\beta)\,\hat{\mathbf{x}} + y_6 b\,\hat{\mathbf{y}} + z_6 c\sin\beta\,\hat{\mathbf{z}}$ | (8$f$) | H III |
| $\mathbf{B}_{18}$ | = | $(-x_6 - y_6)\,\mathbf{a}_1 + (-x_6 + y_6)\,\mathbf{a}_2 + \left(\frac{1}{2} - z_6\right)\mathbf{a}_3$ | = | $\left(\frac{1}{2}c\cos\beta - x_6 a - z_6 c\cos\beta\right)\hat{\mathbf{x}} + y_6 b\,\hat{\mathbf{y}} + \left(\frac{1}{2} - z_6\right)c\sin\beta\,\hat{\mathbf{z}}$ | (8$f$) | H III |
| $\mathbf{B}_{19}$ | = | $(-x_6 + y_6)\,\mathbf{a}_1 + (-x_6 - y_6)\,\mathbf{a}_2 - z_6\,\mathbf{a}_3$ | = | $(-x_6 a - z_6 c\cos\beta)\,\hat{\mathbf{x}} - y_6 b\,\hat{\mathbf{y}} - z_6 c\sin\beta\,\hat{\mathbf{z}}$ | (8$f$) | H III |
| $\mathbf{B}_{20}$ | = | $(x_6 + y_6)\,\mathbf{a}_1 + (x_6 - y_6)\,\mathbf{a}_2 + \left(\frac{1}{2} + z_6\right)\mathbf{a}_3$ | = | $\left(\frac{1}{2}c\cos\beta + x_6 a + z_6 c\cos\beta\right)\hat{\mathbf{x}} - y_6 b\,\hat{\mathbf{y}} + \left(\frac{1}{2} + z_6\right)c\sin\beta\,\hat{\mathbf{z}}$ | (8$f$) | H III |
| $\mathbf{B}_{21}$ | = | $(x_7 - y_7)\,\mathbf{a}_1 + (x_7 + y_7)\,\mathbf{a}_2 + z_7\,\mathbf{a}_3$ | = | $(x_7 a + z_7 c\cos\beta)\,\hat{\mathbf{x}} + y_7 b\,\hat{\mathbf{y}} + z_7 c\sin\beta\,\hat{\mathbf{z}}$ | (8$f$) | H IV |
| $\mathbf{B}_{22}$ | = | $(-x_7 - y_7)\,\mathbf{a}_1 + (-x_7 + y_7)\,\mathbf{a}_2 + \left(\frac{1}{2} - z_7\right)\mathbf{a}_3$ | = | $\left(\frac{1}{2}c\cos\beta - x_7 a - z_7 c\cos\beta\right)\hat{\mathbf{x}} + y_7 b\,\hat{\mathbf{y}} + \left(\frac{1}{2} - z_7\right)c\sin\beta\,\hat{\mathbf{z}}$ | (8$f$) | H IV |
| $\mathbf{B}_{23}$ | = | $(-x_7+y_7)\,\mathbf{a}_1+(-x_7-y_7)\,\mathbf{a}_2-z_7\,\mathbf{a}_3$ | = | $(-x_7a-z_7c\cos\beta)\,\hat{\mathbf{x}}-y_7b\,\hat{\mathbf{y}}-z_7c\sin\beta\,\hat{\mathbf{z}}$ | $(8f)$ | H IV |
| $\mathbf{B}_{24}$ | = | $(x_7+y_7)\,\mathbf{a}_1+(x_7-y_7)\,\mathbf{a}_2+\left(\frac{1}{2}+z_7\right)\mathbf{a}_3$ | = | $\left(\frac{1}{2}c\cos\beta+x_7a+z_7c\cos\beta\right)\hat{\mathbf{x}}-y_7b\,\hat{\mathbf{y}}+\left(\frac{1}{2}+z_7\right)c\sin\beta\,\hat{\mathbf{z}}$ | $(8f)$ | H IV |
| $\mathbf{B}_{25}$ | = | $(x_8-y_8)\,\mathbf{a}_1+(x_8+y_8)\,\mathbf{a}_2+z_8\,\mathbf{a}_3$ | = | $(x_8a+z_8c\cos\beta)\,\hat{\mathbf{x}}+y_8b\,\hat{\mathbf{y}}+z_8c\sin\beta\,\hat{\mathbf{z}}$ | $(8f)$ | H V |
| $\mathbf{B}_{26}$ | = | $(-x_8-y_8)\,\mathbf{a}_1+(-x_8+y_8)\,\mathbf{a}_2+\left(\frac{1}{2}-z_8\right)\mathbf{a}_3$ | = | $\left(\frac{1}{2}c\cos\beta-x_8a-z_8c\cos\beta\right)\hat{\mathbf{x}}+y_8b\,\hat{\mathbf{y}}+\left(\frac{1}{2}-z_8\right)c\sin\beta\,\hat{\mathbf{z}}$ | $(8f)$ | H V |
| $\mathbf{B}_{27}$ | = | $(-x_8+y_8)\,\mathbf{a}_1+(-x_8-y_8)\,\mathbf{a}_2-z_8\,\mathbf{a}_3$ | = | $(-x_8a-z_8c\cos\beta)\,\hat{\mathbf{x}}-y_8b\,\hat{\mathbf{y}}-z_8c\sin\beta\,\hat{\mathbf{z}}$ | $(8f)$ | H V |
| $\mathbf{B}_{28}$ | = | $(x_8+y_8)\,\mathbf{a}_1+(x_8-y_8)\,\mathbf{a}_2+\left(\frac{1}{2}+z_8\right)\mathbf{a}_3$ | = | $\left(\frac{1}{2}c\cos\beta+x_8a+z_8c\cos\beta\right)\hat{\mathbf{x}}-y_8b\,\hat{\mathbf{y}}+\left(\frac{1}{2}+z_8\right)c\sin\beta\,\hat{\mathbf{z}}$ | $(8f)$ | H V |
| $\mathbf{B}_{29}$ | = | $(x_9-y_9)\,\mathbf{a}_1+(x_9+y_9)\,\mathbf{a}_2+z_9\,\mathbf{a}_3$ | = | $(x_9a+z_9c\cos\beta)\,\hat{\mathbf{x}}+y_9b\,\hat{\mathbf{y}}+z_9c\sin\beta\,\hat{\mathbf{z}}$ | $(8f)$ | H VI |
| $\mathbf{B}_{30}$ | = | $(-x_9-y_9)\,\mathbf{a}_1+(-x_9+y_9)\,\mathbf{a}_2+\left(\frac{1}{2}-z_9\right)\mathbf{a}_3$ | = | $\left(\frac{1}{2}c\cos\beta-x_9a-z_9c\cos\beta\right)\hat{\mathbf{x}}+y_9b\,\hat{\mathbf{y}}+\left(\frac{1}{2}-z_9\right)c\sin\beta\,\hat{\mathbf{z}}$ | $(8f)$ | H VI |
| $\mathbf{B}_{31}$ | = | $(-x_9+y_9)\,\mathbf{a}_1+(-x_9-y_9)\,\mathbf{a}_2-z_9\,\mathbf{a}_3$ | = | $(-x_9a-z_9c\cos\beta)\,\hat{\mathbf{x}}-y_9b\,\hat{\mathbf{y}}-z_9c\sin\beta\,\hat{\mathbf{z}}$ | $(8f)$ | H VI |
| $\mathbf{B}_{32}$ | = | $(x_9+y_9)\,\mathbf{a}_1+(x_9-y_9)\,\mathbf{a}_2+\left(\frac{1}{2}+z_9\right)\mathbf{a}_3$ | = | $\left(\frac{1}{2}c\cos\beta+x_9a+z_9c\cos\beta\right)\hat{\mathbf{x}}-y_9b\,\hat{\mathbf{y}}+\left(\frac{1}{2}+z_9\right)c\sin\beta\,\hat{\mathbf{z}}$ | $(8f)$ | H VI |
| $\mathbf{B}_{33}$ | = | $(x_{10}-y_{10})\,\mathbf{a}_1+(x_{10}+y_{10})\,\mathbf{a}_2+z_{10}\,\mathbf{a}_3$ | = | $(x_{10}a+z_{10}c\cos\beta)\,\hat{\mathbf{x}}+y_{10}b\,\hat{\mathbf{y}}+z_{10}c\sin\beta\,\hat{\mathbf{z}}$ | $(8f)$ | H VII |
| $\mathbf{B}_{34}$ | = | $(-x_{10}-y_{10})\,\mathbf{a}_1+(-x_{10}+y_{10})\,\mathbf{a}_2+\left(\frac{1}{2}-z_{10}\right)\mathbf{a}_3$ | = | $\left(\frac{1}{2}c\cos\beta-x_{10}a-z_{10}c\cos\beta\right)\hat{\mathbf{x}}+y_{10}b\,\hat{\mathbf{y}}+\left(\frac{1}{2}-z_{10}\right)c\sin\beta\,\hat{\mathbf{z}}$ | $(8f)$ | H VII |
| $\mathbf{B}_{35}$ | = | $(-x_{10}+y_{10})\,\mathbf{a}_1+(-x_{10}-y_{10})\,\mathbf{a}_2-z_{10}\,\mathbf{a}_3$ | = | $(-x_{10}a-z_{10}c\cos\beta)\,\hat{\mathbf{x}}-y_{10}b\,\hat{\mathbf{y}}-z_{10}c\sin\beta\,\hat{\mathbf{z}}$ | $(8f)$ | H VII |
| $\mathbf{B}_{36}$ | = | $(x_{10}+y_{10})\,\mathbf{a}_1+(x_{10}-y_{10})\,\mathbf{a}_2+\left(\frac{1}{2}+z_{10}\right)\mathbf{a}_3$ | = | $\left(\frac{1}{2}c\cos\beta+x_{10}a+z_{10}c\cos\beta\right)\hat{\mathbf{x}}-y_{10}b\,\hat{\mathbf{y}}+\left(\frac{1}{2}+z_{10}\right)c\sin\beta\,\hat{\mathbf{z}}$ | $(8f)$ | H VII |
| $\mathbf{B}_{37}$ | = | $(x_{11}-y_{11})\,\mathbf{a}_1+(x_{11}+y_{11})\,\mathbf{a}_2+z_{11}\,\mathbf{a}_3$ | = | $(x_{11}a+z_{11}c\cos\beta)\,\hat{\mathbf{x}}+y_{11}b\,\hat{\mathbf{y}}+z_{11}c\sin\beta\,\hat{\mathbf{z}}$ | $(8f)$ | H VIII |
| $\mathbf{B}_{38}$ | = | $(-x_{11}-y_{11})\,\mathbf{a}_1+(-x_{11}+y_{11})\,\mathbf{a}_2+\left(\frac{1}{2}-z_{11}\right)\mathbf{a}_3$ | = | $\left(\frac{1}{2}c\cos\beta-x_{11}a-z_{11}c\cos\beta\right)\hat{\mathbf{x}}+y_{11}b\,\hat{\mathbf{y}}+\left(\frac{1}{2}-z_{11}\right)c\sin\beta\,\hat{\mathbf{z}}$ | $(8f)$ | H VIII |
| $\mathbf{B}_{39}$ | = | $(-x_{11}+y_{11})\,\mathbf{a}_1+(-x_{11}-y_{11})\,\mathbf{a}_2-z_{11}\,\mathbf{a}_3$ | = | $(-x_{11}a-z_{11}c\cos\beta)\,\hat{\mathbf{x}}-y_{11}b\,\hat{\mathbf{y}}-z_{11}c\sin\beta\,\hat{\mathbf{z}}$ | $(8f)$ | H VIII |
| $\mathbf{B}_{40}$ | = | $(x_{11}+y_{11})\,\mathbf{a}_1+(x_{11}-y_{11})\,\mathbf{a}_2+\left(\frac{1}{2}+z_{11}\right)\mathbf{a}_3$ | = | $\left(\frac{1}{2}c\cos\beta+x_{11}a+z_{11}c\cos\beta\right)\hat{\mathbf{x}}-y_{11}b\,\hat{\mathbf{y}}+\left(\frac{1}{2}+z_{11}\right)c\sin\beta\,\hat{\mathbf{z}}$ | $(8f)$ | H VIII |
| $\mathbf{B}_{41}$ | = | $(x_{12}-y_{12})\,\mathbf{a}_1+(x_{12}+y_{12})\,\mathbf{a}_2+z_{12}\,\mathbf{a}_3$ | = | $(x_{12}a+z_{12}c\cos\beta)\,\hat{\mathbf{x}}+y_{12}b\,\hat{\mathbf{y}}+z_{12}c\sin\beta\,\hat{\mathbf{z}}$ | $(8f)$ | O I |
| $\mathbf{B}_{42}$ | = | $(-x_{12}-y_{12})\,\mathbf{a}_1+(-x_{12}+y_{12})\,\mathbf{a}_2+\left(\frac{1}{2}-z_{12}\right)\mathbf{a}_3$ | = | $\left(\frac{1}{2}c\cos\beta-x_{12}a-z_{12}c\cos\beta\right)\hat{\mathbf{x}}+y_{12}b\,\hat{\mathbf{y}}+\left(\frac{1}{2}-z_{12}\right)c\sin\beta\,\hat{\mathbf{z}}$ | $(8f)$ | O I |
| $\mathbf{B}_{43}$ | = | $(-x_{12}+y_{12})\,\mathbf{a}_1+(-x_{12}-y_{12})\,\mathbf{a}_2-z_{12}\,\mathbf{a}_3$ | = | $(-x_{12}a-z_{12}c\cos\beta)\,\hat{\mathbf{x}}-y_{12}b\,\hat{\mathbf{y}}-z_{12}c\sin\beta\,\hat{\mathbf{z}}$ | $(8f)$ | O I |
| $\mathbf{B}_{44}$ | = | $(x_{12}+y_{12})\,\mathbf{a}_1+(x_{12}-y_{12})\,\mathbf{a}_2+\left(\frac{1}{2}+z_{12}\right)\mathbf{a}_3$ | = | $\left(\frac{1}{2}c\cos\beta+x_{12}a+z_{12}c\cos\beta\right)\hat{\mathbf{x}}-y_{12}b\,\hat{\mathbf{y}}+\left(\frac{1}{2}+z_{12}\right)c\sin\beta\,\hat{\mathbf{z}}$ | $(8f)$ | O I |
| $\mathbf{B}_{45}$ | = | $(x_{13}-y_{13})\,\mathbf{a}_1+(x_{13}+y_{13})\,\mathbf{a}_2+z_{13}\,\mathbf{a}_3$ | = | $(x_{13}a+z_{13}c\cos\beta)\,\hat{\mathbf{x}}+y_{13}b\,\hat{\mathbf{y}}+z_{13}c\sin\beta\,\hat{\mathbf{z}}$ | $(8f)$ | O II |
| $\mathbf{B}_{46}$ | = | $(-x_{13}-y_{13})\,\mathbf{a}_1+(-x_{13}+y_{13})\,\mathbf{a}_2+\left(\frac{1}{2}-z_{13}\right)\,\mathbf{a}_3$ | = | $\left(\frac{1}{2}c\cos\beta-x_{13}a-z_{13}c\cos\beta\right)\,\hat{\mathbf{x}}+y_{13}b\,\hat{\mathbf{y}}+\left(\frac{1}{2}-z_{13}\right)c\sin\beta\,\hat{\mathbf{z}}$ | $(8f)$ | O II |
| $\mathbf{B}_{47}$ | = | $(-x_{13}+y_{13})\,\mathbf{a}_1+(-x_{13}-y_{13})\,\mathbf{a}_2-z_{13}\,\mathbf{a}_3$ | = | $(-x_{13}a-z_{13}c\cos\beta)\,\hat{\mathbf{x}}-y_{13}b\,\hat{\mathbf{y}}-z_{13}c\sin\beta\,\hat{\mathbf{z}}$ | $(8f)$ | O II |
| $\mathbf{B}_{48}$ | = | $(x_{13}+y_{13})\,\mathbf{a}_1+(x_{13}-y_{13})\,\mathbf{a}_2+\left(\frac{1}{2}+z_{13}\right)\,\mathbf{a}_3$ | = | $\left(\frac{1}{2}c\cos\beta+x_{13}a+z_{13}c\cos\beta\right)\,\hat{\mathbf{x}}-y_{13}b\,\hat{\mathbf{y}}+\left(\frac{1}{2}+z_{13}\right)c\sin\beta\,\hat{\mathbf{z}}$ | $(8f)$ | O II |
| $\mathbf{B}_{49}$ | = | $(x_{14}-y_{14})\,\mathbf{a}_1+(x_{14}+y_{14})\,\mathbf{a}_2+z_{14}\,\mathbf{a}_3$ | = | $(x_{14}a+z_{14}c\cos\beta)\,\hat{\mathbf{x}}+y_{14}b\,\hat{\mathbf{y}}+z_{14}c\sin\beta\,\hat{\mathbf{z}}$ | $(8f)$ | O III |
| $\mathbf{B}_{50}$ | = | $(-x_{14}-y_{14})\,\mathbf{a}_1+(-x_{14}+y_{14})\,\mathbf{a}_2+\left(\frac{1}{2}-z_{14}\right)\,\mathbf{a}_3$ | = | $\left(\frac{1}{2}c\cos\beta-x_{14}a-z_{14}c\cos\beta\right)\,\hat{\mathbf{x}}+y_{14}b\,\hat{\mathbf{y}}+\left(\frac{1}{2}-z_{14}\right)c\sin\beta\,\hat{\mathbf{z}}$ | $(8f)$ | O III |
| $\mathbf{B}_{51}$ | = | $(-x_{14}+y_{14})\,\mathbf{a}_1+(-x_{14}-y_{14})\,\mathbf{a}_2-z_{14}\,\mathbf{a}_3$ | = | $(-x_{14}a-z_{14}c\cos\beta)\,\hat{\mathbf{x}}-y_{14}b\,\hat{\mathbf{y}}-z_{14}c\sin\beta\,\hat{\mathbf{z}}$ | $(8f)$ | O III |
| $\mathbf{B}_{52}$ | = | $(x_{14}+y_{14})\,\mathbf{a}_1+(x_{14}-y_{14})\,\mathbf{a}_2+\left(\frac{1}{2}+z_{14}\right)\,\mathbf{a}_3$ | = | $\left(\frac{1}{2}c\cos\beta+x_{14}a+z_{14}c\cos\beta\right)\,\hat{\mathbf{x}}-y_{14}b\,\hat{\mathbf{y}}+\left(\frac{1}{2}+z_{14}\right)c\sin\beta\,\hat{\mathbf{z}}$ | $(8f)$ | O III |
| $\mathbf{B}_{53}$ | = | $(x_{15}-y_{15})\,\mathbf{a}_1+(x_{15}+y_{15})\,\mathbf{a}_2+z_{15}\,\mathbf{a}_3$ | = | $(x_{15}a+z_{15}c\cos\beta)\,\hat{\mathbf{x}}+y_{15}b\,\hat{\mathbf{y}}+z_{15}c\sin\beta\,\hat{\mathbf{z}}$ | $(8f)$ | O IV |
| $\mathbf{B}_{54}$ | = | $(-x_{15}-y_{15})\,\mathbf{a}_1+(-x_{15}+y_{15})\,\mathbf{a}_2+\left(\frac{1}{2}-z_{15}\right)\,\mathbf{a}_3$ | = | $\left(\frac{1}{2}c\cos\beta-x_{15}a-z_{15}c\cos\beta\right)\,\hat{\mathbf{x}}+y_{15}b\,\hat{\mathbf{y}}+\left(\frac{1}{2}-z_{15}\right)c\sin\beta\,\hat{\mathbf{z}}$ | $(8f)$ | O IV |
| $\mathbf{B}_{55}$ | = | $(-x_{15}+y_{15})\,\mathbf{a}_1+(-x_{15}-y_{15})\,\mathbf{a}_2-z_{15}\,\mathbf{a}_3$ | = | $(-x_{15}a-z_{15}c\cos\beta)\,\hat{\mathbf{x}}-y_{15}b\,\hat{\mathbf{y}}-z_{15}c\sin\beta\,\hat{\mathbf{z}}$ | $(8f)$ | O IV |
| $\mathbf{B}_{56}$ | = | $(x_{15}+y_{15})\,\mathbf{a}_1+(x_{15}-y_{15})\,\mathbf{a}_2+\left(\frac{1}{2}+z_{15}\right)\,\mathbf{a}_3$ | = | $\left(\frac{1}{2}c\cos\beta+x_{15}a+z_{15}c\cos\beta\right)\,\hat{\mathbf{x}}-y_{15}b\,\hat{\mathbf{y}}+\left(\frac{1}{2}+z_{15}\right)c\sin\beta\,\hat{\mathbf{z}}$ | $(8f)$ | O IV |
| $\mathbf{B}_{57}$ | = | $(x_{16}-y_{16})\,\mathbf{a}_1+(x_{16}+y_{16})\,\mathbf{a}_2+z_{16}\,\mathbf{a}_3$ | = | $(x_{16}a+z_{16}c\cos\beta)\,\hat{\mathbf{x}}+y_{16}b\,\hat{\mathbf{y}}+z_{16}c\sin\beta\,\hat{\mathbf{z}}$ | $(8f)$ | O V |
| $\mathbf{B}_{58}$ | = | $(-x_{16}-y_{16})\,\mathbf{a}_1+(-x_{16}+y_{16})\,\mathbf{a}_2+\left(\frac{1}{2}-z_{16}\right)\,\mathbf{a}_3$ | = | $\left(\frac{1}{2}c\cos\beta-x_{16}a-z_{16}c\cos\beta\right)\,\hat{\mathbf{x}}+y_{16}b\,\hat{\mathbf{y}}+\left(\frac{1}{2}-z_{16}\right)c\sin\beta\,\hat{\mathbf{z}}$ | $(8f)$ | O V |
| $\mathbf{B}_{59}$ | = | $(-x_{16}+y_{16})\,\mathbf{a}_1+(-x_{16}-y_{16})\,\mathbf{a}_2-z_{16}\,\mathbf{a}_3$ | = | $(-x_{16}a-z_{16}c\cos\beta)\,\hat{\mathbf{x}}-y_{16}b\,\hat{\mathbf{y}}-z_{16}c\sin\beta\,\hat{\mathbf{z}}$ | $(8f)$ | O V |
| $\mathbf{B}_{60}$ | = | $(x_{16}+y_{16})\,\mathbf{a}_1+(x_{16}-y_{16})\,\mathbf{a}_2+\left(\frac{1}{2}+z_{16}\right)\,\mathbf{a}_3$ | = | $\left(\frac{1}{2}c\cos\beta+x_{16}a+z_{16}c\cos\beta\right)\,\hat{\mathbf{x}}-y_{16}b\,\hat{\mathbf{y}}+\left(\frac{1}{2}+z_{16}\right)c\sin\beta\,\hat{\mathbf{z}}$ | $(8f)$ | O V |
| $\mathbf{B}_{61}$ | = | $(x_{17}-y_{17})\,\mathbf{a}_1+(x_{17}+y_{17})\,\mathbf{a}_2+z_{17}\,\mathbf{a}_3$ | = | $(x_{17}a+z_{17}c\cos\beta)\,\hat{\mathbf{x}}+y_{17}b\,\hat{\mathbf{y}}+z_{17}c\sin\beta\,\hat{\mathbf{z}}$ | $(8f)$ | O VI |
| $\mathbf{B}_{62}$ | = | $(-x_{17}-y_{17})\,\mathbf{a}_1+(-x_{17}+y_{17})\,\mathbf{a}_2+\left(\frac{1}{2}-z_{17}\right)\,\mathbf{a}_3$ | = | $\left(\frac{1}{2}c\cos\beta-x_{17}a-z_{17}c\cos\beta\right)\,\hat{\mathbf{x}}+y_{17}b\,\hat{\mathbf{y}}+\left(\frac{1}{2}-z_{17}\right)c\sin\beta\,\hat{\mathbf{z}}$ | $(8f)$ | O VI |
| $\mathbf{B}_{63}$ | = | $(-x_{17}+y_{17})\,\mathbf{a}_1+(-x_{17}-y_{17})\,\mathbf{a}_2-z_{17}\,\mathbf{a}_3$ | = | $(-x_{17}a-z_{17}c\cos\beta)\,\hat{\mathbf{x}}-y_{17}b\,\hat{\mathbf{y}}-z_{17}c\sin\beta\,\hat{\mathbf{z}}$ | $(8f)$ | O VI |
| $\mathbf{B}_{64}$ | = | $(x_{17}+y_{17})\,\mathbf{a}_1+(x_{17}-y_{17})\,\mathbf{a}_2+\left(\frac{1}{2}+z_{17}\right)\,\mathbf{a}_3$ | = | $\left(\frac{1}{2}c\cos\beta+x_{17}a+z_{17}c\cos\beta\right)\,\hat{\mathbf{x}}-y_{17}b\,\hat{\mathbf{y}}+\left(\frac{1}{2}+z_{17}\right)c\sin\beta\,\hat{\mathbf{z}}$ | $(8f)$ | O VI |
| $\mathbf{B}_{65}$ | = | $(x_{18}-y_{18})\,\mathbf{a}_1+(x_{18}+y_{18})\,\mathbf{a}_2+z_{18}\,\mathbf{a}_3$ | = | $(x_{18}a+z_{18}c\cos\beta)\,\hat{\mathbf{x}}+y_{18}b\,\hat{\mathbf{y}}+z_{18}c\sin\beta\,\hat{\mathbf{z}}$ | $(8f)$ | O VII |
| $\mathbf{B}_{66}$ | = | $(-x_{18}-y_{18})\,\mathbf{a}_1+(-x_{18}+y_{18})\,\mathbf{a}_2+\left(\frac{1}{2}-z_{18}\right)\,\mathbf{a}_3$ | = | $\left(\frac{1}{2}c\cos\beta-x_{18}a-z_{18}c\cos\beta\right)\,\hat{\mathbf{x}}+y_{18}b\,\hat{\mathbf{y}}+\left(\frac{1}{2}-z_{18}\right)c\sin\beta\,\hat{\mathbf{z}}$ | $(8f)$ | O VII |
| $\mathbf{B}_{67}$ | = | $\left(-x_{18}+y_{18}\right)\,\mathbf{a}_{1}+\left(-x_{18}-y_{18}\right)\,\mathbf{a}_{2}-z_{18}\,\mathbf{a}_{3}$ | = | $\left(-x_{18}a-z_{18}c\cos\beta\right)\,\hat{\mathbf{x}}-y_{18}b\,\hat{\mathbf{y}}-z_{18}c\sin\beta\,\hat{\mathbf{z}}$ | $(8f)$ | O VII |
| $\mathbf{B}_{68}$ | = | $\left(x_{18}+y_{18}\right)\,\mathbf{a}_{1}+\left(x_{18}-y_{18}\right)\,\mathbf{a}_{2}+\left(\frac{1}{2}+z_{18}\right)\,\mathbf{a}_{3}$ | = | $\left(\frac{1}{2}c\cos\beta+x_{18}a+z_{18}c\cos\beta\right)\,\hat{\mathbf{x}}-y_{18}b\,\hat{\mathbf{y}}+\left(\frac{1}{2}+z_{18}\right)c\sin\beta\,\hat{\mathbf{z}}$ | $(8f)$ | O VII |
| $\mathbf{B}_{69}$ | = | $\left(x_{19}-y_{19}\right)\,\mathbf{a}_{1}+\left(x_{19}+y_{19}\right)\,\mathbf{a}_{2}+z_{19}\,\mathbf{a}_{3}$ | = | $\left(x_{19}a+z_{19}c\cos\beta\right)\,\hat{\mathbf{x}}+y_{19}b\,\hat{\mathbf{y}}+z_{19}c\sin\beta\,\hat{\mathbf{z}}$ | $(8f)$ | O VIII |
| $\mathbf{B}_{70}$ | = | $\left(-x_{19}-y_{19}\right)\,\mathbf{a}_{1}+\left(-x_{19}+y_{19}\right)\,\mathbf{a}_{2}+\left(\frac{1}{2}-z_{19}\right)\,\mathbf{a}_{3}$ | = | $\left(\frac{1}{2}c\cos\beta-x_{19}a-z_{19}c\cos\beta\right)\,\hat{\mathbf{x}}+y_{19}b\,\hat{\mathbf{y}}+\left(\frac{1}{2}-z_{19}\right)c\sin\beta\,\hat{\mathbf{z}}$ | $(8f)$ | O VIII |
| $\mathbf{B}_{71}$ | = | $\left(-x_{19}+y_{19}\right)\,\mathbf{a}_{1}+\left(-x_{19}-y_{19}\right)\,\mathbf{a}_{2}-z_{19}\,\mathbf{a}_{3}$ | = | $\left(-x_{19}a-z_{19}c\cos\beta\right)\,\hat{\mathbf{x}}-y_{19}b\,\hat{\mathbf{y}}-z_{19}c\sin\beta\,\hat{\mathbf{z}}$ | $(8f)$ | O VIII |
| $\mathbf{B}_{72}$ | = | $\left(x_{19}+y_{19}\right)\,\mathbf{a}_{1}+\left(x_{19}-y_{19}\right)\,\mathbf{a}_{2}+\left(\frac{1}{2}+z_{19}\right)\,\mathbf{a}_{3}$ | = | $\left(\frac{1}{2}c\cos\beta+x_{19}a+z_{19}c\cos\beta\right)\,\hat{\mathbf{x}}-y_{19}b\,\hat{\mathbf{y}}+\left(\frac{1}{2}+z_{19}\right)c\sin\beta\,\hat{\mathbf{z}}$ | $(8f)$ | O VIII |
| $\mathbf{B}_{73}$ | = | $\left(x_{20}-y_{20}\right)\,\mathbf{a}_{1}+\left(x_{20}+y_{20}\right)\,\mathbf{a}_{2}+z_{20}\,\mathbf{a}_{3}$ | = | $\left(x_{20}a+z_{20}c\cos\beta\right)\,\hat{\mathbf{x}}+y_{20}b\,\hat{\mathbf{y}}+z_{20}c\sin\beta\,\hat{\mathbf{z}}$ | $(8f)$ | O IX |
| $\mathbf{B}_{74}$ | = | $\left(-x_{20}-y_{20}\right)\,\mathbf{a}_{1}+\left(-x_{20}+y_{20}\right)\,\mathbf{a}_{2}+\left(\frac{1}{2}-z_{20}\right)\,\mathbf{a}_{3}$ | = | $\left(\frac{1}{2}c\cos\beta-x_{20}a-z_{20}c\cos\beta\right)\,\hat{\mathbf{x}}+y_{20}b\,\hat{\mathbf{y}}+\left(\frac{1}{2}-z_{20}\right)c\sin\beta\,\hat{\mathbf{z}}$ | $(8f)$ | O IX |
| $\mathbf{B}_{75}$ | = | $\left(-x_{20}+y_{20}\right)\,\mathbf{a}_{1}+\left(-x_{20}-y_{20}\right)\,\mathbf{a}_{2}-z_{20}\,\mathbf{a}_{3}$ | = | $\left(-x_{20}a-z_{20}c\cos\beta\right)\,\hat{\mathbf{x}}-y_{20}b\,\hat{\mathbf{y}}-z_{20}c\sin\beta\,\hat{\mathbf{z}}$ | $(8f)$ | O IX |
| $\mathbf{B}_{76}$ | = | $\left(x_{20}+y_{20}\right)\,\mathbf{a}_{1}+\left(x_{20}-y_{20}\right)\,\mathbf{a}_{2}+\left(\frac{1}{2}+z_{20}\right)\,\mathbf{a}_{3}$ | = | $\left(\frac{1}{2}c\cos\beta+x_{20}a+z_{20}c\cos\beta\right)\,\hat{\mathbf{x}}-y_{20}b\,\hat{\mathbf{y}}+\left(\frac{1}{2}+z_{20}\right)c\sin\beta\,\hat{\mathbf{z}}$ | $(8f)$ | O IX |
| $\mathbf{B}_{77}$ | = | $\left(x_{21}-y_{21}\right)\,\mathbf{a}_{1}+\left(x_{21}+y_{21}\right)\,\mathbf{a}_{2}+z_{21}\,\mathbf{a}_{3}$ | = | $\left(x_{21}a+z_{21}c\cos\beta\right)\,\hat{\mathbf{x}}+y_{21}b\,\hat{\mathbf{y}}+z_{21}c\sin\beta\,\hat{\mathbf{z}}$ | $(8f)$ | O X |
| $\mathbf{B}_{78}$ | = | $\left(-x_{21}-y_{21}\right)\,\mathbf{a}_{1}+\left(-x_{21}+y_{21}\right)\,\mathbf{a}_{2}+\left(\frac{1}{2}-z_{21}\right)\,\mathbf{a}_{3}$ | = | $\left(\frac{1}{2}c\cos\beta-x_{21}a-z_{21}c\cos\beta\right)\,\hat{\mathbf{x}}+y_{21}b\,\hat{\mathbf{y}}+\left(\frac{1}{2}-z_{21}\right)c\sin\beta\,\hat{\mathbf{z}}$ | $(8f)$ | O X |
| $\mathbf{B}_{79}$ | = | $\left(-x_{21}+y_{21}\right)\,\mathbf{a}_{1}+\left(-x_{21}-y_{21}\right)\,\mathbf{a}_{2}-z_{21}\,\mathbf{a}_{3}$ | = | $\left(-x_{21}a-z_{21}c\cos\beta\right)\,\hat{\mathbf{x}}-y_{21}b\,\hat{\mathbf{y}}-z_{21}c\sin\beta\,\hat{\mathbf{z}}$ | $(8f)$ | O X |
| $\mathbf{B}_{80}$ | = | $\left(x_{21}+y_{21}\right)\,\mathbf{a}_{1}+\left(x_{21}-y_{21}\right)\,\mathbf{a}_{2}+\left(\frac{1}{2}+z_{21}\right)\,\mathbf{a}_{3}$ | = | $\left(\frac{1}{2}c\cos\beta+x_{21}a+z_{21}c\cos\beta\right)\,\hat{\mathbf{x}}-y_{21}b\,\hat{\mathbf{y}}+\left(\frac{1}{2}+z_{21}\right)c\sin\beta\,\hat{\mathbf{z}}$ | $(8f)$ | O X |
| $\mathbf{B}_{81}$ | = | $\left(x_{22}-y_{22}\right)\,\mathbf{a}_{1}+\left(x_{22}+y_{22}\right)\,\mathbf{a}_{2}+z_{22}\,\mathbf{a}_{3}$ | = | $\left(x_{22}a+z_{22}c\cos\beta\right)\,\hat{\mathbf{x}}+y_{22}b\,\hat{\mathbf{y}}+z_{22}c\sin\beta\,\hat{\mathbf{z}}$ | $(8f)$ | S II |
| $\mathbf{B}_{82}$ | = | $\left(-x_{22}-y_{22}\right)\,\mathbf{a}_{1}+\left(-x_{22}+y_{22}\right)\,\mathbf{a}_{2}+\left(\frac{1}{2}-z_{22}\right)\,\mathbf{a}_{3}$ | = | $\left(\frac{1}{2}c\cos\beta-x_{22}a-z_{22}c\cos\beta\right)\,\hat{\mathbf{x}}+y_{22}b\,\hat{\mathbf{y}}+\left(\frac{1}{2}-z_{22}\right)c\sin\beta\,\hat{\mathbf{z}}$ | $(8f)$ | S II |
| $\mathbf{B}_{83}$ | = | $\left(-x_{22}+y_{22}\right)\,\mathbf{a}_{1}+\left(-x_{22}-y_{22}\right)\,\mathbf{a}_{2}-z_{22}\,\mathbf{a}_{3}$ | = | $\left(-x_{22}a-z_{22}c\cos\beta\right)\,\hat{\mathbf{x}}-y_{22}b\,\hat{\mathbf{y}}-z_{22}c\sin\beta\,\hat{\mathbf{z}}$ | $(8f)$ | S II |
| $\mathbf{B}_{84}$ | = | $\left(x_{22}+y_{22}\right)\,\mathbf{a}_{1}+\left(x_{22}-y_{22}\right)\,\mathbf{a}_{2}+\left(\frac{1}{2}+z_{22}\right)\,\mathbf{a}_{3}$ | = | $\left(\frac{1}{2}c\cos\beta+x_{22}a+z_{22}c\cos\beta\right)\,\hat{\mathbf{x}}-y_{22}b\,\hat{\mathbf{y}}+\left(\frac{1}{2}+z_{22}\right)c\sin\beta\,\hat{\mathbf{z}}$ | $(8f)$ | S II |

**Geometry files:**

- CIF: pp. 1566
- POSCAR: pp. 1566

# $Al_2Mg_5Si_3O_{10}(OH)_8$ ($S5_5$) Structure: A5B10C8D4_mC108_15_a2ef_5f_4f_2f

http://aflow.org/prototype-encyclopedia/A5B10C8D4_mC108_15_a2ef_5f_4f_2f

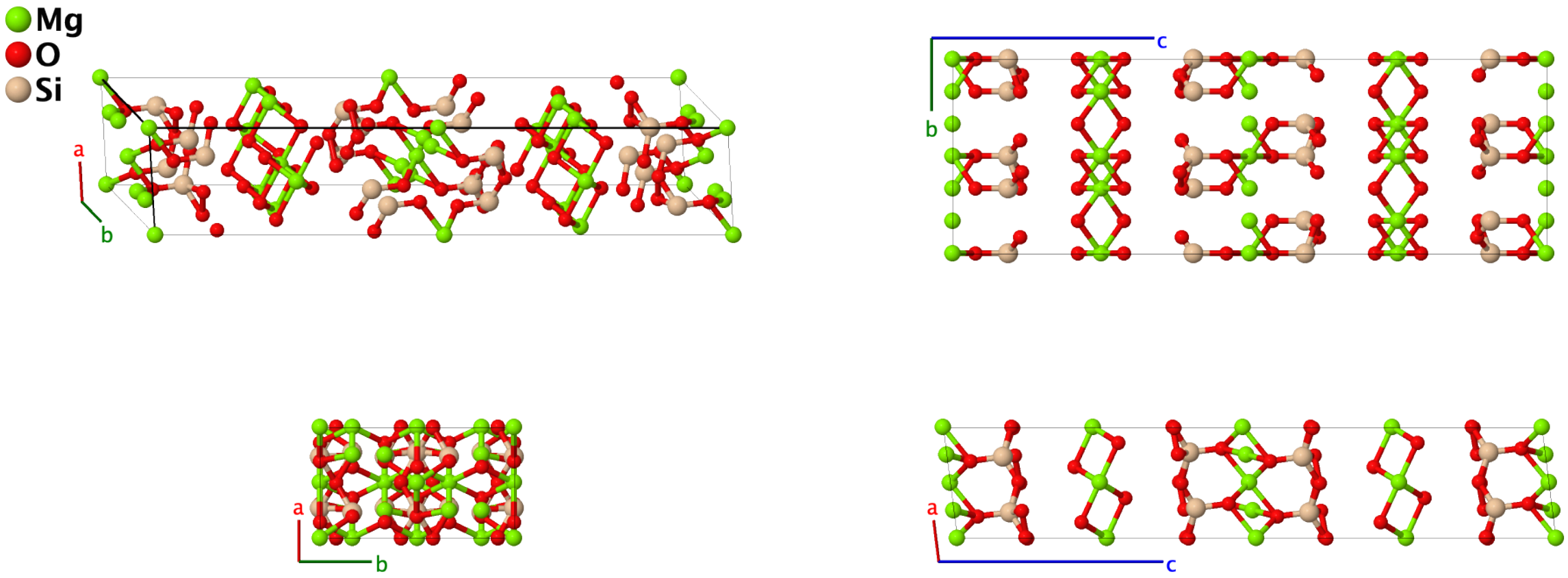

| | | |
|---|---|---|
| **Prototype** | : | $Al_2H_8Mg_5O_{18}Si_3$ |
| **AFLOW prototype label** | : | A5B10C8D4_mC108_15_a2ef_5f_4f_2f |
| ***Strukturbericht* designation** | : | $S5_5$ |
| **Pearson symbol** | : | mC108 |
| **Space group number** | : | 15 |
| **Space group symbol** | : | $C2/c$ |
| **AFLOW prototype command** | : | `aflow --proto=A5B10C8D4_mC108_15_a2ef_5f_4f_2f --params=`$a, b/a, c/a, \beta, y_2, y_3, x_4, y_4, z_4, x_5, y_5, z_5, x_6, y_6, z_6, x_7, y_7, z_7, x_8, y_8, z_8, x_9, y_9, z_9, x_{10}, y_{10}, z_{10}, x_{11}, y_{11}, z_{11}, x_{12}, y_{12}, z_{12}, x_{13}, y_{13}, z_{13}, x_{14}, y_{14}, z_{14}, x_{15}, y_{15}, z_{15}$ |

- This structure was very difficult to reconstruct. (McMurchy, 1934) gives atomic positions with the notation “Four equivalent positions for each atom are listed.” This apparently should be interpreted as “each atom is at one of four equivalent positions,” however if that is the case some of the atoms are listed twice, in two of the equivalent positions. We instead used the Wyckoff positions found in (Gottfried, 1937), which seem to match the positions in McMurchy’s Fig. 5, but these too have problems: the (Mg, Al)II and (Mg, Al)IV sites are listed as being on Wyckoff positions (8f), but they are clearly on (4e) sites.
- The sites we label Mg-I through Mg-IV are actually mixtures of magnesium and aluminum, while the Si-I and Si-II sites are blended silicon and aluminum. Given this, the actual composition of the structure given here should be $(Mg_xAl_{1-x})_5(Si_yAl_{1-y})_4O_{10}(OH)_8$, with appropriate fractional occupations on the “Mg” and “Si” sites. This puts nine atoms on the Mg and Si sites. However, the nominal formula $Al_2Mg_5Si_3O_{10}(OH)_8$ puts ten atoms on those sites. Given this, and our difficulty in reconstructing the unit cell from the original paper, one should treat these positions with some skepticism.

**Base-centered Monoclinic primitive vectors:**

$$\mathbf{a}_1 = \tfrac{1}{2}\,a\,\hat{\mathbf{x}} - \tfrac{1}{2}\,b\,\hat{\mathbf{y}}$$
$$\mathbf{a}_2 = \tfrac{1}{2}\,a\,\hat{\mathbf{x}} + \tfrac{1}{2}\,b\,\hat{\mathbf{y}}$$
$$\mathbf{a}_3 = c\cos\beta\,\hat{\mathbf{x}} + c\sin\beta\,\hat{\mathbf{z}}$$

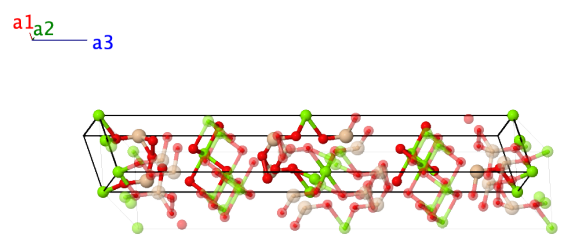

**Basis vectors:**

| | | Lattice Coordinates | | Cartesian Coordinates | Wyckoff Position | Atom Type |
|---|---|---|---|---|---|---|
| $\mathbf{B}_1$ | = | $0\,\mathbf{a}_1 + 0\,\mathbf{a}_2 + 0\,\mathbf{a}_3$ | = | $0\,\hat{\mathbf{x}} + 0\,\hat{\mathbf{y}} + 0\,\hat{\mathbf{z}}$ | (4$a$) | Mg I |
| $\mathbf{B}_2$ | = | $\frac{1}{2}\,\mathbf{a}_3$ | = | $\frac{1}{2}c\cos\beta\,\hat{\mathbf{x}} + \frac{1}{2}c\sin\beta\,\hat{\mathbf{z}}$ | (4$a$) | Mg I |
| $\mathbf{B}_3$ | = | $-y_2\,\mathbf{a}_1 + y_2\,\mathbf{a}_2 + \frac{1}{4}\,\mathbf{a}_3$ | = | $\frac{1}{4}c\cos\beta\,\hat{\mathbf{x}} + y_2b\,\hat{\mathbf{y}} + \frac{1}{4}c\sin\beta\,\hat{\mathbf{z}}$ | (4$e$) | Mg II |
| $\mathbf{B}_4$ | = | $y_2\,\mathbf{a}_1 - y_2\,\mathbf{a}_2 + \frac{3}{4}\,\mathbf{a}_3$ | = | $\frac{3}{4}c\cos\beta\,\hat{\mathbf{x}} - y_2b\,\hat{\mathbf{y}} + \frac{3}{4}c\sin\beta\,\hat{\mathbf{z}}$ | (4$e$) | Mg II |
| $\mathbf{B}_5$ | = | $-y_3\,\mathbf{a}_1 + y_3\,\mathbf{a}_2 + \frac{1}{4}\,\mathbf{a}_3$ | = | $\frac{1}{4}c\cos\beta\,\hat{\mathbf{x}} + y_3b\,\hat{\mathbf{y}} + \frac{1}{4}c\sin\beta\,\hat{\mathbf{z}}$ | (4$e$) | Mg III |
| $\mathbf{B}_6$ | = | $y_3\,\mathbf{a}_1 - y_3\,\mathbf{a}_2 + \frac{3}{4}\,\mathbf{a}_3$ | = | $\frac{3}{4}c\cos\beta\,\hat{\mathbf{x}} - y_3b\,\hat{\mathbf{y}} + \frac{3}{4}c\sin\beta\,\hat{\mathbf{z}}$ | (4$e$) | Mg III |
| $\mathbf{B}_7$ | = | $(x_4 - y_4)\,\mathbf{a}_1 + (x_4 + y_4)\,\mathbf{a}_2 + z_4\,\mathbf{a}_3$ | = | $(x_4a + z_4c\cos\beta)\,\hat{\mathbf{x}} + y_4b\,\hat{\mathbf{y}} + z_4c\sin\beta\,\hat{\mathbf{z}}$ | (8$f$) | Mg IV |
| $\mathbf{B}_8$ | = | $(-x_4 - y_4)\,\mathbf{a}_1 + (-x_4 + y_4)\,\mathbf{a}_2 + \left(\frac{1}{2} - z_4\right)\mathbf{a}_3$ | = | $\left(\frac{1}{2}c\cos\beta - x_4a - z_4c\cos\beta\right)\hat{\mathbf{x}} + y_4b\,\hat{\mathbf{y}} + \left(\frac{1}{2} - z_4\right)c\sin\beta\,\hat{\mathbf{z}}$ | (8$f$) | Mg IV |
| $\mathbf{B}_9$ | = | $(-x_4 + y_4)\,\mathbf{a}_1 + (-x_4 - y_4)\,\mathbf{a}_2 - z_4\,\mathbf{a}_3$ | = | $(-x_4a - z_4c\cos\beta)\,\hat{\mathbf{x}} - y_4b\,\hat{\mathbf{y}} - z_4c\sin\beta\,\hat{\mathbf{z}}$ | (8$f$) | Mg IV |
| $\mathbf{B}_{10}$ | = | $(x_4 + y_4)\,\mathbf{a}_1 + (x_4 - y_4)\,\mathbf{a}_2 + \left(\frac{1}{2} + z_4\right)\mathbf{a}_3$ | = | $\left(\frac{1}{2}c\cos\beta + x_4a + z_4c\cos\beta\right)\hat{\mathbf{x}} - y_4b\,\hat{\mathbf{y}} + \left(\frac{1}{2} + z_4\right)c\sin\beta\,\hat{\mathbf{z}}$ | (8$f$) | Mg IV |
| $\mathbf{B}_{11}$ | = | $(x_5 - y_5)\,\mathbf{a}_1 + (x_5 + y_5)\,\mathbf{a}_2 + z_5\,\mathbf{a}_3$ | = | $(x_5a + z_5c\cos\beta)\,\hat{\mathbf{x}} + y_5b\,\hat{\mathbf{y}} + z_5c\sin\beta\,\hat{\mathbf{z}}$ | (8$f$) | O I |
| $\mathbf{B}_{12}$ | = | $(-x_5 - y_5)\,\mathbf{a}_1 + (-x_5 + y_5)\,\mathbf{a}_2 + \left(\frac{1}{2} - z_5\right)\mathbf{a}_3$ | = | $\left(\frac{1}{2}c\cos\beta - x_5a - z_5c\cos\beta\right)\hat{\mathbf{x}} + y_5b\,\hat{\mathbf{y}} + \left(\frac{1}{2} - z_5\right)c\sin\beta\,\hat{\mathbf{z}}$ | (8$f$) | O I |
| $\mathbf{B}_{13}$ | = | $(-x_5 + y_5)\,\mathbf{a}_1 + (-x_5 - y_5)\,\mathbf{a}_2 - z_5\,\mathbf{a}_3$ | = | $(-x_5a - z_5c\cos\beta)\,\hat{\mathbf{x}} - y_5b\,\hat{\mathbf{y}} - z_5c\sin\beta\,\hat{\mathbf{z}}$ | (8$f$) | O I |
| $\mathbf{B}_{14}$ | = | $(x_5 + y_5)\,\mathbf{a}_1 + (x_5 - y_5)\,\mathbf{a}_2 + \left(\frac{1}{2} + z_5\right)\mathbf{a}_3$ | = | $\left(\frac{1}{2}c\cos\beta + x_5a + z_5c\cos\beta\right)\hat{\mathbf{x}} - y_5b\,\hat{\mathbf{y}} + \left(\frac{1}{2} + z_5\right)c\sin\beta\,\hat{\mathbf{z}}$ | (8$f$) | O I |
| $\mathbf{B}_{15}$ | = | $(x_6 - y_6)\,\mathbf{a}_1 + (x_6 + y_6)\,\mathbf{a}_2 + z_6\,\mathbf{a}_3$ | = | $(x_6a + z_6c\cos\beta)\,\hat{\mathbf{x}} + y_6b\,\hat{\mathbf{y}} + z_6c\sin\beta\,\hat{\mathbf{z}}$ | (8$f$) | O II |
| $\mathbf{B}_{16}$ | = | $(-x_6 - y_6)\,\mathbf{a}_1 + (-x_6 + y_6)\,\mathbf{a}_2 + \left(\frac{1}{2} - z_6\right)\mathbf{a}_3$ | = | $\left(\frac{1}{2}c\cos\beta - x_6a - z_6c\cos\beta\right)\hat{\mathbf{x}} + y_6b\,\hat{\mathbf{y}} + \left(\frac{1}{2} - z_6\right)c\sin\beta\,\hat{\mathbf{z}}$ | (8$f$) | O II |
| $\mathbf{B}_{17}$ | = | $(-x_6 + y_6)\,\mathbf{a}_1 + (-x_6 - y_6)\,\mathbf{a}_2 - z_6\,\mathbf{a}_3$ | = | $(-x_6a - z_6c\cos\beta)\,\hat{\mathbf{x}} - y_6b\,\hat{\mathbf{y}} - z_6c\sin\beta\,\hat{\mathbf{z}}$ | (8$f$) | O II |
| $\mathbf{B}_{18}$ | = | $(x_6 + y_6)\,\mathbf{a}_1 + (x_6 - y_6)\,\mathbf{a}_2 + \left(\frac{1}{2} + z_6\right)\mathbf{a}_3$ | = | $\left(\frac{1}{2}c\cos\beta + x_6a + z_6c\cos\beta\right)\hat{\mathbf{x}} - y_6b\,\hat{\mathbf{y}} + \left(\frac{1}{2} + z_6\right)c\sin\beta\,\hat{\mathbf{z}}$ | (8$f$) | O II |
| $\mathbf{B}_{19}$ | = | $(x_7 - y_7)\,\mathbf{a}_1 + (x_7 + y_7)\,\mathbf{a}_2 + z_7\,\mathbf{a}_3$ | = | $(x_7a + z_7c\cos\beta)\,\hat{\mathbf{x}} + y_7b\,\hat{\mathbf{y}} + z_7c\sin\beta\,\hat{\mathbf{z}}$ | (8$f$) | O III |
| $\mathbf{B}_{20}$ | = | $(-x_7 - y_7)\,\mathbf{a}_1 + (-x_7 + y_7)\,\mathbf{a}_2 + \left(\frac{1}{2} - z_7\right)\mathbf{a}_3$ | = | $\left(\frac{1}{2}c\cos\beta - x_7a - z_7c\cos\beta\right)\hat{\mathbf{x}} + y_7b\,\hat{\mathbf{y}} + \left(\frac{1}{2} - z_7\right)c\sin\beta\,\hat{\mathbf{z}}$ | (8$f$) | O III |
| $\mathbf{B}_{21}$ | = | $(-x_7 + y_7)\,\mathbf{a}_1 + (-x_7 - y_7)\,\mathbf{a}_2 - z_7\,\mathbf{a}_3$ | = | $(-x_7a - z_7c\cos\beta)\,\hat{\mathbf{x}} - y_7b\,\hat{\mathbf{y}} - z_7c\sin\beta\,\hat{\mathbf{z}}$ | (8$f$) | O III |
| $\mathbf{B}_{22}$ | = | $(x_7 + y_7)\,\mathbf{a}_1 + (x_7 - y_7)\,\mathbf{a}_2 + \left(\frac{1}{2} + z_7\right)\mathbf{a}_3$ | = | $\left(\frac{1}{2}c\cos\beta + x_7a + z_7c\cos\beta\right)\hat{\mathbf{x}} - y_7b\,\hat{\mathbf{y}} + \left(\frac{1}{2} + z_7\right)c\sin\beta\,\hat{\mathbf{z}}$ | (8$f$) | O III |
| $\mathbf{B}_{23}$ | = | $(x_8 - y_8)\,\mathbf{a}_1 + (x_8 + y_8)\,\mathbf{a}_2 + z_8\,\mathbf{a}_3$ | = | $(x_8a + z_8c\cos\beta)\,\hat{\mathbf{x}} + y_8b\,\hat{\mathbf{y}} + z_8c\sin\beta\,\hat{\mathbf{z}}$ | (8$f$) | O IV |
| $\mathbf{B}_{24}$ | = | $(-x_8 - y_8)\,\mathbf{a}_1 + (-x_8 + y_8)\,\mathbf{a}_2 + \left(\frac{1}{2} - z_8\right)\mathbf{a}_3$ | = | $\left(\frac{1}{2}c\cos\beta - x_8a - z_8c\cos\beta\right)\hat{\mathbf{x}} + y_8b\,\hat{\mathbf{y}} + \left(\frac{1}{2} - z_8\right)c\sin\beta\,\hat{\mathbf{z}}$ | (8$f$) | O IV |
| $\mathbf{B}_{25}$ | = | $\left(-x_{8}+y_{8}\right)\,\mathbf{a}_{1}+\left(-x_{8}-y_{8}\right)\,\mathbf{a}_{2}-z_{8}\,\mathbf{a}_{3}$ | = | $\left(-x_{8}a-z_{8}c\cos\beta\right)\,\mathbf{\hat{x}}-y_{8}b\,\mathbf{\hat{y}}-z_{8}c\sin\beta\,\mathbf{\hat{z}}$ | $(8f)$ | O IV |
| $\mathbf{B}_{26}$ | = | $\left(x_{8}+y_{8}\right)\,\mathbf{a}_{1}+\left(x_{8}-y_{8}\right)\,\mathbf{a}_{2}+\left(\frac{1}{2}+z_{8}\right)\,\mathbf{a}_{3}$ | = | $\left(\frac{1}{2}c\cos\beta+x_{8}a+z_{8}c\cos\beta\right)\,\mathbf{\hat{x}}-y_{8}b\,\mathbf{\hat{y}}+\left(\frac{1}{2}+z_{8}\right)c\sin\beta\,\mathbf{\hat{z}}$ | $(8f)$ | O IV |
| $\mathbf{B}_{27}$ | = | $\left(x_{9}-y_{9}\right)\,\mathbf{a}_{1}+\left(x_{9}+y_{9}\right)\,\mathbf{a}_{2}+z_{9}\,\mathbf{a}_{3}$ | = | $\left(x_{9}a+z_{9}c\cos\beta\right)\,\mathbf{\hat{x}}+y_{9}b\,\mathbf{\hat{y}}+z_{9}c\sin\beta\,\mathbf{\hat{z}}$ | $(8f)$ | O V |
| $\mathbf{B}_{28}$ | = | $\left(-x_{9}-y_{9}\right)\,\mathbf{a}_{1}+\left(-x_{9}+y_{9}\right)\,\mathbf{a}_{2}+\left(\frac{1}{2}-z_{9}\right)\,\mathbf{a}_{3}$ | = | $\left(\frac{1}{2}c\cos\beta-x_{9}a-z_{9}c\cos\beta\right)\,\mathbf{\hat{x}}+y_{9}b\,\mathbf{\hat{y}}+\left(\frac{1}{2}-z_{9}\right)c\sin\beta\,\mathbf{\hat{z}}$ | $(8f)$ | O V |
| $\mathbf{B}_{29}$ | = | $\left(-x_{9}+y_{9}\right)\,\mathbf{a}_{1}+\left(-x_{9}-y_{9}\right)\,\mathbf{a}_{2}-z_{9}\,\mathbf{a}_{3}$ | = | $\left(-x_{9}a-z_{9}c\cos\beta\right)\,\mathbf{\hat{x}}-y_{9}b\,\mathbf{\hat{y}}-z_{9}c\sin\beta\,\mathbf{\hat{z}}$ | $(8f)$ | O V |
| $\mathbf{B}_{30}$ | = | $\left(x_{9}+y_{9}\right)\,\mathbf{a}_{1}+\left(x_{9}-y_{9}\right)\,\mathbf{a}_{2}+\left(\frac{1}{2}+z_{9}\right)\,\mathbf{a}_{3}$ | = | $\left(\frac{1}{2}c\cos\beta+x_{9}a+z_{9}c\cos\beta\right)\,\mathbf{\hat{x}}-y_{9}b\,\mathbf{\hat{y}}+\left(\frac{1}{2}+z_{9}\right)c\sin\beta\,\mathbf{\hat{z}}$ | $(8f)$ | O V |
| $\mathbf{B}_{31}$ | = | $\left(x_{10}-y_{10}\right)\,\mathbf{a}_{1}+\left(x_{10}+y_{10}\right)\,\mathbf{a}_{2}+z_{10}\,\mathbf{a}_{3}$ | = | $\left(x_{10}a+z_{10}c\cos\beta\right)\,\mathbf{\hat{x}}+y_{10}b\,\mathbf{\hat{y}}+z_{10}c\sin\beta\,\mathbf{\hat{z}}$ | $(8f)$ | OH I |
| $\mathbf{B}_{32}$ | = | $\left(-x_{10}-y_{10}\right)\,\mathbf{a}_{1}+\left(-x_{10}+y_{10}\right)\,\mathbf{a}_{2}+\left(\frac{1}{2}-z_{10}\right)\,\mathbf{a}_{3}$ | = | $\left(\frac{1}{2}c\cos\beta-x_{10}a-z_{10}c\cos\beta\right)\,\mathbf{\hat{x}}+y_{10}b\,\mathbf{\hat{y}}+\left(\frac{1}{2}-z_{10}\right)c\sin\beta\,\mathbf{\hat{z}}$ | $(8f)$ | OH I |
| $\mathbf{B}_{33}$ | = | $\left(-x_{10}+y_{10}\right)\,\mathbf{a}_{1}+\left(-x_{10}-y_{10}\right)\,\mathbf{a}_{2}-z_{10}\,\mathbf{a}_{3}$ | = | $\left(-x_{10}a-z_{10}c\cos\beta\right)\,\mathbf{\hat{x}}-y_{10}b\,\mathbf{\hat{y}}-z_{10}c\sin\beta\,\mathbf{\hat{z}}$ | $(8f)$ | OH I |
| $\mathbf{B}_{34}$ | = | $\left(x_{10}+y_{10}\right)\,\mathbf{a}_{1}+\left(x_{10}-y_{10}\right)\,\mathbf{a}_{2}+\left(\frac{1}{2}+z_{10}\right)\,\mathbf{a}_{3}$ | = | $\left(\frac{1}{2}c\cos\beta+x_{10}a+z_{10}c\cos\beta\right)\,\mathbf{\hat{x}}-y_{10}b\,\mathbf{\hat{y}}+\left(\frac{1}{2}+z_{10}\right)c\sin\beta\,\mathbf{\hat{z}}$ | $(8f)$ | OH I |
| $\mathbf{B}_{35}$ | = | $\left(x_{11}-y_{11}\right)\,\mathbf{a}_{1}+\left(x_{11}+y_{11}\right)\,\mathbf{a}_{2}+z_{11}\,\mathbf{a}_{3}$ | = | $\left(x_{11}a+z_{11}c\cos\beta\right)\,\mathbf{\hat{x}}+y_{11}b\,\mathbf{\hat{y}}+z_{11}c\sin\beta\,\mathbf{\hat{z}}$ | $(8f)$ | OH II |
| $\mathbf{B}_{36}$ | = | $\left(-x_{11}-y_{11}\right)\,\mathbf{a}_{1}+\left(-x_{11}+y_{11}\right)\,\mathbf{a}_{2}+\left(\frac{1}{2}-z_{11}\right)\,\mathbf{a}_{3}$ | = | $\left(\frac{1}{2}c\cos\beta-x_{11}a-z_{11}c\cos\beta\right)\,\mathbf{\hat{x}}+y_{11}b\,\mathbf{\hat{y}}+\left(\frac{1}{2}-z_{11}\right)c\sin\beta\,\mathbf{\hat{z}}$ | $(8f)$ | OH II |
| $\mathbf{B}_{37}$ | = | $\left(-x_{11}+y_{11}\right)\,\mathbf{a}_{1}+\left(-x_{11}-y_{11}\right)\,\mathbf{a}_{2}-z_{11}\,\mathbf{a}_{3}$ | = | $\left(-x_{11}a-z_{11}c\cos\beta\right)\,\mathbf{\hat{x}}-y_{11}b\,\mathbf{\hat{y}}-z_{11}c\sin\beta\,\mathbf{\hat{z}}$ | $(8f)$ | OH II |
| $\mathbf{B}_{38}$ | = | $\left(x_{11}+y_{11}\right)\,\mathbf{a}_{1}+\left(x_{11}-y_{11}\right)\,\mathbf{a}_{2}+\left(\frac{1}{2}+z_{11}\right)\,\mathbf{a}_{3}$ | = | $\left(\frac{1}{2}c\cos\beta+x_{11}a+z_{11}c\cos\beta\right)\,\mathbf{\hat{x}}-y_{11}b\,\mathbf{\hat{y}}+\left(\frac{1}{2}+z_{11}\right)c\sin\beta\,\mathbf{\hat{z}}$ | $(8f)$ | OH II |
| $\mathbf{B}_{39}$ | = | $\left(x_{12}-y_{12}\right)\,\mathbf{a}_{1}+\left(x_{12}+y_{12}\right)\,\mathbf{a}_{2}+z_{12}\,\mathbf{a}_{3}$ | = | $\left(x_{12}a+z_{12}c\cos\beta\right)\,\mathbf{\hat{x}}+y_{12}b\,\mathbf{\hat{y}}+z_{12}c\sin\beta\,\mathbf{\hat{z}}$ | $(8f)$ | OH III |
| $\mathbf{B}_{40}$ | = | $\left(-x_{12}-y_{12}\right)\,\mathbf{a}_{1}+\left(-x_{12}+y_{12}\right)\,\mathbf{a}_{2}+\left(\frac{1}{2}-z_{12}\right)\,\mathbf{a}_{3}$ | = | $\left(\frac{1}{2}c\cos\beta-x_{12}a-z_{12}c\cos\beta\right)\,\mathbf{\hat{x}}+y_{12}b\,\mathbf{\hat{y}}+\left(\frac{1}{2}-z_{12}\right)c\sin\beta\,\mathbf{\hat{z}}$ | $(8f)$ | OH III |
| $\mathbf{B}_{41}$ | = | $\left(-x_{12}+y_{12}\right)\,\mathbf{a}_{1}+\left(-x_{12}-y_{12}\right)\,\mathbf{a}_{2}-z_{12}\,\mathbf{a}_{3}$ | = | $\left(-x_{12}a-z_{12}c\cos\beta\right)\,\mathbf{\hat{x}}-y_{12}b\,\mathbf{\hat{y}}-z_{12}c\sin\beta\,\mathbf{\hat{z}}$ | $(8f)$ | OH III |
| $\mathbf{B}_{42}$ | = | $\left(x_{12}+y_{12}\right)\,\mathbf{a}_{1}+\left(x_{12}-y_{12}\right)\,\mathbf{a}_{2}+\left(\frac{1}{2}+z_{12}\right)\,\mathbf{a}_{3}$ | = | $\left(\frac{1}{2}c\cos\beta+x_{12}a+z_{12}c\cos\beta\right)\,\mathbf{\hat{x}}-y_{12}b\,\mathbf{\hat{y}}+\left(\frac{1}{2}+z_{12}\right)c\sin\beta\,\mathbf{\hat{z}}$ | $(8f)$ | OH III |
| $\mathbf{B}_{43}$ | = | $\left(x_{13}-y_{13}\right)\,\mathbf{a}_{1}+\left(x_{13}+y_{13}\right)\,\mathbf{a}_{2}+z_{13}\,\mathbf{a}_{3}$ | = | $\left(x_{13}a+z_{13}c\cos\beta\right)\,\mathbf{\hat{x}}+y_{13}b\,\mathbf{\hat{y}}+z_{13}c\sin\beta\,\mathbf{\hat{z}}$ | $(8f)$ | OH IV |
| $\mathbf{B}_{44}$ | = | $\left(-x_{13}-y_{13}\right)\,\mathbf{a}_{1}+\left(-x_{13}+y_{13}\right)\,\mathbf{a}_{2}+\left(\frac{1}{2}-z_{13}\right)\,\mathbf{a}_{3}$ | = | $\left(\frac{1}{2}c\cos\beta-x_{13}a-z_{13}c\cos\beta\right)\,\mathbf{\hat{x}}+y_{13}b\,\mathbf{\hat{y}}+\left(\frac{1}{2}-z_{13}\right)c\sin\beta\,\mathbf{\hat{z}}$ | $(8f)$ | OH IV |
| $\mathbf{B}_{45}$ | = | $\left(-x_{13}+y_{13}\right)\,\mathbf{a}_{1}+\left(-x_{13}-y_{13}\right)\,\mathbf{a}_{2}-z_{13}\,\mathbf{a}_{3}$ | = | $\left(-x_{13}a-z_{13}c\cos\beta\right)\,\mathbf{\hat{x}}-y_{13}b\,\mathbf{\hat{y}}-z_{13}c\sin\beta\,\mathbf{\hat{z}}$ | $(8f)$ | OH IV |
| $\mathbf{B}_{46}$ | = | $\left(x_{13}+y_{13}\right)\,\mathbf{a}_{1}+\left(x_{13}-y_{13}\right)\,\mathbf{a}_{2}+\left(\frac{1}{2}+z_{13}\right)\,\mathbf{a}_{3}$ | = | $\left(\frac{1}{2}c\cos\beta+x_{13}a+z_{13}c\cos\beta\right)\,\mathbf{\hat{x}}-y_{13}b\,\mathbf{\hat{y}}+\left(\frac{1}{2}+z_{13}\right)c\sin\beta\,\mathbf{\hat{z}}$ | $(8f)$ | OH IV |
| $\mathbf{B}_{47}$ | = | $(x_{14} - y_{14})\,\mathbf{a}_1 + (x_{14} + y_{14})\,\mathbf{a}_2 + z_{14}\,\mathbf{a}_3$ | = | $(x_{14}a + z_{14}c\cos\beta)\,\hat{\mathbf{x}} + y_{14}b\,\hat{\mathbf{y}} + z_{14}c\sin\beta\,\hat{\mathbf{z}}$ | (8$f$) | Si I |
| $\mathbf{B}_{48}$ | = | $(-x_{14} - y_{14})\,\mathbf{a}_1 + (-x_{14} + y_{14})\,\mathbf{a}_2 + \left(\frac{1}{2} - z_{14}\right)\,\mathbf{a}_3$ | = | $\left(\frac{1}{2}c\cos\beta - x_{14}a - z_{14}c\cos\beta\right)\,\hat{\mathbf{x}} + y_{14}b\,\hat{\mathbf{y}} + \left(\frac{1}{2} - z_{14}\right)c\sin\beta\,\hat{\mathbf{z}}$ | (8$f$) | Si I |
| $\mathbf{B}_{49}$ | = | $(-x_{14} + y_{14})\,\mathbf{a}_1 + (-x_{14} - y_{14})\,\mathbf{a}_2 - z_{14}\,\mathbf{a}_3$ | = | $(-x_{14}a - z_{14}c\cos\beta)\,\hat{\mathbf{x}} - y_{14}b\,\hat{\mathbf{y}} - z_{14}c\sin\beta\,\hat{\mathbf{z}}$ | (8$f$) | Si I |
| $\mathbf{B}_{50}$ | = | $(x_{14} + y_{14})\,\mathbf{a}_1 + (x_{14} - y_{14})\,\mathbf{a}_2 + \left(\frac{1}{2} + z_{14}\right)\,\mathbf{a}_3$ | = | $\left(\frac{1}{2}c\cos\beta + x_{14}a + z_{14}c\cos\beta\right)\,\hat{\mathbf{x}} - y_{14}b\,\hat{\mathbf{y}} + \left(\frac{1}{2} + z_{14}\right)c\sin\beta\,\hat{\mathbf{z}}$ | (8$f$) | Si I |
| $\mathbf{B}_{51}$ | = | $(x_{15} - y_{15})\,\mathbf{a}_1 + (x_{15} + y_{15})\,\mathbf{a}_2 + z_{15}\,\mathbf{a}_3$ | = | $(x_{15}a + z_{15}c\cos\beta)\,\hat{\mathbf{x}} + y_{15}b\,\hat{\mathbf{y}} + z_{15}c\sin\beta\,\hat{\mathbf{z}}$ | (8$f$) | Si II |
| $\mathbf{B}_{52}$ | = | $(-x_{15} - y_{15})\,\mathbf{a}_1 + (-x_{15} + y_{15})\,\mathbf{a}_2 + \left(\frac{1}{2} - z_{15}\right)\,\mathbf{a}_3$ | = | $\left(\frac{1}{2}c\cos\beta - x_{15}a - z_{15}c\cos\beta\right)\,\hat{\mathbf{x}} + y_{15}b\,\hat{\mathbf{y}} + \left(\frac{1}{2} - z_{15}\right)c\sin\beta\,\hat{\mathbf{z}}$ | (8$f$) | Si II |
| $\mathbf{B}_{53}$ | = | $(-x_{15} + y_{15})\,\mathbf{a}_1 + (-x_{15} - y_{15})\,\mathbf{a}_2 - z_{15}\,\mathbf{a}_3$ | = | $(-x_{15}a - z_{15}c\cos\beta)\,\hat{\mathbf{x}} - y_{15}b\,\hat{\mathbf{y}} - z_{15}c\sin\beta\,\hat{\mathbf{z}}$ | (8$f$) | Si II |
| $\mathbf{B}_{54}$ | = | $(x_{15} + y_{15})\,\mathbf{a}_1 + (x_{15} - y_{15})\,\mathbf{a}_2 + \left(\frac{1}{2} + z_{15}\right)\,\mathbf{a}_3$ | = | $\left(\frac{1}{2}c\cos\beta + x_{15}a + z_{15}c\cos\beta\right)\,\hat{\mathbf{x}} - y_{15}b\,\hat{\mathbf{y}} + \left(\frac{1}{2} + z_{15}\right)c\sin\beta\,\hat{\mathbf{z}}$ | (8$f$) | Si II |

**Geometry files:**

- CIF: pp. 1567
- POSCAR: pp. 1567

# $Y_2SiO_5$ ($RE_2SiO_5$ X2) Structure: A5BC2_mC64_15_5f_f_2f

http://aflow.org/prototype-encyclopedia/A5BC2_mC64_15_5f_f_2f

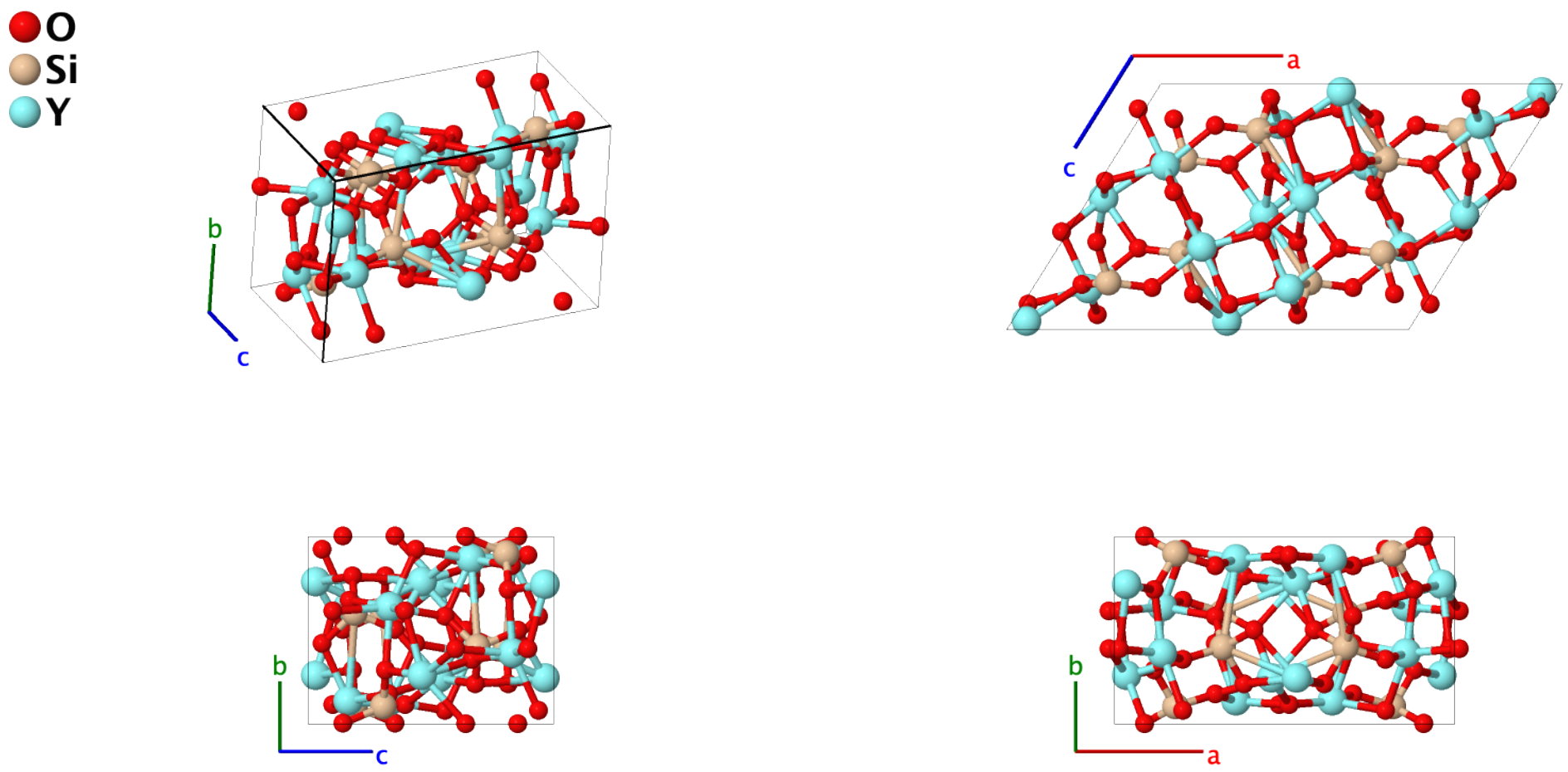

| | | |
|---|---|---|
| **Prototype** | : | $O_5SiY_2$ |
| **AFLOW prototype label** | : | A5BC2_mC64_15_5f_f_2f |
| ***Strukturbericht* designation** | : | None |
| **Pearson symbol** | : | mC64 |
| **Space group number** | : | 15 |
| **Space group symbol** | : | $C2/c$ |
| **AFLOW prototype command** | : | `aflow --proto=A5BC2_mC64_15_5f_f_2f` `--params=`$a, b/a, c/a, \beta, x_1, y_1, z_1, x_2, y_2, z_2, x_3, y_3, z_3, x_4, y_4, z_4, x_5, y_5, z_5, x_6, y_6, z_6, x_7, y_7, z_7, x_8, y_8, z_8$ |

### Other compounds with this structure

- $Dy_2SiO_5$, $Er_2SiO_5$, $Ho_2SiO_5$, $Lu_2SiO_5$, $Tb_2SiO_5$, $Tm_2SiO_5$, and $Yb_2SiO_5$

---

- Compounds of the form $RESiO_5$ ($RE$ = Rare Earth and related elements) crystallize in one of two forms (Wang, 2001 and Tian, 2016): X1, space group $P2_1/c$ #14, for rare earths between lanthanum and gadolinium, and X2, this structure, for the later rare earths, with the dysprosium, yttrium and ytterbium compounds existing in both structures.
- This is the high temperature phase of $Y_2SiO_5$.Below 1190 °C it transforms into the X1 phase.

### Base-centered Monoclinic primitive vectors:

$$\mathbf{a}_1 = \frac{1}{2}\,a\,\hat{\mathbf{x}} - \frac{1}{2}\,b\,\hat{\mathbf{y}}$$
$$\mathbf{a}_2 = \frac{1}{2}\,a\,\hat{\mathbf{x}} + \frac{1}{2}\,b\,\hat{\mathbf{y}}$$
$$\mathbf{a}_3 = c\cos\beta\,\hat{\mathbf{x}} + c\sin\beta\,\hat{\mathbf{z}}$$

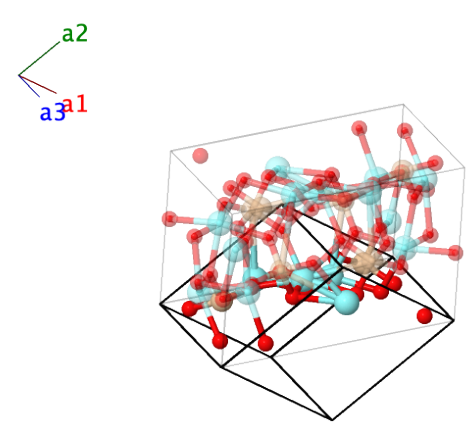

### Basis vectors:

| | | Lattice Coordinates | | Cartesian Coordinates | Wyckoff Position | Atom Type |
|---|---|---|---|---|---|---|
| $\mathbf{B}_1$ | = | $(x_1 - y_1)\,\mathbf{a}_1 + (x_1 + y_1)\,\mathbf{a}_2 + z_1\,\mathbf{a}_3$ | = | $(x_1 a + z_1 c\cos\beta)\,\hat{\mathbf{x}} + y_1 b\,\hat{\mathbf{y}} + z_1 c\sin\beta\,\hat{\mathbf{z}}$ | $(8f)$ | O I |
| $\mathbf{B}_2$ | = | $(-x_1 - y_1)\,\mathbf{a}_1 + (-x_1 + y_1)\,\mathbf{a}_2 + \left(\frac{1}{2} - z_1\right)\mathbf{a}_3$ | = | $\left(\frac{1}{2}c\cos\beta - x_1 a - z_1 c\cos\beta\right)\hat{\mathbf{x}} + y_1 b\,\hat{\mathbf{y}} + \left(\frac{1}{2} - z_1\right)c\sin\beta\,\hat{\mathbf{z}}$ | $(8f)$ | O I |
| $\mathbf{B}_3$ | = | $(-x_1 + y_1)\,\mathbf{a}_1 + (-x_1 - y_1)\,\mathbf{a}_2 - z_1\,\mathbf{a}_3$ | = | $(-x_1 a - z_1 c\cos\beta)\,\hat{\mathbf{x}} - y_1 b\,\hat{\mathbf{y}} - z_1 c\sin\beta\,\hat{\mathbf{z}}$ | $(8f)$ | O I |
| $\mathbf{B}_4$ | = | $(x_1 + y_1)\,\mathbf{a}_1 + (x_1 - y_1)\,\mathbf{a}_2 + \left(\frac{1}{2} + z_1\right)\mathbf{a}_3$ | = | $\left(\frac{1}{2}c\cos\beta + x_1 a + z_1 c\cos\beta\right)\hat{\mathbf{x}} - y_1 b\,\hat{\mathbf{y}} + \left(\frac{1}{2} + z_1\right)c\sin\beta\,\hat{\mathbf{z}}$ | $(8f)$ | O I |
| $\mathbf{B}_5$ | = | $(x_2 - y_2)\,\mathbf{a}_1 + (x_2 + y_2)\,\mathbf{a}_2 + z_2\,\mathbf{a}_3$ | = | $(x_2 a + z_2 c\cos\beta)\,\hat{\mathbf{x}} + y_2 b\,\hat{\mathbf{y}} + z_2 c\sin\beta\,\hat{\mathbf{z}}$ | $(8f)$ | O II |
| $\mathbf{B}_6$ | = | $(-x_2 - y_2)\,\mathbf{a}_1 + (-x_2 + y_2)\,\mathbf{a}_2 + \left(\frac{1}{2} - z_2\right)\mathbf{a}_3$ | = | $\left(\frac{1}{2}c\cos\beta - x_2 a - z_2 c\cos\beta\right)\hat{\mathbf{x}} + y_2 b\,\hat{\mathbf{y}} + \left(\frac{1}{2} - z_2\right)c\sin\beta\,\hat{\mathbf{z}}$ | $(8f)$ | O II |
| $\mathbf{B}_7$ | = | $(-x_2 + y_2)\,\mathbf{a}_1 + (-x_2 - y_2)\,\mathbf{a}_2 - z_2\,\mathbf{a}_3$ | = | $(-x_2 a - z_2 c\cos\beta)\,\hat{\mathbf{x}} - y_2 b\,\hat{\mathbf{y}} - z_2 c\sin\beta\,\hat{\mathbf{z}}$ | $(8f)$ | O II |
| $\mathbf{B}_8$ | = | $(x_2 + y_2)\,\mathbf{a}_1 + (x_2 - y_2)\,\mathbf{a}_2 + \left(\frac{1}{2} + z_2\right)\mathbf{a}_3$ | = | $\left(\frac{1}{2}c\cos\beta + x_2 a + z_2 c\cos\beta\right)\hat{\mathbf{x}} - y_2 b\,\hat{\mathbf{y}} + \left(\frac{1}{2} + z_2\right)c\sin\beta\,\hat{\mathbf{z}}$ | $(8f)$ | O II |
| $\mathbf{B}_9$ | = | $(x_3 - y_3)\,\mathbf{a}_1 + (x_3 + y_3)\,\mathbf{a}_2 + z_3\,\mathbf{a}_3$ | = | $(x_3 a + z_3 c\cos\beta)\,\hat{\mathbf{x}} + y_3 b\,\hat{\mathbf{y}} + z_3 c\sin\beta\,\hat{\mathbf{z}}$ | $(8f)$ | O III |
| $\mathbf{B}_{10}$ | = | $(-x_3 - y_3)\,\mathbf{a}_1 + (-x_3 + y_3)\,\mathbf{a}_2 + \left(\frac{1}{2} - z_3\right)\mathbf{a}_3$ | = | $\left(\frac{1}{2}c\cos\beta - x_3 a - z_3 c\cos\beta\right)\hat{\mathbf{x}} + y_3 b\,\hat{\mathbf{y}} + \left(\frac{1}{2} - z_3\right)c\sin\beta\,\hat{\mathbf{z}}$ | $(8f)$ | O III |
| $\mathbf{B}_{11}$ | = | $(-x_3 + y_3)\,\mathbf{a}_1 + (-x_3 - y_3)\,\mathbf{a}_2 - z_3\,\mathbf{a}_3$ | = | $(-x_3 a - z_3 c\cos\beta)\,\hat{\mathbf{x}} - y_3 b\,\hat{\mathbf{y}} - z_3 c\sin\beta\,\hat{\mathbf{z}}$ | $(8f)$ | O III |
| $\mathbf{B}_{12}$ | = | $(x_3 + y_3)\,\mathbf{a}_1 + (x_3 - y_3)\,\mathbf{a}_2 + \left(\frac{1}{2} + z_3\right)\mathbf{a}_3$ | = | $\left(\frac{1}{2}c\cos\beta + x_3 a + z_3 c\cos\beta\right)\hat{\mathbf{x}} - y_3 b\,\hat{\mathbf{y}} + \left(\frac{1}{2} + z_3\right)c\sin\beta\,\hat{\mathbf{z}}$ | $(8f)$ | O III |
| $\mathbf{B}_{13}$ | = | $(x_4 - y_4)\,\mathbf{a}_1 + (x_4 + y_4)\,\mathbf{a}_2 + z_4\,\mathbf{a}_3$ | = | $(x_4 a + z_4 c\cos\beta)\,\hat{\mathbf{x}} + y_4 b\,\hat{\mathbf{y}} + z_4 c\sin\beta\,\hat{\mathbf{z}}$ | $(8f)$ | O IV |
| $\mathbf{B}_{14}$ | = | $(-x_4 - y_4)\,\mathbf{a}_1 + (-x_4 + y_4)\,\mathbf{a}_2 + \left(\frac{1}{2} - z_4\right)\mathbf{a}_3$ | = | $\left(\frac{1}{2}c\cos\beta - x_4 a - z_4 c\cos\beta\right)\hat{\mathbf{x}} + y_4 b\,\hat{\mathbf{y}} + \left(\frac{1}{2} - z_4\right)c\sin\beta\,\hat{\mathbf{z}}$ | $(8f)$ | O IV |
| $\mathbf{B}_{15}$ | = | $(-x_4 + y_4)\,\mathbf{a}_1 + (-x_4 - y_4)\,\mathbf{a}_2 - z_4\,\mathbf{a}_3$ | = | $(-x_4 a - z_4 c\cos\beta)\,\hat{\mathbf{x}} - y_4 b\,\hat{\mathbf{y}} - z_4 c\sin\beta\,\hat{\mathbf{z}}$ | $(8f)$ | O IV |
| $\mathbf{B}_{16}$ | = | $(x_4 + y_4)\,\mathbf{a}_1 + (x_4 - y_4)\,\mathbf{a}_2 + \left(\frac{1}{2} + z_4\right)\mathbf{a}_3$ | = | $\left(\frac{1}{2}c\cos\beta + x_4 a + z_4 c\cos\beta\right)\hat{\mathbf{x}} - y_4 b\,\hat{\mathbf{y}} + \left(\frac{1}{2} + z_4\right)c\sin\beta\,\hat{\mathbf{z}}$ | $(8f)$ | O IV |
| $\mathbf{B}_{17}$ | = | $(x_5 - y_5)\,\mathbf{a}_1 + (x_5 + y_5)\,\mathbf{a}_2 + z_5\,\mathbf{a}_3$ | = | $(x_5 a + z_5 c\cos\beta)\,\hat{\mathbf{x}} + y_5 b\,\hat{\mathbf{y}} + z_5 c\sin\beta\,\hat{\mathbf{z}}$ | $(8f)$ | O V |
| $\mathbf{B}_{18}$ | = | $(-x_5 - y_5)\,\mathbf{a}_1 + (-x_5 + y_5)\,\mathbf{a}_2 + \left(\frac{1}{2} - z_5\right)\mathbf{a}_3$ | = | $\left(\frac{1}{2}c\cos\beta - x_5 a - z_5 c\cos\beta\right)\hat{\mathbf{x}} + y_5 b\,\hat{\mathbf{y}} + \left(\frac{1}{2} - z_5\right)c\sin\beta\,\hat{\mathbf{z}}$ | $(8f)$ | O V |
| $\mathbf{B}_{19}$ | = | $(-x_5 + y_5)\,\mathbf{a}_1 + (-x_5 - y_5)\,\mathbf{a}_2 - z_5\,\mathbf{a}_3$ | = | $(-x_5 a - z_5 c\cos\beta)\,\hat{\mathbf{x}} - y_5 b\,\hat{\mathbf{y}} - z_5 c\sin\beta\,\hat{\mathbf{z}}$ | $(8f)$ | O V |
| $\mathbf{B}_{20}$ | = | $(x_5 + y_5)\,\mathbf{a}_1 + (x_5 - y_5)\,\mathbf{a}_2 + \left(\frac{1}{2} + z_5\right)\mathbf{a}_3$ | = | $\left(\frac{1}{2}c\cos\beta + x_5 a + z_5 c\cos\beta\right)\hat{\mathbf{x}} - y_5 b\,\hat{\mathbf{y}} + \left(\frac{1}{2} + z_5\right)c\sin\beta\,\hat{\mathbf{z}}$ | $(8f)$ | O V |
| $\mathbf{B}_{21}$ | = | $(x_6 - y_6)\,\mathbf{a}_1 + (x_6 + y_6)\,\mathbf{a}_2 + z_6\,\mathbf{a}_3$ | = | $(x_6 a + z_6 c\cos\beta)\,\hat{\mathbf{x}} + y_6 b\,\hat{\mathbf{y}} + z_6 c\sin\beta\,\hat{\mathbf{z}}$ | $(8f)$ | Si |
| $\mathbf{B}_{22}$ | = | $(-x_6 - y_6)\,\mathbf{a}_1 + (-x_6 + y_6)\,\mathbf{a}_2 + \left(\frac{1}{2} - z_6\right)\,\mathbf{a}_3$ | = | $\left(\frac{1}{2}c\cos\beta - x_6 a - z_6 c\cos\beta\right)\,\hat{\mathbf{x}} + y_6 b\,\hat{\mathbf{y}} + \left(\frac{1}{2} - z_6\right)c\sin\beta\,\hat{\mathbf{z}}$ | (8*f*) | Si |
| $\mathbf{B}_{23}$ | = | $(-x_6 + y_6)\,\mathbf{a}_1 + (-x_6 - y_6)\,\mathbf{a}_2 - z_6\,\mathbf{a}_3$ | = | $(-x_6 a - z_6 c\cos\beta)\,\hat{\mathbf{x}} - y_6 b\,\hat{\mathbf{y}} - z_6 c\sin\beta\,\hat{\mathbf{z}}$ | (8*f*) | Si |
| $\mathbf{B}_{24}$ | = | $(x_6 + y_6)\,\mathbf{a}_1 + (x_6 - y_6)\,\mathbf{a}_2 + \left(\frac{1}{2} + z_6\right)\,\mathbf{a}_3$ | = | $\left(\frac{1}{2}c\cos\beta + x_6 a + z_6 c\cos\beta\right)\,\hat{\mathbf{x}} - y_6 b\,\hat{\mathbf{y}} + \left(\frac{1}{2} + z_6\right)c\sin\beta\,\hat{\mathbf{z}}$ | (8*f*) | Si |
| $\mathbf{B}_{25}$ | = | $(x_7 - y_7)\,\mathbf{a}_1 + (x_7 + y_7)\,\mathbf{a}_2 + z_7\,\mathbf{a}_3$ | = | $(x_7 a + z_7 c\cos\beta)\,\hat{\mathbf{x}} + y_7 b\,\hat{\mathbf{y}} + z_7 c\sin\beta\,\hat{\mathbf{z}}$ | (8*f*) | Y I |
| $\mathbf{B}_{26}$ | = | $(-x_7 - y_7)\,\mathbf{a}_1 + (-x_7 + y_7)\,\mathbf{a}_2 + \left(\frac{1}{2} - z_7\right)\,\mathbf{a}_3$ | = | $\left(\frac{1}{2}c\cos\beta - x_7 a - z_7 c\cos\beta\right)\,\hat{\mathbf{x}} + y_7 b\,\hat{\mathbf{y}} + \left(\frac{1}{2} - z_7\right)c\sin\beta\,\hat{\mathbf{z}}$ | (8*f*) | Y I |
| $\mathbf{B}_{27}$ | = | $(-x_7 + y_7)\,\mathbf{a}_1 + (-x_7 - y_7)\,\mathbf{a}_2 - z_7\,\mathbf{a}_3$ | = | $(-x_7 a - z_7 c\cos\beta)\,\hat{\mathbf{x}} - y_7 b\,\hat{\mathbf{y}} - z_7 c\sin\beta\,\hat{\mathbf{z}}$ | (8*f*) | Y I |
| $\mathbf{B}_{28}$ | = | $(x_7 + y_7)\,\mathbf{a}_1 + (x_7 - y_7)\,\mathbf{a}_2 + \left(\frac{1}{2} + z_7\right)\,\mathbf{a}_3$ | = | $\left(\frac{1}{2}c\cos\beta + x_7 a + z_7 c\cos\beta\right)\,\hat{\mathbf{x}} - y_7 b\,\hat{\mathbf{y}} + \left(\frac{1}{2} + z_7\right)c\sin\beta\,\hat{\mathbf{z}}$ | (8*f*) | Y I |
| $\mathbf{B}_{29}$ | = | $(x_8 - y_8)\,\mathbf{a}_1 + (x_8 + y_8)\,\mathbf{a}_2 + z_8\,\mathbf{a}_3$ | = | $(x_8 a + z_8 c\cos\beta)\,\hat{\mathbf{x}} + y_8 b\,\hat{\mathbf{y}} + z_8 c\sin\beta\,\hat{\mathbf{z}}$ | (8*f*) | Y II |
| $\mathbf{B}_{30}$ | = | $(-x_8 - y_8)\,\mathbf{a}_1 + (-x_8 + y_8)\,\mathbf{a}_2 + \left(\frac{1}{2} - z_8\right)\,\mathbf{a}_3$ | = | $\left(\frac{1}{2}c\cos\beta - x_8 a - z_8 c\cos\beta\right)\,\hat{\mathbf{x}} + y_8 b\,\hat{\mathbf{y}} + \left(\frac{1}{2} - z_8\right)c\sin\beta\,\hat{\mathbf{z}}$ | (8*f*) | Y II |
| $\mathbf{B}_{31}$ | = | $(-x_8 + y_8)\,\mathbf{a}_1 + (-x_8 - y_8)\,\mathbf{a}_2 - z_8\,\mathbf{a}_3$ | = | $(-x_8 a - z_8 c\cos\beta)\,\hat{\mathbf{x}} - y_8 b\,\hat{\mathbf{y}} - z_8 c\sin\beta\,\hat{\mathbf{z}}$ | (8*f*) | Y II |
| $\mathbf{B}_{32}$ | = | $(x_8 + y_8)\,\mathbf{a}_1 + (x_8 - y_8)\,\mathbf{a}_2 + \left(\frac{1}{2} + z_8\right)\,\mathbf{a}_3$ | = | $\left(\frac{1}{2}c\cos\beta + x_8 a + z_8 c\cos\beta\right)\,\hat{\mathbf{x}} - y_8 b\,\hat{\mathbf{y}} + \left(\frac{1}{2} + z_8\right)c\sin\beta\,\hat{\mathbf{z}}$ | (8*f*) | Y II |

**Geometry files:**

- CIF: pp. 1568
- POSCAR: pp. 1568

# $\alpha$-$Zn_2V_2O_7$ Structure: A7B2C2_mC44_15_e3f_f_f

http://aflow.org/prototype-encyclopedia/A7B2C2_mC44_15_e3f_f_f

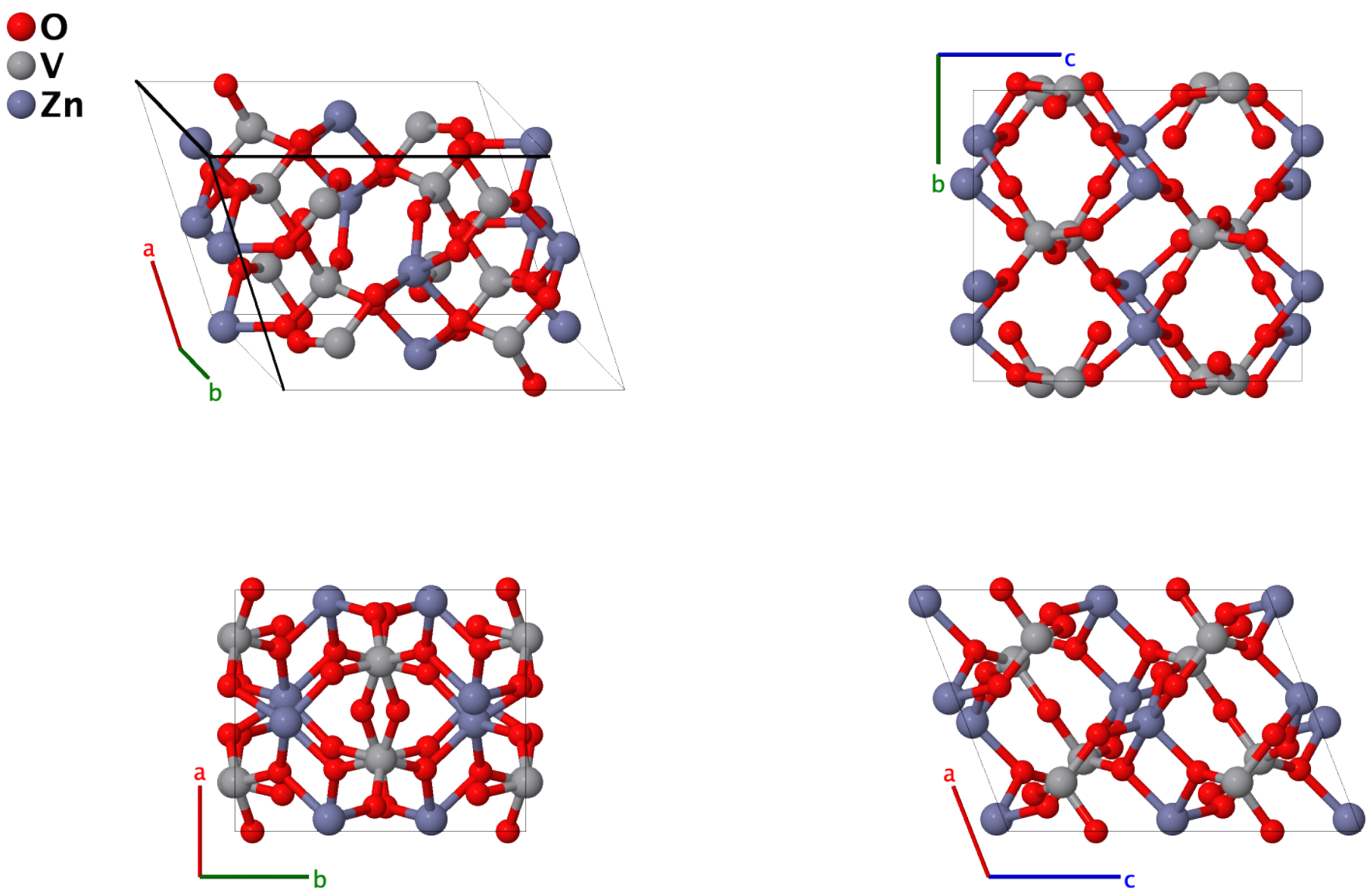

| | | |
|---|---|---|
| **Prototype** | : | $O_7V_2Zn_2$ |
| **AFLOW prototype label** | : | A7B2C2_mC44_15_e3f_f_f |
| ***Strukturbericht* designation** | : | None |
| **Pearson symbol** | : | mC44 |
| **Space group number** | : | 15 |
| **Space group symbol** | : | $C2/c$ |
| **AFLOW prototype command** | : | `aflow --proto=A7B2C2_mC44_15_e3f_f_f` `--params=`$a, b/a, c/a, \beta, y_1, x_2, y_2, z_2, x_3, y_3, z_3, x_4, y_4, z_4, x_5, y_5, z_5, x_6, y_6, z_6$ |

### Other compounds with this structure

- $\beta$-$Cu_2V_2O_7$

### Base-centered Monoclinic primitive vectors:

$$\mathbf{a}_1 = \frac{1}{2}\,a\,\hat{\mathbf{x}} - \frac{1}{2}\,b\,\hat{\mathbf{y}}$$

$$\mathbf{a}_2 = \frac{1}{2}\,a\,\hat{\mathbf{x}} + \frac{1}{2}\,b\,\hat{\mathbf{y}}$$

$$\mathbf{a}_3 = c\cos\beta\,\hat{\mathbf{x}} + c\sin\beta\,\hat{\mathbf{z}}$$

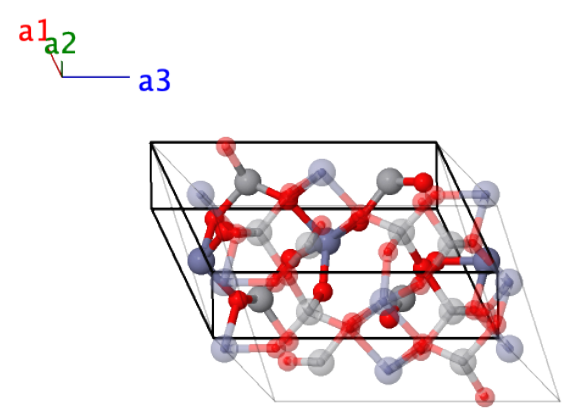

### Basis vectors:

| | | Lattice Coordinates | | Cartesian Coordinates | Wyckoff Position | Atom Type |
|---|---|---|---|---|---|---|
| $\mathbf{B}_1$ | = | $-y_1\,\mathbf{a}_1 + y_1\,\mathbf{a}_2 + \frac{1}{4}\,\mathbf{a}_3$ | = | $\frac{1}{4}c\cos\beta\,\hat{\mathbf{x}} + y_1 b\,\hat{\mathbf{y}} + \frac{1}{4}c\sin\beta\,\hat{\mathbf{z}}$ | (4$e$) | O I |
| $\mathbf{B}_2$ | = | $y_1\,\mathbf{a}_1 - y_1\,\mathbf{a}_2 + \frac{3}{4}\,\mathbf{a}_3$ | = | $\frac{3}{4}c\cos\beta\,\hat{\mathbf{x}} - y_1 b\,\hat{\mathbf{y}} + \frac{3}{4}c\sin\beta\,\hat{\mathbf{z}}$ | (4$e$) | O I |
| $\mathbf{B}_3$ | = | $(x_2 - y_2)\,\mathbf{a}_1 + (x_2 + y_2)\,\mathbf{a}_2 + z_2\,\mathbf{a}_3$ | = | $(x_2 a + z_2 c\cos\beta)\,\hat{\mathbf{x}} + y_2 b\,\hat{\mathbf{y}} + z_2 c\sin\beta\,\hat{\mathbf{z}}$ | (8$f$) | O II |
| $\mathbf{B}_4$ | = | $(-x_2 - y_2)\,\mathbf{a}_1 + (-x_2 + y_2)\,\mathbf{a}_2 + \left(\frac{1}{2} - z_2\right)\,\mathbf{a}_3$ | = | $\left(\frac{1}{2}c\cos\beta - x_2a - z_2c\cos\beta\right)\,\hat{\mathbf{x}} + y_2b\,\hat{\mathbf{y}} + \left(\frac{1}{2} - z_2\right)c\sin\beta\,\hat{\mathbf{z}}$ | $(8f)$ | O II |
| $\mathbf{B}_5$ | = | $(-x_2 + y_2)\,\mathbf{a}_1 + (-x_2 - y_2)\,\mathbf{a}_2 - z_2\,\mathbf{a}_3$ | = | $(-x_2a - z_2c\cos\beta)\,\hat{\mathbf{x}} - y_2b\,\hat{\mathbf{y}} - z_2c\sin\beta\,\hat{\mathbf{z}}$ | $(8f)$ | O II |
| $\mathbf{B}_6$ | = | $(x_2 + y_2)\,\mathbf{a}_1 + (x_2 - y_2)\,\mathbf{a}_2 + \left(\frac{1}{2} + z_2\right)\,\mathbf{a}_3$ | = | $\left(\frac{1}{2}c\cos\beta + x_2a + z_2c\cos\beta\right)\,\hat{\mathbf{x}} - y_2b\,\hat{\mathbf{y}} + \left(\frac{1}{2} + z_2\right)c\sin\beta\,\hat{\mathbf{z}}$ | $(8f)$ | O II |
| $\mathbf{B}_7$ | = | $(x_3 - y_3)\,\mathbf{a}_1 + (x_3 + y_3)\,\mathbf{a}_2 + z_3\,\mathbf{a}_3$ | = | $(x_3a + z_3c\cos\beta)\,\hat{\mathbf{x}} + y_3b\,\hat{\mathbf{y}} + z_3c\sin\beta\,\hat{\mathbf{z}}$ | $(8f)$ | O III |
| $\mathbf{B}_8$ | = | $(-x_3 - y_3)\,\mathbf{a}_1 + (-x_3 + y_3)\,\mathbf{a}_2 + \left(\frac{1}{2} - z_3\right)\,\mathbf{a}_3$ | = | $\left(\frac{1}{2}c\cos\beta - x_3a - z_3c\cos\beta\right)\,\hat{\mathbf{x}} + y_3b\,\hat{\mathbf{y}} + \left(\frac{1}{2} - z_3\right)c\sin\beta\,\hat{\mathbf{z}}$ | $(8f)$ | O III |
| $\mathbf{B}_9$ | = | $(-x_3 + y_3)\,\mathbf{a}_1 + (-x_3 - y_3)\,\mathbf{a}_2 - z_3\,\mathbf{a}_3$ | = | $(-x_3a - z_3c\cos\beta)\,\hat{\mathbf{x}} - y_3b\,\hat{\mathbf{y}} - z_3c\sin\beta\,\hat{\mathbf{z}}$ | $(8f)$ | O III |
| $\mathbf{B}_{10}$ | = | $(x_3 + y_3)\,\mathbf{a}_1 + (x_3 - y_3)\,\mathbf{a}_2 + \left(\frac{1}{2} + z_3\right)\,\mathbf{a}_3$ | = | $\left(\frac{1}{2}c\cos\beta + x_3a + z_3c\cos\beta\right)\,\hat{\mathbf{x}} - y_3b\,\hat{\mathbf{y}} + \left(\frac{1}{2} + z_3\right)c\sin\beta\,\hat{\mathbf{z}}$ | $(8f)$ | O III |
| $\mathbf{B}_{11}$ | = | $(x_4 - y_4)\,\mathbf{a}_1 + (x_4 + y_4)\,\mathbf{a}_2 + z_4\,\mathbf{a}_3$ | = | $(x_4a + z_4c\cos\beta)\,\hat{\mathbf{x}} + y_4b\,\hat{\mathbf{y}} + z_4c\sin\beta\,\hat{\mathbf{z}}$ | $(8f)$ | O IV |
| $\mathbf{B}_{12}$ | = | $(-x_4 - y_4)\,\mathbf{a}_1 + (-x_4 + y_4)\,\mathbf{a}_2 + \left(\frac{1}{2} - z_4\right)\,\mathbf{a}_3$ | = | $\left(\frac{1}{2}c\cos\beta - x_4a - z_4c\cos\beta\right)\,\hat{\mathbf{x}} + y_4b\,\hat{\mathbf{y}} + \left(\frac{1}{2} - z_4\right)c\sin\beta\,\hat{\mathbf{z}}$ | $(8f)$ | O IV |
| $\mathbf{B}_{13}$ | = | $(-x_4 + y_4)\,\mathbf{a}_1 + (-x_4 - y_4)\,\mathbf{a}_2 - z_4\,\mathbf{a}_3$ | = | $(-x_4a - z_4c\cos\beta)\,\hat{\mathbf{x}} - y_4b\,\hat{\mathbf{y}} - z_4c\sin\beta\,\hat{\mathbf{z}}$ | $(8f)$ | O IV |
| $\mathbf{B}_{14}$ | = | $(x_4 + y_4)\,\mathbf{a}_1 + (x_4 - y_4)\,\mathbf{a}_2 + \left(\frac{1}{2} + z_4\right)\,\mathbf{a}_3$ | = | $\left(\frac{1}{2}c\cos\beta + x_4a + z_4c\cos\beta\right)\,\hat{\mathbf{x}} - y_4b\,\hat{\mathbf{y}} + \left(\frac{1}{2} + z_4\right)c\sin\beta\,\hat{\mathbf{z}}$ | $(8f)$ | O IV |
| $\mathbf{B}_{15}$ | = | $(x_5 - y_5)\,\mathbf{a}_1 + (x_5 + y_5)\,\mathbf{a}_2 + z_5\,\mathbf{a}_3$ | = | $(x_5a + z_5c\cos\beta)\,\hat{\mathbf{x}} + y_5b\,\hat{\mathbf{y}} + z_5c\sin\beta\,\hat{\mathbf{z}}$ | $(8f)$ | V |
| $\mathbf{B}_{16}$ | = | $(-x_5 - y_5)\,\mathbf{a}_1 + (-x_5 + y_5)\,\mathbf{a}_2 + \left(\frac{1}{2} - z_5\right)\,\mathbf{a}_3$ | = | $\left(\frac{1}{2}c\cos\beta - x_5a - z_5c\cos\beta\right)\,\hat{\mathbf{x}} + y_5b\,\hat{\mathbf{y}} + \left(\frac{1}{2} - z_5\right)c\sin\beta\,\hat{\mathbf{z}}$ | $(8f)$ | V |
| $\mathbf{B}_{17}$ | = | $(-x_5 + y_5)\,\mathbf{a}_1 + (-x_5 - y_5)\,\mathbf{a}_2 - z_5\,\mathbf{a}_3$ | = | $(-x_5a - z_5c\cos\beta)\,\hat{\mathbf{x}} - y_5b\,\hat{\mathbf{y}} - z_5c\sin\beta\,\hat{\mathbf{z}}$ | $(8f)$ | V |
| $\mathbf{B}_{18}$ | = | $(x_5 + y_5)\,\mathbf{a}_1 + (x_5 - y_5)\,\mathbf{a}_2 + \left(\frac{1}{2} + z_5\right)\,\mathbf{a}_3$ | = | $\left(\frac{1}{2}c\cos\beta + x_5a + z_5c\cos\beta\right)\,\hat{\mathbf{x}} - y_5b\,\hat{\mathbf{y}} + \left(\frac{1}{2} + z_5\right)c\sin\beta\,\hat{\mathbf{z}}$ | $(8f)$ | V |
| $\mathbf{B}_{19}$ | = | $(x_6 - y_6)\,\mathbf{a}_1 + (x_6 + y_6)\,\mathbf{a}_2 + z_6\,\mathbf{a}_3$ | = | $(x_6a + z_6c\cos\beta)\,\hat{\mathbf{x}} + y_6b\,\hat{\mathbf{y}} + z_6c\sin\beta\,\hat{\mathbf{z}}$ | $(8f)$ | Zn |
| $\mathbf{B}_{20}$ | = | $(-x_6 - y_6)\,\mathbf{a}_1 + (-x_6 + y_6)\,\mathbf{a}_2 + \left(\frac{1}{2} - z_6\right)\,\mathbf{a}_3$ | = | $\left(\frac{1}{2}c\cos\beta - x_6a - z_6c\cos\beta\right)\,\hat{\mathbf{x}} + y_6b\,\hat{\mathbf{y}} + \left(\frac{1}{2} - z_6\right)c\sin\beta\,\hat{\mathbf{z}}$ | $(8f)$ | Zn |
| $\mathbf{B}_{21}$ | = | $(-x_6 + y_6)\,\mathbf{a}_1 + (-x_6 - y_6)\,\mathbf{a}_2 - z_6\,\mathbf{a}_3$ | = | $(-x_6a - z_6c\cos\beta)\,\hat{\mathbf{x}} - y_6b\,\hat{\mathbf{y}} - z_6c\sin\beta\,\hat{\mathbf{z}}$ | $(8f)$ | Zn |
| $\mathbf{B}_{22}$ | = | $(x_6 + y_6)\,\mathbf{a}_1 + (x_6 - y_6)\,\mathbf{a}_2 + \left(\frac{1}{2} + z_6\right)\,\mathbf{a}_3$ | = | $\left(\frac{1}{2}c\cos\beta + x_6a + z_6c\cos\beta\right)\,\hat{\mathbf{x}} - y_6b\,\hat{\mathbf{y}} + \left(\frac{1}{2} + z_6\right)c\sin\beta\,\hat{\mathbf{z}}$ | $(8f)$ | Zn |

**Found in:**

**Geometry files:**
- CIF: pp. 1568
- POSCAR: pp. 1569

# Manganese-leonite 185 K [$K_2Mn(SO_4)_2{\cdot}4H_2O$] Structure: A8B2CD12E2_mC200_15_8f_2f_ce_2e11f_2f

http://aflow.org/prototype-encyclopedia/A8B2CD12E2_mC200_15_8f_2f_ce_2e11f_2f

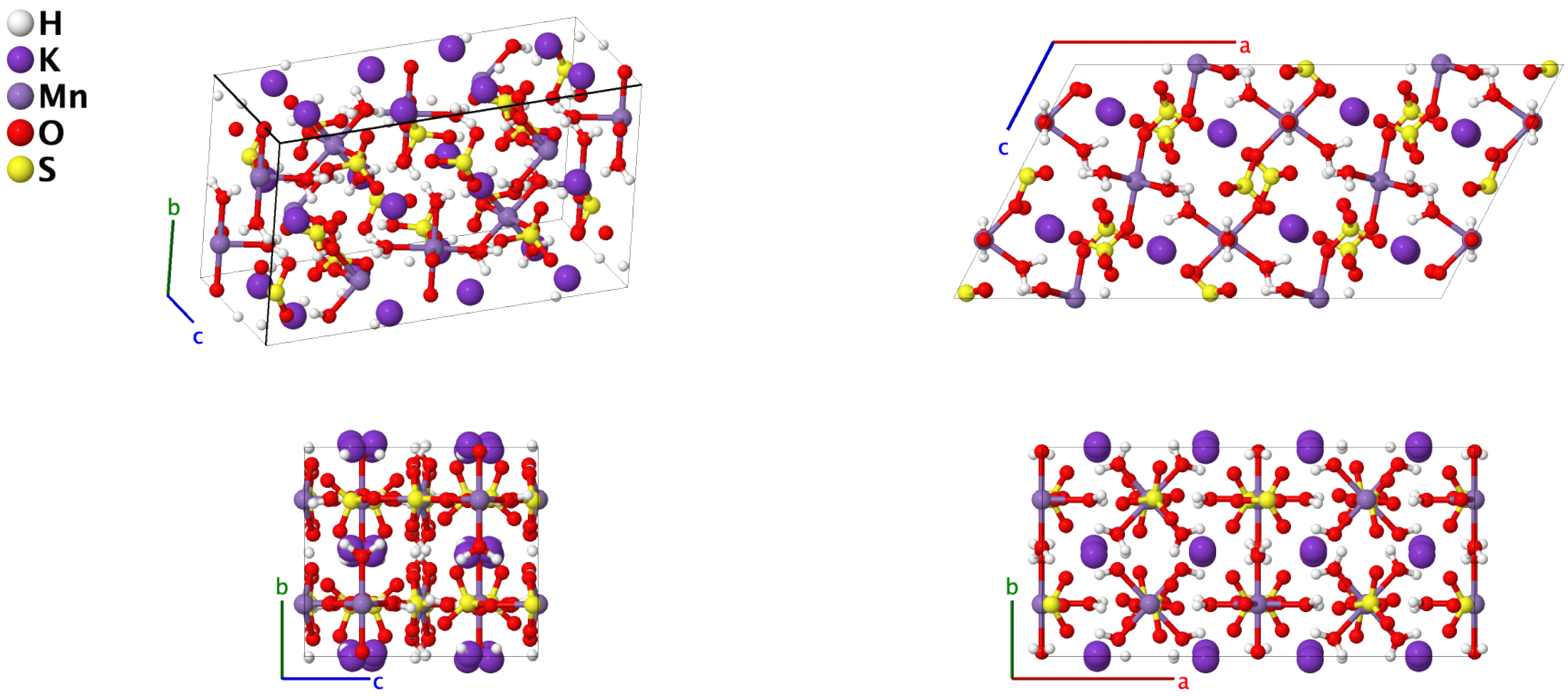

| | | |
|---|---|---|
| **Prototype** | : | $H_8K_2MnO_{12}S_2$ |
| **AFLOW prototype label** | : | A8B2CD12E2_mC200_15_8f_2f_ce_2e11f_2f |
| ***Strukturbericht* designation** | : | None |
| **Pearson symbol** | : | mC200 |
| **Space group number** | : | 15 |
| **Space group symbol** | : | $C2/c$ |
| **AFLOW prototype command** | : | `aflow --proto=A8B2CD12E2_mC200_15_8f_2f_ce_2e11f_2f` `--params=`$a, b/a, c/a, \beta, y_2, y_3, y_4, x_5, y_5, z_5, x_6, y_6, z_6, x_7, y_7, z_7, x_8, y_8, z_8, x_9, y_9, z_9, x_{10}, y_{10}, z_{10}, x_{11}, y_{11}, z_{11}, x_{12}, y_{12}, z_{12}, x_{13}, y_{13}, z_{13}, x_{14}, y_{14}, z_{14}, x_{15}, y_{15}, z_{15}, x_{16}, y_{16}, z_{16}, x_{17}, y_{17}, z_{17}, x_{18}, y_{18}, z_{18}, x_{19}, y_{19}, z_{19}, x_{20}, y_{20}, z_{20}, x_{21}, y_{21}, z_{21}, x_{22}, y_{22}, z_{22}, x_{23}, y_{23}, z_{23}, x_{24}, y_{24}, z_{24}, x_{25}, y_{25}, z_{25}, x_{26}, y_{26}, z_{26}, x_{27}, y_{27}, z_{27}$ |

### Other compounds with this structure

- $K_2Mg(SO_4)_2{\cdot}4H_2O$ (leonite) and $K_2Fe(SO_4)_2{\cdot}4H_2O$ (mereiterite)

- This is the intermediate-low temperature structure of leonite. For Mn-leonite, discussed here, it is stable between 168 K and 205 K, and we show the structure at 185 K.
- The room temperature structure is *Strukturbericht* $H4_{23}$, space group $C2/m$ #13. The current structure doubles the size of the unit cell and orders all of the $SO_4$ radicals.
- The low temperature structure has space group $P2_1/c$ #14.
- (Hertweck, 2001) give crystallographic information of the 185 K phase in the $I2/a$ setting of space group #15, with the origin supposedly shifted by (1/4, 1/4, 1/2) from the -1 point on the $a$-glide plane (the symmetry operations for this setting may be found here). We were unable to use their data to construct a realistic crystal structure. Instead, we used the interpretation of their results by (Villars, 2016) to put the structure in the standard $C2/c$ setting of space group #15.

**Base-centered Monoclinic primitive vectors:**

$$\mathbf{a}_1 = \tfrac{1}{2}\,a\,\hat{\mathbf{x}} - \tfrac{1}{2}\,b\,\hat{\mathbf{y}}$$
$$\mathbf{a}_2 = \tfrac{1}{2}\,a\,\hat{\mathbf{x}} + \tfrac{1}{2}\,b\,\hat{\mathbf{y}}$$
$$\mathbf{a}_3 = c\cos\beta\,\hat{\mathbf{x}} + c\sin\beta\,\hat{\mathbf{z}}$$

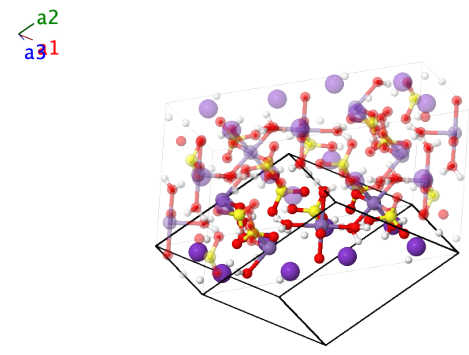

**Basis vectors:**

| | | Lattice Coordinates | | Cartesian Coordinates | Wyckoff Position | Atom Type |
|---|---|---|---|---|---|---|
| $\mathbf{B}_1$ | = | $\frac{1}{2}\,\mathbf{a}_2$ | = | $\frac{1}{4}a\,\hat{\mathbf{x}} + \frac{1}{4}b\,\hat{\mathbf{y}}$ | (4$c$) | Mn I |
| $\mathbf{B}_2$ | = | $\frac{1}{2}\,\mathbf{a}_1 + \frac{1}{2}\,\mathbf{a}_3$ | = | $\left(\frac{1}{4}a + \frac{1}{2}c\cos\beta\right)\,\hat{\mathbf{x}} - \frac{1}{4}b\,\hat{\mathbf{y}} + \frac{1}{2}c\sin\beta\,\hat{\mathbf{z}}$ | (4$c$) | Mn I |
| $\mathbf{B}_3$ | = | $-y_2\,\mathbf{a}_1 + y_2\,\mathbf{a}_2 + \frac{1}{4}\,\mathbf{a}_3$ | = | $\frac{1}{4}c\cos\beta\,\hat{\mathbf{x}} + y_2b\,\hat{\mathbf{y}} + \frac{1}{4}c\sin\beta\,\hat{\mathbf{z}}$ | (4$e$) | Mn II |
| $\mathbf{B}_4$ | = | $y_2\,\mathbf{a}_1 - y_2\,\mathbf{a}_2 + \frac{3}{4}\,\mathbf{a}_3$ | = | $\frac{3}{4}c\cos\beta\,\hat{\mathbf{x}} - y_2b\,\hat{\mathbf{y}} + \frac{3}{4}c\sin\beta\,\hat{\mathbf{z}}$ | (4$e$) | Mn II |
| $\mathbf{B}_5$ | = | $-y_3\,\mathbf{a}_1 + y_3\,\mathbf{a}_2 + \frac{1}{4}\,\mathbf{a}_3$ | = | $\frac{1}{4}c\cos\beta\,\hat{\mathbf{x}} + y_3b\,\hat{\mathbf{y}} + \frac{1}{4}c\sin\beta\,\hat{\mathbf{z}}$ | (4$e$) | O I |
| $\mathbf{B}_6$ | = | $y_3\,\mathbf{a}_1 - y_3\,\mathbf{a}_2 + \frac{3}{4}\,\mathbf{a}_3$ | = | $\frac{3}{4}c\cos\beta\,\hat{\mathbf{x}} - y_3b\,\hat{\mathbf{y}} + \frac{3}{4}c\sin\beta\,\hat{\mathbf{z}}$ | (4$e$) | O I |
| $\mathbf{B}_7$ | = | $-y_4\,\mathbf{a}_1 + y_4\,\mathbf{a}_2 + \frac{1}{4}\,\mathbf{a}_3$ | = | $\frac{1}{4}c\cos\beta\,\hat{\mathbf{x}} + y_4b\,\hat{\mathbf{y}} + \frac{1}{4}c\sin\beta\,\hat{\mathbf{z}}$ | (4$e$) | O II |
| $\mathbf{B}_8$ | = | $y_4\,\mathbf{a}_1 - y_4\,\mathbf{a}_2 + \frac{3}{4}\,\mathbf{a}_3$ | = | $\frac{3}{4}c\cos\beta\,\hat{\mathbf{x}} - y_4b\,\hat{\mathbf{y}} + \frac{3}{4}c\sin\beta\,\hat{\mathbf{z}}$ | (4$e$) | O II |
| $\mathbf{B}_9$ | = | $(x_5 - y_5)\,\mathbf{a}_1 + (x_5 + y_5)\,\mathbf{a}_2 + z_5\,\mathbf{a}_3$ | = | $(x_5a + z_5c\cos\beta)\,\hat{\mathbf{x}} + y_5b\,\hat{\mathbf{y}} + z_5c\sin\beta\,\hat{\mathbf{z}}$ | (8$f$) | H I |
| $\mathbf{B}_{10}$ | = | $(-x_5 - y_5)\,\mathbf{a}_1 + (-x_5 + y_5)\,\mathbf{a}_2 + \left(\frac{1}{2} - z_5\right)\,\mathbf{a}_3$ | = | $\left(\frac{1}{2}c\cos\beta - x_5a - z_5c\cos\beta\right)\,\hat{\mathbf{x}} + y_5b\,\hat{\mathbf{y}} + \left(\frac{1}{2} - z_5\right)c\sin\beta\,\hat{\mathbf{z}}$ | (8$f$) | H I |
| $\mathbf{B}_{11}$ | = | $(-x_5 + y_5)\,\mathbf{a}_1 + (-x_5 - y_5)\,\mathbf{a}_2 - z_5\,\mathbf{a}_3$ | = | $(-x_5a - z_5c\cos\beta)\,\hat{\mathbf{x}} - y_5b\,\hat{\mathbf{y}} - z_5c\sin\beta\,\hat{\mathbf{z}}$ | (8$f$) | H I |
| $\mathbf{B}_{12}$ | = | $(x_5 + y_5)\,\mathbf{a}_1 + (x_5 - y_5)\,\mathbf{a}_2 + \left(\frac{1}{2} + z_5\right)\,\mathbf{a}_3$ | = | $\left(\frac{1}{2}c\cos\beta + x_5a + z_5c\cos\beta\right)\,\hat{\mathbf{x}} - y_5b\,\hat{\mathbf{y}} + \left(\frac{1}{2} + z_5\right)c\sin\beta\,\hat{\mathbf{z}}$ | (8$f$) | H I |
| $\mathbf{B}_{13}$ | = | $(x_6 - y_6)\,\mathbf{a}_1 + (x_6 + y_6)\,\mathbf{a}_2 + z_6\,\mathbf{a}_3$ | = | $(x_6a + z_6c\cos\beta)\,\hat{\mathbf{x}} + y_6b\,\hat{\mathbf{y}} + z_6c\sin\beta\,\hat{\mathbf{z}}$ | (8$f$) | H II |
| $\mathbf{B}_{14}$ | = | $(-x_6 - y_6)\,\mathbf{a}_1 + (-x_6 + y_6)\,\mathbf{a}_2 + \left(\frac{1}{2} - z_6\right)\,\mathbf{a}_3$ | = | $\left(\frac{1}{2}c\cos\beta - x_6a - z_6c\cos\beta\right)\,\hat{\mathbf{x}} + y_6b\,\hat{\mathbf{y}} + \left(\frac{1}{2} - z_6\right)c\sin\beta\,\hat{\mathbf{z}}$ | (8$f$) | H II |
| $\mathbf{B}_{15}$ | = | $(-x_6 + y_6)\,\mathbf{a}_1 + (-x_6 - y_6)\,\mathbf{a}_2 - z_6\,\mathbf{a}_3$ | = | $(-x_6a - z_6c\cos\beta)\,\hat{\mathbf{x}} - y_6b\,\hat{\mathbf{y}} - z_6c\sin\beta\,\hat{\mathbf{z}}$ | (8$f$) | H II |
| $\mathbf{B}_{16}$ | = | $(x_6 + y_6)\,\mathbf{a}_1 + (x_6 - y_6)\,\mathbf{a}_2 + \left(\frac{1}{2} + z_6\right)\,\mathbf{a}_3$ | = | $\left(\frac{1}{2}c\cos\beta + x_6a + z_6c\cos\beta\right)\,\hat{\mathbf{x}} - y_6b\,\hat{\mathbf{y}} + \left(\frac{1}{2} + z_6\right)c\sin\beta\,\hat{\mathbf{z}}$ | (8$f$) | H II |
| $\mathbf{B}_{17}$ | = | $(x_7 - y_7)\,\mathbf{a}_1 + (x_7 + y_7)\,\mathbf{a}_2 + z_7\,\mathbf{a}_3$ | = | $(x_7a + z_7c\cos\beta)\,\hat{\mathbf{x}} + y_7b\,\hat{\mathbf{y}} + z_7c\sin\beta\,\hat{\mathbf{z}}$ | (8$f$) | H III |
| $\mathbf{B}_{18}$ | = | $(-x_7 - y_7)\,\mathbf{a}_1 + (-x_7 + y_7)\,\mathbf{a}_2 + \left(\frac{1}{2} - z_7\right)\,\mathbf{a}_3$ | = | $\left(\frac{1}{2}c\cos\beta - x_7a - z_7c\cos\beta\right)\,\hat{\mathbf{x}} + y_7b\,\hat{\mathbf{y}} + \left(\frac{1}{2} - z_7\right)c\sin\beta\,\hat{\mathbf{z}}$ | (8$f$) | H III |
| $\mathbf{B}_{19}$ | = | $(-x_7 + y_7)\,\mathbf{a}_1 + (-x_7 - y_7)\,\mathbf{a}_2 - z_7\,\mathbf{a}_3$ | = | $(-x_7a - z_7c\cos\beta)\,\hat{\mathbf{x}} - y_7b\,\hat{\mathbf{y}} - z_7c\sin\beta\,\hat{\mathbf{z}}$ | (8$f$) | H III |
| $\mathbf{B}_{20}$ | = | $(x_7 + y_7)\,\mathbf{a}_1 + (x_7 - y_7)\,\mathbf{a}_2 + \left(\frac{1}{2} + z_7\right)\,\mathbf{a}_3$ | = | $\left(\frac{1}{2}c\cos\beta + x_7a + z_7c\cos\beta\right)\,\hat{\mathbf{x}} - y_7b\,\hat{\mathbf{y}} + \left(\frac{1}{2} + z_7\right)c\sin\beta\,\hat{\mathbf{z}}$ | (8$f$) | H III |
| $\mathbf{B}_{21}$ | = | $(x_8 - y_8)\,\mathbf{a}_1 + (x_8 + y_8)\,\mathbf{a}_2 + z_8\,\mathbf{a}_3$ | = | $(x_8a + z_8c\cos\beta)\,\hat{\mathbf{x}} + y_8b\,\hat{\mathbf{y}} + z_8c\sin\beta\,\hat{\mathbf{z}}$ | (8$f$) | H IV |
| $\mathbf{B}_{22}$ | = | $\left(-x_{8}-y_{8}\right)\,\mathbf{a}_{1}+\left(-x_{8}+y_{8}\right)\,\mathbf{a}_{2}+\left(\frac{1}{2}-z_{8}\right)\,\mathbf{a}_{3}$ | = | $\left(\frac{1}{2}c\cos\beta-x_{8}a-z_{8}c\cos\beta\right)\,\mathbf{\hat{x}}+y_{8}b\,\mathbf{\hat{y}}+\left(\frac{1}{2}-z_{8}\right)c\sin\beta\,\mathbf{\hat{z}}$ | (8$f$) | H IV |
| $\mathbf{B}_{23}$ | = | $\left(-x_{8}+y_{8}\right)\,\mathbf{a}_{1}+\left(-x_{8}-y_{8}\right)\,\mathbf{a}_{2}-z_{8}\,\mathbf{a}_{3}$ | = | $\left(-x_{8}a-z_{8}c\cos\beta\right)\,\mathbf{\hat{x}}-y_{8}b\,\mathbf{\hat{y}}-z_{8}c\sin\beta\,\mathbf{\hat{z}}$ | (8$f$) | H IV |
| $\mathbf{B}_{24}$ | = | $\left(x_{8}+y_{8}\right)\,\mathbf{a}_{1}+\left(x_{8}-y_{8}\right)\,\mathbf{a}_{2}+\left(\frac{1}{2}+z_{8}\right)\,\mathbf{a}_{3}$ | = | $\left(\frac{1}{2}c\cos\beta+x_{8}a+z_{8}c\cos\beta\right)\,\mathbf{\hat{x}}-y_{8}b\,\mathbf{\hat{y}}+\left(\frac{1}{2}+z_{8}\right)c\sin\beta\,\mathbf{\hat{z}}$ | (8$f$) | H IV |
| $\mathbf{B}_{25}$ | = | $\left(x_{9}-y_{9}\right)\,\mathbf{a}_{1}+\left(x_{9}+y_{9}\right)\,\mathbf{a}_{2}+z_{9}\,\mathbf{a}_{3}$ | = | $\left(x_{9}a+z_{9}c\cos\beta\right)\,\mathbf{\hat{x}}+y_{9}b\,\mathbf{\hat{y}}+z_{9}c\sin\beta\,\mathbf{\hat{z}}$ | (8$f$) | H V |
| $\mathbf{B}_{26}$ | = | $\left(-x_{9}-y_{9}\right)\,\mathbf{a}_{1}+\left(-x_{9}+y_{9}\right)\,\mathbf{a}_{2}+\left(\frac{1}{2}-z_{9}\right)\,\mathbf{a}_{3}$ | = | $\left(\frac{1}{2}c\cos\beta-x_{9}a-z_{9}c\cos\beta\right)\,\mathbf{\hat{x}}+y_{9}b\,\mathbf{\hat{y}}+\left(\frac{1}{2}-z_{9}\right)c\sin\beta\,\mathbf{\hat{z}}$ | (8$f$) | H V |
| $\mathbf{B}_{27}$ | = | $\left(-x_{9}+y_{9}\right)\,\mathbf{a}_{1}+\left(-x_{9}-y_{9}\right)\,\mathbf{a}_{2}-z_{9}\,\mathbf{a}_{3}$ | = | $\left(-x_{9}a-z_{9}c\cos\beta\right)\,\mathbf{\hat{x}}-y_{9}b\,\mathbf{\hat{y}}-z_{9}c\sin\beta\,\mathbf{\hat{z}}$ | (8$f$) | H V |
| $\mathbf{B}_{28}$ | = | $\left(x_{9}+y_{9}\right)\,\mathbf{a}_{1}+\left(x_{9}-y_{9}\right)\,\mathbf{a}_{2}+\left(\frac{1}{2}+z_{9}\right)\,\mathbf{a}_{3}$ | = | $\left(\frac{1}{2}c\cos\beta+x_{9}a+z_{9}c\cos\beta\right)\,\mathbf{\hat{x}}-y_{9}b\,\mathbf{\hat{y}}+\left(\frac{1}{2}+z_{9}\right)c\sin\beta\,\mathbf{\hat{z}}$ | (8$f$) | H V |
| $\mathbf{B}_{29}$ | = | $\left(x_{10}-y_{10}\right)\,\mathbf{a}_{1}+\left(x_{10}+y_{10}\right)\,\mathbf{a}_{2}+z_{10}\,\mathbf{a}_{3}$ | = | $\left(x_{10}a+z_{10}c\cos\beta\right)\,\mathbf{\hat{x}}+y_{10}b\,\mathbf{\hat{y}}+z_{10}c\sin\beta\,\mathbf{\hat{z}}$ | (8$f$) | H VI |
| $\mathbf{B}_{30}$ | = | $\left(-x_{10}-y_{10}\right)\,\mathbf{a}_{1}+\left(-x_{10}+y_{10}\right)\,\mathbf{a}_{2}+\left(\frac{1}{2}-z_{10}\right)\,\mathbf{a}_{3}$ | = | $\left(\frac{1}{2}c\cos\beta-x_{10}a-z_{10}c\cos\beta\right)\,\mathbf{\hat{x}}+y_{10}b\,\mathbf{\hat{y}}+\left(\frac{1}{2}-z_{10}\right)c\sin\beta\,\mathbf{\hat{z}}$ | (8$f$) | H VI |
| $\mathbf{B}_{31}$ | = | $\left(-x_{10}+y_{10}\right)\,\mathbf{a}_{1}+\left(-x_{10}-y_{10}\right)\,\mathbf{a}_{2}-z_{10}\,\mathbf{a}_{3}$ | = | $\left(-x_{10}a-z_{10}c\cos\beta\right)\,\mathbf{\hat{x}}-y_{10}b\,\mathbf{\hat{y}}-z_{10}c\sin\beta\,\mathbf{\hat{z}}$ | (8$f$) | H VI |
| $\mathbf{B}_{32}$ | = | $\left(x_{10}+y_{10}\right)\,\mathbf{a}_{1}+\left(x_{10}-y_{10}\right)\,\mathbf{a}_{2}+\left(\frac{1}{2}+z_{10}\right)\,\mathbf{a}_{3}$ | = | $\left(\frac{1}{2}c\cos\beta+x_{10}a+z_{10}c\cos\beta\right)\,\mathbf{\hat{x}}-y_{10}b\,\mathbf{\hat{y}}+\left(\frac{1}{2}+z_{10}\right)c\sin\beta\,\mathbf{\hat{z}}$ | (8$f$) | H VI |
| $\mathbf{B}_{33}$ | = | $\left(x_{11}-y_{11}\right)\,\mathbf{a}_{1}+\left(x_{11}+y_{11}\right)\,\mathbf{a}_{2}+z_{11}\,\mathbf{a}_{3}$ | = | $\left(x_{11}a+z_{11}c\cos\beta\right)\,\mathbf{\hat{x}}+y_{11}b\,\mathbf{\hat{y}}+z_{11}c\sin\beta\,\mathbf{\hat{z}}$ | (8$f$) | H VII |
| $\mathbf{B}_{34}$ | = | $\left(-x_{11}-y_{11}\right)\,\mathbf{a}_{1}+\left(-x_{11}+y_{11}\right)\,\mathbf{a}_{2}+\left(\frac{1}{2}-z_{11}\right)\,\mathbf{a}_{3}$ | = | $\left(\frac{1}{2}c\cos\beta-x_{11}a-z_{11}c\cos\beta\right)\,\mathbf{\hat{x}}+y_{11}b\,\mathbf{\hat{y}}+\left(\frac{1}{2}-z_{11}\right)c\sin\beta\,\mathbf{\hat{z}}$ | (8$f$) | H VII |
| $\mathbf{B}_{35}$ | = | $\left(-x_{11}+y_{11}\right)\,\mathbf{a}_{1}+\left(-x_{11}-y_{11}\right)\,\mathbf{a}_{2}-z_{11}\,\mathbf{a}_{3}$ | = | $\left(-x_{11}a-z_{11}c\cos\beta\right)\,\mathbf{\hat{x}}-y_{11}b\,\mathbf{\hat{y}}-z_{11}c\sin\beta\,\mathbf{\hat{z}}$ | (8$f$) | H VII |
| $\mathbf{B}_{36}$ | = | $\left(x_{11}+y_{11}\right)\,\mathbf{a}_{1}+\left(x_{11}-y_{11}\right)\,\mathbf{a}_{2}+\left(\frac{1}{2}+z_{11}\right)\,\mathbf{a}_{3}$ | = | $\left(\frac{1}{2}c\cos\beta+x_{11}a+z_{11}c\cos\beta\right)\,\mathbf{\hat{x}}-y_{11}b\,\mathbf{\hat{y}}+\left(\frac{1}{2}+z_{11}\right)c\sin\beta\,\mathbf{\hat{z}}$ | (8$f$) | H VII |
| $\mathbf{B}_{37}$ | = | $\left(x_{12}-y_{12}\right)\,\mathbf{a}_{1}+\left(x_{12}+y_{12}\right)\,\mathbf{a}_{2}+z_{12}\,\mathbf{a}_{3}$ | = | $\left(x_{12}a+z_{12}c\cos\beta\right)\,\mathbf{\hat{x}}+y_{12}b\,\mathbf{\hat{y}}+z_{12}c\sin\beta\,\mathbf{\hat{z}}$ | (8$f$) | H VIII |
| $\mathbf{B}_{38}$ | = | $\left(-x_{12}-y_{12}\right)\,\mathbf{a}_{1}+\left(-x_{12}+y_{12}\right)\,\mathbf{a}_{2}+\left(\frac{1}{2}-z_{12}\right)\,\mathbf{a}_{3}$ | = | $\left(\frac{1}{2}c\cos\beta-x_{12}a-z_{12}c\cos\beta\right)\,\mathbf{\hat{x}}+y_{12}b\,\mathbf{\hat{y}}+\left(\frac{1}{2}-z_{12}\right)c\sin\beta\,\mathbf{\hat{z}}$ | (8$f$) | H VIII |
| $\mathbf{B}_{39}$ | = | $\left(-x_{12}+y_{12}\right)\,\mathbf{a}_{1}+\left(-x_{12}-y_{12}\right)\,\mathbf{a}_{2}-z_{12}\,\mathbf{a}_{3}$ | = | $\left(-x_{12}a-z_{12}c\cos\beta\right)\,\mathbf{\hat{x}}-y_{12}b\,\mathbf{\hat{y}}-z_{12}c\sin\beta\,\mathbf{\hat{z}}$ | (8$f$) | H VIII |
| $\mathbf{B}_{40}$ | = | $\left(x_{12}+y_{12}\right)\,\mathbf{a}_{1}+\left(x_{12}-y_{12}\right)\,\mathbf{a}_{2}+\left(\frac{1}{2}+z_{12}\right)\,\mathbf{a}_{3}$ | = | $\left(\frac{1}{2}c\cos\beta+x_{12}a+z_{12}c\cos\beta\right)\,\mathbf{\hat{x}}-y_{12}b\,\mathbf{\hat{y}}+\left(\frac{1}{2}+z_{12}\right)c\sin\beta\,\mathbf{\hat{z}}$ | (8$f$) | H VIII |
| $\mathbf{B}_{41}$ | = | $\left(x_{13}-y_{13}\right)\,\mathbf{a}_{1}+\left(x_{13}+y_{13}\right)\,\mathbf{a}_{2}+z_{13}\,\mathbf{a}_{3}$ | = | $\left(x_{13}a+z_{13}c\cos\beta\right)\,\mathbf{\hat{x}}+y_{13}b\,\mathbf{\hat{y}}+z_{13}c\sin\beta\,\mathbf{\hat{z}}$ | (8$f$) | K I |
| $\mathbf{B}_{42}$ | = | $\left(-x_{13}-y_{13}\right)\,\mathbf{a}_{1}+\left(-x_{13}+y_{13}\right)\,\mathbf{a}_{2}+\left(\frac{1}{2}-z_{13}\right)\,\mathbf{a}_{3}$ | = | $\left(\frac{1}{2}c\cos\beta-x_{13}a-z_{13}c\cos\beta\right)\,\mathbf{\hat{x}}+y_{13}b\,\mathbf{\hat{y}}+\left(\frac{1}{2}-z_{13}\right)c\sin\beta\,\mathbf{\hat{z}}$ | (8$f$) | K I |
| $\mathbf{B}_{43}$ | = | $\left(-x_{13}+y_{13}\right)\,\mathbf{a}_{1}+\left(-x_{13}-y_{13}\right)\,\mathbf{a}_{2}-z_{13}\,\mathbf{a}_{3}$ | = | $\left(-x_{13}a-z_{13}c\cos\beta\right)\,\mathbf{\hat{x}}-y_{13}b\,\mathbf{\hat{y}}-z_{13}c\sin\beta\,\mathbf{\hat{z}}$ | (8$f$) | K I |
| $\mathbf{B}_{44}$ | = | $(x_{13}+y_{13})\,\mathbf{a}_1 + (x_{13}-y_{13})\,\mathbf{a}_2 + \left(\frac{1}{2}+z_{13}\right)\mathbf{a}_3$ | = | $\left(\frac{1}{2}c\cos\beta + x_{13}a + z_{13}c\cos\beta\right)\hat{\mathbf{x}} - y_{13}b\,\hat{\mathbf{y}} + \left(\frac{1}{2}+z_{13}\right)c\sin\beta\,\hat{\mathbf{z}}$ | $(8f)$ | K I |
| $\mathbf{B}_{45}$ | = | $(x_{14}-y_{14})\,\mathbf{a}_1 + (x_{14}+y_{14})\,\mathbf{a}_2 + z_{14}\,\mathbf{a}_3$ | = | $(x_{14}a + z_{14}c\cos\beta)\,\hat{\mathbf{x}} + y_{14}b\,\hat{\mathbf{y}} + z_{14}c\sin\beta\,\hat{\mathbf{z}}$ | $(8f)$ | K II |
| $\mathbf{B}_{46}$ | = | $(-x_{14}-y_{14})\,\mathbf{a}_1 + (-x_{14}+y_{14})\,\mathbf{a}_2 + \left(\frac{1}{2}-z_{14}\right)\mathbf{a}_3$ | = | $\left(\frac{1}{2}c\cos\beta - x_{14}a - z_{14}c\cos\beta\right)\hat{\mathbf{x}} + y_{14}b\,\hat{\mathbf{y}} + \left(\frac{1}{2}-z_{14}\right)c\sin\beta\,\hat{\mathbf{z}}$ | $(8f)$ | K II |
| $\mathbf{B}_{47}$ | = | $(-x_{14}+y_{14})\,\mathbf{a}_1 + (-x_{14}-y_{14})\,\mathbf{a}_2 - z_{14}\,\mathbf{a}_3$ | = | $(-x_{14}a - z_{14}c\cos\beta)\,\hat{\mathbf{x}} - y_{14}b\,\hat{\mathbf{y}} - z_{14}c\sin\beta\,\hat{\mathbf{z}}$ | $(8f)$ | K II |
| $\mathbf{B}_{48}$ | = | $(x_{14}+y_{14})\,\mathbf{a}_1 + (x_{14}-y_{14})\,\mathbf{a}_2 + \left(\frac{1}{2}+z_{14}\right)\mathbf{a}_3$ | = | $\left(\frac{1}{2}c\cos\beta + x_{14}a + z_{14}c\cos\beta\right)\hat{\mathbf{x}} - y_{14}b\,\hat{\mathbf{y}} + \left(\frac{1}{2}+z_{14}\right)c\sin\beta\,\hat{\mathbf{z}}$ | $(8f)$ | K II |
| $\mathbf{B}_{49}$ | = | $(x_{15}-y_{15})\,\mathbf{a}_1 + (x_{15}+y_{15})\,\mathbf{a}_2 + z_{15}\,\mathbf{a}_3$ | = | $(x_{15}a + z_{15}c\cos\beta)\,\hat{\mathbf{x}} + y_{15}b\,\hat{\mathbf{y}} + z_{15}c\sin\beta\,\hat{\mathbf{z}}$ | $(8f)$ | O III |
| $\mathbf{B}_{50}$ | = | $(-x_{15}-y_{15})\,\mathbf{a}_1 + (-x_{15}+y_{15})\,\mathbf{a}_2 + \left(\frac{1}{2}-z_{15}\right)\mathbf{a}_3$ | = | $\left(\frac{1}{2}c\cos\beta - x_{15}a - z_{15}c\cos\beta\right)\hat{\mathbf{x}} + y_{15}b\,\hat{\mathbf{y}} + \left(\frac{1}{2}-z_{15}\right)c\sin\beta\,\hat{\mathbf{z}}$ | $(8f)$ | O III |
| $\mathbf{B}_{51}$ | = | $(-x_{15}+y_{15})\,\mathbf{a}_1 + (-x_{15}-y_{15})\,\mathbf{a}_2 - z_{15}\,\mathbf{a}_3$ | = | $(-x_{15}a - z_{15}c\cos\beta)\,\hat{\mathbf{x}} - y_{15}b\,\hat{\mathbf{y}} - z_{15}c\sin\beta\,\hat{\mathbf{z}}$ | $(8f)$ | O III |
| $\mathbf{B}_{52}$ | = | $(x_{15}+y_{15})\,\mathbf{a}_1 + (x_{15}-y_{15})\,\mathbf{a}_2 + \left(\frac{1}{2}+z_{15}\right)\mathbf{a}_3$ | = | $\left(\frac{1}{2}c\cos\beta + x_{15}a + z_{15}c\cos\beta\right)\hat{\mathbf{x}} - y_{15}b\,\hat{\mathbf{y}} + \left(\frac{1}{2}+z_{15}\right)c\sin\beta\,\hat{\mathbf{z}}$ | $(8f)$ | O III |
| $\mathbf{B}_{53}$ | = | $(x_{16}-y_{16})\,\mathbf{a}_1 + (x_{16}+y_{16})\,\mathbf{a}_2 + z_{16}\,\mathbf{a}_3$ | = | $(x_{16}a + z_{16}c\cos\beta)\,\hat{\mathbf{x}} + y_{16}b\,\hat{\mathbf{y}} + z_{16}c\sin\beta\,\hat{\mathbf{z}}$ | $(8f)$ | O IV |
| $\mathbf{B}_{54}$ | = | $(-x_{16}-y_{16})\,\mathbf{a}_1 + (-x_{16}+y_{16})\,\mathbf{a}_2 + \left(\frac{1}{2}-z_{16}\right)\mathbf{a}_3$ | = | $\left(\frac{1}{2}c\cos\beta - x_{16}a - z_{16}c\cos\beta\right)\hat{\mathbf{x}} + y_{16}b\,\hat{\mathbf{y}} + \left(\frac{1}{2}-z_{16}\right)c\sin\beta\,\hat{\mathbf{z}}$ | $(8f)$ | O IV |
| $\mathbf{B}_{55}$ | = | $(-x_{16}+y_{16})\,\mathbf{a}_1 + (-x_{16}-y_{16})\,\mathbf{a}_2 - z_{16}\,\mathbf{a}_3$ | = | $(-x_{16}a - z_{16}c\cos\beta)\,\hat{\mathbf{x}} - y_{16}b\,\hat{\mathbf{y}} - z_{16}c\sin\beta\,\hat{\mathbf{z}}$ | $(8f)$ | O IV |
| $\mathbf{B}_{56}$ | = | $(x_{16}+y_{16})\,\mathbf{a}_1 + (x_{16}-y_{16})\,\mathbf{a}_2 + \left(\frac{1}{2}+z_{16}\right)\mathbf{a}_3$ | = | $\left(\frac{1}{2}c\cos\beta + x_{16}a + z_{16}c\cos\beta\right)\hat{\mathbf{x}} - y_{16}b\,\hat{\mathbf{y}} + \left(\frac{1}{2}+z_{16}\right)c\sin\beta\,\hat{\mathbf{z}}$ | $(8f)$ | O IV |
| $\mathbf{B}_{57}$ | = | $(x_{17}-y_{17})\,\mathbf{a}_1 + (x_{17}+y_{17})\,\mathbf{a}_2 + z_{17}\,\mathbf{a}_3$ | = | $(x_{17}a + z_{17}c\cos\beta)\,\hat{\mathbf{x}} + y_{17}b\,\hat{\mathbf{y}} + z_{17}c\sin\beta\,\hat{\mathbf{z}}$ | $(8f)$ | O V |
| $\mathbf{B}_{58}$ | = | $(-x_{17}-y_{17})\,\mathbf{a}_1 + (-x_{17}+y_{17})\,\mathbf{a}_2 + \left(\frac{1}{2}-z_{17}\right)\mathbf{a}_3$ | = | $\left(\frac{1}{2}c\cos\beta - x_{17}a - z_{17}c\cos\beta\right)\hat{\mathbf{x}} + y_{17}b\,\hat{\mathbf{y}} + \left(\frac{1}{2}-z_{17}\right)c\sin\beta\,\hat{\mathbf{z}}$ | $(8f)$ | O V |
| $\mathbf{B}_{59}$ | = | $(-x_{17}+y_{17})\,\mathbf{a}_1 + (-x_{17}-y_{17})\,\mathbf{a}_2 - z_{17}\,\mathbf{a}_3$ | = | $(-x_{17}a - z_{17}c\cos\beta)\,\hat{\mathbf{x}} - y_{17}b\,\hat{\mathbf{y}} - z_{17}c\sin\beta\,\hat{\mathbf{z}}$ | $(8f)$ | O V |
| $\mathbf{B}_{60}$ | = | $(x_{17}+y_{17})\,\mathbf{a}_1 + (x_{17}-y_{17})\,\mathbf{a}_2 + \left(\frac{1}{2}+z_{17}\right)\mathbf{a}_3$ | = | $\left(\frac{1}{2}c\cos\beta + x_{17}a + z_{17}c\cos\beta\right)\hat{\mathbf{x}} - y_{17}b\,\hat{\mathbf{y}} + \left(\frac{1}{2}+z_{17}\right)c\sin\beta\,\hat{\mathbf{z}}$ | $(8f)$ | O V |
| $\mathbf{B}_{61}$ | = | $(x_{18}-y_{18})\,\mathbf{a}_1 + (x_{18}+y_{18})\,\mathbf{a}_2 + z_{18}\,\mathbf{a}_3$ | = | $(x_{18}a + z_{18}c\cos\beta)\,\hat{\mathbf{x}} + y_{18}b\,\hat{\mathbf{y}} + z_{18}c\sin\beta\,\hat{\mathbf{z}}$ | $(8f)$ | O VI |
| $\mathbf{B}_{62}$ | = | $(-x_{18}-y_{18})\,\mathbf{a}_1 + (-x_{18}+y_{18})\,\mathbf{a}_2 + \left(\frac{1}{2}-z_{18}\right)\mathbf{a}_3$ | = | $\left(\frac{1}{2}c\cos\beta - x_{18}a - z_{18}c\cos\beta\right)\hat{\mathbf{x}} + y_{18}b\,\hat{\mathbf{y}} + \left(\frac{1}{2}-z_{18}\right)c\sin\beta\,\hat{\mathbf{z}}$ | $(8f)$ | O VI |
| $\mathbf{B}_{63}$ | = | $(-x_{18}+y_{18})\,\mathbf{a}_1 + (-x_{18}-y_{18})\,\mathbf{a}_2 - z_{18}\,\mathbf{a}_3$ | = | $(-x_{18}a - z_{18}c\cos\beta)\,\hat{\mathbf{x}} - y_{18}b\,\hat{\mathbf{y}} - z_{18}c\sin\beta\,\hat{\mathbf{z}}$ | $(8f)$ | O VI |
| $\mathbf{B}_{64}$ | = | $(x_{18}+y_{18})\,\mathbf{a}_1 + (x_{18}-y_{18})\,\mathbf{a}_2 + \left(\frac{1}{2}+z_{18}\right)\mathbf{a}_3$ | = | $\left(\frac{1}{2}c\cos\beta + x_{18}a + z_{18}c\cos\beta\right)\hat{\mathbf{x}} - y_{18}b\,\hat{\mathbf{y}} + \left(\frac{1}{2}+z_{18}\right)c\sin\beta\,\hat{\mathbf{z}}$ | $(8f)$ | O VI |
| $\mathbf{B}_{65}$ | = | $(x_{19}-y_{19})\,\mathbf{a}_1 + (x_{19}+y_{19})\,\mathbf{a}_2 + z_{19}\,\mathbf{a}_3$ | = | $(x_{19}a + z_{19}c\cos\beta)\,\hat{\mathbf{x}} + y_{19}b\,\hat{\mathbf{y}} + z_{19}c\sin\beta\,\hat{\mathbf{z}}$ | $(8f)$ | O VII |
| $\mathbf{B}_{66}$ | = | $(-x_{19}-y_{19})\,\mathbf{a}_1 + (-x_{19}+y_{19})\,\mathbf{a}_2 + \left(\frac{1}{2}-z_{19}\right)\mathbf{a}_3$ | = | $\left(\frac{1}{2}c\cos\beta - x_{19}a - z_{19}c\cos\beta\right)\hat{\mathbf{x}} + y_{19}b\,\hat{\mathbf{y}} + \left(\frac{1}{2}-z_{19}\right)c\sin\beta\,\hat{\mathbf{z}}$ | $(8f)$ | O VII |
| $\mathbf{B}_{67}$ | = | $(-x_{19}+y_{19})\,\mathbf{a}_1 + (-x_{19}-y_{19})\,\mathbf{a}_2 - z_{19}\,\mathbf{a}_3$ | = | $(-x_{19}a - z_{19}c\cos\beta)\,\hat{\mathbf{x}} - y_{19}b\,\hat{\mathbf{y}} - z_{19}c\sin\beta\,\hat{\mathbf{z}}$ | $(8f)$ | O VII |
| $\mathbf{B}_{68}$ | = | $(x_{19}+y_{19})\,\mathbf{a}_1 + (x_{19}-y_{19})\,\mathbf{a}_2 + \left(\frac{1}{2}+z_{19}\right)\mathbf{a}_3$ | = | $\left(\frac{1}{2}c\cos\beta + x_{19}a + z_{19}c\cos\beta\right)\hat{\mathbf{x}} - y_{19}b\,\hat{\mathbf{y}} + \left(\frac{1}{2}+z_{19}\right)c\sin\beta\,\hat{\mathbf{z}}$ | $(8f)$ | O VII |
| $\mathbf{B}_{69}$ | = | $(x_{20}-y_{20})\,\mathbf{a}_1 + (x_{20}+y_{20})\,\mathbf{a}_2 + z_{20}\,\mathbf{a}_3$ | = | $(x_{20}a + z_{20}c\cos\beta)\,\hat{\mathbf{x}} + y_{20}b\,\hat{\mathbf{y}} + z_{20}c\sin\beta\,\hat{\mathbf{z}}$ | $(8f)$ | O VIII |
| $\mathbf{B}_{70}$ | = | $(-x_{20}-y_{20})\,\mathbf{a}_1 + (-x_{20}+y_{20})\,\mathbf{a}_2 + \left(\frac{1}{2}-z_{20}\right)\mathbf{a}_3$ | = | $\left(\frac{1}{2}c\cos\beta - x_{20}a - z_{20}c\cos\beta\right)\hat{\mathbf{x}} + y_{20}b\,\hat{\mathbf{y}} + \left(\frac{1}{2}-z_{20}\right)c\sin\beta\,\hat{\mathbf{z}}$ | $(8f)$ | O VIII |
| $\mathbf{B}_{71}$ | = | $(-x_{20}+y_{20})\,\mathbf{a}_1 + (-x_{20}-y_{20})\,\mathbf{a}_2 - z_{20}\,\mathbf{a}_3$ | = | $(-x_{20}a - z_{20}c\cos\beta)\,\hat{\mathbf{x}} - y_{20}b\,\hat{\mathbf{y}} - z_{20}c\sin\beta\,\hat{\mathbf{z}}$ | $(8f)$ | O VIII |
| $\mathbf{B}_{72}$ | = | $(x_{20}+y_{20})\,\mathbf{a}_1 + (x_{20}-y_{20})\,\mathbf{a}_2 + \left(\frac{1}{2}+z_{20}\right)\mathbf{a}_3$ | = | $\left(\frac{1}{2}c\cos\beta + x_{20}a + z_{20}c\cos\beta\right)\hat{\mathbf{x}} - y_{20}b\,\hat{\mathbf{y}} + \left(\frac{1}{2}+z_{20}\right)c\sin\beta\,\hat{\mathbf{z}}$ | $(8f)$ | O VIII |
| $\mathbf{B}_{73}$ | = | $(x_{21}-y_{21})\,\mathbf{a}_1 + (x_{21}+y_{21})\,\mathbf{a}_2 + z_{21}\,\mathbf{a}_3$ | = | $(x_{21}a + z_{21}c\cos\beta)\,\hat{\mathbf{x}} + y_{21}b\,\hat{\mathbf{y}} + z_{21}c\sin\beta\,\hat{\mathbf{z}}$ | $(8f)$ | O IX |
| $\mathbf{B}_{74}$ | = | $(-x_{21}-y_{21})\,\mathbf{a}_1 + (-x_{21}+y_{21})\,\mathbf{a}_2 + \left(\frac{1}{2}-z_{21}\right)\mathbf{a}_3$ | = | $\left(\frac{1}{2}c\cos\beta - x_{21}a - z_{21}c\cos\beta\right)\hat{\mathbf{x}} + y_{21}b\,\hat{\mathbf{y}} + \left(\frac{1}{2}-z_{21}\right)c\sin\beta\,\hat{\mathbf{z}}$ | $(8f)$ | O IX |
| $\mathbf{B}_{75}$ | = | $(-x_{21}+y_{21})\,\mathbf{a}_1 + (-x_{21}-y_{21})\,\mathbf{a}_2 - z_{21}\,\mathbf{a}_3$ | = | $(-x_{21}a - z_{21}c\cos\beta)\,\hat{\mathbf{x}} - y_{21}b\,\hat{\mathbf{y}} - z_{21}c\sin\beta\,\hat{\mathbf{z}}$ | $(8f)$ | O IX |
| $\mathbf{B}_{76}$ | = | $(x_{21}+y_{21})\,\mathbf{a}_1 + (x_{21}-y_{21})\,\mathbf{a}_2 + \left(\frac{1}{2}+z_{21}\right)\mathbf{a}_3$ | = | $\left(\frac{1}{2}c\cos\beta + x_{21}a + z_{21}c\cos\beta\right)\hat{\mathbf{x}} - y_{21}b\,\hat{\mathbf{y}} + \left(\frac{1}{2}+z_{21}\right)c\sin\beta\,\hat{\mathbf{z}}$ | $(8f)$ | O IX |
| $\mathbf{B}_{77}$ | = | $(x_{22}-y_{22})\,\mathbf{a}_1 + (x_{22}+y_{22})\,\mathbf{a}_2 + z_{22}\,\mathbf{a}_3$ | = | $(x_{22}a + z_{22}c\cos\beta)\,\hat{\mathbf{x}} + y_{22}b\,\hat{\mathbf{y}} + z_{22}c\sin\beta\,\hat{\mathbf{z}}$ | $(8f)$ | O X |
| $\mathbf{B}_{78}$ | = | $(-x_{22}-y_{22})\,\mathbf{a}_1 + (-x_{22}+y_{22})\,\mathbf{a}_2 + \left(\frac{1}{2}-z_{22}\right)\mathbf{a}_3$ | = | $\left(\frac{1}{2}c\cos\beta - x_{22}a - z_{22}c\cos\beta\right)\hat{\mathbf{x}} + y_{22}b\,\hat{\mathbf{y}} + \left(\frac{1}{2}-z_{22}\right)c\sin\beta\,\hat{\mathbf{z}}$ | $(8f)$ | O X |
| $\mathbf{B}_{79}$ | = | $(-x_{22}+y_{22})\,\mathbf{a}_1 + (-x_{22}-y_{22})\,\mathbf{a}_2 - z_{22}\,\mathbf{a}_3$ | = | $(-x_{22}a - z_{22}c\cos\beta)\,\hat{\mathbf{x}} - y_{22}b\,\hat{\mathbf{y}} - z_{22}c\sin\beta\,\hat{\mathbf{z}}$ | $(8f)$ | O X |
| $\mathbf{B}_{80}$ | = | $(x_{22}+y_{22})\,\mathbf{a}_1 + (x_{22}-y_{22})\,\mathbf{a}_2 + \left(\frac{1}{2}+z_{22}\right)\mathbf{a}_3$ | = | $\left(\frac{1}{2}c\cos\beta + x_{22}a + z_{22}c\cos\beta\right)\hat{\mathbf{x}} - y_{22}b\,\hat{\mathbf{y}} + \left(\frac{1}{2}+z_{22}\right)c\sin\beta\,\hat{\mathbf{z}}$ | $(8f)$ | O X |
| $\mathbf{B}_{81}$ | = | $(x_{23}-y_{23})\,\mathbf{a}_1 + (x_{23}+y_{23})\,\mathbf{a}_2 + z_{23}\,\mathbf{a}_3$ | = | $(x_{23}a + z_{23}c\cos\beta)\,\hat{\mathbf{x}} + y_{23}b\,\hat{\mathbf{y}} + z_{23}c\sin\beta\,\hat{\mathbf{z}}$ | $(8f)$ | O XI |
| $\mathbf{B}_{82}$ | = | $(-x_{23}-y_{23})\,\mathbf{a}_1 + (-x_{23}+y_{23})\,\mathbf{a}_2 + \left(\frac{1}{2}-z_{23}\right)\mathbf{a}_3$ | = | $\left(\frac{1}{2}c\cos\beta - x_{23}a - z_{23}c\cos\beta\right)\hat{\mathbf{x}} + y_{23}b\,\hat{\mathbf{y}} + \left(\frac{1}{2}-z_{23}\right)c\sin\beta\,\hat{\mathbf{z}}$ | $(8f)$ | O XI |
| $\mathbf{B}_{83}$ | = | $(-x_{23}+y_{23})\,\mathbf{a}_1 + (-x_{23}-y_{23})\,\mathbf{a}_2 - z_{23}\,\mathbf{a}_3$ | = | $(-x_{23}a - z_{23}c\cos\beta)\,\hat{\mathbf{x}} - y_{23}b\,\hat{\mathbf{y}} - z_{23}c\sin\beta\,\hat{\mathbf{z}}$ | $(8f)$ | O XI |
| $\mathbf{B}_{84}$ | = | $(x_{23}+y_{23})\,\mathbf{a}_1 + (x_{23}-y_{23})\,\mathbf{a}_2 + \left(\frac{1}{2}+z_{23}\right)\mathbf{a}_3$ | = | $\left(\frac{1}{2}c\cos\beta + x_{23}a + z_{23}c\cos\beta\right)\hat{\mathbf{x}} - y_{23}b\,\hat{\mathbf{y}} + \left(\frac{1}{2}+z_{23}\right)c\sin\beta\,\hat{\mathbf{z}}$ | $(8f)$ | O XI |
| $\mathbf{B}_{85}$ | = | $(x_{24}-y_{24})\,\mathbf{a}_1 + (x_{24}+y_{24})\,\mathbf{a}_2 + z_{24}\,\mathbf{a}_3$ | = | $(x_{24}a + z_{24}c\cos\beta)\,\hat{\mathbf{x}} + y_{24}b\,\hat{\mathbf{y}} + z_{24}c\sin\beta\,\hat{\mathbf{z}}$ | $(8f)$ | O XII |
| $\mathbf{B}_{86}$ | = | $(-x_{24}-y_{24})\,\mathbf{a}_1 + (-x_{24}+y_{24})\,\mathbf{a}_2 + \left(\frac{1}{2}-z_{24}\right)\mathbf{a}_3$ | = | $\left(\frac{1}{2}c\cos\beta - x_{24}a - z_{24}c\cos\beta\right)\hat{\mathbf{x}} + y_{24}b\,\hat{\mathbf{y}} + \left(\frac{1}{2}-z_{24}\right)c\sin\beta\,\hat{\mathbf{z}}$ | $(8f)$ | O XII |
| $\mathbf{B}_{87}$ | = | $(-x_{24}+y_{24})\,\mathbf{a}_1 + (-x_{24}-y_{24})\,\mathbf{a}_2 - z_{24}\,\mathbf{a}_3$ | = | $(-x_{24}a - z_{24}c\cos\beta)\,\hat{\mathbf{x}} - y_{24}b\,\hat{\mathbf{y}} - z_{24}c\sin\beta\,\hat{\mathbf{z}}$ | $(8f)$ | O XII |
| $\mathbf{B}_{88}$ | = | $(x_{24}+y_{24})\,\mathbf{a}_1+(x_{24}-y_{24})\,\mathbf{a}_2+\left(\frac{1}{2}+z_{24}\right)\mathbf{a}_3$ | = | $\left(\frac{1}{2}c\cos\beta+x_{24}a+z_{24}c\cos\beta\right)\hat{\mathbf{x}}-y_{24}b\,\hat{\mathbf{y}}+\left(\frac{1}{2}+z_{24}\right)c\sin\beta\,\hat{\mathbf{z}}$ | (8*f*) | O XII |
| $\mathbf{B}_{89}$ | = | $(x_{25}-y_{25})\,\mathbf{a}_1+(x_{25}+y_{25})\,\mathbf{a}_2+z_{25}\,\mathbf{a}_3$ | = | $(x_{25}a+z_{25}c\cos\beta)\,\hat{\mathbf{x}}+y_{25}b\,\hat{\mathbf{y}}+z_{25}c\sin\beta\,\hat{\mathbf{z}}$ | (8*f*) | O XIII |
| $\mathbf{B}_{90}$ | = | $(-x_{25}-y_{25})\,\mathbf{a}_1+(-x_{25}+y_{25})\,\mathbf{a}_2+\left(\frac{1}{2}-z_{25}\right)\mathbf{a}_3$ | = | $\left(\frac{1}{2}c\cos\beta-x_{25}a-z_{25}c\cos\beta\right)\hat{\mathbf{x}}+y_{25}b\,\hat{\mathbf{y}}+\left(\frac{1}{2}-z_{25}\right)c\sin\beta\,\hat{\mathbf{z}}$ | (8*f*) | O XIII |
| $\mathbf{B}_{91}$ | = | $(-x_{25}+y_{25})\,\mathbf{a}_1+(-x_{25}-y_{25})\,\mathbf{a}_2-z_{25}\,\mathbf{a}_3$ | = | $(-x_{25}a-z_{25}c\cos\beta)\,\hat{\mathbf{x}}-y_{25}b\,\hat{\mathbf{y}}-z_{25}c\sin\beta\,\hat{\mathbf{z}}$ | (8*f*) | O XIII |
| $\mathbf{B}_{92}$ | = | $(x_{25}+y_{25})\,\mathbf{a}_1+(x_{25}-y_{25})\,\mathbf{a}_2+\left(\frac{1}{2}+z_{25}\right)\mathbf{a}_3$ | = | $\left(\frac{1}{2}c\cos\beta+x_{25}a+z_{25}c\cos\beta\right)\hat{\mathbf{x}}-y_{25}b\,\hat{\mathbf{y}}+\left(\frac{1}{2}+z_{25}\right)c\sin\beta\,\hat{\mathbf{z}}$ | (8*f*) | O XIII |
| $\mathbf{B}_{93}$ | = | $(x_{26}-y_{26})\,\mathbf{a}_1+(x_{26}+y_{26})\,\mathbf{a}_2+z_{26}\,\mathbf{a}_3$ | = | $(x_{26}a+z_{26}c\cos\beta)\,\hat{\mathbf{x}}+y_{26}b\,\hat{\mathbf{y}}+z_{26}c\sin\beta\,\hat{\mathbf{z}}$ | (8*f*) | S I |
| $\mathbf{B}_{94}$ | = | $(-x_{26}-y_{26})\,\mathbf{a}_1+(-x_{26}+y_{26})\,\mathbf{a}_2+\left(\frac{1}{2}-z_{26}\right)\mathbf{a}_3$ | = | $\left(\frac{1}{2}c\cos\beta-x_{26}a-z_{26}c\cos\beta\right)\hat{\mathbf{x}}+y_{26}b\,\hat{\mathbf{y}}+\left(\frac{1}{2}-z_{26}\right)c\sin\beta\,\hat{\mathbf{z}}$ | (8*f*) | S I |
| $\mathbf{B}_{95}$ | = | $(-x_{26}+y_{26})\,\mathbf{a}_1+(-x_{26}-y_{26})\,\mathbf{a}_2-z_{26}\,\mathbf{a}_3$ | = | $(-x_{26}a-z_{26}c\cos\beta)\,\hat{\mathbf{x}}-y_{26}b\,\hat{\mathbf{y}}-z_{26}c\sin\beta\,\hat{\mathbf{z}}$ | (8*f*) | S I |
| $\mathbf{B}_{96}$ | = | $(x_{26}+y_{26})\,\mathbf{a}_1+(x_{26}-y_{26})\,\mathbf{a}_2+\left(\frac{1}{2}+z_{26}\right)\mathbf{a}_3$ | = | $\left(\frac{1}{2}c\cos\beta+x_{26}a+z_{26}c\cos\beta\right)\hat{\mathbf{x}}-y_{26}b\,\hat{\mathbf{y}}+\left(\frac{1}{2}+z_{26}\right)c\sin\beta\,\hat{\mathbf{z}}$ | (8*f*) | S I |
| $\mathbf{B}_{97}$ | = | $(x_{27}-y_{27})\,\mathbf{a}_1+(x_{27}+y_{27})\,\mathbf{a}_2+z_{27}\,\mathbf{a}_3$ | = | $(x_{27}a+z_{27}c\cos\beta)\,\hat{\mathbf{x}}+y_{27}b\,\hat{\mathbf{y}}+z_{27}c\sin\beta\,\hat{\mathbf{z}}$ | (8*f*) | S II |
| $\mathbf{B}_{98}$ | = | $(-x_{27}-y_{27})\,\mathbf{a}_1+(-x_{27}+y_{27})\,\mathbf{a}_2+\left(\frac{1}{2}-z_{27}\right)\mathbf{a}_3$ | = | $\left(\frac{1}{2}c\cos\beta-x_{27}a-z_{27}c\cos\beta\right)\hat{\mathbf{x}}+y_{27}b\,\hat{\mathbf{y}}+\left(\frac{1}{2}-z_{27}\right)c\sin\beta\,\hat{\mathbf{z}}$ | (8*f*) | S II |
| $\mathbf{B}_{99}$ | = | $(-x_{27}+y_{27})\,\mathbf{a}_1+(-x_{27}-y_{27})\,\mathbf{a}_2-z_{27}\,\mathbf{a}_3$ | = | $(-x_{27}a-z_{27}c\cos\beta)\,\hat{\mathbf{x}}-y_{27}b\,\hat{\mathbf{y}}-z_{27}c\sin\beta\,\hat{\mathbf{z}}$ | (8*f*) | S II |
| $\mathbf{B}_{100}$ | = | $(x_{27}+y_{27})\,\mathbf{a}_1+(x_{27}-y_{27})\,\mathbf{a}_2+\left(\frac{1}{2}+z_{27}\right)\mathbf{a}_3$ | = | $\left(\frac{1}{2}c\cos\beta+x_{27}a+z_{27}c\cos\beta\right)\hat{\mathbf{x}}-y_{27}b\,\hat{\mathbf{y}}+\left(\frac{1}{2}+z_{27}\right)c\sin\beta\,\hat{\mathbf{z}}$ | (8*f*) | S II |

**Geometry files:**

- CIF: pp. 1569
- POSCAR: pp. 1569

# $BaNi(CN)_4 \cdot 4H_2O$ ($H4_{22}$) Structure: AB4C4D4E_mC56_15_e_2f_2f_2f_a

http://aflow.org/prototype-encyclopedia/AB4C4D4E_mC56_15_e_2f_2f_2f_a

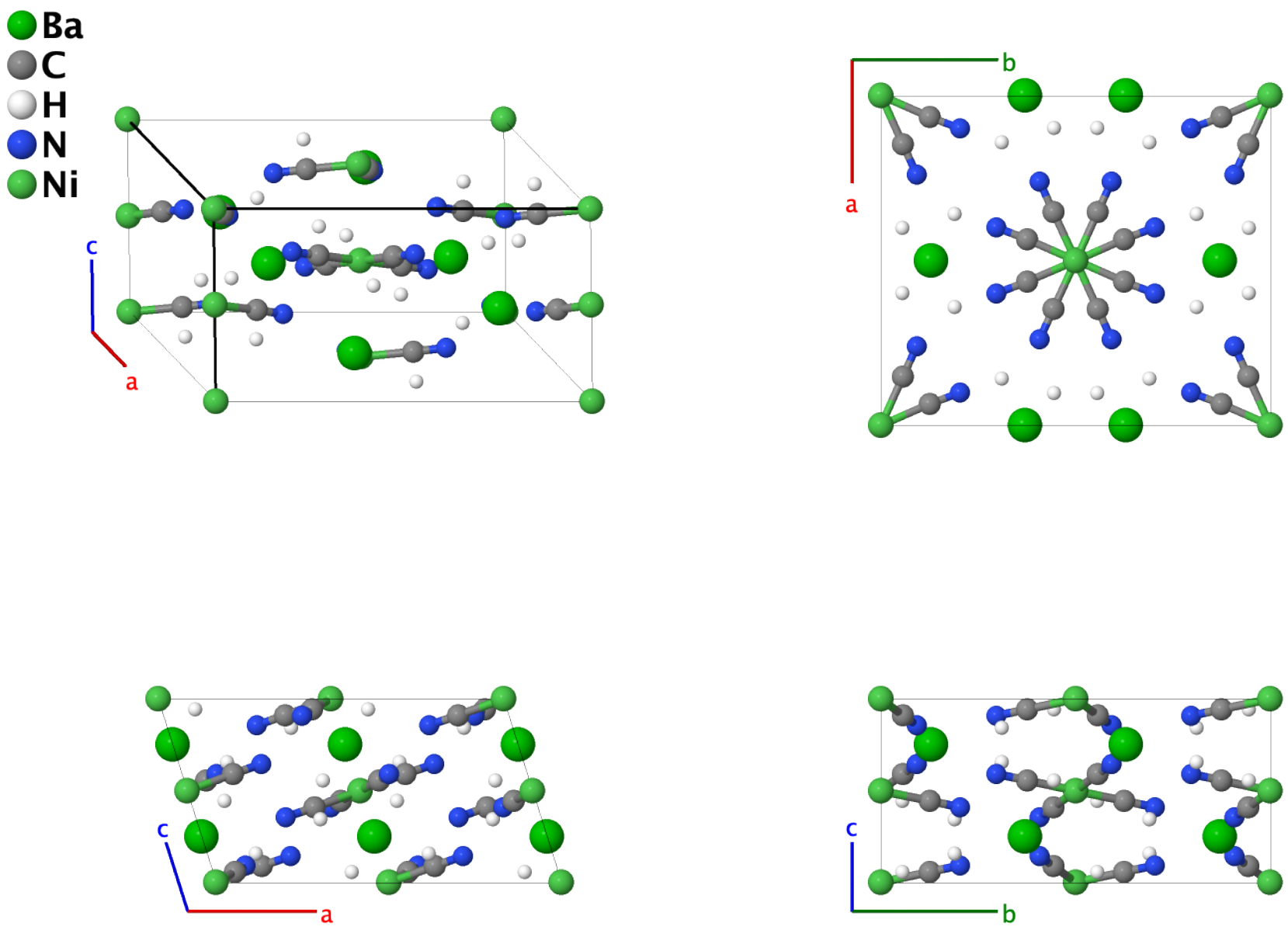

| | | |
|---|---|---|
| **Prototype** | : | $BaC_4(H_2O)_4N_4Ni$ |
| **AFLOW prototype label** | : | AB4C4D4E_mC56_15_e_2f_2f_2f_a |
| ***Strukturbericht* designation** | : | $H4_{22}$ |
| **Pearson symbol** | : | mC56 |
| **Space group number** | : | 15 |
| **Space group symbol** | : | $C2/c$ |
| **AFLOW prototype command** | : | `aflow --proto=AB4C4D4E_mC56_15_e_2f_2f_2f_a --params=`$a, b/a, c/a, \beta, y_2, x_3, y_3, z_3, x_4, y_4, z_4, x_5, y_5, z_5, x_6, y_6, z_6, x_7, y_7, z_7, x_8, y_8, z_8$ |

**Other compounds with this structure**

- $BaC_4N_4Pd \cdot 4H_2O$ and $BaC_4N_4Pt \cdot 4H_2O$

- We use the structure found by (Larsen, 1969). It is very similar to the structure found by (Brasseur, 1938) which was designated $H4_{22}$ by (Herrmann, 1941). The newer structure was able to separate the carbon and nitrogen atoms in the cyanide (CN) radical, and also move the barium atom from the (4*b*) to the (4*e*) Wyckoff position.

**Base-centered Monoclinic primitive vectors:**

$$
\begin{aligned}
\mathbf{a}_1 &= \tfrac{1}{2}\,a\,\hat{\mathbf{x}} - \tfrac{1}{2}\,b\,\hat{\mathbf{y}} \\
\mathbf{a}_2 &= \tfrac{1}{2}\,a\,\hat{\mathbf{x}} + \tfrac{1}{2}\,b\,\hat{\mathbf{y}} \\
\mathbf{a}_3 &= c\cos\beta\,\hat{\mathbf{x}} + c\sin\beta\,\hat{\mathbf{z}}
\end{aligned}
$$

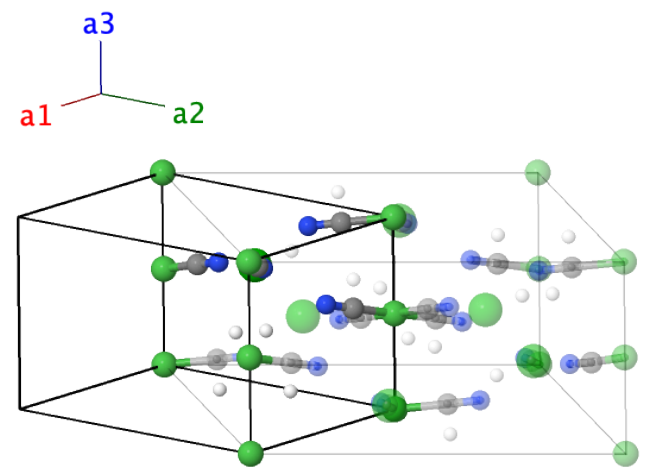

**Basis vectors:**

| | | Lattice Coordinates | | Cartesian Coordinates | Wyckoff Position | Atom Type |
|---|---|---|---|---|---|---|
| $\mathbf{B}_1$ | = | $0\,\mathbf{a}_1 + 0\,\mathbf{a}_2 + 0\,\mathbf{a}_3$ | = | $0\,\hat{\mathbf{x}} + 0\,\hat{\mathbf{y}} + 0\,\hat{\mathbf{z}}$ | (4$a$) | Ni |
| $\mathbf{B}_2$ | = | $\frac{1}{2}\,\mathbf{a}_3$ | = | $\frac{1}{2}c\cos\beta\,\hat{\mathbf{x}} + \frac{1}{2}c\sin\beta\,\hat{\mathbf{z}}$ | (4$a$) | Ni |
| $\mathbf{B}_3$ | = | $-y_2\,\mathbf{a}_1 + y_2\,\mathbf{a}_2 + \frac{1}{4}\,\mathbf{a}_3$ | = | $\frac{1}{4}c\cos\beta\,\hat{\mathbf{x}} + y_2 b\,\hat{\mathbf{y}} + \frac{1}{4}c\sin\beta\,\hat{\mathbf{z}}$ | (4$e$) | Ba |
| $\mathbf{B}_4$ | = | $y_2\,\mathbf{a}_1 - y_2\,\mathbf{a}_2 + \frac{3}{4}\,\mathbf{a}_3$ | = | $\frac{3}{4}c\cos\beta\,\hat{\mathbf{x}} - y_2 b\,\hat{\mathbf{y}} + \frac{3}{4}c\sin\beta\,\hat{\mathbf{z}}$ | (4$e$) | Ba |
| $\mathbf{B}_5$ | = | $(x_3 - y_3)\,\mathbf{a}_1 + (x_3 + y_3)\,\mathbf{a}_2 + z_3\,\mathbf{a}_3$ | = | $(x_3 a + z_3 c\cos\beta)\,\hat{\mathbf{x}} + y_3 b\,\hat{\mathbf{y}} + z_3 c\sin\beta\,\hat{\mathbf{z}}$ | (8$f$) | C I |
| $\mathbf{B}_6$ | = | $(-x_3 - y_3)\,\mathbf{a}_1 + (-x_3 + y_3)\,\mathbf{a}_2 + \left(\frac{1}{2} - z_3\right)\mathbf{a}_3$ | = | $\left(\frac{1}{2}c\cos\beta - x_3 a - z_3 c\cos\beta\right)\hat{\mathbf{x}} + y_3 b\,\hat{\mathbf{y}} + \left(\frac{1}{2} - z_3\right)c\sin\beta\,\hat{\mathbf{z}}$ | (8$f$) | C I |
| $\mathbf{B}_7$ | = | $(-x_3 + y_3)\,\mathbf{a}_1 + (-x_3 - y_3)\,\mathbf{a}_2 - z_3\,\mathbf{a}_3$ | = | $(-x_3 a - z_3 c\cos\beta)\,\hat{\mathbf{x}} - y_3 b\,\hat{\mathbf{y}} - z_3 c\sin\beta\,\hat{\mathbf{z}}$ | (8$f$) | C I |
| $\mathbf{B}_8$ | = | $(x_3 + y_3)\,\mathbf{a}_1 + (x_3 - y_3)\,\mathbf{a}_2 + \left(\frac{1}{2} + z_3\right)\mathbf{a}_3$ | = | $\left(\frac{1}{2}c\cos\beta + x_3 a + z_3 c\cos\beta\right)\hat{\mathbf{x}} - y_3 b\,\hat{\mathbf{y}} + \left(\frac{1}{2} + z_3\right)c\sin\beta\,\hat{\mathbf{z}}$ | (8$f$) | C I |
| $\mathbf{B}_9$ | = | $(x_4 - y_4)\,\mathbf{a}_1 + (x_4 + y_4)\,\mathbf{a}_2 + z_4\,\mathbf{a}_3$ | = | $(x_4 a + z_4 c\cos\beta)\,\hat{\mathbf{x}} + y_4 b\,\hat{\mathbf{y}} + z_4 c\sin\beta\,\hat{\mathbf{z}}$ | (8$f$) | C II |
| $\mathbf{B}_{10}$ | = | $(-x_4 - y_4)\,\mathbf{a}_1 + (-x_4 + y_4)\,\mathbf{a}_2 + \left(\frac{1}{2} - z_4\right)\mathbf{a}_3$ | = | $\left(\frac{1}{2}c\cos\beta - x_4 a - z_4 c\cos\beta\right)\hat{\mathbf{x}} + y_4 b\,\hat{\mathbf{y}} + \left(\frac{1}{2} - z_4\right)c\sin\beta\,\hat{\mathbf{z}}$ | (8$f$) | C II |
| $\mathbf{B}_{11}$ | = | $(-x_4 + y_4)\,\mathbf{a}_1 + (-x_4 - y_4)\,\mathbf{a}_2 - z_4\,\mathbf{a}_3$ | = | $(-x_4 a - z_4 c\cos\beta)\,\hat{\mathbf{x}} - y_4 b\,\hat{\mathbf{y}} - z_4 c\sin\beta\,\hat{\mathbf{z}}$ | (8$f$) | C II |
| $\mathbf{B}_{12}$ | = | $(x_4 + y_4)\,\mathbf{a}_1 + (x_4 - y_4)\,\mathbf{a}_2 + \left(\frac{1}{2} + z_4\right)\mathbf{a}_3$ | = | $\left(\frac{1}{2}c\cos\beta + x_4 a + z_4 c\cos\beta\right)\hat{\mathbf{x}} - y_4 b\,\hat{\mathbf{y}} + \left(\frac{1}{2} + z_4\right)c\sin\beta\,\hat{\mathbf{z}}$ | (8$f$) | C II |
| $\mathbf{B}_{13}$ | = | $(x_5 - y_5)\,\mathbf{a}_1 + (x_5 + y_5)\,\mathbf{a}_2 + z_5\,\mathbf{a}_3$ | = | $(x_5 a + z_5 c\cos\beta)\,\hat{\mathbf{x}} + y_5 b\,\hat{\mathbf{y}} + z_5 c\sin\beta\,\hat{\mathbf{z}}$ | (8$f$) | $H_2O$ I |
| $\mathbf{B}_{14}$ | = | $(-x_5 - y_5)\,\mathbf{a}_1 + (-x_5 + y_5)\,\mathbf{a}_2 + \left(\frac{1}{2} - z_5\right)\mathbf{a}_3$ | = | $\left(\frac{1}{2}c\cos\beta - x_5 a - z_5 c\cos\beta\right)\hat{\mathbf{x}} + y_5 b\,\hat{\mathbf{y}} + \left(\frac{1}{2} - z_5\right)c\sin\beta\,\hat{\mathbf{z}}$ | (8$f$) | $H_2O$ I |
| $\mathbf{B}_{15}$ | = | $(-x_5 + y_5)\,\mathbf{a}_1 + (-x_5 - y_5)\,\mathbf{a}_2 - z_5\,\mathbf{a}_3$ | = | $(-x_5 a - z_5 c\cos\beta)\,\hat{\mathbf{x}} - y_5 b\,\hat{\mathbf{y}} - z_5 c\sin\beta\,\hat{\mathbf{z}}$ | (8$f$) | $H_2O$ I |
| $\mathbf{B}_{16}$ | = | $(x_5 + y_5)\,\mathbf{a}_1 + (x_5 - y_5)\,\mathbf{a}_2 + \left(\frac{1}{2} + z_5\right)\mathbf{a}_3$ | = | $\left(\frac{1}{2}c\cos\beta + x_5 a + z_5 c\cos\beta\right)\hat{\mathbf{x}} - y_5 b\,\hat{\mathbf{y}} + \left(\frac{1}{2} + z_5\right)c\sin\beta\,\hat{\mathbf{z}}$ | (8$f$) | $H_2O$ I |
| $\mathbf{B}_{17}$ | = | $(x_6 - y_6)\,\mathbf{a}_1 + (x_6 + y_6)\,\mathbf{a}_2 + z_6\,\mathbf{a}_3$ | = | $(x_6 a + z_6 c\cos\beta)\,\hat{\mathbf{x}} + y_6 b\,\hat{\mathbf{y}} + z_6 c\sin\beta\,\hat{\mathbf{z}}$ | (8$f$) | $H_2O$ II |
| $\mathbf{B}_{18}$ | = | $(-x_6 - y_6)\,\mathbf{a}_1 + (-x_6 + y_6)\,\mathbf{a}_2 + \left(\frac{1}{2} - z_6\right)\mathbf{a}_3$ | = | $\left(\frac{1}{2}c\cos\beta - x_6 a - z_6 c\cos\beta\right)\hat{\mathbf{x}} + y_6 b\,\hat{\mathbf{y}} + \left(\frac{1}{2} - z_6\right)c\sin\beta\,\hat{\mathbf{z}}$ | (8$f$) | $H_2O$ II |
| $\mathbf{B}_{19}$ | = | $\left(-x_{6}+y_{6}\right)\,\mathbf{a}_{1}+\left(-x_{6}-y_{6}\right)\,\mathbf{a}_{2}-z_{6}\,\mathbf{a}_{3}$ | = | $\left(-x_{6}a-z_{6}c\cos\beta\right)\,\mathbf{\hat{x}}-y_{6}b\,\mathbf{\hat{y}}-z_{6}c\sin\beta\,\mathbf{\hat{z}}$ | $(8f)$ | $H_2O$ II |
| $\mathbf{B}_{20}$ | = | $\left(x_{6}+y_{6}\right)\,\mathbf{a}_{1}+\left(x_{6}-y_{6}\right)\,\mathbf{a}_{2}+\left(\frac{1}{2}+z_{6}\right)\,\mathbf{a}_{3}$ | = | $\left(\frac{1}{2}c\cos\beta+x_{6}a+z_{6}c\cos\beta\right)\,\mathbf{\hat{x}}-y_{6}b\,\mathbf{\hat{y}}+\left(\frac{1}{2}+z_{6}\right)c\sin\beta\,\mathbf{\hat{z}}$ | $(8f)$ | $H_2O$ II |
| $\mathbf{B}_{21}$ | = | $\left(x_{7}-y_{7}\right)\,\mathbf{a}_{1}+\left(x_{7}+y_{7}\right)\,\mathbf{a}_{2}+z_{7}\,\mathbf{a}_{3}$ | = | $\left(x_{7}a+z_{7}c\cos\beta\right)\,\mathbf{\hat{x}}+y_{7}b\,\mathbf{\hat{y}}+z_{7}c\sin\beta\,\mathbf{\hat{z}}$ | $(8f)$ | N I |
| $\mathbf{B}_{22}$ | = | $\left(-x_{7}-y_{7}\right)\,\mathbf{a}_{1}+\left(-x_{7}+y_{7}\right)\,\mathbf{a}_{2}+\left(\frac{1}{2}-z_{7}\right)\,\mathbf{a}_{3}$ | = | $\left(\frac{1}{2}c\cos\beta-x_{7}a-z_{7}c\cos\beta\right)\,\mathbf{\hat{x}}+y_{7}b\,\mathbf{\hat{y}}+\left(\frac{1}{2}-z_{7}\right)c\sin\beta\,\mathbf{\hat{z}}$ | $(8f)$ | N I |
| $\mathbf{B}_{23}$ | = | $\left(-x_{7}+y_{7}\right)\,\mathbf{a}_{1}+\left(-x_{7}-y_{7}\right)\,\mathbf{a}_{2}-z_{7}\,\mathbf{a}_{3}$ | = | $\left(-x_{7}a-z_{7}c\cos\beta\right)\,\mathbf{\hat{x}}-y_{7}b\,\mathbf{\hat{y}}-z_{7}c\sin\beta\,\mathbf{\hat{z}}$ | $(8f)$ | N I |
| $\mathbf{B}_{24}$ | = | $\left(x_{7}+y_{7}\right)\,\mathbf{a}_{1}+\left(x_{7}-y_{7}\right)\,\mathbf{a}_{2}+\left(\frac{1}{2}+z_{7}\right)\,\mathbf{a}_{3}$ | = | $\left(\frac{1}{2}c\cos\beta+x_{7}a+z_{7}c\cos\beta\right)\,\mathbf{\hat{x}}-y_{7}b\,\mathbf{\hat{y}}+\left(\frac{1}{2}+z_{7}\right)c\sin\beta\,\mathbf{\hat{z}}$ | $(8f)$ | N I |
| $\mathbf{B}_{25}$ | = | $\left(x_{8}-y_{8}\right)\,\mathbf{a}_{1}+\left(x_{8}+y_{8}\right)\,\mathbf{a}_{2}+z_{8}\,\mathbf{a}_{3}$ | = | $\left(x_{8}a+z_{8}c\cos\beta\right)\,\mathbf{\hat{x}}+y_{8}b\,\mathbf{\hat{y}}+z_{8}c\sin\beta\,\mathbf{\hat{z}}$ | $(8f)$ | N II |
| $\mathbf{B}_{26}$ | = | $\left(-x_{8}-y_{8}\right)\,\mathbf{a}_{1}+\left(-x_{8}+y_{8}\right)\,\mathbf{a}_{2}+\left(\frac{1}{2}-z_{8}\right)\,\mathbf{a}_{3}$ | = | $\left(\frac{1}{2}c\cos\beta-x_{8}a-z_{8}c\cos\beta\right)\,\mathbf{\hat{x}}+y_{8}b\,\mathbf{\hat{y}}+\left(\frac{1}{2}-z_{8}\right)c\sin\beta\,\mathbf{\hat{z}}$ | $(8f)$ | N II |
| $\mathbf{B}_{27}$ | = | $\left(-x_{8}+y_{8}\right)\,\mathbf{a}_{1}+\left(-x_{8}-y_{8}\right)\,\mathbf{a}_{2}-z_{8}\,\mathbf{a}_{3}$ | = | $\left(-x_{8}a-z_{8}c\cos\beta\right)\,\mathbf{\hat{x}}-y_{8}b\,\mathbf{\hat{y}}-z_{8}c\sin\beta\,\mathbf{\hat{z}}$ | $(8f)$ | N II |
| $\mathbf{B}_{28}$ | = | $\left(x_{8}+y_{8}\right)\,\mathbf{a}_{1}+\left(x_{8}-y_{8}\right)\,\mathbf{a}_{2}+\left(\frac{1}{2}+z_{8}\right)\,\mathbf{a}_{3}$ | = | $\left(\frac{1}{2}c\cos\beta+x_{8}a+z_{8}c\cos\beta\right)\,\mathbf{\hat{x}}-y_{8}b\,\mathbf{\hat{y}}+\left(\frac{1}{2}+z_{8}\right)c\sin\beta\,\mathbf{\hat{z}}$ | $(8f)$ | N II |

---

---

**Geometry files:**

- CIF: pp. 1570
- POSCAR: pp. 1570

# Gypsum ($CaSO_4{\cdot}2H_2O$, $H4_6$) Structure: AB4C6D_mC48_15_e_2f_3f_e

http://aflow.org/prototype-encyclopedia/AB4C6D_mC48_15_e_2f_3f_e

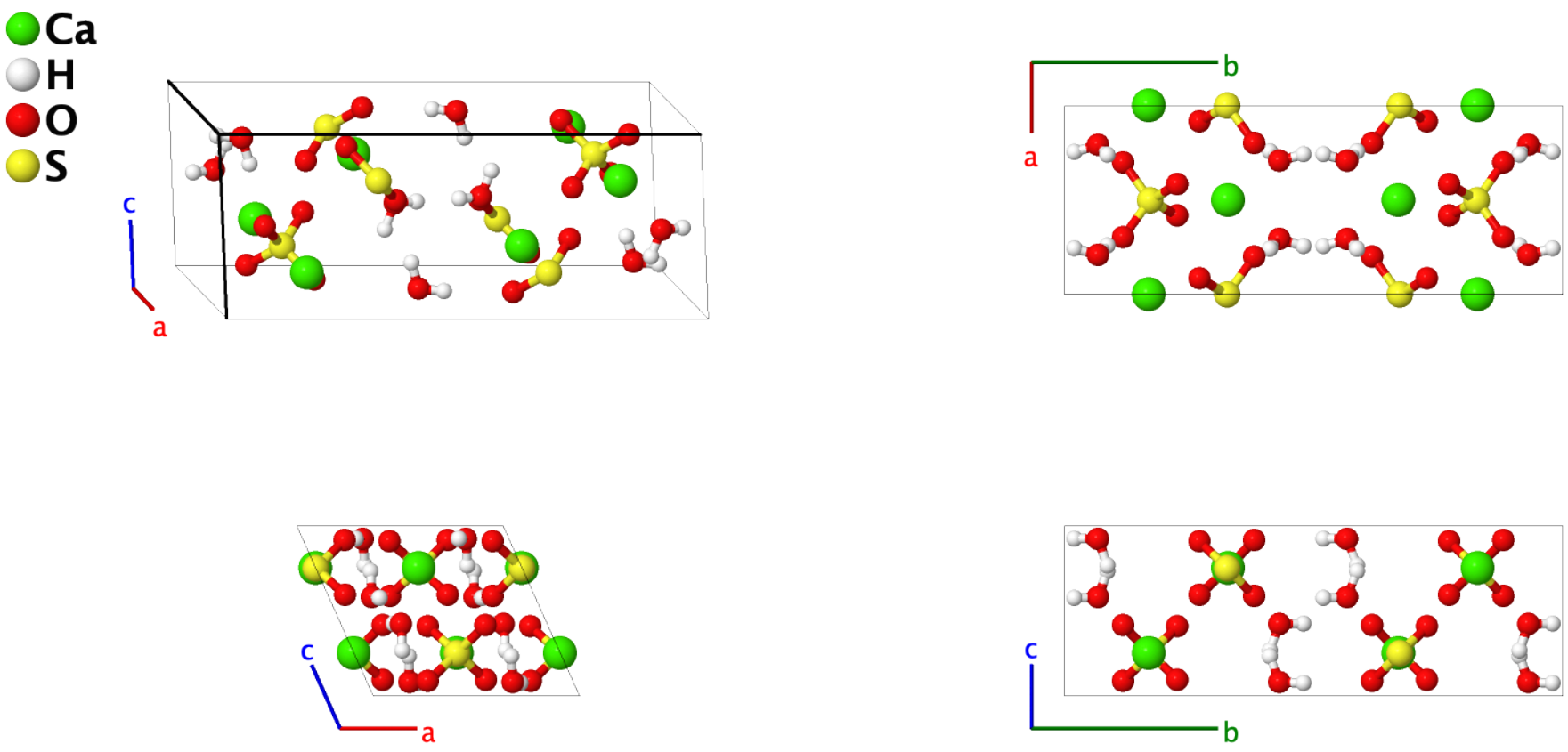

| | | |
|---|---|---|
| **Prototype** | : | $CaH_4O_6S$ |
| **AFLOW prototype label** | : | AB4C6D_mC48_15_e_2f_3f_e |
| ***Strukturbericht* designation** | : | $H4_6$ |
| **Pearson symbol** | : | mC48 |
| **Space group number** | : | 15 |
| **Space group symbol** | : | $C2/c$ |
| **AFLOW prototype command** | : | `aflow --proto=AB4C6D_mC48_15_e_2f_3f_e` `--params=`$a,b/a,c/a,\beta,y_1,y_2,x_3,y_3,z_3,x_4,y_4,z_4,x_5,y_5,z_5,x_6,y_6,z_6,x_7,y_7,z_7$ |

- (Onorato, 1929) determined that gypsum was in space group $C2/m$ #13, and (Hermann, 1937) assigned this to *Strukturbericht* $H4_6$. However, (Wooster, 1936) and others determined that the correct space group was $C2/c$ #15, and (Gottfried, 1938) assigned the $H4_6$ symbol to this structure. We use the 1 atmosphere data of (Comodi, 2008), who also determine the positions of the hydrogen atoms.

**Base-centered Monoclinic primitive vectors:**

$$\mathbf{a}_1 = \tfrac{1}{2}\,a\,\hat{\mathbf{x}} - \tfrac{1}{2}\,b\,\hat{\mathbf{y}}$$
$$\mathbf{a}_2 = \tfrac{1}{2}\,a\,\hat{\mathbf{x}} + \tfrac{1}{2}\,b\,\hat{\mathbf{y}}$$
$$\mathbf{a}_3 = c\cos\beta\,\hat{\mathbf{x}} + c\sin\beta\,\hat{\mathbf{z}}$$

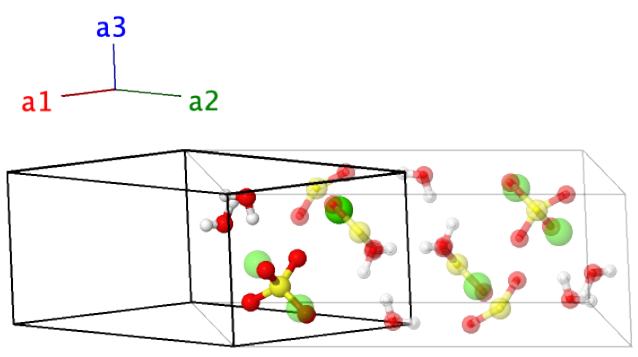

**Basis vectors:**

| | | Lattice Coordinates | | Cartesian Coordinates | Wyckoff Position | Atom Type |
|---|---|---|---|---|---|---|
| $\mathbf{B}_1$ | = | $-y_1\,\mathbf{a}_1 + y_1\,\mathbf{a}_2 + \frac{1}{4}\,\mathbf{a}_3$ | = | $\frac{1}{4}c\cos\beta\,\hat{\mathbf{x}} + y_1 b\,\hat{\mathbf{y}} + \frac{1}{4}c\sin\beta\,\hat{\mathbf{z}}$ | (4$e$) | Ca |
| $\mathbf{B}_2$ | = | $y_1\,\mathbf{a}_1 - y_1\,\mathbf{a}_2 + \frac{3}{4}\,\mathbf{a}_3$ | = | $\frac{3}{4}c\cos\beta\,\hat{\mathbf{x}} - y_1 b\,\hat{\mathbf{y}} + \frac{3}{4}c\sin\beta\,\hat{\mathbf{z}}$ | (4$e$) | Ca |
| $\mathbf{B}_3$ | = | $-y_2\,\mathbf{a}_1 + y_2\,\mathbf{a}_2 + \frac{1}{4}\,\mathbf{a}_3$ | = | $\frac{1}{4}c\cos\beta\,\hat{\mathbf{x}} + y_2 b\,\hat{\mathbf{y}} + \frac{1}{4}c\sin\beta\,\hat{\mathbf{z}}$ | (4$e$) | S |
| $\mathbf{B}_4$ | = | $y_2\,\mathbf{a}_1 - y_2\,\mathbf{a}_2 + \frac{3}{4}\,\mathbf{a}_3$ | = | $\frac{3}{4}c\cos\beta\,\hat{\mathbf{x}} - y_2 b\,\hat{\mathbf{y}} + \frac{3}{4}c\sin\beta\,\hat{\mathbf{z}}$ | (4$e$) | S |
| $\mathbf{B}_5$ | = | $(x_3 - y_3)\,\mathbf{a}_1 + (x_3 + y_3)\,\mathbf{a}_2 + z_3\,\mathbf{a}_3$ | = | $(x_3 a + z_3 c\cos\beta)\,\hat{\mathbf{x}} + y_3 b\,\hat{\mathbf{y}} + z_3 c\sin\beta\,\hat{\mathbf{z}}$ | (8$f$) | H I |
| $\mathbf{B}_6$ | = | $(-x_3 - y_3)\,\mathbf{a}_1 + (-x_3 + y_3)\,\mathbf{a}_2 + \left(\frac{1}{2} - z_3\right)\mathbf{a}_3$ | = | $\left(\frac{1}{2}c\cos\beta - x_3 a - z_3 c\cos\beta\right)\hat{\mathbf{x}} + y_3 b\,\hat{\mathbf{y}} + \left(\frac{1}{2} - z_3\right)c\sin\beta\,\hat{\mathbf{z}}$ | (8$f$) | H I |
| $\mathbf{B}_7$ | = | $(-x_3 + y_3)\,\mathbf{a}_1 + (-x_3 - y_3)\,\mathbf{a}_2 - z_3\,\mathbf{a}_3$ | = | $(-x_3 a - z_3 c\cos\beta)\,\hat{\mathbf{x}} - y_3 b\,\hat{\mathbf{y}} - z_3 c\sin\beta\,\hat{\mathbf{z}}$ | (8$f$) | H I |
| $\mathbf{B}_8$ | = | $(x_3 + y_3)\,\mathbf{a}_1 + (x_3 - y_3)\,\mathbf{a}_2 + \left(\frac{1}{2} + z_3\right)\mathbf{a}_3$ | = | $\left(\frac{1}{2}c\cos\beta + x_3 a + z_3 c\cos\beta\right)\hat{\mathbf{x}} - y_3 b\,\hat{\mathbf{y}} + \left(\frac{1}{2} + z_3\right)c\sin\beta\,\hat{\mathbf{z}}$ | (8$f$) | H I |
| $\mathbf{B}_9$ | = | $(x_4 - y_4)\,\mathbf{a}_1 + (x_4 + y_4)\,\mathbf{a}_2 + z_4\,\mathbf{a}_3$ | = | $(x_4 a + z_4 c\cos\beta)\,\hat{\mathbf{x}} + y_4 b\,\hat{\mathbf{y}} + z_4 c\sin\beta\,\hat{\mathbf{z}}$ | (8$f$) | H II |
| $\mathbf{B}_{10}$ | = | $(-x_4 - y_4)\,\mathbf{a}_1 + (-x_4 + y_4)\,\mathbf{a}_2 + \left(\frac{1}{2} - z_4\right)\mathbf{a}_3$ | = | $\left(\frac{1}{2}c\cos\beta - x_4 a - z_4 c\cos\beta\right)\hat{\mathbf{x}} + y_4 b\,\hat{\mathbf{y}} + \left(\frac{1}{2} - z_4\right)c\sin\beta\,\hat{\mathbf{z}}$ | (8$f$) | H II |
| $\mathbf{B}_{11}$ | = | $(-x_4 + y_4)\,\mathbf{a}_1 + (-x_4 - y_4)\,\mathbf{a}_2 - z_4\,\mathbf{a}_3$ | = | $(-x_4 a - z_4 c\cos\beta)\,\hat{\mathbf{x}} - y_4 b\,\hat{\mathbf{y}} - z_4 c\sin\beta\,\hat{\mathbf{z}}$ | (8$f$) | H II |
| $\mathbf{B}_{12}$ | = | $(x_4 + y_4)\,\mathbf{a}_1 + (x_4 - y_4)\,\mathbf{a}_2 + \left(\frac{1}{2} + z_4\right)\mathbf{a}_3$ | = | $\left(\frac{1}{2}c\cos\beta + x_4 a + z_4 c\cos\beta\right)\hat{\mathbf{x}} - y_4 b\,\hat{\mathbf{y}} + \left(\frac{1}{2} + z_4\right)c\sin\beta\,\hat{\mathbf{z}}$ | (8$f$) | H II |
| $\mathbf{B}_{13}$ | = | $(x_5 - y_5)\,\mathbf{a}_1 + (x_5 + y_5)\,\mathbf{a}_2 + z_5\,\mathbf{a}_3$ | = | $(x_5 a + z_5 c\cos\beta)\,\hat{\mathbf{x}} + y_5 b\,\hat{\mathbf{y}} + z_5 c\sin\beta\,\hat{\mathbf{z}}$ | (8$f$) | O I |
| $\mathbf{B}_{14}$ | = | $(-x_5 - y_5)\,\mathbf{a}_1 + (-x_5 + y_5)\,\mathbf{a}_2 + \left(\frac{1}{2} - z_5\right)\mathbf{a}_3$ | = | $\left(\frac{1}{2}c\cos\beta - x_5 a - z_5 c\cos\beta\right)\hat{\mathbf{x}} + y_5 b\,\hat{\mathbf{y}} + \left(\frac{1}{2} - z_5\right)c\sin\beta\,\hat{\mathbf{z}}$ | (8$f$) | O I |
| $\mathbf{B}_{15}$ | = | $(-x_5 + y_5)\,\mathbf{a}_1 + (-x_5 - y_5)\,\mathbf{a}_2 - z_5\,\mathbf{a}_3$ | = | $(-x_5 a - z_5 c\cos\beta)\,\hat{\mathbf{x}} - y_5 b\,\hat{\mathbf{y}} - z_5 c\sin\beta\,\hat{\mathbf{z}}$ | (8$f$) | O I |
| $\mathbf{B}_{16}$ | = | $(x_5 + y_5)\,\mathbf{a}_1 + (x_5 - y_5)\,\mathbf{a}_2 + \left(\frac{1}{2} + z_5\right)\mathbf{a}_3$ | = | $\left(\frac{1}{2}c\cos\beta + x_5 a + z_5 c\cos\beta\right)\hat{\mathbf{x}} - y_5 b\,\hat{\mathbf{y}} + \left(\frac{1}{2} + z_5\right)c\sin\beta\,\hat{\mathbf{z}}$ | (8$f$) | O I |
| $\mathbf{B}_{17}$ | = | $(x_6 - y_6)\,\mathbf{a}_1 + (x_6 + y_6)\,\mathbf{a}_2 + z_6\,\mathbf{a}_3$ | = | $(x_6 a + z_6 c\cos\beta)\,\hat{\mathbf{x}} + y_6 b\,\hat{\mathbf{y}} + z_6 c\sin\beta\,\hat{\mathbf{z}}$ | (8$f$) | O II |
| $\mathbf{B}_{18}$ | = | $(-x_6 - y_6)\,\mathbf{a}_1 + (-x_6 + y_6)\,\mathbf{a}_2 + \left(\frac{1}{2} - z_6\right)\mathbf{a}_3$ | = | $\left(\frac{1}{2}c\cos\beta - x_6 a - z_6 c\cos\beta\right)\hat{\mathbf{x}} + y_6 b\,\hat{\mathbf{y}} + \left(\frac{1}{2} - z_6\right)c\sin\beta\,\hat{\mathbf{z}}$ | (8$f$) | O II |
| $\mathbf{B}_{19}$ | = | $(-x_6 + y_6)\,\mathbf{a}_1 + (-x_6 - y_6)\,\mathbf{a}_2 - z_6\,\mathbf{a}_3$ | = | $(-x_6 a - z_6 c\cos\beta)\,\hat{\mathbf{x}} - y_6 b\,\hat{\mathbf{y}} - z_6 c\sin\beta\,\hat{\mathbf{z}}$ | (8$f$) | O II |
| $\mathbf{B}_{20}$ | = | $(x_6 + y_6)\,\mathbf{a}_1 + (x_6 - y_6)\,\mathbf{a}_2 + \left(\frac{1}{2} + z_6\right)\mathbf{a}_3$ | = | $\left(\frac{1}{2}c\cos\beta + x_6 a + z_6 c\cos\beta\right)\hat{\mathbf{x}} - y_6 b\,\hat{\mathbf{y}} + \left(\frac{1}{2} + z_6\right)c\sin\beta\,\hat{\mathbf{z}}$ | (8$f$) | O II |
| $\mathbf{B}_{21}$ | = | $(x_7 - y_7)\,\mathbf{a}_1 + (x_7 + y_7)\,\mathbf{a}_2 + z_7\,\mathbf{a}_3$ | = | $(x_7 a + z_7 c\cos\beta)\,\hat{\mathbf{x}} + y_7 b\,\hat{\mathbf{y}} + z_7 c\sin\beta\,\hat{\mathbf{z}}$ | (8$f$) | O III |
| $\mathbf{B}_{22}$ | = | $(-x_7 - y_7)\,\mathbf{a}_1 + (-x_7 + y_7)\,\mathbf{a}_2 + \left(\frac{1}{2} - z_7\right)\mathbf{a}_3$ | = | $\left(\frac{1}{2}c\cos\beta - x_7 a - z_7 c\cos\beta\right)\hat{\mathbf{x}} + y_7 b\,\hat{\mathbf{y}} + \left(\frac{1}{2} - z_7\right)c\sin\beta\,\hat{\mathbf{z}}$ | (8$f$) | O III |
| $\mathbf{B}_{23}$ | = | $(-x_7 + y_7)\,\mathbf{a}_1 + (-x_7 - y_7)\,\mathbf{a}_2 - z_7\,\mathbf{a}_3$ | = | $(-x_7 a - z_7 c\cos\beta)\,\hat{\mathbf{x}} - y_7 b\,\hat{\mathbf{y}} - z_7 c\sin\beta\,\hat{\mathbf{z}}$ | (8$f$) | O III |
| $\mathbf{B}_{24}$ | = | $(x_7 + y_7)\,\mathbf{a}_1 + (x_7 - y_7)\,\mathbf{a}_2 + \left(\frac{1}{2} + z_7\right)\mathbf{a}_3$ | = | $\left(\frac{1}{2}c\cos\beta + x_7 a + z_7 c\cos\beta\right)\hat{\mathbf{x}} - y_7 b\,\hat{\mathbf{y}} + \left(\frac{1}{2} + z_7\right)c\sin\beta\,\hat{\mathbf{z}}$ | (8$f$) | O III |

**Found in:**

---

**Geometry files:**

- CIF: pp. 1570
- POSCAR: pp. 1571

# $Ta_2NiSe_5$ Structure: AB5C2_mC32_15_e_e2f_f

http://aflow.org/prototype-encyclopedia/AB5C2_mC32_15_e_e2f_f

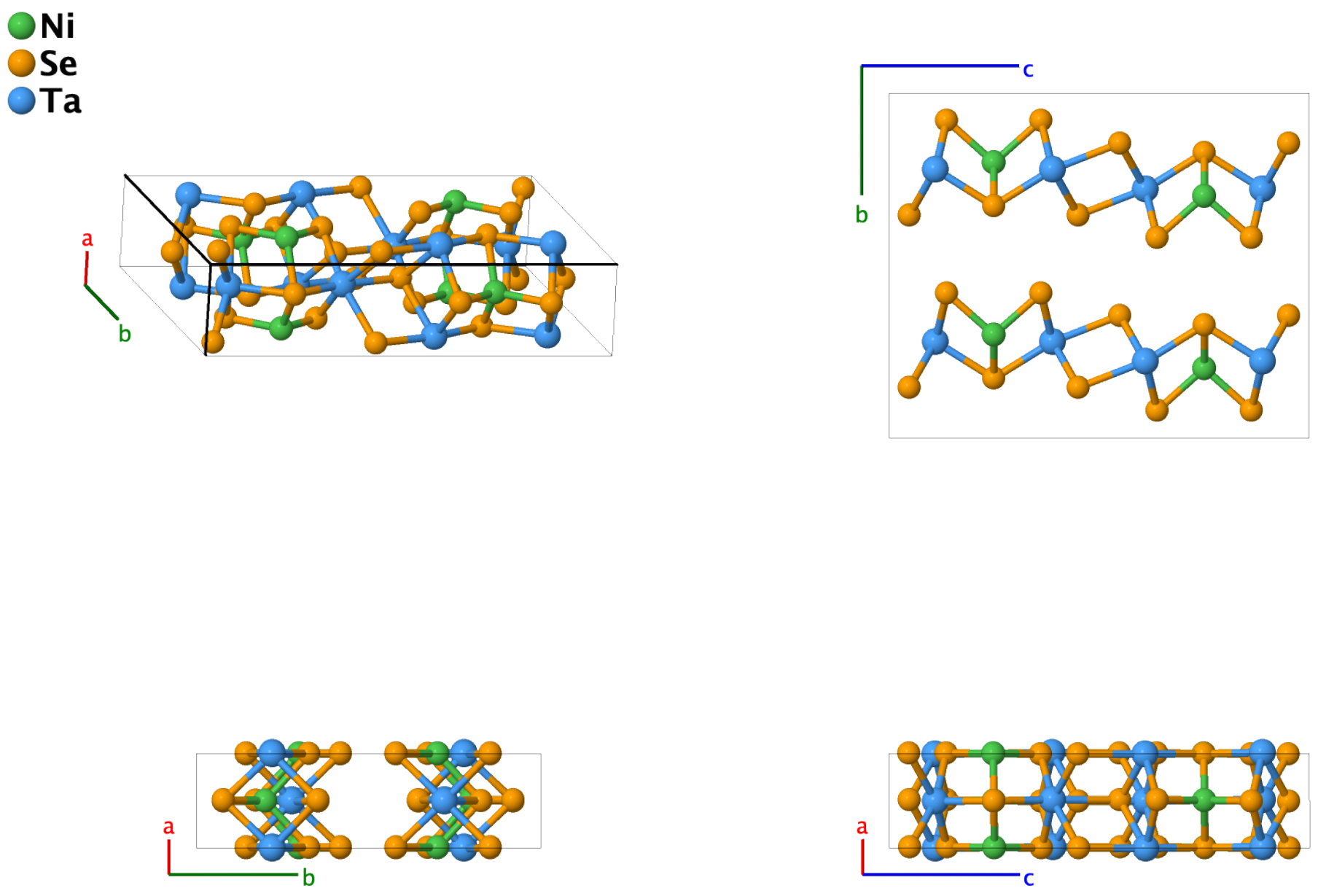

| | | |
|---|---|---|
| **Prototype** | : | $NiSe_5Ta_2$ |
| **AFLOW prototype label** | : | AB5C2_mC32_15_e_e2f_f |
| ***Strukturbericht* designation** | : | None |
| **Pearson symbol** | : | mC32 |
| **Space group number** | : | 15 |
| **Space group symbol** | : | $C2/c$ |
| **AFLOW prototype command** | : | `aflow --proto=AB5C2_mC32_15_e_e2f_f`<br>`--params=`$a, b/a, c/a, \beta, y_1, y_2, x_3, y_3, z_3, x_4, y_4, z_4, x_5, y_5, z_5$ |

- Above 328 K this transforms into the $Ta_2NiS_5$ structure.
- When $\beta = 90^\circ$, $x_3 = 1/2$, and $x_4 = x_5 = 0$, this becomes the $Ta_2NiS_5$ structure.

**Base-centered Monoclinic primitive vectors:**

$$
\begin{aligned}
\mathbf{a}_1 &= \tfrac{1}{2}\,a\,\hat{\mathbf{x}} - \tfrac{1}{2}\,b\,\hat{\mathbf{y}} \\
\mathbf{a}_2 &= \tfrac{1}{2}\,a\,\hat{\mathbf{x}} + \tfrac{1}{2}\,b\,\hat{\mathbf{y}} \\
\mathbf{a}_3 &= c\cos\beta\,\hat{\mathbf{x}} + c\sin\beta\,\hat{\mathbf{z}}
\end{aligned}
$$

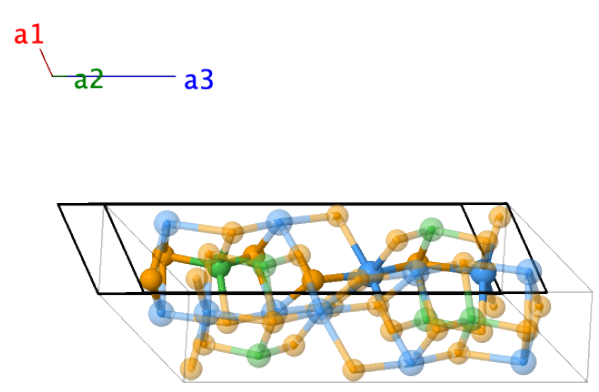

**Basis vectors:**

| | | Lattice Coordinates | | Cartesian Coordinates | Wyckoff Position | Atom Type |
|---|---|---|---|---|---|---|
| $\mathbf{B}_1$ | = | $-y_1\,\mathbf{a}_1 + y_1\,\mathbf{a}_2 + \frac{1}{4}\,\mathbf{a}_3$ | = | $\frac{1}{4}c\cos\beta\,\hat{\mathbf{x}} + y_1 b\,\hat{\mathbf{y}} + \frac{1}{4}c\sin\beta\,\hat{\mathbf{z}}$ | (4$e$) | Ni |
| $\mathbf{B}_2$ | = | $y_1\,\mathbf{a}_1 - y_1\,\mathbf{a}_2 + \frac{3}{4}\,\mathbf{a}_3$ | = | $\frac{3}{4}c\cos\beta\,\hat{\mathbf{x}} - y_1 b\,\hat{\mathbf{y}} + \frac{3}{4}c\sin\beta\,\hat{\mathbf{z}}$ | (4$e$) | Ni |
| $\mathbf{B}_3$ | = | $-y_2\,\mathbf{a}_1 + y_2\,\mathbf{a}_2 + \frac{1}{4}\,\mathbf{a}_3$ | = | $\frac{1}{4}c\cos\beta\,\hat{\mathbf{x}} + y_2 b\,\hat{\mathbf{y}} + \frac{1}{4}c\sin\beta\,\hat{\mathbf{z}}$ | (4$e$) | Se I |
| $\mathbf{B}_4$ | = | $y_2\,\mathbf{a}_1 - y_2\,\mathbf{a}_2 + \frac{3}{4}\,\mathbf{a}_3$ | = | $\frac{3}{4}c\cos\beta\,\hat{\mathbf{x}} - y_2 b\,\hat{\mathbf{y}} + \frac{3}{4}c\sin\beta\,\hat{\mathbf{z}}$ | (4$e$) | Se I |
| $\mathbf{B}_5$ | = | $(x_3 - y_3)\,\mathbf{a}_1 + (x_3 + y_3)\,\mathbf{a}_2 + z_3\,\mathbf{a}_3$ | = | $(x_3 a + z_3 c\cos\beta)\,\hat{\mathbf{x}} + y_3 b\,\hat{\mathbf{y}} + z_3 c\sin\beta\,\hat{\mathbf{z}}$ | (8$f$) | Se II |
| $\mathbf{B}_6$ | = | $(-x_3 - y_3)\,\mathbf{a}_1 + (-x_3 + y_3)\,\mathbf{a}_2 + \left(\frac{1}{2} - z_3\right)\,\mathbf{a}_3$ | = | $\left(\frac{1}{2}c\cos\beta - x_3 a - z_3 c\cos\beta\right)\,\hat{\mathbf{x}} + y_3 b\,\hat{\mathbf{y}} + \left(\frac{1}{2} - z_3\right)c\sin\beta\,\hat{\mathbf{z}}$ | (8$f$) | Se II |
| $\mathbf{B}_7$ | = | $(-x_3 + y_3)\,\mathbf{a}_1 + (-x_3 - y_3)\,\mathbf{a}_2 - z_3\,\mathbf{a}_3$ | = | $(-x_3 a - z_3 c\cos\beta)\,\hat{\mathbf{x}} - y_3 b\,\hat{\mathbf{y}} - z_3 c\sin\beta\,\hat{\mathbf{z}}$ | (8$f$) | Se II |
| $\mathbf{B}_8$ | = | $(x_3 + y_3)\,\mathbf{a}_1 + (x_3 - y_3)\,\mathbf{a}_2 + \left(\frac{1}{2} + z_3\right)\,\mathbf{a}_3$ | = | $\left(\frac{1}{2}c\cos\beta + x_3 a + z_3 c\cos\beta\right)\,\hat{\mathbf{x}} - y_3 b\,\hat{\mathbf{y}} + \left(\frac{1}{2} + z_3\right)c\sin\beta\,\hat{\mathbf{z}}$ | (8$f$) | Se II |
| $\mathbf{B}_9$ | = | $(x_4 - y_4)\,\mathbf{a}_1 + (x_4 + y_4)\,\mathbf{a}_2 + z_4\,\mathbf{a}_3$ | = | $(x_4 a + z_4 c\cos\beta)\,\hat{\mathbf{x}} + y_4 b\,\hat{\mathbf{y}} + z_4 c\sin\beta\,\hat{\mathbf{z}}$ | (8$f$) | Se III |
| $\mathbf{B}_{10}$ | = | $(-x_4 - y_4)\,\mathbf{a}_1 + (-x_4 + y_4)\,\mathbf{a}_2 + \left(\frac{1}{2} - z_4\right)\,\mathbf{a}_3$ | = | $\left(\frac{1}{2}c\cos\beta - x_4 a - z_4 c\cos\beta\right)\,\hat{\mathbf{x}} + y_4 b\,\hat{\mathbf{y}} + \left(\frac{1}{2} - z_4\right)c\sin\beta\,\hat{\mathbf{z}}$ | (8$f$) | Se III |
| $\mathbf{B}_{11}$ | = | $(-x_4 + y_4)\,\mathbf{a}_1 + (-x_4 - y_4)\,\mathbf{a}_2 - z_4\,\mathbf{a}_3$ | = | $(-x_4 a - z_4 c\cos\beta)\,\hat{\mathbf{x}} - y_4 b\,\hat{\mathbf{y}} - z_4 c\sin\beta\,\hat{\mathbf{z}}$ | (8$f$) | Se III |
| $\mathbf{B}_{12}$ | = | $(x_4 + y_4)\,\mathbf{a}_1 + (x_4 - y_4)\,\mathbf{a}_2 + \left(\frac{1}{2} + z_4\right)\,\mathbf{a}_3$ | = | $\left(\frac{1}{2}c\cos\beta + x_4 a + z_4 c\cos\beta\right)\,\hat{\mathbf{x}} - y_4 b\,\hat{\mathbf{y}} + \left(\frac{1}{2} + z_4\right)c\sin\beta\,\hat{\mathbf{z}}$ | (8$f$) | Se III |
| $\mathbf{B}_{13}$ | = | $(x_5 - y_5)\,\mathbf{a}_1 + (x_5 + y_5)\,\mathbf{a}_2 + z_5\,\mathbf{a}_3$ | = | $(x_5 a + z_5 c\cos\beta)\,\hat{\mathbf{x}} + y_5 b\,\hat{\mathbf{y}} + z_5 c\sin\beta\,\hat{\mathbf{z}}$ | (8$f$) | Ta |
| $\mathbf{B}_{14}$ | = | $(-x_5 - y_5)\,\mathbf{a}_1 + (-x_5 + y_5)\,\mathbf{a}_2 + \left(\frac{1}{2} - z_5\right)\,\mathbf{a}_3$ | = | $\left(\frac{1}{2}c\cos\beta - x_5 a - z_5 c\cos\beta\right)\,\hat{\mathbf{x}} + y_5 b\,\hat{\mathbf{y}} + \left(\frac{1}{2} - z_5\right)c\sin\beta\,\hat{\mathbf{z}}$ | (8$f$) | Ta |
| $\mathbf{B}_{15}$ | = | $(-x_5 + y_5)\,\mathbf{a}_1 + (-x_5 - y_5)\,\mathbf{a}_2 - z_5\,\mathbf{a}_3$ | = | $(-x_5 a - z_5 c\cos\beta)\,\hat{\mathbf{x}} - y_5 b\,\hat{\mathbf{y}} - z_5 c\sin\beta\,\hat{\mathbf{z}}$ | (8$f$) | Ta |
| $\mathbf{B}_{16}$ | = | $(x_5 + y_5)\,\mathbf{a}_1 + (x_5 - y_5)\,\mathbf{a}_2 + \left(\frac{1}{2} + z_5\right)\,\mathbf{a}_3$ | = | $\left(\frac{1}{2}c\cos\beta + x_5 a + z_5 c\cos\beta\right)\,\hat{\mathbf{x}} - y_5 b\,\hat{\mathbf{y}} + \left(\frac{1}{2} + z_5\right)c\sin\beta\,\hat{\mathbf{z}}$ | (8$f$) | Ta |

**Geometry files:**

- CIF: pp. 1571
- POSCAR: pp. 1571

# Pyrophyllite [$AlSi_2O_5(OH)$, $S5_6$] Structure: AB5CD2_mC72_15_f_5f_f_2f

http://aflow.org/prototype-encyclopedia/AB5CD2_mC72_15_f_5f_f_2f

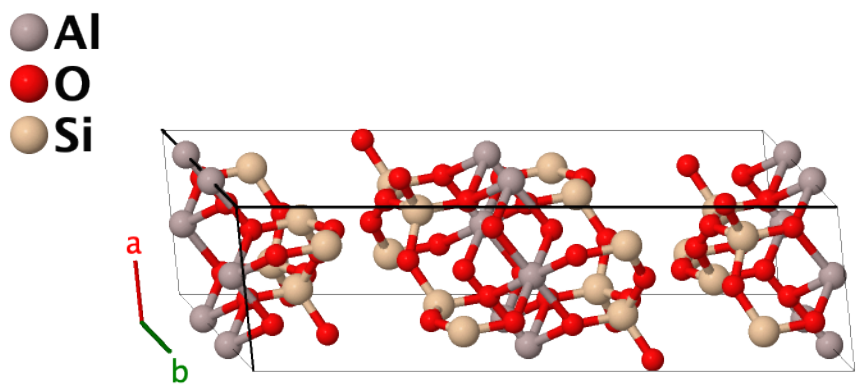

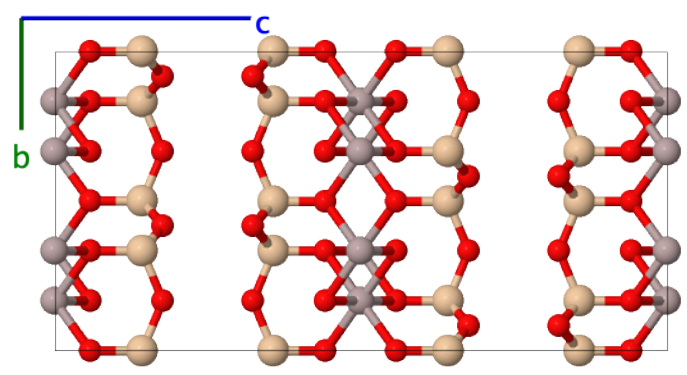

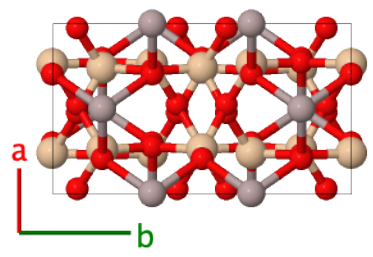

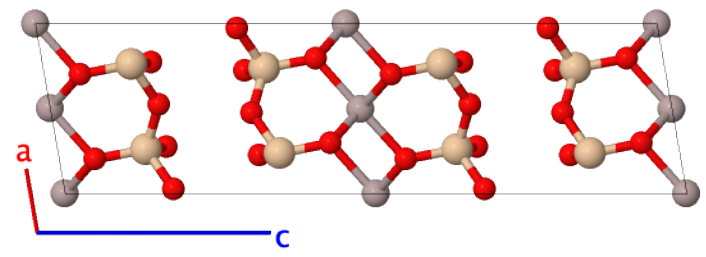

| | | |
|---|---|---|
| **Prototype** | : | $AlO_5(OH)Si_2$ |
| **AFLOW prototype label** | : | AB5CD2_mC72_15_f_5f_f_2f |
| ***Strukturbericht* designation** | : | $S5_6$ |
| **Pearson symbol** | : | mC72 |
| **Space group number** | : | 15 |
| **Space group symbol** | : | $C2/c$ |
| **AFLOW prototype command** | : | `aflow --proto=AB5CD2_mC72_15_f_5f_f_2f` `--params=`$a, b/a, c/a, \beta, x_1, y_1, z_1, x_2, y_2, z_2, x_3, y_3, z_3, x_4, y_4, z_4, x_5, y_5, z_5, x_6, y_6, z_6, x_7, y_7, z_7, x_8, y_8, z_8, x_9, y_9, z_9$ |

### Other compounds with this structure

- $Mg_3Si_4O_{10}(OH)_2$ (talc)

---

- This a double-layer version of the pyrophyllite structure. (Brindley, 1259) has shown that some samples have only one layer, with a corresponding reduction in the $c$ lattice vector. Both Brindley and (Lee, 1981) found pyrophyllite samples with space group $P\overline{1}$ #2.
- Talc ($Mg_3Si_4O_{10}(OH)_2$) is isostructural with pyrophyllite, with magnesium replacing aluminum on the (8f) site and the addition of another magnesium atom on the (4a) Wyckoff site.

### Base-centered Monoclinic primitive vectors:

$$\mathbf{a}_1 = \frac{1}{2}\,a\,\hat{\mathbf{x}} - \frac{1}{2}\,b\,\hat{\mathbf{y}}$$
$$\mathbf{a}_2 = \frac{1}{2}\,a\,\hat{\mathbf{x}} + \frac{1}{2}\,b\,\hat{\mathbf{y}}$$
$$\mathbf{a}_3 = c\cos\beta\,\hat{\mathbf{x}} + c\sin\beta\,\hat{\mathbf{z}}$$

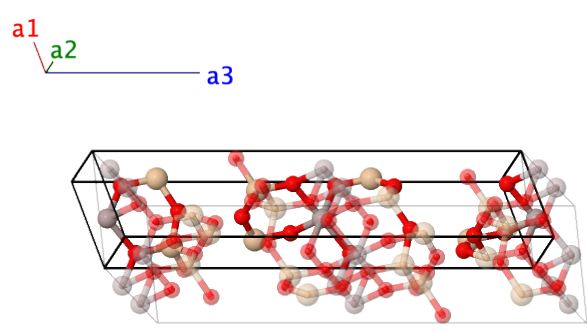

### Basis vectors:

| | | Lattice Coordinates | | Cartesian Coordinates | Wyckoff Position | Atom Type |
|---|---|---|---|---|---|---|
| $\mathbf{B}_1$ | = | $(x_1 - y_1)\,\mathbf{a}_1 + (x_1 + y_1)\,\mathbf{a}_2 + z_1\,\mathbf{a}_3$ | = | $(x_1 a + z_1 c\cos\beta)\,\hat{\mathbf{x}} + y_1 b\,\hat{\mathbf{y}} + z_1 c\sin\beta\,\hat{\mathbf{z}}$ | $(8f)$ | Al |
| $\mathbf{B}_2$ | = | $(-x_1 - y_1)\,\mathbf{a}_1 + (-x_1 + y_1)\,\mathbf{a}_2 + \left(\frac{1}{2} - z_1\right)\,\mathbf{a}_3$ | = | $\left(\frac{1}{2}c\cos\beta - x_1 a - z_1 c\cos\beta\right)\,\hat{\mathbf{x}} + y_1 b\,\hat{\mathbf{y}} + \left(\frac{1}{2} - z_1\right)c\sin\beta\,\hat{\mathbf{z}}$ | $(8f)$ | Al |
| $\mathbf{B}_3$ | = | $(-x_1 + y_1)\,\mathbf{a}_1 + (-x_1 - y_1)\,\mathbf{a}_2 - z_1\,\mathbf{a}_3$ | = | $(-x_1 a - z_1 c\cos\beta)\,\hat{\mathbf{x}} - y_1 b\,\hat{\mathbf{y}} - z_1 c\sin\beta\,\hat{\mathbf{z}}$ | $(8f)$ | Al |
| $\mathbf{B}_4$ | = | $(x_1 + y_1)\,\mathbf{a}_1 + (x_1 - y_1)\,\mathbf{a}_2 + \left(\frac{1}{2} + z_1\right)\,\mathbf{a}_3$ | = | $\left(\frac{1}{2}c\cos\beta + x_1 a + z_1 c\cos\beta\right)\,\hat{\mathbf{x}} - y_1 b\,\hat{\mathbf{y}} + \left(\frac{1}{2} + z_1\right)c\sin\beta\,\hat{\mathbf{z}}$ | $(8f)$ | Al |
| $\mathbf{B}_5$ | = | $(x_2 - y_2)\,\mathbf{a}_1 + (x_2 + y_2)\,\mathbf{a}_2 + z_2\,\mathbf{a}_3$ | = | $(x_2 a + z_2 c\cos\beta)\,\hat{\mathbf{x}} + y_2 b\,\hat{\mathbf{y}} + z_2 c\sin\beta\,\hat{\mathbf{z}}$ | $(8f)$ | O I |
| $\mathbf{B}_6$ | = | $(-x_2 - y_2)\,\mathbf{a}_1 + (-x_2 + y_2)\,\mathbf{a}_2 + \left(\frac{1}{2} - z_2\right)\,\mathbf{a}_3$ | = | $\left(\frac{1}{2}c\cos\beta - x_2 a - z_2 c\cos\beta\right)\,\hat{\mathbf{x}} + y_2 b\,\hat{\mathbf{y}} + \left(\frac{1}{2} - z_2\right)c\sin\beta\,\hat{\mathbf{z}}$ | $(8f)$ | O I |
| $\mathbf{B}_7$ | = | $(-x_2 + y_2)\,\mathbf{a}_1 + (-x_2 - y_2)\,\mathbf{a}_2 - z_2\,\mathbf{a}_3$ | = | $(-x_2 a - z_2 c\cos\beta)\,\hat{\mathbf{x}} - y_2 b\,\hat{\mathbf{y}} - z_2 c\sin\beta\,\hat{\mathbf{z}}$ | $(8f)$ | O I |
| $\mathbf{B}_8$ | = | $(x_2 + y_2)\,\mathbf{a}_1 + (x_2 - y_2)\,\mathbf{a}_2 + \left(\frac{1}{2} + z_2\right)\,\mathbf{a}_3$ | = | $\left(\frac{1}{2}c\cos\beta + x_2 a + z_2 c\cos\beta\right)\,\hat{\mathbf{x}} - y_2 b\,\hat{\mathbf{y}} + \left(\frac{1}{2} + z_2\right)c\sin\beta\,\hat{\mathbf{z}}$ | $(8f)$ | O I |
| $\mathbf{B}_9$ | = | $(x_3 - y_3)\,\mathbf{a}_1 + (x_3 + y_3)\,\mathbf{a}_2 + z_3\,\mathbf{a}_3$ | = | $(x_3 a + z_3 c\cos\beta)\,\hat{\mathbf{x}} + y_3 b\,\hat{\mathbf{y}} + z_3 c\sin\beta\,\hat{\mathbf{z}}$ | $(8f)$ | O II |
| $\mathbf{B}_{10}$ | = | $(-x_3 - y_3)\,\mathbf{a}_1 + (-x_3 + y_3)\,\mathbf{a}_2 + \left(\frac{1}{2} - z_3\right)\,\mathbf{a}_3$ | = | $\left(\frac{1}{2}c\cos\beta - x_3 a - z_3 c\cos\beta\right)\,\hat{\mathbf{x}} + y_3 b\,\hat{\mathbf{y}} + \left(\frac{1}{2} - z_3\right)c\sin\beta\,\hat{\mathbf{z}}$ | $(8f)$ | O II |
| $\mathbf{B}_{11}$ | = | $(-x_3 + y_3)\,\mathbf{a}_1 + (-x_3 - y_3)\,\mathbf{a}_2 - z_3\,\mathbf{a}_3$ | = | $(-x_3 a - z_3 c\cos\beta)\,\hat{\mathbf{x}} - y_3 b\,\hat{\mathbf{y}} - z_3 c\sin\beta\,\hat{\mathbf{z}}$ | $(8f)$ | O II |
| $\mathbf{B}_{12}$ | = | $(x_3 + y_3)\,\mathbf{a}_1 + (x_3 - y_3)\,\mathbf{a}_2 + \left(\frac{1}{2} + z_3\right)\,\mathbf{a}_3$ | = | $\left(\frac{1}{2}c\cos\beta + x_3 a + z_3 c\cos\beta\right)\,\hat{\mathbf{x}} - y_3 b\,\hat{\mathbf{y}} + \left(\frac{1}{2} + z_3\right)c\sin\beta\,\hat{\mathbf{z}}$ | $(8f)$ | O II |
| $\mathbf{B}_{13}$ | = | $(x_4 - y_4)\,\mathbf{a}_1 + (x_4 + y_4)\,\mathbf{a}_2 + z_4\,\mathbf{a}_3$ | = | $(x_4 a + z_4 c\cos\beta)\,\hat{\mathbf{x}} + y_4 b\,\hat{\mathbf{y}} + z_4 c\sin\beta\,\hat{\mathbf{z}}$ | $(8f)$ | O III |
| $\mathbf{B}_{14}$ | = | $(-x_4 - y_4)\,\mathbf{a}_1 + (-x_4 + y_4)\,\mathbf{a}_2 + \left(\frac{1}{2} - z_4\right)\,\mathbf{a}_3$ | = | $\left(\frac{1}{2}c\cos\beta - x_4 a - z_4 c\cos\beta\right)\,\hat{\mathbf{x}} + y_4 b\,\hat{\mathbf{y}} + \left(\frac{1}{2} - z_4\right)c\sin\beta\,\hat{\mathbf{z}}$ | $(8f)$ | O III |
| $\mathbf{B}_{15}$ | = | $(-x_4 + y_4)\,\mathbf{a}_1 + (-x_4 - y_4)\,\mathbf{a}_2 - z_4\,\mathbf{a}_3$ | = | $(-x_4 a - z_4 c\cos\beta)\,\hat{\mathbf{x}} - y_4 b\,\hat{\mathbf{y}} - z_4 c\sin\beta\,\hat{\mathbf{z}}$ | $(8f)$ | O III |
| $\mathbf{B}_{16}$ | = | $(x_4 + y_4)\,\mathbf{a}_1 + (x_4 - y_4)\,\mathbf{a}_2 + \left(\frac{1}{2} + z_4\right)\,\mathbf{a}_3$ | = | $\left(\frac{1}{2}c\cos\beta + x_4 a + z_4 c\cos\beta\right)\,\hat{\mathbf{x}} - y_4 b\,\hat{\mathbf{y}} + \left(\frac{1}{2} + z_4\right)c\sin\beta\,\hat{\mathbf{z}}$ | $(8f)$ | O III |
| $\mathbf{B}_{17}$ | = | $(x_5 - y_5)\,\mathbf{a}_1 + (x_5 + y_5)\,\mathbf{a}_2 + z_5\,\mathbf{a}_3$ | = | $(x_5 a + z_5 c\cos\beta)\,\hat{\mathbf{x}} + y_5 b\,\hat{\mathbf{y}} + z_5 c\sin\beta\,\hat{\mathbf{z}}$ | $(8f)$ | O IV |
| $\mathbf{B}_{18}$ | = | $(-x_5 - y_5)\,\mathbf{a}_1 + (-x_5 + y_5)\,\mathbf{a}_2 + \left(\frac{1}{2} - z_5\right)\,\mathbf{a}_3$ | = | $\left(\frac{1}{2}c\cos\beta - x_5 a - z_5 c\cos\beta\right)\,\hat{\mathbf{x}} + y_5 b\,\hat{\mathbf{y}} + \left(\frac{1}{2} - z_5\right)c\sin\beta\,\hat{\mathbf{z}}$ | $(8f)$ | O IV |
| $\mathbf{B}_{19}$ | = | $(-x_5 + y_5)\,\mathbf{a}_1 + (-x_5 - y_5)\,\mathbf{a}_2 - z_5\,\mathbf{a}_3$ | = | $(-x_5 a - z_5 c\cos\beta)\,\hat{\mathbf{x}} - y_5 b\,\hat{\mathbf{y}} - z_5 c\sin\beta\,\hat{\mathbf{z}}$ | $(8f)$ | O IV |
| $\mathbf{B}_{20}$ | = | $(x_5 + y_5)\,\mathbf{a}_1 + (x_5 - y_5)\,\mathbf{a}_2 + \left(\frac{1}{2} + z_5\right)\,\mathbf{a}_3$ | = | $\left(\frac{1}{2}c\cos\beta + x_5 a + z_5 c\cos\beta\right)\,\hat{\mathbf{x}} - y_5 b\,\hat{\mathbf{y}} + \left(\frac{1}{2} + z_5\right)c\sin\beta\,\hat{\mathbf{z}}$ | $(8f)$ | O IV |
| $\mathbf{B}_{21}$ | = | $(x_6 - y_6)\,\mathbf{a}_1 + (x_6 + y_6)\,\mathbf{a}_2 + z_6\,\mathbf{a}_3$ | = | $(x_6 a + z_6 c\cos\beta)\,\hat{\mathbf{x}} + y_6 b\,\hat{\mathbf{y}} + z_6 c\sin\beta\,\hat{\mathbf{z}}$ | $(8f)$ | O V |
| $\mathbf{B}_{22}$ | = | $\left(-x_6 - y_6\right)\,\mathbf{a}_1 + \left(-x_6 + y_6\right)\,\mathbf{a}_2 + \left(\frac{1}{2} - z_6\right)\,\mathbf{a}_3$ | = | $\left(\frac{1}{2}c\cos\beta - x_6 a - z_6 c\cos\beta\right)\,\mathbf{\hat{x}} + y_6 b\,\mathbf{\hat{y}} + \left(\frac{1}{2} - z_6\right)c\sin\beta\,\mathbf{\hat{z}}$ | (8$f$) | O V |
| $\mathbf{B}_{23}$ | = | $\left(-x_6 + y_6\right)\,\mathbf{a}_1 + \left(-x_6 - y_6\right)\,\mathbf{a}_2 - z_6\,\mathbf{a}_3$ | = | $\left(-x_6 a - z_6 c\cos\beta\right)\,\mathbf{\hat{x}} - y_6 b\,\mathbf{\hat{y}} - z_6 c\sin\beta\,\mathbf{\hat{z}}$ | (8$f$) | O V |
| $\mathbf{B}_{24}$ | = | $\left(x_6 + y_6\right)\,\mathbf{a}_1 + \left(x_6 - y_6\right)\,\mathbf{a}_2 + \left(\frac{1}{2} + z_6\right)\,\mathbf{a}_3$ | = | $\left(\frac{1}{2}c\cos\beta + x_6 a + z_6 c\cos\beta\right)\,\mathbf{\hat{x}} - y_6 b\,\mathbf{\hat{y}} + \left(\frac{1}{2} + z_6\right)c\sin\beta\,\mathbf{\hat{z}}$ | (8$f$) | O V |
| $\mathbf{B}_{25}$ | = | $\left(x_7 - y_7\right)\,\mathbf{a}_1 + \left(x_7 + y_7\right)\,\mathbf{a}_2 + z_7\,\mathbf{a}_3$ | = | $\left(x_7 a + z_7 c\cos\beta\right)\,\mathbf{\hat{x}} + y_7 b\,\mathbf{\hat{y}} + z_7 c\sin\beta\,\mathbf{\hat{z}}$ | (8$f$) | OH |
| $\mathbf{B}_{26}$ | = | $\left(-x_7 - y_7\right)\,\mathbf{a}_1 + \left(-x_7 + y_7\right)\,\mathbf{a}_2 + \left(\frac{1}{2} - z_7\right)\,\mathbf{a}_3$ | = | $\left(\frac{1}{2}c\cos\beta - x_7 a - z_7 c\cos\beta\right)\,\mathbf{\hat{x}} + y_7 b\,\mathbf{\hat{y}} + \left(\frac{1}{2} - z_7\right)c\sin\beta\,\mathbf{\hat{z}}$ | (8$f$) | OH |
| $\mathbf{B}_{27}$ | = | $\left(-x_7 + y_7\right)\,\mathbf{a}_1 + \left(-x_7 - y_7\right)\,\mathbf{a}_2 - z_7\,\mathbf{a}_3$ | = | $\left(-x_7 a - z_7 c\cos\beta\right)\,\mathbf{\hat{x}} - y_7 b\,\mathbf{\hat{y}} - z_7 c\sin\beta\,\mathbf{\hat{z}}$ | (8$f$) | OH |
| $\mathbf{B}_{28}$ | = | $\left(x_7 + y_7\right)\,\mathbf{a}_1 + \left(x_7 - y_7\right)\,\mathbf{a}_2 + \left(\frac{1}{2} + z_7\right)\,\mathbf{a}_3$ | = | $\left(\frac{1}{2}c\cos\beta + x_7 a + z_7 c\cos\beta\right)\,\mathbf{\hat{x}} - y_7 b\,\mathbf{\hat{y}} + \left(\frac{1}{2} + z_7\right)c\sin\beta\,\mathbf{\hat{z}}$ | (8$f$) | OH |
| $\mathbf{B}_{29}$ | = | $\left(x_8 - y_8\right)\,\mathbf{a}_1 + \left(x_8 + y_8\right)\,\mathbf{a}_2 + z_8\,\mathbf{a}_3$ | = | $\left(x_8 a + z_8 c\cos\beta\right)\,\mathbf{\hat{x}} + y_8 b\,\mathbf{\hat{y}} + z_8 c\sin\beta\,\mathbf{\hat{z}}$ | (8$f$) | Si I |
| $\mathbf{B}_{30}$ | = | $\left(-x_8 - y_8\right)\,\mathbf{a}_1 + \left(-x_8 + y_8\right)\,\mathbf{a}_2 + \left(\frac{1}{2} - z_8\right)\,\mathbf{a}_3$ | = | $\left(\frac{1}{2}c\cos\beta - x_8 a - z_8 c\cos\beta\right)\,\mathbf{\hat{x}} + y_8 b\,\mathbf{\hat{y}} + \left(\frac{1}{2} - z_8\right)c\sin\beta\,\mathbf{\hat{z}}$ | (8$f$) | Si I |
| $\mathbf{B}_{31}$ | = | $\left(-x_8 + y_8\right)\,\mathbf{a}_1 + \left(-x_8 - y_8\right)\,\mathbf{a}_2 - z_8\,\mathbf{a}_3$ | = | $\left(-x_8 a - z_8 c\cos\beta\right)\,\mathbf{\hat{x}} - y_8 b\,\mathbf{\hat{y}} - z_8 c\sin\beta\,\mathbf{\hat{z}}$ | (8$f$) | Si I |
| $\mathbf{B}_{32}$ | = | $\left(x_8 + y_8\right)\,\mathbf{a}_1 + \left(x_8 - y_8\right)\,\mathbf{a}_2 + \left(\frac{1}{2} + z_8\right)\,\mathbf{a}_3$ | = | $\left(\frac{1}{2}c\cos\beta + x_8 a + z_8 c\cos\beta\right)\,\mathbf{\hat{x}} - y_8 b\,\mathbf{\hat{y}} + \left(\frac{1}{2} + z_8\right)c\sin\beta\,\mathbf{\hat{z}}$ | (8$f$) | Si I |
| $\mathbf{B}_{33}$ | = | $\left(x_9 - y_9\right)\,\mathbf{a}_1 + \left(x_9 + y_9\right)\,\mathbf{a}_2 + z_9\,\mathbf{a}_3$ | = | $\left(x_9 a + z_9 c\cos\beta\right)\,\mathbf{\hat{x}} + y_9 b\,\mathbf{\hat{y}} + z_9 c\sin\beta\,\mathbf{\hat{z}}$ | (8$f$) | Si II |
| $\mathbf{B}_{34}$ | = | $\left(-x_9 - y_9\right)\,\mathbf{a}_1 + \left(-x_9 + y_9\right)\,\mathbf{a}_2 + \left(\frac{1}{2} - z_9\right)\,\mathbf{a}_3$ | = | $\left(\frac{1}{2}c\cos\beta - x_9 a - z_9 c\cos\beta\right)\,\mathbf{\hat{x}} + y_9 b\,\mathbf{\hat{y}} + \left(\frac{1}{2} - z_9\right)c\sin\beta\,\mathbf{\hat{z}}$ | (8$f$) | Si II |
| $\mathbf{B}_{35}$ | = | $\left(-x_9 + y_9\right)\,\mathbf{a}_1 + \left(-x_9 - y_9\right)\,\mathbf{a}_2 - z_9\,\mathbf{a}_3$ | = | $\left(-x_9 a - z_9 c\cos\beta\right)\,\mathbf{\hat{x}} - y_9 b\,\mathbf{\hat{y}} - z_9 c\sin\beta\,\mathbf{\hat{z}}$ | (8$f$) | Si II |
| $\mathbf{B}_{36}$ | = | $\left(x_9 + y_9\right)\,\mathbf{a}_1 + \left(x_9 - y_9\right)\,\mathbf{a}_2 + \left(\frac{1}{2} + z_9\right)\,\mathbf{a}_3$ | = | $\left(\frac{1}{2}c\cos\beta + x_9 a + z_9 c\cos\beta\right)\,\mathbf{\hat{x}} - y_9 b\,\mathbf{\hat{y}} + \left(\frac{1}{2} + z_9\right)c\sin\beta\,\mathbf{\hat{z}}$ | (8$f$) | Si II |

**Found in:**

**Geometry files:**

- CIF: pp. 1572

- POSCAR: pp. 1572

# Titanite ($CaTiSiO_5$, $S0_6$) Structure: AB5CD_mC32_15_e_e2f_e_b

http://aflow.org/prototype-encyclopedia/AB5CD_mC32_15_e_e2f_e_b

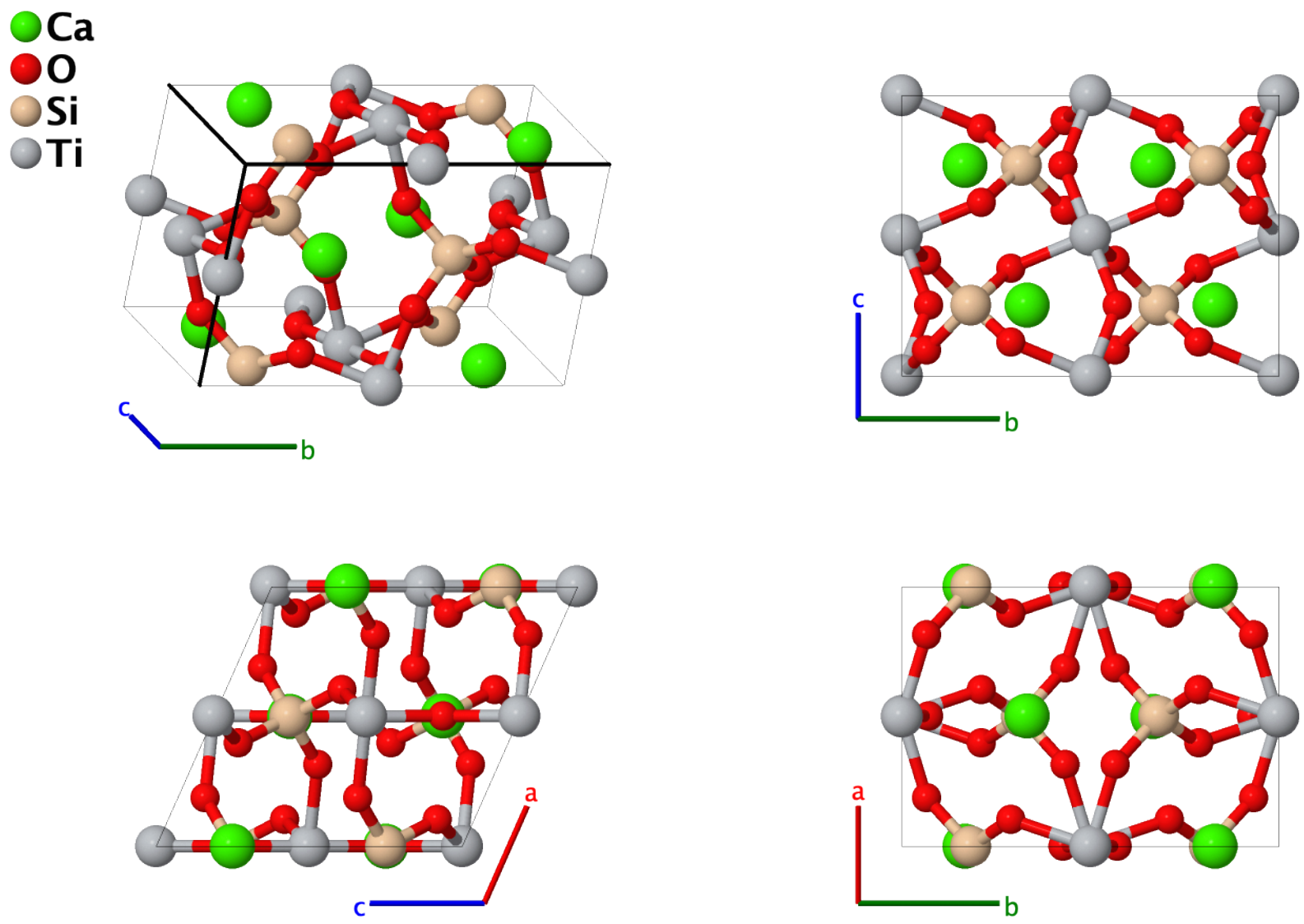

| | | |
|---|---|---|
| **Prototype** | : | $CaO_5SiTi$ |
| **AFLOW prototype label** | : | AB5CD_mC32_15_e_e2f_e_b |
| ***Strukturbericht* designation** | : | $S0_6$ |
| **Pearson symbol** | : | mC32 |
| **Space group number** | : | 15 |
| **Space group symbol** | : | $C2/c$ |
| **AFLOW prototype command** | : | `aflow --proto=AB5CD_mC32_15_e_e2f_e_b` <br> `--params=`$a, b/a, c/a, \beta, y_2, y_3, y_4, x_5, y_5, z_5, x_6, y_6, z_6$ |

- We use the data from sample M28658 to construct our titanite cell.
- (Hermann, 1937) places the titanium atoms at the (4c) Wyckoff position rather than the (4b) position favored by (Hawthorne, 1991). The biggest difference between the two structures is the distance between the silicon and titanium atoms, which is 3.81 Å in Hawthorne and a rather short 2 Å in Hermann. We therefore favor the Hawthorne structure, and, since it does not otherwise substantially change the results, we continue to use $S0_6$ to describe it.
- (Hermann, 1937) also listed this as the $H5_6$ structure in the index.

**Base-centered Monoclinic primitive vectors:**

$$
\begin{aligned}
\mathbf{a}_1 &= \tfrac{1}{2}\,a\,\mathbf{\hat{x}} - \tfrac{1}{2}\,b\,\mathbf{\hat{y}} \\
\mathbf{a}_2 &= \tfrac{1}{2}\,a\,\mathbf{\hat{x}} + \tfrac{1}{2}\,b\,\mathbf{\hat{y}} \\
\mathbf{a}_3 &= c\cos\beta\,\mathbf{\hat{x}} + c\sin\beta\,\mathbf{\hat{z}}
\end{aligned}
$$

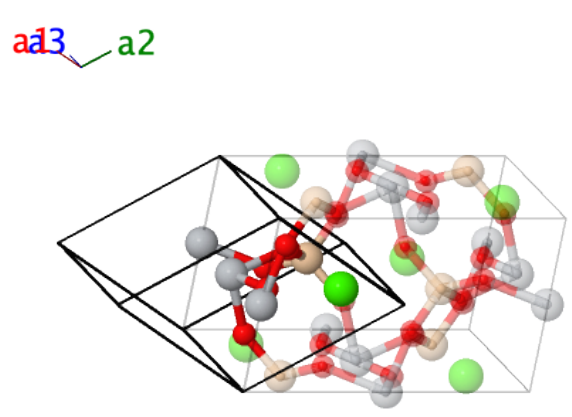

**Basis vectors:**

| | | Lattice Coordinates | | Cartesian Coordinates | Wyckoff Position | Atom Type |
|---|---|---|---|---|---|---|
| $\mathbf{B}_1$ | = | $\frac{1}{2}\,\mathbf{a}_1+\frac{1}{2}\,\mathbf{a}_2$ | = | $\frac{1}{2}a\,\hat{\mathbf{x}}$ | (4$b$) | Ti |
| $\mathbf{B}_2$ | = | $\frac{1}{2}\,\mathbf{a}_1+\frac{1}{2}\,\mathbf{a}_2+\frac{1}{2}\,\mathbf{a}_3$ | = | $\frac{1}{2}\left(a+c\cos\beta\right)\,\hat{\mathbf{x}}+\frac{1}{2}c\sin\beta\,\hat{\mathbf{z}}$ | (4$b$) | Ti |
| $\mathbf{B}_3$ | = | $-y_2\,\mathbf{a}_1+y_2\,\mathbf{a}_2+\frac{1}{4}\,\mathbf{a}_3$ | = | $\frac{1}{4}c\cos\beta\,\hat{\mathbf{x}}+y_2b\,\hat{\mathbf{y}}+\frac{1}{4}c\sin\beta\,\hat{\mathbf{z}}$ | (4$e$) | Ca |
| $\mathbf{B}_4$ | = | $y_2\,\mathbf{a}_1-y_2\,\mathbf{a}_2+\frac{3}{4}\,\mathbf{a}_3$ | = | $\frac{3}{4}c\cos\beta\,\hat{\mathbf{x}}-y_2b\,\hat{\mathbf{y}}+\frac{3}{4}c\sin\beta\,\hat{\mathbf{z}}$ | (4$e$) | Ca |
| $\mathbf{B}_5$ | = | $-y_3\,\mathbf{a}_1+y_3\,\mathbf{a}_2+\frac{1}{4}\,\mathbf{a}_3$ | = | $\frac{1}{4}c\cos\beta\,\hat{\mathbf{x}}+y_3b\,\hat{\mathbf{y}}+\frac{1}{4}c\sin\beta\,\hat{\mathbf{z}}$ | (4$e$) | O I |
| $\mathbf{B}_6$ | = | $y_3\,\mathbf{a}_1-y_3\,\mathbf{a}_2+\frac{3}{4}\,\mathbf{a}_3$ | = | $\frac{3}{4}c\cos\beta\,\hat{\mathbf{x}}-y_3b\,\hat{\mathbf{y}}+\frac{3}{4}c\sin\beta\,\hat{\mathbf{z}}$ | (4$e$) | O I |
| $\mathbf{B}_7$ | = | $-y_4\,\mathbf{a}_1+y_4\,\mathbf{a}_2+\frac{1}{4}\,\mathbf{a}_3$ | = | $\frac{1}{4}c\cos\beta\,\hat{\mathbf{x}}+y_4b\,\hat{\mathbf{y}}+\frac{1}{4}c\sin\beta\,\hat{\mathbf{z}}$ | (4$e$) | Si |
| $\mathbf{B}_8$ | = | $y_4\,\mathbf{a}_1-y_4\,\mathbf{a}_2+\frac{3}{4}\,\mathbf{a}_3$ | = | $\frac{3}{4}c\cos\beta\,\hat{\mathbf{x}}-y_4b\,\hat{\mathbf{y}}+\frac{3}{4}c\sin\beta\,\hat{\mathbf{z}}$ | (4$e$) | Si |
| $\mathbf{B}_9$ | = | $\left(x_5-y_5\right)\,\mathbf{a}_1+\left(x_5+y_5\right)\,\mathbf{a}_2+z_5\,\mathbf{a}_3$ | = | $\left(x_5a+z_5c\cos\beta\right)\,\hat{\mathbf{x}}+y_5b\,\hat{\mathbf{y}}+z_5c\sin\beta\,\hat{\mathbf{z}}$ | (8$f$) | O II |
| $\mathbf{B}_{10}$ | = | $\left(-x_5-y_5\right)\,\mathbf{a}_1+\left(-x_5+y_5\right)\,\mathbf{a}_2+\left(\frac{1}{2}-z_5\right)\,\mathbf{a}_3$ | = | $\left(\frac{1}{2}c\cos\beta-x_5a-z_5c\cos\beta\right)\,\hat{\mathbf{x}}+y_5b\,\hat{\mathbf{y}}+\left(\frac{1}{2}-z_5\right)c\sin\beta\,\hat{\mathbf{z}}$ | (8$f$) | O II |
| $\mathbf{B}_{11}$ | = | $\left(-x_5+y_5\right)\,\mathbf{a}_1+\left(-x_5-y_5\right)\,\mathbf{a}_2-z_5\,\mathbf{a}_3$ | = | $\left(-x_5a-z_5c\cos\beta\right)\,\hat{\mathbf{x}}-y_5b\,\hat{\mathbf{y}}-z_5c\sin\beta\,\hat{\mathbf{z}}$ | (8$f$) | O II |
| $\mathbf{B}_{12}$ | = | $\left(x_5+y_5\right)\,\mathbf{a}_1+\left(x_5-y_5\right)\,\mathbf{a}_2+\left(\frac{1}{2}+z_5\right)\,\mathbf{a}_3$ | = | $\left(\frac{1}{2}c\cos\beta+x_5a+z_5c\cos\beta\right)\,\hat{\mathbf{x}}-y_5b\,\hat{\mathbf{y}}+\left(\frac{1}{2}+z_5\right)c\sin\beta\,\hat{\mathbf{z}}$ | (8$f$) | O II |
| $\mathbf{B}_{13}$ | = | $\left(x_6-y_6\right)\,\mathbf{a}_1+\left(x_6+y_6\right)\,\mathbf{a}_2+z_6\,\mathbf{a}_3$ | = | $\left(x_6a+z_6c\cos\beta\right)\,\hat{\mathbf{x}}+y_6b\,\hat{\mathbf{y}}+z_6c\sin\beta\,\hat{\mathbf{z}}$ | (8$f$) | O III |
| $\mathbf{B}_{14}$ | = | $\left(-x_6-y_6\right)\,\mathbf{a}_1+\left(-x_6+y_6\right)\,\mathbf{a}_2+\left(\frac{1}{2}-z_6\right)\,\mathbf{a}_3$ | = | $\left(\frac{1}{2}c\cos\beta-x_6a-z_6c\cos\beta\right)\,\hat{\mathbf{x}}+y_6b\,\hat{\mathbf{y}}+\left(\frac{1}{2}-z_6\right)c\sin\beta\,\hat{\mathbf{z}}$ | (8$f$) | O III |
| $\mathbf{B}_{15}$ | = | $\left(-x_6+y_6\right)\,\mathbf{a}_1+\left(-x_6-y_6\right)\,\mathbf{a}_2-z_6\,\mathbf{a}_3$ | = | $\left(-x_6a-z_6c\cos\beta\right)\,\hat{\mathbf{x}}-y_6b\,\hat{\mathbf{y}}-z_6c\sin\beta\,\hat{\mathbf{z}}$ | (8$f$) | O III |
| $\mathbf{B}_{16}$ | = | $\left(x_6+y_6\right)\,\mathbf{a}_1+\left(x_6-y_6\right)\,\mathbf{a}_2+\left(\frac{1}{2}+z_6\right)\,\mathbf{a}_3$ | = | $\left(\frac{1}{2}c\cos\beta+x_6a+z_6c\cos\beta\right)\,\hat{\mathbf{x}}-y_6b\,\hat{\mathbf{y}}+\left(\frac{1}{2}+z_6\right)c\sin\beta\,\hat{\mathbf{z}}$ | (8$f$) | O III |

**Geometry files:**

- CIF: pp. 1572
- POSCAR: pp. 1573

# $KFeS_2$ ($F5_a$) Structure: ABC2_mC16_15_e_e_f

http://aflow.org/prototype-encyclopedia/ABC2_mC16_15_e_e_f

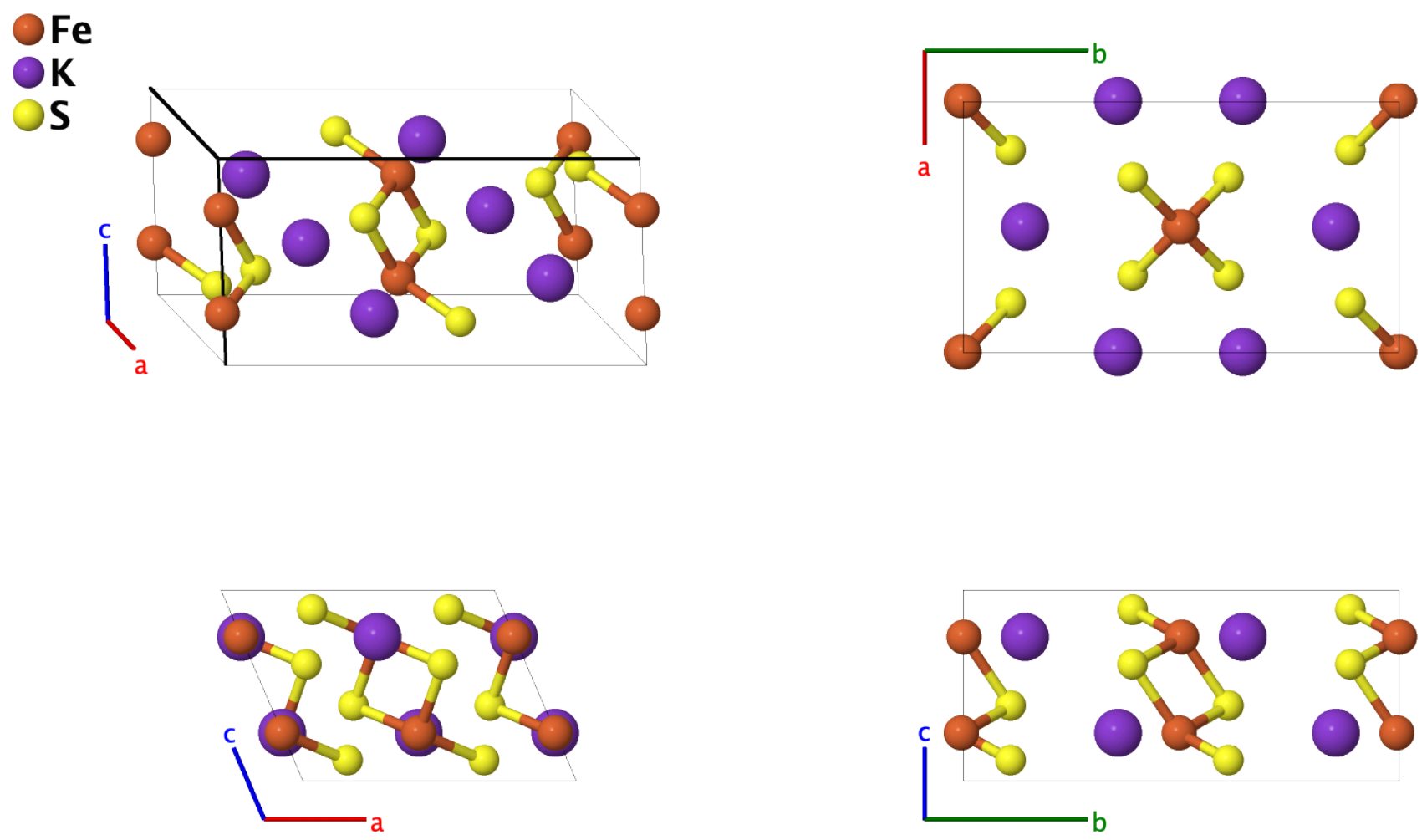

| | | |
|---|---|---|
| **Prototype** | : | $FeKS_2$ |
| **AFLOW prototype label** | : | ABC2_mC16_15_e_e_f |
| ***Strukturbericht* designation** | : | $F5_a$ |
| **Pearson symbol** | : | mC16 |
| **Space group number** | : | 15 |
| **Space group symbol** | : | $C2/c$ |
| **AFLOW prototype command** | : | `aflow --proto=ABC2_mC16_15_e_e_f` <br> `--params=`$a, b/a, c/a, \beta, y_1, y_2, x_3, y_3, z_3$ |

### Other compounds with this structure

- $RbFeS_2$, $KFeSe_2$, $RbFeSe_2$, $CsGaS_2$, and $CsGaSe_2$

- This structure forms cages of Fe atoms surrounded by tetrahedra of S atoms.

### Base-centered Monoclinic primitive vectors:

$$\mathbf{a}_1 = \frac{1}{2}\,a\,\mathbf{\hat{x}} - \frac{1}{2}\,b\,\mathbf{\hat{y}}$$
$$\mathbf{a}_2 = \frac{1}{2}\,a\,\mathbf{\hat{x}} + \frac{1}{2}\,b\,\mathbf{\hat{y}}$$
$$\mathbf{a}_3 = c\cos\beta\,\mathbf{\hat{x}} + c\sin\beta\,\mathbf{\hat{z}}$$

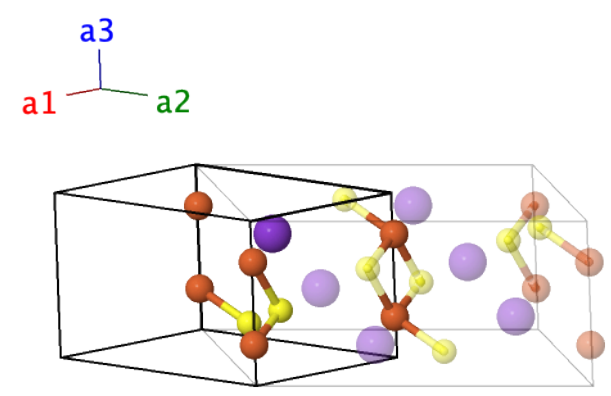

### Basis vectors:

| | | Lattice Coordinates | | Cartesian Coordinates | Wyckoff Position | Atom Type |
|---|---|---|---|---|---|---|
| $\mathbf{B}_1$ | = | $-y_1\,\mathbf{a}_1 + y_1\,\mathbf{a}_2 + \frac{1}{4}\,\mathbf{a}_3$ | = | $\frac{1}{4}c\cos\beta\,\mathbf{\hat{x}} + y_1 b\,\mathbf{\hat{y}} + \frac{1}{4}c\sin\beta\,\mathbf{\hat{z}}$ | (4$e$) | Fe |
| $\mathbf{B}_2$ | = | $y_1\,\mathbf{a}_1 - y_1\,\mathbf{a}_2 + \frac{3}{4}\,\mathbf{a}_3$ | = | $\frac{3}{4}c\cos\beta\,\mathbf{\hat{x}} - y_1 b\,\mathbf{\hat{y}} + \frac{3}{4}c\sin\beta\,\mathbf{\hat{z}}$ | (4$e$) | Fe |
| $\mathbf{B}_3$ | = | $-y_2\,\mathbf{a}_1 + y_2\,\mathbf{a}_2 + \frac{1}{4}\,\mathbf{a}_3$ | = | $\frac{1}{4}c\cos\beta\,\mathbf{\hat{x}} + y_2 b\,\mathbf{\hat{y}} + \frac{1}{4}c\sin\beta\,\mathbf{\hat{z}}$ | (4$e$) | K |
| $\mathbf{B}_4$ | = | $y_2\,\mathbf{a}_1 - y_2\,\mathbf{a}_2 + \frac{3}{4}\,\mathbf{a}_3$ | = | $\frac{3}{4}c\cos\beta\,\hat{\mathbf{x}} - y_2 b\,\hat{\mathbf{y}} + \frac{3}{4}c\sin\beta\,\hat{\mathbf{z}}$ | $(4e)$ | K |
| $\mathbf{B}_5$ | = | $(x_3 - y_3)\,\mathbf{a}_1 + (x_3 + y_3)\,\mathbf{a}_2 + z_3\,\mathbf{a}_3$ | = | $(x_3 a + z_3 c\cos\beta)\,\hat{\mathbf{x}} + y_3 b\,\hat{\mathbf{y}} + z_3 c\sin\beta\,\hat{\mathbf{z}}$ | $(8f)$ | S |
| $\mathbf{B}_6$ | = | $(-x_3 - y_3)\,\mathbf{a}_1 + (-x_3 + y_3)\,\mathbf{a}_2 + \left(\frac{1}{2} - z_3\right)\,\mathbf{a}_3$ | = | $\left(\frac{1}{2}c\cos\beta - x_3 a - z_3 c\cos\beta\right)\,\hat{\mathbf{x}} + y_3 b\,\hat{\mathbf{y}} + \left(\frac{1}{2} - z_3\right)c\sin\beta\,\hat{\mathbf{z}}$ | $(8f)$ | S |
| $\mathbf{B}_7$ | = | $(-x_3 + y_3)\,\mathbf{a}_1 + (-x_3 - y_3)\,\mathbf{a}_2 - z_3\,\mathbf{a}_3$ | = | $(-x_3 a - z_3 c\cos\beta)\,\hat{\mathbf{x}} - y_3 b\,\hat{\mathbf{y}} - z_3 c\sin\beta\,\hat{\mathbf{z}}$ | $(8f)$ | S |
| $\mathbf{B}_8$ | = | $(x_3 + y_3)\,\mathbf{a}_1 + (x_3 - y_3)\,\mathbf{a}_2 + \left(\frac{1}{2} + z_3\right)\,\mathbf{a}_3$ | = | $\left(\frac{1}{2}c\cos\beta + x_3 a + z_3 c\cos\beta\right)\,\hat{\mathbf{x}} - y_3 b\,\hat{\mathbf{y}} + \left(\frac{1}{2} + z_3\right)c\sin\beta\,\hat{\mathbf{z}}$ | $(8f)$ | S |

**Geometry files:**

- CIF: pp. 1573
- POSCAR: pp. 1573

# Diopside [$CaMg(SiO_3)_2$, $S4_1$] Structure: ABC6D2_mC40_15_e_e_3f_f

http://aflow.org/prototype-encyclopedia/ABC6D2_mC40_15_e_e_3f_f.S4_1

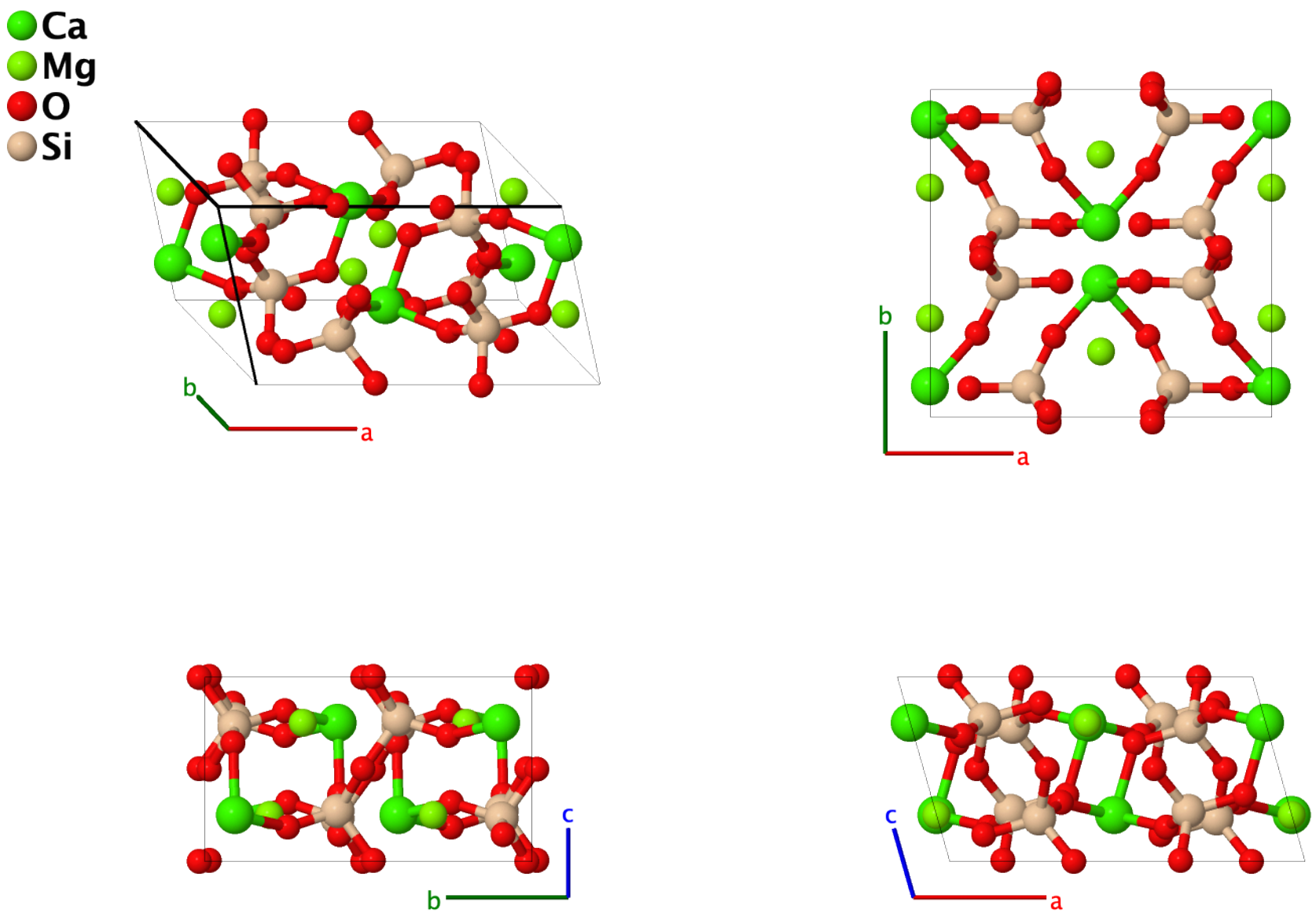

| | | |
|---|---|---|
| **Prototype** | : | $CaMgO_6Si_2$ |
| **AFLOW prototype label** | : | ABC6D2_mC40_15_e_e_3f_f |
| ***Strukturbericht* designation** | : | $S4_1$ |
| **Pearson symbol** | : | mC40 |
| **Space group number** | : | 15 |
| **Space group symbol** | : | $C2/c$ |
| **AFLOW prototype command** | : | `aflow --proto=ABC6D2_mC40_15_e_e_3f_f` `--params=`$a, b/a, c/a, \beta, y_1, y_2, x_3, y_3, z_3, x_4, y_4, z_4, x_5, y_5, z_5, x_6, y_6, z_6$ |

**Other compounds with this structure**

- (Ca,Na)(Mg,Fe,Al,Ti)$(Si,Al)_2O_6$ (augite)

- (Finger, 1976) list the two (4$e$) positions as 'M1' and 'M2', and both are mixtures of calcium and magnesium. For visual clarity we desigated the first (4$e$) position as calcium and the second as magnesium.
- This structure has the same AFLOW label as Esseneite. The structures are generated by the same symmetry operations with different sets of parameters (`--params`) specified in their corresponding CIF files.

**Base-centered Monoclinic primitive vectors:**

$$\mathbf{a}_1 = \frac{1}{2}\,a\,\hat{\mathbf{x}} - \frac{1}{2}\,b\,\hat{\mathbf{y}}$$
$$\mathbf{a}_2 = \frac{1}{2}\,a\,\hat{\mathbf{x}} + \frac{1}{2}\,b\,\hat{\mathbf{y}}$$
$$\mathbf{a}_3 = c\cos\beta\,\hat{\mathbf{x}} + c\sin\beta\,\hat{\mathbf{z}}$$

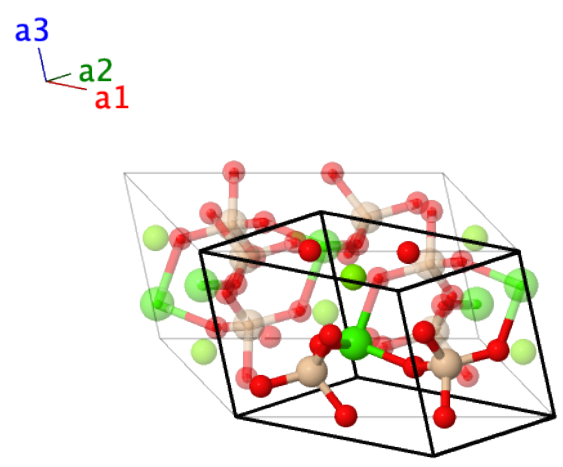

**Basis vectors:**

| | | Lattice Coordinates | | Cartesian Coordinates | Wyckoff Position | Atom Type |
|---|---|---|---|---|---|---|
| $\mathbf{B}_1$ | = | $-y_1\,\mathbf{a}_1 + y_1\,\mathbf{a}_2 + \frac{1}{4}\,\mathbf{a}_3$ | = | $\frac{1}{4}c\cos\beta\,\hat{\mathbf{x}} + y_1 b\,\hat{\mathbf{y}} + \frac{1}{4}c\sin\beta\,\hat{\mathbf{z}}$ | (4$e$) | Ca |
| $\mathbf{B}_2$ | = | $y_1\,\mathbf{a}_1 - y_1\,\mathbf{a}_2 + \frac{3}{4}\,\mathbf{a}_3$ | = | $\frac{3}{4}c\cos\beta\,\hat{\mathbf{x}} - y_1 b\,\hat{\mathbf{y}} + \frac{3}{4}c\sin\beta\,\hat{\mathbf{z}}$ | (4$e$) | Ca |
| $\mathbf{B}_3$ | = | $-y_2\,\mathbf{a}_1 + y_2\,\mathbf{a}_2 + \frac{1}{4}\,\mathbf{a}_3$ | = | $\frac{1}{4}c\cos\beta\,\hat{\mathbf{x}} + y_2 b\,\hat{\mathbf{y}} + \frac{1}{4}c\sin\beta\,\hat{\mathbf{z}}$ | (4$e$) | Mg |
| $\mathbf{B}_4$ | = | $y_2\,\mathbf{a}_1 - y_2\,\mathbf{a}_2 + \frac{3}{4}\,\mathbf{a}_3$ | = | $\frac{3}{4}c\cos\beta\,\hat{\mathbf{x}} - y_2 b\,\hat{\mathbf{y}} + \frac{3}{4}c\sin\beta\,\hat{\mathbf{z}}$ | (4$e$) | Mg |
| $\mathbf{B}_5$ | = | $(x_3 - y_3)\,\mathbf{a}_1 + (x_3 + y_3)\,\mathbf{a}_2 + z_3\,\mathbf{a}_3$ | = | $(x_3 a + z_3 c\cos\beta)\,\hat{\mathbf{x}} + y_3 b\,\hat{\mathbf{y}} + z_3 c\sin\beta\,\hat{\mathbf{z}}$ | (8$f$) | O I |
| $\mathbf{B}_6$ | = | $(-x_3 - y_3)\,\mathbf{a}_1 + (-x_3 + y_3)\,\mathbf{a}_2 + \left(\frac{1}{2} - z_3\right)\mathbf{a}_3$ | = | $\left(\frac{1}{2}c\cos\beta - x_3 a - z_3 c\cos\beta\right)\hat{\mathbf{x}} + y_3 b\,\hat{\mathbf{y}} + \left(\frac{1}{2} - z_3\right)c\sin\beta\,\hat{\mathbf{z}}$ | (8$f$) | O I |
| $\mathbf{B}_7$ | = | $(-x_3 + y_3)\,\mathbf{a}_1 + (-x_3 - y_3)\,\mathbf{a}_2 - z_3\,\mathbf{a}_3$ | = | $(-x_3 a - z_3 c\cos\beta)\,\hat{\mathbf{x}} - y_3 b\,\hat{\mathbf{y}} - z_3 c\sin\beta\,\hat{\mathbf{z}}$ | (8$f$) | O I |
| $\mathbf{B}_8$ | = | $(x_3 + y_3)\,\mathbf{a}_1 + (x_3 - y_3)\,\mathbf{a}_2 + \left(\frac{1}{2} + z_3\right)\mathbf{a}_3$ | = | $\left(\frac{1}{2}c\cos\beta + x_3 a + z_3 c\cos\beta\right)\hat{\mathbf{x}} - y_3 b\,\hat{\mathbf{y}} + \left(\frac{1}{2} + z_3\right)c\sin\beta\,\hat{\mathbf{z}}$ | (8$f$) | O I |
| $\mathbf{B}_9$ | = | $(x_4 - y_4)\,\mathbf{a}_1 + (x_4 + y_4)\,\mathbf{a}_2 + z_4\,\mathbf{a}_3$ | = | $(x_4 a + z_4 c\cos\beta)\,\hat{\mathbf{x}} + y_4 b\,\hat{\mathbf{y}} + z_4 c\sin\beta\,\hat{\mathbf{z}}$ | (8$f$) | O II |
| $\mathbf{B}_{10}$ | = | $(-x_4 - y_4)\,\mathbf{a}_1 + (-x_4 + y_4)\,\mathbf{a}_2 + \left(\frac{1}{2} - z_4\right)\mathbf{a}_3$ | = | $\left(\frac{1}{2}c\cos\beta - x_4 a - z_4 c\cos\beta\right)\hat{\mathbf{x}} + y_4 b\,\hat{\mathbf{y}} + \left(\frac{1}{2} - z_4\right)c\sin\beta\,\hat{\mathbf{z}}$ | (8$f$) | O II |
| $\mathbf{B}_{11}$ | = | $(-x_4 + y_4)\,\mathbf{a}_1 + (-x_4 - y_4)\,\mathbf{a}_2 - z_4\,\mathbf{a}_3$ | = | $(-x_4 a - z_4 c\cos\beta)\,\hat{\mathbf{x}} - y_4 b\,\hat{\mathbf{y}} - z_4 c\sin\beta\,\hat{\mathbf{z}}$ | (8$f$) | O II |
| $\mathbf{B}_{12}$ | = | $(x_4 + y_4)\,\mathbf{a}_1 + (x_4 - y_4)\,\mathbf{a}_2 + \left(\frac{1}{2} + z_4\right)\mathbf{a}_3$ | = | $\left(\frac{1}{2}c\cos\beta + x_4 a + z_4 c\cos\beta\right)\hat{\mathbf{x}} - y_4 b\,\hat{\mathbf{y}} + \left(\frac{1}{2} + z_4\right)c\sin\beta\,\hat{\mathbf{z}}$ | (8$f$) | O II |
| $\mathbf{B}_{13}$ | = | $(x_5 - y_5)\,\mathbf{a}_1 + (x_5 + y_5)\,\mathbf{a}_2 + z_5\,\mathbf{a}_3$ | = | $(x_5 a + z_5 c\cos\beta)\,\hat{\mathbf{x}} + y_5 b\,\hat{\mathbf{y}} + z_5 c\sin\beta\,\hat{\mathbf{z}}$ | (8$f$) | O III |
| $\mathbf{B}_{14}$ | = | $(-x_5 - y_5)\,\mathbf{a}_1 + (-x_5 + y_5)\,\mathbf{a}_2 + \left(\frac{1}{2} - z_5\right)\mathbf{a}_3$ | = | $\left(\frac{1}{2}c\cos\beta - x_5 a - z_5 c\cos\beta\right)\hat{\mathbf{x}} + y_5 b\,\hat{\mathbf{y}} + \left(\frac{1}{2} - z_5\right)c\sin\beta\,\hat{\mathbf{z}}$ | (8$f$) | O III |
| $\mathbf{B}_{15}$ | = | $(-x_5 + y_5)\,\mathbf{a}_1 + (-x_5 - y_5)\,\mathbf{a}_2 - z_5\,\mathbf{a}_3$ | = | $(-x_5 a - z_5 c\cos\beta)\,\hat{\mathbf{x}} - y_5 b\,\hat{\mathbf{y}} - z_5 c\sin\beta\,\hat{\mathbf{z}}$ | (8$f$) | O III |
| $\mathbf{B}_{16}$ | = | $(x_5 + y_5)\,\mathbf{a}_1 + (x_5 - y_5)\,\mathbf{a}_2 + \left(\frac{1}{2} + z_5\right)\mathbf{a}_3$ | = | $\left(\frac{1}{2}c\cos\beta + x_5 a + z_5 c\cos\beta\right)\hat{\mathbf{x}} - y_5 b\,\hat{\mathbf{y}} + \left(\frac{1}{2} + z_5\right)c\sin\beta\,\hat{\mathbf{z}}$ | (8$f$) | O III |
| $\mathbf{B}_{17}$ | = | $(x_6 - y_6)\,\mathbf{a}_1 + (x_6 + y_6)\,\mathbf{a}_2 + z_6\,\mathbf{a}_3$ | = | $(x_6 a + z_6 c\cos\beta)\,\hat{\mathbf{x}} + y_6 b\,\hat{\mathbf{y}} + z_6 c\sin\beta\,\hat{\mathbf{z}}$ | (8$f$) | Si |
| $\mathbf{B}_{18}$ | = | $(-x_6 - y_6)\,\mathbf{a}_1 + (-x_6 + y_6)\,\mathbf{a}_2 + \left(\frac{1}{2} - z_6\right)\mathbf{a}_3$ | = | $\left(\frac{1}{2}c\cos\beta - x_6 a - z_6 c\cos\beta\right)\hat{\mathbf{x}} + y_6 b\,\hat{\mathbf{y}} + \left(\frac{1}{2} - z_6\right)c\sin\beta\,\hat{\mathbf{z}}$ | (8$f$) | Si |
| $\mathbf{B}_{19}$ | = | $(-x_6 + y_6)\,\mathbf{a}_1 + (-x_6 - y_6)\,\mathbf{a}_2 - z_6\,\mathbf{a}_3$ | = | $(-x_6 a - z_6 c\cos\beta)\,\hat{\mathbf{x}} - y_6 b\,\hat{\mathbf{y}} - z_6 c\sin\beta\,\hat{\mathbf{z}}$ | (8$f$) | Si |
| $\mathbf{B}_{20}$ | = | $(x_6 + y_6)\,\mathbf{a}_1 + (x_6 - y_6)\,\mathbf{a}_2 + \left(\frac{1}{2} + z_6\right)\mathbf{a}_3$ | = | $\left(\frac{1}{2}c\cos\beta + x_6 a + z_6 c\cos\beta\right)\hat{\mathbf{x}} - y_6 b\,\hat{\mathbf{y}} + \left(\frac{1}{2} + z_6\right)c\sin\beta\,\hat{\mathbf{z}}$ | (8$f$) | Si |

**Geometry files:**

- CIF: pp. 1573
- POSCAR: pp. 1574

# $\beta$-Ga (*obsolete*) Structure: A_mC4_15_e

http://aflow.org/prototype-encyclopedia/A_mC4_15_e

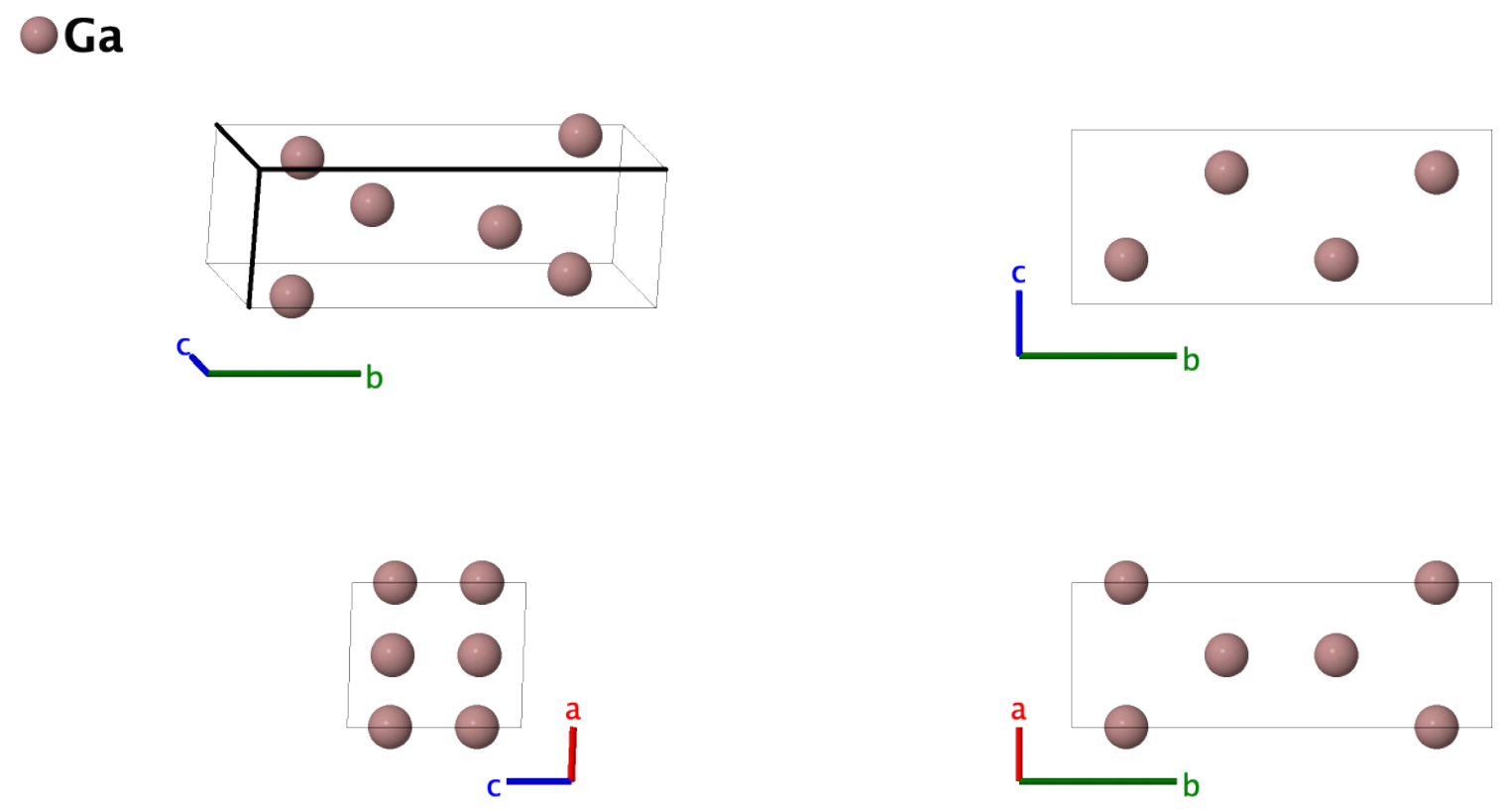

| | | |
|---|---|---|
| **Prototype** | : | Ga |
| **AFLOW prototype label** | : | A_mC4_15_e |
| ***Strukturbericht* designation** | : | None |
| **Pearson symbol** | : | mC4 |
| **Space group number** | : | 15 |
| **Space group symbol** | : | $C2/c$ |
| **AFLOW prototype command** | : | `aflow --proto=A_mC4_15_e` <br> `--params=`$a, b/a, c/a, \beta, y_1$ |

- $\beta$-Ga was proposed as a metastable structure of gallium, visible for short times at atomospheric pressure. The true high-pressure structure of gallium is isostructural with indium (A6).

**Base-centered Monoclinic primitive vectors:**

$$\mathbf{a}_1 = \frac{1}{2}a\,\hat{\mathbf{x}} - \frac{1}{2}b\,\hat{\mathbf{y}}$$
$$\mathbf{a}_2 = \frac{1}{2}a\,\hat{\mathbf{x}} + \frac{1}{2}b\,\hat{\mathbf{y}}$$
$$\mathbf{a}_3 = c\cos\beta\,\hat{\mathbf{x}} + c\sin\beta\,\hat{\mathbf{z}}$$

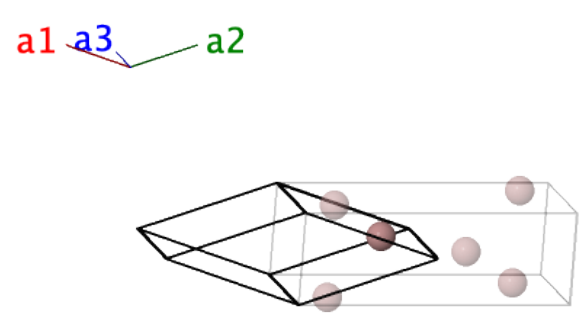

**Basis vectors:**

| | | Lattice Coordinates | | Cartesian Coordinates | Wyckoff Position | Atom Type |
|---|---|---|---|---|---|---|
| $\mathbf{B}_1$ | = | $-y_1\,\mathbf{a}_1 + y_1\,\mathbf{a}_2 + \frac{1}{4}\,\mathbf{a}_3$ | = | $\frac{1}{4}c\cos\beta\,\hat{\mathbf{x}} + y_1 b\,\hat{\mathbf{y}} + \frac{1}{4}c\sin\beta\,\hat{\mathbf{z}}$ | (4*e*) | Ga |
| $\mathbf{B}_2$ | = | $y_1\,\mathbf{a}_1 - y_1\,\mathbf{a}_2 + \frac{3}{4}\,\mathbf{a}_3$ | = | $\frac{3}{4}c\cos\beta\,\hat{\mathbf{x}} - y_1 b\,\hat{\mathbf{y}} + \frac{3}{4}c\sin\beta\,\hat{\mathbf{z}}$ | (4*e*) | Ga |

**Found in:**

---

**Geometry files:**
- CIF: pp. 1574
- POSCAR: pp. 1574

# $NaNbO_3$ Structure: ABC3_oP40_17_abcd_2e_abcd4e

http://aflow.org/prototype-encyclopedia/ABC3_oP40_17_abcd_2e_abcd4e

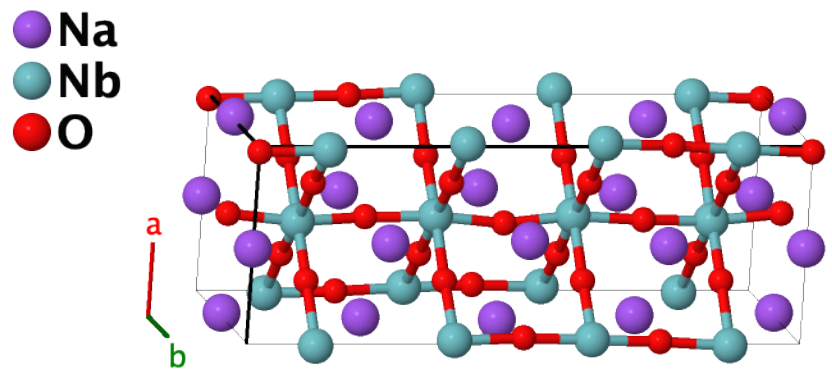

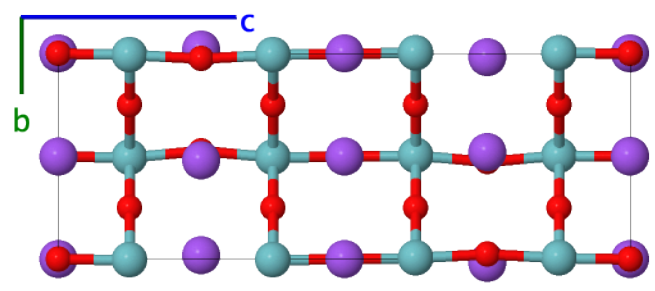

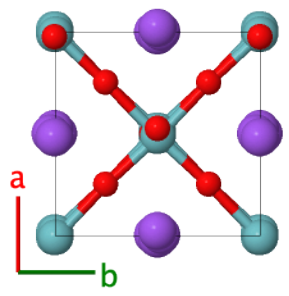

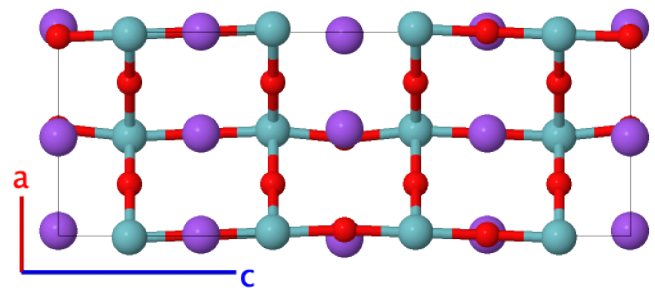

| | | |
|---|---|---|
| **Prototype** | : | $NaNbO_3$ |
| **AFLOW prototype label** | : | ABC3_oP40_17_abcd_2e_abcd4e |
| ***Strukturbericht* designation** | : | None |
| **Pearson symbol** | : | oP40 |
| **Space group number** | : | 17 |
| **Space group symbol** | : | $P222_1$ |
| **AFLOW prototype command** | : | `aflow --proto=ABC3_oP40_17_abcd_2e_abcd4e` `--params=`$a, b/a, c/a, x_1, x_2, x_3, x_4, y_5, y_6, y_7, y_8, x_9, y_9, z_9, x_{10}, y_{10}, z_{10}, x_{11}, y_{11}, z_{11}, x_{12}, y_{12}, z_{12}, x_{13}, y_{13}, z_{13}, x_{14}, y_{14}, z_{14}$ |

- (Downs, 2003) identifies this as a "possible polymorph of Lueshite."
- If the AFLOW parameters are set to `--params=`$a, 1, \sqrt{8}, 1/2, 0, 0, 1/2, 1/2, 0, 0, 1/2, 0, 0, 3/8, 1/2, 1/2, 3/8, 1/4, 1/4, 3/8, 3/4, 1/4, 3/8, 1/4, 3/4, 3/8, 3/4, 3/4, 3/8$ then the structure is equivalent to Cubic Perovskite $E2_1$.

### Simple Orthorhombic primitive vectors:

$$\mathbf{a}_1 = a\,\hat{\mathbf{x}}$$
$$\mathbf{a}_2 = b\,\hat{\mathbf{y}}$$
$$\mathbf{a}_3 = c\,\hat{\mathbf{z}}$$

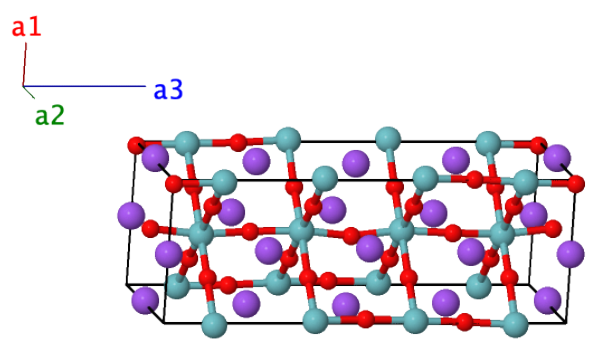

### Basis vectors:

| | | Lattice Coordinates | | Cartesian Coordinates | Wyckoff Position | Atom Type |
|---|---|---|---|---|---|---|
| $\mathbf{B}_1$ | = | $x_1\,\mathbf{a}_1$ | = | $x_1 a\,\hat{\mathbf{x}}$ | (2a) | Na I |
| $\mathbf{B}_2$ | = | $-x_1\,\mathbf{a}_1 + \frac{1}{2}\,\mathbf{a}_3$ | = | $-x_1 a\,\hat{\mathbf{x}} + \frac{1}{2}c\,\hat{\mathbf{z}}$ | (2a) | Na I |
| $\mathbf{B}_3$ | = | $x_2\,\mathbf{a}_1$ | = | $x_2 a\,\hat{\mathbf{x}}$ | (2a) | O I |
| $\mathbf{B}_4$ | = | $-x_2\,\mathbf{a}_1 + \frac{1}{2}\,\mathbf{a}_3$ | = | $-x_2 a\,\hat{\mathbf{x}} + \frac{1}{2}c\,\hat{\mathbf{z}}$ | (2a) | O I |
| $\mathbf{B}_5$ | = | $x_3\,\mathbf{a}_1 + \frac{1}{2}\,\mathbf{a}_2$ | = | $x_3 a\,\hat{\mathbf{x}} + \frac{1}{2}b\,\hat{\mathbf{y}}$ | (2b) | Na II |
| $\mathbf{B}_6$ | = | $-x_3\,\mathbf{a}_1 + \frac{1}{2}\,\mathbf{a}_2 + \frac{1}{2}\,\mathbf{a}_3$ | = | $-x_3 a\,\hat{\mathbf{x}} + \frac{1}{2}b\,\hat{\mathbf{y}} + \frac{1}{2}c\,\hat{\mathbf{z}}$ | (2b) | Na II |
| $\mathbf{B}_{7}$ | = | $x_4\,\mathbf{a}_1+\frac{1}{2}\,\mathbf{a}_2$ | = | $x_4a\,\hat{\mathbf{x}}+\frac{1}{2}b\,\hat{\mathbf{y}}$ | (2$b$) | O II |
| $\mathbf{B}_{8}$ | = | $-x_4\,\mathbf{a}_1+\frac{1}{2}\,\mathbf{a}_2+\frac{1}{2}\,\mathbf{a}_3$ | = | $-x_4a\,\hat{\mathbf{x}}+\frac{1}{2}b\,\hat{\mathbf{y}}+\frac{1}{2}c\,\hat{\mathbf{z}}$ | (2$b$) | O II |
| $\mathbf{B}_{9}$ | = | $y_5\,\mathbf{a}_2+\frac{1}{4}\,\mathbf{a}_3$ | = | $y_5b\,\hat{\mathbf{y}}+\frac{1}{4}c\,\hat{\mathbf{z}}$ | (2$c$) | Na III |
| $\mathbf{B}_{10}$ | = | $-y_5\,\mathbf{a}_2+\frac{3}{4}\,\mathbf{a}_3$ | = | $-y_5b\,\hat{\mathbf{y}}+\frac{3}{4}c\,\hat{\mathbf{z}}$ | (2$c$) | Na III |
| $\mathbf{B}_{11}$ | = | $y_6\,\mathbf{a}_2+\frac{1}{4}\,\mathbf{a}_3$ | = | $y_6b\,\hat{\mathbf{y}}+\frac{1}{4}c\,\hat{\mathbf{z}}$ | (2$c$) | O III |
| $\mathbf{B}_{12}$ | = | $-y_6\,\mathbf{a}_2+\frac{3}{4}\,\mathbf{a}_3$ | = | $-y_6b\,\hat{\mathbf{y}}+\frac{3}{4}c\,\hat{\mathbf{z}}$ | (2$c$) | O III |
| $\mathbf{B}_{13}$ | = | $\frac{1}{2}\,\mathbf{a}_1+y_7\,\mathbf{a}_2+\frac{1}{4}\,\mathbf{a}_3$ | = | $\frac{1}{2}a\,\hat{\mathbf{x}}+y_7b\,\hat{\mathbf{y}}+\frac{1}{4}c\,\hat{\mathbf{z}}$ | (2$d$) | Na IV |
| $\mathbf{B}_{14}$ | = | $\frac{1}{2}\,\mathbf{a}_1-y_7\,\mathbf{a}_2+\frac{3}{4}\,\mathbf{a}_3$ | = | $\frac{1}{2}a\,\hat{\mathbf{x}}-y_7b\,\hat{\mathbf{y}}+\frac{3}{4}c\,\hat{\mathbf{z}}$ | (2$d$) | Na IV |
| $\mathbf{B}_{15}$ | = | $\frac{1}{2}\,\mathbf{a}_1+y_8\,\mathbf{a}_2+\frac{1}{4}\,\mathbf{a}_3$ | = | $\frac{1}{2}a\,\hat{\mathbf{x}}+y_8b\,\hat{\mathbf{y}}+\frac{1}{4}c\,\hat{\mathbf{z}}$ | (2$d$) | O IV |
| $\mathbf{B}_{16}$ | = | $\frac{1}{2}\,\mathbf{a}_1-y_8\,\mathbf{a}_2+\frac{3}{4}\,\mathbf{a}_3$ | = | $\frac{1}{2}a\,\hat{\mathbf{x}}-y_8b\,\hat{\mathbf{y}}+\frac{3}{4}c\,\hat{\mathbf{z}}$ | (2$d$) | O IV |
| $\mathbf{B}_{17}$ | = | $x_9\,\mathbf{a}_1+y_9\,\mathbf{a}_2+z_9\,\mathbf{a}_3$ | = | $x_9a\,\hat{\mathbf{x}}+y_9b\,\hat{\mathbf{y}}+z_9c\,\hat{\mathbf{z}}$ | (4$e$) | Nb I |
| $\mathbf{B}_{18}$ | = | $-x_9\,\mathbf{a}_1-y_9\,\mathbf{a}_2+\left(\frac{1}{2}+z_9\right)\,\mathbf{a}_3$ | = | $-x_9a\,\hat{\mathbf{x}}-y_9b\,\hat{\mathbf{y}}+\left(\frac{1}{2}+z_9\right)c\,\hat{\mathbf{z}}$ | (4$e$) | Nb I |
| $\mathbf{B}_{19}$ | = | $-x_9\,\mathbf{a}_1+y_9\,\mathbf{a}_2+\left(\frac{1}{2}-z_9\right)\,\mathbf{a}_3$ | = | $-x_9a\,\hat{\mathbf{x}}+y_9b\,\hat{\mathbf{y}}+\left(\frac{1}{2}-z_9\right)c\,\hat{\mathbf{z}}$ | (4$e$) | Nb I |
| $\mathbf{B}_{20}$ | = | $x_9\,\mathbf{a}_1-y_9\,\mathbf{a}_2-z_9\,\mathbf{a}_3$ | = | $x_9a\,\hat{\mathbf{x}}-y_9b\,\hat{\mathbf{y}}-z_9c\,\hat{\mathbf{z}}$ | (4$e$) | Nb I |
| $\mathbf{B}_{21}$ | = | $x_{10}\,\mathbf{a}_1+y_{10}\,\mathbf{a}_2+z_{10}\,\mathbf{a}_3$ | = | $x_{10}a\,\hat{\mathbf{x}}+y_{10}b\,\hat{\mathbf{y}}+z_{10}c\,\hat{\mathbf{z}}$ | (4$e$) | Nb II |
| $\mathbf{B}_{22}$ | = | $-x_{10}\,\mathbf{a}_1-y_{10}\,\mathbf{a}_2+\left(\frac{1}{2}+z_{10}\right)\,\mathbf{a}_3$ | = | $-x_{10}a\,\hat{\mathbf{x}}-y_{10}b\,\hat{\mathbf{y}}+\left(\frac{1}{2}+z_{10}\right)c\,\hat{\mathbf{z}}$ | (4$e$) | Nb II |
| $\mathbf{B}_{23}$ | = | $-x_{10}\,\mathbf{a}_1+y_{10}\,\mathbf{a}_2+\left(\frac{1}{2}-z_{10}\right)\,\mathbf{a}_3$ | = | $-x_{10}a\,\hat{\mathbf{x}}+y_{10}b\,\hat{\mathbf{y}}+\left(\frac{1}{2}-z_{10}\right)c\,\hat{\mathbf{z}}$ | (4$e$) | Nb II |
| $\mathbf{B}_{24}$ | = | $x_{10}\,\mathbf{a}_1-y_{10}\,\mathbf{a}_2-z_{10}\,\mathbf{a}_3$ | = | $x_{10}a\,\hat{\mathbf{x}}-y_{10}b\,\hat{\mathbf{y}}-z_{10}c\,\hat{\mathbf{z}}$ | (4$e$) | Nb II |
| $\mathbf{B}_{25}$ | = | $x_{11}\,\mathbf{a}_1+y_{11}\,\mathbf{a}_2+z_{11}\,\mathbf{a}_3$ | = | $x_{11}a\,\hat{\mathbf{x}}+y_{11}b\,\hat{\mathbf{y}}+z_{11}c\,\hat{\mathbf{z}}$ | (4$e$) | O V |
| $\mathbf{B}_{26}$ | = | $-x_{11}\,\mathbf{a}_1-y_{11}\,\mathbf{a}_2+\left(\frac{1}{2}+z_{11}\right)\,\mathbf{a}_3$ | = | $-x_{11}a\,\hat{\mathbf{x}}-y_{11}b\,\hat{\mathbf{y}}+\left(\frac{1}{2}+z_{11}\right)c\,\hat{\mathbf{z}}$ | (4$e$) | O V |
| $\mathbf{B}_{27}$ | = | $-x_{11}\,\mathbf{a}_1+y_{11}\,\mathbf{a}_2+\left(\frac{1}{2}-z_{11}\right)\,\mathbf{a}_3$ | = | $-x_{11}a\,\hat{\mathbf{x}}+y_{11}b\,\hat{\mathbf{y}}+\left(\frac{1}{2}-z_{11}\right)c\,\hat{\mathbf{z}}$ | (4$e$) | O V |
| $\mathbf{B}_{28}$ | = | $x_{11}\,\mathbf{a}_1-y_{11}\,\mathbf{a}_2-z_{11}\,\mathbf{a}_3$ | = | $x_{11}a\,\hat{\mathbf{x}}-y_{11}b\,\hat{\mathbf{y}}-z_{11}c\,\hat{\mathbf{z}}$ | (4$e$) | O V |
| $\mathbf{B}_{29}$ | = | $x_{12}\,\mathbf{a}_1+y_{12}\,\mathbf{a}_2+z_{12}\,\mathbf{a}_3$ | = | $x_{12}a\,\hat{\mathbf{x}}+y_{12}b\,\hat{\mathbf{y}}+z_{12}c\,\hat{\mathbf{z}}$ | (4$e$) | O VI |
| $\mathbf{B}_{30}$ | = | $-x_{12}\,\mathbf{a}_1-y_{12}\,\mathbf{a}_2+\left(\frac{1}{2}+z_{12}\right)\,\mathbf{a}_3$ | = | $-x_{12}a\,\hat{\mathbf{x}}-y_{12}b\,\hat{\mathbf{y}}+\left(\frac{1}{2}+z_{12}\right)c\,\hat{\mathbf{z}}$ | (4$e$) | O VI |
| $\mathbf{B}_{31}$ | = | $-x_{12}\,\mathbf{a}_1+y_{12}\,\mathbf{a}_2+\left(\frac{1}{2}-z_{12}\right)\,\mathbf{a}_3$ | = | $-x_{12}a\,\hat{\mathbf{x}}+y_{12}b\,\hat{\mathbf{y}}+\left(\frac{1}{2}-z_{12}\right)c\,\hat{\mathbf{z}}$ | (4$e$) | O VI |
| $\mathbf{B}_{32}$ | = | $x_{12}\,\mathbf{a}_1-y_{12}\,\mathbf{a}_2-z_{12}\,\mathbf{a}_3$ | = | $x_{12}a\,\hat{\mathbf{x}}-y_{12}b\,\hat{\mathbf{y}}-z_{12}c\,\hat{\mathbf{z}}$ | (4$e$) | O VI |
| $\mathbf{B}_{33}$ | = | $x_{13}\,\mathbf{a}_1+y_{13}\,\mathbf{a}_2+z_{13}\,\mathbf{a}_3$ | = | $x_{13}a\,\hat{\mathbf{x}}+y_{13}b\,\hat{\mathbf{y}}+z_{13}c\,\hat{\mathbf{z}}$ | (4$e$) | O VII |
| $\mathbf{B}_{34}$ | = | $-x_{13}\,\mathbf{a}_1-y_{13}\,\mathbf{a}_2+\left(\frac{1}{2}+z_{13}\right)\,\mathbf{a}_3$ | = | $-x_{13}a\,\hat{\mathbf{x}}-y_{13}b\,\hat{\mathbf{y}}+\left(\frac{1}{2}+z_{13}\right)c\,\hat{\mathbf{z}}$ | (4$e$) | O VII |
| $\mathbf{B}_{35}$ | = | $-x_{13}\,\mathbf{a}_1+y_{13}\,\mathbf{a}_2+\left(\frac{1}{2}-z_{13}\right)\,\mathbf{a}_3$ | = | $-x_{13}a\,\hat{\mathbf{x}}+y_{13}b\,\hat{\mathbf{y}}+\left(\frac{1}{2}-z_{13}\right)c\,\hat{\mathbf{z}}$ | (4$e$) | O VII |
| $\mathbf{B}_{36}$ | = | $x_{13}\,\mathbf{a}_1-y_{13}\,\mathbf{a}_2-z_{13}\,\mathbf{a}_3$ | = | $x_{13}a\,\hat{\mathbf{x}}-y_{13}b\,\hat{\mathbf{y}}-z_{13}c\,\hat{\mathbf{z}}$ | (4$e$) | O VII |
| $\mathbf{B}_{37}$ | = | $x_{14}\,\mathbf{a}_1+y_{14}\,\mathbf{a}_2+z_{14}\,\mathbf{a}_3$ | = | $x_{14}a\,\hat{\mathbf{x}}+y_{14}b\,\hat{\mathbf{y}}+z_{14}c\,\hat{\mathbf{z}}$ | (4$e$) | O VIII |
| $\mathbf{B}_{38}$ | = | $-x_{14}\,\mathbf{a}_1-y_{14}\,\mathbf{a}_2+\left(\frac{1}{2}+z_{14}\right)\,\mathbf{a}_3$ | = | $-x_{14}a\,\hat{\mathbf{x}}-y_{14}b\,\hat{\mathbf{y}}+\left(\frac{1}{2}+z_{14}\right)c\,\hat{\mathbf{z}}$ | (4$e$) | O VIII |
| $\mathbf{B}_{39}$ | = | $-x_{14}\,\mathbf{a}_1+y_{14}\,\mathbf{a}_2+\left(\frac{1}{2}-z_{14}\right)\,\mathbf{a}_3$ | = | $-x_{14}a\,\hat{\mathbf{x}}+y_{14}b\,\hat{\mathbf{y}}+\left(\frac{1}{2}-z_{14}\right)c\,\hat{\mathbf{z}}$ | (4$e$) | O VIII |
| $\mathbf{B}_{40}$ | = | $x_{14}\,\mathbf{a}_1-y_{14}\,\mathbf{a}_2-z_{14}\,\mathbf{a}_3$ | = | $x_{14}a\,\hat{\mathbf{x}}-y_{14}b\,\hat{\mathbf{y}}-z_{14}c\,\hat{\mathbf{z}}$ | (4$e$) | O VIII |

**Found in:**

- R. T. Downs and M. Hall-Wallace, *The American Mineralogist Crystal Structure Database*, Am. Mineral. **88**, 247–250 (2003).

---

**Geometry files:**

- CIF: pp. 1574
- POSCAR: pp. 1575

# $\gamma$-$TeO_2$ Structure: A2B_oP12_18_2c_c

http://aflow.org/prototype-encyclopedia/A2B_oP12_18_2c_c

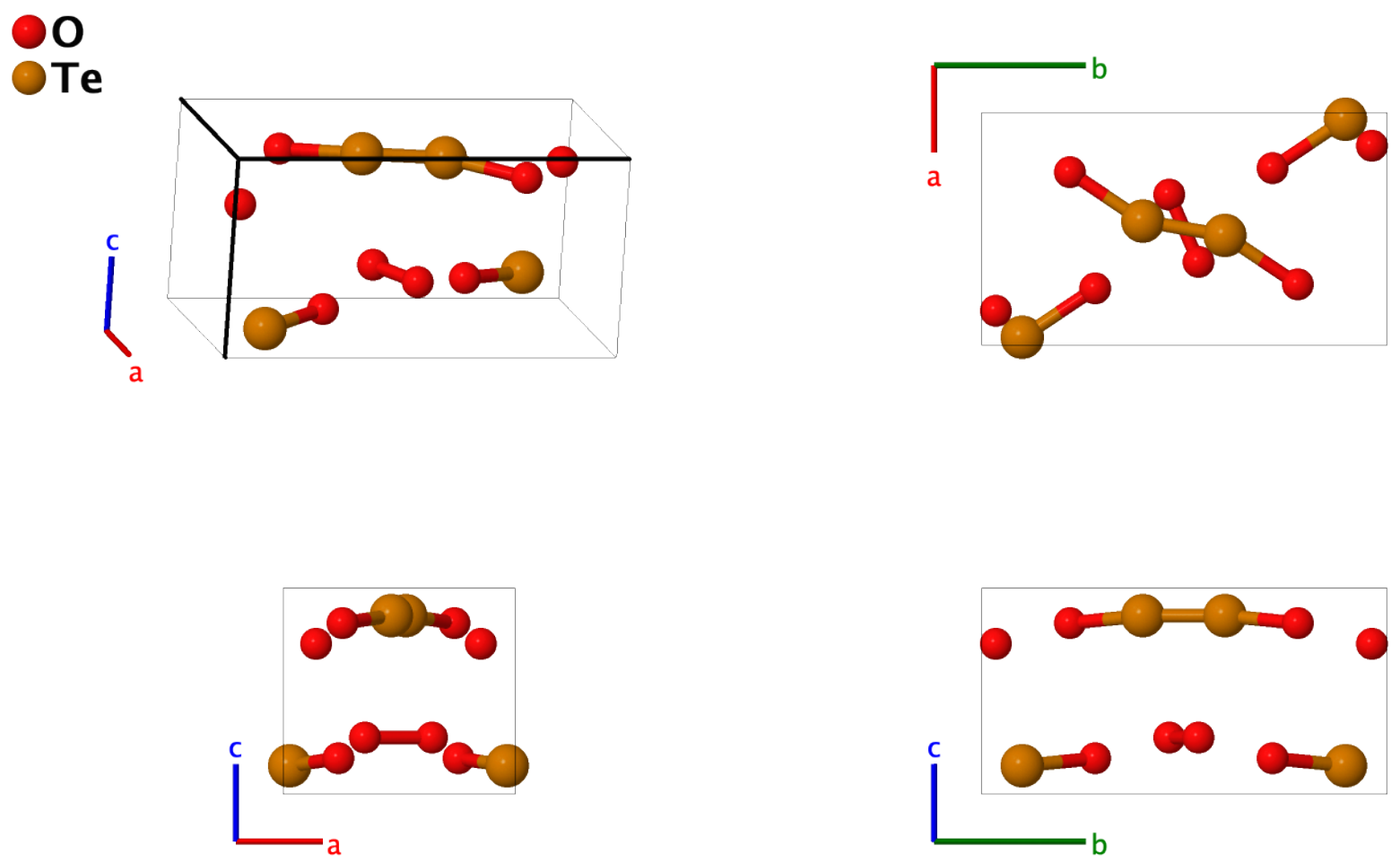

| | | |
|---|---|---|
| **Prototype** | : | $O_2Te$ |
| **AFLOW prototype label** | : | A2B_oP12_18_2c_c |
| ***Strukturbericht* designation** | : | None |
| **Pearson symbol** | : | oP12 |
| **Space group number** | : | 18 |
| **Space group symbol** | : | $P2_12_12$ |
| **AFLOW prototype command** | : | `aflow --proto=A2B_oP12_18_2c_c`<br>`--params=`$a, b/a, c/a, x_1, y_1, z_1, x_2, y_2, z_2, x_3, y_3, z_3$ |

- This is a metastable state of $TeO_2$, similar to $\alpha$-$TeO_2$ paratelluride. (Champarnaud-Mesjard, 2000) labels the atomic positions as being on the '(4$a$)' Wyckoff position, but this does not exist for space group $P2_12_12$ #18. The atoms are actually on the general site for this space group, (4$c$).

**Simple Orthorhombic primitive vectors:**

$$\mathbf{a}_1 = a\,\hat{\mathbf{x}}$$
$$\mathbf{a}_2 = b\,\hat{\mathbf{y}}$$
$$\mathbf{a}_3 = c\,\hat{\mathbf{z}}$$

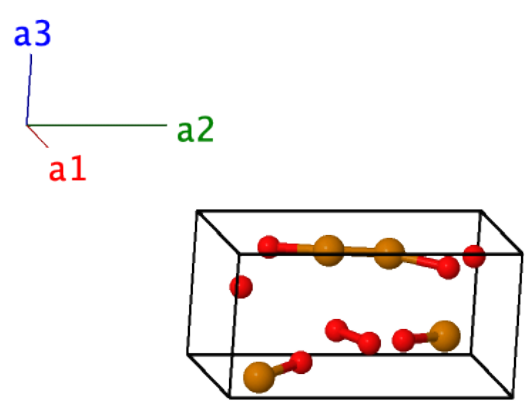

**Basis vectors:**

| | | Lattice Coordinates | | Cartesian Coordinates | Wyckoff Position | Atom Type |
|---|---|---|---|---|---|---|
| $\mathbf{B}_1$ | = | $x_1\,\mathbf{a}_1 + y_1\,\mathbf{a}_2 + z_1\,\mathbf{a}_3$ | = | $x_1 a\,\hat{\mathbf{x}} + y_1 b\,\hat{\mathbf{y}} + z_1 c\,\hat{\mathbf{z}}$ | (4$c$) | O I |
| $\mathbf{B}_2$ | = | $-x_1\,\mathbf{a}_1 - y_1\,\mathbf{a}_2 + z_1\,\mathbf{a}_3$ | = | $-x_1 a\,\hat{\mathbf{x}} - y_1 b\,\hat{\mathbf{y}} + z_1 c\,\hat{\mathbf{z}}$ | (4$c$) | O I |
| $\mathbf{B}_3$ | = | $\left(\frac{1}{2} - x_1\right)\mathbf{a}_1 + \left(\frac{1}{2} + y_1\right)\mathbf{a}_2 - z_1\,\mathbf{a}_3$ | = | $\left(\frac{1}{2} - x_1\right) a\,\hat{\mathbf{x}} + \left(\frac{1}{2} + y_1\right) b\,\hat{\mathbf{y}} - z_1 c\,\hat{\mathbf{z}}$ | (4$c$) | O I |
| $\mathbf{B}_4$ | = | $\left(\frac{1}{2} + x_1\right)\mathbf{a}_1 + \left(\frac{1}{2} - y_1\right)\mathbf{a}_2 - z_1\,\mathbf{a}_3$ | = | $\left(\frac{1}{2} + x_1\right) a\,\hat{\mathbf{x}} + \left(\frac{1}{2} - y_1\right) b\,\hat{\mathbf{y}} - z_1 c\,\hat{\mathbf{z}}$ | (4$c$) | O I |
| $\mathbf{B}_5$ | = | $x_2\,\mathbf{a}_1 + y_2\,\mathbf{a}_2 + z_2\,\mathbf{a}_3$ | = | $x_2 a\,\hat{\mathbf{x}} + y_2 b\,\hat{\mathbf{y}} + z_2 c\,\hat{\mathbf{z}}$ | (4$c$) | O II |
| $\mathbf{B}_6$ | = | $-x_2\,\mathbf{a}_1 - y_2\,\mathbf{a}_2 + z_2\,\mathbf{a}_3$ | = | $-x_2 a\,\hat{\mathbf{x}} - y_2 b\,\hat{\mathbf{y}} + z_2 c\,\hat{\mathbf{z}}$ | (4$c$) | O II |
| $\mathbf{B}_7$ | = | $\left(\frac{1}{2} - x_2\right)\mathbf{a}_1 + \left(\frac{1}{2} + y_2\right)\mathbf{a}_2 - z_2\,\mathbf{a}_3$ | = | $\left(\frac{1}{2} - x_2\right) a\,\hat{\mathbf{x}} + \left(\frac{1}{2} + y_2\right) b\,\hat{\mathbf{y}} - z_2 c\,\hat{\mathbf{z}}$ | (4$c$) | O II |
| $\mathbf{B}_8$ | = | $\left(\frac{1}{2} + x_2\right)\mathbf{a}_1 + \left(\frac{1}{2} - y_2\right)\mathbf{a}_2 - z_2\,\mathbf{a}_3$ | = | $\left(\frac{1}{2} + x_2\right) a\,\hat{\mathbf{x}} + \left(\frac{1}{2} - y_2\right) b\,\hat{\mathbf{y}} - z_2 c\,\hat{\mathbf{z}}$ | (4$c$) | O II |
| $\mathbf{B}_9$ | = | $x_3\,\mathbf{a}_1 + y_3\,\mathbf{a}_2 + z_3\,\mathbf{a}_3$ | = | $x_3 a\,\hat{\mathbf{x}} + y_3 b\,\hat{\mathbf{y}} + z_3 c\,\hat{\mathbf{z}}$ | (4$c$) | Te |
| $\mathbf{B}_{10}$ | = | $-x_3\,\mathbf{a}_1 - y_3\,\mathbf{a}_2 + z_3\,\mathbf{a}_3$ | = | $-x_3 a\,\hat{\mathbf{x}} - y_3 b\,\hat{\mathbf{y}} + z_3 c\,\hat{\mathbf{z}}$ | (4$c$) | Te |
| $\mathbf{B}_{11}$ | = | $\left(\frac{1}{2} - x_3\right)\mathbf{a}_1 + \left(\frac{1}{2} + y_3\right)\mathbf{a}_2 - z_3\,\mathbf{a}_3$ | = | $\left(\frac{1}{2} - x_3\right) a\,\hat{\mathbf{x}} + \left(\frac{1}{2} + y_3\right) b\,\hat{\mathbf{y}} - z_3 c\,\hat{\mathbf{z}}$ | (4$c$) | Te |
| $\mathbf{B}_{12}$ | = | $\left(\frac{1}{2} + x_3\right)\mathbf{a}_1 + \left(\frac{1}{2} - y_3\right)\mathbf{a}_2 - z_3\,\mathbf{a}_3$ | = | $\left(\frac{1}{2} + x_3\right) a\,\hat{\mathbf{x}} + \left(\frac{1}{2} - y_3\right) b\,\hat{\mathbf{y}} - z_3 c\,\hat{\mathbf{z}}$ | (4$c$) | Te |

**Found in:**

**Geometry files:**

- CIF: pp. 1575
- POSCAR: pp. 1575

# Diamminetriamidodizinc Chloride ($[Zn_2(NH_3)_2(NH_2)_3]Cl$) Structure: AB12C5D2_oP40_18_a_6c_b2c_c

http://aflow.org/prototype-encyclopedia/AB12C5D2_oP40_18_a_6c_b2c_c

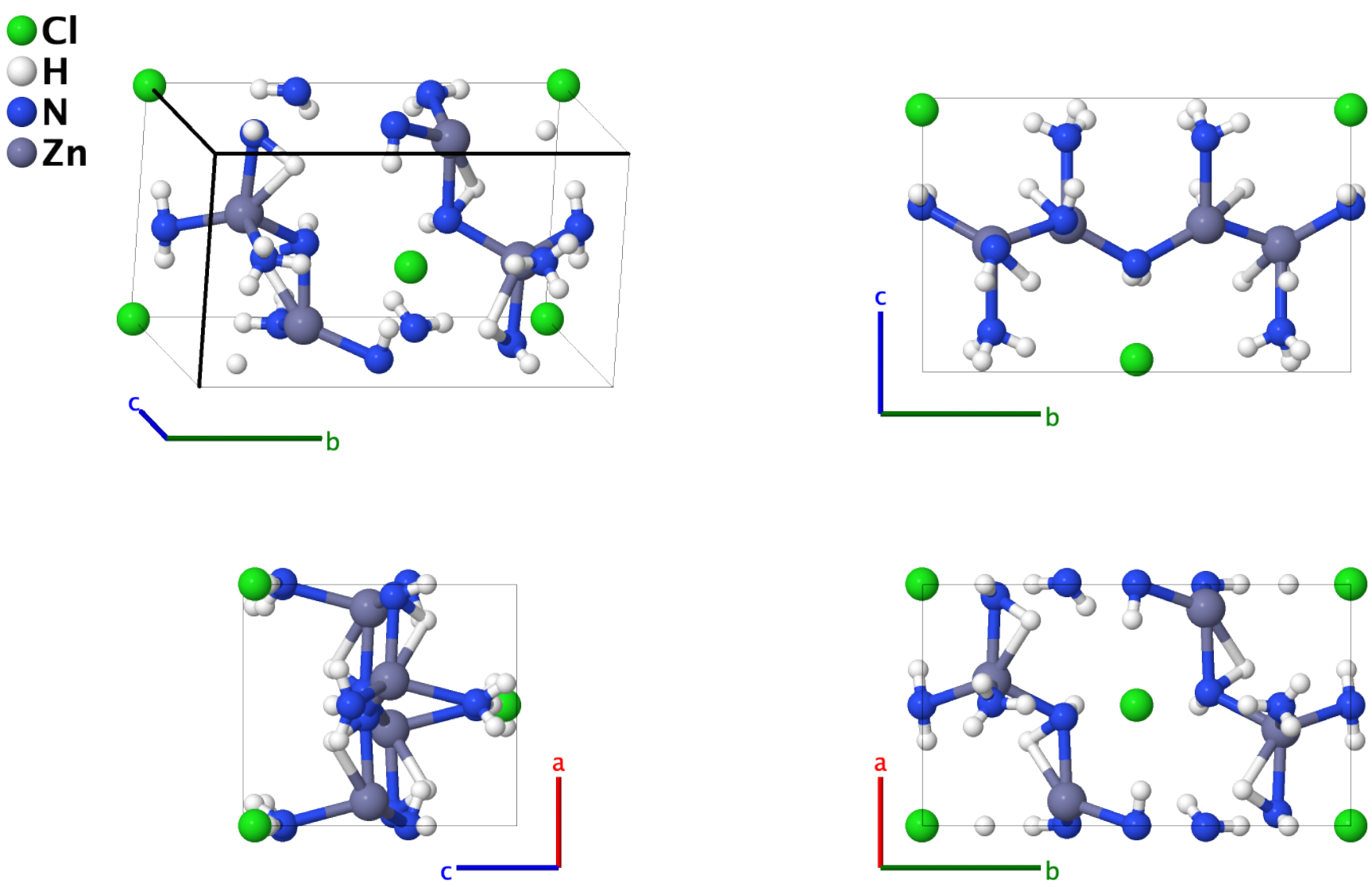

| | | |
|---|---|---|
| **Prototype** | : | $ClH_{12}N_5Zn_2$ |
| **AFLOW prototype label** | : | AB12C5D2_oP40_18_a_6c_b2c_c |
| ***Strukturbericht* designation** | : | None |
| **Pearson symbol** | : | oP40 |
| **Space group number** | : | 18 |
| **Space group symbol** | : | $P2_12_12$ |
| **AFLOW prototype command** | : | `aflow --proto=AB12C5D2_oP40_18_a_6c_b2c_c --params=`$a, b/a, c/a, z_1, z_2, x_3, y_3, z_3, x_4, y_4, z_4, x_5, y_5, z_5, x_6, y_6, z_6, x_7, y_7, z_7, x_8, y_8, z_8, x_9, y_9, z_9, x_{10}, y_{10}, z_{10}, x_{11}, y_{11}, z_{11}$ |

**Simple Orthorhombic primitive vectors:**

$$\mathbf{a}_1 = a\,\hat{\mathbf{x}}$$
$$\mathbf{a}_2 = b\,\hat{\mathbf{y}}$$
$$\mathbf{a}_3 = c\,\hat{\mathbf{z}}$$

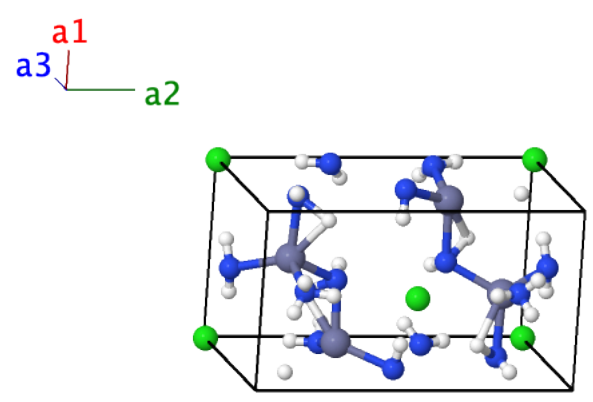

**Basis vectors:**

| | | Lattice Coordinates | | Cartesian Coordinates | Wyckoff Position | Atom Type |
|---|---|---|---|---|---|---|
| $\mathbf{B}_1$ | = | $z_1\,\mathbf{a}_3$ | = | $z_1 c\,\hat{\mathbf{z}}$ | (2$a$) | Cl |
| $\mathbf{B}_2$ | = | $\frac{1}{2}\,\mathbf{a}_1 + \frac{1}{2}\,\mathbf{a}_2 - z_1\,\mathbf{a}_3$ | = | $\frac{1}{2}a\,\hat{\mathbf{x}} + \frac{1}{2}b\,\hat{\mathbf{y}} - z_1 c\,\hat{\mathbf{z}}$ | (2$a$) | Cl |
| $\mathbf{B}_3$ | = | $\frac{1}{2}\,\mathbf{a}_2 + z_2\,\mathbf{a}_3$ | = | $\frac{1}{2}b\,\hat{\mathbf{y}} + z_2 c\,\hat{\mathbf{z}}$ | (2$b$) | N I |
| $\mathbf{B}_4$ | = | $\frac{1}{2}\,\mathbf{a}_1 - z_2\,\mathbf{a}_3$ | = | $\frac{1}{2}a\,\hat{\mathbf{x}} - z_2c\,\hat{\mathbf{z}}$ | (2b) | N I |
| $\mathbf{B}_5$ | = | $x_3\,\mathbf{a}_1 + y_3\,\mathbf{a}_2 + z_3\,\mathbf{a}_3$ | = | $x_3a\,\hat{\mathbf{x}} + y_3b\,\hat{\mathbf{y}} + z_3c\,\hat{\mathbf{z}}$ | (4c) | H I |
| $\mathbf{B}_6$ | = | $-x_3\,\mathbf{a}_1 - y_3\,\mathbf{a}_2 + z_3\,\mathbf{a}_3$ | = | $-x_3a\,\hat{\mathbf{x}} - y_3b\,\hat{\mathbf{y}} + z_3c\,\hat{\mathbf{z}}$ | (4c) | H I |
| $\mathbf{B}_7$ | = | $\left(\frac{1}{2} - x_3\right)\mathbf{a}_1 + \left(\frac{1}{2} + y_3\right)\mathbf{a}_2 - z_3\,\mathbf{a}_3$ | = | $\left(\frac{1}{2} - x_3\right)a\,\hat{\mathbf{x}} + \left(\frac{1}{2} + y_3\right)b\,\hat{\mathbf{y}} - z_3c\,\hat{\mathbf{z}}$ | (4c) | H I |
| $\mathbf{B}_8$ | = | $\left(\frac{1}{2} + x_3\right)\mathbf{a}_1 + \left(\frac{1}{2} - y_3\right)\mathbf{a}_2 - z_3\,\mathbf{a}_3$ | = | $\left(\frac{1}{2} + x_3\right)a\,\hat{\mathbf{x}} + \left(\frac{1}{2} - y_3\right)b\,\hat{\mathbf{y}} - z_3c\,\hat{\mathbf{z}}$ | (4c) | H I |
| $\mathbf{B}_9$ | = | $x_4\,\mathbf{a}_1 + y_4\,\mathbf{a}_2 + z_4\,\mathbf{a}_3$ | = | $x_4a\,\hat{\mathbf{x}} + y_4b\,\hat{\mathbf{y}} + z_4c\,\hat{\mathbf{z}}$ | (4c) | H II |
| $\mathbf{B}_{10}$ | = | $-x_4\,\mathbf{a}_1 - y_4\,\mathbf{a}_2 + z_4\,\mathbf{a}_3$ | = | $-x_4a\,\hat{\mathbf{x}} - y_4b\,\hat{\mathbf{y}} + z_4c\,\hat{\mathbf{z}}$ | (4c) | H II |
| $\mathbf{B}_{11}$ | = | $\left(\frac{1}{2} - x_4\right)\mathbf{a}_1 + \left(\frac{1}{2} + y_4\right)\mathbf{a}_2 - z_4\,\mathbf{a}_3$ | = | $\left(\frac{1}{2} - x_4\right)a\,\hat{\mathbf{x}} + \left(\frac{1}{2} + y_4\right)b\,\hat{\mathbf{y}} - z_4c\,\hat{\mathbf{z}}$ | (4c) | H II |
| $\mathbf{B}_{12}$ | = | $\left(\frac{1}{2} + x_4\right)\mathbf{a}_1 + \left(\frac{1}{2} - y_4\right)\mathbf{a}_2 - z_4\,\mathbf{a}_3$ | = | $\left(\frac{1}{2} + x_4\right)a\,\hat{\mathbf{x}} + \left(\frac{1}{2} - y_4\right)b\,\hat{\mathbf{y}} - z_4c\,\hat{\mathbf{z}}$ | (4c) | H II |
| $\mathbf{B}_{13}$ | = | $x_5\,\mathbf{a}_1 + y_5\,\mathbf{a}_2 + z_5\,\mathbf{a}_3$ | = | $x_5a\,\hat{\mathbf{x}} + y_5b\,\hat{\mathbf{y}} + z_5c\,\hat{\mathbf{z}}$ | (4c) | H III |
| $\mathbf{B}_{14}$ | = | $-x_5\,\mathbf{a}_1 - y_5\,\mathbf{a}_2 + z_5\,\mathbf{a}_3$ | = | $-x_5a\,\hat{\mathbf{x}} - y_5b\,\hat{\mathbf{y}} + z_5c\,\hat{\mathbf{z}}$ | (4c) | H III |
| $\mathbf{B}_{15}$ | = | $\left(\frac{1}{2} - x_5\right)\mathbf{a}_1 + \left(\frac{1}{2} + y_5\right)\mathbf{a}_2 - z_5\,\mathbf{a}_3$ | = | $\left(\frac{1}{2} - x_5\right)a\,\hat{\mathbf{x}} + \left(\frac{1}{2} + y_5\right)b\,\hat{\mathbf{y}} - z_5c\,\hat{\mathbf{z}}$ | (4c) | H III |
| $\mathbf{B}_{16}$ | = | $\left(\frac{1}{2} + x_5\right)\mathbf{a}_1 + \left(\frac{1}{2} - y_5\right)\mathbf{a}_2 - z_5\,\mathbf{a}_3$ | = | $\left(\frac{1}{2} + x_5\right)a\,\hat{\mathbf{x}} + \left(\frac{1}{2} - y_5\right)b\,\hat{\mathbf{y}} - z_5c\,\hat{\mathbf{z}}$ | (4c) | H III |
| $\mathbf{B}_{17}$ | = | $x_6\,\mathbf{a}_1 + y_6\,\mathbf{a}_2 + z_6\,\mathbf{a}_3$ | = | $x_6a\,\hat{\mathbf{x}} + y_6b\,\hat{\mathbf{y}} + z_6c\,\hat{\mathbf{z}}$ | (4c) | H IV |
| $\mathbf{B}_{18}$ | = | $-x_6\,\mathbf{a}_1 - y_6\,\mathbf{a}_2 + z_6\,\mathbf{a}_3$ | = | $-x_6a\,\hat{\mathbf{x}} - y_6b\,\hat{\mathbf{y}} + z_6c\,\hat{\mathbf{z}}$ | (4c) | H IV |
| $\mathbf{B}_{19}$ | = | $\left(\frac{1}{2} - x_6\right)\mathbf{a}_1 + \left(\frac{1}{2} + y_6\right)\mathbf{a}_2 - z_6\,\mathbf{a}_3$ | = | $\left(\frac{1}{2} - x_6\right)a\,\hat{\mathbf{x}} + \left(\frac{1}{2} + y_6\right)b\,\hat{\mathbf{y}} - z_6c\,\hat{\mathbf{z}}$ | (4c) | H IV |
| $\mathbf{B}_{20}$ | = | $\left(\frac{1}{2} + x_6\right)\mathbf{a}_1 + \left(\frac{1}{2} - y_6\right)\mathbf{a}_2 - z_6\,\mathbf{a}_3$ | = | $\left(\frac{1}{2} + x_6\right)a\,\hat{\mathbf{x}} + \left(\frac{1}{2} - y_6\right)b\,\hat{\mathbf{y}} - z_6c\,\hat{\mathbf{z}}$ | (4c) | H IV |
| $\mathbf{B}_{21}$ | = | $x_7\,\mathbf{a}_1 + y_7\,\mathbf{a}_2 + z_7\,\mathbf{a}_3$ | = | $x_7a\,\hat{\mathbf{x}} + y_7b\,\hat{\mathbf{y}} + z_7c\,\hat{\mathbf{z}}$ | (4c) | H V |
| $\mathbf{B}_{22}$ | = | $-x_7\,\mathbf{a}_1 - y_7\,\mathbf{a}_2 + z_7\,\mathbf{a}_3$ | = | $-x_7a\,\hat{\mathbf{x}} - y_7b\,\hat{\mathbf{y}} + z_7c\,\hat{\mathbf{z}}$ | (4c) | H V |
| $\mathbf{B}_{23}$ | = | $\left(\frac{1}{2} - x_7\right)\mathbf{a}_1 + \left(\frac{1}{2} + y_7\right)\mathbf{a}_2 - z_7\,\mathbf{a}_3$ | = | $\left(\frac{1}{2} - x_7\right)a\,\hat{\mathbf{x}} + \left(\frac{1}{2} + y_7\right)b\,\hat{\mathbf{y}} - z_7c\,\hat{\mathbf{z}}$ | (4c) | H V |
| $\mathbf{B}_{24}$ | = | $\left(\frac{1}{2} + x_7\right)\mathbf{a}_1 + \left(\frac{1}{2} - y_7\right)\mathbf{a}_2 - z_7\,\mathbf{a}_3$ | = | $\left(\frac{1}{2} + x_7\right)a\,\hat{\mathbf{x}} + \left(\frac{1}{2} - y_7\right)b\,\hat{\mathbf{y}} - z_7c\,\hat{\mathbf{z}}$ | (4c) | H V |
| $\mathbf{B}_{25}$ | = | $x_8\,\mathbf{a}_1 + y_8\,\mathbf{a}_2 + z_8\,\mathbf{a}_3$ | = | $x_8a\,\hat{\mathbf{x}} + y_8b\,\hat{\mathbf{y}} + z_8c\,\hat{\mathbf{z}}$ | (4c) | H VI |
| $\mathbf{B}_{26}$ | = | $-x_8\,\mathbf{a}_1 - y_8\,\mathbf{a}_2 + z_8\,\mathbf{a}_3$ | = | $-x_8a\,\hat{\mathbf{x}} - y_8b\,\hat{\mathbf{y}} + z_8c\,\hat{\mathbf{z}}$ | (4c) | H VI |
| $\mathbf{B}_{27}$ | = | $\left(\frac{1}{2} - x_8\right)\mathbf{a}_1 + \left(\frac{1}{2} + y_8\right)\mathbf{a}_2 - z_8\,\mathbf{a}_3$ | = | $\left(\frac{1}{2} - x_8\right)a\,\hat{\mathbf{x}} + \left(\frac{1}{2} + y_8\right)b\,\hat{\mathbf{y}} - z_8c\,\hat{\mathbf{z}}$ | (4c) | H VI |
| $\mathbf{B}_{28}$ | = | $\left(\frac{1}{2} + x_8\right)\mathbf{a}_1 + \left(\frac{1}{2} - y_8\right)\mathbf{a}_2 - z_8\,\mathbf{a}_3$ | = | $\left(\frac{1}{2} + x_8\right)a\,\hat{\mathbf{x}} + \left(\frac{1}{2} - y_8\right)b\,\hat{\mathbf{y}} - z_8c\,\hat{\mathbf{z}}$ | (4c) | H VI |
| $\mathbf{B}_{29}$ | = | $x_9\,\mathbf{a}_1 + y_9\,\mathbf{a}_2 + z_9\,\mathbf{a}_3$ | = | $x_9a\,\hat{\mathbf{x}} + y_9b\,\hat{\mathbf{y}} + z_9c\,\hat{\mathbf{z}}$ | (4c) | N II |
| $\mathbf{B}_{30}$ | = | $-x_9\,\mathbf{a}_1 - y_9\,\mathbf{a}_2 + z_9\,\mathbf{a}_3$ | = | $-x_9a\,\hat{\mathbf{x}} - y_9b\,\hat{\mathbf{y}} + z_9c\,\hat{\mathbf{z}}$ | (4c) | N II |
| $\mathbf{B}_{31}$ | = | $\left(\frac{1}{2} - x_9\right)\mathbf{a}_1 + \left(\frac{1}{2} + y_9\right)\mathbf{a}_2 - z_9\,\mathbf{a}_3$ | = | $\left(\frac{1}{2} - x_9\right)a\,\hat{\mathbf{x}} + \left(\frac{1}{2} + y_9\right)b\,\hat{\mathbf{y}} - z_9c\,\hat{\mathbf{z}}$ | (4c) | N II |
| $\mathbf{B}_{32}$ | = | $\left(\frac{1}{2} + x_9\right)\mathbf{a}_1 + \left(\frac{1}{2} - y_9\right)\mathbf{a}_2 - z_9\,\mathbf{a}_3$ | = | $\left(\frac{1}{2} + x_9\right)a\,\hat{\mathbf{x}} + \left(\frac{1}{2} - y_9\right)b\,\hat{\mathbf{y}} - z_9c\,\hat{\mathbf{z}}$ | (4c) | N II |
| $\mathbf{B}_{33}$ | = | $x_{10}\,\mathbf{a}_1 + y_{10}\,\mathbf{a}_2 + z_{10}\,\mathbf{a}_3$ | = | $x_{10}a\,\hat{\mathbf{x}} + y_{10}b\,\hat{\mathbf{y}} + z_{10}c\,\hat{\mathbf{z}}$ | (4c) | N III |
| $\mathbf{B}_{34}$ | = | $-x_{10}\,\mathbf{a}_1 - y_{10}\,\mathbf{a}_2 + z_{10}\,\mathbf{a}_3$ | = | $-x_{10}a\,\hat{\mathbf{x}} - y_{10}b\,\hat{\mathbf{y}} + z_{10}c\,\hat{\mathbf{z}}$ | (4c) | N III |
| $\mathbf{B}_{35}$ | = | $\left(\frac{1}{2} - x_{10}\right)\mathbf{a}_1 + \left(\frac{1}{2} + y_{10}\right)\mathbf{a}_2 - z_{10}\,\mathbf{a}_3$ | = | $\left(\frac{1}{2} - x_{10}\right)a\,\hat{\mathbf{x}} + \left(\frac{1}{2} + y_{10}\right)b\,\hat{\mathbf{y}} - z_{10}c\,\hat{\mathbf{z}}$ | (4c) | N III |
| $\mathbf{B}_{36}$ | = | $\left(\frac{1}{2} + x_{10}\right)\mathbf{a}_1 + \left(\frac{1}{2} - y_{10}\right)\mathbf{a}_2 - z_{10}\,\mathbf{a}_3$ | = | $\left(\frac{1}{2} + x_{10}\right)a\,\hat{\mathbf{x}} + \left(\frac{1}{2} - y_{10}\right)b\,\hat{\mathbf{y}} - z_{10}c\,\hat{\mathbf{z}}$ | (4c) | N III |
| $\mathbf{B}_{37}$ | = | $x_{11}\,\mathbf{a}_1 + y_{11}\,\mathbf{a}_2 + z_{11}\,\mathbf{a}_3$ | = | $x_{11}a\,\hat{\mathbf{x}} + y_{11}b\,\hat{\mathbf{y}} + z_{11}c\,\hat{\mathbf{z}}$ | (4c) | Zn |
| $\mathbf{B}_{38}$ | = | $-x_{11}\,\mathbf{a}_1 - y_{11}\,\mathbf{a}_2 + z_{11}\,\mathbf{a}_3$ | = | $-x_{11}a\,\hat{\mathbf{x}} - y_{11}b\,\hat{\mathbf{y}} + z_{11}c\,\hat{\mathbf{z}}$ | (4c) | Zn |
| $\mathbf{B}_{39}$ | = | $\left(\frac{1}{2} - x_{11}\right)\mathbf{a}_1 + \left(\frac{1}{2} + y_{11}\right)\mathbf{a}_2 - z_{11}\,\mathbf{a}_3$ | = | $\left(\frac{1}{2} - x_{11}\right)a\,\hat{\mathbf{x}} + \left(\frac{1}{2} + y_{11}\right)b\,\hat{\mathbf{y}} - z_{11}c\,\hat{\mathbf{z}}$ | (4c) | Zn |

$$\mathbf{B}_{40} = \left(\frac{1}{2} + x_{11}\right) \mathbf{a}_1 + \left(\frac{1}{2} - y_{11}\right) \mathbf{a}_2 - z_{11}\,\mathbf{a}_3 = \left(\frac{1}{2} + x_{11}\right) a\,\hat{\mathbf{x}} + \left(\frac{1}{2} - y_{11}\right) b\,\hat{\mathbf{y}} - z_{11} c\,\hat{\mathbf{z}} \qquad (4c) \qquad \text{Zn}$$

---

---

**Geometry files:**

- CIF: pp. 1575
- POSCAR: pp. 1576

# Morenosite ($NiSO_4{\cdot}7H_2O$, $H4_{12}$) Structure: A14BC11D_oP108_19_14a_a_11a_a

http://aflow.org/prototype-encyclopedia/A14BC11D_oP108_19_14a_a_11a_a

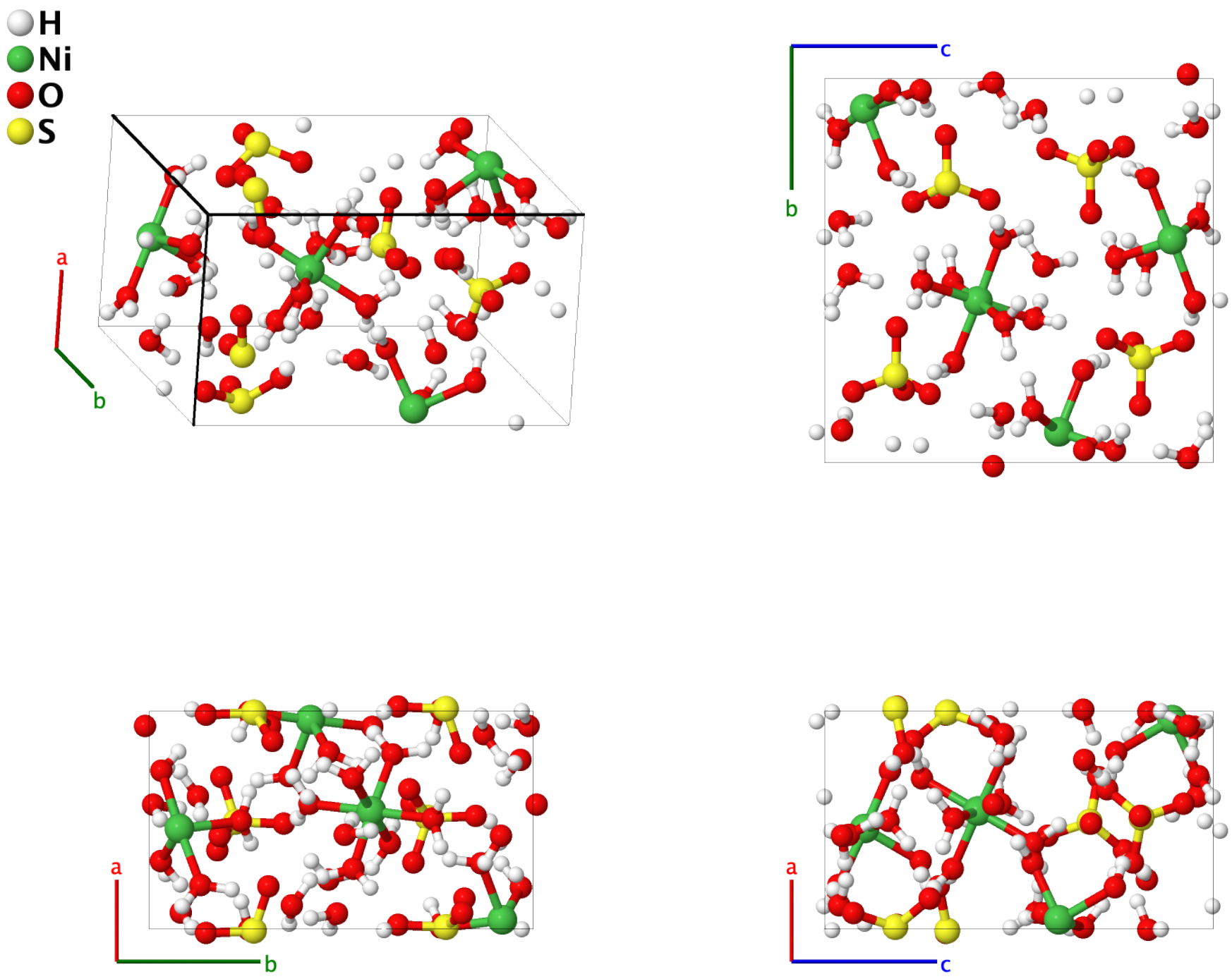

| | | |
|---|---|---|
| **Prototype** | : | $H_{14}NiO_{11}S$ |
| **AFLOW prototype label** | : | A14BC11D_oP108_19_14a_a_11a_a |
| ***Strukturbericht* designation** | : | $H4_{12}$ |
| **Pearson symbol** | : | oP108 |
| **Space group number** | : | 19 |
| **Space group symbol** | : | $P2_12_12_1$ |
| **AFLOW prototype command** | : | `aflow --proto=A14BC11D_oP108_19_14a_a_11a_a` `--params=`$a, b/a, c/a, x_1, y_1, z_1, x_2, y_2, z_2, x_3, y_3, z_3, x_4, y_4, z_4, x_5, y_5, z_5, x_6, y_6, z_6, x_7, y_7, z_7, x_8, y_8, z_8, x_9, y_9, z_9, x_{10}, y_{10}, z_{10}, x_{11}, y_{11}, z_{11}, x_{12}, y_{12}, z_{12}, x_{13}, y_{13}, z_{13}, x_{14}, y_{14}, z_{14}, x_{15}, y_{15}, z_{15}, x_{16}, y_{16}, z_{16}, x_{17}, y_{17}, z_{17}, x_{18}, y_{18}, z_{18}, x_{19}, y_{19}, z_{19}, x_{20}, y_{20}, z_{20}, x_{21}, y_{21}, z_{21}, x_{22}, y_{22}, z_{22}, x_{23}, y_{23}, z_{23}, x_{24}, y_{24}, z_{24}, x_{25}, y_{25}, z_{25}, x_{26}, y_{26}, z_{26}, x_{27}, y_{27}, z_{27}$ |

### Other compounds with this structure

- $MgSO_4{\cdot}7H_2O$ (Epsomite) and $ZnSO_4{\cdot}7H_2O$ (Gosalrite)

---

- We use the structure from the 25 K neutron data taken by (Ptasiewicz-Bak, 1997).

**Simple Orthorhombic primitive vectors:**

$$\mathbf{a}_1 = a\,\hat{\mathbf{x}}$$
$$\mathbf{a}_2 = b\,\hat{\mathbf{y}}$$
$$\mathbf{a}_3 = c\,\hat{\mathbf{z}}$$

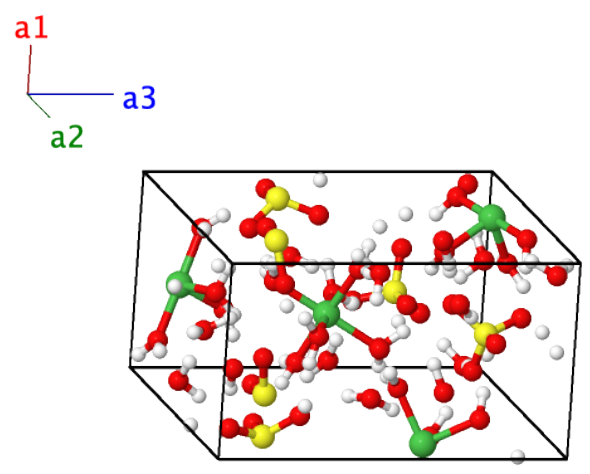

**Basis vectors:**

| | | Lattice Coordinates | | Cartesian Coordinates | Wyckoff Position | Atom Type |
|---|---|---|---|---|---|---|
| $\mathbf{B}_1$ | = | $x_1\,\mathbf{a}_1 + y_1\,\mathbf{a}_2 + z_1\,\mathbf{a}_3$ | = | $x_1 a\,\hat{\mathbf{x}} + y_1 b\,\hat{\mathbf{y}} + z_1 c\,\hat{\mathbf{z}}$ | (4$a$) | H I |
| $\mathbf{B}_2$ | = | $\left(\frac{1}{2} - x_1\right)\mathbf{a}_1 - y_1\,\mathbf{a}_2 + \left(\frac{1}{2} + z_1\right)\mathbf{a}_3$ | = | $\left(\frac{1}{2} - x_1\right) a\,\hat{\mathbf{x}} - y_1 b\,\hat{\mathbf{y}} + \left(\frac{1}{2} + z_1\right) c\,\hat{\mathbf{z}}$ | (4$a$) | H I |
| $\mathbf{B}_3$ | = | $-x_1\,\mathbf{a}_1 + \left(\frac{1}{2} + y_1\right)\mathbf{a}_2 + \left(\frac{1}{2} - z_1\right)\mathbf{a}_3$ | = | $-x_1 a\,\hat{\mathbf{x}} + \left(\frac{1}{2} + y_1\right) b\,\hat{\mathbf{y}} + \left(\frac{1}{2} - z_1\right) c\,\hat{\mathbf{z}}$ | (4$a$) | H I |
| $\mathbf{B}_4$ | = | $\left(\frac{1}{2} + x_1\right)\mathbf{a}_1 + \left(\frac{1}{2} - y_1\right)\mathbf{a}_2 - z_1\,\mathbf{a}_3$ | = | $\left(\frac{1}{2} + x_1\right) a\,\hat{\mathbf{x}} + \left(\frac{1}{2} - y_1\right) b\,\hat{\mathbf{y}} - z_1 c\,\hat{\mathbf{z}}$ | (4$a$) | H I |
| $\mathbf{B}_5$ | = | $x_2\,\mathbf{a}_1 + y_2\,\mathbf{a}_2 + z_2\,\mathbf{a}_3$ | = | $x_2 a\,\hat{\mathbf{x}} + y_2 b\,\hat{\mathbf{y}} + z_2 c\,\hat{\mathbf{z}}$ | (4$a$) | H II |
| $\mathbf{B}_6$ | = | $\left(\frac{1}{2} - x_2\right)\mathbf{a}_1 - y_2\,\mathbf{a}_2 + \left(\frac{1}{2} + z_2\right)\mathbf{a}_3$ | = | $\left(\frac{1}{2} - x_2\right) a\,\hat{\mathbf{x}} - y_2 b\,\hat{\mathbf{y}} + \left(\frac{1}{2} + z_2\right) c\,\hat{\mathbf{z}}$ | (4$a$) | H II |
| $\mathbf{B}_7$ | = | $-x_2\,\mathbf{a}_1 + \left(\frac{1}{2} + y_2\right)\mathbf{a}_2 + \left(\frac{1}{2} - z_2\right)\mathbf{a}_3$ | = | $-x_2 a\,\hat{\mathbf{x}} + \left(\frac{1}{2} + y_2\right) b\,\hat{\mathbf{y}} + \left(\frac{1}{2} - z_2\right) c\,\hat{\mathbf{z}}$ | (4$a$) | H II |
| $\mathbf{B}_8$ | = | $\left(\frac{1}{2} + x_2\right)\mathbf{a}_1 + \left(\frac{1}{2} - y_2\right)\mathbf{a}_2 - z_2\,\mathbf{a}_3$ | = | $\left(\frac{1}{2} + x_2\right) a\,\hat{\mathbf{x}} + \left(\frac{1}{2} - y_2\right) b\,\hat{\mathbf{y}} - z_2 c\,\hat{\mathbf{z}}$ | (4$a$) | H II |
| $\mathbf{B}_9$ | = | $x_3\,\mathbf{a}_1 + y_3\,\mathbf{a}_2 + z_3\,\mathbf{a}_3$ | = | $x_3 a\,\hat{\mathbf{x}} + y_3 b\,\hat{\mathbf{y}} + z_3 c\,\hat{\mathbf{z}}$ | (4$a$) | H III |
| $\mathbf{B}_{10}$ | = | $\left(\frac{1}{2} - x_3\right)\mathbf{a}_1 - y_3\,\mathbf{a}_2 + \left(\frac{1}{2} + z_3\right)\mathbf{a}_3$ | = | $\left(\frac{1}{2} - x_3\right) a\,\hat{\mathbf{x}} - y_3 b\,\hat{\mathbf{y}} + \left(\frac{1}{2} + z_3\right) c\,\hat{\mathbf{z}}$ | (4$a$) | H III |
| $\mathbf{B}_{11}$ | = | $-x_3\,\mathbf{a}_1 + \left(\frac{1}{2} + y_3\right)\mathbf{a}_2 + \left(\frac{1}{2} - z_3\right)\mathbf{a}_3$ | = | $-x_3 a\,\hat{\mathbf{x}} + \left(\frac{1}{2} + y_3\right) b\,\hat{\mathbf{y}} + \left(\frac{1}{2} - z_3\right) c\,\hat{\mathbf{z}}$ | (4$a$) | H III |
| $\mathbf{B}_{12}$ | = | $\left(\frac{1}{2} + x_3\right)\mathbf{a}_1 + \left(\frac{1}{2} - y_3\right)\mathbf{a}_2 - z_3\,\mathbf{a}_3$ | = | $\left(\frac{1}{2} + x_3\right) a\,\hat{\mathbf{x}} + \left(\frac{1}{2} - y_3\right) b\,\hat{\mathbf{y}} - z_3 c\,\hat{\mathbf{z}}$ | (4$a$) | H III |
| $\mathbf{B}_{13}$ | = | $x_4\,\mathbf{a}_1 + y_4\,\mathbf{a}_2 + z_4\,\mathbf{a}_3$ | = | $x_4 a\,\hat{\mathbf{x}} + y_4 b\,\hat{\mathbf{y}} + z_4 c\,\hat{\mathbf{z}}$ | (4$a$) | H IV |
| $\mathbf{B}_{14}$ | = | $\left(\frac{1}{2} - x_4\right)\mathbf{a}_1 - y_4\,\mathbf{a}_2 + \left(\frac{1}{2} + z_4\right)\mathbf{a}_3$ | = | $\left(\frac{1}{2} - x_4\right) a\,\hat{\mathbf{x}} - y_4 b\,\hat{\mathbf{y}} + \left(\frac{1}{2} + z_4\right) c\,\hat{\mathbf{z}}$ | (4$a$) | H IV |
| $\mathbf{B}_{15}$ | = | $-x_4\,\mathbf{a}_1 + \left(\frac{1}{2} + y_4\right)\mathbf{a}_2 + \left(\frac{1}{2} - z_4\right)\mathbf{a}_3$ | = | $-x_4 a\,\hat{\mathbf{x}} + \left(\frac{1}{2} + y_4\right) b\,\hat{\mathbf{y}} + \left(\frac{1}{2} - z_4\right) c\,\hat{\mathbf{z}}$ | (4$a$) | H IV |
| $\mathbf{B}_{16}$ | = | $\left(\frac{1}{2} + x_4\right)\mathbf{a}_1 + \left(\frac{1}{2} - y_4\right)\mathbf{a}_2 - z_4\,\mathbf{a}_3$ | = | $\left(\frac{1}{2} + x_4\right) a\,\hat{\mathbf{x}} + \left(\frac{1}{2} - y_4\right) b\,\hat{\mathbf{y}} - z_4 c\,\hat{\mathbf{z}}$ | (4$a$) | H IV |
| $\mathbf{B}_{17}$ | = | $x_5\,\mathbf{a}_1 + y_5\,\mathbf{a}_2 + z_5\,\mathbf{a}_3$ | = | $x_5 a\,\hat{\mathbf{x}} + y_5 b\,\hat{\mathbf{y}} + z_5 c\,\hat{\mathbf{z}}$ | (4$a$) | H V |
| $\mathbf{B}_{18}$ | = | $\left(\frac{1}{2} - x_5\right)\mathbf{a}_1 - y_5\,\mathbf{a}_2 + \left(\frac{1}{2} + z_5\right)\mathbf{a}_3$ | = | $\left(\frac{1}{2} - x_5\right) a\,\hat{\mathbf{x}} - y_5 b\,\hat{\mathbf{y}} + \left(\frac{1}{2} + z_5\right) c\,\hat{\mathbf{z}}$ | (4$a$) | H V |
| $\mathbf{B}_{19}$ | = | $-x_5\,\mathbf{a}_1 + \left(\frac{1}{2} + y_5\right)\mathbf{a}_2 + \left(\frac{1}{2} - z_5\right)\mathbf{a}_3$ | = | $-x_5 a\,\hat{\mathbf{x}} + \left(\frac{1}{2} + y_5\right) b\,\hat{\mathbf{y}} + \left(\frac{1}{2} - z_5\right) c\,\hat{\mathbf{z}}$ | (4$a$) | H V |
| $\mathbf{B}_{20}$ | = | $\left(\frac{1}{2} + x_5\right)\mathbf{a}_1 + \left(\frac{1}{2} - y_5\right)\mathbf{a}_2 - z_5\,\mathbf{a}_3$ | = | $\left(\frac{1}{2} + x_5\right) a\,\hat{\mathbf{x}} + \left(\frac{1}{2} - y_5\right) b\,\hat{\mathbf{y}} - z_5 c\,\hat{\mathbf{z}}$ | (4$a$) | H V |
| $\mathbf{B}_{21}$ | = | $x_6\,\mathbf{a}_1 + y_6\,\mathbf{a}_2 + z_6\,\mathbf{a}_3$ | = | $x_6 a\,\hat{\mathbf{x}} + y_6 b\,\hat{\mathbf{y}} + z_6 c\,\hat{\mathbf{z}}$ | (4$a$) | H VI |
| $\mathbf{B}_{22}$ | = | $\left(\frac{1}{2} - x_6\right)\mathbf{a}_1 - y_6\,\mathbf{a}_2 + \left(\frac{1}{2} + z_6\right)\mathbf{a}_3$ | = | $\left(\frac{1}{2} - x_6\right) a\,\hat{\mathbf{x}} - y_6 b\,\hat{\mathbf{y}} + \left(\frac{1}{2} + z_6\right) c\,\hat{\mathbf{z}}$ | (4$a$) | H VI |
| $\mathbf{B}_{23}$ | = | $-x_6\,\mathbf{a}_1 + \left(\frac{1}{2} + y_6\right)\mathbf{a}_2 + \left(\frac{1}{2} - z_6\right)\mathbf{a}_3$ | = | $-x_6 a\,\hat{\mathbf{x}} + \left(\frac{1}{2} + y_6\right) b\,\hat{\mathbf{y}} + \left(\frac{1}{2} - z_6\right) c\,\hat{\mathbf{z}}$ | (4$a$) | H VI |
| $\mathbf{B}_{24}$ | = | $\left(\frac{1}{2} + x_6\right)\mathbf{a}_1 + \left(\frac{1}{2} - y_6\right)\mathbf{a}_2 - z_6\,\mathbf{a}_3$ | = | $\left(\frac{1}{2} + x_6\right) a\,\hat{\mathbf{x}} + \left(\frac{1}{2} - y_6\right) b\,\hat{\mathbf{y}} - z_6 c\,\hat{\mathbf{z}}$ | (4$a$) | H VI |
| $\mathbf{B}_{25}$ | = | $x_7\,\mathbf{a}_1 + y_7\,\mathbf{a}_2 + z_7\,\mathbf{a}_3$ | = | $x_7 a\,\hat{\mathbf{x}} + y_7 b\,\hat{\mathbf{y}} + z_7 c\,\hat{\mathbf{z}}$ | (4$a$) | H VII |
| $\mathbf{B}_{26}$ | = | $\left(\frac{1}{2} - x_7\right)\mathbf{a}_1 - y_7\,\mathbf{a}_2 + \left(\frac{1}{2} + z_7\right)\mathbf{a}_3$ | = | $\left(\frac{1}{2} - x_7\right) a\,\hat{\mathbf{x}} - y_7 b\,\hat{\mathbf{y}} + \left(\frac{1}{2} + z_7\right) c\,\hat{\mathbf{z}}$ | (4$a$) | H VII |
| $\mathbf{B}_{27}$ | = | $-x_7\,\mathbf{a}_1 + \left(\frac{1}{2} + y_7\right)\mathbf{a}_2 + \left(\frac{1}{2} - z_7\right)\mathbf{a}_3$ | = | $-x_7 a\,\hat{\mathbf{x}} + \left(\frac{1}{2} + y_7\right) b\,\hat{\mathbf{y}} + \left(\frac{1}{2} - z_7\right) c\,\hat{\mathbf{z}}$ | (4$a$) | H VII |
| $\mathbf{B}_{28}$ | = | $\left(\frac{1}{2}+x_7\right)\mathbf{a}_1+\left(\frac{1}{2}-y_7\right)\mathbf{a}_2-z_7\,\mathbf{a}_3$ | = | $\left(\frac{1}{2}+x_7\right)a\,\hat{\mathbf{x}}+\left(\frac{1}{2}-y_7\right)b\,\hat{\mathbf{y}}-z_7c\,\hat{\mathbf{z}}$ | (4$a$) | H VII |
| $\mathbf{B}_{29}$ | = | $x_8\,\mathbf{a}_1+y_8\,\mathbf{a}_2+z_8\,\mathbf{a}_3$ | = | $x_8a\,\hat{\mathbf{x}}+y_8b\,\hat{\mathbf{y}}+z_8c\,\hat{\mathbf{z}}$ | (4$a$) | H VIII |
| $\mathbf{B}_{30}$ | = | $\left(\frac{1}{2}-x_8\right)\mathbf{a}_1-y_8\,\mathbf{a}_2+\left(\frac{1}{2}+z_8\right)\mathbf{a}_3$ | = | $\left(\frac{1}{2}-x_8\right)a\,\hat{\mathbf{x}}-y_8b\,\hat{\mathbf{y}}+\left(\frac{1}{2}+z_8\right)c\,\hat{\mathbf{z}}$ | (4$a$) | H VIII |
| $\mathbf{B}_{31}$ | = | $-x_8\,\mathbf{a}_1+\left(\frac{1}{2}+y_8\right)\mathbf{a}_2+\left(\frac{1}{2}-z_8\right)\mathbf{a}_3$ | = | $-x_8a\,\hat{\mathbf{x}}+\left(\frac{1}{2}+y_8\right)b\,\hat{\mathbf{y}}+\left(\frac{1}{2}-z_8\right)c\,\hat{\mathbf{z}}$ | (4$a$) | H VIII |
| $\mathbf{B}_{32}$ | = | $\left(\frac{1}{2}+x_8\right)\mathbf{a}_1+\left(\frac{1}{2}-y_8\right)\mathbf{a}_2-z_8\,\mathbf{a}_3$ | = | $\left(\frac{1}{2}+x_8\right)a\,\hat{\mathbf{x}}+\left(\frac{1}{2}-y_8\right)b\,\hat{\mathbf{y}}-z_8c\,\hat{\mathbf{z}}$ | (4$a$) | H VIII |
| $\mathbf{B}_{33}$ | = | $x_9\,\mathbf{a}_1+y_9\,\mathbf{a}_2+z_9\,\mathbf{a}_3$ | = | $x_9a\,\hat{\mathbf{x}}+y_9b\,\hat{\mathbf{y}}+z_9c\,\hat{\mathbf{z}}$ | (4$a$) | H IX |
| $\mathbf{B}_{34}$ | = | $\left(\frac{1}{2}-x_9\right)\mathbf{a}_1-y_9\,\mathbf{a}_2+\left(\frac{1}{2}+z_9\right)\mathbf{a}_3$ | = | $\left(\frac{1}{2}-x_9\right)a\,\hat{\mathbf{x}}-y_9b\,\hat{\mathbf{y}}+\left(\frac{1}{2}+z_9\right)c\,\hat{\mathbf{z}}$ | (4$a$) | H IX |
| $\mathbf{B}_{35}$ | = | $-x_9\,\mathbf{a}_1+\left(\frac{1}{2}+y_9\right)\mathbf{a}_2+\left(\frac{1}{2}-z_9\right)\mathbf{a}_3$ | = | $-x_9a\,\hat{\mathbf{x}}+\left(\frac{1}{2}+y_9\right)b\,\hat{\mathbf{y}}+\left(\frac{1}{2}-z_9\right)c\,\hat{\mathbf{z}}$ | (4$a$) | H IX |
| $\mathbf{B}_{36}$ | = | $\left(\frac{1}{2}+x_9\right)\mathbf{a}_1+\left(\frac{1}{2}-y_9\right)\mathbf{a}_2-z_9\,\mathbf{a}_3$ | = | $\left(\frac{1}{2}+x_9\right)a\,\hat{\mathbf{x}}+\left(\frac{1}{2}-y_9\right)b\,\hat{\mathbf{y}}-z_9c\,\hat{\mathbf{z}}$ | (4$a$) | H IX |
| $\mathbf{B}_{37}$ | = | $x_{10}\,\mathbf{a}_1+y_{10}\,\mathbf{a}_2+z_{10}\,\mathbf{a}_3$ | = | $x_{10}a\,\hat{\mathbf{x}}+y_{10}b\,\hat{\mathbf{y}}+z_{10}c\,\hat{\mathbf{z}}$ | (4$a$) | H X |
| $\mathbf{B}_{38}$ | = | $\left(\frac{1}{2}-x_{10}\right)\mathbf{a}_1-y_{10}\,\mathbf{a}_2+\left(\frac{1}{2}+z_{10}\right)\mathbf{a}_3$ | = | $\left(\frac{1}{2}-x_{10}\right)a\,\hat{\mathbf{x}}-y_{10}b\,\hat{\mathbf{y}}+\left(\frac{1}{2}+z_{10}\right)c\,\hat{\mathbf{z}}$ | (4$a$) | H X |
| $\mathbf{B}_{39}$ | = | $-x_{10}\,\mathbf{a}_1+\left(\frac{1}{2}+y_{10}\right)\mathbf{a}_2+\left(\frac{1}{2}-z_{10}\right)\mathbf{a}_3$ | = | $-x_{10}a\,\hat{\mathbf{x}}+\left(\frac{1}{2}+y_{10}\right)b\,\hat{\mathbf{y}}+\left(\frac{1}{2}-z_{10}\right)c\,\hat{\mathbf{z}}$ | (4$a$) | H X |
| $\mathbf{B}_{40}$ | = | $\left(\frac{1}{2}+x_{10}\right)\mathbf{a}_1+\left(\frac{1}{2}-y_{10}\right)\mathbf{a}_2-z_{10}\,\mathbf{a}_3$ | = | $\left(\frac{1}{2}+x_{10}\right)a\,\hat{\mathbf{x}}+\left(\frac{1}{2}-y_{10}\right)b\,\hat{\mathbf{y}}-z_{10}c\,\hat{\mathbf{z}}$ | (4$a$) | H X |
| $\mathbf{B}_{41}$ | = | $x_{11}\,\mathbf{a}_1+y_{11}\,\mathbf{a}_2+z_{11}\,\mathbf{a}_3$ | = | $x_{11}a\,\hat{\mathbf{x}}+y_{11}b\,\hat{\mathbf{y}}+z_{11}c\,\hat{\mathbf{z}}$ | (4$a$) | H XI |
| $\mathbf{B}_{42}$ | = | $\left(\frac{1}{2}-x_{11}\right)\mathbf{a}_1-y_{11}\,\mathbf{a}_2+\left(\frac{1}{2}+z_{11}\right)\mathbf{a}_3$ | = | $\left(\frac{1}{2}-x_{11}\right)a\,\hat{\mathbf{x}}-y_{11}b\,\hat{\mathbf{y}}+\left(\frac{1}{2}+z_{11}\right)c\,\hat{\mathbf{z}}$ | (4$a$) | H XI |
| $\mathbf{B}_{43}$ | = | $-x_{11}\,\mathbf{a}_1+\left(\frac{1}{2}+y_{11}\right)\mathbf{a}_2+\left(\frac{1}{2}-z_{11}\right)\mathbf{a}_3$ | = | $-x_{11}a\,\hat{\mathbf{x}}+\left(\frac{1}{2}+y_{11}\right)b\,\hat{\mathbf{y}}+\left(\frac{1}{2}-z_{11}\right)c\,\hat{\mathbf{z}}$ | (4$a$) | H XI |
| $\mathbf{B}_{44}$ | = | $\left(\frac{1}{2}+x_{11}\right)\mathbf{a}_1+\left(\frac{1}{2}-y_{11}\right)\mathbf{a}_2-z_{11}\,\mathbf{a}_3$ | = | $\left(\frac{1}{2}+x_{11}\right)a\,\hat{\mathbf{x}}+\left(\frac{1}{2}-y_{11}\right)b\,\hat{\mathbf{y}}-z_{11}c\,\hat{\mathbf{z}}$ | (4$a$) | H XI |
| $\mathbf{B}_{45}$ | = | $x_{12}\,\mathbf{a}_1+y_{12}\,\mathbf{a}_2+z_{12}\,\mathbf{a}_3$ | = | $x_{12}a\,\hat{\mathbf{x}}+y_{12}b\,\hat{\mathbf{y}}+z_{12}c\,\hat{\mathbf{z}}$ | (4$a$) | H XII |
| $\mathbf{B}_{46}$ | = | $\left(\frac{1}{2}-x_{12}\right)\mathbf{a}_1-y_{12}\,\mathbf{a}_2+\left(\frac{1}{2}+z_{12}\right)\mathbf{a}_3$ | = | $\left(\frac{1}{2}-x_{12}\right)a\,\hat{\mathbf{x}}-y_{12}b\,\hat{\mathbf{y}}+\left(\frac{1}{2}+z_{12}\right)c\,\hat{\mathbf{z}}$ | (4$a$) | H XII |
| $\mathbf{B}_{47}$ | = | $-x_{12}\,\mathbf{a}_1+\left(\frac{1}{2}+y_{12}\right)\mathbf{a}_2+\left(\frac{1}{2}-z_{12}\right)\mathbf{a}_3$ | = | $-x_{12}a\,\hat{\mathbf{x}}+\left(\frac{1}{2}+y_{12}\right)b\,\hat{\mathbf{y}}+\left(\frac{1}{2}-z_{12}\right)c\,\hat{\mathbf{z}}$ | (4$a$) | H XII |
| $\mathbf{B}_{48}$ | = | $\left(\frac{1}{2}+x_{12}\right)\mathbf{a}_1+\left(\frac{1}{2}-y_{12}\right)\mathbf{a}_2-z_{12}\,\mathbf{a}_3$ | = | $\left(\frac{1}{2}+x_{12}\right)a\,\hat{\mathbf{x}}+\left(\frac{1}{2}-y_{12}\right)b\,\hat{\mathbf{y}}-z_{12}c\,\hat{\mathbf{z}}$ | (4$a$) | H XII |
| $\mathbf{B}_{49}$ | = | $x_{13}\,\mathbf{a}_1+y_{13}\,\mathbf{a}_2+z_{13}\,\mathbf{a}_3$ | = | $x_{13}a\,\hat{\mathbf{x}}+y_{13}b\,\hat{\mathbf{y}}+z_{13}c\,\hat{\mathbf{z}}$ | (4$a$) | H XIII |
| $\mathbf{B}_{50}$ | = | $\left(\frac{1}{2}-x_{13}\right)\mathbf{a}_1-y_{13}\,\mathbf{a}_2+\left(\frac{1}{2}+z_{13}\right)\mathbf{a}_3$ | = | $\left(\frac{1}{2}-x_{13}\right)a\,\hat{\mathbf{x}}-y_{13}b\,\hat{\mathbf{y}}+\left(\frac{1}{2}+z_{13}\right)c\,\hat{\mathbf{z}}$ | (4$a$) | H XIII |
| $\mathbf{B}_{51}$ | = | $-x_{13}\,\mathbf{a}_1+\left(\frac{1}{2}+y_{13}\right)\mathbf{a}_2+\left(\frac{1}{2}-z_{13}\right)\mathbf{a}_3$ | = | $-x_{13}a\,\hat{\mathbf{x}}+\left(\frac{1}{2}+y_{13}\right)b\,\hat{\mathbf{y}}+\left(\frac{1}{2}-z_{13}\right)c\,\hat{\mathbf{z}}$ | (4$a$) | H XIII |
| $\mathbf{B}_{52}$ | = | $\left(\frac{1}{2}+x_{13}\right)\mathbf{a}_1+\left(\frac{1}{2}-y_{13}\right)\mathbf{a}_2-z_{13}\,\mathbf{a}_3$ | = | $\left(\frac{1}{2}+x_{13}\right)a\,\hat{\mathbf{x}}+\left(\frac{1}{2}-y_{13}\right)b\,\hat{\mathbf{y}}-z_{13}c\,\hat{\mathbf{z}}$ | (4$a$) | H XIII |
| $\mathbf{B}_{53}$ | = | $x_{14}\,\mathbf{a}_1+y_{14}\,\mathbf{a}_2+z_{14}\,\mathbf{a}_3$ | = | $x_{14}a\,\hat{\mathbf{x}}+y_{14}b\,\hat{\mathbf{y}}+z_{14}c\,\hat{\mathbf{z}}$ | (4$a$) | H XIV |
| $\mathbf{B}_{54}$ | = | $\left(\frac{1}{2}-x_{14}\right)\mathbf{a}_1-y_{14}\,\mathbf{a}_2+\left(\frac{1}{2}+z_{14}\right)\mathbf{a}_3$ | = | $\left(\frac{1}{2}-x_{14}\right)a\,\hat{\mathbf{x}}-y_{14}b\,\hat{\mathbf{y}}+\left(\frac{1}{2}+z_{14}\right)c\,\hat{\mathbf{z}}$ | (4$a$) | H XIV |
| $\mathbf{B}_{55}$ | = | $-x_{14}\,\mathbf{a}_1+\left(\frac{1}{2}+y_{14}\right)\mathbf{a}_2+\left(\frac{1}{2}-z_{14}\right)\mathbf{a}_3$ | = | $-x_{14}a\,\hat{\mathbf{x}}+\left(\frac{1}{2}+y_{14}\right)b\,\hat{\mathbf{y}}+\left(\frac{1}{2}-z_{14}\right)c\,\hat{\mathbf{z}}$ | (4$a$) | H XIV |
| $\mathbf{B}_{56}$ | = | $\left(\frac{1}{2}+x_{14}\right)\mathbf{a}_1+\left(\frac{1}{2}-y_{14}\right)\mathbf{a}_2-z_{14}\,\mathbf{a}_3$ | = | $\left(\frac{1}{2}+x_{14}\right)a\,\hat{\mathbf{x}}+\left(\frac{1}{2}-y_{14}\right)b\,\hat{\mathbf{y}}-z_{14}c\,\hat{\mathbf{z}}$ | (4$a$) | H XIV |
| $\mathbf{B}_{57}$ | = | $x_{15}\,\mathbf{a}_1+y_{15}\,\mathbf{a}_2+z_{15}\,\mathbf{a}_3$ | = | $x_{15}a\,\hat{\mathbf{x}}+y_{15}b\,\hat{\mathbf{y}}+z_{15}c\,\hat{\mathbf{z}}$ | (4$a$) | Ni |
| $\mathbf{B}_{58}$ | = | $\left(\frac{1}{2}-x_{15}\right)\mathbf{a}_1-y_{15}\,\mathbf{a}_2+\left(\frac{1}{2}+z_{15}\right)\mathbf{a}_3$ | = | $\left(\frac{1}{2}-x_{15}\right)a\,\hat{\mathbf{x}}-y_{15}b\,\hat{\mathbf{y}}+\left(\frac{1}{2}+z_{15}\right)c\,\hat{\mathbf{z}}$ | (4$a$) | Ni |
| $\mathbf{B}_{59}$ | = | $-x_{15}\,\mathbf{a}_1+\left(\frac{1}{2}+y_{15}\right)\mathbf{a}_2+\left(\frac{1}{2}-z_{15}\right)\mathbf{a}_3$ | = | $-x_{15}a\,\hat{\mathbf{x}}+\left(\frac{1}{2}+y_{15}\right)b\,\hat{\mathbf{y}}+\left(\frac{1}{2}-z_{15}\right)c\,\hat{\mathbf{z}}$ | (4$a$) | Ni |
| $\mathbf{B}_{60}$ | = | $\left(\frac{1}{2} + x_{15}\right)\mathbf{a}_1 + \left(\frac{1}{2} - y_{15}\right)\mathbf{a}_2 - z_{15}\,\mathbf{a}_3$ | = | $\left(\frac{1}{2} + x_{15}\right)a\,\mathbf{\hat{x}} + \left(\frac{1}{2} - y_{15}\right)b\,\mathbf{\hat{y}} - z_{15}c\,\mathbf{\hat{z}}$ | (4$a$) | Ni |
| $\mathbf{B}_{61}$ | = | $x_{16}\,\mathbf{a}_1 + y_{16}\,\mathbf{a}_2 + z_{16}\,\mathbf{a}_3$ | = | $x_{16}a\,\mathbf{\hat{x}} + y_{16}b\,\mathbf{\hat{y}} + z_{16}c\,\mathbf{\hat{z}}$ | (4$a$) | O I |
| $\mathbf{B}_{62}$ | = | $\left(\frac{1}{2} - x_{16}\right)\mathbf{a}_1 - y_{16}\,\mathbf{a}_2 + \left(\frac{1}{2} + z_{16}\right)\mathbf{a}_3$ | = | $\left(\frac{1}{2} - x_{16}\right)a\,\mathbf{\hat{x}} - y_{16}b\,\mathbf{\hat{y}} + \left(\frac{1}{2} + z_{16}\right)c\,\mathbf{\hat{z}}$ | (4$a$) | O I |
| $\mathbf{B}_{63}$ | = | $-x_{16}\,\mathbf{a}_1 + \left(\frac{1}{2} + y_{16}\right)\mathbf{a}_2 + \left(\frac{1}{2} - z_{16}\right)\mathbf{a}_3$ | = | $-x_{16}a\,\mathbf{\hat{x}} + \left(\frac{1}{2} + y_{16}\right)b\,\mathbf{\hat{y}} + \left(\frac{1}{2} - z_{16}\right)c\,\mathbf{\hat{z}}$ | (4$a$) | O I |
| $\mathbf{B}_{64}$ | = | $\left(\frac{1}{2} + x_{16}\right)\mathbf{a}_1 + \left(\frac{1}{2} - y_{16}\right)\mathbf{a}_2 - z_{16}\,\mathbf{a}_3$ | = | $\left(\frac{1}{2} + x_{16}\right)a\,\mathbf{\hat{x}} + \left(\frac{1}{2} - y_{16}\right)b\,\mathbf{\hat{y}} - z_{16}c\,\mathbf{\hat{z}}$ | (4$a$) | O I |
| $\mathbf{B}_{65}$ | = | $x_{17}\,\mathbf{a}_1 + y_{17}\,\mathbf{a}_2 + z_{17}\,\mathbf{a}_3$ | = | $x_{17}a\,\mathbf{\hat{x}} + y_{17}b\,\mathbf{\hat{y}} + z_{17}c\,\mathbf{\hat{z}}$ | (4$a$) | O II |
| $\mathbf{B}_{66}$ | = | $\left(\frac{1}{2} - x_{17}\right)\mathbf{a}_1 - y_{17}\,\mathbf{a}_2 + \left(\frac{1}{2} + z_{17}\right)\mathbf{a}_3$ | = | $\left(\frac{1}{2} - x_{17}\right)a\,\mathbf{\hat{x}} - y_{17}b\,\mathbf{\hat{y}} + \left(\frac{1}{2} + z_{17}\right)c\,\mathbf{\hat{z}}$ | (4$a$) | O II |
| $\mathbf{B}_{67}$ | = | $-x_{17}\,\mathbf{a}_1 + \left(\frac{1}{2} + y_{17}\right)\mathbf{a}_2 + \left(\frac{1}{2} - z_{17}\right)\mathbf{a}_3$ | = | $-x_{17}a\,\mathbf{\hat{x}} + \left(\frac{1}{2} + y_{17}\right)b\,\mathbf{\hat{y}} + \left(\frac{1}{2} - z_{17}\right)c\,\mathbf{\hat{z}}$ | (4$a$) | O II |
| $\mathbf{B}_{68}$ | = | $\left(\frac{1}{2} + x_{17}\right)\mathbf{a}_1 + \left(\frac{1}{2} - y_{17}\right)\mathbf{a}_2 - z_{17}\,\mathbf{a}_3$ | = | $\left(\frac{1}{2} + x_{17}\right)a\,\mathbf{\hat{x}} + \left(\frac{1}{2} - y_{17}\right)b\,\mathbf{\hat{y}} - z_{17}c\,\mathbf{\hat{z}}$ | (4$a$) | O II |
| $\mathbf{B}_{69}$ | = | $x_{18}\,\mathbf{a}_1 + y_{18}\,\mathbf{a}_2 + z_{18}\,\mathbf{a}_3$ | = | $x_{18}a\,\mathbf{\hat{x}} + y_{18}b\,\mathbf{\hat{y}} + z_{18}c\,\mathbf{\hat{z}}$ | (4$a$) | O III |
| $\mathbf{B}_{70}$ | = | $\left(\frac{1}{2} - x_{18}\right)\mathbf{a}_1 - y_{18}\,\mathbf{a}_2 + \left(\frac{1}{2} + z_{18}\right)\mathbf{a}_3$ | = | $\left(\frac{1}{2} - x_{18}\right)a\,\mathbf{\hat{x}} - y_{18}b\,\mathbf{\hat{y}} + \left(\frac{1}{2} + z_{18}\right)c\,\mathbf{\hat{z}}$ | (4$a$) | O III |
| $\mathbf{B}_{71}$ | = | $-x_{18}\,\mathbf{a}_1 + \left(\frac{1}{2} + y_{18}\right)\mathbf{a}_2 + \left(\frac{1}{2} - z_{18}\right)\mathbf{a}_3$ | = | $-x_{18}a\,\mathbf{\hat{x}} + \left(\frac{1}{2} + y_{18}\right)b\,\mathbf{\hat{y}} + \left(\frac{1}{2} - z_{18}\right)c\,\mathbf{\hat{z}}$ | (4$a$) | O III |
| $\mathbf{B}_{72}$ | = | $\left(\frac{1}{2} + x_{18}\right)\mathbf{a}_1 + \left(\frac{1}{2} - y_{18}\right)\mathbf{a}_2 - z_{18}\,\mathbf{a}_3$ | = | $\left(\frac{1}{2} + x_{18}\right)a\,\mathbf{\hat{x}} + \left(\frac{1}{2} - y_{18}\right)b\,\mathbf{\hat{y}} - z_{18}c\,\mathbf{\hat{z}}$ | (4$a$) | O III |
| $\mathbf{B}_{73}$ | = | $x_{19}\,\mathbf{a}_1 + y_{19}\,\mathbf{a}_2 + z_{19}\,\mathbf{a}_3$ | = | $x_{19}a\,\mathbf{\hat{x}} + y_{19}b\,\mathbf{\hat{y}} + z_{19}c\,\mathbf{\hat{z}}$ | (4$a$) | O IV |
| $\mathbf{B}_{74}$ | = | $\left(\frac{1}{2} - x_{19}\right)\mathbf{a}_1 - y_{19}\,\mathbf{a}_2 + \left(\frac{1}{2} + z_{19}\right)\mathbf{a}_3$ | = | $\left(\frac{1}{2} - x_{19}\right)a\,\mathbf{\hat{x}} - y_{19}b\,\mathbf{\hat{y}} + \left(\frac{1}{2} + z_{19}\right)c\,\mathbf{\hat{z}}$ | (4$a$) | O IV |
| $\mathbf{B}_{75}$ | = | $-x_{19}\,\mathbf{a}_1 + \left(\frac{1}{2} + y_{19}\right)\mathbf{a}_2 + \left(\frac{1}{2} - z_{19}\right)\mathbf{a}_3$ | = | $-x_{19}a\,\mathbf{\hat{x}} + \left(\frac{1}{2} + y_{19}\right)b\,\mathbf{\hat{y}} + \left(\frac{1}{2} - z_{19}\right)c\,\mathbf{\hat{z}}$ | (4$a$) | O IV |
| $\mathbf{B}_{76}$ | = | $\left(\frac{1}{2} + x_{19}\right)\mathbf{a}_1 + \left(\frac{1}{2} - y_{19}\right)\mathbf{a}_2 - z_{19}\,\mathbf{a}_3$ | = | $\left(\frac{1}{2} + x_{19}\right)a\,\mathbf{\hat{x}} + \left(\frac{1}{2} - y_{19}\right)b\,\mathbf{\hat{y}} - z_{19}c\,\mathbf{\hat{z}}$ | (4$a$) | O IV |
| $\mathbf{B}_{77}$ | = | $x_{20}\,\mathbf{a}_1 + y_{20}\,\mathbf{a}_2 + z_{20}\,\mathbf{a}_3$ | = | $x_{20}a\,\mathbf{\hat{x}} + y_{20}b\,\mathbf{\hat{y}} + z_{20}c\,\mathbf{\hat{z}}$ | (4$a$) | O V |
| $\mathbf{B}_{78}$ | = | $\left(\frac{1}{2} - x_{20}\right)\mathbf{a}_1 - y_{20}\,\mathbf{a}_2 + \left(\frac{1}{2} + z_{20}\right)\mathbf{a}_3$ | = | $\left(\frac{1}{2} - x_{20}\right)a\,\mathbf{\hat{x}} - y_{20}b\,\mathbf{\hat{y}} + \left(\frac{1}{2} + z_{20}\right)c\,\mathbf{\hat{z}}$ | (4$a$) | O V |
| $\mathbf{B}_{79}$ | = | $-x_{20}\,\mathbf{a}_1 + \left(\frac{1}{2} + y_{20}\right)\mathbf{a}_2 + \left(\frac{1}{2} - z_{20}\right)\mathbf{a}_3$ | = | $-x_{20}a\,\mathbf{\hat{x}} + \left(\frac{1}{2} + y_{20}\right)b\,\mathbf{\hat{y}} + \left(\frac{1}{2} - z_{20}\right)c\,\mathbf{\hat{z}}$ | (4$a$) | O V |
| $\mathbf{B}_{80}$ | = | $\left(\frac{1}{2} + x_{20}\right)\mathbf{a}_1 + \left(\frac{1}{2} - y_{20}\right)\mathbf{a}_2 - z_{20}\,\mathbf{a}_3$ | = | $\left(\frac{1}{2} + x_{20}\right)a\,\mathbf{\hat{x}} + \left(\frac{1}{2} - y_{20}\right)b\,\mathbf{\hat{y}} - z_{20}c\,\mathbf{\hat{z}}$ | (4$a$) | O V |
| $\mathbf{B}_{81}$ | = | $x_{21}\,\mathbf{a}_1 + y_{21}\,\mathbf{a}_2 + z_{21}\,\mathbf{a}_3$ | = | $x_{21}a\,\mathbf{\hat{x}} + y_{21}b\,\mathbf{\hat{y}} + z_{21}c\,\mathbf{\hat{z}}$ | (4$a$) | O VI |
| $\mathbf{B}_{82}$ | = | $\left(\frac{1}{2} - x_{21}\right)\mathbf{a}_1 - y_{21}\,\mathbf{a}_2 + \left(\frac{1}{2} + z_{21}\right)\mathbf{a}_3$ | = | $\left(\frac{1}{2} - x_{21}\right)a\,\mathbf{\hat{x}} - y_{21}b\,\mathbf{\hat{y}} + \left(\frac{1}{2} + z_{21}\right)c\,\mathbf{\hat{z}}$ | (4$a$) | O VI |
| $\mathbf{B}_{83}$ | = | $-x_{21}\,\mathbf{a}_1 + \left(\frac{1}{2} + y_{21}\right)\mathbf{a}_2 + \left(\frac{1}{2} - z_{21}\right)\mathbf{a}_3$ | = | $-x_{21}a\,\mathbf{\hat{x}} + \left(\frac{1}{2} + y_{21}\right)b\,\mathbf{\hat{y}} + \left(\frac{1}{2} - z_{21}\right)c\,\mathbf{\hat{z}}$ | (4$a$) | O VI |
| $\mathbf{B}_{84}$ | = | $\left(\frac{1}{2} + x_{21}\right)\mathbf{a}_1 + \left(\frac{1}{2} - y_{21}\right)\mathbf{a}_2 - z_{21}\,\mathbf{a}_3$ | = | $\left(\frac{1}{2} + x_{21}\right)a\,\mathbf{\hat{x}} + \left(\frac{1}{2} - y_{21}\right)b\,\mathbf{\hat{y}} - z_{21}c\,\mathbf{\hat{z}}$ | (4$a$) | O VI |
| $\mathbf{B}_{85}$ | = | $x_{22}\,\mathbf{a}_1 + y_{22}\,\mathbf{a}_2 + z_{22}\,\mathbf{a}_3$ | = | $x_{22}a\,\mathbf{\hat{x}} + y_{22}b\,\mathbf{\hat{y}} + z_{22}c\,\mathbf{\hat{z}}$ | (4$a$) | O VII |
| $\mathbf{B}_{86}$ | = | $\left(\frac{1}{2} - x_{22}\right)\mathbf{a}_1 - y_{22}\,\mathbf{a}_2 + \left(\frac{1}{2} + z_{22}\right)\mathbf{a}_3$ | = | $\left(\frac{1}{2} - x_{22}\right)a\,\mathbf{\hat{x}} - y_{22}b\,\mathbf{\hat{y}} + \left(\frac{1}{2} + z_{22}\right)c\,\mathbf{\hat{z}}$ | (4$a$) | O VII |
| $\mathbf{B}_{87}$ | = | $-x_{22}\,\mathbf{a}_1 + \left(\frac{1}{2} + y_{22}\right)\mathbf{a}_2 + \left(\frac{1}{2} - z_{22}\right)\mathbf{a}_3$ | = | $-x_{22}a\,\mathbf{\hat{x}} + \left(\frac{1}{2} + y_{22}\right)b\,\mathbf{\hat{y}} + \left(\frac{1}{2} - z_{22}\right)c\,\mathbf{\hat{z}}$ | (4$a$) | O VII |
| $\mathbf{B}_{88}$ | = | $\left(\frac{1}{2} + x_{22}\right)\mathbf{a}_1 + \left(\frac{1}{2} - y_{22}\right)\mathbf{a}_2 - z_{22}\,\mathbf{a}_3$ | = | $\left(\frac{1}{2} + x_{22}\right)a\,\mathbf{\hat{x}} + \left(\frac{1}{2} - y_{22}\right)b\,\mathbf{\hat{y}} - z_{22}c\,\mathbf{\hat{z}}$ | (4$a$) | O VII |
| $\mathbf{B}_{89}$ | = | $x_{23}\,\mathbf{a}_1 + y_{23}\,\mathbf{a}_2 + z_{23}\,\mathbf{a}_3$ | = | $x_{23}a\,\mathbf{\hat{x}} + y_{23}b\,\mathbf{\hat{y}} + z_{23}c\,\mathbf{\hat{z}}$ | (4$a$) | O VIII |
| $\mathbf{B}_{90}$ | = | $\left(\frac{1}{2} - x_{23}\right)\mathbf{a}_1 - y_{23}\,\mathbf{a}_2 + \left(\frac{1}{2} + z_{23}\right)\mathbf{a}_3$ | = | $\left(\frac{1}{2} - x_{23}\right)a\,\mathbf{\hat{x}} - y_{23}b\,\mathbf{\hat{y}} + \left(\frac{1}{2} + z_{23}\right)c\,\mathbf{\hat{z}}$ | (4$a$) | O VIII |
| $\mathbf{B}_{91}$ | = | $-x_{23}\,\mathbf{a}_1+\left(\frac{1}{2}+y_{23}\right)\,\mathbf{a}_2+\left(\frac{1}{2}-z_{23}\right)\,\mathbf{a}_3$ | = | $-x_{23}a\,\mathbf{\hat{x}}+\left(\frac{1}{2}+y_{23}\right)b\,\mathbf{\hat{y}}+\left(\frac{1}{2}-z_{23}\right)c\,\mathbf{\hat{z}}$ | (4$a$) | O VIII |
| $\mathbf{B}_{92}$ | = | $\left(\frac{1}{2}+x_{23}\right)\,\mathbf{a}_1+\left(\frac{1}{2}-y_{23}\right)\,\mathbf{a}_2-z_{23}\,\mathbf{a}_3$ | = | $\left(\frac{1}{2}+x_{23}\right)a\,\mathbf{\hat{x}}+\left(\frac{1}{2}-y_{23}\right)b\,\mathbf{\hat{y}}-z_{23}c\,\mathbf{\hat{z}}$ | (4$a$) | O VIII |
| $\mathbf{B}_{93}$ | = | $x_{24}\,\mathbf{a}_1+y_{24}\,\mathbf{a}_2+z_{24}\,\mathbf{a}_3$ | = | $x_{24}a\,\mathbf{\hat{x}}+y_{24}b\,\mathbf{\hat{y}}+z_{24}c\,\mathbf{\hat{z}}$ | (4$a$) | O IX |
| $\mathbf{B}_{94}$ | = | $\left(\frac{1}{2}-x_{24}\right)\,\mathbf{a}_1-y_{24}\,\mathbf{a}_2+\left(\frac{1}{2}+z_{24}\right)\,\mathbf{a}_3$ | = | $\left(\frac{1}{2}-x_{24}\right)a\,\mathbf{\hat{x}}-y_{24}b\,\mathbf{\hat{y}}+\left(\frac{1}{2}+z_{24}\right)c\,\mathbf{\hat{z}}$ | (4$a$) | O IX |
| $\mathbf{B}_{95}$ | = | $-x_{24}\,\mathbf{a}_1+\left(\frac{1}{2}+y_{24}\right)\,\mathbf{a}_2+\left(\frac{1}{2}-z_{24}\right)\,\mathbf{a}_3$ | = | $-x_{24}a\,\mathbf{\hat{x}}+\left(\frac{1}{2}+y_{24}\right)b\,\mathbf{\hat{y}}+\left(\frac{1}{2}-z_{24}\right)c\,\mathbf{\hat{z}}$ | (4$a$) | O IX |
| $\mathbf{B}_{96}$ | = | $\left(\frac{1}{2}+x_{24}\right)\,\mathbf{a}_1+\left(\frac{1}{2}-y_{24}\right)\,\mathbf{a}_2-z_{24}\,\mathbf{a}_3$ | = | $\left(\frac{1}{2}+x_{24}\right)a\,\mathbf{\hat{x}}+\left(\frac{1}{2}-y_{24}\right)b\,\mathbf{\hat{y}}-z_{24}c\,\mathbf{\hat{z}}$ | (4$a$) | O IX |
| $\mathbf{B}_{97}$ | = | $x_{25}\,\mathbf{a}_1+y_{25}\,\mathbf{a}_2+z_{25}\,\mathbf{a}_3$ | = | $x_{25}a\,\mathbf{\hat{x}}+y_{25}b\,\mathbf{\hat{y}}+z_{25}c\,\mathbf{\hat{z}}$ | (4$a$) | O X |
| $\mathbf{B}_{98}$ | = | $\left(\frac{1}{2}-x_{25}\right)\,\mathbf{a}_1-y_{25}\,\mathbf{a}_2+\left(\frac{1}{2}+z_{25}\right)\,\mathbf{a}_3$ | = | $\left(\frac{1}{2}-x_{25}\right)a\,\mathbf{\hat{x}}-y_{25}b\,\mathbf{\hat{y}}+\left(\frac{1}{2}+z_{25}\right)c\,\mathbf{\hat{z}}$ | (4$a$) | O X |
| $\mathbf{B}_{99}$ | = | $-x_{25}\,\mathbf{a}_1+\left(\frac{1}{2}+y_{25}\right)\,\mathbf{a}_2+\left(\frac{1}{2}-z_{25}\right)\,\mathbf{a}_3$ | = | $-x_{25}a\,\mathbf{\hat{x}}+\left(\frac{1}{2}+y_{25}\right)b\,\mathbf{\hat{y}}+\left(\frac{1}{2}-z_{25}\right)c\,\mathbf{\hat{z}}$ | (4$a$) | O X |
| $\mathbf{B}_{100}$ | = | $\left(\frac{1}{2}+x_{25}\right)\,\mathbf{a}_1+\left(\frac{1}{2}-y_{25}\right)\,\mathbf{a}_2-z_{25}\,\mathbf{a}_3$ | = | $\left(\frac{1}{2}+x_{25}\right)a\,\mathbf{\hat{x}}+\left(\frac{1}{2}-y_{25}\right)b\,\mathbf{\hat{y}}-z_{25}c\,\mathbf{\hat{z}}$ | (4$a$) | O X |
| $\mathbf{B}_{101}$ | = | $x_{26}\,\mathbf{a}_1+y_{26}\,\mathbf{a}_2+z_{26}\,\mathbf{a}_3$ | = | $x_{26}a\,\mathbf{\hat{x}}+y_{26}b\,\mathbf{\hat{y}}+z_{26}c\,\mathbf{\hat{z}}$ | (4$a$) | O XI |
| $\mathbf{B}_{102}$ | = | $\left(\frac{1}{2}-x_{26}\right)\,\mathbf{a}_1-y_{26}\,\mathbf{a}_2+\left(\frac{1}{2}+z_{26}\right)\,\mathbf{a}_3$ | = | $\left(\frac{1}{2}-x_{26}\right)a\,\mathbf{\hat{x}}-y_{26}b\,\mathbf{\hat{y}}+\left(\frac{1}{2}+z_{26}\right)c\,\mathbf{\hat{z}}$ | (4$a$) | O XI |
| $\mathbf{B}_{103}$ | = | $-x_{26}\,\mathbf{a}_1+\left(\frac{1}{2}+y_{26}\right)\,\mathbf{a}_2+\left(\frac{1}{2}-z_{26}\right)\,\mathbf{a}_3$ | = | $-x_{26}a\,\mathbf{\hat{x}}+\left(\frac{1}{2}+y_{26}\right)b\,\mathbf{\hat{y}}+\left(\frac{1}{2}-z_{26}\right)c\,\mathbf{\hat{z}}$ | (4$a$) | O XI |
| $\mathbf{B}_{104}$ | = | $\left(\frac{1}{2}+x_{26}\right)\,\mathbf{a}_1+\left(\frac{1}{2}-y_{26}\right)\,\mathbf{a}_2-z_{26}\,\mathbf{a}_3$ | = | $\left(\frac{1}{2}+x_{26}\right)a\,\mathbf{\hat{x}}+\left(\frac{1}{2}-y_{26}\right)b\,\mathbf{\hat{y}}-z_{26}c\,\mathbf{\hat{z}}$ | (4$a$) | O XI |
| $\mathbf{B}_{105}$ | = | $x_{27}\,\mathbf{a}_1+y_{27}\,\mathbf{a}_2+z_{27}\,\mathbf{a}_3$ | = | $x_{27}a\,\mathbf{\hat{x}}+y_{27}b\,\mathbf{\hat{y}}+z_{27}c\,\mathbf{\hat{z}}$ | (4$a$) | S |
| $\mathbf{B}_{106}$ | = | $\left(\frac{1}{2}-x_{27}\right)\,\mathbf{a}_1-y_{27}\,\mathbf{a}_2+\left(\frac{1}{2}+z_{27}\right)\,\mathbf{a}_3$ | = | $\left(\frac{1}{2}-x_{27}\right)a\,\mathbf{\hat{x}}-y_{27}b\,\mathbf{\hat{y}}+\left(\frac{1}{2}+z_{27}\right)c\,\mathbf{\hat{z}}$ | (4$a$) | S |
| $\mathbf{B}_{107}$ | = | $-x_{27}\,\mathbf{a}_1+\left(\frac{1}{2}+y_{27}\right)\,\mathbf{a}_2+\left(\frac{1}{2}-z_{27}\right)\,\mathbf{a}_3$ | = | $-x_{27}a\,\mathbf{\hat{x}}+\left(\frac{1}{2}+y_{27}\right)b\,\mathbf{\hat{y}}+\left(\frac{1}{2}-z_{27}\right)c\,\mathbf{\hat{z}}$ | (4$a$) | S |
| $\mathbf{B}_{108}$ | = | $\left(\frac{1}{2}+x_{27}\right)\,\mathbf{a}_1+\left(\frac{1}{2}-y_{27}\right)\,\mathbf{a}_2-z_{27}\,\mathbf{a}_3$ | = | $\left(\frac{1}{2}+x_{27}\right)a\,\mathbf{\hat{x}}+\left(\frac{1}{2}-y_{27}\right)b\,\mathbf{\hat{y}}-z_{27}c\,\mathbf{\hat{z}}$ | (4$a$) | S |

**Geometry files:**

- CIF: pp. 1576
- POSCAR: pp. 1577

# Wülfingite ($\epsilon$-Zn(OH)$_2$, $C31$) Structure: A2B2C_oP20_19_2a_2a_a

http://aflow.org/prototype-encyclopedia/A2B2C_oP20_19_2a_2a_a

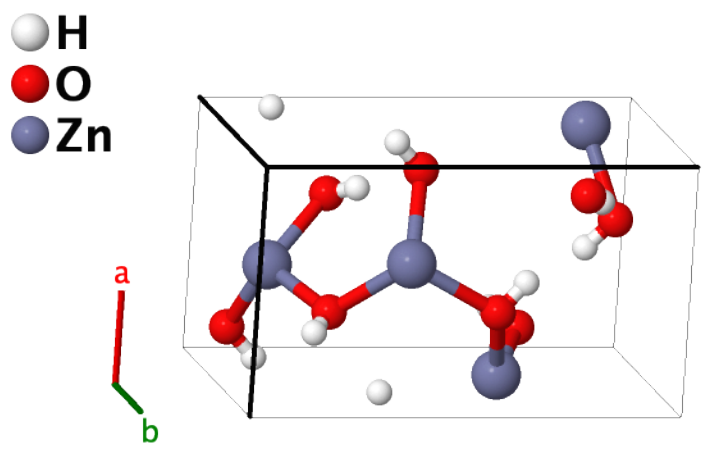

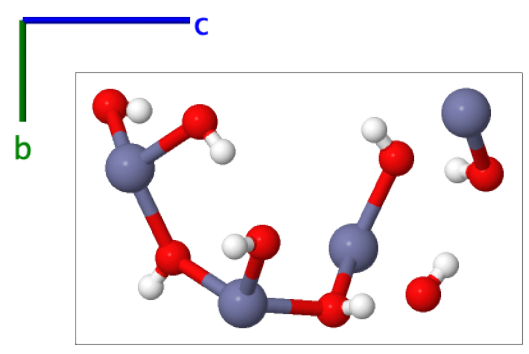

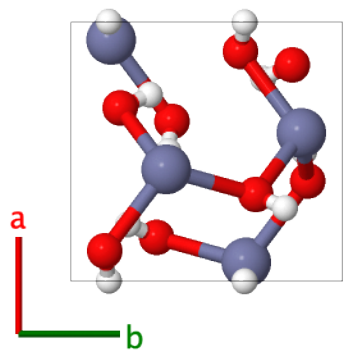

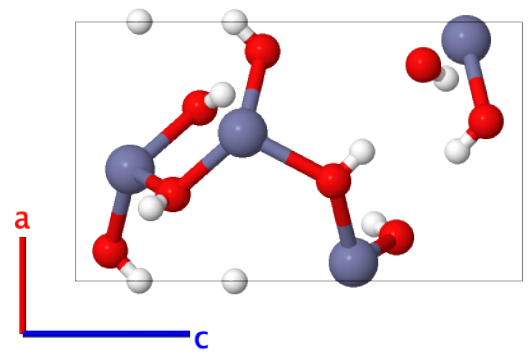

| | | |
|---|---|---|
| **Prototype** | : | $H_2O_2Zn$ |
| **AFLOW prototype label** | : | A2B2C_oP20_19_2a_2a_a |
| ***Strukturbericht* designation** | : | $C31$ |
| **Pearson symbol** | : | oP20 |
| **Space group number** | : | 19 |
| **Space group symbol** | : | $P2_12_12_1$ |
| **AFLOW prototype command** | : | `aflow --proto=A2B2C_oP20_19_2a_2a_a` <br> `--params=`$a, b/a, c/a, x_1, y_1, z_1, x_2, y_2, z_2, x_3, y_3, z_3, x_4, y_4, z_4, x_5, y_5, z_5$ |

### Other compounds with this structure

- $\beta$-Be(OH)$_2$

---

- (Corey, 1933) originally determined the structure of $\epsilon$-Zn(OH)$_2$, but they were unable to locate the hydrogen atoms. This structure was given the *Strukturbericht* designation $C31$ by (Gottfried, 1937). (Stahl, 1998) located the hydrogen atoms, and found that they did not change the space group, so we use the updated structure as our $C31$ prototype.

### Simple Orthorhombic primitive vectors:

$$\mathbf{a}_1 = a\,\hat{\mathbf{x}}$$
$$\mathbf{a}_2 = b\,\hat{\mathbf{y}}$$
$$\mathbf{a}_3 = c\,\hat{\mathbf{z}}$$

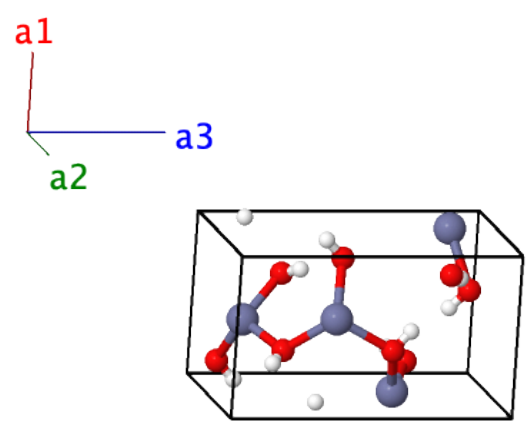

### Basis vectors:

| | | Lattice Coordinates | | Cartesian Coordinates | Wyckoff Position | Atom Type |
|---|---|---|---|---|---|---|
| $\mathbf{B}_1$ | = | $x_1\,\mathbf{a}_1 + y_1\,\mathbf{a}_2 + z_1\,\mathbf{a}_3$ | = | $x_1 a\,\hat{\mathbf{x}} + y_1 b\,\hat{\mathbf{y}} + z_1 c\,\hat{\mathbf{z}}$ | (4$a$) | H I |

$$\begin{array}{lllllll}
\mathbf{B}_2 & = & \left(\frac{1}{2} - x_1\right)\,\mathbf{a}_1 - y_1\,\mathbf{a}_2 + \left(\frac{1}{2} + z_1\right)\,\mathbf{a}_3 & = & \left(\frac{1}{2} - x_1\right)a\,\hat{\mathbf{x}} - y_1 b\,\hat{\mathbf{y}} + \left(\frac{1}{2} + z_1\right)c\,\hat{\mathbf{z}} & (4a) & \text{H I} \\
\mathbf{B}_3 & = & -x_1\,\mathbf{a}_1 + \left(\frac{1}{2} + y_1\right)\,\mathbf{a}_2 + \left(\frac{1}{2} - z_1\right)\,\mathbf{a}_3 & = & -x_1 a\,\hat{\mathbf{x}} + \left(\frac{1}{2} + y_1\right)b\,\hat{\mathbf{y}} + \left(\frac{1}{2} - z_1\right)c\,\hat{\mathbf{z}} & (4a) & \text{H I} \\
\mathbf{B}_4 & = & \left(\frac{1}{2} + x_1\right)\,\mathbf{a}_1 + \left(\frac{1}{2} - y_1\right)\,\mathbf{a}_2 - z_1\,\mathbf{a}_3 & = & \left(\frac{1}{2} + x_1\right)a\,\hat{\mathbf{x}} + \left(\frac{1}{2} - y_1\right)b\,\hat{\mathbf{y}} - z_1 c\,\hat{\mathbf{z}} & (4a) & \text{H I} \\
\mathbf{B}_5 & = & x_2\,\mathbf{a}_1 + y_2\,\mathbf{a}_2 + z_2\,\mathbf{a}_3 & = & x_2 a\,\hat{\mathbf{x}} + y_2 b\,\hat{\mathbf{y}} + z_2 c\,\hat{\mathbf{z}} & (4a) & \text{H II} \\
\mathbf{B}_6 & = & \left(\frac{1}{2} - x_2\right)\,\mathbf{a}_1 - y_2\,\mathbf{a}_2 + \left(\frac{1}{2} + z_2\right)\,\mathbf{a}_3 & = & \left(\frac{1}{2} - x_2\right)a\,\hat{\mathbf{x}} - y_2 b\,\hat{\mathbf{y}} + \left(\frac{1}{2} + z_2\right)c\,\hat{\mathbf{z}} & (4a) & \text{H II} \\
\mathbf{B}_7 & = & -x_2\,\mathbf{a}_1 + \left(\frac{1}{2} + y_2\right)\,\mathbf{a}_2 + \left(\frac{1}{2} - z_2\right)\,\mathbf{a}_3 & = & -x_2 a\,\hat{\mathbf{x}} + \left(\frac{1}{2} + y_2\right)b\,\hat{\mathbf{y}} + \left(\frac{1}{2} - z_2\right)c\,\hat{\mathbf{z}} & (4a) & \text{H II} \\
\mathbf{B}_8 & = & \left(\frac{1}{2} + x_2\right)\,\mathbf{a}_1 + \left(\frac{1}{2} - y_2\right)\,\mathbf{a}_2 - z_2\,\mathbf{a}_3 & = & \left(\frac{1}{2} + x_2\right)a\,\hat{\mathbf{x}} + \left(\frac{1}{2} - y_2\right)b\,\hat{\mathbf{y}} - z_2 c\,\hat{\mathbf{z}} & (4a) & \text{H II} \\
\mathbf{B}_9 & = & x_3\,\mathbf{a}_1 + y_3\,\mathbf{a}_2 + z_3\,\mathbf{a}_3 & = & x_3 a\,\hat{\mathbf{x}} + y_3 b\,\hat{\mathbf{y}} + z_3 c\,\hat{\mathbf{z}} & (4a) & \text{O I} \\
\mathbf{B}_{10} & = & \left(\frac{1}{2} - x_3\right)\,\mathbf{a}_1 - y_3\,\mathbf{a}_2 + \left(\frac{1}{2} + z_3\right)\,\mathbf{a}_3 & = & \left(\frac{1}{2} - x_3\right)a\,\hat{\mathbf{x}} - y_3 b\,\hat{\mathbf{y}} + \left(\frac{1}{2} + z_3\right)c\,\hat{\mathbf{z}} & (4a) & \text{O I} \\
\mathbf{B}_{11} & = & -x_3\,\mathbf{a}_1 + \left(\frac{1}{2} + y_3\right)\,\mathbf{a}_2 + \left(\frac{1}{2} - z_3\right)\,\mathbf{a}_3 & = & -x_3 a\,\hat{\mathbf{x}} + \left(\frac{1}{2} + y_3\right)b\,\hat{\mathbf{y}} + \left(\frac{1}{2} - z_3\right)c\,\hat{\mathbf{z}} & (4a) & \text{O I} \\
\mathbf{B}_{12} & = & \left(\frac{1}{2} + x_3\right)\,\mathbf{a}_1 + \left(\frac{1}{2} - y_3\right)\,\mathbf{a}_2 - z_3\,\mathbf{a}_3 & = & \left(\frac{1}{2} + x_3\right)a\,\hat{\mathbf{x}} + \left(\frac{1}{2} - y_3\right)b\,\hat{\mathbf{y}} - z_3 c\,\hat{\mathbf{z}} & (4a) & \text{O I} \\
\mathbf{B}_{13} & = & x_4\,\mathbf{a}_1 + y_4\,\mathbf{a}_2 + z_4\,\mathbf{a}_3 & = & x_4 a\,\hat{\mathbf{x}} + y_4 b\,\hat{\mathbf{y}} + z_4 c\,\hat{\mathbf{z}} & (4a) & \text{O II} \\
\mathbf{B}_{14} & = & \left(\frac{1}{2} - x_4\right)\,\mathbf{a}_1 - y_4\,\mathbf{a}_2 + \left(\frac{1}{2} + z_4\right)\,\mathbf{a}_3 & = & \left(\frac{1}{2} - x_4\right)a\,\hat{\mathbf{x}} - y_4 b\,\hat{\mathbf{y}} + \left(\frac{1}{2} + z_4\right)c\,\hat{\mathbf{z}} & (4a) & \text{O II} \\
\mathbf{B}_{15} & = & -x_4\,\mathbf{a}_1 + \left(\frac{1}{2} + y_4\right)\,\mathbf{a}_2 + \left(\frac{1}{2} - z_4\right)\,\mathbf{a}_3 & = & -x_4 a\,\hat{\mathbf{x}} + \left(\frac{1}{2} + y_4\right)b\,\hat{\mathbf{y}} + \left(\frac{1}{2} - z_4\right)c\,\hat{\mathbf{z}} & (4a) & \text{O II} \\
\mathbf{B}_{16} & = & \left(\frac{1}{2} + x_4\right)\,\mathbf{a}_1 + \left(\frac{1}{2} - y_4\right)\,\mathbf{a}_2 - z_4\,\mathbf{a}_3 & = & \left(\frac{1}{2} + x_4\right)a\,\hat{\mathbf{x}} + \left(\frac{1}{2} - y_4\right)b\,\hat{\mathbf{y}} - z_4 c\,\hat{\mathbf{z}} & (4a) & \text{O II} \\
\mathbf{B}_{17} & = & x_5\,\mathbf{a}_1 + y_5\,\mathbf{a}_2 + z_5\,\mathbf{a}_3 & = & x_5 a\,\hat{\mathbf{x}} + y_5 b\,\hat{\mathbf{y}} + z_5 c\,\hat{\mathbf{z}} & (4a) & \text{Zn} \\
\mathbf{B}_{18} & = & \left(\frac{1}{2} - x_5\right)\,\mathbf{a}_1 - y_5\,\mathbf{a}_2 + \left(\frac{1}{2} + z_5\right)\,\mathbf{a}_3 & = & \left(\frac{1}{2} - x_5\right)a\,\hat{\mathbf{x}} - y_5 b\,\hat{\mathbf{y}} + \left(\frac{1}{2} + z_5\right)c\,\hat{\mathbf{z}} & (4a) & \text{Zn} \\
\mathbf{B}_{19} & = & -x_5\,\mathbf{a}_1 + \left(\frac{1}{2} + y_5\right)\,\mathbf{a}_2 + \left(\frac{1}{2} - z_5\right)\,\mathbf{a}_3 & = & -x_5 a\,\hat{\mathbf{x}} + \left(\frac{1}{2} + y_5\right)b\,\hat{\mathbf{y}} + \left(\frac{1}{2} - z_5\right)c\,\hat{\mathbf{z}} & (4a) & \text{Zn} \\
\mathbf{B}_{20} & = & \left(\frac{1}{2} + x_5\right)\,\mathbf{a}_1 + \left(\frac{1}{2} - y_5\right)\,\mathbf{a}_2 - z_5\,\mathbf{a}_3 & = & \left(\frac{1}{2} + x_5\right)a\,\hat{\mathbf{x}} + \left(\frac{1}{2} - y_5\right)b\,\hat{\mathbf{y}} - z_5 c\,\hat{\mathbf{z}} & (4a) & \text{Zn}
\end{array}$$

---

**Found in:**

---

**Geometry files:**

- CIF: pp. 1577
- POSCAR: pp. 1578

# Ferroelectric $NH_4H_2PO_4$ Structure: A6BC4D_oP48_19_6a_a_4a_a

http://aflow.org/prototype-encyclopedia/A6BC4D_oP48_19_6a_a_4a_a

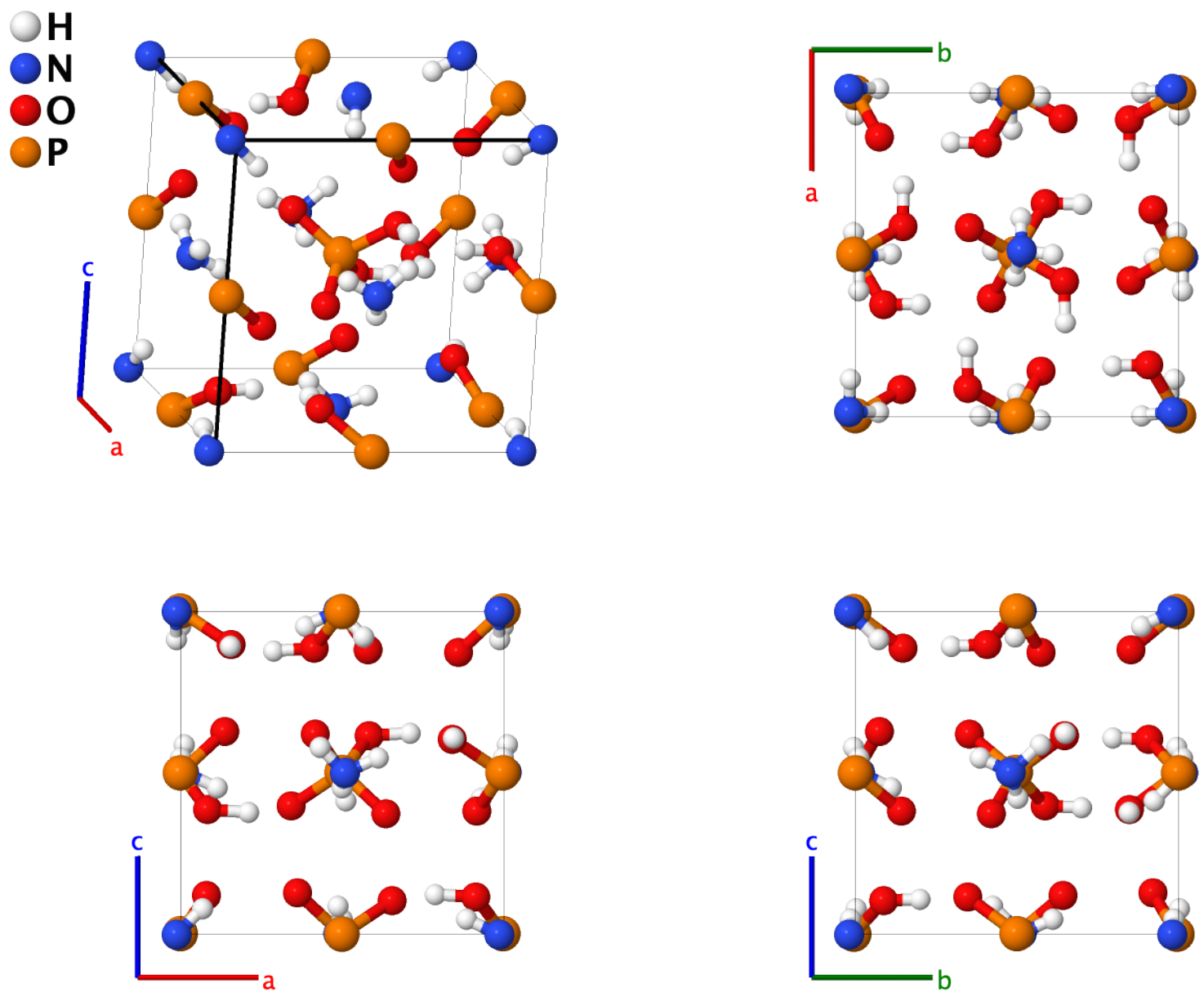

**Prototype** : $H_6NO_4P$

**AFLOW prototype label** : A6BC4D_oP48_19_6a_a_4a_a

***Strukturbericht* designation** : None

**Pearson symbol** : oP48

**Space group number** : 19

**Space group symbol** : $P2_12_12_1$

**AFLOW prototype command** : `aflow --proto=A6BC4D_oP48_19_6a_a_4a_a`
`--params=`$a, b/a, c/a, x_1, y_1, z_1, x_2, y_2, z_2, x_3, y_3, z_3, x_4, y_4, z_4, x_5, y_5, z_5, x_6, y_6, z_6, x_7, y_7, z_7, x_8, y_8, z_8, x_9, y_9, z_9, x_{10}, y_{10}, z_{10}, x_{11}, y_{11}, z_{11}, x_{12}, y_{12}, z_{12}$

### Other compounds with this structure

- $NH_4H_2AsO_4$

---

- Below 148 K, $NH_4H_2PO_4$ undergoes a transition from a tetragonal paraelectric phase to this orthorhombic ferroelectric phase. This seems to occur because the hydrogen atoms associated with the $PO_4$ ions become frozen into place.
- The data for this structure was taken at 143 K.

### Simple Orthorhombic primitive vectors:

$$\mathbf{a}_1 = a\,\hat{\mathbf{x}}$$
$$\mathbf{a}_2 = b\,\hat{\mathbf{y}}$$
$$\mathbf{a}_3 = c\,\hat{\mathbf{z}}$$

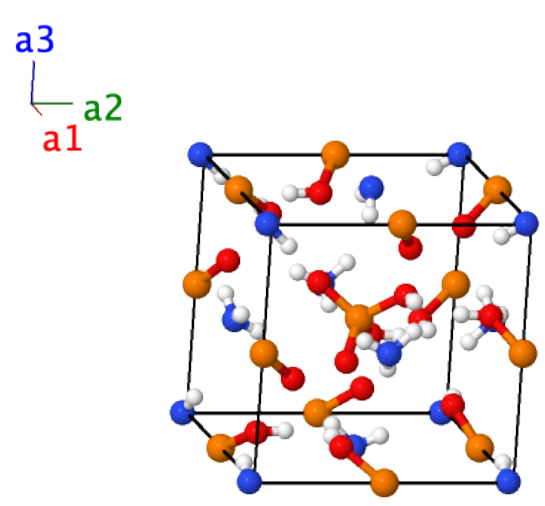

**Basis vectors:**

| | | Lattice Coordinates | | Cartesian Coordinates | Wyckoff Position | Atom Type |
|---|---|---|---|---|---|---|
| $\mathbf{B}_1$ | = | $x_1\,\mathbf{a}_1 + y_1\,\mathbf{a}_2 + z_1\,\mathbf{a}_3$ | = | $x_1 a\,\hat{\mathbf{x}} + y_1 b\,\hat{\mathbf{y}} + z_1 c\,\hat{\mathbf{z}}$ | (4$a$) | H I |
| $\mathbf{B}_2$ | = | $\left(\frac{1}{2} - x_1\right)\mathbf{a}_1 - y_1\,\mathbf{a}_2 + \left(\frac{1}{2} + z_1\right)\mathbf{a}_3$ | = | $\left(\frac{1}{2} - x_1\right) a\,\hat{\mathbf{x}} - y_1 b\,\hat{\mathbf{y}} + \left(\frac{1}{2} + z_1\right) c\,\hat{\mathbf{z}}$ | (4$a$) | H I |
| $\mathbf{B}_3$ | = | $-x_1\,\mathbf{a}_1 + \left(\frac{1}{2} + y_1\right)\mathbf{a}_2 + \left(\frac{1}{2} - z_1\right)\mathbf{a}_3$ | = | $-x_1 a\,\hat{\mathbf{x}} + \left(\frac{1}{2} + y_1\right) b\,\hat{\mathbf{y}} + \left(\frac{1}{2} - z_1\right) c\,\hat{\mathbf{z}}$ | (4$a$) | H I |
| $\mathbf{B}_4$ | = | $\left(\frac{1}{2} + x_1\right)\mathbf{a}_1 + \left(\frac{1}{2} - y_1\right)\mathbf{a}_2 - z_1\,\mathbf{a}_3$ | = | $\left(\frac{1}{2} + x_1\right) a\,\hat{\mathbf{x}} + \left(\frac{1}{2} - y_1\right) b\,\hat{\mathbf{y}} - z_1 c\,\hat{\mathbf{z}}$ | (4$a$) | H I |
| $\mathbf{B}_5$ | = | $x_2\,\mathbf{a}_1 + y_2\,\mathbf{a}_2 + z_2\,\mathbf{a}_3$ | = | $x_2 a\,\hat{\mathbf{x}} + y_2 b\,\hat{\mathbf{y}} + z_2 c\,\hat{\mathbf{z}}$ | (4$a$) | H II |
| $\mathbf{B}_6$ | = | $\left(\frac{1}{2} - x_2\right)\mathbf{a}_1 - y_2\,\mathbf{a}_2 + \left(\frac{1}{2} + z_2\right)\mathbf{a}_3$ | = | $\left(\frac{1}{2} - x_2\right) a\,\hat{\mathbf{x}} - y_2 b\,\hat{\mathbf{y}} + \left(\frac{1}{2} + z_2\right) c\,\hat{\mathbf{z}}$ | (4$a$) | H II |
| $\mathbf{B}_7$ | = | $-x_2\,\mathbf{a}_1 + \left(\frac{1}{2} + y_2\right)\mathbf{a}_2 + \left(\frac{1}{2} - z_2\right)\mathbf{a}_3$ | = | $-x_2 a\,\hat{\mathbf{x}} + \left(\frac{1}{2} + y_2\right) b\,\hat{\mathbf{y}} + \left(\frac{1}{2} - z_2\right) c\,\hat{\mathbf{z}}$ | (4$a$) | H II |
| $\mathbf{B}_8$ | = | $\left(\frac{1}{2} + x_2\right)\mathbf{a}_1 + \left(\frac{1}{2} - y_2\right)\mathbf{a}_2 - z_2\,\mathbf{a}_3$ | = | $\left(\frac{1}{2} + x_2\right) a\,\hat{\mathbf{x}} + \left(\frac{1}{2} - y_2\right) b\,\hat{\mathbf{y}} - z_2 c\,\hat{\mathbf{z}}$ | (4$a$) | H II |
| $\mathbf{B}_9$ | = | $x_3\,\mathbf{a}_1 + y_3\,\mathbf{a}_2 + z_3\,\mathbf{a}_3$ | = | $x_3 a\,\hat{\mathbf{x}} + y_3 b\,\hat{\mathbf{y}} + z_3 c\,\hat{\mathbf{z}}$ | (4$a$) | H III |
| $\mathbf{B}_{10}$ | = | $\left(\frac{1}{2} - x_3\right)\mathbf{a}_1 - y_3\,\mathbf{a}_2 + \left(\frac{1}{2} + z_3\right)\mathbf{a}_3$ | = | $\left(\frac{1}{2} - x_3\right) a\,\hat{\mathbf{x}} - y_3 b\,\hat{\mathbf{y}} + \left(\frac{1}{2} + z_3\right) c\,\hat{\mathbf{z}}$ | (4$a$) | H III |
| $\mathbf{B}_{11}$ | = | $-x_3\,\mathbf{a}_1 + \left(\frac{1}{2} + y_3\right)\mathbf{a}_2 + \left(\frac{1}{2} - z_3\right)\mathbf{a}_3$ | = | $-x_3 a\,\hat{\mathbf{x}} + \left(\frac{1}{2} + y_3\right) b\,\hat{\mathbf{y}} + \left(\frac{1}{2} - z_3\right) c\,\hat{\mathbf{z}}$ | (4$a$) | H III |
| $\mathbf{B}_{12}$ | = | $\left(\frac{1}{2} + x_3\right)\mathbf{a}_1 + \left(\frac{1}{2} - y_3\right)\mathbf{a}_2 - z_3\,\mathbf{a}_3$ | = | $\left(\frac{1}{2} + x_3\right) a\,\hat{\mathbf{x}} + \left(\frac{1}{2} - y_3\right) b\,\hat{\mathbf{y}} - z_3 c\,\hat{\mathbf{z}}$ | (4$a$) | H III |
| $\mathbf{B}_{13}$ | = | $x_4\,\mathbf{a}_1 + y_4\,\mathbf{a}_2 + z_4\,\mathbf{a}_3$ | = | $x_4 a\,\hat{\mathbf{x}} + y_4 b\,\hat{\mathbf{y}} + z_4 c\,\hat{\mathbf{z}}$ | (4$a$) | H IV |
| $\mathbf{B}_{14}$ | = | $\left(\frac{1}{2} - x_4\right)\mathbf{a}_1 - y_4\,\mathbf{a}_2 + \left(\frac{1}{2} + z_4\right)\mathbf{a}_3$ | = | $\left(\frac{1}{2} - x_4\right) a\,\hat{\mathbf{x}} - y_4 b\,\hat{\mathbf{y}} + \left(\frac{1}{2} + z_4\right) c\,\hat{\mathbf{z}}$ | (4$a$) | H IV |
| $\mathbf{B}_{15}$ | = | $-x_4\,\mathbf{a}_1 + \left(\frac{1}{2} + y_4\right)\mathbf{a}_2 + \left(\frac{1}{2} - z_4\right)\mathbf{a}_3$ | = | $-x_4 a\,\hat{\mathbf{x}} + \left(\frac{1}{2} + y_4\right) b\,\hat{\mathbf{y}} + \left(\frac{1}{2} - z_4\right) c\,\hat{\mathbf{z}}$ | (4$a$) | H IV |
| $\mathbf{B}_{16}$ | = | $\left(\frac{1}{2} + x_4\right)\mathbf{a}_1 + \left(\frac{1}{2} - y_4\right)\mathbf{a}_2 - z_4\,\mathbf{a}_3$ | = | $\left(\frac{1}{2} + x_4\right) a\,\hat{\mathbf{x}} + \left(\frac{1}{2} - y_4\right) b\,\hat{\mathbf{y}} - z_4 c\,\hat{\mathbf{z}}$ | (4$a$) | H IV |
| $\mathbf{B}_{17}$ | = | $x_5\,\mathbf{a}_1 + y_5\,\mathbf{a}_2 + z_5\,\mathbf{a}_3$ | = | $x_5 a\,\hat{\mathbf{x}} + y_5 b\,\hat{\mathbf{y}} + z_5 c\,\hat{\mathbf{z}}$ | (4$a$) | H V |
| $\mathbf{B}_{18}$ | = | $\left(\frac{1}{2} - x_5\right)\mathbf{a}_1 - y_5\,\mathbf{a}_2 + \left(\frac{1}{2} + z_5\right)\mathbf{a}_3$ | = | $\left(\frac{1}{2} - x_5\right) a\,\hat{\mathbf{x}} - y_5 b\,\hat{\mathbf{y}} + \left(\frac{1}{2} + z_5\right) c\,\hat{\mathbf{z}}$ | (4$a$) | H V |
| $\mathbf{B}_{19}$ | = | $-x_5\,\mathbf{a}_1 + \left(\frac{1}{2} + y_5\right)\mathbf{a}_2 + \left(\frac{1}{2} - z_5\right)\mathbf{a}_3$ | = | $-x_5 a\,\hat{\mathbf{x}} + \left(\frac{1}{2} + y_5\right) b\,\hat{\mathbf{y}} + \left(\frac{1}{2} - z_5\right) c\,\hat{\mathbf{z}}$ | (4$a$) | H V |
| $\mathbf{B}_{20}$ | = | $\left(\frac{1}{2} + x_5\right)\mathbf{a}_1 + \left(\frac{1}{2} - y_5\right)\mathbf{a}_2 - z_5\,\mathbf{a}_3$ | = | $\left(\frac{1}{2} + x_5\right) a\,\hat{\mathbf{x}} + \left(\frac{1}{2} - y_5\right) b\,\hat{\mathbf{y}} - z_5 c\,\hat{\mathbf{z}}$ | (4$a$) | H V |
| $\mathbf{B}_{21}$ | = | $x_6\,\mathbf{a}_1 + y_6\,\mathbf{a}_2 + z_6\,\mathbf{a}_3$ | = | $x_6 a\,\hat{\mathbf{x}} + y_6 b\,\hat{\mathbf{y}} + z_6 c\,\hat{\mathbf{z}}$ | (4$a$) | H VI |
| $\mathbf{B}_{22}$ | = | $\left(\frac{1}{2} - x_6\right)\mathbf{a}_1 - y_6\,\mathbf{a}_2 + \left(\frac{1}{2} + z_6\right)\mathbf{a}_3$ | = | $\left(\frac{1}{2} - x_6\right) a\,\hat{\mathbf{x}} - y_6 b\,\hat{\mathbf{y}} + \left(\frac{1}{2} + z_6\right) c\,\hat{\mathbf{z}}$ | (4$a$) | H VI |
| $\mathbf{B}_{23}$ | = | $-x_6\,\mathbf{a}_1 + \left(\frac{1}{2} + y_6\right)\mathbf{a}_2 + \left(\frac{1}{2} - z_6\right)\mathbf{a}_3$ | = | $-x_6 a\,\hat{\mathbf{x}} + \left(\frac{1}{2} + y_6\right) b\,\hat{\mathbf{y}} + \left(\frac{1}{2} - z_6\right) c\,\hat{\mathbf{z}}$ | (4$a$) | H VI |
| $\mathbf{B}_{24}$ | = | $\left(\frac{1}{2} + x_6\right)\mathbf{a}_1 + \left(\frac{1}{2} - y_6\right)\mathbf{a}_2 - z_6\,\mathbf{a}_3$ | = | $\left(\frac{1}{2} + x_6\right) a\,\hat{\mathbf{x}} + \left(\frac{1}{2} - y_6\right) b\,\hat{\mathbf{y}} - z_6 c\,\hat{\mathbf{z}}$ | (4$a$) | H VI |
| $\mathbf{B}_{25}$ | = | $x_7\,\mathbf{a}_1 + y_7\,\mathbf{a}_2 + z_7\,\mathbf{a}_3$ | = | $x_7 a\,\hat{\mathbf{x}} + y_7 b\,\hat{\mathbf{y}} + z_7 c\,\hat{\mathbf{z}}$ | (4$a$) | N |
| $\mathbf{B}_{26}$ | = | $\left(\frac{1}{2} - x_7\right)\mathbf{a}_1 - y_7\,\mathbf{a}_2 + \left(\frac{1}{2} + z_7\right)\mathbf{a}_3$ | = | $\left(\frac{1}{2} - x_7\right) a\,\hat{\mathbf{x}} - y_7 b\,\hat{\mathbf{y}} + \left(\frac{1}{2} + z_7\right) c\,\hat{\mathbf{z}}$ | (4$a$) | N |
| $\mathbf{B}_{27}$ | = | $-x_7\,\mathbf{a}_1 + \left(\frac{1}{2} + y_7\right)\mathbf{a}_2 + \left(\frac{1}{2} - z_7\right)\mathbf{a}_3$ | = | $-x_7 a\,\hat{\mathbf{x}} + \left(\frac{1}{2} + y_7\right) b\,\hat{\mathbf{y}} + \left(\frac{1}{2} - z_7\right) c\,\hat{\mathbf{z}}$ | (4$a$) | N |
| $\mathbf{B}_{28}$ | = | $\left(\frac{1}{2} + x_7\right)\mathbf{a}_1 + \left(\frac{1}{2} - y_7\right)\mathbf{a}_2 - z_7\,\mathbf{a}_3$ | = | $\left(\frac{1}{2} + x_7\right) a\,\hat{\mathbf{x}} + \left(\frac{1}{2} - y_7\right) b\,\hat{\mathbf{y}} - z_7 c\,\hat{\mathbf{z}}$ | (4$a$) | N |
| $\mathbf{B}_{29}$ | = | $x_8\,\mathbf{a}_1 + y_8\,\mathbf{a}_2 + z_8\,\mathbf{a}_3$ | = | $x_8 a\,\hat{\mathbf{x}} + y_8 b\,\hat{\mathbf{y}} + z_8 c\,\hat{\mathbf{z}}$ | (4$a$) | O I |
| $\mathbf{B}_{30}$ | = | $\left(\frac{1}{2} - x_8\right)\mathbf{a}_1 - y_8\,\mathbf{a}_2 + \left(\frac{1}{2} + z_8\right)\mathbf{a}_3$ | = | $\left(\frac{1}{2} - x_8\right) a\,\hat{\mathbf{x}} - y_8 b\,\hat{\mathbf{y}} + \left(\frac{1}{2} + z_8\right) c\,\hat{\mathbf{z}}$ | (4$a$) | O I |
| $\mathbf{B}_{31}$ | = | $-x_8\,\mathbf{a}_1 + \left(\frac{1}{2} + y_8\right)\mathbf{a}_2 + \left(\frac{1}{2} - z_8\right)\mathbf{a}_3$ | = | $-x_8 a\,\hat{\mathbf{x}} + \left(\frac{1}{2} + y_8\right) b\,\hat{\mathbf{y}} + \left(\frac{1}{2} - z_8\right) c\,\hat{\mathbf{z}}$ | (4$a$) | O I |
| $\mathbf{B}_{32}$ | = | $\left(\frac{1}{2} + x_8\right)\mathbf{a}_1 + \left(\frac{1}{2} - y_8\right)\mathbf{a}_2 - z_8\,\mathbf{a}_3$ | = | $\left(\frac{1}{2} + x_8\right) a\,\hat{\mathbf{x}} + \left(\frac{1}{2} - y_8\right) b\,\hat{\mathbf{y}} - z_8 c\,\hat{\mathbf{z}}$ | (4$a$) | O I |
| $\mathbf{B}_{33}$ | = | $x_9\,\mathbf{a}_1 + y_9\,\mathbf{a}_2 + z_9\,\mathbf{a}_3$ | = | $x_9 a\,\hat{\mathbf{x}} + y_9 b\,\hat{\mathbf{y}} + z_9 c\,\hat{\mathbf{z}}$ | (4$a$) | O II |
| $\mathbf{B}_{34}$ | = | $\left(\frac{1}{2} - x_9\right)\mathbf{a}_1 - y_9\,\mathbf{a}_2 + \left(\frac{1}{2} + z_9\right)\mathbf{a}_3$ | = | $\left(\frac{1}{2} - x_9\right) a\,\hat{\mathbf{x}} - y_9 b\,\hat{\mathbf{y}} + \left(\frac{1}{2} + z_9\right) c\,\hat{\mathbf{z}}$ | (4$a$) | O II |
| $\mathbf{B}_{35}$ | = | $-x_9\,\mathbf{a}_1 + \left(\frac{1}{2} + y_9\right)\,\mathbf{a}_2 + \left(\frac{1}{2} - z_9\right)\,\mathbf{a}_3$ | = | $-x_9 a\,\hat{\mathbf{x}} + \left(\frac{1}{2} + y_9\right) b\,\hat{\mathbf{y}} + \left(\frac{1}{2} - z_9\right) c\,\hat{\mathbf{z}}$ | (4*a*) | O II |
| $\mathbf{B}_{36}$ | = | $\left(\frac{1}{2} + x_9\right)\,\mathbf{a}_1 + \left(\frac{1}{2} - y_9\right)\,\mathbf{a}_2 - z_9\,\mathbf{a}_3$ | = | $\left(\frac{1}{2} + x_9\right) a\,\hat{\mathbf{x}} + \left(\frac{1}{2} - y_9\right) b\,\hat{\mathbf{y}} - z_9 c\,\hat{\mathbf{z}}$ | (4*a*) | O II |
| $\mathbf{B}_{37}$ | = | $x_{10}\,\mathbf{a}_1 + y_{10}\,\mathbf{a}_2 + z_{10}\,\mathbf{a}_3$ | = | $x_{10} a\,\hat{\mathbf{x}} + y_{10} b\,\hat{\mathbf{y}} + z_{10} c\,\hat{\mathbf{z}}$ | (4*a*) | O III |
| $\mathbf{B}_{38}$ | = | $\left(\frac{1}{2} - x_{10}\right)\,\mathbf{a}_1 - y_{10}\,\mathbf{a}_2 + \left(\frac{1}{2} + z_{10}\right)\,\mathbf{a}_3$ | = | $\left(\frac{1}{2} - x_{10}\right) a\,\hat{\mathbf{x}} - y_{10} b\,\hat{\mathbf{y}} + \left(\frac{1}{2} + z_{10}\right) c\,\hat{\mathbf{z}}$ | (4*a*) | O III |
| $\mathbf{B}_{39}$ | = | $-x_{10}\,\mathbf{a}_1 + \left(\frac{1}{2} + y_{10}\right)\,\mathbf{a}_2 + \left(\frac{1}{2} - z_{10}\right)\,\mathbf{a}_3$ | = | $-x_{10} a\,\hat{\mathbf{x}} + \left(\frac{1}{2} + y_{10}\right) b\,\hat{\mathbf{y}} + \left(\frac{1}{2} - z_{10}\right) c\,\hat{\mathbf{z}}$ | (4*a*) | O III |
| $\mathbf{B}_{40}$ | = | $\left(\frac{1}{2} + x_{10}\right)\,\mathbf{a}_1 + \left(\frac{1}{2} - y_{10}\right)\,\mathbf{a}_2 - z_{10}\,\mathbf{a}_3$ | = | $\left(\frac{1}{2} + x_{10}\right) a\,\hat{\mathbf{x}} + \left(\frac{1}{2} - y_{10}\right) b\,\hat{\mathbf{y}} - z_{10} c\,\hat{\mathbf{z}}$ | (4*a*) | O III |
| $\mathbf{B}_{41}$ | = | $x_{11}\,\mathbf{a}_1 + y_{11}\,\mathbf{a}_2 + z_{11}\,\mathbf{a}_3$ | = | $x_{11} a\,\hat{\mathbf{x}} + y_{11} b\,\hat{\mathbf{y}} + z_{11} c\,\hat{\mathbf{z}}$ | (4*a*) | O IV |
| $\mathbf{B}_{42}$ | = | $\left(\frac{1}{2} - x_{11}\right)\,\mathbf{a}_1 - y_{11}\,\mathbf{a}_2 + \left(\frac{1}{2} + z_{11}\right)\,\mathbf{a}_3$ | = | $\left(\frac{1}{2} - x_{11}\right) a\,\hat{\mathbf{x}} - y_{11} b\,\hat{\mathbf{y}} + \left(\frac{1}{2} + z_{11}\right) c\,\hat{\mathbf{z}}$ | (4*a*) | O IV |
| $\mathbf{B}_{43}$ | = | $-x_{11}\,\mathbf{a}_1 + \left(\frac{1}{2} + y_{11}\right)\,\mathbf{a}_2 + \left(\frac{1}{2} - z_{11}\right)\,\mathbf{a}_3$ | = | $-x_{11} a\,\hat{\mathbf{x}} + \left(\frac{1}{2} + y_{11}\right) b\,\hat{\mathbf{y}} + \left(\frac{1}{2} - z_{11}\right) c\,\hat{\mathbf{z}}$ | (4*a*) | O IV |
| $\mathbf{B}_{44}$ | = | $\left(\frac{1}{2} + x_{11}\right)\,\mathbf{a}_1 + \left(\frac{1}{2} - y_{11}\right)\,\mathbf{a}_2 - z_{11}\,\mathbf{a}_3$ | = | $\left(\frac{1}{2} + x_{11}\right) a\,\hat{\mathbf{x}} + \left(\frac{1}{2} - y_{11}\right) b\,\hat{\mathbf{y}} - z_{11} c\,\hat{\mathbf{z}}$ | (4*a*) | O IV |
| $\mathbf{B}_{45}$ | = | $x_{12}\,\mathbf{a}_1 + y_{12}\,\mathbf{a}_2 + z_{12}\,\mathbf{a}_3$ | = | $x_{12} a\,\hat{\mathbf{x}} + y_{12} b\,\hat{\mathbf{y}} + z_{12} c\,\hat{\mathbf{z}}$ | (4*a*) | P |
| $\mathbf{B}_{46}$ | = | $\left(\frac{1}{2} - x_{12}\right)\,\mathbf{a}_1 - y_{12}\,\mathbf{a}_2 + \left(\frac{1}{2} + z_{12}\right)\,\mathbf{a}_3$ | = | $\left(\frac{1}{2} - x_{12}\right) a\,\hat{\mathbf{x}} - y_{12} b\,\hat{\mathbf{y}} + \left(\frac{1}{2} + z_{12}\right) c\,\hat{\mathbf{z}}$ | (4*a*) | P |
| $\mathbf{B}_{47}$ | = | $-x_{12}\,\mathbf{a}_1 + \left(\frac{1}{2} + y_{12}\right)\,\mathbf{a}_2 + \left(\frac{1}{2} - z_{12}\right)\,\mathbf{a}_3$ | = | $-x_{12} a\,\hat{\mathbf{x}} + \left(\frac{1}{2} + y_{12}\right) b\,\hat{\mathbf{y}} + \left(\frac{1}{2} - z_{12}\right) c\,\hat{\mathbf{z}}$ | (4*a*) | P |
| $\mathbf{B}_{48}$ | = | $\left(\frac{1}{2} + x_{12}\right)\,\mathbf{a}_1 + \left(\frac{1}{2} - y_{12}\right)\,\mathbf{a}_2 - z_{12}\,\mathbf{a}_3$ | = | $\left(\frac{1}{2} + x_{12}\right) a\,\hat{\mathbf{x}} + \left(\frac{1}{2} - y_{12}\right) b\,\hat{\mathbf{y}} - z_{12} c\,\hat{\mathbf{z}}$ | (4*a*) | P |

---

---

**Geometry files:**

- CIF: pp. 1578
- POSCAR: pp. 1578

# $\beta$-Arabinose [$(CH_2O)_{20}$] Structure: AB2C_oP80_19_5a_10a_5a

http://aflow.org/prototype-encyclopedia/AB2C_oP80_19_5a_10a_5a

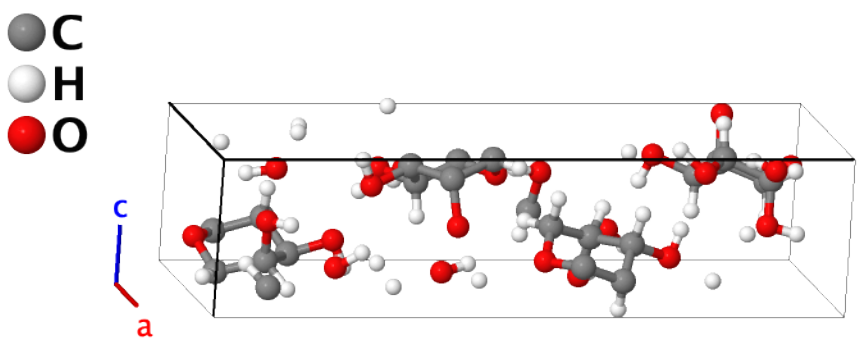

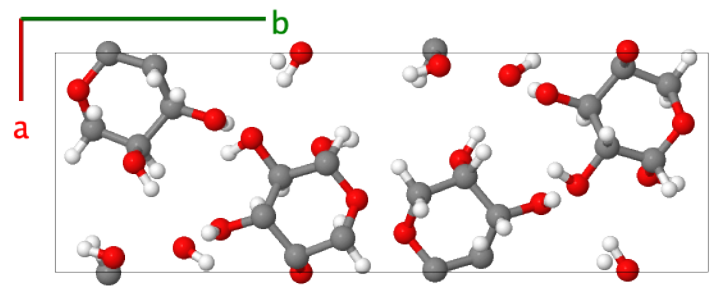

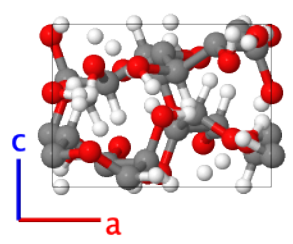

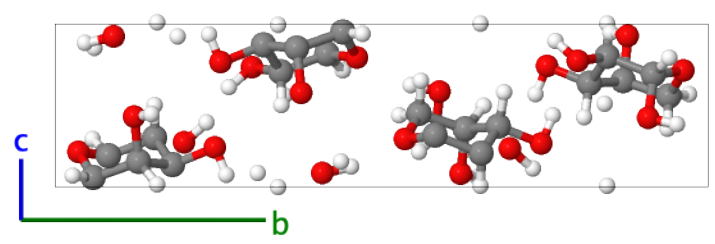

| | | |
|---|---|---|
| **Prototype** | : | $CH_2O$ |
| **AFLOW prototype label** | : | AB2C_oP80_19_5a_10a_5a |
| ***Strukturbericht* designation** | : | None |
| **Pearson symbol** | : | oP80 |
| **Space group number** | : | 19 |
| **Space group symbol** | : | $P2_12_12_1$ |
| **AFLOW prototype command** | : | `aflow --proto=AB2C_oP80_19_5a_10a_5a --params=`$a, b/a, c/a, x_1, y_1, z_1, x_2, y_2, z_2, x_3, y_3, z_3, x_4, y_4, z_4, x_5, y_5, z_5, x_6, y_6, z_6, x_7, y_7, z_7, x_8, y_8, z_8, x_9, y_9, z_9, x_{10}, y_{10}, z_{10}, x_{11}, y_{11}, z_{11}, x_{12}, y_{12}, z_{12}, x_{13}, y_{13}, z_{13}, x_{14}, y_{14}, z_{14}, x_{15}, y_{15}, z_{15}, x_{16}, y_{16}, z_{16}, x_{17}, y_{17}, z_{17}, x_{18}, y_{18}, z_{18}, x_{19}, y_{19}, z_{19}, x_{20}, y_{20}, z_{20}$ |

**Simple Orthorhombic primitive vectors:**

$$\mathbf{a}_1 = a\,\hat{\mathbf{x}}$$
$$\mathbf{a}_2 = b\,\hat{\mathbf{y}}$$
$$\mathbf{a}_3 = c\,\hat{\mathbf{z}}$$

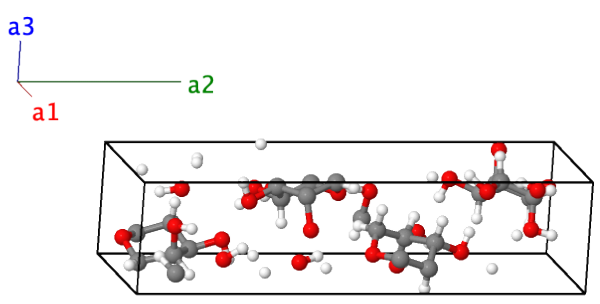

**Basis vectors:**

| | | Lattice Coordinates | | Cartesian Coordinates | Wyckoff Position | Atom Type |
|---|---|---|---|---|---|---|
| $\mathbf{B}_1$ | = | $x_1\,\mathbf{a}_1 + y_1\,\mathbf{a}_2 + z_1\,\mathbf{a}_3$ | = | $x_1 a\,\hat{\mathbf{x}} + y_1 b\,\hat{\mathbf{y}} + z_1 c\,\hat{\mathbf{z}}$ | (4a) | C I |
| $\mathbf{B}_2$ | = | $\left(\frac{1}{2} - x_1\right)\mathbf{a}_1 - y_1\,\mathbf{a}_2 + \left(\frac{1}{2} + z_1\right)\mathbf{a}_3$ | = | $\left(\frac{1}{2} - x_1\right)a\,\hat{\mathbf{x}} - y_1 b\,\hat{\mathbf{y}} + \left(\frac{1}{2} + z_1\right)c\,\hat{\mathbf{z}}$ | (4a) | C I |
| $\mathbf{B}_3$ | = | $-x_1\,\mathbf{a}_1 + \left(\frac{1}{2} + y_1\right)\mathbf{a}_2 + \left(\frac{1}{2} - z_1\right)\mathbf{a}_3$ | = | $-x_1 a\,\hat{\mathbf{x}} + \left(\frac{1}{2} + y_1\right)b\,\hat{\mathbf{y}} + \left(\frac{1}{2} - z_1\right)c\,\hat{\mathbf{z}}$ | (4a) | C I |
| $\mathbf{B}_4$ | = | $\left(\frac{1}{2} + x_1\right)\mathbf{a}_1 + \left(\frac{1}{2} - y_1\right)\mathbf{a}_2 - z_1\,\mathbf{a}_3$ | = | $\left(\frac{1}{2} + x_1\right)a\,\hat{\mathbf{x}} + \left(\frac{1}{2} - y_1\right)b\,\hat{\mathbf{y}} - z_1 c\,\hat{\mathbf{z}}$ | (4a) | C I |
| $\mathbf{B}_5$ | = | $x_2\,\mathbf{a}_1 + y_2\,\mathbf{a}_2 + z_2\,\mathbf{a}_3$ | = | $x_2 a\,\hat{\mathbf{x}} + y_2 b\,\hat{\mathbf{y}} + z_2 c\,\hat{\mathbf{z}}$ | (4a) | C II |
| $\mathbf{B}_6$ | = | $\left(\frac{1}{2} - x_2\right)\mathbf{a}_1 - y_2\,\mathbf{a}_2 + \left(\frac{1}{2} + z_2\right)\mathbf{a}_3$ | = | $\left(\frac{1}{2} - x_2\right)a\,\hat{\mathbf{x}} - y_2 b\,\hat{\mathbf{y}} + \left(\frac{1}{2} + z_2\right)c\,\hat{\mathbf{z}}$ | (4a) | C II |
| $\mathbf{B}_7$ | = | $-x_2\,\mathbf{a}_1 + \left(\frac{1}{2} + y_2\right)\mathbf{a}_2 + \left(\frac{1}{2} - z_2\right)\mathbf{a}_3$ | = | $-x_2 a\,\hat{\mathbf{x}} + \left(\frac{1}{2} + y_2\right)b\,\hat{\mathbf{y}} + \left(\frac{1}{2} - z_2\right)c\,\hat{\mathbf{z}}$ | (4a) | C II |
| $\mathbf{B}_{8}$ | = | $\left(\frac{1}{2}+x_{2}\right)\,\mathbf{a}_{1}+\left(\frac{1}{2}-y_{2}\right)\,\mathbf{a}_{2}-z_{2}\,\mathbf{a}_{3}$ | = | $\left(\frac{1}{2}+x_{2}\right)a\,\mathbf{\hat{x}}+\left(\frac{1}{2}-y_{2}\right)b\,\mathbf{\hat{y}}-z_{2}c\,\mathbf{\hat{z}}$ | (4a) | C II |
| $\mathbf{B}_{9}$ | = | $x_{3}\,\mathbf{a}_{1}+y_{3}\,\mathbf{a}_{2}+z_{3}\,\mathbf{a}_{3}$ | = | $x_{3}a\,\mathbf{\hat{x}}+y_{3}b\,\mathbf{\hat{y}}+z_{3}c\,\mathbf{\hat{z}}$ | (4a) | C III |
| $\mathbf{B}_{10}$ | = | $\left(\frac{1}{2}-x_{3}\right)\,\mathbf{a}_{1}-y_{3}\,\mathbf{a}_{2}+\left(\frac{1}{2}+z_{3}\right)\,\mathbf{a}_{3}$ | = | $\left(\frac{1}{2}-x_{3}\right)a\,\mathbf{\hat{x}}-y_{3}b\,\mathbf{\hat{y}}+\left(\frac{1}{2}+z_{3}\right)c\,\mathbf{\hat{z}}$ | (4a) | C III |
| $\mathbf{B}_{11}$ | = | $-x_{3}\,\mathbf{a}_{1}+\left(\frac{1}{2}+y_{3}\right)\,\mathbf{a}_{2}+\left(\frac{1}{2}-z_{3}\right)\,\mathbf{a}_{3}$ | = | $-x_{3}a\,\mathbf{\hat{x}}+\left(\frac{1}{2}+y_{3}\right)b\,\mathbf{\hat{y}}+\left(\frac{1}{2}-z_{3}\right)c\,\mathbf{\hat{z}}$ | (4a) | C III |
| $\mathbf{B}_{12}$ | = | $\left(\frac{1}{2}+x_{3}\right)\,\mathbf{a}_{1}+\left(\frac{1}{2}-y_{3}\right)\,\mathbf{a}_{2}-z_{3}\,\mathbf{a}_{3}$ | = | $\left(\frac{1}{2}+x_{3}\right)a\,\mathbf{\hat{x}}+\left(\frac{1}{2}-y_{3}\right)b\,\mathbf{\hat{y}}-z_{3}c\,\mathbf{\hat{z}}$ | (4a) | C III |
| $\mathbf{B}_{13}$ | = | $x_{4}\,\mathbf{a}_{1}+y_{4}\,\mathbf{a}_{2}+z_{4}\,\mathbf{a}_{3}$ | = | $x_{4}a\,\mathbf{\hat{x}}+y_{4}b\,\mathbf{\hat{y}}+z_{4}c\,\mathbf{\hat{z}}$ | (4a) | C IV |
| $\mathbf{B}_{14}$ | = | $\left(\frac{1}{2}-x_{4}\right)\,\mathbf{a}_{1}-y_{4}\,\mathbf{a}_{2}+\left(\frac{1}{2}+z_{4}\right)\,\mathbf{a}_{3}$ | = | $\left(\frac{1}{2}-x_{4}\right)a\,\mathbf{\hat{x}}-y_{4}b\,\mathbf{\hat{y}}+\left(\frac{1}{2}+z_{4}\right)c\,\mathbf{\hat{z}}$ | (4a) | C IV |
| $\mathbf{B}_{15}$ | = | $-x_{4}\,\mathbf{a}_{1}+\left(\frac{1}{2}+y_{4}\right)\,\mathbf{a}_{2}+\left(\frac{1}{2}-z_{4}\right)\,\mathbf{a}_{3}$ | = | $-x_{4}a\,\mathbf{\hat{x}}+\left(\frac{1}{2}+y_{4}\right)b\,\mathbf{\hat{y}}+\left(\frac{1}{2}-z_{4}\right)c\,\mathbf{\hat{z}}$ | (4a) | C IV |
| $\mathbf{B}_{16}$ | = | $\left(\frac{1}{2}+x_{4}\right)\,\mathbf{a}_{1}+\left(\frac{1}{2}-y_{4}\right)\,\mathbf{a}_{2}-z_{4}\,\mathbf{a}_{3}$ | = | $\left(\frac{1}{2}+x_{4}\right)a\,\mathbf{\hat{x}}+\left(\frac{1}{2}-y_{4}\right)b\,\mathbf{\hat{y}}-z_{4}c\,\mathbf{\hat{z}}$ | (4a) | C IV |
| $\mathbf{B}_{17}$ | = | $x_{5}\,\mathbf{a}_{1}+y_{5}\,\mathbf{a}_{2}+z_{5}\,\mathbf{a}_{3}$ | = | $x_{5}a\,\mathbf{\hat{x}}+y_{5}b\,\mathbf{\hat{y}}+z_{5}c\,\mathbf{\hat{z}}$ | (4a) | C V |
| $\mathbf{B}_{18}$ | = | $\left(\frac{1}{2}-x_{5}\right)\,\mathbf{a}_{1}-y_{5}\,\mathbf{a}_{2}+\left(\frac{1}{2}+z_{5}\right)\,\mathbf{a}_{3}$ | = | $\left(\frac{1}{2}-x_{5}\right)a\,\mathbf{\hat{x}}-y_{5}b\,\mathbf{\hat{y}}+\left(\frac{1}{2}+z_{5}\right)c\,\mathbf{\hat{z}}$ | (4a) | C V |
| $\mathbf{B}_{19}$ | = | $-x_{5}\,\mathbf{a}_{1}+\left(\frac{1}{2}+y_{5}\right)\,\mathbf{a}_{2}+\left(\frac{1}{2}-z_{5}\right)\,\mathbf{a}_{3}$ | = | $-x_{5}a\,\mathbf{\hat{x}}+\left(\frac{1}{2}+y_{5}\right)b\,\mathbf{\hat{y}}+\left(\frac{1}{2}-z_{5}\right)c\,\mathbf{\hat{z}}$ | (4a) | C V |
| $\mathbf{B}_{20}$ | = | $\left(\frac{1}{2}+x_{5}\right)\,\mathbf{a}_{1}+\left(\frac{1}{2}-y_{5}\right)\,\mathbf{a}_{2}-z_{5}\,\mathbf{a}_{3}$ | = | $\left(\frac{1}{2}+x_{5}\right)a\,\mathbf{\hat{x}}+\left(\frac{1}{2}-y_{5}\right)b\,\mathbf{\hat{y}}-z_{5}c\,\mathbf{\hat{z}}$ | (4a) | C V |
| $\mathbf{B}_{21}$ | = | $x_{6}\,\mathbf{a}_{1}+y_{6}\,\mathbf{a}_{2}+z_{6}\,\mathbf{a}_{3}$ | = | $x_{6}a\,\mathbf{\hat{x}}+y_{6}b\,\mathbf{\hat{y}}+z_{6}c\,\mathbf{\hat{z}}$ | (4a) | H I |
| $\mathbf{B}_{22}$ | = | $\left(\frac{1}{2}-x_{6}\right)\,\mathbf{a}_{1}-y_{6}\,\mathbf{a}_{2}+\left(\frac{1}{2}+z_{6}\right)\,\mathbf{a}_{3}$ | = | $\left(\frac{1}{2}-x_{6}\right)a\,\mathbf{\hat{x}}-y_{6}b\,\mathbf{\hat{y}}+\left(\frac{1}{2}+z_{6}\right)c\,\mathbf{\hat{z}}$ | (4a) | H I |
| $\mathbf{B}_{23}$ | = | $-x_{6}\,\mathbf{a}_{1}+\left(\frac{1}{2}+y_{6}\right)\,\mathbf{a}_{2}+\left(\frac{1}{2}-z_{6}\right)\,\mathbf{a}_{3}$ | = | $-x_{6}a\,\mathbf{\hat{x}}+\left(\frac{1}{2}+y_{6}\right)b\,\mathbf{\hat{y}}+\left(\frac{1}{2}-z_{6}\right)c\,\mathbf{\hat{z}}$ | (4a) | H I |
| $\mathbf{B}_{24}$ | = | $\left(\frac{1}{2}+x_{6}\right)\,\mathbf{a}_{1}+\left(\frac{1}{2}-y_{6}\right)\,\mathbf{a}_{2}-z_{6}\,\mathbf{a}_{3}$ | = | $\left(\frac{1}{2}+x_{6}\right)a\,\mathbf{\hat{x}}+\left(\frac{1}{2}-y_{6}\right)b\,\mathbf{\hat{y}}-z_{6}c\,\mathbf{\hat{z}}$ | (4a) | H I |
| $\mathbf{B}_{25}$ | = | $x_{7}\,\mathbf{a}_{1}+y_{7}\,\mathbf{a}_{2}+z_{7}\,\mathbf{a}_{3}$ | = | $x_{7}a\,\mathbf{\hat{x}}+y_{7}b\,\mathbf{\hat{y}}+z_{7}c\,\mathbf{\hat{z}}$ | (4a) | H II |
| $\mathbf{B}_{26}$ | = | $\left(\frac{1}{2}-x_{7}\right)\,\mathbf{a}_{1}-y_{7}\,\mathbf{a}_{2}+\left(\frac{1}{2}+z_{7}\right)\,\mathbf{a}_{3}$ | = | $\left(\frac{1}{2}-x_{7}\right)a\,\mathbf{\hat{x}}-y_{7}b\,\mathbf{\hat{y}}+\left(\frac{1}{2}+z_{7}\right)c\,\mathbf{\hat{z}}$ | (4a) | H II |
| $\mathbf{B}_{27}$ | = | $-x_{7}\,\mathbf{a}_{1}+\left(\frac{1}{2}+y_{7}\right)\,\mathbf{a}_{2}+\left(\frac{1}{2}-z_{7}\right)\,\mathbf{a}_{3}$ | = | $-x_{7}a\,\mathbf{\hat{x}}+\left(\frac{1}{2}+y_{7}\right)b\,\mathbf{\hat{y}}+\left(\frac{1}{2}-z_{7}\right)c\,\mathbf{\hat{z}}$ | (4a) | H II |
| $\mathbf{B}_{28}$ | = | $\left(\frac{1}{2}+x_{7}\right)\,\mathbf{a}_{1}+\left(\frac{1}{2}-y_{7}\right)\,\mathbf{a}_{2}-z_{7}\,\mathbf{a}_{3}$ | = | $\left(\frac{1}{2}+x_{7}\right)a\,\mathbf{\hat{x}}+\left(\frac{1}{2}-y_{7}\right)b\,\mathbf{\hat{y}}-z_{7}c\,\mathbf{\hat{z}}$ | (4a) | H II |
| $\mathbf{B}_{29}$ | = | $x_{8}\,\mathbf{a}_{1}+y_{8}\,\mathbf{a}_{2}+z_{8}\,\mathbf{a}_{3}$ | = | $x_{8}a\,\mathbf{\hat{x}}+y_{8}b\,\mathbf{\hat{y}}+z_{8}c\,\mathbf{\hat{z}}$ | (4a) | H III |
| $\mathbf{B}_{30}$ | = | $\left(\frac{1}{2}-x_{8}\right)\,\mathbf{a}_{1}-y_{8}\,\mathbf{a}_{2}+\left(\frac{1}{2}+z_{8}\right)\,\mathbf{a}_{3}$ | = | $\left(\frac{1}{2}-x_{8}\right)a\,\mathbf{\hat{x}}-y_{8}b\,\mathbf{\hat{y}}+\left(\frac{1}{2}+z_{8}\right)c\,\mathbf{\hat{z}}$ | (4a) | H III |
| $\mathbf{B}_{31}$ | = | $-x_{8}\,\mathbf{a}_{1}+\left(\frac{1}{2}+y_{8}\right)\,\mathbf{a}_{2}+\left(\frac{1}{2}-z_{8}\right)\,\mathbf{a}_{3}$ | = | $-x_{8}a\,\mathbf{\hat{x}}+\left(\frac{1}{2}+y_{8}\right)b\,\mathbf{\hat{y}}+\left(\frac{1}{2}-z_{8}\right)c\,\mathbf{\hat{z}}$ | (4a) | H III |
| $\mathbf{B}_{32}$ | = | $\left(\frac{1}{2}+x_{8}\right)\,\mathbf{a}_{1}+\left(\frac{1}{2}-y_{8}\right)\,\mathbf{a}_{2}-z_{8}\,\mathbf{a}_{3}$ | = | $\left(\frac{1}{2}+x_{8}\right)a\,\mathbf{\hat{x}}+\left(\frac{1}{2}-y_{8}\right)b\,\mathbf{\hat{y}}-z_{8}c\,\mathbf{\hat{z}}$ | (4a) | H III |
| $\mathbf{B}_{33}$ | = | $x_{9}\,\mathbf{a}_{1}+y_{9}\,\mathbf{a}_{2}+z_{9}\,\mathbf{a}_{3}$ | = | $x_{9}a\,\mathbf{\hat{x}}+y_{9}b\,\mathbf{\hat{y}}+z_{9}c\,\mathbf{\hat{z}}$ | (4a) | H IV |
| $\mathbf{B}_{34}$ | = | $\left(\frac{1}{2}-x_{9}\right)\,\mathbf{a}_{1}-y_{9}\,\mathbf{a}_{2}+\left(\frac{1}{2}+z_{9}\right)\,\mathbf{a}_{3}$ | = | $\left(\frac{1}{2}-x_{9}\right)a\,\mathbf{\hat{x}}-y_{9}b\,\mathbf{\hat{y}}+\left(\frac{1}{2}+z_{9}\right)c\,\mathbf{\hat{z}}$ | (4a) | H IV |
| $\mathbf{B}_{35}$ | = | $-x_{9}\,\mathbf{a}_{1}+\left(\frac{1}{2}+y_{9}\right)\,\mathbf{a}_{2}+\left(\frac{1}{2}-z_{9}\right)\,\mathbf{a}_{3}$ | = | $-x_{9}a\,\mathbf{\hat{x}}+\left(\frac{1}{2}+y_{9}\right)b\,\mathbf{\hat{y}}+\left(\frac{1}{2}-z_{9}\right)c\,\mathbf{\hat{z}}$ | (4a) | H IV |
| $\mathbf{B}_{36}$ | = | $\left(\frac{1}{2}+x_{9}\right)\,\mathbf{a}_{1}+\left(\frac{1}{2}-y_{9}\right)\,\mathbf{a}_{2}-z_{9}\,\mathbf{a}_{3}$ | = | $\left(\frac{1}{2}+x_{9}\right)a\,\mathbf{\hat{x}}+\left(\frac{1}{2}-y_{9}\right)b\,\mathbf{\hat{y}}-z_{9}c\,\mathbf{\hat{z}}$ | (4a) | H IV |
| $\mathbf{B}_{37}$ | = | $x_{10}\,\mathbf{a}_{1}+y_{10}\,\mathbf{a}_{2}+z_{10}\,\mathbf{a}_{3}$ | = | $x_{10}a\,\mathbf{\hat{x}}+y_{10}b\,\mathbf{\hat{y}}+z_{10}c\,\mathbf{\hat{z}}$ | (4a) | H V |
| $\mathbf{B}_{38}$ | = | $\left(\frac{1}{2}-x_{10}\right)\,\mathbf{a}_{1}-y_{10}\,\mathbf{a}_{2}+\left(\frac{1}{2}+z_{10}\right)\,\mathbf{a}_{3}$ | = | $\left(\frac{1}{2}-x_{10}\right)a\,\mathbf{\hat{x}}-y_{10}b\,\mathbf{\hat{y}}+\left(\frac{1}{2}+z_{10}\right)c\,\mathbf{\hat{z}}$ | (4a) | H V |
| $\mathbf{B}_{39}$ | = | $-x_{10}\,\mathbf{a}_{1}+\left(\frac{1}{2}+y_{10}\right)\,\mathbf{a}_{2}+\left(\frac{1}{2}-z_{10}\right)\,\mathbf{a}_{3}$ | = | $-x_{10}a\,\mathbf{\hat{x}}+\left(\frac{1}{2}+y_{10}\right)b\,\mathbf{\hat{y}}+\left(\frac{1}{2}-z_{10}\right)c\,\mathbf{\hat{z}}$ | (4a) | H V |
| $\mathbf{B}_{40}$ | = | $\left(\frac{1}{2}+x_{10}\right)\,\mathbf{a}_{1}+\left(\frac{1}{2}-y_{10}\right)\,\mathbf{a}_{2}-z_{10}\,\mathbf{a}_{3}$ | = | $\left(\frac{1}{2}+x_{10}\right)a\,\mathbf{\hat{x}}+\left(\frac{1}{2}-y_{10}\right)b\,\mathbf{\hat{y}}-z_{10}c\,\mathbf{\hat{z}}$ | (4a) | H V |
| $\mathbf{B}_{41}$ | = | $x_{11}\,\mathbf{a}_{1}+y_{11}\,\mathbf{a}_{2}+z_{11}\,\mathbf{a}_{3}$ | = | $x_{11}a\,\mathbf{\hat{x}}+y_{11}b\,\mathbf{\hat{y}}+z_{11}c\,\mathbf{\hat{z}}$ | (4a) | H VI |
| $\mathbf{B}_{42}$ | = | $\left(\frac{1}{2}-x_{11}\right)\,\mathbf{a}_{1}-y_{11}\,\mathbf{a}_{2}+\left(\frac{1}{2}+z_{11}\right)\,\mathbf{a}_{3}$ | = | $\left(\frac{1}{2}-x_{11}\right)a\,\mathbf{\hat{x}}-y_{11}b\,\mathbf{\hat{y}}+\left(\frac{1}{2}+z_{11}\right)c\,\mathbf{\hat{z}}$ | (4a) | H VI |
| $\mathbf{B}_{43}$ | = | $-x_{11}\,\mathbf{a}_1+\left(\frac{1}{2}+y_{11}\right)\,\mathbf{a}_2+\left(\frac{1}{2}-z_{11}\right)\,\mathbf{a}_3$ | = | $-x_{11}a\,\hat{\mathbf{x}}+\left(\frac{1}{2}+y_{11}\right)b\,\hat{\mathbf{y}}+\left(\frac{1}{2}-z_{11}\right)c\,\hat{\mathbf{z}}$ | (4$a$) | H VI |
| $\mathbf{B}_{44}$ | = | $\left(\frac{1}{2}+x_{11}\right)\,\mathbf{a}_1+\left(\frac{1}{2}-y_{11}\right)\,\mathbf{a}_2-z_{11}\,\mathbf{a}_3$ | = | $\left(\frac{1}{2}+x_{11}\right)a\,\hat{\mathbf{x}}+\left(\frac{1}{2}-y_{11}\right)b\,\hat{\mathbf{y}}-z_{11}c\,\hat{\mathbf{z}}$ | (4$a$) | H VI |
| $\mathbf{B}_{45}$ | = | $x_{12}\,\mathbf{a}_1+y_{12}\,\mathbf{a}_2+z_{12}\,\mathbf{a}_3$ | = | $x_{12}a\,\hat{\mathbf{x}}+y_{12}b\,\hat{\mathbf{y}}+z_{12}c\,\hat{\mathbf{z}}$ | (4$a$) | H VII |
| $\mathbf{B}_{46}$ | = | $\left(\frac{1}{2}-x_{12}\right)\,\mathbf{a}_1-y_{12}\,\mathbf{a}_2+\left(\frac{1}{2}+z_{12}\right)\,\mathbf{a}_3$ | = | $\left(\frac{1}{2}-x_{12}\right)a\,\hat{\mathbf{x}}-y_{12}b\,\hat{\mathbf{y}}+\left(\frac{1}{2}+z_{12}\right)c\,\hat{\mathbf{z}}$ | (4$a$) | H VII |
| $\mathbf{B}_{47}$ | = | $-x_{12}\,\mathbf{a}_1+\left(\frac{1}{2}+y_{12}\right)\,\mathbf{a}_2+\left(\frac{1}{2}-z_{12}\right)\,\mathbf{a}_3$ | = | $-x_{12}a\,\hat{\mathbf{x}}+\left(\frac{1}{2}+y_{12}\right)b\,\hat{\mathbf{y}}+\left(\frac{1}{2}-z_{12}\right)c\,\hat{\mathbf{z}}$ | (4$a$) | H VII |
| $\mathbf{B}_{48}$ | = | $\left(\frac{1}{2}+x_{12}\right)\,\mathbf{a}_1+\left(\frac{1}{2}-y_{12}\right)\,\mathbf{a}_2-z_{12}\,\mathbf{a}_3$ | = | $\left(\frac{1}{2}+x_{12}\right)a\,\hat{\mathbf{x}}+\left(\frac{1}{2}-y_{12}\right)b\,\hat{\mathbf{y}}-z_{12}c\,\hat{\mathbf{z}}$ | (4$a$) | H VII |
| $\mathbf{B}_{49}$ | = | $x_{13}\,\mathbf{a}_1+y_{13}\,\mathbf{a}_2+z_{13}\,\mathbf{a}_3$ | = | $x_{13}a\,\hat{\mathbf{x}}+y_{13}b\,\hat{\mathbf{y}}+z_{13}c\,\hat{\mathbf{z}}$ | (4$a$) | H VIII |
| $\mathbf{B}_{50}$ | = | $\left(\frac{1}{2}-x_{13}\right)\,\mathbf{a}_1-y_{13}\,\mathbf{a}_2+\left(\frac{1}{2}+z_{13}\right)\,\mathbf{a}_3$ | = | $\left(\frac{1}{2}-x_{13}\right)a\,\hat{\mathbf{x}}-y_{13}b\,\hat{\mathbf{y}}+\left(\frac{1}{2}+z_{13}\right)c\,\hat{\mathbf{z}}$ | (4$a$) | H VIII |
| $\mathbf{B}_{51}$ | = | $-x_{13}\,\mathbf{a}_1+\left(\frac{1}{2}+y_{13}\right)\,\mathbf{a}_2+\left(\frac{1}{2}-z_{13}\right)\,\mathbf{a}_3$ | = | $-x_{13}a\,\hat{\mathbf{x}}+\left(\frac{1}{2}+y_{13}\right)b\,\hat{\mathbf{y}}+\left(\frac{1}{2}-z_{13}\right)c\,\hat{\mathbf{z}}$ | (4$a$) | H VIII |
| $\mathbf{B}_{52}$ | = | $\left(\frac{1}{2}+x_{13}\right)\,\mathbf{a}_1+\left(\frac{1}{2}-y_{13}\right)\,\mathbf{a}_2-z_{13}\,\mathbf{a}_3$ | = | $\left(\frac{1}{2}+x_{13}\right)a\,\hat{\mathbf{x}}+\left(\frac{1}{2}-y_{13}\right)b\,\hat{\mathbf{y}}-z_{13}c\,\hat{\mathbf{z}}$ | (4$a$) | H VIII |
| $\mathbf{B}_{53}$ | = | $x_{14}\,\mathbf{a}_1+y_{14}\,\mathbf{a}_2+z_{14}\,\mathbf{a}_3$ | = | $x_{14}a\,\hat{\mathbf{x}}+y_{14}b\,\hat{\mathbf{y}}+z_{14}c\,\hat{\mathbf{z}}$ | (4$a$) | H IX |
| $\mathbf{B}_{54}$ | = | $\left(\frac{1}{2}-x_{14}\right)\,\mathbf{a}_1-y_{14}\,\mathbf{a}_2+\left(\frac{1}{2}+z_{14}\right)\,\mathbf{a}_3$ | = | $\left(\frac{1}{2}-x_{14}\right)a\,\hat{\mathbf{x}}-y_{14}b\,\hat{\mathbf{y}}+\left(\frac{1}{2}+z_{14}\right)c\,\hat{\mathbf{z}}$ | (4$a$) | H IX |
| $\mathbf{B}_{55}$ | = | $-x_{14}\,\mathbf{a}_1+\left(\frac{1}{2}+y_{14}\right)\,\mathbf{a}_2+\left(\frac{1}{2}-z_{14}\right)\,\mathbf{a}_3$ | = | $-x_{14}a\,\hat{\mathbf{x}}+\left(\frac{1}{2}+y_{14}\right)b\,\hat{\mathbf{y}}+\left(\frac{1}{2}-z_{14}\right)c\,\hat{\mathbf{z}}$ | (4$a$) | H IX |
| $\mathbf{B}_{56}$ | = | $\left(\frac{1}{2}+x_{14}\right)\,\mathbf{a}_1+\left(\frac{1}{2}-y_{14}\right)\,\mathbf{a}_2-z_{14}\,\mathbf{a}_3$ | = | $\left(\frac{1}{2}+x_{14}\right)a\,\hat{\mathbf{x}}+\left(\frac{1}{2}-y_{14}\right)b\,\hat{\mathbf{y}}-z_{14}c\,\hat{\mathbf{z}}$ | (4$a$) | H IX |
| $\mathbf{B}_{57}$ | = | $x_{15}\,\mathbf{a}_1+y_{15}\,\mathbf{a}_2+z_{15}\,\mathbf{a}_3$ | = | $x_{15}a\,\hat{\mathbf{x}}+y_{15}b\,\hat{\mathbf{y}}+z_{15}c\,\hat{\mathbf{z}}$ | (4$a$) | H X |
| $\mathbf{B}_{58}$ | = | $\left(\frac{1}{2}-x_{15}\right)\,\mathbf{a}_1-y_{15}\,\mathbf{a}_2+\left(\frac{1}{2}+z_{15}\right)\,\mathbf{a}_3$ | = | $\left(\frac{1}{2}-x_{15}\right)a\,\hat{\mathbf{x}}-y_{15}b\,\hat{\mathbf{y}}+\left(\frac{1}{2}+z_{15}\right)c\,\hat{\mathbf{z}}$ | (4$a$) | H X |
| $\mathbf{B}_{59}$ | = | $-x_{15}\,\mathbf{a}_1+\left(\frac{1}{2}+y_{15}\right)\,\mathbf{a}_2+\left(\frac{1}{2}-z_{15}\right)\,\mathbf{a}_3$ | = | $-x_{15}a\,\hat{\mathbf{x}}+\left(\frac{1}{2}+y_{15}\right)b\,\hat{\mathbf{y}}+\left(\frac{1}{2}-z_{15}\right)c\,\hat{\mathbf{z}}$ | (4$a$) | H X |
| $\mathbf{B}_{60}$ | = | $\left(\frac{1}{2}+x_{15}\right)\,\mathbf{a}_1+\left(\frac{1}{2}-y_{15}\right)\,\mathbf{a}_2-z_{15}\,\mathbf{a}_3$ | = | $\left(\frac{1}{2}+x_{15}\right)a\,\hat{\mathbf{x}}+\left(\frac{1}{2}-y_{15}\right)b\,\hat{\mathbf{y}}-z_{15}c\,\hat{\mathbf{z}}$ | (4$a$) | H X |
| $\mathbf{B}_{61}$ | = | $x_{16}\,\mathbf{a}_1+y_{16}\,\mathbf{a}_2+z_{16}\,\mathbf{a}_3$ | = | $x_{16}a\,\hat{\mathbf{x}}+y_{16}b\,\hat{\mathbf{y}}+z_{16}c\,\hat{\mathbf{z}}$ | (4$a$) | O I |
| $\mathbf{B}_{62}$ | = | $\left(\frac{1}{2}-x_{16}\right)\,\mathbf{a}_1-y_{16}\,\mathbf{a}_2+\left(\frac{1}{2}+z_{16}\right)\,\mathbf{a}_3$ | = | $\left(\frac{1}{2}-x_{16}\right)a\,\hat{\mathbf{x}}-y_{16}b\,\hat{\mathbf{y}}+\left(\frac{1}{2}+z_{16}\right)c\,\hat{\mathbf{z}}$ | (4$a$) | O I |
| $\mathbf{B}_{63}$ | = | $-x_{16}\,\mathbf{a}_1+\left(\frac{1}{2}+y_{16}\right)\,\mathbf{a}_2+\left(\frac{1}{2}-z_{16}\right)\,\mathbf{a}_3$ | = | $-x_{16}a\,\hat{\mathbf{x}}+\left(\frac{1}{2}+y_{16}\right)b\,\hat{\mathbf{y}}+\left(\frac{1}{2}-z_{16}\right)c\,\hat{\mathbf{z}}$ | (4$a$) | O I |
| $\mathbf{B}_{64}$ | = | $\left(\frac{1}{2}+x_{16}\right)\,\mathbf{a}_1+\left(\frac{1}{2}-y_{16}\right)\,\mathbf{a}_2-z_{16}\,\mathbf{a}_3$ | = | $\left(\frac{1}{2}+x_{16}\right)a\,\hat{\mathbf{x}}+\left(\frac{1}{2}-y_{16}\right)b\,\hat{\mathbf{y}}-z_{16}c\,\hat{\mathbf{z}}$ | (4$a$) | O I |
| $\mathbf{B}_{65}$ | = | $x_{17}\,\mathbf{a}_1+y_{17}\,\mathbf{a}_2+z_{17}\,\mathbf{a}_3$ | = | $x_{17}a\,\hat{\mathbf{x}}+y_{17}b\,\hat{\mathbf{y}}+z_{17}c\,\hat{\mathbf{z}}$ | (4$a$) | O II |
| $\mathbf{B}_{66}$ | = | $\left(\frac{1}{2}-x_{17}\right)\,\mathbf{a}_1-y_{17}\,\mathbf{a}_2+\left(\frac{1}{2}+z_{17}\right)\,\mathbf{a}_3$ | = | $\left(\frac{1}{2}-x_{17}\right)a\,\hat{\mathbf{x}}-y_{17}b\,\hat{\mathbf{y}}+\left(\frac{1}{2}+z_{17}\right)c\,\hat{\mathbf{z}}$ | (4$a$) | O II |
| $\mathbf{B}_{67}$ | = | $-x_{17}\,\mathbf{a}_1+\left(\frac{1}{2}+y_{17}\right)\,\mathbf{a}_2+\left(\frac{1}{2}-z_{17}\right)\,\mathbf{a}_3$ | = | $-x_{17}a\,\hat{\mathbf{x}}+\left(\frac{1}{2}+y_{17}\right)b\,\hat{\mathbf{y}}+\left(\frac{1}{2}-z_{17}\right)c\,\hat{\mathbf{z}}$ | (4$a$) | O II |
| $\mathbf{B}_{68}$ | = | $\left(\frac{1}{2}+x_{17}\right)\,\mathbf{a}_1+\left(\frac{1}{2}-y_{17}\right)\,\mathbf{a}_2-z_{17}\,\mathbf{a}_3$ | = | $\left(\frac{1}{2}+x_{17}\right)a\,\hat{\mathbf{x}}+\left(\frac{1}{2}-y_{17}\right)b\,\hat{\mathbf{y}}-z_{17}c\,\hat{\mathbf{z}}$ | (4$a$) | O II |
| $\mathbf{B}_{69}$ | = | $x_{18}\,\mathbf{a}_1+y_{18}\,\mathbf{a}_2+z_{18}\,\mathbf{a}_3$ | = | $x_{18}a\,\hat{\mathbf{x}}+y_{18}b\,\hat{\mathbf{y}}+z_{18}c\,\hat{\mathbf{z}}$ | (4$a$) | O III |
| $\mathbf{B}_{70}$ | = | $\left(\frac{1}{2}-x_{18}\right)\,\mathbf{a}_1-y_{18}\,\mathbf{a}_2+\left(\frac{1}{2}+z_{18}\right)\,\mathbf{a}_3$ | = | $\left(\frac{1}{2}-x_{18}\right)a\,\hat{\mathbf{x}}-y_{18}b\,\hat{\mathbf{y}}+\left(\frac{1}{2}+z_{18}\right)c\,\hat{\mathbf{z}}$ | (4$a$) | O III |
| $\mathbf{B}_{71}$ | = | $-x_{18}\,\mathbf{a}_1+\left(\frac{1}{2}+y_{18}\right)\,\mathbf{a}_2+\left(\frac{1}{2}-z_{18}\right)\,\mathbf{a}_3$ | = | $-x_{18}a\,\hat{\mathbf{x}}+\left(\frac{1}{2}+y_{18}\right)b\,\hat{\mathbf{y}}+\left(\frac{1}{2}-z_{18}\right)c\,\hat{\mathbf{z}}$ | (4$a$) | O III |
| $\mathbf{B}_{72}$ | = | $\left(\frac{1}{2}+x_{18}\right)\,\mathbf{a}_1+\left(\frac{1}{2}-y_{18}\right)\,\mathbf{a}_2-z_{18}\,\mathbf{a}_3$ | = | $\left(\frac{1}{2}+x_{18}\right)a\,\hat{\mathbf{x}}+\left(\frac{1}{2}-y_{18}\right)b\,\hat{\mathbf{y}}-z_{18}c\,\hat{\mathbf{z}}$ | (4$a$) | O III |
| $\mathbf{B}_{73}$ | = | $x_{19}\,\mathbf{a}_1 + y_{19}\,\mathbf{a}_2 + z_{19}\,\mathbf{a}_3$ | = | $x_{19}a\,\hat{\mathbf{x}} + y_{19}b\,\hat{\mathbf{y}} + z_{19}c\,\hat{\mathbf{z}}$ | (4$a$) | O IV |
| $\mathbf{B}_{74}$ | = | $\left(\frac{1}{2} - x_{19}\right)\mathbf{a}_1 - y_{19}\,\mathbf{a}_2 + \left(\frac{1}{2} + z_{19}\right)\mathbf{a}_3$ | = | $\left(\frac{1}{2} - x_{19}\right)a\,\hat{\mathbf{x}} - y_{19}b\,\hat{\mathbf{y}} + \left(\frac{1}{2} + z_{19}\right)c\,\hat{\mathbf{z}}$ | (4$a$) | O IV |
| $\mathbf{B}_{75}$ | = | $-x_{19}\,\mathbf{a}_1 + \left(\frac{1}{2} + y_{19}\right)\mathbf{a}_2 + \left(\frac{1}{2} - z_{19}\right)\mathbf{a}_3$ | = | $-x_{19}a\,\hat{\mathbf{x}} + \left(\frac{1}{2} + y_{19}\right)b\,\hat{\mathbf{y}} + \left(\frac{1}{2} - z_{19}\right)c\,\hat{\mathbf{z}}$ | (4$a$) | O IV |
| $\mathbf{B}_{76}$ | = | $\left(\frac{1}{2} + x_{19}\right)\mathbf{a}_1 + \left(\frac{1}{2} - y_{19}\right)\mathbf{a}_2 - z_{19}\,\mathbf{a}_3$ | = | $\left(\frac{1}{2} + x_{19}\right)a\,\hat{\mathbf{x}} + \left(\frac{1}{2} - y_{19}\right)b\,\hat{\mathbf{y}} - z_{19}c\,\hat{\mathbf{z}}$ | (4$a$) | O IV |
| $\mathbf{B}_{77}$ | = | $x_{20}\,\mathbf{a}_1 + y_{20}\,\mathbf{a}_2 + z_{20}\,\mathbf{a}_3$ | = | $x_{20}a\,\hat{\mathbf{x}} + y_{20}b\,\hat{\mathbf{y}} + z_{20}c\,\hat{\mathbf{z}}$ | (4$a$) | O V |
| $\mathbf{B}_{78}$ | = | $\left(\frac{1}{2} - x_{20}\right)\mathbf{a}_1 - y_{20}\,\mathbf{a}_2 + \left(\frac{1}{2} + z_{20}\right)\mathbf{a}_3$ | = | $\left(\frac{1}{2} - x_{20}\right)a\,\hat{\mathbf{x}} - y_{20}b\,\hat{\mathbf{y}} + \left(\frac{1}{2} + z_{20}\right)c\,\hat{\mathbf{z}}$ | (4$a$) | O V |
| $\mathbf{B}_{79}$ | = | $-x_{20}\,\mathbf{a}_1 + \left(\frac{1}{2} + y_{20}\right)\mathbf{a}_2 + \left(\frac{1}{2} - z_{20}\right)\mathbf{a}_3$ | = | $-x_{20}a\,\hat{\mathbf{x}} + \left(\frac{1}{2} + y_{20}\right)b\,\hat{\mathbf{y}} + \left(\frac{1}{2} - z_{20}\right)c\,\hat{\mathbf{z}}$ | (4$a$) | O V |
| $\mathbf{B}_{80}$ | = | $\left(\frac{1}{2} + x_{20}\right)\mathbf{a}_1 + \left(\frac{1}{2} - y_{20}\right)\mathbf{a}_2 - z_{20}\,\mathbf{a}_3$ | = | $\left(\frac{1}{2} + x_{20}\right)a\,\hat{\mathbf{x}} + \left(\frac{1}{2} - y_{20}\right)b\,\hat{\mathbf{y}} - z_{20}c\,\hat{\mathbf{z}}$ | (4$a$) | O V |

**Geometry files:**

- CIF: pp. 1578
- POSCAR: pp. 1579

# $NaAlCl_4$ Structure: AB4C_oP24_19_a_4a_a

http://aflow.org/prototype-encyclopedia/AB4C_oP24_19_a_4a_a

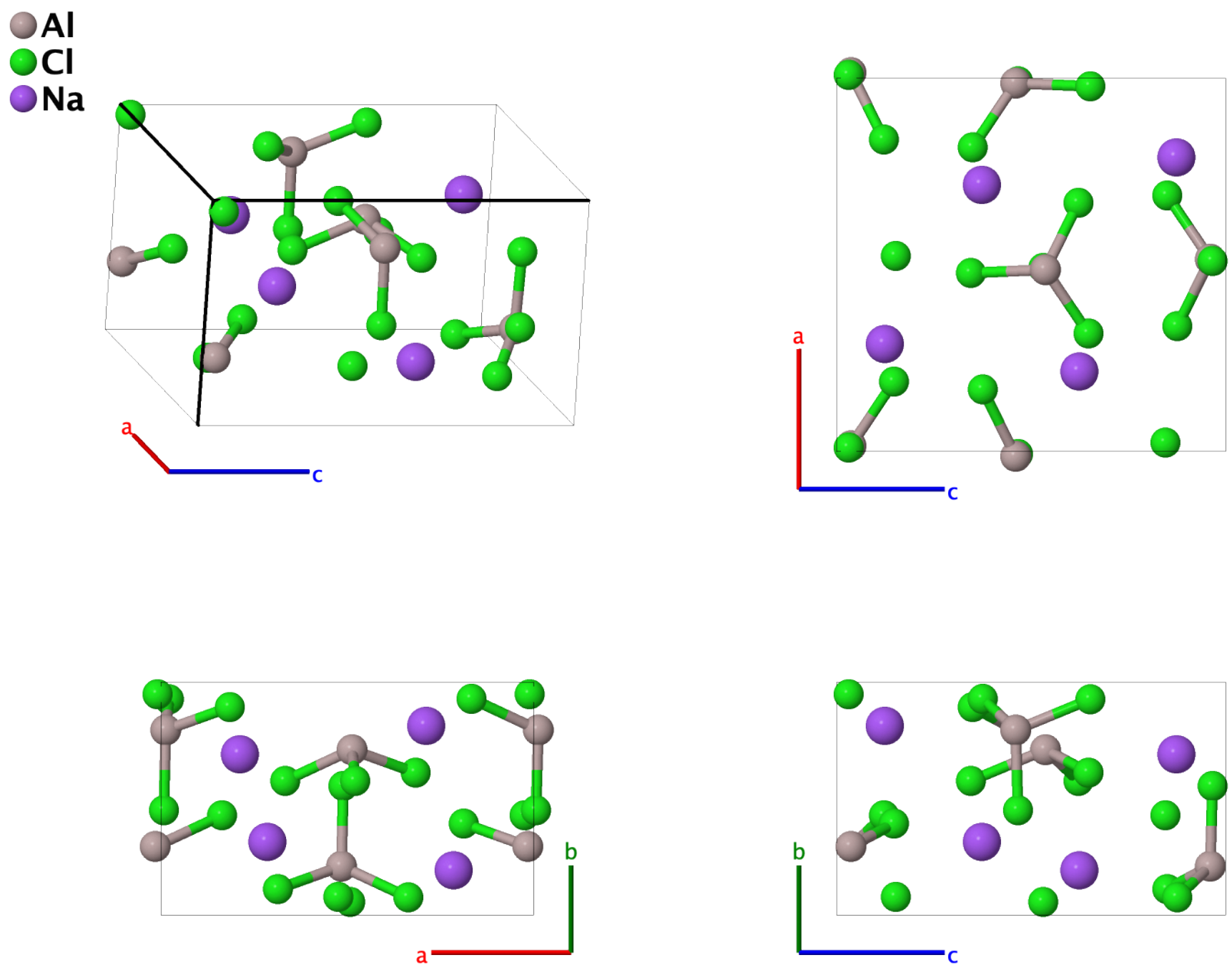

| | | |
|---|---|---|
| **Prototype** | : | $AlCl_4Na$ |
| **AFLOW prototype label** | : | AB4C_oP24_19_a_4a_a |
| ***Strukturbericht* designation** | : | None |
| **Pearson symbol** | : | oP24 |
| **Space group number** | : | 19 |
| **Space group symbol** | : | $P2_12_12_1$ |
| **AFLOW prototype command** | : | `aflow --proto=AB4C_oP24_19_a_4a_a`<br>`--params=`$a, b/a, c/a, x_1, y_1, z_1, x_2, y_2, z_2, x_3, y_3, z_3, x_4, y_4, z_4, x_5, y_5, z_5, x_6, y_6, z_6$ |

**Simple Orthorhombic primitive vectors:**

$$\mathbf{a}_1 = a\,\hat{\mathbf{x}}$$
$$\mathbf{a}_2 = b\,\hat{\mathbf{y}}$$
$$\mathbf{a}_3 = c\,\hat{\mathbf{z}}$$

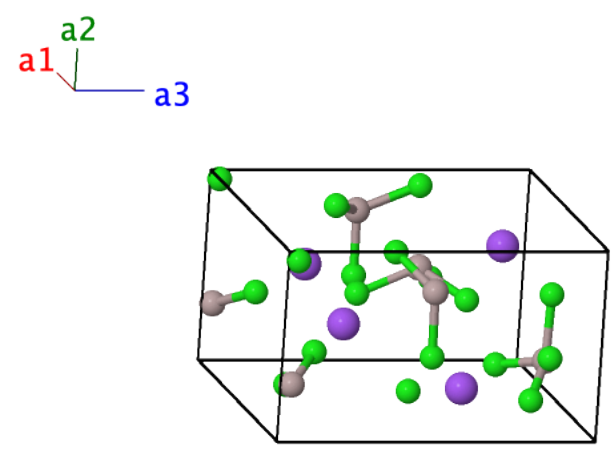

**Basis vectors:**

| | | Lattice Coordinates | | Cartesian Coordinates | Wyckoff Position | Atom Type |
|---|---|---|---|---|---|---|
| $\mathbf{B}_1$ | = | $x_1\,\mathbf{a}_1 + y_1\,\mathbf{a}_2 + z_1\,\mathbf{a}_3$ | = | $x_1\,a\,\hat{\mathbf{x}} + y_1\,b\,\hat{\mathbf{y}} + z_1\,c\,\hat{\mathbf{z}}$ | (4a) | Al |
| $\mathbf{B}_2$ | = | $\left(\frac{1}{2} - x_1\right)\mathbf{a}_1 - y_1\,\mathbf{a}_2 + \left(\frac{1}{2} + z_1\right)\mathbf{a}_3$ | = | $\left(\frac{1}{2} - x_1\right)a\,\hat{\mathbf{x}} - y_1\,b\,\hat{\mathbf{y}} + \left(\frac{1}{2} + z_1\right)c\,\hat{\mathbf{z}}$ | (4a) | Al |
| $\mathbf{B}_3$ | = | $-x_1\,\mathbf{a}_1 + \left(\frac{1}{2}+y_1\right)\mathbf{a}_2 + \left(\frac{1}{2}-z_1\right)\mathbf{a}_3$ | = | $-x_1 a\,\hat{\mathbf{x}} + \left(\frac{1}{2}+y_1\right) b\,\hat{\mathbf{y}} + \left(\frac{1}{2}-z_1\right) c\,\hat{\mathbf{z}}$ | (4$a$) | Al |
| $\mathbf{B}_4$ | = | $\left(\frac{1}{2}+x_1\right)\mathbf{a}_1 + \left(\frac{1}{2}-y_1\right)\mathbf{a}_2 - z_1\,\mathbf{a}_3$ | = | $\left(\frac{1}{2}+x_1\right) a\,\hat{\mathbf{x}} + \left(\frac{1}{2}-y_1\right) b\,\hat{\mathbf{y}} - z_1 c\,\hat{\mathbf{z}}$ | (4$a$) | Al |
| $\mathbf{B}_5$ | = | $x_2\,\mathbf{a}_1 + y_2\,\mathbf{a}_2 + z_2\,\mathbf{a}_3$ | = | $x_2 a\,\hat{\mathbf{x}} + y_2 b\,\hat{\mathbf{y}} + z_2 c\,\hat{\mathbf{z}}$ | (4$a$) | Cl I |
| $\mathbf{B}_6$ | = | $\left(\frac{1}{2}-x_2\right)\mathbf{a}_1 - y_2\,\mathbf{a}_2 + \left(\frac{1}{2}+z_2\right)\mathbf{a}_3$ | = | $\left(\frac{1}{2}-x_2\right) a\,\hat{\mathbf{x}} - y_2 b\,\hat{\mathbf{y}} + \left(\frac{1}{2}+z_2\right) c\,\hat{\mathbf{z}}$ | (4$a$) | Cl I |
| $\mathbf{B}_7$ | = | $-x_2\,\mathbf{a}_1 + \left(\frac{1}{2}+y_2\right)\mathbf{a}_2 + \left(\frac{1}{2}-z_2\right)\mathbf{a}_3$ | = | $-x_2 a\,\hat{\mathbf{x}} + \left(\frac{1}{2}+y_2\right) b\,\hat{\mathbf{y}} + \left(\frac{1}{2}-z_2\right) c\,\hat{\mathbf{z}}$ | (4$a$) | Cl I |
| $\mathbf{B}_8$ | = | $\left(\frac{1}{2}+x_2\right)\mathbf{a}_1 + \left(\frac{1}{2}-y_2\right)\mathbf{a}_2 - z_2\,\mathbf{a}_3$ | = | $\left(\frac{1}{2}+x_2\right) a\,\hat{\mathbf{x}} + \left(\frac{1}{2}-y_2\right) b\,\hat{\mathbf{y}} - z_2 c\,\hat{\mathbf{z}}$ | (4$a$) | Cl I |
| $\mathbf{B}_9$ | = | $x_3\,\mathbf{a}_1 + y_3\,\mathbf{a}_2 + z_3\,\mathbf{a}_3$ | = | $x_3 a\,\hat{\mathbf{x}} + y_3 b\,\hat{\mathbf{y}} + z_3 c\,\hat{\mathbf{z}}$ | (4$a$) | Cl II |
| $\mathbf{B}_{10}$ | = | $\left(\frac{1}{2}-x_3\right)\mathbf{a}_1 - y_3\,\mathbf{a}_2 + \left(\frac{1}{2}+z_3\right)\mathbf{a}_3$ | = | $\left(\frac{1}{2}-x_3\right) a\,\hat{\mathbf{x}} - y_3 b\,\hat{\mathbf{y}} + \left(\frac{1}{2}+z_3\right) c\,\hat{\mathbf{z}}$ | (4$a$) | Cl II |
| $\mathbf{B}_{11}$ | = | $-x_3\,\mathbf{a}_1 + \left(\frac{1}{2}+y_3\right)\mathbf{a}_2 + \left(\frac{1}{2}-z_3\right)\mathbf{a}_3$ | = | $-x_3 a\,\hat{\mathbf{x}} + \left(\frac{1}{2}+y_3\right) b\,\hat{\mathbf{y}} + \left(\frac{1}{2}-z_3\right) c\,\hat{\mathbf{z}}$ | (4$a$) | Cl II |
| $\mathbf{B}_{12}$ | = | $\left(\frac{1}{2}+x_3\right)\mathbf{a}_1 + \left(\frac{1}{2}-y_3\right)\mathbf{a}_2 - z_3\,\mathbf{a}_3$ | = | $\left(\frac{1}{2}+x_3\right) a\,\hat{\mathbf{x}} + \left(\frac{1}{2}-y_3\right) b\,\hat{\mathbf{y}} - z_3 c\,\hat{\mathbf{z}}$ | (4$a$) | Cl II |
| $\mathbf{B}_{13}$ | = | $x_4\,\mathbf{a}_1 + y_4\,\mathbf{a}_2 + z_4\,\mathbf{a}_3$ | = | $x_4 a\,\hat{\mathbf{x}} + y_4 b\,\hat{\mathbf{y}} + z_4 c\,\hat{\mathbf{z}}$ | (4$a$) | Cl III |
| $\mathbf{B}_{14}$ | = | $\left(\frac{1}{2}-x_4\right)\mathbf{a}_1 - y_4\,\mathbf{a}_2 + \left(\frac{1}{2}+z_4\right)\mathbf{a}_3$ | = | $\left(\frac{1}{2}-x_4\right) a\,\hat{\mathbf{x}} - y_4 b\,\hat{\mathbf{y}} + \left(\frac{1}{2}+z_4\right) c\,\hat{\mathbf{z}}$ | (4$a$) | Cl III |
| $\mathbf{B}_{15}$ | = | $-x_4\,\mathbf{a}_1 + \left(\frac{1}{2}+y_4\right)\mathbf{a}_2 + \left(\frac{1}{2}-z_4\right)\mathbf{a}_3$ | = | $-x_4 a\,\hat{\mathbf{x}} + \left(\frac{1}{2}+y_4\right) b\,\hat{\mathbf{y}} + \left(\frac{1}{2}-z_4\right) c\,\hat{\mathbf{z}}$ | (4$a$) | Cl III |
| $\mathbf{B}_{16}$ | = | $\left(\frac{1}{2}+x_4\right)\mathbf{a}_1 + \left(\frac{1}{2}-y_4\right)\mathbf{a}_2 - z_4\,\mathbf{a}_3$ | = | $\left(\frac{1}{2}+x_4\right) a\,\hat{\mathbf{x}} + \left(\frac{1}{2}-y_4\right) b\,\hat{\mathbf{y}} - z_4 c\,\hat{\mathbf{z}}$ | (4$a$) | Cl III |
| $\mathbf{B}_{17}$ | = | $x_5\,\mathbf{a}_1 + y_5\,\mathbf{a}_2 + z_5\,\mathbf{a}_3$ | = | $x_5 a\,\hat{\mathbf{x}} + y_5 b\,\hat{\mathbf{y}} + z_5 c\,\hat{\mathbf{z}}$ | (4$a$) | Cl IV |
| $\mathbf{B}_{18}$ | = | $\left(\frac{1}{2}-x_5\right)\mathbf{a}_1 - y_5\,\mathbf{a}_2 + \left(\frac{1}{2}+z_5\right)\mathbf{a}_3$ | = | $\left(\frac{1}{2}-x_5\right) a\,\hat{\mathbf{x}} - y_5 b\,\hat{\mathbf{y}} + \left(\frac{1}{2}+z_5\right) c\,\hat{\mathbf{z}}$ | (4$a$) | Cl IV |
| $\mathbf{B}_{19}$ | = | $-x_5\,\mathbf{a}_1 + \left(\frac{1}{2}+y_5\right)\mathbf{a}_2 + \left(\frac{1}{2}-z_5\right)\mathbf{a}_3$ | = | $-x_5 a\,\hat{\mathbf{x}} + \left(\frac{1}{2}+y_5\right) b\,\hat{\mathbf{y}} + \left(\frac{1}{2}-z_5\right) c\,\hat{\mathbf{z}}$ | (4$a$) | Cl IV |
| $\mathbf{B}_{20}$ | = | $\left(\frac{1}{2}+x_5\right)\mathbf{a}_1 + \left(\frac{1}{2}-y_5\right)\mathbf{a}_2 - z_5\,\mathbf{a}_3$ | = | $\left(\frac{1}{2}+x_5\right) a\,\hat{\mathbf{x}} + \left(\frac{1}{2}-y_5\right) b\,\hat{\mathbf{y}} - z_5 c\,\hat{\mathbf{z}}$ | (4$a$) | Cl IV |
| $\mathbf{B}_{21}$ | = | $x_6\,\mathbf{a}_1 + y_6\,\mathbf{a}_2 + z_6\,\mathbf{a}_3$ | = | $x_6 a\,\hat{\mathbf{x}} + y_6 b\,\hat{\mathbf{y}} + z_6 c\,\hat{\mathbf{z}}$ | (4$a$) | Na |
| $\mathbf{B}_{22}$ | = | $\left(\frac{1}{2}-x_6\right)\mathbf{a}_1 - y_6\,\mathbf{a}_2 + \left(\frac{1}{2}+z_6\right)\mathbf{a}_3$ | = | $\left(\frac{1}{2}-x_6\right) a\,\hat{\mathbf{x}} - y_6 b\,\hat{\mathbf{y}} + \left(\frac{1}{2}+z_6\right) c\,\hat{\mathbf{z}}$ | (4$a$) | Na |
| $\mathbf{B}_{23}$ | = | $-x_6\,\mathbf{a}_1 + \left(\frac{1}{2}+y_6\right)\mathbf{a}_2 + \left(\frac{1}{2}-z_6\right)\mathbf{a}_3$ | = | $-x_6 a\,\hat{\mathbf{x}} + \left(\frac{1}{2}+y_6\right) b\,\hat{\mathbf{y}} + \left(\frac{1}{2}-z_6\right) c\,\hat{\mathbf{z}}$ | (4$a$) | Na |
| $\mathbf{B}_{24}$ | = | $\left(\frac{1}{2}+x_6\right)\mathbf{a}_1 + \left(\frac{1}{2}-y_6\right)\mathbf{a}_2 - z_6\,\mathbf{a}_3$ | = | $\left(\frac{1}{2}+x_6\right) a\,\hat{\mathbf{x}} + \left(\frac{1}{2}-y_6\right) b\,\hat{\mathbf{y}} - z_6 c\,\hat{\mathbf{z}}$ | (4$a$) | Na |

**Geometry files:**

- CIF: pp. 1579
- POSCAR: pp. 1580

# NaP Structure: AB_oP16_19_2a_2a

http://aflow.org/prototype-encyclopedia/AB_oP16_19_2a_2a

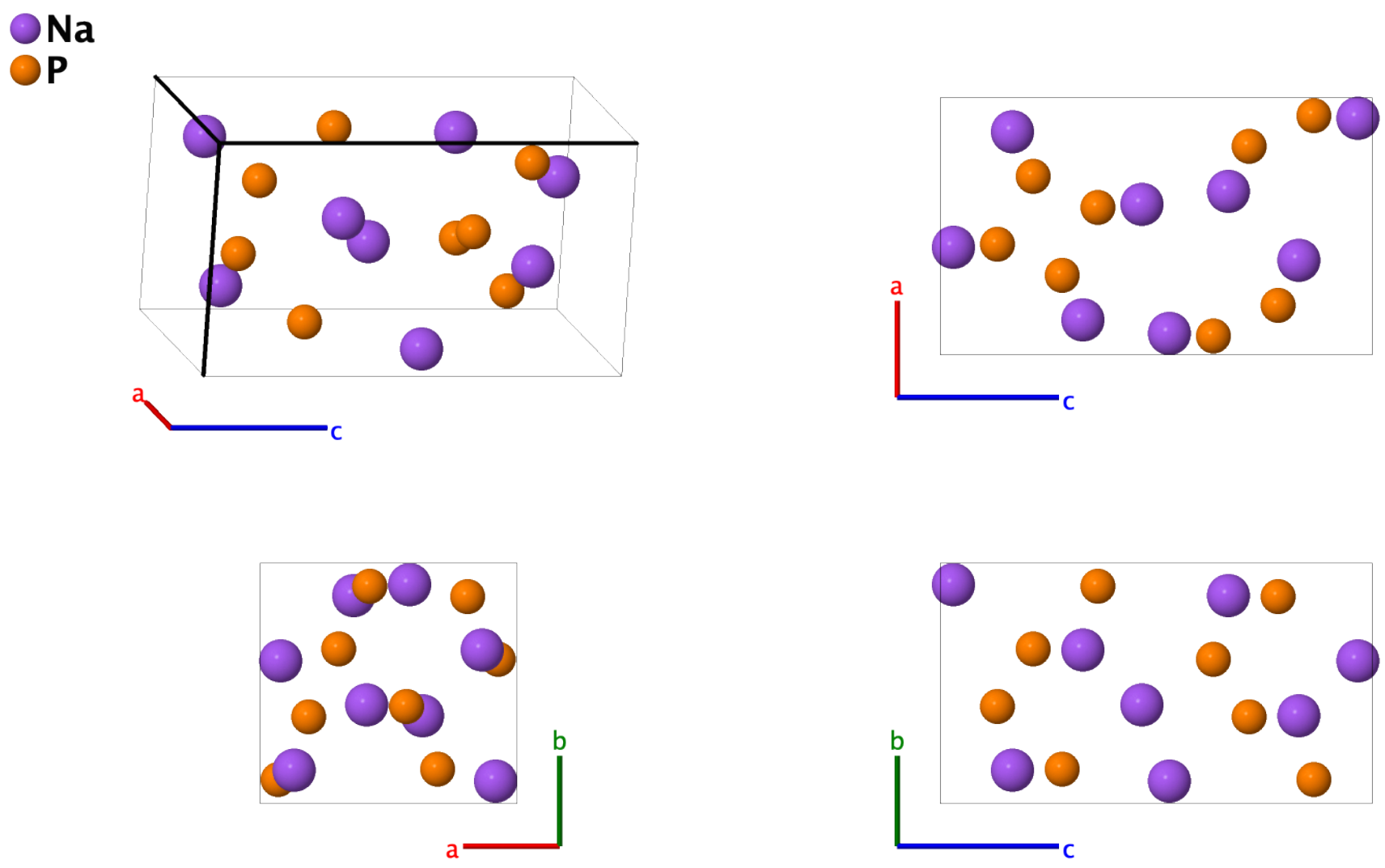

| | | |
|---|---|---|
| **Prototype** | : | NaP |
| **AFLOW prototype label** | : | AB_oP16_19_2a_2a |
| ***Strukturbericht* designation** | : | None |
| **Pearson symbol** | : | oP16 |
| **Space group number** | : | 19 |
| **Space group symbol** | : | $P2_12_12_1$ |
| **AFLOW prototype command** | : | `aflow --proto=AB_oP16_19_2a_2a`<br>`--params=`$a, b/a, c/a, x_1, y_1, z_1, x_2, y_2, z_2, x_3, y_3, z_3, x_4, y_4, z_4$ |

## Other compounds with this structure

- CsSb and KP

## Simple Orthorhombic primitive vectors:

$$\mathbf{a}_1 = a\,\hat{\mathbf{x}}$$
$$\mathbf{a}_2 = b\,\hat{\mathbf{y}}$$
$$\mathbf{a}_3 = c\,\hat{\mathbf{z}}$$

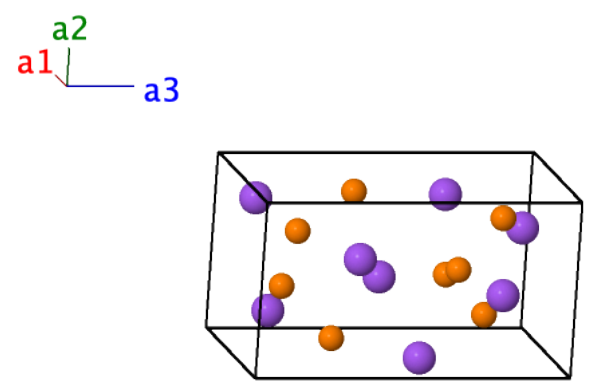

## Basis vectors:

| | | Lattice Coordinates | | Cartesian Coordinates | Wyckoff Position | Atom Type |
|---|---|---|---|---|---|---|
| $\mathbf{B}_1$ | = | $x_1\,\mathbf{a}_1 + y_1\,\mathbf{a}_2 + z_1\,\mathbf{a}_3$ | = | $x_1 a\,\hat{\mathbf{x}} + y_1 b\,\hat{\mathbf{y}} + z_1 c\,\hat{\mathbf{z}}$ | (4$a$) | Na I |
| $\mathbf{B}_2$ | = | $\left(\frac{1}{2} - x_1\right)\mathbf{a}_1 - y_1\,\mathbf{a}_2 + \left(\frac{1}{2} + z_1\right)\mathbf{a}_3$ | = | $\left(\frac{1}{2} - x_1\right) a\,\hat{\mathbf{x}} - y_1 b\,\hat{\mathbf{y}} + \left(\frac{1}{2} + z_1\right) c\,\hat{\mathbf{z}}$ | (4$a$) | Na I |
| $\mathbf{B}_3$ | = | $-x_1\,\mathbf{a}_1 + \left(\frac{1}{2} + y_1\right)\mathbf{a}_2 + \left(\frac{1}{2} - z_1\right)\mathbf{a}_3$ | = | $-x_1 a\,\hat{\mathbf{x}} + \left(\frac{1}{2} + y_1\right) b\,\hat{\mathbf{y}} + \left(\frac{1}{2} - z_1\right) c\,\hat{\mathbf{z}}$ | (4$a$) | Na I |
| $\mathbf{B}_4$ | = | $\left(\frac{1}{2} + x_1\right)\mathbf{a}_1 + \left(\frac{1}{2} - y_1\right)\mathbf{a}_2 - z_1\,\mathbf{a}_3$ | = | $\left(\frac{1}{2} + x_1\right) a\,\hat{\mathbf{x}} + \left(\frac{1}{2} - y_1\right) b\,\hat{\mathbf{y}} - z_1 c\,\hat{\mathbf{z}}$ | (4$a$) | Na I |
| $\mathbf{B}_5$ | = | $x_2\,\mathbf{a}_1 + y_2\,\mathbf{a}_2 + z_2\,\mathbf{a}_3$ | = | $x_2a\,\hat{\mathbf{x}} + y_2b\,\hat{\mathbf{y}} + z_2c\,\hat{\mathbf{z}}$ | (4$a$) | Na II |
| $\mathbf{B}_6$ | = | $\left(\frac{1}{2} - x_2\right)\mathbf{a}_1 - y_2\,\mathbf{a}_2 + \left(\frac{1}{2} + z_2\right)\mathbf{a}_3$ | = | $\left(\frac{1}{2} - x_2\right)a\,\hat{\mathbf{x}} - y_2b\,\hat{\mathbf{y}} + \left(\frac{1}{2} + z_2\right)c\,\hat{\mathbf{z}}$ | (4$a$) | Na II |
| $\mathbf{B}_7$ | = | $-x_2\,\mathbf{a}_1 + \left(\frac{1}{2} + y_2\right)\mathbf{a}_2 + \left(\frac{1}{2} - z_2\right)\mathbf{a}_3$ | = | $-x_2a\,\hat{\mathbf{x}} + \left(\frac{1}{2} + y_2\right)b\,\hat{\mathbf{y}} + \left(\frac{1}{2} - z_2\right)c\,\hat{\mathbf{z}}$ | (4$a$) | Na II |
| $\mathbf{B}_8$ | = | $\left(\frac{1}{2} + x_2\right)\mathbf{a}_1 + \left(\frac{1}{2} - y_2\right)\mathbf{a}_2 - z_2\,\mathbf{a}_3$ | = | $\left(\frac{1}{2} + x_2\right)a\,\hat{\mathbf{x}} + \left(\frac{1}{2} - y_2\right)b\,\hat{\mathbf{y}} - z_2c\,\hat{\mathbf{z}}$ | (4$a$) | Na II |
| $\mathbf{B}_9$ | = | $x_3\,\mathbf{a}_1 + y_3\,\mathbf{a}_2 + z_3\,\mathbf{a}_3$ | = | $x_3a\,\hat{\mathbf{x}} + y_3b\,\hat{\mathbf{y}} + z_3c\,\hat{\mathbf{z}}$ | (4$a$) | P I |
| $\mathbf{B}_{10}$ | = | $\left(\frac{1}{2} - x_3\right)\mathbf{a}_1 - y_3\,\mathbf{a}_2 + \left(\frac{1}{2} + z_3\right)\mathbf{a}_3$ | = | $\left(\frac{1}{2} - x_3\right)a\,\hat{\mathbf{x}} - y_3b\,\hat{\mathbf{y}} + \left(\frac{1}{2} + z_3\right)c\,\hat{\mathbf{z}}$ | (4$a$) | P I |
| $\mathbf{B}_{11}$ | = | $-x_3\,\mathbf{a}_1 + \left(\frac{1}{2} + y_3\right)\mathbf{a}_2 + \left(\frac{1}{2} - z_3\right)\mathbf{a}_3$ | = | $-x_3a\,\hat{\mathbf{x}} + \left(\frac{1}{2} + y_3\right)b\,\hat{\mathbf{y}} + \left(\frac{1}{2} - z_3\right)c\,\hat{\mathbf{z}}$ | (4$a$) | P I |
| $\mathbf{B}_{12}$ | = | $\left(\frac{1}{2} + x_3\right)\mathbf{a}_1 + \left(\frac{1}{2} - y_3\right)\mathbf{a}_2 - z_3\,\mathbf{a}_3$ | = | $\left(\frac{1}{2} + x_3\right)a\,\hat{\mathbf{x}} + \left(\frac{1}{2} - y_3\right)b\,\hat{\mathbf{y}} - z_3c\,\hat{\mathbf{z}}$ | (4$a$) | P I |
| $\mathbf{B}_{13}$ | = | $x_4\,\mathbf{a}_1 + y_4\,\mathbf{a}_2 + z_4\,\mathbf{a}_3$ | = | $x_4a\,\hat{\mathbf{x}} + y_4b\,\hat{\mathbf{y}} + z_4c\,\hat{\mathbf{z}}$ | (4$a$) | P II |
| $\mathbf{B}_{14}$ | = | $\left(\frac{1}{2} - x_4\right)\mathbf{a}_1 - y_4\,\mathbf{a}_2 + \left(\frac{1}{2} + z_4\right)\mathbf{a}_3$ | = | $\left(\frac{1}{2} - x_4\right)a\,\hat{\mathbf{x}} - y_4b\,\hat{\mathbf{y}} + \left(\frac{1}{2} + z_4\right)c\,\hat{\mathbf{z}}$ | (4$a$) | P II |
| $\mathbf{B}_{15}$ | = | $-x_4\,\mathbf{a}_1 + \left(\frac{1}{2} + y_4\right)\mathbf{a}_2 + \left(\frac{1}{2} - z_4\right)\mathbf{a}_3$ | = | $-x_4a\,\hat{\mathbf{x}} + \left(\frac{1}{2} + y_4\right)b\,\hat{\mathbf{y}} + \left(\frac{1}{2} - z_4\right)c\,\hat{\mathbf{z}}$ | (4$a$) | P II |
| $\mathbf{B}_{16}$ | = | $\left(\frac{1}{2} + x_4\right)\mathbf{a}_1 + \left(\frac{1}{2} - y_4\right)\mathbf{a}_2 - z_4\,\mathbf{a}_3$ | = | $\left(\frac{1}{2} + x_4\right)a\,\hat{\mathbf{x}} + \left(\frac{1}{2} - y_4\right)b\,\hat{\mathbf{y}} - z_4c\,\hat{\mathbf{z}}$ | (4$a$) | P II |

---

**Found in:**

---

**Geometry files:**

- CIF: pp. 1580
- POSCAR: pp. 1580

# $D0_7$ ($CrO_3$) (*obsolete*) Structure: AB3_oC16_20_a_bc

http://aflow.org/prototype-encyclopedia/AB3_oC16_20_a_bc

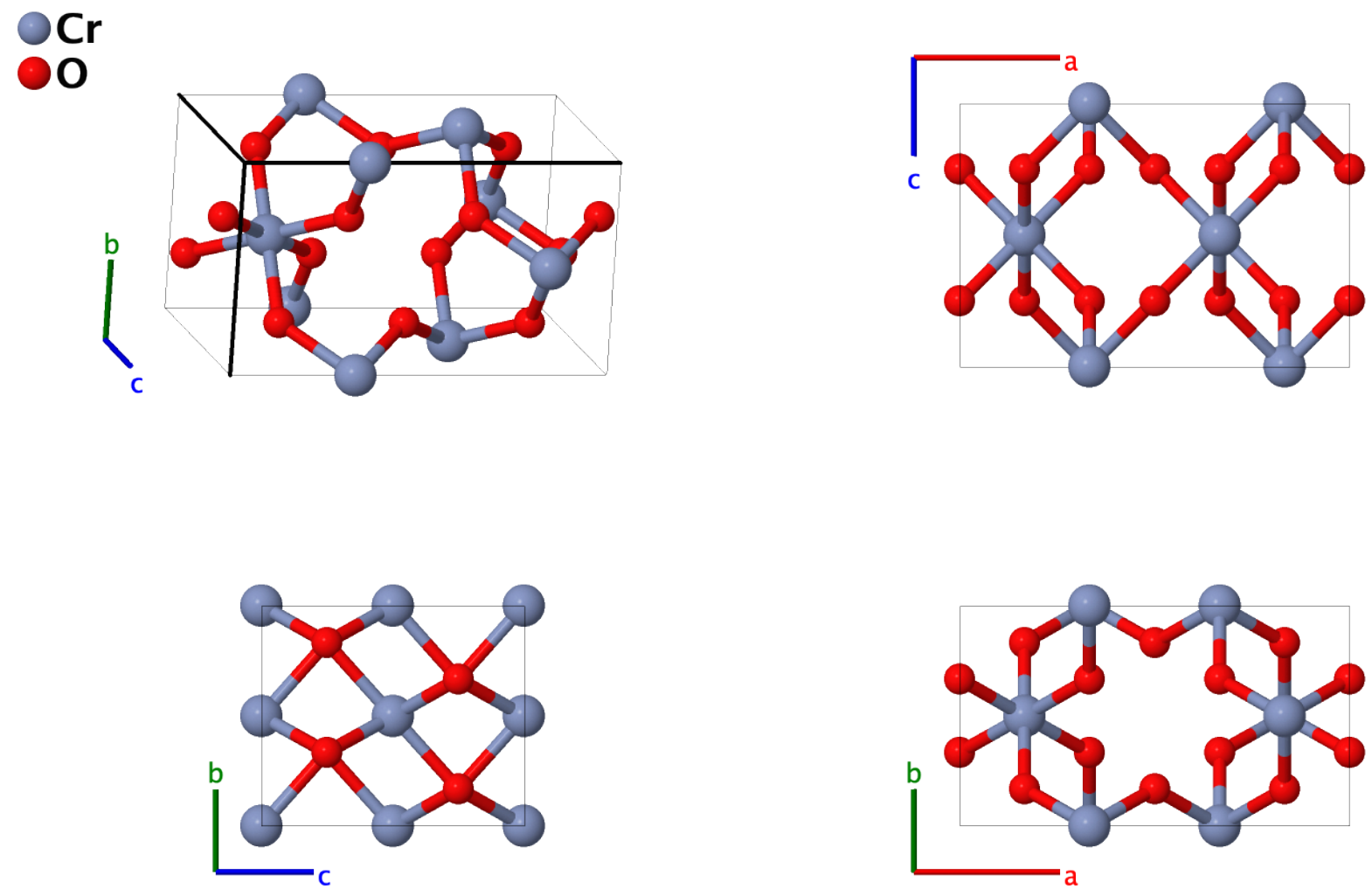

| | | |
|---|---|---|
| **Prototype** | : | $CrO_3$ |
| **AFLOW prototype label** | : | AB3_oC16_20_a_bc |
| ***Strukturbericht* designation** | : | $D0_7$ |
| **Pearson symbol** | : | oC16 |
| **Space group number** | : | 20 |
| **Space group symbol** | : | $C222_1$ |
| **AFLOW prototype command** | : | `aflow --proto=AB3_oC16_20_a_bc --params=`$a, b/a, c/a, x_1, y_2, x_3, y_3, z_3$ |

- This is the determination of the orthorhombic structure of $CrO_3$ given the $D0_7$ designation by (Hermann, 1937). This structure is superceded by the one found by (Byström, 1950), which we call the "orthorhombic $CrO_3$ phase,". The current structure has the Cr atom at the center of a distorted oxygen octahedron, while the newer structure has the Cr atom at center of a distorted tetrahedron.

**Base-centered Orthorhombic primitive vectors:**

$$\mathbf{a}_1 = \frac{1}{2}\,a\,\hat{\mathbf{x}} - \frac{1}{2}\,b\,\hat{\mathbf{y}}$$
$$\mathbf{a}_2 = \frac{1}{2}\,a\,\hat{\mathbf{x}} + \frac{1}{2}\,b\,\hat{\mathbf{y}}$$
$$\mathbf{a}_3 = c\,\hat{\mathbf{z}}$$

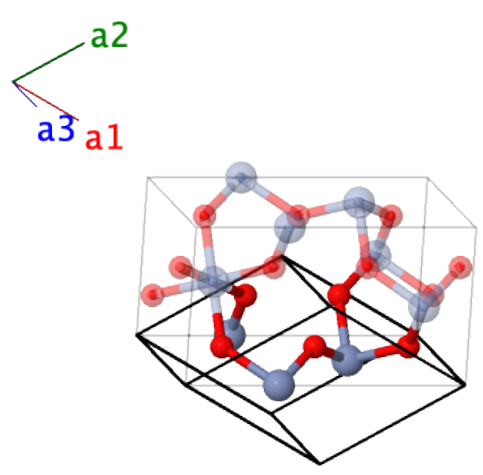

**Basis vectors:**

| | | Lattice Coordinates | | Cartesian Coordinates | Wyckoff Position | Atom Type |
|---|---|---|---|---|---|---|
| $\mathbf{B}_1$ | = | $x_1\,\mathbf{a}_1 + x_1\,\mathbf{a}_2$ | = | $x_1\,a\,\hat{\mathbf{x}}$ | (4$a$) | Cr |
| $\mathbf{B}_2$ | = | $-x_1\,\mathbf{a}_1 - x_1\,\mathbf{a}_2 + \frac{1}{2}\,\mathbf{a}_3$ | = | $-x_1\,a\,\hat{\mathbf{x}} + \frac{1}{2}\,c\,\hat{\mathbf{z}}$ | (4$a$) | Cr |
| $\mathbf{B}_3$ | = | $-y_2\,\mathbf{a}_1 + y_2\,\mathbf{a}_2 + \frac{1}{4}\,\mathbf{a}_3$ | = | $y_2 b\,\hat{\mathbf{y}} + \frac{1}{4} c\,\hat{\mathbf{z}}$ | (4$b$) | O I |
| $\mathbf{B}_4$ | = | $y_2\,\mathbf{a}_1 - y_2\,\mathbf{a}_2 + \frac{3}{4}\,\mathbf{a}_3$ | = | $-y_2 b\,\hat{\mathbf{y}} + \frac{3}{4} c\,\hat{\mathbf{z}}$ | (4$b$) | O I |
| $\mathbf{B}_5$ | = | $(x_3 - y_3)\,\mathbf{a}_1 + (x_3 + y_3)\,\mathbf{a}_2 + z_3\,\mathbf{a}_3$ | = | $x_3 a\,\hat{\mathbf{x}} + y_3 b\,\hat{\mathbf{y}} + z_3 c\,\hat{\mathbf{z}}$ | (8$c$) | O II |
| $\mathbf{B}_6$ | = | $(-x_3 + y_3)\,\mathbf{a}_1 + (-x_3 - y_3)\,\mathbf{a}_2 + \left(\frac{1}{2} + z_3\right)\,\mathbf{a}_3$ | = | $-x_3 a\,\hat{\mathbf{x}} - y_3 b\,\hat{\mathbf{y}} + \left(\frac{1}{2} + z_3\right) c\,\hat{\mathbf{z}}$ | (8$c$) | O II |
| $\mathbf{B}_7$ | = | $(-x_3 - y_3)\,\mathbf{a}_1 + (-x_3 + y_3)\,\mathbf{a}_2 + \left(\frac{1}{2} - z_3\right)\,\mathbf{a}_3$ | = | $-x_3 a\,\hat{\mathbf{x}} + y_3 b\,\hat{\mathbf{y}} + \left(\frac{1}{2} - z_3\right) c\,\hat{\mathbf{z}}$ | (8$c$) | O II |
| $\mathbf{B}_8$ | = | $(x_3 + y_3)\,\mathbf{a}_1 + (x_3 - y_3)\,\mathbf{a}_2 - z_3\,\mathbf{a}_3$ | = | $x_3 a\,\hat{\mathbf{x}} - y_3 b\,\hat{\mathbf{y}} - z_3 c\,\hat{\mathbf{z}}$ | (8$c$) | O II |

---

**Found in:**

---

**Geometry files:**

- CIF: pp. 1580
- POSCAR: pp. 1581

# $AlPO_4$ "low cristobalite type" Structure: AB4C_oC24_20_b_2c_a

http://aflow.org/prototype-encyclopedia/AB4C_oC24_20_b_2c_a

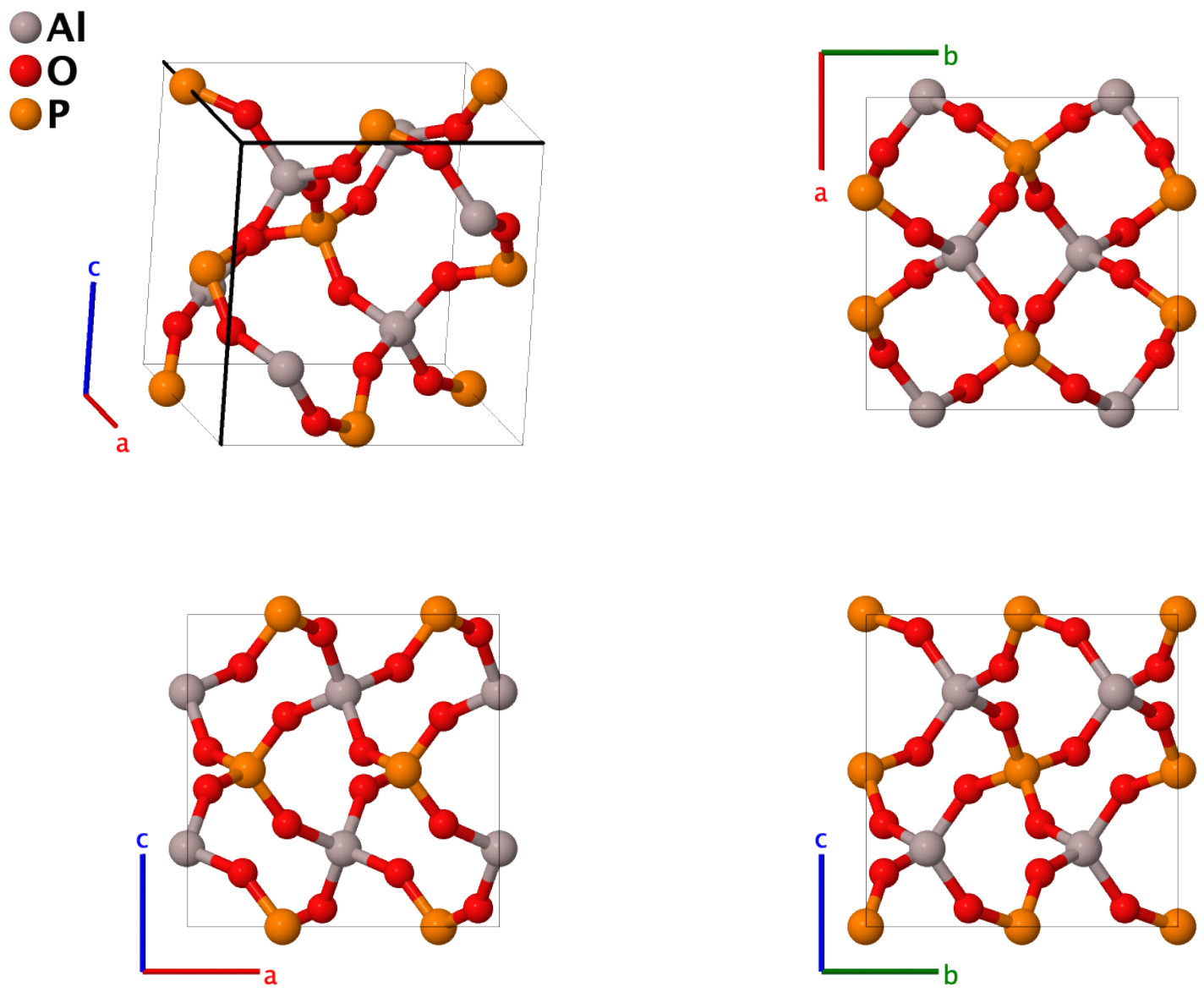

| | | |
|---|---|---|
| **Prototype** | : | $AlO_4P$ |
| **AFLOW prototype label** | : | AB4C_oC24_20_b_2c_a |
| ***Strukturbericht* designation** | : | None |
| **Pearson symbol** | : | oC24 |
| **Space group number** | : | 20 |
| **Space group symbol** | : | $C222_1$ |
| **AFLOW prototype command** | : | `aflow --proto=AB4C_oC24_20_b_2c_a`<br>`--params=`$a, b/a, c/a, x_1, y_2, x_3, y_3, z_3, x_4, y_4, z_4$ |

**Other compounds with this structure**

- $GaPO_4$

- Aluminum and Gallium Phosphate are closely related to the low cristobalite structure of $SiO_2$.
- The change in symmetry from $Fd\bar{3}m$ in $SiO_2$ to $C222_1$ in $AlPO_4$ is caused by the replacement of the silicon atoms by two types of atoms.
- The structure is "pseudo-tetragonal", with $a = b$, but the presence of ordered aluminum and phosphorous atoms reduces the symmetry to orthogonal.

**Base-centered Orthorhombic primitive vectors:**

$$\mathbf{a}_1 = \frac{1}{2}\,a\,\hat{\mathbf{x}} - \frac{1}{2}\,b\,\hat{\mathbf{y}}$$
$$\mathbf{a}_2 = \frac{1}{2}\,a\,\hat{\mathbf{x}} + \frac{1}{2}\,b\,\hat{\mathbf{y}}$$
$$\mathbf{a}_3 = c\,\hat{\mathbf{z}}$$

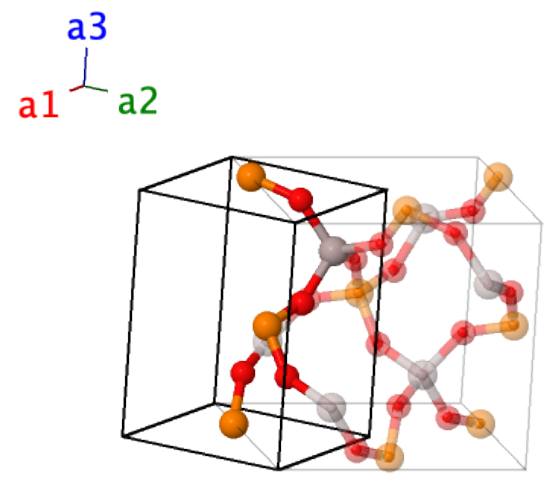

**Basis vectors:**

| | | Lattice Coordinates | | Cartesian Coordinates | Wyckoff Position | Atom Type |
|---|---|---|---|---|---|---|
| $\mathbf{B}_1$ | = | $x_1\,\mathbf{a}_1 + x_1\,\mathbf{a}_2$ | = | $x_1 a\,\hat{\mathbf{x}}$ | (4$a$) | P |
| $\mathbf{B}_2$ | = | $-x_1\,\mathbf{a}_1 - x_1\,\mathbf{a}_2 + \frac{1}{2}\,\mathbf{a}_3$ | = | $-x_1 a\,\hat{\mathbf{x}} + \frac{1}{2}c\,\hat{\mathbf{z}}$ | (4$a$) | P |
| $\mathbf{B}_3$ | = | $-y_2\,\mathbf{a}_1 + y_2\,\mathbf{a}_2 + \frac{1}{4}\,\mathbf{a}_3$ | = | $y_2 b\,\hat{\mathbf{y}} + \frac{1}{4}c\,\hat{\mathbf{z}}$ | (4$b$) | Al |
| $\mathbf{B}_4$ | = | $y_2\,\mathbf{a}_1 - y_2\,\mathbf{a}_2 + \frac{3}{4}\,\mathbf{a}_3$ | = | $-y_2 b\,\hat{\mathbf{y}} + \frac{3}{4}c\,\hat{\mathbf{z}}$ | (4$b$) | Al |
| $\mathbf{B}_5$ | = | $(x_3 - y_3)\,\mathbf{a}_1 + (x_3 + y_3)\,\mathbf{a}_2 + z_3\,\mathbf{a}_3$ | = | $x_3 a\,\hat{\mathbf{x}} + y_3 b\,\hat{\mathbf{y}} + z_3 c\,\hat{\mathbf{z}}$ | (8$c$) | O I |
| $\mathbf{B}_6$ | = | $(-x_3 + y_3)\,\mathbf{a}_1 + (-x_3 - y_3)\,\mathbf{a}_2 + \left(\frac{1}{2} + z_3\right)\,\mathbf{a}_3$ | = | $-x_3 a\,\hat{\mathbf{x}} - y_3 b\,\hat{\mathbf{y}} + \left(\frac{1}{2} + z_3\right) c\,\hat{\mathbf{z}}$ | (8$c$) | O I |
| $\mathbf{B}_7$ | = | $(-x_3 - y_3)\,\mathbf{a}_1 + (-x_3 + y_3)\,\mathbf{a}_2 + \left(\frac{1}{2} - z_3\right)\,\mathbf{a}_3$ | = | $-x_3 a\,\hat{\mathbf{x}} + y_3 b\,\hat{\mathbf{y}} + \left(\frac{1}{2} - z_3\right) c\,\hat{\mathbf{z}}$ | (8$c$) | O I |
| $\mathbf{B}_8$ | = | $(x_3 + y_3)\,\mathbf{a}_1 + (x_3 - y_3)\,\mathbf{a}_2 - z_3\,\mathbf{a}_3$ | = | $x_3 a\,\hat{\mathbf{x}} - y_3 b\,\hat{\mathbf{y}} - z_3 c\,\hat{\mathbf{z}}$ | (8$c$) | O I |
| $\mathbf{B}_9$ | = | $(x_4 - y_4)\,\mathbf{a}_1 + (x_4 + y_4)\,\mathbf{a}_2 + z_4\,\mathbf{a}_3$ | = | $x_4 a\,\hat{\mathbf{x}} + y_4 b\,\hat{\mathbf{y}} + z_4 c\,\hat{\mathbf{z}}$ | (8$c$) | O II |
| $\mathbf{B}_{10}$ | = | $(-x_4 + y_4)\,\mathbf{a}_1 + (-x_4 - y_4)\,\mathbf{a}_2 + \left(\frac{1}{2} + z_4\right)\,\mathbf{a}_3$ | = | $-x_4 a\,\hat{\mathbf{x}} - y_4 b\,\hat{\mathbf{y}} + \left(\frac{1}{2} + z_4\right) c\,\hat{\mathbf{z}}$ | (8$c$) | O II |
| $\mathbf{B}_{11}$ | = | $(-x_4 - y_4)\,\mathbf{a}_1 + (-x_4 + y_4)\,\mathbf{a}_2 + \left(\frac{1}{2} - z_4\right)\,\mathbf{a}_3$ | = | $-x_4 a\,\hat{\mathbf{x}} + y_4 b\,\hat{\mathbf{y}} + \left(\frac{1}{2} - z_4\right) c\,\hat{\mathbf{z}}$ | (8$c$) | O II |
| $\mathbf{B}_{12}$ | = | $(x_4 + y_4)\,\mathbf{a}_1 + (x_4 - y_4)\,\mathbf{a}_2 - z_4\,\mathbf{a}_3$ | = | $x_4 a\,\hat{\mathbf{x}} - y_4 b\,\hat{\mathbf{y}} - z_4 c\,\hat{\mathbf{z}}$ | (8$c$) | O II |

**Found in:**

**Geometry files:**

- CIF: pp. 1581
- POSCAR: pp. 1581

# $Tl_2AlF_5$ ($K3_3$) Structure: AB5C2_oC32_20_b_a2bc_c

http://aflow.org/prototype-encyclopedia/AB5C2_oC32_20_b_a2bc_c

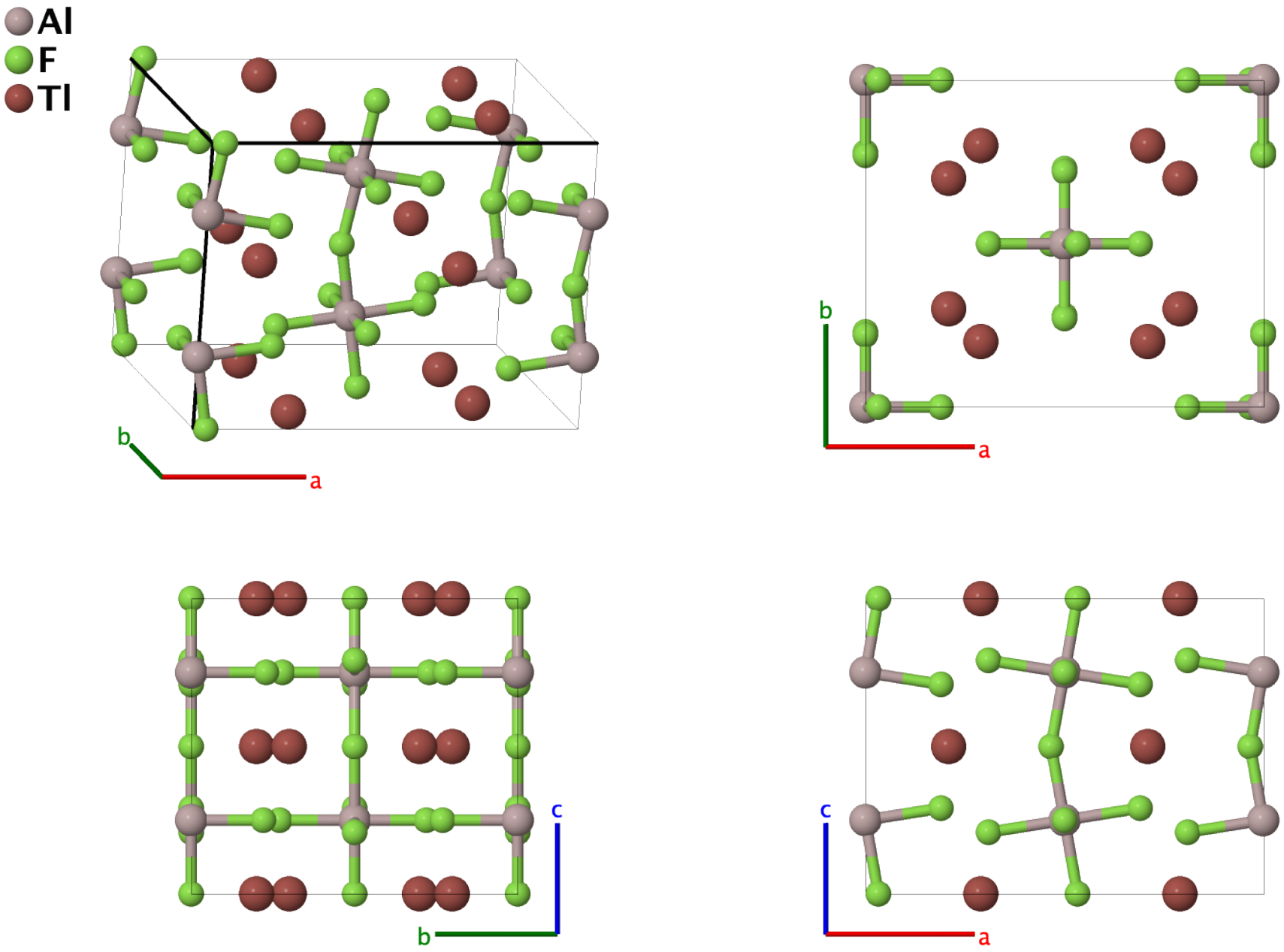

| | | |
|---|---|---|
| **Prototype** | : | $AlF_5Tl_2$ |
| **AFLOW prototype label** | : | AB5C2_oC32_20_b_a2bc_c |
| ***Strukturbericht* designation** | : | None |
| **Pearson symbol** | : | oC32 |
| **Space group number** | : | 20 |
| **Space group symbol** | : | $C222_1$ |
| **AFLOW prototype command** | : | `aflow --proto=AB5C2_oC32_20_b_a2bc_c` `--params=`$a, b/a, c/a, x_1, y_2, y_3, y_4, x_5, y_5, z_5, x_6, y_6, z_6$ |

- There are several problems with this structure:
- First, (Brosset, 1937) gives coordinates $y_3 = -y_2$, which gives the structure an inversion site and makes the space group *Cmcm*. We have adjusted the coordinates of the third atom slightly to avoid this.
- Second, (Molinier, 1993) makes the argument that the structure observed by Brosset is actually $Tl_2AlF_5{\cdot}H_2O$.
- Finally, (Pabst, 1950) indicates that a 1942 work of Brosset refines the structure, but we have not been able to access this paper in any form, nor have we found any reference to this structure in the literature.

**Base-centered Orthorhombic primitive vectors:**

$$\mathbf{a}_1 = \frac{1}{2}\,a\,\mathbf{\hat{x}} - \frac{1}{2}\,b\,\mathbf{\hat{y}}$$
$$\mathbf{a}_2 = \frac{1}{2}\,a\,\mathbf{\hat{x}} + \frac{1}{2}\,b\,\mathbf{\hat{y}}$$
$$\mathbf{a}_3 = c\,\mathbf{\hat{z}}$$

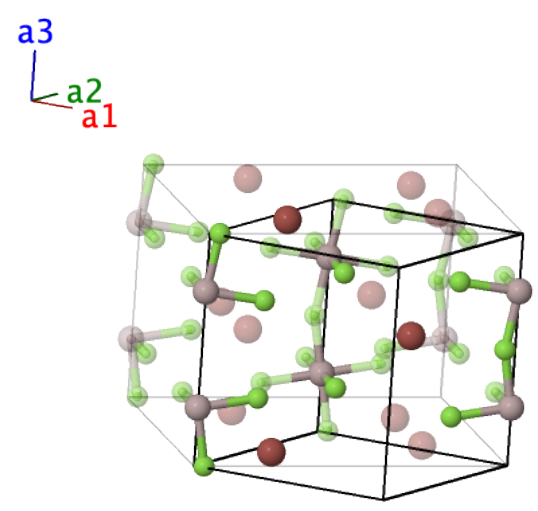

**Basis vectors:**

| | | Lattice Coordinates | | Cartesian Coordinates | Wyckoff Position | Atom Type |
|---|---|---|---|---|---|---|
| $\mathbf{B}_1$ | = | $x_1\,\mathbf{a}_1 + x_1\,\mathbf{a}_2$ | = | $x_1 a\,\hat{\mathbf{x}}$ | (4$a$) | F I |
| $\mathbf{B}_2$ | = | $-x_1\,\mathbf{a}_1 - x_1\,\mathbf{a}_2 + \frac{1}{2}\,\mathbf{a}_3$ | = | $-x_1 a\,\hat{\mathbf{x}} + \frac{1}{2}c\,\hat{\mathbf{z}}$ | (4$a$) | F I |
| $\mathbf{B}_3$ | = | $-y_2\,\mathbf{a}_1 + y_2\,\mathbf{a}_2 + \frac{1}{4}\,\mathbf{a}_3$ | = | $y_2 b\,\hat{\mathbf{y}} + \frac{1}{4}c\,\hat{\mathbf{z}}$ | (4$b$) | Al |
| $\mathbf{B}_4$ | = | $y_2\,\mathbf{a}_1 - y_2\,\mathbf{a}_2 + \frac{3}{4}\,\mathbf{a}_3$ | = | $-y_2 b\,\hat{\mathbf{y}} + \frac{3}{4}c\,\hat{\mathbf{z}}$ | (4$b$) | Al |
| $\mathbf{B}_5$ | = | $-y_3\,\mathbf{a}_1 + y_3\,\mathbf{a}_2 + \frac{1}{4}\,\mathbf{a}_3$ | = | $y_3 b\,\hat{\mathbf{y}} + \frac{1}{4}c\,\hat{\mathbf{z}}$ | (4$b$) | F II |
| $\mathbf{B}_6$ | = | $y_3\,\mathbf{a}_1 - y_3\,\mathbf{a}_2 + \frac{3}{4}\,\mathbf{a}_3$ | = | $-y_3 b\,\hat{\mathbf{y}} + \frac{3}{4}c\,\hat{\mathbf{z}}$ | (4$b$) | F II |
| $\mathbf{B}_7$ | = | $-y_4\,\mathbf{a}_1 + y_4\,\mathbf{a}_2 + \frac{1}{4}\,\mathbf{a}_3$ | = | $y_4 b\,\hat{\mathbf{y}} + \frac{1}{4}c\,\hat{\mathbf{z}}$ | (4$b$) | F III |
| $\mathbf{B}_8$ | = | $y_4\,\mathbf{a}_1 - y_4\,\mathbf{a}_2 + \frac{3}{4}\,\mathbf{a}_3$ | = | $-y_4 b\,\hat{\mathbf{y}} + \frac{3}{4}c\,\hat{\mathbf{z}}$ | (4$b$) | F III |
| $\mathbf{B}_9$ | = | $(x_5 - y_5)\,\mathbf{a}_1 + (x_5 + y_5)\,\mathbf{a}_2 + z_5\,\mathbf{a}_3$ | = | $x_5 a\,\hat{\mathbf{x}} + y_5 b\,\hat{\mathbf{y}} + z_5 c\,\hat{\mathbf{z}}$ | (8$c$) | F IV |
| $\mathbf{B}_{10}$ | = | $(-x_5 + y_5)\,\mathbf{a}_1 + (-x_5 - y_5)\,\mathbf{a}_2 + \left(\frac{1}{2} + z_5\right)\,\mathbf{a}_3$ | = | $-x_5 a\,\hat{\mathbf{x}} - y_5 b\,\hat{\mathbf{y}} + \left(\frac{1}{2} + z_5\right)c\,\hat{\mathbf{z}}$ | (8$c$) | F IV |
| $\mathbf{B}_{11}$ | = | $(-x_5 - y_5)\,\mathbf{a}_1 + (-x_5 + y_5)\,\mathbf{a}_2 + \left(\frac{1}{2} - z_5\right)\,\mathbf{a}_3$ | = | $-x_5 a\,\hat{\mathbf{x}} + y_5 b\,\hat{\mathbf{y}} + \left(\frac{1}{2} - z_5\right)c\,\hat{\mathbf{z}}$ | (8$c$) | F IV |
| $\mathbf{B}_{12}$ | = | $(x_5 + y_5)\,\mathbf{a}_1 + (x_5 - y_5)\,\mathbf{a}_2 - z_5\,\mathbf{a}_3$ | = | $x_5 a\,\hat{\mathbf{x}} - y_5 b\,\hat{\mathbf{y}} - z_5 c\,\hat{\mathbf{z}}$ | (8$c$) | F IV |
| $\mathbf{B}_{13}$ | = | $(x_6 - y_6)\,\mathbf{a}_1 + (x_6 + y_6)\,\mathbf{a}_2 + z_6\,\mathbf{a}_3$ | = | $x_6 a\,\hat{\mathbf{x}} + y_6 b\,\hat{\mathbf{y}} + z_6 c\,\hat{\mathbf{z}}$ | (8$c$) | Tl |
| $\mathbf{B}_{14}$ | = | $(-x_6 + y_6)\,\mathbf{a}_1 + (-x_6 - y_6)\,\mathbf{a}_2 + \left(\frac{1}{2} + z_6\right)\,\mathbf{a}_3$ | = | $-x_6 a\,\hat{\mathbf{x}} - y_6 b\,\hat{\mathbf{y}} + \left(\frac{1}{2} + z_6\right)c\,\hat{\mathbf{z}}$ | (8$c$) | Tl |
| $\mathbf{B}_{15}$ | = | $(-x_6 - y_6)\,\mathbf{a}_1 + (-x_6 + y_6)\,\mathbf{a}_2 + \left(\frac{1}{2} - z_6\right)\,\mathbf{a}_3$ | = | $-x_6 a\,\hat{\mathbf{x}} + y_6 b\,\hat{\mathbf{y}} + \left(\frac{1}{2} - z_6\right)c\,\hat{\mathbf{z}}$ | (8$c$) | Tl |
| $\mathbf{B}_{16}$ | = | $(x_6 + y_6)\,\mathbf{a}_1 + (x_6 - y_6)\,\mathbf{a}_2 - z_6\,\mathbf{a}_3$ | = | $x_6 a\,\hat{\mathbf{x}} - y_6 b\,\hat{\mathbf{y}} - z_6 c\,\hat{\mathbf{z}}$ | (8$c$) | Tl |

**Geometry files:**

- CIF: pp. 1581
- POSCAR: pp. 1582

# $HoSb_2$ Structure: AB2_oC6_21_a_k

http://aflow.org/prototype-encyclopedia/AB2_oC6_21_a_k.HoSb2

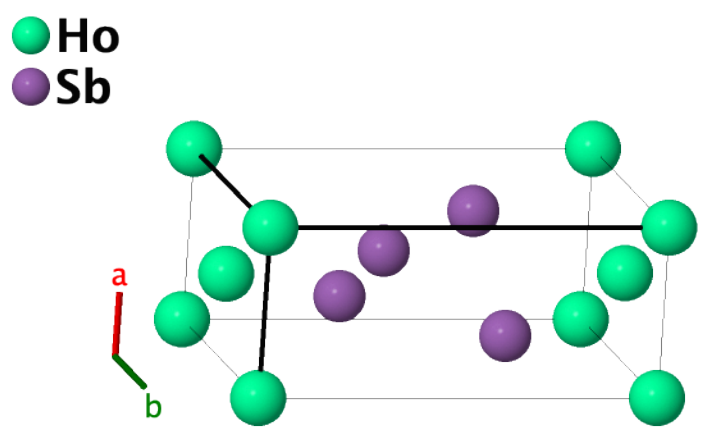

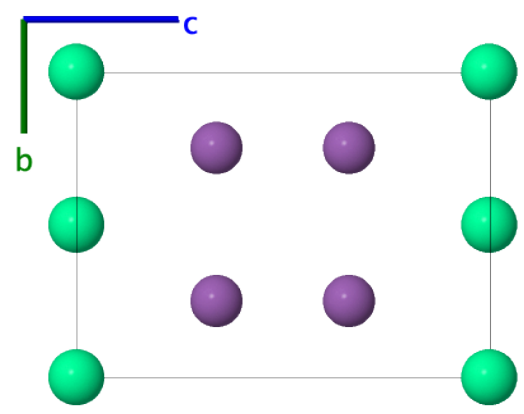

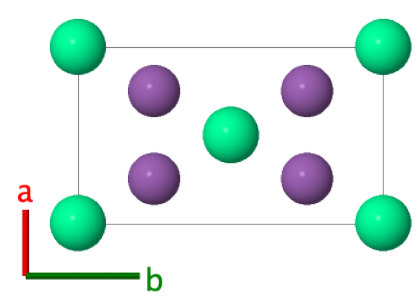

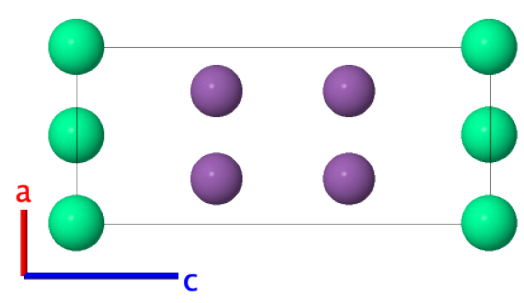

| | | |
|---|---|---|
| **Prototype** | : | $HoSb_2$ |
| **AFLOW prototype label** | : | AB2_oC6_21_a_k |
| ***Strukturbericht* designation** | : | None |
| **Pearson symbol** | : | oC6 |
| **Space group number** | : | 21 |
| **Space group symbol** | : | *C*222 |
| **AFLOW prototype command** | : | `aflow --proto=AB2_oC6_21_a_k --params=`$a, b/a, c/a, z_2$ |

**Other compounds with this structure**

- $LuSb_2$, $YSb_2$, $DySb_2$ (HT), $GdSb_2$ (HT), and $TbSb_2$ (HT)

- Measurements were performed at a pressure of 65 kbar.
- The primitive unit cell is nearly hexagonal.
- The author states "We are well aware of the fact that the structure presented here may indeed be only a subcell of the true structure."
- $HoSb_2$ and $Ta_2H$ share the same AFLOW prototype label, AB2_oC6_21_a_k. The structures are generated by the same symmetry operations with different sets of parameters (`--params`) specified in their corresponding CIF files.

**Base-centered Orthorhombic primitive vectors:**

$$\mathbf{a}_1 = \frac{1}{2}\,a\,\hat{\mathbf{x}} - \frac{1}{2}\,b\,\hat{\mathbf{y}}$$
$$\mathbf{a}_2 = \frac{1}{2}\,a\,\hat{\mathbf{x}} + \frac{1}{2}\,b\,\hat{\mathbf{y}}$$
$$\mathbf{a}_3 = c\,\hat{\mathbf{z}}$$

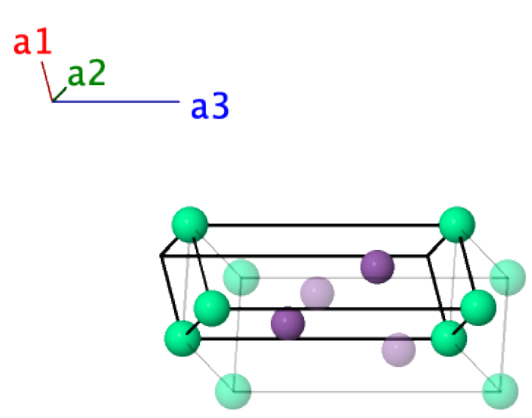

**Basis vectors:**

| | | Lattice Coordinates | | Cartesian Coordinates | Wyckoff Position | Atom Type |
|---|---|---|---|---|---|---|
| $\mathbf{B}_1$ | = | $0\,\mathbf{a}_1 + 0\,\mathbf{a}_2 + 0\,\mathbf{a}_3$ | = | $0\,\hat{\mathbf{x}} + 0\,\hat{\mathbf{y}} + 0\,\hat{\mathbf{z}}$ | $(2a)$ | Ho |
| $\mathbf{B}_2$ | = | $\frac{1}{2}\,\mathbf{a}_2 + z_2\,\mathbf{a}_3$ | = | $\frac{1}{4}a\,\hat{\mathbf{x}} + \frac{1}{4}b\,\hat{\mathbf{y}} + z_2c\,\hat{\mathbf{z}}$ | $(4k)$ | Sb |
| $\mathbf{B}_3$ | = | $\frac{1}{2}\,\mathbf{a}_1 - z_2\,\mathbf{a}_3$ | = | $\frac{1}{4}a\,\hat{\mathbf{x}} - \frac{1}{4}b\,\hat{\mathbf{y}} - z_2c\,\hat{\mathbf{z}}$ | $(4k)$ | Sb |

**Geometry files:**

- CIF: pp. 1582
- POSCAR: pp. 1582

# Predicted Phase IV $Cd_2Re_2O_7$ Structure: A2B7C2_oF88_22_k_bdefghij_k

http://aflow.org/prototype-encyclopedia/A2B7C2_oF88_22_k_bdefghij_k

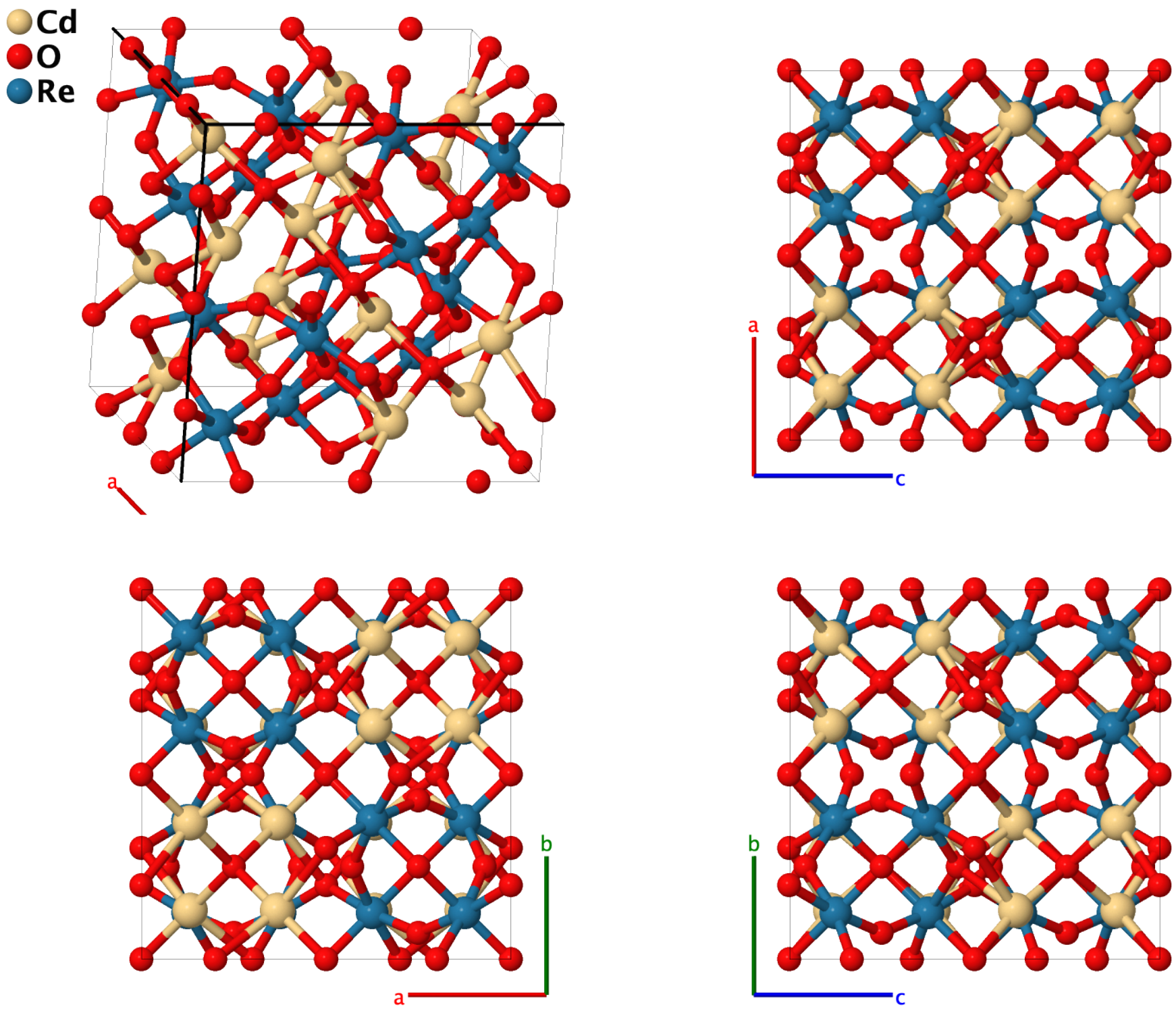

| | | |
|---|---|---|
| **Prototype** | : | $Cd_2O_7Re_2$ |
| **AFLOW prototype label** | : | A2B7C2_oF88_22_k_bdefghij_k |
| ***Strukturbericht* designation** | : | None |
| **Pearson symbol** | : | oF88 |
| **Space group number** | : | 22 |
| **Space group symbol** | : | $F222$ |
| **AFLOW prototype command** | : | `aflow --proto=A2B7C2_oF88_22_k_bdefghij_k` <br> `--params=`$a, b/a, c/a, x_3, y_4, z_5, z_6, y_7, x_8, x_9, y_9, z_9, x_{10}, y_{10}, z_{10}$ |

- $Cd_2Re_2O_7$ exhibits a number of phases. We will use the notation of (Kapcia, 2019) to describe them:
  - Phase I: above 200 K, the system takes on the cubic pyrochlore ($E8_1$) structure.
  - Phase II: in the range 120-200 K the system is in the tetragonal $I\bar{4}m2$ #119 structure.
  - Phase III: in the range 80-120 K the system is in the tetragonal $I4_122$ #98 structure.
  - Phase IV: (Kapcia, 2019) did a first-principles study of this system and found that below 80 K Phase III develops a soft phonon mode which transforms the system into an orthorhombic $F222$ #22 structure. (This structure)
  - (Norman, 2019) points out that both Phase III and Phase IV structures have issues.
- Phase IV is extremely close to Phase II. If AFLOW-SYM and FINDSYM allow symmetry tolerances of 0.002 Å, the orthorhombic phase becomes tetragonal.

## Face-centered Orthorhombic primitive vectors:

$$\mathbf{a}_1 = \frac{1}{2}\,b\,\hat{\mathbf{y}} + \frac{1}{2}\,c\,\hat{\mathbf{z}}$$
$$\mathbf{a}_2 = \frac{1}{2}\,a\,\hat{\mathbf{x}} + \frac{1}{2}\,c\,\hat{\mathbf{z}}$$
$$\mathbf{a}_3 = \frac{1}{2}\,a\,\hat{\mathbf{x}} + \frac{1}{2}\,b\,\hat{\mathbf{y}}$$

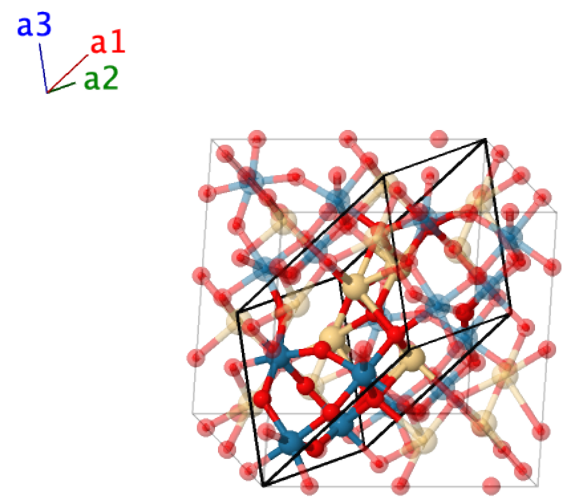

## Basis vectors:

| | | Lattice Coordinates | | Cartesian Coordinates | Wyckoff Position | Atom Type |
|---|---|---|---|---|---|---|
| $\mathbf{B}_1$ | = | $\frac{1}{2}\,\mathbf{a}_1 + \frac{1}{2}\,\mathbf{a}_2 + \frac{1}{2}\,\mathbf{a}_3$ | = | $\frac{1}{2}a\,\hat{\mathbf{x}} + \frac{1}{2}b\,\hat{\mathbf{y}} + \frac{1}{2}c\,\hat{\mathbf{z}}$ | $(4b)$ | O I |
| $\mathbf{B}_2$ | = | $\frac{3}{4}\,\mathbf{a}_1 + \frac{3}{4}\,\mathbf{a}_2 + \frac{3}{4}\,\mathbf{a}_3$ | = | $\frac{3}{4}a\,\hat{\mathbf{x}} + \frac{3}{4}b\,\hat{\mathbf{y}} + \frac{3}{4}c\,\hat{\mathbf{z}}$ | $(4d)$ | O II |
| $\mathbf{B}_3$ | = | $-x_3\,\mathbf{a}_1 + x_3\,\mathbf{a}_2 + x_3\,\mathbf{a}_3$ | = | $x_3a\,\hat{\mathbf{x}}$ | $(8e)$ | O III |
| $\mathbf{B}_4$ | = | $x_3\,\mathbf{a}_1 - x_3\,\mathbf{a}_2 - x_3\,\mathbf{a}_3$ | = | $-x_3a\,\hat{\mathbf{x}}$ | $(8e)$ | O III |
| $\mathbf{B}_5$ | = | $y_4\,\mathbf{a}_1 - y_4\,\mathbf{a}_2 + y_4\,\mathbf{a}_3$ | = | $y_4b\,\hat{\mathbf{y}}$ | $(8f)$ | O IV |
| $\mathbf{B}_6$ | = | $-y_4\,\mathbf{a}_1 + y_4\,\mathbf{a}_2 - y_4\,\mathbf{a}_3$ | = | $-y_4b\,\hat{\mathbf{y}}$ | $(8f)$ | O IV |
| $\mathbf{B}_7$ | = | $z_5\,\mathbf{a}_1 + z_5\,\mathbf{a}_2 - z_5\,\mathbf{a}_3$ | = | $z_5c\,\hat{\mathbf{z}}$ | $(8g)$ | O V |
| $\mathbf{B}_8$ | = | $-z_5\,\mathbf{a}_1 - z_5\,\mathbf{a}_2 + z_5\,\mathbf{a}_3$ | = | $-z_5c\,\hat{\mathbf{z}}$ | $(8g)$ | O V |
| $\mathbf{B}_9$ | = | $z_6\,\mathbf{a}_1 + z_6\,\mathbf{a}_2 + \left(\frac{1}{2} - z_6\right)\,\mathbf{a}_3$ | = | $\frac{1}{4}a\,\hat{\mathbf{x}} + \frac{1}{4}b\,\hat{\mathbf{y}} + z_6c\,\hat{\mathbf{z}}$ | $(8h)$ | O VI |
| $\mathbf{B}_{10}$ | = | $\left(\frac{1}{2} - z_6\right)\,\mathbf{a}_1 + \left(\frac{1}{2} - z_6\right)\,\mathbf{a}_2 + z_6\,\mathbf{a}_3$ | = | $\frac{1}{4}a\,\hat{\mathbf{x}} + \frac{1}{4}b\,\hat{\mathbf{y}} + \left(\frac{1}{2} - z_6\right)c\,\hat{\mathbf{z}}$ | $(8h)$ | O VI |
| $\mathbf{B}_{11}$ | = | $y_7\,\mathbf{a}_1 + \left(\frac{1}{2} - y_7\right)\,\mathbf{a}_2 + y_7\,\mathbf{a}_3$ | = | $\frac{1}{4}a\,\hat{\mathbf{x}} + y_7b\,\hat{\mathbf{y}} + \frac{1}{4}c\,\hat{\mathbf{z}}$ | $(8i)$ | O VII |
| $\mathbf{B}_{12}$ | = | $\left(\frac{1}{2} - y_7\right)\,\mathbf{a}_1 + y_7\,\mathbf{a}_2 + \left(\frac{1}{2} - y_7\right)\,\mathbf{a}_3$ | = | $\frac{1}{4}a\,\hat{\mathbf{x}} + \left(\frac{1}{2} - y_7\right)b\,\hat{\mathbf{y}} + \frac{1}{4}c\,\hat{\mathbf{z}}$ | $(8i)$ | O VII |
| $\mathbf{B}_{13}$ | = | $\left(\frac{1}{2} - x_8\right)\,\mathbf{a}_1 + x_8\,\mathbf{a}_2 + x_8\,\mathbf{a}_3$ | = | $x_8a\,\hat{\mathbf{x}} + \frac{1}{4}b\,\hat{\mathbf{y}} + \frac{1}{4}c\,\hat{\mathbf{z}}$ | $(8j)$ | O VIII |
| $\mathbf{B}_{14}$ | = | $x_8\,\mathbf{a}_1 + \left(\frac{1}{2} - x_8\right)\,\mathbf{a}_2 + \left(\frac{1}{2} - x_8\right)\,\mathbf{a}_3$ | = | $\left(\frac{1}{2} - x_8\right)a\,\hat{\mathbf{x}} + \frac{1}{4}b\,\hat{\mathbf{y}} + \frac{1}{4}c\,\hat{\mathbf{z}}$ | $(8j)$ | O VIII |
| $\mathbf{B}_{15}$ | = | $(-x_9 + y_9 + z_9)\,\mathbf{a}_1 + (x_9 - y_9 + z_9)\,\mathbf{a}_2 + (x_9 + y_9 - z_9)\,\mathbf{a}_3$ | = | $x_9a\,\hat{\mathbf{x}} + y_9b\,\hat{\mathbf{y}} + z_9c\,\hat{\mathbf{z}}$ | $(16k)$ | Cd |
| $\mathbf{B}_{16}$ | = | $(x_9 - y_9 + z_9)\,\mathbf{a}_1 + (-x_9 + y_9 + z_9)\,\mathbf{a}_2 + (-x_9 - y_9 - z_9)\,\mathbf{a}_3$ | = | $-x_9a\,\hat{\mathbf{x}} - y_9b\,\hat{\mathbf{y}} + z_9c\,\hat{\mathbf{z}}$ | $(16k)$ | Cd |
| $\mathbf{B}_{17}$ | = | $(x_9 + y_9 - z_9)\,\mathbf{a}_1 + (-x_9 - y_9 - z_9)\,\mathbf{a}_2 + (-x_9 + y_9 + z_9)\,\mathbf{a}_3$ | = | $-x_9a\,\hat{\mathbf{x}} + y_9b\,\hat{\mathbf{y}} - z_9c\,\hat{\mathbf{z}}$ | $(16k)$ | Cd |
| $\mathbf{B}_{18}$ | = | $(-x_9 - y_9 - z_9)\,\mathbf{a}_1 + (x_9 + y_9 - z_9)\,\mathbf{a}_2 + (x_9 - y_9 + z_9)\,\mathbf{a}_3$ | = | $x_9a\,\hat{\mathbf{x}} - y_9b\,\hat{\mathbf{y}} - z_9c\,\hat{\mathbf{z}}$ | $(16k)$ | Cd |
| $\mathbf{B}_{19}$ | = | $(-x_{10} + y_{10} + z_{10})\,\mathbf{a}_1 + (x_{10} - y_{10} + z_{10})\,\mathbf{a}_2 + (x_{10} + y_{10} - z_{10})\,\mathbf{a}_3$ | = | $x_{10}a\,\hat{\mathbf{x}} + y_{10}b\,\hat{\mathbf{y}} + z_{10}c\,\hat{\mathbf{z}}$ | $(16k)$ | Re |
| $\mathbf{B}_{20}$ | = | $(x_{10} - y_{10} + z_{10})\,\mathbf{a}_1 + (-x_{10} + y_{10} + z_{10})\,\mathbf{a}_2 + (-x_{10} - y_{10} - z_{10})\,\mathbf{a}_3$ | = | $-x_{10}a\,\hat{\mathbf{x}} - y_{10}b\,\hat{\mathbf{y}} + z_{10}c\,\hat{\mathbf{z}}$ | $(16k)$ | Re |
| $\mathbf{B}_{21}$ | = | $(x_{10} + y_{10} - z_{10})\,\mathbf{a}_1 + (-x_{10} - y_{10} - z_{10})\,\mathbf{a}_2 + (-x_{10} + y_{10} + z_{10})\,\mathbf{a}_3$ | = | $-x_{10}a\,\hat{\mathbf{x}} + y_{10}b\,\hat{\mathbf{y}} - z_{10}c\,\hat{\mathbf{z}}$ | $(16k)$ | Re |
| $\mathbf{B}_{22}$ | = | $(-x_{10} - y_{10} - z_{10})\,\mathbf{a}_1 + (x_{10} + y_{10} - z_{10})\,\mathbf{a}_2 + (x_{10} - y_{10} + z_{10})\,\mathbf{a}_3$ | = | $x_{10}a\,\hat{\mathbf{x}} - y_{10}b\,\hat{\mathbf{y}} - z_{10}c\,\hat{\mathbf{z}}$ | $(16k)$ | Re |

**Geometry files:**

- CIF: pp. 1582
- POSCAR: pp. 1583

# Mercury (II) Azide [$Hg(N_3)_2$] Structure: AB6_oP28_29_a_6a

http://aflow.org/prototype-encyclopedia/AB6_oP28_29_a_6a

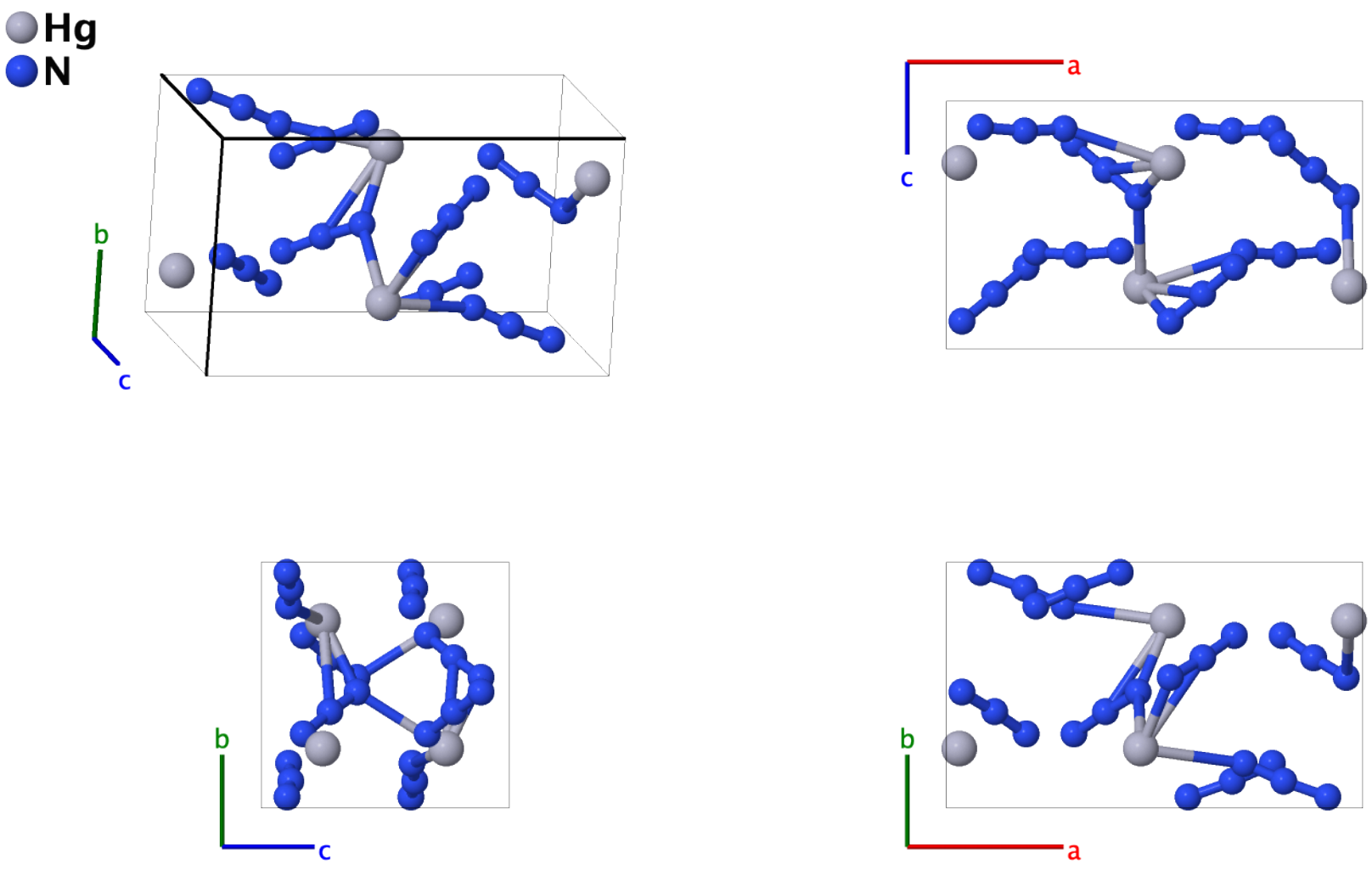

| | | |
|---|---|---|
| **Prototype** | : | $HgN_6$ |
| **AFLOW prototype label** | : | AB6_oP28_29_a_6a |
| ***Strukturbericht* designation** | : | None |
| **Pearson symbol** | : | oP28 |
| **Space group number** | : | 29 |
| **Space group symbol** | : | $Pca2_1$ |
| **AFLOW prototype command** | : | `aflow --proto=AB6_oP28_29_a_6a` `--params=`$a, b/a, c/a, x_1, y_1, z_1, x_2, y_2, z_2, x_3, y_3, z_3, x_4, y_4, z_4, x_5, y_5, z_5, x_6, y_6, z_6, x_7, y_7, z_7$ |

**Simple Orthorhombic primitive vectors:**

$$\mathbf{a}_1 = a\,\hat{\mathbf{x}}$$
$$\mathbf{a}_2 = b\,\hat{\mathbf{y}}$$
$$\mathbf{a}_3 = c\,\hat{\mathbf{z}}$$

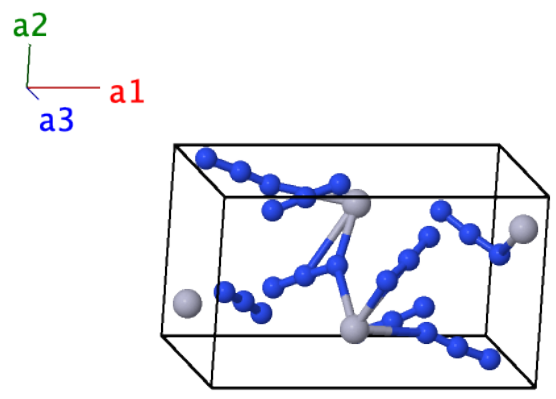

**Basis vectors:**

| | | Lattice Coordinates | | Cartesian Coordinates | Wyckoff Position | Atom Type |
|---|---|---|---|---|---|---|
| $\mathbf{B}_1$ | = | $x_1\,\mathbf{a}_1 + y_1\,\mathbf{a}_2 + z_1\,\mathbf{a}_3$ | = | $x_1 a\,\hat{\mathbf{x}} + y_1 b\,\hat{\mathbf{y}} + z_1 c\,\hat{\mathbf{z}}$ | (4$a$) | Hg |
| $\mathbf{B}_2$ | = | $-x_1\,\mathbf{a}_1 - y_1\,\mathbf{a}_2 + \left(\frac{1}{2} + z_1\right)\mathbf{a}_3$ | = | $-x_1 a\,\hat{\mathbf{x}} - y_1 b\,\hat{\mathbf{y}} + \left(\frac{1}{2} + z_1\right) c\,\hat{\mathbf{z}}$ | (4$a$) | Hg |
| $\mathbf{B}_3$ | = | $\left(\frac{1}{2} + x_1\right)\mathbf{a}_1 - y_1\,\mathbf{a}_2 + z_1\,\mathbf{a}_3$ | = | $\left(\frac{1}{2} + x_1\right) a\,\hat{\mathbf{x}} - y_1 b\,\hat{\mathbf{y}} + z_1 c\,\hat{\mathbf{z}}$ | (4$a$) | Hg |
| $\mathbf{B}_4$ | = | $\left(\frac{1}{2} - x_1\right)\mathbf{a}_1 + y_1\,\mathbf{a}_2 + \left(\frac{1}{2} + z_1\right)\mathbf{a}_3$ | = | $\left(\frac{1}{2} - x_1\right) a\,\hat{\mathbf{x}} + y_1 b\,\hat{\mathbf{y}} + \left(\frac{1}{2} + z_1\right) c\,\hat{\mathbf{z}}$ | (4$a$) | Hg |
| $\mathbf{B}_5$ | = | $x_2\,\mathbf{a}_1 + y_2\,\mathbf{a}_2 + z_2\,\mathbf{a}_3$ | = | $x_2 a\,\hat{\mathbf{x}} + y_2 b\,\hat{\mathbf{y}} + z_2 c\,\hat{\mathbf{z}}$ | (4$a$) | N I |
| $\mathbf{B}_{6}$ | = | $-x_2\,\mathbf{a}_1 - y_2\,\mathbf{a}_2 + \left(\frac{1}{2} + z_2\right)\,\mathbf{a}_3$ | = | $-x_2 a\,\hat{\mathbf{x}} - y_2 b\,\hat{\mathbf{y}} + \left(\frac{1}{2} + z_2\right) c\,\hat{\mathbf{z}}$ | (4$a$) | N I |
| $\mathbf{B}_{7}$ | = | $\left(\frac{1}{2} + x_2\right)\,\mathbf{a}_1 - y_2\,\mathbf{a}_2 + z_2\,\mathbf{a}_3$ | = | $\left(\frac{1}{2} + x_2\right) a\,\hat{\mathbf{x}} - y_2 b\,\hat{\mathbf{y}} + z_2 c\,\hat{\mathbf{z}}$ | (4$a$) | N I |
| $\mathbf{B}_{8}$ | = | $\left(\frac{1}{2} - x_2\right)\,\mathbf{a}_1 + y_2\,\mathbf{a}_2 + \left(\frac{1}{2} + z_2\right)\,\mathbf{a}_3$ | = | $\left(\frac{1}{2} - x_2\right) a\,\hat{\mathbf{x}} + y_2 b\,\hat{\mathbf{y}} + \left(\frac{1}{2} + z_2\right) c\,\hat{\mathbf{z}}$ | (4$a$) | N I |
| $\mathbf{B}_{9}$ | = | $x_3\,\mathbf{a}_1 + y_3\,\mathbf{a}_2 + z_3\,\mathbf{a}_3$ | = | $x_3 a\,\hat{\mathbf{x}} + y_3 b\,\hat{\mathbf{y}} + z_3 c\,\hat{\mathbf{z}}$ | (4$a$) | N II |
| $\mathbf{B}_{10}$ | = | $-x_3\,\mathbf{a}_1 - y_3\,\mathbf{a}_2 + \left(\frac{1}{2} + z_3\right)\,\mathbf{a}_3$ | = | $-x_3 a\,\hat{\mathbf{x}} - y_3 b\,\hat{\mathbf{y}} + \left(\frac{1}{2} + z_3\right) c\,\hat{\mathbf{z}}$ | (4$a$) | N II |
| $\mathbf{B}_{11}$ | = | $\left(\frac{1}{2} + x_3\right)\,\mathbf{a}_1 - y_3\,\mathbf{a}_2 + z_3\,\mathbf{a}_3$ | = | $\left(\frac{1}{2} + x_3\right) a\,\hat{\mathbf{x}} - y_3 b\,\hat{\mathbf{y}} + z_3 c\,\hat{\mathbf{z}}$ | (4$a$) | N II |
| $\mathbf{B}_{12}$ | = | $\left(\frac{1}{2} - x_3\right)\,\mathbf{a}_1 + y_3\,\mathbf{a}_2 + \left(\frac{1}{2} + z_3\right)\,\mathbf{a}_3$ | = | $\left(\frac{1}{2} - x_3\right) a\,\hat{\mathbf{x}} + y_3 b\,\hat{\mathbf{y}} + \left(\frac{1}{2} + z_3\right) c\,\hat{\mathbf{z}}$ | (4$a$) | N II |
| $\mathbf{B}_{13}$ | = | $x_4\,\mathbf{a}_1 + y_4\,\mathbf{a}_2 + z_4\,\mathbf{a}_3$ | = | $x_4 a\,\hat{\mathbf{x}} + y_4 b\,\hat{\mathbf{y}} + z_4 c\,\hat{\mathbf{z}}$ | (4$a$) | N III |
| $\mathbf{B}_{14}$ | = | $-x_4\,\mathbf{a}_1 - y_4\,\mathbf{a}_2 + \left(\frac{1}{2} + z_4\right)\,\mathbf{a}_3$ | = | $-x_4 a\,\hat{\mathbf{x}} - y_4 b\,\hat{\mathbf{y}} + \left(\frac{1}{2} + z_4\right) c\,\hat{\mathbf{z}}$ | (4$a$) | N III |
| $\mathbf{B}_{15}$ | = | $\left(\frac{1}{2} + x_4\right)\,\mathbf{a}_1 - y_4\,\mathbf{a}_2 + z_4\,\mathbf{a}_3$ | = | $\left(\frac{1}{2} + x_4\right) a\,\hat{\mathbf{x}} - y_4 b\,\hat{\mathbf{y}} + z_4 c\,\hat{\mathbf{z}}$ | (4$a$) | N III |
| $\mathbf{B}_{16}$ | = | $\left(\frac{1}{2} - x_4\right)\,\mathbf{a}_1 + y_4\,\mathbf{a}_2 + \left(\frac{1}{2} + z_4\right)\,\mathbf{a}_3$ | = | $\left(\frac{1}{2} - x_4\right) a\,\hat{\mathbf{x}} + y_4 b\,\hat{\mathbf{y}} + \left(\frac{1}{2} + z_4\right) c\,\hat{\mathbf{z}}$ | (4$a$) | N III |
| $\mathbf{B}_{17}$ | = | $x_5\,\mathbf{a}_1 + y_5\,\mathbf{a}_2 + z_5\,\mathbf{a}_3$ | = | $x_5 a\,\hat{\mathbf{x}} + y_5 b\,\hat{\mathbf{y}} + z_5 c\,\hat{\mathbf{z}}$ | (4$a$) | N IV |
| $\mathbf{B}_{18}$ | = | $-x_5\,\mathbf{a}_1 - y_5\,\mathbf{a}_2 + \left(\frac{1}{2} + z_5\right)\,\mathbf{a}_3$ | = | $-x_5 a\,\hat{\mathbf{x}} - y_5 b\,\hat{\mathbf{y}} + \left(\frac{1}{2} + z_5\right) c\,\hat{\mathbf{z}}$ | (4$a$) | N IV |
| $\mathbf{B}_{19}$ | = | $\left(\frac{1}{2} + x_5\right)\,\mathbf{a}_1 - y_5\,\mathbf{a}_2 + z_5\,\mathbf{a}_3$ | = | $\left(\frac{1}{2} + x_5\right) a\,\hat{\mathbf{x}} - y_5 b\,\hat{\mathbf{y}} + z_5 c\,\hat{\mathbf{z}}$ | (4$a$) | N IV |
| $\mathbf{B}_{20}$ | = | $\left(\frac{1}{2} - x_5\right)\,\mathbf{a}_1 + y_5\,\mathbf{a}_2 + \left(\frac{1}{2} + z_5\right)\,\mathbf{a}_3$ | = | $\left(\frac{1}{2} - x_5\right) a\,\hat{\mathbf{x}} + y_5 b\,\hat{\mathbf{y}} + \left(\frac{1}{2} + z_5\right) c\,\hat{\mathbf{z}}$ | (4$a$) | N IV |
| $\mathbf{B}_{21}$ | = | $x_6\,\mathbf{a}_1 + y_6\,\mathbf{a}_2 + z_6\,\mathbf{a}_3$ | = | $x_6 a\,\hat{\mathbf{x}} + y_6 b\,\hat{\mathbf{y}} + z_6 c\,\hat{\mathbf{z}}$ | (4$a$) | N V |
| $\mathbf{B}_{22}$ | = | $-x_6\,\mathbf{a}_1 - y_6\,\mathbf{a}_2 + \left(\frac{1}{2} + z_6\right)\,\mathbf{a}_3$ | = | $-x_6 a\,\hat{\mathbf{x}} - y_6 b\,\hat{\mathbf{y}} + \left(\frac{1}{2} + z_6\right) c\,\hat{\mathbf{z}}$ | (4$a$) | N V |
| $\mathbf{B}_{23}$ | = | $\left(\frac{1}{2} + x_6\right)\,\mathbf{a}_1 - y_6\,\mathbf{a}_2 + z_6\,\mathbf{a}_3$ | = | $\left(\frac{1}{2} + x_6\right) a\,\hat{\mathbf{x}} - y_6 b\,\hat{\mathbf{y}} + z_6 c\,\hat{\mathbf{z}}$ | (4$a$) | N V |
| $\mathbf{B}_{24}$ | = | $\left(\frac{1}{2} - x_6\right)\,\mathbf{a}_1 + y_6\,\mathbf{a}_2 + \left(\frac{1}{2} + z_6\right)\,\mathbf{a}_3$ | = | $\left(\frac{1}{2} - x_6\right) a\,\hat{\mathbf{x}} + y_6 b\,\hat{\mathbf{y}} + \left(\frac{1}{2} + z_6\right) c\,\hat{\mathbf{z}}$ | (4$a$) | N V |
| $\mathbf{B}_{25}$ | = | $x_7\,\mathbf{a}_1 + y_7\,\mathbf{a}_2 + z_7\,\mathbf{a}_3$ | = | $x_7 a\,\hat{\mathbf{x}} + y_7 b\,\hat{\mathbf{y}} + z_7 c\,\hat{\mathbf{z}}$ | (4$a$) | N VI |
| $\mathbf{B}_{26}$ | = | $-x_7\,\mathbf{a}_1 - y_7\,\mathbf{a}_2 + \left(\frac{1}{2} + z_7\right)\,\mathbf{a}_3$ | = | $-x_7 a\,\hat{\mathbf{x}} - y_7 b\,\hat{\mathbf{y}} + \left(\frac{1}{2} + z_7\right) c\,\hat{\mathbf{z}}$ | (4$a$) | N VI |
| $\mathbf{B}_{27}$ | = | $\left(\frac{1}{2} + x_7\right)\,\mathbf{a}_1 - y_7\,\mathbf{a}_2 + z_7\,\mathbf{a}_3$ | = | $\left(\frac{1}{2} + x_7\right) a\,\hat{\mathbf{x}} - y_7 b\,\hat{\mathbf{y}} + z_7 c\,\hat{\mathbf{z}}$ | (4$a$) | N VI |
| $\mathbf{B}_{28}$ | = | $\left(\frac{1}{2} - x_7\right)\,\mathbf{a}_1 + y_7\,\mathbf{a}_2 + \left(\frac{1}{2} + z_7\right)\,\mathbf{a}_3$ | = | $\left(\frac{1}{2} - x_7\right) a\,\hat{\mathbf{x}} + y_7 b\,\hat{\mathbf{y}} + \left(\frac{1}{2} + z_7\right) c\,\hat{\mathbf{z}}$ | (4$a$) | N VI |

**Geometry files:**

- CIF: pp. 1583
- POSCAR: pp. 1583

# Low-Temperature $(NH_3CH_3)Al(SO_4)_2{\cdot}12H_2O$ Structure: ABC30DE20F2_oP220_29_a_a_30a_a_20a_2a

http://aflow.org/prototype-encyclopedia/ABC30DE20F2_oP220_29_a_a_30a_a_20a_2a

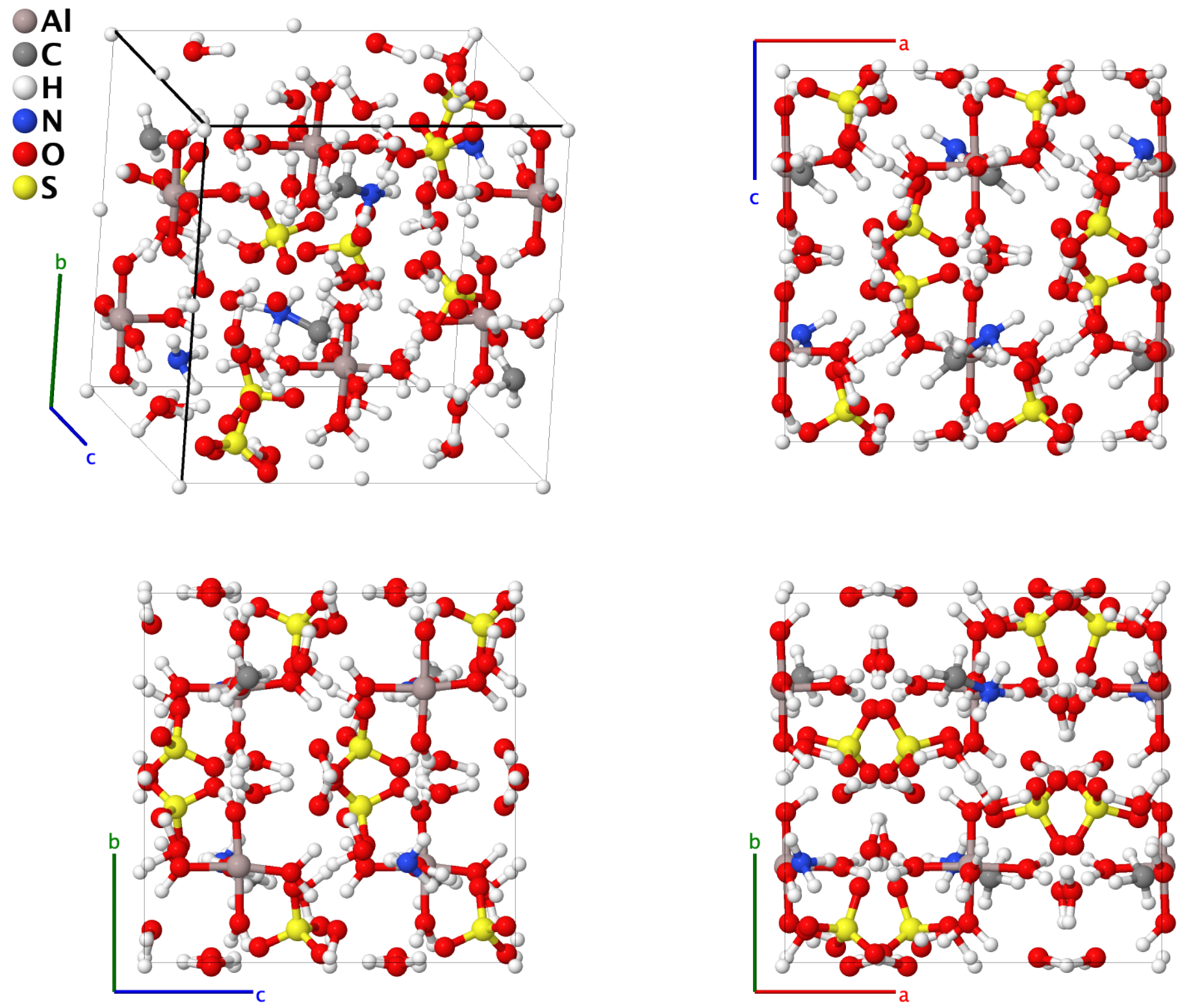

**Prototype** : $AlCH_{30}NO_{20}S_2$

**AFLOW prototype label** : ABC30DE20F2_oP220_29_a_a_30a_a_20a_2a

***Strukturbericht* designation** : None

**Pearson symbol** : oP220

**Space group number** : 29

**Space group symbol** : $Pca2_1$

**AFLOW prototype command** : `aflow --proto=ABC30DE20F2_oP220_29_a_a_30a_a_20a_2a`
`--params=`$a, b/a, c/a, x_1, y_1, z_1, x_2, y_2, z_2, x_3, y_3, z_3, x_4, y_4, z_4, x_5, y_5, z_5, x_6, y_6, z_6, x_7, y_7, z_7, x_8, y_8, z_8, x_9, y_9, z_9, x_{10}, y_{10}, z_{10}, x_{11}, y_{11}, z_{11}, x_{12}, y_{12}, z_{12}, x_{13}, y_{13}, z_{13}, x_{14}, y_{14}, z_{14}, x_{15}, y_{15}, z_{15}, x_{16}, y_{16}, z_{16}, x_{17}, y_{17}, z_{17}, x_{18}, y_{18}, z_{18}, x_{19}, y_{19}, z_{19}, x_{20}, y_{20}, z_{20}, x_{21}, y_{21}, z_{21}, x_{22}, y_{22}, z_{22}, x_{23}, y_{23}, z_{23}, x_{24}, y_{24}, z_{24}, x_{25}, y_{25}, z_{25}, x_{26}, y_{26}, z_{26}, x_{27}, y_{27}, z_{27}, x_{28}, y_{28}, z_{28}, x_{29}, y_{29}, z_{29}, x_{30}, y_{30}, z_{30}, x_{31}, y_{31}, z_{31}, x_{32}, y_{32}, z_{32}, x_{33}, y_{33}, z_{33}, x_{34}, y_{34}, z_{34}, x_{35}, y_{35}, z_{35}, x_{36}, y_{36}, z_{36}, x_{37}, y_{37}, z_{37}, x_{38}, y_{38}, z_{38}, x_{39}, y_{39}, z_{39}, x_{40}, y_{40}, z_{40}, x_{41}, y_{41}, z_{41}, x_{42}, y_{42}, z_{42}, x_{43}, y_{43}, z_{43}, x_{44}, y_{44}, z_{44}, x_{45}, y_{45}, z_{45}, x_{46}, y_{46}, z_{46}, x_{47}, y_{47}, z_{47}, x_{48}, y_{48}, z_{48}, x_{49}, y_{49}, z_{49}, x_{50}, y_{50}, z_{50}, x_{51}, y_{51}, z_{51}, x_{52}, y_{52}, z_{52}, x_{53}, y_{53}, z_{53}, x_{54}, y_{54}, z_{54}, x_{55}, y_{55}, z_{55}$

- The alums have the general formula $AB(XO_4)_2{\cdot}12H_2O$, where $A$ is a monovalent ion, $B$ is a trivalent ion, and $X$ is a chalcogen. In most cases atom $B$ is aluminum and atom $X$ is sulfur, leading to the name alum.
- All alums have their room-temperature form in space group $Pa\bar{3}$ #205, but the bonding between the $A$ and $B$ ions and the $XO_4$ complex can be quite different.

- (Lipson, 1935ab) described three general forms of alum based on the sizes of the monovalent ions. Each of these forms was given a *Strukturbericht* designation by (Gottfried, 1937):
  - $\alpha$-alum, with intermediate sized ions, prototype $KAl(SO_4)_2{\cdot}12H_2O$, $H4_{13}$,
  - $\beta$-alum, with large ions, prototype $(NH_3CH_3)Al(SO_4)_2{\cdot}12H_2O$, $H4_{14}$, and
  - $\gamma$-alum, with small ions, prototype $NaAl(SO_4)_2{\cdot}12H_2O$, $H4_{15}$.
- This classification scheme is not complete, *e.g.*, (Ledsham, 1968) points out that $NaCr(SO_4)_2{\cdot}12H_2O$ does not fit into any of these categories, and that the actual structure depends on the combination of monovalent and trivalent ions.
- As noted above, the $Pa\overline{3}$ structures of alum are the room temperature form. As the temperature decreases the alum structure may transform. For example, in the temperature range 150-170 K, the $\beta$-alum $(NH_3CH_3)Al(SO_4)_2{\cdot}12H_2O$ transforms into this orthorhombic structure with fully ordered $NH_3CH_3$ ions.
- The data presented here was taken at 113 K.

---

**Simple Orthorhombic primitive vectors:**

$$\mathbf{a}_1 = a\,\hat{\mathbf{x}}$$
$$\mathbf{a}_2 = b\,\hat{\mathbf{y}}$$
$$\mathbf{a}_3 = c\,\hat{\mathbf{z}}$$

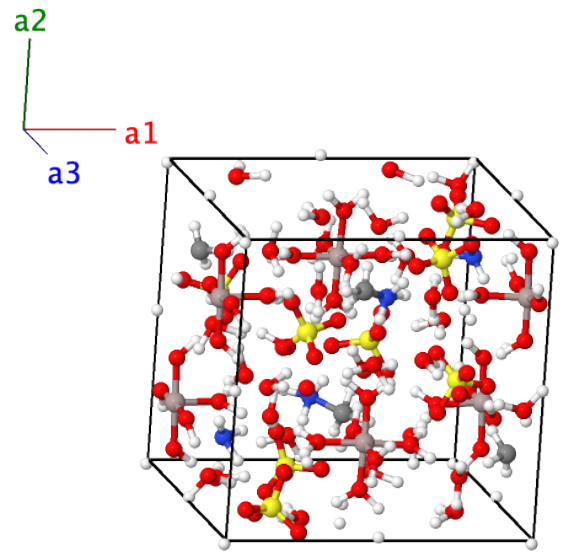

---

**Basis vectors:**

| | | Lattice Coordinates | | Cartesian Coordinates | Wyckoff Position | Atom Type |
|---|---|---|---|---|---|---|
| $\mathbf{B}_1$ | = | $x_1\,\mathbf{a}_1 + y_1\,\mathbf{a}_2 + z_1\,\mathbf{a}_3$ | = | $x_1 a\,\hat{\mathbf{x}} + y_1 b\,\hat{\mathbf{y}} + z_1 c\,\hat{\mathbf{z}}$ | (4$a$) | Al |
| $\mathbf{B}_2$ | = | $-x_1\,\mathbf{a}_1 - y_1\,\mathbf{a}_2 + \left(\frac{1}{2} + z_1\right)\mathbf{a}_3$ | = | $-x_1 a\,\hat{\mathbf{x}} - y_1 b\,\hat{\mathbf{y}} + \left(\frac{1}{2} + z_1\right) c\,\hat{\mathbf{z}}$ | (4$a$) | Al |
| $\mathbf{B}_3$ | = | $\left(\frac{1}{2} + x_1\right)\mathbf{a}_1 - y_1\,\mathbf{a}_2 + z_1\,\mathbf{a}_3$ | = | $\left(\frac{1}{2} + x_1\right) a\,\hat{\mathbf{x}} - y_1 b\,\hat{\mathbf{y}} + z_1 c\,\hat{\mathbf{z}}$ | (4$a$) | Al |
| $\mathbf{B}_4$ | = | $\left(\frac{1}{2} - x_1\right)\mathbf{a}_1 + y_1\,\mathbf{a}_2 + \left(\frac{1}{2} + z_1\right)\mathbf{a}_3$ | = | $\left(\frac{1}{2} - x_1\right) a\,\hat{\mathbf{x}} + y_1 b\,\hat{\mathbf{y}} + \left(\frac{1}{2} + z_1\right) c\,\hat{\mathbf{z}}$ | (4$a$) | Al |
| $\mathbf{B}_5$ | = | $x_2\,\mathbf{a}_1 + y_2\,\mathbf{a}_2 + z_2\,\mathbf{a}_3$ | = | $x_2 a\,\hat{\mathbf{x}} + y_2 b\,\hat{\mathbf{y}} + z_2 c\,\hat{\mathbf{z}}$ | (4$a$) | C |
| $\mathbf{B}_6$ | = | $-x_2\,\mathbf{a}_1 - y_2\,\mathbf{a}_2 + \left(\frac{1}{2} + z_2\right)\mathbf{a}_3$ | = | $-x_2 a\,\hat{\mathbf{x}} - y_2 b\,\hat{\mathbf{y}} + \left(\frac{1}{2} + z_2\right) c\,\hat{\mathbf{z}}$ | (4$a$) | C |
| $\mathbf{B}_7$ | = | $\left(\frac{1}{2} + x_2\right)\mathbf{a}_1 - y_2\,\mathbf{a}_2 + z_2\,\mathbf{a}_3$ | = | $\left(\frac{1}{2} + x_2\right) a\,\hat{\mathbf{x}} - y_2 b\,\hat{\mathbf{y}} + z_2 c\,\hat{\mathbf{z}}$ | (4$a$) | C |
| $\mathbf{B}_8$ | = | $\left(\frac{1}{2} - x_2\right)\mathbf{a}_1 + y_2\,\mathbf{a}_2 + \left(\frac{1}{2} + z_2\right)\mathbf{a}_3$ | = | $\left(\frac{1}{2} - x_2\right) a\,\hat{\mathbf{x}} + y_2 b\,\hat{\mathbf{y}} + \left(\frac{1}{2} + z_2\right) c\,\hat{\mathbf{z}}$ | (4$a$) | C |
| $\mathbf{B}_9$ | = | $x_3\,\mathbf{a}_1 + y_3\,\mathbf{a}_2 + z_3\,\mathbf{a}_3$ | = | $x_3 a\,\hat{\mathbf{x}} + y_3 b\,\hat{\mathbf{y}} + z_3 c\,\hat{\mathbf{z}}$ | (4$a$) | H I |
| $\mathbf{B}_{10}$ | = | $-x_3\,\mathbf{a}_1 - y_3\,\mathbf{a}_2 + \left(\frac{1}{2} + z_3\right)\mathbf{a}_3$ | = | $-x_3 a\,\hat{\mathbf{x}} - y_3 b\,\hat{\mathbf{y}} + \left(\frac{1}{2} + z_3\right) c\,\hat{\mathbf{z}}$ | (4$a$) | H I |
| $\mathbf{B}_{11}$ | = | $\left(\frac{1}{2} + x_3\right)\mathbf{a}_1 - y_3\,\mathbf{a}_2 + z_3\,\mathbf{a}_3$ | = | $\left(\frac{1}{2} + x_3\right) a\,\hat{\mathbf{x}} - y_3 b\,\hat{\mathbf{y}} + z_3 c\,\hat{\mathbf{z}}$ | (4$a$) | H I |
| $\mathbf{B}_{12}$ | = | $\left(\frac{1}{2} - x_3\right)\mathbf{a}_1 + y_3\,\mathbf{a}_2 + \left(\frac{1}{2} + z_3\right)\mathbf{a}_3$ | = | $\left(\frac{1}{2} - x_3\right) a\,\hat{\mathbf{x}} + y_3 b\,\hat{\mathbf{y}} + \left(\frac{1}{2} + z_3\right) c\,\hat{\mathbf{z}}$ | (4$a$) | H I |
| $\mathbf{B}_{13}$ | = | $x_4\,\mathbf{a}_1 + y_4\,\mathbf{a}_2 + z_4\,\mathbf{a}_3$ | = | $x_4 a\,\hat{\mathbf{x}} + y_4 b\,\hat{\mathbf{y}} + z_4 c\,\hat{\mathbf{z}}$ | (4$a$) | H II |
| $\mathbf{B}_{14}$ | = | $-x_4\,\mathbf{a}_1 - y_4\,\mathbf{a}_2 + \left(\frac{1}{2} + z_4\right)\mathbf{a}_3$ | = | $-x_4 a\,\hat{\mathbf{x}} - y_4 b\,\hat{\mathbf{y}} + \left(\frac{1}{2} + z_4\right) c\,\hat{\mathbf{z}}$ | (4$a$) | H II |
| $\mathbf{B}_{15}$ | = | $\left(\frac{1}{2} + x_4\right)\mathbf{a}_1 - y_4\,\mathbf{a}_2 + z_4\,\mathbf{a}_3$ | = | $\left(\frac{1}{2} + x_4\right) a\,\hat{\mathbf{x}} - y_4 b\,\hat{\mathbf{y}} + z_4 c\,\hat{\mathbf{z}}$ | (4$a$) | H II |
| $\mathbf{B}_{16}$ | = | $\left(\frac{1}{2} - x_4\right)\mathbf{a}_1 + y_4\,\mathbf{a}_2 + \left(\frac{1}{2} + z_4\right)\mathbf{a}_3$ | = | $\left(\frac{1}{2} - x_4\right) a\,\hat{\mathbf{x}} + y_4 b\,\hat{\mathbf{y}} + \left(\frac{1}{2} + z_4\right) c\,\hat{\mathbf{z}}$ | (4$a$) | H II |
| $\mathbf{B}_{17}$ | = | $x_5\,\mathbf{a}_1 + y_5\,\mathbf{a}_2 + z_5\,\mathbf{a}_3$ | = | $x_5 a\,\hat{\mathbf{x}} + y_5 b\,\hat{\mathbf{y}} + z_5 c\,\hat{\mathbf{z}}$ | (4$a$) | H III |
| $\mathbf{B}_{18}$ | = | $-x_5\,\mathbf{a}_1 - y_5\,\mathbf{a}_2 + \left(\frac{1}{2}+z_5\right)\,\mathbf{a}_3$ | = | $-x_5 a\,\mathbf{\hat{x}} - y_5 b\,\mathbf{\hat{y}} + \left(\frac{1}{2}+z_5\right) c\,\mathbf{\hat{z}}$ | (4$a$) | H III |
| $\mathbf{B}_{19}$ | = | $\left(\frac{1}{2}+x_5\right)\,\mathbf{a}_1 - y_5\,\mathbf{a}_2 + z_5\,\mathbf{a}_3$ | = | $\left(\frac{1}{2}+x_5\right) a\,\mathbf{\hat{x}} - y_5 b\,\mathbf{\hat{y}} + z_5 c\,\mathbf{\hat{z}}$ | (4$a$) | H III |
| $\mathbf{B}_{20}$ | = | $\left(\frac{1}{2}-x_5\right)\,\mathbf{a}_1 + y_5\,\mathbf{a}_2 + \left(\frac{1}{2}+z_5\right)\,\mathbf{a}_3$ | = | $\left(\frac{1}{2}-x_5\right) a\,\mathbf{\hat{x}} + y_5 b\,\mathbf{\hat{y}} + \left(\frac{1}{2}+z_5\right) c\,\mathbf{\hat{z}}$ | (4$a$) | H III |
| $\mathbf{B}_{21}$ | = | $x_6\,\mathbf{a}_1 + y_6\,\mathbf{a}_2 + z_6\,\mathbf{a}_3$ | = | $x_6 a\,\mathbf{\hat{x}} + y_6 b\,\mathbf{\hat{y}} + z_6 c\,\mathbf{\hat{z}}$ | (4$a$) | H IV |
| $\mathbf{B}_{22}$ | = | $-x_6\,\mathbf{a}_1 - y_6\,\mathbf{a}_2 + \left(\frac{1}{2}+z_6\right)\,\mathbf{a}_3$ | = | $-x_6 a\,\mathbf{\hat{x}} - y_6 b\,\mathbf{\hat{y}} + \left(\frac{1}{2}+z_6\right) c\,\mathbf{\hat{z}}$ | (4$a$) | H IV |
| $\mathbf{B}_{23}$ | = | $\left(\frac{1}{2}+x_6\right)\,\mathbf{a}_1 - y_6\,\mathbf{a}_2 + z_6\,\mathbf{a}_3$ | = | $\left(\frac{1}{2}+x_6\right) a\,\mathbf{\hat{x}} - y_6 b\,\mathbf{\hat{y}} + z_6 c\,\mathbf{\hat{z}}$ | (4$a$) | H IV |
| $\mathbf{B}_{24}$ | = | $\left(\frac{1}{2}-x_6\right)\,\mathbf{a}_1 + y_6\,\mathbf{a}_2 + \left(\frac{1}{2}+z_6\right)\,\mathbf{a}_3$ | = | $\left(\frac{1}{2}-x_6\right) a\,\mathbf{\hat{x}} + y_6 b\,\mathbf{\hat{y}} + \left(\frac{1}{2}+z_6\right) c\,\mathbf{\hat{z}}$ | (4$a$) | H IV |
| $\mathbf{B}_{25}$ | = | $x_7\,\mathbf{a}_1 + y_7\,\mathbf{a}_2 + z_7\,\mathbf{a}_3$ | = | $x_7 a\,\mathbf{\hat{x}} + y_7 b\,\mathbf{\hat{y}} + z_7 c\,\mathbf{\hat{z}}$ | (4$a$) | H V |
| $\mathbf{B}_{26}$ | = | $-x_7\,\mathbf{a}_1 - y_7\,\mathbf{a}_2 + \left(\frac{1}{2}+z_7\right)\,\mathbf{a}_3$ | = | $-x_7 a\,\mathbf{\hat{x}} - y_7 b\,\mathbf{\hat{y}} + \left(\frac{1}{2}+z_7\right) c\,\mathbf{\hat{z}}$ | (4$a$) | H V |
| $\mathbf{B}_{27}$ | = | $\left(\frac{1}{2}+x_7\right)\,\mathbf{a}_1 - y_7\,\mathbf{a}_2 + z_7\,\mathbf{a}_3$ | = | $\left(\frac{1}{2}+x_7\right) a\,\mathbf{\hat{x}} - y_7 b\,\mathbf{\hat{y}} + z_7 c\,\mathbf{\hat{z}}$ | (4$a$) | H V |
| $\mathbf{B}_{28}$ | = | $\left(\frac{1}{2}-x_7\right)\,\mathbf{a}_1 + y_7\,\mathbf{a}_2 + \left(\frac{1}{2}+z_7\right)\,\mathbf{a}_3$ | = | $\left(\frac{1}{2}-x_7\right) a\,\mathbf{\hat{x}} + y_7 b\,\mathbf{\hat{y}} + \left(\frac{1}{2}+z_7\right) c\,\mathbf{\hat{z}}$ | (4$a$) | H V |
| $\mathbf{B}_{29}$ | = | $x_8\,\mathbf{a}_1 + y_8\,\mathbf{a}_2 + z_8\,\mathbf{a}_3$ | = | $x_8 a\,\mathbf{\hat{x}} + y_8 b\,\mathbf{\hat{y}} + z_8 c\,\mathbf{\hat{z}}$ | (4$a$) | H VI |
| $\mathbf{B}_{30}$ | = | $-x_8\,\mathbf{a}_1 - y_8\,\mathbf{a}_2 + \left(\frac{1}{2}+z_8\right)\,\mathbf{a}_3$ | = | $-x_8 a\,\mathbf{\hat{x}} - y_8 b\,\mathbf{\hat{y}} + \left(\frac{1}{2}+z_8\right) c\,\mathbf{\hat{z}}$ | (4$a$) | H VI |
| $\mathbf{B}_{31}$ | = | $\left(\frac{1}{2}+x_8\right)\,\mathbf{a}_1 - y_8\,\mathbf{a}_2 + z_8\,\mathbf{a}_3$ | = | $\left(\frac{1}{2}+x_8\right) a\,\mathbf{\hat{x}} - y_8 b\,\mathbf{\hat{y}} + z_8 c\,\mathbf{\hat{z}}$ | (4$a$) | H VI |
| $\mathbf{B}_{32}$ | = | $\left(\frac{1}{2}-x_8\right)\,\mathbf{a}_1 + y_8\,\mathbf{a}_2 + \left(\frac{1}{2}+z_8\right)\,\mathbf{a}_3$ | = | $\left(\frac{1}{2}-x_8\right) a\,\mathbf{\hat{x}} + y_8 b\,\mathbf{\hat{y}} + \left(\frac{1}{2}+z_8\right) c\,\mathbf{\hat{z}}$ | (4$a$) | H VI |
| $\mathbf{B}_{33}$ | = | $x_9\,\mathbf{a}_1 + y_9\,\mathbf{a}_2 + z_9\,\mathbf{a}_3$ | = | $x_9 a\,\mathbf{\hat{x}} + y_9 b\,\mathbf{\hat{y}} + z_9 c\,\mathbf{\hat{z}}$ | (4$a$) | H VII |
| $\mathbf{B}_{34}$ | = | $-x_9\,\mathbf{a}_1 - y_9\,\mathbf{a}_2 + \left(\frac{1}{2}+z_9\right)\,\mathbf{a}_3$ | = | $-x_9 a\,\mathbf{\hat{x}} - y_9 b\,\mathbf{\hat{y}} + \left(\frac{1}{2}+z_9\right) c\,\mathbf{\hat{z}}$ | (4$a$) | H VII |
| $\mathbf{B}_{35}$ | = | $\left(\frac{1}{2}+x_9\right)\,\mathbf{a}_1 - y_9\,\mathbf{a}_2 + z_9\,\mathbf{a}_3$ | = | $\left(\frac{1}{2}+x_9\right) a\,\mathbf{\hat{x}} - y_9 b\,\mathbf{\hat{y}} + z_9 c\,\mathbf{\hat{z}}$ | (4$a$) | H VII |
| $\mathbf{B}_{36}$ | = | $\left(\frac{1}{2}-x_9\right)\,\mathbf{a}_1 + y_9\,\mathbf{a}_2 + \left(\frac{1}{2}+z_9\right)\,\mathbf{a}_3$ | = | $\left(\frac{1}{2}-x_9\right) a\,\mathbf{\hat{x}} + y_9 b\,\mathbf{\hat{y}} + \left(\frac{1}{2}+z_9\right) c\,\mathbf{\hat{z}}$ | (4$a$) | H VII |
| $\mathbf{B}_{37}$ | = | $x_{10}\,\mathbf{a}_1 + y_{10}\,\mathbf{a}_2 + z_{10}\,\mathbf{a}_3$ | = | $x_{10} a\,\mathbf{\hat{x}} + y_{10} b\,\mathbf{\hat{y}} + z_{10} c\,\mathbf{\hat{z}}$ | (4$a$) | H VIII |
| $\mathbf{B}_{38}$ | = | $-x_{10}\,\mathbf{a}_1 - y_{10}\,\mathbf{a}_2 + \left(\frac{1}{2}+z_{10}\right)\,\mathbf{a}_3$ | = | $-x_{10} a\,\mathbf{\hat{x}} - y_{10} b\,\mathbf{\hat{y}} + \left(\frac{1}{2}+z_{10}\right) c\,\mathbf{\hat{z}}$ | (4$a$) | H VIII |
| $\mathbf{B}_{39}$ | = | $\left(\frac{1}{2}+x_{10}\right)\,\mathbf{a}_1 - y_{10}\,\mathbf{a}_2 + z_{10}\,\mathbf{a}_3$ | = | $\left(\frac{1}{2}+x_{10}\right) a\,\mathbf{\hat{x}} - y_{10} b\,\mathbf{\hat{y}} + z_{10} c\,\mathbf{\hat{z}}$ | (4$a$) | H VIII |
| $\mathbf{B}_{40}$ | = | $\left(\frac{1}{2}-x_{10}\right)\,\mathbf{a}_1 + y_{10}\,\mathbf{a}_2 + \left(\frac{1}{2}+z_{10}\right)\,\mathbf{a}_3$ | = | $\left(\frac{1}{2}-x_{10}\right) a\,\mathbf{\hat{x}} + y_{10} b\,\mathbf{\hat{y}} + \left(\frac{1}{2}+z_{10}\right) c\,\mathbf{\hat{z}}$ | (4$a$) | H VIII |
| $\mathbf{B}_{41}$ | = | $x_{11}\,\mathbf{a}_1 + y_{11}\,\mathbf{a}_2 + z_{11}\,\mathbf{a}_3$ | = | $x_{11} a\,\mathbf{\hat{x}} + y_{11} b\,\mathbf{\hat{y}} + z_{11} c\,\mathbf{\hat{z}}$ | (4$a$) | H IX |
| $\mathbf{B}_{42}$ | = | $-x_{11}\,\mathbf{a}_1 - y_{11}\,\mathbf{a}_2 + \left(\frac{1}{2}+z_{11}\right)\,\mathbf{a}_3$ | = | $-x_{11} a\,\mathbf{\hat{x}} - y_{11} b\,\mathbf{\hat{y}} + \left(\frac{1}{2}+z_{11}\right) c\,\mathbf{\hat{z}}$ | (4$a$) | H IX |
| $\mathbf{B}_{43}$ | = | $\left(\frac{1}{2}+x_{11}\right)\,\mathbf{a}_1 - y_{11}\,\mathbf{a}_2 + z_{11}\,\mathbf{a}_3$ | = | $\left(\frac{1}{2}+x_{11}\right) a\,\mathbf{\hat{x}} - y_{11} b\,\mathbf{\hat{y}} + z_{11} c\,\mathbf{\hat{z}}$ | (4$a$) | H IX |
| $\mathbf{B}_{44}$ | = | $\left(\frac{1}{2}-x_{11}\right)\,\mathbf{a}_1 + y_{11}\,\mathbf{a}_2 + \left(\frac{1}{2}+z_{11}\right)\,\mathbf{a}_3$ | = | $\left(\frac{1}{2}-x_{11}\right) a\,\mathbf{\hat{x}} + y_{11} b\,\mathbf{\hat{y}} + \left(\frac{1}{2}+z_{11}\right) c\,\mathbf{\hat{z}}$ | (4$a$) | H IX |
| $\mathbf{B}_{45}$ | = | $x_{12}\,\mathbf{a}_1 + y_{12}\,\mathbf{a}_2 + z_{12}\,\mathbf{a}_3$ | = | $x_{12} a\,\mathbf{\hat{x}} + y_{12} b\,\mathbf{\hat{y}} + z_{12} c\,\mathbf{\hat{z}}$ | (4$a$) | H X |
| $\mathbf{B}_{46}$ | = | $-x_{12}\,\mathbf{a}_1 - y_{12}\,\mathbf{a}_2 + \left(\frac{1}{2}+z_{12}\right)\,\mathbf{a}_3$ | = | $-x_{12} a\,\mathbf{\hat{x}} - y_{12} b\,\mathbf{\hat{y}} + \left(\frac{1}{2}+z_{12}\right) c\,\mathbf{\hat{z}}$ | (4$a$) | H X |
| $\mathbf{B}_{47}$ | = | $\left(\frac{1}{2}+x_{12}\right)\,\mathbf{a}_1 - y_{12}\,\mathbf{a}_2 + z_{12}\,\mathbf{a}_3$ | = | $\left(\frac{1}{2}+x_{12}\right) a\,\mathbf{\hat{x}} - y_{12} b\,\mathbf{\hat{y}} + z_{12} c\,\mathbf{\hat{z}}$ | (4$a$) | H X |
| $\mathbf{B}_{48}$ | = | $\left(\frac{1}{2}-x_{12}\right)\,\mathbf{a}_1 + y_{12}\,\mathbf{a}_2 + \left(\frac{1}{2}+z_{12}\right)\,\mathbf{a}_3$ | = | $\left(\frac{1}{2}-x_{12}\right) a\,\mathbf{\hat{x}} + y_{12} b\,\mathbf{\hat{y}} + \left(\frac{1}{2}+z_{12}\right) c\,\mathbf{\hat{z}}$ | (4$a$) | H X |
| $\mathbf{B}_{49}$ | = | $x_{13}\,\mathbf{a}_1 + y_{13}\,\mathbf{a}_2 + z_{13}\,\mathbf{a}_3$ | = | $x_{13} a\,\mathbf{\hat{x}} + y_{13} b\,\mathbf{\hat{y}} + z_{13} c\,\mathbf{\hat{z}}$ | (4$a$) | H XI |
| $\mathbf{B}_{50}$ | = | $-x_{13}\,\mathbf{a}_1 - y_{13}\,\mathbf{a}_2 + \left(\frac{1}{2}+z_{13}\right)\,\mathbf{a}_3$ | = | $-x_{13} a\,\mathbf{\hat{x}} - y_{13} b\,\mathbf{\hat{y}} + \left(\frac{1}{2}+z_{13}\right) c\,\mathbf{\hat{z}}$ | (4$a$) | H XI |
| $\mathbf{B}_{51}$ | = | $\left(\frac{1}{2}+x_{13}\right)\,\mathbf{a}_1 - y_{13}\,\mathbf{a}_2 + z_{13}\,\mathbf{a}_3$ | = | $\left(\frac{1}{2}+x_{13}\right) a\,\mathbf{\hat{x}} - y_{13} b\,\mathbf{\hat{y}} + z_{13} c\,\mathbf{\hat{z}}$ | (4$a$) | H XI |
| $\mathbf{B}_{52}$ | = | $\left(\frac{1}{2}-x_{13}\right)\,\mathbf{a}_1 + y_{13}\,\mathbf{a}_2 + \left(\frac{1}{2}+z_{13}\right)\,\mathbf{a}_3$ | = | $\left(\frac{1}{2}-x_{13}\right) a\,\mathbf{\hat{x}} + y_{13} b\,\mathbf{\hat{y}} + \left(\frac{1}{2}+z_{13}\right) c\,\mathbf{\hat{z}}$ | (4$a$) | H XI |
| $\mathbf{B}_{53}$ | = | $x_{14}\,\mathbf{a}_1 + y_{14}\,\mathbf{a}_2 + z_{14}\,\mathbf{a}_3$ | = | $x_{14} a\,\mathbf{\hat{x}} + y_{14} b\,\mathbf{\hat{y}} + z_{14} c\,\mathbf{\hat{z}}$ | (4$a$) | H XII |
| $\mathbf{B}_{54}$ | = | $-x_{14}\,\mathbf{a}_1 - y_{14}\,\mathbf{a}_2 + \left(\frac{1}{2} + z_{14}\right)\mathbf{a}_3$ | = | $-x_{14}a\,\hat{\mathbf{x}} - y_{14}b\,\hat{\mathbf{y}} + \left(\frac{1}{2} + z_{14}\right)c\,\hat{\mathbf{z}}$ | (4$a$) | H XII |
| $\mathbf{B}_{55}$ | = | $\left(\frac{1}{2} + x_{14}\right)\mathbf{a}_1 - y_{14}\,\mathbf{a}_2 + z_{14}\,\mathbf{a}_3$ | = | $\left(\frac{1}{2} + x_{14}\right)a\,\hat{\mathbf{x}} - y_{14}b\,\hat{\mathbf{y}} + z_{14}c\,\hat{\mathbf{z}}$ | (4$a$) | H XII |
| $\mathbf{B}_{56}$ | = | $\left(\frac{1}{2} - x_{14}\right)\mathbf{a}_1 + y_{14}\,\mathbf{a}_2 + \left(\frac{1}{2} + z_{14}\right)\mathbf{a}_3$ | = | $\left(\frac{1}{2} - x_{14}\right)a\,\hat{\mathbf{x}} + y_{14}b\,\hat{\mathbf{y}} + \left(\frac{1}{2} + z_{14}\right)c\,\hat{\mathbf{z}}$ | (4$a$) | H XII |
| $\mathbf{B}_{57}$ | = | $x_{15}\,\mathbf{a}_1 + y_{15}\,\mathbf{a}_2 + z_{15}\,\mathbf{a}_3$ | = | $x_{15}a\,\hat{\mathbf{x}} + y_{15}b\,\hat{\mathbf{y}} + z_{15}c\,\hat{\mathbf{z}}$ | (4$a$) | H XIII |
| $\mathbf{B}_{58}$ | = | $-x_{15}\,\mathbf{a}_1 - y_{15}\,\mathbf{a}_2 + \left(\frac{1}{2} + z_{15}\right)\mathbf{a}_3$ | = | $-x_{15}a\,\hat{\mathbf{x}} - y_{15}b\,\hat{\mathbf{y}} + \left(\frac{1}{2} + z_{15}\right)c\,\hat{\mathbf{z}}$ | (4$a$) | H XIII |
| $\mathbf{B}_{59}$ | = | $\left(\frac{1}{2} + x_{15}\right)\mathbf{a}_1 - y_{15}\,\mathbf{a}_2 + z_{15}\,\mathbf{a}_3$ | = | $\left(\frac{1}{2} + x_{15}\right)a\,\hat{\mathbf{x}} - y_{15}b\,\hat{\mathbf{y}} + z_{15}c\,\hat{\mathbf{z}}$ | (4$a$) | H XIII |
| $\mathbf{B}_{60}$ | = | $\left(\frac{1}{2} - x_{15}\right)\mathbf{a}_1 + y_{15}\,\mathbf{a}_2 + \left(\frac{1}{2} + z_{15}\right)\mathbf{a}_3$ | = | $\left(\frac{1}{2} - x_{15}\right)a\,\hat{\mathbf{x}} + y_{15}b\,\hat{\mathbf{y}} + \left(\frac{1}{2} + z_{15}\right)c\,\hat{\mathbf{z}}$ | (4$a$) | H XIII |
| $\mathbf{B}_{61}$ | = | $x_{16}\,\mathbf{a}_1 + y_{16}\,\mathbf{a}_2 + z_{16}\,\mathbf{a}_3$ | = | $x_{16}a\,\hat{\mathbf{x}} + y_{16}b\,\hat{\mathbf{y}} + z_{16}c\,\hat{\mathbf{z}}$ | (4$a$) | H XIV |
| $\mathbf{B}_{62}$ | = | $-x_{16}\,\mathbf{a}_1 - y_{16}\,\mathbf{a}_2 + \left(\frac{1}{2} + z_{16}\right)\mathbf{a}_3$ | = | $-x_{16}a\,\hat{\mathbf{x}} - y_{16}b\,\hat{\mathbf{y}} + \left(\frac{1}{2} + z_{16}\right)c\,\hat{\mathbf{z}}$ | (4$a$) | H XIV |
| $\mathbf{B}_{63}$ | = | $\left(\frac{1}{2} + x_{16}\right)\mathbf{a}_1 - y_{16}\,\mathbf{a}_2 + z_{16}\,\mathbf{a}_3$ | = | $\left(\frac{1}{2} + x_{16}\right)a\,\hat{\mathbf{x}} - y_{16}b\,\hat{\mathbf{y}} + z_{16}c\,\hat{\mathbf{z}}$ | (4$a$) | H XIV |
| $\mathbf{B}_{64}$ | = | $\left(\frac{1}{2} - x_{16}\right)\mathbf{a}_1 + y_{16}\,\mathbf{a}_2 + \left(\frac{1}{2} + z_{16}\right)\mathbf{a}_3$ | = | $\left(\frac{1}{2} - x_{16}\right)a\,\hat{\mathbf{x}} + y_{16}b\,\hat{\mathbf{y}} + \left(\frac{1}{2} + z_{16}\right)c\,\hat{\mathbf{z}}$ | (4$a$) | H XIV |
| $\mathbf{B}_{65}$ | = | $x_{17}\,\mathbf{a}_1 + y_{17}\,\mathbf{a}_2 + z_{17}\,\mathbf{a}_3$ | = | $x_{17}a\,\hat{\mathbf{x}} + y_{17}b\,\hat{\mathbf{y}} + z_{17}c\,\hat{\mathbf{z}}$ | (4$a$) | H XV |
| $\mathbf{B}_{66}$ | = | $-x_{17}\,\mathbf{a}_1 - y_{17}\,\mathbf{a}_2 + \left(\frac{1}{2} + z_{17}\right)\mathbf{a}_3$ | = | $-x_{17}a\,\hat{\mathbf{x}} - y_{17}b\,\hat{\mathbf{y}} + \left(\frac{1}{2} + z_{17}\right)c\,\hat{\mathbf{z}}$ | (4$a$) | H XV |
| $\mathbf{B}_{67}$ | = | $\left(\frac{1}{2} + x_{17}\right)\mathbf{a}_1 - y_{17}\,\mathbf{a}_2 + z_{17}\,\mathbf{a}_3$ | = | $\left(\frac{1}{2} + x_{17}\right)a\,\hat{\mathbf{x}} - y_{17}b\,\hat{\mathbf{y}} + z_{17}c\,\hat{\mathbf{z}}$ | (4$a$) | H XV |
| $\mathbf{B}_{68}$ | = | $\left(\frac{1}{2} - x_{17}\right)\mathbf{a}_1 + y_{17}\,\mathbf{a}_2 + \left(\frac{1}{2} + z_{17}\right)\mathbf{a}_3$ | = | $\left(\frac{1}{2} - x_{17}\right)a\,\hat{\mathbf{x}} + y_{17}b\,\hat{\mathbf{y}} + \left(\frac{1}{2} + z_{17}\right)c\,\hat{\mathbf{z}}$ | (4$a$) | H XV |
| $\mathbf{B}_{69}$ | = | $x_{18}\,\mathbf{a}_1 + y_{18}\,\mathbf{a}_2 + z_{18}\,\mathbf{a}_3$ | = | $x_{18}a\,\hat{\mathbf{x}} + y_{18}b\,\hat{\mathbf{y}} + z_{18}c\,\hat{\mathbf{z}}$ | (4$a$) | H XVI |
| $\mathbf{B}_{70}$ | = | $-x_{18}\,\mathbf{a}_1 - y_{18}\,\mathbf{a}_2 + \left(\frac{1}{2} + z_{18}\right)\mathbf{a}_3$ | = | $-x_{18}a\,\hat{\mathbf{x}} - y_{18}b\,\hat{\mathbf{y}} + \left(\frac{1}{2} + z_{18}\right)c\,\hat{\mathbf{z}}$ | (4$a$) | H XVI |
| $\mathbf{B}_{71}$ | = | $\left(\frac{1}{2} + x_{18}\right)\mathbf{a}_1 - y_{18}\,\mathbf{a}_2 + z_{18}\,\mathbf{a}_3$ | = | $\left(\frac{1}{2} + x_{18}\right)a\,\hat{\mathbf{x}} - y_{18}b\,\hat{\mathbf{y}} + z_{18}c\,\hat{\mathbf{z}}$ | (4$a$) | H XVI |
| $\mathbf{B}_{72}$ | = | $\left(\frac{1}{2} - x_{18}\right)\mathbf{a}_1 + y_{18}\,\mathbf{a}_2 + \left(\frac{1}{2} + z_{18}\right)\mathbf{a}_3$ | = | $\left(\frac{1}{2} - x_{18}\right)a\,\hat{\mathbf{x}} + y_{18}b\,\hat{\mathbf{y}} + \left(\frac{1}{2} + z_{18}\right)c\,\hat{\mathbf{z}}$ | (4$a$) | H XVI |
| $\mathbf{B}_{73}$ | = | $x_{19}\,\mathbf{a}_1 + y_{19}\,\mathbf{a}_2 + z_{19}\,\mathbf{a}_3$ | = | $x_{19}a\,\hat{\mathbf{x}} + y_{19}b\,\hat{\mathbf{y}} + z_{19}c\,\hat{\mathbf{z}}$ | (4$a$) | H XVII |
| $\mathbf{B}_{74}$ | = | $-x_{19}\,\mathbf{a}_1 - y_{19}\,\mathbf{a}_2 + \left(\frac{1}{2} + z_{19}\right)\mathbf{a}_3$ | = | $-x_{19}a\,\hat{\mathbf{x}} - y_{19}b\,\hat{\mathbf{y}} + \left(\frac{1}{2} + z_{19}\right)c\,\hat{\mathbf{z}}$ | (4$a$) | H XVII |
| $\mathbf{B}_{75}$ | = | $\left(\frac{1}{2} + x_{19}\right)\mathbf{a}_1 - y_{19}\,\mathbf{a}_2 + z_{19}\,\mathbf{a}_3$ | = | $\left(\frac{1}{2} + x_{19}\right)a\,\hat{\mathbf{x}} - y_{19}b\,\hat{\mathbf{y}} + z_{19}c\,\hat{\mathbf{z}}$ | (4$a$) | H XVII |
| $\mathbf{B}_{76}$ | = | $\left(\frac{1}{2} - x_{19}\right)\mathbf{a}_1 + y_{19}\,\mathbf{a}_2 + \left(\frac{1}{2} + z_{19}\right)\mathbf{a}_3$ | = | $\left(\frac{1}{2} - x_{19}\right)a\,\hat{\mathbf{x}} + y_{19}b\,\hat{\mathbf{y}} + \left(\frac{1}{2} + z_{19}\right)c\,\hat{\mathbf{z}}$ | (4$a$) | H XVII |
| $\mathbf{B}_{77}$ | = | $x_{20}\,\mathbf{a}_1 + y_{20}\,\mathbf{a}_2 + z_{20}\,\mathbf{a}_3$ | = | $x_{20}a\,\hat{\mathbf{x}} + y_{20}b\,\hat{\mathbf{y}} + z_{20}c\,\hat{\mathbf{z}}$ | (4$a$) | H XVIII |
| $\mathbf{B}_{78}$ | = | $-x_{20}\,\mathbf{a}_1 - y_{20}\,\mathbf{a}_2 + \left(\frac{1}{2} + z_{20}\right)\mathbf{a}_3$ | = | $-x_{20}a\,\hat{\mathbf{x}} - y_{20}b\,\hat{\mathbf{y}} + \left(\frac{1}{2} + z_{20}\right)c\,\hat{\mathbf{z}}$ | (4$a$) | H XVIII |
| $\mathbf{B}_{79}$ | = | $\left(\frac{1}{2} + x_{20}\right)\mathbf{a}_1 - y_{20}\,\mathbf{a}_2 + z_{20}\,\mathbf{a}_3$ | = | $\left(\frac{1}{2} + x_{20}\right)a\,\hat{\mathbf{x}} - y_{20}b\,\hat{\mathbf{y}} + z_{20}c\,\hat{\mathbf{z}}$ | (4$a$) | H XVIII |
| $\mathbf{B}_{80}$ | = | $\left(\frac{1}{2} - x_{20}\right)\mathbf{a}_1 + y_{20}\,\mathbf{a}_2 + \left(\frac{1}{2} + z_{20}\right)\mathbf{a}_3$ | = | $\left(\frac{1}{2} - x_{20}\right)a\,\hat{\mathbf{x}} + y_{20}b\,\hat{\mathbf{y}} + \left(\frac{1}{2} + z_{20}\right)c\,\hat{\mathbf{z}}$ | (4$a$) | H XVIII |
| $\mathbf{B}_{81}$ | = | $x_{21}\,\mathbf{a}_1 + y_{21}\,\mathbf{a}_2 + z_{21}\,\mathbf{a}_3$ | = | $x_{21}a\,\hat{\mathbf{x}} + y_{21}b\,\hat{\mathbf{y}} + z_{21}c\,\hat{\mathbf{z}}$ | (4$a$) | H XIX |
| $\mathbf{B}_{82}$ | = | $-x_{21}\,\mathbf{a}_1 - y_{21}\,\mathbf{a}_2 + \left(\frac{1}{2} + z_{21}\right)\mathbf{a}_3$ | = | $-x_{21}a\,\hat{\mathbf{x}} - y_{21}b\,\hat{\mathbf{y}} + \left(\frac{1}{2} + z_{21}\right)c\,\hat{\mathbf{z}}$ | (4$a$) | H XIX |
| $\mathbf{B}_{83}$ | = | $\left(\frac{1}{2} + x_{21}\right)\mathbf{a}_1 - y_{21}\,\mathbf{a}_2 + z_{21}\,\mathbf{a}_3$ | = | $\left(\frac{1}{2} + x_{21}\right)a\,\hat{\mathbf{x}} - y_{21}b\,\hat{\mathbf{y}} + z_{21}c\,\hat{\mathbf{z}}$ | (4$a$) | H XIX |
| $\mathbf{B}_{84}$ | = | $\left(\frac{1}{2} - x_{21}\right)\mathbf{a}_1 + y_{21}\,\mathbf{a}_2 + \left(\frac{1}{2} + z_{21}\right)\mathbf{a}_3$ | = | $\left(\frac{1}{2} - x_{21}\right)a\,\hat{\mathbf{x}} + y_{21}b\,\hat{\mathbf{y}} + \left(\frac{1}{2} + z_{21}\right)c\,\hat{\mathbf{z}}$ | (4$a$) | H XIX |
| $\mathbf{B}_{85}$ | = | $x_{22}\,\mathbf{a}_1 + y_{22}\,\mathbf{a}_2 + z_{22}\,\mathbf{a}_3$ | = | $x_{22}a\,\hat{\mathbf{x}} + y_{22}b\,\hat{\mathbf{y}} + z_{22}c\,\hat{\mathbf{z}}$ | (4$a$) | H XX |
| $\mathbf{B}_{86}$ | = | $-x_{22}\,\mathbf{a}_1 - y_{22}\,\mathbf{a}_2 + \left(\frac{1}{2} + z_{22}\right)\mathbf{a}_3$ | = | $-x_{22}a\,\hat{\mathbf{x}} - y_{22}b\,\hat{\mathbf{y}} + \left(\frac{1}{2} + z_{22}\right)c\,\hat{\mathbf{z}}$ | (4$a$) | H XX |
| $\mathbf{B}_{87}$ | = | $\left(\frac{1}{2} + x_{22}\right)\mathbf{a}_1 - y_{22}\,\mathbf{a}_2 + z_{22}\,\mathbf{a}_3$ | = | $\left(\frac{1}{2} + x_{22}\right)a\,\hat{\mathbf{x}} - y_{22}b\,\hat{\mathbf{y}} + z_{22}c\,\hat{\mathbf{z}}$ | (4$a$) | H XX |
| $\mathbf{B}_{88}$ | = | $\left(\frac{1}{2} - x_{22}\right)\mathbf{a}_1 + y_{22}\,\mathbf{a}_2 + \left(\frac{1}{2} + z_{22}\right)\mathbf{a}_3$ | = | $\left(\frac{1}{2} - x_{22}\right)a\,\hat{\mathbf{x}} + y_{22}b\,\hat{\mathbf{y}} + \left(\frac{1}{2} + z_{22}\right)c\,\hat{\mathbf{z}}$ | (4$a$) | H XX |
| $\mathbf{B}_{89}$ | = | $x_{23}\,\mathbf{a}_1 + y_{23}\,\mathbf{a}_2 + z_{23}\,\mathbf{a}_3$ | = | $x_{23}a\,\hat{\mathbf{x}} + y_{23}b\,\hat{\mathbf{y}} + z_{23}c\,\hat{\mathbf{z}}$ | (4$a$) | H XXI |

$$
\begin{array}{llllll}
\mathbf{B}_{90} & = & -x_{23}\,\mathbf{a}_1 - y_{23}\,\mathbf{a}_2 + \left(\frac{1}{2} + z_{23}\right)\,\mathbf{a}_3 & = & -x_{23}a\,\hat{\mathbf{x}} - y_{23}b\,\hat{\mathbf{y}} + \left(\frac{1}{2} + z_{23}\right)c\,\hat{\mathbf{z}} & (4a) & \text{H XXI} \\
\mathbf{B}_{91} & = & \left(\frac{1}{2} + x_{23}\right)\,\mathbf{a}_1 - y_{23}\,\mathbf{a}_2 + z_{23}\,\mathbf{a}_3 & = & \left(\frac{1}{2} + x_{23}\right)a\,\hat{\mathbf{x}} - y_{23}b\,\hat{\mathbf{y}} + z_{23}c\,\hat{\mathbf{z}} & (4a) & \text{H XXI} \\
\mathbf{B}_{92} & = & \left(\frac{1}{2} - x_{23}\right)\,\mathbf{a}_1 + y_{23}\,\mathbf{a}_2 + \left(\frac{1}{2} + z_{23}\right)\,\mathbf{a}_3 & = & \left(\frac{1}{2} - x_{23}\right)a\,\hat{\mathbf{x}} + y_{23}b\,\hat{\mathbf{y}} + \left(\frac{1}{2} + z_{23}\right)c\,\hat{\mathbf{z}} & (4a) & \text{H XXI} \\
\mathbf{B}_{93} & = & x_{24}\,\mathbf{a}_1 + y_{24}\,\mathbf{a}_2 + z_{24}\,\mathbf{a}_3 & = & x_{24}a\,\hat{\mathbf{x}} + y_{24}b\,\hat{\mathbf{y}} + z_{24}c\,\hat{\mathbf{z}} & (4a) & \text{H XXII} \\
\mathbf{B}_{94} & = & -x_{24}\,\mathbf{a}_1 - y_{24}\,\mathbf{a}_2 + \left(\frac{1}{2} + z_{24}\right)\,\mathbf{a}_3 & = & -x_{24}a\,\hat{\mathbf{x}} - y_{24}b\,\hat{\mathbf{y}} + \left(\frac{1}{2} + z_{24}\right)c\,\hat{\mathbf{z}} & (4a) & \text{H XXII} \\
\mathbf{B}_{95} & = & \left(\frac{1}{2} + x_{24}\right)\,\mathbf{a}_1 - y_{24}\,\mathbf{a}_2 + z_{24}\,\mathbf{a}_3 & = & \left(\frac{1}{2} + x_{24}\right)a\,\hat{\mathbf{x}} - y_{24}b\,\hat{\mathbf{y}} + z_{24}c\,\hat{\mathbf{z}} & (4a) & \text{H XXII} \\
\mathbf{B}_{96} & = & \left(\frac{1}{2} - x_{24}\right)\,\mathbf{a}_1 + y_{24}\,\mathbf{a}_2 + \left(\frac{1}{2} + z_{24}\right)\,\mathbf{a}_3 & = & \left(\frac{1}{2} - x_{24}\right)a\,\hat{\mathbf{x}} + y_{24}b\,\hat{\mathbf{y}} + \left(\frac{1}{2} + z_{24}\right)c\,\hat{\mathbf{z}} & (4a) & \text{H XXII} \\
\mathbf{B}_{97} & = & x_{25}\,\mathbf{a}_1 + y_{25}\,\mathbf{a}_2 + z_{25}\,\mathbf{a}_3 & = & x_{25}a\,\hat{\mathbf{x}} + y_{25}b\,\hat{\mathbf{y}} + z_{25}c\,\hat{\mathbf{z}} & (4a) & \text{H XXIII} \\
\mathbf{B}_{98} & = & -x_{25}\,\mathbf{a}_1 - y_{25}\,\mathbf{a}_2 + \left(\frac{1}{2} + z_{25}\right)\,\mathbf{a}_3 & = & -x_{25}a\,\hat{\mathbf{x}} - y_{25}b\,\hat{\mathbf{y}} + \left(\frac{1}{2} + z_{25}\right)c\,\hat{\mathbf{z}} & (4a) & \text{H XXIII} \\
\mathbf{B}_{99} & = & \left(\frac{1}{2} + x_{25}\right)\,\mathbf{a}_1 - y_{25}\,\mathbf{a}_2 + z_{25}\,\mathbf{a}_3 & = & \left(\frac{1}{2} + x_{25}\right)a\,\hat{\mathbf{x}} - y_{25}b\,\hat{\mathbf{y}} + z_{25}c\,\hat{\mathbf{z}} & (4a) & \text{H XXIII} \\
\mathbf{B}_{100} & = & \left(\frac{1}{2} - x_{25}\right)\,\mathbf{a}_1 + y_{25}\,\mathbf{a}_2 + \left(\frac{1}{2} + z_{25}\right)\,\mathbf{a}_3 & = & \left(\frac{1}{2} - x_{25}\right)a\,\hat{\mathbf{x}} + y_{25}b\,\hat{\mathbf{y}} + \left(\frac{1}{2} + z_{25}\right)c\,\hat{\mathbf{z}} & (4a) & \text{H XXIII} \\
\mathbf{B}_{101} & = & x_{26}\,\mathbf{a}_1 + y_{26}\,\mathbf{a}_2 + z_{26}\,\mathbf{a}_3 & = & x_{26}a\,\hat{\mathbf{x}} + y_{26}b\,\hat{\mathbf{y}} + z_{26}c\,\hat{\mathbf{z}} & (4a) & \text{H XXIV} \\
\mathbf{B}_{102} & = & -x_{26}\,\mathbf{a}_1 - y_{26}\,\mathbf{a}_2 + \left(\frac{1}{2} + z_{26}\right)\,\mathbf{a}_3 & = & -x_{26}a\,\hat{\mathbf{x}} - y_{26}b\,\hat{\mathbf{y}} + \left(\frac{1}{2} + z_{26}\right)c\,\hat{\mathbf{z}} & (4a) & \text{H XXIV} \\
\mathbf{B}_{103} & = & \left(\frac{1}{2} + x_{26}\right)\,\mathbf{a}_1 - y_{26}\,\mathbf{a}_2 + z_{26}\,\mathbf{a}_3 & = & \left(\frac{1}{2} + x_{26}\right)a\,\hat{\mathbf{x}} - y_{26}b\,\hat{\mathbf{y}} + z_{26}c\,\hat{\mathbf{z}} & (4a) & \text{H XXIV} \\
\mathbf{B}_{104} & = & \left(\frac{1}{2} - x_{26}\right)\,\mathbf{a}_1 + y_{26}\,\mathbf{a}_2 + \left(\frac{1}{2} + z_{26}\right)\,\mathbf{a}_3 & = & \left(\frac{1}{2} - x_{26}\right)a\,\hat{\mathbf{x}} + y_{26}b\,\hat{\mathbf{y}} + \left(\frac{1}{2} + z_{26}\right)c\,\hat{\mathbf{z}} & (4a) & \text{H XXIV} \\
\mathbf{B}_{105} & = & x_{27}\,\mathbf{a}_1 + y_{27}\,\mathbf{a}_2 + z_{27}\,\mathbf{a}_3 & = & x_{27}a\,\hat{\mathbf{x}} + y_{27}b\,\hat{\mathbf{y}} + z_{27}c\,\hat{\mathbf{z}} & (4a) & \text{H XXV} \\
\mathbf{B}_{106} & = & -x_{27}\,\mathbf{a}_1 - y_{27}\,\mathbf{a}_2 + \left(\frac{1}{2} + z_{27}\right)\,\mathbf{a}_3 & = & -x_{27}a\,\hat{\mathbf{x}} - y_{27}b\,\hat{\mathbf{y}} + \left(\frac{1}{2} + z_{27}\right)c\,\hat{\mathbf{z}} & (4a) & \text{H XXV} \\
\mathbf{B}_{107} & = & \left(\frac{1}{2} + x_{27}\right)\,\mathbf{a}_1 - y_{27}\,\mathbf{a}_2 + z_{27}\,\mathbf{a}_3 & = & \left(\frac{1}{2} + x_{27}\right)a\,\hat{\mathbf{x}} - y_{27}b\,\hat{\mathbf{y}} + z_{27}c\,\hat{\mathbf{z}} & (4a) & \text{H XXV} \\
\mathbf{B}_{108} & = & \left(\frac{1}{2} - x_{27}\right)\,\mathbf{a}_1 + y_{27}\,\mathbf{a}_2 + \left(\frac{1}{2} + z_{27}\right)\,\mathbf{a}_3 & = & \left(\frac{1}{2} - x_{27}\right)a\,\hat{\mathbf{x}} + y_{27}b\,\hat{\mathbf{y}} + \left(\frac{1}{2} + z_{27}\right)c\,\hat{\mathbf{z}} & (4a) & \text{H XXV} \\
\mathbf{B}_{109} & = & x_{28}\,\mathbf{a}_1 + y_{28}\,\mathbf{a}_2 + z_{28}\,\mathbf{a}_3 & = & x_{28}a\,\hat{\mathbf{x}} + y_{28}b\,\hat{\mathbf{y}} + z_{28}c\,\hat{\mathbf{z}} & (4a) & \text{H XXVI} \\
\mathbf{B}_{110} & = & -x_{28}\,\mathbf{a}_1 - y_{28}\,\mathbf{a}_2 + \left(\frac{1}{2} + z_{28}\right)\,\mathbf{a}_3 & = & -x_{28}a\,\hat{\mathbf{x}} - y_{28}b\,\hat{\mathbf{y}} + \left(\frac{1}{2} + z_{28}\right)c\,\hat{\mathbf{z}} & (4a) & \text{H XXVI} \\
\mathbf{B}_{111} & = & \left(\frac{1}{2} + x_{28}\right)\,\mathbf{a}_1 - y_{28}\,\mathbf{a}_2 + z_{28}\,\mathbf{a}_3 & = & \left(\frac{1}{2} + x_{28}\right)a\,\hat{\mathbf{x}} - y_{28}b\,\hat{\mathbf{y}} + z_{28}c\,\hat{\mathbf{z}} & (4a) & \text{H XXVI} \\
\mathbf{B}_{112} & = & \left(\frac{1}{2} - x_{28}\right)\,\mathbf{a}_1 + y_{28}\,\mathbf{a}_2 + \left(\frac{1}{2} + z_{28}\right)\,\mathbf{a}_3 & = & \left(\frac{1}{2} - x_{28}\right)a\,\hat{\mathbf{x}} + y_{28}b\,\hat{\mathbf{y}} + \left(\frac{1}{2} + z_{28}\right)c\,\hat{\mathbf{z}} & (4a) & \text{H XXVI} \\
\mathbf{B}_{113} & = & x_{29}\,\mathbf{a}_1 + y_{29}\,\mathbf{a}_2 + z_{29}\,\mathbf{a}_3 & = & x_{29}a\,\hat{\mathbf{x}} + y_{29}b\,\hat{\mathbf{y}} + z_{29}c\,\hat{\mathbf{z}} & (4a) & \text{H XXVII} \\
\mathbf{B}_{114} & = & -x_{29}\,\mathbf{a}_1 - y_{29}\,\mathbf{a}_2 + \left(\frac{1}{2} + z_{29}\right)\,\mathbf{a}_3 & = & -x_{29}a\,\hat{\mathbf{x}} - y_{29}b\,\hat{\mathbf{y}} + \left(\frac{1}{2} + z_{29}\right)c\,\hat{\mathbf{z}} & (4a) & \text{H XXVII} \\
\mathbf{B}_{115} & = & \left(\frac{1}{2} + x_{29}\right)\,\mathbf{a}_1 - y_{29}\,\mathbf{a}_2 + z_{29}\,\mathbf{a}_3 & = & \left(\frac{1}{2} + x_{29}\right)a\,\hat{\mathbf{x}} - y_{29}b\,\hat{\mathbf{y}} + z_{29}c\,\hat{\mathbf{z}} & (4a) & \text{H XXVII} \\
\mathbf{B}_{116} & = & \left(\frac{1}{2} - x_{29}\right)\,\mathbf{a}_1 + y_{29}\,\mathbf{a}_2 + \left(\frac{1}{2} + z_{29}\right)\,\mathbf{a}_3 & = & \left(\frac{1}{2} - x_{29}\right)a\,\hat{\mathbf{x}} + y_{29}b\,\hat{\mathbf{y}} + \left(\frac{1}{2} + z_{29}\right)c\,\hat{\mathbf{z}} & (4a) & \text{H XXVII} \\
\mathbf{B}_{117} & = & x_{30}\,\mathbf{a}_1 + y_{30}\,\mathbf{a}_2 + z_{30}\,\mathbf{a}_3 & = & x_{30}a\,\hat{\mathbf{x}} + y_{30}b\,\hat{\mathbf{y}} + z_{30}c\,\hat{\mathbf{z}} & (4a) & \text{H XXVIII} \\
\mathbf{B}_{118} & = & -x_{30}\,\mathbf{a}_1 - y_{30}\,\mathbf{a}_2 + \left(\frac{1}{2} + z_{30}\right)\,\mathbf{a}_3 & = & -x_{30}a\,\hat{\mathbf{x}} - y_{30}b\,\hat{\mathbf{y}} + \left(\frac{1}{2} + z_{30}\right)c\,\hat{\mathbf{z}} & (4a) & \text{H XXVIII} \\
\mathbf{B}_{119} & = & \left(\frac{1}{2} + x_{30}\right)\,\mathbf{a}_1 - y_{30}\,\mathbf{a}_2 + z_{30}\,\mathbf{a}_3 & = & \left(\frac{1}{2} + x_{30}\right)a\,\hat{\mathbf{x}} - y_{30}b\,\hat{\mathbf{y}} + z_{30}c\,\hat{\mathbf{z}} & (4a) & \text{H XXVIII} \\
\mathbf{B}_{120} & = & \left(\frac{1}{2} - x_{30}\right)\,\mathbf{a}_1 + y_{30}\,\mathbf{a}_2 + \left(\frac{1}{2} + z_{30}\right)\,\mathbf{a}_3 & = & \left(\frac{1}{2} - x_{30}\right)a\,\hat{\mathbf{x}} + y_{30}b\,\hat{\mathbf{y}} + \left(\frac{1}{2} + z_{30}\right)c\,\hat{\mathbf{z}} & (4a) & \text{H XXVIII} \\
\mathbf{B}_{121} & = & x_{31}\,\mathbf{a}_1 + y_{31}\,\mathbf{a}_2 + z_{31}\,\mathbf{a}_3 & = & x_{31}a\,\hat{\mathbf{x}} + y_{31}b\,\hat{\mathbf{y}} + z_{31}c\,\hat{\mathbf{z}} & (4a) & \text{H XXIX} \\
\mathbf{B}_{122} & = & -x_{31}\,\mathbf{a}_1 - y_{31}\,\mathbf{a}_2 + \left(\frac{1}{2} + z_{31}\right)\,\mathbf{a}_3 & = & -x_{31}a\,\hat{\mathbf{x}} - y_{31}b\,\hat{\mathbf{y}} + \left(\frac{1}{2} + z_{31}\right)c\,\hat{\mathbf{z}} & (4a) & \text{H XXIX} \\
\mathbf{B}_{123} & = & \left(\frac{1}{2} + x_{31}\right)\,\mathbf{a}_1 - y_{31}\,\mathbf{a}_2 + z_{31}\,\mathbf{a}_3 & = & \left(\frac{1}{2} + x_{31}\right)a\,\hat{\mathbf{x}} - y_{31}b\,\hat{\mathbf{y}} + z_{31}c\,\hat{\mathbf{z}} & (4a) & \text{H XXIX} \\
\mathbf{B}_{124} & = & \left(\frac{1}{2} - x_{31}\right)\,\mathbf{a}_1 + y_{31}\,\mathbf{a}_2 + \left(\frac{1}{2} + z_{31}\right)\,\mathbf{a}_3 & = & \left(\frac{1}{2} - x_{31}\right)a\,\hat{\mathbf{x}} + y_{31}b\,\hat{\mathbf{y}} + \left(\frac{1}{2} + z_{31}\right)c\,\hat{\mathbf{z}} & (4a) & \text{H XXIX} \\
\mathbf{B}_{125} & = & x_{32}\,\mathbf{a}_1 + y_{32}\,\mathbf{a}_2 + z_{32}\,\mathbf{a}_3 & = & x_{32}a\,\hat{\mathbf{x}} + y_{32}b\,\hat{\mathbf{y}} + z_{32}c\,\hat{\mathbf{z}} & (4a) & \text{H XXX}
\end{array}
$$

| | | | | | | |
|---|---|---|---|---|---|---|
| $\mathbf{B}_{126}$ | = | $-x_{32}\,\mathbf{a}_1 - y_{32}\,\mathbf{a}_2 + \left(\frac{1}{2} + z_{32}\right)\,\mathbf{a}_3$ | = | $-x_{32}a\,\hat{\mathbf{x}} - y_{32}b\,\hat{\mathbf{y}} + \left(\frac{1}{2} + z_{32}\right)c\,\hat{\mathbf{z}}$ | (4$a$) | H XXX |
| $\mathbf{B}_{127}$ | = | $\left(\frac{1}{2} + x_{32}\right)\,\mathbf{a}_1 - y_{32}\,\mathbf{a}_2 + z_{32}\,\mathbf{a}_3$ | = | $\left(\frac{1}{2} + x_{32}\right)a\,\hat{\mathbf{x}} - y_{32}b\,\hat{\mathbf{y}} + z_{32}c\,\hat{\mathbf{z}}$ | (4$a$) | H XXX |
| $\mathbf{B}_{128}$ | = | $\left(\frac{1}{2} - x_{32}\right)\,\mathbf{a}_1 + y_{32}\,\mathbf{a}_2 + \left(\frac{1}{2} + z_{32}\right)\,\mathbf{a}_3$ | = | $\left(\frac{1}{2} - x_{32}\right)a\,\hat{\mathbf{x}} + y_{32}b\,\hat{\mathbf{y}} + \left(\frac{1}{2} + z_{32}\right)c\,\hat{\mathbf{z}}$ | (4$a$) | H XXX |
| $\mathbf{B}_{129}$ | = | $x_{33}\,\mathbf{a}_1 + y_{33}\,\mathbf{a}_2 + z_{33}\,\mathbf{a}_3$ | = | $x_{33}a\,\hat{\mathbf{x}} + y_{33}b\,\hat{\mathbf{y}} + z_{33}c\,\hat{\mathbf{z}}$ | (4$a$) | N |
| $\mathbf{B}_{130}$ | = | $-x_{33}\,\mathbf{a}_1 - y_{33}\,\mathbf{a}_2 + \left(\frac{1}{2} + z_{33}\right)\,\mathbf{a}_3$ | = | $-x_{33}a\,\hat{\mathbf{x}} - y_{33}b\,\hat{\mathbf{y}} + \left(\frac{1}{2} + z_{33}\right)c\,\hat{\mathbf{z}}$ | (4$a$) | N |
| $\mathbf{B}_{131}$ | = | $\left(\frac{1}{2} + x_{33}\right)\,\mathbf{a}_1 - y_{33}\,\mathbf{a}_2 + z_{33}\,\mathbf{a}_3$ | = | $\left(\frac{1}{2} + x_{33}\right)a\,\hat{\mathbf{x}} - y_{33}b\,\hat{\mathbf{y}} + z_{33}c\,\hat{\mathbf{z}}$ | (4$a$) | N |
| $\mathbf{B}_{132}$ | = | $\left(\frac{1}{2} - x_{33}\right)\,\mathbf{a}_1 + y_{33}\,\mathbf{a}_2 + \left(\frac{1}{2} + z_{33}\right)\,\mathbf{a}_3$ | = | $\left(\frac{1}{2} - x_{33}\right)a\,\hat{\mathbf{x}} + y_{33}b\,\hat{\mathbf{y}} + \left(\frac{1}{2} + z_{33}\right)c\,\hat{\mathbf{z}}$ | (4$a$) | N |
| $\mathbf{B}_{133}$ | = | $x_{34}\,\mathbf{a}_1 + y_{34}\,\mathbf{a}_2 + z_{34}\,\mathbf{a}_3$ | = | $x_{34}a\,\hat{\mathbf{x}} + y_{34}b\,\hat{\mathbf{y}} + z_{34}c\,\hat{\mathbf{z}}$ | (4$a$) | O I |
| $\mathbf{B}_{134}$ | = | $-x_{34}\,\mathbf{a}_1 - y_{34}\,\mathbf{a}_2 + \left(\frac{1}{2} + z_{34}\right)\,\mathbf{a}_3$ | = | $-x_{34}a\,\hat{\mathbf{x}} - y_{34}b\,\hat{\mathbf{y}} + \left(\frac{1}{2} + z_{34}\right)c\,\hat{\mathbf{z}}$ | (4$a$) | O I |
| $\mathbf{B}_{135}$ | = | $\left(\frac{1}{2} + x_{34}\right)\,\mathbf{a}_1 - y_{34}\,\mathbf{a}_2 + z_{34}\,\mathbf{a}_3$ | = | $\left(\frac{1}{2} + x_{34}\right)a\,\hat{\mathbf{x}} - y_{34}b\,\hat{\mathbf{y}} + z_{34}c\,\hat{\mathbf{z}}$ | (4$a$) | O I |
| $\mathbf{B}_{136}$ | = | $\left(\frac{1}{2} - x_{34}\right)\,\mathbf{a}_1 + y_{34}\,\mathbf{a}_2 + \left(\frac{1}{2} + z_{34}\right)\,\mathbf{a}_3$ | = | $\left(\frac{1}{2} - x_{34}\right)a\,\hat{\mathbf{x}} + y_{34}b\,\hat{\mathbf{y}} + \left(\frac{1}{2} + z_{34}\right)c\,\hat{\mathbf{z}}$ | (4$a$) | O I |
| $\mathbf{B}_{137}$ | = | $x_{35}\,\mathbf{a}_1 + y_{35}\,\mathbf{a}_2 + z_{35}\,\mathbf{a}_3$ | = | $x_{35}a\,\hat{\mathbf{x}} + y_{35}b\,\hat{\mathbf{y}} + z_{35}c\,\hat{\mathbf{z}}$ | (4$a$) | O II |
| $\mathbf{B}_{138}$ | = | $-x_{35}\,\mathbf{a}_1 - y_{35}\,\mathbf{a}_2 + \left(\frac{1}{2} + z_{35}\right)\,\mathbf{a}_3$ | = | $-x_{35}a\,\hat{\mathbf{x}} - y_{35}b\,\hat{\mathbf{y}} + \left(\frac{1}{2} + z_{35}\right)c\,\hat{\mathbf{z}}$ | (4$a$) | O II |
| $\mathbf{B}_{139}$ | = | $\left(\frac{1}{2} + x_{35}\right)\,\mathbf{a}_1 - y_{35}\,\mathbf{a}_2 + z_{35}\,\mathbf{a}_3$ | = | $\left(\frac{1}{2} + x_{35}\right)a\,\hat{\mathbf{x}} - y_{35}b\,\hat{\mathbf{y}} + z_{35}c\,\hat{\mathbf{z}}$ | (4$a$) | O II |
| $\mathbf{B}_{140}$ | = | $\left(\frac{1}{2} - x_{35}\right)\,\mathbf{a}_1 + y_{35}\,\mathbf{a}_2 + \left(\frac{1}{2} + z_{35}\right)\,\mathbf{a}_3$ | = | $\left(\frac{1}{2} - x_{35}\right)a\,\hat{\mathbf{x}} + y_{35}b\,\hat{\mathbf{y}} + \left(\frac{1}{2} + z_{35}\right)c\,\hat{\mathbf{z}}$ | (4$a$) | O II |
| $\mathbf{B}_{141}$ | = | $x_{36}\,\mathbf{a}_1 + y_{36}\,\mathbf{a}_2 + z_{36}\,\mathbf{a}_3$ | = | $x_{36}a\,\hat{\mathbf{x}} + y_{36}b\,\hat{\mathbf{y}} + z_{36}c\,\hat{\mathbf{z}}$ | (4$a$) | O III |
| $\mathbf{B}_{142}$ | = | $-x_{36}\,\mathbf{a}_1 - y_{36}\,\mathbf{a}_2 + \left(\frac{1}{2} + z_{36}\right)\,\mathbf{a}_3$ | = | $-x_{36}a\,\hat{\mathbf{x}} - y_{36}b\,\hat{\mathbf{y}} + \left(\frac{1}{2} + z_{36}\right)c\,\hat{\mathbf{z}}$ | (4$a$) | O III |
| $\mathbf{B}_{143}$ | = | $\left(\frac{1}{2} + x_{36}\right)\,\mathbf{a}_1 - y_{36}\,\mathbf{a}_2 + z_{36}\,\mathbf{a}_3$ | = | $\left(\frac{1}{2} + x_{36}\right)a\,\hat{\mathbf{x}} - y_{36}b\,\hat{\mathbf{y}} + z_{36}c\,\hat{\mathbf{z}}$ | (4$a$) | O III |
| $\mathbf{B}_{144}$ | = | $\left(\frac{1}{2} - x_{36}\right)\,\mathbf{a}_1 + y_{36}\,\mathbf{a}_2 + \left(\frac{1}{2} + z_{36}\right)\,\mathbf{a}_3$ | = | $\left(\frac{1}{2} - x_{36}\right)a\,\hat{\mathbf{x}} + y_{36}b\,\hat{\mathbf{y}} + \left(\frac{1}{2} + z_{36}\right)c\,\hat{\mathbf{z}}$ | (4$a$) | O III |
| $\mathbf{B}_{145}$ | = | $x_{37}\,\mathbf{a}_1 + y_{37}\,\mathbf{a}_2 + z_{37}\,\mathbf{a}_3$ | = | $x_{37}a\,\hat{\mathbf{x}} + y_{37}b\,\hat{\mathbf{y}} + z_{37}c\,\hat{\mathbf{z}}$ | (4$a$) | O IV |
| $\mathbf{B}_{146}$ | = | $-x_{37}\,\mathbf{a}_1 - y_{37}\,\mathbf{a}_2 + \left(\frac{1}{2} + z_{37}\right)\,\mathbf{a}_3$ | = | $-x_{37}a\,\hat{\mathbf{x}} - y_{37}b\,\hat{\mathbf{y}} + \left(\frac{1}{2} + z_{37}\right)c\,\hat{\mathbf{z}}$ | (4$a$) | O IV |
| $\mathbf{B}_{147}$ | = | $\left(\frac{1}{2} + x_{37}\right)\,\mathbf{a}_1 - y_{37}\,\mathbf{a}_2 + z_{37}\,\mathbf{a}_3$ | = | $\left(\frac{1}{2} + x_{37}\right)a\,\hat{\mathbf{x}} - y_{37}b\,\hat{\mathbf{y}} + z_{37}c\,\hat{\mathbf{z}}$ | (4$a$) | O IV |
| $\mathbf{B}_{148}$ | = | $\left(\frac{1}{2} - x_{37}\right)\,\mathbf{a}_1 + y_{37}\,\mathbf{a}_2 + \left(\frac{1}{2} + z_{37}\right)\,\mathbf{a}_3$ | = | $\left(\frac{1}{2} - x_{37}\right)a\,\hat{\mathbf{x}} + y_{37}b\,\hat{\mathbf{y}} + \left(\frac{1}{2} + z_{37}\right)c\,\hat{\mathbf{z}}$ | (4$a$) | O IV |
| $\mathbf{B}_{149}$ | = | $x_{38}\,\mathbf{a}_1 + y_{38}\,\mathbf{a}_2 + z_{38}\,\mathbf{a}_3$ | = | $x_{38}a\,\hat{\mathbf{x}} + y_{38}b\,\hat{\mathbf{y}} + z_{38}c\,\hat{\mathbf{z}}$ | (4$a$) | O V |
| $\mathbf{B}_{150}$ | = | $-x_{38}\,\mathbf{a}_1 - y_{38}\,\mathbf{a}_2 + \left(\frac{1}{2} + z_{38}\right)\,\mathbf{a}_3$ | = | $-x_{38}a\,\hat{\mathbf{x}} - y_{38}b\,\hat{\mathbf{y}} + \left(\frac{1}{2} + z_{38}\right)c\,\hat{\mathbf{z}}$ | (4$a$) | O V |
| $\mathbf{B}_{151}$ | = | $\left(\frac{1}{2} + x_{38}\right)\,\mathbf{a}_1 - y_{38}\,\mathbf{a}_2 + z_{38}\,\mathbf{a}_3$ | = | $\left(\frac{1}{2} + x_{38}\right)a\,\hat{\mathbf{x}} - y_{38}b\,\hat{\mathbf{y}} + z_{38}c\,\hat{\mathbf{z}}$ | (4$a$) | O V |
| $\mathbf{B}_{152}$ | = | $\left(\frac{1}{2} - x_{38}\right)\,\mathbf{a}_1 + y_{38}\,\mathbf{a}_2 + \left(\frac{1}{2} + z_{38}\right)\,\mathbf{a}_3$ | = | $\left(\frac{1}{2} - x_{38}\right)a\,\hat{\mathbf{x}} + y_{38}b\,\hat{\mathbf{y}} + \left(\frac{1}{2} + z_{38}\right)c\,\hat{\mathbf{z}}$ | (4$a$) | O V |
| $\mathbf{B}_{153}$ | = | $x_{39}\,\mathbf{a}_1 + y_{39}\,\mathbf{a}_2 + z_{39}\,\mathbf{a}_3$ | = | $x_{39}a\,\hat{\mathbf{x}} + y_{39}b\,\hat{\mathbf{y}} + z_{39}c\,\hat{\mathbf{z}}$ | (4$a$) | O VI |
| $\mathbf{B}_{154}$ | = | $-x_{39}\,\mathbf{a}_1 - y_{39}\,\mathbf{a}_2 + \left(\frac{1}{2} + z_{39}\right)\,\mathbf{a}_3$ | = | $-x_{39}a\,\hat{\mathbf{x}} - y_{39}b\,\hat{\mathbf{y}} + \left(\frac{1}{2} + z_{39}\right)c\,\hat{\mathbf{z}}$ | (4$a$) | O VI |
| $\mathbf{B}_{155}$ | = | $\left(\frac{1}{2} + x_{39}\right)\,\mathbf{a}_1 - y_{39}\,\mathbf{a}_2 + z_{39}\,\mathbf{a}_3$ | = | $\left(\frac{1}{2} + x_{39}\right)a\,\hat{\mathbf{x}} - y_{39}b\,\hat{\mathbf{y}} + z_{39}c\,\hat{\mathbf{z}}$ | (4$a$) | O VI |
| $\mathbf{B}_{156}$ | = | $\left(\frac{1}{2} - x_{39}\right)\,\mathbf{a}_1 + y_{39}\,\mathbf{a}_2 + \left(\frac{1}{2} + z_{39}\right)\,\mathbf{a}_3$ | = | $\left(\frac{1}{2} - x_{39}\right)a\,\hat{\mathbf{x}} + y_{39}b\,\hat{\mathbf{y}} + \left(\frac{1}{2} + z_{39}\right)c\,\hat{\mathbf{z}}$ | (4$a$) | O VI |
| $\mathbf{B}_{157}$ | = | $x_{40}\,\mathbf{a}_1 + y_{40}\,\mathbf{a}_2 + z_{40}\,\mathbf{a}_3$ | = | $x_{40}a\,\hat{\mathbf{x}} + y_{40}b\,\hat{\mathbf{y}} + z_{40}c\,\hat{\mathbf{z}}$ | (4$a$) | O VII |
| $\mathbf{B}_{158}$ | = | $-x_{40}\,\mathbf{a}_1 - y_{40}\,\mathbf{a}_2 + \left(\frac{1}{2} + z_{40}\right)\,\mathbf{a}_3$ | = | $-x_{40}a\,\hat{\mathbf{x}} - y_{40}b\,\hat{\mathbf{y}} + \left(\frac{1}{2} + z_{40}\right)c\,\hat{\mathbf{z}}$ | (4$a$) | O VII |
| $\mathbf{B}_{159}$ | = | $\left(\frac{1}{2} + x_{40}\right)\,\mathbf{a}_1 - y_{40}\,\mathbf{a}_2 + z_{40}\,\mathbf{a}_3$ | = | $\left(\frac{1}{2} + x_{40}\right)a\,\hat{\mathbf{x}} - y_{40}b\,\hat{\mathbf{y}} + z_{40}c\,\hat{\mathbf{z}}$ | (4$a$) | O VII |
| $\mathbf{B}_{160}$ | = | $\left(\frac{1}{2} - x_{40}\right)\,\mathbf{a}_1 + y_{40}\,\mathbf{a}_2 + \left(\frac{1}{2} + z_{40}\right)\,\mathbf{a}_3$ | = | $\left(\frac{1}{2} - x_{40}\right)a\,\hat{\mathbf{x}} + y_{40}b\,\hat{\mathbf{y}} + \left(\frac{1}{2} + z_{40}\right)c\,\hat{\mathbf{z}}$ | (4$a$) | O VII |
| $\mathbf{B}_{161}$ | = | $x_{41}\,\mathbf{a}_1 + y_{41}\,\mathbf{a}_2 + z_{41}\,\mathbf{a}_3$ | = | $x_{41}a\,\hat{\mathbf{x}} + y_{41}b\,\hat{\mathbf{y}} + z_{41}c\,\hat{\mathbf{z}}$ | (4$a$) | O VIII |

$$\mathbf{B}_{162} = -x_{41}\,\mathbf{a}_1 - y_{41}\,\mathbf{a}_2 + \left(\frac{1}{2} + z_{41}\right)\,\mathbf{a}_3 = -x_{41}a\,\hat{\mathbf{x}} - y_{41}b\,\hat{\mathbf{y}} + \left(\frac{1}{2} + z_{41}\right)c\,\hat{\mathbf{z}} \quad (4a) \quad \text{O VIII}$$

$$\mathbf{B}_{163} = \left(\frac{1}{2} + x_{41}\right)\,\mathbf{a}_1 - y_{41}\,\mathbf{a}_2 + z_{41}\,\mathbf{a}_3 = \left(\frac{1}{2} + x_{41}\right)a\,\hat{\mathbf{x}} - y_{41}b\,\hat{\mathbf{y}} + z_{41}c\,\hat{\mathbf{z}} \quad (4a) \quad \text{O VIII}$$

$$\mathbf{B}_{164} = \left(\frac{1}{2} - x_{41}\right)\,\mathbf{a}_1 + y_{41}\,\mathbf{a}_2 + \left(\frac{1}{2} + z_{41}\right)\,\mathbf{a}_3 = \left(\frac{1}{2} - x_{41}\right)a\,\hat{\mathbf{x}} + y_{41}b\,\hat{\mathbf{y}} + \left(\frac{1}{2} + z_{41}\right)c\,\hat{\mathbf{z}} \quad (4a) \quad \text{O VIII}$$

$$\mathbf{B}_{165} = x_{42}\,\mathbf{a}_1 + y_{42}\,\mathbf{a}_2 + z_{42}\,\mathbf{a}_3 = x_{42}a\,\hat{\mathbf{x}} + y_{42}b\,\hat{\mathbf{y}} + z_{42}c\,\hat{\mathbf{z}} \quad (4a) \quad \text{O IX}$$

$$\mathbf{B}_{166} = -x_{42}\,\mathbf{a}_1 - y_{42}\,\mathbf{a}_2 + \left(\frac{1}{2} + z_{42}\right)\,\mathbf{a}_3 = -x_{42}a\,\hat{\mathbf{x}} - y_{42}b\,\hat{\mathbf{y}} + \left(\frac{1}{2} + z_{42}\right)c\,\hat{\mathbf{z}} \quad (4a) \quad \text{O IX}$$

$$\mathbf{B}_{167} = \left(\frac{1}{2} + x_{42}\right)\,\mathbf{a}_1 - y_{42}\,\mathbf{a}_2 + z_{42}\,\mathbf{a}_3 = \left(\frac{1}{2} + x_{42}\right)a\,\hat{\mathbf{x}} - y_{42}b\,\hat{\mathbf{y}} + z_{42}c\,\hat{\mathbf{z}} \quad (4a) \quad \text{O IX}$$

$$\mathbf{B}_{168} = \left(\frac{1}{2} - x_{42}\right)\,\mathbf{a}_1 + y_{42}\,\mathbf{a}_2 + \left(\frac{1}{2} + z_{42}\right)\,\mathbf{a}_3 = \left(\frac{1}{2} - x_{42}\right)a\,\hat{\mathbf{x}} + y_{42}b\,\hat{\mathbf{y}} + \left(\frac{1}{2} + z_{42}\right)c\,\hat{\mathbf{z}} \quad (4a) \quad \text{O IX}$$

$$\mathbf{B}_{169} = x_{43}\,\mathbf{a}_1 + y_{43}\,\mathbf{a}_2 + z_{43}\,\mathbf{a}_3 = x_{43}a\,\hat{\mathbf{x}} + y_{43}b\,\hat{\mathbf{y}} + z_{43}c\,\hat{\mathbf{z}} \quad (4a) \quad \text{O X}$$

$$\mathbf{B}_{170} = -x_{43}\,\mathbf{a}_1 - y_{43}\,\mathbf{a}_2 + \left(\frac{1}{2} + z_{43}\right)\,\mathbf{a}_3 = -x_{43}a\,\hat{\mathbf{x}} - y_{43}b\,\hat{\mathbf{y}} + \left(\frac{1}{2} + z_{43}\right)c\,\hat{\mathbf{z}} \quad (4a) \quad \text{O X}$$

$$\mathbf{B}_{171} = \left(\frac{1}{2} + x_{43}\right)\,\mathbf{a}_1 - y_{43}\,\mathbf{a}_2 + z_{43}\,\mathbf{a}_3 = \left(\frac{1}{2} + x_{43}\right)a\,\hat{\mathbf{x}} - y_{43}b\,\hat{\mathbf{y}} + z_{43}c\,\hat{\mathbf{z}} \quad (4a) \quad \text{O X}$$

$$\mathbf{B}_{172} = \left(\frac{1}{2} - x_{43}\right)\,\mathbf{a}_1 + y_{43}\,\mathbf{a}_2 + \left(\frac{1}{2} + z_{43}\right)\,\mathbf{a}_3 = \left(\frac{1}{2} - x_{43}\right)a\,\hat{\mathbf{x}} + y_{43}b\,\hat{\mathbf{y}} + \left(\frac{1}{2} + z_{43}\right)c\,\hat{\mathbf{z}} \quad (4a) \quad \text{O X}$$

$$\mathbf{B}_{173} = x_{44}\,\mathbf{a}_1 + y_{44}\,\mathbf{a}_2 + z_{44}\,\mathbf{a}_3 = x_{44}a\,\hat{\mathbf{x}} + y_{44}b\,\hat{\mathbf{y}} + z_{44}c\,\hat{\mathbf{z}} \quad (4a) \quad \text{O XI}$$

$$\mathbf{B}_{174} = -x_{44}\,\mathbf{a}_1 - y_{44}\,\mathbf{a}_2 + \left(\frac{1}{2} + z_{44}\right)\,\mathbf{a}_3 = -x_{44}a\,\hat{\mathbf{x}} - y_{44}b\,\hat{\mathbf{y}} + \left(\frac{1}{2} + z_{44}\right)c\,\hat{\mathbf{z}} \quad (4a) \quad \text{O XI}$$

$$\mathbf{B}_{175} = \left(\frac{1}{2} + x_{44}\right)\,\mathbf{a}_1 - y_{44}\,\mathbf{a}_2 + z_{44}\,\mathbf{a}_3 = \left(\frac{1}{2} + x_{44}\right)a\,\hat{\mathbf{x}} - y_{44}b\,\hat{\mathbf{y}} + z_{44}c\,\hat{\mathbf{z}} \quad (4a) \quad \text{O XI}$$

$$\mathbf{B}_{176} = \left(\frac{1}{2} - x_{44}\right)\,\mathbf{a}_1 + y_{44}\,\mathbf{a}_2 + \left(\frac{1}{2} + z_{44}\right)\,\mathbf{a}_3 = \left(\frac{1}{2} - x_{44}\right)a\,\hat{\mathbf{x}} + y_{44}b\,\hat{\mathbf{y}} + \left(\frac{1}{2} + z_{44}\right)c\,\hat{\mathbf{z}} \quad (4a) \quad \text{O XI}$$

$$\mathbf{B}_{177} = x_{45}\,\mathbf{a}_1 + y_{45}\,\mathbf{a}_2 + z_{45}\,\mathbf{a}_3 = x_{45}a\,\hat{\mathbf{x}} + y_{45}b\,\hat{\mathbf{y}} + z_{45}c\,\hat{\mathbf{z}} \quad (4a) \quad \text{O XII}$$

$$\mathbf{B}_{178} = -x_{45}\,\mathbf{a}_1 - y_{45}\,\mathbf{a}_2 + \left(\frac{1}{2} + z_{45}\right)\,\mathbf{a}_3 = -x_{45}a\,\hat{\mathbf{x}} - y_{45}b\,\hat{\mathbf{y}} + \left(\frac{1}{2} + z_{45}\right)c\,\hat{\mathbf{z}} \quad (4a) \quad \text{O XII}$$

$$\mathbf{B}_{179} = \left(\frac{1}{2} + x_{45}\right)\,\mathbf{a}_1 - y_{45}\,\mathbf{a}_2 + z_{45}\,\mathbf{a}_3 = \left(\frac{1}{2} + x_{45}\right)a\,\hat{\mathbf{x}} - y_{45}b\,\hat{\mathbf{y}} + z_{45}c\,\hat{\mathbf{z}} \quad (4a) \quad \text{O XII}$$

$$\mathbf{B}_{180} = \left(\frac{1}{2} - x_{45}\right)\,\mathbf{a}_1 + y_{45}\,\mathbf{a}_2 + \left(\frac{1}{2} + z_{45}\right)\,\mathbf{a}_3 = \left(\frac{1}{2} - x_{45}\right)a\,\hat{\mathbf{x}} + y_{45}b\,\hat{\mathbf{y}} + \left(\frac{1}{2} + z_{45}\right)c\,\hat{\mathbf{z}} \quad (4a) \quad \text{O XII}$$

$$\mathbf{B}_{181} = x_{46}\,\mathbf{a}_1 + y_{46}\,\mathbf{a}_2 + z_{46}\,\mathbf{a}_3 = x_{46}a\,\hat{\mathbf{x}} + y_{46}b\,\hat{\mathbf{y}} + z_{46}c\,\hat{\mathbf{z}} \quad (4a) \quad \text{O XIII}$$

$$\mathbf{B}_{182} = -x_{46}\,\mathbf{a}_1 - y_{46}\,\mathbf{a}_2 + \left(\frac{1}{2} + z_{46}\right)\,\mathbf{a}_3 = -x_{46}a\,\hat{\mathbf{x}} - y_{46}b\,\hat{\mathbf{y}} + \left(\frac{1}{2} + z_{46}\right)c\,\hat{\mathbf{z}} \quad (4a) \quad \text{O XIII}$$

$$\mathbf{B}_{183} = \left(\frac{1}{2} + x_{46}\right)\,\mathbf{a}_1 - y_{46}\,\mathbf{a}_2 + z_{46}\,\mathbf{a}_3 = \left(\frac{1}{2} + x_{46}\right)a\,\hat{\mathbf{x}} - y_{46}b\,\hat{\mathbf{y}} + z_{46}c\,\hat{\mathbf{z}} \quad (4a) \quad \text{O XIII}$$

$$\mathbf{B}_{184} = \left(\frac{1}{2} - x_{46}\right)\,\mathbf{a}_1 + y_{46}\,\mathbf{a}_2 + \left(\frac{1}{2} + z_{46}\right)\,\mathbf{a}_3 = \left(\frac{1}{2} - x_{46}\right)a\,\hat{\mathbf{x}} + y_{46}b\,\hat{\mathbf{y}} + \left(\frac{1}{2} + z_{46}\right)c\,\hat{\mathbf{z}} \quad (4a) \quad \text{O XIII}$$

$$\mathbf{B}_{185} = x_{47}\,\mathbf{a}_1 + y_{47}\,\mathbf{a}_2 + z_{47}\,\mathbf{a}_3 = x_{47}a\,\hat{\mathbf{x}} + y_{47}b\,\hat{\mathbf{y}} + z_{47}c\,\hat{\mathbf{z}} \quad (4a) \quad \text{O XIV}$$

$$\mathbf{B}_{186} = -x_{47}\,\mathbf{a}_1 - y_{47}\,\mathbf{a}_2 + \left(\frac{1}{2} + z_{47}\right)\,\mathbf{a}_3 = -x_{47}a\,\hat{\mathbf{x}} - y_{47}b\,\hat{\mathbf{y}} + \left(\frac{1}{2} + z_{47}\right)c\,\hat{\mathbf{z}} \quad (4a) \quad \text{O XIV}$$

$$\mathbf{B}_{187} = \left(\frac{1}{2} + x_{47}\right)\,\mathbf{a}_1 - y_{47}\,\mathbf{a}_2 + z_{47}\,\mathbf{a}_3 = \left(\frac{1}{2} + x_{47}\right)a\,\hat{\mathbf{x}} - y_{47}b\,\hat{\mathbf{y}} + z_{47}c\,\hat{\mathbf{z}} \quad (4a) \quad \text{O XIV}$$

$$\mathbf{B}_{188} = \left(\frac{1}{2} - x_{47}\right)\,\mathbf{a}_1 + y_{47}\,\mathbf{a}_2 + \left(\frac{1}{2} + z_{47}\right)\,\mathbf{a}_3 = \left(\frac{1}{2} - x_{47}\right)a\,\hat{\mathbf{x}} + y_{47}b\,\hat{\mathbf{y}} + \left(\frac{1}{2} + z_{47}\right)c\,\hat{\mathbf{z}} \quad (4a) \quad \text{O XIV}$$

$$\mathbf{B}_{189} = x_{48}\,\mathbf{a}_1 + y_{48}\,\mathbf{a}_2 + z_{48}\,\mathbf{a}_3 = x_{48}a\,\hat{\mathbf{x}} + y_{48}b\,\hat{\mathbf{y}} + z_{48}c\,\hat{\mathbf{z}} \quad (4a) \quad \text{O XV}$$

$$\mathbf{B}_{190} = -x_{48}\,\mathbf{a}_1 - y_{48}\,\mathbf{a}_2 + \left(\frac{1}{2} + z_{48}\right)\,\mathbf{a}_3 = -x_{48}a\,\hat{\mathbf{x}} - y_{48}b\,\hat{\mathbf{y}} + \left(\frac{1}{2} + z_{48}\right)c\,\hat{\mathbf{z}} \quad (4a) \quad \text{O XV}$$

$$\mathbf{B}_{191} = \left(\frac{1}{2} + x_{48}\right)\,\mathbf{a}_1 - y_{48}\,\mathbf{a}_2 + z_{48}\,\mathbf{a}_3 = \left(\frac{1}{2} + x_{48}\right)a\,\hat{\mathbf{x}} - y_{48}b\,\hat{\mathbf{y}} + z_{48}c\,\hat{\mathbf{z}} \quad (4a) \quad \text{O XV}$$

$$\mathbf{B}_{192} = \left(\frac{1}{2} - x_{48}\right)\,\mathbf{a}_1 + y_{48}\,\mathbf{a}_2 + \left(\frac{1}{2} + z_{48}\right)\,\mathbf{a}_3 = \left(\frac{1}{2} - x_{48}\right)a\,\hat{\mathbf{x}} + y_{48}b\,\hat{\mathbf{y}} + \left(\frac{1}{2} + z_{48}\right)c\,\hat{\mathbf{z}} \quad (4a) \quad \text{O XV}$$

$$\mathbf{B}_{193} = x_{49}\,\mathbf{a}_1 + y_{49}\,\mathbf{a}_2 + z_{49}\,\mathbf{a}_3 = x_{49}a\,\hat{\mathbf{x}} + y_{49}b\,\hat{\mathbf{y}} + z_{49}c\,\hat{\mathbf{z}} \quad (4a) \quad \text{O XVI}$$

$$\mathbf{B}_{194} = -x_{49}\,\mathbf{a}_1 - y_{49}\,\mathbf{a}_2 + \left(\frac{1}{2} + z_{49}\right)\,\mathbf{a}_3 = -x_{49}a\,\hat{\mathbf{x}} - y_{49}b\,\hat{\mathbf{y}} + \left(\frac{1}{2} + z_{49}\right)c\,\hat{\mathbf{z}} \quad (4a) \quad \text{O XVI}$$

$$\mathbf{B}_{195} = \left(\frac{1}{2} + x_{49}\right)\,\mathbf{a}_1 - y_{49}\,\mathbf{a}_2 + z_{49}\,\mathbf{a}_3 = \left(\frac{1}{2} + x_{49}\right)a\,\hat{\mathbf{x}} - y_{49}b\,\hat{\mathbf{y}} + z_{49}c\,\hat{\mathbf{z}} \quad (4a) \quad \text{O XVI}$$

$$\mathbf{B}_{196} = \left(\frac{1}{2} - x_{49}\right)\,\mathbf{a}_1 + y_{49}\,\mathbf{a}_2 + \left(\frac{1}{2} + z_{49}\right)\,\mathbf{a}_3 = \left(\frac{1}{2} - x_{49}\right)a\,\hat{\mathbf{x}} + y_{49}b\,\hat{\mathbf{y}} + \left(\frac{1}{2} + z_{49}\right)c\,\hat{\mathbf{z}} \quad (4a) \quad \text{O XVI}$$

$$\mathbf{B}_{197} = x_{50}\,\mathbf{a}_1 + y_{50}\,\mathbf{a}_2 + z_{50}\,\mathbf{a}_3 = x_{50}a\,\hat{\mathbf{x}} + y_{50}b\,\hat{\mathbf{y}} + z_{50}c\,\hat{\mathbf{z}} \quad (4a) \quad \text{O XVII}$$

| | | | | | | |
|---|---|---|---|---|---|---|
| $\mathbf{B}_{198}$ | = | $-x_{50}\,\mathbf{a}_{1} - y_{50}\,\mathbf{a}_{2} + \left(\frac{1}{2} + z_{50}\right)\,\mathbf{a}_{3}$ | = | $-x_{50}a\,\hat{\mathbf{x}} - y_{50}b\,\hat{\mathbf{y}} + \left(\frac{1}{2} + z_{50}\right)c\,\hat{\mathbf{z}}$ | (4$a$) | O XVII |
| $\mathbf{B}_{199}$ | = | $\left(\frac{1}{2} + x_{50}\right)\,\mathbf{a}_{1} - y_{50}\,\mathbf{a}_{2} + z_{50}\,\mathbf{a}_{3}$ | = | $\left(\frac{1}{2} + x_{50}\right)a\,\hat{\mathbf{x}} - y_{50}b\,\hat{\mathbf{y}} + z_{50}c\,\hat{\mathbf{z}}$ | (4$a$) | O XVII |
| $\mathbf{B}_{200}$ | = | $\left(\frac{1}{2} - x_{50}\right)\,\mathbf{a}_{1} + y_{50}\,\mathbf{a}_{2} + \left(\frac{1}{2} + z_{50}\right)\,\mathbf{a}_{3}$ | = | $\left(\frac{1}{2} - x_{50}\right)a\,\hat{\mathbf{x}} + y_{50}b\,\hat{\mathbf{y}} + \left(\frac{1}{2} + z_{50}\right)c\,\hat{\mathbf{z}}$ | (4$a$) | O XVII |
| $\mathbf{B}_{201}$ | = | $x_{51}\,\mathbf{a}_{1} + y_{51}\,\mathbf{a}_{2} + z_{51}\,\mathbf{a}_{3}$ | = | $x_{51}a\,\hat{\mathbf{x}} + y_{51}b\,\hat{\mathbf{y}} + z_{51}c\,\hat{\mathbf{z}}$ | (4$a$) | O XVIII |
| $\mathbf{B}_{202}$ | = | $-x_{51}\,\mathbf{a}_{1} - y_{51}\,\mathbf{a}_{2} + \left(\frac{1}{2} + z_{51}\right)\,\mathbf{a}_{3}$ | = | $-x_{51}a\,\hat{\mathbf{x}} - y_{51}b\,\hat{\mathbf{y}} + \left(\frac{1}{2} + z_{51}\right)c\,\hat{\mathbf{z}}$ | (4$a$) | O XVIII |
| $\mathbf{B}_{203}$ | = | $\left(\frac{1}{2} + x_{51}\right)\,\mathbf{a}_{1} - y_{51}\,\mathbf{a}_{2} + z_{51}\,\mathbf{a}_{3}$ | = | $\left(\frac{1}{2} + x_{51}\right)a\,\hat{\mathbf{x}} - y_{51}b\,\hat{\mathbf{y}} + z_{51}c\,\hat{\mathbf{z}}$ | (4$a$) | O XVIII |
| $\mathbf{B}_{204}$ | = | $\left(\frac{1}{2} - x_{51}\right)\,\mathbf{a}_{1} + y_{51}\,\mathbf{a}_{2} + \left(\frac{1}{2} + z_{51}\right)\,\mathbf{a}_{3}$ | = | $\left(\frac{1}{2} - x_{51}\right)a\,\hat{\mathbf{x}} + y_{51}b\,\hat{\mathbf{y}} + \left(\frac{1}{2} + z_{51}\right)c\,\hat{\mathbf{z}}$ | (4$a$) | O XVIII |
| $\mathbf{B}_{205}$ | = | $x_{52}\,\mathbf{a}_{1} + y_{52}\,\mathbf{a}_{2} + z_{52}\,\mathbf{a}_{3}$ | = | $x_{52}a\,\hat{\mathbf{x}} + y_{52}b\,\hat{\mathbf{y}} + z_{52}c\,\hat{\mathbf{z}}$ | (4$a$) | O XIX |
| $\mathbf{B}_{206}$ | = | $-x_{52}\,\mathbf{a}_{1} - y_{52}\,\mathbf{a}_{2} + \left(\frac{1}{2} + z_{52}\right)\,\mathbf{a}_{3}$ | = | $-x_{52}a\,\hat{\mathbf{x}} - y_{52}b\,\hat{\mathbf{y}} + \left(\frac{1}{2} + z_{52}\right)c\,\hat{\mathbf{z}}$ | (4$a$) | O XIX |
| $\mathbf{B}_{207}$ | = | $\left(\frac{1}{2} + x_{52}\right)\,\mathbf{a}_{1} - y_{52}\,\mathbf{a}_{2} + z_{52}\,\mathbf{a}_{3}$ | = | $\left(\frac{1}{2} + x_{52}\right)a\,\hat{\mathbf{x}} - y_{52}b\,\hat{\mathbf{y}} + z_{52}c\,\hat{\mathbf{z}}$ | (4$a$) | O XIX |
| $\mathbf{B}_{208}$ | = | $\left(\frac{1}{2} - x_{52}\right)\,\mathbf{a}_{1} + y_{52}\,\mathbf{a}_{2} + \left(\frac{1}{2} + z_{52}\right)\,\mathbf{a}_{3}$ | = | $\left(\frac{1}{2} - x_{52}\right)a\,\hat{\mathbf{x}} + y_{52}b\,\hat{\mathbf{y}} + \left(\frac{1}{2} + z_{52}\right)c\,\hat{\mathbf{z}}$ | (4$a$) | O XIX |
| $\mathbf{B}_{209}$ | = | $x_{53}\,\mathbf{a}_{1} + y_{53}\,\mathbf{a}_{2} + z_{53}\,\mathbf{a}_{3}$ | = | $x_{53}a\,\hat{\mathbf{x}} + y_{53}b\,\hat{\mathbf{y}} + z_{53}c\,\hat{\mathbf{z}}$ | (4$a$) | O XX |
| $\mathbf{B}_{210}$ | = | $-x_{53}\,\mathbf{a}_{1} - y_{53}\,\mathbf{a}_{2} + \left(\frac{1}{2} + z_{53}\right)\,\mathbf{a}_{3}$ | = | $-x_{53}a\,\hat{\mathbf{x}} - y_{53}b\,\hat{\mathbf{y}} + \left(\frac{1}{2} + z_{53}\right)c\,\hat{\mathbf{z}}$ | (4$a$) | O XX |
| $\mathbf{B}_{211}$ | = | $\left(\frac{1}{2} + x_{53}\right)\,\mathbf{a}_{1} - y_{53}\,\mathbf{a}_{2} + z_{53}\,\mathbf{a}_{3}$ | = | $\left(\frac{1}{2} + x_{53}\right)a\,\hat{\mathbf{x}} - y_{53}b\,\hat{\mathbf{y}} + z_{53}c\,\hat{\mathbf{z}}$ | (4$a$) | O XX |
| $\mathbf{B}_{212}$ | = | $\left(\frac{1}{2} - x_{53}\right)\,\mathbf{a}_{1} + y_{53}\,\mathbf{a}_{2} + \left(\frac{1}{2} + z_{53}\right)\,\mathbf{a}_{3}$ | = | $\left(\frac{1}{2} - x_{53}\right)a\,\hat{\mathbf{x}} + y_{53}b\,\hat{\mathbf{y}} + \left(\frac{1}{2} + z_{53}\right)c\,\hat{\mathbf{z}}$ | (4$a$) | O XX |
| $\mathbf{B}_{213}$ | = | $x_{54}\,\mathbf{a}_{1} + y_{54}\,\mathbf{a}_{2} + z_{54}\,\mathbf{a}_{3}$ | = | $x_{54}a\,\hat{\mathbf{x}} + y_{54}b\,\hat{\mathbf{y}} + z_{54}c\,\hat{\mathbf{z}}$ | (4$a$) | S I |
| $\mathbf{B}_{214}$ | = | $-x_{54}\,\mathbf{a}_{1} - y_{54}\,\mathbf{a}_{2} + \left(\frac{1}{2} + z_{54}\right)\,\mathbf{a}_{3}$ | = | $-x_{54}a\,\hat{\mathbf{x}} - y_{54}b\,\hat{\mathbf{y}} + \left(\frac{1}{2} + z_{54}\right)c\,\hat{\mathbf{z}}$ | (4$a$) | S I |
| $\mathbf{B}_{215}$ | = | $\left(\frac{1}{2} + x_{54}\right)\,\mathbf{a}_{1} - y_{54}\,\mathbf{a}_{2} + z_{54}\,\mathbf{a}_{3}$ | = | $\left(\frac{1}{2} + x_{54}\right)a\,\hat{\mathbf{x}} - y_{54}b\,\hat{\mathbf{y}} + z_{54}c\,\hat{\mathbf{z}}$ | (4$a$) | S I |
| $\mathbf{B}_{216}$ | = | $\left(\frac{1}{2} - x_{54}\right)\,\mathbf{a}_{1} + y_{54}\,\mathbf{a}_{2} + \left(\frac{1}{2} + z_{54}\right)\,\mathbf{a}_{3}$ | = | $\left(\frac{1}{2} - x_{54}\right)a\,\hat{\mathbf{x}} + y_{54}b\,\hat{\mathbf{y}} + \left(\frac{1}{2} + z_{54}\right)c\,\hat{\mathbf{z}}$ | (4$a$) | S I |
| $\mathbf{B}_{217}$ | = | $x_{55}\,\mathbf{a}_{1} + y_{55}\,\mathbf{a}_{2} + z_{55}\,\mathbf{a}_{3}$ | = | $x_{55}a\,\hat{\mathbf{x}} + y_{55}b\,\hat{\mathbf{y}} + z_{55}c\,\hat{\mathbf{z}}$ | (4$a$) | S II |
| $\mathbf{B}_{218}$ | = | $-x_{55}\,\mathbf{a}_{1} - y_{55}\,\mathbf{a}_{2} + \left(\frac{1}{2} + z_{55}\right)\,\mathbf{a}_{3}$ | = | $-x_{55}a\,\hat{\mathbf{x}} - y_{55}b\,\hat{\mathbf{y}} + \left(\frac{1}{2} + z_{55}\right)c\,\hat{\mathbf{z}}$ | (4$a$) | S II |
| $\mathbf{B}_{219}$ | = | $\left(\frac{1}{2} + x_{55}\right)\,\mathbf{a}_{1} - y_{55}\,\mathbf{a}_{2} + z_{55}\,\mathbf{a}_{3}$ | = | $\left(\frac{1}{2} + x_{55}\right)a\,\hat{\mathbf{x}} - y_{55}b\,\hat{\mathbf{y}} + z_{55}c\,\hat{\mathbf{z}}$ | (4$a$) | S II |
| $\mathbf{B}_{220}$ | = | $\left(\frac{1}{2} - x_{55}\right)\,\mathbf{a}_{1} + y_{55}\,\mathbf{a}_{2} + \left(\frac{1}{2} + z_{55}\right)\,\mathbf{a}_{3}$ | = | $\left(\frac{1}{2} - x_{55}\right)a\,\hat{\mathbf{x}} + y_{55}b\,\hat{\mathbf{y}} + \left(\frac{1}{2} + z_{55}\right)c\,\hat{\mathbf{z}}$ | (4$a$) | S II |

---

---

**Geometry files:**

- CIF: pp. 1583
- POSCAR: pp. 1584

# Orthorhombic $Co_4Al_{13}$ Structure: A13B4_oP102_31_17a11b_8a2b

http://aflow.org/prototype-encyclopedia/A13B4_oP102_31_17a11b_8a2b

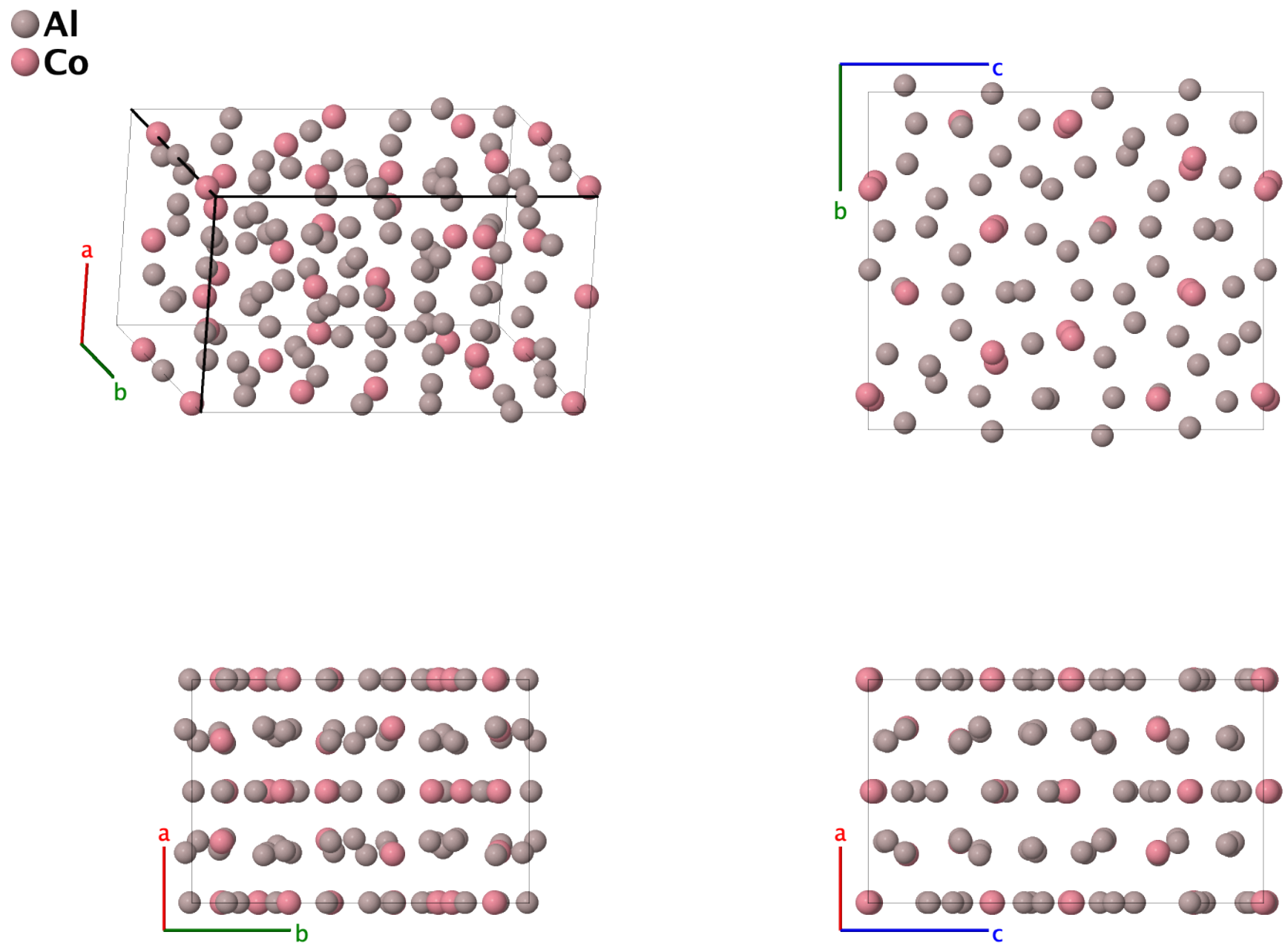

| | | |
|---|---|---|
| **Prototype** | : | $Al_{13}Co_4$ |
| **AFLOW prototype label** | : | A13B4_oP102_31_17a11b_8a2b |
| ***Strukturbericht* designation** | : | None |
| **Pearson symbol** | : | oP102 |
| **Space group number** | : | 31 |
| **Space group symbol** | : | $Pmn2_1$ |
| **AFLOW prototype command** | : | `aflow --proto=A13B4_oP102_31_17a11b_8a2b` `--params=`$a, b/a, c/a, y_1, z_1, y_2, z_2, y_3, z_3, y_4, z_4, y_5, z_5, y_6, z_6, y_7, z_7, y_8, z_8, y_9, z_9, y_{10}, z_{10}, y_{11}, z_{11}, y_{12}, z_{12}, y_{13}, z_{13}, y_{14}, z_{14}, y_{15}, z_{15}, y_{16}, z_{16}, y_{17}, z_{17}, y_{18}, z_{18}, y_{19}, z_{19}, y_{20}, z_{20}, y_{21}, z_{21}, y_{22}, z_{22}, y_{23}, z_{23}, y_{24}, z_{24}, y_{25}, z_{25}, x_{26}, y_{26}, z_{26}, x_{27}, y_{27}, z_{27}, x_{28}, y_{28}, z_{28}, x_{29}, y_{29}, z_{29}, x_{30}, y_{30}, z_{30}, x_{31}, y_{31}, z_{31}, x_{32}, y_{32}, z_{32}, x_{33}, y_{33}, z_{33}, x_{34}, y_{34}, z_{34}, x_{35}, y_{35}, z_{35}, x_{36}, y_{36}, z_{36}, x_{37}, y_{37}, z_{37}, x_{38}, y_{38}, z_{38}$ |

- Space group $Pmn2_1$ #31 allows an arbitrary choice for the origin of the $z$-axis. We follow (Grin, 1994) and set the $z_{25} = 0$.
- If we allow a tolerance of 0.25 Å for AFLOW-SYM and 0.6 Å in the atomic positions for FINDSYM the symmetry is classified as $Pnnm$ #58.

## Simple Orthorhombic primitive vectors:

$$\mathbf{a}_1 = a\,\hat{\mathbf{x}}$$
$$\mathbf{a}_2 = b\,\hat{\mathbf{y}}$$
$$\mathbf{a}_3 = c\,\hat{\mathbf{z}}$$

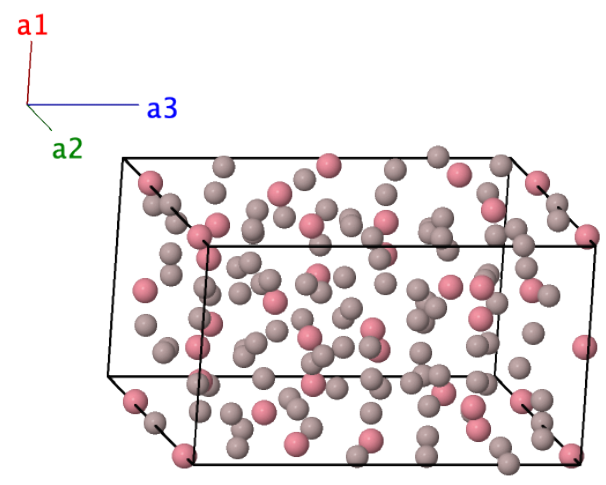

## Basis vectors:

| | | Lattice Coordinates | | Cartesian Coordinates | Wyckoff Position | Atom Type |
|---|---|---|---|---|---|---|
| $\mathbf{B}_1$ | = | $y_1\,\mathbf{a}_2 + z_1\,\mathbf{a}_3$ | = | $y_1 b\,\hat{\mathbf{y}} + z_1 c\,\hat{\mathbf{z}}$ | (2a) | Al I |
| $\mathbf{B}_2$ | = | $\frac{1}{2}\,\mathbf{a}_1 - y_1\,\mathbf{a}_2 + \left(\frac{1}{2} + z_1\right)\,\mathbf{a}_3$ | = | $\frac{1}{2}a\,\hat{\mathbf{x}} - y_1 b\,\hat{\mathbf{y}} + \left(\frac{1}{2} + z_1\right)c\,\hat{\mathbf{z}}$ | (2a) | Al I |
| $\mathbf{B}_3$ | = | $y_2\,\mathbf{a}_2 + z_2\,\mathbf{a}_3$ | = | $y_2 b\,\hat{\mathbf{y}} + z_2 c\,\hat{\mathbf{z}}$ | (2a) | Al II |
| $\mathbf{B}_4$ | = | $\frac{1}{2}\,\mathbf{a}_1 - y_2\,\mathbf{a}_2 + \left(\frac{1}{2} + z_2\right)\,\mathbf{a}_3$ | = | $\frac{1}{2}a\,\hat{\mathbf{x}} - y_2 b\,\hat{\mathbf{y}} + \left(\frac{1}{2} + z_2\right)c\,\hat{\mathbf{z}}$ | (2a) | Al II |
| $\mathbf{B}_5$ | = | $y_3\,\mathbf{a}_2 + z_3\,\mathbf{a}_3$ | = | $y_3 b\,\hat{\mathbf{y}} + z_3 c\,\hat{\mathbf{z}}$ | (2a) | Al III |
| $\mathbf{B}_6$ | = | $\frac{1}{2}\,\mathbf{a}_1 - y_3\,\mathbf{a}_2 + \left(\frac{1}{2} + z_3\right)\,\mathbf{a}_3$ | = | $\frac{1}{2}a\,\hat{\mathbf{x}} - y_3 b\,\hat{\mathbf{y}} + \left(\frac{1}{2} + z_3\right)c\,\hat{\mathbf{z}}$ | (2a) | Al III |
| $\mathbf{B}_7$ | = | $y_4\,\mathbf{a}_2 + z_4\,\mathbf{a}_3$ | = | $y_4 b\,\hat{\mathbf{y}} + z_4 c\,\hat{\mathbf{z}}$ | (2a) | Al IV |
| $\mathbf{B}_8$ | = | $\frac{1}{2}\,\mathbf{a}_1 - y_4\,\mathbf{a}_2 + \left(\frac{1}{2} + z_4\right)\,\mathbf{a}_3$ | = | $\frac{1}{2}a\,\hat{\mathbf{x}} - y_4 b\,\hat{\mathbf{y}} + \left(\frac{1}{2} + z_4\right)c\,\hat{\mathbf{z}}$ | (2a) | Al IV |
| $\mathbf{B}_9$ | = | $y_5\,\mathbf{a}_2 + z_5\,\mathbf{a}_3$ | = | $y_5 b\,\hat{\mathbf{y}} + z_5 c\,\hat{\mathbf{z}}$ | (2a) | Al V |
| $\mathbf{B}_{10}$ | = | $\frac{1}{2}\,\mathbf{a}_1 - y_5\,\mathbf{a}_2 + \left(\frac{1}{2} + z_5\right)\,\mathbf{a}_3$ | = | $\frac{1}{2}a\,\hat{\mathbf{x}} - y_5 b\,\hat{\mathbf{y}} + \left(\frac{1}{2} + z_5\right)c\,\hat{\mathbf{z}}$ | (2a) | Al V |
| $\mathbf{B}_{11}$ | = | $y_6\,\mathbf{a}_2 + z_6\,\mathbf{a}_3$ | = | $y_6 b\,\hat{\mathbf{y}} + z_6 c\,\hat{\mathbf{z}}$ | (2a) | Al VI |
| $\mathbf{B}_{12}$ | = | $\frac{1}{2}\,\mathbf{a}_1 - y_6\,\mathbf{a}_2 + \left(\frac{1}{2} + z_6\right)\,\mathbf{a}_3$ | = | $\frac{1}{2}a\,\hat{\mathbf{x}} - y_6 b\,\hat{\mathbf{y}} + \left(\frac{1}{2} + z_6\right)c\,\hat{\mathbf{z}}$ | (2a) | Al VI |
| $\mathbf{B}_{13}$ | = | $y_7\,\mathbf{a}_2 + z_7\,\mathbf{a}_3$ | = | $y_7 b\,\hat{\mathbf{y}} + z_7 c\,\hat{\mathbf{z}}$ | (2a) | Al VII |
| $\mathbf{B}_{14}$ | = | $\frac{1}{2}\,\mathbf{a}_1 - y_7\,\mathbf{a}_2 + \left(\frac{1}{2} + z_7\right)\,\mathbf{a}_3$ | = | $\frac{1}{2}a\,\hat{\mathbf{x}} - y_7 b\,\hat{\mathbf{y}} + \left(\frac{1}{2} + z_7\right)c\,\hat{\mathbf{z}}$ | (2a) | Al VII |
| $\mathbf{B}_{15}$ | = | $y_8\,\mathbf{a}_2 + z_8\,\mathbf{a}_3$ | = | $y_8 b\,\hat{\mathbf{y}} + z_8 c\,\hat{\mathbf{z}}$ | (2a) | Al VIII |
| $\mathbf{B}_{16}$ | = | $\frac{1}{2}\,\mathbf{a}_1 - y_8\,\mathbf{a}_2 + \left(\frac{1}{2} + z_8\right)\,\mathbf{a}_3$ | = | $\frac{1}{2}a\,\hat{\mathbf{x}} - y_8 b\,\hat{\mathbf{y}} + \left(\frac{1}{2} + z_8\right)c\,\hat{\mathbf{z}}$ | (2a) | Al VIII |
| $\mathbf{B}_{17}$ | = | $y_9\,\mathbf{a}_2 + z_9\,\mathbf{a}_3$ | = | $y_9 b\,\hat{\mathbf{y}} + z_9 c\,\hat{\mathbf{z}}$ | (2a) | Al IX |
| $\mathbf{B}_{18}$ | = | $\frac{1}{2}\,\mathbf{a}_1 - y_9\,\mathbf{a}_2 + \left(\frac{1}{2} + z_9\right)\,\mathbf{a}_3$ | = | $\frac{1}{2}a\,\hat{\mathbf{x}} - y_9 b\,\hat{\mathbf{y}} + \left(\frac{1}{2} + z_9\right)c\,\hat{\mathbf{z}}$ | (2a) | Al IX |
| $\mathbf{B}_{19}$ | = | $y_{10}\,\mathbf{a}_2 + z_{10}\,\mathbf{a}_3$ | = | $y_{10} b\,\hat{\mathbf{y}} + z_{10} c\,\hat{\mathbf{z}}$ | (2a) | Al X |
| $\mathbf{B}_{20}$ | = | $\frac{1}{2}\,\mathbf{a}_1 - y_{10}\,\mathbf{a}_2 + \left(\frac{1}{2} + z_{10}\right)\,\mathbf{a}_3$ | = | $\frac{1}{2}a\,\hat{\mathbf{x}} - y_{10} b\,\hat{\mathbf{y}} + \left(\frac{1}{2} + z_{10}\right)c\,\hat{\mathbf{z}}$ | (2a) | Al X |
| $\mathbf{B}_{21}$ | = | $y_{11}\,\mathbf{a}_2 + z_{11}\,\mathbf{a}_3$ | = | $y_{11} b\,\hat{\mathbf{y}} + z_{11} c\,\hat{\mathbf{z}}$ | (2a) | Al XI |
| $\mathbf{B}_{22}$ | = | $\frac{1}{2}\,\mathbf{a}_1 - y_{11}\,\mathbf{a}_2 + \left(\frac{1}{2} + z_{11}\right)\,\mathbf{a}_3$ | = | $\frac{1}{2}a\,\hat{\mathbf{x}} - y_{11} b\,\hat{\mathbf{y}} + \left(\frac{1}{2} + z_{11}\right)c\,\hat{\mathbf{z}}$ | (2a) | Al XI |
| $\mathbf{B}_{23}$ | = | $y_{12}\,\mathbf{a}_2 + z_{12}\,\mathbf{a}_3$ | = | $y_{12} b\,\hat{\mathbf{y}} + z_{12} c\,\hat{\mathbf{z}}$ | (2a) | Al XII |
| $\mathbf{B}_{24}$ | = | $\frac{1}{2}\,\mathbf{a}_1 - y_{12}\,\mathbf{a}_2 + \left(\frac{1}{2} + z_{12}\right)\,\mathbf{a}_3$ | = | $\frac{1}{2}a\,\hat{\mathbf{x}} - y_{12} b\,\hat{\mathbf{y}} + \left(\frac{1}{2} + z_{12}\right)c\,\hat{\mathbf{z}}$ | (2a) | Al XII |
| $\mathbf{B}_{25}$ | = | $y_{13}\,\mathbf{a}_2 + z_{13}\,\mathbf{a}_3$ | = | $y_{13} b\,\hat{\mathbf{y}} + z_{13} c\,\hat{\mathbf{z}}$ | (2a) | Al XIII |
| $\mathbf{B}_{26}$ | = | $\frac{1}{2}\,\mathbf{a}_1 - y_{13}\,\mathbf{a}_2 + \left(\frac{1}{2} + z_{13}\right)\,\mathbf{a}_3$ | = | $\frac{1}{2}a\,\hat{\mathbf{x}} - y_{13} b\,\hat{\mathbf{y}} + \left(\frac{1}{2} + z_{13}\right)c\,\hat{\mathbf{z}}$ | (2a) | Al XIII |
| $\mathbf{B}_{27}$ | = | $y_{14}\,\mathbf{a}_2 + z_{14}\,\mathbf{a}_3$ | = | $y_{14} b\,\hat{\mathbf{y}} + z_{14} c\,\hat{\mathbf{z}}$ | (2a) | Al XIV |
| $\mathbf{B}_{28}$ | = | $\frac{1}{2}\,\mathbf{a}_1 - y_{14}\,\mathbf{a}_2 + \left(\frac{1}{2} + z_{14}\right)\,\mathbf{a}_3$ | = | $\frac{1}{2}a\,\mathbf{\hat{x}} - y_{14}b\,\mathbf{\hat{y}} + \left(\frac{1}{2} + z_{14}\right)c\,\mathbf{\hat{z}}$ | (2*a*) | Al XIV |
| $\mathbf{B}_{29}$ | = | $y_{15}\,\mathbf{a}_2 + z_{15}\,\mathbf{a}_3$ | = | $y_{15}b\,\mathbf{\hat{y}} + z_{15}c\,\mathbf{\hat{z}}$ | (2*a*) | Al XV |
| $\mathbf{B}_{30}$ | = | $\frac{1}{2}\,\mathbf{a}_1 - y_{15}\,\mathbf{a}_2 + \left(\frac{1}{2} + z_{15}\right)\,\mathbf{a}_3$ | = | $\frac{1}{2}a\,\mathbf{\hat{x}} - y_{15}b\,\mathbf{\hat{y}} + \left(\frac{1}{2} + z_{15}\right)c\,\mathbf{\hat{z}}$ | (2*a*) | Al XV |
| $\mathbf{B}_{31}$ | = | $y_{16}\,\mathbf{a}_2 + z_{16}\,\mathbf{a}_3$ | = | $y_{16}b\,\mathbf{\hat{y}} + z_{16}c\,\mathbf{\hat{z}}$ | (2*a*) | Al XVI |
| $\mathbf{B}_{32}$ | = | $\frac{1}{2}\,\mathbf{a}_1 - y_{16}\,\mathbf{a}_2 + \left(\frac{1}{2} + z_{16}\right)\,\mathbf{a}_3$ | = | $\frac{1}{2}a\,\mathbf{\hat{x}} - y_{16}b\,\mathbf{\hat{y}} + \left(\frac{1}{2} + z_{16}\right)c\,\mathbf{\hat{z}}$ | (2*a*) | Al XVI |
| $\mathbf{B}_{33}$ | = | $y_{17}\,\mathbf{a}_2 + z_{17}\,\mathbf{a}_3$ | = | $y_{17}b\,\mathbf{\hat{y}} + z_{17}c\,\mathbf{\hat{z}}$ | (2*a*) | Al XVII |
| $\mathbf{B}_{34}$ | = | $\frac{1}{2}\,\mathbf{a}_1 - y_{17}\,\mathbf{a}_2 + \left(\frac{1}{2} + z_{17}\right)\,\mathbf{a}_3$ | = | $\frac{1}{2}a\,\mathbf{\hat{x}} - y_{17}b\,\mathbf{\hat{y}} + \left(\frac{1}{2} + z_{17}\right)c\,\mathbf{\hat{z}}$ | (2*a*) | Al XVII |
| $\mathbf{B}_{35}$ | = | $y_{18}\,\mathbf{a}_2 + z_{18}\,\mathbf{a}_3$ | = | $y_{18}b\,\mathbf{\hat{y}} + z_{18}c\,\mathbf{\hat{z}}$ | (2*a*) | Co I |
| $\mathbf{B}_{36}$ | = | $\frac{1}{2}\,\mathbf{a}_1 - y_{18}\,\mathbf{a}_2 + \left(\frac{1}{2} + z_{18}\right)\,\mathbf{a}_3$ | = | $\frac{1}{2}a\,\mathbf{\hat{x}} - y_{18}b\,\mathbf{\hat{y}} + \left(\frac{1}{2} + z_{18}\right)c\,\mathbf{\hat{z}}$ | (2*a*) | Co I |
| $\mathbf{B}_{37}$ | = | $y_{19}\,\mathbf{a}_2 + z_{19}\,\mathbf{a}_3$ | = | $y_{19}b\,\mathbf{\hat{y}} + z_{19}c\,\mathbf{\hat{z}}$ | (2*a*) | Co II |
| $\mathbf{B}_{38}$ | = | $\frac{1}{2}\,\mathbf{a}_1 - y_{19}\,\mathbf{a}_2 + \left(\frac{1}{2} + z_{19}\right)\,\mathbf{a}_3$ | = | $\frac{1}{2}a\,\mathbf{\hat{x}} - y_{19}b\,\mathbf{\hat{y}} + \left(\frac{1}{2} + z_{19}\right)c\,\mathbf{\hat{z}}$ | (2*a*) | Co II |
| $\mathbf{B}_{39}$ | = | $y_{20}\,\mathbf{a}_2 + z_{20}\,\mathbf{a}_3$ | = | $y_{20}b\,\mathbf{\hat{y}} + z_{20}c\,\mathbf{\hat{z}}$ | (2*a*) | Co III |
| $\mathbf{B}_{40}$ | = | $\frac{1}{2}\,\mathbf{a}_1 - y_{20}\,\mathbf{a}_2 + \left(\frac{1}{2} + z_{20}\right)\,\mathbf{a}_3$ | = | $\frac{1}{2}a\,\mathbf{\hat{x}} - y_{20}b\,\mathbf{\hat{y}} + \left(\frac{1}{2} + z_{20}\right)c\,\mathbf{\hat{z}}$ | (2*a*) | Co III |
| $\mathbf{B}_{41}$ | = | $y_{21}\,\mathbf{a}_2 + z_{21}\,\mathbf{a}_3$ | = | $y_{21}b\,\mathbf{\hat{y}} + z_{21}c\,\mathbf{\hat{z}}$ | (2*a*) | Co IV |
| $\mathbf{B}_{42}$ | = | $\frac{1}{2}\,\mathbf{a}_1 - y_{21}\,\mathbf{a}_2 + \left(\frac{1}{2} + z_{21}\right)\,\mathbf{a}_3$ | = | $\frac{1}{2}a\,\mathbf{\hat{x}} - y_{21}b\,\mathbf{\hat{y}} + \left(\frac{1}{2} + z_{21}\right)c\,\mathbf{\hat{z}}$ | (2*a*) | Co IV |
| $\mathbf{B}_{43}$ | = | $y_{22}\,\mathbf{a}_2 + z_{22}\,\mathbf{a}_3$ | = | $y_{22}b\,\mathbf{\hat{y}} + z_{22}c\,\mathbf{\hat{z}}$ | (2*a*) | Co V |
| $\mathbf{B}_{44}$ | = | $\frac{1}{2}\,\mathbf{a}_1 - y_{22}\,\mathbf{a}_2 + \left(\frac{1}{2} + z_{22}\right)\,\mathbf{a}_3$ | = | $\frac{1}{2}a\,\mathbf{\hat{x}} - y_{22}b\,\mathbf{\hat{y}} + \left(\frac{1}{2} + z_{22}\right)c\,\mathbf{\hat{z}}$ | (2*a*) | Co V |
| $\mathbf{B}_{45}$ | = | $y_{23}\,\mathbf{a}_2 + z_{23}\,\mathbf{a}_3$ | = | $y_{23}b\,\mathbf{\hat{y}} + z_{23}c\,\mathbf{\hat{z}}$ | (2*a*) | Co VI |
| $\mathbf{B}_{46}$ | = | $\frac{1}{2}\,\mathbf{a}_1 - y_{23}\,\mathbf{a}_2 + \left(\frac{1}{2} + z_{23}\right)\,\mathbf{a}_3$ | = | $\frac{1}{2}a\,\mathbf{\hat{x}} - y_{23}b\,\mathbf{\hat{y}} + \left(\frac{1}{2} + z_{23}\right)c\,\mathbf{\hat{z}}$ | (2*a*) | Co VI |
| $\mathbf{B}_{47}$ | = | $y_{24}\,\mathbf{a}_2 + z_{24}\,\mathbf{a}_3$ | = | $y_{24}b\,\mathbf{\hat{y}} + z_{24}c\,\mathbf{\hat{z}}$ | (2*a*) | Co VII |
| $\mathbf{B}_{48}$ | = | $\frac{1}{2}\,\mathbf{a}_1 - y_{24}\,\mathbf{a}_2 + \left(\frac{1}{2} + z_{24}\right)\,\mathbf{a}_3$ | = | $\frac{1}{2}a\,\mathbf{\hat{x}} - y_{24}b\,\mathbf{\hat{y}} + \left(\frac{1}{2} + z_{24}\right)c\,\mathbf{\hat{z}}$ | (2*a*) | Co VII |
| $\mathbf{B}_{49}$ | = | $y_{25}\,\mathbf{a}_2 + z_{25}\,\mathbf{a}_3$ | = | $y_{25}b\,\mathbf{\hat{y}} + z_{25}c\,\mathbf{\hat{z}}$ | (2*a*) | Co VIII |
| $\mathbf{B}_{50}$ | = | $\frac{1}{2}\,\mathbf{a}_1 - y_{25}\,\mathbf{a}_2 + \left(\frac{1}{2} + z_{25}\right)\,\mathbf{a}_3$ | = | $\frac{1}{2}a\,\mathbf{\hat{x}} - y_{25}b\,\mathbf{\hat{y}} + \left(\frac{1}{2} + z_{25}\right)c\,\mathbf{\hat{z}}$ | (2*a*) | Co VIII |
| $\mathbf{B}_{51}$ | = | $x_{26}\,\mathbf{a}_1 + y_{26}\,\mathbf{a}_2 + z_{26}\,\mathbf{a}_3$ | = | $x_{26}a\,\mathbf{\hat{x}} + y_{26}b\,\mathbf{\hat{y}} + z_{26}c\,\mathbf{\hat{z}}$ | (4*b*) | Al XVIII |
| $\mathbf{B}_{52}$ | = | $\left(\frac{1}{2} - x_{26}\right)\,\mathbf{a}_1 - y_{26}\,\mathbf{a}_2 + \left(\frac{1}{2} + z_{26}\right)\,\mathbf{a}_3$ | = | $\left(\frac{1}{2} - x_{26}\right)a\,\mathbf{\hat{x}} - y_{26}b\,\mathbf{\hat{y}} + \left(\frac{1}{2} + z_{26}\right)c\,\mathbf{\hat{z}}$ | (4*b*) | Al XVIII |
| $\mathbf{B}_{53}$ | = | $\left(\frac{1}{2} + x_{26}\right)\,\mathbf{a}_1 - y_{26}\,\mathbf{a}_2 + \left(\frac{1}{2} + z_{26}\right)\,\mathbf{a}_3$ | = | $\left(\frac{1}{2} + x_{26}\right)a\,\mathbf{\hat{x}} - y_{26}b\,\mathbf{\hat{y}} + \left(\frac{1}{2} + z_{26}\right)c\,\mathbf{\hat{z}}$ | (4*b*) | Al XVIII |
| $\mathbf{B}_{54}$ | = | $-x_{26}\,\mathbf{a}_1 + y_{26}\,\mathbf{a}_2 + z_{26}\,\mathbf{a}_3$ | = | $-x_{26}a\,\mathbf{\hat{x}} + y_{26}b\,\mathbf{\hat{y}} + z_{26}c\,\mathbf{\hat{z}}$ | (4*b*) | Al XVIII |
| $\mathbf{B}_{55}$ | = | $x_{27}\,\mathbf{a}_1 + y_{27}\,\mathbf{a}_2 + z_{27}\,\mathbf{a}_3$ | = | $x_{27}a\,\mathbf{\hat{x}} + y_{27}b\,\mathbf{\hat{y}} + z_{27}c\,\mathbf{\hat{z}}$ | (4*b*) | Al XIX |
| $\mathbf{B}_{56}$ | = | $\left(\frac{1}{2} - x_{27}\right)\,\mathbf{a}_1 - y_{27}\,\mathbf{a}_2 + \left(\frac{1}{2} + z_{27}\right)\,\mathbf{a}_3$ | = | $\left(\frac{1}{2} - x_{27}\right)a\,\mathbf{\hat{x}} - y_{27}b\,\mathbf{\hat{y}} + \left(\frac{1}{2} + z_{27}\right)c\,\mathbf{\hat{z}}$ | (4*b*) | Al XIX |
| $\mathbf{B}_{57}$ | = | $\left(\frac{1}{2} + x_{27}\right)\,\mathbf{a}_1 - y_{27}\,\mathbf{a}_2 + \left(\frac{1}{2} + z_{27}\right)\,\mathbf{a}_3$ | = | $\left(\frac{1}{2} + x_{27}\right)a\,\mathbf{\hat{x}} - y_{27}b\,\mathbf{\hat{y}} + \left(\frac{1}{2} + z_{27}\right)c\,\mathbf{\hat{z}}$ | (4*b*) | Al XIX |
| $\mathbf{B}_{58}$ | = | $-x_{27}\,\mathbf{a}_1 + y_{27}\,\mathbf{a}_2 + z_{27}\,\mathbf{a}_3$ | = | $-x_{27}a\,\mathbf{\hat{x}} + y_{27}b\,\mathbf{\hat{y}} + z_{27}c\,\mathbf{\hat{z}}$ | (4*b*) | Al XIX |
| $\mathbf{B}_{59}$ | = | $x_{28}\,\mathbf{a}_1 + y_{28}\,\mathbf{a}_2 + z_{28}\,\mathbf{a}_3$ | = | $x_{28}a\,\mathbf{\hat{x}} + y_{28}b\,\mathbf{\hat{y}} + z_{28}c\,\mathbf{\hat{z}}$ | (4*b*) | Al XX |
| $\mathbf{B}_{60}$ | = | $\left(\frac{1}{2} - x_{28}\right)\,\mathbf{a}_1 - y_{28}\,\mathbf{a}_2 + \left(\frac{1}{2} + z_{28}\right)\,\mathbf{a}_3$ | = | $\left(\frac{1}{2} - x_{28}\right)a\,\mathbf{\hat{x}} - y_{28}b\,\mathbf{\hat{y}} + \left(\frac{1}{2} + z_{28}\right)c\,\mathbf{\hat{z}}$ | (4*b*) | Al XX |
| $\mathbf{B}_{61}$ | = | $\left(\frac{1}{2} + x_{28}\right)\,\mathbf{a}_1 - y_{28}\,\mathbf{a}_2 + \left(\frac{1}{2} + z_{28}\right)\,\mathbf{a}_3$ | = | $\left(\frac{1}{2} + x_{28}\right)a\,\mathbf{\hat{x}} - y_{28}b\,\mathbf{\hat{y}} + \left(\frac{1}{2} + z_{28}\right)c\,\mathbf{\hat{z}}$ | (4*b*) | Al XX |
| $\mathbf{B}_{62}$ | = | $-x_{28}\,\mathbf{a}_1 + y_{28}\,\mathbf{a}_2 + z_{28}\,\mathbf{a}_3$ | = | $-x_{28}a\,\mathbf{\hat{x}} + y_{28}b\,\mathbf{\hat{y}} + z_{28}c\,\mathbf{\hat{z}}$ | (4*b*) | Al XX |
| $\mathbf{B}_{63}$ | = | $x_{29}\,\mathbf{a}_1 + y_{29}\,\mathbf{a}_2 + z_{29}\,\mathbf{a}_3$ | = | $x_{29}a\,\mathbf{\hat{x}} + y_{29}b\,\mathbf{\hat{y}} + z_{29}c\,\mathbf{\hat{z}}$ | (4*b*) | Al XXI |
| $\mathbf{B}_{64}$ | = | $\left(\frac{1}{2} - x_{29}\right)\mathbf{a}_1 - y_{29}\,\mathbf{a}_2 + \left(\frac{1}{2} + z_{29}\right)\mathbf{a}_3$ | = | $\left(\frac{1}{2} - x_{29}\right)a\,\hat{\mathbf{x}} - y_{29}b\,\hat{\mathbf{y}} + \left(\frac{1}{2} + z_{29}\right)c\,\hat{\mathbf{z}}$ | (4$b$) | Al XXI |
| $\mathbf{B}_{65}$ | = | $\left(\frac{1}{2} + x_{29}\right)\mathbf{a}_1 - y_{29}\,\mathbf{a}_2 + \left(\frac{1}{2} + z_{29}\right)\mathbf{a}_3$ | = | $\left(\frac{1}{2} + x_{29}\right)a\,\hat{\mathbf{x}} - y_{29}b\,\hat{\mathbf{y}} + \left(\frac{1}{2} + z_{29}\right)c\,\hat{\mathbf{z}}$ | (4$b$) | Al XXI |
| $\mathbf{B}_{66}$ | = | $-x_{29}\,\mathbf{a}_1 + y_{29}\,\mathbf{a}_2 + z_{29}\,\mathbf{a}_3$ | = | $-x_{29}a\,\hat{\mathbf{x}} + y_{29}b\,\hat{\mathbf{y}} + z_{29}c\,\hat{\mathbf{z}}$ | (4$b$) | Al XXI |
| $\mathbf{B}_{67}$ | = | $x_{30}\,\mathbf{a}_1 + y_{30}\,\mathbf{a}_2 + z_{30}\,\mathbf{a}_3$ | = | $x_{30}a\,\hat{\mathbf{x}} + y_{30}b\,\hat{\mathbf{y}} + z_{30}c\,\hat{\mathbf{z}}$ | (4$b$) | Al XXII |
| $\mathbf{B}_{68}$ | = | $\left(\frac{1}{2} - x_{30}\right)\mathbf{a}_1 - y_{30}\,\mathbf{a}_2 + \left(\frac{1}{2} + z_{30}\right)\mathbf{a}_3$ | = | $\left(\frac{1}{2} - x_{30}\right)a\,\hat{\mathbf{x}} - y_{30}b\,\hat{\mathbf{y}} + \left(\frac{1}{2} + z_{30}\right)c\,\hat{\mathbf{z}}$ | (4$b$) | Al XXII |
| $\mathbf{B}_{69}$ | = | $\left(\frac{1}{2} + x_{30}\right)\mathbf{a}_1 - y_{30}\,\mathbf{a}_2 + \left(\frac{1}{2} + z_{30}\right)\mathbf{a}_3$ | = | $\left(\frac{1}{2} + x_{30}\right)a\,\hat{\mathbf{x}} - y_{30}b\,\hat{\mathbf{y}} + \left(\frac{1}{2} + z_{30}\right)c\,\hat{\mathbf{z}}$ | (4$b$) | Al XXII |
| $\mathbf{B}_{70}$ | = | $-x_{30}\,\mathbf{a}_1 + y_{30}\,\mathbf{a}_2 + z_{30}\,\mathbf{a}_3$ | = | $-x_{30}a\,\hat{\mathbf{x}} + y_{30}b\,\hat{\mathbf{y}} + z_{30}c\,\hat{\mathbf{z}}$ | (4$b$) | Al XXII |
| $\mathbf{B}_{71}$ | = | $x_{31}\,\mathbf{a}_1 + y_{31}\,\mathbf{a}_2 + z_{31}\,\mathbf{a}_3$ | = | $x_{31}a\,\hat{\mathbf{x}} + y_{31}b\,\hat{\mathbf{y}} + z_{31}c\,\hat{\mathbf{z}}$ | (4$b$) | Al XXIII |
| $\mathbf{B}_{72}$ | = | $\left(\frac{1}{2} - x_{31}\right)\mathbf{a}_1 - y_{31}\,\mathbf{a}_2 + \left(\frac{1}{2} + z_{31}\right)\mathbf{a}_3$ | = | $\left(\frac{1}{2} - x_{31}\right)a\,\hat{\mathbf{x}} - y_{31}b\,\hat{\mathbf{y}} + \left(\frac{1}{2} + z_{31}\right)c\,\hat{\mathbf{z}}$ | (4$b$) | Al XXIII |
| $\mathbf{B}_{73}$ | = | $\left(\frac{1}{2} + x_{31}\right)\mathbf{a}_1 - y_{31}\,\mathbf{a}_2 + \left(\frac{1}{2} + z_{31}\right)\mathbf{a}_3$ | = | $\left(\frac{1}{2} + x_{31}\right)a\,\hat{\mathbf{x}} - y_{31}b\,\hat{\mathbf{y}} + \left(\frac{1}{2} + z_{31}\right)c\,\hat{\mathbf{z}}$ | (4$b$) | Al XXIII |
| $\mathbf{B}_{74}$ | = | $-x_{31}\,\mathbf{a}_1 + y_{31}\,\mathbf{a}_2 + z_{31}\,\mathbf{a}_3$ | = | $-x_{31}a\,\hat{\mathbf{x}} + y_{31}b\,\hat{\mathbf{y}} + z_{31}c\,\hat{\mathbf{z}}$ | (4$b$) | Al XXIII |
| $\mathbf{B}_{75}$ | = | $x_{32}\,\mathbf{a}_1 + y_{32}\,\mathbf{a}_2 + z_{32}\,\mathbf{a}_3$ | = | $x_{32}a\,\hat{\mathbf{x}} + y_{32}b\,\hat{\mathbf{y}} + z_{32}c\,\hat{\mathbf{z}}$ | (4$b$) | Al XXIV |
| $\mathbf{B}_{76}$ | = | $\left(\frac{1}{2} - x_{32}\right)\mathbf{a}_1 - y_{32}\,\mathbf{a}_2 + \left(\frac{1}{2} + z_{32}\right)\mathbf{a}_3$ | = | $\left(\frac{1}{2} - x_{32}\right)a\,\hat{\mathbf{x}} - y_{32}b\,\hat{\mathbf{y}} + \left(\frac{1}{2} + z_{32}\right)c\,\hat{\mathbf{z}}$ | (4$b$) | Al XXIV |
| $\mathbf{B}_{77}$ | = | $\left(\frac{1}{2} + x_{32}\right)\mathbf{a}_1 - y_{32}\,\mathbf{a}_2 + \left(\frac{1}{2} + z_{32}\right)\mathbf{a}_3$ | = | $\left(\frac{1}{2} + x_{32}\right)a\,\hat{\mathbf{x}} - y_{32}b\,\hat{\mathbf{y}} + \left(\frac{1}{2} + z_{32}\right)c\,\hat{\mathbf{z}}$ | (4$b$) | Al XXIV |
| $\mathbf{B}_{78}$ | = | $-x_{32}\,\mathbf{a}_1 + y_{32}\,\mathbf{a}_2 + z_{32}\,\mathbf{a}_3$ | = | $-x_{32}a\,\hat{\mathbf{x}} + y_{32}b\,\hat{\mathbf{y}} + z_{32}c\,\hat{\mathbf{z}}$ | (4$b$) | Al XXIV |
| $\mathbf{B}_{79}$ | = | $x_{33}\,\mathbf{a}_1 + y_{33}\,\mathbf{a}_2 + z_{33}\,\mathbf{a}_3$ | = | $x_{33}a\,\hat{\mathbf{x}} + y_{33}b\,\hat{\mathbf{y}} + z_{33}c\,\hat{\mathbf{z}}$ | (4$b$) | Al XXV |
| $\mathbf{B}_{80}$ | = | $\left(\frac{1}{2} - x_{33}\right)\mathbf{a}_1 - y_{33}\,\mathbf{a}_2 + \left(\frac{1}{2} + z_{33}\right)\mathbf{a}_3$ | = | $\left(\frac{1}{2} - x_{33}\right)a\,\hat{\mathbf{x}} - y_{33}b\,\hat{\mathbf{y}} + \left(\frac{1}{2} + z_{33}\right)c\,\hat{\mathbf{z}}$ | (4$b$) | Al XXV |
| $\mathbf{B}_{81}$ | = | $\left(\frac{1}{2} + x_{33}\right)\mathbf{a}_1 - y_{33}\,\mathbf{a}_2 + \left(\frac{1}{2} + z_{33}\right)\mathbf{a}_3$ | = | $\left(\frac{1}{2} + x_{33}\right)a\,\hat{\mathbf{x}} - y_{33}b\,\hat{\mathbf{y}} + \left(\frac{1}{2} + z_{33}\right)c\,\hat{\mathbf{z}}$ | (4$b$) | Al XXV |
| $\mathbf{B}_{82}$ | = | $-x_{33}\,\mathbf{a}_1 + y_{33}\,\mathbf{a}_2 + z_{33}\,\mathbf{a}_3$ | = | $-x_{33}a\,\hat{\mathbf{x}} + y_{33}b\,\hat{\mathbf{y}} + z_{33}c\,\hat{\mathbf{z}}$ | (4$b$) | Al XXV |
| $\mathbf{B}_{83}$ | = | $x_{34}\,\mathbf{a}_1 + y_{34}\,\mathbf{a}_2 + z_{34}\,\mathbf{a}_3$ | = | $x_{34}a\,\hat{\mathbf{x}} + y_{34}b\,\hat{\mathbf{y}} + z_{34}c\,\hat{\mathbf{z}}$ | (4$b$) | Al XXVI |
| $\mathbf{B}_{84}$ | = | $\left(\frac{1}{2} - x_{34}\right)\mathbf{a}_1 - y_{34}\,\mathbf{a}_2 + \left(\frac{1}{2} + z_{34}\right)\mathbf{a}_3$ | = | $\left(\frac{1}{2} - x_{34}\right)a\,\hat{\mathbf{x}} - y_{34}b\,\hat{\mathbf{y}} + \left(\frac{1}{2} + z_{34}\right)c\,\hat{\mathbf{z}}$ | (4$b$) | Al XXVI |
| $\mathbf{B}_{85}$ | = | $\left(\frac{1}{2} + x_{34}\right)\mathbf{a}_1 - y_{34}\,\mathbf{a}_2 + \left(\frac{1}{2} + z_{34}\right)\mathbf{a}_3$ | = | $\left(\frac{1}{2} + x_{34}\right)a\,\hat{\mathbf{x}} - y_{34}b\,\hat{\mathbf{y}} + \left(\frac{1}{2} + z_{34}\right)c\,\hat{\mathbf{z}}$ | (4$b$) | Al XXVI |
| $\mathbf{B}_{86}$ | = | $-x_{34}\,\mathbf{a}_1 + y_{34}\,\mathbf{a}_2 + z_{34}\,\mathbf{a}_3$ | = | $-x_{34}a\,\hat{\mathbf{x}} + y_{34}b\,\hat{\mathbf{y}} + z_{34}c\,\hat{\mathbf{z}}$ | (4$b$) | Al XXVI |
| $\mathbf{B}_{87}$ | = | $x_{35}\,\mathbf{a}_1 + y_{35}\,\mathbf{a}_2 + z_{35}\,\mathbf{a}_3$ | = | $x_{35}a\,\hat{\mathbf{x}} + y_{35}b\,\hat{\mathbf{y}} + z_{35}c\,\hat{\mathbf{z}}$ | (4$b$) | Al XXVII |
| $\mathbf{B}_{88}$ | = | $\left(\frac{1}{2} - x_{35}\right)\mathbf{a}_1 - y_{35}\,\mathbf{a}_2 + \left(\frac{1}{2} + z_{35}\right)\mathbf{a}_3$ | = | $\left(\frac{1}{2} - x_{35}\right)a\,\hat{\mathbf{x}} - y_{35}b\,\hat{\mathbf{y}} + \left(\frac{1}{2} + z_{35}\right)c\,\hat{\mathbf{z}}$ | (4$b$) | Al XXVII |
| $\mathbf{B}_{89}$ | = | $\left(\frac{1}{2} + x_{35}\right)\mathbf{a}_1 - y_{35}\,\mathbf{a}_2 + \left(\frac{1}{2} + z_{35}\right)\mathbf{a}_3$ | = | $\left(\frac{1}{2} + x_{35}\right)a\,\hat{\mathbf{x}} - y_{35}b\,\hat{\mathbf{y}} + \left(\frac{1}{2} + z_{35}\right)c\,\hat{\mathbf{z}}$ | (4$b$) | Al XXVII |
| $\mathbf{B}_{90}$ | = | $-x_{35}\,\mathbf{a}_1 + y_{35}\,\mathbf{a}_2 + z_{35}\,\mathbf{a}_3$ | = | $-x_{35}a\,\hat{\mathbf{x}} + y_{35}b\,\hat{\mathbf{y}} + z_{35}c\,\hat{\mathbf{z}}$ | (4$b$) | Al XXVII |
| $\mathbf{B}_{91}$ | = | $x_{36}\,\mathbf{a}_1 + y_{36}\,\mathbf{a}_2 + z_{36}\,\mathbf{a}_3$ | = | $x_{36}a\,\hat{\mathbf{x}} + y_{36}b\,\hat{\mathbf{y}} + z_{36}c\,\hat{\mathbf{z}}$ | (4$b$) | Al XXVIII |
| $\mathbf{B}_{92}$ | = | $\left(\frac{1}{2} - x_{36}\right)\mathbf{a}_1 - y_{36}\,\mathbf{a}_2 + \left(\frac{1}{2} + z_{36}\right)\mathbf{a}_3$ | = | $\left(\frac{1}{2} - x_{36}\right)a\,\hat{\mathbf{x}} - y_{36}b\,\hat{\mathbf{y}} + \left(\frac{1}{2} + z_{36}\right)c\,\hat{\mathbf{z}}$ | (4$b$) | Al XXVIII |
| $\mathbf{B}_{93}$ | = | $\left(\frac{1}{2} + x_{36}\right)\mathbf{a}_1 - y_{36}\,\mathbf{a}_2 + \left(\frac{1}{2} + z_{36}\right)\mathbf{a}_3$ | = | $\left(\frac{1}{2} + x_{36}\right)a\,\hat{\mathbf{x}} - y_{36}b\,\hat{\mathbf{y}} + \left(\frac{1}{2} + z_{36}\right)c\,\hat{\mathbf{z}}$ | (4$b$) | Al XXVIII |
| $\mathbf{B}_{94}$ | = | $-x_{36}\,\mathbf{a}_1 + y_{36}\,\mathbf{a}_2 + z_{36}\,\mathbf{a}_3$ | = | $-x_{36}a\,\hat{\mathbf{x}} + y_{36}b\,\hat{\mathbf{y}} + z_{36}c\,\hat{\mathbf{z}}$ | (4$b$) | Al XXVIII |
| $\mathbf{B}_{95}$ | = | $x_{37}\,\mathbf{a}_1 + y_{37}\,\mathbf{a}_2 + z_{37}\,\mathbf{a}_3$ | = | $x_{37}a\,\hat{\mathbf{x}} + y_{37}b\,\hat{\mathbf{y}} + z_{37}c\,\hat{\mathbf{z}}$ | (4$b$) | Co IX |
| $\mathbf{B}_{96}$ | = | $\left(\frac{1}{2} - x_{37}\right)\mathbf{a}_1 - y_{37}\,\mathbf{a}_2 + \left(\frac{1}{2} + z_{37}\right)\mathbf{a}_3$ | = | $\left(\frac{1}{2} - x_{37}\right)a\,\hat{\mathbf{x}} - y_{37}b\,\hat{\mathbf{y}} + \left(\frac{1}{2} + z_{37}\right)c\,\hat{\mathbf{z}}$ | (4$b$) | Co IX |
| $\mathbf{B}_{97}$ | = | $\left(\frac{1}{2} + x_{37}\right)\mathbf{a}_1 - y_{37}\,\mathbf{a}_2 + \left(\frac{1}{2} + z_{37}\right)\mathbf{a}_3$ | = | $\left(\frac{1}{2} + x_{37}\right)a\,\hat{\mathbf{x}} - y_{37}b\,\hat{\mathbf{y}} + \left(\frac{1}{2} + z_{37}\right)c\,\hat{\mathbf{z}}$ | (4$b$) | Co IX |
| $\mathbf{B}_{98}$ | = | $-x_{37}\,\mathbf{a}_1 + y_{37}\,\mathbf{a}_2 + z_{37}\,\mathbf{a}_3$ | = | $-x_{37}a\,\hat{\mathbf{x}} + y_{37}b\,\hat{\mathbf{y}} + z_{37}c\,\hat{\mathbf{z}}$ | (4$b$) | Co IX |
| $\mathbf{B}_{99}$ | = | $x_{38}\,\mathbf{a}_1 + y_{38}\,\mathbf{a}_2 + z_{38}\,\mathbf{a}_3$ | = | $x_{38}a\,\hat{\mathbf{x}} + y_{38}b\,\hat{\mathbf{y}} + z_{38}c\,\hat{\mathbf{z}}$ | (4$b$) | Co X |
| $\mathbf{B}_{100}$ | = | $\left(\frac{1}{2} - x_{38}\right) \mathbf{a}_1 - y_{38} \mathbf{a}_2 + \left(\frac{1}{2} + z_{38}\right) \mathbf{a}_3$ | = | $\left(\frac{1}{2} - x_{38}\right) a \hat{\mathbf{x}} - y_{38} b \hat{\mathbf{y}} + \left(\frac{1}{2} + z_{38}\right) c \hat{\mathbf{z}}$ | (4$b$) | Co X |
| $\mathbf{B}_{101}$ | = | $\left(\frac{1}{2} + x_{38}\right) \mathbf{a}_1 - y_{38} \mathbf{a}_2 + \left(\frac{1}{2} + z_{38}\right) \mathbf{a}_3$ | = | $\left(\frac{1}{2} + x_{38}\right) a \hat{\mathbf{x}} - y_{38} b \hat{\mathbf{y}} + \left(\frac{1}{2} + z_{38}\right) c \hat{\mathbf{z}}$ | (4$b$) | Co X |
| $\mathbf{B}_{102}$ | = | $-x_{38} \mathbf{a}_1 + y_{38} \mathbf{a}_2 + z_{38} \mathbf{a}_3$ | = | $-x_{38} a \hat{\mathbf{x}} + y_{38} b \hat{\mathbf{y}} + z_{38} c \hat{\mathbf{z}}$ | (4$b$) | Co X |

**Geometry files:**

- CIF: pp. 1585
- POSCAR: pp. 1586

# $Mg(ClO_4)_2{\cdot}6H_2O$ ($H4_{11}$) Structure: A2B6CD8_oP34_31_2a_2a2b_a_4a2b

http://aflow.org/prototype-encyclopedia/A2B6CD8_oP34_31_2a_2a2b_a_4a2b

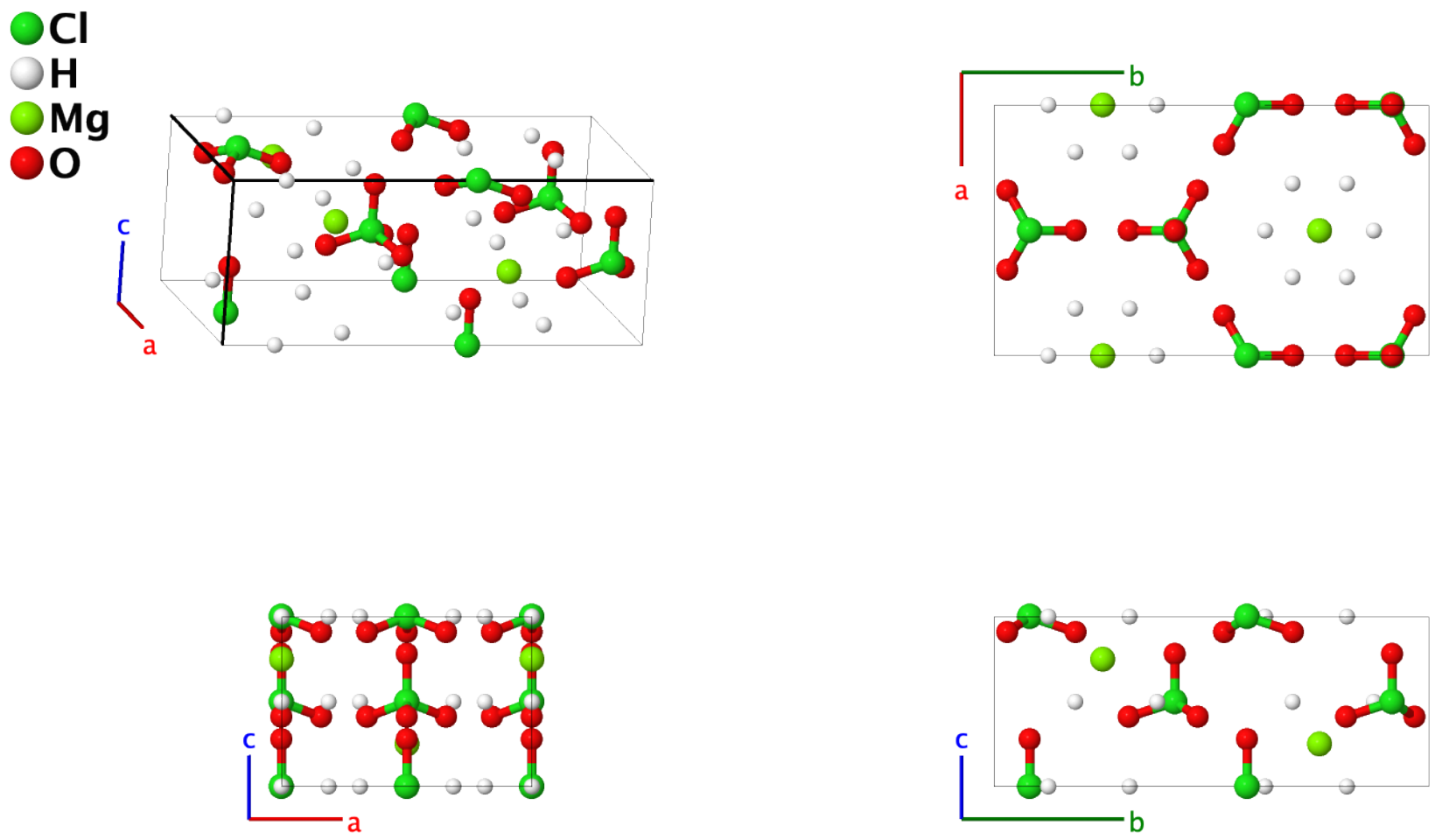

| | | |
|---|---|---|
| **Prototype** | : | $Cl_2(H_2O)_6MgO_8$ |
| **AFLOW prototype label** | : | A2B6CD8_oP34_31_2a_2a2b_a_4a2b |
| ***Strukturbericht* designation** | : | $H4_{11}$ |
| **Pearson symbol** | : | oP34 |
| **Space group number** | : | 31 |
| **Space group symbol** | : | $Pmn2_1$ |
| **AFLOW prototype command** | : | `aflow --proto=A2B6CD8_oP34_31_2a_2a2b_a_4a2b` `--params=`$a, b/a, c/a, y_1, z_1, y_2, z_2, y_3, z_3, y_4, z_4, y_5, z_5, y_6, z_6, y_7, z_7, y_8, z_8, y_9, z_9, x_{10}, y_{10}, z_{10}, x_{11}, y_{11}, z_{11}, x_{12}, y_{12}, z_{12}, x_{13}, y_{13}, z_{13}$ |

### Other compounds with this structure

- $Co(ClO_4)_2{\cdot}6H_2O$, $Fe(ClO_4)_2{\cdot}6H_2O$, $Mn(ClO_4)_2{\cdot}6H_2O$, $Ni(ClO_4)_2{\cdot}6H_2O$, $Zn(ClO_4)_2{\cdot}6H_2O$, $Mg(BF_4)_2{\cdot}6H_2O$, $Co(BF_4)_2{\cdot}6H_2O$, $Fe(BF_4)_2{\cdot}6H_2O$, $Mn(BF_4)_2{\cdot}6H_2O$, $Ni(BF_4)_2{\cdot}6H_2O$, and $Zn(BF_4)_2{\cdot}6H_2O$

---

- (Gottfried, 1937) writes $z_{13} = 0.408$ for the coordinate of the (H2O-II) molecule, but West uses $z_{13} = 0$, which gives a symmetric arrangement of water molecules around the chlorine atom. We use West's value here.

### Simple Orthorhombic primitive vectors:

$$\mathbf{a}_1 = a\,\hat{\mathbf{x}}$$
$$\mathbf{a}_2 = b\,\hat{\mathbf{y}}$$
$$\mathbf{a}_3 = c\,\hat{\mathbf{z}}$$

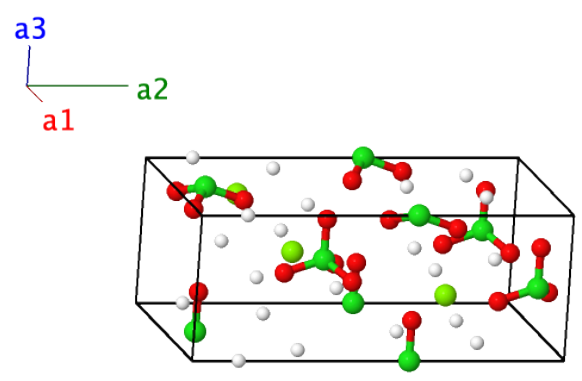

### Basis vectors:

| | | Lattice Coordinates | | Cartesian Coordinates | Wyckoff Position | Atom Type |
|---|---|---|---|---|---|---|
| $\mathbf{B}_1$ | = | $y_1\,\mathbf{a}_2 + z_1\,\mathbf{a}_3$ | = | $y_1 b\,\hat{\mathbf{y}} + z_1 c\,\hat{\mathbf{z}}$ | (2a) | Cl I |
| $\mathbf{B}_2$ | = | $\frac{1}{2}\,\mathbf{a}_1 - y_1\,\mathbf{a}_2 + \left(\frac{1}{2} + z_1\right)\mathbf{a}_3$ | = | $\frac{1}{2}a\,\hat{\mathbf{x}} - y_1 b\,\hat{\mathbf{y}} + \left(\frac{1}{2} + z_1\right)c\,\hat{\mathbf{z}}$ | (2a) | Cl I |
| $\mathbf{B}_3$ | = | $y_2\,\mathbf{a}_2 + z_2\,\mathbf{a}_3$ | = | $y_2 b\,\hat{\mathbf{y}} + z_2 c\,\hat{\mathbf{z}}$ | (2a) | Cl II |
| $\mathbf{B}_4$ | = | $\frac{1}{2}\,\mathbf{a}_1 - y_2\,\mathbf{a}_2 + \left(\frac{1}{2} + z_2\right)\mathbf{a}_3$ | = | $\frac{1}{2}a\,\hat{\mathbf{x}} - y_2 b\,\hat{\mathbf{y}} + \left(\frac{1}{2} + z_2\right)c\,\hat{\mathbf{z}}$ | (2a) | Cl II |
| $\mathbf{B}_5$ | = | $y_3\,\mathbf{a}_2 + z_3\,\mathbf{a}_3$ | = | $y_3 b\,\hat{\mathbf{y}} + z_3 c\,\hat{\mathbf{z}}$ | (2a) | $H_2O$ I |
| $\mathbf{B}_6$ | = | $\frac{1}{2}\,\mathbf{a}_1 - y_3\,\mathbf{a}_2 + \left(\frac{1}{2} + z_3\right)\mathbf{a}_3$ | = | $\frac{1}{2}a\,\hat{\mathbf{x}} - y_3 b\,\hat{\mathbf{y}} + \left(\frac{1}{2} + z_3\right)c\,\hat{\mathbf{z}}$ | (2a) | $H_2O$ I |
| $\mathbf{B}_7$ | = | $y_4\,\mathbf{a}_2 + z_4\,\mathbf{a}_3$ | = | $y_4 b\,\hat{\mathbf{y}} + z_4 c\,\hat{\mathbf{z}}$ | (2a) | $H_2O$ II |
| $\mathbf{B}_8$ | = | $\frac{1}{2}\,\mathbf{a}_1 - y_4\,\mathbf{a}_2 + \left(\frac{1}{2} + z_4\right)\mathbf{a}_3$ | = | $\frac{1}{2}a\,\hat{\mathbf{x}} - y_4 b\,\hat{\mathbf{y}} + \left(\frac{1}{2} + z_4\right)c\,\hat{\mathbf{z}}$ | (2a) | $H_2O$ II |
| $\mathbf{B}_9$ | = | $y_5\,\mathbf{a}_2 + z_5\,\mathbf{a}_3$ | = | $y_5 b\,\hat{\mathbf{y}} + z_5 c\,\hat{\mathbf{z}}$ | (2a) | Mg |
| $\mathbf{B}_{10}$ | = | $\frac{1}{2}\,\mathbf{a}_1 - y_5\,\mathbf{a}_2 + \left(\frac{1}{2} + z_5\right)\mathbf{a}_3$ | = | $\frac{1}{2}a\,\hat{\mathbf{x}} - y_5 b\,\hat{\mathbf{y}} + \left(\frac{1}{2} + z_5\right)c\,\hat{\mathbf{z}}$ | (2a) | Mg |
| $\mathbf{B}_{11}$ | = | $y_6\,\mathbf{a}_2 + z_6\,\mathbf{a}_3$ | = | $y_6 b\,\hat{\mathbf{y}} + z_6 c\,\hat{\mathbf{z}}$ | (2a) | O I |
| $\mathbf{B}_{12}$ | = | $\frac{1}{2}\,\mathbf{a}_1 - y_6\,\mathbf{a}_2 + \left(\frac{1}{2} + z_6\right)\mathbf{a}_3$ | = | $\frac{1}{2}a\,\hat{\mathbf{x}} - y_6 b\,\hat{\mathbf{y}} + \left(\frac{1}{2} + z_6\right)c\,\hat{\mathbf{z}}$ | (2a) | O I |
| $\mathbf{B}_{13}$ | = | $y_7\,\mathbf{a}_2 + z_7\,\mathbf{a}_3$ | = | $y_7 b\,\hat{\mathbf{y}} + z_7 c\,\hat{\mathbf{z}}$ | (2a) | O II |
| $\mathbf{B}_{14}$ | = | $\frac{1}{2}\,\mathbf{a}_1 - y_7\,\mathbf{a}_2 + \left(\frac{1}{2} + z_7\right)\mathbf{a}_3$ | = | $\frac{1}{2}a\,\hat{\mathbf{x}} - y_7 b\,\hat{\mathbf{y}} + \left(\frac{1}{2} + z_7\right)c\,\hat{\mathbf{z}}$ | (2a) | O II |
| $\mathbf{B}_{15}$ | = | $y_8\,\mathbf{a}_2 + z_8\,\mathbf{a}_3$ | = | $y_8 b\,\hat{\mathbf{y}} + z_8 c\,\hat{\mathbf{z}}$ | (2a) | O III |
| $\mathbf{B}_{16}$ | = | $\frac{1}{2}\,\mathbf{a}_1 - y_8\,\mathbf{a}_2 + \left(\frac{1}{2} + z_8\right)\mathbf{a}_3$ | = | $\frac{1}{2}a\,\hat{\mathbf{x}} - y_8 b\,\hat{\mathbf{y}} + \left(\frac{1}{2} + z_8\right)c\,\hat{\mathbf{z}}$ | (2a) | O III |
| $\mathbf{B}_{17}$ | = | $y_9\,\mathbf{a}_2 + z_9\,\mathbf{a}_3$ | = | $y_9 b\,\hat{\mathbf{y}} + z_9 c\,\hat{\mathbf{z}}$ | (2a) | O IV |
| $\mathbf{B}_{18}$ | = | $\frac{1}{2}\,\mathbf{a}_1 - y_9\,\mathbf{a}_2 + \left(\frac{1}{2} + z_9\right)\mathbf{a}_3$ | = | $\frac{1}{2}a\,\hat{\mathbf{x}} - y_9 b\,\hat{\mathbf{y}} + \left(\frac{1}{2} + z_9\right)c\,\hat{\mathbf{z}}$ | (2a) | O IV |
| $\mathbf{B}_{19}$ | = | $x_{10}\,\mathbf{a}_1 + y_{10}\,\mathbf{a}_2 + z_{10}\,\mathbf{a}_3$ | = | $x_{10}a\,\hat{\mathbf{x}} + y_{10}b\,\hat{\mathbf{y}} + z_{10}c\,\hat{\mathbf{z}}$ | (4b) | $H_2O$ III |
| $\mathbf{B}_{20}$ | = | $\left(\frac{1}{2} - x_{10}\right)\mathbf{a}_1 - y_{10}\,\mathbf{a}_2 + \left(\frac{1}{2} + z_{10}\right)\mathbf{a}_3$ | = | $\left(\frac{1}{2} - x_{10}\right)a\,\hat{\mathbf{x}} - y_{10}b\,\hat{\mathbf{y}} + \left(\frac{1}{2} + z_{10}\right)c\,\hat{\mathbf{z}}$ | (4b) | $H_2O$ III |
| $\mathbf{B}_{21}$ | = | $\left(\frac{1}{2} + x_{10}\right)\mathbf{a}_1 - y_{10}\,\mathbf{a}_2 + \left(\frac{1}{2} + z_{10}\right)\mathbf{a}_3$ | = | $\left(\frac{1}{2} + x_{10}\right)a\,\hat{\mathbf{x}} - y_{10}b\,\hat{\mathbf{y}} + \left(\frac{1}{2} + z_{10}\right)c\,\hat{\mathbf{z}}$ | (4b) | $H_2O$ III |
| $\mathbf{B}_{22}$ | = | $-x_{10}\,\mathbf{a}_1 + y_{10}\,\mathbf{a}_2 + z_{10}\,\mathbf{a}_3$ | = | $-x_{10}a\,\hat{\mathbf{x}} + y_{10}b\,\hat{\mathbf{y}} + z_{10}c\,\hat{\mathbf{z}}$ | (4b) | $H_2O$ III |
| $\mathbf{B}_{23}$ | = | $x_{11}\,\mathbf{a}_1 + y_{11}\,\mathbf{a}_2 + z_{11}\,\mathbf{a}_3$ | = | $x_{11}a\,\hat{\mathbf{x}} + y_{11}b\,\hat{\mathbf{y}} + z_{11}c\,\hat{\mathbf{z}}$ | (4b) | $H_2O$ IV |
| $\mathbf{B}_{24}$ | = | $\left(\frac{1}{2} - x_{11}\right)\mathbf{a}_1 - y_{11}\,\mathbf{a}_2 + \left(\frac{1}{2} + z_{11}\right)\mathbf{a}_3$ | = | $\left(\frac{1}{2} - x_{11}\right)a\,\hat{\mathbf{x}} - y_{11}b\,\hat{\mathbf{y}} + \left(\frac{1}{2} + z_{11}\right)c\,\hat{\mathbf{z}}$ | (4b) | $H_2O$ IV |
| $\mathbf{B}_{25}$ | = | $\left(\frac{1}{2} + x_{11}\right)\mathbf{a}_1 - y_{11}\,\mathbf{a}_2 + \left(\frac{1}{2} + z_{11}\right)\mathbf{a}_3$ | = | $\left(\frac{1}{2} + x_{11}\right)a\,\hat{\mathbf{x}} - y_{11}b\,\hat{\mathbf{y}} + \left(\frac{1}{2} + z_{11}\right)c\,\hat{\mathbf{z}}$ | (4b) | $H_2O$ IV |
| $\mathbf{B}_{26}$ | = | $-x_{11}\,\mathbf{a}_1 + y_{11}\,\mathbf{a}_2 + z_{11}\,\mathbf{a}_3$ | = | $-x_{11}a\,\hat{\mathbf{x}} + y_{11}b\,\hat{\mathbf{y}} + z_{11}c\,\hat{\mathbf{z}}$ | (4b) | $H_2O$ IV |
| $\mathbf{B}_{27}$ | = | $x_{12}\,\mathbf{a}_1 + y_{12}\,\mathbf{a}_2 + z_{12}\,\mathbf{a}_3$ | = | $x_{12}a\,\hat{\mathbf{x}} + y_{12}b\,\hat{\mathbf{y}} + z_{12}c\,\hat{\mathbf{z}}$ | (4b) | O V |
| $\mathbf{B}_{28}$ | = | $\left(\frac{1}{2} - x_{12}\right)\mathbf{a}_1 - y_{12}\,\mathbf{a}_2 + \left(\frac{1}{2} + z_{12}\right)\mathbf{a}_3$ | = | $\left(\frac{1}{2} - x_{12}\right)a\,\hat{\mathbf{x}} - y_{12}b\,\hat{\mathbf{y}} + \left(\frac{1}{2} + z_{12}\right)c\,\hat{\mathbf{z}}$ | (4b) | O V |
| $\mathbf{B}_{29}$ | = | $\left(\frac{1}{2} + x_{12}\right)\mathbf{a}_1 - y_{12}\,\mathbf{a}_2 + \left(\frac{1}{2} + z_{12}\right)\mathbf{a}_3$ | = | $\left(\frac{1}{2} + x_{12}\right)a\,\hat{\mathbf{x}} - y_{12}b\,\hat{\mathbf{y}} + \left(\frac{1}{2} + z_{12}\right)c\,\hat{\mathbf{z}}$ | (4b) | O V |
| $\mathbf{B}_{30}$ | = | $-x_{12}\,\mathbf{a}_1 + y_{12}\,\mathbf{a}_2 + z_{12}\,\mathbf{a}_3$ | = | $-x_{12}a\,\hat{\mathbf{x}} + y_{12}b\,\hat{\mathbf{y}} + z_{12}c\,\hat{\mathbf{z}}$ | (4b) | O V |
| $\mathbf{B}_{31}$ | = | $x_{13}\,\mathbf{a}_1 + y_{13}\,\mathbf{a}_2 + z_{13}\,\mathbf{a}_3$ | = | $x_{13}a\,\hat{\mathbf{x}} + y_{13}b\,\hat{\mathbf{y}} + z_{13}c\,\hat{\mathbf{z}}$ | (4b) | O VI |
| $\mathbf{B}_{32}$ | = | $\left(\frac{1}{2} - x_{13}\right)\mathbf{a}_1 - y_{13}\,\mathbf{a}_2 + \left(\frac{1}{2} + z_{13}\right)\mathbf{a}_3$ | = | $\left(\frac{1}{2} - x_{13}\right)a\,\hat{\mathbf{x}} - y_{13}b\,\hat{\mathbf{y}} + \left(\frac{1}{2} + z_{13}\right)c\,\hat{\mathbf{z}}$ | (4b) | O VI |
| $\mathbf{B}_{33}$ | = | $\left(\frac{1}{2} + x_{13}\right)\mathbf{a}_1 - y_{13}\,\mathbf{a}_2 + \left(\frac{1}{2} + z_{13}\right)\mathbf{a}_3$ | = | $\left(\frac{1}{2} + x_{13}\right)a\,\hat{\mathbf{x}} - y_{13}b\,\hat{\mathbf{y}} + \left(\frac{1}{2} + z_{13}\right)c\,\hat{\mathbf{z}}$ | (4b) | O VI |
| $\mathbf{B}_{34}$ | = | $-x_{13}\,\mathbf{a}_1 + y_{13}\,\mathbf{a}_2 + z_{13}\,\mathbf{a}_3$ | = | $-x_{13}a\,\hat{\mathbf{x}} + y_{13}b\,\hat{\mathbf{y}} + z_{13}c\,\hat{\mathbf{z}}$ | (4b) | O VI |

---

**Found in:**

---

**Geometry files:**

- CIF: pp. 1587
- POSCAR: pp. 1587

# $B_4SrO_7$ Structure: A4B7C_oP24_31_2b_a3b_a

http://aflow.org/prototype-encyclopedia/A4B7C_oP24_31_2b_a3b_a

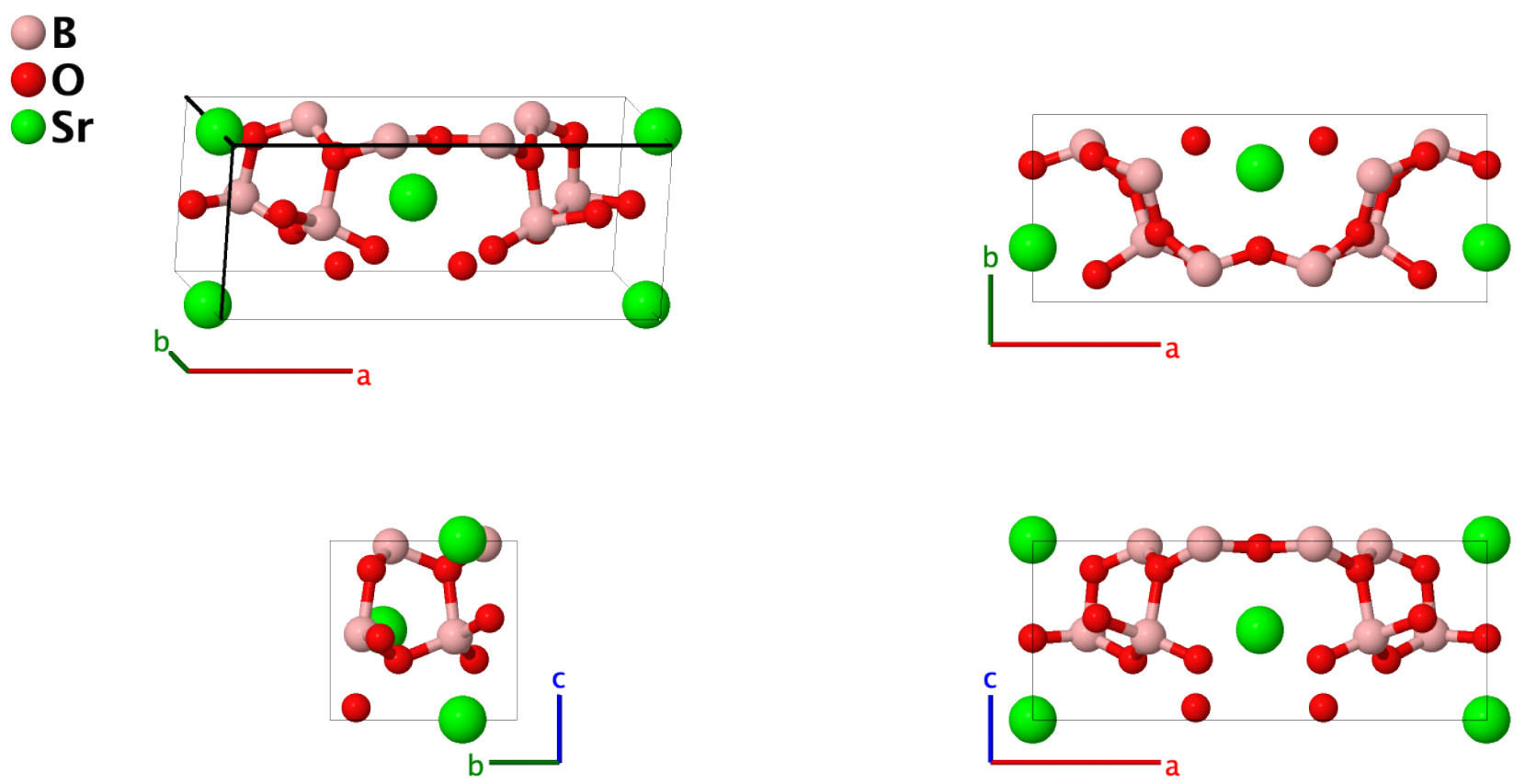

| | | |
|---|---|---|
| **Prototype** | : | $B_4O_7Sr$ |
| **AFLOW prototype label** | : | A4B7C_oP24_31_2b_a3b_a |
| ***Strukturbericht* designation** | : | None |
| **Pearson symbol** | : | oP24 |
| **Space group number** | : | 31 |
| **Space group symbol** | : | $Pmn2_1$ |
| **AFLOW prototype command** | : | `aflow --proto=A4B7C_oP24_31_2b_a3b_a --params=`$a, b/a, c/a, y_1, z_1, y_2, z_2, x_3, y_3, z_3, x_4, y_4, z_4, x_5, y_5, z_5, x_6, y_6, z_6, x_7, y_7, z_7$ |

- Space group $Pmn2_1$ #31 allows an arbitrary choice of the zero of the $z$-axis. Here it is chosen so that $z_1 = 0$ for the strontium atom.

**Simple Orthorhombic primitive vectors:**

$$\mathbf{a}_1 = a\,\hat{\mathbf{x}}$$
$$\mathbf{a}_2 = b\,\hat{\mathbf{y}}$$
$$\mathbf{a}_3 = c\,\hat{\mathbf{z}}$$

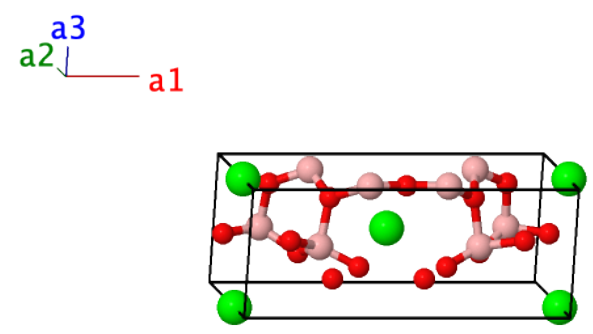

**Basis vectors:**

| | | Lattice Coordinates | | Cartesian Coordinates | Wyckoff Position | Atom Type |
|---|---|---|---|---|---|---|
| $\mathbf{B}_1$ | = | $y_1\,\mathbf{a}_2 + z_1\,\mathbf{a}_3$ | = | $y_1 b\,\hat{\mathbf{y}} + z_1 c\,\hat{\mathbf{z}}$ | (2a) | O I |
| $\mathbf{B}_2$ | = | $\frac{1}{2}\,\mathbf{a}_1 - y_1\,\mathbf{a}_2 + \left(\frac{1}{2} + z_1\right)\mathbf{a}_3$ | = | $\frac{1}{2}a\,\hat{\mathbf{x}} - y_1 b\,\hat{\mathbf{y}} + \left(\frac{1}{2} + z_1\right)c\,\hat{\mathbf{z}}$ | (2a) | O I |
| $\mathbf{B}_3$ | = | $y_2\,\mathbf{a}_2 + z_2\,\mathbf{a}_3$ | = | $y_2 b\,\hat{\mathbf{y}} + z_2 c\,\hat{\mathbf{z}}$ | (2a) | Sr |
| $\mathbf{B}_4$ | = | $\frac{1}{2}\,\mathbf{a}_1 - y_2\,\mathbf{a}_2 + \left(\frac{1}{2} + z_2\right)\mathbf{a}_3$ | = | $\frac{1}{2}a\,\hat{\mathbf{x}} - y_2 b\,\hat{\mathbf{y}} + \left(\frac{1}{2} + z_2\right)c\,\hat{\mathbf{z}}$ | (2a) | Sr |
| $\mathbf{B}_5$ | = | $x_3\,\mathbf{a}_1 + y_3\,\mathbf{a}_2 + z_3\,\mathbf{a}_3$ | = | $x_3 a\,\hat{\mathbf{x}} + y_3 b\,\hat{\mathbf{y}} + z_3 c\,\hat{\mathbf{z}}$ | (4b) | B I |
| $\mathbf{B}_6$ | = | $\left(\frac{1}{2} - x_3\right)\mathbf{a}_1 - y_3\,\mathbf{a}_2 + \left(\frac{1}{2} + z_3\right)\mathbf{a}_3$ | = | $\left(\frac{1}{2} - x_3\right)a\,\hat{\mathbf{x}} - y_3 b\,\hat{\mathbf{y}} + \left(\frac{1}{2} + z_3\right)c\,\hat{\mathbf{z}}$ | (4b) | B I |
| $\mathbf{B}_7$ | = | $\left(\frac{1}{2}+x_3\right)\mathbf{a}_1 - y_3\,\mathbf{a}_2 + \left(\frac{1}{2}+z_3\right)\mathbf{a}_3$ | = | $\left(\frac{1}{2}+x_3\right)a\,\hat{\mathbf{x}} - y_3 b\,\hat{\mathbf{y}} + \left(\frac{1}{2}+z_3\right)c\,\hat{\mathbf{z}}$ | (4$b$) | B I |
| $\mathbf{B}_8$ | = | $-x_3\,\mathbf{a}_1 + y_3\,\mathbf{a}_2 + z_3\,\mathbf{a}_3$ | = | $-x_3 a\,\hat{\mathbf{x}} + y_3 b\,\hat{\mathbf{y}} + z_3 c\,\hat{\mathbf{z}}$ | (4$b$) | B I |
| $\mathbf{B}_9$ | = | $x_4\,\mathbf{a}_1 + y_4\,\mathbf{a}_2 + z_4\,\mathbf{a}_3$ | = | $x_4 a\,\hat{\mathbf{x}} + y_4 b\,\hat{\mathbf{y}} + z_4 c\,\hat{\mathbf{z}}$ | (4$b$) | B II |
| $\mathbf{B}_{10}$ | = | $\left(\frac{1}{2}-x_4\right)\mathbf{a}_1 - y_4\,\mathbf{a}_2 + \left(\frac{1}{2}+z_4\right)\mathbf{a}_3$ | = | $\left(\frac{1}{2}-x_4\right)a\,\hat{\mathbf{x}} - y_4 b\,\hat{\mathbf{y}} + \left(\frac{1}{2}+z_4\right)c\,\hat{\mathbf{z}}$ | (4$b$) | B II |
| $\mathbf{B}_{11}$ | = | $\left(\frac{1}{2}+x_4\right)\mathbf{a}_1 - y_4\,\mathbf{a}_2 + \left(\frac{1}{2}+z_4\right)\mathbf{a}_3$ | = | $\left(\frac{1}{2}+x_4\right)a\,\hat{\mathbf{x}} - y_4 b\,\hat{\mathbf{y}} + \left(\frac{1}{2}+z_4\right)c\,\hat{\mathbf{z}}$ | (4$b$) | B II |
| $\mathbf{B}_{12}$ | = | $-x_4\,\mathbf{a}_1 + y_4\,\mathbf{a}_2 + z_4\,\mathbf{a}_3$ | = | $-x_4 a\,\hat{\mathbf{x}} + y_4 b\,\hat{\mathbf{y}} + z_4 c\,\hat{\mathbf{z}}$ | (4$b$) | B II |
| $\mathbf{B}_{13}$ | = | $x_5\,\mathbf{a}_1 + y_5\,\mathbf{a}_2 + z_5\,\mathbf{a}_3$ | = | $x_5 a\,\hat{\mathbf{x}} + y_5 b\,\hat{\mathbf{y}} + z_5 c\,\hat{\mathbf{z}}$ | (4$b$) | O II |
| $\mathbf{B}_{14}$ | = | $\left(\frac{1}{2}-x_5\right)\mathbf{a}_1 - y_5\,\mathbf{a}_2 + \left(\frac{1}{2}+z_5\right)\mathbf{a}_3$ | = | $\left(\frac{1}{2}-x_5\right)a\,\hat{\mathbf{x}} - y_5 b\,\hat{\mathbf{y}} + \left(\frac{1}{2}+z_5\right)c\,\hat{\mathbf{z}}$ | (4$b$) | O II |
| $\mathbf{B}_{15}$ | = | $\left(\frac{1}{2}+x_5\right)\mathbf{a}_1 - y_5\,\mathbf{a}_2 + \left(\frac{1}{2}+z_5\right)\mathbf{a}_3$ | = | $\left(\frac{1}{2}+x_5\right)a\,\hat{\mathbf{x}} - y_5 b\,\hat{\mathbf{y}} + \left(\frac{1}{2}+z_5\right)c\,\hat{\mathbf{z}}$ | (4$b$) | O II |
| $\mathbf{B}_{16}$ | = | $-x_5\,\mathbf{a}_1 + y_5\,\mathbf{a}_2 + z_5\,\mathbf{a}_3$ | = | $-x_5 a\,\hat{\mathbf{x}} + y_5 b\,\hat{\mathbf{y}} + z_5 c\,\hat{\mathbf{z}}$ | (4$b$) | O II |
| $\mathbf{B}_{17}$ | = | $x_6\,\mathbf{a}_1 + y_6\,\mathbf{a}_2 + z_6\,\mathbf{a}_3$ | = | $x_6 a\,\hat{\mathbf{x}} + y_6 b\,\hat{\mathbf{y}} + z_6 c\,\hat{\mathbf{z}}$ | (4$b$) | O III |
| $\mathbf{B}_{18}$ | = | $\left(\frac{1}{2}-x_6\right)\mathbf{a}_1 - y_6\,\mathbf{a}_2 + \left(\frac{1}{2}+z_6\right)\mathbf{a}_3$ | = | $\left(\frac{1}{2}-x_6\right)a\,\hat{\mathbf{x}} - y_6 b\,\hat{\mathbf{y}} + \left(\frac{1}{2}+z_6\right)c\,\hat{\mathbf{z}}$ | (4$b$) | O III |
| $\mathbf{B}_{19}$ | = | $\left(\frac{1}{2}+x_6\right)\mathbf{a}_1 - y_6\,\mathbf{a}_2 + \left(\frac{1}{2}+z_6\right)\mathbf{a}_3$ | = | $\left(\frac{1}{2}+x_6\right)a\,\hat{\mathbf{x}} - y_6 b\,\hat{\mathbf{y}} + \left(\frac{1}{2}+z_6\right)c\,\hat{\mathbf{z}}$ | (4$b$) | O III |
| $\mathbf{B}_{20}$ | = | $-x_6\,\mathbf{a}_1 + y_6\,\mathbf{a}_2 + z_6\,\mathbf{a}_3$ | = | $-x_6 a\,\hat{\mathbf{x}} + y_6 b\,\hat{\mathbf{y}} + z_6 c\,\hat{\mathbf{z}}$ | (4$b$) | O III |
| $\mathbf{B}_{21}$ | = | $x_7\,\mathbf{a}_1 + y_7\,\mathbf{a}_2 + z_7\,\mathbf{a}_3$ | = | $x_7 a\,\hat{\mathbf{x}} + y_7 b\,\hat{\mathbf{y}} + z_7 c\,\hat{\mathbf{z}}$ | (4$b$) | O IV |
| $\mathbf{B}_{22}$ | = | $\left(\frac{1}{2}-x_7\right)\mathbf{a}_1 - y_7\,\mathbf{a}_2 + \left(\frac{1}{2}+z_7\right)\mathbf{a}_3$ | = | $\left(\frac{1}{2}-x_7\right)a\,\hat{\mathbf{x}} - y_7 b\,\hat{\mathbf{y}} + \left(\frac{1}{2}+z_7\right)c\,\hat{\mathbf{z}}$ | (4$b$) | O IV |
| $\mathbf{B}_{23}$ | = | $\left(\frac{1}{2}+x_7\right)\mathbf{a}_1 - y_7\,\mathbf{a}_2 + \left(\frac{1}{2}+z_7\right)\mathbf{a}_3$ | = | $\left(\frac{1}{2}+x_7\right)a\,\hat{\mathbf{x}} - y_7 b\,\hat{\mathbf{y}} + \left(\frac{1}{2}+z_7\right)c\,\hat{\mathbf{z}}$ | (4$b$) | O IV |
| $\mathbf{B}_{24}$ | = | $-x_7\,\mathbf{a}_1 + y_7\,\mathbf{a}_2 + z_7\,\mathbf{a}_3$ | = | $-x_7 a\,\hat{\mathbf{x}} + y_7 b\,\hat{\mathbf{y}} + z_7 c\,\hat{\mathbf{z}}$ | (4$b$) | O IV |

---

---

**Geometry files:**

- CIF: pp. 1587
- POSCAR: pp. 1588

# $D8_7$ (Shcherbinaite, $V_2O_5$) (*obsolete*) Structure: A5B2_oP14_31_a2b_b

http://aflow.org/prototype-encyclopedia/A5B2_oP14_31_a2b_b

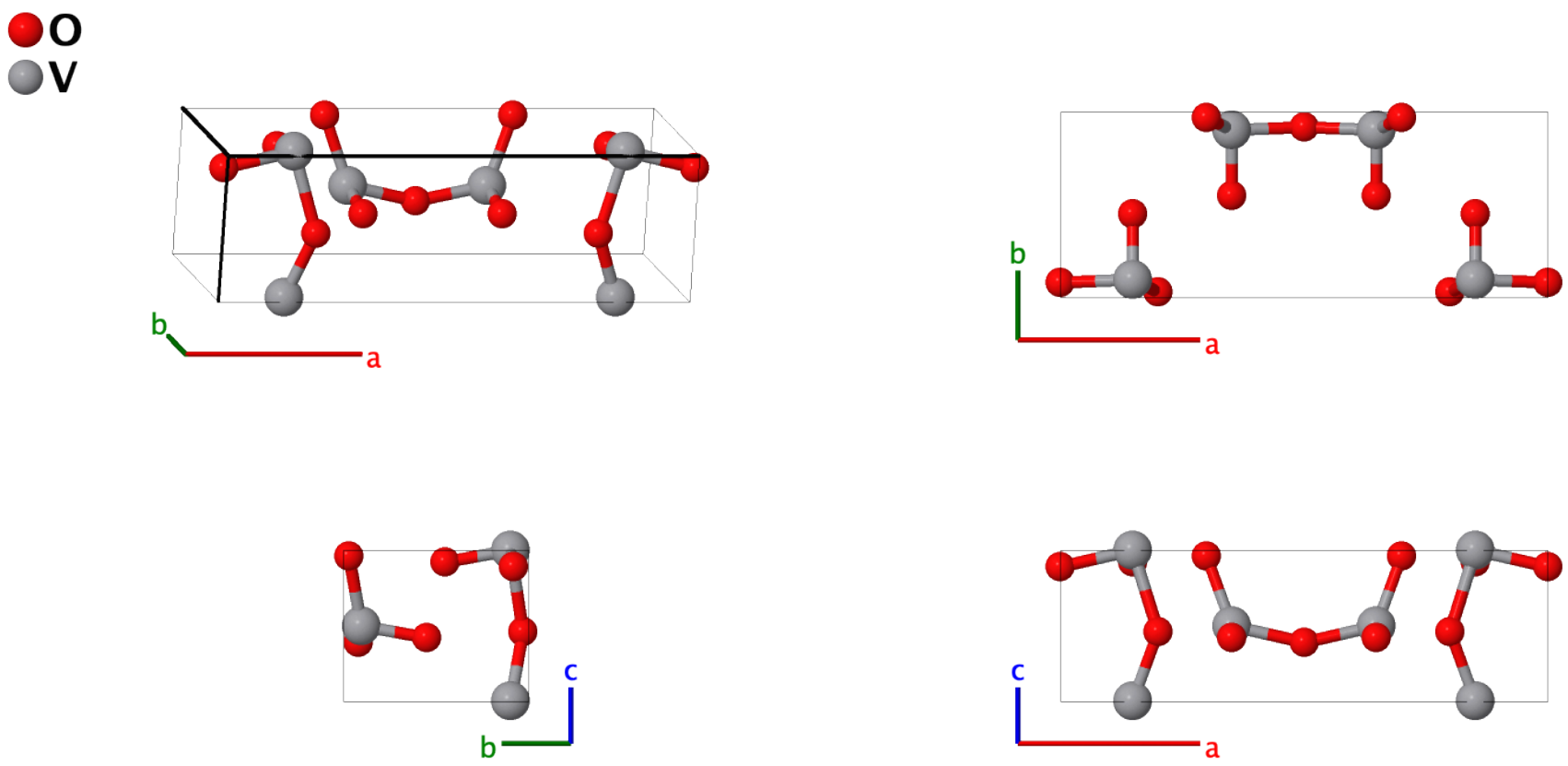

| | | |
|---|---|---|
| **Prototype** | : | $O_5V_2$ |
| **AFLOW prototype label** | : | A5B2_oP14_31_a2b_b |
| ***Strukturbericht* designation** | : | $D8_7$ |
| **Pearson symbol** | : | oP14 |
| **Space group number** | : | 31 |
| **Space group symbol** | : | $Pmn2_1$ |
| **AFLOW prototype command** | : | `aflow --proto=A5B2_oP14_31_a2b_b` `--params=`$a, b/a, c/a, y_1, z_1, x_2, y_2, z_2, x_3, y_3, z_3, x_4, y_4, z_4$ |

- This structure was reinvestigated by (Enjalbert, 1986) and others because of "[dissatisfaction with] some aspects of the crystal structure." They proposed a revised structure in space group *Pmmn*. We list this structure here for the historical record.

**Simple Orthorhombic primitive vectors:**

$$\mathbf{a}_1 = a\,\hat{\mathbf{x}}$$
$$\mathbf{a}_2 = b\,\hat{\mathbf{y}}$$
$$\mathbf{a}_3 = c\,\hat{\mathbf{z}}$$

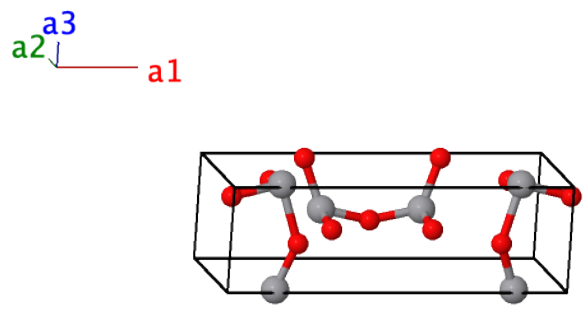

**Basis vectors:**

| | | Lattice Coordinates | | Cartesian Coordinates | Wyckoff Position | Atom Type |
|---|---|---|---|---|---|---|
| $\mathbf{B}_1$ | = | $y_1\,\mathbf{a}_2 + z_1\,\mathbf{a}_3$ | = | $y_1 b\,\hat{\mathbf{y}} + z_1 c\,\hat{\mathbf{z}}$ | (2*a*) | O I |
| $\mathbf{B}_2$ | = | $\frac{1}{2}\,\mathbf{a}_1 - y_1\,\mathbf{a}_2 + \left(\frac{1}{2} + z_1\right)\mathbf{a}_3$ | = | $\frac{1}{2}a\,\hat{\mathbf{x}} - y_1 b\,\hat{\mathbf{y}} + \left(\frac{1}{2} + z_1\right)c\,\hat{\mathbf{z}}$ | (2*a*) | O I |
| $\mathbf{B}_3$ | = | $x_2\,\mathbf{a}_1 + y_2\,\mathbf{a}_2 + z_2\,\mathbf{a}_3$ | = | $x_2 a\,\hat{\mathbf{x}} + y_2 b\,\hat{\mathbf{y}} + z_2 c\,\hat{\mathbf{z}}$ | (4*b*) | O II |
| $\mathbf{B}_4$ | = | $\left(\frac{1}{2} - x_2\right)\mathbf{a}_1 - y_2\,\mathbf{a}_2 + \left(\frac{1}{2} + z_2\right)\mathbf{a}_3$ | = | $\left(\frac{1}{2} - x_2\right)a\,\hat{\mathbf{x}} - y_2 b\,\hat{\mathbf{y}} + \left(\frac{1}{2} + z_2\right)c\,\hat{\mathbf{z}}$ | (4*b*) | O II |
| $\mathbf{B}_5$ | = | $\left(\frac{1}{2} + x_2\right)\mathbf{a}_1 - y_2\,\mathbf{a}_2 + \left(\frac{1}{2} + z_2\right)\mathbf{a}_3$ | = | $\left(\frac{1}{2} + x_2\right)a\,\hat{\mathbf{x}} - y_2 b\,\hat{\mathbf{y}} + \left(\frac{1}{2} + z_2\right)c\,\hat{\mathbf{z}}$ | (4*b*) | O II |
| $\mathbf{B}_6$ | = | $-x_2\,\mathbf{a}_1 + y_2\,\mathbf{a}_2 + z_2\,\mathbf{a}_3$ | = | $-x_2 a\,\hat{\mathbf{x}} + y_2 b\,\hat{\mathbf{y}} + z_2 c\,\hat{\mathbf{z}}$ | $(4b)$ | O II |
| $\mathbf{B}_7$ | = | $x_3\,\mathbf{a}_1 + y_3\,\mathbf{a}_2 + z_3\,\mathbf{a}_3$ | = | $x_3 a\,\hat{\mathbf{x}} + y_3 b\,\hat{\mathbf{y}} + z_3 c\,\hat{\mathbf{z}}$ | $(4b)$ | O III |
| $\mathbf{B}_8$ | = | $\left(\frac{1}{2} - x_3\right)\mathbf{a}_1 - y_3\,\mathbf{a}_2 + \left(\frac{1}{2} + z_3\right)\mathbf{a}_3$ | = | $\left(\frac{1}{2} - x_3\right)a\,\hat{\mathbf{x}} - y_3 b\,\hat{\mathbf{y}} + \left(\frac{1}{2} + z_3\right)c\,\hat{\mathbf{z}}$ | $(4b)$ | O III |
| $\mathbf{B}_9$ | = | $\left(\frac{1}{2} + x_3\right)\mathbf{a}_1 - y_3\,\mathbf{a}_2 + \left(\frac{1}{2} + z_3\right)\mathbf{a}_3$ | = | $\left(\frac{1}{2} + x_3\right)a\,\hat{\mathbf{x}} - y_3 b\,\hat{\mathbf{y}} + \left(\frac{1}{2} + z_3\right)c\,\hat{\mathbf{z}}$ | $(4b)$ | O III |
| $\mathbf{B}_{10}$ | = | $-x_3\,\mathbf{a}_1 + y_3\,\mathbf{a}_2 + z_3\,\mathbf{a}_3$ | = | $-x_3 a\,\hat{\mathbf{x}} + y_3 b\,\hat{\mathbf{y}} + z_3 c\,\hat{\mathbf{z}}$ | $(4b)$ | O III |
| $\mathbf{B}_{11}$ | = | $x_4\,\mathbf{a}_1 + y_4\,\mathbf{a}_2 + z_4\,\mathbf{a}_3$ | = | $x_4 a\,\hat{\mathbf{x}} + y_4 b\,\hat{\mathbf{y}} + z_4 c\,\hat{\mathbf{z}}$ | $(4b)$ | V |
| $\mathbf{B}_{12}$ | = | $\left(\frac{1}{2} - x_4\right)\mathbf{a}_1 - y_4\,\mathbf{a}_2 + \left(\frac{1}{2} + z_4\right)\mathbf{a}_3$ | = | $\left(\frac{1}{2} - x_4\right)a\,\hat{\mathbf{x}} - y_4 b\,\hat{\mathbf{y}} + \left(\frac{1}{2} + z_4\right)c\,\hat{\mathbf{z}}$ | $(4b)$ | V |
| $\mathbf{B}_{13}$ | = | $\left(\frac{1}{2} + x_4\right)\mathbf{a}_1 - y_4\,\mathbf{a}_2 + \left(\frac{1}{2} + z_4\right)\mathbf{a}_3$ | = | $\left(\frac{1}{2} + x_4\right)a\,\hat{\mathbf{x}} - y_4 b\,\hat{\mathbf{y}} + \left(\frac{1}{2} + z_4\right)c\,\hat{\mathbf{z}}$ | $(4b)$ | V |
| $\mathbf{B}_{14}$ | = | $-x_4\,\mathbf{a}_1 + y_4\,\mathbf{a}_2 + z_4\,\mathbf{a}_3$ | = | $-x_4 a\,\hat{\mathbf{x}} + y_4 b\,\hat{\mathbf{y}} + z_4 c\,\hat{\mathbf{z}}$ | $(4b)$ | V |

**Geometry files:**

- CIF: pp. 1588
- POSCAR: pp. 1588

# $Mo_{17}O_{47}$ Structure: A17B47_oP128_32_a8c_a23c

http://aflow.org/prototype-encyclopedia/A17B47_oP128_32_a8c_a23c

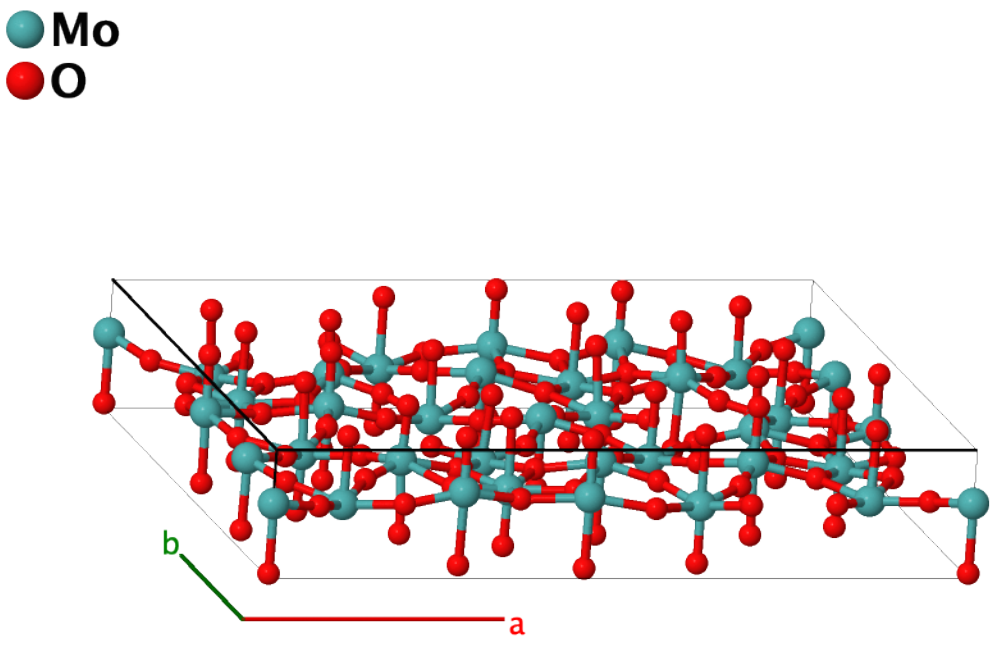

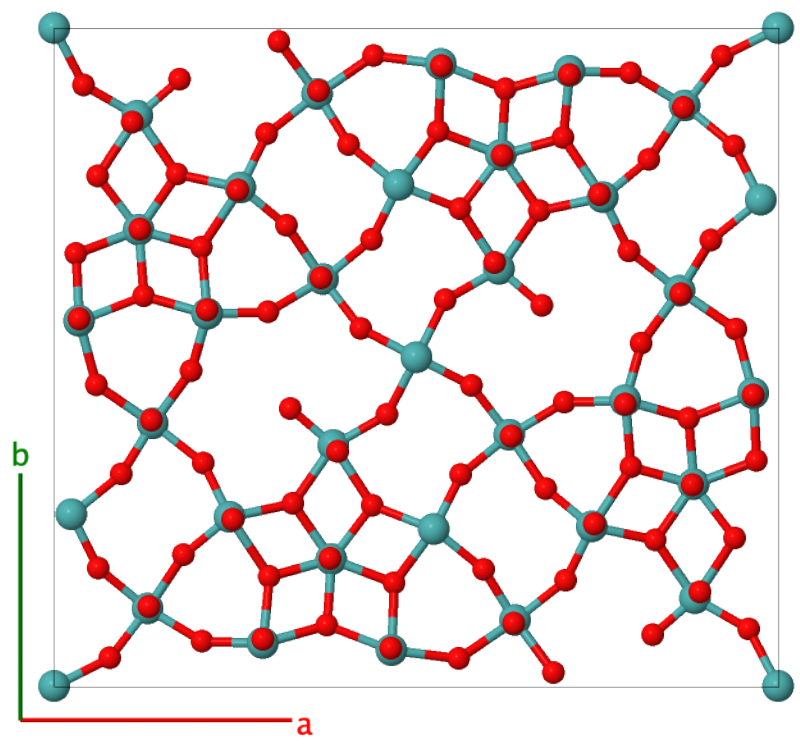

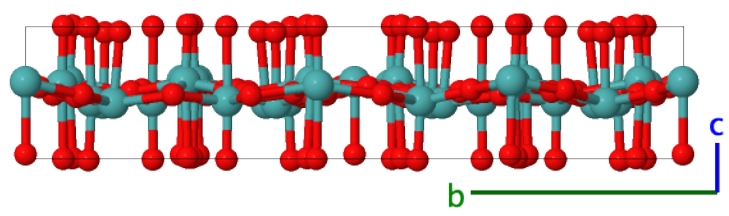

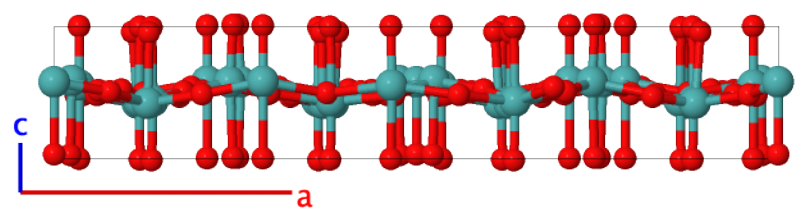

**Prototype** : $Mo_{17}O_{47}$

**AFLOW prototype label** : A17B47_oP128_32_a8c_a23c

***Strukturbericht* designation** : None

**Pearson symbol** : oP128

**Space group number** : 32

**Space group symbol** : *Pba*2

**AFLOW prototype command** : `aflow --proto=A17B47_oP128_32_a8c_a23c`
`--params=`$a, b/a, c/a, z_1, z_2, x_3, y_3, z_3, x_4, y_4, z_4, x_5, y_5, z_5, x_6, y_6, z_6, x_7, y_7, z_7, x_8, y_8, z_8, x_9, y_9, z_9, x_{10}, y_{10}, z_{10}, x_{11}, y_{11}, z_{11}, x_{12}, y_{12}, z_{12}, x_{13}, y_{13}, z_{13}, x_{14}, y_{14}, z_{14}, x_{15}, y_{15}, z_{15}, x_{16}, y_{16}, z_{16}, x_{17}, y_{17}, z_{17}, x_{18}, y_{18}, z_{18}, x_{19}, y_{19}, z_{19}, x_{20}, y_{20}, z_{20}, x_{21}, y_{21}, z_{21}, x_{22}, y_{22}, z_{22}, x_{23}, y_{23}, z_{23}, x_{24}, y_{24}, z_{24}, x_{25}, y_{25}, z_{25}, x_{26}, y_{26}, z_{26}, x_{27}, y_{27}, z_{27}, x_{28}, y_{28}, z_{28}, x_{29}, y_{29}, z_{29}, x_{30}, y_{30}, z_{30}, x_{31}, y_{31}, z_{31}, x_{32}, y_{32}, z_{32}, x_{33}, y_{33}, z_{33}$

## Simple Orthorhombic primitive vectors:

$$\mathbf{a}_1 = a\,\hat{\mathbf{x}}$$
$$\mathbf{a}_2 = b\,\hat{\mathbf{y}}$$
$$\mathbf{a}_3 = c\,\hat{\mathbf{z}}$$

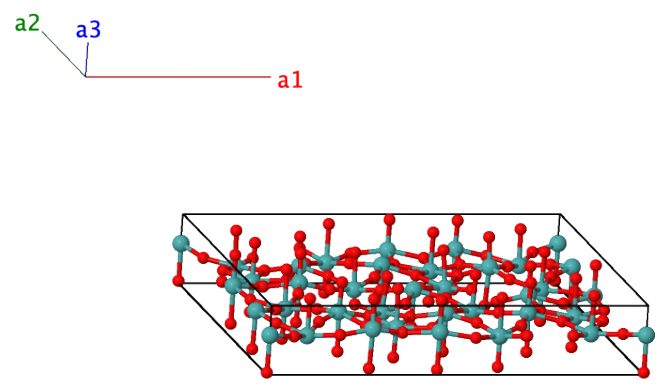

## Basis vectors:

| | | Lattice Coordinates | | Cartesian Coordinates | Wyckoff Position | Atom Type |
|---|---|---|---|---|---|---|
| $\mathbf{B}_1$ | = | $z_1\,\mathbf{a}_3$ | = | $z_1 c\,\hat{\mathbf{z}}$ | (2a) | Mo I |
| $\mathbf{B}_2$ | = | $\frac{1}{2}\,\mathbf{a}_1 + \frac{1}{2}\,\mathbf{a}_2 + z_1\,\mathbf{a}_3$ | = | $\frac{1}{2}a\,\hat{\mathbf{x}} + \frac{1}{2}b\,\hat{\mathbf{y}} + z_1 c\,\hat{\mathbf{z}}$ | (2a) | Mo I |
| $\mathbf{B}_3$ | = | $z_2\,\mathbf{a}_3$ | = | $z_2 c\,\hat{\mathbf{z}}$ | (2a) | O I |
| $\mathbf{B}_4$ | = | $\frac{1}{2}\,\mathbf{a}_1 + \frac{1}{2}\,\mathbf{a}_2 + z_2\,\mathbf{a}_3$ | = | $\frac{1}{2}a\,\hat{\mathbf{x}} + \frac{1}{2}b\,\hat{\mathbf{y}} + z_2 c\,\hat{\mathbf{z}}$ | (2a) | O I |
| $\mathbf{B}_5$ | = | $x_3\,\mathbf{a}_1 + y_3\,\mathbf{a}_2 + z_3\,\mathbf{a}_3$ | = | $x_3 a\,\hat{\mathbf{x}} + y_3 b\,\hat{\mathbf{y}} + z_3 c\,\hat{\mathbf{z}}$ | (4c) | Mo II |
| $\mathbf{B}_6$ | = | $-x_3\,\mathbf{a}_1 - y_3\,\mathbf{a}_2 + z_3\,\mathbf{a}_3$ | = | $-x_3 a\,\hat{\mathbf{x}} - y_3 b\,\hat{\mathbf{y}} + z_3 c\,\hat{\mathbf{z}}$ | (4c) | Mo II |
| $\mathbf{B}_7$ | = | $\left(\frac{1}{2} + x_3\right)\mathbf{a}_1 + \left(\frac{1}{2} - y_3\right)\mathbf{a}_2 + z_3\,\mathbf{a}_3$ | = | $\left(\frac{1}{2} + x_3\right)a\,\hat{\mathbf{x}} + \left(\frac{1}{2} - y_3\right)b\,\hat{\mathbf{y}} + z_3 c\,\hat{\mathbf{z}}$ | (4c) | Mo II |
| $\mathbf{B}_8$ | = | $\left(\frac{1}{2} - x_3\right)\mathbf{a}_1 + \left(\frac{1}{2} + y_3\right)\mathbf{a}_2 + z_3\,\mathbf{a}_3$ | = | $\left(\frac{1}{2} - x_3\right)a\,\hat{\mathbf{x}} + \left(\frac{1}{2} + y_3\right)b\,\hat{\mathbf{y}} + z_3 c\,\hat{\mathbf{z}}$ | (4c) | Mo II |
| $\mathbf{B}_9$ | = | $x_4\,\mathbf{a}_1 + y_4\,\mathbf{a}_2 + z_4\,\mathbf{a}_3$ | = | $x_4 a\,\hat{\mathbf{x}} + y_4 b\,\hat{\mathbf{y}} + z_4 c\,\hat{\mathbf{z}}$ | (4c) | Mo III |
| $\mathbf{B}_{10}$ | = | $-x_4\,\mathbf{a}_1 - y_4\,\mathbf{a}_2 + z_4\,\mathbf{a}_3$ | = | $-x_4 a\,\hat{\mathbf{x}} - y_4 b\,\hat{\mathbf{y}} + z_4 c\,\hat{\mathbf{z}}$ | (4c) | Mo III |
| $\mathbf{B}_{11}$ | = | $\left(\frac{1}{2} + x_4\right)\mathbf{a}_1 + \left(\frac{1}{2} - y_4\right)\mathbf{a}_2 + z_4\,\mathbf{a}_3$ | = | $\left(\frac{1}{2} + x_4\right)a\,\hat{\mathbf{x}} + \left(\frac{1}{2} - y_4\right)b\,\hat{\mathbf{y}} + z_4 c\,\hat{\mathbf{z}}$ | (4c) | Mo III |
| $\mathbf{B}_{12}$ | = | $\left(\frac{1}{2} - x_4\right)\mathbf{a}_1 + \left(\frac{1}{2} + y_4\right)\mathbf{a}_2 + z_4\,\mathbf{a}_3$ | = | $\left(\frac{1}{2} - x_4\right)a\,\hat{\mathbf{x}} + \left(\frac{1}{2} + y_4\right)b\,\hat{\mathbf{y}} + z_4 c\,\hat{\mathbf{z}}$ | (4c) | Mo III |
| $\mathbf{B}_{13}$ | = | $x_5\,\mathbf{a}_1 + y_5\,\mathbf{a}_2 + z_5\,\mathbf{a}_3$ | = | $x_5 a\,\hat{\mathbf{x}} + y_5 b\,\hat{\mathbf{y}} + z_5 c\,\hat{\mathbf{z}}$ | (4c) | Mo IV |
| $\mathbf{B}_{14}$ | = | $-x_5\,\mathbf{a}_1 - y_5\,\mathbf{a}_2 + z_5\,\mathbf{a}_3$ | = | $-x_5 a\,\hat{\mathbf{x}} - y_5 b\,\hat{\mathbf{y}} + z_5 c\,\hat{\mathbf{z}}$ | (4c) | Mo IV |
| $\mathbf{B}_{15}$ | = | $\left(\frac{1}{2} + x_5\right)\mathbf{a}_1 + \left(\frac{1}{2} - y_5\right)\mathbf{a}_2 + z_5\,\mathbf{a}_3$ | = | $\left(\frac{1}{2} + x_5\right)a\,\hat{\mathbf{x}} + \left(\frac{1}{2} - y_5\right)b\,\hat{\mathbf{y}} + z_5 c\,\hat{\mathbf{z}}$ | (4c) | Mo IV |
| $\mathbf{B}_{16}$ | = | $\left(\frac{1}{2} - x_5\right)\mathbf{a}_1 + \left(\frac{1}{2} + y_5\right)\mathbf{a}_2 + z_5\,\mathbf{a}_3$ | = | $\left(\frac{1}{2} - x_5\right)a\,\hat{\mathbf{x}} + \left(\frac{1}{2} + y_5\right)b\,\hat{\mathbf{y}} + z_5 c\,\hat{\mathbf{z}}$ | (4c) | Mo IV |
| $\mathbf{B}_{17}$ | = | $x_6\,\mathbf{a}_1 + y_6\,\mathbf{a}_2 + z_6\,\mathbf{a}_3$ | = | $x_6 a\,\hat{\mathbf{x}} + y_6 b\,\hat{\mathbf{y}} + z_6 c\,\hat{\mathbf{z}}$ | (4c) | Mo V |
| $\mathbf{B}_{18}$ | = | $-x_6\,\mathbf{a}_1 - y_6\,\mathbf{a}_2 + z_6\,\mathbf{a}_3$ | = | $-x_6 a\,\hat{\mathbf{x}} - y_6 b\,\hat{\mathbf{y}} + z_6 c\,\hat{\mathbf{z}}$ | (4c) | Mo V |
| $\mathbf{B}_{19}$ | = | $\left(\frac{1}{2} + x_6\right)\mathbf{a}_1 + \left(\frac{1}{2} - y_6\right)\mathbf{a}_2 + z_6\,\mathbf{a}_3$ | = | $\left(\frac{1}{2} + x_6\right)a\,\hat{\mathbf{x}} + \left(\frac{1}{2} - y_6\right)b\,\hat{\mathbf{y}} + z_6 c\,\hat{\mathbf{z}}$ | (4c) | Mo V |
| $\mathbf{B}_{20}$ | = | $\left(\frac{1}{2} - x_6\right)\mathbf{a}_1 + \left(\frac{1}{2} + y_6\right)\mathbf{a}_2 + z_6\,\mathbf{a}_3$ | = | $\left(\frac{1}{2} - x_6\right)a\,\hat{\mathbf{x}} + \left(\frac{1}{2} + y_6\right)b\,\hat{\mathbf{y}} + z_6 c\,\hat{\mathbf{z}}$ | (4c) | Mo V |
| $\mathbf{B}_{21}$ | = | $x_7\,\mathbf{a}_1 + y_7\,\mathbf{a}_2 + z_7\,\mathbf{a}_3$ | = | $x_7 a\,\hat{\mathbf{x}} + y_7 b\,\hat{\mathbf{y}} + z_7 c\,\hat{\mathbf{z}}$ | (4c) | Mo VI |
| $\mathbf{B}_{22}$ | = | $-x_7\,\mathbf{a}_1 - y_7\,\mathbf{a}_2 + z_7\,\mathbf{a}_3$ | = | $-x_7 a\,\hat{\mathbf{x}} - y_7 b\,\hat{\mathbf{y}} + z_7 c\,\hat{\mathbf{z}}$ | (4c) | Mo VI |
| $\mathbf{B}_{23}$ | = | $\left(\frac{1}{2} + x_7\right)\mathbf{a}_1 + \left(\frac{1}{2} - y_7\right)\mathbf{a}_2 + z_7\,\mathbf{a}_3$ | = | $\left(\frac{1}{2} + x_7\right)a\,\hat{\mathbf{x}} + \left(\frac{1}{2} - y_7\right)b\,\hat{\mathbf{y}} + z_7 c\,\hat{\mathbf{z}}$ | (4c) | Mo VI |
| $\mathbf{B}_{24}$ | = | $\left(\frac{1}{2} - x_7\right)\mathbf{a}_1 + \left(\frac{1}{2} + y_7\right)\mathbf{a}_2 + z_7\,\mathbf{a}_3$ | = | $\left(\frac{1}{2} - x_7\right)a\,\hat{\mathbf{x}} + \left(\frac{1}{2} + y_7\right)b\,\hat{\mathbf{y}} + z_7 c\,\hat{\mathbf{z}}$ | (4c) | Mo VI |
| $\mathbf{B}_{25}$ | = | $x_8\,\mathbf{a}_1 + y_8\,\mathbf{a}_2 + z_8\,\mathbf{a}_3$ | = | $x_8 a\,\hat{\mathbf{x}} + y_8 b\,\hat{\mathbf{y}} + z_8 c\,\hat{\mathbf{z}}$ | (4c) | Mo VII |
| $\mathbf{B}_{26}$ | = | $-x_8\,\mathbf{a}_1 - y_8\,\mathbf{a}_2 + z_8\,\mathbf{a}_3$ | = | $-x_8 a\,\hat{\mathbf{x}} - y_8 b\,\hat{\mathbf{y}} + z_8 c\,\hat{\mathbf{z}}$ | (4c) | Mo VII |
| $\mathbf{B}_{27}$ | = | $\left(\frac{1}{2} + x_8\right)\mathbf{a}_1 + \left(\frac{1}{2} - y_8\right)\mathbf{a}_2 + z_8\,\mathbf{a}_3$ | = | $\left(\frac{1}{2} + x_8\right)a\,\hat{\mathbf{x}} + \left(\frac{1}{2} - y_8\right)b\,\hat{\mathbf{y}} + z_8 c\,\hat{\mathbf{z}}$ | (4c) | Mo VII |
| $\mathbf{B}_{28}$ | = | $\left(\frac{1}{2} - x_8\right) \mathbf{a}_1 + \left(\frac{1}{2} + y_8\right) \mathbf{a}_2 + z_8 \mathbf{a}_3$ | = | $\left(\frac{1}{2} - x_8\right) a \hat{\mathbf{x}} + \left(\frac{1}{2} + y_8\right) b \hat{\mathbf{y}} + z_8 c \hat{\mathbf{z}}$ | (4$c$) | Mo VII |
| $\mathbf{B}_{29}$ | = | $x_9 \mathbf{a}_1 + y_9 \mathbf{a}_2 + z_9 \mathbf{a}_3$ | = | $x_9 a \hat{\mathbf{x}} + y_9 b \hat{\mathbf{y}} + z_9 c \hat{\mathbf{z}}$ | (4$c$) | Mo VIII |
| $\mathbf{B}_{30}$ | = | $-x_9 \mathbf{a}_1 - y_9 \mathbf{a}_2 + z_9 \mathbf{a}_3$ | = | $-x_9 a \hat{\mathbf{x}} - y_9 b \hat{\mathbf{y}} + z_9 c \hat{\mathbf{z}}$ | (4$c$) | Mo VIII |
| $\mathbf{B}_{31}$ | = | $\left(\frac{1}{2} + x_9\right) \mathbf{a}_1 + \left(\frac{1}{2} - y_9\right) \mathbf{a}_2 + z_9 \mathbf{a}_3$ | = | $\left(\frac{1}{2} + x_9\right) a \hat{\mathbf{x}} + \left(\frac{1}{2} - y_9\right) b \hat{\mathbf{y}} + z_9 c \hat{\mathbf{z}}$ | (4$c$) | Mo VIII |
| $\mathbf{B}_{32}$ | = | $\left(\frac{1}{2} - x_9\right) \mathbf{a}_1 + \left(\frac{1}{2} + y_9\right) \mathbf{a}_2 + z_9 \mathbf{a}_3$ | = | $\left(\frac{1}{2} - x_9\right) a \hat{\mathbf{x}} + \left(\frac{1}{2} + y_9\right) b \hat{\mathbf{y}} + z_9 c \hat{\mathbf{z}}$ | (4$c$) | Mo VIII |
| $\mathbf{B}_{33}$ | = | $x_{10} \mathbf{a}_1 + y_{10} \mathbf{a}_2 + z_{10} \mathbf{a}_3$ | = | $x_{10} a \hat{\mathbf{x}} + y_{10} b \hat{\mathbf{y}} + z_{10} c \hat{\mathbf{z}}$ | (4$c$) | Mo IX |
| $\mathbf{B}_{34}$ | = | $-x_{10} \mathbf{a}_1 - y_{10} \mathbf{a}_2 + z_{10} \mathbf{a}_3$ | = | $-x_{10} a \hat{\mathbf{x}} - y_{10} b \hat{\mathbf{y}} + z_{10} c \hat{\mathbf{z}}$ | (4$c$) | Mo IX |
| $\mathbf{B}_{35}$ | = | $\left(\frac{1}{2} + x_{10}\right) \mathbf{a}_1 + \left(\frac{1}{2} - y_{10}\right) \mathbf{a}_2 + z_{10} \mathbf{a}_3$ | = | $\left(\frac{1}{2} + x_{10}\right) a \hat{\mathbf{x}} + \left(\frac{1}{2} - y_{10}\right) b \hat{\mathbf{y}} + z_{10} c \hat{\mathbf{z}}$ | (4$c$) | Mo IX |
| $\mathbf{B}_{36}$ | = | $\left(\frac{1}{2} - x_{10}\right) \mathbf{a}_1 + \left(\frac{1}{2} + y_{10}\right) \mathbf{a}_2 + z_{10} \mathbf{a}_3$ | = | $\left(\frac{1}{2} - x_{10}\right) a \hat{\mathbf{x}} + \left(\frac{1}{2} + y_{10}\right) b \hat{\mathbf{y}} + z_{10} c \hat{\mathbf{z}}$ | (4$c$) | Mo IX |
| $\mathbf{B}_{37}$ | = | $x_{11} \mathbf{a}_1 + y_{11} \mathbf{a}_2 + z_{11} \mathbf{a}_3$ | = | $x_{11} a \hat{\mathbf{x}} + y_{11} b \hat{\mathbf{y}} + z_{11} c \hat{\mathbf{z}}$ | (4$c$) | O II |
| $\mathbf{B}_{38}$ | = | $-x_{11} \mathbf{a}_1 - y_{11} \mathbf{a}_2 + z_{11} \mathbf{a}_3$ | = | $-x_{11} a \hat{\mathbf{x}} - y_{11} b \hat{\mathbf{y}} + z_{11} c \hat{\mathbf{z}}$ | (4$c$) | O II |
| $\mathbf{B}_{39}$ | = | $\left(\frac{1}{2} + x_{11}\right) \mathbf{a}_1 + \left(\frac{1}{2} - y_{11}\right) \mathbf{a}_2 + z_{11} \mathbf{a}_3$ | = | $\left(\frac{1}{2} + x_{11}\right) a \hat{\mathbf{x}} + \left(\frac{1}{2} - y_{11}\right) b \hat{\mathbf{y}} + z_{11} c \hat{\mathbf{z}}$ | (4$c$) | O II |
| $\mathbf{B}_{40}$ | = | $\left(\frac{1}{2} - x_{11}\right) \mathbf{a}_1 + \left(\frac{1}{2} + y_{11}\right) \mathbf{a}_2 + z_{11} \mathbf{a}_3$ | = | $\left(\frac{1}{2} - x_{11}\right) a \hat{\mathbf{x}} + \left(\frac{1}{2} + y_{11}\right) b \hat{\mathbf{y}} + z_{11} c \hat{\mathbf{z}}$ | (4$c$) | O II |
| $\mathbf{B}_{41}$ | = | $x_{12} \mathbf{a}_1 + y_{12} \mathbf{a}_2 + z_{12} \mathbf{a}_3$ | = | $x_{12} a \hat{\mathbf{x}} + y_{12} b \hat{\mathbf{y}} + z_{12} c \hat{\mathbf{z}}$ | (4$c$) | O III |
| $\mathbf{B}_{42}$ | = | $-x_{12} \mathbf{a}_1 - y_{12} \mathbf{a}_2 + z_{12} \mathbf{a}_3$ | = | $-x_{12} a \hat{\mathbf{x}} - y_{12} b \hat{\mathbf{y}} + z_{12} c \hat{\mathbf{z}}$ | (4$c$) | O III |
| $\mathbf{B}_{43}$ | = | $\left(\frac{1}{2} + x_{12}\right) \mathbf{a}_1 + \left(\frac{1}{2} - y_{12}\right) \mathbf{a}_2 + z_{12} \mathbf{a}_3$ | = | $\left(\frac{1}{2} + x_{12}\right) a \hat{\mathbf{x}} + \left(\frac{1}{2} - y_{12}\right) b \hat{\mathbf{y}} + z_{12} c \hat{\mathbf{z}}$ | (4$c$) | O III |
| $\mathbf{B}_{44}$ | = | $\left(\frac{1}{2} - x_{12}\right) \mathbf{a}_1 + \left(\frac{1}{2} + y_{12}\right) \mathbf{a}_2 + z_{12} \mathbf{a}_3$ | = | $\left(\frac{1}{2} - x_{12}\right) a \hat{\mathbf{x}} + \left(\frac{1}{2} + y_{12}\right) b \hat{\mathbf{y}} + z_{12} c \hat{\mathbf{z}}$ | (4$c$) | O III |
| $\mathbf{B}_{45}$ | = | $x_{13} \mathbf{a}_1 + y_{13} \mathbf{a}_2 + z_{13} \mathbf{a}_3$ | = | $x_{13} a \hat{\mathbf{x}} + y_{13} b \hat{\mathbf{y}} + z_{13} c \hat{\mathbf{z}}$ | (4$c$) | O IV |
| $\mathbf{B}_{46}$ | = | $-x_{13} \mathbf{a}_1 - y_{13} \mathbf{a}_2 + z_{13} \mathbf{a}_3$ | = | $-x_{13} a \hat{\mathbf{x}} - y_{13} b \hat{\mathbf{y}} + z_{13} c \hat{\mathbf{z}}$ | (4$c$) | O IV |
| $\mathbf{B}_{47}$ | = | $\left(\frac{1}{2} + x_{13}\right) \mathbf{a}_1 + \left(\frac{1}{2} - y_{13}\right) \mathbf{a}_2 + z_{13} \mathbf{a}_3$ | = | $\left(\frac{1}{2} + x_{13}\right) a \hat{\mathbf{x}} + \left(\frac{1}{2} - y_{13}\right) b \hat{\mathbf{y}} + z_{13} c \hat{\mathbf{z}}$ | (4$c$) | O IV |
| $\mathbf{B}_{48}$ | = | $\left(\frac{1}{2} - x_{13}\right) \mathbf{a}_1 + \left(\frac{1}{2} + y_{13}\right) \mathbf{a}_2 + z_{13} \mathbf{a}_3$ | = | $\left(\frac{1}{2} - x_{13}\right) a \hat{\mathbf{x}} + \left(\frac{1}{2} + y_{13}\right) b \hat{\mathbf{y}} + z_{13} c \hat{\mathbf{z}}$ | (4$c$) | O IV |
| $\mathbf{B}_{49}$ | = | $x_{14} \mathbf{a}_1 + y_{14} \mathbf{a}_2 + z_{14} \mathbf{a}_3$ | = | $x_{14} a \hat{\mathbf{x}} + y_{14} b \hat{\mathbf{y}} + z_{14} c \hat{\mathbf{z}}$ | (4$c$) | O V |
| $\mathbf{B}_{50}$ | = | $-x_{14} \mathbf{a}_1 - y_{14} \mathbf{a}_2 + z_{14} \mathbf{a}_3$ | = | $-x_{14} a \hat{\mathbf{x}} - y_{14} b \hat{\mathbf{y}} + z_{14} c \hat{\mathbf{z}}$ | (4$c$) | O V |
| $\mathbf{B}_{51}$ | = | $\left(\frac{1}{2} + x_{14}\right) \mathbf{a}_1 + \left(\frac{1}{2} - y_{14}\right) \mathbf{a}_2 + z_{14} \mathbf{a}_3$ | = | $\left(\frac{1}{2} + x_{14}\right) a \hat{\mathbf{x}} + \left(\frac{1}{2} - y_{14}\right) b \hat{\mathbf{y}} + z_{14} c \hat{\mathbf{z}}$ | (4$c$) | O V |
| $\mathbf{B}_{52}$ | = | $\left(\frac{1}{2} - x_{14}\right) \mathbf{a}_1 + \left(\frac{1}{2} + y_{14}\right) \mathbf{a}_2 + z_{14} \mathbf{a}_3$ | = | $\left(\frac{1}{2} - x_{14}\right) a \hat{\mathbf{x}} + \left(\frac{1}{2} + y_{14}\right) b \hat{\mathbf{y}} + z_{14} c \hat{\mathbf{z}}$ | (4$c$) | O V |
| $\mathbf{B}_{53}$ | = | $x_{15} \mathbf{a}_1 + y_{15} \mathbf{a}_2 + z_{15} \mathbf{a}_3$ | = | $x_{15} a \hat{\mathbf{x}} + y_{15} b \hat{\mathbf{y}} + z_{15} c \hat{\mathbf{z}}$ | (4$c$) | O VI |
| $\mathbf{B}_{54}$ | = | $-x_{15} \mathbf{a}_1 - y_{15} \mathbf{a}_2 + z_{15} \mathbf{a}_3$ | = | $-x_{15} a \hat{\mathbf{x}} - y_{15} b \hat{\mathbf{y}} + z_{15} c \hat{\mathbf{z}}$ | (4$c$) | O VI |
| $\mathbf{B}_{55}$ | = | $\left(\frac{1}{2} + x_{15}\right) \mathbf{a}_1 + \left(\frac{1}{2} - y_{15}\right) \mathbf{a}_2 + z_{15} \mathbf{a}_3$ | = | $\left(\frac{1}{2} + x_{15}\right) a \hat{\mathbf{x}} + \left(\frac{1}{2} - y_{15}\right) b \hat{\mathbf{y}} + z_{15} c \hat{\mathbf{z}}$ | (4$c$) | O VI |
| $\mathbf{B}_{56}$ | = | $\left(\frac{1}{2} - x_{15}\right) \mathbf{a}_1 + \left(\frac{1}{2} + y_{15}\right) \mathbf{a}_2 + z_{15} \mathbf{a}_3$ | = | $\left(\frac{1}{2} - x_{15}\right) a \hat{\mathbf{x}} + \left(\frac{1}{2} + y_{15}\right) b \hat{\mathbf{y}} + z_{15} c \hat{\mathbf{z}}$ | (4$c$) | O VI |
| $\mathbf{B}_{57}$ | = | $x_{16} \mathbf{a}_1 + y_{16} \mathbf{a}_2 + z_{16} \mathbf{a}_3$ | = | $x_{16} a \hat{\mathbf{x}} + y_{16} b \hat{\mathbf{y}} + z_{16} c \hat{\mathbf{z}}$ | (4$c$) | O VII |
| $\mathbf{B}_{58}$ | = | $-x_{16} \mathbf{a}_1 - y_{16} \mathbf{a}_2 + z_{16} \mathbf{a}_3$ | = | $-x_{16} a \hat{\mathbf{x}} - y_{16} b \hat{\mathbf{y}} + z_{16} c \hat{\mathbf{z}}$ | (4$c$) | O VII |
| $\mathbf{B}_{59}$ | = | $\left(\frac{1}{2} + x_{16}\right) \mathbf{a}_1 + \left(\frac{1}{2} - y_{16}\right) \mathbf{a}_2 + z_{16} \mathbf{a}_3$ | = | $\left(\frac{1}{2} + x_{16}\right) a \hat{\mathbf{x}} + \left(\frac{1}{2} - y_{16}\right) b \hat{\mathbf{y}} + z_{16} c \hat{\mathbf{z}}$ | (4$c$) | O VII |
| $\mathbf{B}_{60}$ | = | $\left(\frac{1}{2} - x_{16}\right) \mathbf{a}_1 + \left(\frac{1}{2} + y_{16}\right) \mathbf{a}_2 + z_{16} \mathbf{a}_3$ | = | $\left(\frac{1}{2} - x_{16}\right) a \hat{\mathbf{x}} + \left(\frac{1}{2} + y_{16}\right) b \hat{\mathbf{y}} + z_{16} c \hat{\mathbf{z}}$ | (4$c$) | O VII |
| $\mathbf{B}_{61}$ | = | $x_{17} \mathbf{a}_1 + y_{17} \mathbf{a}_2 + z_{17} \mathbf{a}_3$ | = | $x_{17} a \hat{\mathbf{x}} + y_{17} b \hat{\mathbf{y}} + z_{17} c \hat{\mathbf{z}}$ | (4$c$) | O VIII |
| $\mathbf{B}_{62}$ | = | $-x_{17} \mathbf{a}_1 - y_{17} \mathbf{a}_2 + z_{17} \mathbf{a}_3$ | = | $-x_{17} a \hat{\mathbf{x}} - y_{17} b \hat{\mathbf{y}} + z_{17} c \hat{\mathbf{z}}$ | (4$c$) | O VIII |
| $\mathbf{B}_{63}$ | = | $\left(\frac{1}{2} + x_{17}\right) \mathbf{a}_1 + \left(\frac{1}{2} - y_{17}\right) \mathbf{a}_2 + z_{17} \mathbf{a}_3$ | = | $\left(\frac{1}{2} + x_{17}\right) a \hat{\mathbf{x}} + \left(\frac{1}{2} - y_{17}\right) b \hat{\mathbf{y}} + z_{17} c \hat{\mathbf{z}}$ | (4$c$) | O VIII |
| $\mathbf{B}_{64}$ | = | $\left(\frac{1}{2} - x_{17}\right) \mathbf{a}_1 + \left(\frac{1}{2} + y_{17}\right) \mathbf{a}_2 + z_{17} \mathbf{a}_3$ | = | $\left(\frac{1}{2} - x_{17}\right) a \hat{\mathbf{x}} + \left(\frac{1}{2} + y_{17}\right) b \hat{\mathbf{y}} + z_{17} c \hat{\mathbf{z}}$ | (4$c$) | O VIII |
| $\mathbf{B}_{65}$ | = | $x_{18} \mathbf{a}_1 + y_{18} \mathbf{a}_2 + z_{18} \mathbf{a}_3$ | = | $x_{18} a \hat{\mathbf{x}} + y_{18} b \hat{\mathbf{y}} + z_{18} c \hat{\mathbf{z}}$ | (4$c$) | O IX |
| $\mathbf{B}_{66}$ | = | $-x_{18} \mathbf{a}_1 - y_{18} \mathbf{a}_2 + z_{18} \mathbf{a}_3$ | = | $-x_{18} a \hat{\mathbf{x}} - y_{18} b \hat{\mathbf{y}} + z_{18} c \hat{\mathbf{z}}$ | (4$c$) | O IX |
| $\mathbf{B}_{67}$ | = | $\left(\frac{1}{2} + x_{18}\right) \mathbf{a}_1 + \left(\frac{1}{2} - y_{18}\right) \mathbf{a}_2 + z_{18} \mathbf{a}_3$ | = | $\left(\frac{1}{2} + x_{18}\right) a \hat{\mathbf{x}} + \left(\frac{1}{2} - y_{18}\right) b \hat{\mathbf{y}} + z_{18} c \hat{\mathbf{z}}$ | (4$c$) | O IX |
| $\mathbf{B}_{68}$ | = | $\left(\frac{1}{2} - x_{18}\right) \mathbf{a}_1 + \left(\frac{1}{2} + y_{18}\right) \mathbf{a}_2 + z_{18} \mathbf{a}_3$ | = | $\left(\frac{1}{2} - x_{18}\right) a \hat{\mathbf{x}} + \left(\frac{1}{2} + y_{18}\right) b \hat{\mathbf{y}} + z_{18} c \hat{\mathbf{z}}$ | (4$c$) | O IX |
| $\mathbf{B}_{69}$ | = | $x_{19} \mathbf{a}_1 + y_{19} \mathbf{a}_2 + z_{19} \mathbf{a}_3$ | = | $x_{19} a \hat{\mathbf{x}} + y_{19} b \hat{\mathbf{y}} + z_{19} c \hat{\mathbf{z}}$ | (4$c$) | O X |
| $\mathbf{B}_{70}$ | = | $-x_{19} \mathbf{a}_1 - y_{19} \mathbf{a}_2 + z_{19} \mathbf{a}_3$ | = | $-x_{19} a \hat{\mathbf{x}} - y_{19} b \hat{\mathbf{y}} + z_{19} c \hat{\mathbf{z}}$ | (4$c$) | O X |
| $\mathbf{B}_{71}$ | = | $\left(\frac{1}{2} + x_{19}\right) \mathbf{a}_1 + \left(\frac{1}{2} - y_{19}\right) \mathbf{a}_2 + z_{19} \mathbf{a}_3$ | = | $\left(\frac{1}{2} + x_{19}\right) a \hat{\mathbf{x}} + \left(\frac{1}{2} - y_{19}\right) b \hat{\mathbf{y}} + z_{19} c \hat{\mathbf{z}}$ | (4$c$) | O X |
| $\mathbf{B}_{72}$ | = | $\left(\frac{1}{2} - x_{19}\right) \mathbf{a}_1 + \left(\frac{1}{2} + y_{19}\right) \mathbf{a}_2 + z_{19} \mathbf{a}_3$ | = | $\left(\frac{1}{2} - x_{19}\right) a \hat{\mathbf{x}} + \left(\frac{1}{2} + y_{19}\right) b \hat{\mathbf{y}} + z_{19} c \hat{\mathbf{z}}$ | (4$c$) | O X |
| $\mathbf{B}_{73}$ | = | $x_{20} \mathbf{a}_1 + y_{20} \mathbf{a}_2 + z_{20} \mathbf{a}_3$ | = | $x_{20} a \hat{\mathbf{x}} + y_{20} b \hat{\mathbf{y}} + z_{20} c \hat{\mathbf{z}}$ | (4$c$) | O XI |
| $\mathbf{B}_{74}$ | = | $-x_{20} \mathbf{a}_1 - y_{20} \mathbf{a}_2 + z_{20} \mathbf{a}_3$ | = | $-x_{20} a \hat{\mathbf{x}} - y_{20} b \hat{\mathbf{y}} + z_{20} c \hat{\mathbf{z}}$ | (4$c$) | O XI |
| $\mathbf{B}_{75}$ | = | $\left(\frac{1}{2} + x_{20}\right) \mathbf{a}_1 + \left(\frac{1}{2} - y_{20}\right) \mathbf{a}_2 + z_{20} \mathbf{a}_3$ | = | $\left(\frac{1}{2} + x_{20}\right) a \hat{\mathbf{x}} + \left(\frac{1}{2} - y_{20}\right) b \hat{\mathbf{y}} + z_{20} c \hat{\mathbf{z}}$ | (4$c$) | O XI |
| $\mathbf{B}_{76}$ | = | $\left(\frac{1}{2} - x_{20}\right) \mathbf{a}_1 + \left(\frac{1}{2} + y_{20}\right) \mathbf{a}_2 + z_{20} \mathbf{a}_3$ | = | $\left(\frac{1}{2} - x_{20}\right) a \hat{\mathbf{x}} + \left(\frac{1}{2} + y_{20}\right) b \hat{\mathbf{y}} + z_{20} c \hat{\mathbf{z}}$ | (4$c$) | O XI |
| $\mathbf{B}_{77}$ | = | $x_{21} \mathbf{a}_1 + y_{21} \mathbf{a}_2 + z_{21} \mathbf{a}_3$ | = | $x_{21} a \hat{\mathbf{x}} + y_{21} b \hat{\mathbf{y}} + z_{21} c \hat{\mathbf{z}}$ | (4$c$) | O XII |
| $\mathbf{B}_{78}$ | = | $-x_{21} \mathbf{a}_1 - y_{21} \mathbf{a}_2 + z_{21} \mathbf{a}_3$ | = | $-x_{21} a \hat{\mathbf{x}} - y_{21} b \hat{\mathbf{y}} + z_{21} c \hat{\mathbf{z}}$ | (4$c$) | O XII |
| $\mathbf{B}_{79}$ | = | $\left(\frac{1}{2} + x_{21}\right) \mathbf{a}_1 + \left(\frac{1}{2} - y_{21}\right) \mathbf{a}_2 + z_{21} \mathbf{a}_3$ | = | $\left(\frac{1}{2} + x_{21}\right) a \hat{\mathbf{x}} + \left(\frac{1}{2} - y_{21}\right) b \hat{\mathbf{y}} + z_{21} c \hat{\mathbf{z}}$ | (4$c$) | O XII |
| $\mathbf{B}_{80}$ | = | $\left(\frac{1}{2} - x_{21}\right) \mathbf{a}_1 + \left(\frac{1}{2} + y_{21}\right) \mathbf{a}_2 + z_{21} \mathbf{a}_3$ | = | $\left(\frac{1}{2} - x_{21}\right) a \hat{\mathbf{x}} + \left(\frac{1}{2} + y_{21}\right) b \hat{\mathbf{y}} + z_{21} c \hat{\mathbf{z}}$ | (4$c$) | O XII |
| $\mathbf{B}_{81}$ | = | $x_{22} \mathbf{a}_1 + y_{22} \mathbf{a}_2 + z_{22} \mathbf{a}_3$ | = | $x_{22} a \hat{\mathbf{x}} + y_{22} b \hat{\mathbf{y}} + z_{22} c \hat{\mathbf{z}}$ | (4$c$) | O XIII |
| $\mathbf{B}_{82}$ | = | $-x_{22} \mathbf{a}_1 - y_{22} \mathbf{a}_2 + z_{22} \mathbf{a}_3$ | = | $-x_{22} a \hat{\mathbf{x}} - y_{22} b \hat{\mathbf{y}} + z_{22} c \hat{\mathbf{z}}$ | (4$c$) | O XIII |
| $\mathbf{B}_{83}$ | = | $\left(\frac{1}{2} + x_{22}\right) \mathbf{a}_1 + \left(\frac{1}{2} - y_{22}\right) \mathbf{a}_2 + z_{22} \mathbf{a}_3$ | = | $\left(\frac{1}{2} + x_{22}\right) a \hat{\mathbf{x}} + \left(\frac{1}{2} - y_{22}\right) b \hat{\mathbf{y}} + z_{22} c \hat{\mathbf{z}}$ | (4$c$) | O XIII |
| $\mathbf{B}_{84}$ | = | $\left(\frac{1}{2} - x_{22}\right) \mathbf{a}_1 + \left(\frac{1}{2} + y_{22}\right) \mathbf{a}_2 + z_{22} \mathbf{a}_3$ | = | $\left(\frac{1}{2} - x_{22}\right) a \hat{\mathbf{x}} + \left(\frac{1}{2} + y_{22}\right) b \hat{\mathbf{y}} + z_{22} c \hat{\mathbf{z}}$ | (4$c$) | O XIII |
| $\mathbf{B}_{85}$ | = | $x_{23} \mathbf{a}_1 + y_{23} \mathbf{a}_2 + z_{23} \mathbf{a}_3$ | = | $x_{23} a \hat{\mathbf{x}} + y_{23} b \hat{\mathbf{y}} + z_{23} c \hat{\mathbf{z}}$ | (4$c$) | O XIV |
| $\mathbf{B}_{86}$ | = | $-x_{23} \mathbf{a}_1 - y_{23} \mathbf{a}_2 + z_{23} \mathbf{a}_3$ | = | $-x_{23} a \hat{\mathbf{x}} - y_{23} b \hat{\mathbf{y}} + z_{23} c \hat{\mathbf{z}}$ | (4$c$) | O XIV |
| $\mathbf{B}_{87}$ | = | $\left(\frac{1}{2} + x_{23}\right) \mathbf{a}_1 + \left(\frac{1}{2} - y_{23}\right) \mathbf{a}_2 + z_{23} \mathbf{a}_3$ | = | $\left(\frac{1}{2} + x_{23}\right) a \hat{\mathbf{x}} + \left(\frac{1}{2} - y_{23}\right) b \hat{\mathbf{y}} + z_{23} c \hat{\mathbf{z}}$ | (4$c$) | O XIV |
| $\mathbf{B}_{88}$ | = | $\left(\frac{1}{2} - x_{23}\right) \mathbf{a}_1 + \left(\frac{1}{2} + y_{23}\right) \mathbf{a}_2 + z_{23} \mathbf{a}_3$ | = | $\left(\frac{1}{2} - x_{23}\right) a \hat{\mathbf{x}} + \left(\frac{1}{2} + y_{23}\right) b \hat{\mathbf{y}} + z_{23} c \hat{\mathbf{z}}$ | (4$c$) | O XIV |
| $\mathbf{B}_{89}$ | = | $x_{24} \mathbf{a}_1 + y_{24} \mathbf{a}_2 + z_{24} \mathbf{a}_3$ | = | $x_{24} a \hat{\mathbf{x}} + y_{24} b \hat{\mathbf{y}} + z_{24} c \hat{\mathbf{z}}$ | (4$c$) | O XV |
| $\mathbf{B}_{90}$ | = | $-x_{24} \mathbf{a}_1 - y_{24} \mathbf{a}_2 + z_{24} \mathbf{a}_3$ | = | $-x_{24} a \hat{\mathbf{x}} - y_{24} b \hat{\mathbf{y}} + z_{24} c \hat{\mathbf{z}}$ | (4$c$) | O XV |
| $\mathbf{B}_{91}$ | = | $\left(\frac{1}{2} + x_{24}\right) \mathbf{a}_1 + \left(\frac{1}{2} - y_{24}\right) \mathbf{a}_2 + z_{24} \mathbf{a}_3$ | = | $\left(\frac{1}{2} + x_{24}\right) a \hat{\mathbf{x}} + \left(\frac{1}{2} - y_{24}\right) b \hat{\mathbf{y}} + z_{24} c \hat{\mathbf{z}}$ | (4$c$) | O XV |
| $\mathbf{B}_{92}$ | = | $\left(\frac{1}{2} - x_{24}\right) \mathbf{a}_1 + \left(\frac{1}{2} + y_{24}\right) \mathbf{a}_2 + z_{24} \mathbf{a}_3$ | = | $\left(\frac{1}{2} - x_{24}\right) a \hat{\mathbf{x}} + \left(\frac{1}{2} + y_{24}\right) b \hat{\mathbf{y}} + z_{24} c \hat{\mathbf{z}}$ | (4$c$) | O XV |
| $\mathbf{B}_{93}$ | = | $x_{25} \mathbf{a}_1 + y_{25} \mathbf{a}_2 + z_{25} \mathbf{a}_3$ | = | $x_{25} a \hat{\mathbf{x}} + y_{25} b \hat{\mathbf{y}} + z_{25} c \hat{\mathbf{z}}$ | (4$c$) | O XVI |
| $\mathbf{B}_{94}$ | = | $-x_{25} \mathbf{a}_1 - y_{25} \mathbf{a}_2 + z_{25} \mathbf{a}_3$ | = | $-x_{25} a \hat{\mathbf{x}} - y_{25} b \hat{\mathbf{y}} + z_{25} c \hat{\mathbf{z}}$ | (4$c$) | O XVI |
| $\mathbf{B}_{95}$ | = | $\left(\frac{1}{2} + x_{25}\right) \mathbf{a}_1 + \left(\frac{1}{2} - y_{25}\right) \mathbf{a}_2 + z_{25} \mathbf{a}_3$ | = | $\left(\frac{1}{2} + x_{25}\right) a \hat{\mathbf{x}} + \left(\frac{1}{2} - y_{25}\right) b \hat{\mathbf{y}} + z_{25} c \hat{\mathbf{z}}$ | (4$c$) | O XVI |
| $\mathbf{B}_{96}$ | = | $\left(\frac{1}{2} - x_{25}\right) \mathbf{a}_1 + \left(\frac{1}{2} + y_{25}\right) \mathbf{a}_2 + z_{25} \mathbf{a}_3$ | = | $\left(\frac{1}{2} - x_{25}\right) a \hat{\mathbf{x}} + \left(\frac{1}{2} + y_{25}\right) b \hat{\mathbf{y}} + z_{25} c \hat{\mathbf{z}}$ | (4$c$) | O XVI |
| $\mathbf{B}_{97}$ | = | $x_{26} \mathbf{a}_1 + y_{26} \mathbf{a}_2 + z_{26} \mathbf{a}_3$ | = | $x_{26} a \hat{\mathbf{x}} + y_{26} b \hat{\mathbf{y}} + z_{26} c \hat{\mathbf{z}}$ | (4$c$) | O XVII |
| $\mathbf{B}_{98}$ | = | $-x_{26} \mathbf{a}_1 - y_{26} \mathbf{a}_2 + z_{26} \mathbf{a}_3$ | = | $-x_{26} a \hat{\mathbf{x}} - y_{26} b \hat{\mathbf{y}} + z_{26} c \hat{\mathbf{z}}$ | (4$c$) | O XVII |
| $\mathbf{B}_{99}$ | = | $\left(\frac{1}{2} + x_{26}\right) \mathbf{a}_1 + \left(\frac{1}{2} - y_{26}\right) \mathbf{a}_2 + z_{26} \mathbf{a}_3$ | = | $\left(\frac{1}{2} + x_{26}\right) a \hat{\mathbf{x}} + \left(\frac{1}{2} - y_{26}\right) b \hat{\mathbf{y}} + z_{26} c \hat{\mathbf{z}}$ | (4$c$) | O XVII |
| $\mathbf{B}_{100}$ | = | $\left(\frac{1}{2} - x_{26}\right)\,\mathbf{a}_1 + \left(\frac{1}{2} + y_{26}\right)\,\mathbf{a}_2 + z_{26}\,\mathbf{a}_3$ | = | $\left(\frac{1}{2} - x_{26}\right)a\,\hat{\mathbf{x}} + \left(\frac{1}{2} + y_{26}\right)b\,\hat{\mathbf{y}} + z_{26}c\,\hat{\mathbf{z}}$ | (4$c$) | O XVII |
| $\mathbf{B}_{101}$ | = | $x_{27}\,\mathbf{a}_1 + y_{27}\,\mathbf{a}_2 + z_{27}\,\mathbf{a}_3$ | = | $x_{27}a\,\hat{\mathbf{x}} + y_{27}b\,\hat{\mathbf{y}} + z_{27}c\,\hat{\mathbf{z}}$ | (4$c$) | O XVIII |
| $\mathbf{B}_{102}$ | = | $-x_{27}\,\mathbf{a}_1 - y_{27}\,\mathbf{a}_2 + z_{27}\,\mathbf{a}_3$ | = | $-x_{27}a\,\hat{\mathbf{x}} - y_{27}b\,\hat{\mathbf{y}} + z_{27}c\,\hat{\mathbf{z}}$ | (4$c$) | O XVIII |
| $\mathbf{B}_{103}$ | = | $\left(\frac{1}{2} + x_{27}\right)\,\mathbf{a}_1 + \left(\frac{1}{2} - y_{27}\right)\,\mathbf{a}_2 + z_{27}\,\mathbf{a}_3$ | = | $\left(\frac{1}{2} + x_{27}\right)a\,\hat{\mathbf{x}} + \left(\frac{1}{2} - y_{27}\right)b\,\hat{\mathbf{y}} + z_{27}c\,\hat{\mathbf{z}}$ | (4$c$) | O XVIII |
| $\mathbf{B}_{104}$ | = | $\left(\frac{1}{2} - x_{27}\right)\,\mathbf{a}_1 + \left(\frac{1}{2} + y_{27}\right)\,\mathbf{a}_2 + z_{27}\,\mathbf{a}_3$ | = | $\left(\frac{1}{2} - x_{27}\right)a\,\hat{\mathbf{x}} + \left(\frac{1}{2} + y_{27}\right)b\,\hat{\mathbf{y}} + z_{27}c\,\hat{\mathbf{z}}$ | (4$c$) | O XVIII |
| $\mathbf{B}_{105}$ | = | $x_{28}\,\mathbf{a}_1 + y_{28}\,\mathbf{a}_2 + z_{28}\,\mathbf{a}_3$ | = | $x_{28}a\,\hat{\mathbf{x}} + y_{28}b\,\hat{\mathbf{y}} + z_{28}c\,\hat{\mathbf{z}}$ | (4$c$) | O XIX |
| $\mathbf{B}_{106}$ | = | $-x_{28}\,\mathbf{a}_1 - y_{28}\,\mathbf{a}_2 + z_{28}\,\mathbf{a}_3$ | = | $-x_{28}a\,\hat{\mathbf{x}} - y_{28}b\,\hat{\mathbf{y}} + z_{28}c\,\hat{\mathbf{z}}$ | (4$c$) | O XIX |
| $\mathbf{B}_{107}$ | = | $\left(\frac{1}{2} + x_{28}\right)\,\mathbf{a}_1 + \left(\frac{1}{2} - y_{28}\right)\,\mathbf{a}_2 + z_{28}\,\mathbf{a}_3$ | = | $\left(\frac{1}{2} + x_{28}\right)a\,\hat{\mathbf{x}} + \left(\frac{1}{2} - y_{28}\right)b\,\hat{\mathbf{y}} + z_{28}c\,\hat{\mathbf{z}}$ | (4$c$) | O XIX |
| $\mathbf{B}_{108}$ | = | $\left(\frac{1}{2} - x_{28}\right)\,\mathbf{a}_1 + \left(\frac{1}{2} + y_{28}\right)\,\mathbf{a}_2 + z_{28}\,\mathbf{a}_3$ | = | $\left(\frac{1}{2} - x_{28}\right)a\,\hat{\mathbf{x}} + \left(\frac{1}{2} + y_{28}\right)b\,\hat{\mathbf{y}} + z_{28}c\,\hat{\mathbf{z}}$ | (4$c$) | O XIX |
| $\mathbf{B}_{109}$ | = | $x_{29}\,\mathbf{a}_1 + y_{29}\,\mathbf{a}_2 + z_{29}\,\mathbf{a}_3$ | = | $x_{29}a\,\hat{\mathbf{x}} + y_{29}b\,\hat{\mathbf{y}} + z_{29}c\,\hat{\mathbf{z}}$ | (4$c$) | O XX |
| $\mathbf{B}_{110}$ | = | $-x_{29}\,\mathbf{a}_1 - y_{29}\,\mathbf{a}_2 + z_{29}\,\mathbf{a}_3$ | = | $-x_{29}a\,\hat{\mathbf{x}} - y_{29}b\,\hat{\mathbf{y}} + z_{29}c\,\hat{\mathbf{z}}$ | (4$c$) | O XX |
| $\mathbf{B}_{111}$ | = | $\left(\frac{1}{2} + x_{29}\right)\,\mathbf{a}_1 + \left(\frac{1}{2} - y_{29}\right)\,\mathbf{a}_2 + z_{29}\,\mathbf{a}_3$ | = | $\left(\frac{1}{2} + x_{29}\right)a\,\hat{\mathbf{x}} + \left(\frac{1}{2} - y_{29}\right)b\,\hat{\mathbf{y}} + z_{29}c\,\hat{\mathbf{z}}$ | (4$c$) | O XX |
| $\mathbf{B}_{112}$ | = | $\left(\frac{1}{2} - x_{29}\right)\,\mathbf{a}_1 + \left(\frac{1}{2} + y_{29}\right)\,\mathbf{a}_2 + z_{29}\,\mathbf{a}_3$ | = | $\left(\frac{1}{2} - x_{29}\right)a\,\hat{\mathbf{x}} + \left(\frac{1}{2} + y_{29}\right)b\,\hat{\mathbf{y}} + z_{29}c\,\hat{\mathbf{z}}$ | (4$c$) | O XX |
| $\mathbf{B}_{113}$ | = | $x_{30}\,\mathbf{a}_1 + y_{30}\,\mathbf{a}_2 + z_{30}\,\mathbf{a}_3$ | = | $x_{30}a\,\hat{\mathbf{x}} + y_{30}b\,\hat{\mathbf{y}} + z_{30}c\,\hat{\mathbf{z}}$ | (4$c$) | O XXI |
| $\mathbf{B}_{114}$ | = | $-x_{30}\,\mathbf{a}_1 - y_{30}\,\mathbf{a}_2 + z_{30}\,\mathbf{a}_3$ | = | $-x_{30}a\,\hat{\mathbf{x}} - y_{30}b\,\hat{\mathbf{y}} + z_{30}c\,\hat{\mathbf{z}}$ | (4$c$) | O XXI |
| $\mathbf{B}_{115}$ | = | $\left(\frac{1}{2} + x_{30}\right)\,\mathbf{a}_1 + \left(\frac{1}{2} - y_{30}\right)\,\mathbf{a}_2 + z_{30}\,\mathbf{a}_3$ | = | $\left(\frac{1}{2} + x_{30}\right)a\,\hat{\mathbf{x}} + \left(\frac{1}{2} - y_{30}\right)b\,\hat{\mathbf{y}} + z_{30}c\,\hat{\mathbf{z}}$ | (4$c$) | O XXI |
| $\mathbf{B}_{116}$ | = | $\left(\frac{1}{2} - x_{30}\right)\,\mathbf{a}_1 + \left(\frac{1}{2} + y_{30}\right)\,\mathbf{a}_2 + z_{30}\,\mathbf{a}_3$ | = | $\left(\frac{1}{2} - x_{30}\right)a\,\hat{\mathbf{x}} + \left(\frac{1}{2} + y_{30}\right)b\,\hat{\mathbf{y}} + z_{30}c\,\hat{\mathbf{z}}$ | (4$c$) | O XXI |
| $\mathbf{B}_{117}$ | = | $x_{31}\,\mathbf{a}_1 + y_{31}\,\mathbf{a}_2 + z_{31}\,\mathbf{a}_3$ | = | $x_{31}a\,\hat{\mathbf{x}} + y_{31}b\,\hat{\mathbf{y}} + z_{31}c\,\hat{\mathbf{z}}$ | (4$c$) | O XXII |
| $\mathbf{B}_{118}$ | = | $-x_{31}\,\mathbf{a}_1 - y_{31}\,\mathbf{a}_2 + z_{31}\,\mathbf{a}_3$ | = | $-x_{31}a\,\hat{\mathbf{x}} - y_{31}b\,\hat{\mathbf{y}} + z_{31}c\,\hat{\mathbf{z}}$ | (4$c$) | O XXII |
| $\mathbf{B}_{119}$ | = | $\left(\frac{1}{2} + x_{31}\right)\,\mathbf{a}_1 + \left(\frac{1}{2} - y_{31}\right)\,\mathbf{a}_2 + z_{31}\,\mathbf{a}_3$ | = | $\left(\frac{1}{2} + x_{31}\right)a\,\hat{\mathbf{x}} + \left(\frac{1}{2} - y_{31}\right)b\,\hat{\mathbf{y}} + z_{31}c\,\hat{\mathbf{z}}$ | (4$c$) | O XXII |
| $\mathbf{B}_{120}$ | = | $\left(\frac{1}{2} - x_{31}\right)\,\mathbf{a}_1 + \left(\frac{1}{2} + y_{31}\right)\,\mathbf{a}_2 + z_{31}\,\mathbf{a}_3$ | = | $\left(\frac{1}{2} - x_{31}\right)a\,\hat{\mathbf{x}} + \left(\frac{1}{2} + y_{31}\right)b\,\hat{\mathbf{y}} + z_{31}c\,\hat{\mathbf{z}}$ | (4$c$) | O XXII |
| $\mathbf{B}_{121}$ | = | $x_{32}\,\mathbf{a}_1 + y_{32}\,\mathbf{a}_2 + z_{32}\,\mathbf{a}_3$ | = | $x_{32}a\,\hat{\mathbf{x}} + y_{32}b\,\hat{\mathbf{y}} + z_{32}c\,\hat{\mathbf{z}}$ | (4$c$) | O XXIII |
| $\mathbf{B}_{122}$ | = | $-x_{32}\,\mathbf{a}_1 - y_{32}\,\mathbf{a}_2 + z_{32}\,\mathbf{a}_3$ | = | $-x_{32}a\,\hat{\mathbf{x}} - y_{32}b\,\hat{\mathbf{y}} + z_{32}c\,\hat{\mathbf{z}}$ | (4$c$) | O XXIII |
| $\mathbf{B}_{123}$ | = | $\left(\frac{1}{2} + x_{32}\right)\,\mathbf{a}_1 + \left(\frac{1}{2} - y_{32}\right)\,\mathbf{a}_2 + z_{32}\,\mathbf{a}_3$ | = | $\left(\frac{1}{2} + x_{32}\right)a\,\hat{\mathbf{x}} + \left(\frac{1}{2} - y_{32}\right)b\,\hat{\mathbf{y}} + z_{32}c\,\hat{\mathbf{z}}$ | (4$c$) | O XXIII |
| $\mathbf{B}_{124}$ | = | $\left(\frac{1}{2} - x_{32}\right)\,\mathbf{a}_1 + \left(\frac{1}{2} + y_{32}\right)\,\mathbf{a}_2 + z_{32}\,\mathbf{a}_3$ | = | $\left(\frac{1}{2} - x_{32}\right)a\,\hat{\mathbf{x}} + \left(\frac{1}{2} + y_{32}\right)b\,\hat{\mathbf{y}} + z_{32}c\,\hat{\mathbf{z}}$ | (4$c$) | O XXIII |
| $\mathbf{B}_{125}$ | = | $x_{33}\,\mathbf{a}_1 + y_{33}\,\mathbf{a}_2 + z_{33}\,\mathbf{a}_3$ | = | $x_{33}a\,\hat{\mathbf{x}} + y_{33}b\,\hat{\mathbf{y}} + z_{33}c\,\hat{\mathbf{z}}$ | (4$c$) | O XXIV |
| $\mathbf{B}_{126}$ | = | $-x_{33}\,\mathbf{a}_1 - y_{33}\,\mathbf{a}_2 + z_{33}\,\mathbf{a}_3$ | = | $-x_{33}a\,\hat{\mathbf{x}} - y_{33}b\,\hat{\mathbf{y}} + z_{33}c\,\hat{\mathbf{z}}$ | (4$c$) | O XXIV |
| $\mathbf{B}_{127}$ | = | $\left(\frac{1}{2} + x_{33}\right)\,\mathbf{a}_1 + \left(\frac{1}{2} - y_{33}\right)\,\mathbf{a}_2 + z_{33}\,\mathbf{a}_3$ | = | $\left(\frac{1}{2} + x_{33}\right)a\,\hat{\mathbf{x}} + \left(\frac{1}{2} - y_{33}\right)b\,\hat{\mathbf{y}} + z_{33}c\,\hat{\mathbf{z}}$ | (4$c$) | O XXIV |
| $\mathbf{B}_{128}$ | = | $\left(\frac{1}{2} - x_{33}\right)\,\mathbf{a}_1 + \left(\frac{1}{2} + y_{33}\right)\,\mathbf{a}_2 + z_{33}\,\mathbf{a}_3$ | = | $\left(\frac{1}{2} - x_{33}\right)a\,\hat{\mathbf{x}} + \left(\frac{1}{2} + y_{33}\right)b\,\hat{\mathbf{y}} + z_{33}c\,\hat{\mathbf{z}}$ | (4$c$) | O XXIV |

**Geometry files:**

- CIF: pp. 1588
- POSCAR: pp. 1589

# Possible $\delta$-$Gd_2Si_2O_7$ Structure: A2B7C2_oP44_33_2a_7a_2a

http://aflow.org/prototype-encyclopedia/A2B7C2_oP44_33_2a_7a_2a

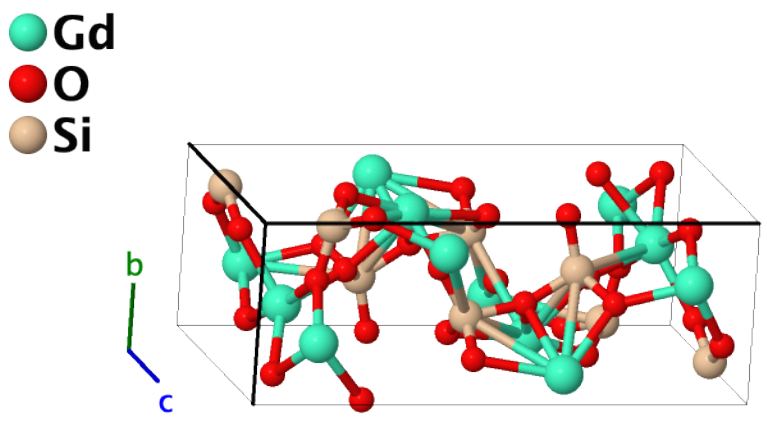

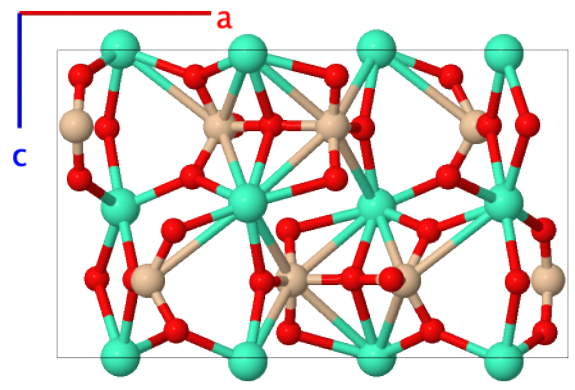

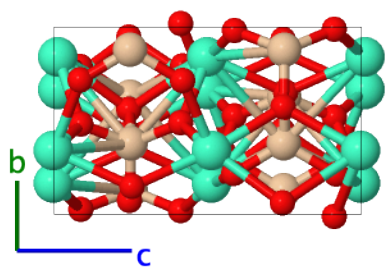

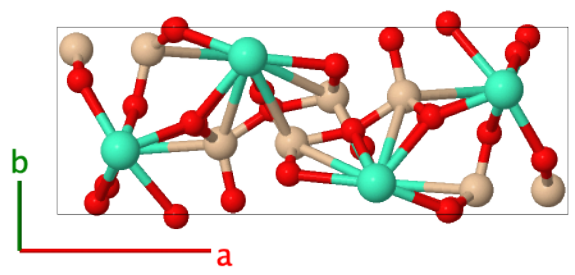

| | | |
|---|---|---|
| **Prototype** | : | $Gd_2O_7Si_2$ |
| **AFLOW prototype label** | : | A2B7C2_oP44_33_2a_7a_2a |
| ***Strukturbericht* designation** | : | None |
| **Pearson symbol** | : | oP44 |
| **Space group number** | : | 33 |
| **Space group symbol** | : | $Pna2_1$ |
| **AFLOW prototype command** | : | `aflow --proto=A2B7C2_oP44_33_2a_7a_2a` `--params=`$a, b/a, c/a, x_1, y_1, z_1, x_2, y_2, z_2, x_3, y_3, z_3, x_4, y_4, z_4, x_5, y_5, z_5, x_6, y_6, z_6, x_7, y_7, z_7, x_8, y_8, z_8, x_9, y_9, z_9, x_{10}, y_{10}, z_{10}, x_{11}, y_{11}, z_{11}$ |

### Other compounds with this structure

- $\delta$-$Ho_2O_7Si_2$, $\delta$-$Dy_2O_7Si_2$, and $\delta$-$Y_2O_7Si_2$

- (Smolin, 1970) found that some structures of $RE_2O_7Si_2$ ($RE$ = Ho, Dy, Gd, Y) were in the orthorhombic $Pna2_1$ #33 space group, in which case this would be the prototype of $\delta$-$RE_2O_7Si_2$ (Becerro, 2004). However, (Dias, 1990) found $\delta$-$Y_2O_7Si_2$ to be in the centro-symmetric *Pnma* #62 space group. This was supported by (Becerro, 2004), who found only one yttrium site in the $\delta$-structure. In addition, if we allow a small amount of uncertainty (0.2 Å) in positions, AFLOW-SYM and FINDSYM place this structure in the *Pnma* group. Nevertheless we have found no work explicitly stating that the structure of (Smolin, 1970) is in error, and indeed (Christensen, 1994) found $\delta$-$Y_2O_7Si_2$ in space group $Pna2_1$. Given this ambiguity, we list $Gd_2O_7Si_2$ as a possible prototype for the $\delta$-phase pyrosilicates.

### Simple Orthorhombic primitive vectors:

$$
\begin{aligned}
\mathbf{a}_1 &= a\,\hat{\mathbf{x}} \\
\mathbf{a}_2 &= b\,\hat{\mathbf{y}} \\
\mathbf{a}_3 &= c\,\hat{\mathbf{z}}
\end{aligned}
$$

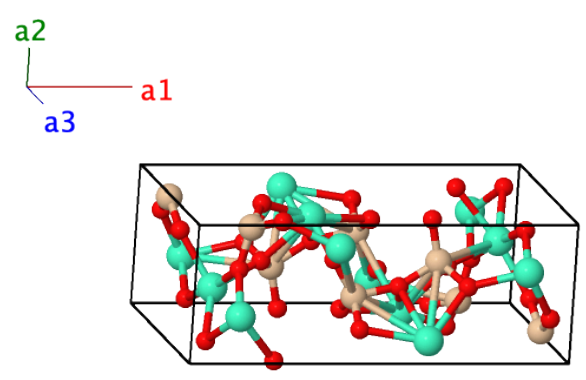

**Basis vectors:**

| | | Lattice Coordinates | | Cartesian Coordinates | Wyckoff Position | Atom Type |
|---|---|---|---|---|---|---|
| $\mathbf{B}_{1}$ | = | $x_{1}\,\mathbf{a}_{1}+y_{1}\,\mathbf{a}_{2}+z_{1}\,\mathbf{a}_{3}$ | = | $x_{1}a\,\hat{\mathbf{x}}+y_{1}b\,\hat{\mathbf{y}}+z_{1}c\,\hat{\mathbf{z}}$ | (4$a$) | Gd I |
| $\mathbf{B}_{2}$ | = | $-x_{1}\,\mathbf{a}_{1}-y_{1}\,\mathbf{a}_{2}+\left(\frac{1}{2}+z_{1}\right)\,\mathbf{a}_{3}$ | = | $-x_{1}a\,\hat{\mathbf{x}}-y_{1}b\,\hat{\mathbf{y}}+\left(\frac{1}{2}+z_{1}\right)c\,\hat{\mathbf{z}}$ | (4$a$) | Gd I |
| $\mathbf{B}_{3}$ | = | $\left(\frac{1}{2}+x_{1}\right)\,\mathbf{a}_{1}+\left(\frac{1}{2}-y_{1}\right)\,\mathbf{a}_{2}+z_{1}\,\mathbf{a}_{3}$ | = | $\left(\frac{1}{2}+x_{1}\right)a\,\hat{\mathbf{x}}+\left(\frac{1}{2}-y_{1}\right)b\,\hat{\mathbf{y}}+z_{1}c\,\hat{\mathbf{z}}$ | (4$a$) | Gd I |
| $\mathbf{B}_{4}$ | = | $\left(\frac{1}{2}-x_{1}\right)\,\mathbf{a}_{1}+\left(\frac{1}{2}+y_{1}\right)\,\mathbf{a}_{2}+\left(\frac{1}{2}+z_{1}\right)\,\mathbf{a}_{3}$ | = | $\left(\frac{1}{2}-x_{1}\right)a\,\hat{\mathbf{x}}+\left(\frac{1}{2}+y_{1}\right)b\,\hat{\mathbf{y}}+\left(\frac{1}{2}+z_{1}\right)c\,\hat{\mathbf{z}}$ | (4$a$) | Gd I |
| $\mathbf{B}_{5}$ | = | $x_{2}\,\mathbf{a}_{1}+y_{2}\,\mathbf{a}_{2}+z_{2}\,\mathbf{a}_{3}$ | = | $x_{2}a\,\hat{\mathbf{x}}+y_{2}b\,\hat{\mathbf{y}}+z_{2}c\,\hat{\mathbf{z}}$ | (4$a$) | Gd II |
| $\mathbf{B}_{6}$ | = | $-x_{2}\,\mathbf{a}_{1}-y_{2}\,\mathbf{a}_{2}+\left(\frac{1}{2}+z_{2}\right)\,\mathbf{a}_{3}$ | = | $-x_{2}a\,\hat{\mathbf{x}}-y_{2}b\,\hat{\mathbf{y}}+\left(\frac{1}{2}+z_{2}\right)c\,\hat{\mathbf{z}}$ | (4$a$) | Gd II |
| $\mathbf{B}_{7}$ | = | $\left(\frac{1}{2}+x_{2}\right)\,\mathbf{a}_{1}+\left(\frac{1}{2}-y_{2}\right)\,\mathbf{a}_{2}+z_{2}\,\mathbf{a}_{3}$ | = | $\left(\frac{1}{2}+x_{2}\right)a\,\hat{\mathbf{x}}+\left(\frac{1}{2}-y_{2}\right)b\,\hat{\mathbf{y}}+z_{2}c\,\hat{\mathbf{z}}$ | (4$a$) | Gd II |
| $\mathbf{B}_{8}$ | = | $\left(\frac{1}{2}-x_{2}\right)\,\mathbf{a}_{1}+\left(\frac{1}{2}+y_{2}\right)\,\mathbf{a}_{2}+\left(\frac{1}{2}+z_{2}\right)\,\mathbf{a}_{3}$ | = | $\left(\frac{1}{2}-x_{2}\right)a\,\hat{\mathbf{x}}+\left(\frac{1}{2}+y_{2}\right)b\,\hat{\mathbf{y}}+\left(\frac{1}{2}+z_{2}\right)c\,\hat{\mathbf{z}}$ | (4$a$) | Gd II |
| $\mathbf{B}_{9}$ | = | $x_{3}\,\mathbf{a}_{1}+y_{3}\,\mathbf{a}_{2}+z_{3}\,\mathbf{a}_{3}$ | = | $x_{3}a\,\hat{\mathbf{x}}+y_{3}b\,\hat{\mathbf{y}}+z_{3}c\,\hat{\mathbf{z}}$ | (4$a$) | O I |
| $\mathbf{B}_{10}$ | = | $-x_{3}\,\mathbf{a}_{1}-y_{3}\,\mathbf{a}_{2}+\left(\frac{1}{2}+z_{3}\right)\,\mathbf{a}_{3}$ | = | $-x_{3}a\,\hat{\mathbf{x}}-y_{3}b\,\hat{\mathbf{y}}+\left(\frac{1}{2}+z_{3}\right)c\,\hat{\mathbf{z}}$ | (4$a$) | O I |
| $\mathbf{B}_{11}$ | = | $\left(\frac{1}{2}+x_{3}\right)\,\mathbf{a}_{1}+\left(\frac{1}{2}-y_{3}\right)\,\mathbf{a}_{2}+z_{3}\,\mathbf{a}_{3}$ | = | $\left(\frac{1}{2}+x_{3}\right)a\,\hat{\mathbf{x}}+\left(\frac{1}{2}-y_{3}\right)b\,\hat{\mathbf{y}}+z_{3}c\,\hat{\mathbf{z}}$ | (4$a$) | O I |
| $\mathbf{B}_{12}$ | = | $\left(\frac{1}{2}-x_{3}\right)\,\mathbf{a}_{1}+\left(\frac{1}{2}+y_{3}\right)\,\mathbf{a}_{2}+\left(\frac{1}{2}+z_{3}\right)\,\mathbf{a}_{3}$ | = | $\left(\frac{1}{2}-x_{3}\right)a\,\hat{\mathbf{x}}+\left(\frac{1}{2}+y_{3}\right)b\,\hat{\mathbf{y}}+\left(\frac{1}{2}+z_{3}\right)c\,\hat{\mathbf{z}}$ | (4$a$) | O I |
| $\mathbf{B}_{13}$ | = | $x_{4}\,\mathbf{a}_{1}+y_{4}\,\mathbf{a}_{2}+z_{4}\,\mathbf{a}_{3}$ | = | $x_{4}a\,\hat{\mathbf{x}}+y_{4}b\,\hat{\mathbf{y}}+z_{4}c\,\hat{\mathbf{z}}$ | (4$a$) | O II |
| $\mathbf{B}_{14}$ | = | $-x_{4}\,\mathbf{a}_{1}-y_{4}\,\mathbf{a}_{2}+\left(\frac{1}{2}+z_{4}\right)\,\mathbf{a}_{3}$ | = | $-x_{4}a\,\hat{\mathbf{x}}-y_{4}b\,\hat{\mathbf{y}}+\left(\frac{1}{2}+z_{4}\right)c\,\hat{\mathbf{z}}$ | (4$a$) | O II |
| $\mathbf{B}_{15}$ | = | $\left(\frac{1}{2}+x_{4}\right)\,\mathbf{a}_{1}+\left(\frac{1}{2}-y_{4}\right)\,\mathbf{a}_{2}+z_{4}\,\mathbf{a}_{3}$ | = | $\left(\frac{1}{2}+x_{4}\right)a\,\hat{\mathbf{x}}+\left(\frac{1}{2}-y_{4}\right)b\,\hat{\mathbf{y}}+z_{4}c\,\hat{\mathbf{z}}$ | (4$a$) | O II |
| $\mathbf{B}_{16}$ | = | $\left(\frac{1}{2}-x_{4}\right)\,\mathbf{a}_{1}+\left(\frac{1}{2}+y_{4}\right)\,\mathbf{a}_{2}+\left(\frac{1}{2}+z_{4}\right)\,\mathbf{a}_{3}$ | = | $\left(\frac{1}{2}-x_{4}\right)a\,\hat{\mathbf{x}}+\left(\frac{1}{2}+y_{4}\right)b\,\hat{\mathbf{y}}+\left(\frac{1}{2}+z_{4}\right)c\,\hat{\mathbf{z}}$ | (4$a$) | O II |
| $\mathbf{B}_{17}$ | = | $x_{5}\,\mathbf{a}_{1}+y_{5}\,\mathbf{a}_{2}+z_{5}\,\mathbf{a}_{3}$ | = | $x_{5}a\,\hat{\mathbf{x}}+y_{5}b\,\hat{\mathbf{y}}+z_{5}c\,\hat{\mathbf{z}}$ | (4$a$) | O III |
| $\mathbf{B}_{18}$ | = | $-x_{5}\,\mathbf{a}_{1}-y_{5}\,\mathbf{a}_{2}+\left(\frac{1}{2}+z_{5}\right)\,\mathbf{a}_{3}$ | = | $-x_{5}a\,\hat{\mathbf{x}}-y_{5}b\,\hat{\mathbf{y}}+\left(\frac{1}{2}+z_{5}\right)c\,\hat{\mathbf{z}}$ | (4$a$) | O III |
| $\mathbf{B}_{19}$ | = | $\left(\frac{1}{2}+x_{5}\right)\,\mathbf{a}_{1}+\left(\frac{1}{2}-y_{5}\right)\,\mathbf{a}_{2}+z_{5}\,\mathbf{a}_{3}$ | = | $\left(\frac{1}{2}+x_{5}\right)a\,\hat{\mathbf{x}}+\left(\frac{1}{2}-y_{5}\right)b\,\hat{\mathbf{y}}+z_{5}c\,\hat{\mathbf{z}}$ | (4$a$) | O III |
| $\mathbf{B}_{20}$ | = | $\left(\frac{1}{2}-x_{5}\right)\,\mathbf{a}_{1}+\left(\frac{1}{2}+y_{5}\right)\,\mathbf{a}_{2}+\left(\frac{1}{2}+z_{5}\right)\,\mathbf{a}_{3}$ | = | $\left(\frac{1}{2}-x_{5}\right)a\,\hat{\mathbf{x}}+\left(\frac{1}{2}+y_{5}\right)b\,\hat{\mathbf{y}}+\left(\frac{1}{2}+z_{5}\right)c\,\hat{\mathbf{z}}$ | (4$a$) | O III |
| $\mathbf{B}_{21}$ | = | $x_{6}\,\mathbf{a}_{1}+y_{6}\,\mathbf{a}_{2}+z_{6}\,\mathbf{a}_{3}$ | = | $x_{6}a\,\hat{\mathbf{x}}+y_{6}b\,\hat{\mathbf{y}}+z_{6}c\,\hat{\mathbf{z}}$ | (4$a$) | O IV |
| $\mathbf{B}_{22}$ | = | $-x_{6}\,\mathbf{a}_{1}-y_{6}\,\mathbf{a}_{2}+\left(\frac{1}{2}+z_{6}\right)\,\mathbf{a}_{3}$ | = | $-x_{6}a\,\hat{\mathbf{x}}-y_{6}b\,\hat{\mathbf{y}}+\left(\frac{1}{2}+z_{6}\right)c\,\hat{\mathbf{z}}$ | (4$a$) | O IV |
| $\mathbf{B}_{23}$ | = | $\left(\frac{1}{2}+x_{6}\right)\,\mathbf{a}_{1}+\left(\frac{1}{2}-y_{6}\right)\,\mathbf{a}_{2}+z_{6}\,\mathbf{a}_{3}$ | = | $\left(\frac{1}{2}+x_{6}\right)a\,\hat{\mathbf{x}}+\left(\frac{1}{2}-y_{6}\right)b\,\hat{\mathbf{y}}+z_{6}c\,\hat{\mathbf{z}}$ | (4$a$) | O IV |
| $\mathbf{B}_{24}$ | = | $\left(\frac{1}{2}-x_{6}\right)\,\mathbf{a}_{1}+\left(\frac{1}{2}+y_{6}\right)\,\mathbf{a}_{2}+\left(\frac{1}{2}+z_{6}\right)\,\mathbf{a}_{3}$ | = | $\left(\frac{1}{2}-x_{6}\right)a\,\hat{\mathbf{x}}+\left(\frac{1}{2}+y_{6}\right)b\,\hat{\mathbf{y}}+\left(\frac{1}{2}+z_{6}\right)c\,\hat{\mathbf{z}}$ | (4$a$) | O IV |
| $\mathbf{B}_{25}$ | = | $x_{7}\,\mathbf{a}_{1}+y_{7}\,\mathbf{a}_{2}+z_{7}\,\mathbf{a}_{3}$ | = | $x_{7}a\,\hat{\mathbf{x}}+y_{7}b\,\hat{\mathbf{y}}+z_{7}c\,\hat{\mathbf{z}}$ | (4$a$) | O V |
| $\mathbf{B}_{26}$ | = | $-x_{7}\,\mathbf{a}_{1}-y_{7}\,\mathbf{a}_{2}+\left(\frac{1}{2}+z_{7}\right)\,\mathbf{a}_{3}$ | = | $-x_{7}a\,\hat{\mathbf{x}}-y_{7}b\,\hat{\mathbf{y}}+\left(\frac{1}{2}+z_{7}\right)c\,\hat{\mathbf{z}}$ | (4$a$) | O V |
| $\mathbf{B}_{27}$ | = | $\left(\frac{1}{2}+x_{7}\right)\,\mathbf{a}_{1}+\left(\frac{1}{2}-y_{7}\right)\,\mathbf{a}_{2}+z_{7}\,\mathbf{a}_{3}$ | = | $\left(\frac{1}{2}+x_{7}\right)a\,\hat{\mathbf{x}}+\left(\frac{1}{2}-y_{7}\right)b\,\hat{\mathbf{y}}+z_{7}c\,\hat{\mathbf{z}}$ | (4$a$) | O V |
| $\mathbf{B}_{28}$ | = | $\left(\frac{1}{2}-x_{7}\right)\,\mathbf{a}_{1}+\left(\frac{1}{2}+y_{7}\right)\,\mathbf{a}_{2}+\left(\frac{1}{2}+z_{7}\right)\,\mathbf{a}_{3}$ | = | $\left(\frac{1}{2}-x_{7}\right)a\,\hat{\mathbf{x}}+\left(\frac{1}{2}+y_{7}\right)b\,\hat{\mathbf{y}}+\left(\frac{1}{2}+z_{7}\right)c\,\hat{\mathbf{z}}$ | (4$a$) | O V |
| $\mathbf{B}_{29}$ | = | $x_8\,\mathbf{a}_1 + y_8\,\mathbf{a}_2 + z_8\,\mathbf{a}_3$ | = | $x_8 a\,\hat{\mathbf{x}} + y_8 b\,\hat{\mathbf{y}} + z_8 c\,\hat{\mathbf{z}}$ | (4a) | O VI |
| $\mathbf{B}_{30}$ | = | $-x_8\,\mathbf{a}_1 - y_8\,\mathbf{a}_2 + \left(\frac{1}{2} + z_8\right)\mathbf{a}_3$ | = | $-x_8 a\,\hat{\mathbf{x}} - y_8 b\,\hat{\mathbf{y}} + \left(\frac{1}{2} + z_8\right) c\,\hat{\mathbf{z}}$ | (4a) | O VI |
| $\mathbf{B}_{31}$ | = | $\left(\frac{1}{2} + x_8\right)\mathbf{a}_1 + \left(\frac{1}{2} - y_8\right)\mathbf{a}_2 + z_8\,\mathbf{a}_3$ | = | $\left(\frac{1}{2} + x_8\right) a\,\hat{\mathbf{x}} + \left(\frac{1}{2} - y_8\right) b\,\hat{\mathbf{y}} + z_8 c\,\hat{\mathbf{z}}$ | (4a) | O VI |
| $\mathbf{B}_{32}$ | = | $\left(\frac{1}{2} - x_8\right)\mathbf{a}_1 + \left(\frac{1}{2} + y_8\right)\mathbf{a}_2 + \left(\frac{1}{2} + z_8\right)\mathbf{a}_3$ | = | $\left(\frac{1}{2} - x_8\right) a\,\hat{\mathbf{x}} + \left(\frac{1}{2} + y_8\right) b\,\hat{\mathbf{y}} + \left(\frac{1}{2} + z_8\right) c\,\hat{\mathbf{z}}$ | (4a) | O VI |
| $\mathbf{B}_{33}$ | = | $x_9\,\mathbf{a}_1 + y_9\,\mathbf{a}_2 + z_9\,\mathbf{a}_3$ | = | $x_9 a\,\hat{\mathbf{x}} + y_9 b\,\hat{\mathbf{y}} + z_9 c\,\hat{\mathbf{z}}$ | (4a) | O VII |
| $\mathbf{B}_{34}$ | = | $-x_9\,\mathbf{a}_1 - y_9\,\mathbf{a}_2 + \left(\frac{1}{2} + z_9\right)\mathbf{a}_3$ | = | $-x_9 a\,\hat{\mathbf{x}} - y_9 b\,\hat{\mathbf{y}} + \left(\frac{1}{2} + z_9\right) c\,\hat{\mathbf{z}}$ | (4a) | O VII |
| $\mathbf{B}_{35}$ | = | $\left(\frac{1}{2} + x_9\right)\mathbf{a}_1 + \left(\frac{1}{2} - y_9\right)\mathbf{a}_2 + z_9\,\mathbf{a}_3$ | = | $\left(\frac{1}{2} + x_9\right) a\,\hat{\mathbf{x}} + \left(\frac{1}{2} - y_9\right) b\,\hat{\mathbf{y}} + z_9 c\,\hat{\mathbf{z}}$ | (4a) | O VII |
| $\mathbf{B}_{36}$ | = | $\left(\frac{1}{2} - x_9\right)\mathbf{a}_1 + \left(\frac{1}{2} + y_9\right)\mathbf{a}_2 + \left(\frac{1}{2} + z_9\right)\mathbf{a}_3$ | = | $\left(\frac{1}{2} - x_9\right) a\,\hat{\mathbf{x}} + \left(\frac{1}{2} + y_9\right) b\,\hat{\mathbf{y}} + \left(\frac{1}{2} + z_9\right) c\,\hat{\mathbf{z}}$ | (4a) | O VII |
| $\mathbf{B}_{37}$ | = | $x_{10}\,\mathbf{a}_1 + y_{10}\,\mathbf{a}_2 + z_{10}\,\mathbf{a}_3$ | = | $x_{10} a\,\hat{\mathbf{x}} + y_{10} b\,\hat{\mathbf{y}} + z_{10} c\,\hat{\mathbf{z}}$ | (4a) | Si I |
| $\mathbf{B}_{38}$ | = | $-x_{10}\,\mathbf{a}_1 - y_{10}\,\mathbf{a}_2 + \left(\frac{1}{2} + z_{10}\right)\mathbf{a}_3$ | = | $-x_{10} a\,\hat{\mathbf{x}} - y_{10} b\,\hat{\mathbf{y}} + \left(\frac{1}{2} + z_{10}\right) c\,\hat{\mathbf{z}}$ | (4a) | Si I |
| $\mathbf{B}_{39}$ | = | $\left(\frac{1}{2} + x_{10}\right)\mathbf{a}_1 + \left(\frac{1}{2} - y_{10}\right)\mathbf{a}_2 + z_{10}\,\mathbf{a}_3$ | = | $\left(\frac{1}{2} + x_{10}\right) a\,\hat{\mathbf{x}} + \left(\frac{1}{2} - y_{10}\right) b\,\hat{\mathbf{y}} + z_{10} c\,\hat{\mathbf{z}}$ | (4a) | Si I |
| $\mathbf{B}_{40}$ | = | $\left(\frac{1}{2} - x_{10}\right)\mathbf{a}_1 + \left(\frac{1}{2} + y_{10}\right)\mathbf{a}_2 + \left(\frac{1}{2} + z_{10}\right)\mathbf{a}_3$ | = | $\left(\frac{1}{2} - x_{10}\right) a\,\hat{\mathbf{x}} + \left(\frac{1}{2} + y_{10}\right) b\,\hat{\mathbf{y}} + \left(\frac{1}{2} + z_{10}\right) c\,\hat{\mathbf{z}}$ | (4a) | Si I |
| $\mathbf{B}_{41}$ | = | $x_{11}\,\mathbf{a}_1 + y_{11}\,\mathbf{a}_2 + z_{11}\,\mathbf{a}_3$ | = | $x_{11} a\,\hat{\mathbf{x}} + y_{11} b\,\hat{\mathbf{y}} + z_{11} c\,\hat{\mathbf{z}}$ | (4a) | Si II |
| $\mathbf{B}_{42}$ | = | $-x_{11}\,\mathbf{a}_1 - y_{11}\,\mathbf{a}_2 + \left(\frac{1}{2} + z_{11}\right)\mathbf{a}_3$ | = | $-x_{11} a\,\hat{\mathbf{x}} - y_{11} b\,\hat{\mathbf{y}} + \left(\frac{1}{2} + z_{11}\right) c\,\hat{\mathbf{z}}$ | (4a) | Si II |
| $\mathbf{B}_{43}$ | = | $\left(\frac{1}{2} + x_{11}\right)\mathbf{a}_1 + \left(\frac{1}{2} - y_{11}\right)\mathbf{a}_2 + z_{11}\,\mathbf{a}_3$ | = | $\left(\frac{1}{2} + x_{11}\right) a\,\hat{\mathbf{x}} + \left(\frac{1}{2} - y_{11}\right) b\,\hat{\mathbf{y}} + z_{11} c\,\hat{\mathbf{z}}$ | (4a) | Si II |
| $\mathbf{B}_{44}$ | = | $\left(\frac{1}{2} - x_{11}\right)\mathbf{a}_1 + \left(\frac{1}{2} + y_{11}\right)\mathbf{a}_2 + \left(\frac{1}{2} + z_{11}\right)\mathbf{a}_3$ | = | $\left(\frac{1}{2} - x_{11}\right) a\,\hat{\mathbf{x}} + \left(\frac{1}{2} + y_{11}\right) b\,\hat{\mathbf{y}} + \left(\frac{1}{2} + z_{11}\right) c\,\hat{\mathbf{z}}$ | (4a) | Si II |

**Found in:**

**Geometry files:**

- CIF: pp. 1590
- POSCAR: pp. 1590

# $CaB_2O_4$ (III) Structure: A2BC4_oP84_33_6a_3a_12a

http://aflow.org/prototype-encyclopedia/A2BC4_oP84_33_6a_3a_12a

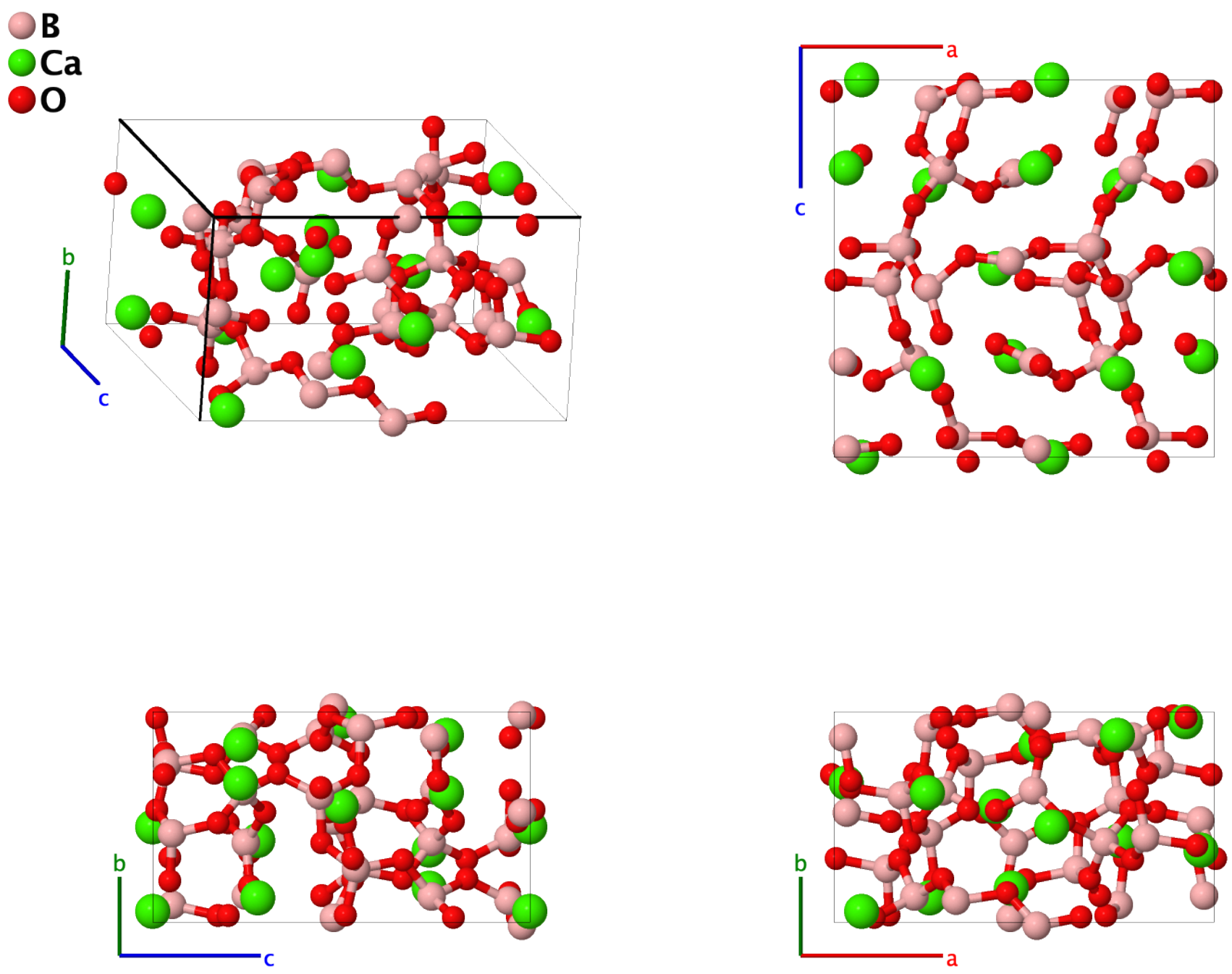

| | | |
|---|---|---|
| **Prototype** | : | $B_2CaO_4$ |
| **AFLOW prototype label** | : | A2BC4_oP84_33_6a_3a_12a |
| ***Strukturbericht* designation** | : | None |
| **Pearson symbol** | : | oP84 |
| **Space group number** | : | 33 |
| **Space group symbol** | : | $Pna2_1$ |
| **AFLOW prototype command** | : | `aflow --proto=A2BC4_oP84_33_6a_3a_12a` `--params=`$a, b/a, c/a, x_1, y_1, z_1, x_2, y_2, z_2, x_3, y_3, z_3, x_4, y_4, z_4, x_5, y_5, z_5, x_6, y_6, z_6, x_7, y_7, z_7, x_8, y_8, z_8, x_9, y_9, z_9, x_{10}, y_{10}, z_{10}, x_{11}, y_{11}, z_{11}, x_{12}, y_{12}, z_{12}, x_{13}, y_{13}, z_{13}, x_{14}, y_{14}, z_{14}, x_{15}, y_{15}, z_{15}, x_{16}, y_{16}, z_{16}, x_{17}, y_{17}, z_{17}, x_{18}, y_{18}, z_{18}, x_{19}, y_{19}, z_{19}, x_{20}, y_{20}, z_{20}, x_{21}, y_{21}, z_{21}$ |

- $CaB_2O_4$ exists in at least four phases (Marezio, 1969):
- I - The ground state, stable below 1.2 GPa, *Strukturbericht* $E3_2$.
- II – Orthorhombic high pressure structure, stable between 1.2 and 1.5 GPa, presumably calciborite.
- III – Orthorhombic high pressure structure, stable between 1.5 and 2.5 GPa (this structure).
- IV – Cubic high pressure structure, stable between 2.5 and 4.0 GPa.

## Simple Orthorhombic primitive vectors:

$$\mathbf{a}_1 = a\,\hat{\mathbf{x}}$$
$$\mathbf{a}_2 = b\,\hat{\mathbf{y}}$$
$$\mathbf{a}_3 = c\,\hat{\mathbf{z}}$$

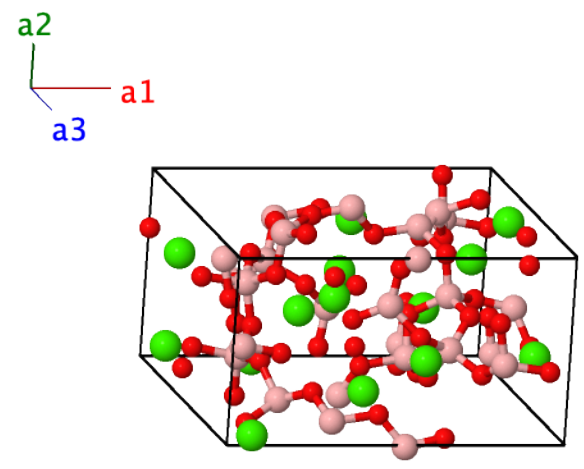

## Basis vectors:

| | | Lattice Coordinates | | Cartesian Coordinates | Wyckoff Position | Atom Type |
|---|---|---|---|---|---|---|
| $\mathbf{B}_1$ | = | $x_1\,\mathbf{a}_1 + y_1\,\mathbf{a}_2 + z_1\,\mathbf{a}_3$ | = | $x_1 a\,\hat{\mathbf{x}} + y_1 b\,\hat{\mathbf{y}} + z_1 c\,\hat{\mathbf{z}}$ | (4$a$) | B I |
| $\mathbf{B}_2$ | = | $-x_1\,\mathbf{a}_1 - y_1\,\mathbf{a}_2 + \left(\frac{1}{2} + z_1\right)\mathbf{a}_3$ | = | $-x_1 a\,\hat{\mathbf{x}} - y_1 b\,\hat{\mathbf{y}} + \left(\frac{1}{2} + z_1\right) c\,\hat{\mathbf{z}}$ | (4$a$) | B I |
| $\mathbf{B}_3$ | = | $\left(\frac{1}{2} + x_1\right)\mathbf{a}_1 + \left(\frac{1}{2} - y_1\right)\mathbf{a}_2 + z_1\,\mathbf{a}_3$ | = | $\left(\frac{1}{2} + x_1\right) a\,\hat{\mathbf{x}} + \left(\frac{1}{2} - y_1\right) b\,\hat{\mathbf{y}} + z_1 c\,\hat{\mathbf{z}}$ | (4$a$) | B I |
| $\mathbf{B}_4$ | = | $\left(\frac{1}{2} - x_1\right)\mathbf{a}_1 + \left(\frac{1}{2} + y_1\right)\mathbf{a}_2 + \left(\frac{1}{2} + z_1\right)\mathbf{a}_3$ | = | $\left(\frac{1}{2} - x_1\right) a\,\hat{\mathbf{x}} + \left(\frac{1}{2} + y_1\right) b\,\hat{\mathbf{y}} + \left(\frac{1}{2} + z_1\right) c\,\hat{\mathbf{z}}$ | (4$a$) | B I |
| $\mathbf{B}_5$ | = | $x_2\,\mathbf{a}_1 + y_2\,\mathbf{a}_2 + z_2\,\mathbf{a}_3$ | = | $x_2 a\,\hat{\mathbf{x}} + y_2 b\,\hat{\mathbf{y}} + z_2 c\,\hat{\mathbf{z}}$ | (4$a$) | B II |
| $\mathbf{B}_6$ | = | $-x_2\,\mathbf{a}_1 - y_2\,\mathbf{a}_2 + \left(\frac{1}{2} + z_2\right)\mathbf{a}_3$ | = | $-x_2 a\,\hat{\mathbf{x}} - y_2 b\,\hat{\mathbf{y}} + \left(\frac{1}{2} + z_2\right) c\,\hat{\mathbf{z}}$ | (4$a$) | B II |
| $\mathbf{B}_7$ | = | $\left(\frac{1}{2} + x_2\right)\mathbf{a}_1 + \left(\frac{1}{2} - y_2\right)\mathbf{a}_2 + z_2\,\mathbf{a}_3$ | = | $\left(\frac{1}{2} + x_2\right) a\,\hat{\mathbf{x}} + \left(\frac{1}{2} - y_2\right) b\,\hat{\mathbf{y}} + z_2 c\,\hat{\mathbf{z}}$ | (4$a$) | B II |
| $\mathbf{B}_8$ | = | $\left(\frac{1}{2} - x_2\right)\mathbf{a}_1 + \left(\frac{1}{2} + y_2\right)\mathbf{a}_2 + \left(\frac{1}{2} + z_2\right)\mathbf{a}_3$ | = | $\left(\frac{1}{2} - x_2\right) a\,\hat{\mathbf{x}} + \left(\frac{1}{2} + y_2\right) b\,\hat{\mathbf{y}} + \left(\frac{1}{2} + z_2\right) c\,\hat{\mathbf{z}}$ | (4$a$) | B II |
| $\mathbf{B}_9$ | = | $x_3\,\mathbf{a}_1 + y_3\,\mathbf{a}_2 + z_3\,\mathbf{a}_3$ | = | $x_3 a\,\hat{\mathbf{x}} + y_3 b\,\hat{\mathbf{y}} + z_3 c\,\hat{\mathbf{z}}$ | (4$a$) | B III |
| $\mathbf{B}_{10}$ | = | $-x_3\,\mathbf{a}_1 - y_3\,\mathbf{a}_2 + \left(\frac{1}{2} + z_3\right)\mathbf{a}_3$ | = | $-x_3 a\,\hat{\mathbf{x}} - y_3 b\,\hat{\mathbf{y}} + \left(\frac{1}{2} + z_3\right) c\,\hat{\mathbf{z}}$ | (4$a$) | B III |
| $\mathbf{B}_{11}$ | = | $\left(\frac{1}{2} + x_3\right)\mathbf{a}_1 + \left(\frac{1}{2} - y_3\right)\mathbf{a}_2 + z_3\,\mathbf{a}_3$ | = | $\left(\frac{1}{2} + x_3\right) a\,\hat{\mathbf{x}} + \left(\frac{1}{2} - y_3\right) b\,\hat{\mathbf{y}} + z_3 c\,\hat{\mathbf{z}}$ | (4$a$) | B III |
| $\mathbf{B}_{12}$ | = | $\left(\frac{1}{2} - x_3\right)\mathbf{a}_1 + \left(\frac{1}{2} + y_3\right)\mathbf{a}_2 + \left(\frac{1}{2} + z_3\right)\mathbf{a}_3$ | = | $\left(\frac{1}{2} - x_3\right) a\,\hat{\mathbf{x}} + \left(\frac{1}{2} + y_3\right) b\,\hat{\mathbf{y}} + \left(\frac{1}{2} + z_3\right) c\,\hat{\mathbf{z}}$ | (4$a$) | B III |
| $\mathbf{B}_{13}$ | = | $x_4\,\mathbf{a}_1 + y_4\,\mathbf{a}_2 + z_4\,\mathbf{a}_3$ | = | $x_4 a\,\hat{\mathbf{x}} + y_4 b\,\hat{\mathbf{y}} + z_4 c\,\hat{\mathbf{z}}$ | (4$a$) | B IV |
| $\mathbf{B}_{14}$ | = | $-x_4\,\mathbf{a}_1 - y_4\,\mathbf{a}_2 + \left(\frac{1}{2} + z_4\right)\mathbf{a}_3$ | = | $-x_4 a\,\hat{\mathbf{x}} - y_4 b\,\hat{\mathbf{y}} + \left(\frac{1}{2} + z_4\right) c\,\hat{\mathbf{z}}$ | (4$a$) | B IV |
| $\mathbf{B}_{15}$ | = | $\left(\frac{1}{2} + x_4\right)\mathbf{a}_1 + \left(\frac{1}{2} - y_4\right)\mathbf{a}_2 + z_4\,\mathbf{a}_3$ | = | $\left(\frac{1}{2} + x_4\right) a\,\hat{\mathbf{x}} + \left(\frac{1}{2} - y_4\right) b\,\hat{\mathbf{y}} + z_4 c\,\hat{\mathbf{z}}$ | (4$a$) | B IV |
| $\mathbf{B}_{16}$ | = | $\left(\frac{1}{2} - x_4\right)\mathbf{a}_1 + \left(\frac{1}{2} + y_4\right)\mathbf{a}_2 + \left(\frac{1}{2} + z_4\right)\mathbf{a}_3$ | = | $\left(\frac{1}{2} - x_4\right) a\,\hat{\mathbf{x}} + \left(\frac{1}{2} + y_4\right) b\,\hat{\mathbf{y}} + \left(\frac{1}{2} + z_4\right) c\,\hat{\mathbf{z}}$ | (4$a$) | B IV |
| $\mathbf{B}_{17}$ | = | $x_5\,\mathbf{a}_1 + y_5\,\mathbf{a}_2 + z_5\,\mathbf{a}_3$ | = | $x_5 a\,\hat{\mathbf{x}} + y_5 b\,\hat{\mathbf{y}} + z_5 c\,\hat{\mathbf{z}}$ | (4$a$) | B V |
| $\mathbf{B}_{18}$ | = | $-x_5\,\mathbf{a}_1 - y_5\,\mathbf{a}_2 + \left(\frac{1}{2} + z_5\right)\mathbf{a}_3$ | = | $-x_5 a\,\hat{\mathbf{x}} - y_5 b\,\hat{\mathbf{y}} + \left(\frac{1}{2} + z_5\right) c\,\hat{\mathbf{z}}$ | (4$a$) | B V |
| $\mathbf{B}_{19}$ | = | $\left(\frac{1}{2} + x_5\right)\mathbf{a}_1 + \left(\frac{1}{2} - y_5\right)\mathbf{a}_2 + z_5\,\mathbf{a}_3$ | = | $\left(\frac{1}{2} + x_5\right) a\,\hat{\mathbf{x}} + \left(\frac{1}{2} - y_5\right) b\,\hat{\mathbf{y}} + z_5 c\,\hat{\mathbf{z}}$ | (4$a$) | B V |
| $\mathbf{B}_{20}$ | = | $\left(\frac{1}{2} - x_5\right)\mathbf{a}_1 + \left(\frac{1}{2} + y_5\right)\mathbf{a}_2 + \left(\frac{1}{2} + z_5\right)\mathbf{a}_3$ | = | $\left(\frac{1}{2} - x_5\right) a\,\hat{\mathbf{x}} + \left(\frac{1}{2} + y_5\right) b\,\hat{\mathbf{y}} + \left(\frac{1}{2} + z_5\right) c\,\hat{\mathbf{z}}$ | (4$a$) | B V |
| $\mathbf{B}_{21}$ | = | $x_6\,\mathbf{a}_1 + y_6\,\mathbf{a}_2 + z_6\,\mathbf{a}_3$ | = | $x_6 a\,\hat{\mathbf{x}} + y_6 b\,\hat{\mathbf{y}} + z_6 c\,\hat{\mathbf{z}}$ | (4$a$) | B VI |
| $\mathbf{B}_{22}$ | = | $-x_6\,\mathbf{a}_1 - y_6\,\mathbf{a}_2 + \left(\frac{1}{2} + z_6\right)\mathbf{a}_3$ | = | $-x_6 a\,\hat{\mathbf{x}} - y_6 b\,\hat{\mathbf{y}} + \left(\frac{1}{2} + z_6\right) c\,\hat{\mathbf{z}}$ | (4$a$) | B VI |
| $\mathbf{B}_{23}$ | = | $\left(\frac{1}{2} + x_6\right)\mathbf{a}_1 + \left(\frac{1}{2} - y_6\right)\mathbf{a}_2 + z_6\,\mathbf{a}_3$ | = | $\left(\frac{1}{2} + x_6\right) a\,\hat{\mathbf{x}} + \left(\frac{1}{2} - y_6\right) b\,\hat{\mathbf{y}} + z_6 c\,\hat{\mathbf{z}}$ | (4$a$) | B VI |

$$\mathbf{B}_{24} = \left(\frac{1}{2} - x_6\right)\mathbf{a}_1 + \left(\frac{1}{2} + y_6\right)\mathbf{a}_2 + \left(\frac{1}{2} + z_6\right)\mathbf{a}_3 = \left(\frac{1}{2} - x_6\right)a\,\hat{\mathbf{x}} + \left(\frac{1}{2} + y_6\right)b\,\hat{\mathbf{y}} + \left(\frac{1}{2} + z_6\right)c\,\hat{\mathbf{z}} \quad (4a) \quad \text{B VI}$$

$$\mathbf{B}_{25} = x_7\,\mathbf{a}_1 + y_7\,\mathbf{a}_2 + z_7\,\mathbf{a}_3 = x_7 a\,\hat{\mathbf{x}} + y_7 b\,\hat{\mathbf{y}} + z_7 c\,\hat{\mathbf{z}} \quad (4a) \quad \text{Ca I}$$

$$\mathbf{B}_{26} = -x_7\,\mathbf{a}_1 - y_7\,\mathbf{a}_2 + \left(\frac{1}{2} + z_7\right)\mathbf{a}_3 = -x_7 a\,\hat{\mathbf{x}} - y_7 b\,\hat{\mathbf{y}} + \left(\frac{1}{2} + z_7\right)c\,\hat{\mathbf{z}} \quad (4a) \quad \text{Ca I}$$

$$\mathbf{B}_{27} = \left(\frac{1}{2} + x_7\right)\mathbf{a}_1 + \left(\frac{1}{2} - y_7\right)\mathbf{a}_2 + z_7\,\mathbf{a}_3 = \left(\frac{1}{2} + x_7\right)a\,\hat{\mathbf{x}} + \left(\frac{1}{2} - y_7\right)b\,\hat{\mathbf{y}} + z_7 c\,\hat{\mathbf{z}} \quad (4a) \quad \text{Ca I}$$

$$\mathbf{B}_{28} = \left(\frac{1}{2} - x_7\right)\mathbf{a}_1 + \left(\frac{1}{2} + y_7\right)\mathbf{a}_2 + \left(\frac{1}{2} + z_7\right)\mathbf{a}_3 = \left(\frac{1}{2} - x_7\right)a\,\hat{\mathbf{x}} + \left(\frac{1}{2} + y_7\right)b\,\hat{\mathbf{y}} + \left(\frac{1}{2} + z_7\right)c\,\hat{\mathbf{z}} \quad (4a) \quad \text{Ca I}$$

$$\mathbf{B}_{29} = x_8\,\mathbf{a}_1 + y_8\,\mathbf{a}_2 + z_8\,\mathbf{a}_3 = x_8 a\,\hat{\mathbf{x}} + y_8 b\,\hat{\mathbf{y}} + z_8 c\,\hat{\mathbf{z}} \quad (4a) \quad \text{Ca II}$$

$$\mathbf{B}_{30} = -x_8\,\mathbf{a}_1 - y_8\,\mathbf{a}_2 + \left(\frac{1}{2} + z_8\right)\mathbf{a}_3 = -x_8 a\,\hat{\mathbf{x}} - y_8 b\,\hat{\mathbf{y}} + \left(\frac{1}{2} + z_8\right)c\,\hat{\mathbf{z}} \quad (4a) \quad \text{Ca II}$$

$$\mathbf{B}_{31} = \left(\frac{1}{2} + x_8\right)\mathbf{a}_1 + \left(\frac{1}{2} - y_8\right)\mathbf{a}_2 + z_8\,\mathbf{a}_3 = \left(\frac{1}{2} + x_8\right)a\,\hat{\mathbf{x}} + \left(\frac{1}{2} - y_8\right)b\,\hat{\mathbf{y}} + z_8 c\,\hat{\mathbf{z}} \quad (4a) \quad \text{Ca II}$$

$$\mathbf{B}_{32} = \left(\frac{1}{2} - x_8\right)\mathbf{a}_1 + \left(\frac{1}{2} + y_8\right)\mathbf{a}_2 + \left(\frac{1}{2} + z_8\right)\mathbf{a}_3 = \left(\frac{1}{2} - x_8\right)a\,\hat{\mathbf{x}} + \left(\frac{1}{2} + y_8\right)b\,\hat{\mathbf{y}} + \left(\frac{1}{2} + z_8\right)c\,\hat{\mathbf{z}} \quad (4a) \quad \text{Ca II}$$

$$\mathbf{B}_{33} = x_9\,\mathbf{a}_1 + y_9\,\mathbf{a}_2 + z_9\,\mathbf{a}_3 = x_9 a\,\hat{\mathbf{x}} + y_9 b\,\hat{\mathbf{y}} + z_9 c\,\hat{\mathbf{z}} \quad (4a) \quad \text{Ca III}$$

$$\mathbf{B}_{34} = -x_9\,\mathbf{a}_1 - y_9\,\mathbf{a}_2 + \left(\frac{1}{2} + z_9\right)\mathbf{a}_3 = -x_9 a\,\hat{\mathbf{x}} - y_9 b\,\hat{\mathbf{y}} + \left(\frac{1}{2} + z_9\right)c\,\hat{\mathbf{z}} \quad (4a) \quad \text{Ca III}$$

$$\mathbf{B}_{35} = \left(\frac{1}{2} + x_9\right)\mathbf{a}_1 + \left(\frac{1}{2} - y_9\right)\mathbf{a}_2 + z_9\,\mathbf{a}_3 = \left(\frac{1}{2} + x_9\right)a\,\hat{\mathbf{x}} + \left(\frac{1}{2} - y_9\right)b\,\hat{\mathbf{y}} + z_9 c\,\hat{\mathbf{z}} \quad (4a) \quad \text{Ca III}$$

$$\mathbf{B}_{36} = \left(\frac{1}{2} - x_9\right)\mathbf{a}_1 + \left(\frac{1}{2} + y_9\right)\mathbf{a}_2 + \left(\frac{1}{2} + z_9\right)\mathbf{a}_3 = \left(\frac{1}{2} - x_9\right)a\,\hat{\mathbf{x}} + \left(\frac{1}{2} + y_9\right)b\,\hat{\mathbf{y}} + \left(\frac{1}{2} + z_9\right)c\,\hat{\mathbf{z}} \quad (4a) \quad \text{Ca III}$$

$$\mathbf{B}_{37} = x_{10}\,\mathbf{a}_1 + y_{10}\,\mathbf{a}_2 + z_{10}\,\mathbf{a}_3 = x_{10} a\,\hat{\mathbf{x}} + y_{10} b\,\hat{\mathbf{y}} + z_{10} c\,\hat{\mathbf{z}} \quad (4a) \quad \text{O I}$$

$$\mathbf{B}_{38} = -x_{10}\,\mathbf{a}_1 - y_{10}\,\mathbf{a}_2 + \left(\frac{1}{2} + z_{10}\right)\mathbf{a}_3 = -x_{10} a\,\hat{\mathbf{x}} - y_{10} b\,\hat{\mathbf{y}} + \left(\frac{1}{2} + z_{10}\right)c\,\hat{\mathbf{z}} \quad (4a) \quad \text{O I}$$

$$\mathbf{B}_{39} = \left(\frac{1}{2} + x_{10}\right)\mathbf{a}_1 + \left(\frac{1}{2} - y_{10}\right)\mathbf{a}_2 + z_{10}\,\mathbf{a}_3 = \left(\frac{1}{2} + x_{10}\right)a\,\hat{\mathbf{x}} + \left(\frac{1}{2} - y_{10}\right)b\,\hat{\mathbf{y}} + z_{10} c\,\hat{\mathbf{z}} \quad (4a) \quad \text{O I}$$

$$\mathbf{B}_{40} = \left(\frac{1}{2} - x_{10}\right)\mathbf{a}_1 + \left(\frac{1}{2} + y_{10}\right)\mathbf{a}_2 + \left(\frac{1}{2} + z_{10}\right)\mathbf{a}_3 = \left(\frac{1}{2} - x_{10}\right)a\,\hat{\mathbf{x}} + \left(\frac{1}{2} + y_{10}\right)b\,\hat{\mathbf{y}} + \left(\frac{1}{2} + z_{10}\right)c\,\hat{\mathbf{z}} \quad (4a) \quad \text{O I}$$

$$\mathbf{B}_{41} = x_{11}\,\mathbf{a}_1 + y_{11}\,\mathbf{a}_2 + z_{11}\,\mathbf{a}_3 = x_{11} a\,\hat{\mathbf{x}} + y_{11} b\,\hat{\mathbf{y}} + z_{11} c\,\hat{\mathbf{z}} \quad (4a) \quad \text{O II}$$

$$\mathbf{B}_{42} = -x_{11}\,\mathbf{a}_1 - y_{11}\,\mathbf{a}_2 + \left(\frac{1}{2} + z_{11}\right)\mathbf{a}_3 = -x_{11} a\,\hat{\mathbf{x}} - y_{11} b\,\hat{\mathbf{y}} + \left(\frac{1}{2} + z_{11}\right)c\,\hat{\mathbf{z}} \quad (4a) \quad \text{O II}$$

$$\mathbf{B}_{43} = \left(\frac{1}{2} + x_{11}\right)\mathbf{a}_1 + \left(\frac{1}{2} - y_{11}\right)\mathbf{a}_2 + z_{11}\,\mathbf{a}_3 = \left(\frac{1}{2} + x_{11}\right)a\,\hat{\mathbf{x}} + \left(\frac{1}{2} - y_{11}\right)b\,\hat{\mathbf{y}} + z_{11} c\,\hat{\mathbf{z}} \quad (4a) \quad \text{O II}$$

$$\mathbf{B}_{44} = \left(\frac{1}{2} - x_{11}\right)\mathbf{a}_1 + \left(\frac{1}{2} + y_{11}\right)\mathbf{a}_2 + \left(\frac{1}{2} + z_{11}\right)\mathbf{a}_3 = \left(\frac{1}{2} - x_{11}\right)a\,\hat{\mathbf{x}} + \left(\frac{1}{2} + y_{11}\right)b\,\hat{\mathbf{y}} + \left(\frac{1}{2} + z_{11}\right)c\,\hat{\mathbf{z}} \quad (4a) \quad \text{O II}$$

$$\mathbf{B}_{45} = x_{12}\,\mathbf{a}_1 + y_{12}\,\mathbf{a}_2 + z_{12}\,\mathbf{a}_3 = x_{12} a\,\hat{\mathbf{x}} + y_{12} b\,\hat{\mathbf{y}} + z_{12} c\,\hat{\mathbf{z}} \quad (4a) \quad \text{O III}$$

$$\mathbf{B}_{46} = -x_{12}\,\mathbf{a}_1 - y_{12}\,\mathbf{a}_2 + \left(\frac{1}{2} + z_{12}\right)\mathbf{a}_3 = -x_{12} a\,\hat{\mathbf{x}} - y_{12} b\,\hat{\mathbf{y}} + \left(\frac{1}{2} + z_{12}\right)c\,\hat{\mathbf{z}} \quad (4a) \quad \text{O III}$$

$$\mathbf{B}_{47} = \left(\frac{1}{2} + x_{12}\right)\mathbf{a}_1 + \left(\frac{1}{2} - y_{12}\right)\mathbf{a}_2 + z_{12}\,\mathbf{a}_3 = \left(\frac{1}{2} + x_{12}\right)a\,\hat{\mathbf{x}} + \left(\frac{1}{2} - y_{12}\right)b\,\hat{\mathbf{y}} + z_{12} c\,\hat{\mathbf{z}} \quad (4a) \quad \text{O III}$$

$$\mathbf{B}_{48} = \left(\frac{1}{2} - x_{12}\right)\mathbf{a}_1 + \left(\frac{1}{2} + y_{12}\right)\mathbf{a}_2 + \left(\frac{1}{2} + z_{12}\right)\mathbf{a}_3 = \left(\frac{1}{2} - x_{12}\right)a\,\hat{\mathbf{x}} + \left(\frac{1}{2} + y_{12}\right)b\,\hat{\mathbf{y}} + \left(\frac{1}{2} + z_{12}\right)c\,\hat{\mathbf{z}} \quad (4a) \quad \text{O III}$$

$$\mathbf{B}_{49} = x_{13}\,\mathbf{a}_1 + y_{13}\,\mathbf{a}_2 + z_{13}\,\mathbf{a}_3 = x_{13} a\,\hat{\mathbf{x}} + y_{13} b\,\hat{\mathbf{y}} + z_{13} c\,\hat{\mathbf{z}} \quad (4a) \quad \text{O IV}$$

$$\mathbf{B}_{50} = -x_{13}\,\mathbf{a}_1 - y_{13}\,\mathbf{a}_2 + \left(\frac{1}{2} + z_{13}\right)\mathbf{a}_3 = -x_{13} a\,\hat{\mathbf{x}} - y_{13} b\,\hat{\mathbf{y}} + \left(\frac{1}{2} + z_{13}\right)c\,\hat{\mathbf{z}} \quad (4a) \quad \text{O IV}$$

$$\mathbf{B}_{51} = \left(\frac{1}{2} + x_{13}\right)\mathbf{a}_1 + \left(\frac{1}{2} - y_{13}\right)\mathbf{a}_2 + z_{13}\,\mathbf{a}_3 = \left(\frac{1}{2} + x_{13}\right)a\,\hat{\mathbf{x}} + \left(\frac{1}{2} - y_{13}\right)b\,\hat{\mathbf{y}} + z_{13} c\,\hat{\mathbf{z}} \quad (4a) \quad \text{O IV}$$

$$\mathbf{B}_{52} = \left(\frac{1}{2} - x_{13}\right)\mathbf{a}_1 + \left(\frac{1}{2} + y_{13}\right)\mathbf{a}_2 + \left(\frac{1}{2} + z_{13}\right)\mathbf{a}_3 = \left(\frac{1}{2} - x_{13}\right)a\,\hat{\mathbf{x}} + \left(\frac{1}{2} + y_{13}\right)b\,\hat{\mathbf{y}} + \left(\frac{1}{2} + z_{13}\right)c\,\hat{\mathbf{z}} \quad (4a) \quad \text{O IV}$$

$$\mathbf{B}_{53} = x_{14}\,\mathbf{a}_1 + y_{14}\,\mathbf{a}_2 + z_{14}\,\mathbf{a}_3 = x_{14} a\,\hat{\mathbf{x}} + y_{14} b\,\hat{\mathbf{y}} + z_{14} c\,\hat{\mathbf{z}} \quad (4a) \quad \text{O V}$$

$$\mathbf{B}_{54} = -x_{14}\,\mathbf{a}_1 - y_{14}\,\mathbf{a}_2 + \left(\frac{1}{2} + z_{14}\right)\mathbf{a}_3 = -x_{14}a\,\hat{\mathbf{x}} - y_{14}b\,\hat{\mathbf{y}} + \left(\frac{1}{2} + z_{14}\right)c\,\hat{\mathbf{z}} \quad (4a) \quad \text{O V}$$

$$\mathbf{B}_{55} = \left(\frac{1}{2} + x_{14}\right)\mathbf{a}_1 + \left(\frac{1}{2} - y_{14}\right)\mathbf{a}_2 + z_{14}\,\mathbf{a}_3 = \left(\frac{1}{2} + x_{14}\right)a\,\hat{\mathbf{x}} + \left(\frac{1}{2} - y_{14}\right)b\,\hat{\mathbf{y}} + z_{14}c\,\hat{\mathbf{z}} \quad (4a) \quad \text{O V}$$

$$\mathbf{B}_{56} = \left(\frac{1}{2} - x_{14}\right)\mathbf{a}_1 + \left(\frac{1}{2} + y_{14}\right)\mathbf{a}_2 + \left(\frac{1}{2} + z_{14}\right)\mathbf{a}_3 = \left(\frac{1}{2} - x_{14}\right)a\,\hat{\mathbf{x}} + \left(\frac{1}{2} + y_{14}\right)b\,\hat{\mathbf{y}} + \left(\frac{1}{2} + z_{14}\right)c\,\hat{\mathbf{z}} \quad (4a) \quad \text{O V}$$

$$\mathbf{B}_{57} = x_{15}\,\mathbf{a}_1 + y_{15}\,\mathbf{a}_2 + z_{15}\,\mathbf{a}_3 = x_{15}a\,\hat{\mathbf{x}} + y_{15}b\,\hat{\mathbf{y}} + z_{15}c\,\hat{\mathbf{z}} \quad (4a) \quad \text{O VI}$$

$$\mathbf{B}_{58} = -x_{15}\,\mathbf{a}_1 - y_{15}\,\mathbf{a}_2 + \left(\frac{1}{2} + z_{15}\right)\mathbf{a}_3 = -x_{15}a\,\hat{\mathbf{x}} - y_{15}b\,\hat{\mathbf{y}} + \left(\frac{1}{2} + z_{15}\right)c\,\hat{\mathbf{z}} \quad (4a) \quad \text{O VI}$$

$$\mathbf{B}_{59} = \left(\frac{1}{2} + x_{15}\right)\mathbf{a}_1 + \left(\frac{1}{2} - y_{15}\right)\mathbf{a}_2 + z_{15}\,\mathbf{a}_3 = \left(\frac{1}{2} + x_{15}\right)a\,\hat{\mathbf{x}} + \left(\frac{1}{2} - y_{15}\right)b\,\hat{\mathbf{y}} + z_{15}c\,\hat{\mathbf{z}} \quad (4a) \quad \text{O VI}$$

$$\mathbf{B}_{60} = \left(\frac{1}{2} - x_{15}\right)\mathbf{a}_1 + \left(\frac{1}{2} + y_{15}\right)\mathbf{a}_2 + \left(\frac{1}{2} + z_{15}\right)\mathbf{a}_3 = \left(\frac{1}{2} - x_{15}\right)a\,\hat{\mathbf{x}} + \left(\frac{1}{2} + y_{15}\right)b\,\hat{\mathbf{y}} + \left(\frac{1}{2} + z_{15}\right)c\,\hat{\mathbf{z}} \quad (4a) \quad \text{O VI}$$

$$\mathbf{B}_{61} = x_{16}\,\mathbf{a}_1 + y_{16}\,\mathbf{a}_2 + z_{16}\,\mathbf{a}_3 = x_{16}a\,\hat{\mathbf{x}} + y_{16}b\,\hat{\mathbf{y}} + z_{16}c\,\hat{\mathbf{z}} \quad (4a) \quad \text{O VII}$$

$$\mathbf{B}_{62} = -x_{16}\,\mathbf{a}_1 - y_{16}\,\mathbf{a}_2 + \left(\frac{1}{2} + z_{16}\right)\mathbf{a}_3 = -x_{16}a\,\hat{\mathbf{x}} - y_{16}b\,\hat{\mathbf{y}} + \left(\frac{1}{2} + z_{16}\right)c\,\hat{\mathbf{z}} \quad (4a) \quad \text{O VII}$$

$$\mathbf{B}_{63} = \left(\frac{1}{2} + x_{16}\right)\mathbf{a}_1 + \left(\frac{1}{2} - y_{16}\right)\mathbf{a}_2 + z_{16}\,\mathbf{a}_3 = \left(\frac{1}{2} + x_{16}\right)a\,\hat{\mathbf{x}} + \left(\frac{1}{2} - y_{16}\right)b\,\hat{\mathbf{y}} + z_{16}c\,\hat{\mathbf{z}} \quad (4a) \quad \text{O VII}$$

$$\mathbf{B}_{64} = \left(\frac{1}{2} - x_{16}\right)\mathbf{a}_1 + \left(\frac{1}{2} + y_{16}\right)\mathbf{a}_2 + \left(\frac{1}{2} + z_{16}\right)\mathbf{a}_3 = \left(\frac{1}{2} - x_{16}\right)a\,\hat{\mathbf{x}} + \left(\frac{1}{2} + y_{16}\right)b\,\hat{\mathbf{y}} + \left(\frac{1}{2} + z_{16}\right)c\,\hat{\mathbf{z}} \quad (4a) \quad \text{O VII}$$

$$\mathbf{B}_{65} = x_{17}\,\mathbf{a}_1 + y_{17}\,\mathbf{a}_2 + z_{17}\,\mathbf{a}_3 = x_{17}a\,\hat{\mathbf{x}} + y_{17}b\,\hat{\mathbf{y}} + z_{17}c\,\hat{\mathbf{z}} \quad (4a) \quad \text{O VIII}$$

$$\mathbf{B}_{66} = -x_{17}\,\mathbf{a}_1 - y_{17}\,\mathbf{a}_2 + \left(\frac{1}{2} + z_{17}\right)\mathbf{a}_3 = -x_{17}a\,\hat{\mathbf{x}} - y_{17}b\,\hat{\mathbf{y}} + \left(\frac{1}{2} + z_{17}\right)c\,\hat{\mathbf{z}} \quad (4a) \quad \text{O VIII}$$

$$\mathbf{B}_{67} = \left(\frac{1}{2} + x_{17}\right)\mathbf{a}_1 + \left(\frac{1}{2} - y_{17}\right)\mathbf{a}_2 + z_{17}\,\mathbf{a}_3 = \left(\frac{1}{2} + x_{17}\right)a\,\hat{\mathbf{x}} + \left(\frac{1}{2} - y_{17}\right)b\,\hat{\mathbf{y}} + z_{17}c\,\hat{\mathbf{z}} \quad (4a) \quad \text{O VIII}$$

$$\mathbf{B}_{68} = \left(\frac{1}{2} - x_{17}\right)\mathbf{a}_1 + \left(\frac{1}{2} + y_{17}\right)\mathbf{a}_2 + \left(\frac{1}{2} + z_{17}\right)\mathbf{a}_3 = \left(\frac{1}{2} - x_{17}\right)a\,\hat{\mathbf{x}} + \left(\frac{1}{2} + y_{17}\right)b\,\hat{\mathbf{y}} + \left(\frac{1}{2} + z_{17}\right)c\,\hat{\mathbf{z}} \quad (4a) \quad \text{O VIII}$$

$$\mathbf{B}_{69} = x_{18}\,\mathbf{a}_1 + y_{18}\,\mathbf{a}_2 + z_{18}\,\mathbf{a}_3 = x_{18}a\,\hat{\mathbf{x}} + y_{18}b\,\hat{\mathbf{y}} + z_{18}c\,\hat{\mathbf{z}} \quad (4a) \quad \text{O IX}$$

$$\mathbf{B}_{70} = -x_{18}\,\mathbf{a}_1 - y_{18}\,\mathbf{a}_2 + \left(\frac{1}{2} + z_{18}\right)\mathbf{a}_3 = -x_{18}a\,\hat{\mathbf{x}} - y_{18}b\,\hat{\mathbf{y}} + \left(\frac{1}{2} + z_{18}\right)c\,\hat{\mathbf{z}} \quad (4a) \quad \text{O IX}$$

$$\mathbf{B}_{71} = \left(\frac{1}{2} + x_{18}\right)\mathbf{a}_1 + \left(\frac{1}{2} - y_{18}\right)\mathbf{a}_2 + z_{18}\,\mathbf{a}_3 = \left(\frac{1}{2} + x_{18}\right)a\,\hat{\mathbf{x}} + \left(\frac{1}{2} - y_{18}\right)b\,\hat{\mathbf{y}} + z_{18}c\,\hat{\mathbf{z}} \quad (4a) \quad \text{O IX}$$

$$\mathbf{B}_{72} = \left(\frac{1}{2} - x_{18}\right)\mathbf{a}_1 + \left(\frac{1}{2} + y_{18}\right)\mathbf{a}_2 + \left(\frac{1}{2} + z_{18}\right)\mathbf{a}_3 = \left(\frac{1}{2} - x_{18}\right)a\,\hat{\mathbf{x}} + \left(\frac{1}{2} + y_{18}\right)b\,\hat{\mathbf{y}} + \left(\frac{1}{2} + z_{18}\right)c\,\hat{\mathbf{z}} \quad (4a) \quad \text{O IX}$$

$$\mathbf{B}_{73} = x_{19}\,\mathbf{a}_1 + y_{19}\,\mathbf{a}_2 + z_{19}\,\mathbf{a}_3 = x_{19}a\,\hat{\mathbf{x}} + y_{19}b\,\hat{\mathbf{y}} + z_{19}c\,\hat{\mathbf{z}} \quad (4a) \quad \text{O X}$$

$$\mathbf{B}_{74} = -x_{19}\,\mathbf{a}_1 - y_{19}\,\mathbf{a}_2 + \left(\frac{1}{2} + z_{19}\right)\mathbf{a}_3 = -x_{19}a\,\hat{\mathbf{x}} - y_{19}b\,\hat{\mathbf{y}} + \left(\frac{1}{2} + z_{19}\right)c\,\hat{\mathbf{z}} \quad (4a) \quad \text{O X}$$

$$\mathbf{B}_{75} = \left(\frac{1}{2} + x_{19}\right)\mathbf{a}_1 + \left(\frac{1}{2} - y_{19}\right)\mathbf{a}_2 + z_{19}\,\mathbf{a}_3 = \left(\frac{1}{2} + x_{19}\right)a\,\hat{\mathbf{x}} + \left(\frac{1}{2} - y_{19}\right)b\,\hat{\mathbf{y}} + z_{19}c\,\hat{\mathbf{z}} \quad (4a) \quad \text{O X}$$

$$\mathbf{B}_{76} = \left(\frac{1}{2} - x_{19}\right)\mathbf{a}_1 + \left(\frac{1}{2} + y_{19}\right)\mathbf{a}_2 + \left(\frac{1}{2} + z_{19}\right)\mathbf{a}_3 = \left(\frac{1}{2} - x_{19}\right)a\,\hat{\mathbf{x}} + \left(\frac{1}{2} + y_{19}\right)b\,\hat{\mathbf{y}} + \left(\frac{1}{2} + z_{19}\right)c\,\hat{\mathbf{z}} \quad (4a) \quad \text{O X}$$

$$\mathbf{B}_{77} = x_{20}\,\mathbf{a}_1 + y_{20}\,\mathbf{a}_2 + z_{20}\,\mathbf{a}_3 = x_{20}a\,\hat{\mathbf{x}} + y_{20}b\,\hat{\mathbf{y}} + z_{20}c\,\hat{\mathbf{z}} \quad (4a) \quad \text{O XI}$$

$$\mathbf{B}_{78} = -x_{20}\,\mathbf{a}_1 - y_{20}\,\mathbf{a}_2 + \left(\frac{1}{2} + z_{20}\right)\mathbf{a}_3 = -x_{20}a\,\hat{\mathbf{x}} - y_{20}b\,\hat{\mathbf{y}} + \left(\frac{1}{2} + z_{20}\right)c\,\hat{\mathbf{z}} \quad (4a) \quad \text{O XI}$$

$$\mathbf{B}_{79} = \left(\frac{1}{2} + x_{20}\right)\mathbf{a}_1 + \left(\frac{1}{2} - y_{20}\right)\mathbf{a}_2 + z_{20}\,\mathbf{a}_3 = \left(\frac{1}{2} + x_{20}\right)a\,\hat{\mathbf{x}} + \left(\frac{1}{2} - y_{20}\right)b\,\hat{\mathbf{y}} + z_{20}c\,\hat{\mathbf{z}} \quad (4a) \quad \text{O XI}$$

$$\mathbf{B}_{80} = \left(\frac{1}{2} - x_{20}\right)\mathbf{a}_1 + \left(\frac{1}{2} + y_{20}\right)\mathbf{a}_2 + \left(\frac{1}{2} + z_{20}\right)\mathbf{a}_3 = \left(\frac{1}{2} - x_{20}\right)a\,\hat{\mathbf{x}} + \left(\frac{1}{2} + y_{20}\right)b\,\hat{\mathbf{y}} + \left(\frac{1}{2} + z_{20}\right)c\,\hat{\mathbf{z}} \quad (4a) \quad \text{O XI}$$

$$\mathbf{B}_{81} = x_{21}\,\mathbf{a}_1 + y_{21}\,\mathbf{a}_2 + z_{21}\,\mathbf{a}_3 = x_{21}a\,\hat{\mathbf{x}} + y_{21}b\,\hat{\mathbf{y}} + z_{21}c\,\hat{\mathbf{z}} \quad (4a) \quad \text{O XII}$$

$$\mathbf{B}_{82} = -x_{21}\,\mathbf{a}_1 - y_{21}\,\mathbf{a}_2 + \left(\frac{1}{2} + z_{21}\right)\mathbf{a}_3 = -x_{21}a\,\hat{\mathbf{x}} - y_{21}b\,\hat{\mathbf{y}} + \left(\frac{1}{2} + z_{21}\right)c\,\hat{\mathbf{z}} \quad (4a) \quad \text{O XII}$$

$$\mathbf{B}_{83} = \left(\frac{1}{2} + x_{21}\right)\mathbf{a}_1 + \left(\frac{1}{2} - y_{21}\right)\mathbf{a}_2 + z_{21}\,\mathbf{a}_3 = \left(\frac{1}{2} + x_{21}\right)a\,\hat{\mathbf{x}} + \left(\frac{1}{2} - y_{21}\right)b\,\hat{\mathbf{y}} + z_{21}c\,\hat{\mathbf{z}} \quad (4a) \quad \text{O XII}$$

$$\mathbf{B}_{84} = \left(\frac{1}{2} - x_{21}\right)\mathbf{a}_1 + \left(\frac{1}{2} + y_{21}\right)\mathbf{a}_2 + \left(\frac{1}{2} + z_{21}\right)\mathbf{a}_3 = \left(\frac{1}{2} - x_{21}\right)a\,\hat{\mathbf{x}} + \left(\frac{1}{2} + y_{21}\right)b\,\hat{\mathbf{y}} + \left(\frac{1}{2} + z_{21}\right)c\,\hat{\mathbf{z}} \quad (4a) \quad \text{O XII}$$

---

---

**Geometry files:**

- CIF: pp. 1590
- POSCAR: pp. 1591

# Cervantite ($\alpha$-$Sb_2O_4$) Structure: A2B_oP24_33_4a_2a

http://aflow.org/prototype-encyclopedia/A2B_oP24_33_4a_2a

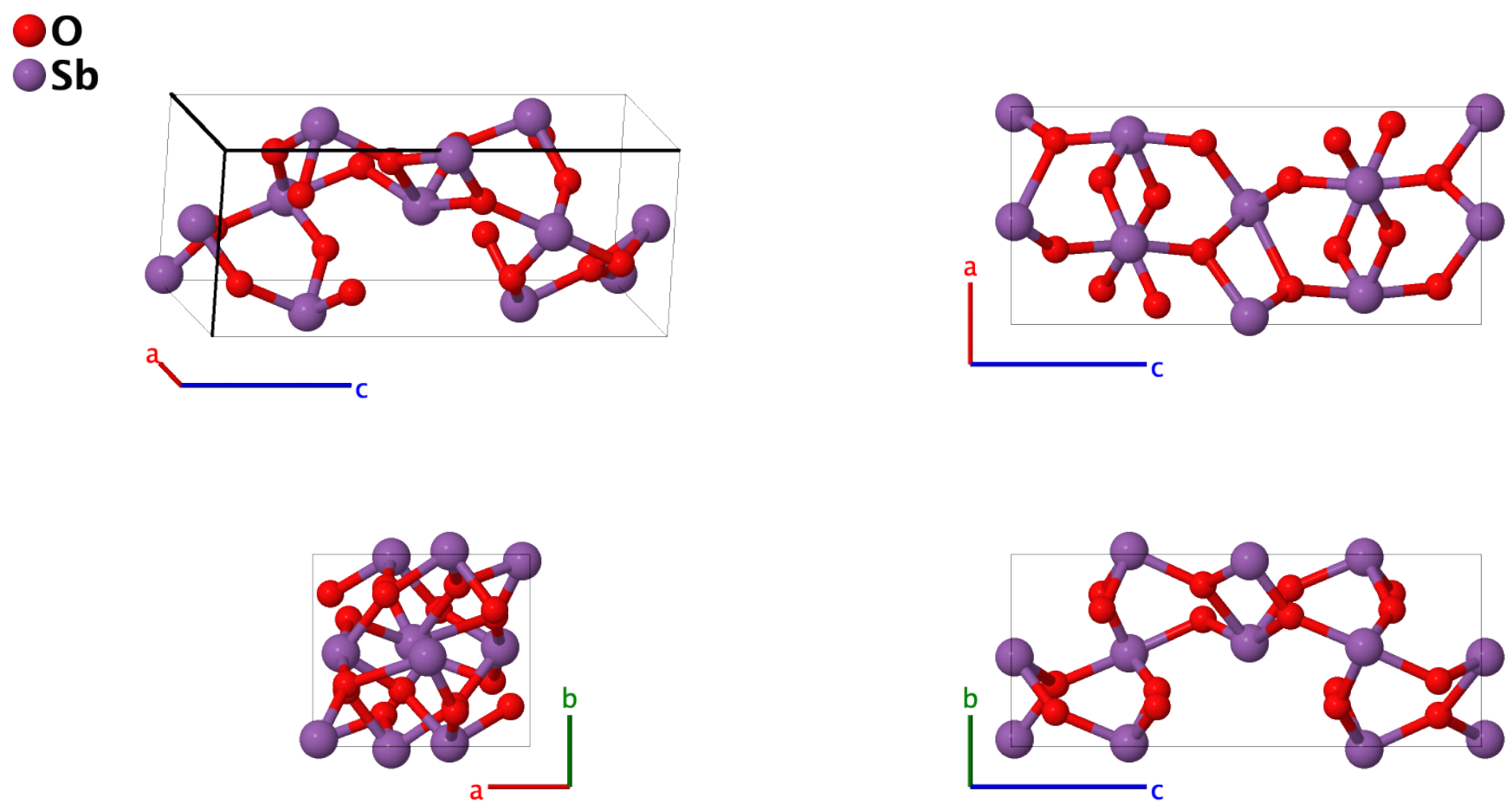

| | | |
|---|---|---|
| **Prototype** | : | $O_2Sb$ |
| **AFLOW prototype label** | : | A2B_oP24_33_4a_2a |
| ***Strukturbericht* designation** | : | None |
| **Pearson symbol** | : | oP24 |
| **Space group number** | : | 33 |
| **Space group symbol** | : | $Pna2_1$ |
| **AFLOW prototype command** | : | `aflow --proto=A2B_oP24_33_4a_2a` `--params=`$a, b/a, c/a, x_1, y_1, z_1, x_2, y_2, z_2, x_3, y_3, z_3, x_4, y_4, z_4, x_5, y_5, z_5, x_6, y_6, z_6$ |

### Other compounds with this structure

- $SbNbO_4$

- This is *not* the $D6_2$ structure of $SbO_2$, which was found to be incorrect (Thornton, 1977). There is also a naturally occurring monoclinic modification of cervantite, clinocervantite ($\beta$-$Sb_2O_4$).

### Simple Orthorhombic primitive vectors:

$$\mathbf{a}_1 = a\,\hat{\mathbf{x}}$$
$$\mathbf{a}_2 = b\,\hat{\mathbf{y}}$$
$$\mathbf{a}_3 = c\,\hat{\mathbf{z}}$$

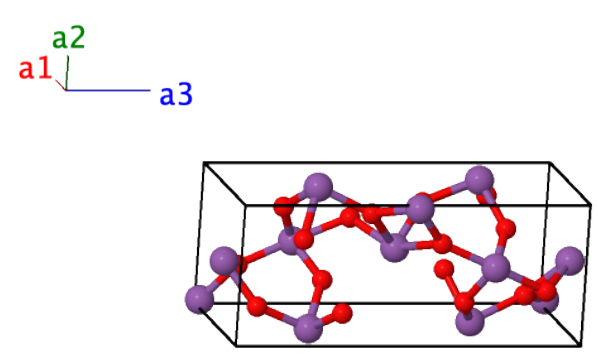

### Basis vectors:

| | | Lattice Coordinates | | Cartesian Coordinates | Wyckoff Position | Atom Type |
|---|---|---|---|---|---|---|
| $\mathbf{B}_1$ | = | $x_1\,\mathbf{a}_1 + y_1\,\mathbf{a}_2 + z_1\,\mathbf{a}_3$ | = | $x_1 a\,\hat{\mathbf{x}} + y_1 b\,\hat{\mathbf{y}} + z_1 c\,\hat{\mathbf{z}}$ | (4$a$) | O I |
| $\mathbf{B}_2$ | = | $-x_1\,\mathbf{a}_1 - y_1\,\mathbf{a}_2 + \left(\frac{1}{2} + z_1\right)\mathbf{a}_3$ | = | $-x_1 a\,\hat{\mathbf{x}} - y_1 b\,\hat{\mathbf{y}} + \left(\frac{1}{2} + z_1\right) c\,\hat{\mathbf{z}}$ | (4$a$) | O I |
| $\mathbf{B}_3$ | = | $\left(\frac{1}{2} + x_1\right)\mathbf{a}_1 + \left(\frac{1}{2} - y_1\right)\mathbf{a}_2 + z_1\,\mathbf{a}_3$ | = | $\left(\frac{1}{2} + x_1\right) a\,\hat{\mathbf{x}} + \left(\frac{1}{2} - y_1\right) b\,\hat{\mathbf{y}} + z_1 c\,\hat{\mathbf{z}}$ | (4$a$) | O I |

$$
\begin{array}{llllll}
\mathbf{B}_{4} & = & \left(\frac{1}{2} - x_{1}\right) \mathbf{a}_{1} + \left(\frac{1}{2} + y_{1}\right) \mathbf{a}_{2} + \left(\frac{1}{2} + z_{1}\right) \mathbf{a}_{3} & = & \left(\frac{1}{2} - x_{1}\right) a \, \hat{\mathbf{x}} + \left(\frac{1}{2} + y_{1}\right) b \, \hat{\mathbf{y}} + \left(\frac{1}{2} + z_{1}\right) c \, \hat{\mathbf{z}} & (4a) & \text{O I} \\
\mathbf{B}_{5} & = & x_{2} \, \mathbf{a}_{1} + y_{2} \, \mathbf{a}_{2} + z_{2} \, \mathbf{a}_{3} & = & x_{2} a \, \hat{\mathbf{x}} + y_{2} b \, \hat{\mathbf{y}} + z_{2} c \, \hat{\mathbf{z}} & (4a) & \text{O II} \\
\mathbf{B}_{6} & = & -x_{2} \, \mathbf{a}_{1} - y_{2} \, \mathbf{a}_{2} + \left(\frac{1}{2} + z_{2}\right) \mathbf{a}_{3} & = & -x_{2} a \, \hat{\mathbf{x}} - y_{2} b \, \hat{\mathbf{y}} + \left(\frac{1}{2} + z_{2}\right) c \, \hat{\mathbf{z}} & (4a) & \text{O II} \\
\mathbf{B}_{7} & = & \left(\frac{1}{2} + x_{2}\right) \mathbf{a}_{1} + \left(\frac{1}{2} - y_{2}\right) \mathbf{a}_{2} + z_{2} \, \mathbf{a}_{3} & = & \left(\frac{1}{2} + x_{2}\right) a \, \hat{\mathbf{x}} + \left(\frac{1}{2} - y_{2}\right) b \, \hat{\mathbf{y}} + z_{2} c \, \hat{\mathbf{z}} & (4a) & \text{O II} \\
\mathbf{B}_{8} & = & \left(\frac{1}{2} - x_{2}\right) \mathbf{a}_{1} + \left(\frac{1}{2} + y_{2}\right) \mathbf{a}_{2} + \left(\frac{1}{2} + z_{2}\right) \mathbf{a}_{3} & = & \left(\frac{1}{2} - x_{2}\right) a \, \hat{\mathbf{x}} + \left(\frac{1}{2} + y_{2}\right) b \, \hat{\mathbf{y}} + \left(\frac{1}{2} + z_{2}\right) c \, \hat{\mathbf{z}} & (4a) & \text{O II} \\
\mathbf{B}_{9} & = & x_{3} \, \mathbf{a}_{1} + y_{3} \, \mathbf{a}_{2} + z_{3} \, \mathbf{a}_{3} & = & x_{3} a \, \hat{\mathbf{x}} + y_{3} b \, \hat{\mathbf{y}} + z_{3} c \, \hat{\mathbf{z}} & (4a) & \text{O III} \\
\mathbf{B}_{10} & = & -x_{3} \, \mathbf{a}_{1} - y_{3} \, \mathbf{a}_{2} + \left(\frac{1}{2} + z_{3}\right) \mathbf{a}_{3} & = & -x_{3} a \, \hat{\mathbf{x}} - y_{3} b \, \hat{\mathbf{y}} + \left(\frac{1}{2} + z_{3}\right) c \, \hat{\mathbf{z}} & (4a) & \text{O III} \\
\mathbf{B}_{11} & = & \left(\frac{1}{2} + x_{3}\right) \mathbf{a}_{1} + \left(\frac{1}{2} - y_{3}\right) \mathbf{a}_{2} + z_{3} \, \mathbf{a}_{3} & = & \left(\frac{1}{2} + x_{3}\right) a \, \hat{\mathbf{x}} + \left(\frac{1}{2} - y_{3}\right) b \, \hat{\mathbf{y}} + z_{3} c \, \hat{\mathbf{z}} & (4a) & \text{O III} \\
\mathbf{B}_{12} & = & \left(\frac{1}{2} - x_{3}\right) \mathbf{a}_{1} + \left(\frac{1}{2} + y_{3}\right) \mathbf{a}_{2} + \left(\frac{1}{2} + z_{3}\right) \mathbf{a}_{3} & = & \left(\frac{1}{2} - x_{3}\right) a \, \hat{\mathbf{x}} + \left(\frac{1}{2} + y_{3}\right) b \, \hat{\mathbf{y}} + \left(\frac{1}{2} + z_{3}\right) c \, \hat{\mathbf{z}} & (4a) & \text{O III} \\
\mathbf{B}_{13} & = & x_{4} \, \mathbf{a}_{1} + y_{4} \, \mathbf{a}_{2} + z_{4} \, \mathbf{a}_{3} & = & x_{4} a \, \hat{\mathbf{x}} + y_{4} b \, \hat{\mathbf{y}} + z_{4} c \, \hat{\mathbf{z}} & (4a) & \text{O IV} \\
\mathbf{B}_{14} & = & -x_{4} \, \mathbf{a}_{1} - y_{4} \, \mathbf{a}_{2} + \left(\frac{1}{2} + z_{4}\right) \mathbf{a}_{3} & = & -x_{4} a \, \hat{\mathbf{x}} - y_{4} b \, \hat{\mathbf{y}} + \left(\frac{1}{2} + z_{4}\right) c \, \hat{\mathbf{z}} & (4a) & \text{O IV} \\
\mathbf{B}_{15} & = & \left(\frac{1}{2} + x_{4}\right) \mathbf{a}_{1} + \left(\frac{1}{2} - y_{4}\right) \mathbf{a}_{2} + z_{4} \, \mathbf{a}_{3} & = & \left(\frac{1}{2} + x_{4}\right) a \, \hat{\mathbf{x}} + \left(\frac{1}{2} - y_{4}\right) b \, \hat{\mathbf{y}} + z_{4} c \, \hat{\mathbf{z}} & (4a) & \text{O IV} \\
\mathbf{B}_{16} & = & \left(\frac{1}{2} - x_{4}\right) \mathbf{a}_{1} + \left(\frac{1}{2} + y_{4}\right) \mathbf{a}_{2} + \left(\frac{1}{2} + z_{4}\right) \mathbf{a}_{3} & = & \left(\frac{1}{2} - x_{4}\right) a \, \hat{\mathbf{x}} + \left(\frac{1}{2} + y_{4}\right) b \, \hat{\mathbf{y}} + \left(\frac{1}{2} + z_{4}\right) c \, \hat{\mathbf{z}} & (4a) & \text{O IV} \\
\mathbf{B}_{17} & = & x_{5} \, \mathbf{a}_{1} + y_{5} \, \mathbf{a}_{2} + z_{5} \, \mathbf{a}_{3} & = & x_{5} a \, \hat{\mathbf{x}} + y_{5} b \, \hat{\mathbf{y}} + z_{5} c \, \hat{\mathbf{z}} & (4a) & \text{Sb I} \\
\mathbf{B}_{18} & = & -x_{5} \, \mathbf{a}_{1} - y_{5} \, \mathbf{a}_{2} + \left(\frac{1}{2} + z_{5}\right) \mathbf{a}_{3} & = & -x_{5} a \, \hat{\mathbf{x}} - y_{5} b \, \hat{\mathbf{y}} + \left(\frac{1}{2} + z_{5}\right) c \, \hat{\mathbf{z}} & (4a) & \text{Sb I} \\
\mathbf{B}_{19} & = & \left(\frac{1}{2} + x_{5}\right) \mathbf{a}_{1} + \left(\frac{1}{2} - y_{5}\right) \mathbf{a}_{2} + z_{5} \, \mathbf{a}_{3} & = & \left(\frac{1}{2} + x_{5}\right) a \, \hat{\mathbf{x}} + \left(\frac{1}{2} - y_{5}\right) b \, \hat{\mathbf{y}} + z_{5} c \, \hat{\mathbf{z}} & (4a) & \text{Sb I} \\
\mathbf{B}_{20} & = & \left(\frac{1}{2} - x_{5}\right) \mathbf{a}_{1} + \left(\frac{1}{2} + y_{5}\right) \mathbf{a}_{2} + \left(\frac{1}{2} + z_{5}\right) \mathbf{a}_{3} & = & \left(\frac{1}{2} - x_{5}\right) a \, \hat{\mathbf{x}} + \left(\frac{1}{2} + y_{5}\right) b \, \hat{\mathbf{y}} + \left(\frac{1}{2} + z_{5}\right) c \, \hat{\mathbf{z}} & (4a) & \text{Sb I} \\
\mathbf{B}_{21} & = & x_{6} \, \mathbf{a}_{1} + y_{6} \, \mathbf{a}_{2} + z_{6} \, \mathbf{a}_{3} & = & x_{6} a \, \hat{\mathbf{x}} + y_{6} b \, \hat{\mathbf{y}} + z_{6} c \, \hat{\mathbf{z}} & (4a) & \text{Sb II} \\
\mathbf{B}_{22} & = & -x_{6} \, \mathbf{a}_{1} - y_{6} \, \mathbf{a}_{2} + \left(\frac{1}{2} + z_{6}\right) \mathbf{a}_{3} & = & -x_{6} a \, \hat{\mathbf{x}} - y_{6} b \, \hat{\mathbf{y}} + \left(\frac{1}{2} + z_{6}\right) c \, \hat{\mathbf{z}} & (4a) & \text{Sb II} \\
\mathbf{B}_{23} & = & \left(\frac{1}{2} + x_{6}\right) \mathbf{a}_{1} + \left(\frac{1}{2} - y_{6}\right) \mathbf{a}_{2} + z_{6} \, \mathbf{a}_{3} & = & \left(\frac{1}{2} + x_{6}\right) a \, \hat{\mathbf{x}} + \left(\frac{1}{2} - y_{6}\right) b \, \hat{\mathbf{y}} + z_{6} c \, \hat{\mathbf{z}} & (4a) & \text{Sb II} \\
\mathbf{B}_{24} & = & \left(\frac{1}{2} - x_{6}\right) \mathbf{a}_{1} + \left(\frac{1}{2} + y_{6}\right) \mathbf{a}_{2} + \left(\frac{1}{2} + z_{6}\right) \mathbf{a}_{3} & = & \left(\frac{1}{2} - x_{6}\right) a \, \hat{\mathbf{x}} + \left(\frac{1}{2} + y_{6}\right) b \, \hat{\mathbf{y}} + \left(\frac{1}{2} + z_{6}\right) c \, \hat{\mathbf{z}} & (4a) & \text{Sb II}
\end{array}
$$

**Geometry files:**

- CIF: pp. 1591
- POSCAR: pp. 1592

# $CsB_4O_6F$ Structure: A4BCD6_oP48_33_4a_a_a_6a

http://aflow.org/prototype-encyclopedia/A4BCD6_oP48_33_4a_a_a_6a

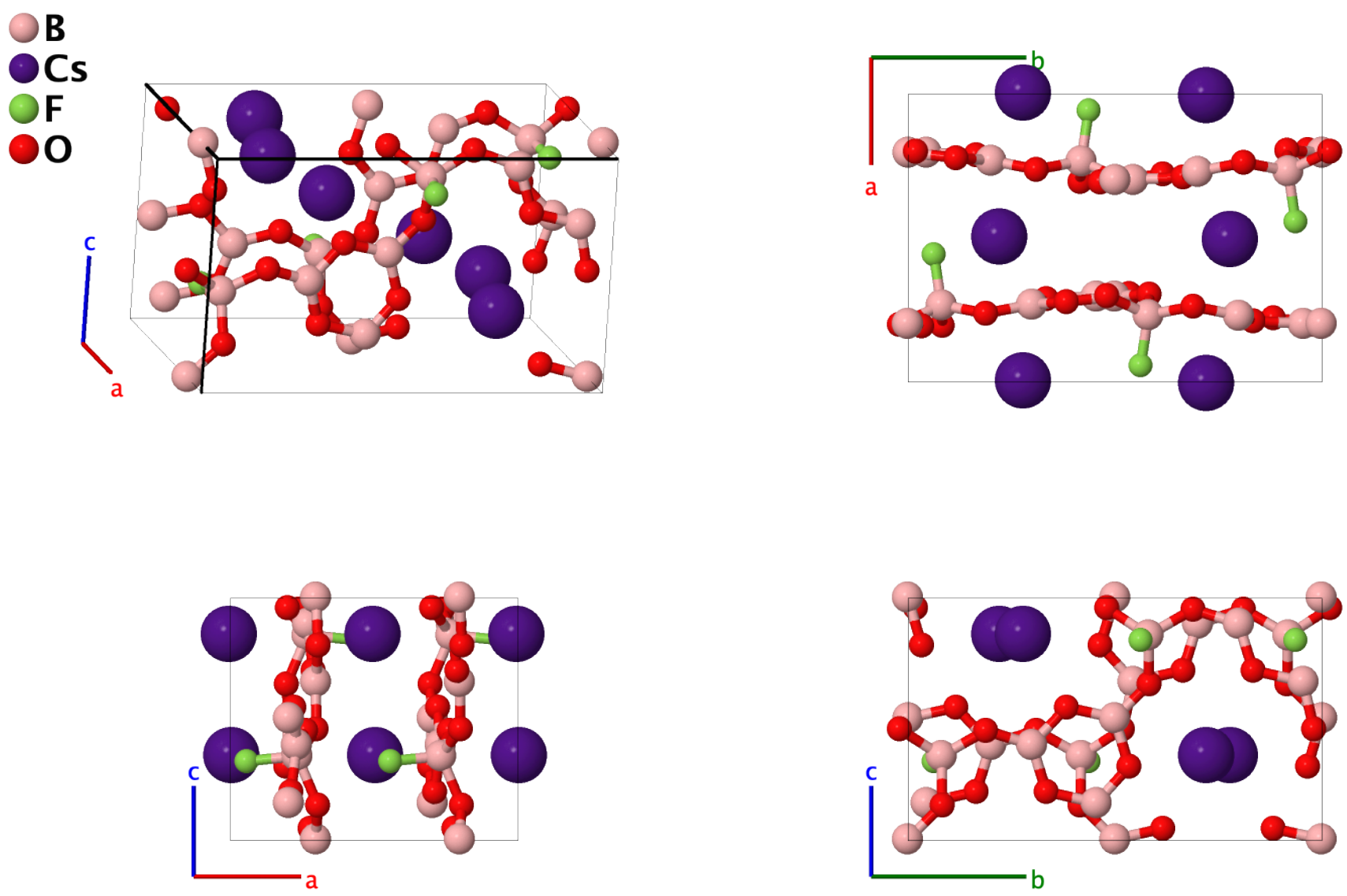

| | | |
|---|---|---|
| **Prototype** | : | $B_4CsFO_6$ |
| **AFLOW prototype label** | : | A4BCD6_oP48_33_4a_a_a_6a |
| ***Strukturbericht* designation** | : | None |
| **Pearson symbol** | : | oP48 |
| **Space group number** | : | 33 |
| **Space group symbol** | : | $Pna2_1$ |
| **AFLOW prototype command** | : | `aflow --proto=A4BCD6_oP48_33_4a_a_a_6a --params=`$a, b/a, c/a, x_1, y_1, z_1, x_2, y_2, z_2, x_3, y_3, z_3, x_4, y_4, z_4, x_5, y_5, z_5, x_6, y_6, z_6, x_7, y_7, z_7, x_8, y_8, z_8, x_9, y_9, z_9, x_{10}, y_{10}, z_{10}, x_{11}, y_{11}, z_{11}, x_{12}, y_{12}, z_{12}$ |

**Simple Orthorhombic primitive vectors:**

$$\mathbf{a}_1 = a\,\hat{\mathbf{x}}$$
$$\mathbf{a}_2 = b\,\hat{\mathbf{y}}$$
$$\mathbf{a}_3 = c\,\hat{\mathbf{z}}$$

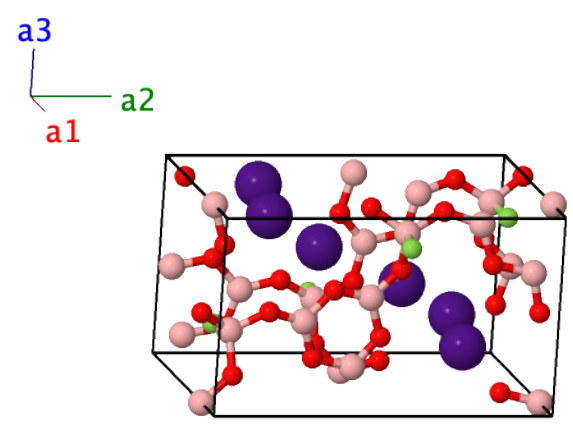

**Basis vectors:**

| | | Lattice Coordinates | | Cartesian Coordinates | Wyckoff Position | Atom Type |
|---|---|---|---|---|---|---|
| $\mathbf{B}_1$ | = | $x_1\,\mathbf{a}_1 + y_1\,\mathbf{a}_2 + z_1\,\mathbf{a}_3$ | = | $x_1 a\,\hat{\mathbf{x}} + y_1 b\,\hat{\mathbf{y}} + z_1 c\,\hat{\mathbf{z}}$ | $(4a)$ | B I |
| $\mathbf{B}_2$ | = | $-x_1\,\mathbf{a}_1 - y_1\,\mathbf{a}_2 + \left(\frac{1}{2} + z_1\right)\mathbf{a}_3$ | = | $-x_1 a\,\hat{\mathbf{x}} - y_1 b\,\hat{\mathbf{y}} + \left(\frac{1}{2} + z_1\right) c\,\hat{\mathbf{z}}$ | $(4a)$ | B I |
| $\mathbf{B}_3$ | = | $\left(\frac{1}{2} + x_1\right)\mathbf{a}_1 + \left(\frac{1}{2} - y_1\right)\mathbf{a}_2 + z_1\,\mathbf{a}_3$ | = | $\left(\frac{1}{2} + x_1\right) a\,\hat{\mathbf{x}} + \left(\frac{1}{2} - y_1\right) b\,\hat{\mathbf{y}} + z_1 c\,\hat{\mathbf{z}}$ | $(4a)$ | B I |
| $\mathbf{B}_4$ | = | $\left(\frac{1}{2} - x_1\right)\mathbf{a}_1 + \left(\frac{1}{2} + y_1\right)\mathbf{a}_2 + \left(\frac{1}{2} + z_1\right)\mathbf{a}_3$ | = | $\left(\frac{1}{2} - x_1\right) a\,\hat{\mathbf{x}} + \left(\frac{1}{2} + y_1\right) b\,\hat{\mathbf{y}} + \left(\frac{1}{2} + z_1\right) c\,\hat{\mathbf{z}}$ | $(4a)$ | B I |
| $\mathbf{B}_{5}$ | = | $x_{2}\,\mathbf{a}_{1}+y_{2}\,\mathbf{a}_{2}+z_{2}\,\mathbf{a}_{3}$ | = | $x_{2}a\,\hat{\mathbf{x}}+y_{2}b\,\hat{\mathbf{y}}+z_{2}c\,\hat{\mathbf{z}}$ | (4a) | B II |
| $\mathbf{B}_{6}$ | = | $-x_{2}\,\mathbf{a}_{1}-y_{2}\,\mathbf{a}_{2}+\left(\frac{1}{2}+z_{2}\right)\,\mathbf{a}_{3}$ | = | $-x_{2}a\,\hat{\mathbf{x}}-y_{2}b\,\hat{\mathbf{y}}+\left(\frac{1}{2}+z_{2}\right)c\,\hat{\mathbf{z}}$ | (4a) | B II |
| $\mathbf{B}_{7}$ | = | $\left(\frac{1}{2}+x_{2}\right)\,\mathbf{a}_{1}+\left(\frac{1}{2}-y_{2}\right)\,\mathbf{a}_{2}+z_{2}\,\mathbf{a}_{3}$ | = | $\left(\frac{1}{2}+x_{2}\right)a\,\hat{\mathbf{x}}+\left(\frac{1}{2}-y_{2}\right)b\,\hat{\mathbf{y}}+z_{2}c\,\hat{\mathbf{z}}$ | (4a) | B II |
| $\mathbf{B}_{8}$ | = | $\left(\frac{1}{2}-x_{2}\right)\,\mathbf{a}_{1}+\left(\frac{1}{2}+y_{2}\right)\,\mathbf{a}_{2}+\left(\frac{1}{2}+z_{2}\right)\,\mathbf{a}_{3}$ | = | $\left(\frac{1}{2}-x_{2}\right)a\,\hat{\mathbf{x}}+\left(\frac{1}{2}+y_{2}\right)b\,\hat{\mathbf{y}}+\left(\frac{1}{2}+z_{2}\right)c\,\hat{\mathbf{z}}$ | (4a) | B II |
| $\mathbf{B}_{9}$ | = | $x_{3}\,\mathbf{a}_{1}+y_{3}\,\mathbf{a}_{2}+z_{3}\,\mathbf{a}_{3}$ | = | $x_{3}a\,\hat{\mathbf{x}}+y_{3}b\,\hat{\mathbf{y}}+z_{3}c\,\hat{\mathbf{z}}$ | (4a) | B III |
| $\mathbf{B}_{10}$ | = | $-x_{3}\,\mathbf{a}_{1}-y_{3}\,\mathbf{a}_{2}+\left(\frac{1}{2}+z_{3}\right)\,\mathbf{a}_{3}$ | = | $-x_{3}a\,\hat{\mathbf{x}}-y_{3}b\,\hat{\mathbf{y}}+\left(\frac{1}{2}+z_{3}\right)c\,\hat{\mathbf{z}}$ | (4a) | B III |
| $\mathbf{B}_{11}$ | = | $\left(\frac{1}{2}+x_{3}\right)\,\mathbf{a}_{1}+\left(\frac{1}{2}-y_{3}\right)\,\mathbf{a}_{2}+z_{3}\,\mathbf{a}_{3}$ | = | $\left(\frac{1}{2}+x_{3}\right)a\,\hat{\mathbf{x}}+\left(\frac{1}{2}-y_{3}\right)b\,\hat{\mathbf{y}}+z_{3}c\,\hat{\mathbf{z}}$ | (4a) | B III |
| $\mathbf{B}_{12}$ | = | $\left(\frac{1}{2}-x_{3}\right)\,\mathbf{a}_{1}+\left(\frac{1}{2}+y_{3}\right)\,\mathbf{a}_{2}+\left(\frac{1}{2}+z_{3}\right)\,\mathbf{a}_{3}$ | = | $\left(\frac{1}{2}-x_{3}\right)a\,\hat{\mathbf{x}}+\left(\frac{1}{2}+y_{3}\right)b\,\hat{\mathbf{y}}+\left(\frac{1}{2}+z_{3}\right)c\,\hat{\mathbf{z}}$ | (4a) | B III |
| $\mathbf{B}_{13}$ | = | $x_{4}\,\mathbf{a}_{1}+y_{4}\,\mathbf{a}_{2}+z_{4}\,\mathbf{a}_{3}$ | = | $x_{4}a\,\hat{\mathbf{x}}+y_{4}b\,\hat{\mathbf{y}}+z_{4}c\,\hat{\mathbf{z}}$ | (4a) | B IV |
| $\mathbf{B}_{14}$ | = | $-x_{4}\,\mathbf{a}_{1}-y_{4}\,\mathbf{a}_{2}+\left(\frac{1}{2}+z_{4}\right)\,\mathbf{a}_{3}$ | = | $-x_{4}a\,\hat{\mathbf{x}}-y_{4}b\,\hat{\mathbf{y}}+\left(\frac{1}{2}+z_{4}\right)c\,\hat{\mathbf{z}}$ | (4a) | B IV |
| $\mathbf{B}_{15}$ | = | $\left(\frac{1}{2}+x_{4}\right)\,\mathbf{a}_{1}+\left(\frac{1}{2}-y_{4}\right)\,\mathbf{a}_{2}+z_{4}\,\mathbf{a}_{3}$ | = | $\left(\frac{1}{2}+x_{4}\right)a\,\hat{\mathbf{x}}+\left(\frac{1}{2}-y_{4}\right)b\,\hat{\mathbf{y}}+z_{4}c\,\hat{\mathbf{z}}$ | (4a) | B IV |
| $\mathbf{B}_{16}$ | = | $\left(\frac{1}{2}-x_{4}\right)\,\mathbf{a}_{1}+\left(\frac{1}{2}+y_{4}\right)\,\mathbf{a}_{2}+\left(\frac{1}{2}+z_{4}\right)\,\mathbf{a}_{3}$ | = | $\left(\frac{1}{2}-x_{4}\right)a\,\hat{\mathbf{x}}+\left(\frac{1}{2}+y_{4}\right)b\,\hat{\mathbf{y}}+\left(\frac{1}{2}+z_{4}\right)c\,\hat{\mathbf{z}}$ | (4a) | B IV |
| $\mathbf{B}_{17}$ | = | $x_{5}\,\mathbf{a}_{1}+y_{5}\,\mathbf{a}_{2}+z_{5}\,\mathbf{a}_{3}$ | = | $x_{5}a\,\hat{\mathbf{x}}+y_{5}b\,\hat{\mathbf{y}}+z_{5}c\,\hat{\mathbf{z}}$ | (4a) | Cs |
| $\mathbf{B}_{18}$ | = | $-x_{5}\,\mathbf{a}_{1}-y_{5}\,\mathbf{a}_{2}+\left(\frac{1}{2}+z_{5}\right)\,\mathbf{a}_{3}$ | = | $-x_{5}a\,\hat{\mathbf{x}}-y_{5}b\,\hat{\mathbf{y}}+\left(\frac{1}{2}+z_{5}\right)c\,\hat{\mathbf{z}}$ | (4a) | Cs |
| $\mathbf{B}_{19}$ | = | $\left(\frac{1}{2}+x_{5}\right)\,\mathbf{a}_{1}+\left(\frac{1}{2}-y_{5}\right)\,\mathbf{a}_{2}+z_{5}\,\mathbf{a}_{3}$ | = | $\left(\frac{1}{2}+x_{5}\right)a\,\hat{\mathbf{x}}+\left(\frac{1}{2}-y_{5}\right)b\,\hat{\mathbf{y}}+z_{5}c\,\hat{\mathbf{z}}$ | (4a) | Cs |
| $\mathbf{B}_{20}$ | = | $\left(\frac{1}{2}-x_{5}\right)\,\mathbf{a}_{1}+\left(\frac{1}{2}+y_{5}\right)\,\mathbf{a}_{2}+\left(\frac{1}{2}+z_{5}\right)\,\mathbf{a}_{3}$ | = | $\left(\frac{1}{2}-x_{5}\right)a\,\hat{\mathbf{x}}+\left(\frac{1}{2}+y_{5}\right)b\,\hat{\mathbf{y}}+\left(\frac{1}{2}+z_{5}\right)c\,\hat{\mathbf{z}}$ | (4a) | Cs |
| $\mathbf{B}_{21}$ | = | $x_{6}\,\mathbf{a}_{1}+y_{6}\,\mathbf{a}_{2}+z_{6}\,\mathbf{a}_{3}$ | = | $x_{6}a\,\hat{\mathbf{x}}+y_{6}b\,\hat{\mathbf{y}}+z_{6}c\,\hat{\mathbf{z}}$ | (4a) | F |
| $\mathbf{B}_{22}$ | = | $-x_{6}\,\mathbf{a}_{1}-y_{6}\,\mathbf{a}_{2}+\left(\frac{1}{2}+z_{6}\right)\,\mathbf{a}_{3}$ | = | $-x_{6}a\,\hat{\mathbf{x}}-y_{6}b\,\hat{\mathbf{y}}+\left(\frac{1}{2}+z_{6}\right)c\,\hat{\mathbf{z}}$ | (4a) | F |
| $\mathbf{B}_{23}$ | = | $\left(\frac{1}{2}+x_{6}\right)\,\mathbf{a}_{1}+\left(\frac{1}{2}-y_{6}\right)\,\mathbf{a}_{2}+z_{6}\,\mathbf{a}_{3}$ | = | $\left(\frac{1}{2}+x_{6}\right)a\,\hat{\mathbf{x}}+\left(\frac{1}{2}-y_{6}\right)b\,\hat{\mathbf{y}}+z_{6}c\,\hat{\mathbf{z}}$ | (4a) | F |
| $\mathbf{B}_{24}$ | = | $\left(\frac{1}{2}-x_{6}\right)\,\mathbf{a}_{1}+\left(\frac{1}{2}+y_{6}\right)\,\mathbf{a}_{2}+\left(\frac{1}{2}+z_{6}\right)\,\mathbf{a}_{3}$ | = | $\left(\frac{1}{2}-x_{6}\right)a\,\hat{\mathbf{x}}+\left(\frac{1}{2}+y_{6}\right)b\,\hat{\mathbf{y}}+\left(\frac{1}{2}+z_{6}\right)c\,\hat{\mathbf{z}}$ | (4a) | F |
| $\mathbf{B}_{25}$ | = | $x_{7}\,\mathbf{a}_{1}+y_{7}\,\mathbf{a}_{2}+z_{7}\,\mathbf{a}_{3}$ | = | $x_{7}a\,\hat{\mathbf{x}}+y_{7}b\,\hat{\mathbf{y}}+z_{7}c\,\hat{\mathbf{z}}$ | (4a) | O I |
| $\mathbf{B}_{26}$ | = | $-x_{7}\,\mathbf{a}_{1}-y_{7}\,\mathbf{a}_{2}+\left(\frac{1}{2}+z_{7}\right)\,\mathbf{a}_{3}$ | = | $-x_{7}a\,\hat{\mathbf{x}}-y_{7}b\,\hat{\mathbf{y}}+\left(\frac{1}{2}+z_{7}\right)c\,\hat{\mathbf{z}}$ | (4a) | O I |
| $\mathbf{B}_{27}$ | = | $\left(\frac{1}{2}+x_{7}\right)\,\mathbf{a}_{1}+\left(\frac{1}{2}-y_{7}\right)\,\mathbf{a}_{2}+z_{7}\,\mathbf{a}_{3}$ | = | $\left(\frac{1}{2}+x_{7}\right)a\,\hat{\mathbf{x}}+\left(\frac{1}{2}-y_{7}\right)b\,\hat{\mathbf{y}}+z_{7}c\,\hat{\mathbf{z}}$ | (4a) | O I |
| $\mathbf{B}_{28}$ | = | $\left(\frac{1}{2}-x_{7}\right)\,\mathbf{a}_{1}+\left(\frac{1}{2}+y_{7}\right)\,\mathbf{a}_{2}+\left(\frac{1}{2}+z_{7}\right)\,\mathbf{a}_{3}$ | = | $\left(\frac{1}{2}-x_{7}\right)a\,\hat{\mathbf{x}}+\left(\frac{1}{2}+y_{7}\right)b\,\hat{\mathbf{y}}+\left(\frac{1}{2}+z_{7}\right)c\,\hat{\mathbf{z}}$ | (4a) | O I |
| $\mathbf{B}_{29}$ | = | $x_{8}\,\mathbf{a}_{1}+y_{8}\,\mathbf{a}_{2}+z_{8}\,\mathbf{a}_{3}$ | = | $x_{8}a\,\hat{\mathbf{x}}+y_{8}b\,\hat{\mathbf{y}}+z_{8}c\,\hat{\mathbf{z}}$ | (4a) | O II |
| $\mathbf{B}_{30}$ | = | $-x_{8}\,\mathbf{a}_{1}-y_{8}\,\mathbf{a}_{2}+\left(\frac{1}{2}+z_{8}\right)\,\mathbf{a}_{3}$ | = | $-x_{8}a\,\hat{\mathbf{x}}-y_{8}b\,\hat{\mathbf{y}}+\left(\frac{1}{2}+z_{8}\right)c\,\hat{\mathbf{z}}$ | (4a) | O II |
| $\mathbf{B}_{31}$ | = | $\left(\frac{1}{2}+x_{8}\right)\,\mathbf{a}_{1}+\left(\frac{1}{2}-y_{8}\right)\,\mathbf{a}_{2}+z_{8}\,\mathbf{a}_{3}$ | = | $\left(\frac{1}{2}+x_{8}\right)a\,\hat{\mathbf{x}}+\left(\frac{1}{2}-y_{8}\right)b\,\hat{\mathbf{y}}+z_{8}c\,\hat{\mathbf{z}}$ | (4a) | O II |
| $\mathbf{B}_{32}$ | = | $\left(\frac{1}{2}-x_{8}\right)\,\mathbf{a}_{1}+\left(\frac{1}{2}+y_{8}\right)\,\mathbf{a}_{2}+\left(\frac{1}{2}+z_{8}\right)\,\mathbf{a}_{3}$ | = | $\left(\frac{1}{2}-x_{8}\right)a\,\hat{\mathbf{x}}+\left(\frac{1}{2}+y_{8}\right)b\,\hat{\mathbf{y}}+\left(\frac{1}{2}+z_{8}\right)c\,\hat{\mathbf{z}}$ | (4a) | O II |
| $\mathbf{B}_{33}$ | = | $x_{9}\,\mathbf{a}_{1}+y_{9}\,\mathbf{a}_{2}+z_{9}\,\mathbf{a}_{3}$ | = | $x_{9}a\,\hat{\mathbf{x}}+y_{9}b\,\hat{\mathbf{y}}+z_{9}c\,\hat{\mathbf{z}}$ | (4a) | O III |
| $\mathbf{B}_{34}$ | = | $-x_{9}\,\mathbf{a}_{1}-y_{9}\,\mathbf{a}_{2}+\left(\frac{1}{2}+z_{9}\right)\,\mathbf{a}_{3}$ | = | $-x_{9}a\,\hat{\mathbf{x}}-y_{9}b\,\hat{\mathbf{y}}+\left(\frac{1}{2}+z_{9}\right)c\,\hat{\mathbf{z}}$ | (4a) | O III |
| $\mathbf{B}_{35}$ | = | $\left(\frac{1}{2}+x_{9}\right)\,\mathbf{a}_{1}+\left(\frac{1}{2}-y_{9}\right)\,\mathbf{a}_{2}+z_{9}\,\mathbf{a}_{3}$ | = | $\left(\frac{1}{2}+x_{9}\right)a\,\hat{\mathbf{x}}+\left(\frac{1}{2}-y_{9}\right)b\,\hat{\mathbf{y}}+z_{9}c\,\hat{\mathbf{z}}$ | (4a) | O III |
| $\mathbf{B}_{36}$ | = | $\left(\frac{1}{2} - x_9\right)\,\mathbf{a}_1 + \left(\frac{1}{2} + y_9\right)\,\mathbf{a}_2 + \left(\frac{1}{2} + z_9\right)\,\mathbf{a}_3$ | = | $\left(\frac{1}{2} - x_9\right)a\,\mathbf{\hat{x}} + \left(\frac{1}{2} + y_9\right)b\,\mathbf{\hat{y}} + \left(\frac{1}{2} + z_9\right)c\,\mathbf{\hat{z}}$ | (4$a$) | O III |
| $\mathbf{B}_{37}$ | = | $x_{10}\,\mathbf{a}_1 + y_{10}\,\mathbf{a}_2 + z_{10}\,\mathbf{a}_3$ | = | $x_{10}a\,\mathbf{\hat{x}} + y_{10}b\,\mathbf{\hat{y}} + z_{10}c\,\mathbf{\hat{z}}$ | (4$a$) | O IV |
| $\mathbf{B}_{38}$ | = | $-x_{10}\,\mathbf{a}_1 - y_{10}\,\mathbf{a}_2 + \left(\frac{1}{2} + z_{10}\right)\,\mathbf{a}_3$ | = | $-x_{10}a\,\mathbf{\hat{x}} - y_{10}b\,\mathbf{\hat{y}} + \left(\frac{1}{2} + z_{10}\right)c\,\mathbf{\hat{z}}$ | (4$a$) | O IV |
| $\mathbf{B}_{39}$ | = | $\left(\frac{1}{2} + x_{10}\right)\,\mathbf{a}_1 + \left(\frac{1}{2} - y_{10}\right)\,\mathbf{a}_2 + z_{10}\,\mathbf{a}_3$ | = | $\left(\frac{1}{2} + x_{10}\right)a\,\mathbf{\hat{x}} + \left(\frac{1}{2} - y_{10}\right)b\,\mathbf{\hat{y}} + z_{10}c\,\mathbf{\hat{z}}$ | (4$a$) | O IV |
| $\mathbf{B}_{40}$ | = | $\left(\frac{1}{2} - x_{10}\right)\,\mathbf{a}_1 + \left(\frac{1}{2} + y_{10}\right)\,\mathbf{a}_2 + \left(\frac{1}{2} + z_{10}\right)\,\mathbf{a}_3$ | = | $\left(\frac{1}{2} - x_{10}\right)a\,\mathbf{\hat{x}} + \left(\frac{1}{2} + y_{10}\right)b\,\mathbf{\hat{y}} + \left(\frac{1}{2} + z_{10}\right)c\,\mathbf{\hat{z}}$ | (4$a$) | O IV |
| $\mathbf{B}_{41}$ | = | $x_{11}\,\mathbf{a}_1 + y_{11}\,\mathbf{a}_2 + z_{11}\,\mathbf{a}_3$ | = | $x_{11}a\,\mathbf{\hat{x}} + y_{11}b\,\mathbf{\hat{y}} + z_{11}c\,\mathbf{\hat{z}}$ | (4$a$) | O V |
| $\mathbf{B}_{42}$ | = | $-x_{11}\,\mathbf{a}_1 - y_{11}\,\mathbf{a}_2 + \left(\frac{1}{2} + z_{11}\right)\,\mathbf{a}_3$ | = | $-x_{11}a\,\mathbf{\hat{x}} - y_{11}b\,\mathbf{\hat{y}} + \left(\frac{1}{2} + z_{11}\right)c\,\mathbf{\hat{z}}$ | (4$a$) | O V |
| $\mathbf{B}_{43}$ | = | $\left(\frac{1}{2} + x_{11}\right)\,\mathbf{a}_1 + \left(\frac{1}{2} - y_{11}\right)\,\mathbf{a}_2 + z_{11}\,\mathbf{a}_3$ | = | $\left(\frac{1}{2} + x_{11}\right)a\,\mathbf{\hat{x}} + \left(\frac{1}{2} - y_{11}\right)b\,\mathbf{\hat{y}} + z_{11}c\,\mathbf{\hat{z}}$ | (4$a$) | O V |
| $\mathbf{B}_{44}$ | = | $\left(\frac{1}{2} - x_{11}\right)\,\mathbf{a}_1 + \left(\frac{1}{2} + y_{11}\right)\,\mathbf{a}_2 + \left(\frac{1}{2} + z_{11}\right)\,\mathbf{a}_3$ | = | $\left(\frac{1}{2} - x_{11}\right)a\,\mathbf{\hat{x}} + \left(\frac{1}{2} + y_{11}\right)b\,\mathbf{\hat{y}} + \left(\frac{1}{2} + z_{11}\right)c\,\mathbf{\hat{z}}$ | (4$a$) | O V |
| $\mathbf{B}_{45}$ | = | $x_{12}\,\mathbf{a}_1 + y_{12}\,\mathbf{a}_2 + z_{12}\,\mathbf{a}_3$ | = | $x_{12}a\,\mathbf{\hat{x}} + y_{12}b\,\mathbf{\hat{y}} + z_{12}c\,\mathbf{\hat{z}}$ | (4$a$) | O VI |
| $\mathbf{B}_{46}$ | = | $-x_{12}\,\mathbf{a}_1 - y_{12}\,\mathbf{a}_2 + \left(\frac{1}{2} + z_{12}\right)\,\mathbf{a}_3$ | = | $-x_{12}a\,\mathbf{\hat{x}} - y_{12}b\,\mathbf{\hat{y}} + \left(\frac{1}{2} + z_{12}\right)c\,\mathbf{\hat{z}}$ | (4$a$) | O VI |
| $\mathbf{B}_{47}$ | = | $\left(\frac{1}{2} + x_{12}\right)\,\mathbf{a}_1 + \left(\frac{1}{2} - y_{12}\right)\,\mathbf{a}_2 + z_{12}\,\mathbf{a}_3$ | = | $\left(\frac{1}{2} + x_{12}\right)a\,\mathbf{\hat{x}} + \left(\frac{1}{2} - y_{12}\right)b\,\mathbf{\hat{y}} + z_{12}c\,\mathbf{\hat{z}}$ | (4$a$) | O VI |
| $\mathbf{B}_{48}$ | = | $\left(\frac{1}{2} - x_{12}\right)\,\mathbf{a}_1 + \left(\frac{1}{2} + y_{12}\right)\,\mathbf{a}_2 + \left(\frac{1}{2} + z_{12}\right)\,\mathbf{a}_3$ | = | $\left(\frac{1}{2} - x_{12}\right)a\,\mathbf{\hat{x}} + \left(\frac{1}{2} + y_{12}\right)b\,\mathbf{\hat{y}} + \left(\frac{1}{2} + z_{12}\right)c\,\mathbf{\hat{z}}$ | (4$a$) | O VI |

**Geometry files:**

- CIF: pp. 1592
- POSCAR: pp. 1592

# $LiGaO_2$ Structure: ABC2_oP16_33_a_a_2a

http://aflow.org/prototype-encyclopedia/ABC2_oP16_33_a_a_2a

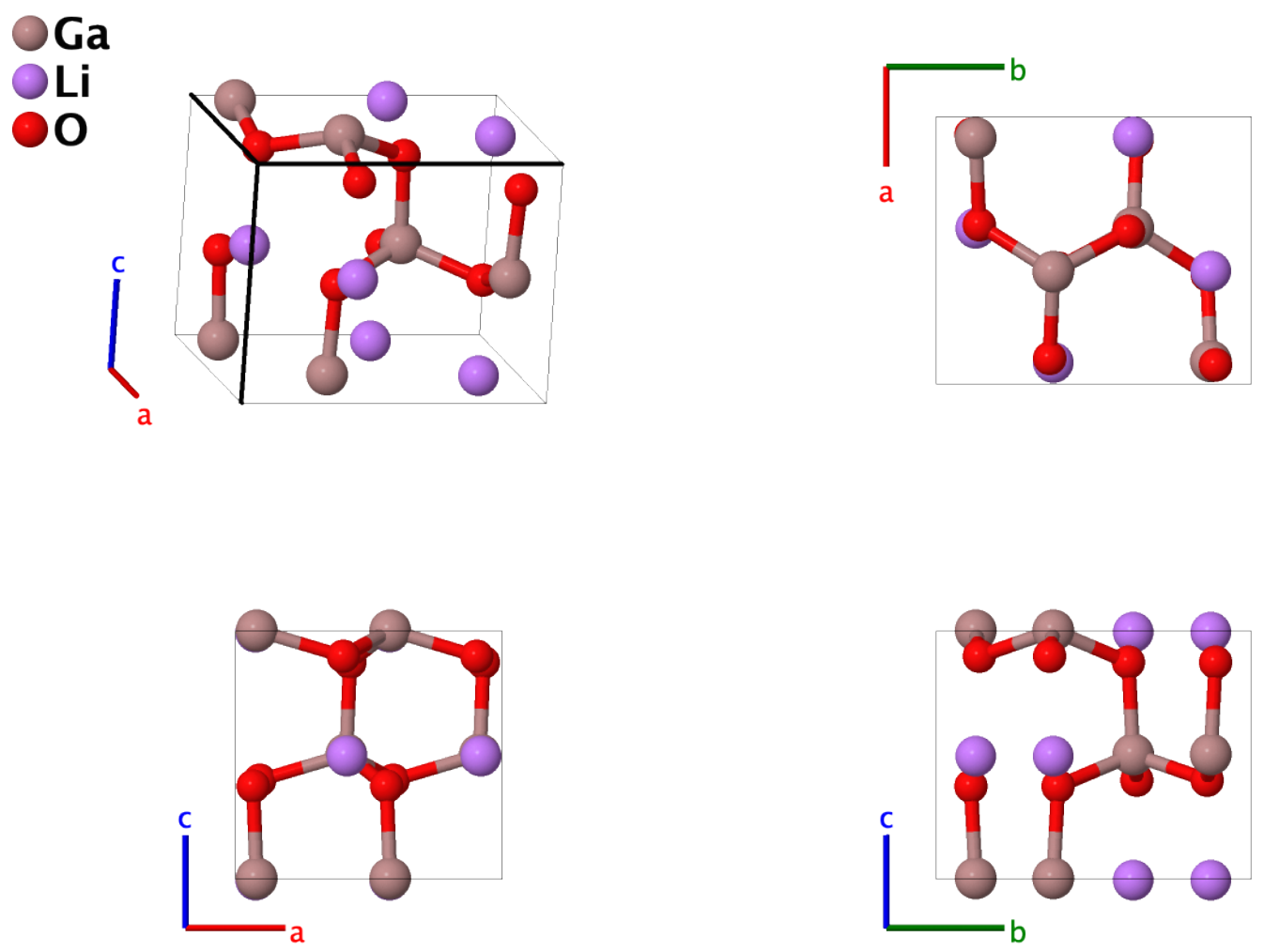

| | | |
|---|---|---|
| **Prototype** | : | $GaLiO_2$ |
| **AFLOW prototype label** | : | ABC2_oP16_33_a_a_2a |
| ***Strukturbericht* designation** | : | None |
| **Pearson symbol** | : | oP16 |
| **Space group number** | : | 33 |
| **Space group symbol** | : | $Pna2_1$ |
| **AFLOW prototype command** | : | `aflow --proto=ABC2_oP16_33_a_a_2a`<br>`--params=`$a, b/a, c/a, x_1, y_1, z_1, x_2, y_2, z_2, x_3, y_3, z_3, x_4, y_4, z_4$ |

- Space group $Pna2_1$ #33 allows an arbitary choice of the zero of the $z$-axis. Here we set $z_1 = 0$ for gallium.

**Simple Orthorhombic primitive vectors:**

$$\mathbf{a}_1 = a\,\hat{\mathbf{x}}$$
$$\mathbf{a}_2 = b\,\hat{\mathbf{y}}$$
$$\mathbf{a}_3 = c\,\hat{\mathbf{z}}$$

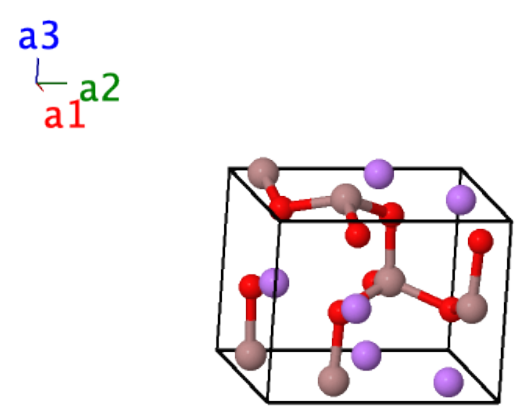

**Basis vectors:**

| | | Lattice Coordinates | | Cartesian Coordinates | Wyckoff Position | Atom Type |
|---|---|---|---|---|---|---|
| $\mathbf{B}_1$ | = | $x_1\,\mathbf{a}_1 + y_1\,\mathbf{a}_2 + z_1\,\mathbf{a}_3$ | = | $x_1 a\,\hat{\mathbf{x}} + y_1 b\,\hat{\mathbf{y}} + z_1 c\,\hat{\mathbf{z}}$ | (4$a$) | Ga |
| $\mathbf{B}_2$ | = | $-x_1\,\mathbf{a}_1 - y_1\,\mathbf{a}_2 + \left(\frac{1}{2} + z_1\right)\mathbf{a}_3$ | = | $-x_1 a\,\hat{\mathbf{x}} - y_1 b\,\hat{\mathbf{y}} + \left(\frac{1}{2} + z_1\right) c\,\hat{\mathbf{z}}$ | (4$a$) | Ga |
| $\mathbf{B}_3$ | = | $\left(\frac{1}{2} + x_1\right)\mathbf{a}_1 + \left(\frac{1}{2} - y_1\right)\mathbf{a}_2 + z_1\,\mathbf{a}_3$ | = | $\left(\frac{1}{2} + x_1\right) a\,\hat{\mathbf{x}} + \left(\frac{1}{2} - y_1\right) b\,\hat{\mathbf{y}} + z_1 c\,\hat{\mathbf{z}}$ | (4$a$) | Ga |
| $\mathbf{B}_4$ | = | $\left(\frac{1}{2} - x_1\right)\mathbf{a}_1 + \left(\frac{1}{2} + y_1\right)\mathbf{a}_2 + \left(\frac{1}{2} + z_1\right)\mathbf{a}_3$ | = | $\left(\frac{1}{2} - x_1\right) a\,\hat{\mathbf{x}} + \left(\frac{1}{2} + y_1\right) b\,\hat{\mathbf{y}} + \left(\frac{1}{2} + z_1\right) c\,\hat{\mathbf{z}}$ | (4$a$) | Ga |
| $\mathbf{B}_{5}$ | = | $x_{2}\,\mathbf{a}_{1} + y_{2}\,\mathbf{a}_{2} + z_{2}\,\mathbf{a}_{3}$ | = | $x_{2}a\,\hat{\mathbf{x}} + y_{2}b\,\hat{\mathbf{y}} + z_{2}c\,\hat{\mathbf{z}}$ | (4a) | Li |
| $\mathbf{B}_{6}$ | = | $-x_{2}\,\mathbf{a}_{1} - y_{2}\,\mathbf{a}_{2} + \left(\frac{1}{2} + z_{2}\right)\,\mathbf{a}_{3}$ | = | $-x_{2}a\,\hat{\mathbf{x}} - y_{2}b\,\hat{\mathbf{y}} + \left(\frac{1}{2} + z_{2}\right)c\,\hat{\mathbf{z}}$ | (4a) | Li |
| $\mathbf{B}_{7}$ | = | $\left(\frac{1}{2} + x_{2}\right)\,\mathbf{a}_{1} + \left(\frac{1}{2} - y_{2}\right)\,\mathbf{a}_{2} + z_{2}\,\mathbf{a}_{3}$ | = | $\left(\frac{1}{2} + x_{2}\right)a\,\hat{\mathbf{x}} + \left(\frac{1}{2} - y_{2}\right)b\,\hat{\mathbf{y}} + z_{2}c\,\hat{\mathbf{z}}$ | (4a) | Li |
| $\mathbf{B}_{8}$ | = | $\left(\frac{1}{2} - x_{2}\right)\,\mathbf{a}_{1} + \left(\frac{1}{2} + y_{2}\right)\,\mathbf{a}_{2} + \left(\frac{1}{2} + z_{2}\right)\,\mathbf{a}_{3}$ | = | $\left(\frac{1}{2} - x_{2}\right)a\,\hat{\mathbf{x}} + \left(\frac{1}{2} + y_{2}\right)b\,\hat{\mathbf{y}} + \left(\frac{1}{2} + z_{2}\right)c\,\hat{\mathbf{z}}$ | (4a) | Li |
| $\mathbf{B}_{9}$ | = | $x_{3}\,\mathbf{a}_{1} + y_{3}\,\mathbf{a}_{2} + z_{3}\,\mathbf{a}_{3}$ | = | $x_{3}a\,\hat{\mathbf{x}} + y_{3}b\,\hat{\mathbf{y}} + z_{3}c\,\hat{\mathbf{z}}$ | (4a) | O I |
| $\mathbf{B}_{10}$ | = | $-x_{3}\,\mathbf{a}_{1} - y_{3}\,\mathbf{a}_{2} + \left(\frac{1}{2} + z_{3}\right)\,\mathbf{a}_{3}$ | = | $-x_{3}a\,\hat{\mathbf{x}} - y_{3}b\,\hat{\mathbf{y}} + \left(\frac{1}{2} + z_{3}\right)c\,\hat{\mathbf{z}}$ | (4a) | O I |
| $\mathbf{B}_{11}$ | = | $\left(\frac{1}{2} + x_{3}\right)\,\mathbf{a}_{1} + \left(\frac{1}{2} - y_{3}\right)\,\mathbf{a}_{2} + z_{3}\,\mathbf{a}_{3}$ | = | $\left(\frac{1}{2} + x_{3}\right)a\,\hat{\mathbf{x}} + \left(\frac{1}{2} - y_{3}\right)b\,\hat{\mathbf{y}} + z_{3}c\,\hat{\mathbf{z}}$ | (4a) | O I |
| $\mathbf{B}_{12}$ | = | $\left(\frac{1}{2} - x_{3}\right)\,\mathbf{a}_{1} + \left(\frac{1}{2} + y_{3}\right)\,\mathbf{a}_{2} + \left(\frac{1}{2} + z_{3}\right)\,\mathbf{a}_{3}$ | = | $\left(\frac{1}{2} - x_{3}\right)a\,\hat{\mathbf{x}} + \left(\frac{1}{2} + y_{3}\right)b\,\hat{\mathbf{y}} + \left(\frac{1}{2} + z_{3}\right)c\,\hat{\mathbf{z}}$ | (4a) | O I |
| $\mathbf{B}_{13}$ | = | $x_{4}\,\mathbf{a}_{1} + y_{4}\,\mathbf{a}_{2} + z_{4}\,\mathbf{a}_{3}$ | = | $x_{4}a\,\hat{\mathbf{x}} + y_{4}b\,\hat{\mathbf{y}} + z_{4}c\,\hat{\mathbf{z}}$ | (4a) | O II |
| $\mathbf{B}_{14}$ | = | $-x_{4}\,\mathbf{a}_{1} - y_{4}\,\mathbf{a}_{2} + \left(\frac{1}{2} + z_{4}\right)\,\mathbf{a}_{3}$ | = | $-x_{4}a\,\hat{\mathbf{x}} - y_{4}b\,\hat{\mathbf{y}} + \left(\frac{1}{2} + z_{4}\right)c\,\hat{\mathbf{z}}$ | (4a) | O II |
| $\mathbf{B}_{15}$ | = | $\left(\frac{1}{2} + x_{4}\right)\,\mathbf{a}_{1} + \left(\frac{1}{2} - y_{4}\right)\,\mathbf{a}_{2} + z_{4}\,\mathbf{a}_{3}$ | = | $\left(\frac{1}{2} + x_{4}\right)a\,\hat{\mathbf{x}} + \left(\frac{1}{2} - y_{4}\right)b\,\hat{\mathbf{y}} + z_{4}c\,\hat{\mathbf{z}}$ | (4a) | O II |
| $\mathbf{B}_{16}$ | = | $\left(\frac{1}{2} - x_{4}\right)\,\mathbf{a}_{1} + \left(\frac{1}{2} + y_{4}\right)\,\mathbf{a}_{2} + \left(\frac{1}{2} + z_{4}\right)\,\mathbf{a}_{3}$ | = | $\left(\frac{1}{2} - x_{4}\right)a\,\hat{\mathbf{x}} + \left(\frac{1}{2} + y_{4}\right)b\,\hat{\mathbf{y}} + \left(\frac{1}{2} + z_{4}\right)c\,\hat{\mathbf{z}}$ | (4a) | O II |

**Geometry files:**

- CIF: pp. 1592
- POSCAR: pp. 1593

# $\gamma$-$LiIO_3$ Structure: ABC3_oP20_33_a_a_3a

http://aflow.org/prototype-encyclopedia/ABC3_oP20_33_a_a_3a

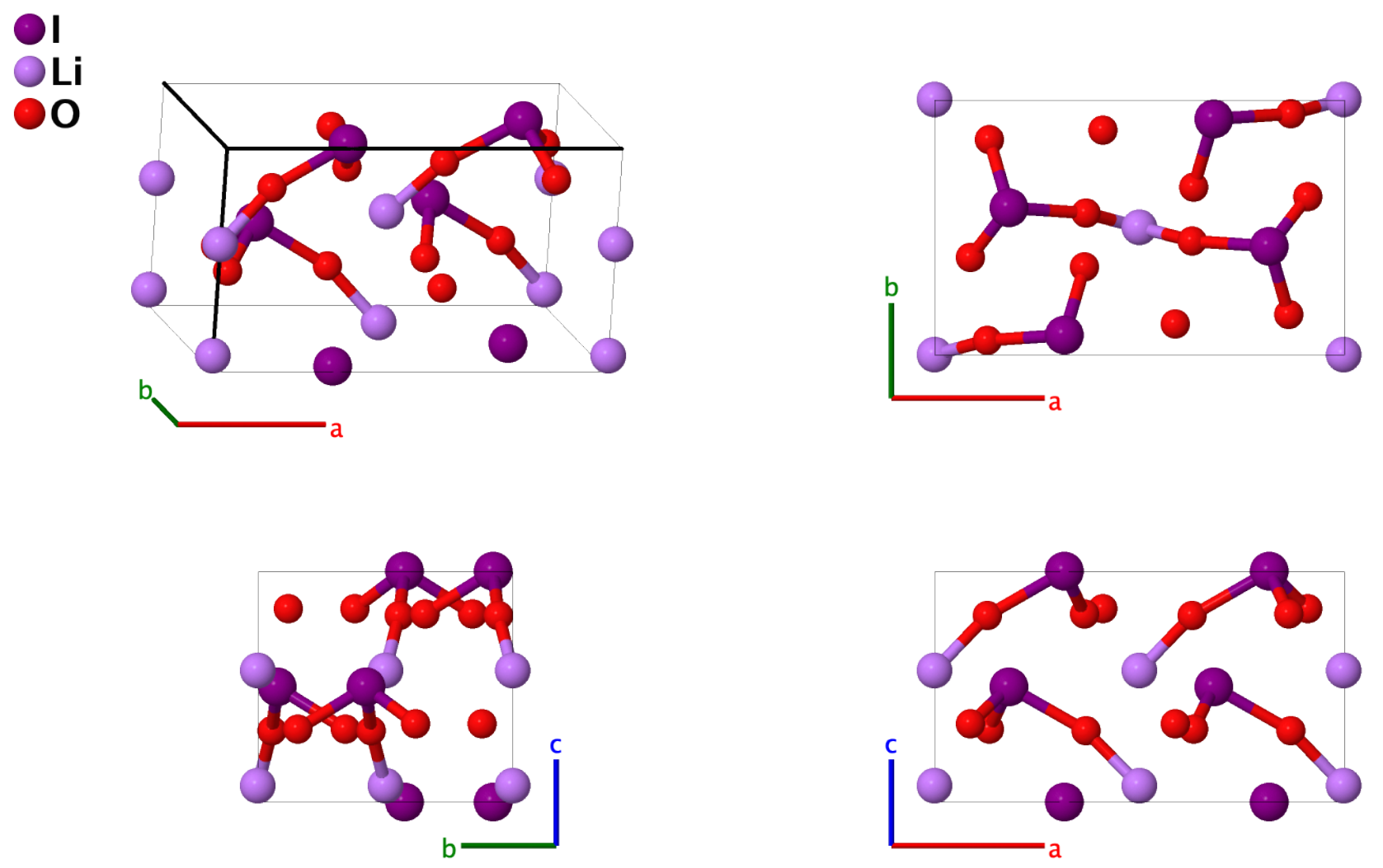

| | | |
|---|---|---|
| **Prototype** | : | $ILiO_3$ |
| **AFLOW prototype label** | : | ABC3_oP20_33_a_a_3a |
| ***Strukturbericht* designation** | : | None |
| **Pearson symbol** | : | oP20 |
| **Space group number** | : | 33 |
| **Space group symbol** | : | $Pna2_1$ |
| **AFLOW prototype command** | : | `aflow --proto=ABC3_oP20_33_a_a_3a`<br>`--params=`$a, b/a, c/a, x_1, y_1, z_1, x_2, y_2, z_2, x_3, y_3, z_3, x_4, y_4, z_4, x_5, y_5, z_5$ |

- $LiIO_3$ is known to exist in three forms:
  - $\alpha$-$LiIO_3$, stable below 473 K: (Zachariasen, 1931) originally determined that the structure of $\alpha$-$LiIO_3$ was in space group $P6_322$ #182, which (Hermann, 1937) designated *Strukturbericht* $E2_3$. (Rosenzweig, 1966) subsequently determined that the true structure was in space group $P6_3$ #173.
  - $\beta$-$LiIO_3$, stable from 573 up to the melting point at 708 K.
  - $\gamma$-$LiIO_3$, stable between the $\alpha$- and $\beta$-phases. Here we show the structure at 515 K.
- Space group $Pna2_1$ #33 allows an arbitary choice of the zero of the $z$-axis. Here we set $z_1 = 0$ for iodine.

**Simple Orthorhombic primitive vectors:**

$$\mathbf{a}_1 = a\,\hat{\mathbf{x}}$$
$$\mathbf{a}_2 = b\,\hat{\mathbf{y}}$$
$$\mathbf{a}_3 = c\,\hat{\mathbf{z}}$$

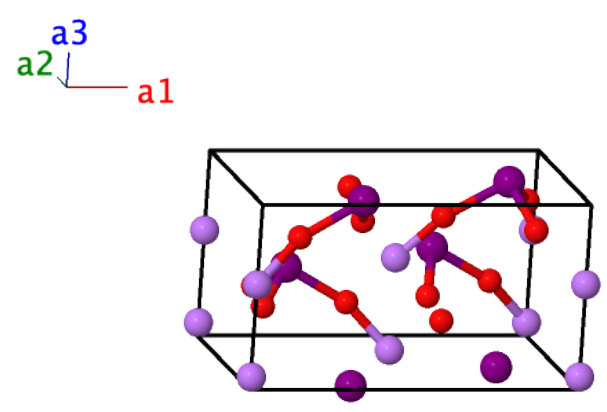

**Basis vectors:**

| | | Lattice Coordinates | | Cartesian Coordinates | Wyckoff Position | Atom Type |
|---|---|---|---|---|---|---|
| $\mathbf{B}_1$ | = | $x_1\,\mathbf{a}_1 + y_1\,\mathbf{a}_2 + z_1\,\mathbf{a}_3$ | = | $x_1 a\,\hat{\mathbf{x}} + y_1 b\,\hat{\mathbf{y}} + z_1 c\,\hat{\mathbf{z}}$ | (4a) | I |
| $\mathbf{B}_2$ | = | $-x_1\,\mathbf{a}_1 - y_1\,\mathbf{a}_2 + \left(\frac{1}{2} + z_1\right)\mathbf{a}_3$ | = | $-x_1 a\,\hat{\mathbf{x}} - y_1 b\,\hat{\mathbf{y}} + \left(\frac{1}{2} + z_1\right) c\,\hat{\mathbf{z}}$ | (4a) | I |
| $\mathbf{B}_3$ | = | $\left(\frac{1}{2} + x_1\right)\mathbf{a}_1 + \left(\frac{1}{2} - y_1\right)\mathbf{a}_2 + z_1\,\mathbf{a}_3$ | = | $\left(\frac{1}{2} + x_1\right) a\,\hat{\mathbf{x}} + \left(\frac{1}{2} - y_1\right) b\,\hat{\mathbf{y}} + z_1 c\,\hat{\mathbf{z}}$ | (4a) | I |
| $\mathbf{B}_4$ | = | $\left(\frac{1}{2} - x_1\right)\mathbf{a}_1 + \left(\frac{1}{2} + y_1\right)\mathbf{a}_2 + \left(\frac{1}{2} + z_1\right)\mathbf{a}_3$ | = | $\left(\frac{1}{2} - x_1\right) a\,\hat{\mathbf{x}} + \left(\frac{1}{2} + y_1\right) b\,\hat{\mathbf{y}} + \left(\frac{1}{2} + z_1\right) c\,\hat{\mathbf{z}}$ | (4a) | I |
| $\mathbf{B}_5$ | = | $x_2\,\mathbf{a}_1 + y_2\,\mathbf{a}_2 + z_2\,\mathbf{a}_3$ | = | $x_2 a\,\hat{\mathbf{x}} + y_2 b\,\hat{\mathbf{y}} + z_2 c\,\hat{\mathbf{z}}$ | (4a) | Li |
| $\mathbf{B}_6$ | = | $-x_2\,\mathbf{a}_1 - y_2\,\mathbf{a}_2 + \left(\frac{1}{2} + z_2\right)\mathbf{a}_3$ | = | $-x_2 a\,\hat{\mathbf{x}} - y_2 b\,\hat{\mathbf{y}} + \left(\frac{1}{2} + z_2\right) c\,\hat{\mathbf{z}}$ | (4a) | Li |
| $\mathbf{B}_7$ | = | $\left(\frac{1}{2} + x_2\right)\mathbf{a}_1 + \left(\frac{1}{2} - y_2\right)\mathbf{a}_2 + z_2\,\mathbf{a}_3$ | = | $\left(\frac{1}{2} + x_2\right) a\,\hat{\mathbf{x}} + \left(\frac{1}{2} - y_2\right) b\,\hat{\mathbf{y}} + z_2 c\,\hat{\mathbf{z}}$ | (4a) | Li |
| $\mathbf{B}_8$ | = | $\left(\frac{1}{2} - x_2\right)\mathbf{a}_1 + \left(\frac{1}{2} + y_2\right)\mathbf{a}_2 + \left(\frac{1}{2} + z_2\right)\mathbf{a}_3$ | = | $\left(\frac{1}{2} - x_2\right) a\,\hat{\mathbf{x}} + \left(\frac{1}{2} + y_2\right) b\,\hat{\mathbf{y}} + \left(\frac{1}{2} + z_2\right) c\,\hat{\mathbf{z}}$ | (4a) | Li |
| $\mathbf{B}_9$ | = | $x_3\,\mathbf{a}_1 + y_3\,\mathbf{a}_2 + z_3\,\mathbf{a}_3$ | = | $x_3 a\,\hat{\mathbf{x}} + y_3 b\,\hat{\mathbf{y}} + z_3 c\,\hat{\mathbf{z}}$ | (4a) | O I |
| $\mathbf{B}_{10}$ | = | $-x_3\,\mathbf{a}_1 - y_3\,\mathbf{a}_2 + \left(\frac{1}{2} + z_3\right)\mathbf{a}_3$ | = | $-x_3 a\,\hat{\mathbf{x}} - y_3 b\,\hat{\mathbf{y}} + \left(\frac{1}{2} + z_3\right) c\,\hat{\mathbf{z}}$ | (4a) | O I |
| $\mathbf{B}_{11}$ | = | $\left(\frac{1}{2} + x_3\right)\mathbf{a}_1 + \left(\frac{1}{2} - y_3\right)\mathbf{a}_2 + z_3\,\mathbf{a}_3$ | = | $\left(\frac{1}{2} + x_3\right) a\,\hat{\mathbf{x}} + \left(\frac{1}{2} - y_3\right) b\,\hat{\mathbf{y}} + z_3 c\,\hat{\mathbf{z}}$ | (4a) | O I |
| $\mathbf{B}_{12}$ | = | $\left(\frac{1}{2} - x_3\right)\mathbf{a}_1 + \left(\frac{1}{2} + y_3\right)\mathbf{a}_2 + \left(\frac{1}{2} + z_3\right)\mathbf{a}_3$ | = | $\left(\frac{1}{2} - x_3\right) a\,\hat{\mathbf{x}} + \left(\frac{1}{2} + y_3\right) b\,\hat{\mathbf{y}} + \left(\frac{1}{2} + z_3\right) c\,\hat{\mathbf{z}}$ | (4a) | O I |
| $\mathbf{B}_{13}$ | = | $x_4\,\mathbf{a}_1 + y_4\,\mathbf{a}_2 + z_4\,\mathbf{a}_3$ | = | $x_4 a\,\hat{\mathbf{x}} + y_4 b\,\hat{\mathbf{y}} + z_4 c\,\hat{\mathbf{z}}$ | (4a) | O II |
| $\mathbf{B}_{14}$ | = | $-x_4\,\mathbf{a}_1 - y_4\,\mathbf{a}_2 + \left(\frac{1}{2} + z_4\right)\mathbf{a}_3$ | = | $-x_4 a\,\hat{\mathbf{x}} - y_4 b\,\hat{\mathbf{y}} + \left(\frac{1}{2} + z_4\right) c\,\hat{\mathbf{z}}$ | (4a) | O II |
| $\mathbf{B}_{15}$ | = | $\left(\frac{1}{2} + x_4\right)\mathbf{a}_1 + \left(\frac{1}{2} - y_4\right)\mathbf{a}_2 + z_4\,\mathbf{a}_3$ | = | $\left(\frac{1}{2} + x_4\right) a\,\hat{\mathbf{x}} + \left(\frac{1}{2} - y_4\right) b\,\hat{\mathbf{y}} + z_4 c\,\hat{\mathbf{z}}$ | (4a) | O II |
| $\mathbf{B}_{16}$ | = | $\left(\frac{1}{2} - x_4\right)\mathbf{a}_1 + \left(\frac{1}{2} + y_4\right)\mathbf{a}_2 + \left(\frac{1}{2} + z_4\right)\mathbf{a}_3$ | = | $\left(\frac{1}{2} - x_4\right) a\,\hat{\mathbf{x}} + \left(\frac{1}{2} + y_4\right) b\,\hat{\mathbf{y}} + \left(\frac{1}{2} + z_4\right) c\,\hat{\mathbf{z}}$ | (4a) | O II |
| $\mathbf{B}_{17}$ | = | $x_5\,\mathbf{a}_1 + y_5\,\mathbf{a}_2 + z_5\,\mathbf{a}_3$ | = | $x_5 a\,\hat{\mathbf{x}} + y_5 b\,\hat{\mathbf{y}} + z_5 c\,\hat{\mathbf{z}}$ | (4a) | O III |
| $\mathbf{B}_{18}$ | = | $-x_5\,\mathbf{a}_1 - y_5\,\mathbf{a}_2 + \left(\frac{1}{2} + z_5\right)\mathbf{a}_3$ | = | $-x_5 a\,\hat{\mathbf{x}} - y_5 b\,\hat{\mathbf{y}} + \left(\frac{1}{2} + z_5\right) c\,\hat{\mathbf{z}}$ | (4a) | O III |
| $\mathbf{B}_{19}$ | = | $\left(\frac{1}{2} + x_5\right)\mathbf{a}_1 + \left(\frac{1}{2} - y_5\right)\mathbf{a}_2 + z_5\,\mathbf{a}_3$ | = | $\left(\frac{1}{2} + x_5\right) a\,\hat{\mathbf{x}} + \left(\frac{1}{2} - y_5\right) b\,\hat{\mathbf{y}} + z_5 c\,\hat{\mathbf{z}}$ | (4a) | O III |
| $\mathbf{B}_{20}$ | = | $\left(\frac{1}{2} - x_5\right)\mathbf{a}_1 + \left(\frac{1}{2} + y_5\right)\mathbf{a}_2 + \left(\frac{1}{2} + z_5\right)\mathbf{a}_3$ | = | $\left(\frac{1}{2} - x_5\right) a\,\hat{\mathbf{x}} + \left(\frac{1}{2} + y_5\right) b\,\hat{\mathbf{y}} + \left(\frac{1}{2} + z_5\right) c\,\hat{\mathbf{z}}$ | (4a) | O III |

**Geometry files:**

- CIF: pp. 1593
- POSCAR: pp. 1593

# $MnF_{2-x}(OH)_x$ Structure: A2B2CD2_oP14_34_c_c_a_c

http://aflow.org/prototype-encyclopedia/A2B2CD2_oP14_34_c_c_a_c

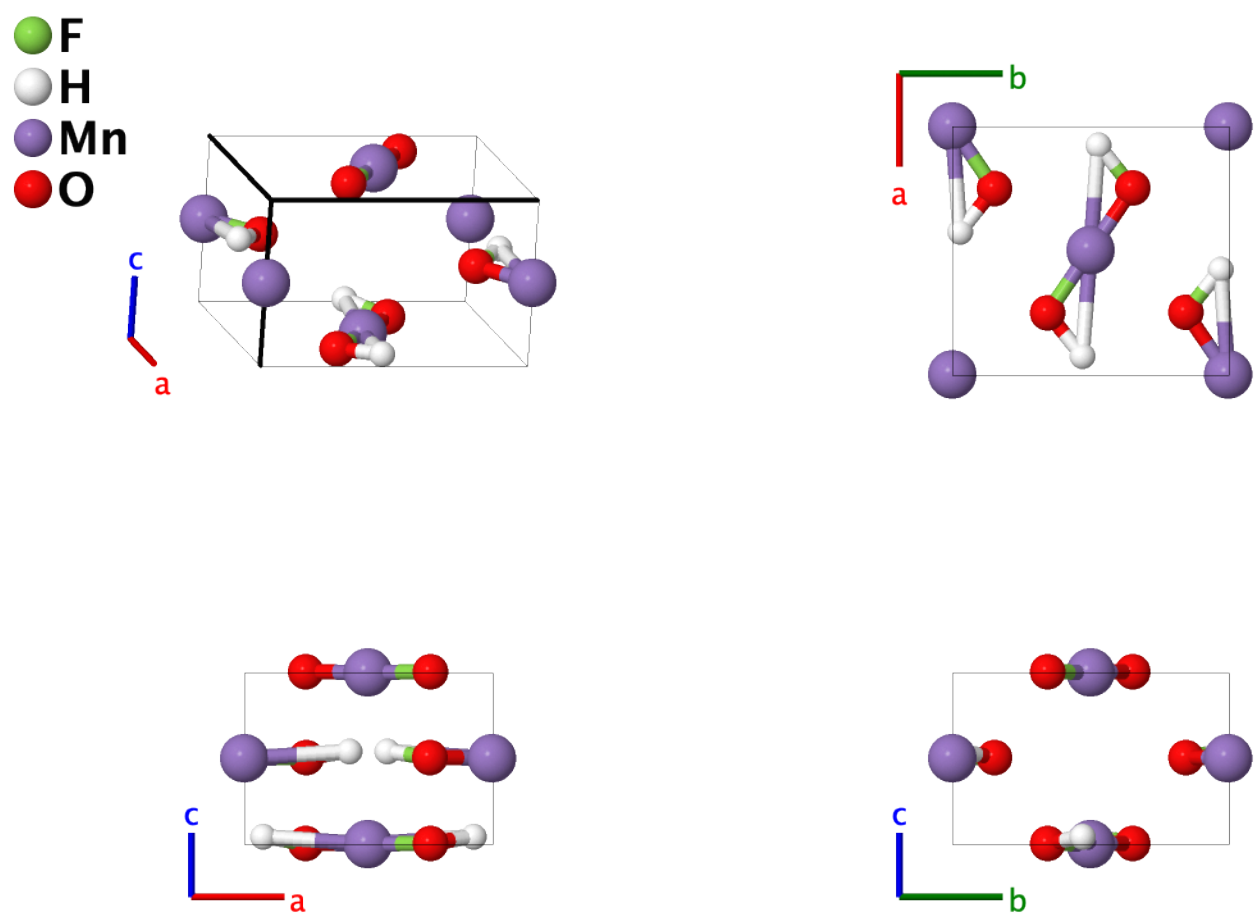

| | | |
|---|---|---|
| **Prototype** | : | $F_{2-x}H_xMnO_x$ |
| **AFLOW prototype label** | : | A2B2CD2_oP14_34_c_c_a_c |
| ***Strukturbericht* designation** | : | None |
| **Pearson symbol** | : | oP14 |
| **Space group number** | : | 34 |
| **Space group symbol** | : | *Pnn*2 |
| **AFLOW prototype command** | : | `aflow --proto=A2B2CD2_oP14_34_c_c_a_c` `--params=`$a, b/a, c/a, z_1, x_2, y_2, z_2, x_3, y_3, z_3, x_4, y_4, z_4$ |

- We use the data (Yahia, 2013) give for the site occupancy of oxygen and hydrogen (0.399). Either the (4*c*) F site is occupied, or the (4*c*) H and O sites are occupied.
- Space group *Pnn*2 #34 allows an arbitrary choice of the origin of the *z*-coordinate. Here we use this to put the manganese atom at the origin, setting $z_1 = 0$.

**Simple Orthorhombic primitive vectors:**

$$\mathbf{a}_1 = a\,\hat{\mathbf{x}}$$
$$\mathbf{a}_2 = b\,\hat{\mathbf{y}}$$
$$\mathbf{a}_3 = c\,\hat{\mathbf{z}}$$

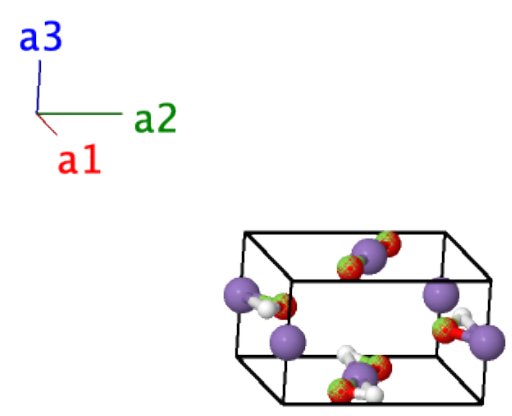

**Basis vectors:**

| | | Lattice Coordinates | | Cartesian Coordinates | Wyckoff Position | Atom Type |
|---|---|---|---|---|---|---|
| $\mathbf{B}_1$ | = | $z_1\,\mathbf{a}_3$ | = | $z_1 c\,\hat{\mathbf{z}}$ | (2*a*) | Mn |
| $\mathbf{B}_2$ | = | $\frac{1}{2}\,\mathbf{a}_1 + \frac{1}{2}\,\mathbf{a}_2 + \left(\frac{1}{2} + z_1\right)\,\mathbf{a}_3$ | = | $\frac{1}{2}a\,\hat{\mathbf{x}} + \frac{1}{2}b\,\hat{\mathbf{y}} + \left(\frac{1}{2} + z_1\right)c\,\hat{\mathbf{z}}$ | (2*a*) | Mn |
| $\mathbf{B}_3$ | = | $x_2\,\mathbf{a}_1 + y_2\,\mathbf{a}_2 + z_2\,\mathbf{a}_3$ | = | $x_2 a\,\hat{\mathbf{x}} + y_2 b\,\hat{\mathbf{y}} + z_2 c\,\hat{\mathbf{z}}$ | (4*c*) | F |
| $\mathbf{B}_4$ | = | $-x_2\,\mathbf{a}_1 - y_2\,\mathbf{a}_2 + z_2\,\mathbf{a}_3$ | = | $-x_2 a\,\hat{\mathbf{x}} - y_2 b\,\hat{\mathbf{y}} + z_2 c\,\hat{\mathbf{z}}$ | (4$c$) | F |
| $\mathbf{B}_5$ | = | $\left(\frac{1}{2} + x_2\right)\mathbf{a}_1 + \left(\frac{1}{2} - y_2\right)\mathbf{a}_2 + \left(\frac{1}{2} + z_2\right)\mathbf{a}_3$ | = | $\left(\frac{1}{2} + x_2\right)a\,\hat{\mathbf{x}} + \left(\frac{1}{2} - y_2\right)b\,\hat{\mathbf{y}} + \left(\frac{1}{2} + z_2\right)c\,\hat{\mathbf{z}}$ | (4$c$) | F |
| $\mathbf{B}_6$ | = | $\left(\frac{1}{2} - x_2\right)\mathbf{a}_1 + \left(\frac{1}{2} + y_2\right)\mathbf{a}_2 + \left(\frac{1}{2} + z_2\right)\mathbf{a}_3$ | = | $\left(\frac{1}{2} - x_2\right)a\,\hat{\mathbf{x}} + \left(\frac{1}{2} + y_2\right)b\,\hat{\mathbf{y}} + \left(\frac{1}{2} + z_2\right)c\,\hat{\mathbf{z}}$ | (4$c$) | F |
| $\mathbf{B}_7$ | = | $x_3\,\mathbf{a}_1 + y_3\,\mathbf{a}_2 + z_3\,\mathbf{a}_3$ | = | $x_3 a\,\hat{\mathbf{x}} + y_3 b\,\hat{\mathbf{y}} + z_3 c\,\hat{\mathbf{z}}$ | (4$c$) | H |
| $\mathbf{B}_8$ | = | $-x_3\,\mathbf{a}_1 - y_3\,\mathbf{a}_2 + z_3\,\mathbf{a}_3$ | = | $-x_3 a\,\hat{\mathbf{x}} - y_3 b\,\hat{\mathbf{y}} + z_3 c\,\hat{\mathbf{z}}$ | (4$c$) | H |
| $\mathbf{B}_9$ | = | $\left(\frac{1}{2} + x_3\right)\mathbf{a}_1 + \left(\frac{1}{2} - y_3\right)\mathbf{a}_2 + \left(\frac{1}{2} + z_3\right)\mathbf{a}_3$ | = | $\left(\frac{1}{2} + x_3\right)a\,\hat{\mathbf{x}} + \left(\frac{1}{2} - y_3\right)b\,\hat{\mathbf{y}} + \left(\frac{1}{2} + z_3\right)c\,\hat{\mathbf{z}}$ | (4$c$) | H |
| $\mathbf{B}_{10}$ | = | $\left(\frac{1}{2} - x_3\right)\mathbf{a}_1 + \left(\frac{1}{2} + y_3\right)\mathbf{a}_2 + \left(\frac{1}{2} + z_3\right)\mathbf{a}_3$ | = | $\left(\frac{1}{2} - x_3\right)a\,\hat{\mathbf{x}} + \left(\frac{1}{2} + y_3\right)b\,\hat{\mathbf{y}} + \left(\frac{1}{2} + z_3\right)c\,\hat{\mathbf{z}}$ | (4$c$) | H |
| $\mathbf{B}_{11}$ | = | $x_4\,\mathbf{a}_1 + y_4\,\mathbf{a}_2 + z_4\,\mathbf{a}_3$ | = | $x_4 a\,\hat{\mathbf{x}} + y_4 b\,\hat{\mathbf{y}} + z_4 c\,\hat{\mathbf{z}}$ | (4$c$) | O |
| $\mathbf{B}_{12}$ | = | $-x_4\,\mathbf{a}_1 - y_4\,\mathbf{a}_2 + z_4\,\mathbf{a}_3$ | = | $-x_4 a\,\hat{\mathbf{x}} - y_4 b\,\hat{\mathbf{y}} + z_4 c\,\hat{\mathbf{z}}$ | (4$c$) | O |
| $\mathbf{B}_{13}$ | = | $\left(\frac{1}{2} + x_4\right)\mathbf{a}_1 + \left(\frac{1}{2} - y_4\right)\mathbf{a}_2 + \left(\frac{1}{2} + z_4\right)\mathbf{a}_3$ | = | $\left(\frac{1}{2} + x_4\right)a\,\hat{\mathbf{x}} + \left(\frac{1}{2} - y_4\right)b\,\hat{\mathbf{y}} + \left(\frac{1}{2} + z_4\right)c\,\hat{\mathbf{z}}$ | (4$c$) | O |
| $\mathbf{B}_{14}$ | = | $\left(\frac{1}{2} - x_4\right)\mathbf{a}_1 + \left(\frac{1}{2} + y_4\right)\mathbf{a}_2 + \left(\frac{1}{2} + z_4\right)\mathbf{a}_3$ | = | $\left(\frac{1}{2} - x_4\right)a\,\hat{\mathbf{x}} + \left(\frac{1}{2} + y_4\right)b\,\hat{\mathbf{y}} + \left(\frac{1}{2} + z_4\right)c\,\hat{\mathbf{z}}$ | (4$c$) | O |

**Geometry files:**

- CIF: pp. 1593
- POSCAR: pp. 1594

# $Si_2N_2O$ Structure: A2BC2_oC20_36_b_a_b

http://aflow.org/prototype-encyclopedia/A2BC2_oC20_36_b_a_b

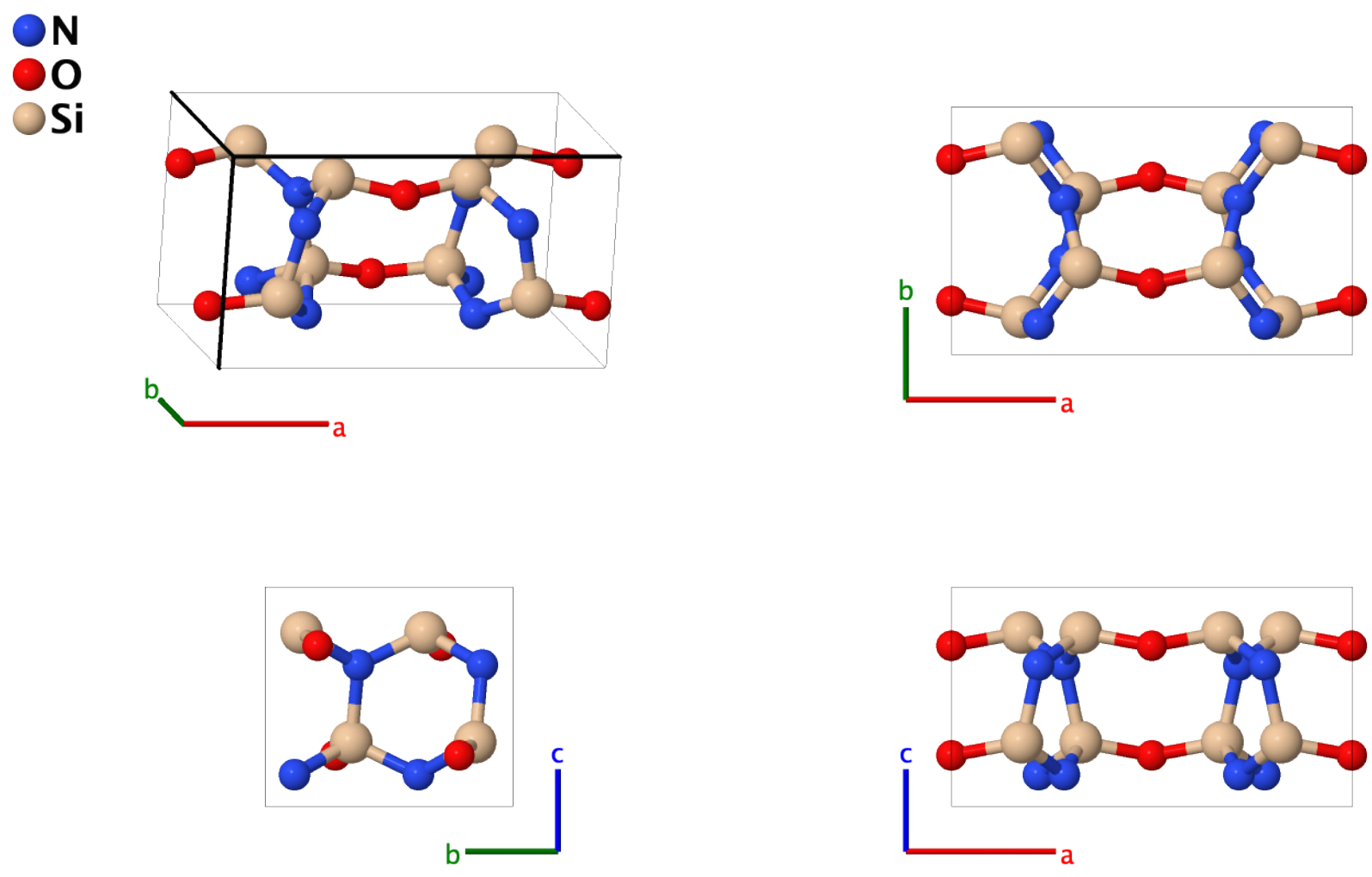

| | | |
|---|---|---|
| **Prototype** | : | $N_2OSi_2$ |
| **AFLOW prototype label** | : | A2BC2_oC20_36_b_a_b |
| ***Strukturbericht* designation** | : | None |
| **Pearson symbol** | : | oC20 |
| **Space group number** | : | 36 |
| **Space group symbol** | : | $Cmc2_1$ |
| **AFLOW prototype command** | : | `aflow --proto=A2BC2_oC20_36_b_a_b` `--params=`$a, b/a, c/a, y_1, z_1, x_2, y_2, z_2, x_3, y_3, z_3$ |

**Base-centered Orthorhombic primitive vectors:**

$$\mathbf{a}_1 = \frac{1}{2}\,a\,\hat{\mathbf{x}} - \frac{1}{2}\,b\,\hat{\mathbf{y}}$$
$$\mathbf{a}_2 = \frac{1}{2}\,a\,\hat{\mathbf{x}} + \frac{1}{2}\,b\,\hat{\mathbf{y}}$$
$$\mathbf{a}_3 = c\,\hat{\mathbf{z}}$$

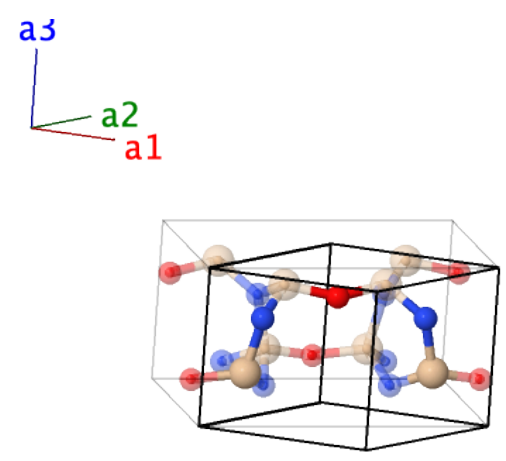

**Basis vectors:**

| | | Lattice Coordinates | | Cartesian Coordinates | Wyckoff Position | Atom Type |
|---|---|---|---|---|---|---|
| $\mathbf{B}_1$ | = | $-y_1\,\mathbf{a}_1 + y_1\,\mathbf{a}_2 + z_1\,\mathbf{a}_3$ | = | $y_1 b\,\hat{\mathbf{y}} + z_1 c\,\hat{\mathbf{z}}$ | (4$a$) | O |
| $\mathbf{B}_2$ | = | $y_1\,\mathbf{a}_1 - y_1\,\mathbf{a}_2 + \left(\frac{1}{2} + z_1\right)\,\mathbf{a}_3$ | = | $-y_1 b\,\hat{\mathbf{y}} + \left(\frac{1}{2} + z_1\right) c\,\hat{\mathbf{z}}$ | (4$a$) | O |
| $\mathbf{B}_3$ | = | $(x_2 - y_2)\,\mathbf{a}_1 + (x_2 + y_2)\,\mathbf{a}_2 + z_2\,\mathbf{a}_3$ | = | $x_2 a\,\hat{\mathbf{x}} + y_2 b\,\hat{\mathbf{y}} + z_2 c\,\hat{\mathbf{z}}$ | (8$b$) | N |
| $\mathbf{B}_4$ | = | $(-x_2 + y_2)\,\mathbf{a}_1 + (-x_2 - y_2)\,\mathbf{a}_2 + \left(\frac{1}{2} + z_2\right)\,\mathbf{a}_3$ | = | $-x_2 a\,\hat{\mathbf{x}} - y_2 b\,\hat{\mathbf{y}} + \left(\frac{1}{2} + z_2\right) c\,\hat{\mathbf{z}}$ | (8$b$) | N |
| $\mathbf{B}_5$ | = | $(x_2 + y_2)\,\mathbf{a}_1 + (x_2 - y_2)\,\mathbf{a}_2 + \left(\frac{1}{2} + z_2\right)\,\mathbf{a}_3$ | = | $x_2 a\,\hat{\mathbf{x}} - y_2 b\,\hat{\mathbf{y}} + \left(\frac{1}{2} + z_2\right) c\,\hat{\mathbf{z}}$ | (8$b$) | N |
| $\mathbf{B}_6$ | = | $(-x_2 - y_2)\,\mathbf{a}_1 + (-x_2 + y_2)\,\mathbf{a}_2 + z_2\,\mathbf{a}_3$ | = | $-x_2 a\,\hat{\mathbf{x}} + y_2 b\,\hat{\mathbf{y}} + z_2 c\,\hat{\mathbf{z}}$ | (8$b$) | N |
| $\mathbf{B}_7$ | = | $(x_3 - y_3)\,\mathbf{a}_1 + (x_3 + y_3)\,\mathbf{a}_2 + z_3\,\mathbf{a}_3$ | = | $x_3 a\,\hat{\mathbf{x}} + y_3 b\,\hat{\mathbf{y}} + z_3 c\,\hat{\mathbf{z}}$ | (8$b$) | Si |
| $\mathbf{B}_8$ | = | $(-x_3 + y_3)\,\mathbf{a}_1 + (-x_3 - y_3)\,\mathbf{a}_2 + \left(\frac{1}{2} + z_3\right)\,\mathbf{a}_3$ | = | $-x_3 a\,\hat{\mathbf{x}} - y_3 b\,\hat{\mathbf{y}} + \left(\frac{1}{2} + z_3\right) c\,\hat{\mathbf{z}}$ | (8$b$) | Si |
| $\mathbf{B}_9$ | = | $(x_3 + y_3)\,\mathbf{a}_1 + (x_3 - y_3)\,\mathbf{a}_2 + \left(\frac{1}{2} + z_3\right)\,\mathbf{a}_3$ | = | $x_3 a\,\hat{\mathbf{x}} - y_3 b\,\hat{\mathbf{y}} + \left(\frac{1}{2} + z_3\right) c\,\hat{\mathbf{z}}$ | (8$b$) | Si |
| $\mathbf{B}_{10}$ | = | $(-x_3 - y_3)\,\mathbf{a}_1 + (-x_3 + y_3)\,\mathbf{a}_2 + z_3\,\mathbf{a}_3$ | = | $-x_3 a\,\hat{\mathbf{x}} + y_3 b\,\hat{\mathbf{y}} + z_3 c\,\hat{\mathbf{z}}$ | (8$b$) | Si |

---

---

**Geometry files:**

- CIF: pp. 1594
- POSCAR: pp. 1594

# $Bi_2GeO_5$ Structure: A2BC5_oC32_36_b_a_a2b

http://aflow.org/prototype-encyclopedia/A2BC5_oC32_36_b_a_a2b

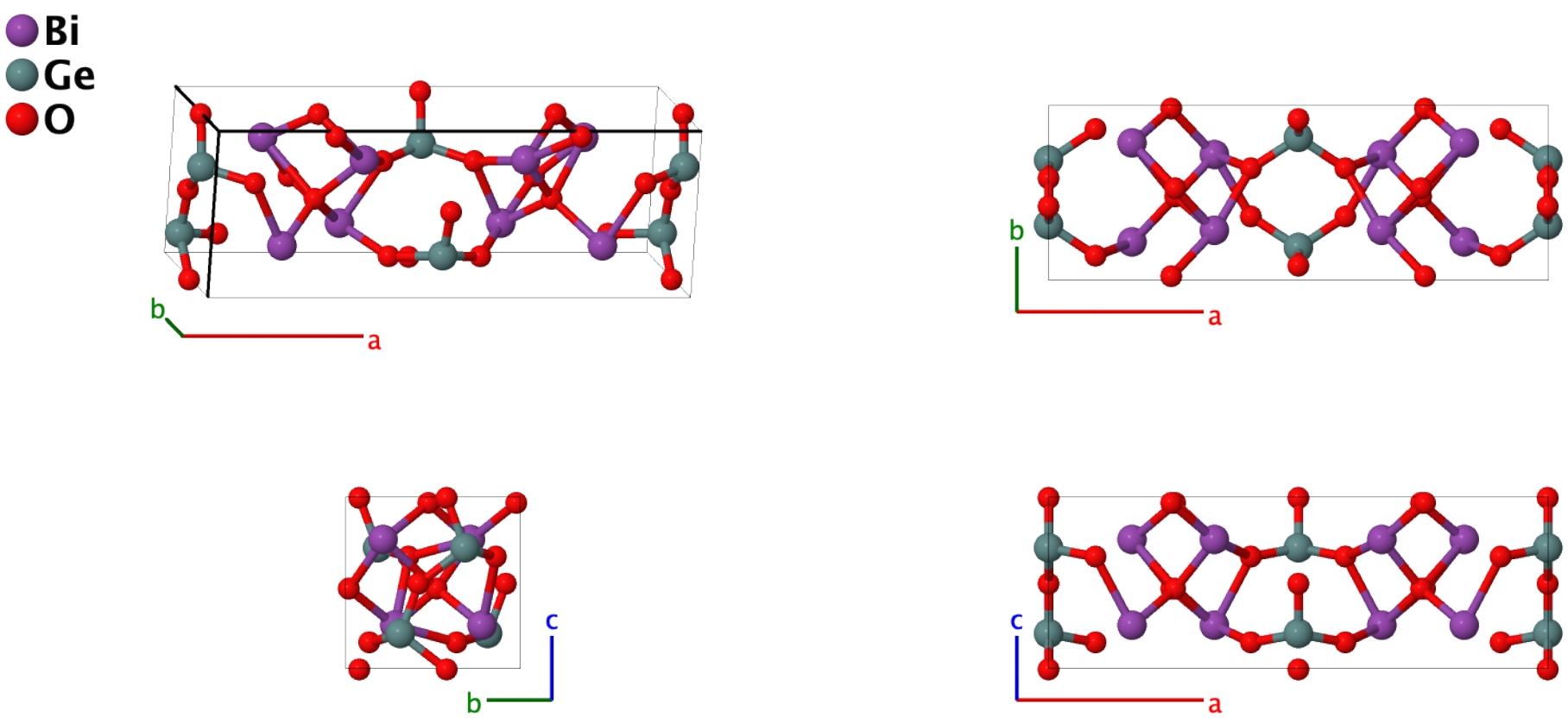

| | | |
|---|---|---|
| **Prototype** | : | $Bi_2GeO_5$ |
| **AFLOW prototype label** | : | A2BC5_oC32_36_b_a_a2b |
| ***Strukturbericht* designation** | : | None |
| **Pearson symbol** | : | oC32 |
| **Space group number** | : | 36 |
| **Space group symbol** | : | $Cmc2_1$ |
| **AFLOW prototype command** | : | `aflow --proto=A2BC5_oC32_36_b_a_a2b` `--params=`$a, b/a, c/a, y_1, z_1, y_2, z_2, x_3, y_3, z_3, x_4, y_4, z_4, x_5, y_5, z_5$ |

### Other compounds with this structure

- $Bi_2SiO_5$

- Space group $Cmc2_1$ #36 allows an arbitrary choice of the zero of the $z$-axis. Here it is chosen so that $z_3 = 0$ for the bismuth atom.

### Base-centered Orthorhombic primitive vectors:

$$\mathbf{a}_1 = \frac{1}{2}\,a\,\mathbf{\hat{x}} - \frac{1}{2}\,b\,\mathbf{\hat{y}}$$
$$\mathbf{a}_2 = \frac{1}{2}\,a\,\mathbf{\hat{x}} + \frac{1}{2}\,b\,\mathbf{\hat{y}}$$
$$\mathbf{a}_3 = c\,\mathbf{\hat{z}}$$

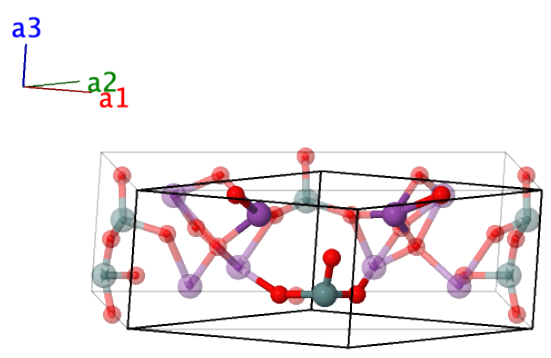

### Basis vectors:

| | | Lattice Coordinates | | Cartesian Coordinates | Wyckoff Position | Atom Type |
|---|---|---|---|---|---|---|
| $\mathbf{B}_1$ | = | $-y_1\,\mathbf{a}_1 + y_1\,\mathbf{a}_2 + z_1\,\mathbf{a}_3$ | = | $y_1 b\,\mathbf{\hat{y}} + z_1 c\,\mathbf{\hat{z}}$ | (4$a$) | Ge |
| $\mathbf{B}_2$ | = | $y_1\,\mathbf{a}_1 - y_1\,\mathbf{a}_2 + \left(\frac{1}{2} + z_1\right)\,\mathbf{a}_3$ | = | $-y_1 b\,\mathbf{\hat{y}} + \left(\frac{1}{2} + z_1\right) c\,\mathbf{\hat{z}}$ | (4$a$) | Ge |
| $\mathbf{B}_3$ | = | $-y_2\,\mathbf{a}_1 + y_2\,\mathbf{a}_2 + z_2\,\mathbf{a}_3$ | = | $y_2 b\,\mathbf{\hat{y}} + z_2 c\,\mathbf{\hat{z}}$ | (4$a$) | O I |
| $\mathbf{B}_4$ | = | $y_2\,\mathbf{a}_1 - y_2\,\mathbf{a}_2 + \left(\frac{1}{2} + z_2\right)\,\mathbf{a}_3$ | = | $-y_2 b\,\mathbf{\hat{y}} + \left(\frac{1}{2} + z_2\right) c\,\mathbf{\hat{z}}$ | (4$a$) | O I |
| $\mathbf{B}_5$ | = | $(x_3 - y_3)\,\mathbf{a}_1 + (x_3 + y_3)\,\mathbf{a}_2 + z_3\,\mathbf{a}_3$ | = | $x_3 a\,\mathbf{\hat{x}} + y_3 b\,\mathbf{\hat{y}} + z_3 c\,\mathbf{\hat{z}}$ | (8$b$) | Bi |
| $\mathbf{B}_6$ | = | $(-x_3 + y_3)\,\mathbf{a}_1 + (-x_3 - y_3)\,\mathbf{a}_2 + \left(\frac{1}{2} + z_3\right)\mathbf{a}_3$ | = | $-x_3 a\,\hat{\mathbf{x}} - y_3 b\,\hat{\mathbf{y}} + \left(\frac{1}{2} + z_3\right) c\,\hat{\mathbf{z}}$ | (8$b$) | Bi |
| $\mathbf{B}_7$ | = | $(x_3 + y_3)\,\mathbf{a}_1 + (x_3 - y_3)\,\mathbf{a}_2 + \left(\frac{1}{2} + z_3\right)\mathbf{a}_3$ | = | $x_3 a\,\hat{\mathbf{x}} - y_3 b\,\hat{\mathbf{y}} + \left(\frac{1}{2} + z_3\right) c\,\hat{\mathbf{z}}$ | (8$b$) | Bi |
| $\mathbf{B}_8$ | = | $(-x_3 - y_3)\,\mathbf{a}_1 + (-x_3 + y_3)\,\mathbf{a}_2 + z_3\,\mathbf{a}_3$ | = | $-x_3 a\,\hat{\mathbf{x}} + y_3 b\,\hat{\mathbf{y}} + z_3 c\,\hat{\mathbf{z}}$ | (8$b$) | Bi |
| $\mathbf{B}_9$ | = | $(x_4 - y_4)\,\mathbf{a}_1 + (x_4 + y_4)\,\mathbf{a}_2 + z_4\,\mathbf{a}_3$ | = | $x_4 a\,\hat{\mathbf{x}} + y_4 b\,\hat{\mathbf{y}} + z_4 c\,\hat{\mathbf{z}}$ | (8$b$) | O II |
| $\mathbf{B}_{10}$ | = | $(-x_4 + y_4)\,\mathbf{a}_1 + (-x_4 - y_4)\,\mathbf{a}_2 + \left(\frac{1}{2} + z_4\right)\mathbf{a}_3$ | = | $-x_4 a\,\hat{\mathbf{x}} - y_4 b\,\hat{\mathbf{y}} + \left(\frac{1}{2} + z_4\right) c\,\hat{\mathbf{z}}$ | (8$b$) | O II |
| $\mathbf{B}_{11}$ | = | $(x_4 + y_4)\,\mathbf{a}_1 + (x_4 - y_4)\,\mathbf{a}_2 + \left(\frac{1}{2} + z_4\right)\mathbf{a}_3$ | = | $x_4 a\,\hat{\mathbf{x}} - y_4 b\,\hat{\mathbf{y}} + \left(\frac{1}{2} + z_4\right) c\,\hat{\mathbf{z}}$ | (8$b$) | O II |
| $\mathbf{B}_{12}$ | = | $(-x_4 - y_4)\,\mathbf{a}_1 + (-x_4 + y_4)\,\mathbf{a}_2 + z_4\,\mathbf{a}_3$ | = | $-x_4 a\,\hat{\mathbf{x}} + y_4 b\,\hat{\mathbf{y}} + z_4 c\,\hat{\mathbf{z}}$ | (8$b$) | O II |
| $\mathbf{B}_{13}$ | = | $(x_5 - y_5)\,\mathbf{a}_1 + (x_5 + y_5)\,\mathbf{a}_2 + z_5\,\mathbf{a}_3$ | = | $x_5 a\,\hat{\mathbf{x}} + y_5 b\,\hat{\mathbf{y}} + z_5 c\,\hat{\mathbf{z}}$ | (8$b$) | O III |
| $\mathbf{B}_{14}$ | = | $(-x_5 + y_5)\,\mathbf{a}_1 + (-x_5 - y_5)\,\mathbf{a}_2 + \left(\frac{1}{2} + z_5\right)\mathbf{a}_3$ | = | $-x_5 a\,\hat{\mathbf{x}} - y_5 b\,\hat{\mathbf{y}} + \left(\frac{1}{2} + z_5\right) c\,\hat{\mathbf{z}}$ | (8$b$) | O III |
| $\mathbf{B}_{15}$ | = | $(x_5 + y_5)\,\mathbf{a}_1 + (x_5 - y_5)\,\mathbf{a}_2 + \left(\frac{1}{2} + z_5\right)\mathbf{a}_3$ | = | $x_5 a\,\hat{\mathbf{x}} - y_5 b\,\hat{\mathbf{y}} + \left(\frac{1}{2} + z_5\right) c\,\hat{\mathbf{z}}$ | (8$b$) | O III |
| $\mathbf{B}_{16}$ | = | $(-x_5 - y_5)\,\mathbf{a}_1 + (-x_5 + y_5)\,\mathbf{a}_2 + z_5\,\mathbf{a}_3$ | = | $-x_5 a\,\hat{\mathbf{x}} + y_5 b\,\hat{\mathbf{y}} + z_5 c\,\hat{\mathbf{z}}$ | (8$b$) | O III |

**Geometry files:**

- CIF: pp. 1594
- POSCAR: pp. 1595

# $Ni_3Si_2$ Structure: A3B2_oC80_36_4a4b_2a3b

http://aflow.org/prototype-encyclopedia/A3B2_oC80_36_4a4b_2a3b

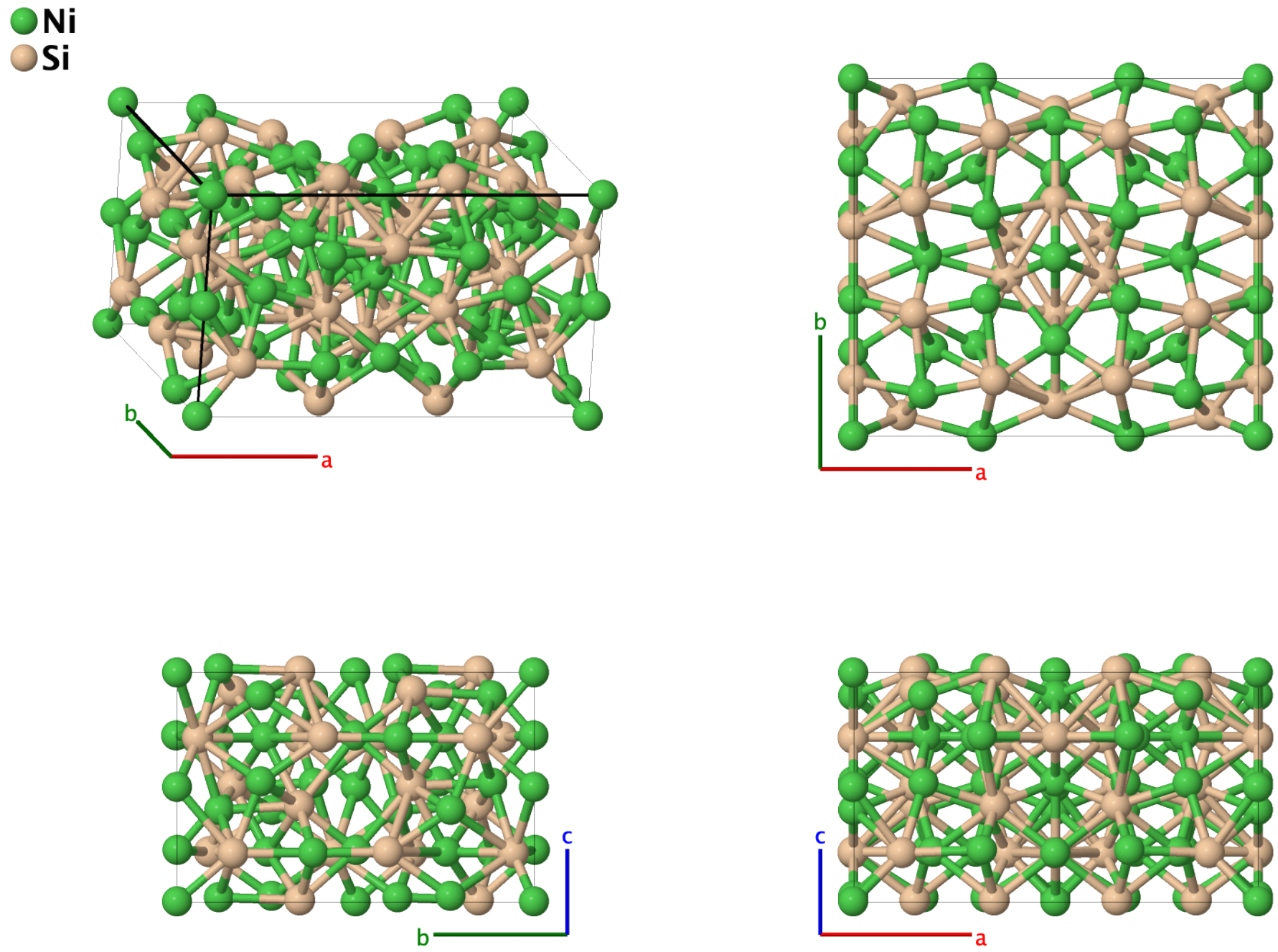

| | | |
|---|---|---|
| **Prototype** | : | $Ni_3Si_2$ |
| **AFLOW prototype label** | : | A3B2_oC80_36_4a4b_2a3b |
| ***Strukturbericht* designation** | : | None |
| **Pearson symbol** | : | oC80 |
| **Space group number** | : | 36 |
| **Space group symbol** | : | $Cmc2_1$ |
| **AFLOW prototype command** | : | `aflow --proto=A3B2_oC80_36_4a4b_2a3b` `--params=`$a, b/a, c/a, y_1, z_1, y_2, z_2, y_3, z_3, y_4, z_4, y_5, z_5, y_6, z_6, x_7, y_7, z_7, x_8, y_8, z_8, x_9, y_9, z_9, x_{10}, y_{10}, z_{10}, x_{11}, y_{11}, z_{11}, x_{12}, y_{12}, z_{12}, x_{13}, y_{13}, z_{13}$ |

**Base-centered Orthorhombic primitive vectors:**

$$\mathbf{a}_1 = \frac{1}{2}\,a\,\hat{\mathbf{x}} - \frac{1}{2}\,b\,\hat{\mathbf{y}}$$
$$\mathbf{a}_2 = \frac{1}{2}\,a\,\hat{\mathbf{x}} + \frac{1}{2}\,b\,\hat{\mathbf{y}}$$
$$\mathbf{a}_3 = c\,\hat{\mathbf{z}}$$

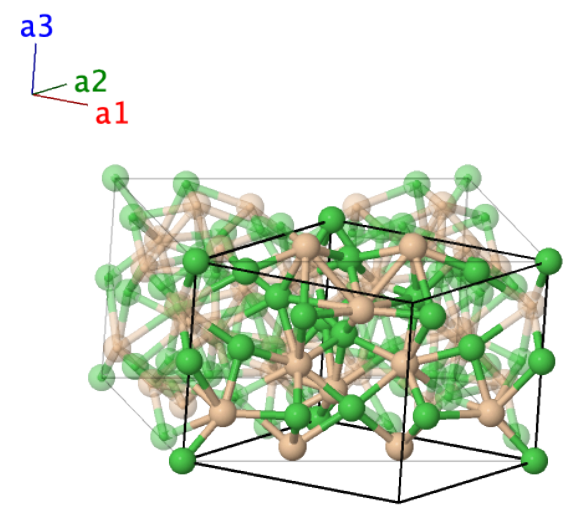

**Basis vectors:**

| | | Lattice Coordinates | | Cartesian Coordinates | Wyckoff Position | Atom Type |
|---|---|---|---|---|---|---|
| $\mathbf{B}_1$ | = | $-y_1\,\mathbf{a}_1 + y_1\,\mathbf{a}_2 + z_1\,\mathbf{a}_3$ | = | $y_1 b\,\hat{\mathbf{y}} + z_1 c\,\hat{\mathbf{z}}$ | (4$a$) | Ni I |
| $\mathbf{B}_2$ | = | $y_1\,\mathbf{a}_1 - y_1\,\mathbf{a}_2 + \left(\frac{1}{2} + z_1\right)\,\mathbf{a}_3$ | = | $-y_1 b\,\hat{\mathbf{y}} + \left(\frac{1}{2} + z_1\right) c\,\hat{\mathbf{z}}$ | (4$a$) | Ni I |
| $\mathbf{B}_{3}$ | = | $-y_{2}\,\mathbf{a}_{1}+y_{2}\,\mathbf{a}_{2}+z_{2}\,\mathbf{a}_{3}$ | = | $y_{2}b\,\mathbf{\hat{y}}+z_{2}c\,\mathbf{\hat{z}}$ | (4$a$) | Ni II |
| $\mathbf{B}_{4}$ | = | $y_{2}\,\mathbf{a}_{1}-y_{2}\,\mathbf{a}_{2}+\left(\frac{1}{2}+z_{2}\right)\mathbf{a}_{3}$ | = | $-y_{2}b\,\mathbf{\hat{y}}+\left(\frac{1}{2}+z_{2}\right)c\,\mathbf{\hat{z}}$ | (4$a$) | Ni II |
| $\mathbf{B}_{5}$ | = | $-y_{3}\,\mathbf{a}_{1}+y_{3}\,\mathbf{a}_{2}+z_{3}\,\mathbf{a}_{3}$ | = | $y_{3}b\,\mathbf{\hat{y}}+z_{3}c\,\mathbf{\hat{z}}$ | (4$a$) | Ni III |
| $\mathbf{B}_{6}$ | = | $y_{3}\,\mathbf{a}_{1}-y_{3}\,\mathbf{a}_{2}+\left(\frac{1}{2}+z_{3}\right)\mathbf{a}_{3}$ | = | $-y_{3}b\,\mathbf{\hat{y}}+\left(\frac{1}{2}+z_{3}\right)c\,\mathbf{\hat{z}}$ | (4$a$) | Ni III |
| $\mathbf{B}_{7}$ | = | $-y_{4}\,\mathbf{a}_{1}+y_{4}\,\mathbf{a}_{2}+z_{4}\,\mathbf{a}_{3}$ | = | $y_{4}b\,\mathbf{\hat{y}}+z_{4}c\,\mathbf{\hat{z}}$ | (4$a$) | Ni IV |
| $\mathbf{B}_{8}$ | = | $y_{4}\,\mathbf{a}_{1}-y_{4}\,\mathbf{a}_{2}+\left(\frac{1}{2}+z_{4}\right)\mathbf{a}_{3}$ | = | $-y_{4}b\,\mathbf{\hat{y}}+\left(\frac{1}{2}+z_{4}\right)c\,\mathbf{\hat{z}}$ | (4$a$) | Ni IV |
| $\mathbf{B}_{9}$ | = | $-y_{5}\,\mathbf{a}_{1}+y_{5}\,\mathbf{a}_{2}+z_{5}\,\mathbf{a}_{3}$ | = | $y_{5}b\,\mathbf{\hat{y}}+z_{5}c\,\mathbf{\hat{z}}$ | (4$a$) | Si I |
| $\mathbf{B}_{10}$ | = | $y_{5}\,\mathbf{a}_{1}-y_{5}\,\mathbf{a}_{2}+\left(\frac{1}{2}+z_{5}\right)\mathbf{a}_{3}$ | = | $-y_{5}b\,\mathbf{\hat{y}}+\left(\frac{1}{2}+z_{5}\right)c\,\mathbf{\hat{z}}$ | (4$a$) | Si I |
| $\mathbf{B}_{11}$ | = | $-y_{6}\,\mathbf{a}_{1}+y_{6}\,\mathbf{a}_{2}+z_{6}\,\mathbf{a}_{3}$ | = | $y_{6}b\,\mathbf{\hat{y}}+z_{6}c\,\mathbf{\hat{z}}$ | (4$a$) | Si II |
| $\mathbf{B}_{12}$ | = | $y_{6}\,\mathbf{a}_{1}-y_{6}\,\mathbf{a}_{2}+\left(\frac{1}{2}+z_{6}\right)\mathbf{a}_{3}$ | = | $-y_{6}b\,\mathbf{\hat{y}}+\left(\frac{1}{2}+z_{6}\right)c\,\mathbf{\hat{z}}$ | (4$a$) | Si II |
| $\mathbf{B}_{13}$ | = | $(x_{7}-y_{7})\,\mathbf{a}_{1}+(x_{7}+y_{7})\,\mathbf{a}_{2}+z_{7}\,\mathbf{a}_{3}$ | = | $x_{7}a\,\mathbf{\hat{x}}+y_{7}b\,\mathbf{\hat{y}}+z_{7}c\,\mathbf{\hat{z}}$ | (8$b$) | Ni V |
| $\mathbf{B}_{14}$ | = | $(-x_{7}+y_{7})\,\mathbf{a}_{1}+(-x_{7}-y_{7})\,\mathbf{a}_{2}+\left(\frac{1}{2}+z_{7}\right)\mathbf{a}_{3}$ | = | $-x_{7}a\,\mathbf{\hat{x}}-y_{7}b\,\mathbf{\hat{y}}+\left(\frac{1}{2}+z_{7}\right)c\,\mathbf{\hat{z}}$ | (8$b$) | Ni V |
| $\mathbf{B}_{15}$ | = | $(x_{7}+y_{7})\,\mathbf{a}_{1}+(x_{7}-y_{7})\,\mathbf{a}_{2}+\left(\frac{1}{2}+z_{7}\right)\mathbf{a}_{3}$ | = | $x_{7}a\,\mathbf{\hat{x}}-y_{7}b\,\mathbf{\hat{y}}+\left(\frac{1}{2}+z_{7}\right)c\,\mathbf{\hat{z}}$ | (8$b$) | Ni V |
| $\mathbf{B}_{16}$ | = | $(-x_{7}-y_{7})\,\mathbf{a}_{1}+(-x_{7}+y_{7})\,\mathbf{a}_{2}+z_{7}\,\mathbf{a}_{3}$ | = | $-x_{7}a\,\mathbf{\hat{x}}+y_{7}b\,\mathbf{\hat{y}}+z_{7}c\,\mathbf{\hat{z}}$ | (8$b$) | Ni V |
| $\mathbf{B}_{17}$ | = | $(x_{8}-y_{8})\,\mathbf{a}_{1}+(x_{8}+y_{8})\,\mathbf{a}_{2}+z_{8}\,\mathbf{a}_{3}$ | = | $x_{8}a\,\mathbf{\hat{x}}+y_{8}b\,\mathbf{\hat{y}}+z_{8}c\,\mathbf{\hat{z}}$ | (8$b$) | Ni VI |
| $\mathbf{B}_{18}$ | = | $(-x_{8}+y_{8})\,\mathbf{a}_{1}+(-x_{8}-y_{8})\,\mathbf{a}_{2}+\left(\frac{1}{2}+z_{8}\right)\mathbf{a}_{3}$ | = | $-x_{8}a\,\mathbf{\hat{x}}-y_{8}b\,\mathbf{\hat{y}}+\left(\frac{1}{2}+z_{8}\right)c\,\mathbf{\hat{z}}$ | (8$b$) | Ni VI |
| $\mathbf{B}_{19}$ | = | $(x_{8}+y_{8})\,\mathbf{a}_{1}+(x_{8}-y_{8})\,\mathbf{a}_{2}+\left(\frac{1}{2}+z_{8}\right)\mathbf{a}_{3}$ | = | $x_{8}a\,\mathbf{\hat{x}}-y_{8}b\,\mathbf{\hat{y}}+\left(\frac{1}{2}+z_{8}\right)c\,\mathbf{\hat{z}}$ | (8$b$) | Ni VI |
| $\mathbf{B}_{20}$ | = | $(-x_{8}-y_{8})\,\mathbf{a}_{1}+(-x_{8}+y_{8})\,\mathbf{a}_{2}+z_{8}\,\mathbf{a}_{3}$ | = | $-x_{8}a\,\mathbf{\hat{x}}+y_{8}b\,\mathbf{\hat{y}}+z_{8}c\,\mathbf{\hat{z}}$ | (8$b$) | Ni VI |
| $\mathbf{B}_{21}$ | = | $(x_{9}-y_{9})\,\mathbf{a}_{1}+(x_{9}+y_{9})\,\mathbf{a}_{2}+z_{9}\,\mathbf{a}_{3}$ | = | $x_{9}a\,\mathbf{\hat{x}}+y_{9}b\,\mathbf{\hat{y}}+z_{9}c\,\mathbf{\hat{z}}$ | (8$b$) | Ni VII |
| $\mathbf{B}_{22}$ | = | $(-x_{9}+y_{9})\,\mathbf{a}_{1}+(-x_{9}-y_{9})\,\mathbf{a}_{2}+\left(\frac{1}{2}+z_{9}\right)\mathbf{a}_{3}$ | = | $-x_{9}a\,\mathbf{\hat{x}}-y_{9}b\,\mathbf{\hat{y}}+\left(\frac{1}{2}+z_{9}\right)c\,\mathbf{\hat{z}}$ | (8$b$) | Ni VII |
| $\mathbf{B}_{23}$ | = | $(x_{9}+y_{9})\,\mathbf{a}_{1}+(x_{9}-y_{9})\,\mathbf{a}_{2}+\left(\frac{1}{2}+z_{9}\right)\mathbf{a}_{3}$ | = | $x_{9}a\,\mathbf{\hat{x}}-y_{9}b\,\mathbf{\hat{y}}+\left(\frac{1}{2}+z_{9}\right)c\,\mathbf{\hat{z}}$ | (8$b$) | Ni VII |
| $\mathbf{B}_{24}$ | = | $(-x_{9}-y_{9})\,\mathbf{a}_{1}+(-x_{9}+y_{9})\,\mathbf{a}_{2}+z_{9}\,\mathbf{a}_{3}$ | = | $-x_{9}a\,\mathbf{\hat{x}}+y_{9}b\,\mathbf{\hat{y}}+z_{9}c\,\mathbf{\hat{z}}$ | (8$b$) | Ni VII |
| $\mathbf{B}_{25}$ | = | $(x_{10}-y_{10})\,\mathbf{a}_{1}+(x_{10}+y_{10})\,\mathbf{a}_{2}+z_{10}\,\mathbf{a}_{3}$ | = | $x_{10}a\,\mathbf{\hat{x}}+y_{10}b\,\mathbf{\hat{y}}+z_{10}c\,\mathbf{\hat{z}}$ | (8$b$) | Ni VIII |
| $\mathbf{B}_{26}$ | = | $(-x_{10}+y_{10})\,\mathbf{a}_{1}+(-x_{10}-y_{10})\,\mathbf{a}_{2}+\left(\frac{1}{2}+z_{10}\right)\mathbf{a}_{3}$ | = | $-x_{10}a\,\mathbf{\hat{x}}-y_{10}b\,\mathbf{\hat{y}}+\left(\frac{1}{2}+z_{10}\right)c\,\mathbf{\hat{z}}$ | (8$b$) | Ni VIII |
| $\mathbf{B}_{27}$ | = | $(x_{10}+y_{10})\,\mathbf{a}_{1}+(x_{10}-y_{10})\,\mathbf{a}_{2}+\left(\frac{1}{2}+z_{10}\right)\mathbf{a}_{3}$ | = | $x_{10}a\,\mathbf{\hat{x}}-y_{10}b\,\mathbf{\hat{y}}+\left(\frac{1}{2}+z_{10}\right)c\,\mathbf{\hat{z}}$ | (8$b$) | Ni VIII |
| $\mathbf{B}_{28}$ | = | $(-x_{10}-y_{10})\,\mathbf{a}_{1}+(-x_{10}+y_{10})\,\mathbf{a}_{2}+z_{10}\,\mathbf{a}_{3}$ | = | $-x_{10}a\,\mathbf{\hat{x}}+y_{10}b\,\mathbf{\hat{y}}+z_{10}c\,\mathbf{\hat{z}}$ | (8$b$) | Ni VIII |
| $\mathbf{B}_{29}$ | = | $(x_{11}-y_{11})\,\mathbf{a}_{1}+(x_{11}+y_{11})\,\mathbf{a}_{2}+z_{11}\,\mathbf{a}_{3}$ | = | $x_{11}a\,\mathbf{\hat{x}}+y_{11}b\,\mathbf{\hat{y}}+z_{11}c\,\mathbf{\hat{z}}$ | (8$b$) | Si III |
| $\mathbf{B}_{30}$ | = | $(-x_{11}+y_{11})\,\mathbf{a}_{1}+(-x_{11}-y_{11})\,\mathbf{a}_{2}+\left(\frac{1}{2}+z_{11}\right)\mathbf{a}_{3}$ | = | $-x_{11}a\,\mathbf{\hat{x}}-y_{11}b\,\mathbf{\hat{y}}+\left(\frac{1}{2}+z_{11}\right)c\,\mathbf{\hat{z}}$ | (8$b$) | Si III |
| $\mathbf{B}_{31}$ | = | $(x_{11}+y_{11})\,\mathbf{a}_{1}+(x_{11}-y_{11})\,\mathbf{a}_{2}+\left(\frac{1}{2}+z_{11}\right)\mathbf{a}_{3}$ | = | $x_{11}a\,\mathbf{\hat{x}}-y_{11}b\,\mathbf{\hat{y}}+\left(\frac{1}{2}+z_{11}\right)c\,\mathbf{\hat{z}}$ | (8$b$) | Si III |
| $\mathbf{B}_{32}$ | = | $(-x_{11}-y_{11})\,\mathbf{a}_1+(-x_{11}+y_{11})\,\mathbf{a}_2+z_{11}\,\mathbf{a}_3$ | = | $-x_{11}a\,\hat{\mathbf{x}}+y_{11}b\,\hat{\mathbf{y}}+z_{11}c\,\hat{\mathbf{z}}$ | (8$b$) | Si III |
| $\mathbf{B}_{33}$ | = | $(x_{12}-y_{12})\,\mathbf{a}_1+(x_{12}+y_{12})\,\mathbf{a}_2+z_{12}\,\mathbf{a}_3$ | = | $x_{12}a\,\hat{\mathbf{x}}+y_{12}b\,\hat{\mathbf{y}}+z_{12}c\,\hat{\mathbf{z}}$ | (8$b$) | Si IV |
| $\mathbf{B}_{34}$ | = | $(-x_{12}+y_{12})\,\mathbf{a}_1+(-x_{12}-y_{12})\,\mathbf{a}_2+\left(\frac{1}{2}+z_{12}\right)\mathbf{a}_3$ | = | $-x_{12}a\,\hat{\mathbf{x}}-y_{12}b\,\hat{\mathbf{y}}+\left(\frac{1}{2}+z_{12}\right)c\,\hat{\mathbf{z}}$ | (8$b$) | Si IV |
| $\mathbf{B}_{35}$ | = | $(x_{12}+y_{12})\,\mathbf{a}_1+(x_{12}-y_{12})\,\mathbf{a}_2+\left(\frac{1}{2}+z_{12}\right)\mathbf{a}_3$ | = | $x_{12}a\,\hat{\mathbf{x}}-y_{12}b\,\hat{\mathbf{y}}+\left(\frac{1}{2}+z_{12}\right)c\,\hat{\mathbf{z}}$ | (8$b$) | Si IV |
| $\mathbf{B}_{36}$ | = | $(-x_{12}-y_{12})\,\mathbf{a}_1+(-x_{12}+y_{12})\,\mathbf{a}_2+z_{12}\,\mathbf{a}_3$ | = | $-x_{12}a\,\hat{\mathbf{x}}+y_{12}b\,\hat{\mathbf{y}}+z_{12}c\,\hat{\mathbf{z}}$ | (8$b$) | Si IV |
| $\mathbf{B}_{37}$ | = | $(x_{13}-y_{13})\,\mathbf{a}_1+(x_{13}+y_{13})\,\mathbf{a}_2+z_{13}\,\mathbf{a}_3$ | = | $x_{13}a\,\hat{\mathbf{x}}+y_{13}b\,\hat{\mathbf{y}}+z_{13}c\,\hat{\mathbf{z}}$ | (8$b$) | Si V |
| $\mathbf{B}_{38}$ | = | $(-x_{13}+y_{13})\,\mathbf{a}_1+(-x_{13}-y_{13})\,\mathbf{a}_2+\left(\frac{1}{2}+z_{13}\right)\mathbf{a}_3$ | = | $-x_{13}a\,\hat{\mathbf{x}}-y_{13}b\,\hat{\mathbf{y}}+\left(\frac{1}{2}+z_{13}\right)c\,\hat{\mathbf{z}}$ | (8$b$) | Si V |
| $\mathbf{B}_{39}$ | = | $(x_{13}+y_{13})\,\mathbf{a}_1+(x_{13}-y_{13})\,\mathbf{a}_2+\left(\frac{1}{2}+z_{13}\right)\mathbf{a}_3$ | = | $x_{13}a\,\hat{\mathbf{x}}-y_{13}b\,\hat{\mathbf{y}}+\left(\frac{1}{2}+z_{13}\right)c\,\hat{\mathbf{z}}$ | (8$b$) | Si V |
| $\mathbf{B}_{40}$ | = | $(-x_{13}-y_{13})\,\mathbf{a}_1+(-x_{13}+y_{13})\,\mathbf{a}_2+z_{13}\,\mathbf{a}_3$ | = | $-x_{13}a\,\hat{\mathbf{x}}+y_{13}b\,\hat{\mathbf{y}}+z_{13}c\,\hat{\mathbf{z}}$ | (8$b$) | Si V |

**Geometry files:**

- CIF: pp. 1595
- POSCAR: pp. 1595

# Bertrandite ($Be_4Si_2O_7(OH)_2$, $S4_6$) Structure: A4B7C2D2_oC60_36_2b_a3b_2a_b

http://aflow.org/prototype-encyclopedia/A4B7C2D2_oC60_36_2b_a3b_2a_b

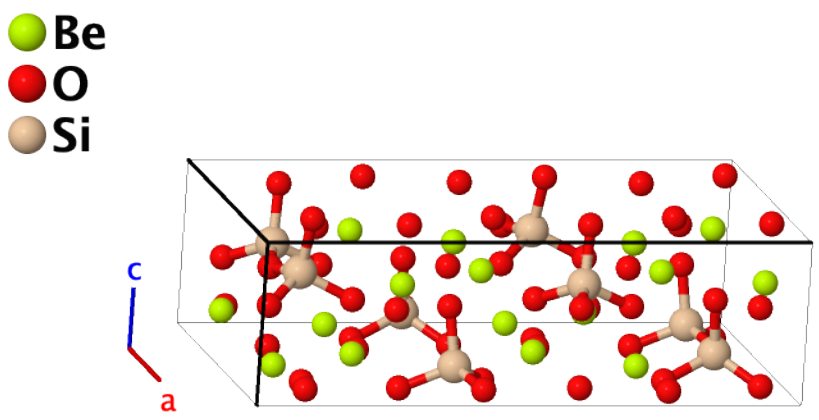

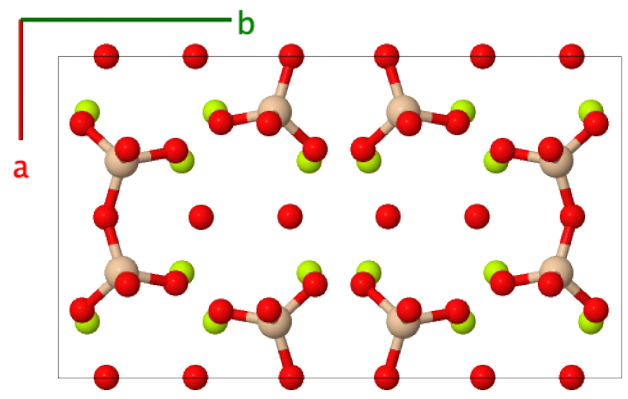

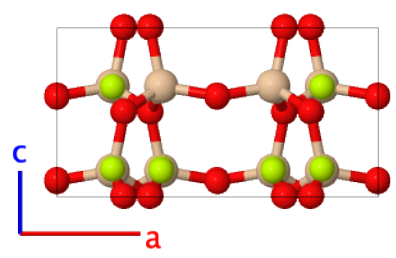

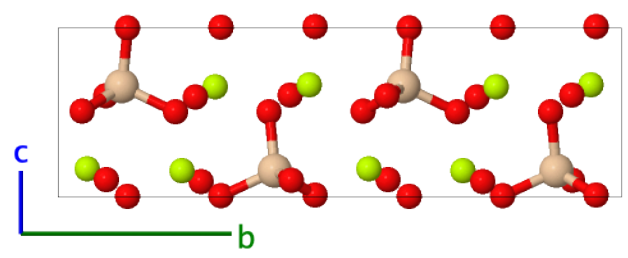

| | | |
|---|---|---|
| **Prototype** | : | $Be_4O_7(OH)_2Si_2$ |
| **AFLOW prototype label** | : | A4B7C2D2_oC60_36_2b_a3b_2a_b |
| ***Strukturbericht* designation** | : | $S4_6$ |
| **Pearson symbol** | : | oC60 |
| **Space group number** | : | 36 |
| **Space group symbol** | : | $Cmc2_1$ |
| **AFLOW prototype command** | : | `aflow --proto=A4B7C2D2_oC60_36_2b_a3b_2a_b` <br> `--params=`$a, b/a, c/a, y_1, z_1, y_2, z_2, y_3, z_3, x_4, y_4, z_4, x_5, y_5, z_5, x_6, y_6, z_6, x_7, y_7, z_7, x_8, y_8, z_8, x_9, y_9, z_9$ |

- All sources agree that the space group for Bertrandite is $Cmc2_1$ #36 but differ slightly in details. We have chosen to use (Hazen, 1986) as the prototype over (Ito, 1932), which was originally designated $S4_6$, and (Solov'eva, 1965), which has unrealistically large Be-O distances.
- No paper locates the hydrogen atoms in the hydroxide ions, which may indicate that these ions can freely rotate.
- We use the data taken at 1 atmosphere.

**Base-centered Orthorhombic primitive vectors:**

$$\mathbf{a}_1 = \frac{1}{2}\,a\,\hat{\mathbf{x}} - \frac{1}{2}\,b\,\hat{\mathbf{y}}$$
$$\mathbf{a}_2 = \frac{1}{2}\,a\,\hat{\mathbf{x}} + \frac{1}{2}\,b\,\hat{\mathbf{y}}$$
$$\mathbf{a}_3 = c\,\hat{\mathbf{z}}$$

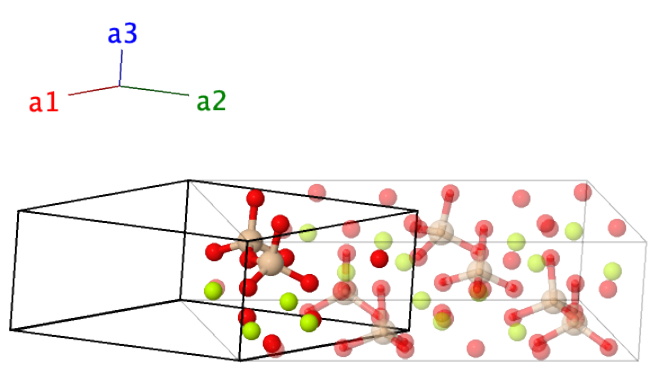

**Basis vectors:**

| | | Lattice Coordinates | | Cartesian Coordinates | Wyckoff Position | Atom Type |
|---|---|---|---|---|---|---|
| $\mathbf{B}_1$ | = | $-y_1\,\mathbf{a}_1 + y_1\,\mathbf{a}_2 + z_1\,\mathbf{a}_3$ | = | $y_1 b\,\hat{\mathbf{y}} + z_1 c\,\hat{\mathbf{z}}$ | (4$a$) | O I |
| $\mathbf{B}_2$ | = | $y_1\,\mathbf{a}_1 - y_1\,\mathbf{a}_2 + \left(\frac{1}{2} + z_1\right)\mathbf{a}_3$ | = | $-y_1 b\,\hat{\mathbf{y}} + \left(\frac{1}{2} + z_1\right) c\,\hat{\mathbf{z}}$ | (4$a$) | O I |
| $\mathbf{B}_3$ | = | $-y_2\,\mathbf{a}_1 + y_2\,\mathbf{a}_2 + z_2\,\mathbf{a}_3$ | = | $y_2 b\,\hat{\mathbf{y}} + z_2 c\,\hat{\mathbf{z}}$ | (4$a$) | OH I |
| $\mathbf{B}_4$ | = | $y_2\,\mathbf{a}_1 - y_2\,\mathbf{a}_2 + \left(\frac{1}{2} + z_2\right)\mathbf{a}_3$ | = | $-y_2 b\,\hat{\mathbf{y}} + \left(\frac{1}{2} + z_2\right) c\,\hat{\mathbf{z}}$ | (4$a$) | OH I |
| $\mathbf{B}_5$ | = | $-y_3\,\mathbf{a}_1 + y_3\,\mathbf{a}_2 + z_3\,\mathbf{a}_3$ | = | $y_3 b\,\hat{\mathbf{y}} + z_3 c\,\hat{\mathbf{z}}$ | (4$a$) | OH II |
| $\mathbf{B}_6$ | = | $y_3\,\mathbf{a}_1 - y_3\,\mathbf{a}_2 + \left(\frac{1}{2} + z_3\right)\mathbf{a}_3$ | = | $-y_3 b\,\hat{\mathbf{y}} + \left(\frac{1}{2} + z_3\right) c\,\hat{\mathbf{z}}$ | (4$a$) | OH II |
| $\mathbf{B}_7$ | = | $(x_4 - y_4)\,\mathbf{a}_1 + (x_4 + y_4)\,\mathbf{a}_2 + z_4\,\mathbf{a}_3$ | = | $x_4 a\,\hat{\mathbf{x}} + y_4 b\,\hat{\mathbf{y}} + z_4 c\,\hat{\mathbf{z}}$ | (8$b$) | Be I |
| $\mathbf{B}_8$ | = | $(-x_4 + y_4)\,\mathbf{a}_1 + (-x_4 - y_4)\,\mathbf{a}_2 + \left(\frac{1}{2} + z_4\right)\mathbf{a}_3$ | = | $-x_4 a\,\hat{\mathbf{x}} - y_4 b\,\hat{\mathbf{y}} + \left(\frac{1}{2} + z_4\right) c\,\hat{\mathbf{z}}$ | (8$b$) | Be I |
| $\mathbf{B}_9$ | = | $(x_4 + y_4)\,\mathbf{a}_1 + (x_4 - y_4)\,\mathbf{a}_2 + \left(\frac{1}{2} + z_4\right)\mathbf{a}_3$ | = | $x_4 a\,\hat{\mathbf{x}} - y_4 b\,\hat{\mathbf{y}} + \left(\frac{1}{2} + z_4\right) c\,\hat{\mathbf{z}}$ | (8$b$) | Be I |
| $\mathbf{B}_{10}$ | = | $(-x_4 - y_4)\,\mathbf{a}_1 + (-x_4 + y_4)\,\mathbf{a}_2 + z_4\,\mathbf{a}_3$ | = | $-x_4 a\,\hat{\mathbf{x}} + y_4 b\,\hat{\mathbf{y}} + z_4 c\,\hat{\mathbf{z}}$ | (8$b$) | Be I |
| $\mathbf{B}_{11}$ | = | $(x_5 - y_5)\,\mathbf{a}_1 + (x_5 + y_5)\,\mathbf{a}_2 + z_5\,\mathbf{a}_3$ | = | $x_5 a\,\hat{\mathbf{x}} + y_5 b\,\hat{\mathbf{y}} + z_5 c\,\hat{\mathbf{z}}$ | (8$b$) | Be II |
| $\mathbf{B}_{12}$ | = | $(-x_5 + y_5)\,\mathbf{a}_1 + (-x_5 - y_5)\,\mathbf{a}_2 + \left(\frac{1}{2} + z_5\right)\mathbf{a}_3$ | = | $-x_5 a\,\hat{\mathbf{x}} - y_5 b\,\hat{\mathbf{y}} + \left(\frac{1}{2} + z_5\right) c\,\hat{\mathbf{z}}$ | (8$b$) | Be II |
| $\mathbf{B}_{13}$ | = | $(x_5 + y_5)\,\mathbf{a}_1 + (x_5 - y_5)\,\mathbf{a}_2 + \left(\frac{1}{2} + z_5\right)\mathbf{a}_3$ | = | $x_5 a\,\hat{\mathbf{x}} - y_5 b\,\hat{\mathbf{y}} + \left(\frac{1}{2} + z_5\right) c\,\hat{\mathbf{z}}$ | (8$b$) | Be II |
| $\mathbf{B}_{14}$ | = | $(-x_5 - y_5)\,\mathbf{a}_1 + (-x_5 + y_5)\,\mathbf{a}_2 + z_5\,\mathbf{a}_3$ | = | $-x_5 a\,\hat{\mathbf{x}} + y_5 b\,\hat{\mathbf{y}} + z_5 c\,\hat{\mathbf{z}}$ | (8$b$) | Be II |
| $\mathbf{B}_{15}$ | = | $(x_6 - y_6)\,\mathbf{a}_1 + (x_6 + y_6)\,\mathbf{a}_2 + z_6\,\mathbf{a}_3$ | = | $x_6 a\,\hat{\mathbf{x}} + y_6 b\,\hat{\mathbf{y}} + z_6 c\,\hat{\mathbf{z}}$ | (8$b$) | O II |
| $\mathbf{B}_{16}$ | = | $(-x_6 + y_6)\,\mathbf{a}_1 + (-x_6 - y_6)\,\mathbf{a}_2 + \left(\frac{1}{2} + z_6\right)\mathbf{a}_3$ | = | $-x_6 a\,\hat{\mathbf{x}} - y_6 b\,\hat{\mathbf{y}} + \left(\frac{1}{2} + z_6\right) c\,\hat{\mathbf{z}}$ | (8$b$) | O II |
| $\mathbf{B}_{17}$ | = | $(x_6 + y_6)\,\mathbf{a}_1 + (x_6 - y_6)\,\mathbf{a}_2 + \left(\frac{1}{2} + z_6\right)\mathbf{a}_3$ | = | $x_6 a\,\hat{\mathbf{x}} - y_6 b\,\hat{\mathbf{y}} + \left(\frac{1}{2} + z_6\right) c\,\hat{\mathbf{z}}$ | (8$b$) | O II |
| $\mathbf{B}_{18}$ | = | $(-x_6 - y_6)\,\mathbf{a}_1 + (-x_6 + y_6)\,\mathbf{a}_2 + z_6\,\mathbf{a}_3$ | = | $-x_6 a\,\hat{\mathbf{x}} + y_6 b\,\hat{\mathbf{y}} + z_6 c\,\hat{\mathbf{z}}$ | (8$b$) | O II |
| $\mathbf{B}_{19}$ | = | $(x_7 - y_7)\,\mathbf{a}_1 + (x_7 + y_7)\,\mathbf{a}_2 + z_7\,\mathbf{a}_3$ | = | $x_7 a\,\hat{\mathbf{x}} + y_7 b\,\hat{\mathbf{y}} + z_7 c\,\hat{\mathbf{z}}$ | (8$b$) | O III |
| $\mathbf{B}_{20}$ | = | $(-x_7 + y_7)\,\mathbf{a}_1 + (-x_7 - y_7)\,\mathbf{a}_2 + \left(\frac{1}{2} + z_7\right)\mathbf{a}_3$ | = | $-x_7 a\,\hat{\mathbf{x}} - y_7 b\,\hat{\mathbf{y}} + \left(\frac{1}{2} + z_7\right) c\,\hat{\mathbf{z}}$ | (8$b$) | O III |
| $\mathbf{B}_{21}$ | = | $(x_7 + y_7)\,\mathbf{a}_1 + (x_7 - y_7)\,\mathbf{a}_2 + \left(\frac{1}{2} + z_7\right)\mathbf{a}_3$ | = | $x_7 a\,\hat{\mathbf{x}} - y_7 b\,\hat{\mathbf{y}} + \left(\frac{1}{2} + z_7\right) c\,\hat{\mathbf{z}}$ | (8$b$) | O III |
| $\mathbf{B}_{22}$ | = | $(-x_7 - y_7)\,\mathbf{a}_1 + (-x_7 + y_7)\,\mathbf{a}_2 + z_7\,\mathbf{a}_3$ | = | $-x_7 a\,\hat{\mathbf{x}} + y_7 b\,\hat{\mathbf{y}} + z_7 c\,\hat{\mathbf{z}}$ | (8$b$) | O III |
| $\mathbf{B}_{23}$ | = | $(x_8 - y_8)\,\mathbf{a}_1 + (x_8 + y_8)\,\mathbf{a}_2 + z_8\,\mathbf{a}_3$ | = | $x_8 a\,\hat{\mathbf{x}} + y_8 b\,\hat{\mathbf{y}} + z_8 c\,\hat{\mathbf{z}}$ | (8$b$) | O IV |
| $\mathbf{B}_{24}$ | = | $(-x_8 + y_8)\,\mathbf{a}_1 + (-x_8 - y_8)\,\mathbf{a}_2 + \left(\frac{1}{2} + z_8\right)\mathbf{a}_3$ | = | $-x_8 a\,\hat{\mathbf{x}} - y_8 b\,\hat{\mathbf{y}} + \left(\frac{1}{2} + z_8\right) c\,\hat{\mathbf{z}}$ | (8$b$) | O IV |
| $\mathbf{B}_{25}$ | = | $(x_8 + y_8)\,\mathbf{a}_1 + (x_8 - y_8)\,\mathbf{a}_2 + \left(\frac{1}{2} + z_8\right)\mathbf{a}_3$ | = | $x_8 a\,\hat{\mathbf{x}} - y_8 b\,\hat{\mathbf{y}} + \left(\frac{1}{2} + z_8\right) c\,\hat{\mathbf{z}}$ | (8$b$) | O IV |
| $\mathbf{B}_{26}$ | = | $(-x_8 - y_8)\,\mathbf{a}_1 + (-x_8 + y_8)\,\mathbf{a}_2 + z_8\,\mathbf{a}_3$ | = | $-x_8 a\,\hat{\mathbf{x}} + y_8 b\,\hat{\mathbf{y}} + z_8 c\,\hat{\mathbf{z}}$ | (8$b$) | O IV |
| $\mathbf{B}_{27}$ | = | $(x_9 - y_9)\,\mathbf{a}_1 + (x_9 + y_9)\,\mathbf{a}_2 + z_9\,\mathbf{a}_3$ | = | $x_9 a\,\hat{\mathbf{x}} + y_9 b\,\hat{\mathbf{y}} + z_9 c\,\hat{\mathbf{z}}$ | (8$b$) | Si |
| $\mathbf{B}_{28}$ | = | $(-x_9 + y_9)\,\mathbf{a}_1 + (-x_9 - y_9)\,\mathbf{a}_2 + \left(\frac{1}{2} + z_9\right)\mathbf{a}_3$ | = | $-x_9 a\,\hat{\mathbf{x}} - y_9 b\,\hat{\mathbf{y}} + \left(\frac{1}{2} + z_9\right) c\,\hat{\mathbf{z}}$ | (8$b$) | Si |
| $\mathbf{B}_{29}$ | = | $(x_9 + y_9)\,\mathbf{a}_1 + (x_9 - y_9)\,\mathbf{a}_2 + \left(\frac{1}{2} + z_9\right)\mathbf{a}_3$ | = | $x_9 a\,\hat{\mathbf{x}} - y_9 b\,\hat{\mathbf{y}} + \left(\frac{1}{2} + z_9\right) c\,\hat{\mathbf{z}}$ | (8$b$) | Si |
| $\mathbf{B}_{30}$ | = | $(-x_9 - y_9)\,\mathbf{a}_1 + (-x_9 + y_9)\,\mathbf{a}_2 + z_9\,\mathbf{a}_3$ | = | $-x_9 a\,\hat{\mathbf{x}} + y_9 b\,\hat{\mathbf{y}} + z_9 c\,\hat{\mathbf{z}}$ | (8$b$) | Si |

**Found in:**

---

**Geometry files:**

- CIF: pp. 1596
- POSCAR: pp. 1596

# $MoP_2$ Structure: AB2_oC12_36_a_2a

http://aflow.org/prototype-encyclopedia/AB2_oC12_36_a_2a

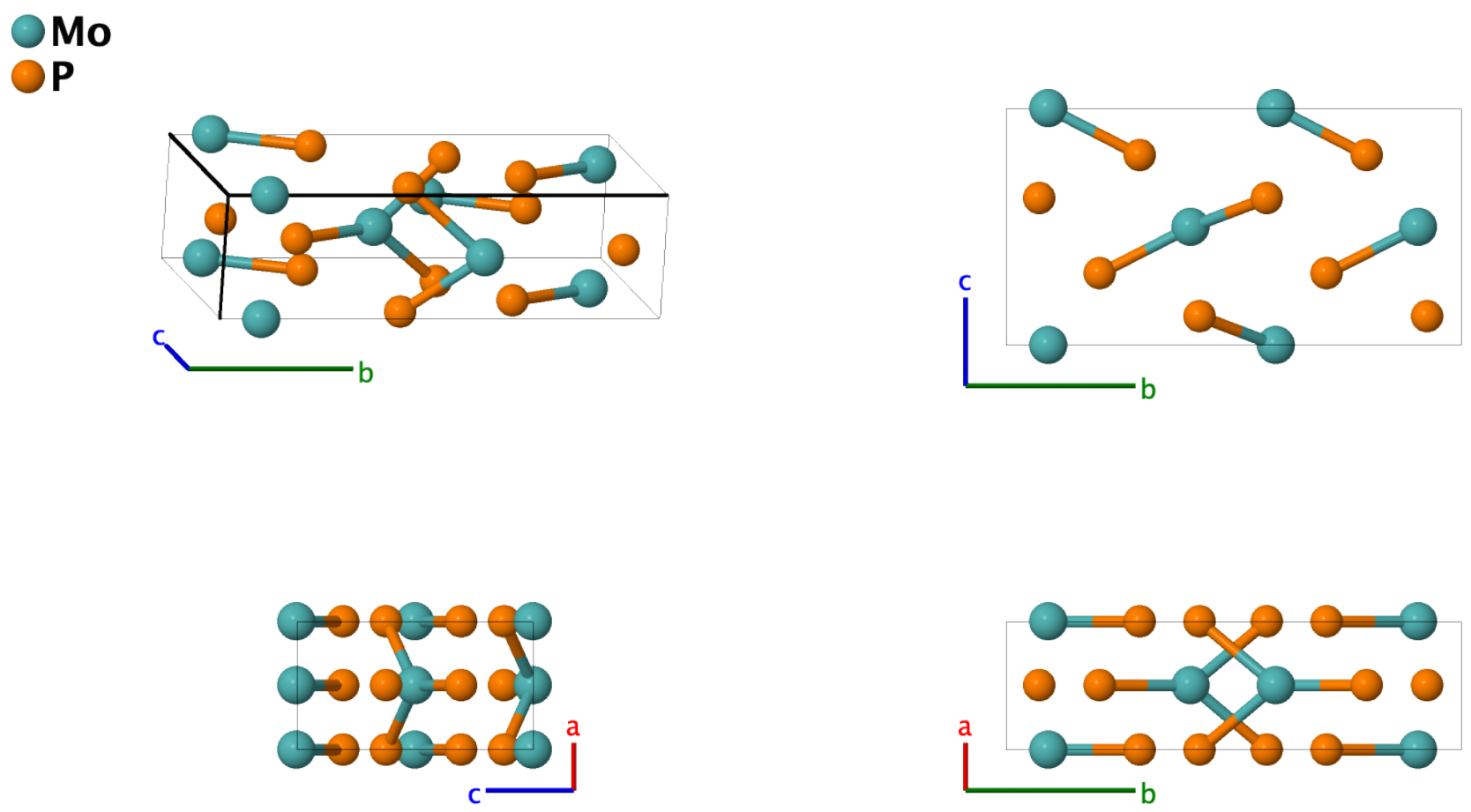

| | | |
|---|---|---|
| **Prototype** | : | $MoP_2$ |
| **AFLOW prototype label** | : | AB2_oC12_36_a_2a |
| ***Strukturbericht* designation** | : | None |
| **Pearson symbol** | : | oC12 |
| **Space group number** | : | 36 |
| **Space group symbol** | : | $Cmc2_1$ |
| **AFLOW prototype command** | : | `aflow --proto=AB2_oC12_36_a_2a`<br>`--params=`$a, b/a, c/a, y_1, z_1, y_2, z_2, y_3, z_3$ |

**Other compounds with this structure**

- $WP_2$

**Base-centered Orthorhombic primitive vectors:**

$$\mathbf{a}_1 = \frac{1}{2}\,a\,\hat{\mathbf{x}} - \frac{1}{2}\,b\,\hat{\mathbf{y}}$$
$$\mathbf{a}_2 = \frac{1}{2}\,a\,\hat{\mathbf{x}} + \frac{1}{2}\,b\,\hat{\mathbf{y}}$$
$$\mathbf{a}_3 = c\,\hat{\mathbf{z}}$$

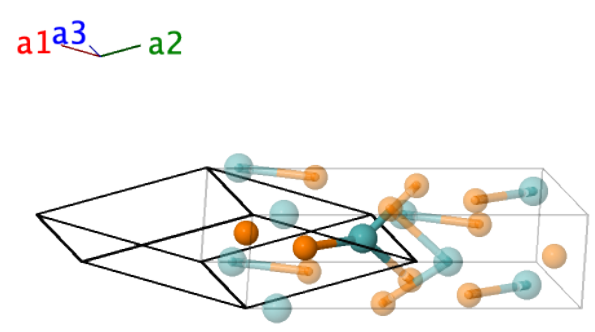

**Basis vectors:**

| | | Lattice Coordinates | | Cartesian Coordinates | Wyckoff Position | Atom Type |
|---|---|---|---|---|---|---|
| $\mathbf{B}_1$ | = | $-y_1\,\mathbf{a}_1 + y_1\,\mathbf{a}_2 + z_1\,\mathbf{a}_3$ | = | $y_1 b\,\hat{\mathbf{y}} + z_1 c\,\hat{\mathbf{z}}$ | (4a) | Mo |
| $\mathbf{B}_2$ | = | $y_1\,\mathbf{a}_1 - y_1\,\mathbf{a}_2 + \left(\frac{1}{2} + z_1\right)\mathbf{a}_3$ | = | $-y_1 b\,\hat{\mathbf{y}} + \left(\frac{1}{2} + z_1\right) c\,\hat{\mathbf{z}}$ | (4a) | Mo |
| $\mathbf{B}_3$ | = | $-y_2\,\mathbf{a}_1 + y_2\,\mathbf{a}_2 + z_2\,\mathbf{a}_3$ | = | $y_2 b\,\hat{\mathbf{y}} + z_2 c\,\hat{\mathbf{z}}$ | (4a) | P I |
| $\mathbf{B}_4$ | = | $y_2\,\mathbf{a}_1 - y_2\,\mathbf{a}_2 + \left(\frac{1}{2} + z_2\right)\mathbf{a}_3$ | = | $-y_2 b\,\hat{\mathbf{y}} + \left(\frac{1}{2} + z_2\right) c\,\hat{\mathbf{z}}$ | (4a) | P I |
| $\mathbf{B}_5$ | = | $-y_3\,\mathbf{a}_1 + y_3\,\mathbf{a}_2 + z_3\,\mathbf{a}_3$ | = | $y_3 b\,\hat{\mathbf{y}} + z_3 c\,\hat{\mathbf{z}}$ | (4a) | P II |
| $\mathbf{B}_6$ | = | $y_3\,\mathbf{a}_1 - y_3\,\mathbf{a}_2 + \left(\frac{1}{2} + z_3\right)\mathbf{a}_3$ | = | $-y_3 b\,\hat{\mathbf{y}} + \left(\frac{1}{2} + z_3\right) c\,\hat{\mathbf{z}}$ | (4a) | P II |

**Geometry files:**

- CIF: pp. 1596
- POSCAR: pp. 1597

# $\alpha$-Potassium Nitrate ($KNO_3$) II Structure: ABC3_oC80_36_2ab_2ab_2a5b

http://aflow.org/prototype-encyclopedia/ABC3_oC80_36_2ab_2ab_2a5b

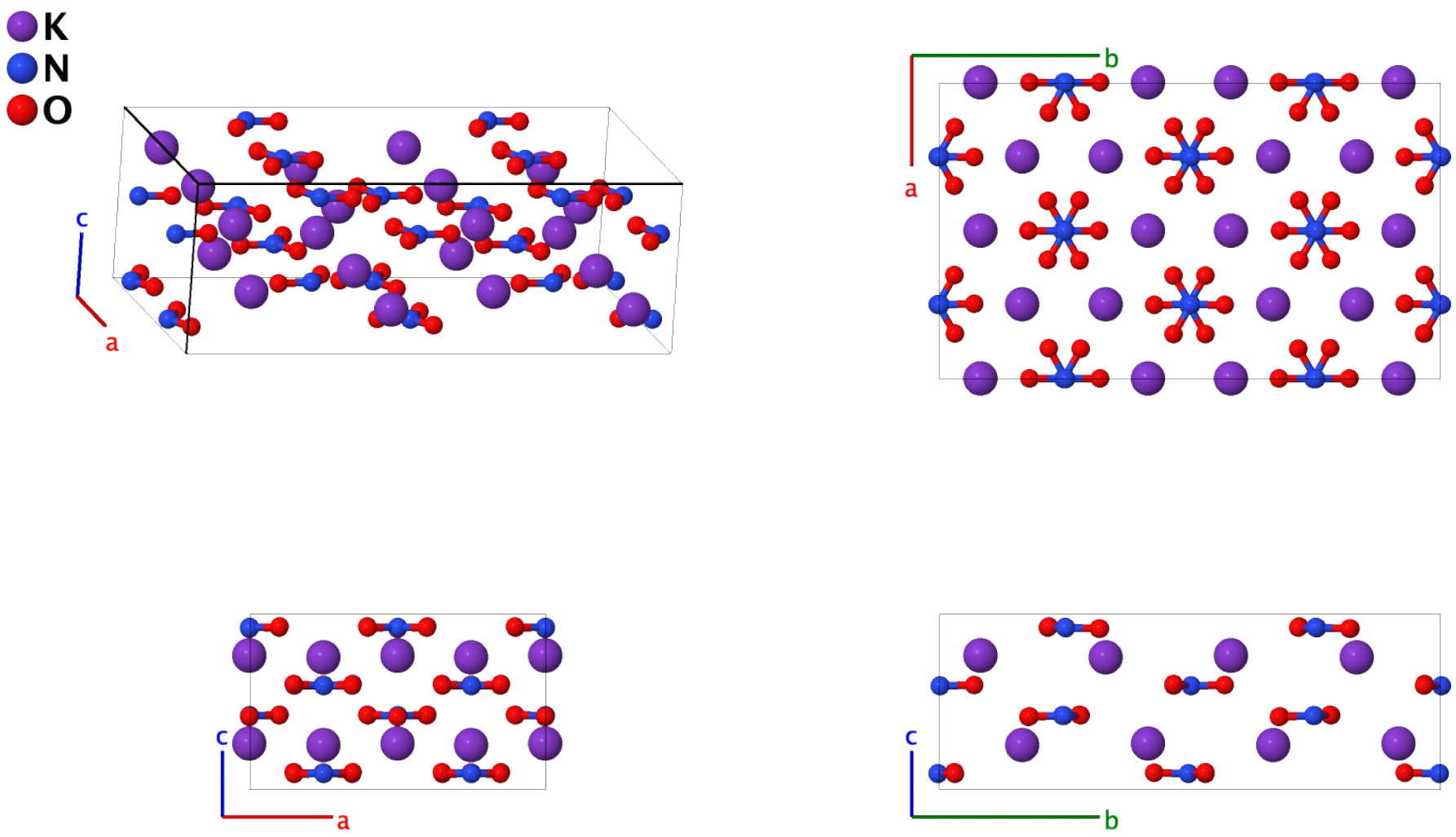

| | | |
|---|---|---|
| **Prototype** | : | $KNO_3$ |
| **AFLOW prototype label** | : | ABC3_oC80_36_2ab_2ab_2a5b |
| ***Strukturbericht* designation** | : | None |
| **Pearson symbol** | : | oC80 |
| **Space group number** | : | 36 |
| **Space group symbol** | : | $Cmc2_1$ |
| **AFLOW prototype command** | : | `aflow --proto=ABC3_oC80_36_2ab_2ab_2a5b`<br>`--params=`$a, b/a, c/a, y_1, z_1, y_2, z_2, y_3, z_3, y_4, z_4, y_5, z_5, y_6, z_6, x_7, y_7, z_7, x_8, y_8, z_8, x_9, y_9, z_9, x_{10}, y_{10}, z_{10}, x_{11}, y_{11}, z_{11}, x_{12}, y_{12}, z_{12}, x_{13}, y_{13}, z_{13}$ |

- Two possible structures have been identified for $\alpha$-$KNO_3$: (Nimmo, 1973) proposed an orthorhombic structure in space group *Pnma* #62, which we call "Structure I". (Adiwidjaja, 2003) found that structure, but also noted that it could be described by a doubling of the unit cell into a superstructure of type I, which we call "Structure II" and present on this page. It is unclear to us which structure is correct, so we present both. However, we do note that if we allow some uncertainty in the atoms of this structure, AFLOW-SYM and FINDSYM show it to be identical to Structure I.
- On heating, $\alpha$-$KNO_3$ transforms into $\beta$-$KNO_3$ at 128 °C. When heated above 200°C and then cooled, the $\beta$ phase transforms into the metastable ferroelectric $\gamma$-$KNO_3$ phase, which remains metastable at room temperature.

**Base-centered Orthorhombic primitive vectors:**

$$\mathbf{a}_1 = \frac{1}{2}\,a\,\hat{\mathbf{x}} - \frac{1}{2}\,b\,\hat{\mathbf{y}}$$
$$\mathbf{a}_2 = \frac{1}{2}\,a\,\hat{\mathbf{x}} + \frac{1}{2}\,b\,\hat{\mathbf{y}}$$
$$\mathbf{a}_3 = c\,\hat{\mathbf{z}}$$

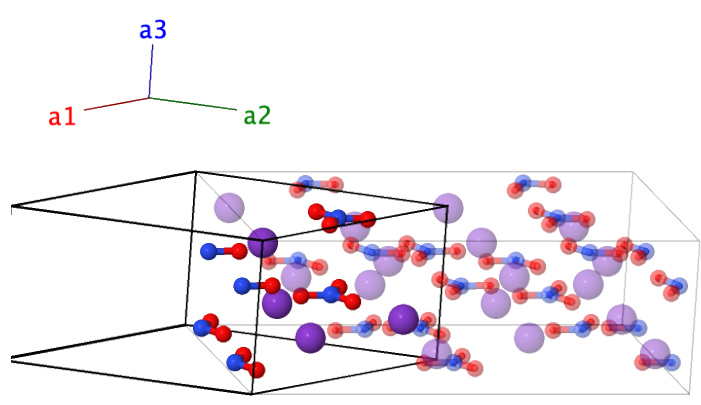

**Basis vectors:**

| | | Lattice Coordinates | | Cartesian Coordinates | Wyckoff Position | Atom Type |
|---|---|---|---|---|---|---|
| $\mathbf{B}_1$ | = | $-y_1\,\mathbf{a}_1 + y_1\,\mathbf{a}_2 + z_1\,\mathbf{a}_3$ | = | $y_1 b\,\hat{\mathbf{y}} + z_1 c\,\hat{\mathbf{z}}$ | (4$a$) | K I |
| $\mathbf{B}_2$ | = | $y_1\,\mathbf{a}_1 - y_1\,\mathbf{a}_2 + \left(\frac{1}{2} + z_1\right)\,\mathbf{a}_3$ | = | $-y_1 b\,\hat{\mathbf{y}} + \left(\frac{1}{2} + z_1\right) c\,\hat{\mathbf{z}}$ | (4$a$) | K I |
| $\mathbf{B}_3$ | = | $-y_2\,\mathbf{a}_1 + y_2\,\mathbf{a}_2 + z_2\,\mathbf{a}_3$ | = | $y_2 b\,\hat{\mathbf{y}} + z_2 c\,\hat{\mathbf{z}}$ | (4$a$) | K II |
| $\mathbf{B}_4$ | = | $y_2\,\mathbf{a}_1 - y_2\,\mathbf{a}_2 + \left(\frac{1}{2} + z_2\right)\,\mathbf{a}_3$ | = | $-y_2 b\,\hat{\mathbf{y}} + \left(\frac{1}{2} + z_2\right) c\,\hat{\mathbf{z}}$ | (4$a$) | K II |
| $\mathbf{B}_5$ | = | $-y_3\,\mathbf{a}_1 + y_3\,\mathbf{a}_2 + z_3\,\mathbf{a}_3$ | = | $y_3 b\,\hat{\mathbf{y}} + z_3 c\,\hat{\mathbf{z}}$ | (4$a$) | N I |
| $\mathbf{B}_6$ | = | $y_3\,\mathbf{a}_1 - y_3\,\mathbf{a}_2 + \left(\frac{1}{2} + z_3\right)\,\mathbf{a}_3$ | = | $-y_3 b\,\hat{\mathbf{y}} + \left(\frac{1}{2} + z_3\right) c\,\hat{\mathbf{z}}$ | (4$a$) | N I |
| $\mathbf{B}_7$ | = | $-y_4\,\mathbf{a}_1 + y_4\,\mathbf{a}_2 + z_4\,\mathbf{a}_3$ | = | $y_4 b\,\hat{\mathbf{y}} + z_4 c\,\hat{\mathbf{z}}$ | (4$a$) | N II |
| $\mathbf{B}_8$ | = | $y_4\,\mathbf{a}_1 - y_4\,\mathbf{a}_2 + \left(\frac{1}{2} + z_4\right)\,\mathbf{a}_3$ | = | $-y_4 b\,\hat{\mathbf{y}} + \left(\frac{1}{2} + z_4\right) c\,\hat{\mathbf{z}}$ | (4$a$) | N II |
| $\mathbf{B}_9$ | = | $-y_5\,\mathbf{a}_1 + y_5\,\mathbf{a}_2 + z_5\,\mathbf{a}_3$ | = | $y_5 b\,\hat{\mathbf{y}} + z_5 c\,\hat{\mathbf{z}}$ | (4$a$) | O I |
| $\mathbf{B}_{10}$ | = | $y_5\,\mathbf{a}_1 - y_5\,\mathbf{a}_2 + \left(\frac{1}{2} + z_5\right)\,\mathbf{a}_3$ | = | $-y_5 b\,\hat{\mathbf{y}} + \left(\frac{1}{2} + z_5\right) c\,\hat{\mathbf{z}}$ | (4$a$) | O I |
| $\mathbf{B}_{11}$ | = | $-y_6\,\mathbf{a}_1 + y_6\,\mathbf{a}_2 + z_6\,\mathbf{a}_3$ | = | $y_6 b\,\hat{\mathbf{y}} + z_6 c\,\hat{\mathbf{z}}$ | (4$a$) | O II |
| $\mathbf{B}_{12}$ | = | $y_6\,\mathbf{a}_1 - y_6\,\mathbf{a}_2 + \left(\frac{1}{2} + z_6\right)\,\mathbf{a}_3$ | = | $-y_6 b\,\hat{\mathbf{y}} + \left(\frac{1}{2} + z_6\right) c\,\hat{\mathbf{z}}$ | (4$a$) | O II |
| $\mathbf{B}_{13}$ | = | $(x_7 - y_7)\,\mathbf{a}_1 + (x_7 + y_7)\,\mathbf{a}_2 + z_7\,\mathbf{a}_3$ | = | $x_7 a\,\hat{\mathbf{x}} + y_7 b\,\hat{\mathbf{y}} + z_7 c\,\hat{\mathbf{z}}$ | (8$b$) | K III |
| $\mathbf{B}_{14}$ | = | $(-x_7 + y_7)\,\mathbf{a}_1 + (-x_7 - y_7)\,\mathbf{a}_2 + \left(\frac{1}{2} + z_7\right)\,\mathbf{a}_3$ | = | $-x_7 a\,\hat{\mathbf{x}} - y_7 b\,\hat{\mathbf{y}} + \left(\frac{1}{2} + z_7\right) c\,\hat{\mathbf{z}}$ | (8$b$) | K III |
| $\mathbf{B}_{15}$ | = | $(x_7 + y_7)\,\mathbf{a}_1 + (x_7 - y_7)\,\mathbf{a}_2 + \left(\frac{1}{2} + z_7\right)\,\mathbf{a}_3$ | = | $x_7 a\,\hat{\mathbf{x}} - y_7 b\,\hat{\mathbf{y}} + \left(\frac{1}{2} + z_7\right) c\,\hat{\mathbf{z}}$ | (8$b$) | K III |
| $\mathbf{B}_{16}$ | = | $(-x_7 - y_7)\,\mathbf{a}_1 + (-x_7 + y_7)\,\mathbf{a}_2 + z_7\,\mathbf{a}_3$ | = | $-x_7 a\,\hat{\mathbf{x}} + y_7 b\,\hat{\mathbf{y}} + z_7 c\,\hat{\mathbf{z}}$ | (8$b$) | K III |
| $\mathbf{B}_{17}$ | = | $(x_8 - y_8)\,\mathbf{a}_1 + (x_8 + y_8)\,\mathbf{a}_2 + z_8\,\mathbf{a}_3$ | = | $x_8 a\,\hat{\mathbf{x}} + y_8 b\,\hat{\mathbf{y}} + z_8 c\,\hat{\mathbf{z}}$ | (8$b$) | N III |
| $\mathbf{B}_{18}$ | = | $(-x_8 + y_8)\,\mathbf{a}_1 + (-x_8 - y_8)\,\mathbf{a}_2 + \left(\frac{1}{2} + z_8\right)\,\mathbf{a}_3$ | = | $-x_8 a\,\hat{\mathbf{x}} - y_8 b\,\hat{\mathbf{y}} + \left(\frac{1}{2} + z_8\right) c\,\hat{\mathbf{z}}$ | (8$b$) | N III |
| $\mathbf{B}_{19}$ | = | $(x_8 + y_8)\,\mathbf{a}_1 + (x_8 - y_8)\,\mathbf{a}_2 + \left(\frac{1}{2} + z_8\right)\,\mathbf{a}_3$ | = | $x_8 a\,\hat{\mathbf{x}} - y_8 b\,\hat{\mathbf{y}} + \left(\frac{1}{2} + z_8\right) c\,\hat{\mathbf{z}}$ | (8$b$) | N III |
| $\mathbf{B}_{20}$ | = | $(-x_8 - y_8)\,\mathbf{a}_1 + (-x_8 + y_8)\,\mathbf{a}_2 + z_8\,\mathbf{a}_3$ | = | $-x_8 a\,\hat{\mathbf{x}} + y_8 b\,\hat{\mathbf{y}} + z_8 c\,\hat{\mathbf{z}}$ | (8$b$) | N III |
| $\mathbf{B}_{21}$ | = | $(x_9 - y_9)\,\mathbf{a}_1 + (x_9 + y_9)\,\mathbf{a}_2 + z_9\,\mathbf{a}_3$ | = | $x_9 a\,\hat{\mathbf{x}} + y_9 b\,\hat{\mathbf{y}} + z_9 c\,\hat{\mathbf{z}}$ | (8$b$) | O III |
| $\mathbf{B}_{22}$ | = | $(-x_9 + y_9)\,\mathbf{a}_1 + (-x_9 - y_9)\,\mathbf{a}_2 + \left(\frac{1}{2} + z_9\right)\,\mathbf{a}_3$ | = | $-x_9 a\,\hat{\mathbf{x}} - y_9 b\,\hat{\mathbf{y}} + \left(\frac{1}{2} + z_9\right) c\,\hat{\mathbf{z}}$ | (8$b$) | O III |
| $\mathbf{B}_{23}$ | = | $(x_9 + y_9)\,\mathbf{a}_1 + (x_9 - y_9)\,\mathbf{a}_2 + \left(\frac{1}{2} + z_9\right)\,\mathbf{a}_3$ | = | $x_9 a\,\hat{\mathbf{x}} - y_9 b\,\hat{\mathbf{y}} + \left(\frac{1}{2} + z_9\right) c\,\hat{\mathbf{z}}$ | (8$b$) | O III |
| $\mathbf{B}_{24}$ | = | $(-x_9 - y_9)\,\mathbf{a}_1 + (-x_9 + y_9)\,\mathbf{a}_2 + z_9\,\mathbf{a}_3$ | = | $-x_9 a\,\hat{\mathbf{x}} + y_9 b\,\hat{\mathbf{y}} + z_9 c\,\hat{\mathbf{z}}$ | (8$b$) | O III |
| $\mathbf{B}_{25}$ | = | $(x_{10} - y_{10})\,\mathbf{a}_1 + (x_{10} + y_{10})\,\mathbf{a}_2 + z_{10}\,\mathbf{a}_3$ | = | $x_{10} a\,\hat{\mathbf{x}} + y_{10} b\,\hat{\mathbf{y}} + z_{10} c\,\hat{\mathbf{z}}$ | (8$b$) | O IV |
| $\mathbf{B}_{26}$ | = | $(-x_{10} + y_{10})\,\mathbf{a}_1 + (-x_{10} - y_{10})\,\mathbf{a}_2 + \left(\frac{1}{2} + z_{10}\right)\,\mathbf{a}_3$ | = | $-x_{10} a\,\hat{\mathbf{x}} - y_{10} b\,\hat{\mathbf{y}} + \left(\frac{1}{2} + z_{10}\right) c\,\hat{\mathbf{z}}$ | (8$b$) | O IV |
| $\mathbf{B}_{27}$ | = | $(x_{10} + y_{10})\,\mathbf{a}_1 + (x_{10} - y_{10})\,\mathbf{a}_2 + \left(\frac{1}{2} + z_{10}\right)\,\mathbf{a}_3$ | = | $x_{10} a\,\hat{\mathbf{x}} - y_{10} b\,\hat{\mathbf{y}} + \left(\frac{1}{2} + z_{10}\right) c\,\hat{\mathbf{z}}$ | (8$b$) | O IV |
| $\mathbf{B}_{28}$ | = | $(-x_{10} - y_{10})\,\mathbf{a}_1 + (-x_{10} + y_{10})\,\mathbf{a}_2 + z_{10}\,\mathbf{a}_3$ | = | $-x_{10} a\,\hat{\mathbf{x}} + y_{10} b\,\hat{\mathbf{y}} + z_{10} c\,\hat{\mathbf{z}}$ | (8$b$) | O IV |
| $\mathbf{B}_{29}$ | = | $(x_{11} - y_{11})\,\mathbf{a}_1 + (x_{11} + y_{11})\,\mathbf{a}_2 + z_{11}\,\mathbf{a}_3$ | = | $x_{11} a\,\hat{\mathbf{x}} + y_{11} b\,\hat{\mathbf{y}} + z_{11} c\,\hat{\mathbf{z}}$ | (8$b$) | O V |
| $\mathbf{B}_{30}$ | = | $(-x_{11}+y_{11})\,\mathbf{a}_1+(-x_{11}-y_{11})\,\mathbf{a}_2+\left(\frac{1}{2}+z_{11}\right)\mathbf{a}_3$ | = | $-x_{11}a\,\hat{\mathbf{x}}-y_{11}b\,\hat{\mathbf{y}}+\left(\frac{1}{2}+z_{11}\right)c\,\hat{\mathbf{z}}$ | (8$b$) | O V |
| $\mathbf{B}_{31}$ | = | $(x_{11}+y_{11})\,\mathbf{a}_1+(x_{11}-y_{11})\,\mathbf{a}_2+\left(\frac{1}{2}+z_{11}\right)\mathbf{a}_3$ | = | $x_{11}a\,\hat{\mathbf{x}}-y_{11}b\,\hat{\mathbf{y}}+\left(\frac{1}{2}+z_{11}\right)c\,\hat{\mathbf{z}}$ | (8$b$) | O V |
| $\mathbf{B}_{32}$ | = | $(-x_{11}-y_{11})\,\mathbf{a}_1+(-x_{11}+y_{11})\,\mathbf{a}_2+z_{11}\,\mathbf{a}_3$ | = | $-x_{11}a\,\hat{\mathbf{x}}+y_{11}b\,\hat{\mathbf{y}}+z_{11}c\,\hat{\mathbf{z}}$ | (8$b$) | O V |
| $\mathbf{B}_{33}$ | = | $(x_{12}-y_{12})\,\mathbf{a}_1+(x_{12}+y_{12})\,\mathbf{a}_2+z_{12}\,\mathbf{a}_3$ | = | $x_{12}a\,\hat{\mathbf{x}}+y_{12}b\,\hat{\mathbf{y}}+z_{12}c\,\hat{\mathbf{z}}$ | (8$b$) | O VI |
| $\mathbf{B}_{34}$ | = | $(-x_{12}+y_{12})\,\mathbf{a}_1+(-x_{12}-y_{12})\,\mathbf{a}_2+\left(\frac{1}{2}+z_{12}\right)\mathbf{a}_3$ | = | $-x_{12}a\,\hat{\mathbf{x}}-y_{12}b\,\hat{\mathbf{y}}+\left(\frac{1}{2}+z_{12}\right)c\,\hat{\mathbf{z}}$ | (8$b$) | O VI |
| $\mathbf{B}_{35}$ | = | $(x_{12}+y_{12})\,\mathbf{a}_1+(x_{12}-y_{12})\,\mathbf{a}_2+\left(\frac{1}{2}+z_{12}\right)\mathbf{a}_3$ | = | $x_{12}a\,\hat{\mathbf{x}}-y_{12}b\,\hat{\mathbf{y}}+\left(\frac{1}{2}+z_{12}\right)c\,\hat{\mathbf{z}}$ | (8$b$) | O VI |
| $\mathbf{B}_{36}$ | = | $(-x_{12}-y_{12})\,\mathbf{a}_1+(-x_{12}+y_{12})\,\mathbf{a}_2+z_{12}\,\mathbf{a}_3$ | = | $-x_{12}a\,\hat{\mathbf{x}}+y_{12}b\,\hat{\mathbf{y}}+z_{12}c\,\hat{\mathbf{z}}$ | (8$b$) | O VI |
| $\mathbf{B}_{37}$ | = | $(x_{13}-y_{13})\,\mathbf{a}_1+(x_{13}+y_{13})\,\mathbf{a}_2+z_{13}\,\mathbf{a}_3$ | = | $x_{13}a\,\hat{\mathbf{x}}+y_{13}b\,\hat{\mathbf{y}}+z_{13}c\,\hat{\mathbf{z}}$ | (8$b$) | O VII |
| $\mathbf{B}_{38}$ | = | $(-x_{13}+y_{13})\,\mathbf{a}_1+(-x_{13}-y_{13})\,\mathbf{a}_2+\left(\frac{1}{2}+z_{13}\right)\mathbf{a}_3$ | = | $-x_{13}a\,\hat{\mathbf{x}}-y_{13}b\,\hat{\mathbf{y}}+\left(\frac{1}{2}+z_{13}\right)c\,\hat{\mathbf{z}}$ | (8$b$) | O VII |
| $\mathbf{B}_{39}$ | = | $(x_{13}+y_{13})\,\mathbf{a}_1+(x_{13}-y_{13})\,\mathbf{a}_2+\left(\frac{1}{2}+z_{13}\right)\mathbf{a}_3$ | = | $x_{13}a\,\hat{\mathbf{x}}-y_{13}b\,\hat{\mathbf{y}}+\left(\frac{1}{2}+z_{13}\right)c\,\hat{\mathbf{z}}$ | (8$b$) | O VII |
| $\mathbf{B}_{40}$ | = | $(-x_{13}-y_{13})\,\mathbf{a}_1+(-x_{13}+y_{13})\,\mathbf{a}_2+z_{13}\,\mathbf{a}_3$ | = | $-x_{13}a\,\hat{\mathbf{x}}+y_{13}b\,\hat{\mathbf{y}}+z_{13}c\,\hat{\mathbf{z}}$ | (8$b$) | O VII |

**Geometry files:**

- CIF: pp. 1597
- POSCAR: pp. 1597

# $Ta_3Ti_{13}$ (BCC SQS-16) Structure: A3B13_oC32_38_ac_a2bcdef

http://aflow.org/prototype-encyclopedia/A3B13_oC32_38_ac_a2bcdef

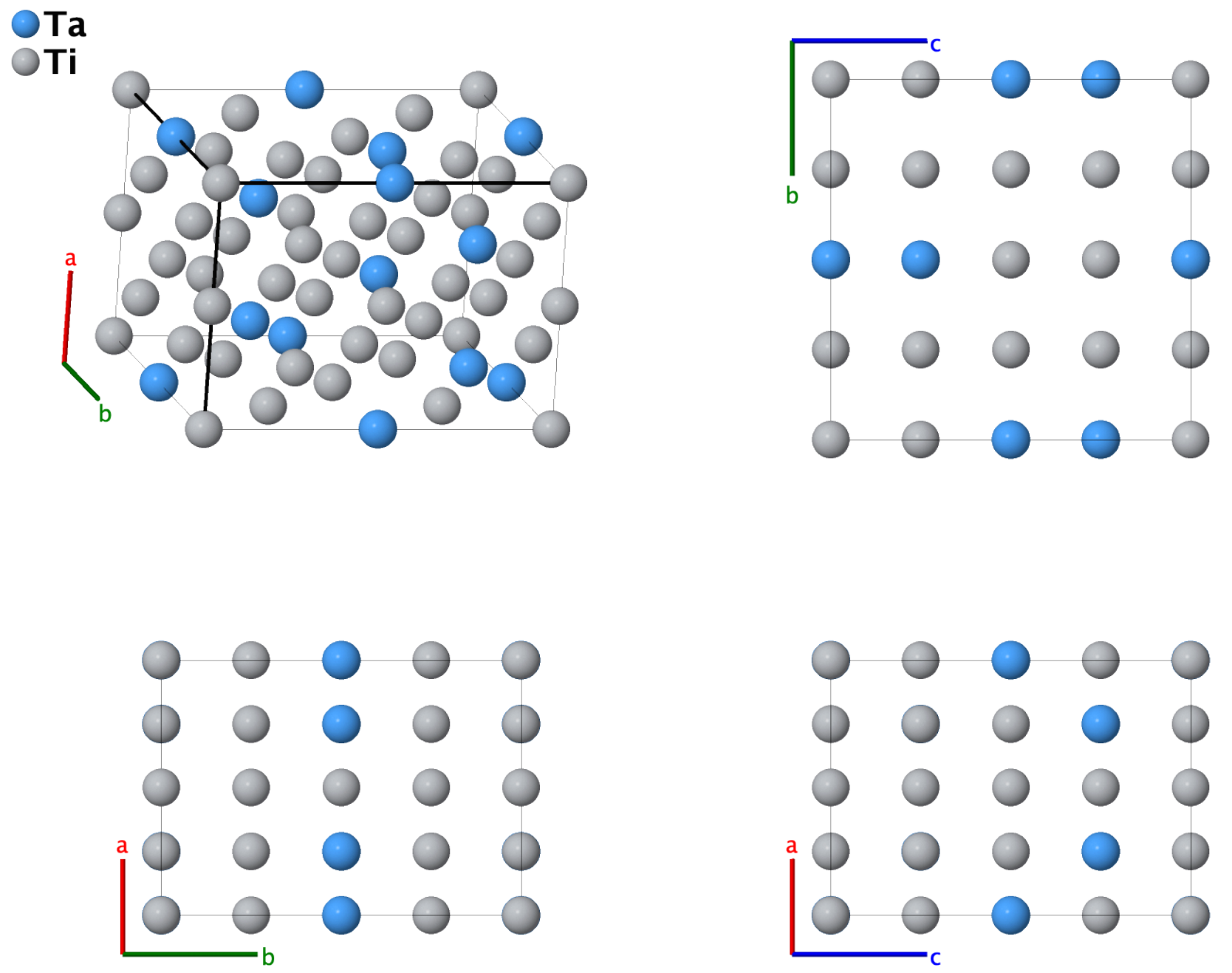

| | | |
|---|---|---|
| **Prototype** | : | $Ta_3Ti_{13}$ |
| **AFLOW prototype label** | : | A3B13_oC32_38_ac_a2bcdef |
| ***Strukturbericht* designation** | : | None |
| **Pearson symbol** | : | oC32 |
| **Space group number** | : | 38 |
| **Space group symbol** | : | *Amm*2 |
| **AFLOW prototype command** | : | `aflow --proto=A3B13_oC32_38_ac_a2bcdef` <br> `--params=`$a, b/a, c/a, z_1, z_2, z_3, z_4, x_5, z_5, x_6, z_6, y_7, z_7, y_8, z_8, x_9, y_9, z_9$ |

- This is a special quasirandom structure with 16 atoms per unit cell (SQS-16) for the $\beta$-phase (high-temperature austenite) bcc substitutional Ti-Ta alloy (Chakraborty, 2016). This protoype contains 18.75% Ta. Prototypes are listed for other Ta-Ti concentrations: 12.5% Ta (AB7_hR16_166_c_c2h), 25% Ta (AB3_mC32_8_4a_4a4b), 31.25% Ta (A5B11_mP16_6_2abc_2a3b3c), and 37.5% Ta (A3B5_oC32_38_abce_abcdf).

**Base-centered Orthorhombic primitive vectors:**

$$
\begin{aligned}
\mathbf{a}_1 &= a\,\hat{\mathbf{x}} \\
\mathbf{a}_2 &= \tfrac{1}{2}\,b\,\hat{\mathbf{y}} - \tfrac{1}{2}\,c\,\hat{\mathbf{z}} \\
\mathbf{a}_3 &= \tfrac{1}{2}\,b\,\hat{\mathbf{y}} + \tfrac{1}{2}\,c\,\hat{\mathbf{z}}
\end{aligned}
$$

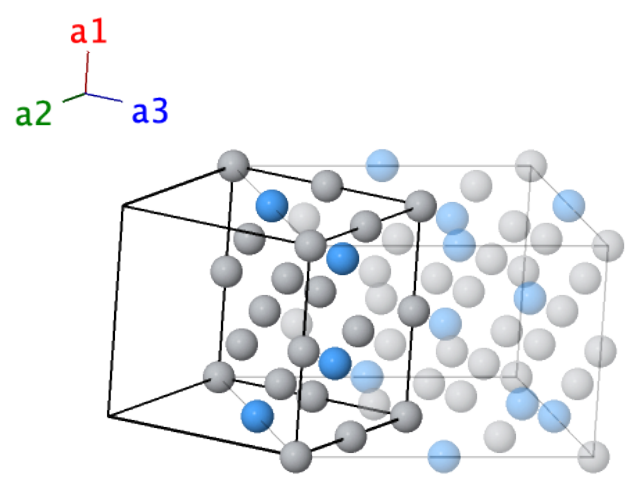

**Basis vectors:**

| | | Lattice Coordinates | | Cartesian Coordinates | Wyckoff Position | Atom Type |
|---|---|---|---|---|---|---|
| $\mathbf{B}_1$ | = | $-z_1\,\mathbf{a}_2 + z_1\,\mathbf{a}_3$ | = | $z_1 c\,\hat{\mathbf{z}}$ | $(2a)$ | Ta I |
| $\mathbf{B}_2$ | = | $-z_2\,\mathbf{a}_2 + z_2\,\mathbf{a}_3$ | = | $z_2 c\,\hat{\mathbf{z}}$ | $(2a)$ | Ti I |
| $\mathbf{B}_3$ | = | $\frac{1}{2}\,\mathbf{a}_1 - z_3\,\mathbf{a}_2 + z_3\,\mathbf{a}_3$ | = | $\frac{1}{2}a\,\hat{\mathbf{x}} + z_3 c\,\hat{\mathbf{z}}$ | $(2b)$ | Ti II |
| $\mathbf{B}_4$ | = | $\frac{1}{2}\,\mathbf{a}_1 - z_4\,\mathbf{a}_2 + z_4\,\mathbf{a}_3$ | = | $\frac{1}{2}a\,\hat{\mathbf{x}} + z_4 c\,\hat{\mathbf{z}}$ | $(2b)$ | Ti III |
| $\mathbf{B}_5$ | = | $x_5\,\mathbf{a}_1 - z_5\,\mathbf{a}_2 + z_5\,\mathbf{a}_3$ | = | $x_5 a\,\hat{\mathbf{x}} + z_5 c\,\hat{\mathbf{z}}$ | $(4c)$ | Ta II |
| $\mathbf{B}_6$ | = | $-x_5\,\mathbf{a}_1 - z_5\,\mathbf{a}_2 + z_5\,\mathbf{a}_3$ | = | $-x_5 a\,\hat{\mathbf{x}} + z_5 c\,\hat{\mathbf{z}}$ | $(4c)$ | Ta II |
| $\mathbf{B}_7$ | = | $x_6\,\mathbf{a}_1 - z_6\,\mathbf{a}_2 + z_6\,\mathbf{a}_3$ | = | $x_6 a\,\hat{\mathbf{x}} + z_6 c\,\hat{\mathbf{z}}$ | $(4c)$ | Ti IV |
| $\mathbf{B}_8$ | = | $-x_6\,\mathbf{a}_1 - z_6\,\mathbf{a}_2 + z_6\,\mathbf{a}_3$ | = | $-x_6 a\,\hat{\mathbf{x}} + z_6 c\,\hat{\mathbf{z}}$ | $(4c)$ | Ti IV |
| $\mathbf{B}_9$ | = | $(y_7 - z_7)\,\mathbf{a}_2 + (y_7 + z_7)\,\mathbf{a}_3$ | = | $y_7 b\,\hat{\mathbf{y}} + z_7 c\,\hat{\mathbf{z}}$ | $(4d)$ | Ti V |
| $\mathbf{B}_{10}$ | = | $(-y_7 - z_7)\,\mathbf{a}_2 + (-y_7 + z_7)\,\mathbf{a}_3$ | = | $-y_7 b\,\hat{\mathbf{y}} + z_7 c\,\hat{\mathbf{z}}$ | $(4d)$ | Ti V |
| $\mathbf{B}_{11}$ | = | $\frac{1}{2}\,\mathbf{a}_1 + (y_8 - z_8)\,\mathbf{a}_2 + (y_8 + z_8)\,\mathbf{a}_3$ | = | $\frac{1}{2}a\,\hat{\mathbf{x}} + y_8 b\,\hat{\mathbf{y}} + z_8 c\,\hat{\mathbf{z}}$ | $(4e)$ | Ti VI |
| $\mathbf{B}_{12}$ | = | $\frac{1}{2}\,\mathbf{a}_1 + (-y_8 - z_8)\,\mathbf{a}_2 + (-y_8 + z_8)\,\mathbf{a}_3$ | = | $\frac{1}{2}a\,\hat{\mathbf{x}} - y_8 b\,\hat{\mathbf{y}} + z_8 c\,\hat{\mathbf{z}}$ | $(4e)$ | Ti VI |
| $\mathbf{B}_{13}$ | = | $x_9\,\mathbf{a}_1 + (y_9 - z_9)\,\mathbf{a}_2 + (y_9 + z_9)\,\mathbf{a}_3$ | = | $x_9 a\,\hat{\mathbf{x}} + y_9 b\,\hat{\mathbf{y}} + z_9 c\,\hat{\mathbf{z}}$ | $(8f)$ | Ti VII |
| $\mathbf{B}_{14}$ | = | $-x_9\,\mathbf{a}_1 + (-y_9 - z_9)\,\mathbf{a}_2 + (-y_9 + z_9)\,\mathbf{a}_3$ | = | $-x_9 a\,\hat{\mathbf{x}} - y_9 b\,\hat{\mathbf{y}} + z_9 c\,\hat{\mathbf{z}}$ | $(8f)$ | Ti VII |
| $\mathbf{B}_{15}$ | = | $x_9\,\mathbf{a}_1 + (-y_9 - z_9)\,\mathbf{a}_2 + (-y_9 + z_9)\,\mathbf{a}_3$ | = | $x_9 a\,\hat{\mathbf{x}} - y_9 b\,\hat{\mathbf{y}} + z_9 c\,\hat{\mathbf{z}}$ | $(8f)$ | Ti VII |
| $\mathbf{B}_{16}$ | = | $-x_9\,\mathbf{a}_1 + (y_9 - z_9)\,\mathbf{a}_2 + (y_9 + z_9)\,\mathbf{a}_3$ | = | $-x_9 a\,\hat{\mathbf{x}} + y_9 b\,\hat{\mathbf{y}} + z_9 c\,\hat{\mathbf{z}}$ | $(8f)$ | Ti VII |

**Geometry files:**

- CIF: pp. 1597
- POSCAR: pp. 1598

# $Ta_3Ti_5$ (BCC SQS-16) Structure: A3B5_oC32_38_abce_abcdf

http://aflow.org/prototype-encyclopedia/A3B5_oC32_38_abce_abcdf

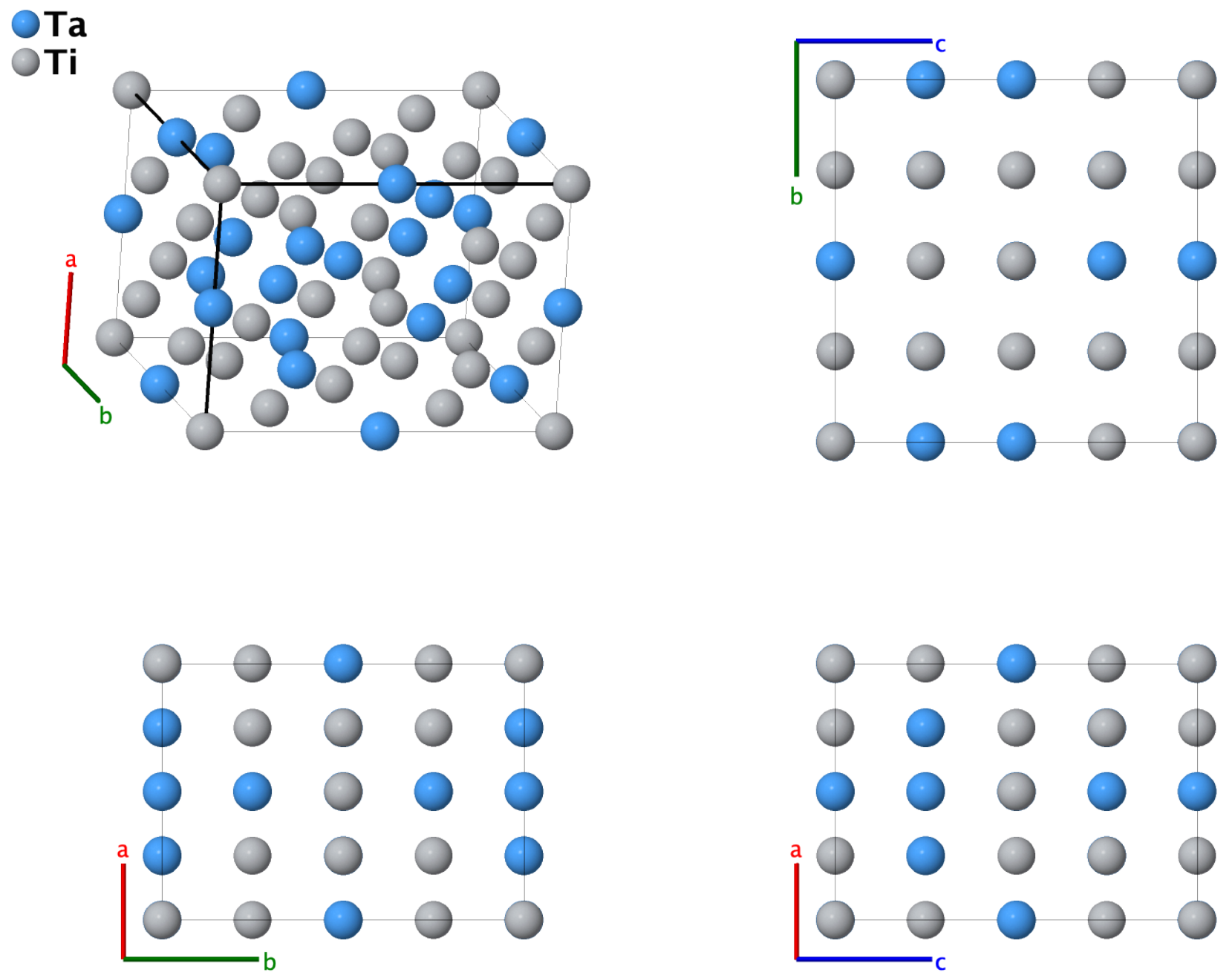

| | | |
|---|---|---|
| **Prototype** | : | $Ta_3Ti_5$ |
| **AFLOW prototype label** | : | A3B5_oC32_38_abce_abcdf |
| ***Strukturbericht* designation** | : | None |
| **Pearson symbol** | : | oC32 |
| **Space group number** | : | 38 |
| **Space group symbol** | : | *Amm*2 |
| **AFLOW prototype command** | : | `aflow --proto=A3B5_oC32_38_abce_abcdf` `--params=`$a, b/a, c/a, z_1, z_2, z_3, z_4, x_5, z_5, x_6, z_6, y_7, z_7, y_8, z_8, x_9, y_9, z_9$ |

- This is a special quasirandom structure with 16 atoms per unit cell (SQS-16) for the $\beta$-phase (high-temperature austenite) bcc substitutional Ti-Ta alloy (Chakraborty, 2016). This protoype contains 37.5% Ta. Prototypes are listed for other Ta-Ti concentrations: 12.5% Ta (AB7_hR16_166_c_c2h), 18.75% Ta (A3B13_oC32_38_ac_a2bcdef), 25% Ta (AB3_mC32_8_4a_4a4b), 31.25% Ta (A5B11_mP16_6_2abc_2a3b3c).

**Base-centered Orthorhombic primitive vectors:**

$$
\begin{aligned}
\mathbf{a}_1 &= a\,\hat{\mathbf{x}} \\
\mathbf{a}_2 &= \tfrac{1}{2}\,b\,\hat{\mathbf{y}} - \tfrac{1}{2}\,c\,\hat{\mathbf{z}} \\
\mathbf{a}_3 &= \tfrac{1}{2}\,b\,\hat{\mathbf{y}} + \tfrac{1}{2}\,c\,\hat{\mathbf{z}}
\end{aligned}
$$

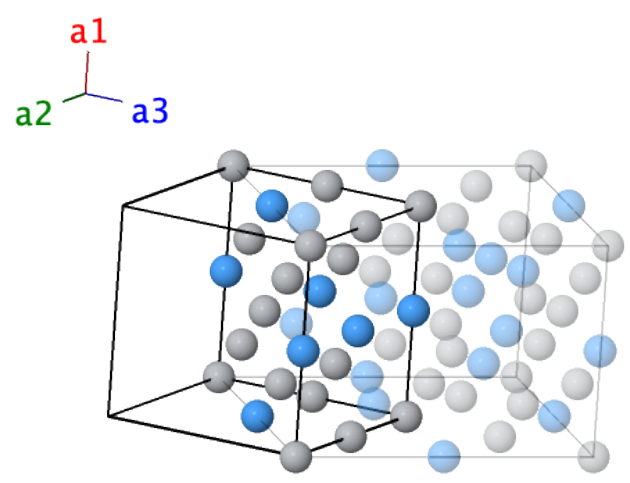

**Basis vectors:**

| | | Lattice Coordinates | | Cartesian Coordinates | Wyckoff Position | Atom Type |
|---|---|---|---|---|---|---|
| $\mathbf{B}_1$ | = | $-z_1\,\mathbf{a}_2 + z_1\,\mathbf{a}_3$ | = | $z_1 c\,\hat{\mathbf{z}}$ | (2$a$) | Ta I |
| $\mathbf{B}_2$ | = | $-z_2\,\mathbf{a}_2 + z_2\,\mathbf{a}_3$ | = | $z_2 c\,\hat{\mathbf{z}}$ | (2$a$) | Ti I |
| $\mathbf{B}_3$ | = | $\frac{1}{2}\,\mathbf{a}_1 - z_3\,\mathbf{a}_2 + z_3\,\mathbf{a}_3$ | = | $\frac{1}{2}a\,\hat{\mathbf{x}} + z_3 c\,\hat{\mathbf{z}}$ | (2$b$) | Ta II |
| $\mathbf{B}_4$ | = | $\frac{1}{2}\,\mathbf{a}_1 - z_4\,\mathbf{a}_2 + z_4\,\mathbf{a}_3$ | = | $\frac{1}{2}a\,\hat{\mathbf{x}} + z_4 c\,\hat{\mathbf{z}}$ | (2$b$) | Ti II |
| $\mathbf{B}_5$ | = | $x_5\,\mathbf{a}_1 - z_5\,\mathbf{a}_2 + z_5\,\mathbf{a}_3$ | = | $x_5 a\,\hat{\mathbf{x}} + z_5 c\,\hat{\mathbf{z}}$ | (4$c$) | Ta III |
| $\mathbf{B}_6$ | = | $-x_5\,\mathbf{a}_1 - z_5\,\mathbf{a}_2 + z_5\,\mathbf{a}_3$ | = | $-x_5 a\,\hat{\mathbf{x}} + z_5 c\,\hat{\mathbf{z}}$ | (4$c$) | Ta III |
| $\mathbf{B}_7$ | = | $x_6\,\mathbf{a}_1 - z_6\,\mathbf{a}_2 + z_6\,\mathbf{a}_3$ | = | $x_6 a\,\hat{\mathbf{x}} + z_6 c\,\hat{\mathbf{z}}$ | (4$c$) | Ti III |
| $\mathbf{B}_8$ | = | $-x_6\,\mathbf{a}_1 - z_6\,\mathbf{a}_2 + z_6\,\mathbf{a}_3$ | = | $-x_6 a\,\hat{\mathbf{x}} + z_6 c\,\hat{\mathbf{z}}$ | (4$c$) | Ti III |
| $\mathbf{B}_9$ | = | $(y_7 - z_7)\,\mathbf{a}_2 + (y_7 + z_7)\,\mathbf{a}_3$ | = | $y_7 b\,\hat{\mathbf{y}} + z_7 c\,\hat{\mathbf{z}}$ | (4$d$) | Ti IV |
| $\mathbf{B}_{10}$ | = | $(-y_7 - z_7)\,\mathbf{a}_2 + (-y_7 + z_7)\,\mathbf{a}_3$ | = | $-y_7 b\,\hat{\mathbf{y}} + z_7 c\,\hat{\mathbf{z}}$ | (4$d$) | Ti IV |
| $\mathbf{B}_{11}$ | = | $\frac{1}{2}\,\mathbf{a}_1 + (y_8 - z_8)\,\mathbf{a}_2 + (y_8 + z_8)\,\mathbf{a}_3$ | = | $\frac{1}{2}a\,\hat{\mathbf{x}} + y_8 b\,\hat{\mathbf{y}} + z_8 c\,\hat{\mathbf{z}}$ | (4$e$) | Ta IV |
| $\mathbf{B}_{12}$ | = | $\frac{1}{2}\,\mathbf{a}_1 + (-y_8 - z_8)\,\mathbf{a}_2 + (-y_8 + z_8)\,\mathbf{a}_3$ | = | $\frac{1}{2}a\,\hat{\mathbf{x}} - y_8 b\,\hat{\mathbf{y}} + z_8 c\,\hat{\mathbf{z}}$ | (4$e$) | Ta IV |
| $\mathbf{B}_{13}$ | = | $x_9\,\mathbf{a}_1 + (y_9 - z_9)\,\mathbf{a}_2 + (y_9 + z_9)\,\mathbf{a}_3$ | = | $x_9 a\,\hat{\mathbf{x}} + y_9 b\,\hat{\mathbf{y}} + z_9 c\,\hat{\mathbf{z}}$ | (8$f$) | Ti V |
| $\mathbf{B}_{14}$ | = | $-x_9\,\mathbf{a}_1 + (-y_9 - z_9)\,\mathbf{a}_2 + (-y_9 + z_9)\,\mathbf{a}_3$ | = | $-x_9 a\,\hat{\mathbf{x}} - y_9 b\,\hat{\mathbf{y}} + z_9 c\,\hat{\mathbf{z}}$ | (8$f$) | Ti V |
| $\mathbf{B}_{15}$ | = | $x_9\,\mathbf{a}_1 + (-y_9 - z_9)\,\mathbf{a}_2 + (-y_9 + z_9)\,\mathbf{a}_3$ | = | $x_9 a\,\hat{\mathbf{x}} - y_9 b\,\hat{\mathbf{y}} + z_9 c\,\hat{\mathbf{z}}$ | (8$f$) | Ti V |
| $\mathbf{B}_{16}$ | = | $-x_9\,\mathbf{a}_1 + (y_9 - z_9)\,\mathbf{a}_2 + (y_9 + z_9)\,\mathbf{a}_3$ | = | $-x_9 a\,\hat{\mathbf{x}} + y_9 b\,\hat{\mathbf{y}} + z_9 c\,\hat{\mathbf{z}}$ | (8$f$) | Ti V |

**Geometry files:**

- CIF: pp. 1598
- POSCAR: pp. 1598

# $NaNb_6O_{15}F$ Structure: ABC6D15_oC46_38_b_b_2a2d_2ab4d2e

http://aflow.org/prototype-encyclopedia/ABC6D15_oC46_38_b_b_2a2d_2ab4d2e

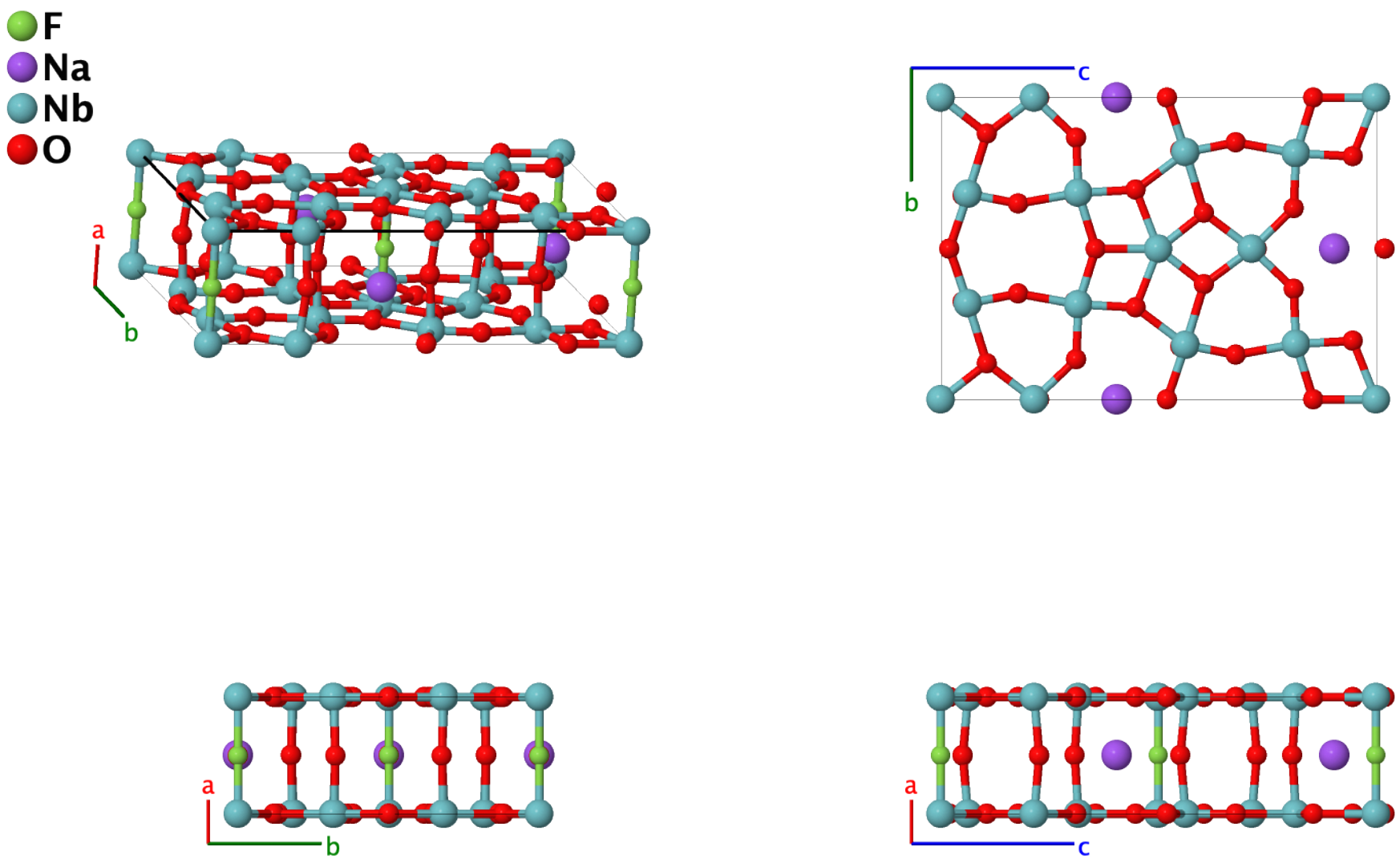

| | | |
|---|---|---|
| **Prototype** | : | $FNaNb_6O_{15}$ |
| **AFLOW prototype label** | : | ABC6D15_oC46_38_b_b_2a2d_2ab4d2e |
| ***Strukturbericht* designation** | : | None |
| **Pearson symbol** | : | oC46 |
| **Space group number** | : | 38 |
| **Space group symbol** | : | $Amm2$ |
| **AFLOW prototype command** | : | `aflow --proto=ABC6D15_oC46_38_b_b_2a2d_2ab4d2e --params=`$a, b/a, c/a, z_1, z_2, z_3, z_4, z_5, z_6, z_7, y_8, z_8, y_9, z_9, y_{10}, z_{10}, y_{11}, z_{11}, y_{12}, z_{12}, y_{13}, z_{13}, y_{14}, z_{14}, y_{15}, z_{15}$ |

**Other compounds with this structure**

- $NaNb_6O_{15}(OH)$

- The X-ray scattering of an $F^-$ ion is almost identical to that of $O^{2-}$, and (Andersson, 1965) was not able to distinguish between them. He arbitrarily labeled the (2*b*) site he designated as O(1) as the location of the fluorine ion and we follow this, but in reality we have no idea if the $F^-$ ions are located on this site, are ordered on another site, or are statistically distributed on the oxygen sites. Presumably the same considerations hold for $NaNb_6O_{15}(OH)$.

- Andersson sets $z_4 = 0.159$ as the coordinate of what we label as O-II and he calls O(10), but this gives an unreasonably short distance between the Nb-II and O-II atoms, and the distances between the O-II atom and the other atoms in the structure do not agree with the distances given his paper. If we assume that the first two digits were transposed when printed, so that $z_4 = 0.519$, then we get within 0.1% of Andersson's distances.

## Base-centered Orthorhombic primitive vectors:

$$
\begin{aligned}
\mathbf{a}_1 &= a\,\mathbf{\hat{x}} \\
\mathbf{a}_2 &= \tfrac{1}{2}\,b\,\mathbf{\hat{y}} - \tfrac{1}{2}\,c\,\mathbf{\hat{z}} \\
\mathbf{a}_3 &= \tfrac{1}{2}\,b\,\mathbf{\hat{y}} + \tfrac{1}{2}\,c\,\mathbf{\hat{z}}
\end{aligned}
$$

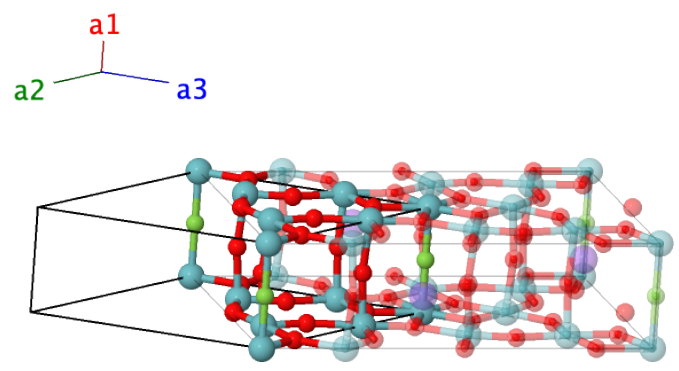

## Basis vectors:

| | | Lattice Coordinates | | Cartesian Coordinates | Wyckoff Position | Atom Type |
|---|---|---|---|---|---|---|
| $\mathbf{B}_1$ | = | $-z_1\,\mathbf{a}_2 + z_1\,\mathbf{a}_3$ | = | $z_1 c\,\mathbf{\hat{z}}$ | $(2a)$ | Nb I |
| $\mathbf{B}_2$ | = | $-z_2\,\mathbf{a}_2 + z_2\,\mathbf{a}_3$ | = | $z_2 c\,\mathbf{\hat{z}}$ | $(2a)$ | Nb II |
| $\mathbf{B}_3$ | = | $-z_3\,\mathbf{a}_2 + z_3\,\mathbf{a}_3$ | = | $z_3 c\,\mathbf{\hat{z}}$ | $(2a)$ | O I |
| $\mathbf{B}_4$ | = | $-z_4\,\mathbf{a}_2 + z_4\,\mathbf{a}_3$ | = | $z_4 c\,\mathbf{\hat{z}}$ | $(2a)$ | O II |
| $\mathbf{B}_5$ | = | $\frac{1}{2}\,\mathbf{a}_1 - z_5\,\mathbf{a}_2 + z_5\,\mathbf{a}_3$ | = | $\frac{1}{2}a\,\mathbf{\hat{x}} + z_5 c\,\mathbf{\hat{z}}$ | $(2b)$ | F |
| $\mathbf{B}_6$ | = | $\frac{1}{2}\,\mathbf{a}_1 - z_6\,\mathbf{a}_2 + z_6\,\mathbf{a}_3$ | = | $\frac{1}{2}a\,\mathbf{\hat{x}} + z_6 c\,\mathbf{\hat{z}}$ | $(2b)$ | Na |
| $\mathbf{B}_7$ | = | $\frac{1}{2}\,\mathbf{a}_1 - z_7\,\mathbf{a}_2 + z_7\,\mathbf{a}_3$ | = | $\frac{1}{2}a\,\mathbf{\hat{x}} + z_7 c\,\mathbf{\hat{z}}$ | $(2b)$ | O III |
| $\mathbf{B}_8$ | = | $(y_8 - z_8)\,\mathbf{a}_2 + (y_8 + z_8)\,\mathbf{a}_3$ | = | $y_8 b\,\mathbf{\hat{y}} + z_8 c\,\mathbf{\hat{z}}$ | $(4d)$ | Nb III |
| $\mathbf{B}_9$ | = | $(-y_8 - z_8)\,\mathbf{a}_2 + (-y_8 + z_8)\,\mathbf{a}_3$ | = | $-y_8 b\,\mathbf{\hat{y}} + z_8 c\,\mathbf{\hat{z}}$ | $(4d)$ | Nb III |
| $\mathbf{B}_{10}$ | = | $(y_9 - z_9)\,\mathbf{a}_2 + (y_9 + z_9)\,\mathbf{a}_3$ | = | $y_9 b\,\mathbf{\hat{y}} + z_9 c\,\mathbf{\hat{z}}$ | $(4d)$ | Nb IV |
| $\mathbf{B}_{11}$ | = | $(-y_9 - z_9)\,\mathbf{a}_2 + (-y_9 + z_9)\,\mathbf{a}_3$ | = | $-y_9 b\,\mathbf{\hat{y}} + z_9 c\,\mathbf{\hat{z}}$ | $(4d)$ | Nb IV |
| $\mathbf{B}_{12}$ | = | $(y_{10} - z_{10})\,\mathbf{a}_2 + (y_{10} + z_{10})\,\mathbf{a}_3$ | = | $y_{10} b\,\mathbf{\hat{y}} + z_{10} c\,\mathbf{\hat{z}}$ | $(4d)$ | O IV |
| $\mathbf{B}_{13}$ | = | $(-y_{10} - z_{10})\,\mathbf{a}_2 + (-y_{10} + z_{10})\,\mathbf{a}_3$ | = | $-y_{10} b\,\mathbf{\hat{y}} + z_{10} c\,\mathbf{\hat{z}}$ | $(4d)$ | O IV |
| $\mathbf{B}_{14}$ | = | $(y_{11} - z_{11})\,\mathbf{a}_2 + (y_{11} + z_{11})\,\mathbf{a}_3$ | = | $y_{11} b\,\mathbf{\hat{y}} + z_{11} c\,\mathbf{\hat{z}}$ | $(4d)$ | O V |
| $\mathbf{B}_{15}$ | = | $(-y_{11} - z_{11})\,\mathbf{a}_2 + (-y_{11} + z_{11})\,\mathbf{a}_3$ | = | $-y_{11} b\,\mathbf{\hat{y}} + z_{11} c\,\mathbf{\hat{z}}$ | $(4d)$ | O V |
| $\mathbf{B}_{16}$ | = | $(y_{12} - z_{12})\,\mathbf{a}_2 + (y_{12} + z_{12})\,\mathbf{a}_3$ | = | $y_{12} b\,\mathbf{\hat{y}} + z_{12} c\,\mathbf{\hat{z}}$ | $(4d)$ | O VI |
| $\mathbf{B}_{17}$ | = | $(-y_{12} - z_{12})\,\mathbf{a}_2 + (-y_{12} + z_{12})\,\mathbf{a}_3$ | = | $-y_{12} b\,\mathbf{\hat{y}} + z_{12} c\,\mathbf{\hat{z}}$ | $(4d)$ | O VI |
| $\mathbf{B}_{18}$ | = | $(y_{13} - z_{13})\,\mathbf{a}_2 + (y_{13} + z_{13})\,\mathbf{a}_3$ | = | $y_{13} b\,\mathbf{\hat{y}} + z_{13} c\,\mathbf{\hat{z}}$ | $(4d)$ | O VII |
| $\mathbf{B}_{19}$ | = | $(-y_{13} - z_{13})\,\mathbf{a}_2 + (-y_{13} + z_{13})\,\mathbf{a}_3$ | = | $-y_{13} b\,\mathbf{\hat{y}} + z_{13} c\,\mathbf{\hat{z}}$ | $(4d)$ | O VII |
| $\mathbf{B}_{20}$ | = | $\frac{1}{2}\,\mathbf{a}_1 + (y_{14} - z_{14})\,\mathbf{a}_2 + (y_{14} + z_{14})\,\mathbf{a}_3$ | = | $\frac{1}{2}a\,\mathbf{\hat{x}} + y_{14} b\,\mathbf{\hat{y}} + z_{14} c\,\mathbf{\hat{z}}$ | $(4e)$ | O VIII |
| $\mathbf{B}_{21}$ | = | $\frac{1}{2}\,\mathbf{a}_1 + (-y_{14} - z_{14})\,\mathbf{a}_2 + (-y_{14} + z_{14})\,\mathbf{a}_3$ | = | $\frac{1}{2}a\,\mathbf{\hat{x}} - y_{14} b\,\mathbf{\hat{y}} + z_{14} c\,\mathbf{\hat{z}}$ | $(4e)$ | O VIII |
| $\mathbf{B}_{22}$ | = | $\frac{1}{2}\,\mathbf{a}_1 + (y_{15} - z_{15})\,\mathbf{a}_2 + (y_{15} + z_{15})\,\mathbf{a}_3$ | = | $\frac{1}{2}a\,\mathbf{\hat{x}} + y_{15} b\,\mathbf{\hat{y}} + z_{15} c\,\mathbf{\hat{z}}$ | $(4e)$ | O IX |
| $\mathbf{B}_{23}$ | = | $\frac{1}{2}\,\mathbf{a}_1 + (-y_{15} - z_{15})\,\mathbf{a}_2 + (-y_{15} + z_{15})\,\mathbf{a}_3$ | = | $\frac{1}{2}a\,\mathbf{\hat{x}} - y_{15} b\,\mathbf{\hat{y}} + z_{15} c\,\mathbf{\hat{z}}$ | $(4e)$ | O IX |

## Geometry files:

- CIF: pp. 1598
- POSCAR: pp. 1599

# $Rb_2Mo_2O_7$ Structure: A2B7C2_oC88_40_abc_2b6c_a3b

http://aflow.org/prototype-encyclopedia/A2B7C2_oC88_40_abc_2b6c_a3b

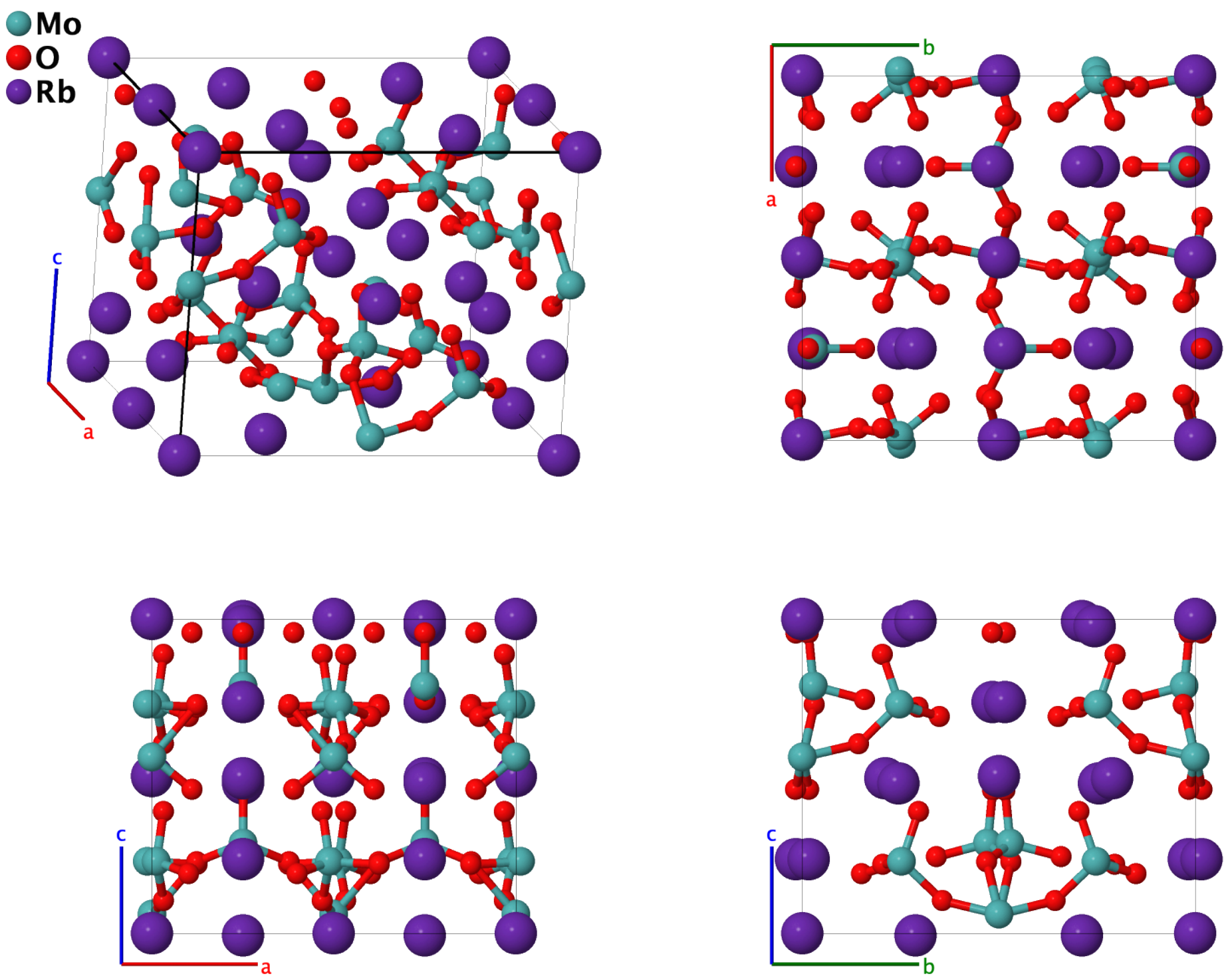

| | | |
|---|---|---|
| **Prototype** | : | $Mo_2O_7Rb_2$ |
| **AFLOW prototype label** | : | A2B7C2_oC88_40_abc_2b6c_a3b |
| ***Strukturbericht* designation** | : | None |
| **Pearson symbol** | : | oC88 |
| **Space group number** | : | 40 |
| **Space group symbol** | : | *Ama*2 |
| **AFLOW prototype command** | : | `aflow --proto=A2B7C2_oC88_40_abc_2b6c_a3b` `--params=`$a, b/a, c/a, z_1, z_2, y_3, z_3, y_4, z_4, y_5, z_5, y_6, z_6, y_7, z_7, y_8, z_8, x_9, y_9, z_9, x_{10}, y_{10}, z_{10}, x_{11}, y_{11}, z_{11}, x_{12}, y_{12}, z_{12}, x_{13}, y_{13}, z_{13}, x_{14}, y_{14}, z_{14}, x_{15}, y_{15}, z_{15}$ |

**Base-centered Orthorhombic primitive vectors:**

$$\mathbf{a}_1 = a\,\hat{\mathbf{x}}$$
$$\mathbf{a}_2 = \frac{1}{2}\,b\,\hat{\mathbf{y}} - \frac{1}{2}\,c\,\hat{\mathbf{z}}$$
$$\mathbf{a}_3 = \frac{1}{2}\,b\,\hat{\mathbf{y}} + \frac{1}{2}\,c\,\hat{\mathbf{z}}$$

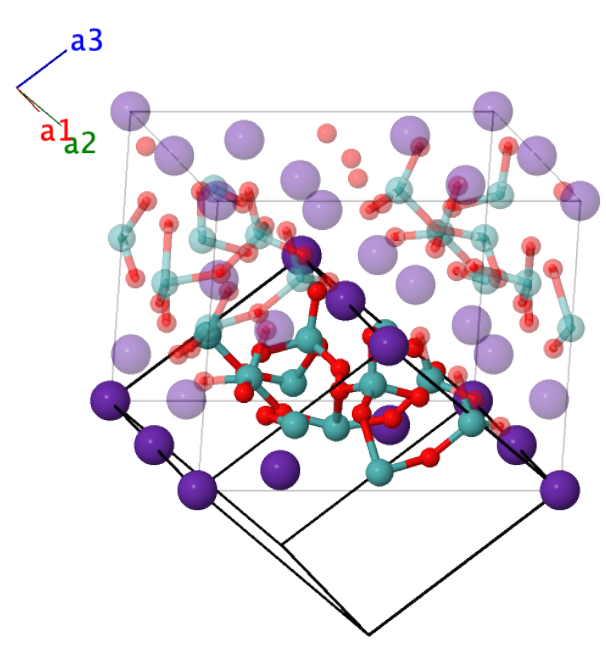

**Basis vectors:**

| | | Lattice Coordinates | | Cartesian Coordinates | Wyckoff Position | Atom Type |
|---|---|---|---|---|---|---|
| $\mathbf{B}_1$ | = | $-z_1\,\mathbf{a}_2 + z_1\,\mathbf{a}_3$ | = | $z_1 c\,\hat{\mathbf{z}}$ | (4*a*) | Mo I |
| $\mathbf{B}_2$ | = | $\frac{1}{2}\,\mathbf{a}_1 - z_1\,\mathbf{a}_2 + z_1\,\mathbf{a}_3$ | = | $\frac{1}{2}a\,\hat{\mathbf{x}} + z_1c\,\hat{\mathbf{z}}$ | (4$a$) | Mo I |
| $\mathbf{B}_3$ | = | $-z_2\,\mathbf{a}_2 + z_2\,\mathbf{a}_3$ | = | $z_2c\,\hat{\mathbf{z}}$ | (4$a$) | Rb I |
| $\mathbf{B}_4$ | = | $\frac{1}{2}\,\mathbf{a}_1 - z_2\,\mathbf{a}_2 + z_2\,\mathbf{a}_3$ | = | $\frac{1}{2}a\,\hat{\mathbf{x}} + z_2c\,\hat{\mathbf{z}}$ | (4$a$) | Rb I |
| $\mathbf{B}_5$ | = | $\frac{1}{4}\,\mathbf{a}_1 + (y_3 - z_3)\,\mathbf{a}_2 + (y_3 + z_3)\,\mathbf{a}_3$ | = | $\frac{1}{4}a\,\hat{\mathbf{x}} + y_3b\,\hat{\mathbf{y}} + z_3c\,\hat{\mathbf{z}}$ | (4$b$) | Mo II |
| $\mathbf{B}_6$ | = | $\frac{3}{4}\,\mathbf{a}_1 + (-y_3 - z_3)\,\mathbf{a}_2 + (-y_3 + z_3)\,\mathbf{a}_3$ | = | $\frac{3}{4}a\,\hat{\mathbf{x}} - y_3b\,\hat{\mathbf{y}} + z_3c\,\hat{\mathbf{z}}$ | (4$b$) | Mo II |
| $\mathbf{B}_7$ | = | $\frac{1}{4}\,\mathbf{a}_1 + (y_4 - z_4)\,\mathbf{a}_2 + (y_4 + z_4)\,\mathbf{a}_3$ | = | $\frac{1}{4}a\,\hat{\mathbf{x}} + y_4b\,\hat{\mathbf{y}} + z_4c\,\hat{\mathbf{z}}$ | (4$b$) | O I |
| $\mathbf{B}_8$ | = | $\frac{3}{4}\,\mathbf{a}_1 + (-y_4 - z_4)\,\mathbf{a}_2 + (-y_4 + z_4)\,\mathbf{a}_3$ | = | $\frac{3}{4}a\,\hat{\mathbf{x}} - y_4b\,\hat{\mathbf{y}} + z_4c\,\hat{\mathbf{z}}$ | (4$b$) | O I |
| $\mathbf{B}_9$ | = | $\frac{1}{4}\,\mathbf{a}_1 + (y_5 - z_5)\,\mathbf{a}_2 + (y_5 + z_5)\,\mathbf{a}_3$ | = | $\frac{1}{4}a\,\hat{\mathbf{x}} + y_5b\,\hat{\mathbf{y}} + z_5c\,\hat{\mathbf{z}}$ | (4$b$) | O II |
| $\mathbf{B}_{10}$ | = | $\frac{3}{4}\,\mathbf{a}_1 + (-y_5 - z_5)\,\mathbf{a}_2 + (-y_5 + z_5)\,\mathbf{a}_3$ | = | $\frac{3}{4}a\,\hat{\mathbf{x}} - y_5b\,\hat{\mathbf{y}} + z_5c\,\hat{\mathbf{z}}$ | (4$b$) | O II |
| $\mathbf{B}_{11}$ | = | $\frac{1}{4}\,\mathbf{a}_1 + (y_6 - z_6)\,\mathbf{a}_2 + (y_6 + z_6)\,\mathbf{a}_3$ | = | $\frac{1}{4}a\,\hat{\mathbf{x}} + y_6b\,\hat{\mathbf{y}} + z_6c\,\hat{\mathbf{z}}$ | (4$b$) | Rb II |
| $\mathbf{B}_{12}$ | = | $\frac{3}{4}\,\mathbf{a}_1 + (-y_6 - z_6)\,\mathbf{a}_2 + (-y_6 + z_6)\,\mathbf{a}_3$ | = | $\frac{3}{4}a\,\hat{\mathbf{x}} - y_6b\,\hat{\mathbf{y}} + z_6c\,\hat{\mathbf{z}}$ | (4$b$) | Rb II |
| $\mathbf{B}_{13}$ | = | $\frac{1}{4}\,\mathbf{a}_1 + (y_7 - z_7)\,\mathbf{a}_2 + (y_7 + z_7)\,\mathbf{a}_3$ | = | $\frac{1}{4}a\,\hat{\mathbf{x}} + y_7b\,\hat{\mathbf{y}} + z_7c\,\hat{\mathbf{z}}$ | (4$b$) | Rb III |
| $\mathbf{B}_{14}$ | = | $\frac{3}{4}\,\mathbf{a}_1 + (-y_7 - z_7)\,\mathbf{a}_2 + (-y_7 + z_7)\,\mathbf{a}_3$ | = | $\frac{3}{4}a\,\hat{\mathbf{x}} - y_7b\,\hat{\mathbf{y}} + z_7c\,\hat{\mathbf{z}}$ | (4$b$) | Rb III |
| $\mathbf{B}_{15}$ | = | $\frac{1}{4}\,\mathbf{a}_1 + (y_8 - z_8)\,\mathbf{a}_2 + (y_8 + z_8)\,\mathbf{a}_3$ | = | $\frac{1}{4}a\,\hat{\mathbf{x}} + y_8b\,\hat{\mathbf{y}} + z_8c\,\hat{\mathbf{z}}$ | (4$b$) | Rb IV |
| $\mathbf{B}_{16}$ | = | $\frac{3}{4}\,\mathbf{a}_1 + (-y_8 - z_8)\,\mathbf{a}_2 + (-y_8 + z_8)\,\mathbf{a}_3$ | = | $\frac{3}{4}a\,\hat{\mathbf{x}} - y_8b\,\hat{\mathbf{y}} + z_8c\,\hat{\mathbf{z}}$ | (4$b$) | Rb IV |
| $\mathbf{B}_{17}$ | = | $x_9\,\mathbf{a}_1 + (y_9 - z_9)\,\mathbf{a}_2 + (y_9 + z_9)\,\mathbf{a}_3$ | = | $x_9a\,\hat{\mathbf{x}} + y_9b\,\hat{\mathbf{y}} + z_9c\,\hat{\mathbf{z}}$ | (8$c$) | Mo III |
| $\mathbf{B}_{18}$ | = | $-x_9\,\mathbf{a}_1 + (-y_9 - z_9)\,\mathbf{a}_2 + (-y_9 + z_9)\,\mathbf{a}_3$ | = | $-x_9a\,\hat{\mathbf{x}} - y_9b\,\hat{\mathbf{y}} + z_9c\,\hat{\mathbf{z}}$ | (8$c$) | Mo III |
| $\mathbf{B}_{19}$ | = | $\left(\frac{1}{2} + x_9\right)\,\mathbf{a}_1 + (-y_9 - z_9)\,\mathbf{a}_2 + (-y_9 + z_9)\,\mathbf{a}_3$ | = | $\left(\frac{1}{2} + x_9\right)a\,\hat{\mathbf{x}} - y_9b\,\hat{\mathbf{y}} + z_9c\,\hat{\mathbf{z}}$ | (8$c$) | Mo III |
| $\mathbf{B}_{20}$ | = | $\left(\frac{1}{2} - x_9\right)\,\mathbf{a}_1 + (y_9 - z_9)\,\mathbf{a}_2 + (y_9 + z_9)\,\mathbf{a}_3$ | = | $\left(\frac{1}{2} - x_9\right)a\,\hat{\mathbf{x}} + y_9b\,\hat{\mathbf{y}} + z_9c\,\hat{\mathbf{z}}$ | (8$c$) | Mo III |
| $\mathbf{B}_{21}$ | = | $x_{10}\,\mathbf{a}_1 + (y_{10} - z_{10})\,\mathbf{a}_2 + (y_{10} + z_{10})\,\mathbf{a}_3$ | = | $x_{10}a\,\hat{\mathbf{x}} + y_{10}b\,\hat{\mathbf{y}} + z_{10}c\,\hat{\mathbf{z}}$ | (8$c$) | O III |
| $\mathbf{B}_{22}$ | = | $-x_{10}\,\mathbf{a}_1 + (-y_{10} - z_{10})\,\mathbf{a}_2 + (-y_{10} + z_{10})\,\mathbf{a}_3$ | = | $-x_{10}a\,\hat{\mathbf{x}} - y_{10}b\,\hat{\mathbf{y}} + z_{10}c\,\hat{\mathbf{z}}$ | (8$c$) | O III |
| $\mathbf{B}_{23}$ | = | $\left(\frac{1}{2} + x_{10}\right)\,\mathbf{a}_1 + (-y_{10} - z_{10})\,\mathbf{a}_2 + (-y_{10} + z_{10})\,\mathbf{a}_3$ | = | $\left(\frac{1}{2} + x_{10}\right)a\,\hat{\mathbf{x}} - y_{10}b\,\hat{\mathbf{y}} + z_{10}c\,\hat{\mathbf{z}}$ | (8$c$) | O III |
| $\mathbf{B}_{24}$ | = | $\left(\frac{1}{2} - x_{10}\right)\,\mathbf{a}_1 + (y_{10} - z_{10})\,\mathbf{a}_2 + (y_{10} + z_{10})\,\mathbf{a}_3$ | = | $\left(\frac{1}{2} - x_{10}\right)a\,\hat{\mathbf{x}} + y_{10}b\,\hat{\mathbf{y}} + z_{10}c\,\hat{\mathbf{z}}$ | (8$c$) | O III |
| $\mathbf{B}_{25}$ | = | $x_{11}\,\mathbf{a}_1 + (y_{11} - z_{11})\,\mathbf{a}_2 + (y_{11} + z_{11})\,\mathbf{a}_3$ | = | $x_{11}a\,\hat{\mathbf{x}} + y_{11}b\,\hat{\mathbf{y}} + z_{11}c\,\hat{\mathbf{z}}$ | (8$c$) | O IV |
| $\mathbf{B}_{26}$ | = | $-x_{11}\,\mathbf{a}_1 + (-y_{11} - z_{11})\,\mathbf{a}_2 + (-y_{11} + z_{11})\,\mathbf{a}_3$ | = | $-x_{11}a\,\hat{\mathbf{x}} - y_{11}b\,\hat{\mathbf{y}} + z_{11}c\,\hat{\mathbf{z}}$ | (8$c$) | O IV |
| $\mathbf{B}_{27}$ | = | $\left(\frac{1}{2} + x_{11}\right)\,\mathbf{a}_1 + (-y_{11} - z_{11})\,\mathbf{a}_2 + (-y_{11} + z_{11})\,\mathbf{a}_3$ | = | $\left(\frac{1}{2} + x_{11}\right)a\,\hat{\mathbf{x}} - y_{11}b\,\hat{\mathbf{y}} + z_{11}c\,\hat{\mathbf{z}}$ | (8$c$) | O IV |
| $\mathbf{B}_{28}$ | = | $\left(\frac{1}{2} - x_{11}\right)\,\mathbf{a}_1 + (y_{11} - z_{11})\,\mathbf{a}_2 + (y_{11} + z_{11})\,\mathbf{a}_3$ | = | $\left(\frac{1}{2} - x_{11}\right)a\,\hat{\mathbf{x}} + y_{11}b\,\hat{\mathbf{y}} + z_{11}c\,\hat{\mathbf{z}}$ | (8$c$) | O IV |
| $\mathbf{B}_{29}$ | = | $x_{12}\,\mathbf{a}_1 + (y_{12} - z_{12})\,\mathbf{a}_2 + (y_{12} + z_{12})\,\mathbf{a}_3$ | = | $x_{12}a\,\hat{\mathbf{x}} + y_{12}b\,\hat{\mathbf{y}} + z_{12}c\,\hat{\mathbf{z}}$ | (8$c$) | O V |
| $\mathbf{B}_{30}$ | = | $-x_{12}\,\mathbf{a}_1 + (-y_{12} - z_{12})\,\mathbf{a}_2 + (-y_{12} + z_{12})\,\mathbf{a}_3$ | = | $-x_{12}a\,\hat{\mathbf{x}} - y_{12}b\,\hat{\mathbf{y}} + z_{12}c\,\hat{\mathbf{z}}$ | (8$c$) | O V |
| $\mathbf{B}_{31}$ | = | $\left(\frac{1}{2} + x_{12}\right)\,\mathbf{a}_1 + (-y_{12} - z_{12})\,\mathbf{a}_2 + (-y_{12} + z_{12})\,\mathbf{a}_3$ | = | $\left(\frac{1}{2} + x_{12}\right)a\,\hat{\mathbf{x}} - y_{12}b\,\hat{\mathbf{y}} + z_{12}c\,\hat{\mathbf{z}}$ | (8$c$) | O V |
| $\mathbf{B}_{32}$ | = | $\left(\frac{1}{2} - x_{12}\right)\,\mathbf{a}_1 + (y_{12} - z_{12})\,\mathbf{a}_2 + (y_{12} + z_{12})\,\mathbf{a}_3$ | = | $\left(\frac{1}{2} - x_{12}\right)a\,\hat{\mathbf{x}} + y_{12}b\,\hat{\mathbf{y}} + z_{12}c\,\hat{\mathbf{z}}$ | (8$c$) | O V |
| $\mathbf{B}_{33}$ | = | $x_{13}\,\mathbf{a}_1 + (y_{13} - z_{13})\,\mathbf{a}_2 + (y_{13} + z_{13})\,\mathbf{a}_3$ | = | $x_{13}a\,\hat{\mathbf{x}} + y_{13}b\,\hat{\mathbf{y}} + z_{13}c\,\hat{\mathbf{z}}$ | (8$c$) | O VI |
| $\mathbf{B}_{34}$ | = | $-x_{13}\,\mathbf{a}_1 + \left(-y_{13} - z_{13}\right)\,\mathbf{a}_2 + \left(-y_{13} + z_{13}\right)\,\mathbf{a}_3$ | = | $-x_{13}a\,\hat{\mathbf{x}} - y_{13}b\,\hat{\mathbf{y}} + z_{13}c\,\hat{\mathbf{z}}$ | (8$c$) | O VI |
| $\mathbf{B}_{35}$ | = | $\left(\frac{1}{2} + x_{13}\right)\,\mathbf{a}_1 + \left(-y_{13} - z_{13}\right)\,\mathbf{a}_2 + \left(-y_{13} + z_{13}\right)\,\mathbf{a}_3$ | = | $\left(\frac{1}{2} + x_{13}\right)a\,\hat{\mathbf{x}} - y_{13}b\,\hat{\mathbf{y}} + z_{13}c\,\hat{\mathbf{z}}$ | (8$c$) | O VI |
| $\mathbf{B}_{36}$ | = | $\left(\frac{1}{2} - x_{13}\right)\,\mathbf{a}_1 + \left(y_{13} - z_{13}\right)\,\mathbf{a}_2 + \left(y_{13} + z_{13}\right)\,\mathbf{a}_3$ | = | $\left(\frac{1}{2} - x_{13}\right)a\,\hat{\mathbf{x}} + y_{13}b\,\hat{\mathbf{y}} + z_{13}c\,\hat{\mathbf{z}}$ | (8$c$) | O VI |
| $\mathbf{B}_{37}$ | = | $x_{14}\,\mathbf{a}_1 + \left(y_{14} - z_{14}\right)\,\mathbf{a}_2 + \left(y_{14} + z_{14}\right)\,\mathbf{a}_3$ | = | $x_{14}a\,\hat{\mathbf{x}} + y_{14}b\,\hat{\mathbf{y}} + z_{14}c\,\hat{\mathbf{z}}$ | (8$c$) | O VII |
| $\mathbf{B}_{38}$ | = | $-x_{14}\,\mathbf{a}_1 + \left(-y_{14} - z_{14}\right)\,\mathbf{a}_2 + \left(-y_{14} + z_{14}\right)\,\mathbf{a}_3$ | = | $-x_{14}a\,\hat{\mathbf{x}} - y_{14}b\,\hat{\mathbf{y}} + z_{14}c\,\hat{\mathbf{z}}$ | (8$c$) | O VII |
| $\mathbf{B}_{39}$ | = | $\left(\frac{1}{2} + x_{14}\right)\,\mathbf{a}_1 + \left(-y_{14} - z_{14}\right)\,\mathbf{a}_2 + \left(-y_{14} + z_{14}\right)\,\mathbf{a}_3$ | = | $\left(\frac{1}{2} + x_{14}\right)a\,\hat{\mathbf{x}} - y_{14}b\,\hat{\mathbf{y}} + z_{14}c\,\hat{\mathbf{z}}$ | (8$c$) | O VII |
| $\mathbf{B}_{40}$ | = | $\left(\frac{1}{2} - x_{14}\right)\,\mathbf{a}_1 + \left(y_{14} - z_{14}\right)\,\mathbf{a}_2 + \left(y_{14} + z_{14}\right)\,\mathbf{a}_3$ | = | $\left(\frac{1}{2} - x_{14}\right)a\,\hat{\mathbf{x}} + y_{14}b\,\hat{\mathbf{y}} + z_{14}c\,\hat{\mathbf{z}}$ | (8$c$) | O VII |
| $\mathbf{B}_{41}$ | = | $x_{15}\,\mathbf{a}_1 + \left(y_{15} - z_{15}\right)\,\mathbf{a}_2 + \left(y_{15} + z_{15}\right)\,\mathbf{a}_3$ | = | $x_{15}a\,\hat{\mathbf{x}} + y_{15}b\,\hat{\mathbf{y}} + z_{15}c\,\hat{\mathbf{z}}$ | (8$c$) | O VIII |
| $\mathbf{B}_{42}$ | = | $-x_{15}\,\mathbf{a}_1 + \left(-y_{15} - z_{15}\right)\,\mathbf{a}_2 + \left(-y_{15} + z_{15}\right)\,\mathbf{a}_3$ | = | $-x_{15}a\,\hat{\mathbf{x}} - y_{15}b\,\hat{\mathbf{y}} + z_{15}c\,\hat{\mathbf{z}}$ | (8$c$) | O VIII |
| $\mathbf{B}_{43}$ | = | $\left(\frac{1}{2} + x_{15}\right)\,\mathbf{a}_1 + \left(-y_{15} - z_{15}\right)\,\mathbf{a}_2 + \left(-y_{15} + z_{15}\right)\,\mathbf{a}_3$ | = | $\left(\frac{1}{2} + x_{15}\right)a\,\hat{\mathbf{x}} - y_{15}b\,\hat{\mathbf{y}} + z_{15}c\,\hat{\mathbf{z}}$ | (8$c$) | O VIII |
| $\mathbf{B}_{44}$ | = | $\left(\frac{1}{2} - x_{15}\right)\,\mathbf{a}_1 + \left(y_{15} - z_{15}\right)\,\mathbf{a}_2 + \left(y_{15} + z_{15}\right)\,\mathbf{a}_3$ | = | $\left(\frac{1}{2} - x_{15}\right)a\,\hat{\mathbf{x}} + y_{15}b\,\hat{\mathbf{y}} + z_{15}c\,\hat{\mathbf{z}}$ | (8$c$) | O VIII |

**Geometry files:**

- CIF: pp. 1599
- POSCAR: pp. 1599

# Orthorhombic $CrO_3$ Structure: AB3_oC16_40_b_a2b

http://aflow.org/prototype-encyclopedia/AB3_oC16_40_b_a2b

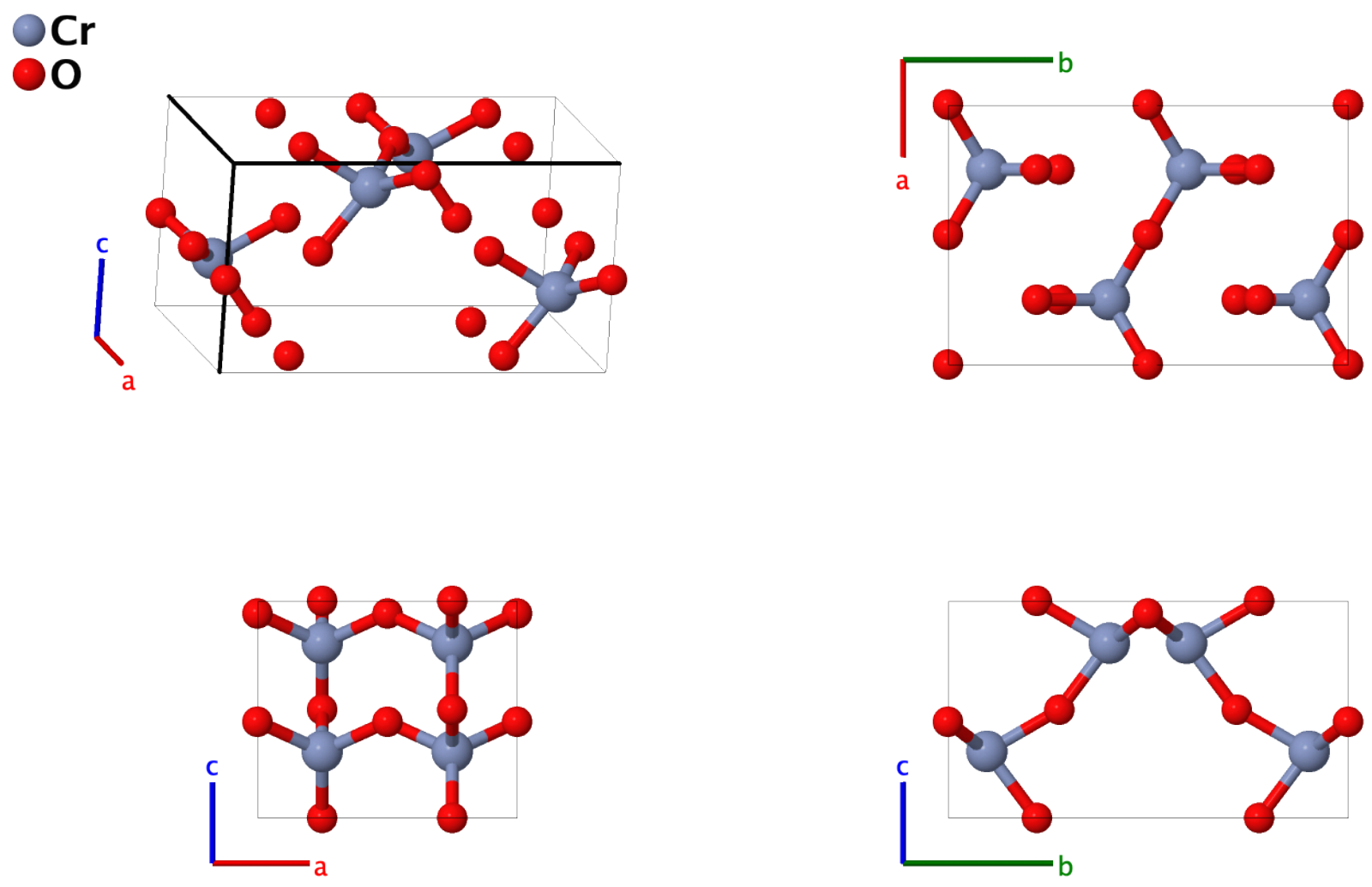

| | | |
|---|---|---|
| **Prototype** | : | $CrO_3$ |
| **AFLOW prototype label** | : | AB3_oC16_40_b_a2b |
| ***Strukturbericht* designation** | : | None |
| **Pearson symbol** | : | oC16 |
| **Space group number** | : | 40 |
| **Space group symbol** | : | *Ama*2 |
| **AFLOW prototype command** | : | `aflow --proto=AB3_oC16_40_b_a2b`<br>`--params=`$a, b/a, c/a, z_1, y_2, z_2, y_3, z_3, y_4, z_4$ |

- This is a more modern version of the $D0_7$ $CrO_3$ structure.
- (Byström, 1950) give this structure in the *C*2*cm* setting of space group #40. We have shifted it to the standard *Ama*2 setting.

**Base-centered Orthorhombic primitive vectors:**

$$
\begin{aligned}
\mathbf{a}_1 &= a\,\hat{\mathbf{x}} \\
\mathbf{a}_2 &= \tfrac{1}{2}\,b\,\hat{\mathbf{y}} - \tfrac{1}{2}\,c\,\hat{\mathbf{z}} \\
\mathbf{a}_3 &= \tfrac{1}{2}\,b\,\hat{\mathbf{y}} + \tfrac{1}{2}\,c\,\hat{\mathbf{z}}
\end{aligned}
$$

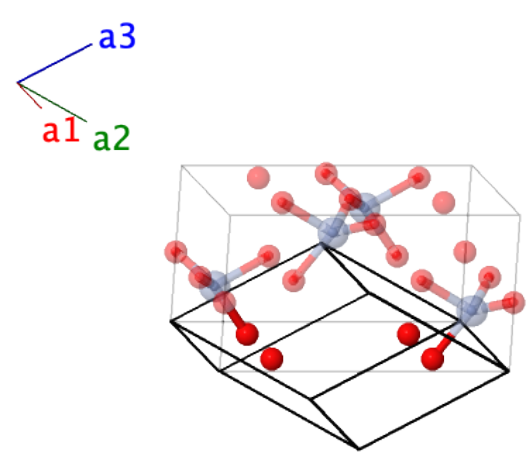

**Basis vectors:**

| | | Lattice Coordinates | | Cartesian Coordinates | Wyckoff Position | Atom Type |
|---|---|---|---|---|---|---|
| $\mathbf{B}_1$ | = | $-z_1\,\mathbf{a}_2 + z_1\,\mathbf{a}_3$ | = | $z_1 c\,\hat{\mathbf{z}}$ | (4*a*) | O I |
| $\mathbf{B}_2$ | = | $\frac{1}{2}\,\mathbf{a}_1 - z_1\,\mathbf{a}_2 + z_1\,\mathbf{a}_3$ | = | $\frac{1}{2}a\,\hat{\mathbf{x}} + z_1 c\,\hat{\mathbf{z}}$ | (4*a*) | O I |
| $\mathbf{B}_3$ | = | $\frac{1}{4}\,\mathbf{a}_1 + (y_2 - z_2)\,\mathbf{a}_2 + (y_2 + z_2)\,\mathbf{a}_3$ | = | $\frac{1}{4}a\,\hat{\mathbf{x}} + y_2 b\,\hat{\mathbf{y}} + z_2 c\,\hat{\mathbf{z}}$ | (4*b*) | Cr |
| $\mathbf{B}_4$ | = | $\frac{3}{4}\,\mathbf{a}_1 + (-y_2 - z_2)\,\mathbf{a}_2 + (-y_2 + z_2)\,\mathbf{a}_3$ | = | $\frac{3}{4}a\,\hat{\mathbf{x}} - y_2 b\,\hat{\mathbf{y}} + z_2 c\,\hat{\mathbf{z}}$ | (4$b$) | Cr |
| $\mathbf{B}_5$ | = | $\frac{1}{4}\,\mathbf{a}_1 + (y_3 - z_3)\,\mathbf{a}_2 + (y_3 + z_3)\,\mathbf{a}_3$ | = | $\frac{1}{4}a\,\hat{\mathbf{x}} + y_3 b\,\hat{\mathbf{y}} + z_3 c\,\hat{\mathbf{z}}$ | (4$b$) | O II |
| $\mathbf{B}_6$ | = | $\frac{3}{4}\,\mathbf{a}_1 + (-y_3 - z_3)\,\mathbf{a}_2 + (-y_3 + z_3)\,\mathbf{a}_3$ | = | $\frac{3}{4}a\,\hat{\mathbf{x}} - y_3 b\,\hat{\mathbf{y}} + z_3 c\,\hat{\mathbf{z}}$ | (4$b$) | O II |
| $\mathbf{B}_7$ | = | $\frac{1}{4}\,\mathbf{a}_1 + (y_4 - z_4)\,\mathbf{a}_2 + (y_4 + z_4)\,\mathbf{a}_3$ | = | $\frac{1}{4}a\,\hat{\mathbf{x}} + y_4 b\,\hat{\mathbf{y}} + z_4 c\,\hat{\mathbf{z}}$ | (4$b$) | O III |
| $\mathbf{B}_8$ | = | $\frac{3}{4}\,\mathbf{a}_1 + (-y_4 - z_4)\,\mathbf{a}_2 + (-y_4 + z_4)\,\mathbf{a}_3$ | = | $\frac{3}{4}a\,\hat{\mathbf{x}} - y_4 b\,\hat{\mathbf{y}} + z_4 c\,\hat{\mathbf{z}}$ | (4$b$) | O III |

**Geometry files:**

- CIF: pp. 1600
- POSCAR: pp. 1600

# Santite ($KB_5O_8 \cdot 4H_2O$, $K3_5$) Structure: A5B8CD12_oC104_41_a2b_4b_a_6b

http://aflow.org/prototype-encyclopedia/A5B8CD12_oC104_41_a2b_4b_a_6b

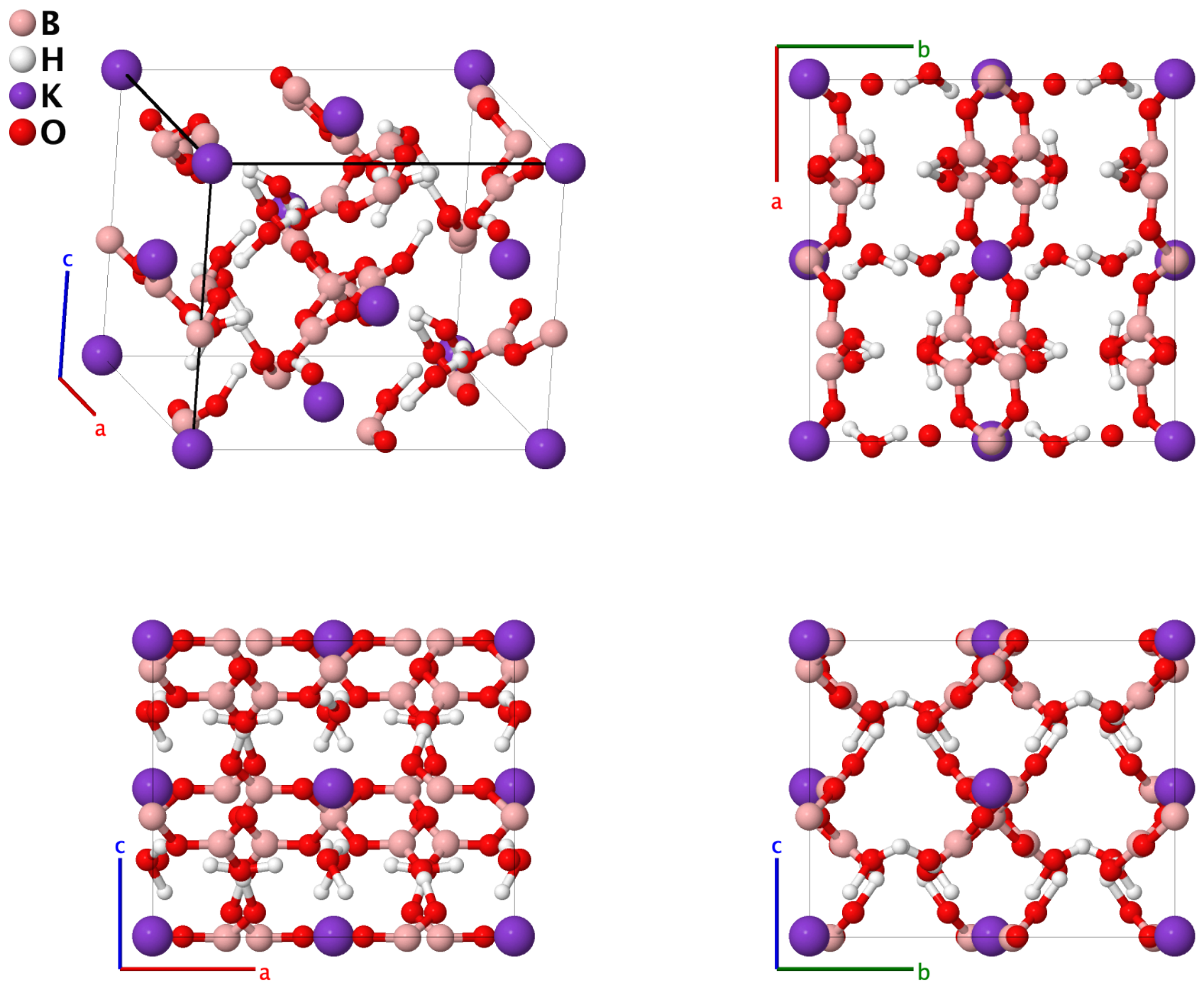

| | | |
|---|---|---|
| **Prototype** | : | $B_5H_8KO_{12}$ |
| **AFLOW prototype label** | : | A5B8CD12_oC104_41_a2b_4b_a_6b |
| ***Strukturbericht* designation** | : | $K3_5$ |
| **Pearson symbol** | : | oC104 |
| **Space group number** | : | 41 |
| **Space group symbol** | : | $Aba2$ |
| **AFLOW prototype command** | : | `aflow --proto=A5B8CD12_oC104_41_a2b_4b_a_6b` `--params=`$a, b/a, c/a, z_1, z_2, x_3, y_3, z_3, x_4, y_4, z_4, x_5, y_5, z_5, x_6, y_6, z_6, x_7, y_7, z_7, x_8, y_8, z_8, x_9, y_9, z_9, x_{10}, y_{10}, z_{10}, x_{11}, y_{11}, z_{11}, x_{12}, y_{12}, z_{12}, x_{13}, y_{13}, z_{13}, x_{14}, y_{14}, z_{14}$ |

- (Zachariasen, 1938) originally determined the structure of this compound, without knowing the positions of the hydrogens, but believed the structure to be correctly written as $KH_2(H_3O)_2B_5O_{10}$. (Gottfried, 1940) gave this the *Strukturbericht* designation $K3_5$. (Zachariasen, 1963) did a refinement of the structure, including the hydrogen positions, noting that the structure should properly be written $K[B_5O_6(OH)_4] \cdot 2H_2O$. As neither the positions of the heavy atoms nor the space group have changed, we retain the $K3_5$ designation, but we give the prototype as $KB_5O_8 \cdot 4H_2O$, which seems to be the standard formula even though it is not structurally correct.

**Base-centered Orthorhombic primitive vectors:**

$$\mathbf{a}_1 = a\,\hat{\mathbf{x}}$$
$$\mathbf{a}_2 = \frac{1}{2}\,b\,\hat{\mathbf{y}} - \frac{1}{2}\,c\,\hat{\mathbf{z}}$$
$$\mathbf{a}_3 = \frac{1}{2}\,b\,\hat{\mathbf{y}} + \frac{1}{2}\,c\,\hat{\mathbf{z}}$$

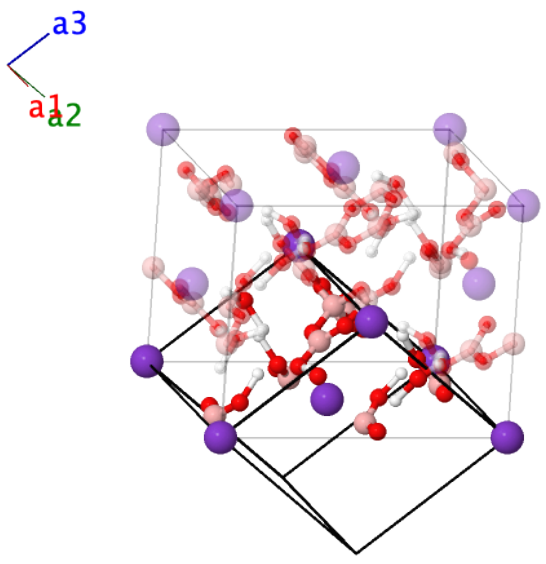

**Basis vectors:**

| | | Lattice Coordinates | | Cartesian Coordinates | Wyckoff Position | Atom Type |
|---|---|---|---|---|---|---|
| $\mathbf{B}_1$ | = | $-z_1\,\mathbf{a}_2 + z_1\,\mathbf{a}_3$ | = | $z_1 c\,\hat{\mathbf{z}}$ | (4$a$) | B I |
| $\mathbf{B}_2$ | = | $\frac{1}{2}\,\mathbf{a}_1 + \left(\frac{1}{2} - z_1\right)\,\mathbf{a}_2 + \left(\frac{1}{2} + z_1\right)\,\mathbf{a}_3$ | = | $\frac{1}{2}a\,\hat{\mathbf{x}} + \frac{1}{2}b\,\hat{\mathbf{y}} + z_1 c\,\hat{\mathbf{z}}$ | (4$a$) | B I |
| $\mathbf{B}_3$ | = | $-z_2\,\mathbf{a}_2 + z_2\,\mathbf{a}_3$ | = | $z_2 c\,\hat{\mathbf{z}}$ | (4$a$) | K |
| $\mathbf{B}_4$ | = | $\frac{1}{2}\,\mathbf{a}_1 + \left(\frac{1}{2} - z_2\right)\,\mathbf{a}_2 + \left(\frac{1}{2} + z_2\right)\,\mathbf{a}_3$ | = | $\frac{1}{2}a\,\hat{\mathbf{x}} + \frac{1}{2}b\,\hat{\mathbf{y}} + z_2 c\,\hat{\mathbf{z}}$ | (4$a$) | K |
| $\mathbf{B}_5$ | = | $x_3\,\mathbf{a}_1 + (y_3 - z_3)\,\mathbf{a}_2 + (y_3 + z_3)\,\mathbf{a}_3$ | = | $x_3 a\,\hat{\mathbf{x}} + y_3 b\,\hat{\mathbf{y}} + z_3 c\,\hat{\mathbf{z}}$ | (8$b$) | B II |
| $\mathbf{B}_6$ | = | $-x_3\,\mathbf{a}_1 + (-y_3 - z_3)\,\mathbf{a}_2 + (-y_3 + z_3)\,\mathbf{a}_3$ | = | $-x_3 a\,\hat{\mathbf{x}} - y_3 b\,\hat{\mathbf{y}} + z_3 c\,\hat{\mathbf{z}}$ | (8$b$) | B II |
| $\mathbf{B}_7$ | = | $\left(\frac{1}{2} + x_3\right)\,\mathbf{a}_1 + \left(\frac{1}{2} - y_3 - z_3\right)\,\mathbf{a}_2 + \left(\frac{1}{2} - y_3 + z_3\right)\,\mathbf{a}_3$ | = | $\left(\frac{1}{2} + x_3\right) a\,\hat{\mathbf{x}} + \left(\frac{1}{2} - y_3\right) b\,\hat{\mathbf{y}} + z_3 c\,\hat{\mathbf{z}}$ | (8$b$) | B II |
| $\mathbf{B}_8$ | = | $\left(\frac{1}{2} - x_3\right)\,\mathbf{a}_1 + \left(\frac{1}{2} + y_3 - z_3\right)\,\mathbf{a}_2 + \left(\frac{1}{2} + y_3 + z_3\right)\,\mathbf{a}_3$ | = | $\left(\frac{1}{2} - x_3\right) a\,\hat{\mathbf{x}} + \left(\frac{1}{2} + y_3\right) b\,\hat{\mathbf{y}} + z_3 c\,\hat{\mathbf{z}}$ | (8$b$) | B II |
| $\mathbf{B}_9$ | = | $x_4\,\mathbf{a}_1 + (y_4 - z_4)\,\mathbf{a}_2 + (y_4 + z_4)\,\mathbf{a}_3$ | = | $x_4 a\,\hat{\mathbf{x}} + y_4 b\,\hat{\mathbf{y}} + z_4 c\,\hat{\mathbf{z}}$ | (8$b$) | B III |
| $\mathbf{B}_{10}$ | = | $-x_4\,\mathbf{a}_1 + (-y_4 - z_4)\,\mathbf{a}_2 + (-y_4 + z_4)\,\mathbf{a}_3$ | = | $-x_4 a\,\hat{\mathbf{x}} - y_4 b\,\hat{\mathbf{y}} + z_4 c\,\hat{\mathbf{z}}$ | (8$b$) | B III |
| $\mathbf{B}_{11}$ | = | $\left(\frac{1}{2} + x_4\right)\,\mathbf{a}_1 + \left(\frac{1}{2} - y_4 - z_4\right)\,\mathbf{a}_2 + \left(\frac{1}{2} - y_4 + z_4\right)\,\mathbf{a}_3$ | = | $\left(\frac{1}{2} + x_4\right) a\,\hat{\mathbf{x}} + \left(\frac{1}{2} - y_4\right) b\,\hat{\mathbf{y}} + z_4 c\,\hat{\mathbf{z}}$ | (8$b$) | B III |
| $\mathbf{B}_{12}$ | = | $\left(\frac{1}{2} - x_4\right)\,\mathbf{a}_1 + \left(\frac{1}{2} + y_4 - z_4\right)\,\mathbf{a}_2 + \left(\frac{1}{2} + y_4 + z_4\right)\,\mathbf{a}_3$ | = | $\left(\frac{1}{2} - x_4\right) a\,\hat{\mathbf{x}} + \left(\frac{1}{2} + y_4\right) b\,\hat{\mathbf{y}} + z_4 c\,\hat{\mathbf{z}}$ | (8$b$) | B III |
| $\mathbf{B}_{13}$ | = | $x_5\,\mathbf{a}_1 + (y_5 - z_5)\,\mathbf{a}_2 + (y_5 + z_5)\,\mathbf{a}_3$ | = | $x_5 a\,\hat{\mathbf{x}} + y_5 b\,\hat{\mathbf{y}} + z_5 c\,\hat{\mathbf{z}}$ | (8$b$) | H I |
| $\mathbf{B}_{14}$ | = | $-x_5\,\mathbf{a}_1 + (-y_5 - z_5)\,\mathbf{a}_2 + (-y_5 + z_5)\,\mathbf{a}_3$ | = | $-x_5 a\,\hat{\mathbf{x}} - y_5 b\,\hat{\mathbf{y}} + z_5 c\,\hat{\mathbf{z}}$ | (8$b$) | H I |
| $\mathbf{B}_{15}$ | = | $\left(\frac{1}{2} + x_5\right)\,\mathbf{a}_1 + \left(\frac{1}{2} - y_5 - z_5\right)\,\mathbf{a}_2 + \left(\frac{1}{2} - y_5 + z_5\right)\,\mathbf{a}_3$ | = | $\left(\frac{1}{2} + x_5\right) a\,\hat{\mathbf{x}} + \left(\frac{1}{2} - y_5\right) b\,\hat{\mathbf{y}} + z_5 c\,\hat{\mathbf{z}}$ | (8$b$) | H I |
| $\mathbf{B}_{16}$ | = | $\left(\frac{1}{2} - x_5\right)\,\mathbf{a}_1 + \left(\frac{1}{2} + y_5 - z_5\right)\,\mathbf{a}_2 + \left(\frac{1}{2} + y_5 + z_5\right)\,\mathbf{a}_3$ | = | $\left(\frac{1}{2} - x_5\right) a\,\hat{\mathbf{x}} + \left(\frac{1}{2} + y_5\right) b\,\hat{\mathbf{y}} + z_5 c\,\hat{\mathbf{z}}$ | (8$b$) | H I |
| $\mathbf{B}_{17}$ | = | $x_6\,\mathbf{a}_1 + (y_6 - z_6)\,\mathbf{a}_2 + (y_6 + z_6)\,\mathbf{a}_3$ | = | $x_6 a\,\hat{\mathbf{x}} + y_6 b\,\hat{\mathbf{y}} + z_6 c\,\hat{\mathbf{z}}$ | (8$b$) | H II |
| $\mathbf{B}_{18}$ | = | $-x_6\,\mathbf{a}_1 + (-y_6 - z_6)\,\mathbf{a}_2 + (-y_6 + z_6)\,\mathbf{a}_3$ | = | $-x_6 a\,\hat{\mathbf{x}} - y_6 b\,\hat{\mathbf{y}} + z_6 c\,\hat{\mathbf{z}}$ | (8$b$) | H II |
| $\mathbf{B}_{19}$ | = | $\left(\frac{1}{2} + x_6\right)\,\mathbf{a}_1 + \left(\frac{1}{2} - y_6 - z_6\right)\,\mathbf{a}_2 + \left(\frac{1}{2} - y_6 + z_6\right)\,\mathbf{a}_3$ | = | $\left(\frac{1}{2} + x_6\right) a\,\hat{\mathbf{x}} + \left(\frac{1}{2} - y_6\right) b\,\hat{\mathbf{y}} + z_6 c\,\hat{\mathbf{z}}$ | (8$b$) | H II |
| $\mathbf{B}_{20}$ | = | $\left(\frac{1}{2} - x_6\right) \mathbf{a}_1 + \left(\frac{1}{2} + y_6 - z_6\right) \mathbf{a}_2 + \left(\frac{1}{2} + y_6 + z_6\right) \mathbf{a}_3$ | = | $\left(\frac{1}{2} - x_6\right) a\,\hat{\mathbf{x}} + \left(\frac{1}{2} + y_6\right) b\,\hat{\mathbf{y}} + z_6 c\,\hat{\mathbf{z}}$ | (8*b*) | H II |
| $\mathbf{B}_{21}$ | = | $x_7\,\mathbf{a}_1 + (y_7 - z_7)\,\mathbf{a}_2 + (y_7 + z_7)\,\mathbf{a}_3$ | = | $x_7 a\,\hat{\mathbf{x}} + y_7 b\,\hat{\mathbf{y}} + z_7 c\,\hat{\mathbf{z}}$ | (8*b*) | H III |
| $\mathbf{B}_{22}$ | = | $-x_7\,\mathbf{a}_1 + (-y_7 - z_7)\,\mathbf{a}_2 + (-y_7 + z_7)\,\mathbf{a}_3$ | = | $-x_7 a\,\hat{\mathbf{x}} - y_7 b\,\hat{\mathbf{y}} + z_7 c\,\hat{\mathbf{z}}$ | (8*b*) | H III |
| $\mathbf{B}_{23}$ | = | $\left(\frac{1}{2} + x_7\right) \mathbf{a}_1 + \left(\frac{1}{2} - y_7 - z_7\right) \mathbf{a}_2 + \left(\frac{1}{2} - y_7 + z_7\right) \mathbf{a}_3$ | = | $\left(\frac{1}{2} + x_7\right) a\,\hat{\mathbf{x}} + \left(\frac{1}{2} - y_7\right) b\,\hat{\mathbf{y}} + z_7 c\,\hat{\mathbf{z}}$ | (8*b*) | H III |
| $\mathbf{B}_{24}$ | = | $\left(\frac{1}{2} - x_7\right) \mathbf{a}_1 + \left(\frac{1}{2} + y_7 - z_7\right) \mathbf{a}_2 + \left(\frac{1}{2} + y_7 + z_7\right) \mathbf{a}_3$ | = | $\left(\frac{1}{2} - x_7\right) a\,\hat{\mathbf{x}} + \left(\frac{1}{2} + y_7\right) b\,\hat{\mathbf{y}} + z_7 c\,\hat{\mathbf{z}}$ | (8*b*) | H III |
| $\mathbf{B}_{25}$ | = | $x_8\,\mathbf{a}_1 + (y_8 - z_8)\,\mathbf{a}_2 + (y_8 + z_8)\,\mathbf{a}_3$ | = | $x_8 a\,\hat{\mathbf{x}} + y_8 b\,\hat{\mathbf{y}} + z_8 c\,\hat{\mathbf{z}}$ | (8*b*) | H IV |
| $\mathbf{B}_{26}$ | = | $-x_8\,\mathbf{a}_1 + (-y_8 - z_8)\,\mathbf{a}_2 + (-y_8 + z_8)\,\mathbf{a}_3$ | = | $-x_8 a\,\hat{\mathbf{x}} - y_8 b\,\hat{\mathbf{y}} + z_8 c\,\hat{\mathbf{z}}$ | (8*b*) | H IV |
| $\mathbf{B}_{27}$ | = | $\left(\frac{1}{2} + x_8\right) \mathbf{a}_1 + \left(\frac{1}{2} - y_8 - z_8\right) \mathbf{a}_2 + \left(\frac{1}{2} - y_8 + z_8\right) \mathbf{a}_3$ | = | $\left(\frac{1}{2} + x_8\right) a\,\hat{\mathbf{x}} + \left(\frac{1}{2} - y_8\right) b\,\hat{\mathbf{y}} + z_8 c\,\hat{\mathbf{z}}$ | (8*b*) | H IV |
| $\mathbf{B}_{28}$ | = | $\left(\frac{1}{2} - x_8\right) \mathbf{a}_1 + \left(\frac{1}{2} + y_8 - z_8\right) \mathbf{a}_2 + \left(\frac{1}{2} + y_8 + z_8\right) \mathbf{a}_3$ | = | $\left(\frac{1}{2} - x_8\right) a\,\hat{\mathbf{x}} + \left(\frac{1}{2} + y_8\right) b\,\hat{\mathbf{y}} + z_8 c\,\hat{\mathbf{z}}$ | (8*b*) | H IV |
| $\mathbf{B}_{29}$ | = | $x_9\,\mathbf{a}_1 + (y_9 - z_9)\,\mathbf{a}_2 + (y_9 + z_9)\,\mathbf{a}_3$ | = | $x_9 a\,\hat{\mathbf{x}} + y_9 b\,\hat{\mathbf{y}} + z_9 c\,\hat{\mathbf{z}}$ | (8*b*) | O I |
| $\mathbf{B}_{30}$ | = | $-x_9\,\mathbf{a}_1 + (-y_9 - z_9)\,\mathbf{a}_2 + (-y_9 + z_9)\,\mathbf{a}_3$ | = | $-x_9 a\,\hat{\mathbf{x}} - y_9 b\,\hat{\mathbf{y}} + z_9 c\,\hat{\mathbf{z}}$ | (8*b*) | O I |
| $\mathbf{B}_{31}$ | = | $\left(\frac{1}{2} + x_9\right) \mathbf{a}_1 + \left(\frac{1}{2} - y_9 - z_9\right) \mathbf{a}_2 + \left(\frac{1}{2} - y_9 + z_9\right) \mathbf{a}_3$ | = | $\left(\frac{1}{2} + x_9\right) a\,\hat{\mathbf{x}} + \left(\frac{1}{2} - y_9\right) b\,\hat{\mathbf{y}} + z_9 c\,\hat{\mathbf{z}}$ | (8*b*) | O I |
| $\mathbf{B}_{32}$ | = | $\left(\frac{1}{2} - x_9\right) \mathbf{a}_1 + \left(\frac{1}{2} + y_9 - z_9\right) \mathbf{a}_2 + \left(\frac{1}{2} + y_9 + z_9\right) \mathbf{a}_3$ | = | $\left(\frac{1}{2} - x_9\right) a\,\hat{\mathbf{x}} + \left(\frac{1}{2} + y_9\right) b\,\hat{\mathbf{y}} + z_9 c\,\hat{\mathbf{z}}$ | (8*b*) | O I |
| $\mathbf{B}_{33}$ | = | $x_{10}\,\mathbf{a}_1 + (y_{10} - z_{10})\,\mathbf{a}_2 + (y_{10} + z_{10})\,\mathbf{a}_3$ | = | $x_{10} a\,\hat{\mathbf{x}} + y_{10} b\,\hat{\mathbf{y}} + z_{10} c\,\hat{\mathbf{z}}$ | (8*b*) | O II |
| $\mathbf{B}_{34}$ | = | $-x_{10}\,\mathbf{a}_1 + (-y_{10} - z_{10})\,\mathbf{a}_2 + (-y_{10} + z_{10})\,\mathbf{a}_3$ | = | $-x_{10} a\,\hat{\mathbf{x}} - y_{10} b\,\hat{\mathbf{y}} + z_{10} c\,\hat{\mathbf{z}}$ | (8*b*) | O II |
| $\mathbf{B}_{35}$ | = | $\left(\frac{1}{2} + x_{10}\right) \mathbf{a}_1 + \left(\frac{1}{2} - y_{10} - z_{10}\right) \mathbf{a}_2 + \left(\frac{1}{2} - y_{10} + z_{10}\right) \mathbf{a}_3$ | = | $\left(\frac{1}{2} + x_{10}\right) a\,\hat{\mathbf{x}} + \left(\frac{1}{2} - y_{10}\right) b\,\hat{\mathbf{y}} + z_{10} c\,\hat{\mathbf{z}}$ | (8*b*) | O II |
| $\mathbf{B}_{36}$ | = | $\left(\frac{1}{2} - x_{10}\right) \mathbf{a}_1 + \left(\frac{1}{2} + y_{10} - z_{10}\right) \mathbf{a}_2 + \left(\frac{1}{2} + y_{10} + z_{10}\right) \mathbf{a}_3$ | = | $\left(\frac{1}{2} - x_{10}\right) a\,\hat{\mathbf{x}} + \left(\frac{1}{2} + y_{10}\right) b\,\hat{\mathbf{y}} + z_{10} c\,\hat{\mathbf{z}}$ | (8*b*) | O II |
| $\mathbf{B}_{37}$ | = | $x_{11}\,\mathbf{a}_1 + (y_{11} - z_{11})\,\mathbf{a}_2 + (y_{11} + z_{11})\,\mathbf{a}_3$ | = | $x_{11} a\,\hat{\mathbf{x}} + y_{11} b\,\hat{\mathbf{y}} + z_{11} c\,\hat{\mathbf{z}}$ | (8*b*) | O III |
| $\mathbf{B}_{38}$ | = | $-x_{11}\,\mathbf{a}_1 + (-y_{11} - z_{11})\,\mathbf{a}_2 + (-y_{11} + z_{11})\,\mathbf{a}_3$ | = | $-x_{11} a\,\hat{\mathbf{x}} - y_{11} b\,\hat{\mathbf{y}} + z_{11} c\,\hat{\mathbf{z}}$ | (8*b*) | O III |
| $\mathbf{B}_{39}$ | = | $\left(\frac{1}{2} + x_{11}\right) \mathbf{a}_1 + \left(\frac{1}{2} - y_{11} - z_{11}\right) \mathbf{a}_2 + \left(\frac{1}{2} - y_{11} + z_{11}\right) \mathbf{a}_3$ | = | $\left(\frac{1}{2} + x_{11}\right) a\,\hat{\mathbf{x}} + \left(\frac{1}{2} - y_{11}\right) b\,\hat{\mathbf{y}} + z_{11} c\,\hat{\mathbf{z}}$ | (8*b*) | O III |
| $\mathbf{B}_{40}$ | = | $\left(\frac{1}{2} - x_{11}\right) \mathbf{a}_1 + \left(\frac{1}{2} + y_{11} - z_{11}\right) \mathbf{a}_2 + \left(\frac{1}{2} + y_{11} + z_{11}\right) \mathbf{a}_3$ | = | $\left(\frac{1}{2} - x_{11}\right) a\,\hat{\mathbf{x}} + \left(\frac{1}{2} + y_{11}\right) b\,\hat{\mathbf{y}} + z_{11} c\,\hat{\mathbf{z}}$ | (8*b*) | O III |
| $\mathbf{B}_{41}$ | = | $x_{12}\,\mathbf{a}_1 + (y_{12} - z_{12})\,\mathbf{a}_2 + (y_{12} + z_{12})\,\mathbf{a}_3$ | = | $x_{12} a\,\hat{\mathbf{x}} + y_{12} b\,\hat{\mathbf{y}} + z_{12} c\,\hat{\mathbf{z}}$ | (8*b*) | O IV |
| $\mathbf{B}_{42}$ | = | $-x_{12}\,\mathbf{a}_1 + (-y_{12} - z_{12})\,\mathbf{a}_2 + (-y_{12} + z_{12})\,\mathbf{a}_3$ | = | $-x_{12} a\,\hat{\mathbf{x}} - y_{12} b\,\hat{\mathbf{y}} + z_{12} c\,\hat{\mathbf{z}}$ | (8*b*) | O IV |
| $\mathbf{B}_{43}$ | = | $\left(\frac{1}{2}+x_{12}\right)\mathbf{a}_1+\left(\frac{1}{2}-y_{12}-z_{12}\right)\mathbf{a}_2+\left(\frac{1}{2}-y_{12}+z_{12}\right)\mathbf{a}_3$ | = | $\left(\frac{1}{2}+x_{12}\right)a\,\hat{\mathbf{x}}+\left(\frac{1}{2}-y_{12}\right)b\,\hat{\mathbf{y}}+z_{12}c\,\hat{\mathbf{z}}$ | (8$b$) | O IV |
| $\mathbf{B}_{44}$ | = | $\left(\frac{1}{2}-x_{12}\right)\mathbf{a}_1+\left(\frac{1}{2}+y_{12}-z_{12}\right)\mathbf{a}_2+\left(\frac{1}{2}+y_{12}+z_{12}\right)\mathbf{a}_3$ | = | $\left(\frac{1}{2}-x_{12}\right)a\,\hat{\mathbf{x}}+\left(\frac{1}{2}+y_{12}\right)b\,\hat{\mathbf{y}}+z_{12}c\,\hat{\mathbf{z}}$ | (8$b$) | O IV |
| $\mathbf{B}_{45}$ | = | $x_{13}\,\mathbf{a}_1+(y_{13}-z_{13})\,\mathbf{a}_2+(y_{13}+z_{13})\,\mathbf{a}_3$ | = | $x_{13}a\,\hat{\mathbf{x}}+y_{13}b\,\hat{\mathbf{y}}+z_{13}c\,\hat{\mathbf{z}}$ | (8$b$) | O V |
| $\mathbf{B}_{46}$ | = | $-x_{13}\,\mathbf{a}_1+(-y_{13}-z_{13})\,\mathbf{a}_2+(-y_{13}+z_{13})\,\mathbf{a}_3$ | = | $-x_{13}a\,\hat{\mathbf{x}}-y_{13}b\,\hat{\mathbf{y}}+z_{13}c\,\hat{\mathbf{z}}$ | (8$b$) | O V |
| $\mathbf{B}_{47}$ | = | $\left(\frac{1}{2}+x_{13}\right)\mathbf{a}_1+\left(\frac{1}{2}-y_{13}-z_{13}\right)\mathbf{a}_2+\left(\frac{1}{2}-y_{13}+z_{13}\right)\mathbf{a}_3$ | = | $\left(\frac{1}{2}+x_{13}\right)a\,\hat{\mathbf{x}}+\left(\frac{1}{2}-y_{13}\right)b\,\hat{\mathbf{y}}+z_{13}c\,\hat{\mathbf{z}}$ | (8$b$) | O V |
| $\mathbf{B}_{48}$ | = | $\left(\frac{1}{2}-x_{13}\right)\mathbf{a}_1+\left(\frac{1}{2}+y_{13}-z_{13}\right)\mathbf{a}_2+\left(\frac{1}{2}+y_{13}+z_{13}\right)\mathbf{a}_3$ | = | $\left(\frac{1}{2}-x_{13}\right)a\,\hat{\mathbf{x}}+\left(\frac{1}{2}+y_{13}\right)b\,\hat{\mathbf{y}}+z_{13}c\,\hat{\mathbf{z}}$ | (8$b$) | O V |
| $\mathbf{B}_{49}$ | = | $x_{14}\,\mathbf{a}_1+(y_{14}-z_{14})\,\mathbf{a}_2+(y_{14}+z_{14})\,\mathbf{a}_3$ | = | $x_{14}a\,\hat{\mathbf{x}}+y_{14}b\,\hat{\mathbf{y}}+z_{14}c\,\hat{\mathbf{z}}$ | (8$b$) | O VI |
| $\mathbf{B}_{50}$ | = | $-x_{14}\,\mathbf{a}_1+(-y_{14}-z_{14})\,\mathbf{a}_2+(-y_{14}+z_{14})\,\mathbf{a}_3$ | = | $-x_{14}a\,\hat{\mathbf{x}}-y_{14}b\,\hat{\mathbf{y}}+z_{14}c\,\hat{\mathbf{z}}$ | (8$b$) | O VI |
| $\mathbf{B}_{51}$ | = | $\left(\frac{1}{2}+x_{14}\right)\mathbf{a}_1+\left(\frac{1}{2}-y_{14}-z_{14}\right)\mathbf{a}_2+\left(\frac{1}{2}-y_{14}+z_{14}\right)\mathbf{a}_3$ | = | $\left(\frac{1}{2}+x_{14}\right)a\,\hat{\mathbf{x}}+\left(\frac{1}{2}-y_{14}\right)b\,\hat{\mathbf{y}}+z_{14}c\,\hat{\mathbf{z}}$ | (8$b$) | O VI |
| $\mathbf{B}_{52}$ | = | $\left(\frac{1}{2}-x_{14}\right)\mathbf{a}_1+\left(\frac{1}{2}+y_{14}-z_{14}\right)\mathbf{a}_2+\left(\frac{1}{2}+y_{14}+z_{14}\right)\mathbf{a}_3$ | = | $\left(\frac{1}{2}-x_{14}\right)a\,\hat{\mathbf{x}}+\left(\frac{1}{2}+y_{14}\right)b\,\hat{\mathbf{y}}+z_{14}c\,\hat{\mathbf{z}}$ | (8$b$) | O VI |

**Geometry files:**

- CIF: pp. 1600
- POSCAR: pp. 1601

# $Ag_2O_3$ Structure: A2B3_oF40_43_b_ab

http://aflow.org/prototype-encyclopedia/A2B3_oF40_43_b_ab

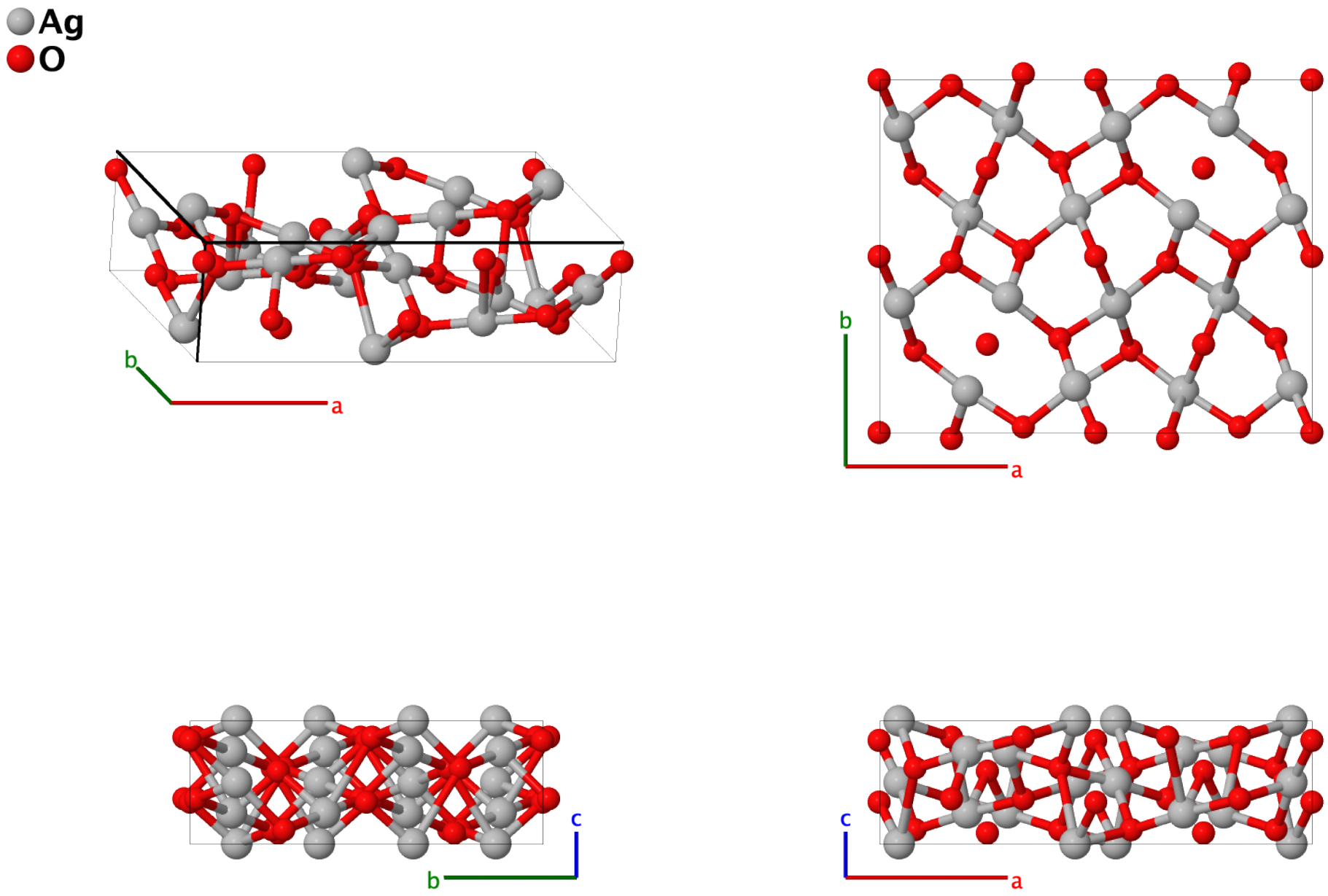

| | | |
|---|---|---|
| **Prototype** | : | $Ag_2O_3$ |
| **AFLOW prototype label** | : | A2B3_oF40_43_b_ab |
| ***Strukturbericht* designation** | : | None |
| **Pearson symbol** | : | oF40 |
| **Space group number** | : | 43 |
| **Space group symbol** | : | $Fdd2$ |
| **AFLOW prototype command** | : | `aflow --proto=A2B3_oF40_43_b_ab` `--params=`$a, b/a, c/a, z_1, x_2, y_2, z_2, x_3, y_3, z_3$ |

- This structure is a distortion of the $D5_5$ $Ag_2O_3$ structure (A3B2_cP10_224_d_b), although (Standke, 1986) does not seem to be aware of the earlier work. This is most likely closer to the correct structure for $Ag_2O_3$ than the $D5_5$ structure is.

**Face-centered Orthorhombic primitive vectors:**

$$\mathbf{a}_1 = \frac{1}{2}\,b\,\hat{\mathbf{y}} + \frac{1}{2}\,c\,\hat{\mathbf{z}}$$
$$\mathbf{a}_2 = \frac{1}{2}\,a\,\hat{\mathbf{x}} + \frac{1}{2}\,c\,\hat{\mathbf{z}}$$
$$\mathbf{a}_3 = \frac{1}{2}\,a\,\hat{\mathbf{x}} + \frac{1}{2}\,b\,\hat{\mathbf{y}}$$

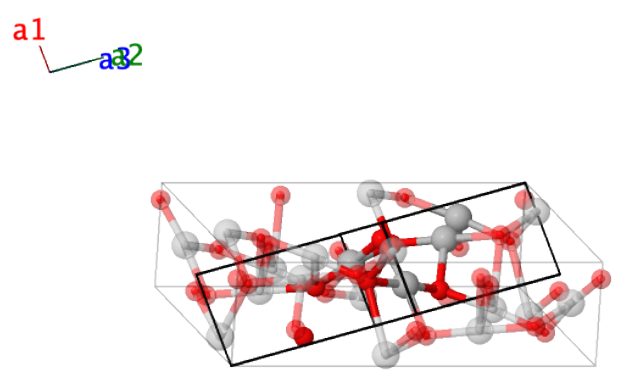

**Basis vectors:**

| | | Lattice Coordinates | | Cartesian Coordinates | Wyckoff Position | Atom Type |
|---|---|---|---|---|---|---|
| $\mathbf{B}_1$ | = | $z_1\,\mathbf{a}_1 + z_1\,\mathbf{a}_2 - z_1\,\mathbf{a}_3$ | = | $z_1 c\,\hat{\mathbf{z}}$ | (8$a$) | O I |
| $\mathbf{B}_2$ | = | $\left(\frac{1}{4}+z_1\right)\mathbf{a}_1 + \left(\frac{1}{4}+z_1\right)\mathbf{a}_2 + \left(\frac{1}{4}-z_1\right)\mathbf{a}_3$ | = | $\frac{1}{4}a\,\hat{\mathbf{x}} + \frac{1}{4}b\,\hat{\mathbf{y}} + \left(\frac{1}{4}+z_1\right)c\,\hat{\mathbf{z}}$ | (8$a$) | O I |
| $\mathbf{B}_3$ | = | $(-x_2+y_2+z_2)\,\mathbf{a}_1 + (x_2-y_2+z_2)\,\mathbf{a}_2 + (x_2+y_2-z_2)\,\mathbf{a}_3$ | = | $x_2 a\,\hat{\mathbf{x}} + y_2 b\,\hat{\mathbf{y}} + z_2 c\,\hat{\mathbf{z}}$ | (16$b$) | Ag |
| $\mathbf{B}_4$ | = | $(x_2-y_2+z_2)\,\mathbf{a}_1 + (-x_2+y_2+z_2)\,\mathbf{a}_2 + (-x_2-y_2-z_2)\,\mathbf{a}_3$ | = | $-x_2 a\,\hat{\mathbf{x}} - y_2 b\,\hat{\mathbf{y}} + z_2 c\,\hat{\mathbf{z}}$ | (16$b$) | Ag |
| $\mathbf{B}_5$ | = | $\left(\frac{1}{4}-x_2-y_2+z_2\right)\mathbf{a}_1 + \left(\frac{1}{4}+x_2+y_2+z_2\right)\mathbf{a}_2 + \left(\frac{1}{4}+x_2-y_2-z_2\right)\mathbf{a}_3$ | = | $\left(\frac{1}{4}+x_2\right)a\,\hat{\mathbf{x}} + \left(\frac{1}{4}-y_2\right)b\,\hat{\mathbf{y}} + \left(\frac{1}{4}+z_2\right)c\,\hat{\mathbf{z}}$ | (16$b$) | Ag |
| $\mathbf{B}_6$ | = | $\left(\frac{1}{4}+x_2+y_2+z_2\right)\mathbf{a}_1 + \left(\frac{1}{4}-x_2-y_2+z_2\right)\mathbf{a}_2 + \left(\frac{1}{4}-x_2+y_2-z_2\right)\mathbf{a}_3$ | = | $\left(\frac{1}{4}-x_2\right)a\,\hat{\mathbf{x}} + \left(\frac{1}{4}+y_2\right)b\,\hat{\mathbf{y}} + \left(\frac{1}{4}+z_2\right)c\,\hat{\mathbf{z}}$ | (16$b$) | Ag |
| $\mathbf{B}_7$ | = | $(-x_3+y_3+z_3)\,\mathbf{a}_1 + (x_3-y_3+z_3)\,\mathbf{a}_2 + (x_3+y_3-z_3)\,\mathbf{a}_3$ | = | $x_3 a\,\hat{\mathbf{x}} + y_3 b\,\hat{\mathbf{y}} + z_3 c\,\hat{\mathbf{z}}$ | (16$b$) | O II |
| $\mathbf{B}_8$ | = | $(x_3-y_3+z_3)\,\mathbf{a}_1 + (-x_3+y_3+z_3)\,\mathbf{a}_2 + (-x_3-y_3-z_3)\,\mathbf{a}_3$ | = | $-x_3 a\,\hat{\mathbf{x}} - y_3 b\,\hat{\mathbf{y}} + z_3 c\,\hat{\mathbf{z}}$ | (16$b$) | O II |
| $\mathbf{B}_9$ | = | $\left(\frac{1}{4}-x_3-y_3+z_3\right)\mathbf{a}_1 + \left(\frac{1}{4}+x_3+y_3+z_3\right)\mathbf{a}_2 + \left(\frac{1}{4}+x_3-y_3-z_3\right)\mathbf{a}_3$ | = | $\left(\frac{1}{4}+x_3\right)a\,\hat{\mathbf{x}} + \left(\frac{1}{4}-y_3\right)b\,\hat{\mathbf{y}} + \left(\frac{1}{4}+z_3\right)c\,\hat{\mathbf{z}}$ | (16$b$) | O II |
| $\mathbf{B}_{10}$ | = | $\left(\frac{1}{4}+x_3+y_3+z_3\right)\mathbf{a}_1 + \left(\frac{1}{4}-x_3-y_3+z_3\right)\mathbf{a}_2 + \left(\frac{1}{4}-x_3+y_3-z_3\right)\mathbf{a}_3$ | = | $\left(\frac{1}{4}-x_3\right)a\,\hat{\mathbf{x}} + \left(\frac{1}{4}+y_3\right)b\,\hat{\mathbf{y}} + \left(\frac{1}{4}+z_3\right)c\,\hat{\mathbf{z}}$ | (16$b$) | O II |

**Geometry files:**

- CIF: pp. 1601
- POSCAR: pp. 1601

# Natrolite ($Na_2Al_2Si_3O_{10}{\cdot}2H_2O$, $S6_{10}$) Structure: A2B4C2D12E3_oF184_43_b_2b_b_6b_ab

http://aflow.org/prototype-encyclopedia/A2B4C2D12E3_oF184_43_b_2b_b_6b_ab

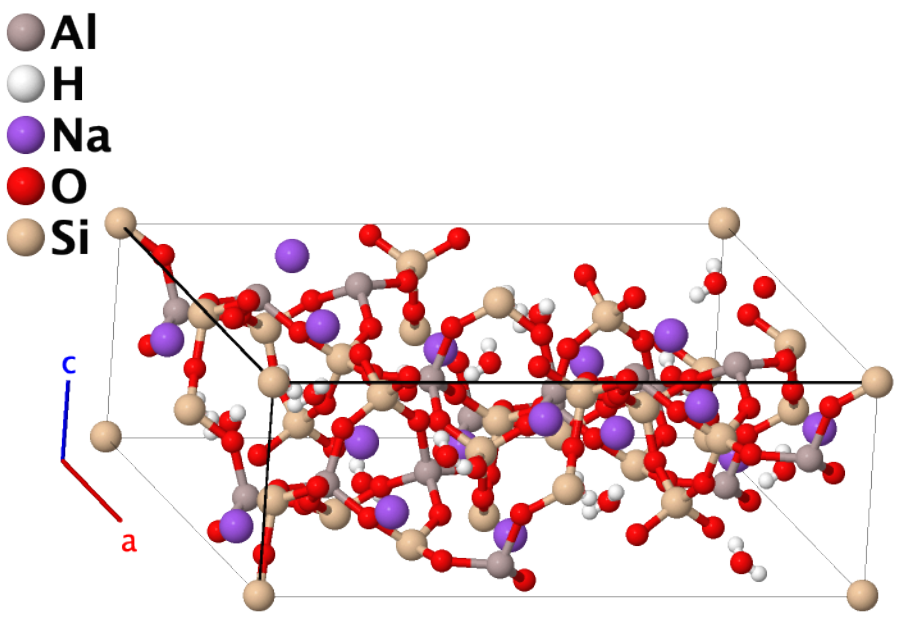

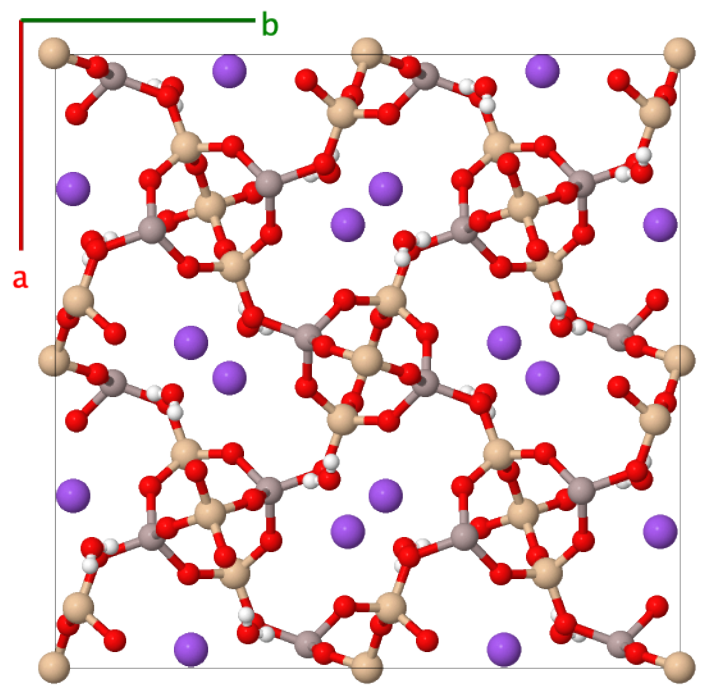

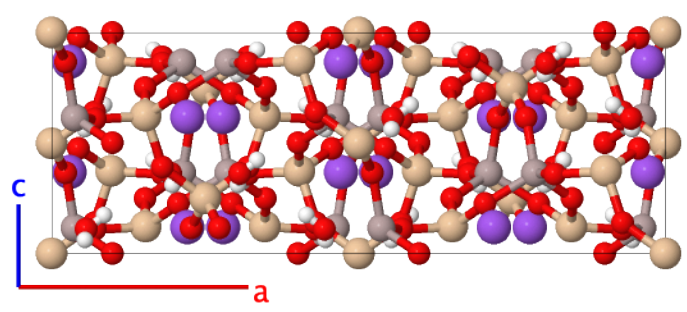

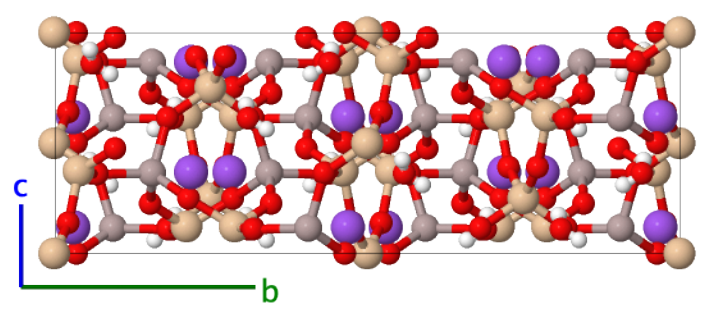

| | | |
|---|---|---|
| **Prototype** | : | $Al_2H_4Na_2O_{12}Si_3$ |
| **AFLOW prototype label** | : | A2B4C2D12E3_oF184_43_b_2b_b_6b_ab |
| ***Strukturbericht* designation** | : | $S6_{10}$ |
| **Pearson symbol** | : | oF184 |
| **Space group number** | : | 43 |
| **Space group symbol** | : | $Fdd2$ |
| **AFLOW prototype command** | : | `aflow --proto=A2B4C2D12E3_oF184_43_b_2b_b_6b_ab` `--params=`$a, b/a, c/a, z_1, x_2, y_2, z_2, x_3, y_3, z_3, x_4, y_4, z_4, x_5, y_5, z_5, x_6, y_6, z_6, x_7, y_7, z_7, x_8, y_8, z_8, x_9, y_9, z_9, x_{10}, y_{10}, z_{10}, x_{11}, y_{11}, z_{11}, x_{12}, y_{12}, z_{12}$ |

- We use the data from the sample that (Kirfel, 1984) call Crystal II. The origin has been arbitrarily set so that $z_1 = 0$.

## Face-centered Orthorhombic primitive vectors:

$$\mathbf{a}_1 = \tfrac{1}{2}\,b\,\hat{\mathbf{y}} + \tfrac{1}{2}\,c\,\hat{\mathbf{z}}$$
$$\mathbf{a}_2 = \tfrac{1}{2}\,a\,\hat{\mathbf{x}} + \tfrac{1}{2}\,c\,\hat{\mathbf{z}}$$
$$\mathbf{a}_3 = \tfrac{1}{2}\,a\,\hat{\mathbf{x}} + \tfrac{1}{2}\,b\,\hat{\mathbf{y}}$$

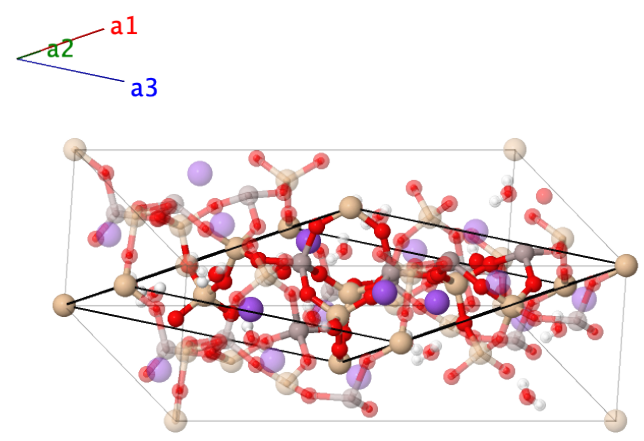

## Basis vectors:

| | | Lattice Coordinates | | Cartesian Coordinates | Wyckoff Position | Atom Type |
|---|---|---|---|---|---|---|
| $\mathbf{B}_1$ | = | $z_1\,\mathbf{a}_1 + z_1\,\mathbf{a}_2 - z_1\,\mathbf{a}_3$ | = | $z_1 c\,\hat{\mathbf{z}}$ | (8$a$) | Si I |
| $\mathbf{B}_2$ | = | $\left(\frac{1}{4} + z_1\right)\mathbf{a}_1 + \left(\frac{1}{4} + z_1\right)\mathbf{a}_2 + \left(\frac{1}{4} - z_1\right)\mathbf{a}_3$ | = | $\frac{1}{4}a\,\hat{\mathbf{x}} + \frac{1}{4}b\,\hat{\mathbf{y}} + \left(\frac{1}{4} + z_1\right)c\,\hat{\mathbf{z}}$ | (8$a$) | Si I |
| $\mathbf{B}_3$ | = | $(-x_2 + y_2 + z_2)\,\mathbf{a}_1 + (x_2 - y_2 + z_2)\,\mathbf{a}_2 + (x_2 + y_2 - z_2)\,\mathbf{a}_3$ | = | $x_2 a\,\hat{\mathbf{x}} + y_2 b\,\hat{\mathbf{y}} + z_2 c\,\hat{\mathbf{z}}$ | (16$b$) | Al |
| $\mathbf{B}_4$ | = | $(x_2 - y_2 + z_2)\,\mathbf{a}_1 + (-x_2 + y_2 + z_2)\,\mathbf{a}_2 + (-x_2 - y_2 - z_2)\,\mathbf{a}_3$ | = | $-x_2 a\,\hat{\mathbf{x}} - y_2 b\,\hat{\mathbf{y}} + z_2 c\,\hat{\mathbf{z}}$ | (16$b$) | Al |
| $\mathbf{B}_5$ | = | $\left(\frac{1}{4} - x_2 - y_2 + z_2\right)\mathbf{a}_1 + \left(\frac{1}{4} + x_2 + y_2 + z_2\right)\mathbf{a}_2 + \left(\frac{1}{4} + x_2 - y_2 - z_2\right)\mathbf{a}_3$ | = | $\left(\frac{1}{4} + x_2\right)a\,\hat{\mathbf{x}} + \left(\frac{1}{4} - y_2\right)b\,\hat{\mathbf{y}} + \left(\frac{1}{4} + z_2\right)c\,\hat{\mathbf{z}}$ | (16$b$) | Al |
| $\mathbf{B}_6$ | = | $\left(\frac{1}{4} + x_2 + y_2 + z_2\right)\mathbf{a}_1 + \left(\frac{1}{4} - x_2 - y_2 + z_2\right)\mathbf{a}_2 + \left(\frac{1}{4} - x_2 + y_2 - z_2\right)\mathbf{a}_3$ | = | $\left(\frac{1}{4} - x_2\right)a\,\hat{\mathbf{x}} + \left(\frac{1}{4} + y_2\right)b\,\hat{\mathbf{y}} + \left(\frac{1}{4} + z_2\right)c\,\hat{\mathbf{z}}$ | (16$b$) | Al |
| $\mathbf{B}_7$ | = | $(-x_3 + y_3 + z_3)\,\mathbf{a}_1 + (x_3 - y_3 + z_3)\,\mathbf{a}_2 + (x_3 + y_3 - z_3)\,\mathbf{a}_3$ | = | $x_3 a\,\hat{\mathbf{x}} + y_3 b\,\hat{\mathbf{y}} + z_3 c\,\hat{\mathbf{z}}$ | (16$b$) | H I |
| $\mathbf{B}_8$ | = | $(x_3 - y_3 + z_3)\,\mathbf{a}_1 + (-x_3 + y_3 + z_3)\,\mathbf{a}_2 + (-x_3 - y_3 - z_3)\,\mathbf{a}_3$ | = | $-x_3 a\,\hat{\mathbf{x}} - y_3 b\,\hat{\mathbf{y}} + z_3 c\,\hat{\mathbf{z}}$ | (16$b$) | H I |
| $\mathbf{B}_9$ | = | $\left(\frac{1}{4} - x_3 - y_3 + z_3\right)\mathbf{a}_1 + \left(\frac{1}{4} + x_3 + y_3 + z_3\right)\mathbf{a}_2 + \left(\frac{1}{4} + x_3 - y_3 - z_3\right)\mathbf{a}_3$ | = | $\left(\frac{1}{4} + x_3\right)a\,\hat{\mathbf{x}} + \left(\frac{1}{4} - y_3\right)b\,\hat{\mathbf{y}} + \left(\frac{1}{4} + z_3\right)c\,\hat{\mathbf{z}}$ | (16$b$) | H I |
| $\mathbf{B}_{10}$ | = | $\left(\frac{1}{4} + x_3 + y_3 + z_3\right)\mathbf{a}_1 + \left(\frac{1}{4} - x_3 - y_3 + z_3\right)\mathbf{a}_2 + \left(\frac{1}{4} - x_3 + y_3 - z_3\right)\mathbf{a}_3$ | = | $\left(\frac{1}{4} - x_3\right)a\,\hat{\mathbf{x}} + \left(\frac{1}{4} + y_3\right)b\,\hat{\mathbf{y}} + \left(\frac{1}{4} + z_3\right)c\,\hat{\mathbf{z}}$ | (16$b$) | H I |
| $\mathbf{B}_{11}$ | = | $(-x_4 + y_4 + z_4)\,\mathbf{a}_1 + (x_4 - y_4 + z_4)\,\mathbf{a}_2 + (x_4 + y_4 - z_4)\,\mathbf{a}_3$ | = | $x_4 a\,\hat{\mathbf{x}} + y_4 b\,\hat{\mathbf{y}} + z_4 c\,\hat{\mathbf{z}}$ | (16$b$) | H II |
| $\mathbf{B}_{12}$ | = | $(x_4 - y_4 + z_4)\,\mathbf{a}_1 + (-x_4 + y_4 + z_4)\,\mathbf{a}_2 + (-x_4 - y_4 - z_4)\,\mathbf{a}_3$ | = | $-x_4 a\,\hat{\mathbf{x}} - y_4 b\,\hat{\mathbf{y}} + z_4 c\,\hat{\mathbf{z}}$ | (16$b$) | H II |
| $\mathbf{B}_{13}$ | = | $\left(\frac{1}{4} - x_4 - y_4 + z_4\right)\,\mathbf{a}_1 + \left(\frac{1}{4} + x_4 + y_4 + z_4\right)\,\mathbf{a}_2 + \left(\frac{1}{4} + x_4 - y_4 - z_4\right)\,\mathbf{a}_3$ | = | $\left(\frac{1}{4} + x_4\right) a\,\hat{\mathbf{x}} + \left(\frac{1}{4} - y_4\right) b\,\hat{\mathbf{y}} + \left(\frac{1}{4} + z_4\right) c\,\hat{\mathbf{z}}$ | (16$b$) | H II |
| $\mathbf{B}_{14}$ | = | $\left(\frac{1}{4} + x_4 + y_4 + z_4\right)\,\mathbf{a}_1 + \left(\frac{1}{4} - x_4 - y_4 + z_4\right)\,\mathbf{a}_2 + \left(\frac{1}{4} - x_4 + y_4 - z_4\right)\,\mathbf{a}_3$ | = | $\left(\frac{1}{4} - x_4\right) a\,\hat{\mathbf{x}} + \left(\frac{1}{4} + y_4\right) b\,\hat{\mathbf{y}} + \left(\frac{1}{4} + z_4\right) c\,\hat{\mathbf{z}}$ | (16$b$) | H II |
| $\mathbf{B}_{15}$ | = | $(-x_5 + y_5 + z_5)\,\mathbf{a}_1 + (x_5 - y_5 + z_5)\,\mathbf{a}_2 + (x_5 + y_5 - z_5)\,\mathbf{a}_3$ | = | $x_5 a\,\hat{\mathbf{x}} + y_5 b\,\hat{\mathbf{y}} + z_5 c\,\hat{\mathbf{z}}$ | (16$b$) | Na |
| $\mathbf{B}_{16}$ | = | $(x_5 - y_5 + z_5)\,\mathbf{a}_1 + (-x_5 + y_5 + z_5)\,\mathbf{a}_2 + (-x_5 - y_5 - z_5)\,\mathbf{a}_3$ | = | $-x_5 a\,\hat{\mathbf{x}} - y_5 b\,\hat{\mathbf{y}} + z_5 c\,\hat{\mathbf{z}}$ | (16$b$) | Na |
| $\mathbf{B}_{17}$ | = | $\left(\frac{1}{4} - x_5 - y_5 + z_5\right)\,\mathbf{a}_1 + \left(\frac{1}{4} + x_5 + y_5 + z_5\right)\,\mathbf{a}_2 + \left(\frac{1}{4} + x_5 - y_5 - z_5\right)\,\mathbf{a}_3$ | = | $\left(\frac{1}{4} + x_5\right) a\,\hat{\mathbf{x}} + \left(\frac{1}{4} - y_5\right) b\,\hat{\mathbf{y}} + \left(\frac{1}{4} + z_5\right) c\,\hat{\mathbf{z}}$ | (16$b$) | Na |
| $\mathbf{B}_{18}$ | = | $\left(\frac{1}{4} + x_5 + y_5 + z_5\right)\,\mathbf{a}_1 + \left(\frac{1}{4} - x_5 - y_5 + z_5\right)\,\mathbf{a}_2 + \left(\frac{1}{4} - x_5 + y_5 - z_5\right)\,\mathbf{a}_3$ | = | $\left(\frac{1}{4} - x_5\right) a\,\hat{\mathbf{x}} + \left(\frac{1}{4} + y_5\right) b\,\hat{\mathbf{y}} + \left(\frac{1}{4} + z_5\right) c\,\hat{\mathbf{z}}$ | (16$b$) | Na |
| $\mathbf{B}_{19}$ | = | $(-x_6 + y_6 + z_6)\,\mathbf{a}_1 + (x_6 - y_6 + z_6)\,\mathbf{a}_2 + (x_6 + y_6 - z_6)\,\mathbf{a}_3$ | = | $x_6 a\,\hat{\mathbf{x}} + y_6 b\,\hat{\mathbf{y}} + z_6 c\,\hat{\mathbf{z}}$ | (16$b$) | O I |
| $\mathbf{B}_{20}$ | = | $(x_6 - y_6 + z_6)\,\mathbf{a}_1 + (-x_6 + y_6 + z_6)\,\mathbf{a}_2 + (-x_6 - y_6 - z_6)\,\mathbf{a}_3$ | = | $-x_6 a\,\hat{\mathbf{x}} - y_6 b\,\hat{\mathbf{y}} + z_6 c\,\hat{\mathbf{z}}$ | (16$b$) | O I |
| $\mathbf{B}_{21}$ | = | $\left(\frac{1}{4} - x_6 - y_6 + z_6\right)\,\mathbf{a}_1 + \left(\frac{1}{4} + x_6 + y_6 + z_6\right)\,\mathbf{a}_2 + \left(\frac{1}{4} + x_6 - y_6 - z_6\right)\,\mathbf{a}_3$ | = | $\left(\frac{1}{4} + x_6\right) a\,\hat{\mathbf{x}} + \left(\frac{1}{4} - y_6\right) b\,\hat{\mathbf{y}} + \left(\frac{1}{4} + z_6\right) c\,\hat{\mathbf{z}}$ | (16$b$) | O I |
| $\mathbf{B}_{22}$ | = | $\left(\frac{1}{4} + x_6 + y_6 + z_6\right)\,\mathbf{a}_1 + \left(\frac{1}{4} - x_6 - y_6 + z_6\right)\,\mathbf{a}_2 + \left(\frac{1}{4} - x_6 + y_6 - z_6\right)\,\mathbf{a}_3$ | = | $\left(\frac{1}{4} - x_6\right) a\,\hat{\mathbf{x}} + \left(\frac{1}{4} + y_6\right) b\,\hat{\mathbf{y}} + \left(\frac{1}{4} + z_6\right) c\,\hat{\mathbf{z}}$ | (16$b$) | O I |
| $\mathbf{B}_{23}$ | = | $(-x_7 + y_7 + z_7)\,\mathbf{a}_1 + (x_7 - y_7 + z_7)\,\mathbf{a}_2 + (x_7 + y_7 - z_7)\,\mathbf{a}_3$ | = | $x_7 a\,\hat{\mathbf{x}} + y_7 b\,\hat{\mathbf{y}} + z_7 c\,\hat{\mathbf{z}}$ | (16$b$) | O II |
| $\mathbf{B}_{24}$ | = | $(x_7 - y_7 + z_7)\,\mathbf{a}_1 + (-x_7 + y_7 + z_7)\,\mathbf{a}_2 + (-x_7 - y_7 - z_7)\,\mathbf{a}_3$ | = | $-x_7 a\,\hat{\mathbf{x}} - y_7 b\,\hat{\mathbf{y}} + z_7 c\,\hat{\mathbf{z}}$ | (16$b$) | O II |
| $\mathbf{B}_{25}$ | = | $\left(\frac{1}{4} - x_7 - y_7 + z_7\right)\,\mathbf{a}_1 + \left(\frac{1}{4} + x_7 + y_7 + z_7\right)\,\mathbf{a}_2 + \left(\frac{1}{4} + x_7 - y_7 - z_7\right)\,\mathbf{a}_3$ | = | $\left(\frac{1}{4} + x_7\right) a\,\hat{\mathbf{x}} + \left(\frac{1}{4} - y_7\right) b\,\hat{\mathbf{y}} + \left(\frac{1}{4} + z_7\right) c\,\hat{\mathbf{z}}$ | (16$b$) | O II |
| $\mathbf{B}_{26}$ | = | $\left(\frac{1}{4} + x_7 + y_7 + z_7\right)\,\mathbf{a}_1 + \left(\frac{1}{4} - x_7 - y_7 + z_7\right)\,\mathbf{a}_2 + \left(\frac{1}{4} - x_7 + y_7 - z_7\right)\,\mathbf{a}_3$ | = | $\left(\frac{1}{4} - x_7\right) a\,\hat{\mathbf{x}} + \left(\frac{1}{4} + y_7\right) b\,\hat{\mathbf{y}} + \left(\frac{1}{4} + z_7\right) c\,\hat{\mathbf{z}}$ | (16$b$) | O II |
| $\mathbf{B}_{27}$ | = | $(-x_8 + y_8 + z_8)\,\mathbf{a}_1 + (x_8 - y_8 + z_8)\,\mathbf{a}_2 + (x_8 + y_8 - z_8)\,\mathbf{a}_3$ | = | $x_8 a\,\hat{\mathbf{x}} + y_8 b\,\hat{\mathbf{y}} + z_8 c\,\hat{\mathbf{z}}$ | (16$b$) | O III |
| $\mathbf{B}_{28}$ | = | $(x_8 - y_8 + z_8)\ \mathbf{a}_1 + (-x_8 + y_8 + z_8)\ \mathbf{a}_2 + (-x_8 - y_8 - z_8)\ \mathbf{a}_3$ | = | $-x_8 a\,\hat{\mathbf{x}} - y_8 b\,\hat{\mathbf{y}} + z_8 c\,\hat{\mathbf{z}}$ | (16*b*) | O III |
| $\mathbf{B}_{29}$ | = | $\left(\frac{1}{4} - x_8 - y_8 + z_8\right)\ \mathbf{a}_1 + \left(\frac{1}{4} + x_8 + y_8 + z_8\right)\ \mathbf{a}_2 + \left(\frac{1}{4} + x_8 - y_8 - z_8\right)\ \mathbf{a}_3$ | = | $\left(\frac{1}{4} + x_8\right) a\,\hat{\mathbf{x}} + \left(\frac{1}{4} - y_8\right) b\,\hat{\mathbf{y}} + \left(\frac{1}{4} + z_8\right) c\,\hat{\mathbf{z}}$ | (16*b*) | O III |
| $\mathbf{B}_{30}$ | = | $\left(\frac{1}{4} + x_8 + y_8 + z_8\right)\ \mathbf{a}_1 + \left(\frac{1}{4} - x_8 - y_8 + z_8\right)\ \mathbf{a}_2 + \left(\frac{1}{4} - x_8 + y_8 - z_8\right)\ \mathbf{a}_3$ | = | $\left(\frac{1}{4} - x_8\right) a\,\hat{\mathbf{x}} + \left(\frac{1}{4} + y_8\right) b\,\hat{\mathbf{y}} + \left(\frac{1}{4} + z_8\right) c\,\hat{\mathbf{z}}$ | (16*b*) | O III |
| $\mathbf{B}_{31}$ | = | $(-x_9 + y_9 + z_9)\ \mathbf{a}_1 + (x_9 - y_9 + z_9)\ \mathbf{a}_2 + (x_9 + y_9 - z_9)\ \mathbf{a}_3$ | = | $x_9 a\,\hat{\mathbf{x}} + y_9 b\,\hat{\mathbf{y}} + z_9 c\,\hat{\mathbf{z}}$ | (16*b*) | O IV |
| $\mathbf{B}_{32}$ | = | $(x_9 - y_9 + z_9)\ \mathbf{a}_1 + (-x_9 + y_9 + z_9)\ \mathbf{a}_2 + (-x_9 - y_9 - z_9)\ \mathbf{a}_3$ | = | $-x_9 a\,\hat{\mathbf{x}} - y_9 b\,\hat{\mathbf{y}} + z_9 c\,\hat{\mathbf{z}}$ | (16*b*) | O IV |
| $\mathbf{B}_{33}$ | = | $\left(\frac{1}{4} - x_9 - y_9 + z_9\right)\ \mathbf{a}_1 + \left(\frac{1}{4} + x_9 + y_9 + z_9\right)\ \mathbf{a}_2 + \left(\frac{1}{4} + x_9 - y_9 - z_9\right)\ \mathbf{a}_3$ | = | $\left(\frac{1}{4} + x_9\right) a\,\hat{\mathbf{x}} + \left(\frac{1}{4} - y_9\right) b\,\hat{\mathbf{y}} + \left(\frac{1}{4} + z_9\right) c\,\hat{\mathbf{z}}$ | (16*b*) | O IV |
| $\mathbf{B}_{34}$ | = | $\left(\frac{1}{4} + x_9 + y_9 + z_9\right)\ \mathbf{a}_1 + \left(\frac{1}{4} - x_9 - y_9 + z_9\right)\ \mathbf{a}_2 + \left(\frac{1}{4} - x_9 + y_9 - z_9\right)\ \mathbf{a}_3$ | = | $\left(\frac{1}{4} - x_9\right) a\,\hat{\mathbf{x}} + \left(\frac{1}{4} + y_9\right) b\,\hat{\mathbf{y}} + \left(\frac{1}{4} + z_9\right) c\,\hat{\mathbf{z}}$ | (16*b*) | O IV |
| $\mathbf{B}_{35}$ | = | $(-x_{10} + y_{10} + z_{10})\ \mathbf{a}_1 + (x_{10} - y_{10} + z_{10})\ \mathbf{a}_2 + (x_{10} + y_{10} - z_{10})\ \mathbf{a}_3$ | = | $x_{10} a\,\hat{\mathbf{x}} + y_{10} b\,\hat{\mathbf{y}} + z_{10} c\,\hat{\mathbf{z}}$ | (16*b*) | O V |
| $\mathbf{B}_{36}$ | = | $(x_{10} - y_{10} + z_{10})\ \mathbf{a}_1 + (-x_{10} + y_{10} + z_{10})\ \mathbf{a}_2 + (-x_{10} - y_{10} - z_{10})\ \mathbf{a}_3$ | = | $-x_{10} a\,\hat{\mathbf{x}} - y_{10} b\,\hat{\mathbf{y}} + z_{10} c\,\hat{\mathbf{z}}$ | (16*b*) | O V |
| $\mathbf{B}_{37}$ | = | $\left(\frac{1}{4} - x_{10} - y_{10} + z_{10}\right)\ \mathbf{a}_1 + \left(\frac{1}{4} + x_{10} + y_{10} + z_{10}\right)\ \mathbf{a}_2 + \left(\frac{1}{4} + x_{10} - y_{10} - z_{10}\right)\ \mathbf{a}_3$ | = | $\left(\frac{1}{4} + x_{10}\right) a\,\hat{\mathbf{x}} + \left(\frac{1}{4} - y_{10}\right) b\,\hat{\mathbf{y}} + \left(\frac{1}{4} + z_{10}\right) c\,\hat{\mathbf{z}}$ | (16*b*) | O V |
| $\mathbf{B}_{38}$ | = | $\left(\frac{1}{4} + x_{10} + y_{10} + z_{10}\right)\ \mathbf{a}_1 + \left(\frac{1}{4} - x_{10} - y_{10} + z_{10}\right)\ \mathbf{a}_2 + \left(\frac{1}{4} - x_{10} + y_{10} - z_{10}\right)\ \mathbf{a}_3$ | = | $\left(\frac{1}{4} - x_{10}\right) a\,\hat{\mathbf{x}} + \left(\frac{1}{4} + y_{10}\right) b\,\hat{\mathbf{y}} + \left(\frac{1}{4} + z_{10}\right) c\,\hat{\mathbf{z}}$ | (16*b*) | O V |
| $\mathbf{B}_{39}$ | = | $(-x_{11} + y_{11} + z_{11})\ \mathbf{a}_1 + (x_{11} - y_{11} + z_{11})\ \mathbf{a}_2 + (x_{11} + y_{11} - z_{11})\ \mathbf{a}_3$ | = | $x_{11} a\,\hat{\mathbf{x}} + y_{11} b\,\hat{\mathbf{y}} + z_{11} c\,\hat{\mathbf{z}}$ | (16*b*) | O VI |
| $\mathbf{B}_{40}$ | = | $(x_{11} - y_{11} + z_{11})\ \mathbf{a}_1 + (-x_{11} + y_{11} + z_{11})\ \mathbf{a}_2 + (-x_{11} - y_{11} - z_{11})\ \mathbf{a}_3$ | = | $-x_{11} a\,\hat{\mathbf{x}} - y_{11} b\,\hat{\mathbf{y}} + z_{11} c\,\hat{\mathbf{z}}$ | (16*b*) | O VI |
| $\mathbf{B}_{41}$ | = | $\left(\frac{1}{4} - x_{11} - y_{11} + z_{11}\right)\ \mathbf{a}_1 + \left(\frac{1}{4} + x_{11} + y_{11} + z_{11}\right)\ \mathbf{a}_2 + \left(\frac{1}{4} + x_{11} - y_{11} - z_{11}\right)\ \mathbf{a}_3$ | = | $\left(\frac{1}{4} + x_{11}\right) a\,\hat{\mathbf{x}} + \left(\frac{1}{4} - y_{11}\right) b\,\hat{\mathbf{y}} + \left(\frac{1}{4} + z_{11}\right) c\,\hat{\mathbf{z}}$ | (16*b*) | O VI |
| $\mathbf{B}_{42}$ | = | $\left(\frac{1}{4} + x_{11} + y_{11} + z_{11}\right)\ \mathbf{a}_1 + \left(\frac{1}{4} - x_{11} - y_{11} + z_{11}\right)\ \mathbf{a}_2 + \left(\frac{1}{4} - x_{11} + y_{11} - z_{11}\right)\ \mathbf{a}_3$ | = | $\left(\frac{1}{4} - x_{11}\right) a\,\hat{\mathbf{x}} + \left(\frac{1}{4} + y_{11}\right) b\,\hat{\mathbf{y}} + \left(\frac{1}{4} + z_{11}\right) c\,\hat{\mathbf{z}}$ | (16*b*) | O VI |
| $\mathbf{B}_{43}$ | = | $\left(-x_{12}+y_{12}+z_{12}\right)\,\mathbf{a}_1 + \left(x_{12}-y_{12}+z_{12}\right)\,\mathbf{a}_2 + \left(x_{12}+y_{12}-z_{12}\right)\,\mathbf{a}_3$ | = | $x_{12}a\,\hat{\mathbf{x}} + y_{12}b\,\hat{\mathbf{y}} + z_{12}c\,\hat{\mathbf{z}}$ | (16$b$) | Si II |
| $\mathbf{B}_{44}$ | = | $\left(x_{12}-y_{12}+z_{12}\right)\,\mathbf{a}_1 + \left(-x_{12}+y_{12}+z_{12}\right)\,\mathbf{a}_2 + \left(-x_{12}-y_{12}-z_{12}\right)\,\mathbf{a}_3$ | = | $-x_{12}a\,\hat{\mathbf{x}} - y_{12}b\,\hat{\mathbf{y}} + z_{12}c\,\hat{\mathbf{z}}$ | (16$b$) | Si II |
| $\mathbf{B}_{45}$ | = | $\left(\frac{1}{4}-x_{12}-y_{12}+z_{12}\right)\,\mathbf{a}_1 + \left(\frac{1}{4}+x_{12}+y_{12}+z_{12}\right)\,\mathbf{a}_2 + \left(\frac{1}{4}+x_{12}-y_{12}-z_{12}\right)\,\mathbf{a}_3$ | = | $\left(\frac{1}{4}+x_{12}\right)a\,\hat{\mathbf{x}} + \left(\frac{1}{4}-y_{12}\right)b\,\hat{\mathbf{y}} + \left(\frac{1}{4}+z_{12}\right)c\,\hat{\mathbf{z}}$ | (16$b$) | Si II |
| $\mathbf{B}_{46}$ | = | $\left(\frac{1}{4}+x_{12}+y_{12}+z_{12}\right)\,\mathbf{a}_1 + \left(\frac{1}{4}-x_{12}-y_{12}+z_{12}\right)\,\mathbf{a}_2 + \left(\frac{1}{4}-x_{12}+y_{12}-z_{12}\right)\,\mathbf{a}_3$ | = | $\left(\frac{1}{4}-x_{12}\right)a\,\hat{\mathbf{x}} + \left(\frac{1}{4}+y_{12}\right)b\,\hat{\mathbf{y}} + \left(\frac{1}{4}+z_{12}\right)c\,\hat{\mathbf{z}}$ | (16$b$) | Si II |

---

---

**Geometry files:**

- CIF: pp. 1601
- POSCAR: pp. 1602

# Blossite ($\alpha$-$Cu_2V_2O_7$) Structure: A2B7C2_oF88_43_b_a3b_b

http://aflow.org/prototype-encyclopedia/A2B7C2_oF88_43_b_a3b_b

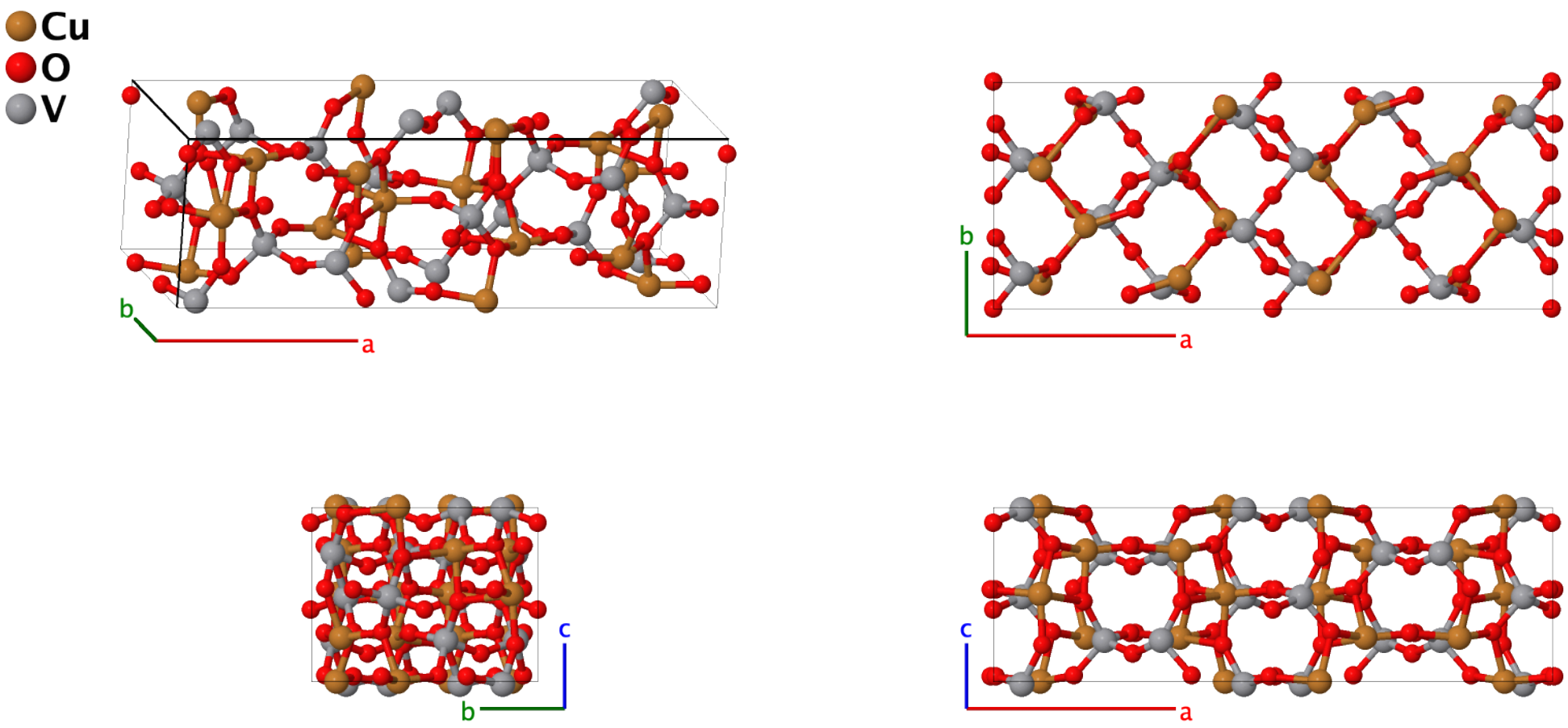

| | | |
|---|---|---|
| **Prototype** | : | $Cu_2O_7V_2$ |
| **AFLOW prototype label** | : | A2B7C2_oF88_43_b_a3b_b |
| ***Strukturbericht* designation** | : | None |
| **Pearson symbol** | : | oF88 |
| **Space group number** | : | 43 |
| **Space group symbol** | : | $Fdd2$ |
| **AFLOW prototype command** | : | `aflow --proto=A2B7C2_oF88_43_b_a3b_b`<br>`--params=`$a, b/a, c/a, z_1, x_2, y_2, z_2, x_3, y_3, z_3, x_4, y_4, z_4, x_5, y_5, z_5, x_6, y_6, z_6$ |

- This structure was given the name "blossite" by (Robinson, 1987).
- Space group $Fdd2$ #43 does not fix the $z = 0$ plane. We do this by setting $z_2 = 3/4$ for the copper atom.

**Face-centered Orthorhombic primitive vectors:**

$$\mathbf{a}_1 = \frac{1}{2}\, b\, \hat{\mathbf{y}} + \frac{1}{2}\, c\, \hat{\mathbf{z}}$$
$$\mathbf{a}_2 = \frac{1}{2}\, a\, \hat{\mathbf{x}} + \frac{1}{2}\, c\, \hat{\mathbf{z}}$$
$$\mathbf{a}_3 = \frac{1}{2}\, a\, \hat{\mathbf{x}} + \frac{1}{2}\, b\, \hat{\mathbf{y}}$$

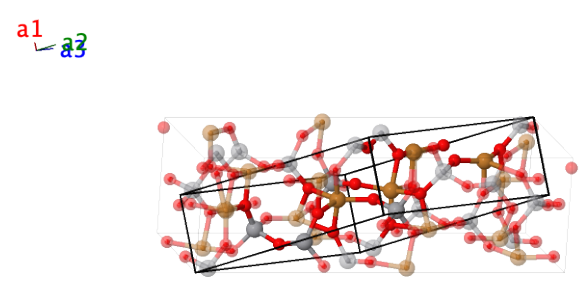

**Basis vectors:**

| | | Lattice Coordinates | | Cartesian Coordinates | Wyckoff Position | Atom Type |
|---|---|---|---|---|---|---|
| $\mathbf{B}_1$ | = | $z_1\,\mathbf{a}_1 + z_1\,\mathbf{a}_2 - z_1\,\mathbf{a}_3$ | = | $z_1 c\,\hat{\mathbf{z}}$ | (8$a$) | O I |
| $\mathbf{B}_2$ | = | $\left(\frac{1}{4} + z_1\right)\mathbf{a}_1 + \left(\frac{1}{4} + z_1\right)\mathbf{a}_2 + \left(\frac{1}{4} - z_1\right)\mathbf{a}_3$ | = | $\frac{1}{4}a\,\hat{\mathbf{x}} + \frac{1}{4}b\,\hat{\mathbf{y}} + \left(\frac{1}{4} + z_1\right)c\,\hat{\mathbf{z}}$ | (8$a$) | O I |
| $\mathbf{B}_3$ | = | $(-x_2 + y_2 + z_2)\,\mathbf{a}_1 + (x_2 - y_2 + z_2)\,\mathbf{a}_2 + (x_2 + y_2 - z_2)\,\mathbf{a}_3$ | = | $x_2 a\,\hat{\mathbf{x}} + y_2 b\,\hat{\mathbf{y}} + z_2 c\,\hat{\mathbf{z}}$ | (16$b$) | Cu |

$$\mathbf{B}_4 = (x_2 - y_2 + z_2)\,\mathbf{a}_1 + (-x_2 + y_2 + z_2)\,\mathbf{a}_2 + (-x_2 - y_2 - z_2)\,\mathbf{a}_3 = -x_2 a\,\hat{\mathbf{x}} - y_2 b\,\hat{\mathbf{y}} + z_2 c\,\hat{\mathbf{z}} \quad (16b) \quad \text{Cu}$$

$$\mathbf{B}_5 = \left(\tfrac{1}{4} - x_2 - y_2 + z_2\right)\mathbf{a}_1 + \left(\tfrac{1}{4} + x_2 + y_2 + z_2\right)\mathbf{a}_2 + \left(\tfrac{1}{4} + x_2 - y_2 - z_2\right)\mathbf{a}_3 = \left(\tfrac{1}{4} + x_2\right)a\,\hat{\mathbf{x}} + \left(\tfrac{1}{4} - y_2\right)b\,\hat{\mathbf{y}} + \left(\tfrac{1}{4} + z_2\right)c\,\hat{\mathbf{z}} \quad (16b) \quad \text{Cu}$$

$$\mathbf{B}_6 = \left(\tfrac{1}{4} + x_2 + y_2 + z_2\right)\mathbf{a}_1 + \left(\tfrac{1}{4} - x_2 - y_2 + z_2\right)\mathbf{a}_2 + \left(\tfrac{1}{4} - x_2 + y_2 - z_2\right)\mathbf{a}_3 = \left(\tfrac{1}{4} - x_2\right)a\,\hat{\mathbf{x}} + \left(\tfrac{1}{4} + y_2\right)b\,\hat{\mathbf{y}} + \left(\tfrac{1}{4} + z_2\right)c\,\hat{\mathbf{z}} \quad (16b) \quad \text{Cu}$$

$$\mathbf{B}_7 = (-x_3 + y_3 + z_3)\,\mathbf{a}_1 + (x_3 - y_3 + z_3)\,\mathbf{a}_2 + (x_3 + y_3 - z_3)\,\mathbf{a}_3 = x_3 a\,\hat{\mathbf{x}} + y_3 b\,\hat{\mathbf{y}} + z_3 c\,\hat{\mathbf{z}} \quad (16b) \quad \text{O II}$$

$$\mathbf{B}_8 = (x_3 - y_3 + z_3)\,\mathbf{a}_1 + (-x_3 + y_3 + z_3)\,\mathbf{a}_2 + (-x_3 - y_3 - z_3)\,\mathbf{a}_3 = -x_3 a\,\hat{\mathbf{x}} - y_3 b\,\hat{\mathbf{y}} + z_3 c\,\hat{\mathbf{z}} \quad (16b) \quad \text{O II}$$

$$\mathbf{B}_9 = \left(\tfrac{1}{4} - x_3 - y_3 + z_3\right)\mathbf{a}_1 + \left(\tfrac{1}{4} + x_3 + y_3 + z_3\right)\mathbf{a}_2 + \left(\tfrac{1}{4} + x_3 - y_3 - z_3\right)\mathbf{a}_3 = \left(\tfrac{1}{4} + x_3\right)a\,\hat{\mathbf{x}} + \left(\tfrac{1}{4} - y_3\right)b\,\hat{\mathbf{y}} + \left(\tfrac{1}{4} + z_3\right)c\,\hat{\mathbf{z}} \quad (16b) \quad \text{O II}$$

$$\mathbf{B}_{10} = \left(\tfrac{1}{4} + x_3 + y_3 + z_3\right)\mathbf{a}_1 + \left(\tfrac{1}{4} - x_3 - y_3 + z_3\right)\mathbf{a}_2 + \left(\tfrac{1}{4} - x_3 + y_3 - z_3\right)\mathbf{a}_3 = \left(\tfrac{1}{4} - x_3\right)a\,\hat{\mathbf{x}} + \left(\tfrac{1}{4} + y_3\right)b\,\hat{\mathbf{y}} + \left(\tfrac{1}{4} + z_3\right)c\,\hat{\mathbf{z}} \quad (16b) \quad \text{O II}$$

$$\mathbf{B}_{11} = (-x_4 + y_4 + z_4)\,\mathbf{a}_1 + (x_4 - y_4 + z_4)\,\mathbf{a}_2 + (x_4 + y_4 - z_4)\,\mathbf{a}_3 = x_4 a\,\hat{\mathbf{x}} + y_4 b\,\hat{\mathbf{y}} + z_4 c\,\hat{\mathbf{z}} \quad (16b) \quad \text{O III}$$

$$\mathbf{B}_{12} = (x_4 - y_4 + z_4)\,\mathbf{a}_1 + (-x_4 + y_4 + z_4)\,\mathbf{a}_2 + (-x_4 - y_4 - z_4)\,\mathbf{a}_3 = -x_4 a\,\hat{\mathbf{x}} - y_4 b\,\hat{\mathbf{y}} + z_4 c\,\hat{\mathbf{z}} \quad (16b) \quad \text{O III}$$

$$\mathbf{B}_{13} = \left(\tfrac{1}{4} - x_4 - y_4 + z_4\right)\mathbf{a}_1 + \left(\tfrac{1}{4} + x_4 + y_4 + z_4\right)\mathbf{a}_2 + \left(\tfrac{1}{4} + x_4 - y_4 - z_4\right)\mathbf{a}_3 = \left(\tfrac{1}{4} + x_4\right)a\,\hat{\mathbf{x}} + \left(\tfrac{1}{4} - y_4\right)b\,\hat{\mathbf{y}} + \left(\tfrac{1}{4} + z_4\right)c\,\hat{\mathbf{z}} \quad (16b) \quad \text{O III}$$

$$\mathbf{B}_{14} = \left(\tfrac{1}{4} + x_4 + y_4 + z_4\right)\mathbf{a}_1 + \left(\tfrac{1}{4} - x_4 - y_4 + z_4\right)\mathbf{a}_2 + \left(\tfrac{1}{4} - x_4 + y_4 - z_4\right)\mathbf{a}_3 = \left(\tfrac{1}{4} - x_4\right)a\,\hat{\mathbf{x}} + \left(\tfrac{1}{4} + y_4\right)b\,\hat{\mathbf{y}} + \left(\tfrac{1}{4} + z_4\right)c\,\hat{\mathbf{z}} \quad (16b) \quad \text{O III}$$

$$\mathbf{B}_{15} = (-x_5 + y_5 + z_5)\,\mathbf{a}_1 + (x_5 - y_5 + z_5)\,\mathbf{a}_2 + (x_5 + y_5 - z_5)\,\mathbf{a}_3 = x_5 a\,\hat{\mathbf{x}} + y_5 b\,\hat{\mathbf{y}} + z_5 c\,\hat{\mathbf{z}} \quad (16b) \quad \text{O IV}$$

$$\mathbf{B}_{16} = (x_5 - y_5 + z_5)\,\mathbf{a}_1 + (-x_5 + y_5 + z_5)\,\mathbf{a}_2 + (-x_5 - y_5 - z_5)\,\mathbf{a}_3 = -x_5 a\,\hat{\mathbf{x}} - y_5 b\,\hat{\mathbf{y}} + z_5 c\,\hat{\mathbf{z}} \quad (16b) \quad \text{O IV}$$

$$\mathbf{B}_{17} = \left(\tfrac{1}{4} - x_5 - y_5 + z_5\right)\mathbf{a}_1 + \left(\tfrac{1}{4} + x_5 + y_5 + z_5\right)\mathbf{a}_2 + \left(\tfrac{1}{4} + x_5 - y_5 - z_5\right)\mathbf{a}_3 = \left(\tfrac{1}{4} + x_5\right)a\,\hat{\mathbf{x}} + \left(\tfrac{1}{4} - y_5\right)b\,\hat{\mathbf{y}} + \left(\tfrac{1}{4} + z_5\right)c\,\hat{\mathbf{z}} \quad (16b) \quad \text{O IV}$$

$$\mathbf{B}_{18} = \left(\tfrac{1}{4} + x_5 + y_5 + z_5\right)\mathbf{a}_1 + \left(\tfrac{1}{4} - x_5 - y_5 + z_5\right)\mathbf{a}_2 + \left(\tfrac{1}{4} - x_5 + y_5 - z_5\right)\mathbf{a}_3 = \left(\tfrac{1}{4} - x_5\right)a\,\hat{\mathbf{x}} + \left(\tfrac{1}{4} + y_5\right)b\,\hat{\mathbf{y}} + \left(\tfrac{1}{4} + z_5\right)c\,\hat{\mathbf{z}} \quad (16b) \quad \text{O IV}$$

| | | | | | | |
|---|---|---|---|---|---|---|
| $\mathbf{B}_{19}$ | = | $(-x_6 + y_6 + z_6)\,\mathbf{a}_1 + (x_6 - y_6 + z_6)\,\mathbf{a}_2 + (x_6 + y_6 - z_6)\,\mathbf{a}_3$ | = | $x_6 a\,\hat{\mathbf{x}} + y_6 b\,\hat{\mathbf{y}} + z_6 c\,\hat{\mathbf{z}}$ | (16$b$) | V |
| $\mathbf{B}_{20}$ | = | $(x_6 - y_6 + z_6)\,\mathbf{a}_1 + (-x_6 + y_6 + z_6)\,\mathbf{a}_2 + (-x_6 - y_6 - z_6)\,\mathbf{a}_3$ | = | $-x_6 a\,\hat{\mathbf{x}} - y_6 b\,\hat{\mathbf{y}} + z_6 c\,\hat{\mathbf{z}}$ | (16$b$) | V |
| $\mathbf{B}_{21}$ | = | $\left(\frac{1}{4} - x_6 - y_6 + z_6\right)\mathbf{a}_1 + \left(\frac{1}{4} + x_6 + y_6 + z_6\right)\mathbf{a}_2 + \left(\frac{1}{4} + x_6 - y_6 - z_6\right)\mathbf{a}_3$ | = | $\left(\frac{1}{4} + x_6\right)a\,\hat{\mathbf{x}} + \left(\frac{1}{4} - y_6\right)b\,\hat{\mathbf{y}} + \left(\frac{1}{4} + z_6\right)c\,\hat{\mathbf{z}}$ | (16$b$) | V |
| $\mathbf{B}_{22}$ | = | $\left(\frac{1}{4} + x_6 + y_6 + z_6\right)\mathbf{a}_1 + \left(\frac{1}{4} - x_6 - y_6 + z_6\right)\mathbf{a}_2 + \left(\frac{1}{4} - x_6 + y_6 - z_6\right)\mathbf{a}_3$ | = | $\left(\frac{1}{4} - x_6\right)a\,\hat{\mathbf{x}} + \left(\frac{1}{4} + y_6\right)b\,\hat{\mathbf{y}} + \left(\frac{1}{4} + z_6\right)c\,\hat{\mathbf{z}}$ | (16$b$) | V |

**Found in:**

**Geometry files:**

- CIF: pp. 1602
- POSCAR: pp. 1603

# Archerite ($KH_2PO_4$) Structure: A2BC4D_oF64_43_b_a_2b_a

http://aflow.org/prototype-encyclopedia/A2BC4D_oF64_43_b_a_2b_a

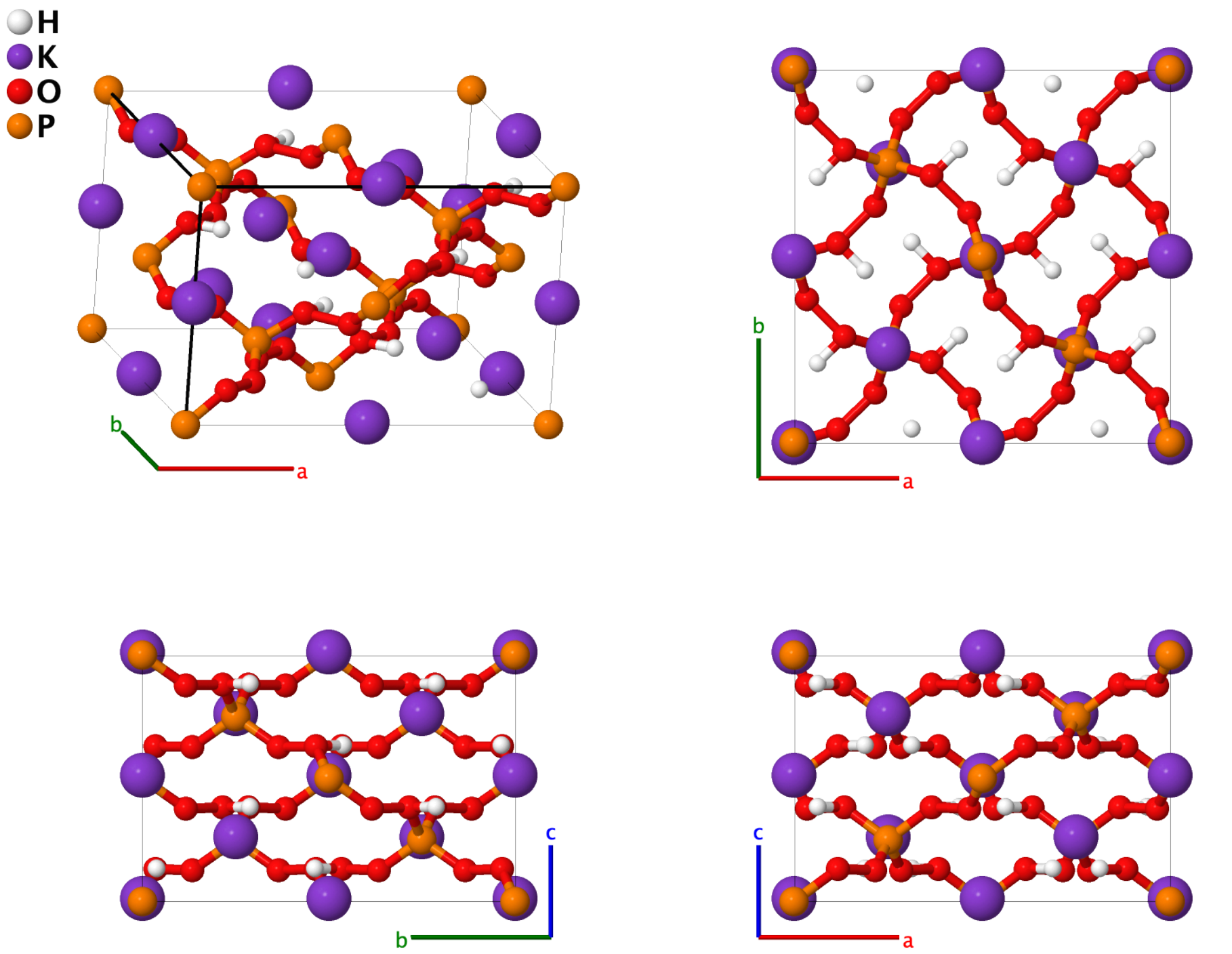

| | | |
|---|---|---|
| **Prototype** | : | $H_2KO_4P$ |
| **AFLOW prototype label** | : | A2BC4D_oF64_43_b_a_2b_a |
| ***Strukturbericht* designation** | : | None |
| **Pearson symbol** | : | oF64 |
| **Space group number** | : | 43 |
| **Space group symbol** | : | $Fdd2$ |
| **AFLOW prototype command** | : | `aflow --proto=A2BC4D_oF64_43_b_a_2b_a --params=`$a, b/a, c/a, z_1, z_2, x_3, y_3, z_3, x_4, y_4, z_4, x_5, y_5, z_5$ |

- This structure is stable below 121 K, and the data was taken at 113 K. The high temperature structure is tetragonal with disordered hydrogen. (Levy, 1954)
- The origin of the $z$-axis is not restricted in space group $Fdd2$ #43. Here it is fixed by putting the phosphorous atom at the origin.

**Face-centered Orthorhombic primitive vectors:**

$$\mathbf{a}_1 = \tfrac{1}{2}\,b\,\hat{\mathbf{y}} + \tfrac{1}{2}\,c\,\hat{\mathbf{z}}$$
$$\mathbf{a}_2 = \tfrac{1}{2}\,a\,\hat{\mathbf{x}} + \tfrac{1}{2}\,c\,\hat{\mathbf{z}}$$
$$\mathbf{a}_3 = \tfrac{1}{2}\,a\,\hat{\mathbf{x}} + \tfrac{1}{2}\,b\,\hat{\mathbf{y}}$$

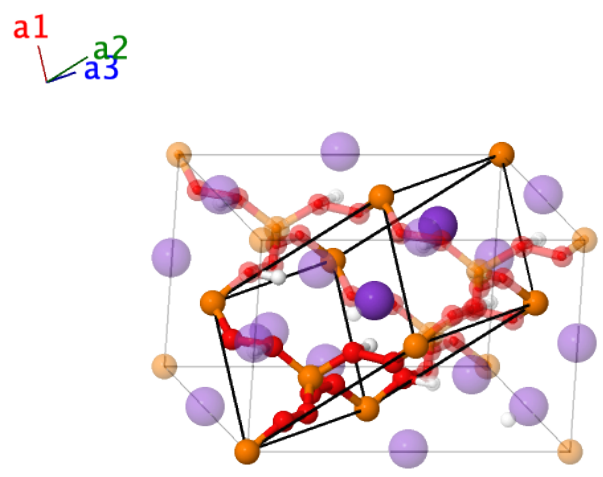

**Basis vectors:**

| | | Lattice Coordinates | | Cartesian Coordinates | Wyckoff Position | Atom Type |
|---|---|---|---|---|---|---|
| $\mathbf{B}_1$ | = | $z_1\,\mathbf{a}_1 + z_1\,\mathbf{a}_2 - z_1\,\mathbf{a}_3$ | = | $z_1 c\,\hat{\mathbf{z}}$ | (8$a$) | K |
| $\mathbf{B}_2$ | = | $\left(\frac{1}{4}+z_1\right)\mathbf{a}_1 + \left(\frac{1}{4}+z_1\right)\mathbf{a}_2 + \left(\frac{1}{4}-z_1\right)\mathbf{a}_3$ | = | $\frac{1}{4}a\,\hat{\mathbf{x}} + \frac{1}{4}b\,\hat{\mathbf{y}} + \left(\frac{1}{4}+z_1\right)c\,\hat{\mathbf{z}}$ | (8$a$) | K |
| $\mathbf{B}_3$ | = | $z_2\,\mathbf{a}_1 + z_2\,\mathbf{a}_2 - z_2\,\mathbf{a}_3$ | = | $z_2 c\,\hat{\mathbf{z}}$ | (8$a$) | P |
| $\mathbf{B}_4$ | = | $\left(\frac{1}{4}+z_2\right)\mathbf{a}_1 + \left(\frac{1}{4}+z_2\right)\mathbf{a}_2 + \left(\frac{1}{4}-z_2\right)\mathbf{a}_3$ | = | $\frac{1}{4}a\,\hat{\mathbf{x}} + \frac{1}{4}b\,\hat{\mathbf{y}} + \left(\frac{1}{4}+z_2\right)c\,\hat{\mathbf{z}}$ | (8$a$) | P |
| $\mathbf{B}_5$ | = | $(-x_3+y_3+z_3)\,\mathbf{a}_1 + (x_3-y_3+z_3)\,\mathbf{a}_2 + (x_3+y_3-z_3)\,\mathbf{a}_3$ | = | $x_3 a\,\hat{\mathbf{x}} + y_3 b\,\hat{\mathbf{y}} + z_3 c\,\hat{\mathbf{z}}$ | (16$b$) | H |
| $\mathbf{B}_6$ | = | $(x_3-y_3+z_3)\,\mathbf{a}_1 + (-x_3+y_3+z_3)\,\mathbf{a}_2 + (-x_3-y_3-z_3)\,\mathbf{a}_3$ | = | $-x_3 a\,\hat{\mathbf{x}} - y_3 b\,\hat{\mathbf{y}} + z_3 c\,\hat{\mathbf{z}}$ | (16$b$) | H |
| $\mathbf{B}_7$ | = | $\left(\frac{1}{4}-x_3-y_3+z_3\right)\mathbf{a}_1 + \left(\frac{1}{4}+x_3+y_3+z_3\right)\mathbf{a}_2 + \left(\frac{1}{4}+x_3-y_3-z_3\right)\mathbf{a}_3$ | = | $\left(\frac{1}{4}+x_3\right)a\,\hat{\mathbf{x}} + \left(\frac{1}{4}-y_3\right)b\,\hat{\mathbf{y}} + \left(\frac{1}{4}+z_3\right)c\,\hat{\mathbf{z}}$ | (16$b$) | H |
| $\mathbf{B}_8$ | = | $\left(\frac{1}{4}+x_3+y_3+z_3\right)\mathbf{a}_1 + \left(\frac{1}{4}-x_3-y_3+z_3\right)\mathbf{a}_2 + \left(\frac{1}{4}-x_3+y_3-z_3\right)\mathbf{a}_3$ | = | $\left(\frac{1}{4}-x_3\right)a\,\hat{\mathbf{x}} + \left(\frac{1}{4}+y_3\right)b\,\hat{\mathbf{y}} + \left(\frac{1}{4}+z_3\right)c\,\hat{\mathbf{z}}$ | (16$b$) | H |
| $\mathbf{B}_9$ | = | $(-x_4+y_4+z_4)\,\mathbf{a}_1 + (x_4-y_4+z_4)\,\mathbf{a}_2 + (x_4+y_4-z_4)\,\mathbf{a}_3$ | = | $x_4 a\,\hat{\mathbf{x}} + y_4 b\,\hat{\mathbf{y}} + z_4 c\,\hat{\mathbf{z}}$ | (16$b$) | O I |
| $\mathbf{B}_{10}$ | = | $(x_4-y_4+z_4)\,\mathbf{a}_1 + (-x_4+y_4+z_4)\,\mathbf{a}_2 + (-x_4-y_4-z_4)\,\mathbf{a}_3$ | = | $-x_4 a\,\hat{\mathbf{x}} - y_4 b\,\hat{\mathbf{y}} + z_4 c\,\hat{\mathbf{z}}$ | (16$b$) | O I |
| $\mathbf{B}_{11}$ | = | $\left(\frac{1}{4}-x_4-y_4+z_4\right)\mathbf{a}_1 + \left(\frac{1}{4}+x_4+y_4+z_4\right)\mathbf{a}_2 + \left(\frac{1}{4}+x_4-y_4-z_4\right)\mathbf{a}_3$ | = | $\left(\frac{1}{4}+x_4\right)a\,\hat{\mathbf{x}} + \left(\frac{1}{4}-y_4\right)b\,\hat{\mathbf{y}} + \left(\frac{1}{4}+z_4\right)c\,\hat{\mathbf{z}}$ | (16$b$) | O I |
| $\mathbf{B}_{12}$ | = | $\left(\frac{1}{4}+x_4+y_4+z_4\right)\mathbf{a}_1 + \left(\frac{1}{4}-x_4-y_4+z_4\right)\mathbf{a}_2 + \left(\frac{1}{4}-x_4+y_4-z_4\right)\mathbf{a}_3$ | = | $\left(\frac{1}{4}-x_4\right)a\,\hat{\mathbf{x}} + \left(\frac{1}{4}+y_4\right)b\,\hat{\mathbf{y}} + \left(\frac{1}{4}+z_4\right)c\,\hat{\mathbf{z}}$ | (16$b$) | O I |
| $\mathbf{B}_{13}$ | = | $(-x_5+y_5+z_5)\,\mathbf{a}_1 + (x_5-y_5+z_5)\,\mathbf{a}_2 + (x_5+y_5-z_5)\,\mathbf{a}_3$ | = | $x_5 a\,\hat{\mathbf{x}} + y_5 b\,\hat{\mathbf{y}} + z_5 c\,\hat{\mathbf{z}}$ | (16$b$) | O II |
| $\mathbf{B}_{14}$ | = | $(x_5 - y_5 + z_5)\,\mathbf{a}_1 + (-x_5 + y_5 + z_5)\,\mathbf{a}_2 + (-x_5 - y_5 - z_5)\,\mathbf{a}_3$ | = | $-x_5 a\,\hat{\mathbf{x}} - y_5 b\,\hat{\mathbf{y}} + z_5 c\,\hat{\mathbf{z}}$ | (16*b*) | O II |
| $\mathbf{B}_{15}$ | = | $\left(\frac{1}{4} - x_5 - y_5 + z_5\right)\mathbf{a}_1 + \left(\frac{1}{4} + x_5 + y_5 + z_5\right)\mathbf{a}_2 + \left(\frac{1}{4} + x_5 - y_5 - z_5\right)\mathbf{a}_3$ | = | $\left(\frac{1}{4} + x_5\right)a\,\hat{\mathbf{x}} + \left(\frac{1}{4} - y_5\right)b\,\hat{\mathbf{y}} + \left(\frac{1}{4} + z_5\right)c\,\hat{\mathbf{z}}$ | (16*b*) | O II |
| $\mathbf{B}_{16}$ | = | $\left(\frac{1}{4} + x_5 + y_5 + z_5\right)\mathbf{a}_1 + \left(\frac{1}{4} - x_5 - y_5 + z_5\right)\mathbf{a}_2 + \left(\frac{1}{4} - x_5 + y_5 - z_5\right)\mathbf{a}_3$ | = | $\left(\frac{1}{4} - x_5\right)a\,\hat{\mathbf{x}} + \left(\frac{1}{4} + y_5\right)b\,\hat{\mathbf{y}} + \left(\frac{1}{4} + z_5\right)c\,\hat{\mathbf{z}}$ | (16*b*) | O II |

---

---

**Geometry files:**

- CIF: pp. 1603
- POSCAR: pp. 1603

# $Cs_2Se$ Structure: A2B_oF24_43_b_a

http://aflow.org/prototype-encyclopedia/A2B_oF24_43_b_a

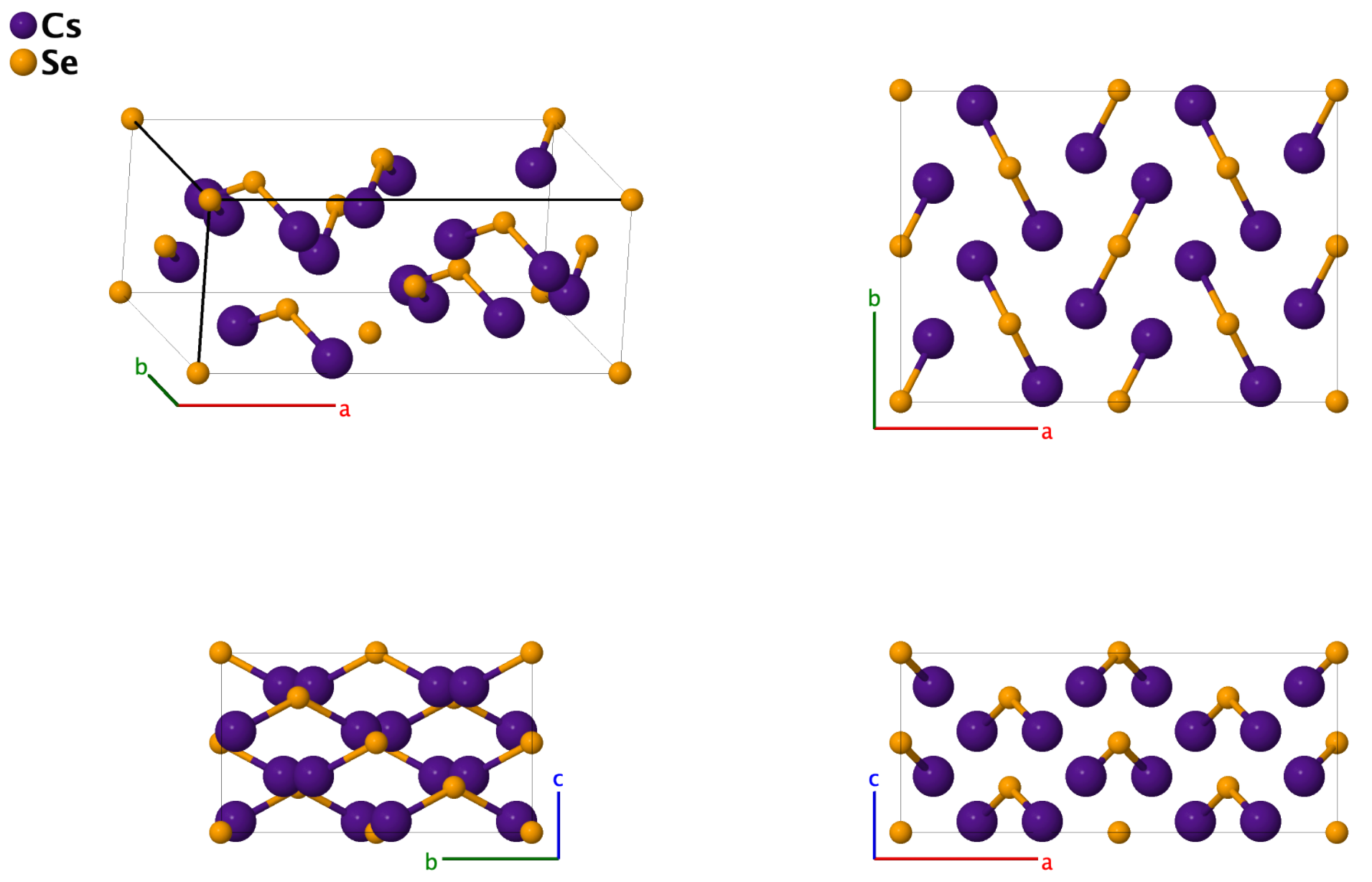

| | | |
|---|---|---|
| **Prototype** | : | $Cs_2Se$ |
| **AFLOW prototype label** | : | A2B_oF24_43_b_a |
| ***Strukturbericht* designation** | : | None |
| **Pearson symbol** | : | oF24 |
| **Space group number** | : | 43 |
| **Space group symbol** | : | $Fdd2$ |
| **AFLOW prototype command** | : | `aflow --proto=A2B_oF24_43_b_a --params=`$a, b/a, c/a, z_1, x_2, y_2, z_2$ |

**Face-centered Orthorhombic primitive vectors:**

$$\mathbf{a}_1 = \frac{1}{2}\,b\,\hat{\mathbf{y}} + \frac{1}{2}\,c\,\hat{\mathbf{z}}$$
$$\mathbf{a}_2 = \frac{1}{2}\,a\,\hat{\mathbf{x}} + \frac{1}{2}\,c\,\hat{\mathbf{z}}$$
$$\mathbf{a}_3 = \frac{1}{2}\,a\,\hat{\mathbf{x}} + \frac{1}{2}\,b\,\hat{\mathbf{y}}$$

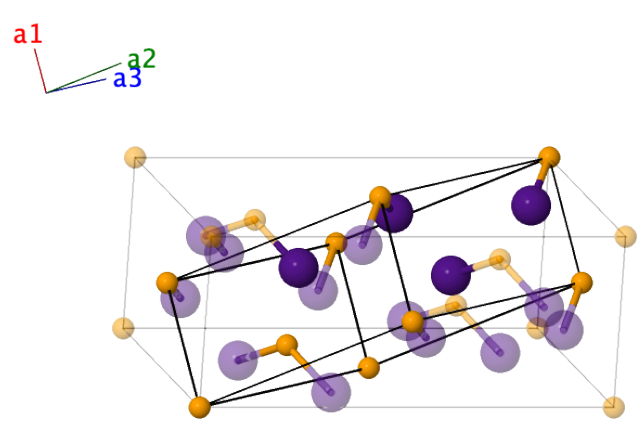

**Basis vectors:**

| | | Lattice Coordinates | | Cartesian Coordinates | Wyckoff Position | Atom Type |
|---|---|---|---|---|---|---|
| $\mathbf{B}_1$ | = | $z_1\,\mathbf{a}_1 + z_1\,\mathbf{a}_2 - z_1\,\mathbf{a}_3$ | = | $z_1\,c\,\hat{\mathbf{z}}$ | (8$a$) | Se |
| $\mathbf{B}_2$ | = | $\left(\frac{1}{4} + z_1\right)\mathbf{a}_1 + \left(\frac{1}{4} + z_1\right)\mathbf{a}_2 + \left(\frac{1}{4} - z_1\right)\mathbf{a}_3$ | = | $\frac{1}{4}a\,\hat{\mathbf{x}} + \frac{1}{4}b\,\hat{\mathbf{y}} + \left(\frac{1}{4} + z_1\right)c\,\hat{\mathbf{z}}$ | (8$a$) | Se |
| $\mathbf{B}_3$ | = | $(-x_2 + y_2 + z_2)\,\mathbf{a}_1 + (x_2 - y_2 + z_2)\,\mathbf{a}_2 + (x_2 + y_2 - z_2)\,\mathbf{a}_3$ | = | $x_2 a\,\hat{\mathbf{x}} + y_2 b\,\hat{\mathbf{y}} + z_2 c\,\hat{\mathbf{z}}$ | (16$b$) | Cs |
| $\mathbf{B}_4$ | = | $(x_2 - y_2 + z_2)\,\mathbf{a}_1 + (-x_2 + y_2 + z_2)\,\mathbf{a}_2 + (-x_2 - y_2 - z_2)\,\mathbf{a}_3$ | = | $-x_2 a\,\hat{\mathbf{x}} - y_2 b\,\hat{\mathbf{y}} + z_2 c\,\hat{\mathbf{z}}$ | (16$b$) | Cs |
| $\mathbf{B}_5$ | = | $\left(\frac{1}{4} - x_2 - y_2 + z_2\right)\mathbf{a}_1 + \left(\frac{1}{4} + x_2 + y_2 + z_2\right)\mathbf{a}_2 + \left(\frac{1}{4} + x_2 - y_2 - z_2\right)\mathbf{a}_3$ | = | $\left(\frac{1}{4} + x_2\right) a\,\hat{\mathbf{x}} + \left(\frac{1}{4} - y_2\right) b\,\hat{\mathbf{y}} + \left(\frac{1}{4} + z_2\right) c\,\hat{\mathbf{z}}$ | (16$b$) | Cs |
| $\mathbf{B}_6$ | = | $\left(\frac{1}{4} + x_2 + y_2 + z_2\right)\mathbf{a}_1 + \left(\frac{1}{4} - x_2 - y_2 + z_2\right)\mathbf{a}_2 + \left(\frac{1}{4} - x_2 + y_2 - z_2\right)\mathbf{a}_3$ | = | $\left(\frac{1}{4} - x_2\right) a\,\hat{\mathbf{x}} + \left(\frac{1}{4} + y_2\right) b\,\hat{\mathbf{y}} + \left(\frac{1}{4} + z_2\right) c\,\hat{\mathbf{z}}$ | (16$b$) | Cs |

---

**Found in:**

---

**Geometry files:**

- CIF: pp. 1603
- POSCAR: pp. 1604

# $Zr_2Al_3$ Structure: A3B2_oF40_43_ab_b

http://aflow.org/prototype-encyclopedia/A3B2_oF40_43_ab_b

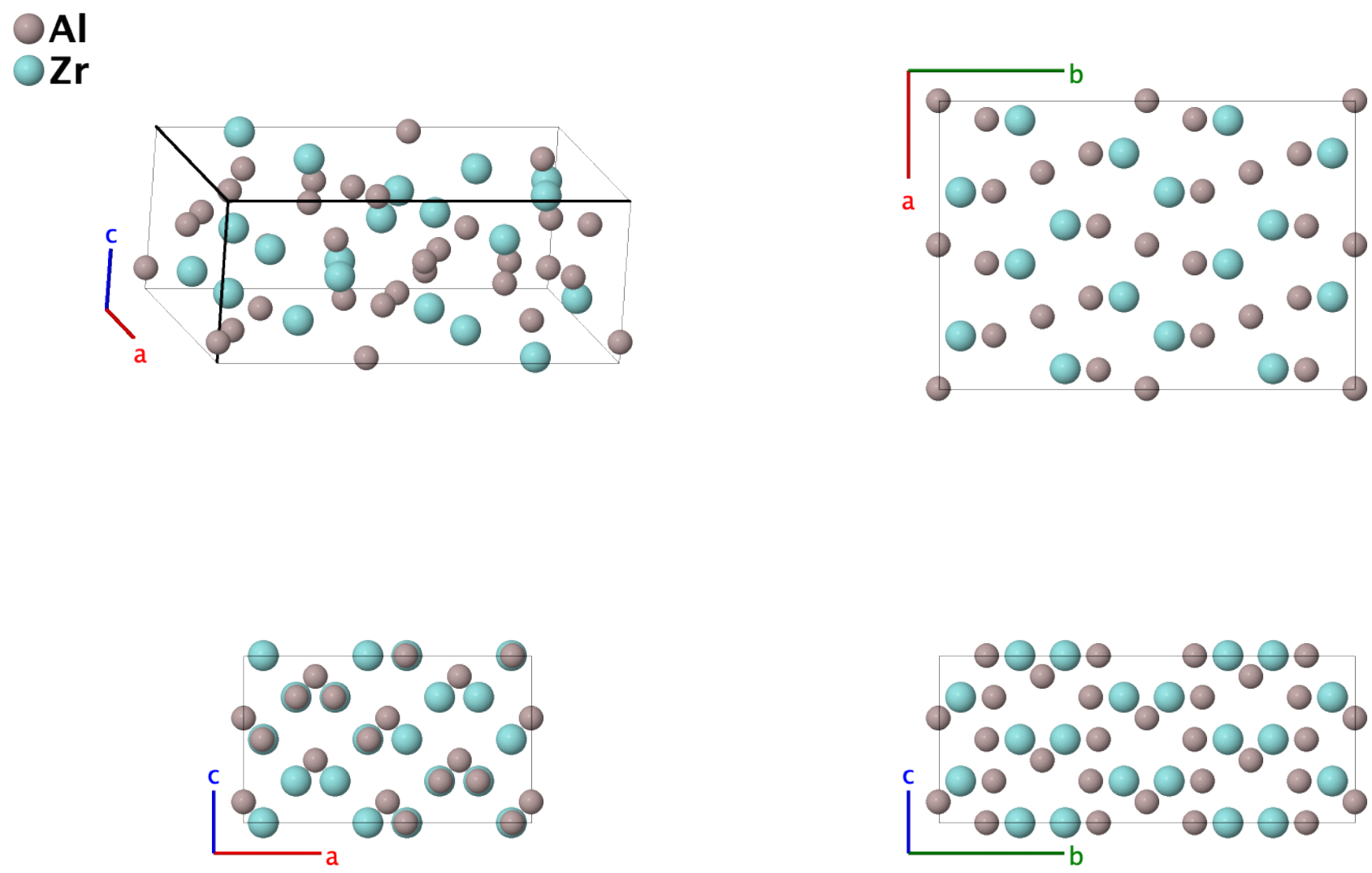

| | | |
|---|---|---|
| **Prototype** | : | $Al_3Zr_2$ |
| **AFLOW prototype label** | : | A3B2_oF40_43_ab_b |
| ***Strukturbericht* designation** | : | None |
| **Pearson symbol** | : | oF40 |
| **Space group number** | : | 43 |
| **Space group symbol** | : | $Fdd2$ |
| **AFLOW prototype command** | : | `aflow --proto=A3B2_oF40_43_ab_b`<br>`--params=`$a, b/a, c/a, z_1, x_2, y_2, z_2, x_3, y_3, z_3$ |

**Other compounds with this structure**

- $Hf_2Al_3$

- The $z = 0$ plane is not fixed in space group $Fdd2$ #43. Here it is arbitrarily chosen so that $z_3 = 0$ for the zirconium atom.

**Face-centered Orthorhombic primitive vectors:**

$$\mathbf{a}_1 = \frac{1}{2}\, b\,\hat{\mathbf{y}} + \frac{1}{2}\, c\,\hat{\mathbf{z}}$$
$$\mathbf{a}_2 = \frac{1}{2}\, a\,\hat{\mathbf{x}} + \frac{1}{2}\, c\,\hat{\mathbf{z}}$$
$$\mathbf{a}_3 = \frac{1}{2}\, a\,\hat{\mathbf{x}} + \frac{1}{2}\, b\,\hat{\mathbf{y}}$$

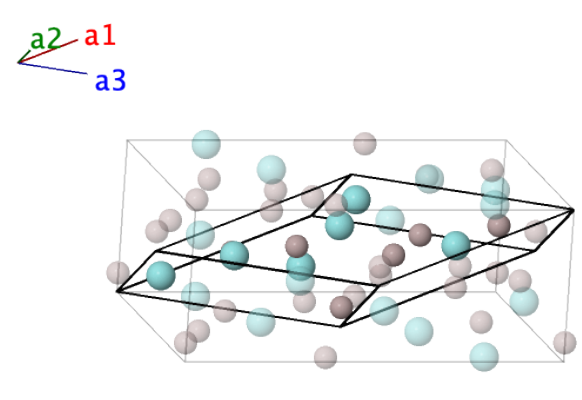

**Basis vectors:**

| | | Lattice Coordinates | | Cartesian Coordinates | Wyckoff Position | Atom Type |
|---|---|---|---|---|---|---|
| $\mathbf{B}_1$ | = | $z_1\,\mathbf{a}_1 + z_1\,\mathbf{a}_2 - z_1\,\mathbf{a}_3$ | = | $z_1\, c\,\hat{\mathbf{z}}$ | (8$a$) | Al I |
| $\mathbf{B}_2$ | = | $\left(\frac{1}{4}+z_1\right)\mathbf{a}_1+\left(\frac{1}{4}+z_1\right)\mathbf{a}_2+\left(\frac{1}{4}-z_1\right)\mathbf{a}_3$ | = | $\frac{1}{4}a\,\hat{\mathbf{x}}+\frac{1}{4}b\,\hat{\mathbf{y}}+\left(\frac{1}{4}+z_1\right)c\,\hat{\mathbf{z}}$ | (8*a*) | Al I |
| $\mathbf{B}_3$ | = | $(-x_2+y_2+z_2)\,\mathbf{a}_1+(x_2-y_2+z_2)\,\mathbf{a}_2+(x_2+y_2-z_2)\,\mathbf{a}_3$ | = | $x_2a\,\hat{\mathbf{x}}+y_2b\,\hat{\mathbf{y}}+z_2c\,\hat{\mathbf{z}}$ | (16*b*) | Al II |
| $\mathbf{B}_4$ | = | $(x_2-y_2+z_2)\,\mathbf{a}_1+(-x_2+y_2+z_2)\,\mathbf{a}_2+(-x_2-y_2-z_2)\,\mathbf{a}_3$ | = | $-x_2a\,\hat{\mathbf{x}}-y_2b\,\hat{\mathbf{y}}+z_2c\,\hat{\mathbf{z}}$ | (16*b*) | Al II |
| $\mathbf{B}_5$ | = | $\left(\frac{1}{4}-x_2-y_2+z_2\right)\mathbf{a}_1+\left(\frac{1}{4}+x_2+y_2+z_2\right)\mathbf{a}_2+\left(\frac{1}{4}+x_2-y_2-z_2\right)\mathbf{a}_3$ | = | $\left(\frac{1}{4}+x_2\right)a\,\hat{\mathbf{x}}+\left(\frac{1}{4}-y_2\right)b\,\hat{\mathbf{y}}+\left(\frac{1}{4}+z_2\right)c\,\hat{\mathbf{z}}$ | (16*b*) | Al II |
| $\mathbf{B}_6$ | = | $\left(\frac{1}{4}+x_2+y_2+z_2\right)\mathbf{a}_1+\left(\frac{1}{4}-x_2-y_2+z_2\right)\mathbf{a}_2+\left(\frac{1}{4}-x_2+y_2-z_2\right)\mathbf{a}_3$ | = | $\left(\frac{1}{4}-x_2\right)a\,\hat{\mathbf{x}}+\left(\frac{1}{4}+y_2\right)b\,\hat{\mathbf{y}}+\left(\frac{1}{4}+z_2\right)c\,\hat{\mathbf{z}}$ | (16*b*) | Al II |
| $\mathbf{B}_7$ | = | $(-x_3+y_3+z_3)\,\mathbf{a}_1+(x_3-y_3+z_3)\,\mathbf{a}_2+(x_3+y_3-z_3)\,\mathbf{a}_3$ | = | $x_3a\,\hat{\mathbf{x}}+y_3b\,\hat{\mathbf{y}}+z_3c\,\hat{\mathbf{z}}$ | (16*b*) | Zr |
| $\mathbf{B}_8$ | = | $(x_3-y_3+z_3)\,\mathbf{a}_1+(-x_3+y_3+z_3)\,\mathbf{a}_2+(-x_3-y_3-z_3)\,\mathbf{a}_3$ | = | $-x_3a\,\hat{\mathbf{x}}-y_3b\,\hat{\mathbf{y}}+z_3c\,\hat{\mathbf{z}}$ | (16*b*) | Zr |
| $\mathbf{B}_9$ | = | $\left(\frac{1}{4}-x_3-y_3+z_3\right)\mathbf{a}_1+\left(\frac{1}{4}+x_3+y_3+z_3\right)\mathbf{a}_2+\left(\frac{1}{4}+x_3-y_3-z_3\right)\mathbf{a}_3$ | = | $\left(\frac{1}{4}+x_3\right)a\,\hat{\mathbf{x}}+\left(\frac{1}{4}-y_3\right)b\,\hat{\mathbf{y}}+\left(\frac{1}{4}+z_3\right)c\,\hat{\mathbf{z}}$ | (16*b*) | Zr |
| $\mathbf{B}_{10}$ | = | $\left(\frac{1}{4}+x_3+y_3+z_3\right)\mathbf{a}_1+\left(\frac{1}{4}-x_3-y_3+z_3\right)\mathbf{a}_2+\left(\frac{1}{4}-x_3+y_3-z_3\right)\mathbf{a}_3$ | = | $\left(\frac{1}{4}-x_3\right)a\,\hat{\mathbf{x}}+\left(\frac{1}{4}+y_3\right)b\,\hat{\mathbf{y}}+\left(\frac{1}{4}+z_3\right)c\,\hat{\mathbf{z}}$ | (16*b*) | Zr |

**Found in:**

**Geometry files:**

- CIF: pp. 1604
- POSCAR: pp. 1604

# Hemimorphite ($Zn_4Si_2O_7(OH)_2 \cdot H_2O$, $S2_2$) Structure: A2B5CD2_oI40_44_2c_abcde_d_e

http://aflow.org/prototype-encyclopedia/A2B5CD2_oI40_44_2c_abcde_d_e

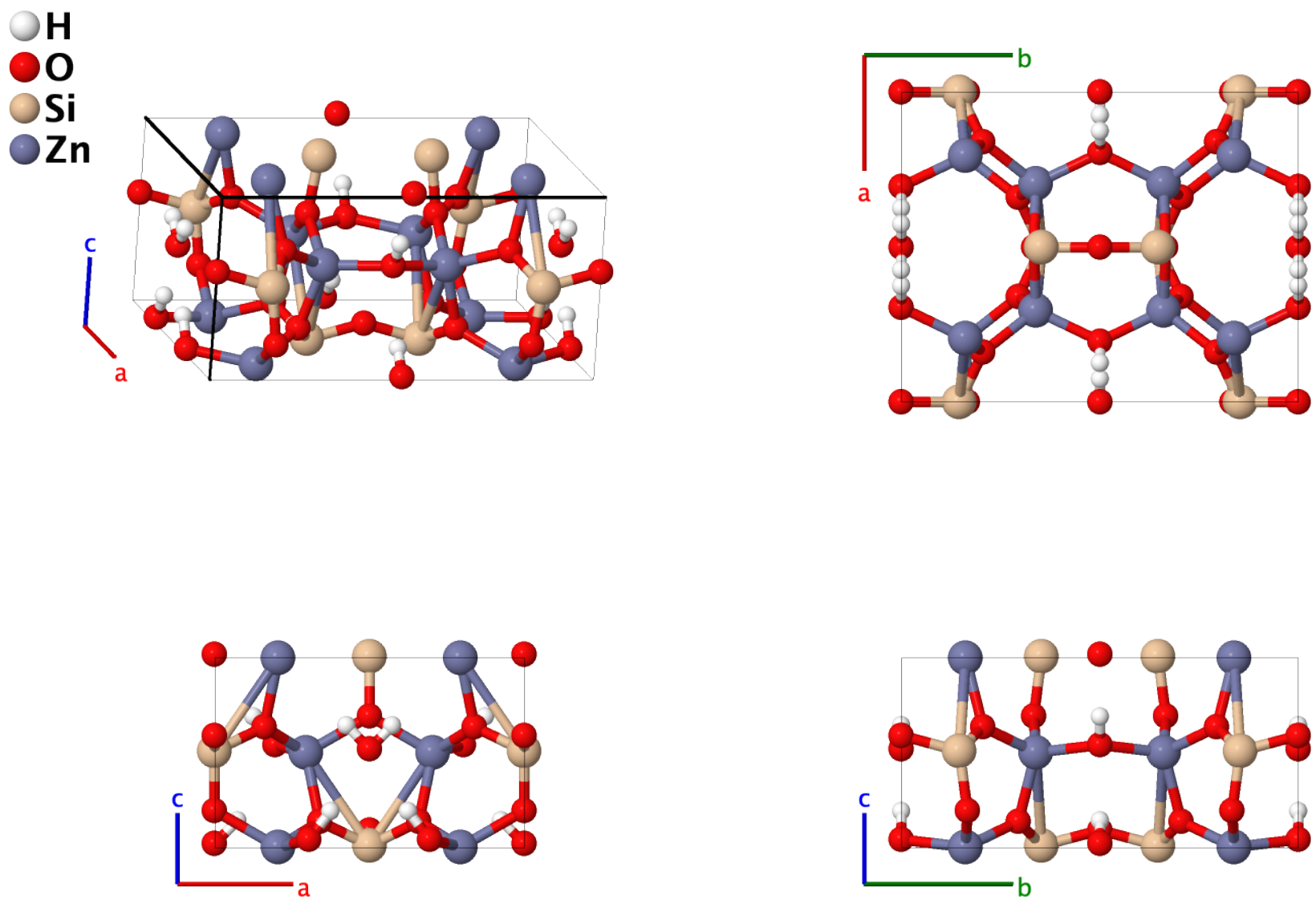

| | | |
|---|---|---|
| **Prototype** | : | $H_2O_5SiZn_2$ |
| **AFLOW prototype label** | : | A2B5CD2_oI40_44_2c_abcde_d_e |
| ***Strukturbericht* designation** | : | $S2_2$ |
| **Pearson symbol** | : | oI40 |
| **Space group number** | : | 44 |
| **Space group symbol** | : | $Imm2$ |
| **AFLOW prototype command** | : | `aflow --proto=A2B5CD2_oI40_44_2c_abcde_d_e`<br>`--params=`$a, b/a, c/a, z_1, z_2, x_3, z_3, x_4, z_4, x_5, z_5, y_6, z_6, y_7, z_7, x_8, y_8, z_8, x_9, y_9, z_9$ |

- The original (Ito, 1932) determination of this structure did not locate the positions of the hydrogen atoms. (Hill, 1977) were able to do this, so we use the updated structure as the prototype.
- (Hill, 1977) gives the $z$ coordinates of the atoms on the (2$c$) sites as 0.0190, 0.0643, and 0.0410, respectively, but this gives unrealistic H-O distances. Examination of the figures and distance tables shows that we should take $z_2$ = 0.190, $z_3$ = 0.643, and $z_4$ = 0.041, a conclusion also reached by (Downs, 2003).

**Body-centered Orthorhombic primitive vectors:**

$$\mathbf{a}_1 = -\frac{1}{2}\,a\,\mathbf{\hat{x}} + \frac{1}{2}\,b\,\mathbf{\hat{y}} + \frac{1}{2}\,c\,\mathbf{\hat{z}}$$
$$\mathbf{a}_2 = \frac{1}{2}\,a\,\mathbf{\hat{x}} - \frac{1}{2}\,b\,\mathbf{\hat{y}} + \frac{1}{2}\,c\,\mathbf{\hat{z}}$$
$$\mathbf{a}_3 = \frac{1}{2}\,a\,\mathbf{\hat{x}} + \frac{1}{2}\,b\,\mathbf{\hat{y}} - \frac{1}{2}\,c\,\mathbf{\hat{z}}$$

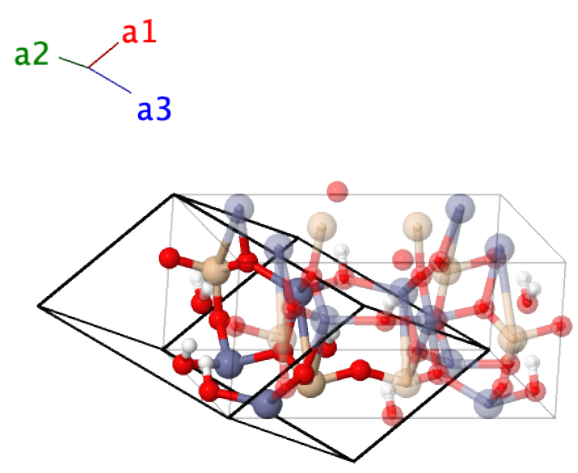

**Basis vectors:**

| | | Lattice Coordinates | | Cartesian Coordinates | Wyckoff Position | Atom Type |
|---|---|---|---|---|---|---|
| $\mathbf{B}_1$ | = | $z_1\,\mathbf{a}_1 + z_1\,\mathbf{a}_2$ | = | $z_1 c\,\hat{\mathbf{z}}$ | (2$a$) | O I |
| $\mathbf{B}_2$ | = | $\left(\frac{1}{2} + z_2\right)\mathbf{a}_1 + z_2\,\mathbf{a}_2 + \frac{1}{2}\,\mathbf{a}_3$ | = | $\frac{1}{2}b\,\hat{\mathbf{y}} + z_2 c\,\hat{\mathbf{z}}$ | (2$b$) | O II |
| $\mathbf{B}_3$ | = | $z_3\,\mathbf{a}_1 + (x_3 + z_3)\,\mathbf{a}_2 + x_3\,\mathbf{a}_3$ | = | $x_3 a\,\hat{\mathbf{x}} + z_3 c\,\hat{\mathbf{z}}$ | (4$c$) | H I |
| $\mathbf{B}_4$ | = | $z_3\,\mathbf{a}_1 + (-x_3 + z_3)\,\mathbf{a}_2 - x_3\,\mathbf{a}_3$ | = | $-x_3 a\,\hat{\mathbf{x}} + z_3 c\,\hat{\mathbf{z}}$ | (4$c$) | H I |
| $\mathbf{B}_5$ | = | $z_4\,\mathbf{a}_1 + (x_4 + z_4)\,\mathbf{a}_2 + x_4\,\mathbf{a}_3$ | = | $x_4 a\,\hat{\mathbf{x}} + z_4 c\,\hat{\mathbf{z}}$ | (4$c$) | H II |
| $\mathbf{B}_6$ | = | $z_4\,\mathbf{a}_1 + (-x_4 + z_4)\,\mathbf{a}_2 - x_4\,\mathbf{a}_3$ | = | $-x_4 a\,\hat{\mathbf{x}} + z_4 c\,\hat{\mathbf{z}}$ | (4$c$) | H II |
| $\mathbf{B}_7$ | = | $z_5\,\mathbf{a}_1 + (x_5 + z_5)\,\mathbf{a}_2 + x_5\,\mathbf{a}_3$ | = | $x_5 a\,\hat{\mathbf{x}} + z_5 c\,\hat{\mathbf{z}}$ | (4$c$) | O III |
| $\mathbf{B}_8$ | = | $z_5\,\mathbf{a}_1 + (-x_5 + z_5)\,\mathbf{a}_2 - x_5\,\mathbf{a}_3$ | = | $-x_5 a\,\hat{\mathbf{x}} + z_5 c\,\hat{\mathbf{z}}$ | (4$c$) | O III |
| $\mathbf{B}_9$ | = | $(y_6 + z_6)\,\mathbf{a}_1 + z_6\,\mathbf{a}_2 + y_6\,\mathbf{a}_3$ | = | $y_6 b\,\hat{\mathbf{y}} + z_6 c\,\hat{\mathbf{z}}$ | (4$d$) | O IV |
| $\mathbf{B}_{10}$ | = | $(-y_6 + z_6)\,\mathbf{a}_1 + z_6\,\mathbf{a}_2 - y_6\,\mathbf{a}_3$ | = | $-y_6 b\,\hat{\mathbf{y}} + z_6 c\,\hat{\mathbf{z}}$ | (4$d$) | O IV |
| $\mathbf{B}_{11}$ | = | $(y_7 + z_7)\,\mathbf{a}_1 + z_7\,\mathbf{a}_2 + y_7\,\mathbf{a}_3$ | = | $y_7 b\,\hat{\mathbf{y}} + z_7 c\,\hat{\mathbf{z}}$ | (4$d$) | Si |
| $\mathbf{B}_{12}$ | = | $(-y_7 + z_7)\,\mathbf{a}_1 + z_7\,\mathbf{a}_2 - y_7\,\mathbf{a}_3$ | = | $-y_7 b\,\hat{\mathbf{y}} + z_7 c\,\hat{\mathbf{z}}$ | (4$d$) | Si |
| $\mathbf{B}_{13}$ | = | $(y_8 + z_8)\,\mathbf{a}_1 + (x_8 + z_8)\,\mathbf{a}_2 + (x_8 + y_8)\,\mathbf{a}_3$ | = | $x_8 a\,\hat{\mathbf{x}} + y_8 b\,\hat{\mathbf{y}} + z_8 c\,\hat{\mathbf{z}}$ | (8$e$) | O V |
| $\mathbf{B}_{14}$ | = | $(-y_8 + z_8)\,\mathbf{a}_1 + (-x_8 + z_8)\,\mathbf{a}_2 + (-x_8 - y_8)\,\mathbf{a}_3$ | = | $-x_8 a\,\hat{\mathbf{x}} - y_8 b\,\hat{\mathbf{y}} + z_8 c\,\hat{\mathbf{z}}$ | (8$e$) | O V |
| $\mathbf{B}_{15}$ | = | $(-y_8 + z_8)\,\mathbf{a}_1 + (x_8 + z_8)\,\mathbf{a}_2 + (x_8 - y_8)\,\mathbf{a}_3$ | = | $x_8 a\,\hat{\mathbf{x}} - y_8 b\,\hat{\mathbf{y}} + z_8 c\,\hat{\mathbf{z}}$ | (8$e$) | O V |
| $\mathbf{B}_{16}$ | = | $(y_8 + z_8)\,\mathbf{a}_1 + (-x_8 + z_8)\,\mathbf{a}_2 + (-x_8 + y_8)\,\mathbf{a}_3$ | = | $-x_8 a\,\hat{\mathbf{x}} + y_8 b\,\hat{\mathbf{y}} + z_8 c\,\hat{\mathbf{z}}$ | (8$e$) | O V |
| $\mathbf{B}_{17}$ | = | $(y_9 + z_9)\,\mathbf{a}_1 + (x_9 + z_9)\,\mathbf{a}_2 + (x_9 + y_9)\,\mathbf{a}_3$ | = | $x_9 a\,\hat{\mathbf{x}} + y_9 b\,\hat{\mathbf{y}} + z_9 c\,\hat{\mathbf{z}}$ | (8$e$) | Zn |
| $\mathbf{B}_{18}$ | = | $(-y_9 + z_9)\,\mathbf{a}_1 + (-x_9 + z_9)\,\mathbf{a}_2 + (-x_9 - y_9)\,\mathbf{a}_3$ | = | $-x_9 a\,\hat{\mathbf{x}} - y_9 b\,\hat{\mathbf{y}} + z_9 c\,\hat{\mathbf{z}}$ | (8$e$) | Zn |
| $\mathbf{B}_{19}$ | = | $(-y_9 + z_9)\,\mathbf{a}_1 + (x_9 + z_9)\,\mathbf{a}_2 + (x_9 - y_9)\,\mathbf{a}_3$ | = | $x_9 a\,\hat{\mathbf{x}} - y_9 b\,\hat{\mathbf{y}} + z_9 c\,\hat{\mathbf{z}}$ | (8$e$) | Zn |
| $\mathbf{B}_{20}$ | = | $(y_9 + z_9)\,\mathbf{a}_1 + (-x_9 + z_9)\,\mathbf{a}_2 + (-x_9 + y_9)\,\mathbf{a}_3$ | = | $-x_9 a\,\hat{\mathbf{x}} + y_9 b\,\hat{\mathbf{y}} + z_9 c\,\hat{\mathbf{z}}$ | (8$e$) | Zn |

**Found in:**

**Geometry files:**

- CIF: pp. 1604
- POSCAR: pp. 1605

# Ferroelectric $NaNO_2$ ($F5_5$) Structure: ABC2_oI8_44_a_a_c

http://aflow.org/prototype-encyclopedia/ABC2_oI8_44_a_a_c

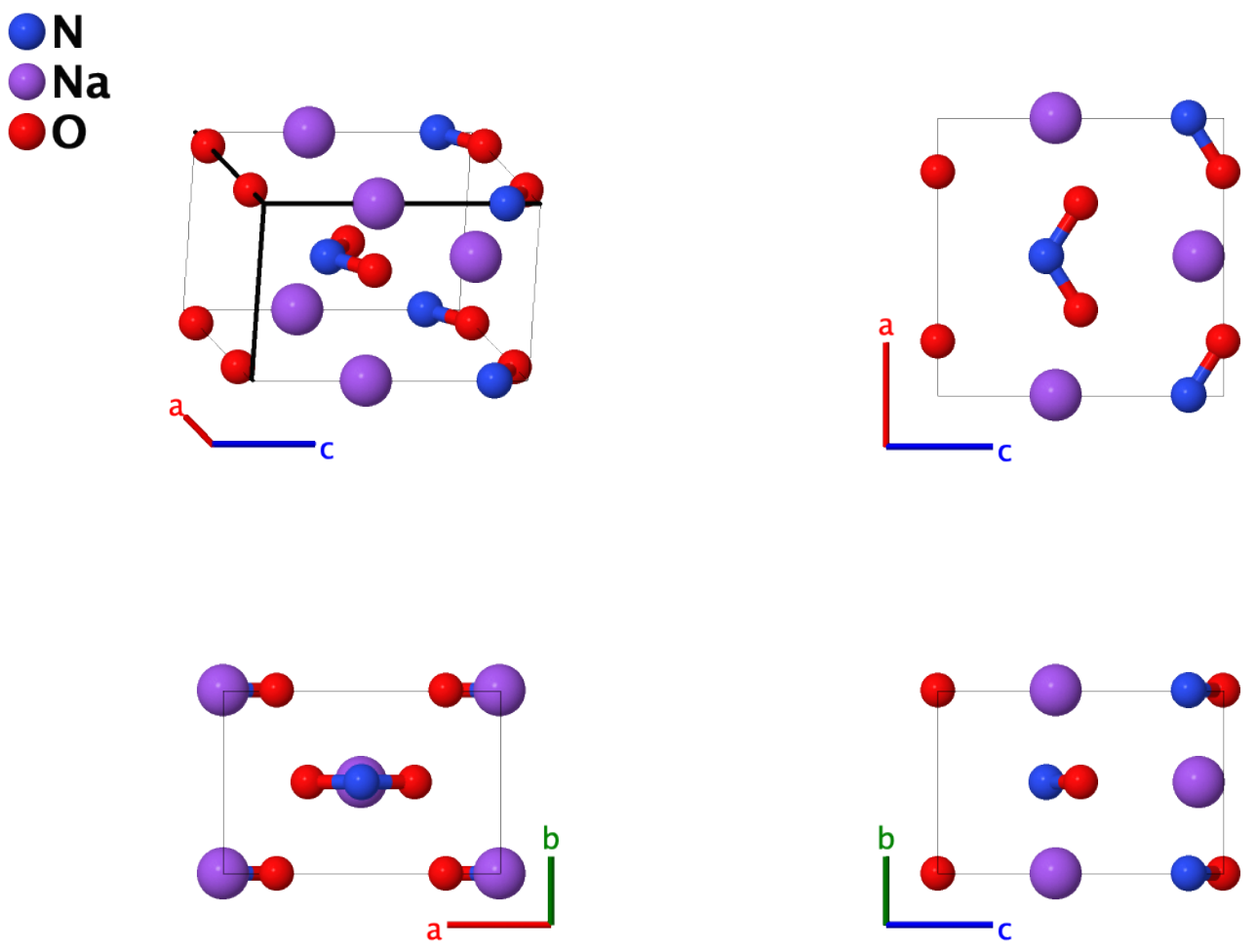

| | | |
|---|---|---|
| **Prototype** | : | $NaNO_2$ |
| **AFLOW prototype label** | : | ABC2_oI8_44_a_a_c |
| ***Strukturbericht* designation** | : | $F5_5$ |
| **Pearson symbol** | : | oI8 |
| **Space group number** | : | 44 |
| **Space group symbol** | : | $Imm2$ |
| **AFLOW prototype command** | : | `aflow --proto=ABC2_oI8_44_a_a_c --params=`$a, b/a, c/a, z_1, z_2, x_3, z_3$ |

**Other compounds with this structure**

- $AgNO_2$ ($F5_{12}$)

- This is the low-temperature phase of $NaNO_2$. It is ferroelectric below 158°C.
- (Kay, 1961) gives the structure in the $Im2m$ setting of space group #44. We have used FINDSYM to put it in the standard $Imm2$ setting.
- This structure is very similar to silver nitrite, $AgNO_2$, $F5_{12}$. We list the two structures as they are listed separately in the *Strukturbericht* volumes.
- The origin of the $z$ coordinate in $Imm2$, or the $y$ coordinate in $Im2m$, is arbitrary, and $z_3$ is set to zero.

### Body-centered Orthorhombic primitive vectors:

$$\mathbf{a}_1 = -\frac{1}{2}\,a\,\hat{\mathbf{x}} + \frac{1}{2}\,b\,\hat{\mathbf{y}} + \frac{1}{2}\,c\,\hat{\mathbf{z}}$$
$$\mathbf{a}_2 = \frac{1}{2}\,a\,\hat{\mathbf{x}} - \frac{1}{2}\,b\,\hat{\mathbf{y}} + \frac{1}{2}\,c\,\hat{\mathbf{z}}$$
$$\mathbf{a}_3 = \frac{1}{2}\,a\,\hat{\mathbf{x}} + \frac{1}{2}\,b\,\hat{\mathbf{y}} - \frac{1}{2}\,c\,\hat{\mathbf{z}}$$

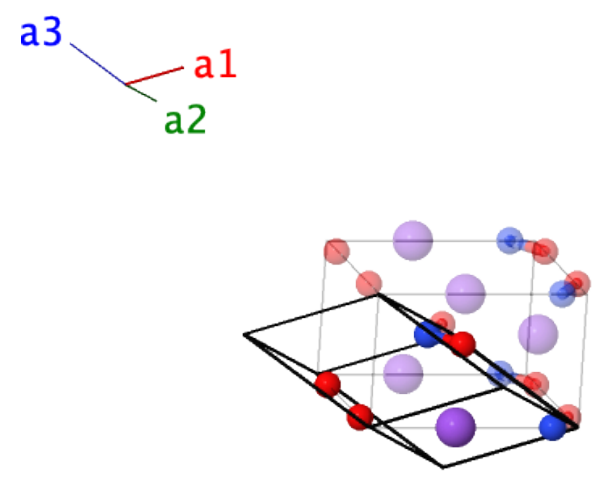

### Basis vectors:

| | | Lattice Coordinates | | Cartesian Coordinates | Wyckoff Position | Atom Type |
|---|---|---|---|---|---|---|
| $\mathbf{B}_1$ | = | $z_1\,\mathbf{a}_1 + z_1\,\mathbf{a}_2$ | = | $z_1c\,\hat{\mathbf{z}}$ | (2$a$) | N |
| $\mathbf{B}_2$ | = | $z_2\,\mathbf{a}_1 + z_2\,\mathbf{a}_2$ | = | $z_2c\,\hat{\mathbf{z}}$ | (2$a$) | Na |
| $\mathbf{B}_3$ | = | $z_3\,\mathbf{a}_1 + (x_3 + z_3)\,\mathbf{a}_2 + x_3\,\mathbf{a}_3$ | = | $x_3a\,\hat{\mathbf{x}} + z_3c\,\hat{\mathbf{z}}$ | (4$c$) | O |
| $\mathbf{B}_4$ | = | $z_3\,\mathbf{a}_1 + (-x_3 + z_3)\,\mathbf{a}_2 - x_3\,\mathbf{a}_3$ | = | $-x_3a\,\hat{\mathbf{x}} + z_3c\,\hat{\mathbf{z}}$ | (4$c$) | O |

### Geometry files:

- CIF: pp. 1605
- POSCAR: pp. 1605

# $AgNO_2$ ($F5_{12}$) Structure: ABC2_oI8_44_a_a_d

http://aflow.org/prototype-encyclopedia/ABC2_oI8_44_a_a_d

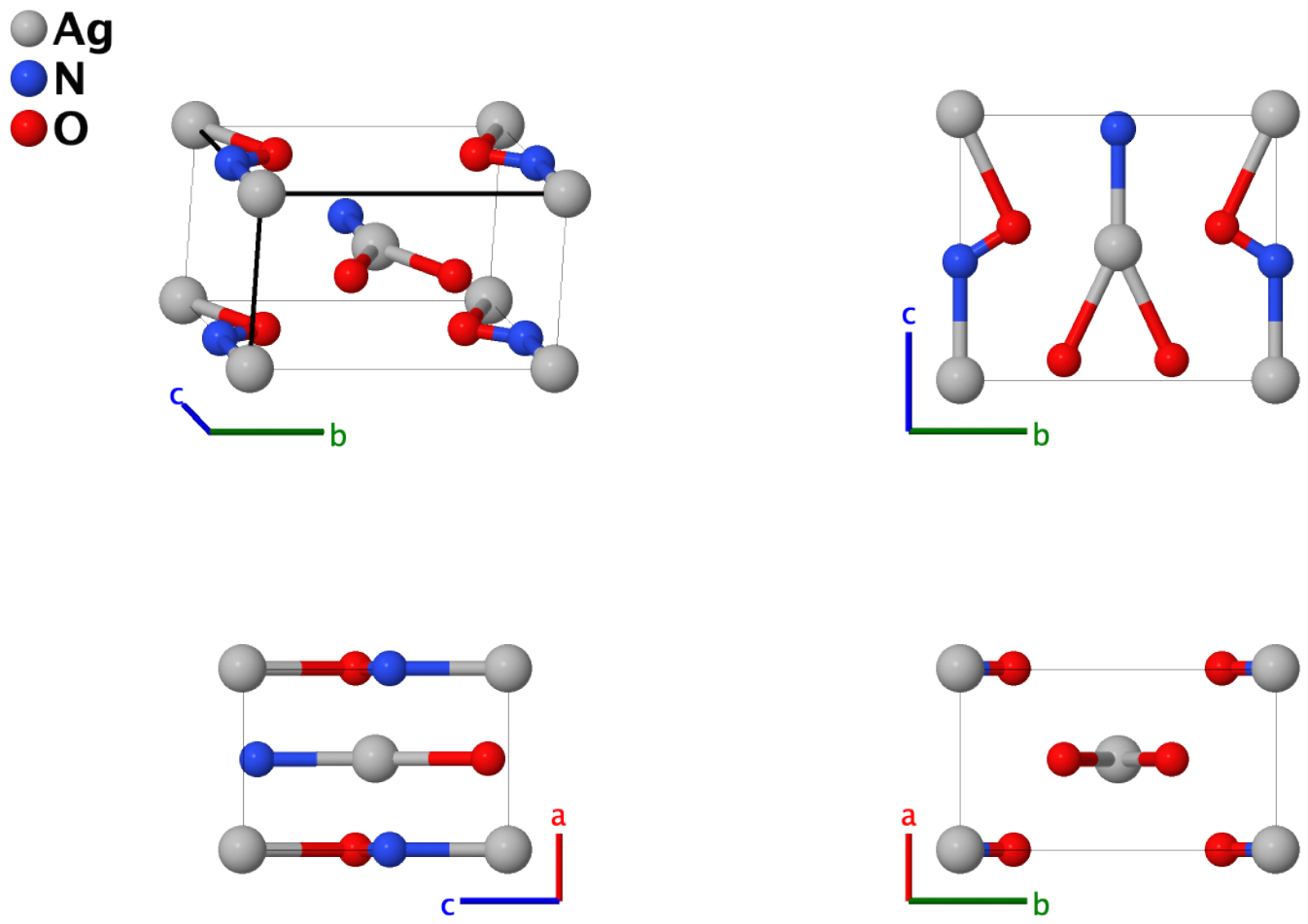

| | | |
|---|---|---|
| **Prototype** | : | $AgNO_2$ |
| **AFLOW prototype label** | : | ABC2_oI8_44_a_a_d |
| ***Strukturbericht* designation** | : | $F5_{12}$ |
| **Pearson symbol** | : | oI8 |
| **Space group number** | : | 44 |
| **Space group symbol** | : | $Imm2$ |
| **AFLOW prototype command** | : | `aflow --proto=ABC2_oI8_44_a_a_d`<br>`--params=`$a, b/a, c/a, z_1, z_2, y_3, z_3$ |

### Other compounds with this structure

- $NaNO_2$ ($F5_5$)

- This structure is very similar to sodium nitrite, $NbNO_2$, $F5_5$. We list the two structures as they are listed separately in the *Strukturbericht* volumes.
- The origin of the $z$ coordinate in $Imm2$ is arbitrary, and we set $z_1$ to zero.

### Body-centered Orthorhombic primitive vectors:

$$
\begin{aligned}
\mathbf{a}_1 &= -\tfrac{1}{2}\,a\,\hat{\mathbf{x}} + \tfrac{1}{2}\,b\,\hat{\mathbf{y}} + \tfrac{1}{2}\,c\,\hat{\mathbf{z}} \\
\mathbf{a}_2 &= \tfrac{1}{2}\,a\,\hat{\mathbf{x}} - \tfrac{1}{2}\,b\,\hat{\mathbf{y}} + \tfrac{1}{2}\,c\,\hat{\mathbf{z}} \\
\mathbf{a}_3 &= \tfrac{1}{2}\,a\,\hat{\mathbf{x}} + \tfrac{1}{2}\,b\,\hat{\mathbf{y}} - \tfrac{1}{2}\,c\,\hat{\mathbf{z}}
\end{aligned}
$$

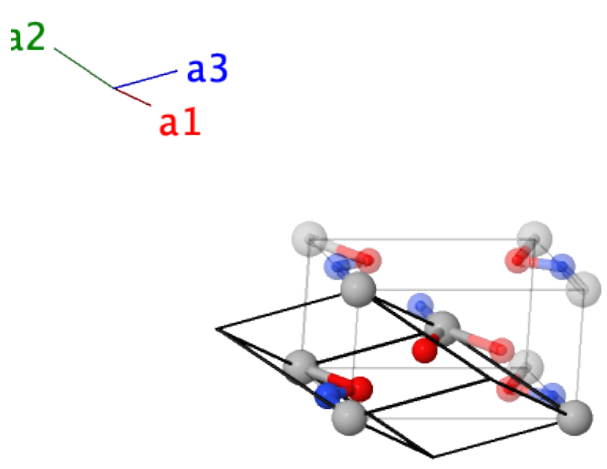

### Basis vectors:

| | | Lattice Coordinates | | Cartesian Coordinates | Wyckoff Position | Atom Type |
|---|---|---|---|---|---|---|
| $\mathbf{B}_1$ | = | $z_1\,\mathbf{a}_1 + z_1\,\mathbf{a}_2$ | = | $z_1 c\,\hat{\mathbf{z}}$ | (2$a$) | Ag |
| $\mathbf{B}_2$ | = | $z_2\,\mathbf{a}_1 + z_2\,\mathbf{a}_2$ | = | $z_2c\,\hat{\mathbf{z}}$ | (2$a$) | N |
| $\mathbf{B}_3$ | = | $(y_3 + z_3)\,\mathbf{a}_1 + z_3\,\mathbf{a}_2 + y_3\,\mathbf{a}_3$ | = | $y_3b\,\hat{\mathbf{y}} + z_3c\,\hat{\mathbf{z}}$ | (4$d$) | O |
| $\mathbf{B}_4$ | = | $(-y_3 + z_3)\,\mathbf{a}_1 + z_3\,\mathbf{a}_2 - y_3\,\mathbf{a}_3$ | = | $-y_3b\,\hat{\mathbf{y}} + z_3c\,\hat{\mathbf{z}}$ | (4$d$) | O |

**Geometry files:**

- CIF: pp. 1605
- POSCAR: pp. 1606

# $B30$ (MgZn?) Structure: AB_oI48_44_6d_ab2cde

http://aflow.org/prototype-encyclopedia/AB_oI48_44_6d_ab2cde

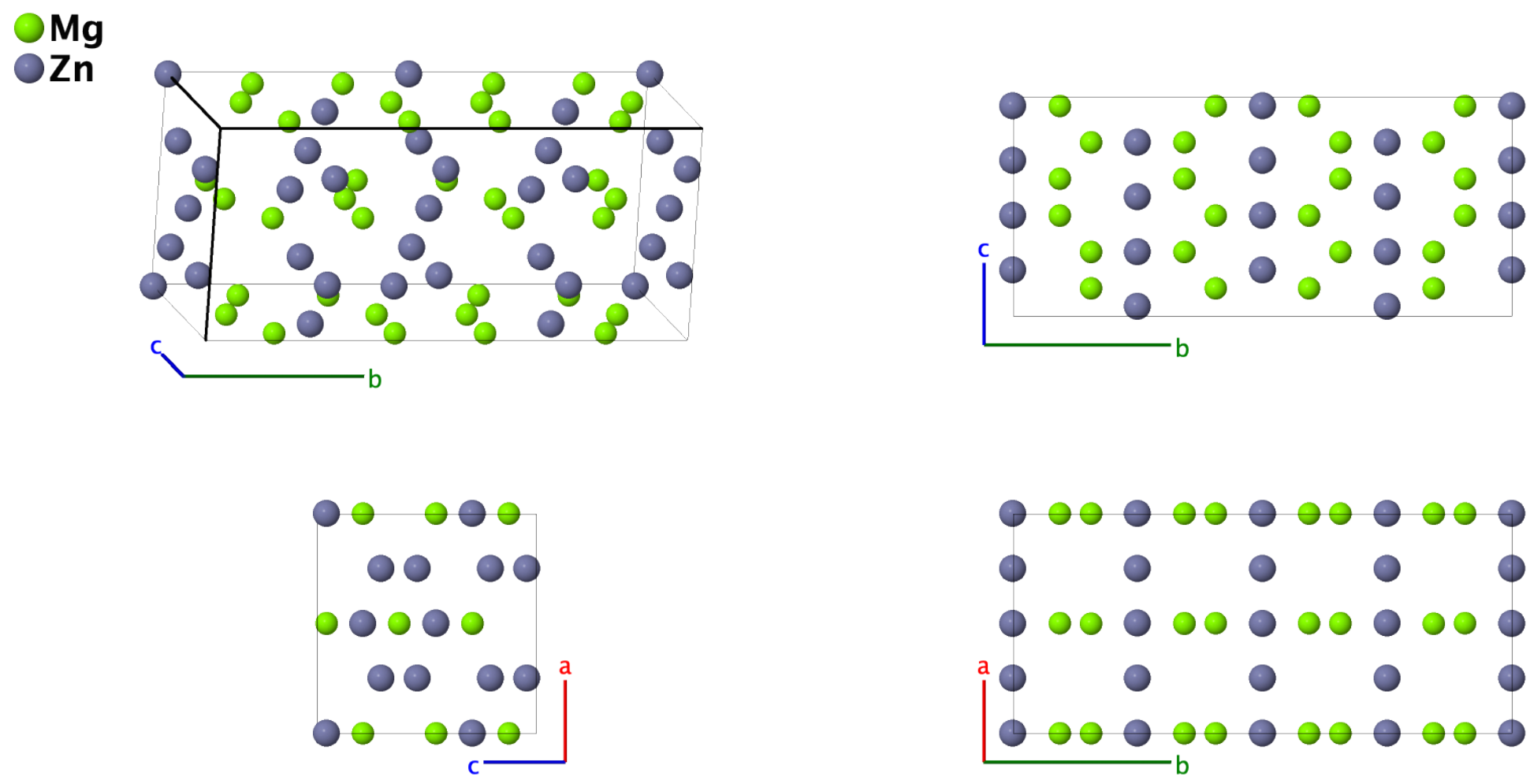

| | | |
|---|---|---|
| **Prototype** | : | MgZn |
| **AFLOW prototype label** | : | AB_oI48_44_6d_ab2cde |
| ***Strukturbericht* designation** | : | $B30$ |
| **Pearson symbol** | : | oI48 |
| **Space group number** | : | 44 |
| **Space group symbol** | : | $Imm2$ |
| **AFLOW prototype command** | : | `aflow --proto=AB_oI48_44_6d_ab2cde` `--params=`$a, b/a, c/a, z_1, z_2, x_3, z_3, x_4, z_4, y_5, z_5, y_6, z_6, y_7, z_7, y_8, z_8, y_9, z_9, y_{10}, z_{10}, y_{11}, z_{11}, x_{12}, y_{12}, z_{12}$ |

- It is rather a mystery why (Hermann, 1937) gave this the *Strukturbericht* designation $B30$, as the structure presented in the literature contradicts itself. (Tarschish, 1933) derived this structure from the hexagonal Laves structure $MgZn_2$ ($C14$) by doubling the unit cell in all directions to obtain a 96 atom unit cell, replacing 16 of the zinc atoms in this structure by magnesium, and shifting the $z$-coordinates of these atoms by $\pm c/16$. He then states that the space group remains $P6_3/mmc$ #194. (McKeehan, 1935) pointed out that this is impossible, as the converted Mg atoms only have a two-fold rotation axis about the $z$-axis. He assigned the structure to space group $Pmm2$ #25. (Hermann, 1937) referenced both papers, giving the space group as $P6_3/mmc$ but listing the atomic coordinates enumerated by McKeehan. In fact, the McKeehan structure has space group $Imm2$ #44, with 48 atoms in the conventional cell, half of the original, and 24 atoms in the primitive cell. This was noted, without reference, by (Parthé, 1993), which is the only comprehensive list of *Strukturbericht* symbols to include the $B30$ structure. We have reproduced this $Imm2$ structure from McKeehan's data. The true structure of MgZn is unclear, as although it is seen in the Mg-Zn binary phase diagram (Massalski, 1990) over a small range of compositions, a complete crystallographic study has never been published. It is possible that the actual structure is off-stoichiometry. There is some evidence of a $Mg_{12}Zn_{13}$ structure (Mezbahul-Islam, 2014), and $Mg_{21}Zn_{25}$ has been determined (Cerný, 2002) to have the $Zr_{21}Re_{25}$ structure. There are similar problems with the $D2_2$ $MgZn_5$ structure, which we discuss on that page.

**Body-centered Orthorhombic primitive vectors:**

$$
\begin{aligned}
\mathbf{a}_1 &= -\tfrac{1}{2}\,a\,\hat{\mathbf{x}} + \tfrac{1}{2}\,b\,\hat{\mathbf{y}} + \tfrac{1}{2}\,c\,\hat{\mathbf{z}} \\
\mathbf{a}_2 &= \tfrac{1}{2}\,a\,\hat{\mathbf{x}} - \tfrac{1}{2}\,b\,\hat{\mathbf{y}} + \tfrac{1}{2}\,c\,\hat{\mathbf{z}} \\
\mathbf{a}_3 &= \tfrac{1}{2}\,a\,\hat{\mathbf{x}} + \tfrac{1}{2}\,b\,\hat{\mathbf{y}} - \tfrac{1}{2}\,c\,\hat{\mathbf{z}}
\end{aligned}
$$

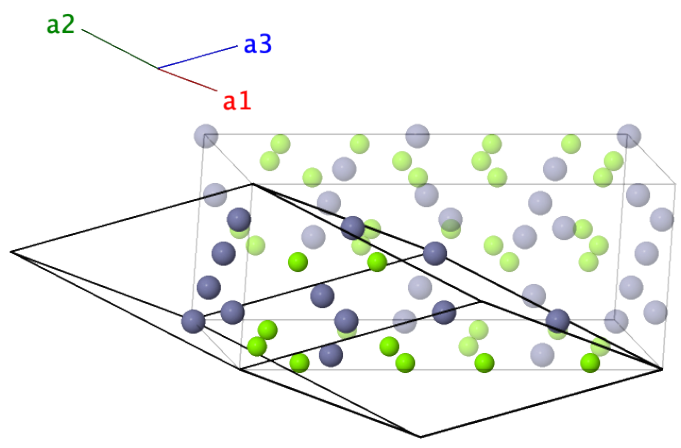

**Basis vectors:**

| | | Lattice Coordinates | | Cartesian Coordinates | Wyckoff Position | Atom Type |
|---|---|---|---|---|---|---|
| $\mathbf{B}_1$ | = | $z_1\,\mathbf{a}_1 + z_1\,\mathbf{a}_2$ | = | $z_1 c\,\hat{\mathbf{z}}$ | (2*a*) | Zn I |
| $\mathbf{B}_2$ | = | $\left(\frac{1}{2} + z_2\right)\mathbf{a}_1 + z_2\,\mathbf{a}_2 + \frac{1}{2}\,\mathbf{a}_3$ | = | $\frac{1}{2} b\,\hat{\mathbf{y}} + z_2 c\,\hat{\mathbf{z}}$ | (2*b*) | Zn II |
| $\mathbf{B}_3$ | = | $z_3\,\mathbf{a}_1 + (x_3 + z_3)\,\mathbf{a}_2 + x_3\,\mathbf{a}_3$ | = | $x_3 a\,\hat{\mathbf{x}} + z_3 c\,\hat{\mathbf{z}}$ | (4*c*) | Zn III |
| $\mathbf{B}_4$ | = | $z_3\,\mathbf{a}_1 + (-x_3 + z_3)\,\mathbf{a}_2 - x_3\,\mathbf{a}_3$ | = | $-x_3 a\,\hat{\mathbf{x}} + z_3 c\,\hat{\mathbf{z}}$ | (4*c*) | Zn III |
| $\mathbf{B}_5$ | = | $z_4\,\mathbf{a}_1 + (x_4 + z_4)\,\mathbf{a}_2 + x_4\,\mathbf{a}_3$ | = | $x_4 a\,\hat{\mathbf{x}} + z_4 c\,\hat{\mathbf{z}}$ | (4*c*) | Zn IV |
| $\mathbf{B}_6$ | = | $z_4\,\mathbf{a}_1 + (-x_4 + z_4)\,\mathbf{a}_2 - x_4\,\mathbf{a}_3$ | = | $-x_4 a\,\hat{\mathbf{x}} + z_4 c\,\hat{\mathbf{z}}$ | (4*c*) | Zn IV |
| $\mathbf{B}_7$ | = | $(y_5 + z_5)\,\mathbf{a}_1 + z_5\,\mathbf{a}_2 + y_5\,\mathbf{a}_3$ | = | $y_5 b\,\hat{\mathbf{y}} + z_5 c\,\hat{\mathbf{z}}$ | (4*d*) | Mg I |
| $\mathbf{B}_8$ | = | $(-y_5 + z_5)\,\mathbf{a}_1 + z_5\,\mathbf{a}_2 - y_5\,\mathbf{a}_3$ | = | $-y_5 b\,\hat{\mathbf{y}} + z_5 c\,\hat{\mathbf{z}}$ | (4*d*) | Mg I |
| $\mathbf{B}_9$ | = | $(y_6 + z_6)\,\mathbf{a}_1 + z_6\,\mathbf{a}_2 + y_6\,\mathbf{a}_3$ | = | $y_6 b\,\hat{\mathbf{y}} + z_6 c\,\hat{\mathbf{z}}$ | (4*d*) | Mg II |
| $\mathbf{B}_{10}$ | = | $(-y_6 + z_6)\,\mathbf{a}_1 + z_6\,\mathbf{a}_2 - y_6\,\mathbf{a}_3$ | = | $-y_6 b\,\hat{\mathbf{y}} + z_6 c\,\hat{\mathbf{z}}$ | (4*d*) | Mg II |
| $\mathbf{B}_{11}$ | = | $(y_7 + z_7)\,\mathbf{a}_1 + z_7\,\mathbf{a}_2 + y_7\,\mathbf{a}_3$ | = | $y_7 b\,\hat{\mathbf{y}} + z_7 c\,\hat{\mathbf{z}}$ | (4*d*) | Mg III |
| $\mathbf{B}_{12}$ | = | $(-y_7 + z_7)\,\mathbf{a}_1 + z_7\,\mathbf{a}_2 - y_7\,\mathbf{a}_3$ | = | $-y_7 b\,\hat{\mathbf{y}} + z_7 c\,\hat{\mathbf{z}}$ | (4*d*) | Mg III |
| $\mathbf{B}_{13}$ | = | $(y_8 + z_8)\,\mathbf{a}_1 + z_8\,\mathbf{a}_2 + y_8\,\mathbf{a}_3$ | = | $y_8 b\,\hat{\mathbf{y}} + z_8 c\,\hat{\mathbf{z}}$ | (4*d*) | Mg IV |
| $\mathbf{B}_{14}$ | = | $(-y_8 + z_8)\,\mathbf{a}_1 + z_8\,\mathbf{a}_2 - y_8\,\mathbf{a}_3$ | = | $-y_8 b\,\hat{\mathbf{y}} + z_8 c\,\hat{\mathbf{z}}$ | (4*d*) | Mg IV |
| $\mathbf{B}_{15}$ | = | $(y_9 + z_9)\,\mathbf{a}_1 + z_9\,\mathbf{a}_2 + y_9\,\mathbf{a}_3$ | = | $y_9 b\,\hat{\mathbf{y}} + z_9 c\,\hat{\mathbf{z}}$ | (4*d*) | Mg V |
| $\mathbf{B}_{16}$ | = | $(-y_9 + z_9)\,\mathbf{a}_1 + z_9\,\mathbf{a}_2 - y_9\,\mathbf{a}_3$ | = | $-y_9 b\,\hat{\mathbf{y}} + z_9 c\,\hat{\mathbf{z}}$ | (4*d*) | Mg V |
| $\mathbf{B}_{17}$ | = | $(y_{10} + z_{10})\,\mathbf{a}_1 + z_{10}\,\mathbf{a}_2 + y_{10}\,\mathbf{a}_3$ | = | $y_{10} b\,\hat{\mathbf{y}} + z_{10} c\,\hat{\mathbf{z}}$ | (4*d*) | Mg VI |
| $\mathbf{B}_{18}$ | = | $(-y_{10} + z_{10})\,\mathbf{a}_1 + z_{10}\,\mathbf{a}_2 - y_{10}\,\mathbf{a}_3$ | = | $-y_{10} b\,\hat{\mathbf{y}} + z_{10} c\,\hat{\mathbf{z}}$ | (4*d*) | Mg VI |
| $\mathbf{B}_{19}$ | = | $(y_{11} + z_{11})\,\mathbf{a}_1 + z_{11}\,\mathbf{a}_2 + y_{11}\,\mathbf{a}_3$ | = | $y_{11} b\,\hat{\mathbf{y}} + z_{11} c\,\hat{\mathbf{z}}$ | (4*d*) | Zn V |
| $\mathbf{B}_{20}$ | = | $(-y_{11} + z_{11})\,\mathbf{a}_1 + z_{11}\,\mathbf{a}_2 - y_{11}\,\mathbf{a}_3$ | = | $-y_{11} b\,\hat{\mathbf{y}} + z_{11} c\,\hat{\mathbf{z}}$ | (4*d*) | Zn V |
| $\mathbf{B}_{21}$ | = | $(y_{12} + z_{12})\,\mathbf{a}_1 + (x_{12} + z_{12})\,\mathbf{a}_2 + (x_{12} + y_{12})\,\mathbf{a}_3$ | = | $x_{12} a\,\hat{\mathbf{x}} + y_{12} b\,\hat{\mathbf{y}} + z_{12} c\,\hat{\mathbf{z}}$ | (8*e*) | Zn VI |
| $\mathbf{B}_{22}$ | = | $(-y_{12} + z_{12})\,\mathbf{a}_1 + (-x_{12} + z_{12})\,\mathbf{a}_2 + (-x_{12} - y_{12})\,\mathbf{a}_3$ | = | $-x_{12} a\,\hat{\mathbf{x}} - y_{12} b\,\hat{\mathbf{y}} + z_{12} c\,\hat{\mathbf{z}}$ | (8*e*) | Zn VI |
| $\mathbf{B}_{23}$ | = | $(-y_{12} + z_{12})\,\mathbf{a}_1 + (x_{12} + z_{12})\,\mathbf{a}_2 + (x_{12} - y_{12})\,\mathbf{a}_3$ | = | $x_{12} a\,\hat{\mathbf{x}} - y_{12} b\,\hat{\mathbf{y}} + z_{12} c\,\hat{\mathbf{z}}$ | (8*e*) | Zn VI |
| $\mathbf{B}_{24}$ | = | $(y_{12} + z_{12})\,\mathbf{a}_1 + (-x_{12} + z_{12})\,\mathbf{a}_2 + (-x_{12} + y_{12})\,\mathbf{a}_3$ | = | $-x_{12} a\,\hat{\mathbf{x}} + y_{12} b\,\hat{\mathbf{y}} + z_{12} c\,\hat{\mathbf{z}}$ | (8*e*) | Zn VI |

---

**Geometry files:**
- CIF: pp. 1606
- POSCAR: pp. 1606

# $Nb_2Zr_6O_{17}$ Structure: A2B17C6_oI100_46_ab_b8c_3c

http://aflow.org/prototype-encyclopedia/A2B17C6_oI100_46_ab_b8c_3c

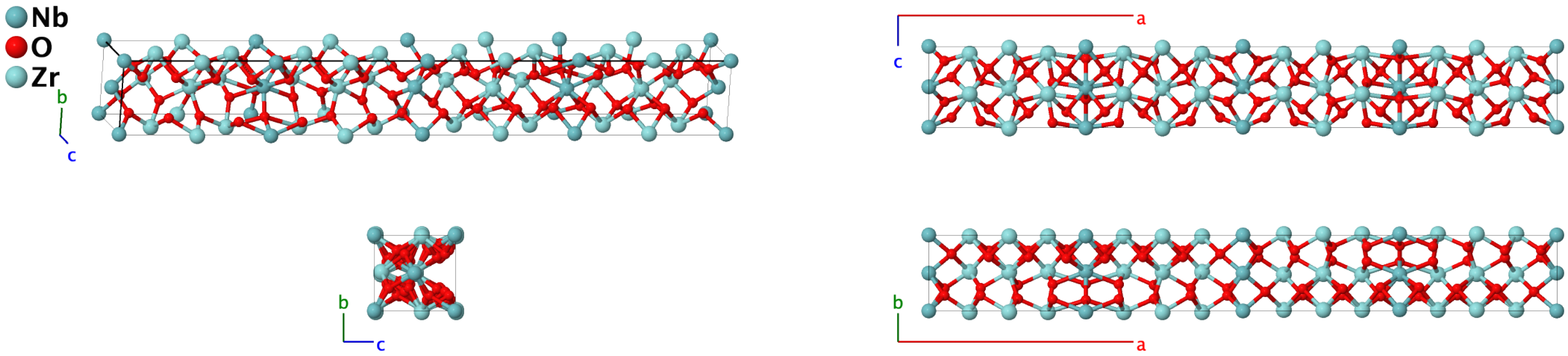

| | | |
|---|---|---|
| **Prototype** | : | $Nb_2O_{17}Zr_6$ |
| **AFLOW prototype label** | : | A2B17C6_oI100_46_ab_b8c_3c |
| ***Strukturbericht* designation** | : | None |
| **Pearson symbol** | : | oI100 |
| **Space group number** | : | 46 |
| **Space group symbol** | : | *Ima*2 |
| **AFLOW prototype command** | : | `aflow --proto=A2B17C6_oI100_46_ab_b8c_3c` `--params=`$a, b/a, c/a, z_1, y_2, z_2, y_3, z_3, x_4, y_4, z_4, x_5, y_5, z_5, x_6, y_6, z_6, x_7, y_7, z_7, x_8, y_8, z_8, x_9, y_9, z_9, x_{10}, y_{10}, z_{10}, x_{11}, y_{11}, z_{11}, x_{12}, y_{12}, z_{12}, x_{13}, y_{13}, z_{13}, x_{14}, y_{14}, z_{14}$ |

### Other compounds with this structure

- $Nb_2Hf_6O_{17}$, $Nb_2Zr_6O_{17}$, $Ta_2Hf_6O_{17}$, and $Ta_2Zr_6O_{17}$

---

- Both (Galy, 1973) and (McCormack, 2019) state that the metallic atom sites are disordered, that is, for the prototype each metallic site has the average composition, $Nb_{0.25}Zr_{0.75}$. We place the niobium atoms on the (2*a*) and (2*b*) sites, and the zirconium on the (4*c*) sites so that the different symmetries are displayed.
- (McCormack, 2019) notes that the metallic composition of these compounds can deviate from the stoichiometry shown here.

### Body-centered Orthorhombic primitive vectors:

$$\mathbf{a}_1 = -\frac{1}{2}\,a\,\hat{\mathbf{x}} + \frac{1}{2}\,b\,\hat{\mathbf{y}} + \frac{1}{2}\,c\,\hat{\mathbf{z}}$$
$$\mathbf{a}_2 = \frac{1}{2}\,a\,\hat{\mathbf{x}} - \frac{1}{2}\,b\,\hat{\mathbf{y}} + \frac{1}{2}\,c\,\hat{\mathbf{z}}$$
$$\mathbf{a}_3 = \frac{1}{2}\,a\,\hat{\mathbf{x}} + \frac{1}{2}\,b\,\hat{\mathbf{y}} - \frac{1}{2}\,c\,\hat{\mathbf{z}}$$

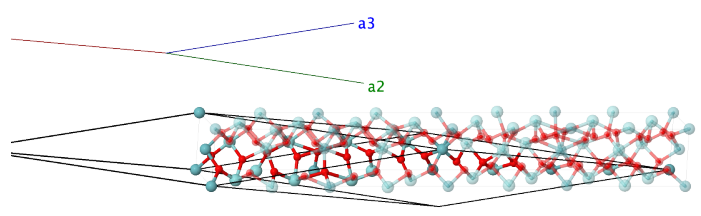

### Basis vectors:

| | | Lattice Coordinates | | Cartesian Coordinates | Wyckoff Position | Atom Type |
|---|---|---|---|---|---|---|
| $\mathbf{B}_1$ | = | $z_1\,\mathbf{a}_1 + z_1\,\mathbf{a}_2$ | = | $z_1 c\,\hat{\mathbf{z}}$ | (4*a*) | Nb I |
| $\mathbf{B}_2$ | = | $z_1\,\mathbf{a}_1 + \left(\frac{1}{2} + z_1\right)\,\mathbf{a}_2 + \frac{1}{2}\,\mathbf{a}_3$ | = | $\frac{1}{2}a\,\hat{\mathbf{x}} + z_1 c\,\hat{\mathbf{z}}$ | (4*a*) | Nb I |
| $\mathbf{B}_3$ | = | $(y_2 + z_2)\,\mathbf{a}_1 + \left(\frac{1}{4} + z_2\right)\,\mathbf{a}_2 + \left(\frac{1}{4} + y_2\right)\,\mathbf{a}_3$ | = | $\frac{1}{4}a\,\hat{\mathbf{x}} + y_2 b\,\hat{\mathbf{y}} + z_2 c\,\hat{\mathbf{z}}$ | (4*b*) | Nb II |
| $\mathbf{B}_4$ | = | $(-y_2 + z_2)\,\mathbf{a}_1 + \left(\frac{3}{4} + z_2\right)\,\mathbf{a}_2 + \left(\frac{3}{4} - y_2\right)\,\mathbf{a}_3$ | = | $\frac{3}{4}a\,\hat{\mathbf{x}} - y_2 b\,\hat{\mathbf{y}} + z_2 c\,\hat{\mathbf{z}}$ | (4*b*) | Nb II |
| $\mathbf{B}_5$ | = | $(y_3 + z_3)\,\mathbf{a}_1 + \left(\frac{1}{4} + z_3\right)\,\mathbf{a}_2 + \left(\frac{1}{4} + y_3\right)\,\mathbf{a}_3$ | = | $\frac{1}{4}a\,\hat{\mathbf{x}} + y_3 b\,\hat{\mathbf{y}} + z_3 c\,\hat{\mathbf{z}}$ | (4*b*) | O I |
| $\mathbf{B}_6$ | = | $(-y_3 + z_3)\,\mathbf{a}_1 + \left(\frac{3}{4} + z_3\right)\,\mathbf{a}_2 + \left(\frac{3}{4} - y_3\right)\,\mathbf{a}_3$ | = | $\frac{3}{4}a\,\hat{\mathbf{x}} - y_3 b\,\hat{\mathbf{y}} + z_3 c\,\hat{\mathbf{z}}$ | (4*b*) | O I |
| $\mathbf{B}_{7}$ | = | $\left(y_{4}+z_{4}\right)\mathbf{a}_{1}+\left(x_{4}+z_{4}\right)\mathbf{a}_{2}+\left(x_{4}+y_{4}\right)\mathbf{a}_{3}$ | = | $x_{4}a\,\hat{\mathbf{x}}+y_{4}b\,\hat{\mathbf{y}}+z_{4}c\,\hat{\mathbf{z}}$ | (8$c$) | O II |
| $\mathbf{B}_{8}$ | = | $\left(-y_{4}+z_{4}\right)\mathbf{a}_{1}+\left(-x_{4}+z_{4}\right)\mathbf{a}_{2}+\left(-x_{4}-y_{4}\right)\mathbf{a}_{3}$ | = | $-x_{4}a\,\hat{\mathbf{x}}-y_{4}b\,\hat{\mathbf{y}}+z_{4}c\,\hat{\mathbf{z}}$ | (8$c$) | O II |
| $\mathbf{B}_{9}$ | = | $\left(-y_{4}+z_{4}\right)\mathbf{a}_{1}+\left(\frac{1}{2}+x_{4}+z_{4}\right)\mathbf{a}_{2}+\left(\frac{1}{2}+x_{4}-y_{4}\right)\mathbf{a}_{3}$ | = | $\left(\frac{1}{2}+x_{4}\right)a\,\hat{\mathbf{x}}-y_{4}b\,\hat{\mathbf{y}}+z_{4}c\,\hat{\mathbf{z}}$ | (8$c$) | O II |
| $\mathbf{B}_{10}$ | = | $\left(y_{4}+z_{4}\right)\mathbf{a}_{1}+\left(\frac{1}{2}-x_{4}+z_{4}\right)\mathbf{a}_{2}+\left(\frac{1}{2}-x_{4}+y_{4}\right)\mathbf{a}_{3}$ | = | $\left(\frac{1}{2}-x_{4}\right)a\,\hat{\mathbf{x}}+y_{4}b\,\hat{\mathbf{y}}+z_{4}c\,\hat{\mathbf{z}}$ | (8$c$) | O II |
| $\mathbf{B}_{11}$ | = | $\left(y_{5}+z_{5}\right)\mathbf{a}_{1}+\left(x_{5}+z_{5}\right)\mathbf{a}_{2}+\left(x_{5}+y_{5}\right)\mathbf{a}_{3}$ | = | $x_{5}a\,\hat{\mathbf{x}}+y_{5}b\,\hat{\mathbf{y}}+z_{5}c\,\hat{\mathbf{z}}$ | (8$c$) | O III |
| $\mathbf{B}_{12}$ | = | $\left(-y_{5}+z_{5}\right)\mathbf{a}_{1}+\left(-x_{5}+z_{5}\right)\mathbf{a}_{2}+\left(-x_{5}-y_{5}\right)\mathbf{a}_{3}$ | = | $-x_{5}a\,\hat{\mathbf{x}}-y_{5}b\,\hat{\mathbf{y}}+z_{5}c\,\hat{\mathbf{z}}$ | (8$c$) | O III |
| $\mathbf{B}_{13}$ | = | $\left(-y_{5}+z_{5}\right)\mathbf{a}_{1}+\left(\frac{1}{2}+x_{5}+z_{5}\right)\mathbf{a}_{2}+\left(\frac{1}{2}+x_{5}-y_{5}\right)\mathbf{a}_{3}$ | = | $\left(\frac{1}{2}+x_{5}\right)a\,\hat{\mathbf{x}}-y_{5}b\,\hat{\mathbf{y}}+z_{5}c\,\hat{\mathbf{z}}$ | (8$c$) | O III |
| $\mathbf{B}_{14}$ | = | $\left(y_{5}+z_{5}\right)\mathbf{a}_{1}+\left(\frac{1}{2}-x_{5}+z_{5}\right)\mathbf{a}_{2}+\left(\frac{1}{2}-x_{5}+y_{5}\right)\mathbf{a}_{3}$ | = | $\left(\frac{1}{2}-x_{5}\right)a\,\hat{\mathbf{x}}+y_{5}b\,\hat{\mathbf{y}}+z_{5}c\,\hat{\mathbf{z}}$ | (8$c$) | O III |
| $\mathbf{B}_{15}$ | = | $\left(y_{6}+z_{6}\right)\mathbf{a}_{1}+\left(x_{6}+z_{6}\right)\mathbf{a}_{2}+\left(x_{6}+y_{6}\right)\mathbf{a}_{3}$ | = | $x_{6}a\,\hat{\mathbf{x}}+y_{6}b\,\hat{\mathbf{y}}+z_{6}c\,\hat{\mathbf{z}}$ | (8$c$) | O IV |
| $\mathbf{B}_{16}$ | = | $\left(-y_{6}+z_{6}\right)\mathbf{a}_{1}+\left(-x_{6}+z_{6}\right)\mathbf{a}_{2}+\left(-x_{6}-y_{6}\right)\mathbf{a}_{3}$ | = | $-x_{6}a\,\hat{\mathbf{x}}-y_{6}b\,\hat{\mathbf{y}}+z_{6}c\,\hat{\mathbf{z}}$ | (8$c$) | O IV |
| $\mathbf{B}_{17}$ | = | $\left(-y_{6}+z_{6}\right)\mathbf{a}_{1}+\left(\frac{1}{2}+x_{6}+z_{6}\right)\mathbf{a}_{2}+\left(\frac{1}{2}+x_{6}-y_{6}\right)\mathbf{a}_{3}$ | = | $\left(\frac{1}{2}+x_{6}\right)a\,\hat{\mathbf{x}}-y_{6}b\,\hat{\mathbf{y}}+z_{6}c\,\hat{\mathbf{z}}$ | (8$c$) | O IV |
| $\mathbf{B}_{18}$ | = | $\left(y_{6}+z_{6}\right)\mathbf{a}_{1}+\left(\frac{1}{2}-x_{6}+z_{6}\right)\mathbf{a}_{2}+\left(\frac{1}{2}-x_{6}+y_{6}\right)\mathbf{a}_{3}$ | = | $\left(\frac{1}{2}-x_{6}\right)a\,\hat{\mathbf{x}}+y_{6}b\,\hat{\mathbf{y}}+z_{6}c\,\hat{\mathbf{z}}$ | (8$c$) | O IV |
| $\mathbf{B}_{19}$ | = | $\left(y_{7}+z_{7}\right)\mathbf{a}_{1}+\left(x_{7}+z_{7}\right)\mathbf{a}_{2}+\left(x_{7}+y_{7}\right)\mathbf{a}_{3}$ | = | $x_{7}a\,\hat{\mathbf{x}}+y_{7}b\,\hat{\mathbf{y}}+z_{7}c\,\hat{\mathbf{z}}$ | (8$c$) | O V |
| $\mathbf{B}_{20}$ | = | $\left(-y_{7}+z_{7}\right)\mathbf{a}_{1}+\left(-x_{7}+z_{7}\right)\mathbf{a}_{2}+\left(-x_{7}-y_{7}\right)\mathbf{a}_{3}$ | = | $-x_{7}a\,\hat{\mathbf{x}}-y_{7}b\,\hat{\mathbf{y}}+z_{7}c\,\hat{\mathbf{z}}$ | (8$c$) | O V |
| $\mathbf{B}_{21}$ | = | $\left(-y_{7}+z_{7}\right)\mathbf{a}_{1}+\left(\frac{1}{2}+x_{7}+z_{7}\right)\mathbf{a}_{2}+\left(\frac{1}{2}+x_{7}-y_{7}\right)\mathbf{a}_{3}$ | = | $\left(\frac{1}{2}+x_{7}\right)a\,\hat{\mathbf{x}}-y_{7}b\,\hat{\mathbf{y}}+z_{7}c\,\hat{\mathbf{z}}$ | (8$c$) | O V |
| $\mathbf{B}_{22}$ | = | $\left(y_{7}+z_{7}\right)\mathbf{a}_{1}+\left(\frac{1}{2}-x_{7}+z_{7}\right)\mathbf{a}_{2}+\left(\frac{1}{2}-x_{7}+y_{7}\right)\mathbf{a}_{3}$ | = | $\left(\frac{1}{2}-x_{7}\right)a\,\hat{\mathbf{x}}+y_{7}b\,\hat{\mathbf{y}}+z_{7}c\,\hat{\mathbf{z}}$ | (8$c$) | O V |
| $\mathbf{B}_{23}$ | = | $\left(y_{8}+z_{8}\right)\mathbf{a}_{1}+\left(x_{8}+z_{8}\right)\mathbf{a}_{2}+\left(x_{8}+y_{8}\right)\mathbf{a}_{3}$ | = | $x_{8}a\,\hat{\mathbf{x}}+y_{8}b\,\hat{\mathbf{y}}+z_{8}c\,\hat{\mathbf{z}}$ | (8$c$) | O VI |
| $\mathbf{B}_{24}$ | = | $\left(-y_{8}+z_{8}\right)\mathbf{a}_{1}+\left(-x_{8}+z_{8}\right)\mathbf{a}_{2}+\left(-x_{8}-y_{8}\right)\mathbf{a}_{3}$ | = | $-x_{8}a\,\hat{\mathbf{x}}-y_{8}b\,\hat{\mathbf{y}}+z_{8}c\,\hat{\mathbf{z}}$ | (8$c$) | O VI |
| $\mathbf{B}_{25}$ | = | $\left(-y_{8}+z_{8}\right)\mathbf{a}_{1}+\left(\frac{1}{2}+x_{8}+z_{8}\right)\mathbf{a}_{2}+\left(\frac{1}{2}+x_{8}-y_{8}\right)\mathbf{a}_{3}$ | = | $\left(\frac{1}{2}+x_{8}\right)a\,\hat{\mathbf{x}}-y_{8}b\,\hat{\mathbf{y}}+z_{8}c\,\hat{\mathbf{z}}$ | (8$c$) | O VI |
| $\mathbf{B}_{26}$ | = | $\left(y_{8}+z_{8}\right)\mathbf{a}_{1}+\left(\frac{1}{2}-x_{8}+z_{8}\right)\mathbf{a}_{2}+\left(\frac{1}{2}-x_{8}+y_{8}\right)\mathbf{a}_{3}$ | = | $\left(\frac{1}{2}-x_{8}\right)a\,\hat{\mathbf{x}}+y_{8}b\,\hat{\mathbf{y}}+z_{8}c\,\hat{\mathbf{z}}$ | (8$c$) | O VI |
| $\mathbf{B}_{27}$ | = | $\left(y_{9}+z_{9}\right)\mathbf{a}_{1}+\left(x_{9}+z_{9}\right)\mathbf{a}_{2}+\left(x_{9}+y_{9}\right)\mathbf{a}_{3}$ | = | $x_{9}a\,\hat{\mathbf{x}}+y_{9}b\,\hat{\mathbf{y}}+z_{9}c\,\hat{\mathbf{z}}$ | (8$c$) | O VII |
| $\mathbf{B}_{28}$ | = | $\left(-y_{9}+z_{9}\right)\mathbf{a}_{1}+\left(-x_{9}+z_{9}\right)\mathbf{a}_{2}+\left(-x_{9}-y_{9}\right)\mathbf{a}_{3}$ | = | $-x_{9}a\,\hat{\mathbf{x}}-y_{9}b\,\hat{\mathbf{y}}+z_{9}c\,\hat{\mathbf{z}}$ | (8$c$) | O VII |
| $\mathbf{B}_{29}$ | = | $\left(-y_{9}+z_{9}\right)\mathbf{a}_{1}+\left(\frac{1}{2}+x_{9}+z_{9}\right)\mathbf{a}_{2}+\left(\frac{1}{2}+x_{9}-y_{9}\right)\mathbf{a}_{3}$ | = | $\left(\frac{1}{2}+x_{9}\right)a\,\hat{\mathbf{x}}-y_{9}b\,\hat{\mathbf{y}}+z_{9}c\,\hat{\mathbf{z}}$ | (8$c$) | O VII |
| $\mathbf{B}_{30}$ | = | $\left(y_{9}+z_{9}\right)\mathbf{a}_{1}+\left(\frac{1}{2}-x_{9}+z_{9}\right)\mathbf{a}_{2}+\left(\frac{1}{2}-x_{9}+y_{9}\right)\mathbf{a}_{3}$ | = | $\left(\frac{1}{2}-x_{9}\right)a\,\hat{\mathbf{x}}+y_{9}b\,\hat{\mathbf{y}}+z_{9}c\,\hat{\mathbf{z}}$ | (8$c$) | O VII |
| $\mathbf{B}_{31}$ | = | $(y_{10}+z_{10})\,\mathbf{a}_1 + (x_{10}+z_{10})\,\mathbf{a}_2 + (x_{10}+y_{10})\,\mathbf{a}_3$ | = | $x_{10}a\,\hat{\mathbf{x}} + y_{10}b\,\hat{\mathbf{y}} + z_{10}c\,\hat{\mathbf{z}}$ | (8c) | O VIII |
| $\mathbf{B}_{32}$ | = | $(-y_{10}+z_{10})\,\mathbf{a}_1 + (-x_{10}+z_{10})\,\mathbf{a}_2 + (-x_{10}-y_{10})\,\mathbf{a}_3$ | = | $-x_{10}a\,\hat{\mathbf{x}} - y_{10}b\,\hat{\mathbf{y}} + z_{10}c\,\hat{\mathbf{z}}$ | (8c) | O VIII |
| $\mathbf{B}_{33}$ | = | $(-y_{10}+z_{10})\,\mathbf{a}_1 + \left(\frac{1}{2}+x_{10}+z_{10}\right)\mathbf{a}_2 + \left(\frac{1}{2}+x_{10}-y_{10}\right)\mathbf{a}_3$ | = | $\left(\frac{1}{2}+x_{10}\right)a\,\hat{\mathbf{x}} - y_{10}b\,\hat{\mathbf{y}} + z_{10}c\,\hat{\mathbf{z}}$ | (8c) | O VIII |
| $\mathbf{B}_{34}$ | = | $(y_{10}+z_{10})\,\mathbf{a}_1 + \left(\frac{1}{2}-x_{10}+z_{10}\right)\mathbf{a}_2 + \left(\frac{1}{2}-x_{10}+y_{10}\right)\mathbf{a}_3$ | = | $\left(\frac{1}{2}-x_{10}\right)a\,\hat{\mathbf{x}} + y_{10}b\,\hat{\mathbf{y}} + z_{10}c\,\hat{\mathbf{z}}$ | (8c) | O VIII |
| $\mathbf{B}_{35}$ | = | $(y_{11}+z_{11})\,\mathbf{a}_1 + (x_{11}+z_{11})\,\mathbf{a}_2 + (x_{11}+y_{11})\,\mathbf{a}_3$ | = | $x_{11}a\,\hat{\mathbf{x}} + y_{11}b\,\hat{\mathbf{y}} + z_{11}c\,\hat{\mathbf{z}}$ | (8c) | O IX |
| $\mathbf{B}_{36}$ | = | $(-y_{11}+z_{11})\,\mathbf{a}_1 + (-x_{11}+z_{11})\,\mathbf{a}_2 + (-x_{11}-y_{11})\,\mathbf{a}_3$ | = | $-x_{11}a\,\hat{\mathbf{x}} - y_{11}b\,\hat{\mathbf{y}} + z_{11}c\,\hat{\mathbf{z}}$ | (8c) | O IX |
| $\mathbf{B}_{37}$ | = | $(-y_{11}+z_{11})\,\mathbf{a}_1 + \left(\frac{1}{2}+x_{11}+z_{11}\right)\mathbf{a}_2 + \left(\frac{1}{2}+x_{11}-y_{11}\right)\mathbf{a}_3$ | = | $\left(\frac{1}{2}+x_{11}\right)a\,\hat{\mathbf{x}} - y_{11}b\,\hat{\mathbf{y}} + z_{11}c\,\hat{\mathbf{z}}$ | (8c) | O IX |
| $\mathbf{B}_{38}$ | = | $(y_{11}+z_{11})\,\mathbf{a}_1 + \left(\frac{1}{2}-x_{11}+z_{11}\right)\mathbf{a}_2 + \left(\frac{1}{2}-x_{11}+y_{11}\right)\mathbf{a}_3$ | = | $\left(\frac{1}{2}-x_{11}\right)a\,\hat{\mathbf{x}} + y_{11}b\,\hat{\mathbf{y}} + z_{11}c\,\hat{\mathbf{z}}$ | (8c) | O IX |
| $\mathbf{B}_{39}$ | = | $(y_{12}+z_{12})\,\mathbf{a}_1 + (x_{12}+z_{12})\,\mathbf{a}_2 + (x_{12}+y_{12})\,\mathbf{a}_3$ | = | $x_{12}a\,\hat{\mathbf{x}} + y_{12}b\,\hat{\mathbf{y}} + z_{12}c\,\hat{\mathbf{z}}$ | (8c) | Zr I |
| $\mathbf{B}_{40}$ | = | $(-y_{12}+z_{12})\,\mathbf{a}_1 + (-x_{12}+z_{12})\,\mathbf{a}_2 + (-x_{12}-y_{12})\,\mathbf{a}_3$ | = | $-x_{12}a\,\hat{\mathbf{x}} - y_{12}b\,\hat{\mathbf{y}} + z_{12}c\,\hat{\mathbf{z}}$ | (8c) | Zr I |
| $\mathbf{B}_{41}$ | = | $(-y_{12}+z_{12})\,\mathbf{a}_1 + \left(\frac{1}{2}+x_{12}+z_{12}\right)\mathbf{a}_2 + \left(\frac{1}{2}+x_{12}-y_{12}\right)\mathbf{a}_3$ | = | $\left(\frac{1}{2}+x_{12}\right)a\,\hat{\mathbf{x}} - y_{12}b\,\hat{\mathbf{y}} + z_{12}c\,\hat{\mathbf{z}}$ | (8c) | Zr I |
| $\mathbf{B}_{42}$ | = | $(y_{12}+z_{12})\,\mathbf{a}_1 + \left(\frac{1}{2}-x_{12}+z_{12}\right)\mathbf{a}_2 + \left(\frac{1}{2}-x_{12}+y_{12}\right)\mathbf{a}_3$ | = | $\left(\frac{1}{2}-x_{12}\right)a\,\hat{\mathbf{x}} + y_{12}b\,\hat{\mathbf{y}} + z_{12}c\,\hat{\mathbf{z}}$ | (8c) | Zr I |
| $\mathbf{B}_{43}$ | = | $(y_{13}+z_{13})\,\mathbf{a}_1 + (x_{13}+z_{13})\,\mathbf{a}_2 + (x_{13}+y_{13})\,\mathbf{a}_3$ | = | $x_{13}a\,\hat{\mathbf{x}} + y_{13}b\,\hat{\mathbf{y}} + z_{13}c\,\hat{\mathbf{z}}$ | (8c) | Zr II |
| $\mathbf{B}_{44}$ | = | $(-y_{13}+z_{13})\,\mathbf{a}_1 + (-x_{13}+z_{13})\,\mathbf{a}_2 + (-x_{13}-y_{13})\,\mathbf{a}_3$ | = | $-x_{13}a\,\hat{\mathbf{x}} - y_{13}b\,\hat{\mathbf{y}} + z_{13}c\,\hat{\mathbf{z}}$ | (8c) | Zr II |
| $\mathbf{B}_{45}$ | = | $(-y_{13}+z_{13})\,\mathbf{a}_1 + \left(\frac{1}{2}+x_{13}+z_{13}\right)\mathbf{a}_2 + \left(\frac{1}{2}+x_{13}-y_{13}\right)\mathbf{a}_3$ | = | $\left(\frac{1}{2}+x_{13}\right)a\,\hat{\mathbf{x}} - y_{13}b\,\hat{\mathbf{y}} + z_{13}c\,\hat{\mathbf{z}}$ | (8c) | Zr II |
| $\mathbf{B}_{46}$ | = | $(y_{13}+z_{13})\,\mathbf{a}_1 + \left(\frac{1}{2}-x_{13}+z_{13}\right)\mathbf{a}_2 + \left(\frac{1}{2}-x_{13}+y_{13}\right)\mathbf{a}_3$ | = | $\left(\frac{1}{2}-x_{13}\right)a\,\hat{\mathbf{x}} + y_{13}b\,\hat{\mathbf{y}} + z_{13}c\,\hat{\mathbf{z}}$ | (8c) | Zr II |
| $\mathbf{B}_{47}$ | = | $(y_{14}+z_{14})\,\mathbf{a}_1 + (x_{14}+z_{14})\,\mathbf{a}_2 + (x_{14}+y_{14})\,\mathbf{a}_3$ | = | $x_{14}a\,\hat{\mathbf{x}} + y_{14}b\,\hat{\mathbf{y}} + z_{14}c\,\hat{\mathbf{z}}$ | (8c) | Zr III |
| $\mathbf{B}_{48}$ | = | $(-y_{14}+z_{14})\,\mathbf{a}_1 + (-x_{14}+z_{14})\,\mathbf{a}_2 + (-x_{14}-y_{14})\,\mathbf{a}_3$ | = | $-x_{14}a\,\hat{\mathbf{x}} - y_{14}b\,\hat{\mathbf{y}} + z_{14}c\,\hat{\mathbf{z}}$ | (8c) | Zr III |
| $\mathbf{B}_{49}$ | = | $(-y_{14}+z_{14})\,\mathbf{a}_1 + \left(\frac{1}{2}+x_{14}+z_{14}\right)\mathbf{a}_2 + \left(\frac{1}{2}+x_{14}-y_{14}\right)\mathbf{a}_3$ | = | $\left(\frac{1}{2}+x_{14}\right)a\,\hat{\mathbf{x}} - y_{14}b\,\hat{\mathbf{y}} + z_{14}c\,\hat{\mathbf{z}}$ | (8c) | Zr III |
| $\mathbf{B}_{50}$ | = | $(y_{14}+z_{14})\,\mathbf{a}_1 + \left(\frac{1}{2}-x_{14}+z_{14}\right)\mathbf{a}_2 + \left(\frac{1}{2}-x_{14}+y_{14}\right)\mathbf{a}_3$ | = | $\left(\frac{1}{2}-x_{14}\right)a\,\hat{\mathbf{x}} + y_{14}b\,\hat{\mathbf{y}} + z_{14}c\,\hat{\mathbf{z}}$ | (8c) | Zr III |

**Found in:**

**Geometry files:**

- CIF: pp. 1606
- POSCAR: pp. 1607

# Parkerite ($Ni_3Bi_2S_2$) Structure: AB2C_oP8_51_e_be_f

http://aflow.org/prototype-encyclopedia/AB2C_oP8_51_e_be_f

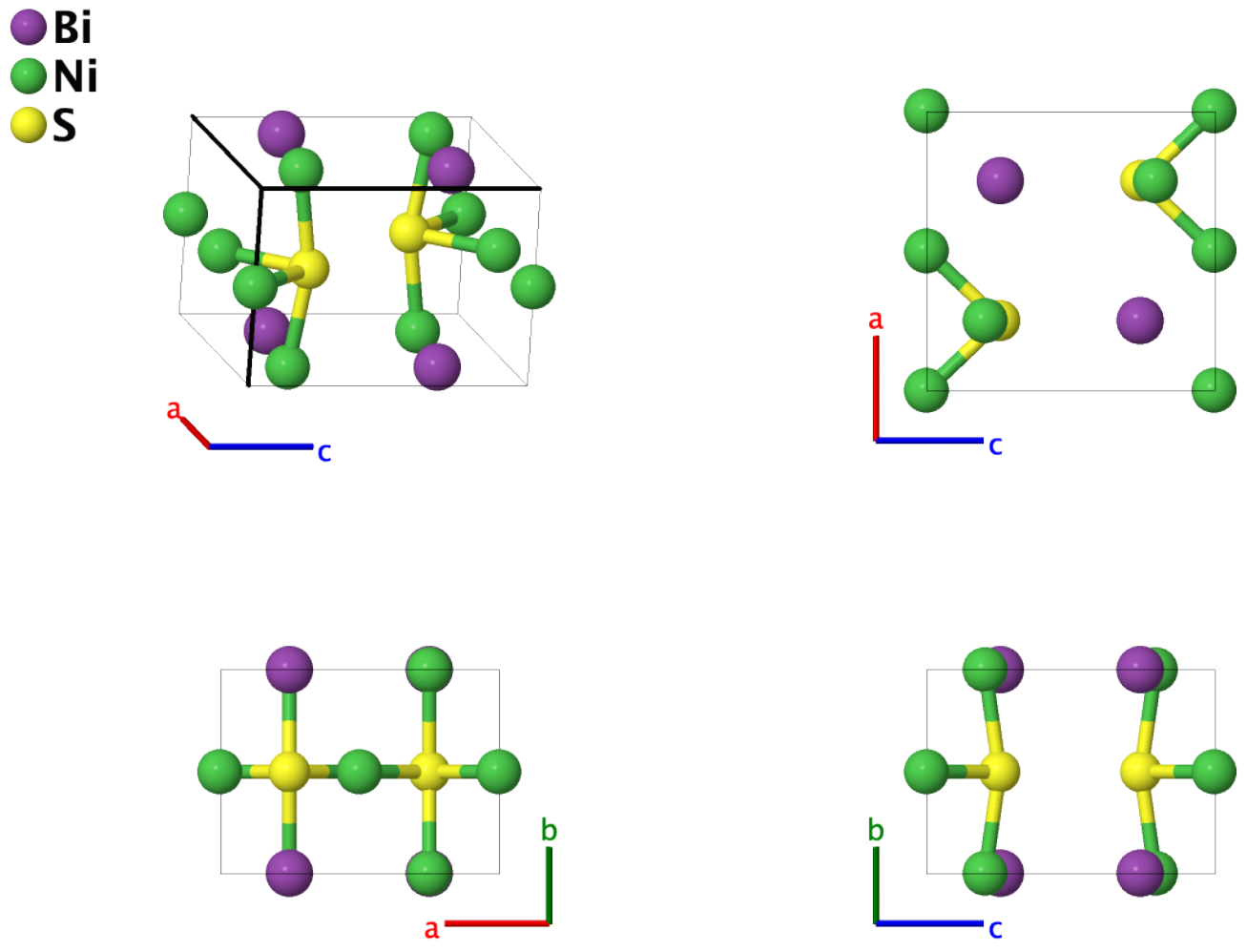

| | | |
|---|---|---|
| **Prototype** | : | $Bi_2Ni_3S_2$ |
| **AFLOW prototype label** | : | AB2C_oP8_51_e_be_f |
| ***Strukturbericht* designation** | : | None |
| **Pearson symbol** | : | oP8 |
| **Space group number** | : | 51 |
| **Space group symbol** | : | *Pmma* |
| **AFLOW prototype command** | : | `aflow --proto=AB2C_oP8_51_e_be_f` `--params=`$a, b/a, c/a, z_2, z_3, z_4$ |

### Other compounds with this structure

- $Ni_3(Bi,Pb)_2S_2$

- (Fleet, 1973) states that parkerite is a derivative of the shandite ($Ni_3Pb_2S_2$) structure, and changes to shandite if more than 4% of the bismuth is replaced by lead.
- The Ni-II (2e) site is occupied 50% of the time, given the observed stoichiometry.
- Earlier sources give parkerite a monoclinic structure. This may be due to an ordering of the nickel atoms at lower temperature. We follow (Downs, 2003) and use the orthorhombic structure.
- (Fleet, 1973) describes the structure in the *Pmam* setting of space group #51. We used FINDSYM to transform it to the standard *Pmma* setting.

### Simple Orthorhombic primitive vectors:

$$\mathbf{a}_1 = a\,\hat{\mathbf{x}}$$
$$\mathbf{a}_2 = b\,\hat{\mathbf{y}}$$
$$\mathbf{a}_3 = c\,\hat{\mathbf{z}}$$

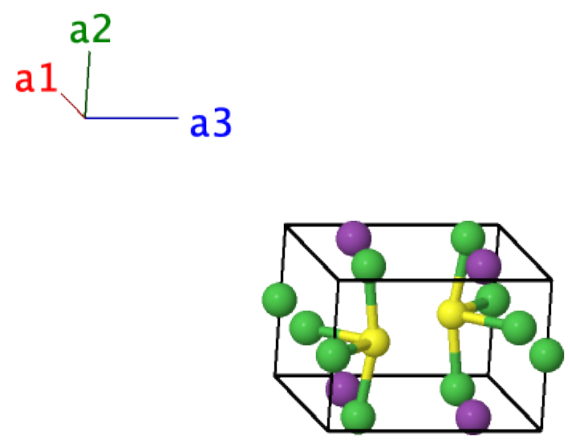

**Basis vectors:**

| | | Lattice Coordinates | | Cartesian Coordinates | Wyckoff Position | Atom Type |
|---|---|---|---|---|---|---|
| $\mathbf{B}_1$ | = | $\frac{1}{2}\,\mathbf{a}_2$ | = | $\frac{1}{2}b\,\hat{\mathbf{y}}$ | (2$b$) | Ni I |
| $\mathbf{B}_2$ | = | $\frac{1}{2}\,\mathbf{a}_1 + \frac{1}{2}\,\mathbf{a}_2$ | = | $\frac{1}{2}a\,\hat{\mathbf{x}} + \frac{1}{2}b\,\hat{\mathbf{y}}$ | (2$b$) | Ni I |
| $\mathbf{B}_3$ | = | $\frac{1}{4}\,\mathbf{a}_1 + z_2\,\mathbf{a}_3$ | = | $\frac{1}{4}a\,\hat{\mathbf{x}} + z_2c\,\hat{\mathbf{z}}$ | (2$e$) | Bi |
| $\mathbf{B}_4$ | = | $\frac{3}{4}\,\mathbf{a}_1 - z_2\,\mathbf{a}_3$ | = | $\frac{3}{4}a\,\hat{\mathbf{x}} - z_2c\,\hat{\mathbf{z}}$ | (2$e$) | Bi |
| $\mathbf{B}_5$ | = | $\frac{1}{4}\,\mathbf{a}_1 + z_3\,\mathbf{a}_3$ | = | $\frac{1}{4}a\,\hat{\mathbf{x}} + z_3c\,\hat{\mathbf{z}}$ | (2$e$) | Ni II |
| $\mathbf{B}_6$ | = | $\frac{3}{4}\,\mathbf{a}_1 - z_3\,\mathbf{a}_3$ | = | $\frac{3}{4}a\,\hat{\mathbf{x}} - z_3c\,\hat{\mathbf{z}}$ | (2$e$) | Ni II |
| $\mathbf{B}_7$ | = | $\frac{1}{4}\,\mathbf{a}_1 + \frac{1}{2}\,\mathbf{a}_2 + z_4\,\mathbf{a}_3$ | = | $\frac{1}{4}a\,\hat{\mathbf{x}} + \frac{1}{2}b\,\hat{\mathbf{y}} + z_4c\,\hat{\mathbf{z}}$ | (2$f$) | S |
| $\mathbf{B}_8$ | = | $\frac{3}{4}\,\mathbf{a}_1 + \frac{1}{2}\,\mathbf{a}_2 - z_4\,\mathbf{a}_3$ | = | $\frac{3}{4}a\,\hat{\mathbf{x}} + \frac{1}{2}b\,\hat{\mathbf{y}} - z_4c\,\hat{\mathbf{z}}$ | (2$f$) | S |

**Geometry files:**

- CIF: pp. 1607
- POSCAR: pp. 1607

# $LiNb_6O_{15}F$ Structure: ABC6D15_oP46_51_f_d_2e2i_aef4i2j

http://aflow.org/prototype-encyclopedia/ABC6D15_oP46_51_f_d_2e2i_aef4i2j

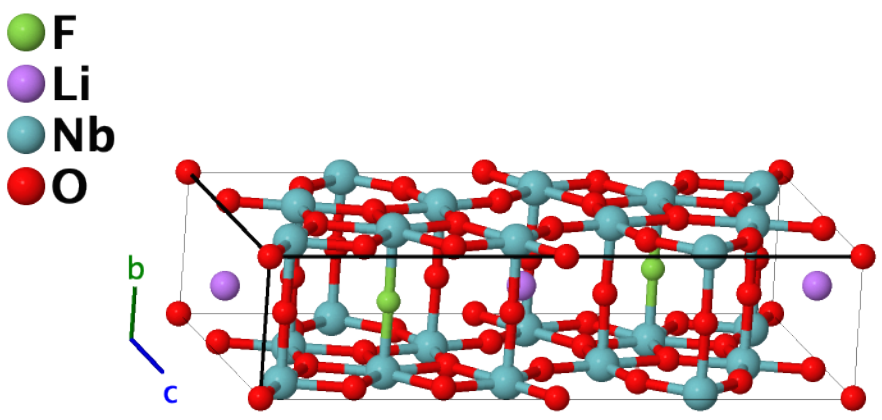

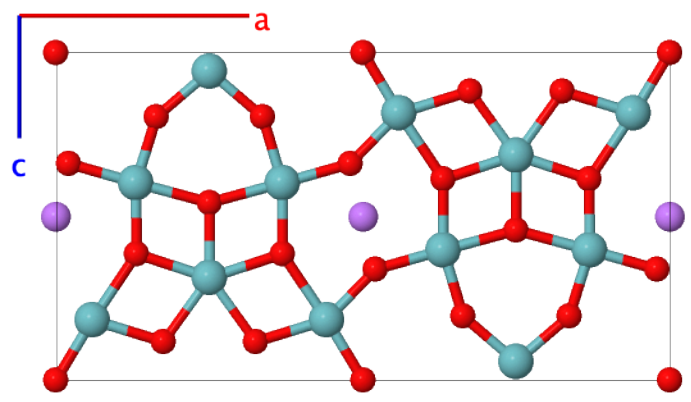

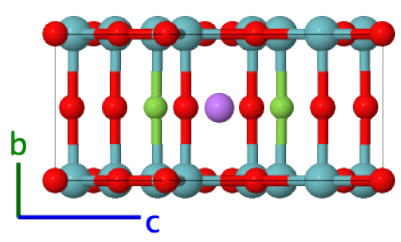

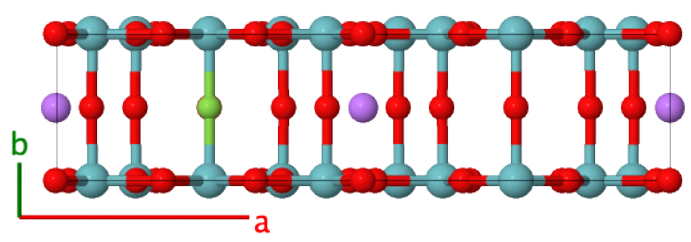

| | | |
|---|---|---|
| **Prototype** | : | $FLiNb_6O_{15}$ |
| **AFLOW prototype label** | : | ABC6D15_oP46_51_f_d_2e2i_aef4i2j |
| ***Strukturbericht* designation** | : | None |
| **Pearson symbol** | : | oP46 |
| **Space group number** | : | 51 |
| **Space group symbol** | : | *Pmma* |
| **AFLOW prototype command** | : | `aflow --proto=ABC6D15_oP46_51_f_d_2e2i_aef4i2j --params=`$a, b/a, c/a, z_3, z_4, z_5, z_6, z_7, x_8, z_8, x_9, z_9, x_{10}, z_{10}, x_{11}, z_{11}, x_{12}, z_{12}, x_{13}, z_{13}, x_{14}, z_{14}, x_{15}, z_{15}$ |

- (Lundberg, 1965) suggests that the lithium atoms are either on the (2*d*) site or are statistically distributed on a (4*f*) site with approximate coordinates (0.08,1/2,0.10). For simplicity we place the atoms on the (2*d*) site.
- The X-ray scattering of a $F^-$ ion is almost identical to that of $O^{2-}$, and Lundberg was not able to distinguish between them. She arbitrarily designated the (2*f*) site she also called $O_4$ as the location of the fluorine ion and we follow this, but in reality we have no idea if the $F^-$ ions are located on this site, are ordered on another site, or are statistically distributed on the oxygen sites.

**Simple Orthorhombic primitive vectors:**

$$\mathbf{a}_1 = a\,\hat{\mathbf{x}}$$
$$\mathbf{a}_2 = b\,\hat{\mathbf{y}}$$
$$\mathbf{a}_3 = c\,\hat{\mathbf{z}}$$

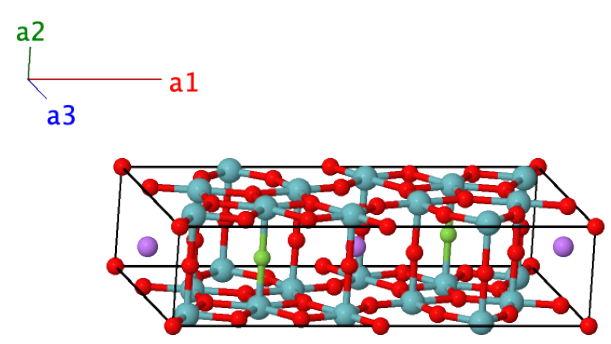

**Basis vectors:**

| | | Lattice Coordinates | | Cartesian Coordinates | Wyckoff Position | Atom Type |
|---|---|---|---|---|---|---|
| $\mathbf{B}_{1}$ | = | $0\,\mathbf{a}_{1}+0\,\mathbf{a}_{2}+0\,\mathbf{a}_{3}$ | = | $0\,\hat{\mathbf{x}}+0\,\hat{\mathbf{y}}+0\,\hat{\mathbf{z}}$ | (2$a$) | O I |
| $\mathbf{B}_{2}$ | = | $\frac{1}{2}\,\mathbf{a}_{1}$ | = | $\frac{1}{2}a\,\hat{\mathbf{x}}$ | (2$a$) | O I |
| $\mathbf{B}_{3}$ | = | $\frac{1}{2}\,\mathbf{a}_{2}+\frac{1}{2}\,\mathbf{a}_{3}$ | = | $\frac{1}{2}b\,\hat{\mathbf{y}}+\frac{1}{2}c\,\hat{\mathbf{z}}$ | (2$d$) | Li |
| $\mathbf{B}_{4}$ | = | $\frac{1}{2}\,\mathbf{a}_{1}+\frac{1}{2}\,\mathbf{a}_{2}+\frac{1}{2}\,\mathbf{a}_{3}$ | = | $\frac{1}{2}a\,\hat{\mathbf{x}}+\frac{1}{2}b\,\hat{\mathbf{y}}+\frac{1}{2}c\,\hat{\mathbf{z}}$ | (2$d$) | Li |
| $\mathbf{B}_{5}$ | = | $\frac{1}{4}\,\mathbf{a}_{1}+z_{3}\,\mathbf{a}_{3}$ | = | $\frac{1}{4}a\,\hat{\mathbf{x}}+z_{3}c\,\hat{\mathbf{z}}$ | (2$e$) | Nb I |
| $\mathbf{B}_{6}$ | = | $\frac{3}{4}\,\mathbf{a}_{1}-z_{3}\,\mathbf{a}_{3}$ | = | $\frac{3}{4}a\,\hat{\mathbf{x}}-z_{3}c\,\hat{\mathbf{z}}$ | (2$e$) | Nb I |
| $\mathbf{B}_{7}$ | = | $\frac{1}{4}\,\mathbf{a}_{1}+z_{4}\,\mathbf{a}_{3}$ | = | $\frac{1}{4}a\,\hat{\mathbf{x}}+z_{4}c\,\hat{\mathbf{z}}$ | (2$e$) | Nb II |
| $\mathbf{B}_{8}$ | = | $\frac{3}{4}\,\mathbf{a}_{1}-z_{4}\,\mathbf{a}_{3}$ | = | $\frac{3}{4}a\,\hat{\mathbf{x}}-z_{4}c\,\hat{\mathbf{z}}$ | (2$e$) | Nb II |
| $\mathbf{B}_{9}$ | = | $\frac{1}{4}\,\mathbf{a}_{1}+z_{5}\,\mathbf{a}_{3}$ | = | $\frac{1}{4}a\,\hat{\mathbf{x}}+z_{5}c\,\hat{\mathbf{z}}$ | (2$e$) | O II |
| $\mathbf{B}_{10}$ | = | $\frac{3}{4}\,\mathbf{a}_{1}-z_{5}\,\mathbf{a}_{3}$ | = | $\frac{3}{4}a\,\hat{\mathbf{x}}-z_{5}c\,\hat{\mathbf{z}}$ | (2$e$) | O II |
| $\mathbf{B}_{11}$ | = | $\frac{1}{4}\,\mathbf{a}_{1}+\frac{1}{2}\,\mathbf{a}_{2}+z_{6}\,\mathbf{a}_{3}$ | = | $\frac{1}{4}a\,\hat{\mathbf{x}}+\frac{1}{2}b\,\hat{\mathbf{y}}+z_{6}c\,\hat{\mathbf{z}}$ | (2$f$) | F |
| $\mathbf{B}_{12}$ | = | $\frac{3}{4}\,\mathbf{a}_{1}+\frac{1}{2}\,\mathbf{a}_{2}-z_{6}\,\mathbf{a}_{3}$ | = | $\frac{3}{4}a\,\hat{\mathbf{x}}+\frac{1}{2}b\,\hat{\mathbf{y}}-z_{6}c\,\hat{\mathbf{z}}$ | (2$f$) | F |
| $\mathbf{B}_{13}$ | = | $\frac{1}{4}\,\mathbf{a}_{1}+\frac{1}{2}\,\mathbf{a}_{2}+z_{7}\,\mathbf{a}_{3}$ | = | $\frac{1}{4}a\,\hat{\mathbf{x}}+\frac{1}{2}b\,\hat{\mathbf{y}}+z_{7}c\,\hat{\mathbf{z}}$ | (2$f$) | O III |
| $\mathbf{B}_{14}$ | = | $\frac{3}{4}\,\mathbf{a}_{1}+\frac{1}{2}\,\mathbf{a}_{2}-z_{7}\,\mathbf{a}_{3}$ | = | $\frac{3}{4}a\,\hat{\mathbf{x}}+\frac{1}{2}b\,\hat{\mathbf{y}}-z_{7}c\,\hat{\mathbf{z}}$ | (2$f$) | O III |
| $\mathbf{B}_{15}$ | = | $x_{8}\,\mathbf{a}_{1}+z_{8}\,\mathbf{a}_{3}$ | = | $x_{8}a\,\hat{\mathbf{x}}+z_{8}c\,\hat{\mathbf{z}}$ | (4$i$) | Nb III |
| $\mathbf{B}_{16}$ | = | $\left(\frac{1}{2}-x_{8}\right)\,\mathbf{a}_{1}+z_{8}\,\mathbf{a}_{3}$ | = | $\left(\frac{1}{2}-x_{8}\right)a\,\hat{\mathbf{x}}+z_{8}c\,\hat{\mathbf{z}}$ | (4$i$) | Nb III |
| $\mathbf{B}_{17}$ | = | $-x_{8}\,\mathbf{a}_{1}-z_{8}\,\mathbf{a}_{3}$ | = | $-x_{8}a\,\hat{\mathbf{x}}-z_{8}c\,\hat{\mathbf{z}}$ | (4$i$) | Nb III |
| $\mathbf{B}_{18}$ | = | $\left(\frac{1}{2}+x_{8}\right)\,\mathbf{a}_{1}-z_{8}\,\mathbf{a}_{3}$ | = | $\left(\frac{1}{2}+x_{8}\right)a\,\hat{\mathbf{x}}-z_{8}c\,\hat{\mathbf{z}}$ | (4$i$) | Nb III |
| $\mathbf{B}_{19}$ | = | $x_{9}\,\mathbf{a}_{1}+z_{9}\,\mathbf{a}_{3}$ | = | $x_{9}a\,\hat{\mathbf{x}}+z_{9}c\,\hat{\mathbf{z}}$ | (4$i$) | Nb IV |
| $\mathbf{B}_{20}$ | = | $\left(\frac{1}{2}-x_{9}\right)\,\mathbf{a}_{1}+z_{9}\,\mathbf{a}_{3}$ | = | $\left(\frac{1}{2}-x_{9}\right)a\,\hat{\mathbf{x}}+z_{9}c\,\hat{\mathbf{z}}$ | (4$i$) | Nb IV |
| $\mathbf{B}_{21}$ | = | $-x_{9}\,\mathbf{a}_{1}-z_{9}\,\mathbf{a}_{3}$ | = | $-x_{9}a\,\hat{\mathbf{x}}-z_{9}c\,\hat{\mathbf{z}}$ | (4$i$) | Nb IV |
| $\mathbf{B}_{22}$ | = | $\left(\frac{1}{2}+x_{9}\right)\,\mathbf{a}_{1}-z_{9}\,\mathbf{a}_{3}$ | = | $\left(\frac{1}{2}+x_{9}\right)a\,\hat{\mathbf{x}}-z_{9}c\,\hat{\mathbf{z}}$ | (4$i$) | Nb IV |
| $\mathbf{B}_{23}$ | = | $x_{10}\,\mathbf{a}_{1}+z_{10}\,\mathbf{a}_{3}$ | = | $x_{10}a\,\hat{\mathbf{x}}+z_{10}c\,\hat{\mathbf{z}}$ | (4$i$) | O IV |
| $\mathbf{B}_{24}$ | = | $\left(\frac{1}{2}-x_{10}\right)\,\mathbf{a}_{1}+z_{10}\,\mathbf{a}_{3}$ | = | $\left(\frac{1}{2}-x_{10}\right)a\,\hat{\mathbf{x}}+z_{10}c\,\hat{\mathbf{z}}$ | (4$i$) | O IV |
| $\mathbf{B}_{25}$ | = | $-x_{10}\,\mathbf{a}_{1}-z_{10}\,\mathbf{a}_{3}$ | = | $-x_{10}a\,\hat{\mathbf{x}}-z_{10}c\,\hat{\mathbf{z}}$ | (4$i$) | O IV |
| $\mathbf{B}_{26}$ | = | $\left(\frac{1}{2}+x_{10}\right)\,\mathbf{a}_{1}-z_{10}\,\mathbf{a}_{3}$ | = | $\left(\frac{1}{2}+x_{10}\right)a\,\hat{\mathbf{x}}-z_{10}c\,\hat{\mathbf{z}}$ | (4$i$) | O IV |
| $\mathbf{B}_{27}$ | = | $x_{11}\,\mathbf{a}_{1}+z_{11}\,\mathbf{a}_{3}$ | = | $x_{11}a\,\hat{\mathbf{x}}+z_{11}c\,\hat{\mathbf{z}}$ | (4$i$) | O V |
| $\mathbf{B}_{28}$ | = | $\left(\frac{1}{2}-x_{11}\right)\,\mathbf{a}_{1}+z_{11}\,\mathbf{a}_{3}$ | = | $\left(\frac{1}{2}-x_{11}\right)a\,\hat{\mathbf{x}}+z_{11}c\,\hat{\mathbf{z}}$ | (4$i$) | O V |
| $\mathbf{B}_{29}$ | = | $-x_{11}\,\mathbf{a}_{1}-z_{11}\,\mathbf{a}_{3}$ | = | $-x_{11}a\,\hat{\mathbf{x}}-z_{11}c\,\hat{\mathbf{z}}$ | (4$i$) | O V |
| $\mathbf{B}_{30}$ | = | $\left(\frac{1}{2}+x_{11}\right)\,\mathbf{a}_{1}-z_{11}\,\mathbf{a}_{3}$ | = | $\left(\frac{1}{2}+x_{11}\right)a\,\hat{\mathbf{x}}-z_{11}c\,\hat{\mathbf{z}}$ | (4$i$) | O V |
| $\mathbf{B}_{31}$ | = | $x_{12}\,\mathbf{a}_{1}+z_{12}\,\mathbf{a}_{3}$ | = | $x_{12}a\,\hat{\mathbf{x}}+z_{12}c\,\hat{\mathbf{z}}$ | (4$i$) | O VI |
| $\mathbf{B}_{32}$ | = | $\left(\frac{1}{2}-x_{12}\right)\,\mathbf{a}_{1}+z_{12}\,\mathbf{a}_{3}$ | = | $\left(\frac{1}{2}-x_{12}\right)a\,\hat{\mathbf{x}}+z_{12}c\,\hat{\mathbf{z}}$ | (4$i$) | O VI |
| $\mathbf{B}_{33}$ | = | $-x_{12}\,\mathbf{a}_{1}-z_{12}\,\mathbf{a}_{3}$ | = | $-x_{12}a\,\hat{\mathbf{x}}-z_{12}c\,\hat{\mathbf{z}}$ | (4$i$) | O VI |
| $\mathbf{B}_{34}$ | = | $\left(\frac{1}{2}+x_{12}\right)\,\mathbf{a}_{1}-z_{12}\,\mathbf{a}_{3}$ | = | $\left(\frac{1}{2}+x_{12}\right)a\,\hat{\mathbf{x}}-z_{12}c\,\hat{\mathbf{z}}$ | (4$i$) | O VI |
| $\mathbf{B}_{35}$ | = | $x_{13}\,\mathbf{a}_{1}+z_{13}\,\mathbf{a}_{3}$ | = | $x_{13}a\,\hat{\mathbf{x}}+z_{13}c\,\hat{\mathbf{z}}$ | (4$i$) | O VII |
| $\mathbf{B}_{36}$ | = | $\left(\frac{1}{2} - x_{13}\right)\, \mathbf{a}_1 + z_{13}\, \mathbf{a}_3$ | = | $\left(\frac{1}{2} - x_{13}\right) a\, \hat{\mathbf{x}} + z_{13} c\, \hat{\mathbf{z}}$ | (4$i$) | O VII |
| $\mathbf{B}_{37}$ | = | $-x_{13}\, \mathbf{a}_1 - z_{13}\, \mathbf{a}_3$ | = | $-x_{13} a\, \hat{\mathbf{x}} - z_{13} c\, \hat{\mathbf{z}}$ | (4$i$) | O VII |
| $\mathbf{B}_{38}$ | = | $\left(\frac{1}{2} + x_{13}\right)\, \mathbf{a}_1 - z_{13}\, \mathbf{a}_3$ | = | $\left(\frac{1}{2} + x_{13}\right) a\, \hat{\mathbf{x}} - z_{13} c\, \hat{\mathbf{z}}$ | (4$i$) | O VII |
| $\mathbf{B}_{39}$ | = | $x_{14}\, \mathbf{a}_1 + \frac{1}{2}\, \mathbf{a}_2 + z_{14}\, \mathbf{a}_3$ | = | $x_{14} a\, \hat{\mathbf{x}} + \frac{1}{2} b\, \hat{\mathbf{y}} + z_{14} c\, \hat{\mathbf{z}}$ | (4$j$) | O VIII |
| $\mathbf{B}_{40}$ | = | $\left(\frac{1}{2} - x_{14}\right)\, \mathbf{a}_1 + \frac{1}{2}\, \mathbf{a}_2 + z_{14}\, \mathbf{a}_3$ | = | $\left(\frac{1}{2} - x_{14}\right) a\, \hat{\mathbf{x}} + \frac{1}{2} b\, \hat{\mathbf{y}} + z_{14} c\, \hat{\mathbf{z}}$ | (4$j$) | O VIII |
| $\mathbf{B}_{41}$ | = | $-x_{14}\, \mathbf{a}_1 + \frac{1}{2}\, \mathbf{a}_2 - z_{14}\, \mathbf{a}_3$ | = | $-x_{14} a\, \hat{\mathbf{x}} + \frac{1}{2} b\, \hat{\mathbf{y}} - z_{14} c\, \hat{\mathbf{z}}$ | (4$j$) | O VIII |
| $\mathbf{B}_{42}$ | = | $\left(\frac{1}{2} + x_{14}\right)\, \mathbf{a}_1 + \frac{1}{2}\, \mathbf{a}_2 - z_{14}\, \mathbf{a}_3$ | = | $\left(\frac{1}{2} + x_{14}\right) a\, \hat{\mathbf{x}} + \frac{1}{2} b\, \hat{\mathbf{y}} - z_{14} c\, \hat{\mathbf{z}}$ | (4$j$) | O VIII |
| $\mathbf{B}_{43}$ | = | $x_{15}\, \mathbf{a}_1 + \frac{1}{2}\, \mathbf{a}_2 + z_{15}\, \mathbf{a}_3$ | = | $x_{15} a\, \hat{\mathbf{x}} + \frac{1}{2} b\, \hat{\mathbf{y}} + z_{15} c\, \hat{\mathbf{z}}$ | (4$j$) | O IX |
| $\mathbf{B}_{44}$ | = | $\left(\frac{1}{2} - x_{15}\right)\, \mathbf{a}_1 + \frac{1}{2}\, \mathbf{a}_2 + z_{15}\, \mathbf{a}_3$ | = | $\left(\frac{1}{2} - x_{15}\right) a\, \hat{\mathbf{x}} + \frac{1}{2} b\, \hat{\mathbf{y}} + z_{15} c\, \hat{\mathbf{z}}$ | (4$j$) | O IX |
| $\mathbf{B}_{45}$ | = | $-x_{15}\, \mathbf{a}_1 + \frac{1}{2}\, \mathbf{a}_2 - z_{15}\, \mathbf{a}_3$ | = | $-x_{15} a\, \hat{\mathbf{x}} + \frac{1}{2} b\, \hat{\mathbf{y}} - z_{15} c\, \hat{\mathbf{z}}$ | (4$j$) | O IX |
| $\mathbf{B}_{46}$ | = | $\left(\frac{1}{2} + x_{15}\right)\, \mathbf{a}_1 + \frac{1}{2}\, \mathbf{a}_2 - z_{15}\, \mathbf{a}_3$ | = | $\left(\frac{1}{2} + x_{15}\right) a\, \hat{\mathbf{x}} + \frac{1}{2} b\, \hat{\mathbf{y}} - z_{15} c\, \hat{\mathbf{z}}$ | (4$j$) | O IX |

**Geometry files:**

- CIF: pp. 1608
- POSCAR: pp. 1608

# Carnallite [$Mg(H_2O)_6KCl_3$] Structure: A3B12CDE6_oP276_52_d4e_18e_ce_de_2d8e

http://aflow.org/prototype-encyclopedia/A3B12CDE6_oP276_52_d4e_18e_ce_de_2d8e

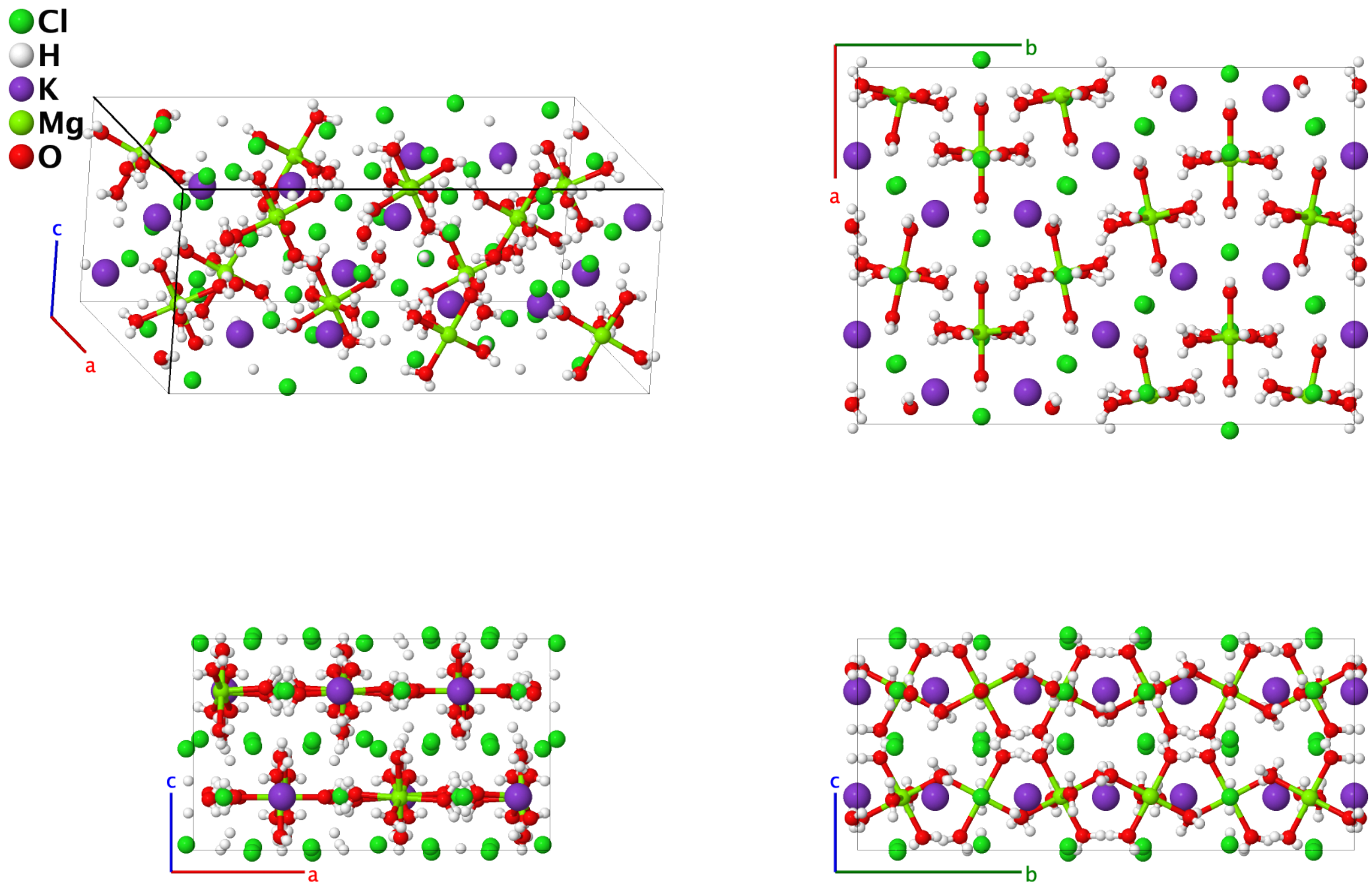

| | | |
|---|---|---|
| **Prototype** | : | $Cl_3H_{12}KMgO_6$ |
| **AFLOW prototype label** | : | A3B12CDE6_oP276_52_d4e_18e_ce_de_2d8e |
| ***Strukturbericht* designation** | : | None |
| **Pearson symbol** | : | oP276 |
| **Space group number** | : | 52 |
| **Space group symbol** | : | *Pnna* |
| **AFLOW prototype command** | : | `aflow --proto=A3B12CDE6_oP276_52_d4e_18e_ce_de_2d8e` `--params=`$a, b/a, c/a, z_1, x_2, x_3, x_4, x_5, x_6, y_6, z_6, x_7, y_7, z_7, x_8, y_8, z_8, x_9, y_9, z_9, x_{10}, y_{10}, z_{10}, x_{11}, y_{11}, z_{11}, x_{12}, y_{12}, z_{12}, x_{13}, y_{13}, z_{13}, x_{14}, y_{14}, z_{14}, x_{15}, y_{15}, z_{15}, x_{16}, y_{16}, z_{16}, x_{17}, y_{17}, z_{17}, x_{18}, y_{18}, z_{18}, x_{19}, y_{19}, z_{19}, x_{20}, y_{20}, z_{20}, x_{21}, y_{21}, z_{21}, x_{22}, y_{22}, z_{22}, x_{23}, y_{23}, z_{23}, x_{24}, y_{24}, z_{24}, x_{25}, y_{25}, z_{25}, x_{26}, y_{26}, z_{26}, x_{27}, y_{27}, z_{27}, x_{28}, y_{28}, z_{28}, x_{29}, y_{29}, z_{29}, x_{30}, y_{30}, z_{30}, x_{31}, y_{31}, z_{31}, x_{32}, y_{32}, z_{32}, x_{33}, y_{33}, z_{33}, x_{34}, y_{34}, z_{34}, x_{35}, y_{35}, z_{35}, x_{36}, y_{36}, z_{36}, x_{37}, y_{37}, z_{37}$ |

### Other compounds with this structure

- $Mg(H_2O)_6K(Cl,Br)_3$

- (Andreß, 1939) determined the crystal structure of brom-carnallite, $Mg(H_2O)_6K(Cl,Br)_3$, finding that it was in space group $P4/n$ #85. (Herrmann, 1939) gave this structure the *Strukturbericht* designation $E2_6$. Later, (Schlemper, 1985), determined that the true space group was *Pnna* #52, and were able to determine the positions of the hydrogen atoms in the water molecules. We present this structure here. For the original structure, see the $E2_6$ page.

**Simple Orthorhombic primitive vectors:**

$$\mathbf{a}_1 = a\,\hat{\mathbf{x}}$$
$$\mathbf{a}_2 = b\,\hat{\mathbf{y}}$$
$$\mathbf{a}_3 = c\,\hat{\mathbf{z}}$$

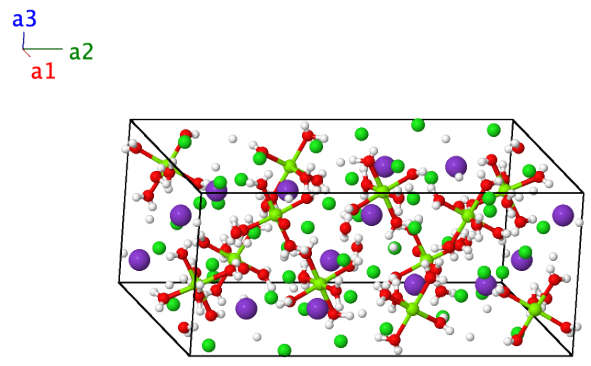

**Basis vectors:**

| | | Lattice Coordinates | | Cartesian Coordinates | Wyckoff Position | Atom Type |
|---|---|---|---|---|---|---|
| $\mathbf{B}_1$ | = | $\frac{1}{4}\,\mathbf{a}_1 + z_1\,\mathbf{a}_3$ | = | $\frac{1}{4}a\,\hat{\mathbf{x}} + z_1 c\,\hat{\mathbf{z}}$ | (4c) | K I |
| $\mathbf{B}_2$ | = | $\frac{1}{4}\,\mathbf{a}_1 + \frac{1}{2}\,\mathbf{a}_2 + \left(\frac{1}{2} - z_1\right)\,\mathbf{a}_3$ | = | $\frac{1}{4}a\,\hat{\mathbf{x}} + \frac{1}{2}b\,\hat{\mathbf{y}} + \left(\frac{1}{2} - z_1\right)c\,\hat{\mathbf{z}}$ | (4c) | K I |
| $\mathbf{B}_3$ | = | $\frac{3}{4}\,\mathbf{a}_1 - z_1\,\mathbf{a}_3$ | = | $\frac{3}{4}a\,\hat{\mathbf{x}} - z_1 c\,\hat{\mathbf{z}}$ | (4c) | K I |
| $\mathbf{B}_4$ | = | $\frac{3}{4}\,\mathbf{a}_1 + \frac{1}{2}\,\mathbf{a}_2 + \left(\frac{1}{2} + z_1\right)\,\mathbf{a}_3$ | = | $\frac{3}{4}a\,\hat{\mathbf{x}} + \frac{1}{2}b\,\hat{\mathbf{y}} + \left(\frac{1}{2} + z_1\right)c\,\hat{\mathbf{z}}$ | (4c) | K I |
| $\mathbf{B}_5$ | = | $x_2\,\mathbf{a}_1 + \frac{1}{4}\,\mathbf{a}_2 + \frac{1}{4}\,\mathbf{a}_3$ | = | $x_2 a\,\hat{\mathbf{x}} + \frac{1}{4}b\,\hat{\mathbf{y}} + \frac{1}{4}c\,\hat{\mathbf{z}}$ | (4d) | Cl I |
| $\mathbf{B}_6$ | = | $\left(\frac{1}{2} - x_2\right)\,\mathbf{a}_1 + \frac{3}{4}\,\mathbf{a}_2 + \frac{1}{4}\,\mathbf{a}_3$ | = | $\left(\frac{1}{2} - x_2\right)a\,\hat{\mathbf{x}} + \frac{3}{4}b\,\hat{\mathbf{y}} + \frac{1}{4}c\,\hat{\mathbf{z}}$ | (4d) | Cl I |
| $\mathbf{B}_7$ | = | $-x_2\,\mathbf{a}_1 + \frac{3}{4}\,\mathbf{a}_2 + \frac{3}{4}\,\mathbf{a}_3$ | = | $-x_2 a\,\hat{\mathbf{x}} + \frac{3}{4}b\,\hat{\mathbf{y}} + \frac{3}{4}c\,\hat{\mathbf{z}}$ | (4d) | Cl I |
| $\mathbf{B}_8$ | = | $\left(\frac{1}{2} + x_2\right)\,\mathbf{a}_1 + \frac{1}{4}\,\mathbf{a}_2 + \frac{3}{4}\,\mathbf{a}_3$ | = | $\left(\frac{1}{2} + x_2\right)a\,\hat{\mathbf{x}} + \frac{1}{4}b\,\hat{\mathbf{y}} + \frac{3}{4}c\,\hat{\mathbf{z}}$ | (4d) | Cl I |
| $\mathbf{B}_9$ | = | $x_3\,\mathbf{a}_1 + \frac{1}{4}\,\mathbf{a}_2 + \frac{1}{4}\,\mathbf{a}_3$ | = | $x_3 a\,\hat{\mathbf{x}} + \frac{1}{4}b\,\hat{\mathbf{y}} + \frac{1}{4}c\,\hat{\mathbf{z}}$ | (4d) | Mg I |
| $\mathbf{B}_{10}$ | = | $\left(\frac{1}{2} - x_3\right)\,\mathbf{a}_1 + \frac{3}{4}\,\mathbf{a}_2 + \frac{1}{4}\,\mathbf{a}_3$ | = | $\left(\frac{1}{2} - x_3\right)a\,\hat{\mathbf{x}} + \frac{3}{4}b\,\hat{\mathbf{y}} + \frac{1}{4}c\,\hat{\mathbf{z}}$ | (4d) | Mg I |
| $\mathbf{B}_{11}$ | = | $-x_3\,\mathbf{a}_1 + \frac{3}{4}\,\mathbf{a}_2 + \frac{3}{4}\,\mathbf{a}_3$ | = | $-x_3 a\,\hat{\mathbf{x}} + \frac{3}{4}b\,\hat{\mathbf{y}} + \frac{3}{4}c\,\hat{\mathbf{z}}$ | (4d) | Mg I |
| $\mathbf{B}_{12}$ | = | $\left(\frac{1}{2} + x_3\right)\,\mathbf{a}_1 + \frac{1}{4}\,\mathbf{a}_2 + \frac{3}{4}\,\mathbf{a}_3$ | = | $\left(\frac{1}{2} + x_3\right)a\,\hat{\mathbf{x}} + \frac{1}{4}b\,\hat{\mathbf{y}} + \frac{3}{4}c\,\hat{\mathbf{z}}$ | (4d) | Mg I |
| $\mathbf{B}_{13}$ | = | $x_4\,\mathbf{a}_1 + \frac{1}{4}\,\mathbf{a}_2 + \frac{1}{4}\,\mathbf{a}_3$ | = | $x_4 a\,\hat{\mathbf{x}} + \frac{1}{4}b\,\hat{\mathbf{y}} + \frac{1}{4}c\,\hat{\mathbf{z}}$ | (4d) | O I |
| $\mathbf{B}_{14}$ | = | $\left(\frac{1}{2} - x_4\right)\,\mathbf{a}_1 + \frac{3}{4}\,\mathbf{a}_2 + \frac{1}{4}\,\mathbf{a}_3$ | = | $\left(\frac{1}{2} - x_4\right)a\,\hat{\mathbf{x}} + \frac{3}{4}b\,\hat{\mathbf{y}} + \frac{1}{4}c\,\hat{\mathbf{z}}$ | (4d) | O I |
| $\mathbf{B}_{15}$ | = | $-x_4\,\mathbf{a}_1 + \frac{3}{4}\,\mathbf{a}_2 + \frac{3}{4}\,\mathbf{a}_3$ | = | $-x_4 a\,\hat{\mathbf{x}} + \frac{3}{4}b\,\hat{\mathbf{y}} + \frac{3}{4}c\,\hat{\mathbf{z}}$ | (4d) | O I |
| $\mathbf{B}_{16}$ | = | $\left(\frac{1}{2} + x_4\right)\,\mathbf{a}_1 + \frac{1}{4}\,\mathbf{a}_2 + \frac{3}{4}\,\mathbf{a}_3$ | = | $\left(\frac{1}{2} + x_4\right)a\,\hat{\mathbf{x}} + \frac{1}{4}b\,\hat{\mathbf{y}} + \frac{3}{4}c\,\hat{\mathbf{z}}$ | (4d) | O I |
| $\mathbf{B}_{17}$ | = | $x_5\,\mathbf{a}_1 + \frac{1}{4}\,\mathbf{a}_2 + \frac{1}{4}\,\mathbf{a}_3$ | = | $x_5 a\,\hat{\mathbf{x}} + \frac{1}{4}b\,\hat{\mathbf{y}} + \frac{1}{4}c\,\hat{\mathbf{z}}$ | (4d) | O II |
| $\mathbf{B}_{18}$ | = | $\left(\frac{1}{2} - x_5\right)\,\mathbf{a}_1 + \frac{3}{4}\,\mathbf{a}_2 + \frac{1}{4}\,\mathbf{a}_3$ | = | $\left(\frac{1}{2} - x_5\right)a\,\hat{\mathbf{x}} + \frac{3}{4}b\,\hat{\mathbf{y}} + \frac{1}{4}c\,\hat{\mathbf{z}}$ | (4d) | O II |
| $\mathbf{B}_{19}$ | = | $-x_5\,\mathbf{a}_1 + \frac{3}{4}\,\mathbf{a}_2 + \frac{3}{4}\,\mathbf{a}_3$ | = | $-x_5 a\,\hat{\mathbf{x}} + \frac{3}{4}b\,\hat{\mathbf{y}} + \frac{3}{4}c\,\hat{\mathbf{z}}$ | (4d) | O II |
| $\mathbf{B}_{20}$ | = | $\left(\frac{1}{2} + x_5\right)\,\mathbf{a}_1 + \frac{1}{4}\,\mathbf{a}_2 + \frac{3}{4}\,\mathbf{a}_3$ | = | $\left(\frac{1}{2} + x_5\right)a\,\hat{\mathbf{x}} + \frac{1}{4}b\,\hat{\mathbf{y}} + \frac{3}{4}c\,\hat{\mathbf{z}}$ | (4d) | O II |
| $\mathbf{B}_{21}$ | = | $x_6\,\mathbf{a}_1 + y_6\,\mathbf{a}_2 + z_6\,\mathbf{a}_3$ | = | $x_6 a\,\hat{\mathbf{x}} + y_6 b\,\hat{\mathbf{y}} + z_6 c\,\hat{\mathbf{z}}$ | (8e) | Cl II |
| $\mathbf{B}_{22}$ | = | $\left(\frac{1}{2} - x_6\right)\,\mathbf{a}_1 - y_6\,\mathbf{a}_2 + z_6\,\mathbf{a}_3$ | = | $\left(\frac{1}{2} - x_6\right)a\,\hat{\mathbf{x}} - y_6 b\,\hat{\mathbf{y}} + z_6 c\,\hat{\mathbf{z}}$ | (8e) | Cl II |
| $\mathbf{B}_{23}$ | = | $\left(\frac{1}{2} - x_6\right)\,\mathbf{a}_1 + \left(\frac{1}{2} + y_6\right)\,\mathbf{a}_2 + \left(\frac{1}{2} - z_6\right)\,\mathbf{a}_3$ | = | $\left(\frac{1}{2} - x_6\right)a\,\hat{\mathbf{x}} + \left(\frac{1}{2} + y_6\right)b\,\hat{\mathbf{y}} + \left(\frac{1}{2} - z_6\right)c\,\hat{\mathbf{z}}$ | (8e) | Cl II |
| $\mathbf{B}_{24}$ | = | $x_6\,\mathbf{a}_1 + \left(\frac{1}{2} - y_6\right)\,\mathbf{a}_2 + \left(\frac{1}{2} - z_6\right)\,\mathbf{a}_3$ | = | $x_6 a\,\hat{\mathbf{x}} + \left(\frac{1}{2} - y_6\right)b\,\hat{\mathbf{y}} + \left(\frac{1}{2} - z_6\right)c\,\hat{\mathbf{z}}$ | (8e) | Cl II |
| $\mathbf{B}_{25}$ | = | $-x_6\,\mathbf{a}_1 - y_6\,\mathbf{a}_2 - z_6\,\mathbf{a}_3$ | = | $-x_6 a\,\hat{\mathbf{x}} - y_6 b\,\hat{\mathbf{y}} - z_6 c\,\hat{\mathbf{z}}$ | (8e) | Cl II |
| $\mathbf{B}_{26}$ | = | $\left(\frac{1}{2} + x_6\right)\,\mathbf{a}_1 + y_6\,\mathbf{a}_2 - z_6\,\mathbf{a}_3$ | = | $\left(\frac{1}{2} + x_6\right)a\,\hat{\mathbf{x}} + y_6 b\,\hat{\mathbf{y}} - z_6 c\,\hat{\mathbf{z}}$ | (8e) | Cl II |
| $\mathbf{B}_{27}$ | = | $\left(\frac{1}{2}+x_6\right)\mathbf{a}_1+\left(\frac{1}{2}-y_6\right)\mathbf{a}_2+\left(\frac{1}{2}+z_6\right)\mathbf{a}_3$ | = | $\left(\frac{1}{2}+x_6\right)a\,\hat{\mathbf{x}}+\left(\frac{1}{2}-y_6\right)b\,\hat{\mathbf{y}}+\left(\frac{1}{2}+z_6\right)c\,\hat{\mathbf{z}}$ | (8*e*) | Cl II |
| $\mathbf{B}_{28}$ | = | $-x_6\,\mathbf{a}_1+\left(\frac{1}{2}+y_6\right)\mathbf{a}_2+\left(\frac{1}{2}+z_6\right)\mathbf{a}_3$ | = | $-x_6a\,\hat{\mathbf{x}}+\left(\frac{1}{2}+y_6\right)b\,\hat{\mathbf{y}}+\left(\frac{1}{2}+z_6\right)c\,\hat{\mathbf{z}}$ | (8*e*) | Cl II |
| $\mathbf{B}_{29}$ | = | $x_7\,\mathbf{a}_1+y_7\,\mathbf{a}_2+z_7\,\mathbf{a}_3$ | = | $x_7a\,\hat{\mathbf{x}}+y_7b\,\hat{\mathbf{y}}+z_7c\,\hat{\mathbf{z}}$ | (8*e*) | Cl III |
| $\mathbf{B}_{30}$ | = | $\left(\frac{1}{2}-x_7\right)\mathbf{a}_1-y_7\,\mathbf{a}_2+z_7\,\mathbf{a}_3$ | = | $\left(\frac{1}{2}-x_7\right)a\,\hat{\mathbf{x}}-y_7b\,\hat{\mathbf{y}}+z_7c\,\hat{\mathbf{z}}$ | (8*e*) | Cl III |
| $\mathbf{B}_{31}$ | = | $\left(\frac{1}{2}-x_7\right)\mathbf{a}_1+\left(\frac{1}{2}+y_7\right)\mathbf{a}_2+\left(\frac{1}{2}-z_7\right)\mathbf{a}_3$ | = | $\left(\frac{1}{2}-x_7\right)a\,\hat{\mathbf{x}}+\left(\frac{1}{2}+y_7\right)b\,\hat{\mathbf{y}}+\left(\frac{1}{2}-z_7\right)c\,\hat{\mathbf{z}}$ | (8*e*) | Cl III |
| $\mathbf{B}_{32}$ | = | $x_7\,\mathbf{a}_1+\left(\frac{1}{2}-y_7\right)\mathbf{a}_2+\left(\frac{1}{2}-z_7\right)\mathbf{a}_3$ | = | $x_7a\,\hat{\mathbf{x}}+\left(\frac{1}{2}-y_7\right)b\,\hat{\mathbf{y}}+\left(\frac{1}{2}-z_7\right)c\,\hat{\mathbf{z}}$ | (8*e*) | Cl III |
| $\mathbf{B}_{33}$ | = | $-x_7\,\mathbf{a}_1-y_7\,\mathbf{a}_2-z_7\,\mathbf{a}_3$ | = | $-x_7a\,\hat{\mathbf{x}}-y_7b\,\hat{\mathbf{y}}-z_7c\,\hat{\mathbf{z}}$ | (8*e*) | Cl III |
| $\mathbf{B}_{34}$ | = | $\left(\frac{1}{2}+x_7\right)\mathbf{a}_1+y_7\,\mathbf{a}_2-z_7\,\mathbf{a}_3$ | = | $\left(\frac{1}{2}+x_7\right)a\,\hat{\mathbf{x}}+y_7b\,\hat{\mathbf{y}}-z_7c\,\hat{\mathbf{z}}$ | (8*e*) | Cl III |
| $\mathbf{B}_{35}$ | = | $\left(\frac{1}{2}+x_7\right)\mathbf{a}_1+\left(\frac{1}{2}-y_7\right)\mathbf{a}_2+\left(\frac{1}{2}+z_7\right)\mathbf{a}_3$ | = | $\left(\frac{1}{2}+x_7\right)a\,\hat{\mathbf{x}}+\left(\frac{1}{2}-y_7\right)b\,\hat{\mathbf{y}}+\left(\frac{1}{2}+z_7\right)c\,\hat{\mathbf{z}}$ | (8*e*) | Cl III |
| $\mathbf{B}_{36}$ | = | $-x_7\,\mathbf{a}_1+\left(\frac{1}{2}+y_7\right)\mathbf{a}_2+\left(\frac{1}{2}+z_7\right)\mathbf{a}_3$ | = | $-x_7a\,\hat{\mathbf{x}}+\left(\frac{1}{2}+y_7\right)b\,\hat{\mathbf{y}}+\left(\frac{1}{2}+z_7\right)c\,\hat{\mathbf{z}}$ | (8*e*) | Cl III |
| $\mathbf{B}_{37}$ | = | $x_8\,\mathbf{a}_1+y_8\,\mathbf{a}_2+z_8\,\mathbf{a}_3$ | = | $x_8a\,\hat{\mathbf{x}}+y_8b\,\hat{\mathbf{y}}+z_8c\,\hat{\mathbf{z}}$ | (8*e*) | Cl IV |
| $\mathbf{B}_{38}$ | = | $\left(\frac{1}{2}-x_8\right)\mathbf{a}_1-y_8\,\mathbf{a}_2+z_8\,\mathbf{a}_3$ | = | $\left(\frac{1}{2}-x_8\right)a\,\hat{\mathbf{x}}-y_8b\,\hat{\mathbf{y}}+z_8c\,\hat{\mathbf{z}}$ | (8*e*) | Cl IV |
| $\mathbf{B}_{39}$ | = | $\left(\frac{1}{2}-x_8\right)\mathbf{a}_1+\left(\frac{1}{2}+y_8\right)\mathbf{a}_2+\left(\frac{1}{2}-z_8\right)\mathbf{a}_3$ | = | $\left(\frac{1}{2}-x_8\right)a\,\hat{\mathbf{x}}+\left(\frac{1}{2}+y_8\right)b\,\hat{\mathbf{y}}+\left(\frac{1}{2}-z_8\right)c\,\hat{\mathbf{z}}$ | (8*e*) | Cl IV |
| $\mathbf{B}_{40}$ | = | $x_8\,\mathbf{a}_1+\left(\frac{1}{2}-y_8\right)\mathbf{a}_2+\left(\frac{1}{2}-z_8\right)\mathbf{a}_3$ | = | $x_8a\,\hat{\mathbf{x}}+\left(\frac{1}{2}-y_8\right)b\,\hat{\mathbf{y}}+\left(\frac{1}{2}-z_8\right)c\,\hat{\mathbf{z}}$ | (8*e*) | Cl IV |
| $\mathbf{B}_{41}$ | = | $-x_8\,\mathbf{a}_1-y_8\,\mathbf{a}_2-z_8\,\mathbf{a}_3$ | = | $-x_8a\,\hat{\mathbf{x}}-y_8b\,\hat{\mathbf{y}}-z_8c\,\hat{\mathbf{z}}$ | (8*e*) | Cl IV |
| $\mathbf{B}_{42}$ | = | $\left(\frac{1}{2}+x_8\right)\mathbf{a}_1+y_8\,\mathbf{a}_2-z_8\,\mathbf{a}_3$ | = | $\left(\frac{1}{2}+x_8\right)a\,\hat{\mathbf{x}}+y_8b\,\hat{\mathbf{y}}-z_8c\,\hat{\mathbf{z}}$ | (8*e*) | Cl IV |
| $\mathbf{B}_{43}$ | = | $\left(\frac{1}{2}+x_8\right)\mathbf{a}_1+\left(\frac{1}{2}-y_8\right)\mathbf{a}_2+\left(\frac{1}{2}+z_8\right)\mathbf{a}_3$ | = | $\left(\frac{1}{2}+x_8\right)a\,\hat{\mathbf{x}}+\left(\frac{1}{2}-y_8\right)b\,\hat{\mathbf{y}}+\left(\frac{1}{2}+z_8\right)c\,\hat{\mathbf{z}}$ | (8*e*) | Cl IV |
| $\mathbf{B}_{44}$ | = | $-x_8\,\mathbf{a}_1+\left(\frac{1}{2}+y_8\right)\mathbf{a}_2+\left(\frac{1}{2}+z_8\right)\mathbf{a}_3$ | = | $-x_8a\,\hat{\mathbf{x}}+\left(\frac{1}{2}+y_8\right)b\,\hat{\mathbf{y}}+\left(\frac{1}{2}+z_8\right)c\,\hat{\mathbf{z}}$ | (8*e*) | Cl IV |
| $\mathbf{B}_{45}$ | = | $x_9\,\mathbf{a}_1+y_9\,\mathbf{a}_2+z_9\,\mathbf{a}_3$ | = | $x_9a\,\hat{\mathbf{x}}+y_9b\,\hat{\mathbf{y}}+z_9c\,\hat{\mathbf{z}}$ | (8*e*) | Cl V |
| $\mathbf{B}_{46}$ | = | $\left(\frac{1}{2}-x_9\right)\mathbf{a}_1-y_9\,\mathbf{a}_2+z_9\,\mathbf{a}_3$ | = | $\left(\frac{1}{2}-x_9\right)a\,\hat{\mathbf{x}}-y_9b\,\hat{\mathbf{y}}+z_9c\,\hat{\mathbf{z}}$ | (8*e*) | Cl V |
| $\mathbf{B}_{47}$ | = | $\left(\frac{1}{2}-x_9\right)\mathbf{a}_1+\left(\frac{1}{2}+y_9\right)\mathbf{a}_2+\left(\frac{1}{2}-z_9\right)\mathbf{a}_3$ | = | $\left(\frac{1}{2}-x_9\right)a\,\hat{\mathbf{x}}+\left(\frac{1}{2}+y_9\right)b\,\hat{\mathbf{y}}+\left(\frac{1}{2}-z_9\right)c\,\hat{\mathbf{z}}$ | (8*e*) | Cl V |
| $\mathbf{B}_{48}$ | = | $x_9\,\mathbf{a}_1+\left(\frac{1}{2}-y_9\right)\mathbf{a}_2+\left(\frac{1}{2}-z_9\right)\mathbf{a}_3$ | = | $x_9a\,\hat{\mathbf{x}}+\left(\frac{1}{2}-y_9\right)b\,\hat{\mathbf{y}}+\left(\frac{1}{2}-z_9\right)c\,\hat{\mathbf{z}}$ | (8*e*) | Cl V |
| $\mathbf{B}_{49}$ | = | $-x_9\,\mathbf{a}_1-y_9\,\mathbf{a}_2-z_9\,\mathbf{a}_3$ | = | $-x_9a\,\hat{\mathbf{x}}-y_9b\,\hat{\mathbf{y}}-z_9c\,\hat{\mathbf{z}}$ | (8*e*) | Cl V |
| $\mathbf{B}_{50}$ | = | $\left(\frac{1}{2}+x_9\right)\mathbf{a}_1+y_9\,\mathbf{a}_2-z_9\,\mathbf{a}_3$ | = | $\left(\frac{1}{2}+x_9\right)a\,\hat{\mathbf{x}}+y_9b\,\hat{\mathbf{y}}-z_9c\,\hat{\mathbf{z}}$ | (8*e*) | Cl V |
| $\mathbf{B}_{51}$ | = | $\left(\frac{1}{2}+x_9\right)\mathbf{a}_1+\left(\frac{1}{2}-y_9\right)\mathbf{a}_2+\left(\frac{1}{2}+z_9\right)\mathbf{a}_3$ | = | $\left(\frac{1}{2}+x_9\right)a\,\hat{\mathbf{x}}+\left(\frac{1}{2}-y_9\right)b\,\hat{\mathbf{y}}+\left(\frac{1}{2}+z_9\right)c\,\hat{\mathbf{z}}$ | (8*e*) | Cl V |
| $\mathbf{B}_{52}$ | = | $-x_9\,\mathbf{a}_1+\left(\frac{1}{2}+y_9\right)\mathbf{a}_2+\left(\frac{1}{2}+z_9\right)\mathbf{a}_3$ | = | $-x_9a\,\hat{\mathbf{x}}+\left(\frac{1}{2}+y_9\right)b\,\hat{\mathbf{y}}+\left(\frac{1}{2}+z_9\right)c\,\hat{\mathbf{z}}$ | (8*e*) | Cl V |
| $\mathbf{B}_{53}$ | = | $x_{10}\,\mathbf{a}_1+y_{10}\,\mathbf{a}_2+z_{10}\,\mathbf{a}_3$ | = | $x_{10}a\,\hat{\mathbf{x}}+y_{10}b\,\hat{\mathbf{y}}+z_{10}c\,\hat{\mathbf{z}}$ | (8*e*) | H I |
| $\mathbf{B}_{54}$ | = | $\left(\frac{1}{2}-x_{10}\right)\mathbf{a}_1-y_{10}\,\mathbf{a}_2+z_{10}\,\mathbf{a}_3$ | = | $\left(\frac{1}{2}-x_{10}\right)a\,\hat{\mathbf{x}}-y_{10}b\,\hat{\mathbf{y}}+z_{10}c\,\hat{\mathbf{z}}$ | (8*e*) | H I |
| $\mathbf{B}_{55}$ | = | $\left(\frac{1}{2}-x_{10}\right)\mathbf{a}_1+\left(\frac{1}{2}+y_{10}\right)\mathbf{a}_2+\left(\frac{1}{2}-z_{10}\right)\mathbf{a}_3$ | = | $\left(\frac{1}{2}-x_{10}\right)a\,\hat{\mathbf{x}}+\left(\frac{1}{2}+y_{10}\right)b\,\hat{\mathbf{y}}+\left(\frac{1}{2}-z_{10}\right)c\,\hat{\mathbf{z}}$ | (8*e*) | H I |
| $\mathbf{B}_{56}$ | = | $x_{10}\,\mathbf{a}_1+\left(\frac{1}{2}-y_{10}\right)\mathbf{a}_2+\left(\frac{1}{2}-z_{10}\right)\mathbf{a}_3$ | = | $x_{10}a\,\hat{\mathbf{x}}+\left(\frac{1}{2}-y_{10}\right)b\,\hat{\mathbf{y}}+\left(\frac{1}{2}-z_{10}\right)c\,\hat{\mathbf{z}}$ | (8*e*) | H I |
| $\mathbf{B}_{57}$ | = | $-x_{10}\,\mathbf{a}_1 - y_{10}\,\mathbf{a}_2 - z_{10}\,\mathbf{a}_3$ | = | $-x_{10}a\,\hat{\mathbf{x}} - y_{10}b\,\hat{\mathbf{y}} - z_{10}c\,\hat{\mathbf{z}}$ | (8*e*) | H I |
| $\mathbf{B}_{58}$ | = | $\left(\frac{1}{2} + x_{10}\right)\mathbf{a}_1 + y_{10}\,\mathbf{a}_2 - z_{10}\,\mathbf{a}_3$ | = | $\left(\frac{1}{2} + x_{10}\right)a\,\hat{\mathbf{x}} + y_{10}b\,\hat{\mathbf{y}} - z_{10}c\,\hat{\mathbf{z}}$ | (8*e*) | H I |
| $\mathbf{B}_{59}$ | = | $\left(\frac{1}{2} + x_{10}\right)\mathbf{a}_1 + \left(\frac{1}{2} - y_{10}\right)\mathbf{a}_2 + \left(\frac{1}{2} + z_{10}\right)\mathbf{a}_3$ | = | $\left(\frac{1}{2} + x_{10}\right)a\,\hat{\mathbf{x}} + \left(\frac{1}{2} - y_{10}\right)b\,\hat{\mathbf{y}} + \left(\frac{1}{2} + z_{10}\right)c\,\hat{\mathbf{z}}$ | (8*e*) | H I |
| $\mathbf{B}_{60}$ | = | $-x_{10}\,\mathbf{a}_1 + \left(\frac{1}{2} + y_{10}\right)\mathbf{a}_2 + \left(\frac{1}{2} + z_{10}\right)\mathbf{a}_3$ | = | $-x_{10}a\,\hat{\mathbf{x}} + \left(\frac{1}{2} + y_{10}\right)b\,\hat{\mathbf{y}} + \left(\frac{1}{2} + z_{10}\right)c\,\hat{\mathbf{z}}$ | (8*e*) | H I |
| $\mathbf{B}_{61}$ | = | $x_{11}\,\mathbf{a}_1 + y_{11}\,\mathbf{a}_2 + z_{11}\,\mathbf{a}_3$ | = | $x_{11}a\,\hat{\mathbf{x}} + y_{11}b\,\hat{\mathbf{y}} + z_{11}c\,\hat{\mathbf{z}}$ | (8*e*) | H II |
| $\mathbf{B}_{62}$ | = | $\left(\frac{1}{2} - x_{11}\right)\mathbf{a}_1 - y_{11}\,\mathbf{a}_2 + z_{11}\,\mathbf{a}_3$ | = | $\left(\frac{1}{2} - x_{11}\right)a\,\hat{\mathbf{x}} - y_{11}b\,\hat{\mathbf{y}} + z_{11}c\,\hat{\mathbf{z}}$ | (8*e*) | H II |
| $\mathbf{B}_{63}$ | = | $\left(\frac{1}{2} - x_{11}\right)\mathbf{a}_1 + \left(\frac{1}{2} + y_{11}\right)\mathbf{a}_2 + \left(\frac{1}{2} - z_{11}\right)\mathbf{a}_3$ | = | $\left(\frac{1}{2} - x_{11}\right)a\,\hat{\mathbf{x}} + \left(\frac{1}{2} + y_{11}\right)b\,\hat{\mathbf{y}} + \left(\frac{1}{2} - z_{11}\right)c\,\hat{\mathbf{z}}$ | (8*e*) | H II |
| $\mathbf{B}_{64}$ | = | $x_{11}\,\mathbf{a}_1 + \left(\frac{1}{2} - y_{11}\right)\mathbf{a}_2 + \left(\frac{1}{2} - z_{11}\right)\mathbf{a}_3$ | = | $x_{11}a\,\hat{\mathbf{x}} + \left(\frac{1}{2} - y_{11}\right)b\,\hat{\mathbf{y}} + \left(\frac{1}{2} - z_{11}\right)c\,\hat{\mathbf{z}}$ | (8*e*) | H II |
| $\mathbf{B}_{65}$ | = | $-x_{11}\,\mathbf{a}_1 - y_{11}\,\mathbf{a}_2 - z_{11}\,\mathbf{a}_3$ | = | $-x_{11}a\,\hat{\mathbf{x}} - y_{11}b\,\hat{\mathbf{y}} - z_{11}c\,\hat{\mathbf{z}}$ | (8*e*) | H II |
| $\mathbf{B}_{66}$ | = | $\left(\frac{1}{2} + x_{11}\right)\mathbf{a}_1 + y_{11}\,\mathbf{a}_2 - z_{11}\,\mathbf{a}_3$ | = | $\left(\frac{1}{2} + x_{11}\right)a\,\hat{\mathbf{x}} + y_{11}b\,\hat{\mathbf{y}} - z_{11}c\,\hat{\mathbf{z}}$ | (8*e*) | H II |
| $\mathbf{B}_{67}$ | = | $\left(\frac{1}{2} + x_{11}\right)\mathbf{a}_1 + \left(\frac{1}{2} - y_{11}\right)\mathbf{a}_2 + \left(\frac{1}{2} + z_{11}\right)\mathbf{a}_3$ | = | $\left(\frac{1}{2} + x_{11}\right)a\,\hat{\mathbf{x}} + \left(\frac{1}{2} - y_{11}\right)b\,\hat{\mathbf{y}} + \left(\frac{1}{2} + z_{11}\right)c\,\hat{\mathbf{z}}$ | (8*e*) | H II |
| $\mathbf{B}_{68}$ | = | $-x_{11}\,\mathbf{a}_1 + \left(\frac{1}{2} + y_{11}\right)\mathbf{a}_2 + \left(\frac{1}{2} + z_{11}\right)\mathbf{a}_3$ | = | $-x_{11}a\,\hat{\mathbf{x}} + \left(\frac{1}{2} + y_{11}\right)b\,\hat{\mathbf{y}} + \left(\frac{1}{2} + z_{11}\right)c\,\hat{\mathbf{z}}$ | (8*e*) | H II |
| $\mathbf{B}_{69}$ | = | $x_{12}\,\mathbf{a}_1 + y_{12}\,\mathbf{a}_2 + z_{12}\,\mathbf{a}_3$ | = | $x_{12}a\,\hat{\mathbf{x}} + y_{12}b\,\hat{\mathbf{y}} + z_{12}c\,\hat{\mathbf{z}}$ | (8*e*) | H III |
| $\mathbf{B}_{70}$ | = | $\left(\frac{1}{2} - x_{12}\right)\mathbf{a}_1 - y_{12}\,\mathbf{a}_2 + z_{12}\,\mathbf{a}_3$ | = | $\left(\frac{1}{2} - x_{12}\right)a\,\hat{\mathbf{x}} - y_{12}b\,\hat{\mathbf{y}} + z_{12}c\,\hat{\mathbf{z}}$ | (8*e*) | H III |
| $\mathbf{B}_{71}$ | = | $\left(\frac{1}{2} - x_{12}\right)\mathbf{a}_1 + \left(\frac{1}{2} + y_{12}\right)\mathbf{a}_2 + \left(\frac{1}{2} - z_{12}\right)\mathbf{a}_3$ | = | $\left(\frac{1}{2} - x_{12}\right)a\,\hat{\mathbf{x}} + \left(\frac{1}{2} + y_{12}\right)b\,\hat{\mathbf{y}} + \left(\frac{1}{2} - z_{12}\right)c\,\hat{\mathbf{z}}$ | (8*e*) | H III |
| $\mathbf{B}_{72}$ | = | $x_{12}\,\mathbf{a}_1 + \left(\frac{1}{2} - y_{12}\right)\mathbf{a}_2 + \left(\frac{1}{2} - z_{12}\right)\mathbf{a}_3$ | = | $x_{12}a\,\hat{\mathbf{x}} + \left(\frac{1}{2} - y_{12}\right)b\,\hat{\mathbf{y}} + \left(\frac{1}{2} - z_{12}\right)c\,\hat{\mathbf{z}}$ | (8*e*) | H III |
| $\mathbf{B}_{73}$ | = | $-x_{12}\,\mathbf{a}_1 - y_{12}\,\mathbf{a}_2 - z_{12}\,\mathbf{a}_3$ | = | $-x_{12}a\,\hat{\mathbf{x}} - y_{12}b\,\hat{\mathbf{y}} - z_{12}c\,\hat{\mathbf{z}}$ | (8*e*) | H III |
| $\mathbf{B}_{74}$ | = | $\left(\frac{1}{2} + x_{12}\right)\mathbf{a}_1 + y_{12}\,\mathbf{a}_2 - z_{12}\,\mathbf{a}_3$ | = | $\left(\frac{1}{2} + x_{12}\right)a\,\hat{\mathbf{x}} + y_{12}b\,\hat{\mathbf{y}} - z_{12}c\,\hat{\mathbf{z}}$ | (8*e*) | H III |
| $\mathbf{B}_{75}$ | = | $\left(\frac{1}{2} + x_{12}\right)\mathbf{a}_1 + \left(\frac{1}{2} - y_{12}\right)\mathbf{a}_2 + \left(\frac{1}{2} + z_{12}\right)\mathbf{a}_3$ | = | $\left(\frac{1}{2} + x_{12}\right)a\,\hat{\mathbf{x}} + \left(\frac{1}{2} - y_{12}\right)b\,\hat{\mathbf{y}} + \left(\frac{1}{2} + z_{12}\right)c\,\hat{\mathbf{z}}$ | (8*e*) | H III |
| $\mathbf{B}_{76}$ | = | $-x_{12}\,\mathbf{a}_1 + \left(\frac{1}{2} + y_{12}\right)\mathbf{a}_2 + \left(\frac{1}{2} + z_{12}\right)\mathbf{a}_3$ | = | $-x_{12}a\,\hat{\mathbf{x}} + \left(\frac{1}{2} + y_{12}\right)b\,\hat{\mathbf{y}} + \left(\frac{1}{2} + z_{12}\right)c\,\hat{\mathbf{z}}$ | (8*e*) | H III |
| $\mathbf{B}_{77}$ | = | $x_{13}\,\mathbf{a}_1 + y_{13}\,\mathbf{a}_2 + z_{13}\,\mathbf{a}_3$ | = | $x_{13}a\,\hat{\mathbf{x}} + y_{13}b\,\hat{\mathbf{y}} + z_{13}c\,\hat{\mathbf{z}}$ | (8*e*) | H IV |
| $\mathbf{B}_{78}$ | = | $\left(\frac{1}{2} - x_{13}\right)\mathbf{a}_1 - y_{13}\,\mathbf{a}_2 + z_{13}\,\mathbf{a}_3$ | = | $\left(\frac{1}{2} - x_{13}\right)a\,\hat{\mathbf{x}} - y_{13}b\,\hat{\mathbf{y}} + z_{13}c\,\hat{\mathbf{z}}$ | (8*e*) | H IV |
| $\mathbf{B}_{79}$ | = | $\left(\frac{1}{2} - x_{13}\right)\mathbf{a}_1 + \left(\frac{1}{2} + y_{13}\right)\mathbf{a}_2 + \left(\frac{1}{2} - z_{13}\right)\mathbf{a}_3$ | = | $\left(\frac{1}{2} - x_{13}\right)a\,\hat{\mathbf{x}} + \left(\frac{1}{2} + y_{13}\right)b\,\hat{\mathbf{y}} + \left(\frac{1}{2} - z_{13}\right)c\,\hat{\mathbf{z}}$ | (8*e*) | H IV |
| $\mathbf{B}_{80}$ | = | $x_{13}\,\mathbf{a}_1 + \left(\frac{1}{2} - y_{13}\right)\mathbf{a}_2 + \left(\frac{1}{2} - z_{13}\right)\mathbf{a}_3$ | = | $x_{13}a\,\hat{\mathbf{x}} + \left(\frac{1}{2} - y_{13}\right)b\,\hat{\mathbf{y}} + \left(\frac{1}{2} - z_{13}\right)c\,\hat{\mathbf{z}}$ | (8*e*) | H IV |
| $\mathbf{B}_{81}$ | = | $-x_{13}\,\mathbf{a}_1 - y_{13}\,\mathbf{a}_2 - z_{13}\,\mathbf{a}_3$ | = | $-x_{13}a\,\hat{\mathbf{x}} - y_{13}b\,\hat{\mathbf{y}} - z_{13}c\,\hat{\mathbf{z}}$ | (8*e*) | H IV |
| $\mathbf{B}_{82}$ | = | $\left(\frac{1}{2} + x_{13}\right)\mathbf{a}_1 + y_{13}\,\mathbf{a}_2 - z_{13}\,\mathbf{a}_3$ | = | $\left(\frac{1}{2} + x_{13}\right)a\,\hat{\mathbf{x}} + y_{13}b\,\hat{\mathbf{y}} - z_{13}c\,\hat{\mathbf{z}}$ | (8*e*) | H IV |
| $\mathbf{B}_{83}$ | = | $\left(\frac{1}{2} + x_{13}\right)\mathbf{a}_1 + \left(\frac{1}{2} - y_{13}\right)\mathbf{a}_2 + \left(\frac{1}{2} + z_{13}\right)\mathbf{a}_3$ | = | $\left(\frac{1}{2} + x_{13}\right)a\,\hat{\mathbf{x}} + \left(\frac{1}{2} - y_{13}\right)b\,\hat{\mathbf{y}} + \left(\frac{1}{2} + z_{13}\right)c\,\hat{\mathbf{z}}$ | (8*e*) | H IV |
| $\mathbf{B}_{84}$ | = | $-x_{13}\,\mathbf{a}_1 + \left(\frac{1}{2} + y_{13}\right)\mathbf{a}_2 + \left(\frac{1}{2} + z_{13}\right)\mathbf{a}_3$ | = | $-x_{13}a\,\hat{\mathbf{x}} + \left(\frac{1}{2} + y_{13}\right)b\,\hat{\mathbf{y}} + \left(\frac{1}{2} + z_{13}\right)c\,\hat{\mathbf{z}}$ | (8*e*) | H IV |
| $\mathbf{B}_{85}$ | = | $x_{14}\,\mathbf{a}_1 + y_{14}\,\mathbf{a}_2 + z_{14}\,\mathbf{a}_3$ | = | $x_{14}a\,\hat{\mathbf{x}} + y_{14}b\,\hat{\mathbf{y}} + z_{14}c\,\hat{\mathbf{z}}$ | (8$e$) | H V |
| $\mathbf{B}_{86}$ | = | $\left(\frac{1}{2} - x_{14}\right)\mathbf{a}_1 - y_{14}\,\mathbf{a}_2 + z_{14}\,\mathbf{a}_3$ | = | $\left(\frac{1}{2} - x_{14}\right)a\,\hat{\mathbf{x}} - y_{14}b\,\hat{\mathbf{y}} + z_{14}c\,\hat{\mathbf{z}}$ | (8$e$) | H V |
| $\mathbf{B}_{87}$ | = | $\left(\frac{1}{2} - x_{14}\right)\mathbf{a}_1 + \left(\frac{1}{2} + y_{14}\right)\mathbf{a}_2 + \left(\frac{1}{2} - z_{14}\right)\mathbf{a}_3$ | = | $\left(\frac{1}{2} - x_{14}\right)a\,\hat{\mathbf{x}} + \left(\frac{1}{2} + y_{14}\right)b\,\hat{\mathbf{y}} + \left(\frac{1}{2} - z_{14}\right)c\,\hat{\mathbf{z}}$ | (8$e$) | H V |
| $\mathbf{B}_{88}$ | = | $x_{14}\,\mathbf{a}_1 + \left(\frac{1}{2} - y_{14}\right)\mathbf{a}_2 + \left(\frac{1}{2} - z_{14}\right)\mathbf{a}_3$ | = | $x_{14}a\,\hat{\mathbf{x}} + \left(\frac{1}{2} - y_{14}\right)b\,\hat{\mathbf{y}} + \left(\frac{1}{2} - z_{14}\right)c\,\hat{\mathbf{z}}$ | (8$e$) | H V |
| $\mathbf{B}_{89}$ | = | $-x_{14}\,\mathbf{a}_1 - y_{14}\,\mathbf{a}_2 - z_{14}\,\mathbf{a}_3$ | = | $-x_{14}a\,\hat{\mathbf{x}} - y_{14}b\,\hat{\mathbf{y}} - z_{14}c\,\hat{\mathbf{z}}$ | (8$e$) | H V |
| $\mathbf{B}_{90}$ | = | $\left(\frac{1}{2} + x_{14}\right)\mathbf{a}_1 + y_{14}\,\mathbf{a}_2 - z_{14}\,\mathbf{a}_3$ | = | $\left(\frac{1}{2} + x_{14}\right)a\,\hat{\mathbf{x}} + y_{14}b\,\hat{\mathbf{y}} - z_{14}c\,\hat{\mathbf{z}}$ | (8$e$) | H V |
| $\mathbf{B}_{91}$ | = | $\left(\frac{1}{2} + x_{14}\right)\mathbf{a}_1 + \left(\frac{1}{2} - y_{14}\right)\mathbf{a}_2 + \left(\frac{1}{2} + z_{14}\right)\mathbf{a}_3$ | = | $\left(\frac{1}{2} + x_{14}\right)a\,\hat{\mathbf{x}} + \left(\frac{1}{2} - y_{14}\right)b\,\hat{\mathbf{y}} + \left(\frac{1}{2} + z_{14}\right)c\,\hat{\mathbf{z}}$ | (8$e$) | H V |
| $\mathbf{B}_{92}$ | = | $-x_{14}\,\mathbf{a}_1 + \left(\frac{1}{2} + y_{14}\right)\mathbf{a}_2 + \left(\frac{1}{2} + z_{14}\right)\mathbf{a}_3$ | = | $-x_{14}a\,\hat{\mathbf{x}} + \left(\frac{1}{2} + y_{14}\right)b\,\hat{\mathbf{y}} + \left(\frac{1}{2} + z_{14}\right)c\,\hat{\mathbf{z}}$ | (8$e$) | H V |
| $\mathbf{B}_{93}$ | = | $x_{15}\,\mathbf{a}_1 + y_{15}\,\mathbf{a}_2 + z_{15}\,\mathbf{a}_3$ | = | $x_{15}a\,\hat{\mathbf{x}} + y_{15}b\,\hat{\mathbf{y}} + z_{15}c\,\hat{\mathbf{z}}$ | (8$e$) | H VI |
| $\mathbf{B}_{94}$ | = | $\left(\frac{1}{2} - x_{15}\right)\mathbf{a}_1 - y_{15}\,\mathbf{a}_2 + z_{15}\,\mathbf{a}_3$ | = | $\left(\frac{1}{2} - x_{15}\right)a\,\hat{\mathbf{x}} - y_{15}b\,\hat{\mathbf{y}} + z_{15}c\,\hat{\mathbf{z}}$ | (8$e$) | H VI |
| $\mathbf{B}_{95}$ | = | $\left(\frac{1}{2} - x_{15}\right)\mathbf{a}_1 + \left(\frac{1}{2} + y_{15}\right)\mathbf{a}_2 + \left(\frac{1}{2} - z_{15}\right)\mathbf{a}_3$ | = | $\left(\frac{1}{2} - x_{15}\right)a\,\hat{\mathbf{x}} + \left(\frac{1}{2} + y_{15}\right)b\,\hat{\mathbf{y}} + \left(\frac{1}{2} - z_{15}\right)c\,\hat{\mathbf{z}}$ | (8$e$) | H VI |
| $\mathbf{B}_{96}$ | = | $x_{15}\,\mathbf{a}_1 + \left(\frac{1}{2} - y_{15}\right)\mathbf{a}_2 + \left(\frac{1}{2} - z_{15}\right)\mathbf{a}_3$ | = | $x_{15}a\,\hat{\mathbf{x}} + \left(\frac{1}{2} - y_{15}\right)b\,\hat{\mathbf{y}} + \left(\frac{1}{2} - z_{15}\right)c\,\hat{\mathbf{z}}$ | (8$e$) | H VI |
| $\mathbf{B}_{97}$ | = | $-x_{15}\,\mathbf{a}_1 - y_{15}\,\mathbf{a}_2 - z_{15}\,\mathbf{a}_3$ | = | $-x_{15}a\,\hat{\mathbf{x}} - y_{15}b\,\hat{\mathbf{y}} - z_{15}c\,\hat{\mathbf{z}}$ | (8$e$) | H VI |
| $\mathbf{B}_{98}$ | = | $\left(\frac{1}{2} + x_{15}\right)\mathbf{a}_1 + y_{15}\,\mathbf{a}_2 - z_{15}\,\mathbf{a}_3$ | = | $\left(\frac{1}{2} + x_{15}\right)a\,\hat{\mathbf{x}} + y_{15}b\,\hat{\mathbf{y}} - z_{15}c\,\hat{\mathbf{z}}$ | (8$e$) | H VI |
| $\mathbf{B}_{99}$ | = | $\left(\frac{1}{2} + x_{15}\right)\mathbf{a}_1 + \left(\frac{1}{2} - y_{15}\right)\mathbf{a}_2 + \left(\frac{1}{2} + z_{15}\right)\mathbf{a}_3$ | = | $\left(\frac{1}{2} + x_{15}\right)a\,\hat{\mathbf{x}} + \left(\frac{1}{2} - y_{15}\right)b\,\hat{\mathbf{y}} + \left(\frac{1}{2} + z_{15}\right)c\,\hat{\mathbf{z}}$ | (8$e$) | H VI |
| $\mathbf{B}_{100}$ | = | $-x_{15}\,\mathbf{a}_1 + \left(\frac{1}{2} + y_{15}\right)\mathbf{a}_2 + \left(\frac{1}{2} + z_{15}\right)\mathbf{a}_3$ | = | $-x_{15}a\,\hat{\mathbf{x}} + \left(\frac{1}{2} + y_{15}\right)b\,\hat{\mathbf{y}} + \left(\frac{1}{2} + z_{15}\right)c\,\hat{\mathbf{z}}$ | (8$e$) | H VI |
| $\mathbf{B}_{101}$ | = | $x_{16}\,\mathbf{a}_1 + y_{16}\,\mathbf{a}_2 + z_{16}\,\mathbf{a}_3$ | = | $x_{16}a\,\hat{\mathbf{x}} + y_{16}b\,\hat{\mathbf{y}} + z_{16}c\,\hat{\mathbf{z}}$ | (8$e$) | H VII |
| $\mathbf{B}_{102}$ | = | $\left(\frac{1}{2} - x_{16}\right)\mathbf{a}_1 - y_{16}\,\mathbf{a}_2 + z_{16}\,\mathbf{a}_3$ | = | $\left(\frac{1}{2} - x_{16}\right)a\,\hat{\mathbf{x}} - y_{16}b\,\hat{\mathbf{y}} + z_{16}c\,\hat{\mathbf{z}}$ | (8$e$) | H VII |
| $\mathbf{B}_{103}$ | = | $\left(\frac{1}{2} - x_{16}\right)\mathbf{a}_1 + \left(\frac{1}{2} + y_{16}\right)\mathbf{a}_2 + \left(\frac{1}{2} - z_{16}\right)\mathbf{a}_3$ | = | $\left(\frac{1}{2} - x_{16}\right)a\,\hat{\mathbf{x}} + \left(\frac{1}{2} + y_{16}\right)b\,\hat{\mathbf{y}} + \left(\frac{1}{2} - z_{16}\right)c\,\hat{\mathbf{z}}$ | (8$e$) | H VII |
| $\mathbf{B}_{104}$ | = | $x_{16}\,\mathbf{a}_1 + \left(\frac{1}{2} - y_{16}\right)\mathbf{a}_2 + \left(\frac{1}{2} - z_{16}\right)\mathbf{a}_3$ | = | $x_{16}a\,\hat{\mathbf{x}} + \left(\frac{1}{2} - y_{16}\right)b\,\hat{\mathbf{y}} + \left(\frac{1}{2} - z_{16}\right)c\,\hat{\mathbf{z}}$ | (8$e$) | H VII |
| $\mathbf{B}_{105}$ | = | $-x_{16}\,\mathbf{a}_1 - y_{16}\,\mathbf{a}_2 - z_{16}\,\mathbf{a}_3$ | = | $-x_{16}a\,\hat{\mathbf{x}} - y_{16}b\,\hat{\mathbf{y}} - z_{16}c\,\hat{\mathbf{z}}$ | (8$e$) | H VII |
| $\mathbf{B}_{106}$ | = | $\left(\frac{1}{2} + x_{16}\right)\mathbf{a}_1 + y_{16}\,\mathbf{a}_2 - z_{16}\,\mathbf{a}_3$ | = | $\left(\frac{1}{2} + x_{16}\right)a\,\hat{\mathbf{x}} + y_{16}b\,\hat{\mathbf{y}} - z_{16}c\,\hat{\mathbf{z}}$ | (8$e$) | H VII |
| $\mathbf{B}_{107}$ | = | $\left(\frac{1}{2} + x_{16}\right)\mathbf{a}_1 + \left(\frac{1}{2} - y_{16}\right)\mathbf{a}_2 + \left(\frac{1}{2} + z_{16}\right)\mathbf{a}_3$ | = | $\left(\frac{1}{2} + x_{16}\right)a\,\hat{\mathbf{x}} + \left(\frac{1}{2} - y_{16}\right)b\,\hat{\mathbf{y}} + \left(\frac{1}{2} + z_{16}\right)c\,\hat{\mathbf{z}}$ | (8$e$) | H VII |
| $\mathbf{B}_{108}$ | = | $-x_{16}\,\mathbf{a}_1 + \left(\frac{1}{2} + y_{16}\right)\mathbf{a}_2 + \left(\frac{1}{2} + z_{16}\right)\mathbf{a}_3$ | = | $-x_{16}a\,\hat{\mathbf{x}} + \left(\frac{1}{2} + y_{16}\right)b\,\hat{\mathbf{y}} + \left(\frac{1}{2} + z_{16}\right)c\,\hat{\mathbf{z}}$ | (8$e$) | H VII |
| $\mathbf{B}_{109}$ | = | $x_{17}\,\mathbf{a}_1 + y_{17}\,\mathbf{a}_2 + z_{17}\,\mathbf{a}_3$ | = | $x_{17}a\,\hat{\mathbf{x}} + y_{17}b\,\hat{\mathbf{y}} + z_{17}c\,\hat{\mathbf{z}}$ | (8$e$) | H VIII |
| $\mathbf{B}_{110}$ | = | $\left(\frac{1}{2} - x_{17}\right)\mathbf{a}_1 - y_{17}\,\mathbf{a}_2 + z_{17}\,\mathbf{a}_3$ | = | $\left(\frac{1}{2} - x_{17}\right)a\,\hat{\mathbf{x}} - y_{17}b\,\hat{\mathbf{y}} + z_{17}c\,\hat{\mathbf{z}}$ | (8$e$) | H VIII |
| $\mathbf{B}_{111}$ | = | $\left(\frac{1}{2} - x_{17}\right)\mathbf{a}_1 + \left(\frac{1}{2} + y_{17}\right)\mathbf{a}_2 + \left(\frac{1}{2} - z_{17}\right)\mathbf{a}_3$ | = | $\left(\frac{1}{2} - x_{17}\right)a\,\hat{\mathbf{x}} + \left(\frac{1}{2} + y_{17}\right)b\,\hat{\mathbf{y}} + \left(\frac{1}{2} - z_{17}\right)c\,\hat{\mathbf{z}}$ | (8$e$) | H VIII |
| $\mathbf{B}_{112}$ | = | $x_{17}\,\mathbf{a}_1 + \left(\frac{1}{2} - y_{17}\right)\mathbf{a}_2 + \left(\frac{1}{2} - z_{17}\right)\mathbf{a}_3$ | = | $x_{17}a\,\hat{\mathbf{x}} + \left(\frac{1}{2} - y_{17}\right)b\,\hat{\mathbf{y}} + \left(\frac{1}{2} - z_{17}\right)c\,\hat{\mathbf{z}}$ | (8$e$) | H VIII |

$$\mathbf{B}_{113} = -x_{17}\,\mathbf{a}_1 - y_{17}\,\mathbf{a}_2 - z_{17}\,\mathbf{a}_3 = -x_{17}a\,\hat{\mathbf{x}} - y_{17}b\,\hat{\mathbf{y}} - z_{17}c\,\hat{\mathbf{z}} \quad (8e) \quad \text{H VIII}$$

$$\mathbf{B}_{114} = \left(\frac{1}{2} + x_{17}\right)\mathbf{a}_1 + y_{17}\,\mathbf{a}_2 - z_{17}\,\mathbf{a}_3 = \left(\frac{1}{2} + x_{17}\right)a\,\hat{\mathbf{x}} + y_{17}b\,\hat{\mathbf{y}} - z_{17}c\,\hat{\mathbf{z}} \quad (8e) \quad \text{H VIII}$$

$$\mathbf{B}_{115} = \left(\frac{1}{2} + x_{17}\right)\mathbf{a}_1 + \left(\frac{1}{2} - y_{17}\right)\mathbf{a}_2 + \left(\frac{1}{2} + z_{17}\right)\mathbf{a}_3 = \left(\frac{1}{2} + x_{17}\right)a\,\hat{\mathbf{x}} + \left(\frac{1}{2} - y_{17}\right)b\,\hat{\mathbf{y}} + \left(\frac{1}{2} + z_{17}\right)c\,\hat{\mathbf{z}} \quad (8e) \quad \text{H VIII}$$

$$\mathbf{B}_{116} = -x_{17}\,\mathbf{a}_1 + \left(\frac{1}{2} + y_{17}\right)\mathbf{a}_2 + \left(\frac{1}{2} + z_{17}\right)\mathbf{a}_3 = -x_{17}a\,\hat{\mathbf{x}} + \left(\frac{1}{2} + y_{17}\right)b\,\hat{\mathbf{y}} + \left(\frac{1}{2} + z_{17}\right)c\,\hat{\mathbf{z}} \quad (8e) \quad \text{H VIII}$$

$$\mathbf{B}_{117} = x_{18}\,\mathbf{a}_1 + y_{18}\,\mathbf{a}_2 + z_{18}\,\mathbf{a}_3 = x_{18}a\,\hat{\mathbf{x}} + y_{18}b\,\hat{\mathbf{y}} + z_{18}c\,\hat{\mathbf{z}} \quad (8e) \quad \text{H IX}$$

$$\mathbf{B}_{118} = \left(\frac{1}{2} - x_{18}\right)\mathbf{a}_1 - y_{18}\,\mathbf{a}_2 + z_{18}\,\mathbf{a}_3 = \left(\frac{1}{2} - x_{18}\right)a\,\hat{\mathbf{x}} - y_{18}b\,\hat{\mathbf{y}} + z_{18}c\,\hat{\mathbf{z}} \quad (8e) \quad \text{H IX}$$

$$\mathbf{B}_{119} = \left(\frac{1}{2} - x_{18}\right)\mathbf{a}_1 + \left(\frac{1}{2} + y_{18}\right)\mathbf{a}_2 + \left(\frac{1}{2} - z_{18}\right)\mathbf{a}_3 = \left(\frac{1}{2} - x_{18}\right)a\,\hat{\mathbf{x}} + \left(\frac{1}{2} + y_{18}\right)b\,\hat{\mathbf{y}} + \left(\frac{1}{2} - z_{18}\right)c\,\hat{\mathbf{z}} \quad (8e) \quad \text{H IX}$$

$$\mathbf{B}_{120} = x_{18}\,\mathbf{a}_1 + \left(\frac{1}{2} - y_{18}\right)\mathbf{a}_2 + \left(\frac{1}{2} - z_{18}\right)\mathbf{a}_3 = x_{18}a\,\hat{\mathbf{x}} + \left(\frac{1}{2} - y_{18}\right)b\,\hat{\mathbf{y}} + \left(\frac{1}{2} - z_{18}\right)c\,\hat{\mathbf{z}} \quad (8e) \quad \text{H IX}$$

$$\mathbf{B}_{121} = -x_{18}\,\mathbf{a}_1 - y_{18}\,\mathbf{a}_2 - z_{18}\,\mathbf{a}_3 = -x_{18}a\,\hat{\mathbf{x}} - y_{18}b\,\hat{\mathbf{y}} - z_{18}c\,\hat{\mathbf{z}} \quad (8e) \quad \text{H IX}$$

$$\mathbf{B}_{122} = \left(\frac{1}{2} + x_{18}\right)\mathbf{a}_1 + y_{18}\,\mathbf{a}_2 - z_{18}\,\mathbf{a}_3 = \left(\frac{1}{2} + x_{18}\right)a\,\hat{\mathbf{x}} + y_{18}b\,\hat{\mathbf{y}} - z_{18}c\,\hat{\mathbf{z}} \quad (8e) \quad \text{H IX}$$

$$\mathbf{B}_{123} = \left(\frac{1}{2} + x_{18}\right)\mathbf{a}_1 + \left(\frac{1}{2} - y_{18}\right)\mathbf{a}_2 + \left(\frac{1}{2} + z_{18}\right)\mathbf{a}_3 = \left(\frac{1}{2} + x_{18}\right)a\,\hat{\mathbf{x}} + \left(\frac{1}{2} - y_{18}\right)b\,\hat{\mathbf{y}} + \left(\frac{1}{2} + z_{18}\right)c\,\hat{\mathbf{z}} \quad (8e) \quad \text{H IX}$$

$$\mathbf{B}_{124} = -x_{18}\,\mathbf{a}_1 + \left(\frac{1}{2} + y_{18}\right)\mathbf{a}_2 + \left(\frac{1}{2} + z_{18}\right)\mathbf{a}_3 = -x_{18}a\,\hat{\mathbf{x}} + \left(\frac{1}{2} + y_{18}\right)b\,\hat{\mathbf{y}} + \left(\frac{1}{2} + z_{18}\right)c\,\hat{\mathbf{z}} \quad (8e) \quad \text{H IX}$$

$$\mathbf{B}_{125} = x_{19}\,\mathbf{a}_1 + y_{19}\,\mathbf{a}_2 + z_{19}\,\mathbf{a}_3 = x_{19}a\,\hat{\mathbf{x}} + y_{19}b\,\hat{\mathbf{y}} + z_{19}c\,\hat{\mathbf{z}} \quad (8e) \quad \text{H X}$$

$$\mathbf{B}_{126} = \left(\frac{1}{2} - x_{19}\right)\mathbf{a}_1 - y_{19}\,\mathbf{a}_2 + z_{19}\,\mathbf{a}_3 = \left(\frac{1}{2} - x_{19}\right)a\,\hat{\mathbf{x}} - y_{19}b\,\hat{\mathbf{y}} + z_{19}c\,\hat{\mathbf{z}} \quad (8e) \quad \text{H X}$$

$$\mathbf{B}_{127} = \left(\frac{1}{2} - x_{19}\right)\mathbf{a}_1 + \left(\frac{1}{2} + y_{19}\right)\mathbf{a}_2 + \left(\frac{1}{2} - z_{19}\right)\mathbf{a}_3 = \left(\frac{1}{2} - x_{19}\right)a\,\hat{\mathbf{x}} + \left(\frac{1}{2} + y_{19}\right)b\,\hat{\mathbf{y}} + \left(\frac{1}{2} - z_{19}\right)c\,\hat{\mathbf{z}} \quad (8e) \quad \text{H X}$$

$$\mathbf{B}_{128} = x_{19}\,\mathbf{a}_1 + \left(\frac{1}{2} - y_{19}\right)\mathbf{a}_2 + \left(\frac{1}{2} - z_{19}\right)\mathbf{a}_3 = x_{19}a\,\hat{\mathbf{x}} + \left(\frac{1}{2} - y_{19}\right)b\,\hat{\mathbf{y}} + \left(\frac{1}{2} - z_{19}\right)c\,\hat{\mathbf{z}} \quad (8e) \quad \text{H X}$$

$$\mathbf{B}_{129} = -x_{19}\,\mathbf{a}_1 - y_{19}\,\mathbf{a}_2 - z_{19}\,\mathbf{a}_3 = -x_{19}a\,\hat{\mathbf{x}} - y_{19}b\,\hat{\mathbf{y}} - z_{19}c\,\hat{\mathbf{z}} \quad (8e) \quad \text{H X}$$

$$\mathbf{B}_{130} = \left(\frac{1}{2} + x_{19}\right)\mathbf{a}_1 + y_{19}\,\mathbf{a}_2 - z_{19}\,\mathbf{a}_3 = \left(\frac{1}{2} + x_{19}\right)a\,\hat{\mathbf{x}} + y_{19}b\,\hat{\mathbf{y}} - z_{19}c\,\hat{\mathbf{z}} \quad (8e) \quad \text{H X}$$

$$\mathbf{B}_{131} = \left(\frac{1}{2} + x_{19}\right)\mathbf{a}_1 + \left(\frac{1}{2} - y_{19}\right)\mathbf{a}_2 + \left(\frac{1}{2} + z_{19}\right)\mathbf{a}_3 = \left(\frac{1}{2} + x_{19}\right)a\,\hat{\mathbf{x}} + \left(\frac{1}{2} - y_{19}\right)b\,\hat{\mathbf{y}} + \left(\frac{1}{2} + z_{19}\right)c\,\hat{\mathbf{z}} \quad (8e) \quad \text{H X}$$

$$\mathbf{B}_{132} = -x_{19}\,\mathbf{a}_1 + \left(\frac{1}{2} + y_{19}\right)\mathbf{a}_2 + \left(\frac{1}{2} + z_{19}\right)\mathbf{a}_3 = -x_{19}a\,\hat{\mathbf{x}} + \left(\frac{1}{2} + y_{19}\right)b\,\hat{\mathbf{y}} + \left(\frac{1}{2} + z_{19}\right)c\,\hat{\mathbf{z}} \quad (8e) \quad \text{H X}$$

$$\mathbf{B}_{133} = x_{20}\,\mathbf{a}_1 + y_{20}\,\mathbf{a}_2 + z_{20}\,\mathbf{a}_3 = x_{20}a\,\hat{\mathbf{x}} + y_{20}b\,\hat{\mathbf{y}} + z_{20}c\,\hat{\mathbf{z}} \quad (8e) \quad \text{H XI}$$

$$\mathbf{B}_{134} = \left(\frac{1}{2} - x_{20}\right)\mathbf{a}_1 - y_{20}\,\mathbf{a}_2 + z_{20}\,\mathbf{a}_3 = \left(\frac{1}{2} - x_{20}\right)a\,\hat{\mathbf{x}} - y_{20}b\,\hat{\mathbf{y}} + z_{20}c\,\hat{\mathbf{z}} \quad (8e) \quad \text{H XI}$$

$$\mathbf{B}_{135} = \left(\frac{1}{2} - x_{20}\right)\mathbf{a}_1 + \left(\frac{1}{2} + y_{20}\right)\mathbf{a}_2 + \left(\frac{1}{2} - z_{20}\right)\mathbf{a}_3 = \left(\frac{1}{2} - x_{20}\right)a\,\hat{\mathbf{x}} + \left(\frac{1}{2} + y_{20}\right)b\,\hat{\mathbf{y}} + \left(\frac{1}{2} - z_{20}\right)c\,\hat{\mathbf{z}} \quad (8e) \quad \text{H XI}$$

$$\mathbf{B}_{136} = x_{20}\,\mathbf{a}_1 + \left(\frac{1}{2} - y_{20}\right)\mathbf{a}_2 + \left(\frac{1}{2} - z_{20}\right)\mathbf{a}_3 = x_{20}a\,\hat{\mathbf{x}} + \left(\frac{1}{2} - y_{20}\right)b\,\hat{\mathbf{y}} + \left(\frac{1}{2} - z_{20}\right)c\,\hat{\mathbf{z}} \quad (8e) \quad \text{H XI}$$

$$\mathbf{B}_{137} = -x_{20}\,\mathbf{a}_1 - y_{20}\,\mathbf{a}_2 - z_{20}\,\mathbf{a}_3 = -x_{20}a\,\hat{\mathbf{x}} - y_{20}b\,\hat{\mathbf{y}} - z_{20}c\,\hat{\mathbf{z}} \quad (8e) \quad \text{H XI}$$

$$\mathbf{B}_{138} = \left(\frac{1}{2} + x_{20}\right)\mathbf{a}_1 + y_{20}\,\mathbf{a}_2 - z_{20}\,\mathbf{a}_3 = \left(\frac{1}{2} + x_{20}\right)a\,\hat{\mathbf{x}} + y_{20}b\,\hat{\mathbf{y}} - z_{20}c\,\hat{\mathbf{z}} \quad (8e) \quad \text{H XI}$$

$$\mathbf{B}_{139} = \left(\frac{1}{2} + x_{20}\right)\mathbf{a}_1 + \left(\frac{1}{2} - y_{20}\right)\mathbf{a}_2 + \left(\frac{1}{2} + z_{20}\right)\mathbf{a}_3 = \left(\frac{1}{2} + x_{20}\right)a\,\hat{\mathbf{x}} + \left(\frac{1}{2} - y_{20}\right)b\,\hat{\mathbf{y}} + \left(\frac{1}{2} + z_{20}\right)c\,\hat{\mathbf{z}} \quad (8e) \quad \text{H XI}$$

$$\mathbf{B}_{140} = -x_{20}\,\mathbf{a}_1 + \left(\frac{1}{2} + y_{20}\right)\mathbf{a}_2 + \left(\frac{1}{2} + z_{20}\right)\mathbf{a}_3 = -x_{20}a\,\hat{\mathbf{x}} + \left(\frac{1}{2} + y_{20}\right)b\,\hat{\mathbf{y}} + \left(\frac{1}{2} + z_{20}\right)c\,\hat{\mathbf{z}} \quad (8e) \quad \text{H XI}$$

| | | | | |
|---|---|---|---|---|
| $\mathbf{B}_{141}$ | $= x_{21}\,\mathbf{a}_1 + y_{21}\,\mathbf{a}_2 + z_{21}\,\mathbf{a}_3$ | $= x_{21}a\,\hat{\mathbf{x}} + y_{21}b\,\hat{\mathbf{y}} + z_{21}c\,\hat{\mathbf{z}}$ | (8*e*) | H XII |
| $\mathbf{B}_{142}$ | $= \left(\frac{1}{2} - x_{21}\right)\mathbf{a}_1 - y_{21}\,\mathbf{a}_2 + z_{21}\,\mathbf{a}_3$ | $= \left(\frac{1}{2} - x_{21}\right)a\,\hat{\mathbf{x}} - y_{21}b\,\hat{\mathbf{y}} + z_{21}c\,\hat{\mathbf{z}}$ | (8*e*) | H XII |
| $\mathbf{B}_{143}$ | $= \left(\frac{1}{2} - x_{21}\right)\mathbf{a}_1 + \left(\frac{1}{2} + y_{21}\right)\mathbf{a}_2 + \left(\frac{1}{2} - z_{21}\right)\mathbf{a}_3$ | $= \left(\frac{1}{2} - x_{21}\right)a\,\hat{\mathbf{x}} + \left(\frac{1}{2} + y_{21}\right)b\,\hat{\mathbf{y}} + \left(\frac{1}{2} - z_{21}\right)c\,\hat{\mathbf{z}}$ | (8*e*) | H XII |
| $\mathbf{B}_{144}$ | $= x_{21}\,\mathbf{a}_1 + \left(\frac{1}{2} - y_{21}\right)\mathbf{a}_2 + \left(\frac{1}{2} - z_{21}\right)\mathbf{a}_3$ | $= x_{21}a\,\hat{\mathbf{x}} + \left(\frac{1}{2} - y_{21}\right)b\,\hat{\mathbf{y}} + \left(\frac{1}{2} - z_{21}\right)c\,\hat{\mathbf{z}}$ | (8*e*) | H XII |
| $\mathbf{B}_{145}$ | $= -x_{21}\,\mathbf{a}_1 - y_{21}\,\mathbf{a}_2 - z_{21}\,\mathbf{a}_3$ | $= -x_{21}a\,\hat{\mathbf{x}} - y_{21}b\,\hat{\mathbf{y}} - z_{21}c\,\hat{\mathbf{z}}$ | (8*e*) | H XII |
| $\mathbf{B}_{146}$ | $= \left(\frac{1}{2} + x_{21}\right)\mathbf{a}_1 + y_{21}\,\mathbf{a}_2 - z_{21}\,\mathbf{a}_3$ | $= \left(\frac{1}{2} + x_{21}\right)a\,\hat{\mathbf{x}} + y_{21}b\,\hat{\mathbf{y}} - z_{21}c\,\hat{\mathbf{z}}$ | (8*e*) | H XII |
| $\mathbf{B}_{147}$ | $= \left(\frac{1}{2} + x_{21}\right)\mathbf{a}_1 + \left(\frac{1}{2} - y_{21}\right)\mathbf{a}_2 + \left(\frac{1}{2} + z_{21}\right)\mathbf{a}_3$ | $= \left(\frac{1}{2} + x_{21}\right)a\,\hat{\mathbf{x}} + \left(\frac{1}{2} - y_{21}\right)b\,\hat{\mathbf{y}} + \left(\frac{1}{2} + z_{21}\right)c\,\hat{\mathbf{z}}$ | (8*e*) | H XII |
| $\mathbf{B}_{148}$ | $= -x_{21}\,\mathbf{a}_1 + \left(\frac{1}{2} + y_{21}\right)\mathbf{a}_2 + \left(\frac{1}{2} + z_{21}\right)\mathbf{a}_3$ | $= -x_{21}a\,\hat{\mathbf{x}} + \left(\frac{1}{2} + y_{21}\right)b\,\hat{\mathbf{y}} + \left(\frac{1}{2} + z_{21}\right)c\,\hat{\mathbf{z}}$ | (8*e*) | H XII |
| $\mathbf{B}_{149}$ | $= x_{22}\,\mathbf{a}_1 + y_{22}\,\mathbf{a}_2 + z_{22}\,\mathbf{a}_3$ | $= x_{22}a\,\hat{\mathbf{x}} + y_{22}b\,\hat{\mathbf{y}} + z_{22}c\,\hat{\mathbf{z}}$ | (8*e*) | H XIII |
| $\mathbf{B}_{150}$ | $= \left(\frac{1}{2} - x_{22}\right)\mathbf{a}_1 - y_{22}\,\mathbf{a}_2 + z_{22}\,\mathbf{a}_3$ | $= \left(\frac{1}{2} - x_{22}\right)a\,\hat{\mathbf{x}} - y_{22}b\,\hat{\mathbf{y}} + z_{22}c\,\hat{\mathbf{z}}$ | (8*e*) | H XIII |
| $\mathbf{B}_{151}$ | $= \left(\frac{1}{2} - x_{22}\right)\mathbf{a}_1 + \left(\frac{1}{2} + y_{22}\right)\mathbf{a}_2 + \left(\frac{1}{2} - z_{22}\right)\mathbf{a}_3$ | $= \left(\frac{1}{2} - x_{22}\right)a\,\hat{\mathbf{x}} + \left(\frac{1}{2} + y_{22}\right)b\,\hat{\mathbf{y}} + \left(\frac{1}{2} - z_{22}\right)c\,\hat{\mathbf{z}}$ | (8*e*) | H XIII |
| $\mathbf{B}_{152}$ | $= x_{22}\,\mathbf{a}_1 + \left(\frac{1}{2} - y_{22}\right)\mathbf{a}_2 + \left(\frac{1}{2} - z_{22}\right)\mathbf{a}_3$ | $= x_{22}a\,\hat{\mathbf{x}} + \left(\frac{1}{2} - y_{22}\right)b\,\hat{\mathbf{y}} + \left(\frac{1}{2} - z_{22}\right)c\,\hat{\mathbf{z}}$ | (8*e*) | H XIII |
| $\mathbf{B}_{153}$ | $= -x_{22}\,\mathbf{a}_1 - y_{22}\,\mathbf{a}_2 - z_{22}\,\mathbf{a}_3$ | $= -x_{22}a\,\hat{\mathbf{x}} - y_{22}b\,\hat{\mathbf{y}} - z_{22}c\,\hat{\mathbf{z}}$ | (8*e*) | H XIII |
| $\mathbf{B}_{154}$ | $= \left(\frac{1}{2} + x_{22}\right)\mathbf{a}_1 + y_{22}\,\mathbf{a}_2 - z_{22}\,\mathbf{a}_3$ | $= \left(\frac{1}{2} + x_{22}\right)a\,\hat{\mathbf{x}} + y_{22}b\,\hat{\mathbf{y}} - z_{22}c\,\hat{\mathbf{z}}$ | (8*e*) | H XIII |
| $\mathbf{B}_{155}$ | $= \left(\frac{1}{2} + x_{22}\right)\mathbf{a}_1 + \left(\frac{1}{2} - y_{22}\right)\mathbf{a}_2 + \left(\frac{1}{2} + z_{22}\right)\mathbf{a}_3$ | $= \left(\frac{1}{2} + x_{22}\right)a\,\hat{\mathbf{x}} + \left(\frac{1}{2} - y_{22}\right)b\,\hat{\mathbf{y}} + \left(\frac{1}{2} + z_{22}\right)c\,\hat{\mathbf{z}}$ | (8*e*) | H XIII |
| $\mathbf{B}_{156}$ | $= -x_{22}\,\mathbf{a}_1 + \left(\frac{1}{2} + y_{22}\right)\mathbf{a}_2 + \left(\frac{1}{2} + z_{22}\right)\mathbf{a}_3$ | $= -x_{22}a\,\hat{\mathbf{x}} + \left(\frac{1}{2} + y_{22}\right)b\,\hat{\mathbf{y}} + \left(\frac{1}{2} + z_{22}\right)c\,\hat{\mathbf{z}}$ | (8*e*) | H XIII |
| $\mathbf{B}_{157}$ | $= x_{23}\,\mathbf{a}_1 + y_{23}\,\mathbf{a}_2 + z_{23}\,\mathbf{a}_3$ | $= x_{23}a\,\hat{\mathbf{x}} + y_{23}b\,\hat{\mathbf{y}} + z_{23}c\,\hat{\mathbf{z}}$ | (8*e*) | H XIV |
| $\mathbf{B}_{158}$ | $= \left(\frac{1}{2} - x_{23}\right)\mathbf{a}_1 - y_{23}\,\mathbf{a}_2 + z_{23}\,\mathbf{a}_3$ | $= \left(\frac{1}{2} - x_{23}\right)a\,\hat{\mathbf{x}} - y_{23}b\,\hat{\mathbf{y}} + z_{23}c\,\hat{\mathbf{z}}$ | (8*e*) | H XIV |
| $\mathbf{B}_{159}$ | $= \left(\frac{1}{2} - x_{23}\right)\mathbf{a}_1 + \left(\frac{1}{2} + y_{23}\right)\mathbf{a}_2 + \left(\frac{1}{2} - z_{23}\right)\mathbf{a}_3$ | $= \left(\frac{1}{2} - x_{23}\right)a\,\hat{\mathbf{x}} + \left(\frac{1}{2} + y_{23}\right)b\,\hat{\mathbf{y}} + \left(\frac{1}{2} - z_{23}\right)c\,\hat{\mathbf{z}}$ | (8*e*) | H XIV |
| $\mathbf{B}_{160}$ | $= x_{23}\,\mathbf{a}_1 + \left(\frac{1}{2} - y_{23}\right)\mathbf{a}_2 + \left(\frac{1}{2} - z_{23}\right)\mathbf{a}_3$ | $= x_{23}a\,\hat{\mathbf{x}} + \left(\frac{1}{2} - y_{23}\right)b\,\hat{\mathbf{y}} + \left(\frac{1}{2} - z_{23}\right)c\,\hat{\mathbf{z}}$ | (8*e*) | H XIV |
| $\mathbf{B}_{161}$ | $= -x_{23}\,\mathbf{a}_1 - y_{23}\,\mathbf{a}_2 - z_{23}\,\mathbf{a}_3$ | $= -x_{23}a\,\hat{\mathbf{x}} - y_{23}b\,\hat{\mathbf{y}} - z_{23}c\,\hat{\mathbf{z}}$ | (8*e*) | H XIV |
| $\mathbf{B}_{162}$ | $= \left(\frac{1}{2} + x_{23}\right)\mathbf{a}_1 + y_{23}\,\mathbf{a}_2 - z_{23}\,\mathbf{a}_3$ | $= \left(\frac{1}{2} + x_{23}\right)a\,\hat{\mathbf{x}} + y_{23}b\,\hat{\mathbf{y}} - z_{23}c\,\hat{\mathbf{z}}$ | (8*e*) | H XIV |
| $\mathbf{B}_{163}$ | $= \left(\frac{1}{2} + x_{23}\right)\mathbf{a}_1 + \left(\frac{1}{2} - y_{23}\right)\mathbf{a}_2 + \left(\frac{1}{2} + z_{23}\right)\mathbf{a}_3$ | $= \left(\frac{1}{2} + x_{23}\right)a\,\hat{\mathbf{x}} + \left(\frac{1}{2} - y_{23}\right)b\,\hat{\mathbf{y}} + \left(\frac{1}{2} + z_{23}\right)c\,\hat{\mathbf{z}}$ | (8*e*) | H XIV |
| $\mathbf{B}_{164}$ | $= -x_{23}\,\mathbf{a}_1 + \left(\frac{1}{2} + y_{23}\right)\mathbf{a}_2 + \left(\frac{1}{2} + z_{23}\right)\mathbf{a}_3$ | $= -x_{23}a\,\hat{\mathbf{x}} + \left(\frac{1}{2} + y_{23}\right)b\,\hat{\mathbf{y}} + \left(\frac{1}{2} + z_{23}\right)c\,\hat{\mathbf{z}}$ | (8*e*) | H XIV |
| $\mathbf{B}_{165}$ | $= x_{24}\,\mathbf{a}_1 + y_{24}\,\mathbf{a}_2 + z_{24}\,\mathbf{a}_3$ | $= x_{24}a\,\hat{\mathbf{x}} + y_{24}b\,\hat{\mathbf{y}} + z_{24}c\,\hat{\mathbf{z}}$ | (8*e*) | H XV |
| $\mathbf{B}_{166}$ | $= \left(\frac{1}{2} - x_{24}\right)\mathbf{a}_1 - y_{24}\,\mathbf{a}_2 + z_{24}\,\mathbf{a}_3$ | $= \left(\frac{1}{2} - x_{24}\right)a\,\hat{\mathbf{x}} - y_{24}b\,\hat{\mathbf{y}} + z_{24}c\,\hat{\mathbf{z}}$ | (8*e*) | H XV |
| $\mathbf{B}_{167}$ | $= \left(\frac{1}{2} - x_{24}\right)\mathbf{a}_1 + \left(\frac{1}{2} + y_{24}\right)\mathbf{a}_2 + \left(\frac{1}{2} - z_{24}\right)\mathbf{a}_3$ | $= \left(\frac{1}{2} - x_{24}\right)a\,\hat{\mathbf{x}} + \left(\frac{1}{2} + y_{24}\right)b\,\hat{\mathbf{y}} + \left(\frac{1}{2} - z_{24}\right)c\,\hat{\mathbf{z}}$ | (8*e*) | H XV |
| $\mathbf{B}_{168}$ | $= x_{24}\,\mathbf{a}_1 + \left(\frac{1}{2} - y_{24}\right)\mathbf{a}_2 + \left(\frac{1}{2} - z_{24}\right)\mathbf{a}_3$ | $= x_{24}a\,\hat{\mathbf{x}} + \left(\frac{1}{2} - y_{24}\right)b\,\hat{\mathbf{y}} + \left(\frac{1}{2} - z_{24}\right)c\,\hat{\mathbf{z}}$ | (8*e*) | H XV |

| | | | | | | |
|---|---|---|---|---|---|---|
| $\mathbf{B}_{169}$ | = | $-x_{24}\,\mathbf{a}_1 - y_{24}\,\mathbf{a}_2 - z_{24}\,\mathbf{a}_3$ | = | $-x_{24}a\,\hat{\mathbf{x}} - y_{24}b\,\hat{\mathbf{y}} - z_{24}c\,\hat{\mathbf{z}}$ | (8*e*) | H XV |
| $\mathbf{B}_{170}$ | = | $\left(\frac{1}{2} + x_{24}\right)\mathbf{a}_1 + y_{24}\,\mathbf{a}_2 - z_{24}\,\mathbf{a}_3$ | = | $\left(\frac{1}{2} + x_{24}\right)a\,\hat{\mathbf{x}} + y_{24}b\,\hat{\mathbf{y}} - z_{24}c\,\hat{\mathbf{z}}$ | (8*e*) | H XV |
| $\mathbf{B}_{171}$ | = | $\left(\frac{1}{2} + x_{24}\right)\mathbf{a}_1 + \left(\frac{1}{2} - y_{24}\right)\mathbf{a}_2 + \left(\frac{1}{2} + z_{24}\right)\mathbf{a}_3$ | = | $\left(\frac{1}{2} + x_{24}\right)a\,\hat{\mathbf{x}} + \left(\frac{1}{2} - y_{24}\right)b\,\hat{\mathbf{y}} + \left(\frac{1}{2} + z_{24}\right)c\,\hat{\mathbf{z}}$ | (8*e*) | H XV |
| $\mathbf{B}_{172}$ | = | $-x_{24}\,\mathbf{a}_1 + \left(\frac{1}{2} + y_{24}\right)\mathbf{a}_2 + \left(\frac{1}{2} + z_{24}\right)\mathbf{a}_3$ | = | $-x_{24}a\,\hat{\mathbf{x}} + \left(\frac{1}{2} + y_{24}\right)b\,\hat{\mathbf{y}} + \left(\frac{1}{2} + z_{24}\right)c\,\hat{\mathbf{z}}$ | (8*e*) | H XV |
| $\mathbf{B}_{173}$ | = | $x_{25}\,\mathbf{a}_1 + y_{25}\,\mathbf{a}_2 + z_{25}\,\mathbf{a}_3$ | = | $x_{25}a\,\hat{\mathbf{x}} + y_{25}b\,\hat{\mathbf{y}} + z_{25}c\,\hat{\mathbf{z}}$ | (8*e*) | H XVI |
| $\mathbf{B}_{174}$ | = | $\left(\frac{1}{2} - x_{25}\right)\mathbf{a}_1 - y_{25}\,\mathbf{a}_2 + z_{25}\,\mathbf{a}_3$ | = | $\left(\frac{1}{2} - x_{25}\right)a\,\hat{\mathbf{x}} - y_{25}b\,\hat{\mathbf{y}} + z_{25}c\,\hat{\mathbf{z}}$ | (8*e*) | H XVI |
| $\mathbf{B}_{175}$ | = | $\left(\frac{1}{2} - x_{25}\right)\mathbf{a}_1 + \left(\frac{1}{2} + y_{25}\right)\mathbf{a}_2 + \left(\frac{1}{2} - z_{25}\right)\mathbf{a}_3$ | = | $\left(\frac{1}{2} - x_{25}\right)a\,\hat{\mathbf{x}} + \left(\frac{1}{2} + y_{25}\right)b\,\hat{\mathbf{y}} + \left(\frac{1}{2} - z_{25}\right)c\,\hat{\mathbf{z}}$ | (8*e*) | H XVI |
| $\mathbf{B}_{176}$ | = | $x_{25}\,\mathbf{a}_1 + \left(\frac{1}{2} - y_{25}\right)\mathbf{a}_2 + \left(\frac{1}{2} - z_{25}\right)\mathbf{a}_3$ | = | $x_{25}a\,\hat{\mathbf{x}} + \left(\frac{1}{2} - y_{25}\right)b\,\hat{\mathbf{y}} + \left(\frac{1}{2} - z_{25}\right)c\,\hat{\mathbf{z}}$ | (8*e*) | H XVI |
| $\mathbf{B}_{177}$ | = | $-x_{25}\,\mathbf{a}_1 - y_{25}\,\mathbf{a}_2 - z_{25}\,\mathbf{a}_3$ | = | $-x_{25}a\,\hat{\mathbf{x}} - y_{25}b\,\hat{\mathbf{y}} - z_{25}c\,\hat{\mathbf{z}}$ | (8*e*) | H XVI |
| $\mathbf{B}_{178}$ | = | $\left(\frac{1}{2} + x_{25}\right)\mathbf{a}_1 + y_{25}\,\mathbf{a}_2 - z_{25}\,\mathbf{a}_3$ | = | $\left(\frac{1}{2} + x_{25}\right)a\,\hat{\mathbf{x}} + y_{25}b\,\hat{\mathbf{y}} - z_{25}c\,\hat{\mathbf{z}}$ | (8*e*) | H XVI |
| $\mathbf{B}_{179}$ | = | $\left(\frac{1}{2} + x_{25}\right)\mathbf{a}_1 + \left(\frac{1}{2} - y_{25}\right)\mathbf{a}_2 + \left(\frac{1}{2} + z_{25}\right)\mathbf{a}_3$ | = | $\left(\frac{1}{2} + x_{25}\right)a\,\hat{\mathbf{x}} + \left(\frac{1}{2} - y_{25}\right)b\,\hat{\mathbf{y}} + \left(\frac{1}{2} + z_{25}\right)c\,\hat{\mathbf{z}}$ | (8*e*) | H XVI |
| $\mathbf{B}_{180}$ | = | $-x_{25}\,\mathbf{a}_1 + \left(\frac{1}{2} + y_{25}\right)\mathbf{a}_2 + \left(\frac{1}{2} + z_{25}\right)\mathbf{a}_3$ | = | $-x_{25}a\,\hat{\mathbf{x}} + \left(\frac{1}{2} + y_{25}\right)b\,\hat{\mathbf{y}} + \left(\frac{1}{2} + z_{25}\right)c\,\hat{\mathbf{z}}$ | (8*e*) | H XVI |
| $\mathbf{B}_{181}$ | = | $x_{26}\,\mathbf{a}_1 + y_{26}\,\mathbf{a}_2 + z_{26}\,\mathbf{a}_3$ | = | $x_{26}a\,\hat{\mathbf{x}} + y_{26}b\,\hat{\mathbf{y}} + z_{26}c\,\hat{\mathbf{z}}$ | (8*e*) | H XVII |
| $\mathbf{B}_{182}$ | = | $\left(\frac{1}{2} - x_{26}\right)\mathbf{a}_1 - y_{26}\,\mathbf{a}_2 + z_{26}\,\mathbf{a}_3$ | = | $\left(\frac{1}{2} - x_{26}\right)a\,\hat{\mathbf{x}} - y_{26}b\,\hat{\mathbf{y}} + z_{26}c\,\hat{\mathbf{z}}$ | (8*e*) | H XVII |
| $\mathbf{B}_{183}$ | = | $\left(\frac{1}{2} - x_{26}\right)\mathbf{a}_1 + \left(\frac{1}{2} + y_{26}\right)\mathbf{a}_2 + \left(\frac{1}{2} - z_{26}\right)\mathbf{a}_3$ | = | $\left(\frac{1}{2} - x_{26}\right)a\,\hat{\mathbf{x}} + \left(\frac{1}{2} + y_{26}\right)b\,\hat{\mathbf{y}} + \left(\frac{1}{2} - z_{26}\right)c\,\hat{\mathbf{z}}$ | (8*e*) | H XVII |
| $\mathbf{B}_{184}$ | = | $x_{26}\,\mathbf{a}_1 + \left(\frac{1}{2} - y_{26}\right)\mathbf{a}_2 + \left(\frac{1}{2} - z_{26}\right)\mathbf{a}_3$ | = | $x_{26}a\,\hat{\mathbf{x}} + \left(\frac{1}{2} - y_{26}\right)b\,\hat{\mathbf{y}} + \left(\frac{1}{2} - z_{26}\right)c\,\hat{\mathbf{z}}$ | (8*e*) | H XVII |
| $\mathbf{B}_{185}$ | = | $-x_{26}\,\mathbf{a}_1 - y_{26}\,\mathbf{a}_2 - z_{26}\,\mathbf{a}_3$ | = | $-x_{26}a\,\hat{\mathbf{x}} - y_{26}b\,\hat{\mathbf{y}} - z_{26}c\,\hat{\mathbf{z}}$ | (8*e*) | H XVII |
| $\mathbf{B}_{186}$ | = | $\left(\frac{1}{2} + x_{26}\right)\mathbf{a}_1 + y_{26}\,\mathbf{a}_2 - z_{26}\,\mathbf{a}_3$ | = | $\left(\frac{1}{2} + x_{26}\right)a\,\hat{\mathbf{x}} + y_{26}b\,\hat{\mathbf{y}} - z_{26}c\,\hat{\mathbf{z}}$ | (8*e*) | H XVII |
| $\mathbf{B}_{187}$ | = | $\left(\frac{1}{2} + x_{26}\right)\mathbf{a}_1 + \left(\frac{1}{2} - y_{26}\right)\mathbf{a}_2 + \left(\frac{1}{2} + z_{26}\right)\mathbf{a}_3$ | = | $\left(\frac{1}{2} + x_{26}\right)a\,\hat{\mathbf{x}} + \left(\frac{1}{2} - y_{26}\right)b\,\hat{\mathbf{y}} + \left(\frac{1}{2} + z_{26}\right)c\,\hat{\mathbf{z}}$ | (8*e*) | H XVII |
| $\mathbf{B}_{188}$ | = | $-x_{26}\,\mathbf{a}_1 + \left(\frac{1}{2} + y_{26}\right)\mathbf{a}_2 + \left(\frac{1}{2} + z_{26}\right)\mathbf{a}_3$ | = | $-x_{26}a\,\hat{\mathbf{x}} + \left(\frac{1}{2} + y_{26}\right)b\,\hat{\mathbf{y}} + \left(\frac{1}{2} + z_{26}\right)c\,\hat{\mathbf{z}}$ | (8*e*) | H XVII |
| $\mathbf{B}_{189}$ | = | $x_{27}\,\mathbf{a}_1 + y_{27}\,\mathbf{a}_2 + z_{27}\,\mathbf{a}_3$ | = | $x_{27}a\,\hat{\mathbf{x}} + y_{27}b\,\hat{\mathbf{y}} + z_{27}c\,\hat{\mathbf{z}}$ | (8*e*) | H XVIII |
| $\mathbf{B}_{190}$ | = | $\left(\frac{1}{2} - x_{27}\right)\mathbf{a}_1 - y_{27}\,\mathbf{a}_2 + z_{27}\,\mathbf{a}_3$ | = | $\left(\frac{1}{2} - x_{27}\right)a\,\hat{\mathbf{x}} - y_{27}b\,\hat{\mathbf{y}} + z_{27}c\,\hat{\mathbf{z}}$ | (8*e*) | H XVIII |
| $\mathbf{B}_{191}$ | = | $\left(\frac{1}{2} - x_{27}\right)\mathbf{a}_1 + \left(\frac{1}{2} + y_{27}\right)\mathbf{a}_2 + \left(\frac{1}{2} - z_{27}\right)\mathbf{a}_3$ | = | $\left(\frac{1}{2} - x_{27}\right)a\,\hat{\mathbf{x}} + \left(\frac{1}{2} + y_{27}\right)b\,\hat{\mathbf{y}} + \left(\frac{1}{2} - z_{27}\right)c\,\hat{\mathbf{z}}$ | (8*e*) | H XVIII |
| $\mathbf{B}_{192}$ | = | $x_{27}\,\mathbf{a}_1 + \left(\frac{1}{2} - y_{27}\right)\mathbf{a}_2 + \left(\frac{1}{2} - z_{27}\right)\mathbf{a}_3$ | = | $x_{27}a\,\hat{\mathbf{x}} + \left(\frac{1}{2} - y_{27}\right)b\,\hat{\mathbf{y}} + \left(\frac{1}{2} - z_{27}\right)c\,\hat{\mathbf{z}}$ | (8*e*) | H XVIII |
| $\mathbf{B}_{193}$ | = | $-x_{27}\,\mathbf{a}_1 - y_{27}\,\mathbf{a}_2 - z_{27}\,\mathbf{a}_3$ | = | $-x_{27}a\,\hat{\mathbf{x}} - y_{27}b\,\hat{\mathbf{y}} - z_{27}c\,\hat{\mathbf{z}}$ | (8*e*) | H XVIII |
| $\mathbf{B}_{194}$ | = | $\left(\frac{1}{2} + x_{27}\right)\mathbf{a}_1 + y_{27}\,\mathbf{a}_2 - z_{27}\,\mathbf{a}_3$ | = | $\left(\frac{1}{2} + x_{27}\right)a\,\hat{\mathbf{x}} + y_{27}b\,\hat{\mathbf{y}} - z_{27}c\,\hat{\mathbf{z}}$ | (8*e*) | H XVIII |
| $\mathbf{B}_{195}$ | = | $\left(\frac{1}{2} + x_{27}\right)\mathbf{a}_1 + \left(\frac{1}{2} - y_{27}\right)\mathbf{a}_2 + \left(\frac{1}{2} + z_{27}\right)\mathbf{a}_3$ | = | $\left(\frac{1}{2} + x_{27}\right)a\,\hat{\mathbf{x}} + \left(\frac{1}{2} - y_{27}\right)b\,\hat{\mathbf{y}} + \left(\frac{1}{2} + z_{27}\right)c\,\hat{\mathbf{z}}$ | (8*e*) | H XVIII |
| $\mathbf{B}_{196}$ | = | $-x_{27}\,\mathbf{a}_1 + \left(\frac{1}{2} + y_{27}\right)\mathbf{a}_2 + \left(\frac{1}{2} + z_{27}\right)\mathbf{a}_3$ | = | $-x_{27}a\,\hat{\mathbf{x}} + \left(\frac{1}{2} + y_{27}\right)b\,\hat{\mathbf{y}} + \left(\frac{1}{2} + z_{27}\right)c\,\hat{\mathbf{z}}$ | (8*e*) | H XVIII |

$$\mathbf{B}_{197} = x_{28}\,\mathbf{a}_{1} + y_{28}\,\mathbf{a}_{2} + z_{28}\,\mathbf{a}_{3} = x_{28}a\,\hat{\mathbf{x}} + y_{28}b\,\hat{\mathbf{y}} + z_{28}c\,\hat{\mathbf{z}} \qquad (8e) \qquad \text{K II}$$

$$\mathbf{B}_{198} = \left(\frac{1}{2} - x_{28}\right)\mathbf{a}_{1} - y_{28}\,\mathbf{a}_{2} + z_{28}\,\mathbf{a}_{3} = \left(\frac{1}{2} - x_{28}\right)a\,\hat{\mathbf{x}} - y_{28}b\,\hat{\mathbf{y}} + z_{28}c\,\hat{\mathbf{z}} \qquad (8e) \qquad \text{K II}$$

$$\mathbf{B}_{199} = \left(\frac{1}{2} - x_{28}\right)\mathbf{a}_{1} + \left(\frac{1}{2} + y_{28}\right)\mathbf{a}_{2} + \left(\frac{1}{2} - z_{28}\right)\mathbf{a}_{3} = \left(\frac{1}{2} - x_{28}\right)a\,\hat{\mathbf{x}} + \left(\frac{1}{2} + y_{28}\right)b\,\hat{\mathbf{y}} + \left(\frac{1}{2} - z_{28}\right)c\,\hat{\mathbf{z}} \qquad (8e) \qquad \text{K II}$$

$$\mathbf{B}_{200} = x_{28}\,\mathbf{a}_{1} + \left(\frac{1}{2} - y_{28}\right)\mathbf{a}_{2} + \left(\frac{1}{2} - z_{28}\right)\mathbf{a}_{3} = x_{28}a\,\hat{\mathbf{x}} + \left(\frac{1}{2} - y_{28}\right)b\,\hat{\mathbf{y}} + \left(\frac{1}{2} - z_{28}\right)c\,\hat{\mathbf{z}} \qquad (8e) \qquad \text{K II}$$

$$\mathbf{B}_{201} = -x_{28}\,\mathbf{a}_{1} - y_{28}\,\mathbf{a}_{2} - z_{28}\,\mathbf{a}_{3} = -x_{28}a\,\hat{\mathbf{x}} - y_{28}b\,\hat{\mathbf{y}} - z_{28}c\,\hat{\mathbf{z}} \qquad (8e) \qquad \text{K II}$$

$$\mathbf{B}_{202} = \left(\frac{1}{2} + x_{28}\right)\mathbf{a}_{1} + y_{28}\,\mathbf{a}_{2} - z_{28}\,\mathbf{a}_{3} = \left(\frac{1}{2} + x_{28}\right)a\,\hat{\mathbf{x}} + y_{28}b\,\hat{\mathbf{y}} - z_{28}c\,\hat{\mathbf{z}} \qquad (8e) \qquad \text{K II}$$

$$\mathbf{B}_{203} = \left(\frac{1}{2} + x_{28}\right)\mathbf{a}_{1} + \left(\frac{1}{2} - y_{28}\right)\mathbf{a}_{2} + \left(\frac{1}{2} + z_{28}\right)\mathbf{a}_{3} = \left(\frac{1}{2} + x_{28}\right)a\,\hat{\mathbf{x}} + \left(\frac{1}{2} - y_{28}\right)b\,\hat{\mathbf{y}} + \left(\frac{1}{2} + z_{28}\right)c\,\hat{\mathbf{z}} \qquad (8e) \qquad \text{K II}$$

$$\mathbf{B}_{204} = -x_{28}\,\mathbf{a}_{1} + \left(\frac{1}{2} + y_{28}\right)\mathbf{a}_{2} + \left(\frac{1}{2} + z_{28}\right)\mathbf{a}_{3} = -x_{28}a\,\hat{\mathbf{x}} + \left(\frac{1}{2} + y_{28}\right)b\,\hat{\mathbf{y}} + \left(\frac{1}{2} + z_{28}\right)c\,\hat{\mathbf{z}} \qquad (8e) \qquad \text{K II}$$

$$\mathbf{B}_{205} = x_{29}\,\mathbf{a}_{1} + y_{29}\,\mathbf{a}_{2} + z_{29}\,\mathbf{a}_{3} = x_{29}a\,\hat{\mathbf{x}} + y_{29}b\,\hat{\mathbf{y}} + z_{29}c\,\hat{\mathbf{z}} \qquad (8e) \qquad \text{Mg II}$$

$$\mathbf{B}_{206} = \left(\frac{1}{2} - x_{29}\right)\mathbf{a}_{1} - y_{29}\,\mathbf{a}_{2} + z_{29}\,\mathbf{a}_{3} = \left(\frac{1}{2} - x_{29}\right)a\,\hat{\mathbf{x}} - y_{29}b\,\hat{\mathbf{y}} + z_{29}c\,\hat{\mathbf{z}} \qquad (8e) \qquad \text{Mg II}$$

$$\mathbf{B}_{207} = \left(\frac{1}{2} - x_{29}\right)\mathbf{a}_{1} + \left(\frac{1}{2} + y_{29}\right)\mathbf{a}_{2} + \left(\frac{1}{2} - z_{29}\right)\mathbf{a}_{3} = \left(\frac{1}{2} - x_{29}\right)a\,\hat{\mathbf{x}} + \left(\frac{1}{2} + y_{29}\right)b\,\hat{\mathbf{y}} + \left(\frac{1}{2} - z_{29}\right)c\,\hat{\mathbf{z}} \qquad (8e) \qquad \text{Mg II}$$

$$\mathbf{B}_{208} = x_{29}\,\mathbf{a}_{1} + \left(\frac{1}{2} - y_{29}\right)\mathbf{a}_{2} + \left(\frac{1}{2} - z_{29}\right)\mathbf{a}_{3} = x_{29}a\,\hat{\mathbf{x}} + \left(\frac{1}{2} - y_{29}\right)b\,\hat{\mathbf{y}} + \left(\frac{1}{2} - z_{29}\right)c\,\hat{\mathbf{z}} \qquad (8e) \qquad \text{Mg II}$$

$$\mathbf{B}_{209} = -x_{29}\,\mathbf{a}_{1} - y_{29}\,\mathbf{a}_{2} - z_{29}\,\mathbf{a}_{3} = -x_{29}a\,\hat{\mathbf{x}} - y_{29}b\,\hat{\mathbf{y}} - z_{29}c\,\hat{\mathbf{z}} \qquad (8e) \qquad \text{Mg II}$$

$$\mathbf{B}_{210} = \left(\frac{1}{2} + x_{29}\right)\mathbf{a}_{1} + y_{29}\,\mathbf{a}_{2} - z_{29}\,\mathbf{a}_{3} = \left(\frac{1}{2} + x_{29}\right)a\,\hat{\mathbf{x}} + y_{29}b\,\hat{\mathbf{y}} - z_{29}c\,\hat{\mathbf{z}} \qquad (8e) \qquad \text{Mg II}$$

$$\mathbf{B}_{211} = \left(\frac{1}{2} + x_{29}\right)\mathbf{a}_{1} + \left(\frac{1}{2} - y_{29}\right)\mathbf{a}_{2} + \left(\frac{1}{2} + z_{29}\right)\mathbf{a}_{3} = \left(\frac{1}{2} + x_{29}\right)a\,\hat{\mathbf{x}} + \left(\frac{1}{2} - y_{29}\right)b\,\hat{\mathbf{y}} + \left(\frac{1}{2} + z_{29}\right)c\,\hat{\mathbf{z}} \qquad (8e) \qquad \text{Mg II}$$

$$\mathbf{B}_{212} = -x_{29}\,\mathbf{a}_{1} + \left(\frac{1}{2} + y_{29}\right)\mathbf{a}_{2} + \left(\frac{1}{2} + z_{29}\right)\mathbf{a}_{3} = -x_{29}a\,\hat{\mathbf{x}} + \left(\frac{1}{2} + y_{29}\right)b\,\hat{\mathbf{y}} + \left(\frac{1}{2} + z_{29}\right)c\,\hat{\mathbf{z}} \qquad (8e) \qquad \text{Mg II}$$

$$\mathbf{B}_{213} = x_{30}\,\mathbf{a}_{1} + y_{30}\,\mathbf{a}_{2} + z_{30}\,\mathbf{a}_{3} = x_{30}a\,\hat{\mathbf{x}} + y_{30}b\,\hat{\mathbf{y}} + z_{30}c\,\hat{\mathbf{z}} \qquad (8e) \qquad \text{O III}$$

$$\mathbf{B}_{214} = \left(\frac{1}{2} - x_{30}\right)\mathbf{a}_{1} - y_{30}\,\mathbf{a}_{2} + z_{30}\,\mathbf{a}_{3} = \left(\frac{1}{2} - x_{30}\right)a\,\hat{\mathbf{x}} - y_{30}b\,\hat{\mathbf{y}} + z_{30}c\,\hat{\mathbf{z}} \qquad (8e) \qquad \text{O III}$$

$$\mathbf{B}_{215} = \left(\frac{1}{2} - x_{30}\right)\mathbf{a}_{1} + \left(\frac{1}{2} + y_{30}\right)\mathbf{a}_{2} + \left(\frac{1}{2} - z_{30}\right)\mathbf{a}_{3} = \left(\frac{1}{2} - x_{30}\right)a\,\hat{\mathbf{x}} + \left(\frac{1}{2} + y_{30}\right)b\,\hat{\mathbf{y}} + \left(\frac{1}{2} - z_{30}\right)c\,\hat{\mathbf{z}} \qquad (8e) \qquad \text{O III}$$

$$\mathbf{B}_{216} = x_{30}\,\mathbf{a}_{1} + \left(\frac{1}{2} - y_{30}\right)\mathbf{a}_{2} + \left(\frac{1}{2} - z_{30}\right)\mathbf{a}_{3} = x_{30}a\,\hat{\mathbf{x}} + \left(\frac{1}{2} - y_{30}\right)b\,\hat{\mathbf{y}} + \left(\frac{1}{2} - z_{30}\right)c\,\hat{\mathbf{z}} \qquad (8e) \qquad \text{O III}$$

$$\mathbf{B}_{217} = -x_{30}\,\mathbf{a}_{1} - y_{30}\,\mathbf{a}_{2} - z_{30}\,\mathbf{a}_{3} = -x_{30}a\,\hat{\mathbf{x}} - y_{30}b\,\hat{\mathbf{y}} - z_{30}c\,\hat{\mathbf{z}} \qquad (8e) \qquad \text{O III}$$

$$\mathbf{B}_{218} = \left(\frac{1}{2} + x_{30}\right)\mathbf{a}_{1} + y_{30}\,\mathbf{a}_{2} - z_{30}\,\mathbf{a}_{3} = \left(\frac{1}{2} + x_{30}\right)a\,\hat{\mathbf{x}} + y_{30}b\,\hat{\mathbf{y}} - z_{30}c\,\hat{\mathbf{z}} \qquad (8e) \qquad \text{O III}$$

$$\mathbf{B}_{219} = \left(\frac{1}{2} + x_{30}\right)\mathbf{a}_{1} + \left(\frac{1}{2} - y_{30}\right)\mathbf{a}_{2} + \left(\frac{1}{2} + z_{30}\right)\mathbf{a}_{3} = \left(\frac{1}{2} + x_{30}\right)a\,\hat{\mathbf{x}} + \left(\frac{1}{2} - y_{30}\right)b\,\hat{\mathbf{y}} + \left(\frac{1}{2} + z_{30}\right)c\,\hat{\mathbf{z}} \qquad (8e) \qquad \text{O III}$$

$$\mathbf{B}_{220} = -x_{30}\,\mathbf{a}_{1} + \left(\frac{1}{2} + y_{30}\right)\mathbf{a}_{2} + \left(\frac{1}{2} + z_{30}\right)\mathbf{a}_{3} = -x_{30}a\,\hat{\mathbf{x}} + \left(\frac{1}{2} + y_{30}\right)b\,\hat{\mathbf{y}} + \left(\frac{1}{2} + z_{30}\right)c\,\hat{\mathbf{z}} \qquad (8e) \qquad \text{O III}$$

$$\mathbf{B}_{221} = x_{31}\,\mathbf{a}_{1} + y_{31}\,\mathbf{a}_{2} + z_{31}\,\mathbf{a}_{3} = x_{31}a\,\hat{\mathbf{x}} + y_{31}b\,\hat{\mathbf{y}} + z_{31}c\,\hat{\mathbf{z}} \qquad (8e) \qquad \text{O IV}$$

$$\mathbf{B}_{222} = \left(\frac{1}{2} - x_{31}\right)\mathbf{a}_{1} - y_{31}\,\mathbf{a}_{2} + z_{31}\,\mathbf{a}_{3} = \left(\frac{1}{2} - x_{31}\right)a\,\hat{\mathbf{x}} - y_{31}b\,\hat{\mathbf{y}} + z_{31}c\,\hat{\mathbf{z}} \qquad (8e) \qquad \text{O IV}$$

$$\mathbf{B}_{223} = \left(\frac{1}{2} - x_{31}\right)\mathbf{a}_{1} + \left(\frac{1}{2} + y_{31}\right)\mathbf{a}_{2} + \left(\frac{1}{2} - z_{31}\right)\mathbf{a}_{3} = \left(\frac{1}{2} - x_{31}\right)a\,\hat{\mathbf{x}} + \left(\frac{1}{2} + y_{31}\right)b\,\hat{\mathbf{y}} + \left(\frac{1}{2} - z_{31}\right)c\,\hat{\mathbf{z}} \qquad (8e) \qquad \text{O IV}$$

$$\mathbf{B}_{224} = x_{31}\,\mathbf{a}_{1} + \left(\frac{1}{2} - y_{31}\right)\mathbf{a}_{2} + \left(\frac{1}{2} - z_{31}\right)\mathbf{a}_{3} = x_{31}a\,\hat{\mathbf{x}} + \left(\frac{1}{2} - y_{31}\right)b\,\hat{\mathbf{y}} + \left(\frac{1}{2} - z_{31}\right)c\,\hat{\mathbf{z}} \qquad (8e) \qquad \text{O IV}$$

| | | | | | | |
|---|---|---|---|---|---|---|
| $\mathbf{B}_{225}$ | = | $-x_{31}\,\mathbf{a}_{1} - y_{31}\,\mathbf{a}_{2} - z_{31}\,\mathbf{a}_{3}$ | = | $-x_{31}a\,\mathbf{\hat{x}} - y_{31}b\,\mathbf{\hat{y}} - z_{31}c\,\mathbf{\hat{z}}$ | (8$e$) | O IV |
| $\mathbf{B}_{226}$ | = | $\left(\frac{1}{2} + x_{31}\right)\,\mathbf{a}_{1} + y_{31}\,\mathbf{a}_{2} - z_{31}\,\mathbf{a}_{3}$ | = | $\left(\frac{1}{2} + x_{31}\right)a\,\mathbf{\hat{x}} + y_{31}b\,\mathbf{\hat{y}} - z_{31}c\,\mathbf{\hat{z}}$ | (8$e$) | O IV |
| $\mathbf{B}_{227}$ | = | $\left(\frac{1}{2} + x_{31}\right)\,\mathbf{a}_{1} + \left(\frac{1}{2} - y_{31}\right)\,\mathbf{a}_{2} + \left(\frac{1}{2} + z_{31}\right)\,\mathbf{a}_{3}$ | = | $\left(\frac{1}{2} + x_{31}\right)a\,\mathbf{\hat{x}} + \left(\frac{1}{2} - y_{31}\right)b\,\mathbf{\hat{y}} + \left(\frac{1}{2} + z_{31}\right)c\,\mathbf{\hat{z}}$ | (8$e$) | O IV |
| $\mathbf{B}_{228}$ | = | $-x_{31}\,\mathbf{a}_{1} + \left(\frac{1}{2} + y_{31}\right)\,\mathbf{a}_{2} + \left(\frac{1}{2} + z_{31}\right)\,\mathbf{a}_{3}$ | = | $-x_{31}a\,\mathbf{\hat{x}} + \left(\frac{1}{2} + y_{31}\right)b\,\mathbf{\hat{y}} + \left(\frac{1}{2} + z_{31}\right)c\,\mathbf{\hat{z}}$ | (8$e$) | O IV |
| $\mathbf{B}_{229}$ | = | $x_{32}\,\mathbf{a}_{1} + y_{32}\,\mathbf{a}_{2} + z_{32}\,\mathbf{a}_{3}$ | = | $x_{32}a\,\mathbf{\hat{x}} + y_{32}b\,\mathbf{\hat{y}} + z_{32}c\,\mathbf{\hat{z}}$ | (8$e$) | O V |
| $\mathbf{B}_{230}$ | = | $\left(\frac{1}{2} - x_{32}\right)\,\mathbf{a}_{1} - y_{32}\,\mathbf{a}_{2} + z_{32}\,\mathbf{a}_{3}$ | = | $\left(\frac{1}{2} - x_{32}\right)a\,\mathbf{\hat{x}} - y_{32}b\,\mathbf{\hat{y}} + z_{32}c\,\mathbf{\hat{z}}$ | (8$e$) | O V |
| $\mathbf{B}_{231}$ | = | $\left(\frac{1}{2} - x_{32}\right)\,\mathbf{a}_{1} + \left(\frac{1}{2} + y_{32}\right)\,\mathbf{a}_{2} + \left(\frac{1}{2} - z_{32}\right)\,\mathbf{a}_{3}$ | = | $\left(\frac{1}{2} - x_{32}\right)a\,\mathbf{\hat{x}} + \left(\frac{1}{2} + y_{32}\right)b\,\mathbf{\hat{y}} + \left(\frac{1}{2} - z_{32}\right)c\,\mathbf{\hat{z}}$ | (8$e$) | O V |
| $\mathbf{B}_{232}$ | = | $x_{32}\,\mathbf{a}_{1} + \left(\frac{1}{2} - y_{32}\right)\,\mathbf{a}_{2} + \left(\frac{1}{2} - z_{32}\right)\,\mathbf{a}_{3}$ | = | $x_{32}a\,\mathbf{\hat{x}} + \left(\frac{1}{2} - y_{32}\right)b\,\mathbf{\hat{y}} + \left(\frac{1}{2} - z_{32}\right)c\,\mathbf{\hat{z}}$ | (8$e$) | O V |
| $\mathbf{B}_{233}$ | = | $-x_{32}\,\mathbf{a}_{1} - y_{32}\,\mathbf{a}_{2} - z_{32}\,\mathbf{a}_{3}$ | = | $-x_{32}a\,\mathbf{\hat{x}} - y_{32}b\,\mathbf{\hat{y}} - z_{32}c\,\mathbf{\hat{z}}$ | (8$e$) | O V |
| $\mathbf{B}_{234}$ | = | $\left(\frac{1}{2} + x_{32}\right)\,\mathbf{a}_{1} + y_{32}\,\mathbf{a}_{2} - z_{32}\,\mathbf{a}_{3}$ | = | $\left(\frac{1}{2} + x_{32}\right)a\,\mathbf{\hat{x}} + y_{32}b\,\mathbf{\hat{y}} - z_{32}c\,\mathbf{\hat{z}}$ | (8$e$) | O V |
| $\mathbf{B}_{235}$ | = | $\left(\frac{1}{2} + x_{32}\right)\,\mathbf{a}_{1} + \left(\frac{1}{2} - y_{32}\right)\,\mathbf{a}_{2} + \left(\frac{1}{2} + z_{32}\right)\,\mathbf{a}_{3}$ | = | $\left(\frac{1}{2} + x_{32}\right)a\,\mathbf{\hat{x}} + \left(\frac{1}{2} - y_{32}\right)b\,\mathbf{\hat{y}} + \left(\frac{1}{2} + z_{32}\right)c\,\mathbf{\hat{z}}$ | (8$e$) | O V |
| $\mathbf{B}_{236}$ | = | $-x_{32}\,\mathbf{a}_{1} + \left(\frac{1}{2} + y_{32}\right)\,\mathbf{a}_{2} + \left(\frac{1}{2} + z_{32}\right)\,\mathbf{a}_{3}$ | = | $-x_{32}a\,\mathbf{\hat{x}} + \left(\frac{1}{2} + y_{32}\right)b\,\mathbf{\hat{y}} + \left(\frac{1}{2} + z_{32}\right)c\,\mathbf{\hat{z}}$ | (8$e$) | O V |
| $\mathbf{B}_{237}$ | = | $x_{33}\,\mathbf{a}_{1} + y_{33}\,\mathbf{a}_{2} + z_{33}\,\mathbf{a}_{3}$ | = | $x_{33}a\,\mathbf{\hat{x}} + y_{33}b\,\mathbf{\hat{y}} + z_{33}c\,\mathbf{\hat{z}}$ | (8$e$) | O VI |
| $\mathbf{B}_{238}$ | = | $\left(\frac{1}{2} - x_{33}\right)\,\mathbf{a}_{1} - y_{33}\,\mathbf{a}_{2} + z_{33}\,\mathbf{a}_{3}$ | = | $\left(\frac{1}{2} - x_{33}\right)a\,\mathbf{\hat{x}} - y_{33}b\,\mathbf{\hat{y}} + z_{33}c\,\mathbf{\hat{z}}$ | (8$e$) | O VI |
| $\mathbf{B}_{239}$ | = | $\left(\frac{1}{2} - x_{33}\right)\,\mathbf{a}_{1} + \left(\frac{1}{2} + y_{33}\right)\,\mathbf{a}_{2} + \left(\frac{1}{2} - z_{33}\right)\,\mathbf{a}_{3}$ | = | $\left(\frac{1}{2} - x_{33}\right)a\,\mathbf{\hat{x}} + \left(\frac{1}{2} + y_{33}\right)b\,\mathbf{\hat{y}} + \left(\frac{1}{2} - z_{33}\right)c\,\mathbf{\hat{z}}$ | (8$e$) | O VI |
| $\mathbf{B}_{240}$ | = | $x_{33}\,\mathbf{a}_{1} + \left(\frac{1}{2} - y_{33}\right)\,\mathbf{a}_{2} + \left(\frac{1}{2} - z_{33}\right)\,\mathbf{a}_{3}$ | = | $x_{33}a\,\mathbf{\hat{x}} + \left(\frac{1}{2} - y_{33}\right)b\,\mathbf{\hat{y}} + \left(\frac{1}{2} - z_{33}\right)c\,\mathbf{\hat{z}}$ | (8$e$) | O VI |
| $\mathbf{B}_{241}$ | = | $-x_{33}\,\mathbf{a}_{1} - y_{33}\,\mathbf{a}_{2} - z_{33}\,\mathbf{a}_{3}$ | = | $-x_{33}a\,\mathbf{\hat{x}} - y_{33}b\,\mathbf{\hat{y}} - z_{33}c\,\mathbf{\hat{z}}$ | (8$e$) | O VI |
| $\mathbf{B}_{242}$ | = | $\left(\frac{1}{2} + x_{33}\right)\,\mathbf{a}_{1} + y_{33}\,\mathbf{a}_{2} - z_{33}\,\mathbf{a}_{3}$ | = | $\left(\frac{1}{2} + x_{33}\right)a\,\mathbf{\hat{x}} + y_{33}b\,\mathbf{\hat{y}} - z_{33}c\,\mathbf{\hat{z}}$ | (8$e$) | O VI |
| $\mathbf{B}_{243}$ | = | $\left(\frac{1}{2} + x_{33}\right)\,\mathbf{a}_{1} + \left(\frac{1}{2} - y_{33}\right)\,\mathbf{a}_{2} + \left(\frac{1}{2} + z_{33}\right)\,\mathbf{a}_{3}$ | = | $\left(\frac{1}{2} + x_{33}\right)a\,\mathbf{\hat{x}} + \left(\frac{1}{2} - y_{33}\right)b\,\mathbf{\hat{y}} + \left(\frac{1}{2} + z_{33}\right)c\,\mathbf{\hat{z}}$ | (8$e$) | O VI |
| $\mathbf{B}_{244}$ | = | $-x_{33}\,\mathbf{a}_{1} + \left(\frac{1}{2} + y_{33}\right)\,\mathbf{a}_{2} + \left(\frac{1}{2} + z_{33}\right)\,\mathbf{a}_{3}$ | = | $-x_{33}a\,\mathbf{\hat{x}} + \left(\frac{1}{2} + y_{33}\right)b\,\mathbf{\hat{y}} + \left(\frac{1}{2} + z_{33}\right)c\,\mathbf{\hat{z}}$ | (8$e$) | O VI |
| $\mathbf{B}_{245}$ | = | $x_{34}\,\mathbf{a}_{1} + y_{34}\,\mathbf{a}_{2} + z_{34}\,\mathbf{a}_{3}$ | = | $x_{34}a\,\mathbf{\hat{x}} + y_{34}b\,\mathbf{\hat{y}} + z_{34}c\,\mathbf{\hat{z}}$ | (8$e$) | O VII |
| $\mathbf{B}_{246}$ | = | $\left(\frac{1}{2} - x_{34}\right)\,\mathbf{a}_{1} - y_{34}\,\mathbf{a}_{2} + z_{34}\,\mathbf{a}_{3}$ | = | $\left(\frac{1}{2} - x_{34}\right)a\,\mathbf{\hat{x}} - y_{34}b\,\mathbf{\hat{y}} + z_{34}c\,\mathbf{\hat{z}}$ | (8$e$) | O VII |
| $\mathbf{B}_{247}$ | = | $\left(\frac{1}{2} - x_{34}\right)\,\mathbf{a}_{1} + \left(\frac{1}{2} + y_{34}\right)\,\mathbf{a}_{2} + \left(\frac{1}{2} - z_{34}\right)\,\mathbf{a}_{3}$ | = | $\left(\frac{1}{2} - x_{34}\right)a\,\mathbf{\hat{x}} + \left(\frac{1}{2} + y_{34}\right)b\,\mathbf{\hat{y}} + \left(\frac{1}{2} - z_{34}\right)c\,\mathbf{\hat{z}}$ | (8$e$) | O VII |
| $\mathbf{B}_{248}$ | = | $x_{34}\,\mathbf{a}_{1} + \left(\frac{1}{2} - y_{34}\right)\,\mathbf{a}_{2} + \left(\frac{1}{2} - z_{34}\right)\,\mathbf{a}_{3}$ | = | $x_{34}a\,\mathbf{\hat{x}} + \left(\frac{1}{2} - y_{34}\right)b\,\mathbf{\hat{y}} + \left(\frac{1}{2} - z_{34}\right)c\,\mathbf{\hat{z}}$ | (8$e$) | O VII |
| $\mathbf{B}_{249}$ | = | $-x_{34}\,\mathbf{a}_{1} - y_{34}\,\mathbf{a}_{2} - z_{34}\,\mathbf{a}_{3}$ | = | $-x_{34}a\,\mathbf{\hat{x}} - y_{34}b\,\mathbf{\hat{y}} - z_{34}c\,\mathbf{\hat{z}}$ | (8$e$) | O VII |
| $\mathbf{B}_{250}$ | = | $\left(\frac{1}{2} + x_{34}\right)\,\mathbf{a}_{1} + y_{34}\,\mathbf{a}_{2} - z_{34}\,\mathbf{a}_{3}$ | = | $\left(\frac{1}{2} + x_{34}\right)a\,\mathbf{\hat{x}} + y_{34}b\,\mathbf{\hat{y}} - z_{34}c\,\mathbf{\hat{z}}$ | (8$e$) | O VII |
| $\mathbf{B}_{251}$ | = | $\left(\frac{1}{2} + x_{34}\right)\,\mathbf{a}_{1} + \left(\frac{1}{2} - y_{34}\right)\,\mathbf{a}_{2} + \left(\frac{1}{2} + z_{34}\right)\,\mathbf{a}_{3}$ | = | $\left(\frac{1}{2} + x_{34}\right)a\,\mathbf{\hat{x}} + \left(\frac{1}{2} - y_{34}\right)b\,\mathbf{\hat{y}} + \left(\frac{1}{2} + z_{34}\right)c\,\mathbf{\hat{z}}$ | (8$e$) | O VII |
| $\mathbf{B}_{252}$ | = | $-x_{34}\,\mathbf{a}_{1} + \left(\frac{1}{2} + y_{34}\right)\,\mathbf{a}_{2} + \left(\frac{1}{2} + z_{34}\right)\,\mathbf{a}_{3}$ | = | $-x_{34}a\,\mathbf{\hat{x}} + \left(\frac{1}{2} + y_{34}\right)b\,\mathbf{\hat{y}} + \left(\frac{1}{2} + z_{34}\right)c\,\mathbf{\hat{z}}$ | (8$e$) | O VII |
| $\mathbf{B}_{253}$ | = | $x_{35}\,\mathbf{a}_1 + y_{35}\,\mathbf{a}_2 + z_{35}\,\mathbf{a}_3$ | = | $x_{35}a\,\hat{\mathbf{x}} + y_{35}b\,\hat{\mathbf{y}} + z_{35}c\,\hat{\mathbf{z}}$ | (8$e$) | O VIII |
| $\mathbf{B}_{254}$ | = | $\left(\frac{1}{2} - x_{35}\right)\mathbf{a}_1 - y_{35}\,\mathbf{a}_2 + z_{35}\,\mathbf{a}_3$ | = | $\left(\frac{1}{2} - x_{35}\right)a\,\hat{\mathbf{x}} - y_{35}b\,\hat{\mathbf{y}} + z_{35}c\,\hat{\mathbf{z}}$ | (8$e$) | O VIII |
| $\mathbf{B}_{255}$ | = | $\left(\frac{1}{2} - x_{35}\right)\mathbf{a}_1 + \left(\frac{1}{2} + y_{35}\right)\mathbf{a}_2 + \left(\frac{1}{2} - z_{35}\right)\mathbf{a}_3$ | = | $\left(\frac{1}{2} - x_{35}\right)a\,\hat{\mathbf{x}} + \left(\frac{1}{2} + y_{35}\right)b\,\hat{\mathbf{y}} + \left(\frac{1}{2} - z_{35}\right)c\,\hat{\mathbf{z}}$ | (8$e$) | O VIII |
| $\mathbf{B}_{256}$ | = | $x_{35}\,\mathbf{a}_1 + \left(\frac{1}{2} - y_{35}\right)\mathbf{a}_2 + \left(\frac{1}{2} - z_{35}\right)\mathbf{a}_3$ | = | $x_{35}a\,\hat{\mathbf{x}} + \left(\frac{1}{2} - y_{35}\right)b\,\hat{\mathbf{y}} + \left(\frac{1}{2} - z_{35}\right)c\,\hat{\mathbf{z}}$ | (8$e$) | O VIII |
| $\mathbf{B}_{257}$ | = | $-x_{35}\,\mathbf{a}_1 - y_{35}\,\mathbf{a}_2 - z_{35}\,\mathbf{a}_3$ | = | $-x_{35}a\,\hat{\mathbf{x}} - y_{35}b\,\hat{\mathbf{y}} - z_{35}c\,\hat{\mathbf{z}}$ | (8$e$) | O VIII |
| $\mathbf{B}_{258}$ | = | $\left(\frac{1}{2} + x_{35}\right)\mathbf{a}_1 + y_{35}\,\mathbf{a}_2 - z_{35}\,\mathbf{a}_3$ | = | $\left(\frac{1}{2} + x_{35}\right)a\,\hat{\mathbf{x}} + y_{35}b\,\hat{\mathbf{y}} - z_{35}c\,\hat{\mathbf{z}}$ | (8$e$) | O VIII |
| $\mathbf{B}_{259}$ | = | $\left(\frac{1}{2} + x_{35}\right)\mathbf{a}_1 + \left(\frac{1}{2} - y_{35}\right)\mathbf{a}_2 + \left(\frac{1}{2} + z_{35}\right)\mathbf{a}_3$ | = | $\left(\frac{1}{2} + x_{35}\right)a\,\hat{\mathbf{x}} + \left(\frac{1}{2} - y_{35}\right)b\,\hat{\mathbf{y}} + \left(\frac{1}{2} + z_{35}\right)c\,\hat{\mathbf{z}}$ | (8$e$) | O VIII |
| $\mathbf{B}_{260}$ | = | $-x_{35}\,\mathbf{a}_1 + \left(\frac{1}{2} + y_{35}\right)\mathbf{a}_2 + \left(\frac{1}{2} + z_{35}\right)\mathbf{a}_3$ | = | $-x_{35}a\,\hat{\mathbf{x}} + \left(\frac{1}{2} + y_{35}\right)b\,\hat{\mathbf{y}} + \left(\frac{1}{2} + z_{35}\right)c\,\hat{\mathbf{z}}$ | (8$e$) | O VIII |
| $\mathbf{B}_{261}$ | = | $x_{36}\,\mathbf{a}_1 + y_{36}\,\mathbf{a}_2 + z_{36}\,\mathbf{a}_3$ | = | $x_{36}a\,\hat{\mathbf{x}} + y_{36}b\,\hat{\mathbf{y}} + z_{36}c\,\hat{\mathbf{z}}$ | (8$e$) | O IX |
| $\mathbf{B}_{262}$ | = | $\left(\frac{1}{2} - x_{36}\right)\mathbf{a}_1 - y_{36}\,\mathbf{a}_2 + z_{36}\,\mathbf{a}_3$ | = | $\left(\frac{1}{2} - x_{36}\right)a\,\hat{\mathbf{x}} - y_{36}b\,\hat{\mathbf{y}} + z_{36}c\,\hat{\mathbf{z}}$ | (8$e$) | O IX |
| $\mathbf{B}_{263}$ | = | $\left(\frac{1}{2} - x_{36}\right)\mathbf{a}_1 + \left(\frac{1}{2} + y_{36}\right)\mathbf{a}_2 + \left(\frac{1}{2} - z_{36}\right)\mathbf{a}_3$ | = | $\left(\frac{1}{2} - x_{36}\right)a\,\hat{\mathbf{x}} + \left(\frac{1}{2} + y_{36}\right)b\,\hat{\mathbf{y}} + \left(\frac{1}{2} - z_{36}\right)c\,\hat{\mathbf{z}}$ | (8$e$) | O IX |
| $\mathbf{B}_{264}$ | = | $x_{36}\,\mathbf{a}_1 + \left(\frac{1}{2} - y_{36}\right)\mathbf{a}_2 + \left(\frac{1}{2} - z_{36}\right)\mathbf{a}_3$ | = | $x_{36}a\,\hat{\mathbf{x}} + \left(\frac{1}{2} - y_{36}\right)b\,\hat{\mathbf{y}} + \left(\frac{1}{2} - z_{36}\right)c\,\hat{\mathbf{z}}$ | (8$e$) | O IX |
| $\mathbf{B}_{265}$ | = | $-x_{36}\,\mathbf{a}_1 - y_{36}\,\mathbf{a}_2 - z_{36}\,\mathbf{a}_3$ | = | $-x_{36}a\,\hat{\mathbf{x}} - y_{36}b\,\hat{\mathbf{y}} - z_{36}c\,\hat{\mathbf{z}}$ | (8$e$) | O IX |
| $\mathbf{B}_{266}$ | = | $\left(\frac{1}{2} + x_{36}\right)\mathbf{a}_1 + y_{36}\,\mathbf{a}_2 - z_{36}\,\mathbf{a}_3$ | = | $\left(\frac{1}{2} + x_{36}\right)a\,\hat{\mathbf{x}} + y_{36}b\,\hat{\mathbf{y}} - z_{36}c\,\hat{\mathbf{z}}$ | (8$e$) | O IX |
| $\mathbf{B}_{267}$ | = | $\left(\frac{1}{2} + x_{36}\right)\mathbf{a}_1 + \left(\frac{1}{2} - y_{36}\right)\mathbf{a}_2 + \left(\frac{1}{2} + z_{36}\right)\mathbf{a}_3$ | = | $\left(\frac{1}{2} + x_{36}\right)a\,\hat{\mathbf{x}} + \left(\frac{1}{2} - y_{36}\right)b\,\hat{\mathbf{y}} + \left(\frac{1}{2} + z_{36}\right)c\,\hat{\mathbf{z}}$ | (8$e$) | O IX |
| $\mathbf{B}_{268}$ | = | $-x_{36}\,\mathbf{a}_1 + \left(\frac{1}{2} + y_{36}\right)\mathbf{a}_2 + \left(\frac{1}{2} + z_{36}\right)\mathbf{a}_3$ | = | $-x_{36}a\,\hat{\mathbf{x}} + \left(\frac{1}{2} + y_{36}\right)b\,\hat{\mathbf{y}} + \left(\frac{1}{2} + z_{36}\right)c\,\hat{\mathbf{z}}$ | (8$e$) | O IX |
| $\mathbf{B}_{269}$ | = | $x_{37}\,\mathbf{a}_1 + y_{37}\,\mathbf{a}_2 + z_{37}\,\mathbf{a}_3$ | = | $x_{37}a\,\hat{\mathbf{x}} + y_{37}b\,\hat{\mathbf{y}} + z_{37}c\,\hat{\mathbf{z}}$ | (8$e$) | O X |
| $\mathbf{B}_{270}$ | = | $\left(\frac{1}{2} - x_{37}\right)\mathbf{a}_1 - y_{37}\,\mathbf{a}_2 + z_{37}\,\mathbf{a}_3$ | = | $\left(\frac{1}{2} - x_{37}\right)a\,\hat{\mathbf{x}} - y_{37}b\,\hat{\mathbf{y}} + z_{37}c\,\hat{\mathbf{z}}$ | (8$e$) | O X |
| $\mathbf{B}_{271}$ | = | $\left(\frac{1}{2} - x_{37}\right)\mathbf{a}_1 + \left(\frac{1}{2} + y_{37}\right)\mathbf{a}_2 + \left(\frac{1}{2} - z_{37}\right)\mathbf{a}_3$ | = | $\left(\frac{1}{2} - x_{37}\right)a\,\hat{\mathbf{x}} + \left(\frac{1}{2} + y_{37}\right)b\,\hat{\mathbf{y}} + \left(\frac{1}{2} - z_{37}\right)c\,\hat{\mathbf{z}}$ | (8$e$) | O X |
| $\mathbf{B}_{272}$ | = | $x_{37}\,\mathbf{a}_1 + \left(\frac{1}{2} - y_{37}\right)\mathbf{a}_2 + \left(\frac{1}{2} - z_{37}\right)\mathbf{a}_3$ | = | $x_{37}a\,\hat{\mathbf{x}} + \left(\frac{1}{2} - y_{37}\right)b\,\hat{\mathbf{y}} + \left(\frac{1}{2} - z_{37}\right)c\,\hat{\mathbf{z}}$ | (8$e$) | O X |
| $\mathbf{B}_{273}$ | = | $-x_{37}\,\mathbf{a}_1 - y_{37}\,\mathbf{a}_2 - z_{37}\,\mathbf{a}_3$ | = | $-x_{37}a\,\hat{\mathbf{x}} - y_{37}b\,\hat{\mathbf{y}} - z_{37}c\,\hat{\mathbf{z}}$ | (8$e$) | O X |
| $\mathbf{B}_{274}$ | = | $\left(\frac{1}{2} + x_{37}\right)\mathbf{a}_1 + y_{37}\,\mathbf{a}_2 - z_{37}\,\mathbf{a}_3$ | = | $\left(\frac{1}{2} + x_{37}\right)a\,\hat{\mathbf{x}} + y_{37}b\,\hat{\mathbf{y}} - z_{37}c\,\hat{\mathbf{z}}$ | (8$e$) | O X |
| $\mathbf{B}_{275}$ | = | $\left(\frac{1}{2} + x_{37}\right)\mathbf{a}_1 + \left(\frac{1}{2} - y_{37}\right)\mathbf{a}_2 + \left(\frac{1}{2} + z_{37}\right)\mathbf{a}_3$ | = | $\left(\frac{1}{2} + x_{37}\right)a\,\hat{\mathbf{x}} + \left(\frac{1}{2} - y_{37}\right)b\,\hat{\mathbf{y}} + \left(\frac{1}{2} + z_{37}\right)c\,\hat{\mathbf{z}}$ | (8$e$) | O X |
| $\mathbf{B}_{276}$ | = | $-x_{37}\,\mathbf{a}_1 + \left(\frac{1}{2} + y_{37}\right)\mathbf{a}_2 + \left(\frac{1}{2} + z_{37}\right)\mathbf{a}_3$ | = | $-x_{37}a\,\hat{\mathbf{x}} + \left(\frac{1}{2} + y_{37}\right)b\,\hat{\mathbf{y}} + \left(\frac{1}{2} + z_{37}\right)c\,\hat{\mathbf{z}}$ | (8$e$) | O X |

**Geometry files:**
- CIF: pp. 1608
- POSCAR: pp. 1609

# Eriochalcite ($CuCl_2 \cdot 2H_2O$, $C45$) Structure: A2BC4D2_oP18_53_h_a_i_e

http://aflow.org/prototype-encyclopedia/A2BC4D2_oP18_53_h_a_i_e

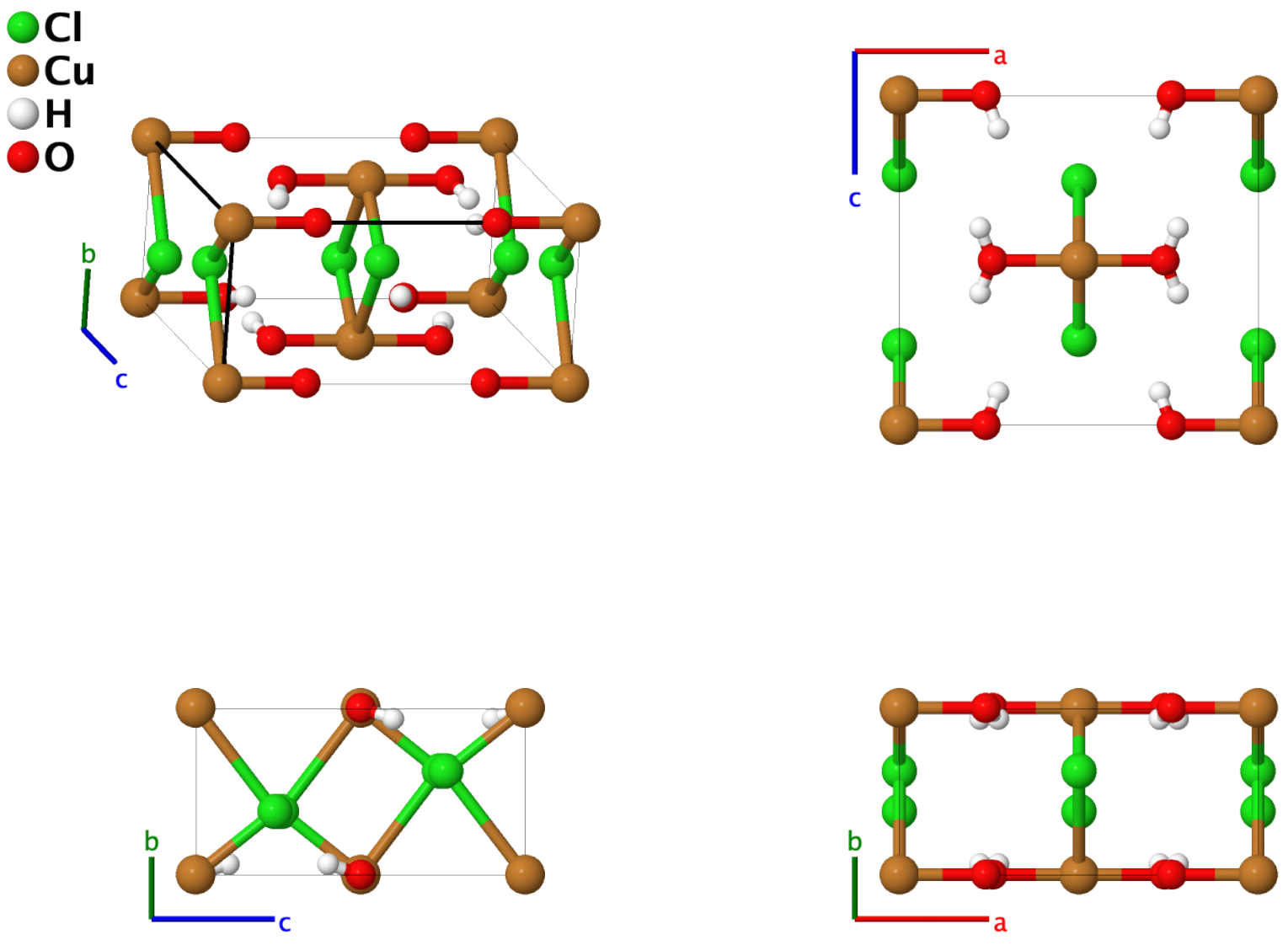

| | | |
|---|---|---|
| **Prototype** | : | $Cl_2CuH_4O_2$ |
| **AFLOW prototype label** | : | A2BC4D2_oP18_53_h_a_i_e |
| ***Strukturbericht* designation** | : | $C45$ |
| **Pearson symbol** | : | oP18 |
| **Space group number** | : | 53 |
| **Space group symbol** | : | $Pmna$ |
| **AFLOW prototype command** | : | `aflow --proto=A2BC4D2_oP18_53_h_a_i_e` <br> `--params=`$a, b/a, c/a, x_2, y_3, z_3, x_4, y_4, z_4$ |

**Simple Orthorhombic primitive vectors:**

$$\mathbf{a}_1 = a\,\hat{\mathbf{x}}$$
$$\mathbf{a}_2 = b\,\hat{\mathbf{y}}$$
$$\mathbf{a}_3 = c\,\hat{\mathbf{z}}$$

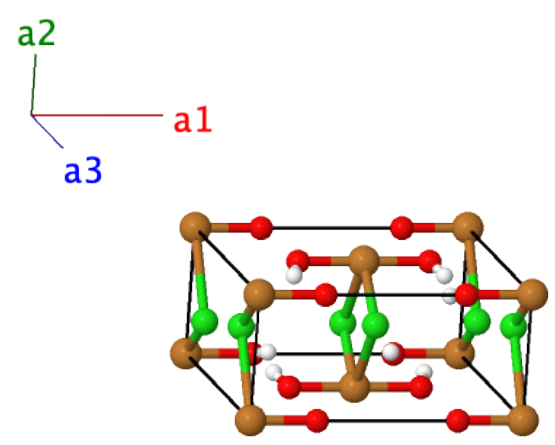

**Basis vectors:**

| | | Lattice Coordinates | | Cartesian Coordinates | Wyckoff Position | Atom Type |
|---|---|---|---|---|---|---|
| $\mathbf{B}_1$ | = | $0\,\mathbf{a}_1 + 0\,\mathbf{a}_2 + 0\,\mathbf{a}_3$ | = | $0\,\hat{\mathbf{x}} + 0\,\hat{\mathbf{y}} + 0\,\hat{\mathbf{z}}$ | $(2a)$ | Cu |
| $\mathbf{B}_2$ | = | $\frac{1}{2}\,\mathbf{a}_1 + \frac{1}{2}\,\mathbf{a}_3$ | = | $\frac{1}{2}a\,\hat{\mathbf{x}} + \frac{1}{2}c\,\hat{\mathbf{z}}$ | $(2a)$ | Cu |
| $\mathbf{B}_3$ | = | $x_2\,\mathbf{a}_1$ | = | $x_2 a\,\hat{\mathbf{x}}$ | $(4e)$ | O |
| $\mathbf{B}_4$ | = | $\left(\frac{1}{2} - x_2\right)\,\mathbf{a}_1 + \frac{1}{2}\,\mathbf{a}_3$ | = | $\left(\frac{1}{2} - x_2\right)a\,\hat{\mathbf{x}} + \frac{1}{2}c\,\hat{\mathbf{z}}$ | (4e) | O |
| $\mathbf{B}_5$ | = | $-x_2\,\mathbf{a}_1$ | = | $-x_2 a\,\hat{\mathbf{x}}$ | (4e) | O |
| $\mathbf{B}_6$ | = | $\left(\frac{1}{2} + x_2\right)\,\mathbf{a}_1 + \frac{1}{2}\,\mathbf{a}_3$ | = | $\left(\frac{1}{2} + x_2\right)a\,\hat{\mathbf{x}} + \frac{1}{2}c\,\hat{\mathbf{z}}$ | (4e) | O |
| $\mathbf{B}_7$ | = | $y_3\,\mathbf{a}_2 + z_3\,\mathbf{a}_3$ | = | $y_3 b\,\hat{\mathbf{y}} + z_3 c\,\hat{\mathbf{z}}$ | (4h) | Cl |
| $\mathbf{B}_8$ | = | $\frac{1}{2}\,\mathbf{a}_1 - y_3\,\mathbf{a}_2 + \left(\frac{1}{2} + z_3\right)\,\mathbf{a}_3$ | = | $\frac{1}{2}a\,\hat{\mathbf{x}} - y_3 b\,\hat{\mathbf{y}} + \left(\frac{1}{2} + z_3\right)c\,\hat{\mathbf{z}}$ | (4h) | Cl |
| $\mathbf{B}_9$ | = | $\frac{1}{2}\,\mathbf{a}_1 + y_3\,\mathbf{a}_2 + \left(\frac{1}{2} - z_3\right)\,\mathbf{a}_3$ | = | $\frac{1}{2}a\,\hat{\mathbf{x}} + y_3 b\,\hat{\mathbf{y}} + \left(\frac{1}{2} - z_3\right)c\,\hat{\mathbf{z}}$ | (4h) | Cl |
| $\mathbf{B}_{10}$ | = | $-y_3\,\mathbf{a}_2 - z_3\,\mathbf{a}_3$ | = | $-y_3 b\,\hat{\mathbf{y}} - z_3 c\,\hat{\mathbf{z}}$ | (4h) | Cl |
| $\mathbf{B}_{11}$ | = | $x_4\,\mathbf{a}_1 + y_4\,\mathbf{a}_2 + z_4\,\mathbf{a}_3$ | = | $x_4 a\,\hat{\mathbf{x}} + y_4 b\,\hat{\mathbf{y}} + z_4 c\,\hat{\mathbf{z}}$ | (8i) | H |
| $\mathbf{B}_{12}$ | = | $\left(\frac{1}{2} - x_4\right)\,\mathbf{a}_1 - y_4\,\mathbf{a}_2 + \left(\frac{1}{2} + z_4\right)\,\mathbf{a}_3$ | = | $\left(\frac{1}{2} - x_4\right)a\,\hat{\mathbf{x}} - y_4 b\,\hat{\mathbf{y}} + \left(\frac{1}{2} + z_4\right)c\,\hat{\mathbf{z}}$ | (8i) | H |
| $\mathbf{B}_{13}$ | = | $\left(\frac{1}{2} - x_4\right)\,\mathbf{a}_1 + y_4\,\mathbf{a}_2 + \left(\frac{1}{2} - z_4\right)\,\mathbf{a}_3$ | = | $\left(\frac{1}{2} - x_4\right)a\,\hat{\mathbf{x}} + y_4 b\,\hat{\mathbf{y}} + \left(\frac{1}{2} - z_4\right)c\,\hat{\mathbf{z}}$ | (8i) | H |
| $\mathbf{B}_{14}$ | = | $x_4\,\mathbf{a}_1 - y_4\,\mathbf{a}_2 - z_4\,\mathbf{a}_3$ | = | $x_4 a\,\hat{\mathbf{x}} - y_4 b\,\hat{\mathbf{y}} - z_4 c\,\hat{\mathbf{z}}$ | (8i) | H |
| $\mathbf{B}_{15}$ | = | $-x_4\,\mathbf{a}_1 - y_4\,\mathbf{a}_2 - z_4\,\mathbf{a}_3$ | = | $-x_4 a\,\hat{\mathbf{x}} - y_4 b\,\hat{\mathbf{y}} - z_4 c\,\hat{\mathbf{z}}$ | (8i) | H |
| $\mathbf{B}_{16}$ | = | $\left(\frac{1}{2} + x_4\right)\,\mathbf{a}_1 + y_4\,\mathbf{a}_2 + \left(\frac{1}{2} - z_4\right)\,\mathbf{a}_3$ | = | $\left(\frac{1}{2} + x_4\right)a\,\hat{\mathbf{x}} + y_4 b\,\hat{\mathbf{y}} + \left(\frac{1}{2} - z_4\right)c\,\hat{\mathbf{z}}$ | (8i) | H |
| $\mathbf{B}_{17}$ | = | $\left(\frac{1}{2} + x_4\right)\,\mathbf{a}_1 - y_4\,\mathbf{a}_2 + \left(\frac{1}{2} + z_4\right)\,\mathbf{a}_3$ | = | $\left(\frac{1}{2} + x_4\right)a\,\hat{\mathbf{x}} - y_4 b\,\hat{\mathbf{y}} + \left(\frac{1}{2} + z_4\right)c\,\hat{\mathbf{z}}$ | (8i) | H |
| $\mathbf{B}_{18}$ | = | $-x_4\,\mathbf{a}_1 + y_4\,\mathbf{a}_2 + z_4\,\mathbf{a}_3$ | = | $-x_4 a\,\hat{\mathbf{x}} + y_4 b\,\hat{\mathbf{y}} + z_4 c\,\hat{\mathbf{z}}$ | (8i) | H |

---

---

**Geometry files:**

- CIF: pp. 1610
- POSCAR: pp. 1611

# $NH_4HF_2$ ($F5_8$) Structure: A2BC_oP16_53_eh_ab_g

http://aflow.org/prototype-encyclopedia/A2BC_oP16_53_eh_ab_g

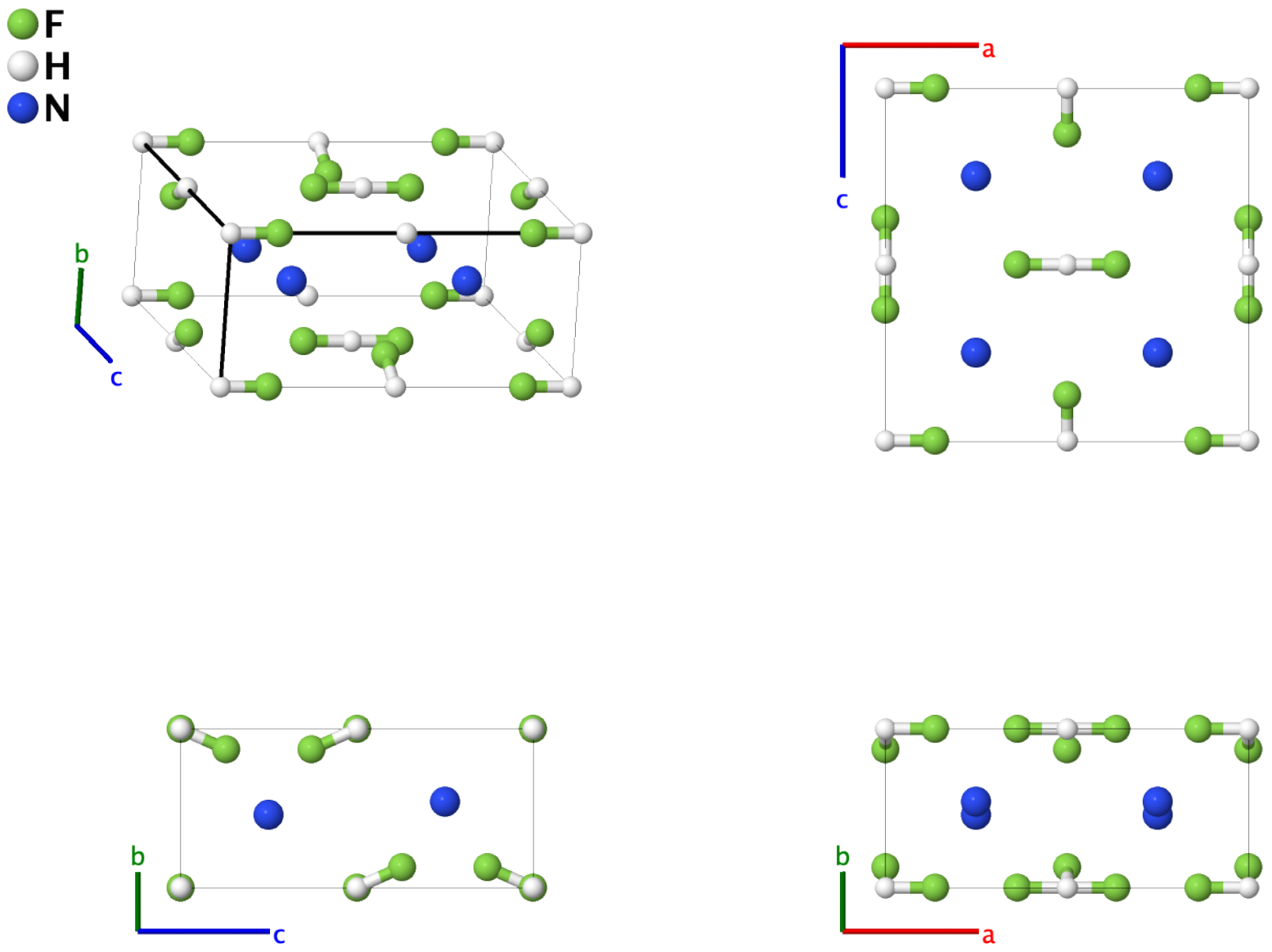

| | | |
|---|---|---|
| **Prototype** | : | $F_2H_5N$ |
| **AFLOW prototype label** | : | A2BC_oP16_53_eh_ab_g |
| ***Strukturbericht* designation** | : | $F5_8$ |
| **Pearson symbol** | : | oP16 |
| **Space group number** | : | 53 |
| **Space group symbol** | : | *Pmna* |
| **AFLOW prototype command** | : | `aflow --proto=A2BC_oP16_53_eh_ab_g` `--params=`$a, b/a, c/a, x_3, y_4, y_5, z_5$ |

- This structure was first investigated by (Pauling, 1933) and assigned *Strukturbericht* designation $F5_8$ by (Gottfried, 1937). It was reinvestigated by (Rogers, 1940). Neither paper notes the positions of the hydrogen atoms, but under the assumption that the structure is similar to $KHF_2$ ($F5_2$), (Downs, 2003) puts some of them the atoms between pairs of fluorine atoms. The remaining hydrogen atoms are part of the $NH_4$ radical.
- The crystal structure was given in the *Pman* setting of space group #53. We used FINDSYM to change it to the standard *Pmna* structure.

**Simple Orthorhombic primitive vectors:**

$$\mathbf{a}_1 = a\,\hat{\mathbf{x}}$$
$$\mathbf{a}_2 = b\,\hat{\mathbf{y}}$$
$$\mathbf{a}_3 = c\,\hat{\mathbf{z}}$$

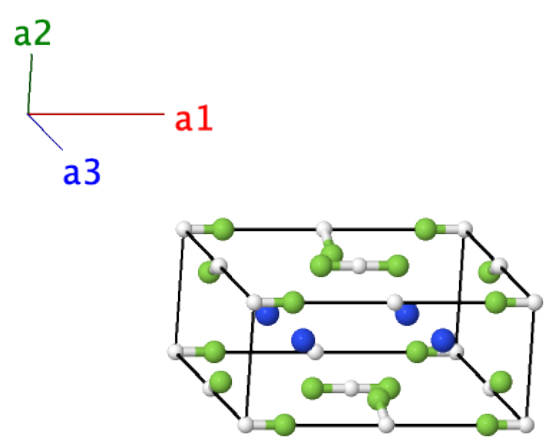

**Basis vectors:**

| | | Lattice Coordinates | | Cartesian Coordinates | Wyckoff Position | Atom Type |
|---|---|---|---|---|---|---|
| $\mathbf{B}_1$ | = | $0\,\mathbf{a}_1 + 0\,\mathbf{a}_2 + 0\,\mathbf{a}_3$ | = | $0\,\hat{\mathbf{x}} + 0\,\hat{\mathbf{y}} + 0\,\hat{\mathbf{z}}$ | $(2a)$ | H I |
| $\mathbf{B}_2$ | = | $\frac{1}{2}\,\mathbf{a}_1 + \frac{1}{2}\,\mathbf{a}_3$ | = | $\frac{1}{2}a\,\hat{\mathbf{x}} + \frac{1}{2}c\,\hat{\mathbf{z}}$ | $(2a)$ | H I |
| $\mathbf{B}_3$ | = | $\frac{1}{2}\,\mathbf{a}_1$ | = | $\frac{1}{2}a\,\hat{\mathbf{x}}$ | $(2b)$ | H II |
| $\mathbf{B}_4$ | = | $\frac{1}{2}\,\mathbf{a}_3$ | = | $\frac{1}{2}c\,\hat{\mathbf{z}}$ | $(2b)$ | H II |
| $\mathbf{B}_5$ | = | $x_3\,\mathbf{a}_1$ | = | $x_3 a\,\hat{\mathbf{x}}$ | $(4e)$ | F I |
| $\mathbf{B}_6$ | = | $\left(\frac{1}{2} - x_3\right)\,\mathbf{a}_1 + \frac{1}{2}\,\mathbf{a}_3$ | = | $\left(\frac{1}{2} - x_3\right)a\,\hat{\mathbf{x}} + \frac{1}{2}c\,\hat{\mathbf{z}}$ | $(4e)$ | F I |
| $\mathbf{B}_7$ | = | $-x_3\,\mathbf{a}_1$ | = | $-x_3 a\,\hat{\mathbf{x}}$ | $(4e)$ | F I |
| $\mathbf{B}_8$ | = | $\left(\frac{1}{2} + x_3\right)\,\mathbf{a}_1 + \frac{1}{2}\,\mathbf{a}_3$ | = | $\left(\frac{1}{2} + x_3\right)a\,\hat{\mathbf{x}} + \frac{1}{2}c\,\hat{\mathbf{z}}$ | $(4e)$ | F I |
| $\mathbf{B}_9$ | = | $\frac{1}{4}\,\mathbf{a}_1 + y_4\,\mathbf{a}_2 + \frac{1}{4}\,\mathbf{a}_3$ | = | $\frac{1}{4}a\,\hat{\mathbf{x}} + y_4 b\,\hat{\mathbf{y}} + \frac{1}{4}c\,\hat{\mathbf{z}}$ | $(4g)$ | $NH_4$ |
| $\mathbf{B}_{10}$ | = | $\frac{1}{4}\,\mathbf{a}_1 - y_4\,\mathbf{a}_2 + \frac{3}{4}\,\mathbf{a}_3$ | = | $\frac{1}{4}a\,\hat{\mathbf{x}} - y_4 b\,\hat{\mathbf{y}} + \frac{3}{4}c\,\hat{\mathbf{z}}$ | $(4g)$ | $NH_4$ |
| $\mathbf{B}_{11}$ | = | $\frac{3}{4}\,\mathbf{a}_1 - y_4\,\mathbf{a}_2 + \frac{3}{4}\,\mathbf{a}_3$ | = | $\frac{3}{4}a\,\hat{\mathbf{x}} - y_4 b\,\hat{\mathbf{y}} + \frac{3}{4}c\,\hat{\mathbf{z}}$ | $(4g)$ | $NH_4$ |
| $\mathbf{B}_{12}$ | = | $\frac{3}{4}\,\mathbf{a}_1 + y_4\,\mathbf{a}_2 + \frac{1}{4}\,\mathbf{a}_3$ | = | $\frac{3}{4}a\,\hat{\mathbf{x}} + y_4 b\,\hat{\mathbf{y}} + \frac{1}{4}c\,\hat{\mathbf{z}}$ | $(4g)$ | $NH_4$ |
| $\mathbf{B}_{13}$ | = | $y_5\,\mathbf{a}_2 + z_5\,\mathbf{a}_3$ | = | $y_5 b\,\hat{\mathbf{y}} + z_5 c\,\hat{\mathbf{z}}$ | $(4h)$ | F II |
| $\mathbf{B}_{14}$ | = | $\frac{1}{2}\,\mathbf{a}_1 - y_5\,\mathbf{a}_2 + \left(\frac{1}{2} + z_5\right)\,\mathbf{a}_3$ | = | $\frac{1}{2}a\,\hat{\mathbf{x}} - y_5 b\,\hat{\mathbf{y}} + \left(\frac{1}{2} + z_5\right)c\,\hat{\mathbf{z}}$ | $(4h)$ | F II |
| $\mathbf{B}_{15}$ | = | $\frac{1}{2}\,\mathbf{a}_1 + y_5\,\mathbf{a}_2 + \left(\frac{1}{2} - z_5\right)\,\mathbf{a}_3$ | = | $\frac{1}{2}a\,\hat{\mathbf{x}} + y_5 b\,\hat{\mathbf{y}} + \left(\frac{1}{2} - z_5\right)c\,\hat{\mathbf{z}}$ | $(4h)$ | F II |
| $\mathbf{B}_{16}$ | = | $-y_5\,\mathbf{a}_2 - z_5\,\mathbf{a}_3$ | = | $-y_5 b\,\hat{\mathbf{y}} - z_5 c\,\hat{\mathbf{z}}$ | $(4h)$ | F II |

**Geometry files:**

- CIF: pp. 1611
- POSCAR: pp. 1611

# Orthorhombic $Sr_4Ru_3O_{10}$ Structure: A10B3C4_oP68_55_2e2fgh2i_adef_2e2f

http://aflow.org/prototype-encyclopedia/A10B3C4_oP68_55_2e2fgh2i_adef_2e2f

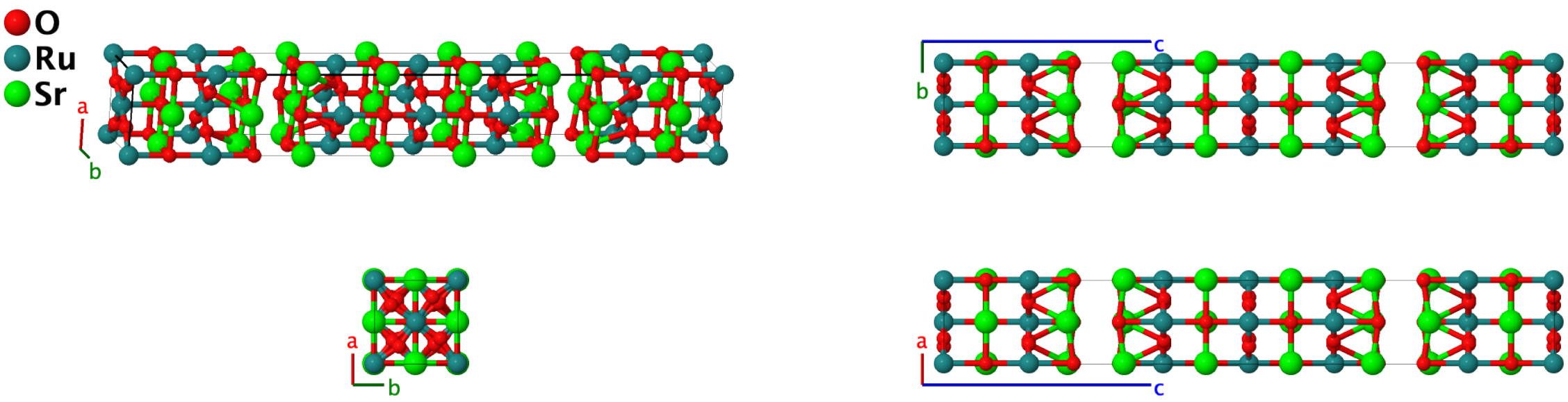

| | | |
|---|---|---|
| **Prototype** | : | $O_{10}Ru_3Sr_4$ |
| **AFLOW prototype label** | : | A10B3C4_oP68_55_2e2fgh2i_adef_2e2f |
| ***Strukturbericht* designation** | : | None |
| **Pearson symbol** | : | oP68 |
| **Space group number** | : | 55 |
| **Space group symbol** | : | *Pbam* |
| **AFLOW prototype command** | : | `aflow --proto=A10B3C4_oP68_55_2e2fgh2i_adef_2e2f --params=`$a, b/a, c/a, z_3, z_4, z_5, z_6, z_7, z_8, z_9, z_{10}, z_{11}, z_{12}, x_{13}, y_{13}, x_{14}, y_{14}, x_{15}, y_{15}, z_{15}, x_{16}, y_{16}, z_{16}$ |

- This structure consists of triple-layer ruthenate structures separated by 2.37 Å from each other. In the *Pbam* #55 space group shown here there are two inequivalent stacks in the orthorhombic cell.
- This cell is very problematic. (Crawford, 2002) note that the x-ray scattering intensities are pseudo body-centered, but found that refining this structure in a body-centered cell with space group *Bbcm* (*Cmca* #64 in our standard orientation) led to non-positive definite thermal parameters. However, if we allow a lattice and atom position uncertainty of only 0.01 Å in the atomic coordinates, AFLOW-SYM and FINDSYM place the structure in base-centered space group *Cmca* #64, oC68. In that case there is only one triple-layer stack in the primitive cell, and the two stacks in the conventional orthorhombic cell are equivalent.

**Simple Orthorhombic primitive vectors:**

$$\mathbf{a}_1 = a\,\hat{\mathbf{x}}$$
$$\mathbf{a}_2 = b\,\hat{\mathbf{y}}$$
$$\mathbf{a}_3 = c\,\hat{\mathbf{z}}$$

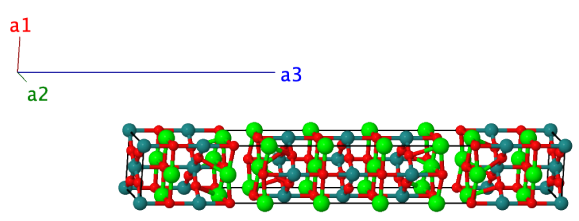

**Basis vectors:**

| | | Lattice Coordinates | | Cartesian Coordinates | Wyckoff Position | Atom Type |
|---|---|---|---|---|---|---|
| $\mathbf{B}_1$ | = | $0\,\mathbf{a}_1 + 0\,\mathbf{a}_2 + 0\,\mathbf{a}_3$ | = | $0\,\hat{\mathbf{x}} + 0\,\hat{\mathbf{y}} + 0\,\hat{\mathbf{z}}$ | (2*a*) | Ru I |
| $\mathbf{B}_2$ | = | $\frac{1}{2}\,\mathbf{a}_1 + \frac{1}{2}\,\mathbf{a}_2$ | = | $\frac{1}{2}a\,\hat{\mathbf{x}} + \frac{1}{2}b\,\hat{\mathbf{y}}$ | (2*a*) | Ru I |
| $\mathbf{B}_3$ | = | $\frac{1}{2}\,\mathbf{a}_2 + \frac{1}{2}\,\mathbf{a}_3$ | = | $\frac{1}{2}b\,\hat{\mathbf{y}} + \frac{1}{2}c\,\hat{\mathbf{z}}$ | (2*d*) | Ru II |
| $\mathbf{B}_4$ | = | $\frac{1}{2}\,\mathbf{a}_1 + \frac{1}{2}\,\mathbf{a}_3$ | = | $\frac{1}{2}a\,\hat{\mathbf{x}} + \frac{1}{2}c\,\hat{\mathbf{z}}$ | (2*d*) | Ru II |
| $\mathbf{B}_5$ | = | $z_3\,\mathbf{a}_3$ | = | $z_3c\,\hat{\mathbf{z}}$ | (4*e*) | O I |
| $\mathbf{B}_{6}$ | = | $\frac{1}{2}\,\mathbf{a}_{1}+\frac{1}{2}\,\mathbf{a}_{2}-z_{3}\,\mathbf{a}_{3}$ | = | $\frac{1}{2}a\,\mathbf{\hat{x}}+\frac{1}{2}b\,\mathbf{\hat{y}}-z_{3}c\,\mathbf{\hat{z}}$ | $(4e)$ | O I |
| $\mathbf{B}_{7}$ | = | $-z_{3}\,\mathbf{a}_{3}$ | = | $-z_{3}c\,\mathbf{\hat{z}}$ | $(4e)$ | O I |
| $\mathbf{B}_{8}$ | = | $\frac{1}{2}\,\mathbf{a}_{1}+\frac{1}{2}\,\mathbf{a}_{2}+z_{3}\,\mathbf{a}_{3}$ | = | $\frac{1}{2}a\,\mathbf{\hat{x}}+\frac{1}{2}b\,\mathbf{\hat{y}}+z_{3}c\,\mathbf{\hat{z}}$ | $(4e)$ | O I |
| $\mathbf{B}_{9}$ | = | $z_{4}\,\mathbf{a}_{3}$ | = | $z_{4}c\,\mathbf{\hat{z}}$ | $(4e)$ | O II |
| $\mathbf{B}_{10}$ | = | $\frac{1}{2}\,\mathbf{a}_{1}+\frac{1}{2}\,\mathbf{a}_{2}-z_{4}\,\mathbf{a}_{3}$ | = | $\frac{1}{2}a\,\mathbf{\hat{x}}+\frac{1}{2}b\,\mathbf{\hat{y}}-z_{4}c\,\mathbf{\hat{z}}$ | $(4e)$ | O II |
| $\mathbf{B}_{11}$ | = | $-z_{4}\,\mathbf{a}_{3}$ | = | $-z_{4}c\,\mathbf{\hat{z}}$ | $(4e)$ | O II |
| $\mathbf{B}_{12}$ | = | $\frac{1}{2}\,\mathbf{a}_{1}+\frac{1}{2}\,\mathbf{a}_{2}+z_{4}\,\mathbf{a}_{3}$ | = | $\frac{1}{2}a\,\mathbf{\hat{x}}+\frac{1}{2}b\,\mathbf{\hat{y}}+z_{4}c\,\mathbf{\hat{z}}$ | $(4e)$ | O II |
| $\mathbf{B}_{13}$ | = | $z_{5}\,\mathbf{a}_{3}$ | = | $z_{5}c\,\mathbf{\hat{z}}$ | $(4e)$ | Ru III |
| $\mathbf{B}_{14}$ | = | $\frac{1}{2}\,\mathbf{a}_{1}+\frac{1}{2}\,\mathbf{a}_{2}-z_{5}\,\mathbf{a}_{3}$ | = | $\frac{1}{2}a\,\mathbf{\hat{x}}+\frac{1}{2}b\,\mathbf{\hat{y}}-z_{5}c\,\mathbf{\hat{z}}$ | $(4e)$ | Ru III |
| $\mathbf{B}_{15}$ | = | $-z_{5}\,\mathbf{a}_{3}$ | = | $-z_{5}c\,\mathbf{\hat{z}}$ | $(4e)$ | Ru III |
| $\mathbf{B}_{16}$ | = | $\frac{1}{2}\,\mathbf{a}_{1}+\frac{1}{2}\,\mathbf{a}_{2}+z_{5}\,\mathbf{a}_{3}$ | = | $\frac{1}{2}a\,\mathbf{\hat{x}}+\frac{1}{2}b\,\mathbf{\hat{y}}+z_{5}c\,\mathbf{\hat{z}}$ | $(4e)$ | Ru III |
| $\mathbf{B}_{17}$ | = | $z_{6}\,\mathbf{a}_{3}$ | = | $z_{6}c\,\mathbf{\hat{z}}$ | $(4e)$ | Sr I |
| $\mathbf{B}_{18}$ | = | $\frac{1}{2}\,\mathbf{a}_{1}+\frac{1}{2}\,\mathbf{a}_{2}-z_{6}\,\mathbf{a}_{3}$ | = | $\frac{1}{2}a\,\mathbf{\hat{x}}+\frac{1}{2}b\,\mathbf{\hat{y}}-z_{6}c\,\mathbf{\hat{z}}$ | $(4e)$ | Sr I |
| $\mathbf{B}_{19}$ | = | $-z_{6}\,\mathbf{a}_{3}$ | = | $-z_{6}c\,\mathbf{\hat{z}}$ | $(4e)$ | Sr I |
| $\mathbf{B}_{20}$ | = | $\frac{1}{2}\,\mathbf{a}_{1}+\frac{1}{2}\,\mathbf{a}_{2}+z_{6}\,\mathbf{a}_{3}$ | = | $\frac{1}{2}a\,\mathbf{\hat{x}}+\frac{1}{2}b\,\mathbf{\hat{y}}+z_{6}c\,\mathbf{\hat{z}}$ | $(4e)$ | Sr I |
| $\mathbf{B}_{21}$ | = | $z_{7}\,\mathbf{a}_{3}$ | = | $z_{7}c\,\mathbf{\hat{z}}$ | $(4e)$ | Sr II |
| $\mathbf{B}_{22}$ | = | $\frac{1}{2}\,\mathbf{a}_{1}+\frac{1}{2}\,\mathbf{a}_{2}-z_{7}\,\mathbf{a}_{3}$ | = | $\frac{1}{2}a\,\mathbf{\hat{x}}+\frac{1}{2}b\,\mathbf{\hat{y}}-z_{7}c\,\mathbf{\hat{z}}$ | $(4e)$ | Sr II |
| $\mathbf{B}_{23}$ | = | $-z_{7}\,\mathbf{a}_{3}$ | = | $-z_{7}c\,\mathbf{\hat{z}}$ | $(4e)$ | Sr II |
| $\mathbf{B}_{24}$ | = | $\frac{1}{2}\,\mathbf{a}_{1}+\frac{1}{2}\,\mathbf{a}_{2}+z_{7}\,\mathbf{a}_{3}$ | = | $\frac{1}{2}a\,\mathbf{\hat{x}}+\frac{1}{2}b\,\mathbf{\hat{y}}+z_{7}c\,\mathbf{\hat{z}}$ | $(4e)$ | Sr II |
| $\mathbf{B}_{25}$ | = | $\frac{1}{2}\,\mathbf{a}_{2}+z_{8}\,\mathbf{a}_{3}$ | = | $\frac{1}{2}b\,\mathbf{\hat{y}}+z_{8}c\,\mathbf{\hat{z}}$ | $(4f)$ | O III |
| $\mathbf{B}_{26}$ | = | $\frac{1}{2}\,\mathbf{a}_{1}-z_{8}\,\mathbf{a}_{3}$ | = | $\frac{1}{2}a\,\mathbf{\hat{x}}-z_{8}c\,\mathbf{\hat{z}}$ | $(4f)$ | O III |
| $\mathbf{B}_{27}$ | = | $\frac{1}{2}\,\mathbf{a}_{2}-z_{8}\,\mathbf{a}_{3}$ | = | $\frac{1}{2}b\,\mathbf{\hat{y}}-z_{8}c\,\mathbf{\hat{z}}$ | $(4f)$ | O III |
| $\mathbf{B}_{28}$ | = | $\frac{1}{2}\,\mathbf{a}_{1}+z_{8}\,\mathbf{a}_{3}$ | = | $\frac{1}{2}a\,\mathbf{\hat{x}}+z_{8}c\,\mathbf{\hat{z}}$ | $(4f)$ | O III |
| $\mathbf{B}_{29}$ | = | $\frac{1}{2}\,\mathbf{a}_{2}+z_{9}\,\mathbf{a}_{3}$ | = | $\frac{1}{2}b\,\mathbf{\hat{y}}+z_{9}c\,\mathbf{\hat{z}}$ | $(4f)$ | O IV |
| $\mathbf{B}_{30}$ | = | $\frac{1}{2}\,\mathbf{a}_{1}-z_{9}\,\mathbf{a}_{3}$ | = | $\frac{1}{2}a\,\mathbf{\hat{x}}-z_{9}c\,\mathbf{\hat{z}}$ | $(4f)$ | O IV |
| $\mathbf{B}_{31}$ | = | $\frac{1}{2}\,\mathbf{a}_{2}-z_{9}\,\mathbf{a}_{3}$ | = | $\frac{1}{2}b\,\mathbf{\hat{y}}-z_{9}c\,\mathbf{\hat{z}}$ | $(4f)$ | O IV |
| $\mathbf{B}_{32}$ | = | $\frac{1}{2}\,\mathbf{a}_{1}+z_{9}\,\mathbf{a}_{3}$ | = | $\frac{1}{2}a\,\mathbf{\hat{x}}+z_{9}c\,\mathbf{\hat{z}}$ | $(4f)$ | O IV |
| $\mathbf{B}_{33}$ | = | $\frac{1}{2}\,\mathbf{a}_{2}+z_{10}\,\mathbf{a}_{3}$ | = | $\frac{1}{2}b\,\mathbf{\hat{y}}+z_{10}c\,\mathbf{\hat{z}}$ | $(4f)$ | Ru IV |
| $\mathbf{B}_{34}$ | = | $\frac{1}{2}\,\mathbf{a}_{1}-z_{10}\,\mathbf{a}_{3}$ | = | $\frac{1}{2}a\,\mathbf{\hat{x}}-z_{10}c\,\mathbf{\hat{z}}$ | $(4f)$ | Ru IV |
| $\mathbf{B}_{35}$ | = | $\frac{1}{2}\,\mathbf{a}_{2}-z_{10}\,\mathbf{a}_{3}$ | = | $\frac{1}{2}b\,\mathbf{\hat{y}}-z_{10}c\,\mathbf{\hat{z}}$ | $(4f)$ | Ru IV |
| $\mathbf{B}_{36}$ | = | $\frac{1}{2}\,\mathbf{a}_{1}+z_{10}\,\mathbf{a}_{3}$ | = | $\frac{1}{2}a\,\mathbf{\hat{x}}+z_{10}c\,\mathbf{\hat{z}}$ | $(4f)$ | Ru IV |
| $\mathbf{B}_{37}$ | = | $\frac{1}{2}\,\mathbf{a}_{2}+z_{11}\,\mathbf{a}_{3}$ | = | $\frac{1}{2}b\,\mathbf{\hat{y}}+z_{11}c\,\mathbf{\hat{z}}$ | $(4f)$ | Sr III |
| $\mathbf{B}_{38}$ | = | $\frac{1}{2}\,\mathbf{a}_{1}-z_{11}\,\mathbf{a}_{3}$ | = | $\frac{1}{2}a\,\mathbf{\hat{x}}-z_{11}c\,\mathbf{\hat{z}}$ | $(4f)$ | Sr III |
| $\mathbf{B}_{39}$ | = | $\frac{1}{2}\,\mathbf{a}_{2}-z_{11}\,\mathbf{a}_{3}$ | = | $\frac{1}{2}b\,\mathbf{\hat{y}}-z_{11}c\,\mathbf{\hat{z}}$ | $(4f)$ | Sr III |
| $\mathbf{B}_{40}$ | = | $\frac{1}{2}\,\mathbf{a}_{1}+z_{11}\,\mathbf{a}_{3}$ | = | $\frac{1}{2}a\,\mathbf{\hat{x}}+z_{11}c\,\mathbf{\hat{z}}$ | $(4f)$ | Sr III |
| $\mathbf{B}_{41}$ | = | $\frac{1}{2}\,\mathbf{a}_{2}+z_{12}\,\mathbf{a}_{3}$ | = | $\frac{1}{2}b\,\mathbf{\hat{y}}+z_{12}c\,\mathbf{\hat{z}}$ | $(4f)$ | Sr IV |
| $\mathbf{B}_{42}$ | = | $\frac{1}{2}\,\mathbf{a}_1 - z_{12}\,\mathbf{a}_3$ | = | $\frac{1}{2}a\,\hat{\mathbf{x}} - z_{12}c\,\hat{\mathbf{z}}$ | (4$f$) | Sr IV |
| $\mathbf{B}_{43}$ | = | $\frac{1}{2}\,\mathbf{a}_2 - z_{12}\,\mathbf{a}_3$ | = | $\frac{1}{2}b\,\hat{\mathbf{y}} - z_{12}c\,\hat{\mathbf{z}}$ | (4$f$) | Sr IV |
| $\mathbf{B}_{44}$ | = | $\frac{1}{2}\,\mathbf{a}_1 + z_{12}\,\mathbf{a}_3$ | = | $\frac{1}{2}a\,\hat{\mathbf{x}} + z_{12}c\,\hat{\mathbf{z}}$ | (4$f$) | Sr IV |
| $\mathbf{B}_{45}$ | = | $x_{13}\,\mathbf{a}_1 + y_{13}\,\mathbf{a}_2$ | = | $x_{13}a\,\hat{\mathbf{x}} + y_{13}b\,\hat{\mathbf{y}}$ | (4$g$) | O V |
| $\mathbf{B}_{46}$ | = | $-x_{13}\,\mathbf{a}_1 - y_{13}\,\mathbf{a}_2$ | = | $-x_{13}a\,\hat{\mathbf{x}} - y_{13}b\,\hat{\mathbf{y}}$ | (4$g$) | O V |
| $\mathbf{B}_{47}$ | = | $\left(\frac{1}{2} - x_{13}\right)\mathbf{a}_1 + \left(\frac{1}{2} + y_{13}\right)\mathbf{a}_2$ | = | $\left(\frac{1}{2} - x_{13}\right)a\,\hat{\mathbf{x}} + \left(\frac{1}{2} + y_{13}\right)b\,\hat{\mathbf{y}}$ | (4$g$) | O V |
| $\mathbf{B}_{48}$ | = | $\left(\frac{1}{2} + x_{13}\right)\mathbf{a}_1 + \left(\frac{1}{2} - y_{13}\right)\mathbf{a}_2$ | = | $\left(\frac{1}{2} + x_{13}\right)a\,\hat{\mathbf{x}} + \left(\frac{1}{2} - y_{13}\right)b\,\hat{\mathbf{y}}$ | (4$g$) | O V |
| $\mathbf{B}_{49}$ | = | $x_{14}\,\mathbf{a}_1 + y_{14}\,\mathbf{a}_2 + \frac{1}{2}\,\mathbf{a}_3$ | = | $x_{14}a\,\hat{\mathbf{x}} + y_{14}b\,\hat{\mathbf{y}} + \frac{1}{2}c\,\hat{\mathbf{z}}$ | (4$h$) | O VI |
| $\mathbf{B}_{50}$ | = | $-x_{14}\,\mathbf{a}_1 - y_{14}\,\mathbf{a}_2 + \frac{1}{2}\,\mathbf{a}_3$ | = | $-x_{14}a\,\hat{\mathbf{x}} - y_{14}b\,\hat{\mathbf{y}} + \frac{1}{2}c\,\hat{\mathbf{z}}$ | (4$h$) | O VI |
| $\mathbf{B}_{51}$ | = | $\left(\frac{1}{2} - x_{14}\right)\mathbf{a}_1 + \left(\frac{1}{2} + y_{14}\right)\mathbf{a}_2 + \frac{1}{2}\,\mathbf{a}_3$ | = | $\left(\frac{1}{2} - x_{14}\right)a\,\hat{\mathbf{x}} + \left(\frac{1}{2} + y_{14}\right)b\,\hat{\mathbf{y}} + \frac{1}{2}c\,\hat{\mathbf{z}}$ | (4$h$) | O VI |
| $\mathbf{B}_{52}$ | = | $\left(\frac{1}{2} + x_{14}\right)\mathbf{a}_1 + \left(\frac{1}{2} - y_{14}\right)\mathbf{a}_2 + \frac{1}{2}\,\mathbf{a}_3$ | = | $\left(\frac{1}{2} + x_{14}\right)a\,\hat{\mathbf{x}} + \left(\frac{1}{2} - y_{14}\right)b\,\hat{\mathbf{y}} + \frac{1}{2}c\,\hat{\mathbf{z}}$ | (4$h$) | O VI |
| $\mathbf{B}_{53}$ | = | $x_{15}\,\mathbf{a}_1 + y_{15}\,\mathbf{a}_2 + z_{15}\,\mathbf{a}_3$ | = | $x_{15}a\,\hat{\mathbf{x}} + y_{15}b\,\hat{\mathbf{y}} + z_{15}c\,\hat{\mathbf{z}}$ | (8$i$) | O VII |
| $\mathbf{B}_{54}$ | = | $-x_{15}\,\mathbf{a}_1 - y_{15}\,\mathbf{a}_2 + z_{15}\,\mathbf{a}_3$ | = | $-x_{15}a\,\hat{\mathbf{x}} - y_{15}b\,\hat{\mathbf{y}} + z_{15}c\,\hat{\mathbf{z}}$ | (8$i$) | O VII |
| $\mathbf{B}_{55}$ | = | $\left(\frac{1}{2} - x_{15}\right)\mathbf{a}_1 + \left(\frac{1}{2} + y_{15}\right)\mathbf{a}_2 - z_{15}\,\mathbf{a}_3$ | = | $\left(\frac{1}{2} - x_{15}\right)a\,\hat{\mathbf{x}} + \left(\frac{1}{2} + y_{15}\right)b\,\hat{\mathbf{y}} - z_{15}c\,\hat{\mathbf{z}}$ | (8$i$) | O VII |
| $\mathbf{B}_{56}$ | = | $\left(\frac{1}{2} + x_{15}\right)\mathbf{a}_1 + \left(\frac{1}{2} - y_{15}\right)\mathbf{a}_2 - z_{15}\,\mathbf{a}_3$ | = | $\left(\frac{1}{2} + x_{15}\right)a\,\hat{\mathbf{x}} + \left(\frac{1}{2} - y_{15}\right)b\,\hat{\mathbf{y}} - z_{15}c\,\hat{\mathbf{z}}$ | (8$i$) | O VII |
| $\mathbf{B}_{57}$ | = | $-x_{15}\,\mathbf{a}_1 - y_{15}\,\mathbf{a}_2 - z_{15}\,\mathbf{a}_3$ | = | $-x_{15}a\,\hat{\mathbf{x}} - y_{15}b\,\hat{\mathbf{y}} - z_{15}c\,\hat{\mathbf{z}}$ | (8$i$) | O VII |
| $\mathbf{B}_{58}$ | = | $x_{15}\,\mathbf{a}_1 + y_{15}\,\mathbf{a}_2 - z_{15}\,\mathbf{a}_3$ | = | $x_{15}a\,\hat{\mathbf{x}} + y_{15}b\,\hat{\mathbf{y}} - z_{15}c\,\hat{\mathbf{z}}$ | (8$i$) | O VII |
| $\mathbf{B}_{59}$ | = | $\left(\frac{1}{2} + x_{15}\right)\mathbf{a}_1 + \left(\frac{1}{2} - y_{15}\right)\mathbf{a}_2 + z_{15}\,\mathbf{a}_3$ | = | $\left(\frac{1}{2} + x_{15}\right)a\,\hat{\mathbf{x}} + \left(\frac{1}{2} - y_{15}\right)b\,\hat{\mathbf{y}} + z_{15}c\,\hat{\mathbf{z}}$ | (8$i$) | O VII |
| $\mathbf{B}_{60}$ | = | $\left(\frac{1}{2} - x_{15}\right)\mathbf{a}_1 + \left(\frac{1}{2} + y_{15}\right)\mathbf{a}_2 + z_{15}\,\mathbf{a}_3$ | = | $\left(\frac{1}{2} - x_{15}\right)a\,\hat{\mathbf{x}} + \left(\frac{1}{2} + y_{15}\right)b\,\hat{\mathbf{y}} + z_{15}c\,\hat{\mathbf{z}}$ | (8$i$) | O VII |
| $\mathbf{B}_{61}$ | = | $x_{16}\,\mathbf{a}_1 + y_{16}\,\mathbf{a}_2 + z_{16}\,\mathbf{a}_3$ | = | $x_{16}a\,\hat{\mathbf{x}} + y_{16}b\,\hat{\mathbf{y}} + z_{16}c\,\hat{\mathbf{z}}$ | (8$i$) | O VIII |
| $\mathbf{B}_{62}$ | = | $-x_{16}\,\mathbf{a}_1 - y_{16}\,\mathbf{a}_2 + z_{16}\,\mathbf{a}_3$ | = | $-x_{16}a\,\hat{\mathbf{x}} - y_{16}b\,\hat{\mathbf{y}} + z_{16}c\,\hat{\mathbf{z}}$ | (8$i$) | O VIII |
| $\mathbf{B}_{63}$ | = | $\left(\frac{1}{2} - x_{16}\right)\mathbf{a}_1 + \left(\frac{1}{2} + y_{16}\right)\mathbf{a}_2 - z_{16}\,\mathbf{a}_3$ | = | $\left(\frac{1}{2} - x_{16}\right)a\,\hat{\mathbf{x}} + \left(\frac{1}{2} + y_{16}\right)b\,\hat{\mathbf{y}} - z_{16}c\,\hat{\mathbf{z}}$ | (8$i$) | O VIII |
| $\mathbf{B}_{64}$ | = | $\left(\frac{1}{2} + x_{16}\right)\mathbf{a}_1 + \left(\frac{1}{2} - y_{16}\right)\mathbf{a}_2 - z_{16}\,\mathbf{a}_3$ | = | $\left(\frac{1}{2} + x_{16}\right)a\,\hat{\mathbf{x}} + \left(\frac{1}{2} - y_{16}\right)b\,\hat{\mathbf{y}} - z_{16}c\,\hat{\mathbf{z}}$ | (8$i$) | O VIII |
| $\mathbf{B}_{65}$ | = | $-x_{16}\,\mathbf{a}_1 - y_{16}\,\mathbf{a}_2 - z_{16}\,\mathbf{a}_3$ | = | $-x_{16}a\,\hat{\mathbf{x}} - y_{16}b\,\hat{\mathbf{y}} - z_{16}c\,\hat{\mathbf{z}}$ | (8$i$) | O VIII |
| $\mathbf{B}_{66}$ | = | $x_{16}\,\mathbf{a}_1 + y_{16}\,\mathbf{a}_2 - z_{16}\,\mathbf{a}_3$ | = | $x_{16}a\,\hat{\mathbf{x}} + y_{16}b\,\hat{\mathbf{y}} - z_{16}c\,\hat{\mathbf{z}}$ | (8$i$) | O VIII |
| $\mathbf{B}_{67}$ | = | $\left(\frac{1}{2} + x_{16}\right)\mathbf{a}_1 + \left(\frac{1}{2} - y_{16}\right)\mathbf{a}_2 + z_{16}\,\mathbf{a}_3$ | = | $\left(\frac{1}{2} + x_{16}\right)a\,\hat{\mathbf{x}} + \left(\frac{1}{2} - y_{16}\right)b\,\hat{\mathbf{y}} + z_{16}c\,\hat{\mathbf{z}}$ | (8$i$) | O VIII |
| $\mathbf{B}_{68}$ | = | $\left(\frac{1}{2} - x_{16}\right)\mathbf{a}_1 + \left(\frac{1}{2} + y_{16}\right)\mathbf{a}_2 + z_{16}\,\mathbf{a}_3$ | = | $\left(\frac{1}{2} - x_{16}\right)a\,\hat{\mathbf{x}} + \left(\frac{1}{2} + y_{16}\right)b\,\hat{\mathbf{y}} + z_{16}c\,\hat{\mathbf{z}}$ | (8$i$) | O VIII |

**Geometry files:**

- CIF: pp. 1611
- POSCAR: pp. 1612

# $Nb_2Pd_3Se_8$ Structure: A2B3C8_oP26_55_h_ag_2g2h

http://aflow.org/prototype-encyclopedia/A2B3C8_oP26_55_h_ag_2g2h

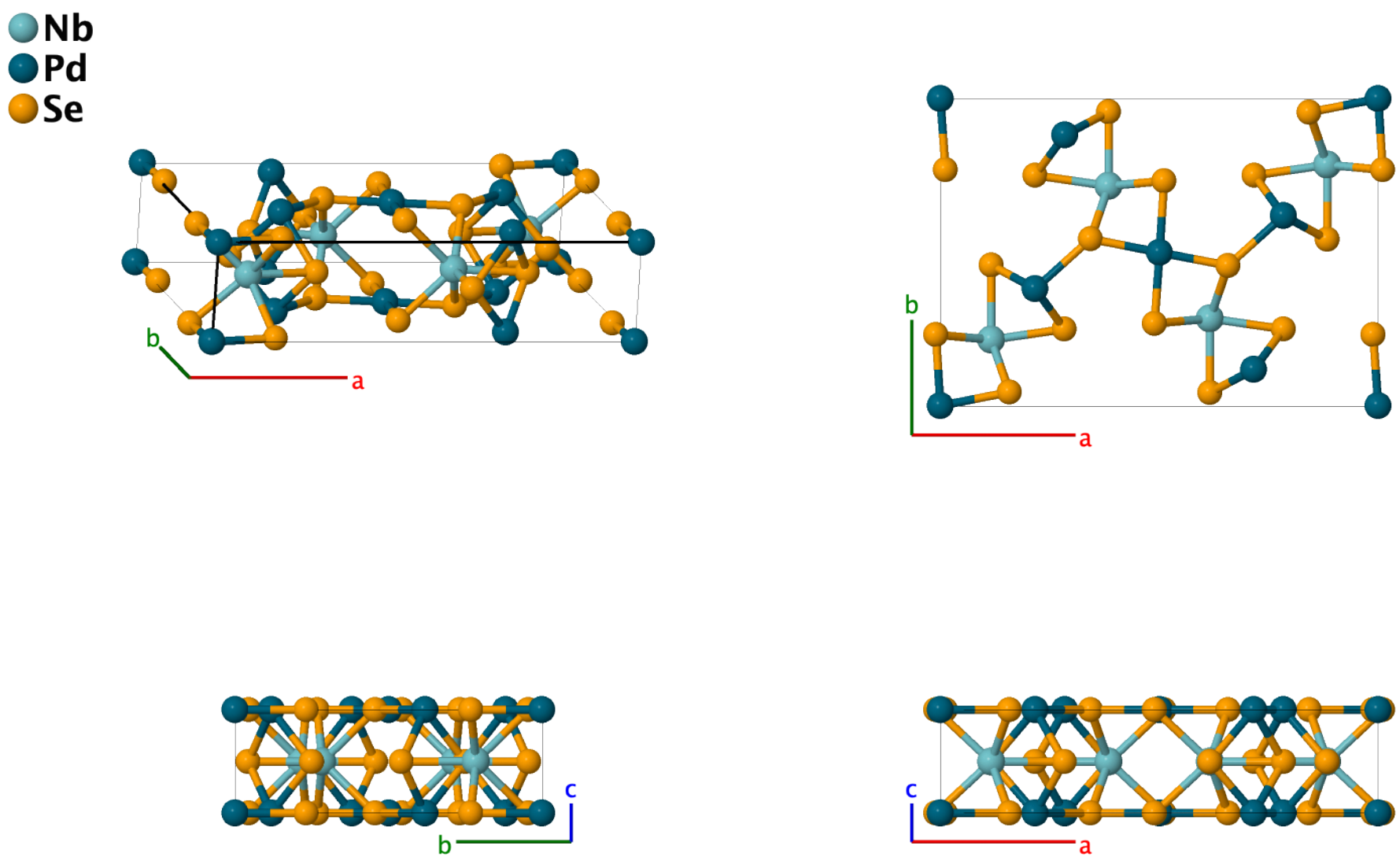

| | | |
|---|---|---|
| **Prototype** | : | $Nb_2Pd_3Se_8$ |
| **AFLOW prototype label** | : | A2B3C8_oP26_55_h_ag_2g2h |
| ***Strukturbericht* designation** | : | None |
| **Pearson symbol** | : | oP26 |
| **Space group number** | : | 55 |
| **Space group symbol** | : | *Pbam* |
| **AFLOW prototype command** | : | `aflow --proto=A2B3C8_oP26_55_h_ag_2g2h` <br> `--params=`$a, b/a, c/a, x_2, y_2, x_3, y_3, x_4, y_4, x_5, y_5, x_6, y_6, x_7, y_7$ |

### Other compounds with this structure

- $Ta_2Pd_3Se_8$

### Simple Orthorhombic primitive vectors:

$$\mathbf{a}_1 = a\,\hat{\mathbf{x}}$$
$$\mathbf{a}_2 = b\,\hat{\mathbf{y}}$$
$$\mathbf{a}_3 = c\,\hat{\mathbf{z}}$$

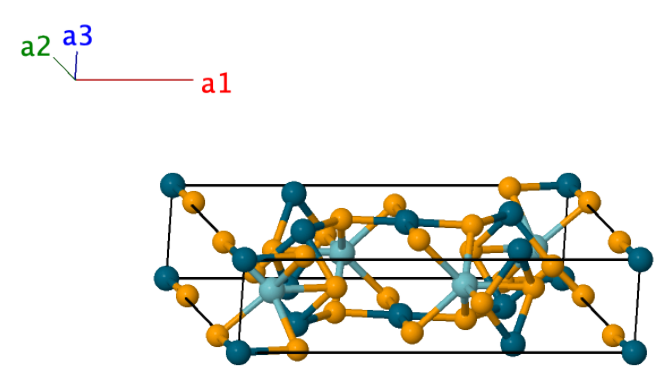

### Basis vectors:

| | | Lattice Coordinates | | Cartesian Coordinates | Wyckoff Position | Atom Type |
|---|---|---|---|---|---|---|
| $\mathbf{B}_1$ | = | $0\,\mathbf{a}_1 + 0\,\mathbf{a}_2 + 0\,\mathbf{a}_3$ | = | $0\,\hat{\mathbf{x}} + 0\,\hat{\mathbf{y}} + 0\,\hat{\mathbf{z}}$ | (2*a*) | Pd I |
| $\mathbf{B}_2$ | = | $\frac{1}{2}\,\mathbf{a}_1 + \frac{1}{2}\,\mathbf{a}_2$ | = | $\frac{1}{2}a\,\hat{\mathbf{x}} + \frac{1}{2}b\,\hat{\mathbf{y}}$ | (2*a*) | Pd I |
| $\mathbf{B}_3$ | = | $x_2\,\mathbf{a}_1 + y_2\,\mathbf{a}_2$ | = | $x_2a\,\hat{\mathbf{x}} + y_2b\,\hat{\mathbf{y}}$ | (4*g*) | Pd II |

$$
\begin{array}{ccccccc}
\mathbf{B}_{4} & = & -x_{2}\,\mathbf{a}_{1}-y_{2}\,\mathbf{a}_{2} & = & -x_{2}a\,\mathbf{\hat{x}}-y_{2}b\,\mathbf{\hat{y}} & (4g) & \text{Pd II} \\
\mathbf{B}_{5} & = & \left(\frac{1}{2}-x_{2}\right)\mathbf{a}_{1}+\left(\frac{1}{2}+y_{2}\right)\mathbf{a}_{2} & = & \left(\frac{1}{2}-x_{2}\right)a\,\mathbf{\hat{x}}+\left(\frac{1}{2}+y_{2}\right)b\,\mathbf{\hat{y}} & (4g) & \text{Pd II} \\
\mathbf{B}_{6} & = & \left(\frac{1}{2}+x_{2}\right)\mathbf{a}_{1}+\left(\frac{1}{2}-y_{2}\right)\mathbf{a}_{2} & = & \left(\frac{1}{2}+x_{2}\right)a\,\mathbf{\hat{x}}+\left(\frac{1}{2}-y_{2}\right)b\,\mathbf{\hat{y}} & (4g) & \text{Pd II} \\
\mathbf{B}_{7} & = & x_{3}\,\mathbf{a}_{1}+y_{3}\,\mathbf{a}_{2} & = & x_{3}a\,\mathbf{\hat{x}}+y_{3}b\,\mathbf{\hat{y}} & (4g) & \text{Se I} \\
\mathbf{B}_{8} & = & -x_{3}\,\mathbf{a}_{1}-y_{3}\,\mathbf{a}_{2} & = & -x_{3}a\,\mathbf{\hat{x}}-y_{3}b\,\mathbf{\hat{y}} & (4g) & \text{Se I} \\
\mathbf{B}_{9} & = & \left(\frac{1}{2}-x_{3}\right)\mathbf{a}_{1}+\left(\frac{1}{2}+y_{3}\right)\mathbf{a}_{2} & = & \left(\frac{1}{2}-x_{3}\right)a\,\mathbf{\hat{x}}+\left(\frac{1}{2}+y_{3}\right)b\,\mathbf{\hat{y}} & (4g) & \text{Se I} \\
\mathbf{B}_{10} & = & \left(\frac{1}{2}+x_{3}\right)\mathbf{a}_{1}+\left(\frac{1}{2}-y_{3}\right)\mathbf{a}_{2} & = & \left(\frac{1}{2}+x_{3}\right)a\,\mathbf{\hat{x}}+\left(\frac{1}{2}-y_{3}\right)b\,\mathbf{\hat{y}} & (4g) & \text{Se I} \\
\mathbf{B}_{11} & = & x_{4}\,\mathbf{a}_{1}+y_{4}\,\mathbf{a}_{2} & = & x_{4}a\,\mathbf{\hat{x}}+y_{4}b\,\mathbf{\hat{y}} & (4g) & \text{Se II} \\
\mathbf{B}_{12} & = & -x_{4}\,\mathbf{a}_{1}-y_{4}\,\mathbf{a}_{2} & = & -x_{4}a\,\mathbf{\hat{x}}-y_{4}b\,\mathbf{\hat{y}} & (4g) & \text{Se II} \\
\mathbf{B}_{13} & = & \left(\frac{1}{2}-x_{4}\right)\mathbf{a}_{1}+\left(\frac{1}{2}+y_{4}\right)\mathbf{a}_{2} & = & \left(\frac{1}{2}-x_{4}\right)a\,\mathbf{\hat{x}}+\left(\frac{1}{2}+y_{4}\right)b\,\mathbf{\hat{y}} & (4g) & \text{Se II} \\
\mathbf{B}_{14} & = & \left(\frac{1}{2}+x_{4}\right)\mathbf{a}_{1}+\left(\frac{1}{2}-y_{4}\right)\mathbf{a}_{2} & = & \left(\frac{1}{2}+x_{4}\right)a\,\mathbf{\hat{x}}+\left(\frac{1}{2}-y_{4}\right)b\,\mathbf{\hat{y}} & (4g) & \text{Se II} \\
\mathbf{B}_{15} & = & x_{5}\,\mathbf{a}_{1}+y_{5}\,\mathbf{a}_{2}+\frac{1}{2}\,\mathbf{a}_{3} & = & x_{5}a\,\mathbf{\hat{x}}+y_{5}b\,\mathbf{\hat{y}}+\frac{1}{2}c\,\mathbf{\hat{z}} & (4h) & \text{Nb} \\
\mathbf{B}_{16} & = & -x_{5}\,\mathbf{a}_{1}-y_{5}\,\mathbf{a}_{2}+\frac{1}{2}\,\mathbf{a}_{3} & = & -x_{5}a\,\mathbf{\hat{x}}-y_{5}b\,\mathbf{\hat{y}}+\frac{1}{2}c\,\mathbf{\hat{z}} & (4h) & \text{Nb} \\
\mathbf{B}_{17} & = & \left(\frac{1}{2}-x_{5}\right)\mathbf{a}_{1}+\left(\frac{1}{2}+y_{5}\right)\mathbf{a}_{2}+\frac{1}{2}\,\mathbf{a}_{3} & = & \left(\frac{1}{2}-x_{5}\right)a\,\mathbf{\hat{x}}+\left(\frac{1}{2}+y_{5}\right)b\,\mathbf{\hat{y}}+\frac{1}{2}c\,\mathbf{\hat{z}} & (4h) & \text{Nb} \\
\mathbf{B}_{18} & = & \left(\frac{1}{2}+x_{5}\right)\mathbf{a}_{1}+\left(\frac{1}{2}-y_{5}\right)\mathbf{a}_{2}+\frac{1}{2}\,\mathbf{a}_{3} & = & \left(\frac{1}{2}+x_{5}\right)a\,\mathbf{\hat{x}}+\left(\frac{1}{2}-y_{5}\right)b\,\mathbf{\hat{y}}+\frac{1}{2}c\,\mathbf{\hat{z}} & (4h) & \text{Nb} \\
\mathbf{B}_{19} & = & x_{6}\,\mathbf{a}_{1}+y_{6}\,\mathbf{a}_{2}+\frac{1}{2}\,\mathbf{a}_{3} & = & x_{6}a\,\mathbf{\hat{x}}+y_{6}b\,\mathbf{\hat{y}}+\frac{1}{2}c\,\mathbf{\hat{z}} & (4h) & \text{Se III} \\
\mathbf{B}_{20} & = & -x_{6}\,\mathbf{a}_{1}-y_{6}\,\mathbf{a}_{2}+\frac{1}{2}\,\mathbf{a}_{3} & = & -x_{6}a\,\mathbf{\hat{x}}-y_{6}b\,\mathbf{\hat{y}}+\frac{1}{2}c\,\mathbf{\hat{z}} & (4h) & \text{Se III} \\
\mathbf{B}_{21} & = & \left(\frac{1}{2}-x_{6}\right)\mathbf{a}_{1}+\left(\frac{1}{2}+y_{6}\right)\mathbf{a}_{2}+\frac{1}{2}\,\mathbf{a}_{3} & = & \left(\frac{1}{2}-x_{6}\right)a\,\mathbf{\hat{x}}+\left(\frac{1}{2}+y_{6}\right)b\,\mathbf{\hat{y}}+\frac{1}{2}c\,\mathbf{\hat{z}} & (4h) & \text{Se III} \\
\mathbf{B}_{22} & = & \left(\frac{1}{2}+x_{6}\right)\mathbf{a}_{1}+\left(\frac{1}{2}-y_{6}\right)\mathbf{a}_{2}+\frac{1}{2}\,\mathbf{a}_{3} & = & \left(\frac{1}{2}+x_{6}\right)a\,\mathbf{\hat{x}}+\left(\frac{1}{2}-y_{6}\right)b\,\mathbf{\hat{y}}+\frac{1}{2}c\,\mathbf{\hat{z}} & (4h) & \text{Se III} \\
\mathbf{B}_{23} & = & x_{7}\,\mathbf{a}_{1}+y_{7}\,\mathbf{a}_{2}+\frac{1}{2}\,\mathbf{a}_{3} & = & x_{7}a\,\mathbf{\hat{x}}+y_{7}b\,\mathbf{\hat{y}}+\frac{1}{2}c\,\mathbf{\hat{z}} & (4h) & \text{Se IV} \\
\mathbf{B}_{24} & = & -x_{7}\,\mathbf{a}_{1}-y_{7}\,\mathbf{a}_{2}+\frac{1}{2}\,\mathbf{a}_{3} & = & -x_{7}a\,\mathbf{\hat{x}}-y_{7}b\,\mathbf{\hat{y}}+\frac{1}{2}c\,\mathbf{\hat{z}} & (4h) & \text{Se IV} \\
\mathbf{B}_{25} & = & \left(\frac{1}{2}-x_{7}\right)\mathbf{a}_{1}+\left(\frac{1}{2}+y_{7}\right)\mathbf{a}_{2}+\frac{1}{2}\,\mathbf{a}_{3} & = & \left(\frac{1}{2}-x_{7}\right)a\,\mathbf{\hat{x}}+\left(\frac{1}{2}+y_{7}\right)b\,\mathbf{\hat{y}}+\frac{1}{2}c\,\mathbf{\hat{z}} & (4h) & \text{Se IV} \\
\mathbf{B}_{26} & = & \left(\frac{1}{2}+x_{7}\right)\mathbf{a}_{1}+\left(\frac{1}{2}-y_{7}\right)\mathbf{a}_{2}+\frac{1}{2}\,\mathbf{a}_{3} & = & \left(\frac{1}{2}+x_{7}\right)a\,\mathbf{\hat{x}}+\left(\frac{1}{2}-y_{7}\right)b\,\mathbf{\hat{y}}+\frac{1}{2}c\,\mathbf{\hat{z}} & (4h) & \text{Se IV}
\end{array}
$$

---

---

**Geometry files:**

- CIF: pp. 1612
- POSCAR: pp. 1612

# $K_2HgCl_4 \cdot H_2O$ ($E3_4$) Structure: A4BCD2_oP32_55_ghi_f_e_gh

http://aflow.org/prototype-encyclopedia/A4BCD2_oP32_55_ghi_f_e_gh

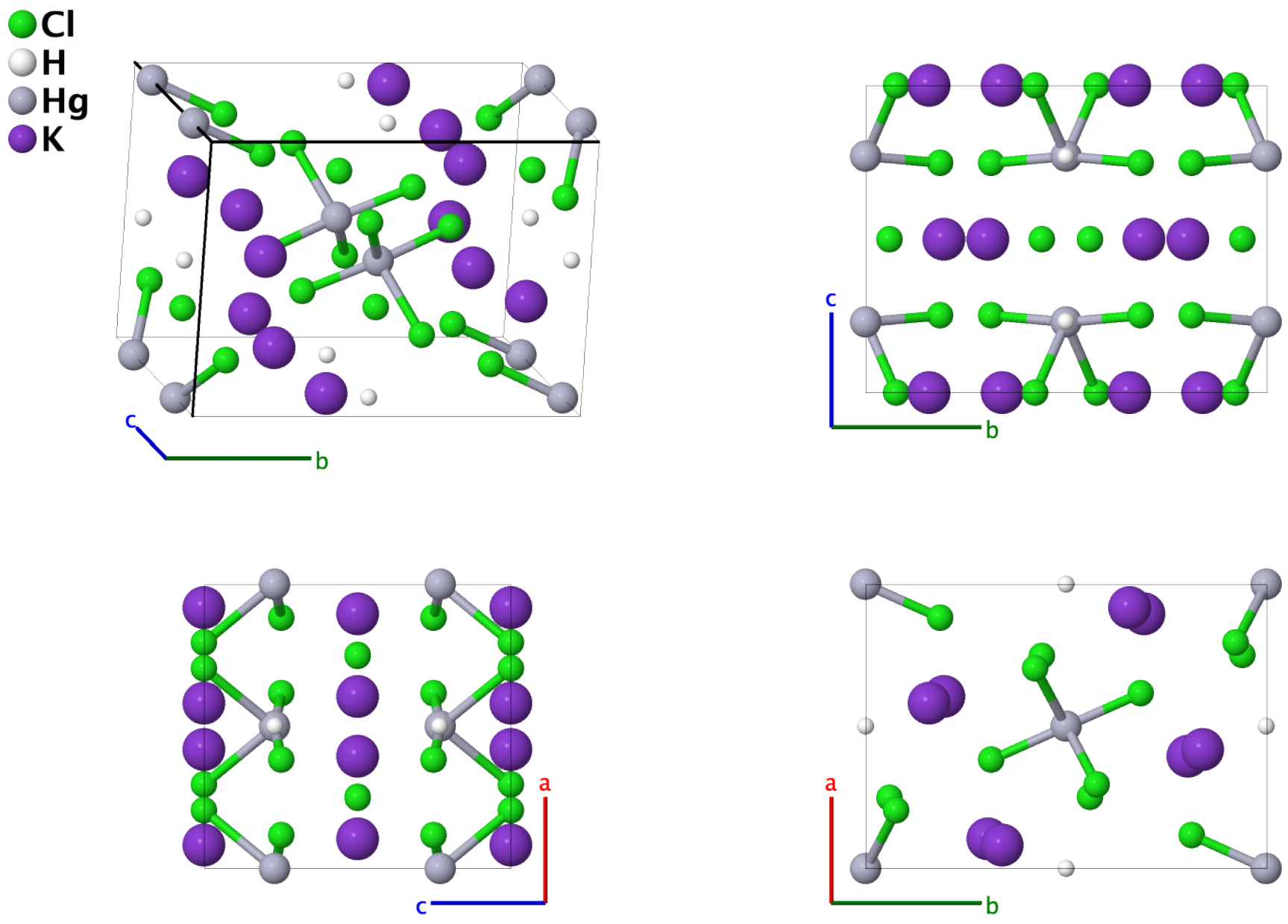

| | | |
|---|---|---|
| **Prototype** | : | $Cl_4(H_2O)HgK_2$ |
| **AFLOW prototype label** | : | A4BCD2_oP32_55_ghi_f_e_gh |
| ***Strukturbericht* designation** | : | $E3_4$ |
| **Pearson symbol** | : | oP32 |
| **Space group number** | : | 55 |
| **Space group symbol** | : | $Pbam$ |
| **AFLOW prototype command** | : | `aflow --proto=A4BCD2_oP32_55_ghi_f_e_gh`<br>`--params=`$a, b/a, c/a, z_1, z_2, x_3, y_3, x_4, y_4, x_5, y_5, x_6, y_6, x_7, y_7, z_7$ |

- The positions of the hydrogen atoms in the water molecules have not been determined.

**Simple Orthorhombic primitive vectors:**

$$\mathbf{a}_1 = a\,\hat{\mathbf{x}}$$
$$\mathbf{a}_2 = b\,\hat{\mathbf{y}}$$
$$\mathbf{a}_3 = c\,\hat{\mathbf{z}}$$

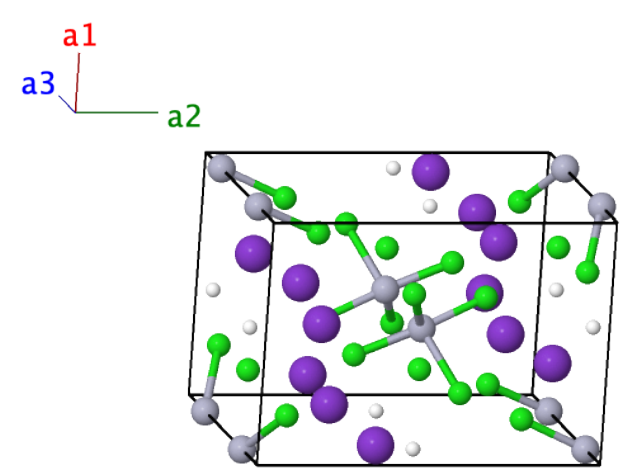

**Basis vectors:**

| | | Lattice Coordinates | | Cartesian Coordinates | Wyckoff Position | Atom Type |
|---|---|---|---|---|---|---|
| $\mathbf{B}_1$ | = | $z_1\,\mathbf{a}_3$ | = | $z_1 c\,\hat{\mathbf{z}}$ | (4$e$) | Hg |
| $\mathbf{B}_2$ | = | $\frac{1}{2}\,\mathbf{a}_1 + \frac{1}{2}\,\mathbf{a}_2 - z_1\,\mathbf{a}_3$ | = | $\frac{1}{2}a\,\hat{\mathbf{x}} + \frac{1}{2}b\,\hat{\mathbf{y}} - z_1 c\,\hat{\mathbf{z}}$ | (4$e$) | Hg |
| $\mathbf{B}_3$ | = | $-z_1\,\mathbf{a}_3$ | = | $-z_1 c\,\hat{\mathbf{z}}$ | (4$e$) | Hg |
| $\mathbf{B}_4$ | = | $\frac{1}{2}\,\mathbf{a}_1+\frac{1}{2}\,\mathbf{a}_2+z_1\,\mathbf{a}_3$ | = | $\frac{1}{2}a\,\hat{\mathbf{x}}+\frac{1}{2}b\,\hat{\mathbf{y}}+z_1 c\,\hat{\mathbf{z}}$ | (4$e$) | Hg |
| $\mathbf{B}_5$ | = | $\frac{1}{2}\,\mathbf{a}_2+z_2\,\mathbf{a}_3$ | = | $\frac{1}{2}b\,\hat{\mathbf{y}}+z_2 c\,\hat{\mathbf{z}}$ | (4$f$) | $H_2O$ |
| $\mathbf{B}_6$ | = | $\frac{1}{2}\,\mathbf{a}_1-z_2\,\mathbf{a}_3$ | = | $\frac{1}{2}a\,\hat{\mathbf{x}}-z_2 c\,\hat{\mathbf{z}}$ | (4$f$) | $H_2O$ |
| $\mathbf{B}_7$ | = | $\frac{1}{2}\,\mathbf{a}_2-z_2\,\mathbf{a}_3$ | = | $\frac{1}{2}b\,\hat{\mathbf{y}}-z_2 c\,\hat{\mathbf{z}}$ | (4$f$) | $H_2O$ |
| $\mathbf{B}_8$ | = | $\frac{1}{2}\,\mathbf{a}_1+z_2\,\mathbf{a}_3$ | = | $\frac{1}{2}a\,\hat{\mathbf{x}}+z_2 c\,\hat{\mathbf{z}}$ | (4$f$) | $H_2O$ |
| $\mathbf{B}_9$ | = | $x_3\,\mathbf{a}_1+y_3\,\mathbf{a}_2$ | = | $x_3 a\,\hat{\mathbf{x}}+y_3 b\,\hat{\mathbf{y}}$ | (4$g$) | Cl I |
| $\mathbf{B}_{10}$ | = | $-x_3\,\mathbf{a}_1-y_3\,\mathbf{a}_2$ | = | $-x_3 a\,\hat{\mathbf{x}}-y_3 b\,\hat{\mathbf{y}}$ | (4$g$) | Cl I |
| $\mathbf{B}_{11}$ | = | $\left(\frac{1}{2}-x_3\right)\mathbf{a}_1+\left(\frac{1}{2}+y_3\right)\mathbf{a}_2$ | = | $\left(\frac{1}{2}-x_3\right)a\,\hat{\mathbf{x}}+\left(\frac{1}{2}+y_3\right)b\,\hat{\mathbf{y}}$ | (4$g$) | Cl I |
| $\mathbf{B}_{12}$ | = | $\left(\frac{1}{2}+x_3\right)\mathbf{a}_1+\left(\frac{1}{2}-y_3\right)\mathbf{a}_2$ | = | $\left(\frac{1}{2}+x_3\right)a\,\hat{\mathbf{x}}+\left(\frac{1}{2}-y_3\right)b\,\hat{\mathbf{y}}$ | (4$g$) | Cl I |
| $\mathbf{B}_{13}$ | = | $x_4\,\mathbf{a}_1+y_4\,\mathbf{a}_2$ | = | $x_4 a\,\hat{\mathbf{x}}+y_4 b\,\hat{\mathbf{y}}$ | (4$g$) | K I |
| $\mathbf{B}_{14}$ | = | $-x_4\,\mathbf{a}_1-y_4\,\mathbf{a}_2$ | = | $-x_4 a\,\hat{\mathbf{x}}-y_4 b\,\hat{\mathbf{y}}$ | (4$g$) | K I |
| $\mathbf{B}_{15}$ | = | $\left(\frac{1}{2}-x_4\right)\mathbf{a}_1+\left(\frac{1}{2}+y_4\right)\mathbf{a}_2$ | = | $\left(\frac{1}{2}-x_4\right)a\,\hat{\mathbf{x}}+\left(\frac{1}{2}+y_4\right)b\,\hat{\mathbf{y}}$ | (4$g$) | K I |
| $\mathbf{B}_{16}$ | = | $\left(\frac{1}{2}+x_4\right)\mathbf{a}_1+\left(\frac{1}{2}-y_4\right)\mathbf{a}_2$ | = | $\left(\frac{1}{2}+x_4\right)a\,\hat{\mathbf{x}}+\left(\frac{1}{2}-y_4\right)b\,\hat{\mathbf{y}}$ | (4$g$) | K I |
| $\mathbf{B}_{17}$ | = | $x_5\,\mathbf{a}_1+y_5\,\mathbf{a}_2+\frac{1}{2}\,\mathbf{a}_3$ | = | $x_5 a\,\hat{\mathbf{x}}+y_5 b\,\hat{\mathbf{y}}+\frac{1}{2}c\,\hat{\mathbf{z}}$ | (4$h$) | Cl II |
| $\mathbf{B}_{18}$ | = | $-x_5\,\mathbf{a}_1-y_5\,\mathbf{a}_2+\frac{1}{2}\,\mathbf{a}_3$ | = | $-x_5 a\,\hat{\mathbf{x}}-y_5 b\,\hat{\mathbf{y}}+\frac{1}{2}c\,\hat{\mathbf{z}}$ | (4$h$) | Cl II |
| $\mathbf{B}_{19}$ | = | $\left(\frac{1}{2}-x_5\right)\mathbf{a}_1+\left(\frac{1}{2}+y_5\right)\mathbf{a}_2+\frac{1}{2}\,\mathbf{a}_3$ | = | $\left(\frac{1}{2}-x_5\right)a\,\hat{\mathbf{x}}+\left(\frac{1}{2}+y_5\right)b\,\hat{\mathbf{y}}+\frac{1}{2}c\,\hat{\mathbf{z}}$ | (4$h$) | Cl II |
| $\mathbf{B}_{20}$ | = | $\left(\frac{1}{2}+x_5\right)\mathbf{a}_1+\left(\frac{1}{2}-y_5\right)\mathbf{a}_2+\frac{1}{2}\,\mathbf{a}_3$ | = | $\left(\frac{1}{2}+x_5\right)a\,\hat{\mathbf{x}}+\left(\frac{1}{2}-y_5\right)b\,\hat{\mathbf{y}}+\frac{1}{2}c\,\hat{\mathbf{z}}$ | (4$h$) | Cl II |
| $\mathbf{B}_{21}$ | = | $x_6\,\mathbf{a}_1+y_6\,\mathbf{a}_2+\frac{1}{2}\,\mathbf{a}_3$ | = | $x_6 a\,\hat{\mathbf{x}}+y_6 b\,\hat{\mathbf{y}}+\frac{1}{2}c\,\hat{\mathbf{z}}$ | (4$h$) | K II |
| $\mathbf{B}_{22}$ | = | $-x_6\,\mathbf{a}_1-y_6\,\mathbf{a}_2+\frac{1}{2}\,\mathbf{a}_3$ | = | $-x_6 a\,\hat{\mathbf{x}}-y_6 b\,\hat{\mathbf{y}}+\frac{1}{2}c\,\hat{\mathbf{z}}$ | (4$h$) | K II |
| $\mathbf{B}_{23}$ | = | $\left(\frac{1}{2}-x_6\right)\mathbf{a}_1+\left(\frac{1}{2}+y_6\right)\mathbf{a}_2+\frac{1}{2}\,\mathbf{a}_3$ | = | $\left(\frac{1}{2}-x_6\right)a\,\hat{\mathbf{x}}+\left(\frac{1}{2}+y_6\right)b\,\hat{\mathbf{y}}+\frac{1}{2}c\,\hat{\mathbf{z}}$ | (4$h$) | K II |
| $\mathbf{B}_{24}$ | = | $\left(\frac{1}{2}+x_6\right)\mathbf{a}_1+\left(\frac{1}{2}-y_6\right)\mathbf{a}_2+\frac{1}{2}\,\mathbf{a}_3$ | = | $\left(\frac{1}{2}+x_6\right)a\,\hat{\mathbf{x}}+\left(\frac{1}{2}-y_6\right)b\,\hat{\mathbf{y}}+\frac{1}{2}c\,\hat{\mathbf{z}}$ | (4$h$) | K II |
| $\mathbf{B}_{25}$ | = | $x_7\,\mathbf{a}_1+y_7\,\mathbf{a}_2+z_7\,\mathbf{a}_3$ | = | $x_7 a\,\hat{\mathbf{x}}+y_7 b\,\hat{\mathbf{y}}+z_7 c\,\hat{\mathbf{z}}$ | (8$i$) | Cl III |
| $\mathbf{B}_{26}$ | = | $-x_7\,\mathbf{a}_1-y_7\,\mathbf{a}_2+z_7\,\mathbf{a}_3$ | = | $-x_7 a\,\hat{\mathbf{x}}-y_7 b\,\hat{\mathbf{y}}+z_7 c\,\hat{\mathbf{z}}$ | (8$i$) | Cl III |
| $\mathbf{B}_{27}$ | = | $\left(\frac{1}{2}-x_7\right)\mathbf{a}_1+\left(\frac{1}{2}+y_7\right)\mathbf{a}_2-z_7\,\mathbf{a}_3$ | = | $\left(\frac{1}{2}-x_7\right)a\,\hat{\mathbf{x}}+\left(\frac{1}{2}+y_7\right)b\,\hat{\mathbf{y}}-z_7 c\,\hat{\mathbf{z}}$ | (8$i$) | Cl III |
| $\mathbf{B}_{28}$ | = | $\left(\frac{1}{2}+x_7\right)\mathbf{a}_1+\left(\frac{1}{2}-y_7\right)\mathbf{a}_2-z_7\,\mathbf{a}_3$ | = | $\left(\frac{1}{2}+x_7\right)a\,\hat{\mathbf{x}}+\left(\frac{1}{2}-y_7\right)b\,\hat{\mathbf{y}}-z_7 c\,\hat{\mathbf{z}}$ | (8$i$) | Cl III |
| $\mathbf{B}_{29}$ | = | $-x_7\,\mathbf{a}_1-y_7\,\mathbf{a}_2-z_7\,\mathbf{a}_3$ | = | $-x_7 a\,\hat{\mathbf{x}}-y_7 b\,\hat{\mathbf{y}}-z_7 c\,\hat{\mathbf{z}}$ | (8$i$) | Cl III |
| $\mathbf{B}_{30}$ | = | $x_7\,\mathbf{a}_1+y_7\,\mathbf{a}_2-z_7\,\mathbf{a}_3$ | = | $x_7 a\,\hat{\mathbf{x}}+y_7 b\,\hat{\mathbf{y}}-z_7 c\,\hat{\mathbf{z}}$ | (8$i$) | Cl III |
| $\mathbf{B}_{31}$ | = | $\left(\frac{1}{2}+x_7\right)\mathbf{a}_1+\left(\frac{1}{2}-y_7\right)\mathbf{a}_2+z_7\,\mathbf{a}_3$ | = | $\left(\frac{1}{2}+x_7\right)a\,\hat{\mathbf{x}}+\left(\frac{1}{2}-y_7\right)b\,\hat{\mathbf{y}}+z_7 c\,\hat{\mathbf{z}}$ | (8$i$) | Cl III |
| $\mathbf{B}_{32}$ | = | $\left(\frac{1}{2}-x_7\right)\mathbf{a}_1+\left(\frac{1}{2}+y_7\right)\mathbf{a}_2+z_7\,\mathbf{a}_3$ | = | $\left(\frac{1}{2}-x_7\right)a\,\hat{\mathbf{x}}+\left(\frac{1}{2}+y_7\right)b\,\hat{\mathbf{y}}+z_7 c\,\hat{\mathbf{z}}$ | (8$i$) | Cl III |

**Geometry files:**

- CIF: pp. 1613
- POSCAR: pp. 1613

# $Ru_{11}B_8$ Structure: A8B11_oP38_55_g3h_a3g2h

http://aflow.org/prototype-encyclopedia/A8B11_oP38_55_g3h_a3g2h

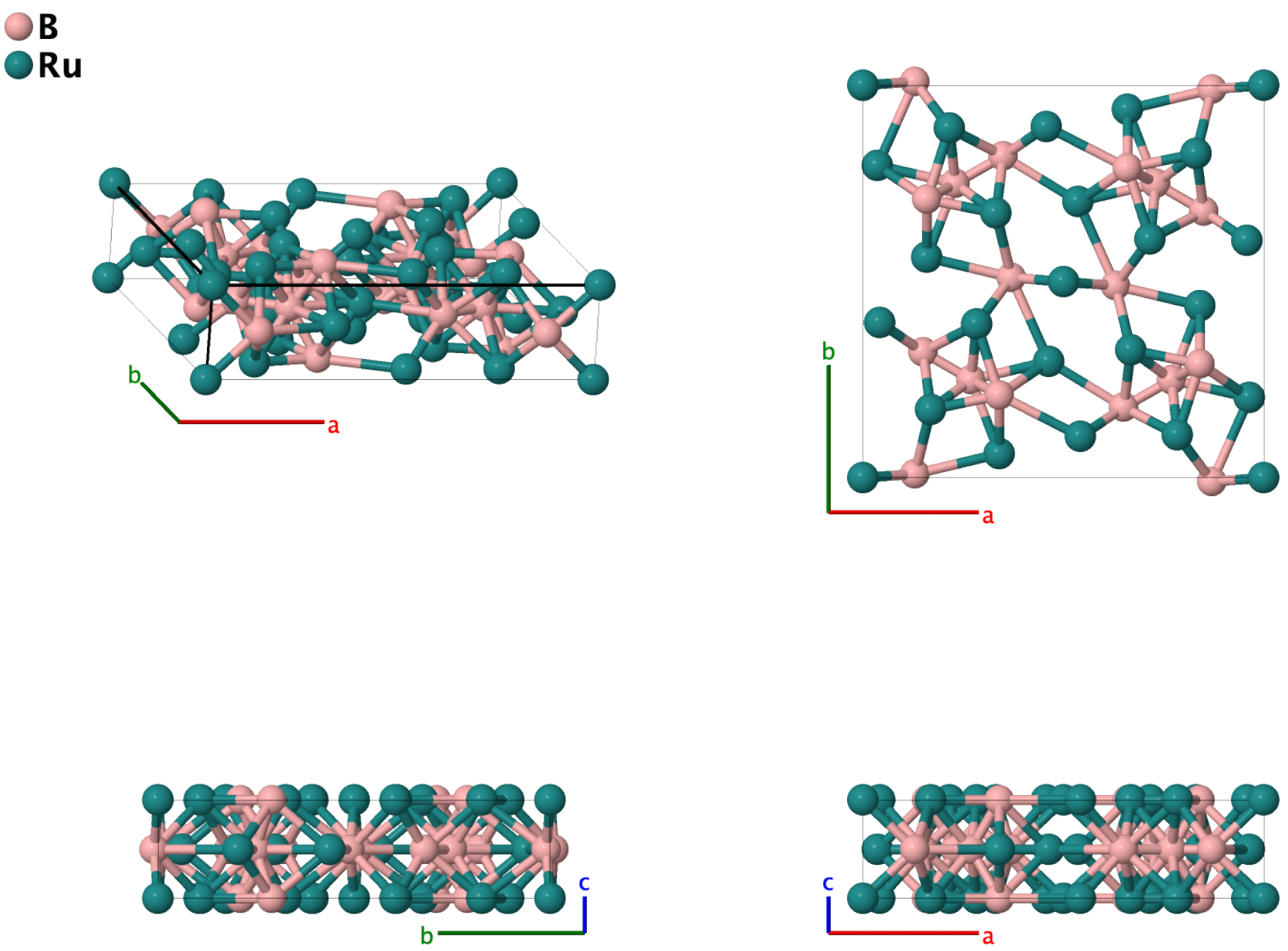

| | | |
|---|---|---|
| **Prototype** | : | $B_8Ru_{11}$ |
| **AFLOW prototype label** | : | A8B11_oP38_55_g3h_a3g2h |
| ***Strukturbericht* designation** | : | None |
| **Pearson symbol** | : | oP38 |
| **Space group number** | : | 55 |
| **Space group symbol** | : | *Pbam* |
| **AFLOW prototype command** | : | `aflow --proto=A8B11_oP38_55_g3h_a3g2h --params=`$a, b/a, c/a, x_2, y_2, x_3, y_3, x_4, y_4, x_5, y_5, x_6, y_6, x_7, y_7, x_8, y_8, x_9, y_9, x_{10}, y_{10}$ |

**Simple Orthorhombic primitive vectors:**

$$\mathbf{a}_1 = a\,\hat{\mathbf{x}}$$
$$\mathbf{a}_2 = b\,\hat{\mathbf{y}}$$
$$\mathbf{a}_3 = c\,\hat{\mathbf{z}}$$

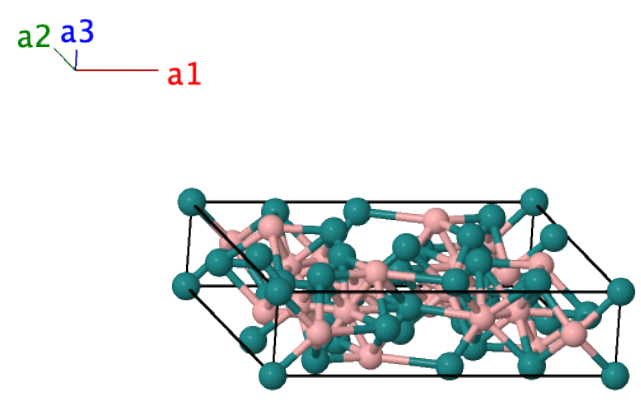

**Basis vectors:**

| | | Lattice Coordinates | | Cartesian Coordinates | Wyckoff Position | Atom Type |
|---|---|---|---|---|---|---|
| $\mathbf{B}_1$ | = | $0\,\mathbf{a}_1 + 0\,\mathbf{a}_2 + 0\,\mathbf{a}_3$ | = | $0\,\hat{\mathbf{x}} + 0\,\hat{\mathbf{y}} + 0\,\hat{\mathbf{z}}$ | (2*a*) | Ru I |
| $\mathbf{B}_{2}$ | = | $\frac{1}{2}\,\mathbf{a}_{1} + \frac{1}{2}\,\mathbf{a}_{2}$ | = | $\frac{1}{2}a\,\mathbf{\hat{x}} + \frac{1}{2}b\,\mathbf{\hat{y}}$ | (2$a$) | Ru I |
| $\mathbf{B}_{3}$ | = | $x_{2}\,\mathbf{a}_{1} + y_{2}\,\mathbf{a}_{2}$ | = | $x_{2}a\,\mathbf{\hat{x}} + y_{2}b\,\mathbf{\hat{y}}$ | (4$g$) | B I |
| $\mathbf{B}_{4}$ | = | $-x_{2}\,\mathbf{a}_{1} - y_{2}\,\mathbf{a}_{2}$ | = | $-x_{2}a\,\mathbf{\hat{x}} - y_{2}b\,\mathbf{\hat{y}}$ | (4$g$) | B I |
| $\mathbf{B}_{5}$ | = | $\left(\frac{1}{2} - x_{2}\right)\,\mathbf{a}_{1} + \left(\frac{1}{2} + y_{2}\right)\,\mathbf{a}_{2}$ | = | $\left(\frac{1}{2} - x_{2}\right)a\,\mathbf{\hat{x}} + \left(\frac{1}{2} + y_{2}\right)b\,\mathbf{\hat{y}}$ | (4$g$) | B I |
| $\mathbf{B}_{6}$ | = | $\left(\frac{1}{2} + x_{2}\right)\,\mathbf{a}_{1} + \left(\frac{1}{2} - y_{2}\right)\,\mathbf{a}_{2}$ | = | $\left(\frac{1}{2} + x_{2}\right)a\,\mathbf{\hat{x}} + \left(\frac{1}{2} - y_{2}\right)b\,\mathbf{\hat{y}}$ | (4$g$) | B I |
| $\mathbf{B}_{7}$ | = | $x_{3}\,\mathbf{a}_{1} + y_{3}\,\mathbf{a}_{2}$ | = | $x_{3}a\,\mathbf{\hat{x}} + y_{3}b\,\mathbf{\hat{y}}$ | (4$g$) | Ru II |
| $\mathbf{B}_{8}$ | = | $-x_{3}\,\mathbf{a}_{1} - y_{3}\,\mathbf{a}_{2}$ | = | $-x_{3}a\,\mathbf{\hat{x}} - y_{3}b\,\mathbf{\hat{y}}$ | (4$g$) | Ru II |
| $\mathbf{B}_{9}$ | = | $\left(\frac{1}{2} - x_{3}\right)\,\mathbf{a}_{1} + \left(\frac{1}{2} + y_{3}\right)\,\mathbf{a}_{2}$ | = | $\left(\frac{1}{2} - x_{3}\right)a\,\mathbf{\hat{x}} + \left(\frac{1}{2} + y_{3}\right)b\,\mathbf{\hat{y}}$ | (4$g$) | Ru II |
| $\mathbf{B}_{10}$ | = | $\left(\frac{1}{2} + x_{3}\right)\,\mathbf{a}_{1} + \left(\frac{1}{2} - y_{3}\right)\,\mathbf{a}_{2}$ | = | $\left(\frac{1}{2} + x_{3}\right)a\,\mathbf{\hat{x}} + \left(\frac{1}{2} - y_{3}\right)b\,\mathbf{\hat{y}}$ | (4$g$) | Ru II |
| $\mathbf{B}_{11}$ | = | $x_{4}\,\mathbf{a}_{1} + y_{4}\,\mathbf{a}_{2}$ | = | $x_{4}a\,\mathbf{\hat{x}} + y_{4}b\,\mathbf{\hat{y}}$ | (4$g$) | Ru III |
| $\mathbf{B}_{12}$ | = | $-x_{4}\,\mathbf{a}_{1} - y_{4}\,\mathbf{a}_{2}$ | = | $-x_{4}a\,\mathbf{\hat{x}} - y_{4}b\,\mathbf{\hat{y}}$ | (4$g$) | Ru III |
| $\mathbf{B}_{13}$ | = | $\left(\frac{1}{2} - x_{4}\right)\,\mathbf{a}_{1} + \left(\frac{1}{2} + y_{4}\right)\,\mathbf{a}_{2}$ | = | $\left(\frac{1}{2} - x_{4}\right)a\,\mathbf{\hat{x}} + \left(\frac{1}{2} + y_{4}\right)b\,\mathbf{\hat{y}}$ | (4$g$) | Ru III |
| $\mathbf{B}_{14}$ | = | $\left(\frac{1}{2} + x_{4}\right)\,\mathbf{a}_{1} + \left(\frac{1}{2} - y_{4}\right)\,\mathbf{a}_{2}$ | = | $\left(\frac{1}{2} + x_{4}\right)a\,\mathbf{\hat{x}} + \left(\frac{1}{2} - y_{4}\right)b\,\mathbf{\hat{y}}$ | (4$g$) | Ru III |
| $\mathbf{B}_{15}$ | = | $x_{5}\,\mathbf{a}_{1} + y_{5}\,\mathbf{a}_{2}$ | = | $x_{5}a\,\mathbf{\hat{x}} + y_{5}b\,\mathbf{\hat{y}}$ | (4$g$) | Ru IV |
| $\mathbf{B}_{16}$ | = | $-x_{5}\,\mathbf{a}_{1} - y_{5}\,\mathbf{a}_{2}$ | = | $-x_{5}a\,\mathbf{\hat{x}} - y_{5}b\,\mathbf{\hat{y}}$ | (4$g$) | Ru IV |
| $\mathbf{B}_{17}$ | = | $\left(\frac{1}{2} - x_{5}\right)\,\mathbf{a}_{1} + \left(\frac{1}{2} + y_{5}\right)\,\mathbf{a}_{2}$ | = | $\left(\frac{1}{2} - x_{5}\right)a\,\mathbf{\hat{x}} + \left(\frac{1}{2} + y_{5}\right)b\,\mathbf{\hat{y}}$ | (4$g$) | Ru IV |
| $\mathbf{B}_{18}$ | = | $\left(\frac{1}{2} + x_{5}\right)\,\mathbf{a}_{1} + \left(\frac{1}{2} - y_{5}\right)\,\mathbf{a}_{2}$ | = | $\left(\frac{1}{2} + x_{5}\right)a\,\mathbf{\hat{x}} + \left(\frac{1}{2} - y_{5}\right)b\,\mathbf{\hat{y}}$ | (4$g$) | Ru IV |
| $\mathbf{B}_{19}$ | = | $x_{6}\,\mathbf{a}_{1} + y_{6}\,\mathbf{a}_{2} + \frac{1}{2}\,\mathbf{a}_{3}$ | = | $x_{6}a\,\mathbf{\hat{x}} + y_{6}b\,\mathbf{\hat{y}} + \frac{1}{2}c\,\mathbf{\hat{z}}$ | (4$h$) | B II |
| $\mathbf{B}_{20}$ | = | $-x_{6}\,\mathbf{a}_{1} - y_{6}\,\mathbf{a}_{2} + \frac{1}{2}\,\mathbf{a}_{3}$ | = | $-x_{6}a\,\mathbf{\hat{x}} - y_{6}b\,\mathbf{\hat{y}} + \frac{1}{2}c\,\mathbf{\hat{z}}$ | (4$h$) | B II |
| $\mathbf{B}_{21}$ | = | $\left(\frac{1}{2} - x_{6}\right)\,\mathbf{a}_{1} + \left(\frac{1}{2} + y_{6}\right)\,\mathbf{a}_{2} + \frac{1}{2}\,\mathbf{a}_{3}$ | = | $\left(\frac{1}{2} - x_{6}\right)a\,\mathbf{\hat{x}} + \left(\frac{1}{2} + y_{6}\right)b\,\mathbf{\hat{y}} + \frac{1}{2}c\,\mathbf{\hat{z}}$ | (4$h$) | B II |
| $\mathbf{B}_{22}$ | = | $\left(\frac{1}{2} + x_{6}\right)\,\mathbf{a}_{1} + \left(\frac{1}{2} - y_{6}\right)\,\mathbf{a}_{2} + \frac{1}{2}\,\mathbf{a}_{3}$ | = | $\left(\frac{1}{2} + x_{6}\right)a\,\mathbf{\hat{x}} + \left(\frac{1}{2} - y_{6}\right)b\,\mathbf{\hat{y}} + \frac{1}{2}c\,\mathbf{\hat{z}}$ | (4$h$) | B II |
| $\mathbf{B}_{23}$ | = | $x_{7}\,\mathbf{a}_{1} + y_{7}\,\mathbf{a}_{2} + \frac{1}{2}\,\mathbf{a}_{3}$ | = | $x_{7}a\,\mathbf{\hat{x}} + y_{7}b\,\mathbf{\hat{y}} + \frac{1}{2}c\,\mathbf{\hat{z}}$ | (4$h$) | B III |
| $\mathbf{B}_{24}$ | = | $-x_{7}\,\mathbf{a}_{1} - y_{7}\,\mathbf{a}_{2} + \frac{1}{2}\,\mathbf{a}_{3}$ | = | $-x_{7}a\,\mathbf{\hat{x}} - y_{7}b\,\mathbf{\hat{y}} + \frac{1}{2}c\,\mathbf{\hat{z}}$ | (4$h$) | B III |
| $\mathbf{B}_{25}$ | = | $\left(\frac{1}{2} - x_{7}\right)\,\mathbf{a}_{1} + \left(\frac{1}{2} + y_{7}\right)\,\mathbf{a}_{2} + \frac{1}{2}\,\mathbf{a}_{3}$ | = | $\left(\frac{1}{2} - x_{7}\right)a\,\mathbf{\hat{x}} + \left(\frac{1}{2} + y_{7}\right)b\,\mathbf{\hat{y}} + \frac{1}{2}c\,\mathbf{\hat{z}}$ | (4$h$) | B III |
| $\mathbf{B}_{26}$ | = | $\left(\frac{1}{2} + x_{7}\right)\,\mathbf{a}_{1} + \left(\frac{1}{2} - y_{7}\right)\,\mathbf{a}_{2} + \frac{1}{2}\,\mathbf{a}_{3}$ | = | $\left(\frac{1}{2} + x_{7}\right)a\,\mathbf{\hat{x}} + \left(\frac{1}{2} - y_{7}\right)b\,\mathbf{\hat{y}} + \frac{1}{2}c\,\mathbf{\hat{z}}$ | (4$h$) | B III |
| $\mathbf{B}_{27}$ | = | $x_{8}\,\mathbf{a}_{1} + y_{8}\,\mathbf{a}_{2} + \frac{1}{2}\,\mathbf{a}_{3}$ | = | $x_{8}a\,\mathbf{\hat{x}} + y_{8}b\,\mathbf{\hat{y}} + \frac{1}{2}c\,\mathbf{\hat{z}}$ | (4$h$) | B IV |
| $\mathbf{B}_{28}$ | = | $-x_{8}\,\mathbf{a}_{1} - y_{8}\,\mathbf{a}_{2} + \frac{1}{2}\,\mathbf{a}_{3}$ | = | $-x_{8}a\,\mathbf{\hat{x}} - y_{8}b\,\mathbf{\hat{y}} + \frac{1}{2}c\,\mathbf{\hat{z}}$ | (4$h$) | B IV |
| $\mathbf{B}_{29}$ | = | $\left(\frac{1}{2} - x_{8}\right)\,\mathbf{a}_{1} + \left(\frac{1}{2} + y_{8}\right)\,\mathbf{a}_{2} + \frac{1}{2}\,\mathbf{a}_{3}$ | = | $\left(\frac{1}{2} - x_{8}\right)a\,\mathbf{\hat{x}} + \left(\frac{1}{2} + y_{8}\right)b\,\mathbf{\hat{y}} + \frac{1}{2}c\,\mathbf{\hat{z}}$ | (4$h$) | B IV |
| $\mathbf{B}_{30}$ | = | $\left(\frac{1}{2} + x_{8}\right)\,\mathbf{a}_{1} + \left(\frac{1}{2} - y_{8}\right)\,\mathbf{a}_{2} + \frac{1}{2}\,\mathbf{a}_{3}$ | = | $\left(\frac{1}{2} + x_{8}\right)a\,\mathbf{\hat{x}} + \left(\frac{1}{2} - y_{8}\right)b\,\mathbf{\hat{y}} + \frac{1}{2}c\,\mathbf{\hat{z}}$ | (4$h$) | B IV |
| $\mathbf{B}_{31}$ | = | $x_{9}\,\mathbf{a}_{1} + y_{9}\,\mathbf{a}_{2} + \frac{1}{2}\,\mathbf{a}_{3}$ | = | $x_{9}a\,\mathbf{\hat{x}} + y_{9}b\,\mathbf{\hat{y}} + \frac{1}{2}c\,\mathbf{\hat{z}}$ | (4$h$) | Ru V |
| $\mathbf{B}_{32}$ | = | $-x_{9}\,\mathbf{a}_{1} - y_{9}\,\mathbf{a}_{2} + \frac{1}{2}\,\mathbf{a}_{3}$ | = | $-x_{9}a\,\mathbf{\hat{x}} - y_{9}b\,\mathbf{\hat{y}} + \frac{1}{2}c\,\mathbf{\hat{z}}$ | (4$h$) | Ru V |
| $\mathbf{B}_{33}$ | = | $\left(\frac{1}{2} - x_{9}\right)\,\mathbf{a}_{1} + \left(\frac{1}{2} + y_{9}\right)\,\mathbf{a}_{2} + \frac{1}{2}\,\mathbf{a}_{3}$ | = | $\left(\frac{1}{2} - x_{9}\right)a\,\mathbf{\hat{x}} + \left(\frac{1}{2} + y_{9}\right)b\,\mathbf{\hat{y}} + \frac{1}{2}c\,\mathbf{\hat{z}}$ | (4$h$) | Ru V |
| $\mathbf{B}_{34}$ | = | $\left(\frac{1}{2} + x_{9}\right)\,\mathbf{a}_{1} + \left(\frac{1}{2} - y_{9}\right)\,\mathbf{a}_{2} + \frac{1}{2}\,\mathbf{a}_{3}$ | = | $\left(\frac{1}{2} + x_{9}\right)a\,\mathbf{\hat{x}} + \left(\frac{1}{2} - y_{9}\right)b\,\mathbf{\hat{y}} + \frac{1}{2}c\,\mathbf{\hat{z}}$ | (4$h$) | Ru V |
| $\mathbf{B}_{35}$ | = | $x_{10}\,\mathbf{a}_{1} + y_{10}\,\mathbf{a}_{2} + \frac{1}{2}\,\mathbf{a}_{3}$ | = | $x_{10}a\,\mathbf{\hat{x}} + y_{10}b\,\mathbf{\hat{y}} + \frac{1}{2}c\,\mathbf{\hat{z}}$ | (4$h$) | Ru VI |
| $\mathbf{B}_{36}$ | = | $-x_{10}\,\mathbf{a}_{1} - y_{10}\,\mathbf{a}_{2} + \frac{1}{2}\,\mathbf{a}_{3}$ | = | $-x_{10}a\,\mathbf{\hat{x}} - y_{10}b\,\mathbf{\hat{y}} + \frac{1}{2}c\,\mathbf{\hat{z}}$ | (4$h$) | Ru VI |
| $\mathbf{B}_{37}$ | = | $\left(\frac{1}{2} - x_{10}\right)\,\mathbf{a}_{1} + \left(\frac{1}{2} + y_{10}\right)\,\mathbf{a}_{2} + \frac{1}{2}\,\mathbf{a}_{3}$ | = | $\left(\frac{1}{2} - x_{10}\right)a\,\mathbf{\hat{x}} + \left(\frac{1}{2} + y_{10}\right)b\,\mathbf{\hat{y}} + \frac{1}{2}c\,\mathbf{\hat{z}}$ | (4$h$) | Ru VI |

$$\mathbf{B}_{38} = \left(\frac{1}{2}+x_{10}\right)\mathbf{a}_1 + \left(\frac{1}{2}-y_{10}\right)\mathbf{a}_2 + \frac{1}{2}\mathbf{a}_3 = \left(\frac{1}{2}+x_{10}\right)a\,\mathbf{\hat{x}} + \left(\frac{1}{2}-y_{10}\right)b\,\mathbf{\hat{y}} + \frac{1}{2}c\,\mathbf{\hat{z}} \qquad (4h) \qquad \text{Ru VI}$$

---

---

**Geometry files:**

- CIF: pp. 1613
- POSCAR: pp. 1614

# $HoMn_2O_5$ Structure: AB2C5_oP32_55_g_fh_eghi

http://aflow.org/prototype-encyclopedia/AB2C5_oP32_55_g_fh_eghi

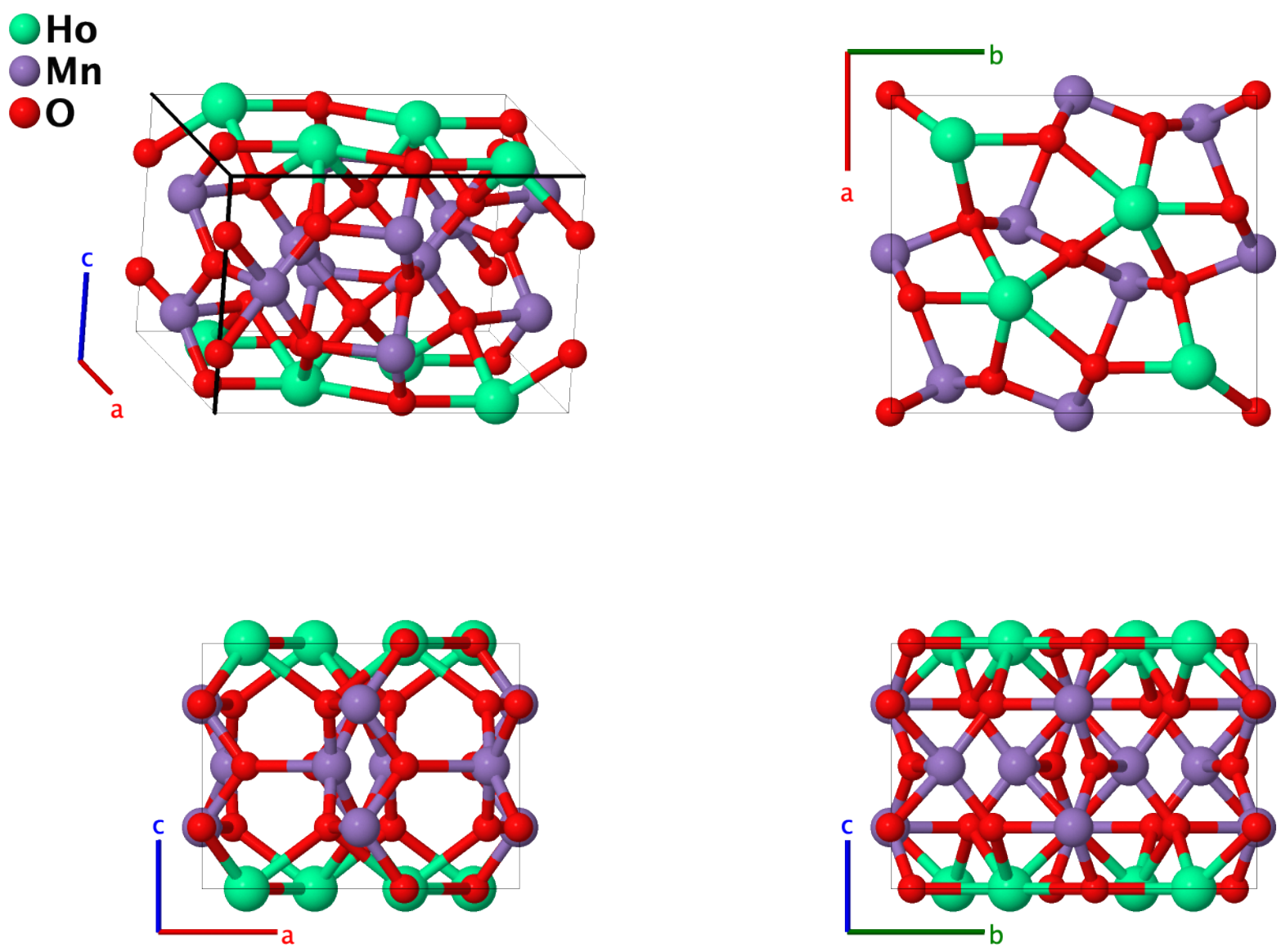

| | | |
|---|---|---|
| **Prototype** | : | $HoMn_2O_5$ |
| **AFLOW prototype label** | : | AB2C5_oP32_55_g_fh_eghi |
| ***Strukturbericht* designation** | : | None |
| **Pearson symbol** | : | oP32 |
| **Space group number** | : | 55 |
| **Space group symbol** | : | *Pbam* |
| **AFLOW prototype command** | : | `aflow --proto=AB2C5_oP32_55_g_fh_eghi` <br> `--params=`$a, b/a, c/a, z_1, z_2, x_3, y_3, x_4, y_4, x_5, y_5, x_6, y_6, x_7, y_7, z_7$ |

**Other compounds with this structure**

- $DyMn_2O_5$, $ErMn_2O_5$, $EuMn_2O_5$, $LaMn_2O_5$, $NdMn_2O_5$, $PrMn_2O_5$, $SmMn_2O_5$, and $TbMn_2O_5$

- We found no definitive definition for the prototype of the structure $XMn_2O_5$, where $X$ is a rare earth metal. (Quezel-Ambrunaz, 1964) has the earliest description of the structure we could find, so we use $HoMn_2O_5$ as the prototype.

**Simple Orthorhombic primitive vectors:**

$$\mathbf{a}_1 = a\,\hat{\mathbf{x}}$$
$$\mathbf{a}_2 = b\,\hat{\mathbf{y}}$$
$$\mathbf{a}_3 = c\,\hat{\mathbf{z}}$$

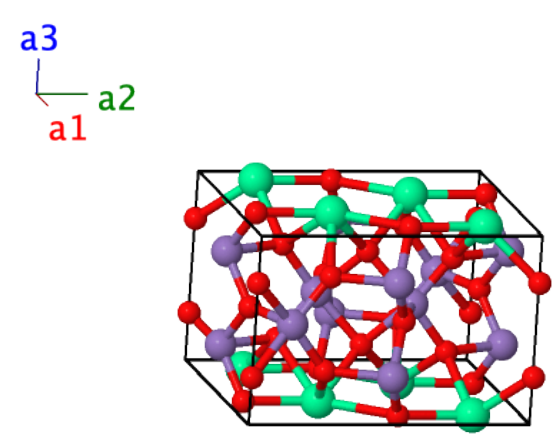

**Basis vectors:**

| | | Lattice Coordinates | | Cartesian Coordinates | Wyckoff Position | Atom Type |
|---|---|---|---|---|---|---|
| $\mathbf{B}_1$ | = | $z_1\,\mathbf{a}_3$ | = | $z_1 c\,\hat{\mathbf{z}}$ | (4e) | O I |
| $\mathbf{B}_2$ | = | $\frac{1}{2}\,\mathbf{a}_1 + \frac{1}{2}\,\mathbf{a}_2 - z_1\,\mathbf{a}_3$ | = | $\frac{1}{2}a\,\hat{\mathbf{x}} + \frac{1}{2}b\,\hat{\mathbf{y}} - z_1 c\,\hat{\mathbf{z}}$ | (4e) | O I |
| $\mathbf{B}_3$ | = | $-z_1\,\mathbf{a}_3$ | = | $-z_1 c\,\hat{\mathbf{z}}$ | (4e) | O I |
| $\mathbf{B}_4$ | = | $\frac{1}{2}\,\mathbf{a}_1 + \frac{1}{2}\,\mathbf{a}_2 + z_1\,\mathbf{a}_3$ | = | $\frac{1}{2}a\,\hat{\mathbf{x}} + \frac{1}{2}b\,\hat{\mathbf{y}} + z_1 c\,\hat{\mathbf{z}}$ | (4e) | O I |
| $\mathbf{B}_5$ | = | $\frac{1}{2}\,\mathbf{a}_2 + z_2\,\mathbf{a}_3$ | = | $\frac{1}{2}b\,\hat{\mathbf{y}} + z_2 c\,\hat{\mathbf{z}}$ | (4f) | Mn I |
| $\mathbf{B}_6$ | = | $\frac{1}{2}\,\mathbf{a}_1 - z_2\,\mathbf{a}_3$ | = | $\frac{1}{2}a\,\hat{\mathbf{x}} - z_2 c\,\hat{\mathbf{z}}$ | (4f) | Mn I |
| $\mathbf{B}_7$ | = | $\frac{1}{2}\,\mathbf{a}_2 - z_2\,\mathbf{a}_3$ | = | $\frac{1}{2}b\,\hat{\mathbf{y}} - z_2 c\,\hat{\mathbf{z}}$ | (4f) | Mn I |
| $\mathbf{B}_8$ | = | $\frac{1}{2}\,\mathbf{a}_1 + z_2\,\mathbf{a}_3$ | = | $\frac{1}{2}a\,\hat{\mathbf{x}} + z_2 c\,\hat{\mathbf{z}}$ | (4f) | Mn I |
| $\mathbf{B}_9$ | = | $x_3\,\mathbf{a}_1 + y_3\,\mathbf{a}_2$ | = | $x_3 a\,\hat{\mathbf{x}} + y_3 b\,\hat{\mathbf{y}}$ | (4g) | Ho |
| $\mathbf{B}_{10}$ | = | $-x_3\,\mathbf{a}_1 - y_3\,\mathbf{a}_2$ | = | $-x_3 a\,\hat{\mathbf{x}} - y_3 b\,\hat{\mathbf{y}}$ | (4g) | Ho |
| $\mathbf{B}_{11}$ | = | $\left(\frac{1}{2} - x_3\right)\,\mathbf{a}_1 + \left(\frac{1}{2} + y_3\right)\,\mathbf{a}_2$ | = | $\left(\frac{1}{2} - x_3\right)a\,\hat{\mathbf{x}} + \left(\frac{1}{2} + y_3\right)b\,\hat{\mathbf{y}}$ | (4g) | Ho |
| $\mathbf{B}_{12}$ | = | $\left(\frac{1}{2} + x_3\right)\,\mathbf{a}_1 + \left(\frac{1}{2} - y_3\right)\,\mathbf{a}_2$ | = | $\left(\frac{1}{2} + x_3\right)a\,\hat{\mathbf{x}} + \left(\frac{1}{2} - y_3\right)b\,\hat{\mathbf{y}}$ | (4g) | Ho |
| $\mathbf{B}_{13}$ | = | $x_4\,\mathbf{a}_1 + y_4\,\mathbf{a}_2$ | = | $x_4 a\,\hat{\mathbf{x}} + y_4 b\,\hat{\mathbf{y}}$ | (4g) | O II |
| $\mathbf{B}_{14}$ | = | $-x_4\,\mathbf{a}_1 - y_4\,\mathbf{a}_2$ | = | $-x_4 a\,\hat{\mathbf{x}} - y_4 b\,\hat{\mathbf{y}}$ | (4g) | O II |
| $\mathbf{B}_{15}$ | = | $\left(\frac{1}{2} - x_4\right)\,\mathbf{a}_1 + \left(\frac{1}{2} + y_4\right)\,\mathbf{a}_2$ | = | $\left(\frac{1}{2} - x_4\right)a\,\hat{\mathbf{x}} + \left(\frac{1}{2} + y_4\right)b\,\hat{\mathbf{y}}$ | (4g) | O II |
| $\mathbf{B}_{16}$ | = | $\left(\frac{1}{2} + x_4\right)\,\mathbf{a}_1 + \left(\frac{1}{2} - y_4\right)\,\mathbf{a}_2$ | = | $\left(\frac{1}{2} + x_4\right)a\,\hat{\mathbf{x}} + \left(\frac{1}{2} - y_4\right)b\,\hat{\mathbf{y}}$ | (4g) | O II |
| $\mathbf{B}_{17}$ | = | $x_5\,\mathbf{a}_1 + y_5\,\mathbf{a}_2 + \frac{1}{2}\,\mathbf{a}_3$ | = | $x_5 a\,\hat{\mathbf{x}} + y_5 b\,\hat{\mathbf{y}} + \frac{1}{2}c\,\hat{\mathbf{z}}$ | (4h) | Mn II |
| $\mathbf{B}_{18}$ | = | $-x_5\,\mathbf{a}_1 - y_5\,\mathbf{a}_2 + \frac{1}{2}\,\mathbf{a}_3$ | = | $-x_5 a\,\hat{\mathbf{x}} - y_5 b\,\hat{\mathbf{y}} + \frac{1}{2}c\,\hat{\mathbf{z}}$ | (4h) | Mn II |
| $\mathbf{B}_{19}$ | = | $\left(\frac{1}{2} - x_5\right)\,\mathbf{a}_1 + \left(\frac{1}{2} + y_5\right)\,\mathbf{a}_2 + \frac{1}{2}\,\mathbf{a}_3$ | = | $\left(\frac{1}{2} - x_5\right)a\,\hat{\mathbf{x}} + \left(\frac{1}{2} + y_5\right)b\,\hat{\mathbf{y}} + \frac{1}{2}c\,\hat{\mathbf{z}}$ | (4h) | Mn II |
| $\mathbf{B}_{20}$ | = | $\left(\frac{1}{2} + x_5\right)\,\mathbf{a}_1 + \left(\frac{1}{2} - y_5\right)\,\mathbf{a}_2 + \frac{1}{2}\,\mathbf{a}_3$ | = | $\left(\frac{1}{2} + x_5\right)a\,\hat{\mathbf{x}} + \left(\frac{1}{2} - y_5\right)b\,\hat{\mathbf{y}} + \frac{1}{2}c\,\hat{\mathbf{z}}$ | (4h) | Mn II |
| $\mathbf{B}_{21}$ | = | $x_6\,\mathbf{a}_1 + y_6\,\mathbf{a}_2 + \frac{1}{2}\,\mathbf{a}_3$ | = | $x_6 a\,\hat{\mathbf{x}} + y_6 b\,\hat{\mathbf{y}} + \frac{1}{2}c\,\hat{\mathbf{z}}$ | (4h) | O III |
| $\mathbf{B}_{22}$ | = | $-x_6\,\mathbf{a}_1 - y_6\,\mathbf{a}_2 + \frac{1}{2}\,\mathbf{a}_3$ | = | $-x_6 a\,\hat{\mathbf{x}} - y_6 b\,\hat{\mathbf{y}} + \frac{1}{2}c\,\hat{\mathbf{z}}$ | (4h) | O III |
| $\mathbf{B}_{23}$ | = | $\left(\frac{1}{2} - x_6\right)\,\mathbf{a}_1 + \left(\frac{1}{2} + y_6\right)\,\mathbf{a}_2 + \frac{1}{2}\,\mathbf{a}_3$ | = | $\left(\frac{1}{2} - x_6\right)a\,\hat{\mathbf{x}} + \left(\frac{1}{2} + y_6\right)b\,\hat{\mathbf{y}} + \frac{1}{2}c\,\hat{\mathbf{z}}$ | (4h) | O III |
| $\mathbf{B}_{24}$ | = | $\left(\frac{1}{2} + x_6\right)\,\mathbf{a}_1 + \left(\frac{1}{2} - y_6\right)\,\mathbf{a}_2 + \frac{1}{2}\,\mathbf{a}_3$ | = | $\left(\frac{1}{2} + x_6\right)a\,\hat{\mathbf{x}} + \left(\frac{1}{2} - y_6\right)b\,\hat{\mathbf{y}} + \frac{1}{2}c\,\hat{\mathbf{z}}$ | (4h) | O III |
| $\mathbf{B}_{25}$ | = | $x_7\,\mathbf{a}_1 + y_7\,\mathbf{a}_2 + z_7\,\mathbf{a}_3$ | = | $x_7 a\,\hat{\mathbf{x}} + y_7 b\,\hat{\mathbf{y}} + z_7 c\,\hat{\mathbf{z}}$ | (8i) | O IV |
| $\mathbf{B}_{26}$ | = | $-x_7\,\mathbf{a}_1 - y_7\,\mathbf{a}_2 + z_7\,\mathbf{a}_3$ | = | $-x_7 a\,\hat{\mathbf{x}} - y_7 b\,\hat{\mathbf{y}} + z_7 c\,\hat{\mathbf{z}}$ | (8i) | O IV |
| $\mathbf{B}_{27}$ | = | $\left(\frac{1}{2} - x_7\right)\,\mathbf{a}_1 + \left(\frac{1}{2} + y_7\right)\,\mathbf{a}_2 - z_7\,\mathbf{a}_3$ | = | $\left(\frac{1}{2} - x_7\right)a\,\hat{\mathbf{x}} + \left(\frac{1}{2} + y_7\right)b\,\hat{\mathbf{y}} - z_7 c\,\hat{\mathbf{z}}$ | (8i) | O IV |
| $\mathbf{B}_{28}$ | = | $\left(\frac{1}{2} + x_7\right)\,\mathbf{a}_1 + \left(\frac{1}{2} - y_7\right)\,\mathbf{a}_2 - z_7\,\mathbf{a}_3$ | = | $\left(\frac{1}{2} + x_7\right)a\,\hat{\mathbf{x}} + \left(\frac{1}{2} - y_7\right)b\,\hat{\mathbf{y}} - z_7 c\,\hat{\mathbf{z}}$ | (8i) | O IV |
| $\mathbf{B}_{29}$ | = | $-x_7\,\mathbf{a}_1 - y_7\,\mathbf{a}_2 - z_7\,\mathbf{a}_3$ | = | $-x_7 a\,\hat{\mathbf{x}} - y_7 b\,\hat{\mathbf{y}} - z_7 c\,\hat{\mathbf{z}}$ | (8i) | O IV |
| $\mathbf{B}_{30}$ | = | $x_7\,\mathbf{a}_1 + y_7\,\mathbf{a}_2 - z_7\,\mathbf{a}_3$ | = | $x_7 a\,\hat{\mathbf{x}} + y_7 b\,\hat{\mathbf{y}} - z_7 c\,\hat{\mathbf{z}}$ | (8i) | O IV |
| $\mathbf{B}_{31}$ | = | $\left(\frac{1}{2} + x_7\right)\,\mathbf{a}_1 + \left(\frac{1}{2} - y_7\right)\,\mathbf{a}_2 + z_7\,\mathbf{a}_3$ | = | $\left(\frac{1}{2} + x_7\right)a\,\hat{\mathbf{x}} + \left(\frac{1}{2} - y_7\right)b\,\hat{\mathbf{y}} + z_7 c\,\hat{\mathbf{z}}$ | (8i) | O IV |
| $\mathbf{B}_{32}$ | = | $\left(\frac{1}{2} - x_7\right)\,\mathbf{a}_1 + \left(\frac{1}{2} + y_7\right)\,\mathbf{a}_2 + z_7\,\mathbf{a}_3$ | = | $\left(\frac{1}{2} - x_7\right)a\,\hat{\mathbf{x}} + \left(\frac{1}{2} + y_7\right)b\,\hat{\mathbf{y}} + z_7 c\,\hat{\mathbf{z}}$ | (8i) | O IV |

http://gallica.bnf.fr/ark:/12148/bpt6k4011c?rk=85837;2.

**Found in:**

- P. Euzen, P. Leone, C. Gueho, and P. Palvadeau, *Structure of* $NdMn_2O_5$, Acta Crystallogr. C **49**, 1875–1877 (1993), doi:10.1107/S0108270193003221.

---

**Geometry files:**

- CIF: pp. 1614
- POSCAR: pp. 1614

# Calciborite ($CaB_2O_4$ II) Structure: A2BC4_oP56_56_2e_e_4e

http://aflow.org/prototype-encyclopedia/A2BC4_oP56_56_2e_e_4e

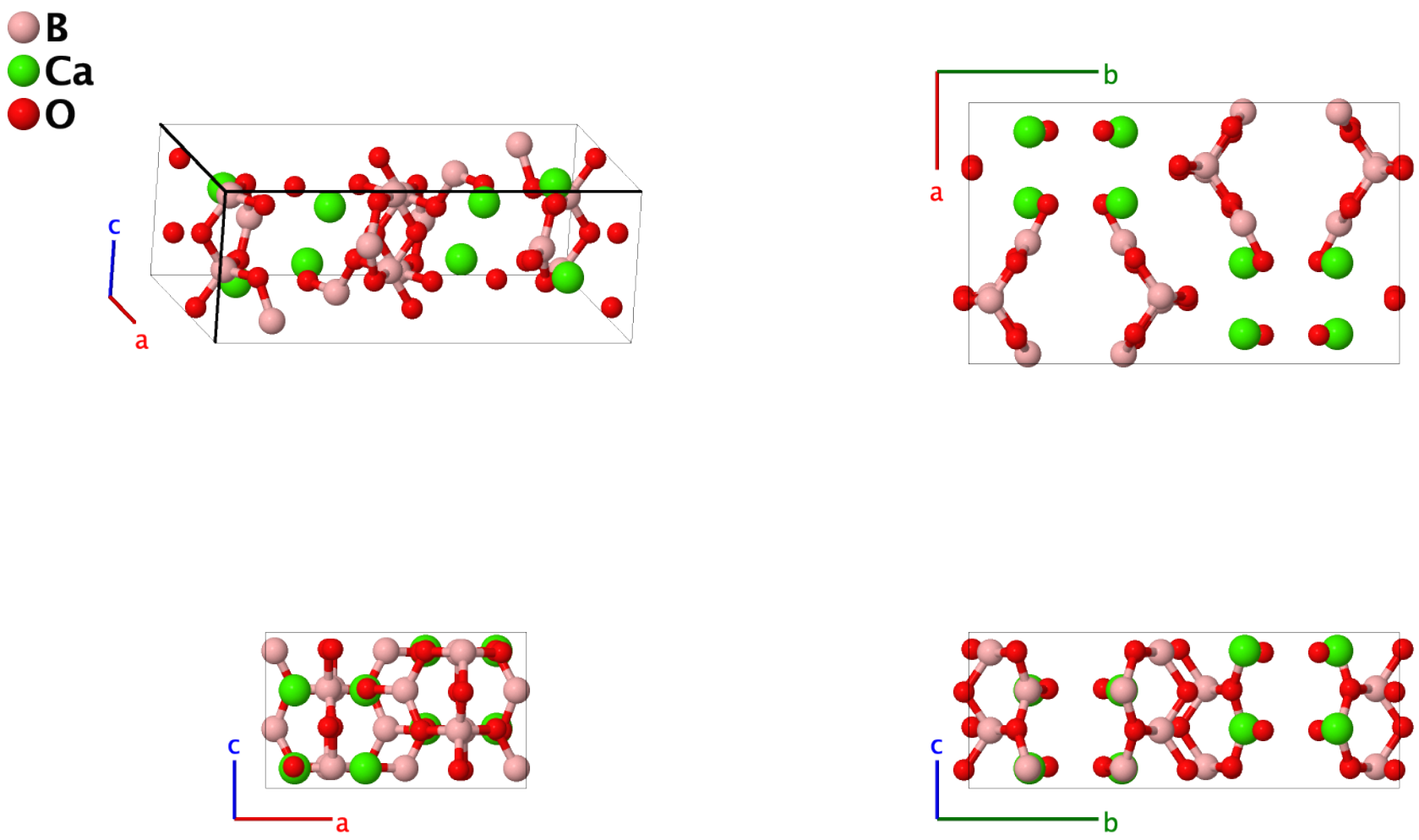

| | | |
|---|---|---|
| **Prototype** | : | $B_2CaO_4$ |
| **AFLOW prototype label** | : | A2BC4_oP56_56_2e_e_4e |
| ***Strukturbericht* designation** | : | None |
| **Pearson symbol** | : | oP56 |
| **Space group number** | : | 56 |
| **Space group symbol** | : | *Pccn* |
| **AFLOW prototype command** | : | `aflow --proto=A2BC4_oP56_56_2e_e_4e` `--params=`$a, b/a, c/a, x_1, y_1, z_1, x_2, y_2, z_2, x_3, y_3, z_3, x_4, y_4, z_4, x_5, y_5, z_5, x_6, y_6, z_6, x_7, y_7, z_7$ |

- $CaB_2O_4$ exists in at least four phases (Marezio, 1969):
- I - The ground state, stable below 1.2 GPa, *Strukturbericht* $E3_2$.
- II – Orthorhombic high pressure structure, stable between 1.2 and 1.5 GPa, presumably calciborite (this structure).
- III – Orthorhombic high pressure structure, stable between 1.5 and 2.5 GPa.
- IV – Cubic high pressure structure, stable between 2.5 and 4.0 GPa.
- This structure has the same orthorhombic lattice constants as the $CaB_2O_4$ II structure identified in (Marezio, 1969ab), however they state that the structure is stable between 1.2 and 1.5 GPa, while this mineral was obtained "From drillcore into a contact metasomatized limestone near a quartz diorite intrusion associated with a copper deposit in skarn." (Downs, 2003).

**Simple Orthorhombic primitive vectors:**

$$\mathbf{a}_1 = a\,\hat{\mathbf{x}}$$
$$\mathbf{a}_2 = b\,\hat{\mathbf{y}}$$
$$\mathbf{a}_3 = c\,\hat{\mathbf{z}}$$

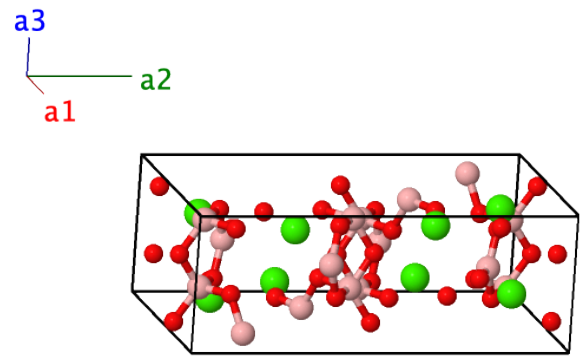

**Basis vectors:**

| | | Lattice Coordinates | | Cartesian Coordinates | Wyckoff Position | Atom Type |
|---|---|---|---|---|---|---|
| $\mathbf{B}_1$ | = | $x_1\,\mathbf{a}_1 + y_1\,\mathbf{a}_2 + z_1\,\mathbf{a}_3$ | = | $x_1 a\,\hat{\mathbf{x}} + y_1 b\,\hat{\mathbf{y}} + z_1 c\,\hat{\mathbf{z}}$ | (8$e$) | B I |
| $\mathbf{B}_2$ | = | $\left(\frac{1}{2} - x_1\right)\mathbf{a}_1 + \left(\frac{1}{2} - y_1\right)\mathbf{a}_2 + z_1\,\mathbf{a}_3$ | = | $\left(\frac{1}{2} - x_1\right)a\,\hat{\mathbf{x}} + \left(\frac{1}{2} - y_1\right)b\,\hat{\mathbf{y}} + z_1 c\,\hat{\mathbf{z}}$ | (8$e$) | B I |
| $\mathbf{B}_3$ | = | $-x_1\,\mathbf{a}_1 + \left(\frac{1}{2} + y_1\right)\mathbf{a}_2 + \left(\frac{1}{2} - z_1\right)\mathbf{a}_3$ | = | $-x_1 a\,\hat{\mathbf{x}} + \left(\frac{1}{2} + y_1\right)b\,\hat{\mathbf{y}} + \left(\frac{1}{2} - z_1\right)c\,\hat{\mathbf{z}}$ | (8$e$) | B I |
| $\mathbf{B}_4$ | = | $\left(\frac{1}{2} + x_1\right)\mathbf{a}_1 - y_1\,\mathbf{a}_2 + \left(\frac{1}{2} - z_1\right)\mathbf{a}_3$ | = | $\left(\frac{1}{2} + x_1\right)a\,\hat{\mathbf{x}} - y_1 b\,\hat{\mathbf{y}} + \left(\frac{1}{2} - z_1\right)c\,\hat{\mathbf{z}}$ | (8$e$) | B I |
| $\mathbf{B}_5$ | = | $-x_1\,\mathbf{a}_1 - y_1\,\mathbf{a}_2 - z_1\,\mathbf{a}_3$ | = | $-x_1 a\,\hat{\mathbf{x}} - y_1 b\,\hat{\mathbf{y}} - z_1 c\,\hat{\mathbf{z}}$ | (8$e$) | B I |
| $\mathbf{B}_6$ | = | $\left(\frac{1}{2} + x_1\right)\mathbf{a}_1 + \left(\frac{1}{2} + y_1\right)\mathbf{a}_2 - z_1\,\mathbf{a}_3$ | = | $\left(\frac{1}{2} + x_1\right)a\,\hat{\mathbf{x}} + \left(\frac{1}{2} + y_1\right)b\,\hat{\mathbf{y}} - z_1 c\,\hat{\mathbf{z}}$ | (8$e$) | B I |
| $\mathbf{B}_7$ | = | $x_1\,\mathbf{a}_1 + \left(\frac{1}{2} - y_1\right)\mathbf{a}_2 + \left(\frac{1}{2} + z_1\right)\mathbf{a}_3$ | = | $x_1 a\,\hat{\mathbf{x}} + \left(\frac{1}{2} - y_1\right)b\,\hat{\mathbf{y}} + \left(\frac{1}{2} + z_1\right)c\,\hat{\mathbf{z}}$ | (8$e$) | B I |
| $\mathbf{B}_8$ | = | $\left(\frac{1}{2} - x_1\right)\mathbf{a}_1 + y_1\,\mathbf{a}_2 + \left(\frac{1}{2} + z_1\right)\mathbf{a}_3$ | = | $\left(\frac{1}{2} - x_1\right)a\,\hat{\mathbf{x}} + y_1 b\,\hat{\mathbf{y}} + \left(\frac{1}{2} + z_1\right)c\,\hat{\mathbf{z}}$ | (8$e$) | B I |
| $\mathbf{B}_9$ | = | $x_2\,\mathbf{a}_1 + y_2\,\mathbf{a}_2 + z_2\,\mathbf{a}_3$ | = | $x_2 a\,\hat{\mathbf{x}} + y_2 b\,\hat{\mathbf{y}} + z_2 c\,\hat{\mathbf{z}}$ | (8$e$) | B II |
| $\mathbf{B}_{10}$ | = | $\left(\frac{1}{2} - x_2\right)\mathbf{a}_1 + \left(\frac{1}{2} - y_2\right)\mathbf{a}_2 + z_2\,\mathbf{a}_3$ | = | $\left(\frac{1}{2} - x_2\right)a\,\hat{\mathbf{x}} + \left(\frac{1}{2} - y_2\right)b\,\hat{\mathbf{y}} + z_2 c\,\hat{\mathbf{z}}$ | (8$e$) | B II |
| $\mathbf{B}_{11}$ | = | $-x_2\,\mathbf{a}_1 + \left(\frac{1}{2} + y_2\right)\mathbf{a}_2 + \left(\frac{1}{2} - z_2\right)\mathbf{a}_3$ | = | $-x_2 a\,\hat{\mathbf{x}} + \left(\frac{1}{2} + y_2\right)b\,\hat{\mathbf{y}} + \left(\frac{1}{2} - z_2\right)c\,\hat{\mathbf{z}}$ | (8$e$) | B II |
| $\mathbf{B}_{12}$ | = | $\left(\frac{1}{2} + x_2\right)\mathbf{a}_1 - y_2\,\mathbf{a}_2 + \left(\frac{1}{2} - z_2\right)\mathbf{a}_3$ | = | $\left(\frac{1}{2} + x_2\right)a\,\hat{\mathbf{x}} - y_2 b\,\hat{\mathbf{y}} + \left(\frac{1}{2} - z_2\right)c\,\hat{\mathbf{z}}$ | (8$e$) | B II |
| $\mathbf{B}_{13}$ | = | $-x_2\,\mathbf{a}_1 - y_2\,\mathbf{a}_2 - z_2\,\mathbf{a}_3$ | = | $-x_2 a\,\hat{\mathbf{x}} - y_2 b\,\hat{\mathbf{y}} - z_2 c\,\hat{\mathbf{z}}$ | (8$e$) | B II |
| $\mathbf{B}_{14}$ | = | $\left(\frac{1}{2} + x_2\right)\mathbf{a}_1 + \left(\frac{1}{2} + y_2\right)\mathbf{a}_2 - z_2\,\mathbf{a}_3$ | = | $\left(\frac{1}{2} + x_2\right)a\,\hat{\mathbf{x}} + \left(\frac{1}{2} + y_2\right)b\,\hat{\mathbf{y}} - z_2 c\,\hat{\mathbf{z}}$ | (8$e$) | B II |
| $\mathbf{B}_{15}$ | = | $x_2\,\mathbf{a}_1 + \left(\frac{1}{2} - y_2\right)\mathbf{a}_2 + \left(\frac{1}{2} + z_2\right)\mathbf{a}_3$ | = | $x_2 a\,\hat{\mathbf{x}} + \left(\frac{1}{2} - y_2\right)b\,\hat{\mathbf{y}} + \left(\frac{1}{2} + z_2\right)c\,\hat{\mathbf{z}}$ | (8$e$) | B II |
| $\mathbf{B}_{16}$ | = | $\left(\frac{1}{2} - x_2\right)\mathbf{a}_1 + y_2\,\mathbf{a}_2 + \left(\frac{1}{2} + z_2\right)\mathbf{a}_3$ | = | $\left(\frac{1}{2} - x_2\right)a\,\hat{\mathbf{x}} + y_2 b\,\hat{\mathbf{y}} + \left(\frac{1}{2} + z_2\right)c\,\hat{\mathbf{z}}$ | (8$e$) | B II |
| $\mathbf{B}_{17}$ | = | $x_3\,\mathbf{a}_1 + y_3\,\mathbf{a}_2 + z_3\,\mathbf{a}_3$ | = | $x_3 a\,\hat{\mathbf{x}} + y_3 b\,\hat{\mathbf{y}} + z_3 c\,\hat{\mathbf{z}}$ | (8$e$) | Ca |
| $\mathbf{B}_{18}$ | = | $\left(\frac{1}{2} - x_3\right)\mathbf{a}_1 + \left(\frac{1}{2} - y_3\right)\mathbf{a}_2 + z_3\,\mathbf{a}_3$ | = | $\left(\frac{1}{2} - x_3\right)a\,\hat{\mathbf{x}} + \left(\frac{1}{2} - y_3\right)b\,\hat{\mathbf{y}} + z_3 c\,\hat{\mathbf{z}}$ | (8$e$) | Ca |
| $\mathbf{B}_{19}$ | = | $-x_3\,\mathbf{a}_1 + \left(\frac{1}{2} + y_3\right)\mathbf{a}_2 + \left(\frac{1}{2} - z_3\right)\mathbf{a}_3$ | = | $-x_3 a\,\hat{\mathbf{x}} + \left(\frac{1}{2} + y_3\right)b\,\hat{\mathbf{y}} + \left(\frac{1}{2} - z_3\right)c\,\hat{\mathbf{z}}$ | (8$e$) | Ca |
| $\mathbf{B}_{20}$ | = | $\left(\frac{1}{2} + x_3\right)\mathbf{a}_1 - y_3\,\mathbf{a}_2 + \left(\frac{1}{2} - z_3\right)\mathbf{a}_3$ | = | $\left(\frac{1}{2} + x_3\right)a\,\hat{\mathbf{x}} - y_3 b\,\hat{\mathbf{y}} + \left(\frac{1}{2} - z_3\right)c\,\hat{\mathbf{z}}$ | (8$e$) | Ca |
| $\mathbf{B}_{21}$ | = | $-x_3\,\mathbf{a}_1 - y_3\,\mathbf{a}_2 - z_3\,\mathbf{a}_3$ | = | $-x_3 a\,\hat{\mathbf{x}} - y_3 b\,\hat{\mathbf{y}} - z_3 c\,\hat{\mathbf{z}}$ | (8$e$) | Ca |
| $\mathbf{B}_{22}$ | = | $\left(\frac{1}{2} + x_3\right)\mathbf{a}_1 + \left(\frac{1}{2} + y_3\right)\mathbf{a}_2 - z_3\,\mathbf{a}_3$ | = | $\left(\frac{1}{2} + x_3\right)a\,\hat{\mathbf{x}} + \left(\frac{1}{2} + y_3\right)b\,\hat{\mathbf{y}} - z_3 c\,\hat{\mathbf{z}}$ | (8$e$) | Ca |
| $\mathbf{B}_{23}$ | = | $x_3\,\mathbf{a}_1 + \left(\frac{1}{2} - y_3\right)\mathbf{a}_2 + \left(\frac{1}{2} + z_3\right)\mathbf{a}_3$ | = | $x_3 a\,\hat{\mathbf{x}} + \left(\frac{1}{2} - y_3\right)b\,\hat{\mathbf{y}} + \left(\frac{1}{2} + z_3\right)c\,\hat{\mathbf{z}}$ | (8$e$) | Ca |
| $\mathbf{B}_{24}$ | = | $\left(\frac{1}{2} - x_3\right)\mathbf{a}_1 + y_3\,\mathbf{a}_2 + \left(\frac{1}{2} + z_3\right)\mathbf{a}_3$ | = | $\left(\frac{1}{2} - x_3\right)a\,\hat{\mathbf{x}} + y_3 b\,\hat{\mathbf{y}} + \left(\frac{1}{2} + z_3\right)c\,\hat{\mathbf{z}}$ | (8$e$) | Ca |
| $\mathbf{B}_{25}$ | = | $x_4\,\mathbf{a}_1 + y_4\,\mathbf{a}_2 + z_4\,\mathbf{a}_3$ | = | $x_4 a\,\hat{\mathbf{x}} + y_4 b\,\hat{\mathbf{y}} + z_4 c\,\hat{\mathbf{z}}$ | (8$e$) | O I |
| $\mathbf{B}_{26}$ | = | $\left(\frac{1}{2} - x_4\right)\mathbf{a}_1 + \left(\frac{1}{2} - y_4\right)\mathbf{a}_2 + z_4\,\mathbf{a}_3$ | = | $\left(\frac{1}{2} - x_4\right)a\,\hat{\mathbf{x}} + \left(\frac{1}{2} - y_4\right)b\,\hat{\mathbf{y}} + z_4 c\,\hat{\mathbf{z}}$ | (8$e$) | O I |
| $\mathbf{B}_{27}$ | = | $-x_4\,\mathbf{a}_1 + \left(\frac{1}{2} + y_4\right)\mathbf{a}_2 + \left(\frac{1}{2} - z_4\right)\mathbf{a}_3$ | = | $-x_4 a\,\hat{\mathbf{x}} + \left(\frac{1}{2} + y_4\right)b\,\hat{\mathbf{y}} + \left(\frac{1}{2} - z_4\right)c\,\hat{\mathbf{z}}$ | (8$e$) | O I |
| $\mathbf{B}_{28}$ | = | $\left(\frac{1}{2} + x_4\right)\mathbf{a}_1 - y_4\,\mathbf{a}_2 + \left(\frac{1}{2} - z_4\right)\mathbf{a}_3$ | = | $\left(\frac{1}{2} + x_4\right)a\,\hat{\mathbf{x}} - y_4 b\,\hat{\mathbf{y}} + \left(\frac{1}{2} - z_4\right)c\,\hat{\mathbf{z}}$ | (8$e$) | O I |
| $\mathbf{B}_{29}$ | = | $-x_4\,\mathbf{a}_1 - y_4\,\mathbf{a}_2 - z_4\,\mathbf{a}_3$ | = | $-x_4 a\,\hat{\mathbf{x}} - y_4 b\,\hat{\mathbf{y}} - z_4 c\,\hat{\mathbf{z}}$ | (8e) | O I |
| $\mathbf{B}_{30}$ | = | $\left(\frac{1}{2}+x_4\right)\mathbf{a}_1 + \left(\frac{1}{2}+y_4\right)\mathbf{a}_2 - z_4\,\mathbf{a}_3$ | = | $\left(\frac{1}{2}+x_4\right)a\,\hat{\mathbf{x}} + \left(\frac{1}{2}+y_4\right)b\,\hat{\mathbf{y}} - z_4 c\,\hat{\mathbf{z}}$ | (8e) | O I |
| $\mathbf{B}_{31}$ | = | $x_4\,\mathbf{a}_1 + \left(\frac{1}{2}-y_4\right)\mathbf{a}_2 + \left(\frac{1}{2}+z_4\right)\mathbf{a}_3$ | = | $x_4 a\,\hat{\mathbf{x}} + \left(\frac{1}{2}-y_4\right)b\,\hat{\mathbf{y}} + \left(\frac{1}{2}+z_4\right)c\,\hat{\mathbf{z}}$ | (8e) | O I |
| $\mathbf{B}_{32}$ | = | $\left(\frac{1}{2}-x_4\right)\mathbf{a}_1 + y_4\,\mathbf{a}_2 + \left(\frac{1}{2}+z_4\right)\mathbf{a}_3$ | = | $\left(\frac{1}{2}-x_4\right)a\,\hat{\mathbf{x}} + y_4 b\,\hat{\mathbf{y}} + \left(\frac{1}{2}+z_4\right)c\,\hat{\mathbf{z}}$ | (8e) | O I |
| $\mathbf{B}_{33}$ | = | $x_5\,\mathbf{a}_1 + y_5\,\mathbf{a}_2 + z_5\,\mathbf{a}_3$ | = | $x_5 a\,\hat{\mathbf{x}} + y_5 b\,\hat{\mathbf{y}} + z_5 c\,\hat{\mathbf{z}}$ | (8e) | O II |
| $\mathbf{B}_{34}$ | = | $\left(\frac{1}{2}-x_5\right)\mathbf{a}_1 + \left(\frac{1}{2}-y_5\right)\mathbf{a}_2 + z_5\,\mathbf{a}_3$ | = | $\left(\frac{1}{2}-x_5\right)a\,\hat{\mathbf{x}} + \left(\frac{1}{2}-y_5\right)b\,\hat{\mathbf{y}} + z_5 c\,\hat{\mathbf{z}}$ | (8e) | O II |
| $\mathbf{B}_{35}$ | = | $-x_5\,\mathbf{a}_1 + \left(\frac{1}{2}+y_5\right)\mathbf{a}_2 + \left(\frac{1}{2}-z_5\right)\mathbf{a}_3$ | = | $-x_5 a\,\hat{\mathbf{x}} + \left(\frac{1}{2}+y_5\right)b\,\hat{\mathbf{y}} + \left(\frac{1}{2}-z_5\right)c\,\hat{\mathbf{z}}$ | (8e) | O II |
| $\mathbf{B}_{36}$ | = | $\left(\frac{1}{2}+x_5\right)\mathbf{a}_1 - y_5\,\mathbf{a}_2 + \left(\frac{1}{2}-z_5\right)\mathbf{a}_3$ | = | $\left(\frac{1}{2}+x_5\right)a\,\hat{\mathbf{x}} - y_5 b\,\hat{\mathbf{y}} + \left(\frac{1}{2}-z_5\right)c\,\hat{\mathbf{z}}$ | (8e) | O II |
| $\mathbf{B}_{37}$ | = | $-x_5\,\mathbf{a}_1 - y_5\,\mathbf{a}_2 - z_5\,\mathbf{a}_3$ | = | $-x_5 a\,\hat{\mathbf{x}} - y_5 b\,\hat{\mathbf{y}} - z_5 c\,\hat{\mathbf{z}}$ | (8e) | O II |
| $\mathbf{B}_{38}$ | = | $\left(\frac{1}{2}+x_5\right)\mathbf{a}_1 + \left(\frac{1}{2}+y_5\right)\mathbf{a}_2 - z_5\,\mathbf{a}_3$ | = | $\left(\frac{1}{2}+x_5\right)a\,\hat{\mathbf{x}} + \left(\frac{1}{2}+y_5\right)b\,\hat{\mathbf{y}} - z_5 c\,\hat{\mathbf{z}}$ | (8e) | O II |
| $\mathbf{B}_{39}$ | = | $x_5\,\mathbf{a}_1 + \left(\frac{1}{2}-y_5\right)\mathbf{a}_2 + \left(\frac{1}{2}+z_5\right)\mathbf{a}_3$ | = | $x_5 a\,\hat{\mathbf{x}} + \left(\frac{1}{2}-y_5\right)b\,\hat{\mathbf{y}} + \left(\frac{1}{2}+z_5\right)c\,\hat{\mathbf{z}}$ | (8e) | O II |
| $\mathbf{B}_{40}$ | = | $\left(\frac{1}{2}-x_5\right)\mathbf{a}_1 + y_5\,\mathbf{a}_2 + \left(\frac{1}{2}+z_5\right)\mathbf{a}_3$ | = | $\left(\frac{1}{2}-x_5\right)a\,\hat{\mathbf{x}} + y_5 b\,\hat{\mathbf{y}} + \left(\frac{1}{2}+z_5\right)c\,\hat{\mathbf{z}}$ | (8e) | O II |
| $\mathbf{B}_{41}$ | = | $x_6\,\mathbf{a}_1 + y_6\,\mathbf{a}_2 + z_6\,\mathbf{a}_3$ | = | $x_6 a\,\hat{\mathbf{x}} + y_6 b\,\hat{\mathbf{y}} + z_6 c\,\hat{\mathbf{z}}$ | (8e) | O III |
| $\mathbf{B}_{42}$ | = | $\left(\frac{1}{2}-x_6\right)\mathbf{a}_1 + \left(\frac{1}{2}-y_6\right)\mathbf{a}_2 + z_6\,\mathbf{a}_3$ | = | $\left(\frac{1}{2}-x_6\right)a\,\hat{\mathbf{x}} + \left(\frac{1}{2}-y_6\right)b\,\hat{\mathbf{y}} + z_6 c\,\hat{\mathbf{z}}$ | (8e) | O III |
| $\mathbf{B}_{43}$ | = | $-x_6\,\mathbf{a}_1 + \left(\frac{1}{2}+y_6\right)\mathbf{a}_2 + \left(\frac{1}{2}-z_6\right)\mathbf{a}_3$ | = | $-x_6 a\,\hat{\mathbf{x}} + \left(\frac{1}{2}+y_6\right)b\,\hat{\mathbf{y}} + \left(\frac{1}{2}-z_6\right)c\,\hat{\mathbf{z}}$ | (8e) | O III |
| $\mathbf{B}_{44}$ | = | $\left(\frac{1}{2}+x_6\right)\mathbf{a}_1 - y_6\,\mathbf{a}_2 + \left(\frac{1}{2}-z_6\right)\mathbf{a}_3$ | = | $\left(\frac{1}{2}+x_6\right)a\,\hat{\mathbf{x}} - y_6 b\,\hat{\mathbf{y}} + \left(\frac{1}{2}-z_6\right)c\,\hat{\mathbf{z}}$ | (8e) | O III |
| $\mathbf{B}_{45}$ | = | $-x_6\,\mathbf{a}_1 - y_6\,\mathbf{a}_2 - z_6\,\mathbf{a}_3$ | = | $-x_6 a\,\hat{\mathbf{x}} - y_6 b\,\hat{\mathbf{y}} - z_6 c\,\hat{\mathbf{z}}$ | (8e) | O III |
| $\mathbf{B}_{46}$ | = | $\left(\frac{1}{2}+x_6\right)\mathbf{a}_1 + \left(\frac{1}{2}+y_6\right)\mathbf{a}_2 - z_6\,\mathbf{a}_3$ | = | $\left(\frac{1}{2}+x_6\right)a\,\hat{\mathbf{x}} + \left(\frac{1}{2}+y_6\right)b\,\hat{\mathbf{y}} - z_6 c\,\hat{\mathbf{z}}$ | (8e) | O III |
| $\mathbf{B}_{47}$ | = | $x_6\,\mathbf{a}_1 + \left(\frac{1}{2}-y_6\right)\mathbf{a}_2 + \left(\frac{1}{2}+z_6\right)\mathbf{a}_3$ | = | $x_6 a\,\hat{\mathbf{x}} + \left(\frac{1}{2}-y_6\right)b\,\hat{\mathbf{y}} + \left(\frac{1}{2}+z_6\right)c\,\hat{\mathbf{z}}$ | (8e) | O III |
| $\mathbf{B}_{48}$ | = | $\left(\frac{1}{2}-x_6\right)\mathbf{a}_1 + y_6\,\mathbf{a}_2 + \left(\frac{1}{2}+z_6\right)\mathbf{a}_3$ | = | $\left(\frac{1}{2}-x_6\right)a\,\hat{\mathbf{x}} + y_6 b\,\hat{\mathbf{y}} + \left(\frac{1}{2}+z_6\right)c\,\hat{\mathbf{z}}$ | (8e) | O III |
| $\mathbf{B}_{49}$ | = | $x_7\,\mathbf{a}_1 + y_7\,\mathbf{a}_2 + z_7\,\mathbf{a}_3$ | = | $x_7 a\,\hat{\mathbf{x}} + y_7 b\,\hat{\mathbf{y}} + z_7 c\,\hat{\mathbf{z}}$ | (8e) | O IV |
| $\mathbf{B}_{50}$ | = | $\left(\frac{1}{2}-x_7\right)\mathbf{a}_1 + \left(\frac{1}{2}-y_7\right)\mathbf{a}_2 + z_7\,\mathbf{a}_3$ | = | $\left(\frac{1}{2}-x_7\right)a\,\hat{\mathbf{x}} + \left(\frac{1}{2}-y_7\right)b\,\hat{\mathbf{y}} + z_7 c\,\hat{\mathbf{z}}$ | (8e) | O IV |
| $\mathbf{B}_{51}$ | = | $-x_7\,\mathbf{a}_1 + \left(\frac{1}{2}+y_7\right)\mathbf{a}_2 + \left(\frac{1}{2}-z_7\right)\mathbf{a}_3$ | = | $-x_7 a\,\hat{\mathbf{x}} + \left(\frac{1}{2}+y_7\right)b\,\hat{\mathbf{y}} + \left(\frac{1}{2}-z_7\right)c\,\hat{\mathbf{z}}$ | (8e) | O IV |
| $\mathbf{B}_{52}$ | = | $\left(\frac{1}{2}+x_7\right)\mathbf{a}_1 - y_7\,\mathbf{a}_2 + \left(\frac{1}{2}-z_7\right)\mathbf{a}_3$ | = | $\left(\frac{1}{2}+x_7\right)a\,\hat{\mathbf{x}} - y_7 b\,\hat{\mathbf{y}} + \left(\frac{1}{2}-z_7\right)c\,\hat{\mathbf{z}}$ | (8e) | O IV |
| $\mathbf{B}_{53}$ | = | $-x_7\,\mathbf{a}_1 - y_7\,\mathbf{a}_2 - z_7\,\mathbf{a}_3$ | = | $-x_7 a\,\hat{\mathbf{x}} - y_7 b\,\hat{\mathbf{y}} - z_7 c\,\hat{\mathbf{z}}$ | (8e) | O IV |
| $\mathbf{B}_{54}$ | = | $\left(\frac{1}{2}+x_7\right)\mathbf{a}_1 + \left(\frac{1}{2}+y_7\right)\mathbf{a}_2 - z_7\,\mathbf{a}_3$ | = | $\left(\frac{1}{2}+x_7\right)a\,\hat{\mathbf{x}} + \left(\frac{1}{2}+y_7\right)b\,\hat{\mathbf{y}} - z_7 c\,\hat{\mathbf{z}}$ | (8e) | O IV |
| $\mathbf{B}_{55}$ | = | $x_7\,\mathbf{a}_1 + \left(\frac{1}{2}-y_7\right)\mathbf{a}_2 + \left(\frac{1}{2}+z_7\right)\mathbf{a}_3$ | = | $x_7 a\,\hat{\mathbf{x}} + \left(\frac{1}{2}-y_7\right)b\,\hat{\mathbf{y}} + \left(\frac{1}{2}+z_7\right)c\,\hat{\mathbf{z}}$ | (8e) | O IV |
| $\mathbf{B}_{56}$ | = | $\left(\frac{1}{2}-x_7\right)\mathbf{a}_1 + y_7\,\mathbf{a}_2 + \left(\frac{1}{2}+z_7\right)\mathbf{a}_3$ | = | $\left(\frac{1}{2}-x_7\right)a\,\hat{\mathbf{x}} + y_7 b\,\hat{\mathbf{y}} + \left(\frac{1}{2}+z_7\right)c\,\hat{\mathbf{z}}$ | (8e) | O IV |

**Geometry files:**

- CIF: pp. 1614
- POSCAR: pp. 1615

# $D0_{10}$ ($WO_3$) (*obsolete*) Structure: A3B_oP16_57_a2d_d

http://aflow.org/prototype-encyclopedia/A3B_oP16_57_a2d_d

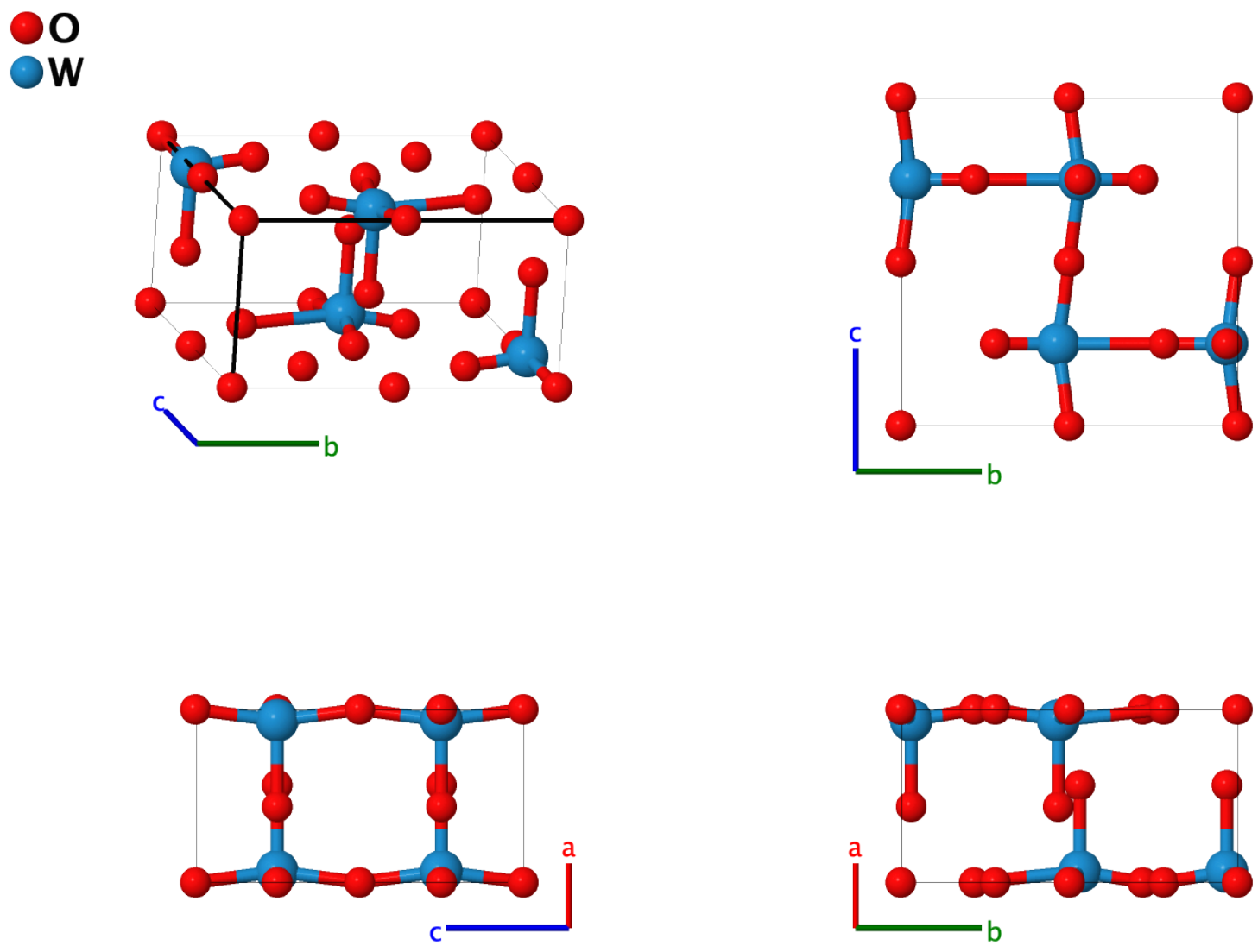

| | | |
|---|---|---|
| **Prototype** | : | $O_3W$ |
| **AFLOW prototype label** | : | A3B_oP16_57_a2d_d |
| ***Strukturbericht* designation** | : | $D0_{10}$ |
| **Pearson symbol** | : | oP16 |
| **Space group number** | : | 57 |
| **Space group symbol** | : | *Pbcm* |
| **AFLOW prototype command** | : | `aflow --proto=A3B_oP16_57_a2d_d` <br> `--params=`$a, b/a, c/a, x_2, y_2, x_3, y_3, x_4, y_4$ |

- All stable phases of $WO_3$ are distortions of the cubic $\alpha$-$ReO_3$ ($D0_9$) phase. (Woodward, 1997 and Vogt, 1999) The known stable phases and their approximate temperature ranges are:
  - $\alpha$-$WO_3$ (1010-1170 K) (Vogt, 1999)
  - $\beta$-$WO_3$ (600-1170 K) (Vogt, 1999)
  - $\gamma$-$WO_3$ (290-600 K) (Vogt, 1999)
  - $\delta$-$WO_3$ (230-290 K) (Diehl, 1978)
  - $\epsilon$-$WO_3$ (below 23 K) (Woodward, 1997)
- In addition, several other structures have been proposed and/or found:
  - The original $D0_{10}$ structure (Bräkken, 1931), (Hermann, 1937), this structure, superseded by $\delta$-$WO_3$
  - Original $\beta$-$WO_3$ (Salje, 1977)
  - Hexagonal $WO_3$ (Gerand, 1979) (metastable)

**Simple Orthorhombic primitive vectors:**

$$
\begin{aligned}
\mathbf{a}_1 &= a\,\hat{\mathbf{x}} \\
\mathbf{a}_2 &= b\,\hat{\mathbf{y}} \\
\mathbf{a}_3 &= c\,\hat{\mathbf{z}}
\end{aligned}
$$

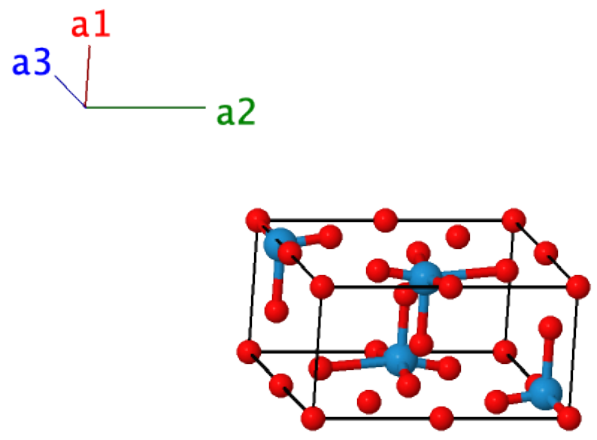

**Basis vectors:**

| | | Lattice Coordinates | | Cartesian Coordinates | Wyckoff Position | Atom Type |
|---|---|---|---|---|---|---|
| $\mathbf{B}_1$ | = | $0\,\mathbf{a}_1 + 0\,\mathbf{a}_2 + 0\,\mathbf{a}_3$ | = | $0\,\hat{\mathbf{x}} + 0\,\hat{\mathbf{y}} + 0\,\hat{\mathbf{z}}$ | (4$a$) | O I |
| $\mathbf{B}_2$ | = | $\frac{1}{2}\,\mathbf{a}_3$ | = | $\frac{1}{2}c\,\hat{\mathbf{z}}$ | (4$a$) | O I |
| $\mathbf{B}_3$ | = | $\frac{1}{2}\,\mathbf{a}_2 + \frac{1}{2}\,\mathbf{a}_3$ | = | $\frac{1}{2}b\,\hat{\mathbf{y}} + \frac{1}{2}c\,\hat{\mathbf{z}}$ | (4$a$) | O I |
| $\mathbf{B}_4$ | = | $\frac{1}{2}\,\mathbf{a}_2$ | = | $\frac{1}{2}b\,\hat{\mathbf{y}}$ | (4$a$) | O I |
| $\mathbf{B}_5$ | = | $x_2\,\mathbf{a}_1 + y_2\,\mathbf{a}_2 + \frac{1}{4}\,\mathbf{a}_3$ | = | $x_2a\,\hat{\mathbf{x}} + y_2b\,\hat{\mathbf{y}} + \frac{1}{4}c\,\hat{\mathbf{z}}$ | (4$d$) | O II |
| $\mathbf{B}_6$ | = | $-x_2\,\mathbf{a}_1 - y_2\,\mathbf{a}_2 + \frac{3}{4}\,\mathbf{a}_3$ | = | $-x_2a\,\hat{\mathbf{x}} - y_2b\,\hat{\mathbf{y}} + \frac{3}{4}c\,\hat{\mathbf{z}}$ | (4$d$) | O II |
| $\mathbf{B}_7$ | = | $-x_2\,\mathbf{a}_1 + \left(\frac{1}{2} + y_2\right)\,\mathbf{a}_2 + \frac{1}{4}\,\mathbf{a}_3$ | = | $-x_2a\,\hat{\mathbf{x}} + \left(\frac{1}{2} + y_2\right)b\,\hat{\mathbf{y}} + \frac{1}{4}c\,\hat{\mathbf{z}}$ | (4$d$) | O II |
| $\mathbf{B}_8$ | = | $x_2\,\mathbf{a}_1 + \left(\frac{1}{2} - y_2\right)\,\mathbf{a}_2 + \frac{3}{4}\,\mathbf{a}_3$ | = | $x_2a\,\hat{\mathbf{x}} + \left(\frac{1}{2} - y_2\right)b\,\hat{\mathbf{y}} + \frac{3}{4}c\,\hat{\mathbf{z}}$ | (4$d$) | O II |
| $\mathbf{B}_9$ | = | $x_3\,\mathbf{a}_1 + y_3\,\mathbf{a}_2 + \frac{1}{4}\,\mathbf{a}_3$ | = | $x_3a\,\hat{\mathbf{x}} + y_3b\,\hat{\mathbf{y}} + \frac{1}{4}c\,\hat{\mathbf{z}}$ | (4$d$) | O III |
| $\mathbf{B}_{10}$ | = | $-x_3\,\mathbf{a}_1 - y_3\,\mathbf{a}_2 + \frac{3}{4}\,\mathbf{a}_3$ | = | $-x_3a\,\hat{\mathbf{x}} - y_3b\,\hat{\mathbf{y}} + \frac{3}{4}c\,\hat{\mathbf{z}}$ | (4$d$) | O III |
| $\mathbf{B}_{11}$ | = | $-x_3\,\mathbf{a}_1 + \left(\frac{1}{2} + y_3\right)\,\mathbf{a}_2 + \frac{1}{4}\,\mathbf{a}_3$ | = | $-x_3a\,\hat{\mathbf{x}} + \left(\frac{1}{2} + y_3\right)b\,\hat{\mathbf{y}} + \frac{1}{4}c\,\hat{\mathbf{z}}$ | (4$d$) | O III |
| $\mathbf{B}_{12}$ | = | $x_3\,\mathbf{a}_1 + \left(\frac{1}{2} - y_3\right)\,\mathbf{a}_2 + \frac{3}{4}\,\mathbf{a}_3$ | = | $x_3a\,\hat{\mathbf{x}} + \left(\frac{1}{2} - y_3\right)b\,\hat{\mathbf{y}} + \frac{3}{4}c\,\hat{\mathbf{z}}$ | (4$d$) | O III |
| $\mathbf{B}_{13}$ | = | $x_4\,\mathbf{a}_1 + y_4\,\mathbf{a}_2 + \frac{1}{4}\,\mathbf{a}_3$ | = | $x_4a\,\hat{\mathbf{x}} + y_4b\,\hat{\mathbf{y}} + \frac{1}{4}c\,\hat{\mathbf{z}}$ | (4$d$) | W |
| $\mathbf{B}_{14}$ | = | $-x_4\,\mathbf{a}_1 - y_4\,\mathbf{a}_2 + \frac{3}{4}\,\mathbf{a}_3$ | = | $-x_4a\,\hat{\mathbf{x}} - y_4b\,\hat{\mathbf{y}} + \frac{3}{4}c\,\hat{\mathbf{z}}$ | (4$d$) | W |
| $\mathbf{B}_{15}$ | = | $-x_4\,\mathbf{a}_1 + \left(\frac{1}{2} + y_4\right)\,\mathbf{a}_2 + \frac{1}{4}\,\mathbf{a}_3$ | = | $-x_4a\,\hat{\mathbf{x}} + \left(\frac{1}{2} + y_4\right)b\,\hat{\mathbf{y}} + \frac{1}{4}c\,\hat{\mathbf{z}}$ | (4$d$) | W |
| $\mathbf{B}_{16}$ | = | $x_4\,\mathbf{a}_1 + \left(\frac{1}{2} - y_4\right)\,\mathbf{a}_2 + \frac{3}{4}\,\mathbf{a}_3$ | = | $x_4a\,\hat{\mathbf{x}} + \left(\frac{1}{2} - y_4\right)b\,\hat{\mathbf{y}} + \frac{3}{4}c\,\hat{\mathbf{z}}$ | (4$d$) | W |

**Geometry files:**
- CIF: pp. 1615
- POSCAR: pp. 1616

# $SrUO_4$ Structure: A4BC_oP24_57_cde_d_a

http://aflow.org/prototype-encyclopedia/A4BC_oP24_57_cde_d_a

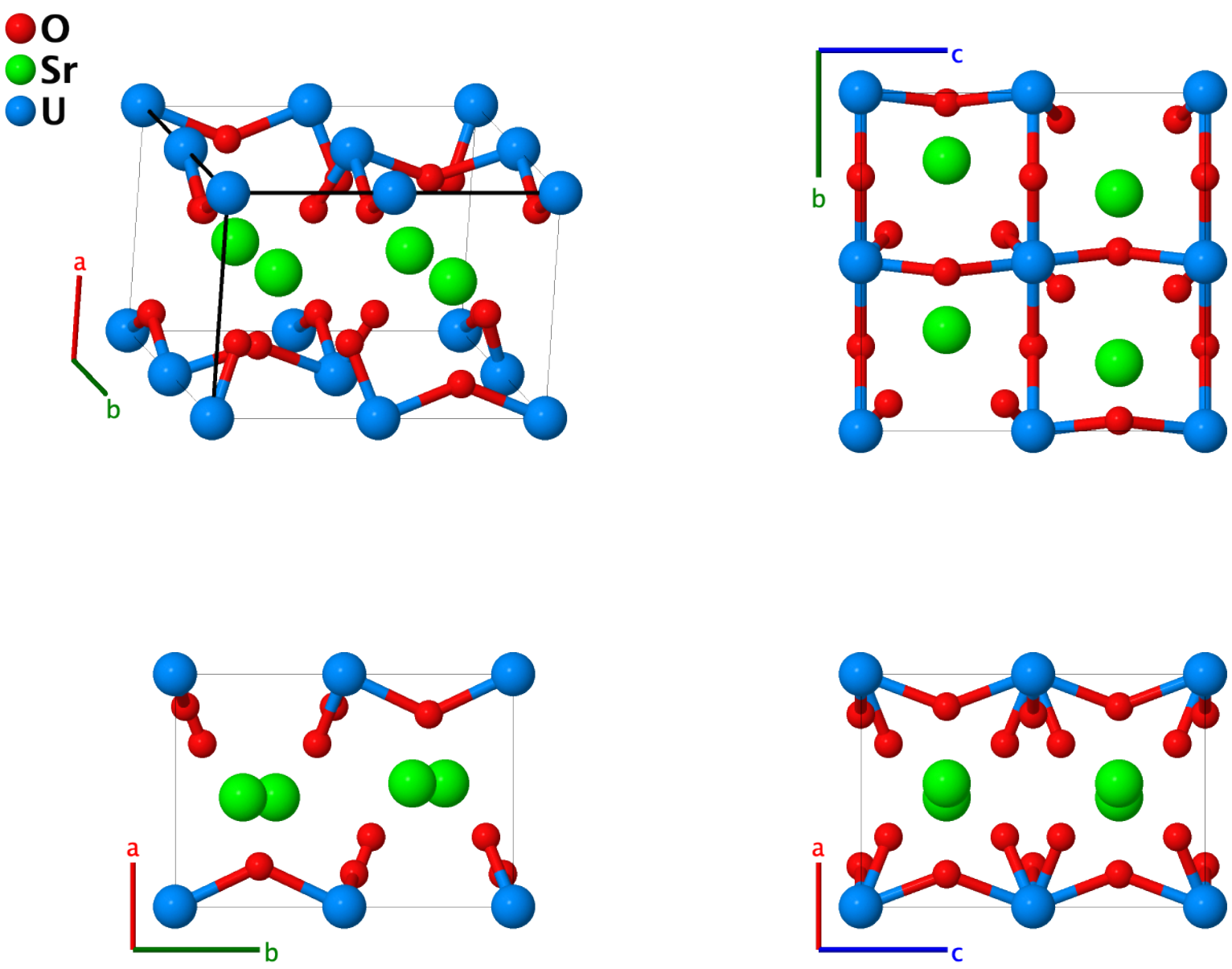

| | | |
|---|---|---|
| **Prototype** | : | $O_4SrU$ |
| **AFLOW prototype label** | : | A4BC_oP24_57_cde_d_a |
| ***Strukturbericht* designation** | : | None |
| **Pearson symbol** | : | oP24 |
| **Space group number** | : | 57 |
| **Space group symbol** | : | *Pbcm* |
| **AFLOW prototype command** | : | `aflow --proto=A4BC_oP24_57_cde_d_a` `--params=`$a, b/a, c/a, x_2, x_3, y_3, x_4, y_4, x_5, y_5, z_5$ |

### Other compounds with this structure

- $BaUO_4$

### Simple Orthorhombic primitive vectors:

$$\mathbf{a}_1 = a\,\hat{\mathbf{x}}$$
$$\mathbf{a}_2 = b\,\hat{\mathbf{y}}$$
$$\mathbf{a}_3 = c\,\hat{\mathbf{z}}$$

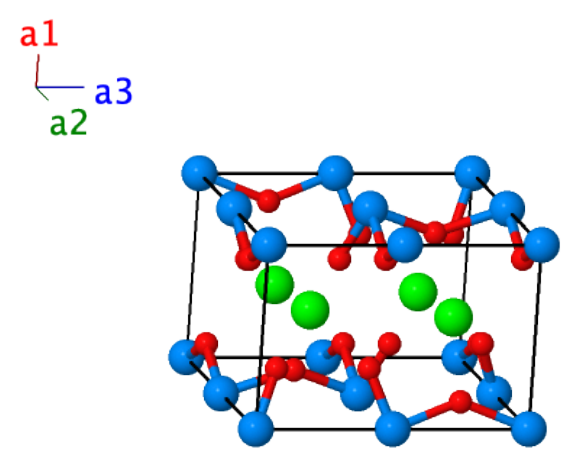

### Basis vectors:

| | | Lattice Coordinates | | Cartesian Coordinates | Wyckoff Position | Atom Type |
|---|---|---|---|---|---|---|
| $\mathbf{B}_1$ | = | $0\,\mathbf{a}_1 + 0\,\mathbf{a}_2 + 0\,\mathbf{a}_3$ | = | $0\,\hat{\mathbf{x}} + 0\,\hat{\mathbf{y}} + 0\,\hat{\mathbf{z}}$ | (4*a*) | U |
| $\mathbf{B}_2$ | = | $\frac{1}{2}\,\mathbf{a}_3$ | = | $\frac{1}{2}c\,\hat{\mathbf{z}}$ | (4*a*) | U |
| $\mathbf{B}_3$ | = | $\frac{1}{2}\,\mathbf{a}_2 + \frac{1}{2}\,\mathbf{a}_3$ | = | $\frac{1}{2}b\,\mathbf{\hat{y}} + \frac{1}{2}c\,\mathbf{\hat{z}}$ | (4*a*) | U |
| $\mathbf{B}_4$ | = | $\frac{1}{2}\,\mathbf{a}_2$ | = | $\frac{1}{2}b\,\mathbf{\hat{y}}$ | (4*a*) | U |
| $\mathbf{B}_5$ | = | $x_2\,\mathbf{a}_1 + \frac{1}{4}\,\mathbf{a}_2$ | = | $x_2a\,\mathbf{\hat{x}} + \frac{1}{4}b\,\mathbf{\hat{y}}$ | (4*c*) | O I |
| $\mathbf{B}_6$ | = | $-x_2\,\mathbf{a}_1 + \frac{3}{4}\,\mathbf{a}_2 + \frac{1}{2}\,\mathbf{a}_3$ | = | $-x_2a\,\mathbf{\hat{x}} + \frac{3}{4}b\,\mathbf{\hat{y}} + \frac{1}{2}c\,\mathbf{\hat{z}}$ | (4*c*) | O I |
| $\mathbf{B}_7$ | = | $-x_2\,\mathbf{a}_1 + \frac{3}{4}\,\mathbf{a}_2$ | = | $-x_2a\,\mathbf{\hat{x}} + \frac{3}{4}b\,\mathbf{\hat{y}}$ | (4*c*) | O I |
| $\mathbf{B}_8$ | = | $x_2\,\mathbf{a}_1 + \frac{1}{4}\,\mathbf{a}_2 + \frac{1}{2}\,\mathbf{a}_3$ | = | $x_2a\,\mathbf{\hat{x}} + \frac{1}{4}b\,\mathbf{\hat{y}} + \frac{1}{2}c\,\mathbf{\hat{z}}$ | (4*c*) | O I |
| $\mathbf{B}_9$ | = | $x_3\,\mathbf{a}_1 + y_3\,\mathbf{a}_2 + \frac{1}{4}\,\mathbf{a}_3$ | = | $x_3a\,\mathbf{\hat{x}} + y_3b\,\mathbf{\hat{y}} + \frac{1}{4}c\,\mathbf{\hat{z}}$ | (4*d*) | O II |
| $\mathbf{B}_{10}$ | = | $-x_3\,\mathbf{a}_1 - y_3\,\mathbf{a}_2 + \frac{3}{4}\,\mathbf{a}_3$ | = | $-x_3a\,\mathbf{\hat{x}} - y_3b\,\mathbf{\hat{y}} + \frac{3}{4}c\,\mathbf{\hat{z}}$ | (4*d*) | O II |
| $\mathbf{B}_{11}$ | = | $-x_3\,\mathbf{a}_1 + \left(\frac{1}{2} + y_3\right)\,\mathbf{a}_2 + \frac{1}{4}\,\mathbf{a}_3$ | = | $-x_3a\,\mathbf{\hat{x}} + \left(\frac{1}{2} + y_3\right)b\,\mathbf{\hat{y}} + \frac{1}{4}c\,\mathbf{\hat{z}}$ | (4*d*) | O II |
| $\mathbf{B}_{12}$ | = | $x_3\,\mathbf{a}_1 + \left(\frac{1}{2} - y_3\right)\,\mathbf{a}_2 + \frac{3}{4}\,\mathbf{a}_3$ | = | $x_3a\,\mathbf{\hat{x}} + \left(\frac{1}{2} - y_3\right)b\,\mathbf{\hat{y}} + \frac{3}{4}c\,\mathbf{\hat{z}}$ | (4*d*) | O II |
| $\mathbf{B}_{13}$ | = | $x_4\,\mathbf{a}_1 + y_4\,\mathbf{a}_2 + \frac{1}{4}\,\mathbf{a}_3$ | = | $x_4a\,\mathbf{\hat{x}} + y_4b\,\mathbf{\hat{y}} + \frac{1}{4}c\,\mathbf{\hat{z}}$ | (4*d*) | Sr |
| $\mathbf{B}_{14}$ | = | $-x_4\,\mathbf{a}_1 - y_4\,\mathbf{a}_2 + \frac{3}{4}\,\mathbf{a}_3$ | = | $-x_4a\,\mathbf{\hat{x}} - y_4b\,\mathbf{\hat{y}} + \frac{3}{4}c\,\mathbf{\hat{z}}$ | (4*d*) | Sr |
| $\mathbf{B}_{15}$ | = | $-x_4\,\mathbf{a}_1 + \left(\frac{1}{2} + y_4\right)\,\mathbf{a}_2 + \frac{1}{4}\,\mathbf{a}_3$ | = | $-x_4a\,\mathbf{\hat{x}} + \left(\frac{1}{2} + y_4\right)b\,\mathbf{\hat{y}} + \frac{1}{4}c\,\mathbf{\hat{z}}$ | (4*d*) | Sr |
| $\mathbf{B}_{16}$ | = | $x_4\,\mathbf{a}_1 + \left(\frac{1}{2} - y_4\right)\,\mathbf{a}_2 + \frac{3}{4}\,\mathbf{a}_3$ | = | $x_4a\,\mathbf{\hat{x}} + \left(\frac{1}{2} - y_4\right)b\,\mathbf{\hat{y}} + \frac{3}{4}c\,\mathbf{\hat{z}}$ | (4*d*) | Sr |
| $\mathbf{B}_{17}$ | = | $x_5\,\mathbf{a}_1 + y_5\,\mathbf{a}_2 + z_5\,\mathbf{a}_3$ | = | $x_5a\,\mathbf{\hat{x}} + y_5b\,\mathbf{\hat{y}} + z_5c\,\mathbf{\hat{z}}$ | (8*e*) | O III |
| $\mathbf{B}_{18}$ | = | $-x_5\,\mathbf{a}_1 - y_5\,\mathbf{a}_2 + \left(\frac{1}{2} + z_5\right)\,\mathbf{a}_3$ | = | $-x_5a\,\mathbf{\hat{x}} - y_5b\,\mathbf{\hat{y}} + \left(\frac{1}{2} + z_5\right)c\,\mathbf{\hat{z}}$ | (8*e*) | O III |
| $\mathbf{B}_{19}$ | = | $-x_5\,\mathbf{a}_1 + \left(\frac{1}{2} + y_5\right)\,\mathbf{a}_2 + \left(\frac{1}{2} - z_5\right)\,\mathbf{a}_3$ | = | $-x_5a\,\mathbf{\hat{x}} + \left(\frac{1}{2} + y_5\right)b\,\mathbf{\hat{y}} + \left(\frac{1}{2} - z_5\right)c\,\mathbf{\hat{z}}$ | (8*e*) | O III |
| $\mathbf{B}_{20}$ | = | $x_5\,\mathbf{a}_1 + \left(\frac{1}{2} - y_5\right)\,\mathbf{a}_2 - z_5\,\mathbf{a}_3$ | = | $x_5a\,\mathbf{\hat{x}} + \left(\frac{1}{2} - y_5\right)b\,\mathbf{\hat{y}} - z_5c\,\mathbf{\hat{z}}$ | (8*e*) | O III |
| $\mathbf{B}_{21}$ | = | $-x_5\,\mathbf{a}_1 - y_5\,\mathbf{a}_2 - z_5\,\mathbf{a}_3$ | = | $-x_5a\,\mathbf{\hat{x}} - y_5b\,\mathbf{\hat{y}} - z_5c\,\mathbf{\hat{z}}$ | (8*e*) | O III |
| $\mathbf{B}_{22}$ | = | $x_5\,\mathbf{a}_1 + y_5\,\mathbf{a}_2 + \left(\frac{1}{2} - z_5\right)\,\mathbf{a}_3$ | = | $x_5a\,\mathbf{\hat{x}} + y_5b\,\mathbf{\hat{y}} + \left(\frac{1}{2} - z_5\right)c\,\mathbf{\hat{z}}$ | (8*e*) | O III |
| $\mathbf{B}_{23}$ | = | $x_5\,\mathbf{a}_1 + \left(\frac{1}{2} - y_5\right)\,\mathbf{a}_2 + \left(\frac{1}{2} + z_5\right)\,\mathbf{a}_3$ | = | $x_5a\,\mathbf{\hat{x}} + \left(\frac{1}{2} - y_5\right)b\,\mathbf{\hat{y}} + \left(\frac{1}{2} + z_5\right)c\,\mathbf{\hat{z}}$ | (8*e*) | O III |
| $\mathbf{B}_{24}$ | = | $-x_5\,\mathbf{a}_1 + \left(\frac{1}{2} + y_5\right)\,\mathbf{a}_2 + z_5\,\mathbf{a}_3$ | = | $-x_5a\,\mathbf{\hat{x}} + \left(\frac{1}{2} + y_5\right)b\,\mathbf{\hat{y}} + z_5c\,\mathbf{\hat{z}}$ | (8*e*) | O III |

**Geometry files:**

- CIF: pp. 1616
- POSCAR: pp. 1616

# Lueshite ($NaNbO_3$) Structure: ABC3_oP40_57_cd_e_cd2e

http://aflow.org/prototype-encyclopedia/ABC3_oP40_57_cd_e_cd2e

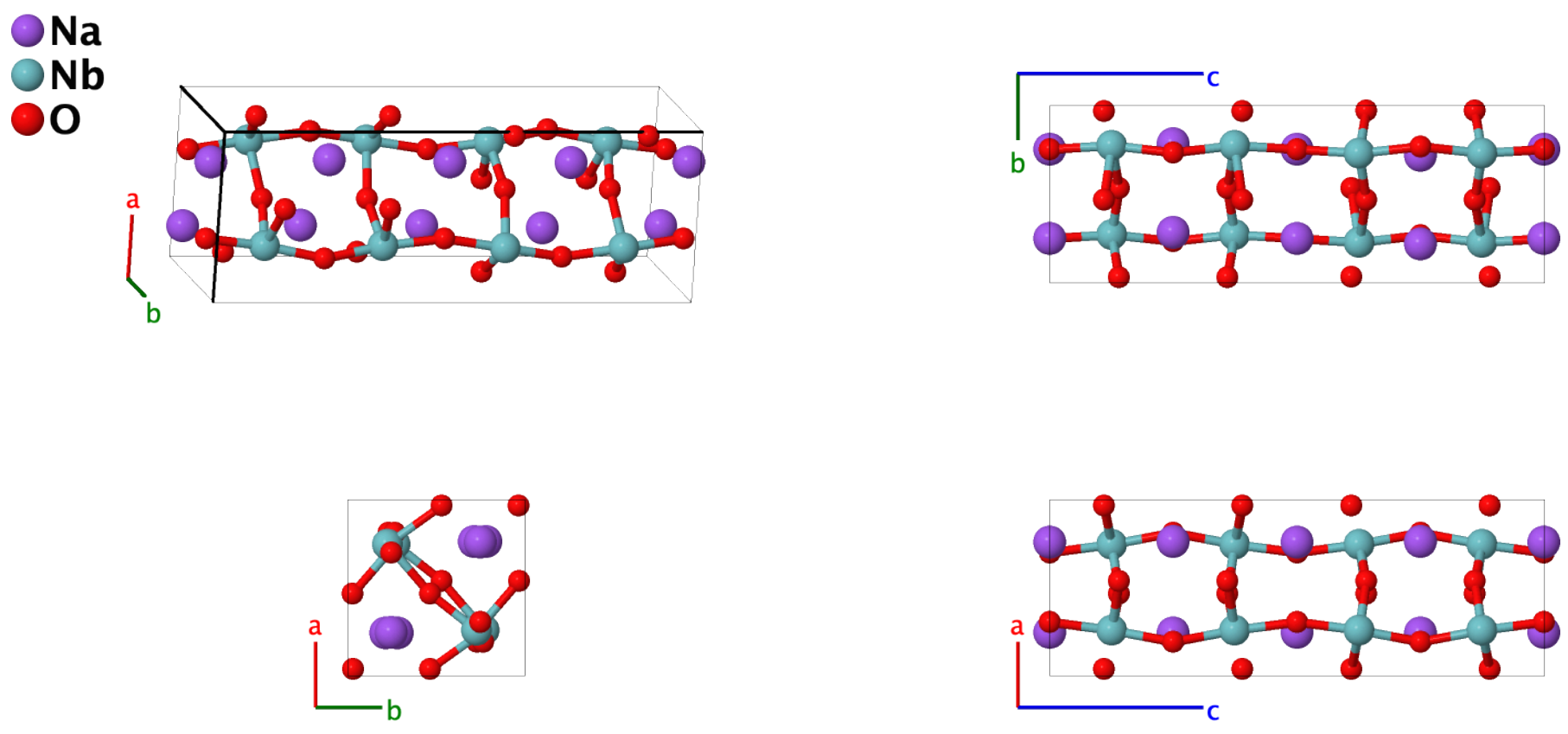

| | | |
|---|---|---|
| **Prototype** | : | $NaNbO_3$ |
| **AFLOW prototype label** | : | ABC3_oP40_57_cd_e_cd2e |
| ***Strukturbericht* designation** | : | None |
| **Pearson symbol** | : | oP40 |
| **Space group number** | : | 57 |
| **Space group symbol** | : | *Pbcm* |
| **AFLOW prototype command** | : | `aflow --proto=ABC3_oP40_57_cd_e_cd2e`<br>`--params=`$a, b/a, c/a, x_1, x_2, x_3, y_3, x_4, y_4, x_5, y_5, z_5, x_6, y_6, z_6, x_7, y_7, z_7$ |

### Other compounds with this structure

- $NaNb_{1-x}Ti_xO_3$

- If the AFLOW parameters are set to `--params=`$a, 1, \sqrt{8}, 1/4, 3/4, 1/4, 1/4, 1/4, 3/4, 1/4, 3/4, 5/8, 1/2, 1/2, 5/8, 0, 0, 5/8$ then the structure is equivalent to Cubic Perovskite $E2_1$.

### Simple Orthorhombic primitive vectors:

$$\mathbf{a}_1 = a\,\hat{\mathbf{x}}$$
$$\mathbf{a}_2 = b\,\hat{\mathbf{y}}$$
$$\mathbf{a}_3 = c\,\hat{\mathbf{z}}$$

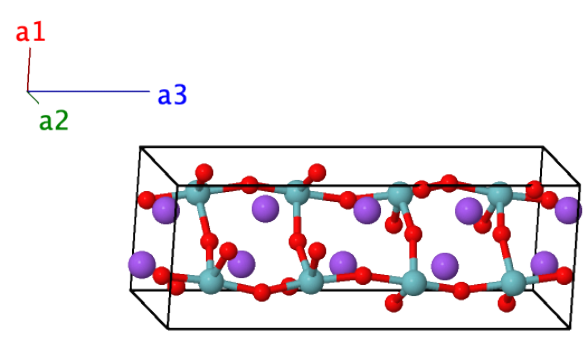

### Basis vectors:

| | | Lattice Coordinates | | Cartesian Coordinates | Wyckoff Position | Atom Type |
|---|---|---|---|---|---|---|
| $\mathbf{B}_1$ | = | $x_1\,\mathbf{a}_1 + \frac{1}{4}\,\mathbf{a}_2$ | = | $x_1 a\,\hat{\mathbf{x}} + \frac{1}{4} b\,\hat{\mathbf{y}}$ | (4*c*) | Na I |
| $\mathbf{B}_2$ | = | $-x_1\,\mathbf{a}_1 + \frac{3}{4}\,\mathbf{a}_2 + \frac{1}{2}\,\mathbf{a}_3$ | = | $-x_1 a\,\hat{\mathbf{x}} + \frac{3}{4} b\,\hat{\mathbf{y}} + \frac{1}{2} c\,\hat{\mathbf{z}}$ | (4*c*) | Na I |
| $\mathbf{B}_3$ | = | $-x_1\,\mathbf{a}_1 + \frac{3}{4}\,\mathbf{a}_2$ | = | $-x_1 a\,\hat{\mathbf{x}} + \frac{3}{4} b\,\hat{\mathbf{y}}$ | (4*c*) | Na I |
| $\mathbf{B}_4$ | = | $x_1\,\mathbf{a}_1 + \frac{1}{4}\,\mathbf{a}_2 + \frac{1}{2}\,\mathbf{a}_3$ | = | $x_1 a\,\hat{\mathbf{x}} + \frac{1}{4} b\,\hat{\mathbf{y}} + \frac{1}{2} c\,\hat{\mathbf{z}}$ | (4*c*) | Na I |
| $\mathbf{B}_5$ | = | $x_2\,\mathbf{a}_1 + \frac{1}{4}\,\mathbf{a}_2$ | = | $x_2 a\,\hat{\mathbf{x}} + \frac{1}{4} b\,\hat{\mathbf{y}}$ | (4*c*) | O I |
| $\mathbf{B}_{6}$ | = | $-x_{2}\,\mathbf{a}_{1}+\frac{3}{4}\,\mathbf{a}_{2}+\frac{1}{2}\,\mathbf{a}_{3}$ | = | $-x_{2}a\,\hat{\mathbf{x}}+\frac{3}{4}b\,\hat{\mathbf{y}}+\frac{1}{2}c\,\hat{\mathbf{z}}$ | (4*c*) | O I |
| $\mathbf{B}_{7}$ | = | $-x_{2}\,\mathbf{a}_{1}+\frac{3}{4}\,\mathbf{a}_{2}$ | = | $-x_{2}a\,\hat{\mathbf{x}}+\frac{3}{4}b\,\hat{\mathbf{y}}$ | (4*c*) | O I |
| $\mathbf{B}_{8}$ | = | $x_{2}\,\mathbf{a}_{1}+\frac{1}{4}\,\mathbf{a}_{2}+\frac{1}{2}\,\mathbf{a}_{3}$ | = | $x_{2}a\,\hat{\mathbf{x}}+\frac{1}{4}b\,\hat{\mathbf{y}}+\frac{1}{2}c\,\hat{\mathbf{z}}$ | (4*c*) | O I |
| $\mathbf{B}_{9}$ | = | $x_{3}\,\mathbf{a}_{1}+y_{3}\,\mathbf{a}_{2}+\frac{1}{4}\,\mathbf{a}_{3}$ | = | $x_{3}a\,\hat{\mathbf{x}}+y_{3}b\,\hat{\mathbf{y}}+\frac{1}{4}c\,\hat{\mathbf{z}}$ | (4*d*) | Na II |
| $\mathbf{B}_{10}$ | = | $-x_{3}\,\mathbf{a}_{1}-y_{3}\,\mathbf{a}_{2}+\frac{3}{4}\,\mathbf{a}_{3}$ | = | $-x_{3}a\,\hat{\mathbf{x}}-y_{3}b\,\hat{\mathbf{y}}+\frac{3}{4}c\,\hat{\mathbf{z}}$ | (4*d*) | Na II |
| $\mathbf{B}_{11}$ | = | $-x_{3}\,\mathbf{a}_{1}+\left(\frac{1}{2}+y_{3}\right)\,\mathbf{a}_{2}+\frac{1}{4}\,\mathbf{a}_{3}$ | = | $-x_{3}a\,\hat{\mathbf{x}}+\left(\frac{1}{2}+y_{3}\right)b\,\hat{\mathbf{y}}+\frac{1}{4}c\,\hat{\mathbf{z}}$ | (4*d*) | Na II |
| $\mathbf{B}_{12}$ | = | $x_{3}\,\mathbf{a}_{1}+\left(\frac{1}{2}-y_{3}\right)\,\mathbf{a}_{2}+\frac{3}{4}\,\mathbf{a}_{3}$ | = | $x_{3}a\,\hat{\mathbf{x}}+\left(\frac{1}{2}-y_{3}\right)b\,\hat{\mathbf{y}}+\frac{3}{4}c\,\hat{\mathbf{z}}$ | (4*d*) | Na II |
| $\mathbf{B}_{13}$ | = | $x_{4}\,\mathbf{a}_{1}+y_{4}\,\mathbf{a}_{2}+\frac{1}{4}\,\mathbf{a}_{3}$ | = | $x_{4}a\,\hat{\mathbf{x}}+y_{4}b\,\hat{\mathbf{y}}+\frac{1}{4}c\,\hat{\mathbf{z}}$ | (4*d*) | O II |
| $\mathbf{B}_{14}$ | = | $-x_{4}\,\mathbf{a}_{1}-y_{4}\,\mathbf{a}_{2}+\frac{3}{4}\,\mathbf{a}_{3}$ | = | $-x_{4}a\,\hat{\mathbf{x}}-y_{4}b\,\hat{\mathbf{y}}+\frac{3}{4}c\,\hat{\mathbf{z}}$ | (4*d*) | O II |
| $\mathbf{B}_{15}$ | = | $-x_{4}\,\mathbf{a}_{1}+\left(\frac{1}{2}+y_{4}\right)\,\mathbf{a}_{2}+\frac{1}{4}\,\mathbf{a}_{3}$ | = | $-x_{4}a\,\hat{\mathbf{x}}+\left(\frac{1}{2}+y_{4}\right)b\,\hat{\mathbf{y}}+\frac{1}{4}c\,\hat{\mathbf{z}}$ | (4*d*) | O II |
| $\mathbf{B}_{16}$ | = | $x_{4}\,\mathbf{a}_{1}+\left(\frac{1}{2}-y_{4}\right)\,\mathbf{a}_{2}+\frac{3}{4}\,\mathbf{a}_{3}$ | = | $x_{4}a\,\hat{\mathbf{x}}+\left(\frac{1}{2}-y_{4}\right)b\,\hat{\mathbf{y}}+\frac{3}{4}c\,\hat{\mathbf{z}}$ | (4*d*) | O II |
| $\mathbf{B}_{17}$ | = | $x_{5}\,\mathbf{a}_{1}+y_{5}\,\mathbf{a}_{2}+z_{5}\,\mathbf{a}_{3}$ | = | $x_{5}a\,\hat{\mathbf{x}}+y_{5}b\,\hat{\mathbf{y}}+z_{5}c\,\hat{\mathbf{z}}$ | (8*e*) | Nb |
| $\mathbf{B}_{18}$ | = | $-x_{5}\,\mathbf{a}_{1}-y_{5}\,\mathbf{a}_{2}+\left(\frac{1}{2}+z_{5}\right)\,\mathbf{a}_{3}$ | = | $-x_{5}a\,\hat{\mathbf{x}}-y_{5}b\,\hat{\mathbf{y}}+\left(\frac{1}{2}+z_{5}\right)c\,\hat{\mathbf{z}}$ | (8*e*) | Nb |
| $\mathbf{B}_{19}$ | = | $-x_{5}\,\mathbf{a}_{1}+\left(\frac{1}{2}+y_{5}\right)\,\mathbf{a}_{2}+\left(\frac{1}{2}-z_{5}\right)\,\mathbf{a}_{3}$ | = | $-x_{5}a\,\hat{\mathbf{x}}+\left(\frac{1}{2}+y_{5}\right)b\,\hat{\mathbf{y}}+\left(\frac{1}{2}-z_{5}\right)c\,\hat{\mathbf{z}}$ | (8*e*) | Nb |
| $\mathbf{B}_{20}$ | = | $x_{5}\,\mathbf{a}_{1}+\left(\frac{1}{2}-y_{5}\right)\,\mathbf{a}_{2}-z_{5}\,\mathbf{a}_{3}$ | = | $x_{5}a\,\hat{\mathbf{x}}+\left(\frac{1}{2}-y_{5}\right)b\,\hat{\mathbf{y}}-z_{5}c\,\hat{\mathbf{z}}$ | (8*e*) | Nb |
| $\mathbf{B}_{21}$ | = | $-x_{5}\,\mathbf{a}_{1}-y_{5}\,\mathbf{a}_{2}-z_{5}\,\mathbf{a}_{3}$ | = | $-x_{5}a\,\hat{\mathbf{x}}-y_{5}b\,\hat{\mathbf{y}}-z_{5}c\,\hat{\mathbf{z}}$ | (8*e*) | Nb |
| $\mathbf{B}_{22}$ | = | $x_{5}\,\mathbf{a}_{1}+y_{5}\,\mathbf{a}_{2}+\left(\frac{1}{2}-z_{5}\right)\,\mathbf{a}_{3}$ | = | $x_{5}a\,\hat{\mathbf{x}}+y_{5}b\,\hat{\mathbf{y}}+\left(\frac{1}{2}-z_{5}\right)c\,\hat{\mathbf{z}}$ | (8*e*) | Nb |
| $\mathbf{B}_{23}$ | = | $x_{5}\,\mathbf{a}_{1}+\left(\frac{1}{2}-y_{5}\right)\,\mathbf{a}_{2}+\left(\frac{1}{2}+z_{5}\right)\,\mathbf{a}_{3}$ | = | $x_{5}a\,\hat{\mathbf{x}}+\left(\frac{1}{2}-y_{5}\right)b\,\hat{\mathbf{y}}+\left(\frac{1}{2}+z_{5}\right)c\,\hat{\mathbf{z}}$ | (8*e*) | Nb |
| $\mathbf{B}_{24}$ | = | $-x_{5}\,\mathbf{a}_{1}+\left(\frac{1}{2}+y_{5}\right)\,\mathbf{a}_{2}+z_{5}\,\mathbf{a}_{3}$ | = | $-x_{5}a\,\hat{\mathbf{x}}+\left(\frac{1}{2}+y_{5}\right)b\,\hat{\mathbf{y}}+z_{5}c\,\hat{\mathbf{z}}$ | (8*e*) | Nb |
| $\mathbf{B}_{25}$ | = | $x_{6}\,\mathbf{a}_{1}+y_{6}\,\mathbf{a}_{2}+z_{6}\,\mathbf{a}_{3}$ | = | $x_{6}a\,\hat{\mathbf{x}}+y_{6}b\,\hat{\mathbf{y}}+z_{6}c\,\hat{\mathbf{z}}$ | (8*e*) | O III |
| $\mathbf{B}_{26}$ | = | $-x_{6}\,\mathbf{a}_{1}-y_{6}\,\mathbf{a}_{2}+\left(\frac{1}{2}+z_{6}\right)\,\mathbf{a}_{3}$ | = | $-x_{6}a\,\hat{\mathbf{x}}-y_{6}b\,\hat{\mathbf{y}}+\left(\frac{1}{2}+z_{6}\right)c\,\hat{\mathbf{z}}$ | (8*e*) | O III |
| $\mathbf{B}_{27}$ | = | $-x_{6}\,\mathbf{a}_{1}+\left(\frac{1}{2}+y_{6}\right)\,\mathbf{a}_{2}+\left(\frac{1}{2}-z_{6}\right)\,\mathbf{a}_{3}$ | = | $-x_{6}a\,\hat{\mathbf{x}}+\left(\frac{1}{2}+y_{6}\right)b\,\hat{\mathbf{y}}+\left(\frac{1}{2}-z_{6}\right)c\,\hat{\mathbf{z}}$ | (8*e*) | O III |
| $\mathbf{B}_{28}$ | = | $x_{6}\,\mathbf{a}_{1}+\left(\frac{1}{2}-y_{6}\right)\,\mathbf{a}_{2}-z_{6}\,\mathbf{a}_{3}$ | = | $x_{6}a\,\hat{\mathbf{x}}+\left(\frac{1}{2}-y_{6}\right)b\,\hat{\mathbf{y}}-z_{6}c\,\hat{\mathbf{z}}$ | (8*e*) | O III |
| $\mathbf{B}_{29}$ | = | $-x_{6}\,\mathbf{a}_{1}-y_{6}\,\mathbf{a}_{2}-z_{6}\,\mathbf{a}_{3}$ | = | $-x_{6}a\,\hat{\mathbf{x}}-y_{6}b\,\hat{\mathbf{y}}-z_{6}c\,\hat{\mathbf{z}}$ | (8*e*) | O III |
| $\mathbf{B}_{30}$ | = | $x_{6}\,\mathbf{a}_{1}+y_{6}\,\mathbf{a}_{2}+\left(\frac{1}{2}-z_{6}\right)\,\mathbf{a}_{3}$ | = | $x_{6}a\,\hat{\mathbf{x}}+y_{6}b\,\hat{\mathbf{y}}+\left(\frac{1}{2}-z_{6}\right)c\,\hat{\mathbf{z}}$ | (8*e*) | O III |
| $\mathbf{B}_{31}$ | = | $x_{6}\,\mathbf{a}_{1}+\left(\frac{1}{2}-y_{6}\right)\,\mathbf{a}_{2}+\left(\frac{1}{2}+z_{6}\right)\,\mathbf{a}_{3}$ | = | $x_{6}a\,\hat{\mathbf{x}}+\left(\frac{1}{2}-y_{6}\right)b\,\hat{\mathbf{y}}+\left(\frac{1}{2}+z_{6}\right)c\,\hat{\mathbf{z}}$ | (8*e*) | O III |
| $\mathbf{B}_{32}$ | = | $-x_{6}\,\mathbf{a}_{1}+\left(\frac{1}{2}+y_{6}\right)\,\mathbf{a}_{2}+z_{6}\,\mathbf{a}_{3}$ | = | $-x_{6}a\,\hat{\mathbf{x}}+\left(\frac{1}{2}+y_{6}\right)b\,\hat{\mathbf{y}}+z_{6}c\,\hat{\mathbf{z}}$ | (8*e*) | O III |
| $\mathbf{B}_{33}$ | = | $x_{7}\,\mathbf{a}_{1}+y_{7}\,\mathbf{a}_{2}+z_{7}\,\mathbf{a}_{3}$ | = | $x_{7}a\,\hat{\mathbf{x}}+y_{7}b\,\hat{\mathbf{y}}+z_{7}c\,\hat{\mathbf{z}}$ | (8*e*) | O IV |
| $\mathbf{B}_{34}$ | = | $-x_{7}\,\mathbf{a}_{1}-y_{7}\,\mathbf{a}_{2}+\left(\frac{1}{2}+z_{7}\right)\,\mathbf{a}_{3}$ | = | $-x_{7}a\,\hat{\mathbf{x}}-y_{7}b\,\hat{\mathbf{y}}+\left(\frac{1}{2}+z_{7}\right)c\,\hat{\mathbf{z}}$ | (8*e*) | O IV |
| $\mathbf{B}_{35}$ | = | $-x_{7}\,\mathbf{a}_{1}+\left(\frac{1}{2}+y_{7}\right)\,\mathbf{a}_{2}+\left(\frac{1}{2}-z_{7}\right)\,\mathbf{a}_{3}$ | = | $-x_{7}a\,\hat{\mathbf{x}}+\left(\frac{1}{2}+y_{7}\right)b\,\hat{\mathbf{y}}+\left(\frac{1}{2}-z_{7}\right)c\,\hat{\mathbf{z}}$ | (8*e*) | O IV |
| $\mathbf{B}_{36}$ | = | $x_{7}\,\mathbf{a}_{1}+\left(\frac{1}{2}-y_{7}\right)\,\mathbf{a}_{2}-z_{7}\,\mathbf{a}_{3}$ | = | $x_{7}a\,\hat{\mathbf{x}}+\left(\frac{1}{2}-y_{7}\right)b\,\hat{\mathbf{y}}-z_{7}c\,\hat{\mathbf{z}}$ | (8*e*) | O IV |
| $\mathbf{B}_{37}$ | = | $-x_{7}\,\mathbf{a}_{1}-y_{7}\,\mathbf{a}_{2}-z_{7}\,\mathbf{a}_{3}$ | = | $-x_{7}a\,\hat{\mathbf{x}}-y_{7}b\,\hat{\mathbf{y}}-z_{7}c\,\hat{\mathbf{z}}$ | (8*e*) | O IV |
| $\mathbf{B}_{38}$ | = | $x_{7}\,\mathbf{a}_{1}+y_{7}\,\mathbf{a}_{2}+\left(\frac{1}{2}-z_{7}\right)\,\mathbf{a}_{3}$ | = | $x_{7}a\,\hat{\mathbf{x}}+y_{7}b\,\hat{\mathbf{y}}+\left(\frac{1}{2}-z_{7}\right)c\,\hat{\mathbf{z}}$ | (8*e*) | O IV |
| $\mathbf{B}_{39}$ | = | $x_{7}\,\mathbf{a}_{1}+\left(\frac{1}{2}-y_{7}\right)\,\mathbf{a}_{2}+\left(\frac{1}{2}+z_{7}\right)\,\mathbf{a}_{3}$ | = | $x_{7}a\,\hat{\mathbf{x}}+\left(\frac{1}{2}-y_{7}\right)b\,\hat{\mathbf{y}}+\left(\frac{1}{2}+z_{7}\right)c\,\hat{\mathbf{z}}$ | (8*e*) | O IV |
| $\mathbf{B}_{40}$ | = | $-x_{7}\,\mathbf{a}_{1}+\left(\frac{1}{2}+y_{7}\right)\,\mathbf{a}_{2}+z_{7}\,\mathbf{a}_{3}$ | = | $-x_{7}a\,\hat{\mathbf{x}}+\left(\frac{1}{2}+y_{7}\right)b\,\hat{\mathbf{y}}+z_{7}c\,\hat{\mathbf{z}}$ | (8*e*) | O IV |

**Found in:**

---

**Geometry files:**

- CIF: pp. 1616
- POSCAR: pp. 1617

# Kotoite ($Mg_3(BO_3)_2$) Structure: A2B3C6_oP22_58_g_af_gh

http://aflow.org/prototype-encyclopedia/A2B3C6_oP22_58_g_af_gh

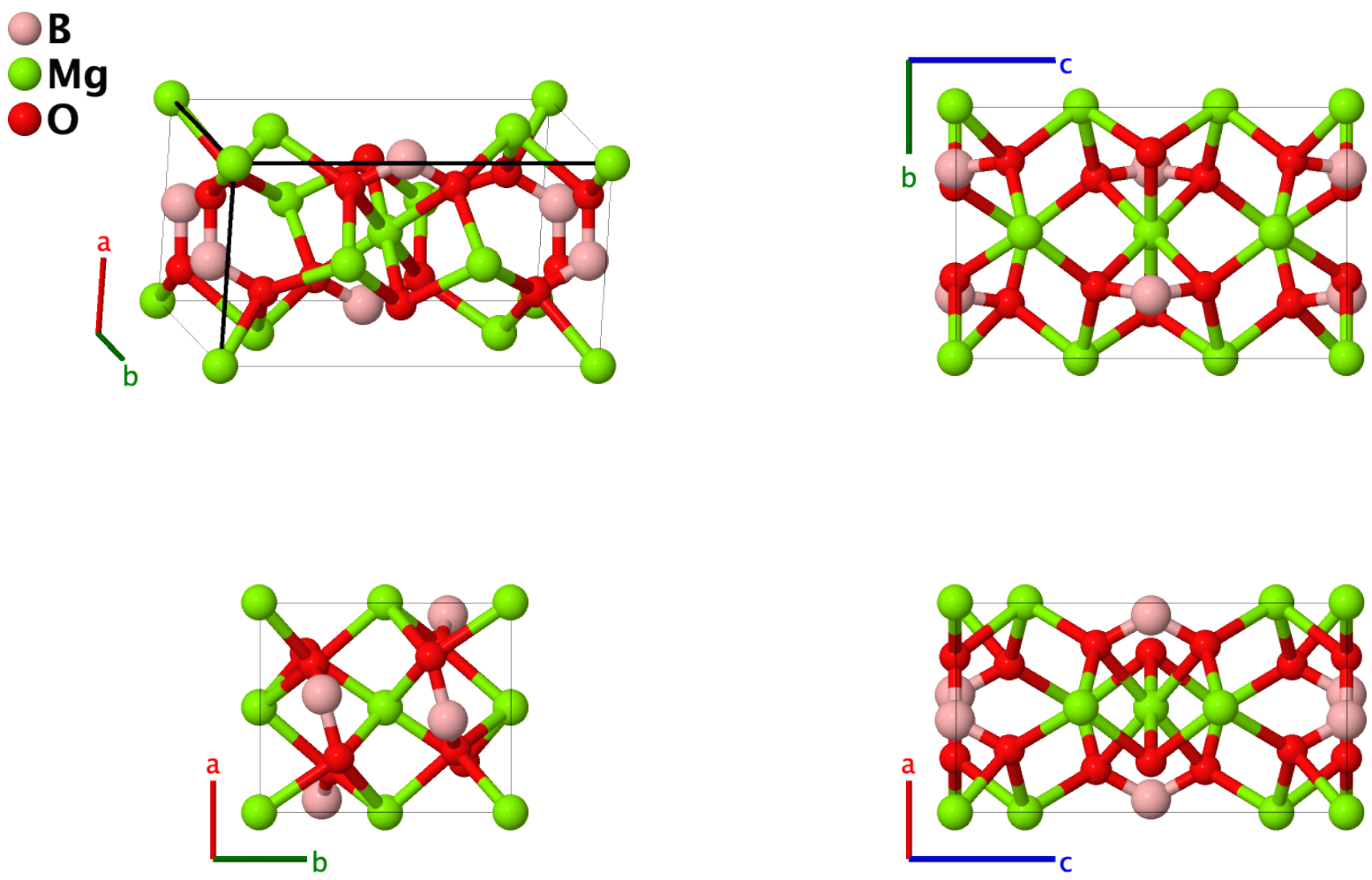

| | | |
|---|---|---|
| **Prototype** | : | $B_2Mg_3O_6$ |
| **AFLOW prototype label** | : | A2B3C6_oP22_58_g_af_gh |
| ***Strukturbericht* designation** | : | None |
| **Pearson symbol** | : | oP22 |
| **Space group number** | : | 58 |
| **Space group symbol** | : | *Pnnm* |
| **AFLOW prototype command** | : | `aflow --proto=A2B3C6_oP22_58_g_af_gh`<br>`--params=`$a, b/a, c/a, z_2, x_3, y_3, x_4, y_4, x_5, y_5, z_5$ |

**Other compounds with this structure**

- $Co_3(BO_3)_2$, $Ni_3(BO_3)_2$, $Mn_3(BO_3)_2$ (Jimboite), $CoNi_2(BO_3)_2$, $KNa_2(BO_3)_2$, and $Eu_3(BO_3)_2$

**Simple Orthorhombic primitive vectors:**

$$\mathbf{a}_1 = a\,\hat{\mathbf{x}}$$
$$\mathbf{a}_2 = b\,\hat{\mathbf{y}}$$
$$\mathbf{a}_3 = c\,\hat{\mathbf{z}}$$

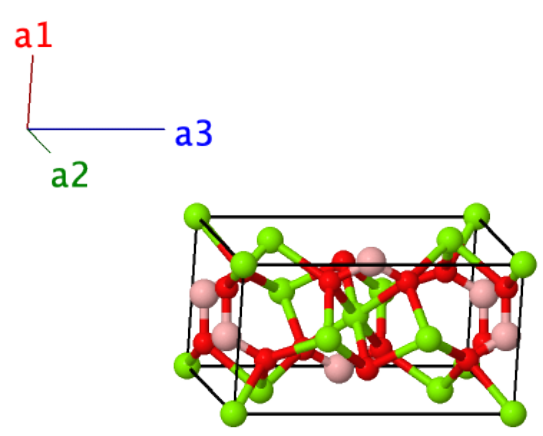

**Basis vectors:**

| | | Lattice Coordinates | | Cartesian Coordinates | Wyckoff Position | Atom Type |
|---|---|---|---|---|---|---|
| $\mathbf{B}_1$ | = | $0\,\mathbf{a}_1 + 0\,\mathbf{a}_2 + 0\,\mathbf{a}_3$ | = | $0\,\hat{\mathbf{x}} + 0\,\hat{\mathbf{y}} + 0\,\hat{\mathbf{z}}$ | (2*a*) | Mg I |
| $\mathbf{B}_2$ | = | $\frac{1}{2}\,\mathbf{a}_1 + \frac{1}{2}\,\mathbf{a}_2 + \frac{1}{2}\,\mathbf{a}_3$ | = | $\frac{1}{2}a\,\hat{\mathbf{x}} + \frac{1}{2}b\,\hat{\mathbf{y}} + \frac{1}{2}c\,\hat{\mathbf{z}}$ | (2*a*) | Mg I |
| $\mathbf{B}_3$ | = | $\frac{1}{2}\,\mathbf{a}_2 + z_2\,\mathbf{a}_3$ | = | $\frac{1}{2}b\,\hat{\mathbf{y}} + z_2c\,\hat{\mathbf{z}}$ | (4*f*) | Mg II |
| $\mathbf{B}_4$ | = | $\frac{1}{2}\,\mathbf{a}_1 + \left(\frac{1}{2} - z_2\right)\,\mathbf{a}_3$ | = | $\frac{1}{2}a\,\mathbf{\hat{x}} + \left(\frac{1}{2} - z_2\right)c\,\mathbf{\hat{z}}$ | (4$f$) | Mg II |
| $\mathbf{B}_5$ | = | $\frac{1}{2}\,\mathbf{a}_2 - z_2\,\mathbf{a}_3$ | = | $\frac{1}{2}b\,\mathbf{\hat{y}} - z_2c\,\mathbf{\hat{z}}$ | (4$f$) | Mg II |
| $\mathbf{B}_6$ | = | $\frac{1}{2}\,\mathbf{a}_1 + \left(\frac{1}{2} + z_2\right)\,\mathbf{a}_3$ | = | $\frac{1}{2}a\,\mathbf{\hat{x}} + \left(\frac{1}{2} + z_2\right)c\,\mathbf{\hat{z}}$ | (4$f$) | Mg II |
| $\mathbf{B}_7$ | = | $x_3\,\mathbf{a}_1 + y_3\,\mathbf{a}_2$ | = | $x_3a\,\mathbf{\hat{x}} + y_3b\,\mathbf{\hat{y}}$ | (4$g$) | B |
| $\mathbf{B}_8$ | = | $-x_3\,\mathbf{a}_1 - y_3\,\mathbf{a}_2$ | = | $-x_3a\,\mathbf{\hat{x}} - y_3b\,\mathbf{\hat{y}}$ | (4$g$) | B |
| $\mathbf{B}_9$ | = | $\left(\frac{1}{2} - x_3\right)\,\mathbf{a}_1 + \left(\frac{1}{2} + y_3\right)\,\mathbf{a}_2 + \frac{1}{2}\,\mathbf{a}_3$ | = | $\left(\frac{1}{2} - x_3\right)a\,\mathbf{\hat{x}} + \left(\frac{1}{2} + y_3\right)b\,\mathbf{\hat{y}} + \frac{1}{2}c\,\mathbf{\hat{z}}$ | (4$g$) | B |
| $\mathbf{B}_{10}$ | = | $\left(\frac{1}{2} + x_3\right)\,\mathbf{a}_1 + \left(\frac{1}{2} - y_3\right)\,\mathbf{a}_2 + \frac{1}{2}\,\mathbf{a}_3$ | = | $\left(\frac{1}{2} + x_3\right)a\,\mathbf{\hat{x}} + \left(\frac{1}{2} - y_3\right)b\,\mathbf{\hat{y}} + \frac{1}{2}c\,\mathbf{\hat{z}}$ | (4$g$) | B |
| $\mathbf{B}_{11}$ | = | $x_4\,\mathbf{a}_1 + y_4\,\mathbf{a}_2$ | = | $x_4a\,\mathbf{\hat{x}} + y_4b\,\mathbf{\hat{y}}$ | (4$g$) | O I |
| $\mathbf{B}_{12}$ | = | $-x_4\,\mathbf{a}_1 - y_4\,\mathbf{a}_2$ | = | $-x_4a\,\mathbf{\hat{x}} - y_4b\,\mathbf{\hat{y}}$ | (4$g$) | O I |
| $\mathbf{B}_{13}$ | = | $\left(\frac{1}{2} - x_4\right)\,\mathbf{a}_1 + \left(\frac{1}{2} + y_4\right)\,\mathbf{a}_2 + \frac{1}{2}\,\mathbf{a}_3$ | = | $\left(\frac{1}{2} - x_4\right)a\,\mathbf{\hat{x}} + \left(\frac{1}{2} + y_4\right)b\,\mathbf{\hat{y}} + \frac{1}{2}c\,\mathbf{\hat{z}}$ | (4$g$) | O I |
| $\mathbf{B}_{14}$ | = | $\left(\frac{1}{2} + x_4\right)\,\mathbf{a}_1 + \left(\frac{1}{2} - y_4\right)\,\mathbf{a}_2 + \frac{1}{2}\,\mathbf{a}_3$ | = | $\left(\frac{1}{2} + x_4\right)a\,\mathbf{\hat{x}} + \left(\frac{1}{2} - y_4\right)b\,\mathbf{\hat{y}} + \frac{1}{2}c\,\mathbf{\hat{z}}$ | (4$g$) | O I |
| $\mathbf{B}_{15}$ | = | $x_5\,\mathbf{a}_1 + y_5\,\mathbf{a}_2 + z_5\,\mathbf{a}_3$ | = | $x_5a\,\mathbf{\hat{x}} + y_5b\,\mathbf{\hat{y}} + z_5c\,\mathbf{\hat{z}}$ | (8$h$) | O II |
| $\mathbf{B}_{16}$ | = | $-x_5\,\mathbf{a}_1 - y_5\,\mathbf{a}_2 + z_5\,\mathbf{a}_3$ | = | $-x_5a\,\mathbf{\hat{x}} - y_5b\,\mathbf{\hat{y}} + z_5c\,\mathbf{\hat{z}}$ | (8$h$) | O II |
| $\mathbf{B}_{17}$ | = | $\left(\frac{1}{2} - x_5\right)\,\mathbf{a}_1 + \left(\frac{1}{2} + y_5\right)\,\mathbf{a}_2 + \left(\frac{1}{2} - z_5\right)\,\mathbf{a}_3$ | = | $\left(\frac{1}{2} - x_5\right)a\,\mathbf{\hat{x}} + \left(\frac{1}{2} + y_5\right)b\,\mathbf{\hat{y}} + \left(\frac{1}{2} - z_5\right)c\,\mathbf{\hat{z}}$ | (8$h$) | O II |
| $\mathbf{B}_{18}$ | = | $\left(\frac{1}{2} + x_5\right)\,\mathbf{a}_1 + \left(\frac{1}{2} - y_5\right)\,\mathbf{a}_2 + \left(\frac{1}{2} - z_5\right)\,\mathbf{a}_3$ | = | $\left(\frac{1}{2} + x_5\right)a\,\mathbf{\hat{x}} + \left(\frac{1}{2} - y_5\right)b\,\mathbf{\hat{y}} + \left(\frac{1}{2} - z_5\right)c\,\mathbf{\hat{z}}$ | (8$h$) | O II |
| $\mathbf{B}_{19}$ | = | $-x_5\,\mathbf{a}_1 - y_5\,\mathbf{a}_2 - z_5\,\mathbf{a}_3$ | = | $-x_5a\,\mathbf{\hat{x}} - y_5b\,\mathbf{\hat{y}} - z_5c\,\mathbf{\hat{z}}$ | (8$h$) | O II |
| $\mathbf{B}_{20}$ | = | $x_5\,\mathbf{a}_1 + y_5\,\mathbf{a}_2 - z_5\,\mathbf{a}_3$ | = | $x_5a\,\mathbf{\hat{x}} + y_5b\,\mathbf{\hat{y}} - z_5c\,\mathbf{\hat{z}}$ | (8$h$) | O II |
| $\mathbf{B}_{21}$ | = | $\left(\frac{1}{2} + x_5\right)\,\mathbf{a}_1 + \left(\frac{1}{2} - y_5\right)\,\mathbf{a}_2 + \left(\frac{1}{2} + z_5\right)\,\mathbf{a}_3$ | = | $\left(\frac{1}{2} + x_5\right)a\,\mathbf{\hat{x}} + \left(\frac{1}{2} - y_5\right)b\,\mathbf{\hat{y}} + \left(\frac{1}{2} + z_5\right)c\,\mathbf{\hat{z}}$ | (8$h$) | O II |
| $\mathbf{B}_{22}$ | = | $\left(\frac{1}{2} - x_5\right)\,\mathbf{a}_1 + \left(\frac{1}{2} + y_5\right)\,\mathbf{a}_2 + \left(\frac{1}{2} + z_5\right)\,\mathbf{a}_3$ | = | $\left(\frac{1}{2} - x_5\right)a\,\mathbf{\hat{x}} + \left(\frac{1}{2} + y_5\right)b\,\mathbf{\hat{y}} + \left(\frac{1}{2} + z_5\right)c\,\mathbf{\hat{z}}$ | (8$h$) | O II |

**Geometry files:**

- CIF: pp. 1617
- POSCAR: pp. 1617

# Andalusite ($Al_2SiO_5$, $S0_2$) Structure: A2B5C_oP32_58_eg_3gh_g

http://aflow.org/prototype-encyclopedia/A2B5C_oP32_58_eg_3gh_g

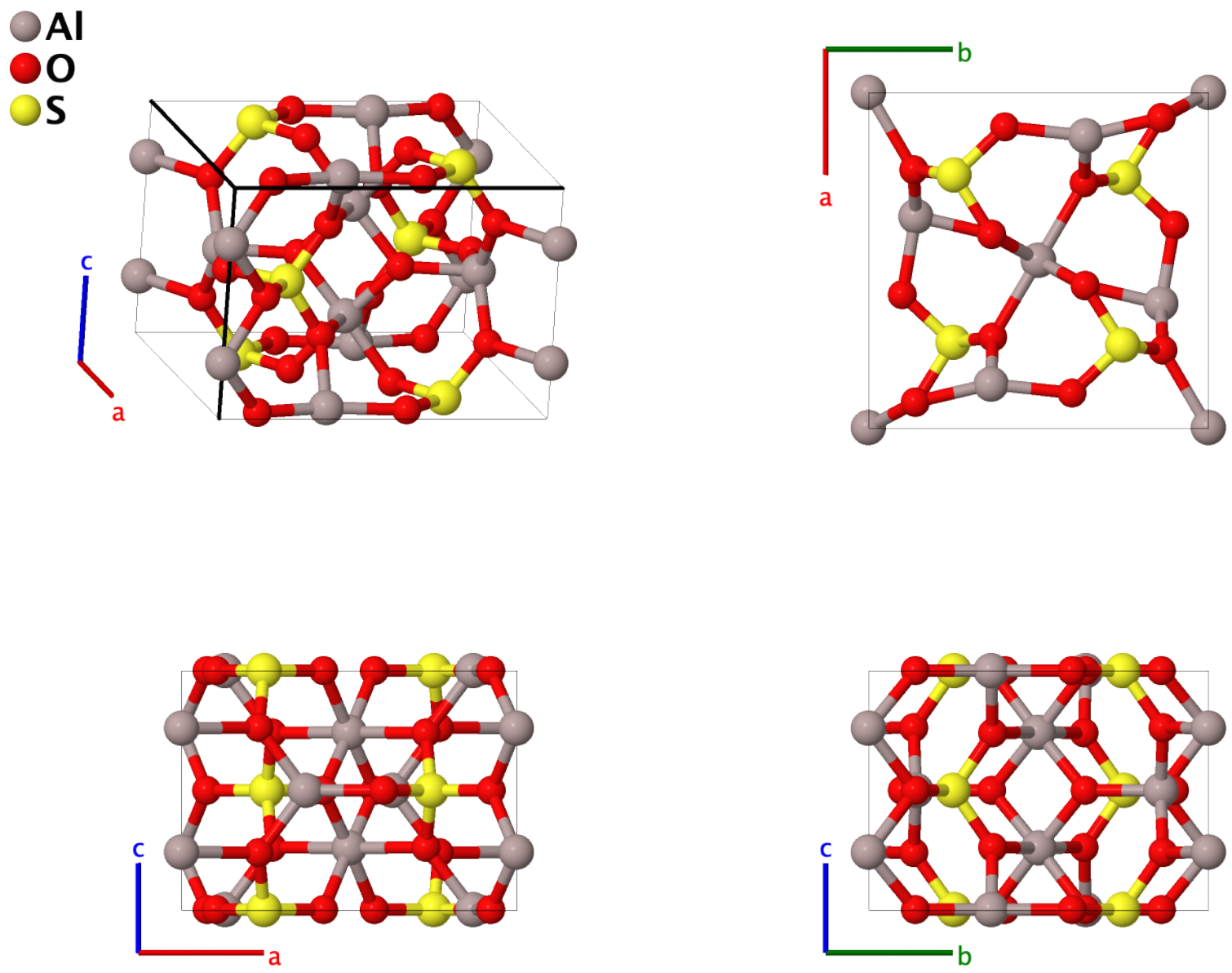

| | | |
|---|---|---|
| **Prototype** | : | $Al_2O_5Si$ |
| **AFLOW prototype label** | : | A2B5C_oP32_58_eg_3gh_g |
| ***Strukturbericht* designation** | : | $S0_2$ |
| **Pearson symbol** | : | oP32 |
| **Space group number** | : | 58 |
| **Space group symbol** | : | *Pnnm* |
| **AFLOW prototype command** | : | `aflow --proto=A2B5C_oP32_58_eg_3gh_g` <br> `--params=`$a, b/a, c/a, z_1, x_2, y_2, x_3, y_3, x_4, y_4, x_5, y_5, x_6, y_6, x_7, y_7, z_7$ |

- Three crystal polymorphs of $Al_2SiO_5$ have been characterized: kyanite ($S0_1$), space group $P\overline{1}$ #2, andalusite ($S0_2$), space group *Pnnm* #58, and sillimanite ($S0_3$), space group *Pnma* #62. All are characterized chains of edge-sharing $SiO_6$ tetrahedra and Al octahedra.
- We use the ambient pressure data of (Yang, 1997).
- (Hermann, 1937) listed this as $S0_2$, but also gave it the $H5_3$ designation in the index.

**Simple Orthorhombic primitive vectors:**

$$\mathbf{a}_1 = a\,\hat{\mathbf{x}}$$
$$\mathbf{a}_2 = b\,\hat{\mathbf{y}}$$
$$\mathbf{a}_3 = c\,\hat{\mathbf{z}}$$

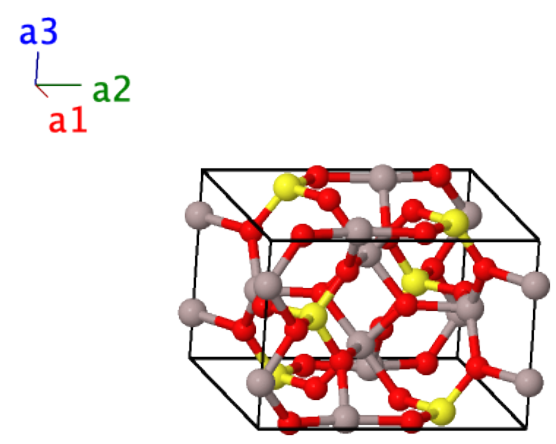

**Basis vectors:**

| | | Lattice Coordinates | | Cartesian Coordinates | Wyckoff Position | Atom Type |
|---|---|---|---|---|---|---|
| $\mathbf{B}_{1}$ | = | $z_{1}\,\mathbf{a}_{3}$ | = | $z_{1}c\,\hat{\mathbf{z}}$ | (4$e$) | Al I |
| $\mathbf{B}_{2}$ | = | $\frac{1}{2}\,\mathbf{a}_{1}+\frac{1}{2}\,\mathbf{a}_{2}+\left(\frac{1}{2}-z_{1}\right)\,\mathbf{a}_{3}$ | = | $\frac{1}{2}a\,\hat{\mathbf{x}}+\frac{1}{2}b\,\hat{\mathbf{y}}+\left(\frac{1}{2}-z_{1}\right)c\,\hat{\mathbf{z}}$ | (4$e$) | Al I |
| $\mathbf{B}_{3}$ | = | $-z_{1}\,\mathbf{a}_{3}$ | = | $-z_{1}c\,\hat{\mathbf{z}}$ | (4$e$) | Al I |
| $\mathbf{B}_{4}$ | = | $\frac{1}{2}\,\mathbf{a}_{1}+\frac{1}{2}\,\mathbf{a}_{2}+\left(\frac{1}{2}+z_{1}\right)\,\mathbf{a}_{3}$ | = | $\frac{1}{2}a\,\hat{\mathbf{x}}+\frac{1}{2}b\,\hat{\mathbf{y}}+\left(\frac{1}{2}+z_{1}\right)c\,\hat{\mathbf{z}}$ | (4$e$) | Al I |
| $\mathbf{B}_{5}$ | = | $x_{2}\,\mathbf{a}_{1}+y_{2}\,\mathbf{a}_{2}$ | = | $x_{2}a\,\hat{\mathbf{x}}+y_{2}b\,\hat{\mathbf{y}}$ | (4$g$) | Al II |
| $\mathbf{B}_{6}$ | = | $-x_{2}\,\mathbf{a}_{1}-y_{2}\,\mathbf{a}_{2}$ | = | $-x_{2}a\,\hat{\mathbf{x}}-y_{2}b\,\hat{\mathbf{y}}$ | (4$g$) | Al II |
| $\mathbf{B}_{7}$ | = | $\left(\frac{1}{2}-x_{2}\right)\,\mathbf{a}_{1}+\left(\frac{1}{2}+y_{2}\right)\,\mathbf{a}_{2}+\frac{1}{2}\,\mathbf{a}_{3}$ | = | $\left(\frac{1}{2}-x_{2}\right)a\,\hat{\mathbf{x}}+\left(\frac{1}{2}+y_{2}\right)b\,\hat{\mathbf{y}}+\frac{1}{2}c\,\hat{\mathbf{z}}$ | (4$g$) | Al II |
| $\mathbf{B}_{8}$ | = | $\left(\frac{1}{2}+x_{2}\right)\,\mathbf{a}_{1}+\left(\frac{1}{2}-y_{2}\right)\,\mathbf{a}_{2}+\frac{1}{2}\,\mathbf{a}_{3}$ | = | $\left(\frac{1}{2}+x_{2}\right)a\,\hat{\mathbf{x}}+\left(\frac{1}{2}-y_{2}\right)b\,\hat{\mathbf{y}}+\frac{1}{2}c\,\hat{\mathbf{z}}$ | (4$g$) | Al II |
| $\mathbf{B}_{9}$ | = | $x_{3}\,\mathbf{a}_{1}+y_{3}\,\mathbf{a}_{2}$ | = | $x_{3}a\,\hat{\mathbf{x}}+y_{3}b\,\hat{\mathbf{y}}$ | (4$g$) | O I |
| $\mathbf{B}_{10}$ | = | $-x_{3}\,\mathbf{a}_{1}-y_{3}\,\mathbf{a}_{2}$ | = | $-x_{3}a\,\hat{\mathbf{x}}-y_{3}b\,\hat{\mathbf{y}}$ | (4$g$) | O I |
| $\mathbf{B}_{11}$ | = | $\left(\frac{1}{2}-x_{3}\right)\,\mathbf{a}_{1}+\left(\frac{1}{2}+y_{3}\right)\,\mathbf{a}_{2}+\frac{1}{2}\,\mathbf{a}_{3}$ | = | $\left(\frac{1}{2}-x_{3}\right)a\,\hat{\mathbf{x}}+\left(\frac{1}{2}+y_{3}\right)b\,\hat{\mathbf{y}}+\frac{1}{2}c\,\hat{\mathbf{z}}$ | (4$g$) | O I |
| $\mathbf{B}_{12}$ | = | $\left(\frac{1}{2}+x_{3}\right)\,\mathbf{a}_{1}+\left(\frac{1}{2}-y_{3}\right)\,\mathbf{a}_{2}+\frac{1}{2}\,\mathbf{a}_{3}$ | = | $\left(\frac{1}{2}+x_{3}\right)a\,\hat{\mathbf{x}}+\left(\frac{1}{2}-y_{3}\right)b\,\hat{\mathbf{y}}+\frac{1}{2}c\,\hat{\mathbf{z}}$ | (4$g$) | O I |
| $\mathbf{B}_{13}$ | = | $x_{4}\,\mathbf{a}_{1}+y_{4}\,\mathbf{a}_{2}$ | = | $x_{4}a\,\hat{\mathbf{x}}+y_{4}b\,\hat{\mathbf{y}}$ | (4$g$) | O II |
| $\mathbf{B}_{14}$ | = | $-x_{4}\,\mathbf{a}_{1}-y_{4}\,\mathbf{a}_{2}$ | = | $-x_{4}a\,\hat{\mathbf{x}}-y_{4}b\,\hat{\mathbf{y}}$ | (4$g$) | O II |
| $\mathbf{B}_{15}$ | = | $\left(\frac{1}{2}-x_{4}\right)\,\mathbf{a}_{1}+\left(\frac{1}{2}+y_{4}\right)\,\mathbf{a}_{2}+\frac{1}{2}\,\mathbf{a}_{3}$ | = | $\left(\frac{1}{2}-x_{4}\right)a\,\hat{\mathbf{x}}+\left(\frac{1}{2}+y_{4}\right)b\,\hat{\mathbf{y}}+\frac{1}{2}c\,\hat{\mathbf{z}}$ | (4$g$) | O II |
| $\mathbf{B}_{16}$ | = | $\left(\frac{1}{2}+x_{4}\right)\,\mathbf{a}_{1}+\left(\frac{1}{2}-y_{4}\right)\,\mathbf{a}_{2}+\frac{1}{2}\,\mathbf{a}_{3}$ | = | $\left(\frac{1}{2}+x_{4}\right)a\,\hat{\mathbf{x}}+\left(\frac{1}{2}-y_{4}\right)b\,\hat{\mathbf{y}}+\frac{1}{2}c\,\hat{\mathbf{z}}$ | (4$g$) | O II |
| $\mathbf{B}_{17}$ | = | $x_{5}\,\mathbf{a}_{1}+y_{5}\,\mathbf{a}_{2}$ | = | $x_{5}a\,\hat{\mathbf{x}}+y_{5}b\,\hat{\mathbf{y}}$ | (4$g$) | O III |
| $\mathbf{B}_{18}$ | = | $-x_{5}\,\mathbf{a}_{1}-y_{5}\,\mathbf{a}_{2}$ | = | $-x_{5}a\,\hat{\mathbf{x}}-y_{5}b\,\hat{\mathbf{y}}$ | (4$g$) | O III |
| $\mathbf{B}_{19}$ | = | $\left(\frac{1}{2}-x_{5}\right)\,\mathbf{a}_{1}+\left(\frac{1}{2}+y_{5}\right)\,\mathbf{a}_{2}+\frac{1}{2}\,\mathbf{a}_{3}$ | = | $\left(\frac{1}{2}-x_{5}\right)a\,\hat{\mathbf{x}}+\left(\frac{1}{2}+y_{5}\right)b\,\hat{\mathbf{y}}+\frac{1}{2}c\,\hat{\mathbf{z}}$ | (4$g$) | O III |
| $\mathbf{B}_{20}$ | = | $\left(\frac{1}{2}+x_{5}\right)\,\mathbf{a}_{1}+\left(\frac{1}{2}-y_{5}\right)\,\mathbf{a}_{2}+\frac{1}{2}\,\mathbf{a}_{3}$ | = | $\left(\frac{1}{2}+x_{5}\right)a\,\hat{\mathbf{x}}+\left(\frac{1}{2}-y_{5}\right)b\,\hat{\mathbf{y}}+\frac{1}{2}c\,\hat{\mathbf{z}}$ | (4$g$) | O III |
| $\mathbf{B}_{21}$ | = | $x_{6}\,\mathbf{a}_{1}+y_{6}\,\mathbf{a}_{2}$ | = | $x_{6}a\,\hat{\mathbf{x}}+y_{6}b\,\hat{\mathbf{y}}$ | (4$g$) | S |
| $\mathbf{B}_{22}$ | = | $-x_{6}\,\mathbf{a}_{1}-y_{6}\,\mathbf{a}_{2}$ | = | $-x_{6}a\,\hat{\mathbf{x}}-y_{6}b\,\hat{\mathbf{y}}$ | (4$g$) | S |
| $\mathbf{B}_{23}$ | = | $\left(\frac{1}{2}-x_{6}\right)\,\mathbf{a}_{1}+\left(\frac{1}{2}+y_{6}\right)\,\mathbf{a}_{2}+\frac{1}{2}\,\mathbf{a}_{3}$ | = | $\left(\frac{1}{2}-x_{6}\right)a\,\hat{\mathbf{x}}+\left(\frac{1}{2}+y_{6}\right)b\,\hat{\mathbf{y}}+\frac{1}{2}c\,\hat{\mathbf{z}}$ | (4$g$) | S |
| $\mathbf{B}_{24}$ | = | $\left(\frac{1}{2}+x_{6}\right)\,\mathbf{a}_{1}+\left(\frac{1}{2}-y_{6}\right)\,\mathbf{a}_{2}+\frac{1}{2}\,\mathbf{a}_{3}$ | = | $\left(\frac{1}{2}+x_{6}\right)a\,\hat{\mathbf{x}}+\left(\frac{1}{2}-y_{6}\right)b\,\hat{\mathbf{y}}+\frac{1}{2}c\,\hat{\mathbf{z}}$ | (4$g$) | S |
| $\mathbf{B}_{25}$ | = | $x_{7}\,\mathbf{a}_{1}+y_{7}\,\mathbf{a}_{2}+z_{7}\,\mathbf{a}_{3}$ | = | $x_{7}a\,\hat{\mathbf{x}}+y_{7}b\,\hat{\mathbf{y}}+z_{7}c\,\hat{\mathbf{z}}$ | (8$h$) | O IV |
| $\mathbf{B}_{26}$ | = | $-x_{7}\,\mathbf{a}_{1}-y_{7}\,\mathbf{a}_{2}+z_{7}\,\mathbf{a}_{3}$ | = | $-x_{7}a\,\hat{\mathbf{x}}-y_{7}b\,\hat{\mathbf{y}}+z_{7}c\,\hat{\mathbf{z}}$ | (8$h$) | O IV |
| $\mathbf{B}_{27}$ | = | $\left(\frac{1}{2}-x_{7}\right)\,\mathbf{a}_{1}+\left(\frac{1}{2}+y_{7}\right)\,\mathbf{a}_{2}+\left(\frac{1}{2}-z_{7}\right)\,\mathbf{a}_{3}$ | = | $\left(\frac{1}{2}-x_{7}\right)a\,\hat{\mathbf{x}}+\left(\frac{1}{2}+y_{7}\right)b\,\hat{\mathbf{y}}+\left(\frac{1}{2}-z_{7}\right)c\,\hat{\mathbf{z}}$ | (8$h$) | O IV |
| $\mathbf{B}_{28}$ | = | $\left(\frac{1}{2}+x_{7}\right)\,\mathbf{a}_{1}+\left(\frac{1}{2}-y_{7}\right)\,\mathbf{a}_{2}+\left(\frac{1}{2}-z_{7}\right)\,\mathbf{a}_{3}$ | = | $\left(\frac{1}{2}+x_{7}\right)a\,\hat{\mathbf{x}}+\left(\frac{1}{2}-y_{7}\right)b\,\hat{\mathbf{y}}+\left(\frac{1}{2}-z_{7}\right)c\,\hat{\mathbf{z}}$ | (8$h$) | O IV |
| $\mathbf{B}_{29}$ | = | $-x_{7}\,\mathbf{a}_{1}-y_{7}\,\mathbf{a}_{2}-z_{7}\,\mathbf{a}_{3}$ | = | $-x_{7}a\,\hat{\mathbf{x}}-y_{7}b\,\hat{\mathbf{y}}-z_{7}c\,\hat{\mathbf{z}}$ | (8$h$) | O IV |
| $\mathbf{B}_{30}$ | = | $x_{7}\,\mathbf{a}_{1}+y_{7}\,\mathbf{a}_{2}-z_{7}\,\mathbf{a}_{3}$ | = | $x_{7}a\,\hat{\mathbf{x}}+y_{7}b\,\hat{\mathbf{y}}-z_{7}c\,\hat{\mathbf{z}}$ | (8$h$) | O IV |
| $\mathbf{B}_{31}$ | = | $\left(\frac{1}{2}+x_{7}\right)\,\mathbf{a}_{1}+\left(\frac{1}{2}-y_{7}\right)\,\mathbf{a}_{2}+\left(\frac{1}{2}+z_{7}\right)\,\mathbf{a}_{3}$ | = | $\left(\frac{1}{2}+x_{7}\right)a\,\hat{\mathbf{x}}+\left(\frac{1}{2}-y_{7}\right)b\,\hat{\mathbf{y}}+\left(\frac{1}{2}+z_{7}\right)c\,\hat{\mathbf{z}}$ | (8$h$) | O IV |
| $\mathbf{B}_{32}$ | = | $\left(\frac{1}{2} - x_7\right) \mathbf{a}_1 + \left(\frac{1}{2} + y_7\right) \mathbf{a}_2 + \left(\frac{1}{2} + z_7\right) \mathbf{a}_3$ | = | $\left(\frac{1}{2} - x_7\right) a \, \hat{\mathbf{x}} + \left(\frac{1}{2} + y_7\right) b \, \hat{\mathbf{y}} + \left(\frac{1}{2} + z_7\right) c \, \hat{\mathbf{z}}$ | (8*h*) | O IV |

---

**Found in:**

---

**Geometry files:**

- CIF: pp. 1617
- POSCAR: pp. 1618

# Protoanthophyllite ($H_2Mg_7Si_8O_{24}$) Structure: A2B7C24D8_oP82_58_g_ae2f_2g5h_2h

http://aflow.org/prototype-encyclopedia/A2B7C24D8_oP82_58_g_ae2f_2g5h_2h

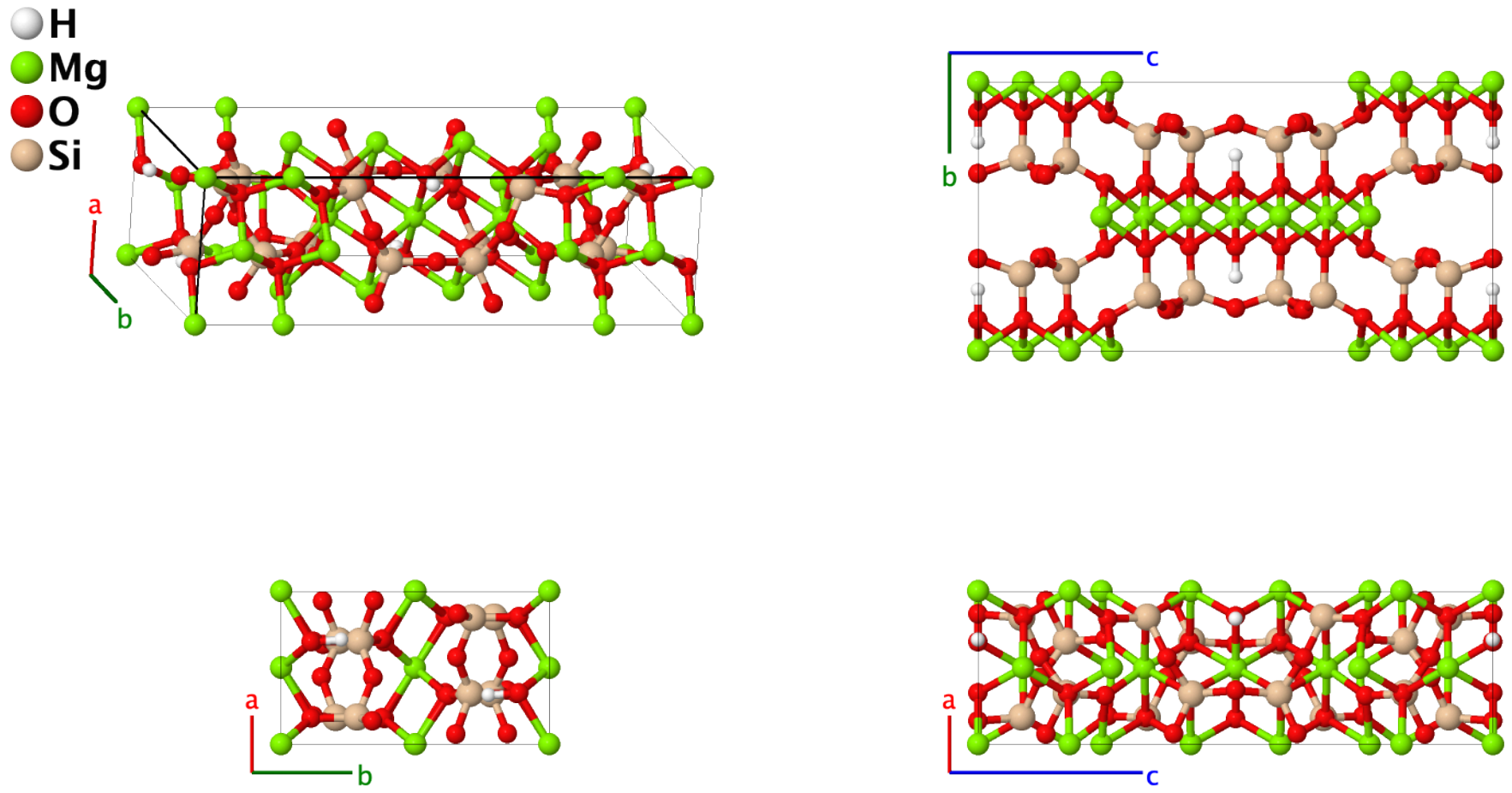

| | | |
|---|---|---|
| **Prototype** | : | $H_2Mg_7O_{24}Si_8$ |
| **AFLOW prototype label** | : | A2B7C24D8_oP82_58_g_ae2f_2g5h_2h |
| ***Strukturbericht* designation** | : | None |
| **Pearson symbol** | : | oP82 |
| **Space group number** | : | 58 |
| **Space group symbol** | : | *Pnnm* |
| **AFLOW prototype command** | : | `aflow --proto=A2B7C24D8_oP82_58_g_ae2f_2g5h_2h` `--params=`$a, b/a, c/a, z_2, z_3, z_4, x_5, y_5, x_6, y_6, x_7, y_7, x_8, y_8, z_8, x_9, y_9, z_9, x_{10}, y_{10}, z_{10}, x_{11}, y_{11}, z_{11}, x_{12}, y_{12}, z_{12}, x_{13}, y_{13}, z_{13}, x_{14}, y_{14}, z_{14}$ |

- This structure is approximately one-half of the unit cell of anthophyllite ($S4_4$).
- Like anthophyllite, iron is sometimes mixed with magnesium. For this sample, (Konishi, 2003) found that the Mg-I (2a) site contains 2.5% iron, Mg-III (4f) contains 2.1%, and Mg-IV (4f) contains 27.7% iron.
- (Konishi, 2003) give the crystal structure in the *Pnmn* setting of space group #58. We used FINDSYM to transform this to the standard *Pnnm* orientation.

**Simple Orthorhombic primitive vectors:**

$$\mathbf{a}_1 = a\,\hat{\mathbf{x}}$$
$$\mathbf{a}_2 = b\,\hat{\mathbf{y}}$$
$$\mathbf{a}_3 = c\,\hat{\mathbf{z}}$$

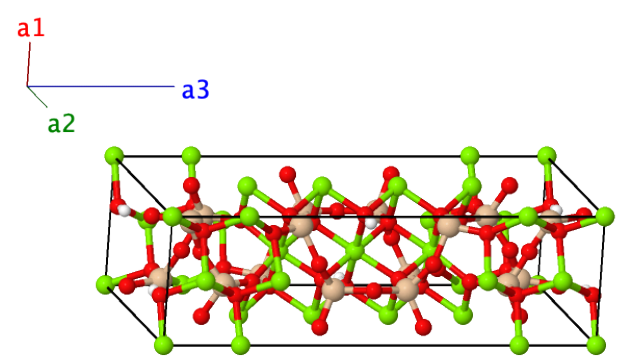

**Basis vectors:**

| | | Lattice Coordinates | | Cartesian Coordinates | Wyckoff Position | Atom Type |
|---|---|---|---|---|---|---|
| $\mathbf{B}_1$ | = | $0\,\mathbf{a}_1 + 0\,\mathbf{a}_2 + 0\,\mathbf{a}_3$ | = | $0\,\hat{\mathbf{x}} + 0\,\hat{\mathbf{y}} + 0\,\hat{\mathbf{z}}$ | (2$a$) | Mg I |
| $\mathbf{B}_{2}$ | = | $\frac{1}{2}\,\mathbf{a}_{1} + \frac{1}{2}\,\mathbf{a}_{2} + \frac{1}{2}\,\mathbf{a}_{3}$ | = | $\frac{1}{2}a\,\hat{\mathbf{x}} + \frac{1}{2}b\,\hat{\mathbf{y}} + \frac{1}{2}c\,\hat{\mathbf{z}}$ | $(2a)$ | Mg I |
| $\mathbf{B}_{3}$ | = | $z_{2}\,\mathbf{a}_{3}$ | = | $z_{2}c\,\hat{\mathbf{z}}$ | $(4e)$ | Mg II |
| $\mathbf{B}_{4}$ | = | $\frac{1}{2}\,\mathbf{a}_{1} + \frac{1}{2}\,\mathbf{a}_{2} + \left(\frac{1}{2} - z_{2}\right)\,\mathbf{a}_{3}$ | = | $\frac{1}{2}a\,\hat{\mathbf{x}} + \frac{1}{2}b\,\hat{\mathbf{y}} + \left(\frac{1}{2} - z_{2}\right)c\,\hat{\mathbf{z}}$ | $(4e)$ | Mg II |
| $\mathbf{B}_{5}$ | = | $-z_{2}\,\mathbf{a}_{3}$ | = | $-z_{2}c\,\hat{\mathbf{z}}$ | $(4e)$ | Mg II |
| $\mathbf{B}_{6}$ | = | $\frac{1}{2}\,\mathbf{a}_{1} + \frac{1}{2}\,\mathbf{a}_{2} + \left(\frac{1}{2} + z_{2}\right)\,\mathbf{a}_{3}$ | = | $\frac{1}{2}a\,\hat{\mathbf{x}} + \frac{1}{2}b\,\hat{\mathbf{y}} + \left(\frac{1}{2} + z_{2}\right)c\,\hat{\mathbf{z}}$ | $(4e)$ | Mg II |
| $\mathbf{B}_{7}$ | = | $\frac{1}{2}\,\mathbf{a}_{2} + z_{3}\,\mathbf{a}_{3}$ | = | $\frac{1}{2}b\,\hat{\mathbf{y}} + z_{3}c\,\hat{\mathbf{z}}$ | $(4f)$ | Mg III |
| $\mathbf{B}_{8}$ | = | $\frac{1}{2}\,\mathbf{a}_{1} + \left(\frac{1}{2} - z_{3}\right)\,\mathbf{a}_{3}$ | = | $\frac{1}{2}a\,\hat{\mathbf{x}} + \left(\frac{1}{2} - z_{3}\right)c\,\hat{\mathbf{z}}$ | $(4f)$ | Mg III |
| $\mathbf{B}_{9}$ | = | $\frac{1}{2}\,\mathbf{a}_{2} - z_{3}\,\mathbf{a}_{3}$ | = | $\frac{1}{2}b\,\hat{\mathbf{y}} - z_{3}c\,\hat{\mathbf{z}}$ | $(4f)$ | Mg III |
| $\mathbf{B}_{10}$ | = | $\frac{1}{2}\,\mathbf{a}_{1} + \left(\frac{1}{2} + z_{3}\right)\,\mathbf{a}_{3}$ | = | $\frac{1}{2}a\,\hat{\mathbf{x}} + \left(\frac{1}{2} + z_{3}\right)c\,\hat{\mathbf{z}}$ | $(4f)$ | Mg III |
| $\mathbf{B}_{11}$ | = | $\frac{1}{2}\,\mathbf{a}_{2} + z_{4}\,\mathbf{a}_{3}$ | = | $\frac{1}{2}b\,\hat{\mathbf{y}} + z_{4}c\,\hat{\mathbf{z}}$ | $(4f)$ | Mg IV |
| $\mathbf{B}_{12}$ | = | $\frac{1}{2}\,\mathbf{a}_{1} + \left(\frac{1}{2} - z_{4}\right)\,\mathbf{a}_{3}$ | = | $\frac{1}{2}a\,\hat{\mathbf{x}} + \left(\frac{1}{2} - z_{4}\right)c\,\hat{\mathbf{z}}$ | $(4f)$ | Mg IV |
| $\mathbf{B}_{13}$ | = | $\frac{1}{2}\,\mathbf{a}_{2} - z_{4}\,\mathbf{a}_{3}$ | = | $\frac{1}{2}b\,\hat{\mathbf{y}} - z_{4}c\,\hat{\mathbf{z}}$ | $(4f)$ | Mg IV |
| $\mathbf{B}_{14}$ | = | $\frac{1}{2}\,\mathbf{a}_{1} + \left(\frac{1}{2} + z_{4}\right)\,\mathbf{a}_{3}$ | = | $\frac{1}{2}a\,\hat{\mathbf{x}} + \left(\frac{1}{2} + z_{4}\right)c\,\hat{\mathbf{z}}$ | $(4f)$ | Mg IV |
| $\mathbf{B}_{15}$ | = | $x_{5}\,\mathbf{a}_{1} + y_{5}\,\mathbf{a}_{2}$ | = | $x_{5}a\,\hat{\mathbf{x}} + y_{5}b\,\hat{\mathbf{y}}$ | $(4g)$ | H |
| $\mathbf{B}_{16}$ | = | $-x_{5}\,\mathbf{a}_{1} - y_{5}\,\mathbf{a}_{2}$ | = | $-x_{5}a\,\hat{\mathbf{x}} - y_{5}b\,\hat{\mathbf{y}}$ | $(4g)$ | H |
| $\mathbf{B}_{17}$ | = | $\left(\frac{1}{2} - x_{5}\right)\,\mathbf{a}_{1} + \left(\frac{1}{2} + y_{5}\right)\,\mathbf{a}_{2} + \frac{1}{2}\,\mathbf{a}_{3}$ | = | $\left(\frac{1}{2} - x_{5}\right)a\,\hat{\mathbf{x}} + \left(\frac{1}{2} + y_{5}\right)b\,\hat{\mathbf{y}} + \frac{1}{2}c\,\hat{\mathbf{z}}$ | $(4g)$ | H |
| $\mathbf{B}_{18}$ | = | $\left(\frac{1}{2} + x_{5}\right)\,\mathbf{a}_{1} + \left(\frac{1}{2} - y_{5}\right)\,\mathbf{a}_{2} + \frac{1}{2}\,\mathbf{a}_{3}$ | = | $\left(\frac{1}{2} + x_{5}\right)a\,\hat{\mathbf{x}} + \left(\frac{1}{2} - y_{5}\right)b\,\hat{\mathbf{y}} + \frac{1}{2}c\,\hat{\mathbf{z}}$ | $(4g)$ | H |
| $\mathbf{B}_{19}$ | = | $x_{6}\,\mathbf{a}_{1} + y_{6}\,\mathbf{a}_{2}$ | = | $x_{6}a\,\hat{\mathbf{x}} + y_{6}b\,\hat{\mathbf{y}}$ | $(4g)$ | O I |
| $\mathbf{B}_{20}$ | = | $-x_{6}\,\mathbf{a}_{1} - y_{6}\,\mathbf{a}_{2}$ | = | $-x_{6}a\,\hat{\mathbf{x}} - y_{6}b\,\hat{\mathbf{y}}$ | $(4g)$ | O I |
| $\mathbf{B}_{21}$ | = | $\left(\frac{1}{2} - x_{6}\right)\,\mathbf{a}_{1} + \left(\frac{1}{2} + y_{6}\right)\,\mathbf{a}_{2} + \frac{1}{2}\,\mathbf{a}_{3}$ | = | $\left(\frac{1}{2} - x_{6}\right)a\,\hat{\mathbf{x}} + \left(\frac{1}{2} + y_{6}\right)b\,\hat{\mathbf{y}} + \frac{1}{2}c\,\hat{\mathbf{z}}$ | $(4g)$ | O I |
| $\mathbf{B}_{22}$ | = | $\left(\frac{1}{2} + x_{6}\right)\,\mathbf{a}_{1} + \left(\frac{1}{2} - y_{6}\right)\,\mathbf{a}_{2} + \frac{1}{2}\,\mathbf{a}_{3}$ | = | $\left(\frac{1}{2} + x_{6}\right)a\,\hat{\mathbf{x}} + \left(\frac{1}{2} - y_{6}\right)b\,\hat{\mathbf{y}} + \frac{1}{2}c\,\hat{\mathbf{z}}$ | $(4g)$ | O I |
| $\mathbf{B}_{23}$ | = | $x_{7}\,\mathbf{a}_{1} + y_{7}\,\mathbf{a}_{2}$ | = | $x_{7}a\,\hat{\mathbf{x}} + y_{7}b\,\hat{\mathbf{y}}$ | $(4g)$ | O II |
| $\mathbf{B}_{24}$ | = | $-x_{7}\,\mathbf{a}_{1} - y_{7}\,\mathbf{a}_{2}$ | = | $-x_{7}a\,\hat{\mathbf{x}} - y_{7}b\,\hat{\mathbf{y}}$ | $(4g)$ | O II |
| $\mathbf{B}_{25}$ | = | $\left(\frac{1}{2} - x_{7}\right)\,\mathbf{a}_{1} + \left(\frac{1}{2} + y_{7}\right)\,\mathbf{a}_{2} + \frac{1}{2}\,\mathbf{a}_{3}$ | = | $\left(\frac{1}{2} - x_{7}\right)a\,\hat{\mathbf{x}} + \left(\frac{1}{2} + y_{7}\right)b\,\hat{\mathbf{y}} + \frac{1}{2}c\,\hat{\mathbf{z}}$ | $(4g)$ | O II |
| $\mathbf{B}_{26}$ | = | $\left(\frac{1}{2} + x_{7}\right)\,\mathbf{a}_{1} + \left(\frac{1}{2} - y_{7}\right)\,\mathbf{a}_{2} + \frac{1}{2}\,\mathbf{a}_{3}$ | = | $\left(\frac{1}{2} + x_{7}\right)a\,\hat{\mathbf{x}} + \left(\frac{1}{2} - y_{7}\right)b\,\hat{\mathbf{y}} + \frac{1}{2}c\,\hat{\mathbf{z}}$ | $(4g)$ | O II |
| $\mathbf{B}_{27}$ | = | $x_{8}\,\mathbf{a}_{1} + y_{8}\,\mathbf{a}_{2} + z_{8}\,\mathbf{a}_{3}$ | = | $x_{8}a\,\hat{\mathbf{x}} + y_{8}b\,\hat{\mathbf{y}} + z_{8}c\,\hat{\mathbf{z}}$ | $(8h)$ | O III |
| $\mathbf{B}_{28}$ | = | $-x_{8}\,\mathbf{a}_{1} - y_{8}\,\mathbf{a}_{2} + z_{8}\,\mathbf{a}_{3}$ | = | $-x_{8}a\,\hat{\mathbf{x}} - y_{8}b\,\hat{\mathbf{y}} + z_{8}c\,\hat{\mathbf{z}}$ | $(8h)$ | O III |
| $\mathbf{B}_{29}$ | = | $\left(\frac{1}{2} - x_{8}\right)\,\mathbf{a}_{1} + \left(\frac{1}{2} + y_{8}\right)\,\mathbf{a}_{2} + \left(\frac{1}{2} - z_{8}\right)\,\mathbf{a}_{3}$ | = | $\left(\frac{1}{2} - x_{8}\right)a\,\hat{\mathbf{x}} + \left(\frac{1}{2} + y_{8}\right)b\,\hat{\mathbf{y}} + \left(\frac{1}{2} - z_{8}\right)c\,\hat{\mathbf{z}}$ | $(8h)$ | O III |
| $\mathbf{B}_{30}$ | = | $\left(\frac{1}{2} + x_{8}\right)\,\mathbf{a}_{1} + \left(\frac{1}{2} - y_{8}\right)\,\mathbf{a}_{2} + \left(\frac{1}{2} - z_{8}\right)\,\mathbf{a}_{3}$ | = | $\left(\frac{1}{2} + x_{8}\right)a\,\hat{\mathbf{x}} + \left(\frac{1}{2} - y_{8}\right)b\,\hat{\mathbf{y}} + \left(\frac{1}{2} - z_{8}\right)c\,\hat{\mathbf{z}}$ | $(8h)$ | O III |
| $\mathbf{B}_{31}$ | = | $-x_{8}\,\mathbf{a}_{1} - y_{8}\,\mathbf{a}_{2} - z_{8}\,\mathbf{a}_{3}$ | = | $-x_{8}a\,\hat{\mathbf{x}} - y_{8}b\,\hat{\mathbf{y}} - z_{8}c\,\hat{\mathbf{z}}$ | $(8h)$ | O III |
| $\mathbf{B}_{32}$ | = | $x_{8}\,\mathbf{a}_{1} + y_{8}\,\mathbf{a}_{2} - z_{8}\,\mathbf{a}_{3}$ | = | $x_{8}a\,\hat{\mathbf{x}} + y_{8}b\,\hat{\mathbf{y}} - z_{8}c\,\hat{\mathbf{z}}$ | $(8h)$ | O III |
| $\mathbf{B}_{33}$ | = | $\left(\frac{1}{2} + x_{8}\right)\,\mathbf{a}_{1} + \left(\frac{1}{2} - y_{8}\right)\,\mathbf{a}_{2} + \left(\frac{1}{2} + z_{8}\right)\,\mathbf{a}_{3}$ | = | $\left(\frac{1}{2} + x_{8}\right)a\,\hat{\mathbf{x}} + \left(\frac{1}{2} - y_{8}\right)b\,\hat{\mathbf{y}} + \left(\frac{1}{2} + z_{8}\right)c\,\hat{\mathbf{z}}$ | $(8h)$ | O III |
| $\mathbf{B}_{34}$ | = | $\left(\frac{1}{2} - x_{8}\right)\,\mathbf{a}_{1} + \left(\frac{1}{2} + y_{8}\right)\,\mathbf{a}_{2} + \left(\frac{1}{2} + z_{8}\right)\,\mathbf{a}_{3}$ | = | $\left(\frac{1}{2} - x_{8}\right)a\,\hat{\mathbf{x}} + \left(\frac{1}{2} + y_{8}\right)b\,\hat{\mathbf{y}} + \left(\frac{1}{2} + z_{8}\right)c\,\hat{\mathbf{z}}$ | $(8h)$ | O III |
| $\mathbf{B}_{35}$ | = | $x_9\,\mathbf{a}_1 + y_9\,\mathbf{a}_2 + z_9\,\mathbf{a}_3$ | = | $x_9 a\,\hat{\mathbf{x}} + y_9 b\,\hat{\mathbf{y}} + z_9 c\,\hat{\mathbf{z}}$ | (8$h$) | O IV |
| $\mathbf{B}_{36}$ | = | $-x_9\,\mathbf{a}_1 - y_9\,\mathbf{a}_2 + z_9\,\mathbf{a}_3$ | = | $-x_9 a\,\hat{\mathbf{x}} - y_9 b\,\hat{\mathbf{y}} + z_9 c\,\hat{\mathbf{z}}$ | (8$h$) | O IV |
| $\mathbf{B}_{37}$ | = | $\left(\frac{1}{2} - x_9\right)\mathbf{a}_1 + \left(\frac{1}{2} + y_9\right)\mathbf{a}_2 + \left(\frac{1}{2} - z_9\right)\mathbf{a}_3$ | = | $\left(\frac{1}{2} - x_9\right) a\,\hat{\mathbf{x}} + \left(\frac{1}{2} + y_9\right) b\,\hat{\mathbf{y}} + \left(\frac{1}{2} - z_9\right) c\,\hat{\mathbf{z}}$ | (8$h$) | O IV |
| $\mathbf{B}_{38}$ | = | $\left(\frac{1}{2} + x_9\right)\mathbf{a}_1 + \left(\frac{1}{2} - y_9\right)\mathbf{a}_2 + \left(\frac{1}{2} - z_9\right)\mathbf{a}_3$ | = | $\left(\frac{1}{2} + x_9\right) a\,\hat{\mathbf{x}} + \left(\frac{1}{2} - y_9\right) b\,\hat{\mathbf{y}} + \left(\frac{1}{2} - z_9\right) c\,\hat{\mathbf{z}}$ | (8$h$) | O IV |
| $\mathbf{B}_{39}$ | = | $-x_9\,\mathbf{a}_1 - y_9\,\mathbf{a}_2 - z_9\,\mathbf{a}_3$ | = | $-x_9 a\,\hat{\mathbf{x}} - y_9 b\,\hat{\mathbf{y}} - z_9 c\,\hat{\mathbf{z}}$ | (8$h$) | O IV |
| $\mathbf{B}_{40}$ | = | $x_9\,\mathbf{a}_1 + y_9\,\mathbf{a}_2 - z_9\,\mathbf{a}_3$ | = | $x_9 a\,\hat{\mathbf{x}} + y_9 b\,\hat{\mathbf{y}} - z_9 c\,\hat{\mathbf{z}}$ | (8$h$) | O IV |
| $\mathbf{B}_{41}$ | = | $\left(\frac{1}{2} + x_9\right)\mathbf{a}_1 + \left(\frac{1}{2} - y_9\right)\mathbf{a}_2 + \left(\frac{1}{2} + z_9\right)\mathbf{a}_3$ | = | $\left(\frac{1}{2} + x_9\right) a\,\hat{\mathbf{x}} + \left(\frac{1}{2} - y_9\right) b\,\hat{\mathbf{y}} + \left(\frac{1}{2} + z_9\right) c\,\hat{\mathbf{z}}$ | (8$h$) | O IV |
| $\mathbf{B}_{42}$ | = | $\left(\frac{1}{2} - x_9\right)\mathbf{a}_1 + \left(\frac{1}{2} + y_9\right)\mathbf{a}_2 + \left(\frac{1}{2} + z_9\right)\mathbf{a}_3$ | = | $\left(\frac{1}{2} - x_9\right) a\,\hat{\mathbf{x}} + \left(\frac{1}{2} + y_9\right) b\,\hat{\mathbf{y}} + \left(\frac{1}{2} + z_9\right) c\,\hat{\mathbf{z}}$ | (8$h$) | O IV |
| $\mathbf{B}_{43}$ | = | $x_{10}\,\mathbf{a}_1 + y_{10}\,\mathbf{a}_2 + z_{10}\,\mathbf{a}_3$ | = | $x_{10} a\,\hat{\mathbf{x}} + y_{10} b\,\hat{\mathbf{y}} + z_{10} c\,\hat{\mathbf{z}}$ | (8$h$) | O V |
| $\mathbf{B}_{44}$ | = | $-x_{10}\,\mathbf{a}_1 - y_{10}\,\mathbf{a}_2 + z_{10}\,\mathbf{a}_3$ | = | $-x_{10} a\,\hat{\mathbf{x}} - y_{10} b\,\hat{\mathbf{y}} + z_{10} c\,\hat{\mathbf{z}}$ | (8$h$) | O V |
| $\mathbf{B}_{45}$ | = | $\left(\frac{1}{2} - x_{10}\right)\mathbf{a}_1 + \left(\frac{1}{2} + y_{10}\right)\mathbf{a}_2 + \left(\frac{1}{2} - z_{10}\right)\mathbf{a}_3$ | = | $\left(\frac{1}{2} - x_{10}\right) a\,\hat{\mathbf{x}} + \left(\frac{1}{2} + y_{10}\right) b\,\hat{\mathbf{y}} + \left(\frac{1}{2} - z_{10}\right) c\,\hat{\mathbf{z}}$ | (8$h$) | O V |
| $\mathbf{B}_{46}$ | = | $\left(\frac{1}{2} + x_{10}\right)\mathbf{a}_1 + \left(\frac{1}{2} - y_{10}\right)\mathbf{a}_2 + \left(\frac{1}{2} - z_{10}\right)\mathbf{a}_3$ | = | $\left(\frac{1}{2} + x_{10}\right) a\,\hat{\mathbf{x}} + \left(\frac{1}{2} - y_{10}\right) b\,\hat{\mathbf{y}} + \left(\frac{1}{2} - z_{10}\right) c\,\hat{\mathbf{z}}$ | (8$h$) | O V |
| $\mathbf{B}_{47}$ | = | $-x_{10}\,\mathbf{a}_1 - y_{10}\,\mathbf{a}_2 - z_{10}\,\mathbf{a}_3$ | = | $-x_{10} a\,\hat{\mathbf{x}} - y_{10} b\,\hat{\mathbf{y}} - z_{10} c\,\hat{\mathbf{z}}$ | (8$h$) | O V |
| $\mathbf{B}_{48}$ | = | $x_{10}\,\mathbf{a}_1 + y_{10}\,\mathbf{a}_2 - z_{10}\,\mathbf{a}_3$ | = | $x_{10} a\,\hat{\mathbf{x}} + y_{10} b\,\hat{\mathbf{y}} - z_{10} c\,\hat{\mathbf{z}}$ | (8$h$) | O V |
| $\mathbf{B}_{49}$ | = | $\left(\frac{1}{2} + x_{10}\right)\mathbf{a}_1 + \left(\frac{1}{2} - y_{10}\right)\mathbf{a}_2 + \left(\frac{1}{2} + z_{10}\right)\mathbf{a}_3$ | = | $\left(\frac{1}{2} + x_{10}\right) a\,\hat{\mathbf{x}} + \left(\frac{1}{2} - y_{10}\right) b\,\hat{\mathbf{y}} + \left(\frac{1}{2} + z_{10}\right) c\,\hat{\mathbf{z}}$ | (8$h$) | O V |
| $\mathbf{B}_{50}$ | = | $\left(\frac{1}{2} - x_{10}\right)\mathbf{a}_1 + \left(\frac{1}{2} + y_{10}\right)\mathbf{a}_2 + \left(\frac{1}{2} + z_{10}\right)\mathbf{a}_3$ | = | $\left(\frac{1}{2} - x_{10}\right) a\,\hat{\mathbf{x}} + \left(\frac{1}{2} + y_{10}\right) b\,\hat{\mathbf{y}} + \left(\frac{1}{2} + z_{10}\right) c\,\hat{\mathbf{z}}$ | (8$h$) | O V |
| $\mathbf{B}_{51}$ | = | $x_{11}\,\mathbf{a}_1 + y_{11}\,\mathbf{a}_2 + z_{11}\,\mathbf{a}_3$ | = | $x_{11} a\,\hat{\mathbf{x}} + y_{11} b\,\hat{\mathbf{y}} + z_{11} c\,\hat{\mathbf{z}}$ | (8$h$) | O VI |
| $\mathbf{B}_{52}$ | = | $-x_{11}\,\mathbf{a}_1 - y_{11}\,\mathbf{a}_2 + z_{11}\,\mathbf{a}_3$ | = | $-x_{11} a\,\hat{\mathbf{x}} - y_{11} b\,\hat{\mathbf{y}} + z_{11} c\,\hat{\mathbf{z}}$ | (8$h$) | O VI |
| $\mathbf{B}_{53}$ | = | $\left(\frac{1}{2} - x_{11}\right)\mathbf{a}_1 + \left(\frac{1}{2} + y_{11}\right)\mathbf{a}_2 + \left(\frac{1}{2} - z_{11}\right)\mathbf{a}_3$ | = | $\left(\frac{1}{2} - x_{11}\right) a\,\hat{\mathbf{x}} + \left(\frac{1}{2} + y_{11}\right) b\,\hat{\mathbf{y}} + \left(\frac{1}{2} - z_{11}\right) c\,\hat{\mathbf{z}}$ | (8$h$) | O VI |
| $\mathbf{B}_{54}$ | = | $\left(\frac{1}{2} + x_{11}\right)\mathbf{a}_1 + \left(\frac{1}{2} - y_{11}\right)\mathbf{a}_2 + \left(\frac{1}{2} - z_{11}\right)\mathbf{a}_3$ | = | $\left(\frac{1}{2} + x_{11}\right) a\,\hat{\mathbf{x}} + \left(\frac{1}{2} - y_{11}\right) b\,\hat{\mathbf{y}} + \left(\frac{1}{2} - z_{11}\right) c\,\hat{\mathbf{z}}$ | (8$h$) | O VI |
| $\mathbf{B}_{55}$ | = | $-x_{11}\,\mathbf{a}_1 - y_{11}\,\mathbf{a}_2 - z_{11}\,\mathbf{a}_3$ | = | $-x_{11} a\,\hat{\mathbf{x}} - y_{11} b\,\hat{\mathbf{y}} - z_{11} c\,\hat{\mathbf{z}}$ | (8$h$) | O VI |
| $\mathbf{B}_{56}$ | = | $x_{11}\,\mathbf{a}_1 + y_{11}\,\mathbf{a}_2 - z_{11}\,\mathbf{a}_3$ | = | $x_{11} a\,\hat{\mathbf{x}} + y_{11} b\,\hat{\mathbf{y}} - z_{11} c\,\hat{\mathbf{z}}$ | (8$h$) | O VI |
| $\mathbf{B}_{57}$ | = | $\left(\frac{1}{2} + x_{11}\right)\mathbf{a}_1 + \left(\frac{1}{2} - y_{11}\right)\mathbf{a}_2 + \left(\frac{1}{2} + z_{11}\right)\mathbf{a}_3$ | = | $\left(\frac{1}{2} + x_{11}\right) a\,\hat{\mathbf{x}} + \left(\frac{1}{2} - y_{11}\right) b\,\hat{\mathbf{y}} + \left(\frac{1}{2} + z_{11}\right) c\,\hat{\mathbf{z}}$ | (8$h$) | O VI |
| $\mathbf{B}_{58}$ | = | $\left(\frac{1}{2} - x_{11}\right)\mathbf{a}_1 + \left(\frac{1}{2} + y_{11}\right)\mathbf{a}_2 + \left(\frac{1}{2} + z_{11}\right)\mathbf{a}_3$ | = | $\left(\frac{1}{2} - x_{11}\right) a\,\hat{\mathbf{x}} + \left(\frac{1}{2} + y_{11}\right) b\,\hat{\mathbf{y}} + \left(\frac{1}{2} + z_{11}\right) c\,\hat{\mathbf{z}}$ | (8$h$) | O VI |
| $\mathbf{B}_{59}$ | = | $x_{12}\,\mathbf{a}_1 + y_{12}\,\mathbf{a}_2 + z_{12}\,\mathbf{a}_3$ | = | $x_{12} a\,\hat{\mathbf{x}} + y_{12} b\,\hat{\mathbf{y}} + z_{12} c\,\hat{\mathbf{z}}$ | (8$h$) | O VII |
| $\mathbf{B}_{60}$ | = | $-x_{12}\,\mathbf{a}_1 - y_{12}\,\mathbf{a}_2 + z_{12}\,\mathbf{a}_3$ | = | $-x_{12} a\,\hat{\mathbf{x}} - y_{12} b\,\hat{\mathbf{y}} + z_{12} c\,\hat{\mathbf{z}}$ | (8$h$) | O VII |
| $\mathbf{B}_{61}$ | = | $\left(\frac{1}{2} - x_{12}\right)\,\mathbf{a}_1 + \left(\frac{1}{2} + y_{12}\right)\,\mathbf{a}_2 + \left(\frac{1}{2} - z_{12}\right)\,\mathbf{a}_3$ | = | $\left(\frac{1}{2} - x_{12}\right)a\,\hat{\mathbf{x}} + \left(\frac{1}{2} + y_{12}\right)b\,\hat{\mathbf{y}} + \left(\frac{1}{2} - z_{12}\right)c\,\hat{\mathbf{z}}$ | (8$h$) | O VII |
| $\mathbf{B}_{62}$ | = | $\left(\frac{1}{2} + x_{12}\right)\,\mathbf{a}_1 + \left(\frac{1}{2} - y_{12}\right)\,\mathbf{a}_2 + \left(\frac{1}{2} - z_{12}\right)\,\mathbf{a}_3$ | = | $\left(\frac{1}{2} + x_{12}\right)a\,\hat{\mathbf{x}} + \left(\frac{1}{2} - y_{12}\right)b\,\hat{\mathbf{y}} + \left(\frac{1}{2} - z_{12}\right)c\,\hat{\mathbf{z}}$ | (8$h$) | O VII |
| $\mathbf{B}_{63}$ | = | $-x_{12}\,\mathbf{a}_1 - y_{12}\,\mathbf{a}_2 - z_{12}\,\mathbf{a}_3$ | = | $-x_{12}a\,\hat{\mathbf{x}} - y_{12}b\,\hat{\mathbf{y}} - z_{12}c\,\hat{\mathbf{z}}$ | (8$h$) | O VII |
| $\mathbf{B}_{64}$ | = | $x_{12}\,\mathbf{a}_1 + y_{12}\,\mathbf{a}_2 - z_{12}\,\mathbf{a}_3$ | = | $x_{12}a\,\hat{\mathbf{x}} + y_{12}b\,\hat{\mathbf{y}} - z_{12}c\,\hat{\mathbf{z}}$ | (8$h$) | O VII |
| $\mathbf{B}_{65}$ | = | $\left(\frac{1}{2} + x_{12}\right)\,\mathbf{a}_1 + \left(\frac{1}{2} - y_{12}\right)\,\mathbf{a}_2 + \left(\frac{1}{2} + z_{12}\right)\,\mathbf{a}_3$ | = | $\left(\frac{1}{2} + x_{12}\right)a\,\hat{\mathbf{x}} + \left(\frac{1}{2} - y_{12}\right)b\,\hat{\mathbf{y}} + \left(\frac{1}{2} + z_{12}\right)c\,\hat{\mathbf{z}}$ | (8$h$) | O VII |
| $\mathbf{B}_{66}$ | = | $\left(\frac{1}{2} - x_{12}\right)\,\mathbf{a}_1 + \left(\frac{1}{2} + y_{12}\right)\,\mathbf{a}_2 + \left(\frac{1}{2} + z_{12}\right)\,\mathbf{a}_3$ | = | $\left(\frac{1}{2} - x_{12}\right)a\,\hat{\mathbf{x}} + \left(\frac{1}{2} + y_{12}\right)b\,\hat{\mathbf{y}} + \left(\frac{1}{2} + z_{12}\right)c\,\hat{\mathbf{z}}$ | (8$h$) | O VII |
| $\mathbf{B}_{67}$ | = | $x_{13}\,\mathbf{a}_1 + y_{13}\,\mathbf{a}_2 + z_{13}\,\mathbf{a}_3$ | = | $x_{13}a\,\hat{\mathbf{x}} + y_{13}b\,\hat{\mathbf{y}} + z_{13}c\,\hat{\mathbf{z}}$ | (8$h$) | Si I |
| $\mathbf{B}_{68}$ | = | $-x_{13}\,\mathbf{a}_1 - y_{13}\,\mathbf{a}_2 + z_{13}\,\mathbf{a}_3$ | = | $-x_{13}a\,\hat{\mathbf{x}} - y_{13}b\,\hat{\mathbf{y}} + z_{13}c\,\hat{\mathbf{z}}$ | (8$h$) | Si I |
| $\mathbf{B}_{69}$ | = | $\left(\frac{1}{2} - x_{13}\right)\,\mathbf{a}_1 + \left(\frac{1}{2} + y_{13}\right)\,\mathbf{a}_2 + \left(\frac{1}{2} - z_{13}\right)\,\mathbf{a}_3$ | = | $\left(\frac{1}{2} - x_{13}\right)a\,\hat{\mathbf{x}} + \left(\frac{1}{2} + y_{13}\right)b\,\hat{\mathbf{y}} + \left(\frac{1}{2} - z_{13}\right)c\,\hat{\mathbf{z}}$ | (8$h$) | Si I |
| $\mathbf{B}_{70}$ | = | $\left(\frac{1}{2} + x_{13}\right)\,\mathbf{a}_1 + \left(\frac{1}{2} - y_{13}\right)\,\mathbf{a}_2 + \left(\frac{1}{2} - z_{13}\right)\,\mathbf{a}_3$ | = | $\left(\frac{1}{2} + x_{13}\right)a\,\hat{\mathbf{x}} + \left(\frac{1}{2} - y_{13}\right)b\,\hat{\mathbf{y}} + \left(\frac{1}{2} - z_{13}\right)c\,\hat{\mathbf{z}}$ | (8$h$) | Si I |
| $\mathbf{B}_{71}$ | = | $-x_{13}\,\mathbf{a}_1 - y_{13}\,\mathbf{a}_2 - z_{13}\,\mathbf{a}_3$ | = | $-x_{13}a\,\hat{\mathbf{x}} - y_{13}b\,\hat{\mathbf{y}} - z_{13}c\,\hat{\mathbf{z}}$ | (8$h$) | Si I |
| $\mathbf{B}_{72}$ | = | $x_{13}\,\mathbf{a}_1 + y_{13}\,\mathbf{a}_2 - z_{13}\,\mathbf{a}_3$ | = | $x_{13}a\,\hat{\mathbf{x}} + y_{13}b\,\hat{\mathbf{y}} - z_{13}c\,\hat{\mathbf{z}}$ | (8$h$) | Si I |
| $\mathbf{B}_{73}$ | = | $\left(\frac{1}{2} + x_{13}\right)\,\mathbf{a}_1 + \left(\frac{1}{2} - y_{13}\right)\,\mathbf{a}_2 + \left(\frac{1}{2} + z_{13}\right)\,\mathbf{a}_3$ | = | $\left(\frac{1}{2} + x_{13}\right)a\,\hat{\mathbf{x}} + \left(\frac{1}{2} - y_{13}\right)b\,\hat{\mathbf{y}} + \left(\frac{1}{2} + z_{13}\right)c\,\hat{\mathbf{z}}$ | (8$h$) | Si I |
| $\mathbf{B}_{74}$ | = | $\left(\frac{1}{2} - x_{13}\right)\,\mathbf{a}_1 + \left(\frac{1}{2} + y_{13}\right)\,\mathbf{a}_2 + \left(\frac{1}{2} + z_{13}\right)\,\mathbf{a}_3$ | = | $\left(\frac{1}{2} - x_{13}\right)a\,\hat{\mathbf{x}} + \left(\frac{1}{2} + y_{13}\right)b\,\hat{\mathbf{y}} + \left(\frac{1}{2} + z_{13}\right)c\,\hat{\mathbf{z}}$ | (8$h$) | Si I |
| $\mathbf{B}_{75}$ | = | $x_{14}\,\mathbf{a}_1 + y_{14}\,\mathbf{a}_2 + z_{14}\,\mathbf{a}_3$ | = | $x_{14}a\,\hat{\mathbf{x}} + y_{14}b\,\hat{\mathbf{y}} + z_{14}c\,\hat{\mathbf{z}}$ | (8$h$) | Si II |
| $\mathbf{B}_{76}$ | = | $-x_{14}\,\mathbf{a}_1 - y_{14}\,\mathbf{a}_2 + z_{14}\,\mathbf{a}_3$ | = | $-x_{14}a\,\hat{\mathbf{x}} - y_{14}b\,\hat{\mathbf{y}} + z_{14}c\,\hat{\mathbf{z}}$ | (8$h$) | Si II |
| $\mathbf{B}_{77}$ | = | $\left(\frac{1}{2} - x_{14}\right)\,\mathbf{a}_1 + \left(\frac{1}{2} + y_{14}\right)\,\mathbf{a}_2 + \left(\frac{1}{2} - z_{14}\right)\,\mathbf{a}_3$ | = | $\left(\frac{1}{2} - x_{14}\right)a\,\hat{\mathbf{x}} + \left(\frac{1}{2} + y_{14}\right)b\,\hat{\mathbf{y}} + \left(\frac{1}{2} - z_{14}\right)c\,\hat{\mathbf{z}}$ | (8$h$) | Si II |
| $\mathbf{B}_{78}$ | = | $\left(\frac{1}{2} + x_{14}\right)\,\mathbf{a}_1 + \left(\frac{1}{2} - y_{14}\right)\,\mathbf{a}_2 + \left(\frac{1}{2} - z_{14}\right)\,\mathbf{a}_3$ | = | $\left(\frac{1}{2} + x_{14}\right)a\,\hat{\mathbf{x}} + \left(\frac{1}{2} - y_{14}\right)b\,\hat{\mathbf{y}} + \left(\frac{1}{2} - z_{14}\right)c\,\hat{\mathbf{z}}$ | (8$h$) | Si II |
| $\mathbf{B}_{79}$ | = | $-x_{14}\,\mathbf{a}_1 - y_{14}\,\mathbf{a}_2 - z_{14}\,\mathbf{a}_3$ | = | $-x_{14}a\,\hat{\mathbf{x}} - y_{14}b\,\hat{\mathbf{y}} - z_{14}c\,\hat{\mathbf{z}}$ | (8$h$) | Si II |
| $\mathbf{B}_{80}$ | = | $x_{14}\,\mathbf{a}_1 + y_{14}\,\mathbf{a}_2 - z_{14}\,\mathbf{a}_3$ | = | $x_{14}a\,\hat{\mathbf{x}} + y_{14}b\,\hat{\mathbf{y}} - z_{14}c\,\hat{\mathbf{z}}$ | (8$h$) | Si II |
| $\mathbf{B}_{81}$ | = | $\left(\frac{1}{2} + x_{14}\right)\,\mathbf{a}_1 + \left(\frac{1}{2} - y_{14}\right)\,\mathbf{a}_2 + \left(\frac{1}{2} + z_{14}\right)\,\mathbf{a}_3$ | = | $\left(\frac{1}{2} + x_{14}\right)a\,\hat{\mathbf{x}} + \left(\frac{1}{2} - y_{14}\right)b\,\hat{\mathbf{y}} + \left(\frac{1}{2} + z_{14}\right)c\,\hat{\mathbf{z}}$ | (8$h$) | Si II |
| $\mathbf{B}_{82}$ | = | $\left(\frac{1}{2} - x_{14}\right)\,\mathbf{a}_1 + \left(\frac{1}{2} + y_{14}\right)\,\mathbf{a}_2 + \left(\frac{1}{2} + z_{14}\right)\,\mathbf{a}_3$ | = | $\left(\frac{1}{2} - x_{14}\right)a\,\hat{\mathbf{x}} + \left(\frac{1}{2} + y_{14}\right)b\,\hat{\mathbf{y}} + \left(\frac{1}{2} + z_{14}\right)c\,\hat{\mathbf{z}}$ | (8$h$) | Si II |

**Geometry files:**

- CIF: pp. 1618
- POSCAR: pp. 1618

# $In_4Se_3$ Structure: A4B3_oP28_58_4g_3g

http://aflow.org/prototype-encyclopedia/A4B3_oP28_58_4g_3g

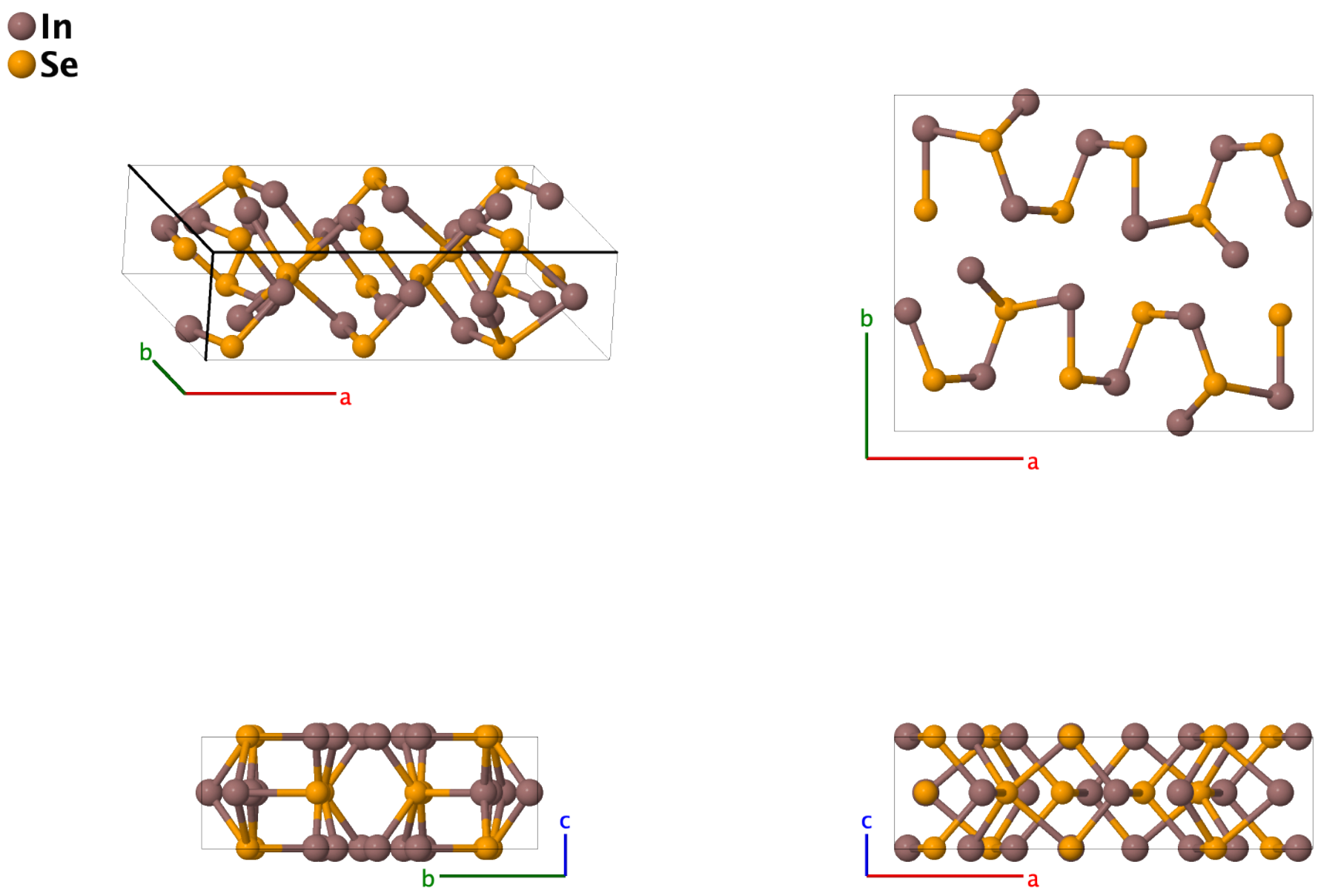

| | | |
|---|---|---|
| **Prototype** | : | $In_4Se_3$ |
| **AFLOW prototype label** | : | A4B3_oP28_58_4g_3g |
| ***Strukturbericht* designation** | : | None |
| **Pearson symbol** | : | oP28 |
| **Space group number** | : | 58 |
| **Space group symbol** | : | *Pnnm* |
| **AFLOW prototype command** | : | `aflow --proto=A4B3_oP28_58_4g_3g` `--params=`$a, b/a, c/a, x_1, y_1, x_2, y_2, x_3, y_3, x_4, y_4, x_5, y_5, x_6, y_6, x_7, y_7$ |

**Simple Orthorhombic primitive vectors:**

$$\mathbf{a}_1 = a\,\hat{\mathbf{x}}$$
$$\mathbf{a}_2 = b\,\hat{\mathbf{y}}$$
$$\mathbf{a}_3 = c\,\hat{\mathbf{z}}$$

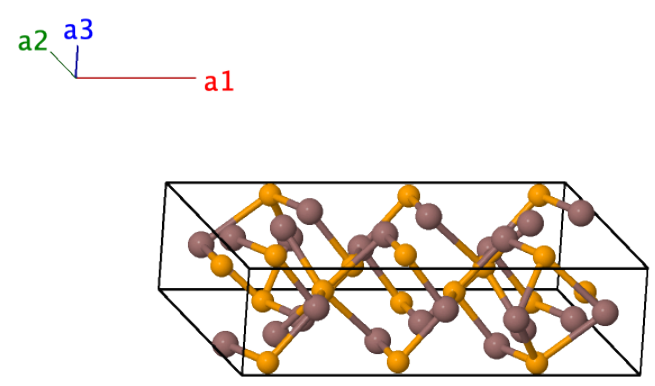

**Basis vectors:**

| | | Lattice Coordinates | | Cartesian Coordinates | Wyckoff Position | Atom Type |
|---|---|---|---|---|---|---|
| $\mathbf{B}_1$ | = | $x_1\,\mathbf{a}_1 + y_1\,\mathbf{a}_2$ | = | $x_1 a\,\hat{\mathbf{x}} + y_1 b\,\hat{\mathbf{y}}$ | (4$g$) | In I |
| $\mathbf{B}_2$ | = | $-x_1\,\mathbf{a}_1 - y_1\,\mathbf{a}_2$ | = | $-x_1 a\,\hat{\mathbf{x}} - y_1 b\,\hat{\mathbf{y}}$ | (4$g$) | In I |
| $\mathbf{B}_3$ | = | $\left(\frac{1}{2} - x_1\right)\mathbf{a}_1 + \left(\frac{1}{2} + y_1\right)\mathbf{a}_2 + \frac{1}{2}\,\mathbf{a}_3$ | = | $\left(\frac{1}{2} - x_1\right)a\,\hat{\mathbf{x}} + \left(\frac{1}{2} + y_1\right)b\,\hat{\mathbf{y}} + \frac{1}{2}c\,\hat{\mathbf{z}}$ | (4$g$) | In I |
| $\mathbf{B}_4$ | = | $\left(\frac{1}{2}+x_1\right)\mathbf{a}_1+\left(\frac{1}{2}-y_1\right)\mathbf{a}_2+\frac{1}{2}\mathbf{a}_3$ | = | $\left(\frac{1}{2}+x_1\right)a\,\hat{\mathbf{x}}+\left(\frac{1}{2}-y_1\right)b\,\hat{\mathbf{y}}+\frac{1}{2}c\,\hat{\mathbf{z}}$ | (4g) | In I |
| $\mathbf{B}_5$ | = | $x_2\,\mathbf{a}_1+y_2\,\mathbf{a}_2$ | = | $x_2a\,\hat{\mathbf{x}}+y_2b\,\hat{\mathbf{y}}$ | (4g) | In II |
| $\mathbf{B}_6$ | = | $-x_2\,\mathbf{a}_1-y_2\,\mathbf{a}_2$ | = | $-x_2a\,\hat{\mathbf{x}}-y_2b\,\hat{\mathbf{y}}$ | (4g) | In II |
| $\mathbf{B}_7$ | = | $\left(\frac{1}{2}-x_2\right)\mathbf{a}_1+\left(\frac{1}{2}+y_2\right)\mathbf{a}_2+\frac{1}{2}\mathbf{a}_3$ | = | $\left(\frac{1}{2}-x_2\right)a\,\hat{\mathbf{x}}+\left(\frac{1}{2}+y_2\right)b\,\hat{\mathbf{y}}+\frac{1}{2}c\,\hat{\mathbf{z}}$ | (4g) | In II |
| $\mathbf{B}_8$ | = | $\left(\frac{1}{2}+x_2\right)\mathbf{a}_1+\left(\frac{1}{2}-y_2\right)\mathbf{a}_2+\frac{1}{2}\mathbf{a}_3$ | = | $\left(\frac{1}{2}+x_2\right)a\,\hat{\mathbf{x}}+\left(\frac{1}{2}-y_2\right)b\,\hat{\mathbf{y}}+\frac{1}{2}c\,\hat{\mathbf{z}}$ | (4g) | In II |
| $\mathbf{B}_9$ | = | $x_3\,\mathbf{a}_1+y_3\,\mathbf{a}_2$ | = | $x_3a\,\hat{\mathbf{x}}+y_3b\,\hat{\mathbf{y}}$ | (4g) | In III |
| $\mathbf{B}_{10}$ | = | $-x_3\,\mathbf{a}_1-y_3\,\mathbf{a}_2$ | = | $-x_3a\,\hat{\mathbf{x}}-y_3b\,\hat{\mathbf{y}}$ | (4g) | In III |
| $\mathbf{B}_{11}$ | = | $\left(\frac{1}{2}-x_3\right)\mathbf{a}_1+\left(\frac{1}{2}+y_3\right)\mathbf{a}_2+\frac{1}{2}\mathbf{a}_3$ | = | $\left(\frac{1}{2}-x_3\right)a\,\hat{\mathbf{x}}+\left(\frac{1}{2}+y_3\right)b\,\hat{\mathbf{y}}+\frac{1}{2}c\,\hat{\mathbf{z}}$ | (4g) | In III |
| $\mathbf{B}_{12}$ | = | $\left(\frac{1}{2}+x_3\right)\mathbf{a}_1+\left(\frac{1}{2}-y_3\right)\mathbf{a}_2+\frac{1}{2}\mathbf{a}_3$ | = | $\left(\frac{1}{2}+x_3\right)a\,\hat{\mathbf{x}}+\left(\frac{1}{2}-y_3\right)b\,\hat{\mathbf{y}}+\frac{1}{2}c\,\hat{\mathbf{z}}$ | (4g) | In III |
| $\mathbf{B}_{13}$ | = | $x_4\,\mathbf{a}_1+y_4\,\mathbf{a}_2$ | = | $x_4a\,\hat{\mathbf{x}}+y_4b\,\hat{\mathbf{y}}$ | (4g) | In IV |
| $\mathbf{B}_{14}$ | = | $-x_4\,\mathbf{a}_1-y_4\,\mathbf{a}_2$ | = | $-x_4a\,\hat{\mathbf{x}}-y_4b\,\hat{\mathbf{y}}$ | (4g) | In IV |
| $\mathbf{B}_{15}$ | = | $\left(\frac{1}{2}-x_4\right)\mathbf{a}_1+\left(\frac{1}{2}+y_4\right)\mathbf{a}_2+\frac{1}{2}\mathbf{a}_3$ | = | $\left(\frac{1}{2}-x_4\right)a\,\hat{\mathbf{x}}+\left(\frac{1}{2}+y_4\right)b\,\hat{\mathbf{y}}+\frac{1}{2}c\,\hat{\mathbf{z}}$ | (4g) | In IV |
| $\mathbf{B}_{16}$ | = | $\left(\frac{1}{2}+x_4\right)\mathbf{a}_1+\left(\frac{1}{2}-y_4\right)\mathbf{a}_2+\frac{1}{2}\mathbf{a}_3$ | = | $\left(\frac{1}{2}+x_4\right)a\,\hat{\mathbf{x}}+\left(\frac{1}{2}-y_4\right)b\,\hat{\mathbf{y}}+\frac{1}{2}c\,\hat{\mathbf{z}}$ | (4g) | In IV |
| $\mathbf{B}_{17}$ | = | $x_5\,\mathbf{a}_1+y_5\,\mathbf{a}_2$ | = | $x_5a\,\hat{\mathbf{x}}+y_5b\,\hat{\mathbf{y}}$ | (4g) | Se I |
| $\mathbf{B}_{18}$ | = | $-x_5\,\mathbf{a}_1-y_5\,\mathbf{a}_2$ | = | $-x_5a\,\hat{\mathbf{x}}-y_5b\,\hat{\mathbf{y}}$ | (4g) | Se I |
| $\mathbf{B}_{19}$ | = | $\left(\frac{1}{2}-x_5\right)\mathbf{a}_1+\left(\frac{1}{2}+y_5\right)\mathbf{a}_2+\frac{1}{2}\mathbf{a}_3$ | = | $\left(\frac{1}{2}-x_5\right)a\,\hat{\mathbf{x}}+\left(\frac{1}{2}+y_5\right)b\,\hat{\mathbf{y}}+\frac{1}{2}c\,\hat{\mathbf{z}}$ | (4g) | Se I |
| $\mathbf{B}_{20}$ | = | $\left(\frac{1}{2}+x_5\right)\mathbf{a}_1+\left(\frac{1}{2}-y_5\right)\mathbf{a}_2+\frac{1}{2}\mathbf{a}_3$ | = | $\left(\frac{1}{2}+x_5\right)a\,\hat{\mathbf{x}}+\left(\frac{1}{2}-y_5\right)b\,\hat{\mathbf{y}}+\frac{1}{2}c\,\hat{\mathbf{z}}$ | (4g) | Se I |
| $\mathbf{B}_{21}$ | = | $x_6\,\mathbf{a}_1+y_6\,\mathbf{a}_2$ | = | $x_6a\,\hat{\mathbf{x}}+y_6b\,\hat{\mathbf{y}}$ | (4g) | Se II |
| $\mathbf{B}_{22}$ | = | $-x_6\,\mathbf{a}_1-y_6\,\mathbf{a}_2$ | = | $-x_6a\,\hat{\mathbf{x}}-y_6b\,\hat{\mathbf{y}}$ | (4g) | Se II |
| $\mathbf{B}_{23}$ | = | $\left(\frac{1}{2}-x_6\right)\mathbf{a}_1+\left(\frac{1}{2}+y_6\right)\mathbf{a}_2+\frac{1}{2}\mathbf{a}_3$ | = | $\left(\frac{1}{2}-x_6\right)a\,\hat{\mathbf{x}}+\left(\frac{1}{2}+y_6\right)b\,\hat{\mathbf{y}}+\frac{1}{2}c\,\hat{\mathbf{z}}$ | (4g) | Se II |
| $\mathbf{B}_{24}$ | = | $\left(\frac{1}{2}+x_6\right)\mathbf{a}_1+\left(\frac{1}{2}-y_6\right)\mathbf{a}_2+\frac{1}{2}\mathbf{a}_3$ | = | $\left(\frac{1}{2}+x_6\right)a\,\hat{\mathbf{x}}+\left(\frac{1}{2}-y_6\right)b\,\hat{\mathbf{y}}+\frac{1}{2}c\,\hat{\mathbf{z}}$ | (4g) | Se II |
| $\mathbf{B}_{25}$ | = | $x_7\,\mathbf{a}_1+y_7\,\mathbf{a}_2$ | = | $x_7a\,\hat{\mathbf{x}}+y_7b\,\hat{\mathbf{y}}$ | (4g) | Se III |
| $\mathbf{B}_{26}$ | = | $-x_7\,\mathbf{a}_1-y_7\,\mathbf{a}_2$ | = | $-x_7a\,\hat{\mathbf{x}}-y_7b\,\hat{\mathbf{y}}$ | (4g) | Se III |
| $\mathbf{B}_{27}$ | = | $\left(\frac{1}{2}-x_7\right)\mathbf{a}_1+\left(\frac{1}{2}+y_7\right)\mathbf{a}_2+\frac{1}{2}\mathbf{a}_3$ | = | $\left(\frac{1}{2}-x_7\right)a\,\hat{\mathbf{x}}+\left(\frac{1}{2}+y_7\right)b\,\hat{\mathbf{y}}+\frac{1}{2}c\,\hat{\mathbf{z}}$ | (4g) | Se III |
| $\mathbf{B}_{28}$ | = | $\left(\frac{1}{2}+x_7\right)\mathbf{a}_1+\left(\frac{1}{2}-y_7\right)\mathbf{a}_2+\frac{1}{2}\mathbf{a}_3$ | = | $\left(\frac{1}{2}+x_7\right)a\,\hat{\mathbf{x}}+\left(\frac{1}{2}-y_7\right)b\,\hat{\mathbf{y}}+\frac{1}{2}c\,\hat{\mathbf{z}}$ | (4g) | Se III |

---

**Found in:**

---

**Geometry files:**

- CIF: pp. 1619
- POSCAR: pp. 1619

# Adamite [$Zn_2(AsO_4)(OH)$, $H2_7$] Structure: ABC5D2_oP36_58_g_g_3gh_eg

http://aflow.org/prototype-encyclopedia/ABC5D2_oP36_58_g_g_3gh_eg

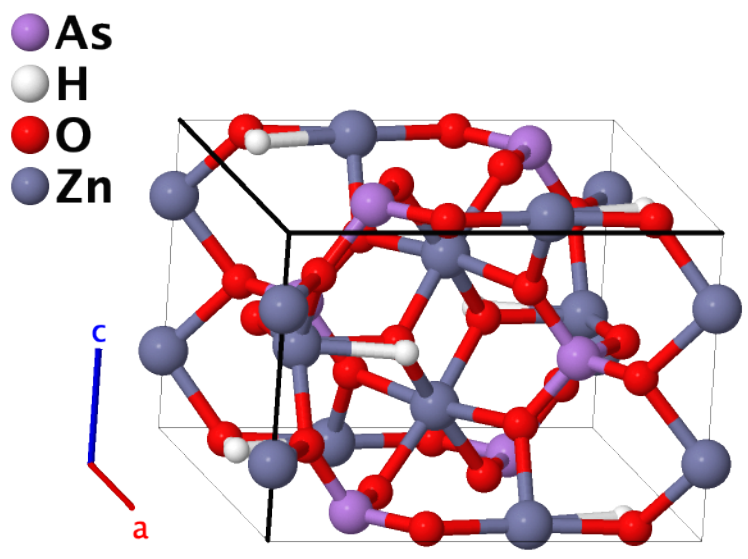

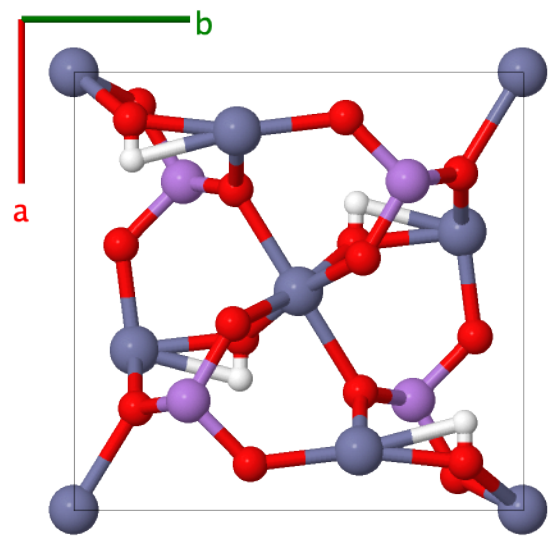

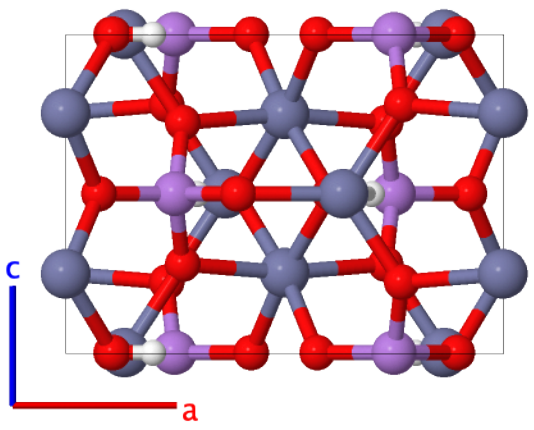

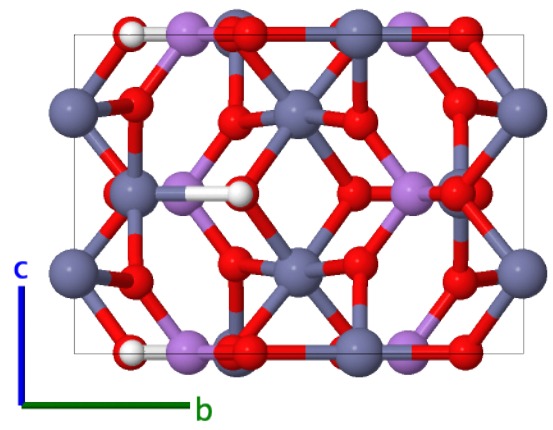

| | | |
|---|---|---|
| **Prototype** | : | $AsHO_5Zn_2$ |
| **AFLOW prototype label** | : | ABC5D2_oP36_58_g_g_3gh_eg |
| ***Strukturbericht* designation** | : | $H2_7$ |
| **Pearson symbol** | : | oP36 |
| **Space group number** | : | 58 |
| **Space group symbol** | : | *Pnnm* |
| **AFLOW prototype command** | : | `aflow --proto=ABC5D2_oP36_58_g_g_3gh_eg` `--params=`$a, b/a, c/a, z_1, x_2, y_2, x_3, y_3, x_4, y_4, x_5, y_5, x_6, y_6, x_7, y_7, x_8, y_8, z_8$ |

**Other compounds with this structure**

- $Cu_2(AsO_4)(OH)$ and $Cu_2(PO_4)(OH)$

- This structure was originally determined by (Kokkoros, 1937) and designated $H2_7$ by (Gottfried, 1940). (Hill, 1976) refined the structure, including the positions of the hydrogen atoms in the OH radical.

**Simple Orthorhombic primitive vectors:**

$$\mathbf{a}_1 = a\,\hat{\mathbf{x}}$$
$$\mathbf{a}_2 = b\,\hat{\mathbf{y}}$$
$$\mathbf{a}_3 = c\,\hat{\mathbf{z}}$$

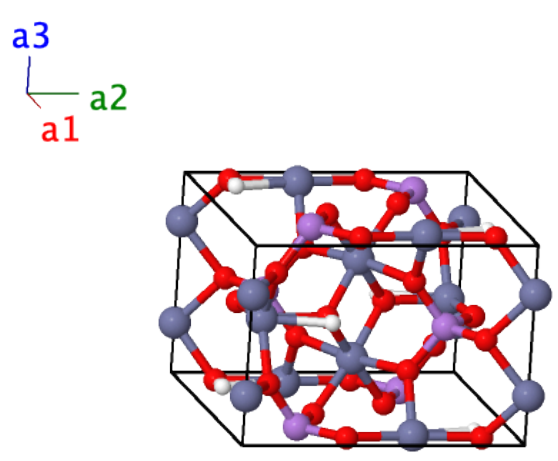

**Basis vectors:**

| | | Lattice Coordinates | | Cartesian Coordinates | Wyckoff Position | Atom Type |
|---|---|---|---|---|---|---|
| $\mathbf{B}_1$ | = | $z_1\,\mathbf{a}_3$ | = | $z_1 c\,\hat{\mathbf{z}}$ | (4*e*) | Zn I |
| $\mathbf{B}_2$ | = | $\frac{1}{2}\,\mathbf{a}_1 + \frac{1}{2}\,\mathbf{a}_2 + \left(\frac{1}{2} - z_1\right)\,\mathbf{a}_3$ | = | $\frac{1}{2}a\,\hat{\mathbf{x}} + \frac{1}{2}b\,\hat{\mathbf{y}} + \left(\frac{1}{2} - z_1\right)c\,\hat{\mathbf{z}}$ | (4*e*) | Zn I |
| $\mathbf{B}_3$ | = | $-z_1\,\mathbf{a}_3$ | = | $-z_1 c\,\hat{\mathbf{z}}$ | (4*e*) | Zn I |
| $\mathbf{B}_4$ | = | $\frac{1}{2}\,\mathbf{a}_1 + \frac{1}{2}\,\mathbf{a}_2 + \left(\frac{1}{2} + z_1\right)\,\mathbf{a}_3$ | = | $\frac{1}{2}a\,\hat{\mathbf{x}} + \frac{1}{2}b\,\hat{\mathbf{y}} + \left(\frac{1}{2} + z_1\right)c\,\hat{\mathbf{z}}$ | (4*e*) | Zn I |
| $\mathbf{B}_5$ | = | $x_2\,\mathbf{a}_1 + y_2\,\mathbf{a}_2$ | = | $x_2 a\,\hat{\mathbf{x}} + y_2 b\,\hat{\mathbf{y}}$ | (4*g*) | As |
| $\mathbf{B}_6$ | = | $-x_2\,\mathbf{a}_1 - y_2\,\mathbf{a}_2$ | = | $-x_2 a\,\hat{\mathbf{x}} - y_2 b\,\hat{\mathbf{y}}$ | (4*g*) | As |
| $\mathbf{B}_7$ | = | $\left(\frac{1}{2} - x_2\right)\,\mathbf{a}_1 + \left(\frac{1}{2} + y_2\right)\,\mathbf{a}_2 + \frac{1}{2}\,\mathbf{a}_3$ | = | $\left(\frac{1}{2} - x_2\right)a\,\hat{\mathbf{x}} + \left(\frac{1}{2} + y_2\right)b\,\hat{\mathbf{y}} + \frac{1}{2}c\,\hat{\mathbf{z}}$ | (4*g*) | As |
| $\mathbf{B}_8$ | = | $\left(\frac{1}{2} + x_2\right)\,\mathbf{a}_1 + \left(\frac{1}{2} - y_2\right)\,\mathbf{a}_2 + \frac{1}{2}\,\mathbf{a}_3$ | = | $\left(\frac{1}{2} + x_2\right)a\,\hat{\mathbf{x}} + \left(\frac{1}{2} - y_2\right)b\,\hat{\mathbf{y}} + \frac{1}{2}c\,\hat{\mathbf{z}}$ | (4*g*) | As |
| $\mathbf{B}_9$ | = | $x_3\,\mathbf{a}_1 + y_3\,\mathbf{a}_2$ | = | $x_3 a\,\hat{\mathbf{x}} + y_3 b\,\hat{\mathbf{y}}$ | (4*g*) | H |
| $\mathbf{B}_{10}$ | = | $-x_3\,\mathbf{a}_1 - y_3\,\mathbf{a}_2$ | = | $-x_3 a\,\hat{\mathbf{x}} - y_3 b\,\hat{\mathbf{y}}$ | (4*g*) | H |
| $\mathbf{B}_{11}$ | = | $\left(\frac{1}{2} - x_3\right)\,\mathbf{a}_1 + \left(\frac{1}{2} + y_3\right)\,\mathbf{a}_2 + \frac{1}{2}\,\mathbf{a}_3$ | = | $\left(\frac{1}{2} - x_3\right)a\,\hat{\mathbf{x}} + \left(\frac{1}{2} + y_3\right)b\,\hat{\mathbf{y}} + \frac{1}{2}c\,\hat{\mathbf{z}}$ | (4*g*) | H |
| $\mathbf{B}_{12}$ | = | $\left(\frac{1}{2} + x_3\right)\,\mathbf{a}_1 + \left(\frac{1}{2} - y_3\right)\,\mathbf{a}_2 + \frac{1}{2}\,\mathbf{a}_3$ | = | $\left(\frac{1}{2} + x_3\right)a\,\hat{\mathbf{x}} + \left(\frac{1}{2} - y_3\right)b\,\hat{\mathbf{y}} + \frac{1}{2}c\,\hat{\mathbf{z}}$ | (4*g*) | H |
| $\mathbf{B}_{13}$ | = | $x_4\,\mathbf{a}_1 + y_4\,\mathbf{a}_2$ | = | $x_4 a\,\hat{\mathbf{x}} + y_4 b\,\hat{\mathbf{y}}$ | (4*g*) | O I |
| $\mathbf{B}_{14}$ | = | $-x_4\,\mathbf{a}_1 - y_4\,\mathbf{a}_2$ | = | $-x_4 a\,\hat{\mathbf{x}} - y_4 b\,\hat{\mathbf{y}}$ | (4*g*) | O I |
| $\mathbf{B}_{15}$ | = | $\left(\frac{1}{2} - x_4\right)\,\mathbf{a}_1 + \left(\frac{1}{2} + y_4\right)\,\mathbf{a}_2 + \frac{1}{2}\,\mathbf{a}_3$ | = | $\left(\frac{1}{2} - x_4\right)a\,\hat{\mathbf{x}} + \left(\frac{1}{2} + y_4\right)b\,\hat{\mathbf{y}} + \frac{1}{2}c\,\hat{\mathbf{z}}$ | (4*g*) | O I |
| $\mathbf{B}_{16}$ | = | $\left(\frac{1}{2} + x_4\right)\,\mathbf{a}_1 + \left(\frac{1}{2} - y_4\right)\,\mathbf{a}_2 + \frac{1}{2}\,\mathbf{a}_3$ | = | $\left(\frac{1}{2} + x_4\right)a\,\hat{\mathbf{x}} + \left(\frac{1}{2} - y_4\right)b\,\hat{\mathbf{y}} + \frac{1}{2}c\,\hat{\mathbf{z}}$ | (4*g*) | O I |
| $\mathbf{B}_{17}$ | = | $x_5\,\mathbf{a}_1 + y_5\,\mathbf{a}_2$ | = | $x_5 a\,\hat{\mathbf{x}} + y_5 b\,\hat{\mathbf{y}}$ | (4*g*) | O II |
| $\mathbf{B}_{18}$ | = | $-x_5\,\mathbf{a}_1 - y_5\,\mathbf{a}_2$ | = | $-x_5 a\,\hat{\mathbf{x}} - y_5 b\,\hat{\mathbf{y}}$ | (4*g*) | O II |
| $\mathbf{B}_{19}$ | = | $\left(\frac{1}{2} - x_5\right)\,\mathbf{a}_1 + \left(\frac{1}{2} + y_5\right)\,\mathbf{a}_2 + \frac{1}{2}\,\mathbf{a}_3$ | = | $\left(\frac{1}{2} - x_5\right)a\,\hat{\mathbf{x}} + \left(\frac{1}{2} + y_5\right)b\,\hat{\mathbf{y}} + \frac{1}{2}c\,\hat{\mathbf{z}}$ | (4*g*) | O II |
| $\mathbf{B}_{20}$ | = | $\left(\frac{1}{2} + x_5\right)\,\mathbf{a}_1 + \left(\frac{1}{2} - y_5\right)\,\mathbf{a}_2 + \frac{1}{2}\,\mathbf{a}_3$ | = | $\left(\frac{1}{2} + x_5\right)a\,\hat{\mathbf{x}} + \left(\frac{1}{2} - y_5\right)b\,\hat{\mathbf{y}} + \frac{1}{2}c\,\hat{\mathbf{z}}$ | (4*g*) | O II |
| $\mathbf{B}_{21}$ | = | $x_6\,\mathbf{a}_1 + y_6\,\mathbf{a}_2$ | = | $x_6 a\,\hat{\mathbf{x}} + y_6 b\,\hat{\mathbf{y}}$ | (4*g*) | O III |
| $\mathbf{B}_{22}$ | = | $-x_6\,\mathbf{a}_1 - y_6\,\mathbf{a}_2$ | = | $-x_6 a\,\hat{\mathbf{x}} - y_6 b\,\hat{\mathbf{y}}$ | (4*g*) | O III |
| $\mathbf{B}_{23}$ | = | $\left(\frac{1}{2} - x_6\right)\,\mathbf{a}_1 + \left(\frac{1}{2} + y_6\right)\,\mathbf{a}_2 + \frac{1}{2}\,\mathbf{a}_3$ | = | $\left(\frac{1}{2} - x_6\right)a\,\hat{\mathbf{x}} + \left(\frac{1}{2} + y_6\right)b\,\hat{\mathbf{y}} + \frac{1}{2}c\,\hat{\mathbf{z}}$ | (4*g*) | O III |
| $\mathbf{B}_{24}$ | = | $\left(\frac{1}{2} + x_6\right)\,\mathbf{a}_1 + \left(\frac{1}{2} - y_6\right)\,\mathbf{a}_2 + \frac{1}{2}\,\mathbf{a}_3$ | = | $\left(\frac{1}{2} + x_6\right)a\,\hat{\mathbf{x}} + \left(\frac{1}{2} - y_6\right)b\,\hat{\mathbf{y}} + \frac{1}{2}c\,\hat{\mathbf{z}}$ | (4*g*) | O III |
| $\mathbf{B}_{25}$ | = | $x_7\,\mathbf{a}_1 + y_7\,\mathbf{a}_2$ | = | $x_7 a\,\hat{\mathbf{x}} + y_7 b\,\hat{\mathbf{y}}$ | (4*g*) | Zn II |
| $\mathbf{B}_{26}$ | = | $-x_7\,\mathbf{a}_1 - y_7\,\mathbf{a}_2$ | = | $-x_7 a\,\hat{\mathbf{x}} - y_7 b\,\hat{\mathbf{y}}$ | (4*g*) | Zn II |
| $\mathbf{B}_{27}$ | = | $\left(\frac{1}{2} - x_7\right)\,\mathbf{a}_1 + \left(\frac{1}{2} + y_7\right)\,\mathbf{a}_2 + \frac{1}{2}\,\mathbf{a}_3$ | = | $\left(\frac{1}{2} - x_7\right)a\,\hat{\mathbf{x}} + \left(\frac{1}{2} + y_7\right)b\,\hat{\mathbf{y}} + \frac{1}{2}c\,\hat{\mathbf{z}}$ | (4*g*) | Zn II |
| $\mathbf{B}_{28}$ | = | $\left(\frac{1}{2} + x_7\right)\,\mathbf{a}_1 + \left(\frac{1}{2} - y_7\right)\,\mathbf{a}_2 + \frac{1}{2}\,\mathbf{a}_3$ | = | $\left(\frac{1}{2} + x_7\right)a\,\hat{\mathbf{x}} + \left(\frac{1}{2} - y_7\right)b\,\hat{\mathbf{y}} + \frac{1}{2}c\,\hat{\mathbf{z}}$ | (4*g*) | Zn II |
| $\mathbf{B}_{29}$ | = | $x_8\,\mathbf{a}_1 + y_8\,\mathbf{a}_2 + z_8\,\mathbf{a}_3$ | = | $x_8 a\,\hat{\mathbf{x}} + y_8 b\,\hat{\mathbf{y}} + z_8 c\,\hat{\mathbf{z}}$ | (8*h*) | O IV |
| $\mathbf{B}_{30}$ | = | $-x_8\,\mathbf{a}_1 - y_8\,\mathbf{a}_2 + z_8\,\mathbf{a}_3$ | = | $-x_8 a\,\hat{\mathbf{x}} - y_8 b\,\hat{\mathbf{y}} + z_8 c\,\hat{\mathbf{z}}$ | (8*h*) | O IV |
| $\mathbf{B}_{31}$ | = | $\left(\frac{1}{2} - x_8\right)\,\mathbf{a}_1 + \left(\frac{1}{2} + y_8\right)\,\mathbf{a}_2 + \left(\frac{1}{2} - z_8\right)\,\mathbf{a}_3$ | = | $\left(\frac{1}{2} - x_8\right)a\,\hat{\mathbf{x}} + \left(\frac{1}{2} + y_8\right)b\,\hat{\mathbf{y}} + \left(\frac{1}{2} - z_8\right)c\,\hat{\mathbf{z}}$ | (8*h*) | O IV |
| $\mathbf{B}_{32}$ | = | $\left(\frac{1}{2} + x_8\right)\,\mathbf{a}_1 + \left(\frac{1}{2} - y_8\right)\,\mathbf{a}_2 + \left(\frac{1}{2} - z_8\right)\,\mathbf{a}_3$ | = | $\left(\frac{1}{2} + x_8\right)a\,\hat{\mathbf{x}} + \left(\frac{1}{2} - y_8\right)b\,\hat{\mathbf{y}} + \left(\frac{1}{2} - z_8\right)c\,\hat{\mathbf{z}}$ | (8*h*) | O IV |
| $\mathbf{B}_{33}$ | = | $-x_8\,\mathbf{a}_1 - y_8\,\mathbf{a}_2 - z_8\,\mathbf{a}_3$ | = | $-x_8 a\,\hat{\mathbf{x}} - y_8 b\,\hat{\mathbf{y}} - z_8 c\,\hat{\mathbf{z}}$ | (8$h$) | O IV |
| $\mathbf{B}_{34}$ | = | $x_8\,\mathbf{a}_1 + y_8\,\mathbf{a}_2 - z_8\,\mathbf{a}_3$ | = | $x_8 a\,\hat{\mathbf{x}} + y_8 b\,\hat{\mathbf{y}} - z_8 c\,\hat{\mathbf{z}}$ | (8$h$) | O IV |
| $\mathbf{B}_{35}$ | = | $\left(\frac{1}{2} + x_8\right)\mathbf{a}_1 + \left(\frac{1}{2} - y_8\right)\mathbf{a}_2 + \left(\frac{1}{2} + z_8\right)\mathbf{a}_3$ | = | $\left(\frac{1}{2} + x_8\right)a\,\hat{\mathbf{x}} + \left(\frac{1}{2} - y_8\right)b\,\hat{\mathbf{y}} + \left(\frac{1}{2} + z_8\right)c\,\hat{\mathbf{z}}$ | (8$h$) | O IV |
| $\mathbf{B}_{36}$ | = | $\left(\frac{1}{2} - x_8\right)\mathbf{a}_1 + \left(\frac{1}{2} + y_8\right)\mathbf{a}_2 + \left(\frac{1}{2} + z_8\right)\mathbf{a}_3$ | = | $\left(\frac{1}{2} - x_8\right)a\,\hat{\mathbf{x}} + \left(\frac{1}{2} + y_8\right)b\,\hat{\mathbf{y}} + \left(\frac{1}{2} + z_8\right)c\,\hat{\mathbf{z}}$ | (8$h$) | O IV |

**Geometry files:**

- CIF: pp. 1619
- POSCAR: pp. 1620

# InS Structure: AB_oP8_58_g_g

http://aflow.org/prototype-encyclopedia/AB_oP8_58_g_g

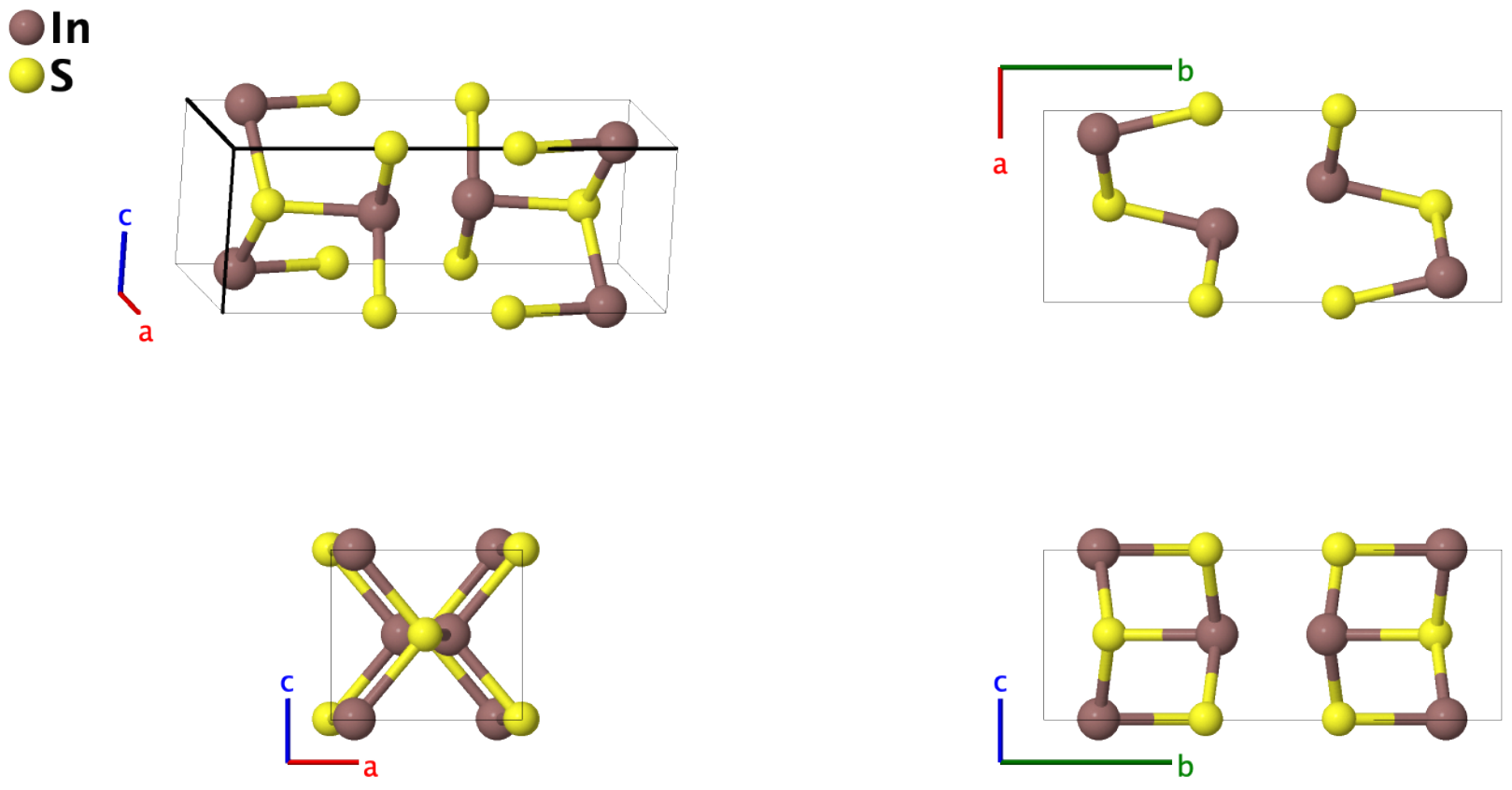

| | | |
|---|---|---|
| **Prototype** | : | InS |
| **AFLOW prototype label** | : | AB_oP8_58_g_g |
| ***Strukturbericht* designation** | : | None |
| **Pearson symbol** | : | oP8 |
| **Space group number** | : | 58 |
| **Space group symbol** | : | *Pnnm* |
| **AFLOW prototype command** | : | `aflow --proto=AB_oP8_58_g_g`<br>`--params=`$a, b/a, c/a, x_1, y_1, x_2, y_2$ |

### Simple Orthorhombic primitive vectors:

$$\mathbf{a}_1 = a\,\hat{\mathbf{x}}$$
$$\mathbf{a}_2 = b\,\hat{\mathbf{y}}$$
$$\mathbf{a}_3 = c\,\hat{\mathbf{z}}$$

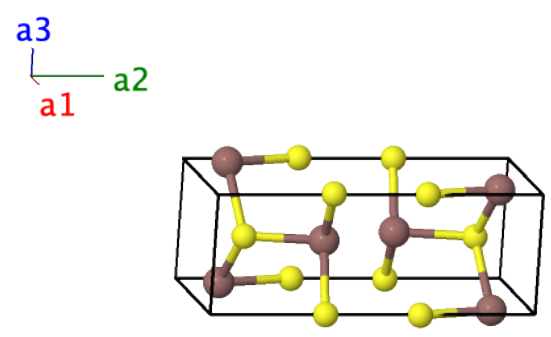

### Basis vectors:

| | | Lattice Coordinates | | Cartesian Coordinates | Wyckoff Position | Atom Type |
|---|---|---|---|---|---|---|
| $\mathbf{B}_1$ | = | $x_1\,\mathbf{a}_1 + y_1\,\mathbf{a}_2$ | = | $x_1 a\,\hat{\mathbf{x}} + y_1 b\,\hat{\mathbf{y}}$ | (4g) | In |
| $\mathbf{B}_2$ | = | $-x_1\,\mathbf{a}_1 - y_1\,\mathbf{a}_2$ | = | $-x_1 a\,\hat{\mathbf{x}} - y_1 b\,\hat{\mathbf{y}}$ | (4g) | In |
| $\mathbf{B}_3$ | = | $\left(\frac{1}{2} - x_1\right)\mathbf{a}_1 + \left(\frac{1}{2} + y_1\right)\mathbf{a}_2 + \frac{1}{2}\,\mathbf{a}_3$ | = | $\left(\frac{1}{2} - x_1\right) a\,\hat{\mathbf{x}} + \left(\frac{1}{2} + y_1\right) b\,\hat{\mathbf{y}} + \frac{1}{2} c\,\hat{\mathbf{z}}$ | (4g) | In |
| $\mathbf{B}_4$ | = | $\left(\frac{1}{2} + x_1\right)\mathbf{a}_1 + \left(\frac{1}{2} - y_1\right)\mathbf{a}_2 + \frac{1}{2}\,\mathbf{a}_3$ | = | $\left(\frac{1}{2} + x_1\right) a\,\hat{\mathbf{x}} + \left(\frac{1}{2} - y_1\right) b\,\hat{\mathbf{y}} + \frac{1}{2} c\,\hat{\mathbf{z}}$ | (4g) | In |
| $\mathbf{B}_5$ | = | $x_2\,\mathbf{a}_1 + y_2\,\mathbf{a}_2$ | = | $x_2 a\,\hat{\mathbf{x}} + y_2 b\,\hat{\mathbf{y}}$ | (4g) | S |
| $\mathbf{B}_6$ | = | $-x_2\,\mathbf{a}_1 - y_2\,\mathbf{a}_2$ | = | $-x_2 a\,\hat{\mathbf{x}} - y_2 b\,\hat{\mathbf{y}}$ | (4g) | S |
| $\mathbf{B}_7$ | = | $\left(\frac{1}{2} - x_2\right)\mathbf{a}_1 + \left(\frac{1}{2} + y_2\right)\mathbf{a}_2 + \frac{1}{2}\,\mathbf{a}_3$ | = | $\left(\frac{1}{2} - x_2\right) a\,\hat{\mathbf{x}} + \left(\frac{1}{2} + y_2\right) b\,\hat{\mathbf{y}} + \frac{1}{2} c\,\hat{\mathbf{z}}$ | (4g) | S |
| $\mathbf{B}_8$ | = | $\left(\frac{1}{2} + x_2\right)\mathbf{a}_1 + \left(\frac{1}{2} - y_2\right)\mathbf{a}_2 + \frac{1}{2}\,\mathbf{a}_3$ | = | $\left(\frac{1}{2} + x_2\right) a\,\hat{\mathbf{x}} + \left(\frac{1}{2} - y_2\right) b\,\hat{\mathbf{y}} + \frac{1}{2} c\,\hat{\mathbf{z}}$ | (4g) | S |

**Found in:**

---

**Geometry files:**

- CIF: pp. 1620
- POSCAR: pp. 1620

# $RuB_2$ Structure: A2B_oP6_59_f_a

http://aflow.org/prototype-encyclopedia/A2B_oP6_59_f_a

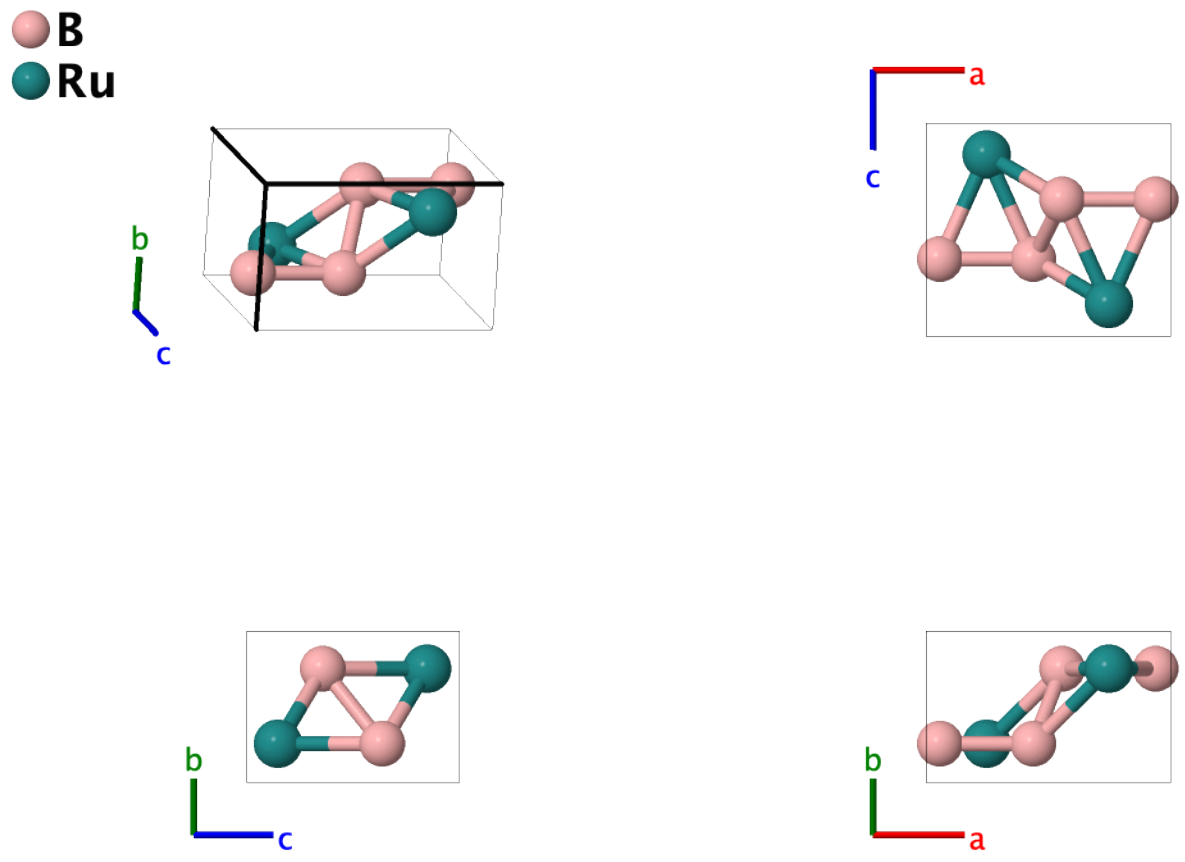

| | | |
|---|---|---|
| **Prototype** | : | $B_2Ru$ |
| **AFLOW prototype label** | : | A2B_oP6_59_f_a |
| ***Strukturbericht* designation** | : | None |
| **Pearson symbol** | : | oP6 |
| **Space group number** | : | 59 |
| **Space group symbol** | : | *Pmmn* |
| **AFLOW prototype command** | : | `aflow --proto=A2B_oP6_59_f_a` <br> `--params=`$a, b/a, c/a, z_1, x_2, z_2$ |

### Other compounds with this structure

- $OsB_2$

### Simple Orthorhombic primitive vectors:

$$\mathbf{a}_1 = a\,\hat{\mathbf{x}}$$
$$\mathbf{a}_2 = b\,\hat{\mathbf{y}}$$
$$\mathbf{a}_3 = c\,\hat{\mathbf{z}}$$

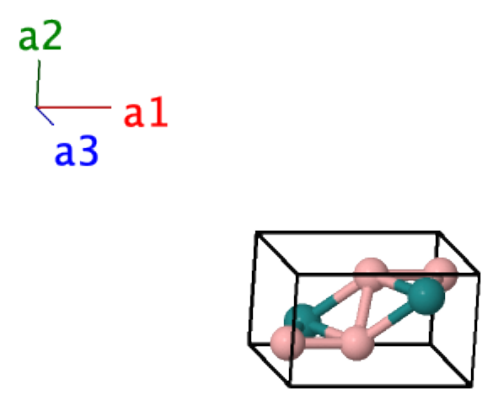

### Basis vectors:

| | | Lattice Coordinates | | Cartesian Coordinates | Wyckoff Position | Atom Type |
|---|---|---|---|---|---|---|
| $\mathbf{B}_1$ | = | $\frac{1}{4}\,\mathbf{a}_1 + \frac{1}{4}\,\mathbf{a}_2 + z_1\,\mathbf{a}_3$ | = | $\frac{1}{4}a\,\hat{\mathbf{x}} + \frac{1}{4}b\,\hat{\mathbf{y}} + z_1 c\,\hat{\mathbf{z}}$ | (2*a*) | Ru |
| $\mathbf{B}_2$ | = | $\frac{3}{4}\,\mathbf{a}_1 + \frac{3}{4}\,\mathbf{a}_2 - z_1\,\mathbf{a}_3$ | = | $\frac{3}{4}a\,\hat{\mathbf{x}} + \frac{3}{4}b\,\hat{\mathbf{y}} - z_1 c\,\hat{\mathbf{z}}$ | (2*a*) | Ru |
| $\mathbf{B}_3$ | = | $x_2\,\mathbf{a}_1 + \frac{1}{4}\,\mathbf{a}_2 + z_2\,\mathbf{a}_3$ | = | $x_2 a\,\hat{\mathbf{x}} + \frac{1}{4}b\,\hat{\mathbf{y}} + z_2 c\,\hat{\mathbf{z}}$ | (4*f*) | B |
| $\mathbf{B}_4$ | = | $\left(\frac{1}{2} - x_2\right)\mathbf{a}_1 + \frac{1}{4}\,\mathbf{a}_2 + z_2\,\mathbf{a}_3$ | = | $\left(\frac{1}{2} - x_2\right)a\,\hat{\mathbf{x}} + \frac{1}{4}b\,\hat{\mathbf{y}} + z_2 c\,\hat{\mathbf{z}}$ | (4*f*) | B |
| $\mathbf{B}_5$ | = | $-x_2\,\mathbf{a}_1 + \frac{3}{4}\,\mathbf{a}_2 - z_2\,\mathbf{a}_3$ | = | $-x_2 a\,\hat{\mathbf{x}} + \frac{3}{4}b\,\hat{\mathbf{y}} - z_2 c\,\hat{\mathbf{z}}$ | (4*f*) | B |
| $\mathbf{B}_6$ | = | $\left(\frac{1}{2} + x_2\right)\mathbf{a}_1 + \frac{3}{4}\,\mathbf{a}_2 - z_2\,\mathbf{a}_3$ | = | $\left(\frac{1}{2} + x_2\right)a\,\hat{\mathbf{x}} + \frac{3}{4}b\,\hat{\mathbf{y}} - z_2 c\,\hat{\mathbf{z}}$ | (4*f*) | B |

**Geometry files:**

- CIF: pp. 1621
- POSCAR: pp. 1621

# $NH_4NO_3$ IV ($G0_{11}$) Structure: A4B2C3_oP18_59_ef_ab_af

http://aflow.org/prototype-encyclopedia/A4B2C3_oP18_59_ef_ab_af

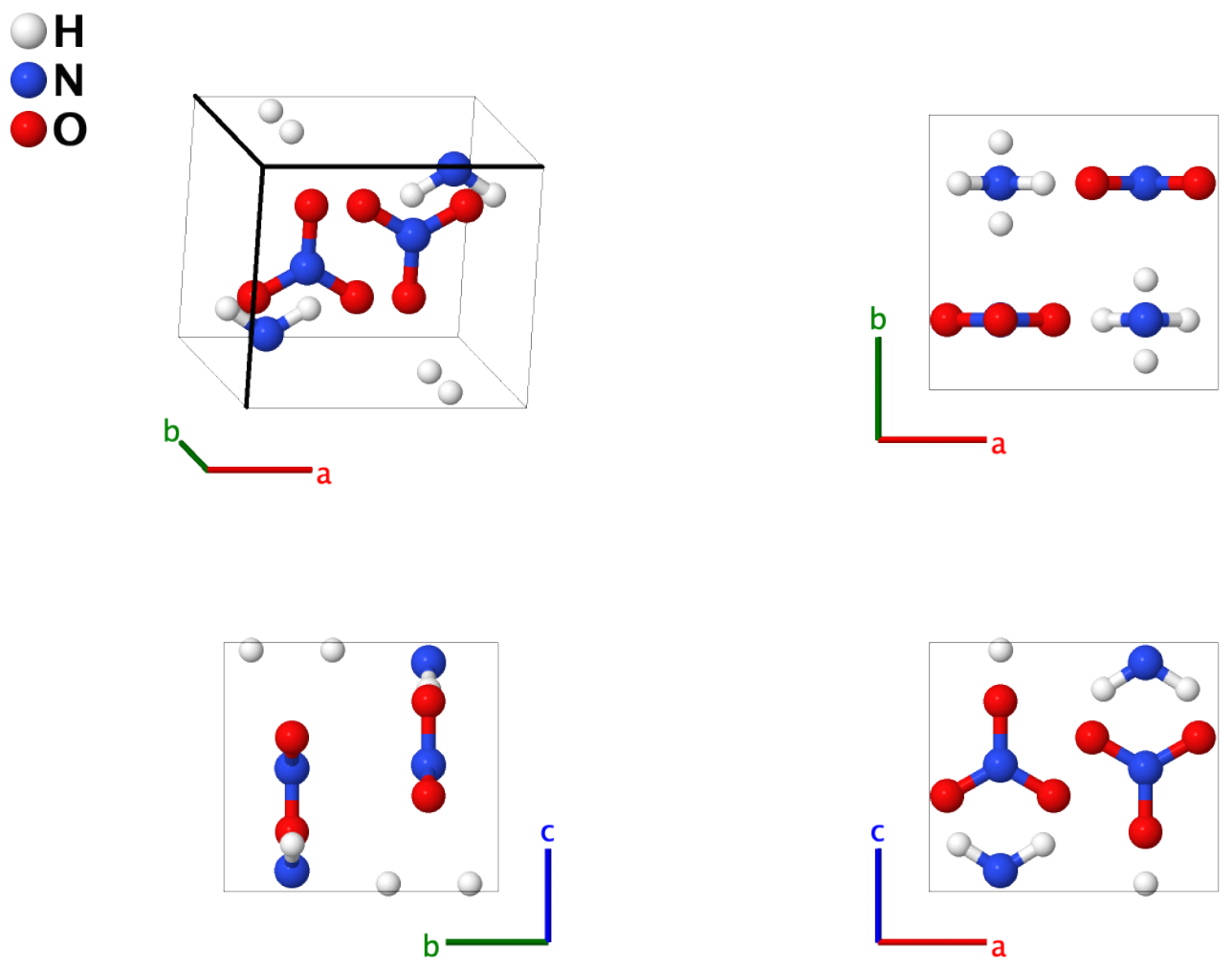

| | | |
|---|---|---|
| **Prototype** | : | $H_4N_2O_3$ |
| **AFLOW prototype label** | : | A4B2C3_oP18_59_ef_ab_af |
| ***Strukturbericht* designation** | : | $G0_{11}$ |
| **Pearson symbol** | : | oP18 |
| **Space group number** | : | 59 |
| **Space group symbol** | : | *Pmmn* |
| **AFLOW prototype command** | : | `aflow --proto=A4B2C3_oP18_59_ef_ab_af` <br> `--params=`$a, b/a, c/a, z_1, z_2, z_3, y_4, z_4, x_5, z_5, x_6, z_6$ |

- Ammonium Nitrate exists in a variety of forms, (Hermann, 1937) depending on the temperature:

| Phase | Temperature °C | *Strukturbericht* | Page | |
|---|---|---|---|---|
| I | 125 – 170 | $G0_8$ | AB_cP2_221_a_b.NH4.NO3 | |
| II | 84 – 125 | $G0_9$ | ABC3_tP10_100_b_a_bc | |
| III | 32 – 84 | $G0_{10}$ | ABC3_oP20_62_c_c_cd.N.NH4.O | |
| IV | -17 – 32 | $G0_{11}$ | A4B2C3_oP18_59_ef_ab_af | (this structure) |
| V | $< -17$ | Gwihabaite | A4B2C3_tP72_77_8d_ab2c2d_6d2 | |

- In the original reference (West, 1932) did not determine the positions of the hydrogen atoms. Since the hydrogen atoms are in the same space group, we continue to designate this the $G0_{11}$ structure.

**Simple Orthorhombic primitive vectors:**

$$\mathbf{a}_1 = a\,\mathbf{\hat{x}}$$
$$\mathbf{a}_2 = b\,\mathbf{\hat{y}}$$
$$\mathbf{a}_3 = c\,\mathbf{\hat{z}}$$

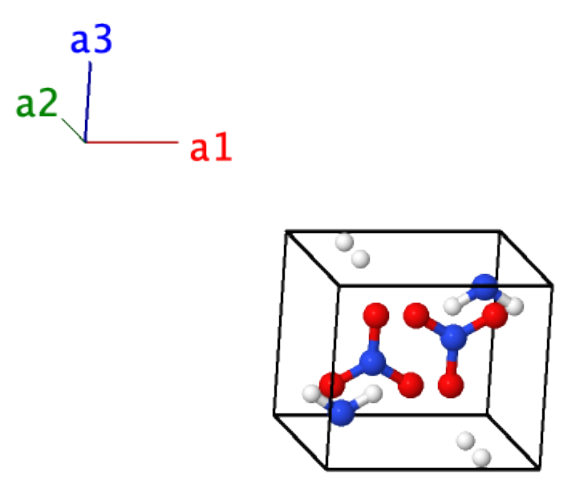

**Basis vectors:**

| | | Lattice Coordinates | | Cartesian Coordinates | Wyckoff Position | Atom Type |
|---|---|---|---|---|---|---|
| $\mathbf{B}_1$ | = | $\frac{1}{4}\,\mathbf{a}_1 + \frac{1}{4}\,\mathbf{a}_2 + z_1\,\mathbf{a}_3$ | = | $\frac{1}{4}a\,\mathbf{\hat{x}} + \frac{1}{4}b\,\mathbf{\hat{y}} + z_1c\,\mathbf{\hat{z}}$ | (2$a$) | N I |
| $\mathbf{B}_2$ | = | $\frac{3}{4}\,\mathbf{a}_1 + \frac{3}{4}\,\mathbf{a}_2 - z_1\,\mathbf{a}_3$ | = | $\frac{3}{4}a\,\mathbf{\hat{x}} + \frac{3}{4}b\,\mathbf{\hat{y}} - z_1c\,\mathbf{\hat{z}}$ | (2$a$) | N I |
| $\mathbf{B}_3$ | = | $\frac{1}{4}\,\mathbf{a}_1 + \frac{1}{4}\,\mathbf{a}_2 + z_2\,\mathbf{a}_3$ | = | $\frac{1}{4}a\,\mathbf{\hat{x}} + \frac{1}{4}b\,\mathbf{\hat{y}} + z_2c\,\mathbf{\hat{z}}$ | (2$a$) | O I |
| $\mathbf{B}_4$ | = | $\frac{3}{4}\,\mathbf{a}_1 + \frac{3}{4}\,\mathbf{a}_2 - z_2\,\mathbf{a}_3$ | = | $\frac{3}{4}a\,\mathbf{\hat{x}} + \frac{3}{4}b\,\mathbf{\hat{y}} - z_2c\,\mathbf{\hat{z}}$ | (2$a$) | O I |
| $\mathbf{B}_5$ | = | $\frac{1}{4}\,\mathbf{a}_1 + \frac{3}{4}\,\mathbf{a}_2 + z_3\,\mathbf{a}_3$ | = | $\frac{1}{4}a\,\mathbf{\hat{x}} + \frac{3}{4}b\,\mathbf{\hat{y}} + z_3c\,\mathbf{\hat{z}}$ | (2$b$) | N II |
| $\mathbf{B}_6$ | = | $\frac{3}{4}\,\mathbf{a}_1 + \frac{1}{4}\,\mathbf{a}_2 - z_3\,\mathbf{a}_3$ | = | $\frac{3}{4}a\,\mathbf{\hat{x}} + \frac{1}{4}b\,\mathbf{\hat{y}} - z_3c\,\mathbf{\hat{z}}$ | (2$b$) | N II |
| $\mathbf{B}_7$ | = | $\frac{1}{4}\,\mathbf{a}_1 + y_4\,\mathbf{a}_2 + z_4\,\mathbf{a}_3$ | = | $\frac{1}{4}a\,\mathbf{\hat{x}} + y_4b\,\mathbf{\hat{y}} + z_4c\,\mathbf{\hat{z}}$ | (4$e$) | H I |
| $\mathbf{B}_8$ | = | $\frac{1}{4}\,\mathbf{a}_1 + \left(\frac{1}{2} - y_4\right)\,\mathbf{a}_2 + z_4\,\mathbf{a}_3$ | = | $\frac{1}{4}a\,\mathbf{\hat{x}} + \left(\frac{1}{2} - y_4\right)b\,\mathbf{\hat{y}} + z_4c\,\mathbf{\hat{z}}$ | (4$e$) | H I |
| $\mathbf{B}_9$ | = | $\frac{3}{4}\,\mathbf{a}_1 + \left(\frac{1}{2} + y_4\right)\,\mathbf{a}_2 - z_4\,\mathbf{a}_3$ | = | $\frac{3}{4}a\,\mathbf{\hat{x}} + \left(\frac{1}{2} + y_4\right)b\,\mathbf{\hat{y}} - z_4c\,\mathbf{\hat{z}}$ | (4$e$) | H I |
| $\mathbf{B}_{10}$ | = | $\frac{3}{4}\,\mathbf{a}_1 - y_4\,\mathbf{a}_2 - z_4\,\mathbf{a}_3$ | = | $\frac{3}{4}a\,\mathbf{\hat{x}} - y_4b\,\mathbf{\hat{y}} - z_4c\,\mathbf{\hat{z}}$ | (4$e$) | H I |
| $\mathbf{B}_{11}$ | = | $x_5\,\mathbf{a}_1 + \frac{1}{4}\,\mathbf{a}_2 + z_5\,\mathbf{a}_3$ | = | $x_5a\,\mathbf{\hat{x}} + \frac{1}{4}b\,\mathbf{\hat{y}} + z_5c\,\mathbf{\hat{z}}$ | (4$f$) | H II |
| $\mathbf{B}_{12}$ | = | $\left(\frac{1}{2} - x_5\right)\,\mathbf{a}_1 + \frac{1}{4}\,\mathbf{a}_2 + z_5\,\mathbf{a}_3$ | = | $\left(\frac{1}{2} - x_5\right)a\,\mathbf{\hat{x}} + \frac{1}{4}b\,\mathbf{\hat{y}} + z_5c\,\mathbf{\hat{z}}$ | (4$f$) | H II |
| $\mathbf{B}_{13}$ | = | $-x_5\,\mathbf{a}_1 + \frac{3}{4}\,\mathbf{a}_2 - z_5\,\mathbf{a}_3$ | = | $-x_5a\,\mathbf{\hat{x}} + \frac{3}{4}b\,\mathbf{\hat{y}} - z_5c\,\mathbf{\hat{z}}$ | (4$f$) | H II |
| $\mathbf{B}_{14}$ | = | $\left(\frac{1}{2} + x_5\right)\,\mathbf{a}_1 + \frac{3}{4}\,\mathbf{a}_2 - z_5\,\mathbf{a}_3$ | = | $\left(\frac{1}{2} + x_5\right)a\,\mathbf{\hat{x}} + \frac{3}{4}b\,\mathbf{\hat{y}} - z_5c\,\mathbf{\hat{z}}$ | (4$f$) | H II |
| $\mathbf{B}_{15}$ | = | $x_6\,\mathbf{a}_1 + \frac{1}{4}\,\mathbf{a}_2 + z_6\,\mathbf{a}_3$ | = | $x_6a\,\mathbf{\hat{x}} + \frac{1}{4}b\,\mathbf{\hat{y}} + z_6c\,\mathbf{\hat{z}}$ | (4$f$) | O II |
| $\mathbf{B}_{16}$ | = | $\left(\frac{1}{2} - x_6\right)\,\mathbf{a}_1 + \frac{1}{4}\,\mathbf{a}_2 + z_6\,\mathbf{a}_3$ | = | $\left(\frac{1}{2} - x_6\right)a\,\mathbf{\hat{x}} + \frac{1}{4}b\,\mathbf{\hat{y}} + z_6c\,\mathbf{\hat{z}}$ | (4$f$) | O II |
| $\mathbf{B}_{17}$ | = | $-x_6\,\mathbf{a}_1 + \frac{3}{4}\,\mathbf{a}_2 - z_6\,\mathbf{a}_3$ | = | $-x_6a\,\mathbf{\hat{x}} + \frac{3}{4}b\,\mathbf{\hat{y}} - z_6c\,\mathbf{\hat{z}}$ | (4$f$) | O II |
| $\mathbf{B}_{18}$ | = | $\left(\frac{1}{2} + x_6\right)\,\mathbf{a}_1 + \frac{3}{4}\,\mathbf{a}_2 - z_6\,\mathbf{a}_3$ | = | $\left(\frac{1}{2} + x_6\right)a\,\mathbf{\hat{x}} + \frac{3}{4}b\,\mathbf{\hat{y}} - z_6c\,\mathbf{\hat{z}}$ | (4$f$) | O II |

**Geometry files:**

- CIF: pp. 1621
- POSCAR: pp. 1621

# Shcherbinaite ($V_2O_5$) (*Revised*) Structure: A5B2_oP14_59_a2f_f

http://aflow.org/prototype-encyclopedia/A5B2_oP14_59_a2f_f

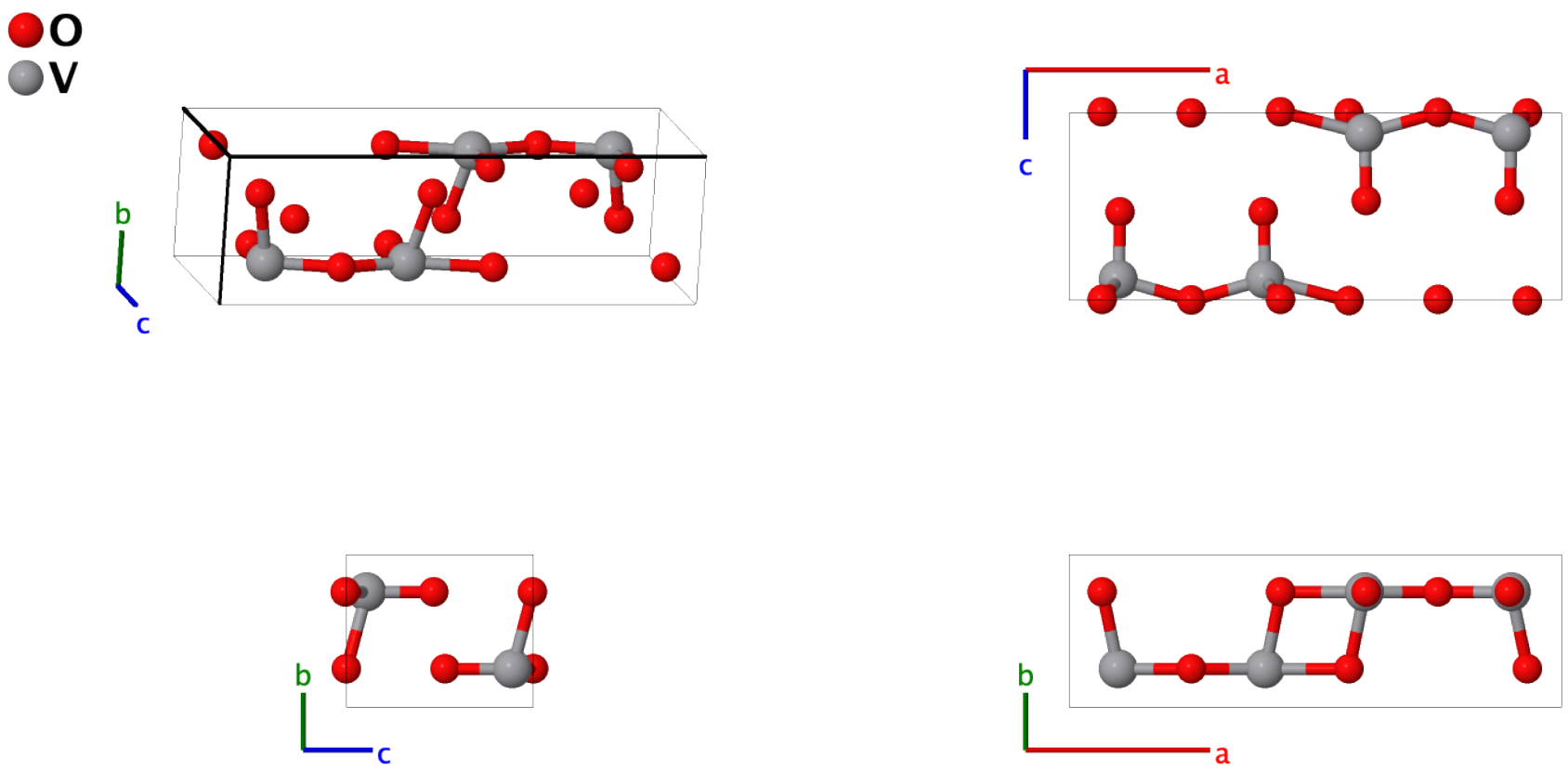

| | | |
|---|---|---|
| **Prototype** | : | $O_5V_2$ |
| **AFLOW prototype label** | : | A5B2_oP14_59_a2f_f |
| ***Strukturbericht* designation** | : | None |
| **Pearson symbol** | : | oP14 |
| **Space group number** | : | 59 |
| **Space group symbol** | : | *Pmmn* |
| **AFLOW prototype command** | : | `aflow --proto=A5B2_oP14_59_a2f_f --params=`$a, b/a, c/a, z_1, x_2, z_2, x_3, z_3, x_4, z_4$ |

- An earlier version of this structure found by (Ketelaar, 1936) was given the $D8_7$ *Strukturbericht* designation in (Gottfried, 1938). It was later realized that this structure had a rather unexpected arrangement of vanadium atoms, and the structure was revised by (Enjalbert, 1986) and others.

**Simple Orthorhombic primitive vectors:**

$$\mathbf{a}_1 = a\,\mathbf{\hat{x}}$$
$$\mathbf{a}_2 = b\,\mathbf{\hat{y}}$$
$$\mathbf{a}_3 = c\,\mathbf{\hat{z}}$$

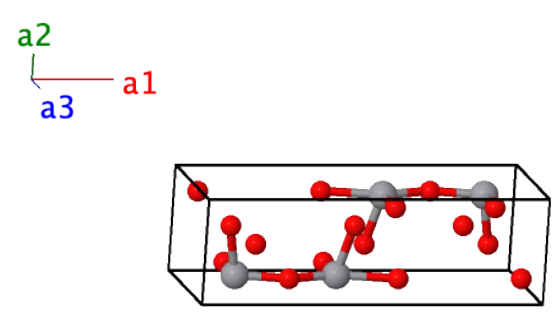

**Basis vectors:**

| | | Lattice Coordinates | | Cartesian Coordinates | Wyckoff Position | Atom Type |
|---|---|---|---|---|---|---|
| $\mathbf{B}_1$ | = | $\frac{1}{4}\,\mathbf{a}_1 + \frac{1}{4}\,\mathbf{a}_2 + z_1\,\mathbf{a}_3$ | = | $\frac{1}{4}a\,\mathbf{\hat{x}} + \frac{1}{4}b\,\mathbf{\hat{y}} + z_1c\,\mathbf{\hat{z}}$ | (2$a$) | O I |
| $\mathbf{B}_2$ | = | $\frac{3}{4}\,\mathbf{a}_1 + \frac{3}{4}\,\mathbf{a}_2 - z_1\,\mathbf{a}_3$ | = | $\frac{3}{4}a\,\mathbf{\hat{x}} + \frac{3}{4}b\,\mathbf{\hat{y}} - z_1c\,\mathbf{\hat{z}}$ | (2$a$) | O I |
| $\mathbf{B}_3$ | = | $x_2\,\mathbf{a}_1 + \frac{1}{4}\,\mathbf{a}_2 + z_2\,\mathbf{a}_3$ | = | $x_2a\,\mathbf{\hat{x}} + \frac{1}{4}b\,\mathbf{\hat{y}} + z_2c\,\mathbf{\hat{z}}$ | (4$f$) | O II |
| $\mathbf{B}_4$ | = | $\left(\frac{1}{2} - x_2\right)\,\mathbf{a}_1 + \frac{1}{4}\,\mathbf{a}_2 + z_2\,\mathbf{a}_3$ | = | $\left(\frac{1}{2} - x_2\right)a\,\mathbf{\hat{x}} + \frac{1}{4}b\,\mathbf{\hat{y}} + z_2c\,\mathbf{\hat{z}}$ | (4$f$) | O II |
| $\mathbf{B}_5$ | = | $-x_2\,\mathbf{a}_1 + \frac{3}{4}\,\mathbf{a}_2 - z_2\,\mathbf{a}_3$ | = | $-x_2a\,\mathbf{\hat{x}} + \frac{3}{4}b\,\mathbf{\hat{y}} - z_2c\,\mathbf{\hat{z}}$ | (4$f$) | O II |
| $\mathbf{B}_6$ | = | $\left(\frac{1}{2}+x_2\right)\mathbf{a}_1+\frac{3}{4}\mathbf{a}_2-z_2\mathbf{a}_3$ | = | $\left(\frac{1}{2}+x_2\right)a\,\hat{\mathbf{x}}+\frac{3}{4}b\,\hat{\mathbf{y}}-z_2c\,\hat{\mathbf{z}}$ | (4$f$) | O II |
| $\mathbf{B}_7$ | = | $x_3\,\mathbf{a}_1+\frac{1}{4}\mathbf{a}_2+z_3\,\mathbf{a}_3$ | = | $x_3a\,\hat{\mathbf{x}}+\frac{1}{4}b\,\hat{\mathbf{y}}+z_3c\,\hat{\mathbf{z}}$ | (4$f$) | O III |
| $\mathbf{B}_8$ | = | $\left(\frac{1}{2}-x_3\right)\mathbf{a}_1+\frac{1}{4}\mathbf{a}_2+z_3\,\mathbf{a}_3$ | = | $\left(\frac{1}{2}-x_3\right)a\,\hat{\mathbf{x}}+\frac{1}{4}b\,\hat{\mathbf{y}}+z_3c\,\hat{\mathbf{z}}$ | (4$f$) | O III |
| $\mathbf{B}_9$ | = | $-x_3\,\mathbf{a}_1+\frac{3}{4}\mathbf{a}_2-z_3\,\mathbf{a}_3$ | = | $-x_3a\,\hat{\mathbf{x}}+\frac{3}{4}b\,\hat{\mathbf{y}}-z_3c\,\hat{\mathbf{z}}$ | (4$f$) | O III |
| $\mathbf{B}_{10}$ | = | $\left(\frac{1}{2}+x_3\right)\mathbf{a}_1+\frac{3}{4}\mathbf{a}_2-z_3\,\mathbf{a}_3$ | = | $\left(\frac{1}{2}+x_3\right)a\,\hat{\mathbf{x}}+\frac{3}{4}b\,\hat{\mathbf{y}}-z_3c\,\hat{\mathbf{z}}$ | (4$f$) | O III |
| $\mathbf{B}_{11}$ | = | $x_4\,\mathbf{a}_1+\frac{1}{4}\mathbf{a}_2+z_4\,\mathbf{a}_3$ | = | $x_4a\,\hat{\mathbf{x}}+\frac{1}{4}b\,\hat{\mathbf{y}}+z_4c\,\hat{\mathbf{z}}$ | (4$f$) | V |
| $\mathbf{B}_{12}$ | = | $\left(\frac{1}{2}-x_4\right)\mathbf{a}_1+\frac{1}{4}\mathbf{a}_2+z_4\,\mathbf{a}_3$ | = | $\left(\frac{1}{2}-x_4\right)a\,\hat{\mathbf{x}}+\frac{1}{4}b\,\hat{\mathbf{y}}+z_4c\,\hat{\mathbf{z}}$ | (4$f$) | V |
| $\mathbf{B}_{13}$ | = | $-x_4\,\mathbf{a}_1+\frac{3}{4}\mathbf{a}_2-z_4\,\mathbf{a}_3$ | = | $-x_4a\,\hat{\mathbf{x}}+\frac{3}{4}b\,\hat{\mathbf{y}}-z_4c\,\hat{\mathbf{z}}$ | (4$f$) | V |
| $\mathbf{B}_{14}$ | = | $\left(\frac{1}{2}+x_4\right)\mathbf{a}_1+\frac{3}{4}\mathbf{a}_2-z_4\,\mathbf{a}_3$ | = | $\left(\frac{1}{2}+x_4\right)a\,\hat{\mathbf{x}}+\frac{3}{4}b\,\hat{\mathbf{y}}-z_4c\,\hat{\mathbf{z}}$ | (4$f$) | V |

**Geometry files:**

- CIF: pp. 1621
- POSCAR: pp. 1622

# $CaB_2O_4$ I ($E3_2$) Structure: A2BC4_oP28_60_d_c_2d

http://aflow.org/prototype-encyclopedia/A2BC4_oP28_60_d_c_2d

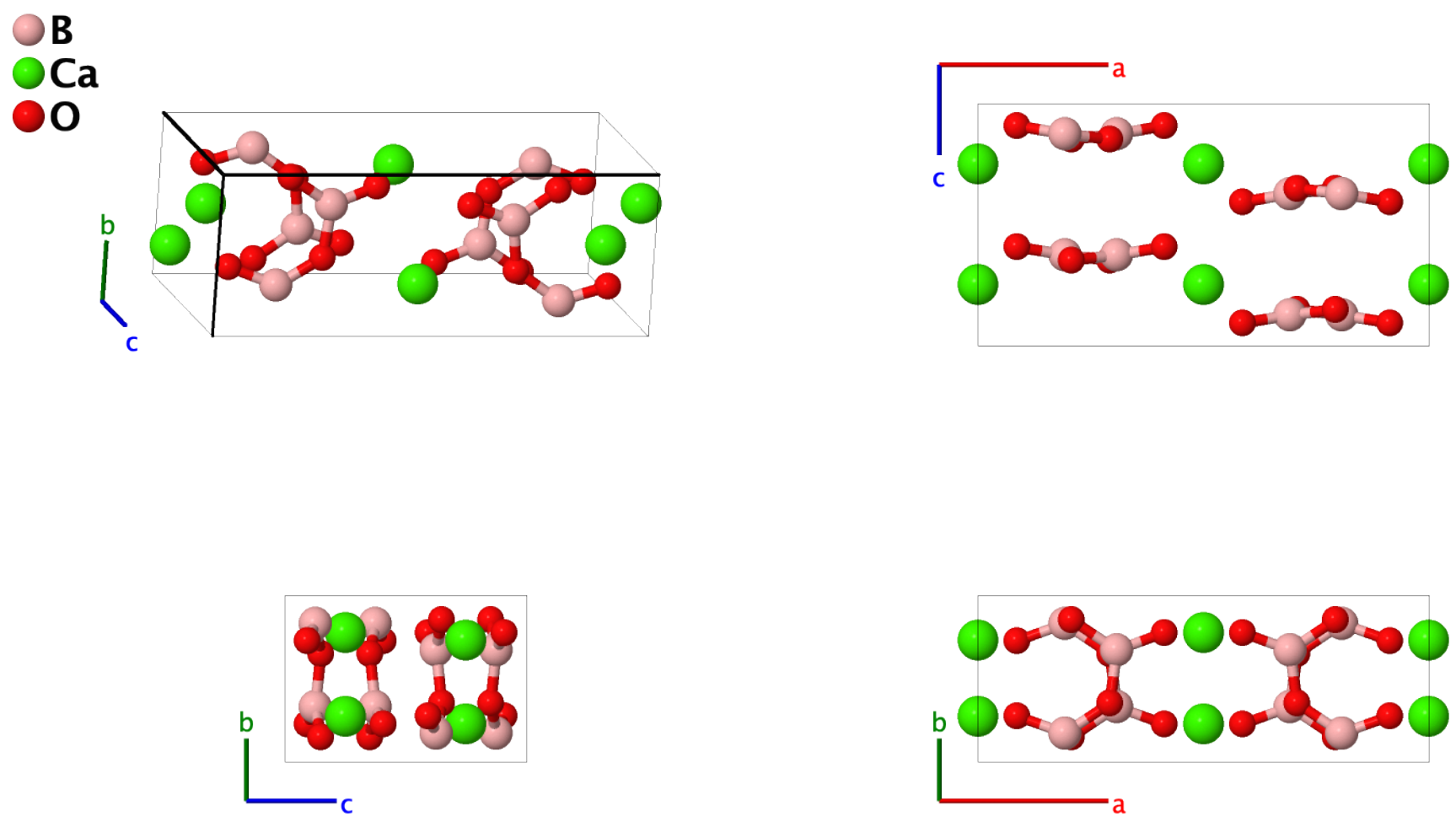

| | | |
|---|---|---|
| **Prototype** | : | $B_2CaO_4$ |
| **AFLOW prototype label** | : | A2BC4_oP28_60_d_c_2d |
| ***Strukturbericht* designation** | : | $E3_2$ |
| **Pearson symbol** | : | oP28 |
| **Space group number** | : | 60 |
| **Space group symbol** | : | *Pbcn* |
| **AFLOW prototype command** | : | `aflow --proto=A2BC4_oP28_60_d_c_2d`<br>`--params=`$a, b/a, c/a, y_1, x_2, y_2, z_2, x_3, y_3, z_3, x_4, y_4, z_4$ |

- $CaB_2O_4$ exists in at least four phases (Marezio, 1969):
- I - The ground state, stable below 1.2 GPa, *Strukturbericht* $E3_2$ (this structure).
- II – Orthorhombic high pressure structure, stable between 1.2 and 1.5 GPa, presumably calciborite.
- III – Orthorhombic high pressure structure, stable between 1.5 and 2.5 GPa.
- IV – Cubic high pressure structure, stable between 2.5 and 4.0 GPa.
- (Marezio, 1963) gives the crystal structure in the *Pnca* setting of space group #60. We have used FINDSYM to transform this to the standard *Pbcn* setting.

**Simple Orthorhombic primitive vectors:**

$$\mathbf{a}_1 = a\,\hat{\mathbf{x}}$$
$$\mathbf{a}_2 = b\,\hat{\mathbf{y}}$$
$$\mathbf{a}_3 = c\,\hat{\mathbf{z}}$$

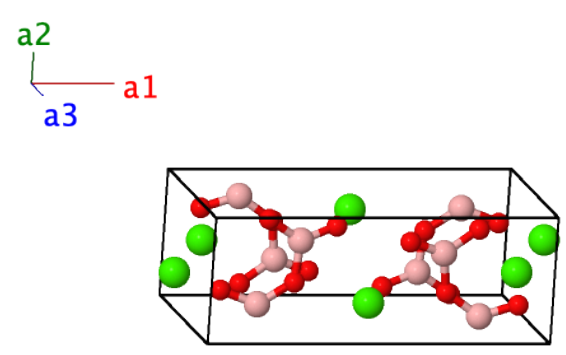

**Basis vectors:**

| | | Lattice Coordinates | | Cartesian Coordinates | Wyckoff Position | Atom Type |
|---|---|---|---|---|---|---|
| $\mathbf{B}_{1}$ | = | $y_{1}\,\mathbf{a}_{2}+\frac{1}{4}\,\mathbf{a}_{3}$ | = | $y_{1}b\,\hat{\mathbf{y}}+\frac{1}{4}c\,\hat{\mathbf{z}}$ | (4$c$) | Ca |
| $\mathbf{B}_{2}$ | = | $\frac{1}{2}\,\mathbf{a}_{1}+\left(\frac{1}{2}-y_{1}\right)\,\mathbf{a}_{2}+\frac{3}{4}\,\mathbf{a}_{3}$ | = | $\frac{1}{2}a\,\hat{\mathbf{x}}+\left(\frac{1}{2}-y_{1}\right)b\,\hat{\mathbf{y}}+\frac{3}{4}c\,\hat{\mathbf{z}}$ | (4$c$) | Ca |
| $\mathbf{B}_{3}$ | = | $-y_{1}\,\mathbf{a}_{2}+\frac{3}{4}\,\mathbf{a}_{3}$ | = | $-y_{1}b\,\hat{\mathbf{y}}+\frac{3}{4}c\,\hat{\mathbf{z}}$ | (4$c$) | Ca |
| $\mathbf{B}_{4}$ | = | $\frac{1}{2}\,\mathbf{a}_{1}+\left(\frac{1}{2}+y_{1}\right)\,\mathbf{a}_{2}+\frac{1}{4}\,\mathbf{a}_{3}$ | = | $\frac{1}{2}a\,\hat{\mathbf{x}}+\left(\frac{1}{2}+y_{1}\right)b\,\hat{\mathbf{y}}+\frac{1}{4}c\,\hat{\mathbf{z}}$ | (4$c$) | Ca |
| $\mathbf{B}_{5}$ | = | $x_{2}\,\mathbf{a}_{1}+y_{2}\,\mathbf{a}_{2}+z_{2}\,\mathbf{a}_{3}$ | = | $x_{2}a\,\hat{\mathbf{x}}+y_{2}b\,\hat{\mathbf{y}}+z_{2}c\,\hat{\mathbf{z}}$ | (8$d$) | B |
| $\mathbf{B}_{6}$ | = | $\left(\frac{1}{2}-x_{2}\right)\,\mathbf{a}_{1}+\left(\frac{1}{2}-y_{2}\right)\,\mathbf{a}_{2}+\left(\frac{1}{2}+z_{2}\right)\,\mathbf{a}_{3}$ | = | $\left(\frac{1}{2}-x_{2}\right)a\,\hat{\mathbf{x}}+\left(\frac{1}{2}-y_{2}\right)b\,\hat{\mathbf{y}}+\left(\frac{1}{2}+z_{2}\right)c\,\hat{\mathbf{z}}$ | (8$d$) | B |
| $\mathbf{B}_{7}$ | = | $-x_{2}\,\mathbf{a}_{1}+y_{2}\,\mathbf{a}_{2}+\left(\frac{1}{2}-z_{2}\right)\,\mathbf{a}_{3}$ | = | $-x_{2}a\,\hat{\mathbf{x}}+y_{2}b\,\hat{\mathbf{y}}+\left(\frac{1}{2}-z_{2}\right)c\,\hat{\mathbf{z}}$ | (8$d$) | B |
| $\mathbf{B}_{8}$ | = | $\left(\frac{1}{2}+x_{2}\right)\,\mathbf{a}_{1}+\left(\frac{1}{2}-y_{2}\right)\,\mathbf{a}_{2}-z_{2}\,\mathbf{a}_{3}$ | = | $\left(\frac{1}{2}+x_{2}\right)a\,\hat{\mathbf{x}}+\left(\frac{1}{2}-y_{2}\right)b\,\hat{\mathbf{y}}-z_{2}c\,\hat{\mathbf{z}}$ | (8$d$) | B |
| $\mathbf{B}_{9}$ | = | $-x_{2}\,\mathbf{a}_{1}-y_{2}\,\mathbf{a}_{2}-z_{2}\,\mathbf{a}_{3}$ | = | $-x_{2}a\,\hat{\mathbf{x}}-y_{2}b\,\hat{\mathbf{y}}-z_{2}c\,\hat{\mathbf{z}}$ | (8$d$) | B |
| $\mathbf{B}_{10}$ | = | $\left(\frac{1}{2}+x_{2}\right)\,\mathbf{a}_{1}+\left(\frac{1}{2}+y_{2}\right)\,\mathbf{a}_{2}+\left(\frac{1}{2}-z_{2}\right)\,\mathbf{a}_{3}$ | = | $\left(\frac{1}{2}+x_{2}\right)a\,\hat{\mathbf{x}}+\left(\frac{1}{2}+y_{2}\right)b\,\hat{\mathbf{y}}+\left(\frac{1}{2}-z_{2}\right)c\,\hat{\mathbf{z}}$ | (8$d$) | B |
| $\mathbf{B}_{11}$ | = | $x_{2}\,\mathbf{a}_{1}-y_{2}\,\mathbf{a}_{2}+\left(\frac{1}{2}+z_{2}\right)\,\mathbf{a}_{3}$ | = | $x_{2}a\,\hat{\mathbf{x}}-y_{2}b\,\hat{\mathbf{y}}+\left(\frac{1}{2}+z_{2}\right)c\,\hat{\mathbf{z}}$ | (8$d$) | B |
| $\mathbf{B}_{12}$ | = | $\left(\frac{1}{2}-x_{2}\right)\,\mathbf{a}_{1}+\left(\frac{1}{2}+y_{2}\right)\,\mathbf{a}_{2}+z_{2}\,\mathbf{a}_{3}$ | = | $\left(\frac{1}{2}-x_{2}\right)a\,\hat{\mathbf{x}}+\left(\frac{1}{2}+y_{2}\right)b\,\hat{\mathbf{y}}+z_{2}c\,\hat{\mathbf{z}}$ | (8$d$) | B |
| $\mathbf{B}_{13}$ | = | $x_{3}\,\mathbf{a}_{1}+y_{3}\,\mathbf{a}_{2}+z_{3}\,\mathbf{a}_{3}$ | = | $x_{3}a\,\hat{\mathbf{x}}+y_{3}b\,\hat{\mathbf{y}}+z_{3}c\,\hat{\mathbf{z}}$ | (8$d$) | O I |
| $\mathbf{B}_{14}$ | = | $\left(\frac{1}{2}-x_{3}\right)\,\mathbf{a}_{1}+\left(\frac{1}{2}-y_{3}\right)\,\mathbf{a}_{2}+\left(\frac{1}{2}+z_{3}\right)\,\mathbf{a}_{3}$ | = | $\left(\frac{1}{2}-x_{3}\right)a\,\hat{\mathbf{x}}+\left(\frac{1}{2}-y_{3}\right)b\,\hat{\mathbf{y}}+\left(\frac{1}{2}+z_{3}\right)c\,\hat{\mathbf{z}}$ | (8$d$) | O I |
| $\mathbf{B}_{15}$ | = | $-x_{3}\,\mathbf{a}_{1}+y_{3}\,\mathbf{a}_{2}+\left(\frac{1}{2}-z_{3}\right)\,\mathbf{a}_{3}$ | = | $-x_{3}a\,\hat{\mathbf{x}}+y_{3}b\,\hat{\mathbf{y}}+\left(\frac{1}{2}-z_{3}\right)c\,\hat{\mathbf{z}}$ | (8$d$) | O I |
| $\mathbf{B}_{16}$ | = | $\left(\frac{1}{2}+x_{3}\right)\,\mathbf{a}_{1}+\left(\frac{1}{2}-y_{3}\right)\,\mathbf{a}_{2}-z_{3}\,\mathbf{a}_{3}$ | = | $\left(\frac{1}{2}+x_{3}\right)a\,\hat{\mathbf{x}}+\left(\frac{1}{2}-y_{3}\right)b\,\hat{\mathbf{y}}-z_{3}c\,\hat{\mathbf{z}}$ | (8$d$) | O I |
| $\mathbf{B}_{17}$ | = | $-x_{3}\,\mathbf{a}_{1}-y_{3}\,\mathbf{a}_{2}-z_{3}\,\mathbf{a}_{3}$ | = | $-x_{3}a\,\hat{\mathbf{x}}-y_{3}b\,\hat{\mathbf{y}}-z_{3}c\,\hat{\mathbf{z}}$ | (8$d$) | O I |
| $\mathbf{B}_{18}$ | = | $\left(\frac{1}{2}+x_{3}\right)\,\mathbf{a}_{1}+\left(\frac{1}{2}+y_{3}\right)\,\mathbf{a}_{2}+\left(\frac{1}{2}-z_{3}\right)\,\mathbf{a}_{3}$ | = | $\left(\frac{1}{2}+x_{3}\right)a\,\hat{\mathbf{x}}+\left(\frac{1}{2}+y_{3}\right)b\,\hat{\mathbf{y}}+\left(\frac{1}{2}-z_{3}\right)c\,\hat{\mathbf{z}}$ | (8$d$) | O I |
| $\mathbf{B}_{19}$ | = | $x_{3}\,\mathbf{a}_{1}-y_{3}\,\mathbf{a}_{2}+\left(\frac{1}{2}+z_{3}\right)\,\mathbf{a}_{3}$ | = | $x_{3}a\,\hat{\mathbf{x}}-y_{3}b\,\hat{\mathbf{y}}+\left(\frac{1}{2}+z_{3}\right)c\,\hat{\mathbf{z}}$ | (8$d$) | O I |
| $\mathbf{B}_{20}$ | = | $\left(\frac{1}{2}-x_{3}\right)\,\mathbf{a}_{1}+\left(\frac{1}{2}+y_{3}\right)\,\mathbf{a}_{2}+z_{3}\,\mathbf{a}_{3}$ | = | $\left(\frac{1}{2}-x_{3}\right)a\,\hat{\mathbf{x}}+\left(\frac{1}{2}+y_{3}\right)b\,\hat{\mathbf{y}}+z_{3}c\,\hat{\mathbf{z}}$ | (8$d$) | O I |
| $\mathbf{B}_{21}$ | = | $x_{4}\,\mathbf{a}_{1}+y_{4}\,\mathbf{a}_{2}+z_{4}\,\mathbf{a}_{3}$ | = | $x_{4}a\,\hat{\mathbf{x}}+y_{4}b\,\hat{\mathbf{y}}+z_{4}c\,\hat{\mathbf{z}}$ | (8$d$) | O II |
| $\mathbf{B}_{22}$ | = | $\left(\frac{1}{2}-x_{4}\right)\,\mathbf{a}_{1}+\left(\frac{1}{2}-y_{4}\right)\,\mathbf{a}_{2}+\left(\frac{1}{2}+z_{4}\right)\,\mathbf{a}_{3}$ | = | $\left(\frac{1}{2}-x_{4}\right)a\,\hat{\mathbf{x}}+\left(\frac{1}{2}-y_{4}\right)b\,\hat{\mathbf{y}}+\left(\frac{1}{2}+z_{4}\right)c\,\hat{\mathbf{z}}$ | (8$d$) | O II |
| $\mathbf{B}_{23}$ | = | $-x_{4}\,\mathbf{a}_{1}+y_{4}\,\mathbf{a}_{2}+\left(\frac{1}{2}-z_{4}\right)\,\mathbf{a}_{3}$ | = | $-x_{4}a\,\hat{\mathbf{x}}+y_{4}b\,\hat{\mathbf{y}}+\left(\frac{1}{2}-z_{4}\right)c\,\hat{\mathbf{z}}$ | (8$d$) | O II |
| $\mathbf{B}_{24}$ | = | $\left(\frac{1}{2}+x_{4}\right)\,\mathbf{a}_{1}+\left(\frac{1}{2}-y_{4}\right)\,\mathbf{a}_{2}-z_{4}\,\mathbf{a}_{3}$ | = | $\left(\frac{1}{2}+x_{4}\right)a\,\hat{\mathbf{x}}+\left(\frac{1}{2}-y_{4}\right)b\,\hat{\mathbf{y}}-z_{4}c\,\hat{\mathbf{z}}$ | (8$d$) | O II |
| $\mathbf{B}_{25}$ | = | $-x_{4}\,\mathbf{a}_{1}-y_{4}\,\mathbf{a}_{2}-z_{4}\,\mathbf{a}_{3}$ | = | $-x_{4}a\,\hat{\mathbf{x}}-y_{4}b\,\hat{\mathbf{y}}-z_{4}c\,\hat{\mathbf{z}}$ | (8$d$) | O II |
| $\mathbf{B}_{26}$ | = | $\left(\frac{1}{2}+x_{4}\right)\,\mathbf{a}_{1}+\left(\frac{1}{2}+y_{4}\right)\,\mathbf{a}_{2}+\left(\frac{1}{2}-z_{4}\right)\,\mathbf{a}_{3}$ | = | $\left(\frac{1}{2}+x_{4}\right)a\,\hat{\mathbf{x}}+\left(\frac{1}{2}+y_{4}\right)b\,\hat{\mathbf{y}}+\left(\frac{1}{2}-z_{4}\right)c\,\hat{\mathbf{z}}$ | (8$d$) | O II |
| $\mathbf{B}_{27}$ | = | $x_{4}\,\mathbf{a}_{1}-y_{4}\,\mathbf{a}_{2}+\left(\frac{1}{2}+z_{4}\right)\,\mathbf{a}_{3}$ | = | $x_{4}a\,\hat{\mathbf{x}}-y_{4}b\,\hat{\mathbf{y}}+\left(\frac{1}{2}+z_{4}\right)c\,\hat{\mathbf{z}}$ | (8$d$) | O II |
| $\mathbf{B}_{28}$ | = | $\left(\frac{1}{2}-x_{4}\right)\,\mathbf{a}_{1}+\left(\frac{1}{2}+y_{4}\right)\,\mathbf{a}_{2}+z_{4}\,\mathbf{a}_{3}$ | = | $\left(\frac{1}{2}-x_{4}\right)a\,\hat{\mathbf{x}}+\left(\frac{1}{2}+y_{4}\right)b\,\hat{\mathbf{y}}+z_{4}c\,\hat{\mathbf{z}}$ | (8$d$) | O II |

**Found in:**

---

**Geometry files:**

- CIF: pp. 1622
- POSCAR: pp. 1622

# $\zeta$-$Fe_2N$ Structure: A2B_oP12_60_d_c

http://aflow.org/prototype-encyclopedia/A2B_oP12_60_d_c.Fe2N

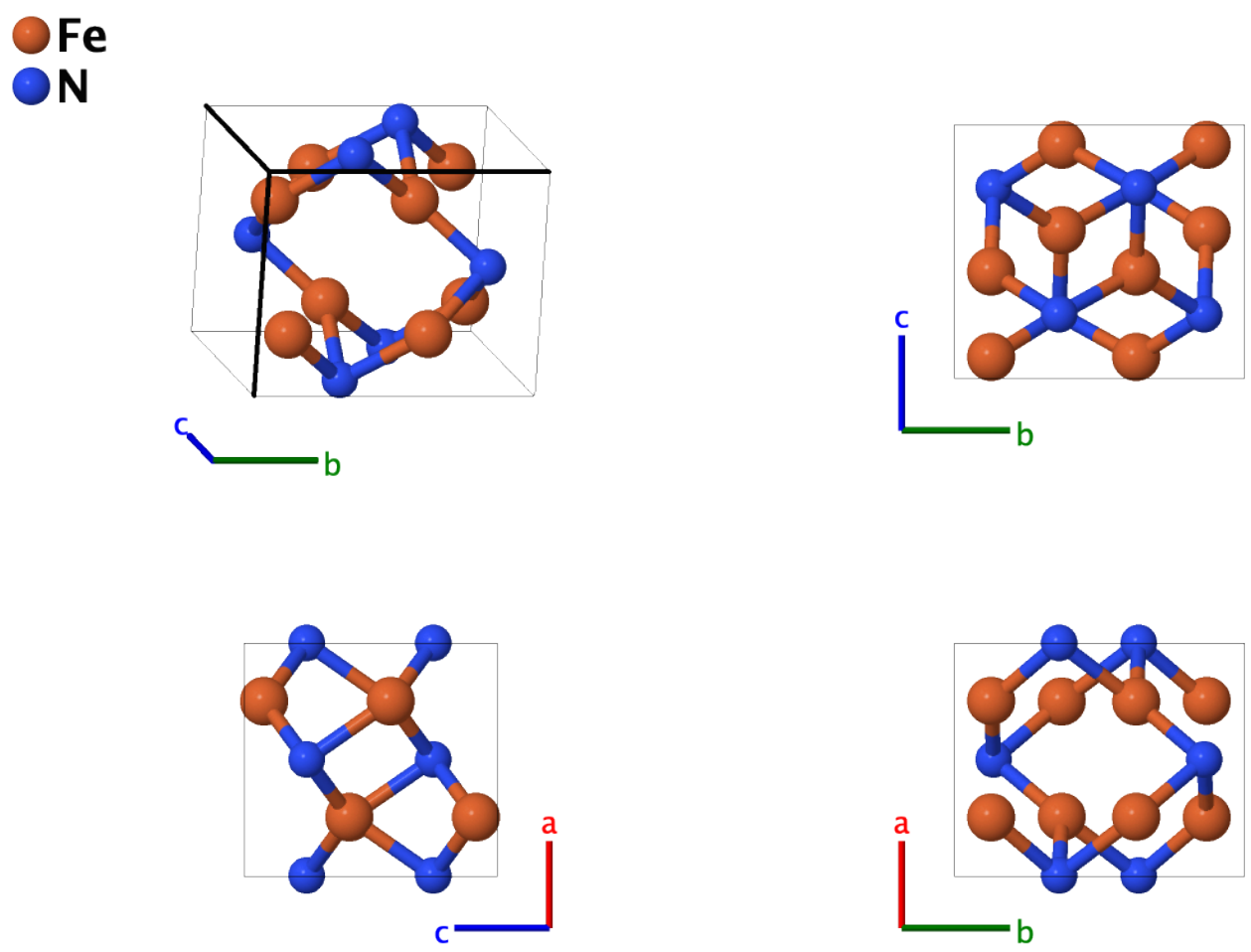

| | | |
|---|---|---|
| **Prototype** | : | $Fe_2N$ |
| **AFLOW prototype label** | : | A2B_oP12_60_d_c |
| ***Strukturbericht* designation** | : | None |
| **Pearson symbol** | : | oP12 |
| **Space group number** | : | 60 |
| **Space group symbol** | : | *Pbcn* |
| **AFLOW prototype command** | : | `aflow --proto=A2B_oP12_60_d_c` `--params=`$a, b/a, c/a, y_1, x_2, y_2, z_2$ |

- This structure has the same AFLOW label as $\alpha$-$PbO_2$, but in that case the lead site is only 49% occupied, so the composition is actually closer to $PbO_4$.
- The structures are generated by the same symmetry operations with different sets of parameters (`--params`) specified in their corresponding CIF files.

**Simple Orthorhombic primitive vectors:**

$$\mathbf{a}_1 = a\,\hat{\mathbf{x}}$$
$$\mathbf{a}_2 = b\,\hat{\mathbf{y}}$$
$$\mathbf{a}_3 = c\,\hat{\mathbf{z}}$$

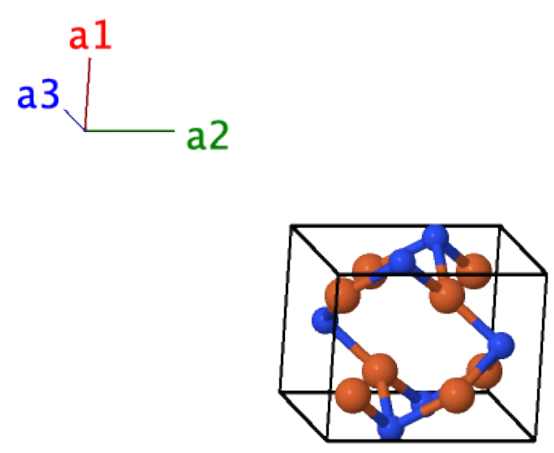

**Basis vectors:**

| | | Lattice Coordinates | | Cartesian Coordinates | Wyckoff Position | Atom Type |
|---|---|---|---|---|---|---|
| $\mathbf{B}_1$ | = | $y_1\,\mathbf{a}_2 + \frac{1}{4}\,\mathbf{a}_3$ | = | $y_1\,b\,\hat{\mathbf{y}} + \frac{1}{4}c\,\hat{\mathbf{z}}$ | (4*c*) | N |
| $\mathbf{B}_2$ | = | $\frac{1}{2}\,\mathbf{a}_1 + \left(\frac{1}{2} - y_1\right)\,\mathbf{a}_2 + \frac{3}{4}\,\mathbf{a}_3$ | = | $\frac{1}{2}a\,\hat{\mathbf{x}} + \left(\frac{1}{2} - y_1\right)\,b\,\hat{\mathbf{y}} + \frac{3}{4}c\,\hat{\mathbf{z}}$ | (4*c*) | N |
| $\mathbf{B}_3$ | = | $-y_1\,\mathbf{a}_2 + \frac{3}{4}\,\mathbf{a}_3$ | = | $-y_1\,b\,\hat{\mathbf{y}} + \frac{3}{4}c\,\hat{\mathbf{z}}$ | (4*c*) | N |
| $\mathbf{B}_4$ | = | $\frac{1}{2}\,\mathbf{a}_1 + \left(\frac{1}{2} + y_1\right)\,\mathbf{a}_2 + \frac{1}{4}\,\mathbf{a}_3$ | = | $\frac{1}{2}a\,\hat{\mathbf{x}} + \left(\frac{1}{2} + y_1\right)b\,\hat{\mathbf{y}} + \frac{1}{4}c\,\hat{\mathbf{z}}$ | (4$c$) | N |
| $\mathbf{B}_5$ | = | $x_2\,\mathbf{a}_1 + y_2\,\mathbf{a}_2 + z_2\,\mathbf{a}_3$ | = | $x_2a\,\hat{\mathbf{x}} + y_2b\,\hat{\mathbf{y}} + z_2c\,\hat{\mathbf{z}}$ | (8$d$) | Fe |
| $\mathbf{B}_6$ | = | $\left(\frac{1}{2} - x_2\right)\,\mathbf{a}_1 + \left(\frac{1}{2} - y_2\right)\,\mathbf{a}_2 + \left(\frac{1}{2} + z_2\right)\,\mathbf{a}_3$ | = | $\left(\frac{1}{2} - x_2\right)a\,\hat{\mathbf{x}} + \left(\frac{1}{2} - y_2\right)b\,\hat{\mathbf{y}} + \left(\frac{1}{2} + z_2\right)c\,\hat{\mathbf{z}}$ | (8$d$) | Fe |
| $\mathbf{B}_7$ | = | $-x_2\,\mathbf{a}_1 + y_2\,\mathbf{a}_2 + \left(\frac{1}{2} - z_2\right)\,\mathbf{a}_3$ | = | $-x_2a\,\hat{\mathbf{x}} + y_2b\,\hat{\mathbf{y}} + \left(\frac{1}{2} - z_2\right)c\,\hat{\mathbf{z}}$ | (8$d$) | Fe |
| $\mathbf{B}_8$ | = | $\left(\frac{1}{2} + x_2\right)\,\mathbf{a}_1 + \left(\frac{1}{2} - y_2\right)\,\mathbf{a}_2 - z_2\,\mathbf{a}_3$ | = | $\left(\frac{1}{2} + x_2\right)a\,\hat{\mathbf{x}} + \left(\frac{1}{2} - y_2\right)b\,\hat{\mathbf{y}} - z_2c\,\hat{\mathbf{z}}$ | (8$d$) | Fe |
| $\mathbf{B}_9$ | = | $-x_2\,\mathbf{a}_1 - y_2\,\mathbf{a}_2 - z_2\,\mathbf{a}_3$ | = | $-x_2a\,\hat{\mathbf{x}} - y_2b\,\hat{\mathbf{y}} - z_2c\,\hat{\mathbf{z}}$ | (8$d$) | Fe |
| $\mathbf{B}_{10}$ | = | $\left(\frac{1}{2} + x_2\right)\,\mathbf{a}_1 + \left(\frac{1}{2} + y_2\right)\,\mathbf{a}_2 + \left(\frac{1}{2} - z_2\right)\,\mathbf{a}_3$ | = | $\left(\frac{1}{2} + x_2\right)a\,\hat{\mathbf{x}} + \left(\frac{1}{2} + y_2\right)b\,\hat{\mathbf{y}} + \left(\frac{1}{2} - z_2\right)c\,\hat{\mathbf{z}}$ | (8$d$) | Fe |
| $\mathbf{B}_{11}$ | = | $x_2\,\mathbf{a}_1 - y_2\,\mathbf{a}_2 + \left(\frac{1}{2} + z_2\right)\,\mathbf{a}_3$ | = | $x_2a\,\hat{\mathbf{x}} - y_2b\,\hat{\mathbf{y}} + \left(\frac{1}{2} + z_2\right)c\,\hat{\mathbf{z}}$ | (8$d$) | Fe |
| $\mathbf{B}_{12}$ | = | $\left(\frac{1}{2} - x_2\right)\,\mathbf{a}_1 + \left(\frac{1}{2} + y_2\right)\,\mathbf{a}_2 + z_2\,\mathbf{a}_3$ | = | $\left(\frac{1}{2} - x_2\right)a\,\hat{\mathbf{x}} + \left(\frac{1}{2} + y_2\right)b\,\hat{\mathbf{y}} + z_2c\,\hat{\mathbf{z}}$ | (8$d$) | Fe |

**Geometry files:**

- CIF: pp. 1623
- POSCAR: pp. 1623

# $\alpha$-$PbO_2$ Structure: A2B_oP12_60_d_c

http://aflow.org/prototype-encyclopedia/A2B_oP12_60_d_c

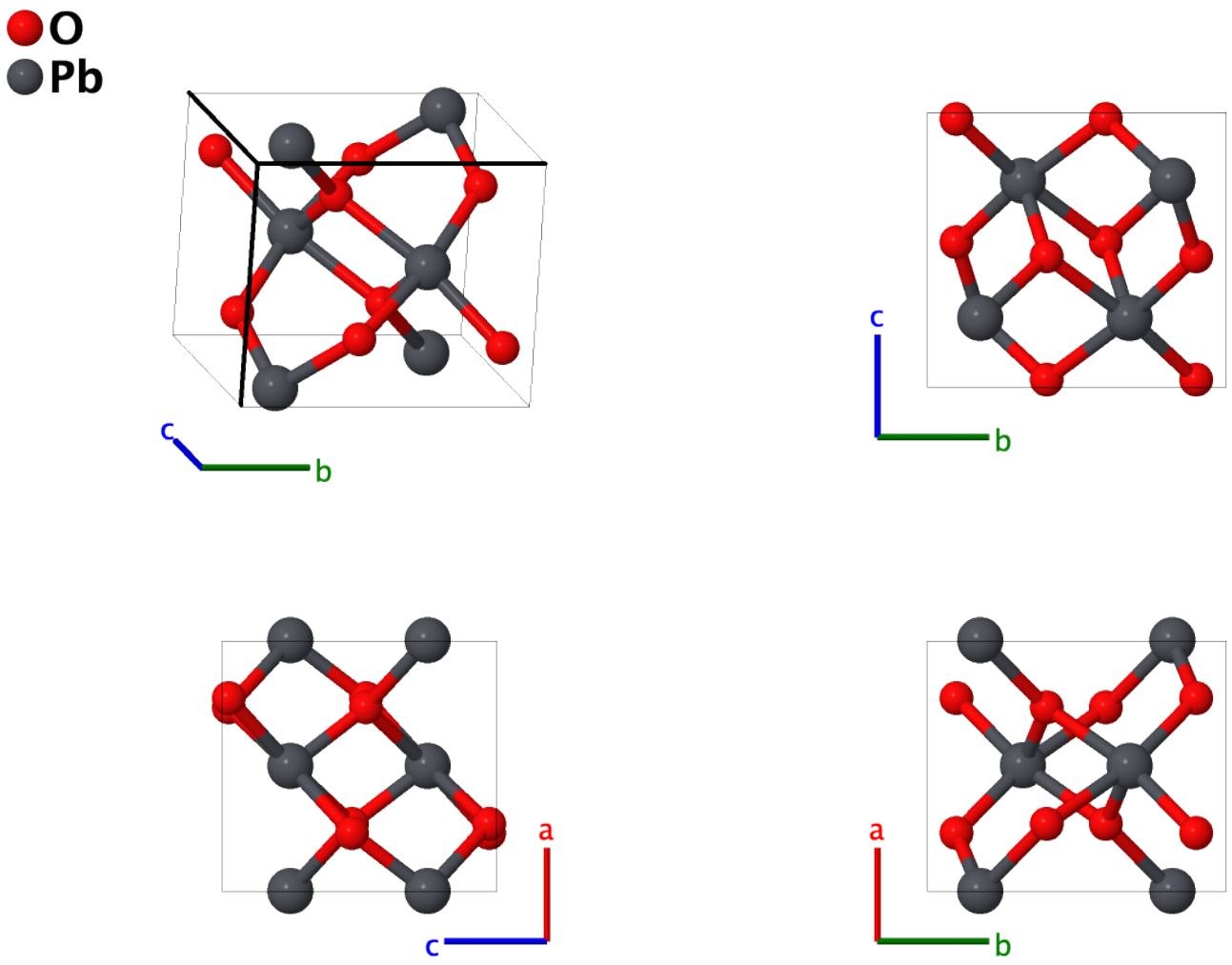

| | | |
|---|---|---|
| **Prototype** | : | $O_2Pb$ |
| **AFLOW prototype label** | : | A2B_oP12_60_d_c |
| ***Strukturbericht* designation** | : | None |
| **Pearson symbol** | : | oP12 |
| **Space group number** | : | 60 |
| **Space group symbol** | : | *Pbcn* |
| **AFLOW prototype command** | : | `aflow --proto=A2B_oP12_60_d_c`<br>`--params=`$a, b/a, c/a, y_1, x_2, y_2, z_2$ |

**Other compounds with this structure**

- (Ti,Zr)$O_2$ (srilankite)

- The experimental evidence shows that the Pb site is only 49% occupied, so stoichiometrically this compound is closer to $PbO_4$.
- This structure has the same AFLOW label as $\zeta$-$Fe_2N$, but in that case the nitrogen site is fully occupied. The structures are generated by the same symmetry operations with different sets of parameters (`--params`) specified in their corresponding CIF files.

**Simple Orthorhombic primitive vectors:**

$$\mathbf{a}_1 = a\,\hat{\mathbf{x}}$$
$$\mathbf{a}_2 = b\,\hat{\mathbf{y}}$$
$$\mathbf{a}_3 = c\,\hat{\mathbf{z}}$$

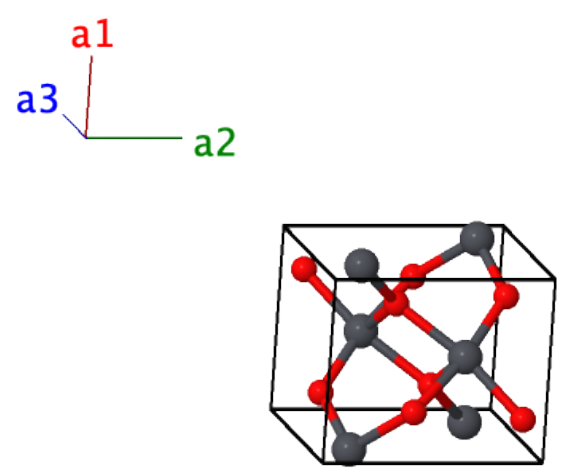

**Basis vectors:**

| | | Lattice Coordinates | | Cartesian Coordinates | Wyckoff Position | Atom Type |
|---|---|---|---|---|---|---|
| $\mathbf{B}_1$ | = | $y_1\,\mathbf{a}_2 + \frac{1}{4}\,\mathbf{a}_3$ | = | $y_1 b\,\hat{\mathbf{y}} + \frac{1}{4} c\,\hat{\mathbf{z}}$ | (4$c$) | Pb |
| $\mathbf{B}_2$ | = | $\frac{1}{2}\,\mathbf{a}_1 + \left(\frac{1}{2} - y_1\right)\,\mathbf{a}_2 + \frac{3}{4}\,\mathbf{a}_3$ | = | $\frac{1}{2} a\,\hat{\mathbf{x}} + \left(\frac{1}{2} - y_1\right) b\,\hat{\mathbf{y}} + \frac{3}{4} c\,\hat{\mathbf{z}}$ | (4$c$) | Pb |
| $\mathbf{B}_3$ | = | $-y_1\,\mathbf{a}_2 + \frac{3}{4}\,\mathbf{a}_3$ | = | $-y_1 b\,\hat{\mathbf{y}} + \frac{3}{4} c\,\hat{\mathbf{z}}$ | (4$c$) | Pb |
| $\mathbf{B}_4$ | = | $\frac{1}{2}\,\mathbf{a}_1 + \left(\frac{1}{2} + y_1\right)\,\mathbf{a}_2 + \frac{1}{4}\,\mathbf{a}_3$ | = | $\frac{1}{2} a\,\hat{\mathbf{x}} + \left(\frac{1}{2} + y_1\right) b\,\hat{\mathbf{y}} + \frac{1}{4} c\,\hat{\mathbf{z}}$ | (4$c$) | Pb |
| $\mathbf{B}_5$ | = | $x_2\,\mathbf{a}_1 + y_2\,\mathbf{a}_2 + z_2\,\mathbf{a}_3$ | = | $x_2 a\,\hat{\mathbf{x}} + y_2 b\,\hat{\mathbf{y}} + z_2 c\,\hat{\mathbf{z}}$ | (8$d$) | O |
| $\mathbf{B}_6$ | = | $\left(\frac{1}{2} - x_2\right)\,\mathbf{a}_1 + \left(\frac{1}{2} - y_2\right)\,\mathbf{a}_2 + \left(\frac{1}{2} + z_2\right)\,\mathbf{a}_3$ | = | $\left(\frac{1}{2} - x_2\right) a\,\hat{\mathbf{x}} + \left(\frac{1}{2} - y_2\right) b\,\hat{\mathbf{y}} + \left(\frac{1}{2} + z_2\right) c\,\hat{\mathbf{z}}$ | (8$d$) | O |
| $\mathbf{B}_7$ | = | $-x_2\,\mathbf{a}_1 + y_2\,\mathbf{a}_2 + \left(\frac{1}{2} - z_2\right)\,\mathbf{a}_3$ | = | $-x_2 a\,\hat{\mathbf{x}} + y_2 b\,\hat{\mathbf{y}} + \left(\frac{1}{2} - z_2\right) c\,\hat{\mathbf{z}}$ | (8$d$) | O |
| $\mathbf{B}_8$ | = | $\left(\frac{1}{2} + x_2\right)\,\mathbf{a}_1 + \left(\frac{1}{2} - y_2\right)\,\mathbf{a}_2 - z_2\,\mathbf{a}_3$ | = | $\left(\frac{1}{2} + x_2\right) a\,\hat{\mathbf{x}} + \left(\frac{1}{2} - y_2\right) b\,\hat{\mathbf{y}} - z_2 c\,\hat{\mathbf{z}}$ | (8$d$) | O |
| $\mathbf{B}_9$ | = | $-x_2\,\mathbf{a}_1 - y_2\,\mathbf{a}_2 - z_2\,\mathbf{a}_3$ | = | $-x_2 a\,\hat{\mathbf{x}} - y_2 b\,\hat{\mathbf{y}} - z_2 c\,\hat{\mathbf{z}}$ | (8$d$) | O |
| $\mathbf{B}_{10}$ | = | $\left(\frac{1}{2} + x_2\right)\,\mathbf{a}_1 + \left(\frac{1}{2} + y_2\right)\,\mathbf{a}_2 + \left(\frac{1}{2} - z_2\right)\,\mathbf{a}_3$ | = | $\left(\frac{1}{2} + x_2\right) a\,\hat{\mathbf{x}} + \left(\frac{1}{2} + y_2\right) b\,\hat{\mathbf{y}} + \left(\frac{1}{2} - z_2\right) c\,\hat{\mathbf{z}}$ | (8$d$) | O |
| $\mathbf{B}_{11}$ | = | $x_2\,\mathbf{a}_1 - y_2\,\mathbf{a}_2 + \left(\frac{1}{2} + z_2\right)\,\mathbf{a}_3$ | = | $x_2 a\,\hat{\mathbf{x}} - y_2 b\,\hat{\mathbf{y}} + \left(\frac{1}{2} + z_2\right) c\,\hat{\mathbf{z}}$ | (8$d$) | O |
| $\mathbf{B}_{12}$ | = | $\left(\frac{1}{2} - x_2\right)\,\mathbf{a}_1 + \left(\frac{1}{2} + y_2\right)\,\mathbf{a}_2 + z_2\,\mathbf{a}_3$ | = | $\left(\frac{1}{2} - x_2\right) a\,\hat{\mathbf{x}} + \left(\frac{1}{2} + y_2\right) b\,\hat{\mathbf{y}} + z_2 c\,\hat{\mathbf{z}}$ | (8$d$) | O |

---

**Found in:**

---

**Geometry files:**

- CIF: pp. 1623
- POSCAR: pp. 1623

# $Cr_5O_{12}$ Structure: A5B12_oP68_60_c2d_6d

http://aflow.org/prototype-encyclopedia/A5B12_oP68_60_c2d_6d

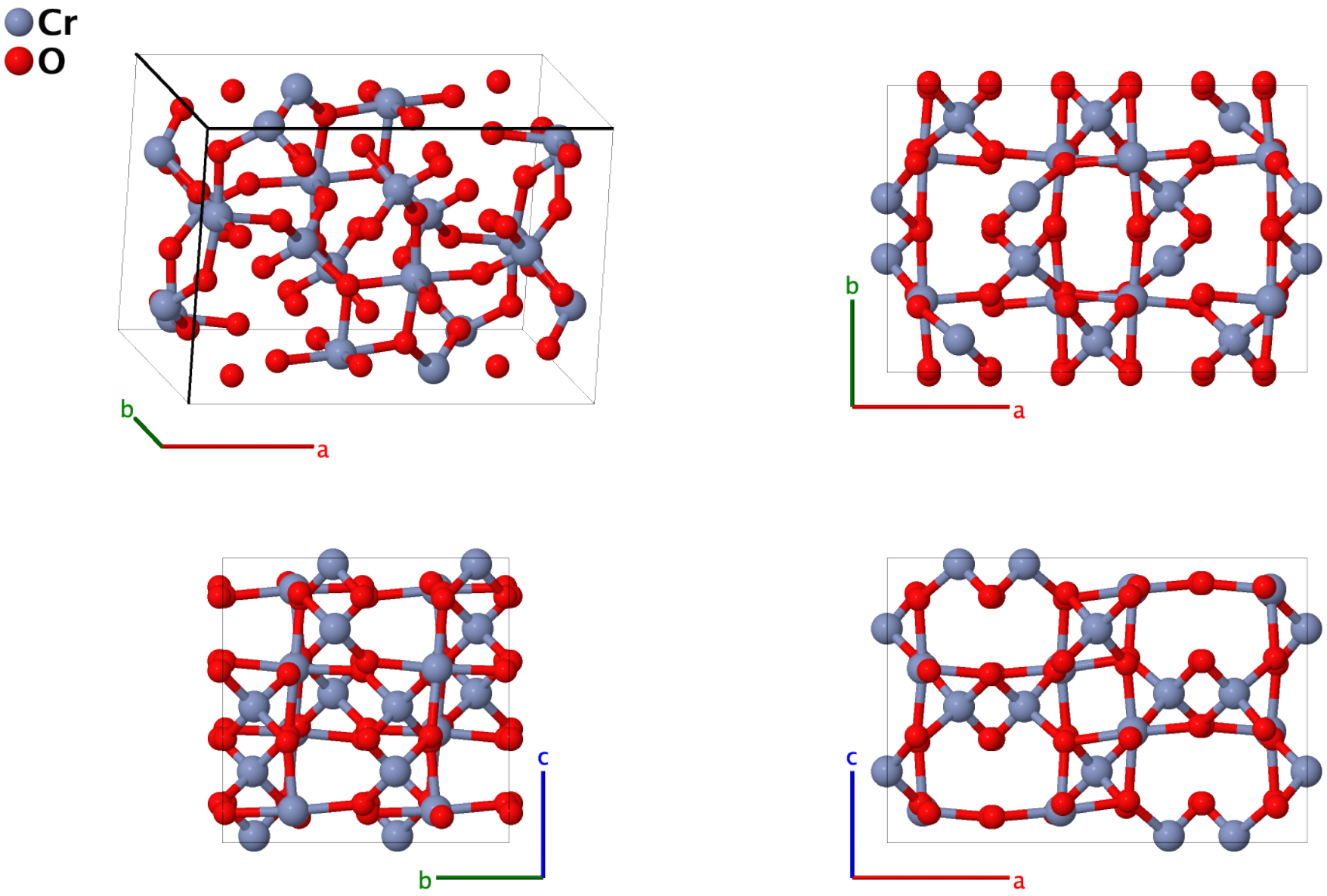

| | | |
|---|---|---|
| **Prototype** | : | $Cr_5O_{12}$ |
| **AFLOW prototype label** | : | A5B12_oP68_60_c2d_6d |
| ***Strukturbericht* designation** | : | None |
| **Pearson symbol** | : | oP68 |
| **Space group number** | : | 60 |
| **Space group symbol** | : | *Pbcn* |
| **AFLOW prototype command** | : | `aflow --proto=A5B12_oP68_60_c2d_6d`<br>`--params=`$a, b/a, c/a, y_1, x_2, y_2, z_2, x_3, y_3, z_3, x_4, y_4, z_4, x_5, y_5, z_5, x_6, y_6, z_6, x_7, y_7, z_7, x_8, y_8, z_8, x_9, y_9, z_9$ |

- (Wilhelmi, 1965) describes this as a distorted face-centered cubic lattice of oxygen atoms with the chromium atoms in interstitial sites. One can very clearly see the $CrO_4$ tetrahedra and $CrO_6$ octahedra, indicating the presence of both $Cr^{2+}$ and $Cr^{3+}$ ions.

**Simple Orthorhombic primitive vectors:**

$$\mathbf{a}_1 = a\,\hat{\mathbf{x}}$$
$$\mathbf{a}_2 = b\,\hat{\mathbf{y}}$$
$$\mathbf{a}_3 = c\,\hat{\mathbf{z}}$$

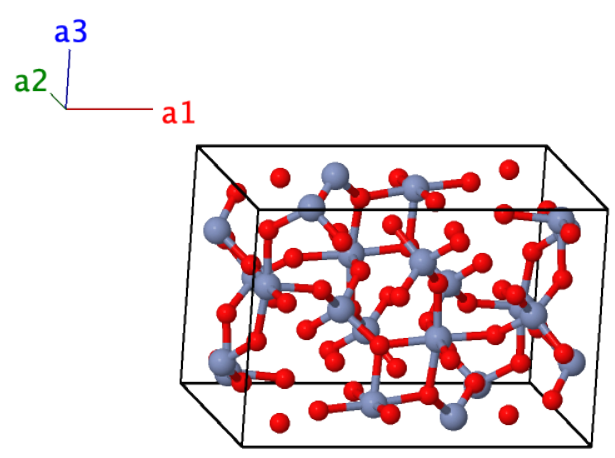

**Basis vectors:**

| | | Lattice Coordinates | | Cartesian Coordinates | Wyckoff Position | Atom Type |
|---|---|---|---|---|---|---|
| $\mathbf{B}_1$ | = | $y_1\,\mathbf{a}_2 + \frac{1}{4}\,\mathbf{a}_3$ | = | $y_1 b\,\hat{\mathbf{y}} + \frac{1}{4}c\,\hat{\mathbf{z}}$ | (4c) | Cr I |
| $\mathbf{B}_2$ | = | $\frac{1}{2}\,\mathbf{a}_1 + \left(\frac{1}{2} - y_1\right)\mathbf{a}_2 + \frac{3}{4}\,\mathbf{a}_3$ | = | $\frac{1}{2}a\,\hat{\mathbf{x}} + \left(\frac{1}{2} - y_1\right)b\,\hat{\mathbf{y}} + \frac{3}{4}c\,\hat{\mathbf{z}}$ | (4c) | Cr I |
| $\mathbf{B}_{3}$ | = | $-y_{1}\,\mathbf{a}_{2}+\frac{3}{4}\,\mathbf{a}_{3}$ | = | $-y_{1}b\,\hat{\mathbf{y}}+\frac{3}{4}c\,\hat{\mathbf{z}}$ | (4$c$) | Cr I |
| $\mathbf{B}_{4}$ | = | $\frac{1}{2}\,\mathbf{a}_{1}+\left(\frac{1}{2}+y_{1}\right)\,\mathbf{a}_{2}+\frac{1}{4}\,\mathbf{a}_{3}$ | = | $\frac{1}{2}a\,\hat{\mathbf{x}}+\left(\frac{1}{2}+y_{1}\right)b\,\hat{\mathbf{y}}+\frac{1}{4}c\,\hat{\mathbf{z}}$ | (4$c$) | Cr I |
| $\mathbf{B}_{5}$ | = | $x_{2}\,\mathbf{a}_{1}+y_{2}\,\mathbf{a}_{2}+z_{2}\,\mathbf{a}_{3}$ | = | $x_{2}a\,\hat{\mathbf{x}}+y_{2}b\,\hat{\mathbf{y}}+z_{2}c\,\hat{\mathbf{z}}$ | (8$d$) | Cr II |
| $\mathbf{B}_{6}$ | = | $\left(\frac{1}{2}-x_{2}\right)\,\mathbf{a}_{1}+\left(\frac{1}{2}-y_{2}\right)\,\mathbf{a}_{2}+\left(\frac{1}{2}+z_{2}\right)\,\mathbf{a}_{3}$ | = | $\left(\frac{1}{2}-x_{2}\right)a\,\hat{\mathbf{x}}+\left(\frac{1}{2}-y_{2}\right)b\,\hat{\mathbf{y}}+\left(\frac{1}{2}+z_{2}\right)c\,\hat{\mathbf{z}}$ | (8$d$) | Cr II |
| $\mathbf{B}_{7}$ | = | $-x_{2}\,\mathbf{a}_{1}+y_{2}\,\mathbf{a}_{2}+\left(\frac{1}{2}-z_{2}\right)\,\mathbf{a}_{3}$ | = | $-x_{2}a\,\hat{\mathbf{x}}+y_{2}b\,\hat{\mathbf{y}}+\left(\frac{1}{2}-z_{2}\right)c\,\hat{\mathbf{z}}$ | (8$d$) | Cr II |
| $\mathbf{B}_{8}$ | = | $\left(\frac{1}{2}+x_{2}\right)\,\mathbf{a}_{1}+\left(\frac{1}{2}-y_{2}\right)\,\mathbf{a}_{2}-z_{2}\,\mathbf{a}_{3}$ | = | $\left(\frac{1}{2}+x_{2}\right)a\,\hat{\mathbf{x}}+\left(\frac{1}{2}-y_{2}\right)b\,\hat{\mathbf{y}}-z_{2}c\,\hat{\mathbf{z}}$ | (8$d$) | Cr II |
| $\mathbf{B}_{9}$ | = | $-x_{2}\,\mathbf{a}_{1}-y_{2}\,\mathbf{a}_{2}-z_{2}\,\mathbf{a}_{3}$ | = | $-x_{2}a\,\hat{\mathbf{x}}-y_{2}b\,\hat{\mathbf{y}}-z_{2}c\,\hat{\mathbf{z}}$ | (8$d$) | Cr II |
| $\mathbf{B}_{10}$ | = | $\left(\frac{1}{2}+x_{2}\right)\,\mathbf{a}_{1}+\left(\frac{1}{2}+y_{2}\right)\,\mathbf{a}_{2}+\left(\frac{1}{2}-z_{2}\right)\,\mathbf{a}_{3}$ | = | $\left(\frac{1}{2}+x_{2}\right)a\,\hat{\mathbf{x}}+\left(\frac{1}{2}+y_{2}\right)b\,\hat{\mathbf{y}}+\left(\frac{1}{2}-z_{2}\right)c\,\hat{\mathbf{z}}$ | (8$d$) | Cr II |
| $\mathbf{B}_{11}$ | = | $x_{2}\,\mathbf{a}_{1}-y_{2}\,\mathbf{a}_{2}+\left(\frac{1}{2}+z_{2}\right)\,\mathbf{a}_{3}$ | = | $x_{2}a\,\hat{\mathbf{x}}-y_{2}b\,\hat{\mathbf{y}}+\left(\frac{1}{2}+z_{2}\right)c\,\hat{\mathbf{z}}$ | (8$d$) | Cr II |
| $\mathbf{B}_{12}$ | = | $\left(\frac{1}{2}-x_{2}\right)\,\mathbf{a}_{1}+\left(\frac{1}{2}+y_{2}\right)\,\mathbf{a}_{2}+z_{2}\,\mathbf{a}_{3}$ | = | $\left(\frac{1}{2}-x_{2}\right)a\,\hat{\mathbf{x}}+\left(\frac{1}{2}+y_{2}\right)b\,\hat{\mathbf{y}}+z_{2}c\,\hat{\mathbf{z}}$ | (8$d$) | Cr II |
| $\mathbf{B}_{13}$ | = | $x_{3}\,\mathbf{a}_{1}+y_{3}\,\mathbf{a}_{2}+z_{3}\,\mathbf{a}_{3}$ | = | $x_{3}a\,\hat{\mathbf{x}}+y_{3}b\,\hat{\mathbf{y}}+z_{3}c\,\hat{\mathbf{z}}$ | (8$d$) | Cr III |
| $\mathbf{B}_{14}$ | = | $\left(\frac{1}{2}-x_{3}\right)\,\mathbf{a}_{1}+\left(\frac{1}{2}-y_{3}\right)\,\mathbf{a}_{2}+\left(\frac{1}{2}+z_{3}\right)\,\mathbf{a}_{3}$ | = | $\left(\frac{1}{2}-x_{3}\right)a\,\hat{\mathbf{x}}+\left(\frac{1}{2}-y_{3}\right)b\,\hat{\mathbf{y}}+\left(\frac{1}{2}+z_{3}\right)c\,\hat{\mathbf{z}}$ | (8$d$) | Cr III |
| $\mathbf{B}_{15}$ | = | $-x_{3}\,\mathbf{a}_{1}+y_{3}\,\mathbf{a}_{2}+\left(\frac{1}{2}-z_{3}\right)\,\mathbf{a}_{3}$ | = | $-x_{3}a\,\hat{\mathbf{x}}+y_{3}b\,\hat{\mathbf{y}}+\left(\frac{1}{2}-z_{3}\right)c\,\hat{\mathbf{z}}$ | (8$d$) | Cr III |
| $\mathbf{B}_{16}$ | = | $\left(\frac{1}{2}+x_{3}\right)\,\mathbf{a}_{1}+\left(\frac{1}{2}-y_{3}\right)\,\mathbf{a}_{2}-z_{3}\,\mathbf{a}_{3}$ | = | $\left(\frac{1}{2}+x_{3}\right)a\,\hat{\mathbf{x}}+\left(\frac{1}{2}-y_{3}\right)b\,\hat{\mathbf{y}}-z_{3}c\,\hat{\mathbf{z}}$ | (8$d$) | Cr III |
| $\mathbf{B}_{17}$ | = | $-x_{3}\,\mathbf{a}_{1}-y_{3}\,\mathbf{a}_{2}-z_{3}\,\mathbf{a}_{3}$ | = | $-x_{3}a\,\hat{\mathbf{x}}-y_{3}b\,\hat{\mathbf{y}}-z_{3}c\,\hat{\mathbf{z}}$ | (8$d$) | Cr III |
| $\mathbf{B}_{18}$ | = | $\left(\frac{1}{2}+x_{3}\right)\,\mathbf{a}_{1}+\left(\frac{1}{2}+y_{3}\right)\,\mathbf{a}_{2}+\left(\frac{1}{2}-z_{3}\right)\,\mathbf{a}_{3}$ | = | $\left(\frac{1}{2}+x_{3}\right)a\,\hat{\mathbf{x}}+\left(\frac{1}{2}+y_{3}\right)b\,\hat{\mathbf{y}}+\left(\frac{1}{2}-z_{3}\right)c\,\hat{\mathbf{z}}$ | (8$d$) | Cr III |
| $\mathbf{B}_{19}$ | = | $x_{3}\,\mathbf{a}_{1}-y_{3}\,\mathbf{a}_{2}+\left(\frac{1}{2}+z_{3}\right)\,\mathbf{a}_{3}$ | = | $x_{3}a\,\hat{\mathbf{x}}-y_{3}b\,\hat{\mathbf{y}}+\left(\frac{1}{2}+z_{3}\right)c\,\hat{\mathbf{z}}$ | (8$d$) | Cr III |
| $\mathbf{B}_{20}$ | = | $\left(\frac{1}{2}-x_{3}\right)\,\mathbf{a}_{1}+\left(\frac{1}{2}+y_{3}\right)\,\mathbf{a}_{2}+z_{3}\,\mathbf{a}_{3}$ | = | $\left(\frac{1}{2}-x_{3}\right)a\,\hat{\mathbf{x}}+\left(\frac{1}{2}+y_{3}\right)b\,\hat{\mathbf{y}}+z_{3}c\,\hat{\mathbf{z}}$ | (8$d$) | Cr III |
| $\mathbf{B}_{21}$ | = | $x_{4}\,\mathbf{a}_{1}+y_{4}\,\mathbf{a}_{2}+z_{4}\,\mathbf{a}_{3}$ | = | $x_{4}a\,\hat{\mathbf{x}}+y_{4}b\,\hat{\mathbf{y}}+z_{4}c\,\hat{\mathbf{z}}$ | (8$d$) | O I |
| $\mathbf{B}_{22}$ | = | $\left(\frac{1}{2}-x_{4}\right)\,\mathbf{a}_{1}+\left(\frac{1}{2}-y_{4}\right)\,\mathbf{a}_{2}+\left(\frac{1}{2}+z_{4}\right)\,\mathbf{a}_{3}$ | = | $\left(\frac{1}{2}-x_{4}\right)a\,\hat{\mathbf{x}}+\left(\frac{1}{2}-y_{4}\right)b\,\hat{\mathbf{y}}+\left(\frac{1}{2}+z_{4}\right)c\,\hat{\mathbf{z}}$ | (8$d$) | O I |
| $\mathbf{B}_{23}$ | = | $-x_{4}\,\mathbf{a}_{1}+y_{4}\,\mathbf{a}_{2}+\left(\frac{1}{2}-z_{4}\right)\,\mathbf{a}_{3}$ | = | $-x_{4}a\,\hat{\mathbf{x}}+y_{4}b\,\hat{\mathbf{y}}+\left(\frac{1}{2}-z_{4}\right)c\,\hat{\mathbf{z}}$ | (8$d$) | O I |
| $\mathbf{B}_{24}$ | = | $\left(\frac{1}{2}+x_{4}\right)\,\mathbf{a}_{1}+\left(\frac{1}{2}-y_{4}\right)\,\mathbf{a}_{2}-z_{4}\,\mathbf{a}_{3}$ | = | $\left(\frac{1}{2}+x_{4}\right)a\,\hat{\mathbf{x}}+\left(\frac{1}{2}-y_{4}\right)b\,\hat{\mathbf{y}}-z_{4}c\,\hat{\mathbf{z}}$ | (8$d$) | O I |
| $\mathbf{B}_{25}$ | = | $-x_{4}\,\mathbf{a}_{1}-y_{4}\,\mathbf{a}_{2}-z_{4}\,\mathbf{a}_{3}$ | = | $-x_{4}a\,\hat{\mathbf{x}}-y_{4}b\,\hat{\mathbf{y}}-z_{4}c\,\hat{\mathbf{z}}$ | (8$d$) | O I |
| $\mathbf{B}_{26}$ | = | $\left(\frac{1}{2}+x_{4}\right)\,\mathbf{a}_{1}+\left(\frac{1}{2}+y_{4}\right)\,\mathbf{a}_{2}+\left(\frac{1}{2}-z_{4}\right)\,\mathbf{a}_{3}$ | = | $\left(\frac{1}{2}+x_{4}\right)a\,\hat{\mathbf{x}}+\left(\frac{1}{2}+y_{4}\right)b\,\hat{\mathbf{y}}+\left(\frac{1}{2}-z_{4}\right)c\,\hat{\mathbf{z}}$ | (8$d$) | O I |
| $\mathbf{B}_{27}$ | = | $x_{4}\,\mathbf{a}_{1}-y_{4}\,\mathbf{a}_{2}+\left(\frac{1}{2}+z_{4}\right)\,\mathbf{a}_{3}$ | = | $x_{4}a\,\hat{\mathbf{x}}-y_{4}b\,\hat{\mathbf{y}}+\left(\frac{1}{2}+z_{4}\right)c\,\hat{\mathbf{z}}$ | (8$d$) | O I |
| $\mathbf{B}_{28}$ | = | $\left(\frac{1}{2}-x_{4}\right)\,\mathbf{a}_{1}+\left(\frac{1}{2}+y_{4}\right)\,\mathbf{a}_{2}+z_{4}\,\mathbf{a}_{3}$ | = | $\left(\frac{1}{2}-x_{4}\right)a\,\hat{\mathbf{x}}+\left(\frac{1}{2}+y_{4}\right)b\,\hat{\mathbf{y}}+z_{4}c\,\hat{\mathbf{z}}$ | (8$d$) | O I |
| $\mathbf{B}_{29}$ | = | $x_{5}\,\mathbf{a}_{1}+y_{5}\,\mathbf{a}_{2}+z_{5}\,\mathbf{a}_{3}$ | = | $x_{5}a\,\hat{\mathbf{x}}+y_{5}b\,\hat{\mathbf{y}}+z_{5}c\,\hat{\mathbf{z}}$ | (8$d$) | O II |
| $\mathbf{B}_{30}$ | = | $\left(\frac{1}{2}-x_{5}\right)\,\mathbf{a}_{1}+\left(\frac{1}{2}-y_{5}\right)\,\mathbf{a}_{2}+\left(\frac{1}{2}+z_{5}\right)\,\mathbf{a}_{3}$ | = | $\left(\frac{1}{2}-x_{5}\right)a\,\hat{\mathbf{x}}+\left(\frac{1}{2}-y_{5}\right)b\,\hat{\mathbf{y}}+\left(\frac{1}{2}+z_{5}\right)c\,\hat{\mathbf{z}}$ | (8$d$) | O II |
| $\mathbf{B}_{31}$ | = | $-x_{5}\,\mathbf{a}_{1}+y_{5}\,\mathbf{a}_{2}+\left(\frac{1}{2}-z_{5}\right)\,\mathbf{a}_{3}$ | = | $-x_{5}a\,\hat{\mathbf{x}}+y_{5}b\,\hat{\mathbf{y}}+\left(\frac{1}{2}-z_{5}\right)c\,\hat{\mathbf{z}}$ | (8$d$) | O II |
| $\mathbf{B}_{32}$ | = | $\left(\frac{1}{2}+x_{5}\right)\,\mathbf{a}_{1}+\left(\frac{1}{2}-y_{5}\right)\,\mathbf{a}_{2}-z_{5}\,\mathbf{a}_{3}$ | = | $\left(\frac{1}{2}+x_{5}\right)a\,\hat{\mathbf{x}}+\left(\frac{1}{2}-y_{5}\right)b\,\hat{\mathbf{y}}-z_{5}c\,\hat{\mathbf{z}}$ | (8$d$) | O II |
| $\mathbf{B}_{33}$ | = | $-x_{5}\,\mathbf{a}_{1}-y_{5}\,\mathbf{a}_{2}-z_{5}\,\mathbf{a}_{3}$ | = | $-x_{5}a\,\hat{\mathbf{x}}-y_{5}b\,\hat{\mathbf{y}}-z_{5}c\,\hat{\mathbf{z}}$ | (8$d$) | O II |
| $\mathbf{B}_{34}$ | = | $\left(\frac{1}{2}+x_5\right)\mathbf{a}_1+\left(\frac{1}{2}+y_5\right)\mathbf{a}_2+\left(\frac{1}{2}-z_5\right)\mathbf{a}_3$ | = | $\left(\frac{1}{2}+x_5\right)a\,\hat{\mathbf{x}}+\left(\frac{1}{2}+y_5\right)b\,\hat{\mathbf{y}}+\left(\frac{1}{2}-z_5\right)c\,\hat{\mathbf{z}}$ | (8$d$) | O II |
| $\mathbf{B}_{35}$ | = | $x_5\,\mathbf{a}_1-y_5\,\mathbf{a}_2+\left(\frac{1}{2}+z_5\right)\mathbf{a}_3$ | = | $x_5a\,\hat{\mathbf{x}}-y_5b\,\hat{\mathbf{y}}+\left(\frac{1}{2}+z_5\right)c\,\hat{\mathbf{z}}$ | (8$d$) | O II |
| $\mathbf{B}_{36}$ | = | $\left(\frac{1}{2}-x_5\right)\mathbf{a}_1+\left(\frac{1}{2}+y_5\right)\mathbf{a}_2+z_5\,\mathbf{a}_3$ | = | $\left(\frac{1}{2}-x_5\right)a\,\hat{\mathbf{x}}+\left(\frac{1}{2}+y_5\right)b\,\hat{\mathbf{y}}+z_5c\,\hat{\mathbf{z}}$ | (8$d$) | O II |
| $\mathbf{B}_{37}$ | = | $x_6\,\mathbf{a}_1+y_6\,\mathbf{a}_2+z_6\,\mathbf{a}_3$ | = | $x_6a\,\hat{\mathbf{x}}+y_6b\,\hat{\mathbf{y}}+z_6c\,\hat{\mathbf{z}}$ | (8$d$) | O III |
| $\mathbf{B}_{38}$ | = | $\left(\frac{1}{2}-x_6\right)\mathbf{a}_1+\left(\frac{1}{2}-y_6\right)\mathbf{a}_2+\left(\frac{1}{2}+z_6\right)\mathbf{a}_3$ | = | $\left(\frac{1}{2}-x_6\right)a\,\hat{\mathbf{x}}+\left(\frac{1}{2}-y_6\right)b\,\hat{\mathbf{y}}+\left(\frac{1}{2}+z_6\right)c\,\hat{\mathbf{z}}$ | (8$d$) | O III |
| $\mathbf{B}_{39}$ | = | $-x_6\,\mathbf{a}_1+y_6\,\mathbf{a}_2+\left(\frac{1}{2}-z_6\right)\mathbf{a}_3$ | = | $-x_6a\,\hat{\mathbf{x}}+y_6b\,\hat{\mathbf{y}}+\left(\frac{1}{2}-z_6\right)c\,\hat{\mathbf{z}}$ | (8$d$) | O III |
| $\mathbf{B}_{40}$ | = | $\left(\frac{1}{2}+x_6\right)\mathbf{a}_1+\left(\frac{1}{2}-y_6\right)\mathbf{a}_2-z_6\,\mathbf{a}_3$ | = | $\left(\frac{1}{2}+x_6\right)a\,\hat{\mathbf{x}}+\left(\frac{1}{2}-y_6\right)b\,\hat{\mathbf{y}}-z_6c\,\hat{\mathbf{z}}$ | (8$d$) | O III |
| $\mathbf{B}_{41}$ | = | $-x_6\,\mathbf{a}_1-y_6\,\mathbf{a}_2-z_6\,\mathbf{a}_3$ | = | $-x_6a\,\hat{\mathbf{x}}-y_6b\,\hat{\mathbf{y}}-z_6c\,\hat{\mathbf{z}}$ | (8$d$) | O III |
| $\mathbf{B}_{42}$ | = | $\left(\frac{1}{2}+x_6\right)\mathbf{a}_1+\left(\frac{1}{2}+y_6\right)\mathbf{a}_2+\left(\frac{1}{2}-z_6\right)\mathbf{a}_3$ | = | $\left(\frac{1}{2}+x_6\right)a\,\hat{\mathbf{x}}+\left(\frac{1}{2}+y_6\right)b\,\hat{\mathbf{y}}+\left(\frac{1}{2}-z_6\right)c\,\hat{\mathbf{z}}$ | (8$d$) | O III |
| $\mathbf{B}_{43}$ | = | $x_6\,\mathbf{a}_1-y_6\,\mathbf{a}_2+\left(\frac{1}{2}+z_6\right)\mathbf{a}_3$ | = | $x_6a\,\hat{\mathbf{x}}-y_6b\,\hat{\mathbf{y}}+\left(\frac{1}{2}+z_6\right)c\,\hat{\mathbf{z}}$ | (8$d$) | O III |
| $\mathbf{B}_{44}$ | = | $\left(\frac{1}{2}-x_6\right)\mathbf{a}_1+\left(\frac{1}{2}+y_6\right)\mathbf{a}_2+z_6\,\mathbf{a}_3$ | = | $\left(\frac{1}{2}-x_6\right)a\,\hat{\mathbf{x}}+\left(\frac{1}{2}+y_6\right)b\,\hat{\mathbf{y}}+z_6c\,\hat{\mathbf{z}}$ | (8$d$) | O III |
| $\mathbf{B}_{45}$ | = | $x_7\,\mathbf{a}_1+y_7\,\mathbf{a}_2+z_7\,\mathbf{a}_3$ | = | $x_7a\,\hat{\mathbf{x}}+y_7b\,\hat{\mathbf{y}}+z_7c\,\hat{\mathbf{z}}$ | (8$d$) | O IV |
| $\mathbf{B}_{46}$ | = | $\left(\frac{1}{2}-x_7\right)\mathbf{a}_1+\left(\frac{1}{2}-y_7\right)\mathbf{a}_2+\left(\frac{1}{2}+z_7\right)\mathbf{a}_3$ | = | $\left(\frac{1}{2}-x_7\right)a\,\hat{\mathbf{x}}+\left(\frac{1}{2}-y_7\right)b\,\hat{\mathbf{y}}+\left(\frac{1}{2}+z_7\right)c\,\hat{\mathbf{z}}$ | (8$d$) | O IV |
| $\mathbf{B}_{47}$ | = | $-x_7\,\mathbf{a}_1+y_7\,\mathbf{a}_2+\left(\frac{1}{2}-z_7\right)\mathbf{a}_3$ | = | $-x_7a\,\hat{\mathbf{x}}+y_7b\,\hat{\mathbf{y}}+\left(\frac{1}{2}-z_7\right)c\,\hat{\mathbf{z}}$ | (8$d$) | O IV |
| $\mathbf{B}_{48}$ | = | $\left(\frac{1}{2}+x_7\right)\mathbf{a}_1+\left(\frac{1}{2}-y_7\right)\mathbf{a}_2-z_7\,\mathbf{a}_3$ | = | $\left(\frac{1}{2}+x_7\right)a\,\hat{\mathbf{x}}+\left(\frac{1}{2}-y_7\right)b\,\hat{\mathbf{y}}-z_7c\,\hat{\mathbf{z}}$ | (8$d$) | O IV |
| $\mathbf{B}_{49}$ | = | $-x_7\,\mathbf{a}_1-y_7\,\mathbf{a}_2-z_7\,\mathbf{a}_3$ | = | $-x_7a\,\hat{\mathbf{x}}-y_7b\,\hat{\mathbf{y}}-z_7c\,\hat{\mathbf{z}}$ | (8$d$) | O IV |
| $\mathbf{B}_{50}$ | = | $\left(\frac{1}{2}+x_7\right)\mathbf{a}_1+\left(\frac{1}{2}+y_7\right)\mathbf{a}_2+\left(\frac{1}{2}-z_7\right)\mathbf{a}_3$ | = | $\left(\frac{1}{2}+x_7\right)a\,\hat{\mathbf{x}}+\left(\frac{1}{2}+y_7\right)b\,\hat{\mathbf{y}}+\left(\frac{1}{2}-z_7\right)c\,\hat{\mathbf{z}}$ | (8$d$) | O IV |
| $\mathbf{B}_{51}$ | = | $x_7\,\mathbf{a}_1-y_7\,\mathbf{a}_2+\left(\frac{1}{2}+z_7\right)\mathbf{a}_3$ | = | $x_7a\,\hat{\mathbf{x}}-y_7b\,\hat{\mathbf{y}}+\left(\frac{1}{2}+z_7\right)c\,\hat{\mathbf{z}}$ | (8$d$) | O IV |
| $\mathbf{B}_{52}$ | = | $\left(\frac{1}{2}-x_7\right)\mathbf{a}_1+\left(\frac{1}{2}+y_7\right)\mathbf{a}_2+z_7\,\mathbf{a}_3$ | = | $\left(\frac{1}{2}-x_7\right)a\,\hat{\mathbf{x}}+\left(\frac{1}{2}+y_7\right)b\,\hat{\mathbf{y}}+z_7c\,\hat{\mathbf{z}}$ | (8$d$) | O IV |
| $\mathbf{B}_{53}$ | = | $x_8\,\mathbf{a}_1+y_8\,\mathbf{a}_2+z_8\,\mathbf{a}_3$ | = | $x_8a\,\hat{\mathbf{x}}+y_8b\,\hat{\mathbf{y}}+z_8c\,\hat{\mathbf{z}}$ | (8$d$) | O V |
| $\mathbf{B}_{54}$ | = | $\left(\frac{1}{2}-x_8\right)\mathbf{a}_1+\left(\frac{1}{2}-y_8\right)\mathbf{a}_2+\left(\frac{1}{2}+z_8\right)\mathbf{a}_3$ | = | $\left(\frac{1}{2}-x_8\right)a\,\hat{\mathbf{x}}+\left(\frac{1}{2}-y_8\right)b\,\hat{\mathbf{y}}+\left(\frac{1}{2}+z_8\right)c\,\hat{\mathbf{z}}$ | (8$d$) | O V |
| $\mathbf{B}_{55}$ | = | $-x_8\,\mathbf{a}_1+y_8\,\mathbf{a}_2+\left(\frac{1}{2}-z_8\right)\mathbf{a}_3$ | = | $-x_8a\,\hat{\mathbf{x}}+y_8b\,\hat{\mathbf{y}}+\left(\frac{1}{2}-z_8\right)c\,\hat{\mathbf{z}}$ | (8$d$) | O V |
| $\mathbf{B}_{56}$ | = | $\left(\frac{1}{2}+x_8\right)\mathbf{a}_1+\left(\frac{1}{2}-y_8\right)\mathbf{a}_2-z_8\,\mathbf{a}_3$ | = | $\left(\frac{1}{2}+x_8\right)a\,\hat{\mathbf{x}}+\left(\frac{1}{2}-y_8\right)b\,\hat{\mathbf{y}}-z_8c\,\hat{\mathbf{z}}$ | (8$d$) | O V |
| $\mathbf{B}_{57}$ | = | $-x_8\,\mathbf{a}_1-y_8\,\mathbf{a}_2-z_8\,\mathbf{a}_3$ | = | $-x_8a\,\hat{\mathbf{x}}-y_8b\,\hat{\mathbf{y}}-z_8c\,\hat{\mathbf{z}}$ | (8$d$) | O V |
| $\mathbf{B}_{58}$ | = | $\left(\frac{1}{2}+x_8\right)\mathbf{a}_1+\left(\frac{1}{2}+y_8\right)\mathbf{a}_2+\left(\frac{1}{2}-z_8\right)\mathbf{a}_3$ | = | $\left(\frac{1}{2}+x_8\right)a\,\hat{\mathbf{x}}+\left(\frac{1}{2}+y_8\right)b\,\hat{\mathbf{y}}+\left(\frac{1}{2}-z_8\right)c\,\hat{\mathbf{z}}$ | (8$d$) | O V |
| $\mathbf{B}_{59}$ | = | $x_8\,\mathbf{a}_1-y_8\,\mathbf{a}_2+\left(\frac{1}{2}+z_8\right)\mathbf{a}_3$ | = | $x_8a\,\hat{\mathbf{x}}-y_8b\,\hat{\mathbf{y}}+\left(\frac{1}{2}+z_8\right)c\,\hat{\mathbf{z}}$ | (8$d$) | O V |
| $\mathbf{B}_{60}$ | = | $\left(\frac{1}{2}-x_8\right)\mathbf{a}_1+\left(\frac{1}{2}+y_8\right)\mathbf{a}_2+z_8\,\mathbf{a}_3$ | = | $\left(\frac{1}{2}-x_8\right)a\,\hat{\mathbf{x}}+\left(\frac{1}{2}+y_8\right)b\,\hat{\mathbf{y}}+z_8c\,\hat{\mathbf{z}}$ | (8$d$) | O V |
| $\mathbf{B}_{61}$ | = | $x_9\,\mathbf{a}_1+y_9\,\mathbf{a}_2+z_9\,\mathbf{a}_3$ | = | $x_9a\,\hat{\mathbf{x}}+y_9b\,\hat{\mathbf{y}}+z_9c\,\hat{\mathbf{z}}$ | (8$d$) | O VI |
| $\mathbf{B}_{62}$ | = | $\left(\frac{1}{2}-x_9\right)\mathbf{a}_1+\left(\frac{1}{2}-y_9\right)\mathbf{a}_2+\left(\frac{1}{2}+z_9\right)\mathbf{a}_3$ | = | $\left(\frac{1}{2}-x_9\right)a\,\hat{\mathbf{x}}+\left(\frac{1}{2}-y_9\right)b\,\hat{\mathbf{y}}+\left(\frac{1}{2}+z_9\right)c\,\hat{\mathbf{z}}$ | (8$d$) | O VI |
| $\mathbf{B}_{63}$ | = | $-x_9\,\mathbf{a}_1+y_9\,\mathbf{a}_2+\left(\frac{1}{2}-z_9\right)\mathbf{a}_3$ | = | $-x_9a\,\hat{\mathbf{x}}+y_9b\,\hat{\mathbf{y}}+\left(\frac{1}{2}-z_9\right)c\,\hat{\mathbf{z}}$ | (8$d$) | O VI |
| $\mathbf{B}_{64}$ | = | $\left(\frac{1}{2}+x_9\right)\mathbf{a}_1+\left(\frac{1}{2}-y_9\right)\mathbf{a}_2-z_9\,\mathbf{a}_3$ | = | $\left(\frac{1}{2}+x_9\right)a\,\hat{\mathbf{x}}+\left(\frac{1}{2}-y_9\right)b\,\hat{\mathbf{y}}-z_9c\,\hat{\mathbf{z}}$ | (8$d$) | O VI |
| $\mathbf{B}_{65}$ | = | $-x_9\,\mathbf{a}_1-y_9\,\mathbf{a}_2-z_9\,\mathbf{a}_3$ | = | $-x_9a\,\hat{\mathbf{x}}-y_9b\,\hat{\mathbf{y}}-z_9c\,\hat{\mathbf{z}}$ | (8$d$) | O VI |
| $\mathbf{B}_{66}$ | = | $\left(\frac{1}{2}+x_9\right)\mathbf{a}_1+\left(\frac{1}{2}+y_9\right)\mathbf{a}_2+\left(\frac{1}{2}-z_9\right)\mathbf{a}_3$ | = | $\left(\frac{1}{2}+x_9\right)a\,\hat{\mathbf{x}}+\left(\frac{1}{2}+y_9\right)b\,\hat{\mathbf{y}}+\left(\frac{1}{2}-z_9\right)c\,\hat{\mathbf{z}}$ | (8$d$) | O VI |
| $\mathbf{B}_{67}$ | = | $x_9\,\mathbf{a}_1-y_9\,\mathbf{a}_2+\left(\frac{1}{2}+z_9\right)\mathbf{a}_3$ | = | $x_9a\,\hat{\mathbf{x}}-y_9b\,\hat{\mathbf{y}}+\left(\frac{1}{2}+z_9\right)c\,\hat{\mathbf{z}}$ | (8$d$) | O VI |
| $\mathbf{B}_{68}$ | = | $\left(\frac{1}{2}-x_9\right)\mathbf{a}_1+\left(\frac{1}{2}+y_9\right)\mathbf{a}_2+z_9\,\mathbf{a}_3$ | = | $\left(\frac{1}{2}-x_9\right)a\,\hat{\mathbf{x}}+\left(\frac{1}{2}+y_9\right)b\,\hat{\mathbf{y}}+z_9c\,\hat{\mathbf{z}}$ | (8$d$) | O VI |

**Geometry files:**

- CIF: pp. 1623
- POSCAR: pp. 1624

# Columbite ($FeNb_2O_4$, $E5_1$) Structure: AB2C6_oP36_60_c_d_3d

http://aflow.org/prototype-encyclopedia/AB2C6_oP36_60_c_d_3d

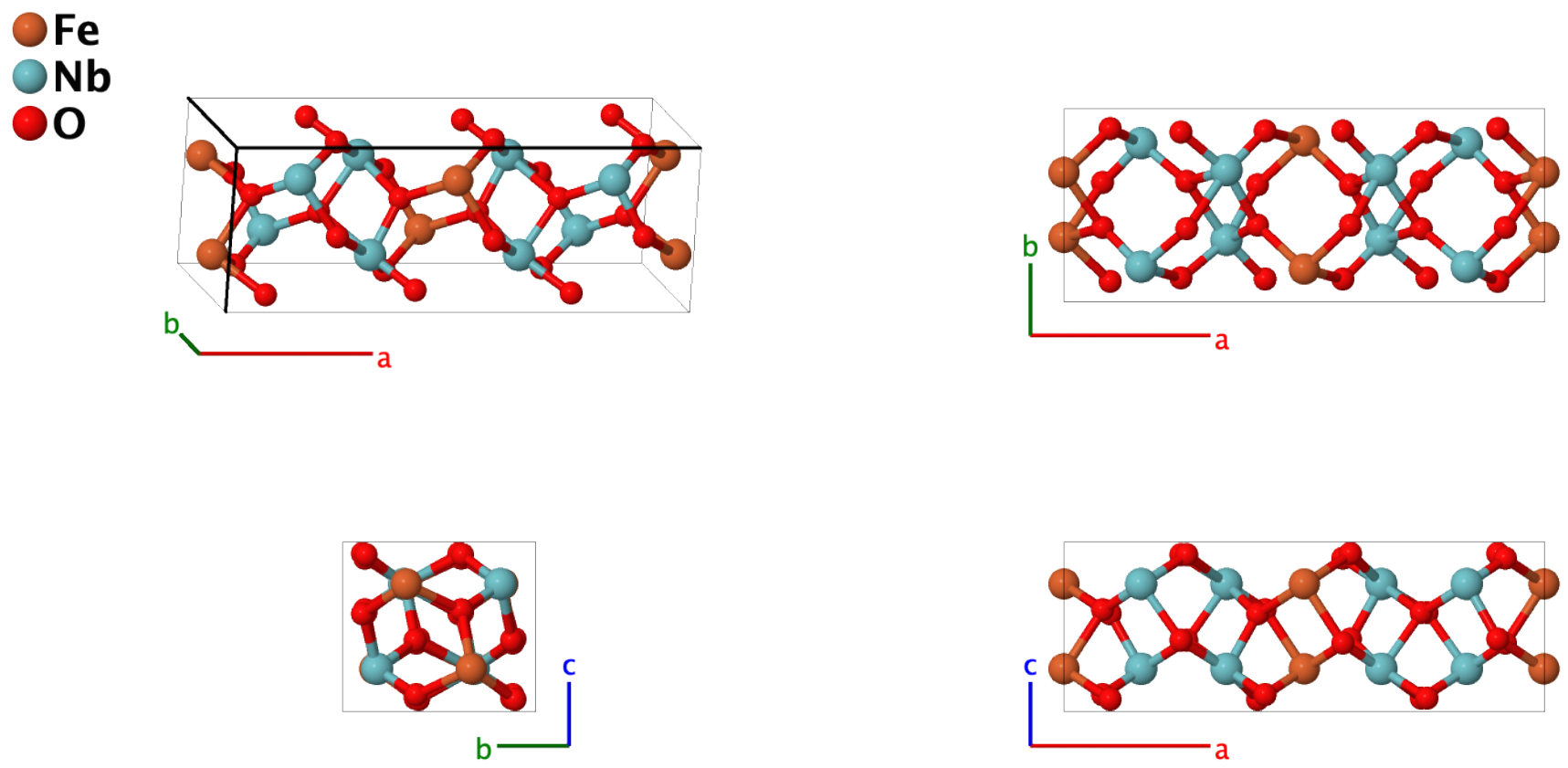

| | | |
|---|---|---|
| **Prototype** | : | $FeNb_2O_6$ |
| **AFLOW prototype label** | : | AB2C6_oP36_60_c_d_3d |
| ***Strukturbericht* designation** | : | $E5_1$ |
| **Pearson symbol** | : | oP36 |
| **Space group number** | : | 60 |
| **Space group symbol** | : | $Pbcn$ |
| **AFLOW prototype command** | : | `aflow --proto=AB2C6_oP36_60_c_d_3d` `--params=`$a, b/a, c/a, y_1, x_2, y_2, z_2, x_3, y_3, z_3, x_4, y_4, z_4, x_5, y_5, z_5$ |

### Other compounds with this structure

- $CaNb_2O_6$ (Fermsmite) and $Zr_5Ti_7O_{24}$

- The original *Strukturbericht* reference uses (Fe,Mn)$Nb_2O_6$ as the prototype for Columbite/$E5_1$.

### Simple Orthorhombic primitive vectors:

$$\mathbf{a}_1 = a\,\hat{\mathbf{x}}$$
$$\mathbf{a}_2 = b\,\hat{\mathbf{y}}$$
$$\mathbf{a}_3 = c\,\hat{\mathbf{z}}$$

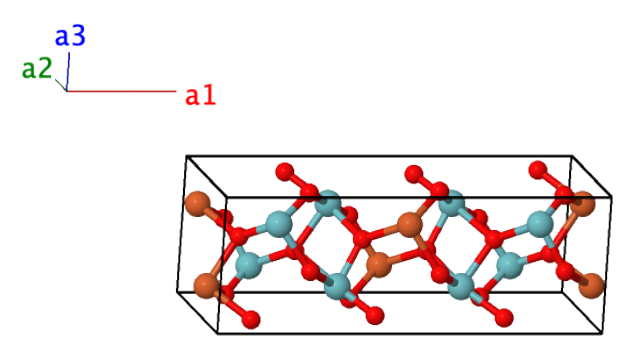

### Basis vectors:

| | | Lattice Coordinates | | Cartesian Coordinates | Wyckoff Position | Atom Type |
|---|---|---|---|---|---|---|
| $\mathbf{B}_1$ | = | $y_1\,\mathbf{a}_2 + \frac{1}{4}\,\mathbf{a}_3$ | = | $y_1 b\,\hat{\mathbf{y}} + \frac{1}{4}c\,\hat{\mathbf{z}}$ | (4$c$) | Fe |
| $\mathbf{B}_2$ | = | $\frac{1}{2}\,\mathbf{a}_1 + \left(\frac{1}{2} - y_1\right)\mathbf{a}_2 + \frac{3}{4}\,\mathbf{a}_3$ | = | $\frac{1}{2}a\,\hat{\mathbf{x}} + \left(\frac{1}{2} - y_1\right)b\,\hat{\mathbf{y}} + \frac{3}{4}c\,\hat{\mathbf{z}}$ | (4$c$) | Fe |
| $\mathbf{B}_3$ | = | $-y_1\,\mathbf{a}_2 + \frac{3}{4}\,\mathbf{a}_3$ | = | $-y_1 b\,\hat{\mathbf{y}} + \frac{3}{4}c\,\hat{\mathbf{z}}$ | (4$c$) | Fe |
| $\mathbf{B}_4$ | = | $\frac{1}{2}\,\mathbf{a}_1 + \left(\frac{1}{2} + y_1\right)\mathbf{a}_2 + \frac{1}{4}\,\mathbf{a}_3$ | = | $\frac{1}{2}a\,\hat{\mathbf{x}} + \left(\frac{1}{2} + y_1\right)b\,\hat{\mathbf{y}} + \frac{1}{4}c\,\hat{\mathbf{z}}$ | (4$c$) | Fe |
| $\mathbf{B}_{5}$ | = | $x_{2}\,\mathbf{a}_{1}+y_{2}\,\mathbf{a}_{2}+z_{2}\,\mathbf{a}_{3}$ | = | $x_{2}a\,\mathbf{\hat{x}}+y_{2}b\,\mathbf{\hat{y}}+z_{2}c\,\mathbf{\hat{z}}$ | (8$d$) | Nb |
| $\mathbf{B}_{6}$ | = | $\left(\frac{1}{2}-x_{2}\right)\mathbf{a}_{1}+\left(\frac{1}{2}-y_{2}\right)\mathbf{a}_{2}+\left(\frac{1}{2}+z_{2}\right)\mathbf{a}_{3}$ | = | $\left(\frac{1}{2}-x_{2}\right)a\,\mathbf{\hat{x}}+\left(\frac{1}{2}-y_{2}\right)b\,\mathbf{\hat{y}}+\left(\frac{1}{2}+z_{2}\right)c\,\mathbf{\hat{z}}$ | (8$d$) | Nb |
| $\mathbf{B}_{7}$ | = | $-x_{2}\,\mathbf{a}_{1}+y_{2}\,\mathbf{a}_{2}+\left(\frac{1}{2}-z_{2}\right)\mathbf{a}_{3}$ | = | $-x_{2}a\,\mathbf{\hat{x}}+y_{2}b\,\mathbf{\hat{y}}+\left(\frac{1}{2}-z_{2}\right)c\,\mathbf{\hat{z}}$ | (8$d$) | Nb |
| $\mathbf{B}_{8}$ | = | $\left(\frac{1}{2}+x_{2}\right)\mathbf{a}_{1}+\left(\frac{1}{2}-y_{2}\right)\mathbf{a}_{2}-z_{2}\,\mathbf{a}_{3}$ | = | $\left(\frac{1}{2}+x_{2}\right)a\,\mathbf{\hat{x}}+\left(\frac{1}{2}-y_{2}\right)b\,\mathbf{\hat{y}}-z_{2}c\,\mathbf{\hat{z}}$ | (8$d$) | Nb |
| $\mathbf{B}_{9}$ | = | $-x_{2}\,\mathbf{a}_{1}-y_{2}\,\mathbf{a}_{2}-z_{2}\,\mathbf{a}_{3}$ | = | $-x_{2}a\,\mathbf{\hat{x}}-y_{2}b\,\mathbf{\hat{y}}-z_{2}c\,\mathbf{\hat{z}}$ | (8$d$) | Nb |
| $\mathbf{B}_{10}$ | = | $\left(\frac{1}{2}+x_{2}\right)\mathbf{a}_{1}+\left(\frac{1}{2}+y_{2}\right)\mathbf{a}_{2}+\left(\frac{1}{2}-z_{2}\right)\mathbf{a}_{3}$ | = | $\left(\frac{1}{2}+x_{2}\right)a\,\mathbf{\hat{x}}+\left(\frac{1}{2}+y_{2}\right)b\,\mathbf{\hat{y}}+\left(\frac{1}{2}-z_{2}\right)c\,\mathbf{\hat{z}}$ | (8$d$) | Nb |
| $\mathbf{B}_{11}$ | = | $x_{2}\,\mathbf{a}_{1}-y_{2}\,\mathbf{a}_{2}+\left(\frac{1}{2}+z_{2}\right)\mathbf{a}_{3}$ | = | $x_{2}a\,\mathbf{\hat{x}}-y_{2}b\,\mathbf{\hat{y}}+\left(\frac{1}{2}+z_{2}\right)c\,\mathbf{\hat{z}}$ | (8$d$) | Nb |
| $\mathbf{B}_{12}$ | = | $\left(\frac{1}{2}-x_{2}\right)\mathbf{a}_{1}+\left(\frac{1}{2}+y_{2}\right)\mathbf{a}_{2}+z_{2}\,\mathbf{a}_{3}$ | = | $\left(\frac{1}{2}-x_{2}\right)a\,\mathbf{\hat{x}}+\left(\frac{1}{2}+y_{2}\right)b\,\mathbf{\hat{y}}+z_{2}c\,\mathbf{\hat{z}}$ | (8$d$) | Nb |
| $\mathbf{B}_{13}$ | = | $x_{3}\,\mathbf{a}_{1}+y_{3}\,\mathbf{a}_{2}+z_{3}\,\mathbf{a}_{3}$ | = | $x_{3}a\,\mathbf{\hat{x}}+y_{3}b\,\mathbf{\hat{y}}+z_{3}c\,\mathbf{\hat{z}}$ | (8$d$) | O I |
| $\mathbf{B}_{14}$ | = | $\left(\frac{1}{2}-x_{3}\right)\mathbf{a}_{1}+\left(\frac{1}{2}-y_{3}\right)\mathbf{a}_{2}+\left(\frac{1}{2}+z_{3}\right)\mathbf{a}_{3}$ | = | $\left(\frac{1}{2}-x_{3}\right)a\,\mathbf{\hat{x}}+\left(\frac{1}{2}-y_{3}\right)b\,\mathbf{\hat{y}}+\left(\frac{1}{2}+z_{3}\right)c\,\mathbf{\hat{z}}$ | (8$d$) | O I |
| $\mathbf{B}_{15}$ | = | $-x_{3}\,\mathbf{a}_{1}+y_{3}\,\mathbf{a}_{2}+\left(\frac{1}{2}-z_{3}\right)\mathbf{a}_{3}$ | = | $-x_{3}a\,\mathbf{\hat{x}}+y_{3}b\,\mathbf{\hat{y}}+\left(\frac{1}{2}-z_{3}\right)c\,\mathbf{\hat{z}}$ | (8$d$) | O I |
| $\mathbf{B}_{16}$ | = | $\left(\frac{1}{2}+x_{3}\right)\mathbf{a}_{1}+\left(\frac{1}{2}-y_{3}\right)\mathbf{a}_{2}-z_{3}\,\mathbf{a}_{3}$ | = | $\left(\frac{1}{2}+x_{3}\right)a\,\mathbf{\hat{x}}+\left(\frac{1}{2}-y_{3}\right)b\,\mathbf{\hat{y}}-z_{3}c\,\mathbf{\hat{z}}$ | (8$d$) | O I |
| $\mathbf{B}_{17}$ | = | $-x_{3}\,\mathbf{a}_{1}-y_{3}\,\mathbf{a}_{2}-z_{3}\,\mathbf{a}_{3}$ | = | $-x_{3}a\,\mathbf{\hat{x}}-y_{3}b\,\mathbf{\hat{y}}-z_{3}c\,\mathbf{\hat{z}}$ | (8$d$) | O I |
| $\mathbf{B}_{18}$ | = | $\left(\frac{1}{2}+x_{3}\right)\mathbf{a}_{1}+\left(\frac{1}{2}+y_{3}\right)\mathbf{a}_{2}+\left(\frac{1}{2}-z_{3}\right)\mathbf{a}_{3}$ | = | $\left(\frac{1}{2}+x_{3}\right)a\,\mathbf{\hat{x}}+\left(\frac{1}{2}+y_{3}\right)b\,\mathbf{\hat{y}}+\left(\frac{1}{2}-z_{3}\right)c\,\mathbf{\hat{z}}$ | (8$d$) | O I |
| $\mathbf{B}_{19}$ | = | $x_{3}\,\mathbf{a}_{1}-y_{3}\,\mathbf{a}_{2}+\left(\frac{1}{2}+z_{3}\right)\mathbf{a}_{3}$ | = | $x_{3}a\,\mathbf{\hat{x}}-y_{3}b\,\mathbf{\hat{y}}+\left(\frac{1}{2}+z_{3}\right)c\,\mathbf{\hat{z}}$ | (8$d$) | O I |
| $\mathbf{B}_{20}$ | = | $\left(\frac{1}{2}-x_{3}\right)\mathbf{a}_{1}+\left(\frac{1}{2}+y_{3}\right)\mathbf{a}_{2}+z_{3}\,\mathbf{a}_{3}$ | = | $\left(\frac{1}{2}-x_{3}\right)a\,\mathbf{\hat{x}}+\left(\frac{1}{2}+y_{3}\right)b\,\mathbf{\hat{y}}+z_{3}c\,\mathbf{\hat{z}}$ | (8$d$) | O I |
| $\mathbf{B}_{21}$ | = | $x_{4}\,\mathbf{a}_{1}+y_{4}\,\mathbf{a}_{2}+z_{4}\,\mathbf{a}_{3}$ | = | $x_{4}a\,\mathbf{\hat{x}}+y_{4}b\,\mathbf{\hat{y}}+z_{4}c\,\mathbf{\hat{z}}$ | (8$d$) | O II |
| $\mathbf{B}_{22}$ | = | $\left(\frac{1}{2}-x_{4}\right)\mathbf{a}_{1}+\left(\frac{1}{2}-y_{4}\right)\mathbf{a}_{2}+\left(\frac{1}{2}+z_{4}\right)\mathbf{a}_{3}$ | = | $\left(\frac{1}{2}-x_{4}\right)a\,\mathbf{\hat{x}}+\left(\frac{1}{2}-y_{4}\right)b\,\mathbf{\hat{y}}+\left(\frac{1}{2}+z_{4}\right)c\,\mathbf{\hat{z}}$ | (8$d$) | O II |
| $\mathbf{B}_{23}$ | = | $-x_{4}\,\mathbf{a}_{1}+y_{4}\,\mathbf{a}_{2}+\left(\frac{1}{2}-z_{4}\right)\mathbf{a}_{3}$ | = | $-x_{4}a\,\mathbf{\hat{x}}+y_{4}b\,\mathbf{\hat{y}}+\left(\frac{1}{2}-z_{4}\right)c\,\mathbf{\hat{z}}$ | (8$d$) | O II |
| $\mathbf{B}_{24}$ | = | $\left(\frac{1}{2}+x_{4}\right)\mathbf{a}_{1}+\left(\frac{1}{2}-y_{4}\right)\mathbf{a}_{2}-z_{4}\,\mathbf{a}_{3}$ | = | $\left(\frac{1}{2}+x_{4}\right)a\,\mathbf{\hat{x}}+\left(\frac{1}{2}-y_{4}\right)b\,\mathbf{\hat{y}}-z_{4}c\,\mathbf{\hat{z}}$ | (8$d$) | O II |
| $\mathbf{B}_{25}$ | = | $-x_{4}\,\mathbf{a}_{1}-y_{4}\,\mathbf{a}_{2}-z_{4}\,\mathbf{a}_{3}$ | = | $-x_{4}a\,\mathbf{\hat{x}}-y_{4}b\,\mathbf{\hat{y}}-z_{4}c\,\mathbf{\hat{z}}$ | (8$d$) | O II |
| $\mathbf{B}_{26}$ | = | $\left(\frac{1}{2}+x_{4}\right)\mathbf{a}_{1}+\left(\frac{1}{2}+y_{4}\right)\mathbf{a}_{2}+\left(\frac{1}{2}-z_{4}\right)\mathbf{a}_{3}$ | = | $\left(\frac{1}{2}+x_{4}\right)a\,\mathbf{\hat{x}}+\left(\frac{1}{2}+y_{4}\right)b\,\mathbf{\hat{y}}+\left(\frac{1}{2}-z_{4}\right)c\,\mathbf{\hat{z}}$ | (8$d$) | O II |
| $\mathbf{B}_{27}$ | = | $x_{4}\,\mathbf{a}_{1}-y_{4}\,\mathbf{a}_{2}+\left(\frac{1}{2}+z_{4}\right)\mathbf{a}_{3}$ | = | $x_{4}a\,\mathbf{\hat{x}}-y_{4}b\,\mathbf{\hat{y}}+\left(\frac{1}{2}+z_{4}\right)c\,\mathbf{\hat{z}}$ | (8$d$) | O II |
| $\mathbf{B}_{28}$ | = | $\left(\frac{1}{2}-x_{4}\right)\mathbf{a}_{1}+\left(\frac{1}{2}+y_{4}\right)\mathbf{a}_{2}+z_{4}\,\mathbf{a}_{3}$ | = | $\left(\frac{1}{2}-x_{4}\right)a\,\mathbf{\hat{x}}+\left(\frac{1}{2}+y_{4}\right)b\,\mathbf{\hat{y}}+z_{4}c\,\mathbf{\hat{z}}$ | (8$d$) | O II |
| $\mathbf{B}_{29}$ | = | $x_{5}\,\mathbf{a}_{1}+y_{5}\,\mathbf{a}_{2}+z_{5}\,\mathbf{a}_{3}$ | = | $x_{5}a\,\mathbf{\hat{x}}+y_{5}b\,\mathbf{\hat{y}}+z_{5}c\,\mathbf{\hat{z}}$ | (8$d$) | O III |
| $\mathbf{B}_{30}$ | = | $\left(\frac{1}{2}-x_{5}\right)\mathbf{a}_{1}+\left(\frac{1}{2}-y_{5}\right)\mathbf{a}_{2}+\left(\frac{1}{2}+z_{5}\right)\mathbf{a}_{3}$ | = | $\left(\frac{1}{2}-x_{5}\right)a\,\mathbf{\hat{x}}+\left(\frac{1}{2}-y_{5}\right)b\,\mathbf{\hat{y}}+\left(\frac{1}{2}+z_{5}\right)c\,\mathbf{\hat{z}}$ | (8$d$) | O III |
| $\mathbf{B}_{31}$ | = | $-x_{5}\,\mathbf{a}_{1}+y_{5}\,\mathbf{a}_{2}+\left(\frac{1}{2}-z_{5}\right)\mathbf{a}_{3}$ | = | $-x_{5}a\,\mathbf{\hat{x}}+y_{5}b\,\mathbf{\hat{y}}+\left(\frac{1}{2}-z_{5}\right)c\,\mathbf{\hat{z}}$ | (8$d$) | O III |
| $\mathbf{B}_{32}$ | = | $\left(\frac{1}{2}+x_{5}\right)\mathbf{a}_{1}+\left(\frac{1}{2}-y_{5}\right)\mathbf{a}_{2}-z_{5}\,\mathbf{a}_{3}$ | = | $\left(\frac{1}{2}+x_{5}\right)a\,\mathbf{\hat{x}}+\left(\frac{1}{2}-y_{5}\right)b\,\mathbf{\hat{y}}-z_{5}c\,\mathbf{\hat{z}}$ | (8$d$) | O III |
| $\mathbf{B}_{33}$ | = | $-x_{5}\,\mathbf{a}_{1}-y_{5}\,\mathbf{a}_{2}-z_{5}\,\mathbf{a}_{3}$ | = | $-x_{5}a\,\mathbf{\hat{x}}-y_{5}b\,\mathbf{\hat{y}}-z_{5}c\,\mathbf{\hat{z}}$ | (8$d$) | O III |
| $\mathbf{B}_{34}$ | = | $\left(\frac{1}{2}+x_{5}\right)\mathbf{a}_{1}+\left(\frac{1}{2}+y_{5}\right)\mathbf{a}_{2}+\left(\frac{1}{2}-z_{5}\right)\mathbf{a}_{3}$ | = | $\left(\frac{1}{2}+x_{5}\right)a\,\mathbf{\hat{x}}+\left(\frac{1}{2}+y_{5}\right)b\,\mathbf{\hat{y}}+\left(\frac{1}{2}-z_{5}\right)c\,\mathbf{\hat{z}}$ | (8$d$) | O III |
| $\mathbf{B}_{35}$ | = | $x_5\,\mathbf{a}_1 - y_5\,\mathbf{a}_2 + \left(\frac{1}{2} + z_5\right)\mathbf{a}_3$ | = | $x_5 a\,\hat{\mathbf{x}} - y_5 b\,\hat{\mathbf{y}} + \left(\frac{1}{2} + z_5\right) c\,\hat{\mathbf{z}}$ | (8*d*) | O III |
| $\mathbf{B}_{36}$ | = | $\left(\frac{1}{2} - x_5\right)\mathbf{a}_1 + \left(\frac{1}{2} + y_5\right)\mathbf{a}_2 + z_5\,\mathbf{a}_3$ | = | $\left(\frac{1}{2} - x_5\right) a\,\hat{\mathbf{x}} + \left(\frac{1}{2} + y_5\right) b\,\hat{\mathbf{y}} + z_5 c\,\hat{\mathbf{z}}$ | (8*d*) | O III |

---

---

**Geometry files:**

- CIF: pp. 1624
- POSCAR: pp. 1625

# $Ca_2RuO_4$ Structure: A2B4C_oP28_61_c_2c_a

http://aflow.org/prototype-encyclopedia/A2B4C_oP28_61_c_2c_a

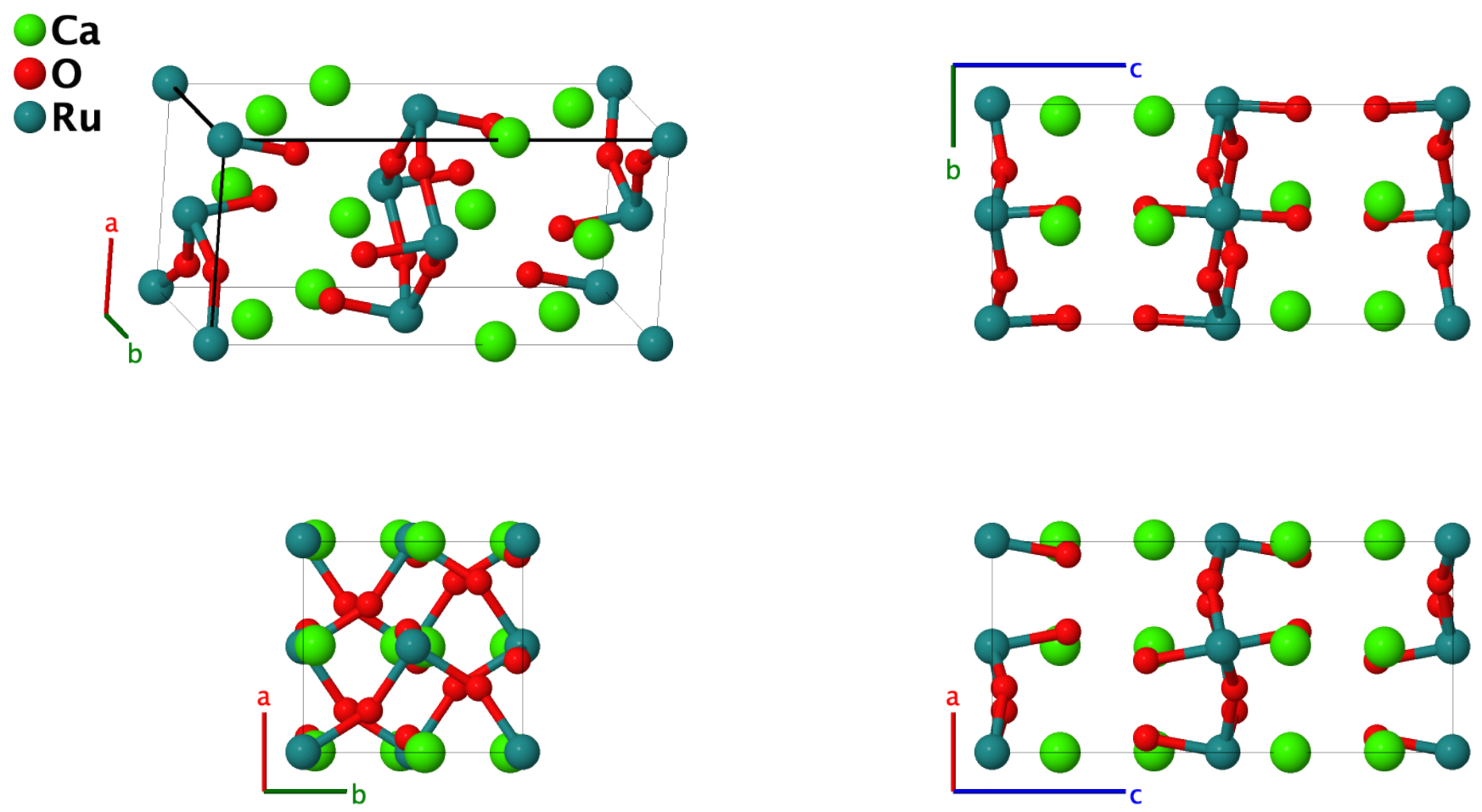

| | | |
|---|---|---|
| **Prototype** | : | $Ca_2O_4Ru$ |
| **AFLOW prototype label** | : | A2B4C_oP28_61_c_2c_a |
| ***Strukturbericht* designation** | : | None |
| **Pearson symbol** | : | oP28 |
| **Space group number** | : | 61 |
| **Space group symbol** | : | *Pbca* |
| **AFLOW prototype command** | : | `aflow --proto=A2B4C_oP28_61_c_2c_a` <br> `--params=`$a, b/a, c/a, x_2, y_2, z_2, x_3, y_3, z_3, x_4, y_4, z_4$ |

### Other compounds with this structure

- $Ca_{2-x}Sr_xRuO_4$

- (Friedt, 2001) never give the positions of the ruthenium atoms in this structure, however the composition of the crystal dictates that they can only be at the (4*a*) or (4*b*) Wyckoff positions, and the Ru-O distances given are consistent with the (4*a*) site.
- The authors identify a low temperature "$S - Pbca$" phase and a high temperature "$L - Pbca$" phase, with a phase transition in the range 350-400 K. The major difference between the two phases is a 4% elongation of the $c$ axis in the $L - Pbca$ phase, accompanied by a 4.5% contraction along the $b$ axis, with the $a$ axis being substantially unchanged. There is also a substantial change in the tilt of the oxygen octahedra surrounding the Ru atoms. Here we show the crystal in the $S - Pbca$ phase using data taken at 180 K.
- (Bertinshaw, 2019) identify $S^*$ and $L^*$ phases that have the same symmetry but stabilized by the application of an electric current.

### Simple Orthorhombic primitive vectors:

$$\mathbf{a}_1 = a\,\hat{\mathbf{x}}$$
$$\mathbf{a}_2 = b\,\hat{\mathbf{y}}$$
$$\mathbf{a}_3 = c\,\hat{\mathbf{z}}$$

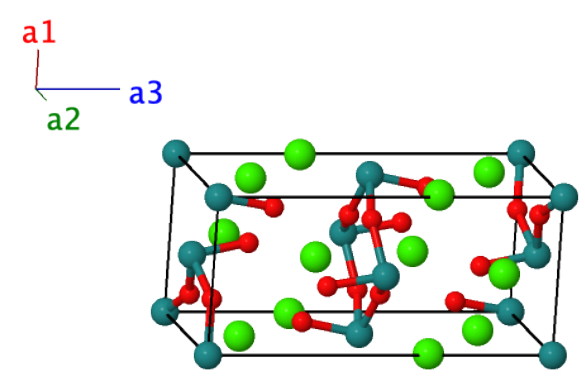

**Basis vectors:**

| | | Lattice Coordinates | | Cartesian Coordinates | Wyckoff Position | Atom Type |
|---|---|---|---|---|---|---|
| $\mathbf{B}_1$ | = | $0\,\mathbf{a}_1 + 0\,\mathbf{a}_2 + 0\,\mathbf{a}_3$ | = | $0\,\hat{\mathbf{x}} + 0\,\hat{\mathbf{y}} + 0\,\hat{\mathbf{z}}$ | (4$a$) | Ru |
| $\mathbf{B}_2$ | = | $\frac{1}{2}\,\mathbf{a}_1 + \frac{1}{2}\,\mathbf{a}_3$ | = | $\frac{1}{2}a\,\hat{\mathbf{x}} + \frac{1}{2}c\,\hat{\mathbf{z}}$ | (4$a$) | Ru |
| $\mathbf{B}_3$ | = | $\frac{1}{2}\,\mathbf{a}_2 + \frac{1}{2}\,\mathbf{a}_3$ | = | $\frac{1}{2}b\,\hat{\mathbf{y}} + \frac{1}{2}c\,\hat{\mathbf{z}}$ | (4$a$) | Ru |
| $\mathbf{B}_4$ | = | $\frac{1}{2}\,\mathbf{a}_1 + \frac{1}{2}\,\mathbf{a}_2$ | = | $\frac{1}{2}a\,\hat{\mathbf{x}} + \frac{1}{2}b\,\hat{\mathbf{y}}$ | (4$a$) | Ru |
| $\mathbf{B}_5$ | = | $x_2\,\mathbf{a}_1 + y_2\,\mathbf{a}_2 + z_2\,\mathbf{a}_3$ | = | $x_2a\,\hat{\mathbf{x}} + y_2b\,\hat{\mathbf{y}} + z_2c\,\hat{\mathbf{z}}$ | (8$c$) | Ca |
| $\mathbf{B}_6$ | = | $\left(\frac{1}{2} - x_2\right)\mathbf{a}_1 - y_2\,\mathbf{a}_2 + \left(\frac{1}{2} + z_2\right)\mathbf{a}_3$ | = | $\left(\frac{1}{2} - x_2\right)a\,\hat{\mathbf{x}} - y_2b\,\hat{\mathbf{y}} + \left(\frac{1}{2} + z_2\right)c\,\hat{\mathbf{z}}$ | (8$c$) | Ca |
| $\mathbf{B}_7$ | = | $-x_2\,\mathbf{a}_1 + \left(\frac{1}{2} + y_2\right)\mathbf{a}_2 + \left(\frac{1}{2} - z_2\right)\mathbf{a}_3$ | = | $-x_2a\,\hat{\mathbf{x}} + \left(\frac{1}{2} + y_2\right)b\,\hat{\mathbf{y}} + \left(\frac{1}{2} - z_2\right)c\,\hat{\mathbf{z}}$ | (8$c$) | Ca |
| $\mathbf{B}_8$ | = | $\left(\frac{1}{2} + x_2\right)\mathbf{a}_1 + \left(\frac{1}{2} - y_2\right)\mathbf{a}_2 - z_2\,\mathbf{a}_3$ | = | $\left(\frac{1}{2} + x_2\right)a\,\hat{\mathbf{x}} + \left(\frac{1}{2} - y_2\right)b\,\hat{\mathbf{y}} - z_2c\,\hat{\mathbf{z}}$ | (8$c$) | Ca |
| $\mathbf{B}_9$ | = | $-x_2\,\mathbf{a}_1 - y_2\,\mathbf{a}_2 - z_2\,\mathbf{a}_3$ | = | $-x_2a\,\hat{\mathbf{x}} - y_2b\,\hat{\mathbf{y}} - z_2c\,\hat{\mathbf{z}}$ | (8$c$) | Ca |
| $\mathbf{B}_{10}$ | = | $\left(\frac{1}{2} + x_2\right)\mathbf{a}_1 + y_2\,\mathbf{a}_2 + \left(\frac{1}{2} - z_2\right)\mathbf{a}_3$ | = | $\left(\frac{1}{2} + x_2\right)a\,\hat{\mathbf{x}} + y_2b\,\hat{\mathbf{y}} + \left(\frac{1}{2} - z_2\right)c\,\hat{\mathbf{z}}$ | (8$c$) | Ca |
| $\mathbf{B}_{11}$ | = | $x_2\,\mathbf{a}_1 + \left(\frac{1}{2} - y_2\right)\mathbf{a}_2 + \left(\frac{1}{2} + z_2\right)\mathbf{a}_3$ | = | $x_2a\,\hat{\mathbf{x}} + \left(\frac{1}{2} - y_2\right)b\,\hat{\mathbf{y}} + \left(\frac{1}{2} + z_2\right)c\,\hat{\mathbf{z}}$ | (8$c$) | Ca |
| $\mathbf{B}_{12}$ | = | $\left(\frac{1}{2} - x_2\right)\mathbf{a}_1 + \left(\frac{1}{2} + y_2\right)\mathbf{a}_2 + z_2\,\mathbf{a}_3$ | = | $\left(\frac{1}{2} - x_2\right)a\,\hat{\mathbf{x}} + \left(\frac{1}{2} + y_2\right)b\,\hat{\mathbf{y}} + z_2c\,\hat{\mathbf{z}}$ | (8$c$) | Ca |
| $\mathbf{B}_{13}$ | = | $x_3\,\mathbf{a}_1 + y_3\,\mathbf{a}_2 + z_3\,\mathbf{a}_3$ | = | $x_3a\,\hat{\mathbf{x}} + y_3b\,\hat{\mathbf{y}} + z_3c\,\hat{\mathbf{z}}$ | (8$c$) | O I |
| $\mathbf{B}_{14}$ | = | $\left(\frac{1}{2} - x_3\right)\mathbf{a}_1 - y_3\,\mathbf{a}_2 + \left(\frac{1}{2} + z_3\right)\mathbf{a}_3$ | = | $\left(\frac{1}{2} - x_3\right)a\,\hat{\mathbf{x}} - y_3b\,\hat{\mathbf{y}} + \left(\frac{1}{2} + z_3\right)c\,\hat{\mathbf{z}}$ | (8$c$) | O I |
| $\mathbf{B}_{15}$ | = | $-x_3\,\mathbf{a}_1 + \left(\frac{1}{2} + y_3\right)\mathbf{a}_2 + \left(\frac{1}{2} - z_3\right)\mathbf{a}_3$ | = | $-x_3a\,\hat{\mathbf{x}} + \left(\frac{1}{2} + y_3\right)b\,\hat{\mathbf{y}} + \left(\frac{1}{2} - z_3\right)c\,\hat{\mathbf{z}}$ | (8$c$) | O I |
| $\mathbf{B}_{16}$ | = | $\left(\frac{1}{2} + x_3\right)\mathbf{a}_1 + \left(\frac{1}{2} - y_3\right)\mathbf{a}_2 - z_3\,\mathbf{a}_3$ | = | $\left(\frac{1}{2} + x_3\right)a\,\hat{\mathbf{x}} + \left(\frac{1}{2} - y_3\right)b\,\hat{\mathbf{y}} - z_3c\,\hat{\mathbf{z}}$ | (8$c$) | O I |
| $\mathbf{B}_{17}$ | = | $-x_3\,\mathbf{a}_1 - y_3\,\mathbf{a}_2 - z_3\,\mathbf{a}_3$ | = | $-x_3a\,\hat{\mathbf{x}} - y_3b\,\hat{\mathbf{y}} - z_3c\,\hat{\mathbf{z}}$ | (8$c$) | O I |
| $\mathbf{B}_{18}$ | = | $\left(\frac{1}{2} + x_3\right)\mathbf{a}_1 + y_3\,\mathbf{a}_2 + \left(\frac{1}{2} - z_3\right)\mathbf{a}_3$ | = | $\left(\frac{1}{2} + x_3\right)a\,\hat{\mathbf{x}} + y_3b\,\hat{\mathbf{y}} + \left(\frac{1}{2} - z_3\right)c\,\hat{\mathbf{z}}$ | (8$c$) | O I |
| $\mathbf{B}_{19}$ | = | $x_3\,\mathbf{a}_1 + \left(\frac{1}{2} - y_3\right)\mathbf{a}_2 + \left(\frac{1}{2} + z_3\right)\mathbf{a}_3$ | = | $x_3a\,\hat{\mathbf{x}} + \left(\frac{1}{2} - y_3\right)b\,\hat{\mathbf{y}} + \left(\frac{1}{2} + z_3\right)c\,\hat{\mathbf{z}}$ | (8$c$) | O I |
| $\mathbf{B}_{20}$ | = | $\left(\frac{1}{2} - x_3\right)\mathbf{a}_1 + \left(\frac{1}{2} + y_3\right)\mathbf{a}_2 + z_3\,\mathbf{a}_3$ | = | $\left(\frac{1}{2} - x_3\right)a\,\hat{\mathbf{x}} + \left(\frac{1}{2} + y_3\right)b\,\hat{\mathbf{y}} + z_3c\,\hat{\mathbf{z}}$ | (8$c$) | O I |
| $\mathbf{B}_{21}$ | = | $x_4\,\mathbf{a}_1 + y_4\,\mathbf{a}_2 + z_4\,\mathbf{a}_3$ | = | $x_4a\,\hat{\mathbf{x}} + y_4b\,\hat{\mathbf{y}} + z_4c\,\hat{\mathbf{z}}$ | (8$c$) | O II |
| $\mathbf{B}_{22}$ | = | $\left(\frac{1}{2} - x_4\right)\mathbf{a}_1 - y_4\,\mathbf{a}_2 + \left(\frac{1}{2} + z_4\right)\mathbf{a}_3$ | = | $\left(\frac{1}{2} - x_4\right)a\,\hat{\mathbf{x}} - y_4b\,\hat{\mathbf{y}} + \left(\frac{1}{2} + z_4\right)c\,\hat{\mathbf{z}}$ | (8$c$) | O II |
| $\mathbf{B}_{23}$ | = | $-x_4\,\mathbf{a}_1 + \left(\frac{1}{2} + y_4\right)\mathbf{a}_2 + \left(\frac{1}{2} - z_4\right)\mathbf{a}_3$ | = | $-x_4a\,\hat{\mathbf{x}} + \left(\frac{1}{2} + y_4\right)b\,\hat{\mathbf{y}} + \left(\frac{1}{2} - z_4\right)c\,\hat{\mathbf{z}}$ | (8$c$) | O II |
| $\mathbf{B}_{24}$ | = | $\left(\frac{1}{2} + x_4\right)\mathbf{a}_1 + \left(\frac{1}{2} - y_4\right)\mathbf{a}_2 - z_4\,\mathbf{a}_3$ | = | $\left(\frac{1}{2} + x_4\right)a\,\hat{\mathbf{x}} + \left(\frac{1}{2} - y_4\right)b\,\hat{\mathbf{y}} - z_4c\,\hat{\mathbf{z}}$ | (8$c$) | O II |
| $\mathbf{B}_{25}$ | = | $-x_4\,\mathbf{a}_1 - y_4\,\mathbf{a}_2 - z_4\,\mathbf{a}_3$ | = | $-x_4a\,\hat{\mathbf{x}} - y_4b\,\hat{\mathbf{y}} - z_4c\,\hat{\mathbf{z}}$ | (8$c$) | O II |
| $\mathbf{B}_{26}$ | = | $\left(\frac{1}{2} + x_4\right)\mathbf{a}_1 + y_4\,\mathbf{a}_2 + \left(\frac{1}{2} - z_4\right)\mathbf{a}_3$ | = | $\left(\frac{1}{2} + x_4\right)a\,\hat{\mathbf{x}} + y_4b\,\hat{\mathbf{y}} + \left(\frac{1}{2} - z_4\right)c\,\hat{\mathbf{z}}$ | (8$c$) | O II |
| $\mathbf{B}_{27}$ | = | $x_4\,\mathbf{a}_1 + \left(\frac{1}{2} - y_4\right)\mathbf{a}_2 + \left(\frac{1}{2} + z_4\right)\mathbf{a}_3$ | = | $x_4a\,\hat{\mathbf{x}} + \left(\frac{1}{2} - y_4\right)b\,\hat{\mathbf{y}} + \left(\frac{1}{2} + z_4\right)c\,\hat{\mathbf{z}}$ | (8$c$) | O II |
| $\mathbf{B}_{28}$ | = | $\left(\frac{1}{2} - x_4\right)\mathbf{a}_1 + \left(\frac{1}{2} + y_4\right)\mathbf{a}_2 + z_4\,\mathbf{a}_3$ | = | $\left(\frac{1}{2} - x_4\right)a\,\hat{\mathbf{x}} + \left(\frac{1}{2} + y_4\right)b\,\hat{\mathbf{y}} + z_4c\,\hat{\mathbf{z}}$ | (8$c$) | O II |

**Geometry files:**

- CIF: pp. 1625
- POSCAR: pp. 1625

# Tellurite ($\beta$-$TeO_2$, $C52$) Structure: A2B_oP24_61_2c_c

http://aflow.org/prototype-encyclopedia/A2B_oP24_61_2c_c.TeO2

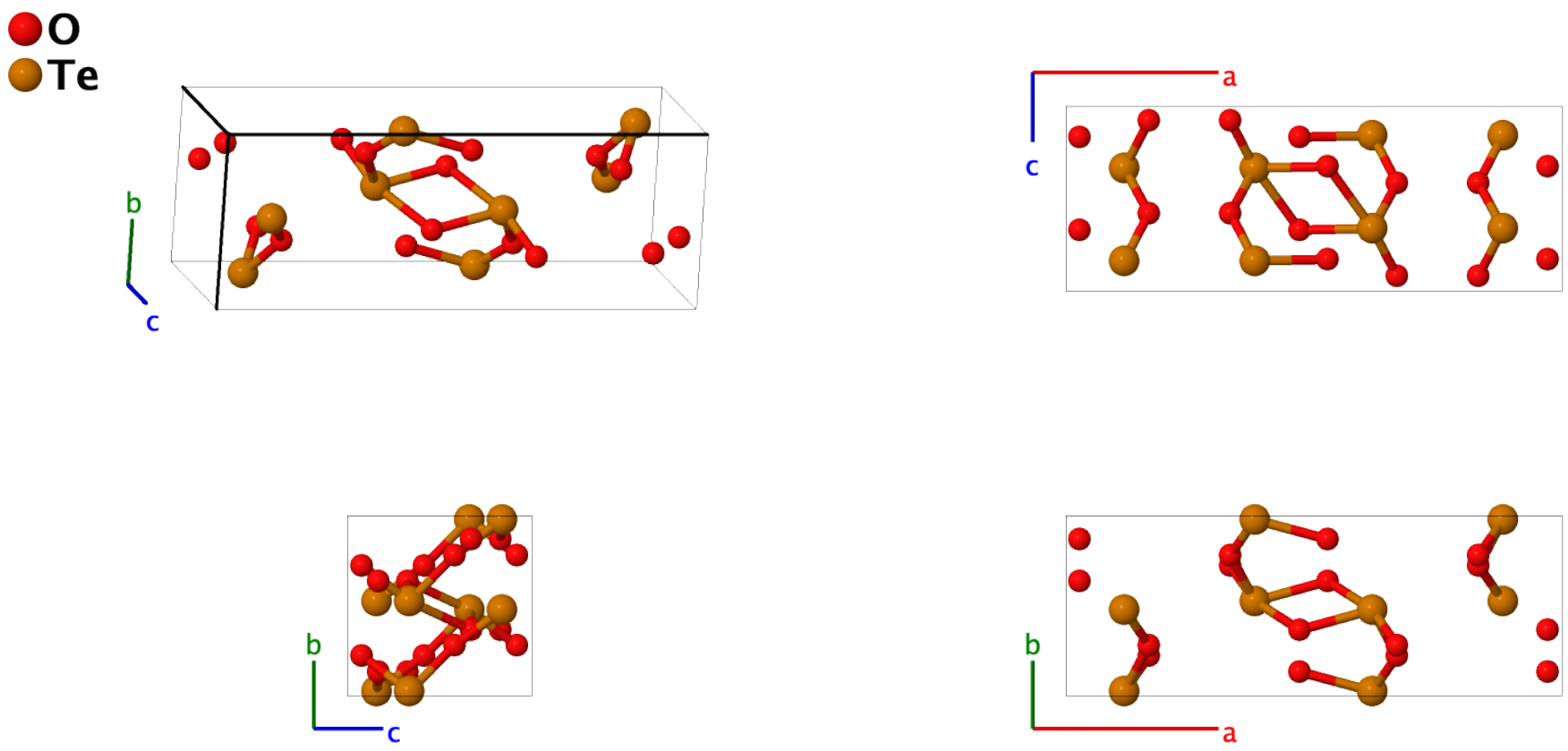

| | | |
|---|---|---|
| **Prototype** | : | $O_2Te$ |
| **AFLOW prototype label** | : | A2B_oP24_61_2c_c |
| ***Strukturbericht* designation** | : | $C52$ |
| **Pearson symbol** | : | oP24 |
| **Space group number** | : | 61 |
| **Space group symbol** | : | $Pbca$ |
| **AFLOW prototype command** | : | `aflow --proto=A2B_oP24_61_2c_c` <br> `--params=`$a, b/a, c/a, x_1, y_1, z_1, x_2, y_2, z_2, x_3, y_3, z_3$ |

**Simple Orthorhombic primitive vectors:**

$$\mathbf{a}_1 = a\,\hat{\mathbf{x}}$$
$$\mathbf{a}_2 = b\,\hat{\mathbf{y}}$$
$$\mathbf{a}_3 = c\,\hat{\mathbf{z}}$$

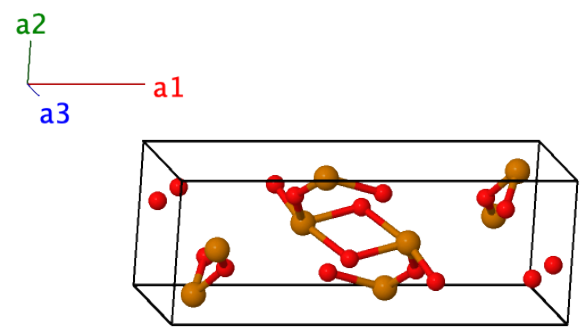

**Basis vectors:**

| | | Lattice Coordinates | | Cartesian Coordinates | Wyckoff Position | Atom Type |
|---|---|---|---|---|---|---|
| $\mathbf{B}_1$ | = | $x_1\,\mathbf{a}_1 + y_1\,\mathbf{a}_2 + z_1\,\mathbf{a}_3$ | = | $x_1 a\,\hat{\mathbf{x}} + y_1 b\,\hat{\mathbf{y}} + z_1 c\,\hat{\mathbf{z}}$ | (8$c$) | O I |
| $\mathbf{B}_2$ | = | $\left(\frac{1}{2} - x_1\right)\mathbf{a}_1 - y_1\,\mathbf{a}_2 + \left(\frac{1}{2} + z_1\right)\mathbf{a}_3$ | = | $\left(\frac{1}{2} - x_1\right)a\,\hat{\mathbf{x}} - y_1 b\,\hat{\mathbf{y}} + \left(\frac{1}{2} + z_1\right)c\,\hat{\mathbf{z}}$ | (8$c$) | O I |
| $\mathbf{B}_3$ | = | $-x_1\,\mathbf{a}_1 + \left(\frac{1}{2} + y_1\right)\mathbf{a}_2 + \left(\frac{1}{2} - z_1\right)\mathbf{a}_3$ | = | $-x_1 a\,\hat{\mathbf{x}} + \left(\frac{1}{2} + y_1\right)b\,\hat{\mathbf{y}} + \left(\frac{1}{2} - z_1\right)c\,\hat{\mathbf{z}}$ | (8$c$) | O I |
| $\mathbf{B}_4$ | = | $\left(\frac{1}{2} + x_1\right)\mathbf{a}_1 + \left(\frac{1}{2} - y_1\right)\mathbf{a}_2 - z_1\,\mathbf{a}_3$ | = | $\left(\frac{1}{2} + x_1\right)a\,\hat{\mathbf{x}} + \left(\frac{1}{2} - y_1\right)b\,\hat{\mathbf{y}} - z_1 c\,\hat{\mathbf{z}}$ | (8$c$) | O I |
| $\mathbf{B}_5$ | = | $-x_1\,\mathbf{a}_1 - y_1\,\mathbf{a}_2 - z_1\,\mathbf{a}_3$ | = | $-x_1 a\,\hat{\mathbf{x}} - y_1 b\,\hat{\mathbf{y}} - z_1 c\,\hat{\mathbf{z}}$ | (8$c$) | O I |
| $\mathbf{B}_6$ | = | $\left(\frac{1}{2} + x_1\right)\mathbf{a}_1 + y_1\,\mathbf{a}_2 + \left(\frac{1}{2} - z_1\right)\mathbf{a}_3$ | = | $\left(\frac{1}{2} + x_1\right)a\,\hat{\mathbf{x}} + y_1 b\,\hat{\mathbf{y}} + \left(\frac{1}{2} - z_1\right)c\,\hat{\mathbf{z}}$ | (8$c$) | O I |
| $\mathbf{B}_7$ | = | $x_1\,\mathbf{a}_1 + \left(\frac{1}{2} - y_1\right)\mathbf{a}_2 + \left(\frac{1}{2} + z_1\right)\mathbf{a}_3$ | = | $x_1 a\,\hat{\mathbf{x}} + \left(\frac{1}{2} - y_1\right)b\,\hat{\mathbf{y}} + \left(\frac{1}{2} + z_1\right)c\,\hat{\mathbf{z}}$ | (8$c$) | O I |
| $\mathbf{B}_8$ | = | $\left(\frac{1}{2} - x_1\right)\mathbf{a}_1 + \left(\frac{1}{2} + y_1\right)\mathbf{a}_2 + z_1\,\mathbf{a}_3$ | = | $\left(\frac{1}{2} - x_1\right)a\,\hat{\mathbf{x}} + \left(\frac{1}{2} + y_1\right)b\,\hat{\mathbf{y}} + z_1 c\,\hat{\mathbf{z}}$ | (8$c$) | O I |
| $\mathbf{B}_9$ | = | $x_2\,\mathbf{a}_1 + y_2\,\mathbf{a}_2 + z_2\,\mathbf{a}_3$ | = | $x_2 a\,\hat{\mathbf{x}} + y_2 b\,\hat{\mathbf{y}} + z_2 c\,\hat{\mathbf{z}}$ | (8$c$) | O II |
| $\mathbf{B}_{10}$ | = | $\left(\frac{1}{2}-x_2\right)\mathbf{a}_1 - y_2\,\mathbf{a}_2 + \left(\frac{1}{2}+z_2\right)\mathbf{a}_3$ | = | $\left(\frac{1}{2}-x_2\right)a\,\hat{\mathbf{x}} - y_2 b\,\hat{\mathbf{y}} + \left(\frac{1}{2}+z_2\right)c\,\hat{\mathbf{z}}$ | (8c) | O II |
| $\mathbf{B}_{11}$ | = | $-x_2\,\mathbf{a}_1 + \left(\frac{1}{2}+y_2\right)\mathbf{a}_2 + \left(\frac{1}{2}-z_2\right)\mathbf{a}_3$ | = | $-x_2 a\,\hat{\mathbf{x}} + \left(\frac{1}{2}+y_2\right)b\,\hat{\mathbf{y}} + \left(\frac{1}{2}-z_2\right)c\,\hat{\mathbf{z}}$ | (8c) | O II |
| $\mathbf{B}_{12}$ | = | $\left(\frac{1}{2}+x_2\right)\mathbf{a}_1 + \left(\frac{1}{2}-y_2\right)\mathbf{a}_2 - z_2\,\mathbf{a}_3$ | = | $\left(\frac{1}{2}+x_2\right)a\,\hat{\mathbf{x}} + \left(\frac{1}{2}-y_2\right)b\,\hat{\mathbf{y}} - z_2 c\,\hat{\mathbf{z}}$ | (8c) | O II |
| $\mathbf{B}_{13}$ | = | $-x_2\,\mathbf{a}_1 - y_2\,\mathbf{a}_2 - z_2\,\mathbf{a}_3$ | = | $-x_2 a\,\hat{\mathbf{x}} - y_2 b\,\hat{\mathbf{y}} - z_2 c\,\hat{\mathbf{z}}$ | (8c) | O II |
| $\mathbf{B}_{14}$ | = | $\left(\frac{1}{2}+x_2\right)\mathbf{a}_1 + y_2\,\mathbf{a}_2 + \left(\frac{1}{2}-z_2\right)\mathbf{a}_3$ | = | $\left(\frac{1}{2}+x_2\right)a\,\hat{\mathbf{x}} + y_2 b\,\hat{\mathbf{y}} + \left(\frac{1}{2}-z_2\right)c\,\hat{\mathbf{z}}$ | (8c) | O II |
| $\mathbf{B}_{15}$ | = | $x_2\,\mathbf{a}_1 + \left(\frac{1}{2}-y_2\right)\mathbf{a}_2 + \left(\frac{1}{2}+z_2\right)\mathbf{a}_3$ | = | $x_2 a\,\hat{\mathbf{x}} + \left(\frac{1}{2}-y_2\right)b\,\hat{\mathbf{y}} + \left(\frac{1}{2}+z_2\right)c\,\hat{\mathbf{z}}$ | (8c) | O II |
| $\mathbf{B}_{16}$ | = | $\left(\frac{1}{2}-x_2\right)\mathbf{a}_1 + \left(\frac{1}{2}+y_2\right)\mathbf{a}_2 + z_2\,\mathbf{a}_3$ | = | $\left(\frac{1}{2}-x_2\right)a\,\hat{\mathbf{x}} + \left(\frac{1}{2}+y_2\right)b\,\hat{\mathbf{y}} + z_2 c\,\hat{\mathbf{z}}$ | (8c) | O II |
| $\mathbf{B}_{17}$ | = | $x_3\,\mathbf{a}_1 + y_3\,\mathbf{a}_2 + z_3\,\mathbf{a}_3$ | = | $x_3 a\,\hat{\mathbf{x}} + y_3 b\,\hat{\mathbf{y}} + z_3 c\,\hat{\mathbf{z}}$ | (8c) | Te |
| $\mathbf{B}_{18}$ | = | $\left(\frac{1}{2}-x_3\right)\mathbf{a}_1 - y_3\,\mathbf{a}_2 + \left(\frac{1}{2}+z_3\right)\mathbf{a}_3$ | = | $\left(\frac{1}{2}-x_3\right)a\,\hat{\mathbf{x}} - y_3 b\,\hat{\mathbf{y}} + \left(\frac{1}{2}+z_3\right)c\,\hat{\mathbf{z}}$ | (8c) | Te |
| $\mathbf{B}_{19}$ | = | $-x_3\,\mathbf{a}_1 + \left(\frac{1}{2}+y_3\right)\mathbf{a}_2 + \left(\frac{1}{2}-z_3\right)\mathbf{a}_3$ | = | $-x_3 a\,\hat{\mathbf{x}} + \left(\frac{1}{2}+y_3\right)b\,\hat{\mathbf{y}} + \left(\frac{1}{2}-z_3\right)c\,\hat{\mathbf{z}}$ | (8c) | Te |
| $\mathbf{B}_{20}$ | = | $\left(\frac{1}{2}+x_3\right)\mathbf{a}_1 + \left(\frac{1}{2}-y_3\right)\mathbf{a}_2 - z_3\,\mathbf{a}_3$ | = | $\left(\frac{1}{2}+x_3\right)a\,\hat{\mathbf{x}} + \left(\frac{1}{2}-y_3\right)b\,\hat{\mathbf{y}} - z_3 c\,\hat{\mathbf{z}}$ | (8c) | Te |
| $\mathbf{B}_{21}$ | = | $-x_3\,\mathbf{a}_1 - y_3\,\mathbf{a}_2 - z_3\,\mathbf{a}_3$ | = | $-x_3 a\,\hat{\mathbf{x}} - y_3 b\,\hat{\mathbf{y}} - z_3 c\,\hat{\mathbf{z}}$ | (8c) | Te |
| $\mathbf{B}_{22}$ | = | $\left(\frac{1}{2}+x_3\right)\mathbf{a}_1 + y_3\,\mathbf{a}_2 + \left(\frac{1}{2}-z_3\right)\mathbf{a}_3$ | = | $\left(\frac{1}{2}+x_3\right)a\,\hat{\mathbf{x}} + y_3 b\,\hat{\mathbf{y}} + \left(\frac{1}{2}-z_3\right)c\,\hat{\mathbf{z}}$ | (8c) | Te |
| $\mathbf{B}_{23}$ | = | $x_3\,\mathbf{a}_1 + \left(\frac{1}{2}-y_3\right)\mathbf{a}_2 + \left(\frac{1}{2}+z_3\right)\mathbf{a}_3$ | = | $x_3 a\,\hat{\mathbf{x}} + \left(\frac{1}{2}-y_3\right)b\,\hat{\mathbf{y}} + \left(\frac{1}{2}+z_3\right)c\,\hat{\mathbf{z}}$ | (8c) | Te |
| $\mathbf{B}_{24}$ | = | $\left(\frac{1}{2}-x_3\right)\mathbf{a}_1 + \left(\frac{1}{2}+y_3\right)\mathbf{a}_2 + z_3\,\mathbf{a}_3$ | = | $\left(\frac{1}{2}-x_3\right)a\,\hat{\mathbf{x}} + \left(\frac{1}{2}+y_3\right)b\,\hat{\mathbf{y}} + z_3 c\,\hat{\mathbf{z}}$ | (8c) | Te |

**Found in:**

**Geometry files:**

- CIF: pp. 1625
- POSCAR: pp. 1626

# $(TiCl_4{\cdot}POCl_3)_2$ Structure: A7BCD_oP80_61_7c_c_c_c

http://aflow.org/prototype-encyclopedia/A7BCD_oP80_61_7c_c_c_c

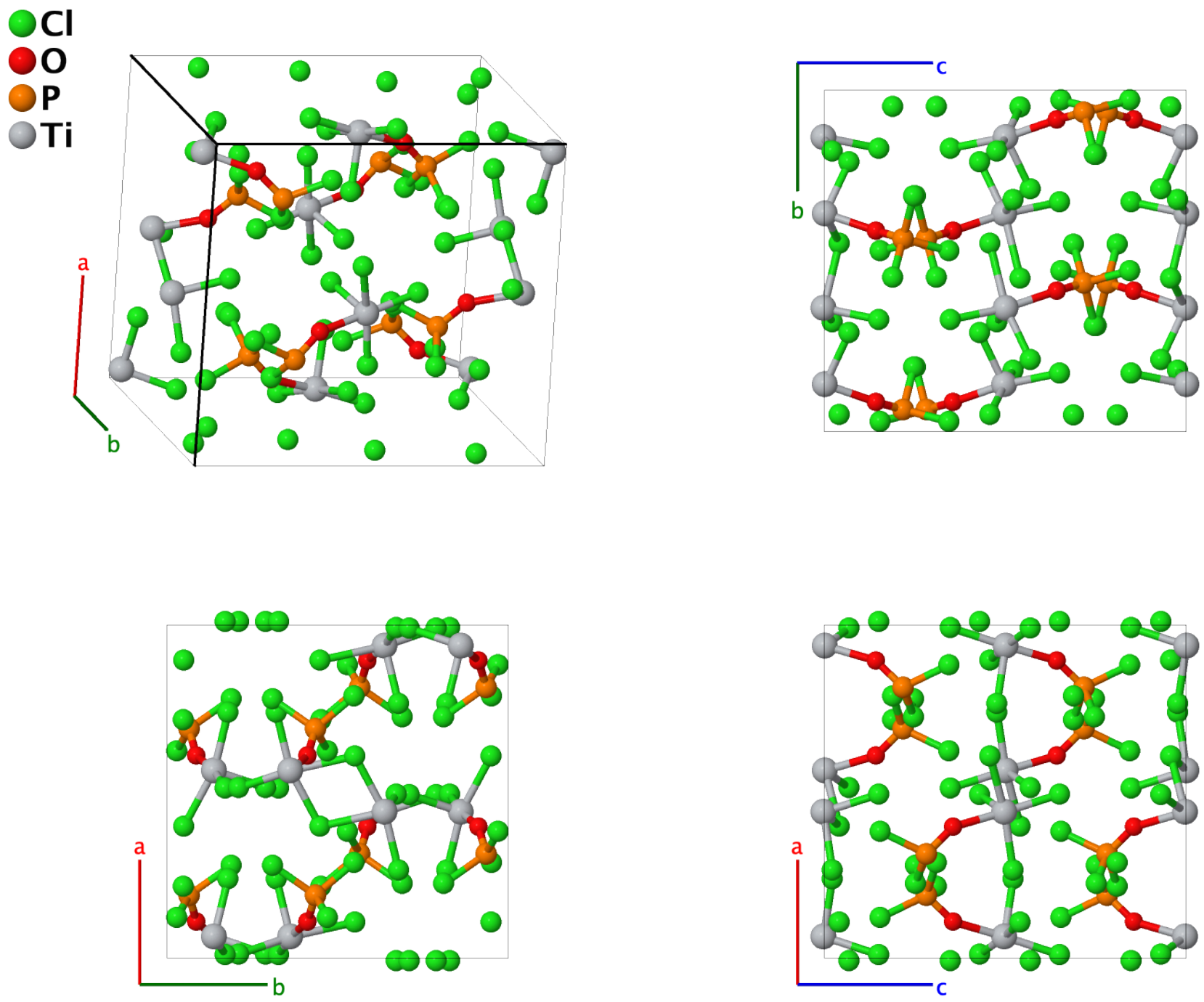

| | | |
|---|---|---|
| **Prototype** | : | $Cl_7OPTi$ |
| **AFLOW prototype label** | : | A7BCD_oP80_61_7c_c_c_c |
| ***Strukturbericht* designation** | : | None |
| **Pearson symbol** | : | oP80 |
| **Space group number** | : | 61 |
| **Space group symbol** | : | *Pbca* |
| **AFLOW prototype command** | : | `aflow --proto=A7BCD_oP80_61_7c_c_c_c --params=`$a, b/a, c/a, x_1, y_1, z_1, x_2, y_2, z_2, x_3, y_3, z_3, x_4, y_4, z_4, x_5, y_5, z_5, x_6, y_6, z_6, x_7, y_7, z_7, x_8, y_8, z_8, x_9, y_9, z_9, x_{10}, y_{10}, z_{10}$ |

**Simple Orthorhombic primitive vectors:**

$$\mathbf{a}_1 = a\,\hat{\mathbf{x}}$$
$$\mathbf{a}_2 = b\,\hat{\mathbf{y}}$$
$$\mathbf{a}_3 = c\,\hat{\mathbf{z}}$$

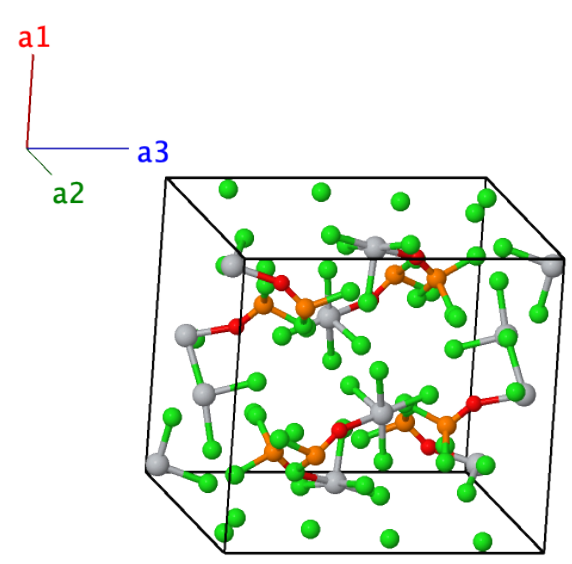

**Basis vectors:**

| | | Lattice Coordinates | | Cartesian Coordinates | Wyckoff Position | Atom Type |
|---|---|---|---|---|---|---|
| $\mathbf{B}_1$ | = | $x_1\,\mathbf{a}_1 + y_1\,\mathbf{a}_2 + z_1\,\mathbf{a}_3$ | = | $x_1 a\,\hat{\mathbf{x}} + y_1 b\,\hat{\mathbf{y}} + z_1 c\,\hat{\mathbf{z}}$ | (8*c*) | Cl I |
| $\mathbf{B}_{2}$ | = | $\left(\frac{1}{2} - x_{1}\right) \mathbf{a}_{1} - y_{1} \mathbf{a}_{2} + \left(\frac{1}{2} + z_{1}\right) \mathbf{a}_{3}$ | = | $\left(\frac{1}{2} - x_{1}\right) a \hat{\mathbf{x}} - y_{1} b \hat{\mathbf{y}} + \left(\frac{1}{2} + z_{1}\right) c \hat{\mathbf{z}}$ | (8*c*) | Cl I |
| $\mathbf{B}_{3}$ | = | $-x_{1} \mathbf{a}_{1} + \left(\frac{1}{2} + y_{1}\right) \mathbf{a}_{2} + \left(\frac{1}{2} - z_{1}\right) \mathbf{a}_{3}$ | = | $-x_{1} a \hat{\mathbf{x}} + \left(\frac{1}{2} + y_{1}\right) b \hat{\mathbf{y}} + \left(\frac{1}{2} - z_{1}\right) c \hat{\mathbf{z}}$ | (8*c*) | Cl I |
| $\mathbf{B}_{4}$ | = | $\left(\frac{1}{2} + x_{1}\right) \mathbf{a}_{1} + \left(\frac{1}{2} - y_{1}\right) \mathbf{a}_{2} - z_{1} \mathbf{a}_{3}$ | = | $\left(\frac{1}{2} + x_{1}\right) a \hat{\mathbf{x}} + \left(\frac{1}{2} - y_{1}\right) b \hat{\mathbf{y}} - z_{1} c \hat{\mathbf{z}}$ | (8*c*) | Cl I |
| $\mathbf{B}_{5}$ | = | $-x_{1} \mathbf{a}_{1} - y_{1} \mathbf{a}_{2} - z_{1} \mathbf{a}_{3}$ | = | $-x_{1} a \hat{\mathbf{x}} - y_{1} b \hat{\mathbf{y}} - z_{1} c \hat{\mathbf{z}}$ | (8*c*) | Cl I |
| $\mathbf{B}_{6}$ | = | $\left(\frac{1}{2} + x_{1}\right) \mathbf{a}_{1} + y_{1} \mathbf{a}_{2} + \left(\frac{1}{2} - z_{1}\right) \mathbf{a}_{3}$ | = | $\left(\frac{1}{2} + x_{1}\right) a \hat{\mathbf{x}} + y_{1} b \hat{\mathbf{y}} + \left(\frac{1}{2} - z_{1}\right) c \hat{\mathbf{z}}$ | (8*c*) | Cl I |
| $\mathbf{B}_{7}$ | = | $x_{1} \mathbf{a}_{1} + \left(\frac{1}{2} - y_{1}\right) \mathbf{a}_{2} + \left(\frac{1}{2} + z_{1}\right) \mathbf{a}_{3}$ | = | $x_{1} a \hat{\mathbf{x}} + \left(\frac{1}{2} - y_{1}\right) b \hat{\mathbf{y}} + \left(\frac{1}{2} + z_{1}\right) c \hat{\mathbf{z}}$ | (8*c*) | Cl I |
| $\mathbf{B}_{8}$ | = | $\left(\frac{1}{2} - x_{1}\right) \mathbf{a}_{1} + \left(\frac{1}{2} + y_{1}\right) \mathbf{a}_{2} + z_{1} \mathbf{a}_{3}$ | = | $\left(\frac{1}{2} - x_{1}\right) a \hat{\mathbf{x}} + \left(\frac{1}{2} + y_{1}\right) b \hat{\mathbf{y}} + z_{1} c \hat{\mathbf{z}}$ | (8*c*) | Cl I |
| $\mathbf{B}_{9}$ | = | $x_{2} \mathbf{a}_{1} + y_{2} \mathbf{a}_{2} + z_{2} \mathbf{a}_{3}$ | = | $x_{2} a \hat{\mathbf{x}} + y_{2} b \hat{\mathbf{y}} + z_{2} c \hat{\mathbf{z}}$ | (8*c*) | Cl II |
| $\mathbf{B}_{10}$ | = | $\left(\frac{1}{2} - x_{2}\right) \mathbf{a}_{1} - y_{2} \mathbf{a}_{2} + \left(\frac{1}{2} + z_{2}\right) \mathbf{a}_{3}$ | = | $\left(\frac{1}{2} - x_{2}\right) a \hat{\mathbf{x}} - y_{2} b \hat{\mathbf{y}} + \left(\frac{1}{2} + z_{2}\right) c \hat{\mathbf{z}}$ | (8*c*) | Cl II |
| $\mathbf{B}_{11}$ | = | $-x_{2} \mathbf{a}_{1} + \left(\frac{1}{2} + y_{2}\right) \mathbf{a}_{2} + \left(\frac{1}{2} - z_{2}\right) \mathbf{a}_{3}$ | = | $-x_{2} a \hat{\mathbf{x}} + \left(\frac{1}{2} + y_{2}\right) b \hat{\mathbf{y}} + \left(\frac{1}{2} - z_{2}\right) c \hat{\mathbf{z}}$ | (8*c*) | Cl II |
| $\mathbf{B}_{12}$ | = | $\left(\frac{1}{2} + x_{2}\right) \mathbf{a}_{1} + \left(\frac{1}{2} - y_{2}\right) \mathbf{a}_{2} - z_{2} \mathbf{a}_{3}$ | = | $\left(\frac{1}{2} + x_{2}\right) a \hat{\mathbf{x}} + \left(\frac{1}{2} - y_{2}\right) b \hat{\mathbf{y}} - z_{2} c \hat{\mathbf{z}}$ | (8*c*) | Cl II |
| $\mathbf{B}_{13}$ | = | $-x_{2} \mathbf{a}_{1} - y_{2} \mathbf{a}_{2} - z_{2} \mathbf{a}_{3}$ | = | $-x_{2} a \hat{\mathbf{x}} - y_{2} b \hat{\mathbf{y}} - z_{2} c \hat{\mathbf{z}}$ | (8*c*) | Cl II |
| $\mathbf{B}_{14}$ | = | $\left(\frac{1}{2} + x_{2}\right) \mathbf{a}_{1} + y_{2} \mathbf{a}_{2} + \left(\frac{1}{2} - z_{2}\right) \mathbf{a}_{3}$ | = | $\left(\frac{1}{2} + x_{2}\right) a \hat{\mathbf{x}} + y_{2} b \hat{\mathbf{y}} + \left(\frac{1}{2} - z_{2}\right) c \hat{\mathbf{z}}$ | (8*c*) | Cl II |
| $\mathbf{B}_{15}$ | = | $x_{2} \mathbf{a}_{1} + \left(\frac{1}{2} - y_{2}\right) \mathbf{a}_{2} + \left(\frac{1}{2} + z_{2}\right) \mathbf{a}_{3}$ | = | $x_{2} a \hat{\mathbf{x}} + \left(\frac{1}{2} - y_{2}\right) b \hat{\mathbf{y}} + \left(\frac{1}{2} + z_{2}\right) c \hat{\mathbf{z}}$ | (8*c*) | Cl II |
| $\mathbf{B}_{16}$ | = | $\left(\frac{1}{2} - x_{2}\right) \mathbf{a}_{1} + \left(\frac{1}{2} + y_{2}\right) \mathbf{a}_{2} + z_{2} \mathbf{a}_{3}$ | = | $\left(\frac{1}{2} - x_{2}\right) a \hat{\mathbf{x}} + \left(\frac{1}{2} + y_{2}\right) b \hat{\mathbf{y}} + z_{2} c \hat{\mathbf{z}}$ | (8*c*) | Cl II |
| $\mathbf{B}_{17}$ | = | $x_{3} \mathbf{a}_{1} + y_{3} \mathbf{a}_{2} + z_{3} \mathbf{a}_{3}$ | = | $x_{3} a \hat{\mathbf{x}} + y_{3} b \hat{\mathbf{y}} + z_{3} c \hat{\mathbf{z}}$ | (8*c*) | Cl III |
| $\mathbf{B}_{18}$ | = | $\left(\frac{1}{2} - x_{3}\right) \mathbf{a}_{1} - y_{3} \mathbf{a}_{2} + \left(\frac{1}{2} + z_{3}\right) \mathbf{a}_{3}$ | = | $\left(\frac{1}{2} - x_{3}\right) a \hat{\mathbf{x}} - y_{3} b \hat{\mathbf{y}} + \left(\frac{1}{2} + z_{3}\right) c \hat{\mathbf{z}}$ | (8*c*) | Cl III |
| $\mathbf{B}_{19}$ | = | $-x_{3} \mathbf{a}_{1} + \left(\frac{1}{2} + y_{3}\right) \mathbf{a}_{2} + \left(\frac{1}{2} - z_{3}\right) \mathbf{a}_{3}$ | = | $-x_{3} a \hat{\mathbf{x}} + \left(\frac{1}{2} + y_{3}\right) b \hat{\mathbf{y}} + \left(\frac{1}{2} - z_{3}\right) c \hat{\mathbf{z}}$ | (8*c*) | Cl III |
| $\mathbf{B}_{20}$ | = | $\left(\frac{1}{2} + x_{3}\right) \mathbf{a}_{1} + \left(\frac{1}{2} - y_{3}\right) \mathbf{a}_{2} - z_{3} \mathbf{a}_{3}$ | = | $\left(\frac{1}{2} + x_{3}\right) a \hat{\mathbf{x}} + \left(\frac{1}{2} - y_{3}\right) b \hat{\mathbf{y}} - z_{3} c \hat{\mathbf{z}}$ | (8*c*) | Cl III |
| $\mathbf{B}_{21}$ | = | $-x_{3} \mathbf{a}_{1} - y_{3} \mathbf{a}_{2} - z_{3} \mathbf{a}_{3}$ | = | $-x_{3} a \hat{\mathbf{x}} - y_{3} b \hat{\mathbf{y}} - z_{3} c \hat{\mathbf{z}}$ | (8*c*) | Cl III |
| $\mathbf{B}_{22}$ | = | $\left(\frac{1}{2} + x_{3}\right) \mathbf{a}_{1} + y_{3} \mathbf{a}_{2} + \left(\frac{1}{2} - z_{3}\right) \mathbf{a}_{3}$ | = | $\left(\frac{1}{2} + x_{3}\right) a \hat{\mathbf{x}} + y_{3} b \hat{\mathbf{y}} + \left(\frac{1}{2} - z_{3}\right) c \hat{\mathbf{z}}$ | (8*c*) | Cl III |
| $\mathbf{B}_{23}$ | = | $x_{3} \mathbf{a}_{1} + \left(\frac{1}{2} - y_{3}\right) \mathbf{a}_{2} + \left(\frac{1}{2} + z_{3}\right) \mathbf{a}_{3}$ | = | $x_{3} a \hat{\mathbf{x}} + \left(\frac{1}{2} - y_{3}\right) b \hat{\mathbf{y}} + \left(\frac{1}{2} + z_{3}\right) c \hat{\mathbf{z}}$ | (8*c*) | Cl III |
| $\mathbf{B}_{24}$ | = | $\left(\frac{1}{2} - x_{3}\right) \mathbf{a}_{1} + \left(\frac{1}{2} + y_{3}\right) \mathbf{a}_{2} + z_{3} \mathbf{a}_{3}$ | = | $\left(\frac{1}{2} - x_{3}\right) a \hat{\mathbf{x}} + \left(\frac{1}{2} + y_{3}\right) b \hat{\mathbf{y}} + z_{3} c \hat{\mathbf{z}}$ | (8*c*) | Cl III |
| $\mathbf{B}_{25}$ | = | $x_{4} \mathbf{a}_{1} + y_{4} \mathbf{a}_{2} + z_{4} \mathbf{a}_{3}$ | = | $x_{4} a \hat{\mathbf{x}} + y_{4} b \hat{\mathbf{y}} + z_{4} c \hat{\mathbf{z}}$ | (8*c*) | Cl IV |
| $\mathbf{B}_{26}$ | = | $\left(\frac{1}{2} - x_{4}\right) \mathbf{a}_{1} - y_{4} \mathbf{a}_{2} + \left(\frac{1}{2} + z_{4}\right) \mathbf{a}_{3}$ | = | $\left(\frac{1}{2} - x_{4}\right) a \hat{\mathbf{x}} - y_{4} b \hat{\mathbf{y}} + \left(\frac{1}{2} + z_{4}\right) c \hat{\mathbf{z}}$ | (8*c*) | Cl IV |
| $\mathbf{B}_{27}$ | = | $-x_{4} \mathbf{a}_{1} + \left(\frac{1}{2} + y_{4}\right) \mathbf{a}_{2} + \left(\frac{1}{2} - z_{4}\right) \mathbf{a}_{3}$ | = | $-x_{4} a \hat{\mathbf{x}} + \left(\frac{1}{2} + y_{4}\right) b \hat{\mathbf{y}} + \left(\frac{1}{2} - z_{4}\right) c \hat{\mathbf{z}}$ | (8*c*) | Cl IV |
| $\mathbf{B}_{28}$ | = | $\left(\frac{1}{2} + x_{4}\right) \mathbf{a}_{1} + \left(\frac{1}{2} - y_{4}\right) \mathbf{a}_{2} - z_{4} \mathbf{a}_{3}$ | = | $\left(\frac{1}{2} + x_{4}\right) a \hat{\mathbf{x}} + \left(\frac{1}{2} - y_{4}\right) b \hat{\mathbf{y}} - z_{4} c \hat{\mathbf{z}}$ | (8*c*) | Cl IV |
| $\mathbf{B}_{29}$ | = | $-x_{4} \mathbf{a}_{1} - y_{4} \mathbf{a}_{2} - z_{4} \mathbf{a}_{3}$ | = | $-x_{4} a \hat{\mathbf{x}} - y_{4} b \hat{\mathbf{y}} - z_{4} c \hat{\mathbf{z}}$ | (8*c*) | Cl IV |
| $\mathbf{B}_{30}$ | = | $\left(\frac{1}{2} + x_{4}\right) \mathbf{a}_{1} + y_{4} \mathbf{a}_{2} + \left(\frac{1}{2} - z_{4}\right) \mathbf{a}_{3}$ | = | $\left(\frac{1}{2} + x_{4}\right) a \hat{\mathbf{x}} + y_{4} b \hat{\mathbf{y}} + \left(\frac{1}{2} - z_{4}\right) c \hat{\mathbf{z}}$ | (8*c*) | Cl IV |
| $\mathbf{B}_{31}$ | = | $x_{4} \mathbf{a}_{1} + \left(\frac{1}{2} - y_{4}\right) \mathbf{a}_{2} + \left(\frac{1}{2} + z_{4}\right) \mathbf{a}_{3}$ | = | $x_{4} a \hat{\mathbf{x}} + \left(\frac{1}{2} - y_{4}\right) b \hat{\mathbf{y}} + \left(\frac{1}{2} + z_{4}\right) c \hat{\mathbf{z}}$ | (8*c*) | Cl IV |
| $\mathbf{B}_{32}$ | = | $\left(\frac{1}{2} - x_{4}\right) \mathbf{a}_{1} + \left(\frac{1}{2} + y_{4}\right) \mathbf{a}_{2} + z_{4} \mathbf{a}_{3}$ | = | $\left(\frac{1}{2} - x_{4}\right) a \hat{\mathbf{x}} + \left(\frac{1}{2} + y_{4}\right) b \hat{\mathbf{y}} + z_{4} c \hat{\mathbf{z}}$ | (8*c*) | Cl IV |
| $\mathbf{B}_{33}$ | = | $x_{5} \mathbf{a}_{1} + y_{5} \mathbf{a}_{2} + z_{5} \mathbf{a}_{3}$ | = | $x_{5} a \hat{\mathbf{x}} + y_{5} b \hat{\mathbf{y}} + z_{5} c \hat{\mathbf{z}}$ | (8*c*) | Cl V |
| $\mathbf{B}_{34}$ | = | $\left(\frac{1}{2} - x_{5}\right) \mathbf{a}_{1} - y_{5} \mathbf{a}_{2} + \left(\frac{1}{2} + z_{5}\right) \mathbf{a}_{3}$ | = | $\left(\frac{1}{2} - x_{5}\right) a \hat{\mathbf{x}} - y_{5} b \hat{\mathbf{y}} + \left(\frac{1}{2} + z_{5}\right) c \hat{\mathbf{z}}$ | (8*c*) | Cl V |
| $\mathbf{B}_{35}$ | = | $-x_{5} \mathbf{a}_{1} + \left(\frac{1}{2} + y_{5}\right) \mathbf{a}_{2} + \left(\frac{1}{2} - z_{5}\right) \mathbf{a}_{3}$ | = | $-x_{5} a \hat{\mathbf{x}} + \left(\frac{1}{2} + y_{5}\right) b \hat{\mathbf{y}} + \left(\frac{1}{2} - z_{5}\right) c \hat{\mathbf{z}}$ | (8*c*) | Cl V |
| $\mathbf{B}_{36}$ | = | $\left(\frac{1}{2} + x_{5}\right) \mathbf{a}_{1} + \left(\frac{1}{2} - y_{5}\right) \mathbf{a}_{2} - z_{5} \mathbf{a}_{3}$ | = | $\left(\frac{1}{2} + x_{5}\right) a \hat{\mathbf{x}} + \left(\frac{1}{2} - y_{5}\right) b \hat{\mathbf{y}} - z_{5} c \hat{\mathbf{z}}$ | (8*c*) | Cl V |
| $\mathbf{B}_{37}$ | = | $-x_{5} \mathbf{a}_{1} - y_{5} \mathbf{a}_{2} - z_{5} \mathbf{a}_{3}$ | = | $-x_{5} a \hat{\mathbf{x}} - y_{5} b \hat{\mathbf{y}} - z_{5} c \hat{\mathbf{z}}$ | (8*c*) | Cl V |
| $\mathbf{B}_{38}$ | = | $\left(\frac{1}{2}+x_5\right)\mathbf{a}_1+y_5\,\mathbf{a}_2+\left(\frac{1}{2}-z_5\right)\mathbf{a}_3$ | = | $\left(\frac{1}{2}+x_5\right)a\,\hat{\mathbf{x}}+y_5b\,\hat{\mathbf{y}}+\left(\frac{1}{2}-z_5\right)c\,\hat{\mathbf{z}}$ | (8$c$) | Cl V |
| $\mathbf{B}_{39}$ | = | $x_5\,\mathbf{a}_1+\left(\frac{1}{2}-y_5\right)\mathbf{a}_2+\left(\frac{1}{2}+z_5\right)\mathbf{a}_3$ | = | $x_5a\,\hat{\mathbf{x}}+\left(\frac{1}{2}-y_5\right)b\,\hat{\mathbf{y}}+\left(\frac{1}{2}+z_5\right)c\,\hat{\mathbf{z}}$ | (8$c$) | Cl V |
| $\mathbf{B}_{40}$ | = | $\left(\frac{1}{2}-x_5\right)\mathbf{a}_1+\left(\frac{1}{2}+y_5\right)\mathbf{a}_2+z_5\,\mathbf{a}_3$ | = | $\left(\frac{1}{2}-x_5\right)a\,\hat{\mathbf{x}}+\left(\frac{1}{2}+y_5\right)b\,\hat{\mathbf{y}}+z_5c\,\hat{\mathbf{z}}$ | (8$c$) | Cl V |
| $\mathbf{B}_{41}$ | = | $x_6\,\mathbf{a}_1+y_6\,\mathbf{a}_2+z_6\,\mathbf{a}_3$ | = | $x_6a\,\hat{\mathbf{x}}+y_6b\,\hat{\mathbf{y}}+z_6c\,\hat{\mathbf{z}}$ | (8$c$) | Cl VI |
| $\mathbf{B}_{42}$ | = | $\left(\frac{1}{2}-x_6\right)\mathbf{a}_1-y_6\,\mathbf{a}_2+\left(\frac{1}{2}+z_6\right)\mathbf{a}_3$ | = | $\left(\frac{1}{2}-x_6\right)a\,\hat{\mathbf{x}}-y_6b\,\hat{\mathbf{y}}+\left(\frac{1}{2}+z_6\right)c\,\hat{\mathbf{z}}$ | (8$c$) | Cl VI |
| $\mathbf{B}_{43}$ | = | $-x_6\,\mathbf{a}_1+\left(\frac{1}{2}+y_6\right)\mathbf{a}_2+\left(\frac{1}{2}-z_6\right)\mathbf{a}_3$ | = | $-x_6a\,\hat{\mathbf{x}}+\left(\frac{1}{2}+y_6\right)b\,\hat{\mathbf{y}}+\left(\frac{1}{2}-z_6\right)c\,\hat{\mathbf{z}}$ | (8$c$) | Cl VI |
| $\mathbf{B}_{44}$ | = | $\left(\frac{1}{2}+x_6\right)\mathbf{a}_1+\left(\frac{1}{2}-y_6\right)\mathbf{a}_2-z_6\,\mathbf{a}_3$ | = | $\left(\frac{1}{2}+x_6\right)a\,\hat{\mathbf{x}}+\left(\frac{1}{2}-y_6\right)b\,\hat{\mathbf{y}}-z_6c\,\hat{\mathbf{z}}$ | (8$c$) | Cl VI |
| $\mathbf{B}_{45}$ | = | $-x_6\,\mathbf{a}_1-y_6\,\mathbf{a}_2-z_6\,\mathbf{a}_3$ | = | $-x_6a\,\hat{\mathbf{x}}-y_6b\,\hat{\mathbf{y}}-z_6c\,\hat{\mathbf{z}}$ | (8$c$) | Cl VI |
| $\mathbf{B}_{46}$ | = | $\left(\frac{1}{2}+x_6\right)\mathbf{a}_1+y_6\,\mathbf{a}_2+\left(\frac{1}{2}-z_6\right)\mathbf{a}_3$ | = | $\left(\frac{1}{2}+x_6\right)a\,\hat{\mathbf{x}}+y_6b\,\hat{\mathbf{y}}+\left(\frac{1}{2}-z_6\right)c\,\hat{\mathbf{z}}$ | (8$c$) | Cl VI |
| $\mathbf{B}_{47}$ | = | $x_6\,\mathbf{a}_1+\left(\frac{1}{2}-y_6\right)\mathbf{a}_2+\left(\frac{1}{2}+z_6\right)\mathbf{a}_3$ | = | $x_6a\,\hat{\mathbf{x}}+\left(\frac{1}{2}-y_6\right)b\,\hat{\mathbf{y}}+\left(\frac{1}{2}+z_6\right)c\,\hat{\mathbf{z}}$ | (8$c$) | Cl VI |
| $\mathbf{B}_{48}$ | = | $\left(\frac{1}{2}-x_6\right)\mathbf{a}_1+\left(\frac{1}{2}+y_6\right)\mathbf{a}_2+z_6\,\mathbf{a}_3$ | = | $\left(\frac{1}{2}-x_6\right)a\,\hat{\mathbf{x}}+\left(\frac{1}{2}+y_6\right)b\,\hat{\mathbf{y}}+z_6c\,\hat{\mathbf{z}}$ | (8$c$) | Cl VI |
| $\mathbf{B}_{49}$ | = | $x_7\,\mathbf{a}_1+y_7\,\mathbf{a}_2+z_7\,\mathbf{a}_3$ | = | $x_7a\,\hat{\mathbf{x}}+y_7b\,\hat{\mathbf{y}}+z_7c\,\hat{\mathbf{z}}$ | (8$c$) | Cl VII |
| $\mathbf{B}_{50}$ | = | $\left(\frac{1}{2}-x_7\right)\mathbf{a}_1-y_7\,\mathbf{a}_2+\left(\frac{1}{2}+z_7\right)\mathbf{a}_3$ | = | $\left(\frac{1}{2}-x_7\right)a\,\hat{\mathbf{x}}-y_7b\,\hat{\mathbf{y}}+\left(\frac{1}{2}+z_7\right)c\,\hat{\mathbf{z}}$ | (8$c$) | Cl VII |
| $\mathbf{B}_{51}$ | = | $-x_7\,\mathbf{a}_1+\left(\frac{1}{2}+y_7\right)\mathbf{a}_2+\left(\frac{1}{2}-z_7\right)\mathbf{a}_3$ | = | $-x_7a\,\hat{\mathbf{x}}+\left(\frac{1}{2}+y_7\right)b\,\hat{\mathbf{y}}+\left(\frac{1}{2}-z_7\right)c\,\hat{\mathbf{z}}$ | (8$c$) | Cl VII |
| $\mathbf{B}_{52}$ | = | $\left(\frac{1}{2}+x_7\right)\mathbf{a}_1+\left(\frac{1}{2}-y_7\right)\mathbf{a}_2-z_7\,\mathbf{a}_3$ | = | $\left(\frac{1}{2}+x_7\right)a\,\hat{\mathbf{x}}+\left(\frac{1}{2}-y_7\right)b\,\hat{\mathbf{y}}-z_7c\,\hat{\mathbf{z}}$ | (8$c$) | Cl VII |
| $\mathbf{B}_{53}$ | = | $-x_7\,\mathbf{a}_1-y_7\,\mathbf{a}_2-z_7\,\mathbf{a}_3$ | = | $-x_7a\,\hat{\mathbf{x}}-y_7b\,\hat{\mathbf{y}}-z_7c\,\hat{\mathbf{z}}$ | (8$c$) | Cl VII |
| $\mathbf{B}_{54}$ | = | $\left(\frac{1}{2}+x_7\right)\mathbf{a}_1+y_7\,\mathbf{a}_2+\left(\frac{1}{2}-z_7\right)\mathbf{a}_3$ | = | $\left(\frac{1}{2}+x_7\right)a\,\hat{\mathbf{x}}+y_7b\,\hat{\mathbf{y}}+\left(\frac{1}{2}-z_7\right)c\,\hat{\mathbf{z}}$ | (8$c$) | Cl VII |
| $\mathbf{B}_{55}$ | = | $x_7\,\mathbf{a}_1+\left(\frac{1}{2}-y_7\right)\mathbf{a}_2+\left(\frac{1}{2}+z_7\right)\mathbf{a}_3$ | = | $x_7a\,\hat{\mathbf{x}}+\left(\frac{1}{2}-y_7\right)b\,\hat{\mathbf{y}}+\left(\frac{1}{2}+z_7\right)c\,\hat{\mathbf{z}}$ | (8$c$) | Cl VII |
| $\mathbf{B}_{56}$ | = | $\left(\frac{1}{2}-x_7\right)\mathbf{a}_1+\left(\frac{1}{2}+y_7\right)\mathbf{a}_2+z_7\,\mathbf{a}_3$ | = | $\left(\frac{1}{2}-x_7\right)a\,\hat{\mathbf{x}}+\left(\frac{1}{2}+y_7\right)b\,\hat{\mathbf{y}}+z_7c\,\hat{\mathbf{z}}$ | (8$c$) | Cl VII |
| $\mathbf{B}_{57}$ | = | $x_8\,\mathbf{a}_1+y_8\,\mathbf{a}_2+z_8\,\mathbf{a}_3$ | = | $x_8a\,\hat{\mathbf{x}}+y_8b\,\hat{\mathbf{y}}+z_8c\,\hat{\mathbf{z}}$ | (8$c$) | O |
| $\mathbf{B}_{58}$ | = | $\left(\frac{1}{2}-x_8\right)\mathbf{a}_1-y_8\,\mathbf{a}_2+\left(\frac{1}{2}+z_8\right)\mathbf{a}_3$ | = | $\left(\frac{1}{2}-x_8\right)a\,\hat{\mathbf{x}}-y_8b\,\hat{\mathbf{y}}+\left(\frac{1}{2}+z_8\right)c\,\hat{\mathbf{z}}$ | (8$c$) | O |
| $\mathbf{B}_{59}$ | = | $-x_8\,\mathbf{a}_1+\left(\frac{1}{2}+y_8\right)\mathbf{a}_2+\left(\frac{1}{2}-z_8\right)\mathbf{a}_3$ | = | $-x_8a\,\hat{\mathbf{x}}+\left(\frac{1}{2}+y_8\right)b\,\hat{\mathbf{y}}+\left(\frac{1}{2}-z_8\right)c\,\hat{\mathbf{z}}$ | (8$c$) | O |
| $\mathbf{B}_{60}$ | = | $\left(\frac{1}{2}+x_8\right)\mathbf{a}_1+\left(\frac{1}{2}-y_8\right)\mathbf{a}_2-z_8\,\mathbf{a}_3$ | = | $\left(\frac{1}{2}+x_8\right)a\,\hat{\mathbf{x}}+\left(\frac{1}{2}-y_8\right)b\,\hat{\mathbf{y}}-z_8c\,\hat{\mathbf{z}}$ | (8$c$) | O |
| $\mathbf{B}_{61}$ | = | $-x_8\,\mathbf{a}_1-y_8\,\mathbf{a}_2-z_8\,\mathbf{a}_3$ | = | $-x_8a\,\hat{\mathbf{x}}-y_8b\,\hat{\mathbf{y}}-z_8c\,\hat{\mathbf{z}}$ | (8$c$) | O |
| $\mathbf{B}_{62}$ | = | $\left(\frac{1}{2}+x_8\right)\mathbf{a}_1+y_8\,\mathbf{a}_2+\left(\frac{1}{2}-z_8\right)\mathbf{a}_3$ | = | $\left(\frac{1}{2}+x_8\right)a\,\hat{\mathbf{x}}+y_8b\,\hat{\mathbf{y}}+\left(\frac{1}{2}-z_8\right)c\,\hat{\mathbf{z}}$ | (8$c$) | O |
| $\mathbf{B}_{63}$ | = | $x_8\,\mathbf{a}_1+\left(\frac{1}{2}-y_8\right)\mathbf{a}_2+\left(\frac{1}{2}+z_8\right)\mathbf{a}_3$ | = | $x_8a\,\hat{\mathbf{x}}+\left(\frac{1}{2}-y_8\right)b\,\hat{\mathbf{y}}+\left(\frac{1}{2}+z_8\right)c\,\hat{\mathbf{z}}$ | (8$c$) | O |
| $\mathbf{B}_{64}$ | = | $\left(\frac{1}{2}-x_8\right)\mathbf{a}_1+\left(\frac{1}{2}+y_8\right)\mathbf{a}_2+z_8\,\mathbf{a}_3$ | = | $\left(\frac{1}{2}-x_8\right)a\,\hat{\mathbf{x}}+\left(\frac{1}{2}+y_8\right)b\,\hat{\mathbf{y}}+z_8c\,\hat{\mathbf{z}}$ | (8$c$) | O |
| $\mathbf{B}_{65}$ | = | $x_9\,\mathbf{a}_1+y_9\,\mathbf{a}_2+z_9\,\mathbf{a}_3$ | = | $x_9a\,\hat{\mathbf{x}}+y_9b\,\hat{\mathbf{y}}+z_9c\,\hat{\mathbf{z}}$ | (8$c$) | P |
| $\mathbf{B}_{66}$ | = | $\left(\frac{1}{2}-x_9\right)\mathbf{a}_1-y_9\,\mathbf{a}_2+\left(\frac{1}{2}+z_9\right)\mathbf{a}_3$ | = | $\left(\frac{1}{2}-x_9\right)a\,\hat{\mathbf{x}}-y_9b\,\hat{\mathbf{y}}+\left(\frac{1}{2}+z_9\right)c\,\hat{\mathbf{z}}$ | (8$c$) | P |
| $\mathbf{B}_{67}$ | = | $-x_9\,\mathbf{a}_1+\left(\frac{1}{2}+y_9\right)\mathbf{a}_2+\left(\frac{1}{2}-z_9\right)\mathbf{a}_3$ | = | $-x_9a\,\hat{\mathbf{x}}+\left(\frac{1}{2}+y_9\right)b\,\hat{\mathbf{y}}+\left(\frac{1}{2}-z_9\right)c\,\hat{\mathbf{z}}$ | (8$c$) | P |
| $\mathbf{B}_{68}$ | = | $\left(\frac{1}{2}+x_9\right)\mathbf{a}_1+\left(\frac{1}{2}-y_9\right)\mathbf{a}_2-z_9\,\mathbf{a}_3$ | = | $\left(\frac{1}{2}+x_9\right)a\,\hat{\mathbf{x}}+\left(\frac{1}{2}-y_9\right)b\,\hat{\mathbf{y}}-z_9c\,\hat{\mathbf{z}}$ | (8$c$) | P |
| $\mathbf{B}_{69}$ | = | $-x_9\,\mathbf{a}_1-y_9\,\mathbf{a}_2-z_9\,\mathbf{a}_3$ | = | $-x_9a\,\hat{\mathbf{x}}-y_9b\,\hat{\mathbf{y}}-z_9c\,\hat{\mathbf{z}}$ | (8$c$) | P |
| $\mathbf{B}_{70}$ | = | $\left(\frac{1}{2}+x_9\right)\mathbf{a}_1+y_9\,\mathbf{a}_2+\left(\frac{1}{2}-z_9\right)\mathbf{a}_3$ | = | $\left(\frac{1}{2}+x_9\right)a\,\hat{\mathbf{x}}+y_9b\,\hat{\mathbf{y}}+\left(\frac{1}{2}-z_9\right)c\,\hat{\mathbf{z}}$ | (8$c$) | P |
| $\mathbf{B}_{71}$ | = | $x_9\,\mathbf{a}_1+\left(\frac{1}{2}-y_9\right)\mathbf{a}_2+\left(\frac{1}{2}+z_9\right)\mathbf{a}_3$ | = | $x_9a\,\hat{\mathbf{x}}+\left(\frac{1}{2}-y_9\right)b\,\hat{\mathbf{y}}+\left(\frac{1}{2}+z_9\right)c\,\hat{\mathbf{z}}$ | (8$c$) | P |
| $\mathbf{B}_{72}$ | = | $\left(\frac{1}{2}-x_9\right)\mathbf{a}_1+\left(\frac{1}{2}+y_9\right)\mathbf{a}_2+z_9\,\mathbf{a}_3$ | = | $\left(\frac{1}{2}-x_9\right)a\,\hat{\mathbf{x}}+\left(\frac{1}{2}+y_9\right)b\,\hat{\mathbf{y}}+z_9c\,\hat{\mathbf{z}}$ | (8$c$) | P |
| $\mathbf{B}_{73}$ | = | $x_{10}\,\mathbf{a}_1+y_{10}\,\mathbf{a}_2+z_{10}\,\mathbf{a}_3$ | = | $x_{10}a\,\hat{\mathbf{x}}+y_{10}b\,\hat{\mathbf{y}}+z_{10}c\,\hat{\mathbf{z}}$ | (8$c$) | Ti |
| $\mathbf{B}_{74}$ | = | $\left(\frac{1}{2} - x_{10}\right) \mathbf{a}_1 - y_{10} \mathbf{a}_2 + \left(\frac{1}{2} + z_{10}\right) \mathbf{a}_3$ | = | $\left(\frac{1}{2} - x_{10}\right) a \hat{\mathbf{x}} - y_{10} b \hat{\mathbf{y}} + \left(\frac{1}{2} + z_{10}\right) c \hat{\mathbf{z}}$ | (8$c$) | Ti |
| $\mathbf{B}_{75}$ | = | $-x_{10} \mathbf{a}_1 + \left(\frac{1}{2} + y_{10}\right) \mathbf{a}_2 + \left(\frac{1}{2} - z_{10}\right) \mathbf{a}_3$ | = | $-x_{10} a \hat{\mathbf{x}} + \left(\frac{1}{2} + y_{10}\right) b \hat{\mathbf{y}} + \left(\frac{1}{2} - z_{10}\right) c \hat{\mathbf{z}}$ | (8$c$) | Ti |
| $\mathbf{B}_{76}$ | = | $\left(\frac{1}{2} + x_{10}\right) \mathbf{a}_1 + \left(\frac{1}{2} - y_{10}\right) \mathbf{a}_2 - z_{10} \mathbf{a}_3$ | = | $\left(\frac{1}{2} + x_{10}\right) a \hat{\mathbf{x}} + \left(\frac{1}{2} - y_{10}\right) b \hat{\mathbf{y}} - z_{10} c \hat{\mathbf{z}}$ | (8$c$) | Ti |
| $\mathbf{B}_{77}$ | = | $-x_{10} \mathbf{a}_1 - y_{10} \mathbf{a}_2 - z_{10} \mathbf{a}_3$ | = | $-x_{10} a \hat{\mathbf{x}} - y_{10} b \hat{\mathbf{y}} - z_{10} c \hat{\mathbf{z}}$ | (8$c$) | Ti |
| $\mathbf{B}_{78}$ | = | $\left(\frac{1}{2} + x_{10}\right) \mathbf{a}_1 + y_{10} \mathbf{a}_2 + \left(\frac{1}{2} - z_{10}\right) \mathbf{a}_3$ | = | $\left(\frac{1}{2} + x_{10}\right) a \hat{\mathbf{x}} + y_{10} b \hat{\mathbf{y}} + \left(\frac{1}{2} - z_{10}\right) c \hat{\mathbf{z}}$ | (8$c$) | Ti |
| $\mathbf{B}_{79}$ | = | $x_{10} \mathbf{a}_1 + \left(\frac{1}{2} - y_{10}\right) \mathbf{a}_2 + \left(\frac{1}{2} + z_{10}\right) \mathbf{a}_3$ | = | $x_{10} a \hat{\mathbf{x}} + \left(\frac{1}{2} - y_{10}\right) b \hat{\mathbf{y}} + \left(\frac{1}{2} + z_{10}\right) c \hat{\mathbf{z}}$ | (8$c$) | Ti |
| $\mathbf{B}_{80}$ | = | $\left(\frac{1}{2} - x_{10}\right) \mathbf{a}_1 + \left(\frac{1}{2} + y_{10}\right) \mathbf{a}_2 + z_{10} \mathbf{a}_3$ | = | $\left(\frac{1}{2} - x_{10}\right) a \hat{\mathbf{x}} + \left(\frac{1}{2} + y_{10}\right) b \hat{\mathbf{y}} + z_{10} c \hat{\mathbf{z}}$ | (8$c$) | Ti |

---

---

**Geometry files:**

- CIF: pp. 1626
- POSCAR: pp. 1626

# Hambergite [$Be_2BO_3(OH)$ ($G7_2$)] Structure: AB2CD4_oP64_61_c_2c_c_4c

http://aflow.org/prototype-encyclopedia/AB2CD4_oP64_61_c_2c_c_4c

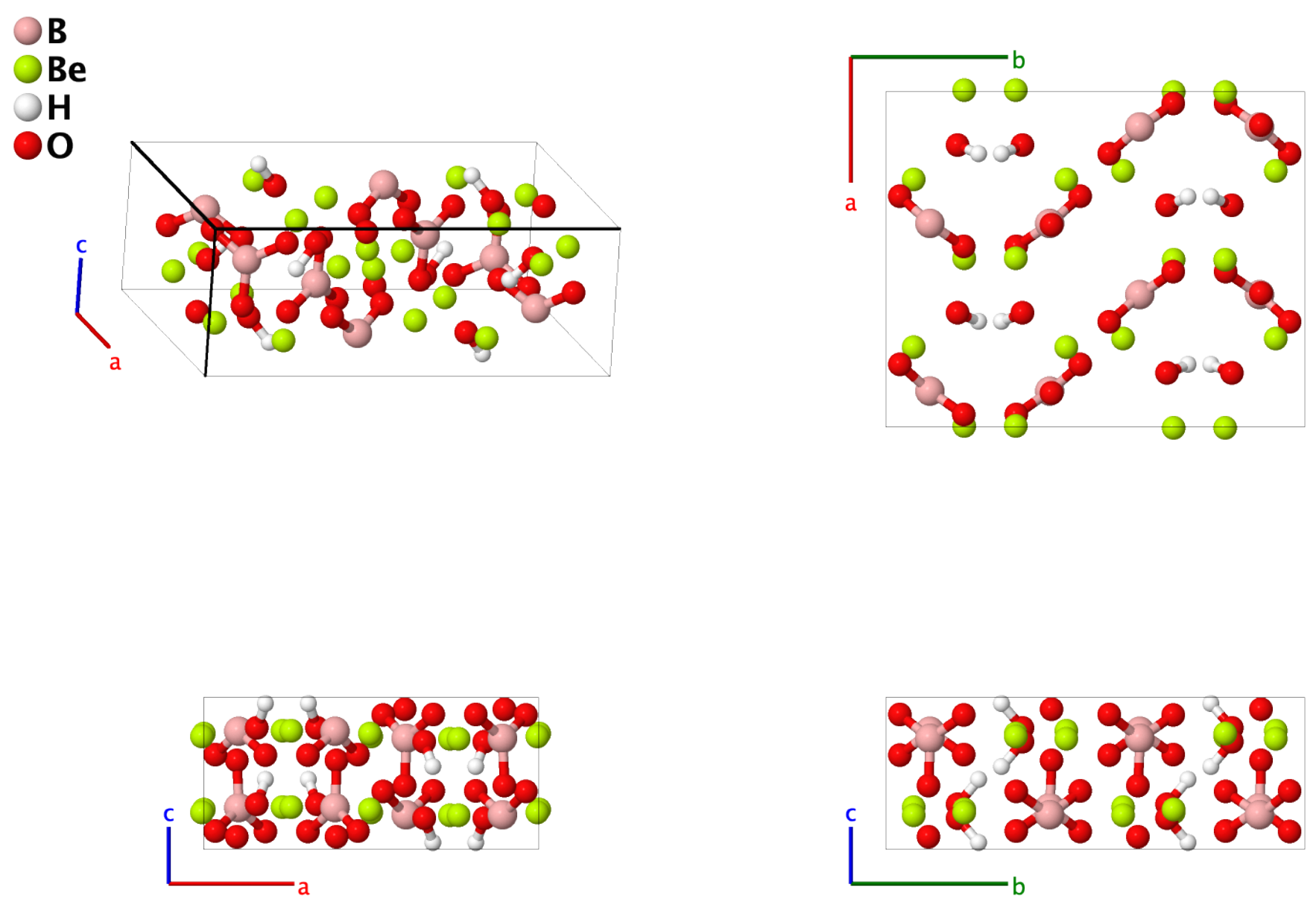

| | | |
|---|---|---|
| **Prototype** | : | $BBe_2HO_4$ |
| **AFLOW prototype label** | : | AB2CD4_oP64_61_c_2c_c_4c |
| ***Strukturbericht* designation** | : | $G7_2$ |
| **Pearson symbol** | : | oP64 |
| **Space group number** | : | 61 |
| **Space group symbol** | : | $Pbca$ |
| **AFLOW prototype command** | : | `aflow --proto=AB2CD4_oP64_61_c_2c_c_4c --params=`$a, b/a, c/a, x_1, y_1, z_1, x_2, y_2, z_2, x_3, y_3, z_3, x_4, y_4, z_4, x_5, y_5, z_5, x_6, y_6, z_6, x_7, y_7, z_7, x_8, y_8, z_8$ |

- The original description of this structure in (Hermann, 1937) did not include the positions of the hydrogen atoms, instead using OH for the O(IV)-H radical.
- The sample actually measured by (Gatta, 2012) has 4% of the OH radicals replaced by fluorine.

**Simple Orthorhombic primitive vectors:**

$$\mathbf{a}_1 = a\,\hat{\mathbf{x}}$$
$$\mathbf{a}_2 = b\,\hat{\mathbf{y}}$$
$$\mathbf{a}_3 = c\,\hat{\mathbf{z}}$$

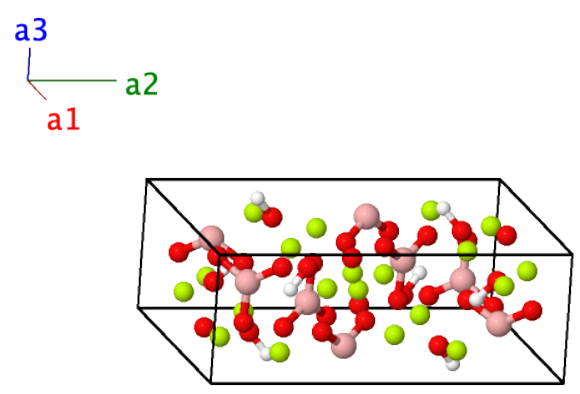

**Basis vectors:**

| | | Lattice Coordinates | | Cartesian Coordinates | Wyckoff Position | Atom Type |
|---|---|---|---|---|---|---|
| $\mathbf{B}_{1}$ | = | $x_{1}\,\mathbf{a}_{1}+y_{1}\,\mathbf{a}_{2}+z_{1}\,\mathbf{a}_{3}$ | = | $x_{1}a\,\hat{\mathbf{x}}+y_{1}b\,\hat{\mathbf{y}}+z_{1}c\,\hat{\mathbf{z}}$ | (8$c$) | B |
| $\mathbf{B}_{2}$ | = | $\left(\frac{1}{2}-x_{1}\right)\mathbf{a}_{1}-y_{1}\,\mathbf{a}_{2}+\left(\frac{1}{2}+z_{1}\right)\mathbf{a}_{3}$ | = | $\left(\frac{1}{2}-x_{1}\right)a\,\hat{\mathbf{x}}-y_{1}b\,\hat{\mathbf{y}}+\left(\frac{1}{2}+z_{1}\right)c\,\hat{\mathbf{z}}$ | (8$c$) | B |
| $\mathbf{B}_{3}$ | = | $-x_{1}\,\mathbf{a}_{1}+\left(\frac{1}{2}+y_{1}\right)\mathbf{a}_{2}+\left(\frac{1}{2}-z_{1}\right)\mathbf{a}_{3}$ | = | $-x_{1}a\,\hat{\mathbf{x}}+\left(\frac{1}{2}+y_{1}\right)b\,\hat{\mathbf{y}}+\left(\frac{1}{2}-z_{1}\right)c\,\hat{\mathbf{z}}$ | (8$c$) | B |
| $\mathbf{B}_{4}$ | = | $\left(\frac{1}{2}+x_{1}\right)\mathbf{a}_{1}+\left(\frac{1}{2}-y_{1}\right)\mathbf{a}_{2}-z_{1}\,\mathbf{a}_{3}$ | = | $\left(\frac{1}{2}+x_{1}\right)a\,\hat{\mathbf{x}}+\left(\frac{1}{2}-y_{1}\right)b\,\hat{\mathbf{y}}-z_{1}c\,\hat{\mathbf{z}}$ | (8$c$) | B |
| $\mathbf{B}_{5}$ | = | $-x_{1}\,\mathbf{a}_{1}-y_{1}\,\mathbf{a}_{2}-z_{1}\,\mathbf{a}_{3}$ | = | $-x_{1}a\,\hat{\mathbf{x}}-y_{1}b\,\hat{\mathbf{y}}-z_{1}c\,\hat{\mathbf{z}}$ | (8$c$) | B |
| $\mathbf{B}_{6}$ | = | $\left(\frac{1}{2}+x_{1}\right)\mathbf{a}_{1}+y_{1}\,\mathbf{a}_{2}+\left(\frac{1}{2}-z_{1}\right)\mathbf{a}_{3}$ | = | $\left(\frac{1}{2}+x_{1}\right)a\,\hat{\mathbf{x}}+y_{1}b\,\hat{\mathbf{y}}+\left(\frac{1}{2}-z_{1}\right)c\,\hat{\mathbf{z}}$ | (8$c$) | B |
| $\mathbf{B}_{7}$ | = | $x_{1}\,\mathbf{a}_{1}+\left(\frac{1}{2}-y_{1}\right)\mathbf{a}_{2}+\left(\frac{1}{2}+z_{1}\right)\mathbf{a}_{3}$ | = | $x_{1}a\,\hat{\mathbf{x}}+\left(\frac{1}{2}-y_{1}\right)b\,\hat{\mathbf{y}}+\left(\frac{1}{2}+z_{1}\right)c\,\hat{\mathbf{z}}$ | (8$c$) | B |
| $\mathbf{B}_{8}$ | = | $\left(\frac{1}{2}-x_{1}\right)\mathbf{a}_{1}+\left(\frac{1}{2}+y_{1}\right)\mathbf{a}_{2}+z_{1}\,\mathbf{a}_{3}$ | = | $\left(\frac{1}{2}-x_{1}\right)a\,\hat{\mathbf{x}}+\left(\frac{1}{2}+y_{1}\right)b\,\hat{\mathbf{y}}+z_{1}c\,\hat{\mathbf{z}}$ | (8$c$) | B |
| $\mathbf{B}_{9}$ | = | $x_{2}\,\mathbf{a}_{1}+y_{2}\,\mathbf{a}_{2}+z_{2}\,\mathbf{a}_{3}$ | = | $x_{2}a\,\hat{\mathbf{x}}+y_{2}b\,\hat{\mathbf{y}}+z_{2}c\,\hat{\mathbf{z}}$ | (8$c$) | Be I |
| $\mathbf{B}_{10}$ | = | $\left(\frac{1}{2}-x_{2}\right)\mathbf{a}_{1}-y_{2}\,\mathbf{a}_{2}+\left(\frac{1}{2}+z_{2}\right)\mathbf{a}_{3}$ | = | $\left(\frac{1}{2}-x_{2}\right)a\,\hat{\mathbf{x}}-y_{2}b\,\hat{\mathbf{y}}+\left(\frac{1}{2}+z_{2}\right)c\,\hat{\mathbf{z}}$ | (8$c$) | Be I |
| $\mathbf{B}_{11}$ | = | $-x_{2}\,\mathbf{a}_{1}+\left(\frac{1}{2}+y_{2}\right)\mathbf{a}_{2}+\left(\frac{1}{2}-z_{2}\right)\mathbf{a}_{3}$ | = | $-x_{2}a\,\hat{\mathbf{x}}+\left(\frac{1}{2}+y_{2}\right)b\,\hat{\mathbf{y}}+\left(\frac{1}{2}-z_{2}\right)c\,\hat{\mathbf{z}}$ | (8$c$) | Be I |
| $\mathbf{B}_{12}$ | = | $\left(\frac{1}{2}+x_{2}\right)\mathbf{a}_{1}+\left(\frac{1}{2}-y_{2}\right)\mathbf{a}_{2}-z_{2}\,\mathbf{a}_{3}$ | = | $\left(\frac{1}{2}+x_{2}\right)a\,\hat{\mathbf{x}}+\left(\frac{1}{2}-y_{2}\right)b\,\hat{\mathbf{y}}-z_{2}c\,\hat{\mathbf{z}}$ | (8$c$) | Be I |
| $\mathbf{B}_{13}$ | = | $-x_{2}\,\mathbf{a}_{1}-y_{2}\,\mathbf{a}_{2}-z_{2}\,\mathbf{a}_{3}$ | = | $-x_{2}a\,\hat{\mathbf{x}}-y_{2}b\,\hat{\mathbf{y}}-z_{2}c\,\hat{\mathbf{z}}$ | (8$c$) | Be I |
| $\mathbf{B}_{14}$ | = | $\left(\frac{1}{2}+x_{2}\right)\mathbf{a}_{1}+y_{2}\,\mathbf{a}_{2}+\left(\frac{1}{2}-z_{2}\right)\mathbf{a}_{3}$ | = | $\left(\frac{1}{2}+x_{2}\right)a\,\hat{\mathbf{x}}+y_{2}b\,\hat{\mathbf{y}}+\left(\frac{1}{2}-z_{2}\right)c\,\hat{\mathbf{z}}$ | (8$c$) | Be I |
| $\mathbf{B}_{15}$ | = | $x_{2}\,\mathbf{a}_{1}+\left(\frac{1}{2}-y_{2}\right)\mathbf{a}_{2}+\left(\frac{1}{2}+z_{2}\right)\mathbf{a}_{3}$ | = | $x_{2}a\,\hat{\mathbf{x}}+\left(\frac{1}{2}-y_{2}\right)b\,\hat{\mathbf{y}}+\left(\frac{1}{2}+z_{2}\right)c\,\hat{\mathbf{z}}$ | (8$c$) | Be I |
| $\mathbf{B}_{16}$ | = | $\left(\frac{1}{2}-x_{2}\right)\mathbf{a}_{1}+\left(\frac{1}{2}+y_{2}\right)\mathbf{a}_{2}+z_{2}\,\mathbf{a}_{3}$ | = | $\left(\frac{1}{2}-x_{2}\right)a\,\hat{\mathbf{x}}+\left(\frac{1}{2}+y_{2}\right)b\,\hat{\mathbf{y}}+z_{2}c\,\hat{\mathbf{z}}$ | (8$c$) | Be I |
| $\mathbf{B}_{17}$ | = | $x_{3}\,\mathbf{a}_{1}+y_{3}\,\mathbf{a}_{2}+z_{3}\,\mathbf{a}_{3}$ | = | $x_{3}a\,\hat{\mathbf{x}}+y_{3}b\,\hat{\mathbf{y}}+z_{3}c\,\hat{\mathbf{z}}$ | (8$c$) | Be II |
| $\mathbf{B}_{18}$ | = | $\left(\frac{1}{2}-x_{3}\right)\mathbf{a}_{1}-y_{3}\,\mathbf{a}_{2}+\left(\frac{1}{2}+z_{3}\right)\mathbf{a}_{3}$ | = | $\left(\frac{1}{2}-x_{3}\right)a\,\hat{\mathbf{x}}-y_{3}b\,\hat{\mathbf{y}}+\left(\frac{1}{2}+z_{3}\right)c\,\hat{\mathbf{z}}$ | (8$c$) | Be II |
| $\mathbf{B}_{19}$ | = | $-x_{3}\,\mathbf{a}_{1}+\left(\frac{1}{2}+y_{3}\right)\mathbf{a}_{2}+\left(\frac{1}{2}-z_{3}\right)\mathbf{a}_{3}$ | = | $-x_{3}a\,\hat{\mathbf{x}}+\left(\frac{1}{2}+y_{3}\right)b\,\hat{\mathbf{y}}+\left(\frac{1}{2}-z_{3}\right)c\,\hat{\mathbf{z}}$ | (8$c$) | Be II |
| $\mathbf{B}_{20}$ | = | $\left(\frac{1}{2}+x_{3}\right)\mathbf{a}_{1}+\left(\frac{1}{2}-y_{3}\right)\mathbf{a}_{2}-z_{3}\,\mathbf{a}_{3}$ | = | $\left(\frac{1}{2}+x_{3}\right)a\,\hat{\mathbf{x}}+\left(\frac{1}{2}-y_{3}\right)b\,\hat{\mathbf{y}}-z_{3}c\,\hat{\mathbf{z}}$ | (8$c$) | Be II |
| $\mathbf{B}_{21}$ | = | $-x_{3}\,\mathbf{a}_{1}-y_{3}\,\mathbf{a}_{2}-z_{3}\,\mathbf{a}_{3}$ | = | $-x_{3}a\,\hat{\mathbf{x}}-y_{3}b\,\hat{\mathbf{y}}-z_{3}c\,\hat{\mathbf{z}}$ | (8$c$) | Be II |
| $\mathbf{B}_{22}$ | = | $\left(\frac{1}{2}+x_{3}\right)\mathbf{a}_{1}+y_{3}\,\mathbf{a}_{2}+\left(\frac{1}{2}-z_{3}\right)\mathbf{a}_{3}$ | = | $\left(\frac{1}{2}+x_{3}\right)a\,\hat{\mathbf{x}}+y_{3}b\,\hat{\mathbf{y}}+\left(\frac{1}{2}-z_{3}\right)c\,\hat{\mathbf{z}}$ | (8$c$) | Be II |
| $\mathbf{B}_{23}$ | = | $x_{3}\,\mathbf{a}_{1}+\left(\frac{1}{2}-y_{3}\right)\mathbf{a}_{2}+\left(\frac{1}{2}+z_{3}\right)\mathbf{a}_{3}$ | = | $x_{3}a\,\hat{\mathbf{x}}+\left(\frac{1}{2}-y_{3}\right)b\,\hat{\mathbf{y}}+\left(\frac{1}{2}+z_{3}\right)c\,\hat{\mathbf{z}}$ | (8$c$) | Be II |
| $\mathbf{B}_{24}$ | = | $\left(\frac{1}{2}-x_{3}\right)\mathbf{a}_{1}+\left(\frac{1}{2}+y_{3}\right)\mathbf{a}_{2}+z_{3}\,\mathbf{a}_{3}$ | = | $\left(\frac{1}{2}-x_{3}\right)a\,\hat{\mathbf{x}}+\left(\frac{1}{2}+y_{3}\right)b\,\hat{\mathbf{y}}+z_{3}c\,\hat{\mathbf{z}}$ | (8$c$) | Be II |
| $\mathbf{B}_{25}$ | = | $x_{4}\,\mathbf{a}_{1}+y_{4}\,\mathbf{a}_{2}+z_{4}\,\mathbf{a}_{3}$ | = | $x_{4}a\,\hat{\mathbf{x}}+y_{4}b\,\hat{\mathbf{y}}+z_{4}c\,\hat{\mathbf{z}}$ | (8$c$) | H |
| $\mathbf{B}_{26}$ | = | $\left(\frac{1}{2}-x_{4}\right)\mathbf{a}_{1}-y_{4}\,\mathbf{a}_{2}+\left(\frac{1}{2}+z_{4}\right)\mathbf{a}_{3}$ | = | $\left(\frac{1}{2}-x_{4}\right)a\,\hat{\mathbf{x}}-y_{4}b\,\hat{\mathbf{y}}+\left(\frac{1}{2}+z_{4}\right)c\,\hat{\mathbf{z}}$ | (8$c$) | H |
| $\mathbf{B}_{27}$ | = | $-x_{4}\,\mathbf{a}_{1}+\left(\frac{1}{2}+y_{4}\right)\mathbf{a}_{2}+\left(\frac{1}{2}-z_{4}\right)\mathbf{a}_{3}$ | = | $-x_{4}a\,\hat{\mathbf{x}}+\left(\frac{1}{2}+y_{4}\right)b\,\hat{\mathbf{y}}+\left(\frac{1}{2}-z_{4}\right)c\,\hat{\mathbf{z}}$ | (8$c$) | H |
| $\mathbf{B}_{28}$ | = | $\left(\frac{1}{2}+x_{4}\right)\mathbf{a}_{1}+\left(\frac{1}{2}-y_{4}\right)\mathbf{a}_{2}-z_{4}\,\mathbf{a}_{3}$ | = | $\left(\frac{1}{2}+x_{4}\right)a\,\hat{\mathbf{x}}+\left(\frac{1}{2}-y_{4}\right)b\,\hat{\mathbf{y}}-z_{4}c\,\hat{\mathbf{z}}$ | (8$c$) | H |
| $\mathbf{B}_{29}$ | = | $-x_{4}\,\mathbf{a}_{1}-y_{4}\,\mathbf{a}_{2}-z_{4}\,\mathbf{a}_{3}$ | = | $-x_{4}a\,\hat{\mathbf{x}}-y_{4}b\,\hat{\mathbf{y}}-z_{4}c\,\hat{\mathbf{z}}$ | (8$c$) | H |
| $\mathbf{B}_{30}$ | = | $\left(\frac{1}{2}+x_{4}\right)\mathbf{a}_{1}+y_{4}\,\mathbf{a}_{2}+\left(\frac{1}{2}-z_{4}\right)\mathbf{a}_{3}$ | = | $\left(\frac{1}{2}+x_{4}\right)a\,\hat{\mathbf{x}}+y_{4}b\,\hat{\mathbf{y}}+\left(\frac{1}{2}-z_{4}\right)c\,\hat{\mathbf{z}}$ | (8$c$) | H |
| $\mathbf{B}_{31}$ | = | $x_{4}\,\mathbf{a}_{1}+\left(\frac{1}{2}-y_{4}\right)\mathbf{a}_{2}+\left(\frac{1}{2}+z_{4}\right)\mathbf{a}_{3}$ | = | $x_{4}a\,\hat{\mathbf{x}}+\left(\frac{1}{2}-y_{4}\right)b\,\hat{\mathbf{y}}+\left(\frac{1}{2}+z_{4}\right)c\,\hat{\mathbf{z}}$ | (8$c$) | H |
| $\mathbf{B}_{32}$ | = | $\left(\frac{1}{2}-x_{4}\right)\mathbf{a}_{1}+\left(\frac{1}{2}+y_{4}\right)\mathbf{a}_{2}+z_{4}\,\mathbf{a}_{3}$ | = | $\left(\frac{1}{2}-x_{4}\right)a\,\hat{\mathbf{x}}+\left(\frac{1}{2}+y_{4}\right)b\,\hat{\mathbf{y}}+z_{4}c\,\hat{\mathbf{z}}$ | (8$c$) | H |
| $\mathbf{B}_{33}$ | = | $x_{5}\,\mathbf{a}_{1}+y_{5}\,\mathbf{a}_{2}+z_{5}\,\mathbf{a}_{3}$ | = | $x_{5}a\,\hat{\mathbf{x}}+y_{5}b\,\hat{\mathbf{y}}+z_{5}c\,\hat{\mathbf{z}}$ | (8$c$) | O I |
| $\mathbf{B}_{34}$ | = | $\left(\frac{1}{2}-x_{5}\right)\mathbf{a}_{1}-y_{5}\,\mathbf{a}_{2}+\left(\frac{1}{2}+z_{5}\right)\mathbf{a}_{3}$ | = | $\left(\frac{1}{2}-x_{5}\right)a\,\hat{\mathbf{x}}-y_{5}b\,\hat{\mathbf{y}}+\left(\frac{1}{2}+z_{5}\right)c\,\hat{\mathbf{z}}$ | (8$c$) | O I |
| $\mathbf{B}_{35}$ | = | $-x_{5}\,\mathbf{a}_{1}+\left(\frac{1}{2}+y_{5}\right)\mathbf{a}_{2}+\left(\frac{1}{2}-z_{5}\right)\mathbf{a}_{3}$ | = | $-x_{5}a\,\hat{\mathbf{x}}+\left(\frac{1}{2}+y_{5}\right)b\,\hat{\mathbf{y}}+\left(\frac{1}{2}-z_{5}\right)c\,\hat{\mathbf{z}}$ | (8$c$) | O I |
| $\mathbf{B}_{36}$ | = | $\left(\frac{1}{2}+x_5\right)\mathbf{a}_1+\left(\frac{1}{2}-y_5\right)\mathbf{a}_2-z_5\,\mathbf{a}_3$ | = | $\left(\frac{1}{2}+x_5\right)a\,\hat{\mathbf{x}}+\left(\frac{1}{2}-y_5\right)b\,\hat{\mathbf{y}}-z_5c\,\hat{\mathbf{z}}$ | (8c) | O I |
| $\mathbf{B}_{37}$ | = | $-x_5\,\mathbf{a}_1-y_5\,\mathbf{a}_2-z_5\,\mathbf{a}_3$ | = | $-x_5a\,\hat{\mathbf{x}}-y_5b\,\hat{\mathbf{y}}-z_5c\,\hat{\mathbf{z}}$ | (8c) | O I |
| $\mathbf{B}_{38}$ | = | $\left(\frac{1}{2}+x_5\right)\mathbf{a}_1+y_5\,\mathbf{a}_2+\left(\frac{1}{2}-z_5\right)\mathbf{a}_3$ | = | $\left(\frac{1}{2}+x_5\right)a\,\hat{\mathbf{x}}+y_5b\,\hat{\mathbf{y}}+\left(\frac{1}{2}-z_5\right)c\,\hat{\mathbf{z}}$ | (8c) | O I |
| $\mathbf{B}_{39}$ | = | $x_5\,\mathbf{a}_1+\left(\frac{1}{2}-y_5\right)\mathbf{a}_2+\left(\frac{1}{2}+z_5\right)\mathbf{a}_3$ | = | $x_5a\,\hat{\mathbf{x}}+\left(\frac{1}{2}-y_5\right)b\,\hat{\mathbf{y}}+\left(\frac{1}{2}+z_5\right)c\,\hat{\mathbf{z}}$ | (8c) | O I |
| $\mathbf{B}_{40}$ | = | $\left(\frac{1}{2}-x_5\right)\mathbf{a}_1+\left(\frac{1}{2}+y_5\right)\mathbf{a}_2+z_5\,\mathbf{a}_3$ | = | $\left(\frac{1}{2}-x_5\right)a\,\hat{\mathbf{x}}+\left(\frac{1}{2}+y_5\right)b\,\hat{\mathbf{y}}+z_5c\,\hat{\mathbf{z}}$ | (8c) | O I |
| $\mathbf{B}_{41}$ | = | $x_6\,\mathbf{a}_1+y_6\,\mathbf{a}_2+z_6\,\mathbf{a}_3$ | = | $x_6a\,\hat{\mathbf{x}}+y_6b\,\hat{\mathbf{y}}+z_6c\,\hat{\mathbf{z}}$ | (8c) | O II |
| $\mathbf{B}_{42}$ | = | $\left(\frac{1}{2}-x_6\right)\mathbf{a}_1-y_6\,\mathbf{a}_2+\left(\frac{1}{2}+z_6\right)\mathbf{a}_3$ | = | $\left(\frac{1}{2}-x_6\right)a\,\hat{\mathbf{x}}-y_6b\,\hat{\mathbf{y}}+\left(\frac{1}{2}+z_6\right)c\,\hat{\mathbf{z}}$ | (8c) | O II |
| $\mathbf{B}_{43}$ | = | $-x_6\,\mathbf{a}_1+\left(\frac{1}{2}+y_6\right)\mathbf{a}_2+\left(\frac{1}{2}-z_6\right)\mathbf{a}_3$ | = | $-x_6a\,\hat{\mathbf{x}}+\left(\frac{1}{2}+y_6\right)b\,\hat{\mathbf{y}}+\left(\frac{1}{2}-z_6\right)c\,\hat{\mathbf{z}}$ | (8c) | O II |
| $\mathbf{B}_{44}$ | = | $\left(\frac{1}{2}+x_6\right)\mathbf{a}_1+\left(\frac{1}{2}-y_6\right)\mathbf{a}_2-z_6\,\mathbf{a}_3$ | = | $\left(\frac{1}{2}+x_6\right)a\,\hat{\mathbf{x}}+\left(\frac{1}{2}-y_6\right)b\,\hat{\mathbf{y}}-z_6c\,\hat{\mathbf{z}}$ | (8c) | O II |
| $\mathbf{B}_{45}$ | = | $-x_6\,\mathbf{a}_1-y_6\,\mathbf{a}_2-z_6\,\mathbf{a}_3$ | = | $-x_6a\,\hat{\mathbf{x}}-y_6b\,\hat{\mathbf{y}}-z_6c\,\hat{\mathbf{z}}$ | (8c) | O II |
| $\mathbf{B}_{46}$ | = | $\left(\frac{1}{2}+x_6\right)\mathbf{a}_1+y_6\,\mathbf{a}_2+\left(\frac{1}{2}-z_6\right)\mathbf{a}_3$ | = | $\left(\frac{1}{2}+x_6\right)a\,\hat{\mathbf{x}}+y_6b\,\hat{\mathbf{y}}+\left(\frac{1}{2}-z_6\right)c\,\hat{\mathbf{z}}$ | (8c) | O II |
| $\mathbf{B}_{47}$ | = | $x_6\,\mathbf{a}_1+\left(\frac{1}{2}-y_6\right)\mathbf{a}_2+\left(\frac{1}{2}+z_6\right)\mathbf{a}_3$ | = | $x_6a\,\hat{\mathbf{x}}+\left(\frac{1}{2}-y_6\right)b\,\hat{\mathbf{y}}+\left(\frac{1}{2}+z_6\right)c\,\hat{\mathbf{z}}$ | (8c) | O II |
| $\mathbf{B}_{48}$ | = | $\left(\frac{1}{2}-x_6\right)\mathbf{a}_1+\left(\frac{1}{2}+y_6\right)\mathbf{a}_2+z_6\,\mathbf{a}_3$ | = | $\left(\frac{1}{2}-x_6\right)a\,\hat{\mathbf{x}}+\left(\frac{1}{2}+y_6\right)b\,\hat{\mathbf{y}}+z_6c\,\hat{\mathbf{z}}$ | (8c) | O II |
| $\mathbf{B}_{49}$ | = | $x_7\,\mathbf{a}_1+y_7\,\mathbf{a}_2+z_7\,\mathbf{a}_3$ | = | $x_7a\,\hat{\mathbf{x}}+y_7b\,\hat{\mathbf{y}}+z_7c\,\hat{\mathbf{z}}$ | (8c) | O III |
| $\mathbf{B}_{50}$ | = | $\left(\frac{1}{2}-x_7\right)\mathbf{a}_1-y_7\,\mathbf{a}_2+\left(\frac{1}{2}+z_7\right)\mathbf{a}_3$ | = | $\left(\frac{1}{2}-x_7\right)a\,\hat{\mathbf{x}}-y_7b\,\hat{\mathbf{y}}+\left(\frac{1}{2}+z_7\right)c\,\hat{\mathbf{z}}$ | (8c) | O III |
| $\mathbf{B}_{51}$ | = | $-x_7\,\mathbf{a}_1+\left(\frac{1}{2}+y_7\right)\mathbf{a}_2+\left(\frac{1}{2}-z_7\right)\mathbf{a}_3$ | = | $-x_7a\,\hat{\mathbf{x}}+\left(\frac{1}{2}+y_7\right)b\,\hat{\mathbf{y}}+\left(\frac{1}{2}-z_7\right)c\,\hat{\mathbf{z}}$ | (8c) | O III |
| $\mathbf{B}_{52}$ | = | $\left(\frac{1}{2}+x_7\right)\mathbf{a}_1+\left(\frac{1}{2}-y_7\right)\mathbf{a}_2-z_7\,\mathbf{a}_3$ | = | $\left(\frac{1}{2}+x_7\right)a\,\hat{\mathbf{x}}+\left(\frac{1}{2}-y_7\right)b\,\hat{\mathbf{y}}-z_7c\,\hat{\mathbf{z}}$ | (8c) | O III |
| $\mathbf{B}_{53}$ | = | $-x_7\,\mathbf{a}_1-y_7\,\mathbf{a}_2-z_7\,\mathbf{a}_3$ | = | $-x_7a\,\hat{\mathbf{x}}-y_7b\,\hat{\mathbf{y}}-z_7c\,\hat{\mathbf{z}}$ | (8c) | O III |
| $\mathbf{B}_{54}$ | = | $\left(\frac{1}{2}+x_7\right)\mathbf{a}_1+y_7\,\mathbf{a}_2+\left(\frac{1}{2}-z_7\right)\mathbf{a}_3$ | = | $\left(\frac{1}{2}+x_7\right)a\,\hat{\mathbf{x}}+y_7b\,\hat{\mathbf{y}}+\left(\frac{1}{2}-z_7\right)c\,\hat{\mathbf{z}}$ | (8c) | O III |
| $\mathbf{B}_{55}$ | = | $x_7\,\mathbf{a}_1+\left(\frac{1}{2}-y_7\right)\mathbf{a}_2+\left(\frac{1}{2}+z_7\right)\mathbf{a}_3$ | = | $x_7a\,\hat{\mathbf{x}}+\left(\frac{1}{2}-y_7\right)b\,\hat{\mathbf{y}}+\left(\frac{1}{2}+z_7\right)c\,\hat{\mathbf{z}}$ | (8c) | O III |
| $\mathbf{B}_{56}$ | = | $\left(\frac{1}{2}-x_7\right)\mathbf{a}_1+\left(\frac{1}{2}+y_7\right)\mathbf{a}_2+z_7\,\mathbf{a}_3$ | = | $\left(\frac{1}{2}-x_7\right)a\,\hat{\mathbf{x}}+\left(\frac{1}{2}+y_7\right)b\,\hat{\mathbf{y}}+z_7c\,\hat{\mathbf{z}}$ | (8c) | O III |
| $\mathbf{B}_{57}$ | = | $x_8\,\mathbf{a}_1+y_8\,\mathbf{a}_2+z_8\,\mathbf{a}_3$ | = | $x_8a\,\hat{\mathbf{x}}+y_8b\,\hat{\mathbf{y}}+z_8c\,\hat{\mathbf{z}}$ | (8c) | O IV |
| $\mathbf{B}_{58}$ | = | $\left(\frac{1}{2}-x_8\right)\mathbf{a}_1-y_8\,\mathbf{a}_2+\left(\frac{1}{2}+z_8\right)\mathbf{a}_3$ | = | $\left(\frac{1}{2}-x_8\right)a\,\hat{\mathbf{x}}-y_8b\,\hat{\mathbf{y}}+\left(\frac{1}{2}+z_8\right)c\,\hat{\mathbf{z}}$ | (8c) | O IV |
| $\mathbf{B}_{59}$ | = | $-x_8\,\mathbf{a}_1+\left(\frac{1}{2}+y_8\right)\mathbf{a}_2+\left(\frac{1}{2}-z_8\right)\mathbf{a}_3$ | = | $-x_8a\,\hat{\mathbf{x}}+\left(\frac{1}{2}+y_8\right)b\,\hat{\mathbf{y}}+\left(\frac{1}{2}-z_8\right)c\,\hat{\mathbf{z}}$ | (8c) | O IV |
| $\mathbf{B}_{60}$ | = | $\left(\frac{1}{2}+x_8\right)\mathbf{a}_1+\left(\frac{1}{2}-y_8\right)\mathbf{a}_2-z_8\,\mathbf{a}_3$ | = | $\left(\frac{1}{2}+x_8\right)a\,\hat{\mathbf{x}}+\left(\frac{1}{2}-y_8\right)b\,\hat{\mathbf{y}}-z_8c\,\hat{\mathbf{z}}$ | (8c) | O IV |
| $\mathbf{B}_{61}$ | = | $-x_8\,\mathbf{a}_1-y_8\,\mathbf{a}_2-z_8\,\mathbf{a}_3$ | = | $-x_8a\,\hat{\mathbf{x}}-y_8b\,\hat{\mathbf{y}}-z_8c\,\hat{\mathbf{z}}$ | (8c) | O IV |
| $\mathbf{B}_{62}$ | = | $\left(\frac{1}{2}+x_8\right)\mathbf{a}_1+y_8\,\mathbf{a}_2+\left(\frac{1}{2}-z_8\right)\mathbf{a}_3$ | = | $\left(\frac{1}{2}+x_8\right)a\,\hat{\mathbf{x}}+y_8b\,\hat{\mathbf{y}}+\left(\frac{1}{2}-z_8\right)c\,\hat{\mathbf{z}}$ | (8c) | O IV |
| $\mathbf{B}_{63}$ | = | $x_8\,\mathbf{a}_1+\left(\frac{1}{2}-y_8\right)\mathbf{a}_2+\left(\frac{1}{2}+z_8\right)\mathbf{a}_3$ | = | $x_8a\,\hat{\mathbf{x}}+\left(\frac{1}{2}-y_8\right)b\,\hat{\mathbf{y}}+\left(\frac{1}{2}+z_8\right)c\,\hat{\mathbf{z}}$ | (8c) | O IV |
| $\mathbf{B}_{64}$ | = | $\left(\frac{1}{2}-x_8\right)\mathbf{a}_1+\left(\frac{1}{2}+y_8\right)\mathbf{a}_2+z_8\,\mathbf{a}_3$ | = | $\left(\frac{1}{2}-x_8\right)a\,\hat{\mathbf{x}}+\left(\frac{1}{2}+y_8\right)b\,\hat{\mathbf{y}}+z_8c\,\hat{\mathbf{z}}$ | (8c) | O IV |

**Geometry files:**

- CIF: pp. 1627

- POSCAR: pp. 1627

# Enstatite ($MgSiO_3$, $S4_3$) Structure: AB3C_oP80_61_2c_6c_2c

http://aflow.org/prototype-encyclopedia/AB3C_oP80_61_2c_6c_2c

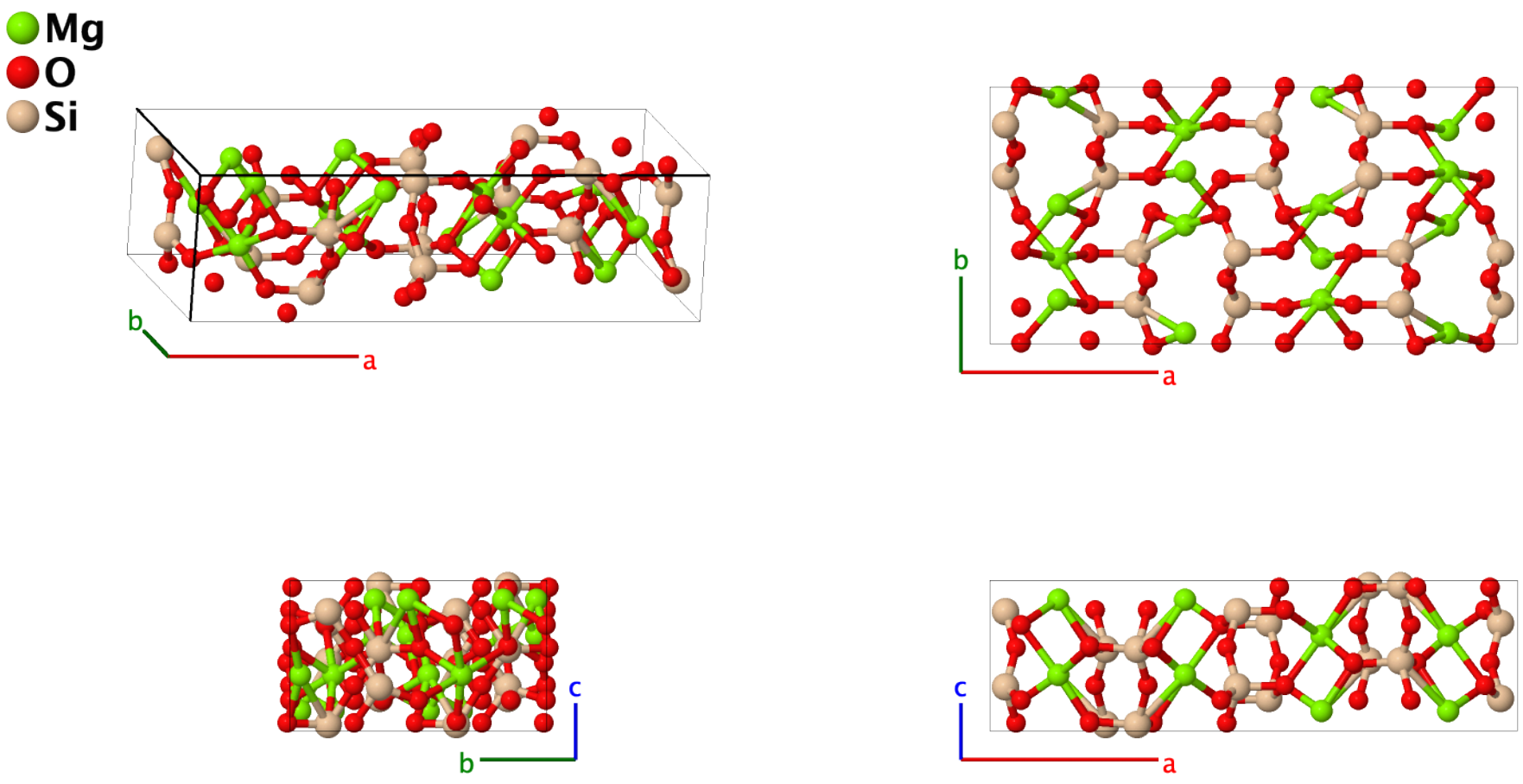

| | | |
|---|---|---|
| **Prototype** | : | $MgO_3Si$ |
| **AFLOW prototype label** | : | AB3C_oP80_61_2c_6c_2c |
| ***Strukturbericht* designation** | : | $S4_3$ |
| **Pearson symbol** | : | oP80 |
| **Space group number** | : | 61 |
| **Space group symbol** | : | *Pbca* |
| **AFLOW prototype command** | : | `aflow --proto=AB3C_oP80_61_2c_6c_2c` `--params=`$a, b/a, c/a, x_1, y_1, z_1, x_2, y_2, z_2, x_3, y_3, z_3, x_4, y_4, z_4, x_5, y_5, z_5, x_6, y_6, z_6, x_7, y_7, z_7, x_8, y_8, z_8, x_9, y_9, z_9, x_{10}, y_{10}, z_{10}$ |

### Other compounds with this structure

- $FeSiO_3$ (ferrosilite) and $CoSiO_3$ (Co-pyroxene)

---

- Enstatite (or properly orthoenstatite) is an end point of orthopyroxene, $XYSi_2O_6$, where $X$ and $Y$ are small-radius divalent cations. If either $X$ or $Y$ is replaced by a transition metal the system becomes a monoclinic clinopyroxene, in space group $C2/c$ #15.

### Simple Orthorhombic primitive vectors:

$$\mathbf{a}_1 = a\,\hat{\mathbf{x}}$$
$$\mathbf{a}_2 = b\,\hat{\mathbf{y}}$$
$$\mathbf{a}_3 = c\,\hat{\mathbf{z}}$$

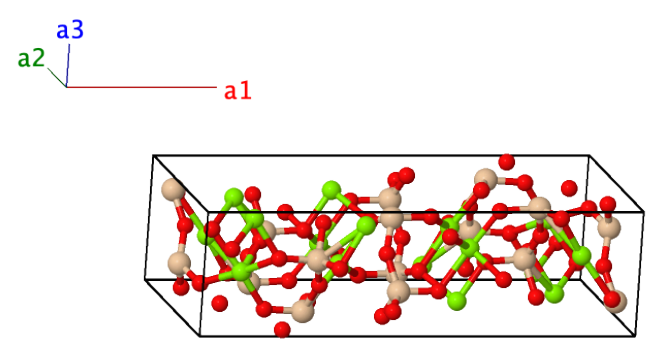

### Basis vectors:

| | | Lattice Coordinates | | Cartesian Coordinates | Wyckoff Position | Atom Type |
|---|---|---|---|---|---|---|
| $\mathbf{B}_1$ | = | $x_1\,\mathbf{a}_1 + y_1\,\mathbf{a}_2 + z_1\,\mathbf{a}_3$ | = | $x_1 a\,\hat{\mathbf{x}} + y_1 b\,\hat{\mathbf{y}} + z_1 c\,\hat{\mathbf{z}}$ | (8c) | Mg I |
| $\mathbf{B}_{2}$ | = | $\left(\frac{1}{2} - x_{1}\right) \mathbf{a}_{1} - y_{1} \mathbf{a}_{2} + \left(\frac{1}{2} + z_{1}\right) \mathbf{a}_{3}$ | = | $\left(\frac{1}{2} - x_{1}\right) a \mathbf{\hat{x}} - y_{1} b \mathbf{\hat{y}} + \left(\frac{1}{2} + z_{1}\right) c \mathbf{\hat{z}}$ | (8*c*) | Mg I |
| $\mathbf{B}_{3}$ | = | $-x_{1} \mathbf{a}_{1} + \left(\frac{1}{2} + y_{1}\right) \mathbf{a}_{2} + \left(\frac{1}{2} - z_{1}\right) \mathbf{a}_{3}$ | = | $-x_{1} a \mathbf{\hat{x}} + \left(\frac{1}{2} + y_{1}\right) b \mathbf{\hat{y}} + \left(\frac{1}{2} - z_{1}\right) c \mathbf{\hat{z}}$ | (8*c*) | Mg I |
| $\mathbf{B}_{4}$ | = | $\left(\frac{1}{2} + x_{1}\right) \mathbf{a}_{1} + \left(\frac{1}{2} - y_{1}\right) \mathbf{a}_{2} - z_{1} \mathbf{a}_{3}$ | = | $\left(\frac{1}{2} + x_{1}\right) a \mathbf{\hat{x}} + \left(\frac{1}{2} - y_{1}\right) b \mathbf{\hat{y}} - z_{1} c \mathbf{\hat{z}}$ | (8*c*) | Mg I |
| $\mathbf{B}_{5}$ | = | $-x_{1} \mathbf{a}_{1} - y_{1} \mathbf{a}_{2} - z_{1} \mathbf{a}_{3}$ | = | $-x_{1} a \mathbf{\hat{x}} - y_{1} b \mathbf{\hat{y}} - z_{1} c \mathbf{\hat{z}}$ | (8*c*) | Mg I |
| $\mathbf{B}_{6}$ | = | $\left(\frac{1}{2} + x_{1}\right) \mathbf{a}_{1} + y_{1} \mathbf{a}_{2} + \left(\frac{1}{2} - z_{1}\right) \mathbf{a}_{3}$ | = | $\left(\frac{1}{2} + x_{1}\right) a \mathbf{\hat{x}} + y_{1} b \mathbf{\hat{y}} + \left(\frac{1}{2} - z_{1}\right) c \mathbf{\hat{z}}$ | (8*c*) | Mg I |
| $\mathbf{B}_{7}$ | = | $x_{1} \mathbf{a}_{1} + \left(\frac{1}{2} - y_{1}\right) \mathbf{a}_{2} + \left(\frac{1}{2} + z_{1}\right) \mathbf{a}_{3}$ | = | $x_{1} a \mathbf{\hat{x}} + \left(\frac{1}{2} - y_{1}\right) b \mathbf{\hat{y}} + \left(\frac{1}{2} + z_{1}\right) c \mathbf{\hat{z}}$ | (8*c*) | Mg I |
| $\mathbf{B}_{8}$ | = | $\left(\frac{1}{2} - x_{1}\right) \mathbf{a}_{1} + \left(\frac{1}{2} + y_{1}\right) \mathbf{a}_{2} + z_{1} \mathbf{a}_{3}$ | = | $\left(\frac{1}{2} - x_{1}\right) a \mathbf{\hat{x}} + \left(\frac{1}{2} + y_{1}\right) b \mathbf{\hat{y}} + z_{1} c \mathbf{\hat{z}}$ | (8*c*) | Mg I |
| $\mathbf{B}_{9}$ | = | $x_{2} \mathbf{a}_{1} + y_{2} \mathbf{a}_{2} + z_{2} \mathbf{a}_{3}$ | = | $x_{2} a \mathbf{\hat{x}} + y_{2} b \mathbf{\hat{y}} + z_{2} c \mathbf{\hat{z}}$ | (8*c*) | Mg II |
| $\mathbf{B}_{10}$ | = | $\left(\frac{1}{2} - x_{2}\right) \mathbf{a}_{1} - y_{2} \mathbf{a}_{2} + \left(\frac{1}{2} + z_{2}\right) \mathbf{a}_{3}$ | = | $\left(\frac{1}{2} - x_{2}\right) a \mathbf{\hat{x}} - y_{2} b \mathbf{\hat{y}} + \left(\frac{1}{2} + z_{2}\right) c \mathbf{\hat{z}}$ | (8*c*) | Mg II |
| $\mathbf{B}_{11}$ | = | $-x_{2} \mathbf{a}_{1} + \left(\frac{1}{2} + y_{2}\right) \mathbf{a}_{2} + \left(\frac{1}{2} - z_{2}\right) \mathbf{a}_{3}$ | = | $-x_{2} a \mathbf{\hat{x}} + \left(\frac{1}{2} + y_{2}\right) b \mathbf{\hat{y}} + \left(\frac{1}{2} - z_{2}\right) c \mathbf{\hat{z}}$ | (8*c*) | Mg II |
| $\mathbf{B}_{12}$ | = | $\left(\frac{1}{2} + x_{2}\right) \mathbf{a}_{1} + \left(\frac{1}{2} - y_{2}\right) \mathbf{a}_{2} - z_{2} \mathbf{a}_{3}$ | = | $\left(\frac{1}{2} + x_{2}\right) a \mathbf{\hat{x}} + \left(\frac{1}{2} - y_{2}\right) b \mathbf{\hat{y}} - z_{2} c \mathbf{\hat{z}}$ | (8*c*) | Mg II |
| $\mathbf{B}_{13}$ | = | $-x_{2} \mathbf{a}_{1} - y_{2} \mathbf{a}_{2} - z_{2} \mathbf{a}_{3}$ | = | $-x_{2} a \mathbf{\hat{x}} - y_{2} b \mathbf{\hat{y}} - z_{2} c \mathbf{\hat{z}}$ | (8*c*) | Mg II |
| $\mathbf{B}_{14}$ | = | $\left(\frac{1}{2} + x_{2}\right) \mathbf{a}_{1} + y_{2} \mathbf{a}_{2} + \left(\frac{1}{2} - z_{2}\right) \mathbf{a}_{3}$ | = | $\left(\frac{1}{2} + x_{2}\right) a \mathbf{\hat{x}} + y_{2} b \mathbf{\hat{y}} + \left(\frac{1}{2} - z_{2}\right) c \mathbf{\hat{z}}$ | (8*c*) | Mg II |
| $\mathbf{B}_{15}$ | = | $x_{2} \mathbf{a}_{1} + \left(\frac{1}{2} - y_{2}\right) \mathbf{a}_{2} + \left(\frac{1}{2} + z_{2}\right) \mathbf{a}_{3}$ | = | $x_{2} a \mathbf{\hat{x}} + \left(\frac{1}{2} - y_{2}\right) b \mathbf{\hat{y}} + \left(\frac{1}{2} + z_{2}\right) c \mathbf{\hat{z}}$ | (8*c*) | Mg II |
| $\mathbf{B}_{16}$ | = | $\left(\frac{1}{2} - x_{2}\right) \mathbf{a}_{1} + \left(\frac{1}{2} + y_{2}\right) \mathbf{a}_{2} + z_{2} \mathbf{a}_{3}$ | = | $\left(\frac{1}{2} - x_{2}\right) a \mathbf{\hat{x}} + \left(\frac{1}{2} + y_{2}\right) b \mathbf{\hat{y}} + z_{2} c \mathbf{\hat{z}}$ | (8*c*) | Mg II |
| $\mathbf{B}_{17}$ | = | $x_{3} \mathbf{a}_{1} + y_{3} \mathbf{a}_{2} + z_{3} \mathbf{a}_{3}$ | = | $x_{3} a \mathbf{\hat{x}} + y_{3} b \mathbf{\hat{y}} + z_{3} c \mathbf{\hat{z}}$ | (8*c*) | O I |
| $\mathbf{B}_{18}$ | = | $\left(\frac{1}{2} - x_{3}\right) \mathbf{a}_{1} - y_{3} \mathbf{a}_{2} + \left(\frac{1}{2} + z_{3}\right) \mathbf{a}_{3}$ | = | $\left(\frac{1}{2} - x_{3}\right) a \mathbf{\hat{x}} - y_{3} b \mathbf{\hat{y}} + \left(\frac{1}{2} + z_{3}\right) c \mathbf{\hat{z}}$ | (8*c*) | O I |
| $\mathbf{B}_{19}$ | = | $-x_{3} \mathbf{a}_{1} + \left(\frac{1}{2} + y_{3}\right) \mathbf{a}_{2} + \left(\frac{1}{2} - z_{3}\right) \mathbf{a}_{3}$ | = | $-x_{3} a \mathbf{\hat{x}} + \left(\frac{1}{2} + y_{3}\right) b \mathbf{\hat{y}} + \left(\frac{1}{2} - z_{3}\right) c \mathbf{\hat{z}}$ | (8*c*) | O I |
| $\mathbf{B}_{20}$ | = | $\left(\frac{1}{2} + x_{3}\right) \mathbf{a}_{1} + \left(\frac{1}{2} - y_{3}\right) \mathbf{a}_{2} - z_{3} \mathbf{a}_{3}$ | = | $\left(\frac{1}{2} + x_{3}\right) a \mathbf{\hat{x}} + \left(\frac{1}{2} - y_{3}\right) b \mathbf{\hat{y}} - z_{3} c \mathbf{\hat{z}}$ | (8*c*) | O I |
| $\mathbf{B}_{21}$ | = | $-x_{3} \mathbf{a}_{1} - y_{3} \mathbf{a}_{2} - z_{3} \mathbf{a}_{3}$ | = | $-x_{3} a \mathbf{\hat{x}} - y_{3} b \mathbf{\hat{y}} - z_{3} c \mathbf{\hat{z}}$ | (8*c*) | O I |
| $\mathbf{B}_{22}$ | = | $\left(\frac{1}{2} + x_{3}\right) \mathbf{a}_{1} + y_{3} \mathbf{a}_{2} + \left(\frac{1}{2} - z_{3}\right) \mathbf{a}_{3}$ | = | $\left(\frac{1}{2} + x_{3}\right) a \mathbf{\hat{x}} + y_{3} b \mathbf{\hat{y}} + \left(\frac{1}{2} - z_{3}\right) c \mathbf{\hat{z}}$ | (8*c*) | O I |
| $\mathbf{B}_{23}$ | = | $x_{3} \mathbf{a}_{1} + \left(\frac{1}{2} - y_{3}\right) \mathbf{a}_{2} + \left(\frac{1}{2} + z_{3}\right) \mathbf{a}_{3}$ | = | $x_{3} a \mathbf{\hat{x}} + \left(\frac{1}{2} - y_{3}\right) b \mathbf{\hat{y}} + \left(\frac{1}{2} + z_{3}\right) c \mathbf{\hat{z}}$ | (8*c*) | O I |
| $\mathbf{B}_{24}$ | = | $\left(\frac{1}{2} - x_{3}\right) \mathbf{a}_{1} + \left(\frac{1}{2} + y_{3}\right) \mathbf{a}_{2} + z_{3} \mathbf{a}_{3}$ | = | $\left(\frac{1}{2} - x_{3}\right) a \mathbf{\hat{x}} + \left(\frac{1}{2} + y_{3}\right) b \mathbf{\hat{y}} + z_{3} c \mathbf{\hat{z}}$ | (8*c*) | O I |
| $\mathbf{B}_{25}$ | = | $x_{4} \mathbf{a}_{1} + y_{4} \mathbf{a}_{2} + z_{4} \mathbf{a}_{3}$ | = | $x_{4} a \mathbf{\hat{x}} + y_{4} b \mathbf{\hat{y}} + z_{4} c \mathbf{\hat{z}}$ | (8*c*) | O II |
| $\mathbf{B}_{26}$ | = | $\left(\frac{1}{2} - x_{4}\right) \mathbf{a}_{1} - y_{4} \mathbf{a}_{2} + \left(\frac{1}{2} + z_{4}\right) \mathbf{a}_{3}$ | = | $\left(\frac{1}{2} - x_{4}\right) a \mathbf{\hat{x}} - y_{4} b \mathbf{\hat{y}} + \left(\frac{1}{2} + z_{4}\right) c \mathbf{\hat{z}}$ | (8*c*) | O II |
| $\mathbf{B}_{27}$ | = | $-x_{4} \mathbf{a}_{1} + \left(\frac{1}{2} + y_{4}\right) \mathbf{a}_{2} + \left(\frac{1}{2} - z_{4}\right) \mathbf{a}_{3}$ | = | $-x_{4} a \mathbf{\hat{x}} + \left(\frac{1}{2} + y_{4}\right) b \mathbf{\hat{y}} + \left(\frac{1}{2} - z_{4}\right) c \mathbf{\hat{z}}$ | (8*c*) | O II |
| $\mathbf{B}_{28}$ | = | $\left(\frac{1}{2} + x_{4}\right) \mathbf{a}_{1} + \left(\frac{1}{2} - y_{4}\right) \mathbf{a}_{2} - z_{4} \mathbf{a}_{3}$ | = | $\left(\frac{1}{2} + x_{4}\right) a \mathbf{\hat{x}} + \left(\frac{1}{2} - y_{4}\right) b \mathbf{\hat{y}} - z_{4} c \mathbf{\hat{z}}$ | (8*c*) | O II |
| $\mathbf{B}_{29}$ | = | $-x_{4} \mathbf{a}_{1} - y_{4} \mathbf{a}_{2} - z_{4} \mathbf{a}_{3}$ | = | $-x_{4} a \mathbf{\hat{x}} - y_{4} b \mathbf{\hat{y}} - z_{4} c \mathbf{\hat{z}}$ | (8*c*) | O II |
| $\mathbf{B}_{30}$ | = | $\left(\frac{1}{2} + x_{4}\right) \mathbf{a}_{1} + y_{4} \mathbf{a}_{2} + \left(\frac{1}{2} - z_{4}\right) \mathbf{a}_{3}$ | = | $\left(\frac{1}{2} + x_{4}\right) a \mathbf{\hat{x}} + y_{4} b \mathbf{\hat{y}} + \left(\frac{1}{2} - z_{4}\right) c \mathbf{\hat{z}}$ | (8*c*) | O II |
| $\mathbf{B}_{31}$ | = | $x_{4} \mathbf{a}_{1} + \left(\frac{1}{2} - y_{4}\right) \mathbf{a}_{2} + \left(\frac{1}{2} + z_{4}\right) \mathbf{a}_{3}$ | = | $x_{4} a \mathbf{\hat{x}} + \left(\frac{1}{2} - y_{4}\right) b \mathbf{\hat{y}} + \left(\frac{1}{2} + z_{4}\right) c \mathbf{\hat{z}}$ | (8*c*) | O II |
| $\mathbf{B}_{32}$ | = | $\left(\frac{1}{2} - x_{4}\right) \mathbf{a}_{1} + \left(\frac{1}{2} + y_{4}\right) \mathbf{a}_{2} + z_{4} \mathbf{a}_{3}$ | = | $\left(\frac{1}{2} - x_{4}\right) a \mathbf{\hat{x}} + \left(\frac{1}{2} + y_{4}\right) b \mathbf{\hat{y}} + z_{4} c \mathbf{\hat{z}}$ | (8*c*) | O II |
| $\mathbf{B}_{33}$ | = | $x_{5} \mathbf{a}_{1} + y_{5} \mathbf{a}_{2} + z_{5} \mathbf{a}_{3}$ | = | $x_{5} a \mathbf{\hat{x}} + y_{5} b \mathbf{\hat{y}} + z_{5} c \mathbf{\hat{z}}$ | (8*c*) | O III |
| $\mathbf{B}_{34}$ | = | $\left(\frac{1}{2} - x_{5}\right) \mathbf{a}_{1} - y_{5} \mathbf{a}_{2} + \left(\frac{1}{2} + z_{5}\right) \mathbf{a}_{3}$ | = | $\left(\frac{1}{2} - x_{5}\right) a \mathbf{\hat{x}} - y_{5} b \mathbf{\hat{y}} + \left(\frac{1}{2} + z_{5}\right) c \mathbf{\hat{z}}$ | (8*c*) | O III |
| $\mathbf{B}_{35}$ | = | $-x_{5} \mathbf{a}_{1} + \left(\frac{1}{2} + y_{5}\right) \mathbf{a}_{2} + \left(\frac{1}{2} - z_{5}\right) \mathbf{a}_{3}$ | = | $-x_{5} a \mathbf{\hat{x}} + \left(\frac{1}{2} + y_{5}\right) b \mathbf{\hat{y}} + \left(\frac{1}{2} - z_{5}\right) c \mathbf{\hat{z}}$ | (8*c*) | O III |
| $\mathbf{B}_{36}$ | = | $\left(\frac{1}{2} + x_{5}\right) \mathbf{a}_{1} + \left(\frac{1}{2} - y_{5}\right) \mathbf{a}_{2} - z_{5} \mathbf{a}_{3}$ | = | $\left(\frac{1}{2} + x_{5}\right) a \mathbf{\hat{x}} + \left(\frac{1}{2} - y_{5}\right) b \mathbf{\hat{y}} - z_{5} c \mathbf{\hat{z}}$ | (8*c*) | O III |
| $\mathbf{B}_{37}$ | = | $-x_{5} \mathbf{a}_{1} - y_{5} \mathbf{a}_{2} - z_{5} \mathbf{a}_{3}$ | = | $-x_{5} a \mathbf{\hat{x}} - y_{5} b \mathbf{\hat{y}} - z_{5} c \mathbf{\hat{z}}$ | (8*c*) | O III |
| $\mathbf{B}_{38}$ | = | $\left(\frac{1}{2}+x_{5}\right)\mathbf{a}_{1}+y_{5}\,\mathbf{a}_{2}+\left(\frac{1}{2}-z_{5}\right)\mathbf{a}_{3}$ | = | $\left(\frac{1}{2}+x_{5}\right)a\,\hat{\mathbf{x}}+y_{5}b\,\hat{\mathbf{y}}+\left(\frac{1}{2}-z_{5}\right)c\,\hat{\mathbf{z}}$ | (8$c$) | O III |
| $\mathbf{B}_{39}$ | = | $x_{5}\,\mathbf{a}_{1}+\left(\frac{1}{2}-y_{5}\right)\mathbf{a}_{2}+\left(\frac{1}{2}+z_{5}\right)\mathbf{a}_{3}$ | = | $x_{5}a\,\hat{\mathbf{x}}+\left(\frac{1}{2}-y_{5}\right)b\,\hat{\mathbf{y}}+\left(\frac{1}{2}+z_{5}\right)c\,\hat{\mathbf{z}}$ | (8$c$) | O III |
| $\mathbf{B}_{40}$ | = | $\left(\frac{1}{2}-x_{5}\right)\mathbf{a}_{1}+\left(\frac{1}{2}+y_{5}\right)\mathbf{a}_{2}+z_{5}\,\mathbf{a}_{3}$ | = | $\left(\frac{1}{2}-x_{5}\right)a\,\hat{\mathbf{x}}+\left(\frac{1}{2}+y_{5}\right)b\,\hat{\mathbf{y}}+z_{5}c\,\hat{\mathbf{z}}$ | (8$c$) | O III |
| $\mathbf{B}_{41}$ | = | $x_{6}\,\mathbf{a}_{1}+y_{6}\,\mathbf{a}_{2}+z_{6}\,\mathbf{a}_{3}$ | = | $x_{6}a\,\hat{\mathbf{x}}+y_{6}b\,\hat{\mathbf{y}}+z_{6}c\,\hat{\mathbf{z}}$ | (8$c$) | O IV |
| $\mathbf{B}_{42}$ | = | $\left(\frac{1}{2}-x_{6}\right)\mathbf{a}_{1}-y_{6}\,\mathbf{a}_{2}+\left(\frac{1}{2}+z_{6}\right)\mathbf{a}_{3}$ | = | $\left(\frac{1}{2}-x_{6}\right)a\,\hat{\mathbf{x}}-y_{6}b\,\hat{\mathbf{y}}+\left(\frac{1}{2}+z_{6}\right)c\,\hat{\mathbf{z}}$ | (8$c$) | O IV |
| $\mathbf{B}_{43}$ | = | $-x_{6}\,\mathbf{a}_{1}+\left(\frac{1}{2}+y_{6}\right)\mathbf{a}_{2}+\left(\frac{1}{2}-z_{6}\right)\mathbf{a}_{3}$ | = | $-x_{6}a\,\hat{\mathbf{x}}+\left(\frac{1}{2}+y_{6}\right)b\,\hat{\mathbf{y}}+\left(\frac{1}{2}-z_{6}\right)c\,\hat{\mathbf{z}}$ | (8$c$) | O IV |
| $\mathbf{B}_{44}$ | = | $\left(\frac{1}{2}+x_{6}\right)\mathbf{a}_{1}+\left(\frac{1}{2}-y_{6}\right)\mathbf{a}_{2}-z_{6}\,\mathbf{a}_{3}$ | = | $\left(\frac{1}{2}+x_{6}\right)a\,\hat{\mathbf{x}}+\left(\frac{1}{2}-y_{6}\right)b\,\hat{\mathbf{y}}-z_{6}c\,\hat{\mathbf{z}}$ | (8$c$) | O IV |
| $\mathbf{B}_{45}$ | = | $-x_{6}\,\mathbf{a}_{1}-y_{6}\,\mathbf{a}_{2}-z_{6}\,\mathbf{a}_{3}$ | = | $-x_{6}a\,\hat{\mathbf{x}}-y_{6}b\,\hat{\mathbf{y}}-z_{6}c\,\hat{\mathbf{z}}$ | (8$c$) | O IV |
| $\mathbf{B}_{46}$ | = | $\left(\frac{1}{2}+x_{6}\right)\mathbf{a}_{1}+y_{6}\,\mathbf{a}_{2}+\left(\frac{1}{2}-z_{6}\right)\mathbf{a}_{3}$ | = | $\left(\frac{1}{2}+x_{6}\right)a\,\hat{\mathbf{x}}+y_{6}b\,\hat{\mathbf{y}}+\left(\frac{1}{2}-z_{6}\right)c\,\hat{\mathbf{z}}$ | (8$c$) | O IV |
| $\mathbf{B}_{47}$ | = | $x_{6}\,\mathbf{a}_{1}+\left(\frac{1}{2}-y_{6}\right)\mathbf{a}_{2}+\left(\frac{1}{2}+z_{6}\right)\mathbf{a}_{3}$ | = | $x_{6}a\,\hat{\mathbf{x}}+\left(\frac{1}{2}-y_{6}\right)b\,\hat{\mathbf{y}}+\left(\frac{1}{2}+z_{6}\right)c\,\hat{\mathbf{z}}$ | (8$c$) | O IV |
| $\mathbf{B}_{48}$ | = | $\left(\frac{1}{2}-x_{6}\right)\mathbf{a}_{1}+\left(\frac{1}{2}+y_{6}\right)\mathbf{a}_{2}+z_{6}\,\mathbf{a}_{3}$ | = | $\left(\frac{1}{2}-x_{6}\right)a\,\hat{\mathbf{x}}+\left(\frac{1}{2}+y_{6}\right)b\,\hat{\mathbf{y}}+z_{6}c\,\hat{\mathbf{z}}$ | (8$c$) | O IV |
| $\mathbf{B}_{49}$ | = | $x_{7}\,\mathbf{a}_{1}+y_{7}\,\mathbf{a}_{2}+z_{7}\,\mathbf{a}_{3}$ | = | $x_{7}a\,\hat{\mathbf{x}}+y_{7}b\,\hat{\mathbf{y}}+z_{7}c\,\hat{\mathbf{z}}$ | (8$c$) | O V |
| $\mathbf{B}_{50}$ | = | $\left(\frac{1}{2}-x_{7}\right)\mathbf{a}_{1}-y_{7}\,\mathbf{a}_{2}+\left(\frac{1}{2}+z_{7}\right)\mathbf{a}_{3}$ | = | $\left(\frac{1}{2}-x_{7}\right)a\,\hat{\mathbf{x}}-y_{7}b\,\hat{\mathbf{y}}+\left(\frac{1}{2}+z_{7}\right)c\,\hat{\mathbf{z}}$ | (8$c$) | O V |
| $\mathbf{B}_{51}$ | = | $-x_{7}\,\mathbf{a}_{1}+\left(\frac{1}{2}+y_{7}\right)\mathbf{a}_{2}+\left(\frac{1}{2}-z_{7}\right)\mathbf{a}_{3}$ | = | $-x_{7}a\,\hat{\mathbf{x}}+\left(\frac{1}{2}+y_{7}\right)b\,\hat{\mathbf{y}}+\left(\frac{1}{2}-z_{7}\right)c\,\hat{\mathbf{z}}$ | (8$c$) | O V |
| $\mathbf{B}_{52}$ | = | $\left(\frac{1}{2}+x_{7}\right)\mathbf{a}_{1}+\left(\frac{1}{2}-y_{7}\right)\mathbf{a}_{2}-z_{7}\,\mathbf{a}_{3}$ | = | $\left(\frac{1}{2}+x_{7}\right)a\,\hat{\mathbf{x}}+\left(\frac{1}{2}-y_{7}\right)b\,\hat{\mathbf{y}}-z_{7}c\,\hat{\mathbf{z}}$ | (8$c$) | O V |
| $\mathbf{B}_{53}$ | = | $-x_{7}\,\mathbf{a}_{1}-y_{7}\,\mathbf{a}_{2}-z_{7}\,\mathbf{a}_{3}$ | = | $-x_{7}a\,\hat{\mathbf{x}}-y_{7}b\,\hat{\mathbf{y}}-z_{7}c\,\hat{\mathbf{z}}$ | (8$c$) | O V |
| $\mathbf{B}_{54}$ | = | $\left(\frac{1}{2}+x_{7}\right)\mathbf{a}_{1}+y_{7}\,\mathbf{a}_{2}+\left(\frac{1}{2}-z_{7}\right)\mathbf{a}_{3}$ | = | $\left(\frac{1}{2}+x_{7}\right)a\,\hat{\mathbf{x}}+y_{7}b\,\hat{\mathbf{y}}+\left(\frac{1}{2}-z_{7}\right)c\,\hat{\mathbf{z}}$ | (8$c$) | O V |
| $\mathbf{B}_{55}$ | = | $x_{7}\,\mathbf{a}_{1}+\left(\frac{1}{2}-y_{7}\right)\mathbf{a}_{2}+\left(\frac{1}{2}+z_{7}\right)\mathbf{a}_{3}$ | = | $x_{7}a\,\hat{\mathbf{x}}+\left(\frac{1}{2}-y_{7}\right)b\,\hat{\mathbf{y}}+\left(\frac{1}{2}+z_{7}\right)c\,\hat{\mathbf{z}}$ | (8$c$) | O V |
| $\mathbf{B}_{56}$ | = | $\left(\frac{1}{2}-x_{7}\right)\mathbf{a}_{1}+\left(\frac{1}{2}+y_{7}\right)\mathbf{a}_{2}+z_{7}\,\mathbf{a}_{3}$ | = | $\left(\frac{1}{2}-x_{7}\right)a\,\hat{\mathbf{x}}+\left(\frac{1}{2}+y_{7}\right)b\,\hat{\mathbf{y}}+z_{7}c\,\hat{\mathbf{z}}$ | (8$c$) | O V |
| $\mathbf{B}_{57}$ | = | $x_{8}\,\mathbf{a}_{1}+y_{8}\,\mathbf{a}_{2}+z_{8}\,\mathbf{a}_{3}$ | = | $x_{8}a\,\hat{\mathbf{x}}+y_{8}b\,\hat{\mathbf{y}}+z_{8}c\,\hat{\mathbf{z}}$ | (8$c$) | O VI |
| $\mathbf{B}_{58}$ | = | $\left(\frac{1}{2}-x_{8}\right)\mathbf{a}_{1}-y_{8}\,\mathbf{a}_{2}+\left(\frac{1}{2}+z_{8}\right)\mathbf{a}_{3}$ | = | $\left(\frac{1}{2}-x_{8}\right)a\,\hat{\mathbf{x}}-y_{8}b\,\hat{\mathbf{y}}+\left(\frac{1}{2}+z_{8}\right)c\,\hat{\mathbf{z}}$ | (8$c$) | O VI |
| $\mathbf{B}_{59}$ | = | $-x_{8}\,\mathbf{a}_{1}+\left(\frac{1}{2}+y_{8}\right)\mathbf{a}_{2}+\left(\frac{1}{2}-z_{8}\right)\mathbf{a}_{3}$ | = | $-x_{8}a\,\hat{\mathbf{x}}+\left(\frac{1}{2}+y_{8}\right)b\,\hat{\mathbf{y}}+\left(\frac{1}{2}-z_{8}\right)c\,\hat{\mathbf{z}}$ | (8$c$) | O VI |
| $\mathbf{B}_{60}$ | = | $\left(\frac{1}{2}+x_{8}\right)\mathbf{a}_{1}+\left(\frac{1}{2}-y_{8}\right)\mathbf{a}_{2}-z_{8}\,\mathbf{a}_{3}$ | = | $\left(\frac{1}{2}+x_{8}\right)a\,\hat{\mathbf{x}}+\left(\frac{1}{2}-y_{8}\right)b\,\hat{\mathbf{y}}-z_{8}c\,\hat{\mathbf{z}}$ | (8$c$) | O VI |
| $\mathbf{B}_{61}$ | = | $-x_{8}\,\mathbf{a}_{1}-y_{8}\,\mathbf{a}_{2}-z_{8}\,\mathbf{a}_{3}$ | = | $-x_{8}a\,\hat{\mathbf{x}}-y_{8}b\,\hat{\mathbf{y}}-z_{8}c\,\hat{\mathbf{z}}$ | (8$c$) | O VI |
| $\mathbf{B}_{62}$ | = | $\left(\frac{1}{2}+x_{8}\right)\mathbf{a}_{1}+y_{8}\,\mathbf{a}_{2}+\left(\frac{1}{2}-z_{8}\right)\mathbf{a}_{3}$ | = | $\left(\frac{1}{2}+x_{8}\right)a\,\hat{\mathbf{x}}+y_{8}b\,\hat{\mathbf{y}}+\left(\frac{1}{2}-z_{8}\right)c\,\hat{\mathbf{z}}$ | (8$c$) | O VI |
| $\mathbf{B}_{63}$ | = | $x_{8}\,\mathbf{a}_{1}+\left(\frac{1}{2}-y_{8}\right)\mathbf{a}_{2}+\left(\frac{1}{2}+z_{8}\right)\mathbf{a}_{3}$ | = | $x_{8}a\,\hat{\mathbf{x}}+\left(\frac{1}{2}-y_{8}\right)b\,\hat{\mathbf{y}}+\left(\frac{1}{2}+z_{8}\right)c\,\hat{\mathbf{z}}$ | (8$c$) | O VI |
| $\mathbf{B}_{64}$ | = | $\left(\frac{1}{2}-x_{8}\right)\mathbf{a}_{1}+\left(\frac{1}{2}+y_{8}\right)\mathbf{a}_{2}+z_{8}\,\mathbf{a}_{3}$ | = | $\left(\frac{1}{2}-x_{8}\right)a\,\hat{\mathbf{x}}+\left(\frac{1}{2}+y_{8}\right)b\,\hat{\mathbf{y}}+z_{8}c\,\hat{\mathbf{z}}$ | (8$c$) | O VI |
| $\mathbf{B}_{65}$ | = | $x_{9}\,\mathbf{a}_{1}+y_{9}\,\mathbf{a}_{2}+z_{9}\,\mathbf{a}_{3}$ | = | $x_{9}a\,\hat{\mathbf{x}}+y_{9}b\,\hat{\mathbf{y}}+z_{9}c\,\hat{\mathbf{z}}$ | (8$c$) | Si I |
| $\mathbf{B}_{66}$ | = | $\left(\frac{1}{2}-x_{9}\right)\mathbf{a}_{1}-y_{9}\,\mathbf{a}_{2}+\left(\frac{1}{2}+z_{9}\right)\mathbf{a}_{3}$ | = | $\left(\frac{1}{2}-x_{9}\right)a\,\hat{\mathbf{x}}-y_{9}b\,\hat{\mathbf{y}}+\left(\frac{1}{2}+z_{9}\right)c\,\hat{\mathbf{z}}$ | (8$c$) | Si I |
| $\mathbf{B}_{67}$ | = | $-x_{9}\,\mathbf{a}_{1}+\left(\frac{1}{2}+y_{9}\right)\mathbf{a}_{2}+\left(\frac{1}{2}-z_{9}\right)\mathbf{a}_{3}$ | = | $-x_{9}a\,\hat{\mathbf{x}}+\left(\frac{1}{2}+y_{9}\right)b\,\hat{\mathbf{y}}+\left(\frac{1}{2}-z_{9}\right)c\,\hat{\mathbf{z}}$ | (8$c$) | Si I |
| $\mathbf{B}_{68}$ | = | $\left(\frac{1}{2}+x_{9}\right)\mathbf{a}_{1}+\left(\frac{1}{2}-y_{9}\right)\mathbf{a}_{2}-z_{9}\,\mathbf{a}_{3}$ | = | $\left(\frac{1}{2}+x_{9}\right)a\,\hat{\mathbf{x}}+\left(\frac{1}{2}-y_{9}\right)b\,\hat{\mathbf{y}}-z_{9}c\,\hat{\mathbf{z}}$ | (8$c$) | Si I |
| $\mathbf{B}_{69}$ | = | $-x_{9}\,\mathbf{a}_{1}-y_{9}\,\mathbf{a}_{2}-z_{9}\,\mathbf{a}_{3}$ | = | $-x_{9}a\,\hat{\mathbf{x}}-y_{9}b\,\hat{\mathbf{y}}-z_{9}c\,\hat{\mathbf{z}}$ | (8$c$) | Si I |
| $\mathbf{B}_{70}$ | = | $\left(\frac{1}{2}+x_{9}\right)\mathbf{a}_{1}+y_{9}\,\mathbf{a}_{2}+\left(\frac{1}{2}-z_{9}\right)\mathbf{a}_{3}$ | = | $\left(\frac{1}{2}+x_{9}\right)a\,\hat{\mathbf{x}}+y_{9}b\,\hat{\mathbf{y}}+\left(\frac{1}{2}-z_{9}\right)c\,\hat{\mathbf{z}}$ | (8$c$) | Si I |
| $\mathbf{B}_{71}$ | = | $x_{9}\,\mathbf{a}_{1}+\left(\frac{1}{2}-y_{9}\right)\mathbf{a}_{2}+\left(\frac{1}{2}+z_{9}\right)\mathbf{a}_{3}$ | = | $x_{9}a\,\hat{\mathbf{x}}+\left(\frac{1}{2}-y_{9}\right)b\,\hat{\mathbf{y}}+\left(\frac{1}{2}+z_{9}\right)c\,\hat{\mathbf{z}}$ | (8$c$) | Si I |
| $\mathbf{B}_{72}$ | = | $\left(\frac{1}{2}-x_{9}\right)\mathbf{a}_{1}+\left(\frac{1}{2}+y_{9}\right)\mathbf{a}_{2}+z_{9}\,\mathbf{a}_{3}$ | = | $\left(\frac{1}{2}-x_{9}\right)a\,\hat{\mathbf{x}}+\left(\frac{1}{2}+y_{9}\right)b\,\hat{\mathbf{y}}+z_{9}c\,\hat{\mathbf{z}}$ | (8$c$) | Si I |
| $\mathbf{B}_{73}$ | = | $x_{10}\,\mathbf{a}_{1}+y_{10}\,\mathbf{a}_{2}+z_{10}\,\mathbf{a}_{3}$ | = | $x_{10}a\,\hat{\mathbf{x}}+y_{10}b\,\hat{\mathbf{y}}+z_{10}c\,\hat{\mathbf{z}}$ | (8$c$) | Si II |
| $\mathbf{B}_{74}$ | = | $\left(\frac{1}{2} - x_{10}\right)\,\mathbf{a}_1 - y_{10}\,\mathbf{a}_2 + \left(\frac{1}{2} + z_{10}\right)\,\mathbf{a}_3$ | = | $\left(\frac{1}{2} - x_{10}\right) a\,\mathbf{\hat{x}} - y_{10} b\,\mathbf{\hat{y}} + \left(\frac{1}{2} + z_{10}\right) c\,\mathbf{\hat{z}}$ | (8$c$) | Si II |
| $\mathbf{B}_{75}$ | = | $-x_{10}\,\mathbf{a}_1 + \left(\frac{1}{2} + y_{10}\right)\,\mathbf{a}_2 + \left(\frac{1}{2} - z_{10}\right)\,\mathbf{a}_3$ | = | $-x_{10} a\,\mathbf{\hat{x}} + \left(\frac{1}{2} + y_{10}\right) b\,\mathbf{\hat{y}} + \left(\frac{1}{2} - z_{10}\right) c\,\mathbf{\hat{z}}$ | (8$c$) | Si II |
| $\mathbf{B}_{76}$ | = | $\left(\frac{1}{2} + x_{10}\right)\,\mathbf{a}_1 + \left(\frac{1}{2} - y_{10}\right)\,\mathbf{a}_2 - z_{10}\,\mathbf{a}_3$ | = | $\left(\frac{1}{2} + x_{10}\right) a\,\mathbf{\hat{x}} + \left(\frac{1}{2} - y_{10}\right) b\,\mathbf{\hat{y}} - z_{10} c\,\mathbf{\hat{z}}$ | (8$c$) | Si II |
| $\mathbf{B}_{77}$ | = | $-x_{10}\,\mathbf{a}_1 - y_{10}\,\mathbf{a}_2 - z_{10}\,\mathbf{a}_3$ | = | $-x_{10} a\,\mathbf{\hat{x}} - y_{10} b\,\mathbf{\hat{y}} - z_{10} c\,\mathbf{\hat{z}}$ | (8$c$) | Si II |
| $\mathbf{B}_{78}$ | = | $\left(\frac{1}{2} + x_{10}\right)\,\mathbf{a}_1 + y_{10}\,\mathbf{a}_2 + \left(\frac{1}{2} - z_{10}\right)\,\mathbf{a}_3$ | = | $\left(\frac{1}{2} + x_{10}\right) a\,\mathbf{\hat{x}} + y_{10} b\,\mathbf{\hat{y}} + \left(\frac{1}{2} - z_{10}\right) c\,\mathbf{\hat{z}}$ | (8$c$) | Si II |
| $\mathbf{B}_{79}$ | = | $x_{10}\,\mathbf{a}_1 + \left(\frac{1}{2} - y_{10}\right)\,\mathbf{a}_2 + \left(\frac{1}{2} + z_{10}\right)\,\mathbf{a}_3$ | = | $x_{10} a\,\mathbf{\hat{x}} + \left(\frac{1}{2} - y_{10}\right) b\,\mathbf{\hat{y}} + \left(\frac{1}{2} + z_{10}\right) c\,\mathbf{\hat{z}}$ | (8$c$) | Si II |
| $\mathbf{B}_{80}$ | = | $\left(\frac{1}{2} - x_{10}\right)\,\mathbf{a}_1 + \left(\frac{1}{2} + y_{10}\right)\,\mathbf{a}_2 + z_{10}\,\mathbf{a}_3$ | = | $\left(\frac{1}{2} - x_{10}\right) a\,\mathbf{\hat{x}} + \left(\frac{1}{2} + y_{10}\right) b\,\mathbf{\hat{y}} + z_{10} c\,\mathbf{\hat{z}}$ | (8$c$) | Si II |

**Geometry files:**

- CIF: pp. 1628
- POSCAR: pp. 1628

# COCl Structure: ABC_oP24_61_c_c_c

http://aflow.org/prototype-encyclopedia/ABC_oP24_61_c_c_c

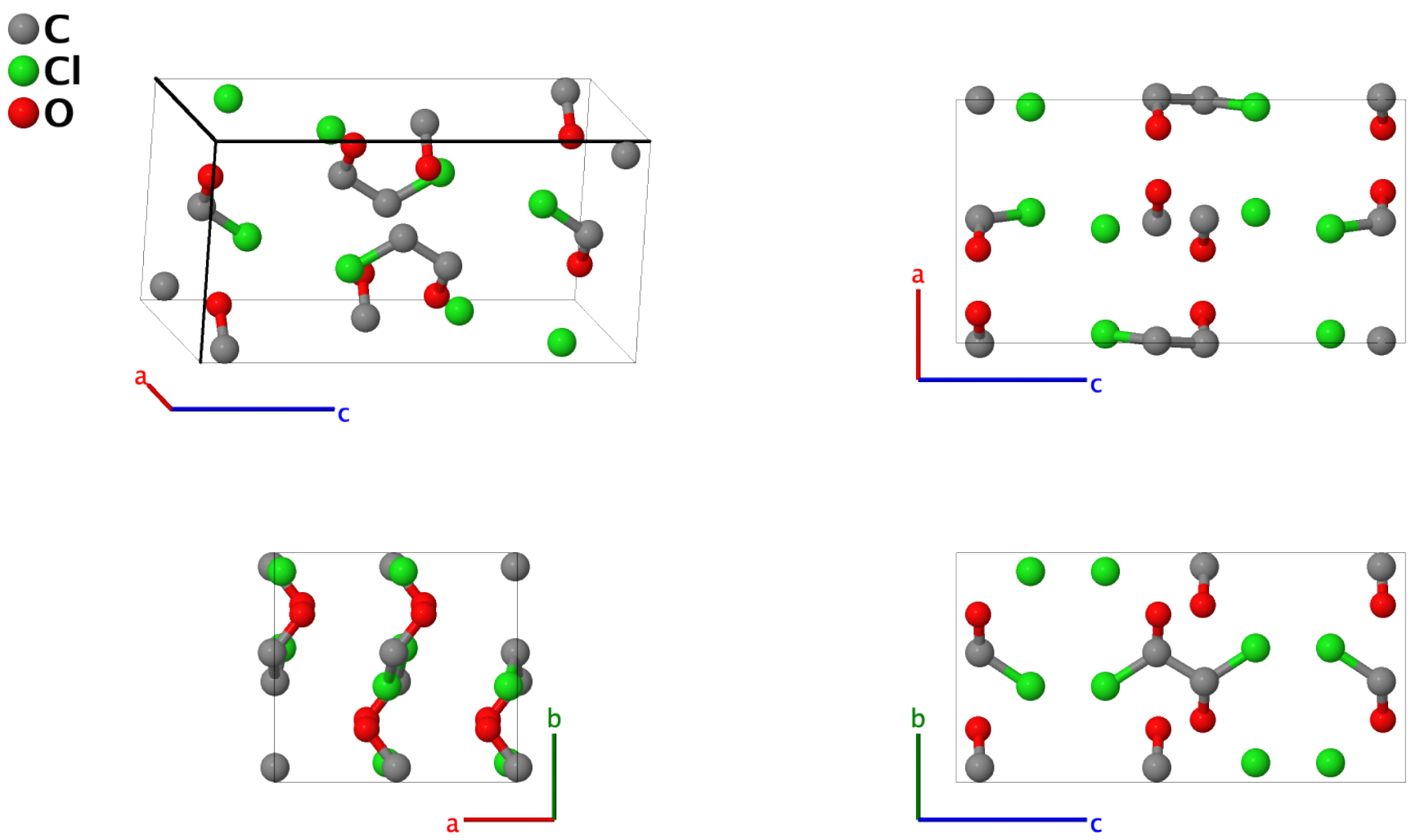

**Prototype** : CClO

**AFLOW prototype label** : ABC_oP24_61_c_c_c

***Strukturbericht* designation** : None

**Pearson symbol** : oP24

**Space group number** : 61

**Space group symbol** : *Pbca*

**AFLOW prototype command** : `aflow --proto=ABC_oP24_61_c_c_c`
`--params=`$a, b/a, c/a, x_1, y_1, z_1, x_2, y_2, z_2, x_3, y_3, z_3$

- (Groth, 1962) are not confident about the atomic positions, saying that "The computed values for the oxalyl chloride molecule ... [are probably] less reliable than those listed for the bromide molecule [COBr]." In fact, we have not been able to reconcile their crystal structure for COBr with the distances between the atoms, so we have left that structure out of the database at this time.

**Simple Orthorhombic primitive vectors:**

$$\mathbf{a}_1 = a\,\hat{\mathbf{x}}$$
$$\mathbf{a}_2 = b\,\hat{\mathbf{y}}$$
$$\mathbf{a}_3 = c\,\hat{\mathbf{z}}$$

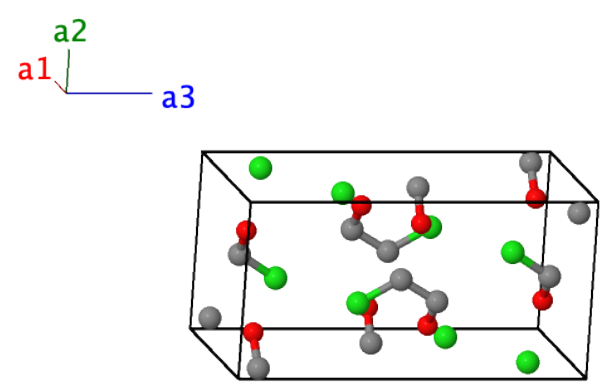

**Basis vectors:**

| | | Lattice Coordinates | | Cartesian Coordinates | Wyckoff Position | Atom Type |
|---|---|---|---|---|---|---|
| $\mathbf{B}_1$ | = | $x_1\,\mathbf{a}_1 + y_1\,\mathbf{a}_2 + z_1\,\mathbf{a}_3$ | = | $x_1 a\,\hat{\mathbf{x}} + y_1 b\,\hat{\mathbf{y}} + z_1 c\,\hat{\mathbf{z}}$ | (8c) | C |
| $\mathbf{B}_2$ | = | $\left(\frac{1}{2} - x_1\right)\mathbf{a}_1 - y_1\,\mathbf{a}_2 + \left(\frac{1}{2} + z_1\right)\mathbf{a}_3$ | = | $\left(\frac{1}{2} - x_1\right)a\,\hat{\mathbf{x}} - y_1 b\,\hat{\mathbf{y}} + \left(\frac{1}{2} + z_1\right)c\,\hat{\mathbf{z}}$ | (8c) | C |
| $\mathbf{B}_3$ | = | $-x_1\,\mathbf{a}_1 + \left(\frac{1}{2} + y_1\right)\mathbf{a}_2 + \left(\frac{1}{2} - z_1\right)\mathbf{a}_3$ | = | $-x_1 a\,\hat{\mathbf{x}} + \left(\frac{1}{2} + y_1\right)b\,\hat{\mathbf{y}} + \left(\frac{1}{2} - z_1\right)c\,\hat{\mathbf{z}}$ | (8c) | C |
| $\mathbf{B}_4$ | = | $\left(\frac{1}{2} + x_1\right)\mathbf{a}_1 + \left(\frac{1}{2} - y_1\right)\mathbf{a}_2 - z_1\,\mathbf{a}_3$ | = | $\left(\frac{1}{2} + x_1\right)a\,\hat{\mathbf{x}} + \left(\frac{1}{2} - y_1\right)b\,\hat{\mathbf{y}} - z_1 c\,\hat{\mathbf{z}}$ | (8c) | C |
| $\mathbf{B}_{5}$ | = | $-x_{1}\,\mathbf{a}_{1}-y_{1}\,\mathbf{a}_{2}-z_{1}\,\mathbf{a}_{3}$ | = | $-x_{1}a\,\hat{\mathbf{x}}-y_{1}b\,\hat{\mathbf{y}}-z_{1}c\,\hat{\mathbf{z}}$ | (8$c$) | C |
| $\mathbf{B}_{6}$ | = | $\left(\frac{1}{2}+x_{1}\right)\,\mathbf{a}_{1}+y_{1}\,\mathbf{a}_{2}+\left(\frac{1}{2}-z_{1}\right)\,\mathbf{a}_{3}$ | = | $\left(\frac{1}{2}+x_{1}\right)a\,\hat{\mathbf{x}}+y_{1}b\,\hat{\mathbf{y}}+\left(\frac{1}{2}-z_{1}\right)c\,\hat{\mathbf{z}}$ | (8$c$) | C |
| $\mathbf{B}_{7}$ | = | $x_{1}\,\mathbf{a}_{1}+\left(\frac{1}{2}-y_{1}\right)\,\mathbf{a}_{2}+\left(\frac{1}{2}+z_{1}\right)\,\mathbf{a}_{3}$ | = | $x_{1}a\,\hat{\mathbf{x}}+\left(\frac{1}{2}-y_{1}\right)b\,\hat{\mathbf{y}}+\left(\frac{1}{2}+z_{1}\right)c\,\hat{\mathbf{z}}$ | (8$c$) | C |
| $\mathbf{B}_{8}$ | = | $\left(\frac{1}{2}-x_{1}\right)\,\mathbf{a}_{1}+\left(\frac{1}{2}+y_{1}\right)\,\mathbf{a}_{2}+z_{1}\,\mathbf{a}_{3}$ | = | $\left(\frac{1}{2}-x_{1}\right)a\,\hat{\mathbf{x}}+\left(\frac{1}{2}+y_{1}\right)b\,\hat{\mathbf{y}}+z_{1}c\,\hat{\mathbf{z}}$ | (8$c$) | C |
| $\mathbf{B}_{9}$ | = | $x_{2}\,\mathbf{a}_{1}+y_{2}\,\mathbf{a}_{2}+z_{2}\,\mathbf{a}_{3}$ | = | $x_{2}a\,\hat{\mathbf{x}}+y_{2}b\,\hat{\mathbf{y}}+z_{2}c\,\hat{\mathbf{z}}$ | (8$c$) | Cl |
| $\mathbf{B}_{10}$ | = | $\left(\frac{1}{2}-x_{2}\right)\,\mathbf{a}_{1}-y_{2}\,\mathbf{a}_{2}+\left(\frac{1}{2}+z_{2}\right)\,\mathbf{a}_{3}$ | = | $\left(\frac{1}{2}-x_{2}\right)a\,\hat{\mathbf{x}}-y_{2}b\,\hat{\mathbf{y}}+\left(\frac{1}{2}+z_{2}\right)c\,\hat{\mathbf{z}}$ | (8$c$) | Cl |
| $\mathbf{B}_{11}$ | = | $-x_{2}\,\mathbf{a}_{1}+\left(\frac{1}{2}+y_{2}\right)\,\mathbf{a}_{2}+\left(\frac{1}{2}-z_{2}\right)\,\mathbf{a}_{3}$ | = | $-x_{2}a\,\hat{\mathbf{x}}+\left(\frac{1}{2}+y_{2}\right)b\,\hat{\mathbf{y}}+\left(\frac{1}{2}-z_{2}\right)c\,\hat{\mathbf{z}}$ | (8$c$) | Cl |
| $\mathbf{B}_{12}$ | = | $\left(\frac{1}{2}+x_{2}\right)\,\mathbf{a}_{1}+\left(\frac{1}{2}-y_{2}\right)\,\mathbf{a}_{2}-z_{2}\,\mathbf{a}_{3}$ | = | $\left(\frac{1}{2}+x_{2}\right)a\,\hat{\mathbf{x}}+\left(\frac{1}{2}-y_{2}\right)b\,\hat{\mathbf{y}}-z_{2}c\,\hat{\mathbf{z}}$ | (8$c$) | Cl |
| $\mathbf{B}_{13}$ | = | $-x_{2}\,\mathbf{a}_{1}-y_{2}\,\mathbf{a}_{2}-z_{2}\,\mathbf{a}_{3}$ | = | $-x_{2}a\,\hat{\mathbf{x}}-y_{2}b\,\hat{\mathbf{y}}-z_{2}c\,\hat{\mathbf{z}}$ | (8$c$) | Cl |
| $\mathbf{B}_{14}$ | = | $\left(\frac{1}{2}+x_{2}\right)\,\mathbf{a}_{1}+y_{2}\,\mathbf{a}_{2}+\left(\frac{1}{2}-z_{2}\right)\,\mathbf{a}_{3}$ | = | $\left(\frac{1}{2}+x_{2}\right)a\,\hat{\mathbf{x}}+y_{2}b\,\hat{\mathbf{y}}+\left(\frac{1}{2}-z_{2}\right)c\,\hat{\mathbf{z}}$ | (8$c$) | Cl |
| $\mathbf{B}_{15}$ | = | $x_{2}\,\mathbf{a}_{1}+\left(\frac{1}{2}-y_{2}\right)\,\mathbf{a}_{2}+\left(\frac{1}{2}+z_{2}\right)\,\mathbf{a}_{3}$ | = | $x_{2}a\,\hat{\mathbf{x}}+\left(\frac{1}{2}-y_{2}\right)b\,\hat{\mathbf{y}}+\left(\frac{1}{2}+z_{2}\right)c\,\hat{\mathbf{z}}$ | (8$c$) | Cl |
| $\mathbf{B}_{16}$ | = | $\left(\frac{1}{2}-x_{2}\right)\,\mathbf{a}_{1}+\left(\frac{1}{2}+y_{2}\right)\,\mathbf{a}_{2}+z_{2}\,\mathbf{a}_{3}$ | = | $\left(\frac{1}{2}-x_{2}\right)a\,\hat{\mathbf{x}}+\left(\frac{1}{2}+y_{2}\right)b\,\hat{\mathbf{y}}+z_{2}c\,\hat{\mathbf{z}}$ | (8$c$) | Cl |
| $\mathbf{B}_{17}$ | = | $x_{3}\,\mathbf{a}_{1}+y_{3}\,\mathbf{a}_{2}+z_{3}\,\mathbf{a}_{3}$ | = | $x_{3}a\,\hat{\mathbf{x}}+y_{3}b\,\hat{\mathbf{y}}+z_{3}c\,\hat{\mathbf{z}}$ | (8$c$) | O |
| $\mathbf{B}_{18}$ | = | $\left(\frac{1}{2}-x_{3}\right)\,\mathbf{a}_{1}-y_{3}\,\mathbf{a}_{2}+\left(\frac{1}{2}+z_{3}\right)\,\mathbf{a}_{3}$ | = | $\left(\frac{1}{2}-x_{3}\right)a\,\hat{\mathbf{x}}-y_{3}b\,\hat{\mathbf{y}}+\left(\frac{1}{2}+z_{3}\right)c\,\hat{\mathbf{z}}$ | (8$c$) | O |
| $\mathbf{B}_{19}$ | = | $-x_{3}\,\mathbf{a}_{1}+\left(\frac{1}{2}+y_{3}\right)\,\mathbf{a}_{2}+\left(\frac{1}{2}-z_{3}\right)\,\mathbf{a}_{3}$ | = | $-x_{3}a\,\hat{\mathbf{x}}+\left(\frac{1}{2}+y_{3}\right)b\,\hat{\mathbf{y}}+\left(\frac{1}{2}-z_{3}\right)c\,\hat{\mathbf{z}}$ | (8$c$) | O |
| $\mathbf{B}_{20}$ | = | $\left(\frac{1}{2}+x_{3}\right)\,\mathbf{a}_{1}+\left(\frac{1}{2}-y_{3}\right)\,\mathbf{a}_{2}-z_{3}\,\mathbf{a}_{3}$ | = | $\left(\frac{1}{2}+x_{3}\right)a\,\hat{\mathbf{x}}+\left(\frac{1}{2}-y_{3}\right)b\,\hat{\mathbf{y}}-z_{3}c\,\hat{\mathbf{z}}$ | (8$c$) | O |
| $\mathbf{B}_{21}$ | = | $-x_{3}\,\mathbf{a}_{1}-y_{3}\,\mathbf{a}_{2}-z_{3}\,\mathbf{a}_{3}$ | = | $-x_{3}a\,\hat{\mathbf{x}}-y_{3}b\,\hat{\mathbf{y}}-z_{3}c\,\hat{\mathbf{z}}$ | (8$c$) | O |
| $\mathbf{B}_{22}$ | = | $\left(\frac{1}{2}+x_{3}\right)\,\mathbf{a}_{1}+y_{3}\,\mathbf{a}_{2}+\left(\frac{1}{2}-z_{3}\right)\,\mathbf{a}_{3}$ | = | $\left(\frac{1}{2}+x_{3}\right)a\,\hat{\mathbf{x}}+y_{3}b\,\hat{\mathbf{y}}+\left(\frac{1}{2}-z_{3}\right)c\,\hat{\mathbf{z}}$ | (8$c$) | O |
| $\mathbf{B}_{23}$ | = | $x_{3}\,\mathbf{a}_{1}+\left(\frac{1}{2}-y_{3}\right)\,\mathbf{a}_{2}+\left(\frac{1}{2}+z_{3}\right)\,\mathbf{a}_{3}$ | = | $x_{3}a\,\hat{\mathbf{x}}+\left(\frac{1}{2}-y_{3}\right)b\,\hat{\mathbf{y}}+\left(\frac{1}{2}+z_{3}\right)c\,\hat{\mathbf{z}}$ | (8$c$) | O |
| $\mathbf{B}_{24}$ | = | $\left(\frac{1}{2}-x_{3}\right)\,\mathbf{a}_{1}+\left(\frac{1}{2}+y_{3}\right)\,\mathbf{a}_{2}+z_{3}\,\mathbf{a}_{3}$ | = | $\left(\frac{1}{2}-x_{3}\right)a\,\hat{\mathbf{x}}+\left(\frac{1}{2}+y_{3}\right)b\,\hat{\mathbf{y}}+z_{3}c\,\hat{\mathbf{z}}$ | (8$c$) | O |

---

---

**Geometry files:**

- CIF: pp. 1628
- POSCAR: pp. 1629

# Topaz ($Al_2SiO_4F_2$, $S0_5$) Structure: A2B2C4D_oP36_62_d_d_2cd_c

http://aflow.org/prototype-encyclopedia/A2B2C4D_oP36_62_d_d_2cd_c

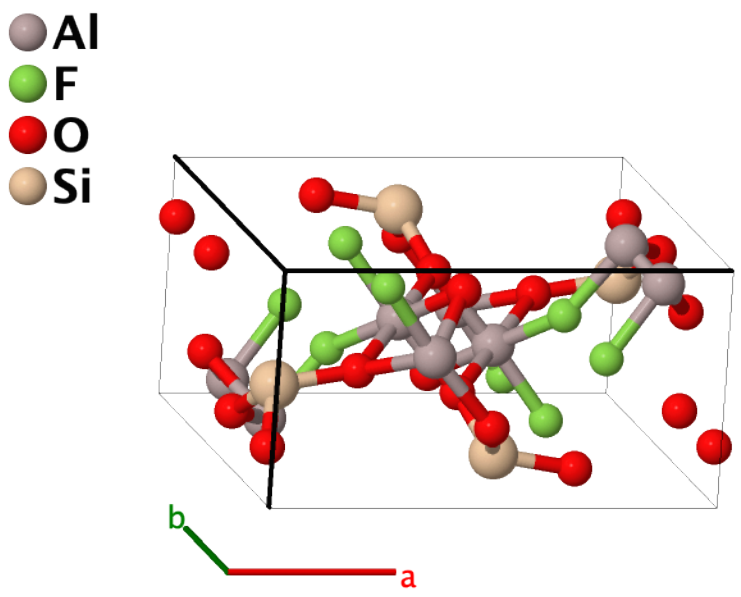

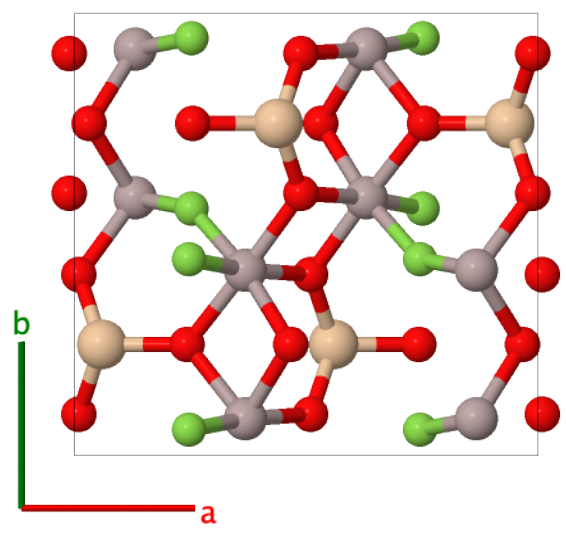

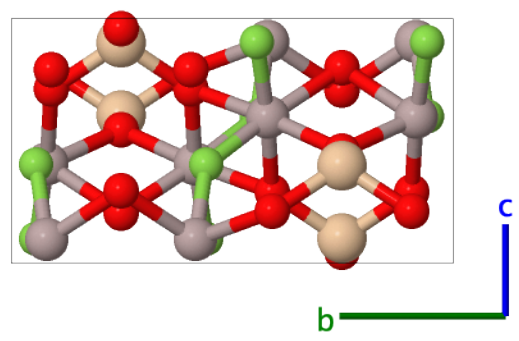

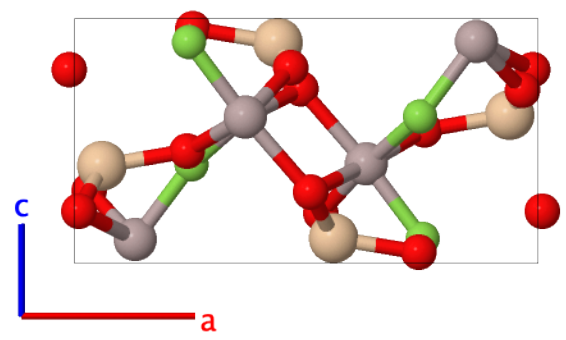

| | | |
|---|---|---|
| **Prototype** | : | $Al_2F_2O_4Si$ |
| **AFLOW prototype label** | : | A2B2C4D_oP36_62_d_d_2cd_c |
| ***Strukturbericht* designation** | : | $S0_5$ |
| **Pearson symbol** | : | oP36 |
| **Space group number** | : | 62 |
| **Space group symbol** | : | *Pnma* |
| **AFLOW prototype command** | : | `aflow --proto=A2B2C4D_oP36_62_d_d_2cd_c` `--params=`$a, b/a, c/a, x_1, z_1, x_2, z_2, x_3, z_3, x_4, y_4, z_4, x_5, y_5, z_5, x_6, y_6, z_6$ |

**Other compounds with this structure**

- $Al_2SiO_4(F,OH)_2$

- The fluorine (8*d*) site in topaz can have a composition $F_{1-x}(OH)_x$, where $x < 0.3$.
- We use the data at $T = 298$ K from the "Pax08" sample in (Komatsu, 2003).
- The data was presented in the *Pbnm* setting of space group #62. We used FINDSYM to transform this into the standard *Pnma* setting.
- (Herman, 1937) gave this the $S0_5$ designation, but co-listed it as $H5_5$ in the index.

**Simple Orthorhombic primitive vectors:**

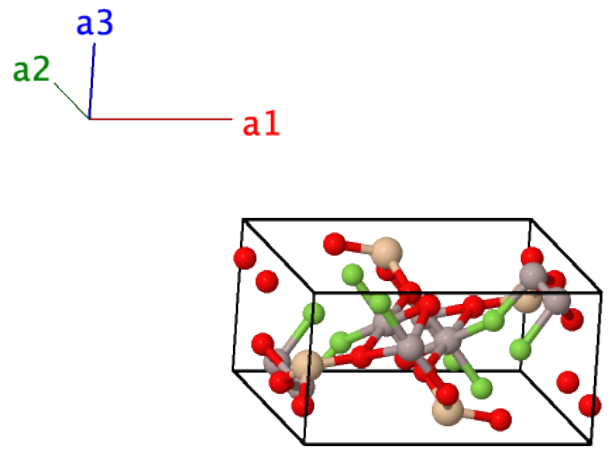

$$\mathbf{a}_1 = a\,\hat{\mathbf{x}}$$
$$\mathbf{a}_2 = b\,\hat{\mathbf{y}}$$
$$\mathbf{a}_3 = c\,\hat{\mathbf{z}}$$

**Basis vectors:**

| | | Lattice Coordinates | | Cartesian Coordinates | Wyckoff Position | Atom Type |
|---|---|---|---|---|---|---|
| $\mathbf{B}_1$ | = | $x_1\,\mathbf{a}_1 + \frac{1}{4}\,\mathbf{a}_2 + z_1\,\mathbf{a}_3$ | = | $x_1 a\,\hat{\mathbf{x}} + \frac{1}{4} b\,\hat{\mathbf{y}} + z_1 c\,\hat{\mathbf{z}}$ | (4$c$) | O I |
| $\mathbf{B}_2$ | = | $\left(\frac{1}{2} - x_1\right)\mathbf{a}_1 + \frac{3}{4}\,\mathbf{a}_2 + \left(\frac{1}{2} + z_1\right)\mathbf{a}_3$ | = | $\left(\frac{1}{2} - x_1\right) a\,\hat{\mathbf{x}} + \frac{3}{4} b\,\hat{\mathbf{y}} + \left(\frac{1}{2} + z_1\right) c\,\hat{\mathbf{z}}$ | (4$c$) | O I |
| $\mathbf{B}_3$ | = | $-x_1\,\mathbf{a}_1 + \frac{3}{4}\,\mathbf{a}_2 - z_1\,\mathbf{a}_3$ | = | $-x_1 a\,\hat{\mathbf{x}} + \frac{3}{4} b\,\hat{\mathbf{y}} - z_1 c\,\hat{\mathbf{z}}$ | (4$c$) | O I |
| $\mathbf{B}_4$ | = | $\left(\frac{1}{2} + x_1\right)\mathbf{a}_1 + \frac{1}{4}\,\mathbf{a}_2 + \left(\frac{1}{2} - z_1\right)\mathbf{a}_3$ | = | $\left(\frac{1}{2} + x_1\right) a\,\hat{\mathbf{x}} + \frac{1}{4} b\,\hat{\mathbf{y}} + \left(\frac{1}{2} - z_1\right) c\,\hat{\mathbf{z}}$ | (4$c$) | O I |
| $\mathbf{B}_5$ | = | $x_2\,\mathbf{a}_1 + \frac{1}{4}\,\mathbf{a}_2 + z_2\,\mathbf{a}_3$ | = | $x_2 a\,\hat{\mathbf{x}} + \frac{1}{4} b\,\hat{\mathbf{y}} + z_2 c\,\hat{\mathbf{z}}$ | (4$c$) | O II |
| $\mathbf{B}_6$ | = | $\left(\frac{1}{2} - x_2\right)\mathbf{a}_1 + \frac{3}{4}\,\mathbf{a}_2 + \left(\frac{1}{2} + z_2\right)\mathbf{a}_3$ | = | $\left(\frac{1}{2} - x_2\right) a\,\hat{\mathbf{x}} + \frac{3}{4} b\,\hat{\mathbf{y}} + \left(\frac{1}{2} + z_2\right) c\,\hat{\mathbf{z}}$ | (4$c$) | O II |
| $\mathbf{B}_7$ | = | $-x_2\,\mathbf{a}_1 + \frac{3}{4}\,\mathbf{a}_2 - z_2\,\mathbf{a}_3$ | = | $-x_2 a\,\hat{\mathbf{x}} + \frac{3}{4} b\,\hat{\mathbf{y}} - z_2 c\,\hat{\mathbf{z}}$ | (4$c$) | O II |
| $\mathbf{B}_8$ | = | $\left(\frac{1}{2} + x_2\right)\mathbf{a}_1 + \frac{1}{4}\,\mathbf{a}_2 + \left(\frac{1}{2} - z_2\right)\mathbf{a}_3$ | = | $\left(\frac{1}{2} + x_2\right) a\,\hat{\mathbf{x}} + \frac{1}{4} b\,\hat{\mathbf{y}} + \left(\frac{1}{2} - z_2\right) c\,\hat{\mathbf{z}}$ | (4$c$) | O II |
| $\mathbf{B}_9$ | = | $x_3\,\mathbf{a}_1 + \frac{1}{4}\,\mathbf{a}_2 + z_3\,\mathbf{a}_3$ | = | $x_3 a\,\hat{\mathbf{x}} + \frac{1}{4} b\,\hat{\mathbf{y}} + z_3 c\,\hat{\mathbf{z}}$ | (4$c$) | Si |
| $\mathbf{B}_{10}$ | = | $\left(\frac{1}{2} - x_3\right)\mathbf{a}_1 + \frac{3}{4}\,\mathbf{a}_2 + \left(\frac{1}{2} + z_3\right)\mathbf{a}_3$ | = | $\left(\frac{1}{2} - x_3\right) a\,\hat{\mathbf{x}} + \frac{3}{4} b\,\hat{\mathbf{y}} + \left(\frac{1}{2} + z_3\right) c\,\hat{\mathbf{z}}$ | (4$c$) | Si |
| $\mathbf{B}_{11}$ | = | $-x_3\,\mathbf{a}_1 + \frac{3}{4}\,\mathbf{a}_2 - z_3\,\mathbf{a}_3$ | = | $-x_3 a\,\hat{\mathbf{x}} + \frac{3}{4} b\,\hat{\mathbf{y}} - z_3 c\,\hat{\mathbf{z}}$ | (4$c$) | Si |
| $\mathbf{B}_{12}$ | = | $\left(\frac{1}{2} + x_3\right)\mathbf{a}_1 + \frac{1}{4}\,\mathbf{a}_2 + \left(\frac{1}{2} - z_3\right)\mathbf{a}_3$ | = | $\left(\frac{1}{2} + x_3\right) a\,\hat{\mathbf{x}} + \frac{1}{4} b\,\hat{\mathbf{y}} + \left(\frac{1}{2} - z_3\right) c\,\hat{\mathbf{z}}$ | (4$c$) | Si |
| $\mathbf{B}_{13}$ | = | $x_4\,\mathbf{a}_1 + y_4\,\mathbf{a}_2 + z_4\,\mathbf{a}_3$ | = | $x_4 a\,\hat{\mathbf{x}} + y_4 b\,\hat{\mathbf{y}} + z_4 c\,\hat{\mathbf{z}}$ | (8$d$) | Al |
| $\mathbf{B}_{14}$ | = | $\left(\frac{1}{2} - x_4\right)\mathbf{a}_1 - y_4\,\mathbf{a}_2 + \left(\frac{1}{2} + z_4\right)\mathbf{a}_3$ | = | $\left(\frac{1}{2} - x_4\right) a\,\hat{\mathbf{x}} - y_4 b\,\hat{\mathbf{y}} + \left(\frac{1}{2} + z_4\right) c\,\hat{\mathbf{z}}$ | (8$d$) | Al |
| $\mathbf{B}_{15}$ | = | $-x_4\,\mathbf{a}_1 + \left(\frac{1}{2} + y_4\right)\mathbf{a}_2 - z_4\,\mathbf{a}_3$ | = | $-x_4 a\,\hat{\mathbf{x}} + \left(\frac{1}{2} + y_4\right) b\,\hat{\mathbf{y}} - z_4 c\,\hat{\mathbf{z}}$ | (8$d$) | Al |
| $\mathbf{B}_{16}$ | = | $\left(\frac{1}{2} + x_4\right)\mathbf{a}_1 + \left(\frac{1}{2} - y_4\right)\mathbf{a}_2 + \left(\frac{1}{2} - z_4\right)\mathbf{a}_3$ | = | $\left(\frac{1}{2} + x_4\right) a\,\hat{\mathbf{x}} + \left(\frac{1}{2} - y_4\right) b\,\hat{\mathbf{y}} + \left(\frac{1}{2} - z_4\right) c\,\hat{\mathbf{z}}$ | (8$d$) | Al |
| $\mathbf{B}_{17}$ | = | $-x_4\,\mathbf{a}_1 - y_4\,\mathbf{a}_2 - z_4\,\mathbf{a}_3$ | = | $-x_4 a\,\hat{\mathbf{x}} - y_4 b\,\hat{\mathbf{y}} - z_4 c\,\hat{\mathbf{z}}$ | (8$d$) | Al |
| $\mathbf{B}_{18}$ | = | $\left(\frac{1}{2} + x_4\right)\mathbf{a}_1 + y_4\,\mathbf{a}_2 + \left(\frac{1}{2} - z_4\right)\mathbf{a}_3$ | = | $\left(\frac{1}{2} + x_4\right) a\,\hat{\mathbf{x}} + y_4 b\,\hat{\mathbf{y}} + \left(\frac{1}{2} - z_4\right) c\,\hat{\mathbf{z}}$ | (8$d$) | Al |
| $\mathbf{B}_{19}$ | = | $x_4\,\mathbf{a}_1 + \left(\frac{1}{2} - y_4\right)\mathbf{a}_2 + z_4\,\mathbf{a}_3$ | = | $x_4 a\,\hat{\mathbf{x}} + \left(\frac{1}{2} - y_4\right) b\,\hat{\mathbf{y}} + z_4 c\,\hat{\mathbf{z}}$ | (8$d$) | Al |
| $\mathbf{B}_{20}$ | = | $\left(\frac{1}{2} - x_4\right)\mathbf{a}_1 + \left(\frac{1}{2} + y_4\right)\mathbf{a}_2 + \left(\frac{1}{2} + z_4\right)\mathbf{a}_3$ | = | $\left(\frac{1}{2} - x_4\right) a\,\hat{\mathbf{x}} + \left(\frac{1}{2} + y_4\right) b\,\hat{\mathbf{y}} + \left(\frac{1}{2} + z_4\right) c\,\hat{\mathbf{z}}$ | (8$d$) | Al |
| $\mathbf{B}_{21}$ | = | $x_5\,\mathbf{a}_1 + y_5\,\mathbf{a}_2 + z_5\,\mathbf{a}_3$ | = | $x_5 a\,\hat{\mathbf{x}} + y_5 b\,\hat{\mathbf{y}} + z_5 c\,\hat{\mathbf{z}}$ | (8$d$) | F |
| $\mathbf{B}_{22}$ | = | $\left(\frac{1}{2} - x_5\right)\mathbf{a}_1 - y_5\,\mathbf{a}_2 + \left(\frac{1}{2} + z_5\right)\mathbf{a}_3$ | = | $\left(\frac{1}{2} - x_5\right) a\,\hat{\mathbf{x}} - y_5 b\,\hat{\mathbf{y}} + \left(\frac{1}{2} + z_5\right) c\,\hat{\mathbf{z}}$ | (8$d$) | F |
| $\mathbf{B}_{23}$ | = | $-x_5\,\mathbf{a}_1 + \left(\frac{1}{2} + y_5\right)\mathbf{a}_2 - z_5\,\mathbf{a}_3$ | = | $-x_5 a\,\hat{\mathbf{x}} + \left(\frac{1}{2} + y_5\right) b\,\hat{\mathbf{y}} - z_5 c\,\hat{\mathbf{z}}$ | (8$d$) | F |
| $\mathbf{B}_{24}$ | = | $\left(\frac{1}{2} + x_5\right)\mathbf{a}_1 + \left(\frac{1}{2} - y_5\right)\mathbf{a}_2 + \left(\frac{1}{2} - z_5\right)\mathbf{a}_3$ | = | $\left(\frac{1}{2} + x_5\right) a\,\hat{\mathbf{x}} + \left(\frac{1}{2} - y_5\right) b\,\hat{\mathbf{y}} + \left(\frac{1}{2} - z_5\right) c\,\hat{\mathbf{z}}$ | (8$d$) | F |
| $\mathbf{B}_{25}$ | = | $-x_5\,\mathbf{a}_1 - y_5\,\mathbf{a}_2 - z_5\,\mathbf{a}_3$ | = | $-x_5 a\,\hat{\mathbf{x}} - y_5 b\,\hat{\mathbf{y}} - z_5 c\,\hat{\mathbf{z}}$ | (8$d$) | F |
| $\mathbf{B}_{26}$ | = | $\left(\frac{1}{2}+x_5\right)\mathbf{a}_1+y_5\,\mathbf{a}_2+\left(\frac{1}{2}-z_5\right)\mathbf{a}_3$ | = | $\left(\frac{1}{2}+x_5\right)a\,\hat{\mathbf{x}}+y_5 b\,\hat{\mathbf{y}}+\left(\frac{1}{2}-z_5\right)c\,\hat{\mathbf{z}}$ | (8*d*) | F |
| $\mathbf{B}_{27}$ | = | $x_5\,\mathbf{a}_1+\left(\frac{1}{2}-y_5\right)\mathbf{a}_2+z_5\,\mathbf{a}_3$ | = | $x_5 a\,\hat{\mathbf{x}}+\left(\frac{1}{2}-y_5\right)b\,\hat{\mathbf{y}}+z_5 c\,\hat{\mathbf{z}}$ | (8*d*) | F |
| $\mathbf{B}_{28}$ | = | $\left(\frac{1}{2}-x_5\right)\mathbf{a}_1+\left(\frac{1}{2}+y_5\right)\mathbf{a}_2+\left(\frac{1}{2}+z_5\right)\mathbf{a}_3$ | = | $\left(\frac{1}{2}-x_5\right)a\,\hat{\mathbf{x}}+\left(\frac{1}{2}+y_5\right)b\,\hat{\mathbf{y}}+\left(\frac{1}{2}+z_5\right)c\,\hat{\mathbf{z}}$ | (8*d*) | F |
| $\mathbf{B}_{29}$ | = | $x_6\,\mathbf{a}_1+y_6\,\mathbf{a}_2+z_6\,\mathbf{a}_3$ | = | $x_6 a\,\hat{\mathbf{x}}+y_6 b\,\hat{\mathbf{y}}+z_6 c\,\hat{\mathbf{z}}$ | (8*d*) | O III |
| $\mathbf{B}_{30}$ | = | $\left(\frac{1}{2}-x_6\right)\mathbf{a}_1-y_6\,\mathbf{a}_2+\left(\frac{1}{2}+z_6\right)\mathbf{a}_3$ | = | $\left(\frac{1}{2}-x_6\right)a\,\hat{\mathbf{x}}-y_6 b\,\hat{\mathbf{y}}+\left(\frac{1}{2}+z_6\right)c\,\hat{\mathbf{z}}$ | (8*d*) | O III |
| $\mathbf{B}_{31}$ | = | $-x_6\,\mathbf{a}_1+\left(\frac{1}{2}+y_6\right)\mathbf{a}_2-z_6\,\mathbf{a}_3$ | = | $-x_6 a\,\hat{\mathbf{x}}+\left(\frac{1}{2}+y_6\right)b\,\hat{\mathbf{y}}-z_6 c\,\hat{\mathbf{z}}$ | (8*d*) | O III |
| $\mathbf{B}_{32}$ | = | $\left(\frac{1}{2}+x_6\right)\mathbf{a}_1+\left(\frac{1}{2}-y_6\right)\mathbf{a}_2+\left(\frac{1}{2}-z_6\right)\mathbf{a}_3$ | = | $\left(\frac{1}{2}+x_6\right)a\,\hat{\mathbf{x}}+\left(\frac{1}{2}-y_6\right)b\,\hat{\mathbf{y}}+\left(\frac{1}{2}-z_6\right)c\,\hat{\mathbf{z}}$ | (8*d*) | O III |
| $\mathbf{B}_{33}$ | = | $-x_6\,\mathbf{a}_1-y_6\,\mathbf{a}_2-z_6\,\mathbf{a}_3$ | = | $-x_6 a\,\hat{\mathbf{x}}-y_6 b\,\hat{\mathbf{y}}-z_6 c\,\hat{\mathbf{z}}$ | (8*d*) | O III |
| $\mathbf{B}_{34}$ | = | $\left(\frac{1}{2}+x_6\right)\mathbf{a}_1+y_6\,\mathbf{a}_2+\left(\frac{1}{2}-z_6\right)\mathbf{a}_3$ | = | $\left(\frac{1}{2}+x_6\right)a\,\hat{\mathbf{x}}+y_6 b\,\hat{\mathbf{y}}+\left(\frac{1}{2}-z_6\right)c\,\hat{\mathbf{z}}$ | (8*d*) | O III |
| $\mathbf{B}_{35}$ | = | $x_6\,\mathbf{a}_1+\left(\frac{1}{2}-y_6\right)\mathbf{a}_2+z_6\,\mathbf{a}_3$ | = | $x_6 a\,\hat{\mathbf{x}}+\left(\frac{1}{2}-y_6\right)b\,\hat{\mathbf{y}}+z_6 c\,\hat{\mathbf{z}}$ | (8*d*) | O III |
| $\mathbf{B}_{36}$ | = | $\left(\frac{1}{2}-x_6\right)\mathbf{a}_1+\left(\frac{1}{2}+y_6\right)\mathbf{a}_2+\left(\frac{1}{2}+z_6\right)\mathbf{a}_3$ | = | $\left(\frac{1}{2}-x_6\right)a\,\hat{\mathbf{x}}+\left(\frac{1}{2}+y_6\right)b\,\hat{\mathbf{y}}+\left(\frac{1}{2}+z_6\right)c\,\hat{\mathbf{z}}$ | (8*d*) | O III |

---

---

**Geometry files:**

- CIF: pp. 1629
- POSCAR: pp. 1629

# Norbergite [$Mg(F,OH)_2 \cdot Mg_2SiO_4$, $S0_7$] Structure: A2B3C4D_oP40_62_d_cd_2cd_c

http://aflow.org/prototype-encyclopedia/A2B3C4D_oP40_62_d_cd_2cd_c

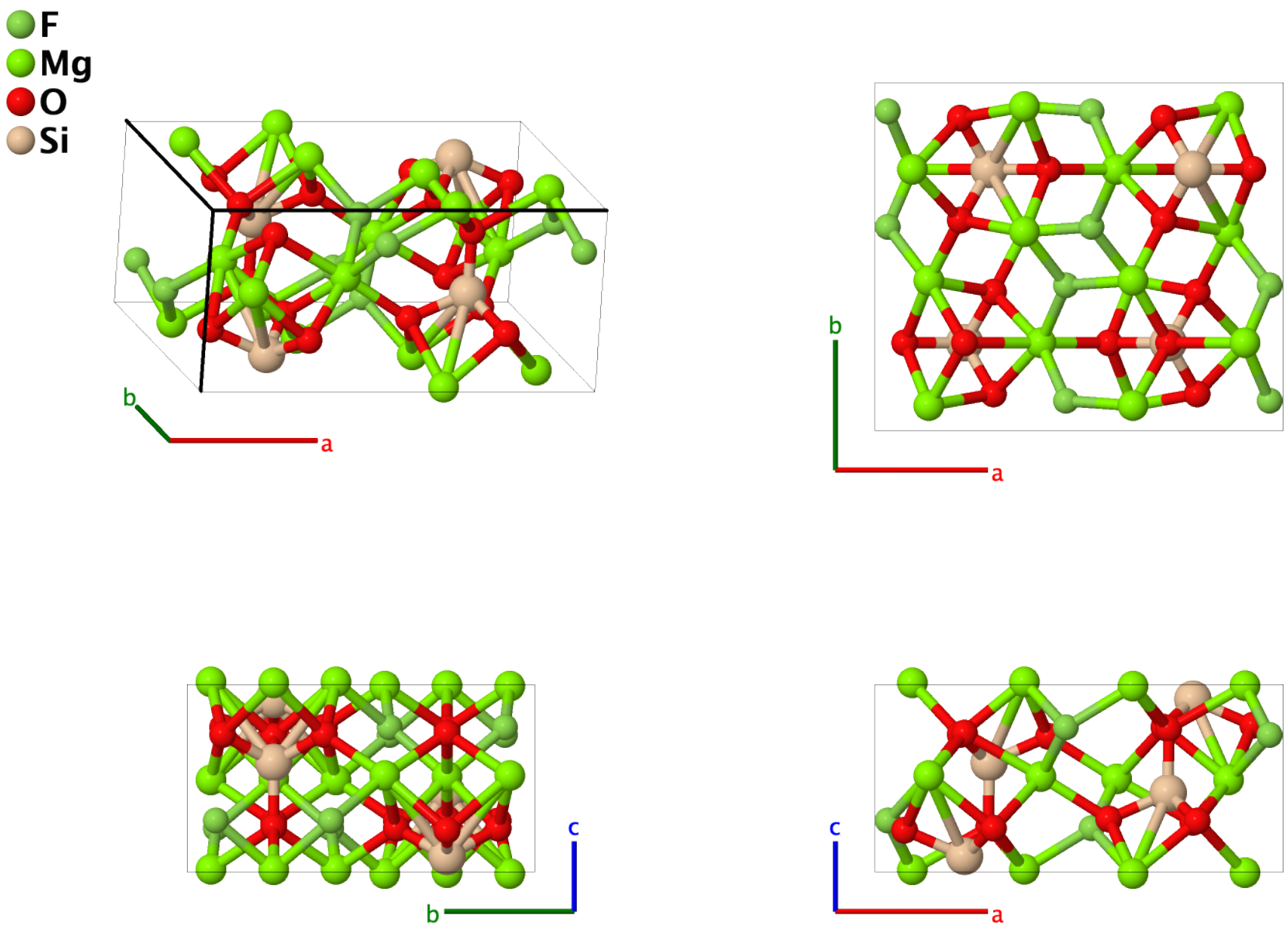

| | | |
|---|---|---|
| **Prototype** | : | $F_2Mg_3O_4Si$ |
| **AFLOW prototype label** | : | A2B3C4D_oP40_62_d_cd_2cd_c |
| ***Strukturbericht* designation** | : | $S0_7$ |
| **Pearson symbol** | : | oP40 |
| **Space group number** | : | 62 |
| **Space group symbol** | : | *Pnma* |
| **AFLOW prototype command** | : | `aflow --proto=A2B3C4D_oP40_62_d_cd_2cd_c` `--params=`$a, b/a, c/a, x_1, z_1, x_2, z_2, x_3, z_3, x_4, z_4, x_5, y_5, z_5, x_6, y_6, z_6, x_7, y_7, z_7$ |

- (Hermann, 1937) define *Strukturbericht* designation $S0_7$ as the general class of humite materials, $Mg(F,OH)_2 \cdot nMg_2SiO_4$. Their major reference for this is (Taylor, 1929), who gave structural data for the $n = 1$ structure, norbergite.
- We use the data from the fluorine-rich sample studied by (Gibbs, 1969), and assume that the (F, OH) site is pure fluorine. (Camara, 1997) studied OH-rich samples, and were able to locate the hydrogen atoms.
- (Gibbs, 1969) give the structure in the *Pbnm* setting of space group #62. We used FINDSYM to change this the the standard *Pnma* setting. This involves a rotation of the axis so that $\hat{x} \rightarrow \hat{z} \rightarrow \hat{y} \rightarrow \hat{x}$.

**Simple Orthorhombic primitive vectors:**

$$\mathbf{a}_1 = a\,\hat{\mathbf{x}}$$
$$\mathbf{a}_2 = b\,\hat{\mathbf{y}}$$
$$\mathbf{a}_3 = c\,\hat{\mathbf{z}}$$

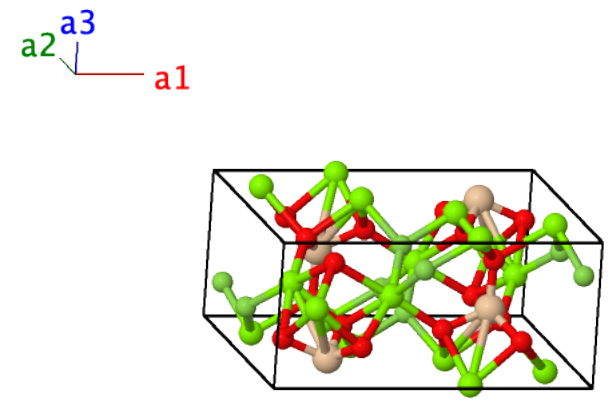

**Basis vectors:**

| | | Lattice Coordinates | | Cartesian Coordinates | Wyckoff Position | Atom Type |
|---|---|---|---|---|---|---|
| $\mathbf{B}_1$ | = | $x_1\,\mathbf{a}_1 + \frac{1}{4}\,\mathbf{a}_2 + z_1\,\mathbf{a}_3$ | = | $x_1 a\,\hat{\mathbf{x}} + \frac{1}{4} b\,\hat{\mathbf{y}} + z_1 c\,\hat{\mathbf{z}}$ | (4$c$) | Mg I |
| $\mathbf{B}_2$ | = | $\left(\frac{1}{2} - x_1\right)\mathbf{a}_1 + \frac{3}{4}\,\mathbf{a}_2 + \left(\frac{1}{2} + z_1\right)\mathbf{a}_3$ | = | $\left(\frac{1}{2} - x_1\right) a\,\hat{\mathbf{x}} + \frac{3}{4} b\,\hat{\mathbf{y}} + \left(\frac{1}{2} + z_1\right) c\,\hat{\mathbf{z}}$ | (4$c$) | Mg I |
| $\mathbf{B}_3$ | = | $-x_1\,\mathbf{a}_1 + \frac{3}{4}\,\mathbf{a}_2 - z_1\,\mathbf{a}_3$ | = | $-x_1 a\,\hat{\mathbf{x}} + \frac{3}{4} b\,\hat{\mathbf{y}} - z_1 c\,\hat{\mathbf{z}}$ | (4$c$) | Mg I |
| $\mathbf{B}_4$ | = | $\left(\frac{1}{2} + x_1\right)\mathbf{a}_1 + \frac{1}{4}\,\mathbf{a}_2 + \left(\frac{1}{2} - z_1\right)\mathbf{a}_3$ | = | $\left(\frac{1}{2} + x_1\right) a\,\hat{\mathbf{x}} + \frac{1}{4} b\,\hat{\mathbf{y}} + \left(\frac{1}{2} - z_1\right) c\,\hat{\mathbf{z}}$ | (4$c$) | Mg I |
| $\mathbf{B}_5$ | = | $x_2\,\mathbf{a}_1 + \frac{1}{4}\,\mathbf{a}_2 + z_2\,\mathbf{a}_3$ | = | $x_2 a\,\hat{\mathbf{x}} + \frac{1}{4} b\,\hat{\mathbf{y}} + z_2 c\,\hat{\mathbf{z}}$ | (4$c$) | O I |
| $\mathbf{B}_6$ | = | $\left(\frac{1}{2} - x_2\right)\mathbf{a}_1 + \frac{3}{4}\,\mathbf{a}_2 + \left(\frac{1}{2} + z_2\right)\mathbf{a}_3$ | = | $\left(\frac{1}{2} - x_2\right) a\,\hat{\mathbf{x}} + \frac{3}{4} b\,\hat{\mathbf{y}} + \left(\frac{1}{2} + z_2\right) c\,\hat{\mathbf{z}}$ | (4$c$) | O I |
| $\mathbf{B}_7$ | = | $-x_2\,\mathbf{a}_1 + \frac{3}{4}\,\mathbf{a}_2 - z_2\,\mathbf{a}_3$ | = | $-x_2 a\,\hat{\mathbf{x}} + \frac{3}{4} b\,\hat{\mathbf{y}} - z_2 c\,\hat{\mathbf{z}}$ | (4$c$) | O I |
| $\mathbf{B}_8$ | = | $\left(\frac{1}{2} + x_2\right)\mathbf{a}_1 + \frac{1}{4}\,\mathbf{a}_2 + \left(\frac{1}{2} - z_2\right)\mathbf{a}_3$ | = | $\left(\frac{1}{2} + x_2\right) a\,\hat{\mathbf{x}} + \frac{1}{4} b\,\hat{\mathbf{y}} + \left(\frac{1}{2} - z_2\right) c\,\hat{\mathbf{z}}$ | (4$c$) | O I |
| $\mathbf{B}_9$ | = | $x_3\,\mathbf{a}_1 + \frac{1}{4}\,\mathbf{a}_2 + z_3\,\mathbf{a}_3$ | = | $x_3 a\,\hat{\mathbf{x}} + \frac{1}{4} b\,\hat{\mathbf{y}} + z_3 c\,\hat{\mathbf{z}}$ | (4$c$) | O II |
| $\mathbf{B}_{10}$ | = | $\left(\frac{1}{2} - x_3\right)\mathbf{a}_1 + \frac{3}{4}\,\mathbf{a}_2 + \left(\frac{1}{2} + z_3\right)\mathbf{a}_3$ | = | $\left(\frac{1}{2} - x_3\right) a\,\hat{\mathbf{x}} + \frac{3}{4} b\,\hat{\mathbf{y}} + \left(\frac{1}{2} + z_3\right) c\,\hat{\mathbf{z}}$ | (4$c$) | O II |
| $\mathbf{B}_{11}$ | = | $-x_3\,\mathbf{a}_1 + \frac{3}{4}\,\mathbf{a}_2 - z_3\,\mathbf{a}_3$ | = | $-x_3 a\,\hat{\mathbf{x}} + \frac{3}{4} b\,\hat{\mathbf{y}} - z_3 c\,\hat{\mathbf{z}}$ | (4$c$) | O II |
| $\mathbf{B}_{12}$ | = | $\left(\frac{1}{2} + x_3\right)\mathbf{a}_1 + \frac{1}{4}\,\mathbf{a}_2 + \left(\frac{1}{2} - z_3\right)\mathbf{a}_3$ | = | $\left(\frac{1}{2} + x_3\right) a\,\hat{\mathbf{x}} + \frac{1}{4} b\,\hat{\mathbf{y}} + \left(\frac{1}{2} - z_3\right) c\,\hat{\mathbf{z}}$ | (4$c$) | O II |
| $\mathbf{B}_{13}$ | = | $x_4\,\mathbf{a}_1 + \frac{1}{4}\,\mathbf{a}_2 + z_4\,\mathbf{a}_3$ | = | $x_4 a\,\hat{\mathbf{x}} + \frac{1}{4} b\,\hat{\mathbf{y}} + z_4 c\,\hat{\mathbf{z}}$ | (4$c$) | Si |
| $\mathbf{B}_{14}$ | = | $\left(\frac{1}{2} - x_4\right)\mathbf{a}_1 + \frac{3}{4}\,\mathbf{a}_2 + \left(\frac{1}{2} + z_4\right)\mathbf{a}_3$ | = | $\left(\frac{1}{2} - x_4\right) a\,\hat{\mathbf{x}} + \frac{3}{4} b\,\hat{\mathbf{y}} + \left(\frac{1}{2} + z_4\right) c\,\hat{\mathbf{z}}$ | (4$c$) | Si |
| $\mathbf{B}_{15}$ | = | $-x_4\,\mathbf{a}_1 + \frac{3}{4}\,\mathbf{a}_2 - z_4\,\mathbf{a}_3$ | = | $-x_4 a\,\hat{\mathbf{x}} + \frac{3}{4} b\,\hat{\mathbf{y}} - z_4 c\,\hat{\mathbf{z}}$ | (4$c$) | Si |
| $\mathbf{B}_{16}$ | = | $\left(\frac{1}{2} + x_4\right)\mathbf{a}_1 + \frac{1}{4}\,\mathbf{a}_2 + \left(\frac{1}{2} - z_4\right)\mathbf{a}_3$ | = | $\left(\frac{1}{2} + x_4\right) a\,\hat{\mathbf{x}} + \frac{1}{4} b\,\hat{\mathbf{y}} + \left(\frac{1}{2} - z_4\right) c\,\hat{\mathbf{z}}$ | (4$c$) | Si |
| $\mathbf{B}_{17}$ | = | $x_5\,\mathbf{a}_1 + y_5\,\mathbf{a}_2 + z_5\,\mathbf{a}_3$ | = | $x_5 a\,\hat{\mathbf{x}} + y_5 b\,\hat{\mathbf{y}} + z_5 c\,\hat{\mathbf{z}}$ | (8$d$) | F |
| $\mathbf{B}_{18}$ | = | $\left(\frac{1}{2} - x_5\right)\mathbf{a}_1 - y_5\,\mathbf{a}_2 + \left(\frac{1}{2} + z_5\right)\mathbf{a}_3$ | = | $\left(\frac{1}{2} - x_5\right) a\,\hat{\mathbf{x}} - y_5 b\,\hat{\mathbf{y}} + \left(\frac{1}{2} + z_5\right) c\,\hat{\mathbf{z}}$ | (8$d$) | F |
| $\mathbf{B}_{19}$ | = | $-x_5\,\mathbf{a}_1 + \left(\frac{1}{2} + y_5\right)\mathbf{a}_2 - z_5\,\mathbf{a}_3$ | = | $-x_5 a\,\hat{\mathbf{x}} + \left(\frac{1}{2} + y_5\right) b\,\hat{\mathbf{y}} - z_5 c\,\hat{\mathbf{z}}$ | (8$d$) | F |
| $\mathbf{B}_{20}$ | = | $\left(\frac{1}{2} + x_5\right)\mathbf{a}_1 + \left(\frac{1}{2} - y_5\right)\mathbf{a}_2 + \left(\frac{1}{2} - z_5\right)\mathbf{a}_3$ | = | $\left(\frac{1}{2} + x_5\right) a\,\hat{\mathbf{x}} + \left(\frac{1}{2} - y_5\right) b\,\hat{\mathbf{y}} + \left(\frac{1}{2} - z_5\right) c\,\hat{\mathbf{z}}$ | (8$d$) | F |
| $\mathbf{B}_{21}$ | = | $-x_5\,\mathbf{a}_1 - y_5\,\mathbf{a}_2 - z_5\,\mathbf{a}_3$ | = | $-x_5 a\,\hat{\mathbf{x}} - y_5 b\,\hat{\mathbf{y}} - z_5 c\,\hat{\mathbf{z}}$ | (8$d$) | F |
| $\mathbf{B}_{22}$ | = | $\left(\frac{1}{2} + x_5\right)\mathbf{a}_1 + y_5\,\mathbf{a}_2 + \left(\frac{1}{2} - z_5\right)\mathbf{a}_3$ | = | $\left(\frac{1}{2} + x_5\right) a\,\hat{\mathbf{x}} + y_5 b\,\hat{\mathbf{y}} + \left(\frac{1}{2} - z_5\right) c\,\hat{\mathbf{z}}$ | (8$d$) | F |
| $\mathbf{B}_{23}$ | = | $x_5\,\mathbf{a}_1 + \left(\frac{1}{2} - y_5\right)\mathbf{a}_2 + z_5\,\mathbf{a}_3$ | = | $x_5 a\,\hat{\mathbf{x}} + \left(\frac{1}{2} - y_5\right) b\,\hat{\mathbf{y}} + z_5 c\,\hat{\mathbf{z}}$ | (8$d$) | F |
| $\mathbf{B}_{24}$ | = | $\left(\frac{1}{2} - x_5\right)\mathbf{a}_1 + \left(\frac{1}{2} + y_5\right)\mathbf{a}_2 + \left(\frac{1}{2} + z_5\right)\mathbf{a}_3$ | = | $\left(\frac{1}{2} - x_5\right) a\,\hat{\mathbf{x}} + \left(\frac{1}{2} + y_5\right) b\,\hat{\mathbf{y}} + \left(\frac{1}{2} + z_5\right) c\,\hat{\mathbf{z}}$ | (8$d$) | F |
| $\mathbf{B}_{25}$ | = | $x_6\,\mathbf{a}_1 + y_6\,\mathbf{a}_2 + z_6\,\mathbf{a}_3$ | = | $x_6 a\,\hat{\mathbf{x}} + y_6 b\,\hat{\mathbf{y}} + z_6 c\,\hat{\mathbf{z}}$ | (8$d$) | Mg II |
| $\mathbf{B}_{26}$ | = | $\left(\frac{1}{2} - x_6\right)\mathbf{a}_1 - y_6\,\mathbf{a}_2 + \left(\frac{1}{2} + z_6\right)\mathbf{a}_3$ | = | $\left(\frac{1}{2} - x_6\right) a\,\hat{\mathbf{x}} - y_6 b\,\hat{\mathbf{y}} + \left(\frac{1}{2} + z_6\right) c\,\hat{\mathbf{z}}$ | (8$d$) | Mg II |

$$\mathbf{B}_{27} = -x_6\,\mathbf{a}_1 + \left(\frac{1}{2} + y_6\right)\,\mathbf{a}_2 - z_6\,\mathbf{a}_3 = -x_6 a\,\hat{\mathbf{x}} + \left(\frac{1}{2} + y_6\right) b\,\hat{\mathbf{y}} - z_6 c\,\hat{\mathbf{z}} \quad (8d) \quad \text{Mg II}$$

$$\mathbf{B}_{28} = \left(\frac{1}{2} + x_6\right)\,\mathbf{a}_1 + \left(\frac{1}{2} - y_6\right)\,\mathbf{a}_2 + \left(\frac{1}{2} - z_6\right)\,\mathbf{a}_3 = \left(\frac{1}{2} + x_6\right) a\,\hat{\mathbf{x}} + \left(\frac{1}{2} - y_6\right) b\,\hat{\mathbf{y}} + \left(\frac{1}{2} - z_6\right) c\,\hat{\mathbf{z}} \quad (8d) \quad \text{Mg II}$$

$$\mathbf{B}_{29} = -x_6\,\mathbf{a}_1 - y_6\,\mathbf{a}_2 - z_6\,\mathbf{a}_3 = -x_6 a\,\hat{\mathbf{x}} - y_6 b\,\hat{\mathbf{y}} - z_6 c\,\hat{\mathbf{z}} \quad (8d) \quad \text{Mg II}$$

$$\mathbf{B}_{30} = \left(\frac{1}{2} + x_6\right)\,\mathbf{a}_1 + y_6\,\mathbf{a}_2 + \left(\frac{1}{2} - z_6\right)\,\mathbf{a}_3 = \left(\frac{1}{2} + x_6\right) a\,\hat{\mathbf{x}} + y_6 b\,\hat{\mathbf{y}} + \left(\frac{1}{2} - z_6\right) c\,\hat{\mathbf{z}} \quad (8d) \quad \text{Mg II}$$

$$\mathbf{B}_{31} = x_6\,\mathbf{a}_1 + \left(\frac{1}{2} - y_6\right)\,\mathbf{a}_2 + z_6\,\mathbf{a}_3 = x_6 a\,\hat{\mathbf{x}} + \left(\frac{1}{2} - y_6\right) b\,\hat{\mathbf{y}} + z_6 c\,\hat{\mathbf{z}} \quad (8d) \quad \text{Mg II}$$

$$\mathbf{B}_{32} = \left(\frac{1}{2} - x_6\right)\,\mathbf{a}_1 + \left(\frac{1}{2} + y_6\right)\,\mathbf{a}_2 + \left(\frac{1}{2} + z_6\right)\,\mathbf{a}_3 = \left(\frac{1}{2} - x_6\right) a\,\hat{\mathbf{x}} + \left(\frac{1}{2} + y_6\right) b\,\hat{\mathbf{y}} + \left(\frac{1}{2} + z_6\right) c\,\hat{\mathbf{z}} \quad (8d) \quad \text{Mg II}$$

$$\mathbf{B}_{33} = x_7\,\mathbf{a}_1 + y_7\,\mathbf{a}_2 + z_7\,\mathbf{a}_3 = x_7 a\,\hat{\mathbf{x}} + y_7 b\,\hat{\mathbf{y}} + z_7 c\,\hat{\mathbf{z}} \quad (8d) \quad \text{O III}$$

$$\mathbf{B}_{34} = \left(\frac{1}{2} - x_7\right)\,\mathbf{a}_1 - y_7\,\mathbf{a}_2 + \left(\frac{1}{2} + z_7\right)\,\mathbf{a}_3 = \left(\frac{1}{2} - x_7\right) a\,\hat{\mathbf{x}} - y_7 b\,\hat{\mathbf{y}} + \left(\frac{1}{2} + z_7\right) c\,\hat{\mathbf{z}} \quad (8d) \quad \text{O III}$$

$$\mathbf{B}_{35} = -x_7\,\mathbf{a}_1 + \left(\frac{1}{2} + y_7\right)\,\mathbf{a}_2 - z_7\,\mathbf{a}_3 = -x_7 a\,\hat{\mathbf{x}} + \left(\frac{1}{2} + y_7\right) b\,\hat{\mathbf{y}} - z_7 c\,\hat{\mathbf{z}} \quad (8d) \quad \text{O III}$$

$$\mathbf{B}_{36} = \left(\frac{1}{2} + x_7\right)\,\mathbf{a}_1 + \left(\frac{1}{2} - y_7\right)\,\mathbf{a}_2 + \left(\frac{1}{2} - z_7\right)\,\mathbf{a}_3 = \left(\frac{1}{2} + x_7\right) a\,\hat{\mathbf{x}} + \left(\frac{1}{2} - y_7\right) b\,\hat{\mathbf{y}} + \left(\frac{1}{2} - z_7\right) c\,\hat{\mathbf{z}} \quad (8d) \quad \text{O III}$$

$$\mathbf{B}_{37} = -x_7\,\mathbf{a}_1 - y_7\,\mathbf{a}_2 - z_7\,\mathbf{a}_3 = -x_7 a\,\hat{\mathbf{x}} - y_7 b\,\hat{\mathbf{y}} - z_7 c\,\hat{\mathbf{z}} \quad (8d) \quad \text{O III}$$

$$\mathbf{B}_{38} = \left(\frac{1}{2} + x_7\right)\,\mathbf{a}_1 + y_7\,\mathbf{a}_2 + \left(\frac{1}{2} - z_7\right)\,\mathbf{a}_3 = \left(\frac{1}{2} + x_7\right) a\,\hat{\mathbf{x}} + y_7 b\,\hat{\mathbf{y}} + \left(\frac{1}{2} - z_7\right) c\,\hat{\mathbf{z}} \quad (8d) \quad \text{O III}$$

$$\mathbf{B}_{39} = x_7\,\mathbf{a}_1 + \left(\frac{1}{2} - y_7\right)\,\mathbf{a}_2 + z_7\,\mathbf{a}_3 = x_7 a\,\hat{\mathbf{x}} + \left(\frac{1}{2} - y_7\right) b\,\hat{\mathbf{y}} + z_7 c\,\hat{\mathbf{z}} \quad (8d) \quad \text{O III}$$

$$\mathbf{B}_{40} = \left(\frac{1}{2} - x_7\right)\,\mathbf{a}_1 + \left(\frac{1}{2} + y_7\right)\,\mathbf{a}_2 + \left(\frac{1}{2} + z_7\right)\,\mathbf{a}_3 = \left(\frac{1}{2} - x_7\right) a\,\hat{\mathbf{x}} + \left(\frac{1}{2} + y_7\right) b\,\hat{\mathbf{y}} + \left(\frac{1}{2} + z_7\right) c\,\hat{\mathbf{z}} \quad (8d) \quad \text{O III}$$

---

---

**Geometry files:**

- CIF: pp. 1630
- POSCAR: pp. 1630

# Arcanite ($K_2SO_4$, $H1_6$) Structure: A2B4C_oP28_62_2c_2cd_c

http://aflow.org/prototype-encyclopedia/A2B4C_oP28_62_2c_2cd_c

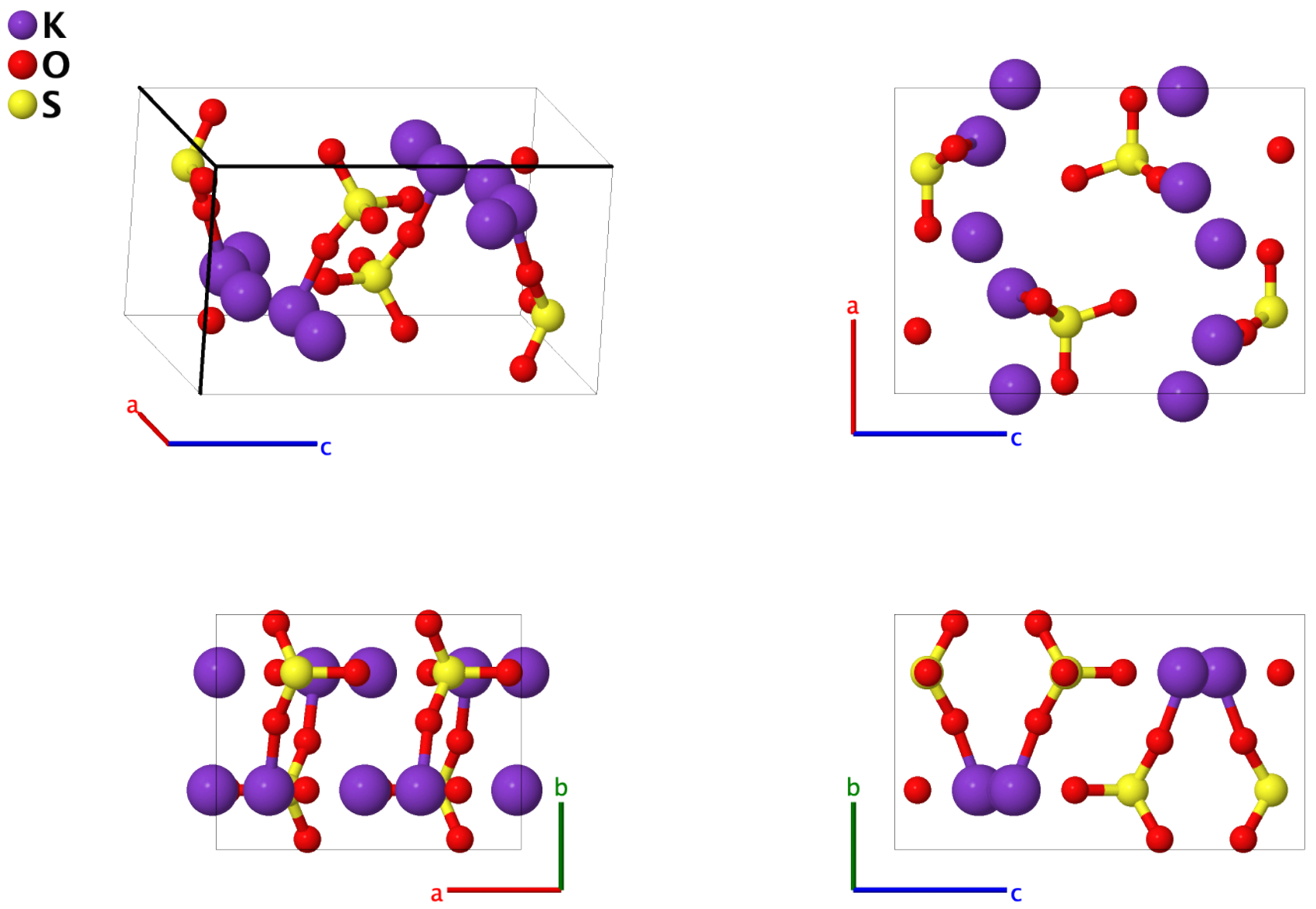

| | | |
|---|---|---|
| **Prototype** | : | $K_2O_4S$ |
| **AFLOW prototype label** | : | A2B4C_oP28_62_2c_2cd_c |
| ***Strukturbericht* designation** | : | $H1_6$ |
| **Pearson symbol** | : | oP28 |
| **Space group number** | : | 62 |
| **Space group symbol** | : | *Pnma* |
| **AFLOW prototype command** | : | `aflow --proto=A2B4C_oP28_62_2c_2cd_c`<br>`--params=`$a, b/a, c/a, x_1, z_1, x_2, z_2, x_3, z_3, x_4, z_4, x_5, z_5, x_6, y_6, z_6$ |

### Other compounds with this structure

- $K_2CrO_4$, $K_2MnO_4$, $K_2SeO_4$, and $Cs_2CuCl_4$

- (McGinnety, 1972) gives the atomic positions using the *Pnam* setting of space group #62. We have switched the $y$- and $z$-axes to put the crystal in the *Pnma* setting.

### Simple Orthorhombic primitive vectors:

$$\mathbf{a}_1 = a\,\hat{\mathbf{x}}$$
$$\mathbf{a}_2 = b\,\hat{\mathbf{y}}$$
$$\mathbf{a}_3 = c\,\hat{\mathbf{z}}$$

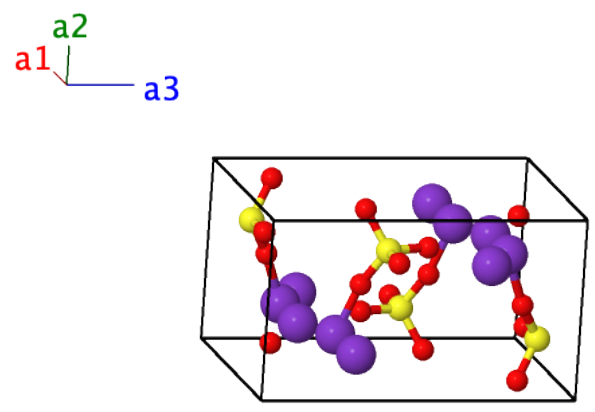

### Basis vectors:

| | | Lattice Coordinates | | Cartesian Coordinates | Wyckoff Position | Atom Type |
|---|---|---|---|---|---|---|
| $\mathbf{B}_{1}$ | = | $x_{1}\,\mathbf{a}_{1}+\frac{1}{4}\,\mathbf{a}_{2}+z_{1}\,\mathbf{a}_{3}$ | = | $x_{1}a\,\hat{\mathbf{x}}+\frac{1}{4}b\,\hat{\mathbf{y}}+z_{1}c\,\hat{\mathbf{z}}$ | (4$c$) | K I |
| $\mathbf{B}_{2}$ | = | $\left(\frac{1}{2}-x_{1}\right)\mathbf{a}_{1}+\frac{3}{4}\,\mathbf{a}_{2}+\left(\frac{1}{2}+z_{1}\right)\mathbf{a}_{3}$ | = | $\left(\frac{1}{2}-x_{1}\right)a\,\hat{\mathbf{x}}+\frac{3}{4}b\,\hat{\mathbf{y}}+\left(\frac{1}{2}+z_{1}\right)c\,\hat{\mathbf{z}}$ | (4$c$) | K I |
| $\mathbf{B}_{3}$ | = | $-x_{1}\,\mathbf{a}_{1}+\frac{3}{4}\,\mathbf{a}_{2}-z_{1}\,\mathbf{a}_{3}$ | = | $-x_{1}a\,\hat{\mathbf{x}}+\frac{3}{4}b\,\hat{\mathbf{y}}-z_{1}c\,\hat{\mathbf{z}}$ | (4$c$) | K I |
| $\mathbf{B}_{4}$ | = | $\left(\frac{1}{2}+x_{1}\right)\mathbf{a}_{1}+\frac{1}{4}\,\mathbf{a}_{2}+\left(\frac{1}{2}-z_{1}\right)\mathbf{a}_{3}$ | = | $\left(\frac{1}{2}+x_{1}\right)a\,\hat{\mathbf{x}}+\frac{1}{4}b\,\hat{\mathbf{y}}+\left(\frac{1}{2}-z_{1}\right)c\,\hat{\mathbf{z}}$ | (4$c$) | K I |
| $\mathbf{B}_{5}$ | = | $x_{2}\,\mathbf{a}_{1}+\frac{1}{4}\,\mathbf{a}_{2}+z_{2}\,\mathbf{a}_{3}$ | = | $x_{2}a\,\hat{\mathbf{x}}+\frac{1}{4}b\,\hat{\mathbf{y}}+z_{2}c\,\hat{\mathbf{z}}$ | (4$c$) | K II |
| $\mathbf{B}_{6}$ | = | $\left(\frac{1}{2}-x_{2}\right)\mathbf{a}_{1}+\frac{3}{4}\,\mathbf{a}_{2}+\left(\frac{1}{2}+z_{2}\right)\mathbf{a}_{3}$ | = | $\left(\frac{1}{2}-x_{2}\right)a\,\hat{\mathbf{x}}+\frac{3}{4}b\,\hat{\mathbf{y}}+\left(\frac{1}{2}+z_{2}\right)c\,\hat{\mathbf{z}}$ | (4$c$) | K II |
| $\mathbf{B}_{7}$ | = | $-x_{2}\,\mathbf{a}_{1}+\frac{3}{4}\,\mathbf{a}_{2}-z_{2}\,\mathbf{a}_{3}$ | = | $-x_{2}a\,\hat{\mathbf{x}}+\frac{3}{4}b\,\hat{\mathbf{y}}-z_{2}c\,\hat{\mathbf{z}}$ | (4$c$) | K II |
| $\mathbf{B}_{8}$ | = | $\left(\frac{1}{2}+x_{2}\right)\mathbf{a}_{1}+\frac{1}{4}\,\mathbf{a}_{2}+\left(\frac{1}{2}-z_{2}\right)\mathbf{a}_{3}$ | = | $\left(\frac{1}{2}+x_{2}\right)a\,\hat{\mathbf{x}}+\frac{1}{4}b\,\hat{\mathbf{y}}+\left(\frac{1}{2}-z_{2}\right)c\,\hat{\mathbf{z}}$ | (4$c$) | K II |
| $\mathbf{B}_{9}$ | = | $x_{3}\,\mathbf{a}_{1}+\frac{1}{4}\,\mathbf{a}_{2}+z_{3}\,\mathbf{a}_{3}$ | = | $x_{3}a\,\hat{\mathbf{x}}+\frac{1}{4}b\,\hat{\mathbf{y}}+z_{3}c\,\hat{\mathbf{z}}$ | (4$c$) | O I |
| $\mathbf{B}_{10}$ | = | $\left(\frac{1}{2}-x_{3}\right)\mathbf{a}_{1}+\frac{3}{4}\,\mathbf{a}_{2}+\left(\frac{1}{2}+z_{3}\right)\mathbf{a}_{3}$ | = | $\left(\frac{1}{2}-x_{3}\right)a\,\hat{\mathbf{x}}+\frac{3}{4}b\,\hat{\mathbf{y}}+\left(\frac{1}{2}+z_{3}\right)c\,\hat{\mathbf{z}}$ | (4$c$) | O I |
| $\mathbf{B}_{11}$ | = | $-x_{3}\,\mathbf{a}_{1}+\frac{3}{4}\,\mathbf{a}_{2}-z_{3}\,\mathbf{a}_{3}$ | = | $-x_{3}a\,\hat{\mathbf{x}}+\frac{3}{4}b\,\hat{\mathbf{y}}-z_{3}c\,\hat{\mathbf{z}}$ | (4$c$) | O I |
| $\mathbf{B}_{12}$ | = | $\left(\frac{1}{2}+x_{3}\right)\mathbf{a}_{1}+\frac{1}{4}\,\mathbf{a}_{2}+\left(\frac{1}{2}-z_{3}\right)\mathbf{a}_{3}$ | = | $\left(\frac{1}{2}+x_{3}\right)a\,\hat{\mathbf{x}}+\frac{1}{4}b\,\hat{\mathbf{y}}+\left(\frac{1}{2}-z_{3}\right)c\,\hat{\mathbf{z}}$ | (4$c$) | O I |
| $\mathbf{B}_{13}$ | = | $x_{4}\,\mathbf{a}_{1}+\frac{1}{4}\,\mathbf{a}_{2}+z_{4}\,\mathbf{a}_{3}$ | = | $x_{4}a\,\hat{\mathbf{x}}+\frac{1}{4}b\,\hat{\mathbf{y}}+z_{4}c\,\hat{\mathbf{z}}$ | (4$c$) | O II |
| $\mathbf{B}_{14}$ | = | $\left(\frac{1}{2}-x_{4}\right)\mathbf{a}_{1}+\frac{3}{4}\,\mathbf{a}_{2}+\left(\frac{1}{2}+z_{4}\right)\mathbf{a}_{3}$ | = | $\left(\frac{1}{2}-x_{4}\right)a\,\hat{\mathbf{x}}+\frac{3}{4}b\,\hat{\mathbf{y}}+\left(\frac{1}{2}+z_{4}\right)c\,\hat{\mathbf{z}}$ | (4$c$) | O II |
| $\mathbf{B}_{15}$ | = | $-x_{4}\,\mathbf{a}_{1}+\frac{3}{4}\,\mathbf{a}_{2}-z_{4}\,\mathbf{a}_{3}$ | = | $-x_{4}a\,\hat{\mathbf{x}}+\frac{3}{4}b\,\hat{\mathbf{y}}-z_{4}c\,\hat{\mathbf{z}}$ | (4$c$) | O II |
| $\mathbf{B}_{16}$ | = | $\left(\frac{1}{2}+x_{4}\right)\mathbf{a}_{1}+\frac{1}{4}\,\mathbf{a}_{2}+\left(\frac{1}{2}-z_{4}\right)\mathbf{a}_{3}$ | = | $\left(\frac{1}{2}+x_{4}\right)a\,\hat{\mathbf{x}}+\frac{1}{4}b\,\hat{\mathbf{y}}+\left(\frac{1}{2}-z_{4}\right)c\,\hat{\mathbf{z}}$ | (4$c$) | O II |
| $\mathbf{B}_{17}$ | = | $x_{5}\,\mathbf{a}_{1}+\frac{1}{4}\,\mathbf{a}_{2}+z_{5}\,\mathbf{a}_{3}$ | = | $x_{5}a\,\hat{\mathbf{x}}+\frac{1}{4}b\,\hat{\mathbf{y}}+z_{5}c\,\hat{\mathbf{z}}$ | (4$c$) | S |
| $\mathbf{B}_{18}$ | = | $\left(\frac{1}{2}-x_{5}\right)\mathbf{a}_{1}+\frac{3}{4}\,\mathbf{a}_{2}+\left(\frac{1}{2}+z_{5}\right)\mathbf{a}_{3}$ | = | $\left(\frac{1}{2}-x_{5}\right)a\,\hat{\mathbf{x}}+\frac{3}{4}b\,\hat{\mathbf{y}}+\left(\frac{1}{2}+z_{5}\right)c\,\hat{\mathbf{z}}$ | (4$c$) | S |
| $\mathbf{B}_{19}$ | = | $-x_{5}\,\mathbf{a}_{1}+\frac{3}{4}\,\mathbf{a}_{2}-z_{5}\,\mathbf{a}_{3}$ | = | $-x_{5}a\,\hat{\mathbf{x}}+\frac{3}{4}b\,\hat{\mathbf{y}}-z_{5}c\,\hat{\mathbf{z}}$ | (4$c$) | S |
| $\mathbf{B}_{20}$ | = | $\left(\frac{1}{2}+x_{5}\right)\mathbf{a}_{1}+\frac{1}{4}\,\mathbf{a}_{2}+\left(\frac{1}{2}-z_{5}\right)\mathbf{a}_{3}$ | = | $\left(\frac{1}{2}+x_{5}\right)a\,\hat{\mathbf{x}}+\frac{1}{4}b\,\hat{\mathbf{y}}+\left(\frac{1}{2}-z_{5}\right)c\,\hat{\mathbf{z}}$ | (4$c$) | S |
| $\mathbf{B}_{21}$ | = | $x_{6}\,\mathbf{a}_{1}+y_{6}\,\mathbf{a}_{2}+z_{6}\,\mathbf{a}_{3}$ | = | $x_{6}a\,\hat{\mathbf{x}}+y_{6}b\,\hat{\mathbf{y}}+z_{6}c\,\hat{\mathbf{z}}$ | (8$d$) | O III |
| $\mathbf{B}_{22}$ | = | $\left(\frac{1}{2}-x_{6}\right)\mathbf{a}_{1}-y_{6}\,\mathbf{a}_{2}+\left(\frac{1}{2}+z_{6}\right)\mathbf{a}_{3}$ | = | $\left(\frac{1}{2}-x_{6}\right)a\,\hat{\mathbf{x}}-y_{6}b\,\hat{\mathbf{y}}+\left(\frac{1}{2}+z_{6}\right)c\,\hat{\mathbf{z}}$ | (8$d$) | O III |
| $\mathbf{B}_{23}$ | = | $-x_{6}\,\mathbf{a}_{1}+\left(\frac{1}{2}+y_{6}\right)\mathbf{a}_{2}-z_{6}\,\mathbf{a}_{3}$ | = | $-x_{6}a\,\hat{\mathbf{x}}+\left(\frac{1}{2}+y_{6}\right)b\,\hat{\mathbf{y}}-z_{6}c\,\hat{\mathbf{z}}$ | (8$d$) | O III |
| $\mathbf{B}_{24}$ | = | $\left(\frac{1}{2}+x_{6}\right)\mathbf{a}_{1}+\left(\frac{1}{2}-y_{6}\right)\mathbf{a}_{2}+\left(\frac{1}{2}-z_{6}\right)\mathbf{a}_{3}$ | = | $\left(\frac{1}{2}+x_{6}\right)a\,\hat{\mathbf{x}}+\left(\frac{1}{2}-y_{6}\right)b\,\hat{\mathbf{y}}+\left(\frac{1}{2}-z_{6}\right)c\,\hat{\mathbf{z}}$ | (8$d$) | O III |
| $\mathbf{B}_{25}$ | = | $-x_{6}\,\mathbf{a}_{1}-y_{6}\,\mathbf{a}_{2}-z_{6}\,\mathbf{a}_{3}$ | = | $-x_{6}a\,\hat{\mathbf{x}}-y_{6}b\,\hat{\mathbf{y}}-z_{6}c\,\hat{\mathbf{z}}$ | (8$d$) | O III |
| $\mathbf{B}_{26}$ | = | $\left(\frac{1}{2}+x_{6}\right)\mathbf{a}_{1}+y_{6}\,\mathbf{a}_{2}+\left(\frac{1}{2}-z_{6}\right)\mathbf{a}_{3}$ | = | $\left(\frac{1}{2}+x_{6}\right)a\,\hat{\mathbf{x}}+y_{6}b\,\hat{\mathbf{y}}+\left(\frac{1}{2}-z_{6}\right)c\,\hat{\mathbf{z}}$ | (8$d$) | O III |
| $\mathbf{B}_{27}$ | = | $x_{6}\,\mathbf{a}_{1}+\left(\frac{1}{2}-y_{6}\right)\mathbf{a}_{2}+z_{6}\,\mathbf{a}_{3}$ | = | $x_{6}a\,\hat{\mathbf{x}}+\left(\frac{1}{2}-y_{6}\right)b\,\hat{\mathbf{y}}+z_{6}c\,\hat{\mathbf{z}}$ | (8$d$) | O III |
| $\mathbf{B}_{28}$ | = | $\left(\frac{1}{2}-x_{6}\right)\mathbf{a}_{1}+\left(\frac{1}{2}+y_{6}\right)\mathbf{a}_{2}+\left(\frac{1}{2}+z_{6}\right)\mathbf{a}_{3}$ | = | $\left(\frac{1}{2}-x_{6}\right)a\,\hat{\mathbf{x}}+\left(\frac{1}{2}+y_{6}\right)b\,\hat{\mathbf{y}}+\left(\frac{1}{2}+z_{6}\right)c\,\hat{\mathbf{z}}$ | (8$d$) | O III |

**Geometry files:**

- CIF: pp. 1630
- POSCAR: pp. 1631

# Anthophyllite ($Mg_5Fe_2Si_8O_{22}(OH)_2$, $S4_4$) Structure: A2B5C22D2E8_oP156_62_d_c2d_2c10d_2c_4d

http://aflow.org/prototype-encyclopedia/A2B5C22D2E8_oP156_62_d_c2d_2c10d_2c_4d

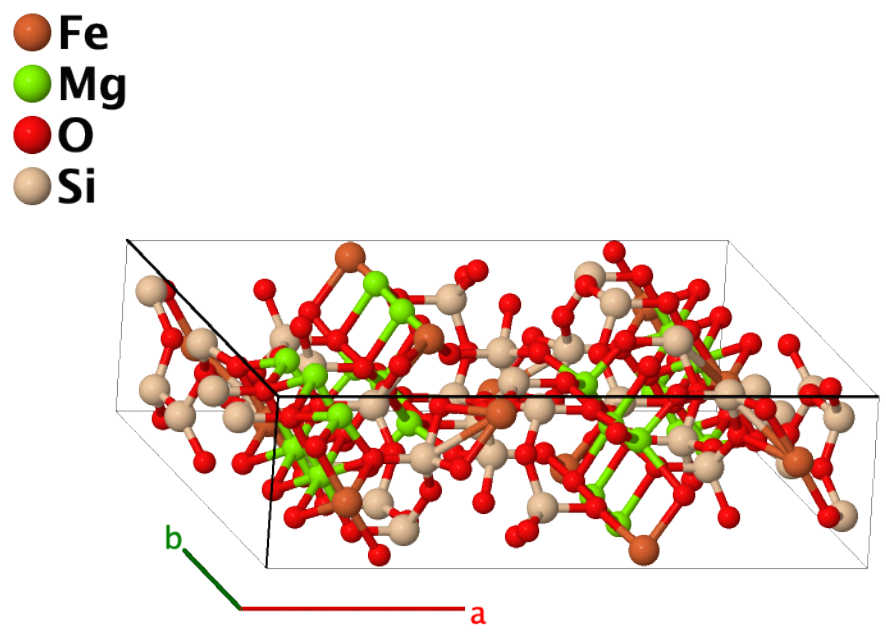

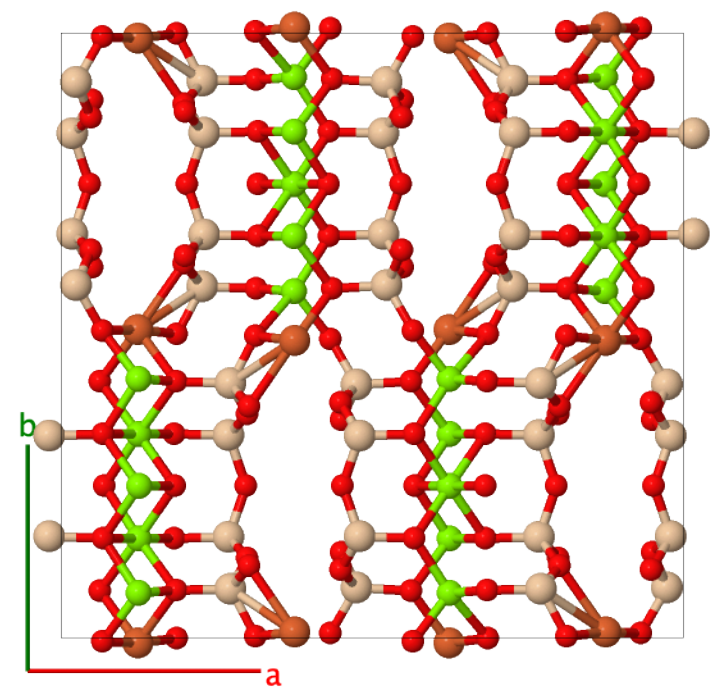

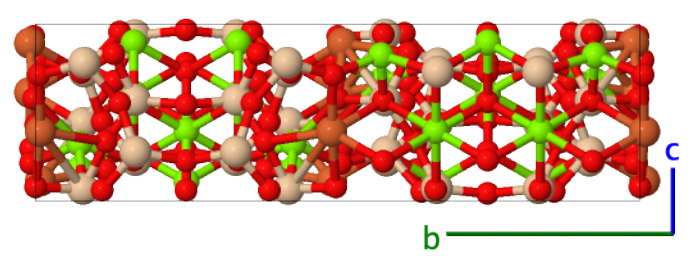

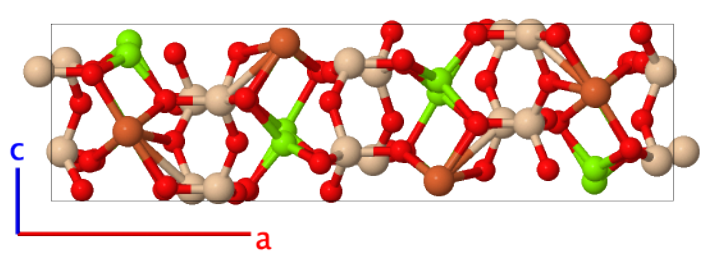

| | | |
|---|---|---|
| **Prototype** | : | $Fe_2Mg_5O_{22}(OH)_2Si_8$ |
| **AFLOW prototype label** | : | A2B5C22D2E8_oP156_62_d_c2d_2c10d_2c_4d |
| ***Strukturbericht* designation** | : | $S4_4$ |
| **Pearson symbol** | : | oP156 |
| **Space group number** | : | 62 |
| **Space group symbol** | : | *Pnma* |
| **AFLOW prototype command** | : | `aflow --proto=A2B5C22D2E8_oP156_62_d_c2d_2c10d_2c_4d` `--params=`$a, b/a, c/a, x_1, z_1, x_2, z_2, x_3, z_3, x_4, z_4, x_5, z_5, x_6, y_6, z_6, x_7, y_7, z_7, x_8, y_8, z_8, x_9, y_9, z_9, x_{10}, y_{10}, z_{10}, x_{11}, y_{11}, z_{11}, x_{12}, y_{12}, z_{12}, x_{13}, y_{13}, z_{13}, x_{14}, y_{14}, z_{14}, x_{15}, y_{15}, z_{15}, x_{16}, y_{16}, z_{16}, x_{17}, y_{17}, z_{17}, x_{18}, y_{18}, z_{18}, x_{19}, y_{19}, z_{19}, x_{20}, y_{20}, z_{20}, x_{21}, y_{21}, z_{21}, x_{22}, y_{22}, z_{22}$ |

### Other compounds with this structure

- $H_2Mg_7Si_8O_{24}$

- (Warren, 1930) analyzed their sample of anthophyllite under the assumption that it was free of iron, and this structure was given the $S4_4$ designation by (Hermann, 1937).
- (Walitzi, 1989) found a more accurate determination of the structure, but their sample included substantial amounts of iron. The Mg-I (4c) site is actually ($Mg_{0.99}Fe_{0.01}$), Mg-II (8d) is ($Mg_{0.98}Fe_{0.02}$), and Fe (8d) is ($Mg_{0.38}Fe_{0.62}$). Trace amounts of calcium, manganese and sodium were found on the magnesium/iron sites, and trace amounts of aluminum were found on the silicon sites. Neither paper was able to determine the positions of the hydrogen atoms, which are included in the OH radicals.

- Anthophyllite can be thought of as an approximation to a doubled unit cell of protoanthophyllite, where the hydrogen atoms have been located.

---

**Simple Orthorhombic primitive vectors:**

$$\mathbf{a}_1 = a\,\hat{\mathbf{x}}$$
$$\mathbf{a}_2 = b\,\hat{\mathbf{y}}$$
$$\mathbf{a}_3 = c\,\hat{\mathbf{z}}$$

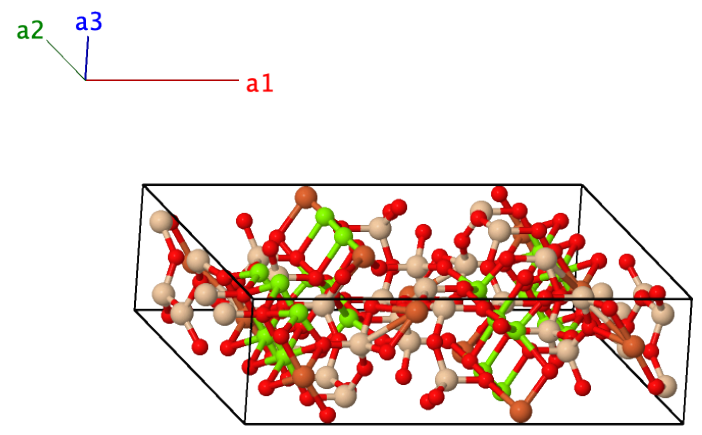

---

**Basis vectors:**

| | | Lattice Coordinates | | Cartesian Coordinates | Wyckoff Position | Atom Type |
|---|---|---|---|---|---|---|
| $\mathbf{B}_1$ | = | $x_1\,\mathbf{a}_1 + \frac{1}{4}\,\mathbf{a}_2 + z_1\,\mathbf{a}_3$ | = | $x_1 a\,\hat{\mathbf{x}} + \frac{1}{4} b\,\hat{\mathbf{y}} + z_1 c\,\hat{\mathbf{z}}$ | (4c) | Mg I |
| $\mathbf{B}_2$ | = | $\left(\frac{1}{2} - x_1\right)\mathbf{a}_1 + \frac{3}{4}\,\mathbf{a}_2 + \left(\frac{1}{2} + z_1\right)\mathbf{a}_3$ | = | $\left(\frac{1}{2} - x_1\right) a\,\hat{\mathbf{x}} + \frac{3}{4} b\,\hat{\mathbf{y}} + \left(\frac{1}{2} + z_1\right) c\,\hat{\mathbf{z}}$ | (4c) | Mg I |
| $\mathbf{B}_3$ | = | $-x_1\,\mathbf{a}_1 + \frac{3}{4}\,\mathbf{a}_2 - z_1\,\mathbf{a}_3$ | = | $-x_1 a\,\hat{\mathbf{x}} + \frac{3}{4} b\,\hat{\mathbf{y}} - z_1 c\,\hat{\mathbf{z}}$ | (4c) | Mg I |
| $\mathbf{B}_4$ | = | $\left(\frac{1}{2} + x_1\right)\mathbf{a}_1 + \frac{1}{4}\,\mathbf{a}_2 + \left(\frac{1}{2} - z_1\right)\mathbf{a}_3$ | = | $\left(\frac{1}{2} + x_1\right) a\,\hat{\mathbf{x}} + \frac{1}{4} b\,\hat{\mathbf{y}} + \left(\frac{1}{2} - z_1\right) c\,\hat{\mathbf{z}}$ | (4c) | Mg I |
| $\mathbf{B}_5$ | = | $x_2\,\mathbf{a}_1 + \frac{1}{4}\,\mathbf{a}_2 + z_2\,\mathbf{a}_3$ | = | $x_2 a\,\hat{\mathbf{x}} + \frac{1}{4} b\,\hat{\mathbf{y}} + z_2 c\,\hat{\mathbf{z}}$ | (4c) | O I |
| $\mathbf{B}_6$ | = | $\left(\frac{1}{2} - x_2\right)\mathbf{a}_1 + \frac{3}{4}\,\mathbf{a}_2 + \left(\frac{1}{2} + z_2\right)\mathbf{a}_3$ | = | $\left(\frac{1}{2} - x_2\right) a\,\hat{\mathbf{x}} + \frac{3}{4} b\,\hat{\mathbf{y}} + \left(\frac{1}{2} + z_2\right) c\,\hat{\mathbf{z}}$ | (4c) | O I |
| $\mathbf{B}_7$ | = | $-x_2\,\mathbf{a}_1 + \frac{3}{4}\,\mathbf{a}_2 - z_2\,\mathbf{a}_3$ | = | $-x_2 a\,\hat{\mathbf{x}} + \frac{3}{4} b\,\hat{\mathbf{y}} - z_2 c\,\hat{\mathbf{z}}$ | (4c) | O I |
| $\mathbf{B}_8$ | = | $\left(\frac{1}{2} + x_2\right)\mathbf{a}_1 + \frac{1}{4}\,\mathbf{a}_2 + \left(\frac{1}{2} - z_2\right)\mathbf{a}_3$ | = | $\left(\frac{1}{2} + x_2\right) a\,\hat{\mathbf{x}} + \frac{1}{4} b\,\hat{\mathbf{y}} + \left(\frac{1}{2} - z_2\right) c\,\hat{\mathbf{z}}$ | (4c) | O I |
| $\mathbf{B}_9$ | = | $x_3\,\mathbf{a}_1 + \frac{1}{4}\,\mathbf{a}_2 + z_3\,\mathbf{a}_3$ | = | $x_3 a\,\hat{\mathbf{x}} + \frac{1}{4} b\,\hat{\mathbf{y}} + z_3 c\,\hat{\mathbf{z}}$ | (4c) | O II |
| $\mathbf{B}_{10}$ | = | $\left(\frac{1}{2} - x_3\right)\mathbf{a}_1 + \frac{3}{4}\,\mathbf{a}_2 + \left(\frac{1}{2} + z_3\right)\mathbf{a}_3$ | = | $\left(\frac{1}{2} - x_3\right) a\,\hat{\mathbf{x}} + \frac{3}{4} b\,\hat{\mathbf{y}} + \left(\frac{1}{2} + z_3\right) c\,\hat{\mathbf{z}}$ | (4c) | O II |
| $\mathbf{B}_{11}$ | = | $-x_3\,\mathbf{a}_1 + \frac{3}{4}\,\mathbf{a}_2 - z_3\,\mathbf{a}_3$ | = | $-x_3 a\,\hat{\mathbf{x}} + \frac{3}{4} b\,\hat{\mathbf{y}} - z_3 c\,\hat{\mathbf{z}}$ | (4c) | O II |
| $\mathbf{B}_{12}$ | = | $\left(\frac{1}{2} + x_3\right)\mathbf{a}_1 + \frac{1}{4}\,\mathbf{a}_2 + \left(\frac{1}{2} - z_3\right)\mathbf{a}_3$ | = | $\left(\frac{1}{2} + x_3\right) a\,\hat{\mathbf{x}} + \frac{1}{4} b\,\hat{\mathbf{y}} + \left(\frac{1}{2} - z_3\right) c\,\hat{\mathbf{z}}$ | (4c) | O II |
| $\mathbf{B}_{13}$ | = | $x_4\,\mathbf{a}_1 + \frac{1}{4}\,\mathbf{a}_2 + z_4\,\mathbf{a}_3$ | = | $x_4 a\,\hat{\mathbf{x}} + \frac{1}{4} b\,\hat{\mathbf{y}} + z_4 c\,\hat{\mathbf{z}}$ | (4c) | OH I |
| $\mathbf{B}_{14}$ | = | $\left(\frac{1}{2} - x_4\right)\mathbf{a}_1 + \frac{3}{4}\,\mathbf{a}_2 + \left(\frac{1}{2} + z_4\right)\mathbf{a}_3$ | = | $\left(\frac{1}{2} - x_4\right) a\,\hat{\mathbf{x}} + \frac{3}{4} b\,\hat{\mathbf{y}} + \left(\frac{1}{2} + z_4\right) c\,\hat{\mathbf{z}}$ | (4c) | OH I |
| $\mathbf{B}_{15}$ | = | $-x_4\,\mathbf{a}_1 + \frac{3}{4}\,\mathbf{a}_2 - z_4\,\mathbf{a}_3$ | = | $-x_4 a\,\hat{\mathbf{x}} + \frac{3}{4} b\,\hat{\mathbf{y}} - z_4 c\,\hat{\mathbf{z}}$ | (4c) | OH I |
| $\mathbf{B}_{16}$ | = | $\left(\frac{1}{2} + x_4\right)\mathbf{a}_1 + \frac{1}{4}\,\mathbf{a}_2 + \left(\frac{1}{2} - z_4\right)\mathbf{a}_3$ | = | $\left(\frac{1}{2} + x_4\right) a\,\hat{\mathbf{x}} + \frac{1}{4} b\,\hat{\mathbf{y}} + \left(\frac{1}{2} - z_4\right) c\,\hat{\mathbf{z}}$ | (4c) | OH I |
| $\mathbf{B}_{17}$ | = | $x_5\,\mathbf{a}_1 + \frac{1}{4}\,\mathbf{a}_2 + z_5\,\mathbf{a}_3$ | = | $x_5 a\,\hat{\mathbf{x}} + \frac{1}{4} b\,\hat{\mathbf{y}} + z_5 c\,\hat{\mathbf{z}}$ | (4c) | OH II |
| $\mathbf{B}_{18}$ | = | $\left(\frac{1}{2} - x_5\right)\mathbf{a}_1 + \frac{3}{4}\,\mathbf{a}_2 + \left(\frac{1}{2} + z_5\right)\mathbf{a}_3$ | = | $\left(\frac{1}{2} - x_5\right) a\,\hat{\mathbf{x}} + \frac{3}{4} b\,\hat{\mathbf{y}} + \left(\frac{1}{2} + z_5\right) c\,\hat{\mathbf{z}}$ | (4c) | OH II |
| $\mathbf{B}_{19}$ | = | $-x_5\,\mathbf{a}_1 + \frac{3}{4}\,\mathbf{a}_2 - z_5\,\mathbf{a}_3$ | = | $-x_5 a\,\hat{\mathbf{x}} + \frac{3}{4} b\,\hat{\mathbf{y}} - z_5 c\,\hat{\mathbf{z}}$ | (4c) | OH II |
| $\mathbf{B}_{20}$ | = | $\left(\frac{1}{2} + x_5\right)\mathbf{a}_1 + \frac{1}{4}\,\mathbf{a}_2 + \left(\frac{1}{2} - z_5\right)\mathbf{a}_3$ | = | $\left(\frac{1}{2} + x_5\right) a\,\hat{\mathbf{x}} + \frac{1}{4} b\,\hat{\mathbf{y}} + \left(\frac{1}{2} - z_5\right) c\,\hat{\mathbf{z}}$ | (4c) | OH II |
| $\mathbf{B}_{21}$ | = | $x_6\,\mathbf{a}_1 + y_6\,\mathbf{a}_2 + z_6\,\mathbf{a}_3$ | = | $x_6 a\,\hat{\mathbf{x}} + y_6 b\,\hat{\mathbf{y}} + z_6 c\,\hat{\mathbf{z}}$ | (8d) | Fe |
| $\mathbf{B}_{22}$ | = | $\left(\frac{1}{2} - x_6\right)\mathbf{a}_1 - y_6\,\mathbf{a}_2 + \left(\frac{1}{2} + z_6\right)\mathbf{a}_3$ | = | $\left(\frac{1}{2} - x_6\right) a\,\hat{\mathbf{x}} - y_6 b\,\hat{\mathbf{y}} + \left(\frac{1}{2} + z_6\right) c\,\hat{\mathbf{z}}$ | (8d) | Fe |
| $\mathbf{B}_{23}$ | = | $-x_6\,\mathbf{a}_1 + \left(\frac{1}{2} + y_6\right)\mathbf{a}_2 - z_6\,\mathbf{a}_3$ | = | $-x_6 a\,\hat{\mathbf{x}} + \left(\frac{1}{2} + y_6\right) b\,\hat{\mathbf{y}} - z_6 c\,\hat{\mathbf{z}}$ | (8d) | Fe |
| $\mathbf{B}_{24}$ | = | $\left(\frac{1}{2} + x_{6}\right) \mathbf{a}_{1} + \left(\frac{1}{2} - y_{6}\right) \mathbf{a}_{2} + \left(\frac{1}{2} - z_{6}\right) \mathbf{a}_{3}$ | = | $\left(\frac{1}{2} + x_{6}\right) a \,\hat{\mathbf{x}} + \left(\frac{1}{2} - y_{6}\right) b \,\hat{\mathbf{y}} + \left(\frac{1}{2} - z_{6}\right) c \,\hat{\mathbf{z}}$ | (8$d$) | Fe |
| $\mathbf{B}_{25}$ | = | $-x_{6} \mathbf{a}_{1} - y_{6} \mathbf{a}_{2} - z_{6} \mathbf{a}_{3}$ | = | $-x_{6}a \,\hat{\mathbf{x}} - y_{6}b \,\hat{\mathbf{y}} - z_{6}c \,\hat{\mathbf{z}}$ | (8$d$) | Fe |
| $\mathbf{B}_{26}$ | = | $\left(\frac{1}{2} + x_{6}\right) \mathbf{a}_{1} + y_{6} \mathbf{a}_{2} + \left(\frac{1}{2} - z_{6}\right) \mathbf{a}_{3}$ | = | $\left(\frac{1}{2} + x_{6}\right) a \,\hat{\mathbf{x}} + y_{6}b \,\hat{\mathbf{y}} + \left(\frac{1}{2} - z_{6}\right) c \,\hat{\mathbf{z}}$ | (8$d$) | Fe |
| $\mathbf{B}_{27}$ | = | $x_{6} \mathbf{a}_{1} + \left(\frac{1}{2} - y_{6}\right) \mathbf{a}_{2} + z_{6} \mathbf{a}_{3}$ | = | $x_{6}a \,\hat{\mathbf{x}} + \left(\frac{1}{2} - y_{6}\right) b \,\hat{\mathbf{y}} + z_{6}c \,\hat{\mathbf{z}}$ | (8$d$) | Fe |
| $\mathbf{B}_{28}$ | = | $\left(\frac{1}{2} - x_{6}\right) \mathbf{a}_{1} + \left(\frac{1}{2} + y_{6}\right) \mathbf{a}_{2} + \left(\frac{1}{2} + z_{6}\right) \mathbf{a}_{3}$ | = | $\left(\frac{1}{2} - x_{6}\right) a \,\hat{\mathbf{x}} + \left(\frac{1}{2} + y_{6}\right) b \,\hat{\mathbf{y}} + \left(\frac{1}{2} + z_{6}\right) c \,\hat{\mathbf{z}}$ | (8$d$) | Fe |
| $\mathbf{B}_{29}$ | = | $x_{7} \mathbf{a}_{1} + y_{7} \mathbf{a}_{2} + z_{7} \mathbf{a}_{3}$ | = | $x_{7}a \,\hat{\mathbf{x}} + y_{7}b \,\hat{\mathbf{y}} + z_{7}c \,\hat{\mathbf{z}}$ | (8$d$) | Mg II |
| $\mathbf{B}_{30}$ | = | $\left(\frac{1}{2} - x_{7}\right) \mathbf{a}_{1} - y_{7} \mathbf{a}_{2} + \left(\frac{1}{2} + z_{7}\right) \mathbf{a}_{3}$ | = | $\left(\frac{1}{2} - x_{7}\right) a \,\hat{\mathbf{x}} - y_{7}b \,\hat{\mathbf{y}} + \left(\frac{1}{2} + z_{7}\right) c \,\hat{\mathbf{z}}$ | (8$d$) | Mg II |
| $\mathbf{B}_{31}$ | = | $-x_{7} \mathbf{a}_{1} + \left(\frac{1}{2} + y_{7}\right) \mathbf{a}_{2} - z_{7} \mathbf{a}_{3}$ | = | $-x_{7}a \,\hat{\mathbf{x}} + \left(\frac{1}{2} + y_{7}\right) b \,\hat{\mathbf{y}} - z_{7}c \,\hat{\mathbf{z}}$ | (8$d$) | Mg II |
| $\mathbf{B}_{32}$ | = | $\left(\frac{1}{2} + x_{7}\right) \mathbf{a}_{1} + \left(\frac{1}{2} - y_{7}\right) \mathbf{a}_{2} + \left(\frac{1}{2} - z_{7}\right) \mathbf{a}_{3}$ | = | $\left(\frac{1}{2} + x_{7}\right) a \,\hat{\mathbf{x}} + \left(\frac{1}{2} - y_{7}\right) b \,\hat{\mathbf{y}} + \left(\frac{1}{2} - z_{7}\right) c \,\hat{\mathbf{z}}$ | (8$d$) | Mg II |
| $\mathbf{B}_{33}$ | = | $-x_{7} \mathbf{a}_{1} - y_{7} \mathbf{a}_{2} - z_{7} \mathbf{a}_{3}$ | = | $-x_{7}a \,\hat{\mathbf{x}} - y_{7}b \,\hat{\mathbf{y}} - z_{7}c \,\hat{\mathbf{z}}$ | (8$d$) | Mg II |
| $\mathbf{B}_{34}$ | = | $\left(\frac{1}{2} + x_{7}\right) \mathbf{a}_{1} + y_{7} \mathbf{a}_{2} + \left(\frac{1}{2} - z_{7}\right) \mathbf{a}_{3}$ | = | $\left(\frac{1}{2} + x_{7}\right) a \,\hat{\mathbf{x}} + y_{7}b \,\hat{\mathbf{y}} + \left(\frac{1}{2} - z_{7}\right) c \,\hat{\mathbf{z}}$ | (8$d$) | Mg II |
| $\mathbf{B}_{35}$ | = | $x_{7} \mathbf{a}_{1} + \left(\frac{1}{2} - y_{7}\right) \mathbf{a}_{2} + z_{7} \mathbf{a}_{3}$ | = | $x_{7}a \,\hat{\mathbf{x}} + \left(\frac{1}{2} - y_{7}\right) b \,\hat{\mathbf{y}} + z_{7}c \,\hat{\mathbf{z}}$ | (8$d$) | Mg II |
| $\mathbf{B}_{36}$ | = | $\left(\frac{1}{2} - x_{7}\right) \mathbf{a}_{1} + \left(\frac{1}{2} + y_{7}\right) \mathbf{a}_{2} + \left(\frac{1}{2} + z_{7}\right) \mathbf{a}_{3}$ | = | $\left(\frac{1}{2} - x_{7}\right) a \,\hat{\mathbf{x}} + \left(\frac{1}{2} + y_{7}\right) b \,\hat{\mathbf{y}} + \left(\frac{1}{2} + z_{7}\right) c \,\hat{\mathbf{z}}$ | (8$d$) | Mg II |
| $\mathbf{B}_{37}$ | = | $x_{8} \mathbf{a}_{1} + y_{8} \mathbf{a}_{2} + z_{8} \mathbf{a}_{3}$ | = | $x_{8}a \,\hat{\mathbf{x}} + y_{8}b \,\hat{\mathbf{y}} + z_{8}c \,\hat{\mathbf{z}}$ | (8$d$) | Mg III |
| $\mathbf{B}_{38}$ | = | $\left(\frac{1}{2} - x_{8}\right) \mathbf{a}_{1} - y_{8} \mathbf{a}_{2} + \left(\frac{1}{2} + z_{8}\right) \mathbf{a}_{3}$ | = | $\left(\frac{1}{2} - x_{8}\right) a \,\hat{\mathbf{x}} - y_{8}b \,\hat{\mathbf{y}} + \left(\frac{1}{2} + z_{8}\right) c \,\hat{\mathbf{z}}$ | (8$d$) | Mg III |
| $\mathbf{B}_{39}$ | = | $-x_{8} \mathbf{a}_{1} + \left(\frac{1}{2} + y_{8}\right) \mathbf{a}_{2} - z_{8} \mathbf{a}_{3}$ | = | $-x_{8}a \,\hat{\mathbf{x}} + \left(\frac{1}{2} + y_{8}\right) b \,\hat{\mathbf{y}} - z_{8}c \,\hat{\mathbf{z}}$ | (8$d$) | Mg III |
| $\mathbf{B}_{40}$ | = | $\left(\frac{1}{2} + x_{8}\right) \mathbf{a}_{1} + \left(\frac{1}{2} - y_{8}\right) \mathbf{a}_{2} + \left(\frac{1}{2} - z_{8}\right) \mathbf{a}_{3}$ | = | $\left(\frac{1}{2} + x_{8}\right) a \,\hat{\mathbf{x}} + \left(\frac{1}{2} - y_{8}\right) b \,\hat{\mathbf{y}} + \left(\frac{1}{2} - z_{8}\right) c \,\hat{\mathbf{z}}$ | (8$d$) | Mg III |
| $\mathbf{B}_{41}$ | = | $-x_{8} \mathbf{a}_{1} - y_{8} \mathbf{a}_{2} - z_{8} \mathbf{a}_{3}$ | = | $-x_{8}a \,\hat{\mathbf{x}} - y_{8}b \,\hat{\mathbf{y}} - z_{8}c \,\hat{\mathbf{z}}$ | (8$d$) | Mg III |
| $\mathbf{B}_{42}$ | = | $\left(\frac{1}{2} + x_{8}\right) \mathbf{a}_{1} + y_{8} \mathbf{a}_{2} + \left(\frac{1}{2} - z_{8}\right) \mathbf{a}_{3}$ | = | $\left(\frac{1}{2} + x_{8}\right) a \,\hat{\mathbf{x}} + y_{8}b \,\hat{\mathbf{y}} + \left(\frac{1}{2} - z_{8}\right) c \,\hat{\mathbf{z}}$ | (8$d$) | Mg III |
| $\mathbf{B}_{43}$ | = | $x_{8} \mathbf{a}_{1} + \left(\frac{1}{2} - y_{8}\right) \mathbf{a}_{2} + z_{8} \mathbf{a}_{3}$ | = | $x_{8}a \,\hat{\mathbf{x}} + \left(\frac{1}{2} - y_{8}\right) b \,\hat{\mathbf{y}} + z_{8}c \,\hat{\mathbf{z}}$ | (8$d$) | Mg III |
| $\mathbf{B}_{44}$ | = | $\left(\frac{1}{2} - x_{8}\right) \mathbf{a}_{1} + \left(\frac{1}{2} + y_{8}\right) \mathbf{a}_{2} + \left(\frac{1}{2} + z_{8}\right) \mathbf{a}_{3}$ | = | $\left(\frac{1}{2} - x_{8}\right) a \,\hat{\mathbf{x}} + \left(\frac{1}{2} + y_{8}\right) b \,\hat{\mathbf{y}} + \left(\frac{1}{2} + z_{8}\right) c \,\hat{\mathbf{z}}$ | (8$d$) | Mg III |
| $\mathbf{B}_{45}$ | = | $x_{9} \mathbf{a}_{1} + y_{9} \mathbf{a}_{2} + z_{9} \mathbf{a}_{3}$ | = | $x_{9}a \,\hat{\mathbf{x}} + y_{9}b \,\hat{\mathbf{y}} + z_{9}c \,\hat{\mathbf{z}}$ | (8$d$) | O III |
| $\mathbf{B}_{46}$ | = | $\left(\frac{1}{2} - x_{9}\right) \mathbf{a}_{1} - y_{9} \mathbf{a}_{2} + \left(\frac{1}{2} + z_{9}\right) \mathbf{a}_{3}$ | = | $\left(\frac{1}{2} - x_{9}\right) a \,\hat{\mathbf{x}} - y_{9}b \,\hat{\mathbf{y}} + \left(\frac{1}{2} + z_{9}\right) c \,\hat{\mathbf{z}}$ | (8$d$) | O III |
| $\mathbf{B}_{47}$ | = | $-x_{9} \mathbf{a}_{1} + \left(\frac{1}{2} + y_{9}\right) \mathbf{a}_{2} - z_{9} \mathbf{a}_{3}$ | = | $-x_{9}a \,\hat{\mathbf{x}} + \left(\frac{1}{2} + y_{9}\right) b \,\hat{\mathbf{y}} - z_{9}c \,\hat{\mathbf{z}}$ | (8$d$) | O III |
| $\mathbf{B}_{48}$ | = | $\left(\frac{1}{2} + x_{9}\right) \mathbf{a}_{1} + \left(\frac{1}{2} - y_{9}\right) \mathbf{a}_{2} + \left(\frac{1}{2} - z_{9}\right) \mathbf{a}_{3}$ | = | $\left(\frac{1}{2} + x_{9}\right) a \,\hat{\mathbf{x}} + \left(\frac{1}{2} - y_{9}\right) b \,\hat{\mathbf{y}} + \left(\frac{1}{2} - z_{9}\right) c \,\hat{\mathbf{z}}$ | (8$d$) | O III |
| $\mathbf{B}_{49}$ | = | $-x_{9} \mathbf{a}_{1} - y_{9} \mathbf{a}_{2} - z_{9} \mathbf{a}_{3}$ | = | $-x_{9}a \,\hat{\mathbf{x}} - y_{9}b \,\hat{\mathbf{y}} - z_{9}c \,\hat{\mathbf{z}}$ | (8$d$) | O III |
| $\mathbf{B}_{50}$ | = | $\left(\frac{1}{2} + x_{9}\right) \mathbf{a}_{1} + y_{9} \mathbf{a}_{2} + \left(\frac{1}{2} - z_{9}\right) \mathbf{a}_{3}$ | = | $\left(\frac{1}{2} + x_{9}\right) a \,\hat{\mathbf{x}} + y_{9}b \,\hat{\mathbf{y}} + \left(\frac{1}{2} - z_{9}\right) c \,\hat{\mathbf{z}}$ | (8$d$) | O III |
| $\mathbf{B}_{51}$ | = | $x_{9} \mathbf{a}_{1} + \left(\frac{1}{2} - y_{9}\right) \mathbf{a}_{2} + z_{9} \mathbf{a}_{3}$ | = | $x_{9}a \,\hat{\mathbf{x}} + \left(\frac{1}{2} - y_{9}\right) b \,\hat{\mathbf{y}} + z_{9}c \,\hat{\mathbf{z}}$ | (8$d$) | O III |
| $\mathbf{B}_{52}$ | = | $\left(\frac{1}{2} - x_{9}\right) \mathbf{a}_{1} + \left(\frac{1}{2} + y_{9}\right) \mathbf{a}_{2} + \left(\frac{1}{2} + z_{9}\right) \mathbf{a}_{3}$ | = | $\left(\frac{1}{2} - x_{9}\right) a \,\hat{\mathbf{x}} + \left(\frac{1}{2} + y_{9}\right) b \,\hat{\mathbf{y}} + \left(\frac{1}{2} + z_{9}\right) c \,\hat{\mathbf{z}}$ | (8$d$) | O III |
| $\mathbf{B}_{53}$ | = | $x_{10} \mathbf{a}_{1} + y_{10} \mathbf{a}_{2} + z_{10} \mathbf{a}_{3}$ | = | $x_{10}a \,\hat{\mathbf{x}} + y_{10}b \,\hat{\mathbf{y}} + z_{10}c \,\hat{\mathbf{z}}$ | (8$d$) | O IV |
| $\mathbf{B}_{54}$ | = | $\left(\frac{1}{2} - x_{10}\right) \mathbf{a}_1 - y_{10} \mathbf{a}_2 + \left(\frac{1}{2} + z_{10}\right) \mathbf{a}_3$ | = | $\left(\frac{1}{2} - x_{10}\right) a \hat{\mathbf{x}} - y_{10} b \hat{\mathbf{y}} + \left(\frac{1}{2} + z_{10}\right) c \hat{\mathbf{z}}$ | (8$d$) | O IV |
| $\mathbf{B}_{55}$ | = | $-x_{10} \mathbf{a}_1 + \left(\frac{1}{2} + y_{10}\right) \mathbf{a}_2 - z_{10} \mathbf{a}_3$ | = | $-x_{10} a \hat{\mathbf{x}} + \left(\frac{1}{2} + y_{10}\right) b \hat{\mathbf{y}} - z_{10} c \hat{\mathbf{z}}$ | (8$d$) | O IV |
| $\mathbf{B}_{56}$ | = | $\left(\frac{1}{2} + x_{10}\right) \mathbf{a}_1 + \left(\frac{1}{2} - y_{10}\right) \mathbf{a}_2 + \left(\frac{1}{2} - z_{10}\right) \mathbf{a}_3$ | = | $\left(\frac{1}{2} + x_{10}\right) a \hat{\mathbf{x}} + \left(\frac{1}{2} - y_{10}\right) b \hat{\mathbf{y}} + \left(\frac{1}{2} - z_{10}\right) c \hat{\mathbf{z}}$ | (8$d$) | O IV |
| $\mathbf{B}_{57}$ | = | $-x_{10} \mathbf{a}_1 - y_{10} \mathbf{a}_2 - z_{10} \mathbf{a}_3$ | = | $-x_{10} a \hat{\mathbf{x}} - y_{10} b \hat{\mathbf{y}} - z_{10} c \hat{\mathbf{z}}$ | (8$d$) | O IV |
| $\mathbf{B}_{58}$ | = | $\left(\frac{1}{2} + x_{10}\right) \mathbf{a}_1 + y_{10} \mathbf{a}_2 + \left(\frac{1}{2} - z_{10}\right) \mathbf{a}_3$ | = | $\left(\frac{1}{2} + x_{10}\right) a \hat{\mathbf{x}} + y_{10} b \hat{\mathbf{y}} + \left(\frac{1}{2} - z_{10}\right) c \hat{\mathbf{z}}$ | (8$d$) | O IV |
| $\mathbf{B}_{59}$ | = | $x_{10} \mathbf{a}_1 + \left(\frac{1}{2} - y_{10}\right) \mathbf{a}_2 + z_{10} \mathbf{a}_3$ | = | $x_{10} a \hat{\mathbf{x}} + \left(\frac{1}{2} - y_{10}\right) b \hat{\mathbf{y}} + z_{10} c \hat{\mathbf{z}}$ | (8$d$) | O IV |
| $\mathbf{B}_{60}$ | = | $\left(\frac{1}{2} - x_{10}\right) \mathbf{a}_1 + \left(\frac{1}{2} + y_{10}\right) \mathbf{a}_2 + \left(\frac{1}{2} + z_{10}\right) \mathbf{a}_3$ | = | $\left(\frac{1}{2} - x_{10}\right) a \hat{\mathbf{x}} + \left(\frac{1}{2} + y_{10}\right) b \hat{\mathbf{y}} + \left(\frac{1}{2} + z_{10}\right) c \hat{\mathbf{z}}$ | (8$d$) | O IV |
| $\mathbf{B}_{61}$ | = | $x_{11} \mathbf{a}_1 + y_{11} \mathbf{a}_2 + z_{11} \mathbf{a}_3$ | = | $x_{11} a \hat{\mathbf{x}} + y_{11} b \hat{\mathbf{y}} + z_{11} c \hat{\mathbf{z}}$ | (8$d$) | O V |
| $\mathbf{B}_{62}$ | = | $\left(\frac{1}{2} - x_{11}\right) \mathbf{a}_1 - y_{11} \mathbf{a}_2 + \left(\frac{1}{2} + z_{11}\right) \mathbf{a}_3$ | = | $\left(\frac{1}{2} - x_{11}\right) a \hat{\mathbf{x}} - y_{11} b \hat{\mathbf{y}} + \left(\frac{1}{2} + z_{11}\right) c \hat{\mathbf{z}}$ | (8$d$) | O V |
| $\mathbf{B}_{63}$ | = | $-x_{11} \mathbf{a}_1 + \left(\frac{1}{2} + y_{11}\right) \mathbf{a}_2 - z_{11} \mathbf{a}_3$ | = | $-x_{11} a \hat{\mathbf{x}} + \left(\frac{1}{2} + y_{11}\right) b \hat{\mathbf{y}} - z_{11} c \hat{\mathbf{z}}$ | (8$d$) | O V |
| $\mathbf{B}_{64}$ | = | $\left(\frac{1}{2} + x_{11}\right) \mathbf{a}_1 + \left(\frac{1}{2} - y_{11}\right) \mathbf{a}_2 + \left(\frac{1}{2} - z_{11}\right) \mathbf{a}_3$ | = | $\left(\frac{1}{2} + x_{11}\right) a \hat{\mathbf{x}} + \left(\frac{1}{2} - y_{11}\right) b \hat{\mathbf{y}} + \left(\frac{1}{2} - z_{11}\right) c \hat{\mathbf{z}}$ | (8$d$) | O V |
| $\mathbf{B}_{65}$ | = | $-x_{11} \mathbf{a}_1 - y_{11} \mathbf{a}_2 - z_{11} \mathbf{a}_3$ | = | $-x_{11} a \hat{\mathbf{x}} - y_{11} b \hat{\mathbf{y}} - z_{11} c \hat{\mathbf{z}}$ | (8$d$) | O V |
| $\mathbf{B}_{66}$ | = | $\left(\frac{1}{2} + x_{11}\right) \mathbf{a}_1 + y_{11} \mathbf{a}_2 + \left(\frac{1}{2} - z_{11}\right) \mathbf{a}_3$ | = | $\left(\frac{1}{2} + x_{11}\right) a \hat{\mathbf{x}} + y_{11} b \hat{\mathbf{y}} + \left(\frac{1}{2} - z_{11}\right) c \hat{\mathbf{z}}$ | (8$d$) | O V |
| $\mathbf{B}_{67}$ | = | $x_{11} \mathbf{a}_1 + \left(\frac{1}{2} - y_{11}\right) \mathbf{a}_2 + z_{11} \mathbf{a}_3$ | = | $x_{11} a \hat{\mathbf{x}} + \left(\frac{1}{2} - y_{11}\right) b \hat{\mathbf{y}} + z_{11} c \hat{\mathbf{z}}$ | (8$d$) | O V |
| $\mathbf{B}_{68}$ | = | $\left(\frac{1}{2} - x_{11}\right) \mathbf{a}_1 + \left(\frac{1}{2} + y_{11}\right) \mathbf{a}_2 + \left(\frac{1}{2} + z_{11}\right) \mathbf{a}_3$ | = | $\left(\frac{1}{2} - x_{11}\right) a \hat{\mathbf{x}} + \left(\frac{1}{2} + y_{11}\right) b \hat{\mathbf{y}} + \left(\frac{1}{2} + z_{11}\right) c \hat{\mathbf{z}}$ | (8$d$) | O V |
| $\mathbf{B}_{69}$ | = | $x_{12} \mathbf{a}_1 + y_{12} \mathbf{a}_2 + z_{12} \mathbf{a}_3$ | = | $x_{12} a \hat{\mathbf{x}} + y_{12} b \hat{\mathbf{y}} + z_{12} c \hat{\mathbf{z}}$ | (8$d$) | O VI |
| $\mathbf{B}_{70}$ | = | $\left(\frac{1}{2} - x_{12}\right) \mathbf{a}_1 - y_{12} \mathbf{a}_2 + \left(\frac{1}{2} + z_{12}\right) \mathbf{a}_3$ | = | $\left(\frac{1}{2} - x_{12}\right) a \hat{\mathbf{x}} - y_{12} b \hat{\mathbf{y}} + \left(\frac{1}{2} + z_{12}\right) c \hat{\mathbf{z}}$ | (8$d$) | O VI |
| $\mathbf{B}_{71}$ | = | $-x_{12} \mathbf{a}_1 + \left(\frac{1}{2} + y_{12}\right) \mathbf{a}_2 - z_{12} \mathbf{a}_3$ | = | $-x_{12} a \hat{\mathbf{x}} + \left(\frac{1}{2} + y_{12}\right) b \hat{\mathbf{y}} - z_{12} c \hat{\mathbf{z}}$ | (8$d$) | O VI |
| $\mathbf{B}_{72}$ | = | $\left(\frac{1}{2} + x_{12}\right) \mathbf{a}_1 + \left(\frac{1}{2} - y_{12}\right) \mathbf{a}_2 + \left(\frac{1}{2} - z_{12}\right) \mathbf{a}_3$ | = | $\left(\frac{1}{2} + x_{12}\right) a \hat{\mathbf{x}} + \left(\frac{1}{2} - y_{12}\right) b \hat{\mathbf{y}} + \left(\frac{1}{2} - z_{12}\right) c \hat{\mathbf{z}}$ | (8$d$) | O VI |
| $\mathbf{B}_{73}$ | = | $-x_{12} \mathbf{a}_1 - y_{12} \mathbf{a}_2 - z_{12} \mathbf{a}_3$ | = | $-x_{12} a \hat{\mathbf{x}} - y_{12} b \hat{\mathbf{y}} - z_{12} c \hat{\mathbf{z}}$ | (8$d$) | O VI |
| $\mathbf{B}_{74}$ | = | $\left(\frac{1}{2} + x_{12}\right) \mathbf{a}_1 + y_{12} \mathbf{a}_2 + \left(\frac{1}{2} - z_{12}\right) \mathbf{a}_3$ | = | $\left(\frac{1}{2} + x_{12}\right) a \hat{\mathbf{x}} + y_{12} b \hat{\mathbf{y}} + \left(\frac{1}{2} - z_{12}\right) c \hat{\mathbf{z}}$ | (8$d$) | O VI |
| $\mathbf{B}_{75}$ | = | $x_{12} \mathbf{a}_1 + \left(\frac{1}{2} - y_{12}\right) \mathbf{a}_2 + z_{12} \mathbf{a}_3$ | = | $x_{12} a \hat{\mathbf{x}} + \left(\frac{1}{2} - y_{12}\right) b \hat{\mathbf{y}} + z_{12} c \hat{\mathbf{z}}$ | (8$d$) | O VI |
| $\mathbf{B}_{76}$ | = | $\left(\frac{1}{2} - x_{12}\right) \mathbf{a}_1 + \left(\frac{1}{2} + y_{12}\right) \mathbf{a}_2 + \left(\frac{1}{2} + z_{12}\right) \mathbf{a}_3$ | = | $\left(\frac{1}{2} - x_{12}\right) a \hat{\mathbf{x}} + \left(\frac{1}{2} + y_{12}\right) b \hat{\mathbf{y}} + \left(\frac{1}{2} + z_{12}\right) c \hat{\mathbf{z}}$ | (8$d$) | O VI |
| $\mathbf{B}_{77}$ | = | $x_{13} \mathbf{a}_1 + y_{13} \mathbf{a}_2 + z_{13} \mathbf{a}_3$ | = | $x_{13} a \hat{\mathbf{x}} + y_{13} b \hat{\mathbf{y}} + z_{13} c \hat{\mathbf{z}}$ | (8$d$) | O VII |
| $\mathbf{B}_{78}$ | = | $\left(\frac{1}{2} - x_{13}\right) \mathbf{a}_1 - y_{13} \mathbf{a}_2 + \left(\frac{1}{2} + z_{13}\right) \mathbf{a}_3$ | = | $\left(\frac{1}{2} - x_{13}\right) a \hat{\mathbf{x}} - y_{13} b \hat{\mathbf{y}} + \left(\frac{1}{2} + z_{13}\right) c \hat{\mathbf{z}}$ | (8$d$) | O VII |
| $\mathbf{B}_{79}$ | = | $-x_{13} \mathbf{a}_1 + \left(\frac{1}{2} + y_{13}\right) \mathbf{a}_2 - z_{13} \mathbf{a}_3$ | = | $-x_{13} a \hat{\mathbf{x}} + \left(\frac{1}{2} + y_{13}\right) b \hat{\mathbf{y}} - z_{13} c \hat{\mathbf{z}}$ | (8$d$) | O VII |
| $\mathbf{B}_{80}$ | = | $\left(\frac{1}{2} + x_{13}\right) \mathbf{a}_1 + \left(\frac{1}{2} - y_{13}\right) \mathbf{a}_2 + \left(\frac{1}{2} - z_{13}\right) \mathbf{a}_3$ | = | $\left(\frac{1}{2} + x_{13}\right) a \hat{\mathbf{x}} + \left(\frac{1}{2} - y_{13}\right) b \hat{\mathbf{y}} + \left(\frac{1}{2} - z_{13}\right) c \hat{\mathbf{z}}$ | (8$d$) | O VII |
| $\mathbf{B}_{81}$ | = | $-x_{13} \mathbf{a}_1 - y_{13} \mathbf{a}_2 - z_{13} \mathbf{a}_3$ | = | $-x_{13} a \hat{\mathbf{x}} - y_{13} b \hat{\mathbf{y}} - z_{13} c \hat{\mathbf{z}}$ | (8$d$) | O VII |
| $\mathbf{B}_{82}$ | = | $\left(\frac{1}{2} + x_{13}\right) \mathbf{a}_1 + y_{13} \mathbf{a}_2 + \left(\frac{1}{2} - z_{13}\right) \mathbf{a}_3$ | = | $\left(\frac{1}{2} + x_{13}\right) a \hat{\mathbf{x}} + y_{13} b \hat{\mathbf{y}} + \left(\frac{1}{2} - z_{13}\right) c \hat{\mathbf{z}}$ | (8$d$) | O VII |
| $\mathbf{B}_{83}$ | = | $x_{13} \mathbf{a}_1 + \left(\frac{1}{2} - y_{13}\right) \mathbf{a}_2 + z_{13} \mathbf{a}_3$ | = | $x_{13} a \hat{\mathbf{x}} + \left(\frac{1}{2} - y_{13}\right) b \hat{\mathbf{y}} + z_{13} c \hat{\mathbf{z}}$ | (8$d$) | O VII |
| $\mathbf{B}_{84}$ | = | $\left(\frac{1}{2} - x_{13}\right) \mathbf{a}_1 + \left(\frac{1}{2} + y_{13}\right) \mathbf{a}_2 + \left(\frac{1}{2} + z_{13}\right) \mathbf{a}_3$ | = | $\left(\frac{1}{2} - x_{13}\right) a \, \hat{\mathbf{x}} + \left(\frac{1}{2} + y_{13}\right) b \, \hat{\mathbf{y}} + \left(\frac{1}{2} + z_{13}\right) c \, \hat{\mathbf{z}}$ | $(8d)$ | O VII |
| $\mathbf{B}_{85}$ | = | $x_{14} \, \mathbf{a}_1 + y_{14} \, \mathbf{a}_2 + z_{14} \, \mathbf{a}_3$ | = | $x_{14} a \, \hat{\mathbf{x}} + y_{14} b \, \hat{\mathbf{y}} + z_{14} c \, \hat{\mathbf{z}}$ | $(8d)$ | O VIII |
| $\mathbf{B}_{86}$ | = | $\left(\frac{1}{2} - x_{14}\right) \mathbf{a}_1 - y_{14} \, \mathbf{a}_2 + \left(\frac{1}{2} + z_{14}\right) \mathbf{a}_3$ | = | $\left(\frac{1}{2} - x_{14}\right) a \, \hat{\mathbf{x}} - y_{14} b \, \hat{\mathbf{y}} + \left(\frac{1}{2} + z_{14}\right) c \, \hat{\mathbf{z}}$ | $(8d)$ | O VIII |
| $\mathbf{B}_{87}$ | = | $-x_{14} \, \mathbf{a}_1 + \left(\frac{1}{2} + y_{14}\right) \mathbf{a}_2 - z_{14} \, \mathbf{a}_3$ | = | $-x_{14} a \, \hat{\mathbf{x}} + \left(\frac{1}{2} + y_{14}\right) b \, \hat{\mathbf{y}} - z_{14} c \, \hat{\mathbf{z}}$ | $(8d)$ | O VIII |
| $\mathbf{B}_{88}$ | = | $\left(\frac{1}{2} + x_{14}\right) \mathbf{a}_1 + \left(\frac{1}{2} - y_{14}\right) \mathbf{a}_2 + \left(\frac{1}{2} - z_{14}\right) \mathbf{a}_3$ | = | $\left(\frac{1}{2} + x_{14}\right) a \, \hat{\mathbf{x}} + \left(\frac{1}{2} - y_{14}\right) b \, \hat{\mathbf{y}} + \left(\frac{1}{2} - z_{14}\right) c \, \hat{\mathbf{z}}$ | $(8d)$ | O VIII |
| $\mathbf{B}_{89}$ | = | $-x_{14} \, \mathbf{a}_1 - y_{14} \, \mathbf{a}_2 - z_{14} \, \mathbf{a}_3$ | = | $-x_{14} a \, \hat{\mathbf{x}} - y_{14} b \, \hat{\mathbf{y}} - z_{14} c \, \hat{\mathbf{z}}$ | $(8d)$ | O VIII |
| $\mathbf{B}_{90}$ | = | $\left(\frac{1}{2} + x_{14}\right) \mathbf{a}_1 + y_{14} \, \mathbf{a}_2 + \left(\frac{1}{2} - z_{14}\right) \mathbf{a}_3$ | = | $\left(\frac{1}{2} + x_{14}\right) a \, \hat{\mathbf{x}} + y_{14} b \, \hat{\mathbf{y}} + \left(\frac{1}{2} - z_{14}\right) c \, \hat{\mathbf{z}}$ | $(8d)$ | O VIII |
| $\mathbf{B}_{91}$ | = | $x_{14} \, \mathbf{a}_1 + \left(\frac{1}{2} - y_{14}\right) \mathbf{a}_2 + z_{14} \, \mathbf{a}_3$ | = | $x_{14} a \, \hat{\mathbf{x}} + \left(\frac{1}{2} - y_{14}\right) b \, \hat{\mathbf{y}} + z_{14} c \, \hat{\mathbf{z}}$ | $(8d)$ | O VIII |
| $\mathbf{B}_{92}$ | = | $\left(\frac{1}{2} - x_{14}\right) \mathbf{a}_1 + \left(\frac{1}{2} + y_{14}\right) \mathbf{a}_2 + \left(\frac{1}{2} + z_{14}\right) \mathbf{a}_3$ | = | $\left(\frac{1}{2} - x_{14}\right) a \, \hat{\mathbf{x}} + \left(\frac{1}{2} + y_{14}\right) b \, \hat{\mathbf{y}} + \left(\frac{1}{2} + z_{14}\right) c \, \hat{\mathbf{z}}$ | $(8d)$ | O VIII |
| $\mathbf{B}_{93}$ | = | $x_{15} \, \mathbf{a}_1 + y_{15} \, \mathbf{a}_2 + z_{15} \, \mathbf{a}_3$ | = | $x_{15} a \, \hat{\mathbf{x}} + y_{15} b \, \hat{\mathbf{y}} + z_{15} c \, \hat{\mathbf{z}}$ | $(8d)$ | O IX |
| $\mathbf{B}_{94}$ | = | $\left(\frac{1}{2} - x_{15}\right) \mathbf{a}_1 - y_{15} \, \mathbf{a}_2 + \left(\frac{1}{2} + z_{15}\right) \mathbf{a}_3$ | = | $\left(\frac{1}{2} - x_{15}\right) a \, \hat{\mathbf{x}} - y_{15} b \, \hat{\mathbf{y}} + \left(\frac{1}{2} + z_{15}\right) c \, \hat{\mathbf{z}}$ | $(8d)$ | O IX |
| $\mathbf{B}_{95}$ | = | $-x_{15} \, \mathbf{a}_1 + \left(\frac{1}{2} + y_{15}\right) \mathbf{a}_2 - z_{15} \, \mathbf{a}_3$ | = | $-x_{15} a \, \hat{\mathbf{x}} + \left(\frac{1}{2} + y_{15}\right) b \, \hat{\mathbf{y}} - z_{15} c \, \hat{\mathbf{z}}$ | $(8d)$ | O IX |
| $\mathbf{B}_{96}$ | = | $\left(\frac{1}{2} + x_{15}\right) \mathbf{a}_1 + \left(\frac{1}{2} - y_{15}\right) \mathbf{a}_2 + \left(\frac{1}{2} - z_{15}\right) \mathbf{a}_3$ | = | $\left(\frac{1}{2} + x_{15}\right) a \, \hat{\mathbf{x}} + \left(\frac{1}{2} - y_{15}\right) b \, \hat{\mathbf{y}} + \left(\frac{1}{2} - z_{15}\right) c \, \hat{\mathbf{z}}$ | $(8d)$ | O IX |
| $\mathbf{B}_{97}$ | = | $-x_{15} \, \mathbf{a}_1 - y_{15} \, \mathbf{a}_2 - z_{15} \, \mathbf{a}_3$ | = | $-x_{15} a \, \hat{\mathbf{x}} - y_{15} b \, \hat{\mathbf{y}} - z_{15} c \, \hat{\mathbf{z}}$ | $(8d)$ | O IX |
| $\mathbf{B}_{98}$ | = | $\left(\frac{1}{2} + x_{15}\right) \mathbf{a}_1 + y_{15} \, \mathbf{a}_2 + \left(\frac{1}{2} - z_{15}\right) \mathbf{a}_3$ | = | $\left(\frac{1}{2} + x_{15}\right) a \, \hat{\mathbf{x}} + y_{15} b \, \hat{\mathbf{y}} + \left(\frac{1}{2} - z_{15}\right) c \, \hat{\mathbf{z}}$ | $(8d)$ | O IX |
| $\mathbf{B}_{99}$ | = | $x_{15} \, \mathbf{a}_1 + \left(\frac{1}{2} - y_{15}\right) \mathbf{a}_2 + z_{15} \, \mathbf{a}_3$ | = | $x_{15} a \, \hat{\mathbf{x}} + \left(\frac{1}{2} - y_{15}\right) b \, \hat{\mathbf{y}} + z_{15} c \, \hat{\mathbf{z}}$ | $(8d)$ | O IX |
| $\mathbf{B}_{100}$ | = | $\left(\frac{1}{2} - x_{15}\right) \mathbf{a}_1 + \left(\frac{1}{2} + y_{15}\right) \mathbf{a}_2 + \left(\frac{1}{2} + z_{15}\right) \mathbf{a}_3$ | = | $\left(\frac{1}{2} - x_{15}\right) a \, \hat{\mathbf{x}} + \left(\frac{1}{2} + y_{15}\right) b \, \hat{\mathbf{y}} + \left(\frac{1}{2} + z_{15}\right) c \, \hat{\mathbf{z}}$ | $(8d)$ | O IX |
| $\mathbf{B}_{101}$ | = | $x_{16} \, \mathbf{a}_1 + y_{16} \, \mathbf{a}_2 + z_{16} \, \mathbf{a}_3$ | = | $x_{16} a \, \hat{\mathbf{x}} + y_{16} b \, \hat{\mathbf{y}} + z_{16} c \, \hat{\mathbf{z}}$ | $(8d)$ | O X |
| $\mathbf{B}_{102}$ | = | $\left(\frac{1}{2} - x_{16}\right) \mathbf{a}_1 - y_{16} \, \mathbf{a}_2 + \left(\frac{1}{2} + z_{16}\right) \mathbf{a}_3$ | = | $\left(\frac{1}{2} - x_{16}\right) a \, \hat{\mathbf{x}} - y_{16} b \, \hat{\mathbf{y}} + \left(\frac{1}{2} + z_{16}\right) c \, \hat{\mathbf{z}}$ | $(8d)$ | O X |
| $\mathbf{B}_{103}$ | = | $-x_{16} \, \mathbf{a}_1 + \left(\frac{1}{2} + y_{16}\right) \mathbf{a}_2 - z_{16} \, \mathbf{a}_3$ | = | $-x_{16} a \, \hat{\mathbf{x}} + \left(\frac{1}{2} + y_{16}\right) b \, \hat{\mathbf{y}} - z_{16} c \, \hat{\mathbf{z}}$ | $(8d)$ | O X |
| $\mathbf{B}_{104}$ | = | $\left(\frac{1}{2} + x_{16}\right) \mathbf{a}_1 + \left(\frac{1}{2} - y_{16}\right) \mathbf{a}_2 + \left(\frac{1}{2} - z_{16}\right) \mathbf{a}_3$ | = | $\left(\frac{1}{2} + x_{16}\right) a \, \hat{\mathbf{x}} + \left(\frac{1}{2} - y_{16}\right) b \, \hat{\mathbf{y}} + \left(\frac{1}{2} - z_{16}\right) c \, \hat{\mathbf{z}}$ | $(8d)$ | O X |
| $\mathbf{B}_{105}$ | = | $-x_{16} \, \mathbf{a}_1 - y_{16} \, \mathbf{a}_2 - z_{16} \, \mathbf{a}_3$ | = | $-x_{16} a \, \hat{\mathbf{x}} - y_{16} b \, \hat{\mathbf{y}} - z_{16} c \, \hat{\mathbf{z}}$ | $(8d)$ | O X |
| $\mathbf{B}_{106}$ | = | $\left(\frac{1}{2} + x_{16}\right) \mathbf{a}_1 + y_{16} \, \mathbf{a}_2 + \left(\frac{1}{2} - z_{16}\right) \mathbf{a}_3$ | = | $\left(\frac{1}{2} + x_{16}\right) a \, \hat{\mathbf{x}} + y_{16} b \, \hat{\mathbf{y}} + \left(\frac{1}{2} - z_{16}\right) c \, \hat{\mathbf{z}}$ | $(8d)$ | O X |
| $\mathbf{B}_{107}$ | = | $x_{16} \, \mathbf{a}_1 + \left(\frac{1}{2} - y_{16}\right) \mathbf{a}_2 + z_{16} \, \mathbf{a}_3$ | = | $x_{16} a \, \hat{\mathbf{x}} + \left(\frac{1}{2} - y_{16}\right) b \, \hat{\mathbf{y}} + z_{16} c \, \hat{\mathbf{z}}$ | $(8d)$ | O X |
| $\mathbf{B}_{108}$ | = | $\left(\frac{1}{2} - x_{16}\right) \mathbf{a}_1 + \left(\frac{1}{2} + y_{16}\right) \mathbf{a}_2 + \left(\frac{1}{2} + z_{16}\right) \mathbf{a}_3$ | = | $\left(\frac{1}{2} - x_{16}\right) a \, \hat{\mathbf{x}} + \left(\frac{1}{2} + y_{16}\right) b \, \hat{\mathbf{y}} + \left(\frac{1}{2} + z_{16}\right) c \, \hat{\mathbf{z}}$ | $(8d)$ | O X |
| $\mathbf{B}_{109}$ | = | $x_{17} \, \mathbf{a}_1 + y_{17} \, \mathbf{a}_2 + z_{17} \, \mathbf{a}_3$ | = | $x_{17} a \, \hat{\mathbf{x}} + y_{17} b \, \hat{\mathbf{y}} + z_{17} c \, \hat{\mathbf{z}}$ | $(8d)$ | O XI |
| $\mathbf{B}_{110}$ | = | $\left(\frac{1}{2} - x_{17}\right) \mathbf{a}_1 - y_{17} \, \mathbf{a}_2 + \left(\frac{1}{2} + z_{17}\right) \mathbf{a}_3$ | = | $\left(\frac{1}{2} - x_{17}\right) a \, \hat{\mathbf{x}} - y_{17} b \, \hat{\mathbf{y}} + \left(\frac{1}{2} + z_{17}\right) c \, \hat{\mathbf{z}}$ | $(8d)$ | O XI |
| $\mathbf{B}_{111}$ | = | $-x_{17} \, \mathbf{a}_1 + \left(\frac{1}{2} + y_{17}\right) \mathbf{a}_2 - z_{17} \, \mathbf{a}_3$ | = | $-x_{17} a \, \hat{\mathbf{x}} + \left(\frac{1}{2} + y_{17}\right) b \, \hat{\mathbf{y}} - z_{17} c \, \hat{\mathbf{z}}$ | $(8d)$ | O XI |
| $\mathbf{B}_{112}$ | = | $\left(\frac{1}{2} + x_{17}\right) \mathbf{a}_1 + \left(\frac{1}{2} - y_{17}\right) \mathbf{a}_2 + \left(\frac{1}{2} - z_{17}\right) \mathbf{a}_3$ | = | $\left(\frac{1}{2} + x_{17}\right) a \, \hat{\mathbf{x}} + \left(\frac{1}{2} - y_{17}\right) b \, \hat{\mathbf{y}} + \left(\frac{1}{2} - z_{17}\right) c \, \hat{\mathbf{z}}$ | $(8d)$ | O XI |
| $\mathbf{B}_{113}$ | = | $-x_{17} \, \mathbf{a}_1 - y_{17} \, \mathbf{a}_2 - z_{17} \, \mathbf{a}_3$ | = | $-x_{17} a \, \hat{\mathbf{x}} - y_{17} b \, \hat{\mathbf{y}} - z_{17} c \, \hat{\mathbf{z}}$ | $(8d)$ | O XI |
| $\mathbf{B}_{114}$ | = | $\left(\frac{1}{2} + x_{17}\right) \mathbf{a}_1 + y_{17} \mathbf{a}_2 + \left(\frac{1}{2} - z_{17}\right) \mathbf{a}_3$ | = | $\left(\frac{1}{2} + x_{17}\right) a \hat{\mathbf{x}} + y_{17} b \hat{\mathbf{y}} + \left(\frac{1}{2} - z_{17}\right) c \hat{\mathbf{z}}$ | (8*d*) | O XI |
| $\mathbf{B}_{115}$ | = | $x_{17} \mathbf{a}_1 + \left(\frac{1}{2} - y_{17}\right) \mathbf{a}_2 + z_{17} \mathbf{a}_3$ | = | $x_{17} a \hat{\mathbf{x}} + \left(\frac{1}{2} - y_{17}\right) b \hat{\mathbf{y}} + z_{17} c \hat{\mathbf{z}}$ | (8*d*) | O XI |
| $\mathbf{B}_{116}$ | = | $\left(\frac{1}{2} - x_{17}\right) \mathbf{a}_1 + \left(\frac{1}{2} + y_{17}\right) \mathbf{a}_2 + \left(\frac{1}{2} + z_{17}\right) \mathbf{a}_3$ | = | $\left(\frac{1}{2} - x_{17}\right) a \hat{\mathbf{x}} + \left(\frac{1}{2} + y_{17}\right) b \hat{\mathbf{y}} + \left(\frac{1}{2} + z_{17}\right) c \hat{\mathbf{z}}$ | (8*d*) | O XI |
| $\mathbf{B}_{117}$ | = | $x_{18} \mathbf{a}_1 + y_{18} \mathbf{a}_2 + z_{18} \mathbf{a}_3$ | = | $x_{18} a \hat{\mathbf{x}} + y_{18} b \hat{\mathbf{y}} + z_{18} c \hat{\mathbf{z}}$ | (8*d*) | O XII |
| $\mathbf{B}_{118}$ | = | $\left(\frac{1}{2} - x_{18}\right) \mathbf{a}_1 - y_{18} \mathbf{a}_2 + \left(\frac{1}{2} + z_{18}\right) \mathbf{a}_3$ | = | $\left(\frac{1}{2} - x_{18}\right) a \hat{\mathbf{x}} - y_{18} b \hat{\mathbf{y}} + \left(\frac{1}{2} + z_{18}\right) c \hat{\mathbf{z}}$ | (8*d*) | O XII |
| $\mathbf{B}_{119}$ | = | $-x_{18} \mathbf{a}_1 + \left(\frac{1}{2} + y_{18}\right) \mathbf{a}_2 - z_{18} \mathbf{a}_3$ | = | $-x_{18} a \hat{\mathbf{x}} + \left(\frac{1}{2} + y_{18}\right) b \hat{\mathbf{y}} - z_{18} c \hat{\mathbf{z}}$ | (8*d*) | O XII |
| $\mathbf{B}_{120}$ | = | $\left(\frac{1}{2} + x_{18}\right) \mathbf{a}_1 + \left(\frac{1}{2} - y_{18}\right) \mathbf{a}_2 + \left(\frac{1}{2} - z_{18}\right) \mathbf{a}_3$ | = | $\left(\frac{1}{2} + x_{18}\right) a \hat{\mathbf{x}} + \left(\frac{1}{2} - y_{18}\right) b \hat{\mathbf{y}} + \left(\frac{1}{2} - z_{18}\right) c \hat{\mathbf{z}}$ | (8*d*) | O XII |
| $\mathbf{B}_{121}$ | = | $-x_{18} \mathbf{a}_1 - y_{18} \mathbf{a}_2 - z_{18} \mathbf{a}_3$ | = | $-x_{18} a \hat{\mathbf{x}} - y_{18} b \hat{\mathbf{y}} - z_{18} c \hat{\mathbf{z}}$ | (8*d*) | O XII |
| $\mathbf{B}_{122}$ | = | $\left(\frac{1}{2} + x_{18}\right) \mathbf{a}_1 + y_{18} \mathbf{a}_2 + \left(\frac{1}{2} - z_{18}\right) \mathbf{a}_3$ | = | $\left(\frac{1}{2} + x_{18}\right) a \hat{\mathbf{x}} + y_{18} b \hat{\mathbf{y}} + \left(\frac{1}{2} - z_{18}\right) c \hat{\mathbf{z}}$ | (8*d*) | O XII |
| $\mathbf{B}_{123}$ | = | $x_{18} \mathbf{a}_1 + \left(\frac{1}{2} - y_{18}\right) \mathbf{a}_2 + z_{18} \mathbf{a}_3$ | = | $x_{18} a \hat{\mathbf{x}} + \left(\frac{1}{2} - y_{18}\right) b \hat{\mathbf{y}} + z_{18} c \hat{\mathbf{z}}$ | (8*d*) | O XII |
| $\mathbf{B}_{124}$ | = | $\left(\frac{1}{2} - x_{18}\right) \mathbf{a}_1 + \left(\frac{1}{2} + y_{18}\right) \mathbf{a}_2 + \left(\frac{1}{2} + z_{18}\right) \mathbf{a}_3$ | = | $\left(\frac{1}{2} - x_{18}\right) a \hat{\mathbf{x}} + \left(\frac{1}{2} + y_{18}\right) b \hat{\mathbf{y}} + \left(\frac{1}{2} + z_{18}\right) c \hat{\mathbf{z}}$ | (8*d*) | O XII |
| $\mathbf{B}_{125}$ | = | $x_{19} \mathbf{a}_1 + y_{19} \mathbf{a}_2 + z_{19} \mathbf{a}_3$ | = | $x_{19} a \hat{\mathbf{x}} + y_{19} b \hat{\mathbf{y}} + z_{19} c \hat{\mathbf{z}}$ | (8*d*) | Si I |
| $\mathbf{B}_{126}$ | = | $\left(\frac{1}{2} - x_{19}\right) \mathbf{a}_1 - y_{19} \mathbf{a}_2 + \left(\frac{1}{2} + z_{19}\right) \mathbf{a}_3$ | = | $\left(\frac{1}{2} - x_{19}\right) a \hat{\mathbf{x}} - y_{19} b \hat{\mathbf{y}} + \left(\frac{1}{2} + z_{19}\right) c \hat{\mathbf{z}}$ | (8*d*) | Si I |
| $\mathbf{B}_{127}$ | = | $-x_{19} \mathbf{a}_1 + \left(\frac{1}{2} + y_{19}\right) \mathbf{a}_2 - z_{19} \mathbf{a}_3$ | = | $-x_{19} a \hat{\mathbf{x}} + \left(\frac{1}{2} + y_{19}\right) b \hat{\mathbf{y}} - z_{19} c \hat{\mathbf{z}}$ | (8*d*) | Si I |
| $\mathbf{B}_{128}$ | = | $\left(\frac{1}{2} + x_{19}\right) \mathbf{a}_1 + \left(\frac{1}{2} - y_{19}\right) \mathbf{a}_2 + \left(\frac{1}{2} - z_{19}\right) \mathbf{a}_3$ | = | $\left(\frac{1}{2} + x_{19}\right) a \hat{\mathbf{x}} + \left(\frac{1}{2} - y_{19}\right) b \hat{\mathbf{y}} + \left(\frac{1}{2} - z_{19}\right) c \hat{\mathbf{z}}$ | (8*d*) | Si I |
| $\mathbf{B}_{129}$ | = | $-x_{19} \mathbf{a}_1 - y_{19} \mathbf{a}_2 - z_{19} \mathbf{a}_3$ | = | $-x_{19} a \hat{\mathbf{x}} - y_{19} b \hat{\mathbf{y}} - z_{19} c \hat{\mathbf{z}}$ | (8*d*) | Si I |
| $\mathbf{B}_{130}$ | = | $\left(\frac{1}{2} + x_{19}\right) \mathbf{a}_1 + y_{19} \mathbf{a}_2 + \left(\frac{1}{2} - z_{19}\right) \mathbf{a}_3$ | = | $\left(\frac{1}{2} + x_{19}\right) a \hat{\mathbf{x}} + y_{19} b \hat{\mathbf{y}} + \left(\frac{1}{2} - z_{19}\right) c \hat{\mathbf{z}}$ | (8*d*) | Si I |
| $\mathbf{B}_{131}$ | = | $x_{19} \mathbf{a}_1 + \left(\frac{1}{2} - y_{19}\right) \mathbf{a}_2 + z_{19} \mathbf{a}_3$ | = | $x_{19} a \hat{\mathbf{x}} + \left(\frac{1}{2} - y_{19}\right) b \hat{\mathbf{y}} + z_{19} c \hat{\mathbf{z}}$ | (8*d*) | Si I |
| $\mathbf{B}_{132}$ | = | $\left(\frac{1}{2} - x_{19}\right) \mathbf{a}_1 + \left(\frac{1}{2} + y_{19}\right) \mathbf{a}_2 + \left(\frac{1}{2} + z_{19}\right) \mathbf{a}_3$ | = | $\left(\frac{1}{2} - x_{19}\right) a \hat{\mathbf{x}} + \left(\frac{1}{2} + y_{19}\right) b \hat{\mathbf{y}} + \left(\frac{1}{2} + z_{19}\right) c \hat{\mathbf{z}}$ | (8*d*) | Si I |
| $\mathbf{B}_{133}$ | = | $x_{20} \mathbf{a}_1 + y_{20} \mathbf{a}_2 + z_{20} \mathbf{a}_3$ | = | $x_{20} a \hat{\mathbf{x}} + y_{20} b \hat{\mathbf{y}} + z_{20} c \hat{\mathbf{z}}$ | (8*d*) | Si II |
| $\mathbf{B}_{134}$ | = | $\left(\frac{1}{2} - x_{20}\right) \mathbf{a}_1 - y_{20} \mathbf{a}_2 + \left(\frac{1}{2} + z_{20}\right) \mathbf{a}_3$ | = | $\left(\frac{1}{2} - x_{20}\right) a \hat{\mathbf{x}} - y_{20} b \hat{\mathbf{y}} + \left(\frac{1}{2} + z_{20}\right) c \hat{\mathbf{z}}$ | (8*d*) | Si II |
| $\mathbf{B}_{135}$ | = | $-x_{20} \mathbf{a}_1 + \left(\frac{1}{2} + y_{20}\right) \mathbf{a}_2 - z_{20} \mathbf{a}_3$ | = | $-x_{20} a \hat{\mathbf{x}} + \left(\frac{1}{2} + y_{20}\right) b \hat{\mathbf{y}} - z_{20} c \hat{\mathbf{z}}$ | (8*d*) | Si II |
| $\mathbf{B}_{136}$ | = | $\left(\frac{1}{2} + x_{20}\right) \mathbf{a}_1 + \left(\frac{1}{2} - y_{20}\right) \mathbf{a}_2 + \left(\frac{1}{2} - z_{20}\right) \mathbf{a}_3$ | = | $\left(\frac{1}{2} + x_{20}\right) a \hat{\mathbf{x}} + \left(\frac{1}{2} - y_{20}\right) b \hat{\mathbf{y}} + \left(\frac{1}{2} - z_{20}\right) c \hat{\mathbf{z}}$ | (8*d*) | Si II |
| $\mathbf{B}_{137}$ | = | $-x_{20} \mathbf{a}_1 - y_{20} \mathbf{a}_2 - z_{20} \mathbf{a}_3$ | = | $-x_{20} a \hat{\mathbf{x}} - y_{20} b \hat{\mathbf{y}} - z_{20} c \hat{\mathbf{z}}$ | (8*d*) | Si II |
| $\mathbf{B}_{138}$ | = | $\left(\frac{1}{2} + x_{20}\right) \mathbf{a}_1 + y_{20} \mathbf{a}_2 + \left(\frac{1}{2} - z_{20}\right) \mathbf{a}_3$ | = | $\left(\frac{1}{2} + x_{20}\right) a \hat{\mathbf{x}} + y_{20} b \hat{\mathbf{y}} + \left(\frac{1}{2} - z_{20}\right) c \hat{\mathbf{z}}$ | (8*d*) | Si II |
| $\mathbf{B}_{139}$ | = | $x_{20} \mathbf{a}_1 + \left(\frac{1}{2} - y_{20}\right) \mathbf{a}_2 + z_{20} \mathbf{a}_3$ | = | $x_{20} a \hat{\mathbf{x}} + \left(\frac{1}{2} - y_{20}\right) b \hat{\mathbf{y}} + z_{20} c \hat{\mathbf{z}}$ | (8*d*) | Si II |
| $\mathbf{B}_{140}$ | = | $\left(\frac{1}{2} - x_{20}\right) \mathbf{a}_1 + \left(\frac{1}{2} + y_{20}\right) \mathbf{a}_2 + \left(\frac{1}{2} + z_{20}\right) \mathbf{a}_3$ | = | $\left(\frac{1}{2} - x_{20}\right) a \hat{\mathbf{x}} + \left(\frac{1}{2} + y_{20}\right) b \hat{\mathbf{y}} + \left(\frac{1}{2} + z_{20}\right) c \hat{\mathbf{z}}$ | (8*d*) | Si II |
| $\mathbf{B}_{141}$ | = | $x_{21} \mathbf{a}_1 + y_{21} \mathbf{a}_2 + z_{21} \mathbf{a}_3$ | = | $x_{21} a \hat{\mathbf{x}} + y_{21} b \hat{\mathbf{y}} + z_{21} c \hat{\mathbf{z}}$ | (8*d*) | Si III |
| $\mathbf{B}_{142}$ | = | $\left(\frac{1}{2} - x_{21}\right) \mathbf{a}_1 - y_{21} \mathbf{a}_2 + \left(\frac{1}{2} + z_{21}\right) \mathbf{a}_3$ | = | $\left(\frac{1}{2} - x_{21}\right) a \hat{\mathbf{x}} - y_{21} b \hat{\mathbf{y}} + \left(\frac{1}{2} + z_{21}\right) c \hat{\mathbf{z}}$ | (8*d*) | Si III |
| $\mathbf{B}_{143}$ | = | $-x_{21} \mathbf{a}_1 + \left(\frac{1}{2} + y_{21}\right) \mathbf{a}_2 - z_{21} \mathbf{a}_3$ | = | $-x_{21} a \hat{\mathbf{x}} + \left(\frac{1}{2} + y_{21}\right) b \hat{\mathbf{y}} - z_{21} c \hat{\mathbf{z}}$ | (8*d*) | Si III |
| $\mathbf{B}_{144}$ | = | $\left(\frac{1}{2}+x_{21}\right)\mathbf{a}_1+\left(\frac{1}{2}-y_{21}\right)\mathbf{a}_2+\left(\frac{1}{2}-z_{21}\right)\mathbf{a}_3$ | = | $\left(\frac{1}{2}+x_{21}\right)a\,\hat{\mathbf{x}}+\left(\frac{1}{2}-y_{21}\right)b\,\hat{\mathbf{y}}+\left(\frac{1}{2}-z_{21}\right)c\,\hat{\mathbf{z}}$ | (8$d$) | Si III |
| $\mathbf{B}_{145}$ | = | $-x_{21}\,\mathbf{a}_1-y_{21}\,\mathbf{a}_2-z_{21}\,\mathbf{a}_3$ | = | $-x_{21}a\,\hat{\mathbf{x}}-y_{21}b\,\hat{\mathbf{y}}-z_{21}c\,\hat{\mathbf{z}}$ | (8$d$) | Si III |
| $\mathbf{B}_{146}$ | = | $\left(\frac{1}{2}+x_{21}\right)\mathbf{a}_1+y_{21}\,\mathbf{a}_2+\left(\frac{1}{2}-z_{21}\right)\mathbf{a}_3$ | = | $\left(\frac{1}{2}+x_{21}\right)a\,\hat{\mathbf{x}}+y_{21}b\,\hat{\mathbf{y}}+\left(\frac{1}{2}-z_{21}\right)c\,\hat{\mathbf{z}}$ | (8$d$) | Si III |
| $\mathbf{B}_{147}$ | = | $x_{21}\,\mathbf{a}_1+\left(\frac{1}{2}-y_{21}\right)\mathbf{a}_2+z_{21}\,\mathbf{a}_3$ | = | $x_{21}a\,\hat{\mathbf{x}}+\left(\frac{1}{2}-y_{21}\right)b\,\hat{\mathbf{y}}+z_{21}c\,\hat{\mathbf{z}}$ | (8$d$) | Si III |
| $\mathbf{B}_{148}$ | = | $\left(\frac{1}{2}-x_{21}\right)\mathbf{a}_1+\left(\frac{1}{2}+y_{21}\right)\mathbf{a}_2+\left(\frac{1}{2}+z_{21}\right)\mathbf{a}_3$ | = | $\left(\frac{1}{2}-x_{21}\right)a\,\hat{\mathbf{x}}+\left(\frac{1}{2}+y_{21}\right)b\,\hat{\mathbf{y}}+\left(\frac{1}{2}+z_{21}\right)c\,\hat{\mathbf{z}}$ | (8$d$) | Si III |
| $\mathbf{B}_{149}$ | = | $x_{22}\,\mathbf{a}_1+y_{22}\,\mathbf{a}_2+z_{22}\,\mathbf{a}_3$ | = | $x_{22}a\,\hat{\mathbf{x}}+y_{22}b\,\hat{\mathbf{y}}+z_{22}c\,\hat{\mathbf{z}}$ | (8$d$) | Si IV |
| $\mathbf{B}_{150}$ | = | $\left(\frac{1}{2}-x_{22}\right)\mathbf{a}_1-y_{22}\,\mathbf{a}_2+\left(\frac{1}{2}+z_{22}\right)\mathbf{a}_3$ | = | $\left(\frac{1}{2}-x_{22}\right)a\,\hat{\mathbf{x}}-y_{22}b\,\hat{\mathbf{y}}+\left(\frac{1}{2}+z_{22}\right)c\,\hat{\mathbf{z}}$ | (8$d$) | Si IV |
| $\mathbf{B}_{151}$ | = | $-x_{22}\,\mathbf{a}_1+\left(\frac{1}{2}+y_{22}\right)\mathbf{a}_2-z_{22}\,\mathbf{a}_3$ | = | $-x_{22}a\,\hat{\mathbf{x}}+\left(\frac{1}{2}+y_{22}\right)b\,\hat{\mathbf{y}}-z_{22}c\,\hat{\mathbf{z}}$ | (8$d$) | Si IV |
| $\mathbf{B}_{152}$ | = | $\left(\frac{1}{2}+x_{22}\right)\mathbf{a}_1+\left(\frac{1}{2}-y_{22}\right)\mathbf{a}_2+\left(\frac{1}{2}-z_{22}\right)\mathbf{a}_3$ | = | $\left(\frac{1}{2}+x_{22}\right)a\,\hat{\mathbf{x}}+\left(\frac{1}{2}-y_{22}\right)b\,\hat{\mathbf{y}}+\left(\frac{1}{2}-z_{22}\right)c\,\hat{\mathbf{z}}$ | (8$d$) | Si IV |
| $\mathbf{B}_{153}$ | = | $-x_{22}\,\mathbf{a}_1-y_{22}\,\mathbf{a}_2-z_{22}\,\mathbf{a}_3$ | = | $-x_{22}a\,\hat{\mathbf{x}}-y_{22}b\,\hat{\mathbf{y}}-z_{22}c\,\hat{\mathbf{z}}$ | (8$d$) | Si IV |
| $\mathbf{B}_{154}$ | = | $\left(\frac{1}{2}+x_{22}\right)\mathbf{a}_1+y_{22}\,\mathbf{a}_2+\left(\frac{1}{2}-z_{22}\right)\mathbf{a}_3$ | = | $\left(\frac{1}{2}+x_{22}\right)a\,\hat{\mathbf{x}}+y_{22}b\,\hat{\mathbf{y}}+\left(\frac{1}{2}-z_{22}\right)c\,\hat{\mathbf{z}}$ | (8$d$) | Si IV |
| $\mathbf{B}_{155}$ | = | $x_{22}\,\mathbf{a}_1+\left(\frac{1}{2}-y_{22}\right)\mathbf{a}_2+z_{22}\,\mathbf{a}_3$ | = | $x_{22}a\,\hat{\mathbf{x}}+\left(\frac{1}{2}-y_{22}\right)b\,\hat{\mathbf{y}}+z_{22}c\,\hat{\mathbf{z}}$ | (8$d$) | Si IV |
| $\mathbf{B}_{156}$ | = | $\left(\frac{1}{2}-x_{22}\right)\mathbf{a}_1+\left(\frac{1}{2}+y_{22}\right)\mathbf{a}_2+\left(\frac{1}{2}+z_{22}\right)\mathbf{a}_3$ | = | $\left(\frac{1}{2}-x_{22}\right)a\,\hat{\mathbf{x}}+\left(\frac{1}{2}+y_{22}\right)b\,\hat{\mathbf{y}}+\left(\frac{1}{2}+z_{22}\right)c\,\hat{\mathbf{z}}$ | (8$d$) | Si IV |

**Found in:**

**Geometry files:**

- CIF: pp. 1631
- POSCAR: pp. 1631

# Sillimanite ($Al_2SiO_5$, $S0_3$) Structure: A2B5C_oP32_62_bc_3cd_c

http://aflow.org/prototype-encyclopedia/A2B5C_oP32_62_bc_3cd_c

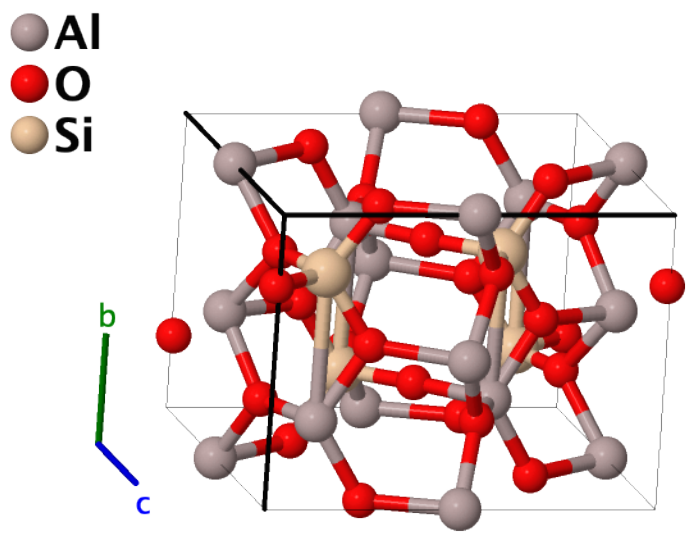

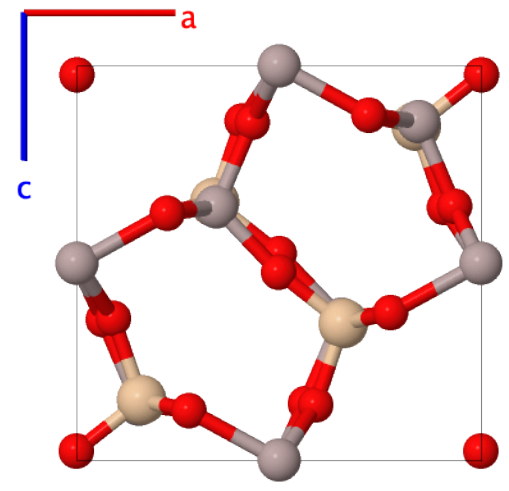

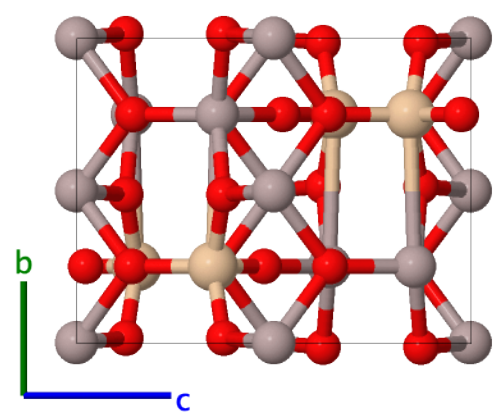

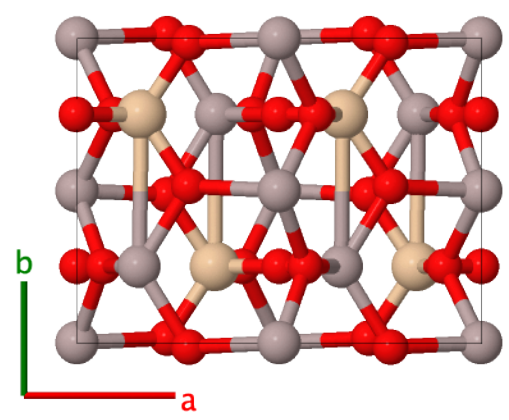

| | | |
|---|---|---|
| **Prototype** | : | $Al_2O_5Si$ |
| **AFLOW prototype label** | : | A2B5C_oP32_62_bc_3cd_c |
| ***Strukturbericht* designation** | : | $S0_3$ |
| **Pearson symbol** | : | oP32 |
| **Space group number** | : | 62 |
| **Space group symbol** | : | *Pnma* |
| **AFLOW prototype command** | : | `aflow --proto=A2B5C_oP32_62_bc_3cd_c` `--params=`$a, b/a, c/a, x_2, z_2, x_3, z_3, x_4, z_4, x_5, z_5, x_6, z_6, x_7, y_7, z_7$ |

- Three crystal polymorphs of $Al_2SiO_5$ have been characterized: kyanite ($S0_1$), space group $P\bar{1}$ #2, andalusite ($S0_2$), space group *Pnnm* #58, and sillimanite ($S0_3$), space group *Pnma* #62. All are characterized chains of edge-sharing $SiO_6$ tetrahedra and Al octahedra.

- We use the ambient pressure data of (Yang, 1997).

- (Yang, 1997) give the Wyckoff positions in the *Pbnm* setting of space group #62. We use FINDSYM to transform this to the standard *Pnma* setting. This involves a rotation of the principle axes and a shift of the set of aluminum atoms from the (4*a*) to the (4*b*) Wyckoff position.

- (Hermann, 1937) defined this as $S0_3$, but also listed it as $H5_2$ in the index.

**Simple Orthorhombic primitive vectors:**

$$\mathbf{a}_1 = a\,\hat{\mathbf{x}}$$
$$\mathbf{a}_2 = b\,\hat{\mathbf{y}}$$
$$\mathbf{a}_3 = c\,\hat{\mathbf{z}}$$

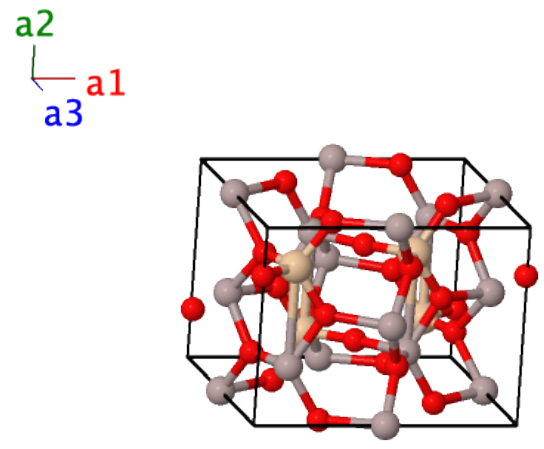

**Basis vectors:**

| | | Lattice Coordinates | | Cartesian Coordinates | Wyckoff Position | Atom Type |
|---|---|---|---|---|---|---|
| $\mathbf{B}_1$ | = | $\frac{1}{2}\,\mathbf{a}_3$ | = | $\frac{1}{2}c\,\hat{\mathbf{z}}$ | (4*b*) | Al I |
| $\mathbf{B}_2$ | = | $\frac{1}{2}\,\mathbf{a}_1$ | = | $\frac{1}{2}a\,\hat{\mathbf{x}}$ | (4*b*) | Al I |
| $\mathbf{B}_3$ | = | $\frac{1}{2}\,\mathbf{a}_2 + \frac{1}{2}\,\mathbf{a}_3$ | = | $\frac{1}{2}b\,\hat{\mathbf{y}} + \frac{1}{2}c\,\hat{\mathbf{z}}$ | (4*b*) | Al I |
| $\mathbf{B}_4$ | = | $\frac{1}{2}\,\mathbf{a}_1 + \frac{1}{2}\,\mathbf{a}_2$ | = | $\frac{1}{2}a\,\hat{\mathbf{x}} + \frac{1}{2}b\,\hat{\mathbf{y}}$ | (4*b*) | Al I |
| $\mathbf{B}_5$ | = | $x_2\,\mathbf{a}_1 + \frac{1}{4}\,\mathbf{a}_2 + z_2\,\mathbf{a}_3$ | = | $x_2a\,\hat{\mathbf{x}} + \frac{1}{4}b\,\hat{\mathbf{y}} + z_2c\,\hat{\mathbf{z}}$ | (4*c*) | Al II |
| $\mathbf{B}_6$ | = | $\left(\frac{1}{2} - x_2\right)\mathbf{a}_1 + \frac{3}{4}\,\mathbf{a}_2 + \left(\frac{1}{2} + z_2\right)\mathbf{a}_3$ | = | $\left(\frac{1}{2} - x_2\right)a\,\hat{\mathbf{x}} + \frac{3}{4}b\,\hat{\mathbf{y}} + \left(\frac{1}{2} + z_2\right)c\,\hat{\mathbf{z}}$ | (4*c*) | Al II |
| $\mathbf{B}_7$ | = | $-x_2\,\mathbf{a}_1 + \frac{3}{4}\,\mathbf{a}_2 - z_2\,\mathbf{a}_3$ | = | $-x_2a\,\hat{\mathbf{x}} + \frac{3}{4}b\,\hat{\mathbf{y}} - z_2c\,\hat{\mathbf{z}}$ | (4*c*) | Al II |
| $\mathbf{B}_8$ | = | $\left(\frac{1}{2} + x_2\right)\mathbf{a}_1 + \frac{1}{4}\,\mathbf{a}_2 + \left(\frac{1}{2} - z_2\right)\mathbf{a}_3$ | = | $\left(\frac{1}{2} + x_2\right)a\,\hat{\mathbf{x}} + \frac{1}{4}b\,\hat{\mathbf{y}} + \left(\frac{1}{2} - z_2\right)c\,\hat{\mathbf{z}}$ | (4*c*) | Al II |
| $\mathbf{B}_9$ | = | $x_3\,\mathbf{a}_1 + \frac{1}{4}\,\mathbf{a}_2 + z_3\,\mathbf{a}_3$ | = | $x_3a\,\hat{\mathbf{x}} + \frac{1}{4}b\,\hat{\mathbf{y}} + z_3c\,\hat{\mathbf{z}}$ | (4*c*) | O I |
| $\mathbf{B}_{10}$ | = | $\left(\frac{1}{2} - x_3\right)\mathbf{a}_1 + \frac{3}{4}\,\mathbf{a}_2 + \left(\frac{1}{2} + z_3\right)\mathbf{a}_3$ | = | $\left(\frac{1}{2} - x_3\right)a\,\hat{\mathbf{x}} + \frac{3}{4}b\,\hat{\mathbf{y}} + \left(\frac{1}{2} + z_3\right)c\,\hat{\mathbf{z}}$ | (4*c*) | O I |
| $\mathbf{B}_{11}$ | = | $-x_3\,\mathbf{a}_1 + \frac{3}{4}\,\mathbf{a}_2 - z_3\,\mathbf{a}_3$ | = | $-x_3a\,\hat{\mathbf{x}} + \frac{3}{4}b\,\hat{\mathbf{y}} - z_3c\,\hat{\mathbf{z}}$ | (4*c*) | O I |
| $\mathbf{B}_{12}$ | = | $\left(\frac{1}{2} + x_3\right)\mathbf{a}_1 + \frac{1}{4}\,\mathbf{a}_2 + \left(\frac{1}{2} - z_3\right)\mathbf{a}_3$ | = | $\left(\frac{1}{2} + x_3\right)a\,\hat{\mathbf{x}} + \frac{1}{4}b\,\hat{\mathbf{y}} + \left(\frac{1}{2} - z_3\right)c\,\hat{\mathbf{z}}$ | (4*c*) | O I |
| $\mathbf{B}_{13}$ | = | $x_4\,\mathbf{a}_1 + \frac{1}{4}\,\mathbf{a}_2 + z_4\,\mathbf{a}_3$ | = | $x_4a\,\hat{\mathbf{x}} + \frac{1}{4}b\,\hat{\mathbf{y}} + z_4c\,\hat{\mathbf{z}}$ | (4*c*) | O II |
| $\mathbf{B}_{14}$ | = | $\left(\frac{1}{2} - x_4\right)\mathbf{a}_1 + \frac{3}{4}\,\mathbf{a}_2 + \left(\frac{1}{2} + z_4\right)\mathbf{a}_3$ | = | $\left(\frac{1}{2} - x_4\right)a\,\hat{\mathbf{x}} + \frac{3}{4}b\,\hat{\mathbf{y}} + \left(\frac{1}{2} + z_4\right)c\,\hat{\mathbf{z}}$ | (4*c*) | O II |
| $\mathbf{B}_{15}$ | = | $-x_4\,\mathbf{a}_1 + \frac{3}{4}\,\mathbf{a}_2 - z_4\,\mathbf{a}_3$ | = | $-x_4a\,\hat{\mathbf{x}} + \frac{3}{4}b\,\hat{\mathbf{y}} - z_4c\,\hat{\mathbf{z}}$ | (4*c*) | O II |
| $\mathbf{B}_{16}$ | = | $\left(\frac{1}{2} + x_4\right)\mathbf{a}_1 + \frac{1}{4}\,\mathbf{a}_2 + \left(\frac{1}{2} - z_4\right)\mathbf{a}_3$ | = | $\left(\frac{1}{2} + x_4\right)a\,\hat{\mathbf{x}} + \frac{1}{4}b\,\hat{\mathbf{y}} + \left(\frac{1}{2} - z_4\right)c\,\hat{\mathbf{z}}$ | (4*c*) | O II |
| $\mathbf{B}_{17}$ | = | $x_5\,\mathbf{a}_1 + \frac{1}{4}\,\mathbf{a}_2 + z_5\,\mathbf{a}_3$ | = | $x_5a\,\hat{\mathbf{x}} + \frac{1}{4}b\,\hat{\mathbf{y}} + z_5c\,\hat{\mathbf{z}}$ | (4*c*) | O III |
| $\mathbf{B}_{18}$ | = | $\left(\frac{1}{2} - x_5\right)\mathbf{a}_1 + \frac{3}{4}\,\mathbf{a}_2 + \left(\frac{1}{2} + z_5\right)\mathbf{a}_3$ | = | $\left(\frac{1}{2} - x_5\right)a\,\hat{\mathbf{x}} + \frac{3}{4}b\,\hat{\mathbf{y}} + \left(\frac{1}{2} + z_5\right)c\,\hat{\mathbf{z}}$ | (4*c*) | O III |
| $\mathbf{B}_{19}$ | = | $-x_5\,\mathbf{a}_1 + \frac{3}{4}\,\mathbf{a}_2 - z_5\,\mathbf{a}_3$ | = | $-x_5a\,\hat{\mathbf{x}} + \frac{3}{4}b\,\hat{\mathbf{y}} - z_5c\,\hat{\mathbf{z}}$ | (4*c*) | O III |
| $\mathbf{B}_{20}$ | = | $\left(\frac{1}{2} + x_5\right)\mathbf{a}_1 + \frac{1}{4}\,\mathbf{a}_2 + \left(\frac{1}{2} - z_5\right)\mathbf{a}_3$ | = | $\left(\frac{1}{2} + x_5\right)a\,\hat{\mathbf{x}} + \frac{1}{4}b\,\hat{\mathbf{y}} + \left(\frac{1}{2} - z_5\right)c\,\hat{\mathbf{z}}$ | (4*c*) | O III |
| $\mathbf{B}_{21}$ | = | $x_6\,\mathbf{a}_1 + \frac{1}{4}\,\mathbf{a}_2 + z_6\,\mathbf{a}_3$ | = | $x_6a\,\hat{\mathbf{x}} + \frac{1}{4}b\,\hat{\mathbf{y}} + z_6c\,\hat{\mathbf{z}}$ | (4*c*) | Si |
| $\mathbf{B}_{22}$ | = | $\left(\frac{1}{2} - x_6\right)\mathbf{a}_1 + \frac{3}{4}\,\mathbf{a}_2 + \left(\frac{1}{2} + z_6\right)\mathbf{a}_3$ | = | $\left(\frac{1}{2} - x_6\right)a\,\hat{\mathbf{x}} + \frac{3}{4}b\,\hat{\mathbf{y}} + \left(\frac{1}{2} + z_6\right)c\,\hat{\mathbf{z}}$ | (4*c*) | Si |
| $\mathbf{B}_{23}$ | = | $-x_6\,\mathbf{a}_1 + \frac{3}{4}\,\mathbf{a}_2 - z_6\,\mathbf{a}_3$ | = | $-x_6a\,\hat{\mathbf{x}} + \frac{3}{4}b\,\hat{\mathbf{y}} - z_6c\,\hat{\mathbf{z}}$ | (4*c*) | Si |
| $\mathbf{B}_{24}$ | = | $\left(\frac{1}{2} + x_6\right)\mathbf{a}_1 + \frac{1}{4}\,\mathbf{a}_2 + \left(\frac{1}{2} - z_6\right)\mathbf{a}_3$ | = | $\left(\frac{1}{2} + x_6\right)a\,\hat{\mathbf{x}} + \frac{1}{4}b\,\hat{\mathbf{y}} + \left(\frac{1}{2} - z_6\right)c\,\hat{\mathbf{z}}$ | (4*c*) | Si |
| $\mathbf{B}_{25}$ | = | $x_7\,\mathbf{a}_1 + y_7\,\mathbf{a}_2 + z_7\,\mathbf{a}_3$ | = | $x_7a\,\hat{\mathbf{x}} + y_7b\,\hat{\mathbf{y}} + z_7c\,\hat{\mathbf{z}}$ | (8*d*) | O IV |
| $\mathbf{B}_{26}$ | = | $\left(\frac{1}{2} - x_7\right)\mathbf{a}_1 - y_7\,\mathbf{a}_2 + \left(\frac{1}{2} + z_7\right)\mathbf{a}_3$ | = | $\left(\frac{1}{2} - x_7\right)a\,\hat{\mathbf{x}} - y_7b\,\hat{\mathbf{y}} + \left(\frac{1}{2} + z_7\right)c\,\hat{\mathbf{z}}$ | (8*d*) | O IV |
| $\mathbf{B}_{27}$ | = | $-x_7\,\mathbf{a}_1 + \left(\frac{1}{2} + y_7\right)\mathbf{a}_2 - z_7\,\mathbf{a}_3$ | = | $-x_7a\,\hat{\mathbf{x}} + \left(\frac{1}{2} + y_7\right)b\,\hat{\mathbf{y}} - z_7c\,\hat{\mathbf{z}}$ | (8*d*) | O IV |
| $\mathbf{B}_{28}$ | = | $\left(\frac{1}{2}+x_7\right)\mathbf{a}_1+\left(\frac{1}{2}-y_7\right)\mathbf{a}_2+\left(\frac{1}{2}-z_7\right)\mathbf{a}_3$ | = | $\left(\frac{1}{2}+x_7\right)a\,\hat{\mathbf{x}}+\left(\frac{1}{2}-y_7\right)b\,\hat{\mathbf{y}}+\left(\frac{1}{2}-z_7\right)c\,\hat{\mathbf{z}}$ | (8*d*) | O IV |
| $\mathbf{B}_{29}$ | = | $-x_7\,\mathbf{a}_1-y_7\,\mathbf{a}_2-z_7\,\mathbf{a}_3$ | = | $-x_7a\,\hat{\mathbf{x}}-y_7b\,\hat{\mathbf{y}}-z_7c\,\hat{\mathbf{z}}$ | (8*d*) | O IV |
| $\mathbf{B}_{30}$ | = | $\left(\frac{1}{2}+x_7\right)\mathbf{a}_1+y_7\,\mathbf{a}_2+\left(\frac{1}{2}-z_7\right)\mathbf{a}_3$ | = | $\left(\frac{1}{2}+x_7\right)a\,\hat{\mathbf{x}}+y_7b\,\hat{\mathbf{y}}+\left(\frac{1}{2}-z_7\right)c\,\hat{\mathbf{z}}$ | (8*d*) | O IV |
| $\mathbf{B}_{31}$ | = | $x_7\,\mathbf{a}_1+\left(\frac{1}{2}-y_7\right)\mathbf{a}_2+z_7\,\mathbf{a}_3$ | = | $x_7a\,\hat{\mathbf{x}}+\left(\frac{1}{2}-y_7\right)b\,\hat{\mathbf{y}}+z_7c\,\hat{\mathbf{z}}$ | (8*d*) | O IV |
| $\mathbf{B}_{32}$ | = | $\left(\frac{1}{2}-x_7\right)\mathbf{a}_1+\left(\frac{1}{2}+y_7\right)\mathbf{a}_2+\left(\frac{1}{2}+z_7\right)\mathbf{a}_3$ | = | $\left(\frac{1}{2}-x_7\right)a\,\hat{\mathbf{x}}+\left(\frac{1}{2}+y_7\right)b\,\hat{\mathbf{y}}+\left(\frac{1}{2}+z_7\right)c\,\hat{\mathbf{z}}$ | (8*d*) | O IV |

**Geometry files:**

- CIF: pp. 1632
- POSCAR: pp. 1633

# $K_2S_3O_6$ ($K5_1$) Structure: A2B6C3_oP44_62_2c_2c2d_3c

http://aflow.org/prototype-encyclopedia/A2B6C3_oP44_62_2c_2c2d_3c

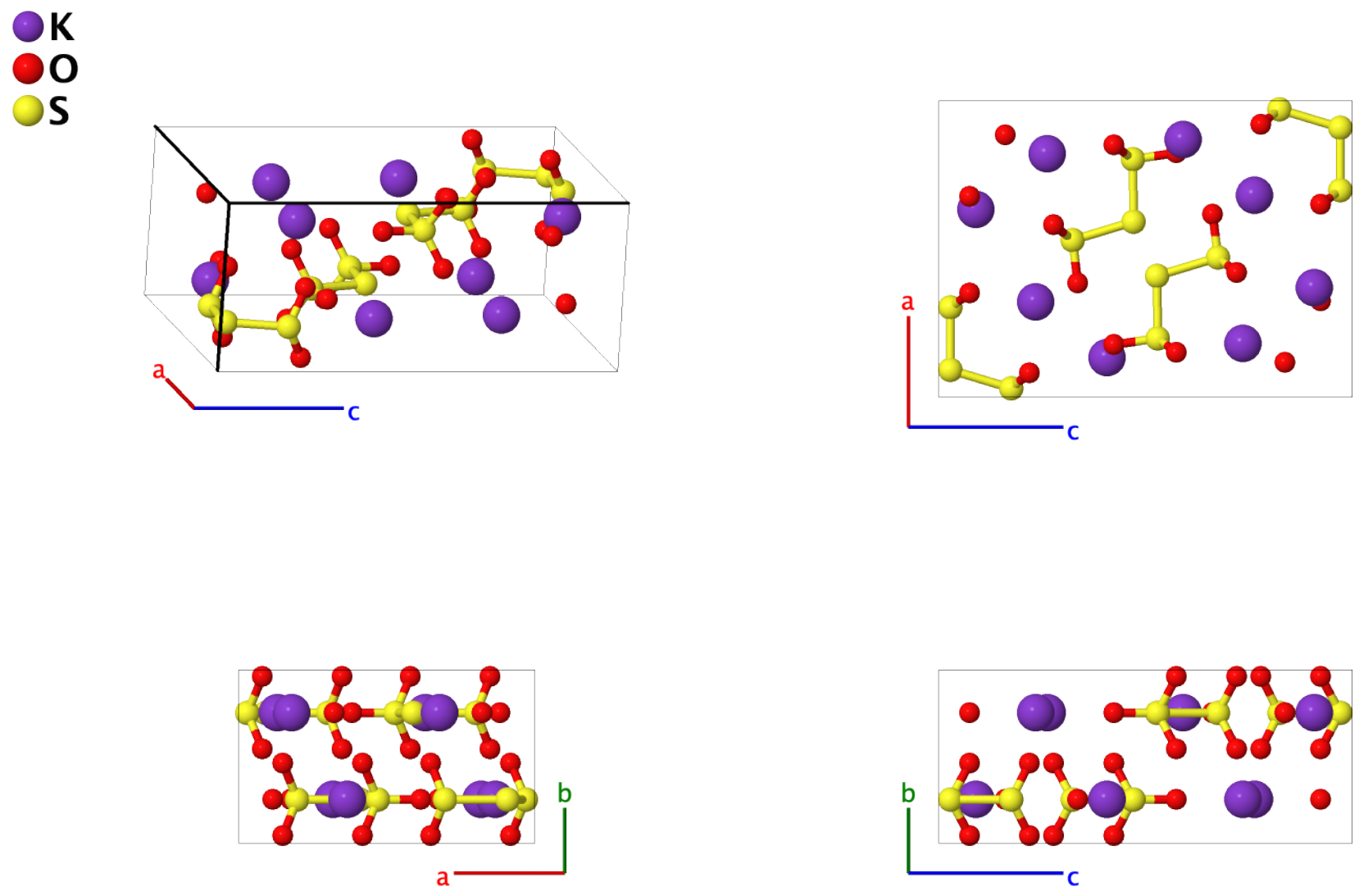

| | | |
|---|---|---|
| **Prototype** | : | $K_2O_6S_3$ |
| **AFLOW prototype label** | : | A2B6C3_oP44_62_2c_2c2d_3c |
| ***Strukturbericht* designation** | : | $K5_1$ |
| **Pearson symbol** | : | oP44 |
| **Space group number** | : | 62 |
| **Space group symbol** | : | *Pnma* |
| **AFLOW prototype command** | : | `aflow --proto=A2B6C3_oP44_62_2c_2c2d_3c` `--params=`$a, b/a, c/a, x_1, z_1, x_2, z_2, x_3, z_3, x_4, z_4, x_5, z_5, x_6, z_6, x_7, z_7, x_8, y_8, z_8, x_9, y_9, z_9$ |

**Other compounds with this structure**

- $Rb_2S_3O_6$

- (Stewart, 1979) give the Wyckoff positions of this structure using the *Pnam* orientation of space group #62. We have used FINDSYM to change this to our standard *Pnma* orientation.

**Simple Orthorhombic primitive vectors:**

$$\mathbf{a}_1 = a\,\hat{\mathbf{x}}$$
$$\mathbf{a}_2 = b\,\hat{\mathbf{y}}$$
$$\mathbf{a}_3 = c\,\hat{\mathbf{z}}$$

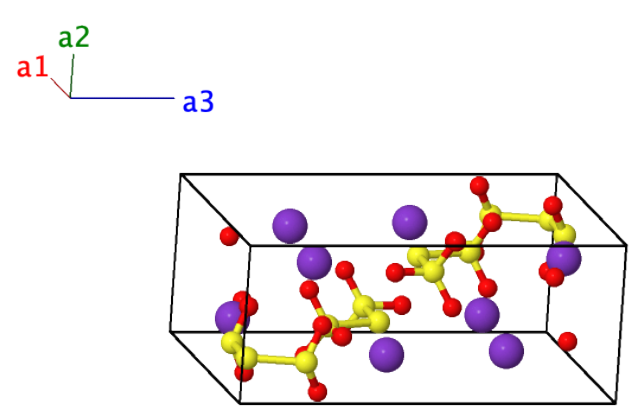

**Basis vectors:**

| | | Lattice Coordinates | | Cartesian Coordinates | Wyckoff Position | Atom Type |
|---|---|---|---|---|---|---|
| $\mathbf{B}_1$ | = | $x_1\,\mathbf{a}_1 + \frac{1}{4}\,\mathbf{a}_2 + z_1\,\mathbf{a}_3$ | = | $x_1 a\,\hat{\mathbf{x}} + \frac{1}{4} b\,\hat{\mathbf{y}} + z_1 c\,\hat{\mathbf{z}}$ | (4$c$) | K I |
| $\mathbf{B}_2$ | = | $\left(\frac{1}{2} - x_1\right)\mathbf{a}_1 + \frac{3}{4}\,\mathbf{a}_2 + \left(\frac{1}{2} + z_1\right)\mathbf{a}_3$ | = | $\left(\frac{1}{2} - x_1\right) a\,\hat{\mathbf{x}} + \frac{3}{4} b\,\hat{\mathbf{y}} + \left(\frac{1}{2} + z_1\right) c\,\hat{\mathbf{z}}$ | (4$c$) | K I |
| $\mathbf{B}_3$ | = | $-x_1\,\mathbf{a}_1 + \frac{3}{4}\,\mathbf{a}_2 - z_1\,\mathbf{a}_3$ | = | $-x_1 a\,\hat{\mathbf{x}} + \frac{3}{4} b\,\hat{\mathbf{y}} - z_1 c\,\hat{\mathbf{z}}$ | (4$c$) | K I |
| $\mathbf{B}_4$ | = | $\left(\frac{1}{2} + x_1\right)\mathbf{a}_1 + \frac{1}{4}\,\mathbf{a}_2 + \left(\frac{1}{2} - z_1\right)\mathbf{a}_3$ | = | $\left(\frac{1}{2} + x_1\right) a\,\hat{\mathbf{x}} + \frac{1}{4} b\,\hat{\mathbf{y}} + \left(\frac{1}{2} - z_1\right) c\,\hat{\mathbf{z}}$ | (4$c$) | K I |
| $\mathbf{B}_5$ | = | $x_2\,\mathbf{a}_1 + \frac{1}{4}\,\mathbf{a}_2 + z_2\,\mathbf{a}_3$ | = | $x_2 a\,\hat{\mathbf{x}} + \frac{1}{4} b\,\hat{\mathbf{y}} + z_2 c\,\hat{\mathbf{z}}$ | (4$c$) | K II |
| $\mathbf{B}_6$ | = | $\left(\frac{1}{2} - x_2\right)\mathbf{a}_1 + \frac{3}{4}\,\mathbf{a}_2 + \left(\frac{1}{2} + z_2\right)\mathbf{a}_3$ | = | $\left(\frac{1}{2} - x_2\right) a\,\hat{\mathbf{x}} + \frac{3}{4} b\,\hat{\mathbf{y}} + \left(\frac{1}{2} + z_2\right) c\,\hat{\mathbf{z}}$ | (4$c$) | K II |
| $\mathbf{B}_7$ | = | $-x_2\,\mathbf{a}_1 + \frac{3}{4}\,\mathbf{a}_2 - z_2\,\mathbf{a}_3$ | = | $-x_2 a\,\hat{\mathbf{x}} + \frac{3}{4} b\,\hat{\mathbf{y}} - z_2 c\,\hat{\mathbf{z}}$ | (4$c$) | K II |
| $\mathbf{B}_8$ | = | $\left(\frac{1}{2} + x_2\right)\mathbf{a}_1 + \frac{1}{4}\,\mathbf{a}_2 + \left(\frac{1}{2} - z_2\right)\mathbf{a}_3$ | = | $\left(\frac{1}{2} + x_2\right) a\,\hat{\mathbf{x}} + \frac{1}{4} b\,\hat{\mathbf{y}} + \left(\frac{1}{2} - z_2\right) c\,\hat{\mathbf{z}}$ | (4$c$) | K II |
| $\mathbf{B}_9$ | = | $x_3\,\mathbf{a}_1 + \frac{1}{4}\,\mathbf{a}_2 + z_3\,\mathbf{a}_3$ | = | $x_3 a\,\hat{\mathbf{x}} + \frac{1}{4} b\,\hat{\mathbf{y}} + z_3 c\,\hat{\mathbf{z}}$ | (4$c$) | O I |
| $\mathbf{B}_{10}$ | = | $\left(\frac{1}{2} - x_3\right)\mathbf{a}_1 + \frac{3}{4}\,\mathbf{a}_2 + \left(\frac{1}{2} + z_3\right)\mathbf{a}_3$ | = | $\left(\frac{1}{2} - x_3\right) a\,\hat{\mathbf{x}} + \frac{3}{4} b\,\hat{\mathbf{y}} + \left(\frac{1}{2} + z_3\right) c\,\hat{\mathbf{z}}$ | (4$c$) | O I |
| $\mathbf{B}_{11}$ | = | $-x_3\,\mathbf{a}_1 + \frac{3}{4}\,\mathbf{a}_2 - z_3\,\mathbf{a}_3$ | = | $-x_3 a\,\hat{\mathbf{x}} + \frac{3}{4} b\,\hat{\mathbf{y}} - z_3 c\,\hat{\mathbf{z}}$ | (4$c$) | O I |
| $\mathbf{B}_{12}$ | = | $\left(\frac{1}{2} + x_3\right)\mathbf{a}_1 + \frac{1}{4}\,\mathbf{a}_2 + \left(\frac{1}{2} - z_3\right)\mathbf{a}_3$ | = | $\left(\frac{1}{2} + x_3\right) a\,\hat{\mathbf{x}} + \frac{1}{4} b\,\hat{\mathbf{y}} + \left(\frac{1}{2} - z_3\right) c\,\hat{\mathbf{z}}$ | (4$c$) | O I |
| $\mathbf{B}_{13}$ | = | $x_4\,\mathbf{a}_1 + \frac{1}{4}\,\mathbf{a}_2 + z_4\,\mathbf{a}_3$ | = | $x_4 a\,\hat{\mathbf{x}} + \frac{1}{4} b\,\hat{\mathbf{y}} + z_4 c\,\hat{\mathbf{z}}$ | (4$c$) | O II |
| $\mathbf{B}_{14}$ | = | $\left(\frac{1}{2} - x_4\right)\mathbf{a}_1 + \frac{3}{4}\,\mathbf{a}_2 + \left(\frac{1}{2} + z_4\right)\mathbf{a}_3$ | = | $\left(\frac{1}{2} - x_4\right) a\,\hat{\mathbf{x}} + \frac{3}{4} b\,\hat{\mathbf{y}} + \left(\frac{1}{2} + z_4\right) c\,\hat{\mathbf{z}}$ | (4$c$) | O II |
| $\mathbf{B}_{15}$ | = | $-x_4\,\mathbf{a}_1 + \frac{3}{4}\,\mathbf{a}_2 - z_4\,\mathbf{a}_3$ | = | $-x_4 a\,\hat{\mathbf{x}} + \frac{3}{4} b\,\hat{\mathbf{y}} - z_4 c\,\hat{\mathbf{z}}$ | (4$c$) | O II |
| $\mathbf{B}_{16}$ | = | $\left(\frac{1}{2} + x_4\right)\mathbf{a}_1 + \frac{1}{4}\,\mathbf{a}_2 + \left(\frac{1}{2} - z_4\right)\mathbf{a}_3$ | = | $\left(\frac{1}{2} + x_4\right) a\,\hat{\mathbf{x}} + \frac{1}{4} b\,\hat{\mathbf{y}} + \left(\frac{1}{2} - z_4\right) c\,\hat{\mathbf{z}}$ | (4$c$) | O II |
| $\mathbf{B}_{17}$ | = | $x_5\,\mathbf{a}_1 + \frac{1}{4}\,\mathbf{a}_2 + z_5\,\mathbf{a}_3$ | = | $x_5 a\,\hat{\mathbf{x}} + \frac{1}{4} b\,\hat{\mathbf{y}} + z_5 c\,\hat{\mathbf{z}}$ | (4$c$) | S I |
| $\mathbf{B}_{18}$ | = | $\left(\frac{1}{2} - x_5\right)\mathbf{a}_1 + \frac{3}{4}\,\mathbf{a}_2 + \left(\frac{1}{2} + z_5\right)\mathbf{a}_3$ | = | $\left(\frac{1}{2} - x_5\right) a\,\hat{\mathbf{x}} + \frac{3}{4} b\,\hat{\mathbf{y}} + \left(\frac{1}{2} + z_5\right) c\,\hat{\mathbf{z}}$ | (4$c$) | S I |
| $\mathbf{B}_{19}$ | = | $-x_5\,\mathbf{a}_1 + \frac{3}{4}\,\mathbf{a}_2 - z_5\,\mathbf{a}_3$ | = | $-x_5 a\,\hat{\mathbf{x}} + \frac{3}{4} b\,\hat{\mathbf{y}} - z_5 c\,\hat{\mathbf{z}}$ | (4$c$) | S I |
| $\mathbf{B}_{20}$ | = | $\left(\frac{1}{2} + x_5\right)\mathbf{a}_1 + \frac{1}{4}\,\mathbf{a}_2 + \left(\frac{1}{2} - z_5\right)\mathbf{a}_3$ | = | $\left(\frac{1}{2} + x_5\right) a\,\hat{\mathbf{x}} + \frac{1}{4} b\,\hat{\mathbf{y}} + \left(\frac{1}{2} - z_5\right) c\,\hat{\mathbf{z}}$ | (4$c$) | S I |
| $\mathbf{B}_{21}$ | = | $x_6\,\mathbf{a}_1 + \frac{1}{4}\,\mathbf{a}_2 + z_6\,\mathbf{a}_3$ | = | $x_6 a\,\hat{\mathbf{x}} + \frac{1}{4} b\,\hat{\mathbf{y}} + z_6 c\,\hat{\mathbf{z}}$ | (4$c$) | S II |
| $\mathbf{B}_{22}$ | = | $\left(\frac{1}{2} - x_6\right)\mathbf{a}_1 + \frac{3}{4}\,\mathbf{a}_2 + \left(\frac{1}{2} + z_6\right)\mathbf{a}_3$ | = | $\left(\frac{1}{2} - x_6\right) a\,\hat{\mathbf{x}} + \frac{3}{4} b\,\hat{\mathbf{y}} + \left(\frac{1}{2} + z_6\right) c\,\hat{\mathbf{z}}$ | (4$c$) | S II |
| $\mathbf{B}_{23}$ | = | $-x_6\,\mathbf{a}_1 + \frac{3}{4}\,\mathbf{a}_2 - z_6\,\mathbf{a}_3$ | = | $-x_6 a\,\hat{\mathbf{x}} + \frac{3}{4} b\,\hat{\mathbf{y}} - z_6 c\,\hat{\mathbf{z}}$ | (4$c$) | S II |
| $\mathbf{B}_{24}$ | = | $\left(\frac{1}{2} + x_6\right)\mathbf{a}_1 + \frac{1}{4}\,\mathbf{a}_2 + \left(\frac{1}{2} - z_6\right)\mathbf{a}_3$ | = | $\left(\frac{1}{2} + x_6\right) a\,\hat{\mathbf{x}} + \frac{1}{4} b\,\hat{\mathbf{y}} + \left(\frac{1}{2} - z_6\right) c\,\hat{\mathbf{z}}$ | (4$c$) | S II |
| $\mathbf{B}_{25}$ | = | $x_7\,\mathbf{a}_1 + \frac{1}{4}\,\mathbf{a}_2 + z_7\,\mathbf{a}_3$ | = | $x_7 a\,\hat{\mathbf{x}} + \frac{1}{4} b\,\hat{\mathbf{y}} + z_7 c\,\hat{\mathbf{z}}$ | (4$c$) | S III |
| $\mathbf{B}_{26}$ | = | $\left(\frac{1}{2} - x_7\right)\mathbf{a}_1 + \frac{3}{4}\,\mathbf{a}_2 + \left(\frac{1}{2} + z_7\right)\mathbf{a}_3$ | = | $\left(\frac{1}{2} - x_7\right) a\,\hat{\mathbf{x}} + \frac{3}{4} b\,\hat{\mathbf{y}} + \left(\frac{1}{2} + z_7\right) c\,\hat{\mathbf{z}}$ | (4$c$) | S III |
| $\mathbf{B}_{27}$ | = | $-x_7\,\mathbf{a}_1 + \frac{3}{4}\,\mathbf{a}_2 - z_7\,\mathbf{a}_3$ | = | $-x_7 a\,\hat{\mathbf{x}} + \frac{3}{4} b\,\hat{\mathbf{y}} - z_7 c\,\hat{\mathbf{z}}$ | (4$c$) | S III |
| $\mathbf{B}_{28}$ | = | $\left(\frac{1}{2} + x_7\right)\mathbf{a}_1 + \frac{1}{4}\,\mathbf{a}_2 + \left(\frac{1}{2} - z_7\right)\mathbf{a}_3$ | = | $\left(\frac{1}{2} + x_7\right) a\,\hat{\mathbf{x}} + \frac{1}{4} b\,\hat{\mathbf{y}} + \left(\frac{1}{2} - z_7\right) c\,\hat{\mathbf{z}}$ | (4$c$) | S III |
| $\mathbf{B}_{29}$ | = | $x_8\,\mathbf{a}_1 + y_8\,\mathbf{a}_2 + z_8\,\mathbf{a}_3$ | = | $x_8 a\,\hat{\mathbf{x}} + y_8 b\,\hat{\mathbf{y}} + z_8 c\,\hat{\mathbf{z}}$ | (8$d$) | O III |
| $\mathbf{B}_{30}$ | = | $\left(\frac{1}{2} - x_8\right)\mathbf{a}_1 - y_8\,\mathbf{a}_2 + \left(\frac{1}{2} + z_8\right)\mathbf{a}_3$ | = | $\left(\frac{1}{2} - x_8\right) a\,\hat{\mathbf{x}} - y_8 b\,\hat{\mathbf{y}} + \left(\frac{1}{2} + z_8\right) c\,\hat{\mathbf{z}}$ | (8$d$) | O III |
| $\mathbf{B}_{31}$ | = | $-x_8\,\mathbf{a}_1 + \left(\frac{1}{2} + y_8\right)\mathbf{a}_2 - z_8\,\mathbf{a}_3$ | = | $-x_8 a\,\hat{\mathbf{x}} + \left(\frac{1}{2} + y_8\right) b\,\hat{\mathbf{y}} - z_8 c\,\hat{\mathbf{z}}$ | (8$d$) | O III |
| $\mathbf{B}_{32}$ | = | $\left(\frac{1}{2} + x_8\right)\mathbf{a}_1 + \left(\frac{1}{2} - y_8\right)\mathbf{a}_2 + \left(\frac{1}{2} - z_8\right)\mathbf{a}_3$ | = | $\left(\frac{1}{2} + x_8\right) a\,\hat{\mathbf{x}} + \left(\frac{1}{2} - y_8\right) b\,\hat{\mathbf{y}} + \left(\frac{1}{2} - z_8\right) c\,\hat{\mathbf{z}}$ | (8$d$) | O III |
| $\mathbf{B}_{33}$ | = | $-x_8\,\mathbf{a}_1 - y_8\,\mathbf{a}_2 - z_8\,\mathbf{a}_3$ | = | $-x_8 a\,\hat{\mathbf{x}} - y_8 b\,\hat{\mathbf{y}} - z_8 c\,\hat{\mathbf{z}}$ | (8$d$) | O III |
| $\mathbf{B}_{34}$ | = | $\left(\frac{1}{2} + x_8\right)\mathbf{a}_1 + y_8\,\mathbf{a}_2 + \left(\frac{1}{2} - z_8\right)\mathbf{a}_3$ | = | $\left(\frac{1}{2} + x_8\right) a\,\hat{\mathbf{x}} + y_8 b\,\hat{\mathbf{y}} + \left(\frac{1}{2} - z_8\right) c\,\hat{\mathbf{z}}$ | (8$d$) | O III |
| $\mathbf{B}_{35}$ | = | $x_8\,\mathbf{a}_1 + \left(\frac{1}{2} - y_8\right)\mathbf{a}_2 + z_8\,\mathbf{a}_3$ | = | $x_8 a\,\hat{\mathbf{x}} + \left(\frac{1}{2} - y_8\right) b\,\hat{\mathbf{y}} + z_8 c\,\hat{\mathbf{z}}$ | (8$d$) | O III |
| $\mathbf{B}_{36}$ | = | $\left(\frac{1}{2} - x_8\right)\mathbf{a}_1 + \left(\frac{1}{2} + y_8\right)\mathbf{a}_2 + \left(\frac{1}{2} + z_8\right)\mathbf{a}_3$ | = | $\left(\frac{1}{2} - x_8\right)a\,\hat{\mathbf{x}} + \left(\frac{1}{2} + y_8\right)b\,\hat{\mathbf{y}} + \left(\frac{1}{2} + z_8\right)c\,\hat{\mathbf{z}}$ | (8$d$) | O III |
| $\mathbf{B}_{37}$ | = | $x_9\,\mathbf{a}_1 + y_9\,\mathbf{a}_2 + z_9\,\mathbf{a}_3$ | = | $x_9a\,\hat{\mathbf{x}} + y_9b\,\hat{\mathbf{y}} + z_9c\,\hat{\mathbf{z}}$ | (8$d$) | O IV |
| $\mathbf{B}_{38}$ | = | $\left(\frac{1}{2} - x_9\right)\mathbf{a}_1 - y_9\,\mathbf{a}_2 + \left(\frac{1}{2} + z_9\right)\mathbf{a}_3$ | = | $\left(\frac{1}{2} - x_9\right)a\,\hat{\mathbf{x}} - y_9b\,\hat{\mathbf{y}} + \left(\frac{1}{2} + z_9\right)c\,\hat{\mathbf{z}}$ | (8$d$) | O IV |
| $\mathbf{B}_{39}$ | = | $-x_9\,\mathbf{a}_1 + \left(\frac{1}{2} + y_9\right)\mathbf{a}_2 - z_9\,\mathbf{a}_3$ | = | $-x_9a\,\hat{\mathbf{x}} + \left(\frac{1}{2} + y_9\right)b\,\hat{\mathbf{y}} - z_9c\,\hat{\mathbf{z}}$ | (8$d$) | O IV |
| $\mathbf{B}_{40}$ | = | $\left(\frac{1}{2} + x_9\right)\mathbf{a}_1 + \left(\frac{1}{2} - y_9\right)\mathbf{a}_2 + \left(\frac{1}{2} - z_9\right)\mathbf{a}_3$ | = | $\left(\frac{1}{2} + x_9\right)a\,\hat{\mathbf{x}} + \left(\frac{1}{2} - y_9\right)b\,\hat{\mathbf{y}} + \left(\frac{1}{2} - z_9\right)c\,\hat{\mathbf{z}}$ | (8$d$) | O IV |
| $\mathbf{B}_{41}$ | = | $-x_9\,\mathbf{a}_1 - y_9\,\mathbf{a}_2 - z_9\,\mathbf{a}_3$ | = | $-x_9a\,\hat{\mathbf{x}} - y_9b\,\hat{\mathbf{y}} - z_9c\,\hat{\mathbf{z}}$ | (8$d$) | O IV |
| $\mathbf{B}_{42}$ | = | $\left(\frac{1}{2} + x_9\right)\mathbf{a}_1 + y_9\,\mathbf{a}_2 + \left(\frac{1}{2} - z_9\right)\mathbf{a}_3$ | = | $\left(\frac{1}{2} + x_9\right)a\,\hat{\mathbf{x}} + y_9b\,\hat{\mathbf{y}} + \left(\frac{1}{2} - z_9\right)c\,\hat{\mathbf{z}}$ | (8$d$) | O IV |
| $\mathbf{B}_{43}$ | = | $x_9\,\mathbf{a}_1 + \left(\frac{1}{2} - y_9\right)\mathbf{a}_2 + z_9\,\mathbf{a}_3$ | = | $x_9a\,\hat{\mathbf{x}} + \left(\frac{1}{2} - y_9\right)b\,\hat{\mathbf{y}} + z_9c\,\hat{\mathbf{z}}$ | (8$d$) | O IV |
| $\mathbf{B}_{44}$ | = | $\left(\frac{1}{2} - x_9\right)\mathbf{a}_1 + \left(\frac{1}{2} + y_9\right)\mathbf{a}_2 + \left(\frac{1}{2} + z_9\right)\mathbf{a}_3$ | = | $\left(\frac{1}{2} - x_9\right)a\,\hat{\mathbf{x}} + \left(\frac{1}{2} + y_9\right)b\,\hat{\mathbf{y}} + \left(\frac{1}{2} + z_9\right)c\,\hat{\mathbf{z}}$ | (8$d$) | O IV |

---

---

**Geometry files:**

- CIF: pp. 1633
- POSCAR: pp. 1633

# Danburite ($CaB_2Si_2O_8$, $S6_3$) Structure: A2BC8D2_oP52_62_d_c_2c3d_d

http://aflow.org/prototype-encyclopedia/A2BC8D2_oP52_62_d_c_2c3d_d

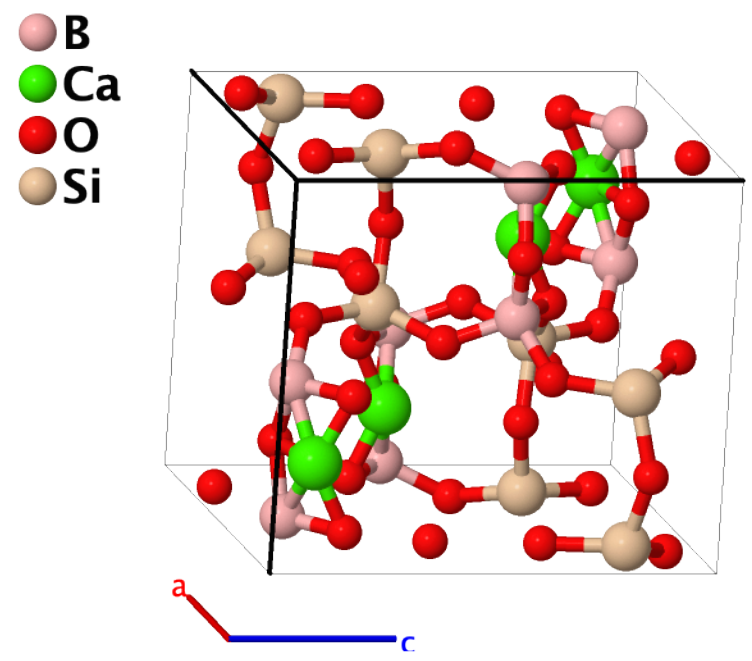

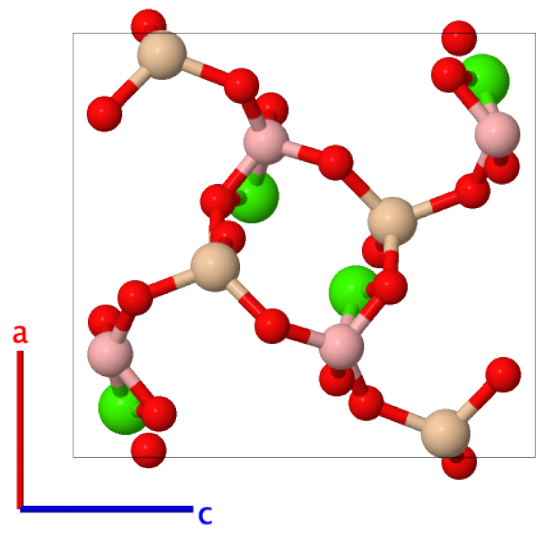

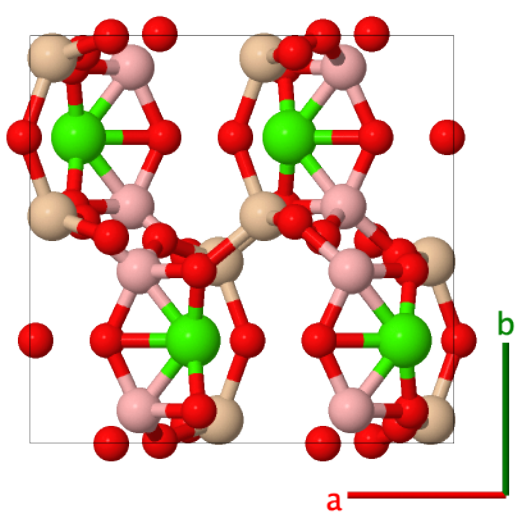

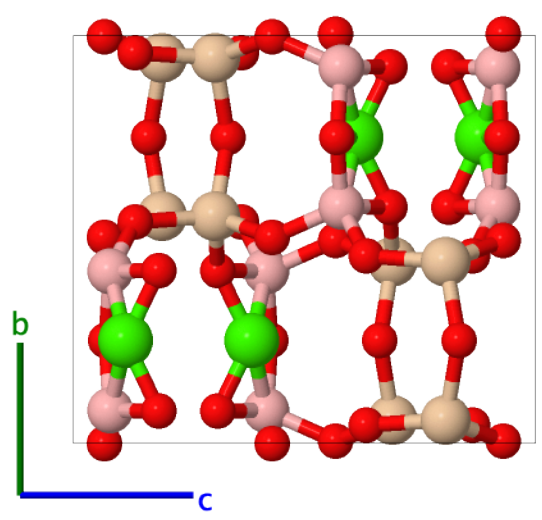

| | | |
|---|---|---|
| **Prototype** | : | $B_2CaO_8Si_2$ |
| **AFLOW prototype label** | : | A2BC8D2_oP52_62_d_c_2c3d_d |
| ***Strukturbericht* designation** | : | $S6_3$ |
| **Pearson symbol** | : | oP52 |
| **Space group number** | : | 62 |
| **Space group symbol** | : | *Pnma* |
| **AFLOW prototype command** | : | `aflow --proto=A2BC8D2_oP52_62_d_c_2c3d_d --params=`$a, b/a, c/a, x_1, z_1, x_2, z_2, x_3, z_3, x_4, y_4, z_4, x_5, y_5, z_5, x_6, y_6, z_6, x_7, y_7, z_7, x_8, y_8, z_8$ |

- We use the 25 °C data from (Sugiyama, 1985). They presented the structure in the *Pnam* orientation of space group #62. We used FINDSYM to transform this to the standard *Pnma* orientation.

**Simple Orthorhombic primitive vectors:**

$$\mathbf{a}_1 = a\,\hat{\mathbf{x}}$$
$$\mathbf{a}_2 = b\,\hat{\mathbf{y}}$$
$$\mathbf{a}_3 = c\,\hat{\mathbf{z}}$$

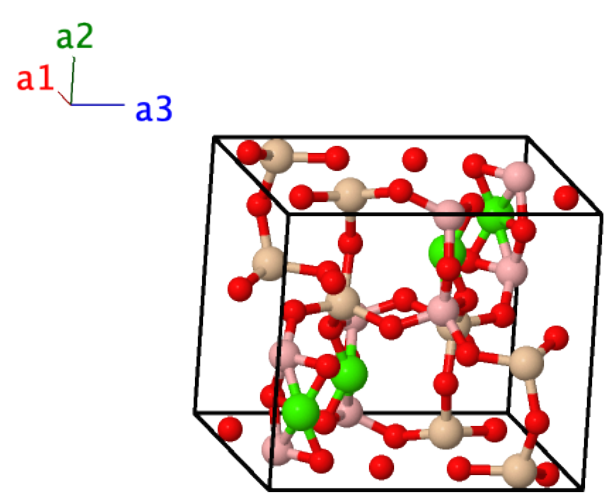

**Basis vectors:**

| | | Lattice Coordinates | | Cartesian Coordinates | Wyckoff Position | Atom Type |
|---|---|---|---|---|---|---|
| $\mathbf{B}_{1}$ | = | $x_{1}\,\mathbf{a}_{1}+\frac{1}{4}\,\mathbf{a}_{2}+z_{1}\,\mathbf{a}_{3}$ | = | $x_{1}a\,\mathbf{\hat{x}}+\frac{1}{4}b\,\mathbf{\hat{y}}+z_{1}c\,\mathbf{\hat{z}}$ | (4$c$) | Ca |
| $\mathbf{B}_{2}$ | = | $\left(\frac{1}{2}-x_{1}\right)\mathbf{a}_{1}+\frac{3}{4}\,\mathbf{a}_{2}+\left(\frac{1}{2}+z_{1}\right)\mathbf{a}_{3}$ | = | $\left(\frac{1}{2}-x_{1}\right)a\,\mathbf{\hat{x}}+\frac{3}{4}b\,\mathbf{\hat{y}}+\left(\frac{1}{2}+z_{1}\right)c\,\mathbf{\hat{z}}$ | (4$c$) | Ca |
| $\mathbf{B}_{3}$ | = | $-x_{1}\,\mathbf{a}_{1}+\frac{3}{4}\,\mathbf{a}_{2}-z_{1}\,\mathbf{a}_{3}$ | = | $-x_{1}a\,\mathbf{\hat{x}}+\frac{3}{4}b\,\mathbf{\hat{y}}-z_{1}c\,\mathbf{\hat{z}}$ | (4$c$) | Ca |
| $\mathbf{B}_{4}$ | = | $\left(\frac{1}{2}+x_{1}\right)\mathbf{a}_{1}+\frac{1}{4}\,\mathbf{a}_{2}+\left(\frac{1}{2}-z_{1}\right)\mathbf{a}_{3}$ | = | $\left(\frac{1}{2}+x_{1}\right)a\,\mathbf{\hat{x}}+\frac{1}{4}b\,\mathbf{\hat{y}}+\left(\frac{1}{2}-z_{1}\right)c\,\mathbf{\hat{z}}$ | (4$c$) | Ca |
| $\mathbf{B}_{5}$ | = | $x_{2}\,\mathbf{a}_{1}+\frac{1}{4}\,\mathbf{a}_{2}+z_{2}\,\mathbf{a}_{3}$ | = | $x_{2}a\,\mathbf{\hat{x}}+\frac{1}{4}b\,\mathbf{\hat{y}}+z_{2}c\,\mathbf{\hat{z}}$ | (4$c$) | O I |
| $\mathbf{B}_{6}$ | = | $\left(\frac{1}{2}-x_{2}\right)\mathbf{a}_{1}+\frac{3}{4}\,\mathbf{a}_{2}+\left(\frac{1}{2}+z_{2}\right)\mathbf{a}_{3}$ | = | $\left(\frac{1}{2}-x_{2}\right)a\,\mathbf{\hat{x}}+\frac{3}{4}b\,\mathbf{\hat{y}}+\left(\frac{1}{2}+z_{2}\right)c\,\mathbf{\hat{z}}$ | (4$c$) | O I |
| $\mathbf{B}_{7}$ | = | $-x_{2}\,\mathbf{a}_{1}+\frac{3}{4}\,\mathbf{a}_{2}-z_{2}\,\mathbf{a}_{3}$ | = | $-x_{2}a\,\mathbf{\hat{x}}+\frac{3}{4}b\,\mathbf{\hat{y}}-z_{2}c\,\mathbf{\hat{z}}$ | (4$c$) | O I |
| $\mathbf{B}_{8}$ | = | $\left(\frac{1}{2}+x_{2}\right)\mathbf{a}_{1}+\frac{1}{4}\,\mathbf{a}_{2}+\left(\frac{1}{2}-z_{2}\right)\mathbf{a}_{3}$ | = | $\left(\frac{1}{2}+x_{2}\right)a\,\mathbf{\hat{x}}+\frac{1}{4}b\,\mathbf{\hat{y}}+\left(\frac{1}{2}-z_{2}\right)c\,\mathbf{\hat{z}}$ | (4$c$) | O I |
| $\mathbf{B}_{9}$ | = | $x_{3}\,\mathbf{a}_{1}+\frac{1}{4}\,\mathbf{a}_{2}+z_{3}\,\mathbf{a}_{3}$ | = | $x_{3}a\,\mathbf{\hat{x}}+\frac{1}{4}b\,\mathbf{\hat{y}}+z_{3}c\,\mathbf{\hat{z}}$ | (4$c$) | O II |
| $\mathbf{B}_{10}$ | = | $\left(\frac{1}{2}-x_{3}\right)\mathbf{a}_{1}+\frac{3}{4}\,\mathbf{a}_{2}+\left(\frac{1}{2}+z_{3}\right)\mathbf{a}_{3}$ | = | $\left(\frac{1}{2}-x_{3}\right)a\,\mathbf{\hat{x}}+\frac{3}{4}b\,\mathbf{\hat{y}}+\left(\frac{1}{2}+z_{3}\right)c\,\mathbf{\hat{z}}$ | (4$c$) | O II |
| $\mathbf{B}_{11}$ | = | $-x_{3}\,\mathbf{a}_{1}+\frac{3}{4}\,\mathbf{a}_{2}-z_{3}\,\mathbf{a}_{3}$ | = | $-x_{3}a\,\mathbf{\hat{x}}+\frac{3}{4}b\,\mathbf{\hat{y}}-z_{3}c\,\mathbf{\hat{z}}$ | (4$c$) | O II |
| $\mathbf{B}_{12}$ | = | $\left(\frac{1}{2}+x_{3}\right)\mathbf{a}_{1}+\frac{1}{4}\,\mathbf{a}_{2}+\left(\frac{1}{2}-z_{3}\right)\mathbf{a}_{3}$ | = | $\left(\frac{1}{2}+x_{3}\right)a\,\mathbf{\hat{x}}+\frac{1}{4}b\,\mathbf{\hat{y}}+\left(\frac{1}{2}-z_{3}\right)c\,\mathbf{\hat{z}}$ | (4$c$) | O II |
| $\mathbf{B}_{13}$ | = | $x_{4}\,\mathbf{a}_{1}+y_{4}\,\mathbf{a}_{2}+z_{4}\,\mathbf{a}_{3}$ | = | $x_{4}a\,\mathbf{\hat{x}}+y_{4}b\,\mathbf{\hat{y}}+z_{4}c\,\mathbf{\hat{z}}$ | (8$d$) | B |
| $\mathbf{B}_{14}$ | = | $\left(\frac{1}{2}-x_{4}\right)\mathbf{a}_{1}-y_{4}\,\mathbf{a}_{2}+\left(\frac{1}{2}+z_{4}\right)\mathbf{a}_{3}$ | = | $\left(\frac{1}{2}-x_{4}\right)a\,\mathbf{\hat{x}}-y_{4}b\,\mathbf{\hat{y}}+\left(\frac{1}{2}+z_{4}\right)c\,\mathbf{\hat{z}}$ | (8$d$) | B |
| $\mathbf{B}_{15}$ | = | $-x_{4}\,\mathbf{a}_{1}+\left(\frac{1}{2}+y_{4}\right)\mathbf{a}_{2}-z_{4}\,\mathbf{a}_{3}$ | = | $-x_{4}a\,\mathbf{\hat{x}}+\left(\frac{1}{2}+y_{4}\right)b\,\mathbf{\hat{y}}-z_{4}c\,\mathbf{\hat{z}}$ | (8$d$) | B |
| $\mathbf{B}_{16}$ | = | $\left(\frac{1}{2}+x_{4}\right)\mathbf{a}_{1}+\left(\frac{1}{2}-y_{4}\right)\mathbf{a}_{2}+\left(\frac{1}{2}-z_{4}\right)\mathbf{a}_{3}$ | = | $\left(\frac{1}{2}+x_{4}\right)a\,\mathbf{\hat{x}}+\left(\frac{1}{2}-y_{4}\right)b\,\mathbf{\hat{y}}+\left(\frac{1}{2}-z_{4}\right)c\,\mathbf{\hat{z}}$ | (8$d$) | B |
| $\mathbf{B}_{17}$ | = | $-x_{4}\,\mathbf{a}_{1}-y_{4}\,\mathbf{a}_{2}-z_{4}\,\mathbf{a}_{3}$ | = | $-x_{4}a\,\mathbf{\hat{x}}-y_{4}b\,\mathbf{\hat{y}}-z_{4}c\,\mathbf{\hat{z}}$ | (8$d$) | B |
| $\mathbf{B}_{18}$ | = | $\left(\frac{1}{2}+x_{4}\right)\mathbf{a}_{1}+y_{4}\,\mathbf{a}_{2}+\left(\frac{1}{2}-z_{4}\right)\mathbf{a}_{3}$ | = | $\left(\frac{1}{2}+x_{4}\right)a\,\mathbf{\hat{x}}+y_{4}b\,\mathbf{\hat{y}}+\left(\frac{1}{2}-z_{4}\right)c\,\mathbf{\hat{z}}$ | (8$d$) | B |
| $\mathbf{B}_{19}$ | = | $x_{4}\,\mathbf{a}_{1}+\left(\frac{1}{2}-y_{4}\right)\mathbf{a}_{2}+z_{4}\,\mathbf{a}_{3}$ | = | $x_{4}a\,\mathbf{\hat{x}}+\left(\frac{1}{2}-y_{4}\right)b\,\mathbf{\hat{y}}+z_{4}c\,\mathbf{\hat{z}}$ | (8$d$) | B |
| $\mathbf{B}_{20}$ | = | $\left(\frac{1}{2}-x_{4}\right)\mathbf{a}_{1}+\left(\frac{1}{2}+y_{4}\right)\mathbf{a}_{2}+\left(\frac{1}{2}+z_{4}\right)\mathbf{a}_{3}$ | = | $\left(\frac{1}{2}-x_{4}\right)a\,\mathbf{\hat{x}}+\left(\frac{1}{2}+y_{4}\right)b\,\mathbf{\hat{y}}+\left(\frac{1}{2}+z_{4}\right)c\,\mathbf{\hat{z}}$ | (8$d$) | B |
| $\mathbf{B}_{21}$ | = | $x_{5}\,\mathbf{a}_{1}+y_{5}\,\mathbf{a}_{2}+z_{5}\,\mathbf{a}_{3}$ | = | $x_{5}a\,\mathbf{\hat{x}}+y_{5}b\,\mathbf{\hat{y}}+z_{5}c\,\mathbf{\hat{z}}$ | (8$d$) | O III |
| $\mathbf{B}_{22}$ | = | $\left(\frac{1}{2}-x_{5}\right)\mathbf{a}_{1}-y_{5}\,\mathbf{a}_{2}+\left(\frac{1}{2}+z_{5}\right)\mathbf{a}_{3}$ | = | $\left(\frac{1}{2}-x_{5}\right)a\,\mathbf{\hat{x}}-y_{5}b\,\mathbf{\hat{y}}+\left(\frac{1}{2}+z_{5}\right)c\,\mathbf{\hat{z}}$ | (8$d$) | O III |
| $\mathbf{B}_{23}$ | = | $-x_{5}\,\mathbf{a}_{1}+\left(\frac{1}{2}+y_{5}\right)\mathbf{a}_{2}-z_{5}\,\mathbf{a}_{3}$ | = | $-x_{5}a\,\mathbf{\hat{x}}+\left(\frac{1}{2}+y_{5}\right)b\,\mathbf{\hat{y}}-z_{5}c\,\mathbf{\hat{z}}$ | (8$d$) | O III |
| $\mathbf{B}_{24}$ | = | $\left(\frac{1}{2}+x_{5}\right)\mathbf{a}_{1}+\left(\frac{1}{2}-y_{5}\right)\mathbf{a}_{2}+\left(\frac{1}{2}-z_{5}\right)\mathbf{a}_{3}$ | = | $\left(\frac{1}{2}+x_{5}\right)a\,\mathbf{\hat{x}}+\left(\frac{1}{2}-y_{5}\right)b\,\mathbf{\hat{y}}+\left(\frac{1}{2}-z_{5}\right)c\,\mathbf{\hat{z}}$ | (8$d$) | O III |
| $\mathbf{B}_{25}$ | = | $-x_{5}\,\mathbf{a}_{1}-y_{5}\,\mathbf{a}_{2}-z_{5}\,\mathbf{a}_{3}$ | = | $-x_{5}a\,\mathbf{\hat{x}}-y_{5}b\,\mathbf{\hat{y}}-z_{5}c\,\mathbf{\hat{z}}$ | (8$d$) | O III |
| $\mathbf{B}_{26}$ | = | $\left(\frac{1}{2}+x_{5}\right)\mathbf{a}_{1}+y_{5}\,\mathbf{a}_{2}+\left(\frac{1}{2}-z_{5}\right)\mathbf{a}_{3}$ | = | $\left(\frac{1}{2}+x_{5}\right)a\,\mathbf{\hat{x}}+y_{5}b\,\mathbf{\hat{y}}+\left(\frac{1}{2}-z_{5}\right)c\,\mathbf{\hat{z}}$ | (8$d$) | O III |
| $\mathbf{B}_{27}$ | = | $x_{5}\,\mathbf{a}_{1}+\left(\frac{1}{2}-y_{5}\right)\mathbf{a}_{2}+z_{5}\,\mathbf{a}_{3}$ | = | $x_{5}a\,\mathbf{\hat{x}}+\left(\frac{1}{2}-y_{5}\right)b\,\mathbf{\hat{y}}+z_{5}c\,\mathbf{\hat{z}}$ | (8$d$) | O III |
| $\mathbf{B}_{28}$ | = | $\left(\frac{1}{2}-x_{5}\right)\mathbf{a}_{1}+\left(\frac{1}{2}+y_{5}\right)\mathbf{a}_{2}+\left(\frac{1}{2}+z_{5}\right)\mathbf{a}_{3}$ | = | $\left(\frac{1}{2}-x_{5}\right)a\,\mathbf{\hat{x}}+\left(\frac{1}{2}+y_{5}\right)b\,\mathbf{\hat{y}}+\left(\frac{1}{2}+z_{5}\right)c\,\mathbf{\hat{z}}$ | (8$d$) | O III |
| $\mathbf{B}_{29}$ | = | $x_{6}\,\mathbf{a}_{1}+y_{6}\,\mathbf{a}_{2}+z_{6}\,\mathbf{a}_{3}$ | = | $x_{6}a\,\mathbf{\hat{x}}+y_{6}b\,\mathbf{\hat{y}}+z_{6}c\,\mathbf{\hat{z}}$ | (8$d$) | O IV |
| $\mathbf{B}_{30}$ | = | $\left(\frac{1}{2}-x_{6}\right)\mathbf{a}_{1}-y_{6}\,\mathbf{a}_{2}+\left(\frac{1}{2}+z_{6}\right)\mathbf{a}_{3}$ | = | $\left(\frac{1}{2}-x_{6}\right)a\,\mathbf{\hat{x}}-y_{6}b\,\mathbf{\hat{y}}+\left(\frac{1}{2}+z_{6}\right)c\,\mathbf{\hat{z}}$ | (8$d$) | O IV |
| $\mathbf{B}_{31}$ | = | $-x_{6}\,\mathbf{a}_{1}+\left(\frac{1}{2}+y_{6}\right)\mathbf{a}_{2}-z_{6}\,\mathbf{a}_{3}$ | = | $-x_{6}a\,\mathbf{\hat{x}}+\left(\frac{1}{2}+y_{6}\right)b\,\mathbf{\hat{y}}-z_{6}c\,\mathbf{\hat{z}}$ | (8$d$) | O IV |
| $\mathbf{B}_{32}$ | = | $\left(\frac{1}{2}+x_6\right)\mathbf{a}_1+\left(\frac{1}{2}-y_6\right)\mathbf{a}_2+\left(\frac{1}{2}-z_6\right)\mathbf{a}_3$ | = | $\left(\frac{1}{2}+x_6\right)a\,\hat{\mathbf{x}}+\left(\frac{1}{2}-y_6\right)b\,\hat{\mathbf{y}}+\left(\frac{1}{2}-z_6\right)c\,\hat{\mathbf{z}}$ | (8*d*) | O IV |
| $\mathbf{B}_{33}$ | = | $-x_6\,\mathbf{a}_1-y_6\,\mathbf{a}_2-z_6\,\mathbf{a}_3$ | = | $-x_6a\,\hat{\mathbf{x}}-y_6b\,\hat{\mathbf{y}}-z_6c\,\hat{\mathbf{z}}$ | (8*d*) | O IV |
| $\mathbf{B}_{34}$ | = | $\left(\frac{1}{2}+x_6\right)\mathbf{a}_1+y_6\,\mathbf{a}_2+\left(\frac{1}{2}-z_6\right)\mathbf{a}_3$ | = | $\left(\frac{1}{2}+x_6\right)a\,\hat{\mathbf{x}}+y_6b\,\hat{\mathbf{y}}+\left(\frac{1}{2}-z_6\right)c\,\hat{\mathbf{z}}$ | (8*d*) | O IV |
| $\mathbf{B}_{35}$ | = | $x_6\,\mathbf{a}_1+\left(\frac{1}{2}-y_6\right)\mathbf{a}_2+z_6\,\mathbf{a}_3$ | = | $x_6a\,\hat{\mathbf{x}}+\left(\frac{1}{2}-y_6\right)b\,\hat{\mathbf{y}}+z_6c\,\hat{\mathbf{z}}$ | (8*d*) | O IV |
| $\mathbf{B}_{36}$ | = | $\left(\frac{1}{2}-x_6\right)\mathbf{a}_1+\left(\frac{1}{2}+y_6\right)\mathbf{a}_2+\left(\frac{1}{2}+z_6\right)\mathbf{a}_3$ | = | $\left(\frac{1}{2}-x_6\right)a\,\hat{\mathbf{x}}+\left(\frac{1}{2}+y_6\right)b\,\hat{\mathbf{y}}+\left(\frac{1}{2}+z_6\right)c\,\hat{\mathbf{z}}$ | (8*d*) | O IV |
| $\mathbf{B}_{37}$ | = | $x_7\,\mathbf{a}_1+y_7\,\mathbf{a}_2+z_7\,\mathbf{a}_3$ | = | $x_7a\,\hat{\mathbf{x}}+y_7b\,\hat{\mathbf{y}}+z_7c\,\hat{\mathbf{z}}$ | (8*d*) | O V |
| $\mathbf{B}_{38}$ | = | $\left(\frac{1}{2}-x_7\right)\mathbf{a}_1-y_7\,\mathbf{a}_2+\left(\frac{1}{2}+z_7\right)\mathbf{a}_3$ | = | $\left(\frac{1}{2}-x_7\right)a\,\hat{\mathbf{x}}-y_7b\,\hat{\mathbf{y}}+\left(\frac{1}{2}+z_7\right)c\,\hat{\mathbf{z}}$ | (8*d*) | O V |
| $\mathbf{B}_{39}$ | = | $-x_7\,\mathbf{a}_1+\left(\frac{1}{2}+y_7\right)\mathbf{a}_2-z_7\,\mathbf{a}_3$ | = | $-x_7a\,\hat{\mathbf{x}}+\left(\frac{1}{2}+y_7\right)b\,\hat{\mathbf{y}}-z_7c\,\hat{\mathbf{z}}$ | (8*d*) | O V |
| $\mathbf{B}_{40}$ | = | $\left(\frac{1}{2}+x_7\right)\mathbf{a}_1+\left(\frac{1}{2}-y_7\right)\mathbf{a}_2+\left(\frac{1}{2}-z_7\right)\mathbf{a}_3$ | = | $\left(\frac{1}{2}+x_7\right)a\,\hat{\mathbf{x}}+\left(\frac{1}{2}-y_7\right)b\,\hat{\mathbf{y}}+\left(\frac{1}{2}-z_7\right)c\,\hat{\mathbf{z}}$ | (8*d*) | O V |
| $\mathbf{B}_{41}$ | = | $-x_7\,\mathbf{a}_1-y_7\,\mathbf{a}_2-z_7\,\mathbf{a}_3$ | = | $-x_7a\,\hat{\mathbf{x}}-y_7b\,\hat{\mathbf{y}}-z_7c\,\hat{\mathbf{z}}$ | (8*d*) | O V |
| $\mathbf{B}_{42}$ | = | $\left(\frac{1}{2}+x_7\right)\mathbf{a}_1+y_7\,\mathbf{a}_2+\left(\frac{1}{2}-z_7\right)\mathbf{a}_3$ | = | $\left(\frac{1}{2}+x_7\right)a\,\hat{\mathbf{x}}+y_7b\,\hat{\mathbf{y}}+\left(\frac{1}{2}-z_7\right)c\,\hat{\mathbf{z}}$ | (8*d*) | O V |
| $\mathbf{B}_{43}$ | = | $x_7\,\mathbf{a}_1+\left(\frac{1}{2}-y_7\right)\mathbf{a}_2+z_7\,\mathbf{a}_3$ | = | $x_7a\,\hat{\mathbf{x}}+\left(\frac{1}{2}-y_7\right)b\,\hat{\mathbf{y}}+z_7c\,\hat{\mathbf{z}}$ | (8*d*) | O V |
| $\mathbf{B}_{44}$ | = | $\left(\frac{1}{2}-x_7\right)\mathbf{a}_1+\left(\frac{1}{2}+y_7\right)\mathbf{a}_2+\left(\frac{1}{2}+z_7\right)\mathbf{a}_3$ | = | $\left(\frac{1}{2}-x_7\right)a\,\hat{\mathbf{x}}+\left(\frac{1}{2}+y_7\right)b\,\hat{\mathbf{y}}+\left(\frac{1}{2}+z_7\right)c\,\hat{\mathbf{z}}$ | (8*d*) | O V |
| $\mathbf{B}_{45}$ | = | $x_8\,\mathbf{a}_1+y_8\,\mathbf{a}_2+z_8\,\mathbf{a}_3$ | = | $x_8a\,\hat{\mathbf{x}}+y_8b\,\hat{\mathbf{y}}+z_8c\,\hat{\mathbf{z}}$ | (8*d*) | Si |
| $\mathbf{B}_{46}$ | = | $\left(\frac{1}{2}-x_8\right)\mathbf{a}_1-y_8\,\mathbf{a}_2+\left(\frac{1}{2}+z_8\right)\mathbf{a}_3$ | = | $\left(\frac{1}{2}-x_8\right)a\,\hat{\mathbf{x}}-y_8b\,\hat{\mathbf{y}}+\left(\frac{1}{2}+z_8\right)c\,\hat{\mathbf{z}}$ | (8*d*) | Si |
| $\mathbf{B}_{47}$ | = | $-x_8\,\mathbf{a}_1+\left(\frac{1}{2}+y_8\right)\mathbf{a}_2-z_8\,\mathbf{a}_3$ | = | $-x_8a\,\hat{\mathbf{x}}+\left(\frac{1}{2}+y_8\right)b\,\hat{\mathbf{y}}-z_8c\,\hat{\mathbf{z}}$ | (8*d*) | Si |
| $\mathbf{B}_{48}$ | = | $\left(\frac{1}{2}+x_8\right)\mathbf{a}_1+\left(\frac{1}{2}-y_8\right)\mathbf{a}_2+\left(\frac{1}{2}-z_8\right)\mathbf{a}_3$ | = | $\left(\frac{1}{2}+x_8\right)a\,\hat{\mathbf{x}}+\left(\frac{1}{2}-y_8\right)b\,\hat{\mathbf{y}}+\left(\frac{1}{2}-z_8\right)c\,\hat{\mathbf{z}}$ | (8*d*) | Si |
| $\mathbf{B}_{49}$ | = | $-x_8\,\mathbf{a}_1-y_8\,\mathbf{a}_2-z_8\,\mathbf{a}_3$ | = | $-x_8a\,\hat{\mathbf{x}}-y_8b\,\hat{\mathbf{y}}-z_8c\,\hat{\mathbf{z}}$ | (8*d*) | Si |
| $\mathbf{B}_{50}$ | = | $\left(\frac{1}{2}+x_8\right)\mathbf{a}_1+y_8\,\mathbf{a}_2+\left(\frac{1}{2}-z_8\right)\mathbf{a}_3$ | = | $\left(\frac{1}{2}+x_8\right)a\,\hat{\mathbf{x}}+y_8b\,\hat{\mathbf{y}}+\left(\frac{1}{2}-z_8\right)c\,\hat{\mathbf{z}}$ | (8*d*) | Si |
| $\mathbf{B}_{51}$ | = | $x_8\,\mathbf{a}_1+\left(\frac{1}{2}-y_8\right)\mathbf{a}_2+z_8\,\mathbf{a}_3$ | = | $x_8a\,\hat{\mathbf{x}}+\left(\frac{1}{2}-y_8\right)b\,\hat{\mathbf{y}}+z_8c\,\hat{\mathbf{z}}$ | (8*d*) | Si |
| $\mathbf{B}_{52}$ | = | $\left(\frac{1}{2}-x_8\right)\mathbf{a}_1+\left(\frac{1}{2}+y_8\right)\mathbf{a}_2+\left(\frac{1}{2}+z_8\right)\mathbf{a}_3$ | = | $\left(\frac{1}{2}-x_8\right)a\,\hat{\mathbf{x}}+\left(\frac{1}{2}+y_8\right)b\,\hat{\mathbf{y}}+\left(\frac{1}{2}+z_8\right)c\,\hat{\mathbf{z}}$ | (8*d*) | Si |

**Found in:**

**Geometry files:**

- CIF: pp. 1633
- POSCAR: pp. 1634

# $C53$ ($SrBr_2$) (*obsolete*) Structure: A2B_oP12_62_2c_c

http://aflow.org/prototype-encyclopedia/A2B_oP12_62_2c_c.SrBr2

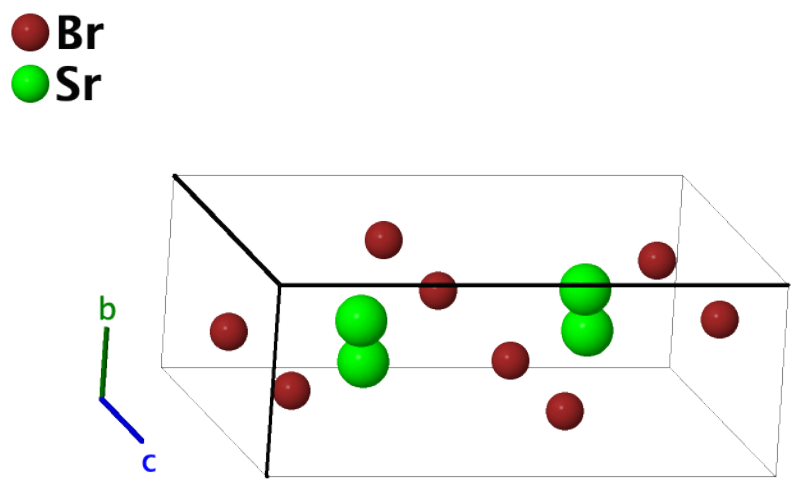

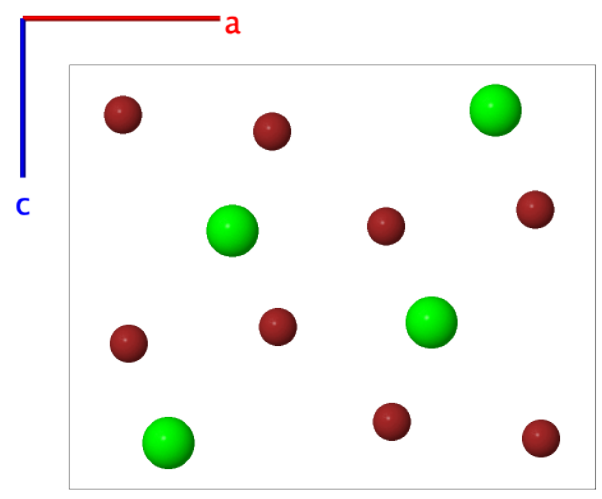

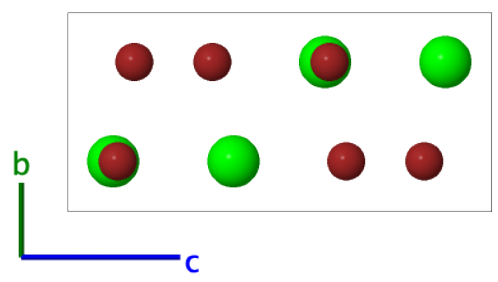

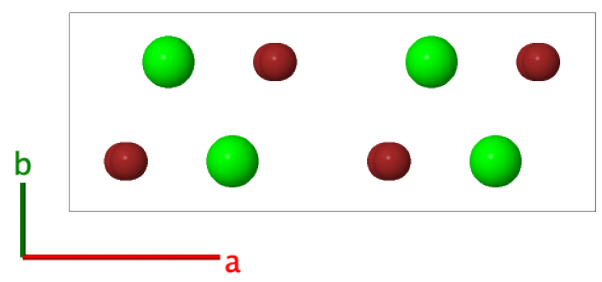

| | | |
|---|---|---|
| **Prototype** | : | $Br_2Sr$ |
| **AFLOW prototype label** | : | A2B_oP12_62_2c_c |
| ***Strukturbericht* designation** | : | $C53$ |
| **Pearson symbol** | : | oP12 |
| **Space group number** | : | 62 |
| **Space group symbol** | : | *Pnma* |
| **AFLOW prototype command** | : | `aflow --proto=A2B_oP12_62_2c_c` `--params=`$a, b/a, c/a, x_1, z_1, x_2, z_2, x_3, z_3$ |

- (Kamermans, 1939) determined that $SrBr_2$ had a distorted $PbCl_2$ structure, and assigned it to space group *Pnma* #62. (Hermann, 1939) gave this the *Strukturbericht* designation $C53$.
- Some years later, (Sass, 1963) showed that this structure was incorrect, and that $\alpha$-$SrBr_2$ was actually in space group $P4/n$ #85, rendering the $C53$ symbol obsolete. We include it here for historical interest.
- (Kamermans, 1939) gave the atomic positions in terms of the *Pbnm* setting of space group #62. We have used FINDSYM to translate this to the standard *Pnma* setting.
- Note that
  - $PbCl_2$ $C23$,
  - $HgCl_2$ $C25$,
  - $SrH_2$ $C29$,
  - $Co_2Si$ $C37$, and
  - $SrBr_2$ $C53$

all have the same AFLOW prototype label. They are generated by the same symmetry operations with different sets of parameters (`--params`) specified in their corresponding CIF files.

---

**Simple Orthorhombic primitive vectors:**

$$\mathbf{a}_1 = a\,\hat{\mathbf{x}}$$
$$\mathbf{a}_2 = b\,\hat{\mathbf{y}}$$
$$\mathbf{a}_3 = c\,\hat{\mathbf{z}}$$

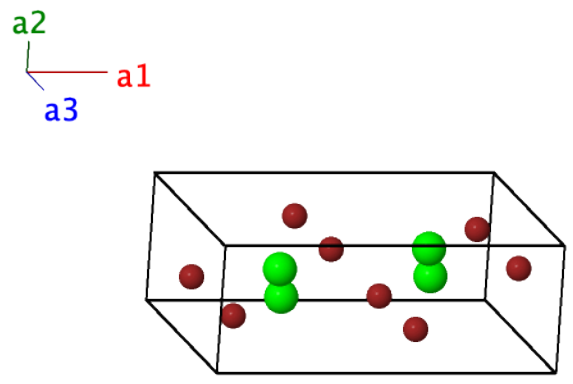

---

**Basis vectors:**

| | | Lattice Coordinates | | Cartesian Coordinates | Wyckoff Position | Atom Type |
|---|---|---|---|---|---|---|
| $\mathbf{B}_1$ | = | $x_1\,\mathbf{a}_1 + \frac{1}{4}\,\mathbf{a}_2 + z_1\,\mathbf{a}_3$ | = | $x_1 a\,\hat{\mathbf{x}} + \frac{1}{4} b\,\hat{\mathbf{y}} + z_1 c\,\hat{\mathbf{z}}$ | (4c) | Br I |
| $\mathbf{B}_2$ | = | $\left(\frac{1}{2} - x_1\right)\mathbf{a}_1 + \frac{3}{4}\,\mathbf{a}_2 + \left(\frac{1}{2} + z_1\right)\mathbf{a}_3$ | = | $\left(\frac{1}{2} - x_1\right) a\,\hat{\mathbf{x}} + \frac{3}{4} b\,\hat{\mathbf{y}} + \left(\frac{1}{2} + z_1\right) c\,\hat{\mathbf{z}}$ | (4c) | Br I |
| $\mathbf{B}_3$ | = | $-x_1\,\mathbf{a}_1 + \frac{3}{4}\,\mathbf{a}_2 - z_1\,\mathbf{a}_3$ | = | $-x_1 a\,\hat{\mathbf{x}} + \frac{3}{4} b\,\hat{\mathbf{y}} - z_1 c\,\hat{\mathbf{z}}$ | (4c) | Br I |
| $\mathbf{B}_4$ | = | $\left(\frac{1}{2} + x_1\right)\mathbf{a}_1 + \frac{1}{4}\,\mathbf{a}_2 + \left(\frac{1}{2} - z_1\right)\mathbf{a}_3$ | = | $\left(\frac{1}{2} + x_1\right) a\,\hat{\mathbf{x}} + \frac{1}{4} b\,\hat{\mathbf{y}} + \left(\frac{1}{2} - z_1\right) c\,\hat{\mathbf{z}}$ | (4c) | Br I |
| $\mathbf{B}_5$ | = | $x_2\,\mathbf{a}_1 + \frac{1}{4}\,\mathbf{a}_2 + z_2\,\mathbf{a}_3$ | = | $x_2 a\,\hat{\mathbf{x}} + \frac{1}{4} b\,\hat{\mathbf{y}} + z_2 c\,\hat{\mathbf{z}}$ | (4c) | Br II |
| $\mathbf{B}_6$ | = | $\left(\frac{1}{2} - x_2\right)\mathbf{a}_1 + \frac{3}{4}\,\mathbf{a}_2 + \left(\frac{1}{2} + z_2\right)\mathbf{a}_3$ | = | $\left(\frac{1}{2} - x_2\right) a\,\hat{\mathbf{x}} + \frac{3}{4} b\,\hat{\mathbf{y}} + \left(\frac{1}{2} + z_2\right) c\,\hat{\mathbf{z}}$ | (4c) | Br II |
| $\mathbf{B}_7$ | = | $-x_2\,\mathbf{a}_1 + \frac{3}{4}\,\mathbf{a}_2 - z_2\,\mathbf{a}_3$ | = | $-x_2 a\,\hat{\mathbf{x}} + \frac{3}{4} b\,\hat{\mathbf{y}} - z_2 c\,\hat{\mathbf{z}}$ | (4c) | Br II |
| $\mathbf{B}_8$ | = | $\left(\frac{1}{2} + x_2\right)\mathbf{a}_1 + \frac{1}{4}\,\mathbf{a}_2 + \left(\frac{1}{2} - z_2\right)\mathbf{a}_3$ | = | $\left(\frac{1}{2} + x_2\right) a\,\hat{\mathbf{x}} + \frac{1}{4} b\,\hat{\mathbf{y}} + \left(\frac{1}{2} - z_2\right) c\,\hat{\mathbf{z}}$ | (4c) | Br II |
| $\mathbf{B}_9$ | = | $x_3\,\mathbf{a}_1 + \frac{1}{4}\,\mathbf{a}_2 + z_3\,\mathbf{a}_3$ | = | $x_3 a\,\hat{\mathbf{x}} + \frac{1}{4} b\,\hat{\mathbf{y}} + z_3 c\,\hat{\mathbf{z}}$ | (4c) | Sr |
| $\mathbf{B}_{10}$ | = | $\left(\frac{1}{2} - x_3\right)\mathbf{a}_1 + \frac{3}{4}\,\mathbf{a}_2 + \left(\frac{1}{2} + z_3\right)\mathbf{a}_3$ | = | $\left(\frac{1}{2} - x_3\right) a\,\hat{\mathbf{x}} + \frac{3}{4} b\,\hat{\mathbf{y}} + \left(\frac{1}{2} + z_3\right) c\,\hat{\mathbf{z}}$ | (4c) | Sr |
| $\mathbf{B}_{11}$ | = | $-x_3\,\mathbf{a}_1 + \frac{3}{4}\,\mathbf{a}_2 - z_3\,\mathbf{a}_3$ | = | $-x_3 a\,\hat{\mathbf{x}} + \frac{3}{4} b\,\hat{\mathbf{y}} - z_3 c\,\hat{\mathbf{z}}$ | (4c) | Sr |
| $\mathbf{B}_{12}$ | = | $\left(\frac{1}{2} + x_3\right)\mathbf{a}_1 + \frac{1}{4}\,\mathbf{a}_2 + \left(\frac{1}{2} - z_3\right)\mathbf{a}_3$ | = | $\left(\frac{1}{2} + x_3\right) a\,\hat{\mathbf{x}} + \frac{1}{4} b\,\hat{\mathbf{y}} + \left(\frac{1}{2} - z_3\right) c\,\hat{\mathbf{z}}$ | (4c) | Sr |

---

**Found in:**

---

**Geometry files:**

- CIF: pp. 1634
- POSCAR: pp. 1635

# $Cs_2Sb$ Structure: A2B_oP24_62_4c_2c

http://aflow.org/prototype-encyclopedia/A2B_oP24_62_4c_2c

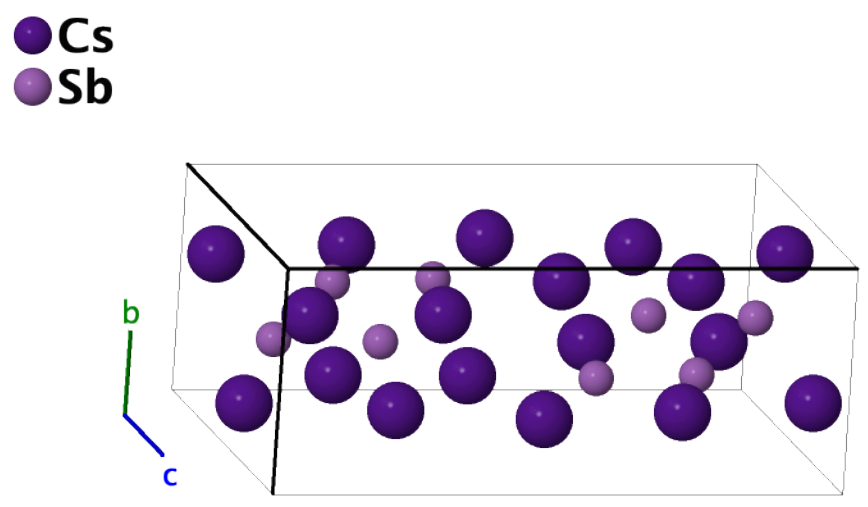

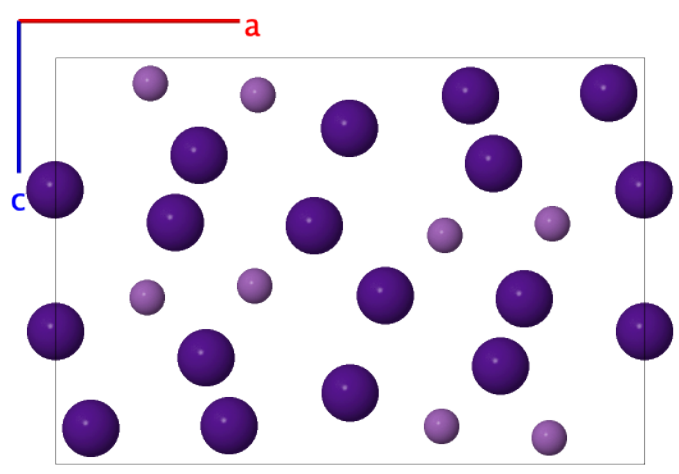

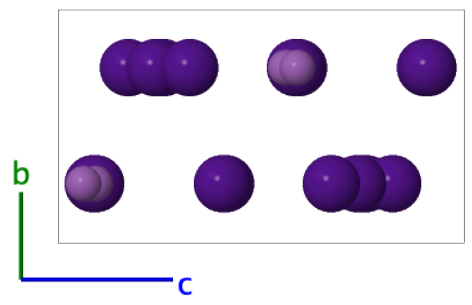

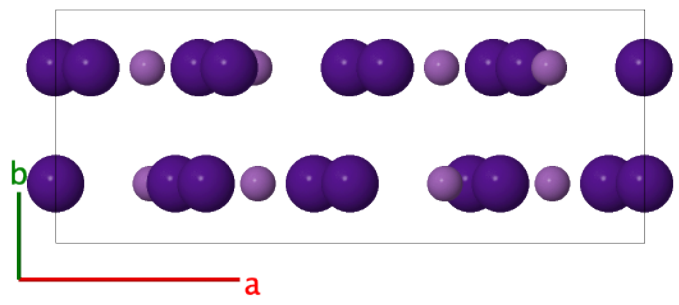

| | | |
|---|---|---|
| **Prototype** | : | $Cs_2Sb$ |
| **AFLOW prototype label** | : | A2B_oP24_62_4c_2c |
| ***Strukturbericht* designation** | : | None |
| **Pearson symbol** | : | oP24 |
| **Space group number** | : | 62 |
| **Space group symbol** | : | *Pnma* |
| **AFLOW prototype command** | : | `aflow --proto=A2B_oP24_62_4c_2c` `--params=`$a, b/a, c/a, x_1, z_1, x_2, z_2, x_3, z_3, x_4, z_4, x_5, z_5, x_6, z_6$ |

**Simple Orthorhombic primitive vectors:**

$$\mathbf{a}_1 = a\,\hat{\mathbf{x}}$$
$$\mathbf{a}_2 = b\,\hat{\mathbf{y}}$$
$$\mathbf{a}_3 = c\,\hat{\mathbf{z}}$$

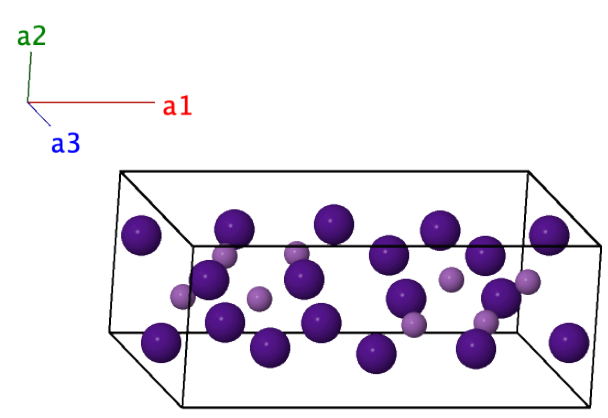

**Basis vectors:**

| | | Lattice Coordinates | | Cartesian Coordinates | Wyckoff Position | Atom Type |
|---|---|---|---|---|---|---|
| $\mathbf{B}_1$ | = | $x_1\,\mathbf{a}_1 + \frac{1}{4}\,\mathbf{a}_2 + z_1\,\mathbf{a}_3$ | = | $x_1 a\,\hat{\mathbf{x}} + \frac{1}{4} b\,\hat{\mathbf{y}} + z_1 c\,\hat{\mathbf{z}}$ | (4*c*) | Cs I |
| $\mathbf{B}_2$ | = | $\left(\frac{1}{2} - x_1\right)\mathbf{a}_1 + \frac{3}{4}\,\mathbf{a}_2 + \left(\frac{1}{2} + z_1\right)\mathbf{a}_3$ | = | $\left(\frac{1}{2} - x_1\right) a\,\hat{\mathbf{x}} + \frac{3}{4} b\,\hat{\mathbf{y}} + \left(\frac{1}{2} + z_1\right) c\,\hat{\mathbf{z}}$ | (4*c*) | Cs I |
| $\mathbf{B}_3$ | = | $-x_1\,\mathbf{a}_1 + \frac{3}{4}\,\mathbf{a}_2 - z_1\,\mathbf{a}_3$ | = | $-x_1 a\,\hat{\mathbf{x}} + \frac{3}{4} b\,\hat{\mathbf{y}} - z_1 c\,\hat{\mathbf{z}}$ | (4*c*) | Cs I |
| $\mathbf{B}_4$ | = | $\left(\frac{1}{2} + x_1\right)\mathbf{a}_1 + \frac{1}{4}\,\mathbf{a}_2 + \left(\frac{1}{2} - z_1\right)\mathbf{a}_3$ | = | $\left(\frac{1}{2} + x_1\right) a\,\hat{\mathbf{x}} + \frac{1}{4} b\,\hat{\mathbf{y}} + \left(\frac{1}{2} - z_1\right) c\,\hat{\mathbf{z}}$ | (4*c*) | Cs I |
| $\mathbf{B}_5$ | = | $x_2\,\mathbf{a}_1 + \frac{1}{4}\,\mathbf{a}_2 + z_2\,\mathbf{a}_3$ | = | $x_2 a\,\hat{\mathbf{x}} + \frac{1}{4} b\,\hat{\mathbf{y}} + z_2 c\,\hat{\mathbf{z}}$ | (4*c*) | Cs II |
| $\mathbf{B}_6$ | = | $\left(\frac{1}{2} - x_2\right) \mathbf{a}_1 + \frac{3}{4} \mathbf{a}_2 + \left(\frac{1}{2} + z_2\right) \mathbf{a}_3$ | = | $\left(\frac{1}{2} - x_2\right) a \, \hat{\mathbf{x}} + \frac{3}{4} b \, \hat{\mathbf{y}} + \left(\frac{1}{2} + z_2\right) c \, \hat{\mathbf{z}}$ | (4c) | Cs II |
| $\mathbf{B}_7$ | = | $-x_2 \mathbf{a}_1 + \frac{3}{4} \mathbf{a}_2 - z_2 \mathbf{a}_3$ | = | $-x_2 a \, \hat{\mathbf{x}} + \frac{3}{4} b \, \hat{\mathbf{y}} - z_2 c \, \hat{\mathbf{z}}$ | (4c) | Cs II |
| $\mathbf{B}_8$ | = | $\left(\frac{1}{2} + x_2\right) \mathbf{a}_1 + \frac{1}{4} \mathbf{a}_2 + \left(\frac{1}{2} - z_2\right) \mathbf{a}_3$ | = | $\left(\frac{1}{2} + x_2\right) a \, \hat{\mathbf{x}} + \frac{1}{4} b \, \hat{\mathbf{y}} + \left(\frac{1}{2} - z_2\right) c \, \hat{\mathbf{z}}$ | (4c) | Cs II |
| $\mathbf{B}_9$ | = | $x_3 \mathbf{a}_1 + \frac{1}{4} \mathbf{a}_2 + z_3 \mathbf{a}_3$ | = | $x_3 a \, \hat{\mathbf{x}} + \frac{1}{4} b \, \hat{\mathbf{y}} + z_3 c \, \hat{\mathbf{z}}$ | (4c) | Cs III |
| $\mathbf{B}_{10}$ | = | $\left(\frac{1}{2} - x_3\right) \mathbf{a}_1 + \frac{3}{4} \mathbf{a}_2 + \left(\frac{1}{2} + z_3\right) \mathbf{a}_3$ | = | $\left(\frac{1}{2} - x_3\right) a \, \hat{\mathbf{x}} + \frac{3}{4} b \, \hat{\mathbf{y}} + \left(\frac{1}{2} + z_3\right) c \, \hat{\mathbf{z}}$ | (4c) | Cs III |
| $\mathbf{B}_{11}$ | = | $-x_3 \mathbf{a}_1 + \frac{3}{4} \mathbf{a}_2 - z_3 \mathbf{a}_3$ | = | $-x_3 a \, \hat{\mathbf{x}} + \frac{3}{4} b \, \hat{\mathbf{y}} - z_3 c \, \hat{\mathbf{z}}$ | (4c) | Cs III |
| $\mathbf{B}_{12}$ | = | $\left(\frac{1}{2} + x_3\right) \mathbf{a}_1 + \frac{1}{4} \mathbf{a}_2 + \left(\frac{1}{2} - z_3\right) \mathbf{a}_3$ | = | $\left(\frac{1}{2} + x_3\right) a \, \hat{\mathbf{x}} + \frac{1}{4} b \, \hat{\mathbf{y}} + \left(\frac{1}{2} - z_3\right) c \, \hat{\mathbf{z}}$ | (4c) | Cs III |
| $\mathbf{B}_{13}$ | = | $x_4 \mathbf{a}_1 + \frac{1}{4} \mathbf{a}_2 + z_4 \mathbf{a}_3$ | = | $x_4 a \, \hat{\mathbf{x}} + \frac{1}{4} b \, \hat{\mathbf{y}} + z_4 c \, \hat{\mathbf{z}}$ | (4c) | Cs IV |
| $\mathbf{B}_{14}$ | = | $\left(\frac{1}{2} - x_4\right) \mathbf{a}_1 + \frac{3}{4} \mathbf{a}_2 + \left(\frac{1}{2} + z_4\right) \mathbf{a}_3$ | = | $\left(\frac{1}{2} - x_4\right) a \, \hat{\mathbf{x}} + \frac{3}{4} b \, \hat{\mathbf{y}} + \left(\frac{1}{2} + z_4\right) c \, \hat{\mathbf{z}}$ | (4c) | Cs IV |
| $\mathbf{B}_{15}$ | = | $-x_4 \mathbf{a}_1 + \frac{3}{4} \mathbf{a}_2 - z_4 \mathbf{a}_3$ | = | $-x_4 a \, \hat{\mathbf{x}} + \frac{3}{4} b \, \hat{\mathbf{y}} - z_4 c \, \hat{\mathbf{z}}$ | (4c) | Cs IV |
| $\mathbf{B}_{16}$ | = | $\left(\frac{1}{2} + x_4\right) \mathbf{a}_1 + \frac{1}{4} \mathbf{a}_2 + \left(\frac{1}{2} - z_4\right) \mathbf{a}_3$ | = | $\left(\frac{1}{2} + x_4\right) a \, \hat{\mathbf{x}} + \frac{1}{4} b \, \hat{\mathbf{y}} + \left(\frac{1}{2} - z_4\right) c \, \hat{\mathbf{z}}$ | (4c) | Cs IV |
| $\mathbf{B}_{17}$ | = | $x_5 \mathbf{a}_1 + \frac{1}{4} \mathbf{a}_2 + z_5 \mathbf{a}_3$ | = | $x_5 a \, \hat{\mathbf{x}} + \frac{1}{4} b \, \hat{\mathbf{y}} + z_5 c \, \hat{\mathbf{z}}$ | (4c) | Sb I |
| $\mathbf{B}_{18}$ | = | $\left(\frac{1}{2} - x_5\right) \mathbf{a}_1 + \frac{3}{4} \mathbf{a}_2 + \left(\frac{1}{2} + z_5\right) \mathbf{a}_3$ | = | $\left(\frac{1}{2} - x_5\right) a \, \hat{\mathbf{x}} + \frac{3}{4} b \, \hat{\mathbf{y}} + \left(\frac{1}{2} + z_5\right) c \, \hat{\mathbf{z}}$ | (4c) | Sb I |
| $\mathbf{B}_{19}$ | = | $-x_5 \mathbf{a}_1 + \frac{3}{4} \mathbf{a}_2 - z_5 \mathbf{a}_3$ | = | $-x_5 a \, \hat{\mathbf{x}} + \frac{3}{4} b \, \hat{\mathbf{y}} - z_5 c \, \hat{\mathbf{z}}$ | (4c) | Sb I |
| $\mathbf{B}_{20}$ | = | $\left(\frac{1}{2} + x_5\right) \mathbf{a}_1 + \frac{1}{4} \mathbf{a}_2 + \left(\frac{1}{2} - z_5\right) \mathbf{a}_3$ | = | $\left(\frac{1}{2} + x_5\right) a \, \hat{\mathbf{x}} + \frac{1}{4} b \, \hat{\mathbf{y}} + \left(\frac{1}{2} - z_5\right) c \, \hat{\mathbf{z}}$ | (4c) | Sb I |
| $\mathbf{B}_{21}$ | = | $x_6 \mathbf{a}_1 + \frac{1}{4} \mathbf{a}_2 + z_6 \mathbf{a}_3$ | = | $x_6 a \, \hat{\mathbf{x}} + \frac{1}{4} b \, \hat{\mathbf{y}} + z_6 c \, \hat{\mathbf{z}}$ | (4c) | Sb II |
| $\mathbf{B}_{22}$ | = | $\left(\frac{1}{2} - x_6\right) \mathbf{a}_1 + \frac{3}{4} \mathbf{a}_2 + \left(\frac{1}{2} + z_6\right) \mathbf{a}_3$ | = | $\left(\frac{1}{2} - x_6\right) a \, \hat{\mathbf{x}} + \frac{3}{4} b \, \hat{\mathbf{y}} + \left(\frac{1}{2} + z_6\right) c \, \hat{\mathbf{z}}$ | (4c) | Sb II |
| $\mathbf{B}_{23}$ | = | $-x_6 \mathbf{a}_1 + \frac{3}{4} \mathbf{a}_2 - z_6 \mathbf{a}_3$ | = | $-x_6 a \, \hat{\mathbf{x}} + \frac{3}{4} b \, \hat{\mathbf{y}} - z_6 c \, \hat{\mathbf{z}}$ | (4c) | Sb II |
| $\mathbf{B}_{24}$ | = | $\left(\frac{1}{2} + x_6\right) \mathbf{a}_1 + \frac{1}{4} \mathbf{a}_2 + \left(\frac{1}{2} - z_6\right) \mathbf{a}_3$ | = | $\left(\frac{1}{2} + x_6\right) a \, \hat{\mathbf{x}} + \frac{1}{4} b \, \hat{\mathbf{y}} + \left(\frac{1}{2} - z_6\right) c \, \hat{\mathbf{z}}$ | (4c) | Sb II |

**Found in:**

**Geometry files:**

- CIF: pp. 1635
- POSCAR: pp. 1635

# $RhCl_2(NH_3)_5Cl$ ($J1_8$) Structure: A3B15C5D_oP96_62_cd_3c6d_3cd_c

http://aflow.org/prototype-encyclopedia/A3B15C5D_oP96_62_cd_3c6d_3cd_c

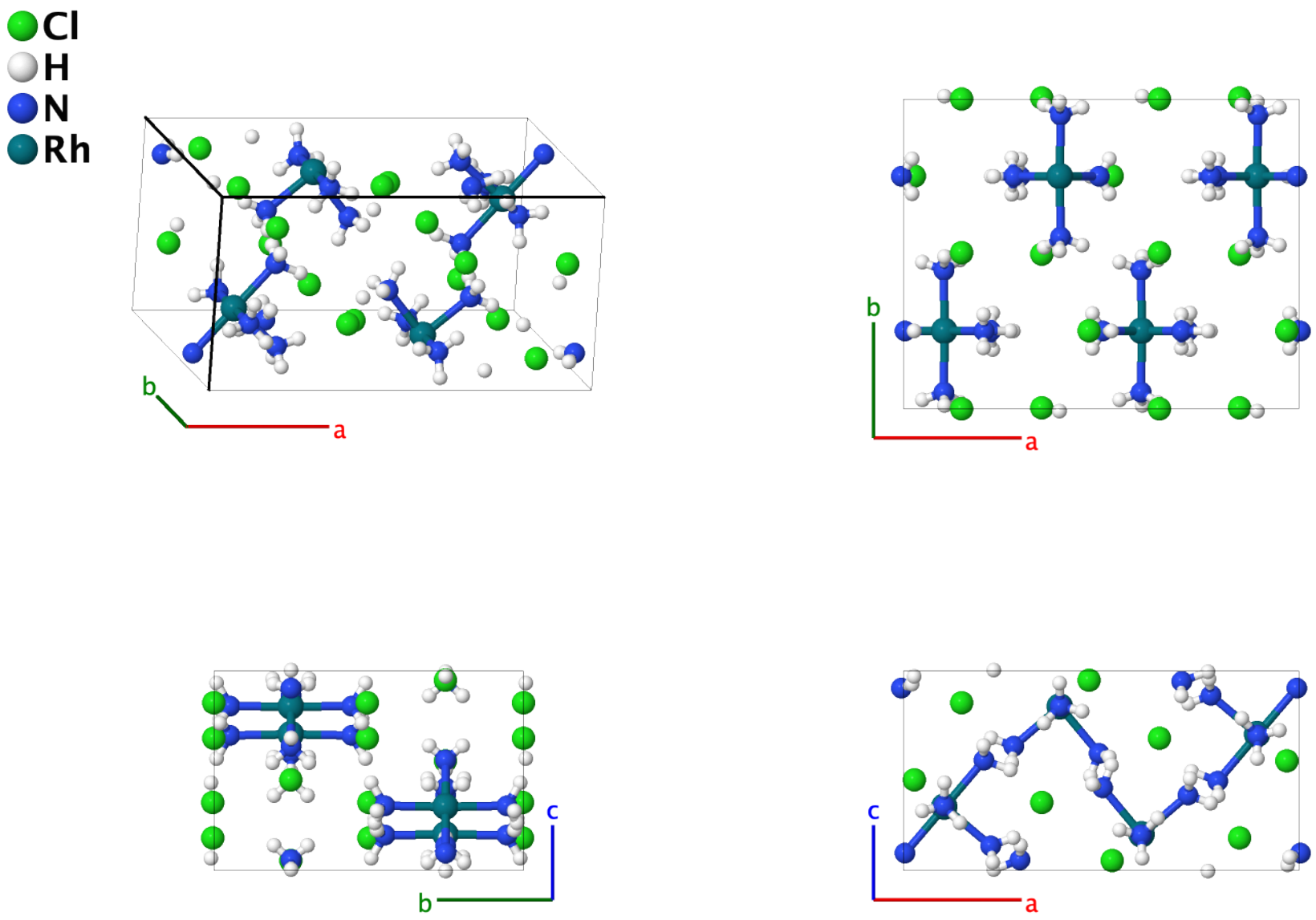

| | | |
|---|---|---|
| **Prototype** | : | $Cl_3H_{15}N_5Rh$ |
| **AFLOW prototype label** | : | A3B15C5D_oP96_62_cd_3c6d_3cd_c |
| ***Strukturbericht* designation** | : | $J1_8$ |
| **Pearson symbol** | : | oP96 |
| **Space group number** | : | 62 |
| **Space group symbol** | : | $Pnma$ |
| **AFLOW prototype command** | : | `aflow --proto=A3B15C5D_oP96_62_cd_3c6d_3cd_c --params=`$a, b/a, c/a, x_1, z_1, x_2, z_2, x_3, z_3, x_4, z_4, x_5, z_5, x_6, z_6, x_7, z_7, x_8, z_8, x_9, y_9, z_9, x_{10}, y_{10}, z_{10}, x_{11}, y_{11}, z_{11}, x_{12}, y_{12}, z_{12}, x_{13}, y_{13}, z_{13}, x_{14}, y_{14}, z_{14}, x_{15}, y_{15}, z_{15}, x_{16}, y_{16}, z_{16}$ |

### Other compounds with this structure

- $CoCl_2(NH_3)_5Cl$, $CoI_2(NH_3)_5Cl$, $CrCl_2(NH_3)_5Cl$, $IrCl_2(NH_3)_5Cl$, $OsCl_2(NH_3)_5Cl$, $RhBr_2(NH_3)_5Br$, $RhCl_2(NH_3)_5Cl$, $RhI_2(NH_3)_5Cl$, and $RuCl_2(NH_3)_5Cl$

### Simple Orthorhombic primitive vectors:

$$\mathbf{a}_1 = a\,\hat{\mathbf{x}}$$
$$\mathbf{a}_2 = b\,\hat{\mathbf{y}}$$
$$\mathbf{a}_3 = c\,\hat{\mathbf{z}}$$

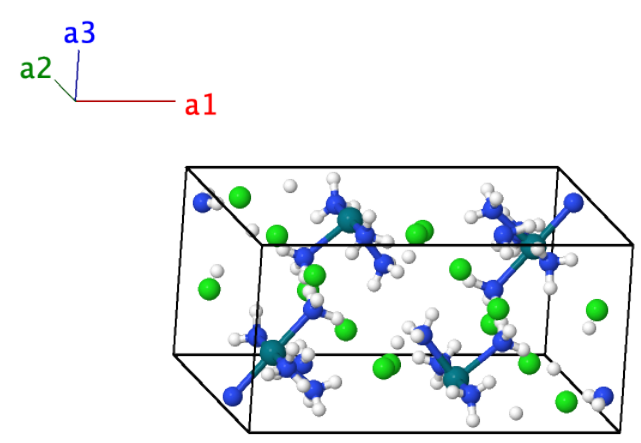

### Basis vectors:

| | | Lattice Coordinates | | Cartesian Coordinates | Wyckoff Position | Atom Type |
|---|---|---|---|---|---|---|
| $\mathbf{B}_{1}$ | = | $x_{1}\,\mathbf{a}_{1}+\frac{1}{4}\,\mathbf{a}_{2}+z_{1}\,\mathbf{a}_{3}$ | = | $x_{1}a\,\mathbf{\hat{x}}+\frac{1}{4}b\,\mathbf{\hat{y}}+z_{1}c\,\mathbf{\hat{z}}$ | (4$c$) | Cl I |
| $\mathbf{B}_{2}$ | = | $\left(\frac{1}{2}-x_{1}\right)\,\mathbf{a}_{1}+\frac{3}{4}\,\mathbf{a}_{2}+\left(\frac{1}{2}+z_{1}\right)\,\mathbf{a}_{3}$ | = | $\left(\frac{1}{2}-x_{1}\right)a\,\mathbf{\hat{x}}+\frac{3}{4}b\,\mathbf{\hat{y}}+\left(\frac{1}{2}+z_{1}\right)c\,\mathbf{\hat{z}}$ | (4$c$) | Cl I |
| $\mathbf{B}_{3}$ | = | $-x_{1}\,\mathbf{a}_{1}+\frac{3}{4}\,\mathbf{a}_{2}-z_{1}\,\mathbf{a}_{3}$ | = | $-x_{1}a\,\mathbf{\hat{x}}+\frac{3}{4}b\,\mathbf{\hat{y}}-z_{1}c\,\mathbf{\hat{z}}$ | (4$c$) | Cl I |
| $\mathbf{B}_{4}$ | = | $\left(\frac{1}{2}+x_{1}\right)\,\mathbf{a}_{1}+\frac{1}{4}\,\mathbf{a}_{2}+\left(\frac{1}{2}-z_{1}\right)\,\mathbf{a}_{3}$ | = | $\left(\frac{1}{2}+x_{1}\right)a\,\mathbf{\hat{x}}+\frac{1}{4}b\,\mathbf{\hat{y}}+\left(\frac{1}{2}-z_{1}\right)c\,\mathbf{\hat{z}}$ | (4$c$) | Cl I |
| $\mathbf{B}_{5}$ | = | $x_{2}\,\mathbf{a}_{1}+\frac{1}{4}\,\mathbf{a}_{2}+z_{2}\,\mathbf{a}_{3}$ | = | $x_{2}a\,\mathbf{\hat{x}}+\frac{1}{4}b\,\mathbf{\hat{y}}+z_{2}c\,\mathbf{\hat{z}}$ | (4$c$) | H I |
| $\mathbf{B}_{6}$ | = | $\left(\frac{1}{2}-x_{2}\right)\,\mathbf{a}_{1}+\frac{3}{4}\,\mathbf{a}_{2}+\left(\frac{1}{2}+z_{2}\right)\,\mathbf{a}_{3}$ | = | $\left(\frac{1}{2}-x_{2}\right)a\,\mathbf{\hat{x}}+\frac{3}{4}b\,\mathbf{\hat{y}}+\left(\frac{1}{2}+z_{2}\right)c\,\mathbf{\hat{z}}$ | (4$c$) | H I |
| $\mathbf{B}_{7}$ | = | $-x_{2}\,\mathbf{a}_{1}+\frac{3}{4}\,\mathbf{a}_{2}-z_{2}\,\mathbf{a}_{3}$ | = | $-x_{2}a\,\mathbf{\hat{x}}+\frac{3}{4}b\,\mathbf{\hat{y}}-z_{2}c\,\mathbf{\hat{z}}$ | (4$c$) | H I |
| $\mathbf{B}_{8}$ | = | $\left(\frac{1}{2}+x_{2}\right)\,\mathbf{a}_{1}+\frac{1}{4}\,\mathbf{a}_{2}+\left(\frac{1}{2}-z_{2}\right)\,\mathbf{a}_{3}$ | = | $\left(\frac{1}{2}+x_{2}\right)a\,\mathbf{\hat{x}}+\frac{1}{4}b\,\mathbf{\hat{y}}+\left(\frac{1}{2}-z_{2}\right)c\,\mathbf{\hat{z}}$ | (4$c$) | H I |
| $\mathbf{B}_{9}$ | = | $x_{3}\,\mathbf{a}_{1}+\frac{1}{4}\,\mathbf{a}_{2}+z_{3}\,\mathbf{a}_{3}$ | = | $x_{3}a\,\mathbf{\hat{x}}+\frac{1}{4}b\,\mathbf{\hat{y}}+z_{3}c\,\mathbf{\hat{z}}$ | (4$c$) | H II |
| $\mathbf{B}_{10}$ | = | $\left(\frac{1}{2}-x_{3}\right)\,\mathbf{a}_{1}+\frac{3}{4}\,\mathbf{a}_{2}+\left(\frac{1}{2}+z_{3}\right)\,\mathbf{a}_{3}$ | = | $\left(\frac{1}{2}-x_{3}\right)a\,\mathbf{\hat{x}}+\frac{3}{4}b\,\mathbf{\hat{y}}+\left(\frac{1}{2}+z_{3}\right)c\,\mathbf{\hat{z}}$ | (4$c$) | H II |
| $\mathbf{B}_{11}$ | = | $-x_{3}\,\mathbf{a}_{1}+\frac{3}{4}\,\mathbf{a}_{2}-z_{3}\,\mathbf{a}_{3}$ | = | $-x_{3}a\,\mathbf{\hat{x}}+\frac{3}{4}b\,\mathbf{\hat{y}}-z_{3}c\,\mathbf{\hat{z}}$ | (4$c$) | H II |
| $\mathbf{B}_{12}$ | = | $\left(\frac{1}{2}+x_{3}\right)\,\mathbf{a}_{1}+\frac{1}{4}\,\mathbf{a}_{2}+\left(\frac{1}{2}-z_{3}\right)\,\mathbf{a}_{3}$ | = | $\left(\frac{1}{2}+x_{3}\right)a\,\mathbf{\hat{x}}+\frac{1}{4}b\,\mathbf{\hat{y}}+\left(\frac{1}{2}-z_{3}\right)c\,\mathbf{\hat{z}}$ | (4$c$) | H II |
| $\mathbf{B}_{13}$ | = | $x_{4}\,\mathbf{a}_{1}+\frac{1}{4}\,\mathbf{a}_{2}+z_{4}\,\mathbf{a}_{3}$ | = | $x_{4}a\,\mathbf{\hat{x}}+\frac{1}{4}b\,\mathbf{\hat{y}}+z_{4}c\,\mathbf{\hat{z}}$ | (4$c$) | H III |
| $\mathbf{B}_{14}$ | = | $\left(\frac{1}{2}-x_{4}\right)\,\mathbf{a}_{1}+\frac{3}{4}\,\mathbf{a}_{2}+\left(\frac{1}{2}+z_{4}\right)\,\mathbf{a}_{3}$ | = | $\left(\frac{1}{2}-x_{4}\right)a\,\mathbf{\hat{x}}+\frac{3}{4}b\,\mathbf{\hat{y}}+\left(\frac{1}{2}+z_{4}\right)c\,\mathbf{\hat{z}}$ | (4$c$) | H III |
| $\mathbf{B}_{15}$ | = | $-x_{4}\,\mathbf{a}_{1}+\frac{3}{4}\,\mathbf{a}_{2}-z_{4}\,\mathbf{a}_{3}$ | = | $-x_{4}a\,\mathbf{\hat{x}}+\frac{3}{4}b\,\mathbf{\hat{y}}-z_{4}c\,\mathbf{\hat{z}}$ | (4$c$) | H III |
| $\mathbf{B}_{16}$ | = | $\left(\frac{1}{2}+x_{4}\right)\,\mathbf{a}_{1}+\frac{1}{4}\,\mathbf{a}_{2}+\left(\frac{1}{2}-z_{4}\right)\,\mathbf{a}_{3}$ | = | $\left(\frac{1}{2}+x_{4}\right)a\,\mathbf{\hat{x}}+\frac{1}{4}b\,\mathbf{\hat{y}}+\left(\frac{1}{2}-z_{4}\right)c\,\mathbf{\hat{z}}$ | (4$c$) | H III |
| $\mathbf{B}_{17}$ | = | $x_{5}\,\mathbf{a}_{1}+\frac{1}{4}\,\mathbf{a}_{2}+z_{5}\,\mathbf{a}_{3}$ | = | $x_{5}a\,\mathbf{\hat{x}}+\frac{1}{4}b\,\mathbf{\hat{y}}+z_{5}c\,\mathbf{\hat{z}}$ | (4$c$) | N I |
| $\mathbf{B}_{18}$ | = | $\left(\frac{1}{2}-x_{5}\right)\,\mathbf{a}_{1}+\frac{3}{4}\,\mathbf{a}_{2}+\left(\frac{1}{2}+z_{5}\right)\,\mathbf{a}_{3}$ | = | $\left(\frac{1}{2}-x_{5}\right)a\,\mathbf{\hat{x}}+\frac{3}{4}b\,\mathbf{\hat{y}}+\left(\frac{1}{2}+z_{5}\right)c\,\mathbf{\hat{z}}$ | (4$c$) | N I |
| $\mathbf{B}_{19}$ | = | $-x_{5}\,\mathbf{a}_{1}+\frac{3}{4}\,\mathbf{a}_{2}-z_{5}\,\mathbf{a}_{3}$ | = | $-x_{5}a\,\mathbf{\hat{x}}+\frac{3}{4}b\,\mathbf{\hat{y}}-z_{5}c\,\mathbf{\hat{z}}$ | (4$c$) | N I |
| $\mathbf{B}_{20}$ | = | $\left(\frac{1}{2}+x_{5}\right)\,\mathbf{a}_{1}+\frac{1}{4}\,\mathbf{a}_{2}+\left(\frac{1}{2}-z_{5}\right)\,\mathbf{a}_{3}$ | = | $\left(\frac{1}{2}+x_{5}\right)a\,\mathbf{\hat{x}}+\frac{1}{4}b\,\mathbf{\hat{y}}+\left(\frac{1}{2}-z_{5}\right)c\,\mathbf{\hat{z}}$ | (4$c$) | N I |
| $\mathbf{B}_{21}$ | = | $x_{6}\,\mathbf{a}_{1}+\frac{1}{4}\,\mathbf{a}_{2}+z_{6}\,\mathbf{a}_{3}$ | = | $x_{6}a\,\mathbf{\hat{x}}+\frac{1}{4}b\,\mathbf{\hat{y}}+z_{6}c\,\mathbf{\hat{z}}$ | (4$c$) | N II |
| $\mathbf{B}_{22}$ | = | $\left(\frac{1}{2}-x_{6}\right)\,\mathbf{a}_{1}+\frac{3}{4}\,\mathbf{a}_{2}+\left(\frac{1}{2}+z_{6}\right)\,\mathbf{a}_{3}$ | = | $\left(\frac{1}{2}-x_{6}\right)a\,\mathbf{\hat{x}}+\frac{3}{4}b\,\mathbf{\hat{y}}+\left(\frac{1}{2}+z_{6}\right)c\,\mathbf{\hat{z}}$ | (4$c$) | N II |
| $\mathbf{B}_{23}$ | = | $-x_{6}\,\mathbf{a}_{1}+\frac{3}{4}\,\mathbf{a}_{2}-z_{6}\,\mathbf{a}_{3}$ | = | $-x_{6}a\,\mathbf{\hat{x}}+\frac{3}{4}b\,\mathbf{\hat{y}}-z_{6}c\,\mathbf{\hat{z}}$ | (4$c$) | N II |
| $\mathbf{B}_{24}$ | = | $\left(\frac{1}{2}+x_{6}\right)\,\mathbf{a}_{1}+\frac{1}{4}\,\mathbf{a}_{2}+\left(\frac{1}{2}-z_{6}\right)\,\mathbf{a}_{3}$ | = | $\left(\frac{1}{2}+x_{6}\right)a\,\mathbf{\hat{x}}+\frac{1}{4}b\,\mathbf{\hat{y}}+\left(\frac{1}{2}-z_{6}\right)c\,\mathbf{\hat{z}}$ | (4$c$) | N II |
| $\mathbf{B}_{25}$ | = | $x_{7}\,\mathbf{a}_{1}+\frac{1}{4}\,\mathbf{a}_{2}+z_{7}\,\mathbf{a}_{3}$ | = | $x_{7}a\,\mathbf{\hat{x}}+\frac{1}{4}b\,\mathbf{\hat{y}}+z_{7}c\,\mathbf{\hat{z}}$ | (4$c$) | N III |
| $\mathbf{B}_{26}$ | = | $\left(\frac{1}{2}-x_{7}\right)\,\mathbf{a}_{1}+\frac{3}{4}\,\mathbf{a}_{2}+\left(\frac{1}{2}+z_{7}\right)\,\mathbf{a}_{3}$ | = | $\left(\frac{1}{2}-x_{7}\right)a\,\mathbf{\hat{x}}+\frac{3}{4}b\,\mathbf{\hat{y}}+\left(\frac{1}{2}+z_{7}\right)c\,\mathbf{\hat{z}}$ | (4$c$) | N III |
| $\mathbf{B}_{27}$ | = | $-x_{7}\,\mathbf{a}_{1}+\frac{3}{4}\,\mathbf{a}_{2}-z_{7}\,\mathbf{a}_{3}$ | = | $-x_{7}a\,\mathbf{\hat{x}}+\frac{3}{4}b\,\mathbf{\hat{y}}-z_{7}c\,\mathbf{\hat{z}}$ | (4$c$) | N III |
| $\mathbf{B}_{28}$ | = | $\left(\frac{1}{2}+x_{7}\right)\,\mathbf{a}_{1}+\frac{1}{4}\,\mathbf{a}_{2}+\left(\frac{1}{2}-z_{7}\right)\,\mathbf{a}_{3}$ | = | $\left(\frac{1}{2}+x_{7}\right)a\,\mathbf{\hat{x}}+\frac{1}{4}b\,\mathbf{\hat{y}}+\left(\frac{1}{2}-z_{7}\right)c\,\mathbf{\hat{z}}$ | (4$c$) | N III |
| $\mathbf{B}_{29}$ | = | $x_{8}\,\mathbf{a}_{1}+\frac{1}{4}\,\mathbf{a}_{2}+z_{8}\,\mathbf{a}_{3}$ | = | $x_{8}a\,\mathbf{\hat{x}}+\frac{1}{4}b\,\mathbf{\hat{y}}+z_{8}c\,\mathbf{\hat{z}}$ | (4$c$) | Rh |
| $\mathbf{B}_{30}$ | = | $\left(\frac{1}{2}-x_{8}\right)\,\mathbf{a}_{1}+\frac{3}{4}\,\mathbf{a}_{2}+\left(\frac{1}{2}+z_{8}\right)\,\mathbf{a}_{3}$ | = | $\left(\frac{1}{2}-x_{8}\right)a\,\mathbf{\hat{x}}+\frac{3}{4}b\,\mathbf{\hat{y}}+\left(\frac{1}{2}+z_{8}\right)c\,\mathbf{\hat{z}}$ | (4$c$) | Rh |
| $\mathbf{B}_{31}$ | = | $-x_{8}\,\mathbf{a}_{1}+\frac{3}{4}\,\mathbf{a}_{2}-z_{8}\,\mathbf{a}_{3}$ | = | $-x_{8}a\,\mathbf{\hat{x}}+\frac{3}{4}b\,\mathbf{\hat{y}}-z_{8}c\,\mathbf{\hat{z}}$ | (4$c$) | Rh |
| $\mathbf{B}_{32}$ | = | $\left(\frac{1}{2}+x_{8}\right)\,\mathbf{a}_{1}+\frac{1}{4}\,\mathbf{a}_{2}+\left(\frac{1}{2}-z_{8}\right)\,\mathbf{a}_{3}$ | = | $\left(\frac{1}{2}+x_{8}\right)a\,\mathbf{\hat{x}}+\frac{1}{4}b\,\mathbf{\hat{y}}+\left(\frac{1}{2}-z_{8}\right)c\,\mathbf{\hat{z}}$ | (4$c$) | Rh |
| $\mathbf{B}_{33}$ | = | $x_{9}\,\mathbf{a}_{1}+y_{9}\,\mathbf{a}_{2}+z_{9}\,\mathbf{a}_{3}$ | = | $x_{9}a\,\mathbf{\hat{x}}+y_{9}b\,\mathbf{\hat{y}}+z_{9}c\,\mathbf{\hat{z}}$ | (8$d$) | Cl II |
| $\mathbf{B}_{34}$ | = | $\left(\frac{1}{2}-x_{9}\right)\,\mathbf{a}_{1}-y_{9}\,\mathbf{a}_{2}+\left(\frac{1}{2}+z_{9}\right)\,\mathbf{a}_{3}$ | = | $\left(\frac{1}{2}-x_{9}\right)a\,\mathbf{\hat{x}}-y_{9}b\,\mathbf{\hat{y}}+\left(\frac{1}{2}+z_{9}\right)c\,\mathbf{\hat{z}}$ | (8$d$) | Cl II |
| $\mathbf{B}_{35}$ | = | $-x_{9}\,\mathbf{a}_{1}+\left(\frac{1}{2}+y_{9}\right)\,\mathbf{a}_{2}-z_{9}\,\mathbf{a}_{3}$ | = | $-x_{9}a\,\mathbf{\hat{x}}+\left(\frac{1}{2}+y_{9}\right)b\,\mathbf{\hat{y}}-z_{9}c\,\mathbf{\hat{z}}$ | (8$d$) | Cl II |
| $\mathbf{B}_{36}$ | = | $\left(\frac{1}{2}+x_{9}\right)\mathbf{a}_{1}+\left(\frac{1}{2}-y_{9}\right)\mathbf{a}_{2}+\left(\frac{1}{2}-z_{9}\right)\mathbf{a}_{3}$ | = | $\left(\frac{1}{2}+x_{9}\right)a\,\hat{\mathbf{x}}+\left(\frac{1}{2}-y_{9}\right)b\,\hat{\mathbf{y}}+\left(\frac{1}{2}-z_{9}\right)c\,\hat{\mathbf{z}}$ | (8$d$) | Cl II |
| $\mathbf{B}_{37}$ | = | $-x_{9}\,\mathbf{a}_{1}-y_{9}\,\mathbf{a}_{2}-z_{9}\,\mathbf{a}_{3}$ | = | $-x_{9}a\,\hat{\mathbf{x}}-y_{9}b\,\hat{\mathbf{y}}-z_{9}c\,\hat{\mathbf{z}}$ | (8$d$) | Cl II |
| $\mathbf{B}_{38}$ | = | $\left(\frac{1}{2}+x_{9}\right)\mathbf{a}_{1}+y_{9}\,\mathbf{a}_{2}+\left(\frac{1}{2}-z_{9}\right)\mathbf{a}_{3}$ | = | $\left(\frac{1}{2}+x_{9}\right)a\,\hat{\mathbf{x}}+y_{9}b\,\hat{\mathbf{y}}+\left(\frac{1}{2}-z_{9}\right)c\,\hat{\mathbf{z}}$ | (8$d$) | Cl II |
| $\mathbf{B}_{39}$ | = | $x_{9}\,\mathbf{a}_{1}+\left(\frac{1}{2}-y_{9}\right)\mathbf{a}_{2}+z_{9}\,\mathbf{a}_{3}$ | = | $x_{9}a\,\hat{\mathbf{x}}+\left(\frac{1}{2}-y_{9}\right)b\,\hat{\mathbf{y}}+z_{9}c\,\hat{\mathbf{z}}$ | (8$d$) | Cl II |
| $\mathbf{B}_{40}$ | = | $\left(\frac{1}{2}-x_{9}\right)\mathbf{a}_{1}+\left(\frac{1}{2}+y_{9}\right)\mathbf{a}_{2}+\left(\frac{1}{2}+z_{9}\right)\mathbf{a}_{3}$ | = | $\left(\frac{1}{2}-x_{9}\right)a\,\hat{\mathbf{x}}+\left(\frac{1}{2}+y_{9}\right)b\,\hat{\mathbf{y}}+\left(\frac{1}{2}+z_{9}\right)c\,\hat{\mathbf{z}}$ | (8$d$) | Cl II |
| $\mathbf{B}_{41}$ | = | $x_{10}\,\mathbf{a}_{1}+y_{10}\,\mathbf{a}_{2}+z_{10}\,\mathbf{a}_{3}$ | = | $x_{10}a\,\hat{\mathbf{x}}+y_{10}b\,\hat{\mathbf{y}}+z_{10}c\,\hat{\mathbf{z}}$ | (8$d$) | H IV |
| $\mathbf{B}_{42}$ | = | $\left(\frac{1}{2}-x_{10}\right)\mathbf{a}_{1}-y_{10}\,\mathbf{a}_{2}+\left(\frac{1}{2}+z_{10}\right)\mathbf{a}_{3}$ | = | $\left(\frac{1}{2}-x_{10}\right)a\,\hat{\mathbf{x}}-y_{10}b\,\hat{\mathbf{y}}+\left(\frac{1}{2}+z_{10}\right)c\,\hat{\mathbf{z}}$ | (8$d$) | H IV |
| $\mathbf{B}_{43}$ | = | $-x_{10}\,\mathbf{a}_{1}+\left(\frac{1}{2}+y_{10}\right)\mathbf{a}_{2}-z_{10}\,\mathbf{a}_{3}$ | = | $-x_{10}a\,\hat{\mathbf{x}}+\left(\frac{1}{2}+y_{10}\right)b\,\hat{\mathbf{y}}-z_{10}c\,\hat{\mathbf{z}}$ | (8$d$) | H IV |
| $\mathbf{B}_{44}$ | = | $\left(\frac{1}{2}+x_{10}\right)\mathbf{a}_{1}+\left(\frac{1}{2}-y_{10}\right)\mathbf{a}_{2}+\left(\frac{1}{2}-z_{10}\right)\mathbf{a}_{3}$ | = | $\left(\frac{1}{2}+x_{10}\right)a\,\hat{\mathbf{x}}+\left(\frac{1}{2}-y_{10}\right)b\,\hat{\mathbf{y}}+\left(\frac{1}{2}-z_{10}\right)c\,\hat{\mathbf{z}}$ | (8$d$) | H IV |
| $\mathbf{B}_{45}$ | = | $-x_{10}\,\mathbf{a}_{1}-y_{10}\,\mathbf{a}_{2}-z_{10}\,\mathbf{a}_{3}$ | = | $-x_{10}a\,\hat{\mathbf{x}}-y_{10}b\,\hat{\mathbf{y}}-z_{10}c\,\hat{\mathbf{z}}$ | (8$d$) | H IV |
| $\mathbf{B}_{46}$ | = | $\left(\frac{1}{2}+x_{10}\right)\mathbf{a}_{1}+y_{10}\,\mathbf{a}_{2}+\left(\frac{1}{2}-z_{10}\right)\mathbf{a}_{3}$ | = | $\left(\frac{1}{2}+x_{10}\right)a\,\hat{\mathbf{x}}+y_{10}b\,\hat{\mathbf{y}}+\left(\frac{1}{2}-z_{10}\right)c\,\hat{\mathbf{z}}$ | (8$d$) | H IV |
| $\mathbf{B}_{47}$ | = | $x_{10}\,\mathbf{a}_{1}+\left(\frac{1}{2}-y_{10}\right)\mathbf{a}_{2}+z_{10}\,\mathbf{a}_{3}$ | = | $x_{10}a\,\hat{\mathbf{x}}+\left(\frac{1}{2}-y_{10}\right)b\,\hat{\mathbf{y}}+z_{10}c\,\hat{\mathbf{z}}$ | (8$d$) | H IV |
| $\mathbf{B}_{48}$ | = | $\left(\frac{1}{2}-x_{10}\right)\mathbf{a}_{1}+\left(\frac{1}{2}+y_{10}\right)\mathbf{a}_{2}+\left(\frac{1}{2}+z_{10}\right)\mathbf{a}_{3}$ | = | $\left(\frac{1}{2}-x_{10}\right)a\,\hat{\mathbf{x}}+\left(\frac{1}{2}+y_{10}\right)b\,\hat{\mathbf{y}}+\left(\frac{1}{2}+z_{10}\right)c\,\hat{\mathbf{z}}$ | (8$d$) | H IV |
| $\mathbf{B}_{49}$ | = | $x_{11}\,\mathbf{a}_{1}+y_{11}\,\mathbf{a}_{2}+z_{11}\,\mathbf{a}_{3}$ | = | $x_{11}a\,\hat{\mathbf{x}}+y_{11}b\,\hat{\mathbf{y}}+z_{11}c\,\hat{\mathbf{z}}$ | (8$d$) | H V |
| $\mathbf{B}_{50}$ | = | $\left(\frac{1}{2}-x_{11}\right)\mathbf{a}_{1}-y_{11}\,\mathbf{a}_{2}+\left(\frac{1}{2}+z_{11}\right)\mathbf{a}_{3}$ | = | $\left(\frac{1}{2}-x_{11}\right)a\,\hat{\mathbf{x}}-y_{11}b\,\hat{\mathbf{y}}+\left(\frac{1}{2}+z_{11}\right)c\,\hat{\mathbf{z}}$ | (8$d$) | H V |
| $\mathbf{B}_{51}$ | = | $-x_{11}\,\mathbf{a}_{1}+\left(\frac{1}{2}+y_{11}\right)\mathbf{a}_{2}-z_{11}\,\mathbf{a}_{3}$ | = | $-x_{11}a\,\hat{\mathbf{x}}+\left(\frac{1}{2}+y_{11}\right)b\,\hat{\mathbf{y}}-z_{11}c\,\hat{\mathbf{z}}$ | (8$d$) | H V |
| $\mathbf{B}_{52}$ | = | $\left(\frac{1}{2}+x_{11}\right)\mathbf{a}_{1}+\left(\frac{1}{2}-y_{11}\right)\mathbf{a}_{2}+\left(\frac{1}{2}-z_{11}\right)\mathbf{a}_{3}$ | = | $\left(\frac{1}{2}+x_{11}\right)a\,\hat{\mathbf{x}}+\left(\frac{1}{2}-y_{11}\right)b\,\hat{\mathbf{y}}+\left(\frac{1}{2}-z_{11}\right)c\,\hat{\mathbf{z}}$ | (8$d$) | H V |
| $\mathbf{B}_{53}$ | = | $-x_{11}\,\mathbf{a}_{1}-y_{11}\,\mathbf{a}_{2}-z_{11}\,\mathbf{a}_{3}$ | = | $-x_{11}a\,\hat{\mathbf{x}}-y_{11}b\,\hat{\mathbf{y}}-z_{11}c\,\hat{\mathbf{z}}$ | (8$d$) | H V |
| $\mathbf{B}_{54}$ | = | $\left(\frac{1}{2}+x_{11}\right)\mathbf{a}_{1}+y_{11}\,\mathbf{a}_{2}+\left(\frac{1}{2}-z_{11}\right)\mathbf{a}_{3}$ | = | $\left(\frac{1}{2}+x_{11}\right)a\,\hat{\mathbf{x}}+y_{11}b\,\hat{\mathbf{y}}+\left(\frac{1}{2}-z_{11}\right)c\,\hat{\mathbf{z}}$ | (8$d$) | H V |
| $\mathbf{B}_{55}$ | = | $x_{11}\,\mathbf{a}_{1}+\left(\frac{1}{2}-y_{11}\right)\mathbf{a}_{2}+z_{11}\,\mathbf{a}_{3}$ | = | $x_{11}a\,\hat{\mathbf{x}}+\left(\frac{1}{2}-y_{11}\right)b\,\hat{\mathbf{y}}+z_{11}c\,\hat{\mathbf{z}}$ | (8$d$) | H V |
| $\mathbf{B}_{56}$ | = | $\left(\frac{1}{2}-x_{11}\right)\mathbf{a}_{1}+\left(\frac{1}{2}+y_{11}\right)\mathbf{a}_{2}+\left(\frac{1}{2}+z_{11}\right)\mathbf{a}_{3}$ | = | $\left(\frac{1}{2}-x_{11}\right)a\,\hat{\mathbf{x}}+\left(\frac{1}{2}+y_{11}\right)b\,\hat{\mathbf{y}}+\left(\frac{1}{2}+z_{11}\right)c\,\hat{\mathbf{z}}$ | (8$d$) | H V |
| $\mathbf{B}_{57}$ | = | $x_{12}\,\mathbf{a}_{1}+y_{12}\,\mathbf{a}_{2}+z_{12}\,\mathbf{a}_{3}$ | = | $x_{12}a\,\hat{\mathbf{x}}+y_{12}b\,\hat{\mathbf{y}}+z_{12}c\,\hat{\mathbf{z}}$ | (8$d$) | H VI |
| $\mathbf{B}_{58}$ | = | $\left(\frac{1}{2}-x_{12}\right)\mathbf{a}_{1}-y_{12}\,\mathbf{a}_{2}+\left(\frac{1}{2}+z_{12}\right)\mathbf{a}_{3}$ | = | $\left(\frac{1}{2}-x_{12}\right)a\,\hat{\mathbf{x}}-y_{12}b\,\hat{\mathbf{y}}+\left(\frac{1}{2}+z_{12}\right)c\,\hat{\mathbf{z}}$ | (8$d$) | H VI |
| $\mathbf{B}_{59}$ | = | $-x_{12}\,\mathbf{a}_{1}+\left(\frac{1}{2}+y_{12}\right)\mathbf{a}_{2}-z_{12}\,\mathbf{a}_{3}$ | = | $-x_{12}a\,\hat{\mathbf{x}}+\left(\frac{1}{2}+y_{12}\right)b\,\hat{\mathbf{y}}-z_{12}c\,\hat{\mathbf{z}}$ | (8$d$) | H VI |
| $\mathbf{B}_{60}$ | = | $\left(\frac{1}{2}+x_{12}\right)\mathbf{a}_{1}+\left(\frac{1}{2}-y_{12}\right)\mathbf{a}_{2}+\left(\frac{1}{2}-z_{12}\right)\mathbf{a}_{3}$ | = | $\left(\frac{1}{2}+x_{12}\right)a\,\hat{\mathbf{x}}+\left(\frac{1}{2}-y_{12}\right)b\,\hat{\mathbf{y}}+\left(\frac{1}{2}-z_{12}\right)c\,\hat{\mathbf{z}}$ | (8$d$) | H VI |
| $\mathbf{B}_{61}$ | = | $-x_{12}\,\mathbf{a}_{1}-y_{12}\,\mathbf{a}_{2}-z_{12}\,\mathbf{a}_{3}$ | = | $-x_{12}a\,\hat{\mathbf{x}}-y_{12}b\,\hat{\mathbf{y}}-z_{12}c\,\hat{\mathbf{z}}$ | (8$d$) | H VI |
| $\mathbf{B}_{62}$ | = | $\left(\frac{1}{2}+x_{12}\right)\mathbf{a}_{1}+y_{12}\,\mathbf{a}_{2}+\left(\frac{1}{2}-z_{12}\right)\mathbf{a}_{3}$ | = | $\left(\frac{1}{2}+x_{12}\right)a\,\hat{\mathbf{x}}+y_{12}b\,\hat{\mathbf{y}}+\left(\frac{1}{2}-z_{12}\right)c\,\hat{\mathbf{z}}$ | (8$d$) | H VI |
| $\mathbf{B}_{63}$ | = | $x_{12}\,\mathbf{a}_{1}+\left(\frac{1}{2}-y_{12}\right)\mathbf{a}_{2}+z_{12}\,\mathbf{a}_{3}$ | = | $x_{12}a\,\hat{\mathbf{x}}+\left(\frac{1}{2}-y_{12}\right)b\,\hat{\mathbf{y}}+z_{12}c\,\hat{\mathbf{z}}$ | (8$d$) | H VI |
| $\mathbf{B}_{64}$ | = | $\left(\frac{1}{2}-x_{12}\right)\mathbf{a}_{1}+\left(\frac{1}{2}+y_{12}\right)\mathbf{a}_{2}+\left(\frac{1}{2}+z_{12}\right)\mathbf{a}_{3}$ | = | $\left(\frac{1}{2}-x_{12}\right)a\,\hat{\mathbf{x}}+\left(\frac{1}{2}+y_{12}\right)b\,\hat{\mathbf{y}}+\left(\frac{1}{2}+z_{12}\right)c\,\hat{\mathbf{z}}$ | (8$d$) | H VI |
| $\mathbf{B}_{65}$ | = | $x_{13}\,\mathbf{a}_{1}+y_{13}\,\mathbf{a}_{2}+z_{13}\,\mathbf{a}_{3}$ | = | $x_{13}a\,\hat{\mathbf{x}}+y_{13}b\,\hat{\mathbf{y}}+z_{13}c\,\hat{\mathbf{z}}$ | (8$d$) | H VII |
| $\mathbf{B}_{66}$ | = | $\left(\frac{1}{2}-x_{13}\right)\mathbf{a}_1-y_{13}\,\mathbf{a}_2+\left(\frac{1}{2}+z_{13}\right)\mathbf{a}_3$ | = | $\left(\frac{1}{2}-x_{13}\right)a\,\hat{\mathbf{x}}-y_{13}b\,\hat{\mathbf{y}}+\left(\frac{1}{2}+z_{13}\right)c\,\hat{\mathbf{z}}$ | (8*d*) | H VII |
| $\mathbf{B}_{67}$ | = | $-x_{13}\,\mathbf{a}_1+\left(\frac{1}{2}+y_{13}\right)\mathbf{a}_2-z_{13}\,\mathbf{a}_3$ | = | $-x_{13}a\,\hat{\mathbf{x}}+\left(\frac{1}{2}+y_{13}\right)b\,\hat{\mathbf{y}}-z_{13}c\,\hat{\mathbf{z}}$ | (8*d*) | H VII |
| $\mathbf{B}_{68}$ | = | $\left(\frac{1}{2}+x_{13}\right)\mathbf{a}_1+\left(\frac{1}{2}-y_{13}\right)\mathbf{a}_2+\left(\frac{1}{2}-z_{13}\right)\mathbf{a}_3$ | = | $\left(\frac{1}{2}+x_{13}\right)a\,\hat{\mathbf{x}}+\left(\frac{1}{2}-y_{13}\right)b\,\hat{\mathbf{y}}+\left(\frac{1}{2}-z_{13}\right)c\,\hat{\mathbf{z}}$ | (8*d*) | H VII |
| $\mathbf{B}_{69}$ | = | $-x_{13}\,\mathbf{a}_1-y_{13}\,\mathbf{a}_2-z_{13}\,\mathbf{a}_3$ | = | $-x_{13}a\,\hat{\mathbf{x}}-y_{13}b\,\hat{\mathbf{y}}-z_{13}c\,\hat{\mathbf{z}}$ | (8*d*) | H VII |
| $\mathbf{B}_{70}$ | = | $\left(\frac{1}{2}+x_{13}\right)\mathbf{a}_1+y_{13}\,\mathbf{a}_2+\left(\frac{1}{2}-z_{13}\right)\mathbf{a}_3$ | = | $\left(\frac{1}{2}+x_{13}\right)a\,\hat{\mathbf{x}}+y_{13}b\,\hat{\mathbf{y}}+\left(\frac{1}{2}-z_{13}\right)c\,\hat{\mathbf{z}}$ | (8*d*) | H VII |
| $\mathbf{B}_{71}$ | = | $x_{13}\,\mathbf{a}_1+\left(\frac{1}{2}-y_{13}\right)\mathbf{a}_2+z_{13}\,\mathbf{a}_3$ | = | $x_{13}a\,\hat{\mathbf{x}}+\left(\frac{1}{2}-y_{13}\right)b\,\hat{\mathbf{y}}+z_{13}c\,\hat{\mathbf{z}}$ | (8*d*) | H VII |
| $\mathbf{B}_{72}$ | = | $\left(\frac{1}{2}-x_{13}\right)\mathbf{a}_1+\left(\frac{1}{2}+y_{13}\right)\mathbf{a}_2+\left(\frac{1}{2}+z_{13}\right)\mathbf{a}_3$ | = | $\left(\frac{1}{2}-x_{13}\right)a\,\hat{\mathbf{x}}+\left(\frac{1}{2}+y_{13}\right)b\,\hat{\mathbf{y}}+\left(\frac{1}{2}+z_{13}\right)c\,\hat{\mathbf{z}}$ | (8*d*) | H VII |
| $\mathbf{B}_{73}$ | = | $x_{14}\,\mathbf{a}_1+y_{14}\,\mathbf{a}_2+z_{14}\,\mathbf{a}_3$ | = | $x_{14}a\,\hat{\mathbf{x}}+y_{14}b\,\hat{\mathbf{y}}+z_{14}c\,\hat{\mathbf{z}}$ | (8*d*) | H VIII |
| $\mathbf{B}_{74}$ | = | $\left(\frac{1}{2}-x_{14}\right)\mathbf{a}_1-y_{14}\,\mathbf{a}_2+\left(\frac{1}{2}+z_{14}\right)\mathbf{a}_3$ | = | $\left(\frac{1}{2}-x_{14}\right)a\,\hat{\mathbf{x}}-y_{14}b\,\hat{\mathbf{y}}+\left(\frac{1}{2}+z_{14}\right)c\,\hat{\mathbf{z}}$ | (8*d*) | H VIII |
| $\mathbf{B}_{75}$ | = | $-x_{14}\,\mathbf{a}_1+\left(\frac{1}{2}+y_{14}\right)\mathbf{a}_2-z_{14}\,\mathbf{a}_3$ | = | $-x_{14}a\,\hat{\mathbf{x}}+\left(\frac{1}{2}+y_{14}\right)b\,\hat{\mathbf{y}}-z_{14}c\,\hat{\mathbf{z}}$ | (8*d*) | H VIII |
| $\mathbf{B}_{76}$ | = | $\left(\frac{1}{2}+x_{14}\right)\mathbf{a}_1+\left(\frac{1}{2}-y_{14}\right)\mathbf{a}_2+\left(\frac{1}{2}-z_{14}\right)\mathbf{a}_3$ | = | $\left(\frac{1}{2}+x_{14}\right)a\,\hat{\mathbf{x}}+\left(\frac{1}{2}-y_{14}\right)b\,\hat{\mathbf{y}}+\left(\frac{1}{2}-z_{14}\right)c\,\hat{\mathbf{z}}$ | (8*d*) | H VIII |
| $\mathbf{B}_{77}$ | = | $-x_{14}\,\mathbf{a}_1-y_{14}\,\mathbf{a}_2-z_{14}\,\mathbf{a}_3$ | = | $-x_{14}a\,\hat{\mathbf{x}}-y_{14}b\,\hat{\mathbf{y}}-z_{14}c\,\hat{\mathbf{z}}$ | (8*d*) | H VIII |
| $\mathbf{B}_{78}$ | = | $\left(\frac{1}{2}+x_{14}\right)\mathbf{a}_1+y_{14}\,\mathbf{a}_2+\left(\frac{1}{2}-z_{14}\right)\mathbf{a}_3$ | = | $\left(\frac{1}{2}+x_{14}\right)a\,\hat{\mathbf{x}}+y_{14}b\,\hat{\mathbf{y}}+\left(\frac{1}{2}-z_{14}\right)c\,\hat{\mathbf{z}}$ | (8*d*) | H VIII |
| $\mathbf{B}_{79}$ | = | $x_{14}\,\mathbf{a}_1+\left(\frac{1}{2}-y_{14}\right)\mathbf{a}_2+z_{14}\,\mathbf{a}_3$ | = | $x_{14}a\,\hat{\mathbf{x}}+\left(\frac{1}{2}-y_{14}\right)b\,\hat{\mathbf{y}}+z_{14}c\,\hat{\mathbf{z}}$ | (8*d*) | H VIII |
| $\mathbf{B}_{80}$ | = | $\left(\frac{1}{2}-x_{14}\right)\mathbf{a}_1+\left(\frac{1}{2}+y_{14}\right)\mathbf{a}_2+\left(\frac{1}{2}+z_{14}\right)\mathbf{a}_3$ | = | $\left(\frac{1}{2}-x_{14}\right)a\,\hat{\mathbf{x}}+\left(\frac{1}{2}+y_{14}\right)b\,\hat{\mathbf{y}}+\left(\frac{1}{2}+z_{14}\right)c\,\hat{\mathbf{z}}$ | (8*d*) | H VIII |
| $\mathbf{B}_{81}$ | = | $x_{15}\,\mathbf{a}_1+y_{15}\,\mathbf{a}_2+z_{15}\,\mathbf{a}_3$ | = | $x_{15}a\,\hat{\mathbf{x}}+y_{15}b\,\hat{\mathbf{y}}+z_{15}c\,\hat{\mathbf{z}}$ | (8*d*) | H IX |
| $\mathbf{B}_{82}$ | = | $\left(\frac{1}{2}-x_{15}\right)\mathbf{a}_1-y_{15}\,\mathbf{a}_2+\left(\frac{1}{2}+z_{15}\right)\mathbf{a}_3$ | = | $\left(\frac{1}{2}-x_{15}\right)a\,\hat{\mathbf{x}}-y_{15}b\,\hat{\mathbf{y}}+\left(\frac{1}{2}+z_{15}\right)c\,\hat{\mathbf{z}}$ | (8*d*) | H IX |
| $\mathbf{B}_{83}$ | = | $-x_{15}\,\mathbf{a}_1+\left(\frac{1}{2}+y_{15}\right)\mathbf{a}_2-z_{15}\,\mathbf{a}_3$ | = | $-x_{15}a\,\hat{\mathbf{x}}+\left(\frac{1}{2}+y_{15}\right)b\,\hat{\mathbf{y}}-z_{15}c\,\hat{\mathbf{z}}$ | (8*d*) | H IX |
| $\mathbf{B}_{84}$ | = | $\left(\frac{1}{2}+x_{15}\right)\mathbf{a}_1+\left(\frac{1}{2}-y_{15}\right)\mathbf{a}_2+\left(\frac{1}{2}-z_{15}\right)\mathbf{a}_3$ | = | $\left(\frac{1}{2}+x_{15}\right)a\,\hat{\mathbf{x}}+\left(\frac{1}{2}-y_{15}\right)b\,\hat{\mathbf{y}}+\left(\frac{1}{2}-z_{15}\right)c\,\hat{\mathbf{z}}$ | (8*d*) | H IX |
| $\mathbf{B}_{85}$ | = | $-x_{15}\,\mathbf{a}_1-y_{15}\,\mathbf{a}_2-z_{15}\,\mathbf{a}_3$ | = | $-x_{15}a\,\hat{\mathbf{x}}-y_{15}b\,\hat{\mathbf{y}}-z_{15}c\,\hat{\mathbf{z}}$ | (8*d*) | H IX |
| $\mathbf{B}_{86}$ | = | $\left(\frac{1}{2}+x_{15}\right)\mathbf{a}_1+y_{15}\,\mathbf{a}_2+\left(\frac{1}{2}-z_{15}\right)\mathbf{a}_3$ | = | $\left(\frac{1}{2}+x_{15}\right)a\,\hat{\mathbf{x}}+y_{15}b\,\hat{\mathbf{y}}+\left(\frac{1}{2}-z_{15}\right)c\,\hat{\mathbf{z}}$ | (8*d*) | H IX |
| $\mathbf{B}_{87}$ | = | $x_{15}\,\mathbf{a}_1+\left(\frac{1}{2}-y_{15}\right)\mathbf{a}_2+z_{15}\,\mathbf{a}_3$ | = | $x_{15}a\,\hat{\mathbf{x}}+\left(\frac{1}{2}-y_{15}\right)b\,\hat{\mathbf{y}}+z_{15}c\,\hat{\mathbf{z}}$ | (8*d*) | H IX |
| $\mathbf{B}_{88}$ | = | $\left(\frac{1}{2}-x_{15}\right)\mathbf{a}_1+\left(\frac{1}{2}+y_{15}\right)\mathbf{a}_2+\left(\frac{1}{2}+z_{15}\right)\mathbf{a}_3$ | = | $\left(\frac{1}{2}-x_{15}\right)a\,\hat{\mathbf{x}}+\left(\frac{1}{2}+y_{15}\right)b\,\hat{\mathbf{y}}+\left(\frac{1}{2}+z_{15}\right)c\,\hat{\mathbf{z}}$ | (8*d*) | H IX |
| $\mathbf{B}_{89}$ | = | $x_{16}\,\mathbf{a}_1+y_{16}\,\mathbf{a}_2+z_{16}\,\mathbf{a}_3$ | = | $x_{16}a\,\hat{\mathbf{x}}+y_{16}b\,\hat{\mathbf{y}}+z_{16}c\,\hat{\mathbf{z}}$ | (8*d*) | N IV |
| $\mathbf{B}_{90}$ | = | $\left(\frac{1}{2}-x_{16}\right)\mathbf{a}_1-y_{16}\,\mathbf{a}_2+\left(\frac{1}{2}+z_{16}\right)\mathbf{a}_3$ | = | $\left(\frac{1}{2}-x_{16}\right)a\,\hat{\mathbf{x}}-y_{16}b\,\hat{\mathbf{y}}+\left(\frac{1}{2}+z_{16}\right)c\,\hat{\mathbf{z}}$ | (8*d*) | N IV |
| $\mathbf{B}_{91}$ | = | $-x_{16}\,\mathbf{a}_1+\left(\frac{1}{2}+y_{16}\right)\mathbf{a}_2-z_{16}\,\mathbf{a}_3$ | = | $-x_{16}a\,\hat{\mathbf{x}}+\left(\frac{1}{2}+y_{16}\right)b\,\hat{\mathbf{y}}-z_{16}c\,\hat{\mathbf{z}}$ | (8*d*) | N IV |
| $\mathbf{B}_{92}$ | = | $\left(\frac{1}{2}+x_{16}\right)\mathbf{a}_1+\left(\frac{1}{2}-y_{16}\right)\mathbf{a}_2+\left(\frac{1}{2}-z_{16}\right)\mathbf{a}_3$ | = | $\left(\frac{1}{2}+x_{16}\right)a\,\hat{\mathbf{x}}+\left(\frac{1}{2}-y_{16}\right)b\,\hat{\mathbf{y}}+\left(\frac{1}{2}-z_{16}\right)c\,\hat{\mathbf{z}}$ | (8*d*) | N IV |
| $\mathbf{B}_{93}$ | = | $-x_{16}\,\mathbf{a}_1-y_{16}\,\mathbf{a}_2-z_{16}\,\mathbf{a}_3$ | = | $-x_{16}a\,\hat{\mathbf{x}}-y_{16}b\,\hat{\mathbf{y}}-z_{16}c\,\hat{\mathbf{z}}$ | (8*d*) | N IV |
| $\mathbf{B}_{94}$ | = | $\left(\frac{1}{2}+x_{16}\right)\mathbf{a}_1+y_{16}\,\mathbf{a}_2+\left(\frac{1}{2}-z_{16}\right)\mathbf{a}_3$ | = | $\left(\frac{1}{2}+x_{16}\right)a\,\hat{\mathbf{x}}+y_{16}b\,\hat{\mathbf{y}}+\left(\frac{1}{2}-z_{16}\right)c\,\hat{\mathbf{z}}$ | (8*d*) | N IV |
| $\mathbf{B}_{95}$ | = | $x_{16}\,\mathbf{a}_1+\left(\frac{1}{2}-y_{16}\right)\mathbf{a}_2+z_{16}\,\mathbf{a}_3$ | = | $x_{16}a\,\hat{\mathbf{x}}+\left(\frac{1}{2}-y_{16}\right)b\,\hat{\mathbf{y}}+z_{16}c\,\hat{\mathbf{z}}$ | (8*d*) | N IV |

| | | | | | |
|---|---|---|---|---|---|
| $\mathbf{B}_{96}$ | = | $\left(\frac{1}{2} - x_{16}\right) \mathbf{a}_1 + \left(\frac{1}{2} + y_{16}\right) \mathbf{a}_2 + \left(\frac{1}{2} + z_{16}\right) \mathbf{a}_3$ | = | $\left(\frac{1}{2} - x_{16}\right) a\, \hat{\mathbf{x}} + \left(\frac{1}{2} + y_{16}\right) b\, \hat{\mathbf{y}} + \left(\frac{1}{2} + z_{16}\right) c\, \hat{\mathbf{z}}$ | (8*d*) | N IV |

**Geometry files:**

- CIF: pp. 1635
- POSCAR: pp. 1636

# $NH_4I_3$ ($D0_{16}$) Structure: A3B_oP16_62_3c_c

http://aflow.org/prototype-encyclopedia/A3B_oP16_62_3c_c

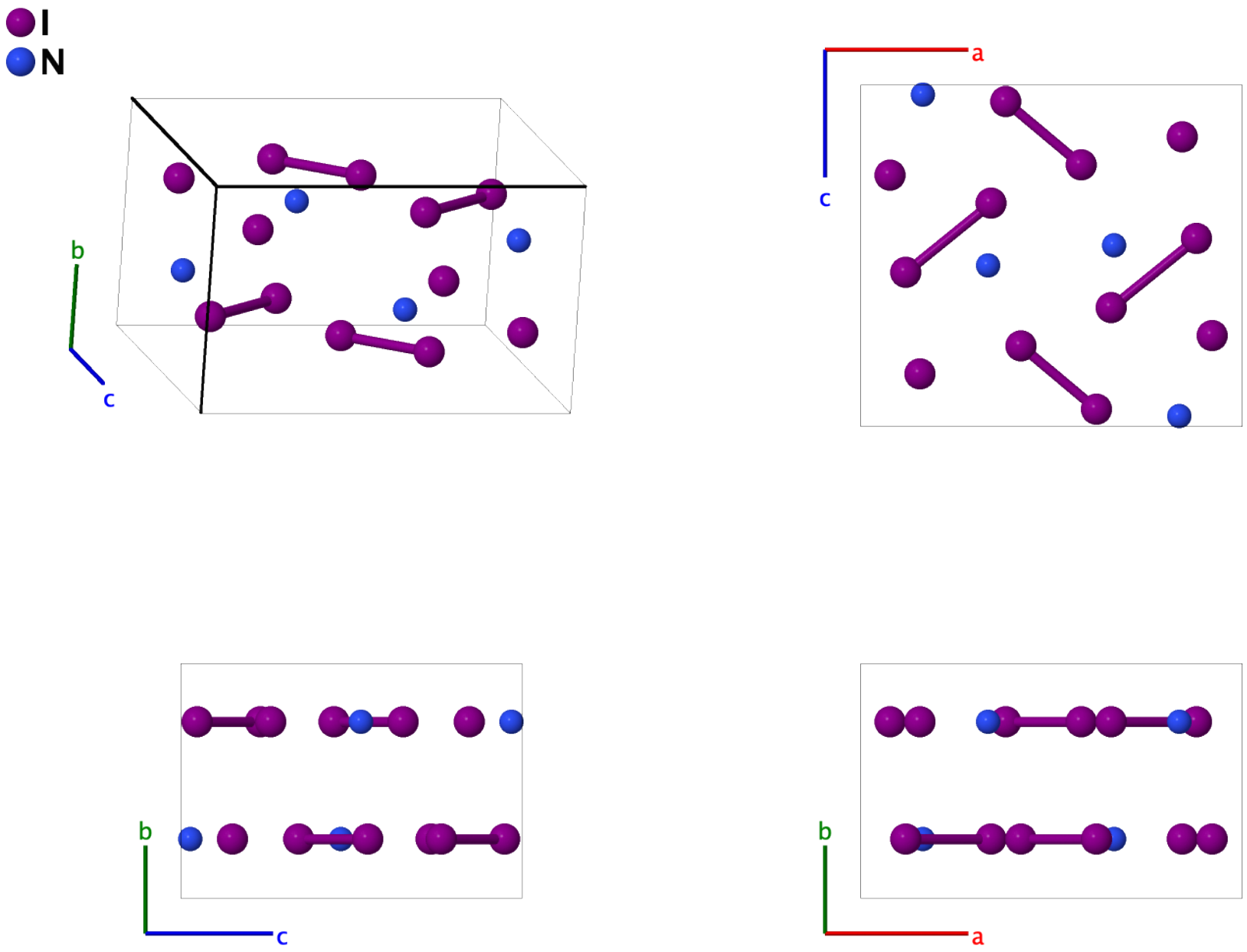

| | | |
|---|---|---|
| **Prototype** | : | $I_3(NH_4)$ |
| **AFLOW prototype label** | : | A3B_oP16_62_3c_c |
| ***Strukturbericht* designation** | : | $D0_{16}$ |
| **Pearson symbol** | : | oP16 |
| **Space group number** | : | 62 |
| **Space group symbol** | : | *Pnma* |
| **AFLOW prototype command** | : | `aflow --proto=A3B_oP16_62_3c_c`<br>`--params=`$a, b/a, c/a, x_1, z_1, x_2, z_2, x_3, z_3, x_4, z_4$ |

### Other compounds with this structure

- $CsI_3$ and $CsBr_3$

---

- (Cheesman, 1970) gives the positions of the atoms in Cartesian coordinates, which we have translated into the standard Wyckoff notation.
- The hydrogen positions are not given, so the $NH_4$ molecules are centered on the (4*c*) Wyckoff position.

### Simple Orthorhombic primitive vectors:

$$\mathbf{a}_1 = a\,\hat{\mathbf{x}}$$
$$\mathbf{a}_2 = b\,\hat{\mathbf{y}}$$
$$\mathbf{a}_3 = c\,\hat{\mathbf{z}}$$

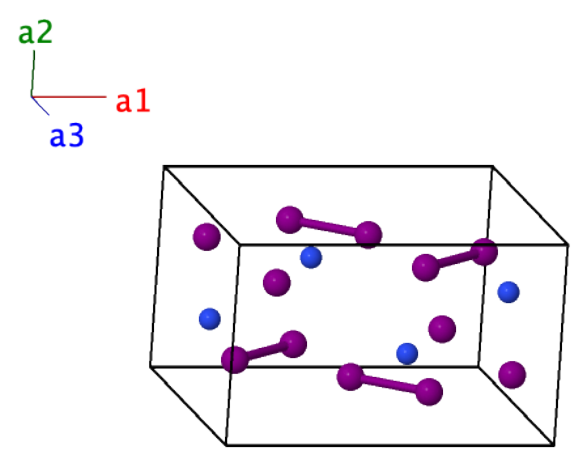

### Basis vectors:

| | | Lattice Coordinates | | Cartesian Coordinates | Wyckoff Position | Atom Type |
|---|---|---|---|---|---|---|
| $\mathbf{B}_1$ | = | $x_1\,\mathbf{a}_1 + \frac{1}{4}\,\mathbf{a}_2 + z_1\,\mathbf{a}_3$ | = | $x_1 a\,\hat{\mathbf{x}} + \frac{1}{4} b\,\hat{\mathbf{y}} + z_1 c\,\hat{\mathbf{z}}$ | (4c) | I I |
| $\mathbf{B}_2$ | = | $\left(\frac{1}{2} - x_1\right)\mathbf{a}_1 + \frac{3}{4}\,\mathbf{a}_2 + \left(\frac{1}{2} + z_1\right)\mathbf{a}_3$ | = | $\left(\frac{1}{2} - x_1\right) a\,\hat{\mathbf{x}} + \frac{3}{4} b\,\hat{\mathbf{y}} + \left(\frac{1}{2} + z_1\right) c\,\hat{\mathbf{z}}$ | (4c) | I I |
| $\mathbf{B}_3$ | = | $-x_1\,\mathbf{a}_1 + \frac{3}{4}\,\mathbf{a}_2 - z_1\,\mathbf{a}_3$ | = | $-x_1 a\,\hat{\mathbf{x}} + \frac{3}{4} b\,\hat{\mathbf{y}} - z_1 c\,\hat{\mathbf{z}}$ | (4c) | I I |
| $\mathbf{B}_4$ | = | $\left(\frac{1}{2} + x_1\right)\mathbf{a}_1 + \frac{1}{4}\,\mathbf{a}_2 + \left(\frac{1}{2} - z_1\right)\mathbf{a}_3$ | = | $\left(\frac{1}{2} + x_1\right) a\,\hat{\mathbf{x}} + \frac{1}{4} b\,\hat{\mathbf{y}} + \left(\frac{1}{2} - z_1\right) c\,\hat{\mathbf{z}}$ | (4c) | I I |
| $\mathbf{B}_5$ | = | $x_2\,\mathbf{a}_1 + \frac{1}{4}\,\mathbf{a}_2 + z_2\,\mathbf{a}_3$ | = | $x_2 a\,\hat{\mathbf{x}} + \frac{1}{4} b\,\hat{\mathbf{y}} + z_2 c\,\hat{\mathbf{z}}$ | (4c) | I II |
| $\mathbf{B}_6$ | = | $\left(\frac{1}{2} - x_2\right)\mathbf{a}_1 + \frac{3}{4}\,\mathbf{a}_2 + \left(\frac{1}{2} + z_2\right)\mathbf{a}_3$ | = | $\left(\frac{1}{2} - x_2\right) a\,\hat{\mathbf{x}} + \frac{3}{4} b\,\hat{\mathbf{y}} + \left(\frac{1}{2} + z_2\right) c\,\hat{\mathbf{z}}$ | (4c) | I II |
| $\mathbf{B}_7$ | = | $-x_2\,\mathbf{a}_1 + \frac{3}{4}\,\mathbf{a}_2 - z_2\,\mathbf{a}_3$ | = | $-x_2 a\,\hat{\mathbf{x}} + \frac{3}{4} b\,\hat{\mathbf{y}} - z_2 c\,\hat{\mathbf{z}}$ | (4c) | I II |
| $\mathbf{B}_8$ | = | $\left(\frac{1}{2} + x_2\right)\mathbf{a}_1 + \frac{1}{4}\,\mathbf{a}_2 + \left(\frac{1}{2} - z_2\right)\mathbf{a}_3$ | = | $\left(\frac{1}{2} + x_2\right) a\,\hat{\mathbf{x}} + \frac{1}{4} b\,\hat{\mathbf{y}} + \left(\frac{1}{2} - z_2\right) c\,\hat{\mathbf{z}}$ | (4c) | I II |
| $\mathbf{B}_9$ | = | $x_3\,\mathbf{a}_1 + \frac{1}{4}\,\mathbf{a}_2 + z_3\,\mathbf{a}_3$ | = | $x_3 a\,\hat{\mathbf{x}} + \frac{1}{4} b\,\hat{\mathbf{y}} + z_3 c\,\hat{\mathbf{z}}$ | (4c) | I III |
| $\mathbf{B}_{10}$ | = | $\left(\frac{1}{2} - x_3\right)\mathbf{a}_1 + \frac{3}{4}\,\mathbf{a}_2 + \left(\frac{1}{2} + z_3\right)\mathbf{a}_3$ | = | $\left(\frac{1}{2} - x_3\right) a\,\hat{\mathbf{x}} + \frac{3}{4} b\,\hat{\mathbf{y}} + \left(\frac{1}{2} + z_3\right) c\,\hat{\mathbf{z}}$ | (4c) | I III |
| $\mathbf{B}_{11}$ | = | $-x_3\,\mathbf{a}_1 + \frac{3}{4}\,\mathbf{a}_2 - z_3\,\mathbf{a}_3$ | = | $-x_3 a\,\hat{\mathbf{x}} + \frac{3}{4} b\,\hat{\mathbf{y}} - z_3 c\,\hat{\mathbf{z}}$ | (4c) | I III |
| $\mathbf{B}_{12}$ | = | $\left(\frac{1}{2} + x_3\right)\mathbf{a}_1 + \frac{1}{4}\,\mathbf{a}_2 + \left(\frac{1}{2} - z_3\right)\mathbf{a}_3$ | = | $\left(\frac{1}{2} + x_3\right) a\,\hat{\mathbf{x}} + \frac{1}{4} b\,\hat{\mathbf{y}} + \left(\frac{1}{2} - z_3\right) c\,\hat{\mathbf{z}}$ | (4c) | I III |
| $\mathbf{B}_{13}$ | = | $x_4\,\mathbf{a}_1 + \frac{1}{4}\,\mathbf{a}_2 + z_4\,\mathbf{a}_3$ | = | $x_4 a\,\hat{\mathbf{x}} + \frac{1}{4} b\,\hat{\mathbf{y}} + z_4 c\,\hat{\mathbf{z}}$ | (4c) | $NH_4$ |
| $\mathbf{B}_{14}$ | = | $\left(\frac{1}{2} - x_4\right)\mathbf{a}_1 + \frac{3}{4}\,\mathbf{a}_2 + \left(\frac{1}{2} + z_4\right)\mathbf{a}_3$ | = | $\left(\frac{1}{2} - x_4\right) a\,\hat{\mathbf{x}} + \frac{3}{4} b\,\hat{\mathbf{y}} + \left(\frac{1}{2} + z_4\right) c\,\hat{\mathbf{z}}$ | (4c) | $NH_4$ |
| $\mathbf{B}_{15}$ | = | $-x_4\,\mathbf{a}_1 + \frac{3}{4}\,\mathbf{a}_2 - z_4\,\mathbf{a}_3$ | = | $-x_4 a\,\hat{\mathbf{x}} + \frac{3}{4} b\,\hat{\mathbf{y}} - z_4 c\,\hat{\mathbf{z}}$ | (4c) | $NH_4$ |
| $\mathbf{B}_{16}$ | = | $\left(\frac{1}{2} + x_4\right)\mathbf{a}_1 + \frac{1}{4}\,\mathbf{a}_2 + \left(\frac{1}{2} - z_4\right)\mathbf{a}_3$ | = | $\left(\frac{1}{2} + x_4\right) a\,\hat{\mathbf{x}} + \frac{1}{4} b\,\hat{\mathbf{y}} + \left(\frac{1}{2} - z_4\right) c\,\hat{\mathbf{z}}$ | (4c) | $NH_4$ |

**Geometry files:**

- CIF: pp. 1636
- POSCAR: pp. 1637

# Original $\beta$-$WO_3$ (*obsolete*) Structure: A3B_oP32_62_ab4c_2c

http://aflow.org/prototype-encyclopedia/A3B_oP32_62_ab4c_2c

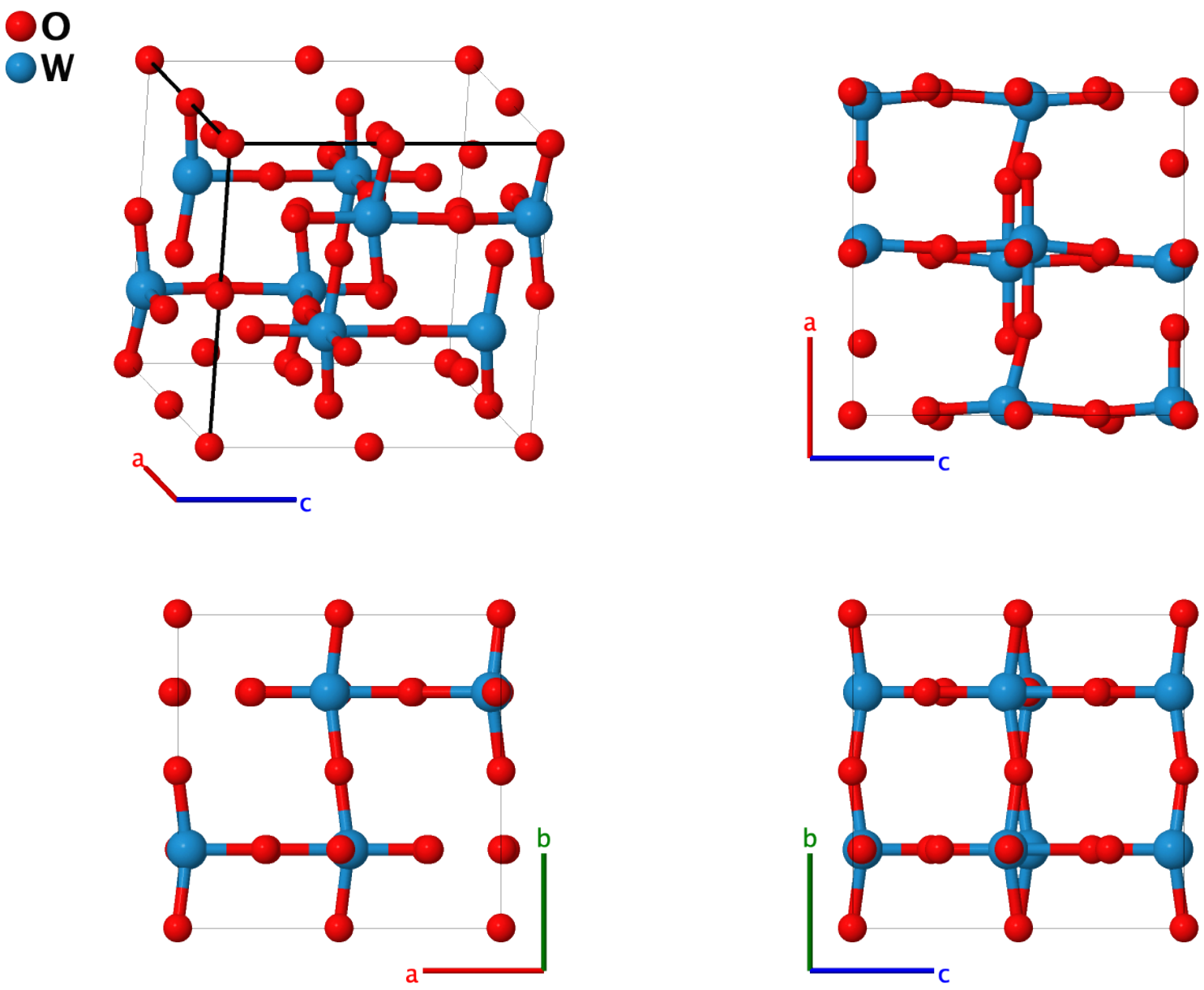

| | | |
|---|---|---|
| **Prototype** | : | $O_3W$ |
| **AFLOW prototype label** | : | A3B_oP32_62_ab4c_2c |
| ***Strukturbericht* designation** | : | None |
| **Pearson symbol** | : | oP32 |
| **Space group number** | : | 62 |
| **Space group symbol** | : | *Pnma* |
| **AFLOW prototype command** | : | `aflow --proto=A3B_oP32_62_ab4c_2c` <br> `--params=`$a, b/a, c/a, x_3, z_3, x_4, z_4, x_5, z_5, x_6, z_6, x_7, z_7, x_8, z_8$ |

- All stable phases of $WO_3$ are distortions of the cubic $\alpha$-$ReO_3$ ($D0_9$) phase. (Woodward, 1997 and Vogt, 1999) The known stable phases and their approximate temperature ranges are:
  - $\alpha$-$WO_3$ (1010-1170 K) (Vogt, 1999)
  - $\beta$-$WO_3$ (600-1170 K) (Vogt, 1999)
  - $\gamma$-$WO_3$ (290-600 K) (Vogt, 1999)
  - $\delta$-$WO_3$ (230-290 K) (Diehl, 1978)
  - $\epsilon$-$WO_3$ (below 23 K) (Woodward, 1997)
- In addition, several other structures have been proposed and/or found:
  - The original $D0_{10}$ structure (Bräkken, 1931), (Hermann, 1937) superseded by $\delta$-$WO_3$
  - Original $\beta$-$WO_3$ (Salje, 1977), this structure
  - Hexagonal $WO_3$ (Gerand, 1979) (metastable)

## Simple Orthorhombic primitive vectors:

$$
\begin{aligned}
\mathbf{a}_1 &= a\,\hat{\mathbf{x}} \\
\mathbf{a}_2 &= b\,\hat{\mathbf{y}} \\
\mathbf{a}_3 &= c\,\hat{\mathbf{z}}
\end{aligned}
$$

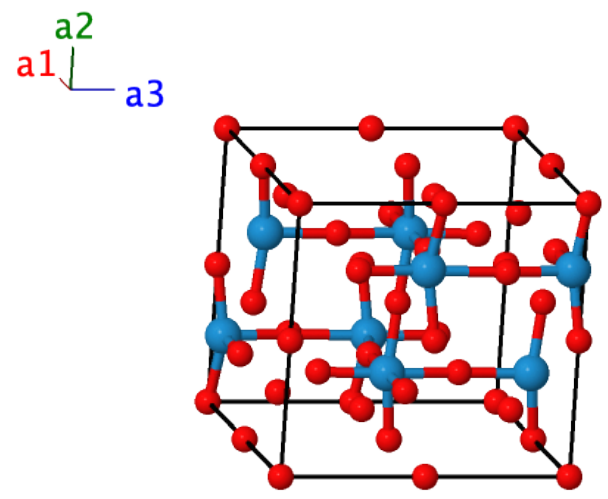

## Basis vectors:

| | | Lattice Coordinates | | Cartesian Coordinates | Wyckoff Position | Atom Type |
|---|---|---|---|---|---|---|
| $\mathbf{B}_1$ | = | $0\,\mathbf{a}_1 + 0\,\mathbf{a}_2 + 0\,\mathbf{a}_3$ | = | $0\,\hat{\mathbf{x}} + 0\,\hat{\mathbf{y}} + 0\,\hat{\mathbf{z}}$ | (4$a$) | O I |
| $\mathbf{B}_2$ | = | $\frac{1}{2}\,\mathbf{a}_1 + \frac{1}{2}\,\mathbf{a}_3$ | = | $\frac{1}{2}a\,\hat{\mathbf{x}} + \frac{1}{2}c\,\hat{\mathbf{z}}$ | (4$a$) | O I |
| $\mathbf{B}_3$ | = | $\frac{1}{2}\,\mathbf{a}_2$ | = | $\frac{1}{2}b\,\hat{\mathbf{y}}$ | (4$a$) | O I |
| $\mathbf{B}_4$ | = | $\frac{1}{2}\,\mathbf{a}_1 + \frac{1}{2}\,\mathbf{a}_2 + \frac{1}{2}\,\mathbf{a}_3$ | = | $\frac{1}{2}a\,\hat{\mathbf{x}} + \frac{1}{2}b\,\hat{\mathbf{y}} + \frac{1}{2}c\,\hat{\mathbf{z}}$ | (4$a$) | O I |
| $\mathbf{B}_5$ | = | $\frac{1}{2}\,\mathbf{a}_3$ | = | $\frac{1}{2}c\,\hat{\mathbf{z}}$ | (4$b$) | O II |
| $\mathbf{B}_6$ | = | $\frac{1}{2}\,\mathbf{a}_1$ | = | $\frac{1}{2}a\,\hat{\mathbf{x}}$ | (4$b$) | O II |
| $\mathbf{B}_7$ | = | $\frac{1}{2}\,\mathbf{a}_2 + \frac{1}{2}\,\mathbf{a}_3$ | = | $\frac{1}{2}b\,\hat{\mathbf{y}} + \frac{1}{2}c\,\hat{\mathbf{z}}$ | (4$b$) | O II |
| $\mathbf{B}_8$ | = | $\frac{1}{2}\,\mathbf{a}_1 + \frac{1}{2}\,\mathbf{a}_2$ | = | $\frac{1}{2}a\,\hat{\mathbf{x}} + \frac{1}{2}b\,\hat{\mathbf{y}}$ | (4$b$) | O II |
| $\mathbf{B}_9$ | = | $x_3\,\mathbf{a}_1 + \frac{1}{4}\,\mathbf{a}_2 + z_3\,\mathbf{a}_3$ | = | $x_3a\,\hat{\mathbf{x}} + \frac{1}{4}b\,\hat{\mathbf{y}} + z_3c\,\hat{\mathbf{z}}$ | (4$c$) | O III |
| $\mathbf{B}_{10}$ | = | $\left(\frac{1}{2} - x_3\right)\,\mathbf{a}_1 + \frac{3}{4}\,\mathbf{a}_2 + \left(\frac{1}{2} + z_3\right)\,\mathbf{a}_3$ | = | $\left(\frac{1}{2} - x_3\right)a\,\hat{\mathbf{x}} + \frac{3}{4}b\,\hat{\mathbf{y}} + \left(\frac{1}{2} + z_3\right)c\,\hat{\mathbf{z}}$ | (4$c$) | O III |
| $\mathbf{B}_{11}$ | = | $-x_3\,\mathbf{a}_1 + \frac{3}{4}\,\mathbf{a}_2 - z_3\,\mathbf{a}_3$ | = | $-x_3a\,\hat{\mathbf{x}} + \frac{3}{4}b\,\hat{\mathbf{y}} - z_3c\,\hat{\mathbf{z}}$ | (4$c$) | O III |
| $\mathbf{B}_{12}$ | = | $\left(\frac{1}{2} + x_3\right)\,\mathbf{a}_1 + \frac{1}{4}\,\mathbf{a}_2 + \left(\frac{1}{2} - z_3\right)\,\mathbf{a}_3$ | = | $\left(\frac{1}{2} + x_3\right)a\,\hat{\mathbf{x}} + \frac{1}{4}b\,\hat{\mathbf{y}} + \left(\frac{1}{2} - z_3\right)c\,\hat{\mathbf{z}}$ | (4$c$) | O III |
| $\mathbf{B}_{13}$ | = | $x_4\,\mathbf{a}_1 + \frac{1}{4}\,\mathbf{a}_2 + z_4\,\mathbf{a}_3$ | = | $x_4a\,\hat{\mathbf{x}} + \frac{1}{4}b\,\hat{\mathbf{y}} + z_4c\,\hat{\mathbf{z}}$ | (4$c$) | O IV |
| $\mathbf{B}_{14}$ | = | $\left(\frac{1}{2} - x_4\right)\,\mathbf{a}_1 + \frac{3}{4}\,\mathbf{a}_2 + \left(\frac{1}{2} + z_4\right)\,\mathbf{a}_3$ | = | $\left(\frac{1}{2} - x_4\right)a\,\hat{\mathbf{x}} + \frac{3}{4}b\,\hat{\mathbf{y}} + \left(\frac{1}{2} + z_4\right)c\,\hat{\mathbf{z}}$ | (4$c$) | O IV |
| $\mathbf{B}_{15}$ | = | $-x_4\,\mathbf{a}_1 + \frac{3}{4}\,\mathbf{a}_2 - z_4\,\mathbf{a}_3$ | = | $-x_4a\,\hat{\mathbf{x}} + \frac{3}{4}b\,\hat{\mathbf{y}} - z_4c\,\hat{\mathbf{z}}$ | (4$c$) | O IV |
| $\mathbf{B}_{16}$ | = | $\left(\frac{1}{2} + x_4\right)\,\mathbf{a}_1 + \frac{1}{4}\,\mathbf{a}_2 + \left(\frac{1}{2} - z_4\right)\,\mathbf{a}_3$ | = | $\left(\frac{1}{2} + x_4\right)a\,\hat{\mathbf{x}} + \frac{1}{4}b\,\hat{\mathbf{y}} + \left(\frac{1}{2} - z_4\right)c\,\hat{\mathbf{z}}$ | (4$c$) | O IV |
| $\mathbf{B}_{17}$ | = | $x_5\,\mathbf{a}_1 + \frac{1}{4}\,\mathbf{a}_2 + z_5\,\mathbf{a}_3$ | = | $x_5a\,\hat{\mathbf{x}} + \frac{1}{4}b\,\hat{\mathbf{y}} + z_5c\,\hat{\mathbf{z}}$ | (4$c$) | O V |
| $\mathbf{B}_{18}$ | = | $\left(\frac{1}{2} - x_5\right)\,\mathbf{a}_1 + \frac{3}{4}\,\mathbf{a}_2 + \left(\frac{1}{2} + z_5\right)\,\mathbf{a}_3$ | = | $\left(\frac{1}{2} - x_5\right)a\,\hat{\mathbf{x}} + \frac{3}{4}b\,\hat{\mathbf{y}} + \left(\frac{1}{2} + z_5\right)c\,\hat{\mathbf{z}}$ | (4$c$) | O V |
| $\mathbf{B}_{19}$ | = | $-x_5\,\mathbf{a}_1 + \frac{3}{4}\,\mathbf{a}_2 - z_5\,\mathbf{a}_3$ | = | $-x_5a\,\hat{\mathbf{x}} + \frac{3}{4}b\,\hat{\mathbf{y}} - z_5c\,\hat{\mathbf{z}}$ | (4$c$) | O V |
| $\mathbf{B}_{20}$ | = | $\left(\frac{1}{2} + x_5\right)\,\mathbf{a}_1 + \frac{1}{4}\,\mathbf{a}_2 + \left(\frac{1}{2} - z_5\right)\,\mathbf{a}_3$ | = | $\left(\frac{1}{2} + x_5\right)a\,\hat{\mathbf{x}} + \frac{1}{4}b\,\hat{\mathbf{y}} + \left(\frac{1}{2} - z_5\right)c\,\hat{\mathbf{z}}$ | (4$c$) | O V |
| $\mathbf{B}_{21}$ | = | $x_6\,\mathbf{a}_1 + \frac{1}{4}\,\mathbf{a}_2 + z_6\,\mathbf{a}_3$ | = | $x_6a\,\hat{\mathbf{x}} + \frac{1}{4}b\,\hat{\mathbf{y}} + z_6c\,\hat{\mathbf{z}}$ | (4$c$) | O VI |
| $\mathbf{B}_{22}$ | = | $\left(\frac{1}{2} - x_6\right)\,\mathbf{a}_1 + \frac{3}{4}\,\mathbf{a}_2 + \left(\frac{1}{2} + z_6\right)\,\mathbf{a}_3$ | = | $\left(\frac{1}{2} - x_6\right)a\,\hat{\mathbf{x}} + \frac{3}{4}b\,\hat{\mathbf{y}} + \left(\frac{1}{2} + z_6\right)c\,\hat{\mathbf{z}}$ | (4$c$) | O VI |
| $\mathbf{B}_{23}$ | = | $-x_6\,\mathbf{a}_1 + \frac{3}{4}\,\mathbf{a}_2 - z_6\,\mathbf{a}_3$ | = | $-x_6a\,\hat{\mathbf{x}} + \frac{3}{4}b\,\hat{\mathbf{y}} - z_6c\,\hat{\mathbf{z}}$ | (4$c$) | O VI |
| $\mathbf{B}_{24}$ | = | $\left(\frac{1}{2} + x_6\right)\,\mathbf{a}_1 + \frac{1}{4}\,\mathbf{a}_2 + \left(\frac{1}{2} - z_6\right)\,\mathbf{a}_3$ | = | $\left(\frac{1}{2} + x_6\right)a\,\hat{\mathbf{x}} + \frac{1}{4}b\,\hat{\mathbf{y}} + \left(\frac{1}{2} - z_6\right)c\,\hat{\mathbf{z}}$ | (4$c$) | O VI |
| $\mathbf{B}_{25}$ | = | $x_7\,\mathbf{a}_1 + \frac{1}{4}\,\mathbf{a}_2 + z_7\,\mathbf{a}_3$ | = | $x_7a\,\hat{\mathbf{x}} + \frac{1}{4}b\,\hat{\mathbf{y}} + z_7c\,\hat{\mathbf{z}}$ | (4$c$) | W I |
| $\mathbf{B}_{26}$ | = | $\left(\frac{1}{2} - x_7\right)\,\mathbf{a}_1 + \frac{3}{4}\,\mathbf{a}_2 + \left(\frac{1}{2} + z_7\right)\,\mathbf{a}_3$ | = | $\left(\frac{1}{2} - x_7\right)a\,\hat{\mathbf{x}} + \frac{3}{4}b\,\hat{\mathbf{y}} + \left(\frac{1}{2} + z_7\right)c\,\hat{\mathbf{z}}$ | (4$c$) | W I |
| $\mathbf{B}_{27}$ | = | $-x_7\,\mathbf{a}_1 + \frac{3}{4}\,\mathbf{a}_2 - z_7\,\mathbf{a}_3$ | = | $-x_7a\,\hat{\mathbf{x}} + \frac{3}{4}b\,\hat{\mathbf{y}} - z_7c\,\hat{\mathbf{z}}$ | (4$c$) | W I |
| $\mathbf{B}_{28}$ | = | $\left(\frac{1}{2}+x_7\right)\mathbf{a}_1+\frac{1}{4}\mathbf{a}_2+\left(\frac{1}{2}-z_7\right)\mathbf{a}_3$ | = | $\left(\frac{1}{2}+x_7\right)a\,\hat{\mathbf{x}}+\frac{1}{4}b\,\hat{\mathbf{y}}+\left(\frac{1}{2}-z_7\right)c\,\hat{\mathbf{z}}$ | (4$c$) | W I |
| $\mathbf{B}_{29}$ | = | $x_8\,\mathbf{a}_1+\frac{1}{4}\mathbf{a}_2+z_8\,\mathbf{a}_3$ | = | $x_8a\,\hat{\mathbf{x}}+\frac{1}{4}b\,\hat{\mathbf{y}}+z_8c\,\hat{\mathbf{z}}$ | (4$c$) | W II |
| $\mathbf{B}_{30}$ | = | $\left(\frac{1}{2}-x_8\right)\mathbf{a}_1+\frac{3}{4}\mathbf{a}_2+\left(\frac{1}{2}+z_8\right)\mathbf{a}_3$ | = | $\left(\frac{1}{2}-x_8\right)a\,\hat{\mathbf{x}}+\frac{3}{4}b\,\hat{\mathbf{y}}+\left(\frac{1}{2}+z_8\right)c\,\hat{\mathbf{z}}$ | (4$c$) | W II |
| $\mathbf{B}_{31}$ | = | $-x_8\,\mathbf{a}_1+\frac{3}{4}\mathbf{a}_2-z_8\,\mathbf{a}_3$ | = | $-x_8a\,\hat{\mathbf{x}}+\frac{3}{4}b\,\hat{\mathbf{y}}-z_8c\,\hat{\mathbf{z}}$ | (4$c$) | W II |
| $\mathbf{B}_{32}$ | = | $\left(\frac{1}{2}+x_8\right)\mathbf{a}_1+\frac{1}{4}\mathbf{a}_2+\left(\frac{1}{2}-z_8\right)\mathbf{a}_3$ | = | $\left(\frac{1}{2}+x_8\right)a\,\hat{\mathbf{x}}+\frac{1}{4}b\,\hat{\mathbf{y}}+\left(\frac{1}{2}-z_8\right)c\,\hat{\mathbf{z}}$ | (4$c$) | W II |

**Geometry files:**

- CIF: pp. 1637
- POSCAR: pp. 1637

# $P_4Se_3$ Structure: A4B3_oP112_62_8c4d_4c4d

http://aflow.org/prototype-encyclopedia/A4B3_oP112_62_8c4d_4c4d

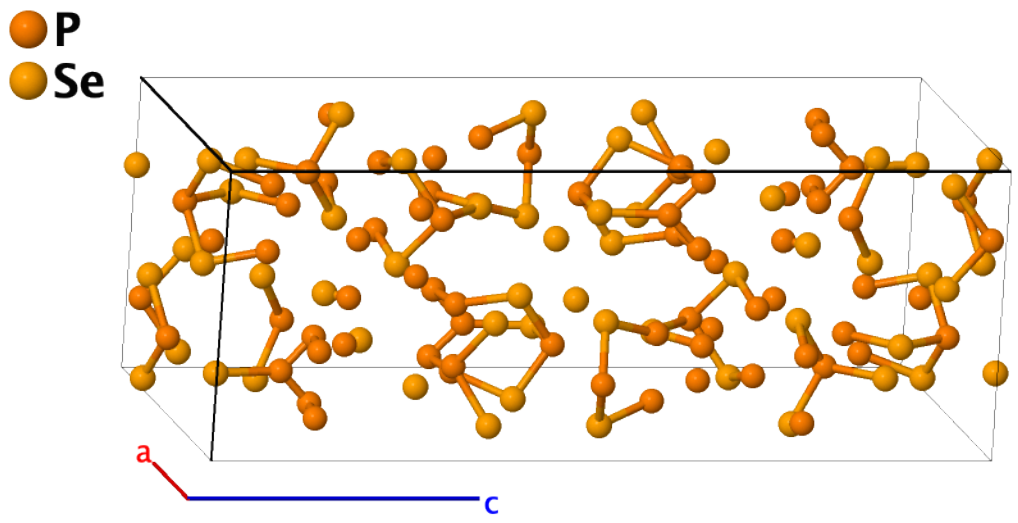

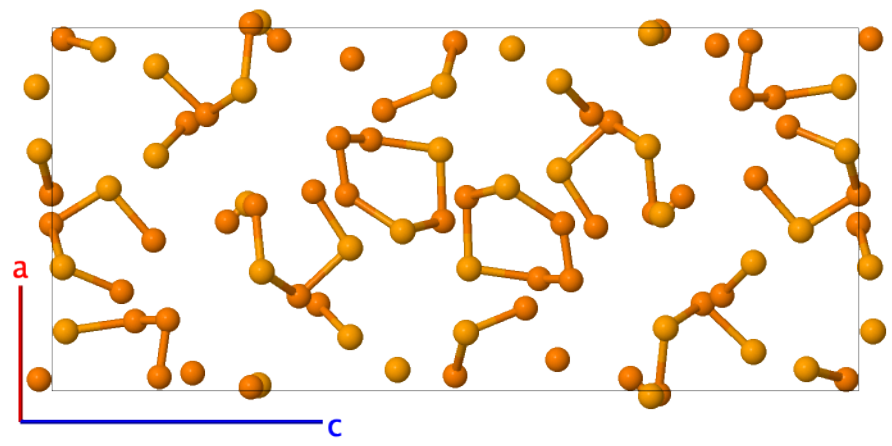

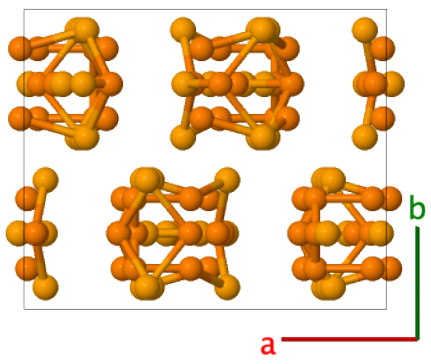

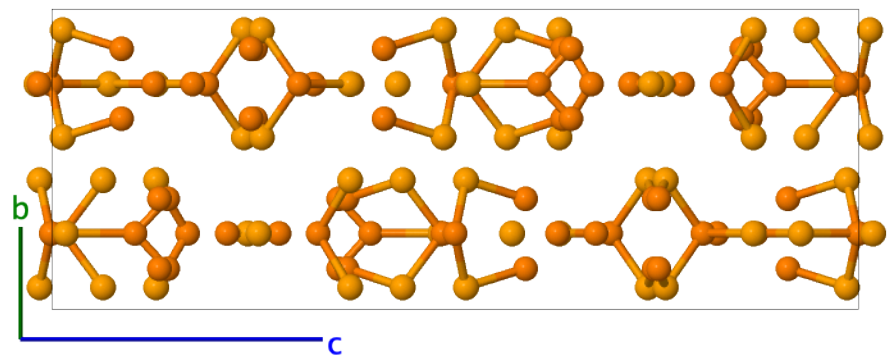

| | | |
|---|---|---|
| **Prototype** | : | $P_4Se_3$ |
| **AFLOW prototype label** | : | A4B3_oP112_62_8c4d_4c4d |
| ***Strukturbericht* designation** | : | None |
| **Pearson symbol** | : | oP112 |
| **Space group number** | : | 62 |
| **Space group symbol** | : | *Pnma* |
| **AFLOW prototype command** | : | `aflow --proto=A4B3_oP112_62_8c4d_4c4d` `--params=`$a, b/a, c/a, x_1, z_1, x_2, z_2, x_3, z_3, x_4, z_4, x_5, z_5, x_6, z_6, x_7, z_7, x_8, z_8, x_9, z_9, x_{10}, z_{10}, x_{11}, z_{11}, x_{12}, z_{12}, x_{13}, y_{13}, z_{13}, x_{14}, y_{14}, z_{14}, x_{15}, y_{15}, z_{15}, x_{16}, y_{16}, z_{16}, x_{17}, y_{17}, z_{17}, x_{18}, y_{18}, z_{18}, x_{19}, y_{19}, z_{19}, x_{20}, y_{20}, z_{20}$ |

- (Keulen, 1959) give this structure in the *Pmnb* setting of space group #62. We used FINDSYM to translate this into the standard *Pnma* setting.

**Simple Orthorhombic primitive vectors:**

$$\mathbf{a}_1 = a\,\hat{\mathbf{x}}$$
$$\mathbf{a}_2 = b\,\hat{\mathbf{y}}$$
$$\mathbf{a}_3 = c\,\hat{\mathbf{z}}$$

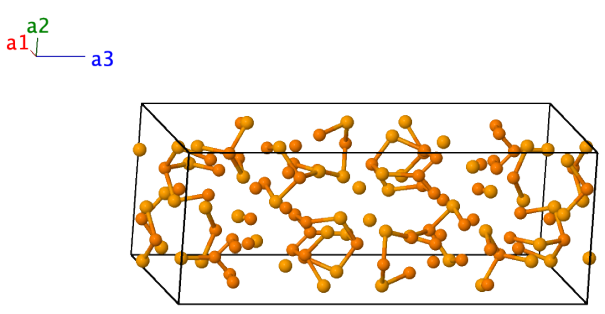

**Basis vectors:**

| | | Lattice Coordinates | | Cartesian Coordinates | Wyckoff Position | Atom Type |
|---|---|---|---|---|---|---|
| $\mathbf{B}_1$ | = | $x_1\,\mathbf{a}_1 + \frac{1}{4}\,\mathbf{a}_2 + z_1\,\mathbf{a}_3$ | = | $x_1 a\,\hat{\mathbf{x}} + \frac{1}{4} b\,\hat{\mathbf{y}} + z_1 c\,\hat{\mathbf{z}}$ | (4c) | P I |
| $\mathbf{B}_2$ | = | $\left(\frac{1}{2} - x_1\right)\mathbf{a}_1 + \frac{3}{4}\,\mathbf{a}_2 + \left(\frac{1}{2} + z_1\right)\mathbf{a}_3$ | = | $\left(\frac{1}{2} - x_1\right) a\,\hat{\mathbf{x}} + \frac{3}{4} b\,\hat{\mathbf{y}} + \left(\frac{1}{2} + z_1\right) c\,\hat{\mathbf{z}}$ | (4c) | P I |
| $\mathbf{B}_3$ | = | $-x_1\,\mathbf{a}_1 + \frac{3}{4}\,\mathbf{a}_2 - z_1\,\mathbf{a}_3$ | = | $-x_1 a\,\hat{\mathbf{x}} + \frac{3}{4} b\,\hat{\mathbf{y}} - z_1 c\,\hat{\mathbf{z}}$ | (4c) | P I |
| $\mathbf{B}_4$ | = | $\left(\frac{1}{2} + x_1\right)\mathbf{a}_1 + \frac{1}{4}\,\mathbf{a}_2 + \left(\frac{1}{2} - z_1\right)\mathbf{a}_3$ | = | $\left(\frac{1}{2} + x_1\right) a\,\hat{\mathbf{x}} + \frac{1}{4} b\,\hat{\mathbf{y}} + \left(\frac{1}{2} - z_1\right) c\,\hat{\mathbf{z}}$ | (4c) | P I |
| $\mathbf{B}_{5}$ | = | $x_{2}\,\mathbf{a}_{1}+\frac{1}{4}\,\mathbf{a}_{2}+z_{2}\,\mathbf{a}_{3}$ | = | $x_{2}a\,\hat{\mathbf{x}}+\frac{1}{4}b\,\hat{\mathbf{y}}+z_{2}c\,\hat{\mathbf{z}}$ | (4$c$) | P II |
| $\mathbf{B}_{6}$ | = | $\left(\frac{1}{2}-x_{2}\right)\mathbf{a}_{1}+\frac{3}{4}\,\mathbf{a}_{2}+\left(\frac{1}{2}+z_{2}\right)\mathbf{a}_{3}$ | = | $\left(\frac{1}{2}-x_{2}\right)a\,\hat{\mathbf{x}}+\frac{3}{4}b\,\hat{\mathbf{y}}+\left(\frac{1}{2}+z_{2}\right)c\,\hat{\mathbf{z}}$ | (4$c$) | P II |
| $\mathbf{B}_{7}$ | = | $-x_{2}\,\mathbf{a}_{1}+\frac{3}{4}\,\mathbf{a}_{2}-z_{2}\,\mathbf{a}_{3}$ | = | $-x_{2}a\,\hat{\mathbf{x}}+\frac{3}{4}b\,\hat{\mathbf{y}}-z_{2}c\,\hat{\mathbf{z}}$ | (4$c$) | P II |
| $\mathbf{B}_{8}$ | = | $\left(\frac{1}{2}+x_{2}\right)\mathbf{a}_{1}+\frac{1}{4}\,\mathbf{a}_{2}+\left(\frac{1}{2}-z_{2}\right)\mathbf{a}_{3}$ | = | $\left(\frac{1}{2}+x_{2}\right)a\,\hat{\mathbf{x}}+\frac{1}{4}b\,\hat{\mathbf{y}}+\left(\frac{1}{2}-z_{2}\right)c\,\hat{\mathbf{z}}$ | (4$c$) | P II |
| $\mathbf{B}_{9}$ | = | $x_{3}\,\mathbf{a}_{1}+\frac{1}{4}\,\mathbf{a}_{2}+z_{3}\,\mathbf{a}_{3}$ | = | $x_{3}a\,\hat{\mathbf{x}}+\frac{1}{4}b\,\hat{\mathbf{y}}+z_{3}c\,\hat{\mathbf{z}}$ | (4$c$) | P III |
| $\mathbf{B}_{10}$ | = | $\left(\frac{1}{2}-x_{3}\right)\mathbf{a}_{1}+\frac{3}{4}\,\mathbf{a}_{2}+\left(\frac{1}{2}+z_{3}\right)\mathbf{a}_{3}$ | = | $\left(\frac{1}{2}-x_{3}\right)a\,\hat{\mathbf{x}}+\frac{3}{4}b\,\hat{\mathbf{y}}+\left(\frac{1}{2}+z_{3}\right)c\,\hat{\mathbf{z}}$ | (4$c$) | P III |
| $\mathbf{B}_{11}$ | = | $-x_{3}\,\mathbf{a}_{1}+\frac{3}{4}\,\mathbf{a}_{2}-z_{3}\,\mathbf{a}_{3}$ | = | $-x_{3}a\,\hat{\mathbf{x}}+\frac{3}{4}b\,\hat{\mathbf{y}}-z_{3}c\,\hat{\mathbf{z}}$ | (4$c$) | P III |
| $\mathbf{B}_{12}$ | = | $\left(\frac{1}{2}+x_{3}\right)\mathbf{a}_{1}+\frac{1}{4}\,\mathbf{a}_{2}+\left(\frac{1}{2}-z_{3}\right)\mathbf{a}_{3}$ | = | $\left(\frac{1}{2}+x_{3}\right)a\,\hat{\mathbf{x}}+\frac{1}{4}b\,\hat{\mathbf{y}}+\left(\frac{1}{2}-z_{3}\right)c\,\hat{\mathbf{z}}$ | (4$c$) | P III |
| $\mathbf{B}_{13}$ | = | $x_{4}\,\mathbf{a}_{1}+\frac{1}{4}\,\mathbf{a}_{2}+z_{4}\,\mathbf{a}_{3}$ | = | $x_{4}a\,\hat{\mathbf{x}}+\frac{1}{4}b\,\hat{\mathbf{y}}+z_{4}c\,\hat{\mathbf{z}}$ | (4$c$) | P IV |
| $\mathbf{B}_{14}$ | = | $\left(\frac{1}{2}-x_{4}\right)\mathbf{a}_{1}+\frac{3}{4}\,\mathbf{a}_{2}+\left(\frac{1}{2}+z_{4}\right)\mathbf{a}_{3}$ | = | $\left(\frac{1}{2}-x_{4}\right)a\,\hat{\mathbf{x}}+\frac{3}{4}b\,\hat{\mathbf{y}}+\left(\frac{1}{2}+z_{4}\right)c\,\hat{\mathbf{z}}$ | (4$c$) | P IV |
| $\mathbf{B}_{15}$ | = | $-x_{4}\,\mathbf{a}_{1}+\frac{3}{4}\,\mathbf{a}_{2}-z_{4}\,\mathbf{a}_{3}$ | = | $-x_{4}a\,\hat{\mathbf{x}}+\frac{3}{4}b\,\hat{\mathbf{y}}-z_{4}c\,\hat{\mathbf{z}}$ | (4$c$) | P IV |
| $\mathbf{B}_{16}$ | = | $\left(\frac{1}{2}+x_{4}\right)\mathbf{a}_{1}+\frac{1}{4}\,\mathbf{a}_{2}+\left(\frac{1}{2}-z_{4}\right)\mathbf{a}_{3}$ | = | $\left(\frac{1}{2}+x_{4}\right)a\,\hat{\mathbf{x}}+\frac{1}{4}b\,\hat{\mathbf{y}}+\left(\frac{1}{2}-z_{4}\right)c\,\hat{\mathbf{z}}$ | (4$c$) | P IV |
| $\mathbf{B}_{17}$ | = | $x_{5}\,\mathbf{a}_{1}+\frac{1}{4}\,\mathbf{a}_{2}+z_{5}\,\mathbf{a}_{3}$ | = | $x_{5}a\,\hat{\mathbf{x}}+\frac{1}{4}b\,\hat{\mathbf{y}}+z_{5}c\,\hat{\mathbf{z}}$ | (4$c$) | P V |
| $\mathbf{B}_{18}$ | = | $\left(\frac{1}{2}-x_{5}\right)\mathbf{a}_{1}+\frac{3}{4}\,\mathbf{a}_{2}+\left(\frac{1}{2}+z_{5}\right)\mathbf{a}_{3}$ | = | $\left(\frac{1}{2}-x_{5}\right)a\,\hat{\mathbf{x}}+\frac{3}{4}b\,\hat{\mathbf{y}}+\left(\frac{1}{2}+z_{5}\right)c\,\hat{\mathbf{z}}$ | (4$c$) | P V |
| $\mathbf{B}_{19}$ | = | $-x_{5}\,\mathbf{a}_{1}+\frac{3}{4}\,\mathbf{a}_{2}-z_{5}\,\mathbf{a}_{3}$ | = | $-x_{5}a\,\hat{\mathbf{x}}+\frac{3}{4}b\,\hat{\mathbf{y}}-z_{5}c\,\hat{\mathbf{z}}$ | (4$c$) | P V |
| $\mathbf{B}_{20}$ | = | $\left(\frac{1}{2}+x_{5}\right)\mathbf{a}_{1}+\frac{1}{4}\,\mathbf{a}_{2}+\left(\frac{1}{2}-z_{5}\right)\mathbf{a}_{3}$ | = | $\left(\frac{1}{2}+x_{5}\right)a\,\hat{\mathbf{x}}+\frac{1}{4}b\,\hat{\mathbf{y}}+\left(\frac{1}{2}-z_{5}\right)c\,\hat{\mathbf{z}}$ | (4$c$) | P V |
| $\mathbf{B}_{21}$ | = | $x_{6}\,\mathbf{a}_{1}+\frac{1}{4}\,\mathbf{a}_{2}+z_{6}\,\mathbf{a}_{3}$ | = | $x_{6}a\,\hat{\mathbf{x}}+\frac{1}{4}b\,\hat{\mathbf{y}}+z_{6}c\,\hat{\mathbf{z}}$ | (4$c$) | P VI |
| $\mathbf{B}_{22}$ | = | $\left(\frac{1}{2}-x_{6}\right)\mathbf{a}_{1}+\frac{3}{4}\,\mathbf{a}_{2}+\left(\frac{1}{2}+z_{6}\right)\mathbf{a}_{3}$ | = | $\left(\frac{1}{2}-x_{6}\right)a\,\hat{\mathbf{x}}+\frac{3}{4}b\,\hat{\mathbf{y}}+\left(\frac{1}{2}+z_{6}\right)c\,\hat{\mathbf{z}}$ | (4$c$) | P VI |
| $\mathbf{B}_{23}$ | = | $-x_{6}\,\mathbf{a}_{1}+\frac{3}{4}\,\mathbf{a}_{2}-z_{6}\,\mathbf{a}_{3}$ | = | $-x_{6}a\,\hat{\mathbf{x}}+\frac{3}{4}b\,\hat{\mathbf{y}}-z_{6}c\,\hat{\mathbf{z}}$ | (4$c$) | P VI |
| $\mathbf{B}_{24}$ | = | $\left(\frac{1}{2}+x_{6}\right)\mathbf{a}_{1}+\frac{1}{4}\,\mathbf{a}_{2}+\left(\frac{1}{2}-z_{6}\right)\mathbf{a}_{3}$ | = | $\left(\frac{1}{2}+x_{6}\right)a\,\hat{\mathbf{x}}+\frac{1}{4}b\,\hat{\mathbf{y}}+\left(\frac{1}{2}-z_{6}\right)c\,\hat{\mathbf{z}}$ | (4$c$) | P VI |
| $\mathbf{B}_{25}$ | = | $x_{7}\,\mathbf{a}_{1}+\frac{1}{4}\,\mathbf{a}_{2}+z_{7}\,\mathbf{a}_{3}$ | = | $x_{7}a\,\hat{\mathbf{x}}+\frac{1}{4}b\,\hat{\mathbf{y}}+z_{7}c\,\hat{\mathbf{z}}$ | (4$c$) | P VII |
| $\mathbf{B}_{26}$ | = | $\left(\frac{1}{2}-x_{7}\right)\mathbf{a}_{1}+\frac{3}{4}\,\mathbf{a}_{2}+\left(\frac{1}{2}+z_{7}\right)\mathbf{a}_{3}$ | = | $\left(\frac{1}{2}-x_{7}\right)a\,\hat{\mathbf{x}}+\frac{3}{4}b\,\hat{\mathbf{y}}+\left(\frac{1}{2}+z_{7}\right)c\,\hat{\mathbf{z}}$ | (4$c$) | P VII |
| $\mathbf{B}_{27}$ | = | $-x_{7}\,\mathbf{a}_{1}+\frac{3}{4}\,\mathbf{a}_{2}-z_{7}\,\mathbf{a}_{3}$ | = | $-x_{7}a\,\hat{\mathbf{x}}+\frac{3}{4}b\,\hat{\mathbf{y}}-z_{7}c\,\hat{\mathbf{z}}$ | (4$c$) | P VII |
| $\mathbf{B}_{28}$ | = | $\left(\frac{1}{2}+x_{7}\right)\mathbf{a}_{1}+\frac{1}{4}\,\mathbf{a}_{2}+\left(\frac{1}{2}-z_{7}\right)\mathbf{a}_{3}$ | = | $\left(\frac{1}{2}+x_{7}\right)a\,\hat{\mathbf{x}}+\frac{1}{4}b\,\hat{\mathbf{y}}+\left(\frac{1}{2}-z_{7}\right)c\,\hat{\mathbf{z}}$ | (4$c$) | P VII |
| $\mathbf{B}_{29}$ | = | $x_{8}\,\mathbf{a}_{1}+\frac{1}{4}\,\mathbf{a}_{2}+z_{8}\,\mathbf{a}_{3}$ | = | $x_{8}a\,\hat{\mathbf{x}}+\frac{1}{4}b\,\hat{\mathbf{y}}+z_{8}c\,\hat{\mathbf{z}}$ | (4$c$) | P VIII |
| $\mathbf{B}_{30}$ | = | $\left(\frac{1}{2}-x_{8}\right)\mathbf{a}_{1}+\frac{3}{4}\,\mathbf{a}_{2}+\left(\frac{1}{2}+z_{8}\right)\mathbf{a}_{3}$ | = | $\left(\frac{1}{2}-x_{8}\right)a\,\hat{\mathbf{x}}+\frac{3}{4}b\,\hat{\mathbf{y}}+\left(\frac{1}{2}+z_{8}\right)c\,\hat{\mathbf{z}}$ | (4$c$) | P VIII |
| $\mathbf{B}_{31}$ | = | $-x_{8}\,\mathbf{a}_{1}+\frac{3}{4}\,\mathbf{a}_{2}-z_{8}\,\mathbf{a}_{3}$ | = | $-x_{8}a\,\hat{\mathbf{x}}+\frac{3}{4}b\,\hat{\mathbf{y}}-z_{8}c\,\hat{\mathbf{z}}$ | (4$c$) | P VIII |
| $\mathbf{B}_{32}$ | = | $\left(\frac{1}{2}+x_{8}\right)\mathbf{a}_{1}+\frac{1}{4}\,\mathbf{a}_{2}+\left(\frac{1}{2}-z_{8}\right)\mathbf{a}_{3}$ | = | $\left(\frac{1}{2}+x_{8}\right)a\,\hat{\mathbf{x}}+\frac{1}{4}b\,\hat{\mathbf{y}}+\left(\frac{1}{2}-z_{8}\right)c\,\hat{\mathbf{z}}$ | (4$c$) | P VIII |
| $\mathbf{B}_{33}$ | = | $x_{9}\,\mathbf{a}_{1}+\frac{1}{4}\,\mathbf{a}_{2}+z_{9}\,\mathbf{a}_{3}$ | = | $x_{9}a\,\hat{\mathbf{x}}+\frac{1}{4}b\,\hat{\mathbf{y}}+z_{9}c\,\hat{\mathbf{z}}$ | (4$c$) | Se I |
| $\mathbf{B}_{34}$ | = | $\left(\frac{1}{2}-x_{9}\right)\mathbf{a}_{1}+\frac{3}{4}\,\mathbf{a}_{2}+\left(\frac{1}{2}+z_{9}\right)\mathbf{a}_{3}$ | = | $\left(\frac{1}{2}-x_{9}\right)a\,\hat{\mathbf{x}}+\frac{3}{4}b\,\hat{\mathbf{y}}+\left(\frac{1}{2}+z_{9}\right)c\,\hat{\mathbf{z}}$ | (4$c$) | Se I |
| $\mathbf{B}_{35}$ | = | $-x_{9}\,\mathbf{a}_{1}+\frac{3}{4}\,\mathbf{a}_{2}-z_{9}\,\mathbf{a}_{3}$ | = | $-x_{9}a\,\hat{\mathbf{x}}+\frac{3}{4}b\,\hat{\mathbf{y}}-z_{9}c\,\hat{\mathbf{z}}$ | (4$c$) | Se I |
| $\mathbf{B}_{36}$ | = | $\left(\frac{1}{2}+x_{9}\right)\mathbf{a}_{1}+\frac{1}{4}\,\mathbf{a}_{2}+\left(\frac{1}{2}-z_{9}\right)\mathbf{a}_{3}$ | = | $\left(\frac{1}{2}+x_{9}\right)a\,\hat{\mathbf{x}}+\frac{1}{4}b\,\hat{\mathbf{y}}+\left(\frac{1}{2}-z_{9}\right)c\,\hat{\mathbf{z}}$ | (4$c$) | Se I |
| $\mathbf{B}_{37}$ | = | $x_{10}\,\mathbf{a}_{1}+\frac{1}{4}\,\mathbf{a}_{2}+z_{10}\,\mathbf{a}_{3}$ | = | $x_{10}a\,\hat{\mathbf{x}}+\frac{1}{4}b\,\hat{\mathbf{y}}+z_{10}c\,\hat{\mathbf{z}}$ | (4$c$) | Se II |
| $\mathbf{B}_{38}$ | = | $\left(\frac{1}{2}-x_{10}\right)\mathbf{a}_{1}+\frac{3}{4}\,\mathbf{a}_{2}+\left(\frac{1}{2}+z_{10}\right)\mathbf{a}_{3}$ | = | $\left(\frac{1}{2}-x_{10}\right)a\,\hat{\mathbf{x}}+\frac{3}{4}b\,\hat{\mathbf{y}}+\left(\frac{1}{2}+z_{10}\right)c\,\hat{\mathbf{z}}$ | (4$c$) | Se II |
| $\mathbf{B}_{39}$ | = | $-x_{10}\,\mathbf{a}_{1}+\frac{3}{4}\,\mathbf{a}_{2}-z_{10}\,\mathbf{a}_{3}$ | = | $-x_{10}a\,\hat{\mathbf{x}}+\frac{3}{4}b\,\hat{\mathbf{y}}-z_{10}c\,\hat{\mathbf{z}}$ | (4$c$) | Se II |
| $\mathbf{B}_{40}$ | = | $\left(\frac{1}{2}+x_{10}\right)\mathbf{a}_{1}+\frac{1}{4}\,\mathbf{a}_{2}+\left(\frac{1}{2}-z_{10}\right)\mathbf{a}_{3}$ | = | $\left(\frac{1}{2}+x_{10}\right)a\,\hat{\mathbf{x}}+\frac{1}{4}b\,\hat{\mathbf{y}}+\left(\frac{1}{2}-z_{10}\right)c\,\hat{\mathbf{z}}$ | (4$c$) | Se II |
| $\mathbf{B}_{41}$ | = | $x_{11}\,\mathbf{a}_1 + \frac{1}{4}\,\mathbf{a}_2 + z_{11}\,\mathbf{a}_3$ | = | $x_{11}a\,\hat{\mathbf{x}} + \frac{1}{4}b\,\hat{\mathbf{y}} + z_{11}c\,\hat{\mathbf{z}}$ | (4$c$) | Se III |
| $\mathbf{B}_{42}$ | = | $\left(\frac{1}{2} - x_{11}\right)\mathbf{a}_1 + \frac{3}{4}\,\mathbf{a}_2 + \left(\frac{1}{2} + z_{11}\right)\mathbf{a}_3$ | = | $\left(\frac{1}{2} - x_{11}\right)a\,\hat{\mathbf{x}} + \frac{3}{4}b\,\hat{\mathbf{y}} + \left(\frac{1}{2} + z_{11}\right)c\,\hat{\mathbf{z}}$ | (4$c$) | Se III |
| $\mathbf{B}_{43}$ | = | $-x_{11}\,\mathbf{a}_1 + \frac{3}{4}\,\mathbf{a}_2 - z_{11}\,\mathbf{a}_3$ | = | $-x_{11}a\,\hat{\mathbf{x}} + \frac{3}{4}b\,\hat{\mathbf{y}} - z_{11}c\,\hat{\mathbf{z}}$ | (4$c$) | Se III |
| $\mathbf{B}_{44}$ | = | $\left(\frac{1}{2} + x_{11}\right)\mathbf{a}_1 + \frac{1}{4}\,\mathbf{a}_2 + \left(\frac{1}{2} - z_{11}\right)\mathbf{a}_3$ | = | $\left(\frac{1}{2} + x_{11}\right)a\,\hat{\mathbf{x}} + \frac{1}{4}b\,\hat{\mathbf{y}} + \left(\frac{1}{2} - z_{11}\right)c\,\hat{\mathbf{z}}$ | (4$c$) | Se III |
| $\mathbf{B}_{45}$ | = | $x_{12}\,\mathbf{a}_1 + \frac{1}{4}\,\mathbf{a}_2 + z_{12}\,\mathbf{a}_3$ | = | $x_{12}a\,\hat{\mathbf{x}} + \frac{1}{4}b\,\hat{\mathbf{y}} + z_{12}c\,\hat{\mathbf{z}}$ | (4$c$) | Se IV |
| $\mathbf{B}_{46}$ | = | $\left(\frac{1}{2} - x_{12}\right)\mathbf{a}_1 + \frac{3}{4}\,\mathbf{a}_2 + \left(\frac{1}{2} + z_{12}\right)\mathbf{a}_3$ | = | $\left(\frac{1}{2} - x_{12}\right)a\,\hat{\mathbf{x}} + \frac{3}{4}b\,\hat{\mathbf{y}} + \left(\frac{1}{2} + z_{12}\right)c\,\hat{\mathbf{z}}$ | (4$c$) | Se IV |
| $\mathbf{B}_{47}$ | = | $-x_{12}\,\mathbf{a}_1 + \frac{3}{4}\,\mathbf{a}_2 - z_{12}\,\mathbf{a}_3$ | = | $-x_{12}a\,\hat{\mathbf{x}} + \frac{3}{4}b\,\hat{\mathbf{y}} - z_{12}c\,\hat{\mathbf{z}}$ | (4$c$) | Se IV |
| $\mathbf{B}_{48}$ | = | $\left(\frac{1}{2} + x_{12}\right)\mathbf{a}_1 + \frac{1}{4}\,\mathbf{a}_2 + \left(\frac{1}{2} - z_{12}\right)\mathbf{a}_3$ | = | $\left(\frac{1}{2} + x_{12}\right)a\,\hat{\mathbf{x}} + \frac{1}{4}b\,\hat{\mathbf{y}} + \left(\frac{1}{2} - z_{12}\right)c\,\hat{\mathbf{z}}$ | (4$c$) | Se IV |
| $\mathbf{B}_{49}$ | = | $x_{13}\,\mathbf{a}_1 + y_{13}\,\mathbf{a}_2 + z_{13}\,\mathbf{a}_3$ | = | $x_{13}a\,\hat{\mathbf{x}} + y_{13}b\,\hat{\mathbf{y}} + z_{13}c\,\hat{\mathbf{z}}$ | (8$d$) | P IX |
| $\mathbf{B}_{50}$ | = | $\left(\frac{1}{2} - x_{13}\right)\mathbf{a}_1 - y_{13}\,\mathbf{a}_2 + \left(\frac{1}{2} + z_{13}\right)\mathbf{a}_3$ | = | $\left(\frac{1}{2} - x_{13}\right)a\,\hat{\mathbf{x}} - y_{13}b\,\hat{\mathbf{y}} + \left(\frac{1}{2} + z_{13}\right)c\,\hat{\mathbf{z}}$ | (8$d$) | P IX |
| $\mathbf{B}_{51}$ | = | $-x_{13}\,\mathbf{a}_1 + \left(\frac{1}{2} + y_{13}\right)\mathbf{a}_2 - z_{13}\,\mathbf{a}_3$ | = | $-x_{13}a\,\hat{\mathbf{x}} + \left(\frac{1}{2} + y_{13}\right)b\,\hat{\mathbf{y}} - z_{13}c\,\hat{\mathbf{z}}$ | (8$d$) | P IX |
| $\mathbf{B}_{52}$ | = | $\left(\frac{1}{2} + x_{13}\right)\mathbf{a}_1 + \left(\frac{1}{2} - y_{13}\right)\mathbf{a}_2 + \left(\frac{1}{2} - z_{13}\right)\mathbf{a}_3$ | = | $\left(\frac{1}{2} + x_{13}\right)a\,\hat{\mathbf{x}} + \left(\frac{1}{2} - y_{13}\right)b\,\hat{\mathbf{y}} + \left(\frac{1}{2} - z_{13}\right)c\,\hat{\mathbf{z}}$ | (8$d$) | P IX |
| $\mathbf{B}_{53}$ | = | $-x_{13}\,\mathbf{a}_1 - y_{13}\,\mathbf{a}_2 - z_{13}\,\mathbf{a}_3$ | = | $-x_{13}a\,\hat{\mathbf{x}} - y_{13}b\,\hat{\mathbf{y}} - z_{13}c\,\hat{\mathbf{z}}$ | (8$d$) | P IX |
| $\mathbf{B}_{54}$ | = | $\left(\frac{1}{2} + x_{13}\right)\mathbf{a}_1 + y_{13}\,\mathbf{a}_2 + \left(\frac{1}{2} - z_{13}\right)\mathbf{a}_3$ | = | $\left(\frac{1}{2} + x_{13}\right)a\,\hat{\mathbf{x}} + y_{13}b\,\hat{\mathbf{y}} + \left(\frac{1}{2} - z_{13}\right)c\,\hat{\mathbf{z}}$ | (8$d$) | P IX |
| $\mathbf{B}_{55}$ | = | $x_{13}\,\mathbf{a}_1 + \left(\frac{1}{2} - y_{13}\right)\mathbf{a}_2 + z_{13}\,\mathbf{a}_3$ | = | $x_{13}a\,\hat{\mathbf{x}} + \left(\frac{1}{2} - y_{13}\right)b\,\hat{\mathbf{y}} + z_{13}c\,\hat{\mathbf{z}}$ | (8$d$) | P IX |
| $\mathbf{B}_{56}$ | = | $\left(\frac{1}{2} - x_{13}\right)\mathbf{a}_1 + \left(\frac{1}{2} + y_{13}\right)\mathbf{a}_2 + \left(\frac{1}{2} + z_{13}\right)\mathbf{a}_3$ | = | $\left(\frac{1}{2} - x_{13}\right)a\,\hat{\mathbf{x}} + \left(\frac{1}{2} + y_{13}\right)b\,\hat{\mathbf{y}} + \left(\frac{1}{2} + z_{13}\right)c\,\hat{\mathbf{z}}$ | (8$d$) | P IX |
| $\mathbf{B}_{57}$ | = | $x_{14}\,\mathbf{a}_1 + y_{14}\,\mathbf{a}_2 + z_{14}\,\mathbf{a}_3$ | = | $x_{14}a\,\hat{\mathbf{x}} + y_{14}b\,\hat{\mathbf{y}} + z_{14}c\,\hat{\mathbf{z}}$ | (8$d$) | P X |
| $\mathbf{B}_{58}$ | = | $\left(\frac{1}{2} - x_{14}\right)\mathbf{a}_1 - y_{14}\,\mathbf{a}_2 + \left(\frac{1}{2} + z_{14}\right)\mathbf{a}_3$ | = | $\left(\frac{1}{2} - x_{14}\right)a\,\hat{\mathbf{x}} - y_{14}b\,\hat{\mathbf{y}} + \left(\frac{1}{2} + z_{14}\right)c\,\hat{\mathbf{z}}$ | (8$d$) | P X |
| $\mathbf{B}_{59}$ | = | $-x_{14}\,\mathbf{a}_1 + \left(\frac{1}{2} + y_{14}\right)\mathbf{a}_2 - z_{14}\,\mathbf{a}_3$ | = | $-x_{14}a\,\hat{\mathbf{x}} + \left(\frac{1}{2} + y_{14}\right)b\,\hat{\mathbf{y}} - z_{14}c\,\hat{\mathbf{z}}$ | (8$d$) | P X |
| $\mathbf{B}_{60}$ | = | $\left(\frac{1}{2} + x_{14}\right)\mathbf{a}_1 + \left(\frac{1}{2} - y_{14}\right)\mathbf{a}_2 + \left(\frac{1}{2} - z_{14}\right)\mathbf{a}_3$ | = | $\left(\frac{1}{2} + x_{14}\right)a\,\hat{\mathbf{x}} + \left(\frac{1}{2} - y_{14}\right)b\,\hat{\mathbf{y}} + \left(\frac{1}{2} - z_{14}\right)c\,\hat{\mathbf{z}}$ | (8$d$) | P X |
| $\mathbf{B}_{61}$ | = | $-x_{14}\,\mathbf{a}_1 - y_{14}\,\mathbf{a}_2 - z_{14}\,\mathbf{a}_3$ | = | $-x_{14}a\,\hat{\mathbf{x}} - y_{14}b\,\hat{\mathbf{y}} - z_{14}c\,\hat{\mathbf{z}}$ | (8$d$) | P X |
| $\mathbf{B}_{62}$ | = | $\left(\frac{1}{2} + x_{14}\right)\mathbf{a}_1 + y_{14}\,\mathbf{a}_2 + \left(\frac{1}{2} - z_{14}\right)\mathbf{a}_3$ | = | $\left(\frac{1}{2} + x_{14}\right)a\,\hat{\mathbf{x}} + y_{14}b\,\hat{\mathbf{y}} + \left(\frac{1}{2} - z_{14}\right)c\,\hat{\mathbf{z}}$ | (8$d$) | P X |
| $\mathbf{B}_{63}$ | = | $x_{14}\,\mathbf{a}_1 + \left(\frac{1}{2} - y_{14}\right)\mathbf{a}_2 + z_{14}\,\mathbf{a}_3$ | = | $x_{14}a\,\hat{\mathbf{x}} + \left(\frac{1}{2} - y_{14}\right)b\,\hat{\mathbf{y}} + z_{14}c\,\hat{\mathbf{z}}$ | (8$d$) | P X |
| $\mathbf{B}_{64}$ | = | $\left(\frac{1}{2} - x_{14}\right)\mathbf{a}_1 + \left(\frac{1}{2} + y_{14}\right)\mathbf{a}_2 + \left(\frac{1}{2} + z_{14}\right)\mathbf{a}_3$ | = | $\left(\frac{1}{2} - x_{14}\right)a\,\hat{\mathbf{x}} + \left(\frac{1}{2} + y_{14}\right)b\,\hat{\mathbf{y}} + \left(\frac{1}{2} + z_{14}\right)c\,\hat{\mathbf{z}}$ | (8$d$) | P X |
| $\mathbf{B}_{65}$ | = | $x_{15}\,\mathbf{a}_1 + y_{15}\,\mathbf{a}_2 + z_{15}\,\mathbf{a}_3$ | = | $x_{15}a\,\hat{\mathbf{x}} + y_{15}b\,\hat{\mathbf{y}} + z_{15}c\,\hat{\mathbf{z}}$ | (8$d$) | P XI |
| $\mathbf{B}_{66}$ | = | $\left(\frac{1}{2} - x_{15}\right)\mathbf{a}_1 - y_{15}\,\mathbf{a}_2 + \left(\frac{1}{2} + z_{15}\right)\mathbf{a}_3$ | = | $\left(\frac{1}{2} - x_{15}\right)a\,\hat{\mathbf{x}} - y_{15}b\,\hat{\mathbf{y}} + \left(\frac{1}{2} + z_{15}\right)c\,\hat{\mathbf{z}}$ | (8$d$) | P XI |
| $\mathbf{B}_{67}$ | = | $-x_{15}\,\mathbf{a}_1 + \left(\frac{1}{2} + y_{15}\right)\mathbf{a}_2 - z_{15}\,\mathbf{a}_3$ | = | $-x_{15}a\,\hat{\mathbf{x}} + \left(\frac{1}{2} + y_{15}\right)b\,\hat{\mathbf{y}} - z_{15}c\,\hat{\mathbf{z}}$ | (8$d$) | P XI |
| $\mathbf{B}_{68}$ | = | $\left(\frac{1}{2} + x_{15}\right)\mathbf{a}_1 + \left(\frac{1}{2} - y_{15}\right)\mathbf{a}_2 + \left(\frac{1}{2} - z_{15}\right)\mathbf{a}_3$ | = | $\left(\frac{1}{2} + x_{15}\right)a\,\hat{\mathbf{x}} + \left(\frac{1}{2} - y_{15}\right)b\,\hat{\mathbf{y}} + \left(\frac{1}{2} - z_{15}\right)c\,\hat{\mathbf{z}}$ | (8$d$) | P XI |
| $\mathbf{B}_{69}$ | = | $-x_{15}\,\mathbf{a}_1 - y_{15}\,\mathbf{a}_2 - z_{15}\,\mathbf{a}_3$ | = | $-x_{15}a\,\hat{\mathbf{x}} - y_{15}b\,\hat{\mathbf{y}} - z_{15}c\,\hat{\mathbf{z}}$ | (8$d$) | P XI |
| $\mathbf{B}_{70}$ | = | $\left(\frac{1}{2} + x_{15}\right)\mathbf{a}_1 + y_{15}\,\mathbf{a}_2 + \left(\frac{1}{2} - z_{15}\right)\mathbf{a}_3$ | = | $\left(\frac{1}{2} + x_{15}\right)a\,\hat{\mathbf{x}} + y_{15}b\,\hat{\mathbf{y}} + \left(\frac{1}{2} - z_{15}\right)c\,\hat{\mathbf{z}}$ | (8$d$) | P XI |
| $\mathbf{B}_{71}$ | = | $x_{15}\,\mathbf{a}_1 + \left(\frac{1}{2} - y_{15}\right)\mathbf{a}_2 + z_{15}\,\mathbf{a}_3$ | = | $x_{15}a\,\hat{\mathbf{x}} + \left(\frac{1}{2} - y_{15}\right)b\,\hat{\mathbf{y}} + z_{15}c\,\hat{\mathbf{z}}$ | (8$d$) | P XI |
| $\mathbf{B}_{72}$ | = | $\left(\frac{1}{2} - x_{15}\right)\mathbf{a}_1 + \left(\frac{1}{2} + y_{15}\right)\mathbf{a}_2 + \left(\frac{1}{2} + z_{15}\right)\mathbf{a}_3$ | = | $\left(\frac{1}{2} - x_{15}\right)a\,\hat{\mathbf{x}} + \left(\frac{1}{2} + y_{15}\right)b\,\hat{\mathbf{y}} + \left(\frac{1}{2} + z_{15}\right)c\,\hat{\mathbf{z}}$ | (8$d$) | P XI |
| $\mathbf{B}_{73}$ | = | $x_{16}\,\mathbf{a}_{1} + y_{16}\,\mathbf{a}_{2} + z_{16}\,\mathbf{a}_{3}$ | = | $x_{16}a\,\mathbf{\hat{x}} + y_{16}b\,\mathbf{\hat{y}} + z_{16}c\,\mathbf{\hat{z}}$ | (8$d$) | P XII |
| $\mathbf{B}_{74}$ | = | $\left(\frac{1}{2} - x_{16}\right)\,\mathbf{a}_{1} - y_{16}\,\mathbf{a}_{2} + \left(\frac{1}{2} + z_{16}\right)\,\mathbf{a}_{3}$ | = | $\left(\frac{1}{2} - x_{16}\right)a\,\mathbf{\hat{x}} - y_{16}b\,\mathbf{\hat{y}} + \left(\frac{1}{2} + z_{16}\right)c\,\mathbf{\hat{z}}$ | (8$d$) | P XII |
| $\mathbf{B}_{75}$ | = | $-x_{16}\,\mathbf{a}_{1} + \left(\frac{1}{2} + y_{16}\right)\,\mathbf{a}_{2} - z_{16}\,\mathbf{a}_{3}$ | = | $-x_{16}a\,\mathbf{\hat{x}} + \left(\frac{1}{2} + y_{16}\right)b\,\mathbf{\hat{y}} - z_{16}c\,\mathbf{\hat{z}}$ | (8$d$) | P XII |
| $\mathbf{B}_{76}$ | = | $\left(\frac{1}{2} + x_{16}\right)\,\mathbf{a}_{1} + \left(\frac{1}{2} - y_{16}\right)\,\mathbf{a}_{2} + \left(\frac{1}{2} - z_{16}\right)\,\mathbf{a}_{3}$ | = | $\left(\frac{1}{2} + x_{16}\right)a\,\mathbf{\hat{x}} + \left(\frac{1}{2} - y_{16}\right)b\,\mathbf{\hat{y}} + \left(\frac{1}{2} - z_{16}\right)c\,\mathbf{\hat{z}}$ | (8$d$) | P XII |
| $\mathbf{B}_{77}$ | = | $-x_{16}\,\mathbf{a}_{1} - y_{16}\,\mathbf{a}_{2} - z_{16}\,\mathbf{a}_{3}$ | = | $-x_{16}a\,\mathbf{\hat{x}} - y_{16}b\,\mathbf{\hat{y}} - z_{16}c\,\mathbf{\hat{z}}$ | (8$d$) | P XII |
| $\mathbf{B}_{78}$ | = | $\left(\frac{1}{2} + x_{16}\right)\,\mathbf{a}_{1} + y_{16}\,\mathbf{a}_{2} + \left(\frac{1}{2} - z_{16}\right)\,\mathbf{a}_{3}$ | = | $\left(\frac{1}{2} + x_{16}\right)a\,\mathbf{\hat{x}} + y_{16}b\,\mathbf{\hat{y}} + \left(\frac{1}{2} - z_{16}\right)c\,\mathbf{\hat{z}}$ | (8$d$) | P XII |
| $\mathbf{B}_{79}$ | = | $x_{16}\,\mathbf{a}_{1} + \left(\frac{1}{2} - y_{16}\right)\,\mathbf{a}_{2} + z_{16}\,\mathbf{a}_{3}$ | = | $x_{16}a\,\mathbf{\hat{x}} + \left(\frac{1}{2} - y_{16}\right)b\,\mathbf{\hat{y}} + z_{16}c\,\mathbf{\hat{z}}$ | (8$d$) | P XII |
| $\mathbf{B}_{80}$ | = | $\left(\frac{1}{2} - x_{16}\right)\,\mathbf{a}_{1} + \left(\frac{1}{2} + y_{16}\right)\,\mathbf{a}_{2} + \left(\frac{1}{2} + z_{16}\right)\,\mathbf{a}_{3}$ | = | $\left(\frac{1}{2} - x_{16}\right)a\,\mathbf{\hat{x}} + \left(\frac{1}{2} + y_{16}\right)b\,\mathbf{\hat{y}} + \left(\frac{1}{2} + z_{16}\right)c\,\mathbf{\hat{z}}$ | (8$d$) | P XII |
| $\mathbf{B}_{81}$ | = | $x_{17}\,\mathbf{a}_{1} + y_{17}\,\mathbf{a}_{2} + z_{17}\,\mathbf{a}_{3}$ | = | $x_{17}a\,\mathbf{\hat{x}} + y_{17}b\,\mathbf{\hat{y}} + z_{17}c\,\mathbf{\hat{z}}$ | (8$d$) | Se V |
| $\mathbf{B}_{82}$ | = | $\left(\frac{1}{2} - x_{17}\right)\,\mathbf{a}_{1} - y_{17}\,\mathbf{a}_{2} + \left(\frac{1}{2} + z_{17}\right)\,\mathbf{a}_{3}$ | = | $\left(\frac{1}{2} - x_{17}\right)a\,\mathbf{\hat{x}} - y_{17}b\,\mathbf{\hat{y}} + \left(\frac{1}{2} + z_{17}\right)c\,\mathbf{\hat{z}}$ | (8$d$) | Se V |
| $\mathbf{B}_{83}$ | = | $-x_{17}\,\mathbf{a}_{1} + \left(\frac{1}{2} + y_{17}\right)\,\mathbf{a}_{2} - z_{17}\,\mathbf{a}_{3}$ | = | $-x_{17}a\,\mathbf{\hat{x}} + \left(\frac{1}{2} + y_{17}\right)b\,\mathbf{\hat{y}} - z_{17}c\,\mathbf{\hat{z}}$ | (8$d$) | Se V |
| $\mathbf{B}_{84}$ | = | $\left(\frac{1}{2} + x_{17}\right)\,\mathbf{a}_{1} + \left(\frac{1}{2} - y_{17}\right)\,\mathbf{a}_{2} + \left(\frac{1}{2} - z_{17}\right)\,\mathbf{a}_{3}$ | = | $\left(\frac{1}{2} + x_{17}\right)a\,\mathbf{\hat{x}} + \left(\frac{1}{2} - y_{17}\right)b\,\mathbf{\hat{y}} + \left(\frac{1}{2} - z_{17}\right)c\,\mathbf{\hat{z}}$ | (8$d$) | Se V |
| $\mathbf{B}_{85}$ | = | $-x_{17}\,\mathbf{a}_{1} - y_{17}\,\mathbf{a}_{2} - z_{17}\,\mathbf{a}_{3}$ | = | $-x_{17}a\,\mathbf{\hat{x}} - y_{17}b\,\mathbf{\hat{y}} - z_{17}c\,\mathbf{\hat{z}}$ | (8$d$) | Se V |
| $\mathbf{B}_{86}$ | = | $\left(\frac{1}{2} + x_{17}\right)\,\mathbf{a}_{1} + y_{17}\,\mathbf{a}_{2} + \left(\frac{1}{2} - z_{17}\right)\,\mathbf{a}_{3}$ | = | $\left(\frac{1}{2} + x_{17}\right)a\,\mathbf{\hat{x}} + y_{17}b\,\mathbf{\hat{y}} + \left(\frac{1}{2} - z_{17}\right)c\,\mathbf{\hat{z}}$ | (8$d$) | Se V |
| $\mathbf{B}_{87}$ | = | $x_{17}\,\mathbf{a}_{1} + \left(\frac{1}{2} - y_{17}\right)\,\mathbf{a}_{2} + z_{17}\,\mathbf{a}_{3}$ | = | $x_{17}a\,\mathbf{\hat{x}} + \left(\frac{1}{2} - y_{17}\right)b\,\mathbf{\hat{y}} + z_{17}c\,\mathbf{\hat{z}}$ | (8$d$) | Se V |
| $\mathbf{B}_{88}$ | = | $\left(\frac{1}{2} - x_{17}\right)\,\mathbf{a}_{1} + \left(\frac{1}{2} + y_{17}\right)\,\mathbf{a}_{2} + \left(\frac{1}{2} + z_{17}\right)\,\mathbf{a}_{3}$ | = | $\left(\frac{1}{2} - x_{17}\right)a\,\mathbf{\hat{x}} + \left(\frac{1}{2} + y_{17}\right)b\,\mathbf{\hat{y}} + \left(\frac{1}{2} + z_{17}\right)c\,\mathbf{\hat{z}}$ | (8$d$) | Se V |
| $\mathbf{B}_{89}$ | = | $x_{18}\,\mathbf{a}_{1} + y_{18}\,\mathbf{a}_{2} + z_{18}\,\mathbf{a}_{3}$ | = | $x_{18}a\,\mathbf{\hat{x}} + y_{18}b\,\mathbf{\hat{y}} + z_{18}c\,\mathbf{\hat{z}}$ | (8$d$) | Se VI |
| $\mathbf{B}_{90}$ | = | $\left(\frac{1}{2} - x_{18}\right)\,\mathbf{a}_{1} - y_{18}\,\mathbf{a}_{2} + \left(\frac{1}{2} + z_{18}\right)\,\mathbf{a}_{3}$ | = | $\left(\frac{1}{2} - x_{18}\right)a\,\mathbf{\hat{x}} - y_{18}b\,\mathbf{\hat{y}} + \left(\frac{1}{2} + z_{18}\right)c\,\mathbf{\hat{z}}$ | (8$d$) | Se VI |
| $\mathbf{B}_{91}$ | = | $-x_{18}\,\mathbf{a}_{1} + \left(\frac{1}{2} + y_{18}\right)\,\mathbf{a}_{2} - z_{18}\,\mathbf{a}_{3}$ | = | $-x_{18}a\,\mathbf{\hat{x}} + \left(\frac{1}{2} + y_{18}\right)b\,\mathbf{\hat{y}} - z_{18}c\,\mathbf{\hat{z}}$ | (8$d$) | Se VI |
| $\mathbf{B}_{92}$ | = | $\left(\frac{1}{2} + x_{18}\right)\,\mathbf{a}_{1} + \left(\frac{1}{2} - y_{18}\right)\,\mathbf{a}_{2} + \left(\frac{1}{2} - z_{18}\right)\,\mathbf{a}_{3}$ | = | $\left(\frac{1}{2} + x_{18}\right)a\,\mathbf{\hat{x}} + \left(\frac{1}{2} - y_{18}\right)b\,\mathbf{\hat{y}} + \left(\frac{1}{2} - z_{18}\right)c\,\mathbf{\hat{z}}$ | (8$d$) | Se VI |
| $\mathbf{B}_{93}$ | = | $-x_{18}\,\mathbf{a}_{1} - y_{18}\,\mathbf{a}_{2} - z_{18}\,\mathbf{a}_{3}$ | = | $-x_{18}a\,\mathbf{\hat{x}} - y_{18}b\,\mathbf{\hat{y}} - z_{18}c\,\mathbf{\hat{z}}$ | (8$d$) | Se VI |
| $\mathbf{B}_{94}$ | = | $\left(\frac{1}{2} + x_{18}\right)\,\mathbf{a}_{1} + y_{18}\,\mathbf{a}_{2} + \left(\frac{1}{2} - z_{18}\right)\,\mathbf{a}_{3}$ | = | $\left(\frac{1}{2} + x_{18}\right)a\,\mathbf{\hat{x}} + y_{18}b\,\mathbf{\hat{y}} + \left(\frac{1}{2} - z_{18}\right)c\,\mathbf{\hat{z}}$ | (8$d$) | Se VI |
| $\mathbf{B}_{95}$ | = | $x_{18}\,\mathbf{a}_{1} + \left(\frac{1}{2} - y_{18}\right)\,\mathbf{a}_{2} + z_{18}\,\mathbf{a}_{3}$ | = | $x_{18}a\,\mathbf{\hat{x}} + \left(\frac{1}{2} - y_{18}\right)b\,\mathbf{\hat{y}} + z_{18}c\,\mathbf{\hat{z}}$ | (8$d$) | Se VI |
| $\mathbf{B}_{96}$ | = | $\left(\frac{1}{2} - x_{18}\right)\,\mathbf{a}_{1} + \left(\frac{1}{2} + y_{18}\right)\,\mathbf{a}_{2} + \left(\frac{1}{2} + z_{18}\right)\,\mathbf{a}_{3}$ | = | $\left(\frac{1}{2} - x_{18}\right)a\,\mathbf{\hat{x}} + \left(\frac{1}{2} + y_{18}\right)b\,\mathbf{\hat{y}} + \left(\frac{1}{2} + z_{18}\right)c\,\mathbf{\hat{z}}$ | (8$d$) | Se VI |
| $\mathbf{B}_{97}$ | = | $x_{19}\,\mathbf{a}_{1} + y_{19}\,\mathbf{a}_{2} + z_{19}\,\mathbf{a}_{3}$ | = | $x_{19}a\,\mathbf{\hat{x}} + y_{19}b\,\mathbf{\hat{y}} + z_{19}c\,\mathbf{\hat{z}}$ | (8$d$) | Se VII |
| $\mathbf{B}_{98}$ | = | $\left(\frac{1}{2} - x_{19}\right)\,\mathbf{a}_{1} - y_{19}\,\mathbf{a}_{2} + \left(\frac{1}{2} + z_{19}\right)\,\mathbf{a}_{3}$ | = | $\left(\frac{1}{2} - x_{19}\right)a\,\mathbf{\hat{x}} - y_{19}b\,\mathbf{\hat{y}} + \left(\frac{1}{2} + z_{19}\right)c\,\mathbf{\hat{z}}$ | (8$d$) | Se VII |
| $\mathbf{B}_{99}$ | = | $-x_{19}\,\mathbf{a}_{1} + \left(\frac{1}{2} + y_{19}\right)\,\mathbf{a}_{2} - z_{19}\,\mathbf{a}_{3}$ | = | $-x_{19}a\,\mathbf{\hat{x}} + \left(\frac{1}{2} + y_{19}\right)b\,\mathbf{\hat{y}} - z_{19}c\,\mathbf{\hat{z}}$ | (8$d$) | Se VII |
| $\mathbf{B}_{100}$ | = | $\left(\frac{1}{2} + x_{19}\right)\,\mathbf{a}_{1} + \left(\frac{1}{2} - y_{19}\right)\,\mathbf{a}_{2} + \left(\frac{1}{2} - z_{19}\right)\,\mathbf{a}_{3}$ | = | $\left(\frac{1}{2} + x_{19}\right)a\,\mathbf{\hat{x}} + \left(\frac{1}{2} - y_{19}\right)b\,\mathbf{\hat{y}} + \left(\frac{1}{2} - z_{19}\right)c\,\mathbf{\hat{z}}$ | (8$d$) | Se VII |
| $\mathbf{B}_{101}$ | = | $-x_{19}\,\mathbf{a}_{1} - y_{19}\,\mathbf{a}_{2} - z_{19}\,\mathbf{a}_{3}$ | = | $-x_{19}a\,\mathbf{\hat{x}} - y_{19}b\,\mathbf{\hat{y}} - z_{19}c\,\mathbf{\hat{z}}$ | (8$d$) | Se VII |
| $\mathbf{B}_{102}$ | = | $\left(\frac{1}{2} + x_{19}\right)\,\mathbf{a}_{1} + y_{19}\,\mathbf{a}_{2} + \left(\frac{1}{2} - z_{19}\right)\,\mathbf{a}_{3}$ | = | $\left(\frac{1}{2} + x_{19}\right)a\,\mathbf{\hat{x}} + y_{19}b\,\mathbf{\hat{y}} + \left(\frac{1}{2} - z_{19}\right)c\,\mathbf{\hat{z}}$ | (8$d$) | Se VII |
| $\mathbf{B}_{103}$ | = | $x_{19}\,\mathbf{a}_{1} + \left(\frac{1}{2} - y_{19}\right)\,\mathbf{a}_{2} + z_{19}\,\mathbf{a}_{3}$ | = | $x_{19}a\,\mathbf{\hat{x}} + \left(\frac{1}{2} - y_{19}\right)b\,\mathbf{\hat{y}} + z_{19}c\,\mathbf{\hat{z}}$ | (8$d$) | Se VII |
| $\mathbf{B}_{104}$ | = | $\left(\frac{1}{2} - x_{19}\right) \mathbf{a}_1 + \left(\frac{1}{2} + y_{19}\right) \mathbf{a}_2 + \left(\frac{1}{2} + z_{19}\right) \mathbf{a}_3$ | = | $\left(\frac{1}{2} - x_{19}\right) a \, \mathbf{\hat{x}} + \left(\frac{1}{2} + y_{19}\right) b \, \mathbf{\hat{y}} + \left(\frac{1}{2} + z_{19}\right) c \, \mathbf{\hat{z}}$ | (8$d$) | Se VII |
| $\mathbf{B}_{105}$ | = | $x_{20} \, \mathbf{a}_1 + y_{20} \, \mathbf{a}_2 + z_{20} \, \mathbf{a}_3$ | = | $x_{20} a \, \mathbf{\hat{x}} + y_{20} b \, \mathbf{\hat{y}} + z_{20} c \, \mathbf{\hat{z}}$ | (8$d$) | Se VIII |
| $\mathbf{B}_{106}$ | = | $\left(\frac{1}{2} - x_{20}\right) \mathbf{a}_1 - y_{20} \, \mathbf{a}_2 + \left(\frac{1}{2} + z_{20}\right) \mathbf{a}_3$ | = | $\left(\frac{1}{2} - x_{20}\right) a \, \mathbf{\hat{x}} - y_{20} b \, \mathbf{\hat{y}} + \left(\frac{1}{2} + z_{20}\right) c \, \mathbf{\hat{z}}$ | (8$d$) | Se VIII |
| $\mathbf{B}_{107}$ | = | $-x_{20} \, \mathbf{a}_1 + \left(\frac{1}{2} + y_{20}\right) \mathbf{a}_2 - z_{20} \, \mathbf{a}_3$ | = | $-x_{20} a \, \mathbf{\hat{x}} + \left(\frac{1}{2} + y_{20}\right) b \, \mathbf{\hat{y}} - z_{20} c \, \mathbf{\hat{z}}$ | (8$d$) | Se VIII |
| $\mathbf{B}_{108}$ | = | $\left(\frac{1}{2} + x_{20}\right) \mathbf{a}_1 + \left(\frac{1}{2} - y_{20}\right) \mathbf{a}_2 + \left(\frac{1}{2} - z_{20}\right) \mathbf{a}_3$ | = | $\left(\frac{1}{2} + x_{20}\right) a \, \mathbf{\hat{x}} + \left(\frac{1}{2} - y_{20}\right) b \, \mathbf{\hat{y}} + \left(\frac{1}{2} - z_{20}\right) c \, \mathbf{\hat{z}}$ | (8$d$) | Se VIII |
| $\mathbf{B}_{109}$ | = | $-x_{20} \, \mathbf{a}_1 - y_{20} \, \mathbf{a}_2 - z_{20} \, \mathbf{a}_3$ | = | $-x_{20} a \, \mathbf{\hat{x}} - y_{20} b \, \mathbf{\hat{y}} - z_{20} c \, \mathbf{\hat{z}}$ | (8$d$) | Se VIII |
| $\mathbf{B}_{110}$ | = | $\left(\frac{1}{2} + x_{20}\right) \mathbf{a}_1 + y_{20} \, \mathbf{a}_2 + \left(\frac{1}{2} - z_{20}\right) \mathbf{a}_3$ | = | $\left(\frac{1}{2} + x_{20}\right) a \, \mathbf{\hat{x}} + y_{20} b \, \mathbf{\hat{y}} + \left(\frac{1}{2} - z_{20}\right) c \, \mathbf{\hat{z}}$ | (8$d$) | Se VIII |
| $\mathbf{B}_{111}$ | = | $x_{20} \, \mathbf{a}_1 + \left(\frac{1}{2} - y_{20}\right) \mathbf{a}_2 + z_{20} \, \mathbf{a}_3$ | = | $x_{20} a \, \mathbf{\hat{x}} + \left(\frac{1}{2} - y_{20}\right) b \, \mathbf{\hat{y}} + z_{20} c \, \mathbf{\hat{z}}$ | (8$d$) | Se VIII |
| $\mathbf{B}_{112}$ | = | $\left(\frac{1}{2} - x_{20}\right) \mathbf{a}_1 + \left(\frac{1}{2} + y_{20}\right) \mathbf{a}_2 + \left(\frac{1}{2} + z_{20}\right) \mathbf{a}_3$ | = | $\left(\frac{1}{2} - x_{20}\right) a \, \mathbf{\hat{x}} + \left(\frac{1}{2} + y_{20}\right) b \, \mathbf{\hat{y}} + \left(\frac{1}{2} + z_{20}\right) c \, \mathbf{\hat{z}}$ | (8$d$) | Se VIII |

**Geometry files:**

- CIF: pp. 1637
- POSCAR: pp. 1638

# $Mo_4P_3$ Structure: A4B3_oP56_62_8c_6c

http://aflow.org/prototype-encyclopedia/A4B3_oP56_62_8c_6c

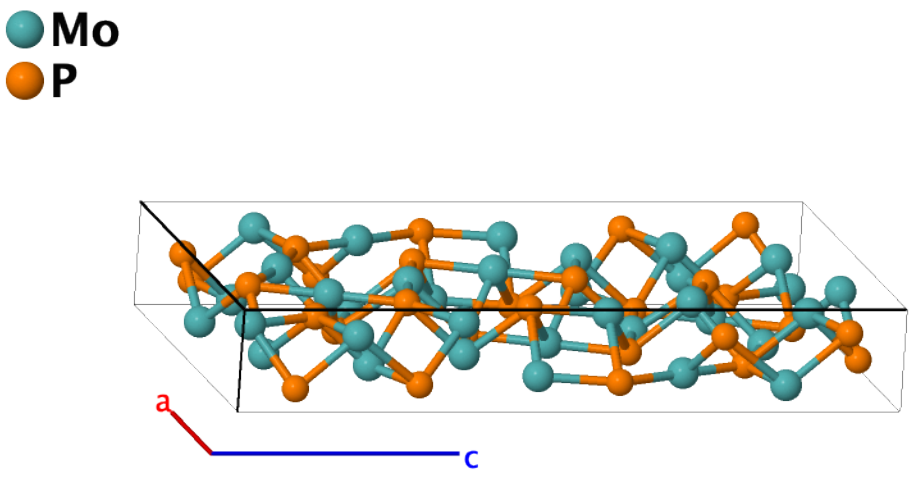

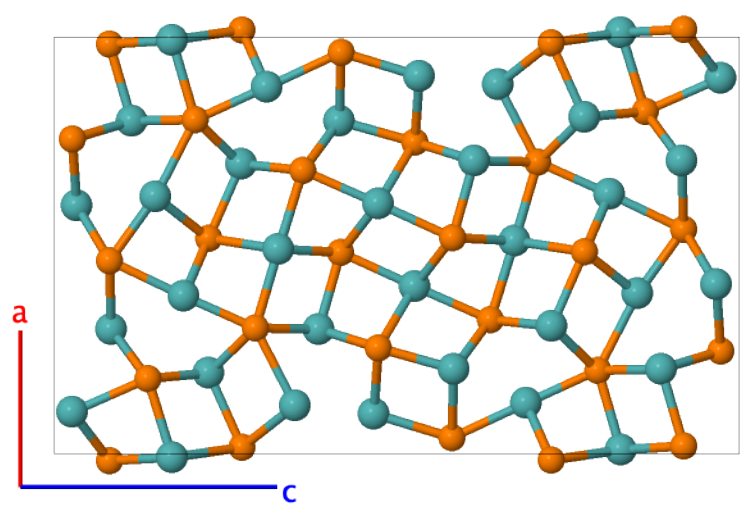

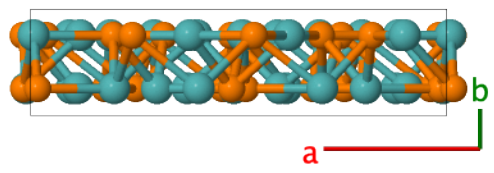

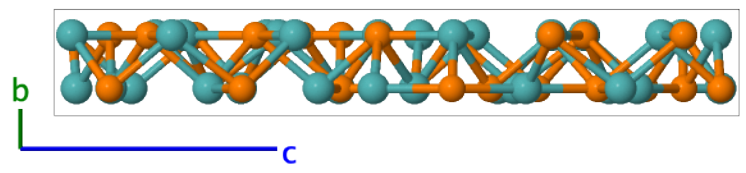

| | | |
|---|---|---|
| **Prototype** | : | $Mo_4P_3$ |
| **AFLOW prototype label** | : | A4B3_oP56_62_8c_6c |
| ***Strukturbericht* designation** | : | None |
| **Pearson symbol** | : | oP56 |
| **Space group number** | : | 62 |
| **Space group symbol** | : | *Pnma* |
| **AFLOW prototype command** | : | `aflow --proto=A4B3_oP56_62_8c_6c` <br> `--params=`$a, b/a, c/a, x_1, z_1, x_2, z_2, x_3, z_3, x_4, z_4, x_5, z_5, x_6, z_6, x_7, z_7, x_8, z_8, x_9, z_9, x_{10}, z_{10}, x_{11}, z_{11}, x_{12}, z_{12}, x_{13}, z_{13}, x_{14}, z_{14}$ |

**Simple Orthorhombic primitive vectors:**

$$\mathbf{a}_1 = a\,\hat{\mathbf{x}}$$
$$\mathbf{a}_2 = b\,\hat{\mathbf{y}}$$
$$\mathbf{a}_3 = c\,\hat{\mathbf{z}}$$

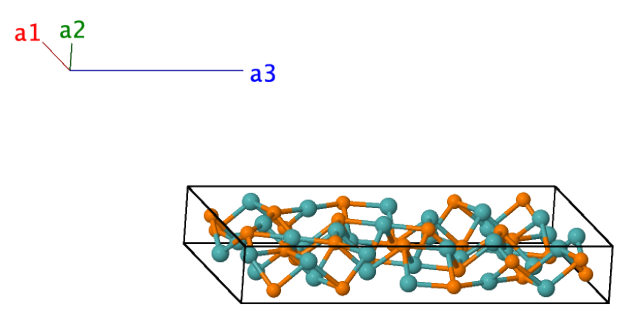

**Basis vectors:**

| | | Lattice Coordinates | | Cartesian Coordinates | Wyckoff Position | Atom Type |
|---|---|---|---|---|---|---|
| $\mathbf{B}_1$ | = | $x_1\,\mathbf{a}_1 + \frac{1}{4}\,\mathbf{a}_2 + z_1\,\mathbf{a}_3$ | = | $x_1 a\,\hat{\mathbf{x}} + \frac{1}{4} b\,\hat{\mathbf{y}} + z_1 c\,\hat{\mathbf{z}}$ | (4$c$) | Mo I |
| $\mathbf{B}_2$ | = | $\left(\frac{1}{2} - x_1\right)\mathbf{a}_1 + \frac{3}{4}\,\mathbf{a}_2 + \left(\frac{1}{2} + z_1\right)\mathbf{a}_3$ | = | $\left(\frac{1}{2} - x_1\right) a\,\hat{\mathbf{x}} + \frac{3}{4} b\,\hat{\mathbf{y}} + \left(\frac{1}{2} + z_1\right) c\,\hat{\mathbf{z}}$ | (4$c$) | Mo I |
| $\mathbf{B}_3$ | = | $-x_1\,\mathbf{a}_1 + \frac{3}{4}\,\mathbf{a}_2 - z_1\,\mathbf{a}_3$ | = | $-x_1 a\,\hat{\mathbf{x}} + \frac{3}{4} b\,\hat{\mathbf{y}} - z_1 c\,\hat{\mathbf{z}}$ | (4$c$) | Mo I |
| $\mathbf{B}_4$ | = | $\left(\frac{1}{2} + x_1\right)\mathbf{a}_1 + \frac{1}{4}\,\mathbf{a}_2 + \left(\frac{1}{2} - z_1\right)\mathbf{a}_3$ | = | $\left(\frac{1}{2} + x_1\right) a\,\hat{\mathbf{x}} + \frac{1}{4} b\,\hat{\mathbf{y}} + \left(\frac{1}{2} - z_1\right) c\,\hat{\mathbf{z}}$ | (4$c$) | Mo I |
| $\mathbf{B}_5$ | = | $x_2\,\mathbf{a}_1 + \frac{1}{4}\,\mathbf{a}_2 + z_2\,\mathbf{a}_3$ | = | $x_2 a\,\hat{\mathbf{x}} + \frac{1}{4} b\,\hat{\mathbf{y}} + z_2 c\,\hat{\mathbf{z}}$ | (4$c$) | Mo II |

$$\begin{array}{llllll}
\mathbf{B}_{6} & = & \left(\frac{1}{2}-x_{2}\right)\,\mathbf{a}_{1}+\frac{3}{4}\,\mathbf{a}_{2}+\left(\frac{1}{2}+z_{2}\right)\,\mathbf{a}_{3} & = & \left(\frac{1}{2}-x_{2}\right)a\,\hat{\mathbf{x}}+\frac{3}{4}b\,\hat{\mathbf{y}}+\left(\frac{1}{2}+z_{2}\right)c\,\hat{\mathbf{z}} & (4c) & \text{Mo II} \\
\mathbf{B}_{7} & = & -x_{2}\,\mathbf{a}_{1}+\frac{3}{4}\,\mathbf{a}_{2}-z_{2}\,\mathbf{a}_{3} & = & -x_{2}a\,\hat{\mathbf{x}}+\frac{3}{4}b\,\hat{\mathbf{y}}-z_{2}c\,\hat{\mathbf{z}} & (4c) & \text{Mo II} \\
\mathbf{B}_{8} & = & \left(\frac{1}{2}+x_{2}\right)\,\mathbf{a}_{1}+\frac{1}{4}\,\mathbf{a}_{2}+\left(\frac{1}{2}-z_{2}\right)\,\mathbf{a}_{3} & = & \left(\frac{1}{2}+x_{2}\right)a\,\hat{\mathbf{x}}+\frac{1}{4}b\,\hat{\mathbf{y}}+\left(\frac{1}{2}-z_{2}\right)c\,\hat{\mathbf{z}} & (4c) & \text{Mo II} \\
\mathbf{B}_{9} & = & x_{3}\,\mathbf{a}_{1}+\frac{1}{4}\,\mathbf{a}_{2}+z_{3}\,\mathbf{a}_{3} & = & x_{3}a\,\hat{\mathbf{x}}+\frac{1}{4}b\,\hat{\mathbf{y}}+z_{3}c\,\hat{\mathbf{z}} & (4c) & \text{Mo III} \\
\mathbf{B}_{10} & = & \left(\frac{1}{2}-x_{3}\right)\,\mathbf{a}_{1}+\frac{3}{4}\,\mathbf{a}_{2}+\left(\frac{1}{2}+z_{3}\right)\,\mathbf{a}_{3} & = & \left(\frac{1}{2}-x_{3}\right)a\,\hat{\mathbf{x}}+\frac{3}{4}b\,\hat{\mathbf{y}}+\left(\frac{1}{2}+z_{3}\right)c\,\hat{\mathbf{z}} & (4c) & \text{Mo III} \\
\mathbf{B}_{11} & = & -x_{3}\,\mathbf{a}_{1}+\frac{3}{4}\,\mathbf{a}_{2}-z_{3}\,\mathbf{a}_{3} & = & -x_{3}a\,\hat{\mathbf{x}}+\frac{3}{4}b\,\hat{\mathbf{y}}-z_{3}c\,\hat{\mathbf{z}} & (4c) & \text{Mo III} \\
\mathbf{B}_{12} & = & \left(\frac{1}{2}+x_{3}\right)\,\mathbf{a}_{1}+\frac{1}{4}\,\mathbf{a}_{2}+\left(\frac{1}{2}-z_{3}\right)\,\mathbf{a}_{3} & = & \left(\frac{1}{2}+x_{3}\right)a\,\hat{\mathbf{x}}+\frac{1}{4}b\,\hat{\mathbf{y}}+\left(\frac{1}{2}-z_{3}\right)c\,\hat{\mathbf{z}} & (4c) & \text{Mo III} \\
\mathbf{B}_{13} & = & x_{4}\,\mathbf{a}_{1}+\frac{1}{4}\,\mathbf{a}_{2}+z_{4}\,\mathbf{a}_{3} & = & x_{4}a\,\hat{\mathbf{x}}+\frac{1}{4}b\,\hat{\mathbf{y}}+z_{4}c\,\hat{\mathbf{z}} & (4c) & \text{Mo IV} \\
\mathbf{B}_{14} & = & \left(\frac{1}{2}-x_{4}\right)\,\mathbf{a}_{1}+\frac{3}{4}\,\mathbf{a}_{2}+\left(\frac{1}{2}+z_{4}\right)\,\mathbf{a}_{3} & = & \left(\frac{1}{2}-x_{4}\right)a\,\hat{\mathbf{x}}+\frac{3}{4}b\,\hat{\mathbf{y}}+\left(\frac{1}{2}+z_{4}\right)c\,\hat{\mathbf{z}} & (4c) & \text{Mo IV} \\
\mathbf{B}_{15} & = & -x_{4}\,\mathbf{a}_{1}+\frac{3}{4}\,\mathbf{a}_{2}-z_{4}\,\mathbf{a}_{3} & = & -x_{4}a\,\hat{\mathbf{x}}+\frac{3}{4}b\,\hat{\mathbf{y}}-z_{4}c\,\hat{\mathbf{z}} & (4c) & \text{Mo IV} \\
\mathbf{B}_{16} & = & \left(\frac{1}{2}+x_{4}\right)\,\mathbf{a}_{1}+\frac{1}{4}\,\mathbf{a}_{2}+\left(\frac{1}{2}-z_{4}\right)\,\mathbf{a}_{3} & = & \left(\frac{1}{2}+x_{4}\right)a\,\hat{\mathbf{x}}+\frac{1}{4}b\,\hat{\mathbf{y}}+\left(\frac{1}{2}-z_{4}\right)c\,\hat{\mathbf{z}} & (4c) & \text{Mo IV} \\
\mathbf{B}_{17} & = & x_{5}\,\mathbf{a}_{1}+\frac{1}{4}\,\mathbf{a}_{2}+z_{5}\,\mathbf{a}_{3} & = & x_{5}a\,\hat{\mathbf{x}}+\frac{1}{4}b\,\hat{\mathbf{y}}+z_{5}c\,\hat{\mathbf{z}} & (4c) & \text{Mo V} \\
\mathbf{B}_{18} & = & \left(\frac{1}{2}-x_{5}\right)\,\mathbf{a}_{1}+\frac{3}{4}\,\mathbf{a}_{2}+\left(\frac{1}{2}+z_{5}\right)\,\mathbf{a}_{3} & = & \left(\frac{1}{2}-x_{5}\right)a\,\hat{\mathbf{x}}+\frac{3}{4}b\,\hat{\mathbf{y}}+\left(\frac{1}{2}+z_{5}\right)c\,\hat{\mathbf{z}} & (4c) & \text{Mo V} \\
\mathbf{B}_{19} & = & -x_{5}\,\mathbf{a}_{1}+\frac{3}{4}\,\mathbf{a}_{2}-z_{5}\,\mathbf{a}_{3} & = & -x_{5}a\,\hat{\mathbf{x}}+\frac{3}{4}b\,\hat{\mathbf{y}}-z_{5}c\,\hat{\mathbf{z}} & (4c) & \text{Mo V} \\
\mathbf{B}_{20} & = & \left(\frac{1}{2}+x_{5}\right)\,\mathbf{a}_{1}+\frac{1}{4}\,\mathbf{a}_{2}+\left(\frac{1}{2}-z_{5}\right)\,\mathbf{a}_{3} & = & \left(\frac{1}{2}+x_{5}\right)a\,\hat{\mathbf{x}}+\frac{1}{4}b\,\hat{\mathbf{y}}+\left(\frac{1}{2}-z_{5}\right)c\,\hat{\mathbf{z}} & (4c) & \text{Mo V} \\
\mathbf{B}_{21} & = & x_{6}\,\mathbf{a}_{1}+\frac{1}{4}\,\mathbf{a}_{2}+z_{6}\,\mathbf{a}_{3} & = & x_{6}a\,\hat{\mathbf{x}}+\frac{1}{4}b\,\hat{\mathbf{y}}+z_{6}c\,\hat{\mathbf{z}} & (4c) & \text{Mo VI} \\
\mathbf{B}_{22} & = & \left(\frac{1}{2}-x_{6}\right)\,\mathbf{a}_{1}+\frac{3}{4}\,\mathbf{a}_{2}+\left(\frac{1}{2}+z_{6}\right)\,\mathbf{a}_{3} & = & \left(\frac{1}{2}-x_{6}\right)a\,\hat{\mathbf{x}}+\frac{3}{4}b\,\hat{\mathbf{y}}+\left(\frac{1}{2}+z_{6}\right)c\,\hat{\mathbf{z}} & (4c) & \text{Mo VI} \\
\mathbf{B}_{23} & = & -x_{6}\,\mathbf{a}_{1}+\frac{3}{4}\,\mathbf{a}_{2}-z_{6}\,\mathbf{a}_{3} & = & -x_{6}a\,\hat{\mathbf{x}}+\frac{3}{4}b\,\hat{\mathbf{y}}-z_{6}c\,\hat{\mathbf{z}} & (4c) & \text{Mo VI} \\
\mathbf{B}_{24} & = & \left(\frac{1}{2}+x_{6}\right)\,\mathbf{a}_{1}+\frac{1}{4}\,\mathbf{a}_{2}+\left(\frac{1}{2}-z_{6}\right)\,\mathbf{a}_{3} & = & \left(\frac{1}{2}+x_{6}\right)a\,\hat{\mathbf{x}}+\frac{1}{4}b\,\hat{\mathbf{y}}+\left(\frac{1}{2}-z_{6}\right)c\,\hat{\mathbf{z}} & (4c) & \text{Mo VI} \\
\mathbf{B}_{25} & = & x_{7}\,\mathbf{a}_{1}+\frac{1}{4}\,\mathbf{a}_{2}+z_{7}\,\mathbf{a}_{3} & = & x_{7}a\,\hat{\mathbf{x}}+\frac{1}{4}b\,\hat{\mathbf{y}}+z_{7}c\,\hat{\mathbf{z}} & (4c) & \text{Mo VII} \\
\mathbf{B}_{26} & = & \left(\frac{1}{2}-x_{7}\right)\,\mathbf{a}_{1}+\frac{3}{4}\,\mathbf{a}_{2}+\left(\frac{1}{2}+z_{7}\right)\,\mathbf{a}_{3} & = & \left(\frac{1}{2}-x_{7}\right)a\,\hat{\mathbf{x}}+\frac{3}{4}b\,\hat{\mathbf{y}}+\left(\frac{1}{2}+z_{7}\right)c\,\hat{\mathbf{z}} & (4c) & \text{Mo VII} \\
\mathbf{B}_{27} & = & -x_{7}\,\mathbf{a}_{1}+\frac{3}{4}\,\mathbf{a}_{2}-z_{7}\,\mathbf{a}_{3} & = & -x_{7}a\,\hat{\mathbf{x}}+\frac{3}{4}b\,\hat{\mathbf{y}}-z_{7}c\,\hat{\mathbf{z}} & (4c) & \text{Mo VII} \\
\mathbf{B}_{28} & = & \left(\frac{1}{2}+x_{7}\right)\,\mathbf{a}_{1}+\frac{1}{4}\,\mathbf{a}_{2}+\left(\frac{1}{2}-z_{7}\right)\,\mathbf{a}_{3} & = & \left(\frac{1}{2}+x_{7}\right)a\,\hat{\mathbf{x}}+\frac{1}{4}b\,\hat{\mathbf{y}}+\left(\frac{1}{2}-z_{7}\right)c\,\hat{\mathbf{z}} & (4c) & \text{Mo VII} \\
\mathbf{B}_{29} & = & x_{8}\,\mathbf{a}_{1}+\frac{1}{4}\,\mathbf{a}_{2}+z_{8}\,\mathbf{a}_{3} & = & x_{8}a\,\hat{\mathbf{x}}+\frac{1}{4}b\,\hat{\mathbf{y}}+z_{8}c\,\hat{\mathbf{z}} & (4c) & \text{Mo VIII} \\
\mathbf{B}_{30} & = & \left(\frac{1}{2}-x_{8}\right)\,\mathbf{a}_{1}+\frac{3}{4}\,\mathbf{a}_{2}+\left(\frac{1}{2}+z_{8}\right)\,\mathbf{a}_{3} & = & \left(\frac{1}{2}-x_{8}\right)a\,\hat{\mathbf{x}}+\frac{3}{4}b\,\hat{\mathbf{y}}+\left(\frac{1}{2}+z_{8}\right)c\,\hat{\mathbf{z}} & (4c) & \text{Mo VIII} \\
\mathbf{B}_{31} & = & -x_{8}\,\mathbf{a}_{1}+\frac{3}{4}\,\mathbf{a}_{2}-z_{8}\,\mathbf{a}_{3} & = & -x_{8}a\,\hat{\mathbf{x}}+\frac{3}{4}b\,\hat{\mathbf{y}}-z_{8}c\,\hat{\mathbf{z}} & (4c) & \text{Mo VIII} \\
\mathbf{B}_{32} & = & \left(\frac{1}{2}+x_{8}\right)\,\mathbf{a}_{1}+\frac{1}{4}\,\mathbf{a}_{2}+\left(\frac{1}{2}-z_{8}\right)\,\mathbf{a}_{3} & = & \left(\frac{1}{2}+x_{8}\right)a\,\hat{\mathbf{x}}+\frac{1}{4}b\,\hat{\mathbf{y}}+\left(\frac{1}{2}-z_{8}\right)c\,\hat{\mathbf{z}} & (4c) & \text{Mo VIII} \\
\mathbf{B}_{33} & = & x_{9}\,\mathbf{a}_{1}+\frac{1}{4}\,\mathbf{a}_{2}+z_{9}\,\mathbf{a}_{3} & = & x_{9}a\,\hat{\mathbf{x}}+\frac{1}{4}b\,\hat{\mathbf{y}}+z_{9}c\,\hat{\mathbf{z}} & (4c) & \text{P I} \\
\mathbf{B}_{34} & = & \left(\frac{1}{2}-x_{9}\right)\,\mathbf{a}_{1}+\frac{3}{4}\,\mathbf{a}_{2}+\left(\frac{1}{2}+z_{9}\right)\,\mathbf{a}_{3} & = & \left(\frac{1}{2}-x_{9}\right)a\,\hat{\mathbf{x}}+\frac{3}{4}b\,\hat{\mathbf{y}}+\left(\frac{1}{2}+z_{9}\right)c\,\hat{\mathbf{z}} & (4c) & \text{P I} \\
\mathbf{B}_{35} & = & -x_{9}\,\mathbf{a}_{1}+\frac{3}{4}\,\mathbf{a}_{2}-z_{9}\,\mathbf{a}_{3} & = & -x_{9}a\,\hat{\mathbf{x}}+\frac{3}{4}b\,\hat{\mathbf{y}}-z_{9}c\,\hat{\mathbf{z}} & (4c) & \text{P I} \\
\mathbf{B}_{36} & = & \left(\frac{1}{2}+x_{9}\right)\,\mathbf{a}_{1}+\frac{1}{4}\,\mathbf{a}_{2}+\left(\frac{1}{2}-z_{9}\right)\,\mathbf{a}_{3} & = & \left(\frac{1}{2}+x_{9}\right)a\,\hat{\mathbf{x}}+\frac{1}{4}b\,\hat{\mathbf{y}}+\left(\frac{1}{2}-z_{9}\right)c\,\hat{\mathbf{z}} & (4c) & \text{P I} \\
\mathbf{B}_{37} & = & x_{10}\,\mathbf{a}_{1}+\frac{1}{4}\,\mathbf{a}_{2}+z_{10}\,\mathbf{a}_{3} & = & x_{10}a\,\hat{\mathbf{x}}+\frac{1}{4}b\,\hat{\mathbf{y}}+z_{10}c\,\hat{\mathbf{z}} & (4c) & \text{P II} \\
\mathbf{B}_{38} & = & \left(\frac{1}{2}-x_{10}\right)\,\mathbf{a}_{1}+\frac{3}{4}\,\mathbf{a}_{2}+\left(\frac{1}{2}+z_{10}\right)\,\mathbf{a}_{3} & = & \left(\frac{1}{2}-x_{10}\right)a\,\hat{\mathbf{x}}+\frac{3}{4}b\,\hat{\mathbf{y}}+\left(\frac{1}{2}+z_{10}\right)c\,\hat{\mathbf{z}} & (4c) & \text{P II} \\
\mathbf{B}_{39} & = & -x_{10}\,\mathbf{a}_{1}+\frac{3}{4}\,\mathbf{a}_{2}-z_{10}\,\mathbf{a}_{3} & = & -x_{10}a\,\hat{\mathbf{x}}+\frac{3}{4}b\,\hat{\mathbf{y}}-z_{10}c\,\hat{\mathbf{z}} & (4c) & \text{P II} \\
\mathbf{B}_{40} & = & \left(\frac{1}{2}+x_{10}\right)\,\mathbf{a}_{1}+\frac{1}{4}\,\mathbf{a}_{2}+\left(\frac{1}{2}-z_{10}\right)\,\mathbf{a}_{3} & = & \left(\frac{1}{2}+x_{10}\right)a\,\hat{\mathbf{x}}+\frac{1}{4}b\,\hat{\mathbf{y}}+\left(\frac{1}{2}-z_{10}\right)c\,\hat{\mathbf{z}} & (4c) & \text{P II} \\
\mathbf{B}_{41} & = & x_{11}\,\mathbf{a}_{1}+\frac{1}{4}\,\mathbf{a}_{2}+z_{11}\,\mathbf{a}_{3} & = & x_{11}a\,\hat{\mathbf{x}}+\frac{1}{4}b\,\hat{\mathbf{y}}+z_{11}c\,\hat{\mathbf{z}} & (4c) & \text{P III}
\end{array}$$

| | | | | | | |
|---|---|---|---|---|---|---|
| $\mathbf{B}_{42}$ | = | $\left(\frac{1}{2} - x_{11}\right) \mathbf{a}_1 + \frac{3}{4} \mathbf{a}_2 + \left(\frac{1}{2} + z_{11}\right) \mathbf{a}_3$ | = | $\left(\frac{1}{2} - x_{11}\right) a \, \mathbf{\hat{x}} + \frac{3}{4} b \, \mathbf{\hat{y}} + \left(\frac{1}{2} + z_{11}\right) c \, \mathbf{\hat{z}}$ | (4$c$) | P III |
| $\mathbf{B}_{43}$ | = | $-x_{11} \mathbf{a}_1 + \frac{3}{4} \mathbf{a}_2 - z_{11} \mathbf{a}_3$ | = | $-x_{11} a \, \mathbf{\hat{x}} + \frac{3}{4} b \, \mathbf{\hat{y}} - z_{11} c \, \mathbf{\hat{z}}$ | (4$c$) | P III |
| $\mathbf{B}_{44}$ | = | $\left(\frac{1}{2} + x_{11}\right) \mathbf{a}_1 + \frac{1}{4} \mathbf{a}_2 + \left(\frac{1}{2} - z_{11}\right) \mathbf{a}_3$ | = | $\left(\frac{1}{2} + x_{11}\right) a \, \mathbf{\hat{x}} + \frac{1}{4} b \, \mathbf{\hat{y}} + \left(\frac{1}{2} - z_{11}\right) c \, \mathbf{\hat{z}}$ | (4$c$) | P III |
| $\mathbf{B}_{45}$ | = | $x_{12} \mathbf{a}_1 + \frac{1}{4} \mathbf{a}_2 + z_{12} \mathbf{a}_3$ | = | $x_{12} a \, \mathbf{\hat{x}} + \frac{1}{4} b \, \mathbf{\hat{y}} + z_{12} c \, \mathbf{\hat{z}}$ | (4$c$) | P IV |
| $\mathbf{B}_{46}$ | = | $\left(\frac{1}{2} - x_{12}\right) \mathbf{a}_1 + \frac{3}{4} \mathbf{a}_2 + \left(\frac{1}{2} + z_{12}\right) \mathbf{a}_3$ | = | $\left(\frac{1}{2} - x_{12}\right) a \, \mathbf{\hat{x}} + \frac{3}{4} b \, \mathbf{\hat{y}} + \left(\frac{1}{2} + z_{12}\right) c \, \mathbf{\hat{z}}$ | (4$c$) | P IV |
| $\mathbf{B}_{47}$ | = | $-x_{12} \mathbf{a}_1 + \frac{3}{4} \mathbf{a}_2 - z_{12} \mathbf{a}_3$ | = | $-x_{12} a \, \mathbf{\hat{x}} + \frac{3}{4} b \, \mathbf{\hat{y}} - z_{12} c \, \mathbf{\hat{z}}$ | (4$c$) | P IV |
| $\mathbf{B}_{48}$ | = | $\left(\frac{1}{2} + x_{12}\right) \mathbf{a}_1 + \frac{1}{4} \mathbf{a}_2 + \left(\frac{1}{2} - z_{12}\right) \mathbf{a}_3$ | = | $\left(\frac{1}{2} + x_{12}\right) a \, \mathbf{\hat{x}} + \frac{1}{4} b \, \mathbf{\hat{y}} + \left(\frac{1}{2} - z_{12}\right) c \, \mathbf{\hat{z}}$ | (4$c$) | P IV |
| $\mathbf{B}_{49}$ | = | $x_{13} \mathbf{a}_1 + \frac{1}{4} \mathbf{a}_2 + z_{13} \mathbf{a}_3$ | = | $x_{13} a \, \mathbf{\hat{x}} + \frac{1}{4} b \, \mathbf{\hat{y}} + z_{13} c \, \mathbf{\hat{z}}$ | (4$c$) | P V |
| $\mathbf{B}_{50}$ | = | $\left(\frac{1}{2} - x_{13}\right) \mathbf{a}_1 + \frac{3}{4} \mathbf{a}_2 + \left(\frac{1}{2} + z_{13}\right) \mathbf{a}_3$ | = | $\left(\frac{1}{2} - x_{13}\right) a \, \mathbf{\hat{x}} + \frac{3}{4} b \, \mathbf{\hat{y}} + \left(\frac{1}{2} + z_{13}\right) c \, \mathbf{\hat{z}}$ | (4$c$) | P V |
| $\mathbf{B}_{51}$ | = | $-x_{13} \mathbf{a}_1 + \frac{3}{4} \mathbf{a}_2 - z_{13} \mathbf{a}_3$ | = | $-x_{13} a \, \mathbf{\hat{x}} + \frac{3}{4} b \, \mathbf{\hat{y}} - z_{13} c \, \mathbf{\hat{z}}$ | (4$c$) | P V |
| $\mathbf{B}_{52}$ | = | $\left(\frac{1}{2} + x_{13}\right) \mathbf{a}_1 + \frac{1}{4} \mathbf{a}_2 + \left(\frac{1}{2} - z_{13}\right) \mathbf{a}_3$ | = | $\left(\frac{1}{2} + x_{13}\right) a \, \mathbf{\hat{x}} + \frac{1}{4} b \, \mathbf{\hat{y}} + \left(\frac{1}{2} - z_{13}\right) c \, \mathbf{\hat{z}}$ | (4$c$) | P V |
| $\mathbf{B}_{53}$ | = | $x_{14} \mathbf{a}_1 + \frac{1}{4} \mathbf{a}_2 + z_{14} \mathbf{a}_3$ | = | $x_{14} a \, \mathbf{\hat{x}} + \frac{1}{4} b \, \mathbf{\hat{y}} + z_{14} c \, \mathbf{\hat{z}}$ | (4$c$) | P VI |
| $\mathbf{B}_{54}$ | = | $\left(\frac{1}{2} - x_{14}\right) \mathbf{a}_1 + \frac{3}{4} \mathbf{a}_2 + \left(\frac{1}{2} + z_{14}\right) \mathbf{a}_3$ | = | $\left(\frac{1}{2} - x_{14}\right) a \, \mathbf{\hat{x}} + \frac{3}{4} b \, \mathbf{\hat{y}} + \left(\frac{1}{2} + z_{14}\right) c \, \mathbf{\hat{z}}$ | (4$c$) | P VI |
| $\mathbf{B}_{55}$ | = | $-x_{14} \mathbf{a}_1 + \frac{3}{4} \mathbf{a}_2 - z_{14} \mathbf{a}_3$ | = | $-x_{14} a \, \mathbf{\hat{x}} + \frac{3}{4} b \, \mathbf{\hat{y}} - z_{14} c \, \mathbf{\hat{z}}$ | (4$c$) | P VI |
| $\mathbf{B}_{56}$ | = | $\left(\frac{1}{2} + x_{14}\right) \mathbf{a}_1 + \frac{1}{4} \mathbf{a}_2 + \left(\frac{1}{2} - z_{14}\right) \mathbf{a}_3$ | = | $\left(\frac{1}{2} + x_{14}\right) a \, \mathbf{\hat{x}} + \frac{1}{4} b \, \mathbf{\hat{y}} + \left(\frac{1}{2} - z_{14}\right) c \, \mathbf{\hat{z}}$ | (4$c$) | P VI |

---

---

**Geometry files:**

- CIF: pp. 1638
- POSCAR: pp. 1639

# $K_2SnCl_4 \cdot H_2O$ ($E3_5$) Structure: A4BC2D_oP32_62_2cd_b_2c_a

http://aflow.org/prototype-encyclopedia/A4BC2D_oP32_62_2cd_b_2c_a

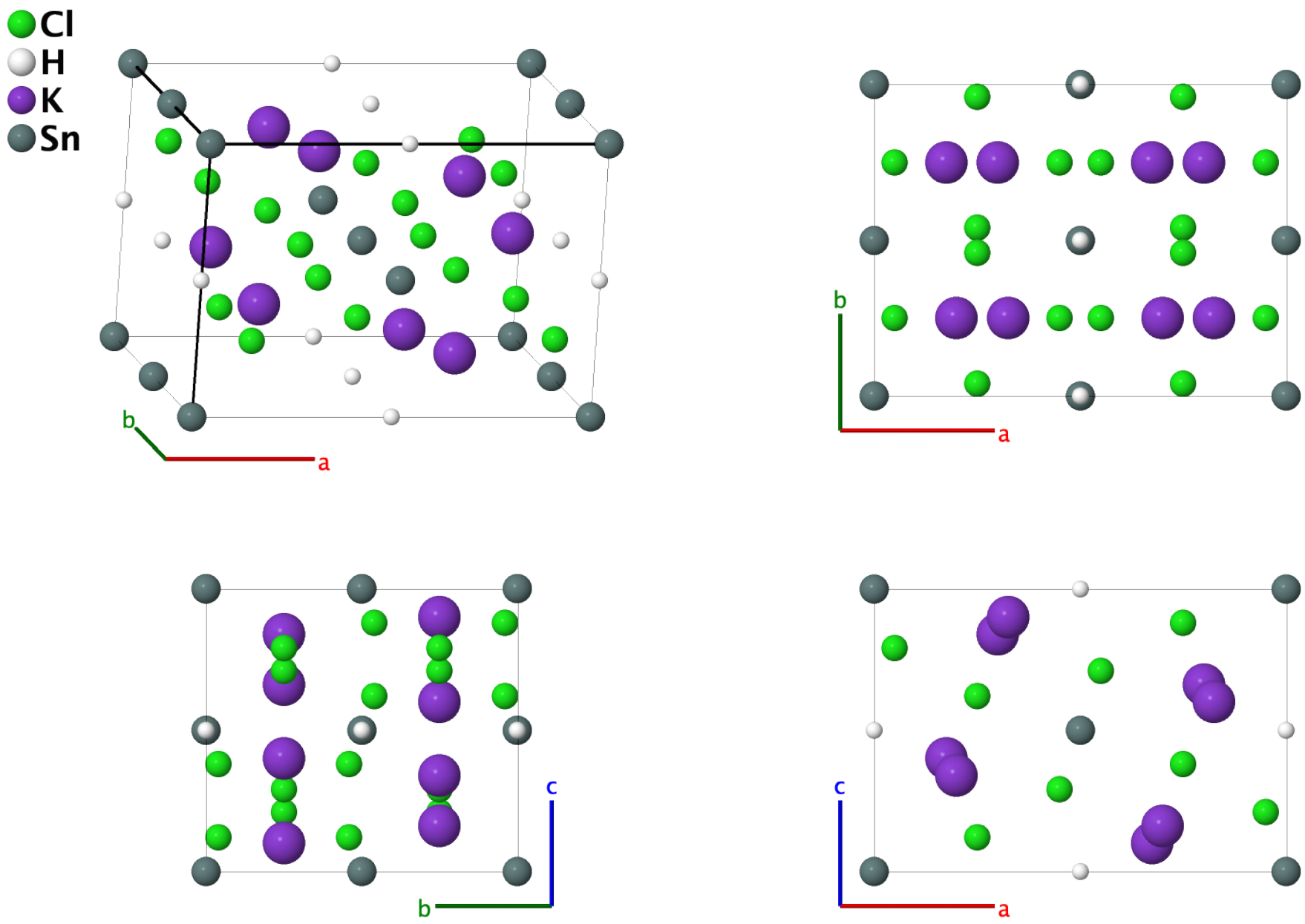

| | | |
|---|---|---|
| **Prototype** | : | $Cl_4(H_2O)K_2Sn$ |
| **AFLOW prototype label** | : | A4BC2D_oP32_62_2cd_b_2c_a |
| ***Strukturbericht* designation** | : | $E3_5$ |
| **Pearson symbol** | : | oP32 |
| **Space group number** | : | 62 |
| **Space group symbol** | : | *Pnma* |
| **AFLOW prototype command** | : | `aflow --proto=A4BC2D_oP32_62_2cd_b_2c_a` `--params=`$a, b/a, c/a, x_3, z_3, x_4, z_4, x_5, z_5, x_6, z_6, x_7, y_7, z_7$ |

- This is the structure listed in (Hermann, 1939) as the $E3_5$ structure.
- (Kamenar, 1962) proposed a somewhat different version of the structure for this crystal. We have no clear indication as to which structure is correct.
- The positions of the hydrogen atoms in the water molecules were not determined, so we only give the positions of the oxygen atoms (labeled as $H_2O$).
- (Brasseur, 1939) gives the crystal structure and atomic positions in the *Pbnm* setting of space group #62. We have used FINDSYM to show the structure in the standard *Pnma* setting.
- (Hermann, 1939) gives incorrect values for the Wyckoff positions of the Cl III atoms. The correct coordinates are ($x_{\rm ClIII} = 0.79$, $y_{\rm ClIII} = +0.05$).

**Simple Orthorhombic primitive vectors:**

$$\mathbf{a}_1 = a\,\hat{\mathbf{x}}$$
$$\mathbf{a}_2 = b\,\hat{\mathbf{y}}$$
$$\mathbf{a}_3 = c\,\hat{\mathbf{z}}$$

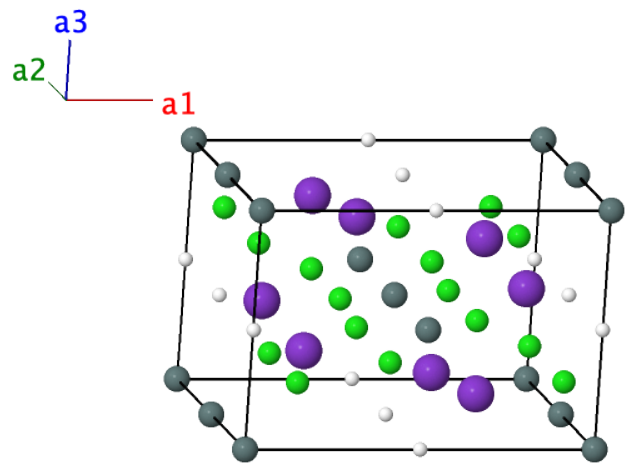

**Basis vectors:**

| | | Lattice Coordinates | | Cartesian Coordinates | Wyckoff Position | Atom Type |
|---|---|---|---|---|---|---|
| $\mathbf{B}_1$ | = | $0\,\mathbf{a}_1 + 0\,\mathbf{a}_2 + 0\,\mathbf{a}_3$ | = | $0\,\hat{\mathbf{x}} + 0\,\hat{\mathbf{y}} + 0\,\hat{\mathbf{z}}$ | (4*a*) | Sn |
| $\mathbf{B}_2$ | = | $\frac{1}{2}\,\mathbf{a}_1 + \frac{1}{2}\,\mathbf{a}_3$ | = | $\frac{1}{2}a\,\hat{\mathbf{x}} + \frac{1}{2}c\,\hat{\mathbf{z}}$ | (4*a*) | Sn |
| $\mathbf{B}_3$ | = | $\frac{1}{2}\,\mathbf{a}_2$ | = | $\frac{1}{2}b\,\hat{\mathbf{y}}$ | (4*a*) | Sn |
| $\mathbf{B}_4$ | = | $\frac{1}{2}\,\mathbf{a}_1 + \frac{1}{2}\,\mathbf{a}_2 + \frac{1}{2}\,\mathbf{a}_3$ | = | $\frac{1}{2}a\,\hat{\mathbf{x}} + \frac{1}{2}b\,\hat{\mathbf{y}} + \frac{1}{2}c\,\hat{\mathbf{z}}$ | (4*a*) | Sn |
| $\mathbf{B}_5$ | = | $\frac{1}{2}\,\mathbf{a}_3$ | = | $\frac{1}{2}c\,\hat{\mathbf{z}}$ | (4*b*) | $H_2O$ |
| $\mathbf{B}_6$ | = | $\frac{1}{2}\,\mathbf{a}_1$ | = | $\frac{1}{2}a\,\hat{\mathbf{x}}$ | (4*b*) | $H_2O$ |
| $\mathbf{B}_7$ | = | $\frac{1}{2}\,\mathbf{a}_2 + \frac{1}{2}\,\mathbf{a}_3$ | = | $\frac{1}{2}b\,\hat{\mathbf{y}} + \frac{1}{2}c\,\hat{\mathbf{z}}$ | (4*b*) | $H_2O$ |
| $\mathbf{B}_8$ | = | $\frac{1}{2}\,\mathbf{a}_1 + \frac{1}{2}\,\mathbf{a}_2$ | = | $\frac{1}{2}a\,\hat{\mathbf{x}} + \frac{1}{2}b\,\hat{\mathbf{y}}$ | (4*b*) | $H_2O$ |
| $\mathbf{B}_9$ | = | $x_3\,\mathbf{a}_1 + \frac{1}{4}\,\mathbf{a}_2 + z_3\,\mathbf{a}_3$ | = | $x_3a\,\hat{\mathbf{x}} + \frac{1}{4}b\,\hat{\mathbf{y}} + z_3c\,\hat{\mathbf{z}}$ | (4*c*) | Cl I |
| $\mathbf{B}_{10}$ | = | $\left(\frac{1}{2} - x_3\right)\mathbf{a}_1 + \frac{3}{4}\,\mathbf{a}_2 + \left(\frac{1}{2} + z_3\right)\mathbf{a}_3$ | = | $\left(\frac{1}{2} - x_3\right)a\,\hat{\mathbf{x}} + \frac{3}{4}b\,\hat{\mathbf{y}} + \left(\frac{1}{2} + z_3\right)c\,\hat{\mathbf{z}}$ | (4*c*) | Cl I |
| $\mathbf{B}_{11}$ | = | $-x_3\,\mathbf{a}_1 + \frac{3}{4}\,\mathbf{a}_2 - z_3\,\mathbf{a}_3$ | = | $-x_3a\,\hat{\mathbf{x}} + \frac{3}{4}b\,\hat{\mathbf{y}} - z_3c\,\hat{\mathbf{z}}$ | (4*c*) | Cl I |
| $\mathbf{B}_{12}$ | = | $\left(\frac{1}{2} + x_3\right)\mathbf{a}_1 + \frac{1}{4}\,\mathbf{a}_2 + \left(\frac{1}{2} - z_3\right)\mathbf{a}_3$ | = | $\left(\frac{1}{2} + x_3\right)a\,\hat{\mathbf{x}} + \frac{1}{4}b\,\hat{\mathbf{y}} + \left(\frac{1}{2} - z_3\right)c\,\hat{\mathbf{z}}$ | (4*c*) | Cl I |
| $\mathbf{B}_{13}$ | = | $x_4\,\mathbf{a}_1 + \frac{1}{4}\,\mathbf{a}_2 + z_4\,\mathbf{a}_3$ | = | $x_4a\,\hat{\mathbf{x}} + \frac{1}{4}b\,\hat{\mathbf{y}} + z_4c\,\hat{\mathbf{z}}$ | (4*c*) | Cl II |
| $\mathbf{B}_{14}$ | = | $\left(\frac{1}{2} - x_4\right)\mathbf{a}_1 + \frac{3}{4}\,\mathbf{a}_2 + \left(\frac{1}{2} + z_4\right)\mathbf{a}_3$ | = | $\left(\frac{1}{2} - x_4\right)a\,\hat{\mathbf{x}} + \frac{3}{4}b\,\hat{\mathbf{y}} + \left(\frac{1}{2} + z_4\right)c\,\hat{\mathbf{z}}$ | (4*c*) | Cl II |
| $\mathbf{B}_{15}$ | = | $-x_4\,\mathbf{a}_1 + \frac{3}{4}\,\mathbf{a}_2 - z_4\,\mathbf{a}_3$ | = | $-x_4a\,\hat{\mathbf{x}} + \frac{3}{4}b\,\hat{\mathbf{y}} - z_4c\,\hat{\mathbf{z}}$ | (4*c*) | Cl II |
| $\mathbf{B}_{16}$ | = | $\left(\frac{1}{2} + x_4\right)\mathbf{a}_1 + \frac{1}{4}\,\mathbf{a}_2 + \left(\frac{1}{2} - z_4\right)\mathbf{a}_3$ | = | $\left(\frac{1}{2} + x_4\right)a\,\hat{\mathbf{x}} + \frac{1}{4}b\,\hat{\mathbf{y}} + \left(\frac{1}{2} - z_4\right)c\,\hat{\mathbf{z}}$ | (4*c*) | Cl II |
| $\mathbf{B}_{17}$ | = | $x_5\,\mathbf{a}_1 + \frac{1}{4}\,\mathbf{a}_2 + z_5\,\mathbf{a}_3$ | = | $x_5a\,\hat{\mathbf{x}} + \frac{1}{4}b\,\hat{\mathbf{y}} + z_5c\,\hat{\mathbf{z}}$ | (4*c*) | K I |
| $\mathbf{B}_{18}$ | = | $\left(\frac{1}{2} - x_5\right)\mathbf{a}_1 + \frac{3}{4}\,\mathbf{a}_2 + \left(\frac{1}{2} + z_5\right)\mathbf{a}_3$ | = | $\left(\frac{1}{2} - x_5\right)a\,\hat{\mathbf{x}} + \frac{3}{4}b\,\hat{\mathbf{y}} + \left(\frac{1}{2} + z_5\right)c\,\hat{\mathbf{z}}$ | (4*c*) | K I |
| $\mathbf{B}_{19}$ | = | $-x_5\,\mathbf{a}_1 + \frac{3}{4}\,\mathbf{a}_2 - z_5\,\mathbf{a}_3$ | = | $-x_5a\,\hat{\mathbf{x}} + \frac{3}{4}b\,\hat{\mathbf{y}} - z_5c\,\hat{\mathbf{z}}$ | (4*c*) | K I |
| $\mathbf{B}_{20}$ | = | $\left(\frac{1}{2} + x_5\right)\mathbf{a}_1 + \frac{1}{4}\,\mathbf{a}_2 + \left(\frac{1}{2} - z_5\right)\mathbf{a}_3$ | = | $\left(\frac{1}{2} + x_5\right)a\,\hat{\mathbf{x}} + \frac{1}{4}b\,\hat{\mathbf{y}} + \left(\frac{1}{2} - z_5\right)c\,\hat{\mathbf{z}}$ | (4*c*) | K I |
| $\mathbf{B}_{21}$ | = | $x_6\,\mathbf{a}_1 + \frac{1}{4}\,\mathbf{a}_2 + z_6\,\mathbf{a}_3$ | = | $x_6a\,\hat{\mathbf{x}} + \frac{1}{4}b\,\hat{\mathbf{y}} + z_6c\,\hat{\mathbf{z}}$ | (4*c*) | K II |
| $\mathbf{B}_{22}$ | = | $\left(\frac{1}{2} - x_6\right)\mathbf{a}_1 + \frac{3}{4}\,\mathbf{a}_2 + \left(\frac{1}{2} + z_6\right)\mathbf{a}_3$ | = | $\left(\frac{1}{2} - x_6\right)a\,\hat{\mathbf{x}} + \frac{3}{4}b\,\hat{\mathbf{y}} + \left(\frac{1}{2} + z_6\right)c\,\hat{\mathbf{z}}$ | (4*c*) | K II |
| $\mathbf{B}_{23}$ | = | $-x_6\,\mathbf{a}_1 + \frac{3}{4}\,\mathbf{a}_2 - z_6\,\mathbf{a}_3$ | = | $-x_6a\,\hat{\mathbf{x}} + \frac{3}{4}b\,\hat{\mathbf{y}} - z_6c\,\hat{\mathbf{z}}$ | (4*c*) | K II |
| $\mathbf{B}_{24}$ | = | $\left(\frac{1}{2} + x_6\right)\mathbf{a}_1 + \frac{1}{4}\,\mathbf{a}_2 + \left(\frac{1}{2} - z_6\right)\mathbf{a}_3$ | = | $\left(\frac{1}{2} + x_6\right)a\,\hat{\mathbf{x}} + \frac{1}{4}b\,\hat{\mathbf{y}} + \left(\frac{1}{2} - z_6\right)c\,\hat{\mathbf{z}}$ | (4*c*) | K II |
| $\mathbf{B}_{25}$ | = | $x_7\,\mathbf{a}_1 + y_7\,\mathbf{a}_2 + z_7\,\mathbf{a}_3$ | = | $x_7a\,\hat{\mathbf{x}} + y_7b\,\hat{\mathbf{y}} + z_7c\,\hat{\mathbf{z}}$ | (8*d*) | Cl III |
| $\mathbf{B}_{26}$ | = | $\left(\frac{1}{2} - x_7\right)\mathbf{a}_1 - y_7\,\mathbf{a}_2 + \left(\frac{1}{2} + z_7\right)\mathbf{a}_3$ | = | $\left(\frac{1}{2} - x_7\right)a\,\hat{\mathbf{x}} - y_7b\,\hat{\mathbf{y}} + \left(\frac{1}{2} + z_7\right)c\,\hat{\mathbf{z}}$ | (8*d*) | Cl III |
| $\mathbf{B}_{27}$ | = | $-x_7\,\mathbf{a}_1 + \left(\frac{1}{2} + y_7\right)\mathbf{a}_2 - z_7\,\mathbf{a}_3$ | = | $-x_7a\,\hat{\mathbf{x}} + \left(\frac{1}{2} + y_7\right)b\,\hat{\mathbf{y}} - z_7c\,\hat{\mathbf{z}}$ | (8*d*) | Cl III |
| $\mathbf{B}_{28}$ | = | $\left(\frac{1}{2}+x_7\right)\mathbf{a}_1+\left(\frac{1}{2}-y_7\right)\mathbf{a}_2+\left(\frac{1}{2}-z_7\right)\mathbf{a}_3$ | = | $\left(\frac{1}{2}+x_7\right)a\,\hat{\mathbf{x}}+\left(\frac{1}{2}-y_7\right)b\,\hat{\mathbf{y}}+\left(\frac{1}{2}-z_7\right)c\,\hat{\mathbf{z}}$ | (8$d$) | Cl III |
| $\mathbf{B}_{29}$ | = | $-x_7\,\mathbf{a}_1-y_7\,\mathbf{a}_2-z_7\,\mathbf{a}_3$ | = | $-x_7a\,\hat{\mathbf{x}}-y_7b\,\hat{\mathbf{y}}-z_7c\,\hat{\mathbf{z}}$ | (8$d$) | Cl III |
| $\mathbf{B}_{30}$ | = | $\left(\frac{1}{2}+x_7\right)\mathbf{a}_1+y_7\,\mathbf{a}_2+\left(\frac{1}{2}-z_7\right)\mathbf{a}_3$ | = | $\left(\frac{1}{2}+x_7\right)a\,\hat{\mathbf{x}}+y_7b\,\hat{\mathbf{y}}+\left(\frac{1}{2}-z_7\right)c\,\hat{\mathbf{z}}$ | (8$d$) | Cl III |
| $\mathbf{B}_{31}$ | = | $x_7\,\mathbf{a}_1+\left(\frac{1}{2}-y_7\right)\mathbf{a}_2+z_7\,\mathbf{a}_3$ | = | $x_7a\,\hat{\mathbf{x}}+\left(\frac{1}{2}-y_7\right)b\,\hat{\mathbf{y}}+z_7c\,\hat{\mathbf{z}}$ | (8$d$) | Cl III |
| $\mathbf{B}_{32}$ | = | $\left(\frac{1}{2}-x_7\right)\mathbf{a}_1+\left(\frac{1}{2}+y_7\right)\mathbf{a}_2+\left(\frac{1}{2}+z_7\right)\mathbf{a}_3$ | = | $\left(\frac{1}{2}-x_7\right)a\,\hat{\mathbf{x}}+\left(\frac{1}{2}+y_7\right)b\,\hat{\mathbf{y}}+\left(\frac{1}{2}+z_7\right)c\,\hat{\mathbf{z}}$ | (8$d$) | Cl III |

**Geometry files:**

- CIF: pp. 1639
- POSCAR: pp. 1640

# $K_2SnCl_4 \cdot H_2O$ Structure: A4BC2D_oP32_62_2cd_c_d_c

http://aflow.org/prototype-encyclopedia/A4BC2D_oP32_62_2cd_c_d_c

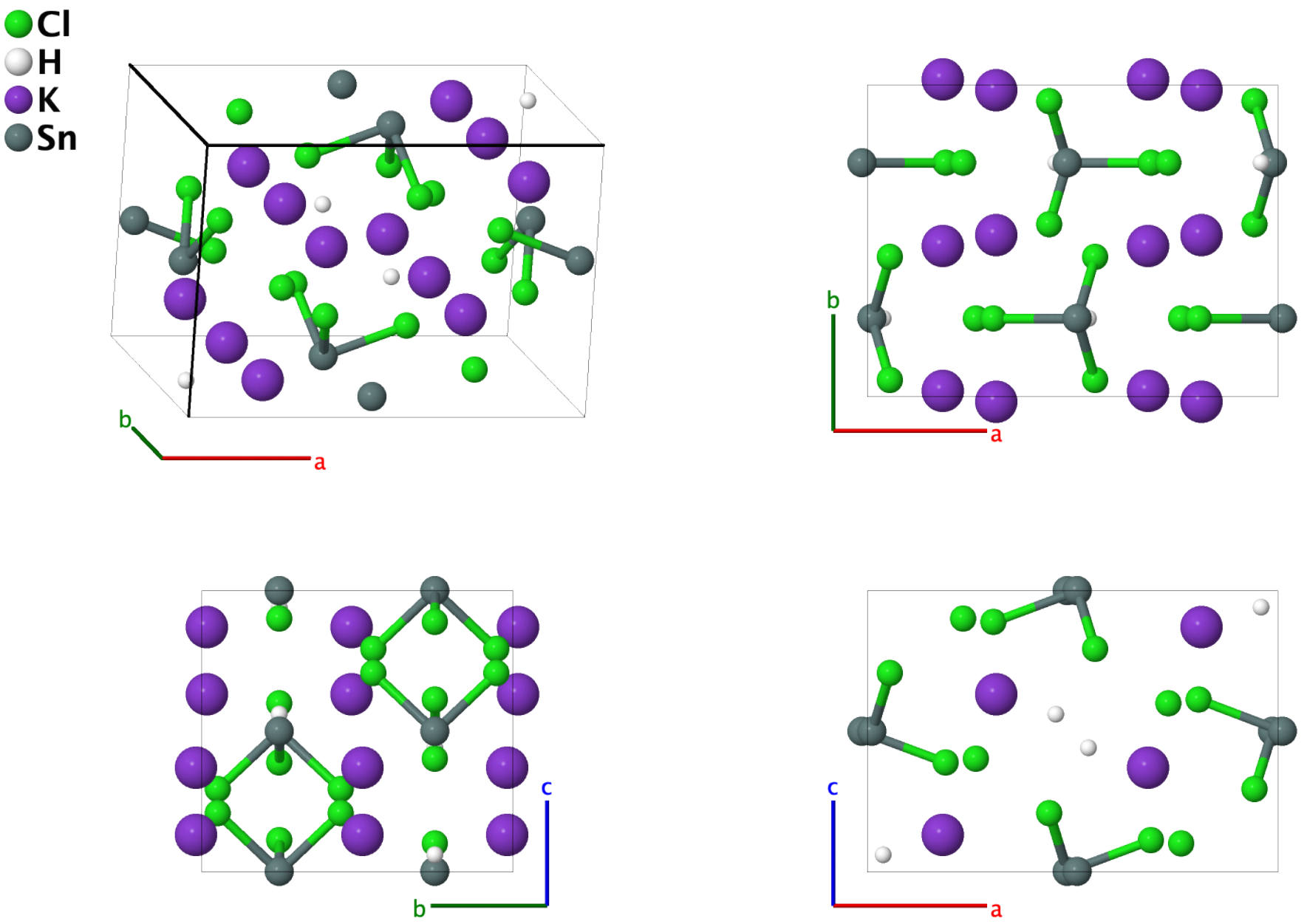

| | | |
|---|---|---|
| **Prototype** | : | $Cl_4(H_2O)K_2Sn$ |
| **AFLOW prototype label** | : | A4BC2D_oP32_62_2cd_c_d_c |
| ***Strukturbericht* designation** | : | None |
| **Pearson symbol** | : | oP32 |
| **Space group number** | : | 62 |
| **Space group symbol** | : | *Pnma* |
| **AFLOW prototype command** | : | `aflow --proto=A4BC2D_oP32_62_2cd_c_d_c` `--params=`$a, b/a, c/a, x_1, z_1, x_2, z_2, x_3, z_3, x_4, z_4, x_5, y_5, z_5, x_6, y_6, z_6$ |

- This is *not* the version of the structure given in (Hermann, 1939) as *Strukturbericht* $E3_3$. That structure was determined by (Brasseur, 1939). We have no clear indication as to which structure is correct.
- The positions of the hydrogen atoms in the water molecules were not determined, so we only give the positions of the oxygen atoms (labeled as $H_2O$).
- (Kamenar, 1962) gives the crystal structure and atomic positions in the *Pbnm* setting of space group #62. We have used FINDSYM to show the structure in the standard *Pnma* setting.

**Simple Orthorhombic primitive vectors:**

$$\mathbf{a}_1 = a\,\hat{\mathbf{x}}$$
$$\mathbf{a}_2 = b\,\hat{\mathbf{y}}$$
$$\mathbf{a}_3 = c\,\hat{\mathbf{z}}$$

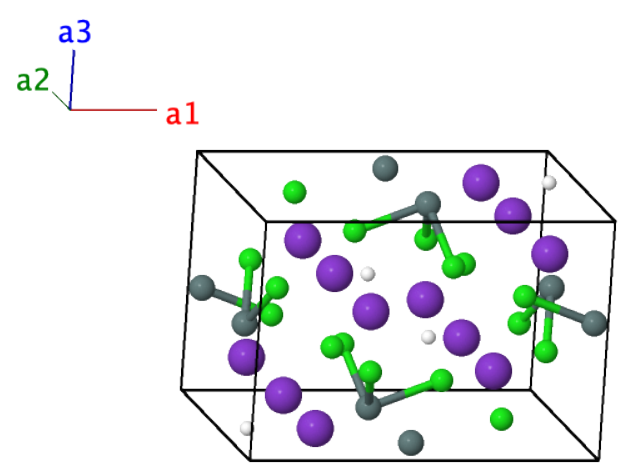

**Basis vectors:**

| | | Lattice Coordinates | | Cartesian Coordinates | Wyckoff Position | Atom Type |
|---|---|---|---|---|---|---|
| $\mathbf{B}_1$ | = | $x_1\,\mathbf{a}_1 + \frac{1}{4}\,\mathbf{a}_2 + z_1\,\mathbf{a}_3$ | = | $x_1 a\,\hat{\mathbf{x}} + \frac{1}{4}b\,\hat{\mathbf{y}} + z_1 c\,\hat{\mathbf{z}}$ | (4$c$) | Cl I |
| $\mathbf{B}_2$ | = | $\left(\frac{1}{2} - x_1\right)\mathbf{a}_1 + \frac{3}{4}\,\mathbf{a}_2 + \left(\frac{1}{2} + z_1\right)\mathbf{a}_3$ | = | $\left(\frac{1}{2} - x_1\right)a\,\hat{\mathbf{x}} + \frac{3}{4}b\,\hat{\mathbf{y}} + \left(\frac{1}{2} + z_1\right)c\,\hat{\mathbf{z}}$ | (4$c$) | Cl I |
| $\mathbf{B}_3$ | = | $-x_1\,\mathbf{a}_1 + \frac{3}{4}\,\mathbf{a}_2 - z_1\,\mathbf{a}_3$ | = | $-x_1 a\,\hat{\mathbf{x}} + \frac{3}{4}b\,\hat{\mathbf{y}} - z_1 c\,\hat{\mathbf{z}}$ | (4$c$) | Cl I |
| $\mathbf{B}_4$ | = | $\left(\frac{1}{2} + x_1\right)\mathbf{a}_1 + \frac{1}{4}\,\mathbf{a}_2 + \left(\frac{1}{2} - z_1\right)\mathbf{a}_3$ | = | $\left(\frac{1}{2} + x_1\right)a\,\hat{\mathbf{x}} + \frac{1}{4}b\,\hat{\mathbf{y}} + \left(\frac{1}{2} - z_1\right)c\,\hat{\mathbf{z}}$ | (4$c$) | Cl I |
| $\mathbf{B}_5$ | = | $x_2\,\mathbf{a}_1 + \frac{1}{4}\,\mathbf{a}_2 + z_2\,\mathbf{a}_3$ | = | $x_2 a\,\hat{\mathbf{x}} + \frac{1}{4}b\,\hat{\mathbf{y}} + z_2 c\,\hat{\mathbf{z}}$ | (4$c$) | Cl II |
| $\mathbf{B}_6$ | = | $\left(\frac{1}{2} - x_2\right)\mathbf{a}_1 + \frac{3}{4}\,\mathbf{a}_2 + \left(\frac{1}{2} + z_2\right)\mathbf{a}_3$ | = | $\left(\frac{1}{2} - x_2\right)a\,\hat{\mathbf{x}} + \frac{3}{4}b\,\hat{\mathbf{y}} + \left(\frac{1}{2} + z_2\right)c\,\hat{\mathbf{z}}$ | (4$c$) | Cl II |
| $\mathbf{B}_7$ | = | $-x_2\,\mathbf{a}_1 + \frac{3}{4}\,\mathbf{a}_2 - z_2\,\mathbf{a}_3$ | = | $-x_2 a\,\hat{\mathbf{x}} + \frac{3}{4}b\,\hat{\mathbf{y}} - z_2 c\,\hat{\mathbf{z}}$ | (4$c$) | Cl II |
| $\mathbf{B}_8$ | = | $\left(\frac{1}{2} + x_2\right)\mathbf{a}_1 + \frac{1}{4}\,\mathbf{a}_2 + \left(\frac{1}{2} - z_2\right)\mathbf{a}_3$ | = | $\left(\frac{1}{2} + x_2\right)a\,\hat{\mathbf{x}} + \frac{1}{4}b\,\hat{\mathbf{y}} + \left(\frac{1}{2} - z_2\right)c\,\hat{\mathbf{z}}$ | (4$c$) | Cl II |
| $\mathbf{B}_9$ | = | $x_3\,\mathbf{a}_1 + \frac{1}{4}\,\mathbf{a}_2 + z_3\,\mathbf{a}_3$ | = | $x_3 a\,\hat{\mathbf{x}} + \frac{1}{4}b\,\hat{\mathbf{y}} + z_3 c\,\hat{\mathbf{z}}$ | (4$c$) | $H_2O$ |
| $\mathbf{B}_{10}$ | = | $\left(\frac{1}{2} - x_3\right)\mathbf{a}_1 + \frac{3}{4}\,\mathbf{a}_2 + \left(\frac{1}{2} + z_3\right)\mathbf{a}_3$ | = | $\left(\frac{1}{2} - x_3\right)a\,\hat{\mathbf{x}} + \frac{3}{4}b\,\hat{\mathbf{y}} + \left(\frac{1}{2} + z_3\right)c\,\hat{\mathbf{z}}$ | (4$c$) | $H_2O$ |
| $\mathbf{B}_{11}$ | = | $-x_3\,\mathbf{a}_1 + \frac{3}{4}\,\mathbf{a}_2 - z_3\,\mathbf{a}_3$ | = | $-x_3 a\,\hat{\mathbf{x}} + \frac{3}{4}b\,\hat{\mathbf{y}} - z_3 c\,\hat{\mathbf{z}}$ | (4$c$) | $H_2O$ |
| $\mathbf{B}_{12}$ | = | $\left(\frac{1}{2} + x_3\right)\mathbf{a}_1 + \frac{1}{4}\,\mathbf{a}_2 + \left(\frac{1}{2} - z_3\right)\mathbf{a}_3$ | = | $\left(\frac{1}{2} + x_3\right)a\,\hat{\mathbf{x}} + \frac{1}{4}b\,\hat{\mathbf{y}} + \left(\frac{1}{2} - z_3\right)c\,\hat{\mathbf{z}}$ | (4$c$) | $H_2O$ |
| $\mathbf{B}_{13}$ | = | $x_4\,\mathbf{a}_1 + \frac{1}{4}\,\mathbf{a}_2 + z_4\,\mathbf{a}_3$ | = | $x_4 a\,\hat{\mathbf{x}} + \frac{1}{4}b\,\hat{\mathbf{y}} + z_4 c\,\hat{\mathbf{z}}$ | (4$c$) | Sn |
| $\mathbf{B}_{14}$ | = | $\left(\frac{1}{2} - x_4\right)\mathbf{a}_1 + \frac{3}{4}\,\mathbf{a}_2 + \left(\frac{1}{2} + z_4\right)\mathbf{a}_3$ | = | $\left(\frac{1}{2} - x_4\right)a\,\hat{\mathbf{x}} + \frac{3}{4}b\,\hat{\mathbf{y}} + \left(\frac{1}{2} + z_4\right)c\,\hat{\mathbf{z}}$ | (4$c$) | Sn |
| $\mathbf{B}_{15}$ | = | $-x_4\,\mathbf{a}_1 + \frac{3}{4}\,\mathbf{a}_2 - z_4\,\mathbf{a}_3$ | = | $-x_4 a\,\hat{\mathbf{x}} + \frac{3}{4}b\,\hat{\mathbf{y}} - z_4 c\,\hat{\mathbf{z}}$ | (4$c$) | Sn |
| $\mathbf{B}_{16}$ | = | $\left(\frac{1}{2} + x_4\right)\mathbf{a}_1 + \frac{1}{4}\,\mathbf{a}_2 + \left(\frac{1}{2} - z_4\right)\mathbf{a}_3$ | = | $\left(\frac{1}{2} + x_4\right)a\,\hat{\mathbf{x}} + \frac{1}{4}b\,\hat{\mathbf{y}} + \left(\frac{1}{2} - z_4\right)c\,\hat{\mathbf{z}}$ | (4$c$) | Sn |
| $\mathbf{B}_{17}$ | = | $x_5\,\mathbf{a}_1 + y_5\,\mathbf{a}_2 + z_5\,\mathbf{a}_3$ | = | $x_5 a\,\hat{\mathbf{x}} + y_5 b\,\hat{\mathbf{y}} + z_5 c\,\hat{\mathbf{z}}$ | (8$d$) | Cl III |
| $\mathbf{B}_{18}$ | = | $\left(\frac{1}{2} - x_5\right)\mathbf{a}_1 - y_5\,\mathbf{a}_2 + \left(\frac{1}{2} + z_5\right)\mathbf{a}_3$ | = | $\left(\frac{1}{2} - x_5\right)a\,\hat{\mathbf{x}} - y_5 b\,\hat{\mathbf{y}} + \left(\frac{1}{2} + z_5\right)c\,\hat{\mathbf{z}}$ | (8$d$) | Cl III |
| $\mathbf{B}_{19}$ | = | $-x_5\,\mathbf{a}_1 + \left(\frac{1}{2} + y_5\right)\mathbf{a}_2 - z_5\,\mathbf{a}_3$ | = | $-x_5 a\,\hat{\mathbf{x}} + \left(\frac{1}{2} + y_5\right)b\,\hat{\mathbf{y}} - z_5 c\,\hat{\mathbf{z}}$ | (8$d$) | Cl III |
| $\mathbf{B}_{20}$ | = | $\left(\frac{1}{2} + x_5\right)\mathbf{a}_1 + \left(\frac{1}{2} - y_5\right)\mathbf{a}_2 + \left(\frac{1}{2} - z_5\right)\mathbf{a}_3$ | = | $\left(\frac{1}{2} + x_5\right)a\,\hat{\mathbf{x}} + \left(\frac{1}{2} - y_5\right)b\,\hat{\mathbf{y}} + \left(\frac{1}{2} - z_5\right)c\,\hat{\mathbf{z}}$ | (8$d$) | Cl III |
| $\mathbf{B}_{21}$ | = | $-x_5\,\mathbf{a}_1 - y_5\,\mathbf{a}_2 - z_5\,\mathbf{a}_3$ | = | $-x_5 a\,\hat{\mathbf{x}} - y_5 b\,\hat{\mathbf{y}} - z_5 c\,\hat{\mathbf{z}}$ | (8$d$) | Cl III |
| $\mathbf{B}_{22}$ | = | $\left(\frac{1}{2} + x_5\right)\mathbf{a}_1 + y_5\,\mathbf{a}_2 + \left(\frac{1}{2} - z_5\right)\mathbf{a}_3$ | = | $\left(\frac{1}{2} + x_5\right)a\,\hat{\mathbf{x}} + y_5 b\,\hat{\mathbf{y}} + \left(\frac{1}{2} - z_5\right)c\,\hat{\mathbf{z}}$ | (8$d$) | Cl III |
| $\mathbf{B}_{23}$ | = | $x_5\,\mathbf{a}_1 + \left(\frac{1}{2} - y_5\right)\mathbf{a}_2 + z_5\,\mathbf{a}_3$ | = | $x_5 a\,\hat{\mathbf{x}} + \left(\frac{1}{2} - y_5\right)b\,\hat{\mathbf{y}} + z_5 c\,\hat{\mathbf{z}}$ | (8$d$) | Cl III |
| $\mathbf{B}_{24}$ | = | $\left(\frac{1}{2} - x_5\right)\mathbf{a}_1 + \left(\frac{1}{2} + y_5\right)\mathbf{a}_2 + \left(\frac{1}{2} + z_5\right)\mathbf{a}_3$ | = | $\left(\frac{1}{2} - x_5\right)a\,\hat{\mathbf{x}} + \left(\frac{1}{2} + y_5\right)b\,\hat{\mathbf{y}} + \left(\frac{1}{2} + z_5\right)c\,\hat{\mathbf{z}}$ | (8$d$) | Cl III |
| $\mathbf{B}_{25}$ | = | $x_6\,\mathbf{a}_1 + y_6\,\mathbf{a}_2 + z_6\,\mathbf{a}_3$ | = | $x_6 a\,\hat{\mathbf{x}} + y_6 b\,\hat{\mathbf{y}} + z_6 c\,\hat{\mathbf{z}}$ | (8$d$) | K |
| $\mathbf{B}_{26}$ | = | $\left(\frac{1}{2} - x_6\right)\mathbf{a}_1 - y_6\,\mathbf{a}_2 + \left(\frac{1}{2} + z_6\right)\mathbf{a}_3$ | = | $\left(\frac{1}{2} - x_6\right)a\,\hat{\mathbf{x}} - y_6 b\,\hat{\mathbf{y}} + \left(\frac{1}{2} + z_6\right)c\,\hat{\mathbf{z}}$ | (8$d$) | K |
| $\mathbf{B}_{27}$ | = | $-x_6\,\mathbf{a}_1 + \left(\frac{1}{2} + y_6\right)\mathbf{a}_2 - z_6\,\mathbf{a}_3$ | = | $-x_6 a\,\hat{\mathbf{x}} + \left(\frac{1}{2} + y_6\right)b\,\hat{\mathbf{y}} - z_6 c\,\hat{\mathbf{z}}$ | (8$d$) | K |
| $\mathbf{B}_{28}$ | = | $\left(\frac{1}{2} + x_6\right)\mathbf{a}_1 + \left(\frac{1}{2} - y_6\right)\mathbf{a}_2 + \left(\frac{1}{2} - z_6\right)\mathbf{a}_3$ | = | $\left(\frac{1}{2} + x_6\right)a\,\hat{\mathbf{x}} + \left(\frac{1}{2} - y_6\right)b\,\hat{\mathbf{y}} + \left(\frac{1}{2} - z_6\right)c\,\hat{\mathbf{z}}$ | (8$d$) | K |
| $\mathbf{B}_{29}$ | = | $-x_6\,\mathbf{a}_1 - y_6\,\mathbf{a}_2 - z_6\,\mathbf{a}_3$ | = | $-x_6 a\,\hat{\mathbf{x}} - y_6 b\,\hat{\mathbf{y}} - z_6 c\,\hat{\mathbf{z}}$ | (8$d$) | K |
| $\mathbf{B}_{30}$ | = | $\left(\frac{1}{2} + x_6\right)\mathbf{a}_1 + y_6\,\mathbf{a}_2 + \left(\frac{1}{2} - z_6\right)\mathbf{a}_3$ | = | $\left(\frac{1}{2} + x_6\right)a\,\hat{\mathbf{x}} + y_6 b\,\hat{\mathbf{y}} + \left(\frac{1}{2} - z_6\right)c\,\hat{\mathbf{z}}$ | (8$d$) | K |
| $\mathbf{B}_{31}$ | = | $x_6\,\mathbf{a}_1 + \left(\frac{1}{2} - y_6\right)\mathbf{a}_2 + z_6\,\mathbf{a}_3$ | = | $x_6 a\,\hat{\mathbf{x}} + \left(\frac{1}{2} - y_6\right)b\,\hat{\mathbf{y}} + z_6 c\,\hat{\mathbf{z}}$ | (8$d$) | K |
| $\mathbf{B}_{32}$ | = | $\left(\frac{1}{2} - x_6\right)\mathbf{a}_1 + \left(\frac{1}{2} + y_6\right)\mathbf{a}_2 + \left(\frac{1}{2} + z_6\right)\mathbf{a}_3$ | = | $\left(\frac{1}{2} - x_6\right)a\,\hat{\mathbf{x}} + \left(\frac{1}{2} + y_6\right)b\,\hat{\mathbf{y}} + \left(\frac{1}{2} + z_6\right)c\,\hat{\mathbf{z}}$ | (8$d$) | K |

---

---

**Geometry files:**

- CIF: pp. 1640
- POSCAR: pp. 1640

# $VOSO_4$ Structure: A5BC_oP28_62_3cd_c_c

http://aflow.org/prototype-encyclopedia/A5BC_oP28_62_3cd_c_c

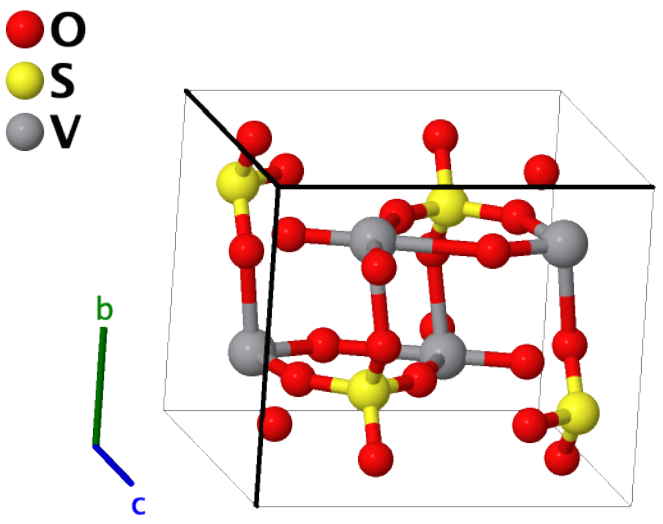

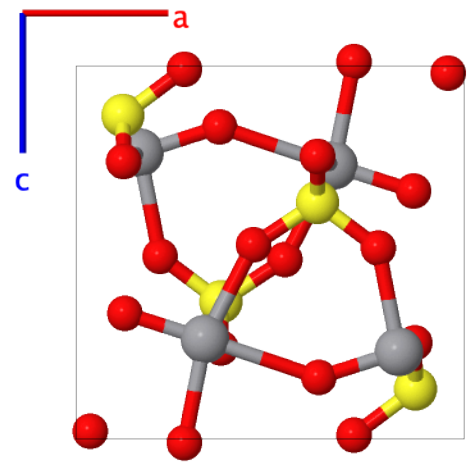

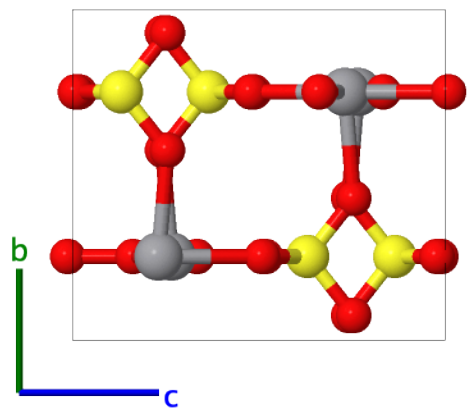

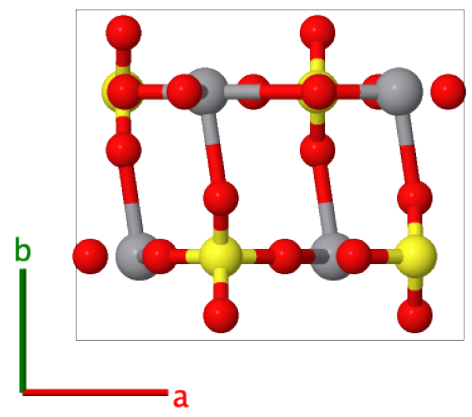

| | | |
|---|---|---|
| **Prototype** | : | $O_5SV$ |
| **AFLOW prototype label** | : | A5BC_oP28_62_3cd_c_c |
| ***Strukturbericht* designation** | : | None |
| **Pearson symbol** | : | oP28 |
| **Space group number** | : | 62 |
| **Space group symbol** | : | *Pnma* |
| **AFLOW prototype command** | : | `aflow --proto=A5BC_oP28_62_3cd_c_c`<br>`--params=`$a, b/a, c/a, x_1, z_1, x_2, z_2, x_3, z_3, x_4, z_4, x_5, z_5, x_6, y_6, z_6$ |

- This structure consists of distorted $SO_4$ tetrahedra linked to distorted $VO_6$ tetrahedra.

**Simple Orthorhombic primitive vectors:**

$$\mathbf{a}_1 = a\,\hat{\mathbf{x}}$$
$$\mathbf{a}_2 = b\,\hat{\mathbf{y}}$$
$$\mathbf{a}_3 = c\,\hat{\mathbf{z}}$$

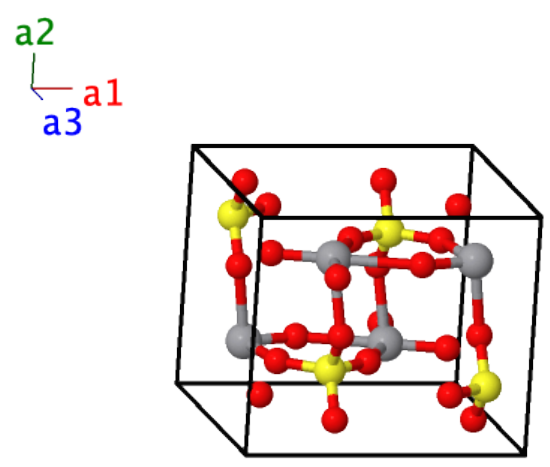

**Basis vectors:**

| | | Lattice Coordinates | | Cartesian Coordinates | Wyckoff Position | Atom Type |
|---|---|---|---|---|---|---|
| $\mathbf{B}_1$ | = | $x_1\,\mathbf{a}_1 + \frac{1}{4}\,\mathbf{a}_2 + z_1\,\mathbf{a}_3$ | = | $x_1\,a\,\hat{\mathbf{x}} + \frac{1}{4}b\,\hat{\mathbf{y}} + z_1\,c\,\hat{\mathbf{z}}$ | (4*c*) | O I |
| $\mathbf{B}_2$ | = | $\left(\frac{1}{2} - x_1\right)\mathbf{a}_1 + \frac{3}{4}\,\mathbf{a}_2 + \left(\frac{1}{2} + z_1\right)\mathbf{a}_3$ | = | $\left(\frac{1}{2} - x_1\right)a\,\hat{\mathbf{x}} + \frac{3}{4}b\,\hat{\mathbf{y}} + \left(\frac{1}{2} + z_1\right)c\,\hat{\mathbf{z}}$ | (4*c*) | O I |
| $\mathbf{B}_3$ | = | $-x_1\,\mathbf{a}_1 + \frac{3}{4}\,\mathbf{a}_2 - z_1\,\mathbf{a}_3$ | = | $-x_1\,a\,\hat{\mathbf{x}} + \frac{3}{4}b\,\hat{\mathbf{y}} - z_1\,c\,\hat{\mathbf{z}}$ | (4*c*) | O I |
| $\mathbf{B}_4$ | = | $\left(\frac{1}{2}+x_1\right)\mathbf{a}_1+\frac{1}{4}\mathbf{a}_2+\left(\frac{1}{2}-z_1\right)\mathbf{a}_3$ | = | $\left(\frac{1}{2}+x_1\right)a\,\hat{\mathbf{x}}+\frac{1}{4}b\,\hat{\mathbf{y}}+\left(\frac{1}{2}-z_1\right)c\,\hat{\mathbf{z}}$ | (4$c$) | O I |
| $\mathbf{B}_5$ | = | $x_2\,\mathbf{a}_1+\frac{1}{4}\mathbf{a}_2+z_2\,\mathbf{a}_3$ | = | $x_2a\,\hat{\mathbf{x}}+\frac{1}{4}b\,\hat{\mathbf{y}}+z_2c\,\hat{\mathbf{z}}$ | (4$c$) | O II |
| $\mathbf{B}_6$ | = | $\left(\frac{1}{2}-x_2\right)\mathbf{a}_1+\frac{3}{4}\mathbf{a}_2+\left(\frac{1}{2}+z_2\right)\mathbf{a}_3$ | = | $\left(\frac{1}{2}-x_2\right)a\,\hat{\mathbf{x}}+\frac{3}{4}b\,\hat{\mathbf{y}}+\left(\frac{1}{2}+z_2\right)c\,\hat{\mathbf{z}}$ | (4$c$) | O II |
| $\mathbf{B}_7$ | = | $-x_2\,\mathbf{a}_1+\frac{3}{4}\mathbf{a}_2-z_2\,\mathbf{a}_3$ | = | $-x_2a\,\hat{\mathbf{x}}+\frac{3}{4}b\,\hat{\mathbf{y}}-z_2c\,\hat{\mathbf{z}}$ | (4$c$) | O II |
| $\mathbf{B}_8$ | = | $\left(\frac{1}{2}+x_2\right)\mathbf{a}_1+\frac{1}{4}\mathbf{a}_2+\left(\frac{1}{2}-z_2\right)\mathbf{a}_3$ | = | $\left(\frac{1}{2}+x_2\right)a\,\hat{\mathbf{x}}+\frac{1}{4}b\,\hat{\mathbf{y}}+\left(\frac{1}{2}-z_2\right)c\,\hat{\mathbf{z}}$ | (4$c$) | O II |
| $\mathbf{B}_9$ | = | $x_3\,\mathbf{a}_1+\frac{1}{4}\mathbf{a}_2+z_3\,\mathbf{a}_3$ | = | $x_3a\,\hat{\mathbf{x}}+\frac{1}{4}b\,\hat{\mathbf{y}}+z_3c\,\hat{\mathbf{z}}$ | (4$c$) | O III |
| $\mathbf{B}_{10}$ | = | $\left(\frac{1}{2}-x_3\right)\mathbf{a}_1+\frac{3}{4}\mathbf{a}_2+\left(\frac{1}{2}+z_3\right)\mathbf{a}_3$ | = | $\left(\frac{1}{2}-x_3\right)a\,\hat{\mathbf{x}}+\frac{3}{4}b\,\hat{\mathbf{y}}+\left(\frac{1}{2}+z_3\right)c\,\hat{\mathbf{z}}$ | (4$c$) | O III |
| $\mathbf{B}_{11}$ | = | $-x_3\,\mathbf{a}_1+\frac{3}{4}\mathbf{a}_2-z_3\,\mathbf{a}_3$ | = | $-x_3a\,\hat{\mathbf{x}}+\frac{3}{4}b\,\hat{\mathbf{y}}-z_3c\,\hat{\mathbf{z}}$ | (4$c$) | O III |
| $\mathbf{B}_{12}$ | = | $\left(\frac{1}{2}+x_3\right)\mathbf{a}_1+\frac{1}{4}\mathbf{a}_2+\left(\frac{1}{2}-z_3\right)\mathbf{a}_3$ | = | $\left(\frac{1}{2}+x_3\right)a\,\hat{\mathbf{x}}+\frac{1}{4}b\,\hat{\mathbf{y}}+\left(\frac{1}{2}-z_3\right)c\,\hat{\mathbf{z}}$ | (4$c$) | O III |
| $\mathbf{B}_{13}$ | = | $x_4\,\mathbf{a}_1+\frac{1}{4}\mathbf{a}_2+z_4\,\mathbf{a}_3$ | = | $x_4a\,\hat{\mathbf{x}}+\frac{1}{4}b\,\hat{\mathbf{y}}+z_4c\,\hat{\mathbf{z}}$ | (4$c$) | S |
| $\mathbf{B}_{14}$ | = | $\left(\frac{1}{2}-x_4\right)\mathbf{a}_1+\frac{3}{4}\mathbf{a}_2+\left(\frac{1}{2}+z_4\right)\mathbf{a}_3$ | = | $\left(\frac{1}{2}-x_4\right)a\,\hat{\mathbf{x}}+\frac{3}{4}b\,\hat{\mathbf{y}}+\left(\frac{1}{2}+z_4\right)c\,\hat{\mathbf{z}}$ | (4$c$) | S |
| $\mathbf{B}_{15}$ | = | $-x_4\,\mathbf{a}_1+\frac{3}{4}\mathbf{a}_2-z_4\,\mathbf{a}_3$ | = | $-x_4a\,\hat{\mathbf{x}}+\frac{3}{4}b\,\hat{\mathbf{y}}-z_4c\,\hat{\mathbf{z}}$ | (4$c$) | S |
| $\mathbf{B}_{16}$ | = | $\left(\frac{1}{2}+x_4\right)\mathbf{a}_1+\frac{1}{4}\mathbf{a}_2+\left(\frac{1}{2}-z_4\right)\mathbf{a}_3$ | = | $\left(\frac{1}{2}+x_4\right)a\,\hat{\mathbf{x}}+\frac{1}{4}b\,\hat{\mathbf{y}}+\left(\frac{1}{2}-z_4\right)c\,\hat{\mathbf{z}}$ | (4$c$) | S |
| $\mathbf{B}_{17}$ | = | $x_5\,\mathbf{a}_1+\frac{1}{4}\mathbf{a}_2+z_5\,\mathbf{a}_3$ | = | $x_5a\,\hat{\mathbf{x}}+\frac{1}{4}b\,\hat{\mathbf{y}}+z_5c\,\hat{\mathbf{z}}$ | (4$c$) | V |
| $\mathbf{B}_{18}$ | = | $\left(\frac{1}{2}-x_5\right)\mathbf{a}_1+\frac{3}{4}\mathbf{a}_2+\left(\frac{1}{2}+z_5\right)\mathbf{a}_3$ | = | $\left(\frac{1}{2}-x_5\right)a\,\hat{\mathbf{x}}+\frac{3}{4}b\,\hat{\mathbf{y}}+\left(\frac{1}{2}+z_5\right)c\,\hat{\mathbf{z}}$ | (4$c$) | V |
| $\mathbf{B}_{19}$ | = | $-x_5\,\mathbf{a}_1+\frac{3}{4}\mathbf{a}_2-z_5\,\mathbf{a}_3$ | = | $-x_5a\,\hat{\mathbf{x}}+\frac{3}{4}b\,\hat{\mathbf{y}}-z_5c\,\hat{\mathbf{z}}$ | (4$c$) | V |
| $\mathbf{B}_{20}$ | = | $\left(\frac{1}{2}+x_5\right)\mathbf{a}_1+\frac{1}{4}\mathbf{a}_2+\left(\frac{1}{2}-z_5\right)\mathbf{a}_3$ | = | $\left(\frac{1}{2}+x_5\right)a\,\hat{\mathbf{x}}+\frac{1}{4}b\,\hat{\mathbf{y}}+\left(\frac{1}{2}-z_5\right)c\,\hat{\mathbf{z}}$ | (4$c$) | V |
| $\mathbf{B}_{21}$ | = | $x_6\,\mathbf{a}_1+y_6\,\mathbf{a}_2+z_6\,\mathbf{a}_3$ | = | $x_6a\,\hat{\mathbf{x}}+y_6b\,\hat{\mathbf{y}}+z_6c\,\hat{\mathbf{z}}$ | (8$d$) | O IV |
| $\mathbf{B}_{22}$ | = | $\left(\frac{1}{2}-x_6\right)\mathbf{a}_1-y_6\,\mathbf{a}_2+\left(\frac{1}{2}+z_6\right)\mathbf{a}_3$ | = | $\left(\frac{1}{2}-x_6\right)a\,\hat{\mathbf{x}}-y_6b\,\hat{\mathbf{y}}+\left(\frac{1}{2}+z_6\right)c\,\hat{\mathbf{z}}$ | (8$d$) | O IV |
| $\mathbf{B}_{23}$ | = | $-x_6\,\mathbf{a}_1+\left(\frac{1}{2}+y_6\right)\mathbf{a}_2-z_6\,\mathbf{a}_3$ | = | $-x_6a\,\hat{\mathbf{x}}+\left(\frac{1}{2}+y_6\right)b\,\hat{\mathbf{y}}-z_6c\,\hat{\mathbf{z}}$ | (8$d$) | O IV |
| $\mathbf{B}_{24}$ | = | $\left(\frac{1}{2}+x_6\right)\mathbf{a}_1+\left(\frac{1}{2}-y_6\right)\mathbf{a}_2+\left(\frac{1}{2}-z_6\right)\mathbf{a}_3$ | = | $\left(\frac{1}{2}+x_6\right)a\,\hat{\mathbf{x}}+\left(\frac{1}{2}-y_6\right)b\,\hat{\mathbf{y}}+\left(\frac{1}{2}-z_6\right)c\,\hat{\mathbf{z}}$ | (8$d$) | O IV |
| $\mathbf{B}_{25}$ | = | $-x_6\,\mathbf{a}_1-y_6\,\mathbf{a}_2-z_6\,\mathbf{a}_3$ | = | $-x_6a\,\hat{\mathbf{x}}-y_6b\,\hat{\mathbf{y}}-z_6c\,\hat{\mathbf{z}}$ | (8$d$) | O IV |
| $\mathbf{B}_{26}$ | = | $\left(\frac{1}{2}+x_6\right)\mathbf{a}_1+y_6\,\mathbf{a}_2+\left(\frac{1}{2}-z_6\right)\mathbf{a}_3$ | = | $\left(\frac{1}{2}+x_6\right)a\,\hat{\mathbf{x}}+y_6b\,\hat{\mathbf{y}}+\left(\frac{1}{2}-z_6\right)c\,\hat{\mathbf{z}}$ | (8$d$) | O IV |
| $\mathbf{B}_{27}$ | = | $x_6\,\mathbf{a}_1+\left(\frac{1}{2}-y_6\right)\mathbf{a}_2+z_6\,\mathbf{a}_3$ | = | $x_6a\,\hat{\mathbf{x}}+\left(\frac{1}{2}-y_6\right)b\,\hat{\mathbf{y}}+z_6c\,\hat{\mathbf{z}}$ | (8$d$) | O IV |
| $\mathbf{B}_{28}$ | = | $\left(\frac{1}{2}-x_6\right)\mathbf{a}_1+\left(\frac{1}{2}+y_6\right)\mathbf{a}_2+\left(\frac{1}{2}+z_6\right)\mathbf{a}_3$ | = | $\left(\frac{1}{2}-x_6\right)a\,\hat{\mathbf{x}}+\left(\frac{1}{2}+y_6\right)b\,\hat{\mathbf{y}}+\left(\frac{1}{2}+z_6\right)c\,\hat{\mathbf{z}}$ | (8$d$) | O IV |

**Geometry files:**

- CIF: pp. 1640
- POSCAR: pp. 1641

# Possible $\delta$-$Y_2Si_2O_7$ Structure: A7B2C2_oP44_62_3c2d_2c_d

http://aflow.org/prototype-encyclopedia/A7B2C2_oP44_62_3c2d_2c_d

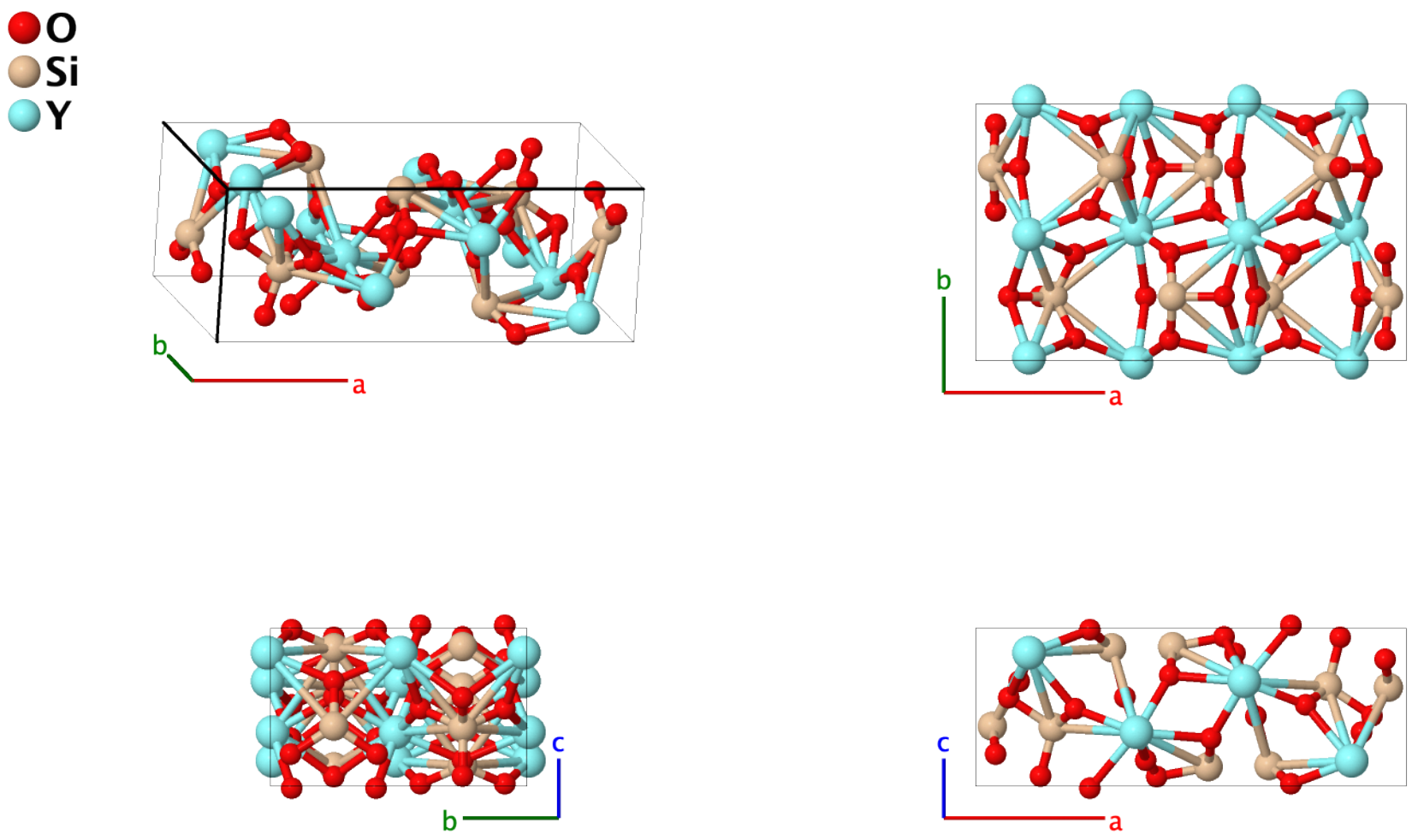

| | | |
|---|---|---|
| **Prototype** | : | $O_7Si_2Y_2$ |
| **AFLOW prototype label** | : | A7B2C2_oP44_62_3c2d_2c_d |
| ***Strukturbericht* designation** | : | None |
| **Pearson symbol** | : | oP44 |
| **Space group number** | : | 62 |
| **Space group symbol** | : | *Pnma* |
| **AFLOW prototype command** | : | `aflow --proto=A7B2C2_oP44_62_3c2d_2c_d` `--params=`$a, b/a, c/a, x_1, z_1, x_2, z_2, x_3, z_3, x_4, z_4, x_5, z_5, x_6, y_6, z_6, x_7, y_7, z_7, x_8, y_8, z_8$ |

### Other compounds with this structure

- $\delta$-$Ho_2O_7Si_2$, $\delta$-$Dy_2O_7Si_2$, $\delta$-$Y_2O_7Si_2$, and $Li_2S_2O_7$

---

- (Dias, 1990) found that some structures of $RE_2Si_2O_7$ ($RE$ = Ho, Dy, Gd, Y) were in the centro-symmetric orthorhombic *Pnma* #62 space group, in which case this would be the prototype of $\delta$-$RE_2Si_2O_7$ (Becerro, 2004). However, (Smolin, 1970) found $\delta$-$Gd_2Si_2O_7$ to be in the non-centro-symmetric $Pna2_1$ #33 space group. (Becerro, 2004) found only one yttrium site in the $\delta$-structure, supporting (Dias, 1990). In addition, if we allow a small amount of uncertainty (0.2 Å) in positions of the $Pna2_1$ structure, AFLOW-SYM and FINDSYM place this structure in the *Pnma* group. Nevertheless we have found no work explicitly stating that the structure of (Smolin, 1970) is in error, and indeed (Christensen, 1994) found $\delta$-$Y_2Si_2O_7$ in space group $Pna2_1$. Given this ambiguity, we list $Y_2Si_2O_7$ only as a possible prototype for the $\delta$-phase pyrosilicates.

**Simple Orthorhombic primitive vectors:**

$$\begin{aligned}\mathbf{a}_1 &= a\,\hat{\mathbf{x}}\\ \mathbf{a}_2 &= b\,\hat{\mathbf{y}}\\ \mathbf{a}_3 &= c\,\hat{\mathbf{z}}\end{aligned}$$

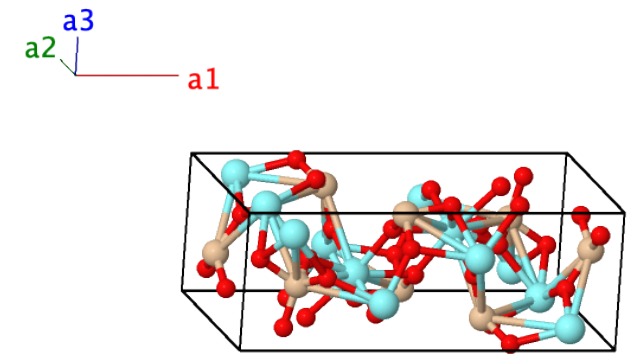

**Basis vectors:**

| | | Lattice Coordinates | | Cartesian Coordinates | Wyckoff Position | Atom Type |
|---|---|---|---|---|---|---|
| $\mathbf{B}_1$ | = | $x_1\,\mathbf{a}_1 + \frac{1}{4}\,\mathbf{a}_2 + z_1\,\mathbf{a}_3$ | = | $x_1 a\,\hat{\mathbf{x}} + \frac{1}{4} b\,\hat{\mathbf{y}} + z_1 c\,\hat{\mathbf{z}}$ | (4$c$) | O I |
| $\mathbf{B}_2$ | = | $\left(\frac{1}{2} - x_1\right)\mathbf{a}_1 + \frac{3}{4}\,\mathbf{a}_2 + \left(\frac{1}{2} + z_1\right)\mathbf{a}_3$ | = | $\left(\frac{1}{2} - x_1\right) a\,\hat{\mathbf{x}} + \frac{3}{4} b\,\hat{\mathbf{y}} + \left(\frac{1}{2} + z_1\right) c\,\hat{\mathbf{z}}$ | (4$c$) | O I |
| $\mathbf{B}_3$ | = | $-x_1\,\mathbf{a}_1 + \frac{3}{4}\,\mathbf{a}_2 - z_1\,\mathbf{a}_3$ | = | $-x_1 a\,\hat{\mathbf{x}} + \frac{3}{4} b\,\hat{\mathbf{y}} - z_1 c\,\hat{\mathbf{z}}$ | (4$c$) | O I |
| $\mathbf{B}_4$ | = | $\left(\frac{1}{2} + x_1\right)\mathbf{a}_1 + \frac{1}{4}\,\mathbf{a}_2 + \left(\frac{1}{2} - z_1\right)\mathbf{a}_3$ | = | $\left(\frac{1}{2} + x_1\right) a\,\hat{\mathbf{x}} + \frac{1}{4} b\,\hat{\mathbf{y}} + \left(\frac{1}{2} - z_1\right) c\,\hat{\mathbf{z}}$ | (4$c$) | O I |
| $\mathbf{B}_5$ | = | $x_2\,\mathbf{a}_1 + \frac{1}{4}\,\mathbf{a}_2 + z_2\,\mathbf{a}_3$ | = | $x_2 a\,\hat{\mathbf{x}} + \frac{1}{4} b\,\hat{\mathbf{y}} + z_2 c\,\hat{\mathbf{z}}$ | (4$c$) | O II |
| $\mathbf{B}_6$ | = | $\left(\frac{1}{2} - x_2\right)\mathbf{a}_1 + \frac{3}{4}\,\mathbf{a}_2 + \left(\frac{1}{2} + z_2\right)\mathbf{a}_3$ | = | $\left(\frac{1}{2} - x_2\right) a\,\hat{\mathbf{x}} + \frac{3}{4} b\,\hat{\mathbf{y}} + \left(\frac{1}{2} + z_2\right) c\,\hat{\mathbf{z}}$ | (4$c$) | O II |
| $\mathbf{B}_7$ | = | $-x_2\,\mathbf{a}_1 + \frac{3}{4}\,\mathbf{a}_2 - z_2\,\mathbf{a}_3$ | = | $-x_2 a\,\hat{\mathbf{x}} + \frac{3}{4} b\,\hat{\mathbf{y}} - z_2 c\,\hat{\mathbf{z}}$ | (4$c$) | O II |
| $\mathbf{B}_8$ | = | $\left(\frac{1}{2} + x_2\right)\mathbf{a}_1 + \frac{1}{4}\,\mathbf{a}_2 + \left(\frac{1}{2} - z_2\right)\mathbf{a}_3$ | = | $\left(\frac{1}{2} + x_2\right) a\,\hat{\mathbf{x}} + \frac{1}{4} b\,\hat{\mathbf{y}} + \left(\frac{1}{2} - z_2\right) c\,\hat{\mathbf{z}}$ | (4$c$) | O II |
| $\mathbf{B}_9$ | = | $x_3\,\mathbf{a}_1 + \frac{1}{4}\,\mathbf{a}_2 + z_3\,\mathbf{a}_3$ | = | $x_3 a\,\hat{\mathbf{x}} + \frac{1}{4} b\,\hat{\mathbf{y}} + z_3 c\,\hat{\mathbf{z}}$ | (4$c$) | O III |
| $\mathbf{B}_{10}$ | = | $\left(\frac{1}{2} - x_3\right)\mathbf{a}_1 + \frac{3}{4}\,\mathbf{a}_2 + \left(\frac{1}{2} + z_3\right)\mathbf{a}_3$ | = | $\left(\frac{1}{2} - x_3\right) a\,\hat{\mathbf{x}} + \frac{3}{4} b\,\hat{\mathbf{y}} + \left(\frac{1}{2} + z_3\right) c\,\hat{\mathbf{z}}$ | (4$c$) | O III |
| $\mathbf{B}_{11}$ | = | $-x_3\,\mathbf{a}_1 + \frac{3}{4}\,\mathbf{a}_2 - z_3\,\mathbf{a}_3$ | = | $-x_3 a\,\hat{\mathbf{x}} + \frac{3}{4} b\,\hat{\mathbf{y}} - z_3 c\,\hat{\mathbf{z}}$ | (4$c$) | O III |
| $\mathbf{B}_{12}$ | = | $\left(\frac{1}{2} + x_3\right)\mathbf{a}_1 + \frac{1}{4}\,\mathbf{a}_2 + \left(\frac{1}{2} - z_3\right)\mathbf{a}_3$ | = | $\left(\frac{1}{2} + x_3\right) a\,\hat{\mathbf{x}} + \frac{1}{4} b\,\hat{\mathbf{y}} + \left(\frac{1}{2} - z_3\right) c\,\hat{\mathbf{z}}$ | (4$c$) | O III |
| $\mathbf{B}_{13}$ | = | $x_4\,\mathbf{a}_1 + \frac{1}{4}\,\mathbf{a}_2 + z_4\,\mathbf{a}_3$ | = | $x_4 a\,\hat{\mathbf{x}} + \frac{1}{4} b\,\hat{\mathbf{y}} + z_4 c\,\hat{\mathbf{z}}$ | (4$c$) | Si I |
| $\mathbf{B}_{14}$ | = | $\left(\frac{1}{2} - x_4\right)\mathbf{a}_1 + \frac{3}{4}\,\mathbf{a}_2 + \left(\frac{1}{2} + z_4\right)\mathbf{a}_3$ | = | $\left(\frac{1}{2} - x_4\right) a\,\hat{\mathbf{x}} + \frac{3}{4} b\,\hat{\mathbf{y}} + \left(\frac{1}{2} + z_4\right) c\,\hat{\mathbf{z}}$ | (4$c$) | Si I |
| $\mathbf{B}_{15}$ | = | $-x_4\,\mathbf{a}_1 + \frac{3}{4}\,\mathbf{a}_2 - z_4\,\mathbf{a}_3$ | = | $-x_4 a\,\hat{\mathbf{x}} + \frac{3}{4} b\,\hat{\mathbf{y}} - z_4 c\,\hat{\mathbf{z}}$ | (4$c$) | Si I |
| $\mathbf{B}_{16}$ | = | $\left(\frac{1}{2} + x_4\right)\mathbf{a}_1 + \frac{1}{4}\,\mathbf{a}_2 + \left(\frac{1}{2} - z_4\right)\mathbf{a}_3$ | = | $\left(\frac{1}{2} + x_4\right) a\,\hat{\mathbf{x}} + \frac{1}{4} b\,\hat{\mathbf{y}} + \left(\frac{1}{2} - z_4\right) c\,\hat{\mathbf{z}}$ | (4$c$) | Si I |
| $\mathbf{B}_{17}$ | = | $x_5\,\mathbf{a}_1 + \frac{1}{4}\,\mathbf{a}_2 + z_5\,\mathbf{a}_3$ | = | $x_5 a\,\hat{\mathbf{x}} + \frac{1}{4} b\,\hat{\mathbf{y}} + z_5 c\,\hat{\mathbf{z}}$ | (4$c$) | Si II |
| $\mathbf{B}_{18}$ | = | $\left(\frac{1}{2} - x_5\right)\mathbf{a}_1 + \frac{3}{4}\,\mathbf{a}_2 + \left(\frac{1}{2} + z_5\right)\mathbf{a}_3$ | = | $\left(\frac{1}{2} - x_5\right) a\,\hat{\mathbf{x}} + \frac{3}{4} b\,\hat{\mathbf{y}} + \left(\frac{1}{2} + z_5\right) c\,\hat{\mathbf{z}}$ | (4$c$) | Si II |
| $\mathbf{B}_{19}$ | = | $-x_5\,\mathbf{a}_1 + \frac{3}{4}\,\mathbf{a}_2 - z_5\,\mathbf{a}_3$ | = | $-x_5 a\,\hat{\mathbf{x}} + \frac{3}{4} b\,\hat{\mathbf{y}} - z_5 c\,\hat{\mathbf{z}}$ | (4$c$) | Si II |
| $\mathbf{B}_{20}$ | = | $\left(\frac{1}{2} + x_5\right)\mathbf{a}_1 + \frac{1}{4}\,\mathbf{a}_2 + \left(\frac{1}{2} - z_5\right)\mathbf{a}_3$ | = | $\left(\frac{1}{2} + x_5\right) a\,\hat{\mathbf{x}} + \frac{1}{4} b\,\hat{\mathbf{y}} + \left(\frac{1}{2} - z_5\right) c\,\hat{\mathbf{z}}$ | (4$c$) | Si II |
| $\mathbf{B}_{21}$ | = | $x_6\,\mathbf{a}_1 + y_6\,\mathbf{a}_2 + z_6\,\mathbf{a}_3$ | = | $x_6 a\,\hat{\mathbf{x}} + y_6 b\,\hat{\mathbf{y}} + z_6 c\,\hat{\mathbf{z}}$ | (8$d$) | O IV |
| $\mathbf{B}_{22}$ | = | $\left(\frac{1}{2} - x_6\right)\mathbf{a}_1 - y_6\,\mathbf{a}_2 + \left(\frac{1}{2} + z_6\right)\mathbf{a}_3$ | = | $\left(\frac{1}{2} - x_6\right) a\,\hat{\mathbf{x}} - y_6 b\,\hat{\mathbf{y}} + \left(\frac{1}{2} + z_6\right) c\,\hat{\mathbf{z}}$ | (8$d$) | O IV |
| $\mathbf{B}_{23}$ | = | $-x_6\,\mathbf{a}_1 + \left(\frac{1}{2} + y_6\right)\mathbf{a}_2 - z_6\,\mathbf{a}_3$ | = | $-x_6 a\,\hat{\mathbf{x}} + \left(\frac{1}{2} + y_6\right) b\,\hat{\mathbf{y}} - z_6 c\,\hat{\mathbf{z}}$ | (8$d$) | O IV |
| $\mathbf{B}_{24}$ | = | $\left(\frac{1}{2} + x_6\right)\mathbf{a}_1 + \left(\frac{1}{2} - y_6\right)\mathbf{a}_2 + \left(\frac{1}{2} - z_6\right)\mathbf{a}_3$ | = | $\left(\frac{1}{2} + x_6\right) a\,\hat{\mathbf{x}} + \left(\frac{1}{2} - y_6\right) b\,\hat{\mathbf{y}} + \left(\frac{1}{2} - z_6\right) c\,\hat{\mathbf{z}}$ | (8$d$) | O IV |
| $\mathbf{B}_{25}$ | = | $-x_6\,\mathbf{a}_1 - y_6\,\mathbf{a}_2 - z_6\,\mathbf{a}_3$ | = | $-x_6 a\,\hat{\mathbf{x}} - y_6 b\,\hat{\mathbf{y}} - z_6 c\,\hat{\mathbf{z}}$ | (8$d$) | O IV |
| $\mathbf{B}_{26}$ | = | $\left(\frac{1}{2} + x_6\right)\mathbf{a}_1 + y_6\,\mathbf{a}_2 + \left(\frac{1}{2} - z_6\right)\mathbf{a}_3$ | = | $\left(\frac{1}{2} + x_6\right) a\,\hat{\mathbf{x}} + y_6 b\,\hat{\mathbf{y}} + \left(\frac{1}{2} - z_6\right) c\,\hat{\mathbf{z}}$ | (8$d$) | O IV |
| $\mathbf{B}_{27}$ | = | $x_6\,\mathbf{a}_1 + \left(\frac{1}{2} - y_6\right)\mathbf{a}_2 + z_6\,\mathbf{a}_3$ | = | $x_6 a\,\hat{\mathbf{x}} + \left(\frac{1}{2} - y_6\right) b\,\hat{\mathbf{y}} + z_6 c\,\hat{\mathbf{z}}$ | (8$d$) | O IV |
| $\mathbf{B}_{28}$ | = | $\left(\frac{1}{2} - x_6\right)\mathbf{a}_1 + \left(\frac{1}{2} + y_6\right)\mathbf{a}_2 + \left(\frac{1}{2} + z_6\right)\mathbf{a}_3$ | = | $\left(\frac{1}{2} - x_6\right)a\,\hat{\mathbf{x}} + \left(\frac{1}{2} + y_6\right)b\,\hat{\mathbf{y}} + \left(\frac{1}{2} + z_6\right)c\,\hat{\mathbf{z}}$ | (8$d$) | O IV |
| $\mathbf{B}_{29}$ | = | $x_7\,\mathbf{a}_1 + y_7\,\mathbf{a}_2 + z_7\,\mathbf{a}_3$ | = | $x_7a\,\hat{\mathbf{x}} + y_7b\,\hat{\mathbf{y}} + z_7c\,\hat{\mathbf{z}}$ | (8$d$) | O V |
| $\mathbf{B}_{30}$ | = | $\left(\frac{1}{2} - x_7\right)\mathbf{a}_1 - y_7\,\mathbf{a}_2 + \left(\frac{1}{2} + z_7\right)\mathbf{a}_3$ | = | $\left(\frac{1}{2} - x_7\right)a\,\hat{\mathbf{x}} - y_7b\,\hat{\mathbf{y}} + \left(\frac{1}{2} + z_7\right)c\,\hat{\mathbf{z}}$ | (8$d$) | O V |
| $\mathbf{B}_{31}$ | = | $-x_7\,\mathbf{a}_1 + \left(\frac{1}{2} + y_7\right)\mathbf{a}_2 - z_7\,\mathbf{a}_3$ | = | $-x_7a\,\hat{\mathbf{x}} + \left(\frac{1}{2} + y_7\right)b\,\hat{\mathbf{y}} - z_7c\,\hat{\mathbf{z}}$ | (8$d$) | O V |
| $\mathbf{B}_{32}$ | = | $\left(\frac{1}{2} + x_7\right)\mathbf{a}_1 + \left(\frac{1}{2} - y_7\right)\mathbf{a}_2 + \left(\frac{1}{2} - z_7\right)\mathbf{a}_3$ | = | $\left(\frac{1}{2} + x_7\right)a\,\hat{\mathbf{x}} + \left(\frac{1}{2} - y_7\right)b\,\hat{\mathbf{y}} + \left(\frac{1}{2} - z_7\right)c\,\hat{\mathbf{z}}$ | (8$d$) | O V |
| $\mathbf{B}_{33}$ | = | $-x_7\,\mathbf{a}_1 - y_7\,\mathbf{a}_2 - z_7\,\mathbf{a}_3$ | = | $-x_7a\,\hat{\mathbf{x}} - y_7b\,\hat{\mathbf{y}} - z_7c\,\hat{\mathbf{z}}$ | (8$d$) | O V |
| $\mathbf{B}_{34}$ | = | $\left(\frac{1}{2} + x_7\right)\mathbf{a}_1 + y_7\,\mathbf{a}_2 + \left(\frac{1}{2} - z_7\right)\mathbf{a}_3$ | = | $\left(\frac{1}{2} + x_7\right)a\,\hat{\mathbf{x}} + y_7b\,\hat{\mathbf{y}} + \left(\frac{1}{2} - z_7\right)c\,\hat{\mathbf{z}}$ | (8$d$) | O V |
| $\mathbf{B}_{35}$ | = | $x_7\,\mathbf{a}_1 + \left(\frac{1}{2} - y_7\right)\mathbf{a}_2 + z_7\,\mathbf{a}_3$ | = | $x_7a\,\hat{\mathbf{x}} + \left(\frac{1}{2} - y_7\right)b\,\hat{\mathbf{y}} + z_7c\,\hat{\mathbf{z}}$ | (8$d$) | O V |
| $\mathbf{B}_{36}$ | = | $\left(\frac{1}{2} - x_7\right)\mathbf{a}_1 + \left(\frac{1}{2} + y_7\right)\mathbf{a}_2 + \left(\frac{1}{2} + z_7\right)\mathbf{a}_3$ | = | $\left(\frac{1}{2} - x_7\right)a\,\hat{\mathbf{x}} + \left(\frac{1}{2} + y_7\right)b\,\hat{\mathbf{y}} + \left(\frac{1}{2} + z_7\right)c\,\hat{\mathbf{z}}$ | (8$d$) | O V |
| $\mathbf{B}_{37}$ | = | $x_8\,\mathbf{a}_1 + y_8\,\mathbf{a}_2 + z_8\,\mathbf{a}_3$ | = | $x_8a\,\hat{\mathbf{x}} + y_8b\,\hat{\mathbf{y}} + z_8c\,\hat{\mathbf{z}}$ | (8$d$) | Y |
| $\mathbf{B}_{38}$ | = | $\left(\frac{1}{2} - x_8\right)\mathbf{a}_1 - y_8\,\mathbf{a}_2 + \left(\frac{1}{2} + z_8\right)\mathbf{a}_3$ | = | $\left(\frac{1}{2} - x_8\right)a\,\hat{\mathbf{x}} - y_8b\,\hat{\mathbf{y}} + \left(\frac{1}{2} + z_8\right)c\,\hat{\mathbf{z}}$ | (8$d$) | Y |
| $\mathbf{B}_{39}$ | = | $-x_8\,\mathbf{a}_1 + \left(\frac{1}{2} + y_8\right)\mathbf{a}_2 - z_8\,\mathbf{a}_3$ | = | $-x_8a\,\hat{\mathbf{x}} + \left(\frac{1}{2} + y_8\right)b\,\hat{\mathbf{y}} - z_8c\,\hat{\mathbf{z}}$ | (8$d$) | Y |
| $\mathbf{B}_{40}$ | = | $\left(\frac{1}{2} + x_8\right)\mathbf{a}_1 + \left(\frac{1}{2} - y_8\right)\mathbf{a}_2 + \left(\frac{1}{2} - z_8\right)\mathbf{a}_3$ | = | $\left(\frac{1}{2} + x_8\right)a\,\hat{\mathbf{x}} + \left(\frac{1}{2} - y_8\right)b\,\hat{\mathbf{y}} + \left(\frac{1}{2} - z_8\right)c\,\hat{\mathbf{z}}$ | (8$d$) | Y |
| $\mathbf{B}_{41}$ | = | $-x_8\,\mathbf{a}_1 - y_8\,\mathbf{a}_2 - z_8\,\mathbf{a}_3$ | = | $-x_8a\,\hat{\mathbf{x}} - y_8b\,\hat{\mathbf{y}} - z_8c\,\hat{\mathbf{z}}$ | (8$d$) | Y |
| $\mathbf{B}_{42}$ | = | $\left(\frac{1}{2} + x_8\right)\mathbf{a}_1 + y_8\,\mathbf{a}_2 + \left(\frac{1}{2} - z_8\right)\mathbf{a}_3$ | = | $\left(\frac{1}{2} + x_8\right)a\,\hat{\mathbf{x}} + y_8b\,\hat{\mathbf{y}} + \left(\frac{1}{2} - z_8\right)c\,\hat{\mathbf{z}}$ | (8$d$) | Y |
| $\mathbf{B}_{43}$ | = | $x_8\,\mathbf{a}_1 + \left(\frac{1}{2} - y_8\right)\mathbf{a}_2 + z_8\,\mathbf{a}_3$ | = | $x_8a\,\hat{\mathbf{x}} + \left(\frac{1}{2} - y_8\right)b\,\hat{\mathbf{y}} + z_8c\,\hat{\mathbf{z}}$ | (8$d$) | Y |
| $\mathbf{B}_{44}$ | = | $\left(\frac{1}{2} - x_8\right)\mathbf{a}_1 + \left(\frac{1}{2} + y_8\right)\mathbf{a}_2 + \left(\frac{1}{2} + z_8\right)\mathbf{a}_3$ | = | $\left(\frac{1}{2} - x_8\right)a\,\hat{\mathbf{x}} + \left(\frac{1}{2} + y_8\right)b\,\hat{\mathbf{y}} + \left(\frac{1}{2} + z_8\right)c\,\hat{\mathbf{z}}$ | (8$d$) | Y |

**Geometry files:**

- CIF: pp. 1641
- POSCAR: pp. 1641

# $K_4[Mo(CN)_8]\cdot 2H_2O$ ($F2_1$) Structure: A8B4C4DE8F2_oP108_62_4c2d_2d_2cd_c_4c2d_d

http://aflow.org/prototype-encyclopedia/A8B4C4DE8F2_oP108_62_4c2d_2d_2cd_c_4c2d_d

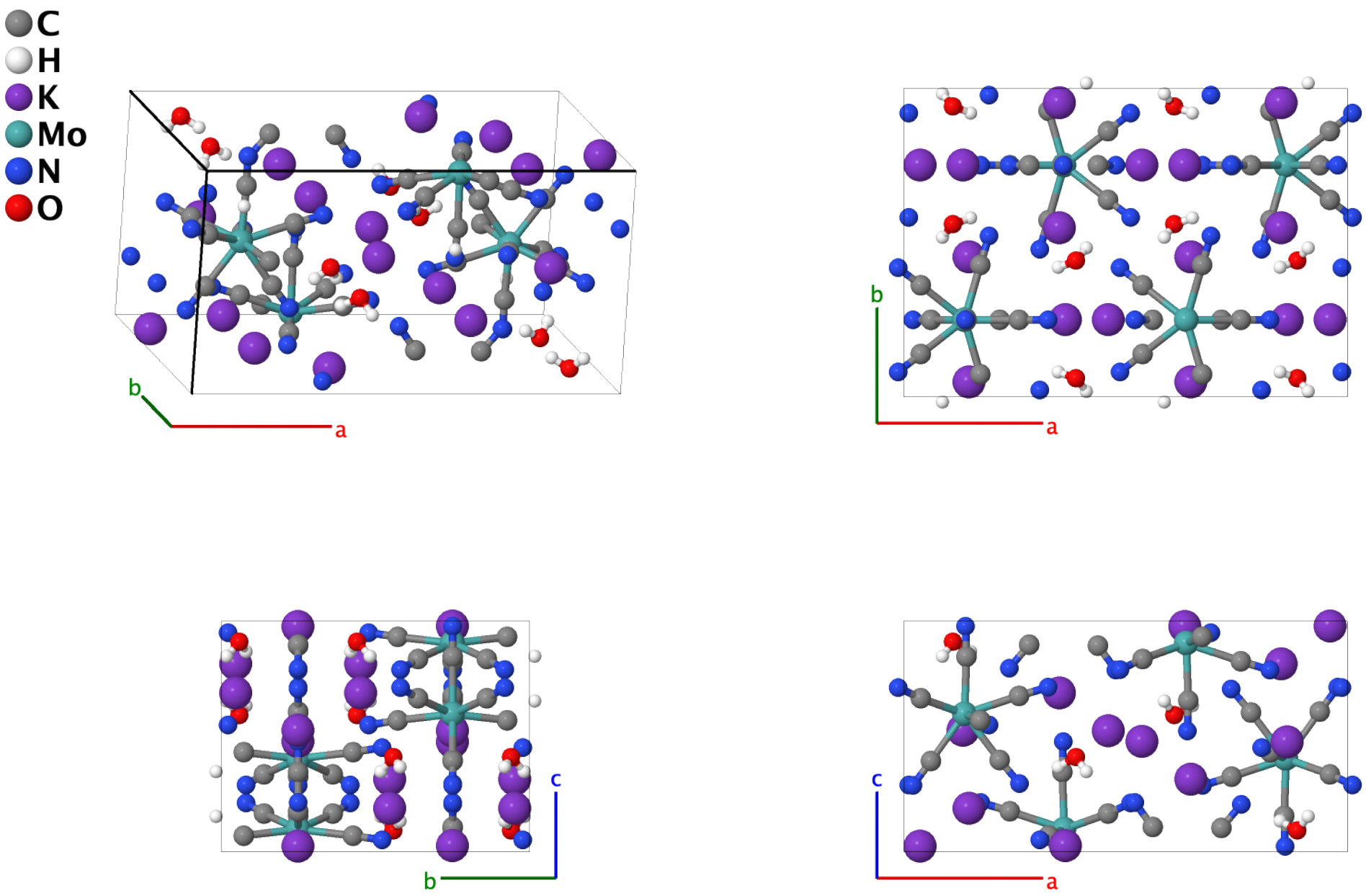

| | | |
|---|---|---|
| **Prototype** | : | $C_8H_4K_4MoN_8O_2$ |
| **AFLOW prototype label** | : | A8B4C4DE8F2_oP108_62_4c2d_2d_2cd_c_4c2d_d |
| ***Strukturbericht* designation** | : | $F2_1$ |
| **Pearson symbol** | : | oP108 |
| **Space group number** | : | 62 |
| **Space group symbol** | : | *Pnma* |
| **AFLOW prototype command** | : | `aflow --proto=A8B4C4DE8F2_oP108_62_4c2d_2d_2cd_c_4c2d_d` `--params=`$a, b/a, c/a, x_1, z_1, x_2, z_2, x_3, z_3, x_4, z_4, x_5, z_5, x_6, z_6, x_7, z_7, x_8, z_8, x_9, z_9, x_{10}, z_{10}, x_{11}, z_{11}, x_{12}, y_{12}, z_{12}, x_{13}, y_{13}, z_{13}, x_{14}, y_{14}, z_{14}, x_{15}, y_{15}, z_{15}, x_{16}, y_{16}, z_{16}, x_{17}, y_{17}, z_{17}, x_{18}, y_{18}, z_{18}, x_{19}, y_{19}, z_{19}$ |

- (Hoard, 1939) originally determined this structure, but were unable to locate the hydrogen atoms. (Herrmann, 1943) gave this the *Struckturbericht* designation $F2_1$. (Typilo, 2010) were able to refine the structure at 173 K, including the hydrogen positions. Since the space group and Wyckoff positions are otherwise unchanged we use the newer structure as our prototype.

**Simple Orthorhombic primitive vectors:**

$$\mathbf{a}_1 = a\,\hat{\mathbf{x}}$$
$$\mathbf{a}_2 = b\,\hat{\mathbf{y}}$$
$$\mathbf{a}_3 = c\,\hat{\mathbf{z}}$$

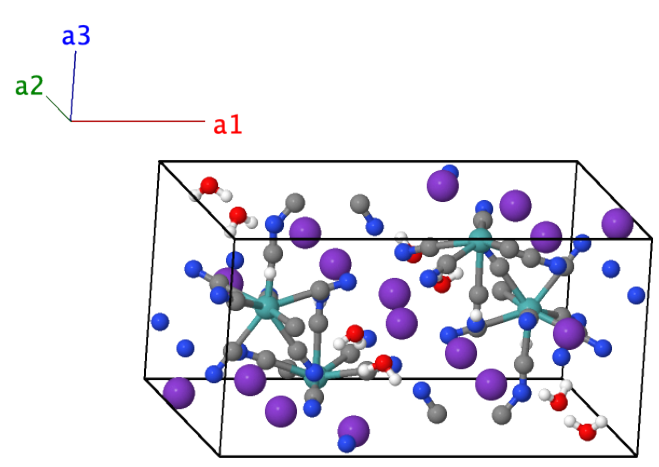

**Basis vectors:**

| | | Lattice Coordinates | | Cartesian Coordinates | Wyckoff Position | Atom Type |
|---|---|---|---|---|---|---|
| $\mathbf{B}_{1}$ | = | $x_{1}\,\mathbf{a}_{1}+\frac{1}{4}\,\mathbf{a}_{2}+z_{1}\,\mathbf{a}_{3}$ | = | $x_{1}a\,\mathbf{\hat{x}}+\frac{1}{4}b\,\mathbf{\hat{y}}+z_{1}c\,\mathbf{\hat{z}}$ | (4$c$) | C I |
| $\mathbf{B}_{2}$ | = | $\left(\frac{1}{2}-x_{1}\right)\,\mathbf{a}_{1}+\frac{3}{4}\,\mathbf{a}_{2}+\left(\frac{1}{2}+z_{1}\right)\,\mathbf{a}_{3}$ | = | $\left(\frac{1}{2}-x_{1}\right)a\,\mathbf{\hat{x}}+\frac{3}{4}b\,\mathbf{\hat{y}}+\left(\frac{1}{2}+z_{1}\right)c\,\mathbf{\hat{z}}$ | (4$c$) | C I |
| $\mathbf{B}_{3}$ | = | $-x_{1}\,\mathbf{a}_{1}+\frac{3}{4}\,\mathbf{a}_{2}-z_{1}\,\mathbf{a}_{3}$ | = | $-x_{1}a\,\mathbf{\hat{x}}+\frac{3}{4}b\,\mathbf{\hat{y}}-z_{1}c\,\mathbf{\hat{z}}$ | (4$c$) | C I |
| $\mathbf{B}_{4}$ | = | $\left(\frac{1}{2}+x_{1}\right)\,\mathbf{a}_{1}+\frac{1}{4}\,\mathbf{a}_{2}+\left(\frac{1}{2}-z_{1}\right)\,\mathbf{a}_{3}$ | = | $\left(\frac{1}{2}+x_{1}\right)a\,\mathbf{\hat{x}}+\frac{1}{4}b\,\mathbf{\hat{y}}+\left(\frac{1}{2}-z_{1}\right)c\,\mathbf{\hat{z}}$ | (4$c$) | C I |
| $\mathbf{B}_{5}$ | = | $x_{2}\,\mathbf{a}_{1}+\frac{1}{4}\,\mathbf{a}_{2}+z_{2}\,\mathbf{a}_{3}$ | = | $x_{2}a\,\mathbf{\hat{x}}+\frac{1}{4}b\,\mathbf{\hat{y}}+z_{2}c\,\mathbf{\hat{z}}$ | (4$c$) | C II |
| $\mathbf{B}_{6}$ | = | $\left(\frac{1}{2}-x_{2}\right)\,\mathbf{a}_{1}+\frac{3}{4}\,\mathbf{a}_{2}+\left(\frac{1}{2}+z_{2}\right)\,\mathbf{a}_{3}$ | = | $\left(\frac{1}{2}-x_{2}\right)a\,\mathbf{\hat{x}}+\frac{3}{4}b\,\mathbf{\hat{y}}+\left(\frac{1}{2}+z_{2}\right)c\,\mathbf{\hat{z}}$ | (4$c$) | C II |
| $\mathbf{B}_{7}$ | = | $-x_{2}\,\mathbf{a}_{1}+\frac{3}{4}\,\mathbf{a}_{2}-z_{2}\,\mathbf{a}_{3}$ | = | $-x_{2}a\,\mathbf{\hat{x}}+\frac{3}{4}b\,\mathbf{\hat{y}}-z_{2}c\,\mathbf{\hat{z}}$ | (4$c$) | C II |
| $\mathbf{B}_{8}$ | = | $\left(\frac{1}{2}+x_{2}\right)\,\mathbf{a}_{1}+\frac{1}{4}\,\mathbf{a}_{2}+\left(\frac{1}{2}-z_{2}\right)\,\mathbf{a}_{3}$ | = | $\left(\frac{1}{2}+x_{2}\right)a\,\mathbf{\hat{x}}+\frac{1}{4}b\,\mathbf{\hat{y}}+\left(\frac{1}{2}-z_{2}\right)c\,\mathbf{\hat{z}}$ | (4$c$) | C II |
| $\mathbf{B}_{9}$ | = | $x_{3}\,\mathbf{a}_{1}+\frac{1}{4}\,\mathbf{a}_{2}+z_{3}\,\mathbf{a}_{3}$ | = | $x_{3}a\,\mathbf{\hat{x}}+\frac{1}{4}b\,\mathbf{\hat{y}}+z_{3}c\,\mathbf{\hat{z}}$ | (4$c$) | C III |
| $\mathbf{B}_{10}$ | = | $\left(\frac{1}{2}-x_{3}\right)\,\mathbf{a}_{1}+\frac{3}{4}\,\mathbf{a}_{2}+\left(\frac{1}{2}+z_{3}\right)\,\mathbf{a}_{3}$ | = | $\left(\frac{1}{2}-x_{3}\right)a\,\mathbf{\hat{x}}+\frac{3}{4}b\,\mathbf{\hat{y}}+\left(\frac{1}{2}+z_{3}\right)c\,\mathbf{\hat{z}}$ | (4$c$) | C III |
| $\mathbf{B}_{11}$ | = | $-x_{3}\,\mathbf{a}_{1}+\frac{3}{4}\,\mathbf{a}_{2}-z_{3}\,\mathbf{a}_{3}$ | = | $-x_{3}a\,\mathbf{\hat{x}}+\frac{3}{4}b\,\mathbf{\hat{y}}-z_{3}c\,\mathbf{\hat{z}}$ | (4$c$) | C III |
| $\mathbf{B}_{12}$ | = | $\left(\frac{1}{2}+x_{3}\right)\,\mathbf{a}_{1}+\frac{1}{4}\,\mathbf{a}_{2}+\left(\frac{1}{2}-z_{3}\right)\,\mathbf{a}_{3}$ | = | $\left(\frac{1}{2}+x_{3}\right)a\,\mathbf{\hat{x}}+\frac{1}{4}b\,\mathbf{\hat{y}}+\left(\frac{1}{2}-z_{3}\right)c\,\mathbf{\hat{z}}$ | (4$c$) | C III |
| $\mathbf{B}_{13}$ | = | $x_{4}\,\mathbf{a}_{1}+\frac{1}{4}\,\mathbf{a}_{2}+z_{4}\,\mathbf{a}_{3}$ | = | $x_{4}a\,\mathbf{\hat{x}}+\frac{1}{4}b\,\mathbf{\hat{y}}+z_{4}c\,\mathbf{\hat{z}}$ | (4$c$) | C IV |
| $\mathbf{B}_{14}$ | = | $\left(\frac{1}{2}-x_{4}\right)\,\mathbf{a}_{1}+\frac{3}{4}\,\mathbf{a}_{2}+\left(\frac{1}{2}+z_{4}\right)\,\mathbf{a}_{3}$ | = | $\left(\frac{1}{2}-x_{4}\right)a\,\mathbf{\hat{x}}+\frac{3}{4}b\,\mathbf{\hat{y}}+\left(\frac{1}{2}+z_{4}\right)c\,\mathbf{\hat{z}}$ | (4$c$) | C IV |
| $\mathbf{B}_{15}$ | = | $-x_{4}\,\mathbf{a}_{1}+\frac{3}{4}\,\mathbf{a}_{2}-z_{4}\,\mathbf{a}_{3}$ | = | $-x_{4}a\,\mathbf{\hat{x}}+\frac{3}{4}b\,\mathbf{\hat{y}}-z_{4}c\,\mathbf{\hat{z}}$ | (4$c$) | C IV |
| $\mathbf{B}_{16}$ | = | $\left(\frac{1}{2}+x_{4}\right)\,\mathbf{a}_{1}+\frac{1}{4}\,\mathbf{a}_{2}+\left(\frac{1}{2}-z_{4}\right)\,\mathbf{a}_{3}$ | = | $\left(\frac{1}{2}+x_{4}\right)a\,\mathbf{\hat{x}}+\frac{1}{4}b\,\mathbf{\hat{y}}+\left(\frac{1}{2}-z_{4}\right)c\,\mathbf{\hat{z}}$ | (4$c$) | C IV |
| $\mathbf{B}_{17}$ | = | $x_{5}\,\mathbf{a}_{1}+\frac{1}{4}\,\mathbf{a}_{2}+z_{5}\,\mathbf{a}_{3}$ | = | $x_{5}a\,\mathbf{\hat{x}}+\frac{1}{4}b\,\mathbf{\hat{y}}+z_{5}c\,\mathbf{\hat{z}}$ | (4$c$) | K I |
| $\mathbf{B}_{18}$ | = | $\left(\frac{1}{2}-x_{5}\right)\,\mathbf{a}_{1}+\frac{3}{4}\,\mathbf{a}_{2}+\left(\frac{1}{2}+z_{5}\right)\,\mathbf{a}_{3}$ | = | $\left(\frac{1}{2}-x_{5}\right)a\,\mathbf{\hat{x}}+\frac{3}{4}b\,\mathbf{\hat{y}}+\left(\frac{1}{2}+z_{5}\right)c\,\mathbf{\hat{z}}$ | (4$c$) | K I |
| $\mathbf{B}_{19}$ | = | $-x_{5}\,\mathbf{a}_{1}+\frac{3}{4}\,\mathbf{a}_{2}-z_{5}\,\mathbf{a}_{3}$ | = | $-x_{5}a\,\mathbf{\hat{x}}+\frac{3}{4}b\,\mathbf{\hat{y}}-z_{5}c\,\mathbf{\hat{z}}$ | (4$c$) | K I |
| $\mathbf{B}_{20}$ | = | $\left(\frac{1}{2}+x_{5}\right)\,\mathbf{a}_{1}+\frac{1}{4}\,\mathbf{a}_{2}+\left(\frac{1}{2}-z_{5}\right)\,\mathbf{a}_{3}$ | = | $\left(\frac{1}{2}+x_{5}\right)a\,\mathbf{\hat{x}}+\frac{1}{4}b\,\mathbf{\hat{y}}+\left(\frac{1}{2}-z_{5}\right)c\,\mathbf{\hat{z}}$ | (4$c$) | K I |
| $\mathbf{B}_{21}$ | = | $x_{6}\,\mathbf{a}_{1}+\frac{1}{4}\,\mathbf{a}_{2}+z_{6}\,\mathbf{a}_{3}$ | = | $x_{6}a\,\mathbf{\hat{x}}+\frac{1}{4}b\,\mathbf{\hat{y}}+z_{6}c\,\mathbf{\hat{z}}$ | (4$c$) | K II |
| $\mathbf{B}_{22}$ | = | $\left(\frac{1}{2}-x_{6}\right)\,\mathbf{a}_{1}+\frac{3}{4}\,\mathbf{a}_{2}+\left(\frac{1}{2}+z_{6}\right)\,\mathbf{a}_{3}$ | = | $\left(\frac{1}{2}-x_{6}\right)a\,\mathbf{\hat{x}}+\frac{3}{4}b\,\mathbf{\hat{y}}+\left(\frac{1}{2}+z_{6}\right)c\,\mathbf{\hat{z}}$ | (4$c$) | K II |
| $\mathbf{B}_{23}$ | = | $-x_{6}\,\mathbf{a}_{1}+\frac{3}{4}\,\mathbf{a}_{2}-z_{6}\,\mathbf{a}_{3}$ | = | $-x_{6}a\,\mathbf{\hat{x}}+\frac{3}{4}b\,\mathbf{\hat{y}}-z_{6}c\,\mathbf{\hat{z}}$ | (4$c$) | K II |
| $\mathbf{B}_{24}$ | = | $\left(\frac{1}{2}+x_{6}\right)\,\mathbf{a}_{1}+\frac{1}{4}\,\mathbf{a}_{2}+\left(\frac{1}{2}-z_{6}\right)\,\mathbf{a}_{3}$ | = | $\left(\frac{1}{2}+x_{6}\right)a\,\mathbf{\hat{x}}+\frac{1}{4}b\,\mathbf{\hat{y}}+\left(\frac{1}{2}-z_{6}\right)c\,\mathbf{\hat{z}}$ | (4$c$) | K II |
| $\mathbf{B}_{25}$ | = | $x_{7}\,\mathbf{a}_{1}+\frac{1}{4}\,\mathbf{a}_{2}+z_{7}\,\mathbf{a}_{3}$ | = | $x_{7}a\,\mathbf{\hat{x}}+\frac{1}{4}b\,\mathbf{\hat{y}}+z_{7}c\,\mathbf{\hat{z}}$ | (4$c$) | Mo |
| $\mathbf{B}_{26}$ | = | $\left(\frac{1}{2}-x_{7}\right)\,\mathbf{a}_{1}+\frac{3}{4}\,\mathbf{a}_{2}+\left(\frac{1}{2}+z_{7}\right)\,\mathbf{a}_{3}$ | = | $\left(\frac{1}{2}-x_{7}\right)a\,\mathbf{\hat{x}}+\frac{3}{4}b\,\mathbf{\hat{y}}+\left(\frac{1}{2}+z_{7}\right)c\,\mathbf{\hat{z}}$ | (4$c$) | Mo |
| $\mathbf{B}_{27}$ | = | $-x_{7}\,\mathbf{a}_{1}+\frac{3}{4}\,\mathbf{a}_{2}-z_{7}\,\mathbf{a}_{3}$ | = | $-x_{7}a\,\mathbf{\hat{x}}+\frac{3}{4}b\,\mathbf{\hat{y}}-z_{7}c\,\mathbf{\hat{z}}$ | (4$c$) | Mo |
| $\mathbf{B}_{28}$ | = | $\left(\frac{1}{2}+x_{7}\right)\,\mathbf{a}_{1}+\frac{1}{4}\,\mathbf{a}_{2}+\left(\frac{1}{2}-z_{7}\right)\,\mathbf{a}_{3}$ | = | $\left(\frac{1}{2}+x_{7}\right)a\,\mathbf{\hat{x}}+\frac{1}{4}b\,\mathbf{\hat{y}}+\left(\frac{1}{2}-z_{7}\right)c\,\mathbf{\hat{z}}$ | (4$c$) | Mo |
| $\mathbf{B}_{29}$ | = | $x_{8}\,\mathbf{a}_{1}+\frac{1}{4}\,\mathbf{a}_{2}+z_{8}\,\mathbf{a}_{3}$ | = | $x_{8}a\,\mathbf{\hat{x}}+\frac{1}{4}b\,\mathbf{\hat{y}}+z_{8}c\,\mathbf{\hat{z}}$ | (4$c$) | N I |
| $\mathbf{B}_{30}$ | = | $\left(\frac{1}{2}-x_{8}\right)\,\mathbf{a}_{1}+\frac{3}{4}\,\mathbf{a}_{2}+\left(\frac{1}{2}+z_{8}\right)\,\mathbf{a}_{3}$ | = | $\left(\frac{1}{2}-x_{8}\right)a\,\mathbf{\hat{x}}+\frac{3}{4}b\,\mathbf{\hat{y}}+\left(\frac{1}{2}+z_{8}\right)c\,\mathbf{\hat{z}}$ | (4$c$) | N I |
| $\mathbf{B}_{31}$ | = | $-x_{8}\,\mathbf{a}_{1}+\frac{3}{4}\,\mathbf{a}_{2}-z_{8}\,\mathbf{a}_{3}$ | = | $-x_{8}a\,\mathbf{\hat{x}}+\frac{3}{4}b\,\mathbf{\hat{y}}-z_{8}c\,\mathbf{\hat{z}}$ | (4$c$) | N I |
| $\mathbf{B}_{32}$ | = | $\left(\frac{1}{2}+x_{8}\right)\,\mathbf{a}_{1}+\frac{1}{4}\,\mathbf{a}_{2}+\left(\frac{1}{2}-z_{8}\right)\,\mathbf{a}_{3}$ | = | $\left(\frac{1}{2}+x_{8}\right)a\,\mathbf{\hat{x}}+\frac{1}{4}b\,\mathbf{\hat{y}}+\left(\frac{1}{2}-z_{8}\right)c\,\mathbf{\hat{z}}$ | (4$c$) | N I |
| $\mathbf{B}_{33}$ | = | $x_{9}\,\mathbf{a}_{1}+\frac{1}{4}\,\mathbf{a}_{2}+z_{9}\,\mathbf{a}_{3}$ | = | $x_{9}a\,\mathbf{\hat{x}}+\frac{1}{4}b\,\mathbf{\hat{y}}+z_{9}c\,\mathbf{\hat{z}}$ | (4$c$) | N II |
| $\mathbf{B}_{34}$ | = | $\left(\frac{1}{2}-x_{9}\right)\,\mathbf{a}_{1}+\frac{3}{4}\,\mathbf{a}_{2}+\left(\frac{1}{2}+z_{9}\right)\,\mathbf{a}_{3}$ | = | $\left(\frac{1}{2}-x_{9}\right)a\,\mathbf{\hat{x}}+\frac{3}{4}b\,\mathbf{\hat{y}}+\left(\frac{1}{2}+z_{9}\right)c\,\mathbf{\hat{z}}$ | (4$c$) | N II |
| $\mathbf{B}_{35}$ | = | $-x_{9}\,\mathbf{a}_{1}+\frac{3}{4}\,\mathbf{a}_{2}-z_{9}\,\mathbf{a}_{3}$ | = | $-x_{9}a\,\hat{\mathbf{x}}+\frac{3}{4}b\,\hat{\mathbf{y}}-z_{9}c\,\hat{\mathbf{z}}$ | (4$c$) | N II |
| $\mathbf{B}_{36}$ | = | $\left(\frac{1}{2}+x_{9}\right)\mathbf{a}_{1}+\frac{1}{4}\,\mathbf{a}_{2}+\left(\frac{1}{2}-z_{9}\right)\mathbf{a}_{3}$ | = | $\left(\frac{1}{2}+x_{9}\right)a\,\hat{\mathbf{x}}+\frac{1}{4}b\,\hat{\mathbf{y}}+\left(\frac{1}{2}-z_{9}\right)c\,\hat{\mathbf{z}}$ | (4$c$) | N II |
| $\mathbf{B}_{37}$ | = | $x_{10}\,\mathbf{a}_{1}+\frac{1}{4}\,\mathbf{a}_{2}+z_{10}\,\mathbf{a}_{3}$ | = | $x_{10}a\,\hat{\mathbf{x}}+\frac{1}{4}b\,\hat{\mathbf{y}}+z_{10}c\,\hat{\mathbf{z}}$ | (4$c$) | N III |
| $\mathbf{B}_{38}$ | = | $\left(\frac{1}{2}-x_{10}\right)\mathbf{a}_{1}+\frac{3}{4}\,\mathbf{a}_{2}+\left(\frac{1}{2}+z_{10}\right)\mathbf{a}_{3}$ | = | $\left(\frac{1}{2}-x_{10}\right)a\,\hat{\mathbf{x}}+\frac{3}{4}b\,\hat{\mathbf{y}}+\left(\frac{1}{2}+z_{10}\right)c\,\hat{\mathbf{z}}$ | (4$c$) | N III |
| $\mathbf{B}_{39}$ | = | $-x_{10}\,\mathbf{a}_{1}+\frac{3}{4}\,\mathbf{a}_{2}-z_{10}\,\mathbf{a}_{3}$ | = | $-x_{10}a\,\hat{\mathbf{x}}+\frac{3}{4}b\,\hat{\mathbf{y}}-z_{10}c\,\hat{\mathbf{z}}$ | (4$c$) | N III |
| $\mathbf{B}_{40}$ | = | $\left(\frac{1}{2}+x_{10}\right)\mathbf{a}_{1}+\frac{1}{4}\,\mathbf{a}_{2}+\left(\frac{1}{2}-z_{10}\right)\mathbf{a}_{3}$ | = | $\left(\frac{1}{2}+x_{10}\right)a\,\hat{\mathbf{x}}+\frac{1}{4}b\,\hat{\mathbf{y}}+\left(\frac{1}{2}-z_{10}\right)c\,\hat{\mathbf{z}}$ | (4$c$) | N III |
| $\mathbf{B}_{41}$ | = | $x_{11}\,\mathbf{a}_{1}+\frac{1}{4}\,\mathbf{a}_{2}+z_{11}\,\mathbf{a}_{3}$ | = | $x_{11}a\,\hat{\mathbf{x}}+\frac{1}{4}b\,\hat{\mathbf{y}}+z_{11}c\,\hat{\mathbf{z}}$ | (4$c$) | N IV |
| $\mathbf{B}_{42}$ | = | $\left(\frac{1}{2}-x_{11}\right)\mathbf{a}_{1}+\frac{3}{4}\,\mathbf{a}_{2}+\left(\frac{1}{2}+z_{11}\right)\mathbf{a}_{3}$ | = | $\left(\frac{1}{2}-x_{11}\right)a\,\hat{\mathbf{x}}+\frac{3}{4}b\,\hat{\mathbf{y}}+\left(\frac{1}{2}+z_{11}\right)c\,\hat{\mathbf{z}}$ | (4$c$) | N IV |
| $\mathbf{B}_{43}$ | = | $-x_{11}\,\mathbf{a}_{1}+\frac{3}{4}\,\mathbf{a}_{2}-z_{11}\,\mathbf{a}_{3}$ | = | $-x_{11}a\,\hat{\mathbf{x}}+\frac{3}{4}b\,\hat{\mathbf{y}}-z_{11}c\,\hat{\mathbf{z}}$ | (4$c$) | N IV |
| $\mathbf{B}_{44}$ | = | $\left(\frac{1}{2}+x_{11}\right)\mathbf{a}_{1}+\frac{1}{4}\,\mathbf{a}_{2}+\left(\frac{1}{2}-z_{11}\right)\mathbf{a}_{3}$ | = | $\left(\frac{1}{2}+x_{11}\right)a\,\hat{\mathbf{x}}+\frac{1}{4}b\,\hat{\mathbf{y}}+\left(\frac{1}{2}-z_{11}\right)c\,\hat{\mathbf{z}}$ | (4$c$) | N IV |
| $\mathbf{B}_{45}$ | = | $x_{12}\,\mathbf{a}_{1}+y_{12}\,\mathbf{a}_{2}+z_{12}\,\mathbf{a}_{3}$ | = | $x_{12}a\,\hat{\mathbf{x}}+y_{12}b\,\hat{\mathbf{y}}+z_{12}c\,\hat{\mathbf{z}}$ | (8$d$) | C V |
| $\mathbf{B}_{46}$ | = | $\left(\frac{1}{2}-x_{12}\right)\mathbf{a}_{1}-y_{12}\,\mathbf{a}_{2}+\left(\frac{1}{2}+z_{12}\right)\mathbf{a}_{3}$ | = | $\left(\frac{1}{2}-x_{12}\right)a\,\hat{\mathbf{x}}-y_{12}b\,\hat{\mathbf{y}}+\left(\frac{1}{2}+z_{12}\right)c\,\hat{\mathbf{z}}$ | (8$d$) | C V |
| $\mathbf{B}_{47}$ | = | $-x_{12}\,\mathbf{a}_{1}+\left(\frac{1}{2}+y_{12}\right)\mathbf{a}_{2}-z_{12}\,\mathbf{a}_{3}$ | = | $-x_{12}a\,\hat{\mathbf{x}}+\left(\frac{1}{2}+y_{12}\right)b\,\hat{\mathbf{y}}-z_{12}c\,\hat{\mathbf{z}}$ | (8$d$) | C V |
| $\mathbf{B}_{48}$ | = | $\left(\frac{1}{2}+x_{12}\right)\mathbf{a}_{1}+\left(\frac{1}{2}-y_{12}\right)\mathbf{a}_{2}+\left(\frac{1}{2}-z_{12}\right)\mathbf{a}_{3}$ | = | $\left(\frac{1}{2}+x_{12}\right)a\,\hat{\mathbf{x}}+\left(\frac{1}{2}-y_{12}\right)b\,\hat{\mathbf{y}}+\left(\frac{1}{2}-z_{12}\right)c\,\hat{\mathbf{z}}$ | (8$d$) | C V |
| $\mathbf{B}_{49}$ | = | $-x_{12}\,\mathbf{a}_{1}-y_{12}\,\mathbf{a}_{2}-z_{12}\,\mathbf{a}_{3}$ | = | $-x_{12}a\,\hat{\mathbf{x}}-y_{12}b\,\hat{\mathbf{y}}-z_{12}c\,\hat{\mathbf{z}}$ | (8$d$) | C V |
| $\mathbf{B}_{50}$ | = | $\left(\frac{1}{2}+x_{12}\right)\mathbf{a}_{1}+y_{12}\,\mathbf{a}_{2}+\left(\frac{1}{2}-z_{12}\right)\mathbf{a}_{3}$ | = | $\left(\frac{1}{2}+x_{12}\right)a\,\hat{\mathbf{x}}+y_{12}b\,\hat{\mathbf{y}}+\left(\frac{1}{2}-z_{12}\right)c\,\hat{\mathbf{z}}$ | (8$d$) | C V |
| $\mathbf{B}_{51}$ | = | $x_{12}\,\mathbf{a}_{1}+\left(\frac{1}{2}-y_{12}\right)\mathbf{a}_{2}+z_{12}\,\mathbf{a}_{3}$ | = | $x_{12}a\,\hat{\mathbf{x}}+\left(\frac{1}{2}-y_{12}\right)b\,\hat{\mathbf{y}}+z_{12}c\,\hat{\mathbf{z}}$ | (8$d$) | C V |
| $\mathbf{B}_{52}$ | = | $\left(\frac{1}{2}-x_{12}\right)\mathbf{a}_{1}+\left(\frac{1}{2}+y_{12}\right)\mathbf{a}_{2}+\left(\frac{1}{2}+z_{12}\right)\mathbf{a}_{3}$ | = | $\left(\frac{1}{2}-x_{12}\right)a\,\hat{\mathbf{x}}+\left(\frac{1}{2}+y_{12}\right)b\,\hat{\mathbf{y}}+\left(\frac{1}{2}+z_{12}\right)c\,\hat{\mathbf{z}}$ | (8$d$) | C V |
| $\mathbf{B}_{53}$ | = | $x_{13}\,\mathbf{a}_{1}+y_{13}\,\mathbf{a}_{2}+z_{13}\,\mathbf{a}_{3}$ | = | $x_{13}a\,\hat{\mathbf{x}}+y_{13}b\,\hat{\mathbf{y}}+z_{13}c\,\hat{\mathbf{z}}$ | (8$d$) | C VI |
| $\mathbf{B}_{54}$ | = | $\left(\frac{1}{2}-x_{13}\right)\mathbf{a}_{1}-y_{13}\,\mathbf{a}_{2}+\left(\frac{1}{2}+z_{13}\right)\mathbf{a}_{3}$ | = | $\left(\frac{1}{2}-x_{13}\right)a\,\hat{\mathbf{x}}-y_{13}b\,\hat{\mathbf{y}}+\left(\frac{1}{2}+z_{13}\right)c\,\hat{\mathbf{z}}$ | (8$d$) | C VI |
| $\mathbf{B}_{55}$ | = | $-x_{13}\,\mathbf{a}_{1}+\left(\frac{1}{2}+y_{13}\right)\mathbf{a}_{2}-z_{13}\,\mathbf{a}_{3}$ | = | $-x_{13}a\,\hat{\mathbf{x}}+\left(\frac{1}{2}+y_{13}\right)b\,\hat{\mathbf{y}}-z_{13}c\,\hat{\mathbf{z}}$ | (8$d$) | C VI |
| $\mathbf{B}_{56}$ | = | $\left(\frac{1}{2}+x_{13}\right)\mathbf{a}_{1}+\left(\frac{1}{2}-y_{13}\right)\mathbf{a}_{2}+\left(\frac{1}{2}-z_{13}\right)\mathbf{a}_{3}$ | = | $\left(\frac{1}{2}+x_{13}\right)a\,\hat{\mathbf{x}}+\left(\frac{1}{2}-y_{13}\right)b\,\hat{\mathbf{y}}+\left(\frac{1}{2}-z_{13}\right)c\,\hat{\mathbf{z}}$ | (8$d$) | C VI |
| $\mathbf{B}_{57}$ | = | $-x_{13}\,\mathbf{a}_{1}-y_{13}\,\mathbf{a}_{2}-z_{13}\,\mathbf{a}_{3}$ | = | $-x_{13}a\,\hat{\mathbf{x}}-y_{13}b\,\hat{\mathbf{y}}-z_{13}c\,\hat{\mathbf{z}}$ | (8$d$) | C VI |
| $\mathbf{B}_{58}$ | = | $\left(\frac{1}{2}+x_{13}\right)\mathbf{a}_{1}+y_{13}\,\mathbf{a}_{2}+\left(\frac{1}{2}-z_{13}\right)\mathbf{a}_{3}$ | = | $\left(\frac{1}{2}+x_{13}\right)a\,\hat{\mathbf{x}}+y_{13}b\,\hat{\mathbf{y}}+\left(\frac{1}{2}-z_{13}\right)c\,\hat{\mathbf{z}}$ | (8$d$) | C VI |
| $\mathbf{B}_{59}$ | = | $x_{13}\,\mathbf{a}_{1}+\left(\frac{1}{2}-y_{13}\right)\mathbf{a}_{2}+z_{13}\,\mathbf{a}_{3}$ | = | $x_{13}a\,\hat{\mathbf{x}}+\left(\frac{1}{2}-y_{13}\right)b\,\hat{\mathbf{y}}+z_{13}c\,\hat{\mathbf{z}}$ | (8$d$) | C VI |
| $\mathbf{B}_{60}$ | = | $\left(\frac{1}{2}-x_{13}\right)\mathbf{a}_{1}+\left(\frac{1}{2}+y_{13}\right)\mathbf{a}_{2}+\left(\frac{1}{2}+z_{13}\right)\mathbf{a}_{3}$ | = | $\left(\frac{1}{2}-x_{13}\right)a\,\hat{\mathbf{x}}+\left(\frac{1}{2}+y_{13}\right)b\,\hat{\mathbf{y}}+\left(\frac{1}{2}+z_{13}\right)c\,\hat{\mathbf{z}}$ | (8$d$) | C VI |
| $\mathbf{B}_{61}$ | = | $x_{14}\,\mathbf{a}_{1}+y_{14}\,\mathbf{a}_{2}+z_{14}\,\mathbf{a}_{3}$ | = | $x_{14}a\,\hat{\mathbf{x}}+y_{14}b\,\hat{\mathbf{y}}+z_{14}c\,\hat{\mathbf{z}}$ | (8$d$) | H I |
| $\mathbf{B}_{62}$ | = | $\left(\frac{1}{2}-x_{14}\right)\mathbf{a}_{1}-y_{14}\,\mathbf{a}_{2}+\left(\frac{1}{2}+z_{14}\right)\mathbf{a}_{3}$ | = | $\left(\frac{1}{2}-x_{14}\right)a\,\hat{\mathbf{x}}-y_{14}b\,\hat{\mathbf{y}}+\left(\frac{1}{2}+z_{14}\right)c\,\hat{\mathbf{z}}$ | (8$d$) | H I |
| $\mathbf{B}_{63}$ | = | $-x_{14}\,\mathbf{a}_{1}+\left(\frac{1}{2}+y_{14}\right)\mathbf{a}_{2}-z_{14}\,\mathbf{a}_{3}$ | = | $-x_{14}a\,\hat{\mathbf{x}}+\left(\frac{1}{2}+y_{14}\right)b\,\hat{\mathbf{y}}-z_{14}c\,\hat{\mathbf{z}}$ | (8$d$) | H I |
| $\mathbf{B}_{64}$ | = | $\left(\frac{1}{2}+x_{14}\right)\mathbf{a}_{1}+\left(\frac{1}{2}-y_{14}\right)\mathbf{a}_{2}+\left(\frac{1}{2}-z_{14}\right)\mathbf{a}_{3}$ | = | $\left(\frac{1}{2}+x_{14}\right)a\,\hat{\mathbf{x}}+\left(\frac{1}{2}-y_{14}\right)b\,\hat{\mathbf{y}}+\left(\frac{1}{2}-z_{14}\right)c\,\hat{\mathbf{z}}$ | (8$d$) | H I |
| $\mathbf{B}_{65}$ | = | $-x_{14}\,\mathbf{a}_{1}-y_{14}\,\mathbf{a}_{2}-z_{14}\,\mathbf{a}_{3}$ | = | $-x_{14}a\,\hat{\mathbf{x}}-y_{14}b\,\hat{\mathbf{y}}-z_{14}c\,\hat{\mathbf{z}}$ | (8$d$) | H I |
| $\mathbf{B}_{66}$ | = | $\left(\frac{1}{2}+x_{14}\right)\mathbf{a}_{1}+y_{14}\,\mathbf{a}_{2}+\left(\frac{1}{2}-z_{14}\right)\mathbf{a}_{3}$ | = | $\left(\frac{1}{2}+x_{14}\right)a\,\hat{\mathbf{x}}+y_{14}b\,\hat{\mathbf{y}}+\left(\frac{1}{2}-z_{14}\right)c\,\hat{\mathbf{z}}$ | (8$d$) | H I |
| $\mathbf{B}_{67}$ | = | $x_{14}\,\mathbf{a}_1 + \left(\frac{1}{2} - y_{14}\right)\mathbf{a}_2 + z_{14}\,\mathbf{a}_3$ | = | $x_{14}a\,\hat{\mathbf{x}} + \left(\frac{1}{2} - y_{14}\right)b\,\hat{\mathbf{y}} + z_{14}c\,\hat{\mathbf{z}}$ | (8$d$) | H I |
| $\mathbf{B}_{68}$ | = | $\left(\frac{1}{2} - x_{14}\right)\mathbf{a}_1 + \left(\frac{1}{2} + y_{14}\right)\mathbf{a}_2 + \left(\frac{1}{2} + z_{14}\right)\mathbf{a}_3$ | = | $\left(\frac{1}{2} - x_{14}\right)a\,\hat{\mathbf{x}} + \left(\frac{1}{2} + y_{14}\right)b\,\hat{\mathbf{y}} + \left(\frac{1}{2} + z_{14}\right)c\,\hat{\mathbf{z}}$ | (8$d$) | H I |
| $\mathbf{B}_{69}$ | = | $x_{15}\,\mathbf{a}_1 + y_{15}\,\mathbf{a}_2 + z_{15}\,\mathbf{a}_3$ | = | $x_{15}a\,\hat{\mathbf{x}} + y_{15}b\,\hat{\mathbf{y}} + z_{15}c\,\hat{\mathbf{z}}$ | (8$d$) | H II |
| $\mathbf{B}_{70}$ | = | $\left(\frac{1}{2} - x_{15}\right)\mathbf{a}_1 - y_{15}\,\mathbf{a}_2 + \left(\frac{1}{2} + z_{15}\right)\mathbf{a}_3$ | = | $\left(\frac{1}{2} - x_{15}\right)a\,\hat{\mathbf{x}} - y_{15}b\,\hat{\mathbf{y}} + \left(\frac{1}{2} + z_{15}\right)c\,\hat{\mathbf{z}}$ | (8$d$) | H II |
| $\mathbf{B}_{71}$ | = | $-x_{15}\,\mathbf{a}_1 + \left(\frac{1}{2} + y_{15}\right)\mathbf{a}_2 - z_{15}\,\mathbf{a}_3$ | = | $-x_{15}a\,\hat{\mathbf{x}} + \left(\frac{1}{2} + y_{15}\right)b\,\hat{\mathbf{y}} - z_{15}c\,\hat{\mathbf{z}}$ | (8$d$) | H II |
| $\mathbf{B}_{72}$ | = | $\left(\frac{1}{2} + x_{15}\right)\mathbf{a}_1 + \left(\frac{1}{2} - y_{15}\right)\mathbf{a}_2 + \left(\frac{1}{2} - z_{15}\right)\mathbf{a}_3$ | = | $\left(\frac{1}{2} + x_{15}\right)a\,\hat{\mathbf{x}} + \left(\frac{1}{2} - y_{15}\right)b\,\hat{\mathbf{y}} + \left(\frac{1}{2} - z_{15}\right)c\,\hat{\mathbf{z}}$ | (8$d$) | H II |
| $\mathbf{B}_{73}$ | = | $-x_{15}\,\mathbf{a}_1 - y_{15}\,\mathbf{a}_2 - z_{15}\,\mathbf{a}_3$ | = | $-x_{15}a\,\hat{\mathbf{x}} - y_{15}b\,\hat{\mathbf{y}} - z_{15}c\,\hat{\mathbf{z}}$ | (8$d$) | H II |
| $\mathbf{B}_{74}$ | = | $\left(\frac{1}{2} + x_{15}\right)\mathbf{a}_1 + y_{15}\,\mathbf{a}_2 + \left(\frac{1}{2} - z_{15}\right)\mathbf{a}_3$ | = | $\left(\frac{1}{2} + x_{15}\right)a\,\hat{\mathbf{x}} + y_{15}b\,\hat{\mathbf{y}} + \left(\frac{1}{2} - z_{15}\right)c\,\hat{\mathbf{z}}$ | (8$d$) | H II |
| $\mathbf{B}_{75}$ | = | $x_{15}\,\mathbf{a}_1 + \left(\frac{1}{2} - y_{15}\right)\mathbf{a}_2 + z_{15}\,\mathbf{a}_3$ | = | $x_{15}a\,\hat{\mathbf{x}} + \left(\frac{1}{2} - y_{15}\right)b\,\hat{\mathbf{y}} + z_{15}c\,\hat{\mathbf{z}}$ | (8$d$) | H II |
| $\mathbf{B}_{76}$ | = | $\left(\frac{1}{2} - x_{15}\right)\mathbf{a}_1 + \left(\frac{1}{2} + y_{15}\right)\mathbf{a}_2 + \left(\frac{1}{2} + z_{15}\right)\mathbf{a}_3$ | = | $\left(\frac{1}{2} - x_{15}\right)a\,\hat{\mathbf{x}} + \left(\frac{1}{2} + y_{15}\right)b\,\hat{\mathbf{y}} + \left(\frac{1}{2} + z_{15}\right)c\,\hat{\mathbf{z}}$ | (8$d$) | H II |
| $\mathbf{B}_{77}$ | = | $x_{16}\,\mathbf{a}_1 + y_{16}\,\mathbf{a}_2 + z_{16}\,\mathbf{a}_3$ | = | $x_{16}a\,\hat{\mathbf{x}} + y_{16}b\,\hat{\mathbf{y}} + z_{16}c\,\hat{\mathbf{z}}$ | (8$d$) | K III |
| $\mathbf{B}_{78}$ | = | $\left(\frac{1}{2} - x_{16}\right)\mathbf{a}_1 - y_{16}\,\mathbf{a}_2 + \left(\frac{1}{2} + z_{16}\right)\mathbf{a}_3$ | = | $\left(\frac{1}{2} - x_{16}\right)a\,\hat{\mathbf{x}} - y_{16}b\,\hat{\mathbf{y}} + \left(\frac{1}{2} + z_{16}\right)c\,\hat{\mathbf{z}}$ | (8$d$) | K III |
| $\mathbf{B}_{79}$ | = | $-x_{16}\,\mathbf{a}_1 + \left(\frac{1}{2} + y_{16}\right)\mathbf{a}_2 - z_{16}\,\mathbf{a}_3$ | = | $-x_{16}a\,\hat{\mathbf{x}} + \left(\frac{1}{2} + y_{16}\right)b\,\hat{\mathbf{y}} - z_{16}c\,\hat{\mathbf{z}}$ | (8$d$) | K III |
| $\mathbf{B}_{80}$ | = | $\left(\frac{1}{2} + x_{16}\right)\mathbf{a}_1 + \left(\frac{1}{2} - y_{16}\right)\mathbf{a}_2 + \left(\frac{1}{2} - z_{16}\right)\mathbf{a}_3$ | = | $\left(\frac{1}{2} + x_{16}\right)a\,\hat{\mathbf{x}} + \left(\frac{1}{2} - y_{16}\right)b\,\hat{\mathbf{y}} + \left(\frac{1}{2} - z_{16}\right)c\,\hat{\mathbf{z}}$ | (8$d$) | K III |
| $\mathbf{B}_{81}$ | = | $-x_{16}\,\mathbf{a}_1 - y_{16}\,\mathbf{a}_2 - z_{16}\,\mathbf{a}_3$ | = | $-x_{16}a\,\hat{\mathbf{x}} - y_{16}b\,\hat{\mathbf{y}} - z_{16}c\,\hat{\mathbf{z}}$ | (8$d$) | K III |
| $\mathbf{B}_{82}$ | = | $\left(\frac{1}{2} + x_{16}\right)\mathbf{a}_1 + y_{16}\,\mathbf{a}_2 + \left(\frac{1}{2} - z_{16}\right)\mathbf{a}_3$ | = | $\left(\frac{1}{2} + x_{16}\right)a\,\hat{\mathbf{x}} + y_{16}b\,\hat{\mathbf{y}} + \left(\frac{1}{2} - z_{16}\right)c\,\hat{\mathbf{z}}$ | (8$d$) | K III |
| $\mathbf{B}_{83}$ | = | $x_{16}\,\mathbf{a}_1 + \left(\frac{1}{2} - y_{16}\right)\mathbf{a}_2 + z_{16}\,\mathbf{a}_3$ | = | $x_{16}a\,\hat{\mathbf{x}} + \left(\frac{1}{2} - y_{16}\right)b\,\hat{\mathbf{y}} + z_{16}c\,\hat{\mathbf{z}}$ | (8$d$) | K III |
| $\mathbf{B}_{84}$ | = | $\left(\frac{1}{2} - x_{16}\right)\mathbf{a}_1 + \left(\frac{1}{2} + y_{16}\right)\mathbf{a}_2 + \left(\frac{1}{2} + z_{16}\right)\mathbf{a}_3$ | = | $\left(\frac{1}{2} - x_{16}\right)a\,\hat{\mathbf{x}} + \left(\frac{1}{2} + y_{16}\right)b\,\hat{\mathbf{y}} + \left(\frac{1}{2} + z_{16}\right)c\,\hat{\mathbf{z}}$ | (8$d$) | K III |
| $\mathbf{B}_{85}$ | = | $x_{17}\,\mathbf{a}_1 + y_{17}\,\mathbf{a}_2 + z_{17}\,\mathbf{a}_3$ | = | $x_{17}a\,\hat{\mathbf{x}} + y_{17}b\,\hat{\mathbf{y}} + z_{17}c\,\hat{\mathbf{z}}$ | (8$d$) | N V |
| $\mathbf{B}_{86}$ | = | $\left(\frac{1}{2} - x_{17}\right)\mathbf{a}_1 - y_{17}\,\mathbf{a}_2 + \left(\frac{1}{2} + z_{17}\right)\mathbf{a}_3$ | = | $\left(\frac{1}{2} - x_{17}\right)a\,\hat{\mathbf{x}} - y_{17}b\,\hat{\mathbf{y}} + \left(\frac{1}{2} + z_{17}\right)c\,\hat{\mathbf{z}}$ | (8$d$) | N V |
| $\mathbf{B}_{87}$ | = | $-x_{17}\,\mathbf{a}_1 + \left(\frac{1}{2} + y_{17}\right)\mathbf{a}_2 - z_{17}\,\mathbf{a}_3$ | = | $-x_{17}a\,\hat{\mathbf{x}} + \left(\frac{1}{2} + y_{17}\right)b\,\hat{\mathbf{y}} - z_{17}c\,\hat{\mathbf{z}}$ | (8$d$) | N V |
| $\mathbf{B}_{88}$ | = | $\left(\frac{1}{2} + x_{17}\right)\mathbf{a}_1 + \left(\frac{1}{2} - y_{17}\right)\mathbf{a}_2 + \left(\frac{1}{2} - z_{17}\right)\mathbf{a}_3$ | = | $\left(\frac{1}{2} + x_{17}\right)a\,\hat{\mathbf{x}} + \left(\frac{1}{2} - y_{17}\right)b\,\hat{\mathbf{y}} + \left(\frac{1}{2} - z_{17}\right)c\,\hat{\mathbf{z}}$ | (8$d$) | N V |
| $\mathbf{B}_{89}$ | = | $-x_{17}\,\mathbf{a}_1 - y_{17}\,\mathbf{a}_2 - z_{17}\,\mathbf{a}_3$ | = | $-x_{17}a\,\hat{\mathbf{x}} - y_{17}b\,\hat{\mathbf{y}} - z_{17}c\,\hat{\mathbf{z}}$ | (8$d$) | N V |
| $\mathbf{B}_{90}$ | = | $\left(\frac{1}{2} + x_{17}\right)\mathbf{a}_1 + y_{17}\,\mathbf{a}_2 + \left(\frac{1}{2} - z_{17}\right)\mathbf{a}_3$ | = | $\left(\frac{1}{2} + x_{17}\right)a\,\hat{\mathbf{x}} + y_{17}b\,\hat{\mathbf{y}} + \left(\frac{1}{2} - z_{17}\right)c\,\hat{\mathbf{z}}$ | (8$d$) | N V |
| $\mathbf{B}_{91}$ | = | $x_{17}\,\mathbf{a}_1 + \left(\frac{1}{2} - y_{17}\right)\mathbf{a}_2 + z_{17}\,\mathbf{a}_3$ | = | $x_{17}a\,\hat{\mathbf{x}} + \left(\frac{1}{2} - y_{17}\right)b\,\hat{\mathbf{y}} + z_{17}c\,\hat{\mathbf{z}}$ | (8$d$) | N V |
| $\mathbf{B}_{92}$ | = | $\left(\frac{1}{2} - x_{17}\right)\mathbf{a}_1 + \left(\frac{1}{2} + y_{17}\right)\mathbf{a}_2 + \left(\frac{1}{2} + z_{17}\right)\mathbf{a}_3$ | = | $\left(\frac{1}{2} - x_{17}\right)a\,\hat{\mathbf{x}} + \left(\frac{1}{2} + y_{17}\right)b\,\hat{\mathbf{y}} + \left(\frac{1}{2} + z_{17}\right)c\,\hat{\mathbf{z}}$ | (8$d$) | N V |
| $\mathbf{B}_{93}$ | = | $x_{18}\,\mathbf{a}_1 + y_{18}\,\mathbf{a}_2 + z_{18}\,\mathbf{a}_3$ | = | $x_{18}a\,\hat{\mathbf{x}} + y_{18}b\,\hat{\mathbf{y}} + z_{18}c\,\hat{\mathbf{z}}$ | (8$d$) | N VI |
| $\mathbf{B}_{94}$ | = | $\left(\frac{1}{2} - x_{18}\right)\mathbf{a}_1 - y_{18}\,\mathbf{a}_2 + \left(\frac{1}{2} + z_{18}\right)\mathbf{a}_3$ | = | $\left(\frac{1}{2} - x_{18}\right)a\,\hat{\mathbf{x}} - y_{18}b\,\hat{\mathbf{y}} + \left(\frac{1}{2} + z_{18}\right)c\,\hat{\mathbf{z}}$ | (8$d$) | N VI |
| $\mathbf{B}_{95}$ | = | $-x_{18}\,\mathbf{a}_1 + \left(\frac{1}{2} + y_{18}\right)\mathbf{a}_2 - z_{18}\,\mathbf{a}_3$ | = | $-x_{18}a\,\hat{\mathbf{x}} + \left(\frac{1}{2} + y_{18}\right)b\,\hat{\mathbf{y}} - z_{18}c\,\hat{\mathbf{z}}$ | (8$d$) | N VI |
| $\mathbf{B}_{96}$ | = | $\left(\frac{1}{2} + x_{18}\right)\mathbf{a}_1 + \left(\frac{1}{2} - y_{18}\right)\mathbf{a}_2 + \left(\frac{1}{2} - z_{18}\right)\mathbf{a}_3$ | = | $\left(\frac{1}{2} + x_{18}\right)a\,\hat{\mathbf{x}} + \left(\frac{1}{2} - y_{18}\right)b\,\hat{\mathbf{y}} + \left(\frac{1}{2} - z_{18}\right)c\,\hat{\mathbf{z}}$ | (8$d$) | N VI |
| $\mathbf{B}_{97}$ | = | $-x_{18}\,\mathbf{a}_1 - y_{18}\,\mathbf{a}_2 - z_{18}\,\mathbf{a}_3$ | = | $-x_{18}a\,\mathbf{\hat{x}} - y_{18}b\,\mathbf{\hat{y}} - z_{18}c\,\mathbf{\hat{z}}$ | (8*d*) | N VI |
| $\mathbf{B}_{98}$ | = | $\left(\frac{1}{2} + x_{18}\right)\mathbf{a}_1 + y_{18}\,\mathbf{a}_2 + \left(\frac{1}{2} - z_{18}\right)\mathbf{a}_3$ | = | $\left(\frac{1}{2} + x_{18}\right)a\,\mathbf{\hat{x}} + y_{18}b\,\mathbf{\hat{y}} + \left(\frac{1}{2} - z_{18}\right)c\,\mathbf{\hat{z}}$ | (8*d*) | N VI |
| $\mathbf{B}_{99}$ | = | $x_{18}\,\mathbf{a}_1 + \left(\frac{1}{2} - y_{18}\right)\mathbf{a}_2 + z_{18}\,\mathbf{a}_3$ | = | $x_{18}a\,\mathbf{\hat{x}} + \left(\frac{1}{2} - y_{18}\right)b\,\mathbf{\hat{y}} + z_{18}c\,\mathbf{\hat{z}}$ | (8*d*) | N VI |
| $\mathbf{B}_{100}$ | = | $\left(\frac{1}{2} - x_{18}\right)\mathbf{a}_1 + \left(\frac{1}{2} + y_{18}\right)\mathbf{a}_2 + \left(\frac{1}{2} + z_{18}\right)\mathbf{a}_3$ | = | $\left(\frac{1}{2} - x_{18}\right)a\,\mathbf{\hat{x}} + \left(\frac{1}{2} + y_{18}\right)b\,\mathbf{\hat{y}} + \left(\frac{1}{2} + z_{18}\right)c\,\mathbf{\hat{z}}$ | (8*d*) | N VI |
| $\mathbf{B}_{101}$ | = | $x_{19}\,\mathbf{a}_1 + y_{19}\,\mathbf{a}_2 + z_{19}\,\mathbf{a}_3$ | = | $x_{19}a\,\mathbf{\hat{x}} + y_{19}b\,\mathbf{\hat{y}} + z_{19}c\,\mathbf{\hat{z}}$ | (8*d*) | O |
| $\mathbf{B}_{102}$ | = | $\left(\frac{1}{2} - x_{19}\right)\mathbf{a}_1 - y_{19}\,\mathbf{a}_2 + \left(\frac{1}{2} + z_{19}\right)\mathbf{a}_3$ | = | $\left(\frac{1}{2} - x_{19}\right)a\,\mathbf{\hat{x}} - y_{19}b\,\mathbf{\hat{y}} + \left(\frac{1}{2} + z_{19}\right)c\,\mathbf{\hat{z}}$ | (8*d*) | O |
| $\mathbf{B}_{103}$ | = | $-x_{19}\,\mathbf{a}_1 + \left(\frac{1}{2} + y_{19}\right)\mathbf{a}_2 - z_{19}\,\mathbf{a}_3$ | = | $-x_{19}a\,\mathbf{\hat{x}} + \left(\frac{1}{2} + y_{19}\right)b\,\mathbf{\hat{y}} - z_{19}c\,\mathbf{\hat{z}}$ | (8*d*) | O |
| $\mathbf{B}_{104}$ | = | $\left(\frac{1}{2} + x_{19}\right)\mathbf{a}_1 + \left(\frac{1}{2} - y_{19}\right)\mathbf{a}_2 + \left(\frac{1}{2} - z_{19}\right)\mathbf{a}_3$ | = | $\left(\frac{1}{2} + x_{19}\right)a\,\mathbf{\hat{x}} + \left(\frac{1}{2} - y_{19}\right)b\,\mathbf{\hat{y}} + \left(\frac{1}{2} - z_{19}\right)c\,\mathbf{\hat{z}}$ | (8*d*) | O |
| $\mathbf{B}_{105}$ | = | $-x_{19}\,\mathbf{a}_1 - y_{19}\,\mathbf{a}_2 - z_{19}\,\mathbf{a}_3$ | = | $-x_{19}a\,\mathbf{\hat{x}} - y_{19}b\,\mathbf{\hat{y}} - z_{19}c\,\mathbf{\hat{z}}$ | (8*d*) | O |
| $\mathbf{B}_{106}$ | = | $\left(\frac{1}{2} + x_{19}\right)\mathbf{a}_1 + y_{19}\,\mathbf{a}_2 + \left(\frac{1}{2} - z_{19}\right)\mathbf{a}_3$ | = | $\left(\frac{1}{2} + x_{19}\right)a\,\mathbf{\hat{x}} + y_{19}b\,\mathbf{\hat{y}} + \left(\frac{1}{2} - z_{19}\right)c\,\mathbf{\hat{z}}$ | (8*d*) | O |
| $\mathbf{B}_{107}$ | = | $x_{19}\,\mathbf{a}_1 + \left(\frac{1}{2} - y_{19}\right)\mathbf{a}_2 + z_{19}\,\mathbf{a}_3$ | = | $x_{19}a\,\mathbf{\hat{x}} + \left(\frac{1}{2} - y_{19}\right)b\,\mathbf{\hat{y}} + z_{19}c\,\mathbf{\hat{z}}$ | (8*d*) | O |
| $\mathbf{B}_{108}$ | = | $\left(\frac{1}{2} - x_{19}\right)\mathbf{a}_1 + \left(\frac{1}{2} + y_{19}\right)\mathbf{a}_2 + \left(\frac{1}{2} + z_{19}\right)\mathbf{a}_3$ | = | $\left(\frac{1}{2} - x_{19}\right)a\,\mathbf{\hat{x}} + \left(\frac{1}{2} + y_{19}\right)b\,\mathbf{\hat{y}} + \left(\frac{1}{2} + z_{19}\right)c\,\mathbf{\hat{z}}$ | (8*d*) | O |

**Geometry files:**

- CIF: pp. 1642
- POSCAR: pp. 1642

# $SbCl_5 \cdot POCl_3$ Structure: A8BCD_oP44_62_4c2d_c_c_c

http://aflow.org/prototype-encyclopedia/A8BCD_oP44_62_4c2d_c_c_c

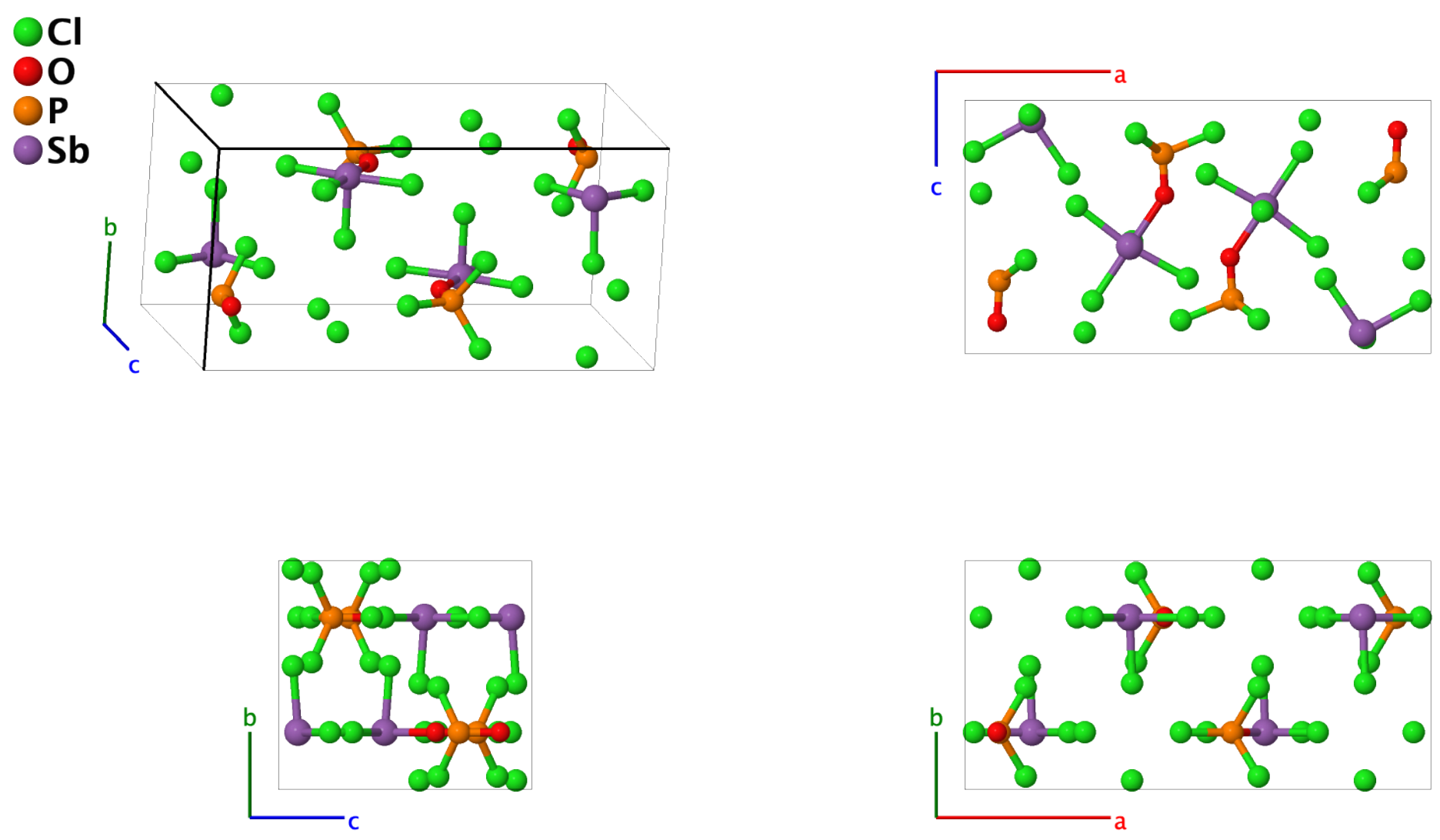

| | | |
|---|---|---|
| **Prototype** | : | $Cl_8OPSb$ |
| **AFLOW prototype label** | : | A8BCD_oP44_62_4c2d_c_c_c |
| ***Strukturbericht* designation** | : | None |
| **Pearson symbol** | : | oP44 |
| **Space group number** | : | 62 |
| **Space group symbol** | : | *Pnma* |
| **AFLOW prototype command** | : | `aflow --proto=A8BCD_oP44_62_4c2d_c_c_c` `--params=`$a, b/a, c/a, x_1, z_1, x_2, z_2, x_3, z_3, x_4, z_4, x_5, z_5, x_6, z_6, x_7, z_7, x_8, y_8, z_8, x_9, y_9, z_9$ |

**Other compounds with this structure**

- $SbCl_5 \cdot POCl_3(CH_3)_3$

- (Lindqvist, 1959) give the data for this structure in setting *Pmnb* of space group #62. We used FINDSYM to exchange the *a*- and *b*-axes so that the structure is now in the standard *Pnma* orientation.

**Simple Orthorhombic primitive vectors:**

$$\mathbf{a}_1 = a\,\hat{\mathbf{x}}$$
$$\mathbf{a}_2 = b\,\hat{\mathbf{y}}$$
$$\mathbf{a}_3 = c\,\hat{\mathbf{z}}$$

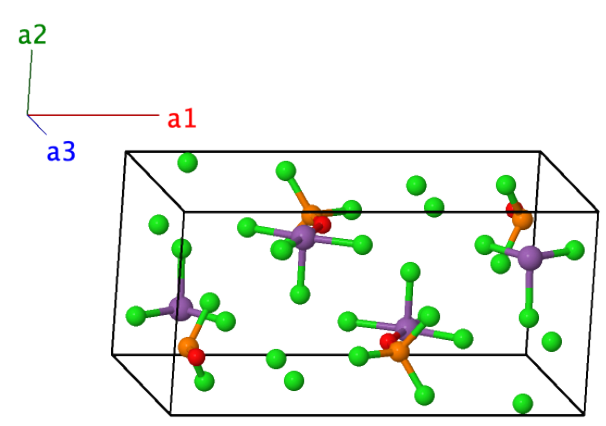

**Basis vectors:**

| | | Lattice Coordinates | | Cartesian Coordinates | Wyckoff Position | Atom Type |
|---|---|---|---|---|---|---|
| $\mathbf{B}_{1}$ | = | $x_1\,\mathbf{a}_1 + \frac{1}{4}\,\mathbf{a}_2 + z_1\,\mathbf{a}_3$ | = | $x_1 a\,\hat{\mathbf{x}} + \frac{1}{4} b\,\hat{\mathbf{y}} + z_1 c\,\hat{\mathbf{z}}$ | (4$c$) | Cl I |
| $\mathbf{B}_{2}$ | = | $\left(\frac{1}{2} - x_1\right)\mathbf{a}_1 + \frac{3}{4}\,\mathbf{a}_2 + \left(\frac{1}{2} + z_1\right)\mathbf{a}_3$ | = | $\left(\frac{1}{2} - x_1\right) a\,\hat{\mathbf{x}} + \frac{3}{4} b\,\hat{\mathbf{y}} + \left(\frac{1}{2} + z_1\right) c\,\hat{\mathbf{z}}$ | (4$c$) | Cl I |
| $\mathbf{B}_{3}$ | = | $-x_1\,\mathbf{a}_1 + \frac{3}{4}\,\mathbf{a}_2 - z_1\,\mathbf{a}_3$ | = | $-x_1 a\,\hat{\mathbf{x}} + \frac{3}{4} b\,\hat{\mathbf{y}} - z_1 c\,\hat{\mathbf{z}}$ | (4$c$) | Cl I |
| $\mathbf{B}_{4}$ | = | $\left(\frac{1}{2} + x_1\right)\mathbf{a}_1 + \frac{1}{4}\,\mathbf{a}_2 + \left(\frac{1}{2} - z_1\right)\mathbf{a}_3$ | = | $\left(\frac{1}{2} + x_1\right) a\,\hat{\mathbf{x}} + \frac{1}{4} b\,\hat{\mathbf{y}} + \left(\frac{1}{2} - z_1\right) c\,\hat{\mathbf{z}}$ | (4$c$) | Cl I |
| $\mathbf{B}_{5}$ | = | $x_2\,\mathbf{a}_1 + \frac{1}{4}\,\mathbf{a}_2 + z_2\,\mathbf{a}_3$ | = | $x_2 a\,\hat{\mathbf{x}} + \frac{1}{4} b\,\hat{\mathbf{y}} + z_2 c\,\hat{\mathbf{z}}$ | (4$c$) | Cl II |
| $\mathbf{B}_{6}$ | = | $\left(\frac{1}{2} - x_2\right)\mathbf{a}_1 + \frac{3}{4}\,\mathbf{a}_2 + \left(\frac{1}{2} + z_2\right)\mathbf{a}_3$ | = | $\left(\frac{1}{2} - x_2\right) a\,\hat{\mathbf{x}} + \frac{3}{4} b\,\hat{\mathbf{y}} + \left(\frac{1}{2} + z_2\right) c\,\hat{\mathbf{z}}$ | (4$c$) | Cl II |
| $\mathbf{B}_{7}$ | = | $-x_2\,\mathbf{a}_1 + \frac{3}{4}\,\mathbf{a}_2 - z_2\,\mathbf{a}_3$ | = | $-x_2 a\,\hat{\mathbf{x}} + \frac{3}{4} b\,\hat{\mathbf{y}} - z_2 c\,\hat{\mathbf{z}}$ | (4$c$) | Cl II |
| $\mathbf{B}_{8}$ | = | $\left(\frac{1}{2} + x_2\right)\mathbf{a}_1 + \frac{1}{4}\,\mathbf{a}_2 + \left(\frac{1}{2} - z_2\right)\mathbf{a}_3$ | = | $\left(\frac{1}{2} + x_2\right) a\,\hat{\mathbf{x}} + \frac{1}{4} b\,\hat{\mathbf{y}} + \left(\frac{1}{2} - z_2\right) c\,\hat{\mathbf{z}}$ | (4$c$) | Cl II |
| $\mathbf{B}_{9}$ | = | $x_3\,\mathbf{a}_1 + \frac{1}{4}\,\mathbf{a}_2 + z_3\,\mathbf{a}_3$ | = | $x_3 a\,\hat{\mathbf{x}} + \frac{1}{4} b\,\hat{\mathbf{y}} + z_3 c\,\hat{\mathbf{z}}$ | (4$c$) | Cl III |
| $\mathbf{B}_{10}$ | = | $\left(\frac{1}{2} - x_3\right)\mathbf{a}_1 + \frac{3}{4}\,\mathbf{a}_2 + \left(\frac{1}{2} + z_3\right)\mathbf{a}_3$ | = | $\left(\frac{1}{2} - x_3\right) a\,\hat{\mathbf{x}} + \frac{3}{4} b\,\hat{\mathbf{y}} + \left(\frac{1}{2} + z_3\right) c\,\hat{\mathbf{z}}$ | (4$c$) | Cl III |
| $\mathbf{B}_{11}$ | = | $-x_3\,\mathbf{a}_1 + \frac{3}{4}\,\mathbf{a}_2 - z_3\,\mathbf{a}_3$ | = | $-x_3 a\,\hat{\mathbf{x}} + \frac{3}{4} b\,\hat{\mathbf{y}} - z_3 c\,\hat{\mathbf{z}}$ | (4$c$) | Cl III |
| $\mathbf{B}_{12}$ | = | $\left(\frac{1}{2} + x_3\right)\mathbf{a}_1 + \frac{1}{4}\,\mathbf{a}_2 + \left(\frac{1}{2} - z_3\right)\mathbf{a}_3$ | = | $\left(\frac{1}{2} + x_3\right) a\,\hat{\mathbf{x}} + \frac{1}{4} b\,\hat{\mathbf{y}} + \left(\frac{1}{2} - z_3\right) c\,\hat{\mathbf{z}}$ | (4$c$) | Cl III |
| $\mathbf{B}_{13}$ | = | $x_4\,\mathbf{a}_1 + \frac{1}{4}\,\mathbf{a}_2 + z_4\,\mathbf{a}_3$ | = | $x_4 a\,\hat{\mathbf{x}} + \frac{1}{4} b\,\hat{\mathbf{y}} + z_4 c\,\hat{\mathbf{z}}$ | (4$c$) | Cl IV |
| $\mathbf{B}_{14}$ | = | $\left(\frac{1}{2} - x_4\right)\mathbf{a}_1 + \frac{3}{4}\,\mathbf{a}_2 + \left(\frac{1}{2} + z_4\right)\mathbf{a}_3$ | = | $\left(\frac{1}{2} - x_4\right) a\,\hat{\mathbf{x}} + \frac{3}{4} b\,\hat{\mathbf{y}} + \left(\frac{1}{2} + z_4\right) c\,\hat{\mathbf{z}}$ | (4$c$) | Cl IV |
| $\mathbf{B}_{15}$ | = | $-x_4\,\mathbf{a}_1 + \frac{3}{4}\,\mathbf{a}_2 - z_4\,\mathbf{a}_3$ | = | $-x_4 a\,\hat{\mathbf{x}} + \frac{3}{4} b\,\hat{\mathbf{y}} - z_4 c\,\hat{\mathbf{z}}$ | (4$c$) | Cl IV |
| $\mathbf{B}_{16}$ | = | $\left(\frac{1}{2} + x_4\right)\mathbf{a}_1 + \frac{1}{4}\,\mathbf{a}_2 + \left(\frac{1}{2} - z_4\right)\mathbf{a}_3$ | = | $\left(\frac{1}{2} + x_4\right) a\,\hat{\mathbf{x}} + \frac{1}{4} b\,\hat{\mathbf{y}} + \left(\frac{1}{2} - z_4\right) c\,\hat{\mathbf{z}}$ | (4$c$) | Cl IV |
| $\mathbf{B}_{17}$ | = | $x_5\,\mathbf{a}_1 + \frac{1}{4}\,\mathbf{a}_2 + z_5\,\mathbf{a}_3$ | = | $x_5 a\,\hat{\mathbf{x}} + \frac{1}{4} b\,\hat{\mathbf{y}} + z_5 c\,\hat{\mathbf{z}}$ | (4$c$) | O |
| $\mathbf{B}_{18}$ | = | $\left(\frac{1}{2} - x_5\right)\mathbf{a}_1 + \frac{3}{4}\,\mathbf{a}_2 + \left(\frac{1}{2} + z_5\right)\mathbf{a}_3$ | = | $\left(\frac{1}{2} - x_5\right) a\,\hat{\mathbf{x}} + \frac{3}{4} b\,\hat{\mathbf{y}} + \left(\frac{1}{2} + z_5\right) c\,\hat{\mathbf{z}}$ | (4$c$) | O |
| $\mathbf{B}_{19}$ | = | $-x_5\,\mathbf{a}_1 + \frac{3}{4}\,\mathbf{a}_2 - z_5\,\mathbf{a}_3$ | = | $-x_5 a\,\hat{\mathbf{x}} + \frac{3}{4} b\,\hat{\mathbf{y}} - z_5 c\,\hat{\mathbf{z}}$ | (4$c$) | O |
| $\mathbf{B}_{20}$ | = | $\left(\frac{1}{2} + x_5\right)\mathbf{a}_1 + \frac{1}{4}\,\mathbf{a}_2 + \left(\frac{1}{2} - z_5\right)\mathbf{a}_3$ | = | $\left(\frac{1}{2} + x_5\right) a\,\hat{\mathbf{x}} + \frac{1}{4} b\,\hat{\mathbf{y}} + \left(\frac{1}{2} - z_5\right) c\,\hat{\mathbf{z}}$ | (4$c$) | O |
| $\mathbf{B}_{21}$ | = | $x_6\,\mathbf{a}_1 + \frac{1}{4}\,\mathbf{a}_2 + z_6\,\mathbf{a}_3$ | = | $x_6 a\,\hat{\mathbf{x}} + \frac{1}{4} b\,\hat{\mathbf{y}} + z_6 c\,\hat{\mathbf{z}}$ | (4$c$) | P |
| $\mathbf{B}_{22}$ | = | $\left(\frac{1}{2} - x_6\right)\mathbf{a}_1 + \frac{3}{4}\,\mathbf{a}_2 + \left(\frac{1}{2} + z_6\right)\mathbf{a}_3$ | = | $\left(\frac{1}{2} - x_6\right) a\,\hat{\mathbf{x}} + \frac{3}{4} b\,\hat{\mathbf{y}} + \left(\frac{1}{2} + z_6\right) c\,\hat{\mathbf{z}}$ | (4$c$) | P |
| $\mathbf{B}_{23}$ | = | $-x_6\,\mathbf{a}_1 + \frac{3}{4}\,\mathbf{a}_2 - z_6\,\mathbf{a}_3$ | = | $-x_6 a\,\hat{\mathbf{x}} + \frac{3}{4} b\,\hat{\mathbf{y}} - z_6 c\,\hat{\mathbf{z}}$ | (4$c$) | P |
| $\mathbf{B}_{24}$ | = | $\left(\frac{1}{2} + x_6\right)\mathbf{a}_1 + \frac{1}{4}\,\mathbf{a}_2 + \left(\frac{1}{2} - z_6\right)\mathbf{a}_3$ | = | $\left(\frac{1}{2} + x_6\right) a\,\hat{\mathbf{x}} + \frac{1}{4} b\,\hat{\mathbf{y}} + \left(\frac{1}{2} - z_6\right) c\,\hat{\mathbf{z}}$ | (4$c$) | P |
| $\mathbf{B}_{25}$ | = | $x_7\,\mathbf{a}_1 + \frac{1}{4}\,\mathbf{a}_2 + z_7\,\mathbf{a}_3$ | = | $x_7 a\,\hat{\mathbf{x}} + \frac{1}{4} b\,\hat{\mathbf{y}} + z_7 c\,\hat{\mathbf{z}}$ | (4$c$) | Sb |
| $\mathbf{B}_{26}$ | = | $\left(\frac{1}{2} - x_7\right)\mathbf{a}_1 + \frac{3}{4}\,\mathbf{a}_2 + \left(\frac{1}{2} + z_7\right)\mathbf{a}_3$ | = | $\left(\frac{1}{2} - x_7\right) a\,\hat{\mathbf{x}} + \frac{3}{4} b\,\hat{\mathbf{y}} + \left(\frac{1}{2} + z_7\right) c\,\hat{\mathbf{z}}$ | (4$c$) | Sb |
| $\mathbf{B}_{27}$ | = | $-x_7\,\mathbf{a}_1 + \frac{3}{4}\,\mathbf{a}_2 - z_7\,\mathbf{a}_3$ | = | $-x_7 a\,\hat{\mathbf{x}} + \frac{3}{4} b\,\hat{\mathbf{y}} - z_7 c\,\hat{\mathbf{z}}$ | (4$c$) | Sb |
| $\mathbf{B}_{28}$ | = | $\left(\frac{1}{2} + x_7\right)\mathbf{a}_1 + \frac{1}{4}\,\mathbf{a}_2 + \left(\frac{1}{2} - z_7\right)\mathbf{a}_3$ | = | $\left(\frac{1}{2} + x_7\right) a\,\hat{\mathbf{x}} + \frac{1}{4} b\,\hat{\mathbf{y}} + \left(\frac{1}{2} - z_7\right) c\,\hat{\mathbf{z}}$ | (4$c$) | Sb |
| $\mathbf{B}_{29}$ | = | $x_8\,\mathbf{a}_1 + y_8\,\mathbf{a}_2 + z_8\,\mathbf{a}_3$ | = | $x_8 a\,\hat{\mathbf{x}} + y_8 b\,\hat{\mathbf{y}} + z_8 c\,\hat{\mathbf{z}}$ | (8$d$) | Cl V |
| $\mathbf{B}_{30}$ | = | $\left(\frac{1}{2} - x_8\right)\mathbf{a}_1 - y_8\,\mathbf{a}_2 + \left(\frac{1}{2} + z_8\right)\mathbf{a}_3$ | = | $\left(\frac{1}{2} - x_8\right) a\,\hat{\mathbf{x}} - y_8 b\,\hat{\mathbf{y}} + \left(\frac{1}{2} + z_8\right) c\,\hat{\mathbf{z}}$ | (8$d$) | Cl V |
| $\mathbf{B}_{31}$ | = | $-x_8\,\mathbf{a}_1 + \left(\frac{1}{2} + y_8\right)\mathbf{a}_2 - z_8\,\mathbf{a}_3$ | = | $-x_8 a\,\hat{\mathbf{x}} + \left(\frac{1}{2} + y_8\right) b\,\hat{\mathbf{y}} - z_8 c\,\hat{\mathbf{z}}$ | (8$d$) | Cl V |
| $\mathbf{B}_{32}$ | = | $\left(\frac{1}{2} + x_8\right)\mathbf{a}_1 + \left(\frac{1}{2} - y_8\right)\mathbf{a}_2 + \left(\frac{1}{2} - z_8\right)\mathbf{a}_3$ | = | $\left(\frac{1}{2} + x_8\right) a\,\hat{\mathbf{x}} + \left(\frac{1}{2} - y_8\right) b\,\hat{\mathbf{y}} + \left(\frac{1}{2} - z_8\right) c\,\hat{\mathbf{z}}$ | (8$d$) | Cl V |
| $\mathbf{B}_{33}$ | = | $-x_8\,\mathbf{a}_1 - y_8\,\mathbf{a}_2 - z_8\,\mathbf{a}_3$ | = | $-x_8 a\,\hat{\mathbf{x}} - y_8 b\,\hat{\mathbf{y}} - z_8 c\,\hat{\mathbf{z}}$ | (8$d$) | Cl V |
| $\mathbf{B}_{34}$ | = | $\left(\frac{1}{2} + x_8\right)\mathbf{a}_1 + y_8\,\mathbf{a}_2 + \left(\frac{1}{2} - z_8\right)\mathbf{a}_3$ | = | $\left(\frac{1}{2} + x_8\right) a\,\hat{\mathbf{x}} + y_8 b\,\hat{\mathbf{y}} + \left(\frac{1}{2} - z_8\right) c\,\hat{\mathbf{z}}$ | (8$d$) | Cl V |
| $\mathbf{B}_{35}$ | = | $x_8\,\mathbf{a}_1 + \left(\frac{1}{2} - y_8\right)\mathbf{a}_2 + z_8\,\mathbf{a}_3$ | = | $x_8 a\,\hat{\mathbf{x}} + \left(\frac{1}{2} - y_8\right) b\,\hat{\mathbf{y}} + z_8 c\,\hat{\mathbf{z}}$ | (8$d$) | Cl V |
| $\mathbf{B}_{36}$ | = | $\left(\frac{1}{2} - x_8\right) \mathbf{a}_1 + \left(\frac{1}{2} + y_8\right) \mathbf{a}_2 + \left(\frac{1}{2} + z_8\right) \mathbf{a}_3$ | = | $\left(\frac{1}{2} - x_8\right) a \, \hat{\mathbf{x}} + \left(\frac{1}{2} + y_8\right) b \, \hat{\mathbf{y}} + \left(\frac{1}{2} + z_8\right) c \, \hat{\mathbf{z}}$ | (8$d$) | Cl V |
| $\mathbf{B}_{37}$ | = | $x_9 \, \mathbf{a}_1 + y_9 \, \mathbf{a}_2 + z_9 \, \mathbf{a}_3$ | = | $x_9 a \, \hat{\mathbf{x}} + y_9 b \, \hat{\mathbf{y}} + z_9 c \, \hat{\mathbf{z}}$ | (8$d$) | Cl VI |
| $\mathbf{B}_{38}$ | = | $\left(\frac{1}{2} - x_9\right) \mathbf{a}_1 - y_9 \, \mathbf{a}_2 + \left(\frac{1}{2} + z_9\right) \mathbf{a}_3$ | = | $\left(\frac{1}{2} - x_9\right) a \, \hat{\mathbf{x}} - y_9 b \, \hat{\mathbf{y}} + \left(\frac{1}{2} + z_9\right) c \, \hat{\mathbf{z}}$ | (8$d$) | Cl VI |
| $\mathbf{B}_{39}$ | = | $-x_9 \, \mathbf{a}_1 + \left(\frac{1}{2} + y_9\right) \mathbf{a}_2 - z_9 \, \mathbf{a}_3$ | = | $-x_9 a \, \hat{\mathbf{x}} + \left(\frac{1}{2} + y_9\right) b \, \hat{\mathbf{y}} - z_9 c \, \hat{\mathbf{z}}$ | (8$d$) | Cl VI |
| $\mathbf{B}_{40}$ | = | $\left(\frac{1}{2} + x_9\right) \mathbf{a}_1 + \left(\frac{1}{2} - y_9\right) \mathbf{a}_2 + \left(\frac{1}{2} - z_9\right) \mathbf{a}_3$ | = | $\left(\frac{1}{2} + x_9\right) a \, \hat{\mathbf{x}} + \left(\frac{1}{2} - y_9\right) b \, \hat{\mathbf{y}} + \left(\frac{1}{2} - z_9\right) c \, \hat{\mathbf{z}}$ | (8$d$) | Cl VI |
| $\mathbf{B}_{41}$ | = | $-x_9 \, \mathbf{a}_1 - y_9 \, \mathbf{a}_2 - z_9 \, \mathbf{a}_3$ | = | $-x_9 a \, \hat{\mathbf{x}} - y_9 b \, \hat{\mathbf{y}} - z_9 c \, \hat{\mathbf{z}}$ | (8$d$) | Cl VI |
| $\mathbf{B}_{42}$ | = | $\left(\frac{1}{2} + x_9\right) \mathbf{a}_1 + y_9 \, \mathbf{a}_2 + \left(\frac{1}{2} - z_9\right) \mathbf{a}_3$ | = | $\left(\frac{1}{2} + x_9\right) a \, \hat{\mathbf{x}} + y_9 b \, \hat{\mathbf{y}} + \left(\frac{1}{2} - z_9\right) c \, \hat{\mathbf{z}}$ | (8$d$) | Cl VI |
| $\mathbf{B}_{43}$ | = | $x_9 \, \mathbf{a}_1 + \left(\frac{1}{2} - y_9\right) \mathbf{a}_2 + z_9 \, \mathbf{a}_3$ | = | $x_9 a \, \hat{\mathbf{x}} + \left(\frac{1}{2} - y_9\right) b \, \hat{\mathbf{y}} + z_9 c \, \hat{\mathbf{z}}$ | (8$d$) | Cl VI |
| $\mathbf{B}_{44}$ | = | $\left(\frac{1}{2} - x_9\right) \mathbf{a}_1 + \left(\frac{1}{2} + y_9\right) \mathbf{a}_2 + \left(\frac{1}{2} + z_9\right) \mathbf{a}_3$ | = | $\left(\frac{1}{2} - x_9\right) a \, \hat{\mathbf{x}} + \left(\frac{1}{2} + y_9\right) b \, \hat{\mathbf{y}} + \left(\frac{1}{2} + z_9\right) c \, \hat{\mathbf{z}}$ | (8$d$) | Cl VI |

**Geometry files:**

- CIF: pp. 1643
- POSCAR: pp. 1643

# Autunite $\{Ca[(UO_2)(PO_4)]_2(H_2O)_{11}\}$ Structure: AB22C23D2E2_oP200_62_c_11d_3c10d_d_d

http://aflow.org/prototype-encyclopedia/AB22C23D2E2_oP200_62_c_11d_3c10d_d_d

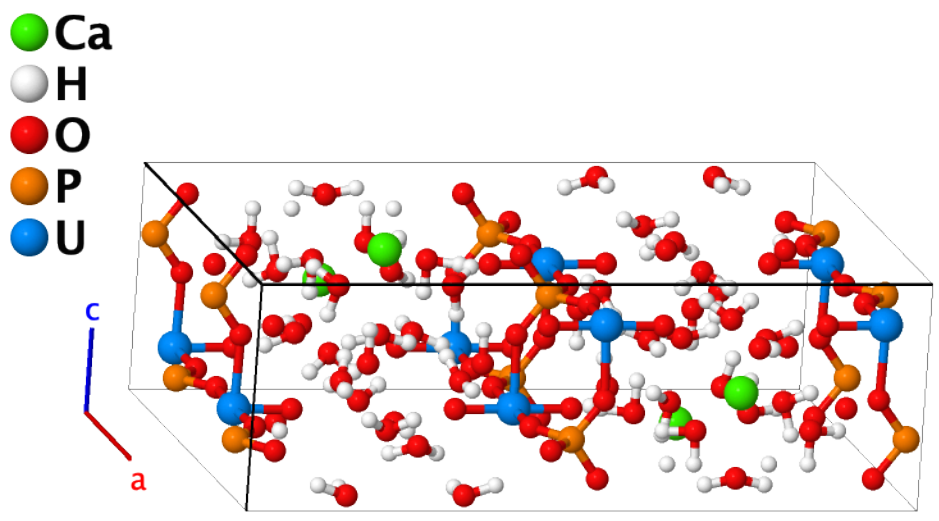

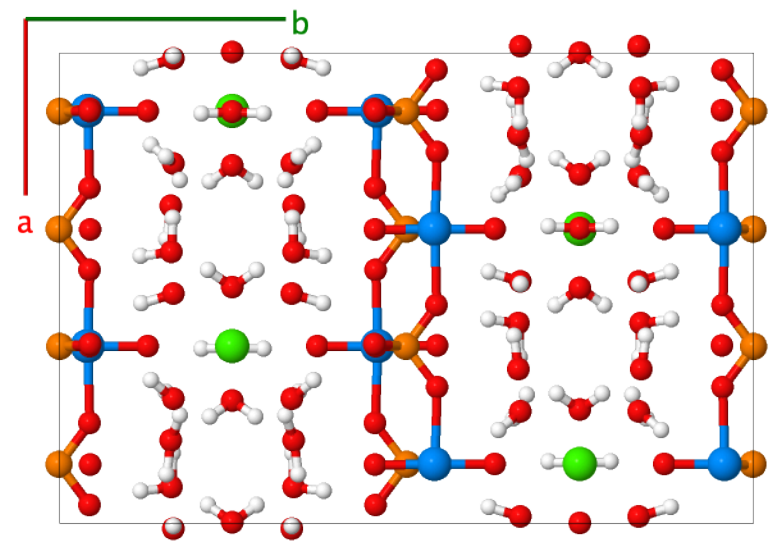

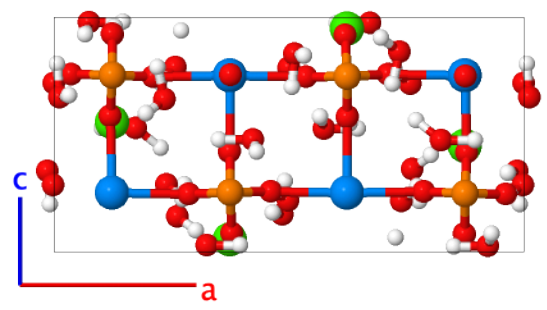

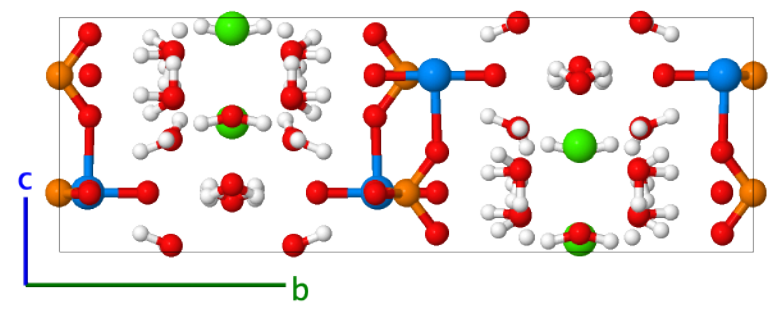

| | | |
|---|---|---|
| **Prototype** | : | $CaH_{22}O_{23}P_2U_2$ |
| **AFLOW prototype label** | : | AB22C23D2E2_oP200_62_c_11d_3c10d_d_d |
| ***Strukturbericht* designation** | : | None |
| **Pearson symbol** | : | oP200 |
| **Space group number** | : | 62 |
| **Space group symbol** | : | *Pnma* |
| **AFLOW prototype command** | : | `aflow --proto=AB22C23D2E2_oP200_62_c_11d_3c10d_d_d` `--params=`$a, b/a, c/a, x_1, z_1, x_2, z_2, x_3, z_3, x_4, z_4, x_5, y_5, z_5, x_6, y_6, z_6, x_7, y_7, z_7, x_8, y_8, z_8, x_9, y_9, z_9, x_{10}, y_{10}, z_{10}, x_{11}, y_{11}, z_{11}, x_{12}, y_{12}, z_{12}, x_{13}, y_{13}, z_{13}, x_{14}, y_{14}, z_{14}, x_{15}, y_{15}, z_{15}, x_{16}, y_{16}, z_{16}, x_{17}, y_{17}, z_{17}, x_{18}, y_{18}, z_{18}, x_{19}, y_{19}, z_{19}, x_{20}, y_{20}, z_{20}, x_{21}, y_{21}, z_{21}, x_{22}, y_{22}, z_{22}, x_{23}, y_{23}, z_{23}, x_{24}, y_{24}, z_{24}, x_{25}, y_{25}, z_{25}, x_{26}, y_{26}, z_{26}, x_{27}, y_{27}, z_{27}$ |

- Autunite, $Ca(UO_2)_2(PO_4)_2{\cdot}nH_2O$, is found in three varieties: naturally occurring Autunite, with $n \gtrsim 10$, and meta-autunite (I), which is partially dehydrated, $6 \gtrsim n \gtrsim 10$. Further dehydration in the laboratory produces meta-autunite (II).
- The original determination of the autunite structure designated *H*5$_9$ by (Herrmann, 1941) was a tetragonal structure, none of the positions of the oxygen atom or water molecules was determined. (Locock, 2003) find a pseudo-tetragonal ($a \approx 2c$) unit cell which doubles the size of the original cell. They were able to locate all of the atoms in the structure.
- (Locock, 2003) found the (4*c*) calcium site to be occupied only 86% of the time.

## Simple Orthorhombic primitive vectors:

$$\mathbf{a}_1 = a\,\hat{\mathbf{x}}$$
$$\mathbf{a}_2 = b\,\hat{\mathbf{y}}$$
$$\mathbf{a}_3 = c\,\hat{\mathbf{z}}$$

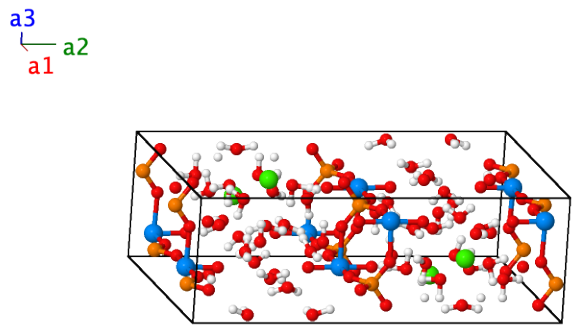

## Basis vectors:

| | | Lattice Coordinates | | Cartesian Coordinates | Wyckoff Position | Atom Type |
|---|---|---|---|---|---|---|
| $\mathbf{B}_1$ | = | $x_1\,\mathbf{a}_1 + \frac{1}{4}\,\mathbf{a}_2 + z_1\,\mathbf{a}_3$ | = | $x_1 a\,\hat{\mathbf{x}} + \frac{1}{4} b\,\hat{\mathbf{y}} + z_1 c\,\hat{\mathbf{z}}$ | (4$c$) | Ca |
| $\mathbf{B}_2$ | = | $\left(\frac{1}{2} - x_1\right)\mathbf{a}_1 + \frac{3}{4}\,\mathbf{a}_2 + \left(\frac{1}{2} + z_1\right)\mathbf{a}_3$ | = | $\left(\frac{1}{2} - x_1\right) a\,\hat{\mathbf{x}} + \frac{3}{4} b\,\hat{\mathbf{y}} + \left(\frac{1}{2} + z_1\right) c\,\hat{\mathbf{z}}$ | (4$c$) | Ca |
| $\mathbf{B}_3$ | = | $-x_1\,\mathbf{a}_1 + \frac{3}{4}\,\mathbf{a}_2 - z_1\,\mathbf{a}_3$ | = | $-x_1 a\,\hat{\mathbf{x}} + \frac{3}{4} b\,\hat{\mathbf{y}} - z_1 c\,\hat{\mathbf{z}}$ | (4$c$) | Ca |
| $\mathbf{B}_4$ | = | $\left(\frac{1}{2} + x_1\right)\mathbf{a}_1 + \frac{1}{4}\,\mathbf{a}_2 + \left(\frac{1}{2} - z_1\right)\mathbf{a}_3$ | = | $\left(\frac{1}{2} + x_1\right) a\,\hat{\mathbf{x}} + \frac{1}{4} b\,\hat{\mathbf{y}} + \left(\frac{1}{2} - z_1\right) c\,\hat{\mathbf{z}}$ | (4$c$) | Ca |
| $\mathbf{B}_5$ | = | $x_2\,\mathbf{a}_1 + \frac{1}{4}\,\mathbf{a}_2 + z_2\,\mathbf{a}_3$ | = | $x_2 a\,\hat{\mathbf{x}} + \frac{1}{4} b\,\hat{\mathbf{y}} + z_2 c\,\hat{\mathbf{z}}$ | (4$c$) | O I |
| $\mathbf{B}_6$ | = | $\left(\frac{1}{2} - x_2\right)\mathbf{a}_1 + \frac{3}{4}\,\mathbf{a}_2 + \left(\frac{1}{2} + z_2\right)\mathbf{a}_3$ | = | $\left(\frac{1}{2} - x_2\right) a\,\hat{\mathbf{x}} + \frac{3}{4} b\,\hat{\mathbf{y}} + \left(\frac{1}{2} + z_2\right) c\,\hat{\mathbf{z}}$ | (4$c$) | O I |
| $\mathbf{B}_7$ | = | $-x_2\,\mathbf{a}_1 + \frac{3}{4}\,\mathbf{a}_2 - z_2\,\mathbf{a}_3$ | = | $-x_2 a\,\hat{\mathbf{x}} + \frac{3}{4} b\,\hat{\mathbf{y}} - z_2 c\,\hat{\mathbf{z}}$ | (4$c$) | O I |
| $\mathbf{B}_8$ | = | $\left(\frac{1}{2} + x_2\right)\mathbf{a}_1 + \frac{1}{4}\,\mathbf{a}_2 + \left(\frac{1}{2} - z_2\right)\mathbf{a}_3$ | = | $\left(\frac{1}{2} + x_2\right) a\,\hat{\mathbf{x}} + \frac{1}{4} b\,\hat{\mathbf{y}} + \left(\frac{1}{2} - z_2\right) c\,\hat{\mathbf{z}}$ | (4$c$) | O I |
| $\mathbf{B}_9$ | = | $x_3\,\mathbf{a}_1 + \frac{1}{4}\,\mathbf{a}_2 + z_3\,\mathbf{a}_3$ | = | $x_3 a\,\hat{\mathbf{x}} + \frac{1}{4} b\,\hat{\mathbf{y}} + z_3 c\,\hat{\mathbf{z}}$ | (4$c$) | O II |
| $\mathbf{B}_{10}$ | = | $\left(\frac{1}{2} - x_3\right)\mathbf{a}_1 + \frac{3}{4}\,\mathbf{a}_2 + \left(\frac{1}{2} + z_3\right)\mathbf{a}_3$ | = | $\left(\frac{1}{2} - x_3\right) a\,\hat{\mathbf{x}} + \frac{3}{4} b\,\hat{\mathbf{y}} + \left(\frac{1}{2} + z_3\right) c\,\hat{\mathbf{z}}$ | (4$c$) | O II |
| $\mathbf{B}_{11}$ | = | $-x_3\,\mathbf{a}_1 + \frac{3}{4}\,\mathbf{a}_2 - z_3\,\mathbf{a}_3$ | = | $-x_3 a\,\hat{\mathbf{x}} + \frac{3}{4} b\,\hat{\mathbf{y}} - z_3 c\,\hat{\mathbf{z}}$ | (4$c$) | O II |
| $\mathbf{B}_{12}$ | = | $\left(\frac{1}{2} + x_3\right)\mathbf{a}_1 + \frac{1}{4}\,\mathbf{a}_2 + \left(\frac{1}{2} - z_3\right)\mathbf{a}_3$ | = | $\left(\frac{1}{2} + x_3\right) a\,\hat{\mathbf{x}} + \frac{1}{4} b\,\hat{\mathbf{y}} + \left(\frac{1}{2} - z_3\right) c\,\hat{\mathbf{z}}$ | (4$c$) | O II |
| $\mathbf{B}_{13}$ | = | $x_4\,\mathbf{a}_1 + \frac{1}{4}\,\mathbf{a}_2 + z_4\,\mathbf{a}_3$ | = | $x_4 a\,\hat{\mathbf{x}} + \frac{1}{4} b\,\hat{\mathbf{y}} + z_4 c\,\hat{\mathbf{z}}$ | (4$c$) | O III |
| $\mathbf{B}_{14}$ | = | $\left(\frac{1}{2} - x_4\right)\mathbf{a}_1 + \frac{3}{4}\,\mathbf{a}_2 + \left(\frac{1}{2} + z_4\right)\mathbf{a}_3$ | = | $\left(\frac{1}{2} - x_4\right) a\,\hat{\mathbf{x}} + \frac{3}{4} b\,\hat{\mathbf{y}} + \left(\frac{1}{2} + z_4\right) c\,\hat{\mathbf{z}}$ | (4$c$) | O III |
| $\mathbf{B}_{15}$ | = | $-x_4\,\mathbf{a}_1 + \frac{3}{4}\,\mathbf{a}_2 - z_4\,\mathbf{a}_3$ | = | $-x_4 a\,\hat{\mathbf{x}} + \frac{3}{4} b\,\hat{\mathbf{y}} - z_4 c\,\hat{\mathbf{z}}$ | (4$c$) | O III |
| $\mathbf{B}_{16}$ | = | $\left(\frac{1}{2} + x_4\right)\mathbf{a}_1 + \frac{1}{4}\,\mathbf{a}_2 + \left(\frac{1}{2} - z_4\right)\mathbf{a}_3$ | = | $\left(\frac{1}{2} + x_4\right) a\,\hat{\mathbf{x}} + \frac{1}{4} b\,\hat{\mathbf{y}} + \left(\frac{1}{2} - z_4\right) c\,\hat{\mathbf{z}}$ | (4$c$) | O III |
| $\mathbf{B}_{17}$ | = | $x_5\,\mathbf{a}_1 + y_5\,\mathbf{a}_2 + z_5\,\mathbf{a}_3$ | = | $x_5 a\,\hat{\mathbf{x}} + y_5 b\,\hat{\mathbf{y}} + z_5 c\,\hat{\mathbf{z}}$ | (8$d$) | H I |
| $\mathbf{B}_{18}$ | = | $\left(\frac{1}{2} - x_5\right)\mathbf{a}_1 - y_5\,\mathbf{a}_2 + \left(\frac{1}{2} + z_5\right)\mathbf{a}_3$ | = | $\left(\frac{1}{2} - x_5\right) a\,\hat{\mathbf{x}} - y_5 b\,\hat{\mathbf{y}} + \left(\frac{1}{2} + z_5\right) c\,\hat{\mathbf{z}}$ | (8$d$) | H I |
| $\mathbf{B}_{19}$ | = | $-x_5\,\mathbf{a}_1 + \left(\frac{1}{2} + y_5\right)\mathbf{a}_2 - z_5\,\mathbf{a}_3$ | = | $-x_5 a\,\hat{\mathbf{x}} + \left(\frac{1}{2} + y_5\right) b\,\hat{\mathbf{y}} - z_5 c\,\hat{\mathbf{z}}$ | (8$d$) | H I |
| $\mathbf{B}_{20}$ | = | $\left(\frac{1}{2} + x_5\right)\mathbf{a}_1 + \left(\frac{1}{2} - y_5\right)\mathbf{a}_2 + \left(\frac{1}{2} - z_5\right)\mathbf{a}_3$ | = | $\left(\frac{1}{2} + x_5\right) a\,\hat{\mathbf{x}} + \left(\frac{1}{2} - y_5\right) b\,\hat{\mathbf{y}} + \left(\frac{1}{2} - z_5\right) c\,\hat{\mathbf{z}}$ | (8$d$) | H I |
| $\mathbf{B}_{21}$ | = | $-x_5\,\mathbf{a}_1 - y_5\,\mathbf{a}_2 - z_5\,\mathbf{a}_3$ | = | $-x_5 a\,\hat{\mathbf{x}} - y_5 b\,\hat{\mathbf{y}} - z_5 c\,\hat{\mathbf{z}}$ | (8$d$) | H I |
| $\mathbf{B}_{22}$ | = | $\left(\frac{1}{2} + x_5\right)\mathbf{a}_1 + y_5\,\mathbf{a}_2 + \left(\frac{1}{2} - z_5\right)\mathbf{a}_3$ | = | $\left(\frac{1}{2} + x_5\right) a\,\hat{\mathbf{x}} + y_5 b\,\hat{\mathbf{y}} + \left(\frac{1}{2} - z_5\right) c\,\hat{\mathbf{z}}$ | (8$d$) | H I |
| $\mathbf{B}_{23}$ | = | $x_5\,\mathbf{a}_1 + \left(\frac{1}{2} - y_5\right)\mathbf{a}_2 + z_5\,\mathbf{a}_3$ | = | $x_5 a\,\hat{\mathbf{x}} + \left(\frac{1}{2} - y_5\right) b\,\hat{\mathbf{y}} + z_5 c\,\hat{\mathbf{z}}$ | (8$d$) | H I |
| $\mathbf{B}_{24}$ | = | $\left(\frac{1}{2} - x_5\right)\mathbf{a}_1 + \left(\frac{1}{2} + y_5\right)\mathbf{a}_2 + \left(\frac{1}{2} + z_5\right)\mathbf{a}_3$ | = | $\left(\frac{1}{2} - x_5\right) a\,\hat{\mathbf{x}} + \left(\frac{1}{2} + y_5\right) b\,\hat{\mathbf{y}} + \left(\frac{1}{2} + z_5\right) c\,\hat{\mathbf{z}}$ | (8$d$) | H I |
| $\mathbf{B}_{25}$ | = | $x_6\,\mathbf{a}_1 + y_6\,\mathbf{a}_2 + z_6\,\mathbf{a}_3$ | = | $x_6 a\,\hat{\mathbf{x}} + y_6 b\,\hat{\mathbf{y}} + z_6 c\,\hat{\mathbf{z}}$ | (8$d$) | H II |
| $\mathbf{B}_{26}$ | = | $\left(\frac{1}{2} - x_6\right)\mathbf{a}_1 - y_6\,\mathbf{a}_2 + \left(\frac{1}{2} + z_6\right)\mathbf{a}_3$ | = | $\left(\frac{1}{2} - x_6\right) a\,\hat{\mathbf{x}} - y_6 b\,\hat{\mathbf{y}} + \left(\frac{1}{2} + z_6\right) c\,\hat{\mathbf{z}}$ | (8$d$) | H II |

$$\mathbf{B}_{27} = -x_6\,\mathbf{a}_1 + \left(\frac{1}{2} + y_6\right)\mathbf{a}_2 - z_6\,\mathbf{a}_3 = -x_6 a\,\hat{\mathbf{x}} + \left(\frac{1}{2} + y_6\right) b\,\hat{\mathbf{y}} - z_6 c\,\hat{\mathbf{z}} \quad (8d) \quad \text{H II}$$

$$\mathbf{B}_{28} = \left(\frac{1}{2} + x_6\right)\mathbf{a}_1 + \left(\frac{1}{2} - y_6\right)\mathbf{a}_2 + \left(\frac{1}{2} - z_6\right)\mathbf{a}_3 = \left(\frac{1}{2} + x_6\right) a\,\hat{\mathbf{x}} + \left(\frac{1}{2} - y_6\right) b\,\hat{\mathbf{y}} + \left(\frac{1}{2} - z_6\right) c\,\hat{\mathbf{z}} \quad (8d) \quad \text{H II}$$

$$\mathbf{B}_{29} = -x_6\,\mathbf{a}_1 - y_6\,\mathbf{a}_2 - z_6\,\mathbf{a}_3 = -x_6 a\,\hat{\mathbf{x}} - y_6 b\,\hat{\mathbf{y}} - z_6 c\,\hat{\mathbf{z}} \quad (8d) \quad \text{H II}$$

$$\mathbf{B}_{30} = \left(\frac{1}{2} + x_6\right)\mathbf{a}_1 + y_6\,\mathbf{a}_2 + \left(\frac{1}{2} - z_6\right)\mathbf{a}_3 = \left(\frac{1}{2} + x_6\right) a\,\hat{\mathbf{x}} + y_6 b\,\hat{\mathbf{y}} + \left(\frac{1}{2} - z_6\right) c\,\hat{\mathbf{z}} \quad (8d) \quad \text{H II}$$

$$\mathbf{B}_{31} = x_6\,\mathbf{a}_1 + \left(\frac{1}{2} - y_6\right)\mathbf{a}_2 + z_6\,\mathbf{a}_3 = x_6 a\,\hat{\mathbf{x}} + \left(\frac{1}{2} - y_6\right) b\,\hat{\mathbf{y}} + z_6 c\,\hat{\mathbf{z}} \quad (8d) \quad \text{H II}$$

$$\mathbf{B}_{32} = \left(\frac{1}{2} - x_6\right)\mathbf{a}_1 + \left(\frac{1}{2} + y_6\right)\mathbf{a}_2 + \left(\frac{1}{2} + z_6\right)\mathbf{a}_3 = \left(\frac{1}{2} - x_6\right) a\,\hat{\mathbf{x}} + \left(\frac{1}{2} + y_6\right) b\,\hat{\mathbf{y}} + \left(\frac{1}{2} + z_6\right) c\,\hat{\mathbf{z}} \quad (8d) \quad \text{H II}$$

$$\mathbf{B}_{33} = x_7\,\mathbf{a}_1 + y_7\,\mathbf{a}_2 + z_7\,\mathbf{a}_3 = x_7 a\,\hat{\mathbf{x}} + y_7 b\,\hat{\mathbf{y}} + z_7 c\,\hat{\mathbf{z}} \quad (8d) \quad \text{H III}$$

$$\mathbf{B}_{34} = \left(\frac{1}{2} - x_7\right)\mathbf{a}_1 - y_7\,\mathbf{a}_2 + \left(\frac{1}{2} + z_7\right)\mathbf{a}_3 = \left(\frac{1}{2} - x_7\right) a\,\hat{\mathbf{x}} - y_7 b\,\hat{\mathbf{y}} + \left(\frac{1}{2} + z_7\right) c\,\hat{\mathbf{z}} \quad (8d) \quad \text{H III}$$

$$\mathbf{B}_{35} = -x_7\,\mathbf{a}_1 + \left(\frac{1}{2} + y_7\right)\mathbf{a}_2 - z_7\,\mathbf{a}_3 = -x_7 a\,\hat{\mathbf{x}} + \left(\frac{1}{2} + y_7\right) b\,\hat{\mathbf{y}} - z_7 c\,\hat{\mathbf{z}} \quad (8d) \quad \text{H III}$$

$$\mathbf{B}_{36} = \left(\frac{1}{2} + x_7\right)\mathbf{a}_1 + \left(\frac{1}{2} - y_7\right)\mathbf{a}_2 + \left(\frac{1}{2} - z_7\right)\mathbf{a}_3 = \left(\frac{1}{2} + x_7\right) a\,\hat{\mathbf{x}} + \left(\frac{1}{2} - y_7\right) b\,\hat{\mathbf{y}} + \left(\frac{1}{2} - z_7\right) c\,\hat{\mathbf{z}} \quad (8d) \quad \text{H III}$$

$$\mathbf{B}_{37} = -x_7\,\mathbf{a}_1 - y_7\,\mathbf{a}_2 - z_7\,\mathbf{a}_3 = -x_7 a\,\hat{\mathbf{x}} - y_7 b\,\hat{\mathbf{y}} - z_7 c\,\hat{\mathbf{z}} \quad (8d) \quad \text{H III}$$

$$\mathbf{B}_{38} = \left(\frac{1}{2} + x_7\right)\mathbf{a}_1 + y_7\,\mathbf{a}_2 + \left(\frac{1}{2} - z_7\right)\mathbf{a}_3 = \left(\frac{1}{2} + x_7\right) a\,\hat{\mathbf{x}} + y_7 b\,\hat{\mathbf{y}} + \left(\frac{1}{2} - z_7\right) c\,\hat{\mathbf{z}} \quad (8d) \quad \text{H III}$$

$$\mathbf{B}_{39} = x_7\,\mathbf{a}_1 + \left(\frac{1}{2} - y_7\right)\mathbf{a}_2 + z_7\,\mathbf{a}_3 = x_7 a\,\hat{\mathbf{x}} + \left(\frac{1}{2} - y_7\right) b\,\hat{\mathbf{y}} + z_7 c\,\hat{\mathbf{z}} \quad (8d) \quad \text{H III}$$

$$\mathbf{B}_{40} = \left(\frac{1}{2} - x_7\right)\mathbf{a}_1 + \left(\frac{1}{2} + y_7\right)\mathbf{a}_2 + \left(\frac{1}{2} + z_7\right)\mathbf{a}_3 = \left(\frac{1}{2} - x_7\right) a\,\hat{\mathbf{x}} + \left(\frac{1}{2} + y_7\right) b\,\hat{\mathbf{y}} + \left(\frac{1}{2} + z_7\right) c\,\hat{\mathbf{z}} \quad (8d) \quad \text{H III}$$

$$\mathbf{B}_{41} = x_8\,\mathbf{a}_1 + y_8\,\mathbf{a}_2 + z_8\,\mathbf{a}_3 = x_8 a\,\hat{\mathbf{x}} + y_8 b\,\hat{\mathbf{y}} + z_8 c\,\hat{\mathbf{z}} \quad (8d) \quad \text{H IV}$$

$$\mathbf{B}_{42} = \left(\frac{1}{2} - x_8\right)\mathbf{a}_1 - y_8\,\mathbf{a}_2 + \left(\frac{1}{2} + z_8\right)\mathbf{a}_3 = \left(\frac{1}{2} - x_8\right) a\,\hat{\mathbf{x}} - y_8 b\,\hat{\mathbf{y}} + \left(\frac{1}{2} + z_8\right) c\,\hat{\mathbf{z}} \quad (8d) \quad \text{H IV}$$

$$\mathbf{B}_{43} = -x_8\,\mathbf{a}_1 + \left(\frac{1}{2} + y_8\right)\mathbf{a}_2 - z_8\,\mathbf{a}_3 = -x_8 a\,\hat{\mathbf{x}} + \left(\frac{1}{2} + y_8\right) b\,\hat{\mathbf{y}} - z_8 c\,\hat{\mathbf{z}} \quad (8d) \quad \text{H IV}$$

$$\mathbf{B}_{44} = \left(\frac{1}{2} + x_8\right)\mathbf{a}_1 + \left(\frac{1}{2} - y_8\right)\mathbf{a}_2 + \left(\frac{1}{2} - z_8\right)\mathbf{a}_3 = \left(\frac{1}{2} + x_8\right) a\,\hat{\mathbf{x}} + \left(\frac{1}{2} - y_8\right) b\,\hat{\mathbf{y}} + \left(\frac{1}{2} - z_8\right) c\,\hat{\mathbf{z}} \quad (8d) \quad \text{H IV}$$

$$\mathbf{B}_{45} = -x_8\,\mathbf{a}_1 - y_8\,\mathbf{a}_2 - z_8\,\mathbf{a}_3 = -x_8 a\,\hat{\mathbf{x}} - y_8 b\,\hat{\mathbf{y}} - z_8 c\,\hat{\mathbf{z}} \quad (8d) \quad \text{H IV}$$

$$\mathbf{B}_{46} = \left(\frac{1}{2} + x_8\right)\mathbf{a}_1 + y_8\,\mathbf{a}_2 + \left(\frac{1}{2} - z_8\right)\mathbf{a}_3 = \left(\frac{1}{2} + x_8\right) a\,\hat{\mathbf{x}} + y_8 b\,\hat{\mathbf{y}} + \left(\frac{1}{2} - z_8\right) c\,\hat{\mathbf{z}} \quad (8d) \quad \text{H IV}$$

$$\mathbf{B}_{47} = x_8\,\mathbf{a}_1 + \left(\frac{1}{2} - y_8\right)\mathbf{a}_2 + z_8\,\mathbf{a}_3 = x_8 a\,\hat{\mathbf{x}} + \left(\frac{1}{2} - y_8\right) b\,\hat{\mathbf{y}} + z_8 c\,\hat{\mathbf{z}} \quad (8d) \quad \text{H IV}$$

$$\mathbf{B}_{48} = \left(\frac{1}{2} - x_8\right)\mathbf{a}_1 + \left(\frac{1}{2} + y_8\right)\mathbf{a}_2 + \left(\frac{1}{2} + z_8\right)\mathbf{a}_3 = \left(\frac{1}{2} - x_8\right) a\,\hat{\mathbf{x}} + \left(\frac{1}{2} + y_8\right) b\,\hat{\mathbf{y}} + \left(\frac{1}{2} + z_8\right) c\,\hat{\mathbf{z}} \quad (8d) \quad \text{H IV}$$

$$\mathbf{B}_{49} = x_9\,\mathbf{a}_1 + y_9\,\mathbf{a}_2 + z_9\,\mathbf{a}_3 = x_9 a\,\hat{\mathbf{x}} + y_9 b\,\hat{\mathbf{y}} + z_9 c\,\hat{\mathbf{z}} \quad (8d) \quad \text{H V}$$

$$\mathbf{B}_{50} = \left(\frac{1}{2} - x_9\right)\mathbf{a}_1 - y_9\,\mathbf{a}_2 + \left(\frac{1}{2} + z_9\right)\mathbf{a}_3 = \left(\frac{1}{2} - x_9\right) a\,\hat{\mathbf{x}} - y_9 b\,\hat{\mathbf{y}} + \left(\frac{1}{2} + z_9\right) c\,\hat{\mathbf{z}} \quad (8d) \quad \text{H V}$$

$$\mathbf{B}_{51} = -x_9\,\mathbf{a}_1 + \left(\frac{1}{2} + y_9\right)\mathbf{a}_2 - z_9\,\mathbf{a}_3 = -x_9 a\,\hat{\mathbf{x}} + \left(\frac{1}{2} + y_9\right) b\,\hat{\mathbf{y}} - z_9 c\,\hat{\mathbf{z}} \quad (8d) \quad \text{H V}$$

$$\mathbf{B}_{52} = \left(\frac{1}{2} + x_9\right)\mathbf{a}_1 + \left(\frac{1}{2} - y_9\right)\mathbf{a}_2 + \left(\frac{1}{2} - z_9\right)\mathbf{a}_3 = \left(\frac{1}{2} + x_9\right) a\,\hat{\mathbf{x}} + \left(\frac{1}{2} - y_9\right) b\,\hat{\mathbf{y}} + \left(\frac{1}{2} - z_9\right) c\,\hat{\mathbf{z}} \quad (8d) \quad \text{H V}$$

$$\mathbf{B}_{53} = -x_9\,\mathbf{a}_1 - y_9\,\mathbf{a}_2 - z_9\,\mathbf{a}_3 = -x_9 a\,\hat{\mathbf{x}} - y_9 b\,\hat{\mathbf{y}} - z_9 c\,\hat{\mathbf{z}} \quad (8d) \quad \text{H V}$$

$$\mathbf{B}_{54} = \left(\frac{1}{2} + x_9\right)\mathbf{a}_1 + y_9\,\mathbf{a}_2 + \left(\frac{1}{2} - z_9\right)\mathbf{a}_3 = \left(\frac{1}{2} + x_9\right) a\,\hat{\mathbf{x}} + y_9 b\,\hat{\mathbf{y}} + \left(\frac{1}{2} - z_9\right) c\,\hat{\mathbf{z}} \quad (8d) \quad \text{H V}$$

$$\mathbf{B}_{55} = x_9\,\mathbf{a}_1 + \left(\frac{1}{2} - y_9\right)\mathbf{a}_2 + z_9\,\mathbf{a}_3 = x_9 a\,\hat{\mathbf{x}} + \left(\frac{1}{2} - y_9\right) b\,\hat{\mathbf{y}} + z_9 c\,\hat{\mathbf{z}} \quad (8d) \quad \text{H V}$$

$$\mathbf{B}_{56} = \left(\frac{1}{2} - x_9\right)\mathbf{a}_1 + \left(\frac{1}{2} + y_9\right)\mathbf{a}_2 + \left(\frac{1}{2} + z_9\right)\mathbf{a}_3 = \left(\frac{1}{2} - x_9\right) a\,\hat{\mathbf{x}} + \left(\frac{1}{2} + y_9\right) b\,\hat{\mathbf{y}} + \left(\frac{1}{2} + z_9\right) c\,\hat{\mathbf{z}} \quad (8d) \quad \text{H V}$$

| | | | | | | |
|---|---|---|---|---|---|---|
| $\mathbf{B}_{57}$ | = | $x_{10}\,\mathbf{a}_1 + y_{10}\,\mathbf{a}_2 + z_{10}\,\mathbf{a}_3$ | = | $x_{10}a\,\hat{\mathbf{x}} + y_{10}b\,\hat{\mathbf{y}} + z_{10}c\,\hat{\mathbf{z}}$ | (8*d*) | H VI |
| $\mathbf{B}_{58}$ | = | $\left(\frac{1}{2} - x_{10}\right)\mathbf{a}_1 - y_{10}\,\mathbf{a}_2 + \left(\frac{1}{2} + z_{10}\right)\mathbf{a}_3$ | = | $\left(\frac{1}{2} - x_{10}\right)a\,\hat{\mathbf{x}} - y_{10}b\,\hat{\mathbf{y}} + \left(\frac{1}{2} + z_{10}\right)c\,\hat{\mathbf{z}}$ | (8*d*) | H VI |
| $\mathbf{B}_{59}$ | = | $-x_{10}\,\mathbf{a}_1 + \left(\frac{1}{2} + y_{10}\right)\mathbf{a}_2 - z_{10}\,\mathbf{a}_3$ | = | $-x_{10}a\,\hat{\mathbf{x}} + \left(\frac{1}{2} + y_{10}\right)b\,\hat{\mathbf{y}} - z_{10}c\,\hat{\mathbf{z}}$ | (8*d*) | H VI |
| $\mathbf{B}_{60}$ | = | $\left(\frac{1}{2} + x_{10}\right)\mathbf{a}_1 + \left(\frac{1}{2} - y_{10}\right)\mathbf{a}_2 + \left(\frac{1}{2} - z_{10}\right)\mathbf{a}_3$ | = | $\left(\frac{1}{2} + x_{10}\right)a\,\hat{\mathbf{x}} + \left(\frac{1}{2} - y_{10}\right)b\,\hat{\mathbf{y}} + \left(\frac{1}{2} - z_{10}\right)c\,\hat{\mathbf{z}}$ | (8*d*) | H VI |
| $\mathbf{B}_{61}$ | = | $-x_{10}\,\mathbf{a}_1 - y_{10}\,\mathbf{a}_2 - z_{10}\,\mathbf{a}_3$ | = | $-x_{10}a\,\hat{\mathbf{x}} - y_{10}b\,\hat{\mathbf{y}} - z_{10}c\,\hat{\mathbf{z}}$ | (8*d*) | H VI |
| $\mathbf{B}_{62}$ | = | $\left(\frac{1}{2} + x_{10}\right)\mathbf{a}_1 + y_{10}\,\mathbf{a}_2 + \left(\frac{1}{2} - z_{10}\right)\mathbf{a}_3$ | = | $\left(\frac{1}{2} + x_{10}\right)a\,\hat{\mathbf{x}} + y_{10}b\,\hat{\mathbf{y}} + \left(\frac{1}{2} - z_{10}\right)c\,\hat{\mathbf{z}}$ | (8*d*) | H VI |
| $\mathbf{B}_{63}$ | = | $x_{10}\,\mathbf{a}_1 + \left(\frac{1}{2} - y_{10}\right)\mathbf{a}_2 + z_{10}\,\mathbf{a}_3$ | = | $x_{10}a\,\hat{\mathbf{x}} + \left(\frac{1}{2} - y_{10}\right)b\,\hat{\mathbf{y}} + z_{10}c\,\hat{\mathbf{z}}$ | (8*d*) | H VI |
| $\mathbf{B}_{64}$ | = | $\left(\frac{1}{2} - x_{10}\right)\mathbf{a}_1 + \left(\frac{1}{2} + y_{10}\right)\mathbf{a}_2 + \left(\frac{1}{2} + z_{10}\right)\mathbf{a}_3$ | = | $\left(\frac{1}{2} - x_{10}\right)a\,\hat{\mathbf{x}} + \left(\frac{1}{2} + y_{10}\right)b\,\hat{\mathbf{y}} + \left(\frac{1}{2} + z_{10}\right)c\,\hat{\mathbf{z}}$ | (8*d*) | H VI |
| $\mathbf{B}_{65}$ | = | $x_{11}\,\mathbf{a}_1 + y_{11}\,\mathbf{a}_2 + z_{11}\,\mathbf{a}_3$ | = | $x_{11}a\,\hat{\mathbf{x}} + y_{11}b\,\hat{\mathbf{y}} + z_{11}c\,\hat{\mathbf{z}}$ | (8*d*) | H VII |
| $\mathbf{B}_{66}$ | = | $\left(\frac{1}{2} - x_{11}\right)\mathbf{a}_1 - y_{11}\,\mathbf{a}_2 + \left(\frac{1}{2} + z_{11}\right)\mathbf{a}_3$ | = | $\left(\frac{1}{2} - x_{11}\right)a\,\hat{\mathbf{x}} - y_{11}b\,\hat{\mathbf{y}} + \left(\frac{1}{2} + z_{11}\right)c\,\hat{\mathbf{z}}$ | (8*d*) | H VII |
| $\mathbf{B}_{67}$ | = | $-x_{11}\,\mathbf{a}_1 + \left(\frac{1}{2} + y_{11}\right)\mathbf{a}_2 - z_{11}\,\mathbf{a}_3$ | = | $-x_{11}a\,\hat{\mathbf{x}} + \left(\frac{1}{2} + y_{11}\right)b\,\hat{\mathbf{y}} - z_{11}c\,\hat{\mathbf{z}}$ | (8*d*) | H VII |
| $\mathbf{B}_{68}$ | = | $\left(\frac{1}{2} + x_{11}\right)\mathbf{a}_1 + \left(\frac{1}{2} - y_{11}\right)\mathbf{a}_2 + \left(\frac{1}{2} - z_{11}\right)\mathbf{a}_3$ | = | $\left(\frac{1}{2} + x_{11}\right)a\,\hat{\mathbf{x}} + \left(\frac{1}{2} - y_{11}\right)b\,\hat{\mathbf{y}} + \left(\frac{1}{2} - z_{11}\right)c\,\hat{\mathbf{z}}$ | (8*d*) | H VII |
| $\mathbf{B}_{69}$ | = | $-x_{11}\,\mathbf{a}_1 - y_{11}\,\mathbf{a}_2 - z_{11}\,\mathbf{a}_3$ | = | $-x_{11}a\,\hat{\mathbf{x}} - y_{11}b\,\hat{\mathbf{y}} - z_{11}c\,\hat{\mathbf{z}}$ | (8*d*) | H VII |
| $\mathbf{B}_{70}$ | = | $\left(\frac{1}{2} + x_{11}\right)\mathbf{a}_1 + y_{11}\,\mathbf{a}_2 + \left(\frac{1}{2} - z_{11}\right)\mathbf{a}_3$ | = | $\left(\frac{1}{2} + x_{11}\right)a\,\hat{\mathbf{x}} + y_{11}b\,\hat{\mathbf{y}} + \left(\frac{1}{2} - z_{11}\right)c\,\hat{\mathbf{z}}$ | (8*d*) | H VII |
| $\mathbf{B}_{71}$ | = | $x_{11}\,\mathbf{a}_1 + \left(\frac{1}{2} - y_{11}\right)\mathbf{a}_2 + z_{11}\,\mathbf{a}_3$ | = | $x_{11}a\,\hat{\mathbf{x}} + \left(\frac{1}{2} - y_{11}\right)b\,\hat{\mathbf{y}} + z_{11}c\,\hat{\mathbf{z}}$ | (8*d*) | H VII |
| $\mathbf{B}_{72}$ | = | $\left(\frac{1}{2} - x_{11}\right)\mathbf{a}_1 + \left(\frac{1}{2} + y_{11}\right)\mathbf{a}_2 + \left(\frac{1}{2} + z_{11}\right)\mathbf{a}_3$ | = | $\left(\frac{1}{2} - x_{11}\right)a\,\hat{\mathbf{x}} + \left(\frac{1}{2} + y_{11}\right)b\,\hat{\mathbf{y}} + \left(\frac{1}{2} + z_{11}\right)c\,\hat{\mathbf{z}}$ | (8*d*) | H VII |
| $\mathbf{B}_{73}$ | = | $x_{12}\,\mathbf{a}_1 + y_{12}\,\mathbf{a}_2 + z_{12}\,\mathbf{a}_3$ | = | $x_{12}a\,\hat{\mathbf{x}} + y_{12}b\,\hat{\mathbf{y}} + z_{12}c\,\hat{\mathbf{z}}$ | (8*d*) | H VIII |
| $\mathbf{B}_{74}$ | = | $\left(\frac{1}{2} - x_{12}\right)\mathbf{a}_1 - y_{12}\,\mathbf{a}_2 + \left(\frac{1}{2} + z_{12}\right)\mathbf{a}_3$ | = | $\left(\frac{1}{2} - x_{12}\right)a\,\hat{\mathbf{x}} - y_{12}b\,\hat{\mathbf{y}} + \left(\frac{1}{2} + z_{12}\right)c\,\hat{\mathbf{z}}$ | (8*d*) | H VIII |
| $\mathbf{B}_{75}$ | = | $-x_{12}\,\mathbf{a}_1 + \left(\frac{1}{2} + y_{12}\right)\mathbf{a}_2 - z_{12}\,\mathbf{a}_3$ | = | $-x_{12}a\,\hat{\mathbf{x}} + \left(\frac{1}{2} + y_{12}\right)b\,\hat{\mathbf{y}} - z_{12}c\,\hat{\mathbf{z}}$ | (8*d*) | H VIII |
| $\mathbf{B}_{76}$ | = | $\left(\frac{1}{2} + x_{12}\right)\mathbf{a}_1 + \left(\frac{1}{2} - y_{12}\right)\mathbf{a}_2 + \left(\frac{1}{2} - z_{12}\right)\mathbf{a}_3$ | = | $\left(\frac{1}{2} + x_{12}\right)a\,\hat{\mathbf{x}} + \left(\frac{1}{2} - y_{12}\right)b\,\hat{\mathbf{y}} + \left(\frac{1}{2} - z_{12}\right)c\,\hat{\mathbf{z}}$ | (8*d*) | H VIII |
| $\mathbf{B}_{77}$ | = | $-x_{12}\,\mathbf{a}_1 - y_{12}\,\mathbf{a}_2 - z_{12}\,\mathbf{a}_3$ | = | $-x_{12}a\,\hat{\mathbf{x}} - y_{12}b\,\hat{\mathbf{y}} - z_{12}c\,\hat{\mathbf{z}}$ | (8*d*) | H VIII |
| $\mathbf{B}_{78}$ | = | $\left(\frac{1}{2} + x_{12}\right)\mathbf{a}_1 + y_{12}\,\mathbf{a}_2 + \left(\frac{1}{2} - z_{12}\right)\mathbf{a}_3$ | = | $\left(\frac{1}{2} + x_{12}\right)a\,\hat{\mathbf{x}} + y_{12}b\,\hat{\mathbf{y}} + \left(\frac{1}{2} - z_{12}\right)c\,\hat{\mathbf{z}}$ | (8*d*) | H VIII |
| $\mathbf{B}_{79}$ | = | $x_{12}\,\mathbf{a}_1 + \left(\frac{1}{2} - y_{12}\right)\mathbf{a}_2 + z_{12}\,\mathbf{a}_3$ | = | $x_{12}a\,\hat{\mathbf{x}} + \left(\frac{1}{2} - y_{12}\right)b\,\hat{\mathbf{y}} + z_{12}c\,\hat{\mathbf{z}}$ | (8*d*) | H VIII |
| $\mathbf{B}_{80}$ | = | $\left(\frac{1}{2} - x_{12}\right)\mathbf{a}_1 + \left(\frac{1}{2} + y_{12}\right)\mathbf{a}_2 + \left(\frac{1}{2} + z_{12}\right)\mathbf{a}_3$ | = | $\left(\frac{1}{2} - x_{12}\right)a\,\hat{\mathbf{x}} + \left(\frac{1}{2} + y_{12}\right)b\,\hat{\mathbf{y}} + \left(\frac{1}{2} + z_{12}\right)c\,\hat{\mathbf{z}}$ | (8*d*) | H VIII |
| $\mathbf{B}_{81}$ | = | $x_{13}\,\mathbf{a}_1 + y_{13}\,\mathbf{a}_2 + z_{13}\,\mathbf{a}_3$ | = | $x_{13}a\,\hat{\mathbf{x}} + y_{13}b\,\hat{\mathbf{y}} + z_{13}c\,\hat{\mathbf{z}}$ | (8*d*) | H IX |
| $\mathbf{B}_{82}$ | = | $\left(\frac{1}{2} - x_{13}\right)\mathbf{a}_1 - y_{13}\,\mathbf{a}_2 + \left(\frac{1}{2} + z_{13}\right)\mathbf{a}_3$ | = | $\left(\frac{1}{2} - x_{13}\right)a\,\hat{\mathbf{x}} - y_{13}b\,\hat{\mathbf{y}} + \left(\frac{1}{2} + z_{13}\right)c\,\hat{\mathbf{z}}$ | (8*d*) | H IX |
| $\mathbf{B}_{83}$ | = | $-x_{13}\,\mathbf{a}_1 + \left(\frac{1}{2} + y_{13}\right)\mathbf{a}_2 - z_{13}\,\mathbf{a}_3$ | = | $-x_{13}a\,\hat{\mathbf{x}} + \left(\frac{1}{2} + y_{13}\right)b\,\hat{\mathbf{y}} - z_{13}c\,\hat{\mathbf{z}}$ | (8*d*) | H IX |
| $\mathbf{B}_{84}$ | = | $\left(\frac{1}{2} + x_{13}\right)\mathbf{a}_1 + \left(\frac{1}{2} - y_{13}\right)\mathbf{a}_2 + \left(\frac{1}{2} - z_{13}\right)\mathbf{a}_3$ | = | $\left(\frac{1}{2} + x_{13}\right)a\,\hat{\mathbf{x}} + \left(\frac{1}{2} - y_{13}\right)b\,\hat{\mathbf{y}} + \left(\frac{1}{2} - z_{13}\right)c\,\hat{\mathbf{z}}$ | (8*d*) | H IX |
| $\mathbf{B}_{85}$ | = | $-x_{13}\,\mathbf{a}_1 - y_{13}\,\mathbf{a}_2 - z_{13}\,\mathbf{a}_3$ | = | $-x_{13}a\,\hat{\mathbf{x}} - y_{13}b\,\hat{\mathbf{y}} - z_{13}c\,\hat{\mathbf{z}}$ | (8*d*) | H IX |
| $\mathbf{B}_{86}$ | = | $\left(\frac{1}{2} + x_{13}\right)\mathbf{a}_1 + y_{13}\,\mathbf{a}_2 + \left(\frac{1}{2} - z_{13}\right)\mathbf{a}_3$ | = | $\left(\frac{1}{2} + x_{13}\right)a\,\hat{\mathbf{x}} + y_{13}b\,\hat{\mathbf{y}} + \left(\frac{1}{2} - z_{13}\right)c\,\hat{\mathbf{z}}$ | (8*d*) | H IX |
| $\mathbf{B}_{87}$ | = | $x_{13}\,\mathbf{a}_1 + \left(\frac{1}{2} - y_{13}\right)\mathbf{a}_2 + z_{13}\,\mathbf{a}_3$ | = | $x_{13}a\,\hat{\mathbf{x}} + \left(\frac{1}{2} - y_{13}\right)b\,\hat{\mathbf{y}} + z_{13}c\,\hat{\mathbf{z}}$ | (8*d*) | H IX |
| $\mathbf{B}_{88}$ | = | $\left(\frac{1}{2} - x_{13}\right) \mathbf{a}_1 + \left(\frac{1}{2} + y_{13}\right) \mathbf{a}_2 + \left(\frac{1}{2} + z_{13}\right) \mathbf{a}_3$ | = | $\left(\frac{1}{2} - x_{13}\right) a \, \hat{\mathbf{x}} + \left(\frac{1}{2} + y_{13}\right) b \, \hat{\mathbf{y}} + \left(\frac{1}{2} + z_{13}\right) c \, \hat{\mathbf{z}}$ | (8$d$) | H IX |
| $\mathbf{B}_{89}$ | = | $x_{14} \, \mathbf{a}_1 + y_{14} \, \mathbf{a}_2 + z_{14} \, \mathbf{a}_3$ | = | $x_{14} a \, \hat{\mathbf{x}} + y_{14} b \, \hat{\mathbf{y}} + z_{14} c \, \hat{\mathbf{z}}$ | (8$d$) | H X |
| $\mathbf{B}_{90}$ | = | $\left(\frac{1}{2} - x_{14}\right) \mathbf{a}_1 - y_{14} \, \mathbf{a}_2 + \left(\frac{1}{2} + z_{14}\right) \mathbf{a}_3$ | = | $\left(\frac{1}{2} - x_{14}\right) a \, \hat{\mathbf{x}} - y_{14} b \, \hat{\mathbf{y}} + \left(\frac{1}{2} + z_{14}\right) c \, \hat{\mathbf{z}}$ | (8$d$) | H X |
| $\mathbf{B}_{91}$ | = | $-x_{14} \, \mathbf{a}_1 + \left(\frac{1}{2} + y_{14}\right) \mathbf{a}_2 - z_{14} \, \mathbf{a}_3$ | = | $-x_{14} a \, \hat{\mathbf{x}} + \left(\frac{1}{2} + y_{14}\right) b \, \hat{\mathbf{y}} - z_{14} c \, \hat{\mathbf{z}}$ | (8$d$) | H X |
| $\mathbf{B}_{92}$ | = | $\left(\frac{1}{2} + x_{14}\right) \mathbf{a}_1 + \left(\frac{1}{2} - y_{14}\right) \mathbf{a}_2 + \left(\frac{1}{2} - z_{14}\right) \mathbf{a}_3$ | = | $\left(\frac{1}{2} + x_{14}\right) a \, \hat{\mathbf{x}} + \left(\frac{1}{2} - y_{14}\right) b \, \hat{\mathbf{y}} + \left(\frac{1}{2} - z_{14}\right) c \, \hat{\mathbf{z}}$ | (8$d$) | H X |
| $\mathbf{B}_{93}$ | = | $-x_{14} \, \mathbf{a}_1 - y_{14} \, \mathbf{a}_2 - z_{14} \, \mathbf{a}_3$ | = | $-x_{14} a \, \hat{\mathbf{x}} - y_{14} b \, \hat{\mathbf{y}} - z_{14} c \, \hat{\mathbf{z}}$ | (8$d$) | H X |
| $\mathbf{B}_{94}$ | = | $\left(\frac{1}{2} + x_{14}\right) \mathbf{a}_1 + y_{14} \, \mathbf{a}_2 + \left(\frac{1}{2} - z_{14}\right) \mathbf{a}_3$ | = | $\left(\frac{1}{2} + x_{14}\right) a \, \hat{\mathbf{x}} + y_{14} b \, \hat{\mathbf{y}} + \left(\frac{1}{2} - z_{14}\right) c \, \hat{\mathbf{z}}$ | (8$d$) | H X |
| $\mathbf{B}_{95}$ | = | $x_{14} \, \mathbf{a}_1 + \left(\frac{1}{2} - y_{14}\right) \mathbf{a}_2 + z_{14} \, \mathbf{a}_3$ | = | $x_{14} a \, \hat{\mathbf{x}} + \left(\frac{1}{2} - y_{14}\right) b \, \hat{\mathbf{y}} + z_{14} c \, \hat{\mathbf{z}}$ | (8$d$) | H X |
| $\mathbf{B}_{96}$ | = | $\left(\frac{1}{2} - x_{14}\right) \mathbf{a}_1 + \left(\frac{1}{2} + y_{14}\right) \mathbf{a}_2 + \left(\frac{1}{2} + z_{14}\right) \mathbf{a}_3$ | = | $\left(\frac{1}{2} - x_{14}\right) a \, \hat{\mathbf{x}} + \left(\frac{1}{2} + y_{14}\right) b \, \hat{\mathbf{y}} + \left(\frac{1}{2} + z_{14}\right) c \, \hat{\mathbf{z}}$ | (8$d$) | H X |
| $\mathbf{B}_{97}$ | = | $x_{15} \, \mathbf{a}_1 + y_{15} \, \mathbf{a}_2 + z_{15} \, \mathbf{a}_3$ | = | $x_{15} a \, \hat{\mathbf{x}} + y_{15} b \, \hat{\mathbf{y}} + z_{15} c \, \hat{\mathbf{z}}$ | (8$d$) | H XI |
| $\mathbf{B}_{98}$ | = | $\left(\frac{1}{2} - x_{15}\right) \mathbf{a}_1 - y_{15} \, \mathbf{a}_2 + \left(\frac{1}{2} + z_{15}\right) \mathbf{a}_3$ | = | $\left(\frac{1}{2} - x_{15}\right) a \, \hat{\mathbf{x}} - y_{15} b \, \hat{\mathbf{y}} + \left(\frac{1}{2} + z_{15}\right) c \, \hat{\mathbf{z}}$ | (8$d$) | H XI |
| $\mathbf{B}_{99}$ | = | $-x_{15} \, \mathbf{a}_1 + \left(\frac{1}{2} + y_{15}\right) \mathbf{a}_2 - z_{15} \, \mathbf{a}_3$ | = | $-x_{15} a \, \hat{\mathbf{x}} + \left(\frac{1}{2} + y_{15}\right) b \, \hat{\mathbf{y}} - z_{15} c \, \hat{\mathbf{z}}$ | (8$d$) | H XI |
| $\mathbf{B}_{100}$ | = | $\left(\frac{1}{2} + x_{15}\right) \mathbf{a}_1 + \left(\frac{1}{2} - y_{15}\right) \mathbf{a}_2 + \left(\frac{1}{2} - z_{15}\right) \mathbf{a}_3$ | = | $\left(\frac{1}{2} + x_{15}\right) a \, \hat{\mathbf{x}} + \left(\frac{1}{2} - y_{15}\right) b \, \hat{\mathbf{y}} + \left(\frac{1}{2} - z_{15}\right) c \, \hat{\mathbf{z}}$ | (8$d$) | H XI |
| $\mathbf{B}_{101}$ | = | $-x_{15} \, \mathbf{a}_1 - y_{15} \, \mathbf{a}_2 - z_{15} \, \mathbf{a}_3$ | = | $-x_{15} a \, \hat{\mathbf{x}} - y_{15} b \, \hat{\mathbf{y}} - z_{15} c \, \hat{\mathbf{z}}$ | (8$d$) | H XI |
| $\mathbf{B}_{102}$ | = | $\left(\frac{1}{2} + x_{15}\right) \mathbf{a}_1 + y_{15} \, \mathbf{a}_2 + \left(\frac{1}{2} - z_{15}\right) \mathbf{a}_3$ | = | $\left(\frac{1}{2} + x_{15}\right) a \, \hat{\mathbf{x}} + y_{15} b \, \hat{\mathbf{y}} + \left(\frac{1}{2} - z_{15}\right) c \, \hat{\mathbf{z}}$ | (8$d$) | H XI |
| $\mathbf{B}_{103}$ | = | $x_{15} \, \mathbf{a}_1 + \left(\frac{1}{2} - y_{15}\right) \mathbf{a}_2 + z_{15} \, \mathbf{a}_3$ | = | $x_{15} a \, \hat{\mathbf{x}} + \left(\frac{1}{2} - y_{15}\right) b \, \hat{\mathbf{y}} + z_{15} c \, \hat{\mathbf{z}}$ | (8$d$) | H XI |
| $\mathbf{B}_{104}$ | = | $\left(\frac{1}{2} - x_{15}\right) \mathbf{a}_1 + \left(\frac{1}{2} + y_{15}\right) \mathbf{a}_2 + \left(\frac{1}{2} + z_{15}\right) \mathbf{a}_3$ | = | $\left(\frac{1}{2} - x_{15}\right) a \, \hat{\mathbf{x}} + \left(\frac{1}{2} + y_{15}\right) b \, \hat{\mathbf{y}} + \left(\frac{1}{2} + z_{15}\right) c \, \hat{\mathbf{z}}$ | (8$d$) | H XI |
| $\mathbf{B}_{105}$ | = | $x_{16} \, \mathbf{a}_1 + y_{16} \, \mathbf{a}_2 + z_{16} \, \mathbf{a}_3$ | = | $x_{16} a \, \hat{\mathbf{x}} + y_{16} b \, \hat{\mathbf{y}} + z_{16} c \, \hat{\mathbf{z}}$ | (8$d$) | O IV |
| $\mathbf{B}_{106}$ | = | $\left(\frac{1}{2} - x_{16}\right) \mathbf{a}_1 - y_{16} \, \mathbf{a}_2 + \left(\frac{1}{2} + z_{16}\right) \mathbf{a}_3$ | = | $\left(\frac{1}{2} - x_{16}\right) a \, \hat{\mathbf{x}} - y_{16} b \, \hat{\mathbf{y}} + \left(\frac{1}{2} + z_{16}\right) c \, \hat{\mathbf{z}}$ | (8$d$) | O IV |
| $\mathbf{B}_{107}$ | = | $-x_{16} \, \mathbf{a}_1 + \left(\frac{1}{2} + y_{16}\right) \mathbf{a}_2 - z_{16} \, \mathbf{a}_3$ | = | $-x_{16} a \, \hat{\mathbf{x}} + \left(\frac{1}{2} + y_{16}\right) b \, \hat{\mathbf{y}} - z_{16} c \, \hat{\mathbf{z}}$ | (8$d$) | O IV |
| $\mathbf{B}_{108}$ | = | $\left(\frac{1}{2} + x_{16}\right) \mathbf{a}_1 + \left(\frac{1}{2} - y_{16}\right) \mathbf{a}_2 + \left(\frac{1}{2} - z_{16}\right) \mathbf{a}_3$ | = | $\left(\frac{1}{2} + x_{16}\right) a \, \hat{\mathbf{x}} + \left(\frac{1}{2} - y_{16}\right) b \, \hat{\mathbf{y}} + \left(\frac{1}{2} - z_{16}\right) c \, \hat{\mathbf{z}}$ | (8$d$) | O IV |
| $\mathbf{B}_{109}$ | = | $-x_{16} \, \mathbf{a}_1 - y_{16} \, \mathbf{a}_2 - z_{16} \, \mathbf{a}_3$ | = | $-x_{16} a \, \hat{\mathbf{x}} - y_{16} b \, \hat{\mathbf{y}} - z_{16} c \, \hat{\mathbf{z}}$ | (8$d$) | O IV |
| $\mathbf{B}_{110}$ | = | $\left(\frac{1}{2} + x_{16}\right) \mathbf{a}_1 + y_{16} \, \mathbf{a}_2 + \left(\frac{1}{2} - z_{16}\right) \mathbf{a}_3$ | = | $\left(\frac{1}{2} + x_{16}\right) a \, \hat{\mathbf{x}} + y_{16} b \, \hat{\mathbf{y}} + \left(\frac{1}{2} - z_{16}\right) c \, \hat{\mathbf{z}}$ | (8$d$) | O IV |
| $\mathbf{B}_{111}$ | = | $x_{16} \, \mathbf{a}_1 + \left(\frac{1}{2} - y_{16}\right) \mathbf{a}_2 + z_{16} \, \mathbf{a}_3$ | = | $x_{16} a \, \hat{\mathbf{x}} + \left(\frac{1}{2} - y_{16}\right) b \, \hat{\mathbf{y}} + z_{16} c \, \hat{\mathbf{z}}$ | (8$d$) | O IV |
| $\mathbf{B}_{112}$ | = | $\left(\frac{1}{2} - x_{16}\right) \mathbf{a}_1 + \left(\frac{1}{2} + y_{16}\right) \mathbf{a}_2 + \left(\frac{1}{2} + z_{16}\right) \mathbf{a}_3$ | = | $\left(\frac{1}{2} - x_{16}\right) a \, \hat{\mathbf{x}} + \left(\frac{1}{2} + y_{16}\right) b \, \hat{\mathbf{y}} + \left(\frac{1}{2} + z_{16}\right) c \, \hat{\mathbf{z}}$ | (8$d$) | O IV |
| $\mathbf{B}_{113}$ | = | $x_{17} \, \mathbf{a}_1 + y_{17} \, \mathbf{a}_2 + z_{17} \, \mathbf{a}_3$ | = | $x_{17} a \, \hat{\mathbf{x}} + y_{17} b \, \hat{\mathbf{y}} + z_{17} c \, \hat{\mathbf{z}}$ | (8$d$) | O V |
| $\mathbf{B}_{114}$ | = | $\left(\frac{1}{2} - x_{17}\right) \mathbf{a}_1 - y_{17} \, \mathbf{a}_2 + \left(\frac{1}{2} + z_{17}\right) \mathbf{a}_3$ | = | $\left(\frac{1}{2} - x_{17}\right) a \, \hat{\mathbf{x}} - y_{17} b \, \hat{\mathbf{y}} + \left(\frac{1}{2} + z_{17}\right) c \, \hat{\mathbf{z}}$ | (8$d$) | O V |
| $\mathbf{B}_{115}$ | = | $-x_{17} \, \mathbf{a}_1 + \left(\frac{1}{2} + y_{17}\right) \mathbf{a}_2 - z_{17} \, \mathbf{a}_3$ | = | $-x_{17} a \, \hat{\mathbf{x}} + \left(\frac{1}{2} + y_{17}\right) b \, \hat{\mathbf{y}} - z_{17} c \, \hat{\mathbf{z}}$ | (8$d$) | O V |
| $\mathbf{B}_{116}$ | = | $\left(\frac{1}{2} + x_{17}\right) \mathbf{a}_1 + \left(\frac{1}{2} - y_{17}\right) \mathbf{a}_2 + \left(\frac{1}{2} - z_{17}\right) \mathbf{a}_3$ | = | $\left(\frac{1}{2} + x_{17}\right) a \, \hat{\mathbf{x}} + \left(\frac{1}{2} - y_{17}\right) b \, \hat{\mathbf{y}} + \left(\frac{1}{2} - z_{17}\right) c \, \hat{\mathbf{z}}$ | (8$d$) | O V |
| $\mathbf{B}_{117}$ | = | $-x_{17} \, \mathbf{a}_1 - y_{17} \, \mathbf{a}_2 - z_{17} \, \mathbf{a}_3$ | = | $-x_{17} a \, \hat{\mathbf{x}} - y_{17} b \, \hat{\mathbf{y}} - z_{17} c \, \hat{\mathbf{z}}$ | (8$d$) | O V |
| $\mathbf{B}_{118}$ | = | $\left(\frac{1}{2}+x_{17}\right)\mathbf{a}_1+y_{17}\,\mathbf{a}_2+\left(\frac{1}{2}-z_{17}\right)\mathbf{a}_3$ | = | $\left(\frac{1}{2}+x_{17}\right)a\,\hat{\mathbf{x}}+y_{17}b\,\hat{\mathbf{y}}+\left(\frac{1}{2}-z_{17}\right)c\,\hat{\mathbf{z}}$ | (8*d*) | O V |
| $\mathbf{B}_{119}$ | = | $x_{17}\,\mathbf{a}_1+\left(\frac{1}{2}-y_{17}\right)\mathbf{a}_2+z_{17}\,\mathbf{a}_3$ | = | $x_{17}a\,\hat{\mathbf{x}}+\left(\frac{1}{2}-y_{17}\right)b\,\hat{\mathbf{y}}+z_{17}c\,\hat{\mathbf{z}}$ | (8*d*) | O V |
| $\mathbf{B}_{120}$ | = | $\left(\frac{1}{2}-x_{17}\right)\mathbf{a}_1+\left(\frac{1}{2}+y_{17}\right)\mathbf{a}_2+\left(\frac{1}{2}+z_{17}\right)\mathbf{a}_3$ | = | $\left(\frac{1}{2}-x_{17}\right)a\,\hat{\mathbf{x}}+\left(\frac{1}{2}+y_{17}\right)b\,\hat{\mathbf{y}}+\left(\frac{1}{2}+z_{17}\right)c\,\hat{\mathbf{z}}$ | (8*d*) | O V |
| $\mathbf{B}_{121}$ | = | $x_{18}\,\mathbf{a}_1+y_{18}\,\mathbf{a}_2+z_{18}\,\mathbf{a}_3$ | = | $x_{18}a\,\hat{\mathbf{x}}+y_{18}b\,\hat{\mathbf{y}}+z_{18}c\,\hat{\mathbf{z}}$ | (8*d*) | O VI |
| $\mathbf{B}_{122}$ | = | $\left(\frac{1}{2}-x_{18}\right)\mathbf{a}_1-y_{18}\,\mathbf{a}_2+\left(\frac{1}{2}+z_{18}\right)\mathbf{a}_3$ | = | $\left(\frac{1}{2}-x_{18}\right)a\,\hat{\mathbf{x}}-y_{18}b\,\hat{\mathbf{y}}+\left(\frac{1}{2}+z_{18}\right)c\,\hat{\mathbf{z}}$ | (8*d*) | O VI |
| $\mathbf{B}_{123}$ | = | $-x_{18}\,\mathbf{a}_1+\left(\frac{1}{2}+y_{18}\right)\mathbf{a}_2-z_{18}\,\mathbf{a}_3$ | = | $-x_{18}a\,\hat{\mathbf{x}}+\left(\frac{1}{2}+y_{18}\right)b\,\hat{\mathbf{y}}-z_{18}c\,\hat{\mathbf{z}}$ | (8*d*) | O VI |
| $\mathbf{B}_{124}$ | = | $\left(\frac{1}{2}+x_{18}\right)\mathbf{a}_1+\left(\frac{1}{2}-y_{18}\right)\mathbf{a}_2+\left(\frac{1}{2}-z_{18}\right)\mathbf{a}_3$ | = | $\left(\frac{1}{2}+x_{18}\right)a\,\hat{\mathbf{x}}+\left(\frac{1}{2}-y_{18}\right)b\,\hat{\mathbf{y}}+\left(\frac{1}{2}-z_{18}\right)c\,\hat{\mathbf{z}}$ | (8*d*) | O VI |
| $\mathbf{B}_{125}$ | = | $-x_{18}\,\mathbf{a}_1-y_{18}\,\mathbf{a}_2-z_{18}\,\mathbf{a}_3$ | = | $-x_{18}a\,\hat{\mathbf{x}}-y_{18}b\,\hat{\mathbf{y}}-z_{18}c\,\hat{\mathbf{z}}$ | (8*d*) | O VI |
| $\mathbf{B}_{126}$ | = | $\left(\frac{1}{2}+x_{18}\right)\mathbf{a}_1+y_{18}\,\mathbf{a}_2+\left(\frac{1}{2}-z_{18}\right)\mathbf{a}_3$ | = | $\left(\frac{1}{2}+x_{18}\right)a\,\hat{\mathbf{x}}+y_{18}b\,\hat{\mathbf{y}}+\left(\frac{1}{2}-z_{18}\right)c\,\hat{\mathbf{z}}$ | (8*d*) | O VI |
| $\mathbf{B}_{127}$ | = | $x_{18}\,\mathbf{a}_1+\left(\frac{1}{2}-y_{18}\right)\mathbf{a}_2+z_{18}\,\mathbf{a}_3$ | = | $x_{18}a\,\hat{\mathbf{x}}+\left(\frac{1}{2}-y_{18}\right)b\,\hat{\mathbf{y}}+z_{18}c\,\hat{\mathbf{z}}$ | (8*d*) | O VI |
| $\mathbf{B}_{128}$ | = | $\left(\frac{1}{2}-x_{18}\right)\mathbf{a}_1+\left(\frac{1}{2}+y_{18}\right)\mathbf{a}_2+\left(\frac{1}{2}+z_{18}\right)\mathbf{a}_3$ | = | $\left(\frac{1}{2}-x_{18}\right)a\,\hat{\mathbf{x}}+\left(\frac{1}{2}+y_{18}\right)b\,\hat{\mathbf{y}}+\left(\frac{1}{2}+z_{18}\right)c\,\hat{\mathbf{z}}$ | (8*d*) | O VI |
| $\mathbf{B}_{129}$ | = | $x_{19}\,\mathbf{a}_1+y_{19}\,\mathbf{a}_2+z_{19}\,\mathbf{a}_3$ | = | $x_{19}a\,\hat{\mathbf{x}}+y_{19}b\,\hat{\mathbf{y}}+z_{19}c\,\hat{\mathbf{z}}$ | (8*d*) | O VII |
| $\mathbf{B}_{130}$ | = | $\left(\frac{1}{2}-x_{19}\right)\mathbf{a}_1-y_{19}\,\mathbf{a}_2+\left(\frac{1}{2}+z_{19}\right)\mathbf{a}_3$ | = | $\left(\frac{1}{2}-x_{19}\right)a\,\hat{\mathbf{x}}-y_{19}b\,\hat{\mathbf{y}}+\left(\frac{1}{2}+z_{19}\right)c\,\hat{\mathbf{z}}$ | (8*d*) | O VII |
| $\mathbf{B}_{131}$ | = | $-x_{19}\,\mathbf{a}_1+\left(\frac{1}{2}+y_{19}\right)\mathbf{a}_2-z_{19}\,\mathbf{a}_3$ | = | $-x_{19}a\,\hat{\mathbf{x}}+\left(\frac{1}{2}+y_{19}\right)b\,\hat{\mathbf{y}}-z_{19}c\,\hat{\mathbf{z}}$ | (8*d*) | O VII |
| $\mathbf{B}_{132}$ | = | $\left(\frac{1}{2}+x_{19}\right)\mathbf{a}_1+\left(\frac{1}{2}-y_{19}\right)\mathbf{a}_2+\left(\frac{1}{2}-z_{19}\right)\mathbf{a}_3$ | = | $\left(\frac{1}{2}+x_{19}\right)a\,\hat{\mathbf{x}}+\left(\frac{1}{2}-y_{19}\right)b\,\hat{\mathbf{y}}+\left(\frac{1}{2}-z_{19}\right)c\,\hat{\mathbf{z}}$ | (8*d*) | O VII |
| $\mathbf{B}_{133}$ | = | $-x_{19}\,\mathbf{a}_1-y_{19}\,\mathbf{a}_2-z_{19}\,\mathbf{a}_3$ | = | $-x_{19}a\,\hat{\mathbf{x}}-y_{19}b\,\hat{\mathbf{y}}-z_{19}c\,\hat{\mathbf{z}}$ | (8*d*) | O VII |
| $\mathbf{B}_{134}$ | = | $\left(\frac{1}{2}+x_{19}\right)\mathbf{a}_1+y_{19}\,\mathbf{a}_2+\left(\frac{1}{2}-z_{19}\right)\mathbf{a}_3$ | = | $\left(\frac{1}{2}+x_{19}\right)a\,\hat{\mathbf{x}}+y_{19}b\,\hat{\mathbf{y}}+\left(\frac{1}{2}-z_{19}\right)c\,\hat{\mathbf{z}}$ | (8*d*) | O VII |
| $\mathbf{B}_{135}$ | = | $x_{19}\,\mathbf{a}_1+\left(\frac{1}{2}-y_{19}\right)\mathbf{a}_2+z_{19}\,\mathbf{a}_3$ | = | $x_{19}a\,\hat{\mathbf{x}}+\left(\frac{1}{2}-y_{19}\right)b\,\hat{\mathbf{y}}+z_{19}c\,\hat{\mathbf{z}}$ | (8*d*) | O VII |
| $\mathbf{B}_{136}$ | = | $\left(\frac{1}{2}-x_{19}\right)\mathbf{a}_1+\left(\frac{1}{2}+y_{19}\right)\mathbf{a}_2+\left(\frac{1}{2}+z_{19}\right)\mathbf{a}_3$ | = | $\left(\frac{1}{2}-x_{19}\right)a\,\hat{\mathbf{x}}+\left(\frac{1}{2}+y_{19}\right)b\,\hat{\mathbf{y}}+\left(\frac{1}{2}+z_{19}\right)c\,\hat{\mathbf{z}}$ | (8*d*) | O VII |
| $\mathbf{B}_{137}$ | = | $x_{20}\,\mathbf{a}_1+y_{20}\,\mathbf{a}_2+z_{20}\,\mathbf{a}_3$ | = | $x_{20}a\,\hat{\mathbf{x}}+y_{20}b\,\hat{\mathbf{y}}+z_{20}c\,\hat{\mathbf{z}}$ | (8*d*) | O VIII |
| $\mathbf{B}_{138}$ | = | $\left(\frac{1}{2}-x_{20}\right)\mathbf{a}_1-y_{20}\,\mathbf{a}_2+\left(\frac{1}{2}+z_{20}\right)\mathbf{a}_3$ | = | $\left(\frac{1}{2}-x_{20}\right)a\,\hat{\mathbf{x}}-y_{20}b\,\hat{\mathbf{y}}+\left(\frac{1}{2}+z_{20}\right)c\,\hat{\mathbf{z}}$ | (8*d*) | O VIII |
| $\mathbf{B}_{139}$ | = | $-x_{20}\,\mathbf{a}_1+\left(\frac{1}{2}+y_{20}\right)\mathbf{a}_2-z_{20}\,\mathbf{a}_3$ | = | $-x_{20}a\,\hat{\mathbf{x}}+\left(\frac{1}{2}+y_{20}\right)b\,\hat{\mathbf{y}}-z_{20}c\,\hat{\mathbf{z}}$ | (8*d*) | O VIII |
| $\mathbf{B}_{140}$ | = | $\left(\frac{1}{2}+x_{20}\right)\mathbf{a}_1+\left(\frac{1}{2}-y_{20}\right)\mathbf{a}_2+\left(\frac{1}{2}-z_{20}\right)\mathbf{a}_3$ | = | $\left(\frac{1}{2}+x_{20}\right)a\,\hat{\mathbf{x}}+\left(\frac{1}{2}-y_{20}\right)b\,\hat{\mathbf{y}}+\left(\frac{1}{2}-z_{20}\right)c\,\hat{\mathbf{z}}$ | (8*d*) | O VIII |
| $\mathbf{B}_{141}$ | = | $-x_{20}\,\mathbf{a}_1-y_{20}\,\mathbf{a}_2-z_{20}\,\mathbf{a}_3$ | = | $-x_{20}a\,\hat{\mathbf{x}}-y_{20}b\,\hat{\mathbf{y}}-z_{20}c\,\hat{\mathbf{z}}$ | (8*d*) | O VIII |
| $\mathbf{B}_{142}$ | = | $\left(\frac{1}{2}+x_{20}\right)\mathbf{a}_1+y_{20}\,\mathbf{a}_2+\left(\frac{1}{2}-z_{20}\right)\mathbf{a}_3$ | = | $\left(\frac{1}{2}+x_{20}\right)a\,\hat{\mathbf{x}}+y_{20}b\,\hat{\mathbf{y}}+\left(\frac{1}{2}-z_{20}\right)c\,\hat{\mathbf{z}}$ | (8*d*) | O VIII |
| $\mathbf{B}_{143}$ | = | $x_{20}\,\mathbf{a}_1+\left(\frac{1}{2}-y_{20}\right)\mathbf{a}_2+z_{20}\,\mathbf{a}_3$ | = | $x_{20}a\,\hat{\mathbf{x}}+\left(\frac{1}{2}-y_{20}\right)b\,\hat{\mathbf{y}}+z_{20}c\,\hat{\mathbf{z}}$ | (8*d*) | O VIII |
| $\mathbf{B}_{144}$ | = | $\left(\frac{1}{2}-x_{20}\right)\mathbf{a}_1+\left(\frac{1}{2}+y_{20}\right)\mathbf{a}_2+\left(\frac{1}{2}+z_{20}\right)\mathbf{a}_3$ | = | $\left(\frac{1}{2}-x_{20}\right)a\,\hat{\mathbf{x}}+\left(\frac{1}{2}+y_{20}\right)b\,\hat{\mathbf{y}}+\left(\frac{1}{2}+z_{20}\right)c\,\hat{\mathbf{z}}$ | (8*d*) | O VIII |
| $\mathbf{B}_{145}$ | = | $x_{21}\,\mathbf{a}_1+y_{21}\,\mathbf{a}_2+z_{21}\,\mathbf{a}_3$ | = | $x_{21}a\,\hat{\mathbf{x}}+y_{21}b\,\hat{\mathbf{y}}+z_{21}c\,\hat{\mathbf{z}}$ | (8*d*) | O IX |
| $\mathbf{B}_{146}$ | = | $\left(\frac{1}{2}-x_{21}\right)\mathbf{a}_1-y_{21}\,\mathbf{a}_2+\left(\frac{1}{2}+z_{21}\right)\mathbf{a}_3$ | = | $\left(\frac{1}{2}-x_{21}\right)a\,\hat{\mathbf{x}}-y_{21}b\,\hat{\mathbf{y}}+\left(\frac{1}{2}+z_{21}\right)c\,\hat{\mathbf{z}}$ | (8*d*) | O IX |
| $\mathbf{B}_{147}$ | = | $-x_{21}\,\mathbf{a}_1+\left(\frac{1}{2}+y_{21}\right)\mathbf{a}_2-z_{21}\,\mathbf{a}_3$ | = | $-x_{21}a\,\hat{\mathbf{x}}+\left(\frac{1}{2}+y_{21}\right)b\,\hat{\mathbf{y}}-z_{21}c\,\hat{\mathbf{z}}$ | (8*d*) | O IX |
| $\mathbf{B}_{148}$ | = | $\left(\frac{1}{2}+x_{21}\right)\mathbf{a}_1+\left(\frac{1}{2}-y_{21}\right)\mathbf{a}_2+\left(\frac{1}{2}-z_{21}\right)\mathbf{a}_3$ | = | $\left(\frac{1}{2}+x_{21}\right)a\,\mathbf{\hat{x}}+\left(\frac{1}{2}-y_{21}\right)b\,\mathbf{\hat{y}}+\left(\frac{1}{2}-z_{21}\right)c\,\mathbf{\hat{z}}$ | (8$d$) | O IX |
| $\mathbf{B}_{149}$ | = | $-x_{21}\,\mathbf{a}_1-y_{21}\,\mathbf{a}_2-z_{21}\,\mathbf{a}_3$ | = | $-x_{21}a\,\mathbf{\hat{x}}-y_{21}b\,\mathbf{\hat{y}}-z_{21}c\,\mathbf{\hat{z}}$ | (8$d$) | O IX |
| $\mathbf{B}_{150}$ | = | $\left(\frac{1}{2}+x_{21}\right)\mathbf{a}_1+y_{21}\,\mathbf{a}_2+\left(\frac{1}{2}-z_{21}\right)\mathbf{a}_3$ | = | $\left(\frac{1}{2}+x_{21}\right)a\,\mathbf{\hat{x}}+y_{21}b\,\mathbf{\hat{y}}+\left(\frac{1}{2}-z_{21}\right)c\,\mathbf{\hat{z}}$ | (8$d$) | O IX |
| $\mathbf{B}_{151}$ | = | $x_{21}\,\mathbf{a}_1+\left(\frac{1}{2}-y_{21}\right)\mathbf{a}_2+z_{21}\,\mathbf{a}_3$ | = | $x_{21}a\,\mathbf{\hat{x}}+\left(\frac{1}{2}-y_{21}\right)b\,\mathbf{\hat{y}}+z_{21}c\,\mathbf{\hat{z}}$ | (8$d$) | O IX |
| $\mathbf{B}_{152}$ | = | $\left(\frac{1}{2}-x_{21}\right)\mathbf{a}_1+\left(\frac{1}{2}+y_{21}\right)\mathbf{a}_2+\left(\frac{1}{2}+z_{21}\right)\mathbf{a}_3$ | = | $\left(\frac{1}{2}-x_{21}\right)a\,\mathbf{\hat{x}}+\left(\frac{1}{2}+y_{21}\right)b\,\mathbf{\hat{y}}+\left(\frac{1}{2}+z_{21}\right)c\,\mathbf{\hat{z}}$ | (8$d$) | O IX |
| $\mathbf{B}_{153}$ | = | $x_{22}\,\mathbf{a}_1+y_{22}\,\mathbf{a}_2+z_{22}\,\mathbf{a}_3$ | = | $x_{22}a\,\mathbf{\hat{x}}+y_{22}b\,\mathbf{\hat{y}}+z_{22}c\,\mathbf{\hat{z}}$ | (8$d$) | O X |
| $\mathbf{B}_{154}$ | = | $\left(\frac{1}{2}-x_{22}\right)\mathbf{a}_1-y_{22}\,\mathbf{a}_2+\left(\frac{1}{2}+z_{22}\right)\mathbf{a}_3$ | = | $\left(\frac{1}{2}-x_{22}\right)a\,\mathbf{\hat{x}}-y_{22}b\,\mathbf{\hat{y}}+\left(\frac{1}{2}+z_{22}\right)c\,\mathbf{\hat{z}}$ | (8$d$) | O X |
| $\mathbf{B}_{155}$ | = | $-x_{22}\,\mathbf{a}_1+\left(\frac{1}{2}+y_{22}\right)\mathbf{a}_2-z_{22}\,\mathbf{a}_3$ | = | $-x_{22}a\,\mathbf{\hat{x}}+\left(\frac{1}{2}+y_{22}\right)b\,\mathbf{\hat{y}}-z_{22}c\,\mathbf{\hat{z}}$ | (8$d$) | O X |
| $\mathbf{B}_{156}$ | = | $\left(\frac{1}{2}+x_{22}\right)\mathbf{a}_1+\left(\frac{1}{2}-y_{22}\right)\mathbf{a}_2+\left(\frac{1}{2}-z_{22}\right)\mathbf{a}_3$ | = | $\left(\frac{1}{2}+x_{22}\right)a\,\mathbf{\hat{x}}+\left(\frac{1}{2}-y_{22}\right)b\,\mathbf{\hat{y}}+\left(\frac{1}{2}-z_{22}\right)c\,\mathbf{\hat{z}}$ | (8$d$) | O X |
| $\mathbf{B}_{157}$ | = | $-x_{22}\,\mathbf{a}_1-y_{22}\,\mathbf{a}_2-z_{22}\,\mathbf{a}_3$ | = | $-x_{22}a\,\mathbf{\hat{x}}-y_{22}b\,\mathbf{\hat{y}}-z_{22}c\,\mathbf{\hat{z}}$ | (8$d$) | O X |
| $\mathbf{B}_{158}$ | = | $\left(\frac{1}{2}+x_{22}\right)\mathbf{a}_1+y_{22}\,\mathbf{a}_2+\left(\frac{1}{2}-z_{22}\right)\mathbf{a}_3$ | = | $\left(\frac{1}{2}+x_{22}\right)a\,\mathbf{\hat{x}}+y_{22}b\,\mathbf{\hat{y}}+\left(\frac{1}{2}-z_{22}\right)c\,\mathbf{\hat{z}}$ | (8$d$) | O X |
| $\mathbf{B}_{159}$ | = | $x_{22}\,\mathbf{a}_1+\left(\frac{1}{2}-y_{22}\right)\mathbf{a}_2+z_{22}\,\mathbf{a}_3$ | = | $x_{22}a\,\mathbf{\hat{x}}+\left(\frac{1}{2}-y_{22}\right)b\,\mathbf{\hat{y}}+z_{22}c\,\mathbf{\hat{z}}$ | (8$d$) | O X |
| $\mathbf{B}_{160}$ | = | $\left(\frac{1}{2}-x_{22}\right)\mathbf{a}_1+\left(\frac{1}{2}+y_{22}\right)\mathbf{a}_2+\left(\frac{1}{2}+z_{22}\right)\mathbf{a}_3$ | = | $\left(\frac{1}{2}-x_{22}\right)a\,\mathbf{\hat{x}}+\left(\frac{1}{2}+y_{22}\right)b\,\mathbf{\hat{y}}+\left(\frac{1}{2}+z_{22}\right)c\,\mathbf{\hat{z}}$ | (8$d$) | O X |
| $\mathbf{B}_{161}$ | = | $x_{23}\,\mathbf{a}_1+y_{23}\,\mathbf{a}_2+z_{23}\,\mathbf{a}_3$ | = | $x_{23}a\,\mathbf{\hat{x}}+y_{23}b\,\mathbf{\hat{y}}+z_{23}c\,\mathbf{\hat{z}}$ | (8$d$) | O XI |
| $\mathbf{B}_{162}$ | = | $\left(\frac{1}{2}-x_{23}\right)\mathbf{a}_1-y_{23}\,\mathbf{a}_2+\left(\frac{1}{2}+z_{23}\right)\mathbf{a}_3$ | = | $\left(\frac{1}{2}-x_{23}\right)a\,\mathbf{\hat{x}}-y_{23}b\,\mathbf{\hat{y}}+\left(\frac{1}{2}+z_{23}\right)c\,\mathbf{\hat{z}}$ | (8$d$) | O XI |
| $\mathbf{B}_{163}$ | = | $-x_{23}\,\mathbf{a}_1+\left(\frac{1}{2}+y_{23}\right)\mathbf{a}_2-z_{23}\,\mathbf{a}_3$ | = | $-x_{23}a\,\mathbf{\hat{x}}+\left(\frac{1}{2}+y_{23}\right)b\,\mathbf{\hat{y}}-z_{23}c\,\mathbf{\hat{z}}$ | (8$d$) | O XI |
| $\mathbf{B}_{164}$ | = | $\left(\frac{1}{2}+x_{23}\right)\mathbf{a}_1+\left(\frac{1}{2}-y_{23}\right)\mathbf{a}_2+\left(\frac{1}{2}-z_{23}\right)\mathbf{a}_3$ | = | $\left(\frac{1}{2}+x_{23}\right)a\,\mathbf{\hat{x}}+\left(\frac{1}{2}-y_{23}\right)b\,\mathbf{\hat{y}}+\left(\frac{1}{2}-z_{23}\right)c\,\mathbf{\hat{z}}$ | (8$d$) | O XI |
| $\mathbf{B}_{165}$ | = | $-x_{23}\,\mathbf{a}_1-y_{23}\,\mathbf{a}_2-z_{23}\,\mathbf{a}_3$ | = | $-x_{23}a\,\mathbf{\hat{x}}-y_{23}b\,\mathbf{\hat{y}}-z_{23}c\,\mathbf{\hat{z}}$ | (8$d$) | O XI |
| $\mathbf{B}_{166}$ | = | $\left(\frac{1}{2}+x_{23}\right)\mathbf{a}_1+y_{23}\,\mathbf{a}_2+\left(\frac{1}{2}-z_{23}\right)\mathbf{a}_3$ | = | $\left(\frac{1}{2}+x_{23}\right)a\,\mathbf{\hat{x}}+y_{23}b\,\mathbf{\hat{y}}+\left(\frac{1}{2}-z_{23}\right)c\,\mathbf{\hat{z}}$ | (8$d$) | O XI |
| $\mathbf{B}_{167}$ | = | $x_{23}\,\mathbf{a}_1+\left(\frac{1}{2}-y_{23}\right)\mathbf{a}_2+z_{23}\,\mathbf{a}_3$ | = | $x_{23}a\,\mathbf{\hat{x}}+\left(\frac{1}{2}-y_{23}\right)b\,\mathbf{\hat{y}}+z_{23}c\,\mathbf{\hat{z}}$ | (8$d$) | O XI |
| $\mathbf{B}_{168}$ | = | $\left(\frac{1}{2}-x_{23}\right)\mathbf{a}_1+\left(\frac{1}{2}+y_{23}\right)\mathbf{a}_2+\left(\frac{1}{2}+z_{23}\right)\mathbf{a}_3$ | = | $\left(\frac{1}{2}-x_{23}\right)a\,\mathbf{\hat{x}}+\left(\frac{1}{2}+y_{23}\right)b\,\mathbf{\hat{y}}+\left(\frac{1}{2}+z_{23}\right)c\,\mathbf{\hat{z}}$ | (8$d$) | O XI |
| $\mathbf{B}_{169}$ | = | $x_{24}\,\mathbf{a}_1+y_{24}\,\mathbf{a}_2+z_{24}\,\mathbf{a}_3$ | = | $x_{24}a\,\mathbf{\hat{x}}+y_{24}b\,\mathbf{\hat{y}}+z_{24}c\,\mathbf{\hat{z}}$ | (8$d$) | O XII |
| $\mathbf{B}_{170}$ | = | $\left(\frac{1}{2}-x_{24}\right)\mathbf{a}_1-y_{24}\,\mathbf{a}_2+\left(\frac{1}{2}+z_{24}\right)\mathbf{a}_3$ | = | $\left(\frac{1}{2}-x_{24}\right)a\,\mathbf{\hat{x}}-y_{24}b\,\mathbf{\hat{y}}+\left(\frac{1}{2}+z_{24}\right)c\,\mathbf{\hat{z}}$ | (8$d$) | O XII |
| $\mathbf{B}_{171}$ | = | $-x_{24}\,\mathbf{a}_1+\left(\frac{1}{2}+y_{24}\right)\mathbf{a}_2-z_{24}\,\mathbf{a}_3$ | = | $-x_{24}a\,\mathbf{\hat{x}}+\left(\frac{1}{2}+y_{24}\right)b\,\mathbf{\hat{y}}-z_{24}c\,\mathbf{\hat{z}}$ | (8$d$) | O XII |
| $\mathbf{B}_{172}$ | = | $\left(\frac{1}{2}+x_{24}\right)\mathbf{a}_1+\left(\frac{1}{2}-y_{24}\right)\mathbf{a}_2+\left(\frac{1}{2}-z_{24}\right)\mathbf{a}_3$ | = | $\left(\frac{1}{2}+x_{24}\right)a\,\mathbf{\hat{x}}+\left(\frac{1}{2}-y_{24}\right)b\,\mathbf{\hat{y}}+\left(\frac{1}{2}-z_{24}\right)c\,\mathbf{\hat{z}}$ | (8$d$) | O XII |
| $\mathbf{B}_{173}$ | = | $-x_{24}\,\mathbf{a}_1-y_{24}\,\mathbf{a}_2-z_{24}\,\mathbf{a}_3$ | = | $-x_{24}a\,\mathbf{\hat{x}}-y_{24}b\,\mathbf{\hat{y}}-z_{24}c\,\mathbf{\hat{z}}$ | (8$d$) | O XII |
| $\mathbf{B}_{174}$ | = | $\left(\frac{1}{2}+x_{24}\right)\mathbf{a}_1+y_{24}\,\mathbf{a}_2+\left(\frac{1}{2}-z_{24}\right)\mathbf{a}_3$ | = | $\left(\frac{1}{2}+x_{24}\right)a\,\mathbf{\hat{x}}+y_{24}b\,\mathbf{\hat{y}}+\left(\frac{1}{2}-z_{24}\right)c\,\mathbf{\hat{z}}$ | (8$d$) | O XII |
| $\mathbf{B}_{175}$ | = | $x_{24}\,\mathbf{a}_1+\left(\frac{1}{2}-y_{24}\right)\mathbf{a}_2+z_{24}\,\mathbf{a}_3$ | = | $x_{24}a\,\mathbf{\hat{x}}+\left(\frac{1}{2}-y_{24}\right)b\,\mathbf{\hat{y}}+z_{24}c\,\mathbf{\hat{z}}$ | (8$d$) | O XII |
| $\mathbf{B}_{176}$ | = | $\left(\frac{1}{2}-x_{24}\right)\mathbf{a}_1+\left(\frac{1}{2}+y_{24}\right)\mathbf{a}_2+\left(\frac{1}{2}+z_{24}\right)\mathbf{a}_3$ | = | $\left(\frac{1}{2}-x_{24}\right)a\,\mathbf{\hat{x}}+\left(\frac{1}{2}+y_{24}\right)b\,\mathbf{\hat{y}}+\left(\frac{1}{2}+z_{24}\right)c\,\mathbf{\hat{z}}$ | (8$d$) | O XII |
| $\mathbf{B}_{177}$ | = | $x_{25}\,\mathbf{a}_1+y_{25}\,\mathbf{a}_2+z_{25}\,\mathbf{a}_3$ | = | $x_{25}a\,\mathbf{\hat{x}}+y_{25}b\,\mathbf{\hat{y}}+z_{25}c\,\mathbf{\hat{z}}$ | (8$d$) | O XIII |
| $\mathbf{B}_{178}$ | = | $\left(\frac{1}{2}-x_{25}\right)\mathbf{a}_1 - y_{25}\,\mathbf{a}_2 + \left(\frac{1}{2}+z_{25}\right)\mathbf{a}_3$ | = | $\left(\frac{1}{2}-x_{25}\right)a\,\hat{\mathbf{x}} - y_{25}b\,\hat{\mathbf{y}} + \left(\frac{1}{2}+z_{25}\right)c\,\hat{\mathbf{z}}$ | (8$d$) | O XIII |
| $\mathbf{B}_{179}$ | = | $-x_{25}\,\mathbf{a}_1 + \left(\frac{1}{2}+y_{25}\right)\mathbf{a}_2 - z_{25}\,\mathbf{a}_3$ | = | $-x_{25}a\,\hat{\mathbf{x}} + \left(\frac{1}{2}+y_{25}\right)b\,\hat{\mathbf{y}} - z_{25}c\,\hat{\mathbf{z}}$ | (8$d$) | O XIII |
| $\mathbf{B}_{180}$ | = | $\left(\frac{1}{2}+x_{25}\right)\mathbf{a}_1 + \left(\frac{1}{2}-y_{25}\right)\mathbf{a}_2 + \left(\frac{1}{2}-z_{25}\right)\mathbf{a}_3$ | = | $\left(\frac{1}{2}+x_{25}\right)a\,\hat{\mathbf{x}} + \left(\frac{1}{2}-y_{25}\right)b\,\hat{\mathbf{y}} + \left(\frac{1}{2}-z_{25}\right)c\,\hat{\mathbf{z}}$ | (8$d$) | O XIII |
| $\mathbf{B}_{181}$ | = | $-x_{25}\,\mathbf{a}_1 - y_{25}\,\mathbf{a}_2 - z_{25}\,\mathbf{a}_3$ | = | $-x_{25}a\,\hat{\mathbf{x}} - y_{25}b\,\hat{\mathbf{y}} - z_{25}c\,\hat{\mathbf{z}}$ | (8$d$) | O XIII |
| $\mathbf{B}_{182}$ | = | $\left(\frac{1}{2}+x_{25}\right)\mathbf{a}_1 + y_{25}\,\mathbf{a}_2 + \left(\frac{1}{2}-z_{25}\right)\mathbf{a}_3$ | = | $\left(\frac{1}{2}+x_{25}\right)a\,\hat{\mathbf{x}} + y_{25}b\,\hat{\mathbf{y}} + \left(\frac{1}{2}-z_{25}\right)c\,\hat{\mathbf{z}}$ | (8$d$) | O XIII |
| $\mathbf{B}_{183}$ | = | $x_{25}\,\mathbf{a}_1 + \left(\frac{1}{2}-y_{25}\right)\mathbf{a}_2 + z_{25}\,\mathbf{a}_3$ | = | $x_{25}a\,\hat{\mathbf{x}} + \left(\frac{1}{2}-y_{25}\right)b\,\hat{\mathbf{y}} + z_{25}c\,\hat{\mathbf{z}}$ | (8$d$) | O XIII |
| $\mathbf{B}_{184}$ | = | $\left(\frac{1}{2}-x_{25}\right)\mathbf{a}_1 + \left(\frac{1}{2}+y_{25}\right)\mathbf{a}_2 + \left(\frac{1}{2}+z_{25}\right)\mathbf{a}_3$ | = | $\left(\frac{1}{2}-x_{25}\right)a\,\hat{\mathbf{x}} + \left(\frac{1}{2}+y_{25}\right)b\,\hat{\mathbf{y}} + \left(\frac{1}{2}+z_{25}\right)c\,\hat{\mathbf{z}}$ | (8$d$) | O XIII |
| $\mathbf{B}_{185}$ | = | $x_{26}\,\mathbf{a}_1 + y_{26}\,\mathbf{a}_2 + z_{26}\,\mathbf{a}_3$ | = | $x_{26}a\,\hat{\mathbf{x}} + y_{26}b\,\hat{\mathbf{y}} + z_{26}c\,\hat{\mathbf{z}}$ | (8$d$) | P |
| $\mathbf{B}_{186}$ | = | $\left(\frac{1}{2}-x_{26}\right)\mathbf{a}_1 - y_{26}\,\mathbf{a}_2 + \left(\frac{1}{2}+z_{26}\right)\mathbf{a}_3$ | = | $\left(\frac{1}{2}-x_{26}\right)a\,\hat{\mathbf{x}} - y_{26}b\,\hat{\mathbf{y}} + \left(\frac{1}{2}+z_{26}\right)c\,\hat{\mathbf{z}}$ | (8$d$) | P |
| $\mathbf{B}_{187}$ | = | $-x_{26}\,\mathbf{a}_1 + \left(\frac{1}{2}+y_{26}\right)\mathbf{a}_2 - z_{26}\,\mathbf{a}_3$ | = | $-x_{26}a\,\hat{\mathbf{x}} + \left(\frac{1}{2}+y_{26}\right)b\,\hat{\mathbf{y}} - z_{26}c\,\hat{\mathbf{z}}$ | (8$d$) | P |
| $\mathbf{B}_{188}$ | = | $\left(\frac{1}{2}+x_{26}\right)\mathbf{a}_1 + \left(\frac{1}{2}-y_{26}\right)\mathbf{a}_2 + \left(\frac{1}{2}-z_{26}\right)\mathbf{a}_3$ | = | $\left(\frac{1}{2}+x_{26}\right)a\,\hat{\mathbf{x}} + \left(\frac{1}{2}-y_{26}\right)b\,\hat{\mathbf{y}} + \left(\frac{1}{2}-z_{26}\right)c\,\hat{\mathbf{z}}$ | (8$d$) | P |
| $\mathbf{B}_{189}$ | = | $-x_{26}\,\mathbf{a}_1 - y_{26}\,\mathbf{a}_2 - z_{26}\,\mathbf{a}_3$ | = | $-x_{26}a\,\hat{\mathbf{x}} - y_{26}b\,\hat{\mathbf{y}} - z_{26}c\,\hat{\mathbf{z}}$ | (8$d$) | P |
| $\mathbf{B}_{190}$ | = | $\left(\frac{1}{2}+x_{26}\right)\mathbf{a}_1 + y_{26}\,\mathbf{a}_2 + \left(\frac{1}{2}-z_{26}\right)\mathbf{a}_3$ | = | $\left(\frac{1}{2}+x_{26}\right)a\,\hat{\mathbf{x}} + y_{26}b\,\hat{\mathbf{y}} + \left(\frac{1}{2}-z_{26}\right)c\,\hat{\mathbf{z}}$ | (8$d$) | P |
| $\mathbf{B}_{191}$ | = | $x_{26}\,\mathbf{a}_1 + \left(\frac{1}{2}-y_{26}\right)\mathbf{a}_2 + z_{26}\,\mathbf{a}_3$ | = | $x_{26}a\,\hat{\mathbf{x}} + \left(\frac{1}{2}-y_{26}\right)b\,\hat{\mathbf{y}} + z_{26}c\,\hat{\mathbf{z}}$ | (8$d$) | P |
| $\mathbf{B}_{192}$ | = | $\left(\frac{1}{2}-x_{26}\right)\mathbf{a}_1 + \left(\frac{1}{2}+y_{26}\right)\mathbf{a}_2 + \left(\frac{1}{2}+z_{26}\right)\mathbf{a}_3$ | = | $\left(\frac{1}{2}-x_{26}\right)a\,\hat{\mathbf{x}} + \left(\frac{1}{2}+y_{26}\right)b\,\hat{\mathbf{y}} + \left(\frac{1}{2}+z_{26}\right)c\,\hat{\mathbf{z}}$ | (8$d$) | P |
| $\mathbf{B}_{193}$ | = | $x_{27}\,\mathbf{a}_1 + y_{27}\,\mathbf{a}_2 + z_{27}\,\mathbf{a}_3$ | = | $x_{27}a\,\hat{\mathbf{x}} + y_{27}b\,\hat{\mathbf{y}} + z_{27}c\,\hat{\mathbf{z}}$ | (8$d$) | U |
| $\mathbf{B}_{194}$ | = | $\left(\frac{1}{2}-x_{27}\right)\mathbf{a}_1 - y_{27}\,\mathbf{a}_2 + \left(\frac{1}{2}+z_{27}\right)\mathbf{a}_3$ | = | $\left(\frac{1}{2}-x_{27}\right)a\,\hat{\mathbf{x}} - y_{27}b\,\hat{\mathbf{y}} + \left(\frac{1}{2}+z_{27}\right)c\,\hat{\mathbf{z}}$ | (8$d$) | U |
| $\mathbf{B}_{195}$ | = | $-x_{27}\,\mathbf{a}_1 + \left(\frac{1}{2}+y_{27}\right)\mathbf{a}_2 - z_{27}\,\mathbf{a}_3$ | = | $-x_{27}a\,\hat{\mathbf{x}} + \left(\frac{1}{2}+y_{27}\right)b\,\hat{\mathbf{y}} - z_{27}c\,\hat{\mathbf{z}}$ | (8$d$) | U |
| $\mathbf{B}_{196}$ | = | $\left(\frac{1}{2}+x_{27}\right)\mathbf{a}_1 + \left(\frac{1}{2}-y_{27}\right)\mathbf{a}_2 + \left(\frac{1}{2}-z_{27}\right)\mathbf{a}_3$ | = | $\left(\frac{1}{2}+x_{27}\right)a\,\hat{\mathbf{x}} + \left(\frac{1}{2}-y_{27}\right)b\,\hat{\mathbf{y}} + \left(\frac{1}{2}-z_{27}\right)c\,\hat{\mathbf{z}}$ | (8$d$) | U |
| $\mathbf{B}_{197}$ | = | $-x_{27}\,\mathbf{a}_1 - y_{27}\,\mathbf{a}_2 - z_{27}\,\mathbf{a}_3$ | = | $-x_{27}a\,\hat{\mathbf{x}} - y_{27}b\,\hat{\mathbf{y}} - z_{27}c\,\hat{\mathbf{z}}$ | (8$d$) | U |
| $\mathbf{B}_{198}$ | = | $\left(\frac{1}{2}+x_{27}\right)\mathbf{a}_1 + y_{27}\,\mathbf{a}_2 + \left(\frac{1}{2}-z_{27}\right)\mathbf{a}_3$ | = | $\left(\frac{1}{2}+x_{27}\right)a\,\hat{\mathbf{x}} + y_{27}b\,\hat{\mathbf{y}} + \left(\frac{1}{2}-z_{27}\right)c\,\hat{\mathbf{z}}$ | (8$d$) | U |
| $\mathbf{B}_{199}$ | = | $x_{27}\,\mathbf{a}_1 + \left(\frac{1}{2}-y_{27}\right)\mathbf{a}_2 + z_{27}\,\mathbf{a}_3$ | = | $x_{27}a\,\hat{\mathbf{x}} + \left(\frac{1}{2}-y_{27}\right)b\,\hat{\mathbf{y}} + z_{27}c\,\hat{\mathbf{z}}$ | (8$d$) | U |
| $\mathbf{B}_{200}$ | = | $\left(\frac{1}{2}-x_{27}\right)\mathbf{a}_1 + \left(\frac{1}{2}+y_{27}\right)\mathbf{a}_2 + \left(\frac{1}{2}+z_{27}\right)\mathbf{a}_3$ | = | $\left(\frac{1}{2}-x_{27}\right)a\,\hat{\mathbf{x}} + \left(\frac{1}{2}+y_{27}\right)b\,\hat{\mathbf{y}} + \left(\frac{1}{2}+z_{27}\right)c\,\hat{\mathbf{z}}$ | (8$d$) | U |

**Found in:**

**Geometry files:**

- CIF: pp. 1643

- POSCAR: pp. 1644

# Atacamite ($Cu_2(OH)_3Cl$) Structure: AB2C3D3_oP36_62_c_ac_cd_cd

http://aflow.org/prototype-encyclopedia/AB2C3D3_oP36_62_c_ac_cd_cd

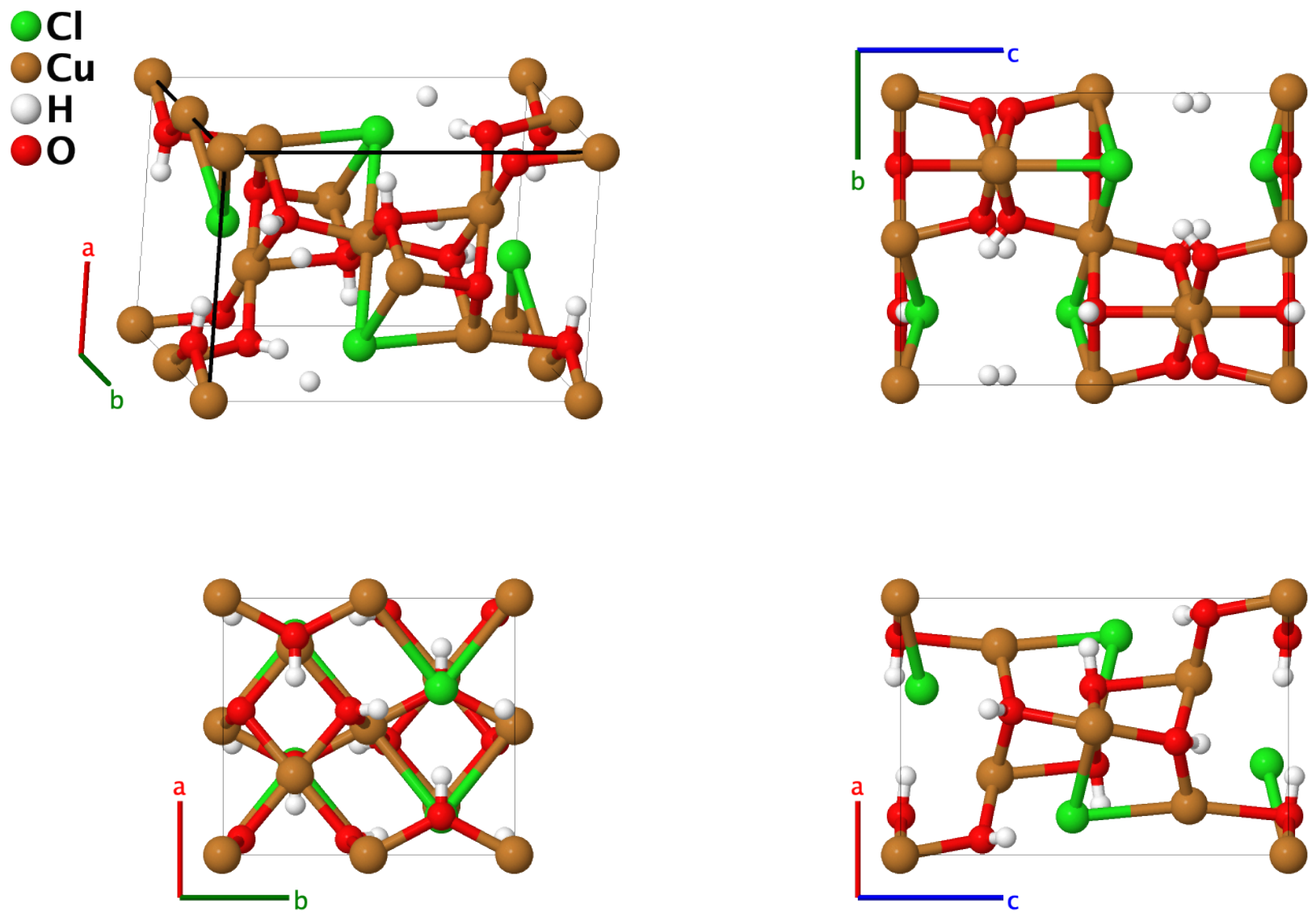

| | | |
|---|---|---|
| **Prototype** | : | $ClCu_2H_3O_3$ |
| **AFLOW prototype label** | : | AB2C3D3_oP36_62_c_ac_cd_cd |
| ***Strukturbericht* designation** | : | None |
| **Pearson symbol** | : | oP36 |
| **Space group number** | : | 62 |
| **Space group symbol** | : | *Pnma* |
| **AFLOW prototype command** | : | `aflow --proto=AB2C3D3_oP36_62_c_ac_cd_cd` `--params=`$a, b/a, c/a, x_2, z_2, x_3, z_3, x_4, z_4, x_5, z_5, x_6, y_6, z_6, x_7, y_7, z_7$ |

**Simple Orthorhombic primitive vectors:**

$$\mathbf{a}_1 = a\,\hat{\mathbf{x}}$$
$$\mathbf{a}_2 = b\,\hat{\mathbf{y}}$$
$$\mathbf{a}_3 = c\,\hat{\mathbf{z}}$$

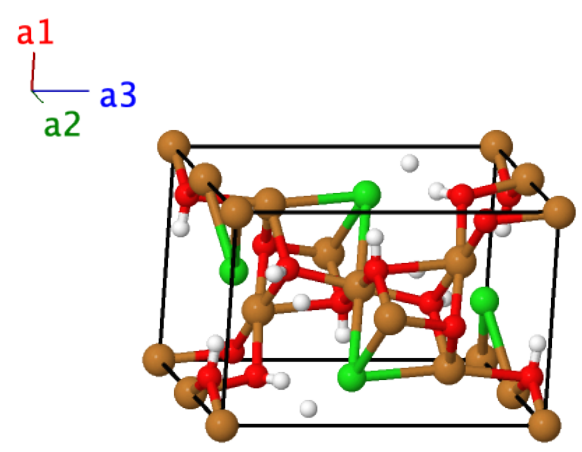

**Basis vectors:**

| | | Lattice Coordinates | | Cartesian Coordinates | Wyckoff Position | Atom Type |
|---|---|---|---|---|---|---|
| $\mathbf{B}_1$ | = | $0\,\mathbf{a}_1 + 0\,\mathbf{a}_2 + 0\,\mathbf{a}_3$ | = | $0\,\hat{\mathbf{x}} + 0\,\hat{\mathbf{y}} + 0\,\hat{\mathbf{z}}$ | (4*a*) | Cu I |
| $\mathbf{B}_2$ | = | $\frac{1}{2}\,\mathbf{a}_1 + \frac{1}{2}\,\mathbf{a}_3$ | = | $\frac{1}{2}a\,\hat{\mathbf{x}} + \frac{1}{2}c\,\hat{\mathbf{z}}$ | (4*a*) | Cu I |
| $\mathbf{B}_3$ | = | $\frac{1}{2}\,\mathbf{a}_2$ | = | $\frac{1}{2}b\,\hat{\mathbf{y}}$ | (4*a*) | Cu I |
| $\mathbf{B}_4$ | = | $\frac{1}{2}\,\mathbf{a}_1 + \frac{1}{2}\,\mathbf{a}_2 + \frac{1}{2}\,\mathbf{a}_3$ | = | $\frac{1}{2}a\,\hat{\mathbf{x}} + \frac{1}{2}b\,\hat{\mathbf{y}} + \frac{1}{2}c\,\hat{\mathbf{z}}$ | (4*a*) | Cu I |
| $\mathbf{B}_{5}$ | = | $x_2\,\mathbf{a}_1+\frac{1}{4}\,\mathbf{a}_2+z_2\,\mathbf{a}_3$ | = | $x_2a\,\hat{\mathbf{x}}+\frac{1}{4}b\,\hat{\mathbf{y}}+z_2c\,\hat{\mathbf{z}}$ | (4$c$) | Cl |
| $\mathbf{B}_{6}$ | = | $\left(\frac{1}{2}-x_2\right)\mathbf{a}_1+\frac{3}{4}\,\mathbf{a}_2+\left(\frac{1}{2}+z_2\right)\mathbf{a}_3$ | = | $\left(\frac{1}{2}-x_2\right)a\,\hat{\mathbf{x}}+\frac{3}{4}b\,\hat{\mathbf{y}}+\left(\frac{1}{2}+z_2\right)c\,\hat{\mathbf{z}}$ | (4$c$) | Cl |
| $\mathbf{B}_{7}$ | = | $-x_2\,\mathbf{a}_1+\frac{3}{4}\,\mathbf{a}_2-z_2\,\mathbf{a}_3$ | = | $-x_2a\,\hat{\mathbf{x}}+\frac{3}{4}b\,\hat{\mathbf{y}}-z_2c\,\hat{\mathbf{z}}$ | (4$c$) | Cl |
| $\mathbf{B}_{8}$ | = | $\left(\frac{1}{2}+x_2\right)\mathbf{a}_1+\frac{1}{4}\,\mathbf{a}_2+\left(\frac{1}{2}-z_2\right)\mathbf{a}_3$ | = | $\left(\frac{1}{2}+x_2\right)a\,\hat{\mathbf{x}}+\frac{1}{4}b\,\hat{\mathbf{y}}+\left(\frac{1}{2}-z_2\right)c\,\hat{\mathbf{z}}$ | (4$c$) | Cl |
| $\mathbf{B}_{9}$ | = | $x_3\,\mathbf{a}_1+\frac{1}{4}\,\mathbf{a}_2+z_3\,\mathbf{a}_3$ | = | $x_3a\,\hat{\mathbf{x}}+\frac{1}{4}b\,\hat{\mathbf{y}}+z_3c\,\hat{\mathbf{z}}$ | (4$c$) | Cu II |
| $\mathbf{B}_{10}$ | = | $\left(\frac{1}{2}-x_3\right)\mathbf{a}_1+\frac{3}{4}\,\mathbf{a}_2+\left(\frac{1}{2}+z_3\right)\mathbf{a}_3$ | = | $\left(\frac{1}{2}-x_3\right)a\,\hat{\mathbf{x}}+\frac{3}{4}b\,\hat{\mathbf{y}}+\left(\frac{1}{2}+z_3\right)c\,\hat{\mathbf{z}}$ | (4$c$) | Cu II |
| $\mathbf{B}_{11}$ | = | $-x_3\,\mathbf{a}_1+\frac{3}{4}\,\mathbf{a}_2-z_3\,\mathbf{a}_3$ | = | $-x_3a\,\hat{\mathbf{x}}+\frac{3}{4}b\,\hat{\mathbf{y}}-z_3c\,\hat{\mathbf{z}}$ | (4$c$) | Cu II |
| $\mathbf{B}_{12}$ | = | $\left(\frac{1}{2}+x_3\right)\mathbf{a}_1+\frac{1}{4}\,\mathbf{a}_2+\left(\frac{1}{2}-z_3\right)\mathbf{a}_3$ | = | $\left(\frac{1}{2}+x_3\right)a\,\hat{\mathbf{x}}+\frac{1}{4}b\,\hat{\mathbf{y}}+\left(\frac{1}{2}-z_3\right)c\,\hat{\mathbf{z}}$ | (4$c$) | Cu II |
| $\mathbf{B}_{13}$ | = | $x_4\,\mathbf{a}_1+\frac{1}{4}\,\mathbf{a}_2+z_4\,\mathbf{a}_3$ | = | $x_4a\,\hat{\mathbf{x}}+\frac{1}{4}b\,\hat{\mathbf{y}}+z_4c\,\hat{\mathbf{z}}$ | (4$c$) | H I |
| $\mathbf{B}_{14}$ | = | $\left(\frac{1}{2}-x_4\right)\mathbf{a}_1+\frac{3}{4}\,\mathbf{a}_2+\left(\frac{1}{2}+z_4\right)\mathbf{a}_3$ | = | $\left(\frac{1}{2}-x_4\right)a\,\hat{\mathbf{x}}+\frac{3}{4}b\,\hat{\mathbf{y}}+\left(\frac{1}{2}+z_4\right)c\,\hat{\mathbf{z}}$ | (4$c$) | H I |
| $\mathbf{B}_{15}$ | = | $-x_4\,\mathbf{a}_1+\frac{3}{4}\,\mathbf{a}_2-z_4\,\mathbf{a}_3$ | = | $-x_4a\,\hat{\mathbf{x}}+\frac{3}{4}b\,\hat{\mathbf{y}}-z_4c\,\hat{\mathbf{z}}$ | (4$c$) | H I |
| $\mathbf{B}_{16}$ | = | $\left(\frac{1}{2}+x_4\right)\mathbf{a}_1+\frac{1}{4}\,\mathbf{a}_2+\left(\frac{1}{2}-z_4\right)\mathbf{a}_3$ | = | $\left(\frac{1}{2}+x_4\right)a\,\hat{\mathbf{x}}+\frac{1}{4}b\,\hat{\mathbf{y}}+\left(\frac{1}{2}-z_4\right)c\,\hat{\mathbf{z}}$ | (4$c$) | H I |
| $\mathbf{B}_{17}$ | = | $x_5\,\mathbf{a}_1+\frac{1}{4}\,\mathbf{a}_2+z_5\,\mathbf{a}_3$ | = | $x_5a\,\hat{\mathbf{x}}+\frac{1}{4}b\,\hat{\mathbf{y}}+z_5c\,\hat{\mathbf{z}}$ | (4$c$) | O I |
| $\mathbf{B}_{18}$ | = | $\left(\frac{1}{2}-x_5\right)\mathbf{a}_1+\frac{3}{4}\,\mathbf{a}_2+\left(\frac{1}{2}+z_5\right)\mathbf{a}_3$ | = | $\left(\frac{1}{2}-x_5\right)a\,\hat{\mathbf{x}}+\frac{3}{4}b\,\hat{\mathbf{y}}+\left(\frac{1}{2}+z_5\right)c\,\hat{\mathbf{z}}$ | (4$c$) | O I |
| $\mathbf{B}_{19}$ | = | $-x_5\,\mathbf{a}_1+\frac{3}{4}\,\mathbf{a}_2-z_5\,\mathbf{a}_3$ | = | $-x_5a\,\hat{\mathbf{x}}+\frac{3}{4}b\,\hat{\mathbf{y}}-z_5c\,\hat{\mathbf{z}}$ | (4$c$) | O I |
| $\mathbf{B}_{20}$ | = | $\left(\frac{1}{2}+x_5\right)\mathbf{a}_1+\frac{1}{4}\,\mathbf{a}_2+\left(\frac{1}{2}-z_5\right)\mathbf{a}_3$ | = | $\left(\frac{1}{2}+x_5\right)a\,\hat{\mathbf{x}}+\frac{1}{4}b\,\hat{\mathbf{y}}+\left(\frac{1}{2}-z_5\right)c\,\hat{\mathbf{z}}$ | (4$c$) | O I |
| $\mathbf{B}_{21}$ | = | $x_6\,\mathbf{a}_1+y_6\,\mathbf{a}_2+z_6\,\mathbf{a}_3$ | = | $x_6a\,\hat{\mathbf{x}}+y_6b\,\hat{\mathbf{y}}+z_6c\,\hat{\mathbf{z}}$ | (8$d$) | H II |
| $\mathbf{B}_{22}$ | = | $\left(\frac{1}{2}-x_6\right)\mathbf{a}_1-y_6\,\mathbf{a}_2+\left(\frac{1}{2}+z_6\right)\mathbf{a}_3$ | = | $\left(\frac{1}{2}-x_6\right)a\,\hat{\mathbf{x}}-y_6b\,\hat{\mathbf{y}}+\left(\frac{1}{2}+z_6\right)c\,\hat{\mathbf{z}}$ | (8$d$) | H II |
| $\mathbf{B}_{23}$ | = | $-x_6\,\mathbf{a}_1+\left(\frac{1}{2}+y_6\right)\mathbf{a}_2-z_6\,\mathbf{a}_3$ | = | $-x_6a\,\hat{\mathbf{x}}+\left(\frac{1}{2}+y_6\right)b\,\hat{\mathbf{y}}-z_6c\,\hat{\mathbf{z}}$ | (8$d$) | H II |
| $\mathbf{B}_{24}$ | = | $\left(\frac{1}{2}+x_6\right)\mathbf{a}_1+\left(\frac{1}{2}-y_6\right)\mathbf{a}_2+\left(\frac{1}{2}-z_6\right)\mathbf{a}_3$ | = | $\left(\frac{1}{2}+x_6\right)a\,\hat{\mathbf{x}}+\left(\frac{1}{2}-y_6\right)b\,\hat{\mathbf{y}}+\left(\frac{1}{2}-z_6\right)c\,\hat{\mathbf{z}}$ | (8$d$) | H II |
| $\mathbf{B}_{25}$ | = | $-x_6\,\mathbf{a}_1-y_6\,\mathbf{a}_2-z_6\,\mathbf{a}_3$ | = | $-x_6a\,\hat{\mathbf{x}}-y_6b\,\hat{\mathbf{y}}-z_6c\,\hat{\mathbf{z}}$ | (8$d$) | H II |
| $\mathbf{B}_{26}$ | = | $\left(\frac{1}{2}+x_6\right)\mathbf{a}_1+y_6\,\mathbf{a}_2+\left(\frac{1}{2}-z_6\right)\mathbf{a}_3$ | = | $\left(\frac{1}{2}+x_6\right)a\,\hat{\mathbf{x}}+y_6b\,\hat{\mathbf{y}}+\left(\frac{1}{2}-z_6\right)c\,\hat{\mathbf{z}}$ | (8$d$) | H II |
| $\mathbf{B}_{27}$ | = | $x_6\,\mathbf{a}_1+\left(\frac{1}{2}-y_6\right)\mathbf{a}_2+z_6\,\mathbf{a}_3$ | = | $x_6a\,\hat{\mathbf{x}}+\left(\frac{1}{2}-y_6\right)b\,\hat{\mathbf{y}}+z_6c\,\hat{\mathbf{z}}$ | (8$d$) | H II |
| $\mathbf{B}_{28}$ | = | $\left(\frac{1}{2}-x_6\right)\mathbf{a}_1+\left(\frac{1}{2}+y_6\right)\mathbf{a}_2+\left(\frac{1}{2}+z_6\right)\mathbf{a}_3$ | = | $\left(\frac{1}{2}-x_6\right)a\,\hat{\mathbf{x}}+\left(\frac{1}{2}+y_6\right)b\,\hat{\mathbf{y}}+\left(\frac{1}{2}+z_6\right)c\,\hat{\mathbf{z}}$ | (8$d$) | H II |
| $\mathbf{B}_{29}$ | = | $x_7\,\mathbf{a}_1+y_7\,\mathbf{a}_2+z_7\,\mathbf{a}_3$ | = | $x_7a\,\hat{\mathbf{x}}+y_7b\,\hat{\mathbf{y}}+z_7c\,\hat{\mathbf{z}}$ | (8$d$) | O II |
| $\mathbf{B}_{30}$ | = | $\left(\frac{1}{2}-x_7\right)\mathbf{a}_1-y_7\,\mathbf{a}_2+\left(\frac{1}{2}+z_7\right)\mathbf{a}_3$ | = | $\left(\frac{1}{2}-x_7\right)a\,\hat{\mathbf{x}}-y_7b\,\hat{\mathbf{y}}+\left(\frac{1}{2}+z_7\right)c\,\hat{\mathbf{z}}$ | (8$d$) | O II |
| $\mathbf{B}_{31}$ | = | $-x_7\,\mathbf{a}_1+\left(\frac{1}{2}+y_7\right)\mathbf{a}_2-z_7\,\mathbf{a}_3$ | = | $-x_7a\,\hat{\mathbf{x}}+\left(\frac{1}{2}+y_7\right)b\,\hat{\mathbf{y}}-z_7c\,\hat{\mathbf{z}}$ | (8$d$) | O II |
| $\mathbf{B}_{32}$ | = | $\left(\frac{1}{2}+x_7\right)\mathbf{a}_1+\left(\frac{1}{2}-y_7\right)\mathbf{a}_2+\left(\frac{1}{2}-z_7\right)\mathbf{a}_3$ | = | $\left(\frac{1}{2}+x_7\right)a\,\hat{\mathbf{x}}+\left(\frac{1}{2}-y_7\right)b\,\hat{\mathbf{y}}+\left(\frac{1}{2}-z_7\right)c\,\hat{\mathbf{z}}$ | (8$d$) | O II |
| $\mathbf{B}_{33}$ | = | $-x_7\,\mathbf{a}_1-y_7\,\mathbf{a}_2-z_7\,\mathbf{a}_3$ | = | $-x_7a\,\hat{\mathbf{x}}-y_7b\,\hat{\mathbf{y}}-z_7c\,\hat{\mathbf{z}}$ | (8$d$) | O II |
| $\mathbf{B}_{34}$ | = | $\left(\frac{1}{2}+x_7\right)\mathbf{a}_1+y_7\,\mathbf{a}_2+\left(\frac{1}{2}-z_7\right)\mathbf{a}_3$ | = | $\left(\frac{1}{2}+x_7\right)a\,\hat{\mathbf{x}}+y_7b\,\hat{\mathbf{y}}+\left(\frac{1}{2}-z_7\right)c\,\hat{\mathbf{z}}$ | (8$d$) | O II |
| $\mathbf{B}_{35}$ | = | $x_7\,\mathbf{a}_1+\left(\frac{1}{2}-y_7\right)\mathbf{a}_2+z_7\,\mathbf{a}_3$ | = | $x_7a\,\hat{\mathbf{x}}+\left(\frac{1}{2}-y_7\right)b\,\hat{\mathbf{y}}+z_7c\,\hat{\mathbf{z}}$ | (8$d$) | O II |
| $\mathbf{B}_{36}$ | = | $\left(\frac{1}{2}-x_7\right)\mathbf{a}_1+\left(\frac{1}{2}+y_7\right)\mathbf{a}_2+\left(\frac{1}{2}+z_7\right)\mathbf{a}_3$ | = | $\left(\frac{1}{2}-x_7\right)a\,\hat{\mathbf{x}}+\left(\frac{1}{2}+y_7\right)b\,\hat{\mathbf{y}}+\left(\frac{1}{2}+z_7\right)c\,\hat{\mathbf{z}}$ | (8$d$) | O II |

**Found in:**

---

**Geometry files:**
- CIF: pp. 1645
- POSCAR: pp. 1645

# $NH_4CdCl_3$ ($E2_4$) Structure: AB3C_oP20_62_c_3c_c

http://aflow.org/prototype-encyclopedia/AB3C_oP20_62_c_3c_c

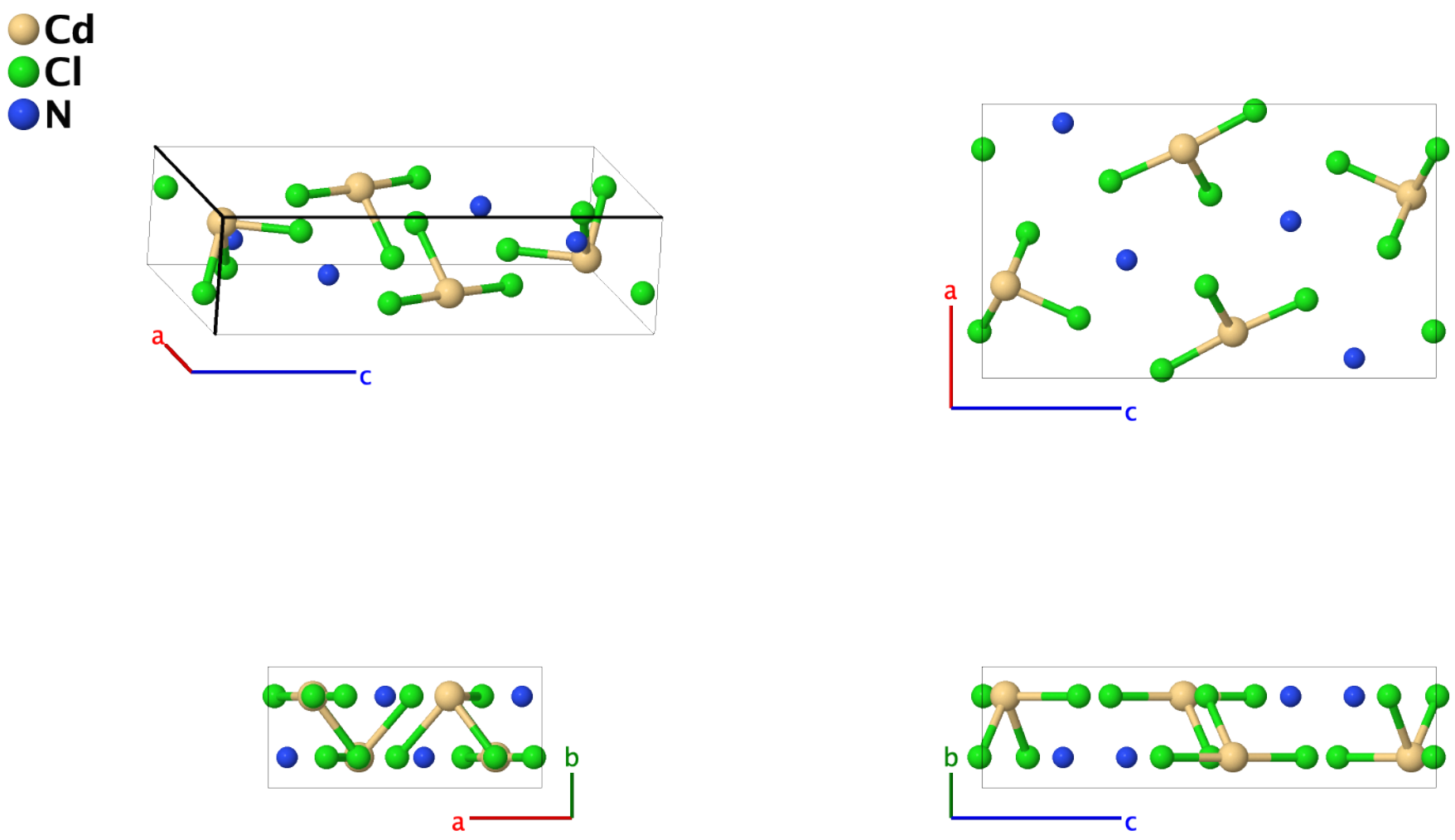

| | | |
|---|---|---|
| **Prototype** | : | $CdCl_3(NH_4)$ |
| **AFLOW prototype label** | : | AB3C_oP20_62_c_3c_c |
| ***Strukturbericht* designation** | : | $E2_4$ |
| **Pearson symbol** | : | oP20 |
| **Space group number** | : | 62 |
| **Space group symbol** | : | *Pnma* |
| **AFLOW prototype command** | : | `aflow --proto=AB3C_oP20_62_c_3c_c` `--params=`$a, b/a, c/a, x_1, z_1, x_2, z_2, x_3, z_3, x_4, z_4, x_5, z_5$ |

### Other compounds with this structure

- $LaCrS_3$, $CeCrS_3$, $SmCr S_3$, $NdCrS_3$, $CeCrSe_3$, $PbSnS_3$, $RbCdBr_3$, $RbCdCl_3$, and $(Ti,Sn)_2S_3$

---

- (Brasseur, 1938) gave the Wyckoff positions in the *Pnam* setting of space group #62. We used FINDSYM to rotate the structure and put it in the standard *Pnma* setting.
- The hydrogen atoms for the $NH_4$ ions were not determined, so we only give the positions of the nitrogen atoms (labeled as $NH_4$).

### Simple Orthorhombic primitive vectors:

$$\mathbf{a}_1 = a\,\hat{\mathbf{x}}$$
$$\mathbf{a}_2 = b\,\hat{\mathbf{y}}$$
$$\mathbf{a}_3 = c\,\hat{\mathbf{z}}$$

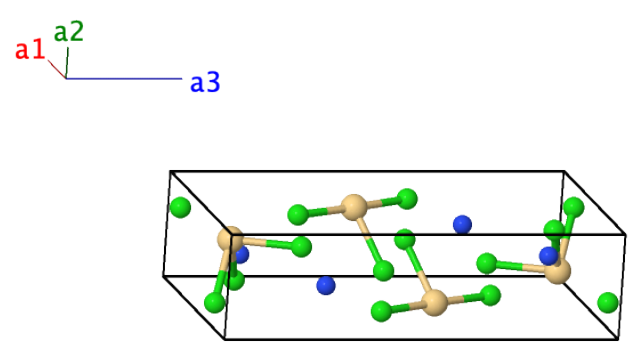

### Basis vectors:

| | | Lattice Coordinates | | Cartesian Coordinates | Wyckoff Position | Atom Type |
|---|---|---|---|---|---|---|
| $\mathbf{B}_1$ | = | $x_1\,\mathbf{a}_1 + \frac{1}{4}\,\mathbf{a}_2 + z_1\,\mathbf{a}_3$ | = | $x_1 a\,\hat{\mathbf{x}} + \frac{1}{4} b\,\hat{\mathbf{y}} + z_1 c\,\hat{\mathbf{z}}$ | (4c) | Cd |
| $\mathbf{B}_2$ | = | $\left(\frac{1}{2} - x_1\right)\mathbf{a}_1 + \frac{3}{4}\,\mathbf{a}_2 + \left(\frac{1}{2} + z_1\right)\mathbf{a}_3$ | = | $\left(\frac{1}{2} - x_1\right) a\,\hat{\mathbf{x}} + \frac{3}{4} b\,\hat{\mathbf{y}} + \left(\frac{1}{2} + z_1\right) c\,\hat{\mathbf{z}}$ | (4c) | Cd |
| $\mathbf{B}_3$ | = | $-x_1\,\mathbf{a}_1 + \frac{3}{4}\,\mathbf{a}_2 - z_1\,\mathbf{a}_3$ | = | $-x_1 a\,\hat{\mathbf{x}} + \frac{3}{4} b\,\hat{\mathbf{y}} - z_1 c\,\hat{\mathbf{z}}$ | (4c) | Cd |
| $\mathbf{B}_4$ | = | $\left(\frac{1}{2} + x_1\right)\mathbf{a}_1 + \frac{1}{4}\,\mathbf{a}_2 + \left(\frac{1}{2} - z_1\right)\mathbf{a}_3$ | = | $\left(\frac{1}{2} + x_1\right) a\,\hat{\mathbf{x}} + \frac{1}{4} b\,\hat{\mathbf{y}} + \left(\frac{1}{2} - z_1\right) c\,\hat{\mathbf{z}}$ | (4c) | Cd |
| $\mathbf{B}_5$ | = | $x_2\,\mathbf{a}_1 + \frac{1}{4}\,\mathbf{a}_2 + z_2\,\mathbf{a}_3$ | = | $x_2 a\,\hat{\mathbf{x}} + \frac{1}{4} b\,\hat{\mathbf{y}} + z_2 c\,\hat{\mathbf{z}}$ | (4c) | Cl I |
| $\mathbf{B}_6$ | = | $\left(\frac{1}{2} - x_2\right)\mathbf{a}_1 + \frac{3}{4}\,\mathbf{a}_2 + \left(\frac{1}{2} + z_2\right)\mathbf{a}_3$ | = | $\left(\frac{1}{2} - x_2\right) a\,\hat{\mathbf{x}} + \frac{3}{4} b\,\hat{\mathbf{y}} + \left(\frac{1}{2} + z_2\right) c\,\hat{\mathbf{z}}$ | (4c) | Cl I |
| $\mathbf{B}_7$ | = | $-x_2\,\mathbf{a}_1 + \frac{3}{4}\,\mathbf{a}_2 - z_2\,\mathbf{a}_3$ | = | $-x_2 a\,\hat{\mathbf{x}} + \frac{3}{4} b\,\hat{\mathbf{y}} - z_2 c\,\hat{\mathbf{z}}$ | (4c) | Cl I |
| $\mathbf{B}_8$ | = | $\left(\frac{1}{2} + x_2\right)\mathbf{a}_1 + \frac{1}{4}\,\mathbf{a}_2 + \left(\frac{1}{2} - z_2\right)\mathbf{a}_3$ | = | $\left(\frac{1}{2} + x_2\right) a\,\hat{\mathbf{x}} + \frac{1}{4} b\,\hat{\mathbf{y}} + \left(\frac{1}{2} - z_2\right) c\,\hat{\mathbf{z}}$ | (4c) | Cl I |
| $\mathbf{B}_9$ | = | $x_3\,\mathbf{a}_1 + \frac{1}{4}\,\mathbf{a}_2 + z_3\,\mathbf{a}_3$ | = | $x_3 a\,\hat{\mathbf{x}} + \frac{1}{4} b\,\hat{\mathbf{y}} + z_3 c\,\hat{\mathbf{z}}$ | (4c) | Cl II |
| $\mathbf{B}_{10}$ | = | $\left(\frac{1}{2} - x_3\right)\mathbf{a}_1 + \frac{3}{4}\,\mathbf{a}_2 + \left(\frac{1}{2} + z_3\right)\mathbf{a}_3$ | = | $\left(\frac{1}{2} - x_3\right) a\,\hat{\mathbf{x}} + \frac{3}{4} b\,\hat{\mathbf{y}} + \left(\frac{1}{2} + z_3\right) c\,\hat{\mathbf{z}}$ | (4c) | Cl II |
| $\mathbf{B}_{11}$ | = | $-x_3\,\mathbf{a}_1 + \frac{3}{4}\,\mathbf{a}_2 - z_3\,\mathbf{a}_3$ | = | $-x_3 a\,\hat{\mathbf{x}} + \frac{3}{4} b\,\hat{\mathbf{y}} - z_3 c\,\hat{\mathbf{z}}$ | (4c) | Cl II |
| $\mathbf{B}_{12}$ | = | $\left(\frac{1}{2} + x_3\right)\mathbf{a}_1 + \frac{1}{4}\,\mathbf{a}_2 + \left(\frac{1}{2} - z_3\right)\mathbf{a}_3$ | = | $\left(\frac{1}{2} + x_3\right) a\,\hat{\mathbf{x}} + \frac{1}{4} b\,\hat{\mathbf{y}} + \left(\frac{1}{2} - z_3\right) c\,\hat{\mathbf{z}}$ | (4c) | Cl II |
| $\mathbf{B}_{13}$ | = | $x_4\,\mathbf{a}_1 + \frac{1}{4}\,\mathbf{a}_2 + z_4\,\mathbf{a}_3$ | = | $x_4 a\,\hat{\mathbf{x}} + \frac{1}{4} b\,\hat{\mathbf{y}} + z_4 c\,\hat{\mathbf{z}}$ | (4c) | Cl III |
| $\mathbf{B}_{14}$ | = | $\left(\frac{1}{2} - x_4\right)\mathbf{a}_1 + \frac{3}{4}\,\mathbf{a}_2 + \left(\frac{1}{2} + z_4\right)\mathbf{a}_3$ | = | $\left(\frac{1}{2} - x_4\right) a\,\hat{\mathbf{x}} + \frac{3}{4} b\,\hat{\mathbf{y}} + \left(\frac{1}{2} + z_4\right) c\,\hat{\mathbf{z}}$ | (4c) | Cl III |
| $\mathbf{B}_{15}$ | = | $-x_4\,\mathbf{a}_1 + \frac{3}{4}\,\mathbf{a}_2 - z_4\,\mathbf{a}_3$ | = | $-x_4 a\,\hat{\mathbf{x}} + \frac{3}{4} b\,\hat{\mathbf{y}} - z_4 c\,\hat{\mathbf{z}}$ | (4c) | Cl III |
| $\mathbf{B}_{16}$ | = | $\left(\frac{1}{2} + x_4\right)\mathbf{a}_1 + \frac{1}{4}\,\mathbf{a}_2 + \left(\frac{1}{2} - z_4\right)\mathbf{a}_3$ | = | $\left(\frac{1}{2} + x_4\right) a\,\hat{\mathbf{x}} + \frac{1}{4} b\,\hat{\mathbf{y}} + \left(\frac{1}{2} - z_4\right) c\,\hat{\mathbf{z}}$ | (4c) | Cl III |
| $\mathbf{B}_{17}$ | = | $x_5\,\mathbf{a}_1 + \frac{1}{4}\,\mathbf{a}_2 + z_5\,\mathbf{a}_3$ | = | $x_5 a\,\hat{\mathbf{x}} + \frac{1}{4} b\,\hat{\mathbf{y}} + z_5 c\,\hat{\mathbf{z}}$ | (4c) | $NH_4$ |
| $\mathbf{B}_{18}$ | = | $\left(\frac{1}{2} - x_5\right)\mathbf{a}_1 + \frac{3}{4}\,\mathbf{a}_2 + \left(\frac{1}{2} + z_5\right)\mathbf{a}_3$ | = | $\left(\frac{1}{2} - x_5\right) a\,\hat{\mathbf{x}} + \frac{3}{4} b\,\hat{\mathbf{y}} + \left(\frac{1}{2} + z_5\right) c\,\hat{\mathbf{z}}$ | (4c) | $NH_4$ |
| $\mathbf{B}_{19}$ | = | $-x_5\,\mathbf{a}_1 + \frac{3}{4}\,\mathbf{a}_2 - z_5\,\mathbf{a}_3$ | = | $-x_5 a\,\hat{\mathbf{x}} + \frac{3}{4} b\,\hat{\mathbf{y}} - z_5 c\,\hat{\mathbf{z}}$ | (4c) | $NH_4$ |
| $\mathbf{B}_{20}$ | = | $\left(\frac{1}{2} + x_5\right)\mathbf{a}_1 + \frac{1}{4}\,\mathbf{a}_2 + \left(\frac{1}{2} - z_5\right)\mathbf{a}_3$ | = | $\left(\frac{1}{2} + x_5\right) a\,\hat{\mathbf{x}} + \frac{1}{4} b\,\hat{\mathbf{y}} + \left(\frac{1}{2} - z_5\right) c\,\hat{\mathbf{z}}$ | (4c) | $NH_4$ |

**Geometry files:**

- CIF: pp. 1646
- POSCAR: pp. 1646

# Berthierite ($FeSb_2S_4$, $E3_3$) Structure: AB4C2_oP28_62_c_4c_2c

http://aflow.org/prototype-encyclopedia/AB4C2_oP28_62_c_4c_2c

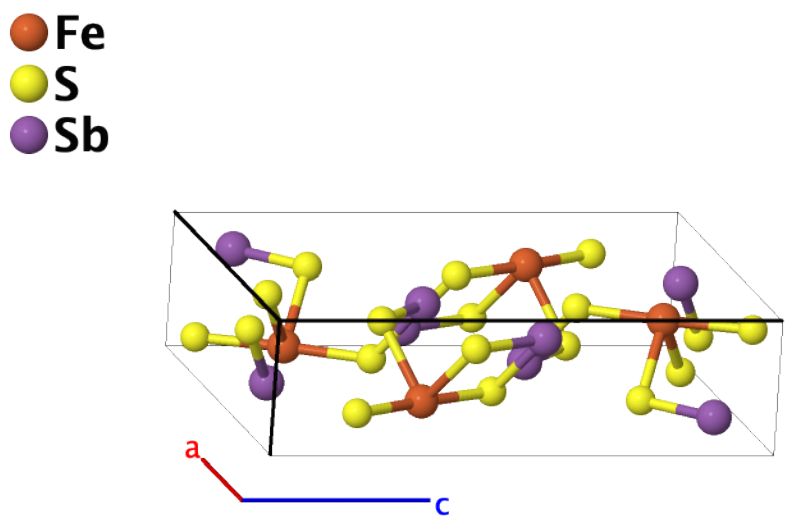

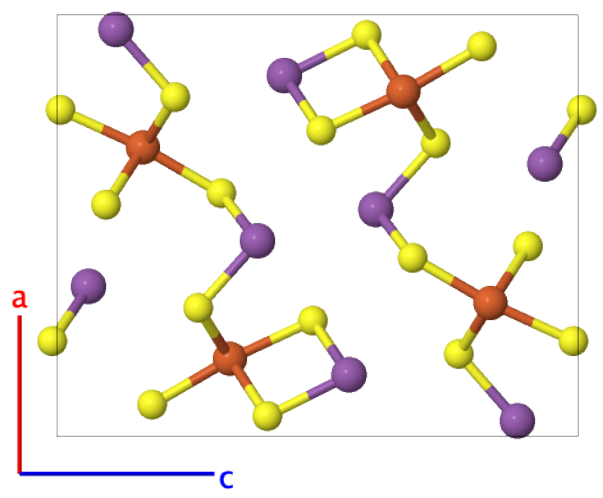

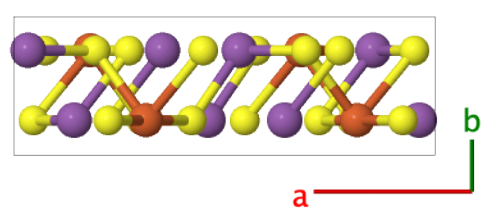

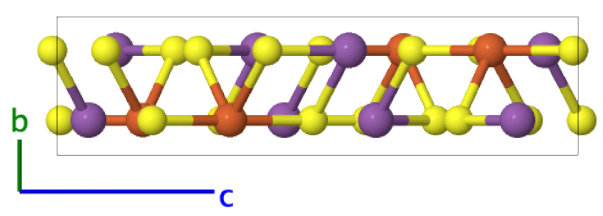

| | | |
|---|---|---|
| **Prototype** | : | $FeS_4Sb_2$ |
| **AFLOW prototype label** | : | AB4C2_oP28_62_c_4c_2c |
| ***Strukturbericht* designation** | : | $E3_3$ |
| **Pearson symbol** | : | oP28 |
| **Space group number** | : | 62 |
| **Space group symbol** | : | *Pnma* |
| **AFLOW prototype command** | : | `aflow --proto=AB4C2_oP28_62_c_4c_2c` `--params=`$a, b/a, c/a, x_1, z_1, x_2, z_2, x_3, z_3, x_4, z_4, x_5, z_5, x_6, z_6, x_7, z_7$ |

- The data for this structure was given in the *Pnam* orientation of space group #62. We used FINDSYM to place it in the standard *Pnma* orientation.

**Simple Orthorhombic primitive vectors:**

$$\mathbf{a}_1 = a\,\hat{\mathbf{x}}$$
$$\mathbf{a}_2 = b\,\hat{\mathbf{y}}$$
$$\mathbf{a}_3 = c\,\hat{\mathbf{z}}$$

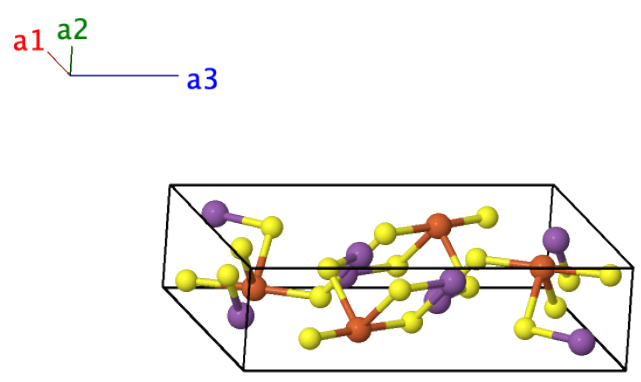

**Basis vectors:**

| Lattice Coordinates | Cartesian Coordinates | Wyckoff Position | Atom Type |
|---|---|---|---|

$$
\begin{array}{llllll}
\mathbf{B}_{1} & = & x_{1}\,\mathbf{a}_{1}+\frac{1}{4}\,\mathbf{a}_{2}+z_{1}\,\mathbf{a}_{3} & = & x_{1}a\,\mathbf{\hat{x}}+\frac{1}{4}b\,\mathbf{\hat{y}}+z_{1}c\,\mathbf{\hat{z}} & (4c) & \text{Fe} \\
\mathbf{B}_{2} & = & \left(\frac{1}{2}-x_{1}\right)\mathbf{a}_{1}+\frac{3}{4}\,\mathbf{a}_{2}+\left(\frac{1}{2}+z_{1}\right)\mathbf{a}_{3} & = & \left(\frac{1}{2}-x_{1}\right)a\,\mathbf{\hat{x}}+\frac{3}{4}b\,\mathbf{\hat{y}}+\left(\frac{1}{2}+z_{1}\right)c\,\mathbf{\hat{z}} & (4c) & \text{Fe} \\
\mathbf{B}_{3} & = & -x_{1}\,\mathbf{a}_{1}+\frac{3}{4}\,\mathbf{a}_{2}-z_{1}\,\mathbf{a}_{3} & = & -x_{1}a\,\mathbf{\hat{x}}+\frac{3}{4}b\,\mathbf{\hat{y}}-z_{1}c\,\mathbf{\hat{z}} & (4c) & \text{Fe} \\
\mathbf{B}_{4} & = & \left(\frac{1}{2}+x_{1}\right)\mathbf{a}_{1}+\frac{1}{4}\,\mathbf{a}_{2}+\left(\frac{1}{2}-z_{1}\right)\mathbf{a}_{3} & = & \left(\frac{1}{2}+x_{1}\right)a\,\mathbf{\hat{x}}+\frac{1}{4}b\,\mathbf{\hat{y}}+\left(\frac{1}{2}-z_{1}\right)c\,\mathbf{\hat{z}} & (4c) & \text{Fe} \\
\mathbf{B}_{5} & = & x_{2}\,\mathbf{a}_{1}+\frac{1}{4}\,\mathbf{a}_{2}+z_{2}\,\mathbf{a}_{3} & = & x_{2}a\,\mathbf{\hat{x}}+\frac{1}{4}b\,\mathbf{\hat{y}}+z_{2}c\,\mathbf{\hat{z}} & (4c) & \text{S I} \\
\mathbf{B}_{6} & = & \left(\frac{1}{2}-x_{2}\right)\mathbf{a}_{1}+\frac{3}{4}\,\mathbf{a}_{2}+\left(\frac{1}{2}+z_{2}\right)\mathbf{a}_{3} & = & \left(\frac{1}{2}-x_{2}\right)a\,\mathbf{\hat{x}}+\frac{3}{4}b\,\mathbf{\hat{y}}+\left(\frac{1}{2}+z_{2}\right)c\,\mathbf{\hat{z}} & (4c) & \text{S I} \\
\mathbf{B}_{7} & = & -x_{2}\,\mathbf{a}_{1}+\frac{3}{4}\,\mathbf{a}_{2}-z_{2}\,\mathbf{a}_{3} & = & -x_{2}a\,\mathbf{\hat{x}}+\frac{3}{4}b\,\mathbf{\hat{y}}-z_{2}c\,\mathbf{\hat{z}} & (4c) & \text{S I} \\
\mathbf{B}_{8} & = & \left(\frac{1}{2}+x_{2}\right)\mathbf{a}_{1}+\frac{1}{4}\,\mathbf{a}_{2}+\left(\frac{1}{2}-z_{2}\right)\mathbf{a}_{3} & = & \left(\frac{1}{2}+x_{2}\right)a\,\mathbf{\hat{x}}+\frac{1}{4}b\,\mathbf{\hat{y}}+\left(\frac{1}{2}-z_{2}\right)c\,\mathbf{\hat{z}} & (4c) & \text{S I} \\
\mathbf{B}_{9} & = & x_{3}\,\mathbf{a}_{1}+\frac{1}{4}\,\mathbf{a}_{2}+z_{3}\,\mathbf{a}_{3} & = & x_{3}a\,\mathbf{\hat{x}}+\frac{1}{4}b\,\mathbf{\hat{y}}+z_{3}c\,\mathbf{\hat{z}} & (4c) & \text{S II} \\
\mathbf{B}_{10} & = & \left(\frac{1}{2}-x_{3}\right)\mathbf{a}_{1}+\frac{3}{4}\,\mathbf{a}_{2}+\left(\frac{1}{2}+z_{3}\right)\mathbf{a}_{3} & = & \left(\frac{1}{2}-x_{3}\right)a\,\mathbf{\hat{x}}+\frac{3}{4}b\,\mathbf{\hat{y}}+\left(\frac{1}{2}+z_{3}\right)c\,\mathbf{\hat{z}} & (4c) & \text{S II} \\
\mathbf{B}_{11} & = & -x_{3}\,\mathbf{a}_{1}+\frac{3}{4}\,\mathbf{a}_{2}-z_{3}\,\mathbf{a}_{3} & = & -x_{3}a\,\mathbf{\hat{x}}+\frac{3}{4}b\,\mathbf{\hat{y}}-z_{3}c\,\mathbf{\hat{z}} & (4c) & \text{S II} \\
\mathbf{B}_{12} & = & \left(\frac{1}{2}+x_{3}\right)\mathbf{a}_{1}+\frac{1}{4}\,\mathbf{a}_{2}+\left(\frac{1}{2}-z_{3}\right)\mathbf{a}_{3} & = & \left(\frac{1}{2}+x_{3}\right)a\,\mathbf{\hat{x}}+\frac{1}{4}b\,\mathbf{\hat{y}}+\left(\frac{1}{2}-z_{3}\right)c\,\mathbf{\hat{z}} & (4c) & \text{S II} \\
\mathbf{B}_{13} & = & x_{4}\,\mathbf{a}_{1}+\frac{1}{4}\,\mathbf{a}_{2}+z_{4}\,\mathbf{a}_{3} & = & x_{4}a\,\mathbf{\hat{x}}+\frac{1}{4}b\,\mathbf{\hat{y}}+z_{4}c\,\mathbf{\hat{z}} & (4c) & \text{S III} \\
\mathbf{B}_{14} & = & \left(\frac{1}{2}-x_{4}\right)\mathbf{a}_{1}+\frac{3}{4}\,\mathbf{a}_{2}+\left(\frac{1}{2}+z_{4}\right)\mathbf{a}_{3} & = & \left(\frac{1}{2}-x_{4}\right)a\,\mathbf{\hat{x}}+\frac{3}{4}b\,\mathbf{\hat{y}}+\left(\frac{1}{2}+z_{4}\right)c\,\mathbf{\hat{z}} & (4c) & \text{S III} \\
\mathbf{B}_{15} & = & -x_{4}\,\mathbf{a}_{1}+\frac{3}{4}\,\mathbf{a}_{2}-z_{4}\,\mathbf{a}_{3} & = & -x_{4}a\,\mathbf{\hat{x}}+\frac{3}{4}b\,\mathbf{\hat{y}}-z_{4}c\,\mathbf{\hat{z}} & (4c) & \text{S III} \\
\mathbf{B}_{16} & = & \left(\frac{1}{2}+x_{4}\right)\mathbf{a}_{1}+\frac{1}{4}\,\mathbf{a}_{2}+\left(\frac{1}{2}-z_{4}\right)\mathbf{a}_{3} & = & \left(\frac{1}{2}+x_{4}\right)a\,\mathbf{\hat{x}}+\frac{1}{4}b\,\mathbf{\hat{y}}+\left(\frac{1}{2}-z_{4}\right)c\,\mathbf{\hat{z}} & (4c) & \text{S III} \\
\mathbf{B}_{17} & = & x_{5}\,\mathbf{a}_{1}+\frac{1}{4}\,\mathbf{a}_{2}+z_{5}\,\mathbf{a}_{3} & = & x_{5}a\,\mathbf{\hat{x}}+\frac{1}{4}b\,\mathbf{\hat{y}}+z_{5}c\,\mathbf{\hat{z}} & (4c) & \text{S IV} \\
\mathbf{B}_{18} & = & \left(\frac{1}{2}-x_{5}\right)\mathbf{a}_{1}+\frac{3}{4}\,\mathbf{a}_{2}+\left(\frac{1}{2}+z_{5}\right)\mathbf{a}_{3} & = & \left(\frac{1}{2}-x_{5}\right)a\,\mathbf{\hat{x}}+\frac{3}{4}b\,\mathbf{\hat{y}}+\left(\frac{1}{2}+z_{5}\right)c\,\mathbf{\hat{z}} & (4c) & \text{S IV} \\
\mathbf{B}_{19} & = & -x_{5}\,\mathbf{a}_{1}+\frac{3}{4}\,\mathbf{a}_{2}-z_{5}\,\mathbf{a}_{3} & = & -x_{5}a\,\mathbf{\hat{x}}+\frac{3}{4}b\,\mathbf{\hat{y}}-z_{5}c\,\mathbf{\hat{z}} & (4c) & \text{S IV} \\
\mathbf{B}_{20} & = & \left(\frac{1}{2}+x_{5}\right)\mathbf{a}_{1}+\frac{1}{4}\,\mathbf{a}_{2}+\left(\frac{1}{2}-z_{5}\right)\mathbf{a}_{3} & = & \left(\frac{1}{2}+x_{5}\right)a\,\mathbf{\hat{x}}+\frac{1}{4}b\,\mathbf{\hat{y}}+\left(\frac{1}{2}-z_{5}\right)c\,\mathbf{\hat{z}} & (4c) & \text{S IV} \\
\mathbf{B}_{21} & = & x_{6}\,\mathbf{a}_{1}+\frac{1}{4}\,\mathbf{a}_{2}+z_{6}\,\mathbf{a}_{3} & = & x_{6}a\,\mathbf{\hat{x}}+\frac{1}{4}b\,\mathbf{\hat{y}}+z_{6}c\,\mathbf{\hat{z}} & (4c) & \text{Sb I} \\
\mathbf{B}_{22} & = & \left(\frac{1}{2}-x_{6}\right)\mathbf{a}_{1}+\frac{3}{4}\,\mathbf{a}_{2}+\left(\frac{1}{2}+z_{6}\right)\mathbf{a}_{3} & = & \left(\frac{1}{2}-x_{6}\right)a\,\mathbf{\hat{x}}+\frac{3}{4}b\,\mathbf{\hat{y}}+\left(\frac{1}{2}+z_{6}\right)c\,\mathbf{\hat{z}} & (4c) & \text{Sb I} \\
\mathbf{B}_{23} & = & -x_{6}\,\mathbf{a}_{1}+\frac{3}{4}\,\mathbf{a}_{2}-z_{6}\,\mathbf{a}_{3} & = & -x_{6}a\,\mathbf{\hat{x}}+\frac{3}{4}b\,\mathbf{\hat{y}}-z_{6}c\,\mathbf{\hat{z}} & (4c) & \text{Sb I} \\
\mathbf{B}_{24} & = & \left(\frac{1}{2}+x_{6}\right)\mathbf{a}_{1}+\frac{1}{4}\,\mathbf{a}_{2}+\left(\frac{1}{2}-z_{6}\right)\mathbf{a}_{3} & = & \left(\frac{1}{2}+x_{6}\right)a\,\mathbf{\hat{x}}+\frac{1}{4}b\,\mathbf{\hat{y}}+\left(\frac{1}{2}-z_{6}\right)c\,\mathbf{\hat{z}} & (4c) & \text{Sb I} \\
\mathbf{B}_{25} & = & x_{7}\,\mathbf{a}_{1}+\frac{1}{4}\,\mathbf{a}_{2}+z_{7}\,\mathbf{a}_{3} & = & x_{7}a\,\mathbf{\hat{x}}+\frac{1}{4}b\,\mathbf{\hat{y}}+z_{7}c\,\mathbf{\hat{z}} & (4c) & \text{Sb II} \\
\mathbf{B}_{26} & = & \left(\frac{1}{2}-x_{7}\right)\mathbf{a}_{1}+\frac{3}{4}\,\mathbf{a}_{2}+\left(\frac{1}{2}+z_{7}\right)\mathbf{a}_{3} & = & \left(\frac{1}{2}-x_{7}\right)a\,\mathbf{\hat{x}}+\frac{3}{4}b\,\mathbf{\hat{y}}+\left(\frac{1}{2}+z_{7}\right)c\,\mathbf{\hat{z}} & (4c) & \text{Sb II} \\
\mathbf{B}_{27} & = & -x_{7}\,\mathbf{a}_{1}+\frac{3}{4}\,\mathbf{a}_{2}-z_{7}\,\mathbf{a}_{3} & = & -x_{7}a\,\mathbf{\hat{x}}+\frac{3}{4}b\,\mathbf{\hat{y}}-z_{7}c\,\mathbf{\hat{z}} & (4c) & \text{Sb II} \\
\mathbf{B}_{28} & = & \left(\frac{1}{2}+x_{7}\right)\mathbf{a}_{1}+\frac{1}{4}\,\mathbf{a}_{2}+\left(\frac{1}{2}-z_{7}\right)\mathbf{a}_{3} & = & \left(\frac{1}{2}+x_{7}\right)a\,\mathbf{\hat{x}}+\frac{1}{4}b\,\mathbf{\hat{y}}+\left(\frac{1}{2}-z_{7}\right)c\,\mathbf{\hat{z}} & (4c) & \text{Sb II}
\end{array}
$$

**Geometry files:**

- CIF: pp. 1646
- POSCAR: pp. 1646

# Chalcocyanite ($CuSO_4$) Structure: AB4C_oP24_62_a_2cd_c

http://aflow.org/prototype-encyclopedia/AB4C_oP24_62_a_2cd_c

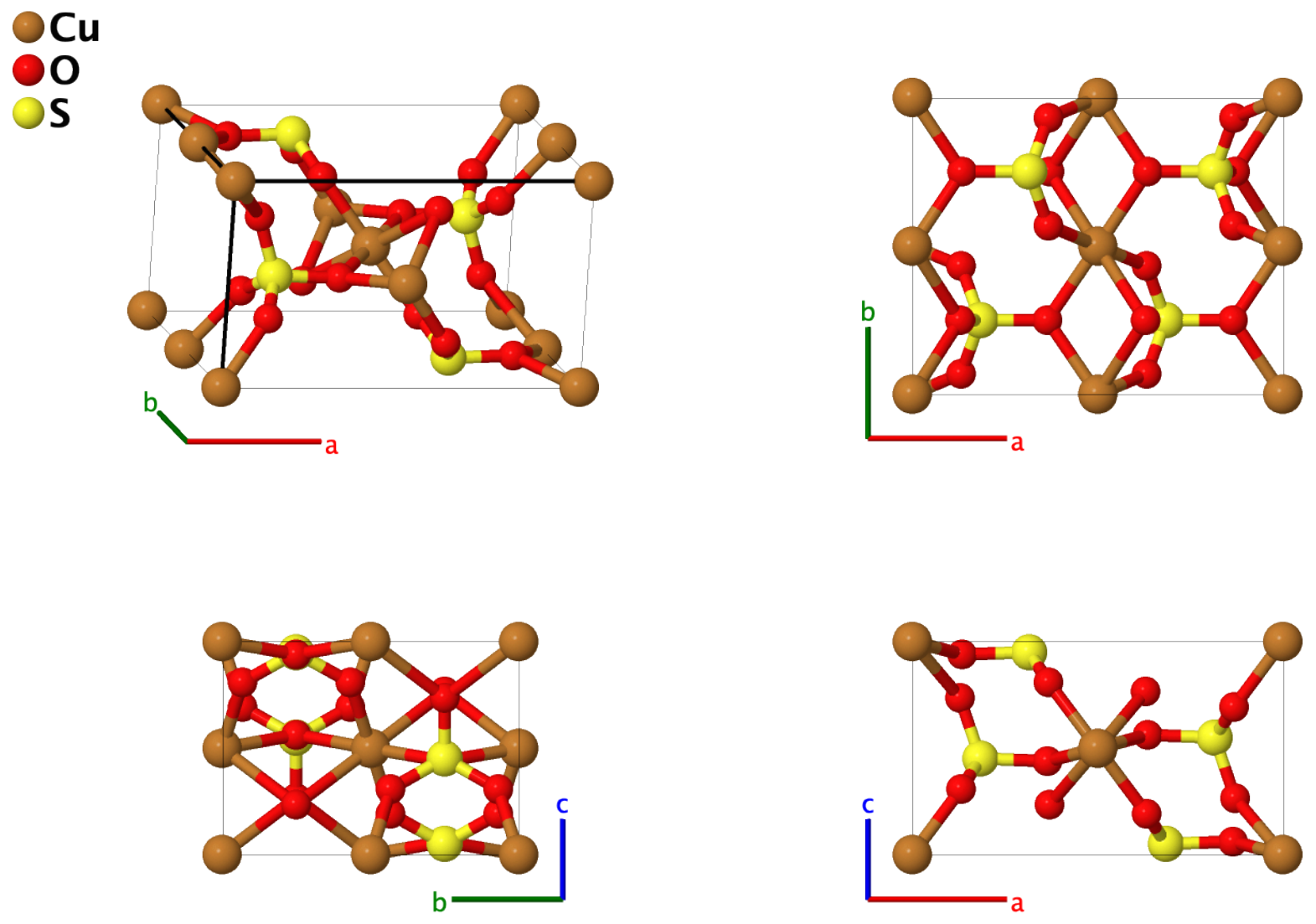

| | | |
|---|---|---|
| **Prototype** | : | $CuO_4S$ |
| **AFLOW prototype label** | : | AB4C_oP24_62_a_2cd_c |
| ***Strukturbericht* designation** | : | None |
| **Pearson symbol** | : | oP24 |
| **Space group number** | : | 62 |
| **Space group symbol** | : | *Pnma* |
| **AFLOW prototype command** | : | `aflow --proto=AB4C_oP24_62_a_2cd_c` `--params=`$a, b/a, c/a, x_2, z_2, x_3, z_3, x_4, z_4, x_5, y_5, z_5$ |

### Other compounds with this structure

- $ZnSO_4$ (Zincosite)

### Simple Orthorhombic primitive vectors:

$$\mathbf{a}_1 = a\,\hat{\mathbf{x}}$$
$$\mathbf{a}_2 = b\,\hat{\mathbf{y}}$$
$$\mathbf{a}_3 = c\,\hat{\mathbf{z}}$$

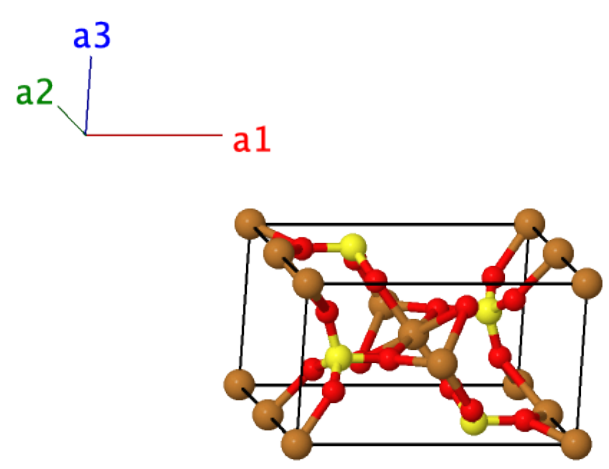

### Basis vectors:

| | | Lattice Coordinates | | Cartesian Coordinates | Wyckoff Position | Atom Type |
|---|---|---|---|---|---|---|
| $\mathbf{B}_1$ | = | $0\,\mathbf{a}_1 + 0\,\mathbf{a}_2 + 0\,\mathbf{a}_3$ | = | $0\,\hat{\mathbf{x}} + 0\,\hat{\mathbf{y}} + 0\,\hat{\mathbf{z}}$ | (4*a*) | Cu |
| $\mathbf{B}_2$ | = | $\frac{1}{2}\,\mathbf{a}_1 + \frac{1}{2}\,\mathbf{a}_3$ | = | $\frac{1}{2}a\,\hat{\mathbf{x}} + \frac{1}{2}c\,\hat{\mathbf{z}}$ | (4*a*) | Cu |
| $\mathbf{B}_3$ | = | $\frac{1}{2}\,\mathbf{a}_2$ | = | $\frac{1}{2}b\,\hat{\mathbf{y}}$ | (4$a$) | Cu |
| $\mathbf{B}_4$ | = | $\frac{1}{2}\,\mathbf{a}_1 + \frac{1}{2}\,\mathbf{a}_2 + \frac{1}{2}\,\mathbf{a}_3$ | = | $\frac{1}{2}a\,\hat{\mathbf{x}} + \frac{1}{2}b\,\hat{\mathbf{y}} + \frac{1}{2}c\,\hat{\mathbf{z}}$ | (4$a$) | Cu |
| $\mathbf{B}_5$ | = | $x_2\,\mathbf{a}_1 + \frac{1}{4}\,\mathbf{a}_2 + z_2\,\mathbf{a}_3$ | = | $x_2a\,\hat{\mathbf{x}} + \frac{1}{4}b\,\hat{\mathbf{y}} + z_2c\,\hat{\mathbf{z}}$ | (4$c$) | O I |
| $\mathbf{B}_6$ | = | $\left(\frac{1}{2} - x_2\right)\,\mathbf{a}_1 + \frac{3}{4}\,\mathbf{a}_2 + \left(\frac{1}{2} + z_2\right)\,\mathbf{a}_3$ | = | $\left(\frac{1}{2} - x_2\right)a\,\hat{\mathbf{x}} + \frac{3}{4}b\,\hat{\mathbf{y}} + \left(\frac{1}{2} + z_2\right)c\,\hat{\mathbf{z}}$ | (4$c$) | O I |
| $\mathbf{B}_7$ | = | $-x_2\,\mathbf{a}_1 + \frac{3}{4}\,\mathbf{a}_2 - z_2\,\mathbf{a}_3$ | = | $-x_2a\,\hat{\mathbf{x}} + \frac{3}{4}b\,\hat{\mathbf{y}} - z_2c\,\hat{\mathbf{z}}$ | (4$c$) | O I |
| $\mathbf{B}_8$ | = | $\left(\frac{1}{2} + x_2\right)\,\mathbf{a}_1 + \frac{1}{4}\,\mathbf{a}_2 + \left(\frac{1}{2} - z_2\right)\,\mathbf{a}_3$ | = | $\left(\frac{1}{2} + x_2\right)a\,\hat{\mathbf{x}} + \frac{1}{4}b\,\hat{\mathbf{y}} + \left(\frac{1}{2} - z_2\right)c\,\hat{\mathbf{z}}$ | (4$c$) | O I |
| $\mathbf{B}_9$ | = | $x_3\,\mathbf{a}_1 + \frac{1}{4}\,\mathbf{a}_2 + z_3\,\mathbf{a}_3$ | = | $x_3a\,\hat{\mathbf{x}} + \frac{1}{4}b\,\hat{\mathbf{y}} + z_3c\,\hat{\mathbf{z}}$ | (4$c$) | O II |
| $\mathbf{B}_{10}$ | = | $\left(\frac{1}{2} - x_3\right)\,\mathbf{a}_1 + \frac{3}{4}\,\mathbf{a}_2 + \left(\frac{1}{2} + z_3\right)\,\mathbf{a}_3$ | = | $\left(\frac{1}{2} - x_3\right)a\,\hat{\mathbf{x}} + \frac{3}{4}b\,\hat{\mathbf{y}} + \left(\frac{1}{2} + z_3\right)c\,\hat{\mathbf{z}}$ | (4$c$) | O II |
| $\mathbf{B}_{11}$ | = | $-x_3\,\mathbf{a}_1 + \frac{3}{4}\,\mathbf{a}_2 - z_3\,\mathbf{a}_3$ | = | $-x_3a\,\hat{\mathbf{x}} + \frac{3}{4}b\,\hat{\mathbf{y}} - z_3c\,\hat{\mathbf{z}}$ | (4$c$) | O II |
| $\mathbf{B}_{12}$ | = | $\left(\frac{1}{2} + x_3\right)\,\mathbf{a}_1 + \frac{1}{4}\,\mathbf{a}_2 + \left(\frac{1}{2} - z_3\right)\,\mathbf{a}_3$ | = | $\left(\frac{1}{2} + x_3\right)a\,\hat{\mathbf{x}} + \frac{1}{4}b\,\hat{\mathbf{y}} + \left(\frac{1}{2} - z_3\right)c\,\hat{\mathbf{z}}$ | (4$c$) | O II |
| $\mathbf{B}_{13}$ | = | $x_4\,\mathbf{a}_1 + \frac{1}{4}\,\mathbf{a}_2 + z_4\,\mathbf{a}_3$ | = | $x_4a\,\hat{\mathbf{x}} + \frac{1}{4}b\,\hat{\mathbf{y}} + z_4c\,\hat{\mathbf{z}}$ | (4$c$) | S |
| $\mathbf{B}_{14}$ | = | $\left(\frac{1}{2} - x_4\right)\,\mathbf{a}_1 + \frac{3}{4}\,\mathbf{a}_2 + \left(\frac{1}{2} + z_4\right)\,\mathbf{a}_3$ | = | $\left(\frac{1}{2} - x_4\right)a\,\hat{\mathbf{x}} + \frac{3}{4}b\,\hat{\mathbf{y}} + \left(\frac{1}{2} + z_4\right)c\,\hat{\mathbf{z}}$ | (4$c$) | S |
| $\mathbf{B}_{15}$ | = | $-x_4\,\mathbf{a}_1 + \frac{3}{4}\,\mathbf{a}_2 - z_4\,\mathbf{a}_3$ | = | $-x_4a\,\hat{\mathbf{x}} + \frac{3}{4}b\,\hat{\mathbf{y}} - z_4c\,\hat{\mathbf{z}}$ | (4$c$) | S |
| $\mathbf{B}_{16}$ | = | $\left(\frac{1}{2} + x_4\right)\,\mathbf{a}_1 + \frac{1}{4}\,\mathbf{a}_2 + \left(\frac{1}{2} - z_4\right)\,\mathbf{a}_3$ | = | $\left(\frac{1}{2} + x_4\right)a\,\hat{\mathbf{x}} + \frac{1}{4}b\,\hat{\mathbf{y}} + \left(\frac{1}{2} - z_4\right)c\,\hat{\mathbf{z}}$ | (4$c$) | S |
| $\mathbf{B}_{17}$ | = | $x_5\,\mathbf{a}_1 + y_5\,\mathbf{a}_2 + z_5\,\mathbf{a}_3$ | = | $x_5a\,\hat{\mathbf{x}} + y_5b\,\hat{\mathbf{y}} + z_5c\,\hat{\mathbf{z}}$ | (8$d$) | O III |
| $\mathbf{B}_{18}$ | = | $\left(\frac{1}{2} - x_5\right)\,\mathbf{a}_1 - y_5\,\mathbf{a}_2 + \left(\frac{1}{2} + z_5\right)\,\mathbf{a}_3$ | = | $\left(\frac{1}{2} - x_5\right)a\,\hat{\mathbf{x}} - y_5b\,\hat{\mathbf{y}} + \left(\frac{1}{2} + z_5\right)c\,\hat{\mathbf{z}}$ | (8$d$) | O III |
| $\mathbf{B}_{19}$ | = | $-x_5\,\mathbf{a}_1 + \left(\frac{1}{2} + y_5\right)\,\mathbf{a}_2 - z_5\,\mathbf{a}_3$ | = | $-x_5a\,\hat{\mathbf{x}} + \left(\frac{1}{2} + y_5\right)b\,\hat{\mathbf{y}} - z_5c\,\hat{\mathbf{z}}$ | (8$d$) | O III |
| $\mathbf{B}_{20}$ | = | $\left(\frac{1}{2} + x_5\right)\,\mathbf{a}_1 + \left(\frac{1}{2} - y_5\right)\,\mathbf{a}_2 + \left(\frac{1}{2} - z_5\right)\,\mathbf{a}_3$ | = | $\left(\frac{1}{2} + x_5\right)a\,\hat{\mathbf{x}} + \left(\frac{1}{2} - y_5\right)b\,\hat{\mathbf{y}} + \left(\frac{1}{2} - z_5\right)c\,\hat{\mathbf{z}}$ | (8$d$) | O III |
| $\mathbf{B}_{21}$ | = | $-x_5\,\mathbf{a}_1 - y_5\,\mathbf{a}_2 - z_5\,\mathbf{a}_3$ | = | $-x_5a\,\hat{\mathbf{x}} - y_5b\,\hat{\mathbf{y}} - z_5c\,\hat{\mathbf{z}}$ | (8$d$) | O III |
| $\mathbf{B}_{22}$ | = | $\left(\frac{1}{2} + x_5\right)\,\mathbf{a}_1 + y_5\,\mathbf{a}_2 + \left(\frac{1}{2} - z_5\right)\,\mathbf{a}_3$ | = | $\left(\frac{1}{2} + x_5\right)a\,\hat{\mathbf{x}} + y_5b\,\hat{\mathbf{y}} + \left(\frac{1}{2} - z_5\right)c\,\hat{\mathbf{z}}$ | (8$d$) | O III |
| $\mathbf{B}_{23}$ | = | $x_5\,\mathbf{a}_1 + \left(\frac{1}{2} - y_5\right)\,\mathbf{a}_2 + z_5\,\mathbf{a}_3$ | = | $x_5a\,\hat{\mathbf{x}} + \left(\frac{1}{2} - y_5\right)b\,\hat{\mathbf{y}} + z_5c\,\hat{\mathbf{z}}$ | (8$d$) | O III |
| $\mathbf{B}_{24}$ | = | $\left(\frac{1}{2} - x_5\right)\,\mathbf{a}_1 + \left(\frac{1}{2} + y_5\right)\,\mathbf{a}_2 + \left(\frac{1}{2} + z_5\right)\,\mathbf{a}_3$ | = | $\left(\frac{1}{2} - x_5\right)a\,\hat{\mathbf{x}} + \left(\frac{1}{2} + y_5\right)b\,\hat{\mathbf{y}} + \left(\frac{1}{2} + z_5\right)c\,\hat{\mathbf{z}}$ | (8$d$) | O III |

**Geometry files:**

- CIF: pp. 1647
- POSCAR: pp. 1647

# Rynersonite (Orthorhombic $CaTa_2O_6$) Structure: AB6C2_oP36_62_c_2c2d_d

http://aflow.org/prototype-encyclopedia/AB6C2_oP36_62_c_2c2d_d

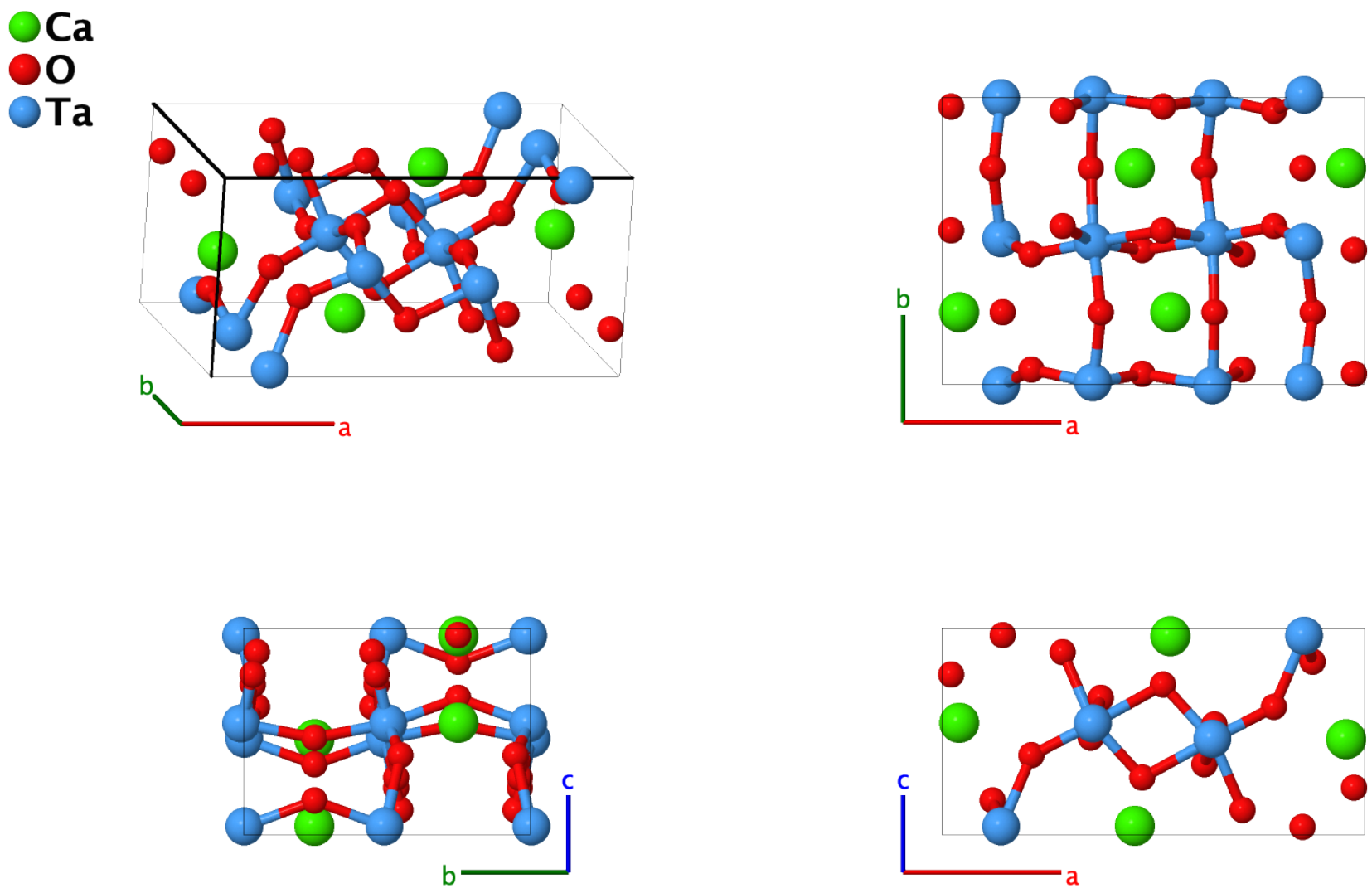

| | | |
|---|---|---|
| **Prototype** | : | $CaO_6Ta_2$ |
| **AFLOW prototype label** | : | AB6C2_oP36_62_c_2c2d_d |
| ***Strukturbericht* designation** | : | None |
| **Pearson symbol** | : | oP36 |
| **Space group number** | : | 62 |
| **Space group symbol** | : | *Pnma* |
| **AFLOW prototype command** | : | `aflow --proto=AB6C2_oP36_62_c_2c2d_d` `--params=`$a, b/a, c/a, x_1, z_1, x_2, z_2, x_3, z_3, x_4, y_4, z_4, x_5, y_5, z_5, x_6, y_6, z_6$ |

**Other compounds with this structure**

- Ca(Ta,Ti)$O_6$, $CaTa_2O_6$, $EuTa_2O_6$, $GdTa_2O_6$, $GdTi_2O_6$, La(Nb,Ti)$O_6$, $LaTa_2O_6$, $LaTi_2O_6$, Nd(Ta,Ti)$O_6$, Pr(Ta,Ti)$O_6$, $SrNb_2O_6$, Tb(Ta,Ti)$O_6$, $YTa_2O_6$, and $(Sr_{0.7}La_{0.3})Nb_2O_6$

- This is the room-temperature structure of $CaTa_2O_6$. At 500 °C, this takes on the cubic perovskite structure with a half-filled calcium site.

**Simple Orthorhombic primitive vectors:**

$$\mathbf{a}_1 = a\,\hat{\mathbf{x}}$$
$$\mathbf{a}_2 = b\,\hat{\mathbf{y}}$$
$$\mathbf{a}_3 = c\,\hat{\mathbf{z}}$$

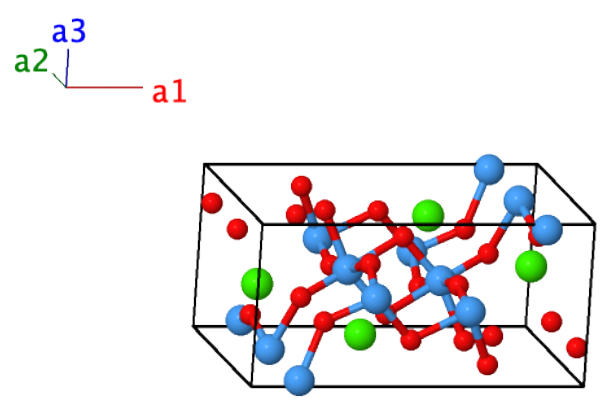

**Basis vectors:**

| | | Lattice Coordinates | | Cartesian Coordinates | Wyckoff Position | Atom Type |
|---|---|---|---|---|---|---|
| $\mathbf{B}_1$ | = | $x_1\,\mathbf{a}_1 + \frac{1}{4}\,\mathbf{a}_2 + z_1\,\mathbf{a}_3$ | = | $x_1 a\,\mathbf{\hat{x}} + \frac{1}{4} b\,\mathbf{\hat{y}} + z_1 c\,\mathbf{\hat{z}}$ | (4$c$) | Ca |
| $\mathbf{B}_2$ | = | $\left(\frac{1}{2} - x_1\right)\mathbf{a}_1 + \frac{3}{4}\,\mathbf{a}_2 + \left(\frac{1}{2} + z_1\right)\mathbf{a}_3$ | = | $\left(\frac{1}{2} - x_1\right) a\,\mathbf{\hat{x}} + \frac{3}{4} b\,\mathbf{\hat{y}} + \left(\frac{1}{2} + z_1\right) c\,\mathbf{\hat{z}}$ | (4$c$) | Ca |
| $\mathbf{B}_3$ | = | $-x_1\,\mathbf{a}_1 + \frac{3}{4}\,\mathbf{a}_2 - z_1\,\mathbf{a}_3$ | = | $-x_1 a\,\mathbf{\hat{x}} + \frac{3}{4} b\,\mathbf{\hat{y}} - z_1 c\,\mathbf{\hat{z}}$ | (4$c$) | Ca |
| $\mathbf{B}_4$ | = | $\left(\frac{1}{2} + x_1\right)\mathbf{a}_1 + \frac{1}{4}\,\mathbf{a}_2 + \left(\frac{1}{2} - z_1\right)\mathbf{a}_3$ | = | $\left(\frac{1}{2} + x_1\right) a\,\mathbf{\hat{x}} + \frac{1}{4} b\,\mathbf{\hat{y}} + \left(\frac{1}{2} - z_1\right) c\,\mathbf{\hat{z}}$ | (4$c$) | Ca |
| $\mathbf{B}_5$ | = | $x_2\,\mathbf{a}_1 + \frac{1}{4}\,\mathbf{a}_2 + z_2\,\mathbf{a}_3$ | = | $x_2 a\,\mathbf{\hat{x}} + \frac{1}{4} b\,\mathbf{\hat{y}} + z_2 c\,\mathbf{\hat{z}}$ | (4$c$) | O I |
| $\mathbf{B}_6$ | = | $\left(\frac{1}{2} - x_2\right)\mathbf{a}_1 + \frac{3}{4}\,\mathbf{a}_2 + \left(\frac{1}{2} + z_2\right)\mathbf{a}_3$ | = | $\left(\frac{1}{2} - x_2\right) a\,\mathbf{\hat{x}} + \frac{3}{4} b\,\mathbf{\hat{y}} + \left(\frac{1}{2} + z_2\right) c\,\mathbf{\hat{z}}$ | (4$c$) | O I |
| $\mathbf{B}_7$ | = | $-x_2\,\mathbf{a}_1 + \frac{3}{4}\,\mathbf{a}_2 - z_2\,\mathbf{a}_3$ | = | $-x_2 a\,\mathbf{\hat{x}} + \frac{3}{4} b\,\mathbf{\hat{y}} - z_2 c\,\mathbf{\hat{z}}$ | (4$c$) | O I |
| $\mathbf{B}_8$ | = | $\left(\frac{1}{2} + x_2\right)\mathbf{a}_1 + \frac{1}{4}\,\mathbf{a}_2 + \left(\frac{1}{2} - z_2\right)\mathbf{a}_3$ | = | $\left(\frac{1}{2} + x_2\right) a\,\mathbf{\hat{x}} + \frac{1}{4} b\,\mathbf{\hat{y}} + \left(\frac{1}{2} - z_2\right) c\,\mathbf{\hat{z}}$ | (4$c$) | O I |
| $\mathbf{B}_9$ | = | $x_3\,\mathbf{a}_1 + \frac{1}{4}\,\mathbf{a}_2 + z_3\,\mathbf{a}_3$ | = | $x_3 a\,\mathbf{\hat{x}} + \frac{1}{4} b\,\mathbf{\hat{y}} + z_3 c\,\mathbf{\hat{z}}$ | (4$c$) | O II |
| $\mathbf{B}_{10}$ | = | $\left(\frac{1}{2} - x_3\right)\mathbf{a}_1 + \frac{3}{4}\,\mathbf{a}_2 + \left(\frac{1}{2} + z_3\right)\mathbf{a}_3$ | = | $\left(\frac{1}{2} - x_3\right) a\,\mathbf{\hat{x}} + \frac{3}{4} b\,\mathbf{\hat{y}} + \left(\frac{1}{2} + z_3\right) c\,\mathbf{\hat{z}}$ | (4$c$) | O II |
| $\mathbf{B}_{11}$ | = | $-x_3\,\mathbf{a}_1 + \frac{3}{4}\,\mathbf{a}_2 - z_3\,\mathbf{a}_3$ | = | $-x_3 a\,\mathbf{\hat{x}} + \frac{3}{4} b\,\mathbf{\hat{y}} - z_3 c\,\mathbf{\hat{z}}$ | (4$c$) | O II |
| $\mathbf{B}_{12}$ | = | $\left(\frac{1}{2} + x_3\right)\mathbf{a}_1 + \frac{1}{4}\,\mathbf{a}_2 + \left(\frac{1}{2} - z_3\right)\mathbf{a}_3$ | = | $\left(\frac{1}{2} + x_3\right) a\,\mathbf{\hat{x}} + \frac{1}{4} b\,\mathbf{\hat{y}} + \left(\frac{1}{2} - z_3\right) c\,\mathbf{\hat{z}}$ | (4$c$) | O II |
| $\mathbf{B}_{13}$ | = | $x_4\,\mathbf{a}_1 + y_4\,\mathbf{a}_2 + z_4\,\mathbf{a}_3$ | = | $x_4 a\,\mathbf{\hat{x}} + y_4 b\,\mathbf{\hat{y}} + z_4 c\,\mathbf{\hat{z}}$ | (8$d$) | O III |
| $\mathbf{B}_{14}$ | = | $\left(\frac{1}{2} - x_4\right)\mathbf{a}_1 - y_4\,\mathbf{a}_2 + \left(\frac{1}{2} + z_4\right)\mathbf{a}_3$ | = | $\left(\frac{1}{2} - x_4\right) a\,\mathbf{\hat{x}} - y_4 b\,\mathbf{\hat{y}} + \left(\frac{1}{2} + z_4\right) c\,\mathbf{\hat{z}}$ | (8$d$) | O III |
| $\mathbf{B}_{15}$ | = | $-x_4\,\mathbf{a}_1 + \left(\frac{1}{2} + y_4\right)\mathbf{a}_2 - z_4\,\mathbf{a}_3$ | = | $-x_4 a\,\mathbf{\hat{x}} + \left(\frac{1}{2} + y_4\right) b\,\mathbf{\hat{y}} - z_4 c\,\mathbf{\hat{z}}$ | (8$d$) | O III |
| $\mathbf{B}_{16}$ | = | $\left(\frac{1}{2} + x_4\right)\mathbf{a}_1 + \left(\frac{1}{2} - y_4\right)\mathbf{a}_2 + \left(\frac{1}{2} - z_4\right)\mathbf{a}_3$ | = | $\left(\frac{1}{2} + x_4\right) a\,\mathbf{\hat{x}} + \left(\frac{1}{2} - y_4\right) b\,\mathbf{\hat{y}} + \left(\frac{1}{2} - z_4\right) c\,\mathbf{\hat{z}}$ | (8$d$) | O III |
| $\mathbf{B}_{17}$ | = | $-x_4\,\mathbf{a}_1 - y_4\,\mathbf{a}_2 - z_4\,\mathbf{a}_3$ | = | $-x_4 a\,\mathbf{\hat{x}} - y_4 b\,\mathbf{\hat{y}} - z_4 c\,\mathbf{\hat{z}}$ | (8$d$) | O III |
| $\mathbf{B}_{18}$ | = | $\left(\frac{1}{2} + x_4\right)\mathbf{a}_1 + y_4\,\mathbf{a}_2 + \left(\frac{1}{2} - z_4\right)\mathbf{a}_3$ | = | $\left(\frac{1}{2} + x_4\right) a\,\mathbf{\hat{x}} + y_4 b\,\mathbf{\hat{y}} + \left(\frac{1}{2} - z_4\right) c\,\mathbf{\hat{z}}$ | (8$d$) | O III |
| $\mathbf{B}_{19}$ | = | $x_4\,\mathbf{a}_1 + \left(\frac{1}{2} - y_4\right)\mathbf{a}_2 + z_4\,\mathbf{a}_3$ | = | $x_4 a\,\mathbf{\hat{x}} + \left(\frac{1}{2} - y_4\right) b\,\mathbf{\hat{y}} + z_4 c\,\mathbf{\hat{z}}$ | (8$d$) | O III |
| $\mathbf{B}_{20}$ | = | $\left(\frac{1}{2} - x_4\right)\mathbf{a}_1 + \left(\frac{1}{2} + y_4\right)\mathbf{a}_2 + \left(\frac{1}{2} + z_4\right)\mathbf{a}_3$ | = | $\left(\frac{1}{2} - x_4\right) a\,\mathbf{\hat{x}} + \left(\frac{1}{2} + y_4\right) b\,\mathbf{\hat{y}} + \left(\frac{1}{2} + z_4\right) c\,\mathbf{\hat{z}}$ | (8$d$) | O III |
| $\mathbf{B}_{21}$ | = | $x_5\,\mathbf{a}_1 + y_5\,\mathbf{a}_2 + z_5\,\mathbf{a}_3$ | = | $x_5 a\,\mathbf{\hat{x}} + y_5 b\,\mathbf{\hat{y}} + z_5 c\,\mathbf{\hat{z}}$ | (8$d$) | O IV |
| $\mathbf{B}_{22}$ | = | $\left(\frac{1}{2} - x_5\right)\mathbf{a}_1 - y_5\,\mathbf{a}_2 + \left(\frac{1}{2} + z_5\right)\mathbf{a}_3$ | = | $\left(\frac{1}{2} - x_5\right) a\,\mathbf{\hat{x}} - y_5 b\,\mathbf{\hat{y}} + \left(\frac{1}{2} + z_5\right) c\,\mathbf{\hat{z}}$ | (8$d$) | O IV |
| $\mathbf{B}_{23}$ | = | $-x_5\,\mathbf{a}_1 + \left(\frac{1}{2} + y_5\right)\mathbf{a}_2 - z_5\,\mathbf{a}_3$ | = | $-x_5 a\,\mathbf{\hat{x}} + \left(\frac{1}{2} + y_5\right) b\,\mathbf{\hat{y}} - z_5 c\,\mathbf{\hat{z}}$ | (8$d$) | O IV |
| $\mathbf{B}_{24}$ | = | $\left(\frac{1}{2} + x_5\right)\mathbf{a}_1 + \left(\frac{1}{2} - y_5\right)\mathbf{a}_2 + \left(\frac{1}{2} - z_5\right)\mathbf{a}_3$ | = | $\left(\frac{1}{2} + x_5\right) a\,\mathbf{\hat{x}} + \left(\frac{1}{2} - y_5\right) b\,\mathbf{\hat{y}} + \left(\frac{1}{2} - z_5\right) c\,\mathbf{\hat{z}}$ | (8$d$) | O IV |
| $\mathbf{B}_{25}$ | = | $-x_5\,\mathbf{a}_1 - y_5\,\mathbf{a}_2 - z_5\,\mathbf{a}_3$ | = | $-x_5 a\,\mathbf{\hat{x}} - y_5 b\,\mathbf{\hat{y}} - z_5 c\,\mathbf{\hat{z}}$ | (8$d$) | O IV |
| $\mathbf{B}_{26}$ | = | $\left(\frac{1}{2} + x_5\right)\mathbf{a}_1 + y_5\,\mathbf{a}_2 + \left(\frac{1}{2} - z_5\right)\mathbf{a}_3$ | = | $\left(\frac{1}{2} + x_5\right) a\,\mathbf{\hat{x}} + y_5 b\,\mathbf{\hat{y}} + \left(\frac{1}{2} - z_5\right) c\,\mathbf{\hat{z}}$ | (8$d$) | O IV |
| $\mathbf{B}_{27}$ | = | $x_5\,\mathbf{a}_1 + \left(\frac{1}{2} - y_5\right)\mathbf{a}_2 + z_5\,\mathbf{a}_3$ | = | $x_5 a\,\mathbf{\hat{x}} + \left(\frac{1}{2} - y_5\right) b\,\mathbf{\hat{y}} + z_5 c\,\mathbf{\hat{z}}$ | (8$d$) | O IV |
| $\mathbf{B}_{28}$ | = | $\left(\frac{1}{2} - x_5\right)\mathbf{a}_1 + \left(\frac{1}{2} + y_5\right)\mathbf{a}_2 + \left(\frac{1}{2} + z_5\right)\mathbf{a}_3$ | = | $\left(\frac{1}{2} - x_5\right) a\,\mathbf{\hat{x}} + \left(\frac{1}{2} + y_5\right) b\,\mathbf{\hat{y}} + \left(\frac{1}{2} + z_5\right) c\,\mathbf{\hat{z}}$ | (8$d$) | O IV |
| $\mathbf{B}_{29}$ | = | $x_6\,\mathbf{a}_1 + y_6\,\mathbf{a}_2 + z_6\,\mathbf{a}_3$ | = | $x_6 a\,\mathbf{\hat{x}} + y_6 b\,\mathbf{\hat{y}} + z_6 c\,\mathbf{\hat{z}}$ | (8$d$) | Ta |
| $\mathbf{B}_{30}$ | = | $\left(\frac{1}{2} - x_6\right)\mathbf{a}_1 - y_6\,\mathbf{a}_2 + \left(\frac{1}{2} + z_6\right)\mathbf{a}_3$ | = | $\left(\frac{1}{2} - x_6\right) a\,\mathbf{\hat{x}} - y_6 b\,\mathbf{\hat{y}} + \left(\frac{1}{2} + z_6\right) c\,\mathbf{\hat{z}}$ | (8$d$) | Ta |
| $\mathbf{B}_{31}$ | = | $-x_6\,\mathbf{a}_1 + \left(\frac{1}{2} + y_6\right)\mathbf{a}_2 - z_6\,\mathbf{a}_3$ | = | $-x_6 a\,\mathbf{\hat{x}} + \left(\frac{1}{2} + y_6\right) b\,\mathbf{\hat{y}} - z_6 c\,\mathbf{\hat{z}}$ | (8$d$) | Ta |
| $\mathbf{B}_{32}$ | = | $\left(\frac{1}{2}+x_6\right)\mathbf{a}_1+\left(\frac{1}{2}-y_6\right)\mathbf{a}_2+\left(\frac{1}{2}-z_6\right)\mathbf{a}_3$ | = | $\left(\frac{1}{2}+x_6\right)a\,\hat{\mathbf{x}}+\left(\frac{1}{2}-y_6\right)b\,\hat{\mathbf{y}}+\left(\frac{1}{2}-z_6\right)c\,\hat{\mathbf{z}}$ | (8*d*) | Ta |
| $\mathbf{B}_{33}$ | = | $-x_6\,\mathbf{a}_1-y_6\,\mathbf{a}_2-z_6\,\mathbf{a}_3$ | = | $-x_6a\,\hat{\mathbf{x}}-y_6b\,\hat{\mathbf{y}}-z_6c\,\hat{\mathbf{z}}$ | (8*d*) | Ta |
| $\mathbf{B}_{34}$ | = | $\left(\frac{1}{2}+x_6\right)\mathbf{a}_1+y_6\,\mathbf{a}_2+\left(\frac{1}{2}-z_6\right)\mathbf{a}_3$ | = | $\left(\frac{1}{2}+x_6\right)a\,\hat{\mathbf{x}}+y_6b\,\hat{\mathbf{y}}+\left(\frac{1}{2}-z_6\right)c\,\hat{\mathbf{z}}$ | (8*d*) | Ta |
| $\mathbf{B}_{35}$ | = | $x_6\,\mathbf{a}_1+\left(\frac{1}{2}-y_6\right)\mathbf{a}_2+z_6\,\mathbf{a}_3$ | = | $x_6a\,\hat{\mathbf{x}}+\left(\frac{1}{2}-y_6\right)b\,\hat{\mathbf{y}}+z_6c\,\hat{\mathbf{z}}$ | (8*d*) | Ta |
| $\mathbf{B}_{36}$ | = | $\left(\frac{1}{2}-x_6\right)\mathbf{a}_1+\left(\frac{1}{2}+y_6\right)\mathbf{a}_2+\left(\frac{1}{2}+z_6\right)\mathbf{a}_3$ | = | $\left(\frac{1}{2}-x_6\right)a\,\hat{\mathbf{x}}+\left(\frac{1}{2}+y_6\right)b\,\hat{\mathbf{y}}+\left(\frac{1}{2}+z_6\right)c\,\hat{\mathbf{z}}$ | (8*d*) | Ta |

**Geometry files:**

- CIF: pp. 1647
- POSCAR: pp. 1648

# Copper (II) Azide [$Cu(N_3)_2$] Structure: AB6_oP28_62_c_6c

http://aflow.org/prototype-encyclopedia/AB6_oP28_62_c_6c

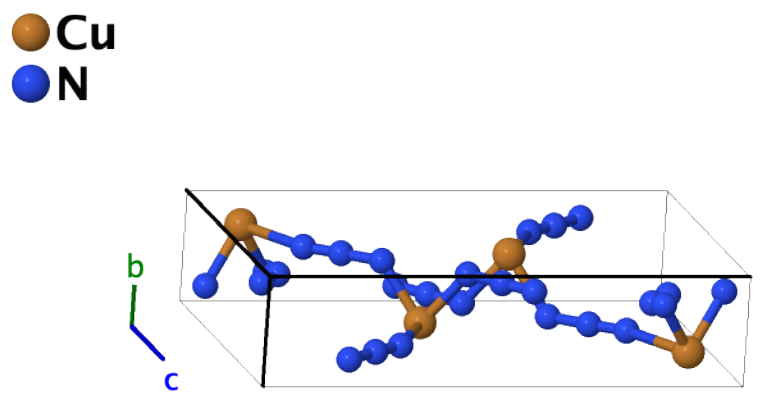

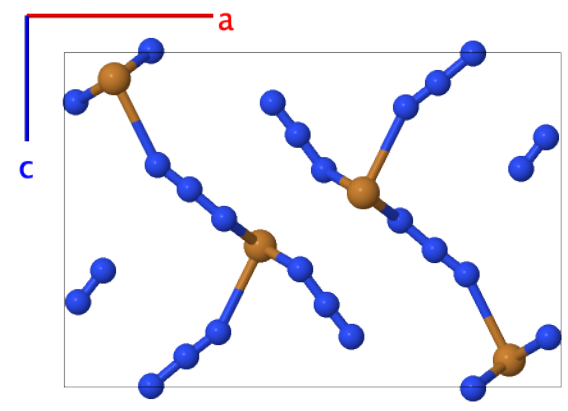

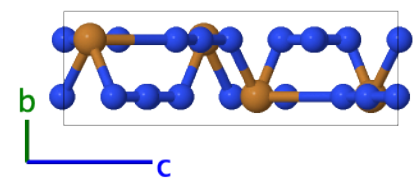

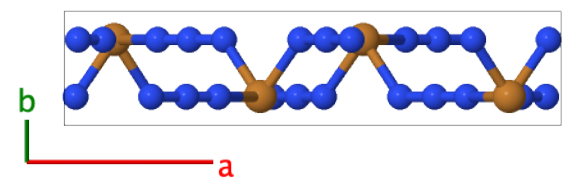

| | | |
|---|---|---|
| **Prototype** | : | $CuN_6$ |
| **AFLOW prototype label** | : | AB6_oP28_62_c_6c |
| ***Strukturbericht* designation** | : | None |
| **Pearson symbol** | : | oP28 |
| **Space group number** | : | 62 |
| **Space group symbol** | : | *Pnma* |
| **AFLOW prototype command** | : | `aflow --proto=AB6_oP28_62_c_6c`<br>`--params=`$a, b/a, c/a, x_1, z_1, x_2, z_2, x_3, z_3, x_4, z_4, x_5, z_5, x_6, z_6, x_7, z_7$ |

- Not to be confused with Copper (I) Azide, $CuN_3$.

**Simple Orthorhombic primitive vectors:**

$$\mathbf{a}_1 = a\,\hat{\mathbf{x}}$$
$$\mathbf{a}_2 = b\,\hat{\mathbf{y}}$$
$$\mathbf{a}_3 = c\,\hat{\mathbf{z}}$$

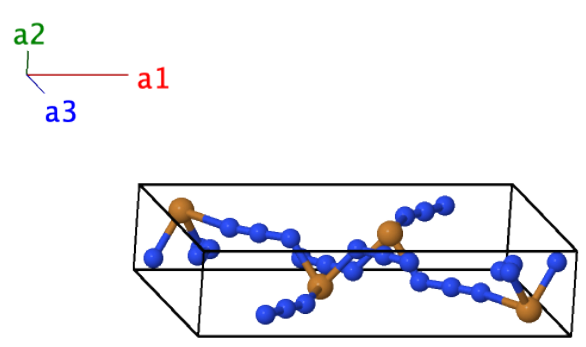

**Basis vectors:**

| | | Lattice Coordinates | | Cartesian Coordinates | Wyckoff Position | Atom Type |
|---|---|---|---|---|---|---|
| $\mathbf{B}_1$ | = | $x_1\,\mathbf{a}_1 + \frac{1}{4}\,\mathbf{a}_2 + z_1\,\mathbf{a}_3$ | = | $x_1 a\,\hat{\mathbf{x}} + \frac{1}{4} b\,\hat{\mathbf{y}} + z_1 c\,\hat{\mathbf{z}}$ | (4*c*) | Cu |
| $\mathbf{B}_2$ | = | $\left(\frac{1}{2} - x_1\right)\mathbf{a}_1 + \frac{3}{4}\,\mathbf{a}_2 + \left(\frac{1}{2} + z_1\right)\mathbf{a}_3$ | = | $\left(\frac{1}{2} - x_1\right) a\,\hat{\mathbf{x}} + \frac{3}{4} b\,\hat{\mathbf{y}} + \left(\frac{1}{2} + z_1\right) c\,\hat{\mathbf{z}}$ | (4*c*) | Cu |
| $\mathbf{B}_3$ | = | $-x_1\,\mathbf{a}_1 + \frac{3}{4}\,\mathbf{a}_2 - z_1\,\mathbf{a}_3$ | = | $-x_1 a\,\hat{\mathbf{x}} + \frac{3}{4} b\,\hat{\mathbf{y}} - z_1 c\,\hat{\mathbf{z}}$ | (4*c*) | Cu |
| $\mathbf{B}_4$ | = | $\left(\frac{1}{2}+x_1\right)\mathbf{a}_1+\frac{1}{4}\mathbf{a}_2+\left(\frac{1}{2}-z_1\right)\mathbf{a}_3$ | = | $\left(\frac{1}{2}+x_1\right)a\,\hat{\mathbf{x}}+\frac{1}{4}b\,\hat{\mathbf{y}}+\left(\frac{1}{2}-z_1\right)c\,\hat{\mathbf{z}}$ | (4c) | Cu |
| $\mathbf{B}_5$ | = | $x_2\,\mathbf{a}_1+\frac{1}{4}\mathbf{a}_2+z_2\,\mathbf{a}_3$ | = | $x_2a\,\hat{\mathbf{x}}+\frac{1}{4}b\,\hat{\mathbf{y}}+z_2c\,\hat{\mathbf{z}}$ | (4c) | N I |
| $\mathbf{B}_6$ | = | $\left(\frac{1}{2}-x_2\right)\mathbf{a}_1+\frac{3}{4}\mathbf{a}_2+\left(\frac{1}{2}+z_2\right)\mathbf{a}_3$ | = | $\left(\frac{1}{2}-x_2\right)a\,\hat{\mathbf{x}}+\frac{3}{4}b\,\hat{\mathbf{y}}+\left(\frac{1}{2}+z_2\right)c\,\hat{\mathbf{z}}$ | (4c) | N I |
| $\mathbf{B}_7$ | = | $-x_2\,\mathbf{a}_1+\frac{3}{4}\mathbf{a}_2-z_2\,\mathbf{a}_3$ | = | $-x_2a\,\hat{\mathbf{x}}+\frac{3}{4}b\,\hat{\mathbf{y}}-z_2c\,\hat{\mathbf{z}}$ | (4c) | N I |
| $\mathbf{B}_8$ | = | $\left(\frac{1}{2}+x_2\right)\mathbf{a}_1+\frac{1}{4}\mathbf{a}_2+\left(\frac{1}{2}-z_2\right)\mathbf{a}_3$ | = | $\left(\frac{1}{2}+x_2\right)a\,\hat{\mathbf{x}}+\frac{1}{4}b\,\hat{\mathbf{y}}+\left(\frac{1}{2}-z_2\right)c\,\hat{\mathbf{z}}$ | (4c) | N I |
| $\mathbf{B}_9$ | = | $x_3\,\mathbf{a}_1+\frac{1}{4}\mathbf{a}_2+z_3\,\mathbf{a}_3$ | = | $x_3a\,\hat{\mathbf{x}}+\frac{1}{4}b\,\hat{\mathbf{y}}+z_3c\,\hat{\mathbf{z}}$ | (4c) | N II |
| $\mathbf{B}_{10}$ | = | $\left(\frac{1}{2}-x_3\right)\mathbf{a}_1+\frac{3}{4}\mathbf{a}_2+\left(\frac{1}{2}+z_3\right)\mathbf{a}_3$ | = | $\left(\frac{1}{2}-x_3\right)a\,\hat{\mathbf{x}}+\frac{3}{4}b\,\hat{\mathbf{y}}+\left(\frac{1}{2}+z_3\right)c\,\hat{\mathbf{z}}$ | (4c) | N II |
| $\mathbf{B}_{11}$ | = | $-x_3\,\mathbf{a}_1+\frac{3}{4}\mathbf{a}_2-z_3\,\mathbf{a}_3$ | = | $-x_3a\,\hat{\mathbf{x}}+\frac{3}{4}b\,\hat{\mathbf{y}}-z_3c\,\hat{\mathbf{z}}$ | (4c) | N II |
| $\mathbf{B}_{12}$ | = | $\left(\frac{1}{2}+x_3\right)\mathbf{a}_1+\frac{1}{4}\mathbf{a}_2+\left(\frac{1}{2}-z_3\right)\mathbf{a}_3$ | = | $\left(\frac{1}{2}+x_3\right)a\,\hat{\mathbf{x}}+\frac{1}{4}b\,\hat{\mathbf{y}}+\left(\frac{1}{2}-z_3\right)c\,\hat{\mathbf{z}}$ | (4c) | N II |
| $\mathbf{B}_{13}$ | = | $x_4\,\mathbf{a}_1+\frac{1}{4}\mathbf{a}_2+z_4\,\mathbf{a}_3$ | = | $x_4a\,\hat{\mathbf{x}}+\frac{1}{4}b\,\hat{\mathbf{y}}+z_4c\,\hat{\mathbf{z}}$ | (4c) | N III |
| $\mathbf{B}_{14}$ | = | $\left(\frac{1}{2}-x_4\right)\mathbf{a}_1+\frac{3}{4}\mathbf{a}_2+\left(\frac{1}{2}+z_4\right)\mathbf{a}_3$ | = | $\left(\frac{1}{2}-x_4\right)a\,\hat{\mathbf{x}}+\frac{3}{4}b\,\hat{\mathbf{y}}+\left(\frac{1}{2}+z_4\right)c\,\hat{\mathbf{z}}$ | (4c) | N III |
| $\mathbf{B}_{15}$ | = | $-x_4\,\mathbf{a}_1+\frac{3}{4}\mathbf{a}_2-z_4\,\mathbf{a}_3$ | = | $-x_4a\,\hat{\mathbf{x}}+\frac{3}{4}b\,\hat{\mathbf{y}}-z_4c\,\hat{\mathbf{z}}$ | (4c) | N III |
| $\mathbf{B}_{16}$ | = | $\left(\frac{1}{2}+x_4\right)\mathbf{a}_1+\frac{1}{4}\mathbf{a}_2+\left(\frac{1}{2}-z_4\right)\mathbf{a}_3$ | = | $\left(\frac{1}{2}+x_4\right)a\,\hat{\mathbf{x}}+\frac{1}{4}b\,\hat{\mathbf{y}}+\left(\frac{1}{2}-z_4\right)c\,\hat{\mathbf{z}}$ | (4c) | N III |
| $\mathbf{B}_{17}$ | = | $x_5\,\mathbf{a}_1+\frac{1}{4}\mathbf{a}_2+z_5\,\mathbf{a}_3$ | = | $x_5a\,\hat{\mathbf{x}}+\frac{1}{4}b\,\hat{\mathbf{y}}+z_5c\,\hat{\mathbf{z}}$ | (4c) | N IV |
| $\mathbf{B}_{18}$ | = | $\left(\frac{1}{2}-x_5\right)\mathbf{a}_1+\frac{3}{4}\mathbf{a}_2+\left(\frac{1}{2}+z_5\right)\mathbf{a}_3$ | = | $\left(\frac{1}{2}-x_5\right)a\,\hat{\mathbf{x}}+\frac{3}{4}b\,\hat{\mathbf{y}}+\left(\frac{1}{2}+z_5\right)c\,\hat{\mathbf{z}}$ | (4c) | N IV |
| $\mathbf{B}_{19}$ | = | $-x_5\,\mathbf{a}_1+\frac{3}{4}\mathbf{a}_2-z_5\,\mathbf{a}_3$ | = | $-x_5a\,\hat{\mathbf{x}}+\frac{3}{4}b\,\hat{\mathbf{y}}-z_5c\,\hat{\mathbf{z}}$ | (4c) | N IV |
| $\mathbf{B}_{20}$ | = | $\left(\frac{1}{2}+x_5\right)\mathbf{a}_1+\frac{1}{4}\mathbf{a}_2+\left(\frac{1}{2}-z_5\right)\mathbf{a}_3$ | = | $\left(\frac{1}{2}+x_5\right)a\,\hat{\mathbf{x}}+\frac{1}{4}b\,\hat{\mathbf{y}}+\left(\frac{1}{2}-z_5\right)c\,\hat{\mathbf{z}}$ | (4c) | N IV |
| $\mathbf{B}_{21}$ | = | $x_6\,\mathbf{a}_1+\frac{1}{4}\mathbf{a}_2+z_6\,\mathbf{a}_3$ | = | $x_6a\,\hat{\mathbf{x}}+\frac{1}{4}b\,\hat{\mathbf{y}}+z_6c\,\hat{\mathbf{z}}$ | (4c) | N V |
| $\mathbf{B}_{22}$ | = | $\left(\frac{1}{2}-x_6\right)\mathbf{a}_1+\frac{3}{4}\mathbf{a}_2+\left(\frac{1}{2}+z_6\right)\mathbf{a}_3$ | = | $\left(\frac{1}{2}-x_6\right)a\,\hat{\mathbf{x}}+\frac{3}{4}b\,\hat{\mathbf{y}}+\left(\frac{1}{2}+z_6\right)c\,\hat{\mathbf{z}}$ | (4c) | N V |
| $\mathbf{B}_{23}$ | = | $-x_6\,\mathbf{a}_1+\frac{3}{4}\mathbf{a}_2-z_6\,\mathbf{a}_3$ | = | $-x_6a\,\hat{\mathbf{x}}+\frac{3}{4}b\,\hat{\mathbf{y}}-z_6c\,\hat{\mathbf{z}}$ | (4c) | N V |
| $\mathbf{B}_{24}$ | = | $\left(\frac{1}{2}+x_6\right)\mathbf{a}_1+\frac{1}{4}\mathbf{a}_2+\left(\frac{1}{2}-z_6\right)\mathbf{a}_3$ | = | $\left(\frac{1}{2}+x_6\right)a\,\hat{\mathbf{x}}+\frac{1}{4}b\,\hat{\mathbf{y}}+\left(\frac{1}{2}-z_6\right)c\,\hat{\mathbf{z}}$ | (4c) | N V |
| $\mathbf{B}_{25}$ | = | $x_7\,\mathbf{a}_1+\frac{1}{4}\mathbf{a}_2+z_7\,\mathbf{a}_3$ | = | $x_7a\,\hat{\mathbf{x}}+\frac{1}{4}b\,\hat{\mathbf{y}}+z_7c\,\hat{\mathbf{z}}$ | (4c) | N VI |
| $\mathbf{B}_{26}$ | = | $\left(\frac{1}{2}-x_7\right)\mathbf{a}_1+\frac{3}{4}\mathbf{a}_2+\left(\frac{1}{2}+z_7\right)\mathbf{a}_3$ | = | $\left(\frac{1}{2}-x_7\right)a\,\hat{\mathbf{x}}+\frac{3}{4}b\,\hat{\mathbf{y}}+\left(\frac{1}{2}+z_7\right)c\,\hat{\mathbf{z}}$ | (4c) | N VI |
| $\mathbf{B}_{27}$ | = | $-x_7\,\mathbf{a}_1+\frac{3}{4}\mathbf{a}_2-z_7\,\mathbf{a}_3$ | = | $-x_7a\,\hat{\mathbf{x}}+\frac{3}{4}b\,\hat{\mathbf{y}}-z_7c\,\hat{\mathbf{z}}$ | (4c) | N VI |
| $\mathbf{B}_{28}$ | = | $\left(\frac{1}{2}+x_7\right)\mathbf{a}_1+\frac{1}{4}\mathbf{a}_2+\left(\frac{1}{2}-z_7\right)\mathbf{a}_3$ | = | $\left(\frac{1}{2}+x_7\right)a\,\hat{\mathbf{x}}+\frac{1}{4}b\,\hat{\mathbf{y}}+\left(\frac{1}{2}-z_7\right)c\,\hat{\mathbf{z}}$ | (4c) | N VI |

**Geometry files:**

- CIF: pp. 1648
- POSCAR: pp. 1648

# Diaspore (AlOOH, $E0_2$) Structure: ABC2_oP16_62_c_c_2c

http://aflow.org/prototype-encyclopedia/ABC2_oP16_62_c_c_2c

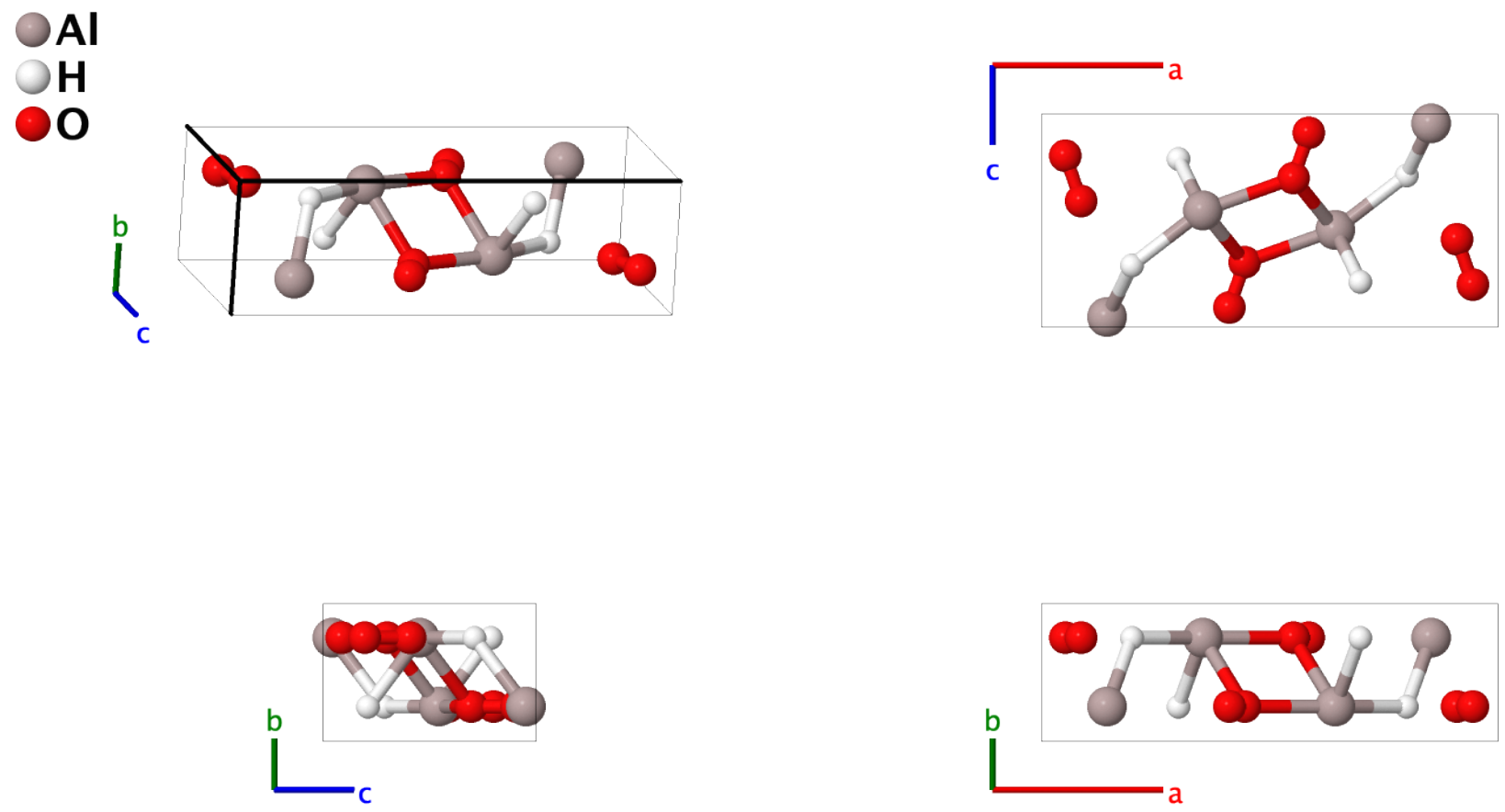

| | | |
|---|---|---|
| **Prototype** | : | $AlHO_2$ |
| **AFLOW prototype label** | : | ABC2_oP16_62_c_c_2c |
| ***Strukturbericht* designation** | : | $E0_2$ |
| **Pearson symbol** | : | oP16 |
| **Space group number** | : | 62 |
| **Space group symbol** | : | $Pnma$ |
| **AFLOW prototype command** | : | `aflow --proto=ABC2_oP16_62_c_c_2c` <br> `--params=`$a, b/a, c/a, x_1, z_1, x_2, z_2, x_3, z_3, x_4, z_4$ |

**Other compounds with this structure**

- FeO(OH) (Goethite) and MnO(OH) (Groutite)

- (Hermann, 1937), following (Verway, 1935) designated this as *Strukturbericht* $E0_2$, but had no knowledge of the positions of the hydrogen atoms, which we take from (Hill, 1979) and which do not change the space group. Hill gave the Wyckoff positions of the atoms in terms of the *Pbnm* setting of space group #62. We used FINDSYM to transform this structure to the standard *Pnma* setting.

**Simple Orthorhombic primitive vectors:**

$$\mathbf{a}_1 = a\,\hat{\mathbf{x}}$$
$$\mathbf{a}_2 = b\,\hat{\mathbf{y}}$$
$$\mathbf{a}_3 = c\,\hat{\mathbf{z}}$$

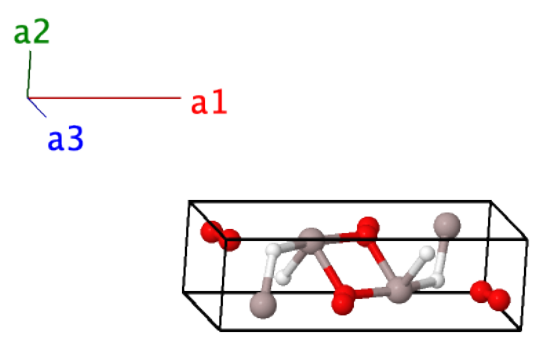

**Basis vectors:**

| | | Lattice Coordinates | | Cartesian Coordinates | Wyckoff Position | Atom Type |
|---|---|---|---|---|---|---|
| $\mathbf{B}_1$ | = | $x_1\,\mathbf{a}_1 + \frac{1}{4}\,\mathbf{a}_2 + z_1\,\mathbf{a}_3$ | = | $x_1 a\,\hat{\mathbf{x}} + \frac{1}{4} b\,\hat{\mathbf{y}} + z_1 c\,\hat{\mathbf{z}}$ | (4$c$) | Al |

$$\mathbf{B}_2 = \left(\frac{1}{2} - x_1\right) \mathbf{a}_1 + \frac{3}{4} \mathbf{a}_2 + \left(\frac{1}{2} + z_1\right) \mathbf{a}_3 = \left(\frac{1}{2} - x_1\right) a \, \hat{\mathbf{x}} + \frac{3}{4} b \, \hat{\mathbf{y}} + \left(\frac{1}{2} + z_1\right) c \, \hat{\mathbf{z}} \quad (4c) \quad \text{Al}$$

$$\mathbf{B}_3 = -x_1 \, \mathbf{a}_1 + \frac{3}{4} \mathbf{a}_2 - z_1 \, \mathbf{a}_3 = -x_1 a \, \hat{\mathbf{x}} + \frac{3}{4} b \, \hat{\mathbf{y}} - z_1 c \, \hat{\mathbf{z}} \quad (4c) \quad \text{Al}$$

$$\mathbf{B}_4 = \left(\frac{1}{2} + x_1\right) \mathbf{a}_1 + \frac{1}{4} \mathbf{a}_2 + \left(\frac{1}{2} - z_1\right) \mathbf{a}_3 = \left(\frac{1}{2} + x_1\right) a \, \hat{\mathbf{x}} + \frac{1}{4} b \, \hat{\mathbf{y}} + \left(\frac{1}{2} - z_1\right) c \, \hat{\mathbf{z}} \quad (4c) \quad \text{Al}$$

$$\mathbf{B}_5 = x_2 \, \mathbf{a}_1 + \frac{1}{4} \mathbf{a}_2 + z_2 \, \mathbf{a}_3 = x_2 a \, \hat{\mathbf{x}} + \frac{1}{4} b \, \hat{\mathbf{y}} + z_2 c \, \hat{\mathbf{z}} \quad (4c) \quad \text{H}$$

$$\mathbf{B}_6 = \left(\frac{1}{2} - x_2\right) \mathbf{a}_1 + \frac{3}{4} \mathbf{a}_2 + \left(\frac{1}{2} + z_2\right) \mathbf{a}_3 = \left(\frac{1}{2} - x_2\right) a \, \hat{\mathbf{x}} + \frac{3}{4} b \, \hat{\mathbf{y}} + \left(\frac{1}{2} + z_2\right) c \, \hat{\mathbf{z}} \quad (4c) \quad \text{H}$$

$$\mathbf{B}_7 = -x_2 \, \mathbf{a}_1 + \frac{3}{4} \mathbf{a}_2 - z_2 \, \mathbf{a}_3 = -x_2 a \, \hat{\mathbf{x}} + \frac{3}{4} b \, \hat{\mathbf{y}} - z_2 c \, \hat{\mathbf{z}} \quad (4c) \quad \text{H}$$

$$\mathbf{B}_8 = \left(\frac{1}{2} + x_2\right) \mathbf{a}_1 + \frac{1}{4} \mathbf{a}_2 + \left(\frac{1}{2} - z_2\right) \mathbf{a}_3 = \left(\frac{1}{2} + x_2\right) a \, \hat{\mathbf{x}} + \frac{1}{4} b \, \hat{\mathbf{y}} + \left(\frac{1}{2} - z_2\right) c \, \hat{\mathbf{z}} \quad (4c) \quad \text{H}$$

$$\mathbf{B}_9 = x_3 \, \mathbf{a}_1 + \frac{1}{4} \mathbf{a}_2 + z_3 \, \mathbf{a}_3 = x_3 a \, \hat{\mathbf{x}} + \frac{1}{4} b \, \hat{\mathbf{y}} + z_3 c \, \hat{\mathbf{z}} \quad (4c) \quad \text{O I}$$

$$\mathbf{B}_{10} = \left(\frac{1}{2} - x_3\right) \mathbf{a}_1 + \frac{3}{4} \mathbf{a}_2 + \left(\frac{1}{2} + z_3\right) \mathbf{a}_3 = \left(\frac{1}{2} - x_3\right) a \, \hat{\mathbf{x}} + \frac{3}{4} b \, \hat{\mathbf{y}} + \left(\frac{1}{2} + z_3\right) c \, \hat{\mathbf{z}} \quad (4c) \quad \text{O I}$$

$$\mathbf{B}_{11} = -x_3 \, \mathbf{a}_1 + \frac{3}{4} \mathbf{a}_2 - z_3 \, \mathbf{a}_3 = -x_3 a \, \hat{\mathbf{x}} + \frac{3}{4} b \, \hat{\mathbf{y}} - z_3 c \, \hat{\mathbf{z}} \quad (4c) \quad \text{O I}$$

$$\mathbf{B}_{12} = \left(\frac{1}{2} + x_3\right) \mathbf{a}_1 + \frac{1}{4} \mathbf{a}_2 + \left(\frac{1}{2} - z_3\right) \mathbf{a}_3 = \left(\frac{1}{2} + x_3\right) a \, \hat{\mathbf{x}} + \frac{1}{4} b \, \hat{\mathbf{y}} + \left(\frac{1}{2} - z_3\right) c \, \hat{\mathbf{z}} \quad (4c) \quad \text{O I}$$

$$\mathbf{B}_{13} = x_4 \, \mathbf{a}_1 + \frac{1}{4} \mathbf{a}_2 + z_4 \, \mathbf{a}_3 = x_4 a \, \hat{\mathbf{x}} + \frac{1}{4} b \, \hat{\mathbf{y}} + z_4 c \, \hat{\mathbf{z}} \quad (4c) \quad \text{O II}$$

$$\mathbf{B}_{14} = \left(\frac{1}{2} - x_4\right) \mathbf{a}_1 + \frac{3}{4} \mathbf{a}_2 + \left(\frac{1}{2} + z_4\right) \mathbf{a}_3 = \left(\frac{1}{2} - x_4\right) a \, \hat{\mathbf{x}} + \frac{3}{4} b \, \hat{\mathbf{y}} + \left(\frac{1}{2} + z_4\right) c \, \hat{\mathbf{z}} \quad (4c) \quad \text{O II}$$

$$\mathbf{B}_{15} = -x_4 \, \mathbf{a}_1 + \frac{3}{4} \mathbf{a}_2 - z_4 \, \mathbf{a}_3 = -x_4 a \, \hat{\mathbf{x}} + \frac{3}{4} b \, \hat{\mathbf{y}} - z_4 c \, \hat{\mathbf{z}} \quad (4c) \quad \text{O II}$$

$$\mathbf{B}_{16} = \left(\frac{1}{2} + x_4\right) \mathbf{a}_1 + \frac{1}{4} \mathbf{a}_2 + \left(\frac{1}{2} - z_4\right) \mathbf{a}_3 = \left(\frac{1}{2} + x_4\right) a \, \hat{\mathbf{x}} + \frac{1}{4} b \, \hat{\mathbf{y}} + \left(\frac{1}{2} - z_4\right) c \, \hat{\mathbf{z}} \quad (4c) \quad \text{O II}$$

---

**Found in:**

---

**Geometry files:**

- CIF: pp. 1648
- POSCAR: pp. 1649

# $\alpha$-Potassium Nitrate ($KNO_3$) I Structure: ABC3_oP20_62_c_c_cd

http://aflow.org/prototype-encyclopedia/ABC3_oP20_62_c_c_cd.K.N.O3

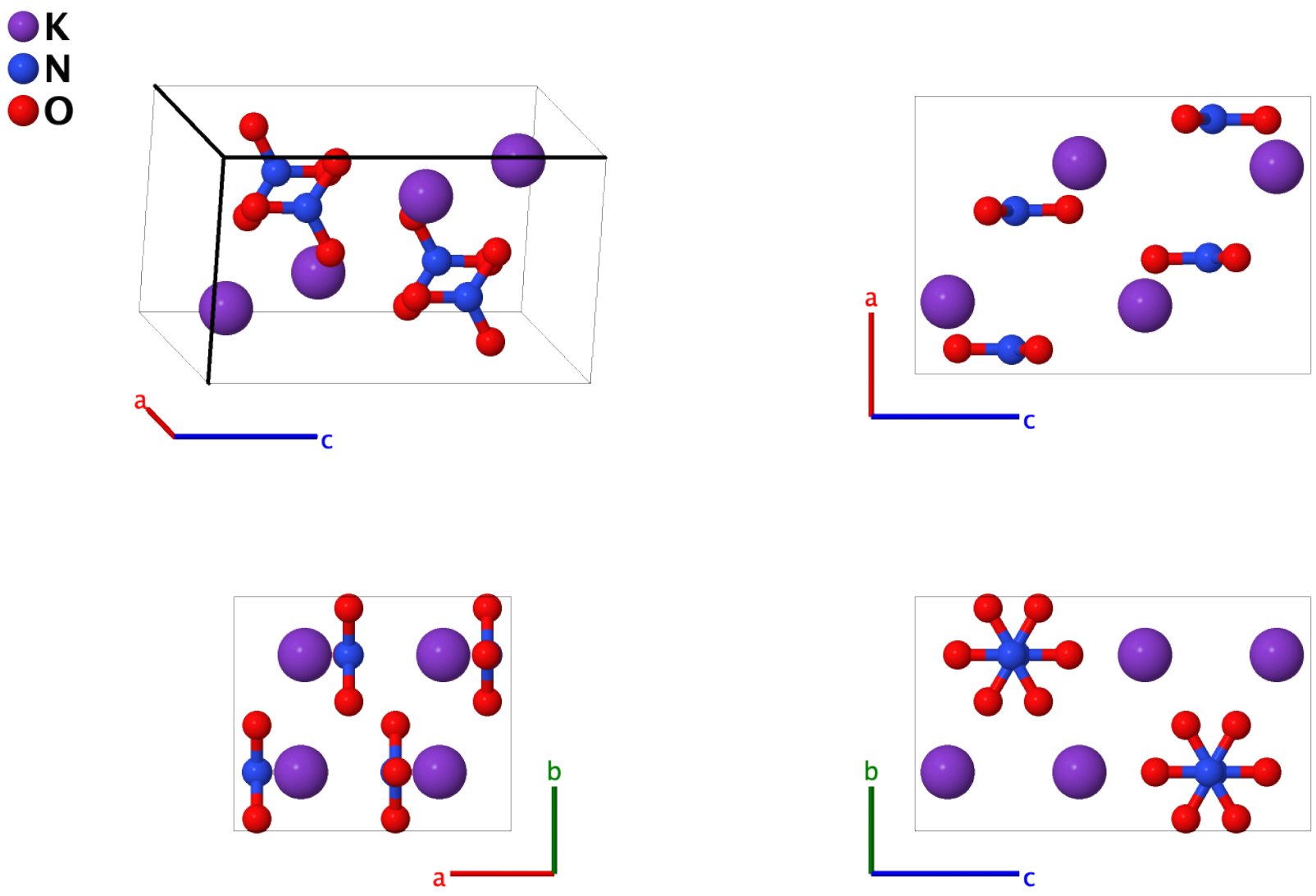

| | | |
|---|---|---|
| **Prototype** | : | $KNO_3$ |
| **AFLOW prototype label** | : | ABC3_oP20_62_c_c_cd |
| ***Strukturbericht* designation** | : | None |
| **Pearson symbol** | : | oP20 |
| **Space group number** | : | 62 |
| **Space group symbol** | : | *Pnma* |
| **AFLOW prototype command** | : | `aflow --proto=ABC3_oP20_62_c_c_cd` `--params=`$a, b/a, c/a, x_1, z_1, x_2, z_2, x_3, z_3, x_4, y_4, z_4$ |

- Two possible structures have been identified for $\alpha$-$KNO_3$: (Nimmo, 1973) proposed the current structure, which we call "Structure I", in space group *Pnma* #62. (Adiwidjaja, 2003) found this structure, but also noted that it could be described by a doubling of the current unit cell into a superstructure with space group $Cmc2_1$ #36, which we call "Structure II". It is unclear to us which structure is correct, so we present both of them.
- (Nimmo, 1973) determined the lattice constants and Wyckoff positions of the type I structure at 25 °C in the *Pmcn* setting of space group #62. We present the data in the standard *Pnma* setting.
- This structure has the same space group and Wyckoff positions as the $MgB_4$ (A4B_oP20_62_2cd_c) structure and the $G0_{10}$ $(NH_4)NO_3$ (ABC3_oP20_62_c_c_cd.N.NH4.O) structure, but the structures are all different. We give the previous two structures the noted AFLOW designations, and give this one the designation ABC3_oP20_62_c_c_cd.K.N.O3.
- On heating, $\alpha$-$KNO_3$ transforms into $\beta$-$KNO_3$ at 128 °C. When heated above 200 °C and then cooled, the $\beta$-phase transforms into the metastable ferroelectric $\gamma$-$KNO_3$ phase, which can remain down to room temperature.

## Simple Orthorhombic primitive vectors:

$$\mathbf{a}_1 = a\,\hat{\mathbf{x}}$$
$$\mathbf{a}_2 = b\,\hat{\mathbf{y}}$$
$$\mathbf{a}_3 = c\,\hat{\mathbf{z}}$$

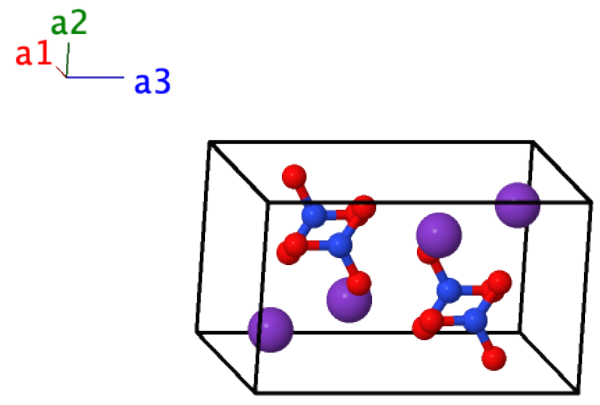

## Basis vectors:

| | | Lattice Coordinates | | Cartesian Coordinates | Wyckoff Position | Atom Type |
|---|---|---|---|---|---|---|
| $\mathbf{B}_1$ | = | $x_1\,\mathbf{a}_1 + \frac{1}{4}\,\mathbf{a}_2 + z_1\,\mathbf{a}_3$ | = | $x_1 a\,\hat{\mathbf{x}} + \frac{1}{4} b\,\hat{\mathbf{y}} + z_1 c\,\hat{\mathbf{z}}$ | (4c) | K |
| $\mathbf{B}_2$ | = | $\left(\frac{1}{2} - x_1\right)\mathbf{a}_1 + \frac{3}{4}\,\mathbf{a}_2 + \left(\frac{1}{2} + z_1\right)\mathbf{a}_3$ | = | $\left(\frac{1}{2} - x_1\right) a\,\hat{\mathbf{x}} + \frac{3}{4} b\,\hat{\mathbf{y}} + \left(\frac{1}{2} + z_1\right) c\,\hat{\mathbf{z}}$ | (4c) | K |
| $\mathbf{B}_3$ | = | $-x_1\,\mathbf{a}_1 + \frac{3}{4}\,\mathbf{a}_2 - z_1\,\mathbf{a}_3$ | = | $-x_1 a\,\hat{\mathbf{x}} + \frac{3}{4} b\,\hat{\mathbf{y}} - z_1 c\,\hat{\mathbf{z}}$ | (4c) | K |
| $\mathbf{B}_4$ | = | $\left(\frac{1}{2} + x_1\right)\mathbf{a}_1 + \frac{1}{4}\,\mathbf{a}_2 + \left(\frac{1}{2} - z_1\right)\mathbf{a}_3$ | = | $\left(\frac{1}{2} + x_1\right) a\,\hat{\mathbf{x}} + \frac{1}{4} b\,\hat{\mathbf{y}} + \left(\frac{1}{2} - z_1\right) c\,\hat{\mathbf{z}}$ | (4c) | K |
| $\mathbf{B}_5$ | = | $x_2\,\mathbf{a}_1 + \frac{1}{4}\,\mathbf{a}_2 + z_2\,\mathbf{a}_3$ | = | $x_2 a\,\hat{\mathbf{x}} + \frac{1}{4} b\,\hat{\mathbf{y}} + z_2 c\,\hat{\mathbf{z}}$ | (4c) | N |
| $\mathbf{B}_6$ | = | $\left(\frac{1}{2} - x_2\right)\mathbf{a}_1 + \frac{3}{4}\,\mathbf{a}_2 + \left(\frac{1}{2} + z_2\right)\mathbf{a}_3$ | = | $\left(\frac{1}{2} - x_2\right) a\,\hat{\mathbf{x}} + \frac{3}{4} b\,\hat{\mathbf{y}} + \left(\frac{1}{2} + z_2\right) c\,\hat{\mathbf{z}}$ | (4c) | N |
| $\mathbf{B}_7$ | = | $-x_2\,\mathbf{a}_1 + \frac{3}{4}\,\mathbf{a}_2 - z_2\,\mathbf{a}_3$ | = | $-x_2 a\,\hat{\mathbf{x}} + \frac{3}{4} b\,\hat{\mathbf{y}} - z_2 c\,\hat{\mathbf{z}}$ | (4c) | N |
| $\mathbf{B}_8$ | = | $\left(\frac{1}{2} + x_2\right)\mathbf{a}_1 + \frac{1}{4}\,\mathbf{a}_2 + \left(\frac{1}{2} - z_2\right)\mathbf{a}_3$ | = | $\left(\frac{1}{2} + x_2\right) a\,\hat{\mathbf{x}} + \frac{1}{4} b\,\hat{\mathbf{y}} + \left(\frac{1}{2} - z_2\right) c\,\hat{\mathbf{z}}$ | (4c) | N |
| $\mathbf{B}_9$ | = | $x_3\,\mathbf{a}_1 + \frac{1}{4}\,\mathbf{a}_2 + z_3\,\mathbf{a}_3$ | = | $x_3 a\,\hat{\mathbf{x}} + \frac{1}{4} b\,\hat{\mathbf{y}} + z_3 c\,\hat{\mathbf{z}}$ | (4c) | O I |
| $\mathbf{B}_{10}$ | = | $\left(\frac{1}{2} - x_3\right)\mathbf{a}_1 + \frac{3}{4}\,\mathbf{a}_2 + \left(\frac{1}{2} + z_3\right)\mathbf{a}_3$ | = | $\left(\frac{1}{2} - x_3\right) a\,\hat{\mathbf{x}} + \frac{3}{4} b\,\hat{\mathbf{y}} + \left(\frac{1}{2} + z_3\right) c\,\hat{\mathbf{z}}$ | (4c) | O I |
| $\mathbf{B}_{11}$ | = | $-x_3\,\mathbf{a}_1 + \frac{3}{4}\,\mathbf{a}_2 - z_3\,\mathbf{a}_3$ | = | $-x_3 a\,\hat{\mathbf{x}} + \frac{3}{4} b\,\hat{\mathbf{y}} - z_3 c\,\hat{\mathbf{z}}$ | (4c) | O I |
| $\mathbf{B}_{12}$ | = | $\left(\frac{1}{2} + x_3\right)\mathbf{a}_1 + \frac{1}{4}\,\mathbf{a}_2 + \left(\frac{1}{2} - z_3\right)\mathbf{a}_3$ | = | $\left(\frac{1}{2} + x_3\right) a\,\hat{\mathbf{x}} + \frac{1}{4} b\,\hat{\mathbf{y}} + \left(\frac{1}{2} - z_3\right) c\,\hat{\mathbf{z}}$ | (4c) | O I |
| $\mathbf{B}_{13}$ | = | $x_4\,\mathbf{a}_1 + y_4\,\mathbf{a}_2 + z_4\,\mathbf{a}_3$ | = | $x_4 a\,\hat{\mathbf{x}} + y_4 b\,\hat{\mathbf{y}} + z_4 c\,\hat{\mathbf{z}}$ | (8d) | O II |
| $\mathbf{B}_{14}$ | = | $\left(\frac{1}{2} - x_4\right)\mathbf{a}_1 - y_4\,\mathbf{a}_2 + \left(\frac{1}{2} + z_4\right)\mathbf{a}_3$ | = | $\left(\frac{1}{2} - x_4\right) a\,\hat{\mathbf{x}} - y_4 b\,\hat{\mathbf{y}} + \left(\frac{1}{2} + z_4\right) c\,\hat{\mathbf{z}}$ | (8d) | O II |
| $\mathbf{B}_{15}$ | = | $-x_4\,\mathbf{a}_1 + \left(\frac{1}{2} + y_4\right)\mathbf{a}_2 - z_4\,\mathbf{a}_3$ | = | $-x_4 a\,\hat{\mathbf{x}} + \left(\frac{1}{2} + y_4\right) b\,\hat{\mathbf{y}} - z_4 c\,\hat{\mathbf{z}}$ | (8d) | O II |
| $\mathbf{B}_{16}$ | = | $\left(\frac{1}{2} + x_4\right)\mathbf{a}_1 + \left(\frac{1}{2} - y_4\right)\mathbf{a}_2 + \left(\frac{1}{2} - z_4\right)\mathbf{a}_3$ | = | $\left(\frac{1}{2} + x_4\right) a\,\hat{\mathbf{x}} + \left(\frac{1}{2} - y_4\right) b\,\hat{\mathbf{y}} + \left(\frac{1}{2} - z_4\right) c\,\hat{\mathbf{z}}$ | (8d) | O II |
| $\mathbf{B}_{17}$ | = | $-x_4\,\mathbf{a}_1 - y_4\,\mathbf{a}_2 - z_4\,\mathbf{a}_3$ | = | $-x_4 a\,\hat{\mathbf{x}} - y_4 b\,\hat{\mathbf{y}} - z_4 c\,\hat{\mathbf{z}}$ | (8d) | O II |
| $\mathbf{B}_{18}$ | = | $\left(\frac{1}{2} + x_4\right)\mathbf{a}_1 + y_4\,\mathbf{a}_2 + \left(\frac{1}{2} - z_4\right)\mathbf{a}_3$ | = | $\left(\frac{1}{2} + x_4\right) a\,\hat{\mathbf{x}} + y_4 b\,\hat{\mathbf{y}} + \left(\frac{1}{2} - z_4\right) c\,\hat{\mathbf{z}}$ | (8d) | O II |
| $\mathbf{B}_{19}$ | = | $x_4\,\mathbf{a}_1 + \left(\frac{1}{2} - y_4\right)\mathbf{a}_2 + z_4\,\mathbf{a}_3$ | = | $x_4 a\,\hat{\mathbf{x}} + \left(\frac{1}{2} - y_4\right) b\,\hat{\mathbf{y}} + z_4 c\,\hat{\mathbf{z}}$ | (8d) | O II |
| $\mathbf{B}_{20}$ | = | $\left(\frac{1}{2} - x_4\right)\mathbf{a}_1 + \left(\frac{1}{2} + y_4\right)\mathbf{a}_2 + \left(\frac{1}{2} + z_4\right)\mathbf{a}_3$ | = | $\left(\frac{1}{2} - x_4\right) a\,\hat{\mathbf{x}} + \left(\frac{1}{2} + y_4\right) b\,\hat{\mathbf{y}} + \left(\frac{1}{2} + z_4\right) c\,\hat{\mathbf{z}}$ | (8d) | O II |

## Found in:

---

**Geometry files:**

- CIF: pp. 1649
- POSCAR: pp. 1649

# $NH_4NO_3$ III ($G0_{10}$) Structure: ABC3_oP20_62_c_c_cd

http://aflow.org/prototype-encyclopedia/ABC3_oP20_62_c_c_cd.N.NH4.O

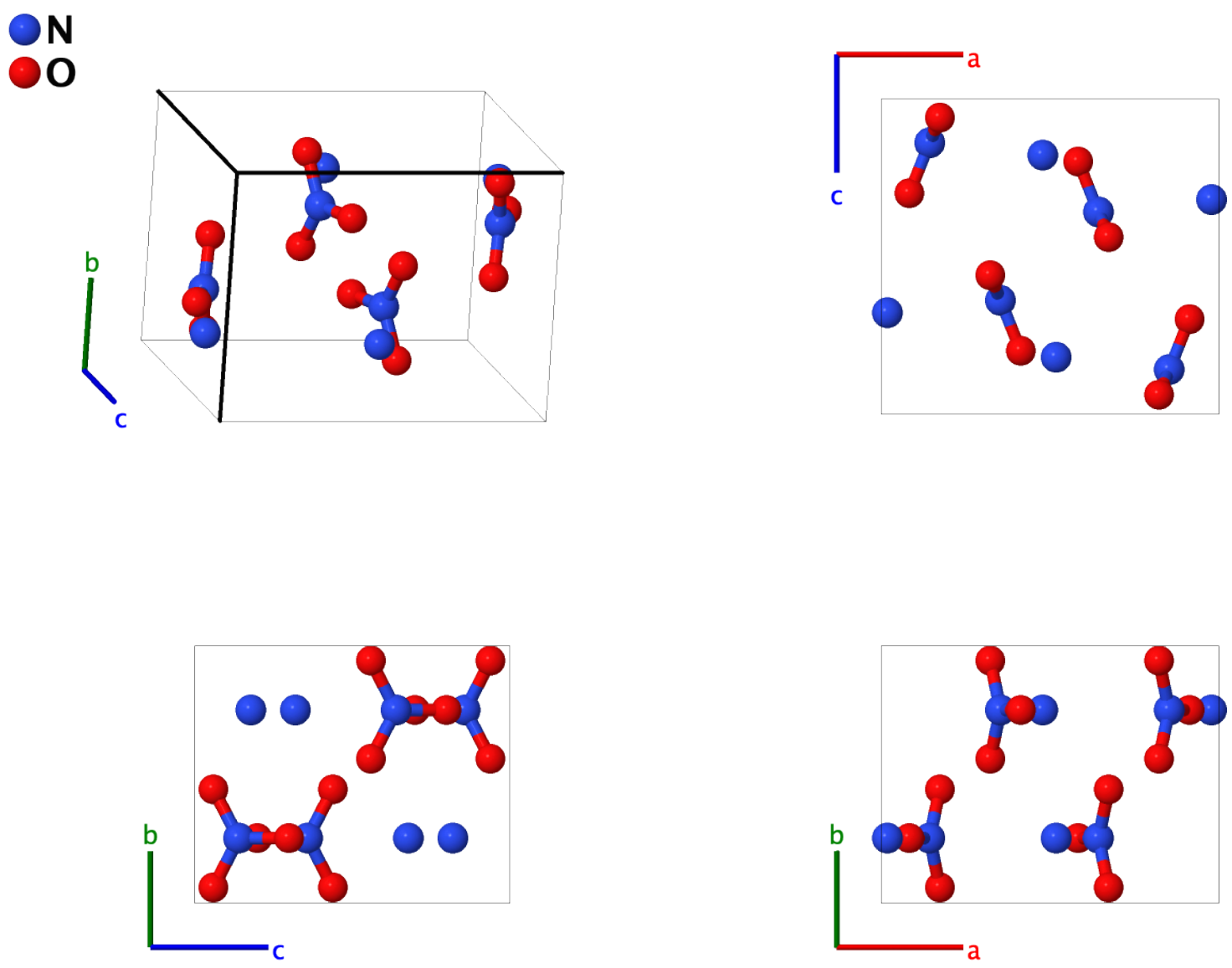

| | | |
|---|---|---|
| **Prototype** | : | $N(NH_4)O_3$ |
| **AFLOW prototype label** | : | ABC3_oP20_62_c_c_cd |
| ***Strukturbericht* designation** | : | $G0_{10}$ |
| **Pearson symbol** | : | oP20 |
| **Space group number** | : | 62 |
| **Space group symbol** | : | *Pnma* |
| **AFLOW prototype command** | : | `aflow --proto=ABC3_oP20_62_c_c_cd`<br>`--params=`$a, b/a, c/a, x_1, z_1, x_2, z_2, x_3, z_3, x_4, y_4, z_4$ |

- Ammonium Nitrate exists in a variety of forms, (Hermann, 1937) depending on the temperature:

| Phase | Temperature °C | *Strukturbericht* | Page | |
|---|---|---|---|---|
| I | 125 – 170 | $G0_8$ | AB_cP2_221_a_b.NH4.NO3 | |
| II | 84 – 125 | $G0_9$ | ABC3_tP10_100_b_a_bc | |
| III | 32 – 84 | $G0_{10}$ | ABC3_oP20_62_c_c_cd.N.NH4.O | (this structure) |
| IV | -17 – 32 | $G0_{11}$ | A4B2C3_oP18_59_ef_ab_af | |
| V | $< -17$ | Gwihabaite | A4B2C3_tP72_77_8d_ab2c2d_6d2 | |

- Data for this structure was taken at 46 °C.
- It is likely that the $NH_4$ ions are free to rotate (Kracek, 1937).
- The positions of the hydrogen atoms in the $NH_4$ ions were not determined, so we only provide the positions of the nitrogen atoms (labeled as $NH_4$).
- (Goodwin, 1947) gave the structure in the *Pbnm* setting of space group #62. We used FINDSYM to change this to the standard *Pnma* setting.

## Simple Orthorhombic primitive vectors:

$$\mathbf{a}_1 = a\,\hat{\mathbf{x}}$$
$$\mathbf{a}_2 = b\,\hat{\mathbf{y}}$$
$$\mathbf{a}_3 = c\,\hat{\mathbf{z}}$$

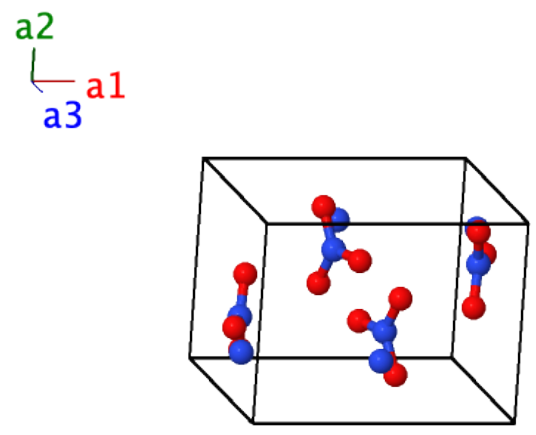

## Basis vectors:

| | | Lattice Coordinates | | Cartesian Coordinates | Wyckoff Position | Atom Type |
|---|---|---|---|---|---|---|
| $\mathbf{B}_1$ | = | $x_1\,\mathbf{a}_1 + \frac{1}{4}\,\mathbf{a}_2 + z_1\,\mathbf{a}_3$ | = | $x_1 a\,\hat{\mathbf{x}} + \frac{1}{4} b\,\hat{\mathbf{y}} + z_1 c\,\hat{\mathbf{z}}$ | (4$c$) | N |
| $\mathbf{B}_2$ | = | $\left(\frac{1}{2} - x_1\right)\mathbf{a}_1 + \frac{3}{4}\,\mathbf{a}_2 + \left(\frac{1}{2} + z_1\right)\mathbf{a}_3$ | = | $\left(\frac{1}{2} - x_1\right) a\,\hat{\mathbf{x}} + \frac{3}{4} b\,\hat{\mathbf{y}} + \left(\frac{1}{2} + z_1\right) c\,\hat{\mathbf{z}}$ | (4$c$) | N |
| $\mathbf{B}_3$ | = | $-x_1\,\mathbf{a}_1 + \frac{3}{4}\,\mathbf{a}_2 - z_1\,\mathbf{a}_3$ | = | $-x_1 a\,\hat{\mathbf{x}} + \frac{3}{4} b\,\hat{\mathbf{y}} - z_1 c\,\hat{\mathbf{z}}$ | (4$c$) | N |
| $\mathbf{B}_4$ | = | $\left(\frac{1}{2} + x_1\right)\mathbf{a}_1 + \frac{1}{4}\,\mathbf{a}_2 + \left(\frac{1}{2} - z_1\right)\mathbf{a}_3$ | = | $\left(\frac{1}{2} + x_1\right) a\,\hat{\mathbf{x}} + \frac{1}{4} b\,\hat{\mathbf{y}} + \left(\frac{1}{2} - z_1\right) c\,\hat{\mathbf{z}}$ | (4$c$) | N |
| $\mathbf{B}_5$ | = | $x_2\,\mathbf{a}_1 + \frac{1}{4}\,\mathbf{a}_2 + z_2\,\mathbf{a}_3$ | = | $x_2 a\,\hat{\mathbf{x}} + \frac{1}{4} b\,\hat{\mathbf{y}} + z_2 c\,\hat{\mathbf{z}}$ | (4$c$) | $NH_4$ |
| $\mathbf{B}_6$ | = | $\left(\frac{1}{2} - x_2\right)\mathbf{a}_1 + \frac{3}{4}\,\mathbf{a}_2 + \left(\frac{1}{2} + z_2\right)\mathbf{a}_3$ | = | $\left(\frac{1}{2} - x_2\right) a\,\hat{\mathbf{x}} + \frac{3}{4} b\,\hat{\mathbf{y}} + \left(\frac{1}{2} + z_2\right) c\,\hat{\mathbf{z}}$ | (4$c$) | $NH_4$ |
| $\mathbf{B}_7$ | = | $-x_2\,\mathbf{a}_1 + \frac{3}{4}\,\mathbf{a}_2 - z_2\,\mathbf{a}_3$ | = | $-x_2 a\,\hat{\mathbf{x}} + \frac{3}{4} b\,\hat{\mathbf{y}} - z_2 c\,\hat{\mathbf{z}}$ | (4$c$) | $NH_4$ |
| $\mathbf{B}_8$ | = | $\left(\frac{1}{2} + x_2\right)\mathbf{a}_1 + \frac{1}{4}\,\mathbf{a}_2 + \left(\frac{1}{2} - z_2\right)\mathbf{a}_3$ | = | $\left(\frac{1}{2} + x_2\right) a\,\hat{\mathbf{x}} + \frac{1}{4} b\,\hat{\mathbf{y}} + \left(\frac{1}{2} - z_2\right) c\,\hat{\mathbf{z}}$ | (4$c$) | $NH_4$ |
| $\mathbf{B}_9$ | = | $x_3\,\mathbf{a}_1 + \frac{1}{4}\,\mathbf{a}_2 + z_3\,\mathbf{a}_3$ | = | $x_3 a\,\hat{\mathbf{x}} + \frac{1}{4} b\,\hat{\mathbf{y}} + z_3 c\,\hat{\mathbf{z}}$ | (4$c$) | O I |
| $\mathbf{B}_{10}$ | = | $\left(\frac{1}{2} - x_3\right)\mathbf{a}_1 + \frac{3}{4}\,\mathbf{a}_2 + \left(\frac{1}{2} + z_3\right)\mathbf{a}_3$ | = | $\left(\frac{1}{2} - x_3\right) a\,\hat{\mathbf{x}} + \frac{3}{4} b\,\hat{\mathbf{y}} + \left(\frac{1}{2} + z_3\right) c\,\hat{\mathbf{z}}$ | (4$c$) | O I |
| $\mathbf{B}_{11}$ | = | $-x_3\,\mathbf{a}_1 + \frac{3}{4}\,\mathbf{a}_2 - z_3\,\mathbf{a}_3$ | = | $-x_3 a\,\hat{\mathbf{x}} + \frac{3}{4} b\,\hat{\mathbf{y}} - z_3 c\,\hat{\mathbf{z}}$ | (4$c$) | O I |
| $\mathbf{B}_{12}$ | = | $\left(\frac{1}{2} + x_3\right)\mathbf{a}_1 + \frac{1}{4}\,\mathbf{a}_2 + \left(\frac{1}{2} - z_3\right)\mathbf{a}_3$ | = | $\left(\frac{1}{2} + x_3\right) a\,\hat{\mathbf{x}} + \frac{1}{4} b\,\hat{\mathbf{y}} + \left(\frac{1}{2} - z_3\right) c\,\hat{\mathbf{z}}$ | (4$c$) | O I |
| $\mathbf{B}_{13}$ | = | $x_4\,\mathbf{a}_1 + y_4\,\mathbf{a}_2 + z_4\,\mathbf{a}_3$ | = | $x_4 a\,\hat{\mathbf{x}} + y_4 b\,\hat{\mathbf{y}} + z_4 c\,\hat{\mathbf{z}}$ | (8$d$) | O II |
| $\mathbf{B}_{14}$ | = | $\left(\frac{1}{2} - x_4\right)\mathbf{a}_1 - y_4\,\mathbf{a}_2 + \left(\frac{1}{2} + z_4\right)\mathbf{a}_3$ | = | $\left(\frac{1}{2} - x_4\right) a\,\hat{\mathbf{x}} - y_4 b\,\hat{\mathbf{y}} + \left(\frac{1}{2} + z_4\right) c\,\hat{\mathbf{z}}$ | (8$d$) | O II |
| $\mathbf{B}_{15}$ | = | $-x_4\,\mathbf{a}_1 + \left(\frac{1}{2} + y_4\right)\mathbf{a}_2 - z_4\,\mathbf{a}_3$ | = | $-x_4 a\,\hat{\mathbf{x}} + \left(\frac{1}{2} + y_4\right) b\,\hat{\mathbf{y}} - z_4 c\,\hat{\mathbf{z}}$ | (8$d$) | O II |
| $\mathbf{B}_{16}$ | = | $\left(\frac{1}{2} + x_4\right)\mathbf{a}_1 + \left(\frac{1}{2} - y_4\right)\mathbf{a}_2 + \left(\frac{1}{2} - z_4\right)\mathbf{a}_3$ | = | $\left(\frac{1}{2} + x_4\right) a\,\hat{\mathbf{x}} + \left(\frac{1}{2} - y_4\right) b\,\hat{\mathbf{y}} + \left(\frac{1}{2} - z_4\right) c\,\hat{\mathbf{z}}$ | (8$d$) | O II |
| $\mathbf{B}_{17}$ | = | $-x_4\,\mathbf{a}_1 - y_4\,\mathbf{a}_2 - z_4\,\mathbf{a}_3$ | = | $-x_4 a\,\hat{\mathbf{x}} - y_4 b\,\hat{\mathbf{y}} - z_4 c\,\hat{\mathbf{z}}$ | (8$d$) | O II |
| $\mathbf{B}_{18}$ | = | $\left(\frac{1}{2} + x_4\right)\mathbf{a}_1 + y_4\,\mathbf{a}_2 + \left(\frac{1}{2} - z_4\right)\mathbf{a}_3$ | = | $\left(\frac{1}{2} + x_4\right) a\,\hat{\mathbf{x}} + y_4 b\,\hat{\mathbf{y}} + \left(\frac{1}{2} - z_4\right) c\,\hat{\mathbf{z}}$ | (8$d$) | O II |
| $\mathbf{B}_{19}$ | = | $x_4\,\mathbf{a}_1 + \left(\frac{1}{2} - y_4\right)\mathbf{a}_2 + z_4\,\mathbf{a}_3$ | = | $x_4 a\,\hat{\mathbf{x}} + \left(\frac{1}{2} - y_4\right) b\,\hat{\mathbf{y}} + z_4 c\,\hat{\mathbf{z}}$ | (8$d$) | O II |
| $\mathbf{B}_{20}$ | = | $\left(\frac{1}{2} - x_4\right)\mathbf{a}_1 + \left(\frac{1}{2} + y_4\right)\mathbf{a}_2 + \left(\frac{1}{2} + z_4\right)\mathbf{a}_3$ | = | $\left(\frac{1}{2} - x_4\right) a\,\hat{\mathbf{x}} + \left(\frac{1}{2} + y_4\right) b\,\hat{\mathbf{y}} + \left(\frac{1}{2} + z_4\right) c\,\hat{\mathbf{z}}$ | (8$d$) | O II |

**Geometry files:**

- CIF: pp. 1649
- POSCAR: pp. 1650

# Aragonite ($CaCO_3$, $G0_2$) Structure: ABC3_oP20_62_c_c_cd

http://aflow.org/prototype-encyclopedia/ABC3_oP20_62_c_c_cd

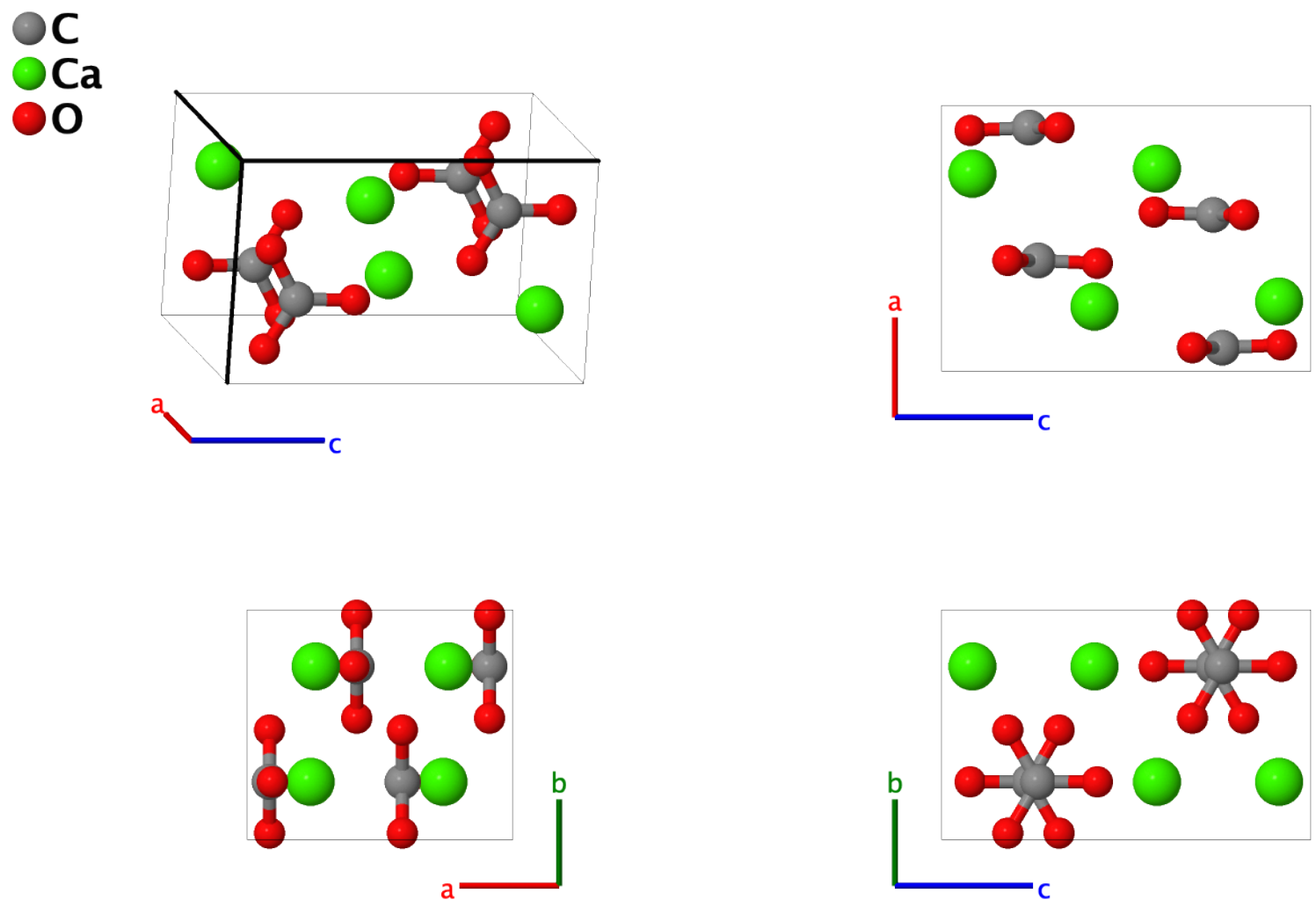

**Prototype** : $CaCO_3$

**AFLOW prototype label** : ABC3_oP20_62_c_c_cd

***Strukturbericht* designation** : $G0_2$

**Pearson symbol** : oP20

**Space group number** : 62

**Space group symbol** : $Pnma$

**AFLOW prototype command** : `aflow --proto=ABC3_oP20_62_c_c_cd`
`--params=`$a, b/a, c/a, x_1, z_1, x_2, z_2, x_3, z_3, x_4, y_4, z_4$

---

**Other compounds with this structure**

- $BaCO_3$ (Witherite), (Ba,Ca)$CO_3$ (Bromlite), (Cu,Mg,Mn)$CO_3$ (Manganocalcite), $MnCO_3$ (Rhodochrosite), $PbCO_3$ (Cerussite), and $SrCO_3$ (Strontianite)

---

- (Ewald, 1931) originally listed this as *Strukturbericht* $G2$, but it was designated $G0_2$ in (Gottfried, 1937).

---

**Simple Orthorhombic primitive vectors:**

$$\mathbf{a}_1 = a\,\hat{\mathbf{x}}$$
$$\mathbf{a}_2 = b\,\hat{\mathbf{y}}$$
$$\mathbf{a}_3 = c\,\hat{\mathbf{z}}$$

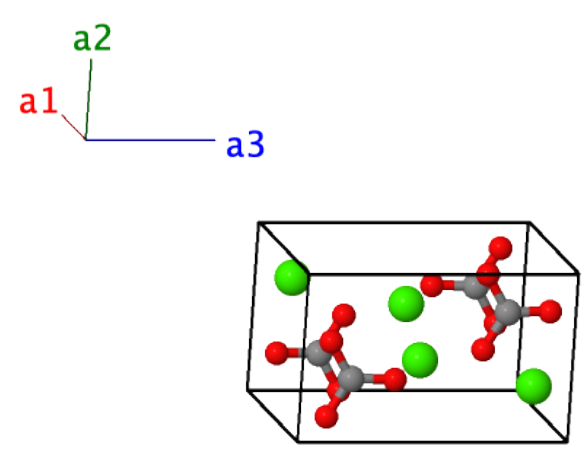

---

**Basis vectors:**

Lattice Coordinates | Cartesian Coordinates | Wyckoff Position | Atom Type

| | | | | | | |
|---|---|---|---|---|---|---|
| $\mathbf{B}_1$ | = | $x_1\,\mathbf{a}_1 + \frac{1}{4}\,\mathbf{a}_2 + z_1\,\mathbf{a}_3$ | = | $x_1 a\,\hat{\mathbf{x}} + \frac{1}{4} b\,\hat{\mathbf{y}} + z_1 c\,\hat{\mathbf{z}}$ | (4c) | C |
| $\mathbf{B}_2$ | = | $\left(\frac{1}{2} - x_1\right)\mathbf{a}_1 + \frac{3}{4}\,\mathbf{a}_2 + \left(\frac{1}{2} + z_1\right)\mathbf{a}_3$ | = | $\left(\frac{1}{2} - x_1\right) a\,\hat{\mathbf{x}} + \frac{3}{4} b\,\hat{\mathbf{y}} + \left(\frac{1}{2} + z_1\right) c\,\hat{\mathbf{z}}$ | (4c) | C |
| $\mathbf{B}_3$ | = | $-x_1\,\mathbf{a}_1 + \frac{3}{4}\,\mathbf{a}_2 - z_1\,\mathbf{a}_3$ | = | $-x_1 a\,\hat{\mathbf{x}} + \frac{3}{4} b\,\hat{\mathbf{y}} - z_1 c\,\hat{\mathbf{z}}$ | (4c) | C |
| $\mathbf{B}_4$ | = | $\left(\frac{1}{2} + x_1\right)\mathbf{a}_1 + \frac{1}{4}\,\mathbf{a}_2 + \left(\frac{1}{2} - z_1\right)\mathbf{a}_3$ | = | $\left(\frac{1}{2} + x_1\right) a\,\hat{\mathbf{x}} + \frac{1}{4} b\,\hat{\mathbf{y}} + \left(\frac{1}{2} - z_1\right) c\,\hat{\mathbf{z}}$ | (4c) | C |
| $\mathbf{B}_5$ | = | $x_2\,\mathbf{a}_1 + \frac{1}{4}\,\mathbf{a}_2 + z_2\,\mathbf{a}_3$ | = | $x_2 a\,\hat{\mathbf{x}} + \frac{1}{4} b\,\hat{\mathbf{y}} + z_2 c\,\hat{\mathbf{z}}$ | (4c) | Ca |
| $\mathbf{B}_6$ | = | $\left(\frac{1}{2} - x_2\right)\mathbf{a}_1 + \frac{3}{4}\,\mathbf{a}_2 + \left(\frac{1}{2} + z_2\right)\mathbf{a}_3$ | = | $\left(\frac{1}{2} - x_2\right) a\,\hat{\mathbf{x}} + \frac{3}{4} b\,\hat{\mathbf{y}} + \left(\frac{1}{2} + z_2\right) c\,\hat{\mathbf{z}}$ | (4c) | Ca |
| $\mathbf{B}_7$ | = | $-x_2\,\mathbf{a}_1 + \frac{3}{4}\,\mathbf{a}_2 - z_2\,\mathbf{a}_3$ | = | $-x_2 a\,\hat{\mathbf{x}} + \frac{3}{4} b\,\hat{\mathbf{y}} - z_2 c\,\hat{\mathbf{z}}$ | (4c) | Ca |
| $\mathbf{B}_8$ | = | $\left(\frac{1}{2} + x_2\right)\mathbf{a}_1 + \frac{1}{4}\,\mathbf{a}_2 + \left(\frac{1}{2} - z_2\right)\mathbf{a}_3$ | = | $\left(\frac{1}{2} + x_2\right) a\,\hat{\mathbf{x}} + \frac{1}{4} b\,\hat{\mathbf{y}} + \left(\frac{1}{2} - z_2\right) c\,\hat{\mathbf{z}}$ | (4c) | Ca |
| $\mathbf{B}_9$ | = | $x_3\,\mathbf{a}_1 + \frac{1}{4}\,\mathbf{a}_2 + z_3\,\mathbf{a}_3$ | = | $x_3 a\,\hat{\mathbf{x}} + \frac{1}{4} b\,\hat{\mathbf{y}} + z_3 c\,\hat{\mathbf{z}}$ | (4c) | O I |
| $\mathbf{B}_{10}$ | = | $\left(\frac{1}{2} - x_3\right)\mathbf{a}_1 + \frac{3}{4}\,\mathbf{a}_2 + \left(\frac{1}{2} + z_3\right)\mathbf{a}_3$ | = | $\left(\frac{1}{2} - x_3\right) a\,\hat{\mathbf{x}} + \frac{3}{4} b\,\hat{\mathbf{y}} + \left(\frac{1}{2} + z_3\right) c\,\hat{\mathbf{z}}$ | (4c) | O I |
| $\mathbf{B}_{11}$ | = | $-x_3\,\mathbf{a}_1 + \frac{3}{4}\,\mathbf{a}_2 - z_3\,\mathbf{a}_3$ | = | $-x_3 a\,\hat{\mathbf{x}} + \frac{3}{4} b\,\hat{\mathbf{y}} - z_3 c\,\hat{\mathbf{z}}$ | (4c) | O I |
| $\mathbf{B}_{12}$ | = | $\left(\frac{1}{2} + x_3\right)\mathbf{a}_1 + \frac{1}{4}\,\mathbf{a}_2 + \left(\frac{1}{2} - z_3\right)\mathbf{a}_3$ | = | $\left(\frac{1}{2} + x_3\right) a\,\hat{\mathbf{x}} + \frac{1}{4} b\,\hat{\mathbf{y}} + \left(\frac{1}{2} - z_3\right) c\,\hat{\mathbf{z}}$ | (4c) | O I |
| $\mathbf{B}_{13}$ | = | $x_4\,\mathbf{a}_1 + y_4\,\mathbf{a}_2 + z_4\,\mathbf{a}_3$ | = | $x_4 a\,\hat{\mathbf{x}} + y_4 b\,\hat{\mathbf{y}} + z_4 c\,\hat{\mathbf{z}}$ | (8d) | O II |
| $\mathbf{B}_{14}$ | = | $\left(\frac{1}{2} - x_4\right)\mathbf{a}_1 - y_4\,\mathbf{a}_2 + \left(\frac{1}{2} + z_4\right)\mathbf{a}_3$ | = | $\left(\frac{1}{2} - x_4\right) a\,\hat{\mathbf{x}} - y_4 b\,\hat{\mathbf{y}} + \left(\frac{1}{2} + z_4\right) c\,\hat{\mathbf{z}}$ | (8d) | O II |
| $\mathbf{B}_{15}$ | = | $-x_4\,\mathbf{a}_1 + \left(\frac{1}{2} + y_4\right)\mathbf{a}_2 - z_4\,\mathbf{a}_3$ | = | $-x_4 a\,\hat{\mathbf{x}} + \left(\frac{1}{2} + y_4\right) b\,\hat{\mathbf{y}} - z_4 c\,\hat{\mathbf{z}}$ | (8d) | O II |
| $\mathbf{B}_{16}$ | = | $\left(\frac{1}{2} + x_4\right)\mathbf{a}_1 + \left(\frac{1}{2} - y_4\right)\mathbf{a}_2 + \left(\frac{1}{2} - z_4\right)\mathbf{a}_3$ | = | $\left(\frac{1}{2} + x_4\right) a\,\hat{\mathbf{x}} + \left(\frac{1}{2} - y_4\right) b\,\hat{\mathbf{y}} + \left(\frac{1}{2} - z_4\right) c\,\hat{\mathbf{z}}$ | (8d) | O II |
| $\mathbf{B}_{17}$ | = | $-x_4\,\mathbf{a}_1 - y_4\,\mathbf{a}_2 - z_4\,\mathbf{a}_3$ | = | $-x_4 a\,\hat{\mathbf{x}} - y_4 b\,\hat{\mathbf{y}} - z_4 c\,\hat{\mathbf{z}}$ | (8d) | O II |
| $\mathbf{B}_{18}$ | = | $\left(\frac{1}{2} + x_4\right)\mathbf{a}_1 + y_4\,\mathbf{a}_2 + \left(\frac{1}{2} - z_4\right)\mathbf{a}_3$ | = | $\left(\frac{1}{2} + x_4\right) a\,\hat{\mathbf{x}} + y_4 b\,\hat{\mathbf{y}} + \left(\frac{1}{2} - z_4\right) c\,\hat{\mathbf{z}}$ | (8d) | O II |
| $\mathbf{B}_{19}$ | = | $x_4\,\mathbf{a}_1 + \left(\frac{1}{2} - y_4\right)\mathbf{a}_2 + z_4\,\mathbf{a}_3$ | = | $x_4 a\,\hat{\mathbf{x}} + \left(\frac{1}{2} - y_4\right) b\,\hat{\mathbf{y}} + z_4 c\,\hat{\mathbf{z}}$ | (8d) | O II |
| $\mathbf{B}_{20}$ | = | $\left(\frac{1}{2} - x_4\right)\mathbf{a}_1 + \left(\frac{1}{2} + y_4\right)\mathbf{a}_2 + \left(\frac{1}{2} + z_4\right)\mathbf{a}_3$ | = | $\left(\frac{1}{2} - x_4\right) a\,\hat{\mathbf{x}} + \left(\frac{1}{2} + y_4\right) b\,\hat{\mathbf{y}} + \left(\frac{1}{2} + z_4\right) c\,\hat{\mathbf{z}}$ | (8d) | O II |

**Geometry files:**

- CIF: pp. 1650
- POSCAR: pp. 1650

# Epididymite ($BeHNaO_8Si_3$, $S4_7$) Structure: ABCD8E3_oP112_62_d_2c_d_4c6d_3d

http://aflow.org/prototype-encyclopedia/ABCD8E3_oP112_62_d_2c_d_4c6d_3d

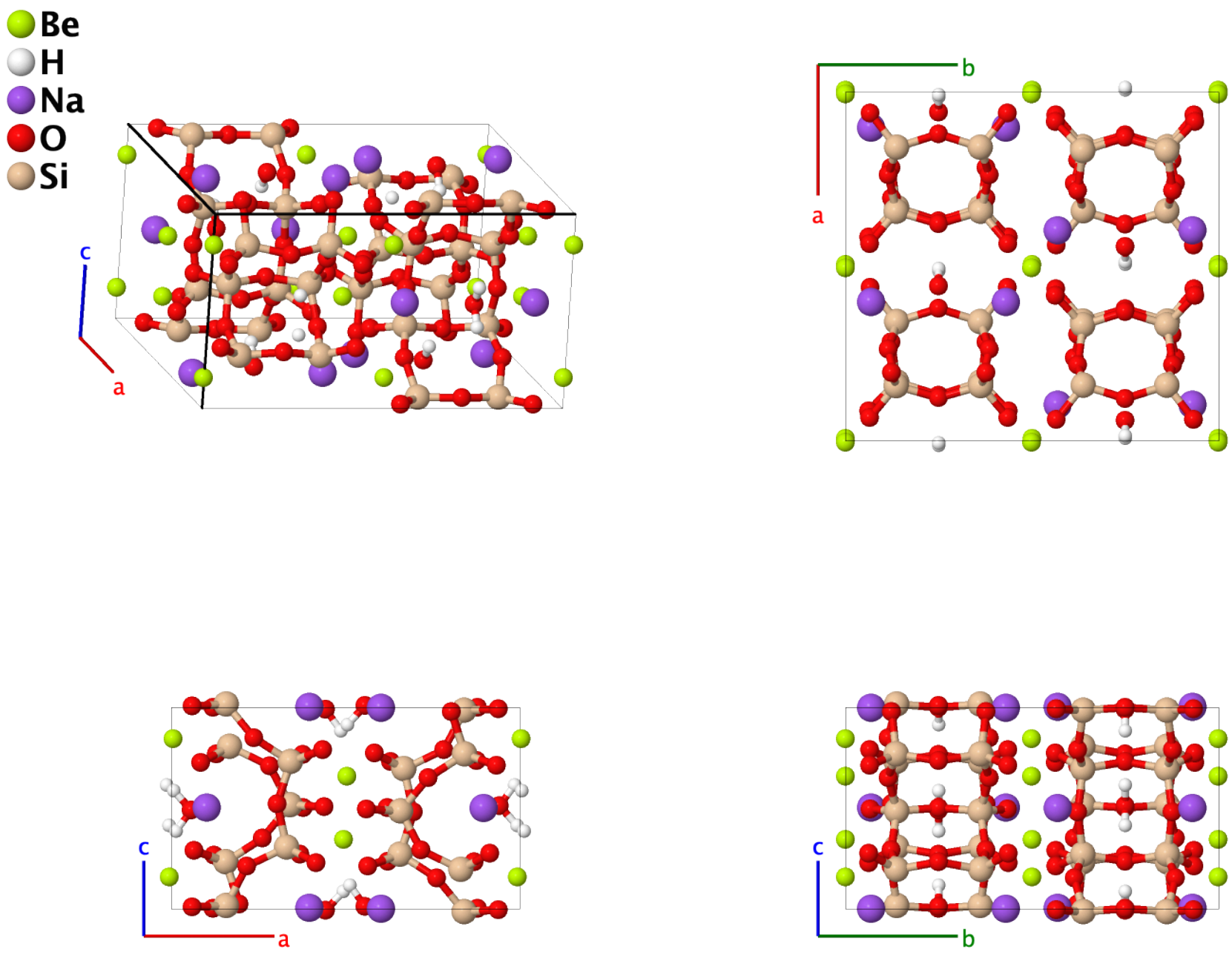

| | | |
|---|---|---|
| **Prototype** | : | $BeHNaO_8Si_3$ |
| **AFLOW prototype label** | : | ABCD8E3_oP112_62_d_2c_d_4c6d_3d |
| ***Strukturbericht* designation** | : | $S4_7$ |
| **Pearson symbol** | : | oP112 |
| **Space group number** | : | 62 |
| **Space group symbol** | : | *Pnma* |
| **AFLOW prototype command** | : | `aflow --proto=ABCD8E3_oP112_62_d_2c_d_4c6d_3d --params=`$a, b/a, c/a, x_1, z_1, x_2, z_2, x_3, z_3, x_4, z_4, x_5, z_5, x_6, z_6, x_7, y_7, z_7, x_8, y_8, z_8, x_9, y_9, z_9, x_{10}, y_{10}, z_{10}, x_{11}, y_{11}, z_{11}, x_{12}, y_{12}, z_{12}, x_{13}, y_{13}, z_{13}, x_{14}, y_{14}, z_{14}, x_{15}, y_{15}, z_{15}, x_{16}, y_{16}, z_{16}, x_{17}, y_{17}, z_{17}$ |

- (Ito, 1934)'s original determination of the epididymite structure, designated as $S4_7$ by (Gottfried, 1937), was flawed, even in the author's estimation. In particular, both the Si-O and Si-Be distances are very small, with Si-Be less than 1Å. In addition, Ito and many later authors assumed that the hydrogen atoms formed OH radicals. (Diego Gatta, 2008) found that the hydrogen atoms were actually a part of a water molecule. Since the initial structure is not even suitable for starting first-principles calculations, while the new structure maintains the *Pnma* #62 space group, we have dropped the former and designate the later as $S4_7$.

- Epididymite and its dimorph, eudidymite, are two forms of hydrated sodium beryllium silicate which are stable under ambient conditions. (Diego Gatta, 2008)

## Simple Orthorhombic primitive vectors:

$$
\begin{aligned}
\mathbf{a}_1 &= a\,\mathbf{\hat{x}} \\
\mathbf{a}_2 &= b\,\mathbf{\hat{y}} \\
\mathbf{a}_3 &= c\,\mathbf{\hat{z}}
\end{aligned}
$$

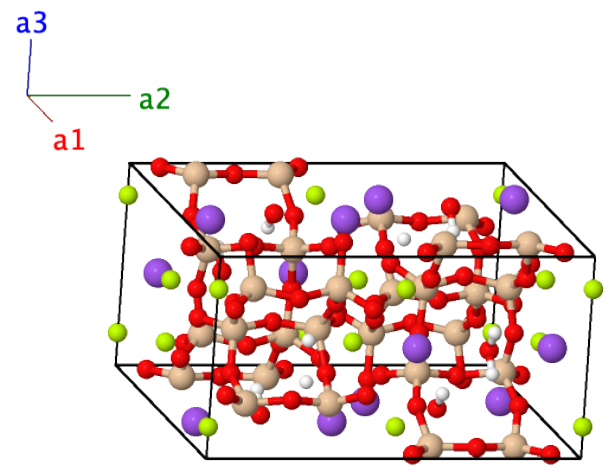

## Basis vectors:

| | | Lattice Coordinates | | Cartesian Coordinates | Wyckoff Position | Atom Type |
|---|---|---|---|---|---|---|
| $\mathbf{B}_1$ | = | $x_1\,\mathbf{a}_1 + \frac{1}{4}\,\mathbf{a}_2 + z_1\,\mathbf{a}_3$ | = | $x_1 a\,\mathbf{\hat{x}} + \frac{1}{4} b\,\mathbf{\hat{y}} + z_1 c\,\mathbf{\hat{z}}$ | (4$c$) | H I |
| $\mathbf{B}_2$ | = | $\left(\frac{1}{2} - x_1\right)\mathbf{a}_1 + \frac{3}{4}\,\mathbf{a}_2 + \left(\frac{1}{2} + z_1\right)\mathbf{a}_3$ | = | $\left(\frac{1}{2} - x_1\right) a\,\mathbf{\hat{x}} + \frac{3}{4} b\,\mathbf{\hat{y}} + \left(\frac{1}{2} + z_1\right) c\,\mathbf{\hat{z}}$ | (4$c$) | H I |
| $\mathbf{B}_3$ | = | $-x_1\,\mathbf{a}_1 + \frac{3}{4}\,\mathbf{a}_2 - z_1\,\mathbf{a}_3$ | = | $-x_1 a\,\mathbf{\hat{x}} + \frac{3}{4} b\,\mathbf{\hat{y}} - z_1 c\,\mathbf{\hat{z}}$ | (4$c$) | H I |
| $\mathbf{B}_4$ | = | $\left(\frac{1}{2} + x_1\right)\mathbf{a}_1 + \frac{1}{4}\,\mathbf{a}_2 + \left(\frac{1}{2} - z_1\right)\mathbf{a}_3$ | = | $\left(\frac{1}{2} + x_1\right) a\,\mathbf{\hat{x}} + \frac{1}{4} b\,\mathbf{\hat{y}} + \left(\frac{1}{2} - z_1\right) c\,\mathbf{\hat{z}}$ | (4$c$) | H I |
| $\mathbf{B}_5$ | = | $x_2\,\mathbf{a}_1 + \frac{1}{4}\,\mathbf{a}_2 + z_2\,\mathbf{a}_3$ | = | $x_2 a\,\mathbf{\hat{x}} + \frac{1}{4} b\,\mathbf{\hat{y}} + z_2 c\,\mathbf{\hat{z}}$ | (4$c$) | H II |
| $\mathbf{B}_6$ | = | $\left(\frac{1}{2} - x_2\right)\mathbf{a}_1 + \frac{3}{4}\,\mathbf{a}_2 + \left(\frac{1}{2} + z_2\right)\mathbf{a}_3$ | = | $\left(\frac{1}{2} - x_2\right) a\,\mathbf{\hat{x}} + \frac{3}{4} b\,\mathbf{\hat{y}} + \left(\frac{1}{2} + z_2\right) c\,\mathbf{\hat{z}}$ | (4$c$) | H II |
| $\mathbf{B}_7$ | = | $-x_2\,\mathbf{a}_1 + \frac{3}{4}\,\mathbf{a}_2 - z_2\,\mathbf{a}_3$ | = | $-x_2 a\,\mathbf{\hat{x}} + \frac{3}{4} b\,\mathbf{\hat{y}} - z_2 c\,\mathbf{\hat{z}}$ | (4$c$) | H II |
| $\mathbf{B}_8$ | = | $\left(\frac{1}{2} + x_2\right)\mathbf{a}_1 + \frac{1}{4}\,\mathbf{a}_2 + \left(\frac{1}{2} - z_2\right)\mathbf{a}_3$ | = | $\left(\frac{1}{2} + x_2\right) a\,\mathbf{\hat{x}} + \frac{1}{4} b\,\mathbf{\hat{y}} + \left(\frac{1}{2} - z_2\right) c\,\mathbf{\hat{z}}$ | (4$c$) | H II |
| $\mathbf{B}_9$ | = | $x_3\,\mathbf{a}_1 + \frac{1}{4}\,\mathbf{a}_2 + z_3\,\mathbf{a}_3$ | = | $x_3 a\,\mathbf{\hat{x}} + \frac{1}{4} b\,\mathbf{\hat{y}} + z_3 c\,\mathbf{\hat{z}}$ | (4$c$) | O I |
| $\mathbf{B}_{10}$ | = | $\left(\frac{1}{2} - x_3\right)\mathbf{a}_1 + \frac{3}{4}\,\mathbf{a}_2 + \left(\frac{1}{2} + z_3\right)\mathbf{a}_3$ | = | $\left(\frac{1}{2} - x_3\right) a\,\mathbf{\hat{x}} + \frac{3}{4} b\,\mathbf{\hat{y}} + \left(\frac{1}{2} + z_3\right) c\,\mathbf{\hat{z}}$ | (4$c$) | O I |
| $\mathbf{B}_{11}$ | = | $-x_3\,\mathbf{a}_1 + \frac{3}{4}\,\mathbf{a}_2 - z_3\,\mathbf{a}_3$ | = | $-x_3 a\,\mathbf{\hat{x}} + \frac{3}{4} b\,\mathbf{\hat{y}} - z_3 c\,\mathbf{\hat{z}}$ | (4$c$) | O I |
| $\mathbf{B}_{12}$ | = | $\left(\frac{1}{2} + x_3\right)\mathbf{a}_1 + \frac{1}{4}\,\mathbf{a}_2 + \left(\frac{1}{2} - z_3\right)\mathbf{a}_3$ | = | $\left(\frac{1}{2} + x_3\right) a\,\mathbf{\hat{x}} + \frac{1}{4} b\,\mathbf{\hat{y}} + \left(\frac{1}{2} - z_3\right) c\,\mathbf{\hat{z}}$ | (4$c$) | O I |
| $\mathbf{B}_{13}$ | = | $x_4\,\mathbf{a}_1 + \frac{1}{4}\,\mathbf{a}_2 + z_4\,\mathbf{a}_3$ | = | $x_4 a\,\mathbf{\hat{x}} + \frac{1}{4} b\,\mathbf{\hat{y}} + z_4 c\,\mathbf{\hat{z}}$ | (4$c$) | O II |
| $\mathbf{B}_{14}$ | = | $\left(\frac{1}{2} - x_4\right)\mathbf{a}_1 + \frac{3}{4}\,\mathbf{a}_2 + \left(\frac{1}{2} + z_4\right)\mathbf{a}_3$ | = | $\left(\frac{1}{2} - x_4\right) a\,\mathbf{\hat{x}} + \frac{3}{4} b\,\mathbf{\hat{y}} + \left(\frac{1}{2} + z_4\right) c\,\mathbf{\hat{z}}$ | (4$c$) | O II |
| $\mathbf{B}_{15}$ | = | $-x_4\,\mathbf{a}_1 + \frac{3}{4}\,\mathbf{a}_2 - z_4\,\mathbf{a}_3$ | = | $-x_4 a\,\mathbf{\hat{x}} + \frac{3}{4} b\,\mathbf{\hat{y}} - z_4 c\,\mathbf{\hat{z}}$ | (4$c$) | O II |
| $\mathbf{B}_{16}$ | = | $\left(\frac{1}{2} + x_4\right)\mathbf{a}_1 + \frac{1}{4}\,\mathbf{a}_2 + \left(\frac{1}{2} - z_4\right)\mathbf{a}_3$ | = | $\left(\frac{1}{2} + x_4\right) a\,\mathbf{\hat{x}} + \frac{1}{4} b\,\mathbf{\hat{y}} + \left(\frac{1}{2} - z_4\right) c\,\mathbf{\hat{z}}$ | (4$c$) | O II |
| $\mathbf{B}_{17}$ | = | $x_5\,\mathbf{a}_1 + \frac{1}{4}\,\mathbf{a}_2 + z_5\,\mathbf{a}_3$ | = | $x_5 a\,\mathbf{\hat{x}} + \frac{1}{4} b\,\mathbf{\hat{y}} + z_5 c\,\mathbf{\hat{z}}$ | (4$c$) | O III |
| $\mathbf{B}_{18}$ | = | $\left(\frac{1}{2} - x_5\right)\mathbf{a}_1 + \frac{3}{4}\,\mathbf{a}_2 + \left(\frac{1}{2} + z_5\right)\mathbf{a}_3$ | = | $\left(\frac{1}{2} - x_5\right) a\,\mathbf{\hat{x}} + \frac{3}{4} b\,\mathbf{\hat{y}} + \left(\frac{1}{2} + z_5\right) c\,\mathbf{\hat{z}}$ | (4$c$) | O III |
| $\mathbf{B}_{19}$ | = | $-x_5\,\mathbf{a}_1 + \frac{3}{4}\,\mathbf{a}_2 - z_5\,\mathbf{a}_3$ | = | $-x_5 a\,\mathbf{\hat{x}} + \frac{3}{4} b\,\mathbf{\hat{y}} - z_5 c\,\mathbf{\hat{z}}$ | (4$c$) | O III |
| $\mathbf{B}_{20}$ | = | $\left(\frac{1}{2} + x_5\right)\mathbf{a}_1 + \frac{1}{4}\,\mathbf{a}_2 + \left(\frac{1}{2} - z_5\right)\mathbf{a}_3$ | = | $\left(\frac{1}{2} + x_5\right) a\,\mathbf{\hat{x}} + \frac{1}{4} b\,\mathbf{\hat{y}} + \left(\frac{1}{2} - z_5\right) c\,\mathbf{\hat{z}}$ | (4$c$) | O III |
| $\mathbf{B}_{21}$ | = | $x_6\,\mathbf{a}_1 + \frac{1}{4}\,\mathbf{a}_2 + z_6\,\mathbf{a}_3$ | = | $x_6 a\,\mathbf{\hat{x}} + \frac{1}{4} b\,\mathbf{\hat{y}} + z_6 c\,\mathbf{\hat{z}}$ | (4$c$) | O IV |
| $\mathbf{B}_{22}$ | = | $\left(\frac{1}{2} - x_6\right)\mathbf{a}_1 + \frac{3}{4}\,\mathbf{a}_2 + \left(\frac{1}{2} + z_6\right)\mathbf{a}_3$ | = | $\left(\frac{1}{2} - x_6\right) a\,\mathbf{\hat{x}} + \frac{3}{4} b\,\mathbf{\hat{y}} + \left(\frac{1}{2} + z_6\right) c\,\mathbf{\hat{z}}$ | (4$c$) | O IV |
| $\mathbf{B}_{23}$ | = | $-x_6\,\mathbf{a}_1 + \frac{3}{4}\,\mathbf{a}_2 - z_6\,\mathbf{a}_3$ | = | $-x_6 a\,\mathbf{\hat{x}} + \frac{3}{4} b\,\mathbf{\hat{y}} - z_6 c\,\mathbf{\hat{z}}$ | (4$c$) | O IV |
| $\mathbf{B}_{24}$ | = | $\left(\frac{1}{2} + x_6\right)\mathbf{a}_1 + \frac{1}{4}\,\mathbf{a}_2 + \left(\frac{1}{2} - z_6\right)\mathbf{a}_3$ | = | $\left(\frac{1}{2} + x_6\right) a\,\mathbf{\hat{x}} + \frac{1}{4} b\,\mathbf{\hat{y}} + \left(\frac{1}{2} - z_6\right) c\,\mathbf{\hat{z}}$ | (4$c$) | O IV |
| $\mathbf{B}_{25}$ | = | $x_7\,\mathbf{a}_1 + y_7\,\mathbf{a}_2 + z_7\,\mathbf{a}_3$ | = | $x_7 a\,\mathbf{\hat{x}} + y_7 b\,\mathbf{\hat{y}} + z_7 c\,\mathbf{\hat{z}}$ | (8$d$) | Be |
| $\mathbf{B}_{26}$ | = | $\left(\frac{1}{2} - x_7\right)\mathbf{a}_1 - y_7\,\mathbf{a}_2 + \left(\frac{1}{2} + z_7\right)\mathbf{a}_3$ | = | $\left(\frac{1}{2} - x_7\right) a\,\mathbf{\hat{x}} - y_7 b\,\mathbf{\hat{y}} + \left(\frac{1}{2} + z_7\right) c\,\mathbf{\hat{z}}$ | (8$d$) | Be |
| $\mathbf{B}_{27}$ | = | $-x_7\,\mathbf{a}_1 + \left(\frac{1}{2} + y_7\right)\mathbf{a}_2 - z_7\,\mathbf{a}_3$ | = | $-x_7 a\,\mathbf{\hat{x}} + \left(\frac{1}{2} + y_7\right) b\,\mathbf{\hat{y}} - z_7 c\,\mathbf{\hat{z}}$ | (8$d$) | Be |
| $\mathbf{B}_{28}$ | = | $\left(\frac{1}{2}+x_{7}\right)\mathbf{a}_{1}+\left(\frac{1}{2}-y_{7}\right)\mathbf{a}_{2}+\left(\frac{1}{2}-z_{7}\right)\mathbf{a}_{3}$ | = | $\left(\frac{1}{2}+x_{7}\right)a\,\mathbf{\hat{x}}+\left(\frac{1}{2}-y_{7}\right)b\,\mathbf{\hat{y}}+\left(\frac{1}{2}-z_{7}\right)c\,\mathbf{\hat{z}}$ | (8$d$) | Be |
| $\mathbf{B}_{29}$ | = | $-x_{7}\,\mathbf{a}_{1}-y_{7}\,\mathbf{a}_{2}-z_{7}\,\mathbf{a}_{3}$ | = | $-x_{7}a\,\mathbf{\hat{x}}-y_{7}b\,\mathbf{\hat{y}}-z_{7}c\,\mathbf{\hat{z}}$ | (8$d$) | Be |
| $\mathbf{B}_{30}$ | = | $\left(\frac{1}{2}+x_{7}\right)\mathbf{a}_{1}+y_{7}\,\mathbf{a}_{2}+\left(\frac{1}{2}-z_{7}\right)\mathbf{a}_{3}$ | = | $\left(\frac{1}{2}+x_{7}\right)a\,\mathbf{\hat{x}}+y_{7}b\,\mathbf{\hat{y}}+\left(\frac{1}{2}-z_{7}\right)c\,\mathbf{\hat{z}}$ | (8$d$) | Be |
| $\mathbf{B}_{31}$ | = | $x_{7}\,\mathbf{a}_{1}+\left(\frac{1}{2}-y_{7}\right)\mathbf{a}_{2}+z_{7}\,\mathbf{a}_{3}$ | = | $x_{7}a\,\mathbf{\hat{x}}+\left(\frac{1}{2}-y_{7}\right)b\,\mathbf{\hat{y}}+z_{7}c\,\mathbf{\hat{z}}$ | (8$d$) | Be |
| $\mathbf{B}_{32}$ | = | $\left(\frac{1}{2}-x_{7}\right)\mathbf{a}_{1}+\left(\frac{1}{2}+y_{7}\right)\mathbf{a}_{2}+\left(\frac{1}{2}+z_{7}\right)\mathbf{a}_{3}$ | = | $\left(\frac{1}{2}-x_{7}\right)a\,\mathbf{\hat{x}}+\left(\frac{1}{2}+y_{7}\right)b\,\mathbf{\hat{y}}+\left(\frac{1}{2}+z_{7}\right)c\,\mathbf{\hat{z}}$ | (8$d$) | Be |
| $\mathbf{B}_{33}$ | = | $x_{8}\,\mathbf{a}_{1}+y_{8}\,\mathbf{a}_{2}+z_{8}\,\mathbf{a}_{3}$ | = | $x_{8}a\,\mathbf{\hat{x}}+y_{8}b\,\mathbf{\hat{y}}+z_{8}c\,\mathbf{\hat{z}}$ | (8$d$) | Na |
| $\mathbf{B}_{34}$ | = | $\left(\frac{1}{2}-x_{8}\right)\mathbf{a}_{1}-y_{8}\,\mathbf{a}_{2}+\left(\frac{1}{2}+z_{8}\right)\mathbf{a}_{3}$ | = | $\left(\frac{1}{2}-x_{8}\right)a\,\mathbf{\hat{x}}-y_{8}b\,\mathbf{\hat{y}}+\left(\frac{1}{2}+z_{8}\right)c\,\mathbf{\hat{z}}$ | (8$d$) | Na |
| $\mathbf{B}_{35}$ | = | $-x_{8}\,\mathbf{a}_{1}+\left(\frac{1}{2}+y_{8}\right)\mathbf{a}_{2}-z_{8}\,\mathbf{a}_{3}$ | = | $-x_{8}a\,\mathbf{\hat{x}}+\left(\frac{1}{2}+y_{8}\right)b\,\mathbf{\hat{y}}-z_{8}c\,\mathbf{\hat{z}}$ | (8$d$) | Na |
| $\mathbf{B}_{36}$ | = | $\left(\frac{1}{2}+x_{8}\right)\mathbf{a}_{1}+\left(\frac{1}{2}-y_{8}\right)\mathbf{a}_{2}+\left(\frac{1}{2}-z_{8}\right)\mathbf{a}_{3}$ | = | $\left(\frac{1}{2}+x_{8}\right)a\,\mathbf{\hat{x}}+\left(\frac{1}{2}-y_{8}\right)b\,\mathbf{\hat{y}}+\left(\frac{1}{2}-z_{8}\right)c\,\mathbf{\hat{z}}$ | (8$d$) | Na |
| $\mathbf{B}_{37}$ | = | $-x_{8}\,\mathbf{a}_{1}-y_{8}\,\mathbf{a}_{2}-z_{8}\,\mathbf{a}_{3}$ | = | $-x_{8}a\,\mathbf{\hat{x}}-y_{8}b\,\mathbf{\hat{y}}-z_{8}c\,\mathbf{\hat{z}}$ | (8$d$) | Na |
| $\mathbf{B}_{38}$ | = | $\left(\frac{1}{2}+x_{8}\right)\mathbf{a}_{1}+y_{8}\,\mathbf{a}_{2}+\left(\frac{1}{2}-z_{8}\right)\mathbf{a}_{3}$ | = | $\left(\frac{1}{2}+x_{8}\right)a\,\mathbf{\hat{x}}+y_{8}b\,\mathbf{\hat{y}}+\left(\frac{1}{2}-z_{8}\right)c\,\mathbf{\hat{z}}$ | (8$d$) | Na |
| $\mathbf{B}_{39}$ | = | $x_{8}\,\mathbf{a}_{1}+\left(\frac{1}{2}-y_{8}\right)\mathbf{a}_{2}+z_{8}\,\mathbf{a}_{3}$ | = | $x_{8}a\,\mathbf{\hat{x}}+\left(\frac{1}{2}-y_{8}\right)b\,\mathbf{\hat{y}}+z_{8}c\,\mathbf{\hat{z}}$ | (8$d$) | Na |
| $\mathbf{B}_{40}$ | = | $\left(\frac{1}{2}-x_{8}\right)\mathbf{a}_{1}+\left(\frac{1}{2}+y_{8}\right)\mathbf{a}_{2}+\left(\frac{1}{2}+z_{8}\right)\mathbf{a}_{3}$ | = | $\left(\frac{1}{2}-x_{8}\right)a\,\mathbf{\hat{x}}+\left(\frac{1}{2}+y_{8}\right)b\,\mathbf{\hat{y}}+\left(\frac{1}{2}+z_{8}\right)c\,\mathbf{\hat{z}}$ | (8$d$) | Na |
| $\mathbf{B}_{41}$ | = | $x_{9}\,\mathbf{a}_{1}+y_{9}\,\mathbf{a}_{2}+z_{9}\,\mathbf{a}_{3}$ | = | $x_{9}a\,\mathbf{\hat{x}}+y_{9}b\,\mathbf{\hat{y}}+z_{9}c\,\mathbf{\hat{z}}$ | (8$d$) | O V |
| $\mathbf{B}_{42}$ | = | $\left(\frac{1}{2}-x_{9}\right)\mathbf{a}_{1}-y_{9}\,\mathbf{a}_{2}+\left(\frac{1}{2}+z_{9}\right)\mathbf{a}_{3}$ | = | $\left(\frac{1}{2}-x_{9}\right)a\,\mathbf{\hat{x}}-y_{9}b\,\mathbf{\hat{y}}+\left(\frac{1}{2}+z_{9}\right)c\,\mathbf{\hat{z}}$ | (8$d$) | O V |
| $\mathbf{B}_{43}$ | = | $-x_{9}\,\mathbf{a}_{1}+\left(\frac{1}{2}+y_{9}\right)\mathbf{a}_{2}-z_{9}\,\mathbf{a}_{3}$ | = | $-x_{9}a\,\mathbf{\hat{x}}+\left(\frac{1}{2}+y_{9}\right)b\,\mathbf{\hat{y}}-z_{9}c\,\mathbf{\hat{z}}$ | (8$d$) | O V |
| $\mathbf{B}_{44}$ | = | $\left(\frac{1}{2}+x_{9}\right)\mathbf{a}_{1}+\left(\frac{1}{2}-y_{9}\right)\mathbf{a}_{2}+\left(\frac{1}{2}-z_{9}\right)\mathbf{a}_{3}$ | = | $\left(\frac{1}{2}+x_{9}\right)a\,\mathbf{\hat{x}}+\left(\frac{1}{2}-y_{9}\right)b\,\mathbf{\hat{y}}+\left(\frac{1}{2}-z_{9}\right)c\,\mathbf{\hat{z}}$ | (8$d$) | O V |
| $\mathbf{B}_{45}$ | = | $-x_{9}\,\mathbf{a}_{1}-y_{9}\,\mathbf{a}_{2}-z_{9}\,\mathbf{a}_{3}$ | = | $-x_{9}a\,\mathbf{\hat{x}}-y_{9}b\,\mathbf{\hat{y}}-z_{9}c\,\mathbf{\hat{z}}$ | (8$d$) | O V |
| $\mathbf{B}_{46}$ | = | $\left(\frac{1}{2}+x_{9}\right)\mathbf{a}_{1}+y_{9}\,\mathbf{a}_{2}+\left(\frac{1}{2}-z_{9}\right)\mathbf{a}_{3}$ | = | $\left(\frac{1}{2}+x_{9}\right)a\,\mathbf{\hat{x}}+y_{9}b\,\mathbf{\hat{y}}+\left(\frac{1}{2}-z_{9}\right)c\,\mathbf{\hat{z}}$ | (8$d$) | O V |
| $\mathbf{B}_{47}$ | = | $x_{9}\,\mathbf{a}_{1}+\left(\frac{1}{2}-y_{9}\right)\mathbf{a}_{2}+z_{9}\,\mathbf{a}_{3}$ | = | $x_{9}a\,\mathbf{\hat{x}}+\left(\frac{1}{2}-y_{9}\right)b\,\mathbf{\hat{y}}+z_{9}c\,\mathbf{\hat{z}}$ | (8$d$) | O V |
| $\mathbf{B}_{48}$ | = | $\left(\frac{1}{2}-x_{9}\right)\mathbf{a}_{1}+\left(\frac{1}{2}+y_{9}\right)\mathbf{a}_{2}+\left(\frac{1}{2}+z_{9}\right)\mathbf{a}_{3}$ | = | $\left(\frac{1}{2}-x_{9}\right)a\,\mathbf{\hat{x}}+\left(\frac{1}{2}+y_{9}\right)b\,\mathbf{\hat{y}}+\left(\frac{1}{2}+z_{9}\right)c\,\mathbf{\hat{z}}$ | (8$d$) | O V |
| $\mathbf{B}_{49}$ | = | $x_{10}\,\mathbf{a}_{1}+y_{10}\,\mathbf{a}_{2}+z_{10}\,\mathbf{a}_{3}$ | = | $x_{10}a\,\mathbf{\hat{x}}+y_{10}b\,\mathbf{\hat{y}}+z_{10}c\,\mathbf{\hat{z}}$ | (8$d$) | O VI |
| $\mathbf{B}_{50}$ | = | $\left(\frac{1}{2}-x_{10}\right)\mathbf{a}_{1}-y_{10}\,\mathbf{a}_{2}+\left(\frac{1}{2}+z_{10}\right)\mathbf{a}_{3}$ | = | $\left(\frac{1}{2}-x_{10}\right)a\,\mathbf{\hat{x}}-y_{10}b\,\mathbf{\hat{y}}+\left(\frac{1}{2}+z_{10}\right)c\,\mathbf{\hat{z}}$ | (8$d$) | O VI |
| $\mathbf{B}_{51}$ | = | $-x_{10}\,\mathbf{a}_{1}+\left(\frac{1}{2}+y_{10}\right)\mathbf{a}_{2}-z_{10}\,\mathbf{a}_{3}$ | = | $-x_{10}a\,\mathbf{\hat{x}}+\left(\frac{1}{2}+y_{10}\right)b\,\mathbf{\hat{y}}-z_{10}c\,\mathbf{\hat{z}}$ | (8$d$) | O VI |
| $\mathbf{B}_{52}$ | = | $\left(\frac{1}{2}+x_{10}\right)\mathbf{a}_{1}+\left(\frac{1}{2}-y_{10}\right)\mathbf{a}_{2}+\left(\frac{1}{2}-z_{10}\right)\mathbf{a}_{3}$ | = | $\left(\frac{1}{2}+x_{10}\right)a\,\mathbf{\hat{x}}+\left(\frac{1}{2}-y_{10}\right)b\,\mathbf{\hat{y}}+\left(\frac{1}{2}-z_{10}\right)c\,\mathbf{\hat{z}}$ | (8$d$) | O VI |
| $\mathbf{B}_{53}$ | = | $-x_{10}\,\mathbf{a}_{1}-y_{10}\,\mathbf{a}_{2}-z_{10}\,\mathbf{a}_{3}$ | = | $-x_{10}a\,\mathbf{\hat{x}}-y_{10}b\,\mathbf{\hat{y}}-z_{10}c\,\mathbf{\hat{z}}$ | (8$d$) | O VI |
| $\mathbf{B}_{54}$ | = | $\left(\frac{1}{2}+x_{10}\right)\mathbf{a}_{1}+y_{10}\,\mathbf{a}_{2}+\left(\frac{1}{2}-z_{10}\right)\mathbf{a}_{3}$ | = | $\left(\frac{1}{2}+x_{10}\right)a\,\mathbf{\hat{x}}+y_{10}b\,\mathbf{\hat{y}}+\left(\frac{1}{2}-z_{10}\right)c\,\mathbf{\hat{z}}$ | (8$d$) | O VI |
| $\mathbf{B}_{55}$ | = | $x_{10}\,\mathbf{a}_{1}+\left(\frac{1}{2}-y_{10}\right)\mathbf{a}_{2}+z_{10}\,\mathbf{a}_{3}$ | = | $x_{10}a\,\mathbf{\hat{x}}+\left(\frac{1}{2}-y_{10}\right)b\,\mathbf{\hat{y}}+z_{10}c\,\mathbf{\hat{z}}$ | (8$d$) | O VI |
| $\mathbf{B}_{56}$ | = | $\left(\frac{1}{2}-x_{10}\right)\mathbf{a}_{1}+\left(\frac{1}{2}+y_{10}\right)\mathbf{a}_{2}+\left(\frac{1}{2}+z_{10}\right)\mathbf{a}_{3}$ | = | $\left(\frac{1}{2}-x_{10}\right)a\,\mathbf{\hat{x}}+\left(\frac{1}{2}+y_{10}\right)b\,\mathbf{\hat{y}}+\left(\frac{1}{2}+z_{10}\right)c\,\mathbf{\hat{z}}$ | (8$d$) | O VI |
| $\mathbf{B}_{57}$ | = | $x_{11}\,\mathbf{a}_{1}+y_{11}\,\mathbf{a}_{2}+z_{11}\,\mathbf{a}_{3}$ | = | $x_{11}a\,\mathbf{\hat{x}}+y_{11}b\,\mathbf{\hat{y}}+z_{11}c\,\mathbf{\hat{z}}$ | (8$d$) | O VII |
| $\mathbf{B}_{58}$ | = | $\left(\frac{1}{2}-x_{11}\right)\mathbf{a}_1 - y_{11}\,\mathbf{a}_2 + \left(\frac{1}{2}+z_{11}\right)\mathbf{a}_3$ | = | $\left(\frac{1}{2}-x_{11}\right)a\,\hat{\mathbf{x}} - y_{11}b\,\hat{\mathbf{y}} + \left(\frac{1}{2}+z_{11}\right)c\,\hat{\mathbf{z}}$ | $(8d)$ | O VII |
| $\mathbf{B}_{59}$ | = | $-x_{11}\,\mathbf{a}_1 + \left(\frac{1}{2}+y_{11}\right)\mathbf{a}_2 - z_{11}\,\mathbf{a}_3$ | = | $-x_{11}a\,\hat{\mathbf{x}} + \left(\frac{1}{2}+y_{11}\right)b\,\hat{\mathbf{y}} - z_{11}c\,\hat{\mathbf{z}}$ | $(8d)$ | O VII |
| $\mathbf{B}_{60}$ | = | $\left(\frac{1}{2}+x_{11}\right)\mathbf{a}_1 + \left(\frac{1}{2}-y_{11}\right)\mathbf{a}_2 + \left(\frac{1}{2}-z_{11}\right)\mathbf{a}_3$ | = | $\left(\frac{1}{2}+x_{11}\right)a\,\hat{\mathbf{x}} + \left(\frac{1}{2}-y_{11}\right)b\,\hat{\mathbf{y}} + \left(\frac{1}{2}-z_{11}\right)c\,\hat{\mathbf{z}}$ | $(8d)$ | O VII |
| $\mathbf{B}_{61}$ | = | $-x_{11}\,\mathbf{a}_1 - y_{11}\,\mathbf{a}_2 - z_{11}\,\mathbf{a}_3$ | = | $-x_{11}a\,\hat{\mathbf{x}} - y_{11}b\,\hat{\mathbf{y}} - z_{11}c\,\hat{\mathbf{z}}$ | $(8d)$ | O VII |
| $\mathbf{B}_{62}$ | = | $\left(\frac{1}{2}+x_{11}\right)\mathbf{a}_1 + y_{11}\,\mathbf{a}_2 + \left(\frac{1}{2}-z_{11}\right)\mathbf{a}_3$ | = | $\left(\frac{1}{2}+x_{11}\right)a\,\hat{\mathbf{x}} + y_{11}b\,\hat{\mathbf{y}} + \left(\frac{1}{2}-z_{11}\right)c\,\hat{\mathbf{z}}$ | $(8d)$ | O VII |
| $\mathbf{B}_{63}$ | = | $x_{11}\,\mathbf{a}_1 + \left(\frac{1}{2}-y_{11}\right)\mathbf{a}_2 + z_{11}\,\mathbf{a}_3$ | = | $x_{11}a\,\hat{\mathbf{x}} + \left(\frac{1}{2}-y_{11}\right)b\,\hat{\mathbf{y}} + z_{11}c\,\hat{\mathbf{z}}$ | $(8d)$ | O VII |
| $\mathbf{B}_{64}$ | = | $\left(\frac{1}{2}-x_{11}\right)\mathbf{a}_1 + \left(\frac{1}{2}+y_{11}\right)\mathbf{a}_2 + \left(\frac{1}{2}+z_{11}\right)\mathbf{a}_3$ | = | $\left(\frac{1}{2}-x_{11}\right)a\,\hat{\mathbf{x}} + \left(\frac{1}{2}+y_{11}\right)b\,\hat{\mathbf{y}} + \left(\frac{1}{2}+z_{11}\right)c\,\hat{\mathbf{z}}$ | $(8d)$ | O VII |
| $\mathbf{B}_{65}$ | = | $x_{12}\,\mathbf{a}_1 + y_{12}\,\mathbf{a}_2 + z_{12}\,\mathbf{a}_3$ | = | $x_{12}a\,\hat{\mathbf{x}} + y_{12}b\,\hat{\mathbf{y}} + z_{12}c\,\hat{\mathbf{z}}$ | $(8d)$ | O VIII |
| $\mathbf{B}_{66}$ | = | $\left(\frac{1}{2}-x_{12}\right)\mathbf{a}_1 - y_{12}\,\mathbf{a}_2 + \left(\frac{1}{2}+z_{12}\right)\mathbf{a}_3$ | = | $\left(\frac{1}{2}-x_{12}\right)a\,\hat{\mathbf{x}} - y_{12}b\,\hat{\mathbf{y}} + \left(\frac{1}{2}+z_{12}\right)c\,\hat{\mathbf{z}}$ | $(8d)$ | O VIII |
| $\mathbf{B}_{67}$ | = | $-x_{12}\,\mathbf{a}_1 + \left(\frac{1}{2}+y_{12}\right)\mathbf{a}_2 - z_{12}\,\mathbf{a}_3$ | = | $-x_{12}a\,\hat{\mathbf{x}} + \left(\frac{1}{2}+y_{12}\right)b\,\hat{\mathbf{y}} - z_{12}c\,\hat{\mathbf{z}}$ | $(8d)$ | O VIII |
| $\mathbf{B}_{68}$ | = | $\left(\frac{1}{2}+x_{12}\right)\mathbf{a}_1 + \left(\frac{1}{2}-y_{12}\right)\mathbf{a}_2 + \left(\frac{1}{2}-z_{12}\right)\mathbf{a}_3$ | = | $\left(\frac{1}{2}+x_{12}\right)a\,\hat{\mathbf{x}} + \left(\frac{1}{2}-y_{12}\right)b\,\hat{\mathbf{y}} + \left(\frac{1}{2}-z_{12}\right)c\,\hat{\mathbf{z}}$ | $(8d)$ | O VIII |
| $\mathbf{B}_{69}$ | = | $-x_{12}\,\mathbf{a}_1 - y_{12}\,\mathbf{a}_2 - z_{12}\,\mathbf{a}_3$ | = | $-x_{12}a\,\hat{\mathbf{x}} - y_{12}b\,\hat{\mathbf{y}} - z_{12}c\,\hat{\mathbf{z}}$ | $(8d)$ | O VIII |
| $\mathbf{B}_{70}$ | = | $\left(\frac{1}{2}+x_{12}\right)\mathbf{a}_1 + y_{12}\,\mathbf{a}_2 + \left(\frac{1}{2}-z_{12}\right)\mathbf{a}_3$ | = | $\left(\frac{1}{2}+x_{12}\right)a\,\hat{\mathbf{x}} + y_{12}b\,\hat{\mathbf{y}} + \left(\frac{1}{2}-z_{12}\right)c\,\hat{\mathbf{z}}$ | $(8d)$ | O VIII |
| $\mathbf{B}_{71}$ | = | $x_{12}\,\mathbf{a}_1 + \left(\frac{1}{2}-y_{12}\right)\mathbf{a}_2 + z_{12}\,\mathbf{a}_3$ | = | $x_{12}a\,\hat{\mathbf{x}} + \left(\frac{1}{2}-y_{12}\right)b\,\hat{\mathbf{y}} + z_{12}c\,\hat{\mathbf{z}}$ | $(8d)$ | O VIII |
| $\mathbf{B}_{72}$ | = | $\left(\frac{1}{2}-x_{12}\right)\mathbf{a}_1 + \left(\frac{1}{2}+y_{12}\right)\mathbf{a}_2 + \left(\frac{1}{2}+z_{12}\right)\mathbf{a}_3$ | = | $\left(\frac{1}{2}-x_{12}\right)a\,\hat{\mathbf{x}} + \left(\frac{1}{2}+y_{12}\right)b\,\hat{\mathbf{y}} + \left(\frac{1}{2}+z_{12}\right)c\,\hat{\mathbf{z}}$ | $(8d)$ | O VIII |
| $\mathbf{B}_{73}$ | = | $x_{13}\,\mathbf{a}_1 + y_{13}\,\mathbf{a}_2 + z_{13}\,\mathbf{a}_3$ | = | $x_{13}a\,\hat{\mathbf{x}} + y_{13}b\,\hat{\mathbf{y}} + z_{13}c\,\hat{\mathbf{z}}$ | $(8d)$ | O IX |
| $\mathbf{B}_{74}$ | = | $\left(\frac{1}{2}-x_{13}\right)\mathbf{a}_1 - y_{13}\,\mathbf{a}_2 + \left(\frac{1}{2}+z_{13}\right)\mathbf{a}_3$ | = | $\left(\frac{1}{2}-x_{13}\right)a\,\hat{\mathbf{x}} - y_{13}b\,\hat{\mathbf{y}} + \left(\frac{1}{2}+z_{13}\right)c\,\hat{\mathbf{z}}$ | $(8d)$ | O IX |
| $\mathbf{B}_{75}$ | = | $-x_{13}\,\mathbf{a}_1 + \left(\frac{1}{2}+y_{13}\right)\mathbf{a}_2 - z_{13}\,\mathbf{a}_3$ | = | $-x_{13}a\,\hat{\mathbf{x}} + \left(\frac{1}{2}+y_{13}\right)b\,\hat{\mathbf{y}} - z_{13}c\,\hat{\mathbf{z}}$ | $(8d)$ | O IX |
| $\mathbf{B}_{76}$ | = | $\left(\frac{1}{2}+x_{13}\right)\mathbf{a}_1 + \left(\frac{1}{2}-y_{13}\right)\mathbf{a}_2 + \left(\frac{1}{2}-z_{13}\right)\mathbf{a}_3$ | = | $\left(\frac{1}{2}+x_{13}\right)a\,\hat{\mathbf{x}} + \left(\frac{1}{2}-y_{13}\right)b\,\hat{\mathbf{y}} + \left(\frac{1}{2}-z_{13}\right)c\,\hat{\mathbf{z}}$ | $(8d)$ | O IX |
| $\mathbf{B}_{77}$ | = | $-x_{13}\,\mathbf{a}_1 - y_{13}\,\mathbf{a}_2 - z_{13}\,\mathbf{a}_3$ | = | $-x_{13}a\,\hat{\mathbf{x}} - y_{13}b\,\hat{\mathbf{y}} - z_{13}c\,\hat{\mathbf{z}}$ | $(8d)$ | O IX |
| $\mathbf{B}_{78}$ | = | $\left(\frac{1}{2}+x_{13}\right)\mathbf{a}_1 + y_{13}\,\mathbf{a}_2 + \left(\frac{1}{2}-z_{13}\right)\mathbf{a}_3$ | = | $\left(\frac{1}{2}+x_{13}\right)a\,\hat{\mathbf{x}} + y_{13}b\,\hat{\mathbf{y}} + \left(\frac{1}{2}-z_{13}\right)c\,\hat{\mathbf{z}}$ | $(8d)$ | O IX |
| $\mathbf{B}_{79}$ | = | $x_{13}\,\mathbf{a}_1 + \left(\frac{1}{2}-y_{13}\right)\mathbf{a}_2 + z_{13}\,\mathbf{a}_3$ | = | $x_{13}a\,\hat{\mathbf{x}} + \left(\frac{1}{2}-y_{13}\right)b\,\hat{\mathbf{y}} + z_{13}c\,\hat{\mathbf{z}}$ | $(8d)$ | O IX |
| $\mathbf{B}_{80}$ | = | $\left(\frac{1}{2}-x_{13}\right)\mathbf{a}_1 + \left(\frac{1}{2}+y_{13}\right)\mathbf{a}_2 + \left(\frac{1}{2}+z_{13}\right)\mathbf{a}_3$ | = | $\left(\frac{1}{2}-x_{13}\right)a\,\hat{\mathbf{x}} + \left(\frac{1}{2}+y_{13}\right)b\,\hat{\mathbf{y}} + \left(\frac{1}{2}+z_{13}\right)c\,\hat{\mathbf{z}}$ | $(8d)$ | O IX |
| $\mathbf{B}_{81}$ | = | $x_{14}\,\mathbf{a}_1 + y_{14}\,\mathbf{a}_2 + z_{14}\,\mathbf{a}_3$ | = | $x_{14}a\,\hat{\mathbf{x}} + y_{14}b\,\hat{\mathbf{y}} + z_{14}c\,\hat{\mathbf{z}}$ | $(8d)$ | O X |
| $\mathbf{B}_{82}$ | = | $\left(\frac{1}{2}-x_{14}\right)\mathbf{a}_1 - y_{14}\,\mathbf{a}_2 + \left(\frac{1}{2}+z_{14}\right)\mathbf{a}_3$ | = | $\left(\frac{1}{2}-x_{14}\right)a\,\hat{\mathbf{x}} - y_{14}b\,\hat{\mathbf{y}} + \left(\frac{1}{2}+z_{14}\right)c\,\hat{\mathbf{z}}$ | $(8d)$ | O X |
| $\mathbf{B}_{83}$ | = | $-x_{14}\,\mathbf{a}_1 + \left(\frac{1}{2}+y_{14}\right)\mathbf{a}_2 - z_{14}\,\mathbf{a}_3$ | = | $-x_{14}a\,\hat{\mathbf{x}} + \left(\frac{1}{2}+y_{14}\right)b\,\hat{\mathbf{y}} - z_{14}c\,\hat{\mathbf{z}}$ | $(8d)$ | O X |
| $\mathbf{B}_{84}$ | = | $\left(\frac{1}{2}+x_{14}\right)\mathbf{a}_1 + \left(\frac{1}{2}-y_{14}\right)\mathbf{a}_2 + \left(\frac{1}{2}-z_{14}\right)\mathbf{a}_3$ | = | $\left(\frac{1}{2}+x_{14}\right)a\,\hat{\mathbf{x}} + \left(\frac{1}{2}-y_{14}\right)b\,\hat{\mathbf{y}} + \left(\frac{1}{2}-z_{14}\right)c\,\hat{\mathbf{z}}$ | $(8d)$ | O X |
| $\mathbf{B}_{85}$ | = | $-x_{14}\,\mathbf{a}_1 - y_{14}\,\mathbf{a}_2 - z_{14}\,\mathbf{a}_3$ | = | $-x_{14}a\,\hat{\mathbf{x}} - y_{14}b\,\hat{\mathbf{y}} - z_{14}c\,\hat{\mathbf{z}}$ | $(8d)$ | O X |
| $\mathbf{B}_{86}$ | = | $\left(\frac{1}{2}+x_{14}\right)\mathbf{a}_1 + y_{14}\,\mathbf{a}_2 + \left(\frac{1}{2}-z_{14}\right)\mathbf{a}_3$ | = | $\left(\frac{1}{2}+x_{14}\right)a\,\hat{\mathbf{x}} + y_{14}b\,\hat{\mathbf{y}} + \left(\frac{1}{2}-z_{14}\right)c\,\hat{\mathbf{z}}$ | $(8d)$ | O X |
| $\mathbf{B}_{87}$ | = | $x_{14}\,\mathbf{a}_1 + \left(\frac{1}{2}-y_{14}\right)\mathbf{a}_2 + z_{14}\,\mathbf{a}_3$ | = | $x_{14}a\,\hat{\mathbf{x}} + \left(\frac{1}{2}-y_{14}\right)b\,\hat{\mathbf{y}} + z_{14}c\,\hat{\mathbf{z}}$ | $(8d)$ | O X |
| $\mathbf{B}_{88}$ | = | $\left(\frac{1}{2} - x_{14}\right) \mathbf{a}_1 + \left(\frac{1}{2} + y_{14}\right) \mathbf{a}_2 + \left(\frac{1}{2} + z_{14}\right) \mathbf{a}_3$ | = | $\left(\frac{1}{2} - x_{14}\right) a\,\mathbf{\hat{x}} + \left(\frac{1}{2} + y_{14}\right) b\,\mathbf{\hat{y}} + \left(\frac{1}{2} + z_{14}\right) c\,\mathbf{\hat{z}}$ | (8*d*) | O X |
| $\mathbf{B}_{89}$ | = | $x_{15}\,\mathbf{a}_1 + y_{15}\,\mathbf{a}_2 + z_{15}\,\mathbf{a}_3$ | = | $x_{15}a\,\mathbf{\hat{x}} + y_{15}b\,\mathbf{\hat{y}} + z_{15}c\,\mathbf{\hat{z}}$ | (8*d*) | Si I |
| $\mathbf{B}_{90}$ | = | $\left(\frac{1}{2} - x_{15}\right) \mathbf{a}_1 - y_{15}\,\mathbf{a}_2 + \left(\frac{1}{2} + z_{15}\right) \mathbf{a}_3$ | = | $\left(\frac{1}{2} - x_{15}\right) a\,\mathbf{\hat{x}} - y_{15}b\,\mathbf{\hat{y}} + \left(\frac{1}{2} + z_{15}\right) c\,\mathbf{\hat{z}}$ | (8*d*) | Si I |
| $\mathbf{B}_{91}$ | = | $-x_{15}\,\mathbf{a}_1 + \left(\frac{1}{2} + y_{15}\right) \mathbf{a}_2 - z_{15}\,\mathbf{a}_3$ | = | $-x_{15}a\,\mathbf{\hat{x}} + \left(\frac{1}{2} + y_{15}\right) b\,\mathbf{\hat{y}} - z_{15}c\,\mathbf{\hat{z}}$ | (8*d*) | Si I |
| $\mathbf{B}_{92}$ | = | $\left(\frac{1}{2} + x_{15}\right) \mathbf{a}_1 + \left(\frac{1}{2} - y_{15}\right) \mathbf{a}_2 + \left(\frac{1}{2} - z_{15}\right) \mathbf{a}_3$ | = | $\left(\frac{1}{2} + x_{15}\right) a\,\mathbf{\hat{x}} + \left(\frac{1}{2} - y_{15}\right) b\,\mathbf{\hat{y}} + \left(\frac{1}{2} - z_{15}\right) c\,\mathbf{\hat{z}}$ | (8*d*) | Si I |
| $\mathbf{B}_{93}$ | = | $-x_{15}\,\mathbf{a}_1 - y_{15}\,\mathbf{a}_2 - z_{15}\,\mathbf{a}_3$ | = | $-x_{15}a\,\mathbf{\hat{x}} - y_{15}b\,\mathbf{\hat{y}} - z_{15}c\,\mathbf{\hat{z}}$ | (8*d*) | Si I |
| $\mathbf{B}_{94}$ | = | $\left(\frac{1}{2} + x_{15}\right) \mathbf{a}_1 + y_{15}\,\mathbf{a}_2 + \left(\frac{1}{2} - z_{15}\right) \mathbf{a}_3$ | = | $\left(\frac{1}{2} + x_{15}\right) a\,\mathbf{\hat{x}} + y_{15}b\,\mathbf{\hat{y}} + \left(\frac{1}{2} - z_{15}\right) c\,\mathbf{\hat{z}}$ | (8*d*) | Si I |
| $\mathbf{B}_{95}$ | = | $x_{15}\,\mathbf{a}_1 + \left(\frac{1}{2} - y_{15}\right) \mathbf{a}_2 + z_{15}\,\mathbf{a}_3$ | = | $x_{15}a\,\mathbf{\hat{x}} + \left(\frac{1}{2} - y_{15}\right) b\,\mathbf{\hat{y}} + z_{15}c\,\mathbf{\hat{z}}$ | (8*d*) | Si I |
| $\mathbf{B}_{96}$ | = | $\left(\frac{1}{2} - x_{15}\right) \mathbf{a}_1 + \left(\frac{1}{2} + y_{15}\right) \mathbf{a}_2 + \left(\frac{1}{2} + z_{15}\right) \mathbf{a}_3$ | = | $\left(\frac{1}{2} - x_{15}\right) a\,\mathbf{\hat{x}} + \left(\frac{1}{2} + y_{15}\right) b\,\mathbf{\hat{y}} + \left(\frac{1}{2} + z_{15}\right) c\,\mathbf{\hat{z}}$ | (8*d*) | Si I |
| $\mathbf{B}_{97}$ | = | $x_{16}\,\mathbf{a}_1 + y_{16}\,\mathbf{a}_2 + z_{16}\,\mathbf{a}_3$ | = | $x_{16}a\,\mathbf{\hat{x}} + y_{16}b\,\mathbf{\hat{y}} + z_{16}c\,\mathbf{\hat{z}}$ | (8*d*) | Si II |
| $\mathbf{B}_{98}$ | = | $\left(\frac{1}{2} - x_{16}\right) \mathbf{a}_1 - y_{16}\,\mathbf{a}_2 + \left(\frac{1}{2} + z_{16}\right) \mathbf{a}_3$ | = | $\left(\frac{1}{2} - x_{16}\right) a\,\mathbf{\hat{x}} - y_{16}b\,\mathbf{\hat{y}} + \left(\frac{1}{2} + z_{16}\right) c\,\mathbf{\hat{z}}$ | (8*d*) | Si II |
| $\mathbf{B}_{99}$ | = | $-x_{16}\,\mathbf{a}_1 + \left(\frac{1}{2} + y_{16}\right) \mathbf{a}_2 - z_{16}\,\mathbf{a}_3$ | = | $-x_{16}a\,\mathbf{\hat{x}} + \left(\frac{1}{2} + y_{16}\right) b\,\mathbf{\hat{y}} - z_{16}c\,\mathbf{\hat{z}}$ | (8*d*) | Si II |
| $\mathbf{B}_{100}$ | = | $\left(\frac{1}{2} + x_{16}\right) \mathbf{a}_1 + \left(\frac{1}{2} - y_{16}\right) \mathbf{a}_2 + \left(\frac{1}{2} - z_{16}\right) \mathbf{a}_3$ | = | $\left(\frac{1}{2} + x_{16}\right) a\,\mathbf{\hat{x}} + \left(\frac{1}{2} - y_{16}\right) b\,\mathbf{\hat{y}} + \left(\frac{1}{2} - z_{16}\right) c\,\mathbf{\hat{z}}$ | (8*d*) | Si II |
| $\mathbf{B}_{101}$ | = | $-x_{16}\,\mathbf{a}_1 - y_{16}\,\mathbf{a}_2 - z_{16}\,\mathbf{a}_3$ | = | $-x_{16}a\,\mathbf{\hat{x}} - y_{16}b\,\mathbf{\hat{y}} - z_{16}c\,\mathbf{\hat{z}}$ | (8*d*) | Si II |
| $\mathbf{B}_{102}$ | = | $\left(\frac{1}{2} + x_{16}\right) \mathbf{a}_1 + y_{16}\,\mathbf{a}_2 + \left(\frac{1}{2} - z_{16}\right) \mathbf{a}_3$ | = | $\left(\frac{1}{2} + x_{16}\right) a\,\mathbf{\hat{x}} + y_{16}b\,\mathbf{\hat{y}} + \left(\frac{1}{2} - z_{16}\right) c\,\mathbf{\hat{z}}$ | (8*d*) | Si II |
| $\mathbf{B}_{103}$ | = | $x_{16}\,\mathbf{a}_1 + \left(\frac{1}{2} - y_{16}\right) \mathbf{a}_2 + z_{16}\,\mathbf{a}_3$ | = | $x_{16}a\,\mathbf{\hat{x}} + \left(\frac{1}{2} - y_{16}\right) b\,\mathbf{\hat{y}} + z_{16}c\,\mathbf{\hat{z}}$ | (8*d*) | Si II |
| $\mathbf{B}_{104}$ | = | $\left(\frac{1}{2} - x_{16}\right) \mathbf{a}_1 + \left(\frac{1}{2} + y_{16}\right) \mathbf{a}_2 + \left(\frac{1}{2} + z_{16}\right) \mathbf{a}_3$ | = | $\left(\frac{1}{2} - x_{16}\right) a\,\mathbf{\hat{x}} + \left(\frac{1}{2} + y_{16}\right) b\,\mathbf{\hat{y}} + \left(\frac{1}{2} + z_{16}\right) c\,\mathbf{\hat{z}}$ | (8*d*) | Si II |
| $\mathbf{B}_{105}$ | = | $x_{17}\,\mathbf{a}_1 + y_{17}\,\mathbf{a}_2 + z_{17}\,\mathbf{a}_3$ | = | $x_{17}a\,\mathbf{\hat{x}} + y_{17}b\,\mathbf{\hat{y}} + z_{17}c\,\mathbf{\hat{z}}$ | (8*d*) | Si III |
| $\mathbf{B}_{106}$ | = | $\left(\frac{1}{2} - x_{17}\right) \mathbf{a}_1 - y_{17}\,\mathbf{a}_2 + \left(\frac{1}{2} + z_{17}\right) \mathbf{a}_3$ | = | $\left(\frac{1}{2} - x_{17}\right) a\,\mathbf{\hat{x}} - y_{17}b\,\mathbf{\hat{y}} + \left(\frac{1}{2} + z_{17}\right) c\,\mathbf{\hat{z}}$ | (8*d*) | Si III |
| $\mathbf{B}_{107}$ | = | $-x_{17}\,\mathbf{a}_1 + \left(\frac{1}{2} + y_{17}\right) \mathbf{a}_2 - z_{17}\,\mathbf{a}_3$ | = | $-x_{17}a\,\mathbf{\hat{x}} + \left(\frac{1}{2} + y_{17}\right) b\,\mathbf{\hat{y}} - z_{17}c\,\mathbf{\hat{z}}$ | (8*d*) | Si III |
| $\mathbf{B}_{108}$ | = | $\left(\frac{1}{2} + x_{17}\right) \mathbf{a}_1 + \left(\frac{1}{2} - y_{17}\right) \mathbf{a}_2 + \left(\frac{1}{2} - z_{17}\right) \mathbf{a}_3$ | = | $\left(\frac{1}{2} + x_{17}\right) a\,\mathbf{\hat{x}} + \left(\frac{1}{2} - y_{17}\right) b\,\mathbf{\hat{y}} + \left(\frac{1}{2} - z_{17}\right) c\,\mathbf{\hat{z}}$ | (8*d*) | Si III |
| $\mathbf{B}_{109}$ | = | $-x_{17}\,\mathbf{a}_1 - y_{17}\,\mathbf{a}_2 - z_{17}\,\mathbf{a}_3$ | = | $-x_{17}a\,\mathbf{\hat{x}} - y_{17}b\,\mathbf{\hat{y}} - z_{17}c\,\mathbf{\hat{z}}$ | (8*d*) | Si III |
| $\mathbf{B}_{110}$ | = | $\left(\frac{1}{2} + x_{17}\right) \mathbf{a}_1 + y_{17}\,\mathbf{a}_2 + \left(\frac{1}{2} - z_{17}\right) \mathbf{a}_3$ | = | $\left(\frac{1}{2} + x_{17}\right) a\,\mathbf{\hat{x}} + y_{17}b\,\mathbf{\hat{y}} + \left(\frac{1}{2} - z_{17}\right) c\,\mathbf{\hat{z}}$ | (8*d*) | Si III |
| $\mathbf{B}_{111}$ | = | $x_{17}\,\mathbf{a}_1 + \left(\frac{1}{2} - y_{17}\right) \mathbf{a}_2 + z_{17}\,\mathbf{a}_3$ | = | $x_{17}a\,\mathbf{\hat{x}} + \left(\frac{1}{2} - y_{17}\right) b\,\mathbf{\hat{y}} + z_{17}c\,\mathbf{\hat{z}}$ | (8*d*) | Si III |
| $\mathbf{B}_{112}$ | = | $\left(\frac{1}{2} - x_{17}\right) \mathbf{a}_1 + \left(\frac{1}{2} + y_{17}\right) \mathbf{a}_2 + \left(\frac{1}{2} + z_{17}\right) \mathbf{a}_3$ | = | $\left(\frac{1}{2} - x_{17}\right) a\,\mathbf{\hat{x}} + \left(\frac{1}{2} + y_{17}\right) b\,\mathbf{\hat{y}} + \left(\frac{1}{2} + z_{17}\right) c\,\mathbf{\hat{z}}$ | (8*d*) | Si III |

Leipzig, 1937).

**Found in:**

- R. T. Downs and M. Hall-Wallace, *The American Mineralogist Crystal Structure Database*, Am. Mineral. **88**, 247–250 (2003).

---

**Geometry files:**

- CIF: pp. 1650
- POSCAR: pp. 1651

# $NH_4ClBrI$ ($F5_{14}$) Structure: ABCD_oP16_62_c_c_c_c

http://aflow.org/prototype-encyclopedia/ABCD_oP16_62_c_c_c_c

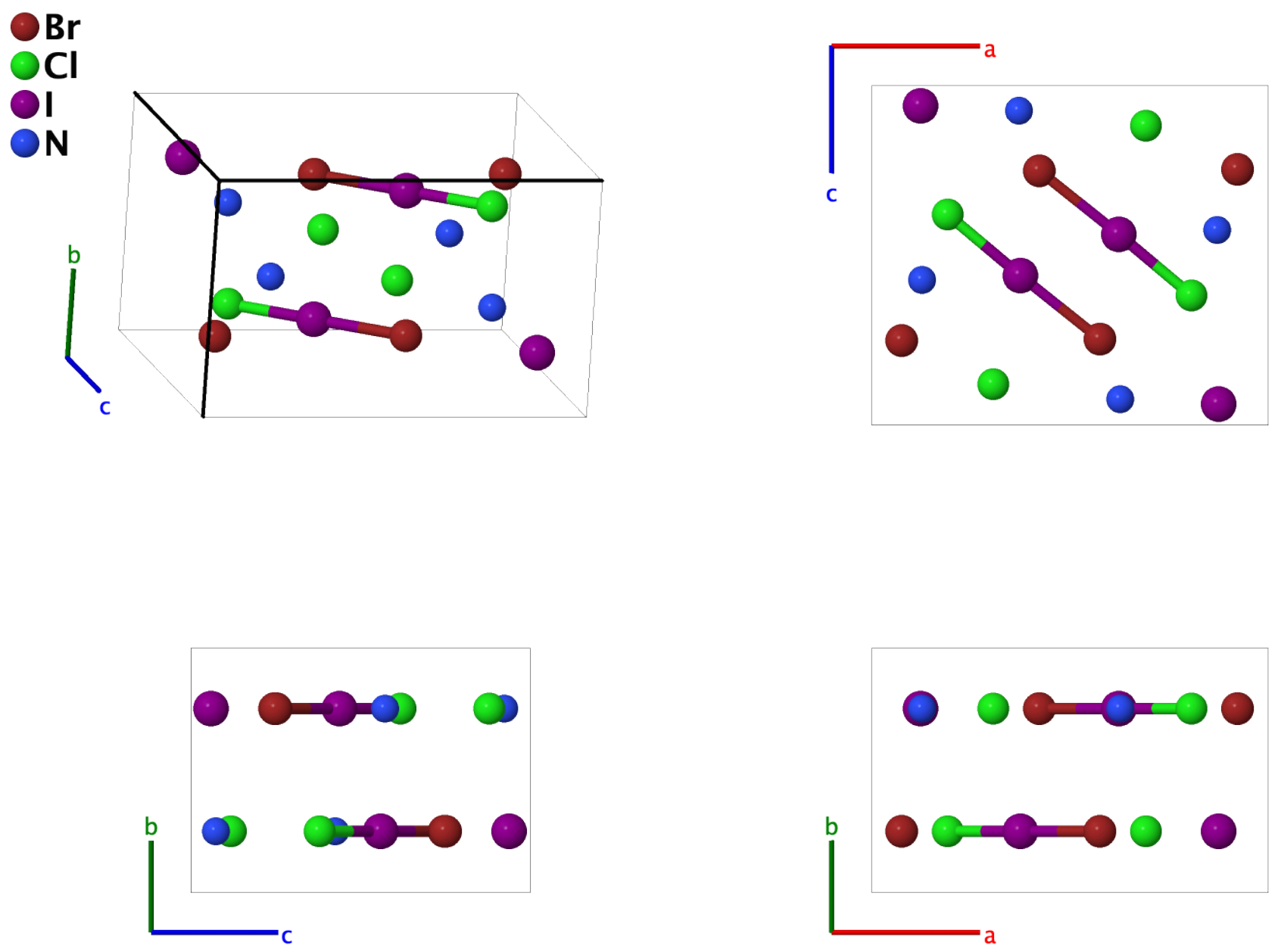

| | | |
|---|---|---|
| **Prototype** | : | $ClBrI(NH_4)$ |
| **AFLOW prototype label** | : | ABCD_oP16_62_c_c_c_c |
| ***Strukturbericht* designation** | : | $F5_{14}$ |
| **Pearson symbol** | : | oP16 |
| **Space group number** | : | 62 |
| **Space group symbol** | : | *Pnma* |
| **AFLOW prototype command** | : | `aflow --proto=ABCD_oP16_62_c_c_c_c`<br>`--params=`$a, b/a, c/a, x_1, z_1, x_2, z_2, x_3, z_3, x_4, z_4$ |

- (Gottfried, 1937) originally described $NH_4ClBrI$ as isostructural with $NH_4I_3$, $D0_{16}$, based on the early work of (Mooney, 1935). However, (Gottfried, 1940) used the newer work of (Mooney, 1937) to create a separate *Strukturbericht* designation for $NH_4ClBrI$, $F5_{14}$, even though this structure is still obviously related to $NH_4I_3$.
- The positions of the hydrogen atoms in the $NH_4$ were not determined, so we only provide the positions of the nitrogen atoms (labeled as $NH_4$).
- The original structure is given in the *Pmcn* setting of space group #62, which was transformed to the standard *Pnma* setting by FINDSYM. However, note that the coordinate transformations for the (4*c*) Wyckoff positions in (Gottfried, 1940) are erroneous, they give a structure in space group *Pmmn* #59. (Gottfried, 1937) has the correct tranformations in the $D0_{16}$ section.

## Simple Orthorhombic primitive vectors:

$$\mathbf{a}_1 = a\,\hat{\mathbf{x}}$$
$$\mathbf{a}_2 = b\,\hat{\mathbf{y}}$$
$$\mathbf{a}_3 = c\,\hat{\mathbf{z}}$$

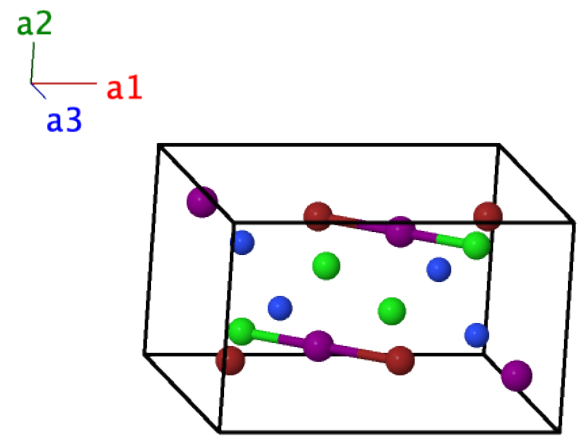

## Basis vectors:

| | | Lattice Coordinates | | Cartesian Coordinates | Wyckoff Position | Atom Type |
|---|---|---|---|---|---|---|
| $\mathbf{B}_1$ | = | $x_1\,\mathbf{a}_1 + \frac{1}{4}\,\mathbf{a}_2 + z_1\,\mathbf{a}_3$ | = | $x_1 a\,\hat{\mathbf{x}} + \frac{1}{4} b\,\hat{\mathbf{y}} + z_1 c\,\hat{\mathbf{z}}$ | (4c) | Br |
| $\mathbf{B}_2$ | = | $\left(\frac{1}{2} - x_1\right)\mathbf{a}_1 + \frac{3}{4}\,\mathbf{a}_2 + \left(\frac{1}{2} + z_1\right)\mathbf{a}_3$ | = | $\left(\frac{1}{2} - x_1\right) a\,\hat{\mathbf{x}} + \frac{3}{4} b\,\hat{\mathbf{y}} + \left(\frac{1}{2} + z_1\right) c\,\hat{\mathbf{z}}$ | (4c) | Br |
| $\mathbf{B}_3$ | = | $-x_1\,\mathbf{a}_1 + \frac{3}{4}\,\mathbf{a}_2 - z_1\,\mathbf{a}_3$ | = | $-x_1 a\,\hat{\mathbf{x}} + \frac{3}{4} b\,\hat{\mathbf{y}} - z_1 c\,\hat{\mathbf{z}}$ | (4c) | Br |
| $\mathbf{B}_4$ | = | $\left(\frac{1}{2} + x_1\right)\mathbf{a}_1 + \frac{1}{4}\,\mathbf{a}_2 + \left(\frac{1}{2} - z_1\right)\mathbf{a}_3$ | = | $\left(\frac{1}{2} + x_1\right) a\,\hat{\mathbf{x}} + \frac{1}{4} b\,\hat{\mathbf{y}} + \left(\frac{1}{2} - z_1\right) c\,\hat{\mathbf{z}}$ | (4c) | Br |
| $\mathbf{B}_5$ | = | $x_2\,\mathbf{a}_1 + \frac{1}{4}\,\mathbf{a}_2 + z_2\,\mathbf{a}_3$ | = | $x_2 a\,\hat{\mathbf{x}} + \frac{1}{4} b\,\hat{\mathbf{y}} + z_2 c\,\hat{\mathbf{z}}$ | (4c) | Cl |
| $\mathbf{B}_6$ | = | $\left(\frac{1}{2} - x_2\right)\mathbf{a}_1 + \frac{3}{4}\,\mathbf{a}_2 + \left(\frac{1}{2} + z_2\right)\mathbf{a}_3$ | = | $\left(\frac{1}{2} - x_2\right) a\,\hat{\mathbf{x}} + \frac{3}{4} b\,\hat{\mathbf{y}} + \left(\frac{1}{2} + z_2\right) c\,\hat{\mathbf{z}}$ | (4c) | Cl |
| $\mathbf{B}_7$ | = | $-x_2\,\mathbf{a}_1 + \frac{3}{4}\,\mathbf{a}_2 - z_2\,\mathbf{a}_3$ | = | $-x_2 a\,\hat{\mathbf{x}} + \frac{3}{4} b\,\hat{\mathbf{y}} - z_2 c\,\hat{\mathbf{z}}$ | (4c) | Cl |
| $\mathbf{B}_8$ | = | $\left(\frac{1}{2} + x_2\right)\mathbf{a}_1 + \frac{1}{4}\,\mathbf{a}_2 + \left(\frac{1}{2} - z_2\right)\mathbf{a}_3$ | = | $\left(\frac{1}{2} + x_2\right) a\,\hat{\mathbf{x}} + \frac{1}{4} b\,\hat{\mathbf{y}} + \left(\frac{1}{2} - z_2\right) c\,\hat{\mathbf{z}}$ | (4c) | Cl |
| $\mathbf{B}_9$ | = | $x_3\,\mathbf{a}_1 + \frac{1}{4}\,\mathbf{a}_2 + z_3\,\mathbf{a}_3$ | = | $x_3 a\,\hat{\mathbf{x}} + \frac{1}{4} b\,\hat{\mathbf{y}} + z_3 c\,\hat{\mathbf{z}}$ | (4c) | I |
| $\mathbf{B}_{10}$ | = | $\left(\frac{1}{2} - x_3\right)\mathbf{a}_1 + \frac{3}{4}\,\mathbf{a}_2 + \left(\frac{1}{2} + z_3\right)\mathbf{a}_3$ | = | $\left(\frac{1}{2} - x_3\right) a\,\hat{\mathbf{x}} + \frac{3}{4} b\,\hat{\mathbf{y}} + \left(\frac{1}{2} + z_3\right) c\,\hat{\mathbf{z}}$ | (4c) | I |
| $\mathbf{B}_{11}$ | = | $-x_3\,\mathbf{a}_1 + \frac{3}{4}\,\mathbf{a}_2 - z_3\,\mathbf{a}_3$ | = | $-x_3 a\,\hat{\mathbf{x}} + \frac{3}{4} b\,\hat{\mathbf{y}} - z_3 c\,\hat{\mathbf{z}}$ | (4c) | I |
| $\mathbf{B}_{12}$ | = | $\left(\frac{1}{2} + x_3\right)\mathbf{a}_1 + \frac{1}{4}\,\mathbf{a}_2 + \left(\frac{1}{2} - z_3\right)\mathbf{a}_3$ | = | $\left(\frac{1}{2} + x_3\right) a\,\hat{\mathbf{x}} + \frac{1}{4} b\,\hat{\mathbf{y}} + \left(\frac{1}{2} - z_3\right) c\,\hat{\mathbf{z}}$ | (4c) | I |
| $\mathbf{B}_{13}$ | = | $x_4\,\mathbf{a}_1 + \frac{1}{4}\,\mathbf{a}_2 + z_4\,\mathbf{a}_3$ | = | $x_4 a\,\hat{\mathbf{x}} + \frac{1}{4} b\,\hat{\mathbf{y}} + z_4 c\,\hat{\mathbf{z}}$ | (4c) | $NH_4$ |
| $\mathbf{B}_{14}$ | = | $\left(\frac{1}{2} - x_4\right)\mathbf{a}_1 + \frac{3}{4}\,\mathbf{a}_2 + \left(\frac{1}{2} + z_4\right)\mathbf{a}_3$ | = | $\left(\frac{1}{2} - x_4\right) a\,\hat{\mathbf{x}} + \frac{3}{4} b\,\hat{\mathbf{y}} + \left(\frac{1}{2} + z_4\right) c\,\hat{\mathbf{z}}$ | (4c) | $NH_4$ |
| $\mathbf{B}_{15}$ | = | $-x_4\,\mathbf{a}_1 + \frac{3}{4}\,\mathbf{a}_2 - z_4\,\mathbf{a}_3$ | = | $-x_4 a\,\hat{\mathbf{x}} + \frac{3}{4} b\,\hat{\mathbf{y}} - z_4 c\,\hat{\mathbf{z}}$ | (4c) | $NH_4$ |
| $\mathbf{B}_{16}$ | = | $\left(\frac{1}{2} + x_4\right)\mathbf{a}_1 + \frac{1}{4}\,\mathbf{a}_2 + \left(\frac{1}{2} - z_4\right)\mathbf{a}_3$ | = | $\left(\frac{1}{2} + x_4\right) a\,\hat{\mathbf{x}} + \frac{1}{4} b\,\hat{\mathbf{y}} + \left(\frac{1}{2} - z_4\right) c\,\hat{\mathbf{z}}$ | (4c) | $NH_4$ |

## Found in:

## Geometry files:

- CIF: pp. 1651
- POSCAR: pp. 1652

# MnCuP Structure: ABC_oP12_62_c_c_c

http://aflow.org/prototype-encyclopedia/ABC_oP12_62_c_c_c

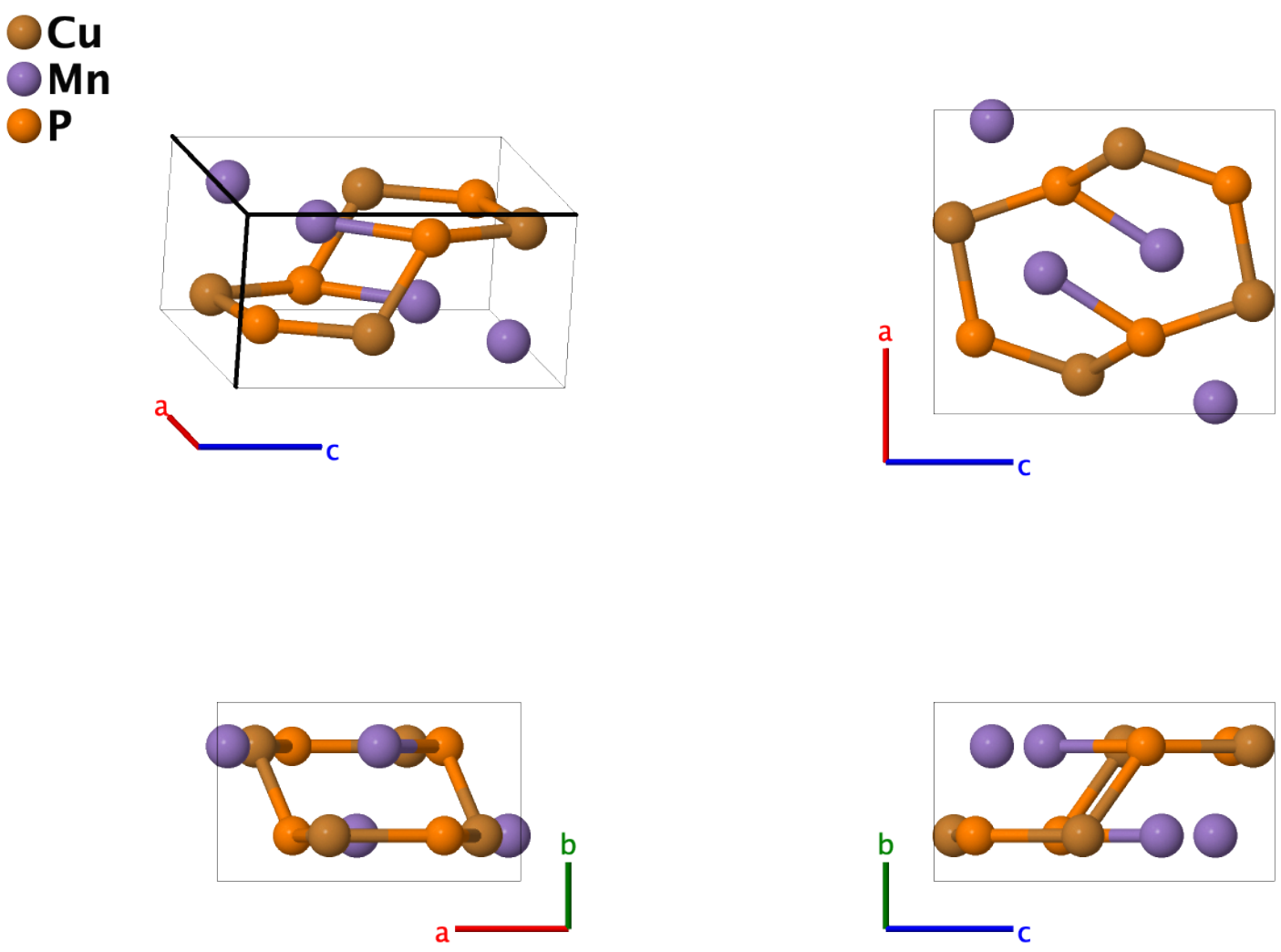

| | | |
|---|---|---|
| **Prototype** | : | CuMnP |
| **AFLOW prototype label** | : | ABC_oP12_62_c_c_c |
| ***Strukturbericht* designation** | : | None |
| **Pearson symbol** | : | oP12 |
| **Space group number** | : | 62 |
| **Space group symbol** | : | *Pnma* |
| **AFLOW prototype command** | : | `aflow --proto=ABC_oP12_62_c_c_c`<br>`--params=`$a, b/a, c/a, x_1, z_1, x_2, z_2, x_3, z_3$ |

### Other compounds with this structure

- CuMnAs and $\text{CuMnP}_x\text{As}_{1-x}$

### Simple Orthorhombic primitive vectors:

$$\mathbf{a}_1 = a\,\hat{\mathbf{x}}$$
$$\mathbf{a}_2 = b\,\hat{\mathbf{y}}$$
$$\mathbf{a}_3 = c\,\hat{\mathbf{z}}$$

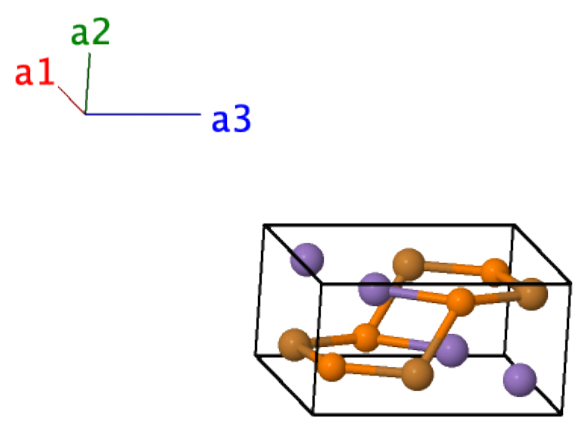

### Basis vectors:

| | | Lattice Coordinates | | Cartesian Coordinates | Wyckoff Position | Atom Type |
|---|---|---|---|---|---|---|
| $\mathbf{B}_1$ | = | $x_1\,\mathbf{a}_1 + \frac{1}{4}\,\mathbf{a}_2 + z_1\,\mathbf{a}_3$ | = | $x_1 a\,\hat{\mathbf{x}} + \frac{1}{4} b\,\hat{\mathbf{y}} + z_1 c\,\hat{\mathbf{z}}$ | (4$c$) | Cu |
| $\mathbf{B}_2$ | = | $\left(\frac{1}{2} - x_1\right)\mathbf{a}_1 + \frac{3}{4}\,\mathbf{a}_2 + \left(\frac{1}{2} + z_1\right)\mathbf{a}_3$ | = | $\left(\frac{1}{2} - x_1\right) a\,\hat{\mathbf{x}} + \frac{3}{4} b\,\hat{\mathbf{y}} + \left(\frac{1}{2} + z_1\right) c\,\hat{\mathbf{z}}$ | (4$c$) | Cu |
| $\mathbf{B}_3$ | = | $-x_1\,\mathbf{a}_1 + \frac{3}{4}\,\mathbf{a}_2 - z_1\,\mathbf{a}_3$ | = | $-x_1 a\,\hat{\mathbf{x}} + \frac{3}{4} b\,\hat{\mathbf{y}} - z_1 c\,\hat{\mathbf{z}}$ | (4$c$) | Cu |

$$\mathbf{B}_4 = \left(\frac{1}{2}+x_1\right)\mathbf{a}_1 + \frac{1}{4}\mathbf{a}_2 + \left(\frac{1}{2}-z_1\right)\mathbf{a}_3 = \left(\frac{1}{2}+x_1\right)a\,\hat{\mathbf{x}} + \frac{1}{4}b\,\hat{\mathbf{y}} + \left(\frac{1}{2}-z_1\right)c\,\hat{\mathbf{z}} \quad (4c) \quad \text{Cu}$$

$$\mathbf{B}_5 = x_2\,\mathbf{a}_1 + \frac{1}{4}\mathbf{a}_2 + z_2\,\mathbf{a}_3 = x_2 a\,\hat{\mathbf{x}} + \frac{1}{4}b\,\hat{\mathbf{y}} + z_2 c\,\hat{\mathbf{z}} \quad (4c) \quad \text{Mn}$$

$$\mathbf{B}_6 = \left(\frac{1}{2}-x_2\right)\mathbf{a}_1 + \frac{3}{4}\mathbf{a}_2 + \left(\frac{1}{2}+z_2\right)\mathbf{a}_3 = \left(\frac{1}{2}-x_2\right)a\,\hat{\mathbf{x}} + \frac{3}{4}b\,\hat{\mathbf{y}} + \left(\frac{1}{2}+z_2\right)c\,\hat{\mathbf{z}} \quad (4c) \quad \text{Mn}$$

$$\mathbf{B}_7 = -x_2\,\mathbf{a}_1 + \frac{3}{4}\mathbf{a}_2 - z_2\,\mathbf{a}_3 = -x_2 a\,\hat{\mathbf{x}} + \frac{3}{4}b\,\hat{\mathbf{y}} - z_2 c\,\hat{\mathbf{z}} \quad (4c) \quad \text{Mn}$$

$$\mathbf{B}_8 = \left(\frac{1}{2}+x_2\right)\mathbf{a}_1 + \frac{1}{4}\mathbf{a}_2 + \left(\frac{1}{2}-z_2\right)\mathbf{a}_3 = \left(\frac{1}{2}+x_2\right)a\,\hat{\mathbf{x}} + \frac{1}{4}b\,\hat{\mathbf{y}} + \left(\frac{1}{2}-z_2\right)c\,\hat{\mathbf{z}} \quad (4c) \quad \text{Mn}$$

$$\mathbf{B}_9 = x_3\,\mathbf{a}_1 + \frac{1}{4}\mathbf{a}_2 + z_3\,\mathbf{a}_3 = x_3 a\,\hat{\mathbf{x}} + \frac{1}{4}b\,\hat{\mathbf{y}} + z_3 c\,\hat{\mathbf{z}} \quad (4c) \quad \text{P}$$

$$\mathbf{B}_{10} = \left(\frac{1}{2}-x_3\right)\mathbf{a}_1 + \frac{3}{4}\mathbf{a}_2 + \left(\frac{1}{2}+z_3\right)\mathbf{a}_3 = \left(\frac{1}{2}-x_3\right)a\,\hat{\mathbf{x}} + \frac{3}{4}b\,\hat{\mathbf{y}} + \left(\frac{1}{2}+z_3\right)c\,\hat{\mathbf{z}} \quad (4c) \quad \text{P}$$

$$\mathbf{B}_{11} = -x_3\,\mathbf{a}_1 + \frac{3}{4}\mathbf{a}_2 - z_3\,\mathbf{a}_3 = -x_3 a\,\hat{\mathbf{x}} + \frac{3}{4}b\,\hat{\mathbf{y}} - z_3 c\,\hat{\mathbf{z}} \quad (4c) \quad \text{P}$$

$$\mathbf{B}_{12} = \left(\frac{1}{2}+x_3\right)\mathbf{a}_1 + \frac{1}{4}\mathbf{a}_2 + \left(\frac{1}{2}-z_3\right)\mathbf{a}_3 = \left(\frac{1}{2}+x_3\right)a\,\hat{\mathbf{x}} + \frac{1}{4}b\,\hat{\mathbf{y}} + \left(\frac{1}{2}-z_3\right)c\,\hat{\mathbf{z}} \quad (4c) \quad \text{P}$$

---

**Found in:**

---

**Geometry files:**

- CIF: pp. 1652
- POSCAR: pp. 1652

# $\eta$-NiSi ($B_d$) Structure: AB_oP8_62_c_c

http://aflow.org/prototype-encyclopedia/AB_oP8_62_c_c.NiSi

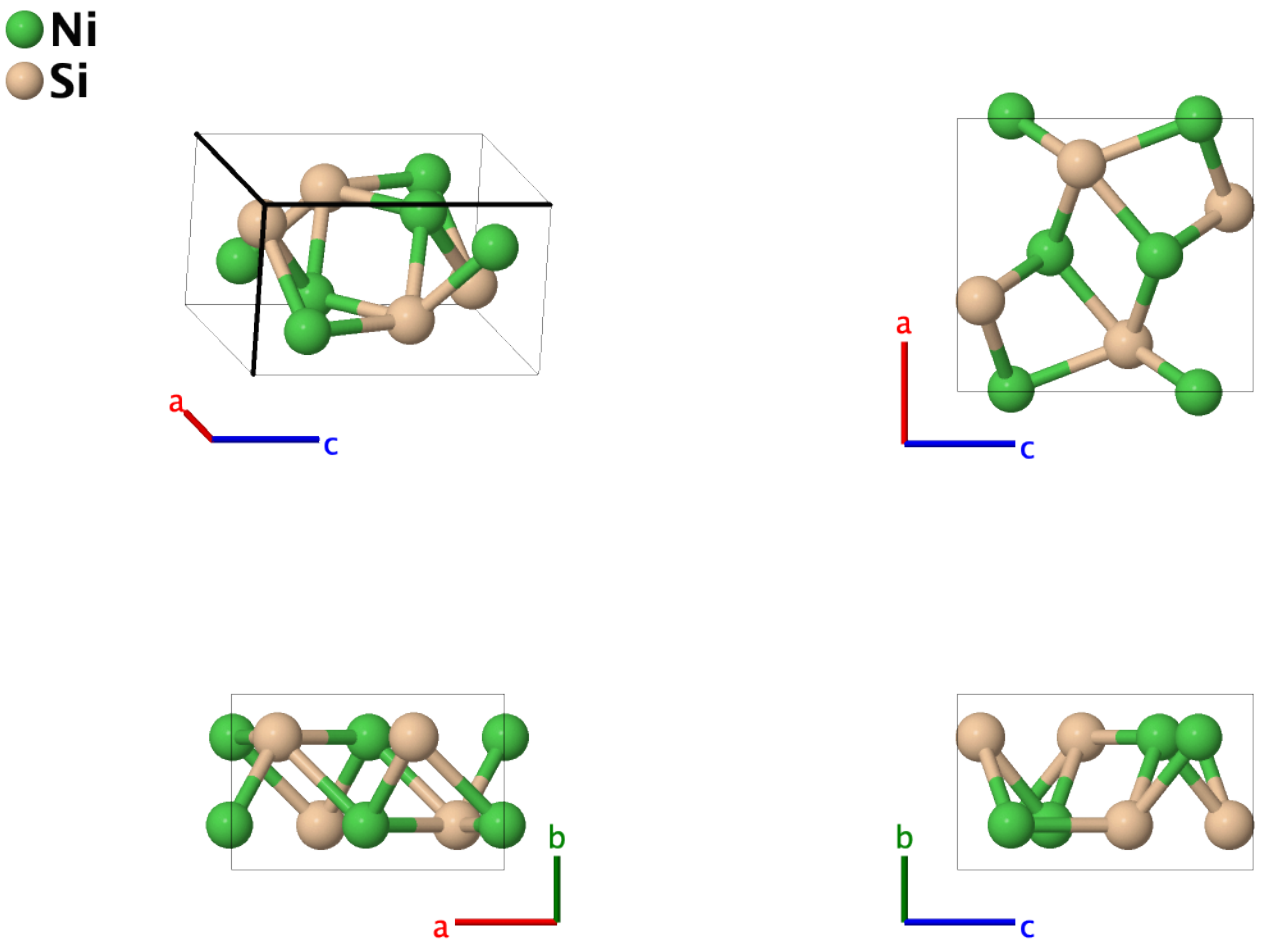

**Prototype** : NiSi

**AFLOW prototype label** : AB_oP8_62_c_c

***Strukturbericht* designation** : $B_d$

**Pearson symbol** : oP8

**Space group number** : 62

**Space group symbol** : *Pnma*

**AFLOW prototype command** : `aflow --proto=AB_oP8_62_c_c` `--params=`$a, b/a, c/a, x_1, z_1, x_2, z_2$

- (Toman, 1951) described this in the *Pbnm* setting of space group #62. We have used FINDSYM to transform this to the standard *Pnma* setting. Note that GeS (AB_oP8_62_c_c, GeS), MnP (AB_oP8_62_c_c, MnP), FeB (AB_oP8_62_c_c, FeB), and SnS (AB_oP8_62_c_c, SnS) have the same AFLOW prototype label. They are generated by the same symmetry operations with different sets of parameters (`--params`) specified in their corresponding CIF files.

**Simple Orthorhombic primitive vectors:**

$$\mathbf{a}_1 = a\,\hat{\mathbf{x}}$$
$$\mathbf{a}_2 = b\,\hat{\mathbf{y}}$$
$$\mathbf{a}_3 = c\,\hat{\mathbf{z}}$$

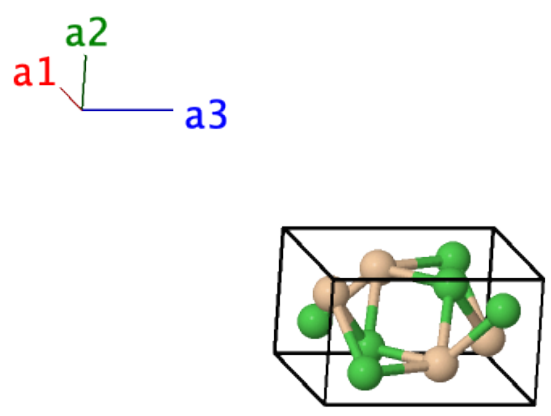

**Basis vectors:**

| | | Lattice Coordinates | | Cartesian Coordinates | Wyckoff Position | Atom Type |
|---|---|---|---|---|---|---|
| $\mathbf{B}_1$ | = | $x_1\,\mathbf{a}_1 + \frac{1}{4}\,\mathbf{a}_2 + z_1\,\mathbf{a}_3$ | = | $x_1 a\,\hat{\mathbf{x}} + \frac{1}{4}b\,\hat{\mathbf{y}} + z_1 c\,\hat{\mathbf{z}}$ | (4c) | Ni |
| $\mathbf{B}_2$ | = | $\left(\frac{1}{2} - x_1\right)\mathbf{a}_1 + \frac{3}{4}\,\mathbf{a}_2 + \left(\frac{1}{2} + z_1\right)\mathbf{a}_3$ | = | $\left(\frac{1}{2} - x_1\right)a\,\hat{\mathbf{x}} + \frac{3}{4}b\,\hat{\mathbf{y}} + \left(\frac{1}{2} + z_1\right)c\,\hat{\mathbf{z}}$ | (4c) | Ni |
| $\mathbf{B}_3$ | = | $-x_1\,\mathbf{a}_1 + \frac{3}{4}\,\mathbf{a}_2 - z_1\,\mathbf{a}_3$ | = | $-x_1 a\,\hat{\mathbf{x}} + \frac{3}{4}b\,\hat{\mathbf{y}} - z_1 c\,\hat{\mathbf{z}}$ | (4c) | Ni |

$$\mathbf{B}_4 = \left(\frac{1}{2}+x_1\right)\mathbf{a}_1 + \frac{1}{4}\mathbf{a}_2 + \left(\frac{1}{2}-z_1\right)\mathbf{a}_3 = \left(\frac{1}{2}+x_1\right)a\,\hat{\mathbf{x}} + \frac{1}{4}b\,\hat{\mathbf{y}} + \left(\frac{1}{2}-z_1\right)c\,\hat{\mathbf{z}} \quad (4c) \quad \text{Ni}$$

$$\mathbf{B}_5 = x_2\,\mathbf{a}_1 + \frac{1}{4}\mathbf{a}_2 + z_2\,\mathbf{a}_3 = x_2a\,\hat{\mathbf{x}} + \frac{1}{4}b\,\hat{\mathbf{y}} + z_2c\,\hat{\mathbf{z}} \quad (4c) \quad \text{Si}$$

$$\mathbf{B}_6 = \left(\frac{1}{2}-x_2\right)\mathbf{a}_1 + \frac{3}{4}\mathbf{a}_2 + \left(\frac{1}{2}+z_2\right)\mathbf{a}_3 = \left(\frac{1}{2}-x_2\right)a\,\hat{\mathbf{x}} + \frac{3}{4}b\,\hat{\mathbf{y}} + \left(\frac{1}{2}+z_2\right)c\,\hat{\mathbf{z}} \quad (4c) \quad \text{Si}$$

$$\mathbf{B}_7 = -x_2\,\mathbf{a}_1 + \frac{3}{4}\mathbf{a}_2 - z_2\,\mathbf{a}_3 = -x_2a\,\hat{\mathbf{x}} + \frac{3}{4}b\,\hat{\mathbf{y}} - z_2c\,\hat{\mathbf{z}} \quad (4c) \quad \text{Si}$$

$$\mathbf{B}_8 = \left(\frac{1}{2}+x_2\right)\mathbf{a}_1 + \frac{1}{4}\mathbf{a}_2 + \left(\frac{1}{2}-z_2\right)\mathbf{a}_3 = \left(\frac{1}{2}+x_2\right)a\,\hat{\mathbf{x}} + \frac{1}{4}b\,\hat{\mathbf{y}} + \left(\frac{1}{2}-z_2\right)c\,\hat{\mathbf{z}} \quad (4c) \quad \text{Si}$$

---

---

**Geometry files:**

- CIF: pp. 1652
- POSCAR: pp. 1653

# $Cu_2Pb(SeO_3)_2Br_2$ Structure: A2B2C6DE2_oC52_63_g_e_fh_c_f

http://aflow.org/prototype-encyclopedia/A2B2C6DE2_oC52_63_g_e_fh_c_f

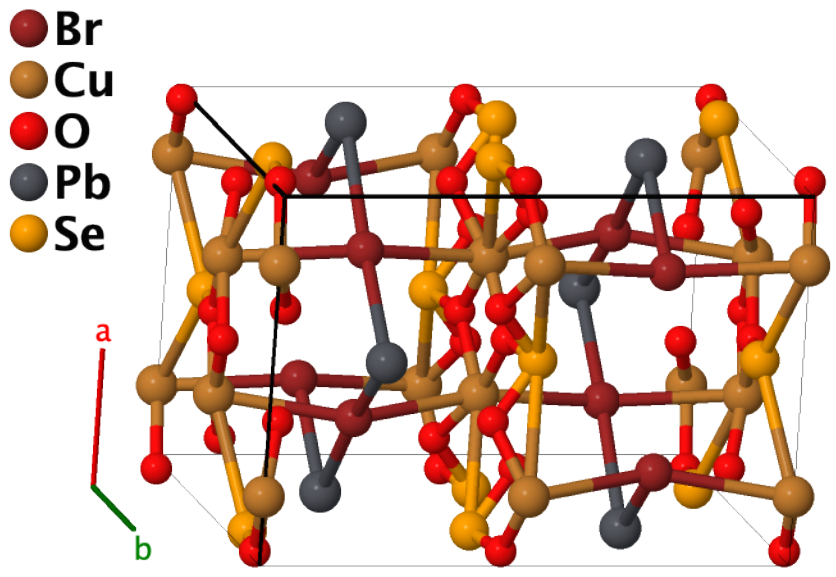

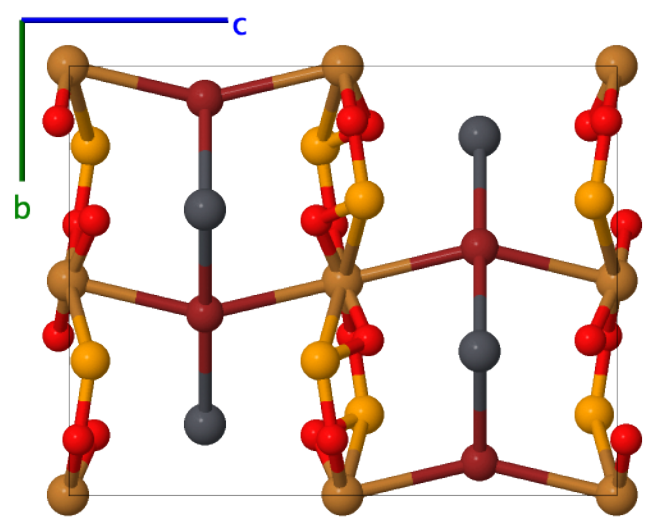

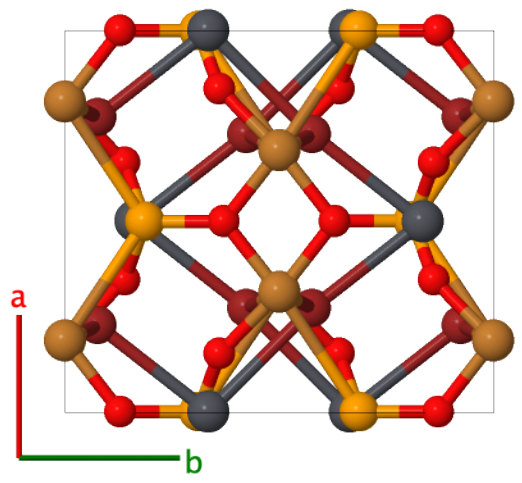

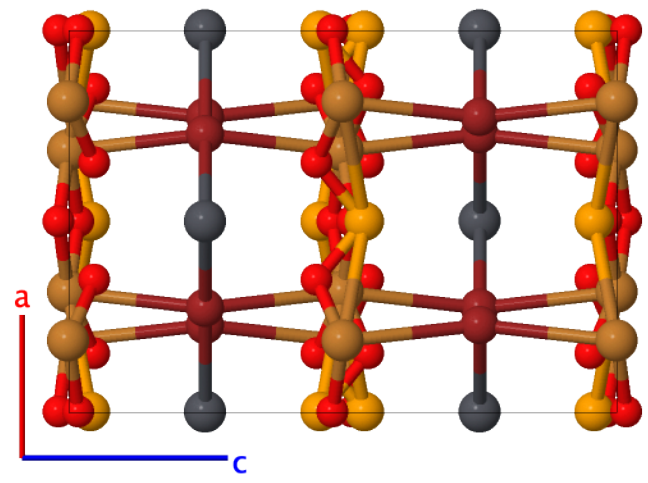

| | | |
|---|---|---|
| **Prototype** | : | $Br_2Cu_2O_6PbSe_2$ |
| **AFLOW prototype label** | : | A2B2C6DE2_oC52_63_g_e_fh_c_f |
| ***Strukturbericht* designation** | : | None |
| **Pearson symbol** | : | oC52 |
| **Space group number** | : | 63 |
| **Space group symbol** | : | *Cmcm* |
| **AFLOW prototype command** | : | `aflow --proto=A2B2C6DE2_oC52_63_g_e_fh_c_f` `--params=`$a, b/a, c/a, y_1, x_2, y_3, z_3, y_4, z_4, x_5, y_5, x_6, y_6, z_6$ |

**Base-centered Orthorhombic primitive vectors:**

$$\mathbf{a}_1 = \frac{1}{2}\,a\,\hat{\mathbf{x}} - \frac{1}{2}\,b\,\hat{\mathbf{y}}$$
$$\mathbf{a}_2 = \frac{1}{2}\,a\,\hat{\mathbf{x}} + \frac{1}{2}\,b\,\hat{\mathbf{y}}$$
$$\mathbf{a}_3 = c\,\hat{\mathbf{z}}$$

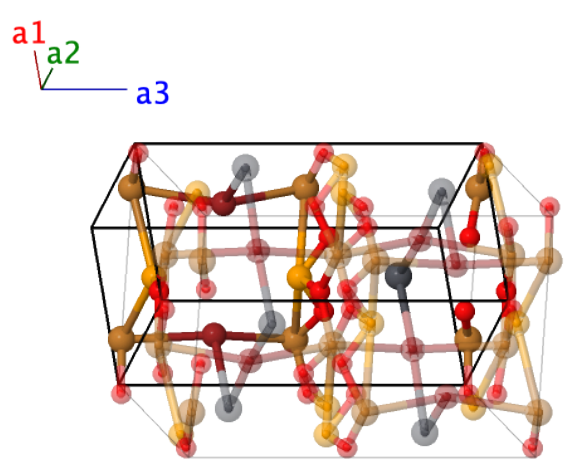

**Basis vectors:**

| | | Lattice Coordinates | | Cartesian Coordinates | Wyckoff Position | Atom Type |
|---|---|---|---|---|---|---|
| $\mathbf{B}_1$ | = | $-y_1\,\mathbf{a}_1 + y_1\,\mathbf{a}_2 + \frac{1}{4}\,\mathbf{a}_3$ | = | $y_1\,b\,\hat{\mathbf{y}} + \frac{1}{4}c\,\hat{\mathbf{z}}$ | (4*c*) | Pb |
| $\mathbf{B}_2$ | = | $y_1\,\mathbf{a}_1 - y_1\,\mathbf{a}_2 + \frac{3}{4}\,\mathbf{a}_3$ | = | $-y_1\,b\,\hat{\mathbf{y}} + \frac{3}{4}c\,\hat{\mathbf{z}}$ | (4*c*) | Pb |
| $\mathbf{B}_3$ | = | $x_2\,\mathbf{a}_1 + x_2\,\mathbf{a}_2$ | = | $x_2\,a\,\hat{\mathbf{x}}$ | (8*e*) | Cu |
| $\mathbf{B}_4$ | = | $-x_2\,\mathbf{a}_1 - x_2\,\mathbf{a}_2 + \frac{1}{2}\,\mathbf{a}_3$ | = | $-x_2 a\,\hat{\mathbf{x}} + \frac{1}{2} c\,\hat{\mathbf{z}}$ | (8$e$) | Cu |
| $\mathbf{B}_5$ | = | $-x_2\,\mathbf{a}_1 - x_2\,\mathbf{a}_2$ | = | $-x_2 a\,\hat{\mathbf{x}}$ | (8$e$) | Cu |
| $\mathbf{B}_6$ | = | $x_2\,\mathbf{a}_1 + x_2\,\mathbf{a}_2 + \frac{1}{2}\,\mathbf{a}_3$ | = | $x_2 a\,\hat{\mathbf{x}} + \frac{1}{2} c\,\hat{\mathbf{z}}$ | (8$e$) | Cu |
| $\mathbf{B}_7$ | = | $-y_3\,\mathbf{a}_1 + y_3\,\mathbf{a}_2 + z_3\,\mathbf{a}_3$ | = | $y_3 b\,\hat{\mathbf{y}} + z_3 c\,\hat{\mathbf{z}}$ | (8$f$) | O I |
| $\mathbf{B}_8$ | = | $y_3\,\mathbf{a}_1 - y_3\,\mathbf{a}_2 + \left(\frac{1}{2} + z_3\right)\,\mathbf{a}_3$ | = | $-y_3 b\,\hat{\mathbf{y}} + \left(\frac{1}{2} + z_3\right) c\,\hat{\mathbf{z}}$ | (8$f$) | O I |
| $\mathbf{B}_9$ | = | $-y_3\,\mathbf{a}_1 + y_3\,\mathbf{a}_2 + \left(\frac{1}{2} - z_3\right)\,\mathbf{a}_3$ | = | $y_3 b\,\hat{\mathbf{y}} + \left(\frac{1}{2} - z_3\right) c\,\hat{\mathbf{z}}$ | (8$f$) | O I |
| $\mathbf{B}_{10}$ | = | $y_3\,\mathbf{a}_1 - y_3\,\mathbf{a}_2 - z_3\,\mathbf{a}_3$ | = | $-y_3 b\,\hat{\mathbf{y}} - z_3 c\,\hat{\mathbf{z}}$ | (8$f$) | O I |
| $\mathbf{B}_{11}$ | = | $-y_4\,\mathbf{a}_1 + y_4\,\mathbf{a}_2 + z_4\,\mathbf{a}_3$ | = | $y_4 b\,\hat{\mathbf{y}} + z_4 c\,\hat{\mathbf{z}}$ | (8$f$) | Se |
| $\mathbf{B}_{12}$ | = | $y_4\,\mathbf{a}_1 - y_4\,\mathbf{a}_2 + \left(\frac{1}{2} + z_4\right)\,\mathbf{a}_3$ | = | $-y_4 b\,\hat{\mathbf{y}} + \left(\frac{1}{2} + z_4\right) c\,\hat{\mathbf{z}}$ | (8$f$) | Se |
| $\mathbf{B}_{13}$ | = | $-y_4\,\mathbf{a}_1 + y_4\,\mathbf{a}_2 + \left(\frac{1}{2} - z_4\right)\,\mathbf{a}_3$ | = | $y_4 b\,\hat{\mathbf{y}} + \left(\frac{1}{2} - z_4\right) c\,\hat{\mathbf{z}}$ | (8$f$) | Se |
| $\mathbf{B}_{14}$ | = | $y_4\,\mathbf{a}_1 - y_4\,\mathbf{a}_2 - z_4\,\mathbf{a}_3$ | = | $-y_4 b\,\hat{\mathbf{y}} - z_4 c\,\hat{\mathbf{z}}$ | (8$f$) | Se |
| $\mathbf{B}_{15}$ | = | $(x_5 - y_5)\,\mathbf{a}_1 + (x_5 + y_5)\,\mathbf{a}_2 + \frac{1}{4}\,\mathbf{a}_3$ | = | $x_5 a\,\hat{\mathbf{x}} + y_5 b\,\hat{\mathbf{y}} + \frac{1}{4} c\,\hat{\mathbf{z}}$ | (8$g$) | Br |
| $\mathbf{B}_{16}$ | = | $(-x_5 + y_5)\,\mathbf{a}_1 + (-x_5 - y_5)\,\mathbf{a}_2 + \frac{3}{4}\,\mathbf{a}_3$ | = | $-x_5 a\,\hat{\mathbf{x}} - y_5 b\,\hat{\mathbf{y}} + \frac{3}{4} c\,\hat{\mathbf{z}}$ | (8$g$) | Br |
| $\mathbf{B}_{17}$ | = | $(-x_5 - y_5)\,\mathbf{a}_1 + (-x_5 + y_5)\,\mathbf{a}_2 + \frac{1}{4}\,\mathbf{a}_3$ | = | $-x_5 a\,\hat{\mathbf{x}} + y_5 b\,\hat{\mathbf{y}} + \frac{1}{4} c\,\hat{\mathbf{z}}$ | (8$g$) | Br |
| $\mathbf{B}_{18}$ | = | $(x_5 + y_5)\,\mathbf{a}_1 + (x_5 - y_5)\,\mathbf{a}_2 + \frac{3}{4}\,\mathbf{a}_3$ | = | $x_5 a\,\hat{\mathbf{x}} - y_5 b\,\hat{\mathbf{y}} + \frac{3}{4} c\,\hat{\mathbf{z}}$ | (8$g$) | Br |
| $\mathbf{B}_{19}$ | = | $(x_6 - y_6)\,\mathbf{a}_1 + (x_6 + y_6)\,\mathbf{a}_2 + z_6\,\mathbf{a}_3$ | = | $x_6 a\,\hat{\mathbf{x}} + y_6 b\,\hat{\mathbf{y}} + z_6 c\,\hat{\mathbf{z}}$ | (16$h$) | O II |
| $\mathbf{B}_{20}$ | = | $(-x_6 + y_6)\,\mathbf{a}_1 + (-x_6 - y_6)\,\mathbf{a}_2 + \left(\frac{1}{2} + z_6\right)\,\mathbf{a}_3$ | = | $-x_6 a\,\hat{\mathbf{x}} - y_6 b\,\hat{\mathbf{y}} + \left(\frac{1}{2} + z_6\right) c\,\hat{\mathbf{z}}$ | (16$h$) | O II |
| $\mathbf{B}_{21}$ | = | $(-x_6 - y_6)\,\mathbf{a}_1 + (-x_6 + y_6)\,\mathbf{a}_2 + \left(\frac{1}{2} - z_6\right)\,\mathbf{a}_3$ | = | $-x_6 a\,\hat{\mathbf{x}} + y_6 b\,\hat{\mathbf{y}} + \left(\frac{1}{2} - z_6\right) c\,\hat{\mathbf{z}}$ | (16$h$) | O II |
| $\mathbf{B}_{22}$ | = | $(x_6 + y_6)\,\mathbf{a}_1 + (x_6 - y_6)\,\mathbf{a}_2 - z_6\,\mathbf{a}_3$ | = | $x_6 a\,\hat{\mathbf{x}} - y_6 b\,\hat{\mathbf{y}} - z_6 c\,\hat{\mathbf{z}}$ | (16$h$) | O II |
| $\mathbf{B}_{23}$ | = | $(-x_6 + y_6)\,\mathbf{a}_1 + (-x_6 - y_6)\,\mathbf{a}_2 - z_6\,\mathbf{a}_3$ | = | $-x_6 a\,\hat{\mathbf{x}} - y_6 b\,\hat{\mathbf{y}} - z_6 c\,\hat{\mathbf{z}}$ | (16$h$) | O II |
| $\mathbf{B}_{24}$ | = | $(x_6 - y_6)\,\mathbf{a}_1 + (x_6 + y_6)\,\mathbf{a}_2 + \left(\frac{1}{2} - z_6\right)\,\mathbf{a}_3$ | = | $x_6 a\,\hat{\mathbf{x}} + y_6 b\,\hat{\mathbf{y}} + \left(\frac{1}{2} - z_6\right) c\,\hat{\mathbf{z}}$ | (16$h$) | O II |
| $\mathbf{B}_{25}$ | = | $(x_6 + y_6)\,\mathbf{a}_1 + (x_6 - y_6)\,\mathbf{a}_2 + \left(\frac{1}{2} + z_6\right)\,\mathbf{a}_3$ | = | $x_6 a\,\hat{\mathbf{x}} - y_6 b\,\hat{\mathbf{y}} + \left(\frac{1}{2} + z_6\right) c\,\hat{\mathbf{z}}$ | (16$h$) | O II |
| $\mathbf{B}_{26}$ | = | $(-x_6 - y_6)\,\mathbf{a}_1 + (-x_6 + y_6)\,\mathbf{a}_2 + z_6\,\mathbf{a}_3$ | = | $-x_6 a\,\hat{\mathbf{x}} + y_6 b\,\hat{\mathbf{y}} + z_6 c\,\hat{\mathbf{z}}$ | (16$h$) | O II |

**Geometry files:**

- CIF: pp. 1653
- POSCAR: pp. 1653

# Pseudobrookite ($Fe_2TiO_5$, $E4_1$) Structure: A2B5C_oC32_63_f_c2f_c

http://aflow.org/prototype-encyclopedia/A2B5C_oC32_63_f_c2f_c

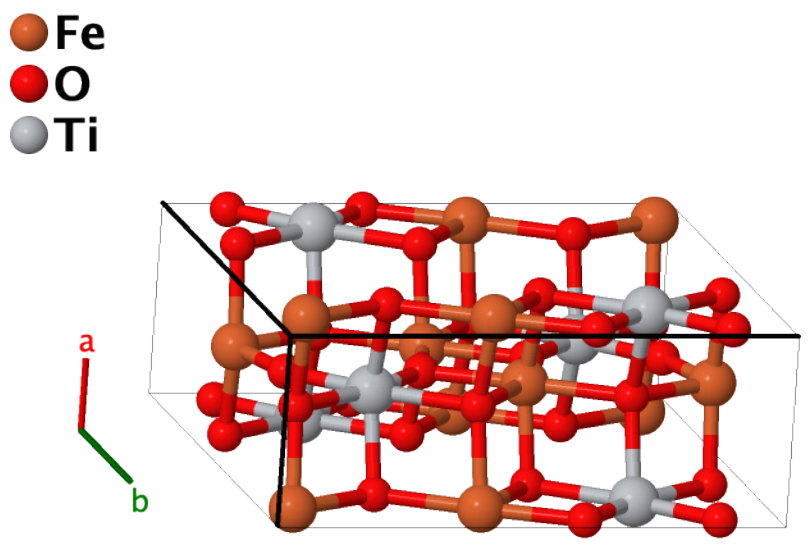

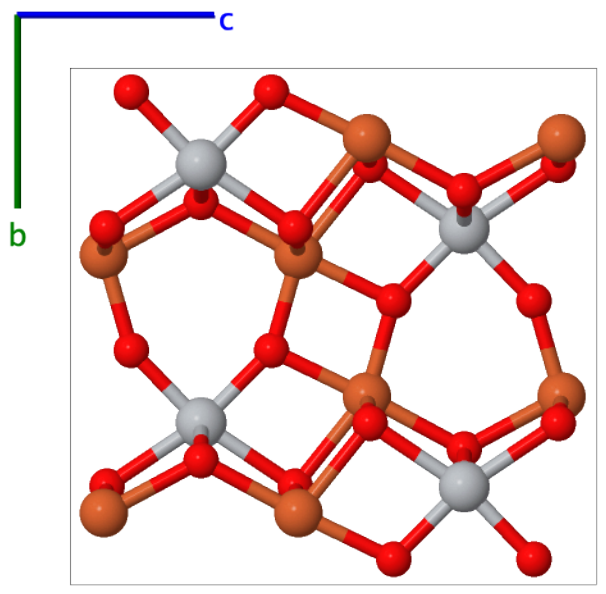

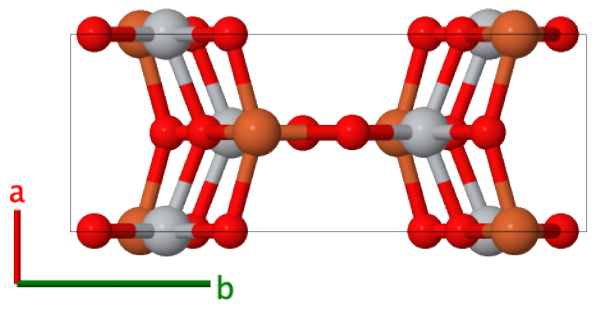

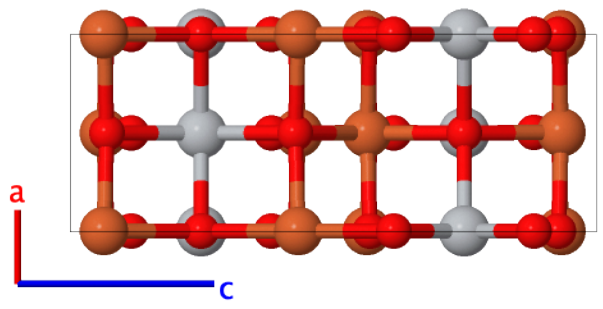

| | | |
|---|---|---|
| **Prototype** | : | $Fe_2O_5Ti$ |
| **AFLOW prototype label** | : | A2B5C_oC32_63_f_c2f_c |
| ***Strukturbericht* designation** | : | $E4_1$ |
| **Pearson symbol** | : | oC32 |
| **Space group number** | : | 63 |
| **Space group symbol** | : | *Cmcm* |
| **AFLOW prototype command** | : | `aflow --proto=A2B5C_oC32_63_f_c2f_c` `--params=`$a, b/a, c/a, y_1, y_2, y_3, z_3, y_4, z_4, y_5, z_5$ |

### Other compounds with this structure

- $Fe_{1+x}Ti_{2-x}O_5$ and $Ti_2MgO_5$

### Base-centered Orthorhombic primitive vectors:

$$\mathbf{a}_1 = \tfrac{1}{2}\,a\,\mathbf{\hat{x}} - \tfrac{1}{2}\,b\,\mathbf{\hat{y}}$$
$$\mathbf{a}_2 = \tfrac{1}{2}\,a\,\mathbf{\hat{x}} + \tfrac{1}{2}\,b\,\mathbf{\hat{y}}$$
$$\mathbf{a}_3 = c\,\mathbf{\hat{z}}$$

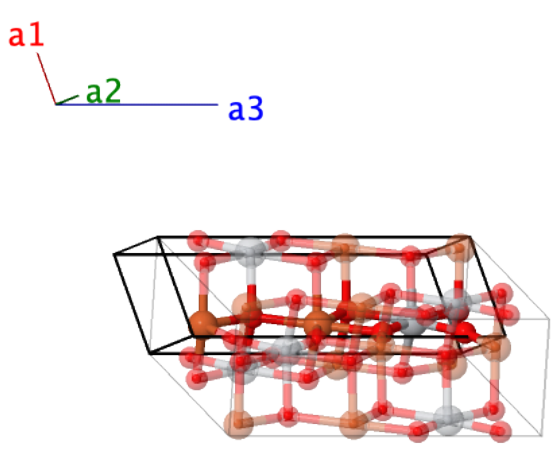

### Basis vectors:

| | | Lattice Coordinates | | Cartesian Coordinates | Wyckoff Position | Atom Type |
|---|---|---|---|---|---|---|
| $\mathbf{B}_1$ | = | $-y_1\,\mathbf{a}_1 + y_1\,\mathbf{a}_2 + \frac{1}{4}\,\mathbf{a}_3$ | = | $y_1 b\,\hat{\mathbf{y}} + \frac{1}{4} c\,\hat{\mathbf{z}}$ | $(4c)$ | O I |
| $\mathbf{B}_2$ | = | $y_1\,\mathbf{a}_1 - y_1\,\mathbf{a}_2 + \frac{3}{4}\,\mathbf{a}_3$ | = | $-y_1 b\,\hat{\mathbf{y}} + \frac{3}{4} c\,\hat{\mathbf{z}}$ | $(4c)$ | O I |
| $\mathbf{B}_3$ | = | $-y_2\,\mathbf{a}_1 + y_2\,\mathbf{a}_2 + \frac{1}{4}\,\mathbf{a}_3$ | = | $y_2 b\,\hat{\mathbf{y}} + \frac{1}{4} c\,\hat{\mathbf{z}}$ | $(4c)$ | Ti |
| $\mathbf{B}_4$ | = | $y_2\,\mathbf{a}_1 - y_2\,\mathbf{a}_2 + \frac{3}{4}\,\mathbf{a}_3$ | = | $-y_2 b\,\hat{\mathbf{y}} + \frac{3}{4} c\,\hat{\mathbf{z}}$ | $(4c)$ | Ti |
| $\mathbf{B}_5$ | = | $-y_3\,\mathbf{a}_1 + y_3\,\mathbf{a}_2 + z_3\,\mathbf{a}_3$ | = | $y_3 b\,\hat{\mathbf{y}} + z_3 c\,\hat{\mathbf{z}}$ | $(8f)$ | Fe |
| $\mathbf{B}_6$ | = | $y_3\,\mathbf{a}_1 - y_3\,\mathbf{a}_2 + \left(\frac{1}{2} + z_3\right)\,\mathbf{a}_3$ | = | $-y_3 b\,\hat{\mathbf{y}} + \left(\frac{1}{2} + z_3\right) c\,\hat{\mathbf{z}}$ | $(8f)$ | Fe |
| $\mathbf{B}_7$ | = | $-y_3\,\mathbf{a}_1 + y_3\,\mathbf{a}_2 + \left(\frac{1}{2} - z_3\right)\,\mathbf{a}_3$ | = | $y_3 b\,\hat{\mathbf{y}} + \left(\frac{1}{2} - z_3\right) c\,\hat{\mathbf{z}}$ | $(8f)$ | Fe |
| $\mathbf{B}_8$ | = | $y_3\,\mathbf{a}_1 - y_3\,\mathbf{a}_2 - z_3\,\mathbf{a}_3$ | = | $-y_3 b\,\hat{\mathbf{y}} - z_3 c\,\hat{\mathbf{z}}$ | $(8f)$ | Fe |
| $\mathbf{B}_9$ | = | $-y_4\,\mathbf{a}_1 + y_4\,\mathbf{a}_2 + z_4\,\mathbf{a}_3$ | = | $y_4 b\,\hat{\mathbf{y}} + z_4 c\,\hat{\mathbf{z}}$ | $(8f)$ | O II |
| $\mathbf{B}_{10}$ | = | $y_4\,\mathbf{a}_1 - y_4\,\mathbf{a}_2 + \left(\frac{1}{2} + z_4\right)\,\mathbf{a}_3$ | = | $-y_4 b\,\hat{\mathbf{y}} + \left(\frac{1}{2} + z_4\right) c\,\hat{\mathbf{z}}$ | $(8f)$ | O II |
| $\mathbf{B}_{11}$ | = | $-y_4\,\mathbf{a}_1 + y_4\,\mathbf{a}_2 + \left(\frac{1}{2} - z_4\right)\,\mathbf{a}_3$ | = | $y_4 b\,\hat{\mathbf{y}} + \left(\frac{1}{2} - z_4\right) c\,\hat{\mathbf{z}}$ | $(8f)$ | O II |
| $\mathbf{B}_{12}$ | = | $y_4\,\mathbf{a}_1 - y_4\,\mathbf{a}_2 - z_4\,\mathbf{a}_3$ | = | $-y_4 b\,\hat{\mathbf{y}} - z_4 c\,\hat{\mathbf{z}}$ | $(8f)$ | O II |
| $\mathbf{B}_{13}$ | = | $-y_5\,\mathbf{a}_1 + y_5\,\mathbf{a}_2 + z_5\,\mathbf{a}_3$ | = | $y_5 b\,\hat{\mathbf{y}} + z_5 c\,\hat{\mathbf{z}}$ | $(8f)$ | O III |
| $\mathbf{B}_{14}$ | = | $y_5\,\mathbf{a}_1 - y_5\,\mathbf{a}_2 + \left(\frac{1}{2} + z_5\right)\,\mathbf{a}_3$ | = | $-y_5 b\,\hat{\mathbf{y}} + \left(\frac{1}{2} + z_5\right) c\,\hat{\mathbf{z}}$ | $(8f)$ | O III |
| $\mathbf{B}_{15}$ | = | $-y_5\,\mathbf{a}_1 + y_5\,\mathbf{a}_2 + \left(\frac{1}{2} - z_5\right)\,\mathbf{a}_3$ | = | $y_5 b\,\hat{\mathbf{y}} + \left(\frac{1}{2} - z_5\right) c\,\hat{\mathbf{z}}$ | $(8f)$ | O III |
| $\mathbf{B}_{16}$ | = | $y_5\,\mathbf{a}_1 - y_5\,\mathbf{a}_2 - z_5\,\mathbf{a}_3$ | = | $-y_5 b\,\hat{\mathbf{y}} - z_5 c\,\hat{\mathbf{z}}$ | $(8f)$ | O III |

**Geometry files:**

- CIF: pp. 1653
- POSCAR: pp. 1654

# $MgCuAl_2$ ($E1_a$) Structure: A2BC_oC16_63_f_c_c

http://aflow.org/prototype-encyclopedia/A2BC_oC16_63_f_c_c

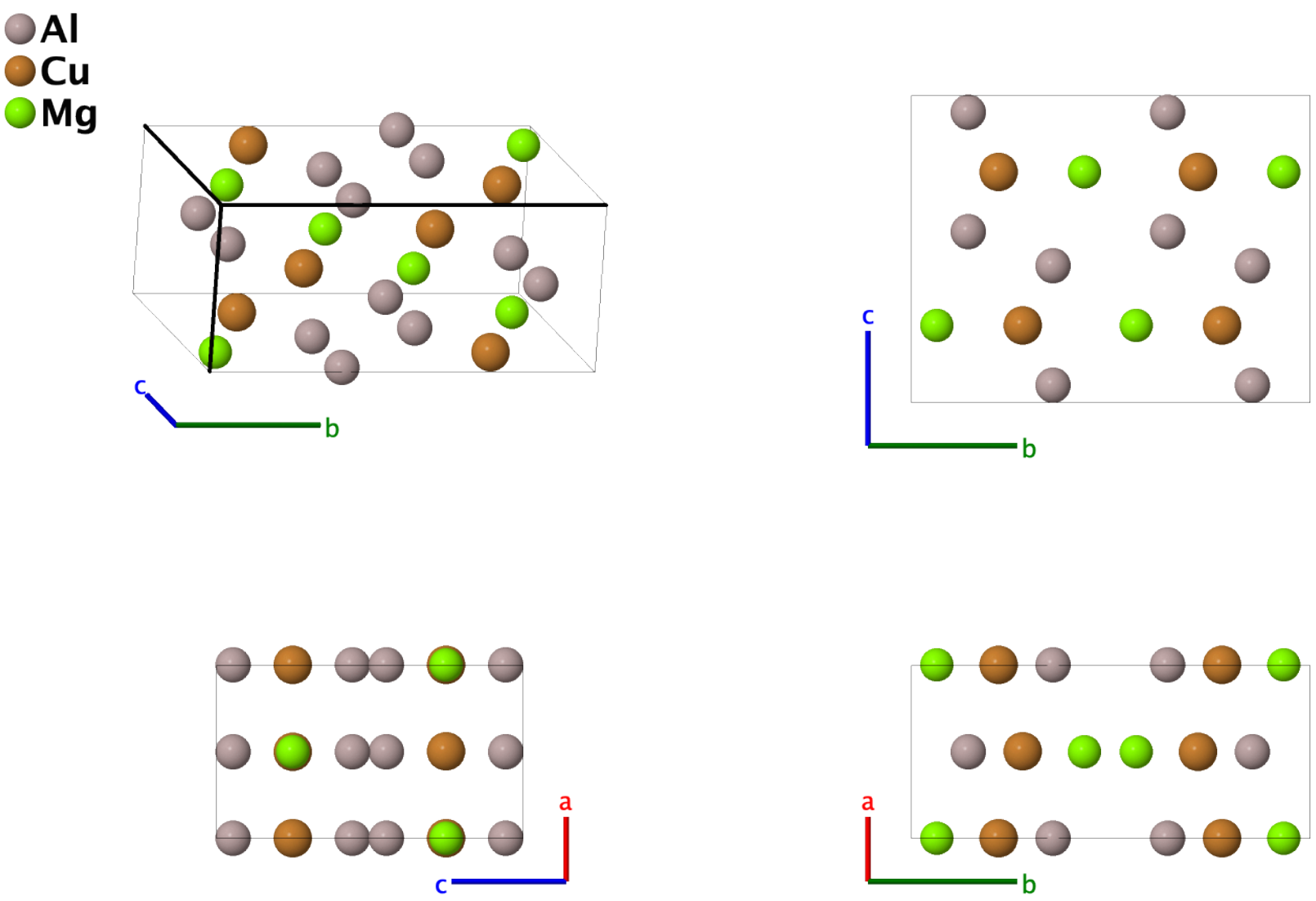

| | | |
|---|---|---|
| **Prototype** | : | $Al_2CuMg$ |
| **AFLOW prototype label** | : | A2BC_oC16_63_f_c_c |
| ***Strukturbericht* designation** | : | None |
| **Pearson symbol** | : | oC16 |
| **Space group number** | : | 63 |
| **Space group symbol** | : | *Cmcm* |
| **AFLOW prototype command** | : | `aflow --proto=A2BC_oC16_63_f_c_c`<br>`--params=`$a, b/a, c/a, y_1, y_2, y_3, z_3$ |

**Other compounds with this structure**

- $YNiAl_2$, $TaBCo_2$, $ScNiAl_2$, $LaPdIn_2$, *RE*-$CuPd_2$ (*RE* = Ce, Pr, Nd, Sm), $CaNiGa_2$, $PPdNi_2$, $EuIrSn_2$, $EuPdIn_2$, $EuPdSn_2$, $YbAuIn_2$, and $YbPdIn_2$

- This is often referred to as an 'S-phase precipitate'. It can be considered as the ternary version of the $Re_3B$ structure.

**Base-centered Orthorhombic primitive vectors:**

$$\mathbf{a}_1 = \frac{1}{2}\,a\,\hat{\mathbf{x}} - \frac{1}{2}\,b\,\hat{\mathbf{y}}$$
$$\mathbf{a}_2 = \frac{1}{2}\,a\,\hat{\mathbf{x}} + \frac{1}{2}\,b\,\hat{\mathbf{y}}$$
$$\mathbf{a}_3 = c\,\hat{\mathbf{z}}$$

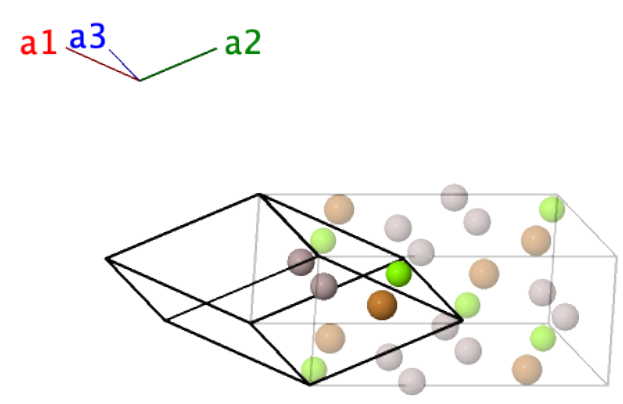

**Basis vectors:**

| Lattice Coordinates | Cartesian Coordinates | Wyckoff Position | Atom Type |
|---|---|---|---|

| | | | | | | |
|---|---|---|---|---|---|---|
| $\mathbf{B}_1$ | = | $-y_1\,\mathbf{a}_1 + y_1\,\mathbf{a}_2 + \frac{1}{4}\,\mathbf{a}_3$ | = | $y_1 b\,\hat{\mathbf{y}} + \frac{1}{4} c\,\hat{\mathbf{z}}$ | (4$c$) | Cu |
| $\mathbf{B}_2$ | = | $y_1\,\mathbf{a}_1 - y_1\,\mathbf{a}_2 + \frac{3}{4}\,\mathbf{a}_3$ | = | $-y_1 b\,\hat{\mathbf{y}} + \frac{3}{4} c\,\hat{\mathbf{z}}$ | (4$c$) | Cu |
| $\mathbf{B}_3$ | = | $-y_2\,\mathbf{a}_1 + y_2\,\mathbf{a}_2 + \frac{1}{4}\,\mathbf{a}_3$ | = | $y_2 b\,\hat{\mathbf{y}} + \frac{1}{4} c\,\hat{\mathbf{z}}$ | (4$c$) | Mg |
| $\mathbf{B}_4$ | = | $y_2\,\mathbf{a}_1 - y_2\,\mathbf{a}_2 + \frac{3}{4}\,\mathbf{a}_3$ | = | $-y_2 b\,\hat{\mathbf{y}} + \frac{3}{4} c\,\hat{\mathbf{z}}$ | (4$c$) | Mg |
| $\mathbf{B}_5$ | = | $-y_3\,\mathbf{a}_1 + y_3\,\mathbf{a}_2 + z_3\,\mathbf{a}_3$ | = | $y_3 b\,\hat{\mathbf{y}} + z_3 c\,\hat{\mathbf{z}}$ | (8$f$) | Al |
| $\mathbf{B}_6$ | = | $y_3\,\mathbf{a}_1 - y_3\,\mathbf{a}_2 + \left(\frac{1}{2} + z_3\right)\,\mathbf{a}_3$ | = | $-y_3 b\,\hat{\mathbf{y}} + \left(\frac{1}{2} + z_3\right) c\,\hat{\mathbf{z}}$ | (8$f$) | Al |
| $\mathbf{B}_7$ | = | $-y_3\,\mathbf{a}_1 + y_3\,\mathbf{a}_2 + \left(\frac{1}{2} - z_3\right)\,\mathbf{a}_3$ | = | $y_3 b\,\hat{\mathbf{y}} + \left(\frac{1}{2} - z_3\right) c\,\hat{\mathbf{z}}$ | (8$f$) | Al |
| $\mathbf{B}_8$ | = | $y_3\,\mathbf{a}_1 - y_3\,\mathbf{a}_2 - z_3\,\mathbf{a}_3$ | = | $-y_3 b\,\hat{\mathbf{y}} - z_3 c\,\hat{\mathbf{z}}$ | (8$f$) | Al |

**Geometry files:**

- CIF: pp. 1654
- POSCAR: pp. 1654

# $S0_4$ (Staurolite, $Fe(OH)_2Al_4Si_2O_{10}$) (*obsolete*) Structure: A4BC12D2_oC76_63_eg_c_f3gh_g

http://aflow.org/prototype-encyclopedia/A4BC12D2_oC76_63_eg_c_f3gh_g

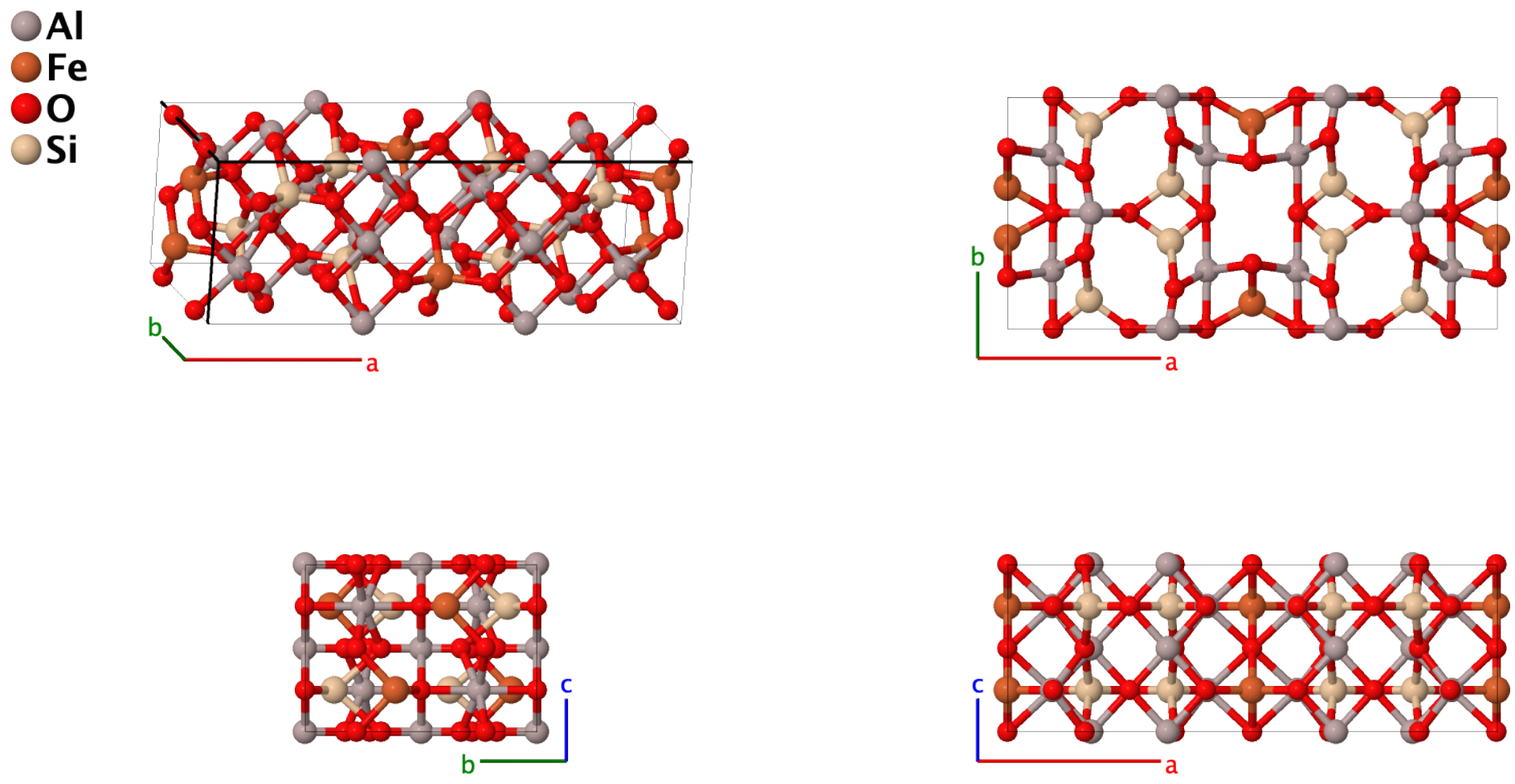

| | | |
|---|---|---|
| **Prototype** | : | $Al_4FeO_{12}Si_2$ |
| **AFLOW prototype label** | : | A4BC12D2_oC76_63_eg_c_f3gh_g |
| ***Strukturbericht* designation** | : | $S0_4$ |
| **Pearson symbol** | : | oC76 |
| **Space group number** | : | 63 |
| **Space group symbol** | : | *Cmcm* |
| **AFLOW prototype command** | : | `aflow --proto=A4BC12D2_oC76_63_eg_c_f3gh_g` `--params=`$a, b/a, c/a, y_1, x_2, y_3, z_3, x_4, y_4, x_5, y_5, x_6, y_6, x_7, y_7, x_8, y_8, x_9, y_9, z_9$ |

- This orthorhombic structure of staurolite determined by (Náray-Szabó, 1929) was given the *Strukturbericht* designation $S0_4$ by (Hermann, 1937). (Smith, 1968) showed that the structure is actually monoclinic with $\beta \approx 90^\circ$, rather than orthorhombic and corrected the chemical composition of the mineral. Accordingly, we have marked this version of the structure as obsolete, but retain it for historical interest.
- One sixth of the oxygen sites listed here should actually be OH ions, but the positions of the hydrogen atoms were not given in (Náray-Szabó, 1929) and are labeled as O.
- We take our data from (Hermann, 1937), who presented it in the *Ccmm* representation of space group #63. We used FINDSYM to transform the structure to the standard *Cmcm* representation.
- (Hermann, 1937) also gave this the label $H5_4$ in the index.

**Base-centered Orthorhombic primitive vectors:**

$$\mathbf{a}_1 = \frac{1}{2}\,a\,\hat{\mathbf{x}} - \frac{1}{2}\,b\,\hat{\mathbf{y}}$$
$$\mathbf{a}_2 = \frac{1}{2}\,a\,\hat{\mathbf{x}} + \frac{1}{2}\,b\,\hat{\mathbf{y}}$$
$$\mathbf{a}_3 = c\,\hat{\mathbf{z}}$$

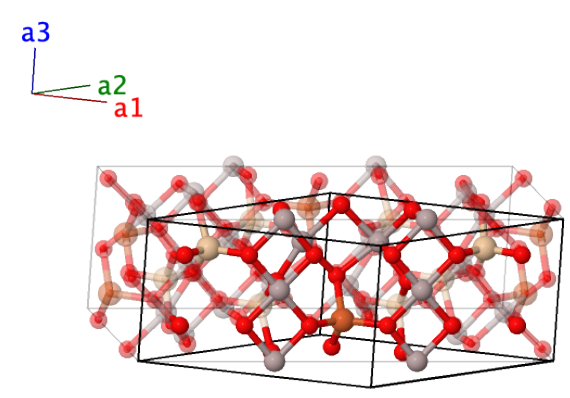

**Basis vectors:**

| | | Lattice Coordinates | | Cartesian Coordinates | Wyckoff Position | Atom Type |
|---|---|---|---|---|---|---|
| $\mathbf{B}_1$ | = | $-y_1\,\mathbf{a}_1 + y_1\,\mathbf{a}_2 + \frac{1}{4}\,\mathbf{a}_3$ | = | $y_1 b\,\hat{\mathbf{y}} + \frac{1}{4} c\,\hat{\mathbf{z}}$ | (4$c$) | Fe |
| $\mathbf{B}_2$ | = | $y_1\,\mathbf{a}_1 - y_1\,\mathbf{a}_2 + \frac{3}{4}\,\mathbf{a}_3$ | = | $-y_1 b\,\hat{\mathbf{y}} + \frac{3}{4} c\,\hat{\mathbf{z}}$ | (4$c$) | Fe |
| $\mathbf{B}_3$ | = | $x_2\,\mathbf{a}_1 + x_2\,\mathbf{a}_2$ | = | $x_2 a\,\hat{\mathbf{x}}$ | (8$e$) | Al I |
| $\mathbf{B}_4$ | = | $-x_2\,\mathbf{a}_1 - x_2\,\mathbf{a}_2 + \frac{1}{2}\,\mathbf{a}_3$ | = | $-x_2 a\,\hat{\mathbf{x}} + \frac{1}{2} c\,\hat{\mathbf{z}}$ | (8$e$) | Al I |
| $\mathbf{B}_5$ | = | $-x_2\,\mathbf{a}_1 - x_2\,\mathbf{a}_2$ | = | $-x_2 a\,\hat{\mathbf{x}}$ | (8$e$) | Al I |
| $\mathbf{B}_6$ | = | $x_2\,\mathbf{a}_1 + x_2\,\mathbf{a}_2 + \frac{1}{2}\,\mathbf{a}_3$ | = | $x_2 a\,\hat{\mathbf{x}} + \frac{1}{2} c\,\hat{\mathbf{z}}$ | (8$e$) | Al I |
| $\mathbf{B}_7$ | = | $-y_3\,\mathbf{a}_1 + y_3\,\mathbf{a}_2 + z_3\,\mathbf{a}_3$ | = | $y_3 b\,\hat{\mathbf{y}} + z_3 c\,\hat{\mathbf{z}}$ | (8$f$) | O I |
| $\mathbf{B}_8$ | = | $y_3\,\mathbf{a}_1 - y_3\,\mathbf{a}_2 + \left(\frac{1}{2} + z_3\right)\,\mathbf{a}_3$ | = | $-y_3 b\,\hat{\mathbf{y}} + \left(\frac{1}{2} + z_3\right) c\,\hat{\mathbf{z}}$ | (8$f$) | O I |
| $\mathbf{B}_9$ | = | $-y_3\,\mathbf{a}_1 + y_3\,\mathbf{a}_2 + \left(\frac{1}{2} - z_3\right)\,\mathbf{a}_3$ | = | $y_3 b\,\hat{\mathbf{y}} + \left(\frac{1}{2} - z_3\right) c\,\hat{\mathbf{z}}$ | (8$f$) | O I |
| $\mathbf{B}_{10}$ | = | $y_3\,\mathbf{a}_1 - y_3\,\mathbf{a}_2 - z_3\,\mathbf{a}_3$ | = | $-y_3 b\,\hat{\mathbf{y}} - z_3 c\,\hat{\mathbf{z}}$ | (8$f$) | O I |
| $\mathbf{B}_{11}$ | = | $(x_4 - y_4)\,\mathbf{a}_1 + (x_4 + y_4)\,\mathbf{a}_2 + \frac{1}{4}\,\mathbf{a}_3$ | = | $x_4 a\,\hat{\mathbf{x}} + y_4 b\,\hat{\mathbf{y}} + \frac{1}{4} c\,\hat{\mathbf{z}}$ | (8$g$) | Al II |
| $\mathbf{B}_{12}$ | = | $(-x_4 + y_4)\,\mathbf{a}_1 + (-x_4 - y_4)\,\mathbf{a}_2 + \frac{3}{4}\,\mathbf{a}_3$ | = | $-x_4 a\,\hat{\mathbf{x}} - y_4 b\,\hat{\mathbf{y}} + \frac{3}{4} c\,\hat{\mathbf{z}}$ | (8$g$) | Al II |
| $\mathbf{B}_{13}$ | = | $(-x_4 - y_4)\,\mathbf{a}_1 + (-x_4 + y_4)\,\mathbf{a}_2 + \frac{1}{4}\,\mathbf{a}_3$ | = | $-x_4 a\,\hat{\mathbf{x}} + y_4 b\,\hat{\mathbf{y}} + \frac{1}{4} c\,\hat{\mathbf{z}}$ | (8$g$) | Al II |
| $\mathbf{B}_{14}$ | = | $(x_4 + y_4)\,\mathbf{a}_1 + (x_4 - y_4)\,\mathbf{a}_2 + \frac{3}{4}\,\mathbf{a}_3$ | = | $x_4 a\,\hat{\mathbf{x}} - y_4 b\,\hat{\mathbf{y}} + \frac{3}{4} c\,\hat{\mathbf{z}}$ | (8$g$) | Al II |
| $\mathbf{B}_{15}$ | = | $(x_5 - y_5)\,\mathbf{a}_1 + (x_5 + y_5)\,\mathbf{a}_2 + \frac{1}{4}\,\mathbf{a}_3$ | = | $x_5 a\,\hat{\mathbf{x}} + y_5 b\,\hat{\mathbf{y}} + \frac{1}{4} c\,\hat{\mathbf{z}}$ | (8$g$) | O II |
| $\mathbf{B}_{16}$ | = | $(-x_5 + y_5)\,\mathbf{a}_1 + (-x_5 - y_5)\,\mathbf{a}_2 + \frac{3}{4}\,\mathbf{a}_3$ | = | $-x_5 a\,\hat{\mathbf{x}} - y_5 b\,\hat{\mathbf{y}} + \frac{3}{4} c\,\hat{\mathbf{z}}$ | (8$g$) | O II |
| $\mathbf{B}_{17}$ | = | $(-x_5 - y_5)\,\mathbf{a}_1 + (-x_5 + y_5)\,\mathbf{a}_2 + \frac{1}{4}\,\mathbf{a}_3$ | = | $-x_5 a\,\hat{\mathbf{x}} + y_5 b\,\hat{\mathbf{y}} + \frac{1}{4} c\,\hat{\mathbf{z}}$ | (8$g$) | O II |
| $\mathbf{B}_{18}$ | = | $(x_5 + y_5)\,\mathbf{a}_1 + (x_5 - y_5)\,\mathbf{a}_2 + \frac{3}{4}\,\mathbf{a}_3$ | = | $x_5 a\,\hat{\mathbf{x}} - y_5 b\,\hat{\mathbf{y}} + \frac{3}{4} c\,\hat{\mathbf{z}}$ | (8$g$) | O II |
| $\mathbf{B}_{19}$ | = | $(x_6 - y_6)\,\mathbf{a}_1 + (x_6 + y_6)\,\mathbf{a}_2 + \frac{1}{4}\,\mathbf{a}_3$ | = | $x_6 a\,\hat{\mathbf{x}} + y_6 b\,\hat{\mathbf{y}} + \frac{1}{4} c\,\hat{\mathbf{z}}$ | (8$g$) | O III |
| $\mathbf{B}_{20}$ | = | $(-x_6 + y_6)\,\mathbf{a}_1 + (-x_6 - y_6)\,\mathbf{a}_2 + \frac{3}{4}\,\mathbf{a}_3$ | = | $-x_6 a\,\hat{\mathbf{x}} - y_6 b\,\hat{\mathbf{y}} + \frac{3}{4} c\,\hat{\mathbf{z}}$ | (8$g$) | O III |
| $\mathbf{B}_{21}$ | = | $(-x_6 - y_6)\,\mathbf{a}_1 + (-x_6 + y_6)\,\mathbf{a}_2 + \frac{1}{4}\,\mathbf{a}_3$ | = | $-x_6 a\,\hat{\mathbf{x}} + y_6 b\,\hat{\mathbf{y}} + \frac{1}{4} c\,\hat{\mathbf{z}}$ | (8$g$) | O III |
| $\mathbf{B}_{22}$ | = | $(x_6 + y_6)\,\mathbf{a}_1 + (x_6 - y_6)\,\mathbf{a}_2 + \frac{3}{4}\,\mathbf{a}_3$ | = | $x_6 a\,\hat{\mathbf{x}} - y_6 b\,\hat{\mathbf{y}} + \frac{3}{4} c\,\hat{\mathbf{z}}$ | (8$g$) | O III |
| $\mathbf{B}_{23}$ | = | $(x_7 - y_7)\,\mathbf{a}_1 + (x_7 + y_7)\,\mathbf{a}_2 + \frac{1}{4}\,\mathbf{a}_3$ | = | $x_7 a\,\hat{\mathbf{x}} + y_7 b\,\hat{\mathbf{y}} + \frac{1}{4} c\,\hat{\mathbf{z}}$ | (8$g$) | O IV |
| $\mathbf{B}_{24}$ | = | $(-x_7 + y_7)\,\mathbf{a}_1 + (-x_7 - y_7)\,\mathbf{a}_2 + \frac{3}{4}\,\mathbf{a}_3$ | = | $-x_7 a\,\hat{\mathbf{x}} - y_7 b\,\hat{\mathbf{y}} + \frac{3}{4} c\,\hat{\mathbf{z}}$ | (8$g$) | O IV |
| $\mathbf{B}_{25}$ | = | $(-x_7 - y_7)\,\mathbf{a}_1 + (-x_7 + y_7)\,\mathbf{a}_2 + \frac{1}{4}\,\mathbf{a}_3$ | = | $-x_7 a\,\hat{\mathbf{x}} + y_7 b\,\hat{\mathbf{y}} + \frac{1}{4} c\,\hat{\mathbf{z}}$ | (8$g$) | O IV |
| $\mathbf{B}_{26}$ | = | $(x_7 + y_7)\,\mathbf{a}_1 + (x_7 - y_7)\,\mathbf{a}_2 + \frac{3}{4}\,\mathbf{a}_3$ | = | $x_7 a\,\hat{\mathbf{x}} - y_7 b\,\hat{\mathbf{y}} + \frac{3}{4} c\,\hat{\mathbf{z}}$ | (8$g$) | O IV |
| $\mathbf{B}_{27}$ | = | $(x_8 - y_8)\,\mathbf{a}_1 + (x_8 + y_8)\,\mathbf{a}_2 + \frac{1}{4}\,\mathbf{a}_3$ | = | $x_8 a\,\hat{\mathbf{x}} + y_8 b\,\hat{\mathbf{y}} + \frac{1}{4} c\,\hat{\mathbf{z}}$ | (8$g$) | Si |
| $\mathbf{B}_{28}$ | = | $(-x_8 + y_8)\,\mathbf{a}_1 + (-x_8 - y_8)\,\mathbf{a}_2 + \frac{3}{4}\,\mathbf{a}_3$ | = | $-x_8 a\,\hat{\mathbf{x}} - y_8 b\,\hat{\mathbf{y}} + \frac{3}{4} c\,\hat{\mathbf{z}}$ | (8$g$) | Si |
| $\mathbf{B}_{29}$ | = | $(-x_8 - y_8)\,\mathbf{a}_1 + (-x_8 + y_8)\,\mathbf{a}_2 + \frac{1}{4}\,\mathbf{a}_3$ | = | $-x_8 a\,\hat{\mathbf{x}} + y_8 b\,\hat{\mathbf{y}} + \frac{1}{4} c\,\hat{\mathbf{z}}$ | (8$g$) | Si |
| $\mathbf{B}_{30}$ | = | $(x_8 + y_8)\,\mathbf{a}_1 + (x_8 - y_8)\,\mathbf{a}_2 + \frac{3}{4}\,\mathbf{a}_3$ | = | $x_8 a\,\hat{\mathbf{x}} - y_8 b\,\hat{\mathbf{y}} + \frac{3}{4} c\,\hat{\mathbf{z}}$ | (8$g$) | Si |
| $\mathbf{B}_{31}$ | = | $(x_9 - y_9)\,\mathbf{a}_1 + (x_9 + y_9)\,\mathbf{a}_2 + z_9\,\mathbf{a}_3$ | = | $x_9 a\,\hat{\mathbf{x}} + y_9 b\,\hat{\mathbf{y}} + z_9 c\,\hat{\mathbf{z}}$ | (16$h$) | O V |
| $\mathbf{B}_{32}$ | = | $(-x_9 + y_9)\,\mathbf{a}_1 + (-x_9 - y_9)\,\mathbf{a}_2 + \left(\frac{1}{2} + z_9\right)\,\mathbf{a}_3$ | = | $-x_9 a\,\hat{\mathbf{x}} - y_9 b\,\hat{\mathbf{y}} + \left(\frac{1}{2} + z_9\right) c\,\hat{\mathbf{z}}$ | (16$h$) | O V |
| $\mathbf{B}_{33}$ | = | $(-x_9 - y_9)\,\mathbf{a}_1 + (-x_9 + y_9)\,\mathbf{a}_2 + \left(\frac{1}{2} - z_9\right)\,\mathbf{a}_3$ | = | $-x_9 a\,\hat{\mathbf{x}} + y_9 b\,\hat{\mathbf{y}} + \left(\frac{1}{2} - z_9\right) c\,\hat{\mathbf{z}}$ | (16$h$) | O V |
| $\mathbf{B}_{34}$ | = | $(x_9 + y_9)\,\mathbf{a}_1 + (x_9 - y_9)\,\mathbf{a}_2 - z_9\,\mathbf{a}_3$ | = | $x_9 a\,\hat{\mathbf{x}} - y_9 b\,\hat{\mathbf{y}} - z_9 c\,\hat{\mathbf{z}}$ | (16$h$) | O V |
| $\mathbf{B}_{35}$ | = | $(-x_9 + y_9)\,\mathbf{a}_1 + (-x_9 - y_9)\,\mathbf{a}_2 - z_9\,\mathbf{a}_3$ | = | $-x_9 a\,\hat{\mathbf{x}} - y_9 b\,\hat{\mathbf{y}} - z_9 c\,\hat{\mathbf{z}}$ | (16$h$) | O V |
| $\mathbf{B}_{36}$ | = | $(x_9 - y_9)\,\mathbf{a}_1 + (x_9 + y_9)\,\mathbf{a}_2 + \left(\frac{1}{2} - z_9\right)\mathbf{a}_3$ | = | $x_9 a\,\hat{\mathbf{x}} + y_9 b\,\hat{\mathbf{y}} + \left(\frac{1}{2} - z_9\right) c\,\hat{\mathbf{z}}$ | (16$h$) | O V |
| $\mathbf{B}_{37}$ | = | $(x_9 + y_9)\,\mathbf{a}_1 + (x_9 - y_9)\,\mathbf{a}_2 + \left(\frac{1}{2} + z_9\right)\mathbf{a}_3$ | = | $x_9 a\,\hat{\mathbf{x}} - y_9 b\,\hat{\mathbf{y}} + \left(\frac{1}{2} + z_9\right) c\,\hat{\mathbf{z}}$ | (16$h$) | O V |
| $\mathbf{B}_{38}$ | = | $(-x_9 - y_9)\,\mathbf{a}_1 + (-x_9 + y_9)\,\mathbf{a}_2 + z_9\,\mathbf{a}_3$ | = | $-x_9 a\,\hat{\mathbf{x}} + y_9 b\,\hat{\mathbf{y}} + z_9 c\,\hat{\mathbf{z}}$ | (16$h$) | O V |

**Geometry files:**

- CIF: pp. 1655
- POSCAR: pp. 1655

# $Pd_5Pu_3$ Structure: A5B3_oC32_63_cfg_ce

http://aflow.org/prototype-encyclopedia/A5B3_oC32_63_cfg_ce

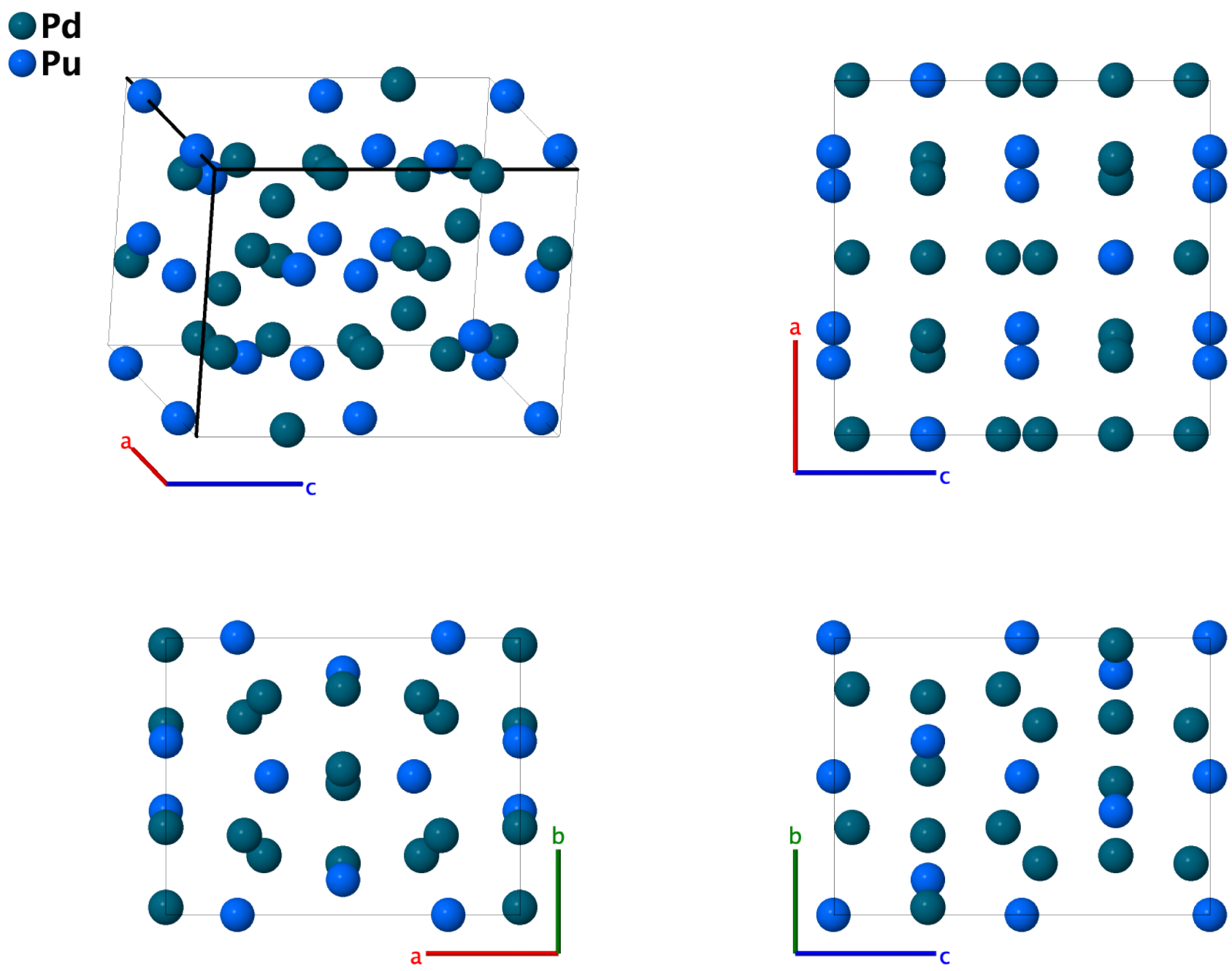

| | | |
|---|---|---|
| **Prototype** | : | $Pd_5Pu_3$ |
| **AFLOW prototype label** | : | A5B3_oC32_63_cfg_ce |
| ***Strukturbericht* designation** | : | None |
| **Pearson symbol** | : | oC32 |
| **Space group number** | : | 63 |
| **Space group symbol** | : | *Cmcm* |
| **AFLOW prototype command** | : | `aflow --proto=A5B3_oC32_63_cfg_ce` <br> `--params=`$a, b/a, c/a, y_1, y_2, x_3, y_4, z_4, x_5, y_5$ |

### Other compounds with this structure

- $Ga_5Zr_3$, $In_5R_3$ (R = Ce,Dy,Er,Gd,Ho,La,Lu,Nd,Pr,Sm,Tb,Th,Y), $Pb_5Ba_3$, $Pd_5R_3$ (R = Sc,Y,Gd-Lu), $Rh_5Zr_3$, $Ga_5U_3$, $(Mg_xSn_{1-x})_5La_3$, $Sn_5La_3$, $Sn_5Sr_3$, and $Sn_5Yb_3$

- Although (Massalski, 1990) lists $Pd_5Pu_3$ as the prototype for many structures, it is not shown in the assessed Pd-Pu phase diagram, which is based on data from 1967.
- (Cromer, 1976) states that this phase may be isostructural with $Ga_5Zr_3$, but at the time of publication the exact structure of that phase had not been solved.

**Base-centered Orthorhombic primitive vectors:**

$$\begin{aligned}
\mathbf{a}_1 &= \tfrac{1}{2}\,a\,\hat{\mathbf{x}} - \tfrac{1}{2}\,b\,\hat{\mathbf{y}} \\
\mathbf{a}_2 &= \tfrac{1}{2}\,a\,\hat{\mathbf{x}} + \tfrac{1}{2}\,b\,\hat{\mathbf{y}} \\
\mathbf{a}_3 &= c\,\hat{\mathbf{z}}
\end{aligned}$$

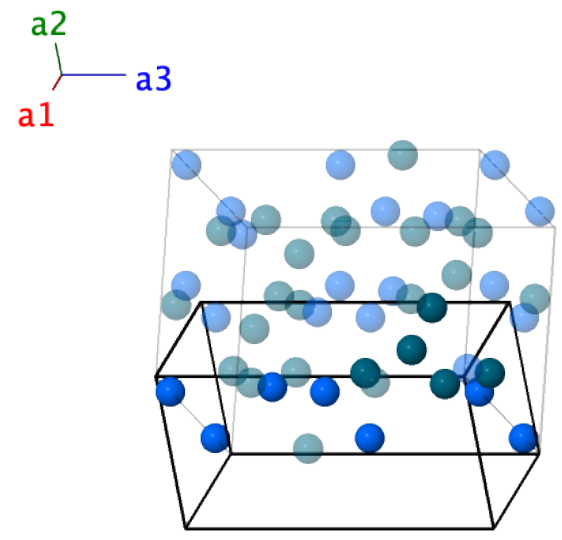

**Basis vectors:**

| | | Lattice Coordinates | | Cartesian Coordinates | Wyckoff Position | Atom Type |
|---|---|---|---|---|---|---|
| $\mathbf{B}_1$ | = | $-y_1\,\mathbf{a}_1 + y_1\,\mathbf{a}_2 + \frac{1}{4}\,\mathbf{a}_3$ | = | $y_1 b\,\hat{\mathbf{y}} + \frac{1}{4} c\,\hat{\mathbf{z}}$ | (4$c$) | Pd I |
| $\mathbf{B}_2$ | = | $y_1\,\mathbf{a}_1 - y_1\,\mathbf{a}_2 + \frac{3}{4}\,\mathbf{a}_3$ | = | $-y_1 b\,\hat{\mathbf{y}} + \frac{3}{4} c\,\hat{\mathbf{z}}$ | (4$c$) | Pd I |
| $\mathbf{B}_3$ | = | $-y_2\,\mathbf{a}_1 + y_2\,\mathbf{a}_2 + \frac{1}{4}\,\mathbf{a}_3$ | = | $y_2 b\,\hat{\mathbf{y}} + \frac{1}{4} c\,\hat{\mathbf{z}}$ | (4$c$) | Pu I |
| $\mathbf{B}_4$ | = | $y_2\,\mathbf{a}_1 - y_2\,\mathbf{a}_2 + \frac{3}{4}\,\mathbf{a}_3$ | = | $-y_2 b\,\hat{\mathbf{y}} + \frac{3}{4} c\,\hat{\mathbf{z}}$ | (4$c$) | Pu I |
| $\mathbf{B}_5$ | = | $x_3\,\mathbf{a}_1 + x_3\,\mathbf{a}_2$ | = | $x_3 a\,\hat{\mathbf{x}}$ | (8$e$) | Pu II |
| $\mathbf{B}_6$ | = | $-x_3\,\mathbf{a}_1 - x_3\,\mathbf{a}_2 + \frac{1}{2}\,\mathbf{a}_3$ | = | $-x_3 a\,\hat{\mathbf{x}} + \frac{1}{2} c\,\hat{\mathbf{z}}$ | (8$e$) | Pu II |
| $\mathbf{B}_7$ | = | $-x_3\,\mathbf{a}_1 - x_3\,\mathbf{a}_2$ | = | $-x_3 a\,\hat{\mathbf{x}}$ | (8$e$) | Pu II |
| $\mathbf{B}_8$ | = | $x_3\,\mathbf{a}_1 + x_3\,\mathbf{a}_2 + \frac{1}{2}\,\mathbf{a}_3$ | = | $x_3 a\,\hat{\mathbf{x}} + \frac{1}{2} c\,\hat{\mathbf{z}}$ | (8$e$) | Pu II |
| $\mathbf{B}_9$ | = | $-y_4\,\mathbf{a}_1 + y_4\,\mathbf{a}_2 + z_4\,\mathbf{a}_3$ | = | $y_4 b\,\hat{\mathbf{y}} + z_4 c\,\hat{\mathbf{z}}$ | (8$f$) | Pd II |
| $\mathbf{B}_{10}$ | = | $y_4\,\mathbf{a}_1 - y_4\,\mathbf{a}_2 + \left(\frac{1}{2} + z_4\right)\mathbf{a}_3$ | = | $-y_4 b\,\hat{\mathbf{y}} + \left(\frac{1}{2} + z_4\right) c\,\hat{\mathbf{z}}$ | (8$f$) | Pd II |
| $\mathbf{B}_{11}$ | = | $-y_4\,\mathbf{a}_1 + y_4\,\mathbf{a}_2 + \left(\frac{1}{2} - z_4\right)\mathbf{a}_3$ | = | $y_4 b\,\hat{\mathbf{y}} + \left(\frac{1}{2} - z_4\right) c\,\hat{\mathbf{z}}$ | (8$f$) | Pd II |
| $\mathbf{B}_{12}$ | = | $y_4\,\mathbf{a}_1 - y_4\,\mathbf{a}_2 - z_4\,\mathbf{a}_3$ | = | $-y_4 b\,\hat{\mathbf{y}} - z_4 c\,\hat{\mathbf{z}}$ | (8$f$) | Pd II |
| $\mathbf{B}_{13}$ | = | $(x_5 - y_5)\,\mathbf{a}_1 + (x_5 + y_5)\,\mathbf{a}_2 + \frac{1}{4}\,\mathbf{a}_3$ | = | $x_5 a\,\hat{\mathbf{x}} + y_5 b\,\hat{\mathbf{y}} + \frac{1}{4} c\,\hat{\mathbf{z}}$ | (8$g$) | Pd III |
| $\mathbf{B}_{14}$ | = | $(-x_5 + y_5)\,\mathbf{a}_1 + (-x_5 - y_5)\,\mathbf{a}_2 + \frac{3}{4}\,\mathbf{a}_3$ | = | $-x_5 a\,\hat{\mathbf{x}} - y_5 b\,\hat{\mathbf{y}} + \frac{3}{4} c\,\hat{\mathbf{z}}$ | (8$g$) | Pd III |
| $\mathbf{B}_{15}$ | = | $(-x_5 - y_5)\,\mathbf{a}_1 + (-x_5 + y_5)\,\mathbf{a}_2 + \frac{1}{4}\,\mathbf{a}_3$ | = | $-x_5 a\,\hat{\mathbf{x}} + y_5 b\,\hat{\mathbf{y}} + \frac{1}{4} c\,\hat{\mathbf{z}}$ | (8$g$) | Pd III |
| $\mathbf{B}_{16}$ | = | $(x_5 + y_5)\,\mathbf{a}_1 + (x_5 - y_5)\,\mathbf{a}_2 + \frac{3}{4}\,\mathbf{a}_3$ | = | $x_5 a\,\hat{\mathbf{x}} - y_5 b\,\hat{\mathbf{y}} + \frac{3}{4} c\,\hat{\mathbf{z}}$ | (8$g$) | Pd III |

**Geometry files:**
- CIF: pp. 1655
- POSCAR: pp. 1656

# $ZrTe_5$ Structure: A5B_oC24_63_c2f_c

http://aflow.org/prototype-encyclopedia/A5B_oC24_63_c2f_c

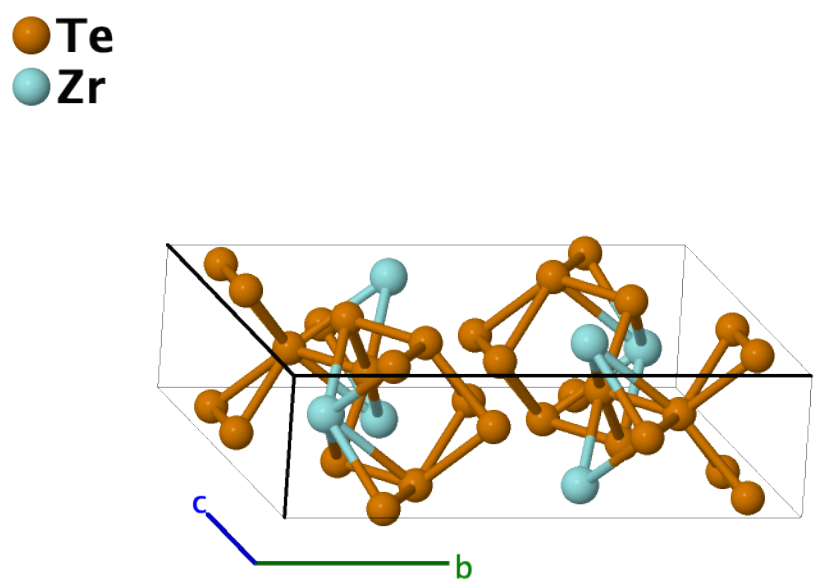

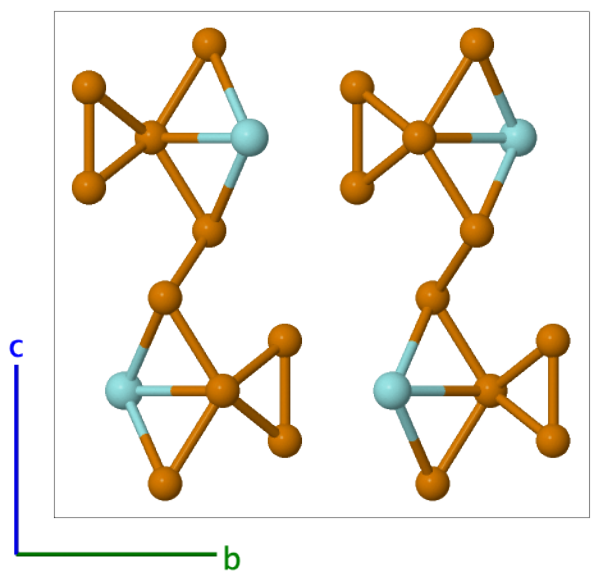

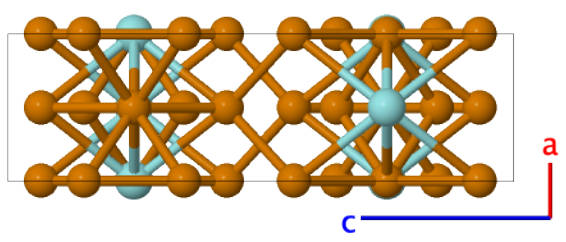

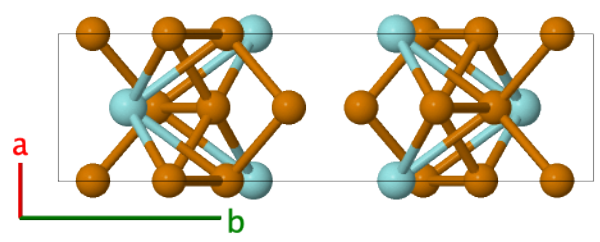

| | | |
|---|---|---|
| **Prototype** | : | $Te_5Zr$ |
| **AFLOW prototype label** | : | A5B_oC24_63_c2f_c |
| ***Strukturbericht* designation** | : | None |
| **Pearson symbol** | : | oC24 |
| **Space group number** | : | 63 |
| **Space group symbol** | : | *Cmcm* |
| **AFLOW prototype command** | : | `aflow --proto=A5B_oC24_63_c2f_c` <br> `--params=`$a, b/a, c/a, y_1, y_2, y_3, z_3, y_4, z_4$ |

### Other compounds with this structure

- $HfTe_5$

- We use the data taken at room temperature.

### Base-centered Orthorhombic primitive vectors:

$$\mathbf{a}_1 = \frac{1}{2}\,a\,\hat{\mathbf{x}} - \frac{1}{2}\,b\,\hat{\mathbf{y}}$$
$$\mathbf{a}_2 = \frac{1}{2}\,a\,\hat{\mathbf{x}} + \frac{1}{2}\,b\,\hat{\mathbf{y}}$$
$$\mathbf{a}_3 = c\,\hat{\mathbf{z}}$$

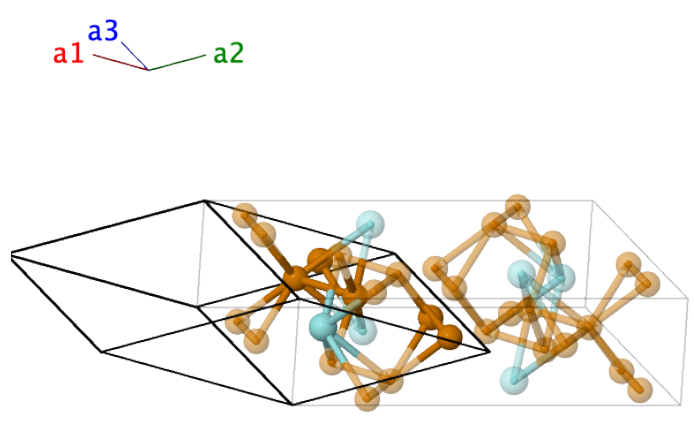

**Basis vectors:**

| | | Lattice Coordinates | | Cartesian Coordinates | Wyckoff Position | Atom Type |
|---|---|---|---|---|---|---|
| $\mathbf{B}_1$ | = | $-y_1\,\mathbf{a}_1 + y_1\,\mathbf{a}_2 + \frac{1}{4}\,\mathbf{a}_3$ | = | $y_1 b\,\hat{\mathbf{y}} + \frac{1}{4} c\,\hat{\mathbf{z}}$ | (4$c$) | Te I |
| $\mathbf{B}_2$ | = | $y_1\,\mathbf{a}_1 - y_1\,\mathbf{a}_2 + \frac{3}{4}\,\mathbf{a}_3$ | = | $-y_1 b\,\hat{\mathbf{y}} + \frac{3}{4} c\,\hat{\mathbf{z}}$ | (4$c$) | Te I |
| $\mathbf{B}_3$ | = | $-y_2\,\mathbf{a}_1 + y_2\,\mathbf{a}_2 + \frac{1}{4}\,\mathbf{a}_3$ | = | $y_2 b\,\hat{\mathbf{y}} + \frac{1}{4} c\,\hat{\mathbf{z}}$ | (4$c$) | Zr |
| $\mathbf{B}_4$ | = | $y_2\,\mathbf{a}_1 - y_2\,\mathbf{a}_2 + \frac{3}{4}\,\mathbf{a}_3$ | = | $-y_2 b\,\hat{\mathbf{y}} + \frac{3}{4} c\,\hat{\mathbf{z}}$ | (4$c$) | Zr |
| $\mathbf{B}_5$ | = | $-y_3\,\mathbf{a}_1 + y_3\,\mathbf{a}_2 + z_3\,\mathbf{a}_3$ | = | $y_3 b\,\hat{\mathbf{y}} + z_3 c\,\hat{\mathbf{z}}$ | (8$f$) | Te II |
| $\mathbf{B}_6$ | = | $y_3\,\mathbf{a}_1 - y_3\,\mathbf{a}_2 + \left(\frac{1}{2} + z_3\right)\mathbf{a}_3$ | = | $-y_3 b\,\hat{\mathbf{y}} + \left(\frac{1}{2} + z_3\right) c\,\hat{\mathbf{z}}$ | (8$f$) | Te II |
| $\mathbf{B}_7$ | = | $-y_3\,\mathbf{a}_1 + y_3\,\mathbf{a}_2 + \left(\frac{1}{2} - z_3\right)\mathbf{a}_3$ | = | $y_3 b\,\hat{\mathbf{y}} + \left(\frac{1}{2} - z_3\right) c\,\hat{\mathbf{z}}$ | (8$f$) | Te II |
| $\mathbf{B}_8$ | = | $y_3\,\mathbf{a}_1 - y_3\,\mathbf{a}_2 - z_3\,\mathbf{a}_3$ | = | $-y_3 b\,\hat{\mathbf{y}} - z_3 c\,\hat{\mathbf{z}}$ | (8$f$) | Te II |
| $\mathbf{B}_9$ | = | $-y_4\,\mathbf{a}_1 + y_4\,\mathbf{a}_2 + z_4\,\mathbf{a}_3$ | = | $y_4 b\,\hat{\mathbf{y}} + z_4 c\,\hat{\mathbf{z}}$ | (8$f$) | Te III |
| $\mathbf{B}_{10}$ | = | $y_4\,\mathbf{a}_1 - y_4\,\mathbf{a}_2 + \left(\frac{1}{2} + z_4\right)\mathbf{a}_3$ | = | $-y_4 b\,\hat{\mathbf{y}} + \left(\frac{1}{2} + z_4\right) c\,\hat{\mathbf{z}}$ | (8$f$) | Te III |
| $\mathbf{B}_{11}$ | = | $-y_4\,\mathbf{a}_1 + y_4\,\mathbf{a}_2 + \left(\frac{1}{2} - z_4\right)\mathbf{a}_3$ | = | $y_4 b\,\hat{\mathbf{y}} + \left(\frac{1}{2} - z_4\right) c\,\hat{\mathbf{z}}$ | (8$f$) | Te III |
| $\mathbf{B}_{12}$ | = | $y_4\,\mathbf{a}_1 - y_4\,\mathbf{a}_2 - z_4\,\mathbf{a}_3$ | = | $-y_4 b\,\hat{\mathbf{y}} - z_4 c\,\hat{\mathbf{z}}$ | (8$f$) | Te III |

**Geometry files:**

- CIF: pp. 1656
- POSCAR: pp. 1656

# Lepidocrocite ($\gamma$-FeO(OH), $E0_4$) Structure: AB2C2_oC20_63_c_f_2c

http://aflow.org/prototype-encyclopedia/AB2C2_oC20_63_c_f_2c

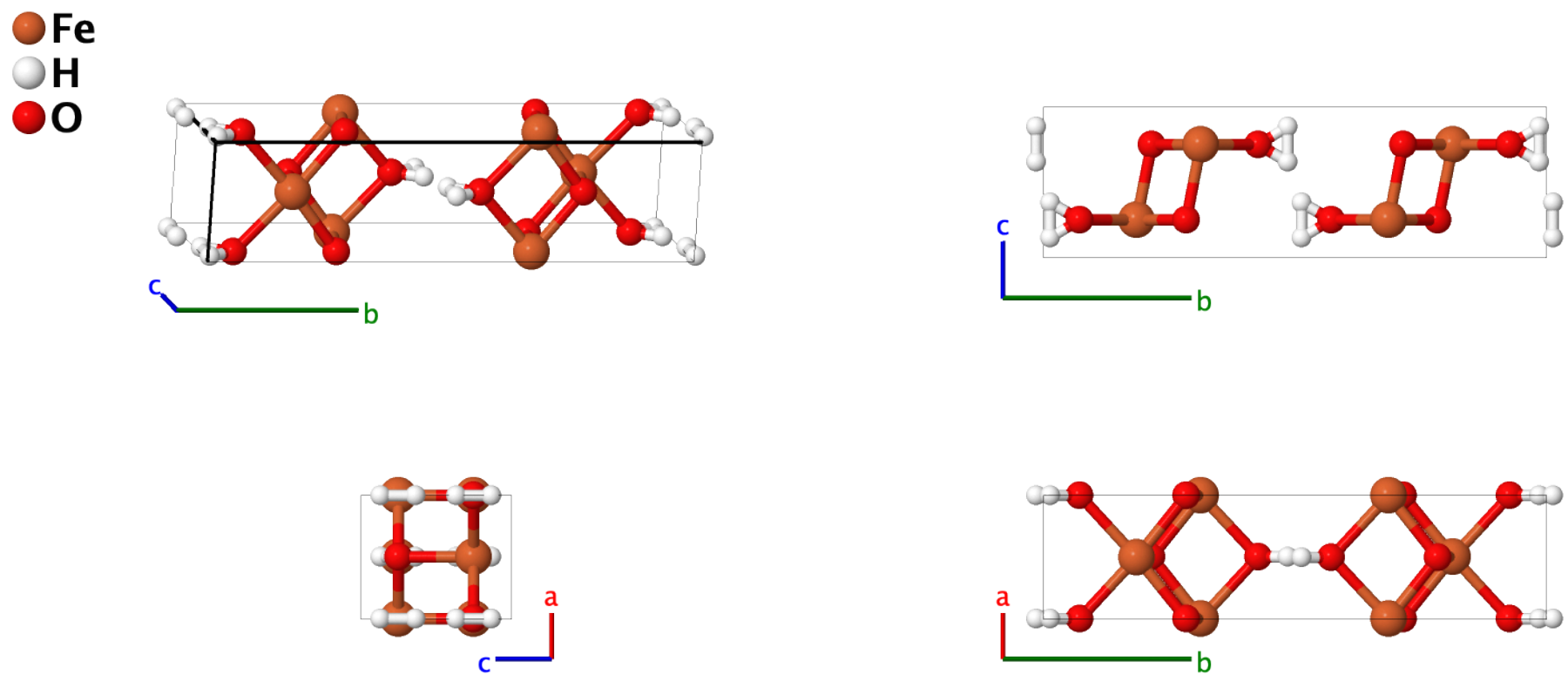

| | | |
|---|---|---|
| **Prototype** | : | $FeHO_2$ |
| **AFLOW prototype label** | : | AB2C2_oC20_63_c_f_2c |
| ***Strukturbericht* designation** | : | $E0_4$ |
| **Pearson symbol** | : | oC20 |
| **Space group number** | : | 63 |
| **Space group symbol** | : | $Cmcm$ |
| **AFLOW prototype command** | : | `aflow --proto=AB2C2_oC20_63_c_f_2c --params=`$a, b/a, c/a, y_1, y_2, y_3, y_4, z_4$ |

- (Gottfried, 1937) gave $\gamma$-FeO(OH) the *Strukturbericht* designation $E0_4$, but did not determine the positions of the hydrogens. Using a deuterated form of FeO(OH) (Christensen, 1982) found that the hydrogen atoms are located on the (8f) Wyckoff sites, but these sites are only 41.7% occupied. We use this structure as the prototype.

**Base-centered Orthorhombic primitive vectors:**

$$\mathbf{a}_1 = \frac{1}{2}\,a\,\hat{\mathbf{x}} - \frac{1}{2}\,b\,\hat{\mathbf{y}}$$
$$\mathbf{a}_2 = \frac{1}{2}\,a\,\hat{\mathbf{x}} + \frac{1}{2}\,b\,\hat{\mathbf{y}}$$
$$\mathbf{a}_3 = c\,\hat{\mathbf{z}}$$

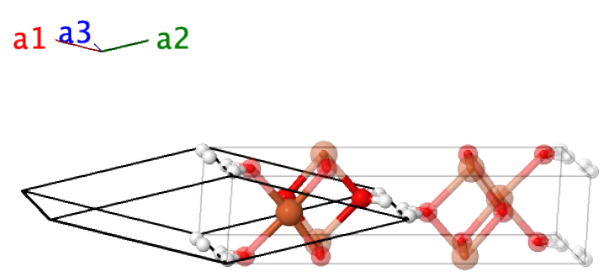

**Basis vectors:**

| | | Lattice Coordinates | | Cartesian Coordinates | Wyckoff Position | Atom Type |
|---|---|---|---|---|---|---|
| $\mathbf{B}_1$ | = | $-y_1\,\mathbf{a}_1 + y_1\,\mathbf{a}_2 + \frac{1}{4}\,\mathbf{a}_3$ | = | $y_1 b\,\hat{\mathbf{y}} + \frac{1}{4}c\,\hat{\mathbf{z}}$ | (4$c$) | Fe |
| $\mathbf{B}_2$ | = | $y_1\,\mathbf{a}_1 - y_1\,\mathbf{a}_2 + \frac{3}{4}\,\mathbf{a}_3$ | = | $-y_1 b\,\hat{\mathbf{y}} + \frac{3}{4}c\,\hat{\mathbf{z}}$ | (4$c$) | Fe |
| $\mathbf{B}_3$ | = | $-y_2\,\mathbf{a}_1 + y_2\,\mathbf{a}_2 + \frac{1}{4}\,\mathbf{a}_3$ | = | $y_2 b\,\hat{\mathbf{y}} + \frac{1}{4}c\,\hat{\mathbf{z}}$ | (4$c$) | O I |
| $\mathbf{B}_4$ | = | $y_2\,\mathbf{a}_1 - y_2\,\mathbf{a}_2 + \frac{3}{4}\,\mathbf{a}_3$ | = | $-y_2 b\,\hat{\mathbf{y}} + \frac{3}{4}c\,\hat{\mathbf{z}}$ | (4$c$) | O I |
| $\mathbf{B}_5$ | = | $-y_3\,\mathbf{a}_1 + y_3\,\mathbf{a}_2 + \frac{1}{4}\,\mathbf{a}_3$ | = | $y_3 b\,\hat{\mathbf{y}} + \frac{1}{4}c\,\hat{\mathbf{z}}$ | (4$c$) | O II |
| $\mathbf{B}_6$ | = | $y_3\,\mathbf{a}_1 - y_3\,\mathbf{a}_2 + \frac{3}{4}\,\mathbf{a}_3$ | = | $-y_3 b\,\hat{\mathbf{y}} + \frac{3}{4}c\,\hat{\mathbf{z}}$ | (4$c$) | O II |

$$\mathbf{B}_7 = -y_4\,\mathbf{a}_1 + y_4\,\mathbf{a}_2 + z_4\,\mathbf{a}_3 = y_4 b\,\hat{\mathbf{y}} + z_4 c\,\hat{\mathbf{z}} \qquad (8f) \qquad \text{H}$$

$$\mathbf{B}_8 = y_4\,\mathbf{a}_1 - y_4\,\mathbf{a}_2 + \left(\frac{1}{2} + z_4\right)\mathbf{a}_3 = -y_4 b\,\hat{\mathbf{y}} + \left(\frac{1}{2} + z_4\right) c\,\hat{\mathbf{z}} \qquad (8f) \qquad \text{H}$$

$$\mathbf{B}_9 = -y_4\,\mathbf{a}_1 + y_4\,\mathbf{a}_2 + \left(\frac{1}{2} - z_4\right)\mathbf{a}_3 = y_4 b\,\hat{\mathbf{y}} + \left(\frac{1}{2} - z_4\right) c\,\hat{\mathbf{z}} \qquad (8f) \qquad \text{H}$$

$$\mathbf{B}_{10} = y_4\,\mathbf{a}_1 - y_4\,\mathbf{a}_2 - z_4\,\mathbf{a}_3 = -y_4 b\,\hat{\mathbf{y}} - z_4 c\,\hat{\mathbf{z}} \qquad (8f) \qquad \text{H}$$

---

---

**Geometry files:**

- CIF: pp. 1656
- POSCAR: pp. 1657

# $Na_2CrO_4$ ($H1_8$) Structure: AB2C4_oC28_63_c_bc_fg

http://aflow.org/prototype-encyclopedia/AB2C4_oC28_63_c_bc_fg

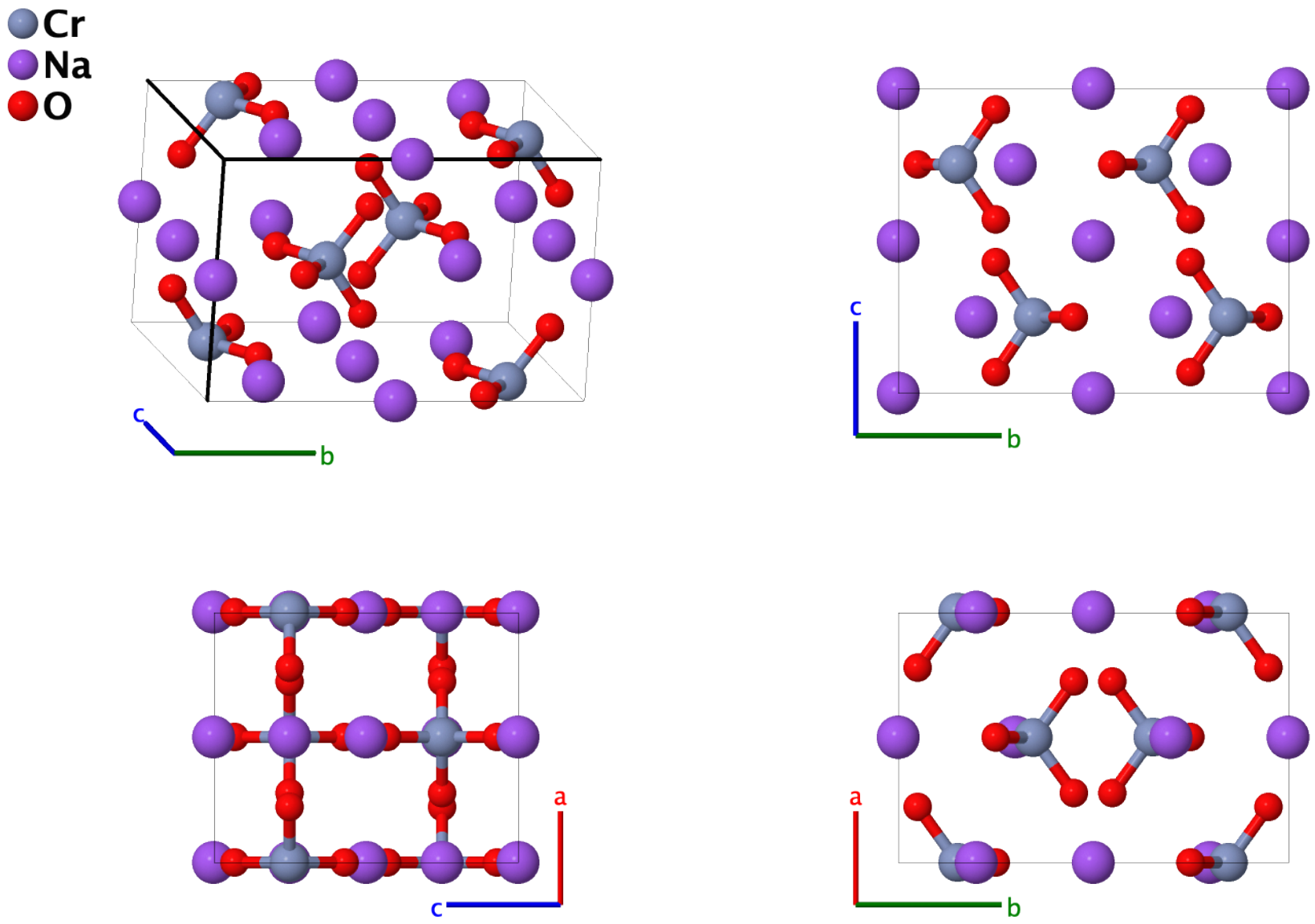

| | | |
|---|---|---|
| **Prototype** | : | $CrNa_2O_4$ |
| **AFLOW prototype label** | : | AB2C4_oC28_63_c_bc_fg |
| ***Strukturbericht* designation** | : | $H1_8$ |
| **Pearson symbol** | : | oC28 |
| **Space group number** | : | 63 |
| **Space group symbol** | : | *Cmcm* |
| **AFLOW prototype command** | : | `aflow --proto=AB2C4_oC28_63_c_bc_fg`<br>`--params=`$a, b/a, c/a, y_2, y_3, y_4, z_4, x_5, y_5$ |

### Other compounds with this structure

- $Li_2SO_4$, $LiFeP_4$, $Na_2FeO_4$, $Na_2SO_4$ (III), $NaCaVO_4$, $NaMnPO_4$, $NaVCdO_4$, and $Tl_2SeO_4$

---

- This structure was originally determined by (Miller, 1936), who placed it in space group *Pnna* #52, and (Gottfried, 1938) uses this data for $H1_8$. Subsequently (Niggli, 1954) rather acerbically pointed out that Miller's coordinates were consistent with the more compact *Cmcm* #63 space group. This does not change the positions of the atoms in the conventional cell, so we use the compact structure as our prototype for *Strukturbericht* designation $H1_8$.
- This structure is stable up to 413 °C. (Amirathanlingam, 1982)

### Base-centered Orthorhombic primitive vectors:

$$\mathbf{a}_1 = \frac{1}{2}\,a\,\hat{\mathbf{x}} - \frac{1}{2}\,b\,\hat{\mathbf{y}}$$
$$\mathbf{a}_2 = \frac{1}{2}\,a\,\hat{\mathbf{x}} + \frac{1}{2}\,b\,\hat{\mathbf{y}}$$
$$\mathbf{a}_3 = c\,\hat{\mathbf{z}}$$

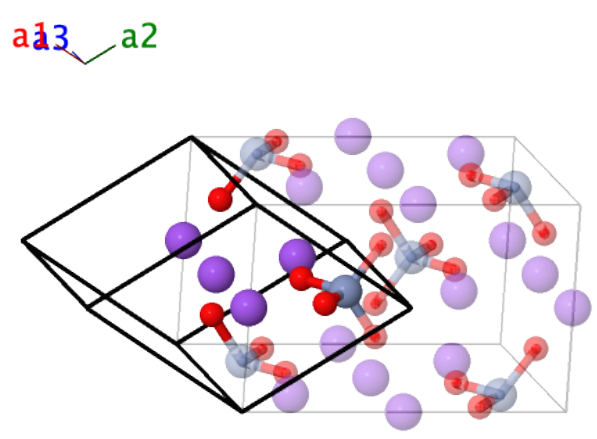

### Basis vectors:

| | | Lattice Coordinates | | Cartesian Coordinates | Wyckoff Position | Atom Type |
|---|---|---|---|---|---|---|
| $\mathbf{B}_1$ | = | $\frac{1}{2}\,\mathbf{a}_1 + \frac{1}{2}\,\mathbf{a}_2$ | = | $\frac{1}{2}a\,\hat{\mathbf{x}}$ | (4b) | Na I |
| $\mathbf{B}_2$ | = | $\frac{1}{2}\,\mathbf{a}_1 + \frac{1}{2}\,\mathbf{a}_2 + \frac{1}{2}\,\mathbf{a}_3$ | = | $\frac{1}{2}a\,\hat{\mathbf{x}} + \frac{1}{2}c\,\hat{\mathbf{z}}$ | (4b) | Na I |
| $\mathbf{B}_3$ | = | $-y_2\,\mathbf{a}_1 + y_2\,\mathbf{a}_2 + \frac{1}{4}\,\mathbf{a}_3$ | = | $y_2b\,\hat{\mathbf{y}} + \frac{1}{4}c\,\hat{\mathbf{z}}$ | (4c) | Cr |
| $\mathbf{B}_4$ | = | $y_2\,\mathbf{a}_1 - y_2\,\mathbf{a}_2 + \frac{3}{4}\,\mathbf{a}_3$ | = | $-y_2b\,\hat{\mathbf{y}} + \frac{3}{4}c\,\hat{\mathbf{z}}$ | (4c) | Cr |
| $\mathbf{B}_5$ | = | $-y_3\,\mathbf{a}_1 + y_3\,\mathbf{a}_2 + \frac{1}{4}\,\mathbf{a}_3$ | = | $y_3b\,\hat{\mathbf{y}} + \frac{1}{4}c\,\hat{\mathbf{z}}$ | (4c) | Na II |
| $\mathbf{B}_6$ | = | $y_3\,\mathbf{a}_1 - y_3\,\mathbf{a}_2 + \frac{3}{4}\,\mathbf{a}_3$ | = | $-y_3b\,\hat{\mathbf{y}} + \frac{3}{4}c\,\hat{\mathbf{z}}$ | (4c) | Na II |
| $\mathbf{B}_7$ | = | $-y_4\,\mathbf{a}_1 + y_4\,\mathbf{a}_2 + z_4\,\mathbf{a}_3$ | = | $y_4b\,\hat{\mathbf{y}} + z_4c\,\hat{\mathbf{z}}$ | (8f) | O I |
| $\mathbf{B}_8$ | = | $y_4\,\mathbf{a}_1 - y_4\,\mathbf{a}_2 + \left(\frac{1}{2} + z_4\right)\,\mathbf{a}_3$ | = | $-y_4b\,\hat{\mathbf{y}} + \left(\frac{1}{2} + z_4\right)c\,\hat{\mathbf{z}}$ | (8f) | O I |
| $\mathbf{B}_9$ | = | $-y_4\,\mathbf{a}_1 + y_4\,\mathbf{a}_2 + \left(\frac{1}{2} - z_4\right)\,\mathbf{a}_3$ | = | $y_4b\,\hat{\mathbf{y}} + \left(\frac{1}{2} - z_4\right)c\,\hat{\mathbf{z}}$ | (8f) | O I |
| $\mathbf{B}_{10}$ | = | $y_4\,\mathbf{a}_1 - y_4\,\mathbf{a}_2 - z_4\,\mathbf{a}_3$ | = | $-y_4b\,\hat{\mathbf{y}} - z_4c\,\hat{\mathbf{z}}$ | (8f) | O I |
| $\mathbf{B}_{11}$ | = | $(x_5 - y_5)\,\mathbf{a}_1 + (x_5 + y_5)\,\mathbf{a}_2 + \frac{1}{4}\,\mathbf{a}_3$ | = | $x_5a\,\hat{\mathbf{x}} + y_5b\,\hat{\mathbf{y}} + \frac{1}{4}c\,\hat{\mathbf{z}}$ | (8g) | O II |
| $\mathbf{B}_{12}$ | = | $(-x_5 + y_5)\,\mathbf{a}_1 + (-x_5 - y_5)\,\mathbf{a}_2 + \frac{3}{4}\,\mathbf{a}_3$ | = | $-x_5a\,\hat{\mathbf{x}} - y_5b\,\hat{\mathbf{y}} + \frac{3}{4}c\,\hat{\mathbf{z}}$ | (8g) | O II |
| $\mathbf{B}_{13}$ | = | $(-x_5 - y_5)\,\mathbf{a}_1 + (-x_5 + y_5)\,\mathbf{a}_2 + \frac{1}{4}\,\mathbf{a}_3$ | = | $-x_5a\,\hat{\mathbf{x}} + y_5b\,\hat{\mathbf{y}} + \frac{1}{4}c\,\hat{\mathbf{z}}$ | (8g) | O II |
| $\mathbf{B}_{14}$ | = | $(x_5 + y_5)\,\mathbf{a}_1 + (x_5 - y_5)\,\mathbf{a}_2 + \frac{3}{4}\,\mathbf{a}_3$ | = | $x_5a\,\hat{\mathbf{x}} - y_5b\,\hat{\mathbf{y}} + \frac{3}{4}c\,\hat{\mathbf{z}}$ | (8g) | O II |

**Geometry files:**

- CIF: pp. 1657
- POSCAR: pp. 1657

# $ThFe_2SiC$ Structure: AB2CD_oC20_63_b_f_c_c

http://aflow.org/prototype-encyclopedia/AB2CD_oC20_63_b_f_c_c

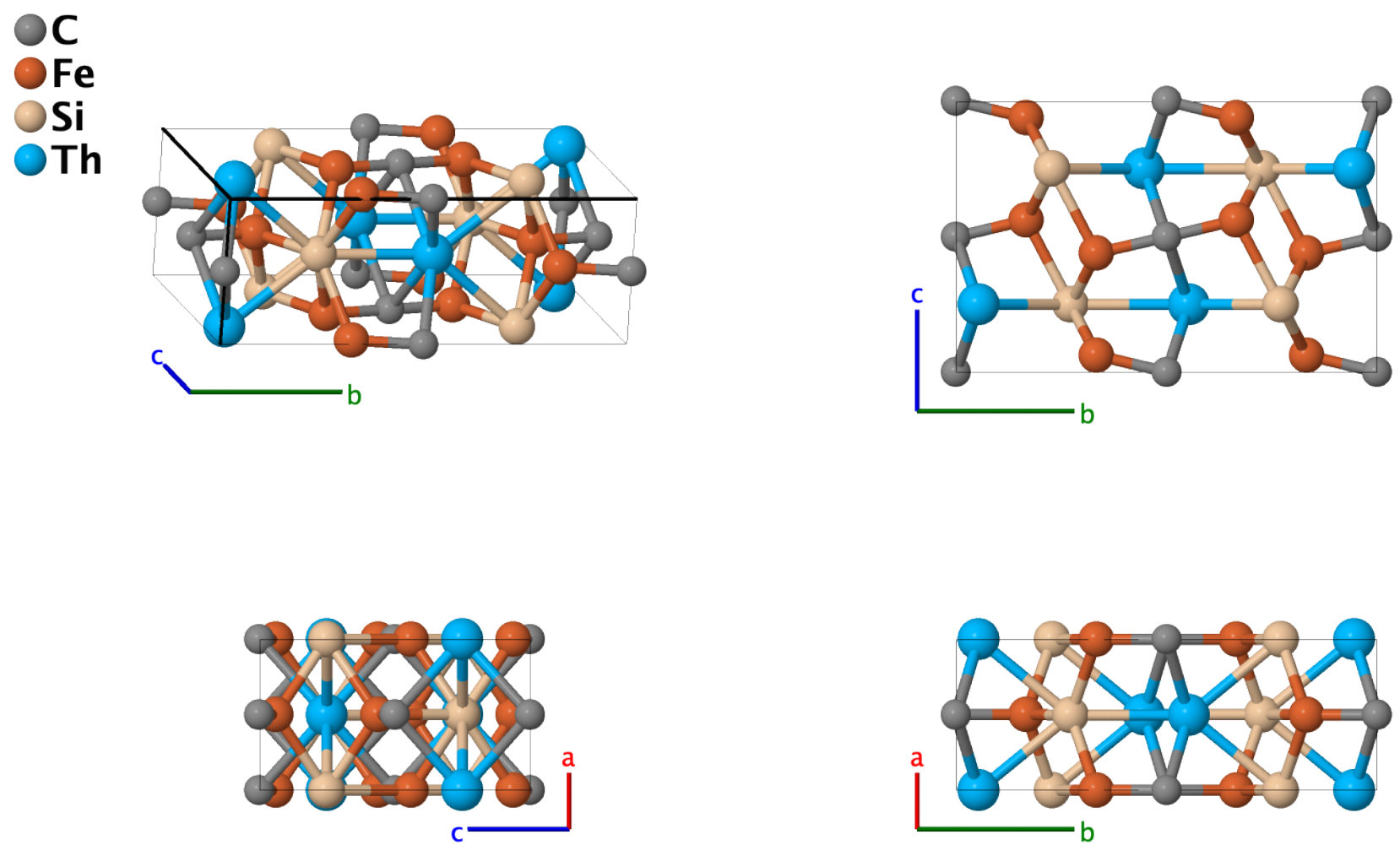

| | | |
|---|---|---|
| **Prototype** | : | $CFe_2SiTh$ |
| **AFLOW prototype label** | : | AB2CD_oC20_63_b_f_c_c |
| ***Strukturbericht* designation** | : | None |
| **Pearson symbol** | : | oC20 |
| **Space group number** | : | 63 |
| **Space group symbol** | : | *Cmcm* |
| **AFLOW prototype command** | : | `aflow --proto=AB2CD_oC20_63_b_f_c_c`<br>`--params=`$a, b/a, c/a, y_2, y_3, y_4, z_4$ |

**Other compounds with this structure**

- $A Fe_2SiC$, ($A$ = Y, Sm, Gd, Tb, Ho, Er, Tm, Lu, U

---

- This structure is a "filled" version of the $Re_3B$ structure, with carbon atoms sitting in the (4*b*) Wyckoff positions. This is the quaternary version of the structure. The ternary version, where one of the (4*c*) Wyckoff positions has the same atom type as the (8*f*) site, is designated the $V_3AsC$ structure.
- (Witte, 1994) does not designate a prototype for this structure, but as they only give full crystallographic data for the $ThFe_2SiC$ structure we will use this as the prototype.
- (Witte, 1994) lists the occupation of each site as Th (100.2%), Fe (97.7%), Si (96.0%), and C (97.0%), indicating that there are vacancies on the Fe, Si, and C sites, and that some of the Th atoms are either interstitial or occuping the other (4*c*) site or the (8*f*) site.

**Base-centered Orthorhombic primitive vectors:**

$$\mathbf{a}_1 = \frac{1}{2} a \hat{\mathbf{x}} - \frac{1}{2} b \hat{\mathbf{y}}$$
$$\mathbf{a}_2 = \frac{1}{2} a \hat{\mathbf{x}} + \frac{1}{2} b \hat{\mathbf{y}}$$
$$\mathbf{a}_3 = c \hat{\mathbf{z}}$$

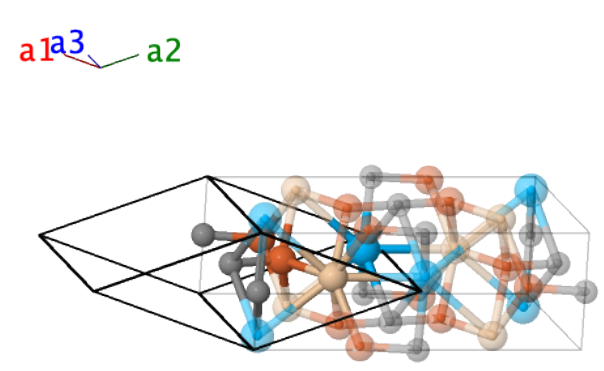

**Basis vectors:**

| | | Lattice Coordinates | | Cartesian Coordinates | Wyckoff Position | Atom Type |
|---|---|---|---|---|---|---|
| $\mathbf{B}_1$ | = | $\frac{1}{2}\,\mathbf{a}_1 + \frac{1}{2}\,\mathbf{a}_2$ | = | $\frac{1}{2}a\,\hat{\mathbf{x}}$ | (4$b$) | C |
| $\mathbf{B}_2$ | = | $\frac{1}{2}\,\mathbf{a}_1 + \frac{1}{2}\,\mathbf{a}_2 + \frac{1}{2}\,\mathbf{a}_3$ | = | $\frac{1}{2}a\,\hat{\mathbf{x}} + \frac{1}{2}c\,\hat{\mathbf{z}}$ | (4$b$) | C |
| $\mathbf{B}_3$ | = | $-y_2\,\mathbf{a}_1 + y_2\,\mathbf{a}_2 + \frac{1}{4}\,\mathbf{a}_3$ | = | $y_2b\,\hat{\mathbf{y}} + \frac{1}{4}c\,\hat{\mathbf{z}}$ | (4$c$) | Si |
| $\mathbf{B}_4$ | = | $y_2\,\mathbf{a}_1 - y_2\,\mathbf{a}_2 + \frac{3}{4}\,\mathbf{a}_3$ | = | $-y_2b\,\hat{\mathbf{y}} + \frac{3}{4}c\,\hat{\mathbf{z}}$ | (4$c$) | Si |
| $\mathbf{B}_5$ | = | $-y_3\,\mathbf{a}_1 + y_3\,\mathbf{a}_2 + \frac{1}{4}\,\mathbf{a}_3$ | = | $y_3b\,\hat{\mathbf{y}} + \frac{1}{4}c\,\hat{\mathbf{z}}$ | (4$c$) | Th |
| $\mathbf{B}_6$ | = | $y_3\,\mathbf{a}_1 - y_3\,\mathbf{a}_2 + \frac{3}{4}\,\mathbf{a}_3$ | = | $-y_3b\,\hat{\mathbf{y}} + \frac{3}{4}c\,\hat{\mathbf{z}}$ | (4$c$) | Th |
| $\mathbf{B}_7$ | = | $-y_4\,\mathbf{a}_1 + y_4\,\mathbf{a}_2 + z_4\,\mathbf{a}_3$ | = | $y_4b\,\hat{\mathbf{y}} + z_4c\,\hat{\mathbf{z}}$ | (8$f$) | Fe |
| $\mathbf{B}_8$ | = | $y_4\,\mathbf{a}_1 - y_4\,\mathbf{a}_2 + \left(\frac{1}{2} + z_4\right)\,\mathbf{a}_3$ | = | $-y_4b\,\hat{\mathbf{y}} + \left(\frac{1}{2} + z_4\right)c\,\hat{\mathbf{z}}$ | (8$f$) | Fe |
| $\mathbf{B}_9$ | = | $-y_4\,\mathbf{a}_1 + y_4\,\mathbf{a}_2 + \left(\frac{1}{2} - z_4\right)\,\mathbf{a}_3$ | = | $y_4b\,\hat{\mathbf{y}} + \left(\frac{1}{2} - z_4\right)c\,\hat{\mathbf{z}}$ | (8$f$) | Fe |
| $\mathbf{B}_{10}$ | = | $y_4\,\mathbf{a}_1 - y_4\,\mathbf{a}_2 - z_4\,\mathbf{a}_3$ | = | $-y_4b\,\hat{\mathbf{y}} - z_4c\,\hat{\mathbf{z}}$ | (8$f$) | Fe |

**Geometry files:**

- CIF: pp. 1657
- POSCAR: pp. 1658

# $Mn_3As$ ($D0_d$) Structure: AB3_oC16_63_c_3c

http://aflow.org/prototype-encyclopedia/AB3_oC16_63_c_3c

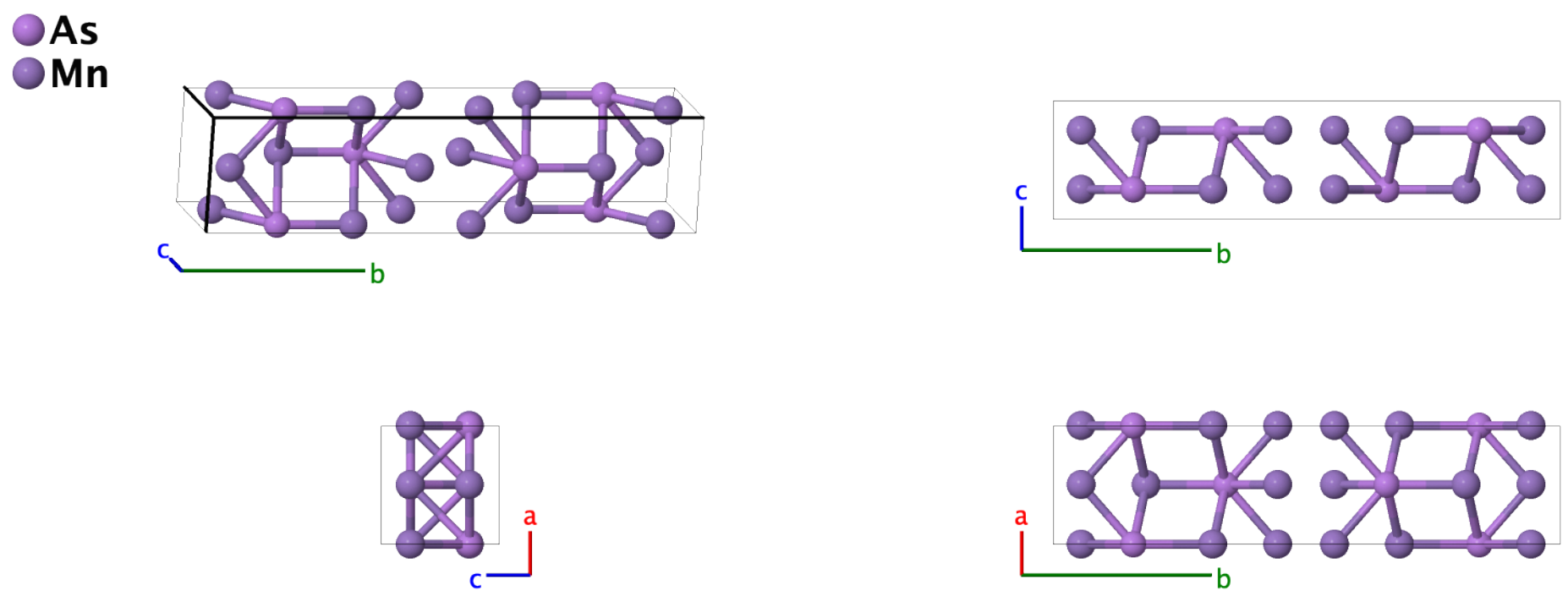

| | | |
|---|---|---|
| **Prototype** | : | $AsMn_3$ |
| **AFLOW prototype label** | : | AB3_oC16_63_c_3c |
| ***Strukturbericht* designation** | : | $D0_d$ |
| **Pearson symbol** | : | oC16 |
| **Space group number** | : | 63 |
| **Space group symbol** | : | *Cmcm* |
| **AFLOW prototype command** | : | `aflow --proto=AB3_oC16_63_c_3c` `--params=`$a, b/a, c/a, y_1, y_2, y_3, y_4$ |

### Other compounds with this structure

- $Te_3Nd$, $DyGe_3$

- (Nowotny, 1951) set the structure of $Mn_3As$ in space group *Pmmn* #59. This was repeated by (Villars, 1991) and (Brandes, 1992). However, (Carrillo-Cabrera, 1983) showed that the structure actually reduces to space group *Cmcm* #63, and this was recognized by (Parthé, 1993). We follow that latter two works and assign the $D0_d$ structure to space group *Cmcm*.
- (Carrillo-Cabrera, 1983) placed the structure in setting *Bmmb* of space group #63, but we have shifted it to the standard *Cmcm* setting.

### Base-centered Orthorhombic primitive vectors:

$$\mathbf{a}_1 = \frac{1}{2}\,a\,\hat{\mathbf{x}} - \frac{1}{2}\,b\,\hat{\mathbf{y}}$$
$$\mathbf{a}_2 = \frac{1}{2}\,a\,\hat{\mathbf{x}} + \frac{1}{2}\,b\,\hat{\mathbf{y}}$$
$$\mathbf{a}_3 = c\,\hat{\mathbf{z}}$$

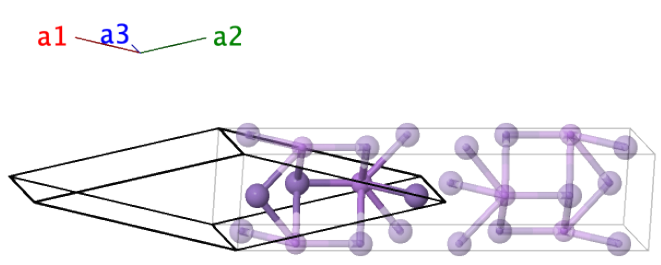

### Basis vectors:

| | | Lattice Coordinates | | Cartesian Coordinates | Wyckoff Position | Atom Type |
|---|---|---|---|---|---|---|
| $\mathbf{B}_1$ | = | $-y_1\,\mathbf{a}_1 + y_1\,\mathbf{a}_2 + \frac{1}{4}\,\mathbf{a}_3$ | = | $y_1 b\,\hat{\mathbf{y}} + \frac{1}{4}c\,\hat{\mathbf{z}}$ | (4*c*) | As |
| $\mathbf{B}_2$ | = | $y_1\,\mathbf{a}_1 - y_1\,\mathbf{a}_2 + \frac{3}{4}\,\mathbf{a}_3$ | = | $-y_1 b\,\hat{\mathbf{y}} + \frac{3}{4}c\,\hat{\mathbf{z}}$ | (4*c*) | As |
| $\mathbf{B}_3$ | = | $-y_2\,\mathbf{a}_1 + y_2\,\mathbf{a}_2 + \frac{1}{4}\,\mathbf{a}_3$ | = | $y_2 b\,\hat{\mathbf{y}} + \frac{1}{4}c\,\hat{\mathbf{z}}$ | (4*c*) | Mn I |
| $\mathbf{B}_4$ | = | $y_2\,\mathbf{a}_1 - y_2\,\mathbf{a}_2 + \frac{3}{4}\,\mathbf{a}_3$ | = | $-y_2 b\,\hat{\mathbf{y}} + \frac{3}{4}c\,\hat{\mathbf{z}}$ | (4*c*) | Mn I |
| $\mathbf{B}_5$ | = | $-y_3\,\mathbf{a}_1 + y_3\,\mathbf{a}_2 + \frac{1}{4}\,\mathbf{a}_3$ | = | $y_3 b\,\hat{\mathbf{y}} + \frac{1}{4} c\,\hat{\mathbf{z}}$ | (4$c$) | Mn II |
| $\mathbf{B}_6$ | = | $y_3\,\mathbf{a}_1 - y_3\,\mathbf{a}_2 + \frac{3}{4}\,\mathbf{a}_3$ | = | $-y_3 b\,\hat{\mathbf{y}} + \frac{3}{4} c\,\hat{\mathbf{z}}$ | (4$c$) | Mn II |
| $\mathbf{B}_7$ | = | $-y_4\,\mathbf{a}_1 + y_4\,\mathbf{a}_2 + \frac{1}{4}\,\mathbf{a}_3$ | = | $y_4 b\,\hat{\mathbf{y}} + \frac{1}{4} c\,\hat{\mathbf{z}}$ | (4$c$) | Mn III |
| $\mathbf{B}_8$ | = | $y_4\,\mathbf{a}_1 - y_4\,\mathbf{a}_2 + \frac{3}{4}\,\mathbf{a}_3$ | = | $-y_4 b\,\hat{\mathbf{y}} + \frac{3}{4} c\,\hat{\mathbf{z}}$ | (4$c$) | Mn III |

**Geometry files:**

- CIF: pp. 1658
- POSCAR: pp. 1658

# $Re_3B$ Structure: AB3_oC16_63_c_cf

http://aflow.org/prototype-encyclopedia/AB3_oC16_63_c_cf

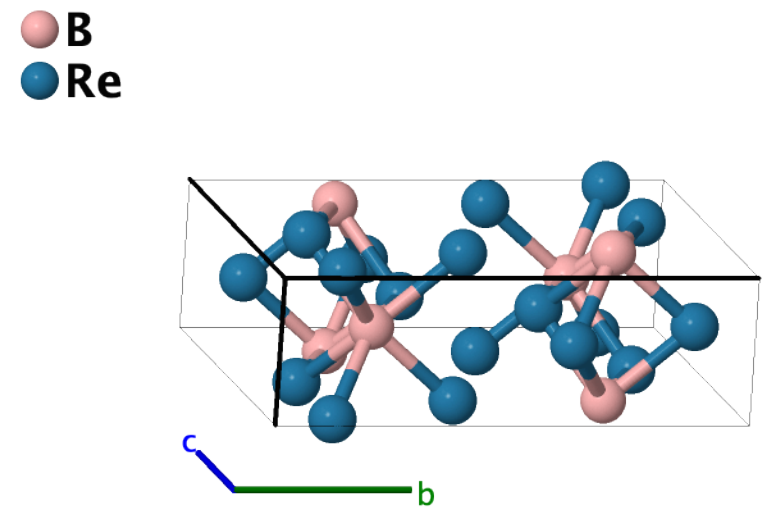

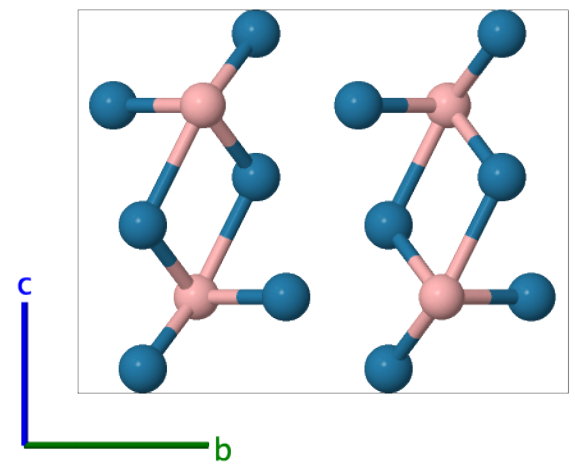

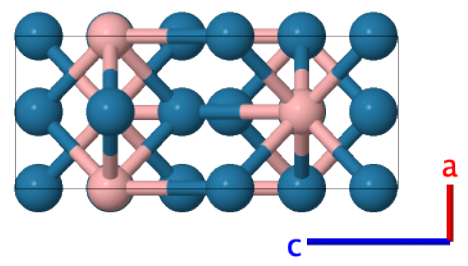

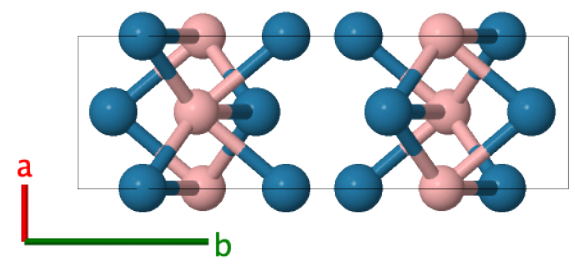

| | | |
|---|---|---|
| **Prototype** | : | $BRe_3$ |
| **AFLOW prototype label** | : | AB3_oC16_63_c_cf |
| ***Strukturbericht* designation** | : | None |
| **Pearson symbol** | : | oC16 |
| **Space group number** | : | 63 |
| **Space group symbol** | : | *Cmcm* |
| **AFLOW prototype command** | : | `aflow --proto=AB3_oC16_63_c_cf`<br>`--params=`$a, b/a, c/a, y_1, y_2, y_3, z_3$ |

- This is the parent structure of the ternary $MgCuAl_2$ ($E1_a$) structure. There are also “filled” versions of this structure, $V_3AsC$ and $ThFe_2SiC$.

**Base-centered Orthorhombic primitive vectors:**

$$\mathbf{a}_1 = \frac{1}{2}\,a\,\hat{\mathbf{x}} - \frac{1}{2}\,b\,\hat{\mathbf{y}}$$
$$\mathbf{a}_2 = \frac{1}{2}\,a\,\hat{\mathbf{x}} + \frac{1}{2}\,b\,\hat{\mathbf{y}}$$
$$\mathbf{a}_3 = c\,\hat{\mathbf{z}}$$

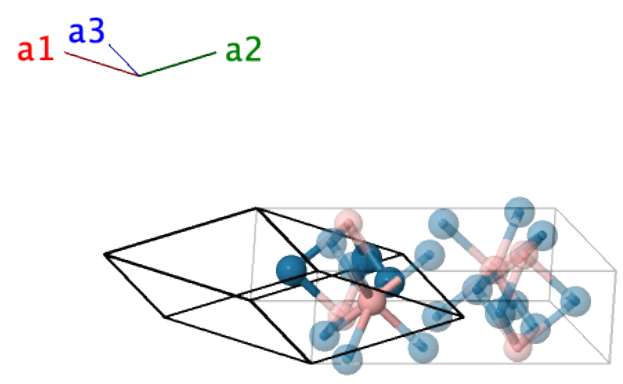

**Basis vectors:**

| | | Lattice Coordinates | | Cartesian Coordinates | Wyckoff Position | Atom Type |
|---|---|---|---|---|---|---|
| $\mathbf{B}_1$ | = | $-y_1\,\mathbf{a}_1 + y_1\,\mathbf{a}_2 + \frac{1}{4}\,\mathbf{a}_3$ | = | $y_1 b\,\hat{\mathbf{y}} + \frac{1}{4} c\,\hat{\mathbf{z}}$ | (4$c$) | B |
| $\mathbf{B}_2$ | = | $y_1\,\mathbf{a}_1 - y_1\,\mathbf{a}_2 + \frac{3}{4}\,\mathbf{a}_3$ | = | $-y_1 b\,\hat{\mathbf{y}} + \frac{3}{4} c\,\hat{\mathbf{z}}$ | (4$c$) | B |
| $\mathbf{B}_3$ | = | $-y_2\,\mathbf{a}_1 + y_2\,\mathbf{a}_2 + \frac{1}{4}\,\mathbf{a}_3$ | = | $y_2 b\,\hat{\mathbf{y}} + \frac{1}{4} c\,\hat{\mathbf{z}}$ | (4$c$) | Re I |
| $\mathbf{B}_4$ | = | $y_2\,\mathbf{a}_1 - y_2\,\mathbf{a}_2 + \frac{3}{4}\,\mathbf{a}_3$ | = | $-y_2 b\,\hat{\mathbf{y}} + \frac{3}{4} c\,\hat{\mathbf{z}}$ | (4$c$) | Re I |
| $\mathbf{B}_5$ | = | $-y_3\,\mathbf{a}_1 + y_3\,\mathbf{a}_2 + z_3\,\mathbf{a}_3$ | = | $y_3 b\,\hat{\mathbf{y}} + z_3 c\,\hat{\mathbf{z}}$ | (8$f$) | Re II |
| $\mathbf{B}_6$ | = | $y_3\,\mathbf{a}_1 - y_3\,\mathbf{a}_2 + \left(\frac{1}{2} + z_3\right)\,\mathbf{a}_3$ | = | $-y_3 b\,\hat{\mathbf{y}} + \left(\frac{1}{2} + z_3\right) c\,\hat{\mathbf{z}}$ | (8$f$) | Re II |
| $\mathbf{B}_7$ | = | $-y_3\,\mathbf{a}_1 + y_3\,\mathbf{a}_2 + \left(\frac{1}{2} - z_3\right)\,\mathbf{a}_3$ | = | $y_3 b\,\hat{\mathbf{y}} + \left(\frac{1}{2} - z_3\right) c\,\hat{\mathbf{z}}$ | (8$f$) | Re II |
| $\mathbf{B}_8$ | = | $y_3\,\mathbf{a}_1 - y_3\,\mathbf{a}_2 - z_3\,\mathbf{a}_3$ | = | $-y_3 b\,\hat{\mathbf{y}} - z_3 c\,\hat{\mathbf{z}}$ | (8$f$) | Re II |

**Geometry files:**

- CIF: pp. 1658
- POSCAR: pp. 1659

# $Ta_2NiS_5$ Structure: AB5C2_oC32_63_c_c2f_f

http://aflow.org/prototype-encyclopedia/AB5C2_oC32_63_c_c2f_f

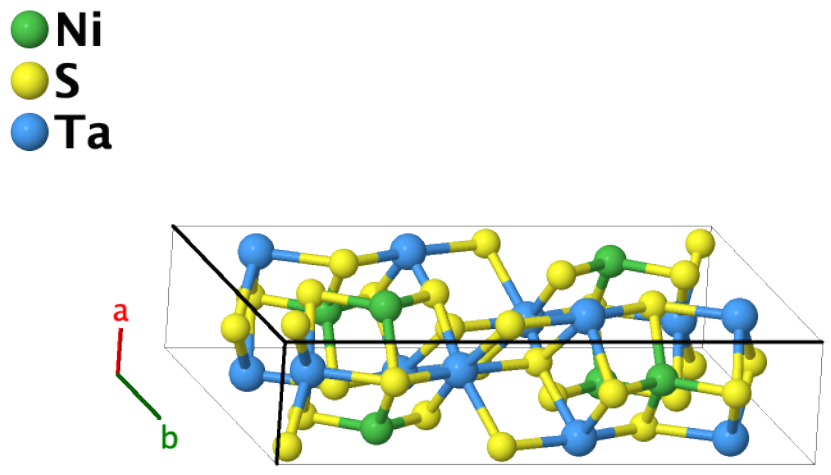

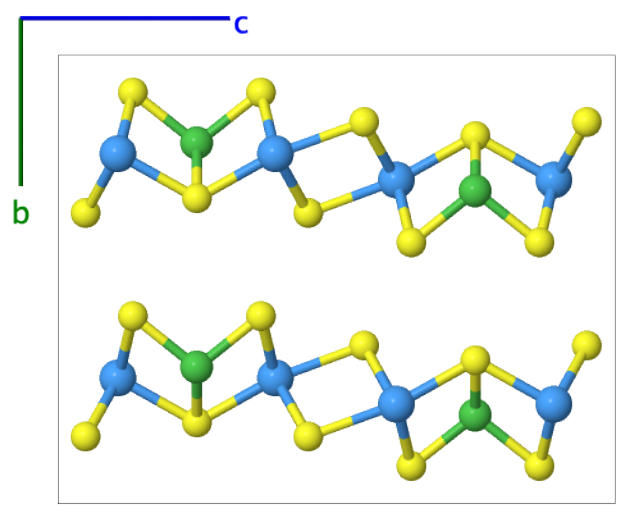

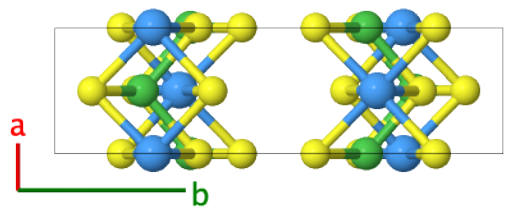

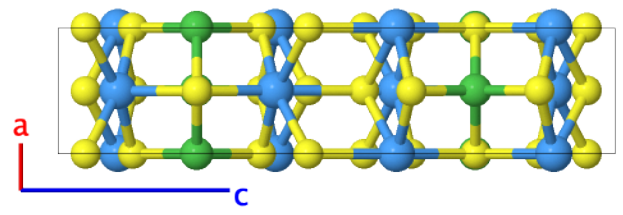

| | | |
|---|---|---|
| **Prototype** | : | $NiS_5Ta_2$ |
| **AFLOW prototype label** | : | AB5C2_oC32_63_c_c2f_f |
| ***Strukturbericht* designation** | : | None |
| **Pearson symbol** | : | oC32 |
| **Space group number** | : | 63 |
| **Space group symbol** | : | *Cmcm* |
| **AFLOW prototype command** | : | `aflow --proto=AB5C2_oC32_63_c_c2f_f` <br> `--params=`$a, b/a, c/a, y_1, y_2, y_3, z_3, y_4, z_4, y_5, z_5$ |

### Other compounds with this structure

- $Ta_2NiSe_5$ (above 328 K)

### Base-centered Orthorhombic primitive vectors:

$$\mathbf{a}_1 = \tfrac{1}{2}\,a\,\hat{\mathbf{x}} - \tfrac{1}{2}\,b\,\hat{\mathbf{y}}$$
$$\mathbf{a}_2 = \tfrac{1}{2}\,a\,\hat{\mathbf{x}} + \tfrac{1}{2}\,b\,\hat{\mathbf{y}}$$
$$\mathbf{a}_3 = c\,\hat{\mathbf{z}}$$

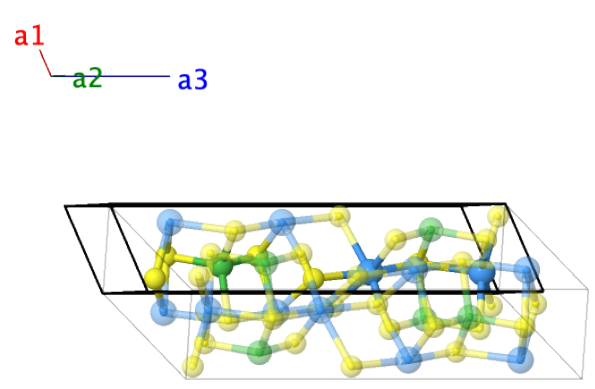

### Basis vectors:

| | | Lattice Coordinates | | Cartesian Coordinates | Wyckoff Position | Atom Type |
|---|---|---|---|---|---|---|
| $\mathbf{B}_1$ | = | $-y_1\,\mathbf{a}_1 + y_1\,\mathbf{a}_2 + \frac{1}{4}\,\mathbf{a}_3$ | = | $y_1 b\,\hat{\mathbf{y}} + \frac{1}{4} c\,\hat{\mathbf{z}}$ | (4*c*) | Ni |
| $\mathbf{B}_2$ | = | $y_1\,\mathbf{a}_1 - y_1\,\mathbf{a}_2 + \frac{3}{4}\,\mathbf{a}_3$ | = | $-y_1 b\,\hat{\mathbf{y}} + \frac{3}{4} c\,\hat{\mathbf{z}}$ | (4$c$) | Ni |
| $\mathbf{B}_3$ | = | $-y_2\,\mathbf{a}_1 + y_2\,\mathbf{a}_2 + \frac{1}{4}\,\mathbf{a}_3$ | = | $y_2 b\,\hat{\mathbf{y}} + \frac{1}{4} c\,\hat{\mathbf{z}}$ | (4$c$) | S I |
| $\mathbf{B}_4$ | = | $y_2\,\mathbf{a}_1 - y_2\,\mathbf{a}_2 + \frac{3}{4}\,\mathbf{a}_3$ | = | $-y_2 b\,\hat{\mathbf{y}} + \frac{3}{4} c\,\hat{\mathbf{z}}$ | (4$c$) | S I |
| $\mathbf{B}_5$ | = | $-y_3\,\mathbf{a}_1 + y_3\,\mathbf{a}_2 + z_3\,\mathbf{a}_3$ | = | $y_3 b\,\hat{\mathbf{y}} + z_3 c\,\hat{\mathbf{z}}$ | (8$f$) | S II |
| $\mathbf{B}_6$ | = | $y_3\,\mathbf{a}_1 - y_3\,\mathbf{a}_2 + \left(\frac{1}{2} + z_3\right)\,\mathbf{a}_3$ | = | $-y_3 b\,\hat{\mathbf{y}} + \left(\frac{1}{2} + z_3\right) c\,\hat{\mathbf{z}}$ | (8$f$) | S II |
| $\mathbf{B}_7$ | = | $-y_3\,\mathbf{a}_1 + y_3\,\mathbf{a}_2 + \left(\frac{1}{2} - z_3\right)\,\mathbf{a}_3$ | = | $y_3 b\,\hat{\mathbf{y}} + \left(\frac{1}{2} - z_3\right) c\,\hat{\mathbf{z}}$ | (8$f$) | S II |
| $\mathbf{B}_8$ | = | $y_3\,\mathbf{a}_1 - y_3\,\mathbf{a}_2 - z_3\,\mathbf{a}_3$ | = | $-y_3 b\,\hat{\mathbf{y}} - z_3 c\,\hat{\mathbf{z}}$ | (8$f$) | S II |
| $\mathbf{B}_9$ | = | $-y_4\,\mathbf{a}_1 + y_4\,\mathbf{a}_2 + z_4\,\mathbf{a}_3$ | = | $y_4 b\,\hat{\mathbf{y}} + z_4 c\,\hat{\mathbf{z}}$ | (8$f$) | S III |
| $\mathbf{B}_{10}$ | = | $y_4\,\mathbf{a}_1 - y_4\,\mathbf{a}_2 + \left(\frac{1}{2} + z_4\right)\,\mathbf{a}_3$ | = | $-y_4 b\,\hat{\mathbf{y}} + \left(\frac{1}{2} + z_4\right) c\,\hat{\mathbf{z}}$ | (8$f$) | S III |
| $\mathbf{B}_{11}$ | = | $-y_4\,\mathbf{a}_1 + y_4\,\mathbf{a}_2 + \left(\frac{1}{2} - z_4\right)\,\mathbf{a}_3$ | = | $y_4 b\,\hat{\mathbf{y}} + \left(\frac{1}{2} - z_4\right) c\,\hat{\mathbf{z}}$ | (8$f$) | S III |
| $\mathbf{B}_{12}$ | = | $y_4\,\mathbf{a}_1 - y_4\,\mathbf{a}_2 - z_4\,\mathbf{a}_3$ | = | $-y_4 b\,\hat{\mathbf{y}} - z_4 c\,\hat{\mathbf{z}}$ | (8$f$) | S III |
| $\mathbf{B}_{13}$ | = | $-y_5\,\mathbf{a}_1 + y_5\,\mathbf{a}_2 + z_5\,\mathbf{a}_3$ | = | $y_5 b\,\hat{\mathbf{y}} + z_5 c\,\hat{\mathbf{z}}$ | (8$f$) | Ta |
| $\mathbf{B}_{14}$ | = | $y_5\,\mathbf{a}_1 - y_5\,\mathbf{a}_2 + \left(\frac{1}{2} + z_5\right)\,\mathbf{a}_3$ | = | $-y_5 b\,\hat{\mathbf{y}} + \left(\frac{1}{2} + z_5\right) c\,\hat{\mathbf{z}}$ | (8$f$) | Ta |
| $\mathbf{B}_{15}$ | = | $-y_5\,\mathbf{a}_1 + y_5\,\mathbf{a}_2 + \left(\frac{1}{2} - z_5\right)\,\mathbf{a}_3$ | = | $y_5 b\,\hat{\mathbf{y}} + \left(\frac{1}{2} - z_5\right) c\,\hat{\mathbf{z}}$ | (8$f$) | Ta |
| $\mathbf{B}_{16}$ | = | $y_5\,\mathbf{a}_1 - y_5\,\mathbf{a}_2 - z_5\,\mathbf{a}_3$ | = | $-y_5 b\,\hat{\mathbf{y}} - z_5 c\,\hat{\mathbf{z}}$ | (8$f$) | Ta |

**Geometry files:**

- CIF: pp. 1659
- POSCAR: pp. 1659

# $V_3AsC$ Structure: ABC3_oC20_63_c_b_cf

http://aflow.org/prototype-encyclopedia/ABC3_oC20_63_c_b_cf

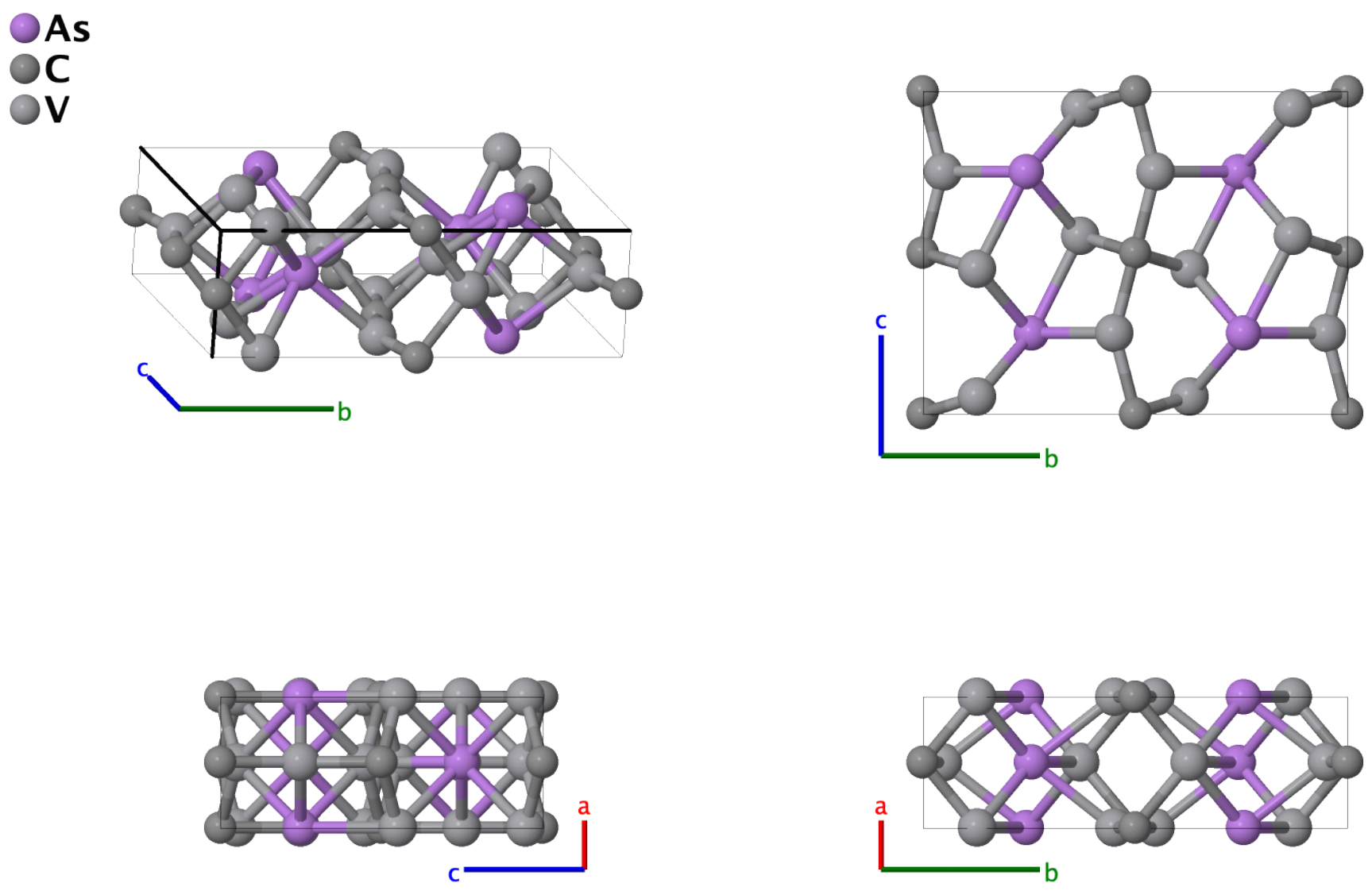

| | | |
|---|---|---|
| **Prototype** | : | $AsCV_3$ |
| **AFLOW prototype label** | : | ABC3_oC20_63_c_b_cf |
| ***Strukturbericht* designation** | : | None |
| **Pearson symbol** | : | oC20 |
| **Space group number** | : | 63 |
| **Space group symbol** | : | *Cmcm* |
| **AFLOW prototype command** | : | `aflow --proto=ABC3_oC20_63_c_b_cf --params=`$a, b/a, c/a, y_2, y_3, y_4, z_4$ |

**Other compounds with this structure**

- $V_3PC$, $V_3PN$, $V_3AsN$, $Cr_3PC$, $Cr_3AsC$, $Zr_3AlN$, and $UScS_3$

- This structure is a "filled" version of the $Re_3B$ structure, with carbon atoms sitting in the (4*b*) Wyckoff positions. This is the ternary version of the structure. The quaternary version, where all Wyckoff positions contain different species of atoms, is listed as the $ThFe_2SiC$ structure.

**Base-centered Orthorhombic primitive vectors:**

$$\mathbf{a}_1 = \frac{1}{2}\,a\,\hat{\mathbf{x}} - \frac{1}{2}\,b\,\hat{\mathbf{y}}$$
$$\mathbf{a}_2 = \frac{1}{2}\,a\,\hat{\mathbf{x}} + \frac{1}{2}\,b\,\hat{\mathbf{y}}$$
$$\mathbf{a}_3 = c\,\hat{\mathbf{z}}$$

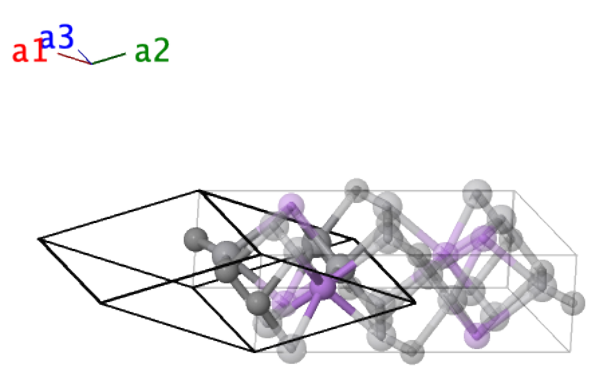

**Basis vectors:**

| | | Lattice Coordinates | | Cartesian Coordinates | Wyckoff Position | Atom Type |
|---|---|---|---|---|---|---|
| $\mathbf{B}_1$ | = | $\frac{1}{2}\,\mathbf{a}_1 + \frac{1}{2}\,\mathbf{a}_2$ | = | $\frac{1}{2}a\,\hat{\mathbf{x}}$ | (4$b$) | C |
| $\mathbf{B}_2$ | = | $\frac{1}{2}\,\mathbf{a}_1 + \frac{1}{2}\,\mathbf{a}_2 + \frac{1}{2}\,\mathbf{a}_3$ | = | $\frac{1}{2}a\,\hat{\mathbf{x}} + \frac{1}{2}c\,\hat{\mathbf{z}}$ | (4$b$) | C |
| $\mathbf{B}_3$ | = | $-y_2\,\mathbf{a}_1 + y_2\,\mathbf{a}_2 + \frac{1}{4}\,\mathbf{a}_3$ | = | $y_2b\,\hat{\mathbf{y}} + \frac{1}{4}c\,\hat{\mathbf{z}}$ | (4$c$) | As |
| $\mathbf{B}_4$ | = | $y_2\,\mathbf{a}_1 - y_2\,\mathbf{a}_2 + \frac{3}{4}\,\mathbf{a}_3$ | = | $-y_2b\,\hat{\mathbf{y}} + \frac{3}{4}c\,\hat{\mathbf{z}}$ | (4$c$) | As |
| $\mathbf{B}_5$ | = | $-y_3\,\mathbf{a}_1 + y_3\,\mathbf{a}_2 + \frac{1}{4}\,\mathbf{a}_3$ | = | $y_3b\,\hat{\mathbf{y}} + \frac{1}{4}c\,\hat{\mathbf{z}}$ | (4$c$) | V I |
| $\mathbf{B}_6$ | = | $y_3\,\mathbf{a}_1 - y_3\,\mathbf{a}_2 + \frac{3}{4}\,\mathbf{a}_3$ | = | $-y_3b\,\hat{\mathbf{y}} + \frac{3}{4}c\,\hat{\mathbf{z}}$ | (4$c$) | V I |
| $\mathbf{B}_7$ | = | $-y_4\,\mathbf{a}_1 + y_4\,\mathbf{a}_2 + z_4\,\mathbf{a}_3$ | = | $y_4b\,\hat{\mathbf{y}} + z_4c\,\hat{\mathbf{z}}$ | (8$f$) | V II |
| $\mathbf{B}_8$ | = | $y_4\,\mathbf{a}_1 - y_4\,\mathbf{a}_2 + \left(\frac{1}{2} + z_4\right)\,\mathbf{a}_3$ | = | $-y_4b\,\hat{\mathbf{y}} + \left(\frac{1}{2} + z_4\right)c\,\hat{\mathbf{z}}$ | (8$f$) | V II |
| $\mathbf{B}_9$ | = | $-y_4\,\mathbf{a}_1 + y_4\,\mathbf{a}_2 + \left(\frac{1}{2} - z_4\right)\,\mathbf{a}_3$ | = | $y_4b\,\hat{\mathbf{y}} + \left(\frac{1}{2} - z_4\right)c\,\hat{\mathbf{z}}$ | (8$f$) | V II |
| $\mathbf{B}_{10}$ | = | $y_4\,\mathbf{a}_1 - y_4\,\mathbf{a}_2 - z_4\,\mathbf{a}_3$ | = | $-y_4b\,\hat{\mathbf{y}} - z_4c\,\hat{\mathbf{z}}$ | (8$f$) | V II |

**Geometry files:**

- CIF: pp. 1659
- POSCAR: pp. 1660

# $Si_{24}$ Clathrate Structure: A_oC24_63_3f

http://aflow.org/prototype-encyclopedia/A_oC24_63_3f

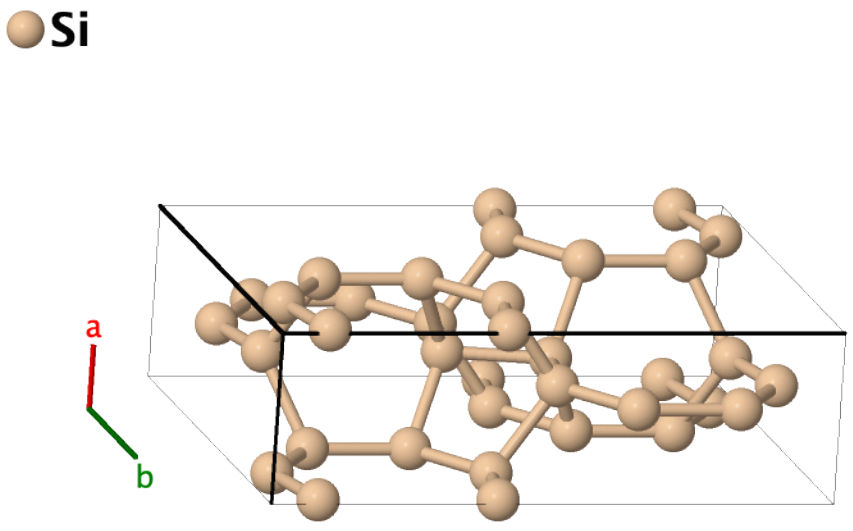

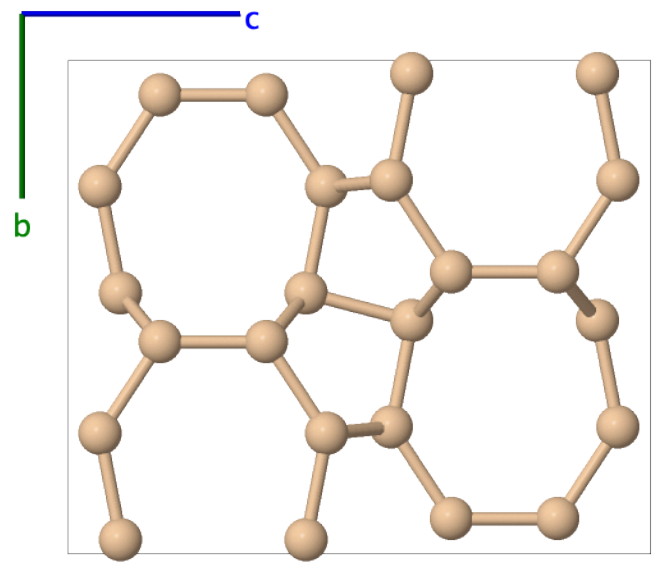

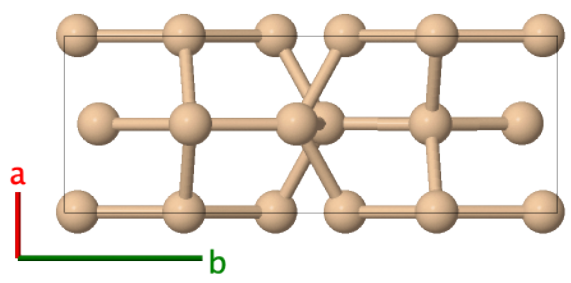

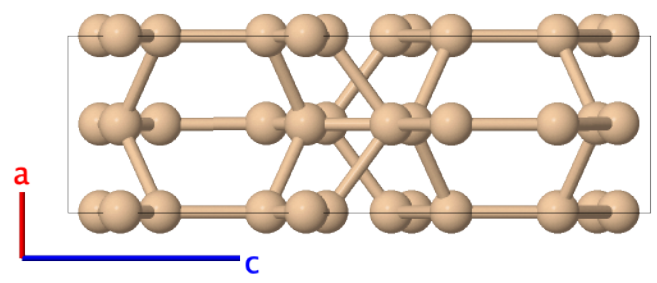

| | | |
|---|---|---|
| **Prototype** | : | $Si_{24}$ |
| **AFLOW prototype label** | : | A_oC24_63_3f |
| ***Strukturbericht* designation** | : | None |
| **Pearson symbol** | : | oC24 |
| **Space group number** | : | 63 |
| **Space group symbol** | : | *Cmcm* |
| **AFLOW prototype command** | : | `aflow --proto=A_oC24_63_3f`<br>`--params=`$a, b/a, c/a, y_1, z_1, y_2, z_2, y_3, z_3$ |

- Unlike the $Si_{34}$ and $Si_{46}$ clathrates, this is an experimentally determined structure, prepared by removing sodium atoms from an $Na_4Si_{24}$ predecessor.
- There is no consistency in the naming of these clathrate structures. $Si_{34}$ and $Si_{46}$ were named by their authors based on the size of the primitive unit cell. Here, (Kim, 2015) has chosen to name the structure based on the size of the conventional cell. For now, at least, we will follow the authors' naming schemes for these structures.

**Base-centered Orthorhombic primitive vectors:**

$$\mathbf{a}_1 = \frac{1}{2}\,a\,\hat{\mathbf{x}} - \frac{1}{2}\,b\,\hat{\mathbf{y}}$$
$$\mathbf{a}_2 = \frac{1}{2}\,a\,\hat{\mathbf{x}} + \frac{1}{2}\,b\,\hat{\mathbf{y}}$$
$$\mathbf{a}_3 = c\,\hat{\mathbf{z}}$$

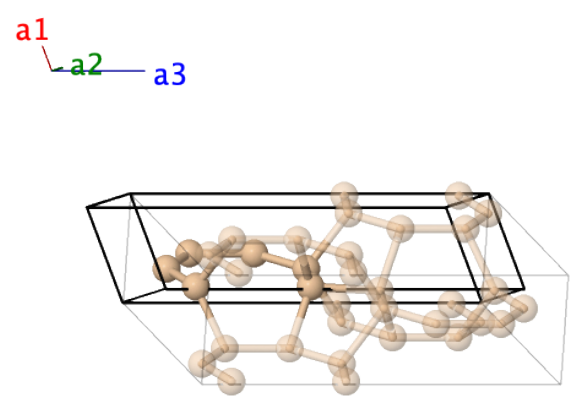

**Basis vectors:**

| | | Lattice Coordinates | | Cartesian Coordinates | Wyckoff Position | Atom Type |
|---|---|---|---|---|---|---|
| $\mathbf{B}_1$ | = | $-y_1\,\mathbf{a}_1 + y_1\,\mathbf{a}_2 + z_1\,\mathbf{a}_3$ | = | $y_1 b\,\hat{\mathbf{y}} + z_1 c\,\hat{\mathbf{z}}$ | $(8f)$ | Si I |
| $\mathbf{B}_2$ | = | $y_1\,\mathbf{a}_1 - y_1\,\mathbf{a}_2 + \left(\frac{1}{2} + z_1\right)\mathbf{a}_3$ | = | $-y_1 b\,\hat{\mathbf{y}} + \left(\frac{1}{2} + z_1\right) c\,\hat{\mathbf{z}}$ | $(8f)$ | Si I |
| $\mathbf{B}_3$ | = | $-y_1\,\mathbf{a}_1 + y_1\,\mathbf{a}_2 + \left(\frac{1}{2} - z_1\right)\mathbf{a}_3$ | = | $y_1 b\,\hat{\mathbf{y}} + \left(\frac{1}{2} - z_1\right) c\,\hat{\mathbf{z}}$ | $(8f)$ | Si I |
| $\mathbf{B}_4$ | = | $y_1\,\mathbf{a}_1 - y_1\,\mathbf{a}_2 - z_1\,\mathbf{a}_3$ | = | $-y_1 b\,\hat{\mathbf{y}} - z_1 c\,\hat{\mathbf{z}}$ | $(8f)$ | Si I |
| $\mathbf{B}_5$ | = | $-y_2\,\mathbf{a}_1 + y_2\,\mathbf{a}_2 + z_2\,\mathbf{a}_3$ | = | $y_2 b\,\hat{\mathbf{y}} + z_2 c\,\hat{\mathbf{z}}$ | $(8f)$ | Si II |
| $\mathbf{B}_6$ | = | $y_2\,\mathbf{a}_1 - y_2\,\mathbf{a}_2 + \left(\frac{1}{2} + z_2\right)\mathbf{a}_3$ | = | $-y_2 b\,\hat{\mathbf{y}} + \left(\frac{1}{2} + z_2\right) c\,\hat{\mathbf{z}}$ | $(8f)$ | Si II |
| $\mathbf{B}_7$ | = | $-y_2\,\mathbf{a}_1 + y_2\,\mathbf{a}_2 + \left(\frac{1}{2} - z_2\right)\mathbf{a}_3$ | = | $y_2 b\,\hat{\mathbf{y}} + \left(\frac{1}{2} - z_2\right) c\,\hat{\mathbf{z}}$ | $(8f)$ | Si II |
| $\mathbf{B}_8$ | = | $y_2\,\mathbf{a}_1 - y_2\,\mathbf{a}_2 - z_2\,\mathbf{a}_3$ | = | $-y_2 b\,\hat{\mathbf{y}} - z_2 c\,\hat{\mathbf{z}}$ | $(8f)$ | Si II |
| $\mathbf{B}_9$ | = | $-y_3\,\mathbf{a}_1 + y_3\,\mathbf{a}_2 + z_3\,\mathbf{a}_3$ | = | $y_3 b\,\hat{\mathbf{y}} + z_3 c\,\hat{\mathbf{z}}$ | $(8f)$ | Si III |
| $\mathbf{B}_{10}$ | = | $y_3\,\mathbf{a}_1 - y_3\,\mathbf{a}_2 + \left(\frac{1}{2} + z_3\right)\mathbf{a}_3$ | = | $-y_3 b\,\hat{\mathbf{y}} + \left(\frac{1}{2} + z_3\right) c\,\hat{\mathbf{z}}$ | $(8f)$ | Si III |
| $\mathbf{B}_{11}$ | = | $-y_3\,\mathbf{a}_1 + y_3\,\mathbf{a}_2 + \left(\frac{1}{2} - z_3\right)\mathbf{a}_3$ | = | $y_3 b\,\hat{\mathbf{y}} + \left(\frac{1}{2} - z_3\right) c\,\hat{\mathbf{z}}$ | $(8f)$ | Si III |
| $\mathbf{B}_{12}$ | = | $y_3\,\mathbf{a}_1 - y_3\,\mathbf{a}_2 - z_3\,\mathbf{a}_3$ | = | $-y_3 b\,\hat{\mathbf{y}} - z_3 c\,\hat{\mathbf{z}}$ | $(8f)$ | Si III |

**Geometry files:**

- CIF: pp. 1660
- POSCAR: pp. 1660

# Base-centered orthorhombic $Sr_4Ru_3O_{10}$ Structure: A10B3C4_oC68_64_2dfg_ad_2d

http://aflow.org/prototype-encyclopedia/A10B3C4_oC68_64_2dfg_ad_2d

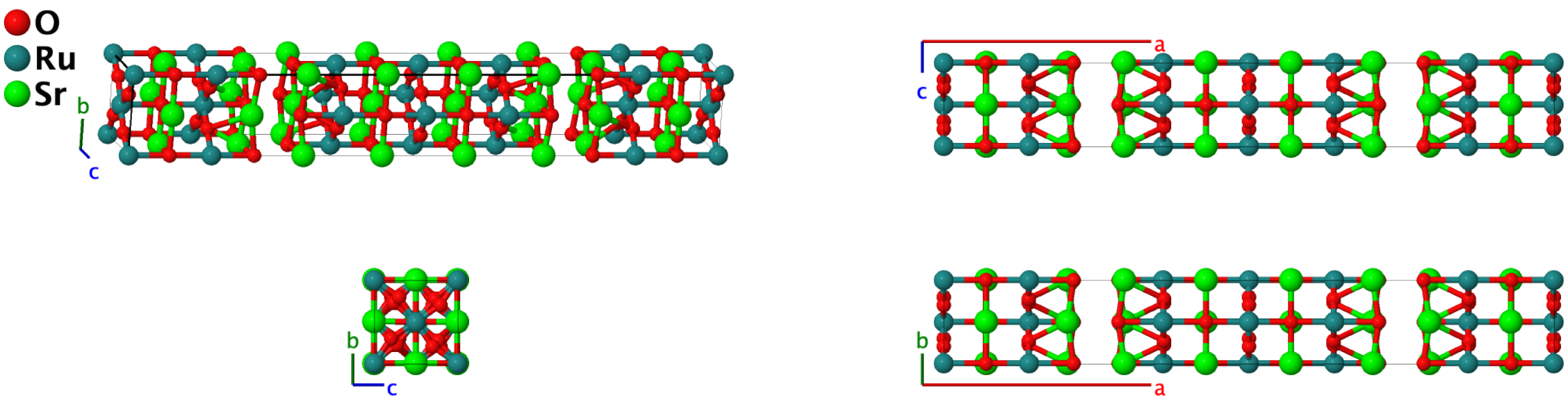

| | | |
|---|---|---|
| **Prototype** | : | $O_{10}Ru_3Sr_4$ |
| **AFLOW prototype label** | : | A10B3C4_oC68_64_2dfg_ad_2d |
| ***Strukturbericht* designation** | : | None |
| **Pearson symbol** | : | oC68 |
| **Space group number** | : | 64 |
| **Space group symbol** | : | *Cmca* |
| **AFLOW prototype command** | : | `aflow --proto=A10B3C4_oC68_64_2dfg_ad_2d --params=`$a, b/a, c/a, x_2, x_3, x_4, x_5, x_6, y_7, z_7, x_8, y_8, z_8$ |

- (Crawford, 2002) placed $Sr_4Ru_3O_{10}$ in space group *Pbam* #55, oP68, but found that it was very close to the current structure. Indeed, we derived this structure from the original structure by allowing a small amount of uncertainty in the atomic positions. In both cases, the structure consists of triple-layer ruthenate separated by 2.37 Å. In the original structure, there are two of these layers in the primitive cell which are not equivalent. In the current structure there is only one triple-layer in the *primitive base-centered cell*, and so the layers in the *conventional* cell are equivalent.

**Base-centered Orthorhombic primitive vectors:**

$$\mathbf{a}_1 = \frac{1}{2}\,a\,\hat{\mathbf{x}} - \frac{1}{2}\,b\,\hat{\mathbf{y}}$$
$$\mathbf{a}_2 = \frac{1}{2}\,a\,\hat{\mathbf{x}} + \frac{1}{2}\,b\,\hat{\mathbf{y}}$$
$$\mathbf{a}_3 = c\,\hat{\mathbf{z}}$$

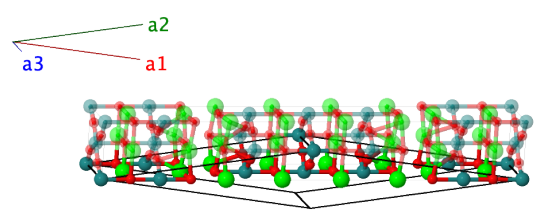

**Basis vectors:**

| | | Lattice Coordinates | | Cartesian Coordinates | Wyckoff Position | Atom Type |
|---|---|---|---|---|---|---|
| $\mathbf{B}_1$ | = | $0\,\mathbf{a}_1 + 0\,\mathbf{a}_2 + 0\,\mathbf{a}_3$ | = | $0\,\hat{\mathbf{x}} + 0\,\hat{\mathbf{y}} + 0\,\hat{\mathbf{z}}$ | (4*a*) | Ru I |
| $\mathbf{B}_2$ | = | $\frac{1}{2}\,\mathbf{a}_1 + \frac{1}{2}\,\mathbf{a}_2 + \frac{1}{2}\,\mathbf{a}_3$ | = | $\frac{1}{2}a\,\hat{\mathbf{x}} + \frac{1}{2}c\,\hat{\mathbf{z}}$ | (4*a*) | Ru I |
| $\mathbf{B}_3$ | = | $x_2\,\mathbf{a}_1 + x_2\,\mathbf{a}_2$ | = | $x_2 a\,\hat{\mathbf{x}}$ | (8*d*) | O I |
| $\mathbf{B}_4$ | = | $\left(\frac{1}{2} - x_2\right)\mathbf{a}_1 + \left(\frac{1}{2} - x_2\right)\mathbf{a}_2 + \frac{1}{2}\,\mathbf{a}_3$ | = | $\left(\frac{1}{2} - x_2\right)a\,\hat{\mathbf{x}} + \frac{1}{2}c\,\hat{\mathbf{z}}$ | (8*d*) | O I |
| $\mathbf{B}_5$ | = | $-x_2\,\mathbf{a}_1 - x_2\,\mathbf{a}_2$ | = | $-x_2 a\,\hat{\mathbf{x}}$ | (8*d*) | O I |
| $\mathbf{B}_6$ | = | $\left(\frac{1}{2} + x_2\right)\mathbf{a}_1 + \left(\frac{1}{2} + x_2\right)\mathbf{a}_2 + \frac{1}{2}\,\mathbf{a}_3$ | = | $\left(\frac{1}{2} + x_2\right)a\,\hat{\mathbf{x}} + \frac{1}{2}c\,\hat{\mathbf{z}}$ | (8*d*) | O I |
| $\mathbf{B}_7$ | = | $x_3\,\mathbf{a}_1 + x_3\,\mathbf{a}_2$ | = | $x_3 a\,\hat{\mathbf{x}}$ | (8*d*) | O II |
| $\mathbf{B}_8$ | = | $\left(\frac{1}{2} - x_3\right)\mathbf{a}_1 + \left(\frac{1}{2} - x_3\right)\mathbf{a}_2 + \frac{1}{2}\,\mathbf{a}_3$ | = | $\left(\frac{1}{2} - x_3\right)a\,\hat{\mathbf{x}} + \frac{1}{2}c\,\hat{\mathbf{z}}$ | (8*d*) | O II |
| $\mathbf{B}_{9}$ | = | $-x_3\,\mathbf{a}_1 - x_3\,\mathbf{a}_2$ | = | $-x_3 a\,\hat{\mathbf{x}}$ | (8$d$) | O II |
| $\mathbf{B}_{10}$ | = | $\left(\frac{1}{2}+x_3\right)\mathbf{a}_1 + \left(\frac{1}{2}+x_3\right)\mathbf{a}_2 + \frac{1}{2}\,\mathbf{a}_3$ | = | $\left(\frac{1}{2}+x_3\right)a\,\hat{\mathbf{x}} + \frac{1}{2}c\,\hat{\mathbf{z}}$ | (8$d$) | O II |
| $\mathbf{B}_{11}$ | = | $x_4\,\mathbf{a}_1 + x_4\,\mathbf{a}_2$ | = | $x_4 a\,\hat{\mathbf{x}}$ | (8$d$) | Ru II |
| $\mathbf{B}_{12}$ | = | $\left(\frac{1}{2}-x_4\right)\mathbf{a}_1 + \left(\frac{1}{2}-x_4\right)\mathbf{a}_2 + \frac{1}{2}\,\mathbf{a}_3$ | = | $\left(\frac{1}{2}-x_4\right)a\,\hat{\mathbf{x}} + \frac{1}{2}c\,\hat{\mathbf{z}}$ | (8$d$) | Ru II |
| $\mathbf{B}_{13}$ | = | $-x_4\,\mathbf{a}_1 - x_4\,\mathbf{a}_2$ | = | $-x_4 a\,\hat{\mathbf{x}}$ | (8$d$) | Ru II |
| $\mathbf{B}_{14}$ | = | $\left(\frac{1}{2}+x_4\right)\mathbf{a}_1 + \left(\frac{1}{2}+x_4\right)\mathbf{a}_2 + \frac{1}{2}\,\mathbf{a}_3$ | = | $\left(\frac{1}{2}+x_4\right)a\,\hat{\mathbf{x}} + \frac{1}{2}c\,\hat{\mathbf{z}}$ | (8$d$) | Ru II |
| $\mathbf{B}_{15}$ | = | $x_5\,\mathbf{a}_1 + x_5\,\mathbf{a}_2$ | = | $x_5 a\,\hat{\mathbf{x}}$ | (8$d$) | Sr I |
| $\mathbf{B}_{16}$ | = | $\left(\frac{1}{2}-x_5\right)\mathbf{a}_1 + \left(\frac{1}{2}-x_5\right)\mathbf{a}_2 + \frac{1}{2}\,\mathbf{a}_3$ | = | $\left(\frac{1}{2}-x_5\right)a\,\hat{\mathbf{x}} + \frac{1}{2}c\,\hat{\mathbf{z}}$ | (8$d$) | Sr I |
| $\mathbf{B}_{17}$ | = | $-x_5\,\mathbf{a}_1 - x_5\,\mathbf{a}_2$ | = | $-x_5 a\,\hat{\mathbf{x}}$ | (8$d$) | Sr I |
| $\mathbf{B}_{18}$ | = | $\left(\frac{1}{2}+x_5\right)\mathbf{a}_1 + \left(\frac{1}{2}+x_5\right)\mathbf{a}_2 + \frac{1}{2}\,\mathbf{a}_3$ | = | $\left(\frac{1}{2}+x_5\right)a\,\hat{\mathbf{x}} + \frac{1}{2}c\,\hat{\mathbf{z}}$ | (8$d$) | Sr I |
| $\mathbf{B}_{19}$ | = | $x_6\,\mathbf{a}_1 + x_6\,\mathbf{a}_2$ | = | $x_6 a\,\hat{\mathbf{x}}$ | (8$d$) | Sr II |
| $\mathbf{B}_{20}$ | = | $\left(\frac{1}{2}-x_6\right)\mathbf{a}_1 + \left(\frac{1}{2}-x_6\right)\mathbf{a}_2 + \frac{1}{2}\,\mathbf{a}_3$ | = | $\left(\frac{1}{2}-x_6\right)a\,\hat{\mathbf{x}} + \frac{1}{2}c\,\hat{\mathbf{z}}$ | (8$d$) | Sr II |
| $\mathbf{B}_{21}$ | = | $-x_6\,\mathbf{a}_1 - x_6\,\mathbf{a}_2$ | = | $-x_6 a\,\hat{\mathbf{x}}$ | (8$d$) | Sr II |
| $\mathbf{B}_{22}$ | = | $\left(\frac{1}{2}+x_6\right)\mathbf{a}_1 + \left(\frac{1}{2}+x_6\right)\mathbf{a}_2 + \frac{1}{2}\,\mathbf{a}_3$ | = | $\left(\frac{1}{2}+x_6\right)a\,\hat{\mathbf{x}} + \frac{1}{2}c\,\hat{\mathbf{z}}$ | (8$d$) | Sr II |
| $\mathbf{B}_{23}$ | = | $-y_7\,\mathbf{a}_1 + y_7\,\mathbf{a}_2 + z_7\,\mathbf{a}_3$ | = | $y_7 b\,\hat{\mathbf{y}} + z_7 c\,\hat{\mathbf{z}}$ | (8$f$) | O III |
| $\mathbf{B}_{24}$ | = | $\left(\frac{1}{2}+y_7\right)\mathbf{a}_1 + \left(\frac{1}{2}-y_7\right)\mathbf{a}_2 + \left(\frac{1}{2}+z_7\right)\mathbf{a}_3$ | = | $\frac{1}{2}a\,\hat{\mathbf{x}} - y_7 b\,\hat{\mathbf{y}} + \left(\frac{1}{2}+z_7\right)c\,\hat{\mathbf{z}}$ | (8$f$) | O III |
| $\mathbf{B}_{25}$ | = | $\left(\frac{1}{2}-y_7\right)\mathbf{a}_1 + \left(\frac{1}{2}+y_7\right)\mathbf{a}_2 + \left(\frac{1}{2}-z_7\right)\mathbf{a}_3$ | = | $\frac{1}{2}a\,\hat{\mathbf{x}} + y_7 b\,\hat{\mathbf{y}} + \left(\frac{1}{2}-z_7\right)c\,\hat{\mathbf{z}}$ | (8$f$) | O III |
| $\mathbf{B}_{26}$ | = | $y_7\,\mathbf{a}_1 - y_7\,\mathbf{a}_2 - z_7\,\mathbf{a}_3$ | = | $-y_7 b\,\hat{\mathbf{y}} - z_7 c\,\hat{\mathbf{z}}$ | (8$f$) | O III |
| $\mathbf{B}_{27}$ | = | $(x_8-y_8)\,\mathbf{a}_1 + (x_8+y_8)\,\mathbf{a}_2 + z_8\,\mathbf{a}_3$ | = | $x_8 a\,\hat{\mathbf{x}} + y_8 b\,\hat{\mathbf{y}} + z_8 c\,\hat{\mathbf{z}}$ | (16$g$) | O IV |
| $\mathbf{B}_{28}$ | = | $\left(\frac{1}{2}-x_8+y_8\right)\mathbf{a}_1 + \left(\frac{1}{2}-x_8-y_8\right)\mathbf{a}_2 + \left(\frac{1}{2}+z_8\right)\mathbf{a}_3$ | = | $\left(\frac{1}{2}-x_8\right)a\,\hat{\mathbf{x}} - y_8 b\,\hat{\mathbf{y}} + \left(\frac{1}{2}+z_8\right)c\,\hat{\mathbf{z}}$ | (16$g$) | O IV |
| $\mathbf{B}_{29}$ | = | $\left(\frac{1}{2}-x_8-y_8\right)\mathbf{a}_1 + \left(\frac{1}{2}-x_8+y_8\right)\mathbf{a}_2 + \left(\frac{1}{2}-z_8\right)\mathbf{a}_3$ | = | $\left(\frac{1}{2}-x_8\right)a\,\hat{\mathbf{x}} + y_8 b\,\hat{\mathbf{y}} + \left(\frac{1}{2}-z_8\right)c\,\hat{\mathbf{z}}$ | (16$g$) | O IV |
| $\mathbf{B}_{30}$ | = | $(x_8+y_8)\,\mathbf{a}_1 + (x_8-y_8)\,\mathbf{a}_2 - z_8\,\mathbf{a}_3$ | = | $x_8 a\,\hat{\mathbf{x}} - y_8 b\,\hat{\mathbf{y}} - z_8 c\,\hat{\mathbf{z}}$ | (16$g$) | O IV |
| $\mathbf{B}_{31}$ | = | $(-x_8+y_8)\,\mathbf{a}_1 + (-x_8-y_8)\,\mathbf{a}_2 - z_8\,\mathbf{a}_3$ | = | $-x_8 a\,\hat{\mathbf{x}} - y_8 b\,\hat{\mathbf{y}} - z_8 c\,\hat{\mathbf{z}}$ | (16$g$) | O IV |
| $\mathbf{B}_{32}$ | = | $\left(\frac{1}{2}+x_8-y_8\right)\mathbf{a}_1 + \left(\frac{1}{2}+x_8+y_8\right)\mathbf{a}_2 + \left(\frac{1}{2}-z_8\right)\mathbf{a}_3$ | = | $\left(\frac{1}{2}+x_8\right)a\,\hat{\mathbf{x}} + y_8 b\,\hat{\mathbf{y}} + \left(\frac{1}{2}-z_8\right)c\,\hat{\mathbf{z}}$ | (16$g$) | O IV |
| $\mathbf{B}_{33}$ | = | $\left(\frac{1}{2}+x_8+y_8\right)\mathbf{a}_1 + \left(\frac{1}{2}+x_8-y_8\right)\mathbf{a}_2 + \left(\frac{1}{2}+z_8\right)\mathbf{a}_3$ | = | $\left(\frac{1}{2}+x_8\right)a\,\hat{\mathbf{x}} - y_8 b\,\hat{\mathbf{y}} + \left(\frac{1}{2}+z_8\right)c\,\hat{\mathbf{z}}$ | (16$g$) | O IV |
| $\mathbf{B}_{34}$ | = | $(-x_8-y_8)\,\mathbf{a}_1 + (-x_8+y_8)\,\mathbf{a}_2 + z_8\,\mathbf{a}_3$ | = | $-x_8 a\,\hat{\mathbf{x}} + y_8 b\,\hat{\mathbf{y}} + z_8 c\,\hat{\mathbf{z}}$ | (16$g$) | O IV |

**Geometry files:**

- CIF: pp. 1660
- POSCAR: pp. 1661

# $Na_2Mo_2O_7$ Structure: A2B2C7_oC88_64_ef_df_3f2g

http://aflow.org/prototype-encyclopedia/A2B2C7_oC88_64_ef_df_3f2g

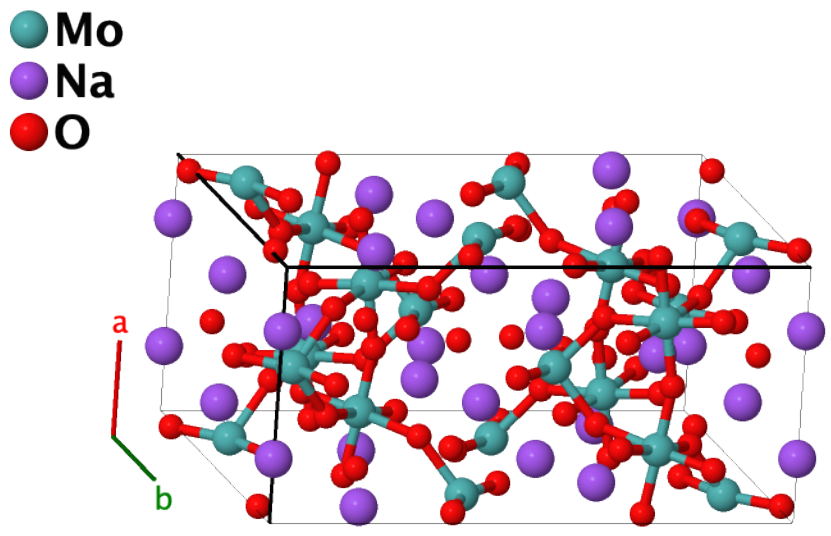

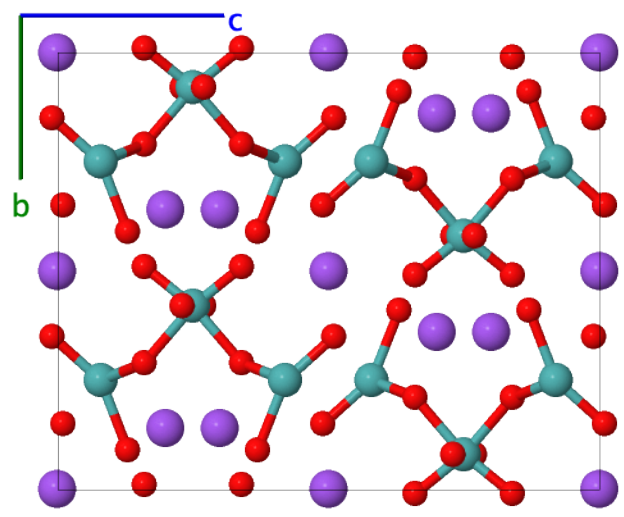

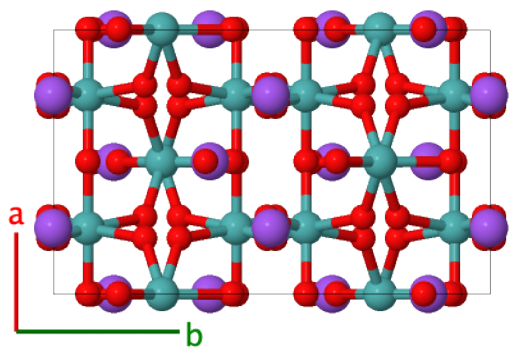

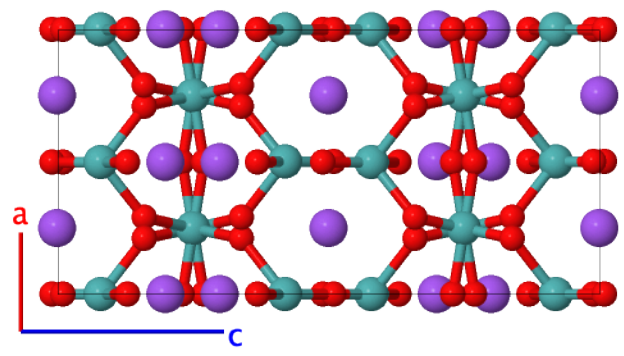

| | | |
|---|---|---|
| **Prototype** | : | $Mo_2Na_2O_7$ |
| **AFLOW prototype label** | : | A2B2C7_oC88_64_ef_df_3f2g |
| ***Strukturbericht* designation** | : | None |
| **Pearson symbol** | : | oC88 |
| **Space group number** | : | 64 |
| **Space group symbol** | : | *Cmca* |
| **AFLOW prototype command** | : | `aflow --proto=A2B2C7_oC88_64_ef_df_3f2g` `--params=`$a, b/a, c/a, x_1, y_2, y_3, z_3, y_4, z_4, y_5, z_5, y_6, z_6, y_7, z_7, x_8, y_8, z_8, x_9, y_9, z_9$ |

### Other compounds with this structure

- $Na_2W_2O_7$

### Base-centered Orthorhombic primitive vectors:

$$\mathbf{a}_1 = \frac{1}{2}\,a\,\hat{\mathbf{x}} - \frac{1}{2}\,b\,\hat{\mathbf{y}}$$
$$\mathbf{a}_2 = \frac{1}{2}\,a\,\hat{\mathbf{x}} + \frac{1}{2}\,b\,\hat{\mathbf{y}}$$
$$\mathbf{a}_3 = c\,\hat{\mathbf{z}}$$

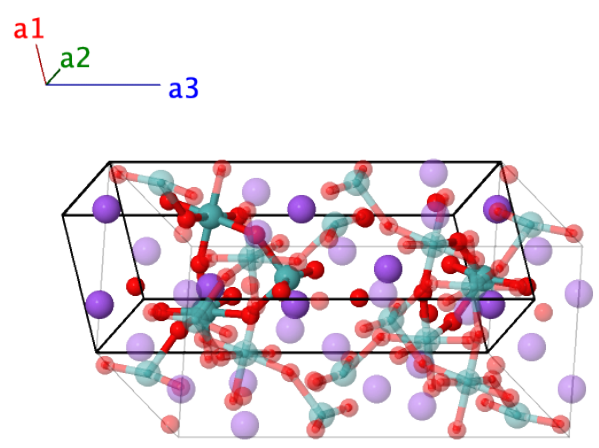

### Basis vectors:

| | | Lattice Coordinates | | Cartesian Coordinates | Wyckoff Position | Atom Type |
|---|---|---|---|---|---|---|
| $\mathbf{B}_1$ | = | $x_1\,\mathbf{a}_1 + x_1\,\mathbf{a}_2$ | = | $x_1\,a\,\hat{\mathbf{x}}$ | (8*d*) | Na I |
| $\mathbf{B}_2$ | = | $\left(\frac{1}{2} - x_1\right)\mathbf{a}_1 + \left(\frac{1}{2} - x_1\right)\mathbf{a}_2 + \frac{1}{2}\mathbf{a}_3$ | = | $\left(\frac{1}{2} - x_1\right)a\,\hat{\mathbf{x}} + \frac{1}{2}c\,\hat{\mathbf{z}}$ | (8$d$) | Na I |
| $\mathbf{B}_3$ | = | $-x_1\,\mathbf{a}_1 - x_1\,\mathbf{a}_2$ | = | $-x_1 a\,\hat{\mathbf{x}}$ | (8$d$) | Na I |
| $\mathbf{B}_4$ | = | $\left(\frac{1}{2} + x_1\right)\mathbf{a}_1 + \left(\frac{1}{2} + x_1\right)\mathbf{a}_2 + \frac{1}{2}\mathbf{a}_3$ | = | $\left(\frac{1}{2} + x_1\right)a\,\hat{\mathbf{x}} + \frac{1}{2}c\,\hat{\mathbf{z}}$ | (8$d$) | Na I |
| $\mathbf{B}_5$ | = | $\left(\frac{1}{4} - y_2\right)\mathbf{a}_1 + \left(\frac{1}{4} + y_2\right)\mathbf{a}_2 + \frac{1}{4}\mathbf{a}_3$ | = | $\frac{1}{4}a\,\hat{\mathbf{x}} + y_2 b\,\hat{\mathbf{y}} + \frac{1}{4}c\,\hat{\mathbf{z}}$ | (8$e$) | Mo I |
| $\mathbf{B}_6$ | = | $\left(\frac{1}{4} + y_2\right)\mathbf{a}_1 + \left(\frac{1}{4} - y_2\right)\mathbf{a}_2 + \frac{3}{4}\mathbf{a}_3$ | = | $\frac{1}{4}a\,\hat{\mathbf{x}} - y_2 b\,\hat{\mathbf{y}} + \frac{3}{4}c\,\hat{\mathbf{z}}$ | (8$e$) | Mo I |
| $\mathbf{B}_7$ | = | $\left(\frac{3}{4} + y_2\right)\mathbf{a}_1 + \left(\frac{3}{4} - y_2\right)\mathbf{a}_2 + \frac{3}{4}\mathbf{a}_3$ | = | $\frac{3}{4}a\,\hat{\mathbf{x}} - y_2 b\,\hat{\mathbf{y}} + \frac{3}{4}c\,\hat{\mathbf{z}}$ | (8$e$) | Mo I |
| $\mathbf{B}_8$ | = | $\left(\frac{3}{4} - y_2\right)\mathbf{a}_1 + \left(\frac{3}{4} + y_2\right)\mathbf{a}_2 + \frac{1}{4}\mathbf{a}_3$ | = | $\frac{3}{4}a\,\hat{\mathbf{x}} + y_2 b\,\hat{\mathbf{y}} + \frac{1}{4}c\,\hat{\mathbf{z}}$ | (8$e$) | Mo I |
| $\mathbf{B}_9$ | = | $-y_3\,\mathbf{a}_1 + y_3\,\mathbf{a}_2 + z_3\,\mathbf{a}_3$ | = | $y_3 b\,\hat{\mathbf{y}} + z_3 c\,\hat{\mathbf{z}}$ | (8$f$) | Mo II |
| $\mathbf{B}_{10}$ | = | $\left(\frac{1}{2} + y_3\right)\mathbf{a}_1 + \left(\frac{1}{2} - y_3\right)\mathbf{a}_2 + \left(\frac{1}{2} + z_3\right)\mathbf{a}_3$ | = | $\frac{1}{2}a\,\hat{\mathbf{x}} - y_3 b\,\hat{\mathbf{y}} + \left(\frac{1}{2} + z_3\right)c\,\hat{\mathbf{z}}$ | (8$f$) | Mo II |
| $\mathbf{B}_{11}$ | = | $\left(\frac{1}{2} - y_3\right)\mathbf{a}_1 + \left(\frac{1}{2} + y_3\right)\mathbf{a}_2 + \left(\frac{1}{2} - z_3\right)\mathbf{a}_3$ | = | $\frac{1}{2}a\,\hat{\mathbf{x}} + y_3 b\,\hat{\mathbf{y}} + \left(\frac{1}{2} - z_3\right)c\,\hat{\mathbf{z}}$ | (8$f$) | Mo II |
| $\mathbf{B}_{12}$ | = | $y_3\,\mathbf{a}_1 - y_3\,\mathbf{a}_2 - z_3\,\mathbf{a}_3$ | = | $-y_3 b\,\hat{\mathbf{y}} - z_3 c\,\hat{\mathbf{z}}$ | (8$f$) | Mo II |
| $\mathbf{B}_{13}$ | = | $-y_4\,\mathbf{a}_1 + y_4\,\mathbf{a}_2 + z_4\,\mathbf{a}_3$ | = | $y_4 b\,\hat{\mathbf{y}} + z_4 c\,\hat{\mathbf{z}}$ | (8$f$) | Na II |
| $\mathbf{B}_{14}$ | = | $\left(\frac{1}{2} + y_4\right)\mathbf{a}_1 + \left(\frac{1}{2} - y_4\right)\mathbf{a}_2 + \left(\frac{1}{2} + z_4\right)\mathbf{a}_3$ | = | $\frac{1}{2}a\,\hat{\mathbf{x}} - y_4 b\,\hat{\mathbf{y}} + \left(\frac{1}{2} + z_4\right)c\,\hat{\mathbf{z}}$ | (8$f$) | Na II |
| $\mathbf{B}_{15}$ | = | $\left(\frac{1}{2} - y_4\right)\mathbf{a}_1 + \left(\frac{1}{2} + y_4\right)\mathbf{a}_2 + \left(\frac{1}{2} - z_4\right)\mathbf{a}_3$ | = | $\frac{1}{2}a\,\hat{\mathbf{x}} + y_4 b\,\hat{\mathbf{y}} + \left(\frac{1}{2} - z_4\right)c\,\hat{\mathbf{z}}$ | (8$f$) | Na II |
| $\mathbf{B}_{16}$ | = | $y_4\,\mathbf{a}_1 - y_4\,\mathbf{a}_2 - z_4\,\mathbf{a}_3$ | = | $-y_4 b\,\hat{\mathbf{y}} - z_4 c\,\hat{\mathbf{z}}$ | (8$f$) | Na II |
| $\mathbf{B}_{17}$ | = | $-y_5\,\mathbf{a}_1 + y_5\,\mathbf{a}_2 + z_5\,\mathbf{a}_3$ | = | $y_5 b\,\hat{\mathbf{y}} + z_5 c\,\hat{\mathbf{z}}$ | (8$f$) | O I |
| $\mathbf{B}_{18}$ | = | $\left(\frac{1}{2} + y_5\right)\mathbf{a}_1 + \left(\frac{1}{2} - y_5\right)\mathbf{a}_2 + \left(\frac{1}{2} + z_5\right)\mathbf{a}_3$ | = | $\frac{1}{2}a\,\hat{\mathbf{x}} - y_5 b\,\hat{\mathbf{y}} + \left(\frac{1}{2} + z_5\right)c\,\hat{\mathbf{z}}$ | (8$f$) | O I |
| $\mathbf{B}_{19}$ | = | $\left(\frac{1}{2} - y_5\right)\mathbf{a}_1 + \left(\frac{1}{2} + y_5\right)\mathbf{a}_2 + \left(\frac{1}{2} - z_5\right)\mathbf{a}_3$ | = | $\frac{1}{2}a\,\hat{\mathbf{x}} + y_5 b\,\hat{\mathbf{y}} + \left(\frac{1}{2} - z_5\right)c\,\hat{\mathbf{z}}$ | (8$f$) | O I |
| $\mathbf{B}_{20}$ | = | $y_5\,\mathbf{a}_1 - y_5\,\mathbf{a}_2 - z_5\,\mathbf{a}_3$ | = | $-y_5 b\,\hat{\mathbf{y}} - z_5 c\,\hat{\mathbf{z}}$ | (8$f$) | O I |
| $\mathbf{B}_{21}$ | = | $-y_6\,\mathbf{a}_1 + y_6\,\mathbf{a}_2 + z_6\,\mathbf{a}_3$ | = | $y_6 b\,\hat{\mathbf{y}} + z_6 c\,\hat{\mathbf{z}}$ | (8$f$) | O II |
| $\mathbf{B}_{22}$ | = | $\left(\frac{1}{2} + y_6\right)\mathbf{a}_1 + \left(\frac{1}{2} - y_6\right)\mathbf{a}_2 + \left(\frac{1}{2} + z_6\right)\mathbf{a}_3$ | = | $\frac{1}{2}a\,\hat{\mathbf{x}} - y_6 b\,\hat{\mathbf{y}} + \left(\frac{1}{2} + z_6\right)c\,\hat{\mathbf{z}}$ | (8$f$) | O II |
| $\mathbf{B}_{23}$ | = | $\left(\frac{1}{2} - y_6\right)\mathbf{a}_1 + \left(\frac{1}{2} + y_6\right)\mathbf{a}_2 + \left(\frac{1}{2} - z_6\right)\mathbf{a}_3$ | = | $\frac{1}{2}a\,\hat{\mathbf{x}} + y_6 b\,\hat{\mathbf{y}} + \left(\frac{1}{2} - z_6\right)c\,\hat{\mathbf{z}}$ | (8$f$) | O II |
| $\mathbf{B}_{24}$ | = | $y_6\,\mathbf{a}_1 - y_6\,\mathbf{a}_2 - z_6\,\mathbf{a}_3$ | = | $-y_6 b\,\hat{\mathbf{y}} - z_6 c\,\hat{\mathbf{z}}$ | (8$f$) | O II |
| $\mathbf{B}_{25}$ | = | $-y_7\,\mathbf{a}_1 + y_7\,\mathbf{a}_2 + z_7\,\mathbf{a}_3$ | = | $y_7 b\,\hat{\mathbf{y}} + z_7 c\,\hat{\mathbf{z}}$ | (8$f$) | O III |
| $\mathbf{B}_{26}$ | = | $\left(\frac{1}{2} + y_7\right)\mathbf{a}_1 + \left(\frac{1}{2} - y_7\right)\mathbf{a}_2 + \left(\frac{1}{2} + z_7\right)\mathbf{a}_3$ | = | $\frac{1}{2}a\,\hat{\mathbf{x}} - y_7 b\,\hat{\mathbf{y}} + \left(\frac{1}{2} + z_7\right)c\,\hat{\mathbf{z}}$ | (8$f$) | O III |
| $\mathbf{B}_{27}$ | = | $\left(\frac{1}{2} - y_7\right)\mathbf{a}_1 + \left(\frac{1}{2} + y_7\right)\mathbf{a}_2 + \left(\frac{1}{2} - z_7\right)\mathbf{a}_3$ | = | $\frac{1}{2}a\,\hat{\mathbf{x}} + y_7 b\,\hat{\mathbf{y}} + \left(\frac{1}{2} - z_7\right)c\,\hat{\mathbf{z}}$ | (8$f$) | O III |
| $\mathbf{B}_{28}$ | = | $y_7\,\mathbf{a}_1 - y_7\,\mathbf{a}_2 - z_7\,\mathbf{a}_3$ | = | $-y_7 b\,\hat{\mathbf{y}} - z_7 c\,\hat{\mathbf{z}}$ | (8$f$) | O III |
| $\mathbf{B}_{29}$ | = | $(x_8 - y_8)\,\mathbf{a}_1 + (x_8 + y_8)\,\mathbf{a}_2 + z_8\,\mathbf{a}_3$ | = | $x_8 a\,\hat{\mathbf{x}} + y_8 b\,\hat{\mathbf{y}} + z_8 c\,\hat{\mathbf{z}}$ | (16$g$) | O IV |
| $\mathbf{B}_{30}$ | = | $\left(\frac{1}{2} - x_8 + y_8\right)\,\mathbf{a}_1 + \left(\frac{1}{2} - x_8 - y_8\right)\,\mathbf{a}_2 + \left(\frac{1}{2} + z_8\right)\,\mathbf{a}_3$ | = | $\left(\frac{1}{2} - x_8\right)a\,\hat{\mathbf{x}} - y_8 b\,\hat{\mathbf{y}} + \left(\frac{1}{2} + z_8\right)c\,\hat{\mathbf{z}}$ | (16$g$) | O IV |
| $\mathbf{B}_{31}$ | = | $\left(\frac{1}{2} - x_8 - y_8\right)\,\mathbf{a}_1 + \left(\frac{1}{2} - x_8 + y_8\right)\,\mathbf{a}_2 + \left(\frac{1}{2} - z_8\right)\,\mathbf{a}_3$ | = | $\left(\frac{1}{2} - x_8\right)a\,\hat{\mathbf{x}} + y_8 b\,\hat{\mathbf{y}} + \left(\frac{1}{2} - z_8\right)c\,\hat{\mathbf{z}}$ | (16$g$) | O IV |
| $\mathbf{B}_{32}$ | = | $(x_8 + y_8)\,\mathbf{a}_1 + (x_8 - y_8)\,\mathbf{a}_2 - z_8\,\mathbf{a}_3$ | = | $x_8 a\,\hat{\mathbf{x}} - y_8 b\,\hat{\mathbf{y}} - z_8 c\,\hat{\mathbf{z}}$ | (16$g$) | O IV |
| $\mathbf{B}_{33}$ | = | $(-x_8 + y_8)\,\mathbf{a}_1 + (-x_8 - y_8)\,\mathbf{a}_2 - z_8\,\mathbf{a}_3$ | = | $-x_8 a\,\hat{\mathbf{x}} - y_8 b\,\hat{\mathbf{y}} - z_8 c\,\hat{\mathbf{z}}$ | (16$g$) | O IV |
| $\mathbf{B}_{34}$ | = | $\left(\frac{1}{2} + x_8 - y_8\right)\,\mathbf{a}_1 + \left(\frac{1}{2} + x_8 + y_8\right)\,\mathbf{a}_2 + \left(\frac{1}{2} - z_8\right)\,\mathbf{a}_3$ | = | $\left(\frac{1}{2} + x_8\right)a\,\hat{\mathbf{x}} + y_8 b\,\hat{\mathbf{y}} + \left(\frac{1}{2} - z_8\right)c\,\hat{\mathbf{z}}$ | (16$g$) | O IV |
| $\mathbf{B}_{35}$ | = | $\left(\frac{1}{2} + x_8 + y_8\right)\,\mathbf{a}_1 + \left(\frac{1}{2} + x_8 - y_8\right)\,\mathbf{a}_2 + \left(\frac{1}{2} + z_8\right)\,\mathbf{a}_3$ | = | $\left(\frac{1}{2} + x_8\right)a\,\hat{\mathbf{x}} - y_8 b\,\hat{\mathbf{y}} + \left(\frac{1}{2} + z_8\right)c\,\hat{\mathbf{z}}$ | (16$g$) | O IV |
| $\mathbf{B}_{36}$ | = | $(-x_8 - y_8)\,\mathbf{a}_1 + (-x_8 + y_8)\,\mathbf{a}_2 + z_8\,\mathbf{a}_3$ | = | $-x_8 a\,\hat{\mathbf{x}} + y_8 b\,\hat{\mathbf{y}} + z_8 c\,\hat{\mathbf{z}}$ | (16$g$) | O IV |
| $\mathbf{B}_{37}$ | = | $(x_9 - y_9)\,\mathbf{a}_1 + (x_9 + y_9)\,\mathbf{a}_2 + z_9\,\mathbf{a}_3$ | = | $x_9 a\,\hat{\mathbf{x}} + y_9 b\,\hat{\mathbf{y}} + z_9 c\,\hat{\mathbf{z}}$ | (16$g$) | O V |
| $\mathbf{B}_{38}$ | = | $\left(\frac{1}{2} - x_9 + y_9\right)\,\mathbf{a}_1 + \left(\frac{1}{2} - x_9 - y_9\right)\,\mathbf{a}_2 + \left(\frac{1}{2} + z_9\right)\,\mathbf{a}_3$ | = | $\left(\frac{1}{2} - x_9\right)a\,\hat{\mathbf{x}} - y_9 b\,\hat{\mathbf{y}} + \left(\frac{1}{2} + z_9\right)c\,\hat{\mathbf{z}}$ | (16$g$) | O V |
| $\mathbf{B}_{39}$ | = | $\left(\frac{1}{2} - x_9 - y_9\right)\,\mathbf{a}_1 + \left(\frac{1}{2} - x_9 + y_9\right)\,\mathbf{a}_2 + \left(\frac{1}{2} - z_9\right)\,\mathbf{a}_3$ | = | $\left(\frac{1}{2} - x_9\right)a\,\hat{\mathbf{x}} + y_9 b\,\hat{\mathbf{y}} + \left(\frac{1}{2} - z_9\right)c\,\hat{\mathbf{z}}$ | (16$g$) | O V |
| $\mathbf{B}_{40}$ | = | $(x_9 + y_9)\,\mathbf{a}_1 + (x_9 - y_9)\,\mathbf{a}_2 - z_9\,\mathbf{a}_3$ | = | $x_9 a\,\hat{\mathbf{x}} - y_9 b\,\hat{\mathbf{y}} - z_9 c\,\hat{\mathbf{z}}$ | (16$g$) | O V |
| $\mathbf{B}_{41}$ | = | $(-x_9 + y_9)\,\mathbf{a}_1 + (-x_9 - y_9)\,\mathbf{a}_2 - z_9\,\mathbf{a}_3$ | = | $-x_9 a\,\hat{\mathbf{x}} - y_9 b\,\hat{\mathbf{y}} - z_9 c\,\hat{\mathbf{z}}$ | (16$g$) | O V |
| $\mathbf{B}_{42}$ | = | $\left(\frac{1}{2} + x_9 - y_9\right)\,\mathbf{a}_1 + \left(\frac{1}{2} + x_9 + y_9\right)\,\mathbf{a}_2 + \left(\frac{1}{2} - z_9\right)\,\mathbf{a}_3$ | = | $\left(\frac{1}{2} + x_9\right)a\,\hat{\mathbf{x}} + y_9 b\,\hat{\mathbf{y}} + \left(\frac{1}{2} - z_9\right)c\,\hat{\mathbf{z}}$ | (16$g$) | O V |
| $\mathbf{B}_{43}$ | = | $\left(\frac{1}{2} + x_9 + y_9\right)\,\mathbf{a}_1 + \left(\frac{1}{2} + x_9 - y_9\right)\,\mathbf{a}_2 + \left(\frac{1}{2} + z_9\right)\,\mathbf{a}_3$ | = | $\left(\frac{1}{2} + x_9\right)a\,\hat{\mathbf{x}} - y_9 b\,\hat{\mathbf{y}} + \left(\frac{1}{2} + z_9\right)c\,\hat{\mathbf{z}}$ | (16$g$) | O V |
| $\mathbf{B}_{44}$ | = | $(-x_9 - y_9)\,\mathbf{a}_1 + (-x_9 + y_9)\,\mathbf{a}_2 + z_9\,\mathbf{a}_3$ | = | $-x_9 a\,\hat{\mathbf{x}} + y_9 b\,\hat{\mathbf{y}} + z_9 c\,\hat{\mathbf{z}}$ | (16$g$) | O V |

---

---

**Geometry files:**

- CIF: pp. 1661
- POSCAR: pp. 1661

# $Li_2PrO_3$ Structure: A2B3C_oC12_65_h_bh_a

http://aflow.org/prototype-encyclopedia/A2B3C_oC12_65_h_bh_a

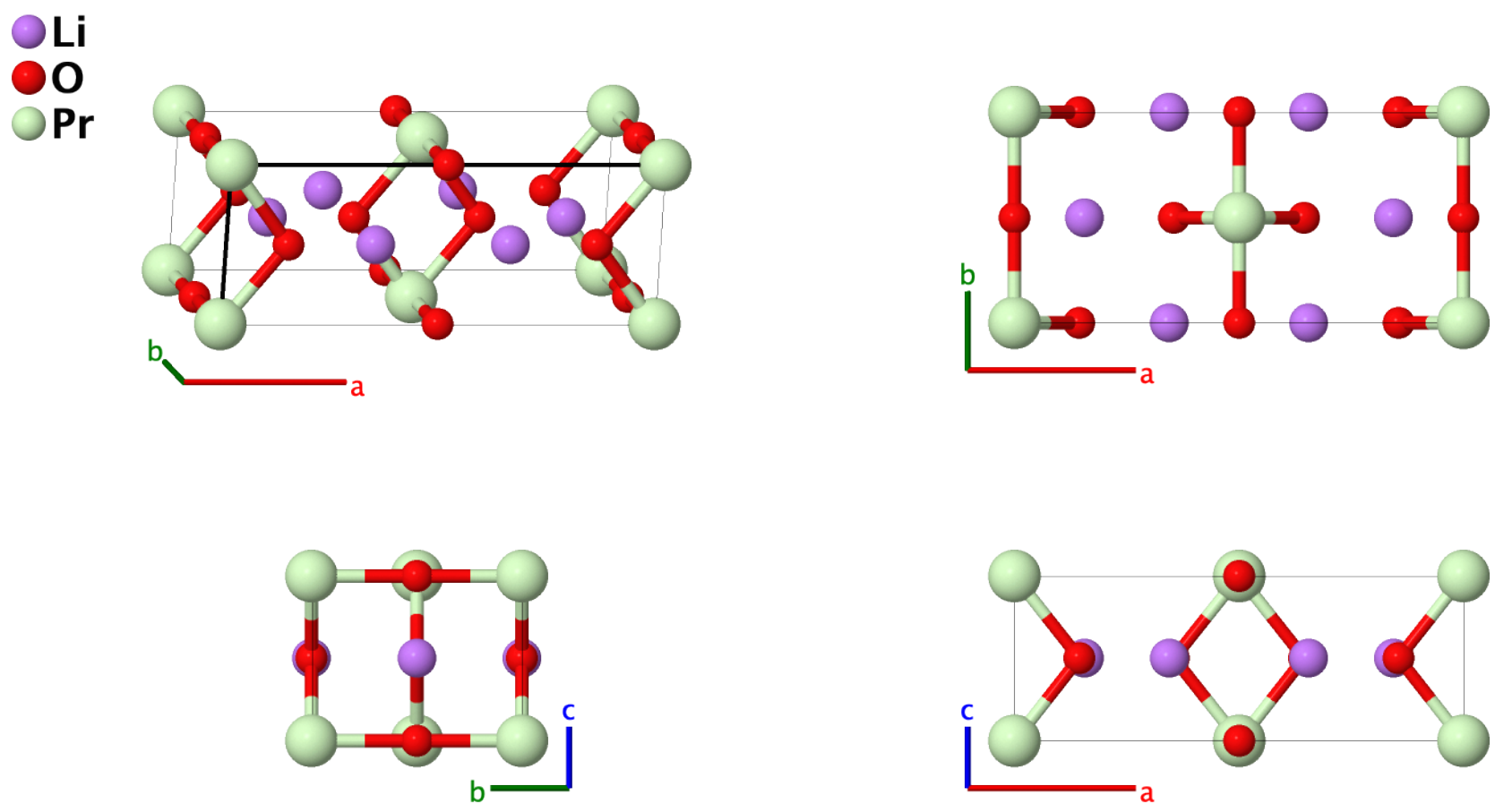

| | | |
|---|---|---|
| **Prototype** | : | $Li_2O_3Pr$ |
| **AFLOW prototype label** | : | A2B3C_oC12_65_h_bh_a |
| ***Strukturbericht* designation** | : | None |
| **Pearson symbol** | : | oC12 |
| **Space group number** | : | 65 |
| **Space group symbol** | : | *Cmmm* |
| **AFLOW prototype command** | : | `aflow --proto=A2B3C_oC12_65_h_bh_a`<br>`--params=`$a, b/a, c/a, x_3, x_4$ |

### Other compounds with this structure

- $Na_2PrO_3$

- Data was taken at 20K.

### Base-centered Orthorhombic primitive vectors:

$$\mathbf{a}_1 = \frac{1}{2}\,a\,\hat{\mathbf{x}} - \frac{1}{2}\,b\,\hat{\mathbf{y}}$$
$$\mathbf{a}_2 = \frac{1}{2}\,a\,\hat{\mathbf{x}} + \frac{1}{2}\,b\,\hat{\mathbf{y}}$$
$$\mathbf{a}_3 = c\,\hat{\mathbf{z}}$$

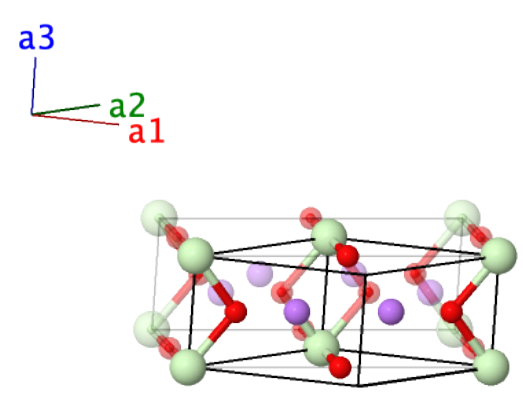

### Basis vectors:

| | | Lattice Coordinates | | Cartesian Coordinates | Wyckoff Position | Atom Type |
|---|---|---|---|---|---|---|
| $\mathbf{B}_1$ | = | $0\,\mathbf{a}_1 + 0\,\mathbf{a}_2 + 0\,\mathbf{a}_3$ | = | $0\,\hat{\mathbf{x}} + 0\,\hat{\mathbf{y}} + 0\,\hat{\mathbf{z}}$ | (2*a*) | Pr |
| $\mathbf{B}_2$ | = | $\frac{1}{2}\,\mathbf{a}_1 + \frac{1}{2}\,\mathbf{a}_2$ | = | $\frac{1}{2}\,a\,\hat{\mathbf{x}}$ | (2*b*) | O I |
| $\mathbf{B}_3$ | = | $x_3\,\mathbf{a}_1 + x_3\,\mathbf{a}_2 + \frac{1}{2}\,\mathbf{a}_3$ | = | $x_3 a\,\hat{\mathbf{x}} + \frac{1}{2} c\,\hat{\mathbf{z}}$ | (4*h*) | Li |
| $\mathbf{B}_4$ | = | $-x_3\,\mathbf{a}_1 - x_3\,\mathbf{a}_2 + \frac{1}{2}\,\mathbf{a}_3$ | = | $-x_3 a\,\hat{\mathbf{x}} + \frac{1}{2} c\,\hat{\mathbf{z}}$ | (4*h*) | Li |
| $\mathbf{B}_5$ | = | $x_4\,\mathbf{a}_1 + x_4\,\mathbf{a}_2 + \frac{1}{2}\,\mathbf{a}_3$ | = | $x_4 a\,\hat{\mathbf{x}} + \frac{1}{2} c\,\hat{\mathbf{z}}$ | (4$h$) | O II |
| $\mathbf{B}_6$ | = | $-x_4\,\mathbf{a}_1 - x_4\,\mathbf{a}_2 + \frac{1}{2}\,\mathbf{a}_3$ | = | $-x_4 a\,\hat{\mathbf{x}} + \frac{1}{2} c\,\hat{\mathbf{z}}$ | (4$h$) | O II |

---

---

**Geometry files:**

- CIF: pp. 1662
- POSCAR: pp. 1662

# $Mg(NH_3)_2Cl_2$ ($E1_3$) Structure: A2B8CD2_oC26_65_h_r_a_i

http://aflow.org/prototype-encyclopedia/A2B8CD2_oC26_65_h_r_a_i

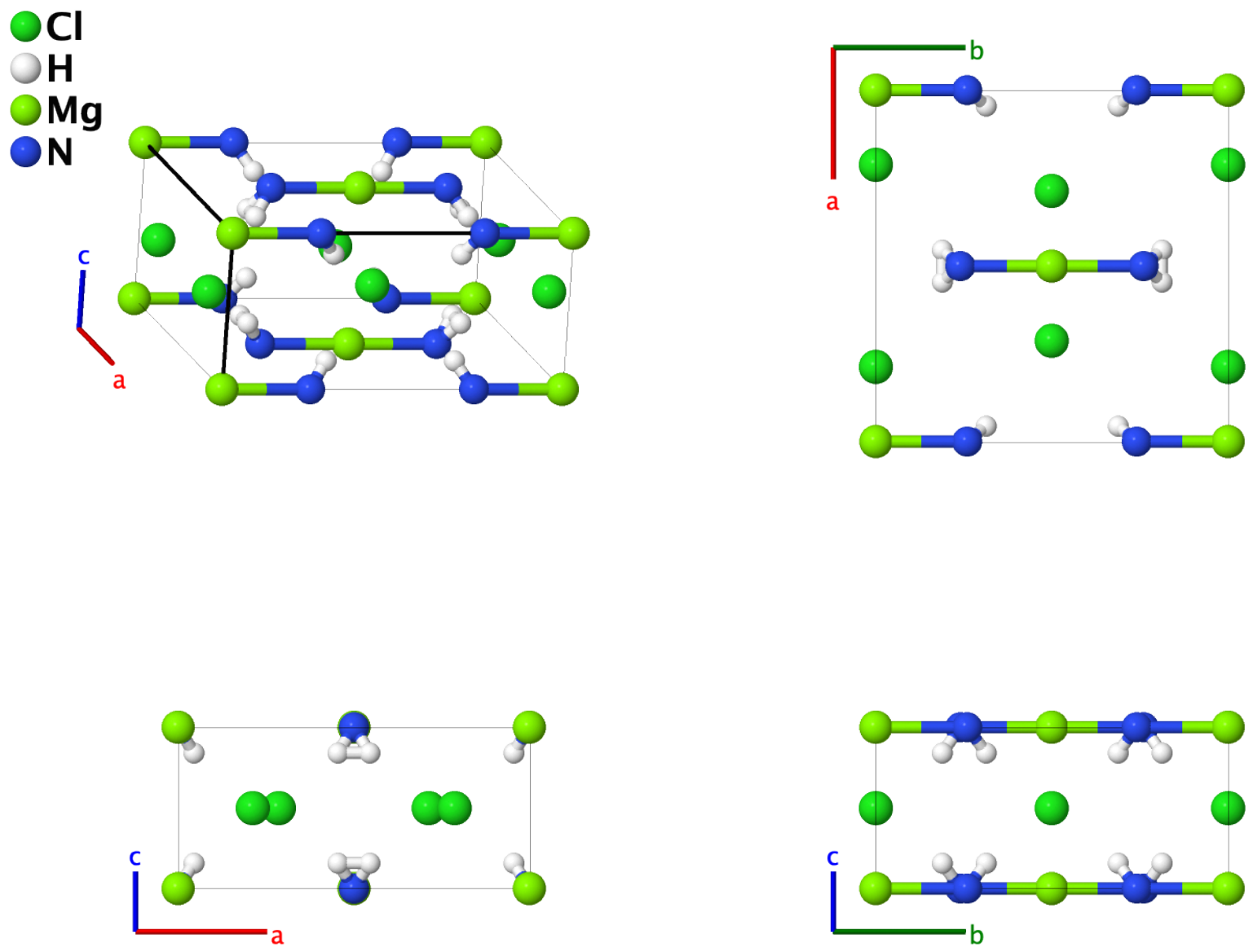

| | | |
|---|---|---|
| **Prototype** | : | $Cl_2H_6MgN_2$ |
| **AFLOW prototype label** | : | A2B8CD2_oC26_65_h_r_a_i |
| ***Strukturbericht* designation** | : | $E1_3$ |
| **Pearson symbol** | : | oC26 |
| **Space group number** | : | 65 |
| **Space group symbol** | : | *Cmmm* |
| **AFLOW prototype command** | : | `aflow --proto=A2B8CD2_oC26_65_h_r_a_i` <br> `--params=`$a, b/a, c/a, x_2, y_3, x_4, y_4, z_4$ |

### Other compounds with this structure

- $Cd(NH_3)_2Cl_2$, $Mg(NH_3)_2Cl_2$, $Hg(NH_3)_2Cl_2$, $Ni(NH_3)_2Br_2$, $Ni(NH_3)_2Cl_2$, and $Ni(NH_3)_2I_2$

---

- (Gottfried, 1938) gave the $E1_3$ designation to $Cd(NH_3)_2Cl_2$, and gave coordinates in the *Cmm*2 #35 space group. However, the cited reference, (MacGillavry, 1936) noted that their coordinates allowed several different space groups, with *Cmmm* #65 having the highest symmetry. We therefore follow most authors and use the *Cmmm* representation.
- (MacGillavry, 1936) could not determine the positions of the hydrogen atoms, but (Leineweber, 1999) was able to do this using the isostructural compound $Mg(NH_3)_2Cl_2$. Accordingly, we use $Mg(NH_3)_2Cl_2$ for the prototype of $E1_3$.
- Twelve hydrogen atoms are statistically distributed among the (16r) positions, so each site has a 75% probability of being occupied.

## Base-centered Orthorhombic primitive vectors:

$$\mathbf{a}_1 = \frac{1}{2}\,a\,\hat{\mathbf{x}} - \frac{1}{2}\,b\,\hat{\mathbf{y}}$$
$$\mathbf{a}_2 = \frac{1}{2}\,a\,\hat{\mathbf{x}} + \frac{1}{2}\,b\,\hat{\mathbf{y}}$$
$$\mathbf{a}_3 = c\,\hat{\mathbf{z}}$$

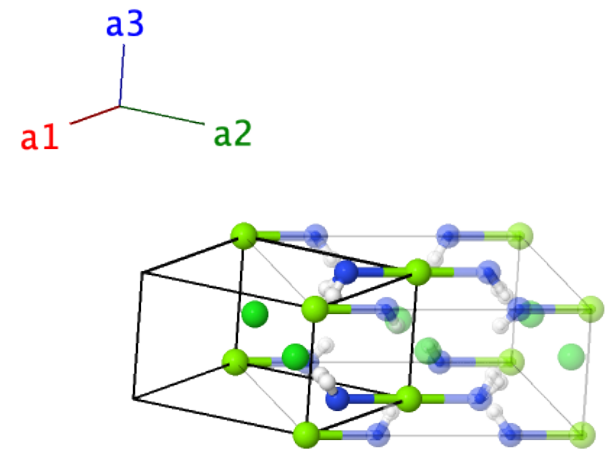

## Basis vectors:

| | | Lattice Coordinates | | Cartesian Coordinates | Wyckoff Position | Atom Type |
|---|---|---|---|---|---|---|
| $\mathbf{B}_1$ | = | $0\,\mathbf{a}_1 + 0\,\mathbf{a}_2 + 0\,\mathbf{a}_3$ | = | $0\,\hat{\mathbf{x}} + 0\,\hat{\mathbf{y}} + 0\,\hat{\mathbf{z}}$ | $(2a)$ | Mg |
| $\mathbf{B}_2$ | = | $x_2\,\mathbf{a}_1 + x_2\,\mathbf{a}_2 + \frac{1}{2}\,\mathbf{a}_3$ | = | $x_2 a\,\hat{\mathbf{x}} + \frac{1}{2} c\,\hat{\mathbf{z}}$ | $(4h)$ | Cl |
| $\mathbf{B}_3$ | = | $-x_2\,\mathbf{a}_1 - x_2\,\mathbf{a}_2 + \frac{1}{2}\,\mathbf{a}_3$ | = | $-x_2 a\,\hat{\mathbf{x}} + \frac{1}{2} c\,\hat{\mathbf{z}}$ | $(4h)$ | Cl |
| $\mathbf{B}_4$ | = | $-y_3\,\mathbf{a}_1 + y_3\,\mathbf{a}_2$ | = | $y_3 b\,\hat{\mathbf{y}}$ | $(4i)$ | N |
| $\mathbf{B}_5$ | = | $y_3\,\mathbf{a}_1 - y_3\,\mathbf{a}_2$ | = | $-y_3 b\,\hat{\mathbf{y}}$ | $(4i)$ | N |
| $\mathbf{B}_6$ | = | $(x_4 - y_4)\,\mathbf{a}_1 + (x_4 + y_4)\,\mathbf{a}_2 + z_4\,\mathbf{a}_3$ | = | $x_4 a\,\hat{\mathbf{x}} + y_4 b\,\hat{\mathbf{y}} + z_4 c\,\hat{\mathbf{z}}$ | $(16r)$ | H |
| $\mathbf{B}_7$ | = | $(-x_4 + y_4)\,\mathbf{a}_1 + (-x_4 - y_4)\,\mathbf{a}_2 + z_4\,\mathbf{a}_3$ | = | $-x_4 a\,\hat{\mathbf{x}} - y_4 b\,\hat{\mathbf{y}} + z_4 c\,\hat{\mathbf{z}}$ | $(16r)$ | H |
| $\mathbf{B}_8$ | = | $(-x_4 - y_4)\,\mathbf{a}_1 + (-x_4 + y_4)\,\mathbf{a}_2 - z_4\,\mathbf{a}_3$ | = | $-x_4 a\,\hat{\mathbf{x}} + y_4 b\,\hat{\mathbf{y}} - z_4 c\,\hat{\mathbf{z}}$ | $(16r)$ | H |
| $\mathbf{B}_9$ | = | $(x_4 + y_4)\,\mathbf{a}_1 + (x_4 - y_4)\,\mathbf{a}_2 - z_4\,\mathbf{a}_3$ | = | $x_4 a\,\hat{\mathbf{x}} - y_4 b\,\hat{\mathbf{y}} - z_4 c\,\hat{\mathbf{z}}$ | $(16r)$ | H |
| $\mathbf{B}_{10}$ | = | $(-x_4 + y_4)\,\mathbf{a}_1 + (-x_4 - y_4)\,\mathbf{a}_2 - z_4\,\mathbf{a}_3$ | = | $-x_4 a\,\hat{\mathbf{x}} - y_4 b\,\hat{\mathbf{y}} - z_4 c\,\hat{\mathbf{z}}$ | $(16r)$ | H |
| $\mathbf{B}_{11}$ | = | $(x_4 - y_4)\,\mathbf{a}_1 + (x_4 + y_4)\,\mathbf{a}_2 - z_4\,\mathbf{a}_3$ | = | $x_4 a\,\hat{\mathbf{x}} + y_4 b\,\hat{\mathbf{y}} - z_4 c\,\hat{\mathbf{z}}$ | $(16r)$ | H |
| $\mathbf{B}_{12}$ | = | $(x_4 + y_4)\,\mathbf{a}_1 + (x_4 - y_4)\,\mathbf{a}_2 + z_4\,\mathbf{a}_3$ | = | $x_4 a\,\hat{\mathbf{x}} - y_4 b\,\hat{\mathbf{y}} + z_4 c\,\hat{\mathbf{z}}$ | $(16r)$ | H |
| $\mathbf{B}_{13}$ | = | $(-x_4 - y_4)\,\mathbf{a}_1 + (-x_4 + y_4)\,\mathbf{a}_2 + z_4\,\mathbf{a}_3$ | = | $-x_4 a\,\hat{\mathbf{x}} + y_4 b\,\hat{\mathbf{y}} + z_4 c\,\hat{\mathbf{z}}$ | $(16r)$ | H |

## Geometry files:

- CIF: pp. 1662
- POSCAR: pp. 1662

# $Nb_3O_7F$ Structure: A3B8_oC22_65_ag_bd2gh

http://aflow.org/prototype-encyclopedia/A3B8_oC22_65_ag_bd2gh

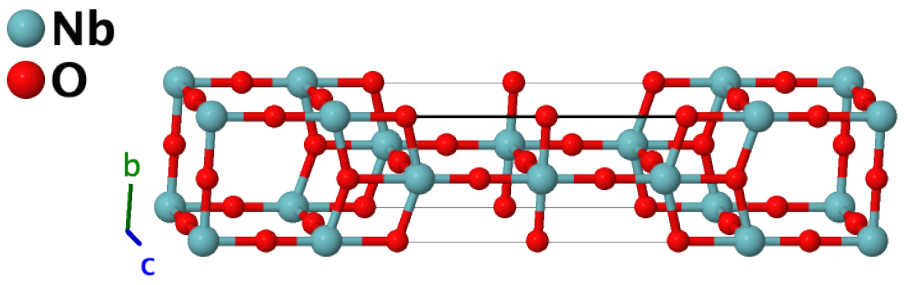

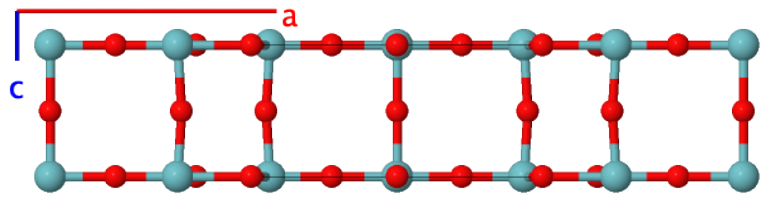

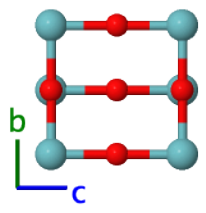

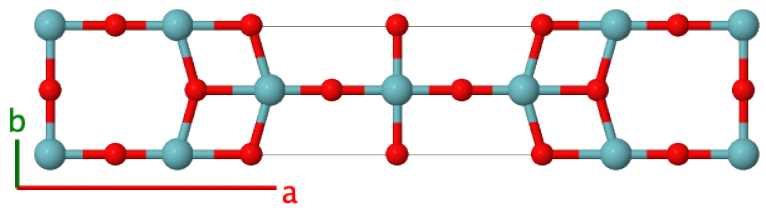

| | | |
|---|---|---|
| **Prototype** | : | $FNb_3O_7$ |
| **AFLOW prototype label** | : | A3B8_oC22_65_ag_bd2gh |
| ***Strukturbericht* designation** | : | None |
| **Pearson symbol** | : | oC22 |
| **Space group number** | : | 65 |
| **Space group symbol** | : | *Cmmm* |
| **AFLOW prototype command** | : | `aflow --proto=A3B8_oC22_65_ag_bd2gh`<br>`--params=`$a, b/a, c/a, x_4, x_5, x_6, x_7$ |

### Other compounds with this structure

- $Nb_3O_7(OH)$

- (Andersson, 1964) was not able to distinguish between the oxygen and fluorine atoms, so it is assumed that the fluorine is distributed randomly on the oxygen sites.

### Base-centered Orthorhombic primitive vectors:

$$\mathbf{a}_1 = \frac{1}{2}\,a\,\hat{\mathbf{x}} - \frac{1}{2}\,b\,\hat{\mathbf{y}}$$
$$\mathbf{a}_2 = \frac{1}{2}\,a\,\hat{\mathbf{x}} + \frac{1}{2}\,b\,\hat{\mathbf{y}}$$
$$\mathbf{a}_3 = c\,\hat{\mathbf{z}}$$

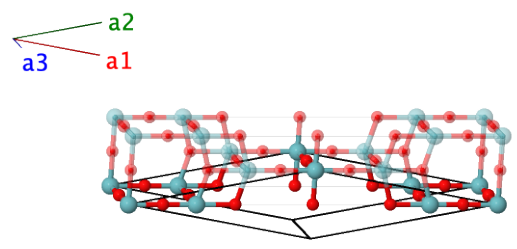

### Basis vectors:

| | | Lattice Coordinates | | Cartesian Coordinates | Wyckoff Position | Atom Type |
|---|---|---|---|---|---|---|
| $\mathbf{B}_1$ | = | $0\,\mathbf{a}_1 + 0\,\mathbf{a}_2 + 0\,\mathbf{a}_3$ | = | $0\,\hat{\mathbf{x}} + 0\,\hat{\mathbf{y}} + 0\,\hat{\mathbf{z}}$ | (2*a*) | Nb I |
| $\mathbf{B}_2$ | = | $\frac{1}{2}\,\mathbf{a}_1 + \frac{1}{2}\,\mathbf{a}_2$ | = | $\frac{1}{2}a\,\hat{\mathbf{x}}$ | (2*b*) | O I |
| $\mathbf{B}_3$ | = | $\frac{1}{2}\,\mathbf{a}_3$ | = | $\frac{1}{2}c\,\hat{\mathbf{z}}$ | (2*d*) | O II |
| $\mathbf{B}_4$ | = | $x_4\,\mathbf{a}_1 + x_4\,\mathbf{a}_2$ | = | $x_4 a\,\hat{\mathbf{x}}$ | (4*g*) | Nb II |
| $\mathbf{B}_5$ | = | $-x_4\,\mathbf{a}_1 - x_4\,\mathbf{a}_2$ | = | $-x_4 a\,\hat{\mathbf{x}}$ | (4*g*) | Nb II |
| $\mathbf{B}_6$ | = | $x_5\,\mathbf{a}_1 + x_5\,\mathbf{a}_2$ | = | $x_5 a\,\hat{\mathbf{x}}$ | (4*g*) | O III |
| $\mathbf{B}_7$ | = | $-x_5\,\mathbf{a}_1 - x_5\,\mathbf{a}_2$ | = | $-x_5 a\,\hat{\mathbf{x}}$ | (4*g*) | O III |
| $\mathbf{B}_8$ | = | $x_6\,\mathbf{a}_1 + x_6\,\mathbf{a}_2$ | = | $x_6 a\,\hat{\mathbf{x}}$ | (4*g*) | O IV |
| $\mathbf{B}_9$ | = | $-x_6\,\mathbf{a}_1 - x_6\,\mathbf{a}_2$ | = | $-x_6 a\,\hat{\mathbf{x}}$ | (4$g$) | O IV |
| $\mathbf{B}_{10}$ | = | $x_7\,\mathbf{a}_1 + x_7\,\mathbf{a}_2 + \frac{1}{2}\,\mathbf{a}_3$ | = | $x_7 a\,\hat{\mathbf{x}} + \frac{1}{2}c\,\hat{\mathbf{z}}$ | (4$h$) | O V |
| $\mathbf{B}_{11}$ | = | $-x_7\,\mathbf{a}_1 - x_7\,\mathbf{a}_2 + \frac{1}{2}\,\mathbf{a}_3$ | = | $-x_7 a\,\hat{\mathbf{x}} + \frac{1}{2}c\,\hat{\mathbf{z}}$ | (4$h$) | O V |

**Geometry files:**

- CIF: pp. 1663
- POSCAR: pp. 1663

# $NH_4H_2PO_2$ ($F5_7$) Structure: A2BC2D_oC24_67_m_a_n_g

http://aflow.org/prototype-encyclopedia/A2BC2D_oC24_67_m_a_n_g

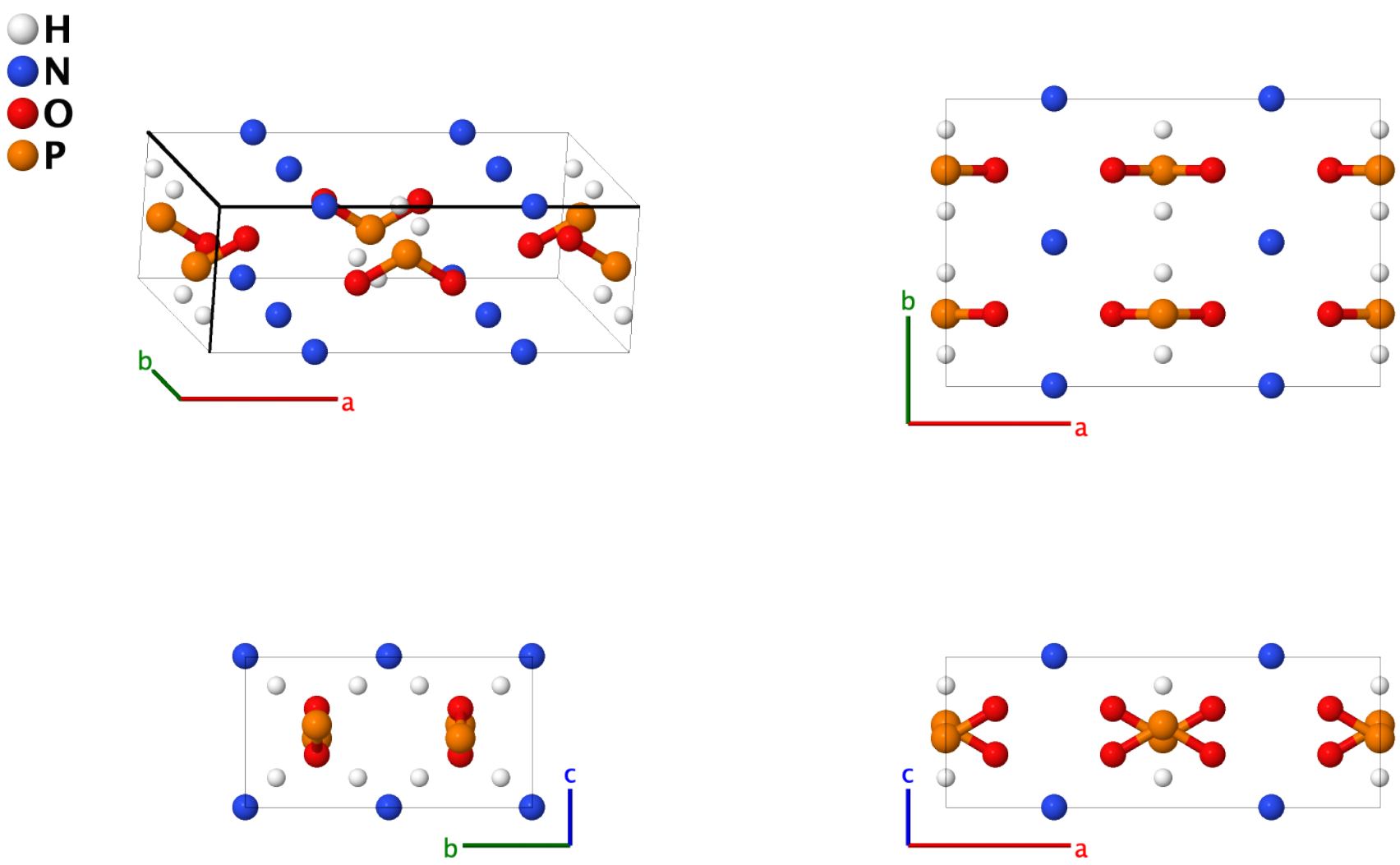

| | | |
|---|---|---|
| **Prototype** | : | $H_2(NH_4)O_2P$ |
| **AFLOW prototype label** | : | A2BC2D_oC24_67_m_a_n_g |
| ***Strukturbericht* designation** | : | $F5_7$ |
| **Pearson symbol** | : | oC24 |
| **Space group number** | : | 67 |
| **Space group symbol** | : | *Cmma* |
| **AFLOW prototype command** | : | `aflow --proto=A2BC2D_oC24_67_m_a_n_g` <br> `--params=`$a, b/a, c/a, z_2, y_3, z_3, x_4, z_4$ |

- (Zachriasen, 1934) state that the H atoms in the ammonium ion must be along the lines between the nitrogen and oxygen atoms, but give no further information.
- The positions of the hydrogen atoms in the $NH_4$ ions were not determined, so we only provide the position of the nitrogen atoms (labeled as $NH_4$).
- The data for this structure was presented in the *Acmm* setting of space group #67. We transformed this to the standard *Cmma* setting using FINDSYM.

**Base-centered Orthorhombic primitive vectors:**

$$\mathbf{a}_1 = \frac{1}{2}\,a\,\hat{\mathbf{x}} - \frac{1}{2}\,b\,\hat{\mathbf{y}}$$
$$\mathbf{a}_2 = \frac{1}{2}\,a\,\hat{\mathbf{x}} + \frac{1}{2}\,b\,\hat{\mathbf{y}}$$
$$\mathbf{a}_3 = c\,\hat{\mathbf{z}}$$

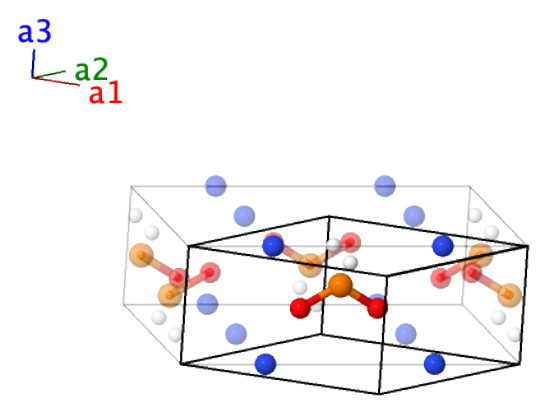

**Basis vectors:**

| | | Lattice Coordinates | | Cartesian Coordinates | Wyckoff Position | Atom Type |
|---|---|---|---|---|---|---|
| $\mathbf{B}_1$ | = | $\frac{1}{4}\,\mathbf{a}_1 + \frac{1}{4}\,\mathbf{a}_2$ | = | $\frac{1}{4}a\,\hat{\mathbf{x}}$ | $(4a)$ | $NH_4$ |
| $\mathbf{B}_2$ | = | $\frac{3}{4}\,\mathbf{a}_1 + \frac{3}{4}\,\mathbf{a}_2$ | = | $\frac{3}{4}a\,\hat{\mathbf{x}}$ | $(4a)$ | $NH_4$ |
| $\mathbf{B}_3$ | = | $\frac{3}{4}\,\mathbf{a}_1 + \frac{1}{4}\,\mathbf{a}_2 + z_2\,\mathbf{a}_3$ | = | $\frac{1}{2}a\,\hat{\mathbf{x}} - \frac{1}{4}b\,\hat{\mathbf{y}} + z_2c\,\hat{\mathbf{z}}$ | $(4g)$ | P |
| $\mathbf{B}_4$ | = | $\frac{1}{4}\,\mathbf{a}_1 + \frac{3}{4}\,\mathbf{a}_2 - z_2\,\mathbf{a}_3$ | = | $\frac{1}{2}a\,\hat{\mathbf{x}} + \frac{1}{4}b\,\hat{\mathbf{y}} - z_2c\,\hat{\mathbf{z}}$ | $(4g)$ | P |
| $\mathbf{B}_5$ | = | $-y_3\,\mathbf{a}_1 + y_3\,\mathbf{a}_2 + z_3\,\mathbf{a}_3$ | = | $y_3b\,\hat{\mathbf{y}} + z_3c\,\hat{\mathbf{z}}$ | $(8m)$ | H |
| $\mathbf{B}_6$ | = | $\left(\frac{1}{2} + y_3\right)\,\mathbf{a}_1 + \left(\frac{1}{2} - y_3\right)\,\mathbf{a}_2 + z_3\,\mathbf{a}_3$ | = | $\frac{1}{2}a\,\hat{\mathbf{x}} - y_3b\,\hat{\mathbf{y}} + z_3c\,\hat{\mathbf{z}}$ | $(8m)$ | H |
| $\mathbf{B}_7$ | = | $\left(\frac{1}{2} - y_3\right)\,\mathbf{a}_1 + \left(\frac{1}{2} + y_3\right)\,\mathbf{a}_2 - z_3\,\mathbf{a}_3$ | = | $\frac{1}{2}a\,\hat{\mathbf{x}} + y_3b\,\hat{\mathbf{y}} - z_3c\,\hat{\mathbf{z}}$ | $(8m)$ | H |
| $\mathbf{B}_8$ | = | $y_3\,\mathbf{a}_1 - y_3\,\mathbf{a}_2 - z_3\,\mathbf{a}_3$ | = | $-y_3b\,\hat{\mathbf{y}} - z_3c\,\hat{\mathbf{z}}$ | $(8m)$ | H |
| $\mathbf{B}_9$ | = | $\left(\frac{3}{4} + x_4\right)\,\mathbf{a}_1 + \left(\frac{1}{4} + x_4\right)\,\mathbf{a}_2 + z_4\,\mathbf{a}_3$ | = | $\left(\frac{1}{2} + x_4\right)a\,\hat{\mathbf{x}} - \frac{1}{4}b\,\hat{\mathbf{y}} + z_4c\,\hat{\mathbf{z}}$ | $(8n)$ | O |
| $\mathbf{B}_{10}$ | = | $\left(\frac{3}{4} - x_4\right)\,\mathbf{a}_1 + \left(\frac{1}{4} - x_4\right)\,\mathbf{a}_2 + z_4\,\mathbf{a}_3$ | = | $\left(\frac{1}{2} - x_4\right)a\,\hat{\mathbf{x}} - \frac{1}{4}b\,\hat{\mathbf{y}} + z_4c\,\hat{\mathbf{z}}$ | $(8n)$ | O |
| $\mathbf{B}_{11}$ | = | $\left(\frac{1}{4} - x_4\right)\,\mathbf{a}_1 + \left(\frac{3}{4} - x_4\right)\,\mathbf{a}_2 - z_4\,\mathbf{a}_3$ | = | $\left(\frac{1}{2} - x_4\right)a\,\hat{\mathbf{x}} + \frac{1}{4}b\,\hat{\mathbf{y}} - z_4c\,\hat{\mathbf{z}}$ | $(8n)$ | O |
| $\mathbf{B}_{12}$ | = | $\left(\frac{1}{4} + x_4\right)\,\mathbf{a}_1 + \left(\frac{3}{4} + x_4\right)\,\mathbf{a}_2 - z_4\,\mathbf{a}_3$ | = | $\left(\frac{1}{2} + x_4\right)a\,\hat{\mathbf{x}} + \frac{1}{4}b\,\hat{\mathbf{y}} - z_4c\,\hat{\mathbf{z}}$ | $(8n)$ | O |

**Geometry files:**

- CIF: pp. 1663
- POSCAR: pp. 1663

# Thenardite [$Na_2SO_4$ (V), $H1_7$] Structure: A2B4C_oF56_70_g_h_a

http://aflow.org/prototype-encyclopedia/A2B4C_oF56_70_g_h_a

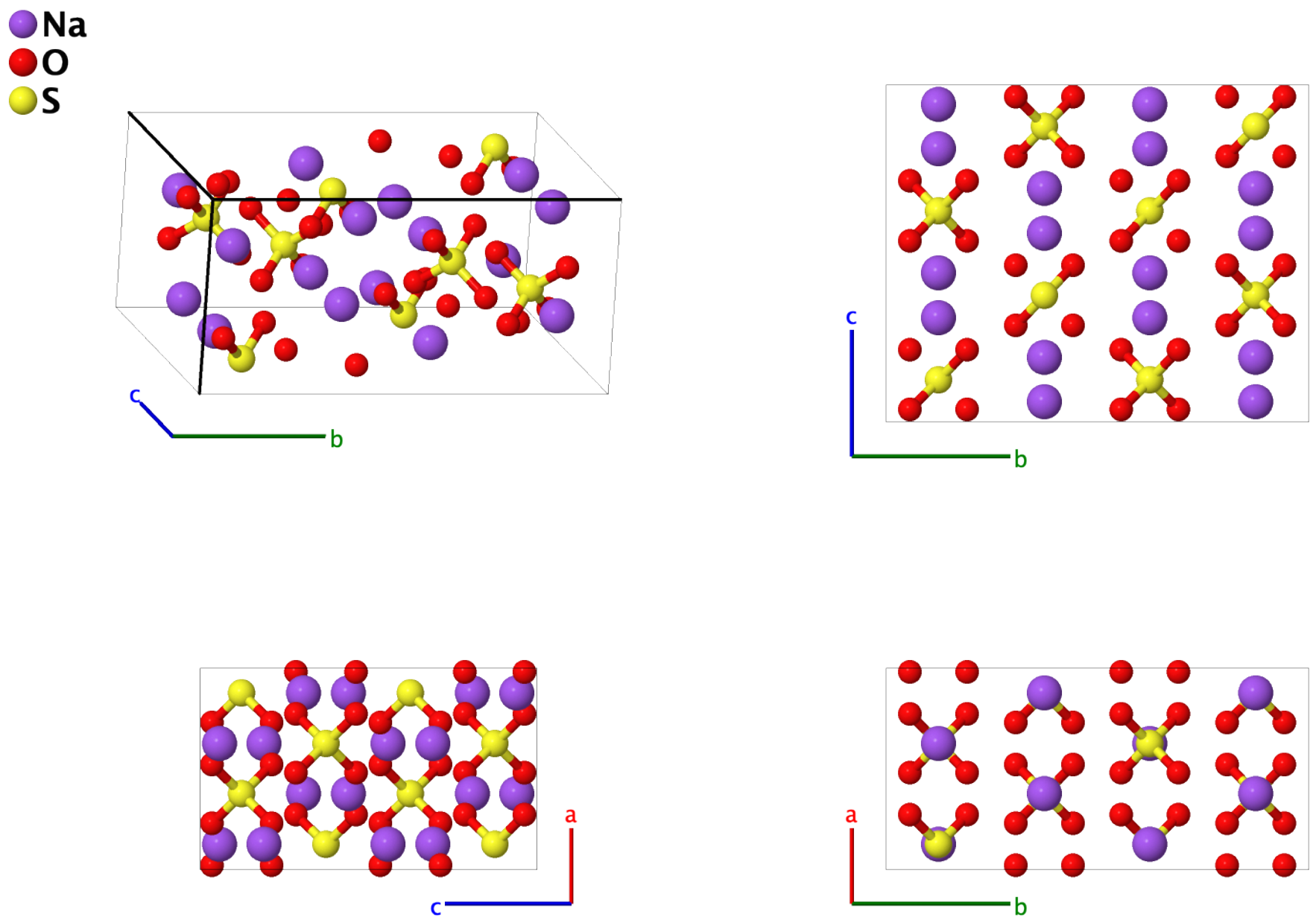

| | | |
|---|---|---|
| **Prototype** | : | $Na_2O_4S$ |
| **AFLOW prototype label** | : | A2B4C_oF56_70_g_h_a |
| ***Strukturbericht* designation** | : | $H1_7$ |
| **Pearson symbol** | : | oF56 |
| **Space group number** | : | 70 |
| **Space group symbol** | : | $Fddd$ |
| **AFLOW prototype command** | : | `aflow --proto=A2B4C_oF56_70_g_h_a` `--params=`$a, b/a, c/a, z_2, x_3, y_3, z_3$ |

**Other compounds with this structure**

- $Ag_2SO_4$ and $Cr_2SO_4$

- $Na_2SO_4$ has eight known anhydrous phases. The thenardite phase is "reported to be stable between 32 °C and about 180 °C" (Nord, 1973), but the data reported here was taken on synthetic thenardite at 25 °C.

**Face-centered Orthorhombic primitive vectors:**

$$\mathbf{a}_1 = \frac{1}{2}\,b\,\hat{\mathbf{y}} + \frac{1}{2}\,c\,\hat{\mathbf{z}}$$
$$\mathbf{a}_2 = \frac{1}{2}\,a\,\hat{\mathbf{x}} + \frac{1}{2}\,c\,\hat{\mathbf{z}}$$
$$\mathbf{a}_3 = \frac{1}{2}\,a\,\hat{\mathbf{x}} + \frac{1}{2}\,b\,\hat{\mathbf{y}}$$

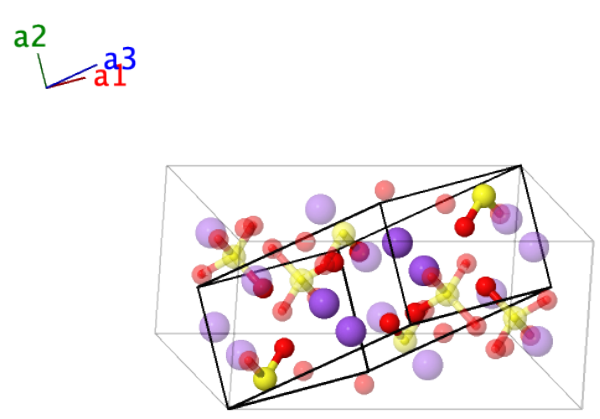

**Basis vectors:**

| | | Lattice Coordinates | | Cartesian Coordinates | Wyckoff Position | Atom Type |
|---|---|---|---|---|---|---|
| $\mathbf{B}_1$ | = | $\frac{1}{8}\,\mathbf{a}_1 + \frac{1}{8}\,\mathbf{a}_2 + \frac{1}{8}\,\mathbf{a}_3$ | = | $\frac{1}{8}a\,\mathbf{\hat{x}} + \frac{1}{8}b\,\mathbf{\hat{y}} + \frac{1}{8}c\,\mathbf{\hat{z}}$ | (8$a$) | S |
| $\mathbf{B}_2$ | = | $\frac{7}{8}\,\mathbf{a}_1 + \frac{7}{8}\,\mathbf{a}_2 + \frac{7}{8}\,\mathbf{a}_3$ | = | $\frac{7}{8}a\,\mathbf{\hat{x}} + \frac{7}{8}b\,\mathbf{\hat{y}} + \frac{7}{8}c\,\mathbf{\hat{z}}$ | (8$a$) | S |
| $\mathbf{B}_3$ | = | $z_2\,\mathbf{a}_1 + z_2\,\mathbf{a}_2 + \left(\frac{1}{4} - z_2\right)\,\mathbf{a}_3$ | = | $\frac{1}{8}a\,\mathbf{\hat{x}} + \frac{1}{8}b\,\mathbf{\hat{y}} + z_2c\,\mathbf{\hat{z}}$ | (16$g$) | Na |
| $\mathbf{B}_4$ | = | $\left(\frac{1}{4} - z_2\right)\,\mathbf{a}_1 + \left(\frac{1}{4} - z_2\right)\,\mathbf{a}_2 + z_2\,\mathbf{a}_3$ | = | $\frac{1}{8}a\,\mathbf{\hat{x}} + \frac{1}{8}b\,\mathbf{\hat{y}} + \left(\frac{1}{4} - z_2\right)c\,\mathbf{\hat{z}}$ | (16$g$) | Na |
| $\mathbf{B}_5$ | = | $-z_2\,\mathbf{a}_1 - z_2\,\mathbf{a}_2 + \left(\frac{3}{4} + z_2\right)\,\mathbf{a}_3$ | = | $\frac{3}{8}a\,\mathbf{\hat{x}} + \frac{3}{8}b\,\mathbf{\hat{y}} - z_2c\,\mathbf{\hat{z}}$ | (16$g$) | Na |
| $\mathbf{B}_6$ | = | $\left(\frac{3}{4} + z_2\right)\,\mathbf{a}_1 + \left(\frac{3}{4} + z_2\right)\,\mathbf{a}_2 - z_2\,\mathbf{a}_3$ | = | $\frac{3}{8}a\,\mathbf{\hat{x}} + \frac{3}{8}b\,\mathbf{\hat{y}} + \left(\frac{3}{4} + z_2\right)c\,\mathbf{\hat{z}}$ | (16$g$) | Na |
| $\mathbf{B}_7$ | = | $(-x_3 + y_3 + z_3)\,\mathbf{a}_1 + (x_3 - y_3 + z_3)\,\mathbf{a}_2 + (x_3 + y_3 - z_3)\,\mathbf{a}_3$ | = | $x_3a\,\mathbf{\hat{x}} + y_3b\,\mathbf{\hat{y}} + z_3c\,\mathbf{\hat{z}}$ | (32$h$) | O |
| $\mathbf{B}_8$ | = | $(x_3 - y_3 + z_3)\,\mathbf{a}_1 + (-x_3 + y_3 + z_3)\,\mathbf{a}_2 + \left(\frac{1}{2} - x_3 - y_3 - z_3\right)\,\mathbf{a}_3$ | = | $\left(\frac{1}{4} - x_3\right)a\,\mathbf{\hat{x}} + \left(\frac{1}{4} - y_3\right)b\,\mathbf{\hat{y}} + z_3c\,\mathbf{\hat{z}}$ | (32$h$) | O |
| $\mathbf{B}_9$ | = | $(x_3 + y_3 - z_3)\,\mathbf{a}_1 + \left(\frac{1}{2} - x_3 - y_3 - z_3\right)\,\mathbf{a}_2 + (-x_3 + y_3 + z_3)\,\mathbf{a}_3$ | = | $\left(\frac{1}{4} - x_3\right)a\,\mathbf{\hat{x}} + y_3b\,\mathbf{\hat{y}} + \left(\frac{1}{4} - z_3\right)c\,\mathbf{\hat{z}}$ | (32$h$) | O |
| $\mathbf{B}_{10}$ | = | $\left(\frac{1}{2} - x_3 - y_3 - z_3\right)\,\mathbf{a}_1 + (x_3 + y_3 - z_3)\,\mathbf{a}_2 + (x_3 - y_3 + z_3)\,\mathbf{a}_3$ | = | $x_3a\,\mathbf{\hat{x}} + \left(\frac{1}{4} - y_3\right)b\,\mathbf{\hat{y}} + \left(\frac{1}{4} - z_3\right)c\,\mathbf{\hat{z}}$ | (32$h$) | O |
| $\mathbf{B}_{11}$ | = | $(x_3 - y_3 - z_3)\,\mathbf{a}_1 + (-x_3 + y_3 - z_3)\,\mathbf{a}_2 + (-x_3 - y_3 + z_3)\,\mathbf{a}_3$ | = | $-x_3a\,\mathbf{\hat{x}} - y_3b\,\mathbf{\hat{y}} - z_3c\,\mathbf{\hat{z}}$ | (32$h$) | O |
| $\mathbf{B}_{12}$ | = | $(-x_3 + y_3 - z_3)\,\mathbf{a}_1 + (x_3 - y_3 - z_3)\,\mathbf{a}_2 + \left(\frac{1}{2} + x_3 + y_3 + z_3\right)\,\mathbf{a}_3$ | = | $\left(\frac{1}{4} + x_3\right)a\,\mathbf{\hat{x}} + \left(\frac{1}{4} + y_3\right)b\,\mathbf{\hat{y}} - z_3c\,\mathbf{\hat{z}}$ | (32$h$) | O |
| $\mathbf{B}_{13}$ | = | $(-x_3 - y_3 + z_3)\,\mathbf{a}_1 + \left(\frac{1}{2} + x_3 + y_3 + z_3\right)\,\mathbf{a}_2 + (x_3 - y_3 - z_3)\,\mathbf{a}_3$ | = | $\left(\frac{1}{4} + x_3\right)a\,\mathbf{\hat{x}} - y_3b\,\mathbf{\hat{y}} + \left(\frac{1}{4} + z_3\right)c\,\mathbf{\hat{z}}$ | (32$h$) | O |
| $\mathbf{B}_{14}$ | = | $\left(\frac{1}{2} + x_3 + y_3 + z_3\right)\,\mathbf{a}_1 + (-x_3 - y_3 + z_3)\,\mathbf{a}_2 + (-x_3 + y_3 - z_3)\,\mathbf{a}_3$ | = | $-x_3a\,\mathbf{\hat{x}} + \left(\frac{1}{4} + y_3\right)b\,\mathbf{\hat{y}} + \left(\frac{1}{4} + z_3\right)c\,\mathbf{\hat{z}}$ | (32$h$) | O |

---

---

**Geometry files:**

- CIF: pp. 1664
- POSCAR: pp. 1664

# $Mg_2Cu$ ($C_b$) Structure: AB2_oF48_70_g_fg

http://aflow.org/prototype-encyclopedia/AB2_oF48_70_g_fg

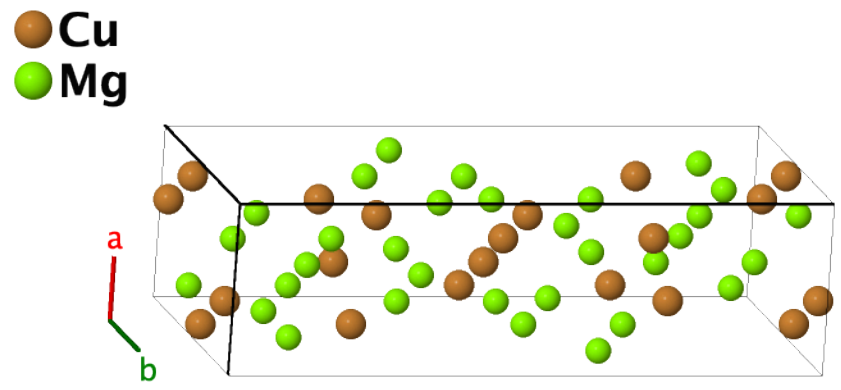

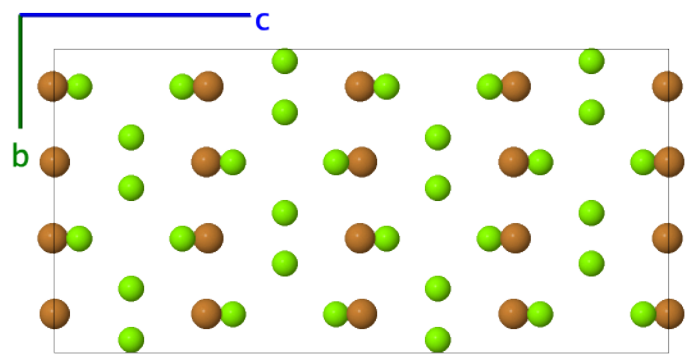

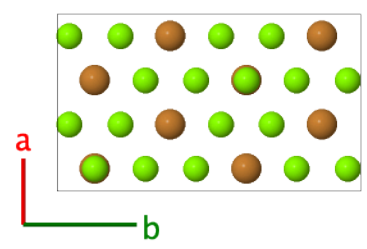

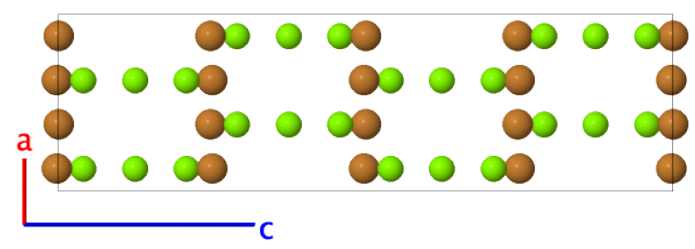

| | | |
|---|---|---|
| **Prototype** | : | $CuMg_2$ |
| **AFLOW prototype label** | : | AB2_oF48_70_g_fg |
| ***Strukturbericht* designation** | : | $C_b$ |
| **Pearson symbol** | : | oF48 |
| **Space group number** | : | 70 |
| **Space group symbol** | : | $Fddd$ |
| **AFLOW prototype command** | : | `aflow --proto=AB2_oF48_70_g_fg --params=`$a, b/a, c/a, y_1, z_2, z_3$ |

### Other compounds with this structure

- $In_2Ir$, $NbSn_2$, $TiBi_2$, FeGaGe, and SbSnTi

### Face-centered Orthorhombic primitive vectors:

$$\mathbf{a}_1 = \frac{1}{2}\,b\,\hat{\mathbf{y}} + \frac{1}{2}\,c\,\hat{\mathbf{z}}$$
$$\mathbf{a}_2 = \frac{1}{2}\,a\,\hat{\mathbf{x}} + \frac{1}{2}\,c\,\hat{\mathbf{z}}$$
$$\mathbf{a}_3 = \frac{1}{2}\,a\,\hat{\mathbf{x}} + \frac{1}{2}\,b\,\hat{\mathbf{y}}$$

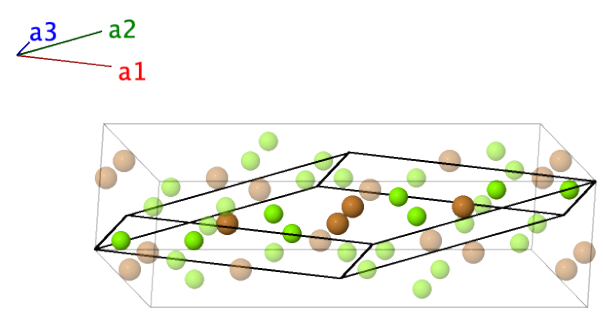

### Basis vectors:

| | | Lattice Coordinates | | Cartesian Coordinates | Wyckoff Position | Atom Type |
|---|---|---|---|---|---|---|
| $\mathbf{B}_1$ | = | $y_1\,\mathbf{a}_1 + \left(\frac{1}{4} - y_1\right)\mathbf{a}_2 + y_1\,\mathbf{a}_3$ | = | $\frac{1}{8}a\,\hat{\mathbf{x}} + y_1 b\,\hat{\mathbf{y}} + \frac{1}{8}c\,\hat{\mathbf{z}}$ | (16$f$) | Mg I |
| $\mathbf{B}_2$ | = | $\left(\frac{1}{4} - y_1\right)\mathbf{a}_1 + y_1\,\mathbf{a}_2 + \left(\frac{1}{4} - y_1\right)\mathbf{a}_3$ | = | $\frac{1}{8}a\,\hat{\mathbf{x}} + \left(\frac{1}{4} - y_1\right)b\,\hat{\mathbf{y}} + \frac{1}{8}c\,\hat{\mathbf{z}}$ | (16$f$) | Mg I |
| $\mathbf{B}_3$ | = | $-y_1\,\mathbf{a}_1 + \left(\frac{3}{4} + y_1\right)\mathbf{a}_2 - y_1\,\mathbf{a}_3$ | = | $\frac{3}{8}a\,\hat{\mathbf{x}} - y_1 b\,\hat{\mathbf{y}} + \frac{3}{8}c\,\hat{\mathbf{z}}$ | (16$f$) | Mg I |
| $\mathbf{B}_4$ | = | $\left(\frac{3}{4} + y_1\right)\mathbf{a}_1 - y_1\,\mathbf{a}_2 + \left(\frac{3}{4} + y_1\right)\mathbf{a}_3$ | = | $\frac{3}{8}a\,\hat{\mathbf{x}} + \left(\frac{3}{4} + y_1\right)b\,\hat{\mathbf{y}} + \frac{3}{8}c\,\hat{\mathbf{z}}$ | (16$f$) | Mg I |
| $\mathbf{B}_5$ | = | $z_2\,\mathbf{a}_1 + z_2\,\mathbf{a}_2 + \left(\frac{1}{4} - z_2\right)\mathbf{a}_3$ | = | $\frac{1}{8}a\,\hat{\mathbf{x}} + \frac{1}{8}b\,\hat{\mathbf{y}} + z_2 c\,\hat{\mathbf{z}}$ | (16$g$) | Cu |
| $\mathbf{B}_6$ | = | $\left(\frac{1}{4} - z_2\right)\mathbf{a}_1 + \left(\frac{1}{4} - z_2\right)\mathbf{a}_2 + z_2\,\mathbf{a}_3$ | = | $\frac{1}{8}a\,\hat{\mathbf{x}} + \frac{1}{8}b\,\hat{\mathbf{y}} + \left(\frac{1}{4} - z_2\right)c\,\hat{\mathbf{z}}$ | (16$g$) | Cu |
| $\mathbf{B}_{7}$ | = | $-z_2\,\mathbf{a}_1 - z_2\,\mathbf{a}_2 + \left(\frac{3}{4} + z_2\right)\,\mathbf{a}_3$ | = | $\frac{3}{8}a\,\hat{\mathbf{x}} + \frac{3}{8}b\,\hat{\mathbf{y}} - z_2c\,\hat{\mathbf{z}}$ | (16$g$) | Cu |
| $\mathbf{B}_{8}$ | = | $\left(\frac{3}{4} + z_2\right)\,\mathbf{a}_1 + \left(\frac{3}{4} + z_2\right)\,\mathbf{a}_2 - z_2\,\mathbf{a}_3$ | = | $\frac{3}{8}a\,\hat{\mathbf{x}} + \frac{3}{8}b\,\hat{\mathbf{y}} + \left(\frac{3}{4} + z_2\right)c\,\hat{\mathbf{z}}$ | (16$g$) | Cu |
| $\mathbf{B}_{9}$ | = | $z_3\,\mathbf{a}_1 + z_3\,\mathbf{a}_2 + \left(\frac{1}{4} - z_3\right)\,\mathbf{a}_3$ | = | $\frac{1}{8}a\,\hat{\mathbf{x}} + \frac{1}{8}b\,\hat{\mathbf{y}} + z_3c\,\hat{\mathbf{z}}$ | (16$g$) | Mg II |
| $\mathbf{B}_{10}$ | = | $\left(\frac{1}{4} - z_3\right)\,\mathbf{a}_1 + \left(\frac{1}{4} - z_3\right)\,\mathbf{a}_2 + z_3\,\mathbf{a}_3$ | = | $\frac{1}{8}a\,\hat{\mathbf{x}} + \frac{1}{8}b\,\hat{\mathbf{y}} + \left(\frac{1}{4} - z_3\right)c\,\hat{\mathbf{z}}$ | (16$g$) | Mg II |
| $\mathbf{B}_{11}$ | = | $-z_3\,\mathbf{a}_1 - z_3\,\mathbf{a}_2 + \left(\frac{3}{4} + z_3\right)\,\mathbf{a}_3$ | = | $\frac{3}{8}a\,\hat{\mathbf{x}} + \frac{3}{8}b\,\hat{\mathbf{y}} - z_3c\,\hat{\mathbf{z}}$ | (16$g$) | Mg II |
| $\mathbf{B}_{12}$ | = | $\left(\frac{3}{4} + z_3\right)\,\mathbf{a}_1 + \left(\frac{3}{4} + z_3\right)\,\mathbf{a}_2 - z_3\,\mathbf{a}_3$ | = | $\frac{3}{8}a\,\hat{\mathbf{x}} + \frac{3}{8}b\,\hat{\mathbf{y}} + \left(\frac{3}{4} + z_3\right)c\,\hat{\mathbf{z}}$ | (16$g$) | Mg II |

**Geometry files:**

- CIF: pp. 1664
- POSCAR: pp. 1665

# High-Temperature Cryolite ($Na_3AlF_6$) Structure: AB6C3_oI20_71_a_in_cj

http://aflow.org/prototype-encyclopedia/AB6C3_oI20_71_a_in_cj

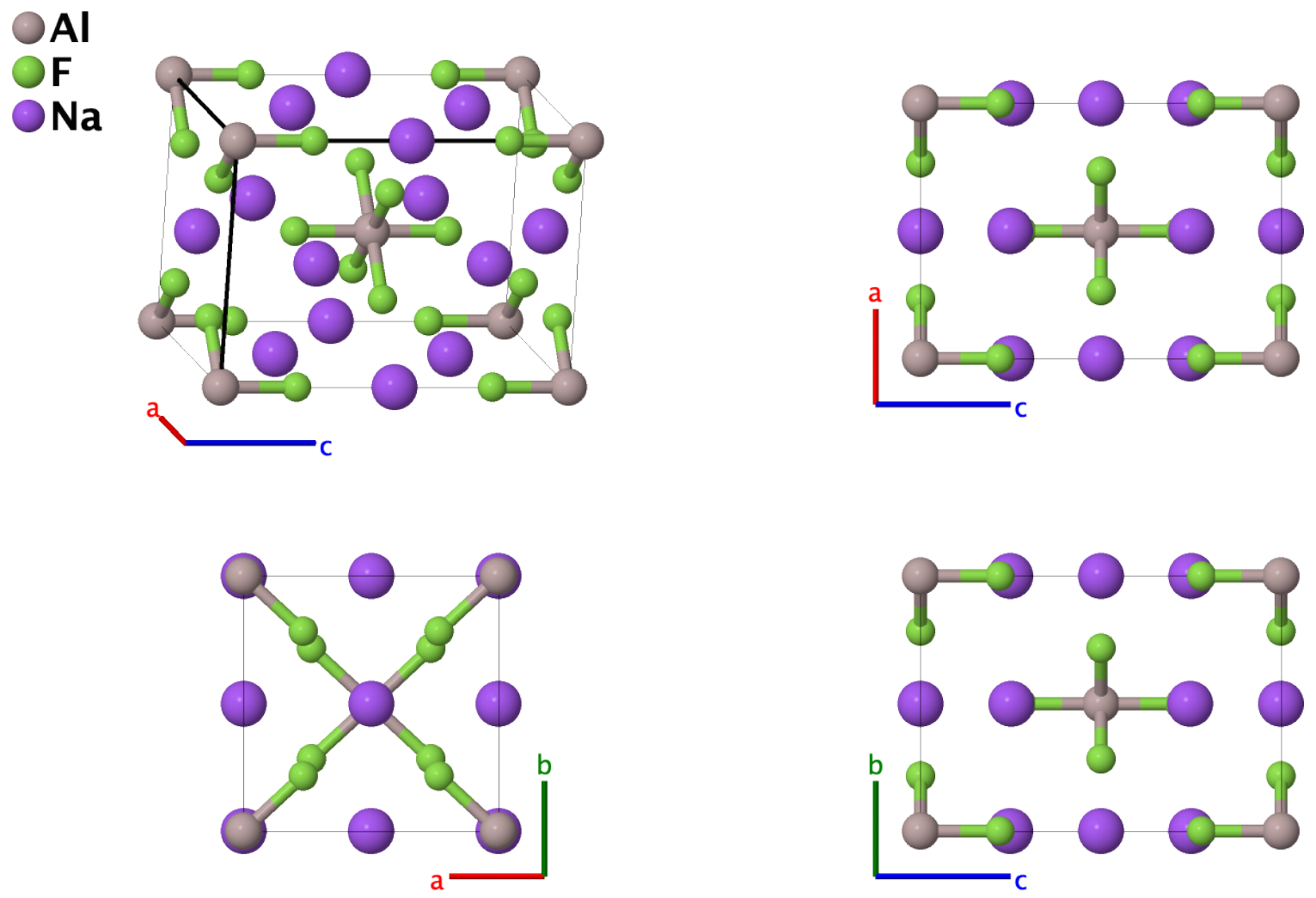

| | | |
|---|---|---|
| **Prototype** | : | $AlF_6Na_3$ |
| **AFLOW prototype label** | : | AB6C3_oI20_71_a_in_cj |
| ***Strukturbericht* designation** | : | None |
| **Pearson symbol** | : | oI20 |
| **Space group number** | : | 71 |
| **Space group symbol** | : | *Immm* |
| **AFLOW prototype command** | : | `aflow --proto=AB6C3_oI20_71_a_in_cj --params=`$a, b/a, c/a, z_3, z_4, x_5, y_5$ |

- Cryolite undergoes a phase transition from the monoclinic $P2_1/c$ $J2_6$ phase to this orthorhombic phase at 890 K. We show structural data taken at 900 K.

**Body-centered Orthorhombic primitive vectors:**

$$\mathbf{a}_1 = -\frac{1}{2}\,a\,\hat{\mathbf{x}} + \frac{1}{2}\,b\,\hat{\mathbf{y}} + \frac{1}{2}\,c\,\hat{\mathbf{z}}$$
$$\mathbf{a}_2 = \frac{1}{2}\,a\,\hat{\mathbf{x}} - \frac{1}{2}\,b\,\hat{\mathbf{y}} + \frac{1}{2}\,c\,\hat{\mathbf{z}}$$
$$\mathbf{a}_3 = \frac{1}{2}\,a\,\hat{\mathbf{x}} + \frac{1}{2}\,b\,\hat{\mathbf{y}} - \frac{1}{2}\,c\,\hat{\mathbf{z}}$$

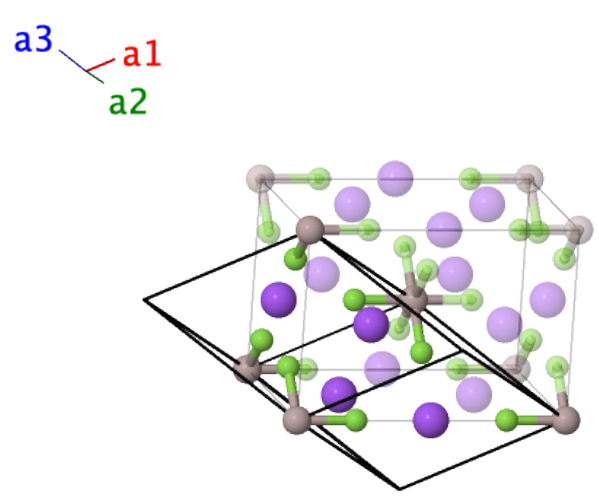

**Basis vectors:**

| | | Lattice Coordinates | | Cartesian Coordinates | Wyckoff Position | Atom Type |
|---|---|---|---|---|---|---|
| $\mathbf{B}_1$ | = | $0\,\mathbf{a}_1 + 0\,\mathbf{a}_2 + 0\,\mathbf{a}_3$ | = | $0\,\hat{\mathbf{x}} + 0\,\hat{\mathbf{y}} + 0\,\hat{\mathbf{z}}$ | (2*a*) | Al |
| $\mathbf{B}_2$ | = | $\frac{1}{2}\,\mathbf{a}_1 + \frac{1}{2}\,\mathbf{a}_2$ | = | $\frac{1}{2}c\,\hat{\mathbf{z}}$ | (2*c*) | Na I |
| $\mathbf{B}_3$ | = | $z_3\,\mathbf{a}_1 + z_3\,\mathbf{a}_2$ | = | $z_3 c\,\hat{\mathbf{z}}$ | (4*i*) | F I |
| $\mathbf{B}_4$ | = | $-z_3\,\mathbf{a}_1 - z_3\,\mathbf{a}_2$ | = | $-z_3 c\,\hat{\mathbf{z}}$ | (4*i*) | F I |
| $\mathbf{B}_5$ | = | $z_4\,\mathbf{a}_1 + \left(\frac{1}{2} + z_4\right)\,\mathbf{a}_2 + \frac{1}{2}\,\mathbf{a}_3$ | = | $\frac{1}{2}a\,\hat{\mathbf{x}} + z_4 c\,\hat{\mathbf{z}}$ | (4*j*) | Na II |
| $\mathbf{B}_6$ | = | $-z_4\,\mathbf{a}_1 + \left(\frac{1}{2} - z_4\right)\,\mathbf{a}_2 + \frac{1}{2}\,\mathbf{a}_3$ | = | $\frac{1}{2}a\,\hat{\mathbf{x}} - z_4 c\,\hat{\mathbf{z}}$ | (4*j*) | Na II |
| $\mathbf{B}_7$ | = | $y_5\,\mathbf{a}_1 + x_5\,\mathbf{a}_2 + (x_5 + y_5)\,\mathbf{a}_3$ | = | $x_5 a\,\hat{\mathbf{x}} + y_5 b\,\hat{\mathbf{y}}$ | (8*n*) | F II |
| $\mathbf{B}_8$ | = | $-y_5\,\mathbf{a}_1 - x_5\,\mathbf{a}_2 + (-x_5 - y_5)\,\mathbf{a}_3$ | = | $-x_5 a\,\hat{\mathbf{x}} - y_5 b\,\hat{\mathbf{y}}$ | (8*n*) | F II |
| $\mathbf{B}_9$ | = | $y_5\,\mathbf{a}_1 - x_5\,\mathbf{a}_2 + (-x_5 + y_5)\,\mathbf{a}_3$ | = | $-x_5 a\,\hat{\mathbf{x}} + y_5 b\,\hat{\mathbf{y}}$ | (8*n*) | F II |
| $\mathbf{B}_{10}$ | = | $-y_5\,\mathbf{a}_1 + x_5\,\mathbf{a}_2 + (x_5 - y_5)\,\mathbf{a}_3$ | = | $x_5 a\,\hat{\mathbf{x}} - y_5 b\,\hat{\mathbf{y}}$ | (8*n*) | F II |

**Geometry files:**

- CIF: pp. 1665
- POSCAR: pp. 1665

# $CsFeS_2$ (100 K) Structure: ABC2_oI16_71_g_i_eh

http://aflow.org/prototype-encyclopedia/ABC2_oI16_71_g_i_eh

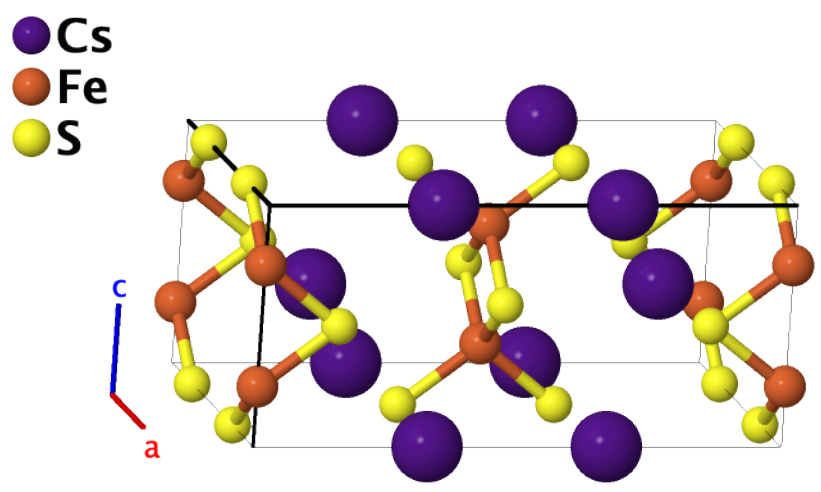

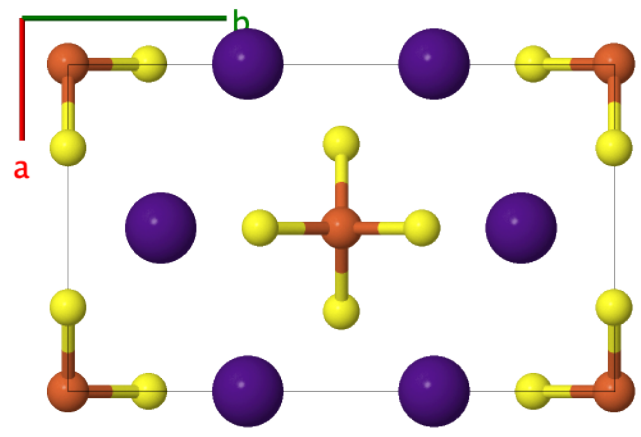

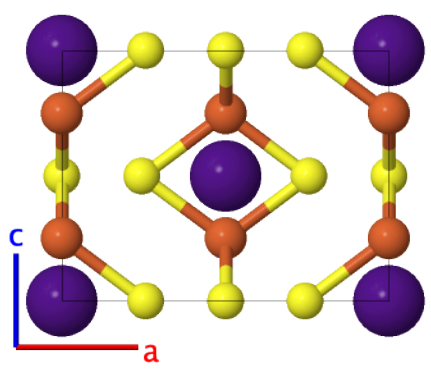

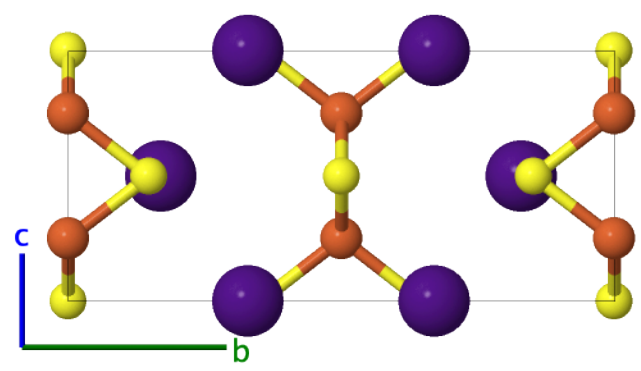

| | | |
|---|---|---|
| **Prototype** | : | $CsFeS_2$ |
| **AFLOW prototype label** | : | ABC2_oI16_71_g_i_eh |
| ***Strukturbericht* designation** | : | None |
| **Pearson symbol** | : | oI16 |
| **Space group number** | : | 71 |
| **Space group symbol** | : | *Immm* |
| **AFLOW prototype command** | : | `aflow --proto=ABC2_oI16_71_g_i_eh` `--params=`$a, b/a, c/a, x_1, y_2, y_3, z_4$ |

### Other compounds with this structure

- $RbFeS_2$

- This structure is stable at 100 K and above. At 40 K the structure is triclinic, with as yet undetermined atomic positions. (Ito, 1985)

### Body-centered Orthorhombic primitive vectors:

$$\mathbf{a}_1 = -\frac{1}{2}\,a\,\hat{\mathbf{x}} + \frac{1}{2}\,b\,\hat{\mathbf{y}} + \frac{1}{2}\,c\,\hat{\mathbf{z}}$$
$$\mathbf{a}_2 = \frac{1}{2}\,a\,\hat{\mathbf{x}} - \frac{1}{2}\,b\,\hat{\mathbf{y}} + \frac{1}{2}\,c\,\hat{\mathbf{z}}$$
$$\mathbf{a}_3 = \frac{1}{2}\,a\,\hat{\mathbf{x}} + \frac{1}{2}\,b\,\hat{\mathbf{y}} - \frac{1}{2}\,c\,\hat{\mathbf{z}}$$

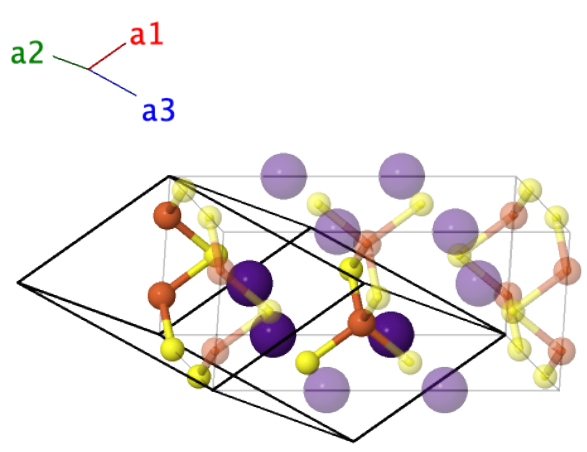

### Basis vectors:

| | | Lattice Coordinates | | Cartesian Coordinates | Wyckoff Position | Atom Type |
|---|---|---|---|---|---|---|
| $\mathbf{B}_1$ | = | $x_1\,\mathbf{a}_2 + x_1\,\mathbf{a}_3$ | = | $x_1 a\,\hat{\mathbf{x}}$ | (4*e*) | S I |
| $\mathbf{B}_2$ | = | $-x_1\,\mathbf{a}_2 - x_1\,\mathbf{a}_3$ | = | $-x_1 a\,\hat{\mathbf{x}}$ | (4*e*) | S I |
| $\mathbf{B}_3$ | = | $y_2\,\mathbf{a}_1 + y_2\,\mathbf{a}_3$ | = | $y_2 b\,\hat{\mathbf{y}}$ | (4$g$) | Cs |
| $\mathbf{B}_4$ | = | $-y_2\,\mathbf{a}_1 - y_2\,\mathbf{a}_3$ | = | $-y_2 b\,\hat{\mathbf{y}}$ | (4$g$) | Cs |
| $\mathbf{B}_5$ | = | $\left(\frac{1}{2} + y_3\right)\mathbf{a}_1 + \frac{1}{2}\,\mathbf{a}_2 + y_3\,\mathbf{a}_3$ | = | $y_3 b\,\hat{\mathbf{y}} + \frac{1}{2}c\,\hat{\mathbf{z}}$ | (4$h$) | S II |
| $\mathbf{B}_6$ | = | $\left(\frac{1}{2} - y_3\right)\mathbf{a}_1 + \frac{1}{2}\,\mathbf{a}_2 - y_3\,\mathbf{a}_3$ | = | $-y_3 b\,\hat{\mathbf{y}} + \frac{1}{2}c\,\hat{\mathbf{z}}$ | (4$h$) | S II |
| $\mathbf{B}_7$ | = | $z_4\,\mathbf{a}_1 + z_4\,\mathbf{a}_2$ | = | $z_4 c\,\hat{\mathbf{z}}$ | (4$i$) | Fe |
| $\mathbf{B}_8$ | = | $-z_4\,\mathbf{a}_1 - z_4\,\mathbf{a}_2$ | = | $-z_4 c\,\hat{\mathbf{z}}$ | (4$i$) | Fe |

---

**Found in:**

---

**Geometry files:**

- CIF: pp. 1665
- POSCAR: pp. 1666

# CsO Structure: AB_oI8_71_g_i

http://aflow.org/prototype-encyclopedia/AB_oI8_71_g_i

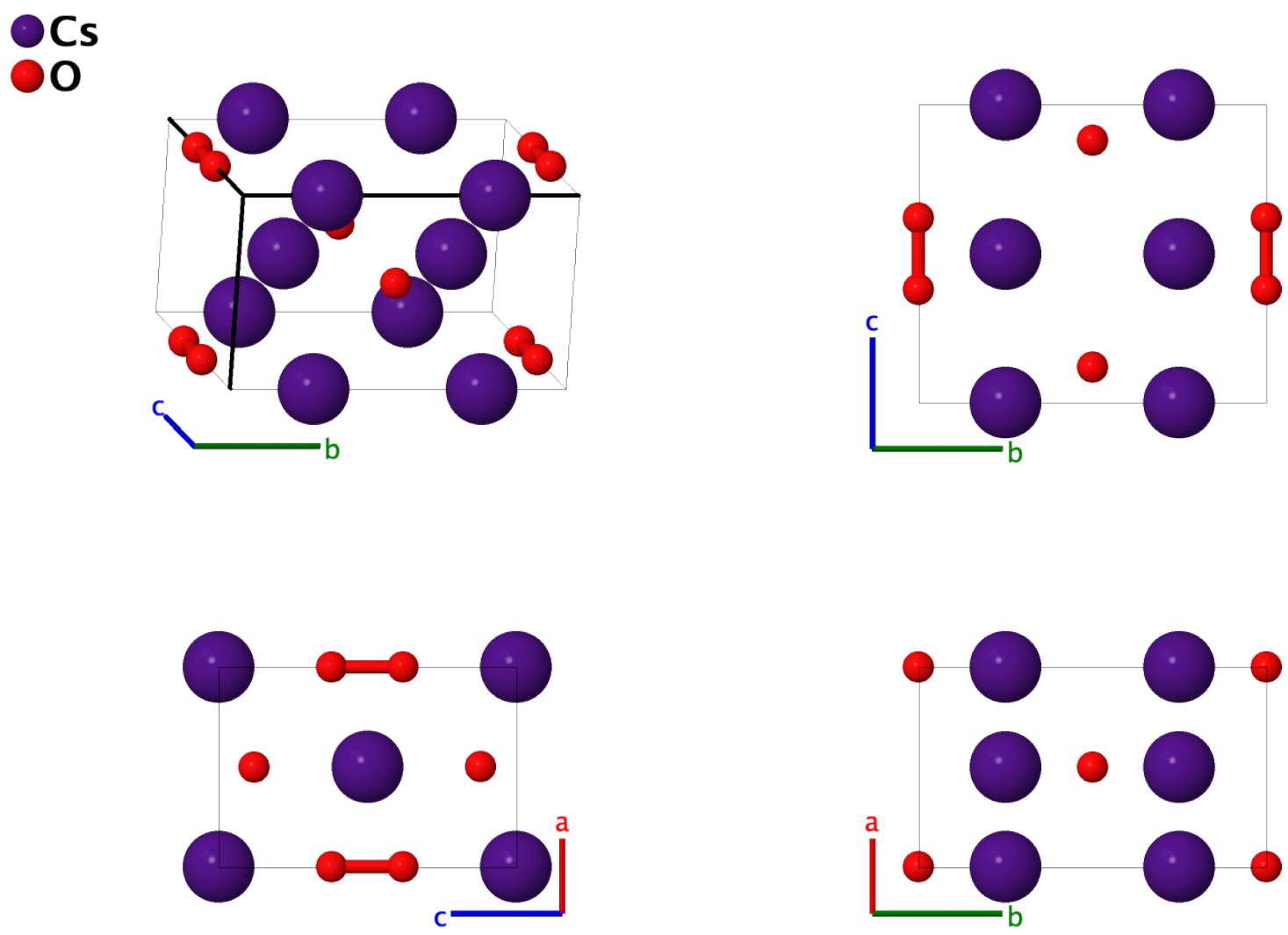

| | | |
|---|---|---|
| **Prototype** | : | CsO |
| **AFLOW prototype label** | : | AB_oI8_71_g_i |
| ***Strukturbericht* designation** | : | None |
| **Pearson symbol** | : | oI8 |
| **Space group number** | : | 71 |
| **Space group symbol** | : | *Immm* |
| **AFLOW prototype command** | : | `aflow --proto=AB_oI8_71_g_i` `--params=`$a, b/a, c/a, y_1, z_2$ |

- (Massalski, 1990) credits the discovery of this structure to (Rengade, 1909), but we have been unable to obtain a copy of this reference. (Downs, 2003) quotes (Wyckoff, 1963) giving an *Immm*-oI8 structure for CsO. Since this is the same space group and Pearson symbol as found in Massalski, we use Wyckoff's data.

**Body-centered Orthorhombic primitive vectors:**

$$
\begin{aligned}
\mathbf{a}_1 &= -\tfrac{1}{2}\,a\,\hat{\mathbf{x}} + \tfrac{1}{2}\,b\,\hat{\mathbf{y}} + \tfrac{1}{2}\,c\,\hat{\mathbf{z}} \\
\mathbf{a}_2 &= \tfrac{1}{2}\,a\,\hat{\mathbf{x}} - \tfrac{1}{2}\,b\,\hat{\mathbf{y}} + \tfrac{1}{2}\,c\,\hat{\mathbf{z}} \\
\mathbf{a}_3 &= \tfrac{1}{2}\,a\,\hat{\mathbf{x}} + \tfrac{1}{2}\,b\,\hat{\mathbf{y}} - \tfrac{1}{2}\,c\,\hat{\mathbf{z}}
\end{aligned}
$$

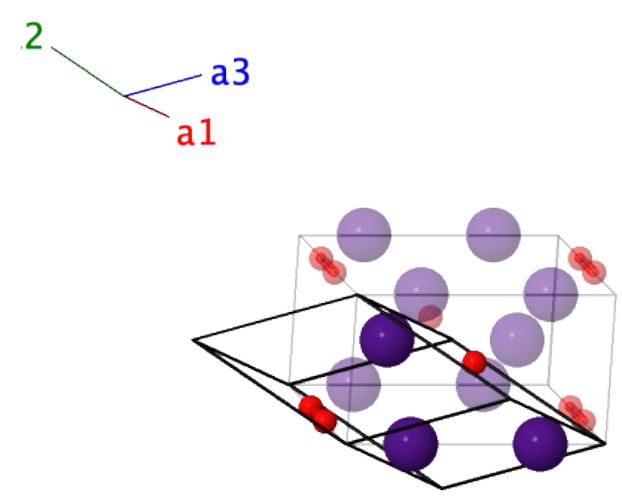

**Basis vectors:**

| | | Lattice Coordinates | | Cartesian Coordinates | Wyckoff Position | Atom Type |
|---|---|---|---|---|---|---|
| $\mathbf{B}_1$ | = | $y_1\,\mathbf{a}_1 + y_1\,\mathbf{a}_3$ | = | $y_1 b\,\hat{\mathbf{y}}$ | (4*g*) | Cs |
| $\mathbf{B}_2$ | = | $-y_1\,\mathbf{a}_1 - y_1\,\mathbf{a}_3$ | = | $-y_1 b\,\hat{\mathbf{y}}$ | (4*g*) | Cs |
| $\mathbf{B}_3$ | = | $z_2\,\mathbf{a}_1 + z_2\,\mathbf{a}_2$ | = | $z_2 c\,\hat{\mathbf{z}}$ | (4*i*) | O |
| $\mathbf{B}_4$ | = | $-z_2\,\mathbf{a}_1 - z_2\,\mathbf{a}_2$ | = | $-z_2 c\,\hat{\mathbf{z}}$ | (4$i$) | O |

---

**Found in:**

---

**Geometry files:**

- CIF: pp. 1666
- POSCAR: pp. 1666

# $Ga_2Mg_5$ ($D8_g$) Structure: A2B5_oI28_72_j_bfj

http://aflow.org/prototype-encyclopedia/A2B5_oI28_72_j_bfj

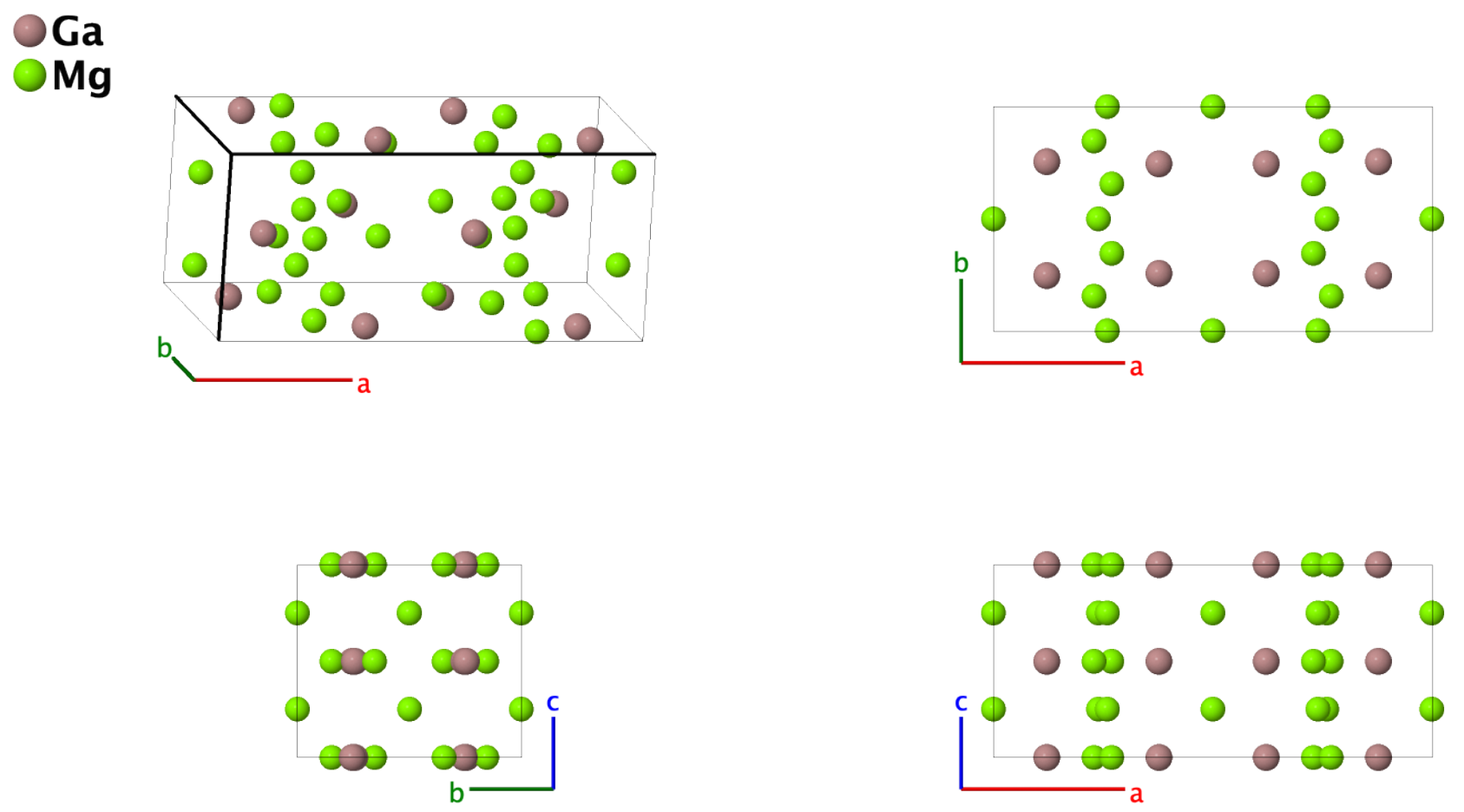

| | | |
|---|---|---|
| **Prototype** | : | $Ga_2Mg_5$ |
| **AFLOW prototype label** | : | A2B5_oI28_72_j_bfj |
| ***Strukturbericht* designation** | : | $D8_g$ |
| **Pearson symbol** | : | oI28 |
| **Space group number** | : | 72 |
| **Space group symbol** | : | *Ibam* |
| **AFLOW prototype command** | : | `aflow --proto=A2B5_oI28_72_j_bfj --params=`$a, b/a, c/a, x_2, x_3, y_3, x_4, y_4$ |

### Other compounds with this structure

- $As_2Cu_5$, $Tl_2Mg_5$, and $Mn_2Ge_5$

### Body-centered Orthorhombic primitive vectors:

$$\mathbf{a}_1 = -\frac{1}{2}\,a\,\hat{\mathbf{x}} + \frac{1}{2}\,b\,\hat{\mathbf{y}} + \frac{1}{2}\,c\,\hat{\mathbf{z}}$$
$$\mathbf{a}_2 = \frac{1}{2}\,a\,\hat{\mathbf{x}} - \frac{1}{2}\,b\,\hat{\mathbf{y}} + \frac{1}{2}\,c\,\hat{\mathbf{z}}$$
$$\mathbf{a}_3 = \frac{1}{2}\,a\,\hat{\mathbf{x}} + \frac{1}{2}\,b\,\hat{\mathbf{y}} - \frac{1}{2}\,c\,\hat{\mathbf{z}}$$

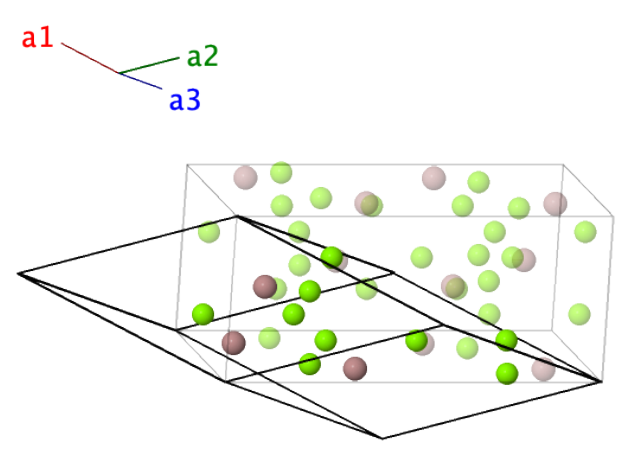

### Basis vectors:

| | | Lattice Coordinates | | Cartesian Coordinates | Wyckoff Position | Atom Type |
|---|---|---|---|---|---|---|
| $\mathbf{B}_1$ | = | $\frac{1}{4}\,\mathbf{a}_1 + \frac{3}{4}\,\mathbf{a}_2 + \frac{1}{2}\,\mathbf{a}_3$ | = | $\frac{1}{2}a\,\hat{\mathbf{x}} + \frac{1}{4}c\,\hat{\mathbf{z}}$ | (4*b*) | Mg I |
| $\mathbf{B}_2$ | = | $\frac{3}{4}\,\mathbf{a}_1 + \frac{1}{4}\,\mathbf{a}_2 + \frac{1}{2}\,\mathbf{a}_3$ | = | $\frac{1}{2}b\,\hat{\mathbf{y}} + \frac{1}{4}c\,\hat{\mathbf{z}}$ | (4*b*) | Mg I |
| $\mathbf{B}_3$ | = | $\frac{1}{4}\,\mathbf{a}_1 + \left(\frac{1}{4} + x_2\right)\,\mathbf{a}_2 + x_2\,\mathbf{a}_3$ | = | $x_2 a\,\hat{\mathbf{x}} + \frac{1}{4}c\,\hat{\mathbf{z}}$ | (8*f*) | Mg II |
| $\mathbf{B}_4$ | = | $\frac{1}{4}\,\mathbf{a}_1 + \left(\frac{1}{4} - x_2\right)\,\mathbf{a}_2 - x_2\,\mathbf{a}_3$ | = | $-x_2 a\,\hat{\mathbf{x}} + \frac{1}{4}c\,\hat{\mathbf{z}}$ | (8*f*) | Mg II |
| $\mathbf{B}_5$ | = | $\frac{3}{4}\,\mathbf{a}_1 + \left(\frac{3}{4} - x_2\right)\,\mathbf{a}_2 - x_2\,\mathbf{a}_3$ | = | $-x_2 a\,\hat{\mathbf{x}} + \frac{3}{4}c\,\hat{\mathbf{z}}$ | (8*f*) | Mg II |
| $\mathbf{B}_6$ | = | $\frac{3}{4}\,\mathbf{a}_1 + \left(\frac{3}{4} + x_2\right)\,\mathbf{a}_2 + x_2\,\mathbf{a}_3$ | = | $x_2 a\,\hat{\mathbf{x}} + \frac{3}{4}c\,\hat{\mathbf{z}}$ | (8*f*) | Mg II |
| $\mathbf{B}_7$ | = | $y_3\,\mathbf{a}_1 + x_3\,\mathbf{a}_2 + (x_3 + y_3)\,\mathbf{a}_3$ | = | $x_3 a\,\hat{\mathbf{x}} + y_3 b\,\hat{\mathbf{y}}$ | $(8j)$ | Ga |
| $\mathbf{B}_8$ | = | $-y_3\,\mathbf{a}_1 - x_3\,\mathbf{a}_2 + (-x_3 - y_3)\,\mathbf{a}_3$ | = | $-x_3 a\,\hat{\mathbf{x}} - y_3 b\,\hat{\mathbf{y}}$ | $(8j)$ | Ga |
| $\mathbf{B}_9$ | = | $\left(\frac{1}{2} + y_3\right)\mathbf{a}_1 + \left(\frac{1}{2} - x_3\right)\mathbf{a}_2 + (-x_3 + y_3)\,\mathbf{a}_3$ | = | $-x_3 a\,\hat{\mathbf{x}} + y_3 b\,\hat{\mathbf{y}} + \frac{1}{2}c\,\hat{\mathbf{z}}$ | $(8j)$ | Ga |
| $\mathbf{B}_{10}$ | = | $\left(\frac{1}{2} - y_3\right)\mathbf{a}_1 + \left(\frac{1}{2} + x_3\right)\mathbf{a}_2 + (x_3 - y_3)\,\mathbf{a}_3$ | = | $x_3 a\,\hat{\mathbf{x}} - y_3 b\,\hat{\mathbf{y}} + \frac{1}{2}c\,\hat{\mathbf{z}}$ | $(8j)$ | Ga |
| $\mathbf{B}_{11}$ | = | $y_4\,\mathbf{a}_1 + x_4\,\mathbf{a}_2 + (x_4 + y_4)\,\mathbf{a}_3$ | = | $x_4 a\,\hat{\mathbf{x}} + y_4 b\,\hat{\mathbf{y}}$ | $(8j)$ | Mg III |
| $\mathbf{B}_{12}$ | = | $-y_4\,\mathbf{a}_1 - x_4\,\mathbf{a}_2 + (-x_4 - y_4)\,\mathbf{a}_3$ | = | $-x_4 a\,\hat{\mathbf{x}} - y_4 b\,\hat{\mathbf{y}}$ | $(8j)$ | Mg III |
| $\mathbf{B}_{13}$ | = | $\left(\frac{1}{2} + y_4\right)\mathbf{a}_1 + \left(\frac{1}{2} - x_4\right)\mathbf{a}_2 + (-x_4 + y_4)\,\mathbf{a}_3$ | = | $-x_4 a\,\hat{\mathbf{x}} + y_4 b\,\hat{\mathbf{y}} + \frac{1}{2}c\,\hat{\mathbf{z}}$ | $(8j)$ | Mg III |
| $\mathbf{B}_{14}$ | = | $\left(\frac{1}{2} - y_4\right)\mathbf{a}_1 + \left(\frac{1}{2} + x_4\right)\mathbf{a}_2 + (x_4 - y_4)\,\mathbf{a}_3$ | = | $x_4 a\,\hat{\mathbf{x}} - y_4 b\,\hat{\mathbf{y}} + \frac{1}{2}c\,\hat{\mathbf{z}}$ | $(8j)$ | Mg III |

---

**Found in:**

---

**Geometry files:**

- CIF: pp. 1666
- POSCAR: pp. 1667

# $Zn(NH_3)_2Cl_2$ ($E1_2$) Structure: A2B6C2D_oI44_74_h_ij_i_e

http://aflow.org/prototype-encyclopedia/A2B6C2D_oI44_74_h_ij_i_e

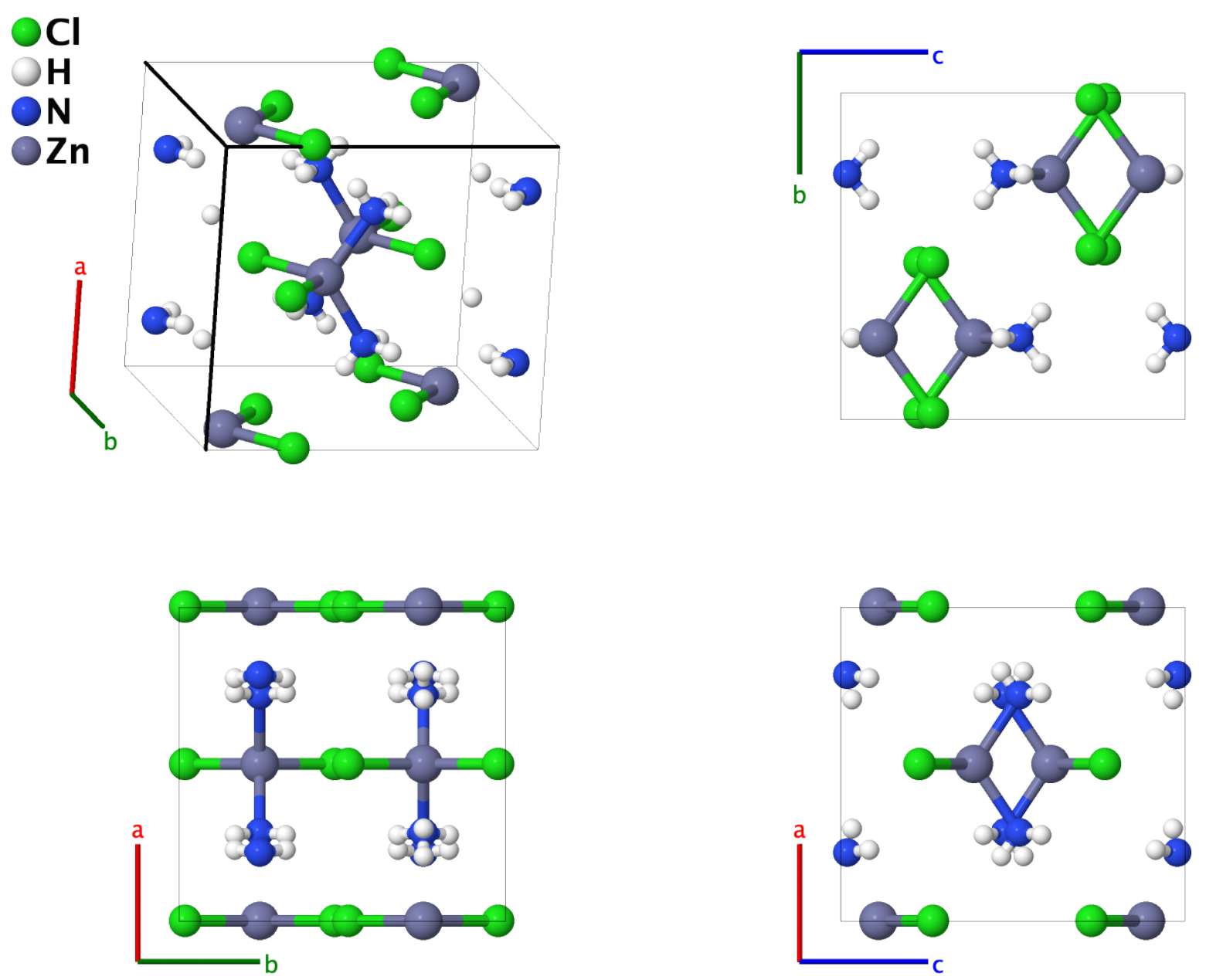

| | | |
|---|---|---|
| **Prototype** | : | $Cl_2H_6N_2Zn$ |
| **AFLOW prototype label** | : | A2B6C2D_oI44_74_h_ij_i_e |
| ***Strukturbericht* designation** | : | $E1_2$ |
| **Pearson symbol** | : | oI44 |
| **Space group number** | : | 74 |
| **Space group symbol** | : | *Imma* |
| **AFLOW prototype command** | : | `aflow --proto=A2B6C2D_oI44_74_h_ij_i_e` `--params=`$a, b/a, c/a, z_1, y_2, z_2, x_3, z_3, x_4, z_4, x_5, y_5, z_5$ |

### Other compounds with this structure

- $Zn(NH_3)_2Br_2$

- The recent work of (Ivšić, 2019) studied this system at 100 K and were able to locate the hydrogen atoms. The positions of the other atoms are similar to those in earlier works such as (Yamaguchi, 1981) and the space group is unchanged, so we use this as the prototype for the $E1_2$ label.

### Body-centered Orthorhombic primitive vectors:

$$
\begin{aligned}
\mathbf{a}_1 &= -\tfrac{1}{2}\,a\,\hat{\mathbf{x}} + \tfrac{1}{2}\,b\,\hat{\mathbf{y}} + \tfrac{1}{2}\,c\,\hat{\mathbf{z}} \\
\mathbf{a}_2 &= \tfrac{1}{2}\,a\,\hat{\mathbf{x}} - \tfrac{1}{2}\,b\,\hat{\mathbf{y}} + \tfrac{1}{2}\,c\,\hat{\mathbf{z}} \\
\mathbf{a}_3 &= \tfrac{1}{2}\,a\,\hat{\mathbf{x}} + \tfrac{1}{2}\,b\,\hat{\mathbf{y}} - \tfrac{1}{2}\,c\,\hat{\mathbf{z}}
\end{aligned}
$$

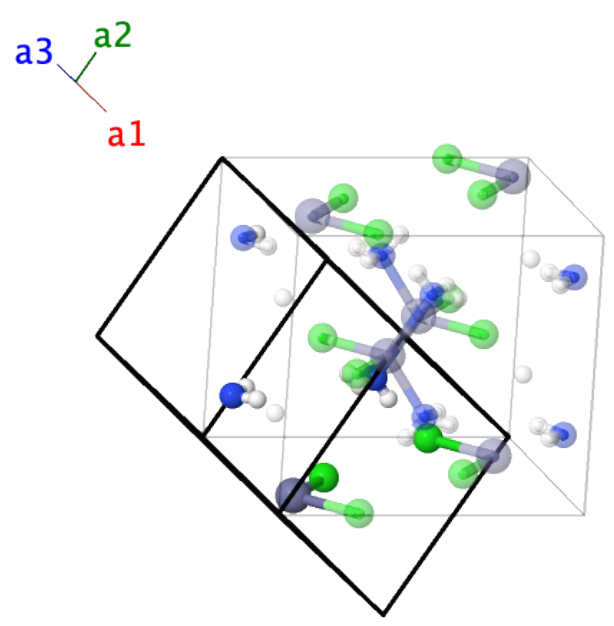

**Basis vectors:**

| | | Lattice Coordinates | | Cartesian Coordinates | Wyckoff Position | Atom Type |
|---|---|---|---|---|---|---|
| $\mathbf{B}_1$ | = | $\left(\frac{1}{4}+z_1\right)\mathbf{a}_1 + z_1\,\mathbf{a}_2 + \frac{1}{4}\,\mathbf{a}_3$ | = | $\frac{1}{4}b\,\hat{\mathbf{y}} + z_1 c\,\hat{\mathbf{z}}$ | (4$e$) | Zn |
| $\mathbf{B}_2$ | = | $\left(\frac{3}{4}-z_1\right)\mathbf{a}_1 - z_1\,\mathbf{a}_2 + \frac{3}{4}\,\mathbf{a}_3$ | = | $\frac{3}{4}b\,\hat{\mathbf{y}} - z_1 c\,\hat{\mathbf{z}}$ | (4$e$) | Zn |
| $\mathbf{B}_3$ | = | $(y_2+z_2)\,\mathbf{a}_1 + z_2\,\mathbf{a}_2 + y_2\,\mathbf{a}_3$ | = | $y_2 b\,\hat{\mathbf{y}} + z_2 c\,\hat{\mathbf{z}}$ | (8$h$) | Cl |
| $\mathbf{B}_4$ | = | $\left(\frac{1}{2}-y_2+z_2\right)\mathbf{a}_1 + z_2\,\mathbf{a}_2 + \left(\frac{1}{2}-y_2\right)\mathbf{a}_3$ | = | $\left(\frac{1}{2}-y_2\right)b\,\hat{\mathbf{y}} + z_2 c\,\hat{\mathbf{z}}$ | (8$h$) | Cl |
| $\mathbf{B}_5$ | = | $\left(\frac{1}{2}+y_2-z_2\right)\mathbf{a}_1 - z_2\,\mathbf{a}_2 + \left(\frac{1}{2}+y_2\right)\mathbf{a}_3$ | = | $\left(\frac{1}{2}+y_2\right)b\,\hat{\mathbf{y}} - z_2 c\,\hat{\mathbf{z}}$ | (8$h$) | Cl |
| $\mathbf{B}_6$ | = | $(-y_2-z_2)\,\mathbf{a}_1 - z_2\,\mathbf{a}_2 - y_2\,\mathbf{a}_3$ | = | $-y_2 b\,\hat{\mathbf{y}} - z_2 c\,\hat{\mathbf{z}}$ | (8$h$) | Cl |
| $\mathbf{B}_7$ | = | $\left(\frac{1}{4}+z_3\right)\mathbf{a}_1 + (x_3+z_3)\,\mathbf{a}_2 + \left(\frac{1}{4}+x_3\right)\mathbf{a}_3$ | = | $x_3 a\,\hat{\mathbf{x}} + \frac{1}{4}b\,\hat{\mathbf{y}} + z_3 c\,\hat{\mathbf{z}}$ | (8$i$) | H I |
| $\mathbf{B}_8$ | = | $\left(\frac{1}{4}+z_3\right)\mathbf{a}_1 + (-x_3+z_3)\,\mathbf{a}_2 + \left(\frac{1}{4}-x_3\right)\mathbf{a}_3$ | = | $-x_3 a\,\hat{\mathbf{x}} + \frac{1}{4}b\,\hat{\mathbf{y}} + z_3 c\,\hat{\mathbf{z}}$ | (8$i$) | H I |
| $\mathbf{B}_9$ | = | $\left(\frac{3}{4}-z_3\right)\mathbf{a}_1 + (-x_3-z_3)\,\mathbf{a}_2 + \left(\frac{3}{4}-x_3\right)\mathbf{a}_3$ | = | $-x_3 a\,\hat{\mathbf{x}} + \frac{3}{4}b\,\hat{\mathbf{y}} - z_3 c\,\hat{\mathbf{z}}$ | (8$i$) | H I |
| $\mathbf{B}_{10}$ | = | $\left(\frac{3}{4}-z_3\right)\mathbf{a}_1 + (x_3-z_3)\,\mathbf{a}_2 + \left(\frac{3}{4}+x_3\right)\mathbf{a}_3$ | = | $x_3 a\,\hat{\mathbf{x}} + \frac{3}{4}b\,\hat{\mathbf{y}} - z_3 c\,\hat{\mathbf{z}}$ | (8$i$) | H I |
| $\mathbf{B}_{11}$ | = | $\left(\frac{1}{4}+z_4\right)\mathbf{a}_1 + (x_4+z_4)\,\mathbf{a}_2 + \left(\frac{1}{4}+x_4\right)\mathbf{a}_3$ | = | $x_4 a\,\hat{\mathbf{x}} + \frac{1}{4}b\,\hat{\mathbf{y}} + z_4 c\,\hat{\mathbf{z}}$ | (8$i$) | N |
| $\mathbf{B}_{12}$ | = | $\left(\frac{1}{4}+z_4\right)\mathbf{a}_1 + (-x_4+z_4)\,\mathbf{a}_2 + \left(\frac{1}{4}-x_4\right)\mathbf{a}_3$ | = | $-x_4 a\,\hat{\mathbf{x}} + \frac{1}{4}b\,\hat{\mathbf{y}} + z_4 c\,\hat{\mathbf{z}}$ | (8$i$) | N |
| $\mathbf{B}_{13}$ | = | $\left(\frac{3}{4}-z_4\right)\mathbf{a}_1 + (-x_4-z_4)\,\mathbf{a}_2 + \left(\frac{3}{4}-x_4\right)\mathbf{a}_3$ | = | $-x_4 a\,\hat{\mathbf{x}} + \frac{3}{4}b\,\hat{\mathbf{y}} - z_4 c\,\hat{\mathbf{z}}$ | (8$i$) | N |
| $\mathbf{B}_{14}$ | = | $\left(\frac{3}{4}-z_4\right)\mathbf{a}_1 + (x_4-z_4)\,\mathbf{a}_2 + \left(\frac{3}{4}+x_4\right)\mathbf{a}_3$ | = | $x_4 a\,\hat{\mathbf{x}} + \frac{3}{4}b\,\hat{\mathbf{y}} - z_4 c\,\hat{\mathbf{z}}$ | (8$i$) | N |
| $\mathbf{B}_{15}$ | = | $(y_5+z_5)\,\mathbf{a}_1 + (x_5+z_5)\,\mathbf{a}_2 + (x_5+y_5)\,\mathbf{a}_3$ | = | $x_5 a\,\hat{\mathbf{x}} + y_5 b\,\hat{\mathbf{y}} + z_5 c\,\hat{\mathbf{z}}$ | (16$j$) | H II |
| $\mathbf{B}_{16}$ | = | $\left(\frac{1}{2}-y_5+z_5\right)\mathbf{a}_1 + (-x_5+z_5)\,\mathbf{a}_2 + \left(\frac{1}{2}-x_5-y_5\right)\mathbf{a}_3$ | = | $-x_5 a\,\hat{\mathbf{x}} + \left(\frac{1}{2}-y_5\right)b\,\hat{\mathbf{y}} + z_5 c\,\hat{\mathbf{z}}$ | (16$j$) | H II |
| $\mathbf{B}_{17}$ | = | $\left(\frac{1}{2}+y_5-z_5\right)\mathbf{a}_1 + (-x_5-z_5)\,\mathbf{a}_2 + \left(\frac{1}{2}-x_5+y_5\right)\mathbf{a}_3$ | = | $-x_5 a\,\hat{\mathbf{x}} + \left(\frac{1}{2}+y_5\right)b\,\hat{\mathbf{y}} - z_5 c\,\hat{\mathbf{z}}$ | (16$j$) | H II |
| $\mathbf{B}_{18}$ | = | $(-y_5-z_5)\,\mathbf{a}_1 + (x_5-z_5)\,\mathbf{a}_2 + (x_5-y_5)\,\mathbf{a}_3$ | = | $x_5 a\,\hat{\mathbf{x}} - y_5 b\,\hat{\mathbf{y}} - z_5 c\,\hat{\mathbf{z}}$ | (16$j$) | H II |
| $\mathbf{B}_{19}$ | = | $(-y_5-z_5)\,\mathbf{a}_1 + (-x_5-z_5)\,\mathbf{a}_2 + (-x_5-y_5)\,\mathbf{a}_3$ | = | $-x_5 a\,\hat{\mathbf{x}} - y_5 b\,\hat{\mathbf{y}} - z_5 c\,\hat{\mathbf{z}}$ | (16$j$) | H II |
| $\mathbf{B}_{20}$ | = | $\left(\frac{1}{2}+y_5-z_5\right)\mathbf{a}_1 + (x_5-z_5)\,\mathbf{a}_2 + \left(\frac{1}{2}+x_5+y_5\right)\mathbf{a}_3$ | = | $x_5 a\,\hat{\mathbf{x}} + \left(\frac{1}{2}+y_5\right)b\,\hat{\mathbf{y}} - z_5 c\,\hat{\mathbf{z}}$ | (16$j$) | H II |
| $\mathbf{B}_{21}$ | = | $\left(\frac{1}{2}-y_5+z_5\right)\mathbf{a}_1 + (x_5+z_5)\,\mathbf{a}_2 + \left(\frac{1}{2}+x_5-y_5\right)\mathbf{a}_3$ | = | $x_5 a\,\hat{\mathbf{x}} + \left(\frac{1}{2}-y_5\right)b\,\hat{\mathbf{y}} + z_5 c\,\hat{\mathbf{z}}$ | (16$j$) | H II |
| $\mathbf{B}_{22}$ | = | $(y_5+z_5)\,\mathbf{a}_1 + (-x_5+z_5)\,\mathbf{a}_2 + (-x_5+y_5)\,\mathbf{a}_3$ | = | $-x_5 a\,\hat{\mathbf{x}} + y_5 b\,\hat{\mathbf{y}} + z_5 c\,\hat{\mathbf{z}}$ | (16$j$) | H II |

**Geometry files:**

- CIF: pp. 1667

- POSCAR: pp. 1667

# $CeCu_2$ Structure: AB2_oI12_74_e_h

http://aflow.org/prototype-encyclopedia/AB2_oI12_74_e_h

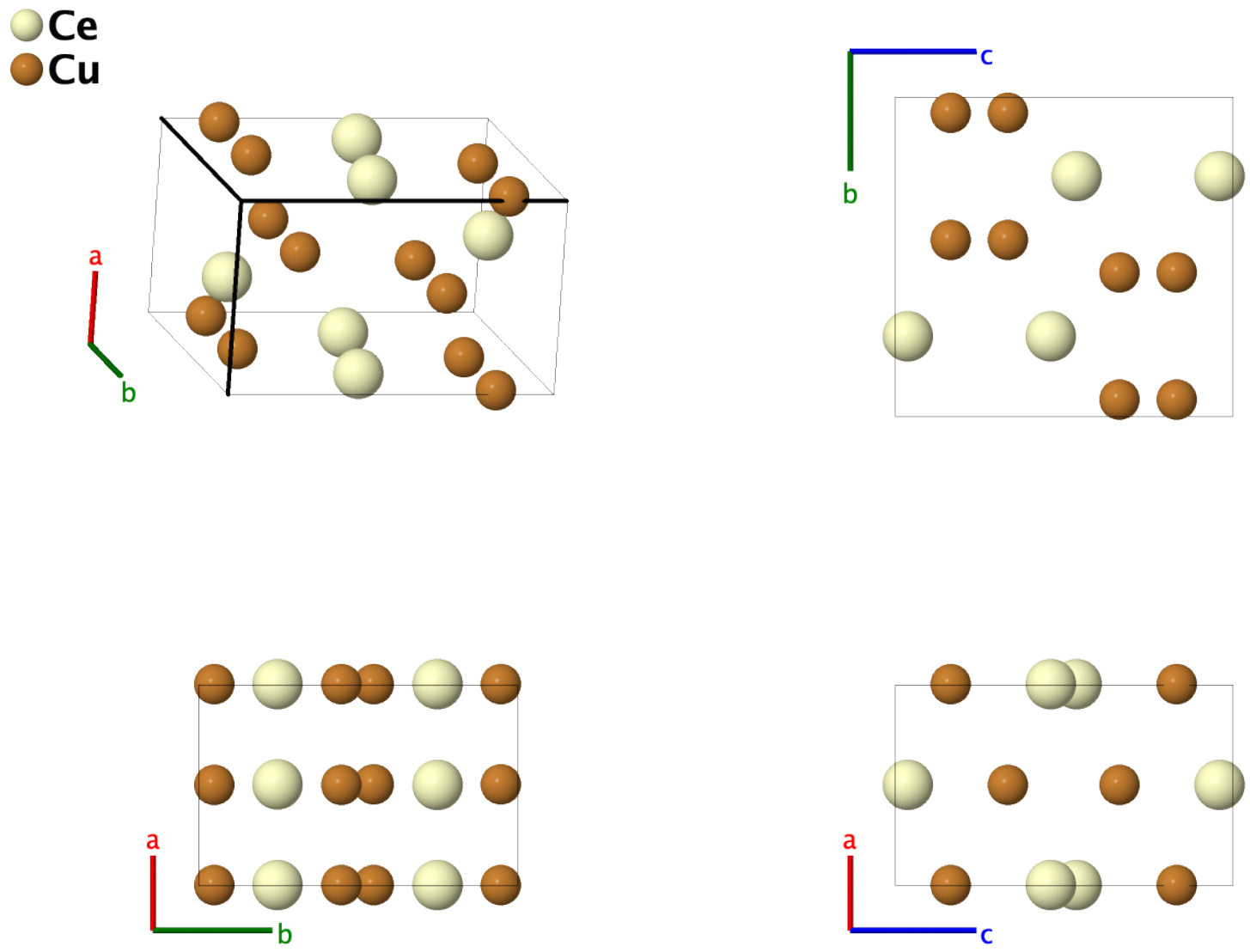

| | | |
|---|---|---|
| **Prototype** | : | $CeCu_2$ |
| **AFLOW prototype label** | : | AB2_oI12_74_e_h |
| ***Strukturbericht* designation** | : | None |
| **Pearson symbol** | : | oI12 |
| **Space group number** | : | 74 |
| **Space group symbol** | : | *Imma* |
| **AFLOW prototype command** | : | `aflow --proto=AB2_oI12_74_e_h`<br>`--params=`$a, b/a, c/a, z_1, y_2, z_2$ |

### Other compounds with this structure

- $CeCu_2$, $PrCu_2$, $NdCu_2$, $SmCu_2$, $EuCu_2$, $GdCu_2$, $TbCu_2$, $DyCu_2$, $HoCu_2$, $ErCu_2$, $YbCu_2$, $LuCu_2$, $LaZn_2$, $CeZn_2$, $PrZn_2$, $NdZn_2$, $SmZn_2$, $EuZn_2$, $GdZn_2$, $ThZn_2$, $DyZn_2$, $HoZn_2$, $ErZn_2$, $TmZn_2$, $YbZn_2$, $LuZn_2$, $LaAg_2$, $CeAg_2$, $PrAg_2$, $NdAg_2$, $EuAg_2$, $YbAg_2$, $LaAu_2$, $CeAu_2$, and $EuAu_2$

### Body-centered Orthorhombic primitive vectors:

$$\mathbf{a}_1 = -\frac{1}{2}\,a\,\hat{\mathbf{x}} + \frac{1}{2}\,b\,\hat{\mathbf{y}} + \frac{1}{2}\,c\,\hat{\mathbf{z}}$$
$$\mathbf{a}_2 = \frac{1}{2}\,a\,\hat{\mathbf{x}} - \frac{1}{2}\,b\,\hat{\mathbf{y}} + \frac{1}{2}\,c\,\hat{\mathbf{z}}$$
$$\mathbf{a}_3 = \frac{1}{2}\,a\,\hat{\mathbf{x}} + \frac{1}{2}\,b\,\hat{\mathbf{y}} - \frac{1}{2}\,c\,\hat{\mathbf{z}}$$

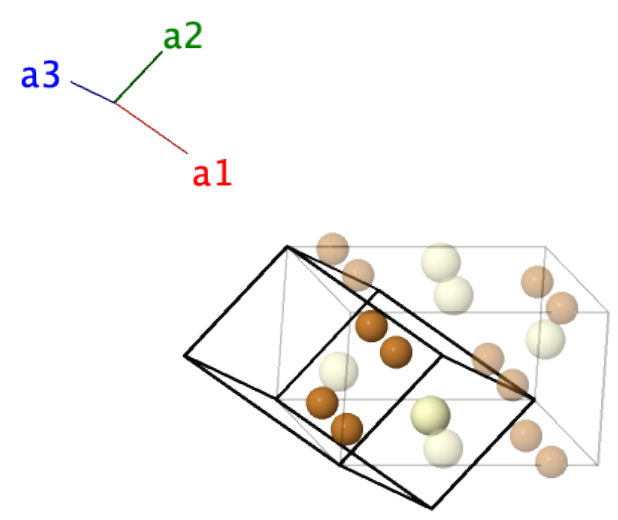

### Basis vectors:

| | | Lattice Coordinates | | Cartesian Coordinates | Wyckoff Position | Atom Type |
|---|---|---|---|---|---|---|
| $\mathbf{B}_1$ | = | $\left(\frac{1}{4} + z_1\right)\mathbf{a}_1 + z_1\,\mathbf{a}_2 + \frac{1}{4}\,\mathbf{a}_3$ | = | $\frac{1}{4}b\,\hat{\mathbf{y}} + z_1 c\,\hat{\mathbf{z}}$ | (4*e*) | Ce |
| $\mathbf{B}_2$ | = | $\left(\frac{3}{4} - z_1\right)\,\mathbf{a}_1 - z_1\,\mathbf{a}_2 + \frac{3}{4}\,\mathbf{a}_3$ | = | $\frac{3}{4}b\,\hat{\mathbf{y}} - z_1 c\,\hat{\mathbf{z}}$ | (4$e$) | Ce |
| $\mathbf{B}_3$ | = | $(y_2 + z_2)\,\mathbf{a}_1 + z_2\,\mathbf{a}_2 + y_2\,\mathbf{a}_3$ | = | $y_2 b\,\hat{\mathbf{y}} + z_2 c\,\hat{\mathbf{z}}$ | (8$h$) | Cu |
| $\mathbf{B}_4$ | = | $\left(\frac{1}{2} - y_2 + z_2\right)\,\mathbf{a}_1 + z_2\,\mathbf{a}_2 + \left(\frac{1}{2} - y_2\right)\,\mathbf{a}_3$ | = | $\left(\frac{1}{2} - y_2\right) b\,\hat{\mathbf{y}} + z_2 c\,\hat{\mathbf{z}}$ | (8$h$) | Cu |
| $\mathbf{B}_5$ | = | $\left(\frac{1}{2} + y_2 - z_2\right)\,\mathbf{a}_1 - z_2\,\mathbf{a}_2 + \left(\frac{1}{2} + y_2\right)\,\mathbf{a}_3$ | = | $\left(\frac{1}{2} + y_2\right) b\,\hat{\mathbf{y}} - z_2 c\,\hat{\mathbf{z}}$ | (8$h$) | Cu |
| $\mathbf{B}_6$ | = | $(-y_2 - z_2)\,\mathbf{a}_1 - z_2\,\mathbf{a}_2 - y_2\,\mathbf{a}_3$ | = | $-y_2 b\,\hat{\mathbf{y}} - z_2 c\,\hat{\mathbf{z}}$ | (8$h$) | Cu |

**Geometry files:**

- CIF: pp. 1667
- POSCAR: pp. 1668

# $LiCuVO_4$ Structure: ABC4D_oI28_74_a_d_hi_e

http://aflow.org/prototype-encyclopedia/ABC4D_oI28_74_a_d_hi_e

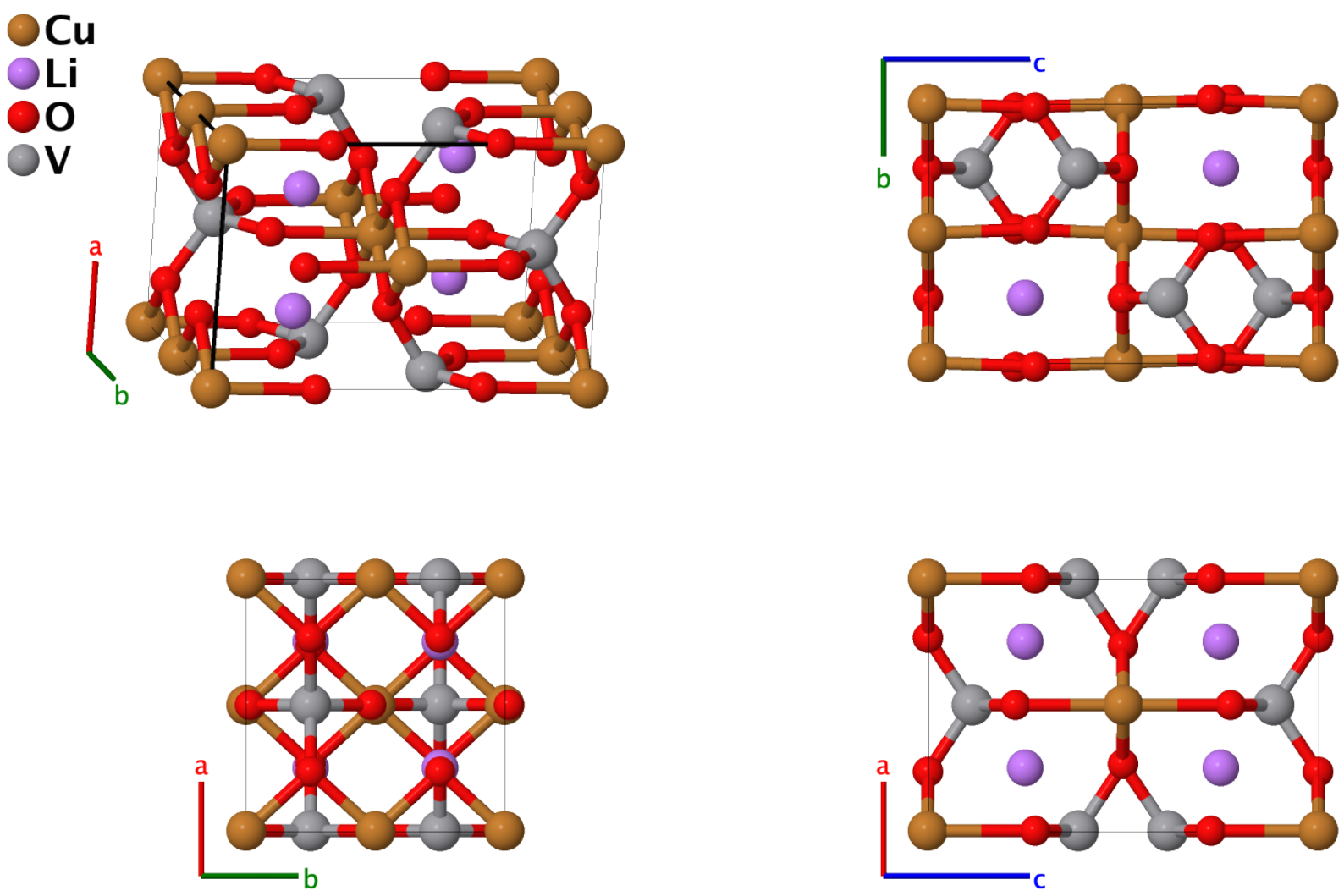

**Prototype** : $CuLiO_4V$

**AFLOW prototype label** : ABC4D_oI28_74_a_d_hi_e

***Strukturbericht* designation** : None

**Pearson symbol** : oI28

**Space group number** : 74

**Space group symbol** : *Imma*

**AFLOW prototype command** : `aflow --proto=ABC4D_oI28_74_a_d_hi_e`
`--params=`$a, b/a, c/a, z_3, y_4, z_4, x_5, z_5$

---

**Body-centered Orthorhombic primitive vectors:**

$$\mathbf{a}_1 = -\frac{1}{2}\,a\,\hat{\mathbf{x}} + \frac{1}{2}\,b\,\hat{\mathbf{y}} + \frac{1}{2}\,c\,\hat{\mathbf{z}}$$
$$\mathbf{a}_2 = \frac{1}{2}\,a\,\hat{\mathbf{x}} - \frac{1}{2}\,b\,\hat{\mathbf{y}} + \frac{1}{2}\,c\,\hat{\mathbf{z}}$$
$$\mathbf{a}_3 = \frac{1}{2}\,a\,\hat{\mathbf{x}} + \frac{1}{2}\,b\,\hat{\mathbf{y}} - \frac{1}{2}\,c\,\hat{\mathbf{z}}$$

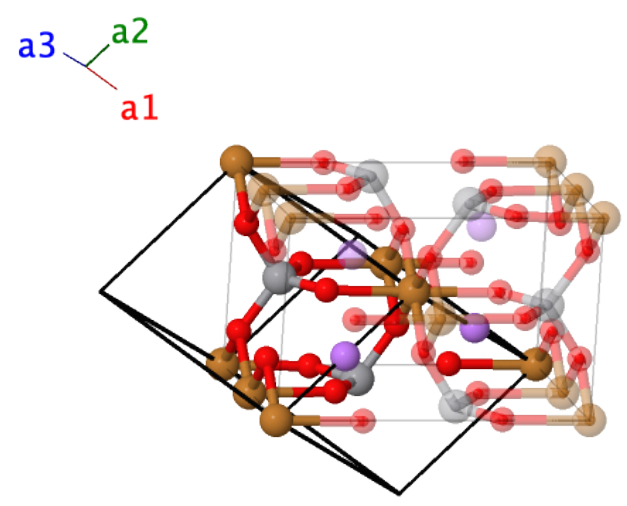

---

**Basis vectors:**

| | | Lattice Coordinates | | Cartesian Coordinates | Wyckoff Position | Atom Type |
|---|---|---|---|---|---|---|
| $\mathbf{B}_1$ | = | $0\,\mathbf{a}_1 + 0\,\mathbf{a}_2 + 0\,\mathbf{a}_3$ | = | $0\,\hat{\mathbf{x}} + 0\,\hat{\mathbf{y}} + 0\,\hat{\mathbf{z}}$ | (4*a*) | Cu |
| $\mathbf{B}_2$ | = | $\frac{1}{2}\,\mathbf{a}_1 + \frac{1}{2}\,\mathbf{a}_3$ | = | $\frac{1}{2}b\,\hat{\mathbf{y}}$ | (4*a*) | Cu |
| $\mathbf{B}_3$ | = | $\frac{1}{2}\,\mathbf{a}_3$ | = | $\frac{1}{4}a\,\hat{\mathbf{x}} + \frac{1}{4}b\,\hat{\mathbf{y}} + \frac{3}{4}c\,\hat{\mathbf{z}}$ | (4*d*) | Li |
| $\mathbf{B}_4$ | = | $\frac{1}{2}\,\mathbf{a}_2$ | = | $\frac{1}{4}a\,\hat{\mathbf{x}} - \frac{1}{4}b\,\hat{\mathbf{y}} + \frac{1}{4}c\,\hat{\mathbf{z}}$ | (4*d*) | Li |
| $\mathbf{B}_5$ | = | $\left(\frac{1}{4} + z_3\right)\,\mathbf{a}_1 + z_3\,\mathbf{a}_2 + \frac{1}{4}\,\mathbf{a}_3$ | = | $\frac{1}{4}b\,\hat{\mathbf{y}} + z_3c\,\hat{\mathbf{z}}$ | (4*e*) | V |
| $\mathbf{B}_6$ | = | $\left(\frac{3}{4} - z_3\right)\,\mathbf{a}_1 - z_3\,\mathbf{a}_2 + \frac{3}{4}\,\mathbf{a}_3$ | = | $\frac{3}{4}b\,\hat{\mathbf{y}} - z_3c\,\hat{\mathbf{z}}$ | (4*e*) | V |
| $\mathbf{B}_7$ | = | $(y_4 + z_4)\,\mathbf{a}_1 + z_4\,\mathbf{a}_2 + y_4\,\mathbf{a}_3$ | = | $y_4b\,\hat{\mathbf{y}} + z_4c\,\hat{\mathbf{z}}$ | (8*h*) | O I |
| $\mathbf{B}_{8}$ | = | $\left(\frac{1}{2} - y_4 + z_4\right)\,\mathbf{a}_{1} + z_4\,\mathbf{a}_{2} + \left(\frac{1}{2} - y_4\right)\,\mathbf{a}_{3}$ | = | $\left(\frac{1}{2} - y_4\right) b\,\mathbf{\hat{y}} + z_4 c\,\mathbf{\hat{z}}$ | (8$h$) | O I |
| $\mathbf{B}_{9}$ | = | $\left(\frac{1}{2} + y_4 - z_4\right)\,\mathbf{a}_{1} - z_4\,\mathbf{a}_{2} + \left(\frac{1}{2} + y_4\right)\,\mathbf{a}_{3}$ | = | $\left(\frac{1}{2} + y_4\right) b\,\mathbf{\hat{y}} - z_4 c\,\mathbf{\hat{z}}$ | (8$h$) | O I |
| $\mathbf{B}_{10}$ | = | $\left(-y_4 - z_4\right)\,\mathbf{a}_{1} - z_4\,\mathbf{a}_{2} - y_4\,\mathbf{a}_{3}$ | = | $-y_4 b\,\mathbf{\hat{y}} - z_4 c\,\mathbf{\hat{z}}$ | (8$h$) | O I |
| $\mathbf{B}_{11}$ | = | $\left(\frac{1}{4} + z_5\right)\,\mathbf{a}_{1} + \left(x_5 + z_5\right)\,\mathbf{a}_{2} + \left(\frac{1}{4} + x_5\right)\,\mathbf{a}_{3}$ | = | $x_5 a\,\mathbf{\hat{x}} + \frac{1}{4} b\,\mathbf{\hat{y}} + z_5 c\,\mathbf{\hat{z}}$ | (8$i$) | O II |
| $\mathbf{B}_{12}$ | = | $\left(\frac{1}{4} + z_5\right)\,\mathbf{a}_{1} + \left(-x_5 + z_5\right)\,\mathbf{a}_{2} + \left(\frac{1}{4} - x_5\right)\,\mathbf{a}_{3}$ | = | $-x_5 a\,\mathbf{\hat{x}} + \frac{1}{4} b\,\mathbf{\hat{y}} + z_5 c\,\mathbf{\hat{z}}$ | (8$i$) | O II |
| $\mathbf{B}_{13}$ | = | $\left(\frac{3}{4} - z_5\right)\,\mathbf{a}_{1} + \left(-x_5 - z_5\right)\,\mathbf{a}_{2} + \left(\frac{3}{4} - x_5\right)\,\mathbf{a}_{3}$ | = | $-x_5 a\,\mathbf{\hat{x}} + \frac{3}{4} b\,\mathbf{\hat{y}} - z_5 c\,\mathbf{\hat{z}}$ | (8$i$) | O II |
| $\mathbf{B}_{14}$ | = | $\left(\frac{3}{4} - z_5\right)\,\mathbf{a}_{1} + \left(x_5 - z_5\right)\,\mathbf{a}_{2} + \left(\frac{3}{4} + x_5\right)\,\mathbf{a}_{3}$ | = | $x_5 a\,\mathbf{\hat{x}} + \frac{3}{4} b\,\mathbf{\hat{y}} - z_5 c\,\mathbf{\hat{z}}$ | (8$i$) | O II |

**Geometry files:**

- CIF: pp. 1668
- POSCAR: pp. 1668

# Gwihabaite [$NH_4NO_3$ (V)] Structure: A4B2C3_tP72_77_8d_ab2c2d_6d

http://aflow.org/prototype-encyclopedia/A4B2C3_tP72_77_8d_ab2c2d_6d

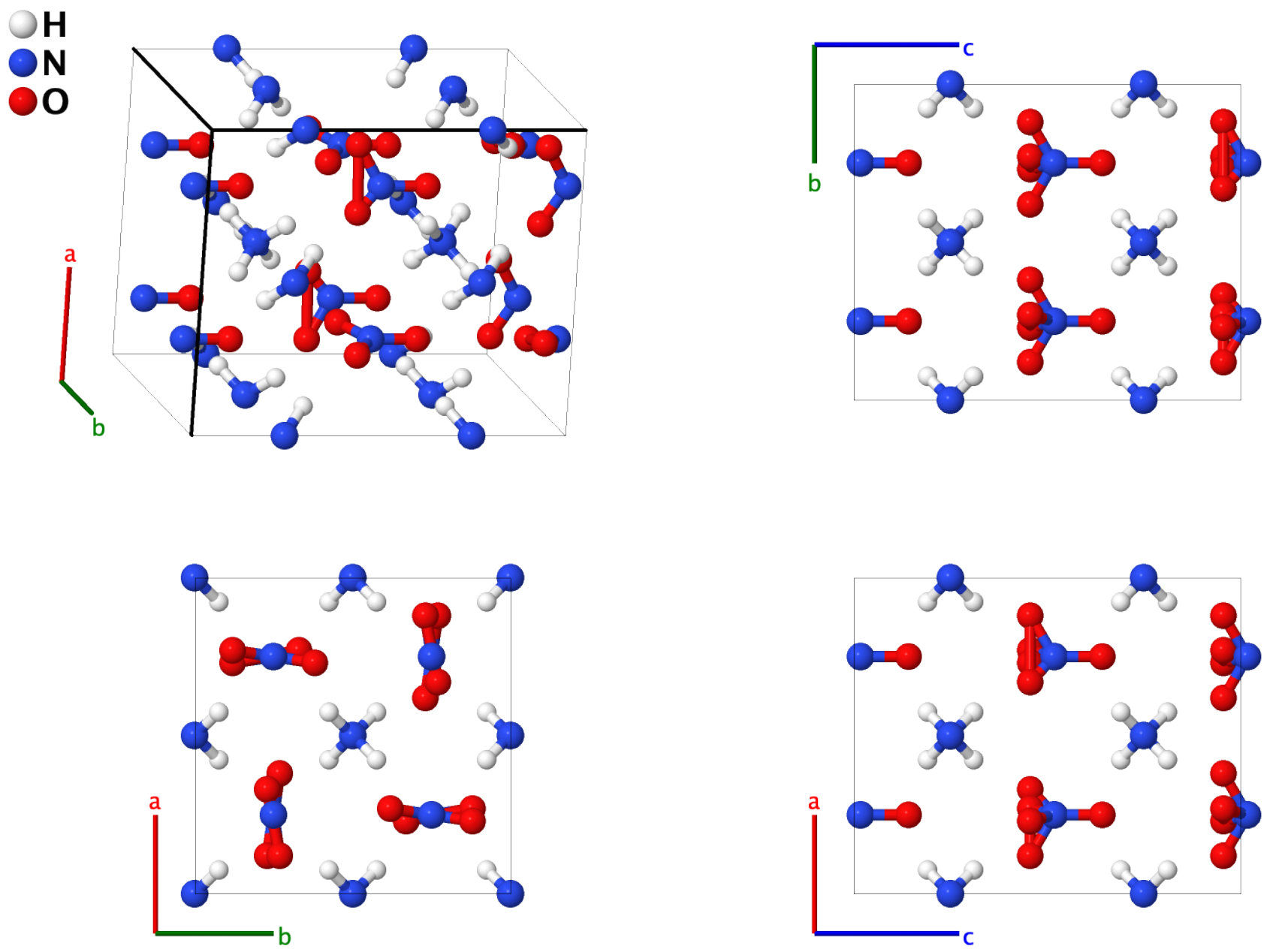

| | | |
|---|---|---|
| **Prototype** | : | $H_4N_2O_3$ |
| **AFLOW prototype label** | : | A4B2C3_tP72_77_8d_ab2c2d_6d |
| ***Strukturbericht* designation** | : | None |
| **Pearson symbol** | : | tP72 |
| **Space group number** | : | 77 |
| **Space group symbol** | : | $P4_2$ |
| **AFLOW prototype command** | : | `aflow --proto=A4B2C3_tP72_77_8d_ab2c2d_6d` `--params=`$a, c/a, z_1, z_2, z_3, z_4, x_5, y_5, z_5, x_6, y_6, z_6, x_7, y_7, z_7, x_8, y_8, z_8, x_9, y_9, z_9, x_{10}, y_{10}, z_{10}, x_{11}, y_{11}, z_{11}, x_{12}, y_{12}, z_{12}, x_{13}, y_{13}, z_{13}, x_{14}, y_{14}, z_{14}, x_{15}, y_{15}, z_{15}, x_{16}, y_{16}, z_{16}, x_{17}, y_{17}, z_{17}, x_{18}, y_{18}, z_{18}, x_{19}, y_{19}, z_{19}, x_{20}, y_{20}, z_{20}$ |

- Ammonium Nitrate exists in a variety of forms, (Hermann, 1937) depending on the temperature:

| Phase | Temperature °C | *Strukturbericht* | Page |
|---|---|---|---|
| I | 125 – 170 | $G0_8$ | AB_cP2_221_a_b.NH4.NO3 |
| II | 84 – 125 | $G0_9$ | ABC3_tP10_100_b_a_bc |
| III | 32 – 84 | $G0_{10}$ | ABC3_oP20_62_c_c_cd.N.NH4.O |
| IV | -17 – 32 | $G0_{11}$ | A4B2C3_oP18_59_ef_ab_af |
| V | $< -17$ | Gwihabaite | A4B2C3_tP72_77_8d_ab2c2d_6d2 |

- Data for this structure was taken at $-150$ °C.

## Simple Tetragonal primitive vectors:

$$\mathbf{a}_1 = a\,\hat{\mathbf{x}}$$
$$\mathbf{a}_2 = a\,\hat{\mathbf{y}}$$
$$\mathbf{a}_3 = c\,\hat{\mathbf{z}}$$

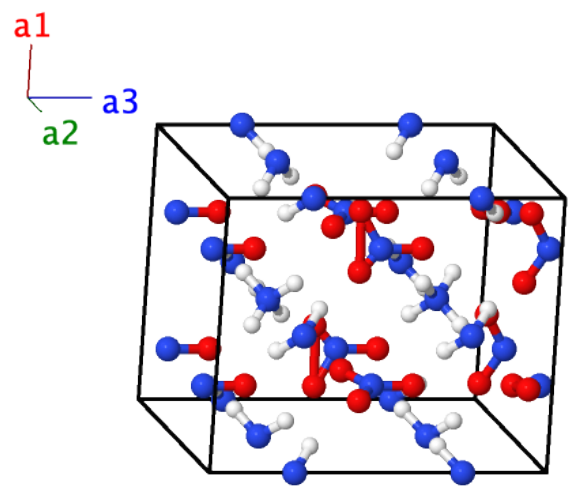

## Basis vectors:

| | | Lattice Coordinates | | Cartesian Coordinates | Wyckoff Position | Atom Type |
|---|---|---|---|---|---|---|
| $\mathbf{B}_1$ | = | $z_1\,\mathbf{a}_3$ | = | $z_1 c\,\hat{\mathbf{z}}$ | (2a) | N I |
| $\mathbf{B}_2$ | = | $\left(\frac{1}{2}+z_1\right)\mathbf{a}_3$ | = | $\left(\frac{1}{2}+z_1\right)c\,\hat{\mathbf{z}}$ | (2a) | N I |
| $\mathbf{B}_3$ | = | $\frac{1}{2}\,\mathbf{a}_1+\frac{1}{2}\,\mathbf{a}_2+z_2\,\mathbf{a}_3$ | = | $\frac{1}{2}a\,\hat{\mathbf{x}}+\frac{1}{2}a\,\hat{\mathbf{y}}+z_2 c\,\hat{\mathbf{z}}$ | (2b) | N II |
| $\mathbf{B}_4$ | = | $\frac{1}{2}\,\mathbf{a}_1+\frac{1}{2}\,\mathbf{a}_2+\left(\frac{1}{2}+z_2\right)\mathbf{a}_3$ | = | $\frac{1}{2}a\,\hat{\mathbf{x}}+\frac{1}{2}a\,\hat{\mathbf{y}}+\left(\frac{1}{2}+z_2\right)c\,\hat{\mathbf{z}}$ | (2b) | N II |
| $\mathbf{B}_5$ | = | $\frac{1}{2}\,\mathbf{a}_2+z_3\,\mathbf{a}_3$ | = | $\frac{1}{2}a\,\hat{\mathbf{y}}+z_3 c\,\hat{\mathbf{z}}$ | (2c) | N III |
| $\mathbf{B}_6$ | = | $\frac{1}{2}\,\mathbf{a}_1+\left(\frac{1}{2}+z_3\right)\mathbf{a}_3$ | = | $\frac{1}{2}a\,\hat{\mathbf{x}}+\left(\frac{1}{2}+z_3\right)c\,\hat{\mathbf{z}}$ | (2c) | N III |
| $\mathbf{B}_7$ | = | $\frac{1}{2}\,\mathbf{a}_2+z_4\,\mathbf{a}_3$ | = | $\frac{1}{2}a\,\hat{\mathbf{y}}+z_4 c\,\hat{\mathbf{z}}$ | (2c) | N IV |
| $\mathbf{B}_8$ | = | $\frac{1}{2}\,\mathbf{a}_1+\left(\frac{1}{2}+z_4\right)\mathbf{a}_3$ | = | $\frac{1}{2}a\,\hat{\mathbf{x}}+\left(\frac{1}{2}+z_4\right)c\,\hat{\mathbf{z}}$ | (2c) | N IV |
| $\mathbf{B}_9$ | = | $x_5\,\mathbf{a}_1+y_5\,\mathbf{a}_2+z_5\,\mathbf{a}_3$ | = | $x_5 a\,\hat{\mathbf{x}}+y_5 a\,\hat{\mathbf{y}}+z_5 c\,\hat{\mathbf{z}}$ | (4d) | H I |
| $\mathbf{B}_{10}$ | = | $-x_5\,\mathbf{a}_1-y_5\,\mathbf{a}_2+z_5\,\mathbf{a}_3$ | = | $-x_5 a\,\hat{\mathbf{x}}-y_5 a\,\hat{\mathbf{y}}+z_5 c\,\hat{\mathbf{z}}$ | (4d) | H I |
| $\mathbf{B}_{11}$ | = | $-y_5\,\mathbf{a}_1+x_5\,\mathbf{a}_2+\left(\frac{1}{2}+z_5\right)\mathbf{a}_3$ | = | $-y_5 a\,\hat{\mathbf{x}}+x_5 a\,\hat{\mathbf{y}}+\left(\frac{1}{2}+z_5\right)c\,\hat{\mathbf{z}}$ | (4d) | H I |
| $\mathbf{B}_{12}$ | = | $y_5\,\mathbf{a}_1-x_5\,\mathbf{a}_2+\left(\frac{1}{2}+z_5\right)\mathbf{a}_3$ | = | $y_5 a\,\hat{\mathbf{x}}-x_5 a\,\hat{\mathbf{y}}+\left(\frac{1}{2}+z_5\right)c\,\hat{\mathbf{z}}$ | (4d) | H I |
| $\mathbf{B}_{13}$ | = | $x_6\,\mathbf{a}_1+y_6\,\mathbf{a}_2+z_6\,\mathbf{a}_3$ | = | $x_6 a\,\hat{\mathbf{x}}+y_6 a\,\hat{\mathbf{y}}+z_6 c\,\hat{\mathbf{z}}$ | (4d) | H II |
| $\mathbf{B}_{14}$ | = | $-x_6\,\mathbf{a}_1-y_6\,\mathbf{a}_2+z_6\,\mathbf{a}_3$ | = | $-x_6 a\,\hat{\mathbf{x}}-y_6 a\,\hat{\mathbf{y}}+z_6 c\,\hat{\mathbf{z}}$ | (4d) | H II |
| $\mathbf{B}_{15}$ | = | $-y_6\,\mathbf{a}_1+x_6\,\mathbf{a}_2+\left(\frac{1}{2}+z_6\right)\mathbf{a}_3$ | = | $-y_6 a\,\hat{\mathbf{x}}+x_6 a\,\hat{\mathbf{y}}+\left(\frac{1}{2}+z_6\right)c\,\hat{\mathbf{z}}$ | (4d) | H II |
| $\mathbf{B}_{16}$ | = | $y_6\,\mathbf{a}_1-x_6\,\mathbf{a}_2+\left(\frac{1}{2}+z_6\right)\mathbf{a}_3$ | = | $y_6 a\,\hat{\mathbf{x}}-x_6 a\,\hat{\mathbf{y}}+\left(\frac{1}{2}+z_6\right)c\,\hat{\mathbf{z}}$ | (4d) | H II |
| $\mathbf{B}_{17}$ | = | $x_7\,\mathbf{a}_1+y_7\,\mathbf{a}_2+z_7\,\mathbf{a}_3$ | = | $x_7 a\,\hat{\mathbf{x}}+y_7 a\,\hat{\mathbf{y}}+z_7 c\,\hat{\mathbf{z}}$ | (4d) | H III |
| $\mathbf{B}_{18}$ | = | $-x_7\,\mathbf{a}_1-y_7\,\mathbf{a}_2+z_7\,\mathbf{a}_3$ | = | $-x_7 a\,\hat{\mathbf{x}}-y_7 a\,\hat{\mathbf{y}}+z_7 c\,\hat{\mathbf{z}}$ | (4d) | H III |
| $\mathbf{B}_{19}$ | = | $-y_7\,\mathbf{a}_1+x_7\,\mathbf{a}_2+\left(\frac{1}{2}+z_7\right)\mathbf{a}_3$ | = | $-y_7 a\,\hat{\mathbf{x}}+x_7 a\,\hat{\mathbf{y}}+\left(\frac{1}{2}+z_7\right)c\,\hat{\mathbf{z}}$ | (4d) | H III |
| $\mathbf{B}_{20}$ | = | $y_7\,\mathbf{a}_1-x_7\,\mathbf{a}_2+\left(\frac{1}{2}+z_7\right)\mathbf{a}_3$ | = | $y_7 a\,\hat{\mathbf{x}}-x_7 a\,\hat{\mathbf{y}}+\left(\frac{1}{2}+z_7\right)c\,\hat{\mathbf{z}}$ | (4d) | H III |
| $\mathbf{B}_{21}$ | = | $x_8\,\mathbf{a}_1+y_8\,\mathbf{a}_2+z_8\,\mathbf{a}_3$ | = | $x_8 a\,\hat{\mathbf{x}}+y_8 a\,\hat{\mathbf{y}}+z_8 c\,\hat{\mathbf{z}}$ | (4d) | H IV |
| $\mathbf{B}_{22}$ | = | $-x_8\,\mathbf{a}_1-y_8\,\mathbf{a}_2+z_8\,\mathbf{a}_3$ | = | $-x_8 a\,\hat{\mathbf{x}}-y_8 a\,\hat{\mathbf{y}}+z_8 c\,\hat{\mathbf{z}}$ | (4d) | H IV |
| $\mathbf{B}_{23}$ | = | $-y_8\,\mathbf{a}_1+x_8\,\mathbf{a}_2+\left(\frac{1}{2}+z_8\right)\mathbf{a}_3$ | = | $-y_8 a\,\hat{\mathbf{x}}+x_8 a\,\hat{\mathbf{y}}+\left(\frac{1}{2}+z_8\right)c\,\hat{\mathbf{z}}$ | (4d) | H IV |
| $\mathbf{B}_{24}$ | = | $y_8\,\mathbf{a}_1-x_8\,\mathbf{a}_2+\left(\frac{1}{2}+z_8\right)\mathbf{a}_3$ | = | $y_8 a\,\hat{\mathbf{x}}-x_8 a\,\hat{\mathbf{y}}+\left(\frac{1}{2}+z_8\right)c\,\hat{\mathbf{z}}$ | (4d) | H IV |
| $\mathbf{B}_{25}$ | = | $x_9\,\mathbf{a}_1+y_9\,\mathbf{a}_2+z_9\,\mathbf{a}_3$ | = | $x_9 a\,\hat{\mathbf{x}}+y_9 a\,\hat{\mathbf{y}}+z_9 c\,\hat{\mathbf{z}}$ | (4d) | H V |
| $\mathbf{B}_{26}$ | = | $-x_9\,\mathbf{a}_1-y_9\,\mathbf{a}_2+z_9\,\mathbf{a}_3$ | = | $-x_9 a\,\hat{\mathbf{x}}-y_9 a\,\hat{\mathbf{y}}+z_9 c\,\hat{\mathbf{z}}$ | (4d) | H V |
| $\mathbf{B}_{27}$ | = | $-y_9\,\mathbf{a}_1+x_9\,\mathbf{a}_2+\left(\frac{1}{2}+z_9\right)\mathbf{a}_3$ | = | $-y_9 a\,\hat{\mathbf{x}}+x_9 a\,\hat{\mathbf{y}}+\left(\frac{1}{2}+z_9\right)c\,\hat{\mathbf{z}}$ | (4d) | H V |
| $\mathbf{B}_{28}$ | = | $y_{9}\,\mathbf{a}_{1} - x_{9}\,\mathbf{a}_{2} + \left(\frac{1}{2} + z_{9}\right)\,\mathbf{a}_{3}$ | = | $y_{9}a\,\mathbf{\hat{x}} - x_{9}a\,\mathbf{\hat{y}} + \left(\frac{1}{2} + z_{9}\right)c\,\mathbf{\hat{z}}$ | $(4d)$ | H V |
| $\mathbf{B}_{29}$ | = | $x_{10}\,\mathbf{a}_{1} + y_{10}\,\mathbf{a}_{2} + z_{10}\,\mathbf{a}_{3}$ | = | $x_{10}a\,\mathbf{\hat{x}} + y_{10}a\,\mathbf{\hat{y}} + z_{10}c\,\mathbf{\hat{z}}$ | $(4d)$ | H VI |
| $\mathbf{B}_{30}$ | = | $-x_{10}\,\mathbf{a}_{1} - y_{10}\,\mathbf{a}_{2} + z_{10}\,\mathbf{a}_{3}$ | = | $-x_{10}a\,\mathbf{\hat{x}} - y_{10}a\,\mathbf{\hat{y}} + z_{10}c\,\mathbf{\hat{z}}$ | $(4d)$ | H VI |
| $\mathbf{B}_{31}$ | = | $-y_{10}\,\mathbf{a}_{1} + x_{10}\,\mathbf{a}_{2} + \left(\frac{1}{2} + z_{10}\right)\,\mathbf{a}_{3}$ | = | $-y_{10}a\,\mathbf{\hat{x}} + x_{10}a\,\mathbf{\hat{y}} + \left(\frac{1}{2} + z_{10}\right)c\,\mathbf{\hat{z}}$ | $(4d)$ | H VI |
| $\mathbf{B}_{32}$ | = | $y_{10}\,\mathbf{a}_{1} - x_{10}\,\mathbf{a}_{2} + \left(\frac{1}{2} + z_{10}\right)\,\mathbf{a}_{3}$ | = | $y_{10}a\,\mathbf{\hat{x}} - x_{10}a\,\mathbf{\hat{y}} + \left(\frac{1}{2} + z_{10}\right)c\,\mathbf{\hat{z}}$ | $(4d)$ | H VI |
| $\mathbf{B}_{33}$ | = | $x_{11}\,\mathbf{a}_{1} + y_{11}\,\mathbf{a}_{2} + z_{11}\,\mathbf{a}_{3}$ | = | $x_{11}a\,\mathbf{\hat{x}} + y_{11}a\,\mathbf{\hat{y}} + z_{11}c\,\mathbf{\hat{z}}$ | $(4d)$ | H VII |
| $\mathbf{B}_{34}$ | = | $-x_{11}\,\mathbf{a}_{1} - y_{11}\,\mathbf{a}_{2} + z_{11}\,\mathbf{a}_{3}$ | = | $-x_{11}a\,\mathbf{\hat{x}} - y_{11}a\,\mathbf{\hat{y}} + z_{11}c\,\mathbf{\hat{z}}$ | $(4d)$ | H VII |
| $\mathbf{B}_{35}$ | = | $-y_{11}\,\mathbf{a}_{1} + x_{11}\,\mathbf{a}_{2} + \left(\frac{1}{2} + z_{11}\right)\,\mathbf{a}_{3}$ | = | $-y_{11}a\,\mathbf{\hat{x}} + x_{11}a\,\mathbf{\hat{y}} + \left(\frac{1}{2} + z_{11}\right)c\,\mathbf{\hat{z}}$ | $(4d)$ | H VII |
| $\mathbf{B}_{36}$ | = | $y_{11}\,\mathbf{a}_{1} - x_{11}\,\mathbf{a}_{2} + \left(\frac{1}{2} + z_{11}\right)\,\mathbf{a}_{3}$ | = | $y_{11}a\,\mathbf{\hat{x}} - x_{11}a\,\mathbf{\hat{y}} + \left(\frac{1}{2} + z_{11}\right)c\,\mathbf{\hat{z}}$ | $(4d)$ | H VII |
| $\mathbf{B}_{37}$ | = | $x_{12}\,\mathbf{a}_{1} + y_{12}\,\mathbf{a}_{2} + z_{12}\,\mathbf{a}_{3}$ | = | $x_{12}a\,\mathbf{\hat{x}} + y_{12}a\,\mathbf{\hat{y}} + z_{12}c\,\mathbf{\hat{z}}$ | $(4d)$ | H VIII |
| $\mathbf{B}_{38}$ | = | $-x_{12}\,\mathbf{a}_{1} - y_{12}\,\mathbf{a}_{2} + z_{12}\,\mathbf{a}_{3}$ | = | $-x_{12}a\,\mathbf{\hat{x}} - y_{12}a\,\mathbf{\hat{y}} + z_{12}c\,\mathbf{\hat{z}}$ | $(4d)$ | H VIII |
| $\mathbf{B}_{39}$ | = | $-y_{12}\,\mathbf{a}_{1} + x_{12}\,\mathbf{a}_{2} + \left(\frac{1}{2} + z_{12}\right)\,\mathbf{a}_{3}$ | = | $-y_{12}a\,\mathbf{\hat{x}} + x_{12}a\,\mathbf{\hat{y}} + \left(\frac{1}{2} + z_{12}\right)c\,\mathbf{\hat{z}}$ | $(4d)$ | H VIII |
| $\mathbf{B}_{40}$ | = | $y_{12}\,\mathbf{a}_{1} - x_{12}\,\mathbf{a}_{2} + \left(\frac{1}{2} + z_{12}\right)\,\mathbf{a}_{3}$ | = | $y_{12}a\,\mathbf{\hat{x}} - x_{12}a\,\mathbf{\hat{y}} + \left(\frac{1}{2} + z_{12}\right)c\,\mathbf{\hat{z}}$ | $(4d)$ | H VIII |
| $\mathbf{B}_{41}$ | = | $x_{13}\,\mathbf{a}_{1} + y_{13}\,\mathbf{a}_{2} + z_{13}\,\mathbf{a}_{3}$ | = | $x_{13}a\,\mathbf{\hat{x}} + y_{13}a\,\mathbf{\hat{y}} + z_{13}c\,\mathbf{\hat{z}}$ | $(4d)$ | N V |
| $\mathbf{B}_{42}$ | = | $-x_{13}\,\mathbf{a}_{1} - y_{13}\,\mathbf{a}_{2} + z_{13}\,\mathbf{a}_{3}$ | = | $-x_{13}a\,\mathbf{\hat{x}} - y_{13}a\,\mathbf{\hat{y}} + z_{13}c\,\mathbf{\hat{z}}$ | $(4d)$ | N V |
| $\mathbf{B}_{43}$ | = | $-y_{13}\,\mathbf{a}_{1} + x_{13}\,\mathbf{a}_{2} + \left(\frac{1}{2} + z_{13}\right)\,\mathbf{a}_{3}$ | = | $-y_{13}a\,\mathbf{\hat{x}} + x_{13}a\,\mathbf{\hat{y}} + \left(\frac{1}{2} + z_{13}\right)c\,\mathbf{\hat{z}}$ | $(4d)$ | N V |
| $\mathbf{B}_{44}$ | = | $y_{13}\,\mathbf{a}_{1} - x_{13}\,\mathbf{a}_{2} + \left(\frac{1}{2} + z_{13}\right)\,\mathbf{a}_{3}$ | = | $y_{13}a\,\mathbf{\hat{x}} - x_{13}a\,\mathbf{\hat{y}} + \left(\frac{1}{2} + z_{13}\right)c\,\mathbf{\hat{z}}$ | $(4d)$ | N V |
| $\mathbf{B}_{45}$ | = | $x_{14}\,\mathbf{a}_{1} + y_{14}\,\mathbf{a}_{2} + z_{14}\,\mathbf{a}_{3}$ | = | $x_{14}a\,\mathbf{\hat{x}} + y_{14}a\,\mathbf{\hat{y}} + z_{14}c\,\mathbf{\hat{z}}$ | $(4d)$ | N VI |
| $\mathbf{B}_{46}$ | = | $-x_{14}\,\mathbf{a}_{1} - y_{14}\,\mathbf{a}_{2} + z_{14}\,\mathbf{a}_{3}$ | = | $-x_{14}a\,\mathbf{\hat{x}} - y_{14}a\,\mathbf{\hat{y}} + z_{14}c\,\mathbf{\hat{z}}$ | $(4d)$ | N VI |
| $\mathbf{B}_{47}$ | = | $-y_{14}\,\mathbf{a}_{1} + x_{14}\,\mathbf{a}_{2} + \left(\frac{1}{2} + z_{14}\right)\,\mathbf{a}_{3}$ | = | $-y_{14}a\,\mathbf{\hat{x}} + x_{14}a\,\mathbf{\hat{y}} + \left(\frac{1}{2} + z_{14}\right)c\,\mathbf{\hat{z}}$ | $(4d)$ | N VI |
| $\mathbf{B}_{48}$ | = | $y_{14}\,\mathbf{a}_{1} - x_{14}\,\mathbf{a}_{2} + \left(\frac{1}{2} + z_{14}\right)\,\mathbf{a}_{3}$ | = | $y_{14}a\,\mathbf{\hat{x}} - x_{14}a\,\mathbf{\hat{y}} + \left(\frac{1}{2} + z_{14}\right)c\,\mathbf{\hat{z}}$ | $(4d)$ | N VI |
| $\mathbf{B}_{49}$ | = | $x_{15}\,\mathbf{a}_{1} + y_{15}\,\mathbf{a}_{2} + z_{15}\,\mathbf{a}_{3}$ | = | $x_{15}a\,\mathbf{\hat{x}} + y_{15}a\,\mathbf{\hat{y}} + z_{15}c\,\mathbf{\hat{z}}$ | $(4d)$ | O I |
| $\mathbf{B}_{50}$ | = | $-x_{15}\,\mathbf{a}_{1} - y_{15}\,\mathbf{a}_{2} + z_{15}\,\mathbf{a}_{3}$ | = | $-x_{15}a\,\mathbf{\hat{x}} - y_{15}a\,\mathbf{\hat{y}} + z_{15}c\,\mathbf{\hat{z}}$ | $(4d)$ | O I |
| $\mathbf{B}_{51}$ | = | $-y_{15}\,\mathbf{a}_{1} + x_{15}\,\mathbf{a}_{2} + \left(\frac{1}{2} + z_{15}\right)\,\mathbf{a}_{3}$ | = | $-y_{15}a\,\mathbf{\hat{x}} + x_{15}a\,\mathbf{\hat{y}} + \left(\frac{1}{2} + z_{15}\right)c\,\mathbf{\hat{z}}$ | $(4d)$ | O I |
| $\mathbf{B}_{52}$ | = | $y_{15}\,\mathbf{a}_{1} - x_{15}\,\mathbf{a}_{2} + \left(\frac{1}{2} + z_{15}\right)\,\mathbf{a}_{3}$ | = | $y_{15}a\,\mathbf{\hat{x}} - x_{15}a\,\mathbf{\hat{y}} + \left(\frac{1}{2} + z_{15}\right)c\,\mathbf{\hat{z}}$ | $(4d)$ | O I |
| $\mathbf{B}_{53}$ | = | $x_{16}\,\mathbf{a}_{1} + y_{16}\,\mathbf{a}_{2} + z_{16}\,\mathbf{a}_{3}$ | = | $x_{16}a\,\mathbf{\hat{x}} + y_{16}a\,\mathbf{\hat{y}} + z_{16}c\,\mathbf{\hat{z}}$ | $(4d)$ | O II |
| $\mathbf{B}_{54}$ | = | $-x_{16}\,\mathbf{a}_{1} - y_{16}\,\mathbf{a}_{2} + z_{16}\,\mathbf{a}_{3}$ | = | $-x_{16}a\,\mathbf{\hat{x}} - y_{16}a\,\mathbf{\hat{y}} + z_{16}c\,\mathbf{\hat{z}}$ | $(4d)$ | O II |
| $\mathbf{B}_{55}$ | = | $-y_{16}\,\mathbf{a}_{1} + x_{16}\,\mathbf{a}_{2} + \left(\frac{1}{2} + z_{16}\right)\,\mathbf{a}_{3}$ | = | $-y_{16}a\,\mathbf{\hat{x}} + x_{16}a\,\mathbf{\hat{y}} + \left(\frac{1}{2} + z_{16}\right)c\,\mathbf{\hat{z}}$ | $(4d)$ | O II |
| $\mathbf{B}_{56}$ | = | $y_{16}\,\mathbf{a}_{1} - x_{16}\,\mathbf{a}_{2} + \left(\frac{1}{2} + z_{16}\right)\,\mathbf{a}_{3}$ | = | $y_{16}a\,\mathbf{\hat{x}} - x_{16}a\,\mathbf{\hat{y}} + \left(\frac{1}{2} + z_{16}\right)c\,\mathbf{\hat{z}}$ | $(4d)$ | O II |
| $\mathbf{B}_{57}$ | = | $x_{17}\,\mathbf{a}_{1} + y_{17}\,\mathbf{a}_{2} + z_{17}\,\mathbf{a}_{3}$ | = | $x_{17}a\,\mathbf{\hat{x}} + y_{17}a\,\mathbf{\hat{y}} + z_{17}c\,\mathbf{\hat{z}}$ | $(4d)$ | O III |
| $\mathbf{B}_{58}$ | = | $-x_{17}\,\mathbf{a}_{1} - y_{17}\,\mathbf{a}_{2} + z_{17}\,\mathbf{a}_{3}$ | = | $-x_{17}a\,\mathbf{\hat{x}} - y_{17}a\,\mathbf{\hat{y}} + z_{17}c\,\mathbf{\hat{z}}$ | $(4d)$ | O III |
| $\mathbf{B}_{59}$ | = | $-y_{17}\,\mathbf{a}_{1} + x_{17}\,\mathbf{a}_{2} + \left(\frac{1}{2} + z_{17}\right)\,\mathbf{a}_{3}$ | = | $-y_{17}a\,\mathbf{\hat{x}} + x_{17}a\,\mathbf{\hat{y}} + \left(\frac{1}{2} + z_{17}\right)c\,\mathbf{\hat{z}}$ | $(4d)$ | O III |
| $\mathbf{B}_{60}$ | = | $y_{17}\,\mathbf{a}_{1} - x_{17}\,\mathbf{a}_{2} + \left(\frac{1}{2} + z_{17}\right)\,\mathbf{a}_{3}$ | = | $y_{17}a\,\mathbf{\hat{x}} - x_{17}a\,\mathbf{\hat{y}} + \left(\frac{1}{2} + z_{17}\right)c\,\mathbf{\hat{z}}$ | $(4d)$ | O III |
| $\mathbf{B}_{61}$ | = | $x_{18}\,\mathbf{a}_{1} + y_{18}\,\mathbf{a}_{2} + z_{18}\,\mathbf{a}_{3}$ | = | $x_{18}a\,\mathbf{\hat{x}} + y_{18}a\,\mathbf{\hat{y}} + z_{18}c\,\mathbf{\hat{z}}$ | $(4d)$ | O IV |
| $\mathbf{B}_{62}$ | = | $-x_{18}\,\mathbf{a}_{1} - y_{18}\,\mathbf{a}_{2} + z_{18}\,\mathbf{a}_{3}$ | = | $-x_{18}a\,\mathbf{\hat{x}} - y_{18}a\,\mathbf{\hat{y}} + z_{18}c\,\mathbf{\hat{z}}$ | $(4d)$ | O IV |
| $\mathbf{B}_{63}$ | = | $-y_{18}\,\mathbf{a}_{1} + x_{18}\,\mathbf{a}_{2} + \left(\frac{1}{2} + z_{18}\right)\,\mathbf{a}_{3}$ | = | $-y_{18}a\,\mathbf{\hat{x}} + x_{18}a\,\mathbf{\hat{y}} + \left(\frac{1}{2} + z_{18}\right)c\,\mathbf{\hat{z}}$ | $(4d)$ | O IV |
| $\mathbf{B}_{64}$ | = | $y_{18}\,\mathbf{a}_1 - x_{18}\,\mathbf{a}_2 + \left(\frac{1}{2} + z_{18}\right)\,\mathbf{a}_3$ | = | $y_{18}a\,\hat{\mathbf{x}} - x_{18}a\,\hat{\mathbf{y}} + \left(\frac{1}{2} + z_{18}\right)c\,\hat{\mathbf{z}}$ | (4$d$) | O IV |
| $\mathbf{B}_{65}$ | = | $x_{19}\,\mathbf{a}_1 + y_{19}\,\mathbf{a}_2 + z_{19}\,\mathbf{a}_3$ | = | $x_{19}a\,\hat{\mathbf{x}} + y_{19}a\,\hat{\mathbf{y}} + z_{19}c\,\hat{\mathbf{z}}$ | (4$d$) | O V |
| $\mathbf{B}_{66}$ | = | $-x_{19}\,\mathbf{a}_1 - y_{19}\,\mathbf{a}_2 + z_{19}\,\mathbf{a}_3$ | = | $-x_{19}a\,\hat{\mathbf{x}} - y_{19}a\,\hat{\mathbf{y}} + z_{19}c\,\hat{\mathbf{z}}$ | (4$d$) | O V |
| $\mathbf{B}_{67}$ | = | $-y_{19}\,\mathbf{a}_1 + x_{19}\,\mathbf{a}_2 + \left(\frac{1}{2} + z_{19}\right)\,\mathbf{a}_3$ | = | $-y_{19}a\,\hat{\mathbf{x}} + x_{19}a\,\hat{\mathbf{y}} + \left(\frac{1}{2} + z_{19}\right)c\,\hat{\mathbf{z}}$ | (4$d$) | O V |
| $\mathbf{B}_{68}$ | = | $y_{19}\,\mathbf{a}_1 - x_{19}\,\mathbf{a}_2 + \left(\frac{1}{2} + z_{19}\right)\,\mathbf{a}_3$ | = | $y_{19}a\,\hat{\mathbf{x}} - x_{19}a\,\hat{\mathbf{y}} + \left(\frac{1}{2} + z_{19}\right)c\,\hat{\mathbf{z}}$ | (4$d$) | O V |
| $\mathbf{B}_{69}$ | = | $x_{20}\,\mathbf{a}_1 + y_{20}\,\mathbf{a}_2 + z_{20}\,\mathbf{a}_3$ | = | $x_{20}a\,\hat{\mathbf{x}} + y_{20}a\,\hat{\mathbf{y}} + z_{20}c\,\hat{\mathbf{z}}$ | (4$d$) | O VI |
| $\mathbf{B}_{70}$ | = | $-x_{20}\,\mathbf{a}_1 - y_{20}\,\mathbf{a}_2 + z_{20}\,\mathbf{a}_3$ | = | $-x_{20}a\,\hat{\mathbf{x}} - y_{20}a\,\hat{\mathbf{y}} + z_{20}c\,\hat{\mathbf{z}}$ | (4$d$) | O VI |
| $\mathbf{B}_{71}$ | = | $-y_{20}\,\mathbf{a}_1 + x_{20}\,\mathbf{a}_2 + \left(\frac{1}{2} + z_{20}\right)\,\mathbf{a}_3$ | = | $-y_{20}a\,\hat{\mathbf{x}} + x_{20}a\,\hat{\mathbf{y}} + \left(\frac{1}{2} + z_{20}\right)c\,\hat{\mathbf{z}}$ | (4$d$) | O VI |
| $\mathbf{B}_{72}$ | = | $y_{20}\,\mathbf{a}_1 - x_{20}\,\mathbf{a}_2 + \left(\frac{1}{2} + z_{20}\right)\,\mathbf{a}_3$ | = | $y_{20}a\,\hat{\mathbf{x}} - x_{20}a\,\hat{\mathbf{y}} + \left(\frac{1}{2} + z_{20}\right)c\,\hat{\mathbf{z}}$ | (4$d$) | O VI |

**Found in:**

**Geometry files:**

- CIF: pp. 1668
- POSCAR: pp. 1669

# Kesterite [$Cu_2$(Zn,Fe)$SnS_4$] Structure: A2BCD4_tI16_82_ac_b_d_g

http://aflow.org/prototype-encyclopedia/A2BCD4_tI16_82_ac_b_d_g

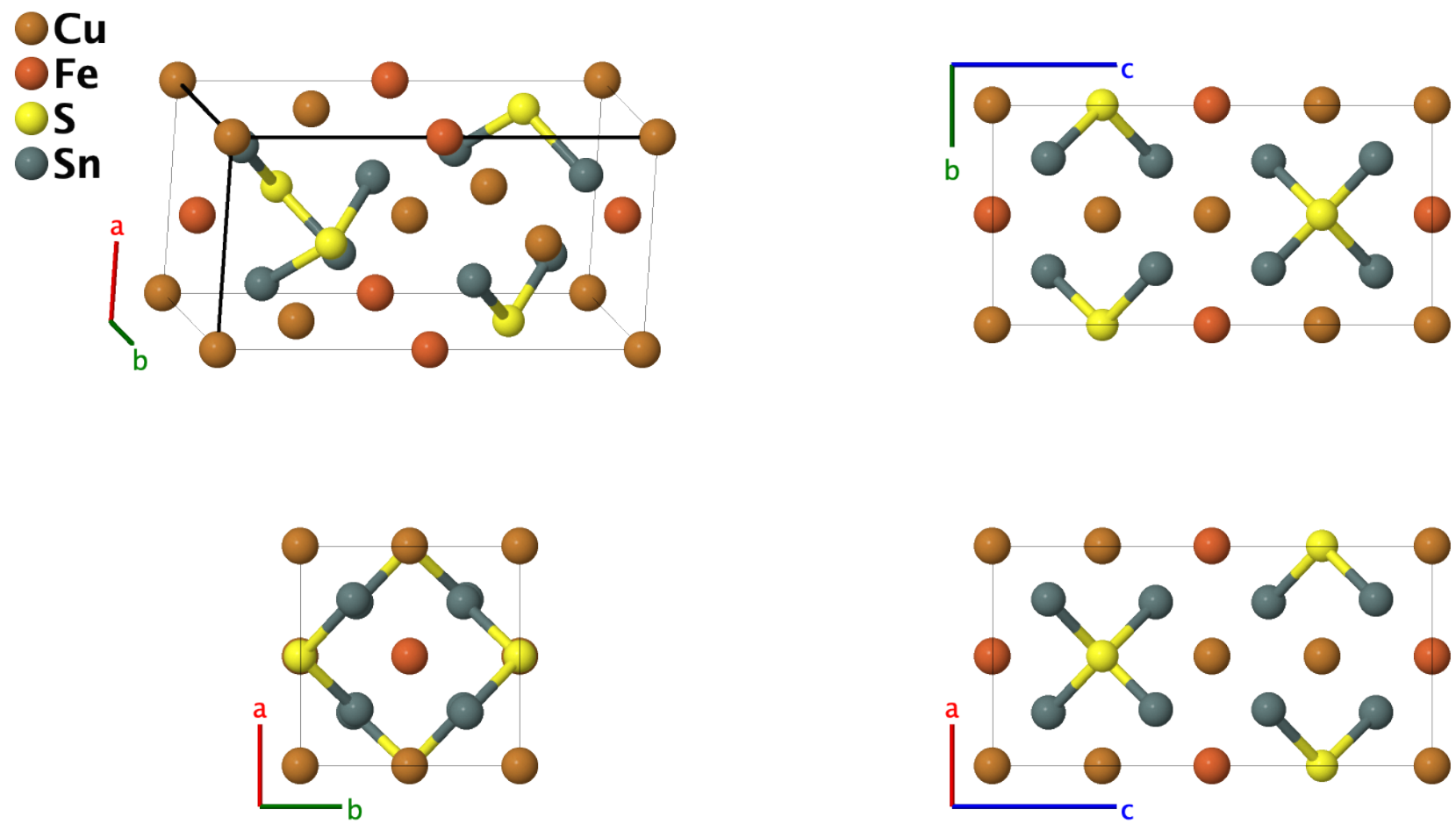

| | | |
|---|---|---|
| **Prototype** | : | $Cu_2$(Fe,Zn)$SnS_4$ |
| **AFLOW prototype label** | : | A2BCD4_tI16_82_ac_b_d_g |
| ***Strukturbericht* designation** | : | None |
| **Pearson symbol** | : | tI16 |
| **Space group number** | : | 82 |
| **Space group symbol** | : | $I\bar{4}$ |
| **AFLOW prototype command** | : | `aflow --proto=A2BCD4_tI16_82_ac_b_d_g --params=`$a, c/a, x_5, y_5, z_5$ |

- The kesterite structure is related to the stannite structure, with differences occurring due to the positioning of the Cu atoms (Hall, 1978). The (2*d*) Wyckoff position is partially occupied with Zn and Fe; however, we decorate it with Fe only.

**Body-centered Tetragonal primitive vectors:**

$$\mathbf{a}_1 = -\frac{1}{2}\,a\,\hat{\mathbf{x}} + \frac{1}{2}\,a\,\hat{\mathbf{y}} + \frac{1}{2}\,c\,\hat{\mathbf{z}}$$
$$\mathbf{a}_2 = \frac{1}{2}\,a\,\hat{\mathbf{x}} - \frac{1}{2}\,a\,\hat{\mathbf{y}} + \frac{1}{2}\,c\,\hat{\mathbf{z}}$$
$$\mathbf{a}_3 = \frac{1}{2}\,a\,\hat{\mathbf{x}} + \frac{1}{2}\,a\,\hat{\mathbf{y}} - \frac{1}{2}\,c\,\hat{\mathbf{z}}$$

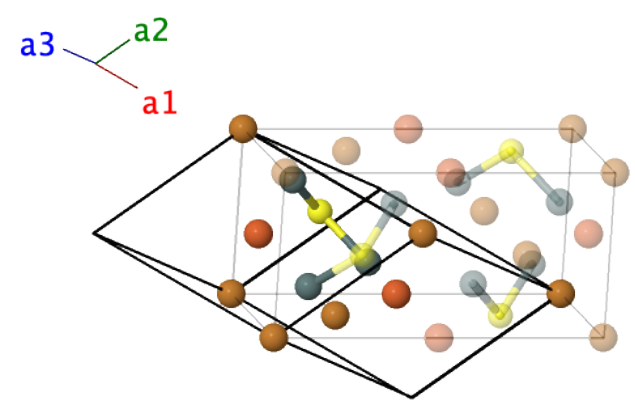

**Basis vectors:**

| | | Lattice Coordinates | | Cartesian Coordinates | Wyckoff Position | Atom Type |
|---|---|---|---|---|---|---|
| $\mathbf{B}_1$ | = | $0\,\mathbf{a}_1 + 0\,\mathbf{a}_2 + 0\,\mathbf{a}_3$ | = | $0\,\hat{\mathbf{x}} + 0\,\hat{\mathbf{y}} + 0\,\hat{\mathbf{z}}$ | (2*a*) | Cu I |
| $\mathbf{B}_2$ | = | $\frac{1}{2}\,\mathbf{a}_1 + \frac{1}{2}\,\mathbf{a}_2$ | = | $\frac{1}{2}c\,\hat{\mathbf{z}}$ | (2*b*) | Fe |
| $\mathbf{B}_3$ | = | $\frac{3}{4}\,\mathbf{a}_1 + \frac{1}{4}\,\mathbf{a}_2 + \frac{1}{2}\,\mathbf{a}_3$ | = | $\frac{1}{2}a\,\hat{\mathbf{y}} + \frac{1}{4}c\,\hat{\mathbf{z}}$ | (2*c*) | Cu II |
| $\mathbf{B}_4$ | = | $\frac{1}{4}\,\mathbf{a}_1 + \frac{3}{4}\,\mathbf{a}_2 + \frac{1}{2}\,\mathbf{a}_3$ | = | $\frac{1}{2}a\,\hat{\mathbf{x}} + \frac{1}{4}c\,\hat{\mathbf{z}}$ | (2*d*) | S |
| $\mathbf{B}_5$ | = | $(y_5 + z_5)\,\mathbf{a}_1 + (x_5 + z_5)\,\mathbf{a}_2 + (x_5 + y_5)\,\mathbf{a}_3$ | = | $x_5 a\,\hat{\mathbf{x}} + y_5 a\,\hat{\mathbf{y}} + z_5 c\,\hat{\mathbf{z}}$ | (8$g$) | Sn |
| $\mathbf{B}_6$ | = | $(-y_5 + z_5)\,\mathbf{a}_1 + (-x_5 + z_5)\,\mathbf{a}_2 + (-x_5 - y_5)\,\mathbf{a}_3$ | = | $-x_5 a\,\hat{\mathbf{x}} - y_5 a\,\hat{\mathbf{y}} + z_5 c\,\hat{\mathbf{z}}$ | (8$g$) | Sn |
| $\mathbf{B}_7$ | = | $(-x_5 - z_5)\,\mathbf{a}_1 + (y_5 - z_5)\,\mathbf{a}_2 + (-x_5 + y_5)\,\mathbf{a}_3$ | = | $y_5 a\,\hat{\mathbf{x}} - x_5 a\,\hat{\mathbf{y}} - z_5 c\,\hat{\mathbf{z}}$ | (8$g$) | Sn |
| $\mathbf{B}_8$ | = | $(x_5 - z_5)\,\mathbf{a}_1 + (-y_5 - z_5)\,\mathbf{a}_2 + (x_5 - y_5)\,\mathbf{a}_3$ | = | $-y_5 a\,\hat{\mathbf{x}} + x_5 a\,\hat{\mathbf{y}} - z_5 c\,\hat{\mathbf{z}}$ | (8$g$) | Sn |

---

---

**Geometry files:**

- CIF: pp. 1669
- POSCAR: pp. 1670

# Bromocarnallite ($KMg(H_2O)_6(Cl,Br)_3$, $E2_6$) Structure: A3B6CD_tP44_85_bcg_3g_ac_e

http://aflow.org/prototype-encyclopedia/A3B6CD_tP44_85_bcg_3g_ac_e

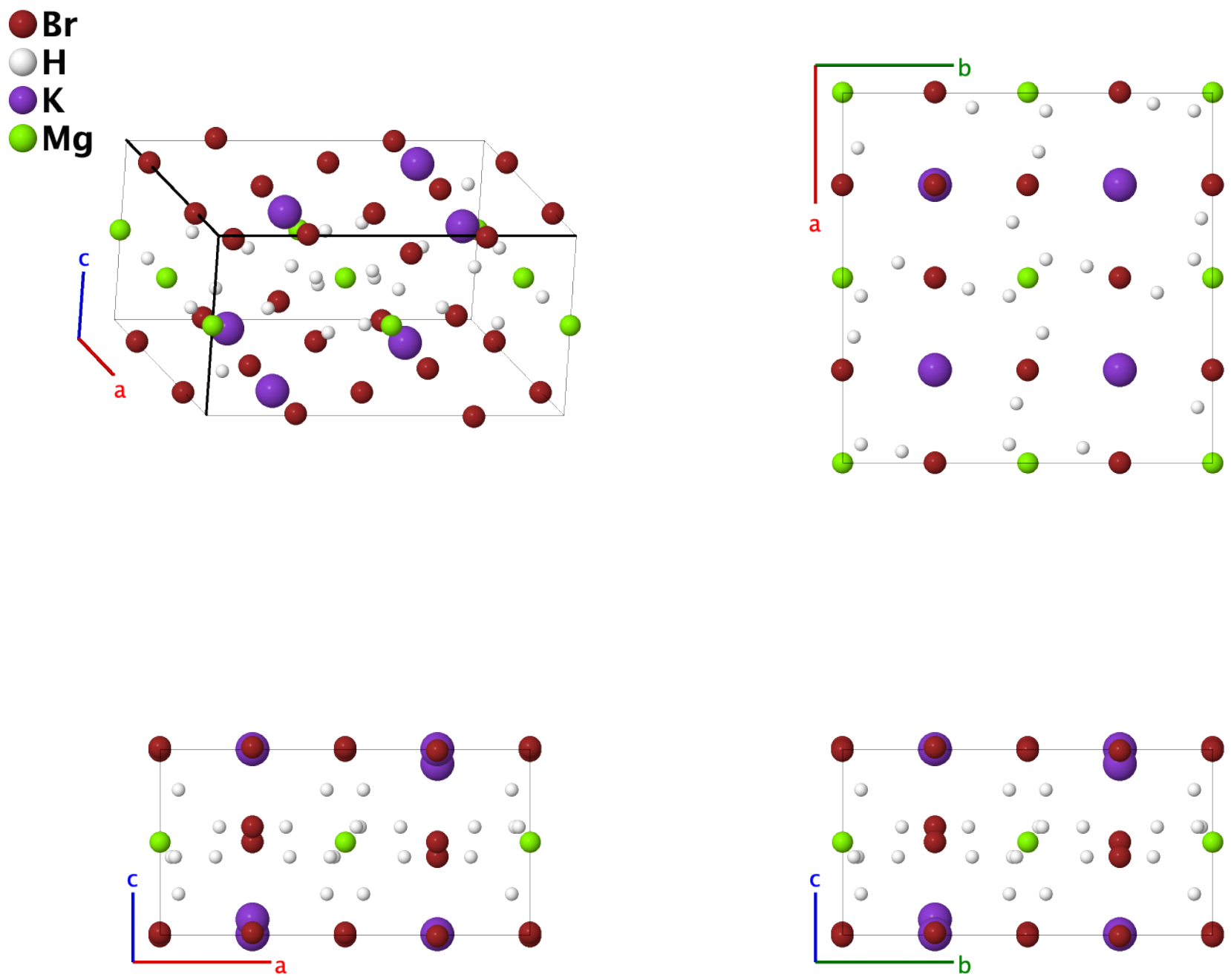

| | | |
|---|---|---|
| **Prototype** | : | $(Br,Cl)_3(H_2O)_6KMg$ |
| **AFLOW prototype label** | : | A3B6CD_tP44_85_bcg_3g_ac_e |
| ***Strukturbericht* designation** | : | $E2_6$ |
| **Pearson symbol** | : | tP44 |
| **Space group number** | : | 85 |
| **Space group symbol** | : | $P4/n$ |
| **AFLOW prototype command** | : | `aflow --proto=A3B6CD_tP44_85_bcg_3g_ac_e` `--params=`$a, c/a, z_3, z_4, x_6, y_6, z_6, x_7, y_7, z_7, x_8, y_8, z_8, x_9, y_9, z_9$ |

- (Andreß, 1939) first determined the structure of Bromocarnallite, using a sample with 75% bromine on the halide site. They were unable to locate the hydrogen atoms, and placed the compound in space group $P4/n$ #85. The data was given by (Hermann, 1939) in setting 1 of this group, but we used FINDSYM to place it in the standard setting 2.
- (Hermann, 1939) assigned this compound the *Strukturbericht* symbol $E2_6$.
- (Schlemper, 1985) re-examined the structure for the pure chlorine version, Carnallite. They located the hydrogen atoms and placed the system in space group *Pnna* #52. We present this structure in the Carnallite structure page.

## Simple Tetragonal primitive vectors:

$$\mathbf{a}_1 = a\,\hat{\mathbf{x}}$$
$$\mathbf{a}_2 = a\,\hat{\mathbf{y}}$$
$$\mathbf{a}_3 = c\,\hat{\mathbf{z}}$$

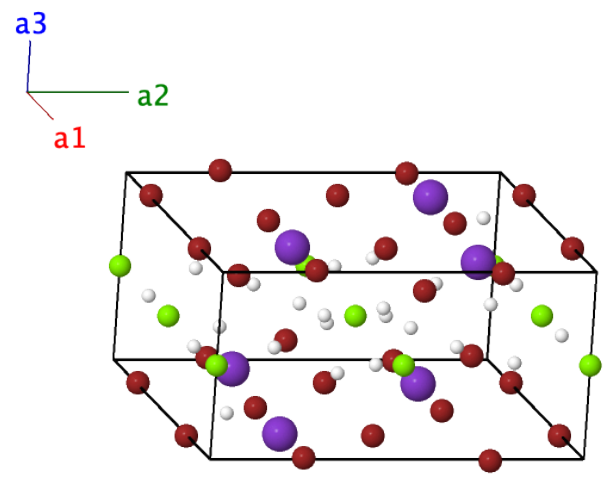

## Basis vectors:

| | | Lattice Coordinates | | Cartesian Coordinates | Wyckoff Position | Atom Type |
|---|---|---|---|---|---|---|
| $\mathbf{B}_1$ | = | $\frac{1}{4}\,\mathbf{a}_1 + \frac{3}{4}\,\mathbf{a}_2$ | = | $\frac{1}{4}a\,\hat{\mathbf{x}} + \frac{3}{4}a\,\hat{\mathbf{y}}$ | (2$a$) | K I |
| $\mathbf{B}_2$ | = | $\frac{3}{4}\,\mathbf{a}_1 + \frac{1}{4}\,\mathbf{a}_2$ | = | $\frac{3}{4}a\,\hat{\mathbf{x}} + \frac{1}{4}a\,\hat{\mathbf{y}}$ | (2$a$) | K I |
| $\mathbf{B}_3$ | = | $\frac{1}{4}\,\mathbf{a}_1 + \frac{3}{4}\,\mathbf{a}_2 + \frac{1}{2}\,\mathbf{a}_3$ | = | $\frac{1}{4}a\,\hat{\mathbf{x}} + \frac{3}{4}a\,\hat{\mathbf{y}} + \frac{1}{2}c\,\hat{\mathbf{z}}$ | (2$b$) | Br I |
| $\mathbf{B}_4$ | = | $\frac{3}{4}\,\mathbf{a}_1 + \frac{1}{4}\,\mathbf{a}_2 + \frac{1}{2}\,\mathbf{a}_3$ | = | $\frac{3}{4}a\,\hat{\mathbf{x}} + \frac{1}{4}a\,\hat{\mathbf{y}} + \frac{1}{2}c\,\hat{\mathbf{z}}$ | (2$b$) | Br I |
| $\mathbf{B}_5$ | = | $\frac{1}{4}\,\mathbf{a}_1 + \frac{1}{4}\,\mathbf{a}_2 + z_3\,\mathbf{a}_3$ | = | $\frac{1}{4}a\,\hat{\mathbf{x}} + \frac{1}{4}a\,\hat{\mathbf{y}} + z_3c\,\hat{\mathbf{z}}$ | (2$c$) | Br II |
| $\mathbf{B}_6$ | = | $\frac{3}{4}\,\mathbf{a}_1 + \frac{3}{4}\,\mathbf{a}_2 - z_3\,\mathbf{a}_3$ | = | $\frac{3}{4}a\,\hat{\mathbf{x}} + \frac{3}{4}a\,\hat{\mathbf{y}} - z_3c\,\hat{\mathbf{z}}$ | (2$c$) | Br II |
| $\mathbf{B}_7$ | = | $\frac{1}{4}\,\mathbf{a}_1 + \frac{1}{4}\,\mathbf{a}_2 + z_4\,\mathbf{a}_3$ | = | $\frac{1}{4}a\,\hat{\mathbf{x}} + \frac{1}{4}a\,\hat{\mathbf{y}} + z_4c\,\hat{\mathbf{z}}$ | (2$c$) | K II |
| $\mathbf{B}_8$ | = | $\frac{3}{4}\,\mathbf{a}_1 + \frac{3}{4}\,\mathbf{a}_2 - z_4\,\mathbf{a}_3$ | = | $\frac{3}{4}a\,\hat{\mathbf{x}} + \frac{3}{4}a\,\hat{\mathbf{y}} - z_4c\,\hat{\mathbf{z}}$ | (2$c$) | K II |
| $\mathbf{B}_9$ | = | $\frac{1}{2}\,\mathbf{a}_3$ | = | $\frac{1}{2}c\,\hat{\mathbf{z}}$ | (4$e$) | Mg |
| $\mathbf{B}_{10}$ | = | $\frac{1}{2}\,\mathbf{a}_1 + \frac{1}{2}\,\mathbf{a}_2 + \frac{1}{2}\,\mathbf{a}_3$ | = | $\frac{1}{2}a\,\hat{\mathbf{x}} + \frac{1}{2}a\,\hat{\mathbf{y}} + \frac{1}{2}c\,\hat{\mathbf{z}}$ | (4$e$) | Mg |
| $\mathbf{B}_{11}$ | = | $\frac{1}{2}\,\mathbf{a}_1 + \frac{1}{2}\,\mathbf{a}_3$ | = | $\frac{1}{2}a\,\hat{\mathbf{x}} + \frac{1}{2}c\,\hat{\mathbf{z}}$ | (4$e$) | Mg |
| $\mathbf{B}_{12}$ | = | $\frac{1}{2}\,\mathbf{a}_2 + \frac{1}{2}\,\mathbf{a}_3$ | = | $\frac{1}{2}a\,\hat{\mathbf{y}} + \frac{1}{2}c\,\hat{\mathbf{z}}$ | (4$e$) | Mg |
| $\mathbf{B}_{13}$ | = | $x_6\,\mathbf{a}_1 + y_6\,\mathbf{a}_2 + z_6\,\mathbf{a}_3$ | = | $x_6a\,\hat{\mathbf{x}} + y_6a\,\hat{\mathbf{y}} + z_6c\,\hat{\mathbf{z}}$ | (8$g$) | Br III |
| $\mathbf{B}_{14}$ | = | $\left(\frac{1}{2} - x_6\right)\,\mathbf{a}_1 + \left(\frac{1}{2} - y_6\right)\,\mathbf{a}_2 + z_6\,\mathbf{a}_3$ | = | $\left(\frac{1}{2} - x_6\right)a\,\hat{\mathbf{x}} + \left(\frac{1}{2} - y_6\right)a\,\hat{\mathbf{y}} + z_6c\,\hat{\mathbf{z}}$ | (8$g$) | Br III |
| $\mathbf{B}_{15}$ | = | $\left(\frac{1}{2} - y_6\right)\,\mathbf{a}_1 + x_6\,\mathbf{a}_2 + z_6\,\mathbf{a}_3$ | = | $\left(\frac{1}{2} - y_6\right)a\,\hat{\mathbf{x}} + x_6a\,\hat{\mathbf{y}} + z_6c\,\hat{\mathbf{z}}$ | (8$g$) | Br III |
| $\mathbf{B}_{16}$ | = | $y_6\,\mathbf{a}_1 + \left(\frac{1}{2} - x_6\right)\,\mathbf{a}_2 + z_6\,\mathbf{a}_3$ | = | $y_6a\,\hat{\mathbf{x}} + \left(\frac{1}{2} - x_6\right)a\,\hat{\mathbf{y}} + z_6c\,\hat{\mathbf{z}}$ | (8$g$) | Br III |
| $\mathbf{B}_{17}$ | = | $-x_6\,\mathbf{a}_1 - y_6\,\mathbf{a}_2 - z_6\,\mathbf{a}_3$ | = | $-x_6a\,\hat{\mathbf{x}} - y_6a\,\hat{\mathbf{y}} - z_6c\,\hat{\mathbf{z}}$ | (8$g$) | Br III |
| $\mathbf{B}_{18}$ | = | $\left(\frac{1}{2} + x_6\right)\,\mathbf{a}_1 + \left(\frac{1}{2} + y_6\right)\,\mathbf{a}_2 - z_6\,\mathbf{a}_3$ | = | $\left(\frac{1}{2} + x_6\right)a\,\hat{\mathbf{x}} + \left(\frac{1}{2} + y_6\right)a\,\hat{\mathbf{y}} - z_6c\,\hat{\mathbf{z}}$ | (8$g$) | Br III |
| $\mathbf{B}_{19}$ | = | $\left(\frac{1}{2} + y_6\right)\,\mathbf{a}_1 - x_6\,\mathbf{a}_2 - z_6\,\mathbf{a}_3$ | = | $\left(\frac{1}{2} + y_6\right)a\,\hat{\mathbf{x}} - x_6a\,\hat{\mathbf{y}} - z_6c\,\hat{\mathbf{z}}$ | (8$g$) | Br III |
| $\mathbf{B}_{20}$ | = | $-y_6\,\mathbf{a}_1 + \left(\frac{1}{2} + x_6\right)\,\mathbf{a}_2 - z_6\,\mathbf{a}_3$ | = | $-y_6a\,\hat{\mathbf{x}} + \left(\frac{1}{2} + x_6\right)a\,\hat{\mathbf{y}} - z_6c\,\hat{\mathbf{z}}$ | (8$g$) | Br III |
| $\mathbf{B}_{21}$ | = | $x_7\,\mathbf{a}_1 + y_7\,\mathbf{a}_2 + z_7\,\mathbf{a}_3$ | = | $x_7a\,\hat{\mathbf{x}} + y_7a\,\hat{\mathbf{y}} + z_7c\,\hat{\mathbf{z}}$ | (8$g$) | $H_2O$ I |
| $\mathbf{B}_{22}$ | = | $\left(\frac{1}{2} - x_7\right)\,\mathbf{a}_1 + \left(\frac{1}{2} - y_7\right)\,\mathbf{a}_2 + z_7\,\mathbf{a}_3$ | = | $\left(\frac{1}{2} - x_7\right)a\,\hat{\mathbf{x}} + \left(\frac{1}{2} - y_7\right)a\,\hat{\mathbf{y}} + z_7c\,\hat{\mathbf{z}}$ | (8$g$) | $H_2O$ I |
| $\mathbf{B}_{23}$ | = | $\left(\frac{1}{2} - y_7\right)\,\mathbf{a}_1 + x_7\,\mathbf{a}_2 + z_7\,\mathbf{a}_3$ | = | $\left(\frac{1}{2} - y_7\right)a\,\hat{\mathbf{x}} + x_7a\,\hat{\mathbf{y}} + z_7c\,\hat{\mathbf{z}}$ | (8$g$) | $H_2O$ I |
| $\mathbf{B}_{24}$ | = | $y_7\,\mathbf{a}_1 + \left(\frac{1}{2} - x_7\right)\,\mathbf{a}_2 + z_7\,\mathbf{a}_3$ | = | $y_7a\,\hat{\mathbf{x}} + \left(\frac{1}{2} - x_7\right)a\,\hat{\mathbf{y}} + z_7c\,\hat{\mathbf{z}}$ | (8$g$) | $H_2O$ I |
| $\mathbf{B}_{25}$ | = | $-x_7\,\mathbf{a}_1 - y_7\,\mathbf{a}_2 - z_7\,\mathbf{a}_3$ | = | $-x_7a\,\hat{\mathbf{x}} - y_7a\,\hat{\mathbf{y}} - z_7c\,\hat{\mathbf{z}}$ | (8$g$) | $H_2O$ I |
| $\mathbf{B}_{26}$ | = | $\left(\frac{1}{2} + x_7\right)\,\mathbf{a}_1 + \left(\frac{1}{2} + y_7\right)\,\mathbf{a}_2 - z_7\,\mathbf{a}_3$ | = | $\left(\frac{1}{2} + x_7\right)a\,\hat{\mathbf{x}} + \left(\frac{1}{2} + y_7\right)a\,\hat{\mathbf{y}} - z_7c\,\hat{\mathbf{z}}$ | (8$g$) | $H_2O$ I |
| $\mathbf{B}_{27}$ | = | $\left(\frac{1}{2} + y_7\right)\,\mathbf{a}_1 - x_7\,\mathbf{a}_2 - z_7\,\mathbf{a}_3$ | = | $\left(\frac{1}{2} + y_7\right)a\,\hat{\mathbf{x}} - x_7a\,\hat{\mathbf{y}} - z_7c\,\hat{\mathbf{z}}$ | (8$g$) | $H_2O$ I |
| $\mathbf{B}_{28}$ | = | $-y_7\,\mathbf{a}_1+\left(\frac{1}{2}+x_7\right)\,\mathbf{a}_2-z_7\,\mathbf{a}_3$ | = | $-y_7a\,\hat{\mathbf{x}}+\left(\frac{1}{2}+x_7\right)a\,\hat{\mathbf{y}}-z_7c\,\hat{\mathbf{z}}$ | (8g) | $H_2O$ I |
| $\mathbf{B}_{29}$ | = | $x_8\,\mathbf{a}_1+y_8\,\mathbf{a}_2+z_8\,\mathbf{a}_3$ | = | $x_8a\,\hat{\mathbf{x}}+y_8a\,\hat{\mathbf{y}}+z_8c\,\hat{\mathbf{z}}$ | (8g) | $H_2O$ II |
| $\mathbf{B}_{30}$ | = | $\left(\frac{1}{2}-x_8\right)\,\mathbf{a}_1+\left(\frac{1}{2}-y_8\right)\,\mathbf{a}_2+z_8\,\mathbf{a}_3$ | = | $\left(\frac{1}{2}-x_8\right)a\,\hat{\mathbf{x}}+\left(\frac{1}{2}-y_8\right)a\,\hat{\mathbf{y}}+z_8c\,\hat{\mathbf{z}}$ | (8g) | $H_2O$ II |
| $\mathbf{B}_{31}$ | = | $\left(\frac{1}{2}-y_8\right)\,\mathbf{a}_1+x_8\,\mathbf{a}_2+z_8\,\mathbf{a}_3$ | = | $\left(\frac{1}{2}-y_8\right)a\,\hat{\mathbf{x}}+x_8a\,\hat{\mathbf{y}}+z_8c\,\hat{\mathbf{z}}$ | (8g) | $H_2O$ II |
| $\mathbf{B}_{32}$ | = | $y_8\,\mathbf{a}_1+\left(\frac{1}{2}-x_8\right)\,\mathbf{a}_2+z_8\,\mathbf{a}_3$ | = | $y_8a\,\hat{\mathbf{x}}+\left(\frac{1}{2}-x_8\right)a\,\hat{\mathbf{y}}+z_8c\,\hat{\mathbf{z}}$ | (8g) | $H_2O$ II |
| $\mathbf{B}_{33}$ | = | $-x_8\,\mathbf{a}_1-y_8\,\mathbf{a}_2-z_8\,\mathbf{a}_3$ | = | $-x_8a\,\hat{\mathbf{x}}-y_8a\,\hat{\mathbf{y}}-z_8c\,\hat{\mathbf{z}}$ | (8g) | $H_2O$ II |
| $\mathbf{B}_{34}$ | = | $\left(\frac{1}{2}+x_8\right)\,\mathbf{a}_1+\left(\frac{1}{2}+y_8\right)\,\mathbf{a}_2-z_8\,\mathbf{a}_3$ | = | $\left(\frac{1}{2}+x_8\right)a\,\hat{\mathbf{x}}+\left(\frac{1}{2}+y_8\right)a\,\hat{\mathbf{y}}-z_8c\,\hat{\mathbf{z}}$ | (8g) | $H_2O$ II |
| $\mathbf{B}_{35}$ | = | $\left(\frac{1}{2}+y_8\right)\,\mathbf{a}_1-x_8\,\mathbf{a}_2-z_8\,\mathbf{a}_3$ | = | $\left(\frac{1}{2}+y_8\right)a\,\hat{\mathbf{x}}-x_8a\,\hat{\mathbf{y}}-z_8c\,\hat{\mathbf{z}}$ | (8g) | $H_2O$ II |
| $\mathbf{B}_{36}$ | = | $-y_8\,\mathbf{a}_1+\left(\frac{1}{2}+x_8\right)\,\mathbf{a}_2-z_8\,\mathbf{a}_3$ | = | $-y_8a\,\hat{\mathbf{x}}+\left(\frac{1}{2}+x_8\right)a\,\hat{\mathbf{y}}-z_8c\,\hat{\mathbf{z}}$ | (8g) | $H_2O$ II |
| $\mathbf{B}_{37}$ | = | $x_9\,\mathbf{a}_1+y_9\,\mathbf{a}_2+z_9\,\mathbf{a}_3$ | = | $x_9a\,\hat{\mathbf{x}}+y_9a\,\hat{\mathbf{y}}+z_9c\,\hat{\mathbf{z}}$ | (8g) | $H_2O$ III |
| $\mathbf{B}_{38}$ | = | $\left(\frac{1}{2}-x_9\right)\,\mathbf{a}_1+\left(\frac{1}{2}-y_9\right)\,\mathbf{a}_2+z_9\,\mathbf{a}_3$ | = | $\left(\frac{1}{2}-x_9\right)a\,\hat{\mathbf{x}}+\left(\frac{1}{2}-y_9\right)a\,\hat{\mathbf{y}}+z_9c\,\hat{\mathbf{z}}$ | (8g) | $H_2O$ III |
| $\mathbf{B}_{39}$ | = | $\left(\frac{1}{2}-y_9\right)\,\mathbf{a}_1+x_9\,\mathbf{a}_2+z_9\,\mathbf{a}_3$ | = | $\left(\frac{1}{2}-y_9\right)a\,\hat{\mathbf{x}}+x_9a\,\hat{\mathbf{y}}+z_9c\,\hat{\mathbf{z}}$ | (8g) | $H_2O$ III |
| $\mathbf{B}_{40}$ | = | $y_9\,\mathbf{a}_1+\left(\frac{1}{2}-x_9\right)\,\mathbf{a}_2+z_9\,\mathbf{a}_3$ | = | $y_9a\,\hat{\mathbf{x}}+\left(\frac{1}{2}-x_9\right)a\,\hat{\mathbf{y}}+z_9c\,\hat{\mathbf{z}}$ | (8g) | $H_2O$ III |
| $\mathbf{B}_{41}$ | = | $-x_9\,\mathbf{a}_1-y_9\,\mathbf{a}_2-z_9\,\mathbf{a}_3$ | = | $-x_9a\,\hat{\mathbf{x}}-y_9a\,\hat{\mathbf{y}}-z_9c\,\hat{\mathbf{z}}$ | (8g) | $H_2O$ III |
| $\mathbf{B}_{42}$ | = | $\left(\frac{1}{2}+x_9\right)\,\mathbf{a}_1+\left(\frac{1}{2}+y_9\right)\,\mathbf{a}_2-z_9\,\mathbf{a}_3$ | = | $\left(\frac{1}{2}+x_9\right)a\,\hat{\mathbf{x}}+\left(\frac{1}{2}+y_9\right)a\,\hat{\mathbf{y}}-z_9c\,\hat{\mathbf{z}}$ | (8g) | $H_2O$ III |
| $\mathbf{B}_{43}$ | = | $\left(\frac{1}{2}+y_9\right)\,\mathbf{a}_1-x_9\,\mathbf{a}_2-z_9\,\mathbf{a}_3$ | = | $\left(\frac{1}{2}+y_9\right)a\,\hat{\mathbf{x}}-x_9a\,\hat{\mathbf{y}}-z_9c\,\hat{\mathbf{z}}$ | (8g) | $H_2O$ III |
| $\mathbf{B}_{44}$ | = | $-y_9\,\mathbf{a}_1+\left(\frac{1}{2}+x_9\right)\,\mathbf{a}_2-z_9\,\mathbf{a}_3$ | = | $-y_9a\,\hat{\mathbf{x}}+\left(\frac{1}{2}+x_9\right)a\,\hat{\mathbf{y}}-z_9c\,\hat{\mathbf{z}}$ | (8g) | $H_2O$ III |

**Found in:**

**Geometry files:**

- CIF: pp. 1670
- POSCAR: pp. 1670

# $MoPO_5$ Structure: AB5C_tP14_85_c_cg_b

http://aflow.org/prototype-encyclopedia/AB5C_tP14_85_c_cg_b

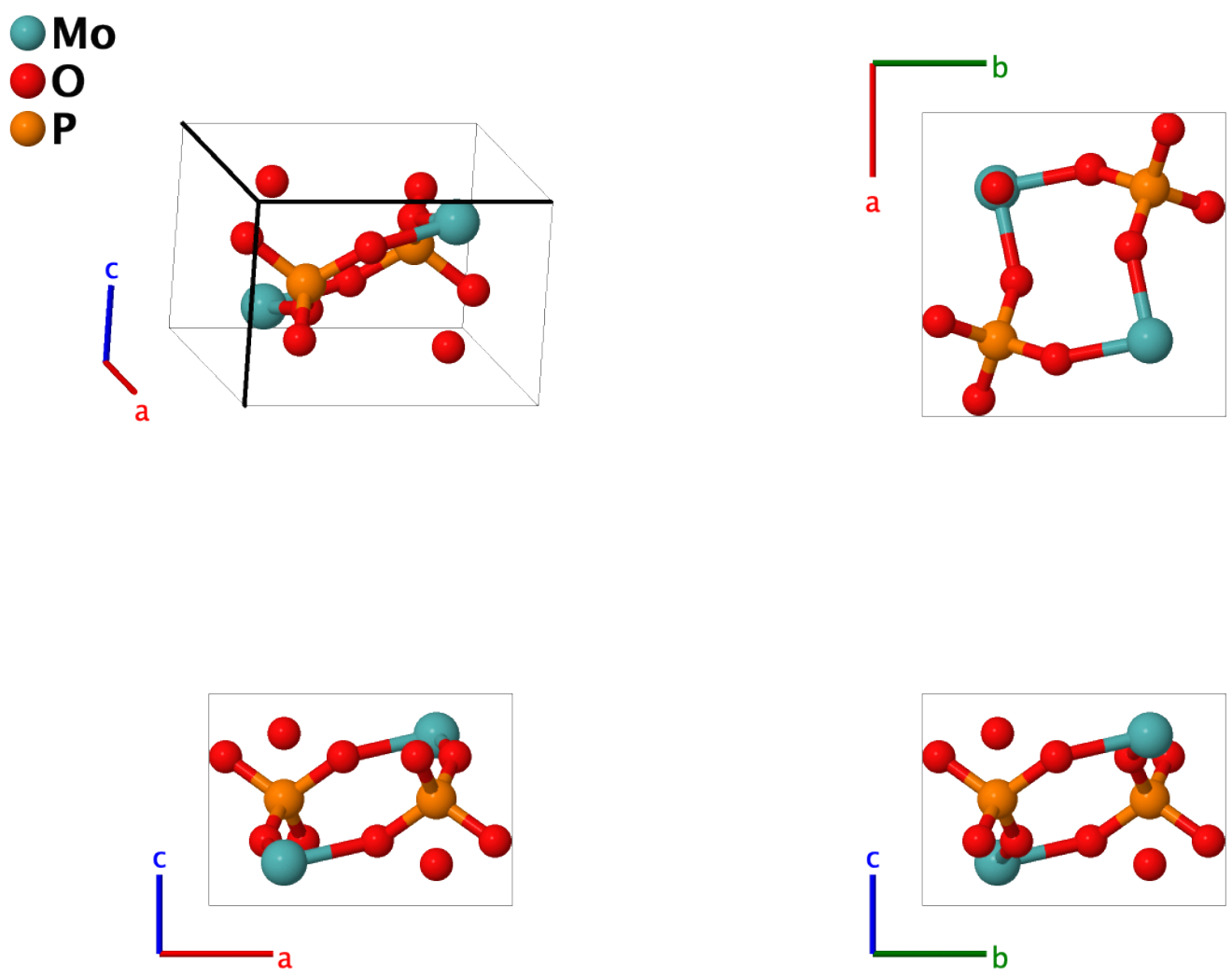

| | | |
|---|---|---|
| **Prototype** | : | $MoO_5P$ |
| **AFLOW prototype label** | : | AB5C_tP14_85_c_cg_b |
| ***Strukturbericht* designation** | : | None |
| **Pearson symbol** | : | tP14 |
| **Space group number** | : | 85 |
| **Space group symbol** | : | $P4/n$ |
| **AFLOW prototype command** | : | `aflow --proto=AB5C_tP14_85_c_cg_b --params=`$a, c/a, z_2, z_3, x_4, y_4, z_4$ |

- (Kierkegaard, 1964) gave this structure in setting 1 of space group $P4/n$ #85. We used FINDSYM to shift this to the standard setting 2.

**Simple Tetragonal primitive vectors:**

$$\mathbf{a}_1 = a\,\hat{\mathbf{x}}$$
$$\mathbf{a}_2 = a\,\hat{\mathbf{y}}$$
$$\mathbf{a}_3 = c\,\hat{\mathbf{z}}$$

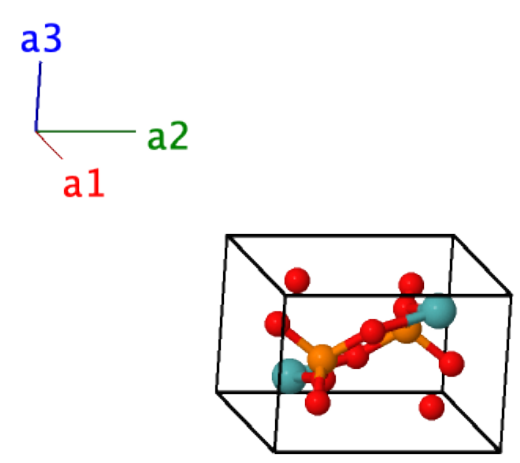

**Basis vectors:**

| | | Lattice Coordinates | | Cartesian Coordinates | Wyckoff Position | Atom Type |
|---|---|---|---|---|---|---|
| $\mathbf{B}_1$ | = | $\frac{1}{4}\,\mathbf{a}_1 + \frac{3}{4}\,\mathbf{a}_2 + \frac{1}{2}\,\mathbf{a}_3$ | = | $\frac{1}{4}a\,\hat{\mathbf{x}} + \frac{3}{4}a\,\hat{\mathbf{y}} + \frac{1}{2}c\,\hat{\mathbf{z}}$ | (2$b$) | P |
| $\mathbf{B}_2$ | = | $\frac{3}{4}\,\mathbf{a}_1 + \frac{1}{4}\,\mathbf{a}_2 + \frac{1}{2}\,\mathbf{a}_3$ | = | $\frac{3}{4}a\,\hat{\mathbf{x}} + \frac{1}{4}a\,\hat{\mathbf{y}} + \frac{1}{2}c\,\hat{\mathbf{z}}$ | (2$b$) | P |
| $\mathbf{B}_3$ | = | $\frac{1}{4}\,\mathbf{a}_1 + \frac{1}{4}\,\mathbf{a}_2 + z_2\,\mathbf{a}_3$ | = | $\frac{1}{4}a\,\hat{\mathbf{x}} + \frac{1}{4}a\,\hat{\mathbf{y}} + z_2c\,\hat{\mathbf{z}}$ | (2$c$) | Mo |
| $\mathbf{B}_4$ | = | $\frac{3}{4}\,\mathbf{a}_1 + \frac{3}{4}\,\mathbf{a}_2 - z_2\,\mathbf{a}_3$ | = | $\frac{3}{4}a\,\hat{\mathbf{x}} + \frac{3}{4}a\,\hat{\mathbf{y}} - z_2c\,\hat{\mathbf{z}}$ | (2c) | Mo |
| $\mathbf{B}_5$ | = | $\frac{1}{4}\,\mathbf{a}_1 + \frac{1}{4}\,\mathbf{a}_2 + z_3\,\mathbf{a}_3$ | = | $\frac{1}{4}a\,\hat{\mathbf{x}} + \frac{1}{4}a\,\hat{\mathbf{y}} + z_3c\,\hat{\mathbf{z}}$ | (2c) | O I |
| $\mathbf{B}_6$ | = | $\frac{3}{4}\,\mathbf{a}_1 + \frac{3}{4}\,\mathbf{a}_2 - z_3\,\mathbf{a}_3$ | = | $\frac{3}{4}a\,\hat{\mathbf{x}} + \frac{3}{4}a\,\hat{\mathbf{y}} - z_3c\,\hat{\mathbf{z}}$ | (2c) | O I |
| $\mathbf{B}_7$ | = | $x_4\,\mathbf{a}_1 + y_4\,\mathbf{a}_2 + z_4\,\mathbf{a}_3$ | = | $x_4a\,\hat{\mathbf{x}} + y_4a\,\hat{\mathbf{y}} + z_4c\,\hat{\mathbf{z}}$ | (8g) | O II |
| $\mathbf{B}_8$ | = | $\left(\frac{1}{2} - x_4\right)\,\mathbf{a}_1 + \left(\frac{1}{2} - y_4\right)\,\mathbf{a}_2 + z_4\,\mathbf{a}_3$ | = | $\left(\frac{1}{2} - x_4\right)a\,\hat{\mathbf{x}} + \left(\frac{1}{2} - y_4\right)a\,\hat{\mathbf{y}} + z_4c\,\hat{\mathbf{z}}$ | (8g) | O II |
| $\mathbf{B}_9$ | = | $\left(\frac{1}{2} - y_4\right)\,\mathbf{a}_1 + x_4\,\mathbf{a}_2 + z_4\,\mathbf{a}_3$ | = | $\left(\frac{1}{2} - y_4\right)a\,\hat{\mathbf{x}} + x_4a\,\hat{\mathbf{y}} + z_4c\,\hat{\mathbf{z}}$ | (8g) | O II |
| $\mathbf{B}_{10}$ | = | $y_4\,\mathbf{a}_1 + \left(\frac{1}{2} - x_4\right)\,\mathbf{a}_2 + z_4\,\mathbf{a}_3$ | = | $y_4a\,\hat{\mathbf{x}} + \left(\frac{1}{2} - x_4\right)a\,\hat{\mathbf{y}} + z_4c\,\hat{\mathbf{z}}$ | (8g) | O II |
| $\mathbf{B}_{11}$ | = | $-x_4\,\mathbf{a}_1 - y_4\,\mathbf{a}_2 - z_4\,\mathbf{a}_3$ | = | $-x_4a\,\hat{\mathbf{x}} - y_4a\,\hat{\mathbf{y}} - z_4c\,\hat{\mathbf{z}}$ | (8g) | O II |
| $\mathbf{B}_{12}$ | = | $\left(\frac{1}{2} + x_4\right)\,\mathbf{a}_1 + \left(\frac{1}{2} + y_4\right)\,\mathbf{a}_2 - z_4\,\mathbf{a}_3$ | = | $\left(\frac{1}{2} + x_4\right)a\,\hat{\mathbf{x}} + \left(\frac{1}{2} + y_4\right)a\,\hat{\mathbf{y}} - z_4c\,\hat{\mathbf{z}}$ | (8g) | O II |
| $\mathbf{B}_{13}$ | = | $\left(\frac{1}{2} + y_4\right)\,\mathbf{a}_1 - x_4\,\mathbf{a}_2 - z_4\,\mathbf{a}_3$ | = | $\left(\frac{1}{2} + y_4\right)a\,\hat{\mathbf{x}} - x_4a\,\hat{\mathbf{y}} - z_4c\,\hat{\mathbf{z}}$ | (8g) | O II |
| $\mathbf{B}_{14}$ | = | $-y_4\,\mathbf{a}_1 + \left(\frac{1}{2} + x_4\right)\,\mathbf{a}_2 - z_4\,\mathbf{a}_3$ | = | $-y_4a\,\hat{\mathbf{x}} + \left(\frac{1}{2} + x_4\right)a\,\hat{\mathbf{y}} - z_4c\,\hat{\mathbf{z}}$ | (8g) | O II |

---

---

**Geometry files:**

- CIF: pp. 1670
- POSCAR: pp. 1671

# $PNCl_2$ ($E1_4$) Structure: A2BC_tP32_86_2g_g_g

http://aflow.org/prototype-encyclopedia/A2BC_tP32_86_2g_g_g

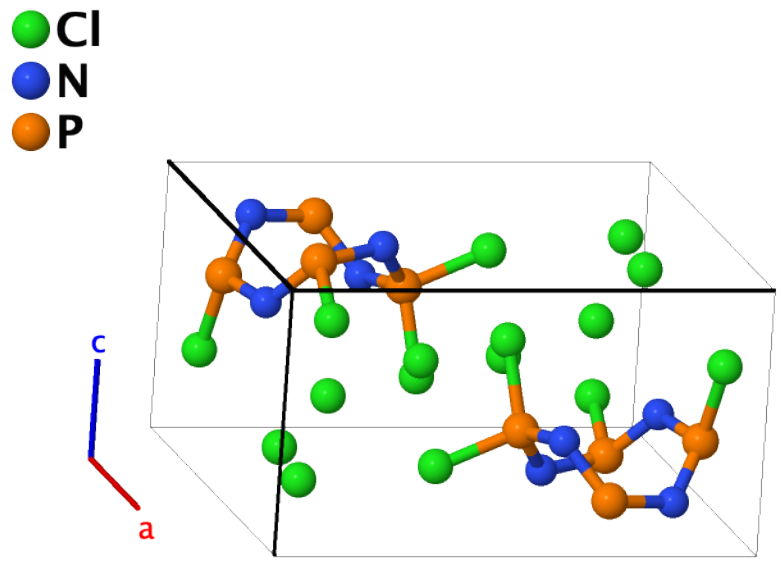

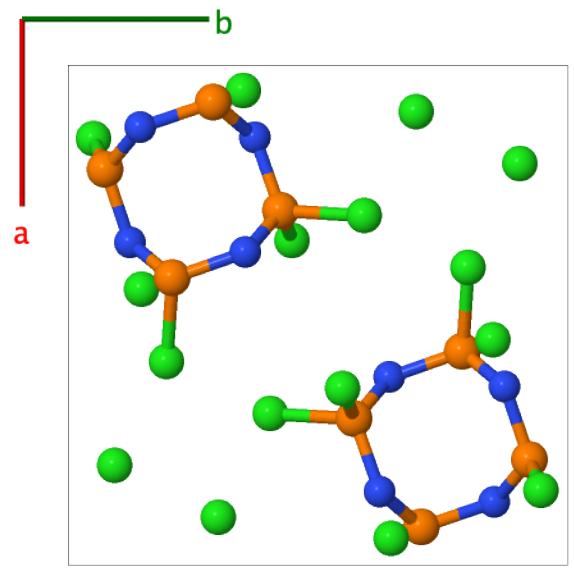

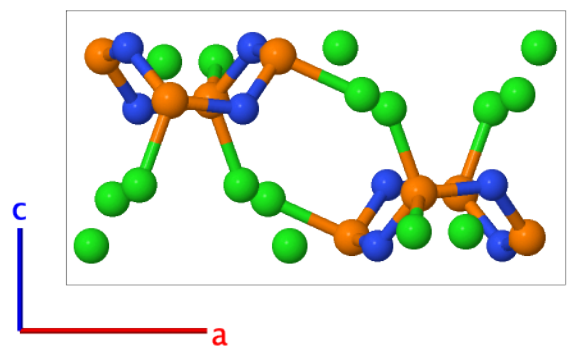

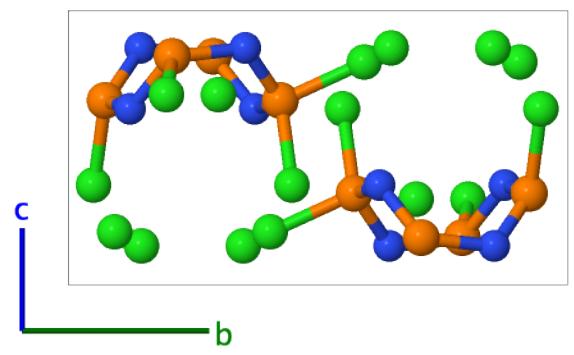

| | | |
|---|---|---|
| **Prototype** | : | $Cl_2NP$ |
| **AFLOW prototype label** | : | A2BC_tP32_86_2g_g_g |
| ***Strukturbericht* designation** | : | $E1_4$ |
| **Pearson symbol** | : | tP32 |
| **Space group number** | : | 86 |
| **Space group symbol** | : | $P4_2/n$ |
| **AFLOW prototype command** | : | `aflow --proto=A2BC_tP32_86_2g_g_g` <br> `--params=`$a, c/a, x_1, y_1, z_1, x_2, y_2, z_2, x_3, y_3, z_3, x_4, y_4, z_4$ |

- (Ketellar, 1939) gave the atomic positions in Setting 1 of space group #86. We have shifted the origin to the inversion site for this lattice and present the atomic positions in setting 2.

**Simple Tetragonal primitive vectors:**

$$\mathbf{a}_1 = a\,\hat{\mathbf{x}}$$
$$\mathbf{a}_2 = a\,\hat{\mathbf{y}}$$
$$\mathbf{a}_3 = c\,\hat{\mathbf{z}}$$

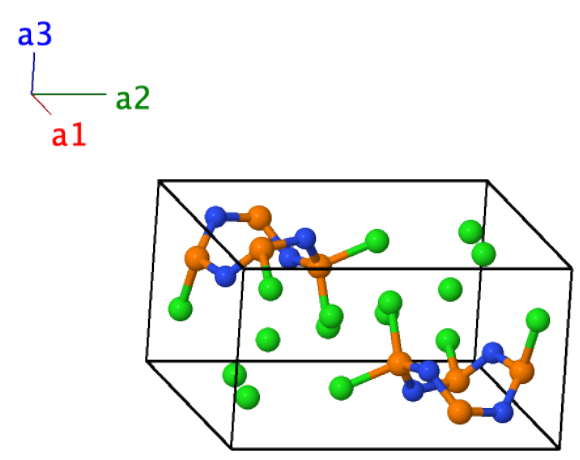

**Basis vectors:**

| | | Lattice Coordinates | | Cartesian Coordinates | Wyckoff Position | Atom Type |
|---|---|---|---|---|---|---|
| $\mathbf{B}_1$ | = | $x_1\,\mathbf{a}_1 + y_1\,\mathbf{a}_2 + z_1\,\mathbf{a}_3$ | = | $x_1 a\,\hat{\mathbf{x}} + y_1 a\,\hat{\mathbf{y}} + z_1 c\,\hat{\mathbf{z}}$ | (8g) | Cl I |
| $\mathbf{B}_2$ | = | $\left(\frac{1}{2} - x_1\right)\mathbf{a}_1 + \left(\frac{1}{2} - y_1\right)\mathbf{a}_2 + z_1\,\mathbf{a}_3$ | = | $\left(\frac{1}{2} - x_1\right) a\,\hat{\mathbf{x}} + \left(\frac{1}{2} - y_1\right) a\,\hat{\mathbf{y}} + z_1 c\,\hat{\mathbf{z}}$ | (8g) | Cl I |
| $\mathbf{B}_3$ | = | $-y_1\,\mathbf{a}_1 + \left(\frac{1}{2} + x_1\right)\mathbf{a}_2 + \left(\frac{1}{2} + z_1\right)\mathbf{a}_3$ | = | $-y_1 a\,\hat{\mathbf{x}} + \left(\frac{1}{2} + x_1\right) a\,\hat{\mathbf{y}} + \left(\frac{1}{2} + z_1\right) c\,\hat{\mathbf{z}}$ | (8g) | Cl I |
| $\mathbf{B}_4$ | = | $\left(\frac{1}{2} + y_1\right)\mathbf{a}_1 - x_1\,\mathbf{a}_2 + \left(\frac{1}{2} + z_1\right)\mathbf{a}_3$ | = | $\left(\frac{1}{2} + y_1\right) a\,\hat{\mathbf{x}} - x_1 a\,\hat{\mathbf{y}} + \left(\frac{1}{2} + z_1\right) c\,\hat{\mathbf{z}}$ | (8g) | Cl I |
| $\mathbf{B}_5$ | = | $-x_1\,\mathbf{a}_1 - y_1\,\mathbf{a}_2 - z_1\,\mathbf{a}_3$ | = | $-x_1 a\,\hat{\mathbf{x}} - y_1 a\,\hat{\mathbf{y}} - z_1 c\,\hat{\mathbf{z}}$ | (8g) | Cl I |
| $\mathbf{B}_6$ | = | $\left(\frac{1}{2} + x_1\right)\mathbf{a}_1 + \left(\frac{1}{2} + y_1\right)\mathbf{a}_2 - z_1\,\mathbf{a}_3$ | = | $\left(\frac{1}{2} + x_1\right) a\,\hat{\mathbf{x}} + \left(\frac{1}{2} + y_1\right) a\,\hat{\mathbf{y}} - z_1 c\,\hat{\mathbf{z}}$ | (8g) | Cl I |
| $\mathbf{B}_7$ | = | $y_1\,\mathbf{a}_1 + \left(\frac{1}{2} - x_1\right)\mathbf{a}_2 + \left(\frac{1}{2} - z_1\right)\mathbf{a}_3$ | = | $y_1 a\,\hat{\mathbf{x}} + \left(\frac{1}{2} - x_1\right) a\,\hat{\mathbf{y}} + \left(\frac{1}{2} - z_1\right) c\,\hat{\mathbf{z}}$ | (8g) | Cl I |
| $\mathbf{B}_8$ | = | $\left(\frac{1}{2} - y_1\right)\mathbf{a}_1 + x_1\,\mathbf{a}_2 + \left(\frac{1}{2} - z_1\right)\mathbf{a}_3$ | = | $\left(\frac{1}{2} - y_1\right) a\,\hat{\mathbf{x}} + x_1 a\,\hat{\mathbf{y}} + \left(\frac{1}{2} - z_1\right) c\,\hat{\mathbf{z}}$ | (8g) | Cl I |
| $\mathbf{B}_9$ | = | $x_2\,\mathbf{a}_1 + y_2\,\mathbf{a}_2 + z_2\,\mathbf{a}_3$ | = | $x_2 a\,\hat{\mathbf{x}} + y_2 a\,\hat{\mathbf{y}} + z_2 c\,\hat{\mathbf{z}}$ | (8g) | Cl II |
| $\mathbf{B}_{10}$ | = | $\left(\frac{1}{2} - x_2\right)\mathbf{a}_1 + \left(\frac{1}{2} - y_2\right)\mathbf{a}_2 + z_2\,\mathbf{a}_3$ | = | $\left(\frac{1}{2} - x_2\right) a\,\hat{\mathbf{x}} + \left(\frac{1}{2} - y_2\right) a\,\hat{\mathbf{y}} + z_2 c\,\hat{\mathbf{z}}$ | (8g) | Cl II |
| $\mathbf{B}_{11}$ | = | $-y_2\,\mathbf{a}_1 + \left(\frac{1}{2} + x_2\right)\mathbf{a}_2 + \left(\frac{1}{2} + z_2\right)\mathbf{a}_3$ | = | $-y_2 a\,\hat{\mathbf{x}} + \left(\frac{1}{2} + x_2\right) a\,\hat{\mathbf{y}} + \left(\frac{1}{2} + z_2\right) c\,\hat{\mathbf{z}}$ | (8g) | Cl II |
| $\mathbf{B}_{12}$ | = | $\left(\frac{1}{2} + y_2\right)\mathbf{a}_1 - x_2\,\mathbf{a}_2 + \left(\frac{1}{2} + z_2\right)\mathbf{a}_3$ | = | $\left(\frac{1}{2} + y_2\right) a\,\hat{\mathbf{x}} - x_2 a\,\hat{\mathbf{y}} + \left(\frac{1}{2} + z_2\right) c\,\hat{\mathbf{z}}$ | (8g) | Cl II |
| $\mathbf{B}_{13}$ | = | $-x_2\,\mathbf{a}_1 - y_2\,\mathbf{a}_2 - z_2\,\mathbf{a}_3$ | = | $-x_2 a\,\hat{\mathbf{x}} - y_2 a\,\hat{\mathbf{y}} - z_2 c\,\hat{\mathbf{z}}$ | (8g) | Cl II |
| $\mathbf{B}_{14}$ | = | $\left(\frac{1}{2} + x_2\right)\mathbf{a}_1 + \left(\frac{1}{2} + y_2\right)\mathbf{a}_2 - z_2\,\mathbf{a}_3$ | = | $\left(\frac{1}{2} + x_2\right) a\,\hat{\mathbf{x}} + \left(\frac{1}{2} + y_2\right) a\,\hat{\mathbf{y}} - z_2 c\,\hat{\mathbf{z}}$ | (8g) | Cl II |
| $\mathbf{B}_{15}$ | = | $y_2\,\mathbf{a}_1 + \left(\frac{1}{2} - x_2\right)\mathbf{a}_2 + \left(\frac{1}{2} - z_2\right)\mathbf{a}_3$ | = | $y_2 a\,\hat{\mathbf{x}} + \left(\frac{1}{2} - x_2\right) a\,\hat{\mathbf{y}} + \left(\frac{1}{2} - z_2\right) c\,\hat{\mathbf{z}}$ | (8g) | Cl II |
| $\mathbf{B}_{16}$ | = | $\left(\frac{1}{2} - y_2\right)\mathbf{a}_1 + x_2\,\mathbf{a}_2 + \left(\frac{1}{2} - z_2\right)\mathbf{a}_3$ | = | $\left(\frac{1}{2} - y_2\right) a\,\hat{\mathbf{x}} + x_2 a\,\hat{\mathbf{y}} + \left(\frac{1}{2} - z_2\right) c\,\hat{\mathbf{z}}$ | (8g) | Cl II |
| $\mathbf{B}_{17}$ | = | $x_3\,\mathbf{a}_1 + y_3\,\mathbf{a}_2 + z_3\,\mathbf{a}_3$ | = | $x_3 a\,\hat{\mathbf{x}} + y_3 a\,\hat{\mathbf{y}} + z_3 c\,\hat{\mathbf{z}}$ | (8g) | N |
| $\mathbf{B}_{18}$ | = | $\left(\frac{1}{2} - x_3\right)\mathbf{a}_1 + \left(\frac{1}{2} - y_3\right)\mathbf{a}_2 + z_3\,\mathbf{a}_3$ | = | $\left(\frac{1}{2} - x_3\right) a\,\hat{\mathbf{x}} + \left(\frac{1}{2} - y_3\right) a\,\hat{\mathbf{y}} + z_3 c\,\hat{\mathbf{z}}$ | (8g) | N |
| $\mathbf{B}_{19}$ | = | $-y_3\,\mathbf{a}_1 + \left(\frac{1}{2} + x_3\right)\mathbf{a}_2 + \left(\frac{1}{2} + z_3\right)\mathbf{a}_3$ | = | $-y_3 a\,\hat{\mathbf{x}} + \left(\frac{1}{2} + x_3\right) a\,\hat{\mathbf{y}} + \left(\frac{1}{2} + z_3\right) c\,\hat{\mathbf{z}}$ | (8g) | N |
| $\mathbf{B}_{20}$ | = | $\left(\frac{1}{2} + y_3\right)\mathbf{a}_1 - x_3\,\mathbf{a}_2 + \left(\frac{1}{2} + z_3\right)\mathbf{a}_3$ | = | $\left(\frac{1}{2} + y_3\right) a\,\hat{\mathbf{x}} - x_3 a\,\hat{\mathbf{y}} + \left(\frac{1}{2} + z_3\right) c\,\hat{\mathbf{z}}$ | (8g) | N |
| $\mathbf{B}_{21}$ | = | $-x_3\,\mathbf{a}_1 - y_3\,\mathbf{a}_2 - z_3\,\mathbf{a}_3$ | = | $-x_3 a\,\hat{\mathbf{x}} - y_3 a\,\hat{\mathbf{y}} - z_3 c\,\hat{\mathbf{z}}$ | (8g) | N |
| $\mathbf{B}_{22}$ | = | $\left(\frac{1}{2} + x_3\right)\mathbf{a}_1 + \left(\frac{1}{2} + y_3\right)\mathbf{a}_2 - z_3\,\mathbf{a}_3$ | = | $\left(\frac{1}{2} + x_3\right) a\,\hat{\mathbf{x}} + \left(\frac{1}{2} + y_3\right) a\,\hat{\mathbf{y}} - z_3 c\,\hat{\mathbf{z}}$ | (8g) | N |
| $\mathbf{B}_{23}$ | = | $y_3\,\mathbf{a}_1 + \left(\frac{1}{2} - x_3\right)\mathbf{a}_2 + \left(\frac{1}{2} - z_3\right)\mathbf{a}_3$ | = | $y_3 a\,\hat{\mathbf{x}} + \left(\frac{1}{2} - x_3\right) a\,\hat{\mathbf{y}} + \left(\frac{1}{2} - z_3\right) c\,\hat{\mathbf{z}}$ | (8g) | N |
| $\mathbf{B}_{24}$ | = | $\left(\frac{1}{2} - y_3\right)\mathbf{a}_1 + x_3\,\mathbf{a}_2 + \left(\frac{1}{2} - z_3\right)\mathbf{a}_3$ | = | $\left(\frac{1}{2} - y_3\right) a\,\hat{\mathbf{x}} + x_3 a\,\hat{\mathbf{y}} + \left(\frac{1}{2} - z_3\right) c\,\hat{\mathbf{z}}$ | (8g) | N |
| $\mathbf{B}_{25}$ | = | $x_4\,\mathbf{a}_1 + y_4\,\mathbf{a}_2 + z_4\,\mathbf{a}_3$ | = | $x_4 a\,\hat{\mathbf{x}} + y_4 a\,\hat{\mathbf{y}} + z_4 c\,\hat{\mathbf{z}}$ | (8g) | P |
| $\mathbf{B}_{26}$ | = | $\left(\frac{1}{2} - x_4\right)\mathbf{a}_1 + \left(\frac{1}{2} - y_4\right)\mathbf{a}_2 + z_4\,\mathbf{a}_3$ | = | $\left(\frac{1}{2} - x_4\right) a\,\hat{\mathbf{x}} + \left(\frac{1}{2} - y_4\right) a\,\hat{\mathbf{y}} + z_4 c\,\hat{\mathbf{z}}$ | (8g) | P |
| $\mathbf{B}_{27}$ | = | $-y_4\,\mathbf{a}_1 + \left(\frac{1}{2} + x_4\right)\mathbf{a}_2 + \left(\frac{1}{2} + z_4\right)\mathbf{a}_3$ | = | $-y_4 a\,\hat{\mathbf{x}} + \left(\frac{1}{2} + x_4\right) a\,\hat{\mathbf{y}} + \left(\frac{1}{2} + z_4\right) c\,\hat{\mathbf{z}}$ | (8g) | P |
| $\mathbf{B}_{28}$ | = | $\left(\frac{1}{2} + y_4\right)\mathbf{a}_1 - x_4\,\mathbf{a}_2 + \left(\frac{1}{2} + z_4\right)\mathbf{a}_3$ | = | $\left(\frac{1}{2} + y_4\right) a\,\hat{\mathbf{x}} - x_4 a\,\hat{\mathbf{y}} + \left(\frac{1}{2} + z_4\right) c\,\hat{\mathbf{z}}$ | (8g) | P |
| $\mathbf{B}_{29}$ | = | $-x_4\,\mathbf{a}_1 - y_4\,\mathbf{a}_2 - z_4\,\mathbf{a}_3$ | = | $-x_4 a\,\hat{\mathbf{x}} - y_4 a\,\hat{\mathbf{y}} - z_4 c\,\hat{\mathbf{z}}$ | (8g) | P |
| $\mathbf{B}_{30}$ | = | $\left(\frac{1}{2} + x_4\right)\mathbf{a}_1 + \left(\frac{1}{2} + y_4\right)\mathbf{a}_2 - z_4\,\mathbf{a}_3$ | = | $\left(\frac{1}{2} + x_4\right) a\,\hat{\mathbf{x}} + \left(\frac{1}{2} + y_4\right) a\,\hat{\mathbf{y}} - z_4 c\,\hat{\mathbf{z}}$ | (8g) | P |
| $\mathbf{B}_{31}$ | = | $y_4\,\mathbf{a}_1 + \left(\frac{1}{2} - x_4\right)\mathbf{a}_2 + \left(\frac{1}{2} - z_4\right)\mathbf{a}_3$ | = | $y_4 a\,\hat{\mathbf{x}} + \left(\frac{1}{2} - x_4\right) a\,\hat{\mathbf{y}} + \left(\frac{1}{2} - z_4\right) c\,\hat{\mathbf{z}}$ | (8g) | P |
| $\mathbf{B}_{32}$ | = | $\left(\frac{1}{2} - y_4\right)\mathbf{a}_1 + x_4\,\mathbf{a}_2 + \left(\frac{1}{2} - z_4\right)\mathbf{a}_3$ | = | $\left(\frac{1}{2} - y_4\right) a\,\hat{\mathbf{x}} + x_4 a\,\hat{\mathbf{y}} + \left(\frac{1}{2} - z_4\right) c\,\hat{\mathbf{z}}$ | (8g) | P |

**Geometry files:**

- CIF: pp. 1671
- POSCAR: pp. 1671

# $Nd_4Re_2O_{11}$ Structure: A4B11C2_tP68_86_2g_ab5g_g

http://aflow.org/prototype-encyclopedia/A4B11C2_tP68_86_2g_ab5g_g

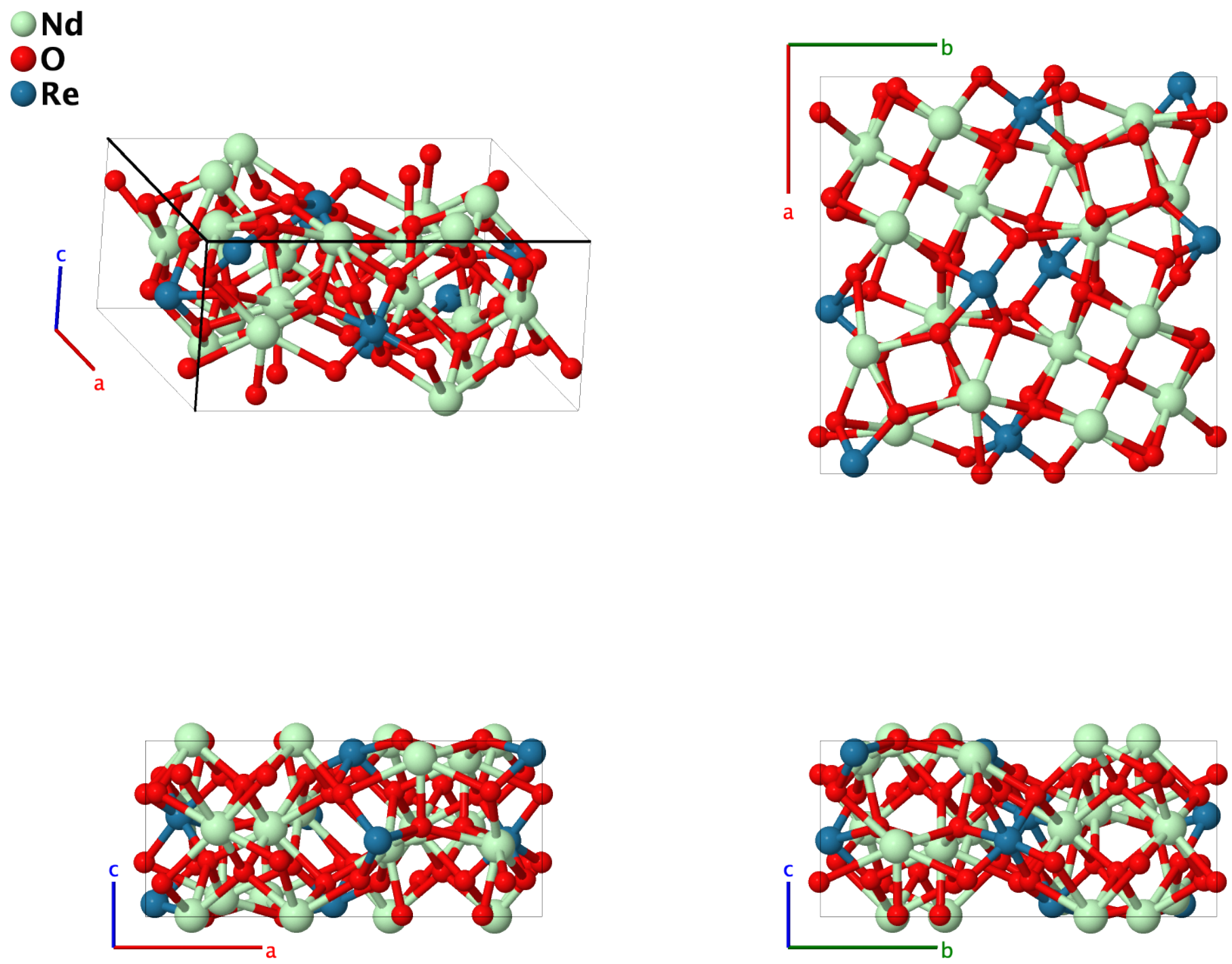

| | | |
|---|---|---|
| **Prototype** | : | $Nd_4O_{11}Re_2$ |
| **AFLOW prototype label** | : | A4B11C2_tP68_86_2g_ab5g_g |
| ***Strukturbericht* designation** | : | None |
| **Pearson symbol** | : | tP68 |
| **Space group number** | : | 86 |
| **Space group symbol** | : | $P4_2/n$ |
| **AFLOW prototype command** | : | `aflow --proto=A4B11C2_tP68_86_2g_ab5g_g` `--params=`$a, c/a, x_3, y_3, z_3, x_4, y_4, z_4, x_5, y_5, z_5, x_6, y_6, z_6, x_7, y_7, z_7, x_8, y_8, z_8, x_9, y_9, z_9, x_{10}, y_{10}, z_{10}$ |

- (Wilhelmi, 1970) has a misprint for the Wyckoff position of the Nd II atom, although the nearest-neighbor distances are correct. (Downs, 2003) corrects the position to be consistent with those distances, and we use their value.

**Simple Tetragonal primitive vectors:**

$$\mathbf{a}_1 = a\,\hat{\mathbf{x}}$$
$$\mathbf{a}_2 = a\,\hat{\mathbf{y}}$$
$$\mathbf{a}_3 = c\,\hat{\mathbf{z}}$$

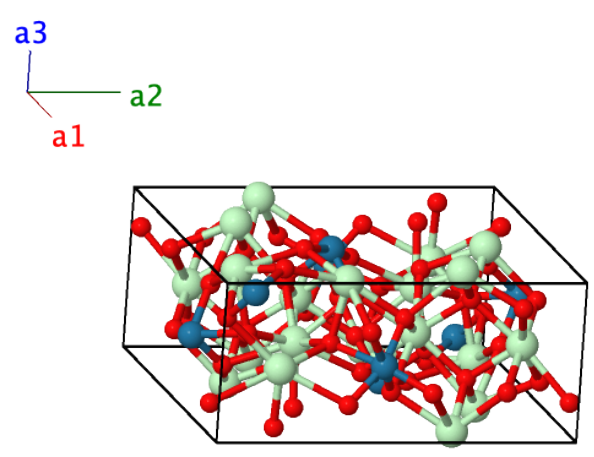

**Basis vectors:**

| | | Lattice Coordinates | | Cartesian Coordinates | Wyckoff Position | Atom Type |
|---|---|---|---|---|---|---|
| $\mathbf{B}_1$ | = | $\frac{1}{4}\,\mathbf{a}_1+\frac{1}{4}\,\mathbf{a}_2+\frac{1}{4}\,\mathbf{a}_3$ | = | $\frac{1}{4}a\,\hat{\mathbf{x}}+\frac{1}{4}a\,\hat{\mathbf{y}}+\frac{1}{4}c\,\hat{\mathbf{z}}$ | (2$a$) | O I |
| $\mathbf{B}_2$ | = | $\frac{3}{4}\,\mathbf{a}_1+\frac{3}{4}\,\mathbf{a}_2+\frac{3}{4}\,\mathbf{a}_3$ | = | $\frac{3}{4}a\,\hat{\mathbf{x}}+\frac{3}{4}a\,\hat{\mathbf{y}}+\frac{3}{4}c\,\hat{\mathbf{z}}$ | (2$a$) | O I |
| $\mathbf{B}_3$ | = | $\frac{1}{4}\,\mathbf{a}_1+\frac{1}{4}\,\mathbf{a}_2+\frac{3}{4}\,\mathbf{a}_3$ | = | $\frac{1}{4}a\,\hat{\mathbf{x}}+\frac{1}{4}a\,\hat{\mathbf{y}}+\frac{3}{4}c\,\hat{\mathbf{z}}$ | (2$b$) | O II |
| $\mathbf{B}_4$ | = | $\frac{3}{4}\,\mathbf{a}_1+\frac{3}{4}\,\mathbf{a}_2+\frac{1}{4}\,\mathbf{a}_3$ | = | $\frac{3}{4}a\,\hat{\mathbf{x}}+\frac{3}{4}a\,\hat{\mathbf{y}}+\frac{1}{4}c\,\hat{\mathbf{z}}$ | (2$b$) | O II |
| $\mathbf{B}_5$ | = | $x_3\,\mathbf{a}_1+y_3\,\mathbf{a}_2+z_3\,\mathbf{a}_3$ | = | $x_3a\,\hat{\mathbf{x}}+y_3a\,\hat{\mathbf{y}}+z_3c\,\hat{\mathbf{z}}$ | (8$g$) | Nd I |
| $\mathbf{B}_6$ | = | $\left(\frac{1}{2}-x_3\right)\,\mathbf{a}_1+\left(\frac{1}{2}-y_3\right)\,\mathbf{a}_2+z_3\,\mathbf{a}_3$ | = | $\left(\frac{1}{2}-x_3\right)a\,\hat{\mathbf{x}}+\left(\frac{1}{2}-y_3\right)a\,\hat{\mathbf{y}}+z_3c\,\hat{\mathbf{z}}$ | (8$g$) | Nd I |
| $\mathbf{B}_7$ | = | $-y_3\,\mathbf{a}_1+\left(\frac{1}{2}+x_3\right)\,\mathbf{a}_2+\left(\frac{1}{2}+z_3\right)\,\mathbf{a}_3$ | = | $-y_3a\,\hat{\mathbf{x}}+\left(\frac{1}{2}+x_3\right)a\,\hat{\mathbf{y}}+\left(\frac{1}{2}+z_3\right)c\,\hat{\mathbf{z}}$ | (8$g$) | Nd I |
| $\mathbf{B}_8$ | = | $\left(\frac{1}{2}+y_3\right)\,\mathbf{a}_1-x_3\,\mathbf{a}_2+\left(\frac{1}{2}+z_3\right)\,\mathbf{a}_3$ | = | $\left(\frac{1}{2}+y_3\right)a\,\hat{\mathbf{x}}-x_3a\,\hat{\mathbf{y}}+\left(\frac{1}{2}+z_3\right)c\,\hat{\mathbf{z}}$ | (8$g$) | Nd I |
| $\mathbf{B}_9$ | = | $-x_3\,\mathbf{a}_1-y_3\,\mathbf{a}_2-z_3\,\mathbf{a}_3$ | = | $-x_3a\,\hat{\mathbf{x}}-y_3a\,\hat{\mathbf{y}}-z_3c\,\hat{\mathbf{z}}$ | (8$g$) | Nd I |
| $\mathbf{B}_{10}$ | = | $\left(\frac{1}{2}+x_3\right)\,\mathbf{a}_1+\left(\frac{1}{2}+y_3\right)\,\mathbf{a}_2-z_3\,\mathbf{a}_3$ | = | $\left(\frac{1}{2}+x_3\right)a\,\hat{\mathbf{x}}+\left(\frac{1}{2}+y_3\right)a\,\hat{\mathbf{y}}-z_3c\,\hat{\mathbf{z}}$ | (8$g$) | Nd I |
| $\mathbf{B}_{11}$ | = | $y_3\,\mathbf{a}_1+\left(\frac{1}{2}-x_3\right)\,\mathbf{a}_2+\left(\frac{1}{2}-z_3\right)\,\mathbf{a}_3$ | = | $y_3a\,\hat{\mathbf{x}}+\left(\frac{1}{2}-x_3\right)a\,\hat{\mathbf{y}}+\left(\frac{1}{2}-z_3\right)c\,\hat{\mathbf{z}}$ | (8$g$) | Nd I |
| $\mathbf{B}_{12}$ | = | $\left(\frac{1}{2}-y_3\right)\,\mathbf{a}_1+x_3\,\mathbf{a}_2+\left(\frac{1}{2}-z_3\right)\,\mathbf{a}_3$ | = | $\left(\frac{1}{2}-y_3\right)a\,\hat{\mathbf{x}}+x_3a\,\hat{\mathbf{y}}+\left(\frac{1}{2}-z_3\right)c\,\hat{\mathbf{z}}$ | (8$g$) | Nd I |
| $\mathbf{B}_{13}$ | = | $x_4\,\mathbf{a}_1+y_4\,\mathbf{a}_2+z_4\,\mathbf{a}_3$ | = | $x_4a\,\hat{\mathbf{x}}+y_4a\,\hat{\mathbf{y}}+z_4c\,\hat{\mathbf{z}}$ | (8$g$) | Nd II |
| $\mathbf{B}_{14}$ | = | $\left(\frac{1}{2}-x_4\right)\,\mathbf{a}_1+\left(\frac{1}{2}-y_4\right)\,\mathbf{a}_2+z_4\,\mathbf{a}_3$ | = | $\left(\frac{1}{2}-x_4\right)a\,\hat{\mathbf{x}}+\left(\frac{1}{2}-y_4\right)a\,\hat{\mathbf{y}}+z_4c\,\hat{\mathbf{z}}$ | (8$g$) | Nd II |
| $\mathbf{B}_{15}$ | = | $-y_4\,\mathbf{a}_1+\left(\frac{1}{2}+x_4\right)\,\mathbf{a}_2+\left(\frac{1}{2}+z_4\right)\,\mathbf{a}_3$ | = | $-y_4a\,\hat{\mathbf{x}}+\left(\frac{1}{2}+x_4\right)a\,\hat{\mathbf{y}}+\left(\frac{1}{2}+z_4\right)c\,\hat{\mathbf{z}}$ | (8$g$) | Nd II |
| $\mathbf{B}_{16}$ | = | $\left(\frac{1}{2}+y_4\right)\,\mathbf{a}_1-x_4\,\mathbf{a}_2+\left(\frac{1}{2}+z_4\right)\,\mathbf{a}_3$ | = | $\left(\frac{1}{2}+y_4\right)a\,\hat{\mathbf{x}}-x_4a\,\hat{\mathbf{y}}+\left(\frac{1}{2}+z_4\right)c\,\hat{\mathbf{z}}$ | (8$g$) | Nd II |
| $\mathbf{B}_{17}$ | = | $-x_4\,\mathbf{a}_1-y_4\,\mathbf{a}_2-z_4\,\mathbf{a}_3$ | = | $-x_4a\,\hat{\mathbf{x}}-y_4a\,\hat{\mathbf{y}}-z_4c\,\hat{\mathbf{z}}$ | (8$g$) | Nd II |
| $\mathbf{B}_{18}$ | = | $\left(\frac{1}{2}+x_4\right)\,\mathbf{a}_1+\left(\frac{1}{2}+y_4\right)\,\mathbf{a}_2-z_4\,\mathbf{a}_3$ | = | $\left(\frac{1}{2}+x_4\right)a\,\hat{\mathbf{x}}+\left(\frac{1}{2}+y_4\right)a\,\hat{\mathbf{y}}-z_4c\,\hat{\mathbf{z}}$ | (8$g$) | Nd II |
| $\mathbf{B}_{19}$ | = | $y_4\,\mathbf{a}_1+\left(\frac{1}{2}-x_4\right)\,\mathbf{a}_2+\left(\frac{1}{2}-z_4\right)\,\mathbf{a}_3$ | = | $y_4a\,\hat{\mathbf{x}}+\left(\frac{1}{2}-x_4\right)a\,\hat{\mathbf{y}}+\left(\frac{1}{2}-z_4\right)c\,\hat{\mathbf{z}}$ | (8$g$) | Nd II |
| $\mathbf{B}_{20}$ | = | $\left(\frac{1}{2}-y_4\right)\,\mathbf{a}_1+x_4\,\mathbf{a}_2+\left(\frac{1}{2}-z_4\right)\,\mathbf{a}_3$ | = | $\left(\frac{1}{2}-y_4\right)a\,\hat{\mathbf{x}}+x_4a\,\hat{\mathbf{y}}+\left(\frac{1}{2}-z_4\right)c\,\hat{\mathbf{z}}$ | (8$g$) | Nd II |
| $\mathbf{B}_{21}$ | = | $x_5\,\mathbf{a}_1+y_5\,\mathbf{a}_2+z_5\,\mathbf{a}_3$ | = | $x_5a\,\hat{\mathbf{x}}+y_5a\,\hat{\mathbf{y}}+z_5c\,\hat{\mathbf{z}}$ | (8$g$) | O III |
| $\mathbf{B}_{22}$ | = | $\left(\frac{1}{2}-x_5\right)\,\mathbf{a}_1+\left(\frac{1}{2}-y_5\right)\,\mathbf{a}_2+z_5\,\mathbf{a}_3$ | = | $\left(\frac{1}{2}-x_5\right)a\,\hat{\mathbf{x}}+\left(\frac{1}{2}-y_5\right)a\,\hat{\mathbf{y}}+z_5c\,\hat{\mathbf{z}}$ | (8$g$) | O III |
| $\mathbf{B}_{23}$ | = | $-y_5\,\mathbf{a}_1+\left(\frac{1}{2}+x_5\right)\,\mathbf{a}_2+\left(\frac{1}{2}+z_5\right)\,\mathbf{a}_3$ | = | $-y_5a\,\hat{\mathbf{x}}+\left(\frac{1}{2}+x_5\right)a\,\hat{\mathbf{y}}+\left(\frac{1}{2}+z_5\right)c\,\hat{\mathbf{z}}$ | (8$g$) | O III |
| $\mathbf{B}_{24}$ | = | $\left(\frac{1}{2}+y_5\right)\,\mathbf{a}_1-x_5\,\mathbf{a}_2+\left(\frac{1}{2}+z_5\right)\,\mathbf{a}_3$ | = | $\left(\frac{1}{2}+y_5\right)a\,\hat{\mathbf{x}}-x_5a\,\hat{\mathbf{y}}+\left(\frac{1}{2}+z_5\right)c\,\hat{\mathbf{z}}$ | (8$g$) | O III |
| $\mathbf{B}_{25}$ | = | $-x_5\,\mathbf{a}_1-y_5\,\mathbf{a}_2-z_5\,\mathbf{a}_3$ | = | $-x_5a\,\hat{\mathbf{x}}-y_5a\,\hat{\mathbf{y}}-z_5c\,\hat{\mathbf{z}}$ | (8$g$) | O III |
| $\mathbf{B}_{26}$ | = | $\left(\frac{1}{2}+x_5\right)\,\mathbf{a}_1+\left(\frac{1}{2}+y_5\right)\,\mathbf{a}_2-z_5\,\mathbf{a}_3$ | = | $\left(\frac{1}{2}+x_5\right)a\,\hat{\mathbf{x}}+\left(\frac{1}{2}+y_5\right)a\,\hat{\mathbf{y}}-z_5c\,\hat{\mathbf{z}}$ | (8$g$) | O III |
| $\mathbf{B}_{27}$ | = | $y_5\,\mathbf{a}_1+\left(\frac{1}{2}-x_5\right)\,\mathbf{a}_2+\left(\frac{1}{2}-z_5\right)\,\mathbf{a}_3$ | = | $y_5a\,\hat{\mathbf{x}}+\left(\frac{1}{2}-x_5\right)a\,\hat{\mathbf{y}}+\left(\frac{1}{2}-z_5\right)c\,\hat{\mathbf{z}}$ | (8$g$) | O III |
| $\mathbf{B}_{28}$ | = | $\left(\frac{1}{2}-y_5\right)\,\mathbf{a}_1+x_5\,\mathbf{a}_2+\left(\frac{1}{2}-z_5\right)\,\mathbf{a}_3$ | = | $\left(\frac{1}{2}-y_5\right)a\,\hat{\mathbf{x}}+x_5a\,\hat{\mathbf{y}}+\left(\frac{1}{2}-z_5\right)c\,\hat{\mathbf{z}}$ | (8$g$) | O III |
| $\mathbf{B}_{29}$ | = | $x_6\,\mathbf{a}_1+y_6\,\mathbf{a}_2+z_6\,\mathbf{a}_3$ | = | $x_6a\,\hat{\mathbf{x}}+y_6a\,\hat{\mathbf{y}}+z_6c\,\hat{\mathbf{z}}$ | (8$g$) | O IV |
| $\mathbf{B}_{30}$ | = | $\left(\frac{1}{2}-x_6\right)\,\mathbf{a}_1+\left(\frac{1}{2}-y_6\right)\,\mathbf{a}_2+z_6\,\mathbf{a}_3$ | = | $\left(\frac{1}{2}-x_6\right)a\,\hat{\mathbf{x}}+\left(\frac{1}{2}-y_6\right)a\,\hat{\mathbf{y}}+z_6c\,\hat{\mathbf{z}}$ | (8$g$) | O IV |
| $\mathbf{B}_{31}$ | = | $-y_6\,\mathbf{a}_1+\left(\frac{1}{2}+x_6\right)\,\mathbf{a}_2+\left(\frac{1}{2}+z_6\right)\,\mathbf{a}_3$ | = | $-y_6a\,\hat{\mathbf{x}}+\left(\frac{1}{2}+x_6\right)a\,\hat{\mathbf{y}}+\left(\frac{1}{2}+z_6\right)c\,\hat{\mathbf{z}}$ | (8$g$) | O IV |
| $\mathbf{B}_{32}$ | = | $\left(\frac{1}{2}+y_6\right)\,\mathbf{a}_1-x_6\,\mathbf{a}_2+\left(\frac{1}{2}+z_6\right)\,\mathbf{a}_3$ | = | $\left(\frac{1}{2}+y_6\right)a\,\hat{\mathbf{x}}-x_6a\,\hat{\mathbf{y}}+\left(\frac{1}{2}+z_6\right)c\,\hat{\mathbf{z}}$ | (8$g$) | O IV |
| $\mathbf{B}_{33}$ | = | $-x_6\,\mathbf{a}_1-y_6\,\mathbf{a}_2-z_6\,\mathbf{a}_3$ | = | $-x_6a\,\hat{\mathbf{x}}-y_6a\,\hat{\mathbf{y}}-z_6c\,\hat{\mathbf{z}}$ | (8$g$) | O IV |
| $\mathbf{B}_{34}$ | = | $\left(\frac{1}{2}+x_6\right)\,\mathbf{a}_1+\left(\frac{1}{2}+y_6\right)\,\mathbf{a}_2-z_6\,\mathbf{a}_3$ | = | $\left(\frac{1}{2}+x_6\right)a\,\hat{\mathbf{x}}+\left(\frac{1}{2}+y_6\right)a\,\hat{\mathbf{y}}-z_6c\,\hat{\mathbf{z}}$ | (8$g$) | O IV |
| $\mathbf{B}_{35}$ | = | $y_{6}\,\mathbf{a}_{1} + \left(\frac{1}{2} - x_{6}\right)\,\mathbf{a}_{2} + \left(\frac{1}{2} - z_{6}\right)\,\mathbf{a}_{3}$ | = | $y_{6}a\,\mathbf{\hat{x}} + \left(\frac{1}{2} - x_{6}\right)a\,\mathbf{\hat{y}} + \left(\frac{1}{2} - z_{6}\right)c\,\mathbf{\hat{z}}$ | (8$g$) | O IV |
| $\mathbf{B}_{36}$ | = | $\left(\frac{1}{2} - y_{6}\right)\,\mathbf{a}_{1} + x_{6}\,\mathbf{a}_{2} + \left(\frac{1}{2} - z_{6}\right)\,\mathbf{a}_{3}$ | = | $\left(\frac{1}{2} - y_{6}\right)a\,\mathbf{\hat{x}} + x_{6}a\,\mathbf{\hat{y}} + \left(\frac{1}{2} - z_{6}\right)c\,\mathbf{\hat{z}}$ | (8$g$) | O IV |
| $\mathbf{B}_{37}$ | = | $x_{7}\,\mathbf{a}_{1} + y_{7}\,\mathbf{a}_{2} + z_{7}\,\mathbf{a}_{3}$ | = | $x_{7}a\,\mathbf{\hat{x}} + y_{7}a\,\mathbf{\hat{y}} + z_{7}c\,\mathbf{\hat{z}}$ | (8$g$) | O V |
| $\mathbf{B}_{38}$ | = | $\left(\frac{1}{2} - x_{7}\right)\,\mathbf{a}_{1} + \left(\frac{1}{2} - y_{7}\right)\,\mathbf{a}_{2} + z_{7}\,\mathbf{a}_{3}$ | = | $\left(\frac{1}{2} - x_{7}\right)a\,\mathbf{\hat{x}} + \left(\frac{1}{2} - y_{7}\right)a\,\mathbf{\hat{y}} + z_{7}c\,\mathbf{\hat{z}}$ | (8$g$) | O V |
| $\mathbf{B}_{39}$ | = | $-y_{7}\,\mathbf{a}_{1} + \left(\frac{1}{2} + x_{7}\right)\,\mathbf{a}_{2} + \left(\frac{1}{2} + z_{7}\right)\,\mathbf{a}_{3}$ | = | $-y_{7}a\,\mathbf{\hat{x}} + \left(\frac{1}{2} + x_{7}\right)a\,\mathbf{\hat{y}} + \left(\frac{1}{2} + z_{7}\right)c\,\mathbf{\hat{z}}$ | (8$g$) | O V |
| $\mathbf{B}_{40}$ | = | $\left(\frac{1}{2} + y_{7}\right)\,\mathbf{a}_{1} - x_{7}\,\mathbf{a}_{2} + \left(\frac{1}{2} + z_{7}\right)\,\mathbf{a}_{3}$ | = | $\left(\frac{1}{2} + y_{7}\right)a\,\mathbf{\hat{x}} - x_{7}a\,\mathbf{\hat{y}} + \left(\frac{1}{2} + z_{7}\right)c\,\mathbf{\hat{z}}$ | (8$g$) | O V |
| $\mathbf{B}_{41}$ | = | $-x_{7}\,\mathbf{a}_{1} - y_{7}\,\mathbf{a}_{2} - z_{7}\,\mathbf{a}_{3}$ | = | $-x_{7}a\,\mathbf{\hat{x}} - y_{7}a\,\mathbf{\hat{y}} - z_{7}c\,\mathbf{\hat{z}}$ | (8$g$) | O V |
| $\mathbf{B}_{42}$ | = | $\left(\frac{1}{2} + x_{7}\right)\,\mathbf{a}_{1} + \left(\frac{1}{2} + y_{7}\right)\,\mathbf{a}_{2} - z_{7}\,\mathbf{a}_{3}$ | = | $\left(\frac{1}{2} + x_{7}\right)a\,\mathbf{\hat{x}} + \left(\frac{1}{2} + y_{7}\right)a\,\mathbf{\hat{y}} - z_{7}c\,\mathbf{\hat{z}}$ | (8$g$) | O V |
| $\mathbf{B}_{43}$ | = | $y_{7}\,\mathbf{a}_{1} + \left(\frac{1}{2} - x_{7}\right)\,\mathbf{a}_{2} + \left(\frac{1}{2} - z_{7}\right)\,\mathbf{a}_{3}$ | = | $y_{7}a\,\mathbf{\hat{x}} + \left(\frac{1}{2} - x_{7}\right)a\,\mathbf{\hat{y}} + \left(\frac{1}{2} - z_{7}\right)c\,\mathbf{\hat{z}}$ | (8$g$) | O V |
| $\mathbf{B}_{44}$ | = | $\left(\frac{1}{2} - y_{7}\right)\,\mathbf{a}_{1} + x_{7}\,\mathbf{a}_{2} + \left(\frac{1}{2} - z_{7}\right)\,\mathbf{a}_{3}$ | = | $\left(\frac{1}{2} - y_{7}\right)a\,\mathbf{\hat{x}} + x_{7}a\,\mathbf{\hat{y}} + \left(\frac{1}{2} - z_{7}\right)c\,\mathbf{\hat{z}}$ | (8$g$) | O V |
| $\mathbf{B}_{45}$ | = | $x_{8}\,\mathbf{a}_{1} + y_{8}\,\mathbf{a}_{2} + z_{8}\,\mathbf{a}_{3}$ | = | $x_{8}a\,\mathbf{\hat{x}} + y_{8}a\,\mathbf{\hat{y}} + z_{8}c\,\mathbf{\hat{z}}$ | (8$g$) | O VI |
| $\mathbf{B}_{46}$ | = | $\left(\frac{1}{2} - x_{8}\right)\,\mathbf{a}_{1} + \left(\frac{1}{2} - y_{8}\right)\,\mathbf{a}_{2} + z_{8}\,\mathbf{a}_{3}$ | = | $\left(\frac{1}{2} - x_{8}\right)a\,\mathbf{\hat{x}} + \left(\frac{1}{2} - y_{8}\right)a\,\mathbf{\hat{y}} + z_{8}c\,\mathbf{\hat{z}}$ | (8$g$) | O VI |
| $\mathbf{B}_{47}$ | = | $-y_{8}\,\mathbf{a}_{1} + \left(\frac{1}{2} + x_{8}\right)\,\mathbf{a}_{2} + \left(\frac{1}{2} + z_{8}\right)\,\mathbf{a}_{3}$ | = | $-y_{8}a\,\mathbf{\hat{x}} + \left(\frac{1}{2} + x_{8}\right)a\,\mathbf{\hat{y}} + \left(\frac{1}{2} + z_{8}\right)c\,\mathbf{\hat{z}}$ | (8$g$) | O VI |
| $\mathbf{B}_{48}$ | = | $\left(\frac{1}{2} + y_{8}\right)\,\mathbf{a}_{1} - x_{8}\,\mathbf{a}_{2} + \left(\frac{1}{2} + z_{8}\right)\,\mathbf{a}_{3}$ | = | $\left(\frac{1}{2} + y_{8}\right)a\,\mathbf{\hat{x}} - x_{8}a\,\mathbf{\hat{y}} + \left(\frac{1}{2} + z_{8}\right)c\,\mathbf{\hat{z}}$ | (8$g$) | O VI |
| $\mathbf{B}_{49}$ | = | $-x_{8}\,\mathbf{a}_{1} - y_{8}\,\mathbf{a}_{2} - z_{8}\,\mathbf{a}_{3}$ | = | $-x_{8}a\,\mathbf{\hat{x}} - y_{8}a\,\mathbf{\hat{y}} - z_{8}c\,\mathbf{\hat{z}}$ | (8$g$) | O VI |
| $\mathbf{B}_{50}$ | = | $\left(\frac{1}{2} + x_{8}\right)\,\mathbf{a}_{1} + \left(\frac{1}{2} + y_{8}\right)\,\mathbf{a}_{2} - z_{8}\,\mathbf{a}_{3}$ | = | $\left(\frac{1}{2} + x_{8}\right)a\,\mathbf{\hat{x}} + \left(\frac{1}{2} + y_{8}\right)a\,\mathbf{\hat{y}} - z_{8}c\,\mathbf{\hat{z}}$ | (8$g$) | O VI |
| $\mathbf{B}_{51}$ | = | $y_{8}\,\mathbf{a}_{1} + \left(\frac{1}{2} - x_{8}\right)\,\mathbf{a}_{2} + \left(\frac{1}{2} - z_{8}\right)\,\mathbf{a}_{3}$ | = | $y_{8}a\,\mathbf{\hat{x}} + \left(\frac{1}{2} - x_{8}\right)a\,\mathbf{\hat{y}} + \left(\frac{1}{2} - z_{8}\right)c\,\mathbf{\hat{z}}$ | (8$g$) | O VI |
| $\mathbf{B}_{52}$ | = | $\left(\frac{1}{2} - y_{8}\right)\,\mathbf{a}_{1} + x_{8}\,\mathbf{a}_{2} + \left(\frac{1}{2} - z_{8}\right)\,\mathbf{a}_{3}$ | = | $\left(\frac{1}{2} - y_{8}\right)a\,\mathbf{\hat{x}} + x_{8}a\,\mathbf{\hat{y}} + \left(\frac{1}{2} - z_{8}\right)c\,\mathbf{\hat{z}}$ | (8$g$) | O VI |
| $\mathbf{B}_{53}$ | = | $x_{9}\,\mathbf{a}_{1} + y_{9}\,\mathbf{a}_{2} + z_{9}\,\mathbf{a}_{3}$ | = | $x_{9}a\,\mathbf{\hat{x}} + y_{9}a\,\mathbf{\hat{y}} + z_{9}c\,\mathbf{\hat{z}}$ | (8$g$) | O VII |
| $\mathbf{B}_{54}$ | = | $\left(\frac{1}{2} - x_{9}\right)\,\mathbf{a}_{1} + \left(\frac{1}{2} - y_{9}\right)\,\mathbf{a}_{2} + z_{9}\,\mathbf{a}_{3}$ | = | $\left(\frac{1}{2} - x_{9}\right)a\,\mathbf{\hat{x}} + \left(\frac{1}{2} - y_{9}\right)a\,\mathbf{\hat{y}} + z_{9}c\,\mathbf{\hat{z}}$ | (8$g$) | O VII |
| $\mathbf{B}_{55}$ | = | $-y_{9}\,\mathbf{a}_{1} + \left(\frac{1}{2} + x_{9}\right)\,\mathbf{a}_{2} + \left(\frac{1}{2} + z_{9}\right)\,\mathbf{a}_{3}$ | = | $-y_{9}a\,\mathbf{\hat{x}} + \left(\frac{1}{2} + x_{9}\right)a\,\mathbf{\hat{y}} + \left(\frac{1}{2} + z_{9}\right)c\,\mathbf{\hat{z}}$ | (8$g$) | O VII |
| $\mathbf{B}_{56}$ | = | $\left(\frac{1}{2} + y_{9}\right)\,\mathbf{a}_{1} - x_{9}\,\mathbf{a}_{2} + \left(\frac{1}{2} + z_{9}\right)\,\mathbf{a}_{3}$ | = | $\left(\frac{1}{2} + y_{9}\right)a\,\mathbf{\hat{x}} - x_{9}a\,\mathbf{\hat{y}} + \left(\frac{1}{2} + z_{9}\right)c\,\mathbf{\hat{z}}$ | (8$g$) | O VII |
| $\mathbf{B}_{57}$ | = | $-x_{9}\,\mathbf{a}_{1} - y_{9}\,\mathbf{a}_{2} - z_{9}\,\mathbf{a}_{3}$ | = | $-x_{9}a\,\mathbf{\hat{x}} - y_{9}a\,\mathbf{\hat{y}} - z_{9}c\,\mathbf{\hat{z}}$ | (8$g$) | O VII |
| $\mathbf{B}_{58}$ | = | $\left(\frac{1}{2} + x_{9}\right)\,\mathbf{a}_{1} + \left(\frac{1}{2} + y_{9}\right)\,\mathbf{a}_{2} - z_{9}\,\mathbf{a}_{3}$ | = | $\left(\frac{1}{2} + x_{9}\right)a\,\mathbf{\hat{x}} + \left(\frac{1}{2} + y_{9}\right)a\,\mathbf{\hat{y}} - z_{9}c\,\mathbf{\hat{z}}$ | (8$g$) | O VII |
| $\mathbf{B}_{59}$ | = | $y_{9}\,\mathbf{a}_{1} + \left(\frac{1}{2} - x_{9}\right)\,\mathbf{a}_{2} + \left(\frac{1}{2} - z_{9}\right)\,\mathbf{a}_{3}$ | = | $y_{9}a\,\mathbf{\hat{x}} + \left(\frac{1}{2} - x_{9}\right)a\,\mathbf{\hat{y}} + \left(\frac{1}{2} - z_{9}\right)c\,\mathbf{\hat{z}}$ | (8$g$) | O VII |
| $\mathbf{B}_{60}$ | = | $\left(\frac{1}{2} - y_{9}\right)\,\mathbf{a}_{1} + x_{9}\,\mathbf{a}_{2} + \left(\frac{1}{2} - z_{9}\right)\,\mathbf{a}_{3}$ | = | $\left(\frac{1}{2} - y_{9}\right)a\,\mathbf{\hat{x}} + x_{9}a\,\mathbf{\hat{y}} + \left(\frac{1}{2} - z_{9}\right)c\,\mathbf{\hat{z}}$ | (8$g$) | O VII |
| $\mathbf{B}_{61}$ | = | $x_{10}\,\mathbf{a}_{1} + y_{10}\,\mathbf{a}_{2} + z_{10}\,\mathbf{a}_{3}$ | = | $x_{10}a\,\mathbf{\hat{x}} + y_{10}a\,\mathbf{\hat{y}} + z_{10}c\,\mathbf{\hat{z}}$ | (8$g$) | Re |
| $\mathbf{B}_{62}$ | = | $\left(\frac{1}{2} - x_{10}\right)\,\mathbf{a}_{1} + \left(\frac{1}{2} - y_{10}\right)\,\mathbf{a}_{2} + z_{10}\,\mathbf{a}_{3}$ | = | $\left(\frac{1}{2} - x_{10}\right)a\,\mathbf{\hat{x}} + \left(\frac{1}{2} - y_{10}\right)a\,\mathbf{\hat{y}} + z_{10}c\,\mathbf{\hat{z}}$ | (8$g$) | Re |
| $\mathbf{B}_{63}$ | = | $-y_{10}\,\mathbf{a}_{1} + \left(\frac{1}{2} + x_{10}\right)\,\mathbf{a}_{2} + \left(\frac{1}{2} + z_{10}\right)\,\mathbf{a}_{3}$ | = | $-y_{10}a\,\mathbf{\hat{x}} + \left(\frac{1}{2} + x_{10}\right)a\,\mathbf{\hat{y}} + \left(\frac{1}{2} + z_{10}\right)c\,\mathbf{\hat{z}}$ | (8$g$) | Re |
| $\mathbf{B}_{64}$ | = | $\left(\frac{1}{2} + y_{10}\right)\,\mathbf{a}_{1} - x_{10}\,\mathbf{a}_{2} + \left(\frac{1}{2} + z_{10}\right)\,\mathbf{a}_{3}$ | = | $\left(\frac{1}{2} + y_{10}\right)a\,\mathbf{\hat{x}} - x_{10}a\,\mathbf{\hat{y}} + \left(\frac{1}{2} + z_{10}\right)c\,\mathbf{\hat{z}}$ | (8$g$) | Re |
| $\mathbf{B}_{65}$ | = | $-x_{10}\,\mathbf{a}_{1} - y_{10}\,\mathbf{a}_{2} - z_{10}\,\mathbf{a}_{3}$ | = | $-x_{10}a\,\mathbf{\hat{x}} - y_{10}a\,\mathbf{\hat{y}} - z_{10}c\,\mathbf{\hat{z}}$ | (8$g$) | Re |
| $\mathbf{B}_{66}$ | = | $\left(\frac{1}{2} + x_{10}\right)\,\mathbf{a}_{1} + \left(\frac{1}{2} + y_{10}\right)\,\mathbf{a}_{2} - z_{10}\,\mathbf{a}_{3}$ | = | $\left(\frac{1}{2} + x_{10}\right)a\,\mathbf{\hat{x}} + \left(\frac{1}{2} + y_{10}\right)a\,\mathbf{\hat{y}} - z_{10}c\,\mathbf{\hat{z}}$ | (8$g$) | Re |
| $\mathbf{B}_{67}$ | = | $y_{10}\,\mathbf{a}_{1} + \left(\frac{1}{2} - x_{10}\right)\,\mathbf{a}_{2} + \left(\frac{1}{2} - z_{10}\right)\,\mathbf{a}_{3}$ | = | $y_{10}a\,\mathbf{\hat{x}} + \left(\frac{1}{2} - x_{10}\right)a\,\mathbf{\hat{y}} + \left(\frac{1}{2} - z_{10}\right)c\,\mathbf{\hat{z}}$ | (8$g$) | Re |
| $\mathbf{B}_{68}$ | = | $\left(\frac{1}{2} - y_{10}\right)\,\mathbf{a}_{1} + x_{10}\,\mathbf{a}_{2} + \left(\frac{1}{2} - z_{10}\right)\,\mathbf{a}_{3}$ | = | $\left(\frac{1}{2} - y_{10}\right)a\,\mathbf{\hat{x}} + x_{10}a\,\mathbf{\hat{y}} + \left(\frac{1}{2} - z_{10}\right)c\,\mathbf{\hat{z}}$ | (8$g$) | Re |

**Found in:**

---

**Geometry files:**

- CIF: pp. 1671
- POSCAR: pp. 1672

# $NaSb(OH)_6$ ($J1_{11}$) Structure: AB6C_tP32_86_d_3g_c

http://aflow.org/prototype-encyclopedia/AB6C_tP32_86_d_3g_c

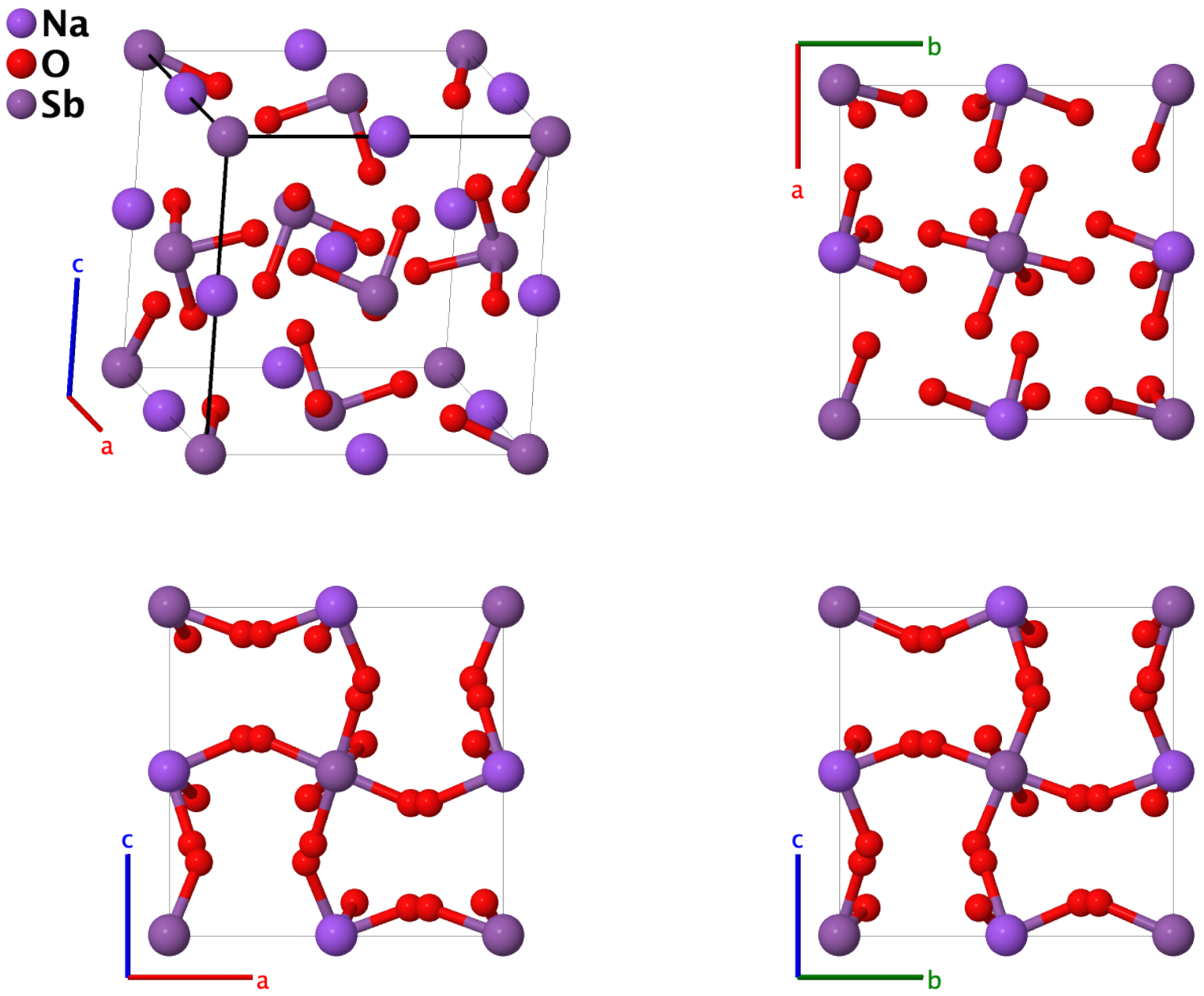

| | | |
|---|---|---|
| **Prototype** | : | $Na(OH)_6Sb$ |
| **AFLOW prototype label** | : | AB6C_tP32_86_d_3g_c |
| ***Strukturbericht* designation** | : | $J1_{11}$ |
| **Pearson symbol** | : | tP32 |
| **Space group number** | : | 86 |
| **Space group symbol** | : | $P4_2/n$ |
| **AFLOW prototype command** | : | `aflow --proto=AB6C_tP32_86_d_3g_c` `--params=`$a, c/a, x_3, y_3, z_3, x_4, y_4, z_4, x_5, y_5, z_5$ |

**Other compounds with this structure**

- $AgSb(OH)_6$

- The atomic positions were originally determined using setting 1 of space group #86. We used FINDSYM to change the origin to setting 2.
- The positions of the hydrogen atoms in the OH ion were not determined, so we only provide the positions of the oxygen atoms (labeled as OH).
- Although the replacement of fluorine by OH does not affect the shape of the Sb-(F,OH)$_6$ ions, it has a profound effect on the structure, as can be seen by looking at $NaSbF_6$ and $NaSbF_4(OH)_2$ ($J1_{12}$).

## Simple Tetragonal primitive vectors:

$$\mathbf{a}_1 = a\,\hat{\mathbf{x}}$$
$$\mathbf{a}_2 = a\,\hat{\mathbf{y}}$$
$$\mathbf{a}_3 = c\,\hat{\mathbf{z}}$$

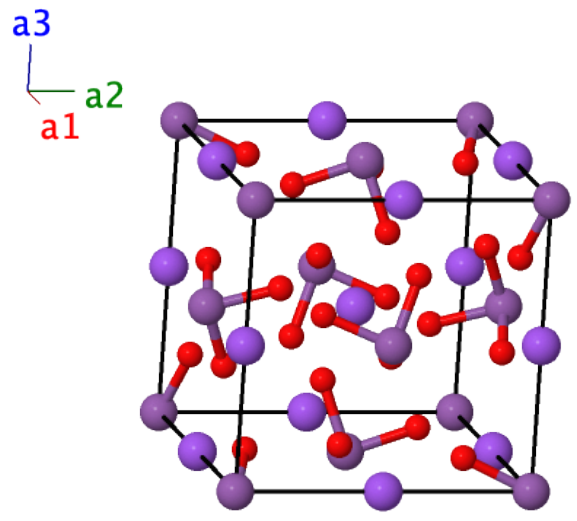

## Basis vectors:

| | | Lattice Coordinates | | Cartesian Coordinates | Wyckoff Position | Atom Type |
|---|---|---|---|---|---|---|
| $\mathbf{B}_1$ | = | $0\,\mathbf{a}_1 + 0\,\mathbf{a}_2 + 0\,\mathbf{a}_3$ | = | $0\,\hat{\mathbf{x}} + 0\,\hat{\mathbf{y}} + 0\,\hat{\mathbf{z}}$ | (4$c$) | Sb |
| $\mathbf{B}_2$ | = | $\frac{1}{2}\,\mathbf{a}_1 + \frac{1}{2}\,\mathbf{a}_2$ | = | $\frac{1}{2}a\,\hat{\mathbf{x}} + \frac{1}{2}a\,\hat{\mathbf{y}}$ | (4$c$) | Sb |
| $\mathbf{B}_3$ | = | $\frac{1}{2}\,\mathbf{a}_2 + \frac{1}{2}\,\mathbf{a}_3$ | = | $\frac{1}{2}a\,\hat{\mathbf{y}} + \frac{1}{2}c\,\hat{\mathbf{z}}$ | (4$c$) | Sb |
| $\mathbf{B}_4$ | = | $\frac{1}{2}\,\mathbf{a}_1 + \frac{1}{2}\,\mathbf{a}_3$ | = | $\frac{1}{2}a\,\hat{\mathbf{x}} + \frac{1}{2}c\,\hat{\mathbf{z}}$ | (4$c$) | Sb |
| $\mathbf{B}_5$ | = | $\frac{1}{2}\,\mathbf{a}_3$ | = | $\frac{1}{2}c\,\hat{\mathbf{z}}$ | (4$d$) | Na |
| $\mathbf{B}_6$ | = | $\frac{1}{2}\,\mathbf{a}_1 + \frac{1}{2}\,\mathbf{a}_2 + \frac{1}{2}\,\mathbf{a}_3$ | = | $\frac{1}{2}a\,\hat{\mathbf{x}} + \frac{1}{2}a\,\hat{\mathbf{y}} + \frac{1}{2}c\,\hat{\mathbf{z}}$ | (4$d$) | Na |
| $\mathbf{B}_7$ | = | $\frac{1}{2}\,\mathbf{a}_2$ | = | $\frac{1}{2}a\,\hat{\mathbf{y}}$ | (4$d$) | Na |
| $\mathbf{B}_8$ | = | $\frac{1}{2}\,\mathbf{a}_1$ | = | $\frac{1}{2}a\,\hat{\mathbf{x}}$ | (4$d$) | Na |
| $\mathbf{B}_9$ | = | $x_3\,\mathbf{a}_1 + y_3\,\mathbf{a}_2 + z_3\,\mathbf{a}_3$ | = | $x_3a\,\hat{\mathbf{x}} + y_3a\,\hat{\mathbf{y}} + z_3c\,\hat{\mathbf{z}}$ | (8$g$) | O I |
| $\mathbf{B}_{10}$ | = | $\left(\frac{1}{2} - x_3\right)\,\mathbf{a}_1 + \left(\frac{1}{2} - y_3\right)\,\mathbf{a}_2 + z_3\,\mathbf{a}_3$ | = | $\left(\frac{1}{2} - x_3\right)a\,\hat{\mathbf{x}} + \left(\frac{1}{2} - y_3\right)a\,\hat{\mathbf{y}} + z_3c\,\hat{\mathbf{z}}$ | (8$g$) | O I |
| $\mathbf{B}_{11}$ | = | $-y_3\,\mathbf{a}_1 + \left(\frac{1}{2} + x_3\right)\,\mathbf{a}_2 + \left(\frac{1}{2} + z_3\right)\,\mathbf{a}_3$ | = | $-y_3a\,\hat{\mathbf{x}} + \left(\frac{1}{2} + x_3\right)a\,\hat{\mathbf{y}} + \left(\frac{1}{2} + z_3\right)c\,\hat{\mathbf{z}}$ | (8$g$) | O I |
| $\mathbf{B}_{12}$ | = | $\left(\frac{1}{2} + y_3\right)\,\mathbf{a}_1 - x_3\,\mathbf{a}_2 + \left(\frac{1}{2} + z_3\right)\,\mathbf{a}_3$ | = | $\left(\frac{1}{2} + y_3\right)a\,\hat{\mathbf{x}} - x_3a\,\hat{\mathbf{y}} + \left(\frac{1}{2} + z_3\right)c\,\hat{\mathbf{z}}$ | (8$g$) | O I |
| $\mathbf{B}_{13}$ | = | $-x_3\,\mathbf{a}_1 - y_3\,\mathbf{a}_2 - z_3\,\mathbf{a}_3$ | = | $-x_3a\,\hat{\mathbf{x}} - y_3a\,\hat{\mathbf{y}} - z_3c\,\hat{\mathbf{z}}$ | (8$g$) | O I |
| $\mathbf{B}_{14}$ | = | $\left(\frac{1}{2} + x_3\right)\,\mathbf{a}_1 + \left(\frac{1}{2} + y_3\right)\,\mathbf{a}_2 - z_3\,\mathbf{a}_3$ | = | $\left(\frac{1}{2} + x_3\right)a\,\hat{\mathbf{x}} + \left(\frac{1}{2} + y_3\right)a\,\hat{\mathbf{y}} - z_3c\,\hat{\mathbf{z}}$ | (8$g$) | O I |
| $\mathbf{B}_{15}$ | = | $y_3\,\mathbf{a}_1 + \left(\frac{1}{2} - x_3\right)\,\mathbf{a}_2 + \left(\frac{1}{2} - z_3\right)\,\mathbf{a}_3$ | = | $y_3a\,\hat{\mathbf{x}} + \left(\frac{1}{2} - x_3\right)a\,\hat{\mathbf{y}} + \left(\frac{1}{2} - z_3\right)c\,\hat{\mathbf{z}}$ | (8$g$) | O I |
| $\mathbf{B}_{16}$ | = | $\left(\frac{1}{2} - y_3\right)\,\mathbf{a}_1 + x_3\,\mathbf{a}_2 + \left(\frac{1}{2} - z_3\right)\,\mathbf{a}_3$ | = | $\left(\frac{1}{2} - y_3\right)a\,\hat{\mathbf{x}} + x_3a\,\hat{\mathbf{y}} + \left(\frac{1}{2} - z_3\right)c\,\hat{\mathbf{z}}$ | (8$g$) | O I |
| $\mathbf{B}_{17}$ | = | $x_4\,\mathbf{a}_1 + y_4\,\mathbf{a}_2 + z_4\,\mathbf{a}_3$ | = | $x_4a\,\hat{\mathbf{x}} + y_4a\,\hat{\mathbf{y}} + z_4c\,\hat{\mathbf{z}}$ | (8$g$) | O II |
| $\mathbf{B}_{18}$ | = | $\left(\frac{1}{2} - x_4\right)\,\mathbf{a}_1 + \left(\frac{1}{2} - y_4\right)\,\mathbf{a}_2 + z_4\,\mathbf{a}_3$ | = | $\left(\frac{1}{2} - x_4\right)a\,\hat{\mathbf{x}} + \left(\frac{1}{2} - y_4\right)a\,\hat{\mathbf{y}} + z_4c\,\hat{\mathbf{z}}$ | (8$g$) | O II |
| $\mathbf{B}_{19}$ | = | $-y_4\,\mathbf{a}_1 + \left(\frac{1}{2} + x_4\right)\,\mathbf{a}_2 + \left(\frac{1}{2} + z_4\right)\,\mathbf{a}_3$ | = | $-y_4a\,\hat{\mathbf{x}} + \left(\frac{1}{2} + x_4\right)a\,\hat{\mathbf{y}} + \left(\frac{1}{2} + z_4\right)c\,\hat{\mathbf{z}}$ | (8$g$) | O II |
| $\mathbf{B}_{20}$ | = | $\left(\frac{1}{2} + y_4\right)\,\mathbf{a}_1 - x_4\,\mathbf{a}_2 + \left(\frac{1}{2} + z_4\right)\,\mathbf{a}_3$ | = | $\left(\frac{1}{2} + y_4\right)a\,\hat{\mathbf{x}} - x_4a\,\hat{\mathbf{y}} + \left(\frac{1}{2} + z_4\right)c\,\hat{\mathbf{z}}$ | (8$g$) | O II |
| $\mathbf{B}_{21}$ | = | $-x_4\,\mathbf{a}_1 - y_4\,\mathbf{a}_2 - z_4\,\mathbf{a}_3$ | = | $-x_4a\,\hat{\mathbf{x}} - y_4a\,\hat{\mathbf{y}} - z_4c\,\hat{\mathbf{z}}$ | (8$g$) | O II |
| $\mathbf{B}_{22}$ | = | $\left(\frac{1}{2} + x_4\right)\,\mathbf{a}_1 + \left(\frac{1}{2} + y_4\right)\,\mathbf{a}_2 - z_4\,\mathbf{a}_3$ | = | $\left(\frac{1}{2} + x_4\right)a\,\hat{\mathbf{x}} + \left(\frac{1}{2} + y_4\right)a\,\hat{\mathbf{y}} - z_4c\,\hat{\mathbf{z}}$ | (8$g$) | O II |
| $\mathbf{B}_{23}$ | = | $y_4\,\mathbf{a}_1 + \left(\frac{1}{2} - x_4\right)\,\mathbf{a}_2 + \left(\frac{1}{2} - z_4\right)\,\mathbf{a}_3$ | = | $y_4a\,\hat{\mathbf{x}} + \left(\frac{1}{2} - x_4\right)a\,\hat{\mathbf{y}} + \left(\frac{1}{2} - z_4\right)c\,\hat{\mathbf{z}}$ | (8$g$) | O II |
| $\mathbf{B}_{24}$ | = | $\left(\frac{1}{2} - y_4\right)\,\mathbf{a}_1 + x_4\,\mathbf{a}_2 + \left(\frac{1}{2} - z_4\right)\,\mathbf{a}_3$ | = | $\left(\frac{1}{2} - y_4\right)a\,\hat{\mathbf{x}} + x_4a\,\hat{\mathbf{y}} + \left(\frac{1}{2} - z_4\right)c\,\hat{\mathbf{z}}$ | (8$g$) | O II |
| $\mathbf{B}_{25}$ | = | $x_5\,\mathbf{a}_1 + y_5\,\mathbf{a}_2 + z_5\,\mathbf{a}_3$ | = | $x_5a\,\hat{\mathbf{x}} + y_5a\,\hat{\mathbf{y}} + z_5c\,\hat{\mathbf{z}}$ | (8$g$) | O III |
| $\mathbf{B}_{26}$ | = | $\left(\frac{1}{2} - x_5\right)\,\mathbf{a}_1 + \left(\frac{1}{2} - y_5\right)\,\mathbf{a}_2 + z_5\,\mathbf{a}_3$ | = | $\left(\frac{1}{2} - x_5\right)a\,\hat{\mathbf{x}} + \left(\frac{1}{2} - y_5\right)a\,\hat{\mathbf{y}} + z_5c\,\hat{\mathbf{z}}$ | (8$g$) | O III |
| $\mathbf{B}_{27}$ | = | $-y_5\,\mathbf{a}_1 + \left(\frac{1}{2} + x_5\right)\,\mathbf{a}_2 + \left(\frac{1}{2} + z_5\right)\,\mathbf{a}_3$ | = | $-y_5a\,\hat{\mathbf{x}} + \left(\frac{1}{2} + x_5\right)a\,\hat{\mathbf{y}} + \left(\frac{1}{2} + z_5\right)c\,\hat{\mathbf{z}}$ | (8$g$) | O III |
| $\mathbf{B}_{28}$ | = | $\left(\frac{1}{2}+y_5\right)\mathbf{a}_1 - x_5\,\mathbf{a}_2 + \left(\frac{1}{2}+z_5\right)\mathbf{a}_3$ | = | $\left(\frac{1}{2}+y_5\right)a\,\hat{\mathbf{x}} - x_5 a\,\hat{\mathbf{y}} + \left(\frac{1}{2}+z_5\right)c\,\hat{\mathbf{z}}$ | (8$g$) | O III |
| $\mathbf{B}_{29}$ | = | $-x_5\,\mathbf{a}_1 - y_5\,\mathbf{a}_2 - z_5\,\mathbf{a}_3$ | = | $-x_5 a\,\hat{\mathbf{x}} - y_5 a\,\hat{\mathbf{y}} - z_5 c\,\hat{\mathbf{z}}$ | (8$g$) | O III |
| $\mathbf{B}_{30}$ | = | $\left(\frac{1}{2}+x_5\right)\mathbf{a}_1 + \left(\frac{1}{2}+y_5\right)\mathbf{a}_2 - z_5\,\mathbf{a}_3$ | = | $\left(\frac{1}{2}+x_5\right)a\,\hat{\mathbf{x}} + \left(\frac{1}{2}+y_5\right)a\,\hat{\mathbf{y}} - z_5 c\,\hat{\mathbf{z}}$ | (8$g$) | O III |
| $\mathbf{B}_{31}$ | = | $y_5\,\mathbf{a}_1 + \left(\frac{1}{2}-x_5\right)\mathbf{a}_2 + \left(\frac{1}{2}-z_5\right)\mathbf{a}_3$ | = | $y_5 a\,\hat{\mathbf{x}} + \left(\frac{1}{2}-x_5\right)a\,\hat{\mathbf{y}} + \left(\frac{1}{2}-z_5\right)c\,\hat{\mathbf{z}}$ | (8$g$) | O III |
| $\mathbf{B}_{32}$ | = | $\left(\frac{1}{2}-y_5\right)\mathbf{a}_1 + x_5\,\mathbf{a}_2 + \left(\frac{1}{2}-z_5\right)\mathbf{a}_3$ | = | $\left(\frac{1}{2}-y_5\right)a\,\hat{\mathbf{x}} + x_5 a\,\hat{\mathbf{y}} + \left(\frac{1}{2}-z_5\right)c\,\hat{\mathbf{z}}$ | (8$g$) | O III |

**Geometry files:**

- CIF: pp. 1672
- POSCAR: pp. 1673

# $\beta$-$LiIO_3$ Structure: ABC3_tP40_86_g_g_3g

http://aflow.org/prototype-encyclopedia/ABC3_tP40_86_g_g_3g

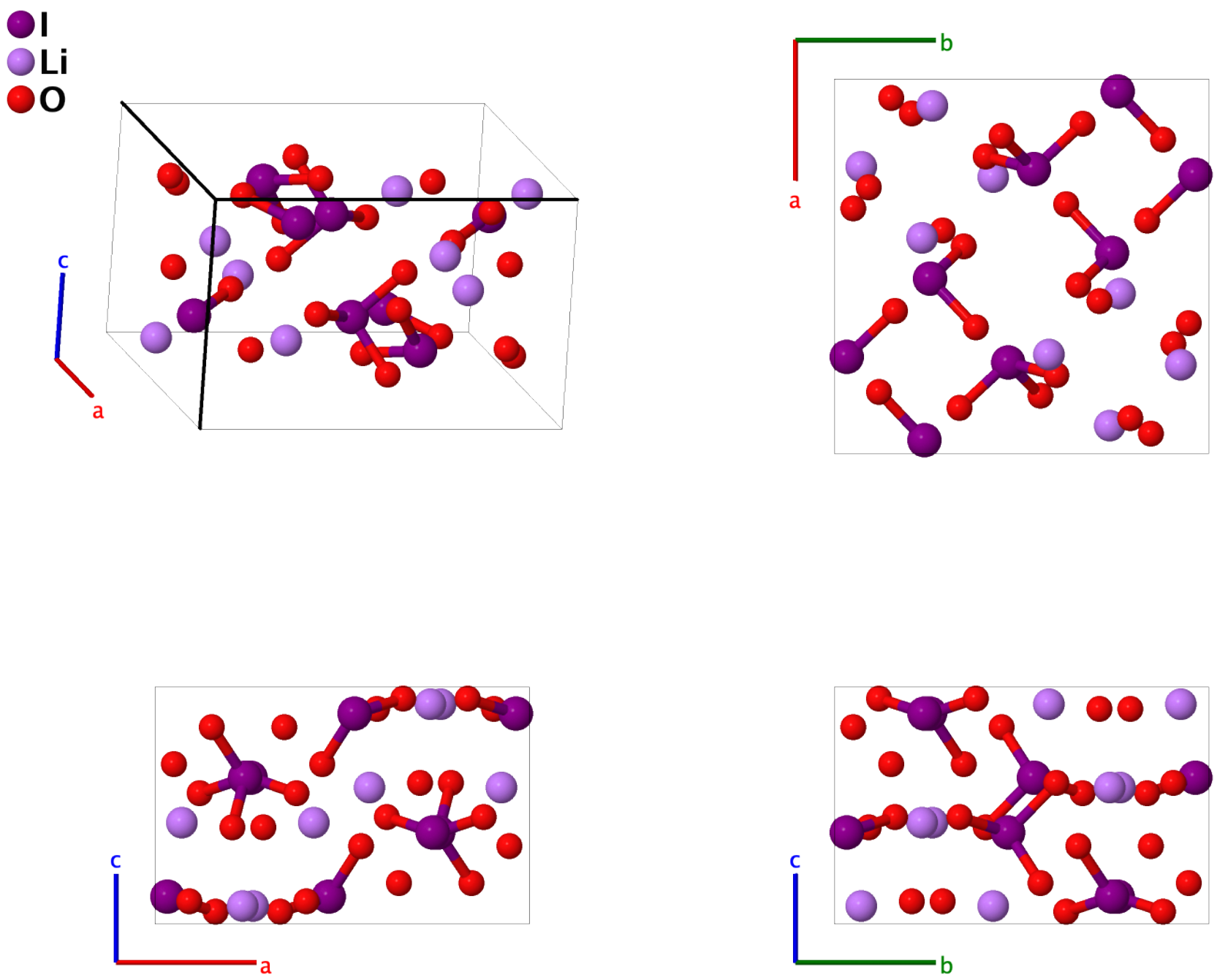

| | | |
|---|---|---|
| **Prototype** | : | $ILiO_3$ |
| **AFLOW prototype label** | : | ABC3_tP40_86_g_g_3g |
| ***Strukturbericht* designation** | : | None |
| **Pearson symbol** | : | tP40 |
| **Space group number** | : | 86 |
| **Space group symbol** | : | $P4_2/n$ |
| **AFLOW prototype command** | : | `aflow --proto=ABC3_tP40_86_g_g_3g` `--params=`$a, c/a, x_1, y_1, z_1, x_2, y_2, z_2, x_3, y_3, z_3, x_4, y_4, z_4, x_5, y_5, z_5$ |

- $LiIO_3$ is known to exist in three forms:
  - $\alpha$-$LiIO_3$, stable below 473 K: (Zachariasen, 1931) originally determined that the structure of $\alpha$-$LiIO_3$ was in space group $P6_322$ #182, which (Hermann, 1937) designated *Strukturbericht* $E2_3$. (Rosenzweig, 1966) subsequently determined that the true structure was in space group $P6_3$ #173.
  - $\beta$-$LiIO_3$, stable from 573 K up to the melting point at 708 K (this Structure).
  - $\gamma$-$LiIO_3$, stable between the $\alpha$- and $\beta$-phases, with an orthorhombic structure in space group $Pna2_1$ #33.

**Simple Tetragonal primitive vectors:**

$$\mathbf{a}_1 = a\,\hat{\mathbf{x}}$$
$$\mathbf{a}_2 = a\,\hat{\mathbf{y}}$$
$$\mathbf{a}_3 = c\,\hat{\mathbf{z}}$$

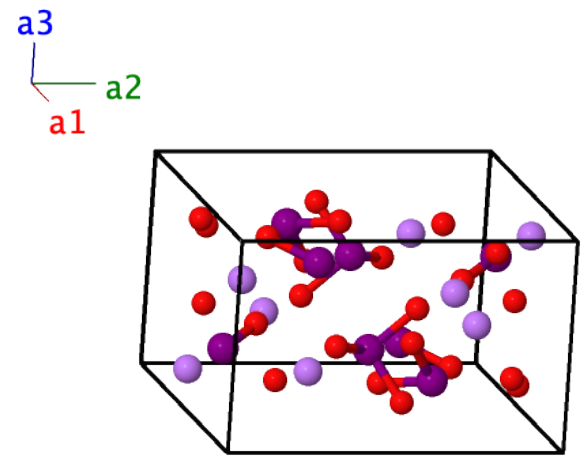

**Basis vectors:**

| | | Lattice Coordinates | | Cartesian Coordinates | Wyckoff Position | Atom Type |
|---|---|---|---|---|---|---|
| $\mathbf{B}_1$ | = | $x_1\,\mathbf{a}_1 + y_1\,\mathbf{a}_2 + z_1\,\mathbf{a}_3$ | = | $x_1 a\,\hat{\mathbf{x}} + y_1 a\,\hat{\mathbf{y}} + z_1 c\,\hat{\mathbf{z}}$ | (8g) | I |
| $\mathbf{B}_2$ | = | $\left(\frac{1}{2} - x_1\right)\mathbf{a}_1 + \left(\frac{1}{2} - y_1\right)\mathbf{a}_2 + z_1\,\mathbf{a}_3$ | = | $\left(\frac{1}{2} - x_1\right)a\,\hat{\mathbf{x}} + \left(\frac{1}{2} - y_1\right)a\,\hat{\mathbf{y}} + z_1 c\,\hat{\mathbf{z}}$ | (8g) | I |
| $\mathbf{B}_3$ | = | $-y_1\,\mathbf{a}_1 + \left(\frac{1}{2} + x_1\right)\mathbf{a}_2 + \left(\frac{1}{2} + z_1\right)\mathbf{a}_3$ | = | $-y_1 a\,\hat{\mathbf{x}} + \left(\frac{1}{2} + x_1\right)a\,\hat{\mathbf{y}} + \left(\frac{1}{2} + z_1\right)c\,\hat{\mathbf{z}}$ | (8g) | I |
| $\mathbf{B}_4$ | = | $\left(\frac{1}{2} + y_1\right)\mathbf{a}_1 - x_1\,\mathbf{a}_2 + \left(\frac{1}{2} + z_1\right)\mathbf{a}_3$ | = | $\left(\frac{1}{2} + y_1\right)a\,\hat{\mathbf{x}} - x_1 a\,\hat{\mathbf{y}} + \left(\frac{1}{2} + z_1\right)c\,\hat{\mathbf{z}}$ | (8g) | I |
| $\mathbf{B}_5$ | = | $-x_1\,\mathbf{a}_1 - y_1\,\mathbf{a}_2 - z_1\,\mathbf{a}_3$ | = | $-x_1 a\,\hat{\mathbf{x}} - y_1 a\,\hat{\mathbf{y}} - z_1 c\,\hat{\mathbf{z}}$ | (8g) | I |
| $\mathbf{B}_6$ | = | $\left(\frac{1}{2} + x_1\right)\mathbf{a}_1 + \left(\frac{1}{2} + y_1\right)\mathbf{a}_2 - z_1\,\mathbf{a}_3$ | = | $\left(\frac{1}{2} + x_1\right)a\,\hat{\mathbf{x}} + \left(\frac{1}{2} + y_1\right)a\,\hat{\mathbf{y}} - z_1 c\,\hat{\mathbf{z}}$ | (8g) | I |
| $\mathbf{B}_7$ | = | $y_1\,\mathbf{a}_1 + \left(\frac{1}{2} - x_1\right)\mathbf{a}_2 + \left(\frac{1}{2} - z_1\right)\mathbf{a}_3$ | = | $y_1 a\,\hat{\mathbf{x}} + \left(\frac{1}{2} - x_1\right)a\,\hat{\mathbf{y}} + \left(\frac{1}{2} - z_1\right)c\,\hat{\mathbf{z}}$ | (8g) | I |
| $\mathbf{B}_8$ | = | $\left(\frac{1}{2} - y_1\right)\mathbf{a}_1 + x_1\,\mathbf{a}_2 + \left(\frac{1}{2} - z_1\right)\mathbf{a}_3$ | = | $\left(\frac{1}{2} - y_1\right)a\,\hat{\mathbf{x}} + x_1 a\,\hat{\mathbf{y}} + \left(\frac{1}{2} - z_1\right)c\,\hat{\mathbf{z}}$ | (8g) | I |
| $\mathbf{B}_9$ | = | $x_2\,\mathbf{a}_1 + y_2\,\mathbf{a}_2 + z_2\,\mathbf{a}_3$ | = | $x_2 a\,\hat{\mathbf{x}} + y_2 a\,\hat{\mathbf{y}} + z_2 c\,\hat{\mathbf{z}}$ | (8g) | Li |
| $\mathbf{B}_{10}$ | = | $\left(\frac{1}{2} - x_2\right)\mathbf{a}_1 + \left(\frac{1}{2} - y_2\right)\mathbf{a}_2 + z_2\,\mathbf{a}_3$ | = | $\left(\frac{1}{2} - x_2\right)a\,\hat{\mathbf{x}} + \left(\frac{1}{2} - y_2\right)a\,\hat{\mathbf{y}} + z_2 c\,\hat{\mathbf{z}}$ | (8g) | Li |
| $\mathbf{B}_{11}$ | = | $-y_2\,\mathbf{a}_1 + \left(\frac{1}{2} + x_2\right)\mathbf{a}_2 + \left(\frac{1}{2} + z_2\right)\mathbf{a}_3$ | = | $-y_2 a\,\hat{\mathbf{x}} + \left(\frac{1}{2} + x_2\right)a\,\hat{\mathbf{y}} + \left(\frac{1}{2} + z_2\right)c\,\hat{\mathbf{z}}$ | (8g) | Li |
| $\mathbf{B}_{12}$ | = | $\left(\frac{1}{2} + y_2\right)\mathbf{a}_1 - x_2\,\mathbf{a}_2 + \left(\frac{1}{2} + z_2\right)\mathbf{a}_3$ | = | $\left(\frac{1}{2} + y_2\right)a\,\hat{\mathbf{x}} - x_2 a\,\hat{\mathbf{y}} + \left(\frac{1}{2} + z_2\right)c\,\hat{\mathbf{z}}$ | (8g) | Li |
| $\mathbf{B}_{13}$ | = | $-x_2\,\mathbf{a}_1 - y_2\,\mathbf{a}_2 - z_2\,\mathbf{a}_3$ | = | $-x_2 a\,\hat{\mathbf{x}} - y_2 a\,\hat{\mathbf{y}} - z_2 c\,\hat{\mathbf{z}}$ | (8g) | Li |
| $\mathbf{B}_{14}$ | = | $\left(\frac{1}{2} + x_2\right)\mathbf{a}_1 + \left(\frac{1}{2} + y_2\right)\mathbf{a}_2 - z_2\,\mathbf{a}_3$ | = | $\left(\frac{1}{2} + x_2\right)a\,\hat{\mathbf{x}} + \left(\frac{1}{2} + y_2\right)a\,\hat{\mathbf{y}} - z_2 c\,\hat{\mathbf{z}}$ | (8g) | Li |
| $\mathbf{B}_{15}$ | = | $y_2\,\mathbf{a}_1 + \left(\frac{1}{2} - x_2\right)\mathbf{a}_2 + \left(\frac{1}{2} - z_2\right)\mathbf{a}_3$ | = | $y_2 a\,\hat{\mathbf{x}} + \left(\frac{1}{2} - x_2\right)a\,\hat{\mathbf{y}} + \left(\frac{1}{2} - z_2\right)c\,\hat{\mathbf{z}}$ | (8g) | Li |
| $\mathbf{B}_{16}$ | = | $\left(\frac{1}{2} - y_2\right)\mathbf{a}_1 + x_2\,\mathbf{a}_2 + \left(\frac{1}{2} - z_2\right)\mathbf{a}_3$ | = | $\left(\frac{1}{2} - y_2\right)a\,\hat{\mathbf{x}} + x_2 a\,\hat{\mathbf{y}} + \left(\frac{1}{2} - z_2\right)c\,\hat{\mathbf{z}}$ | (8g) | Li |
| $\mathbf{B}_{17}$ | = | $x_3\,\mathbf{a}_1 + y_3\,\mathbf{a}_2 + z_3\,\mathbf{a}_3$ | = | $x_3 a\,\hat{\mathbf{x}} + y_3 a\,\hat{\mathbf{y}} + z_3 c\,\hat{\mathbf{z}}$ | (8g) | O I |
| $\mathbf{B}_{18}$ | = | $\left(\frac{1}{2} - x_3\right)\mathbf{a}_1 + \left(\frac{1}{2} - y_3\right)\mathbf{a}_2 + z_3\,\mathbf{a}_3$ | = | $\left(\frac{1}{2} - x_3\right)a\,\hat{\mathbf{x}} + \left(\frac{1}{2} - y_3\right)a\,\hat{\mathbf{y}} + z_3 c\,\hat{\mathbf{z}}$ | (8g) | O I |
| $\mathbf{B}_{19}$ | = | $-y_3\,\mathbf{a}_1 + \left(\frac{1}{2} + x_3\right)\mathbf{a}_2 + \left(\frac{1}{2} + z_3\right)\mathbf{a}_3$ | = | $-y_3 a\,\hat{\mathbf{x}} + \left(\frac{1}{2} + x_3\right)a\,\hat{\mathbf{y}} + \left(\frac{1}{2} + z_3\right)c\,\hat{\mathbf{z}}$ | (8g) | O I |
| $\mathbf{B}_{20}$ | = | $\left(\frac{1}{2} + y_3\right)\mathbf{a}_1 - x_3\,\mathbf{a}_2 + \left(\frac{1}{2} + z_3\right)\mathbf{a}_3$ | = | $\left(\frac{1}{2} + y_3\right)a\,\hat{\mathbf{x}} - x_3 a\,\hat{\mathbf{y}} + \left(\frac{1}{2} + z_3\right)c\,\hat{\mathbf{z}}$ | (8g) | O I |
| $\mathbf{B}_{21}$ | = | $-x_3\,\mathbf{a}_1 - y_3\,\mathbf{a}_2 - z_3\,\mathbf{a}_3$ | = | $-x_3 a\,\hat{\mathbf{x}} - y_3 a\,\hat{\mathbf{y}} - z_3 c\,\hat{\mathbf{z}}$ | (8g) | O I |
| $\mathbf{B}_{22}$ | = | $\left(\frac{1}{2} + x_3\right)\mathbf{a}_1 + \left(\frac{1}{2} + y_3\right)\mathbf{a}_2 - z_3\,\mathbf{a}_3$ | = | $\left(\frac{1}{2} + x_3\right)a\,\hat{\mathbf{x}} + \left(\frac{1}{2} + y_3\right)a\,\hat{\mathbf{y}} - z_3 c\,\hat{\mathbf{z}}$ | (8g) | O I |
| $\mathbf{B}_{23}$ | = | $y_3\,\mathbf{a}_1 + \left(\frac{1}{2} - x_3\right)\mathbf{a}_2 + \left(\frac{1}{2} - z_3\right)\mathbf{a}_3$ | = | $y_3 a\,\hat{\mathbf{x}} + \left(\frac{1}{2} - x_3\right)a\,\hat{\mathbf{y}} + \left(\frac{1}{2} - z_3\right)c\,\hat{\mathbf{z}}$ | (8g) | O I |
| $\mathbf{B}_{24}$ | = | $\left(\frac{1}{2} - y_3\right)\mathbf{a}_1 + x_3\,\mathbf{a}_2 + \left(\frac{1}{2} - z_3\right)\mathbf{a}_3$ | = | $\left(\frac{1}{2} - y_3\right)a\,\hat{\mathbf{x}} + x_3 a\,\hat{\mathbf{y}} + \left(\frac{1}{2} - z_3\right)c\,\hat{\mathbf{z}}$ | (8g) | O I |
| $\mathbf{B}_{25}$ | = | $x_4\,\mathbf{a}_1 + y_4\,\mathbf{a}_2 + z_4\,\mathbf{a}_3$ | = | $x_4 a\,\hat{\mathbf{x}} + y_4 a\,\hat{\mathbf{y}} + z_4 c\,\hat{\mathbf{z}}$ | (8g) | O II |
| $\mathbf{B}_{26}$ | = | $\left(\frac{1}{2} - x_4\right)\mathbf{a}_1 + \left(\frac{1}{2} - y_4\right)\mathbf{a}_2 + z_4\,\mathbf{a}_3$ | = | $\left(\frac{1}{2} - x_4\right)a\,\hat{\mathbf{x}} + \left(\frac{1}{2} - y_4\right)a\,\hat{\mathbf{y}} + z_4 c\,\hat{\mathbf{z}}$ | (8g) | O II |
| $\mathbf{B}_{27}$ | = | $-y_4\,\mathbf{a}_1 + \left(\frac{1}{2} + x_4\right)\mathbf{a}_2 + \left(\frac{1}{2} + z_4\right)\mathbf{a}_3$ | = | $-y_4 a\,\hat{\mathbf{x}} + \left(\frac{1}{2} + x_4\right)a\,\hat{\mathbf{y}} + \left(\frac{1}{2} + z_4\right)c\,\hat{\mathbf{z}}$ | (8g) | O II |
| $\mathbf{B}_{28}$ | = | $\left(\frac{1}{2}+y_4\right)\mathbf{a}_1 - x_4\,\mathbf{a}_2 + \left(\frac{1}{2}+z_4\right)\mathbf{a}_3$ | = | $\left(\frac{1}{2}+y_4\right)a\,\hat{\mathbf{x}} - x_4 a\,\hat{\mathbf{y}} + \left(\frac{1}{2}+z_4\right)c\,\hat{\mathbf{z}}$ | (8$g$) | O II |
| $\mathbf{B}_{29}$ | = | $-x_4\,\mathbf{a}_1 - y_4\,\mathbf{a}_2 - z_4\,\mathbf{a}_3$ | = | $-x_4 a\,\hat{\mathbf{x}} - y_4 a\,\hat{\mathbf{y}} - z_4 c\,\hat{\mathbf{z}}$ | (8$g$) | O II |
| $\mathbf{B}_{30}$ | = | $\left(\frac{1}{2}+x_4\right)\mathbf{a}_1 + \left(\frac{1}{2}+y_4\right)\mathbf{a}_2 - z_4\,\mathbf{a}_3$ | = | $\left(\frac{1}{2}+x_4\right)a\,\hat{\mathbf{x}} + \left(\frac{1}{2}+y_4\right)a\,\hat{\mathbf{y}} - z_4 c\,\hat{\mathbf{z}}$ | (8$g$) | O II |
| $\mathbf{B}_{31}$ | = | $y_4\,\mathbf{a}_1 + \left(\frac{1}{2}-x_4\right)\mathbf{a}_2 + \left(\frac{1}{2}-z_4\right)\mathbf{a}_3$ | = | $y_4 a\,\hat{\mathbf{x}} + \left(\frac{1}{2}-x_4\right)a\,\hat{\mathbf{y}} + \left(\frac{1}{2}-z_4\right)c\,\hat{\mathbf{z}}$ | (8$g$) | O II |
| $\mathbf{B}_{32}$ | = | $\left(\frac{1}{2}-y_4\right)\mathbf{a}_1 + x_4\,\mathbf{a}_2 + \left(\frac{1}{2}-z_4\right)\mathbf{a}_3$ | = | $\left(\frac{1}{2}-y_4\right)a\,\hat{\mathbf{x}} + x_4 a\,\hat{\mathbf{y}} + \left(\frac{1}{2}-z_4\right)c\,\hat{\mathbf{z}}$ | (8$g$) | O II |
| $\mathbf{B}_{33}$ | = | $x_5\,\mathbf{a}_1 + y_5\,\mathbf{a}_2 + z_5\,\mathbf{a}_3$ | = | $x_5 a\,\hat{\mathbf{x}} + y_5 a\,\hat{\mathbf{y}} + z_5 c\,\hat{\mathbf{z}}$ | (8$g$) | O III |
| $\mathbf{B}_{34}$ | = | $\left(\frac{1}{2}-x_5\right)\mathbf{a}_1 + \left(\frac{1}{2}-y_5\right)\mathbf{a}_2 + z_5\,\mathbf{a}_3$ | = | $\left(\frac{1}{2}-x_5\right)a\,\hat{\mathbf{x}} + \left(\frac{1}{2}-y_5\right)a\,\hat{\mathbf{y}} + z_5 c\,\hat{\mathbf{z}}$ | (8$g$) | O III |
| $\mathbf{B}_{35}$ | = | $-y_5\,\mathbf{a}_1 + \left(\frac{1}{2}+x_5\right)\mathbf{a}_2 + \left(\frac{1}{2}+z_5\right)\mathbf{a}_3$ | = | $-y_5 a\,\hat{\mathbf{x}} + \left(\frac{1}{2}+x_5\right)a\,\hat{\mathbf{y}} + \left(\frac{1}{2}+z_5\right)c\,\hat{\mathbf{z}}$ | (8$g$) | O III |
| $\mathbf{B}_{36}$ | = | $\left(\frac{1}{2}+y_5\right)\mathbf{a}_1 - x_5\,\mathbf{a}_2 + \left(\frac{1}{2}+z_5\right)\mathbf{a}_3$ | = | $\left(\frac{1}{2}+y_5\right)a\,\hat{\mathbf{x}} - x_5 a\,\hat{\mathbf{y}} + \left(\frac{1}{2}+z_5\right)c\,\hat{\mathbf{z}}$ | (8$g$) | O III |
| $\mathbf{B}_{37}$ | = | $-x_5\,\mathbf{a}_1 - y_5\,\mathbf{a}_2 - z_5\,\mathbf{a}_3$ | = | $-x_5 a\,\hat{\mathbf{x}} - y_5 a\,\hat{\mathbf{y}} - z_5 c\,\hat{\mathbf{z}}$ | (8$g$) | O III |
| $\mathbf{B}_{38}$ | = | $\left(\frac{1}{2}+x_5\right)\mathbf{a}_1 + \left(\frac{1}{2}+y_5\right)\mathbf{a}_2 - z_5\,\mathbf{a}_3$ | = | $\left(\frac{1}{2}+x_5\right)a\,\hat{\mathbf{x}} + \left(\frac{1}{2}+y_5\right)a\,\hat{\mathbf{y}} - z_5 c\,\hat{\mathbf{z}}$ | (8$g$) | O III |
| $\mathbf{B}_{39}$ | = | $y_5\,\mathbf{a}_1 + \left(\frac{1}{2}-x_5\right)\mathbf{a}_2 + \left(\frac{1}{2}-z_5\right)\mathbf{a}_3$ | = | $y_5 a\,\hat{\mathbf{x}} + \left(\frac{1}{2}-x_5\right)a\,\hat{\mathbf{y}} + \left(\frac{1}{2}-z_5\right)c\,\hat{\mathbf{z}}$ | (8$g$) | O III |
| $\mathbf{B}_{40}$ | = | $\left(\frac{1}{2}-y_5\right)\mathbf{a}_1 + x_5\,\mathbf{a}_2 + \left(\frac{1}{2}-z_5\right)\mathbf{a}_3$ | = | $\left(\frac{1}{2}-y_5\right)a\,\hat{\mathbf{x}} + x_5 a\,\hat{\mathbf{y}} + \left(\frac{1}{2}-z_5\right)c\,\hat{\mathbf{z}}$ | (8$g$) | O III |

---

---

**Geometry files:**

- CIF: pp. 1673
- POSCAR: pp. 1673

# Marialite Scapolite [$Na_4Cl(AlSi_3)_3O_{24}$, $S6_4$] Structure: AB4C24D12_tI82_87_a_h_2h2i_hi

http://aflow.org/prototype-encyclopedia/AB4C24D12_tI82_87_a_h_2h2i_hi

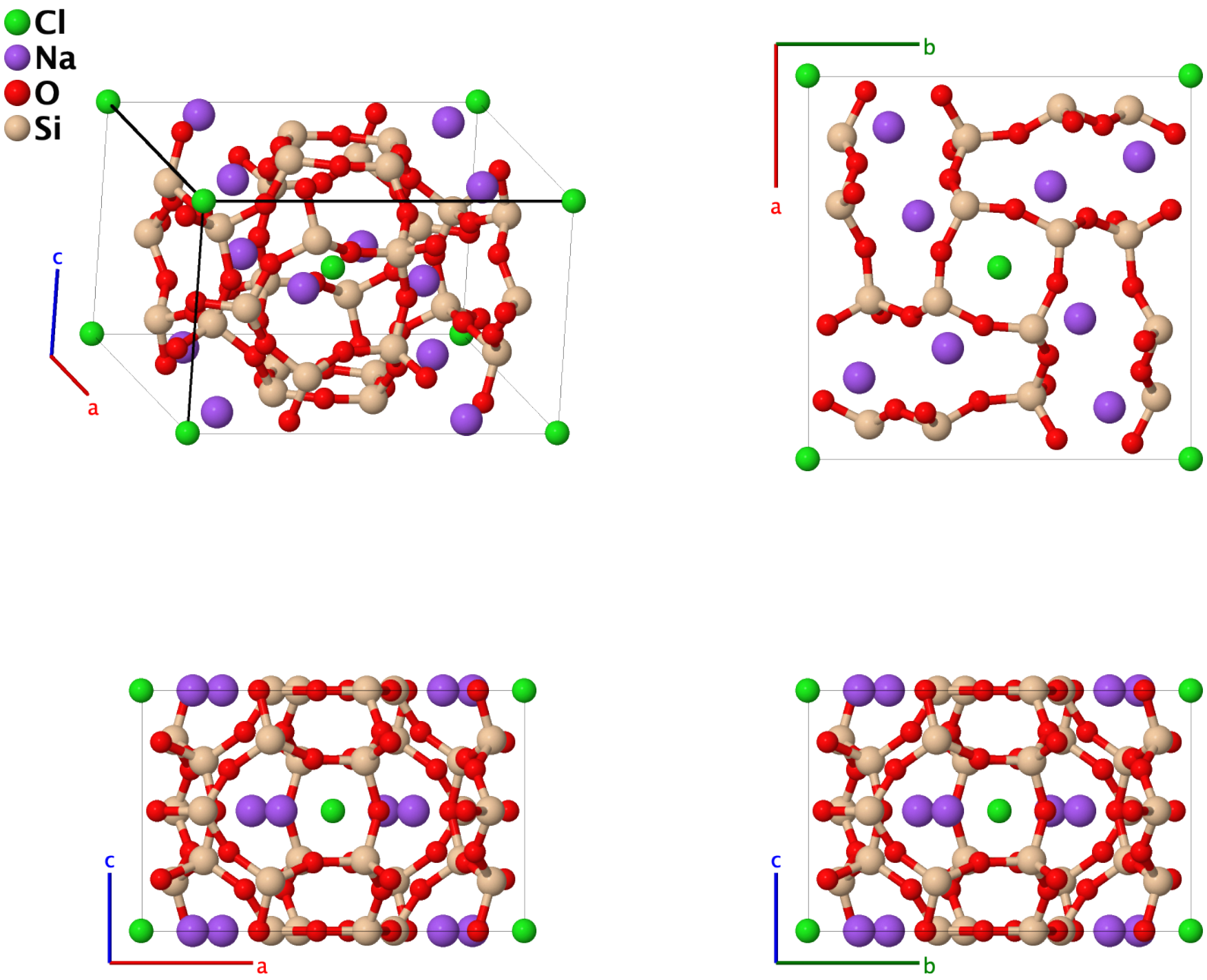

| | | |
|---|---|---|
| **Prototype** | : | $ClNa_4O_{24}(Al_3Si_9)$ |
| **AFLOW prototype label** | : | AB4C24D12_tI82_87_a_h_2h2i_hi |
| ***Strukturbericht* designation** | : | $S6_4$ |
| **Pearson symbol** | : | tI82 |
| **Space group number** | : | 87 |
| **Space group symbol** | : | $I4/m$ |
| **AFLOW prototype command** | : | `aflow --proto=AB4C24D12_tI82_87_a_h_2h2i_hi --params=`$a, c/a, x_2, y_2, x_3, y_3, x_4, y_4, x_5, y_5, x_6, y_6, z_6, x_7, y_7, z_7, x_8, y_8, z_8$ |

- (Papike, 1965) found that the composition of the Si-II (16h) site was actually $Al_{0.458}Si_{0.542}$. This is richer in aluminum than assumed by Pauling, who gave the total aluminum/silicon composition as $AlSi_3$ (Pauling, 1930). If we assume that the Si-I (8h) site is only filled by silicon atoms, then Pauling's composition for the Si-II site is $Al_{0.375}Si_{0.625}$. The name "marialite" scapolite distinguishes this from meionite scapolite, which replaces the sodium atoms by calcium but also includes $SiO_4$ and $CO_3$, which replace the chlorine atoms. According to Pauling, "The minerals of the scapolite group can be considered as solid solution of two end-members, marialite, $Na_4Al_3Si_9O_{24}Cl$, and meionite, $Ca_4Al_6Si_6O_{24}(SO_4,CO_3)$, in various proportions."
- Since the Al-Si sites are partially occupied with a higher concentration of Si, the positions are labeled as Si.

## Body-centered Tetragonal primitive vectors:

$$\mathbf{a}_1 = -\frac{1}{2}\,a\,\hat{\mathbf{x}} + \frac{1}{2}\,a\,\hat{\mathbf{y}} + \frac{1}{2}\,c\,\hat{\mathbf{z}}$$
$$\mathbf{a}_2 = \frac{1}{2}\,a\,\hat{\mathbf{x}} - \frac{1}{2}\,a\,\hat{\mathbf{y}} + \frac{1}{2}\,c\,\hat{\mathbf{z}}$$
$$\mathbf{a}_3 = \frac{1}{2}\,a\,\hat{\mathbf{x}} + \frac{1}{2}\,a\,\hat{\mathbf{y}} - \frac{1}{2}\,c\,\hat{\mathbf{z}}$$

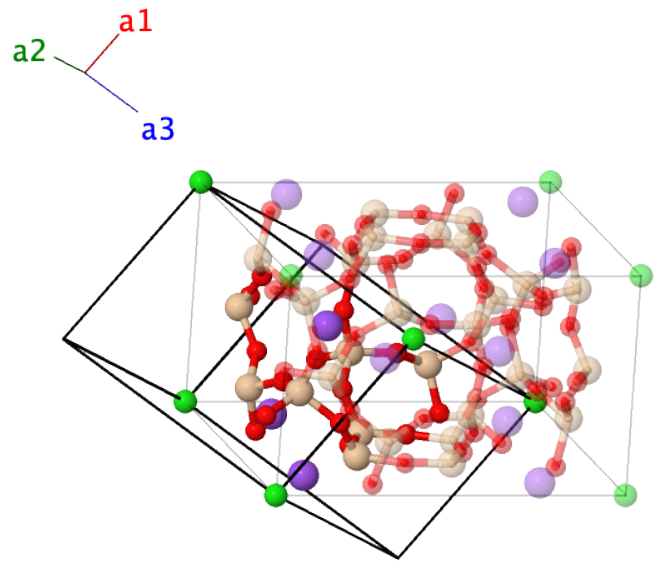

## Basis vectors:

| | | Lattice Coordinates | | Cartesian Coordinates | Wyckoff Position | Atom Type |
|---|---|---|---|---|---|---|
| $\mathbf{B}_1$ | = | $0\,\mathbf{a}_1 + 0\,\mathbf{a}_2 + 0\,\mathbf{a}_3$ | = | $0\,\hat{\mathbf{x}} + 0\,\hat{\mathbf{y}} + 0\,\hat{\mathbf{z}}$ | (2$a$) | Cl |
| $\mathbf{B}_2$ | = | $y_2\,\mathbf{a}_1 + x_2\,\mathbf{a}_2 + (x_2 + y_2)\,\mathbf{a}_3$ | = | $x_2 a\,\hat{\mathbf{x}} + y_2 a\,\hat{\mathbf{y}}$ | (8$h$) | Na |
| $\mathbf{B}_3$ | = | $-y_2\,\mathbf{a}_1 - x_2\,\mathbf{a}_2 + (-x_2 - y_2)\,\mathbf{a}_3$ | = | $-x_2 a\,\hat{\mathbf{x}} - y_2 a\,\hat{\mathbf{y}}$ | (8$h$) | Na |
| $\mathbf{B}_4$ | = | $x_2\,\mathbf{a}_1 - y_2\,\mathbf{a}_2 + (x_2 - y_2)\,\mathbf{a}_3$ | = | $-y_2 a\,\hat{\mathbf{x}} + x_2 a\,\hat{\mathbf{y}}$ | (8$h$) | Na |
| $\mathbf{B}_5$ | = | $-x_2\,\mathbf{a}_1 + y_2\,\mathbf{a}_2 + (-x_2 + y_2)\,\mathbf{a}_3$ | = | $y_2 a\,\hat{\mathbf{x}} - x_2 a\,\hat{\mathbf{y}}$ | (8$h$) | Na |
| $\mathbf{B}_6$ | = | $y_3\,\mathbf{a}_1 + x_3\,\mathbf{a}_2 + (x_3 + y_3)\,\mathbf{a}_3$ | = | $x_3 a\,\hat{\mathbf{x}} + y_3 a\,\hat{\mathbf{y}}$ | (8$h$) | O I |
| $\mathbf{B}_7$ | = | $-y_3\,\mathbf{a}_1 - x_3\,\mathbf{a}_2 + (-x_3 - y_3)\,\mathbf{a}_3$ | = | $-x_3 a\,\hat{\mathbf{x}} - y_3 a\,\hat{\mathbf{y}}$ | (8$h$) | O I |
| $\mathbf{B}_8$ | = | $x_3\,\mathbf{a}_1 - y_3\,\mathbf{a}_2 + (x_3 - y_3)\,\mathbf{a}_3$ | = | $-y_3 a\,\hat{\mathbf{x}} + x_3 a\,\hat{\mathbf{y}}$ | (8$h$) | O I |
| $\mathbf{B}_9$ | = | $-x_3\,\mathbf{a}_1 + y_3\,\mathbf{a}_2 + (-x_3 + y_3)\,\mathbf{a}_3$ | = | $y_3 a\,\hat{\mathbf{x}} - x_3 a\,\hat{\mathbf{y}}$ | (8$h$) | O I |
| $\mathbf{B}_{10}$ | = | $y_4\,\mathbf{a}_1 + x_4\,\mathbf{a}_2 + (x_4 + y_4)\,\mathbf{a}_3$ | = | $x_4 a\,\hat{\mathbf{x}} + y_4 a\,\hat{\mathbf{y}}$ | (8$h$) | O II |
| $\mathbf{B}_{11}$ | = | $-y_4\,\mathbf{a}_1 - x_4\,\mathbf{a}_2 + (-x_4 - y_4)\,\mathbf{a}_3$ | = | $-x_4 a\,\hat{\mathbf{x}} - y_4 a\,\hat{\mathbf{y}}$ | (8$h$) | O II |
| $\mathbf{B}_{12}$ | = | $x_4\,\mathbf{a}_1 - y_4\,\mathbf{a}_2 + (x_4 - y_4)\,\mathbf{a}_3$ | = | $-y_4 a\,\hat{\mathbf{x}} + x_4 a\,\hat{\mathbf{y}}$ | (8$h$) | O II |
| $\mathbf{B}_{13}$ | = | $-x_4\,\mathbf{a}_1 + y_4\,\mathbf{a}_2 + (-x_4 + y_4)\,\mathbf{a}_3$ | = | $y_4 a\,\hat{\mathbf{x}} - x_4 a\,\hat{\mathbf{y}}$ | (8$h$) | O II |
| $\mathbf{B}_{14}$ | = | $y_5\,\mathbf{a}_1 + x_5\,\mathbf{a}_2 + (x_5 + y_5)\,\mathbf{a}_3$ | = | $x_5 a\,\hat{\mathbf{x}} + y_5 a\,\hat{\mathbf{y}}$ | (8$h$) | Si I |
| $\mathbf{B}_{15}$ | = | $-y_5\,\mathbf{a}_1 - x_5\,\mathbf{a}_2 + (-x_5 - y_5)\,\mathbf{a}_3$ | = | $-x_5 a\,\hat{\mathbf{x}} - y_5 a\,\hat{\mathbf{y}}$ | (8$h$) | Si I |
| $\mathbf{B}_{16}$ | = | $x_5\,\mathbf{a}_1 - y_5\,\mathbf{a}_2 + (x_5 - y_5)\,\mathbf{a}_3$ | = | $-y_5 a\,\hat{\mathbf{x}} + x_5 a\,\hat{\mathbf{y}}$ | (8$h$) | Si I |
| $\mathbf{B}_{17}$ | = | $-x_5\,\mathbf{a}_1 + y_5\,\mathbf{a}_2 + (-x_5 + y_5)\,\mathbf{a}_3$ | = | $y_5 a\,\hat{\mathbf{x}} - x_5 a\,\hat{\mathbf{y}}$ | (8$h$) | Si I |
| $\mathbf{B}_{18}$ | = | $(y_6 + z_6)\,\mathbf{a}_1 + (x_6 + z_6)\,\mathbf{a}_2 + (x_6 + y_6)\,\mathbf{a}_3$ | = | $x_6 a\,\hat{\mathbf{x}} + y_6 a\,\hat{\mathbf{y}} + z_6 c\,\hat{\mathbf{z}}$ | (16$i$) | O III |
| $\mathbf{B}_{19}$ | = | $(-y_6 + z_6)\,\mathbf{a}_1 + (-x_6 + z_6)\,\mathbf{a}_2 + (-x_6 - y_6)\,\mathbf{a}_3$ | = | $-x_6 a\,\hat{\mathbf{x}} - y_6 a\,\hat{\mathbf{y}} + z_6 c\,\hat{\mathbf{z}}$ | (16$i$) | O III |
| $\mathbf{B}_{20}$ | = | $(x_6 + z_6)\,\mathbf{a}_1 + (-y_6 + z_6)\,\mathbf{a}_2 + (x_6 - y_6)\,\mathbf{a}_3$ | = | $-y_6 a\,\hat{\mathbf{x}} + x_6 a\,\hat{\mathbf{y}} + z_6 c\,\hat{\mathbf{z}}$ | (16$i$) | O III |
| $\mathbf{B}_{21}$ | = | $(-x_6 + z_6)\,\mathbf{a}_1 + (y_6 + z_6)\,\mathbf{a}_2 + (-x_6 + y_6)\,\mathbf{a}_3$ | = | $y_6 a\,\hat{\mathbf{x}} - x_6 a\,\hat{\mathbf{y}} + z_6 c\,\hat{\mathbf{z}}$ | (16$i$) | O III |
| $\mathbf{B}_{22}$ | = | $(-y_6 - z_6)\,\mathbf{a}_1 + (-x_6 - z_6)\,\mathbf{a}_2 + (-x_6 - y_6)\,\mathbf{a}_3$ | = | $-x_6 a\,\hat{\mathbf{x}} - y_6 a\,\hat{\mathbf{y}} - z_6 c\,\hat{\mathbf{z}}$ | (16$i$) | O III |
| $\mathbf{B}_{23}$ | = | $(y_6 - z_6)\,\mathbf{a}_1 + (x_6 - z_6)\,\mathbf{a}_2 + (x_6 + y_6)\,\mathbf{a}_3$ | = | $x_6 a\,\hat{\mathbf{x}} + y_6 a\,\hat{\mathbf{y}} - z_6 c\,\hat{\mathbf{z}}$ | (16$i$) | O III |
| $\mathbf{B}_{24}$ | = | $(-x_6 - z_6)\,\mathbf{a}_1 + (y_6 - z_6)\,\mathbf{a}_2 + (-x_6 + y_6)\,\mathbf{a}_3$ | = | $y_6 a\,\hat{\mathbf{x}} - x_6 a\,\hat{\mathbf{y}} - z_6 c\,\hat{\mathbf{z}}$ | (16$i$) | O III |
| $\mathbf{B}_{25}$ | = | $(x_6 - z_6)\,\mathbf{a}_1 + (-y_6 - z_6)\,\mathbf{a}_2 + (x_6 - y_6)\,\mathbf{a}_3$ | = | $-y_6 a\,\hat{\mathbf{x}} + x_6 a\,\hat{\mathbf{y}} - z_6 c\,\hat{\mathbf{z}}$ | (16$i$) | O III |
| $\mathbf{B}_{26}$ | = | $(y_7 + z_7)\,\mathbf{a}_1 + (x_7 + z_7)\,\mathbf{a}_2 + (x_7 + y_7)\,\mathbf{a}_3$ | = | $x_7 a\,\hat{\mathbf{x}} + y_7 a\,\hat{\mathbf{y}} + z_7 c\,\hat{\mathbf{z}}$ | (16$i$) | O IV |
| $\mathbf{B}_{27}$ | = | $(-y_7 + z_7)\,\mathbf{a}_1 + (-x_7 + z_7)\,\mathbf{a}_2 + (-x_7 - y_7)\,\mathbf{a}_3$ | = | $-x_7 a\,\hat{\mathbf{x}} - y_7 a\,\hat{\mathbf{y}} + z_7 c\,\hat{\mathbf{z}}$ | (16$i$) | O IV |
| $\mathbf{B}_{28}$ | = | $(x_7 + z_7)\,\mathbf{a}_1 + (-y_7 + z_7)\,\mathbf{a}_2 + (x_7 - y_7)\,\mathbf{a}_3$ | = | $-y_7 a\,\hat{\mathbf{x}} + x_7 a\,\hat{\mathbf{y}} + z_7 c\,\hat{\mathbf{z}}$ | (16*i*) | O IV |
| $\mathbf{B}_{29}$ | = | $(-x_7 + z_7)\,\mathbf{a}_1 + (y_7 + z_7)\,\mathbf{a}_2 + (-x_7 + y_7)\,\mathbf{a}_3$ | = | $y_7 a\,\hat{\mathbf{x}} - x_7 a\,\hat{\mathbf{y}} + z_7 c\,\hat{\mathbf{z}}$ | (16*i*) | O IV |
| $\mathbf{B}_{30}$ | = | $(-y_7 - z_7)\,\mathbf{a}_1 + (-x_7 - z_7)\,\mathbf{a}_2 + (-x_7 - y_7)\,\mathbf{a}_3$ | = | $-x_7 a\,\hat{\mathbf{x}} - y_7 a\,\hat{\mathbf{y}} - z_7 c\,\hat{\mathbf{z}}$ | (16*i*) | O IV |
| $\mathbf{B}_{31}$ | = | $(y_7 - z_7)\,\mathbf{a}_1 + (x_7 - z_7)\,\mathbf{a}_2 + (x_7 + y_7)\,\mathbf{a}_3$ | = | $x_7 a\,\hat{\mathbf{x}} + y_7 a\,\hat{\mathbf{y}} - z_7 c\,\hat{\mathbf{z}}$ | (16*i*) | O IV |
| $\mathbf{B}_{32}$ | = | $(-x_7 - z_7)\,\mathbf{a}_1 + (y_7 - z_7)\,\mathbf{a}_2 + (-x_7 + y_7)\,\mathbf{a}_3$ | = | $y_7 a\,\hat{\mathbf{x}} - x_7 a\,\hat{\mathbf{y}} - z_7 c\,\hat{\mathbf{z}}$ | (16*i*) | O IV |
| $\mathbf{B}_{33}$ | = | $(x_7 - z_7)\,\mathbf{a}_1 + (-y_7 - z_7)\,\mathbf{a}_2 + (x_7 - y_7)\,\mathbf{a}_3$ | = | $-y_7 a\,\hat{\mathbf{x}} + x_7 a\,\hat{\mathbf{y}} - z_7 c\,\hat{\mathbf{z}}$ | (16*i*) | O IV |
| $\mathbf{B}_{34}$ | = | $(y_8 + z_8)\,\mathbf{a}_1 + (x_8 + z_8)\,\mathbf{a}_2 + (x_8 + y_8)\,\mathbf{a}_3$ | = | $x_8 a\,\hat{\mathbf{x}} + y_8 a\,\hat{\mathbf{y}} + z_8 c\,\hat{\mathbf{z}}$ | (16*i*) | Si II |
| $\mathbf{B}_{35}$ | = | $(-y_8 + z_8)\,\mathbf{a}_1 + (-x_8 + z_8)\,\mathbf{a}_2 + (-x_8 - y_8)\,\mathbf{a}_3$ | = | $-x_8 a\,\hat{\mathbf{x}} - y_8 a\,\hat{\mathbf{y}} + z_8 c\,\hat{\mathbf{z}}$ | (16*i*) | Si II |
| $\mathbf{B}_{36}$ | = | $(x_8 + z_8)\,\mathbf{a}_1 + (-y_8 + z_8)\,\mathbf{a}_2 + (x_8 - y_8)\,\mathbf{a}_3$ | = | $-y_8 a\,\hat{\mathbf{x}} + x_8 a\,\hat{\mathbf{y}} + z_8 c\,\hat{\mathbf{z}}$ | (16*i*) | Si II |
| $\mathbf{B}_{37}$ | = | $(-x_8 + z_8)\,\mathbf{a}_1 + (y_8 + z_8)\,\mathbf{a}_2 + (-x_8 + y_8)\,\mathbf{a}_3$ | = | $y_8 a\,\hat{\mathbf{x}} - x_8 a\,\hat{\mathbf{y}} + z_8 c\,\hat{\mathbf{z}}$ | (16*i*) | Si II |
| $\mathbf{B}_{38}$ | = | $(-y_8 - z_8)\,\mathbf{a}_1 + (-x_8 - z_8)\,\mathbf{a}_2 + (-x_8 - y_8)\,\mathbf{a}_3$ | = | $-x_8 a\,\hat{\mathbf{x}} - y_8 a\,\hat{\mathbf{y}} - z_8 c\,\hat{\mathbf{z}}$ | (16*i*) | Si II |
| $\mathbf{B}_{39}$ | = | $(y_8 - z_8)\,\mathbf{a}_1 + (x_8 - z_8)\,\mathbf{a}_2 + (x_8 + y_8)\,\mathbf{a}_3$ | = | $x_8 a\,\hat{\mathbf{x}} + y_8 a\,\hat{\mathbf{y}} - z_8 c\,\hat{\mathbf{z}}$ | (16*i*) | Si II |
| $\mathbf{B}_{40}$ | = | $(-x_8 - z_8)\,\mathbf{a}_1 + (y_8 - z_8)\,\mathbf{a}_2 + (-x_8 + y_8)\,\mathbf{a}_3$ | = | $y_8 a\,\hat{\mathbf{x}} - x_8 a\,\hat{\mathbf{y}} - z_8 c\,\hat{\mathbf{z}}$ | (16*i*) | Si II |
| $\mathbf{B}_{41}$ | = | $(x_8 - z_8)\,\mathbf{a}_1 + (-y_8 - z_8)\,\mathbf{a}_2 + (x_8 - y_8)\,\mathbf{a}_3$ | = | $-y_8 a\,\hat{\mathbf{x}} + x_8 a\,\hat{\mathbf{y}} - z_8 c\,\hat{\mathbf{z}}$ | (16*i*) | Si II |

**Geometry files:**

- CIF: pp. 1673
- POSCAR: pp. 1674

# $Sr_2NiWO_6$ Structure: AB6C2D_tI20_87_a_eh_d_b

http://aflow.org/prototype-encyclopedia/AB6C2D_tI20_87_a_eh_d_b

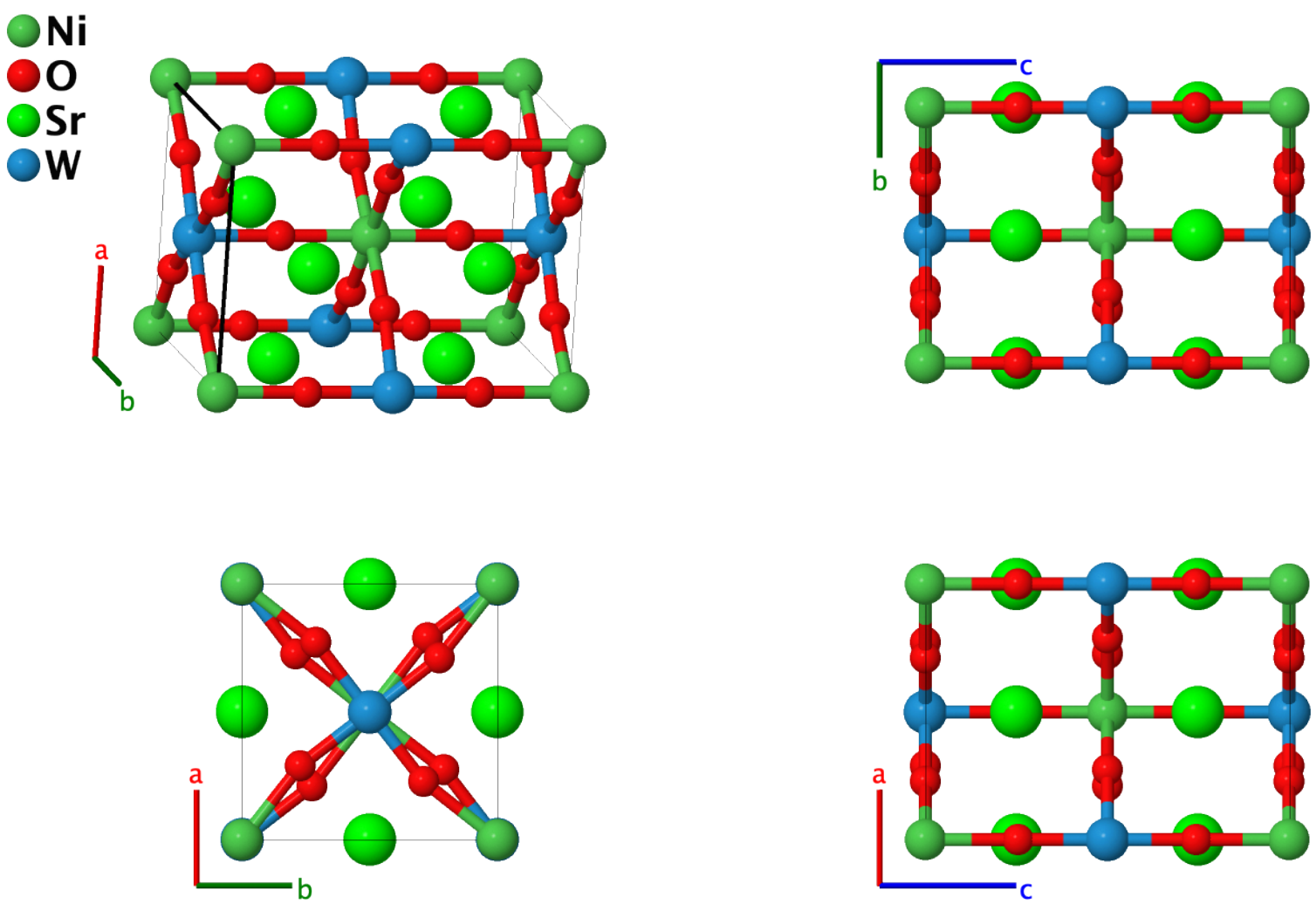

| | | |
|---|---|---|
| **Prototype** | : | $NiO_6Sr_2W$ |
| **AFLOW prototype label** | : | AB6C2D_tI20_87_a_eh_d_b |
| ***Strukturbericht* designation** | : | None |
| **Pearson symbol** | : | tI20 |
| **Space group number** | : | 87 |
| **Space group symbol** | : | $I4/m$ |
| **AFLOW prototype command** | : | `aflow --proto=AB6C2D_tI20_87_a_eh_d_b` `--params=`$a, c/a, z_4, x_5, y_5$ |

### Other compounds with this structure

- $Sr_2MgWO_6$

---

- This double perovskite crystal is the ground state structure of $Sr_2NiWO_6$. Above 300 °C, it transforms into the cubic perovskite $E2_1$ structure. (Iwanga, 2000) places the strontium atoms on the (4$c$) Wyckoff position, but gives the coordinates for the (4$d$) site. The interatomic distances they give with this structure are consistent with the (4d) site.

### Body-centered Tetragonal primitive vectors:

$$\mathbf{a}_1 = -\frac{1}{2}\,a\,\hat{\mathbf{x}} + \frac{1}{2}\,a\,\hat{\mathbf{y}} + \frac{1}{2}\,c\,\hat{\mathbf{z}}$$
$$\mathbf{a}_2 = \frac{1}{2}\,a\,\hat{\mathbf{x}} - \frac{1}{2}\,a\,\hat{\mathbf{y}} + \frac{1}{2}\,c\,\hat{\mathbf{z}}$$
$$\mathbf{a}_3 = \frac{1}{2}\,a\,\hat{\mathbf{x}} + \frac{1}{2}\,a\,\hat{\mathbf{y}} - \frac{1}{2}\,c\,\hat{\mathbf{z}}$$

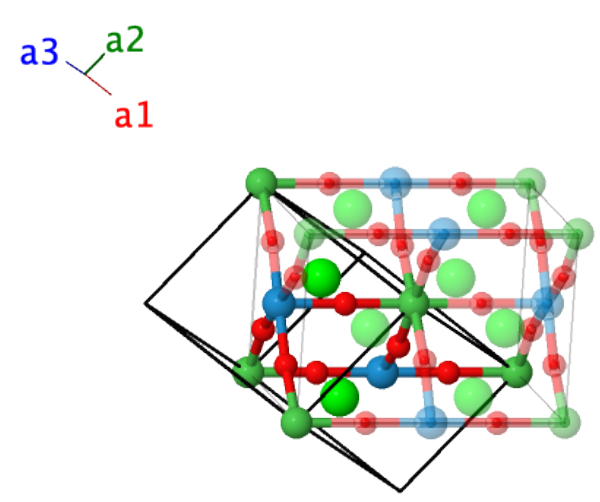

### Basis vectors:

| | | Lattice Coordinates | | Cartesian Coordinates | Wyckoff Position | Atom Type |
|---|---|---|---|---|---|---|
| $\mathbf{B}_1$ | = | $0\,\mathbf{a}_1 + 0\,\mathbf{a}_2 + 0\,\mathbf{a}_3$ | = | $0\,\hat{\mathbf{x}} + 0\,\hat{\mathbf{y}} + 0\,\hat{\mathbf{z}}$ | (2$a$) | Ni |
| $\mathbf{B}_2$ | = | $\frac{1}{2}\,\mathbf{a}_1 + \frac{1}{2}\,\mathbf{a}_2$ | = | $\frac{1}{2}c\,\hat{\mathbf{z}}$ | (2$b$) | W |
| $\mathbf{B}_3$ | = | $\frac{3}{4}\,\mathbf{a}_1 + \frac{1}{4}\,\mathbf{a}_2 + \frac{1}{2}\,\mathbf{a}_3$ | = | $\frac{1}{2}a\,\hat{\mathbf{y}} + \frac{1}{4}c\,\hat{\mathbf{z}}$ | (4$d$) | Sr |
| $\mathbf{B}_4$ | = | $\frac{1}{4}\,\mathbf{a}_1 + \frac{3}{4}\,\mathbf{a}_2 + \frac{1}{2}\,\mathbf{a}_3$ | = | $\frac{1}{2}a\,\hat{\mathbf{x}} + \frac{1}{4}c\,\hat{\mathbf{z}}$ | (4$d$) | Sr |
| $\mathbf{B}_5$ | = | $z_4\,\mathbf{a}_1 + z_4\,\mathbf{a}_2$ | = | $z_4c\,\hat{\mathbf{z}}$ | (4$e$) | O I |
| $\mathbf{B}_6$ | = | $-z_4\,\mathbf{a}_1 - z_4\,\mathbf{a}_2$ | = | $-z_4c\,\hat{\mathbf{z}}$ | (4$e$) | O I |
| $\mathbf{B}_7$ | = | $y_5\,\mathbf{a}_1 + x_5\,\mathbf{a}_2 + (x_5 + y_5)\,\mathbf{a}_3$ | = | $x_5a\,\hat{\mathbf{x}} + y_5a\,\hat{\mathbf{y}}$ | (8$h$) | O II |
| $\mathbf{B}_8$ | = | $-y_5\,\mathbf{a}_1 - x_5\,\mathbf{a}_2 + (-x_5 - y_5)\,\mathbf{a}_3$ | = | $-x_5a\,\hat{\mathbf{x}} - y_5a\,\hat{\mathbf{y}}$ | (8$h$) | O II |
| $\mathbf{B}_9$ | = | $x_5\,\mathbf{a}_1 - y_5\,\mathbf{a}_2 + (x_5 - y_5)\,\mathbf{a}_3$ | = | $-y_5a\,\hat{\mathbf{x}} + x_5a\,\hat{\mathbf{y}}$ | (8$h$) | O II |
| $\mathbf{B}_{10}$ | = | $-x_5\,\mathbf{a}_1 + y_5\,\mathbf{a}_2 + (-x_5 + y_5)\,\mathbf{a}_3$ | = | $y_5a\,\hat{\mathbf{x}} - x_5a\,\hat{\mathbf{y}}$ | (8$h$) | O II |

**Geometry files:**

- CIF: pp. 1674
- POSCAR: pp. 1674

# $Na_4Ge_9O_{20}$ Structure: A9B4C20_tI132_88_a2f_f_5f

http://aflow.org/prototype-encyclopedia/A9B4C20_tI132_88_a2f_f_5f

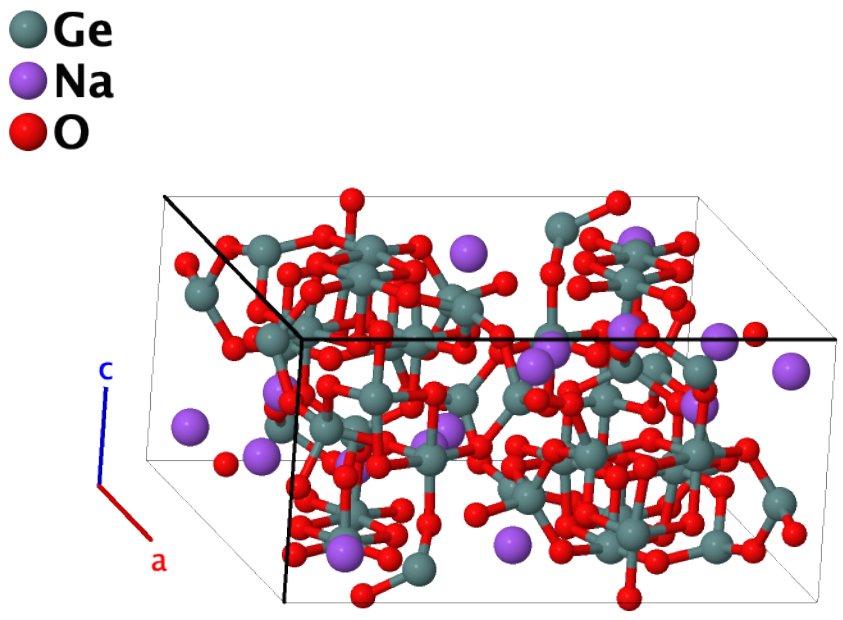

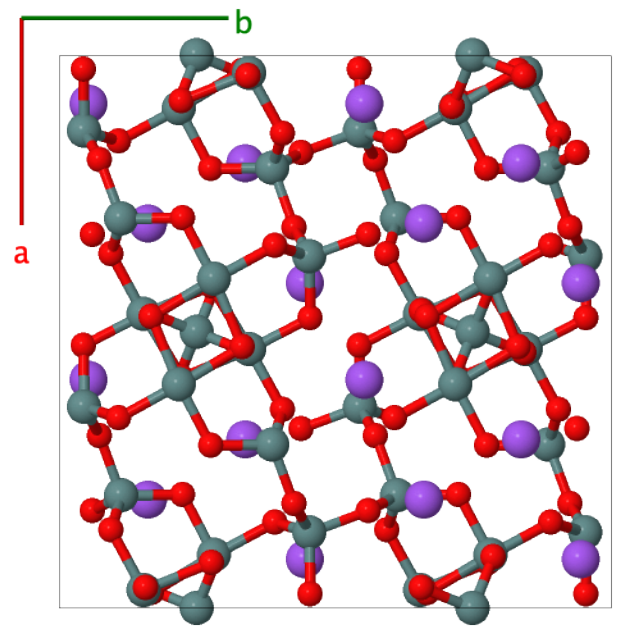

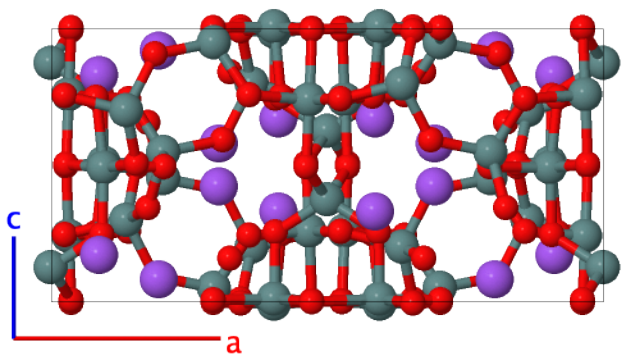

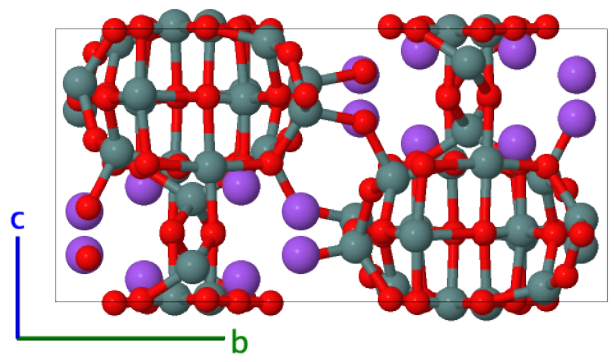

| | | |
|---|---|---|
| **Prototype** | : | $Ge_9Na_4O_{20}$ |
| **AFLOW prototype label** | : | A9B4C20_tI132_88_a2f_f_5f |
| ***Strukturbericht* designation** | : | None |
| **Pearson symbol** | : | tI132 |
| **Space group number** | : | 88 |
| **Space group symbol** | : | $I4_1/a$ |
| **AFLOW prototype command** | : | `aflow --proto=A9B4C20_tI132_88_a2f_f_5f --params=`$a, c/a, x_2, y_2, z_2, x_3, y_3, z_3, x_4, y_4, z_4, x_5, y_5, z_5, x_6, y_6, z_6, x_7, y_7, z_7, x_8, y_8, z_8, x_9, y_9, z_9$ |

**Body-centered Tetragonal primitive vectors:**

$$\mathbf{a}_1 = -\frac{1}{2}\,a\,\hat{\mathbf{x}} + \frac{1}{2}\,a\,\hat{\mathbf{y}} + \frac{1}{2}\,c\,\hat{\mathbf{z}}$$
$$\mathbf{a}_2 = \frac{1}{2}\,a\,\hat{\mathbf{x}} - \frac{1}{2}\,a\,\hat{\mathbf{y}} + \frac{1}{2}\,c\,\hat{\mathbf{z}}$$
$$\mathbf{a}_3 = \frac{1}{2}\,a\,\hat{\mathbf{x}} + \frac{1}{2}\,a\,\hat{\mathbf{y}} - \frac{1}{2}\,c\,\hat{\mathbf{z}}$$

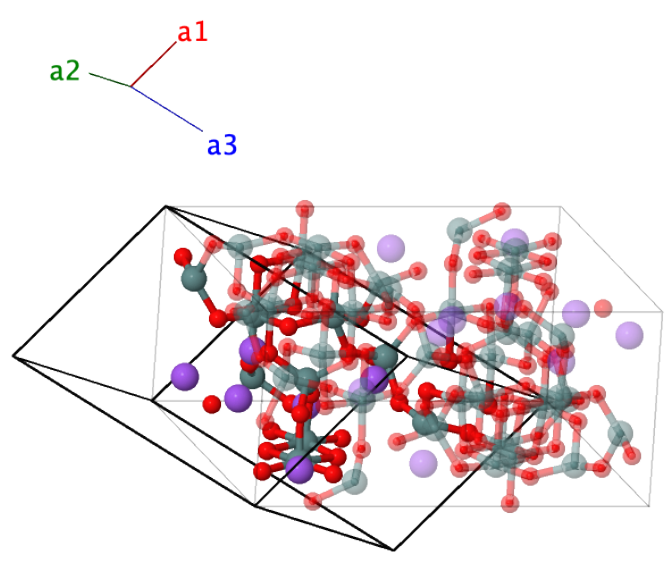

**Basis vectors:**

| | | Lattice Coordinates | | Cartesian Coordinates | Wyckoff Position | Atom Type |
|---|---|---|---|---|---|---|
| $\mathbf{B}_1$ | = | $\frac{3}{8}\,\mathbf{a}_1 + \frac{1}{8}\,\mathbf{a}_2 + \frac{1}{4}\,\mathbf{a}_3$ | = | $\frac{1}{4}a\,\hat{\mathbf{y}} + \frac{1}{8}c\,\hat{\mathbf{z}}$ | (4$a$) | Ge I |
| $\mathbf{B}_2$ | = | $\frac{5}{8}\,\mathbf{a}_1 + \frac{7}{8}\,\mathbf{a}_2 + \frac{3}{4}\,\mathbf{a}_3$ | = | $\frac{1}{2}a\,\hat{\mathbf{x}} + \frac{1}{4}a\,\hat{\mathbf{y}} + \frac{3}{8}c\,\hat{\mathbf{z}}$ | (4$a$) | Ge I |
| $\mathbf{B}_3$ | = | $(y_2 + z_2)\,\mathbf{a}_1 + (x_2 + z_2)\,\mathbf{a}_2 + (x_2 + y_2)\,\mathbf{a}_3$ | = | $x_2a\,\hat{\mathbf{x}} + y_2a\,\hat{\mathbf{y}} + z_2c\,\hat{\mathbf{z}}$ | (16$f$) | Ge II |
| $\mathbf{B}_4$ | = | $\left(\frac{1}{2} - y_2 + z_2\right)\mathbf{a}_1 + (-x_2 + z_2)\,\mathbf{a}_2 + \left(\frac{1}{2} - x_2 - y_2\right)\mathbf{a}_3$ | = | $-x_2a\,\hat{\mathbf{x}} + \left(\frac{1}{2} - y_2\right)a\,\hat{\mathbf{y}} + z_2c\,\hat{\mathbf{z}}$ | (16$f$) | Ge II |
| $\mathbf{B}_5$ | = | $\left(\frac{1}{2} + x_2 + z_2\right)\mathbf{a}_1 + (-y_2 + z_2)\,\mathbf{a}_2 + (x_2 - y_2)\,\mathbf{a}_3$ | = | $\left(\frac{3}{4} - y_2\right)a\,\hat{\mathbf{x}} + \left(\frac{1}{4} + x_2\right)a\,\hat{\mathbf{y}} + \left(\frac{1}{4} + z_2\right)c\,\hat{\mathbf{z}}$ | (16$f$) | Ge II |
| $\mathbf{B}_6$ | = | $\left(\frac{1}{2} - x_2 + z_2\right)\mathbf{a}_1 + \left(\frac{1}{2} + y_2 + z_2\right)\mathbf{a}_2 + \left(\frac{1}{2} - x_2 + y_2\right)\mathbf{a}_3$ | = | $\left(\frac{1}{4} + y_2\right)a\,\hat{\mathbf{x}} + \left(\frac{1}{4} - x_2\right)a\,\hat{\mathbf{y}} + \left(\frac{1}{4} + z_2\right)c\,\hat{\mathbf{z}}$ | (16$f$) | Ge II |
| $\mathbf{B}_7$ | = | $(-y_2 - z_2)\,\mathbf{a}_1 + (-x_2 - z_2)\,\mathbf{a}_2 + (-x_2 - y_2)\,\mathbf{a}_3$ | = | $-x_2a\,\hat{\mathbf{x}} - y_2a\,\hat{\mathbf{y}} - z_2c\,\hat{\mathbf{z}}$ | (16$f$) | Ge II |
| $\mathbf{B}_8$ | = | $\left(\frac{1}{2} + y_2 - z_2\right)\mathbf{a}_1 + (x_2 - z_2)\,\mathbf{a}_2 + \left(\frac{1}{2} + x_2 + y_2\right)\mathbf{a}_3$ | = | $x_2a\,\hat{\mathbf{x}} + \left(\frac{1}{2} + y_2\right)a\,\hat{\mathbf{y}} - z_2c\,\hat{\mathbf{z}}$ | (16$f$) | Ge II |
| $\mathbf{B}_9$ | = | $\left(\frac{1}{2} - x_2 - z_2\right)\mathbf{a}_1 + (y_2 - z_2)\,\mathbf{a}_2 + (-x_2 + y_2)\,\mathbf{a}_3$ | = | $\left(-\frac{1}{4} + y_2\right)a\,\hat{\mathbf{x}} + \left(\frac{1}{4} - x_2\right)a\,\hat{\mathbf{y}} + \left(\frac{1}{4} - z_2\right)c\,\hat{\mathbf{z}}$ | (16$f$) | Ge II |
| $\mathbf{B}_{10}$ | = | $\left(\frac{1}{2} + x_2 - z_2\right)\mathbf{a}_1 + \left(\frac{1}{2} - y_2 - z_2\right)\mathbf{a}_2 + \left(\frac{1}{2} + x_2 - y_2\right)\mathbf{a}_3$ | = | $\left(\frac{1}{4} - y_2\right)a\,\hat{\mathbf{x}} + \left(\frac{1}{4} + x_2\right)a\,\hat{\mathbf{y}} + \left(\frac{1}{4} - z_2\right)c\,\hat{\mathbf{z}}$ | (16$f$) | Ge II |
| $\mathbf{B}_{11}$ | = | $(y_3 + z_3)\,\mathbf{a}_1 + (x_3 + z_3)\,\mathbf{a}_2 + (x_3 + y_3)\,\mathbf{a}_3$ | = | $x_3a\,\hat{\mathbf{x}} + y_3a\,\hat{\mathbf{y}} + z_3c\,\hat{\mathbf{z}}$ | (16$f$) | Ge III |
| $\mathbf{B}_{12}$ | = | $\left(\frac{1}{2} - y_3 + z_3\right)\mathbf{a}_1 + (-x_3 + z_3)\,\mathbf{a}_2 + \left(\frac{1}{2} - x_3 - y_3\right)\mathbf{a}_3$ | = | $-x_3a\,\hat{\mathbf{x}} + \left(\frac{1}{2} - y_3\right)a\,\hat{\mathbf{y}} + z_3c\,\hat{\mathbf{z}}$ | (16$f$) | Ge III |
| $\mathbf{B}_{13}$ | = | $\left(\frac{1}{2} + x_3 + z_3\right)\mathbf{a}_1 + (-y_3 + z_3)\,\mathbf{a}_2 + (x_3 - y_3)\,\mathbf{a}_3$ | = | $\left(\frac{3}{4} - y_3\right)a\,\hat{\mathbf{x}} + \left(\frac{1}{4} + x_3\right)a\,\hat{\mathbf{y}} + \left(\frac{1}{4} + z_3\right)c\,\hat{\mathbf{z}}$ | (16$f$) | Ge III |
| $\mathbf{B}_{14}$ | = | $\left(\frac{1}{2} - x_3 + z_3\right)\mathbf{a}_1 + \left(\frac{1}{2} + y_3 + z_3\right)\mathbf{a}_2 + \left(\frac{1}{2} - x_3 + y_3\right)\mathbf{a}_3$ | = | $\left(\frac{1}{4} + y_3\right)a\,\hat{\mathbf{x}} + \left(\frac{1}{4} - x_3\right)a\,\hat{\mathbf{y}} + \left(\frac{1}{4} + z_3\right)c\,\hat{\mathbf{z}}$ | (16$f$) | Ge III |
| $\mathbf{B}_{15}$ | = | $(-y_3 - z_3)\,\mathbf{a}_1 + (-x_3 - z_3)\,\mathbf{a}_2 + (-x_3 - y_3)\,\mathbf{a}_3$ | = | $-x_3a\,\hat{\mathbf{x}} - y_3a\,\hat{\mathbf{y}} - z_3c\,\hat{\mathbf{z}}$ | (16$f$) | Ge III |
| $\mathbf{B}_{16}$ | = | $\left(\frac{1}{2} + y_3 - z_3\right)\mathbf{a}_1 + (x_3 - z_3)\,\mathbf{a}_2 + \left(\frac{1}{2} + x_3 + y_3\right)\mathbf{a}_3$ | = | $x_3a\,\hat{\mathbf{x}} + \left(\frac{1}{2} + y_3\right)a\,\hat{\mathbf{y}} - z_3c\,\hat{\mathbf{z}}$ | (16$f$) | Ge III |
| $\mathbf{B}_{17}$ | = | $\left(\frac{1}{2} - x_3 - z_3\right)\mathbf{a}_1 + (y_3 - z_3)\,\mathbf{a}_2 + (-x_3 + y_3)\,\mathbf{a}_3$ | = | $\left(-\frac{1}{4} + y_3\right)a\,\hat{\mathbf{x}} + \left(\frac{1}{4} - x_3\right)a\,\hat{\mathbf{y}} + \left(\frac{1}{4} - z_3\right)c\,\hat{\mathbf{z}}$ | (16$f$) | Ge III |
| $\mathbf{B}_{18}$ | = | $\left(\frac{1}{2} + x_3 - z_3\right)\mathbf{a}_1 + \left(\frac{1}{2} - y_3 - z_3\right)\mathbf{a}_2 + \left(\frac{1}{2} + x_3 - y_3\right)\mathbf{a}_3$ | = | $\left(\frac{1}{4} - y_3\right)a\,\hat{\mathbf{x}} + \left(\frac{1}{4} + x_3\right)a\,\hat{\mathbf{y}} + \left(\frac{1}{4} - z_3\right)c\,\hat{\mathbf{z}}$ | (16$f$) | Ge III |
| $\mathbf{B}_{19}$ | = | $(y_4 + z_4)\,\mathbf{a}_1 + (x_4 + z_4)\,\mathbf{a}_2 + (x_4 + y_4)\,\mathbf{a}_3$ | = | $x_4a\,\hat{\mathbf{x}} + y_4a\,\hat{\mathbf{y}} + z_4c\,\hat{\mathbf{z}}$ | (16$f$) | Na |
| $\mathbf{B}_{20}$ | = | $\left(\frac{1}{2} - y_4 + z_4\right)\mathbf{a}_1 + (-x_4 + z_4)\,\mathbf{a}_2 + \left(\frac{1}{2} - x_4 - y_4\right)\mathbf{a}_3$ | = | $-x_4a\,\hat{\mathbf{x}} + \left(\frac{1}{2} - y_4\right)a\,\hat{\mathbf{y}} + z_4c\,\hat{\mathbf{z}}$ | (16$f$) | Na |
| $\mathbf{B}_{21}$ | = | $\left(\frac{1}{2} + x_4 + z_4\right)\mathbf{a}_1 + (-y_4 + z_4)\,\mathbf{a}_2 + (x_4 - y_4)\,\mathbf{a}_3$ | = | $\left(\frac{3}{4} - y_4\right)a\,\hat{\mathbf{x}} + \left(\frac{1}{4} + x_4\right)a\,\hat{\mathbf{y}} + \left(\frac{1}{4} + z_4\right)c\,\hat{\mathbf{z}}$ | (16$f$) | Na |
| $\mathbf{B}_{22}$ | = | $\left(\frac{1}{2} - x_4 + z_4\right) \mathbf{a}_1 + \left(\frac{1}{2} + y_4 + z_4\right) \mathbf{a}_2 + \left(\frac{1}{2} - x_4 + y_4\right) \mathbf{a}_3$ | = | $\left(\frac{1}{4} + y_4\right) a \, \hat{\mathbf{x}} + \left(\frac{1}{4} - x_4\right) a \, \hat{\mathbf{y}} + \left(\frac{1}{4} + z_4\right) c \, \hat{\mathbf{z}}$ | $(16f)$ | Na |
| $\mathbf{B}_{23}$ | = | $(-y_4 - z_4) \, \mathbf{a}_1 + (-x_4 - z_4) \, \mathbf{a}_2 + (-x_4 - y_4) \, \mathbf{a}_3$ | = | $-x_4 a \, \hat{\mathbf{x}} - y_4 a \, \hat{\mathbf{y}} - z_4 c \, \hat{\mathbf{z}}$ | $(16f)$ | Na |
| $\mathbf{B}_{24}$ | = | $\left(\frac{1}{2} + y_4 - z_4\right) \mathbf{a}_1 + (x_4 - z_4) \, \mathbf{a}_2 + \left(\frac{1}{2} + x_4 + y_4\right) \mathbf{a}_3$ | = | $x_4 a \, \hat{\mathbf{x}} + \left(\frac{1}{2} + y_4\right) a \, \hat{\mathbf{y}} - z_4 c \, \hat{\mathbf{z}}$ | $(16f)$ | Na |
| $\mathbf{B}_{25}$ | = | $\left(\frac{1}{2} - x_4 - z_4\right) \mathbf{a}_1 + (y_4 - z_4) \, \mathbf{a}_2 + (-x_4 + y_4) \, \mathbf{a}_3$ | = | $\left(-\frac{1}{4} + y_4\right) a \, \hat{\mathbf{x}} + \left(\frac{1}{4} - x_4\right) a \, \hat{\mathbf{y}} + \left(\frac{1}{4} - z_4\right) c \, \hat{\mathbf{z}}$ | $(16f)$ | Na |
| $\mathbf{B}_{26}$ | = | $\left(\frac{1}{2} + x_4 - z_4\right) \mathbf{a}_1 + \left(\frac{1}{2} - y_4 - z_4\right) \mathbf{a}_2 + \left(\frac{1}{2} + x_4 - y_4\right) \mathbf{a}_3$ | = | $\left(\frac{1}{4} - y_4\right) a \, \hat{\mathbf{x}} + \left(\frac{1}{4} + x_4\right) a \, \hat{\mathbf{y}} + \left(\frac{1}{4} - z_4\right) c \, \hat{\mathbf{z}}$ | $(16f)$ | Na |
| $\mathbf{B}_{27}$ | = | $(y_5 + z_5) \, \mathbf{a}_1 + (x_5 + z_5) \, \mathbf{a}_2 + (x_5 + y_5) \, \mathbf{a}_3$ | = | $x_5 a \, \hat{\mathbf{x}} + y_5 a \, \hat{\mathbf{y}} + z_5 c \, \hat{\mathbf{z}}$ | $(16f)$ | O I |
| $\mathbf{B}_{28}$ | = | $\left(\frac{1}{2} - y_5 + z_5\right) \mathbf{a}_1 + (-x_5 + z_5) \, \mathbf{a}_2 + \left(\frac{1}{2} - x_5 - y_5\right) \mathbf{a}_3$ | = | $-x_5 a \, \hat{\mathbf{x}} + \left(\frac{1}{2} - y_5\right) a \, \hat{\mathbf{y}} + z_5 c \, \hat{\mathbf{z}}$ | $(16f)$ | O I |
| $\mathbf{B}_{29}$ | = | $\left(\frac{1}{2} + x_5 + z_5\right) \mathbf{a}_1 + (-y_5 + z_5) \, \mathbf{a}_2 + (x_5 - y_5) \, \mathbf{a}_3$ | = | $\left(\frac{3}{4} - y_5\right) a \, \hat{\mathbf{x}} + \left(\frac{1}{4} + x_5\right) a \, \hat{\mathbf{y}} + \left(\frac{1}{4} + z_5\right) c \, \hat{\mathbf{z}}$ | $(16f)$ | O I |
| $\mathbf{B}_{30}$ | = | $\left(\frac{1}{2} - x_5 + z_5\right) \mathbf{a}_1 + \left(\frac{1}{2} + y_5 + z_5\right) \mathbf{a}_2 + \left(\frac{1}{2} - x_5 + y_5\right) \mathbf{a}_3$ | = | $\left(\frac{1}{4} + y_5\right) a \, \hat{\mathbf{x}} + \left(\frac{1}{4} - x_5\right) a \, \hat{\mathbf{y}} + \left(\frac{1}{4} + z_5\right) c \, \hat{\mathbf{z}}$ | $(16f)$ | O I |
| $\mathbf{B}_{31}$ | = | $(-y_5 - z_5) \, \mathbf{a}_1 + (-x_5 - z_5) \, \mathbf{a}_2 + (-x_5 - y_5) \, \mathbf{a}_3$ | = | $-x_5 a \, \hat{\mathbf{x}} - y_5 a \, \hat{\mathbf{y}} - z_5 c \, \hat{\mathbf{z}}$ | $(16f)$ | O I |
| $\mathbf{B}_{32}$ | = | $\left(\frac{1}{2} + y_5 - z_5\right) \mathbf{a}_1 + (x_5 - z_5) \, \mathbf{a}_2 + \left(\frac{1}{2} + x_5 + y_5\right) \mathbf{a}_3$ | = | $x_5 a \, \hat{\mathbf{x}} + \left(\frac{1}{2} + y_5\right) a \, \hat{\mathbf{y}} - z_5 c \, \hat{\mathbf{z}}$ | $(16f)$ | O I |
| $\mathbf{B}_{33}$ | = | $\left(\frac{1}{2} - x_5 - z_5\right) \mathbf{a}_1 + (y_5 - z_5) \, \mathbf{a}_2 + (-x_5 + y_5) \, \mathbf{a}_3$ | = | $\left(-\frac{1}{4} + y_5\right) a \, \hat{\mathbf{x}} + \left(\frac{1}{4} - x_5\right) a \, \hat{\mathbf{y}} + \left(\frac{1}{4} - z_5\right) c \, \hat{\mathbf{z}}$ | $(16f)$ | O I |
| $\mathbf{B}_{34}$ | = | $\left(\frac{1}{2} + x_5 - z_5\right) \mathbf{a}_1 + \left(\frac{1}{2} - y_5 - z_5\right) \mathbf{a}_2 + \left(\frac{1}{2} + x_5 - y_5\right) \mathbf{a}_3$ | = | $\left(\frac{1}{4} - y_5\right) a \, \hat{\mathbf{x}} + \left(\frac{1}{4} + x_5\right) a \, \hat{\mathbf{y}} + \left(\frac{1}{4} - z_5\right) c \, \hat{\mathbf{z}}$ | $(16f)$ | O I |
| $\mathbf{B}_{35}$ | = | $(y_6 + z_6) \, \mathbf{a}_1 + (x_6 + z_6) \, \mathbf{a}_2 + (x_6 + y_6) \, \mathbf{a}_3$ | = | $x_6 a \, \hat{\mathbf{x}} + y_6 a \, \hat{\mathbf{y}} + z_6 c \, \hat{\mathbf{z}}$ | $(16f)$ | O II |
| $\mathbf{B}_{36}$ | = | $\left(\frac{1}{2} - y_6 + z_6\right) \mathbf{a}_1 + (-x_6 + z_6) \, \mathbf{a}_2 + \left(\frac{1}{2} - x_6 - y_6\right) \mathbf{a}_3$ | = | $-x_6 a \, \hat{\mathbf{x}} + \left(\frac{1}{2} - y_6\right) a \, \hat{\mathbf{y}} + z_6 c \, \hat{\mathbf{z}}$ | $(16f)$ | O II |
| $\mathbf{B}_{37}$ | = | $\left(\frac{1}{2} + x_6 + z_6\right) \mathbf{a}_1 + (-y_6 + z_6) \, \mathbf{a}_2 + (x_6 - y_6) \, \mathbf{a}_3$ | = | $\left(\frac{3}{4} - y_6\right) a \, \hat{\mathbf{x}} + \left(\frac{1}{4} + x_6\right) a \, \hat{\mathbf{y}} + \left(\frac{1}{4} + z_6\right) c \, \hat{\mathbf{z}}$ | $(16f)$ | O II |
| $\mathbf{B}_{38}$ | = | $\left(\frac{1}{2} - x_6 + z_6\right) \mathbf{a}_1 + \left(\frac{1}{2} + y_6 + z_6\right) \mathbf{a}_2 + \left(\frac{1}{2} - x_6 + y_6\right) \mathbf{a}_3$ | = | $\left(\frac{1}{4} + y_6\right) a \, \hat{\mathbf{x}} + \left(\frac{1}{4} - x_6\right) a \, \hat{\mathbf{y}} + \left(\frac{1}{4} + z_6\right) c \, \hat{\mathbf{z}}$ | $(16f)$ | O II |
| $\mathbf{B}_{39}$ | = | $(-y_6 - z_6) \, \mathbf{a}_1 + (-x_6 - z_6) \, \mathbf{a}_2 + (-x_6 - y_6) \, \mathbf{a}_3$ | = | $-x_6 a \, \hat{\mathbf{x}} - y_6 a \, \hat{\mathbf{y}} - z_6 c \, \hat{\mathbf{z}}$ | $(16f)$ | O II |
| $\mathbf{B}_{40}$ | = | $\left(\frac{1}{2} + y_6 - z_6\right) \mathbf{a}_1 + (x_6 - z_6) \, \mathbf{a}_2 + \left(\frac{1}{2} + x_6 + y_6\right) \mathbf{a}_3$ | = | $x_6 a \, \hat{\mathbf{x}} + \left(\frac{1}{2} + y_6\right) a \, \hat{\mathbf{y}} - z_6 c \, \hat{\mathbf{z}}$ | $(16f)$ | O II |
| $\mathbf{B}_{41}$ | = | $\left(\frac{1}{2} - x_6 - z_6\right) \mathbf{a}_1 + (y_6 - z_6) \, \mathbf{a}_2 + (-x_6 + y_6) \, \mathbf{a}_3$ | = | $\left(-\frac{1}{4} + y_6\right) a \, \hat{\mathbf{x}} + \left(\frac{1}{4} - x_6\right) a \, \hat{\mathbf{y}} + \left(\frac{1}{4} - z_6\right) c \, \hat{\mathbf{z}}$ | $(16f)$ | O II |
| $\mathbf{B}_{42}$ | = | $\left(\frac{1}{2} + x_6 - z_6\right) \mathbf{a}_1 + \left(\frac{1}{2} - y_6 - z_6\right) \mathbf{a}_2 + \left(\frac{1}{2} + x_6 - y_6\right) \mathbf{a}_3$ | = | $\left(\frac{1}{4} - y_6\right) a \, \hat{\mathbf{x}} + \left(\frac{1}{4} + x_6\right) a \, \hat{\mathbf{y}} + \left(\frac{1}{4} - z_6\right) c \, \hat{\mathbf{z}}$ | $(16f)$ | O II |
| $\mathbf{B}_{43}$ | = | $(y_7 + z_7)\,\mathbf{a}_1 + (x_7 + z_7)\,\mathbf{a}_2 + (x_7 + y_7)\,\mathbf{a}_3$ | = | $x_7 a\,\hat{\mathbf{x}} + y_7 a\,\hat{\mathbf{y}} + z_7 c\,\hat{\mathbf{z}}$ | $(16f)$ | O III |
| $\mathbf{B}_{44}$ | = | $\left(\frac{1}{2} - y_7 + z_7\right)\mathbf{a}_1 + (-x_7 + z_7)\,\mathbf{a}_2 + \left(\frac{1}{2} - x_7 - y_7\right)\mathbf{a}_3$ | = | $-x_7 a\,\hat{\mathbf{x}} + \left(\frac{1}{2} - y_7\right) a\,\hat{\mathbf{y}} + z_7 c\,\hat{\mathbf{z}}$ | $(16f)$ | O III |
| $\mathbf{B}_{45}$ | = | $\left(\frac{1}{2} + x_7 + z_7\right)\mathbf{a}_1 + (-y_7 + z_7)\,\mathbf{a}_2 + (x_7 - y_7)\,\mathbf{a}_3$ | = | $\left(\frac{3}{4} - y_7\right) a\,\hat{\mathbf{x}} + \left(\frac{1}{4} + x_7\right) a\,\hat{\mathbf{y}} + \left(\frac{1}{4} + z_7\right) c\,\hat{\mathbf{z}}$ | $(16f)$ | O III |
| $\mathbf{B}_{46}$ | = | $\left(\frac{1}{2} - x_7 + z_7\right)\mathbf{a}_1 + \left(\frac{1}{2} + y_7 + z_7\right)\mathbf{a}_2 + \left(\frac{1}{2} - x_7 + y_7\right)\mathbf{a}_3$ | = | $\left(\frac{1}{4} + y_7\right) a\,\hat{\mathbf{x}} + \left(\frac{1}{4} - x_7\right) a\,\hat{\mathbf{y}} + \left(\frac{1}{4} + z_7\right) c\,\hat{\mathbf{z}}$ | $(16f)$ | O III |
| $\mathbf{B}_{47}$ | = | $(-y_7 - z_7)\,\mathbf{a}_1 + (-x_7 - z_7)\,\mathbf{a}_2 + (-x_7 - y_7)\,\mathbf{a}_3$ | = | $-x_7 a\,\hat{\mathbf{x}} - y_7 a\,\hat{\mathbf{y}} - z_7 c\,\hat{\mathbf{z}}$ | $(16f)$ | O III |
| $\mathbf{B}_{48}$ | = | $\left(\frac{1}{2} + y_7 - z_7\right)\mathbf{a}_1 + (x_7 - z_7)\,\mathbf{a}_2 + \left(\frac{1}{2} + x_7 + y_7\right)\mathbf{a}_3$ | = | $x_7 a\,\hat{\mathbf{x}} + \left(\frac{1}{2} + y_7\right) a\,\hat{\mathbf{y}} - z_7 c\,\hat{\mathbf{z}}$ | $(16f)$ | O III |
| $\mathbf{B}_{49}$ | = | $\left(\frac{1}{2} - x_7 - z_7\right)\mathbf{a}_1 + (y_7 - z_7)\,\mathbf{a}_2 + (-x_7 + y_7)\,\mathbf{a}_3$ | = | $\left(-\frac{1}{4} + y_7\right) a\,\hat{\mathbf{x}} + \left(\frac{1}{4} - x_7\right) a\,\hat{\mathbf{y}} + \left(\frac{1}{4} - z_7\right) c\,\hat{\mathbf{z}}$ | $(16f)$ | O III |
| $\mathbf{B}_{50}$ | = | $\left(\frac{1}{2} + x_7 - z_7\right)\mathbf{a}_1 + \left(\frac{1}{2} - y_7 - z_7\right)\mathbf{a}_2 + \left(\frac{1}{2} + x_7 - y_7\right)\mathbf{a}_3$ | = | $\left(\frac{1}{4} - y_7\right) a\,\hat{\mathbf{x}} + \left(\frac{1}{4} + x_7\right) a\,\hat{\mathbf{y}} + \left(\frac{1}{4} - z_7\right) c\,\hat{\mathbf{z}}$ | $(16f)$ | O III |
| $\mathbf{B}_{51}$ | = | $(y_8 + z_8)\,\mathbf{a}_1 + (x_8 + z_8)\,\mathbf{a}_2 + (x_8 + y_8)\,\mathbf{a}_3$ | = | $x_8 a\,\hat{\mathbf{x}} + y_8 a\,\hat{\mathbf{y}} + z_8 c\,\hat{\mathbf{z}}$ | $(16f)$ | O IV |
| $\mathbf{B}_{52}$ | = | $\left(\frac{1}{2} - y_8 + z_8\right)\mathbf{a}_1 + (-x_8 + z_8)\,\mathbf{a}_2 + \left(\frac{1}{2} - x_8 - y_8\right)\mathbf{a}_3$ | = | $-x_8 a\,\hat{\mathbf{x}} + \left(\frac{1}{2} - y_8\right) a\,\hat{\mathbf{y}} + z_8 c\,\hat{\mathbf{z}}$ | $(16f)$ | O IV |
| $\mathbf{B}_{53}$ | = | $\left(\frac{1}{2} + x_8 + z_8\right)\mathbf{a}_1 + (-y_8 + z_8)\,\mathbf{a}_2 + (x_8 - y_8)\,\mathbf{a}_3$ | = | $\left(\frac{3}{4} - y_8\right) a\,\hat{\mathbf{x}} + \left(\frac{1}{4} + x_8\right) a\,\hat{\mathbf{y}} + \left(\frac{1}{4} + z_8\right) c\,\hat{\mathbf{z}}$ | $(16f)$ | O IV |
| $\mathbf{B}_{54}$ | = | $\left(\frac{1}{2} - x_8 + z_8\right)\mathbf{a}_1 + \left(\frac{1}{2} + y_8 + z_8\right)\mathbf{a}_2 + \left(\frac{1}{2} - x_8 + y_8\right)\mathbf{a}_3$ | = | $\left(\frac{1}{4} + y_8\right) a\,\hat{\mathbf{x}} + \left(\frac{1}{4} - x_8\right) a\,\hat{\mathbf{y}} + \left(\frac{1}{4} + z_8\right) c\,\hat{\mathbf{z}}$ | $(16f)$ | O IV |
| $\mathbf{B}_{55}$ | = | $(-y_8 - z_8)\,\mathbf{a}_1 + (-x_8 - z_8)\,\mathbf{a}_2 + (-x_8 - y_8)\,\mathbf{a}_3$ | = | $-x_8 a\,\hat{\mathbf{x}} - y_8 a\,\hat{\mathbf{y}} - z_8 c\,\hat{\mathbf{z}}$ | $(16f)$ | O IV |
| $\mathbf{B}_{56}$ | = | $\left(\frac{1}{2} + y_8 - z_8\right)\mathbf{a}_1 + (x_8 - z_8)\,\mathbf{a}_2 + \left(\frac{1}{2} + x_8 + y_8\right)\mathbf{a}_3$ | = | $x_8 a\,\hat{\mathbf{x}} + \left(\frac{1}{2} + y_8\right) a\,\hat{\mathbf{y}} - z_8 c\,\hat{\mathbf{z}}$ | $(16f)$ | O IV |
| $\mathbf{B}_{57}$ | = | $\left(\frac{1}{2} - x_8 - z_8\right)\mathbf{a}_1 + (y_8 - z_8)\,\mathbf{a}_2 + (-x_8 + y_8)\,\mathbf{a}_3$ | = | $\left(-\frac{1}{4} + y_8\right) a\,\hat{\mathbf{x}} + \left(\frac{1}{4} - x_8\right) a\,\hat{\mathbf{y}} + \left(\frac{1}{4} - z_8\right) c\,\hat{\mathbf{z}}$ | $(16f)$ | O IV |
| $\mathbf{B}_{58}$ | = | $\left(\frac{1}{2} + x_8 - z_8\right)\mathbf{a}_1 + \left(\frac{1}{2} - y_8 - z_8\right)\mathbf{a}_2 + \left(\frac{1}{2} + x_8 - y_8\right)\mathbf{a}_3$ | = | $\left(\frac{1}{4} - y_8\right) a\,\hat{\mathbf{x}} + \left(\frac{1}{4} + x_8\right) a\,\hat{\mathbf{y}} + \left(\frac{1}{4} - z_8\right) c\,\hat{\mathbf{z}}$ | $(16f)$ | O IV |
| $\mathbf{B}_{59}$ | = | $(y_9 + z_9)\,\mathbf{a}_1 + (x_9 + z_9)\,\mathbf{a}_2 + (x_9 + y_9)\,\mathbf{a}_3$ | = | $x_9 a\,\hat{\mathbf{x}} + y_9 a\,\hat{\mathbf{y}} + z_9 c\,\hat{\mathbf{z}}$ | $(16f)$ | O V |
| $\mathbf{B}_{60}$ | = | $\left(\frac{1}{2} - y_9 + z_9\right)\mathbf{a}_1 + (-x_9 + z_9)\,\mathbf{a}_2 + \left(\frac{1}{2} - x_9 - y_9\right)\mathbf{a}_3$ | = | $-x_9 a\,\hat{\mathbf{x}} + \left(\frac{1}{2} - y_9\right) a\,\hat{\mathbf{y}} + z_9 c\,\hat{\mathbf{z}}$ | $(16f)$ | O V |
| $\mathbf{B}_{61}$ | = | $\left(\frac{1}{2} + x_9 + z_9\right)\mathbf{a}_1 + (-y_9 + z_9)\,\mathbf{a}_2 + (x_9 - y_9)\,\mathbf{a}_3$ | = | $\left(\frac{3}{4} - y_9\right) a\,\hat{\mathbf{x}} + \left(\frac{1}{4} + x_9\right) a\,\hat{\mathbf{y}} + \left(\frac{1}{4} + z_9\right) c\,\hat{\mathbf{z}}$ | $(16f)$ | O V |
| $\mathbf{B}_{62}$ | = | $\left(\frac{1}{2} - x_9 + z_9\right)\mathbf{a}_1 + \left(\frac{1}{2} + y_9 + z_9\right)\mathbf{a}_2 + \left(\frac{1}{2} - x_9 + y_9\right)\mathbf{a}_3$ | = | $\left(\frac{1}{4} + y_9\right) a\,\hat{\mathbf{x}} + \left(\frac{1}{4} - x_9\right) a\,\hat{\mathbf{y}} + \left(\frac{1}{4} + z_9\right) c\,\hat{\mathbf{z}}$ | $(16f)$ | O V |
| $\mathbf{B}_{63}$ | = | $(-y_9 - z_9)\,\mathbf{a}_1 + (-x_9 - z_9)\,\mathbf{a}_2 + (-x_9 - y_9)\,\mathbf{a}_3$ | = | $-x_9 a\,\hat{\mathbf{x}} - y_9 a\,\hat{\mathbf{y}} - z_9 c\,\hat{\mathbf{z}}$ | $(16f)$ | O V |
| $\mathbf{B}_{64}$ | = | $\left(\frac{1}{2} + y_9 - z_9\right) \mathbf{a}_1 + (x_9 - z_9) \mathbf{a}_2 + \left(\frac{1}{2} + x_9 + y_9\right) \mathbf{a}_3$ | = | $x_9 a \, \hat{\mathbf{x}} + \left(\frac{1}{2} + y_9\right) a \, \hat{\mathbf{y}} - z_9 c \, \hat{\mathbf{z}}$ | (16$f$) | O V |
| $\mathbf{B}_{65}$ | = | $\left(\frac{1}{2} - x_9 - z_9\right) \mathbf{a}_1 + (y_9 - z_9) \mathbf{a}_2 + (-x_9 + y_9) \mathbf{a}_3$ | = | $\left(-\frac{1}{4} + y_9\right) a \, \hat{\mathbf{x}} + \left(\frac{1}{4} - x_9\right) a \, \hat{\mathbf{y}} + \left(\frac{1}{4} - z_9\right) c \, \hat{\mathbf{z}}$ | (16$f$) | O V |
| $\mathbf{B}_{66}$ | = | $\left(\frac{1}{2} + x_9 - z_9\right) \mathbf{a}_1 + \left(\frac{1}{2} - y_9 - z_9\right) \mathbf{a}_2 + \left(\frac{1}{2} + x_9 - y_9\right) \mathbf{a}_3$ | = | $\left(\frac{1}{4} - y_9\right) a \, \hat{\mathbf{x}} + \left(\frac{1}{4} + x_9\right) a \, \hat{\mathbf{y}} + \left(\frac{1}{4} - z_9\right) c \, \hat{\mathbf{z}}$ | (16$f$) | O V |

---

---

**Geometry files:**

- CIF: pp. 1675
- POSCAR: pp. 1675

# Copper (I) Azide ($CuN_3$) Structure: AB3_tI32_88_d_cf

http://aflow.org/prototype-encyclopedia/AB3_tI32_88_d_cf

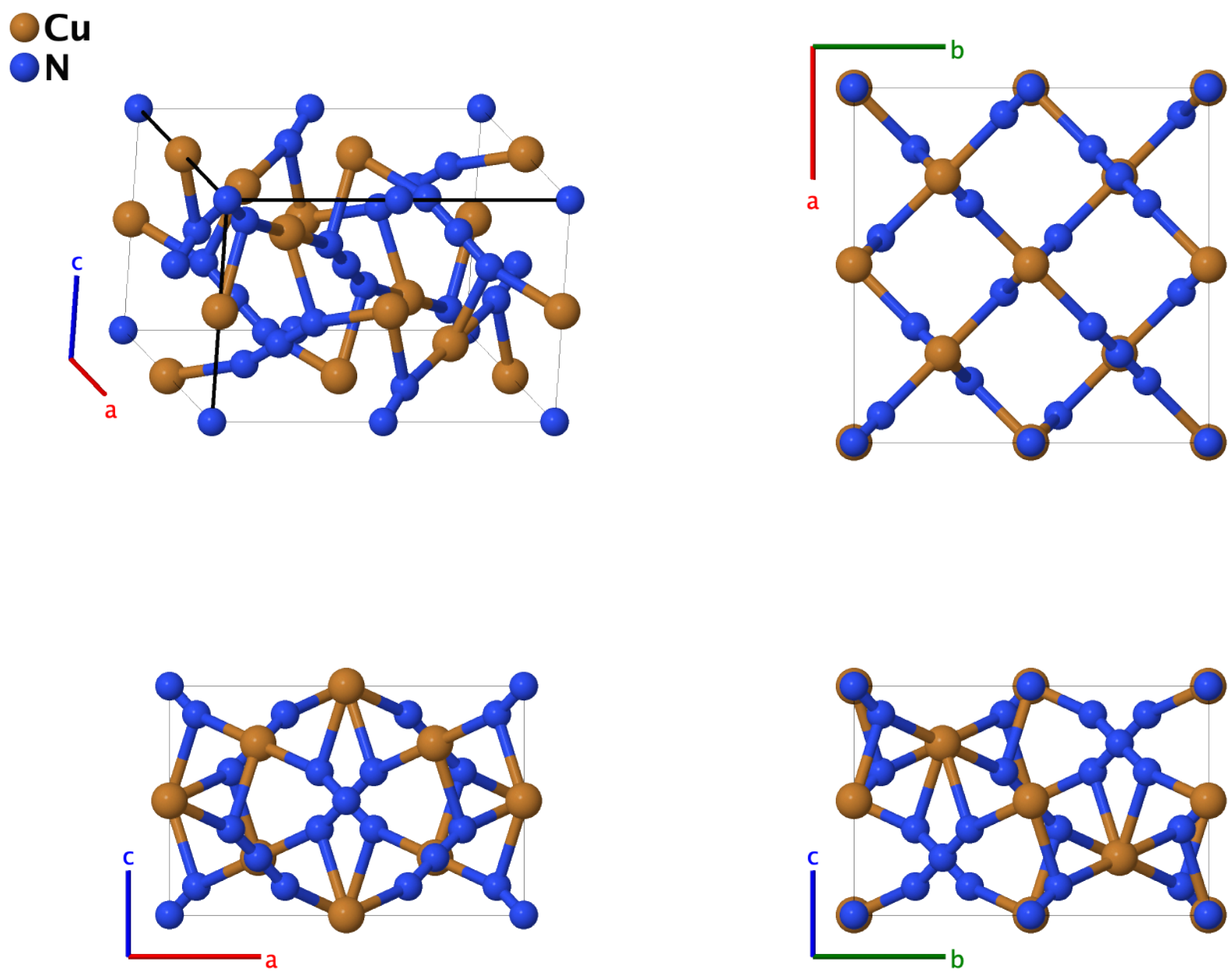

| | | |
|---|---|---|
| **Prototype** | : | $CuN_3$ |
| **AFLOW prototype label** | : | AB3_tI32_88_d_cf |
| ***Strukturbericht* designation** | : | None |
| **Pearson symbol** | : | tI32 |
| **Space group number** | : | 88 |
| **Space group symbol** | : | $I4_1/a$ |
| **AFLOW prototype command** | : | `aflow --proto=AB3_tI32_88_d_cf`<br>`--params=`$a, c/a, x_3, y_3, z_3$ |

### Other compounds with this structure

- $AgN_3$ and $TiN_3$

---

- Not to be confused with Copper (II) Azide, $Cu(N_3)_2$, an explosive.
- (Wilsdorf, 1948) gave the Wyckoff positions in setting 1 of space group #88. We used FINDSYM to translate this to the standard setting 2.

### Body-centered Tetragonal primitive vectors:

$$\mathbf{a}_1 = -\frac{1}{2}\,a\,\mathbf{\hat{x}} + \frac{1}{2}\,a\,\mathbf{\hat{y}} + \frac{1}{2}\,c\,\mathbf{\hat{z}}$$
$$\mathbf{a}_2 = \frac{1}{2}\,a\,\mathbf{\hat{x}} - \frac{1}{2}\,a\,\mathbf{\hat{y}} + \frac{1}{2}\,c\,\mathbf{\hat{z}}$$
$$\mathbf{a}_3 = \frac{1}{2}\,a\,\mathbf{\hat{x}} + \frac{1}{2}\,a\,\mathbf{\hat{y}} - \frac{1}{2}\,c\,\mathbf{\hat{z}}$$

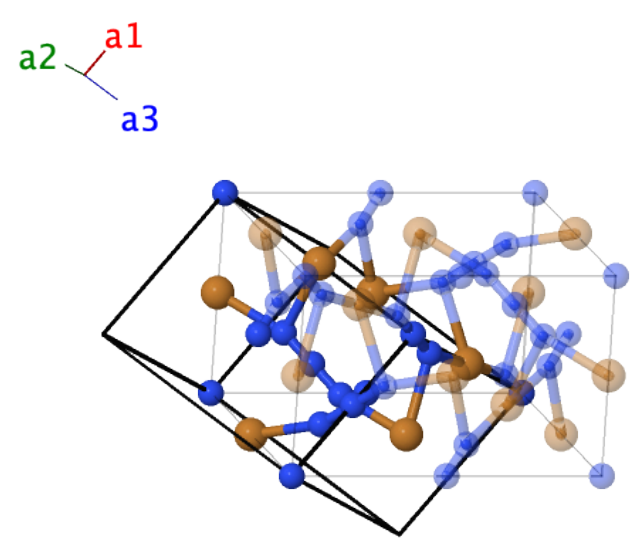

**Basis vectors:**

| | | Lattice Coordinates | | Cartesian Coordinates | Wyckoff Position | Atom Type |
|---|---|---|---|---|---|---|
| $\mathbf{B}_1$ | = | $0\,\mathbf{a}_1 + 0\,\mathbf{a}_2 + 0\,\mathbf{a}_3$ | = | $0\,\hat{\mathbf{x}} + 0\,\hat{\mathbf{y}} + 0\,\hat{\mathbf{z}}$ | (8*c*) | N I |
| $\mathbf{B}_2$ | = | $\frac{1}{2}\,\mathbf{a}_1 + \frac{1}{2}\,\mathbf{a}_3$ | = | $\frac{1}{2}a\,\hat{\mathbf{y}}$ | (8*c*) | N I |
| $\mathbf{B}_3$ | = | $\frac{1}{2}\,\mathbf{a}_1$ | = | $\frac{3}{4}a\,\hat{\mathbf{x}} + \frac{1}{4}a\,\hat{\mathbf{y}} + \frac{1}{4}c\,\hat{\mathbf{z}}$ | (8*c*) | N I |
| $\mathbf{B}_4$ | = | $\frac{1}{2}\,\mathbf{a}_1 + \frac{1}{2}\,\mathbf{a}_2 + \frac{1}{2}\,\mathbf{a}_3$ | = | $\frac{1}{4}a\,\hat{\mathbf{x}} + \frac{1}{4}a\,\hat{\mathbf{y}} + \frac{1}{4}c\,\hat{\mathbf{z}}$ | (8*c*) | N I |
| $\mathbf{B}_5$ | = | $\frac{1}{2}\,\mathbf{a}_1 + \frac{1}{2}\,\mathbf{a}_2$ | = | $\frac{1}{2}c\,\hat{\mathbf{z}}$ | (8*d*) | Cu |
| $\mathbf{B}_6$ | = | $\frac{1}{2}\,\mathbf{a}_2 + \frac{1}{2}\,\mathbf{a}_3$ | = | $\frac{1}{2}a\,\hat{\mathbf{x}}$ | (8*d*) | Cu |
| $\mathbf{B}_7$ | = | $\frac{1}{2}\,\mathbf{a}_2$ | = | $\frac{1}{4}a\,\hat{\mathbf{x}} - \frac{1}{4}a\,\hat{\mathbf{y}} + \frac{1}{4}c\,\hat{\mathbf{z}}$ | (8*d*) | Cu |
| $\mathbf{B}_8$ | = | $\frac{1}{2}\,\mathbf{a}_3$ | = | $\frac{1}{4}a\,\hat{\mathbf{x}} + \frac{1}{4}a\,\hat{\mathbf{y}} - \frac{1}{4}c\,\hat{\mathbf{z}}$ | (8*d*) | Cu |
| $\mathbf{B}_9$ | = | $(y_3 + z_3)\,\mathbf{a}_1 + (x_3 + z_3)\,\mathbf{a}_2 + (x_3 + y_3)\,\mathbf{a}_3$ | = | $x_3a\,\hat{\mathbf{x}} + y_3a\,\hat{\mathbf{y}} + z_3c\,\hat{\mathbf{z}}$ | (16*f*) | N II |
| $\mathbf{B}_{10}$ | = | $\left(\frac{1}{2} - y_3 + z_3\right)\mathbf{a}_1 + (-x_3 + z_3)\,\mathbf{a}_2 + \left(\frac{1}{2} - x_3 - y_3\right)\mathbf{a}_3$ | = | $-x_3a\,\hat{\mathbf{x}} + \left(\frac{1}{2} - y_3\right)a\,\hat{\mathbf{y}} + z_3c\,\hat{\mathbf{z}}$ | (16*f*) | N II |
| $\mathbf{B}_{11}$ | = | $\left(\frac{1}{2} + x_3 + z_3\right)\mathbf{a}_1 + (-y_3 + z_3)\,\mathbf{a}_2 + (x_3 - y_3)\,\mathbf{a}_3$ | = | $\left(\frac{3}{4} - y_3\right)a\,\hat{\mathbf{x}} + \left(\frac{1}{4} + x_3\right)a\,\hat{\mathbf{y}} + \left(\frac{1}{4} + z_3\right)c\,\hat{\mathbf{z}}$ | (16*f*) | N II |
| $\mathbf{B}_{12}$ | = | $\left(\frac{1}{2} - x_3 + z_3\right)\mathbf{a}_1 + \left(\frac{1}{2} + y_3 + z_3\right)\mathbf{a}_2 + \left(\frac{1}{2} - x_3 + y_3\right)\mathbf{a}_3$ | = | $\left(\frac{1}{4} + y_3\right)a\,\hat{\mathbf{x}} + \left(\frac{1}{4} - x_3\right)a\,\hat{\mathbf{y}} + \left(\frac{1}{4} + z_3\right)c\,\hat{\mathbf{z}}$ | (16*f*) | N II |
| $\mathbf{B}_{13}$ | = | $(-y_3 - z_3)\,\mathbf{a}_1 + (-x_3 - z_3)\,\mathbf{a}_2 + (-x_3 - y_3)\,\mathbf{a}_3$ | = | $-x_3a\,\hat{\mathbf{x}} - y_3a\,\hat{\mathbf{y}} - z_3c\,\hat{\mathbf{z}}$ | (16*f*) | N II |
| $\mathbf{B}_{14}$ | = | $\left(\frac{1}{2} + y_3 - z_3\right)\mathbf{a}_1 + (x_3 - z_3)\,\mathbf{a}_2 + \left(\frac{1}{2} + x_3 + y_3\right)\mathbf{a}_3$ | = | $x_3a\,\hat{\mathbf{x}} + \left(\frac{1}{2} + y_3\right)a\,\hat{\mathbf{y}} - z_3c\,\hat{\mathbf{z}}$ | (16*f*) | N II |
| $\mathbf{B}_{15}$ | = | $\left(\frac{1}{2} - x_3 - z_3\right)\mathbf{a}_1 + (y_3 - z_3)\,\mathbf{a}_2 + (-x_3 + y_3)\,\mathbf{a}_3$ | = | $\left(-\frac{1}{4} + y_3\right)a\,\hat{\mathbf{x}} + \left(\frac{1}{4} - x_3\right)a\,\hat{\mathbf{y}} + \left(\frac{1}{4} - z_3\right)c\,\hat{\mathbf{z}}$ | (16*f*) | N II |
| $\mathbf{B}_{16}$ | = | $\left(\frac{1}{2} + x_3 - z_3\right)\mathbf{a}_1 + \left(\frac{1}{2} - y_3 - z_3\right)\mathbf{a}_2 + \left(\frac{1}{2} + x_3 - y_3\right)\mathbf{a}_3$ | = | $\left(\frac{1}{4} - y_3\right)a\,\hat{\mathbf{x}} + \left(\frac{1}{4} + x_3\right)a\,\hat{\mathbf{y}} + \left(\frac{1}{4} - z_3\right)c\,\hat{\mathbf{z}}$ | (16*f*) | N II |

**Found in:**

**Geometry files:**

- CIF: pp. 1675
- POSCAR: pp. 1676

# Scheelite ($CaWO_4$, $H0_4$) Structure: AB4C_tI24_88_b_f_a

http://aflow.org/prototype-encyclopedia/AB4C_tI24_88_b_f_a

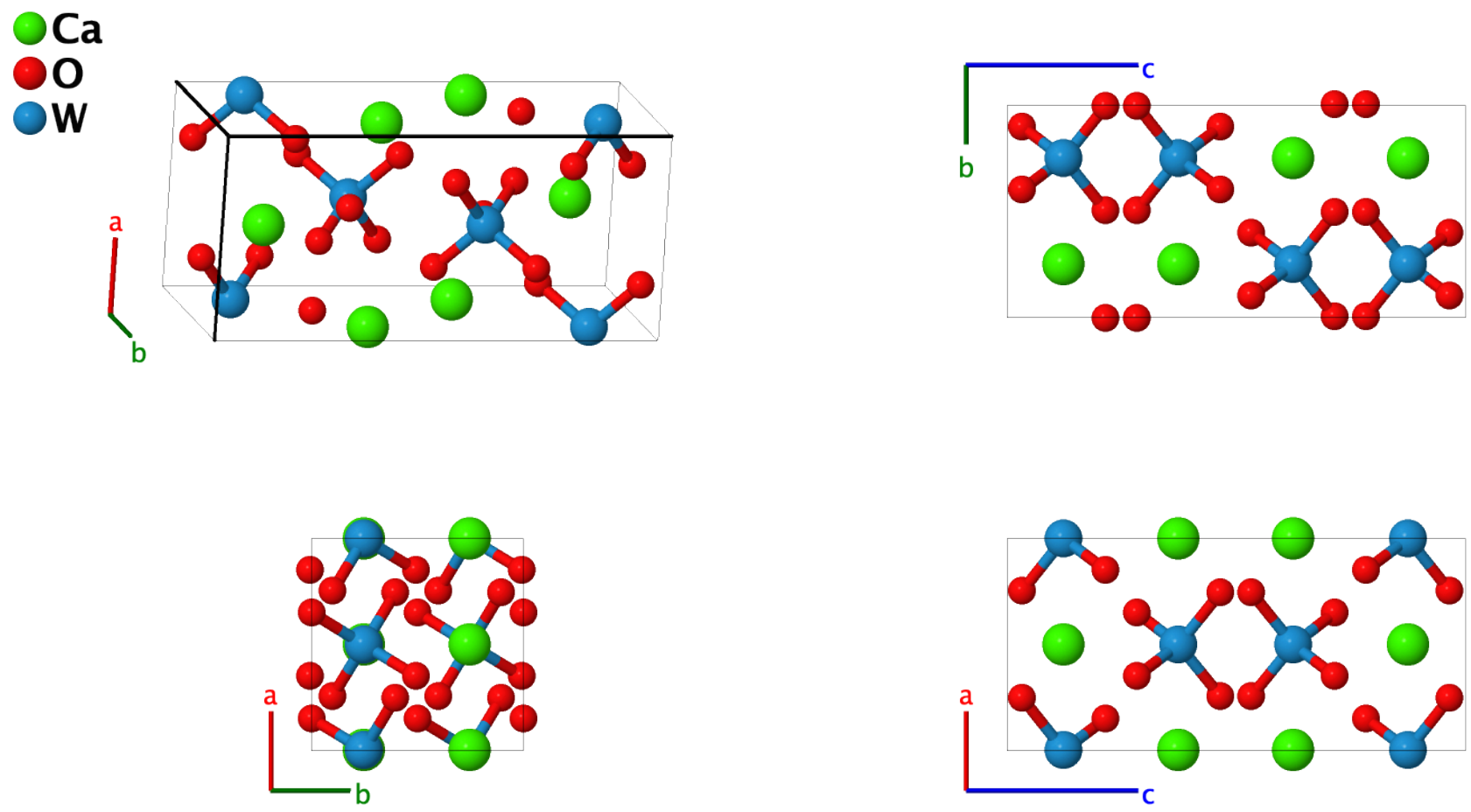

| | | |
|---|---|---|
| **Prototype** | : | $CaO_4W$ |
| **AFLOW prototype label** | : | AB4C_tI24_88_b_f_a |
| ***Strukturbericht* designation** | : | $H0_4$ |
| **Pearson symbol** | : | tI24 |
| **Space group number** | : | 88 |
| **Space group symbol** | : | $I4_1/a$ |
| **AFLOW prototype command** | : | `aflow --proto=AB4C_tI24_88_b_f_a --params=`$a, c/a, x_3, y_3, z_3$ |

### Other compounds with this structure

- $ZrSiO_4$, $LaNbO_4$, $YNbO_4$, (Y,RE)$NbO_4$ (fergusonite), $YVO_4$, $BiVO_4$, $BaWO_4$, $PbWO_4$ (wulfenite), $SrWO_4$, $EuWO_4$, $PbMoO_4$ (stolzite), $SrMoO_4$, $CaMoO_4$ (powellite), $CdMoO_4$, $KReO_4$, $TlReO_4$, $AgReO_4$, and $NaAlH_4$

- (Ewald, 1931) originally gave this structure the *Strukturbericht* designation $H4$, but this was changed to $H0_4$ in (Gottfried, 1937).

### Body-centered Tetragonal primitive vectors:

$$\mathbf{a}_1 = -\frac{1}{2}\,a\,\hat{\mathbf{x}} + \frac{1}{2}\,a\,\hat{\mathbf{y}} + \frac{1}{2}\,c\,\hat{\mathbf{z}}$$
$$\mathbf{a}_2 = \frac{1}{2}\,a\,\hat{\mathbf{x}} - \frac{1}{2}\,a\,\hat{\mathbf{y}} + \frac{1}{2}\,c\,\hat{\mathbf{z}}$$
$$\mathbf{a}_3 = \frac{1}{2}\,a\,\hat{\mathbf{x}} + \frac{1}{2}\,a\,\hat{\mathbf{y}} - \frac{1}{2}\,c\,\hat{\mathbf{z}}$$

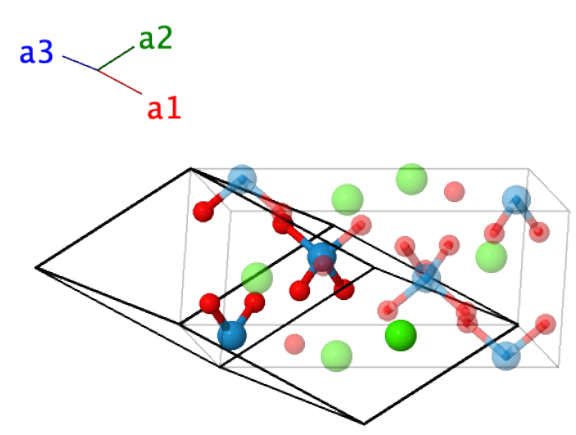

### Basis vectors:

| | | Lattice Coordinates | | Cartesian Coordinates | Wyckoff Position | Atom Type |
|---|---|---|---|---|---|---|
| $\mathbf{B}_1$ | = | $\frac{3}{8}\,\mathbf{a}_1 + \frac{1}{8}\,\mathbf{a}_2 + \frac{1}{4}\,\mathbf{a}_3$ | = | $\frac{1}{4}a\,\hat{\mathbf{y}} + \frac{1}{8}c\,\hat{\mathbf{z}}$ | (4$a$) | W |
| $\mathbf{B}_2$ | = | $\frac{5}{8}\,\mathbf{a}_1 + \frac{7}{8}\,\mathbf{a}_2 + \frac{3}{4}\,\mathbf{a}_3$ | = | $\frac{1}{2}a\,\hat{\mathbf{x}} + \frac{1}{4}a\,\hat{\mathbf{y}} + \frac{3}{8}c\,\hat{\mathbf{z}}$ | (4$a$) | W |
| $\mathbf{B}_3$ | = | $\frac{7}{8}\,\mathbf{a}_1 + \frac{5}{8}\,\mathbf{a}_2 + \frac{1}{4}\,\mathbf{a}_3$ | = | $\frac{1}{4}a\,\mathbf{\hat{y}} + \frac{5}{8}c\,\mathbf{\hat{z}}$ | (4$b$) | Ca |
| $\mathbf{B}_4$ | = | $\frac{1}{8}\,\mathbf{a}_1 + \frac{3}{8}\,\mathbf{a}_2 + \frac{3}{4}\,\mathbf{a}_3$ | = | $\frac{1}{2}a\,\mathbf{\hat{x}} + \frac{1}{4}a\,\mathbf{\hat{y}} + \frac{7}{8}c\,\mathbf{\hat{z}}$ | (4$b$) | Ca |
| $\mathbf{B}_5$ | = | $(y_3 + z_3)\,\mathbf{a}_1 + (x_3 + z_3)\,\mathbf{a}_2 + (x_3 + y_3)\,\mathbf{a}_3$ | = | $x_3 a\,\mathbf{\hat{x}} + y_3 a\,\mathbf{\hat{y}} + z_3 c\,\mathbf{\hat{z}}$ | (16$f$) | O |
| $\mathbf{B}_6$ | = | $\left(\frac{1}{2} - y_3 + z_3\right)\mathbf{a}_1 + (-x_3 + z_3)\,\mathbf{a}_2 + \left(\frac{1}{2} - x_3 - y_3\right)\mathbf{a}_3$ | = | $-x_3 a\,\mathbf{\hat{x}} + \left(\frac{1}{2} - y_3\right)a\,\mathbf{\hat{y}} + z_3 c\,\mathbf{\hat{z}}$ | (16$f$) | O |
| $\mathbf{B}_7$ | = | $\left(\frac{1}{2} + x_3 + z_3\right)\mathbf{a}_1 + (-y_3 + z_3)\,\mathbf{a}_2 + (x_3 - y_3)\,\mathbf{a}_3$ | = | $\left(\frac{3}{4} - y_3\right)a\,\mathbf{\hat{x}} + \left(\frac{1}{4} + x_3\right)a\,\mathbf{\hat{y}} + \left(\frac{1}{4} + z_3\right)c\,\mathbf{\hat{z}}$ | (16$f$) | O |
| $\mathbf{B}_8$ | = | $\left(\frac{1}{2} - x_3 + z_3\right)\mathbf{a}_1 + \left(\frac{1}{2} + y_3 + z_3\right)\mathbf{a}_2 + \left(\frac{1}{2} - x_3 + y_3\right)\mathbf{a}_3$ | = | $\left(\frac{1}{4} + y_3\right)a\,\mathbf{\hat{x}} + \left(\frac{1}{4} - x_3\right)a\,\mathbf{\hat{y}} + \left(\frac{1}{4} + z_3\right)c\,\mathbf{\hat{z}}$ | (16$f$) | O |
| $\mathbf{B}_9$ | = | $(-y_3 - z_3)\,\mathbf{a}_1 + (-x_3 - z_3)\,\mathbf{a}_2 + (-x_3 - y_3)\,\mathbf{a}_3$ | = | $-x_3 a\,\mathbf{\hat{x}} - y_3 a\,\mathbf{\hat{y}} - z_3 c\,\mathbf{\hat{z}}$ | (16$f$) | O |
| $\mathbf{B}_{10}$ | = | $\left(\frac{1}{2} + y_3 - z_3\right)\mathbf{a}_1 + (x_3 - z_3)\,\mathbf{a}_2 + \left(\frac{1}{2} + x_3 + y_3\right)\mathbf{a}_3$ | = | $x_3 a\,\mathbf{\hat{x}} + \left(\frac{1}{2} + y_3\right)a\,\mathbf{\hat{y}} - z_3 c\,\mathbf{\hat{z}}$ | (16$f$) | O |
| $\mathbf{B}_{11}$ | = | $\left(\frac{1}{2} - x_3 - z_3\right)\mathbf{a}_1 + (y_3 - z_3)\,\mathbf{a}_2 + (-x_3 + y_3)\,\mathbf{a}_3$ | = | $\left(-\frac{1}{4} + y_3\right)a\,\mathbf{\hat{x}} + \left(\frac{1}{4} - x_3\right)a\,\mathbf{\hat{y}} + \left(\frac{1}{4} - z_3\right)c\,\mathbf{\hat{z}}$ | (16$f$) | O |
| $\mathbf{B}_{12}$ | = | $\left(\frac{1}{2} + x_3 - z_3\right)\mathbf{a}_1 + \left(\frac{1}{2} - y_3 - z_3\right)\mathbf{a}_2 + \left(\frac{1}{2} + x_3 - y_3\right)\mathbf{a}_3$ | = | $\left(\frac{1}{4} - y_3\right)a\,\mathbf{\hat{x}} + \left(\frac{1}{4} + x_3\right)a\,\mathbf{\hat{y}} + \left(\frac{1}{4} - z_3\right)c\,\mathbf{\hat{z}}$ | (16$f$) | O |

**Geometry files:**

- CIF: pp. 1676
- POSCAR: pp. 1676

# $G7_5$ ($PbCO_3 \cdot PbCl_2$, Phosgenite) (*obsolete*) Structure: AB2C3D2_tP16_90_c_f_ce_e

http://aflow.org/prototype-encyclopedia/AB2C3D2_tP16_90_c_f_ce_e

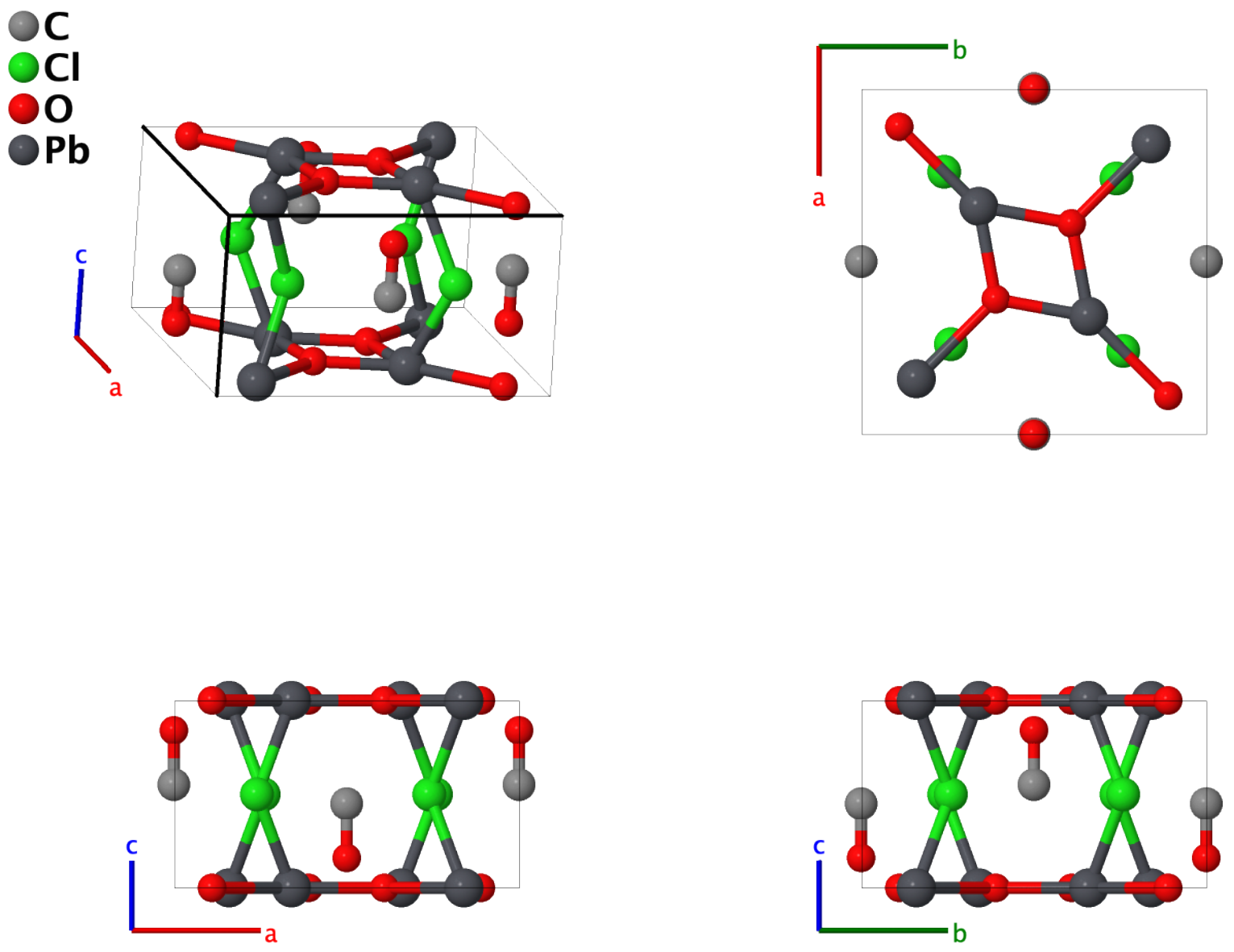

| | | |
|---|---|---|
| **Prototype** | : | $CCl_2O_3Pb_2$ |
| **AFLOW prototype label** | : | AB2C3D2_tP16_90_c_f_ce_e |
| ***Strukturbericht* designation** | : | $G7_5$ |
| **Pearson symbol** | : | tP16 |
| **Space group number** | : | 90 |
| **Space group symbol** | : | $P42_12$ |
| **AFLOW prototype command** | : | `aflow --proto=AB2C3D2_tP16_90_c_f_ce_e` <br> `--params=`$a, c/a, z_1, z_2, x_3, x_4, x_5$ |

- (Onorato, 1934) made an early determination of the structure of phosgenite, and (Gottfried, 1937) assigned it the *Strukturbericht* designation $G7_5$. However, the "proposed structure [was] based on photographic data and partly on steric considerations" (Giuseppetti, 1974). Subsequent investigations showed that the true phosgenite structure was in space group $P4/mbm$ #127. We list the original structure here as part of the historical record.

**Simple Tetragonal primitive vectors:**

$$\mathbf{a}_1 = a\,\hat{\mathbf{x}}$$
$$\mathbf{a}_2 = a\,\hat{\mathbf{y}}$$
$$\mathbf{a}_3 = c\,\hat{\mathbf{z}}$$

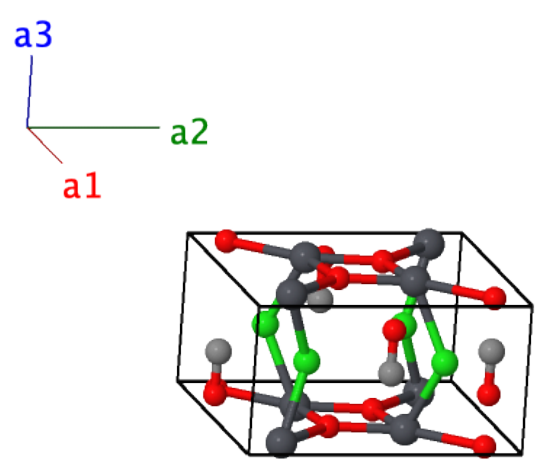

**Basis vectors:**

| | | Lattice Coordinates | | Cartesian Coordinates | Wyckoff Position | Atom Type |
|---|---|---|---|---|---|---|
| $\mathbf{B}_1$ | = | $\frac{1}{2}\,\mathbf{a}_2 + z_1\,\mathbf{a}_3$ | = | $\frac{1}{2}a\,\hat{\mathbf{y}} + z_1c\,\hat{\mathbf{z}}$ | (2c) | C |
| $\mathbf{B}_2$ | = | $\frac{1}{2}\,\mathbf{a}_1 - z_1\,\mathbf{a}_3$ | = | $\frac{1}{2}a\,\hat{\mathbf{x}} - z_1c\,\hat{\mathbf{z}}$ | (2c) | C |
| $\mathbf{B}_3$ | = | $\frac{1}{2}\,\mathbf{a}_2 + z_2\,\mathbf{a}_3$ | = | $\frac{1}{2}a\,\hat{\mathbf{y}} + z_2c\,\hat{\mathbf{z}}$ | (2c) | O I |
| $\mathbf{B}_4$ | = | $\frac{1}{2}\,\mathbf{a}_1 - z_2\,\mathbf{a}_3$ | = | $\frac{1}{2}a\,\hat{\mathbf{x}} - z_2c\,\hat{\mathbf{z}}$ | (2c) | O I |
| $\mathbf{B}_5$ | = | $x_3\,\mathbf{a}_1 + x_3\,\mathbf{a}_2$ | = | $x_3a\,\hat{\mathbf{x}} + x_3a\,\hat{\mathbf{y}}$ | (4e) | O II |
| $\mathbf{B}_6$ | = | $-x_3\,\mathbf{a}_1 - x_3\,\mathbf{a}_2$ | = | $-x_3a\,\hat{\mathbf{x}} - x_3a\,\hat{\mathbf{y}}$ | (4e) | O II |
| $\mathbf{B}_7$ | = | $\left(\frac{1}{2} - x_3\right)\,\mathbf{a}_1 + \left(\frac{1}{2} + x_3\right)\,\mathbf{a}_2$ | = | $\left(\frac{1}{2} - x_3\right)a\,\hat{\mathbf{x}} + \left(\frac{1}{2} + x_3\right)a\,\hat{\mathbf{y}}$ | (4e) | O II |
| $\mathbf{B}_8$ | = | $\left(\frac{1}{2} + x_3\right)\,\mathbf{a}_1 + \left(\frac{1}{2} - x_3\right)\,\mathbf{a}_2$ | = | $\left(\frac{1}{2} + x_3\right)a\,\hat{\mathbf{x}} + \left(\frac{1}{2} - x_3\right)a\,\hat{\mathbf{y}}$ | (4e) | O II |
| $\mathbf{B}_9$ | = | $x_4\,\mathbf{a}_1 + x_4\,\mathbf{a}_2$ | = | $x_4a\,\hat{\mathbf{x}} + x_4a\,\hat{\mathbf{y}}$ | (4e) | Pb |
| $\mathbf{B}_{10}$ | = | $-x_4\,\mathbf{a}_1 - x_4\,\mathbf{a}_2$ | = | $-x_4a\,\hat{\mathbf{x}} - x_4a\,\hat{\mathbf{y}}$ | (4e) | Pb |
| $\mathbf{B}_{11}$ | = | $\left(\frac{1}{2} - x_4\right)\,\mathbf{a}_1 + \left(\frac{1}{2} + x_4\right)\,\mathbf{a}_2$ | = | $\left(\frac{1}{2} - x_4\right)a\,\hat{\mathbf{x}} + \left(\frac{1}{2} + x_4\right)a\,\hat{\mathbf{y}}$ | (4e) | Pb |
| $\mathbf{B}_{12}$ | = | $\left(\frac{1}{2} + x_4\right)\,\mathbf{a}_1 + \left(\frac{1}{2} - x_4\right)\,\mathbf{a}_2$ | = | $\left(\frac{1}{2} + x_4\right)a\,\hat{\mathbf{x}} + \left(\frac{1}{2} - x_4\right)a\,\hat{\mathbf{y}}$ | (4e) | Pb |
| $\mathbf{B}_{13}$ | = | $x_5\,\mathbf{a}_1 + x_5\,\mathbf{a}_2 + \frac{1}{2}\,\mathbf{a}_3$ | = | $x_5a\,\hat{\mathbf{x}} + x_5a\,\hat{\mathbf{y}} + \frac{1}{2}c\,\hat{\mathbf{z}}$ | (4f) | Cl |
| $\mathbf{B}_{14}$ | = | $-x_5\,\mathbf{a}_1 - x_5\,\mathbf{a}_2 + \frac{1}{2}\,\mathbf{a}_3$ | = | $-x_5a\,\hat{\mathbf{x}} - x_5a\,\hat{\mathbf{y}} + \frac{1}{2}c\,\hat{\mathbf{z}}$ | (4f) | Cl |
| $\mathbf{B}_{15}$ | = | $\left(\frac{1}{2} - x_5\right)\,\mathbf{a}_1 + \left(\frac{1}{2} + x_5\right)\,\mathbf{a}_2 + \frac{1}{2}\,\mathbf{a}_3$ | = | $\left(\frac{1}{2} - x_5\right)a\,\hat{\mathbf{x}} + \left(\frac{1}{2} + x_5\right)a\,\hat{\mathbf{y}} + \frac{1}{2}c\,\hat{\mathbf{z}}$ | (4f) | Cl |
| $\mathbf{B}_{16}$ | = | $\left(\frac{1}{2} + x_5\right)\,\mathbf{a}_1 + \left(\frac{1}{2} - x_5\right)\,\mathbf{a}_2 + \frac{1}{2}\,\mathbf{a}_3$ | = | $\left(\frac{1}{2} + x_5\right)a\,\hat{\mathbf{x}} + \left(\frac{1}{2} - x_5\right)a\,\hat{\mathbf{y}} + \frac{1}{2}c\,\hat{\mathbf{z}}$ | (4f) | Cl |

**Geometry files:**

- CIF: pp. 1676
- POSCAR: pp. 1677

# Retgersite ($\alpha$-$NiSO_4 \cdot 6H_2O$, $H4_5$) Structure: A12BC10D_tP96_92_6b_a_5b_a

http://aflow.org/prototype-encyclopedia/A12BC10D_tP96_92_6b_a_5b_a

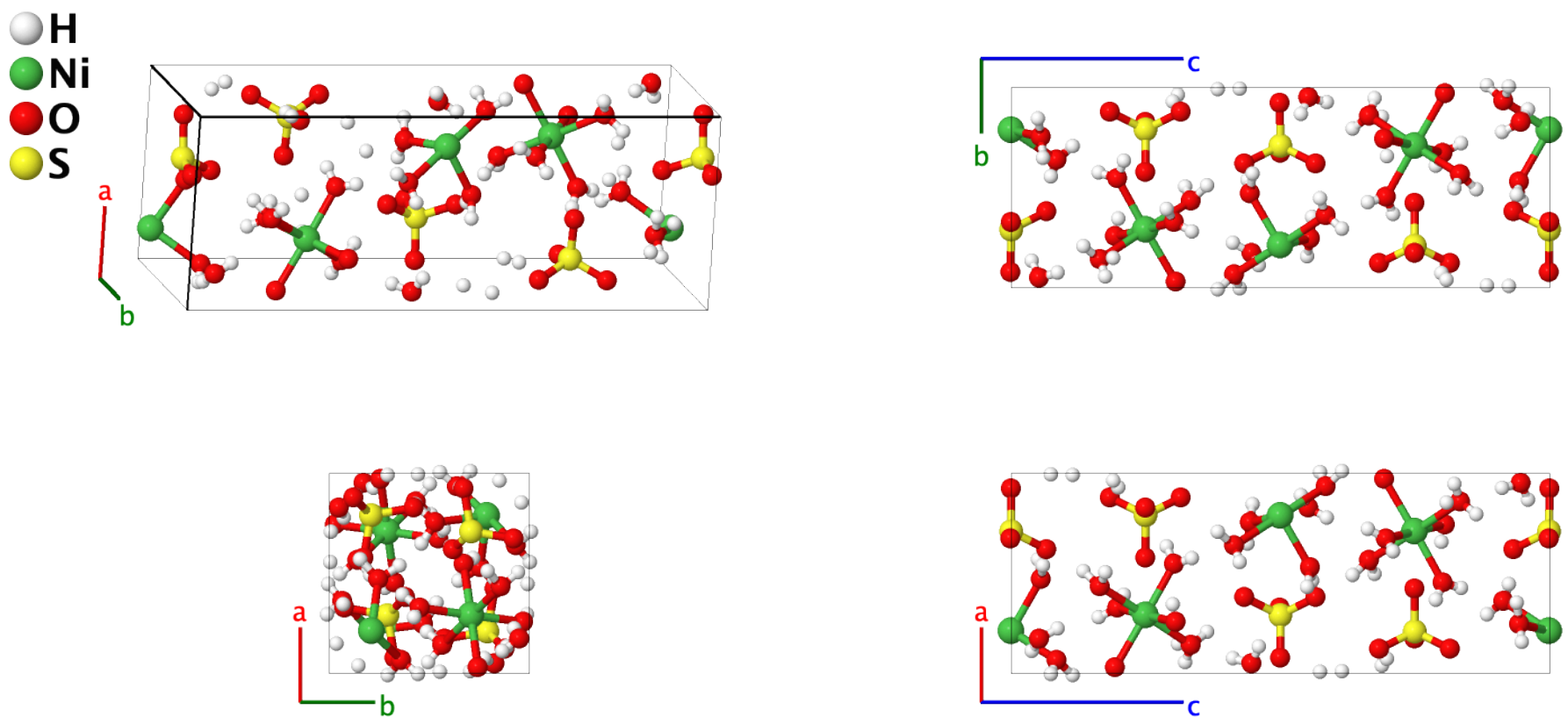

| | | |
|---|---|---|
| **Prototype** | : | $H_{12}NiO_{10}S$ |
| **AFLOW prototype label** | : | A12BC10D_tP96_92_6b_a_5b_a |
| ***Strukturbericht* designation** | : | $H4_5$ |
| **Pearson symbol** | : | tP96 |
| **Space group number** | : | 92 |
| **Space group symbol** | : | $P4_12_12$ |
| **AFLOW prototype command** | : | `aflow --proto=A12BC10D_tP96_92_6b_a_5b_a --params=`$a, c/a, x_1, x_2, x_3, y_3, z_3, x_4, y_4, z_4, x_5, y_5, z_5, x_6, y_6, z_6, x_7, y_7, z_7, x_8, y_8, z_8, x_9, y_9, z_9, x_{10}, y_{10}, z_{10}, x_{11}, y_{11}, z_{11}, x_{12}, y_{12}, z_{12}, x_{13}, y_{13}, z_{13}$ |

- This structure is nearly identical to the one presented in (Hermann, 1937) as $H4_5$, but now includes the positions of the hydrogen atoms.
- This compound can also be found in the enantiomorphic space group $P4_32_12$ #98.

**Simple Tetragonal primitive vectors:**

$$\mathbf{a}_1 = a\,\hat{\mathbf{x}}$$
$$\mathbf{a}_2 = a\,\hat{\mathbf{y}}$$
$$\mathbf{a}_3 = c\,\hat{\mathbf{z}}$$

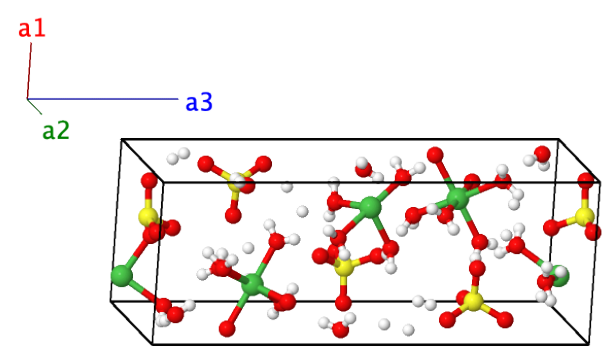

**Basis vectors:**

| | | Lattice Coordinates | | Cartesian Coordinates | Wyckoff Position | Atom Type |
|---|---|---|---|---|---|---|
| $\mathbf{B}_1$ | = | $x_1\,\mathbf{a}_1 + x_1\,\mathbf{a}_2$ | = | $x_1 a\,\hat{\mathbf{x}} + x_1 a\,\hat{\mathbf{y}}$ | (4$a$) | Ni |
| $\mathbf{B}_2$ | = | $-x_1\,\mathbf{a}_1 - x_1\,\mathbf{a}_2 + \frac{1}{2}\,\mathbf{a}_3$ | = | $-x_1 a\,\hat{\mathbf{x}} - x_1 a\,\hat{\mathbf{y}} + \frac{1}{2}c\,\hat{\mathbf{z}}$ | (4$a$) | Ni |
| $\mathbf{B}_3$ | = | $\left(\frac{1}{2} - x_1\right)\mathbf{a}_1 + \left(\frac{1}{2} + x_1\right)\mathbf{a}_2 + \frac{1}{4}\,\mathbf{a}_3$ | = | $\left(\frac{1}{2} - x_1\right)a\,\hat{\mathbf{x}} + \left(\frac{1}{2} + x_1\right)a\,\hat{\mathbf{y}} + \frac{1}{4}c\,\hat{\mathbf{z}}$ | (4$a$) | Ni |
| $\mathbf{B}_4$ | = | $\left(\frac{1}{2} + x_1\right)\mathbf{a}_1 + \left(\frac{1}{2} - x_1\right)\mathbf{a}_2 + \frac{3}{4}\,\mathbf{a}_3$ | = | $\left(\frac{1}{2} + x_1\right)a\,\hat{\mathbf{x}} + \left(\frac{1}{2} - x_1\right)a\,\hat{\mathbf{y}} + \frac{3}{4}c\,\hat{\mathbf{z}}$ | (4$a$) | Ni |
| $\mathbf{B}_5$ | = | $x_2\,\mathbf{a}_1 + x_2\,\mathbf{a}_2$ | = | $x_2 a\,\hat{\mathbf{x}} + x_2 a\,\hat{\mathbf{y}}$ | (4$a$) | S |
| $\mathbf{B}_6$ | = | $-x_2\,\mathbf{a}_1 - x_2\,\mathbf{a}_2 + \frac{1}{2}\,\mathbf{a}_3$ | = | $-x_2 a\,\hat{\mathbf{x}} - x_2 a\,\hat{\mathbf{y}} + \frac{1}{2} c\,\hat{\mathbf{z}}$ | (4$a$) | S |
| $\mathbf{B}_7$ | = | $\left(\frac{1}{2} - x_2\right)\mathbf{a}_1 + \left(\frac{1}{2} + x_2\right)\mathbf{a}_2 + \frac{1}{4}\,\mathbf{a}_3$ | = | $\left(\frac{1}{2} - x_2\right) a\,\hat{\mathbf{x}} + \left(\frac{1}{2} + x_2\right) a\,\hat{\mathbf{y}} + \frac{1}{4} c\,\hat{\mathbf{z}}$ | (4$a$) | S |
| $\mathbf{B}_8$ | = | $\left(\frac{1}{2} + x_2\right)\mathbf{a}_1 + \left(\frac{1}{2} - x_2\right)\mathbf{a}_2 + \frac{3}{4}\,\mathbf{a}_3$ | = | $\left(\frac{1}{2} + x_2\right) a\,\hat{\mathbf{x}} + \left(\frac{1}{2} - x_2\right) a\,\hat{\mathbf{y}} + \frac{3}{4} c\,\hat{\mathbf{z}}$ | (4$a$) | S |
| $\mathbf{B}_9$ | = | $x_3\,\mathbf{a}_1 + y_3\,\mathbf{a}_2 + z_3\,\mathbf{a}_3$ | = | $x_3 a\,\hat{\mathbf{x}} + y_3 a\,\hat{\mathbf{y}} + z_3 c\,\hat{\mathbf{z}}$ | (8$b$) | H I |
| $\mathbf{B}_{10}$ | = | $-x_3\,\mathbf{a}_1 - y_3\,\mathbf{a}_2 + \left(\frac{1}{2} + z_3\right)\mathbf{a}_3$ | = | $-x_3 a\,\hat{\mathbf{x}} - y_3 a\,\hat{\mathbf{y}} + \left(\frac{1}{2} + z_3\right) c\,\hat{\mathbf{z}}$ | (8$b$) | H I |
| $\mathbf{B}_{11}$ | = | $\left(\frac{1}{2} - y_3\right)\mathbf{a}_1 + \left(\frac{1}{2} + x_3\right)\mathbf{a}_2 + \left(\frac{1}{4} + z_3\right)\mathbf{a}_3$ | = | $\left(\frac{1}{2} - y_3\right) a\,\hat{\mathbf{x}} + \left(\frac{1}{2} + x_3\right) a\,\hat{\mathbf{y}} + \left(\frac{1}{4} + z_3\right) c\,\hat{\mathbf{z}}$ | (8$b$) | H I |
| $\mathbf{B}_{12}$ | = | $\left(\frac{1}{2} + y_3\right)\mathbf{a}_1 + \left(\frac{1}{2} - x_3\right)\mathbf{a}_2 + \left(\frac{3}{4} + z_3\right)\mathbf{a}_3$ | = | $\left(\frac{1}{2} + y_3\right) a\,\hat{\mathbf{x}} + \left(\frac{1}{2} - x_3\right) a\,\hat{\mathbf{y}} + \left(\frac{3}{4} + z_3\right) c\,\hat{\mathbf{z}}$ | (8$b$) | H I |
| $\mathbf{B}_{13}$ | = | $\left(\frac{1}{2} - x_3\right)\mathbf{a}_1 + \left(\frac{1}{2} + y_3\right)\mathbf{a}_2 + \left(\frac{1}{4} - z_3\right)\mathbf{a}_3$ | = | $\left(\frac{1}{2} - x_3\right) a\,\hat{\mathbf{x}} + \left(\frac{1}{2} + y_3\right) a\,\hat{\mathbf{y}} + \left(\frac{1}{4} - z_3\right) c\,\hat{\mathbf{z}}$ | (8$b$) | H I |
| $\mathbf{B}_{14}$ | = | $\left(\frac{1}{2} + x_3\right)\mathbf{a}_1 + \left(\frac{1}{2} - y_3\right)\mathbf{a}_2 + \left(\frac{3}{4} - z_3\right)\mathbf{a}_3$ | = | $\left(\frac{1}{2} + x_3\right) a\,\hat{\mathbf{x}} + \left(\frac{1}{2} - y_3\right) a\,\hat{\mathbf{y}} + \left(\frac{3}{4} - z_3\right) c\,\hat{\mathbf{z}}$ | (8$b$) | H I |
| $\mathbf{B}_{15}$ | = | $y_3\,\mathbf{a}_1 + x_3\,\mathbf{a}_2 - z_3\,\mathbf{a}_3$ | = | $y_3 a\,\hat{\mathbf{x}} + x_3 a\,\hat{\mathbf{y}} - z_3 c\,\hat{\mathbf{z}}$ | (8$b$) | H I |
| $\mathbf{B}_{16}$ | = | $-y_3\,\mathbf{a}_1 - x_3\,\mathbf{a}_2 + \left(\frac{1}{2} - z_3\right)\mathbf{a}_3$ | = | $-y_3 a\,\hat{\mathbf{x}} - x_3 a\,\hat{\mathbf{y}} + \left(\frac{1}{2} - z_3\right) c\,\hat{\mathbf{z}}$ | (8$b$) | H I |
| $\mathbf{B}_{17}$ | = | $x_4\,\mathbf{a}_1 + y_4\,\mathbf{a}_2 + z_4\,\mathbf{a}_3$ | = | $x_4 a\,\hat{\mathbf{x}} + y_4 a\,\hat{\mathbf{y}} + z_4 c\,\hat{\mathbf{z}}$ | (8$b$) | H II |
| $\mathbf{B}_{18}$ | = | $-x_4\,\mathbf{a}_1 - y_4\,\mathbf{a}_2 + \left(\frac{1}{2} + z_4\right)\mathbf{a}_3$ | = | $-x_4 a\,\hat{\mathbf{x}} - y_4 a\,\hat{\mathbf{y}} + \left(\frac{1}{2} + z_4\right) c\,\hat{\mathbf{z}}$ | (8$b$) | H II |
| $\mathbf{B}_{19}$ | = | $\left(\frac{1}{2} - y_4\right)\mathbf{a}_1 + \left(\frac{1}{2} + x_4\right)\mathbf{a}_2 + \left(\frac{1}{4} + z_4\right)\mathbf{a}_3$ | = | $\left(\frac{1}{2} - y_4\right) a\,\hat{\mathbf{x}} + \left(\frac{1}{2} + x_4\right) a\,\hat{\mathbf{y}} + \left(\frac{1}{4} + z_4\right) c\,\hat{\mathbf{z}}$ | (8$b$) | H II |
| $\mathbf{B}_{20}$ | = | $\left(\frac{1}{2} + y_4\right)\mathbf{a}_1 + \left(\frac{1}{2} - x_4\right)\mathbf{a}_2 + \left(\frac{3}{4} + z_4\right)\mathbf{a}_3$ | = | $\left(\frac{1}{2} + y_4\right) a\,\hat{\mathbf{x}} + \left(\frac{1}{2} - x_4\right) a\,\hat{\mathbf{y}} + \left(\frac{3}{4} + z_4\right) c\,\hat{\mathbf{z}}$ | (8$b$) | H II |
| $\mathbf{B}_{21}$ | = | $\left(\frac{1}{2} - x_4\right)\mathbf{a}_1 + \left(\frac{1}{2} + y_4\right)\mathbf{a}_2 + \left(\frac{1}{4} - z_4\right)\mathbf{a}_3$ | = | $\left(\frac{1}{2} - x_4\right) a\,\hat{\mathbf{x}} + \left(\frac{1}{2} + y_4\right) a\,\hat{\mathbf{y}} + \left(\frac{1}{4} - z_4\right) c\,\hat{\mathbf{z}}$ | (8$b$) | H II |
| $\mathbf{B}_{22}$ | = | $\left(\frac{1}{2} + x_4\right)\mathbf{a}_1 + \left(\frac{1}{2} - y_4\right)\mathbf{a}_2 + \left(\frac{3}{4} - z_4\right)\mathbf{a}_3$ | = | $\left(\frac{1}{2} + x_4\right) a\,\hat{\mathbf{x}} + \left(\frac{1}{2} - y_4\right) a\,\hat{\mathbf{y}} + \left(\frac{3}{4} - z_4\right) c\,\hat{\mathbf{z}}$ | (8$b$) | H II |
| $\mathbf{B}_{23}$ | = | $y_4\,\mathbf{a}_1 + x_4\,\mathbf{a}_2 - z_4\,\mathbf{a}_3$ | = | $y_4 a\,\hat{\mathbf{x}} + x_4 a\,\hat{\mathbf{y}} - z_4 c\,\hat{\mathbf{z}}$ | (8$b$) | H II |
| $\mathbf{B}_{24}$ | = | $-y_4\,\mathbf{a}_1 - x_4\,\mathbf{a}_2 + \left(\frac{1}{2} - z_4\right)\mathbf{a}_3$ | = | $-y_4 a\,\hat{\mathbf{x}} - x_4 a\,\hat{\mathbf{y}} + \left(\frac{1}{2} - z_4\right) c\,\hat{\mathbf{z}}$ | (8$b$) | H II |
| $\mathbf{B}_{25}$ | = | $x_5\,\mathbf{a}_1 + y_5\,\mathbf{a}_2 + z_5\,\mathbf{a}_3$ | = | $x_5 a\,\hat{\mathbf{x}} + y_5 a\,\hat{\mathbf{y}} + z_5 c\,\hat{\mathbf{z}}$ | (8$b$) | H III |
| $\mathbf{B}_{26}$ | = | $-x_5\,\mathbf{a}_1 - y_5\,\mathbf{a}_2 + \left(\frac{1}{2} + z_5\right)\mathbf{a}_3$ | = | $-x_5 a\,\hat{\mathbf{x}} - y_5 a\,\hat{\mathbf{y}} + \left(\frac{1}{2} + z_5\right) c\,\hat{\mathbf{z}}$ | (8$b$) | H III |
| $\mathbf{B}_{27}$ | = | $\left(\frac{1}{2} - y_5\right)\mathbf{a}_1 + \left(\frac{1}{2} + x_5\right)\mathbf{a}_2 + \left(\frac{1}{4} + z_5\right)\mathbf{a}_3$ | = | $\left(\frac{1}{2} - y_5\right) a\,\hat{\mathbf{x}} + \left(\frac{1}{2} + x_5\right) a\,\hat{\mathbf{y}} + \left(\frac{1}{4} + z_5\right) c\,\hat{\mathbf{z}}$ | (8$b$) | H III |
| $\mathbf{B}_{28}$ | = | $\left(\frac{1}{2} + y_5\right)\mathbf{a}_1 + \left(\frac{1}{2} - x_5\right)\mathbf{a}_2 + \left(\frac{3}{4} + z_5\right)\mathbf{a}_3$ | = | $\left(\frac{1}{2} + y_5\right) a\,\hat{\mathbf{x}} + \left(\frac{1}{2} - x_5\right) a\,\hat{\mathbf{y}} + \left(\frac{3}{4} + z_5\right) c\,\hat{\mathbf{z}}$ | (8$b$) | H III |
| $\mathbf{B}_{29}$ | = | $\left(\frac{1}{2} - x_5\right)\mathbf{a}_1 + \left(\frac{1}{2} + y_5\right)\mathbf{a}_2 + \left(\frac{1}{4} - z_5\right)\mathbf{a}_3$ | = | $\left(\frac{1}{2} - x_5\right) a\,\hat{\mathbf{x}} + \left(\frac{1}{2} + y_5\right) a\,\hat{\mathbf{y}} + \left(\frac{1}{4} - z_5\right) c\,\hat{\mathbf{z}}$ | (8$b$) | H III |
| $\mathbf{B}_{30}$ | = | $\left(\frac{1}{2} + x_5\right)\mathbf{a}_1 + \left(\frac{1}{2} - y_5\right)\mathbf{a}_2 + \left(\frac{3}{4} - z_5\right)\mathbf{a}_3$ | = | $\left(\frac{1}{2} + x_5\right) a\,\hat{\mathbf{x}} + \left(\frac{1}{2} - y_5\right) a\,\hat{\mathbf{y}} + \left(\frac{3}{4} - z_5\right) c\,\hat{\mathbf{z}}$ | (8$b$) | H III |
| $\mathbf{B}_{31}$ | = | $y_5\,\mathbf{a}_1 + x_5\,\mathbf{a}_2 - z_5\,\mathbf{a}_3$ | = | $y_5 a\,\hat{\mathbf{x}} + x_5 a\,\hat{\mathbf{y}} - z_5 c\,\hat{\mathbf{z}}$ | (8$b$) | H III |
| $\mathbf{B}_{32}$ | = | $-y_5\,\mathbf{a}_1 - x_5\,\mathbf{a}_2 + \left(\frac{1}{2} - z_5\right)\mathbf{a}_3$ | = | $-y_5 a\,\hat{\mathbf{x}} - x_5 a\,\hat{\mathbf{y}} + \left(\frac{1}{2} - z_5\right) c\,\hat{\mathbf{z}}$ | (8$b$) | H III |
| $\mathbf{B}_{33}$ | = | $x_6\,\mathbf{a}_1 + y_6\,\mathbf{a}_2 + z_6\,\mathbf{a}_3$ | = | $x_6 a\,\hat{\mathbf{x}} + y_6 a\,\hat{\mathbf{y}} + z_6 c\,\hat{\mathbf{z}}$ | (8*b*) | H IV |
| $\mathbf{B}_{34}$ | = | $-x_6\,\mathbf{a}_1 - y_6\,\mathbf{a}_2 + \left(\frac{1}{2} + z_6\right)\mathbf{a}_3$ | = | $-x_6 a\,\hat{\mathbf{x}} - y_6 a\,\hat{\mathbf{y}} + \left(\frac{1}{2} + z_6\right) c\,\hat{\mathbf{z}}$ | (8*b*) | H IV |
| $\mathbf{B}_{35}$ | = | $\left(\frac{1}{2} - y_6\right)\mathbf{a}_1 + \left(\frac{1}{2} + x_6\right)\mathbf{a}_2 + \left(\frac{1}{4} + z_6\right)\mathbf{a}_3$ | = | $\left(\frac{1}{2} - y_6\right) a\,\hat{\mathbf{x}} + \left(\frac{1}{2} + x_6\right) a\,\hat{\mathbf{y}} + \left(\frac{1}{4} + z_6\right) c\,\hat{\mathbf{z}}$ | (8*b*) | H IV |
| $\mathbf{B}_{36}$ | = | $\left(\frac{1}{2} + y_6\right)\mathbf{a}_1 + \left(\frac{1}{2} - x_6\right)\mathbf{a}_2 + \left(\frac{3}{4} + z_6\right)\mathbf{a}_3$ | = | $\left(\frac{1}{2} + y_6\right) a\,\hat{\mathbf{x}} + \left(\frac{1}{2} - x_6\right) a\,\hat{\mathbf{y}} + \left(\frac{3}{4} + z_6\right) c\,\hat{\mathbf{z}}$ | (8*b*) | H IV |
| $\mathbf{B}_{37}$ | = | $\left(\frac{1}{2} - x_6\right)\mathbf{a}_1 + \left(\frac{1}{2} + y_6\right)\mathbf{a}_2 + \left(\frac{1}{4} - z_6\right)\mathbf{a}_3$ | = | $\left(\frac{1}{2} - x_6\right) a\,\hat{\mathbf{x}} + \left(\frac{1}{2} + y_6\right) a\,\hat{\mathbf{y}} + \left(\frac{1}{4} - z_6\right) c\,\hat{\mathbf{z}}$ | (8*b*) | H IV |
| $\mathbf{B}_{38}$ | = | $\left(\frac{1}{2} + x_6\right)\mathbf{a}_1 + \left(\frac{1}{2} - y_6\right)\mathbf{a}_2 + \left(\frac{3}{4} - z_6\right)\mathbf{a}_3$ | = | $\left(\frac{1}{2} + x_6\right) a\,\hat{\mathbf{x}} + \left(\frac{1}{2} - y_6\right) a\,\hat{\mathbf{y}} + \left(\frac{3}{4} - z_6\right) c\,\hat{\mathbf{z}}$ | (8*b*) | H IV |
| $\mathbf{B}_{39}$ | = | $y_6\,\mathbf{a}_1 + x_6\,\mathbf{a}_2 - z_6\,\mathbf{a}_3$ | = | $y_6 a\,\hat{\mathbf{x}} + x_6 a\,\hat{\mathbf{y}} - z_6 c\,\hat{\mathbf{z}}$ | (8*b*) | H IV |
| $\mathbf{B}_{40}$ | = | $-y_6\,\mathbf{a}_1 - x_6\,\mathbf{a}_2 + \left(\frac{1}{2} - z_6\right)\mathbf{a}_3$ | = | $-y_6 a\,\hat{\mathbf{x}} - x_6 a\,\hat{\mathbf{y}} + \left(\frac{1}{2} - z_6\right) c\,\hat{\mathbf{z}}$ | (8*b*) | H IV |
| $\mathbf{B}_{41}$ | = | $x_7\,\mathbf{a}_1 + y_7\,\mathbf{a}_2 + z_7\,\mathbf{a}_3$ | = | $x_7 a\,\hat{\mathbf{x}} + y_7 a\,\hat{\mathbf{y}} + z_7 c\,\hat{\mathbf{z}}$ | (8*b*) | H V |
| $\mathbf{B}_{42}$ | = | $-x_7\,\mathbf{a}_1 - y_7\,\mathbf{a}_2 + \left(\frac{1}{2} + z_7\right)\mathbf{a}_3$ | = | $-x_7 a\,\hat{\mathbf{x}} - y_7 a\,\hat{\mathbf{y}} + \left(\frac{1}{2} + z_7\right) c\,\hat{\mathbf{z}}$ | (8*b*) | H V |
| $\mathbf{B}_{43}$ | = | $\left(\frac{1}{2} - y_7\right)\mathbf{a}_1 + \left(\frac{1}{2} + x_7\right)\mathbf{a}_2 + \left(\frac{1}{4} + z_7\right)\mathbf{a}_3$ | = | $\left(\frac{1}{2} - y_7\right) a\,\hat{\mathbf{x}} + \left(\frac{1}{2} + x_7\right) a\,\hat{\mathbf{y}} + \left(\frac{1}{4} + z_7\right) c\,\hat{\mathbf{z}}$ | (8*b*) | H V |
| $\mathbf{B}_{44}$ | = | $\left(\frac{1}{2} + y_7\right)\mathbf{a}_1 + \left(\frac{1}{2} - x_7\right)\mathbf{a}_2 + \left(\frac{3}{4} + z_7\right)\mathbf{a}_3$ | = | $\left(\frac{1}{2} + y_7\right) a\,\hat{\mathbf{x}} + \left(\frac{1}{2} - x_7\right) a\,\hat{\mathbf{y}} + \left(\frac{3}{4} + z_7\right) c\,\hat{\mathbf{z}}$ | (8*b*) | H V |
| $\mathbf{B}_{45}$ | = | $\left(\frac{1}{2} - x_7\right)\mathbf{a}_1 + \left(\frac{1}{2} + y_7\right)\mathbf{a}_2 + \left(\frac{1}{4} - z_7\right)\mathbf{a}_3$ | = | $\left(\frac{1}{2} - x_7\right) a\,\hat{\mathbf{x}} + \left(\frac{1}{2} + y_7\right) a\,\hat{\mathbf{y}} + \left(\frac{1}{4} - z_7\right) c\,\hat{\mathbf{z}}$ | (8*b*) | H V |
| $\mathbf{B}_{46}$ | = | $\left(\frac{1}{2} + x_7\right)\mathbf{a}_1 + \left(\frac{1}{2} - y_7\right)\mathbf{a}_2 + \left(\frac{3}{4} - z_7\right)\mathbf{a}_3$ | = | $\left(\frac{1}{2} + x_7\right) a\,\hat{\mathbf{x}} + \left(\frac{1}{2} - y_7\right) a\,\hat{\mathbf{y}} + \left(\frac{3}{4} - z_7\right) c\,\hat{\mathbf{z}}$ | (8*b*) | H V |
| $\mathbf{B}_{47}$ | = | $y_7\,\mathbf{a}_1 + x_7\,\mathbf{a}_2 - z_7\,\mathbf{a}_3$ | = | $y_7 a\,\hat{\mathbf{x}} + x_7 a\,\hat{\mathbf{y}} - z_7 c\,\hat{\mathbf{z}}$ | (8*b*) | H V |
| $\mathbf{B}_{48}$ | = | $-y_7\,\mathbf{a}_1 - x_7\,\mathbf{a}_2 + \left(\frac{1}{2} - z_7\right)\mathbf{a}_3$ | = | $-y_7 a\,\hat{\mathbf{x}} - x_7 a\,\hat{\mathbf{y}} + \left(\frac{1}{2} - z_7\right) c\,\hat{\mathbf{z}}$ | (8*b*) | H V |
| $\mathbf{B}_{49}$ | = | $x_8\,\mathbf{a}_1 + y_8\,\mathbf{a}_2 + z_8\,\mathbf{a}_3$ | = | $x_8 a\,\hat{\mathbf{x}} + y_8 a\,\hat{\mathbf{y}} + z_8 c\,\hat{\mathbf{z}}$ | (8*b*) | H VI |
| $\mathbf{B}_{50}$ | = | $-x_8\,\mathbf{a}_1 - y_8\,\mathbf{a}_2 + \left(\frac{1}{2} + z_8\right)\mathbf{a}_3$ | = | $-x_8 a\,\hat{\mathbf{x}} - y_8 a\,\hat{\mathbf{y}} + \left(\frac{1}{2} + z_8\right) c\,\hat{\mathbf{z}}$ | (8*b*) | H VI |
| $\mathbf{B}_{51}$ | = | $\left(\frac{1}{2} - y_8\right)\mathbf{a}_1 + \left(\frac{1}{2} + x_8\right)\mathbf{a}_2 + \left(\frac{1}{4} + z_8\right)\mathbf{a}_3$ | = | $\left(\frac{1}{2} - y_8\right) a\,\hat{\mathbf{x}} + \left(\frac{1}{2} + x_8\right) a\,\hat{\mathbf{y}} + \left(\frac{1}{4} + z_8\right) c\,\hat{\mathbf{z}}$ | (8*b*) | H VI |
| $\mathbf{B}_{52}$ | = | $\left(\frac{1}{2} + y_8\right)\mathbf{a}_1 + \left(\frac{1}{2} - x_8\right)\mathbf{a}_2 + \left(\frac{3}{4} + z_8\right)\mathbf{a}_3$ | = | $\left(\frac{1}{2} + y_8\right) a\,\hat{\mathbf{x}} + \left(\frac{1}{2} - x_8\right) a\,\hat{\mathbf{y}} + \left(\frac{3}{4} + z_8\right) c\,\hat{\mathbf{z}}$ | (8*b*) | H VI |
| $\mathbf{B}_{53}$ | = | $\left(\frac{1}{2} - x_8\right)\mathbf{a}_1 + \left(\frac{1}{2} + y_8\right)\mathbf{a}_2 + \left(\frac{1}{4} - z_8\right)\mathbf{a}_3$ | = | $\left(\frac{1}{2} - x_8\right) a\,\hat{\mathbf{x}} + \left(\frac{1}{2} + y_8\right) a\,\hat{\mathbf{y}} + \left(\frac{1}{4} - z_8\right) c\,\hat{\mathbf{z}}$ | (8*b*) | H VI |
| $\mathbf{B}_{54}$ | = | $\left(\frac{1}{2} + x_8\right)\mathbf{a}_1 + \left(\frac{1}{2} - y_8\right)\mathbf{a}_2 + \left(\frac{3}{4} - z_8\right)\mathbf{a}_3$ | = | $\left(\frac{1}{2} + x_8\right) a\,\hat{\mathbf{x}} + \left(\frac{1}{2} - y_8\right) a\,\hat{\mathbf{y}} + \left(\frac{3}{4} - z_8\right) c\,\hat{\mathbf{z}}$ | (8*b*) | H VI |
| $\mathbf{B}_{55}$ | = | $y_8\,\mathbf{a}_1 + x_8\,\mathbf{a}_2 - z_8\,\mathbf{a}_3$ | = | $y_8 a\,\hat{\mathbf{x}} + x_8 a\,\hat{\mathbf{y}} - z_8 c\,\hat{\mathbf{z}}$ | (8*b*) | H VI |
| $\mathbf{B}_{56}$ | = | $-y_8\,\mathbf{a}_1 - x_8\,\mathbf{a}_2 + \left(\frac{1}{2} - z_8\right)\mathbf{a}_3$ | = | $-y_8 a\,\hat{\mathbf{x}} - x_8 a\,\hat{\mathbf{y}} + \left(\frac{1}{2} - z_8\right) c\,\hat{\mathbf{z}}$ | (8*b*) | H VI |
| $\mathbf{B}_{57}$ | = | $x_9\,\mathbf{a}_1 + y_9\,\mathbf{a}_2 + z_9\,\mathbf{a}_3$ | = | $x_9 a\,\hat{\mathbf{x}} + y_9 a\,\hat{\mathbf{y}} + z_9 c\,\hat{\mathbf{z}}$ | (8*b*) | O I |
| $\mathbf{B}_{58}$ | = | $-x_9\,\mathbf{a}_1 - y_9\,\mathbf{a}_2 + \left(\frac{1}{2} + z_9\right)\mathbf{a}_3$ | = | $-x_9 a\,\hat{\mathbf{x}} - y_9 a\,\hat{\mathbf{y}} + \left(\frac{1}{2} + z_9\right) c\,\hat{\mathbf{z}}$ | (8*b*) | O I |
| $\mathbf{B}_{59}$ | = | $\left(\frac{1}{2} - y_9\right) \mathbf{a}_1 + \left(\frac{1}{2} + x_9\right) \mathbf{a}_2 + \left(\frac{1}{4} + z_9\right) \mathbf{a}_3$ | = | $\left(\frac{1}{2} - y_9\right) a\,\hat{\mathbf{x}} + \left(\frac{1}{2} + x_9\right) a\,\hat{\mathbf{y}} + \left(\frac{1}{4} + z_9\right) c\,\hat{\mathbf{z}}$ | (8*b*) | O I |
| $\mathbf{B}_{60}$ | = | $\left(\frac{1}{2} + y_9\right) \mathbf{a}_1 + \left(\frac{1}{2} - x_9\right) \mathbf{a}_2 + \left(\frac{3}{4} + z_9\right) \mathbf{a}_3$ | = | $\left(\frac{1}{2} + y_9\right) a\,\hat{\mathbf{x}} + \left(\frac{1}{2} - x_9\right) a\,\hat{\mathbf{y}} + \left(\frac{3}{4} + z_9\right) c\,\hat{\mathbf{z}}$ | (8*b*) | O I |
| $\mathbf{B}_{61}$ | = | $\left(\frac{1}{2} - x_9\right) \mathbf{a}_1 + \left(\frac{1}{2} + y_9\right) \mathbf{a}_2 + \left(\frac{1}{4} - z_9\right) \mathbf{a}_3$ | = | $\left(\frac{1}{2} - x_9\right) a\,\hat{\mathbf{x}} + \left(\frac{1}{2} + y_9\right) a\,\hat{\mathbf{y}} + \left(\frac{1}{4} - z_9\right) c\,\hat{\mathbf{z}}$ | (8*b*) | O I |
| $\mathbf{B}_{62}$ | = | $\left(\frac{1}{2} + x_9\right) \mathbf{a}_1 + \left(\frac{1}{2} - y_9\right) \mathbf{a}_2 + \left(\frac{3}{4} - z_9\right) \mathbf{a}_3$ | = | $\left(\frac{1}{2} + x_9\right) a\,\hat{\mathbf{x}} + \left(\frac{1}{2} - y_9\right) a\,\hat{\mathbf{y}} + \left(\frac{3}{4} - z_9\right) c\,\hat{\mathbf{z}}$ | (8*b*) | O I |
| $\mathbf{B}_{63}$ | = | $y_9\,\mathbf{a}_1 + x_9\,\mathbf{a}_2 - z_9\,\mathbf{a}_3$ | = | $y_9 a\,\hat{\mathbf{x}} + x_9 a\,\hat{\mathbf{y}} - z_9 c\,\hat{\mathbf{z}}$ | (8*b*) | O I |
| $\mathbf{B}_{64}$ | = | $-y_9\,\mathbf{a}_1 - x_9\,\mathbf{a}_2 + \left(\frac{1}{2} - z_9\right) \mathbf{a}_3$ | = | $-y_9 a\,\hat{\mathbf{x}} - x_9 a\,\hat{\mathbf{y}} + \left(\frac{1}{2} - z_9\right) c\,\hat{\mathbf{z}}$ | (8*b*) | O I |
| $\mathbf{B}_{65}$ | = | $x_{10}\,\mathbf{a}_1 + y_{10}\,\mathbf{a}_2 + z_{10}\,\mathbf{a}_3$ | = | $x_{10} a\,\hat{\mathbf{x}} + y_{10} a\,\hat{\mathbf{y}} + z_{10} c\,\hat{\mathbf{z}}$ | (8*b*) | O II |
| $\mathbf{B}_{66}$ | = | $-x_{10}\,\mathbf{a}_1 - y_{10}\,\mathbf{a}_2 + \left(\frac{1}{2} + z_{10}\right) \mathbf{a}_3$ | = | $-x_{10} a\,\hat{\mathbf{x}} - y_{10} a\,\hat{\mathbf{y}} + \left(\frac{1}{2} + z_{10}\right) c\,\hat{\mathbf{z}}$ | (8*b*) | O II |
| $\mathbf{B}_{67}$ | = | $\left(\frac{1}{2} - y_{10}\right) \mathbf{a}_1 + \left(\frac{1}{2} + x_{10}\right) \mathbf{a}_2 + \left(\frac{1}{4} + z_{10}\right) \mathbf{a}_3$ | = | $\left(\frac{1}{2} - y_{10}\right) a\,\hat{\mathbf{x}} + \left(\frac{1}{2} + x_{10}\right) a\,\hat{\mathbf{y}} + \left(\frac{1}{4} + z_{10}\right) c\,\hat{\mathbf{z}}$ | (8*b*) | O II |
| $\mathbf{B}_{68}$ | = | $\left(\frac{1}{2} + y_{10}\right) \mathbf{a}_1 + \left(\frac{1}{2} - x_{10}\right) \mathbf{a}_2 + \left(\frac{3}{4} + z_{10}\right) \mathbf{a}_3$ | = | $\left(\frac{1}{2} + y_{10}\right) a\,\hat{\mathbf{x}} + \left(\frac{1}{2} - x_{10}\right) a\,\hat{\mathbf{y}} + \left(\frac{3}{4} + z_{10}\right) c\,\hat{\mathbf{z}}$ | (8*b*) | O II |
| $\mathbf{B}_{69}$ | = | $\left(\frac{1}{2} - x_{10}\right) \mathbf{a}_1 + \left(\frac{1}{2} + y_{10}\right) \mathbf{a}_2 + \left(\frac{1}{4} - z_{10}\right) \mathbf{a}_3$ | = | $\left(\frac{1}{2} - x_{10}\right) a\,\hat{\mathbf{x}} + \left(\frac{1}{2} + y_{10}\right) a\,\hat{\mathbf{y}} + \left(\frac{1}{4} - z_{10}\right) c\,\hat{\mathbf{z}}$ | (8*b*) | O II |
| $\mathbf{B}_{70}$ | = | $\left(\frac{1}{2} + x_{10}\right) \mathbf{a}_1 + \left(\frac{1}{2} - y_{10}\right) \mathbf{a}_2 + \left(\frac{3}{4} - z_{10}\right) \mathbf{a}_3$ | = | $\left(\frac{1}{2} + x_{10}\right) a\,\hat{\mathbf{x}} + \left(\frac{1}{2} - y_{10}\right) a\,\hat{\mathbf{y}} + \left(\frac{3}{4} - z_{10}\right) c\,\hat{\mathbf{z}}$ | (8*b*) | O II |
| $\mathbf{B}_{71}$ | = | $y_{10}\,\mathbf{a}_1 + x_{10}\,\mathbf{a}_2 - z_{10}\,\mathbf{a}_3$ | = | $y_{10} a\,\hat{\mathbf{x}} + x_{10} a\,\hat{\mathbf{y}} - z_{10} c\,\hat{\mathbf{z}}$ | (8*b*) | O II |
| $\mathbf{B}_{72}$ | = | $-y_{10}\,\mathbf{a}_1 - x_{10}\,\mathbf{a}_2 + \left(\frac{1}{2} - z_{10}\right) \mathbf{a}_3$ | = | $-y_{10} a\,\hat{\mathbf{x}} - x_{10} a\,\hat{\mathbf{y}} + \left(\frac{1}{2} - z_{10}\right) c\,\hat{\mathbf{z}}$ | (8*b*) | O II |
| $\mathbf{B}_{73}$ | = | $x_{11}\,\mathbf{a}_1 + y_{11}\,\mathbf{a}_2 + z_{11}\,\mathbf{a}_3$ | = | $x_{11} a\,\hat{\mathbf{x}} + y_{11} a\,\hat{\mathbf{y}} + z_{11} c\,\hat{\mathbf{z}}$ | (8*b*) | O III |
| $\mathbf{B}_{74}$ | = | $-x_{11}\,\mathbf{a}_1 - y_{11}\,\mathbf{a}_2 + \left(\frac{1}{2} + z_{11}\right) \mathbf{a}_3$ | = | $-x_{11} a\,\hat{\mathbf{x}} - y_{11} a\,\hat{\mathbf{y}} + \left(\frac{1}{2} + z_{11}\right) c\,\hat{\mathbf{z}}$ | (8*b*) | O III |
| $\mathbf{B}_{75}$ | = | $\left(\frac{1}{2} - y_{11}\right) \mathbf{a}_1 + \left(\frac{1}{2} + x_{11}\right) \mathbf{a}_2 + \left(\frac{1}{4} + z_{11}\right) \mathbf{a}_3$ | = | $\left(\frac{1}{2} - y_{11}\right) a\,\hat{\mathbf{x}} + \left(\frac{1}{2} + x_{11}\right) a\,\hat{\mathbf{y}} + \left(\frac{1}{4} + z_{11}\right) c\,\hat{\mathbf{z}}$ | (8*b*) | O III |
| $\mathbf{B}_{76}$ | = | $\left(\frac{1}{2} + y_{11}\right) \mathbf{a}_1 + \left(\frac{1}{2} - x_{11}\right) \mathbf{a}_2 + \left(\frac{3}{4} + z_{11}\right) \mathbf{a}_3$ | = | $\left(\frac{1}{2} + y_{11}\right) a\,\hat{\mathbf{x}} + \left(\frac{1}{2} - x_{11}\right) a\,\hat{\mathbf{y}} + \left(\frac{3}{4} + z_{11}\right) c\,\hat{\mathbf{z}}$ | (8*b*) | O III |
| $\mathbf{B}_{77}$ | = | $\left(\frac{1}{2} - x_{11}\right) \mathbf{a}_1 + \left(\frac{1}{2} + y_{11}\right) \mathbf{a}_2 + \left(\frac{1}{4} - z_{11}\right) \mathbf{a}_3$ | = | $\left(\frac{1}{2} - x_{11}\right) a\,\hat{\mathbf{x}} + \left(\frac{1}{2} + y_{11}\right) a\,\hat{\mathbf{y}} + \left(\frac{1}{4} - z_{11}\right) c\,\hat{\mathbf{z}}$ | (8*b*) | O III |
| $\mathbf{B}_{78}$ | = | $\left(\frac{1}{2} + x_{11}\right) \mathbf{a}_1 + \left(\frac{1}{2} - y_{11}\right) \mathbf{a}_2 + \left(\frac{3}{4} - z_{11}\right) \mathbf{a}_3$ | = | $\left(\frac{1}{2} + x_{11}\right) a\,\hat{\mathbf{x}} + \left(\frac{1}{2} - y_{11}\right) a\,\hat{\mathbf{y}} + \left(\frac{3}{4} - z_{11}\right) c\,\hat{\mathbf{z}}$ | (8*b*) | O III |
| $\mathbf{B}_{79}$ | = | $y_{11}\,\mathbf{a}_1 + x_{11}\,\mathbf{a}_2 - z_{11}\,\mathbf{a}_3$ | = | $y_{11} a\,\hat{\mathbf{x}} + x_{11} a\,\hat{\mathbf{y}} - z_{11} c\,\hat{\mathbf{z}}$ | (8*b*) | O III |
| $\mathbf{B}_{80}$ | = | $-y_{11}\,\mathbf{a}_1 - x_{11}\,\mathbf{a}_2 + \left(\frac{1}{2} - z_{11}\right) \mathbf{a}_3$ | = | $-y_{11} a\,\hat{\mathbf{x}} - x_{11} a\,\hat{\mathbf{y}} + \left(\frac{1}{2} - z_{11}\right) c\,\hat{\mathbf{z}}$ | (8*b*) | O III |
| $\mathbf{B}_{81}$ | = | $x_{12}\,\mathbf{a}_1 + y_{12}\,\mathbf{a}_2 + z_{12}\,\mathbf{a}_3$ | = | $x_{12} a\,\hat{\mathbf{x}} + y_{12} a\,\hat{\mathbf{y}} + z_{12} c\,\hat{\mathbf{z}}$ | (8*b*) | O IV |
| $\mathbf{B}_{82}$ | = | $-x_{12}\,\mathbf{a}_1 - y_{12}\,\mathbf{a}_2 + \left(\frac{1}{2} + z_{12}\right) \mathbf{a}_3$ | = | $-x_{12} a\,\hat{\mathbf{x}} - y_{12} a\,\hat{\mathbf{y}} + \left(\frac{1}{2} + z_{12}\right) c\,\hat{\mathbf{z}}$ | (8*b*) | O IV |
| $\mathbf{B}_{83}$ | = | $\left(\frac{1}{2} - y_{12}\right) \mathbf{a}_1 + \left(\frac{1}{2} + x_{12}\right) \mathbf{a}_2 + \left(\frac{1}{4} + z_{12}\right) \mathbf{a}_3$ | = | $\left(\frac{1}{2} - y_{12}\right) a\,\hat{\mathbf{x}} + \left(\frac{1}{2} + x_{12}\right) a\,\hat{\mathbf{y}} + \left(\frac{1}{4} + z_{12}\right) c\,\hat{\mathbf{z}}$ | (8*b*) | O IV |
| $\mathbf{B}_{84}$ | = | $\left(\frac{1}{2}+y_{12}\right)\mathbf{a}_1+\left(\frac{1}{2}-x_{12}\right)\mathbf{a}_2+\left(\frac{3}{4}+z_{12}\right)\mathbf{a}_3$ | = | $\left(\frac{1}{2}+y_{12}\right)a\,\hat{\mathbf{x}}+\left(\frac{1}{2}-x_{12}\right)a\,\hat{\mathbf{y}}+\left(\frac{3}{4}+z_{12}\right)c\,\hat{\mathbf{z}}$ | (8$b$) | O IV |
| $\mathbf{B}_{85}$ | = | $\left(\frac{1}{2}-x_{12}\right)\mathbf{a}_1+\left(\frac{1}{2}+y_{12}\right)\mathbf{a}_2+\left(\frac{1}{4}-z_{12}\right)\mathbf{a}_3$ | = | $\left(\frac{1}{2}-x_{12}\right)a\,\hat{\mathbf{x}}+\left(\frac{1}{2}+y_{12}\right)a\,\hat{\mathbf{y}}+\left(\frac{1}{4}-z_{12}\right)c\,\hat{\mathbf{z}}$ | (8$b$) | O IV |
| $\mathbf{B}_{86}$ | = | $\left(\frac{1}{2}+x_{12}\right)\mathbf{a}_1+\left(\frac{1}{2}-y_{12}\right)\mathbf{a}_2+\left(\frac{3}{4}-z_{12}\right)\mathbf{a}_3$ | = | $\left(\frac{1}{2}+x_{12}\right)a\,\hat{\mathbf{x}}+\left(\frac{1}{2}-y_{12}\right)a\,\hat{\mathbf{y}}+\left(\frac{3}{4}-z_{12}\right)c\,\hat{\mathbf{z}}$ | (8$b$) | O IV |
| $\mathbf{B}_{87}$ | = | $y_{12}\,\mathbf{a}_1+x_{12}\,\mathbf{a}_2-z_{12}\,\mathbf{a}_3$ | = | $y_{12}a\,\hat{\mathbf{x}}+x_{12}a\,\hat{\mathbf{y}}-z_{12}c\,\hat{\mathbf{z}}$ | (8$b$) | O IV |
| $\mathbf{B}_{88}$ | = | $-y_{12}\,\mathbf{a}_1-x_{12}\,\mathbf{a}_2+\left(\frac{1}{2}-z_{12}\right)\mathbf{a}_3$ | = | $-y_{12}a\,\hat{\mathbf{x}}-x_{12}a\,\hat{\mathbf{y}}+\left(\frac{1}{2}-z_{12}\right)c\,\hat{\mathbf{z}}$ | (8$b$) | O IV |
| $\mathbf{B}_{89}$ | = | $x_{13}\,\mathbf{a}_1+y_{13}\,\mathbf{a}_2+z_{13}\,\mathbf{a}_3$ | = | $x_{13}a\,\hat{\mathbf{x}}+y_{13}a\,\hat{\mathbf{y}}+z_{13}c\,\hat{\mathbf{z}}$ | (8$b$) | O V |
| $\mathbf{B}_{90}$ | = | $-x_{13}\,\mathbf{a}_1-y_{13}\,\mathbf{a}_2+\left(\frac{1}{2}+z_{13}\right)\mathbf{a}_3$ | = | $-x_{13}a\,\hat{\mathbf{x}}-y_{13}a\,\hat{\mathbf{y}}+\left(\frac{1}{2}+z_{13}\right)c\,\hat{\mathbf{z}}$ | (8$b$) | O V |
| $\mathbf{B}_{91}$ | = | $\left(\frac{1}{2}-y_{13}\right)\mathbf{a}_1+\left(\frac{1}{2}+x_{13}\right)\mathbf{a}_2+\left(\frac{1}{4}+z_{13}\right)\mathbf{a}_3$ | = | $\left(\frac{1}{2}-y_{13}\right)a\,\hat{\mathbf{x}}+\left(\frac{1}{2}+x_{13}\right)a\,\hat{\mathbf{y}}+\left(\frac{1}{4}+z_{13}\right)c\,\hat{\mathbf{z}}$ | (8$b$) | O V |
| $\mathbf{B}_{92}$ | = | $\left(\frac{1}{2}+y_{13}\right)\mathbf{a}_1+\left(\frac{1}{2}-x_{13}\right)\mathbf{a}_2+\left(\frac{3}{4}+z_{13}\right)\mathbf{a}_3$ | = | $\left(\frac{1}{2}+y_{13}\right)a\,\hat{\mathbf{x}}+\left(\frac{1}{2}-x_{13}\right)a\,\hat{\mathbf{y}}+\left(\frac{3}{4}+z_{13}\right)c\,\hat{\mathbf{z}}$ | (8$b$) | O V |
| $\mathbf{B}_{93}$ | = | $\left(\frac{1}{2}-x_{13}\right)\mathbf{a}_1+\left(\frac{1}{2}+y_{13}\right)\mathbf{a}_2+\left(\frac{1}{4}-z_{13}\right)\mathbf{a}_3$ | = | $\left(\frac{1}{2}-x_{13}\right)a\,\hat{\mathbf{x}}+\left(\frac{1}{2}+y_{13}\right)a\,\hat{\mathbf{y}}+\left(\frac{1}{4}-z_{13}\right)c\,\hat{\mathbf{z}}$ | (8$b$) | O V |
| $\mathbf{B}_{94}$ | = | $\left(\frac{1}{2}+x_{13}\right)\mathbf{a}_1+\left(\frac{1}{2}-y_{13}\right)\mathbf{a}_2+\left(\frac{3}{4}-z_{13}\right)\mathbf{a}_3$ | = | $\left(\frac{1}{2}+x_{13}\right)a\,\hat{\mathbf{x}}+\left(\frac{1}{2}-y_{13}\right)a\,\hat{\mathbf{y}}+\left(\frac{3}{4}-z_{13}\right)c\,\hat{\mathbf{z}}$ | (8$b$) | O V |
| $\mathbf{B}_{95}$ | = | $y_{13}\,\mathbf{a}_1+x_{13}\,\mathbf{a}_2-z_{13}\,\mathbf{a}_3$ | = | $y_{13}a\,\hat{\mathbf{x}}+x_{13}a\,\hat{\mathbf{y}}-z_{13}c\,\hat{\mathbf{z}}$ | (8$b$) | O V |
| $\mathbf{B}_{96}$ | = | $-y_{13}\,\mathbf{a}_1-x_{13}\,\mathbf{a}_2+\left(\frac{1}{2}-z_{13}\right)\mathbf{a}_3$ | = | $-y_{13}a\,\hat{\mathbf{x}}-x_{13}a\,\hat{\mathbf{y}}+\left(\frac{1}{2}-z_{13}\right)c\,\hat{\mathbf{z}}$ | (8$b$) | O V |

**Found in:**

**Geometry files:**

- CIF: pp. 1677
- POSCAR: pp. 1677

# Paratellurite ($\alpha$-$TeO_2$) Structure: A2B_tP12_92_b_a

http://aflow.org/prototype-encyclopedia/A2B_tP12_92_b_a.TeO2

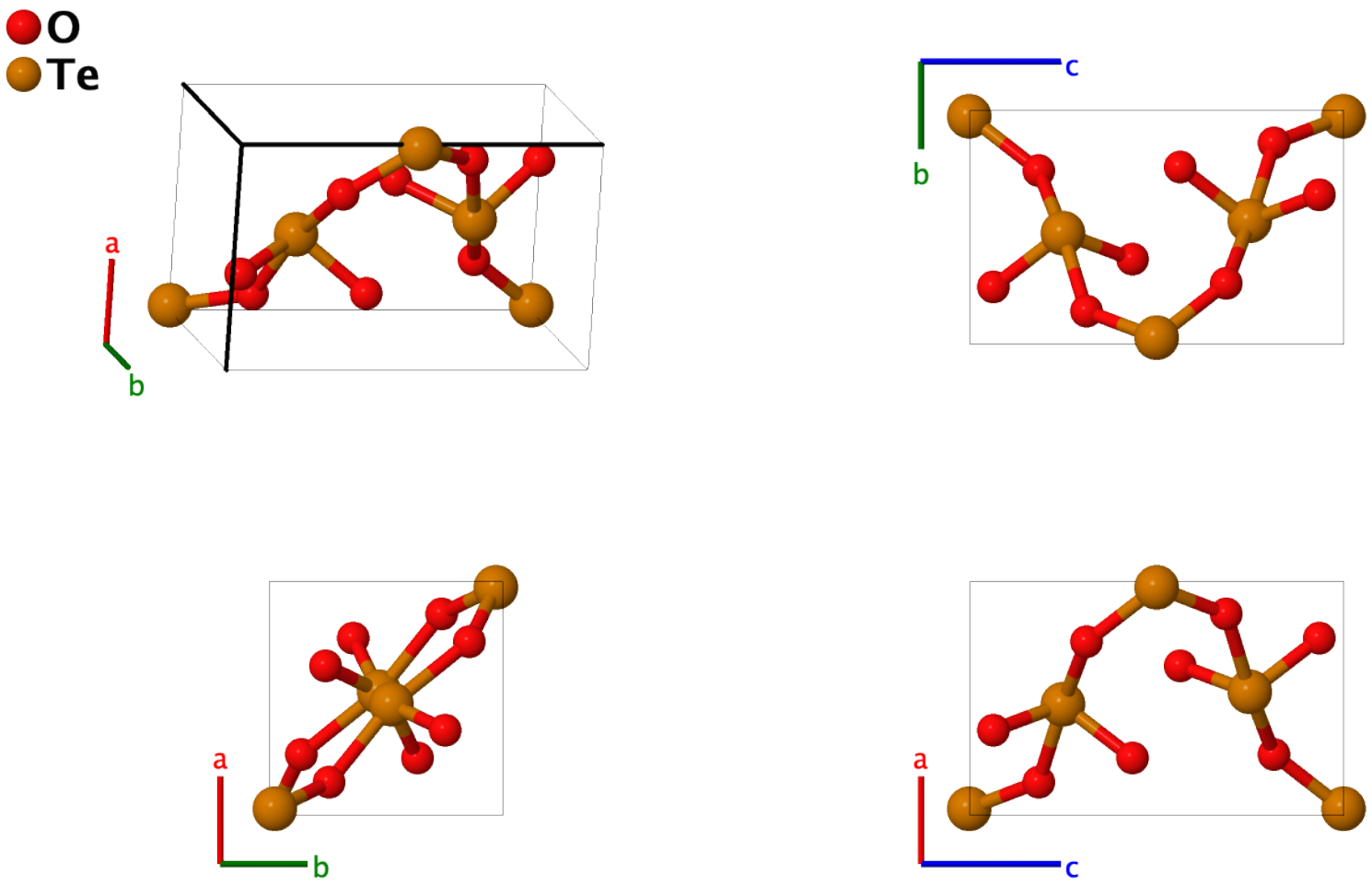

| | | |
|---|---|---|
| **Prototype** | : | $O_2Te$ |
| **AFLOW prototype label** | : | A2B_tP12_92_b_a |
| ***Strukturbericht* designation** | : | None |
| **Pearson symbol** | : | tP12 |
| **Space group number** | : | 92 |
| **Space group symbol** | : | $P4_12_12$ |
| **AFLOW prototype command** | : | `aflow --proto=A2B_tP12_92_b_a --params=`$a, c/a, x_1, x_2, y_2, z_2$ |

- Although this as the same space group and Wyckoff positions as $\alpha$-cristobalite ($C30$), the actual positions of the atoms are substantially different, so we have given the two compounds separate entries in the database.

**Simple Tetragonal primitive vectors:**

$$\mathbf{a}_1 = a\,\hat{\mathbf{x}}$$
$$\mathbf{a}_2 = a\,\hat{\mathbf{y}}$$
$$\mathbf{a}_3 = c\,\hat{\mathbf{z}}$$

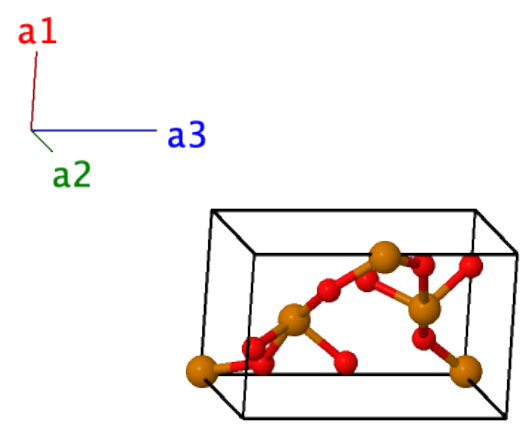

**Basis vectors:**

| | | Lattice Coordinates | | Cartesian Coordinates | Wyckoff Position | Atom Type |
|---|---|---|---|---|---|---|
| $\mathbf{B}_1$ | = | $x_1\,\mathbf{a}_1 + x_1\,\mathbf{a}_2$ | = | $x_1 a\,\hat{\mathbf{x}} + x_1 a\,\hat{\mathbf{y}}$ | (4a) | Te |
| $\mathbf{B}_2$ | = | $-x_1\,\mathbf{a}_1 - x_1\,\mathbf{a}_2 + \frac{1}{2}\,\mathbf{a}_3$ | = | $-x_1 a\,\hat{\mathbf{x}} - x_1 a\,\hat{\mathbf{y}} + \frac{1}{2}c\,\hat{\mathbf{z}}$ | (4a) | Te |
| $\mathbf{B}_3$ | = | $\left(\frac{1}{2} - x_1\right)\mathbf{a}_1 + \left(\frac{1}{2} + x_1\right)\mathbf{a}_2 + \frac{1}{4}\,\mathbf{a}_3$ | = | $\left(\frac{1}{2} - x_1\right)a\,\hat{\mathbf{x}} + \left(\frac{1}{2} + x_1\right)a\,\hat{\mathbf{y}} + \frac{1}{4}c\,\hat{\mathbf{z}}$ | (4a) | Te |
| $\mathbf{B}_4$ | = | $\left(\frac{1}{2} + x_1\right)\mathbf{a}_1 + \left(\frac{1}{2} - x_1\right)\mathbf{a}_2 + \frac{3}{4}\,\mathbf{a}_3$ | = | $\left(\frac{1}{2} + x_1\right)a\,\hat{\mathbf{x}} + \left(\frac{1}{2} - x_1\right)a\,\hat{\mathbf{y}} + \frac{3}{4}c\,\hat{\mathbf{z}}$ | (4a) | Te |
| $\mathbf{B}_5$ | = | $x_2\,\mathbf{a}_1 + y_2\,\mathbf{a}_2 + z_2\,\mathbf{a}_3$ | = | $x_2 a\,\hat{\mathbf{x}} + y_2 a\,\hat{\mathbf{y}} + z_2 c\,\hat{\mathbf{z}}$ | (8b) | O |
| $\mathbf{B}_{6}$ | = | $-x_{2}\,\mathbf{a}_{1} - y_{2}\,\mathbf{a}_{2} + \left(\frac{1}{2} + z_{2}\right)\,\mathbf{a}_{3}$ | = | $-x_{2}a\,\mathbf{\hat{x}} - y_{2}a\,\mathbf{\hat{y}} + \left(\frac{1}{2} + z_{2}\right)c\,\mathbf{\hat{z}}$ | (8$b$) | O |
| $\mathbf{B}_{7}$ | = | $\left(\frac{1}{2} - y_{2}\right)\,\mathbf{a}_{1} + \left(\frac{1}{2} + x_{2}\right)\,\mathbf{a}_{2} + \left(\frac{1}{4} + z_{2}\right)\,\mathbf{a}_{3}$ | = | $\left(\frac{1}{2} - y_{2}\right)a\,\mathbf{\hat{x}} + \left(\frac{1}{2} + x_{2}\right)a\,\mathbf{\hat{y}} + \left(\frac{1}{4} + z_{2}\right)c\,\mathbf{\hat{z}}$ | (8$b$) | O |
| $\mathbf{B}_{8}$ | = | $\left(\frac{1}{2} + y_{2}\right)\,\mathbf{a}_{1} + \left(\frac{1}{2} - x_{2}\right)\,\mathbf{a}_{2} + \left(\frac{3}{4} + z_{2}\right)\,\mathbf{a}_{3}$ | = | $\left(\frac{1}{2} + y_{2}\right)a\,\mathbf{\hat{x}} + \left(\frac{1}{2} - x_{2}\right)a\,\mathbf{\hat{y}} + \left(\frac{3}{4} + z_{2}\right)c\,\mathbf{\hat{z}}$ | (8$b$) | O |
| $\mathbf{B}_{9}$ | = | $\left(\frac{1}{2} - x_{2}\right)\,\mathbf{a}_{1} + \left(\frac{1}{2} + y_{2}\right)\,\mathbf{a}_{2} + \left(\frac{1}{4} - z_{2}\right)\,\mathbf{a}_{3}$ | = | $\left(\frac{1}{2} - x_{2}\right)a\,\mathbf{\hat{x}} + \left(\frac{1}{2} + y_{2}\right)a\,\mathbf{\hat{y}} + \left(\frac{1}{4} - z_{2}\right)c\,\mathbf{\hat{z}}$ | (8$b$) | O |
| $\mathbf{B}_{10}$ | = | $\left(\frac{1}{2} + x_{2}\right)\,\mathbf{a}_{1} + \left(\frac{1}{2} - y_{2}\right)\,\mathbf{a}_{2} + \left(\frac{3}{4} - z_{2}\right)\,\mathbf{a}_{3}$ | = | $\left(\frac{1}{2} + x_{2}\right)a\,\mathbf{\hat{x}} + \left(\frac{1}{2} - y_{2}\right)a\,\mathbf{\hat{y}} + \left(\frac{3}{4} - z_{2}\right)c\,\mathbf{\hat{z}}$ | (8$b$) | O |
| $\mathbf{B}_{11}$ | = | $y_{2}\,\mathbf{a}_{1} + x_{2}\,\mathbf{a}_{2} - z_{2}\,\mathbf{a}_{3}$ | = | $y_{2}a\,\mathbf{\hat{x}} + x_{2}a\,\mathbf{\hat{y}} - z_{2}c\,\mathbf{\hat{z}}$ | (8$b$) | O |
| $\mathbf{B}_{12}$ | = | $-y_{2}\,\mathbf{a}_{1} - x_{2}\,\mathbf{a}_{2} + \left(\frac{1}{2} - z_{2}\right)\,\mathbf{a}_{3}$ | = | $-y_{2}a\,\mathbf{\hat{x}} - x_{2}a\,\mathbf{\hat{y}} + \left(\frac{1}{2} - z_{2}\right)c\,\mathbf{\hat{z}}$ | (8$b$) | O |

**Found in:**

**Geometry files:**

- CIF: pp. 1678
- POSCAR: pp. 1678

# Phase III $Cd_2Re_2O_7$ Structure: A2B7C2_tI44_98_f_bcde_f

http://aflow.org/prototype-encyclopedia/A2B7C2_tI44_98_f_bcde_f

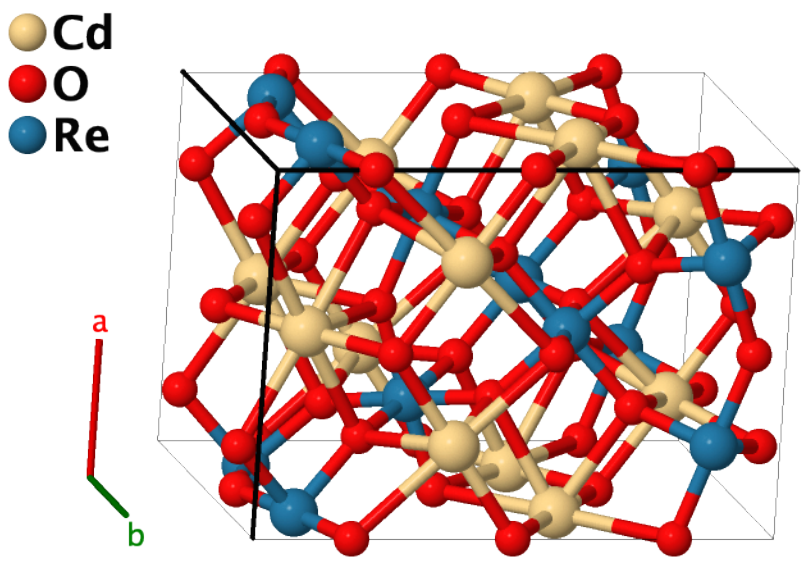

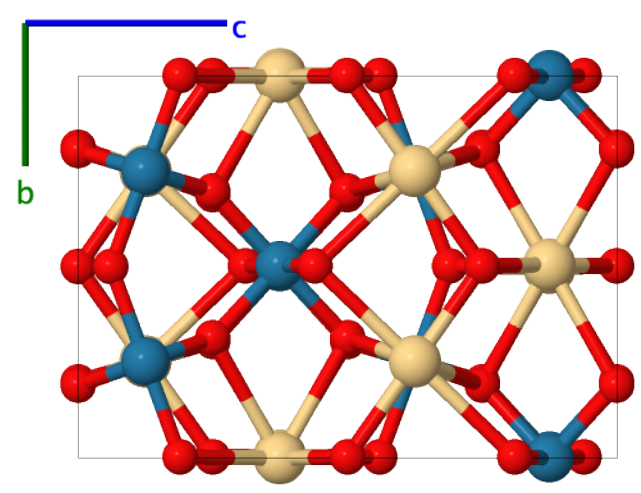

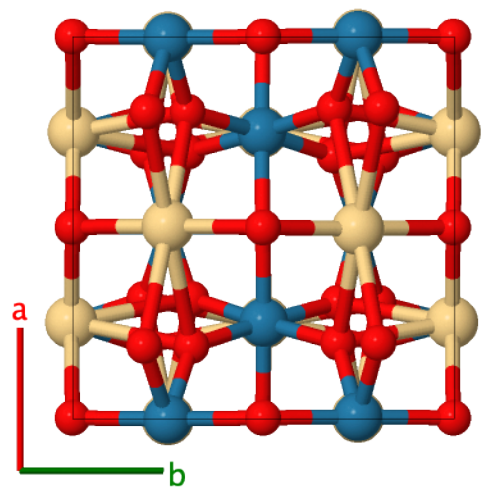

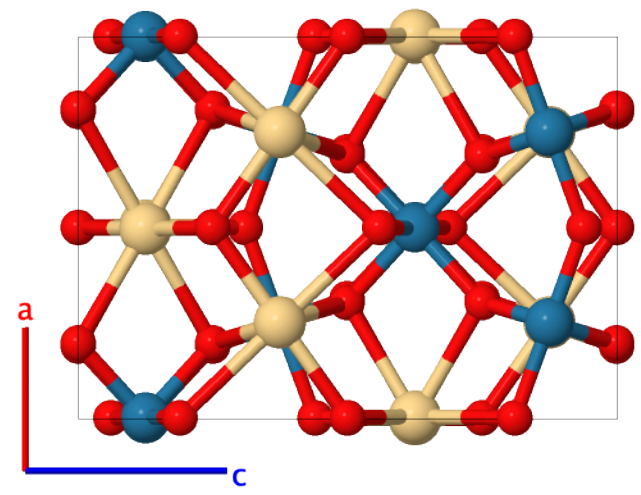

| | | |
|---|---|---|
| **Prototype** | : | $Cd_2O_7Re_2$ |
| **AFLOW prototype label** | : | A2B7C2_tI44_98_f_bcde_f |
| ***Strukturbericht* designation** | : | None |
| **Pearson symbol** | : | tI44 |
| **Space group number** | : | 98 |
| **Space group symbol** | : | $I4_122$ |
| **AFLOW prototype command** | : | `aflow --proto=A2B7C2_tI44_98_f_bcde_f` `--params=`$a, c/a, z_2, x_3, x_4, x_5, x_6$ |

- $Cd_2Re_2O_7$ exhibits a number of phases. We will use the notation of (Kapcia, 2019) to describe them:
-
  - Phase I: above 200 K, the system takes on the cubic pyrochlore ($E8_1$) structure.
  - Phase II: in the range 120-200 K the system is in the tetragonal $I\bar{4}m2$ #119 structure.
  - Phase III: in the range 80-120 K the system is in the tetragonal $I4_122$ #98 structure. (This structure)
  - Phase IV: (Kapcia, 2019) did a first-principles study of this system and found that below 80 K Phase III develops a soft phonon mode which transforms the system into an orthorhombic $F222$ #22 structure.
  - (Norman, 2019) points out that both Phase III and Phase IV structures have issues.
- Both Phase II and Phase III are extremely close to Phase I. If AFLOW-SYM and FINDSYM allow a 0.2 Å uncertainty in lattice vectors and atomic positions both of the tetragonal phases become cubic.
- Phase IV is extremely close to Phase II. If AFLOW-SYM and FINDSYM allow a 0.002 Å uncertainty in the lattice vectors and atomic positions the orthorhombic phase becomes tetragonal.
- Data for the Phase III structure was taken at 90 K.

**Body-centered Tetragonal primitive vectors:**

$$\mathbf{a}_1 = -\frac{1}{2}\,a\,\hat{\mathbf{x}} + \frac{1}{2}\,a\,\hat{\mathbf{y}} + \frac{1}{2}\,c\,\hat{\mathbf{z}}$$
$$\mathbf{a}_2 = \frac{1}{2}\,a\,\hat{\mathbf{x}} - \frac{1}{2}\,a\,\hat{\mathbf{y}} + \frac{1}{2}\,c\,\hat{\mathbf{z}}$$
$$\mathbf{a}_3 = \frac{1}{2}\,a\,\hat{\mathbf{x}} + \frac{1}{2}\,a\,\hat{\mathbf{y}} - \frac{1}{2}\,c\,\hat{\mathbf{z}}$$

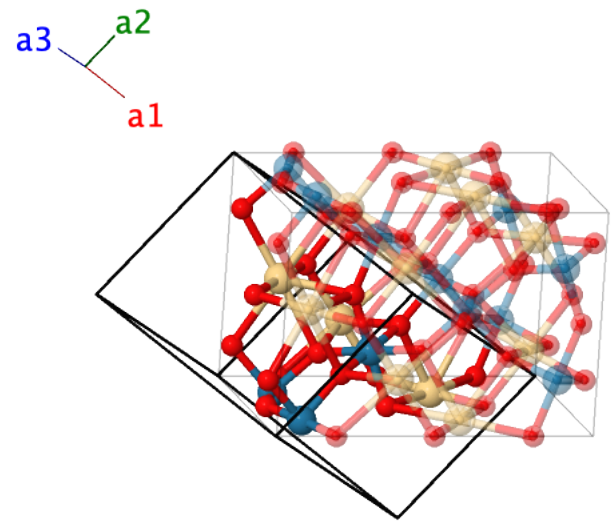

**Basis vectors:**

| | | Lattice Coordinates | | Cartesian Coordinates | Wyckoff Position | Atom Type |
|---|---|---|---|---|---|---|
| $\mathbf{B}_1$ | = | $\frac{1}{2}\,\mathbf{a}_1 + \frac{1}{2}\,\mathbf{a}_2$ | = | $\frac{1}{2}c\,\hat{\mathbf{z}}$ | (4$b$) | O I |
| $\mathbf{B}_2$ | = | $\frac{1}{4}\,\mathbf{a}_1 + \frac{3}{4}\,\mathbf{a}_2 + \frac{1}{2}\,\mathbf{a}_3$ | = | $\frac{1}{2}a\,\hat{\mathbf{x}} + \frac{1}{4}c\,\hat{\mathbf{z}}$ | (4$b$) | O I |
| $\mathbf{B}_3$ | = | $z_2\,\mathbf{a}_1 + z_2\,\mathbf{a}_2$ | = | $z_2c\,\hat{\mathbf{z}}$ | (8$c$) | O II |
| $\mathbf{B}_4$ | = | $\left(\frac{3}{4} + z_2\right)\mathbf{a}_1 + \left(\frac{1}{4} + z_2\right)\mathbf{a}_2 + \frac{1}{2}\,\mathbf{a}_3$ | = | $\frac{1}{2}a\,\hat{\mathbf{y}} + \left(\frac{1}{4} + z_2\right)c\,\hat{\mathbf{z}}$ | (8$c$) | O II |
| $\mathbf{B}_5$ | = | $\left(\frac{3}{4} - z_2\right)\mathbf{a}_1 + \left(\frac{1}{4} - z_2\right)\mathbf{a}_2 + \frac{1}{2}\,\mathbf{a}_3$ | = | $\frac{1}{2}a\,\hat{\mathbf{y}} + \left(\frac{1}{4} - z_2\right)c\,\hat{\mathbf{z}}$ | (8$c$) | O II |
| $\mathbf{B}_6$ | = | $-z_2\,\mathbf{a}_1 - z_2\,\mathbf{a}_2$ | = | $-z_2c\,\hat{\mathbf{z}}$ | (8$c$) | O II |
| $\mathbf{B}_7$ | = | $x_3\,\mathbf{a}_1 + x_3\,\mathbf{a}_2 + 2x_3\,\mathbf{a}_3$ | = | $x_3a\,\hat{\mathbf{x}} + x_3a\,\hat{\mathbf{y}}$ | (8$d$) | O III |
| $\mathbf{B}_8$ | = | $-x_3\,\mathbf{a}_1 - x_3\,\mathbf{a}_2 - 2x_3\,\mathbf{a}_3$ | = | $-x_3a\,\hat{\mathbf{x}} - x_3a\,\hat{\mathbf{y}}$ | (8$d$) | O III |
| $\mathbf{B}_9$ | = | $\left(\frac{3}{4} + x_3\right)\mathbf{a}_1 + \left(\frac{1}{4} - x_3\right)\mathbf{a}_2 + \frac{1}{2}\,\mathbf{a}_3$ | = | $-x_3a\,\hat{\mathbf{x}} + \left(\frac{1}{2} + x_3\right)a\,\hat{\mathbf{y}} + \frac{1}{4}c\,\hat{\mathbf{z}}$ | (8$d$) | O III |
| $\mathbf{B}_{10}$ | = | $\left(\frac{3}{4} - x_3\right)\mathbf{a}_1 + \left(\frac{1}{4} + x_3\right)\mathbf{a}_2 + \frac{1}{2}\,\mathbf{a}_3$ | = | $x_3a\,\hat{\mathbf{x}} + \left(\frac{1}{2} - x_3\right)a\,\hat{\mathbf{y}} + \frac{1}{4}c\,\hat{\mathbf{z}}$ | (8$d$) | O III |
| $\mathbf{B}_{11}$ | = | $x_4\,\mathbf{a}_1 - x_4\,\mathbf{a}_2$ | = | $-x_4a\,\hat{\mathbf{x}} + x_4a\,\hat{\mathbf{y}}$ | (8$e$) | O IV |
| $\mathbf{B}_{12}$ | = | $-x_4\,\mathbf{a}_1 + x_4\,\mathbf{a}_2$ | = | $x_4a\,\hat{\mathbf{x}} - x_4a\,\hat{\mathbf{y}}$ | (8$e$) | O IV |
| $\mathbf{B}_{13}$ | = | $\left(\frac{3}{4} - x_4\right)\mathbf{a}_1 + \left(\frac{1}{4} - x_4\right)\mathbf{a}_2 + \left(\frac{1}{2} - 2x_4\right)\mathbf{a}_3$ | = | $-x_4a\,\hat{\mathbf{x}} + \left(\frac{1}{2} - x_4\right)a\,\hat{\mathbf{y}} + \frac{1}{4}c\,\hat{\mathbf{z}}$ | (8$e$) | O IV |
| $\mathbf{B}_{14}$ | = | $\left(\frac{3}{4} + x_4\right)\mathbf{a}_1 + \left(\frac{1}{4} + x_4\right)\mathbf{a}_2 + \left(\frac{1}{2} + 2x_4\right)\mathbf{a}_3$ | = | $x_4a\,\hat{\mathbf{x}} + \left(\frac{1}{2} + x_4\right)a\,\hat{\mathbf{y}} + \frac{1}{4}c\,\hat{\mathbf{z}}$ | (8$e$) | O IV |
| $\mathbf{B}_{15}$ | = | $\frac{3}{8}\,\mathbf{a}_1 + \left(\frac{1}{8} + x_5\right)\mathbf{a}_2 + \left(\frac{1}{4} + x_5\right)\mathbf{a}_3$ | = | $x_5a\,\hat{\mathbf{x}} + \frac{1}{4}a\,\hat{\mathbf{y}} + \frac{1}{8}c\,\hat{\mathbf{z}}$ | (8$f$) | Cd |
| $\mathbf{B}_{16}$ | = | $\frac{7}{8}\,\mathbf{a}_1 + \left(\frac{1}{8} - x_5\right)\mathbf{a}_2 + \left(\frac{3}{4} - x_5\right)\mathbf{a}_3$ | = | $-x_5a\,\hat{\mathbf{x}} + \frac{3}{4}a\,\hat{\mathbf{y}} + \frac{1}{8}c\,\hat{\mathbf{z}}$ | (8$f$) | Cd |
| $\mathbf{B}_{17}$ | = | $\left(\frac{7}{8} + x_5\right)\mathbf{a}_1 + \frac{1}{8}\,\mathbf{a}_2 + \left(\frac{1}{4} + x_5\right)\mathbf{a}_3$ | = | $\frac{3}{4}a\,\hat{\mathbf{x}} + \left(\frac{1}{2} + x_5\right)a\,\hat{\mathbf{y}} + \frac{3}{8}c\,\hat{\mathbf{z}}$ | (8$f$) | Cd |
| $\mathbf{B}_{18}$ | = | $\left(\frac{7}{8} - x_5\right)\mathbf{a}_1 + \frac{5}{8}\,\mathbf{a}_2 + \left(\frac{3}{4} - x_5\right)\mathbf{a}_3$ | = | $\frac{1}{4}a\,\hat{\mathbf{x}} + \left(\frac{1}{2} - x_5\right)a\,\hat{\mathbf{y}} + \frac{3}{8}c\,\hat{\mathbf{z}}$ | (8$f$) | Cd |
| $\mathbf{B}_{19}$ | = | $\frac{3}{8}\,\mathbf{a}_1 + \left(\frac{1}{8} + x_6\right)\mathbf{a}_2 + \left(\frac{1}{4} + x_6\right)\mathbf{a}_3$ | = | $x_6a\,\hat{\mathbf{x}} + \frac{1}{4}a\,\hat{\mathbf{y}} + \frac{1}{8}c\,\hat{\mathbf{z}}$ | (8$f$) | Re |
| $\mathbf{B}_{20}$ | = | $\frac{7}{8}\,\mathbf{a}_1 + \left(\frac{1}{8} - x_6\right)\mathbf{a}_2 + \left(\frac{3}{4} - x_6\right)\mathbf{a}_3$ | = | $-x_6a\,\hat{\mathbf{x}} + \frac{3}{4}a\,\hat{\mathbf{y}} + \frac{1}{8}c\,\hat{\mathbf{z}}$ | (8$f$) | Re |
| $\mathbf{B}_{21}$ | = | $\left(\frac{7}{8} + x_6\right)\mathbf{a}_1 + \frac{1}{8}\,\mathbf{a}_2 + \left(\frac{1}{4} + x_6\right)\mathbf{a}_3$ | = | $\frac{3}{4}a\,\hat{\mathbf{x}} + \left(\frac{1}{2} + x_6\right)a\,\hat{\mathbf{y}} + \frac{3}{8}c\,\hat{\mathbf{z}}$ | (8$f$) | Re |
| $\mathbf{B}_{22}$ | = | $\left(\frac{7}{8} - x_6\right)\mathbf{a}_1 + \frac{5}{8}\,\mathbf{a}_2 + \left(\frac{3}{4} - x_6\right)\mathbf{a}_3$ | = | $\frac{1}{4}a\,\hat{\mathbf{x}} + \left(\frac{1}{2} - x_6\right)a\,\hat{\mathbf{y}} + \frac{3}{8}c\,\hat{\mathbf{z}}$ | (8$f$) | Re |

**Found in:**

- M. R. Norman, *The crystal structure of the inversion breaking metal $Cd_2Re_2O_7$*, Phys. Rev. B **101**, 045117 (2020), doi:10.1103/PhysRevB.101.045117.

---

**Geometry files:**

- CIF: pp. 1678
- POSCAR: pp. 1679

# $F5_4$ ($NH_4ClO_2$) (*obsolete*) Structure: ABC2_tP8_100_b_a_c

http://aflow.org/prototype-encyclopedia/ABC2_tP8_100_b_a_c

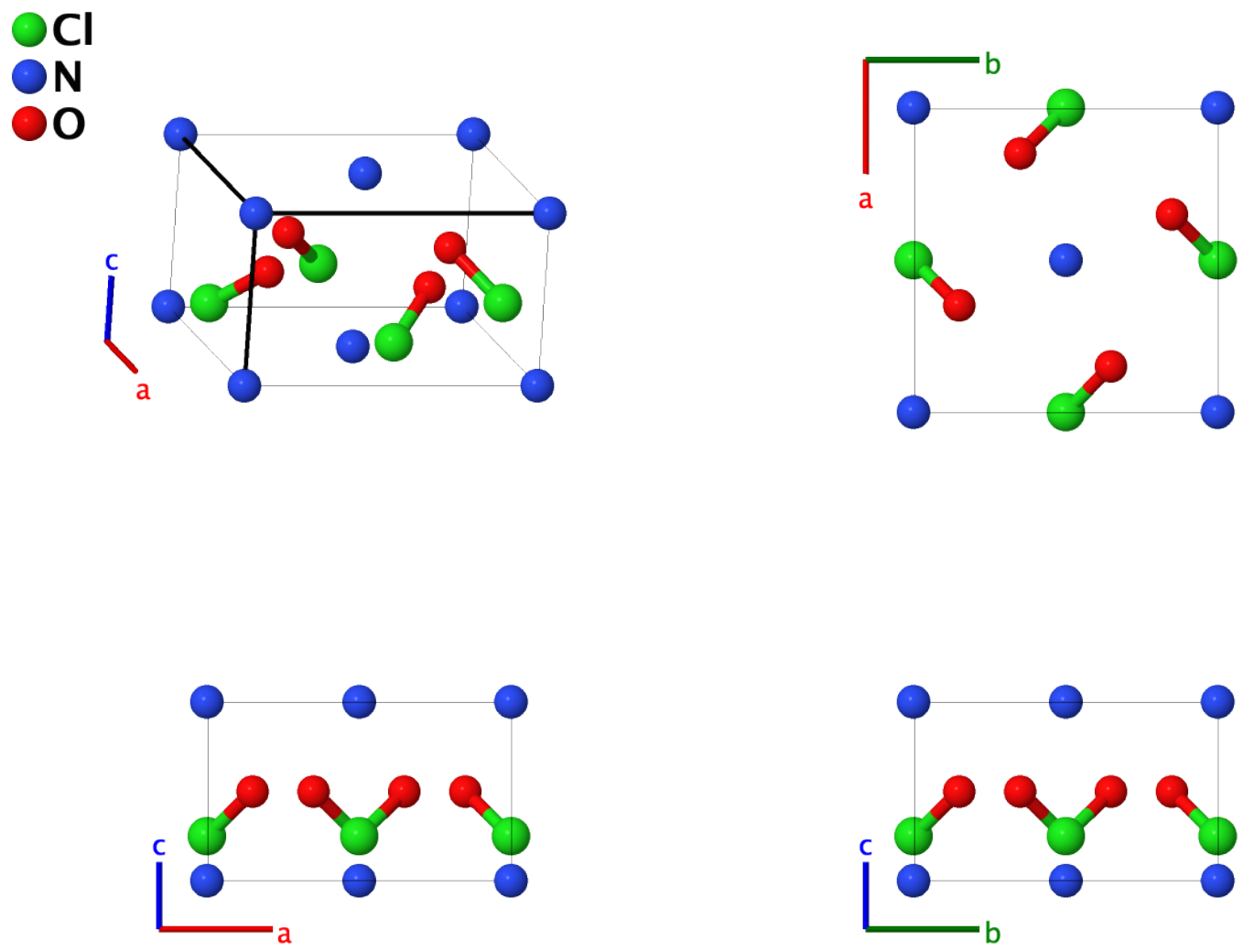

| | | |
|---|---|---|
| **Prototype** | : | $Cl(NH_4)O_2$ |
| **AFLOW prototype label** | : | ABC2_tP8_100_b_a_c |
| ***Strukturbericht* designation** | : | $F5_4$ |
| **Pearson symbol** | : | tP8 |
| **Space group number** | : | 100 |
| **Space group symbol** | : | $P4bm$ |
| **AFLOW prototype command** | : | `aflow --proto=ABC2_tP8_100_b_a_c` `--params=`$a, c/a, z_1, z_2, x_3, z_3$ |

- (Levi, 1931) first detemined this structure, but they were actually looking at a combination of ammonium chlorite, $NH_4ClO_2$ and ammonium chlorate, $NH_4ClO_3$, which takes on the $\gamma$-$KNO_3$ structure. In addition, they were unable to determine the positions of the hydrogen atoms, and it appears that space group $P4bm$ #100 is incompatible with having four hydrogen atoms in a tetrahedral arrangement about the nitrogen atom. (Smolentsev, 2005) determined the positions of the hydrogen atoms, and placed this structure in space group $P\bar{4}2_1m$ #113, making the structure of (Levi, 1931) obsolete. We present it here for historical interest.
- Since the position of the hydrogen atoms in the $NH_4$ ions were not determined, we only provide the nitrogen atom positions (labeled as $NH_4$).

**Simple Tetragonal primitive vectors:**

$$\mathbf{a}_1 = a\,\hat{\mathbf{x}}$$
$$\mathbf{a}_2 = a\,\hat{\mathbf{y}}$$
$$\mathbf{a}_3 = c\,\hat{\mathbf{z}}$$

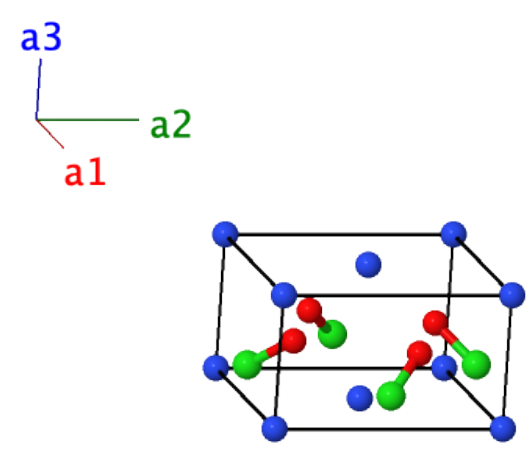

**Basis vectors:**

| | | Lattice Coordinates | | Cartesian Coordinates | Wyckoff Position | Atom Type |
|---|---|---|---|---|---|---|
| $\mathbf{B}_1$ | = | $z_1\,\mathbf{a}_3$ | = | $z_1 c\,\hat{\mathbf{z}}$ | (2$a$) | $NH_4$ |
| $\mathbf{B}_2$ | = | $\frac{1}{2}\,\mathbf{a}_1 + \frac{1}{2}\,\mathbf{a}_2 + z_1\,\mathbf{a}_3$ | = | $\frac{1}{2}a\,\hat{\mathbf{x}} + \frac{1}{2}a\,\hat{\mathbf{y}} + z_1 c\,\hat{\mathbf{z}}$ | (2$a$) | $NH_4$ |
| $\mathbf{B}_3$ | = | $\frac{1}{2}\,\mathbf{a}_1 + z_2\,\mathbf{a}_3$ | = | $\frac{1}{2}a\,\hat{\mathbf{x}} + z_2 c\,\hat{\mathbf{z}}$ | (2$b$) | Cl |
| $\mathbf{B}_4$ | = | $\frac{1}{2}\,\mathbf{a}_2 + z_2\,\mathbf{a}_3$ | = | $\frac{1}{2}a\,\hat{\mathbf{y}} + z_2 c\,\hat{\mathbf{z}}$ | (2$b$) | Cl |
| $\mathbf{B}_5$ | = | $x_3\,\mathbf{a}_1 + \left(\frac{1}{2} + x_3\right)\,\mathbf{a}_2 + z_3\,\mathbf{a}_3$ | = | $x_3 a\,\hat{\mathbf{x}} + \left(\frac{1}{2} + x_3\right) a\,\hat{\mathbf{y}} + z_3 c\,\hat{\mathbf{z}}$ | (4$c$) | O |
| $\mathbf{B}_6$ | = | $-x_3\,\mathbf{a}_1 + \left(\frac{1}{2} - x_3\right)\,\mathbf{a}_2 + z_3\,\mathbf{a}_3$ | = | $-x_3 a\,\hat{\mathbf{x}} + \left(\frac{1}{2} - x_3\right) a\,\hat{\mathbf{y}} + z_3 c\,\hat{\mathbf{z}}$ | (4$c$) | O |
| $\mathbf{B}_7$ | = | $\left(\frac{1}{2} - x_3\right)\,\mathbf{a}_1 + x_3\,\mathbf{a}_2 + z_3\,\mathbf{a}_3$ | = | $\left(\frac{1}{2} - x_3\right) a\,\hat{\mathbf{x}} + x_3 a\,\hat{\mathbf{y}} + z_3 c\,\hat{\mathbf{z}}$ | (4$c$) | O |
| $\mathbf{B}_8$ | = | $\left(\frac{1}{2} + x_3\right)\,\mathbf{a}_1 - x_3\,\mathbf{a}_2 + z_3\,\mathbf{a}_3$ | = | $\left(\frac{1}{2} + x_3\right) a\,\hat{\mathbf{x}} - x_3 a\,\hat{\mathbf{y}} + z_3 c\,\hat{\mathbf{z}}$ | (4$c$) | O |

**Found in:**

**Geometry files:**

- CIF: pp. 1679
- POSCAR: pp. 1679

# $NH_4NO_3$ II ($G0_9$) Structure: ABC3_tP10_100_b_a_bc

http://aflow.org/prototype-encyclopedia/ABC3_tP10_100_b_a_bc

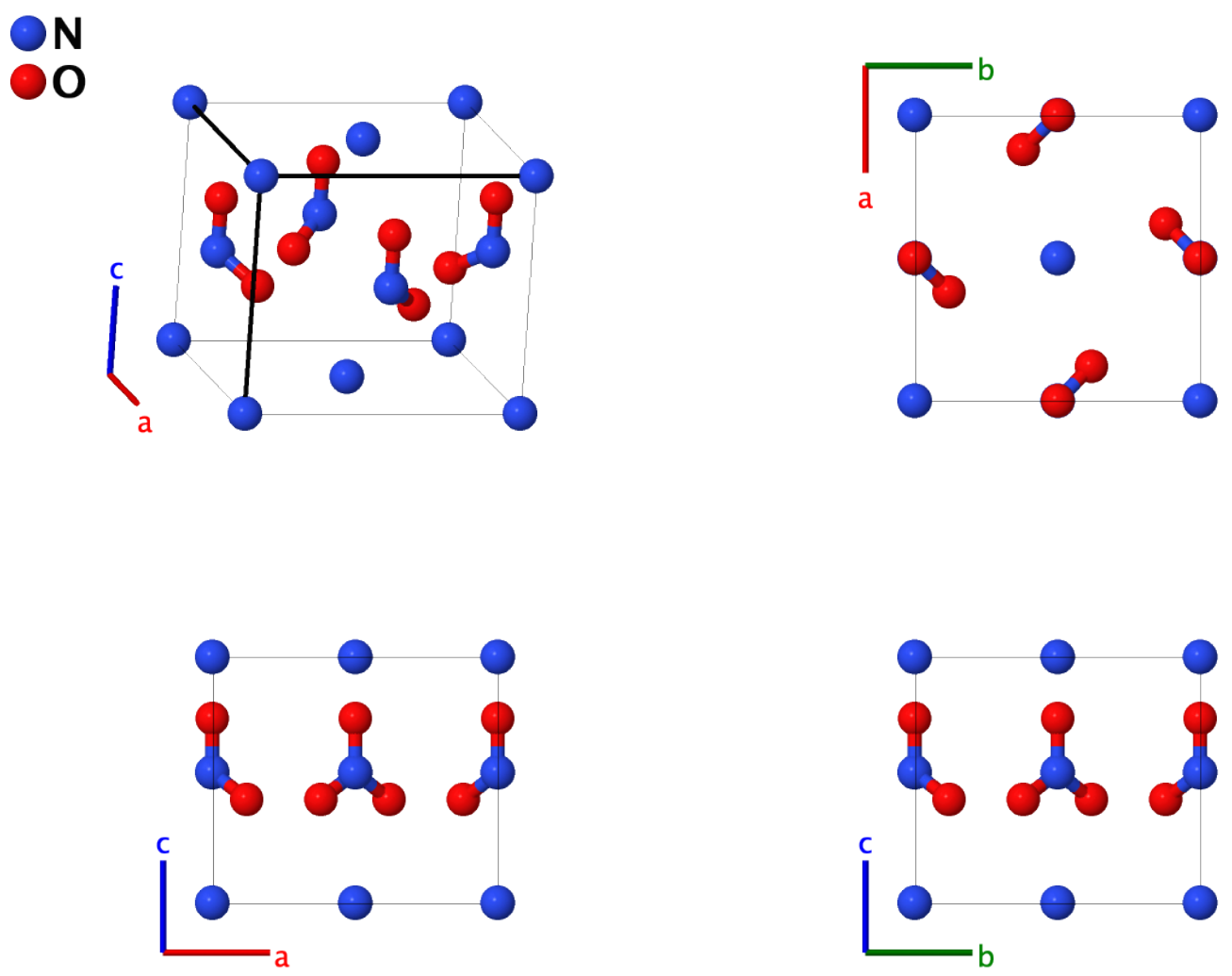

| | | |
|---|---|---|
| **Prototype** | : | $N(NH_4)O_3$ |
| **AFLOW prototype label** | : | ABC3_tP10_100_b_a_bc |
| ***Strukturbericht* designation** | : | $G0_9$ |
| **Pearson symbol** | : | tP10 |
| **Space group number** | : | 100 |
| **Space group symbol** | : | $P4bm$ |
| **AFLOW prototype command** | : | `aflow --proto=ABC3_tP10_100_b_a_bc` <br> `--params=`$a, c/a, z_1, z_2, z_3, x_4, z_4$ |

- Ammonium Nitrate exists in a variety of forms, (Hermann, 1937) depending on the temperature:

| Phase | Temperature °C | *Strukturbericht* | Page | |
|---|---|---|---|---|
| I | 125 – 170 | $G0_8$ | AB_cP2_221_a_b.NH4.NO3 | |
| II | 84 – 125 | $G0_9$ | ABC3_tP10_100_b_a_bc | (this structure) |
| III | 32 – 84 | $G0_{10}$ | ABC3_oP20_62_c_c_cd.N.NH4.O | |
| IV | -17 – 32 | $G0_{11}$ | A4B2C3_oP18_59_ef_ab_af | |
| V | $< -17$ | Gwihabaite | A4B2C3_tP72_77_8d_ab2c2d_6d2 | |

- Data for this structure was taken at 60 °C.
- The positions of the hydrogen atoms were not determined. The isolated nitrogen atoms in this structure's visualization are surrounded by four hydrogen atoms in an approximately tetrahedral arrangement. It is likely that the $NH_4$ ions are free to rotate (Kracek, 1937).
- Both (Shinnaka, 1956) and (Hermann, 1937) state that the available X-ray diffraction data supports a space group of either $P4bm$ (#100) or $P\bar{4}2_1m$ (#113). The atomic positions found by Shinnaka agree with space group $P4bm$.
- (Shinnaka, 1956) states that the $NO_3$ nitrate groups are rotating, but this rotation "is almost bound in two orientations (in opposite directions)." He then gives two possible orientations for the nitrate. We present the first orientation here. The second orientation is obtained by taking $z_3 \rightarrow -z_3$ and $z_4 \rightarrow -z_4$.
- Another way of presenting this information would be to add a second nitrate group to the primitive cell, and set the occupation of all the atoms in the nitrates at 50%. This would give a structure in space group $P4/mbm$ (#127), which might be useful as a pictorial representation but does not correctly represent the physics of the crystal, as the nitrogen and oxygen atoms in an individual nitrate ion must remain together.

- The N–O distances in this structure are about 10% smaller than the distances found in the other phases of $NH_4NO_3$. This suggests that the structure should be reevaluated.

- The positions of the hydrogen atoms in the $NH_4$ ion were not determined, so we only provide the positions of the nitrogen atoms (labeled as $NH_4$).

---

**Simple Tetragonal primitive vectors:**

$$\mathbf{a}_1 = a\,\hat{\mathbf{x}}$$
$$\mathbf{a}_2 = a\,\hat{\mathbf{y}}$$
$$\mathbf{a}_3 = c\,\hat{\mathbf{z}}$$

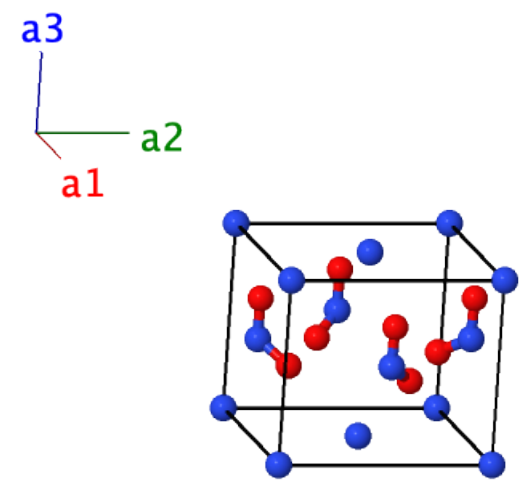

---

**Basis vectors:**

| | | Lattice Coordinates | | Cartesian Coordinates | Wyckoff Position | Atom Type |
|---|---|---|---|---|---|---|
| $\mathbf{B}_1$ | = | $z_1\,\mathbf{a}_3$ | = | $z_1 c\,\hat{\mathbf{z}}$ | (2$a$) | $NH_4$ |
| $\mathbf{B}_2$ | = | $\frac{1}{2}\,\mathbf{a}_1 + \frac{1}{2}\,\mathbf{a}_2 + z_1\,\mathbf{a}_3$ | = | $\frac{1}{2}a\,\hat{\mathbf{x}} + \frac{1}{2}a\,\hat{\mathbf{y}} + z_1 c\,\hat{\mathbf{z}}$ | (2$a$) | $NH_4$ |
| $\mathbf{B}_3$ | = | $\frac{1}{2}\,\mathbf{a}_1 + z_2\,\mathbf{a}_3$ | = | $\frac{1}{2}a\,\hat{\mathbf{x}} + z_2 c\,\hat{\mathbf{z}}$ | (2$b$) | N |
| $\mathbf{B}_4$ | = | $\frac{1}{2}\,\mathbf{a}_2 + z_2\,\mathbf{a}_3$ | = | $\frac{1}{2}a\,\hat{\mathbf{y}} + z_2 c\,\hat{\mathbf{z}}$ | (2$b$) | N |
| $\mathbf{B}_5$ | = | $\frac{1}{2}\,\mathbf{a}_1 + z_3\,\mathbf{a}_3$ | = | $\frac{1}{2}a\,\hat{\mathbf{x}} + z_3 c\,\hat{\mathbf{z}}$ | (2$b$) | O I |
| $\mathbf{B}_6$ | = | $\frac{1}{2}\,\mathbf{a}_2 + z_3\,\mathbf{a}_3$ | = | $\frac{1}{2}a\,\hat{\mathbf{y}} + z_3 c\,\hat{\mathbf{z}}$ | (2$b$) | O I |
| $\mathbf{B}_7$ | = | $x_4\,\mathbf{a}_1 + \left(\frac{1}{2} + x_4\right)\,\mathbf{a}_2 + z_4\,\mathbf{a}_3$ | = | $x_4 a\,\hat{\mathbf{x}} + \left(\frac{1}{2} + x_4\right)a\,\hat{\mathbf{y}} + z_4 c\,\hat{\mathbf{z}}$ | (4$c$) | O II |
| $\mathbf{B}_8$ | = | $-x_4\,\mathbf{a}_1 + \left(\frac{1}{2} - x_4\right)\,\mathbf{a}_2 + z_4\,\mathbf{a}_3$ | = | $-x_4 a\,\hat{\mathbf{x}} + \left(\frac{1}{2} - x_4\right)a\,\hat{\mathbf{y}} + z_4 c\,\hat{\mathbf{z}}$ | (4$c$) | O II |
| $\mathbf{B}_9$ | = | $\left(\frac{1}{2} - x_4\right)\,\mathbf{a}_1 + x_4\,\mathbf{a}_2 + z_4\,\mathbf{a}_3$ | = | $\left(\frac{1}{2} - x_4\right)a\,\hat{\mathbf{x}} + x_4 a\,\hat{\mathbf{y}} + z_4 c\,\hat{\mathbf{z}}$ | (4$c$) | O II |
| $\mathbf{B}_{10}$ | = | $\left(\frac{1}{2} + x_4\right)\,\mathbf{a}_1 - x_4\,\mathbf{a}_2 + z_4\,\mathbf{a}_3$ | = | $\left(\frac{1}{2} + x_4\right)a\,\hat{\mathbf{x}} - x_4 a\,\hat{\mathbf{y}} + z_4 c\,\hat{\mathbf{z}}$ | (4$c$) | O II |

---

---

**Geometry files:**

- CIF: pp. 1679
- POSCAR: pp. 1680

# $VSe_2O_6$ Structure: A6B2C_tP72_103_abc5d_2d_abc

http://aflow.org/prototype-encyclopedia/A6B2C_tP72_103_abc5d_2d_abc

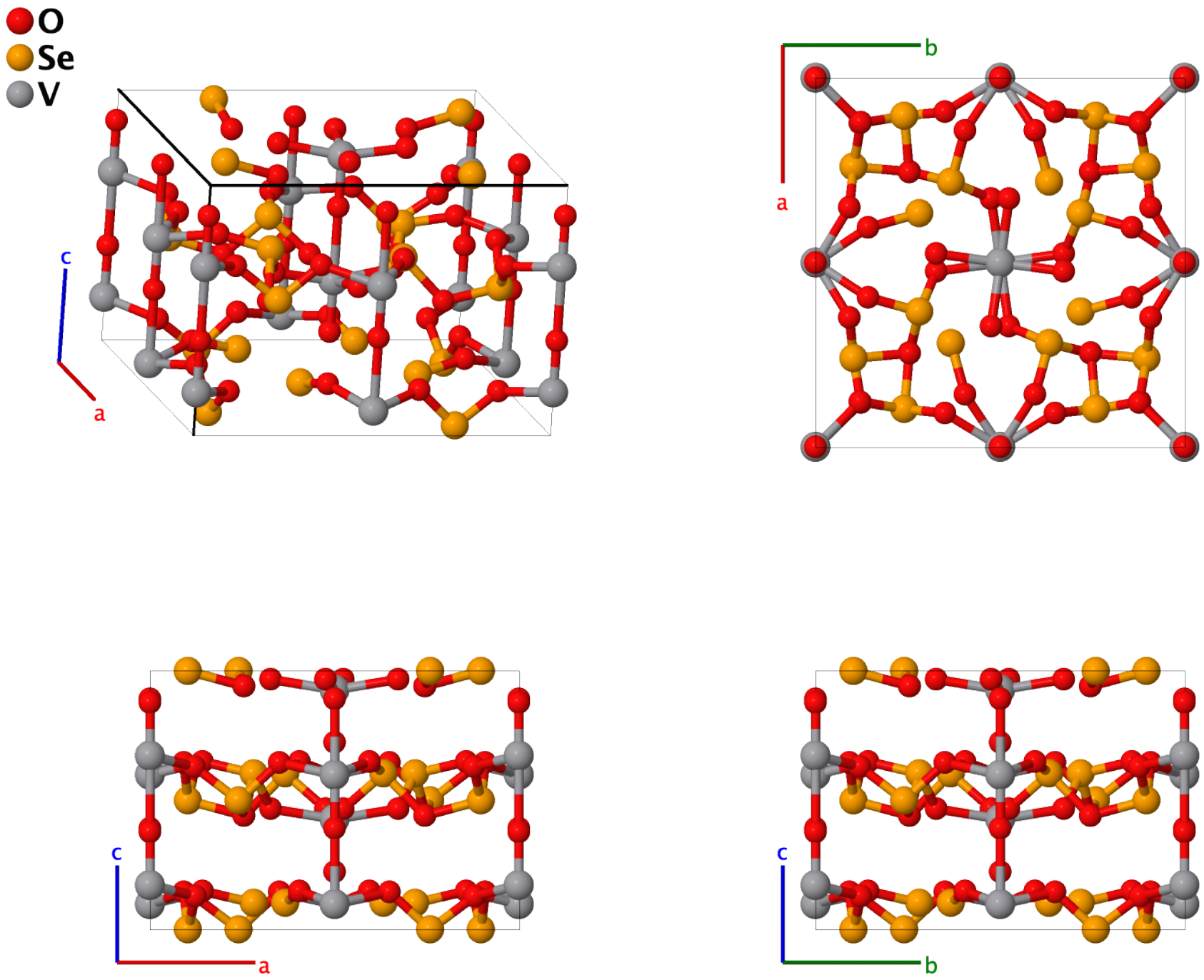

| | | |
|---|---|---|
| **Prototype** | : | $O_6Se_2V$ |
| **AFLOW prototype label** | : | A6B2C_tP72_103_abc5d_2d_abc |
| ***Strukturbericht* designation** | : | None |
| **Pearson symbol** | : | tP72 |
| **Space group number** | : | 103 |
| **Space group symbol** | : | *P4cc* |
| **AFLOW prototype command** | : | `aflow --proto=A6B2C_tP72_103_abc5d_2d_abc` `--params=`$a, c/a, z_1, z_2, z_3, z_4, z_5, z_6, x_7, y_7, z_7, x_8, y_8, z_8, x_9, y_9, z_9, x_{10}, y_{10}, z_{10}, x_{11}, y_{11}, z_{11}, x_{12}, y_{12}, z_{12}, x_{13}, y_{13}, z_{13}$ |

**Simple Tetragonal primitive vectors:**

$$\mathbf{a}_1 = a\,\hat{\mathbf{x}}$$
$$\mathbf{a}_2 = a\,\hat{\mathbf{y}}$$
$$\mathbf{a}_3 = c\,\hat{\mathbf{z}}$$

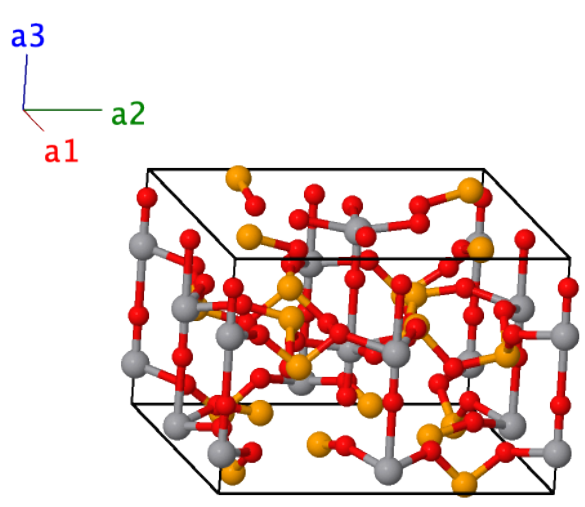

**Basis vectors:**

| | | Lattice Coordinates | | Cartesian Coordinates | Wyckoff Position | Atom Type |
|---|---|---|---|---|---|---|
| $\mathbf{B}_1$ | = | $z_1\,\mathbf{a}_3$ | = | $z_1 c\,\hat{\mathbf{z}}$ | (2a) | O I |
| $\mathbf{B}_{2}$ | = | $\left(\frac{1}{2}+z_{1}\right)\mathbf{a}_{3}$ | = | $\left(\frac{1}{2}+z_{1}\right)c\,\hat{\mathbf{z}}$ | $(2a)$ | O I |
| $\mathbf{B}_{3}$ | = | $z_{2}\,\mathbf{a}_{3}$ | = | $z_{2}c\,\hat{\mathbf{z}}$ | $(2a)$ | V I |
| $\mathbf{B}_{4}$ | = | $\left(\frac{1}{2}+z_{2}\right)\mathbf{a}_{3}$ | = | $\left(\frac{1}{2}+z_{2}\right)c\,\hat{\mathbf{z}}$ | $(2a)$ | V I |
| $\mathbf{B}_{5}$ | = | $\frac{1}{2}\,\mathbf{a}_{1}+\frac{1}{2}\,\mathbf{a}_{2}+z_{3}\,\mathbf{a}_{3}$ | = | $\frac{1}{2}a\,\hat{\mathbf{x}}+\frac{1}{2}a\,\hat{\mathbf{y}}+z_{3}c\,\hat{\mathbf{z}}$ | $(2b)$ | O II |
| $\mathbf{B}_{6}$ | = | $\frac{1}{2}\,\mathbf{a}_{1}+\frac{1}{2}\,\mathbf{a}_{2}+\left(\frac{1}{2}+z_{3}\right)\mathbf{a}_{3}$ | = | $\frac{1}{2}a\,\hat{\mathbf{x}}+\frac{1}{2}a\,\hat{\mathbf{y}}+\left(\frac{1}{2}+z_{3}\right)c\,\hat{\mathbf{z}}$ | $(2b)$ | O II |
| $\mathbf{B}_{7}$ | = | $\frac{1}{2}\,\mathbf{a}_{1}+\frac{1}{2}\,\mathbf{a}_{2}+z_{4}\,\mathbf{a}_{3}$ | = | $\frac{1}{2}a\,\hat{\mathbf{x}}+\frac{1}{2}a\,\hat{\mathbf{y}}+z_{4}c\,\hat{\mathbf{z}}$ | $(2b)$ | V II |
| $\mathbf{B}_{8}$ | = | $\frac{1}{2}\,\mathbf{a}_{1}+\frac{1}{2}\,\mathbf{a}_{2}+\left(\frac{1}{2}+z_{4}\right)\mathbf{a}_{3}$ | = | $\frac{1}{2}a\,\hat{\mathbf{x}}+\frac{1}{2}a\,\hat{\mathbf{y}}+\left(\frac{1}{2}+z_{4}\right)c\,\hat{\mathbf{z}}$ | $(2b)$ | V II |
| $\mathbf{B}_{9}$ | = | $\frac{1}{2}\,\mathbf{a}_{2}+z_{5}\,\mathbf{a}_{3}$ | = | $\frac{1}{2}a\,\hat{\mathbf{y}}+z_{5}c\,\hat{\mathbf{z}}$ | $(4c)$ | O III |
| $\mathbf{B}_{10}$ | = | $\frac{1}{2}\,\mathbf{a}_{1}+z_{5}\,\mathbf{a}_{3}$ | = | $\frac{1}{2}a\,\hat{\mathbf{x}}+z_{5}c\,\hat{\mathbf{z}}$ | $(4c)$ | O III |
| $\mathbf{B}_{11}$ | = | $\frac{1}{2}\,\mathbf{a}_{2}+\left(\frac{1}{2}+z_{5}\right)\mathbf{a}_{3}$ | = | $\frac{1}{2}a\,\hat{\mathbf{y}}+\left(\frac{1}{2}+z_{5}\right)c\,\hat{\mathbf{z}}$ | $(4c)$ | O III |
| $\mathbf{B}_{12}$ | = | $\frac{1}{2}\,\mathbf{a}_{1}+\left(\frac{1}{2}+z_{5}\right)\mathbf{a}_{3}$ | = | $\frac{1}{2}a\,\hat{\mathbf{x}}+\left(\frac{1}{2}+z_{5}\right)c\,\hat{\mathbf{z}}$ | $(4c)$ | O III |
| $\mathbf{B}_{13}$ | = | $\frac{1}{2}\,\mathbf{a}_{2}+z_{6}\,\mathbf{a}_{3}$ | = | $\frac{1}{2}a\,\hat{\mathbf{y}}+z_{6}c\,\hat{\mathbf{z}}$ | $(4c)$ | V III |
| $\mathbf{B}_{14}$ | = | $\frac{1}{2}\,\mathbf{a}_{1}+z_{6}\,\mathbf{a}_{3}$ | = | $\frac{1}{2}a\,\hat{\mathbf{x}}+z_{6}c\,\hat{\mathbf{z}}$ | $(4c)$ | V III |
| $\mathbf{B}_{15}$ | = | $\frac{1}{2}\,\mathbf{a}_{2}+\left(\frac{1}{2}+z_{6}\right)\mathbf{a}_{3}$ | = | $\frac{1}{2}a\,\hat{\mathbf{y}}+\left(\frac{1}{2}+z_{6}\right)c\,\hat{\mathbf{z}}$ | $(4c)$ | V III |
| $\mathbf{B}_{16}$ | = | $\frac{1}{2}\,\mathbf{a}_{1}+\left(\frac{1}{2}+z_{6}\right)\mathbf{a}_{3}$ | = | $\frac{1}{2}a\,\hat{\mathbf{x}}+\left(\frac{1}{2}+z_{6}\right)c\,\hat{\mathbf{z}}$ | $(4c)$ | V III |
| $\mathbf{B}_{17}$ | = | $x_{7}\,\mathbf{a}_{1}+y_{7}\,\mathbf{a}_{2}+z_{7}\,\mathbf{a}_{3}$ | = | $x_{7}a\,\hat{\mathbf{x}}+y_{7}a\,\hat{\mathbf{y}}+z_{7}c\,\hat{\mathbf{z}}$ | $(8d)$ | O IV |
| $\mathbf{B}_{18}$ | = | $-x_{7}\,\mathbf{a}_{1}-y_{7}\,\mathbf{a}_{2}+z_{7}\,\mathbf{a}_{3}$ | = | $-x_{7}a\,\hat{\mathbf{x}}-y_{7}a\,\hat{\mathbf{y}}+z_{7}c\,\hat{\mathbf{z}}$ | $(8d)$ | O IV |
| $\mathbf{B}_{19}$ | = | $-y_{7}\,\mathbf{a}_{1}+x_{7}\,\mathbf{a}_{2}+z_{7}\,\mathbf{a}_{3}$ | = | $-y_{7}a\,\hat{\mathbf{x}}+x_{7}a\,\hat{\mathbf{y}}+z_{7}c\,\hat{\mathbf{z}}$ | $(8d)$ | O IV |
| $\mathbf{B}_{20}$ | = | $y_{7}\,\mathbf{a}_{1}-x_{7}\,\mathbf{a}_{2}+z_{7}\,\mathbf{a}_{3}$ | = | $y_{7}a\,\hat{\mathbf{x}}-x_{7}a\,\hat{\mathbf{y}}+z_{7}c\,\hat{\mathbf{z}}$ | $(8d)$ | O IV |
| $\mathbf{B}_{21}$ | = | $x_{7}\,\mathbf{a}_{1}-y_{7}\,\mathbf{a}_{2}+\left(\frac{1}{2}+z_{7}\right)\mathbf{a}_{3}$ | = | $x_{7}a\,\hat{\mathbf{x}}-y_{7}a\,\hat{\mathbf{y}}+\left(\frac{1}{2}+z_{7}\right)c\,\hat{\mathbf{z}}$ | $(8d)$ | O IV |
| $\mathbf{B}_{22}$ | = | $-x_{7}\,\mathbf{a}_{1}+y_{7}\,\mathbf{a}_{2}+\left(\frac{1}{2}+z_{7}\right)\mathbf{a}_{3}$ | = | $-x_{7}a\,\hat{\mathbf{x}}+y_{7}a\,\hat{\mathbf{y}}+\left(\frac{1}{2}+z_{7}\right)c\,\hat{\mathbf{z}}$ | $(8d)$ | O IV |
| $\mathbf{B}_{23}$ | = | $-y_{7}\,\mathbf{a}_{1}-x_{7}\,\mathbf{a}_{2}+\left(\frac{1}{2}+z_{7}\right)\mathbf{a}_{3}$ | = | $-y_{7}a\,\hat{\mathbf{x}}-x_{7}a\,\hat{\mathbf{y}}+\left(\frac{1}{2}+z_{7}\right)c\,\hat{\mathbf{z}}$ | $(8d)$ | O IV |
| $\mathbf{B}_{24}$ | = | $y_{7}\,\mathbf{a}_{1}+x_{7}\,\mathbf{a}_{2}+\left(\frac{1}{2}+z_{7}\right)\mathbf{a}_{3}$ | = | $y_{7}a\,\hat{\mathbf{x}}+x_{7}a\,\hat{\mathbf{y}}+\left(\frac{1}{2}+z_{7}\right)c\,\hat{\mathbf{z}}$ | $(8d)$ | O IV |
| $\mathbf{B}_{25}$ | = | $x_{8}\,\mathbf{a}_{1}+y_{8}\,\mathbf{a}_{2}+z_{8}\,\mathbf{a}_{3}$ | = | $x_{8}a\,\hat{\mathbf{x}}+y_{8}a\,\hat{\mathbf{y}}+z_{8}c\,\hat{\mathbf{z}}$ | $(8d)$ | O V |
| $\mathbf{B}_{26}$ | = | $-x_{8}\,\mathbf{a}_{1}-y_{8}\,\mathbf{a}_{2}+z_{8}\,\mathbf{a}_{3}$ | = | $-x_{8}a\,\hat{\mathbf{x}}-y_{8}a\,\hat{\mathbf{y}}+z_{8}c\,\hat{\mathbf{z}}$ | $(8d)$ | O V |
| $\mathbf{B}_{27}$ | = | $-y_{8}\,\mathbf{a}_{1}+x_{8}\,\mathbf{a}_{2}+z_{8}\,\mathbf{a}_{3}$ | = | $-y_{8}a\,\hat{\mathbf{x}}+x_{8}a\,\hat{\mathbf{y}}+z_{8}c\,\hat{\mathbf{z}}$ | $(8d)$ | O V |
| $\mathbf{B}_{28}$ | = | $y_{8}\,\mathbf{a}_{1}-x_{8}\,\mathbf{a}_{2}+z_{8}\,\mathbf{a}_{3}$ | = | $y_{8}a\,\hat{\mathbf{x}}-x_{8}a\,\hat{\mathbf{y}}+z_{8}c\,\hat{\mathbf{z}}$ | $(8d)$ | O V |
| $\mathbf{B}_{29}$ | = | $x_{8}\,\mathbf{a}_{1}-y_{8}\,\mathbf{a}_{2}+\left(\frac{1}{2}+z_{8}\right)\mathbf{a}_{3}$ | = | $x_{8}a\,\hat{\mathbf{x}}-y_{8}a\,\hat{\mathbf{y}}+\left(\frac{1}{2}+z_{8}\right)c\,\hat{\mathbf{z}}$ | $(8d)$ | O V |
| $\mathbf{B}_{30}$ | = | $-x_{8}\,\mathbf{a}_{1}+y_{8}\,\mathbf{a}_{2}+\left(\frac{1}{2}+z_{8}\right)\mathbf{a}_{3}$ | = | $-x_{8}a\,\hat{\mathbf{x}}+y_{8}a\,\hat{\mathbf{y}}+\left(\frac{1}{2}+z_{8}\right)c\,\hat{\mathbf{z}}$ | $(8d)$ | O V |
| $\mathbf{B}_{31}$ | = | $-y_{8}\,\mathbf{a}_{1}-x_{8}\,\mathbf{a}_{2}+\left(\frac{1}{2}+z_{8}\right)\mathbf{a}_{3}$ | = | $-y_{8}a\,\hat{\mathbf{x}}-x_{8}a\,\hat{\mathbf{y}}+\left(\frac{1}{2}+z_{8}\right)c\,\hat{\mathbf{z}}$ | $(8d)$ | O V |
| $\mathbf{B}_{32}$ | = | $y_{8}\,\mathbf{a}_{1}+x_{8}\,\mathbf{a}_{2}+\left(\frac{1}{2}+z_{8}\right)\mathbf{a}_{3}$ | = | $y_{8}a\,\hat{\mathbf{x}}+x_{8}a\,\hat{\mathbf{y}}+\left(\frac{1}{2}+z_{8}\right)c\,\hat{\mathbf{z}}$ | $(8d)$ | O V |
| $\mathbf{B}_{33}$ | = | $x_{9}\,\mathbf{a}_{1}+y_{9}\,\mathbf{a}_{2}+z_{9}\,\mathbf{a}_{3}$ | = | $x_{9}a\,\hat{\mathbf{x}}+y_{9}a\,\hat{\mathbf{y}}+z_{9}c\,\hat{\mathbf{z}}$ | $(8d)$ | O VI |
| $\mathbf{B}_{34}$ | = | $-x_{9}\,\mathbf{a}_{1}-y_{9}\,\mathbf{a}_{2}+z_{9}\,\mathbf{a}_{3}$ | = | $-x_{9}a\,\hat{\mathbf{x}}-y_{9}a\,\hat{\mathbf{y}}+z_{9}c\,\hat{\mathbf{z}}$ | $(8d)$ | O VI |
| $\mathbf{B}_{35}$ | = | $-y_{9}\,\mathbf{a}_{1}+x_{9}\,\mathbf{a}_{2}+z_{9}\,\mathbf{a}_{3}$ | = | $-y_{9}a\,\hat{\mathbf{x}}+x_{9}a\,\hat{\mathbf{y}}+z_{9}c\,\hat{\mathbf{z}}$ | $(8d)$ | O VI |
| $\mathbf{B}_{36}$ | = | $y_{9}\,\mathbf{a}_{1}-x_{9}\,\mathbf{a}_{2}+z_{9}\,\mathbf{a}_{3}$ | = | $y_{9}a\,\hat{\mathbf{x}}-x_{9}a\,\hat{\mathbf{y}}+z_{9}c\,\hat{\mathbf{z}}$ | $(8d)$ | O VI |
| $\mathbf{B}_{37}$ | = | $x_{9}\,\mathbf{a}_{1}-y_{9}\,\mathbf{a}_{2}+\left(\frac{1}{2}+z_{9}\right)\mathbf{a}_{3}$ | = | $x_{9}a\,\hat{\mathbf{x}}-y_{9}a\,\hat{\mathbf{y}}+\left(\frac{1}{2}+z_{9}\right)c\,\hat{\mathbf{z}}$ | $(8d)$ | O VI |
| $\mathbf{B}_{38}$ | = | $-x_{9}\,\mathbf{a}_{1} + y_{9}\,\mathbf{a}_{2} + \left(\frac{1}{2} + z_{9}\right)\,\mathbf{a}_{3}$ | = | $-x_{9}a\,\mathbf{\hat{x}} + y_{9}a\,\mathbf{\hat{y}} + \left(\frac{1}{2} + z_{9}\right)c\,\mathbf{\hat{z}}$ | $(8d)$ | O VI |
| $\mathbf{B}_{39}$ | = | $-y_{9}\,\mathbf{a}_{1} - x_{9}\,\mathbf{a}_{2} + \left(\frac{1}{2} + z_{9}\right)\,\mathbf{a}_{3}$ | = | $-y_{9}a\,\mathbf{\hat{x}} - x_{9}a\,\mathbf{\hat{y}} + \left(\frac{1}{2} + z_{9}\right)c\,\mathbf{\hat{z}}$ | $(8d)$ | O VI |
| $\mathbf{B}_{40}$ | = | $y_{9}\,\mathbf{a}_{1} + x_{9}\,\mathbf{a}_{2} + \left(\frac{1}{2} + z_{9}\right)\,\mathbf{a}_{3}$ | = | $y_{9}a\,\mathbf{\hat{x}} + x_{9}a\,\mathbf{\hat{y}} + \left(\frac{1}{2} + z_{9}\right)c\,\mathbf{\hat{z}}$ | $(8d)$ | O VI |
| $\mathbf{B}_{41}$ | = | $x_{10}\,\mathbf{a}_{1} + y_{10}\,\mathbf{a}_{2} + z_{10}\,\mathbf{a}_{3}$ | = | $x_{10}a\,\mathbf{\hat{x}} + y_{10}a\,\mathbf{\hat{y}} + z_{10}c\,\mathbf{\hat{z}}$ | $(8d)$ | O VII |
| $\mathbf{B}_{42}$ | = | $-x_{10}\,\mathbf{a}_{1} - y_{10}\,\mathbf{a}_{2} + z_{10}\,\mathbf{a}_{3}$ | = | $-x_{10}a\,\mathbf{\hat{x}} - y_{10}a\,\mathbf{\hat{y}} + z_{10}c\,\mathbf{\hat{z}}$ | $(8d)$ | O VII |
| $\mathbf{B}_{43}$ | = | $-y_{10}\,\mathbf{a}_{1} + x_{10}\,\mathbf{a}_{2} + z_{10}\,\mathbf{a}_{3}$ | = | $-y_{10}a\,\mathbf{\hat{x}} + x_{10}a\,\mathbf{\hat{y}} + z_{10}c\,\mathbf{\hat{z}}$ | $(8d)$ | O VII |
| $\mathbf{B}_{44}$ | = | $y_{10}\,\mathbf{a}_{1} - x_{10}\,\mathbf{a}_{2} + z_{10}\,\mathbf{a}_{3}$ | = | $y_{10}a\,\mathbf{\hat{x}} - x_{10}a\,\mathbf{\hat{y}} + z_{10}c\,\mathbf{\hat{z}}$ | $(8d)$ | O VII |
| $\mathbf{B}_{45}$ | = | $x_{10}\,\mathbf{a}_{1} - y_{10}\,\mathbf{a}_{2} + \left(\frac{1}{2} + z_{10}\right)\,\mathbf{a}_{3}$ | = | $x_{10}a\,\mathbf{\hat{x}} - y_{10}a\,\mathbf{\hat{y}} + \left(\frac{1}{2} + z_{10}\right)c\,\mathbf{\hat{z}}$ | $(8d)$ | O VII |
| $\mathbf{B}_{46}$ | = | $-x_{10}\,\mathbf{a}_{1} + y_{10}\,\mathbf{a}_{2} + \left(\frac{1}{2} + z_{10}\right)\,\mathbf{a}_{3}$ | = | $-x_{10}a\,\mathbf{\hat{x}} + y_{10}a\,\mathbf{\hat{y}} + \left(\frac{1}{2} + z_{10}\right)c\,\mathbf{\hat{z}}$ | $(8d)$ | O VII |
| $\mathbf{B}_{47}$ | = | $-y_{10}\,\mathbf{a}_{1} - x_{10}\,\mathbf{a}_{2} + \left(\frac{1}{2} + z_{10}\right)\,\mathbf{a}_{3}$ | = | $-y_{10}a\,\mathbf{\hat{x}} - x_{10}a\,\mathbf{\hat{y}} + \left(\frac{1}{2} + z_{10}\right)c\,\mathbf{\hat{z}}$ | $(8d)$ | O VII |
| $\mathbf{B}_{48}$ | = | $y_{10}\,\mathbf{a}_{1} + x_{10}\,\mathbf{a}_{2} + \left(\frac{1}{2} + z_{10}\right)\,\mathbf{a}_{3}$ | = | $y_{10}a\,\mathbf{\hat{x}} + x_{10}a\,\mathbf{\hat{y}} + \left(\frac{1}{2} + z_{10}\right)c\,\mathbf{\hat{z}}$ | $(8d)$ | O VII |
| $\mathbf{B}_{49}$ | = | $x_{11}\,\mathbf{a}_{1} + y_{11}\,\mathbf{a}_{2} + z_{11}\,\mathbf{a}_{3}$ | = | $x_{11}a\,\mathbf{\hat{x}} + y_{11}a\,\mathbf{\hat{y}} + z_{11}c\,\mathbf{\hat{z}}$ | $(8d)$ | O VIII |
| $\mathbf{B}_{50}$ | = | $-x_{11}\,\mathbf{a}_{1} - y_{11}\,\mathbf{a}_{2} + z_{11}\,\mathbf{a}_{3}$ | = | $-x_{11}a\,\mathbf{\hat{x}} - y_{11}a\,\mathbf{\hat{y}} + z_{11}c\,\mathbf{\hat{z}}$ | $(8d)$ | O VIII |
| $\mathbf{B}_{51}$ | = | $-y_{11}\,\mathbf{a}_{1} + x_{11}\,\mathbf{a}_{2} + z_{11}\,\mathbf{a}_{3}$ | = | $-y_{11}a\,\mathbf{\hat{x}} + x_{11}a\,\mathbf{\hat{y}} + z_{11}c\,\mathbf{\hat{z}}$ | $(8d)$ | O VIII |
| $\mathbf{B}_{52}$ | = | $y_{11}\,\mathbf{a}_{1} - x_{11}\,\mathbf{a}_{2} + z_{11}\,\mathbf{a}_{3}$ | = | $y_{11}a\,\mathbf{\hat{x}} - x_{11}a\,\mathbf{\hat{y}} + z_{11}c\,\mathbf{\hat{z}}$ | $(8d)$ | O VIII |
| $\mathbf{B}_{53}$ | = | $x_{11}\,\mathbf{a}_{1} - y_{11}\,\mathbf{a}_{2} + \left(\frac{1}{2} + z_{11}\right)\,\mathbf{a}_{3}$ | = | $x_{11}a\,\mathbf{\hat{x}} - y_{11}a\,\mathbf{\hat{y}} + \left(\frac{1}{2} + z_{11}\right)c\,\mathbf{\hat{z}}$ | $(8d)$ | O VIII |
| $\mathbf{B}_{54}$ | = | $-x_{11}\,\mathbf{a}_{1} + y_{11}\,\mathbf{a}_{2} + \left(\frac{1}{2} + z_{11}\right)\,\mathbf{a}_{3}$ | = | $-x_{11}a\,\mathbf{\hat{x}} + y_{11}a\,\mathbf{\hat{y}} + \left(\frac{1}{2} + z_{11}\right)c\,\mathbf{\hat{z}}$ | $(8d)$ | O VIII |
| $\mathbf{B}_{55}$ | = | $-y_{11}\,\mathbf{a}_{1} - x_{11}\,\mathbf{a}_{2} + \left(\frac{1}{2} + z_{11}\right)\,\mathbf{a}_{3}$ | = | $-y_{11}a\,\mathbf{\hat{x}} - x_{11}a\,\mathbf{\hat{y}} + \left(\frac{1}{2} + z_{11}\right)c\,\mathbf{\hat{z}}$ | $(8d)$ | O VIII |
| $\mathbf{B}_{56}$ | = | $y_{11}\,\mathbf{a}_{1} + x_{11}\,\mathbf{a}_{2} + \left(\frac{1}{2} + z_{11}\right)\,\mathbf{a}_{3}$ | = | $y_{11}a\,\mathbf{\hat{x}} + x_{11}a\,\mathbf{\hat{y}} + \left(\frac{1}{2} + z_{11}\right)c\,\mathbf{\hat{z}}$ | $(8d)$ | O VIII |
| $\mathbf{B}_{57}$ | = | $x_{12}\,\mathbf{a}_{1} + y_{12}\,\mathbf{a}_{2} + z_{12}\,\mathbf{a}_{3}$ | = | $x_{12}a\,\mathbf{\hat{x}} + y_{12}a\,\mathbf{\hat{y}} + z_{12}c\,\mathbf{\hat{z}}$ | $(8d)$ | Se I |
| $\mathbf{B}_{58}$ | = | $-x_{12}\,\mathbf{a}_{1} - y_{12}\,\mathbf{a}_{2} + z_{12}\,\mathbf{a}_{3}$ | = | $-x_{12}a\,\mathbf{\hat{x}} - y_{12}a\,\mathbf{\hat{y}} + z_{12}c\,\mathbf{\hat{z}}$ | $(8d)$ | Se I |
| $\mathbf{B}_{59}$ | = | $-y_{12}\,\mathbf{a}_{1} + x_{12}\,\mathbf{a}_{2} + z_{12}\,\mathbf{a}_{3}$ | = | $-y_{12}a\,\mathbf{\hat{x}} + x_{12}a\,\mathbf{\hat{y}} + z_{12}c\,\mathbf{\hat{z}}$ | $(8d)$ | Se I |
| $\mathbf{B}_{60}$ | = | $y_{12}\,\mathbf{a}_{1} - x_{12}\,\mathbf{a}_{2} + z_{12}\,\mathbf{a}_{3}$ | = | $y_{12}a\,\mathbf{\hat{x}} - x_{12}a\,\mathbf{\hat{y}} + z_{12}c\,\mathbf{\hat{z}}$ | $(8d)$ | Se I |
| $\mathbf{B}_{61}$ | = | $x_{12}\,\mathbf{a}_{1} - y_{12}\,\mathbf{a}_{2} + \left(\frac{1}{2} + z_{12}\right)\,\mathbf{a}_{3}$ | = | $x_{12}a\,\mathbf{\hat{x}} - y_{12}a\,\mathbf{\hat{y}} + \left(\frac{1}{2} + z_{12}\right)c\,\mathbf{\hat{z}}$ | $(8d)$ | Se I |
| $\mathbf{B}_{62}$ | = | $-x_{12}\,\mathbf{a}_{1} + y_{12}\,\mathbf{a}_{2} + \left(\frac{1}{2} + z_{12}\right)\,\mathbf{a}_{3}$ | = | $-x_{12}a\,\mathbf{\hat{x}} + y_{12}a\,\mathbf{\hat{y}} + \left(\frac{1}{2} + z_{12}\right)c\,\mathbf{\hat{z}}$ | $(8d)$ | Se I |
| $\mathbf{B}_{63}$ | = | $-y_{12}\,\mathbf{a}_{1} - x_{12}\,\mathbf{a}_{2} + \left(\frac{1}{2} + z_{12}\right)\,\mathbf{a}_{3}$ | = | $-y_{12}a\,\mathbf{\hat{x}} - x_{12}a\,\mathbf{\hat{y}} + \left(\frac{1}{2} + z_{12}\right)c\,\mathbf{\hat{z}}$ | $(8d)$ | Se I |
| $\mathbf{B}_{64}$ | = | $y_{12}\,\mathbf{a}_{1} + x_{12}\,\mathbf{a}_{2} + \left(\frac{1}{2} + z_{12}\right)\,\mathbf{a}_{3}$ | = | $y_{12}a\,\mathbf{\hat{x}} + x_{12}a\,\mathbf{\hat{y}} + \left(\frac{1}{2} + z_{12}\right)c\,\mathbf{\hat{z}}$ | $(8d)$ | Se I |
| $\mathbf{B}_{65}$ | = | $x_{13}\,\mathbf{a}_{1} + y_{13}\,\mathbf{a}_{2} + z_{13}\,\mathbf{a}_{3}$ | = | $x_{13}a\,\mathbf{\hat{x}} + y_{13}a\,\mathbf{\hat{y}} + z_{13}c\,\mathbf{\hat{z}}$ | $(8d)$ | Se II |
| $\mathbf{B}_{66}$ | = | $-x_{13}\,\mathbf{a}_{1} - y_{13}\,\mathbf{a}_{2} + z_{13}\,\mathbf{a}_{3}$ | = | $-x_{13}a\,\mathbf{\hat{x}} - y_{13}a\,\mathbf{\hat{y}} + z_{13}c\,\mathbf{\hat{z}}$ | $(8d)$ | Se II |
| $\mathbf{B}_{67}$ | = | $-y_{13}\,\mathbf{a}_{1} + x_{13}\,\mathbf{a}_{2} + z_{13}\,\mathbf{a}_{3}$ | = | $-y_{13}a\,\mathbf{\hat{x}} + x_{13}a\,\mathbf{\hat{y}} + z_{13}c\,\mathbf{\hat{z}}$ | $(8d)$ | Se II |
| $\mathbf{B}_{68}$ | = | $y_{13}\,\mathbf{a}_{1} - x_{13}\,\mathbf{a}_{2} + z_{13}\,\mathbf{a}_{3}$ | = | $y_{13}a\,\mathbf{\hat{x}} - x_{13}a\,\mathbf{\hat{y}} + z_{13}c\,\mathbf{\hat{z}}$ | $(8d)$ | Se II |
| $\mathbf{B}_{69}$ | = | $x_{13}\,\mathbf{a}_{1} - y_{13}\,\mathbf{a}_{2} + \left(\frac{1}{2} + z_{13}\right)\,\mathbf{a}_{3}$ | = | $x_{13}a\,\mathbf{\hat{x}} - y_{13}a\,\mathbf{\hat{y}} + \left(\frac{1}{2} + z_{13}\right)c\,\mathbf{\hat{z}}$ | $(8d)$ | Se II |
| $\mathbf{B}_{70}$ | = | $-x_{13}\,\mathbf{a}_{1} + y_{13}\,\mathbf{a}_{2} + \left(\frac{1}{2} + z_{13}\right)\,\mathbf{a}_{3}$ | = | $-x_{13}a\,\mathbf{\hat{x}} + y_{13}a\,\mathbf{\hat{y}} + \left(\frac{1}{2} + z_{13}\right)c\,\mathbf{\hat{z}}$ | $(8d)$ | Se II |
| $\mathbf{B}_{71}$ | = | $-y_{13}\,\mathbf{a}_{1} - x_{13}\,\mathbf{a}_{2} + \left(\frac{1}{2} + z_{13}\right)\,\mathbf{a}_{3}$ | = | $-y_{13}a\,\mathbf{\hat{x}} - x_{13}a\,\mathbf{\hat{y}} + \left(\frac{1}{2} + z_{13}\right)c\,\mathbf{\hat{z}}$ | $(8d)$ | Se II |
| $\mathbf{B}_{72}$ | = | $y_{13}\,\mathbf{a}_{1} + x_{13}\,\mathbf{a}_{2} + \left(\frac{1}{2} + z_{13}\right)\,\mathbf{a}_{3}$ | = | $y_{13}a\,\mathbf{\hat{x}} + x_{13}a\,\mathbf{\hat{y}} + \left(\frac{1}{2} + z_{13}\right)c\,\mathbf{\hat{z}}$ | $(8d)$ | Se II |

---

**Geometry files:**

- CIF: pp. 1680
- POSCAR: pp. 1680

# $BaNiSn_3$ Structure: ABC3_tI10_107_a_a_ab

http://aflow.org/prototype-encyclopedia/ABC3_tI10_107_a_a_ab

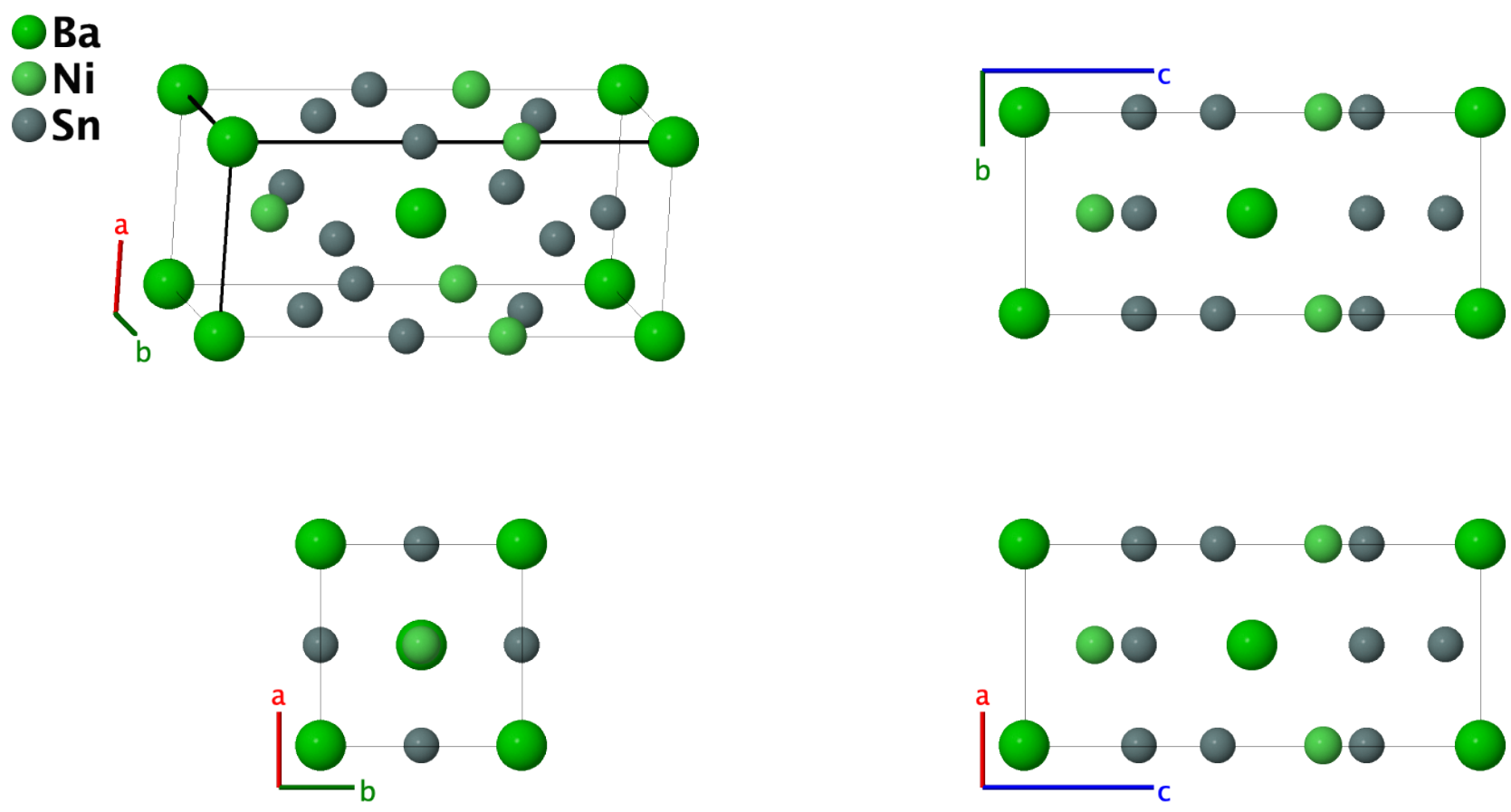

| | | |
|---|---|---|
| **Prototype** | : | $BaNiSn_3$ |
| **AFLOW prototype label** | : | ABC3_tI10_107_a_a_ab |
| ***Strukturbericht* designation** | : | None |
| **Pearson symbol** | : | tI10 |
| **Space group number** | : | 107 |
| **Space group symbol** | : | *I*4*mm* |
| **AFLOW prototype command** | : | `aflow --proto=ABC3_tI10_107_a_a_ab --params=`$a, c/a, z_1, z_2, z_3, z_4$ |

### Other compounds with this structure

- $BaPdSn_3$, $CeCuAl_3$, $EuPdGe_3$, $LaOsSb_3$, $SrIrAl_3$, $SrNiSn_3$, $SrPdAl_3$, and $SrPtAl_3$

---

- This is a ternary form of the $D1_3$ ($BaAl_4$) structure. The atomic positions in both conventional cells are approximately the same, but the distribution of the atoms on those sites and the resulting relaxation leads to a different structure.
- Although (Dörrscheidt, 1978) give the structural information for $BaPtSn_3$ before that of $BaNiSn_3$, (Shatruk, 2019) and others list $BaNiSn_3$ as the prototype for this structure.
- Space group *I*4*mm* #107 does not specify the origin of the $z$-axis. (Dörrscheidt, 1978) places the barium atom at the origin.

### Body-centered Tetragonal primitive vectors:

$$
\begin{aligned}
\mathbf{a}_1 &= -\tfrac{1}{2}\,a\,\hat{\mathbf{x}} + \tfrac{1}{2}\,a\,\hat{\mathbf{y}} + \tfrac{1}{2}\,c\,\hat{\mathbf{z}} \\
\mathbf{a}_2 &= \tfrac{1}{2}\,a\,\hat{\mathbf{x}} - \tfrac{1}{2}\,a\,\hat{\mathbf{y}} + \tfrac{1}{2}\,c\,\hat{\mathbf{z}} \\
\mathbf{a}_3 &= \tfrac{1}{2}\,a\,\hat{\mathbf{x}} + \tfrac{1}{2}\,a\,\hat{\mathbf{y}} - \tfrac{1}{2}\,c\,\hat{\mathbf{z}}
\end{aligned}
$$

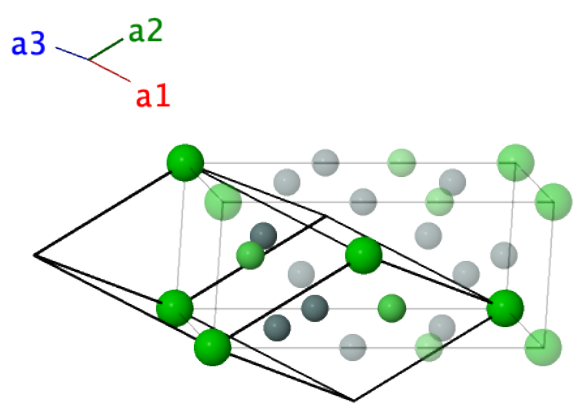

### Basis vectors:

| Lattice Coordinates | Cartesian Coordinates | Wyckoff Position | Atom Type |
|---|---|---|---|

| | | | | | | |
|---|---|---|---|---|---|---|
| $\mathbf{B}_1$ | = | $z_1\,\mathbf{a}_1 + z_1\,\mathbf{a}_2$ | = | $z_1 c\,\hat{\mathbf{z}}$ | (2$a$) | Ba |
| $\mathbf{B}_2$ | = | $z_2\,\mathbf{a}_1 + z_2\,\mathbf{a}_2$ | = | $z_2 c\,\hat{\mathbf{z}}$ | (2$a$) | Ni |
| $\mathbf{B}_3$ | = | $z_3\,\mathbf{a}_1 + z_3\,\mathbf{a}_2$ | = | $z_3 c\,\hat{\mathbf{z}}$ | (2$a$) | Sn I |
| $\mathbf{B}_4$ | = | $\left(\frac{1}{2} + z_4\right)\mathbf{a}_1 + z_4\,\mathbf{a}_2 + \frac{1}{2}\,\mathbf{a}_3$ | = | $\frac{1}{2}a\,\hat{\mathbf{y}} + z_4 c\,\hat{\mathbf{z}}$ | (4$b$) | Sn II |
| $\mathbf{B}_5$ | = | $z_4\,\mathbf{a}_1 + \left(\frac{1}{2} + z_4\right)\mathbf{a}_2 + \frac{1}{2}\,\mathbf{a}_3$ | = | $\frac{1}{2}a\,\hat{\mathbf{x}} + z_4 c\,\hat{\mathbf{z}}$ | (4$b$) | Sn II |

---

---

**Geometry files:**

- CIF: pp. 1681
- POSCAR: pp. 1681

# $E3_1$ ($\beta$-$Ag_2HgI_4$) (*obsolete*) Structure: A2BC4_tP7_111_f_a_n

http://aflow.org/prototype-encyclopedia/A2BC4_tP7_111_f_a_n

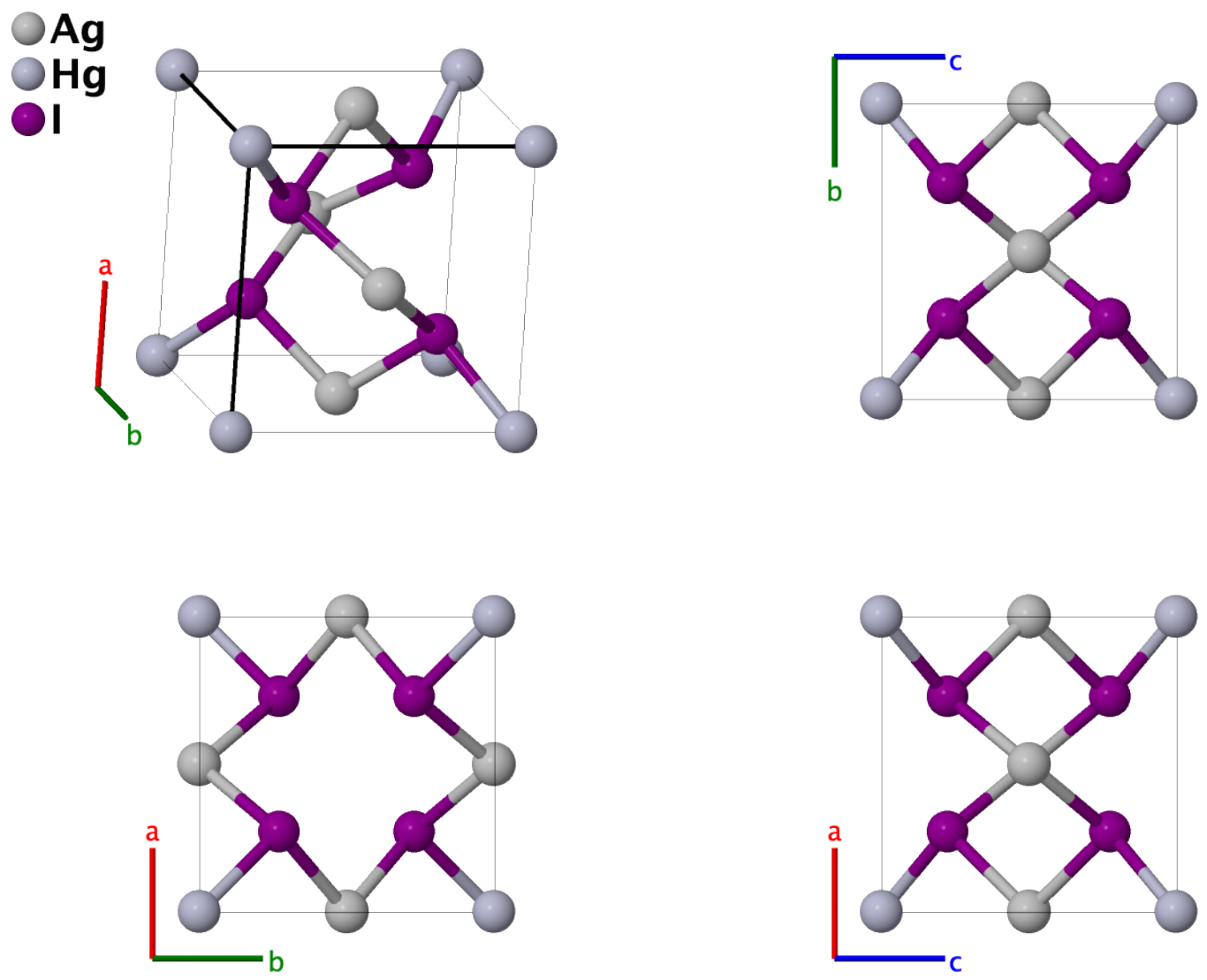

| | | |
|---|---|---|
| **Prototype** | : | $Ag_2HgI_4$ |
| **AFLOW prototype label** | : | A2BC4_tP7_111_f_a_n |
| ***Strukturbericht* designation** | : | $E3_1$ |
| **Pearson symbol** | : | tP7 |
| **Space group number** | : | 111 |
| **Space group symbol** | : | $P\bar{4}2m$ |
| **AFLOW prototype command** | : | `aflow --proto=A2BC4_tP7_111_f_a_n --params=`$a, c/a, x_3, z_3$ |

- (Ketelaar, 1931) determined this pseudo-cubic structure for $\beta$-$Ag_2HgI_4$, and (Hermann, 1937) assigned it the *Strukturbericht* symbol $E3_1$. Later, (Browall, 1974) showed that $\beta$-$Ag_2HgI_4$ takes the $Al_2CdS_4$ structure, which some authorities give as *Strukturbericht* $E3$.

**Simple Tetragonal primitive vectors:**

$$\mathbf{a}_1 = a\,\hat{\mathbf{x}}$$
$$\mathbf{a}_2 = a\,\hat{\mathbf{y}}$$
$$\mathbf{a}_3 = c\,\hat{\mathbf{z}}$$

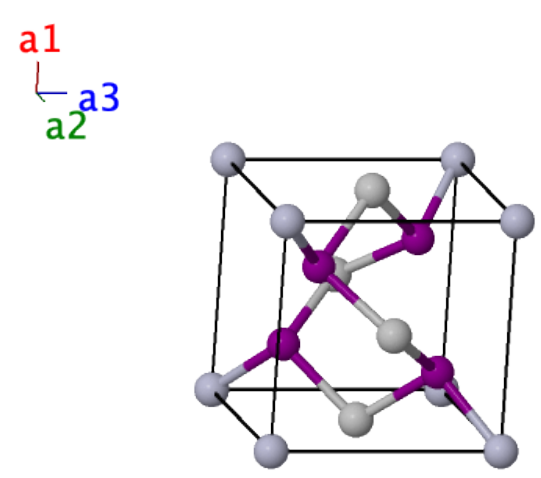

**Basis vectors:**

| | | Lattice Coordinates | | Cartesian Coordinates | Wyckoff Position | Atom Type |
|---|---|---|---|---|---|---|
| $\mathbf{B}_1$ | = | $0\,\mathbf{a}_1 + 0\,\mathbf{a}_2 + 0\,\mathbf{a}_3$ | = | $0\,\hat{\mathbf{x}} + 0\,\hat{\mathbf{y}} + 0\,\hat{\mathbf{z}}$ | (1$a$) | Hg |
| $\mathbf{B}_2$ | = | $\frac{1}{2}\,\mathbf{a}_1 + \frac{1}{2}\,\mathbf{a}_3$ | = | $\frac{1}{2}a\,\hat{\mathbf{x}} + \frac{1}{2}c\,\hat{\mathbf{z}}$ | $(2f)$ | Ag |
| $\mathbf{B}_3$ | = | $\frac{1}{2}\,\mathbf{a}_2 + \frac{1}{2}\,\mathbf{a}_3$ | = | $\frac{1}{2}a\,\hat{\mathbf{y}} + \frac{1}{2}c\,\hat{\mathbf{z}}$ | $(2f)$ | Ag |
| $\mathbf{B}_4$ | = | $x_3\,\mathbf{a}_1 + x_3\,\mathbf{a}_2 + z_3\,\mathbf{a}_3$ | = | $x_3a\,\hat{\mathbf{x}} + x_3a\,\hat{\mathbf{y}} + z_3c\,\hat{\mathbf{z}}$ | $(4n)$ | I |
| $\mathbf{B}_5$ | = | $-x_3\,\mathbf{a}_1 - x_3\,\mathbf{a}_2 + z_3\,\mathbf{a}_3$ | = | $-x_3a\,\hat{\mathbf{x}} - x_3a\,\hat{\mathbf{y}} + z_3c\,\hat{\mathbf{z}}$ | $(4n)$ | I |
| $\mathbf{B}_6$ | = | $x_3\,\mathbf{a}_1 - x_3\,\mathbf{a}_2 - z_3\,\mathbf{a}_3$ | = | $x_3a\,\hat{\mathbf{x}} - x_3a\,\hat{\mathbf{y}} - z_3c\,\hat{\mathbf{z}}$ | $(4n)$ | I |
| $\mathbf{B}_7$ | = | $-x_3\,\mathbf{a}_1 + x_3\,\mathbf{a}_2 - z_3\,\mathbf{a}_3$ | = | $-x_3a\,\hat{\mathbf{x}} + x_3a\,\hat{\mathbf{y}} - z_3c\,\hat{\mathbf{z}}$ | $(4n)$ | I |

**Geometry files:**

- CIF: pp. 1681
- POSCAR: pp. 1681

# Ammonium Chlorite ($NH_4ClO_2$) Structure: AB4CD2_tP16_113_c_f_a_e

http://aflow.org/prototype-encyclopedia/AB4CD2_tP16_113_c_f_a_e

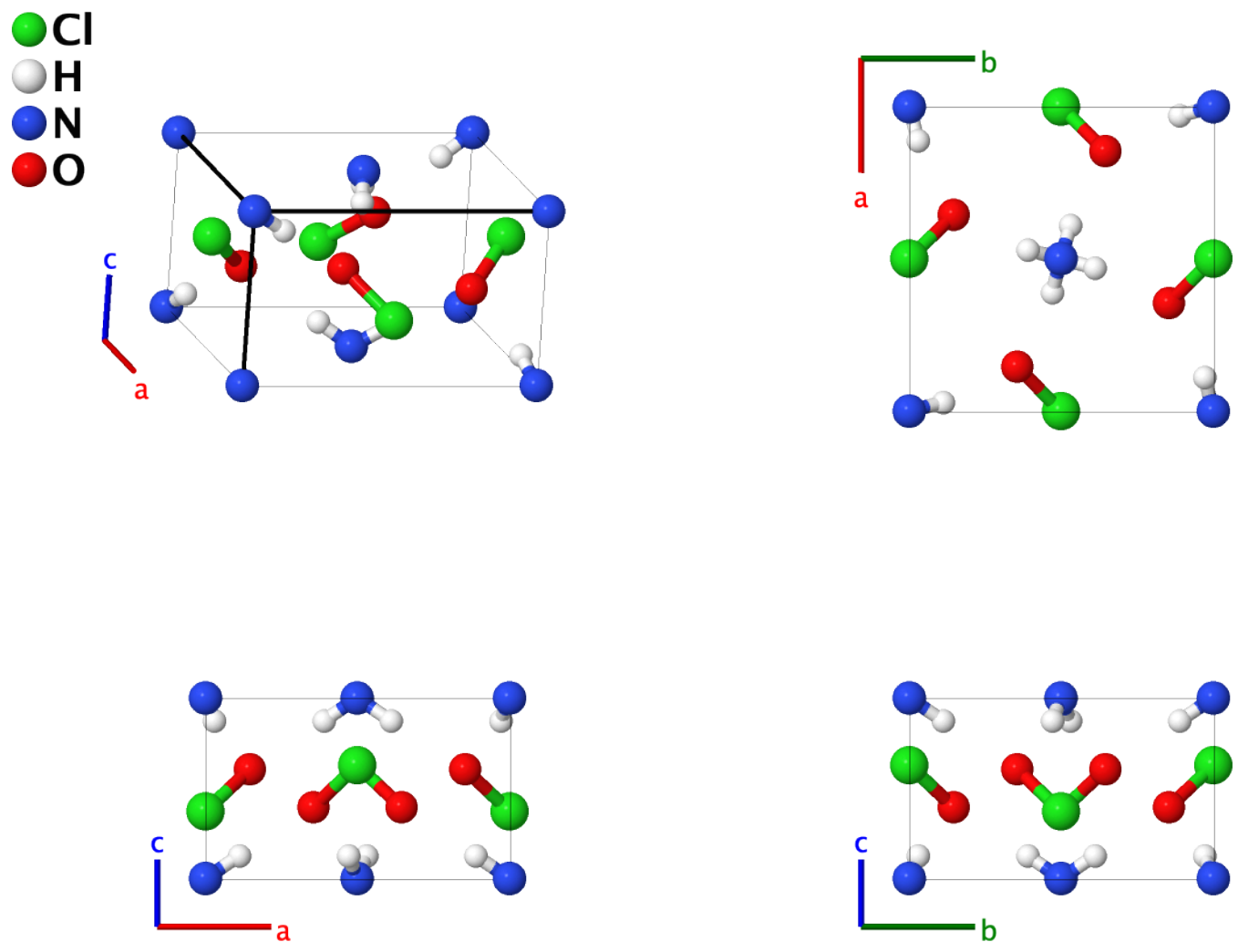

| | | |
|---|---|---|
| **Prototype** | : | $ClH_4NO_2$ |
| **AFLOW prototype label** | : | AB4CD2_tP16_113_c_f_a_e |
| ***Strukturbericht* designation** | : | None |
| **Pearson symbol** | : | tP16 |
| **Space group number** | : | 113 |
| **Space group symbol** | : | $P\bar{4}2_1m$ |
| **AFLOW prototype command** | : | `aflow --proto=AB4CD2_tP16_113_c_f_a_e` `--params=`$a, c/a, z_2, x_3, z_3, x_4, y_4, z_4$ |

- This is the updated version of the $F5_4$ structure, including the positions of the hydrogen atoms.
- The data for this structure was taken at 150 K.
- Not to be confused with ammonium chlorate, $NH_4ClO_3$, which takes on the $KNO_3$ structure, or $NH_4ClO_4$, which has the $H0_5$ structure.

**Simple Tetragonal primitive vectors:**

$$\mathbf{a}_1 = a\,\hat{\mathbf{x}}$$
$$\mathbf{a}_2 = a\,\hat{\mathbf{y}}$$
$$\mathbf{a}_3 = c\,\hat{\mathbf{z}}$$

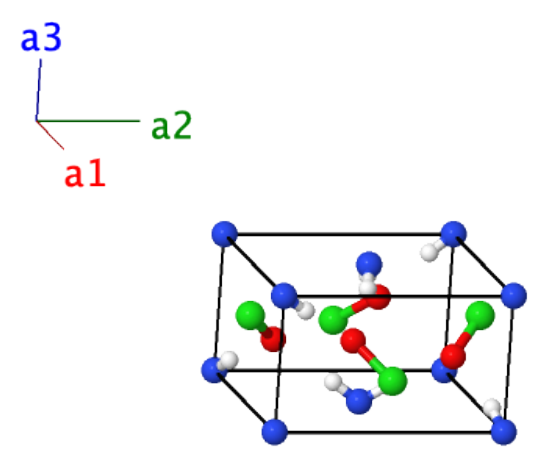

**Basis vectors:**

| | | Lattice Coordinates | | Cartesian Coordinates | Wyckoff Position | Atom Type |
|---|---|---|---|---|---|---|
| $\mathbf{B}_1$ | = | $0\,\mathbf{a}_1 + 0\,\mathbf{a}_2 + 0\,\mathbf{a}_3$ | = | $0\,\hat{\mathbf{x}} + 0\,\hat{\mathbf{y}} + 0\,\hat{\mathbf{z}}$ | (2$a$) | N |
| $\mathbf{B}_2$ | = | $\frac{1}{2}\,\mathbf{a}_1 + \frac{1}{2}\,\mathbf{a}_2$ | = | $\frac{1}{2}a\,\hat{\mathbf{x}} + \frac{1}{2}a\,\hat{\mathbf{y}}$ | (2$a$) | N |
| $\mathbf{B}_3$ | = | $\frac{1}{2}\,\mathbf{a}_2 + z_2\,\mathbf{a}_3$ | = | $\frac{1}{2}a\,\hat{\mathbf{y}} + z_2c\,\hat{\mathbf{z}}$ | (2$c$) | Cl |
| $\mathbf{B}_4$ | = | $\frac{1}{2}\,\mathbf{a}_1 - z_2\,\mathbf{a}_3$ | = | $\frac{1}{2}a\,\hat{\mathbf{x}} - z_2c\,\hat{\mathbf{z}}$ | (2$c$) | Cl |
| $\mathbf{B}_5$ | = | $x_3\,\mathbf{a}_1 + \left(\frac{1}{2} + x_3\right)\,\mathbf{a}_2 + z_3\,\mathbf{a}_3$ | = | $x_3a\,\hat{\mathbf{x}} + \left(\frac{1}{2} + x_3\right)a\,\hat{\mathbf{y}} + z_3c\,\hat{\mathbf{z}}$ | (4$e$) | O |
| $\mathbf{B}_6$ | = | $-x_3\,\mathbf{a}_1 + \left(\frac{1}{2} - x_3\right)\,\mathbf{a}_2 + z_3\,\mathbf{a}_3$ | = | $-x_3a\,\hat{\mathbf{x}} + \left(\frac{1}{2} - x_3\right)a\,\hat{\mathbf{y}} + z_3c\,\hat{\mathbf{z}}$ | (4$e$) | O |
| $\mathbf{B}_7$ | = | $\left(\frac{1}{2} + x_3\right)\,\mathbf{a}_1 - x_3\,\mathbf{a}_2 - z_3\,\mathbf{a}_3$ | = | $\left(\frac{1}{2} + x_3\right)a\,\hat{\mathbf{x}} - x_3a\,\hat{\mathbf{y}} - z_3c\,\hat{\mathbf{z}}$ | (4$e$) | O |
| $\mathbf{B}_8$ | = | $\left(\frac{1}{2} - x_3\right)\,\mathbf{a}_1 + x_3\,\mathbf{a}_2 - z_3\,\mathbf{a}_3$ | = | $\left(\frac{1}{2} - x_3\right)a\,\hat{\mathbf{x}} + x_3a\,\hat{\mathbf{y}} - z_3c\,\hat{\mathbf{z}}$ | (4$e$) | O |
| $\mathbf{B}_9$ | = | $x_4\,\mathbf{a}_1 + y_4\,\mathbf{a}_2 + z_4\,\mathbf{a}_3$ | = | $x_4a\,\hat{\mathbf{x}} + y_4a\,\hat{\mathbf{y}} + z_4c\,\hat{\mathbf{z}}$ | (8$f$) | H |
| $\mathbf{B}_{10}$ | = | $-x_4\,\mathbf{a}_1 - y_4\,\mathbf{a}_2 + z_4\,\mathbf{a}_3$ | = | $-x_4a\,\hat{\mathbf{x}} - y_4a\,\hat{\mathbf{y}} + z_4c\,\hat{\mathbf{z}}$ | (8$f$) | H |
| $\mathbf{B}_{11}$ | = | $y_4\,\mathbf{a}_1 - x_4\,\mathbf{a}_2 - z_4\,\mathbf{a}_3$ | = | $y_4a\,\hat{\mathbf{x}} - x_4a\,\hat{\mathbf{y}} - z_4c\,\hat{\mathbf{z}}$ | (8$f$) | H |
| $\mathbf{B}_{12}$ | = | $-y_4\,\mathbf{a}_1 + x_4\,\mathbf{a}_2 - z_4\,\mathbf{a}_3$ | = | $-y_4a\,\hat{\mathbf{x}} + x_4a\,\hat{\mathbf{y}} - z_4c\,\hat{\mathbf{z}}$ | (8$f$) | H |
| $\mathbf{B}_{13}$ | = | $\left(\frac{1}{2} - x_4\right)\,\mathbf{a}_1 + \left(\frac{1}{2} + y_4\right)\,\mathbf{a}_2 - z_4\,\mathbf{a}_3$ | = | $\left(\frac{1}{2} - x_4\right)a\,\hat{\mathbf{x}} + \left(\frac{1}{2} + y_4\right)a\,\hat{\mathbf{y}} - z_4c\,\hat{\mathbf{z}}$ | (8$f$) | H |
| $\mathbf{B}_{14}$ | = | $\left(\frac{1}{2} + x_4\right)\,\mathbf{a}_1 + \left(\frac{1}{2} - y_4\right)\,\mathbf{a}_2 - z_4\,\mathbf{a}_3$ | = | $\left(\frac{1}{2} + x_4\right)a\,\hat{\mathbf{x}} + \left(\frac{1}{2} - y_4\right)a\,\hat{\mathbf{y}} - z_4c\,\hat{\mathbf{z}}$ | (8$f$) | H |
| $\mathbf{B}_{15}$ | = | $\left(\frac{1}{2} - y_4\right)\,\mathbf{a}_1 + \left(\frac{1}{2} - x_4\right)\,\mathbf{a}_2 + z_4\,\mathbf{a}_3$ | = | $\left(\frac{1}{2} - y_4\right)a\,\hat{\mathbf{x}} + \left(\frac{1}{2} - x_4\right)a\,\hat{\mathbf{y}} + z_4c\,\hat{\mathbf{z}}$ | (8$f$) | H |
| $\mathbf{B}_{16}$ | = | $\left(\frac{1}{2} + y_4\right)\,\mathbf{a}_1 + \left(\frac{1}{2} + x_4\right)\,\mathbf{a}_2 + z_4\,\mathbf{a}_3$ | = | $\left(\frac{1}{2} + y_4\right)a\,\hat{\mathbf{x}} + \left(\frac{1}{2} + x_4\right)a\,\hat{\mathbf{y}} + z_4c\,\hat{\mathbf{z}}$ | (8$f$) | H |

**Geometry files:**

- CIF: pp. 1682
- POSCAR: pp. 1682

# $C_{19}Sc_{15}$ Structure: A19B15_tP68_114_bc4e_ac3e

http://aflow.org/prototype-encyclopedia/A19B15_tP68_114_bc4e_ac3e

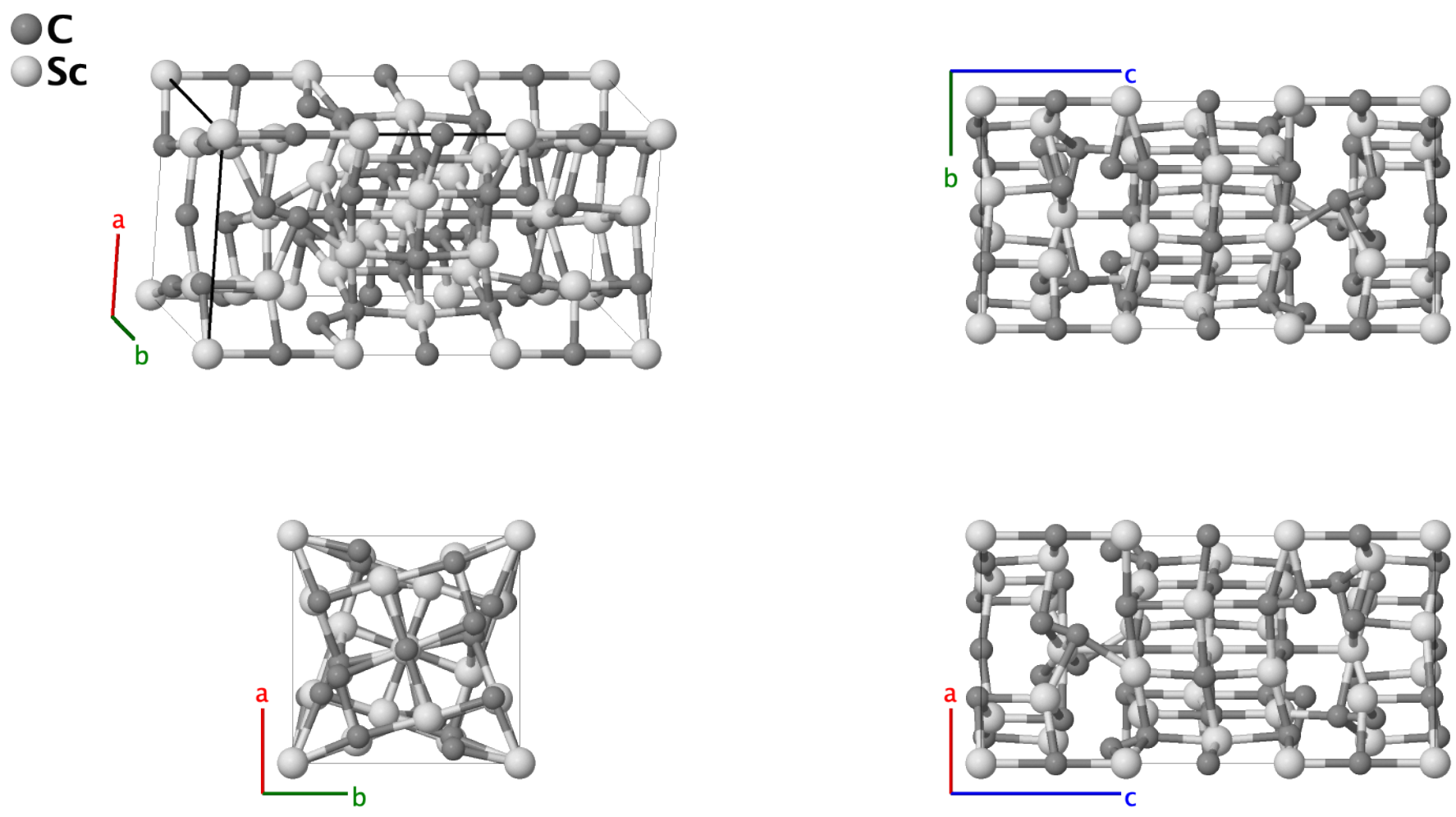

| | | |
|---|---|---|
| **Prototype** | : | $C_{19}Sc_{15}$ |
| **AFLOW prototype label** | : | A19B15_tP68_114_bc4e_ac3e |
| ***Strukturbericht* designation** | : | None |
| **Pearson symbol** | : | tP68 |
| **Space group number** | : | 114 |
| **Space group symbol** | : | $P\bar{4}2_1c$ |
| **AFLOW prototype command** | : | `aflow --proto=A19B15_tP68_114_bc4e_ac3e` `--params=`$a, c/a, z_3, z_4, x_5, y_5, z_5, x_6, y_6, z_6, x_7, y_7, z_7, x_8, y_8, z_8, x_9, y_9, z_9, x_{10}, y_{10}, z_{10}, x_{11}, y_{11}, z_{11}$ |

### Other compounds with this structure

- $C_{19}Er_{15}$, $C_{19}Lu_{15}$, $C_{19}Tm_{15}$, $C_{19}Y_{15}$, and $C_{19}Yb_{15}$

### Simple Tetragonal primitive vectors:

$$\mathbf{a}_1 = a\,\hat{\mathbf{x}}$$
$$\mathbf{a}_2 = a\,\hat{\mathbf{y}}$$
$$\mathbf{a}_3 = c\,\hat{\mathbf{z}}$$

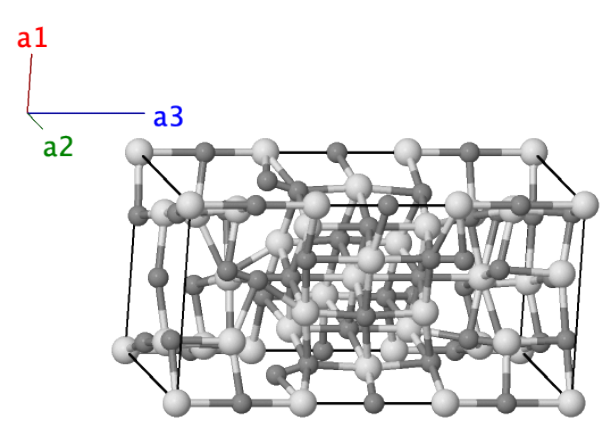

### Basis vectors:

| | | Lattice Coordinates | | Cartesian Coordinates | Wyckoff Position | Atom Type |
|---|---|---|---|---|---|---|
| $\mathbf{B}_1$ | = | $0\,\mathbf{a}_1 + 0\,\mathbf{a}_2 + 0\,\mathbf{a}_3$ | = | $0\,\hat{\mathbf{x}} + 0\,\hat{\mathbf{y}} + 0\,\hat{\mathbf{z}}$ | (2*a*) | Sc I |
| $\mathbf{B}_2$ | = | $\frac{1}{2}\,\mathbf{a}_1 + \frac{1}{2}\,\mathbf{a}_2 + \frac{1}{2}\,\mathbf{a}_3$ | = | $\frac{1}{2}a\,\hat{\mathbf{x}} + \frac{1}{2}a\,\hat{\mathbf{y}} + \frac{1}{2}c\,\hat{\mathbf{z}}$ | (2*a*) | Sc I |
| $\mathbf{B}_3$ | = | $\frac{1}{2}\,\mathbf{a}_3$ | = | $\frac{1}{2}c\,\hat{\mathbf{z}}$ | (2*b*) | C I |
| $\mathbf{B}_4$ | = | $\frac{1}{2}\,\mathbf{a}_1 + \frac{1}{2}\,\mathbf{a}_2$ | = | $\frac{1}{2}a\,\hat{\mathbf{x}} + \frac{1}{2}a\,\hat{\mathbf{y}}$ | (2*b*) | C I |
| $\mathbf{B}_5$ | = | $z_3\,\mathbf{a}_3$ | = | $z_3c\,\hat{\mathbf{z}}$ | (4*c*) | C II |
| $\mathbf{B}_{6}$ | = | $-z_3\,\mathbf{a}_3$ | = | $-z_3 c\,\hat{\mathbf{z}}$ | (4*c*) | C II |
| $\mathbf{B}_{7}$ | = | $\frac{1}{2}\,\mathbf{a}_1 + \frac{1}{2}\,\mathbf{a}_2 + \left(\frac{1}{2} - z_3\right)\,\mathbf{a}_3$ | = | $\frac{1}{2}a\,\hat{\mathbf{x}} + \frac{1}{2}a\,\hat{\mathbf{y}} + \left(\frac{1}{2} - z_3\right)c\,\hat{\mathbf{z}}$ | (4*c*) | C II |
| $\mathbf{B}_{8}$ | = | $\frac{1}{2}\,\mathbf{a}_1 + \frac{1}{2}\,\mathbf{a}_2 + \left(\frac{1}{2} + z_3\right)\,\mathbf{a}_3$ | = | $\frac{1}{2}a\,\hat{\mathbf{x}} + \frac{1}{2}a\,\hat{\mathbf{y}} + \left(\frac{1}{2} + z_3\right)c\,\hat{\mathbf{z}}$ | (4*c*) | C II |
| $\mathbf{B}_{9}$ | = | $z_4\,\mathbf{a}_3$ | = | $z_4 c\,\hat{\mathbf{z}}$ | (4*c*) | Sc II |
| $\mathbf{B}_{10}$ | = | $-z_4\,\mathbf{a}_3$ | = | $-z_4 c\,\hat{\mathbf{z}}$ | (4*c*) | Sc II |
| $\mathbf{B}_{11}$ | = | $\frac{1}{2}\,\mathbf{a}_1 + \frac{1}{2}\,\mathbf{a}_2 + \left(\frac{1}{2} - z_4\right)\,\mathbf{a}_3$ | = | $\frac{1}{2}a\,\hat{\mathbf{x}} + \frac{1}{2}a\,\hat{\mathbf{y}} + \left(\frac{1}{2} - z_4\right)c\,\hat{\mathbf{z}}$ | (4*c*) | Sc II |
| $\mathbf{B}_{12}$ | = | $\frac{1}{2}\,\mathbf{a}_1 + \frac{1}{2}\,\mathbf{a}_2 + \left(\frac{1}{2} + z_4\right)\,\mathbf{a}_3$ | = | $\frac{1}{2}a\,\hat{\mathbf{x}} + \frac{1}{2}a\,\hat{\mathbf{y}} + \left(\frac{1}{2} + z_4\right)c\,\hat{\mathbf{z}}$ | (4*c*) | Sc II |
| $\mathbf{B}_{13}$ | = | $x_5\,\mathbf{a}_1 + y_5\,\mathbf{a}_2 + z_5\,\mathbf{a}_3$ | = | $x_5 a\,\hat{\mathbf{x}} + y_5 a\,\hat{\mathbf{y}} + z_5 c\,\hat{\mathbf{z}}$ | (8*e*) | C III |
| $\mathbf{B}_{14}$ | = | $-x_5\,\mathbf{a}_1 - y_5\,\mathbf{a}_2 + z_5\,\mathbf{a}_3$ | = | $-x_5 a\,\hat{\mathbf{x}} - y_5 a\,\hat{\mathbf{y}} + z_5 c\,\hat{\mathbf{z}}$ | (8*e*) | C III |
| $\mathbf{B}_{15}$ | = | $y_5\,\mathbf{a}_1 - x_5\,\mathbf{a}_2 - z_5\,\mathbf{a}_3$ | = | $y_5 a\,\hat{\mathbf{x}} - x_5 a\,\hat{\mathbf{y}} - z_5 c\,\hat{\mathbf{z}}$ | (8*e*) | C III |
| $\mathbf{B}_{16}$ | = | $-y_5\,\mathbf{a}_1 + x_5\,\mathbf{a}_2 - z_5\,\mathbf{a}_3$ | = | $-y_5 a\,\hat{\mathbf{x}} + x_5 a\,\hat{\mathbf{y}} - z_5 c\,\hat{\mathbf{z}}$ | (8*e*) | C III |
| $\mathbf{B}_{17}$ | = | $\left(\frac{1}{2} - x_5\right)\,\mathbf{a}_1 + \left(\frac{1}{2} + y_5\right)\,\mathbf{a}_2 + \left(\frac{1}{2} - z_5\right)\,\mathbf{a}_3$ | = | $\left(\frac{1}{2} - x_5\right)a\,\hat{\mathbf{x}} + \left(\frac{1}{2} + y_5\right)a\,\hat{\mathbf{y}} + \left(\frac{1}{2} - z_5\right)c\,\hat{\mathbf{z}}$ | (8*e*) | C III |
| $\mathbf{B}_{18}$ | = | $\left(\frac{1}{2} + x_5\right)\,\mathbf{a}_1 + \left(\frac{1}{2} - y_5\right)\,\mathbf{a}_2 + \left(\frac{1}{2} - z_5\right)\,\mathbf{a}_3$ | = | $\left(\frac{1}{2} + x_5\right)a\,\hat{\mathbf{x}} + \left(\frac{1}{2} - y_5\right)a\,\hat{\mathbf{y}} + \left(\frac{1}{2} - z_5\right)c\,\hat{\mathbf{z}}$ | (8*e*) | C III |
| $\mathbf{B}_{19}$ | = | $\left(\frac{1}{2} - y_5\right)\,\mathbf{a}_1 + \left(\frac{1}{2} - x_5\right)\,\mathbf{a}_2 + \left(\frac{1}{2} + z_5\right)\,\mathbf{a}_3$ | = | $\left(\frac{1}{2} - y_5\right)a\,\hat{\mathbf{x}} + \left(\frac{1}{2} - x_5\right)a\,\hat{\mathbf{y}} + \left(\frac{1}{2} + z_5\right)c\,\hat{\mathbf{z}}$ | (8*e*) | C III |
| $\mathbf{B}_{20}$ | = | $\left(\frac{1}{2} + y_5\right)\,\mathbf{a}_1 + \left(\frac{1}{2} + x_5\right)\,\mathbf{a}_2 + \left(\frac{1}{2} + z_5\right)\,\mathbf{a}_3$ | = | $\left(\frac{1}{2} + y_5\right)a\,\hat{\mathbf{x}} + \left(\frac{1}{2} + x_5\right)a\,\hat{\mathbf{y}} + \left(\frac{1}{2} + z_5\right)c\,\hat{\mathbf{z}}$ | (8*e*) | C III |
| $\mathbf{B}_{21}$ | = | $x_6\,\mathbf{a}_1 + y_6\,\mathbf{a}_2 + z_6\,\mathbf{a}_3$ | = | $x_6 a\,\hat{\mathbf{x}} + y_6 a\,\hat{\mathbf{y}} + z_6 c\,\hat{\mathbf{z}}$ | (8*e*) | C IV |
| $\mathbf{B}_{22}$ | = | $-x_6\,\mathbf{a}_1 - y_6\,\mathbf{a}_2 + z_6\,\mathbf{a}_3$ | = | $-x_6 a\,\hat{\mathbf{x}} - y_6 a\,\hat{\mathbf{y}} + z_6 c\,\hat{\mathbf{z}}$ | (8*e*) | C IV |
| $\mathbf{B}_{23}$ | = | $y_6\,\mathbf{a}_1 - x_6\,\mathbf{a}_2 - z_6\,\mathbf{a}_3$ | = | $y_6 a\,\hat{\mathbf{x}} - x_6 a\,\hat{\mathbf{y}} - z_6 c\,\hat{\mathbf{z}}$ | (8*e*) | C IV |
| $\mathbf{B}_{24}$ | = | $-y_6\,\mathbf{a}_1 + x_6\,\mathbf{a}_2 - z_6\,\mathbf{a}_3$ | = | $-y_6 a\,\hat{\mathbf{x}} + x_6 a\,\hat{\mathbf{y}} - z_6 c\,\hat{\mathbf{z}}$ | (8*e*) | C IV |
| $\mathbf{B}_{25}$ | = | $\left(\frac{1}{2} - x_6\right)\,\mathbf{a}_1 + \left(\frac{1}{2} + y_6\right)\,\mathbf{a}_2 + \left(\frac{1}{2} - z_6\right)\,\mathbf{a}_3$ | = | $\left(\frac{1}{2} - x_6\right)a\,\hat{\mathbf{x}} + \left(\frac{1}{2} + y_6\right)a\,\hat{\mathbf{y}} + \left(\frac{1}{2} - z_6\right)c\,\hat{\mathbf{z}}$ | (8*e*) | C IV |
| $\mathbf{B}_{26}$ | = | $\left(\frac{1}{2} + x_6\right)\,\mathbf{a}_1 + \left(\frac{1}{2} - y_6\right)\,\mathbf{a}_2 + \left(\frac{1}{2} - z_6\right)\,\mathbf{a}_3$ | = | $\left(\frac{1}{2} + x_6\right)a\,\hat{\mathbf{x}} + \left(\frac{1}{2} - y_6\right)a\,\hat{\mathbf{y}} + \left(\frac{1}{2} - z_6\right)c\,\hat{\mathbf{z}}$ | (8*e*) | C IV |
| $\mathbf{B}_{27}$ | = | $\left(\frac{1}{2} - y_6\right)\,\mathbf{a}_1 + \left(\frac{1}{2} - x_6\right)\,\mathbf{a}_2 + \left(\frac{1}{2} + z_6\right)\,\mathbf{a}_3$ | = | $\left(\frac{1}{2} - y_6\right)a\,\hat{\mathbf{x}} + \left(\frac{1}{2} - x_6\right)a\,\hat{\mathbf{y}} + \left(\frac{1}{2} + z_6\right)c\,\hat{\mathbf{z}}$ | (8*e*) | C IV |
| $\mathbf{B}_{28}$ | = | $\left(\frac{1}{2} + y_6\right)\,\mathbf{a}_1 + \left(\frac{1}{2} + x_6\right)\,\mathbf{a}_2 + \left(\frac{1}{2} + z_6\right)\,\mathbf{a}_3$ | = | $\left(\frac{1}{2} + y_6\right)a\,\hat{\mathbf{x}} + \left(\frac{1}{2} + x_6\right)a\,\hat{\mathbf{y}} + \left(\frac{1}{2} + z_6\right)c\,\hat{\mathbf{z}}$ | (8*e*) | C IV |
| $\mathbf{B}_{29}$ | = | $x_7\,\mathbf{a}_1 + y_7\,\mathbf{a}_2 + z_7\,\mathbf{a}_3$ | = | $x_7 a\,\hat{\mathbf{x}} + y_7 a\,\hat{\mathbf{y}} + z_7 c\,\hat{\mathbf{z}}$ | (8*e*) | C V |
| $\mathbf{B}_{30}$ | = | $-x_7\,\mathbf{a}_1 - y_7\,\mathbf{a}_2 + z_7\,\mathbf{a}_3$ | = | $-x_7 a\,\hat{\mathbf{x}} - y_7 a\,\hat{\mathbf{y}} + z_7 c\,\hat{\mathbf{z}}$ | (8*e*) | C V |
| $\mathbf{B}_{31}$ | = | $y_7\,\mathbf{a}_1 - x_7\,\mathbf{a}_2 - z_7\,\mathbf{a}_3$ | = | $y_7 a\,\hat{\mathbf{x}} - x_7 a\,\hat{\mathbf{y}} - z_7 c\,\hat{\mathbf{z}}$ | (8*e*) | C V |
| $\mathbf{B}_{32}$ | = | $-y_7\,\mathbf{a}_1 + x_7\,\mathbf{a}_2 - z_7\,\mathbf{a}_3$ | = | $-y_7 a\,\hat{\mathbf{x}} + x_7 a\,\hat{\mathbf{y}} - z_7 c\,\hat{\mathbf{z}}$ | (8*e*) | C V |
| $\mathbf{B}_{33}$ | = | $\left(\frac{1}{2} - x_7\right)\,\mathbf{a}_1 + \left(\frac{1}{2} + y_7\right)\,\mathbf{a}_2 + \left(\frac{1}{2} - z_7\right)\,\mathbf{a}_3$ | = | $\left(\frac{1}{2} - x_7\right)a\,\hat{\mathbf{x}} + \left(\frac{1}{2} + y_7\right)a\,\hat{\mathbf{y}} + \left(\frac{1}{2} - z_7\right)c\,\hat{\mathbf{z}}$ | (8*e*) | C V |
| $\mathbf{B}_{34}$ | = | $\left(\frac{1}{2}+x_7\right)\mathbf{a}_1+\left(\frac{1}{2}-y_7\right)\mathbf{a}_2+\left(\frac{1}{2}-z_7\right)\mathbf{a}_3$ | = | $\left(\frac{1}{2}+x_7\right)a\,\mathbf{\hat{x}}+\left(\frac{1}{2}-y_7\right)a\,\mathbf{\hat{y}}+\left(\frac{1}{2}-z_7\right)c\,\mathbf{\hat{z}}$ | (8$e$) | C V |
| $\mathbf{B}_{35}$ | = | $\left(\frac{1}{2}-y_7\right)\mathbf{a}_1+\left(\frac{1}{2}-x_7\right)\mathbf{a}_2+\left(\frac{1}{2}+z_7\right)\mathbf{a}_3$ | = | $\left(\frac{1}{2}-y_7\right)a\,\mathbf{\hat{x}}+\left(\frac{1}{2}-x_7\right)a\,\mathbf{\hat{y}}+\left(\frac{1}{2}+z_7\right)c\,\mathbf{\hat{z}}$ | (8$e$) | C V |
| $\mathbf{B}_{36}$ | = | $\left(\frac{1}{2}+y_7\right)\mathbf{a}_1+\left(\frac{1}{2}+x_7\right)\mathbf{a}_2+\left(\frac{1}{2}+z_7\right)\mathbf{a}_3$ | = | $\left(\frac{1}{2}+y_7\right)a\,\mathbf{\hat{x}}+\left(\frac{1}{2}+x_7\right)a\,\mathbf{\hat{y}}+\left(\frac{1}{2}+z_7\right)c\,\mathbf{\hat{z}}$ | (8$e$) | C V |
| $\mathbf{B}_{37}$ | = | $x_8\,\mathbf{a}_1+y_8\,\mathbf{a}_2+z_8\,\mathbf{a}_3$ | = | $x_8a\,\mathbf{\hat{x}}+y_8a\,\mathbf{\hat{y}}+z_8c\,\mathbf{\hat{z}}$ | (8$e$) | C VI |
| $\mathbf{B}_{38}$ | = | $-x_8\,\mathbf{a}_1-y_8\,\mathbf{a}_2+z_8\,\mathbf{a}_3$ | = | $-x_8a\,\mathbf{\hat{x}}-y_8a\,\mathbf{\hat{y}}+z_8c\,\mathbf{\hat{z}}$ | (8$e$) | C VI |
| $\mathbf{B}_{39}$ | = | $y_8\,\mathbf{a}_1-x_8\,\mathbf{a}_2-z_8\,\mathbf{a}_3$ | = | $y_8a\,\mathbf{\hat{x}}-x_8a\,\mathbf{\hat{y}}-z_8c\,\mathbf{\hat{z}}$ | (8$e$) | C VI |
| $\mathbf{B}_{40}$ | = | $-y_8\,\mathbf{a}_1+x_8\,\mathbf{a}_2-z_8\,\mathbf{a}_3$ | = | $-y_8a\,\mathbf{\hat{x}}+x_8a\,\mathbf{\hat{y}}-z_8c\,\mathbf{\hat{z}}$ | (8$e$) | C VI |
| $\mathbf{B}_{41}$ | = | $\left(\frac{1}{2}-x_8\right)\mathbf{a}_1+\left(\frac{1}{2}+y_8\right)\mathbf{a}_2+\left(\frac{1}{2}-z_8\right)\mathbf{a}_3$ | = | $\left(\frac{1}{2}-x_8\right)a\,\mathbf{\hat{x}}+\left(\frac{1}{2}+y_8\right)a\,\mathbf{\hat{y}}+\left(\frac{1}{2}-z_8\right)c\,\mathbf{\hat{z}}$ | (8$e$) | C VI |
| $\mathbf{B}_{42}$ | = | $\left(\frac{1}{2}+x_8\right)\mathbf{a}_1+\left(\frac{1}{2}-y_8\right)\mathbf{a}_2+\left(\frac{1}{2}-z_8\right)\mathbf{a}_3$ | = | $\left(\frac{1}{2}+x_8\right)a\,\mathbf{\hat{x}}+\left(\frac{1}{2}-y_8\right)a\,\mathbf{\hat{y}}+\left(\frac{1}{2}-z_8\right)c\,\mathbf{\hat{z}}$ | (8$e$) | C VI |
| $\mathbf{B}_{43}$ | = | $\left(\frac{1}{2}-y_8\right)\mathbf{a}_1+\left(\frac{1}{2}-x_8\right)\mathbf{a}_2+\left(\frac{1}{2}+z_8\right)\mathbf{a}_3$ | = | $\left(\frac{1}{2}-y_8\right)a\,\mathbf{\hat{x}}+\left(\frac{1}{2}-x_8\right)a\,\mathbf{\hat{y}}+\left(\frac{1}{2}+z_8\right)c\,\mathbf{\hat{z}}$ | (8$e$) | C VI |
| $\mathbf{B}_{44}$ | = | $\left(\frac{1}{2}+y_8\right)\mathbf{a}_1+\left(\frac{1}{2}+x_8\right)\mathbf{a}_2+\left(\frac{1}{2}+z_8\right)\mathbf{a}_3$ | = | $\left(\frac{1}{2}+y_8\right)a\,\mathbf{\hat{x}}+\left(\frac{1}{2}+x_8\right)a\,\mathbf{\hat{y}}+\left(\frac{1}{2}+z_8\right)c\,\mathbf{\hat{z}}$ | (8$e$) | C VI |
| $\mathbf{B}_{45}$ | = | $x_9\,\mathbf{a}_1+y_9\,\mathbf{a}_2+z_9\,\mathbf{a}_3$ | = | $x_9a\,\mathbf{\hat{x}}+y_9a\,\mathbf{\hat{y}}+z_9c\,\mathbf{\hat{z}}$ | (8$e$) | Sc III |
| $\mathbf{B}_{46}$ | = | $-x_9\,\mathbf{a}_1-y_9\,\mathbf{a}_2+z_9\,\mathbf{a}_3$ | = | $-x_9a\,\mathbf{\hat{x}}-y_9a\,\mathbf{\hat{y}}+z_9c\,\mathbf{\hat{z}}$ | (8$e$) | Sc III |
| $\mathbf{B}_{47}$ | = | $y_9\,\mathbf{a}_1-x_9\,\mathbf{a}_2-z_9\,\mathbf{a}_3$ | = | $y_9a\,\mathbf{\hat{x}}-x_9a\,\mathbf{\hat{y}}-z_9c\,\mathbf{\hat{z}}$ | (8$e$) | Sc III |
| $\mathbf{B}_{48}$ | = | $-y_9\,\mathbf{a}_1+x_9\,\mathbf{a}_2-z_9\,\mathbf{a}_3$ | = | $-y_9a\,\mathbf{\hat{x}}+x_9a\,\mathbf{\hat{y}}-z_9c\,\mathbf{\hat{z}}$ | (8$e$) | Sc III |
| $\mathbf{B}_{49}$ | = | $\left(\frac{1}{2}-x_9\right)\mathbf{a}_1+\left(\frac{1}{2}+y_9\right)\mathbf{a}_2+\left(\frac{1}{2}-z_9\right)\mathbf{a}_3$ | = | $\left(\frac{1}{2}-x_9\right)a\,\mathbf{\hat{x}}+\left(\frac{1}{2}+y_9\right)a\,\mathbf{\hat{y}}+\left(\frac{1}{2}-z_9\right)c\,\mathbf{\hat{z}}$ | (8$e$) | Sc III |
| $\mathbf{B}_{50}$ | = | $\left(\frac{1}{2}+x_9\right)\mathbf{a}_1+\left(\frac{1}{2}-y_9\right)\mathbf{a}_2+\left(\frac{1}{2}-z_9\right)\mathbf{a}_3$ | = | $\left(\frac{1}{2}+x_9\right)a\,\mathbf{\hat{x}}+\left(\frac{1}{2}-y_9\right)a\,\mathbf{\hat{y}}+\left(\frac{1}{2}-z_9\right)c\,\mathbf{\hat{z}}$ | (8$e$) | Sc III |
| $\mathbf{B}_{51}$ | = | $\left(\frac{1}{2}-y_9\right)\mathbf{a}_1+\left(\frac{1}{2}-x_9\right)\mathbf{a}_2+\left(\frac{1}{2}+z_9\right)\mathbf{a}_3$ | = | $\left(\frac{1}{2}-y_9\right)a\,\mathbf{\hat{x}}+\left(\frac{1}{2}-x_9\right)a\,\mathbf{\hat{y}}+\left(\frac{1}{2}+z_9\right)c\,\mathbf{\hat{z}}$ | (8$e$) | Sc III |
| $\mathbf{B}_{52}$ | = | $\left(\frac{1}{2}+y_9\right)\mathbf{a}_1+\left(\frac{1}{2}+x_9\right)\mathbf{a}_2+\left(\frac{1}{2}+z_9\right)\mathbf{a}_3$ | = | $\left(\frac{1}{2}+y_9\right)a\,\mathbf{\hat{x}}+\left(\frac{1}{2}+x_9\right)a\,\mathbf{\hat{y}}+\left(\frac{1}{2}+z_9\right)c\,\mathbf{\hat{z}}$ | (8$e$) | Sc III |
| $\mathbf{B}_{53}$ | = | $x_{10}\,\mathbf{a}_1+y_{10}\,\mathbf{a}_2+z_{10}\,\mathbf{a}_3$ | = | $x_{10}a\,\mathbf{\hat{x}}+y_{10}a\,\mathbf{\hat{y}}+z_{10}c\,\mathbf{\hat{z}}$ | (8$e$) | Sc IV |
| $\mathbf{B}_{54}$ | = | $-x_{10}\,\mathbf{a}_1-y_{10}\,\mathbf{a}_2+z_{10}\,\mathbf{a}_3$ | = | $-x_{10}a\,\mathbf{\hat{x}}-y_{10}a\,\mathbf{\hat{y}}+z_{10}c\,\mathbf{\hat{z}}$ | (8$e$) | Sc IV |
| $\mathbf{B}_{55}$ | = | $y_{10}\,\mathbf{a}_1-x_{10}\,\mathbf{a}_2-z_{10}\,\mathbf{a}_3$ | = | $y_{10}a\,\mathbf{\hat{x}}-x_{10}a\,\mathbf{\hat{y}}-z_{10}c\,\mathbf{\hat{z}}$ | (8$e$) | Sc IV |
| $\mathbf{B}_{56}$ | = | $-y_{10}\,\mathbf{a}_1+x_{10}\,\mathbf{a}_2-z_{10}\,\mathbf{a}_3$ | = | $-y_{10}a\,\mathbf{\hat{x}}+x_{10}a\,\mathbf{\hat{y}}-z_{10}c\,\mathbf{\hat{z}}$ | (8$e$) | Sc IV |
| $\mathbf{B}_{57}$ | = | $\left(\frac{1}{2}-x_{10}\right)\mathbf{a}_1+\left(\frac{1}{2}+y_{10}\right)\mathbf{a}_2+\left(\frac{1}{2}-z_{10}\right)\mathbf{a}_3$ | = | $\left(\frac{1}{2}-x_{10}\right)a\,\mathbf{\hat{x}}+\left(\frac{1}{2}+y_{10}\right)a\,\mathbf{\hat{y}}+\left(\frac{1}{2}-z_{10}\right)c\,\mathbf{\hat{z}}$ | (8$e$) | Sc IV |
| $\mathbf{B}_{58}$ | = | $\left(\frac{1}{2}+x_{10}\right)\mathbf{a}_1+\left(\frac{1}{2}-y_{10}\right)\mathbf{a}_2+\left(\frac{1}{2}-z_{10}\right)\mathbf{a}_3$ | = | $\left(\frac{1}{2}+x_{10}\right)a\,\mathbf{\hat{x}}+\left(\frac{1}{2}-y_{10}\right)a\,\mathbf{\hat{y}}+\left(\frac{1}{2}-z_{10}\right)c\,\mathbf{\hat{z}}$ | (8$e$) | Sc IV |
| $\mathbf{B}_{59}$ | = | $\left(\frac{1}{2} - y_{10}\right)\mathbf{a}_1 + \left(\frac{1}{2} - x_{10}\right)\mathbf{a}_2 + \left(\frac{1}{2} + z_{10}\right)\mathbf{a}_3$ | = | $\left(\frac{1}{2} - y_{10}\right)a\,\hat{\mathbf{x}} + \left(\frac{1}{2} - x_{10}\right)a\,\hat{\mathbf{y}} + \left(\frac{1}{2} + z_{10}\right)c\,\hat{\mathbf{z}}$ | (8$e$) | Sc IV |
| $\mathbf{B}_{60}$ | = | $\left(\frac{1}{2} + y_{10}\right)\mathbf{a}_1 + \left(\frac{1}{2} + x_{10}\right)\mathbf{a}_2 + \left(\frac{1}{2} + z_{10}\right)\mathbf{a}_3$ | = | $\left(\frac{1}{2} + y_{10}\right)a\,\hat{\mathbf{x}} + \left(\frac{1}{2} + x_{10}\right)a\,\hat{\mathbf{y}} + \left(\frac{1}{2} + z_{10}\right)c\,\hat{\mathbf{z}}$ | (8$e$) | Sc IV |
| $\mathbf{B}_{61}$ | = | $x_{11}\,\mathbf{a}_1 + y_{11}\,\mathbf{a}_2 + z_{11}\,\mathbf{a}_3$ | = | $x_{11}a\,\hat{\mathbf{x}} + y_{11}a\,\hat{\mathbf{y}} + z_{11}c\,\hat{\mathbf{z}}$ | (8$e$) | Sc V |
| $\mathbf{B}_{62}$ | = | $-x_{11}\,\mathbf{a}_1 - y_{11}\,\mathbf{a}_2 + z_{11}\,\mathbf{a}_3$ | = | $-x_{11}a\,\hat{\mathbf{x}} - y_{11}a\,\hat{\mathbf{y}} + z_{11}c\,\hat{\mathbf{z}}$ | (8$e$) | Sc V |
| $\mathbf{B}_{63}$ | = | $y_{11}\,\mathbf{a}_1 - x_{11}\,\mathbf{a}_2 - z_{11}\,\mathbf{a}_3$ | = | $y_{11}a\,\hat{\mathbf{x}} - x_{11}a\,\hat{\mathbf{y}} - z_{11}c\,\hat{\mathbf{z}}$ | (8$e$) | Sc V |
| $\mathbf{B}_{64}$ | = | $-y_{11}\,\mathbf{a}_1 + x_{11}\,\mathbf{a}_2 - z_{11}\,\mathbf{a}_3$ | = | $-y_{11}a\,\hat{\mathbf{x}} + x_{11}a\,\hat{\mathbf{y}} - z_{11}c\,\hat{\mathbf{z}}$ | (8$e$) | Sc V |
| $\mathbf{B}_{65}$ | = | $\left(\frac{1}{2} - x_{11}\right)\mathbf{a}_1 + \left(\frac{1}{2} + y_{11}\right)\mathbf{a}_2 + \left(\frac{1}{2} - z_{11}\right)\mathbf{a}_3$ | = | $\left(\frac{1}{2} - x_{11}\right)a\,\hat{\mathbf{x}} + \left(\frac{1}{2} + y_{11}\right)a\,\hat{\mathbf{y}} + \left(\frac{1}{2} - z_{11}\right)c\,\hat{\mathbf{z}}$ | (8$e$) | Sc V |
| $\mathbf{B}_{66}$ | = | $\left(\frac{1}{2} + x_{11}\right)\mathbf{a}_1 + \left(\frac{1}{2} - y_{11}\right)\mathbf{a}_2 + \left(\frac{1}{2} - z_{11}\right)\mathbf{a}_3$ | = | $\left(\frac{1}{2} + x_{11}\right)a\,\hat{\mathbf{x}} + \left(\frac{1}{2} - y_{11}\right)a\,\hat{\mathbf{y}} + \left(\frac{1}{2} - z_{11}\right)c\,\hat{\mathbf{z}}$ | (8$e$) | Sc V |
| $\mathbf{B}_{67}$ | = | $\left(\frac{1}{2} - y_{11}\right)\mathbf{a}_1 + \left(\frac{1}{2} - x_{11}\right)\mathbf{a}_2 + \left(\frac{1}{2} + z_{11}\right)\mathbf{a}_3$ | = | $\left(\frac{1}{2} - y_{11}\right)a\,\hat{\mathbf{x}} + \left(\frac{1}{2} - x_{11}\right)a\,\hat{\mathbf{y}} + \left(\frac{1}{2} + z_{11}\right)c\,\hat{\mathbf{z}}$ | (8$e$) | Sc V |
| $\mathbf{B}_{68}$ | = | $\left(\frac{1}{2} + y_{11}\right)\mathbf{a}_1 + \left(\frac{1}{2} + x_{11}\right)\mathbf{a}_2 + \left(\frac{1}{2} + z_{11}\right)\mathbf{a}_3$ | = | $\left(\frac{1}{2} + y_{11}\right)a\,\hat{\mathbf{x}} + \left(\frac{1}{2} + x_{11}\right)a\,\hat{\mathbf{y}} + \left(\frac{1}{2} + z_{11}\right)c\,\hat{\mathbf{z}}$ | (8$e$) | Sc V |

---

---

**Geometry files:**

- CIF: pp. 1682
- POSCAR: pp. 1682

# $Ag_2SO_4 \cdot 4NH_3$ ($H4_{17}$) Structure: A2B12C4D4E_tP46_114_d_3e_e_e_a

http://aflow.org/prototype-encyclopedia/A2B12C4D4E_tP46_114_d_3e_e_e_a

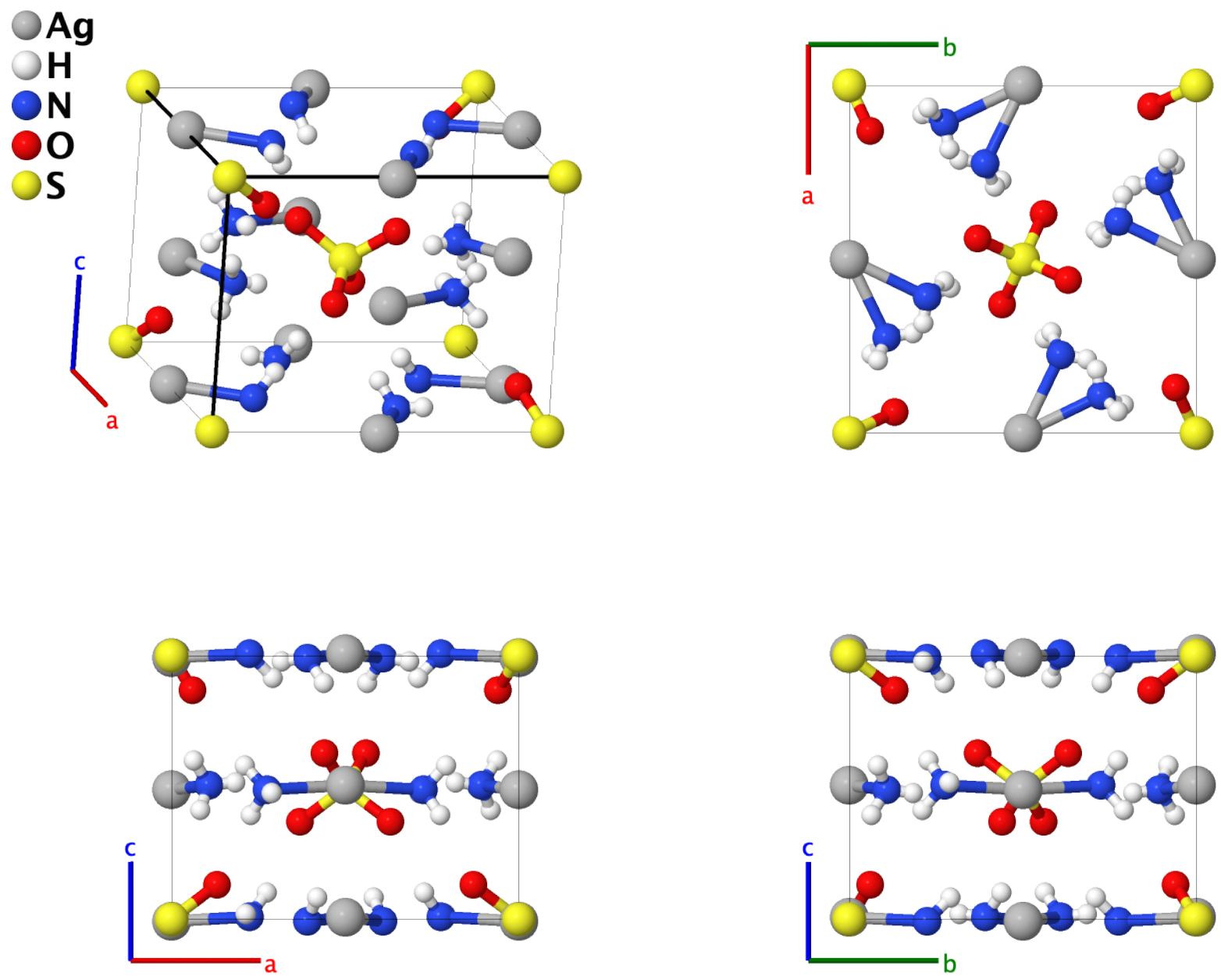

| | | |
|---|---|---|
| **Prototype** | : | $Ag_2H_{12}N_4O_4S$ |
| **AFLOW prototype label** | : | A2B12C4D4E_tP46_114_d_3e_e_e_a |
| ***Strukturbericht* designation** | : | $H4_{17}$ |
| **Pearson symbol** | : | tP46 |
| **Space group number** | : | 114 |
| **Space group symbol** | : | $P\bar{4}2_1c$ |
| **AFLOW prototype command** | : | `aflow --proto=A2B12C4D4E_tP46_114_d_3e_e_e_a` `--params=`$a, c/a, z_2, x_3, y_3, z_3, x_4, y_4, z_4, x_5, y_5, z_5, x_6, y_6, z_6, x_7, y_7, z_7$ |

- We use the data from (Zachwieja, 1992), which refines the data of the original $H4_{17}$ structure of (Corey, 1934) and locates the hydrogen atoms.

**Simple Tetragonal primitive vectors:**

$$\mathbf{a}_1 = a\,\hat{\mathbf{x}}$$
$$\mathbf{a}_2 = a\,\hat{\mathbf{y}}$$
$$\mathbf{a}_3 = c\,\hat{\mathbf{z}}$$

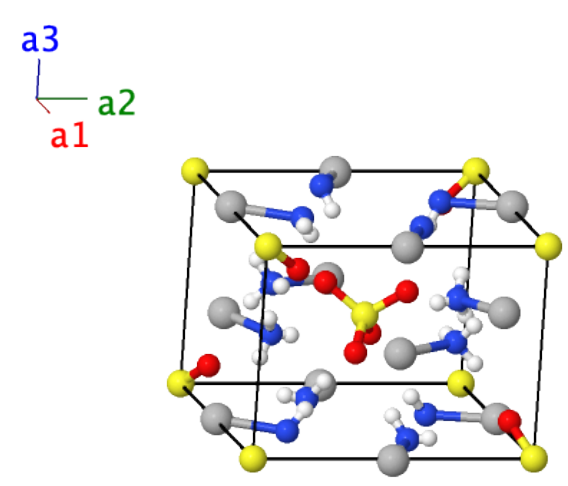

**Basis vectors:**

| | | Lattice Coordinates | | Cartesian Coordinates | Wyckoff Position | Atom Type |
|---|---|---|---|---|---|---|
| $\mathbf{B}_1$ | = | $0\,\mathbf{a}_1 + 0\,\mathbf{a}_2 + 0\,\mathbf{a}_3$ | = | $0\,\hat{\mathbf{x}} + 0\,\hat{\mathbf{y}} + 0\,\hat{\mathbf{z}}$ | (2$a$) | S |
| $\mathbf{B}_2$ | = | $\frac{1}{2}\,\mathbf{a}_1 + \frac{1}{2}\,\mathbf{a}_2 + \frac{1}{2}\,\mathbf{a}_3$ | = | $\frac{1}{2}a\,\hat{\mathbf{x}} + \frac{1}{2}a\,\hat{\mathbf{y}} + \frac{1}{2}c\,\hat{\mathbf{z}}$ | (2$a$) | S |
| $\mathbf{B}_3$ | = | $\frac{1}{2}\,\mathbf{a}_2 + z_2\,\mathbf{a}_3$ | = | $\frac{1}{2}a\,\hat{\mathbf{y}} + z_2c\,\hat{\mathbf{z}}$ | (4$d$) | Ag |
| $\mathbf{B}_4$ | = | $\frac{1}{2}\,\mathbf{a}_1 - z_2\,\mathbf{a}_3$ | = | $\frac{1}{2}a\,\hat{\mathbf{x}} - z_2c\,\hat{\mathbf{z}}$ | (4$d$) | Ag |
| $\mathbf{B}_5$ | = | $\frac{1}{2}\,\mathbf{a}_1 + \left(\frac{1}{2} - z_2\right)\,\mathbf{a}_3$ | = | $\frac{1}{2}a\,\hat{\mathbf{x}} + \left(\frac{1}{2} - z_2\right)c\,\hat{\mathbf{z}}$ | (4$d$) | Ag |
| $\mathbf{B}_6$ | = | $\frac{1}{2}\,\mathbf{a}_2 + \left(\frac{1}{2} + z_2\right)\,\mathbf{a}_3$ | = | $\frac{1}{2}a\,\hat{\mathbf{y}} + \left(\frac{1}{2} + z_2\right)c\,\hat{\mathbf{z}}$ | (4$d$) | Ag |
| $\mathbf{B}_7$ | = | $x_3\,\mathbf{a}_1 + y_3\,\mathbf{a}_2 + z_3\,\mathbf{a}_3$ | = | $x_3a\,\hat{\mathbf{x}} + y_3a\,\hat{\mathbf{y}} + z_3c\,\hat{\mathbf{z}}$ | (8$e$) | H I |
| $\mathbf{B}_8$ | = | $-x_3\,\mathbf{a}_1 - y_3\,\mathbf{a}_2 + z_3\,\mathbf{a}_3$ | = | $-x_3a\,\hat{\mathbf{x}} - y_3a\,\hat{\mathbf{y}} + z_3c\,\hat{\mathbf{z}}$ | (8$e$) | H I |
| $\mathbf{B}_9$ | = | $y_3\,\mathbf{a}_1 - x_3\,\mathbf{a}_2 - z_3\,\mathbf{a}_3$ | = | $y_3a\,\hat{\mathbf{x}} - x_3a\,\hat{\mathbf{y}} - z_3c\,\hat{\mathbf{z}}$ | (8$e$) | H I |
| $\mathbf{B}_{10}$ | = | $-y_3\,\mathbf{a}_1 + x_3\,\mathbf{a}_2 - z_3\,\mathbf{a}_3$ | = | $-y_3a\,\hat{\mathbf{x}} + x_3a\,\hat{\mathbf{y}} - z_3c\,\hat{\mathbf{z}}$ | (8$e$) | H I |
| $\mathbf{B}_{11}$ | = | $\left(\frac{1}{2} - x_3\right)\,\mathbf{a}_1 + \left(\frac{1}{2} + y_3\right)\,\mathbf{a}_2 + \left(\frac{1}{2} - z_3\right)\,\mathbf{a}_3$ | = | $\left(\frac{1}{2} - x_3\right)a\,\hat{\mathbf{x}} + \left(\frac{1}{2} + y_3\right)a\,\hat{\mathbf{y}} + \left(\frac{1}{2} - z_3\right)c\,\hat{\mathbf{z}}$ | (8$e$) | H I |
| $\mathbf{B}_{12}$ | = | $\left(\frac{1}{2} + x_3\right)\,\mathbf{a}_1 + \left(\frac{1}{2} - y_3\right)\,\mathbf{a}_2 + \left(\frac{1}{2} - z_3\right)\,\mathbf{a}_3$ | = | $\left(\frac{1}{2} + x_3\right)a\,\hat{\mathbf{x}} + \left(\frac{1}{2} - y_3\right)a\,\hat{\mathbf{y}} + \left(\frac{1}{2} - z_3\right)c\,\hat{\mathbf{z}}$ | (8$e$) | H I |
| $\mathbf{B}_{13}$ | = | $\left(\frac{1}{2} - y_3\right)\,\mathbf{a}_1 + \left(\frac{1}{2} - x_3\right)\,\mathbf{a}_2 + \left(\frac{1}{2} + z_3\right)\,\mathbf{a}_3$ | = | $\left(\frac{1}{2} - y_3\right)a\,\hat{\mathbf{x}} + \left(\frac{1}{2} - x_3\right)a\,\hat{\mathbf{y}} + \left(\frac{1}{2} + z_3\right)c\,\hat{\mathbf{z}}$ | (8$e$) | H I |
| $\mathbf{B}_{14}$ | = | $\left(\frac{1}{2} + y_3\right)\,\mathbf{a}_1 + \left(\frac{1}{2} + x_3\right)\,\mathbf{a}_2 + \left(\frac{1}{2} + z_3\right)\,\mathbf{a}_3$ | = | $\left(\frac{1}{2} + y_3\right)a\,\hat{\mathbf{x}} + \left(\frac{1}{2} + x_3\right)a\,\hat{\mathbf{y}} + \left(\frac{1}{2} + z_3\right)c\,\hat{\mathbf{z}}$ | (8$e$) | H I |
| $\mathbf{B}_{15}$ | = | $x_4\,\mathbf{a}_1 + y_4\,\mathbf{a}_2 + z_4\,\mathbf{a}_3$ | = | $x_4a\,\hat{\mathbf{x}} + y_4a\,\hat{\mathbf{y}} + z_4c\,\hat{\mathbf{z}}$ | (8$e$) | H II |
| $\mathbf{B}_{16}$ | = | $-x_4\,\mathbf{a}_1 - y_4\,\mathbf{a}_2 + z_4\,\mathbf{a}_3$ | = | $-x_4a\,\hat{\mathbf{x}} - y_4a\,\hat{\mathbf{y}} + z_4c\,\hat{\mathbf{z}}$ | (8$e$) | H II |
| $\mathbf{B}_{17}$ | = | $y_4\,\mathbf{a}_1 - x_4\,\mathbf{a}_2 - z_4\,\mathbf{a}_3$ | = | $y_4a\,\hat{\mathbf{x}} - x_4a\,\hat{\mathbf{y}} - z_4c\,\hat{\mathbf{z}}$ | (8$e$) | H II |
| $\mathbf{B}_{18}$ | = | $-y_4\,\mathbf{a}_1 + x_4\,\mathbf{a}_2 - z_4\,\mathbf{a}_3$ | = | $-y_4a\,\hat{\mathbf{x}} + x_4a\,\hat{\mathbf{y}} - z_4c\,\hat{\mathbf{z}}$ | (8$e$) | H II |
| $\mathbf{B}_{19}$ | = | $\left(\frac{1}{2} - x_4\right)\,\mathbf{a}_1 + \left(\frac{1}{2} + y_4\right)\,\mathbf{a}_2 + \left(\frac{1}{2} - z_4\right)\,\mathbf{a}_3$ | = | $\left(\frac{1}{2} - x_4\right)a\,\hat{\mathbf{x}} + \left(\frac{1}{2} + y_4\right)a\,\hat{\mathbf{y}} + \left(\frac{1}{2} - z_4\right)c\,\hat{\mathbf{z}}$ | (8$e$) | H II |
| $\mathbf{B}_{20}$ | = | $\left(\frac{1}{2} + x_4\right)\,\mathbf{a}_1 + \left(\frac{1}{2} - y_4\right)\,\mathbf{a}_2 + \left(\frac{1}{2} - z_4\right)\,\mathbf{a}_3$ | = | $\left(\frac{1}{2} + x_4\right)a\,\hat{\mathbf{x}} + \left(\frac{1}{2} - y_4\right)a\,\hat{\mathbf{y}} + \left(\frac{1}{2} - z_4\right)c\,\hat{\mathbf{z}}$ | (8$e$) | H II |
| $\mathbf{B}_{21}$ | = | $\left(\frac{1}{2} - y_4\right)\,\mathbf{a}_1 + \left(\frac{1}{2} - x_4\right)\,\mathbf{a}_2 + \left(\frac{1}{2} + z_4\right)\,\mathbf{a}_3$ | = | $\left(\frac{1}{2} - y_4\right)a\,\hat{\mathbf{x}} + \left(\frac{1}{2} - x_4\right)a\,\hat{\mathbf{y}} + \left(\frac{1}{2} + z_4\right)c\,\hat{\mathbf{z}}$ | (8$e$) | H II |
| $\mathbf{B}_{22}$ | = | $\left(\frac{1}{2} + y_4\right)\,\mathbf{a}_1 + \left(\frac{1}{2} + x_4\right)\,\mathbf{a}_2 + \left(\frac{1}{2} + z_4\right)\,\mathbf{a}_3$ | = | $\left(\frac{1}{2} + y_4\right)a\,\hat{\mathbf{x}} + \left(\frac{1}{2} + x_4\right)a\,\hat{\mathbf{y}} + \left(\frac{1}{2} + z_4\right)c\,\hat{\mathbf{z}}$ | (8$e$) | H II |
| $\mathbf{B}_{23}$ | = | $x_5\,\mathbf{a}_1 + y_5\,\mathbf{a}_2 + z_5\,\mathbf{a}_3$ | = | $x_5a\,\hat{\mathbf{x}} + y_5a\,\hat{\mathbf{y}} + z_5c\,\hat{\mathbf{z}}$ | (8$e$) | H III |
| $\mathbf{B}_{24}$ | = | $-x_5\,\mathbf{a}_1 - y_5\,\mathbf{a}_2 + z_5\,\mathbf{a}_3$ | = | $-x_5a\,\hat{\mathbf{x}} - y_5a\,\hat{\mathbf{y}} + z_5c\,\hat{\mathbf{z}}$ | (8$e$) | H III |
| $\mathbf{B}_{25}$ | = | $y_5\,\mathbf{a}_1 - x_5\,\mathbf{a}_2 - z_5\,\mathbf{a}_3$ | = | $y_5a\,\hat{\mathbf{x}} - x_5a\,\hat{\mathbf{y}} - z_5c\,\hat{\mathbf{z}}$ | (8$e$) | H III |
| $\mathbf{B}_{26}$ | = | $-y_5\,\mathbf{a}_1 + x_5\,\mathbf{a}_2 - z_5\,\mathbf{a}_3$ | = | $-y_5a\,\hat{\mathbf{x}} + x_5a\,\hat{\mathbf{y}} - z_5c\,\hat{\mathbf{z}}$ | (8$e$) | H III |
| $\mathbf{B}_{27}$ | = | $\left(\frac{1}{2} - x_5\right)\,\mathbf{a}_1 + \left(\frac{1}{2} + y_5\right)\,\mathbf{a}_2 + \left(\frac{1}{2} - z_5\right)\,\mathbf{a}_3$ | = | $\left(\frac{1}{2} - x_5\right)a\,\hat{\mathbf{x}} + \left(\frac{1}{2} + y_5\right)a\,\hat{\mathbf{y}} + \left(\frac{1}{2} - z_5\right)c\,\hat{\mathbf{z}}$ | (8$e$) | H III |
| $\mathbf{B}_{28}$ | = | $\left(\frac{1}{2}+x_5\right)\mathbf{a}_1+\left(\frac{1}{2}-y_5\right)\mathbf{a}_2+\left(\frac{1}{2}-z_5\right)\mathbf{a}_3$ | = | $\left(\frac{1}{2}+x_5\right)a\,\hat{\mathbf{x}}+\left(\frac{1}{2}-y_5\right)a\,\hat{\mathbf{y}}+\left(\frac{1}{2}-z_5\right)c\,\hat{\mathbf{z}}$ | (8e) | H III |
| $\mathbf{B}_{29}$ | = | $\left(\frac{1}{2}-y_5\right)\mathbf{a}_1+\left(\frac{1}{2}-x_5\right)\mathbf{a}_2+\left(\frac{1}{2}+z_5\right)\mathbf{a}_3$ | = | $\left(\frac{1}{2}-y_5\right)a\,\hat{\mathbf{x}}+\left(\frac{1}{2}-x_5\right)a\,\hat{\mathbf{y}}+\left(\frac{1}{2}+z_5\right)c\,\hat{\mathbf{z}}$ | (8e) | H III |
| $\mathbf{B}_{30}$ | = | $\left(\frac{1}{2}+y_5\right)\mathbf{a}_1+\left(\frac{1}{2}+x_5\right)\mathbf{a}_2+\left(\frac{1}{2}+z_5\right)\mathbf{a}_3$ | = | $\left(\frac{1}{2}+y_5\right)a\,\hat{\mathbf{x}}+\left(\frac{1}{2}+x_5\right)a\,\hat{\mathbf{y}}+\left(\frac{1}{2}+z_5\right)c\,\hat{\mathbf{z}}$ | (8e) | H III |
| $\mathbf{B}_{31}$ | = | $x_6\,\mathbf{a}_1+y_6\,\mathbf{a}_2+z_6\,\mathbf{a}_3$ | = | $x_6a\,\hat{\mathbf{x}}+y_6a\,\hat{\mathbf{y}}+z_6c\,\hat{\mathbf{z}}$ | (8e) | N |
| $\mathbf{B}_{32}$ | = | $-x_6\,\mathbf{a}_1-y_6\,\mathbf{a}_2+z_6\,\mathbf{a}_3$ | = | $-x_6a\,\hat{\mathbf{x}}-y_6a\,\hat{\mathbf{y}}+z_6c\,\hat{\mathbf{z}}$ | (8e) | N |
| $\mathbf{B}_{33}$ | = | $y_6\,\mathbf{a}_1-x_6\,\mathbf{a}_2-z_6\,\mathbf{a}_3$ | = | $y_6a\,\hat{\mathbf{x}}-x_6a\,\hat{\mathbf{y}}-z_6c\,\hat{\mathbf{z}}$ | (8e) | N |
| $\mathbf{B}_{34}$ | = | $-y_6\,\mathbf{a}_1+x_6\,\mathbf{a}_2-z_6\,\mathbf{a}_3$ | = | $-y_6a\,\hat{\mathbf{x}}+x_6a\,\hat{\mathbf{y}}-z_6c\,\hat{\mathbf{z}}$ | (8e) | N |
| $\mathbf{B}_{35}$ | = | $\left(\frac{1}{2}-x_6\right)\mathbf{a}_1+\left(\frac{1}{2}+y_6\right)\mathbf{a}_2+\left(\frac{1}{2}-z_6\right)\mathbf{a}_3$ | = | $\left(\frac{1}{2}-x_6\right)a\,\hat{\mathbf{x}}+\left(\frac{1}{2}+y_6\right)a\,\hat{\mathbf{y}}+\left(\frac{1}{2}-z_6\right)c\,\hat{\mathbf{z}}$ | (8e) | N |
| $\mathbf{B}_{36}$ | = | $\left(\frac{1}{2}+x_6\right)\mathbf{a}_1+\left(\frac{1}{2}-y_6\right)\mathbf{a}_2+\left(\frac{1}{2}-z_6\right)\mathbf{a}_3$ | = | $\left(\frac{1}{2}+x_6\right)a\,\hat{\mathbf{x}}+\left(\frac{1}{2}-y_6\right)a\,\hat{\mathbf{y}}+\left(\frac{1}{2}-z_6\right)c\,\hat{\mathbf{z}}$ | (8e) | N |
| $\mathbf{B}_{37}$ | = | $\left(\frac{1}{2}-y_6\right)\mathbf{a}_1+\left(\frac{1}{2}-x_6\right)\mathbf{a}_2+\left(\frac{1}{2}+z_6\right)\mathbf{a}_3$ | = | $\left(\frac{1}{2}-y_6\right)a\,\hat{\mathbf{x}}+\left(\frac{1}{2}-x_6\right)a\,\hat{\mathbf{y}}+\left(\frac{1}{2}+z_6\right)c\,\hat{\mathbf{z}}$ | (8e) | N |
| $\mathbf{B}_{38}$ | = | $\left(\frac{1}{2}+y_6\right)\mathbf{a}_1+\left(\frac{1}{2}+x_6\right)\mathbf{a}_2+\left(\frac{1}{2}+z_6\right)\mathbf{a}_3$ | = | $\left(\frac{1}{2}+y_6\right)a\,\hat{\mathbf{x}}+\left(\frac{1}{2}+x_6\right)a\,\hat{\mathbf{y}}+\left(\frac{1}{2}+z_6\right)c\,\hat{\mathbf{z}}$ | (8e) | N |
| $\mathbf{B}_{39}$ | = | $x_7\,\mathbf{a}_1+y_7\,\mathbf{a}_2+z_7\,\mathbf{a}_3$ | = | $x_7a\,\hat{\mathbf{x}}+y_7a\,\hat{\mathbf{y}}+z_7c\,\hat{\mathbf{z}}$ | (8e) | O |
| $\mathbf{B}_{40}$ | = | $-x_7\,\mathbf{a}_1-y_7\,\mathbf{a}_2+z_7\,\mathbf{a}_3$ | = | $-x_7a\,\hat{\mathbf{x}}-y_7a\,\hat{\mathbf{y}}+z_7c\,\hat{\mathbf{z}}$ | (8e) | O |
| $\mathbf{B}_{41}$ | = | $y_7\,\mathbf{a}_1-x_7\,\mathbf{a}_2-z_7\,\mathbf{a}_3$ | = | $y_7a\,\hat{\mathbf{x}}-x_7a\,\hat{\mathbf{y}}-z_7c\,\hat{\mathbf{z}}$ | (8e) | O |
| $\mathbf{B}_{42}$ | = | $-y_7\,\mathbf{a}_1+x_7\,\mathbf{a}_2-z_7\,\mathbf{a}_3$ | = | $-y_7a\,\hat{\mathbf{x}}+x_7a\,\hat{\mathbf{y}}-z_7c\,\hat{\mathbf{z}}$ | (8e) | O |
| $\mathbf{B}_{43}$ | = | $\left(\frac{1}{2}-x_7\right)\mathbf{a}_1+\left(\frac{1}{2}+y_7\right)\mathbf{a}_2+\left(\frac{1}{2}-z_7\right)\mathbf{a}_3$ | = | $\left(\frac{1}{2}-x_7\right)a\,\hat{\mathbf{x}}+\left(\frac{1}{2}+y_7\right)a\,\hat{\mathbf{y}}+\left(\frac{1}{2}-z_7\right)c\,\hat{\mathbf{z}}$ | (8e) | O |
| $\mathbf{B}_{44}$ | = | $\left(\frac{1}{2}+x_7\right)\mathbf{a}_1+\left(\frac{1}{2}-y_7\right)\mathbf{a}_2+\left(\frac{1}{2}-z_7\right)\mathbf{a}_3$ | = | $\left(\frac{1}{2}+x_7\right)a\,\hat{\mathbf{x}}+\left(\frac{1}{2}-y_7\right)a\,\hat{\mathbf{y}}+\left(\frac{1}{2}-z_7\right)c\,\hat{\mathbf{z}}$ | (8e) | O |
| $\mathbf{B}_{45}$ | = | $\left(\frac{1}{2}-y_7\right)\mathbf{a}_1+\left(\frac{1}{2}-x_7\right)\mathbf{a}_2+\left(\frac{1}{2}+z_7\right)\mathbf{a}_3$ | = | $\left(\frac{1}{2}-y_7\right)a\,\hat{\mathbf{x}}+\left(\frac{1}{2}-x_7\right)a\,\hat{\mathbf{y}}+\left(\frac{1}{2}+z_7\right)c\,\hat{\mathbf{z}}$ | (8e) | O |
| $\mathbf{B}_{46}$ | = | $\left(\frac{1}{2}+y_7\right)\mathbf{a}_1+\left(\frac{1}{2}+x_7\right)\mathbf{a}_2+\left(\frac{1}{2}+z_7\right)\mathbf{a}_3$ | = | $\left(\frac{1}{2}+y_7\right)a\,\hat{\mathbf{x}}+\left(\frac{1}{2}+x_7\right)a\,\hat{\mathbf{y}}+\left(\frac{1}{2}+z_7\right)c\,\hat{\mathbf{z}}$ | (8e) | O |

**Geometry files:**

- CIF: pp. 1683
- POSCAR: pp. 1683

# $F6_1$ (Chalcopyrite, $CuFeS_2$) (*obsolete*) Structure: ABC2_tP4_115_a_c_g

http://aflow.org/prototype-encyclopedia/ABC2_tP4_115_a_c_g

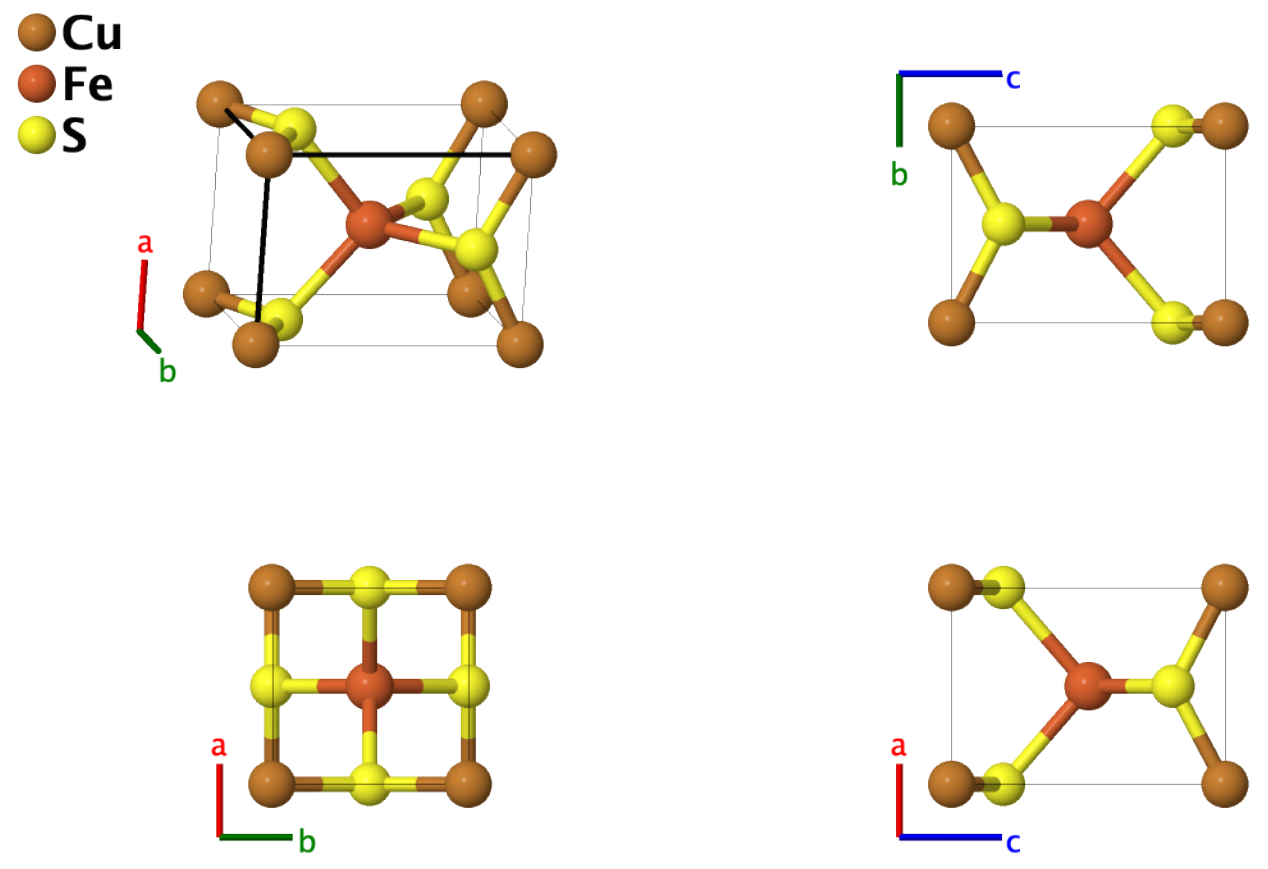

| | | |
|---|---|---|
| **Prototype** | : | $CuFeS_2$ |
| **AFLOW prototype label** | : | ABC2_tP4_115_a_c_g |
| ***Strukturbericht* designation** | : | $F6_1$ |
| **Pearson symbol** | : | tP4 |
| **Space group number** | : | 115 |
| **Space group symbol** | : | $P\bar{4}m2$ |
| **AFLOW prototype command** | : | `aflow --proto=ABC2_tP4_115_a_c_g --params=`$a, c/a, z_3$ |

- This structure was presented as the Chalcopyrite structure and given the *Strukturbericht* designatiion $F6_1$ in (Ewald, 1931). It was subsequently replaced with the $E1_1$ (ABC2_tI16_122_a_b_d) structure, which is now the accepted structure for Chalcopyrite and similar compounds. We include the $F6_1$ structure as part of this historical record.

**Simple Tetragonal primitive vectors:**

$$\mathbf{a}_1 = a\,\hat{\mathbf{x}}$$
$$\mathbf{a}_2 = a\,\hat{\mathbf{y}}$$
$$\mathbf{a}_3 = c\,\hat{\mathbf{z}}$$

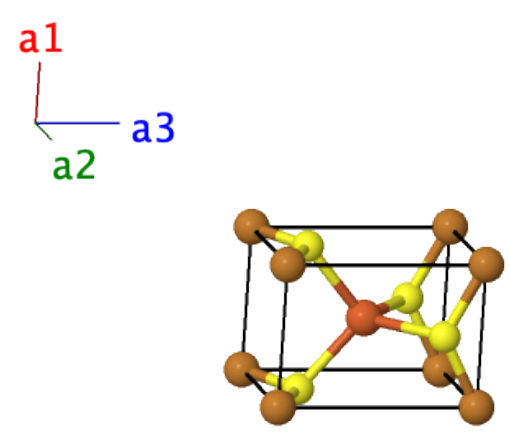

**Basis vectors:**

| | | Lattice Coordinates | | Cartesian Coordinates | Wyckoff Position | Atom Type |
|---|---|---|---|---|---|---|
| $\mathbf{B}_1$ | = | $0\,\mathbf{a}_1 + 0\,\mathbf{a}_2 + 0\,\mathbf{a}_3$ | = | $0\,\hat{\mathbf{x}} + 0\,\hat{\mathbf{y}} + 0\,\hat{\mathbf{z}}$ | (1*a*) | Cu |
| $\mathbf{B}_2$ | = | $\frac{1}{2}\,\mathbf{a}_1 + \frac{1}{2}\,\mathbf{a}_2 + \frac{1}{2}\,\mathbf{a}_3$ | = | $\frac{1}{2}a\,\hat{\mathbf{x}} + \frac{1}{2}a\,\hat{\mathbf{y}} + \frac{1}{2}c\,\hat{\mathbf{z}}$ | (1*c*) | Fe |
| $\mathbf{B}_3$ | = | $\frac{1}{2}\,\mathbf{a}_2 + z_3\,\mathbf{a}_3$ | = | $\frac{1}{2}a\,\hat{\mathbf{y}} + z_3 c\,\hat{\mathbf{z}}$ | (2*g*) | S |
| $\mathbf{B}_4$ | = | $\frac{1}{2}\,\mathbf{a}_1 - z_3\,\mathbf{a}_3$ | = | $\frac{1}{2}a\,\hat{\mathbf{x}} - z_3 c\,\hat{\mathbf{z}}$ | (2*g*) | S |

**Found in:**

---

**Geometry files:**

- CIF: pp. 1684
- POSCAR: pp. 1684

# Phase II $Cd_2Re_2O_7$ Structure: A2B7C2_tI44_119_i_bdefgh_i

http://aflow.org/prototype-encyclopedia/A2B7C2_tI44_119_i_bdefgh_i

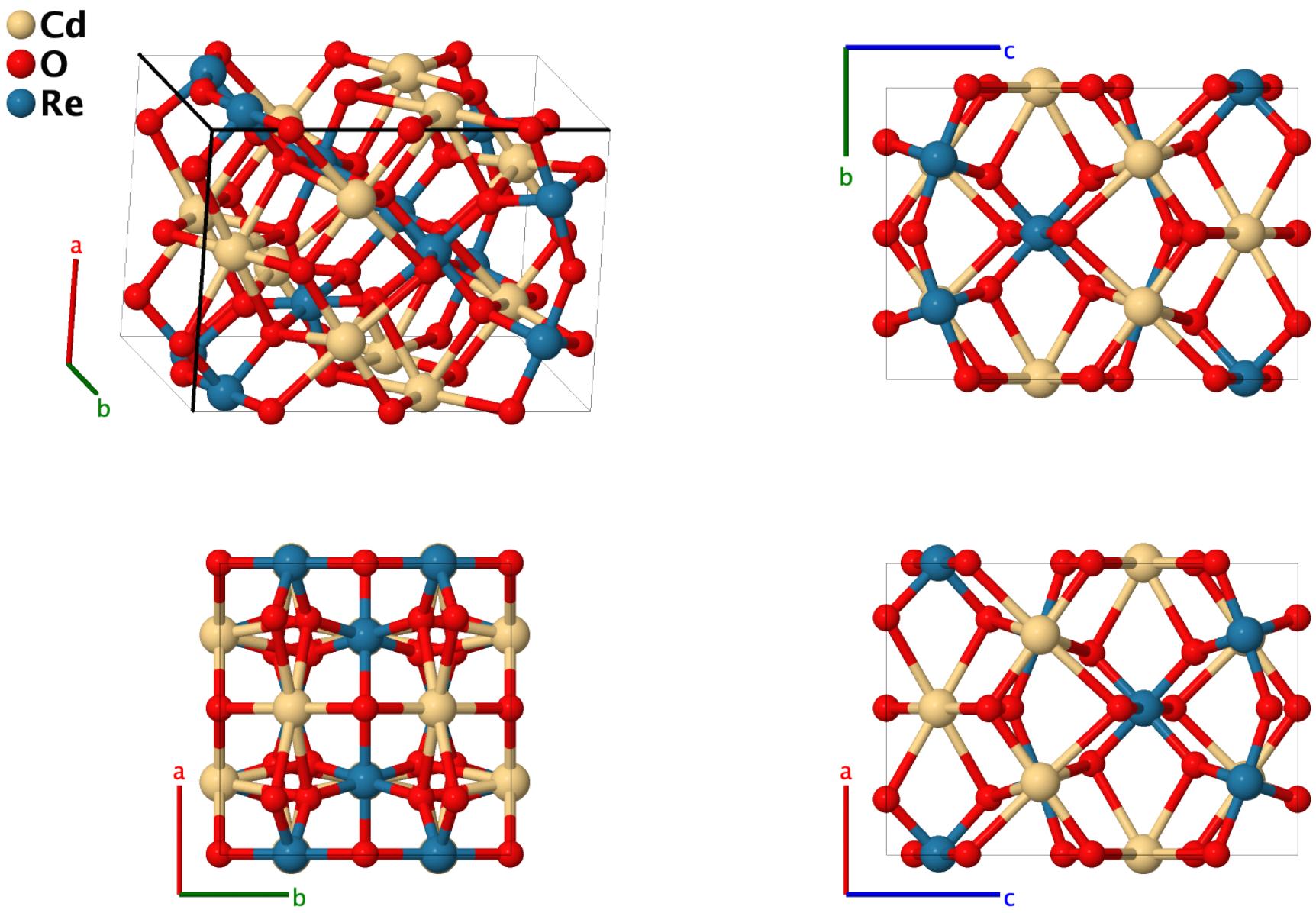

| | | |
|---|---|---|
| **Prototype** | : | $Cd_2O_7Re_2$ |
| **AFLOW prototype label** | : | A2B7C2_tI44_119_i_bdefgh_i |
| ***Strukturbericht* designation** | : | None |
| **Pearson symbol** | : | tI44 |
| **Space group number** | : | 119 |
| **Space group symbol** | : | $I\bar{4}m2$ |
| **AFLOW prototype command** | : | `aflow --proto=A2B7C2_tI44_119_i_bdefgh_i` `--params=`$a, c/a, z_3, z_4, x_5, x_6, x_7, z_7, x_8, z_8$ |

- $Cd_2Re_2O_7$ exhibits a number of phases. We will use the notation of (Kapcia, 2019) to describe them:
  - Phase I: above 200 K, the system takes on the cubic pyrochlore ($E8_1$) structure.
  - Phase II: in the range 120-200 K the system is in the tetragonal $I\bar{4}m2$ #119 structure. (This structure)
  - Phase III: in the range 80-120 K the system is in the tetragonal $I4_122$ #98 structure.
  - Phase IV: (Kapcia, 2019) did a first-principles study of this system and found that below 80 K Phase III develops a soft phonon mode which transforms the system into an orthorhombic $F222$ #22 structure.
  - (Norman, 2019) points out that both Phase III and Phase IV structures have issues.
- Both Phase II and Phase III are extremely close to Phase I. If AFLOW-SYM allows a tolerance of 0.2 Å and FINDSYM allows a tolerance of 0.2 Å for both the lattice vectors and atomic positions both of the tetragonal phases become cubic.
- Phase IV is extremely close to Phase II. If AFLOW-SYM allows a tolerance of 0.002 Å and FINDSYM allows a tolerance 0.002 Å for both the lattice vectors and atomic positions the orthorhombic phase becomes tetragonal.
- Data for the Phase II structure was taken at 160 K.

### Body-centered Tetragonal primitive vectors:

$$\mathbf{a}_1 = -\frac{1}{2}\,a\,\hat{\mathbf{x}} + \frac{1}{2}\,a\,\hat{\mathbf{y}} + \frac{1}{2}\,c\,\hat{\mathbf{z}}$$
$$\mathbf{a}_2 = \frac{1}{2}\,a\,\hat{\mathbf{x}} - \frac{1}{2}\,a\,\hat{\mathbf{y}} + \frac{1}{2}\,c\,\hat{\mathbf{z}}$$
$$\mathbf{a}_3 = \frac{1}{2}\,a\,\hat{\mathbf{x}} + \frac{1}{2}\,a\,\hat{\mathbf{y}} - \frac{1}{2}\,c\,\hat{\mathbf{z}}$$

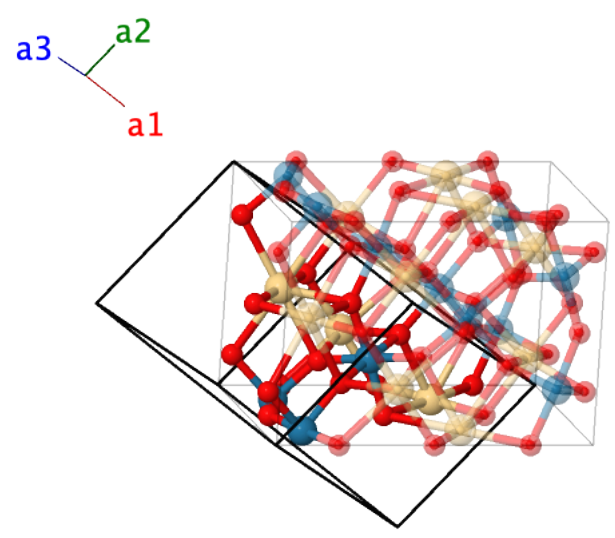

### Basis vectors:

| | | Lattice Coordinates | | Cartesian Coordinates | Wyckoff Position | Atom Type |
|---|---|---|---|---|---|---|
| $\mathbf{B}_1$ | = | $\frac{1}{2}\,\mathbf{a}_1 + \frac{1}{2}\,\mathbf{a}_2$ | = | $\frac{1}{2}c\,\hat{\mathbf{z}}$ | (2$b$) | O I |
| $\mathbf{B}_2$ | = | $\frac{1}{4}\,\mathbf{a}_1 + \frac{3}{4}\,\mathbf{a}_2 + \frac{1}{2}\,\mathbf{a}_3$ | = | $\frac{1}{2}a\,\hat{\mathbf{x}} + \frac{1}{4}c\,\hat{\mathbf{z}}$ | (2$d$) | O II |
| $\mathbf{B}_3$ | = | $z_3\,\mathbf{a}_1 + z_3\,\mathbf{a}_2$ | = | $z_3c\,\hat{\mathbf{z}}$ | (4$e$) | O III |
| $\mathbf{B}_4$ | = | $-z_3\,\mathbf{a}_1 - z_3\,\mathbf{a}_2$ | = | $-z_3c\,\hat{\mathbf{z}}$ | (4$e$) | O III |
| $\mathbf{B}_5$ | = | $\left(\frac{1}{2} + z_4\right)\,\mathbf{a}_1 + z_4\,\mathbf{a}_2 + \frac{1}{2}\,\mathbf{a}_3$ | = | $\frac{1}{2}a\,\hat{\mathbf{y}} + z_4c\,\hat{\mathbf{z}}$ | (4$f$) | O IV |
| $\mathbf{B}_6$ | = | $-z_4\,\mathbf{a}_1 + \left(\frac{1}{2} - z_4\right)\,\mathbf{a}_2 + \frac{1}{2}\,\mathbf{a}_3$ | = | $\frac{1}{2}a\,\hat{\mathbf{x}} - z_4c\,\hat{\mathbf{z}}$ | (4$f$) | O IV |
| $\mathbf{B}_7$ | = | $x_5\,\mathbf{a}_1 + x_5\,\mathbf{a}_2 + 2x_5\,\mathbf{a}_3$ | = | $x_5a\,\hat{\mathbf{x}} + x_5a\,\hat{\mathbf{y}}$ | (8$g$) | O V |
| $\mathbf{B}_8$ | = | $-x_5\,\mathbf{a}_1 - x_5\,\mathbf{a}_2 - 2x_5\,\mathbf{a}_3$ | = | $-x_5a\,\hat{\mathbf{x}} - x_5a\,\hat{\mathbf{y}}$ | (8$g$) | O V |
| $\mathbf{B}_9$ | = | $-x_5\,\mathbf{a}_1 + x_5\,\mathbf{a}_2$ | = | $x_5a\,\hat{\mathbf{x}} - x_5a\,\hat{\mathbf{y}}$ | (8$g$) | O V |
| $\mathbf{B}_{10}$ | = | $x_5\,\mathbf{a}_1 - x_5\,\mathbf{a}_2$ | = | $-x_5a\,\hat{\mathbf{x}} + x_5a\,\hat{\mathbf{y}}$ | (8$g$) | O V |
| $\mathbf{B}_{11}$ | = | $\left(\frac{3}{4} + x_6\right)\,\mathbf{a}_1 + \left(\frac{1}{4} + x_6\right)\,\mathbf{a}_2 + \left(\frac{1}{2} + 2x_6\right)\,\mathbf{a}_3$ | = | $x_6a\,\hat{\mathbf{x}} + \left(\frac{1}{2} + x_6\right)a\,\hat{\mathbf{y}} + \frac{1}{4}c\,\hat{\mathbf{z}}$ | (8$h$) | O VI |
| $\mathbf{B}_{12}$ | = | $\left(\frac{3}{4} - x_6\right)\,\mathbf{a}_1 + \left(\frac{1}{4} - x_6\right)\,\mathbf{a}_2 + \left(\frac{1}{2} - 2x_6\right)\,\mathbf{a}_3$ | = | $-x_6a\,\hat{\mathbf{x}} + \left(\frac{1}{2} - x_6\right)a\,\hat{\mathbf{y}} + \frac{1}{4}c\,\hat{\mathbf{z}}$ | (8$h$) | O VI |
| $\mathbf{B}_{13}$ | = | $\left(\frac{3}{4} - x_6\right)\,\mathbf{a}_1 + \left(\frac{1}{4} + x_6\right)\,\mathbf{a}_2 + \frac{1}{2}\,\mathbf{a}_3$ | = | $x_6a\,\hat{\mathbf{x}} + \left(\frac{1}{2} - x_6\right)a\,\hat{\mathbf{y}} + \frac{1}{4}c\,\hat{\mathbf{z}}$ | (8$h$) | O VI |
| $\mathbf{B}_{14}$ | = | $\left(\frac{3}{4} + x_6\right)\,\mathbf{a}_1 + \left(\frac{1}{4} - x_6\right)\,\mathbf{a}_2 + \frac{1}{2}\,\mathbf{a}_3$ | = | $-x_6a\,\hat{\mathbf{x}} + \left(\frac{1}{2} + x_6\right)a\,\hat{\mathbf{y}} + \frac{1}{4}c\,\hat{\mathbf{z}}$ | (8$h$) | O VI |
| $\mathbf{B}_{15}$ | = | $z_7\,\mathbf{a}_1 + (x_7 + z_7)\,\mathbf{a}_2 + x_7\,\mathbf{a}_3$ | = | $x_7a\,\hat{\mathbf{x}} + z_7c\,\hat{\mathbf{z}}$ | (8$i$) | Cd |
| $\mathbf{B}_{16}$ | = | $z_7\,\mathbf{a}_1 + (-x_7 + z_7)\,\mathbf{a}_2 - x_7\,\mathbf{a}_3$ | = | $-x_7a\,\hat{\mathbf{x}} + z_7c\,\hat{\mathbf{z}}$ | (8$i$) | Cd |
| $\mathbf{B}_{17}$ | = | $(-x_7 - z_7)\,\mathbf{a}_1 - z_7\,\mathbf{a}_2 - x_7\,\mathbf{a}_3$ | = | $-x_7a\,\hat{\mathbf{y}} - z_7c\,\hat{\mathbf{z}}$ | (8$i$) | Cd |
| $\mathbf{B}_{18}$ | = | $(x_7 - z_7)\,\mathbf{a}_1 - z_7\,\mathbf{a}_2 + x_7\,\mathbf{a}_3$ | = | $x_7a\,\hat{\mathbf{y}} - z_7c\,\hat{\mathbf{z}}$ | (8$i$) | Cd |
| $\mathbf{B}_{19}$ | = | $z_8\,\mathbf{a}_1 + (x_8 + z_8)\,\mathbf{a}_2 + x_8\,\mathbf{a}_3$ | = | $x_8a\,\hat{\mathbf{x}} + z_8c\,\hat{\mathbf{z}}$ | (8$i$) | Re |
| $\mathbf{B}_{20}$ | = | $z_8\,\mathbf{a}_1 + (-x_8 + z_8)\,\mathbf{a}_2 - x_8\,\mathbf{a}_3$ | = | $-x_8a\,\hat{\mathbf{x}} + z_8c\,\hat{\mathbf{z}}$ | (8$i$) | Re |
| $\mathbf{B}_{21}$ | = | $(-x_8 - z_8)\,\mathbf{a}_1 - z_8\,\mathbf{a}_2 - x_8\,\mathbf{a}_3$ | = | $-x_8a\,\hat{\mathbf{y}} - z_8c\,\hat{\mathbf{z}}$ | (8$i$) | Re |
| $\mathbf{B}_{22}$ | = | $(x_8 - z_8)\,\mathbf{a}_1 - z_8\,\mathbf{a}_2 + x_8\,\mathbf{a}_3$ | = | $x_8a\,\hat{\mathbf{y}} - z_8c\,\hat{\mathbf{z}}$ | (8$i$) | Re |

### Found in:

- M. R. Norman, *The crystal structure of the inversion breaking metal $Cd_2Re_2O_7$*, Phys. Rev. B **101**, 045117 (2020), doi:10.1103/PhysRevB.101.045117.

---

**Geometry files:**
- CIF: pp. 1684
- POSCAR: pp. 1684

# Tetragonal $TlFeS_2$ Structure: AB2C_tI8_119_c_e_a

http://aflow.org/prototype-encyclopedia/AB2C_tI8_119_c_e_a

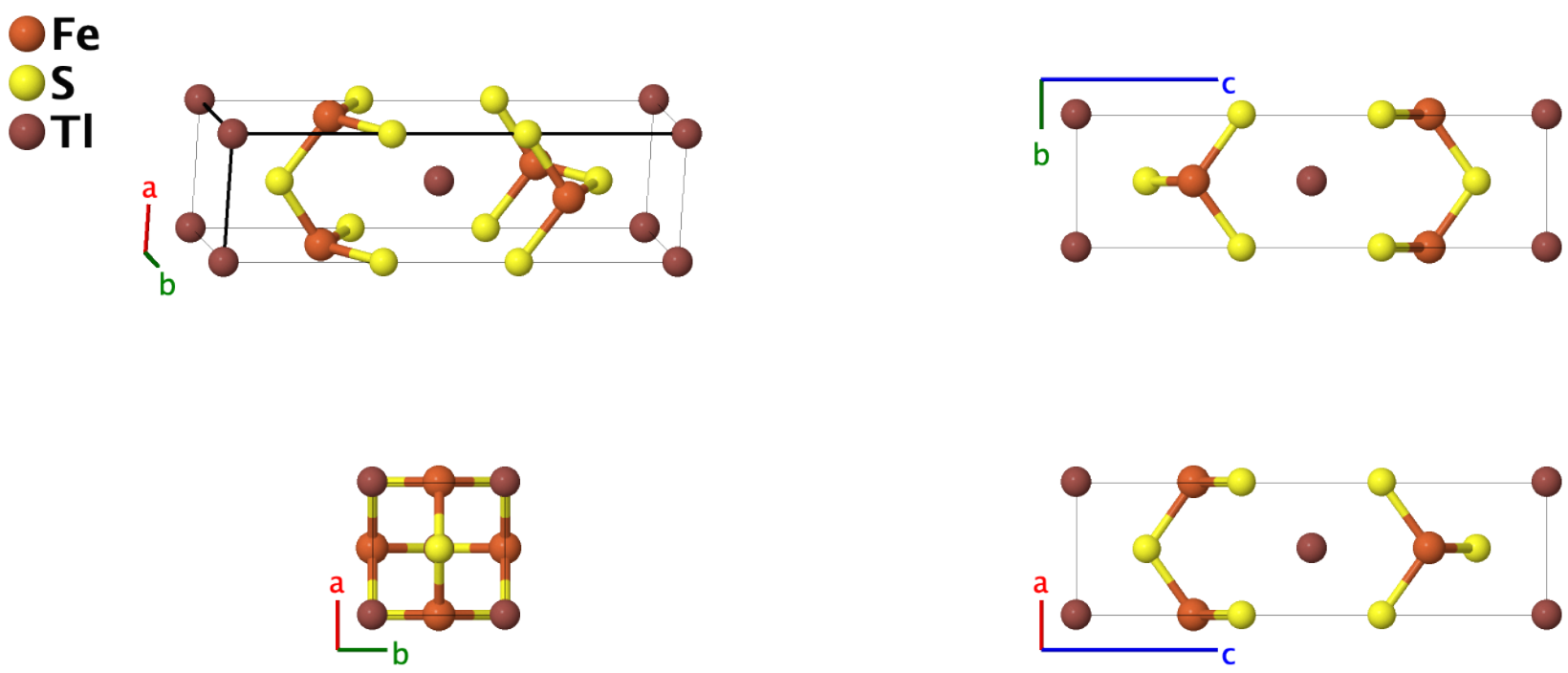

| | | |
|---|---|---|
| **Prototype** | : | $FeS_2Tl$ |
| **AFLOW prototype label** | : | AB2C_tI8_119_c_e_a |
| ***Strukturbericht* designation** | : | None |
| **Pearson symbol** | : | tI8 |
| **Space group number** | : | 119 |
| **Space group symbol** | : | $I\bar{4}m2$ |
| **AFLOW prototype command** | : | `aflow --proto=AB2C_tI8_119_c_e_a --params=`$a, c/a, z_3$ |

### Other compounds with this structure

- $CsFeSe_2$, $KFeSe_2$, $RbFeSe_2$, and $TlFeSe_2$

---

- This is the high-temperature form of $TlFeS_2$/$TlFeSe_2$, formed above 300 °C. Below that temperature, it transforms into the monoclinic $TlFeSe_2$ structure.

### Body-centered Tetragonal primitive vectors:

$$\mathbf{a}_1 = -\frac{1}{2}\,a\,\hat{\mathbf{x}} + \frac{1}{2}\,a\,\hat{\mathbf{y}} + \frac{1}{2}\,c\,\hat{\mathbf{z}}$$
$$\mathbf{a}_2 = \frac{1}{2}\,a\,\hat{\mathbf{x}} - \frac{1}{2}\,a\,\hat{\mathbf{y}} + \frac{1}{2}\,c\,\hat{\mathbf{z}}$$
$$\mathbf{a}_3 = \frac{1}{2}\,a\,\hat{\mathbf{x}} + \frac{1}{2}\,a\,\hat{\mathbf{y}} - \frac{1}{2}\,c\,\hat{\mathbf{z}}$$

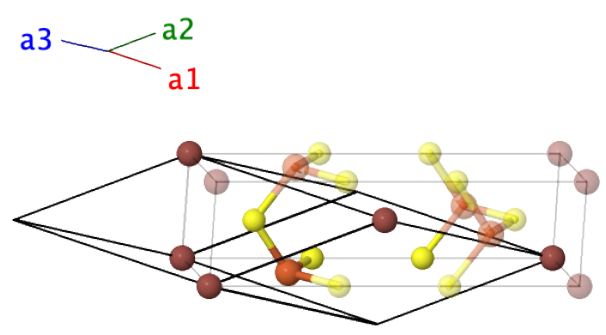

### Basis vectors:

| | | Lattice Coordinates | | Cartesian Coordinates | Wyckoff Position | Atom Type |
|---|---|---|---|---|---|---|
| $\mathbf{B}_1$ | = | $0\,\mathbf{a}_1 + 0\,\mathbf{a}_2 + 0\,\mathbf{a}_3$ | = | $0\,\hat{\mathbf{x}} + 0\,\hat{\mathbf{y}} + 0\,\hat{\mathbf{z}}$ | (2$a$) | Tl |
| $\mathbf{B}_2$ | = | $\frac{3}{4}\,\mathbf{a}_1 + \frac{1}{4}\,\mathbf{a}_2 + \frac{1}{2}\,\mathbf{a}_3$ | = | $\frac{1}{2}a\,\hat{\mathbf{y}} + \frac{1}{4}c\,\hat{\mathbf{z}}$ | (2$c$) | Fe |
| $\mathbf{B}_3$ | = | $z_3\,\mathbf{a}_1 + z_3\,\mathbf{a}_2$ | = | $z_3c\,\hat{\mathbf{z}}$ | (4$e$) | S |
| $\mathbf{B}_4$ | = | $-z_3\,\mathbf{a}_1 - z_3\,\mathbf{a}_2$ | = | $-z_3c\,\hat{\mathbf{z}}$ | (4$e$) | S |

**Found in:**

---

**Geometry files:**

- CIF: pp. 1685
- POSCAR: pp. 1685

# $BeSO_4{\cdot}4H_2O$ ($H4_3$) Structure: AB8C8D_tI72_120_c_2i_2i_b

http://aflow.org/prototype-encyclopedia/AB8C8D_tI72_120_c_2i_2i_b

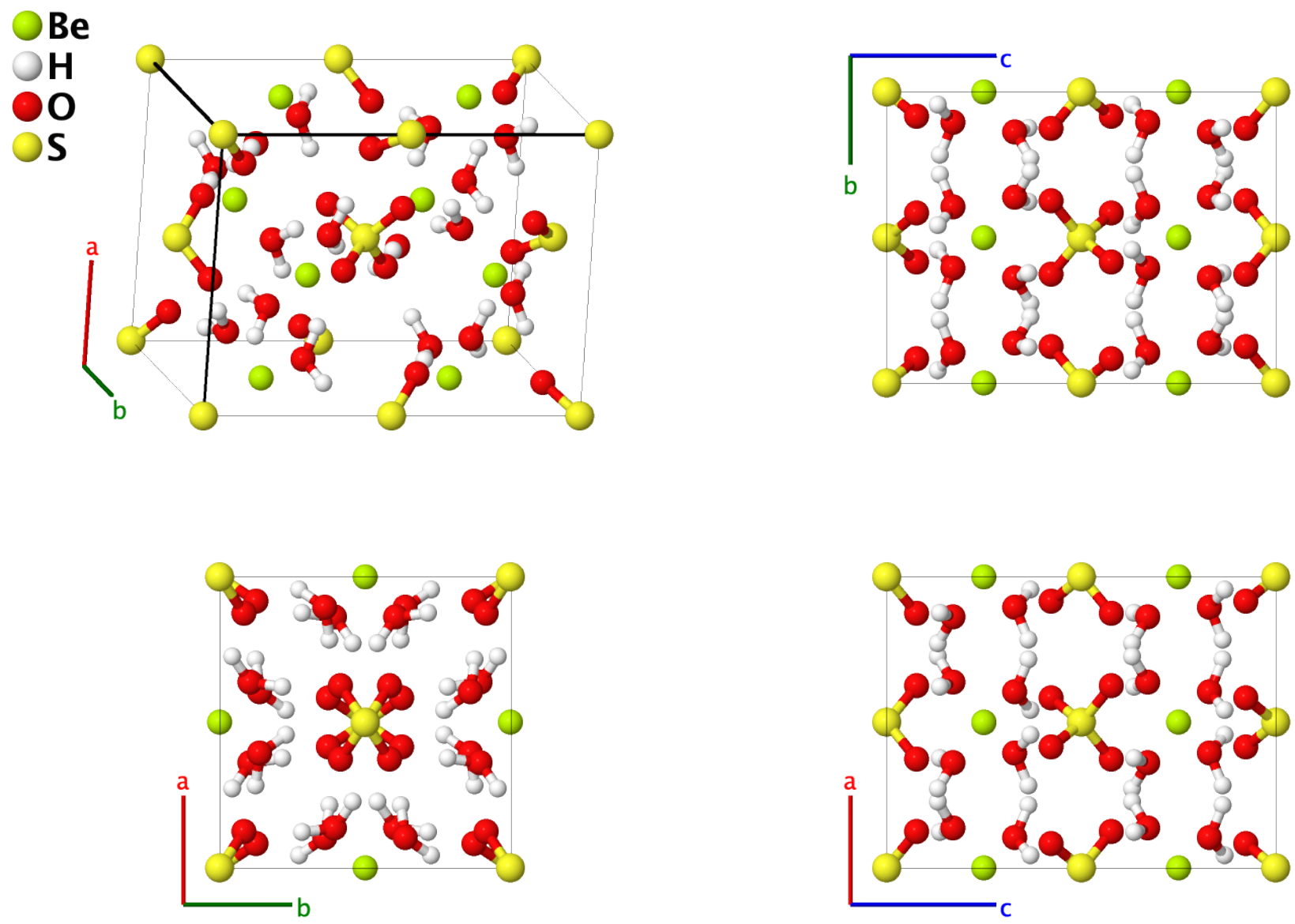

| | | |
|---|---|---|
| **Prototype** | : | $BeH_8O_8S$ |
| **AFLOW prototype label** | : | AB8C8D_tI72_120_c_2i_2i_b |
| ***Strukturbericht* designation** | : | $H4_3$ |
| **Pearson symbol** | : | tI72 |
| **Space group number** | : | 120 |
| **Space group symbol** | : | $I\bar{4}c2$ |
| **AFLOW prototype command** | : | `aflow --proto=AB8C8D_tI72_120_c_2i_2i_b`<br>`--params=`$a, c/a, x_3, y_3, z_3, x_4, y_4, z_4, x_5, y_5, z_5, x_6, y_6, z_6$ |

- The original determination of the $H4_3$ structure did not determine the positions of the hydrogen atoms. Since (Sikka, 1969) showed that the placement of the hydrogen atoms did not substantially affect the positions of the other atoms in the primitive cell, nor change the space group, we retain the original *Strukturbericht* designation for the improved structure.

**Body-centered Tetragonal primitive vectors:**

$$\mathbf{a}_1 = -\frac{1}{2}\,a\,\hat{\mathbf{x}} + \frac{1}{2}\,a\,\hat{\mathbf{y}} + \frac{1}{2}\,c\,\hat{\mathbf{z}}$$
$$\mathbf{a}_2 = \frac{1}{2}\,a\,\hat{\mathbf{x}} - \frac{1}{2}\,a\,\hat{\mathbf{y}} + \frac{1}{2}\,c\,\hat{\mathbf{z}}$$
$$\mathbf{a}_3 = \frac{1}{2}\,a\,\hat{\mathbf{x}} + \frac{1}{2}\,a\,\hat{\mathbf{y}} - \frac{1}{2}\,c\,\hat{\mathbf{z}}$$

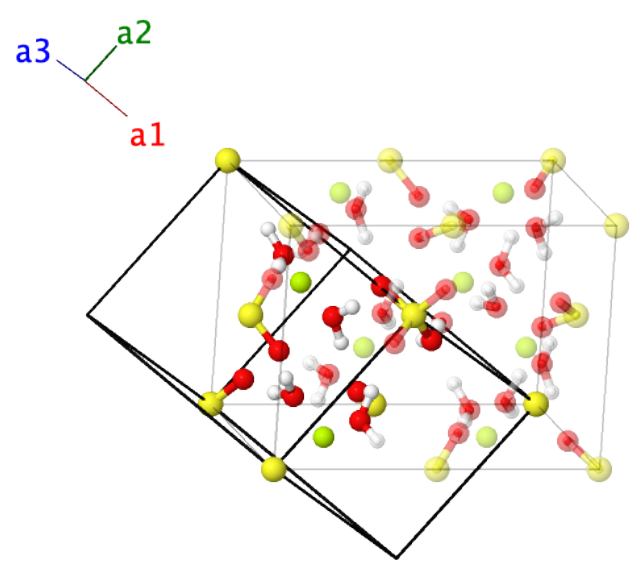

**Basis vectors:**

| Lattice Coordinates | Cartesian Coordinates | Wyckoff Position | Atom Type |
|---|---|---|---|

| | | | | | | |
|---|---|---|---|---|---|---|
| $\mathbf{B}_1$ | = | $0\,\mathbf{a}_1 + 0\,\mathbf{a}_2 + 0\,\mathbf{a}_3$ | = | $0\,\hat{\mathbf{x}} + 0\,\hat{\mathbf{y}} + 0\,\hat{\mathbf{z}}$ | (4*b*) | S |
| $\mathbf{B}_2$ | = | $\frac{1}{2}\,\mathbf{a}_1 + \frac{1}{2}\,\mathbf{a}_2$ | = | $\frac{1}{2}c\,\hat{\mathbf{z}}$ | (4*b*) | S |
| $\mathbf{B}_3$ | = | $\frac{3}{4}\,\mathbf{a}_1 + \frac{1}{4}\,\mathbf{a}_2 + \frac{1}{2}\,\mathbf{a}_3$ | = | $\frac{1}{2}a\,\hat{\mathbf{y}} + \frac{1}{4}c\,\hat{\mathbf{z}}$ | (4*c*) | Be |
| $\mathbf{B}_4$ | = | $\frac{1}{4}\,\mathbf{a}_1 + \frac{3}{4}\,\mathbf{a}_2 + \frac{1}{2}\,\mathbf{a}_3$ | = | $\frac{1}{2}a\,\hat{\mathbf{x}} + \frac{1}{4}c\,\hat{\mathbf{z}}$ | (4*c*) | Be |
| $\mathbf{B}_5$ | = | $(y_3+z_3)\,\mathbf{a}_1 + (x_3+z_3)\,\mathbf{a}_2 + (x_3+y_3)\,\mathbf{a}_3$ | = | $x_3a\,\hat{\mathbf{x}} + y_3a\,\hat{\mathbf{y}} + z_3c\,\hat{\mathbf{z}}$ | (16*i*) | H I |
| $\mathbf{B}_6$ | = | $(-y_3+z_3)\,\mathbf{a}_1 + (-x_3+z_3)\,\mathbf{a}_2 + (-x_3-y_3)\,\mathbf{a}_3$ | = | $-x_3a\,\hat{\mathbf{x}} - y_3a\,\hat{\mathbf{y}} + z_3c\,\hat{\mathbf{z}}$ | (16*i*) | H I |
| $\mathbf{B}_7$ | = | $(-x_3-z_3)\,\mathbf{a}_1 + (y_3-z_3)\,\mathbf{a}_2 + (-x_3+y_3)\,\mathbf{a}_3$ | = | $y_3a\,\hat{\mathbf{x}} - x_3a\,\hat{\mathbf{y}} - z_3c\,\hat{\mathbf{z}}$ | (16*i*) | H I |
| $\mathbf{B}_8$ | = | $(x_3-z_3)\,\mathbf{a}_1 + (-y_3-z_3)\,\mathbf{a}_2 + (x_3-y_3)\,\mathbf{a}_3$ | = | $-y_3a\,\hat{\mathbf{x}} + x_3a\,\hat{\mathbf{y}} - z_3c\,\hat{\mathbf{z}}$ | (16*i*) | H I |
| $\mathbf{B}_9$ | = | $\left(\frac{1}{2}-y_3+z_3\right)\mathbf{a}_1 + \left(\frac{1}{2}+x_3+z_3\right)\mathbf{a}_2 + (x_3-y_3)\,\mathbf{a}_3$ | = | $x_3a\,\hat{\mathbf{x}} - y_3a\,\hat{\mathbf{y}} + \left(\frac{1}{2}+z_3\right)c\,\hat{\mathbf{z}}$ | (16*i*) | H I |
| $\mathbf{B}_{10}$ | = | $\left(\frac{1}{2}+y_3+z_3\right)\mathbf{a}_1 + \left(\frac{1}{2}-x_3+z_3\right)\mathbf{a}_2 + (-x_3+y_3)\,\mathbf{a}_3$ | = | $-x_3a\,\hat{\mathbf{x}} + y_3a\,\hat{\mathbf{y}} + \left(\frac{1}{2}+z_3\right)c\,\hat{\mathbf{z}}$ | (16*i*) | H I |
| $\mathbf{B}_{11}$ | = | $\left(\frac{1}{2}+x_3-z_3\right)\mathbf{a}_1 + \left(\frac{1}{2}+y_3-z_3\right)\mathbf{a}_2 + (x_3+y_3)\,\mathbf{a}_3$ | = | $y_3a\,\hat{\mathbf{x}} + x_3a\,\hat{\mathbf{y}} + \left(\frac{1}{2}-z_3\right)c\,\hat{\mathbf{z}}$ | (16*i*) | H I |
| $\mathbf{B}_{12}$ | = | $\left(\frac{1}{2}-x_3-z_3\right)\mathbf{a}_1 + \left(\frac{1}{2}-y_3-z_3\right)\mathbf{a}_2 + (-x_3-y_3)\,\mathbf{a}_3$ | = | $-y_3a\,\hat{\mathbf{x}} - x_3a\,\hat{\mathbf{y}} + \left(\frac{1}{2}-z_3\right)c\,\hat{\mathbf{z}}$ | (16*i*) | H I |
| $\mathbf{B}_{13}$ | = | $(y_4+z_4)\,\mathbf{a}_1 + (x_4+z_4)\,\mathbf{a}_2 + (x_4+y_4)\,\mathbf{a}_3$ | = | $x_4a\,\hat{\mathbf{x}} + y_4a\,\hat{\mathbf{y}} + z_4c\,\hat{\mathbf{z}}$ | (16*i*) | H II |
| $\mathbf{B}_{14}$ | = | $(-y_4+z_4)\,\mathbf{a}_1 + (-x_4+z_4)\,\mathbf{a}_2 + (-x_4-y_4)\,\mathbf{a}_3$ | = | $-x_4a\,\hat{\mathbf{x}} - y_4a\,\hat{\mathbf{y}} + z_4c\,\hat{\mathbf{z}}$ | (16*i*) | H II |
| $\mathbf{B}_{15}$ | = | $(-x_4-z_4)\,\mathbf{a}_1 + (y_4-z_4)\,\mathbf{a}_2 + (-x_4+y_4)\,\mathbf{a}_3$ | = | $y_4a\,\hat{\mathbf{x}} - x_4a\,\hat{\mathbf{y}} - z_4c\,\hat{\mathbf{z}}$ | (16*i*) | H II |
| $\mathbf{B}_{16}$ | = | $(x_4-z_4)\,\mathbf{a}_1 + (-y_4-z_4)\,\mathbf{a}_2 + (x_4-y_4)\,\mathbf{a}_3$ | = | $-y_4a\,\hat{\mathbf{x}} + x_4a\,\hat{\mathbf{y}} - z_4c\,\hat{\mathbf{z}}$ | (16*i*) | H II |
| $\mathbf{B}_{17}$ | = | $\left(\frac{1}{2}-y_4+z_4\right)\mathbf{a}_1 + \left(\frac{1}{2}+x_4+z_4\right)\mathbf{a}_2 + (x_4-y_4)\,\mathbf{a}_3$ | = | $x_4a\,\hat{\mathbf{x}} - y_4a\,\hat{\mathbf{y}} + \left(\frac{1}{2}+z_4\right)c\,\hat{\mathbf{z}}$ | (16*i*) | H II |
| $\mathbf{B}_{18}$ | = | $\left(\frac{1}{2}+y_4+z_4\right)\mathbf{a}_1 + \left(\frac{1}{2}-x_4+z_4\right)\mathbf{a}_2 + (-x_4+y_4)\,\mathbf{a}_3$ | = | $-x_4a\,\hat{\mathbf{x}} + y_4a\,\hat{\mathbf{y}} + \left(\frac{1}{2}+z_4\right)c\,\hat{\mathbf{z}}$ | (16*i*) | H II |
| $\mathbf{B}_{19}$ | = | $\left(\frac{1}{2}+x_4-z_4\right)\mathbf{a}_1 + \left(\frac{1}{2}+y_4-z_4\right)\mathbf{a}_2 + (x_4+y_4)\,\mathbf{a}_3$ | = | $y_4a\,\hat{\mathbf{x}} + x_4a\,\hat{\mathbf{y}} + \left(\frac{1}{2}-z_4\right)c\,\hat{\mathbf{z}}$ | (16*i*) | H II |
| $\mathbf{B}_{20}$ | = | $\left(\frac{1}{2}-x_4-z_4\right)\mathbf{a}_1 + \left(\frac{1}{2}-y_4-z_4\right)\mathbf{a}_2 + (-x_4-y_4)\,\mathbf{a}_3$ | = | $-y_4a\,\hat{\mathbf{x}} - x_4a\,\hat{\mathbf{y}} + \left(\frac{1}{2}-z_4\right)c\,\hat{\mathbf{z}}$ | (16*i*) | H II |
| $\mathbf{B}_{21}$ | = | $(y_5+z_5)\,\mathbf{a}_1 + (x_5+z_5)\,\mathbf{a}_2 + (x_5+y_5)\,\mathbf{a}_3$ | = | $x_5a\,\hat{\mathbf{x}} + y_5a\,\hat{\mathbf{y}} + z_5c\,\hat{\mathbf{z}}$ | (16*i*) | O I |
| $\mathbf{B}_{22}$ | = | $(-y_5+z_5)\,\mathbf{a}_1 + (-x_5+z_5)\,\mathbf{a}_2 + (-x_5-y_5)\,\mathbf{a}_3$ | = | $-x_5a\,\hat{\mathbf{x}} - y_5a\,\hat{\mathbf{y}} + z_5c\,\hat{\mathbf{z}}$ | (16*i*) | O I |
| $\mathbf{B}_{23}$ | = | $(-x_5-z_5)\,\mathbf{a}_1 + (y_5-z_5)\,\mathbf{a}_2 + (-x_5+y_5)\,\mathbf{a}_3$ | = | $y_5a\,\hat{\mathbf{x}} - x_5a\,\hat{\mathbf{y}} - z_5c\,\hat{\mathbf{z}}$ | (16*i*) | O I |
| $\mathbf{B}_{24}$ | = | $(x_5-z_5)\,\mathbf{a}_1 + (-y_5-z_5)\,\mathbf{a}_2 + (x_5-y_5)\,\mathbf{a}_3$ | = | $-y_5a\,\hat{\mathbf{x}} + x_5a\,\hat{\mathbf{y}} - z_5c\,\hat{\mathbf{z}}$ | (16*i*) | O I |
| $\mathbf{B}_{25}$ | = | $\left(\frac{1}{2}-y_5+z_5\right)\mathbf{a}_1 + \left(\frac{1}{2}+x_5+z_5\right)\mathbf{a}_2 + (x_5-y_5)\,\mathbf{a}_3$ | = | $x_5a\,\hat{\mathbf{x}} - y_5a\,\hat{\mathbf{y}} + \left(\frac{1}{2}+z_5\right)c\,\hat{\mathbf{z}}$ | (16*i*) | O I |
| $\mathbf{B}_{26}$ | = | $\left(\frac{1}{2}+y_5+z_5\right)\mathbf{a}_1 + \left(\frac{1}{2}-x_5+z_5\right)\mathbf{a}_2 + (-x_5+y_5)\,\mathbf{a}_3$ | = | $-x_5a\,\hat{\mathbf{x}} + y_5a\,\hat{\mathbf{y}} + \left(\frac{1}{2}+z_5\right)c\,\hat{\mathbf{z}}$ | (16*i*) | O I |
| $\mathbf{B}_{27}$ | = | $\left(\frac{1}{2}+x_5-z_5\right)\mathbf{a}_1 + \left(\frac{1}{2}+y_5-z_5\right)\mathbf{a}_2 + (x_5+y_5)\,\mathbf{a}_3$ | = | $y_5a\,\hat{\mathbf{x}} + x_5a\,\hat{\mathbf{y}} + \left(\frac{1}{2}-z_5\right)c\,\hat{\mathbf{z}}$ | (16*i*) | O I |
| $\mathbf{B}_{28}$ | = | $\left(\frac{1}{2}-x_5-z_5\right)\mathbf{a}_1 + \left(\frac{1}{2}-y_5-z_5\right)\mathbf{a}_2 + (-x_5-y_5)\,\mathbf{a}_3$ | = | $-y_5a\,\hat{\mathbf{x}} - x_5a\,\hat{\mathbf{y}} + \left(\frac{1}{2}-z_5\right)c\,\hat{\mathbf{z}}$ | (16*i*) | O I |
| $\mathbf{B}_{29}$ | = | $(y_6 + z_6)\,\mathbf{a}_1 + (x_6 + z_6)\,\mathbf{a}_2 + (x_6 + y_6)\,\mathbf{a}_3$ | = | $x_6 a\,\hat{\mathbf{x}} + y_6 a\,\hat{\mathbf{y}} + z_6 c\,\hat{\mathbf{z}}$ | (16$i$) | O II |
| $\mathbf{B}_{30}$ | = | $(-y_6 + z_6)\,\mathbf{a}_1 + (-x_6 + z_6)\,\mathbf{a}_2 + (-x_6 - y_6)\,\mathbf{a}_3$ | = | $-x_6 a\,\hat{\mathbf{x}} - y_6 a\,\hat{\mathbf{y}} + z_6 c\,\hat{\mathbf{z}}$ | (16$i$) | O II |
| $\mathbf{B}_{31}$ | = | $(-x_6 - z_6)\,\mathbf{a}_1 + (y_6 - z_6)\,\mathbf{a}_2 + (-x_6 + y_6)\,\mathbf{a}_3$ | = | $y_6 a\,\hat{\mathbf{x}} - x_6 a\,\hat{\mathbf{y}} - z_6 c\,\hat{\mathbf{z}}$ | (16$i$) | O II |
| $\mathbf{B}_{32}$ | = | $(x_6 - z_6)\,\mathbf{a}_1 + (-y_6 - z_6)\,\mathbf{a}_2 + (x_6 - y_6)\,\mathbf{a}_3$ | = | $-y_6 a\,\hat{\mathbf{x}} + x_6 a\,\hat{\mathbf{y}} - z_6 c\,\hat{\mathbf{z}}$ | (16$i$) | O II |
| $\mathbf{B}_{33}$ | = | $\left(\frac{1}{2} - y_6 + z_6\right)\mathbf{a}_1 + \left(\frac{1}{2} + x_6 + z_6\right)\mathbf{a}_2 + (x_6 - y_6)\,\mathbf{a}_3$ | = | $x_6 a\,\hat{\mathbf{x}} - y_6 a\,\hat{\mathbf{y}} + \left(\frac{1}{2} + z_6\right) c\,\hat{\mathbf{z}}$ | (16$i$) | O II |
| $\mathbf{B}_{34}$ | = | $\left(\frac{1}{2} + y_6 + z_6\right)\mathbf{a}_1 + \left(\frac{1}{2} - x_6 + z_6\right)\mathbf{a}_2 + (-x_6 + y_6)\,\mathbf{a}_3$ | = | $-x_6 a\,\hat{\mathbf{x}} + y_6 a\,\hat{\mathbf{y}} + \left(\frac{1}{2} + z_6\right) c\,\hat{\mathbf{z}}$ | (16$i$) | O II |
| $\mathbf{B}_{35}$ | = | $\left(\frac{1}{2} + x_6 - z_6\right)\mathbf{a}_1 + \left(\frac{1}{2} + y_6 - z_6\right)\mathbf{a}_2 + (x_6 + y_6)\,\mathbf{a}_3$ | = | $y_6 a\,\hat{\mathbf{x}} + x_6 a\,\hat{\mathbf{y}} + \left(\frac{1}{2} - z_6\right) c\,\hat{\mathbf{z}}$ | (16$i$) | O II |
| $\mathbf{B}_{36}$ | = | $\left(\frac{1}{2} - x_6 - z_6\right)\mathbf{a}_1 + \left(\frac{1}{2} - y_6 - z_6\right)\mathbf{a}_2 + (-x_6 - y_6)\,\mathbf{a}_3$ | = | $-y_6 a\,\hat{\mathbf{x}} - x_6 a\,\hat{\mathbf{y}} + \left(\frac{1}{2} - z_6\right) c\,\hat{\mathbf{z}}$ | (16$i$) | O II |

**Geometry files:**

- CIF: pp. 1685
- POSCAR: pp. 1685

# $SrCu_2(BO_3)_2$ Structure: A2B2C6D_tI44_121_i_i_ij_c

http://aflow.org/prototype-encyclopedia/A2B2C6D_tI44_121_i_i_ij_c

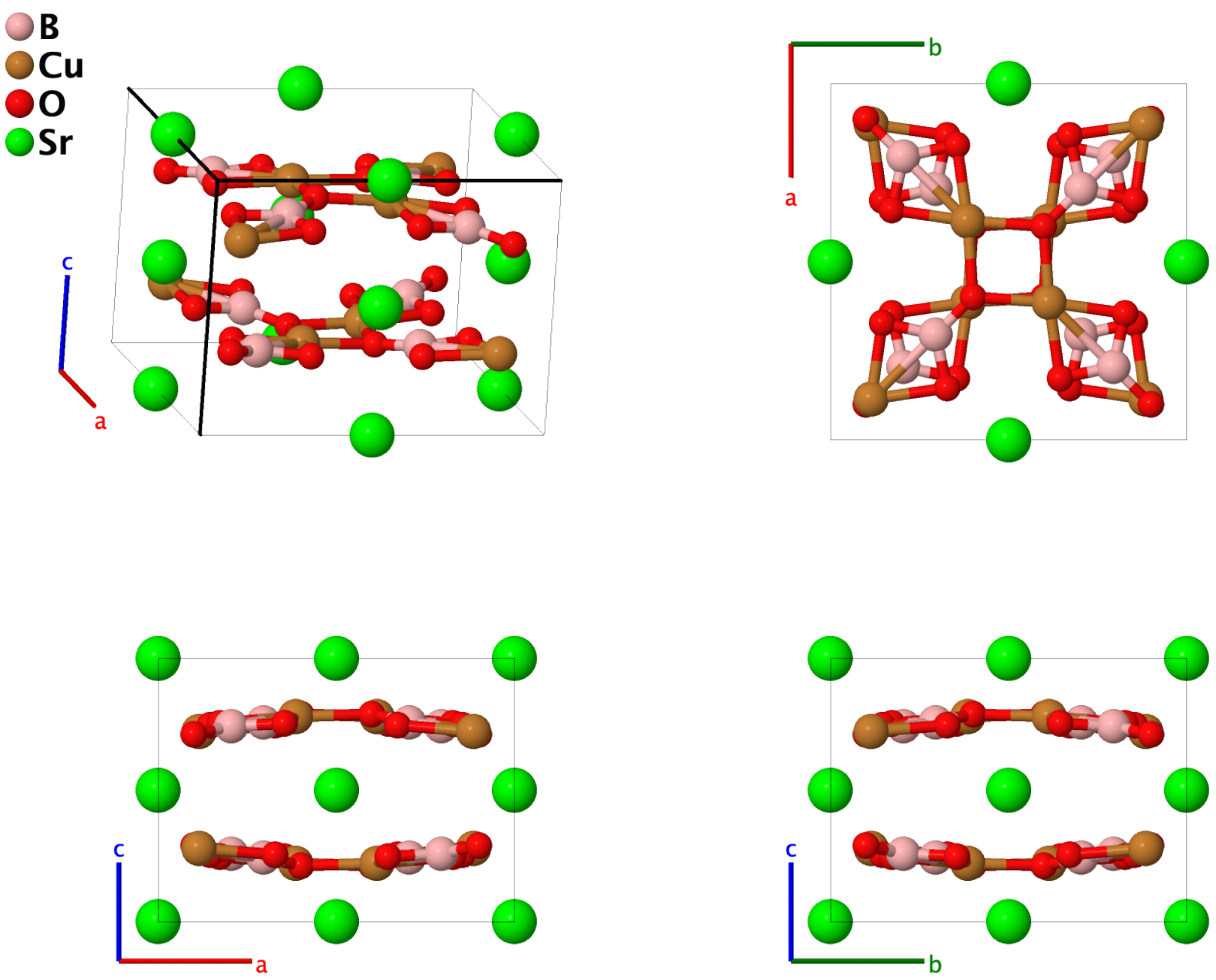

| | | |
|---|---|---|
| **Prototype** | : | $B_2Cu_2O_6Sr$ |
| **AFLOW prototype label** | : | A2B2C6D_tI44_121_i_i_ij_c |
| ***Strukturbericht* designation** | : | None |
| **Pearson symbol** | : | tI44 |
| **Space group number** | : | 121 |
| **Space group symbol** | : | $I\bar{4}2m$ |
| **AFLOW prototype command** | : | `aflow --proto=A2B2C6D_tI44_121_i_i_ij_c --params=`$a, c/a, x_2, z_2, x_3, z_3, x_4, z_4, x_5, y_5, z_5$ |

**Body-centered Tetragonal primitive vectors:**

$$\mathbf{a}_1 = -\frac{1}{2}\,a\,\hat{\mathbf{x}} + \frac{1}{2}\,a\,\hat{\mathbf{y}} + \frac{1}{2}\,c\,\hat{\mathbf{z}}$$

$$\mathbf{a}_2 = \frac{1}{2}\,a\,\hat{\mathbf{x}} - \frac{1}{2}\,a\,\hat{\mathbf{y}} + \frac{1}{2}\,c\,\hat{\mathbf{z}}$$

$$\mathbf{a}_3 = \frac{1}{2}\,a\,\hat{\mathbf{x}} + \frac{1}{2}\,a\,\hat{\mathbf{y}} - \frac{1}{2}\,c\,\hat{\mathbf{z}}$$

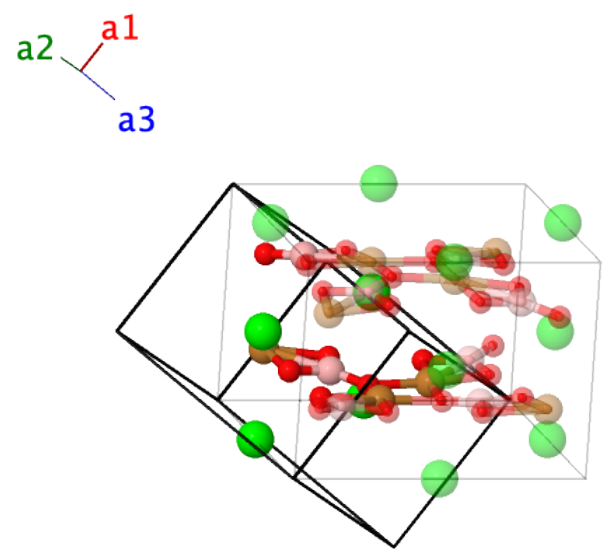

**Basis vectors:**

| | | Lattice Coordinates | | Cartesian Coordinates | Wyckoff Position | Atom Type |
|---|---|---|---|---|---|---|
| $\mathbf{B}_1$ | = | $\frac{1}{2}\,\mathbf{a}_1 + \frac{1}{2}\,\mathbf{a}_3$ | = | $\frac{1}{2}a\,\hat{\mathbf{y}}$ | (4c) | Sr |
| $\mathbf{B}_2$ | = | $\frac{1}{2}\,\mathbf{a}_2 + \frac{1}{2}\,\mathbf{a}_3$ | = | $\frac{1}{2}a\,\hat{\mathbf{x}}$ | (4c) | Sr |
| $\mathbf{B}_3$ | = | $(x_2 + z_2)\,\mathbf{a}_1 + (x_2 + z_2)\,\mathbf{a}_2 + 2x_2\,\mathbf{a}_3$ | = | $x_2a\,\hat{\mathbf{x}} + x_2a\,\hat{\mathbf{y}} + z_2c\,\hat{\mathbf{z}}$ | (8i) | B |
| $\mathbf{B}_4$ | = | $(-x_2+z_2)\,\mathbf{a}_1+(-x_2+z_2)\,\mathbf{a}_2-2x_2\,\mathbf{a}_3$ | = | $-x_2a\,\hat{\mathbf{x}}-x_2a\,\hat{\mathbf{y}}+z_2c\,\hat{\mathbf{z}}$ | (8$i$) | B |
| $\mathbf{B}_5$ | = | $(-x_2-z_2)\,\mathbf{a}_1+(x_2-z_2)\,\mathbf{a}_2$ | = | $x_2a\,\hat{\mathbf{x}}-x_2a\,\hat{\mathbf{y}}-z_2c\,\hat{\mathbf{z}}$ | (8$i$) | B |
| $\mathbf{B}_6$ | = | $(x_2-z_2)\,\mathbf{a}_1+(-x_2-z_2)\,\mathbf{a}_2$ | = | $-x_2a\,\hat{\mathbf{x}}+x_2a\,\hat{\mathbf{y}}-z_2c\,\hat{\mathbf{z}}$ | (8$i$) | B |
| $\mathbf{B}_7$ | = | $(x_3+z_3)\,\mathbf{a}_1+(x_3+z_3)\,\mathbf{a}_2+2x_3\,\mathbf{a}_3$ | = | $x_3a\,\hat{\mathbf{x}}+x_3a\,\hat{\mathbf{y}}+z_3c\,\hat{\mathbf{z}}$ | (8$i$) | Cu |
| $\mathbf{B}_8$ | = | $(-x_3+z_3)\,\mathbf{a}_1+(-x_3+z_3)\,\mathbf{a}_2-2x_3\,\mathbf{a}_3$ | = | $-x_3a\,\hat{\mathbf{x}}-x_3a\,\hat{\mathbf{y}}+z_3c\,\hat{\mathbf{z}}$ | (8$i$) | Cu |
| $\mathbf{B}_9$ | = | $(-x_3-z_3)\,\mathbf{a}_1+(x_3-z_3)\,\mathbf{a}_2$ | = | $x_3a\,\hat{\mathbf{x}}-x_3a\,\hat{\mathbf{y}}-z_3c\,\hat{\mathbf{z}}$ | (8$i$) | Cu |
| $\mathbf{B}_{10}$ | = | $(x_3-z_3)\,\mathbf{a}_1+(-x_3-z_3)\,\mathbf{a}_2$ | = | $-x_3a\,\hat{\mathbf{x}}+x_3a\,\hat{\mathbf{y}}-z_3c\,\hat{\mathbf{z}}$ | (8$i$) | Cu |
| $\mathbf{B}_{11}$ | = | $(x_4+z_4)\,\mathbf{a}_1+(x_4+z_4)\,\mathbf{a}_2+2x_4\,\mathbf{a}_3$ | = | $x_4a\,\hat{\mathbf{x}}+x_4a\,\hat{\mathbf{y}}+z_4c\,\hat{\mathbf{z}}$ | (8$i$) | O I |
| $\mathbf{B}_{12}$ | = | $(-x_4+z_4)\,\mathbf{a}_1+(-x_4+z_4)\,\mathbf{a}_2-2x_4\,\mathbf{a}_3$ | = | $-x_4a\,\hat{\mathbf{x}}-x_4a\,\hat{\mathbf{y}}+z_4c\,\hat{\mathbf{z}}$ | (8$i$) | O I |
| $\mathbf{B}_{13}$ | = | $(-x_4-z_4)\,\mathbf{a}_1+(x_4-z_4)\,\mathbf{a}_2$ | = | $x_4a\,\hat{\mathbf{x}}-x_4a\,\hat{\mathbf{y}}-z_4c\,\hat{\mathbf{z}}$ | (8$i$) | O I |
| $\mathbf{B}_{14}$ | = | $(x_4-z_4)\,\mathbf{a}_1+(-x_4-z_4)\,\mathbf{a}_2$ | = | $-x_4a\,\hat{\mathbf{x}}+x_4a\,\hat{\mathbf{y}}-z_4c\,\hat{\mathbf{z}}$ | (8$i$) | O I |
| $\mathbf{B}_{15}$ | = | $(y_5+z_5)\,\mathbf{a}_1+(x_5+z_5)\,\mathbf{a}_2+(x_5+y_5)\,\mathbf{a}_3$ | = | $x_5a\,\hat{\mathbf{x}}+y_5a\,\hat{\mathbf{y}}+z_5c\,\hat{\mathbf{z}}$ | (16$j$) | O II |
| $\mathbf{B}_{16}$ | = | $(-y_5+z_5)\,\mathbf{a}_1+(-x_5+z_5)\,\mathbf{a}_2+(-x_5-y_5)\,\mathbf{a}_3$ | = | $-x_5a\,\hat{\mathbf{x}}-y_5a\,\hat{\mathbf{y}}+z_5c\,\hat{\mathbf{z}}$ | (16$j$) | O II |
| $\mathbf{B}_{17}$ | = | $(-x_5-z_5)\,\mathbf{a}_1+(y_5-z_5)\,\mathbf{a}_2+(-x_5+y_5)\,\mathbf{a}_3$ | = | $y_5a\,\hat{\mathbf{x}}-x_5a\,\hat{\mathbf{y}}-z_5c\,\hat{\mathbf{z}}$ | (16$j$) | O II |
| $\mathbf{B}_{18}$ | = | $(x_5-z_5)\,\mathbf{a}_1+(-y_5-z_5)\,\mathbf{a}_2+(x_5-y_5)\,\mathbf{a}_3$ | = | $-y_5a\,\hat{\mathbf{x}}+x_5a\,\hat{\mathbf{y}}-z_5c\,\hat{\mathbf{z}}$ | (16$j$) | O II |
| $\mathbf{B}_{19}$ | = | $(y_5-z_5)\,\mathbf{a}_1+(-x_5-z_5)\,\mathbf{a}_2+(-x_5+y_5)\,\mathbf{a}_3$ | = | $-x_5a\,\hat{\mathbf{x}}+y_5a\,\hat{\mathbf{y}}-z_5c\,\hat{\mathbf{z}}$ | (16$j$) | O II |
| $\mathbf{B}_{20}$ | = | $(-y_5-z_5)\,\mathbf{a}_1+(x_5-z_5)\,\mathbf{a}_2+(x_5-y_5)\,\mathbf{a}_3$ | = | $x_5a\,\hat{\mathbf{x}}-y_5a\,\hat{\mathbf{y}}-z_5c\,\hat{\mathbf{z}}$ | (16$j$) | O II |
| $\mathbf{B}_{21}$ | = | $(-x_5+z_5)\,\mathbf{a}_1+(-y_5+z_5)\,\mathbf{a}_2+(-x_5-y_5)\,\mathbf{a}_3$ | = | $-y_5a\,\hat{\mathbf{x}}-x_5a\,\hat{\mathbf{y}}+z_5c\,\hat{\mathbf{z}}$ | (16$j$) | O II |
| $\mathbf{B}_{22}$ | = | $(x_5+z_5)\,\mathbf{a}_1+(y_5+z_5)\,\mathbf{a}_2+(x_5+y_5)\,\mathbf{a}_3$ | = | $y_5a\,\hat{\mathbf{x}}+x_5a\,\hat{\mathbf{y}}+z_5c\,\hat{\mathbf{z}}$ | (16$j$) | O II |

**Geometry files:**

- CIF: pp. 1686
- POSCAR: pp. 1686

# $C17$ ($Fe_2B$) (*obsolete*) Structure: AB2_tI12_121_ab_i

http://aflow.org/prototype-encyclopedia/AB2_tI12_121_ab_i

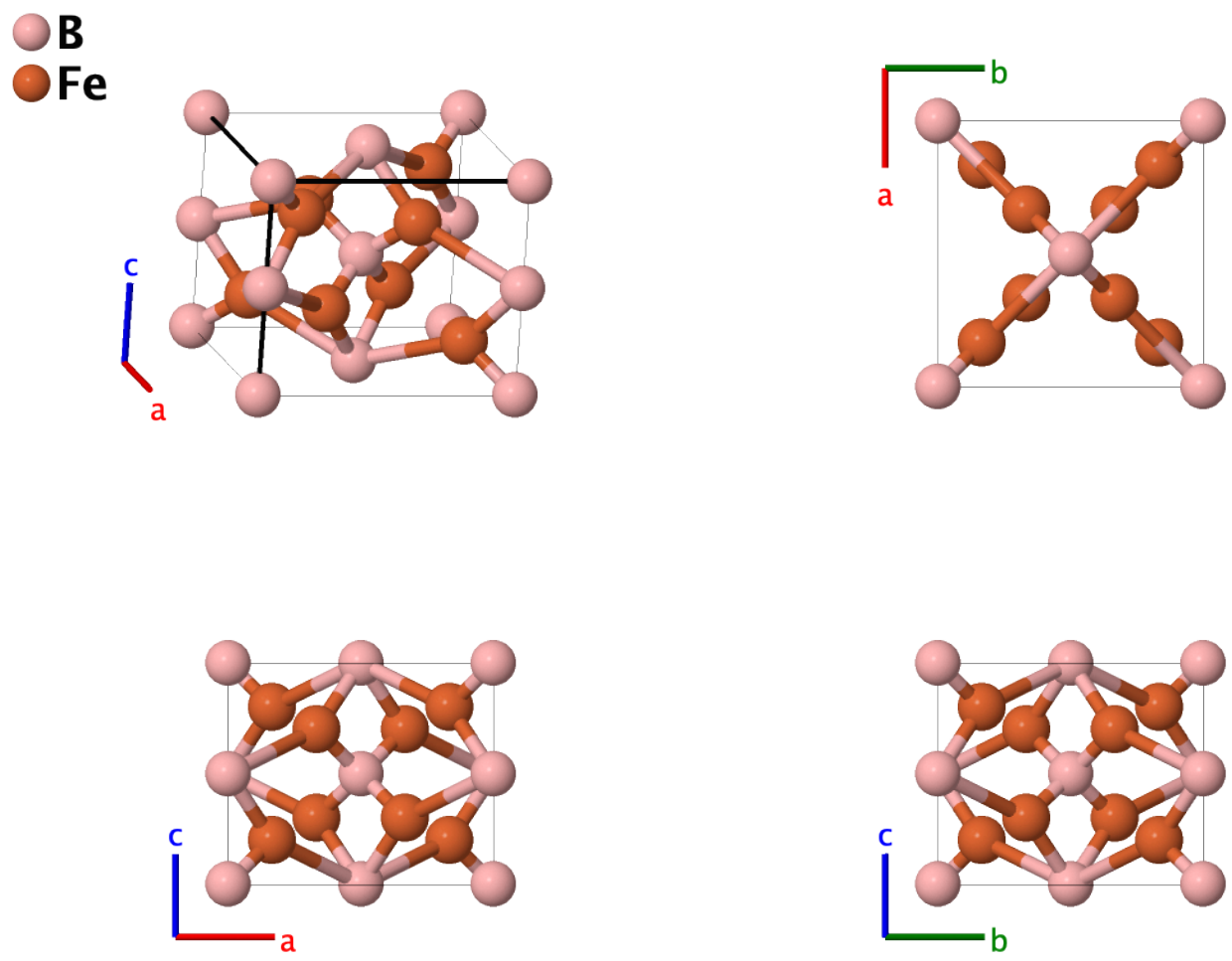

| | | |
|---|---|---|
| **Prototype** | : | $BFe_2$ |
| **AFLOW prototype label** | : | AB2_tI12_121_ab_i |
| ***Strukturbericht* designation** | : | $C17$ |
| **Pearson symbol** | : | tI12 |
| **Space group number** | : | 121 |
| **Space group symbol** | : | $I\bar{4}2m$ |
| **AFLOW prototype command** | : | `aflow --proto=AB2_tI12_121_ab_i` `--params=`$a, c/a, x_3, z_3$ |

- (Wever, 1930) placed this structure in space group #121, and this was used to designate the structure as $C17$ in (Ewald, 1931). However, even then it was recognized that changing $z_3$ from the value of 0.2 assigned by Wever to 0.25 would add an inversion symmetry to the crystal, change the space group to "$I4/mcm$ #140", and make the structure look very much like $Al_2Cu$, *Strukturbericht* $C16$. This has become the accepted structure, as in (Kapfenberger, 2006), and so $C17$ is an obsolete designation. We retain it here for historical interest.

**Body-centered Tetragonal primitive vectors:**

$$\mathbf{a}_1 = -\frac{1}{2}\,a\,\hat{\mathbf{x}} + \frac{1}{2}\,a\,\hat{\mathbf{y}} + \frac{1}{2}\,c\,\hat{\mathbf{z}}$$
$$\mathbf{a}_2 = \frac{1}{2}\,a\,\hat{\mathbf{x}} - \frac{1}{2}\,a\,\hat{\mathbf{y}} + \frac{1}{2}\,c\,\hat{\mathbf{z}}$$
$$\mathbf{a}_3 = \frac{1}{2}\,a\,\hat{\mathbf{x}} + \frac{1}{2}\,a\,\hat{\mathbf{y}} - \frac{1}{2}\,c\,\hat{\mathbf{z}}$$

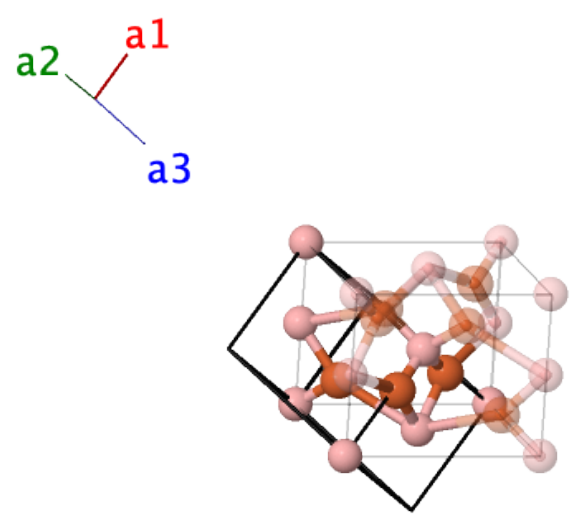

**Basis vectors:**

| | | Lattice Coordinates | | Cartesian Coordinates | Wyckoff Position | Atom Type |
|---|---|---|---|---|---|---|
| $\mathbf{B}_1$ | = | $0\,\mathbf{a}_1 + 0\,\mathbf{a}_2 + 0\,\mathbf{a}_3$ | = | $0\,\hat{\mathbf{x}} + 0\,\hat{\mathbf{y}} + 0\,\hat{\mathbf{z}}$ | $(2a)$ | B I |
| $\mathbf{B}_2$ | = | $\frac{1}{2}\,\mathbf{a}_1 + \frac{1}{2}\,\mathbf{a}_2$ | = | $\frac{1}{2}c\,\hat{\mathbf{z}}$ | (2$b$) | B II |
| $\mathbf{B}_3$ | = | $(x_3+z_3)\,\mathbf{a}_1 + (x_3+z_3)\,\mathbf{a}_2 + 2x_3\,\mathbf{a}_3$ | = | $x_3a\,\hat{\mathbf{x}} + x_3a\,\hat{\mathbf{y}} + z_3c\,\hat{\mathbf{z}}$ | (8$i$) | Fe |
| $\mathbf{B}_4$ | = | $(-x_3+z_3)\,\mathbf{a}_1 + (-x_3+z_3)\,\mathbf{a}_2 - 2x_3\,\mathbf{a}_3$ | = | $-x_3a\,\hat{\mathbf{x}} - x_3a\,\hat{\mathbf{y}} + z_3c\,\hat{\mathbf{z}}$ | (8$i$) | Fe |
| $\mathbf{B}_5$ | = | $(-x_3-z_3)\,\mathbf{a}_1 + (x_3-z_3)\,\mathbf{a}_2$ | = | $x_3a\,\hat{\mathbf{x}} - x_3a\,\hat{\mathbf{y}} - z_3c\,\hat{\mathbf{z}}$ | (8$i$) | Fe |
| $\mathbf{B}_6$ | = | $(x_3-z_3)\,\mathbf{a}_1 + (-x_3-z_3)\,\mathbf{a}_2$ | = | $-x_3a\,\hat{\mathbf{x}} + x_3a\,\hat{\mathbf{y}} - z_3c\,\hat{\mathbf{z}}$ | (8$i$) | Fe |

**Geometry files:**

- CIF: pp. 1686
- POSCAR: pp. 1687

# $K_3CrO_8$ Structure: AB3C8_tI24_121_a_bd_2i

http://aflow.org/prototype-encyclopedia/AB3C8_tI24_121_a_bd_2i

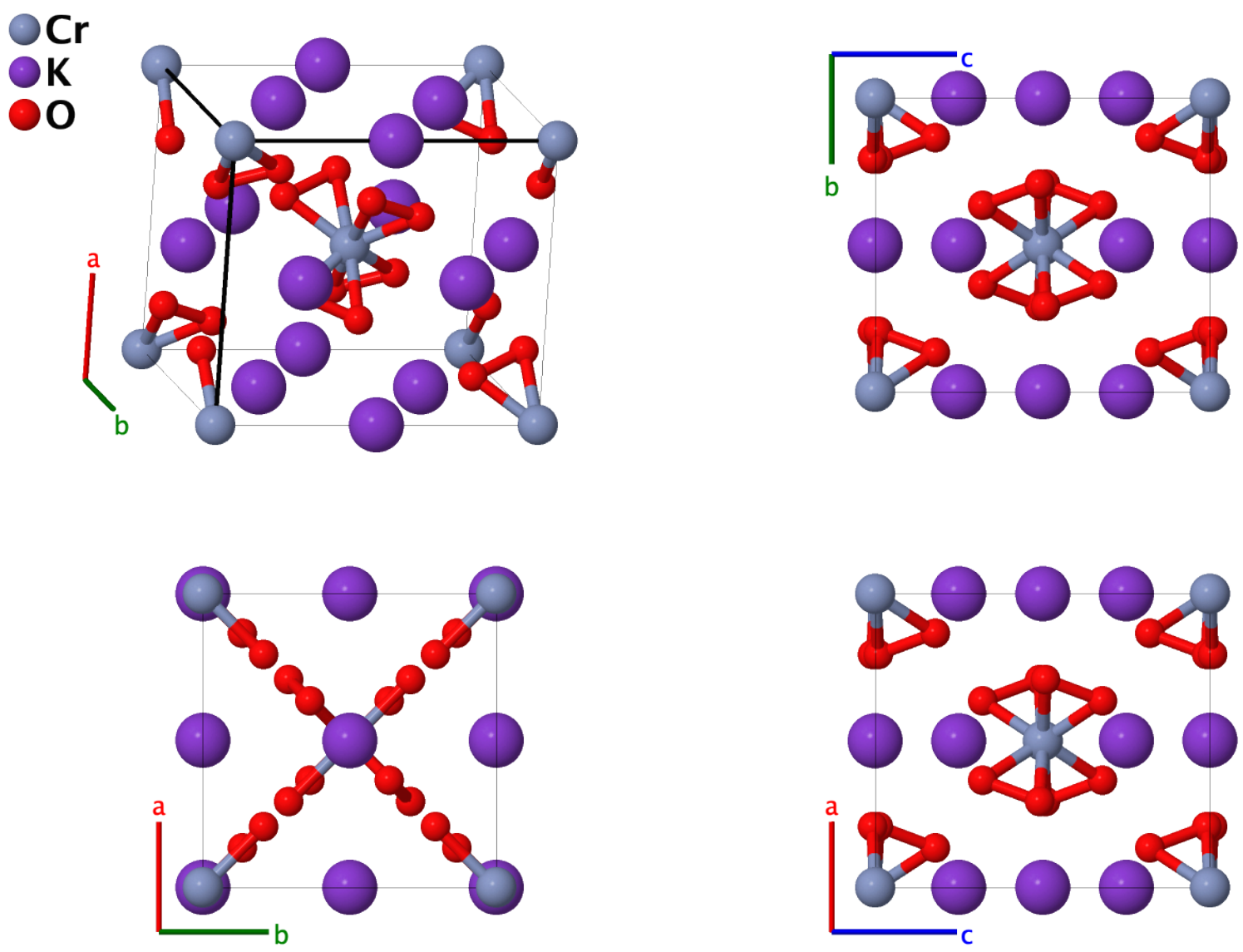

| | | |
|---|---|---|
| **Prototype** | : | $CrK_3O_8$ |
| **AFLOW prototype label** | : | AB3C8_tI24_121_a_bd_2i |
| ***Strukturbericht* designation** | : | None |
| **Pearson symbol** | : | tI24 |
| **Space group number** | : | 121 |
| **Space group symbol** | : | $I\bar{4}2m$ |
| **AFLOW prototype command** | : | `aflow --proto=AB3C8_tI24_121_a_bd_2i` <br> `--params=`$a, c/a, x_4, z_4, x_5, z_5$ |

### Other compounds with this structure

- $Zr_3GeO_8$

### Body-centered Tetragonal primitive vectors:

$$\mathbf{a}_1 = -\frac{1}{2}\,a\,\hat{\mathbf{x}} + \frac{1}{2}\,a\,\hat{\mathbf{y}} + \frac{1}{2}\,c\,\hat{\mathbf{z}}$$
$$\mathbf{a}_2 = \frac{1}{2}\,a\,\hat{\mathbf{x}} - \frac{1}{2}\,a\,\hat{\mathbf{y}} + \frac{1}{2}\,c\,\hat{\mathbf{z}}$$
$$\mathbf{a}_3 = \frac{1}{2}\,a\,\hat{\mathbf{x}} + \frac{1}{2}\,a\,\hat{\mathbf{y}} - \frac{1}{2}\,c\,\hat{\mathbf{z}}$$

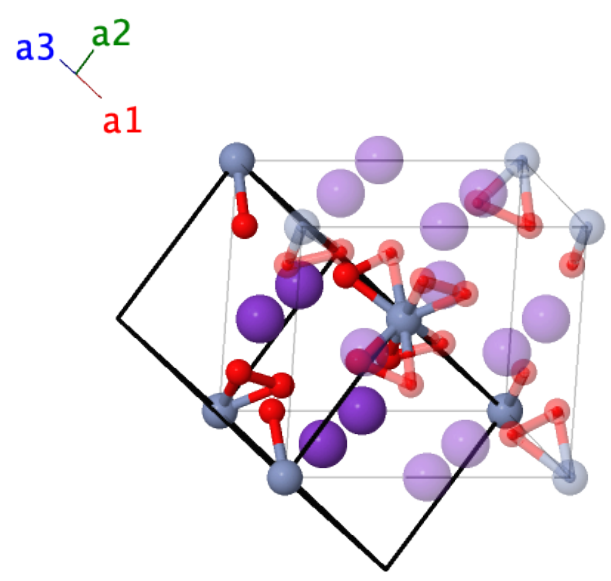

### Basis vectors:

| | | Lattice Coordinates | | Cartesian Coordinates | Wyckoff Position | Atom Type |
|---|---|---|---|---|---|---|
| $\mathbf{B}_1$ | = | $0\,\mathbf{a}_1 + 0\,\mathbf{a}_2 + 0\,\mathbf{a}_3$ | = | $0\,\hat{\mathbf{x}} + 0\,\hat{\mathbf{y}} + 0\,\hat{\mathbf{z}}$ | (2$a$) | Cr |
| $\mathbf{B}_2$ | = | $\frac{1}{2}\,\mathbf{a}_1 + \frac{1}{2}\,\mathbf{a}_2$ | = | $\frac{1}{2}c\,\hat{\mathbf{z}}$ | (2$b$) | K I |
| $\mathbf{B}_3$ | = | $\frac{3}{4}\,\mathbf{a}_1 + \frac{1}{4}\,\mathbf{a}_2 + \frac{1}{2}\,\mathbf{a}_3$ | = | $\frac{1}{2}a\,\hat{\mathbf{y}} + \frac{1}{4}c\,\hat{\mathbf{z}}$ | (4$d$) | K II |
| $\mathbf{B}_4$ | = | $\frac{1}{4}\,\mathbf{a}_1 + \frac{3}{4}\,\mathbf{a}_2 + \frac{1}{2}\,\mathbf{a}_3$ | = | $\frac{1}{2}a\,\hat{\mathbf{x}} + \frac{1}{4}c\,\hat{\mathbf{z}}$ | (4$d$) | K II |
| $\mathbf{B}_5$ | = | $(x_4 + z_4)\,\mathbf{a}_1 + (x_4 + z_4)\,\mathbf{a}_2 + 2x_4\,\mathbf{a}_3$ | = | $x_4 a\,\hat{\mathbf{x}} + x_4 a\,\hat{\mathbf{y}} + z_4 c\,\hat{\mathbf{z}}$ | (8$i$) | O I |
| $\mathbf{B}_6$ | = | $(-x_4 + z_4)\,\mathbf{a}_1 + (-x_4 + z_4)\,\mathbf{a}_2 - 2x_4\,\mathbf{a}_3$ | = | $-x_4 a\,\hat{\mathbf{x}} - x_4 a\,\hat{\mathbf{y}} + z_4 c\,\hat{\mathbf{z}}$ | (8$i$) | O I |
| $\mathbf{B}_7$ | = | $(-x_4 - z_4)\,\mathbf{a}_1 + (x_4 - z_4)\,\mathbf{a}_2$ | = | $x_4 a\,\hat{\mathbf{x}} - x_4 a\,\hat{\mathbf{y}} - z_4 c\,\hat{\mathbf{z}}$ | (8$i$) | O I |
| $\mathbf{B}_8$ | = | $(x_4 - z_4)\,\mathbf{a}_1 + (-x_4 - z_4)\,\mathbf{a}_2$ | = | $-x_4 a\,\hat{\mathbf{x}} + x_4 a\,\hat{\mathbf{y}} - z_4 c\,\hat{\mathbf{z}}$ | (8$i$) | O I |
| $\mathbf{B}_9$ | = | $(x_5 + z_5)\,\mathbf{a}_1 + (x_5 + z_5)\,\mathbf{a}_2 + 2x_5\,\mathbf{a}_3$ | = | $x_5 a\,\hat{\mathbf{x}} + x_5 a\,\hat{\mathbf{y}} + z_5 c\,\hat{\mathbf{z}}$ | (8$i$) | O II |
| $\mathbf{B}_{10}$ | = | $(-x_5 + z_5)\,\mathbf{a}_1 + (-x_5 + z_5)\,\mathbf{a}_2 - 2x_5\,\mathbf{a}_3$ | = | $-x_5 a\,\hat{\mathbf{x}} - x_5 a\,\hat{\mathbf{y}} + z_5 c\,\hat{\mathbf{z}}$ | (8$i$) | O II |
| $\mathbf{B}_{11}$ | = | $(-x_5 - z_5)\,\mathbf{a}_1 + (x_5 - z_5)\,\mathbf{a}_2$ | = | $x_5 a\,\hat{\mathbf{x}} - x_5 a\,\hat{\mathbf{y}} - z_5 c\,\hat{\mathbf{z}}$ | (8$i$) | O II |
| $\mathbf{B}_{12}$ | = | $(x_5 - z_5)\,\mathbf{a}_1 + (-x_5 - z_5)\,\mathbf{a}_2$ | = | $-x_5 a\,\hat{\mathbf{x}} + x_5 a\,\hat{\mathbf{y}} - z_5 c\,\hat{\mathbf{z}}$ | (8$i$) | O II |

**Geometry files:**

- CIF: pp. 1687
- POSCAR: pp. 1687

# $\alpha$-$V_3S$ Structure: AB3_tI32_121_g_f2i

http://aflow.org/prototype-encyclopedia/AB3_tI32_121_g_f2i

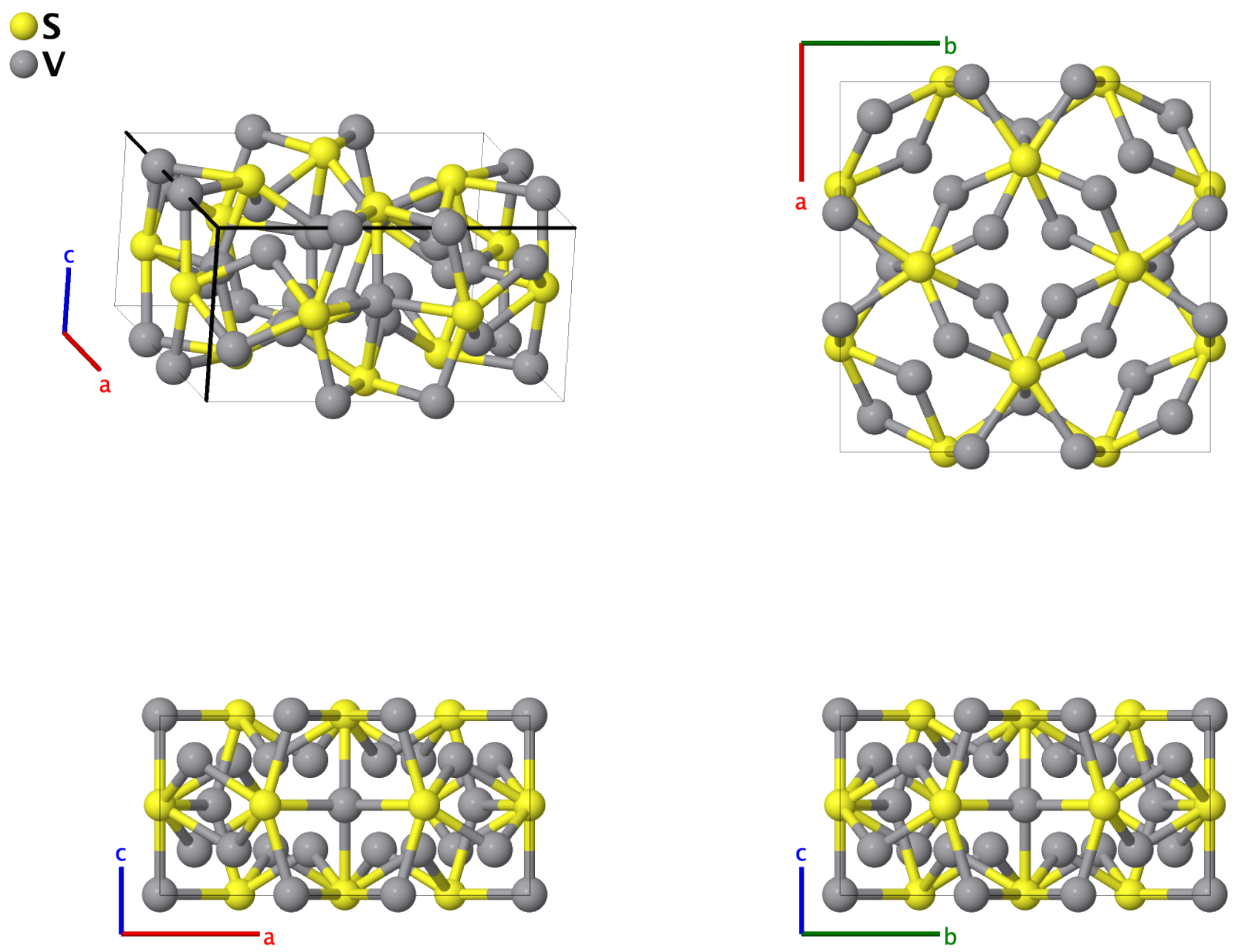

| | | |
|---|---|---|
| **Prototype** | : | $SV_3$ |
| **AFLOW prototype label** | : | AB3_tI32_121_g_f2i |
| ***Strukturbericht* designation** | : | None |
| **Pearson symbol** | : | tI32 |
| **Space group number** | : | 121 |
| **Space group symbol** | : | $I\bar{4}2m$ |
| **AFLOW prototype command** | : | `aflow --proto=AB3_tI32_121_g_f2i` <br> `--params=`$a, c/a, x_1, x_2, x_3, z_3, x_4, z_4$ |

### Other compounds with this structure

- $Zr_3Ir$

- $\alpha$-$V_3S$ is stable above 950 °C, but metastable at 25 °C, where this data was taken. Below 825 °C, the system transforms to the $\beta$-$V_3S$ structure.

### Body-centered Tetragonal primitive vectors:

$$\mathbf{a}_1 = -\frac{1}{2}\,a\,\hat{\mathbf{x}} + \frac{1}{2}\,a\,\hat{\mathbf{y}} + \frac{1}{2}\,c\,\hat{\mathbf{z}}$$
$$\mathbf{a}_2 = \frac{1}{2}\,a\,\hat{\mathbf{x}} - \frac{1}{2}\,a\,\hat{\mathbf{y}} + \frac{1}{2}\,c\,\hat{\mathbf{z}}$$
$$\mathbf{a}_3 = \frac{1}{2}\,a\,\hat{\mathbf{x}} + \frac{1}{2}\,a\,\hat{\mathbf{y}} - \frac{1}{2}\,c\,\hat{\mathbf{z}}$$

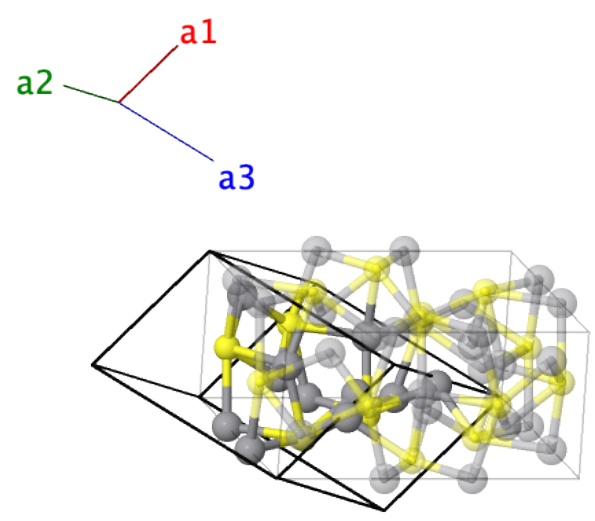

### Basis vectors:

| | | Lattice Coordinates | | Cartesian Coordinates | Wyckoff Position | Atom Type |
|---|---|---|---|---|---|---|
| $\mathbf{B}_1$ | = | $x_1\,\mathbf{a}_2 + x_1\,\mathbf{a}_3$ | = | $x_1 a\,\hat{\mathbf{x}}$ | $(8f)$ | V I |
| $\mathbf{B}_2$ | = | $-x_1\,\mathbf{a}_2 - x_1\,\mathbf{a}_3$ | = | $-x_1 a\,\hat{\mathbf{x}}$ | $(8f)$ | V I |
| $\mathbf{B}_3$ | = | $-x_1\,\mathbf{a}_1 - x_1\,\mathbf{a}_3$ | = | $-x_1 a\,\hat{\mathbf{y}}$ | $(8f)$ | V I |
| $\mathbf{B}_4$ | = | $x_1\,\mathbf{a}_1 + x_1\,\mathbf{a}_3$ | = | $x_1 a\,\hat{\mathbf{y}}$ | $(8f)$ | V I |
| $\mathbf{B}_5$ | = | $\frac{1}{2}\,\mathbf{a}_1 + \left(\frac{1}{2} + x_2\right)\,\mathbf{a}_2 + x_2\,\mathbf{a}_3$ | = | $x_2 a\,\hat{\mathbf{x}} + \frac{1}{2}c\,\hat{\mathbf{z}}$ | $(8g)$ | S |
| $\mathbf{B}_6$ | = | $\frac{1}{2}\,\mathbf{a}_1 + \left(\frac{1}{2} - x_2\right)\,\mathbf{a}_2 - x_2\,\mathbf{a}_3$ | = | $-x_2 a\,\hat{\mathbf{x}} + \frac{1}{2}c\,\hat{\mathbf{z}}$ | $(8g)$ | S |
| $\mathbf{B}_7$ | = | $\left(\frac{1}{2} - x_2\right)\,\mathbf{a}_1 + \frac{1}{2}\,\mathbf{a}_2 - x_2\,\mathbf{a}_3$ | = | $-x_2 a\,\hat{\mathbf{y}} + \frac{1}{2}c\,\hat{\mathbf{z}}$ | $(8g)$ | S |
| $\mathbf{B}_8$ | = | $\left(\frac{1}{2} + x_2\right)\,\mathbf{a}_1 + \frac{1}{2}\,\mathbf{a}_2 + x_2\,\mathbf{a}_3$ | = | $x_2 a\,\hat{\mathbf{y}} + \frac{1}{2}c\,\hat{\mathbf{z}}$ | $(8g)$ | S |
| $\mathbf{B}_9$ | = | $(x_3 + z_3)\,\mathbf{a}_1 + (x_3 + z_3)\,\mathbf{a}_2 + 2x_3\,\mathbf{a}_3$ | = | $x_3 a\,\hat{\mathbf{x}} + x_3 a\,\hat{\mathbf{y}} + z_3 c\,\hat{\mathbf{z}}$ | $(8i)$ | V II |
| $\mathbf{B}_{10}$ | = | $(-x_3 + z_3)\,\mathbf{a}_1 + (-x_3 + z_3)\,\mathbf{a}_2 - 2x_3\,\mathbf{a}_3$ | = | $-x_3 a\,\hat{\mathbf{x}} - x_3 a\,\hat{\mathbf{y}} + z_3 c\,\hat{\mathbf{z}}$ | $(8i)$ | V II |
| $\mathbf{B}_{11}$ | = | $(-x_3 - z_3)\,\mathbf{a}_1 + (x_3 - z_3)\,\mathbf{a}_2$ | = | $x_3 a\,\hat{\mathbf{x}} - x_3 a\,\hat{\mathbf{y}} - z_3 c\,\hat{\mathbf{z}}$ | $(8i)$ | V II |
| $\mathbf{B}_{12}$ | = | $(x_3 - z_3)\,\mathbf{a}_1 + (-x_3 - z_3)\,\mathbf{a}_2$ | = | $-x_3 a\,\hat{\mathbf{x}} + x_3 a\,\hat{\mathbf{y}} - z_3 c\,\hat{\mathbf{z}}$ | $(8i)$ | V II |
| $\mathbf{B}_{13}$ | = | $(x_4 + z_4)\,\mathbf{a}_1 + (x_4 + z_4)\,\mathbf{a}_2 + 2x_4\,\mathbf{a}_3$ | = | $x_4 a\,\hat{\mathbf{x}} + x_4 a\,\hat{\mathbf{y}} + z_4 c\,\hat{\mathbf{z}}$ | $(8i)$ | V III |
| $\mathbf{B}_{14}$ | = | $(-x_4 + z_4)\,\mathbf{a}_1 + (-x_4 + z_4)\,\mathbf{a}_2 - 2x_4\,\mathbf{a}_3$ | = | $-x_4 a\,\hat{\mathbf{x}} - x_4 a\,\hat{\mathbf{y}} + z_4 c\,\hat{\mathbf{z}}$ | $(8i)$ | V III |
| $\mathbf{B}_{15}$ | = | $(-x_4 - z_4)\,\mathbf{a}_1 + (x_4 - z_4)\,\mathbf{a}_2$ | = | $x_4 a\,\hat{\mathbf{x}} - x_4 a\,\hat{\mathbf{y}} - z_4 c\,\hat{\mathbf{z}}$ | $(8i)$ | V III |
| $\mathbf{B}_{16}$ | = | $(x_4 - z_4)\,\mathbf{a}_1 + (-x_4 - z_4)\,\mathbf{a}_2$ | = | $-x_4 a\,\hat{\mathbf{x}} + x_4 a\,\hat{\mathbf{y}} - z_4 c\,\hat{\mathbf{z}}$ | $(8i)$ | V III |

**Geometry files:**

- CIF: pp. 1687
- POSCAR: pp. 1688

# Mercury Cyanide [$Hg(CN)_2$, $F1_1$] Structure: A2BC2_tI40_122_e_d_e

http://aflow.org/prototype-encyclopedia/A2BC2_tI40_122_e_d_e

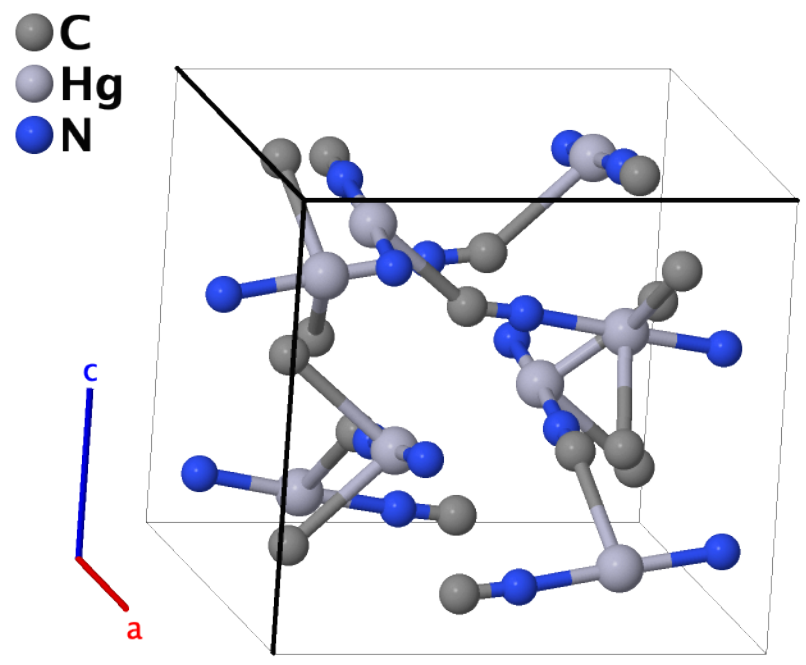

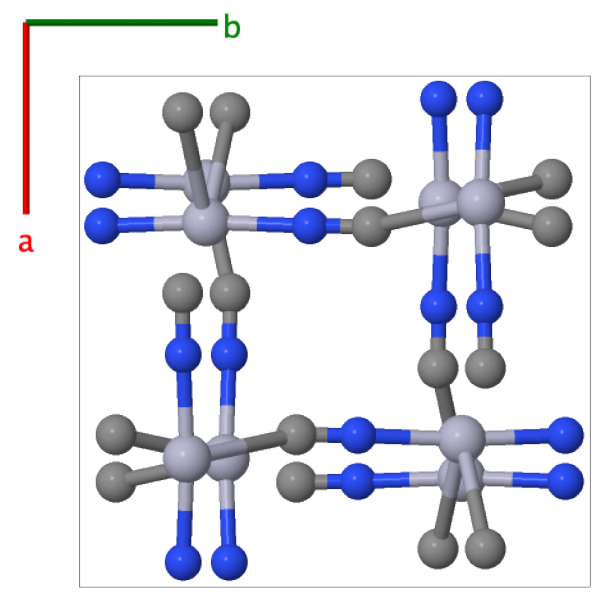

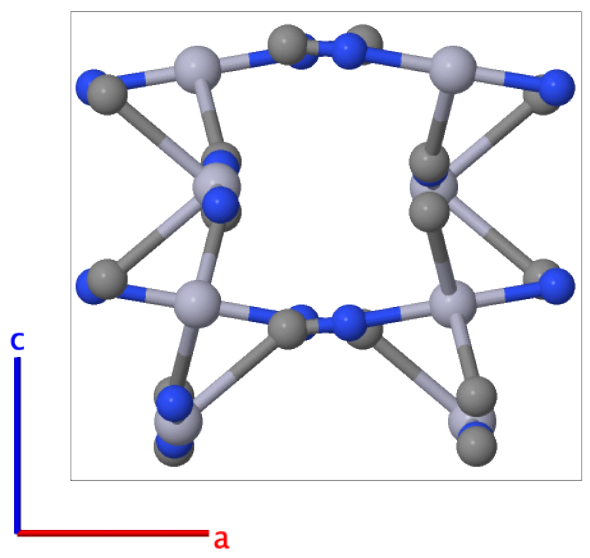

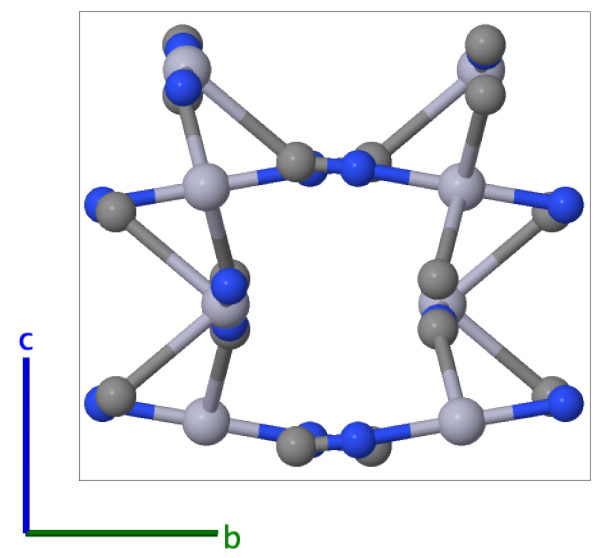

| | | |
|---|---|---|
| **Prototype** | : | $C_2HgN_2$ |
| **AFLOW prototype label** | : | A2BC2_tI40_122_e_d_e |
| ***Strukturbericht* designation** | : | $F1_1$ |
| **Pearson symbol** | : | tI40 |
| **Space group number** | : | 122 |
| **Space group symbol** | : | $I\bar{4}2d$ |
| **AFLOW prototype command** | : | `aflow --proto=A2BC2_tI40_122_e_d_e --params=`$a, c/a, x_1, x_2, y_2, z_2, x_3, y_3, z_3$ |

**Body-centered Tetragonal primitive vectors:**

$$\mathbf{a}_1 = -\frac{1}{2}\,a\,\hat{\mathbf{x}} + \frac{1}{2}\,a\,\hat{\mathbf{y}} + \frac{1}{2}\,c\,\hat{\mathbf{z}}$$
$$\mathbf{a}_2 = \frac{1}{2}\,a\,\hat{\mathbf{x}} - \frac{1}{2}\,a\,\hat{\mathbf{y}} + \frac{1}{2}\,c\,\hat{\mathbf{z}}$$
$$\mathbf{a}_3 = \frac{1}{2}\,a\,\hat{\mathbf{x}} + \frac{1}{2}\,a\,\hat{\mathbf{y}} - \frac{1}{2}\,c\,\hat{\mathbf{z}}$$

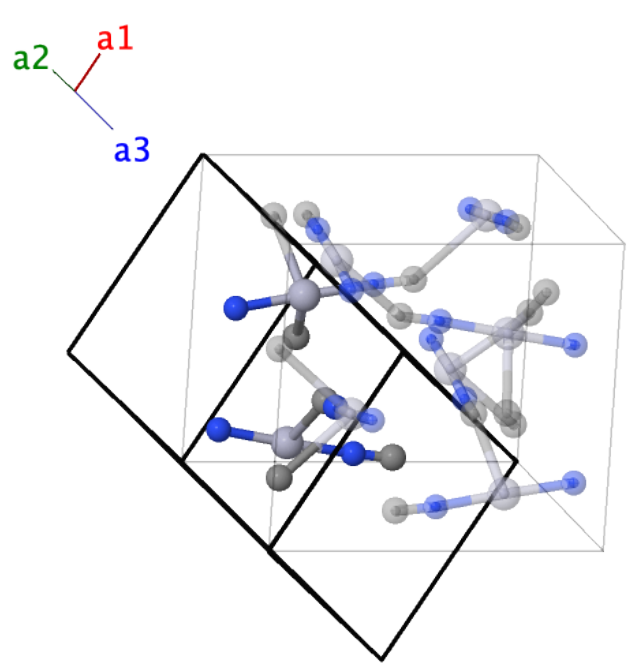

**Basis vectors:**

| | | Lattice Coordinates | | Cartesian Coordinates | Wyckoff Position | Atom Type |
|---|---|---|---|---|---|---|
| $\mathbf{B}_1$ | = | $\frac{3}{8}\,\mathbf{a}_1 + \left(\frac{1}{8} + x_1\right)\mathbf{a}_2 + \left(\frac{1}{4} + x_1\right)\mathbf{a}_3$ | = | $x_1 a\,\hat{\mathbf{x}} + \frac{1}{4} a\,\hat{\mathbf{y}} + \frac{1}{8} c\,\hat{\mathbf{z}}$ | (8$d$) | Hg |

$$\mathbf{B}_2 = \frac{7}{8}\,\mathbf{a}_1 + \left(\frac{1}{8} - x_1\right)\,\mathbf{a}_2 + \left(\frac{3}{4} - x_1\right)\,\mathbf{a}_3 = -x_1 a\,\hat{\mathbf{x}} + \frac{3}{4} a\,\hat{\mathbf{y}} + \frac{1}{8} c\,\hat{\mathbf{z}} \quad (8d) \quad \text{Hg}$$

$$\mathbf{B}_3 = \left(\frac{7}{8} - x_1\right)\,\mathbf{a}_1 + \frac{1}{8}\,\mathbf{a}_2 + \left(\frac{1}{4} - x_1\right)\,\mathbf{a}_3 = -\frac{1}{4} a\,\hat{\mathbf{x}} + \left(\frac{1}{2} - x_1\right) a\,\hat{\mathbf{y}} + \frac{3}{8} c\,\hat{\mathbf{z}} \quad (8d) \quad \text{Hg}$$

$$\mathbf{B}_4 = \left(\frac{7}{8} + x_1\right)\,\mathbf{a}_1 + \frac{5}{8}\,\mathbf{a}_2 + \left(\frac{3}{4} + x_1\right)\,\mathbf{a}_3 = \frac{1}{4} a\,\hat{\mathbf{x}} + \left(\frac{1}{2} + x_1\right) a\,\hat{\mathbf{y}} + \frac{3}{8} c\,\hat{\mathbf{z}} \quad (8d) \quad \text{Hg}$$

$$\mathbf{B}_5 = (y_2 + z_2)\,\mathbf{a}_1 + (x_2 + z_2)\,\mathbf{a}_2 + (x_2 + y_2)\,\mathbf{a}_3 = x_2 a\,\hat{\mathbf{x}} + y_2 a\,\hat{\mathbf{y}} + z_2 c\,\hat{\mathbf{z}} \quad (16e) \quad \text{C}$$

$$\mathbf{B}_6 = (-y_2 + z_2)\,\mathbf{a}_1 + (-x_2 + z_2)\,\mathbf{a}_2 + (-x_2 - y_2)\,\mathbf{a}_3 = -x_2 a\,\hat{\mathbf{x}} - y_2 a\,\hat{\mathbf{y}} + z_2 c\,\hat{\mathbf{z}} \quad (16e) \quad \text{C}$$

$$\mathbf{B}_7 = (-x_2 - z_2)\,\mathbf{a}_1 + (y_2 - z_2)\,\mathbf{a}_2 + (-x_2 + y_2)\,\mathbf{a}_3 = y_2 a\,\hat{\mathbf{x}} - x_2 a\,\hat{\mathbf{y}} - z_2 c\,\hat{\mathbf{z}} \quad (16e) \quad \text{C}$$

$$\mathbf{B}_8 = (x_2 - z_2)\,\mathbf{a}_1 + (-y_2 - z_2)\,\mathbf{a}_2 + (x_2 - y_2)\,\mathbf{a}_3 = -y_2 a\,\hat{\mathbf{x}} + x_2 a\,\hat{\mathbf{y}} - z_2 c\,\hat{\mathbf{z}} \quad (16e) \quad \text{C}$$

$$\mathbf{B}_9 = \left(\frac{3}{4} + y_2 - z_2\right)\,\mathbf{a}_1 + \left(\frac{1}{4} - x_2 - z_2\right)\,\mathbf{a}_2 + \left(\frac{1}{2} - x_2 + y_2\right)\,\mathbf{a}_3 = -x_2 a\,\hat{\mathbf{x}} + \left(\frac{1}{2} + y_2\right) a\,\hat{\mathbf{y}} + \left(\frac{1}{4} - z_2\right) c\,\hat{\mathbf{z}} \quad (16e) \quad \text{C}$$

$$\mathbf{B}_{10} = \left(\frac{3}{4} - y_2 - z_2\right)\,\mathbf{a}_1 + \left(\frac{1}{4} + x_2 - z_2\right)\,\mathbf{a}_2 + \left(\frac{1}{2} + x_2 - y_2\right)\,\mathbf{a}_3 = x_2 a\,\hat{\mathbf{x}} + \left(\frac{1}{2} - y_2\right) a\,\hat{\mathbf{y}} + \left(\frac{1}{4} - z_2\right) c\,\hat{\mathbf{z}} \quad (16e) \quad \text{C}$$

$$\mathbf{B}_{11} = \left(\frac{3}{4} - x_2 + z_2\right)\,\mathbf{a}_1 + \left(\frac{1}{4} - y_2 + z_2\right)\,\mathbf{a}_2 + \left(\frac{1}{2} - x_2 - y_2\right)\,\mathbf{a}_3 = -y_2 a\,\hat{\mathbf{x}} + \left(\frac{1}{2} - x_2\right) a\,\hat{\mathbf{y}} + \left(\frac{1}{4} + z_2\right) c\,\hat{\mathbf{z}} \quad (16e) \quad \text{C}$$

$$\mathbf{B}_{12} = \left(\frac{3}{4} + x_2 + z_2\right)\,\mathbf{a}_1 + \left(\frac{1}{4} + y_2 + z_2\right)\,\mathbf{a}_2 + \left(\frac{1}{2} + x_2 + y_2\right)\,\mathbf{a}_3 = y_2 a\,\hat{\mathbf{x}} + \left(\frac{1}{2} + x_2\right) a\,\hat{\mathbf{y}} + \left(\frac{1}{4} + z_2\right) c\,\hat{\mathbf{z}} \quad (16e) \quad \text{C}$$

$$\mathbf{B}_{13} = (y_3 + z_3)\,\mathbf{a}_1 + (x_3 + z_3)\,\mathbf{a}_2 + (x_3 + y_3)\,\mathbf{a}_3 = x_3 a\,\hat{\mathbf{x}} + y_3 a\,\hat{\mathbf{y}} + z_3 c\,\hat{\mathbf{z}} \quad (16e) \quad \text{N}$$

$$\mathbf{B}_{14} = (-y_3 + z_3)\,\mathbf{a}_1 + (-x_3 + z_3)\,\mathbf{a}_2 + (-x_3 - y_3)\,\mathbf{a}_3 = -x_3 a\,\hat{\mathbf{x}} - y_3 a\,\hat{\mathbf{y}} + z_3 c\,\hat{\mathbf{z}} \quad (16e) \quad \text{N}$$

$$\mathbf{B}_{15} = (-x_3 - z_3)\,\mathbf{a}_1 + (y_3 - z_3)\,\mathbf{a}_2 + (-x_3 + y_3)\,\mathbf{a}_3 = y_3 a\,\hat{\mathbf{x}} - x_3 a\,\hat{\mathbf{y}} - z_3 c\,\hat{\mathbf{z}} \quad (16e) \quad \text{N}$$

$$\mathbf{B}_{16} = (x_3 - z_3)\,\mathbf{a}_1 + (-y_3 - z_3)\,\mathbf{a}_2 + (x_3 - y_3)\,\mathbf{a}_3 = -y_3 a\,\hat{\mathbf{x}} + x_3 a\,\hat{\mathbf{y}} - z_3 c\,\hat{\mathbf{z}} \quad (16e) \quad \text{N}$$

$$\mathbf{B}_{17} = \left(\frac{3}{4} + y_3 - z_3\right)\,\mathbf{a}_1 + \left(\frac{1}{4} - x_3 - z_3\right)\,\mathbf{a}_2 + \left(\frac{1}{2} - x_3 + y_3\right)\,\mathbf{a}_3 = -x_3 a\,\hat{\mathbf{x}} + \left(\frac{1}{2} + y_3\right) a\,\hat{\mathbf{y}} + \left(\frac{1}{4} - z_3\right) c\,\hat{\mathbf{z}} \quad (16e) \quad \text{N}$$

$$\mathbf{B}_{18} = \left(\frac{3}{4} - y_3 - z_3\right)\,\mathbf{a}_1 + \left(\frac{1}{4} + x_3 - z_3\right)\,\mathbf{a}_2 + \left(\frac{1}{2} + x_3 - y_3\right)\,\mathbf{a}_3 = x_3 a\,\hat{\mathbf{x}} + \left(\frac{1}{2} - y_3\right) a\,\hat{\mathbf{y}} + \left(\frac{1}{4} - z_3\right) c\,\hat{\mathbf{z}} \quad (16e) \quad \text{N}$$

$$\mathbf{B}_{19} = \left(\frac{3}{4} - x_3 + z_3\right)\,\mathbf{a}_1 + \left(\frac{1}{4} - y_3 + z_3\right)\,\mathbf{a}_2 + \left(\frac{1}{2} - x_3 - y_3\right)\,\mathbf{a}_3 = -y_3 a\,\hat{\mathbf{x}} + \left(\frac{1}{2} - x_3\right) a\,\hat{\mathbf{y}} + \left(\frac{1}{4} + z_3\right) c\,\hat{\mathbf{z}} \quad (16e) \quad \text{N}$$

$$\mathbf{B}_{20} = \left(\frac{3}{4} + x_3 + z_3\right)\,\mathbf{a}_1 + \left(\frac{1}{4} + y_3 + z_3\right)\,\mathbf{a}_2 + \left(\frac{1}{2} + x_3 + y_3\right)\,\mathbf{a}_3 = y_3 a\,\hat{\mathbf{x}} + \left(\frac{1}{2} + x_3\right) a\,\hat{\mathbf{y}} + \left(\frac{1}{4} + z_3\right) c\,\hat{\mathbf{z}} \quad (16e) \quad \text{N}$$

---

**Found in:**

**Geometry files:**

- CIF: pp. 1688
- POSCAR: pp. 1688

# $KH_2PO_4$ ($H2_2$) Structure: A4BC4D_tI40_122_e_b_e_a

http://aflow.org/prototype-encyclopedia/A4BC4D_tI40_122_e_b_e_a

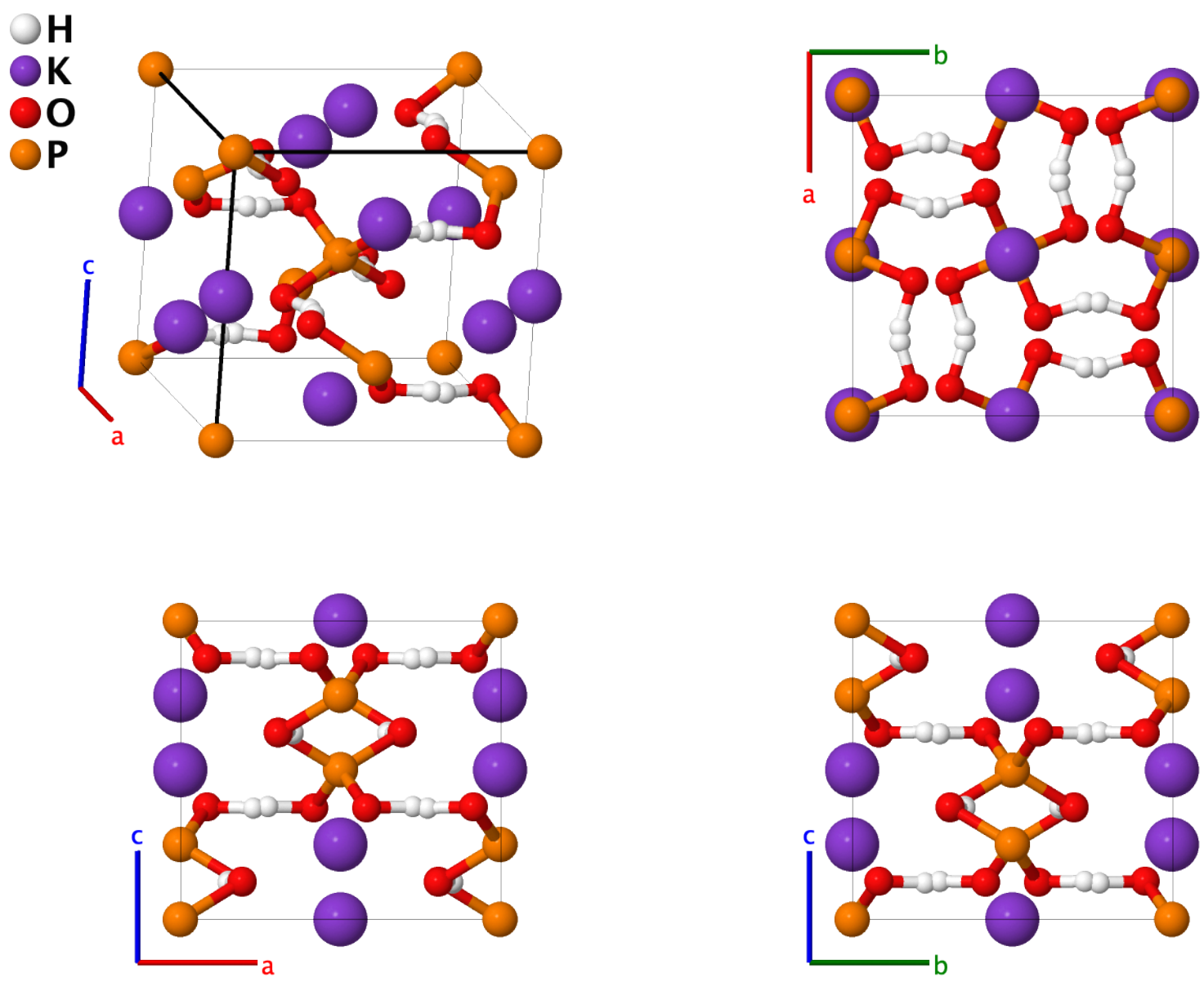

| | | |
|---|---|---|
| **Prototype** | : | $H_2KO_4P$ |
| **AFLOW prototype label** | : | A4BC4D_tI40_122_e_b_e_a |
| ***Strukturbericht* designation** | : | $H2_2$ |
| **Pearson symbol** | : | tI40 |
| **Space group number** | : | 122 |
| **Space group symbol** | : | $I\bar{4}2d$ |
| **AFLOW prototype command** | : | `aflow --proto=A4BC4D_tI40_122_e_b_e_a` <br> `--params=`$a, c/a, x_3, y_3, z_3, x_4, y_4, z_4$ |

**Other compounds with this structure**

- $RbH_2PO_4$, $KH_2AsO_4$, $RbH_2AsO_4$, $CsH_2AsO_4$, $NH_4H_2PO_4$, and $NH_4H_2AsO_4$

- The hydrogen (16*e*) sites are half-occupied. Given the closeness of pairs of hydrogen positions, presumably only one site in each pair is ever occupied.
- This partial occupancy gives a structure that differs slightly from the $H2_2$ structure described by (Ewald, 1928). There, the hydrogen atoms are at their averaged positions, Wyckoff position (8*d*).

**Body-centered Tetragonal primitive vectors:**

$$\mathbf{a}_1 = -\frac{1}{2}\,a\,\hat{\mathbf{x}} + \frac{1}{2}\,a\,\hat{\mathbf{y}} + \frac{1}{2}\,c\,\hat{\mathbf{z}}$$
$$\mathbf{a}_2 = \frac{1}{2}\,a\,\hat{\mathbf{x}} - \frac{1}{2}\,a\,\hat{\mathbf{y}} + \frac{1}{2}\,c\,\hat{\mathbf{z}}$$
$$\mathbf{a}_3 = \frac{1}{2}\,a\,\hat{\mathbf{x}} + \frac{1}{2}\,a\,\hat{\mathbf{y}} - \frac{1}{2}\,c\,\hat{\mathbf{z}}$$

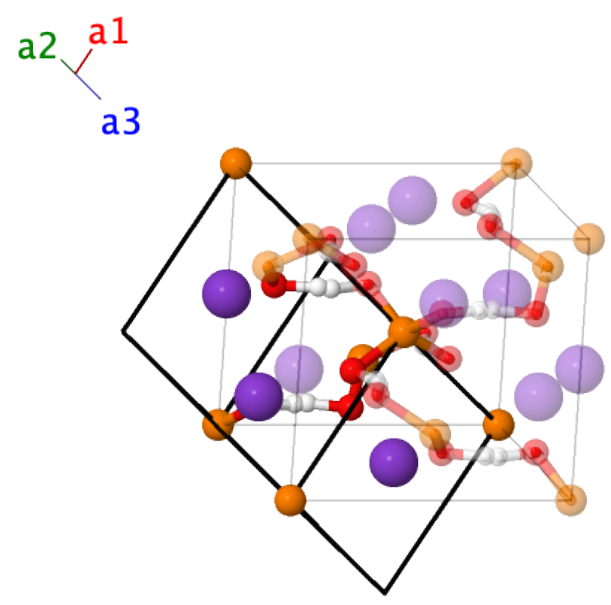

**Basis vectors:**

| | | Lattice Coordinates | | Cartesian Coordinates | Wyckoff Position | Atom Type |
|---|---|---|---|---|---|---|
| $\mathbf{B}_1$ | = | $0\,\mathbf{a}_1 + 0\,\mathbf{a}_2 + 0\,\mathbf{a}_3$ | = | $0\,\hat{\mathbf{x}} + 0\,\hat{\mathbf{y}} + 0\,\hat{\mathbf{z}}$ | (4$a$) | P |
| $\mathbf{B}_2$ | = | $\frac{3}{4}\,\mathbf{a}_1 + \frac{1}{4}\,\mathbf{a}_2 + \frac{1}{2}\,\mathbf{a}_3$ | = | $\frac{1}{2}a\,\hat{\mathbf{y}} + \frac{1}{4}c\,\hat{\mathbf{z}}$ | (4$a$) | P |
| $\mathbf{B}_3$ | = | $\frac{1}{2}\,\mathbf{a}_1 + \frac{1}{2}\,\mathbf{a}_2$ | = | $\frac{1}{2}c\,\hat{\mathbf{z}}$ | (4$b$) | K |
| $\mathbf{B}_4$ | = | $\frac{1}{4}\,\mathbf{a}_1 + \frac{3}{4}\,\mathbf{a}_2 + \frac{1}{2}\,\mathbf{a}_3$ | = | $\frac{1}{2}a\,\hat{\mathbf{x}} + \frac{1}{4}c\,\hat{\mathbf{z}}$ | (4$b$) | K |
| $\mathbf{B}_5$ | = | $(y_3 + z_3)\,\mathbf{a}_1 + (x_3 + z_3)\,\mathbf{a}_2 + (x_3 + y_3)\,\mathbf{a}_3$ | = | $x_3a\,\hat{\mathbf{x}} + y_3a\,\hat{\mathbf{y}} + z_3c\,\hat{\mathbf{z}}$ | (16$e$) | H |
| $\mathbf{B}_6$ | = | $(-y_3 + z_3)\,\mathbf{a}_1 + (-x_3 + z_3)\,\mathbf{a}_2 + (-x_3 - y_3)\,\mathbf{a}_3$ | = | $-x_3a\,\hat{\mathbf{x}} - y_3a\,\hat{\mathbf{y}} + z_3c\,\hat{\mathbf{z}}$ | (16$e$) | H |
| $\mathbf{B}_7$ | = | $(-x_3 - z_3)\,\mathbf{a}_1 + (y_3 - z_3)\,\mathbf{a}_2 + (-x_3 + y_3)\,\mathbf{a}_3$ | = | $y_3a\,\hat{\mathbf{x}} - x_3a\,\hat{\mathbf{y}} - z_3c\,\hat{\mathbf{z}}$ | (16$e$) | H |
| $\mathbf{B}_8$ | = | $(x_3 - z_3)\,\mathbf{a}_1 + (-y_3 - z_3)\,\mathbf{a}_2 + (x_3 - y_3)\,\mathbf{a}_3$ | = | $-y_3a\,\hat{\mathbf{x}} + x_3a\,\hat{\mathbf{y}} - z_3c\,\hat{\mathbf{z}}$ | (16$e$) | H |
| $\mathbf{B}_9$ | = | $\left(\frac{3}{4} + y_3 - z_3\right)\,\mathbf{a}_1 + \left(\frac{1}{4} - x_3 - z_3\right)\,\mathbf{a}_2 + \left(\frac{1}{2} - x_3 + y_3\right)\,\mathbf{a}_3$ | = | $-x_3a\,\hat{\mathbf{x}} + \left(\frac{1}{2} + y_3\right)a\,\hat{\mathbf{y}} + \left(\frac{1}{4} - z_3\right)c\,\hat{\mathbf{z}}$ | (16$e$) | H |
| $\mathbf{B}_{10}$ | = | $\left(\frac{3}{4} - y_3 - z_3\right)\,\mathbf{a}_1 + \left(\frac{1}{4} + x_3 - z_3\right)\,\mathbf{a}_2 + \left(\frac{1}{2} + x_3 - y_3\right)\,\mathbf{a}_3$ | = | $x_3a\,\hat{\mathbf{x}} + \left(\frac{1}{2} - y_3\right)a\,\hat{\mathbf{y}} + \left(\frac{1}{4} - z_3\right)c\,\hat{\mathbf{z}}$ | (16$e$) | H |
| $\mathbf{B}_{11}$ | = | $\left(\frac{3}{4} - x_3 + z_3\right)\,\mathbf{a}_1 + \left(\frac{1}{4} - y_3 + z_3\right)\,\mathbf{a}_2 + \left(\frac{1}{2} - x_3 - y_3\right)\,\mathbf{a}_3$ | = | $-y_3a\,\hat{\mathbf{x}} + \left(\frac{1}{2} - x_3\right)a\,\hat{\mathbf{y}} + \left(\frac{1}{4} + z_3\right)c\,\hat{\mathbf{z}}$ | (16$e$) | H |
| $\mathbf{B}_{12}$ | = | $\left(\frac{3}{4} + x_3 + z_3\right)\,\mathbf{a}_1 + \left(\frac{1}{4} + y_3 + z_3\right)\,\mathbf{a}_2 + \left(\frac{1}{2} + x_3 + y_3\right)\,\mathbf{a}_3$ | = | $y_3a\,\hat{\mathbf{x}} + \left(\frac{1}{2} + x_3\right)a\,\hat{\mathbf{y}} + \left(\frac{1}{4} + z_3\right)c\,\hat{\mathbf{z}}$ | (16$e$) | H |
| $\mathbf{B}_{13}$ | = | $(y_4 + z_4)\,\mathbf{a}_1 + (x_4 + z_4)\,\mathbf{a}_2 + (x_4 + y_4)\,\mathbf{a}_3$ | = | $x_4a\,\hat{\mathbf{x}} + y_4a\,\hat{\mathbf{y}} + z_4c\,\hat{\mathbf{z}}$ | (16$e$) | O |
| $\mathbf{B}_{14}$ | = | $(-y_4 + z_4)\,\mathbf{a}_1 + (-x_4 + z_4)\,\mathbf{a}_2 + (-x_4 - y_4)\,\mathbf{a}_3$ | = | $-x_4a\,\hat{\mathbf{x}} - y_4a\,\hat{\mathbf{y}} + z_4c\,\hat{\mathbf{z}}$ | (16$e$) | O |
| $\mathbf{B}_{15}$ | = | $(-x_4 - z_4)\,\mathbf{a}_1 + (y_4 - z_4)\,\mathbf{a}_2 + (-x_4 + y_4)\,\mathbf{a}_3$ | = | $y_4a\,\hat{\mathbf{x}} - x_4a\,\hat{\mathbf{y}} - z_4c\,\hat{\mathbf{z}}$ | (16$e$) | O |
| $\mathbf{B}_{16}$ | = | $(x_4 - z_4)\,\mathbf{a}_1 + (-y_4 - z_4)\,\mathbf{a}_2 + (x_4 - y_4)\,\mathbf{a}_3$ | = | $-y_4a\,\hat{\mathbf{x}} + x_4a\,\hat{\mathbf{y}} - z_4c\,\hat{\mathbf{z}}$ | (16$e$) | O |
| $\mathbf{B}_{17}$ | = | $\left(\frac{3}{4} + y_4 - z_4\right)\,\mathbf{a}_1 + \left(\frac{1}{4} - x_4 - z_4\right)\,\mathbf{a}_2 + \left(\frac{1}{2} - x_4 + y_4\right)\,\mathbf{a}_3$ | = | $-x_4a\,\hat{\mathbf{x}} + \left(\frac{1}{2} + y_4\right)a\,\hat{\mathbf{y}} + \left(\frac{1}{4} - z_4\right)c\,\hat{\mathbf{z}}$ | (16$e$) | O |
| $\mathbf{B}_{18}$ | = | $\left(\frac{3}{4} - y_4 - z_4\right)\,\mathbf{a}_1 + \left(\frac{1}{4} + x_4 - z_4\right)\,\mathbf{a}_2 + \left(\frac{1}{2} + x_4 - y_4\right)\,\mathbf{a}_3$ | = | $x_4a\,\hat{\mathbf{x}} + \left(\frac{1}{2} - y_4\right)a\,\hat{\mathbf{y}} + \left(\frac{1}{4} - z_4\right)c\,\hat{\mathbf{z}}$ | (16$e$) | O |
| $\mathbf{B}_{19}$ | = | $\left(\frac{3}{4} - x_4 + z_4\right)\,\mathbf{a}_1 + \left(\frac{1}{4} - y_4 + z_4\right)\,\mathbf{a}_2 + \left(\frac{1}{2} - x_4 - y_4\right)\,\mathbf{a}_3$ | = | $-y_4a\,\hat{\mathbf{x}} + \left(\frac{1}{2} - x_4\right)a\,\hat{\mathbf{y}} + \left(\frac{1}{4} + z_4\right)c\,\hat{\mathbf{z}}$ | (16$e$) | O |
| $\mathbf{B}_{20}$ | = | $\left(\frac{3}{4} + x_4 + z_4\right)\,\mathbf{a}_1 + \left(\frac{1}{4} + y_4 + z_4\right)\,\mathbf{a}_2 + \left(\frac{1}{2} + x_4 + y_4\right)\,\mathbf{a}_3$ | = | $y_4a\,\hat{\mathbf{x}} + \left(\frac{1}{2} + x_4\right)a\,\hat{\mathbf{y}} + \left(\frac{1}{4} + z_4\right)c\,\hat{\mathbf{z}}$ | (16$e$) | O |

---

---

**Geometry files:**

- CIF: pp. 1688
- POSCAR: pp. 1689

# $NH_4H_2PO_4$ Structure: A8BC4D_tI56_122_2e_b_e_a

http://aflow.org/prototype-encyclopedia/A8BC4D_tI56_122_2e_b_e_a

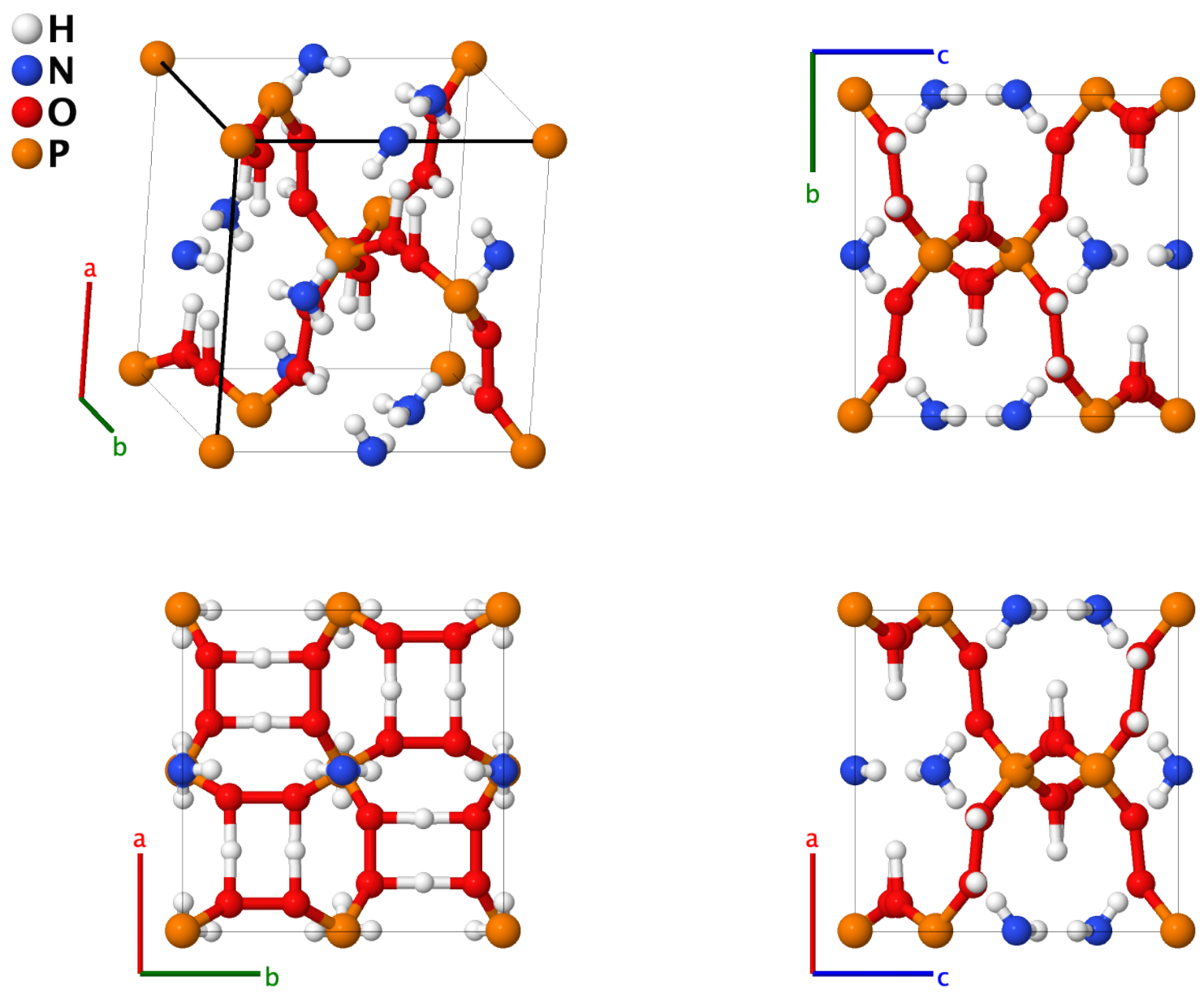

| | | |
|---|---|---|
| **Prototype** | : | $H_6NO_4P$ |
| **AFLOW prototype label** | : | A8BC4D_tI56_122_2e_b_e_a |
| ***Strukturbericht* designation** | : | None |
| **Pearson symbol** | : | tI56 |
| **Space group number** | : | 122 |
| **Space group symbol** | : | $I\bar{4}2d$ |
| **AFLOW prototype command** | : | `aflow --proto=A8BC4D_tI56_122_2e_b_e_a`<br>`--params=`$a, c/a, x_3, y_3, z_3, x_4, y_4, z_4, x_5, y_5, z_5$ |

**Other compounds with this structure**

- $NH_4H_2AsO_4$

---

- $NH_4H_2PO_4$ and $NH_4H_2AsO_4$ are usually considered to be isomorphous with the *H*2$_2$ $KH_2PO_4$ structure, but (Khan, 1973) and (Fukami, 1987) were able to locate the hydrogen atoms in the $NH_4$ ion, so we include this as a new structure.
- As in $KH_2PO_4$, the H-I site, which is associated with the $PO_4$ ion, is 50% occupied.
- Below 148 K the H-I atoms become locked in place, and $NH_4H_2PO_4$ distorts in to a orthorhombic ferroelectric phase.

**Body-centered Tetragonal primitive vectors:**

$$\mathbf{a}_1 = -\frac{1}{2}\,a\,\hat{\mathbf{x}} + \frac{1}{2}\,a\,\hat{\mathbf{y}} + \frac{1}{2}\,c\,\hat{\mathbf{z}}$$
$$\mathbf{a}_2 = \frac{1}{2}\,a\,\hat{\mathbf{x}} - \frac{1}{2}\,a\,\hat{\mathbf{y}} + \frac{1}{2}\,c\,\hat{\mathbf{z}}$$
$$\mathbf{a}_3 = \frac{1}{2}\,a\,\hat{\mathbf{x}} + \frac{1}{2}\,a\,\hat{\mathbf{y}} - \frac{1}{2}\,c\,\hat{\mathbf{z}}$$

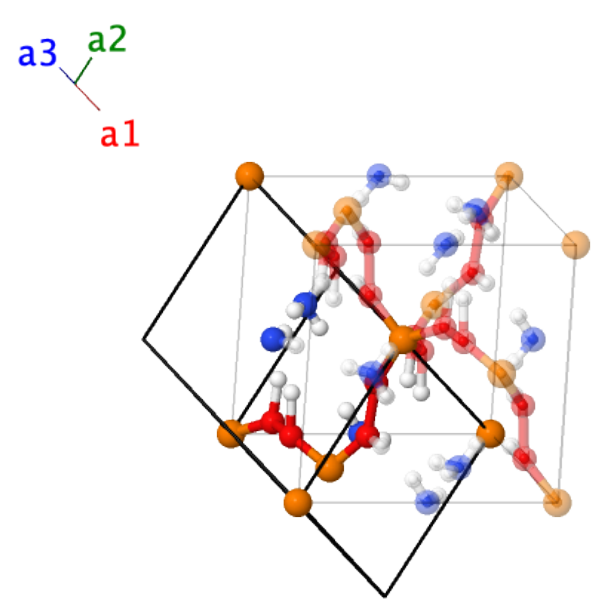

**Basis vectors:**

| | | Lattice Coordinates | | Cartesian Coordinates | Wyckoff Position | Atom Type |
|---|---|---|---|---|---|---|
| $\mathbf{B}_1$ | = | $0\,\mathbf{a}_1 + 0\,\mathbf{a}_2 + 0\,\mathbf{a}_3$ | = | $0\,\hat{\mathbf{x}} + 0\,\hat{\mathbf{y}} + 0\,\hat{\mathbf{z}}$ | (4*a*) | P |
| $\mathbf{B}_2$ | = | $\frac{3}{4}\,\mathbf{a}_1 + \frac{1}{4}\,\mathbf{a}_2 + \frac{1}{2}\,\mathbf{a}_3$ | = | $\frac{1}{2}a\,\hat{\mathbf{y}} + \frac{1}{4}c\,\hat{\mathbf{z}}$ | (4*a*) | P |
| $\mathbf{B}_3$ | = | $\frac{1}{2}\,\mathbf{a}_1 + \frac{1}{2}\,\mathbf{a}_2$ | = | $\frac{1}{2}c\,\hat{\mathbf{z}}$ | (4*b*) | N |
| $\mathbf{B}_4$ | = | $\frac{1}{4}\,\mathbf{a}_1 + \frac{3}{4}\,\mathbf{a}_2 + \frac{1}{2}\,\mathbf{a}_3$ | = | $\frac{1}{2}a\,\hat{\mathbf{x}} + \frac{1}{4}c\,\hat{\mathbf{z}}$ | (4*b*) | N |
| $\mathbf{B}_5$ | = | $(y_3 + z_3)\,\mathbf{a}_1 + (x_3 + z_3)\,\mathbf{a}_2 + (x_3 + y_3)\,\mathbf{a}_3$ | = | $x_3a\,\hat{\mathbf{x}} + y_3a\,\hat{\mathbf{y}} + z_3c\,\hat{\mathbf{z}}$ | (16*e*) | H I |
| $\mathbf{B}_6$ | = | $(-y_3 + z_3)\,\mathbf{a}_1 + (-x_3 + z_3)\,\mathbf{a}_2 + (-x_3 - y_3)\,\mathbf{a}_3$ | = | $-x_3a\,\hat{\mathbf{x}} - y_3a\,\hat{\mathbf{y}} + z_3c\,\hat{\mathbf{z}}$ | (16*e*) | H I |
| $\mathbf{B}_7$ | = | $(-x_3 - z_3)\,\mathbf{a}_1 + (y_3 - z_3)\,\mathbf{a}_2 + (-x_3 + y_3)\,\mathbf{a}_3$ | = | $y_3a\,\hat{\mathbf{x}} - x_3a\,\hat{\mathbf{y}} - z_3c\,\hat{\mathbf{z}}$ | (16*e*) | H I |
| $\mathbf{B}_8$ | = | $(x_3 - z_3)\,\mathbf{a}_1 + (-y_3 - z_3)\,\mathbf{a}_2 + (x_3 - y_3)\,\mathbf{a}_3$ | = | $-y_3a\,\hat{\mathbf{x}} + x_3a\,\hat{\mathbf{y}} - z_3c\,\hat{\mathbf{z}}$ | (16*e*) | H I |
| $\mathbf{B}_9$ | = | $\left(\frac{3}{4} + y_3 - z_3\right)\,\mathbf{a}_1 + \left(\frac{1}{4} - x_3 - z_3\right)\,\mathbf{a}_2 + \left(\frac{1}{2} - x_3 + y_3\right)\,\mathbf{a}_3$ | = | $-x_3a\,\hat{\mathbf{x}} + \left(\frac{1}{2} + y_3\right)a\,\hat{\mathbf{y}} + \left(\frac{1}{4} - z_3\right)c\,\hat{\mathbf{z}}$ | (16*e*) | H I |
| $\mathbf{B}_{10}$ | = | $\left(\frac{3}{4} - y_3 - z_3\right)\,\mathbf{a}_1 + \left(\frac{1}{4} + x_3 - z_3\right)\,\mathbf{a}_2 + \left(\frac{1}{2} + x_3 - y_3\right)\,\mathbf{a}_3$ | = | $x_3a\,\hat{\mathbf{x}} + \left(\frac{1}{2} - y_3\right)a\,\hat{\mathbf{y}} + \left(\frac{1}{4} - z_3\right)c\,\hat{\mathbf{z}}$ | (16*e*) | H I |
| $\mathbf{B}_{11}$ | = | $\left(\frac{3}{4} - x_3 + z_3\right)\,\mathbf{a}_1 + \left(\frac{1}{4} - y_3 + z_3\right)\,\mathbf{a}_2 + \left(\frac{1}{2} - x_3 - y_3\right)\,\mathbf{a}_3$ | = | $-y_3a\,\hat{\mathbf{x}} + \left(\frac{1}{2} - x_3\right)a\,\hat{\mathbf{y}} + \left(\frac{1}{4} + z_3\right)c\,\hat{\mathbf{z}}$ | (16*e*) | H I |
| $\mathbf{B}_{12}$ | = | $\left(\frac{3}{4} + x_3 + z_3\right)\,\mathbf{a}_1 + \left(\frac{1}{4} + y_3 + z_3\right)\,\mathbf{a}_2 + \left(\frac{1}{2} + x_3 + y_3\right)\,\mathbf{a}_3$ | = | $y_3a\,\hat{\mathbf{x}} + \left(\frac{1}{2} + x_3\right)a\,\hat{\mathbf{y}} + \left(\frac{1}{4} + z_3\right)c\,\hat{\mathbf{z}}$ | (16*e*) | H I |
| $\mathbf{B}_{13}$ | = | $(y_4 + z_4)\,\mathbf{a}_1 + (x_4 + z_4)\,\mathbf{a}_2 + (x_4 + y_4)\,\mathbf{a}_3$ | = | $x_4a\,\hat{\mathbf{x}} + y_4a\,\hat{\mathbf{y}} + z_4c\,\hat{\mathbf{z}}$ | (16*e*) | H II |
| $\mathbf{B}_{14}$ | = | $(-y_4 + z_4)\,\mathbf{a}_1 + (-x_4 + z_4)\,\mathbf{a}_2 + (-x_4 - y_4)\,\mathbf{a}_3$ | = | $-x_4a\,\hat{\mathbf{x}} - y_4a\,\hat{\mathbf{y}} + z_4c\,\hat{\mathbf{z}}$ | (16*e*) | H II |
| $\mathbf{B}_{15}$ | = | $(-x_4 - z_4)\,\mathbf{a}_1 + (y_4 - z_4)\,\mathbf{a}_2 + (-x_4 + y_4)\,\mathbf{a}_3$ | = | $y_4a\,\hat{\mathbf{x}} - x_4a\,\hat{\mathbf{y}} - z_4c\,\hat{\mathbf{z}}$ | (16*e*) | H II |
| $\mathbf{B}_{16}$ | = | $(x_4 - z_4)\,\mathbf{a}_1 + (-y_4 - z_4)\,\mathbf{a}_2 + (x_4 - y_4)\,\mathbf{a}_3$ | = | $-y_4a\,\hat{\mathbf{x}} + x_4a\,\hat{\mathbf{y}} - z_4c\,\hat{\mathbf{z}}$ | (16*e*) | H II |
| $\mathbf{B}_{17}$ | = | $\left(\frac{3}{4} + y_4 - z_4\right)\,\mathbf{a}_1 + \left(\frac{1}{4} - x_4 - z_4\right)\,\mathbf{a}_2 + \left(\frac{1}{2} - x_4 + y_4\right)\,\mathbf{a}_3$ | = | $-x_4a\,\hat{\mathbf{x}} + \left(\frac{1}{2} + y_4\right)a\,\hat{\mathbf{y}} + \left(\frac{1}{4} - z_4\right)c\,\hat{\mathbf{z}}$ | (16*e*) | H II |
| $\mathbf{B}_{18}$ | = | $\left(\frac{3}{4} - y_4 - z_4\right)\,\mathbf{a}_1 + \left(\frac{1}{4} + x_4 - z_4\right)\,\mathbf{a}_2 + \left(\frac{1}{2} + x_4 - y_4\right)\,\mathbf{a}_3$ | = | $x_4a\,\hat{\mathbf{x}} + \left(\frac{1}{2} - y_4\right)a\,\hat{\mathbf{y}} + \left(\frac{1}{4} - z_4\right)c\,\hat{\mathbf{z}}$ | (16*e*) | H II |
| $\mathbf{B}_{19}$ | = | $\left(\frac{3}{4} - x_4 + z_4\right)\,\mathbf{a}_1 + \left(\frac{1}{4} - y_4 + z_4\right)\,\mathbf{a}_2 + \left(\frac{1}{2} - x_4 - y_4\right)\,\mathbf{a}_3$ | = | $-y_4a\,\hat{\mathbf{x}} + \left(\frac{1}{2} - x_4\right)a\,\hat{\mathbf{y}} + \left(\frac{1}{4} + z_4\right)c\,\hat{\mathbf{z}}$ | (16*e*) | H II |
| $\mathbf{B}_{20}$ | = | $\left(\frac{3}{4} + x_4 + z_4\right)\,\mathbf{a}_1 + \left(\frac{1}{4} + y_4 + z_4\right)\,\mathbf{a}_2 + \left(\frac{1}{2} + x_4 + y_4\right)\,\mathbf{a}_3$ | = | $y_4a\,\hat{\mathbf{x}} + \left(\frac{1}{2} + x_4\right)a\,\hat{\mathbf{y}} + \left(\frac{1}{4} + z_4\right)c\,\hat{\mathbf{z}}$ | (16*e*) | H II |
| $\mathbf{B}_{21}$ | = | $(y_5 + z_5)\,\mathbf{a}_1 + (x_5 + z_5)\,\mathbf{a}_2 + (x_5 + y_5)\,\mathbf{a}_3$ | = | $x_5a\,\hat{\mathbf{x}} + y_5a\,\hat{\mathbf{y}} + z_5c\,\hat{\mathbf{z}}$ | (16*e*) | O |
| $\mathbf{B}_{22}$ | = | $(-y_5 + z_5)\,\mathbf{a}_1 + (-x_5 + z_5)\,\mathbf{a}_2 + (-x_5 - y_5)\,\mathbf{a}_3$ | = | $-x_5a\,\hat{\mathbf{x}} - y_5a\,\hat{\mathbf{y}} + z_5c\,\hat{\mathbf{z}}$ | (16*e*) | O |
| $\mathbf{B}_{23}$ | = | $(-x_5 - z_5)\,\mathbf{a}_1 + (y_5 - z_5)\,\mathbf{a}_2 + (-x_5 + y_5)\,\mathbf{a}_3$ | = | $y_5 a\,\hat{\mathbf{x}} - x_5 a\,\hat{\mathbf{y}} - z_5 c\,\hat{\mathbf{z}}$ | (16$e$) | O |
| $\mathbf{B}_{24}$ | = | $(x_5 - z_5)\,\mathbf{a}_1 + (-y_5 - z_5)\,\mathbf{a}_2 + (x_5 - y_5)\,\mathbf{a}_3$ | = | $-y_5 a\,\hat{\mathbf{x}} + x_5 a\,\hat{\mathbf{y}} - z_5 c\,\hat{\mathbf{z}}$ | (16$e$) | O |
| $\mathbf{B}_{25}$ | = | $\left(\frac{3}{4} + y_5 - z_5\right)\mathbf{a}_1 + \left(\frac{1}{4} - x_5 - z_5\right)\mathbf{a}_2 + \left(\frac{1}{2} - x_5 + y_5\right)\mathbf{a}_3$ | = | $-x_5 a\,\hat{\mathbf{x}} + \left(\frac{1}{2} + y_5\right) a\,\hat{\mathbf{y}} + \left(\frac{1}{4} - z_5\right) c\,\hat{\mathbf{z}}$ | (16$e$) | O |
| $\mathbf{B}_{26}$ | = | $\left(\frac{3}{4} - y_5 - z_5\right)\mathbf{a}_1 + \left(\frac{1}{4} + x_5 - z_5\right)\mathbf{a}_2 + \left(\frac{1}{2} + x_5 - y_5\right)\mathbf{a}_3$ | = | $x_5 a\,\hat{\mathbf{x}} + \left(\frac{1}{2} - y_5\right) a\,\hat{\mathbf{y}} + \left(\frac{1}{4} - z_5\right) c\,\hat{\mathbf{z}}$ | (16$e$) | O |
| $\mathbf{B}_{27}$ | = | $\left(\frac{3}{4} - x_5 + z_5\right)\mathbf{a}_1 + \left(\frac{1}{4} - y_5 + z_5\right)\mathbf{a}_2 + \left(\frac{1}{2} - x_5 - y_5\right)\mathbf{a}_3$ | = | $-y_5 a\,\hat{\mathbf{x}} + \left(\frac{1}{2} - x_5\right) a\,\hat{\mathbf{y}} + \left(\frac{1}{4} + z_5\right) c\,\hat{\mathbf{z}}$ | (16$e$) | O |
| $\mathbf{B}_{28}$ | = | $\left(\frac{3}{4} + x_5 + z_5\right)\mathbf{a}_1 + \left(\frac{1}{4} + y_5 + z_5\right)\mathbf{a}_2 + \left(\frac{1}{2} + x_5 + y_5\right)\mathbf{a}_3$ | = | $y_5 a\,\hat{\mathbf{x}} + \left(\frac{1}{2} + x_5\right) a\,\hat{\mathbf{y}} + \left(\frac{1}{4} + z_5\right) c\,\hat{\mathbf{z}}$ | (16$e$) | O |

**Geometry files:**

- CIF: pp. 1689
- POSCAR: pp. 1689

# $NaS_2$ Structure: AB2_tI48_122_cd_2e

http://aflow.org/prototype-encyclopedia/AB2_tI48_122_cd_2e

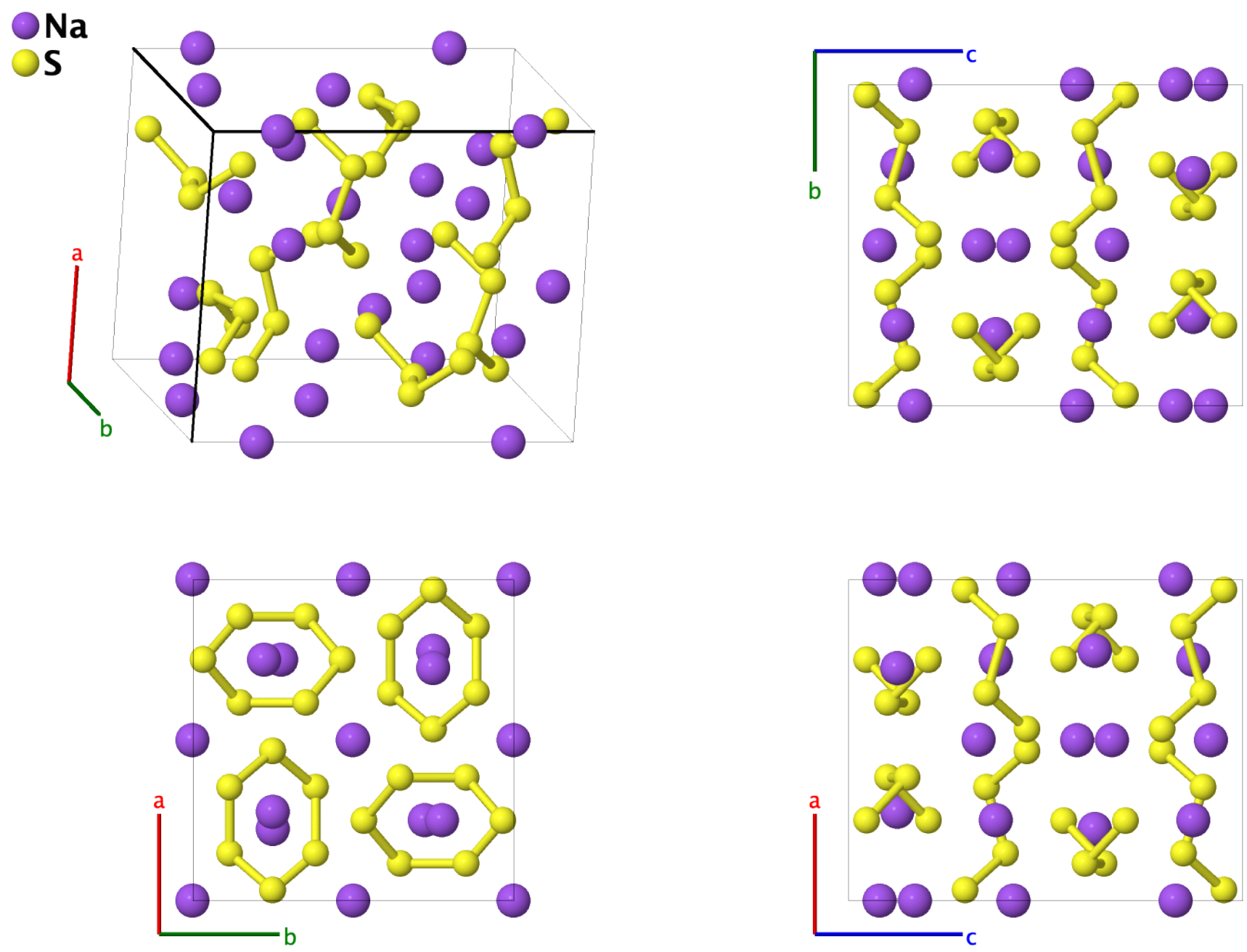

**Prototype** : $NaS_2$

**AFLOW prototype label** : AB2_tI48_122_cd_2e

***Strukturbericht* designation** : None

**Pearson symbol** : tI48

**Space group number** : 122

**Space group symbol** : $I\bar{4}2d$

**AFLOW prototype command** : `aflow --proto=AB2_tI48_122_cd_2e`
`--params=`$a, c/a, z_1, x_2, x_3, y_3, z_3, x_4, y_4, z_4$

**Body-centered Tetragonal primitive vectors:**

$$\mathbf{a}_1 = -\frac{1}{2}\,a\,\hat{\mathbf{x}} + \frac{1}{2}\,a\,\hat{\mathbf{y}} + \frac{1}{2}\,c\,\hat{\mathbf{z}}$$
$$\mathbf{a}_2 = \frac{1}{2}\,a\,\hat{\mathbf{x}} - \frac{1}{2}\,a\,\hat{\mathbf{y}} + \frac{1}{2}\,c\,\hat{\mathbf{z}}$$
$$\mathbf{a}_3 = \frac{1}{2}\,a\,\hat{\mathbf{x}} + \frac{1}{2}\,a\,\hat{\mathbf{y}} - \frac{1}{2}\,c\,\hat{\mathbf{z}}$$

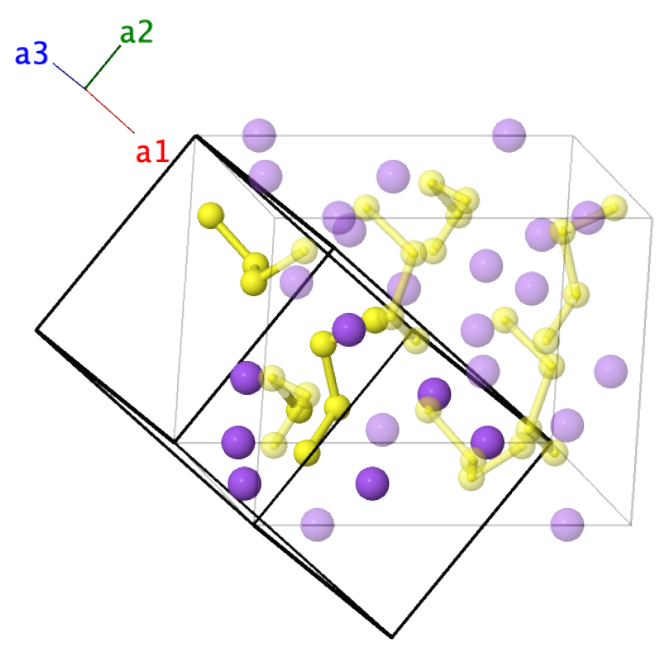

**Basis vectors:**

| | | Lattice Coordinates | | Cartesian Coordinates | Wyckoff Position | Atom Type |
|---|---|---|---|---|---|---|
| $\mathbf{B}_1$ | = | $z_1\,\mathbf{a}_1 + z_1\,\mathbf{a}_2$ | = | $z_1\,c\,\hat{\mathbf{z}}$ | (8$c$) | Na I |
| $\mathbf{B}_2$ | = | $-z_1\,\mathbf{a}_1 - z_1\,\mathbf{a}_2$ | = | $-z_1\,c\,\hat{\mathbf{z}}$ | (8$c$) | Na I |
| $\mathbf{B}_3$ | = | $\left(\frac{3}{4} - z_1\right)\mathbf{a}_1 + \left(\frac{1}{4} - z_1\right)\mathbf{a}_2 + \frac{1}{2}\,\mathbf{a}_3$ | = | $\frac{1}{2}a\,\hat{\mathbf{y}} + \left(\frac{1}{4} - z_1\right)c\,\hat{\mathbf{z}}$ | (8$c$) | Na I |
| $\mathbf{B}_4$ | = | $\left(\frac{3}{4} + z_1\right)\mathbf{a}_1 + \left(\frac{1}{4} + z_1\right)\mathbf{a}_2 + \frac{1}{2}\,\mathbf{a}_3$ | = | $\frac{1}{2}a\,\hat{\mathbf{y}} + \left(\frac{1}{4} + z_1\right)c\,\hat{\mathbf{z}}$ | (8$c$) | Na I |
| $\mathbf{B}_5$ | = | $\frac{3}{8}\,\mathbf{a}_1 + \left(\frac{1}{8}+x_2\right)\,\mathbf{a}_2 + \left(\frac{1}{4}+x_2\right)\,\mathbf{a}_3$ | = | $x_2 a\,\hat{\mathbf{x}} + \frac{1}{4} a\,\hat{\mathbf{y}} + \frac{1}{8} c\,\hat{\mathbf{z}}$ | (8*d*) | Na II |
| $\mathbf{B}_6$ | = | $\frac{7}{8}\,\mathbf{a}_1 + \left(\frac{1}{8}-x_2\right)\,\mathbf{a}_2 + \left(\frac{3}{4}-x_2\right)\,\mathbf{a}_3$ | = | $-x_2 a\,\hat{\mathbf{x}} + \frac{3}{4} a\,\hat{\mathbf{y}} + \frac{1}{8} c\,\hat{\mathbf{z}}$ | (8*d*) | Na II |
| $\mathbf{B}_7$ | = | $\left(\frac{7}{8}-x_2\right)\,\mathbf{a}_1 + \frac{1}{8}\,\mathbf{a}_2 + \left(\frac{1}{4}-x_2\right)\,\mathbf{a}_3$ | = | $-\frac{1}{4} a\,\hat{\mathbf{x}} + \left(\frac{1}{2}-x_2\right) a\,\hat{\mathbf{y}} + \frac{3}{8} c\,\hat{\mathbf{z}}$ | (8*d*) | Na II |
| $\mathbf{B}_8$ | = | $\left(\frac{7}{8}+x_2\right)\,\mathbf{a}_1 + \frac{5}{8}\,\mathbf{a}_2 + \left(\frac{3}{4}+x_2\right)\,\mathbf{a}_3$ | = | $\frac{1}{4} a\,\hat{\mathbf{x}} + \left(\frac{1}{2}+x_2\right) a\,\hat{\mathbf{y}} + \frac{3}{8} c\,\hat{\mathbf{z}}$ | (8*d*) | Na II |
| $\mathbf{B}_9$ | = | $(y_3+z_3)\,\mathbf{a}_1 + (x_3+z_3)\,\mathbf{a}_2 + (x_3+y_3)\,\mathbf{a}_3$ | = | $x_3 a\,\hat{\mathbf{x}} + y_3 a\,\hat{\mathbf{y}} + z_3 c\,\hat{\mathbf{z}}$ | (16*e*) | S I |
| $\mathbf{B}_{10}$ | = | $(-y_3+z_3)\,\mathbf{a}_1 + (-x_3+z_3)\,\mathbf{a}_2 + (-x_3-y_3)\,\mathbf{a}_3$ | = | $-x_3 a\,\hat{\mathbf{x}} - y_3 a\,\hat{\mathbf{y}} + z_3 c\,\hat{\mathbf{z}}$ | (16*e*) | S I |
| $\mathbf{B}_{11}$ | = | $(-x_3-z_3)\,\mathbf{a}_1 + (y_3-z_3)\,\mathbf{a}_2 + (-x_3+y_3)\,\mathbf{a}_3$ | = | $y_3 a\,\hat{\mathbf{x}} - x_3 a\,\hat{\mathbf{y}} - z_3 c\,\hat{\mathbf{z}}$ | (16*e*) | S I |
| $\mathbf{B}_{12}$ | = | $(x_3-z_3)\,\mathbf{a}_1 + (-y_3-z_3)\,\mathbf{a}_2 + (x_3-y_3)\,\mathbf{a}_3$ | = | $-y_3 a\,\hat{\mathbf{x}} + x_3 a\,\hat{\mathbf{y}} - z_3 c\,\hat{\mathbf{z}}$ | (16*e*) | S I |
| $\mathbf{B}_{13}$ | = | $\left(\frac{3}{4}+y_3-z_3\right)\,\mathbf{a}_1 + \left(\frac{1}{4}-x_3-z_3\right)\,\mathbf{a}_2 + \left(\frac{1}{2}-x_3+y_3\right)\,\mathbf{a}_3$ | = | $-x_3 a\,\hat{\mathbf{x}} + \left(\frac{1}{2}+y_3\right) a\,\hat{\mathbf{y}} + \left(\frac{1}{4}-z_3\right) c\,\hat{\mathbf{z}}$ | (16*e*) | S I |
| $\mathbf{B}_{14}$ | = | $\left(\frac{3}{4}-y_3-z_3\right)\,\mathbf{a}_1 + \left(\frac{1}{4}+x_3-z_3\right)\,\mathbf{a}_2 + \left(\frac{1}{2}+x_3-y_3\right)\,\mathbf{a}_3$ | = | $x_3 a\,\hat{\mathbf{x}} + \left(\frac{1}{2}-y_3\right) a\,\hat{\mathbf{y}} + \left(\frac{1}{4}-z_3\right) c\,\hat{\mathbf{z}}$ | (16*e*) | S I |
| $\mathbf{B}_{15}$ | = | $\left(\frac{3}{4}-x_3+z_3\right)\,\mathbf{a}_1 + \left(\frac{1}{4}-y_3+z_3\right)\,\mathbf{a}_2 + \left(\frac{1}{2}-x_3-y_3\right)\,\mathbf{a}_3$ | = | $-y_3 a\,\hat{\mathbf{x}} + \left(\frac{1}{2}-x_3\right) a\,\hat{\mathbf{y}} + \left(\frac{1}{4}+z_3\right) c\,\hat{\mathbf{z}}$ | (16*e*) | S I |
| $\mathbf{B}_{16}$ | = | $\left(\frac{3}{4}+x_3+z_3\right)\,\mathbf{a}_1 + \left(\frac{1}{4}+y_3+z_3\right)\,\mathbf{a}_2 + \left(\frac{1}{2}+x_3+y_3\right)\,\mathbf{a}_3$ | = | $y_3 a\,\hat{\mathbf{x}} + \left(\frac{1}{2}+x_3\right) a\,\hat{\mathbf{y}} + \left(\frac{1}{4}+z_3\right) c\,\hat{\mathbf{z}}$ | (16*e*) | S I |
| $\mathbf{B}_{17}$ | = | $(y_4+z_4)\,\mathbf{a}_1 + (x_4+z_4)\,\mathbf{a}_2 + (x_4+y_4)\,\mathbf{a}_3$ | = | $x_4 a\,\hat{\mathbf{x}} + y_4 a\,\hat{\mathbf{y}} + z_4 c\,\hat{\mathbf{z}}$ | (16*e*) | S II |
| $\mathbf{B}_{18}$ | = | $(-y_4+z_4)\,\mathbf{a}_1 + (-x_4+z_4)\,\mathbf{a}_2 + (-x_4-y_4)\,\mathbf{a}_3$ | = | $-x_4 a\,\hat{\mathbf{x}} - y_4 a\,\hat{\mathbf{y}} + z_4 c\,\hat{\mathbf{z}}$ | (16*e*) | S II |
| $\mathbf{B}_{19}$ | = | $(-x_4-z_4)\,\mathbf{a}_1 + (y_4-z_4)\,\mathbf{a}_2 + (-x_4+y_4)\,\mathbf{a}_3$ | = | $y_4 a\,\hat{\mathbf{x}} - x_4 a\,\hat{\mathbf{y}} - z_4 c\,\hat{\mathbf{z}}$ | (16*e*) | S II |
| $\mathbf{B}_{20}$ | = | $(x_4-z_4)\,\mathbf{a}_1 + (-y_4-z_4)\,\mathbf{a}_2 + (x_4-y_4)\,\mathbf{a}_3$ | = | $-y_4 a\,\hat{\mathbf{x}} + x_4 a\,\hat{\mathbf{y}} - z_4 c\,\hat{\mathbf{z}}$ | (16*e*) | S II |
| $\mathbf{B}_{21}$ | = | $\left(\frac{3}{4}+y_4-z_4\right)\,\mathbf{a}_1 + \left(\frac{1}{4}-x_4-z_4\right)\,\mathbf{a}_2 + \left(\frac{1}{2}-x_4+y_4\right)\,\mathbf{a}_3$ | = | $-x_4 a\,\hat{\mathbf{x}} + \left(\frac{1}{2}+y_4\right) a\,\hat{\mathbf{y}} + \left(\frac{1}{4}-z_4\right) c\,\hat{\mathbf{z}}$ | (16*e*) | S II |
| $\mathbf{B}_{22}$ | = | $\left(\frac{3}{4}-y_4-z_4\right)\,\mathbf{a}_1 + \left(\frac{1}{4}+x_4-z_4\right)\,\mathbf{a}_2 + \left(\frac{1}{2}+x_4-y_4\right)\,\mathbf{a}_3$ | = | $x_4 a\,\hat{\mathbf{x}} + \left(\frac{1}{2}-y_4\right) a\,\hat{\mathbf{y}} + \left(\frac{1}{4}-z_4\right) c\,\hat{\mathbf{z}}$ | (16*e*) | S II |
| $\mathbf{B}_{23}$ | = | $\left(\frac{3}{4}-x_4+z_4\right)\,\mathbf{a}_1 + \left(\frac{1}{4}-y_4+z_4\right)\,\mathbf{a}_2 + \left(\frac{1}{2}-x_4-y_4\right)\,\mathbf{a}_3$ | = | $-y_4 a\,\hat{\mathbf{x}} + \left(\frac{1}{2}-x_4\right) a\,\hat{\mathbf{y}} + \left(\frac{1}{4}+z_4\right) c\,\hat{\mathbf{z}}$ | (16*e*) | S II |
| $\mathbf{B}_{24}$ | = | $\left(\frac{3}{4}+x_4+z_4\right)\,\mathbf{a}_1 + \left(\frac{1}{4}+y_4+z_4\right)\,\mathbf{a}_2 + \left(\frac{1}{2}+x_4+y_4\right)\,\mathbf{a}_3$ | = | $y_4 a\,\hat{\mathbf{x}} + \left(\frac{1}{2}+x_4\right) a\,\hat{\mathbf{y}} + \left(\frac{1}{4}+z_4\right) c\,\hat{\mathbf{z}}$ | (16*e*) | S II |

**Geometry files:**
- CIF: pp. 1689
- POSCAR: pp. 1690

# $NH_4HgCl_3$ ($E2_5$) Structure: A3BC_tP5_123_cg_a_d

http://aflow.org/prototype-encyclopedia/A3BC_tP5_123_cg_a_d

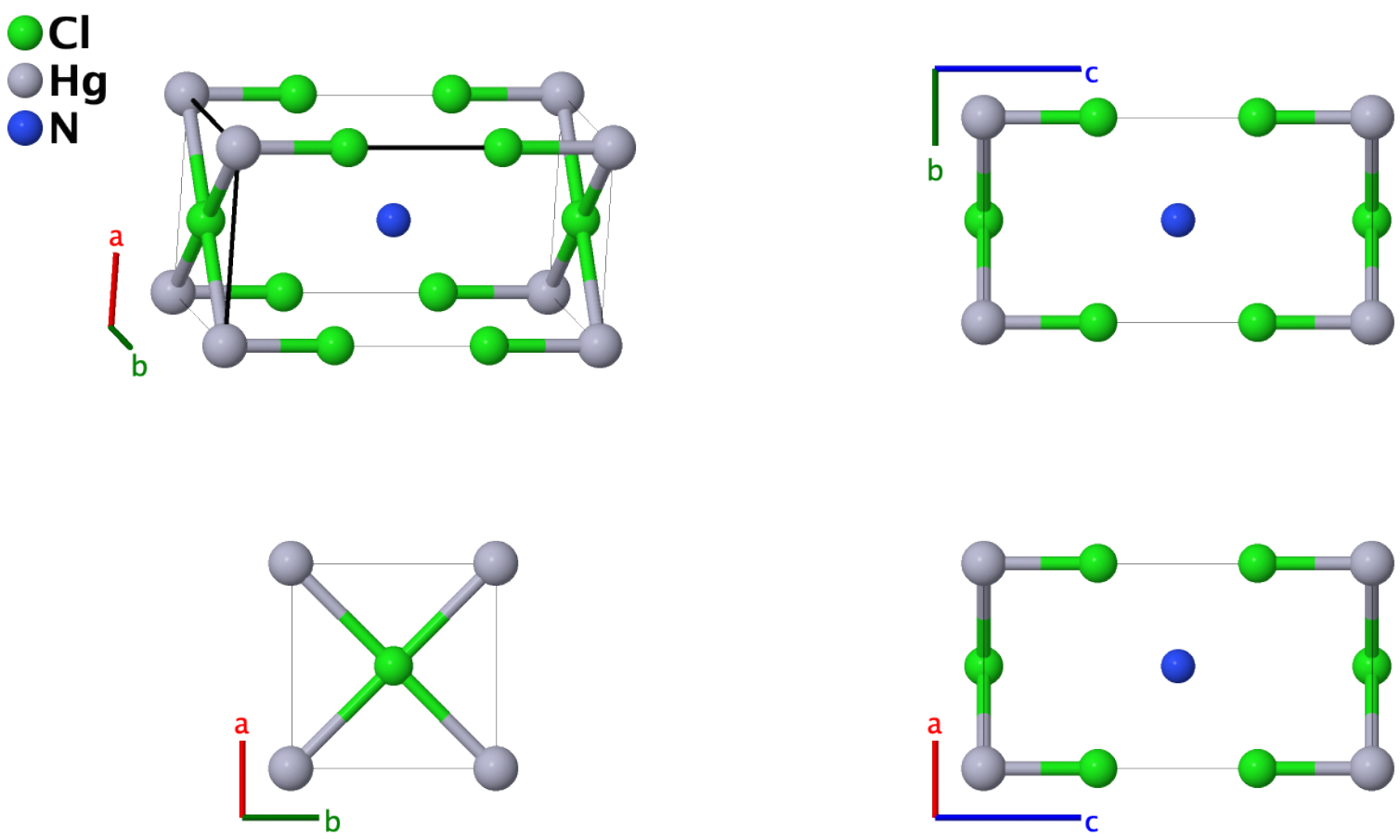

| | | |
|---|---|---|
| **Prototype** | : | $Cl_3Hg(NH_4)$ |
| **AFLOW prototype label** | : | A3BC_tP5_123_cg_a_d |
| ***Strukturbericht* designation** | : | $E2_5$ |
| **Pearson symbol** | : | tP5 |
| **Space group number** | : | 123 |
| **Space group symbol** | : | $P4/mmm$ |
| **AFLOW prototype command** | : | `aflow --proto=A3BC_tP5_123_cg_a_d --params=`$a, c/a, z_4$ |

- The positions of the hydrogen atoms in the $NH_4$ ion are not given, so we only provide the nitrogen positions (labeled as $NH_4$). It is likely that the hydrogen atoms are freely rotating around the nitrogen, as any reasonable fixed positions would destroy both the inversion symmetry and the four-fold rotation axis exhibited by space group $P4/mmm$.
- (Harmsen, 1939) and (Herrmann, 1943) give multiple possible space groups for this structure. We have chosen the highest symmetry representation, space group $P4/mmm$ #123.

**Simple Tetragonal primitive vectors:**

$$\mathbf{a}_1 = a\,\hat{\mathbf{x}}$$
$$\mathbf{a}_2 = a\,\hat{\mathbf{y}}$$
$$\mathbf{a}_3 = c\,\hat{\mathbf{z}}$$

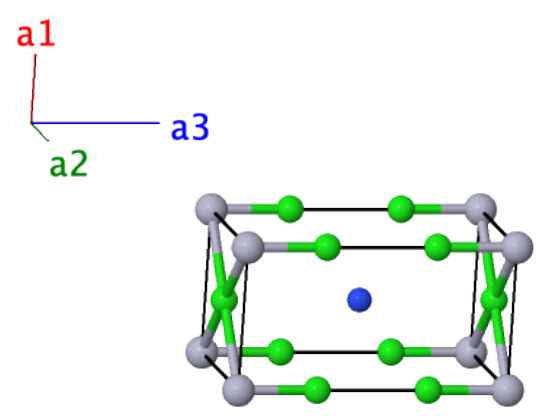

**Basis vectors:**

| | | Lattice Coordinates | | Cartesian Coordinates | Wyckoff Position | Atom Type |
|---|---|---|---|---|---|---|
| $\mathbf{B}_1$ | = | $0\,\mathbf{a}_1 + 0\,\mathbf{a}_2 + 0\,\mathbf{a}_3$ | = | $0\,\hat{\mathbf{x}} + 0\,\hat{\mathbf{y}} + 0\,\hat{\mathbf{z}}$ | $(1a)$ | Hg |
| $\mathbf{B}_2$ | = | $\frac{1}{2}\,\mathbf{a}_1 + \frac{1}{2}\,\mathbf{a}_2$ | = | $\frac{1}{2}a\,\hat{\mathbf{x}} + \frac{1}{2}a\,\hat{\mathbf{y}}$ | $(1c)$ | Cl I |
| $\mathbf{B}_3$ | = | $\frac{1}{2}\,\mathbf{a}_1 + \frac{1}{2}\,\mathbf{a}_2 + \frac{1}{2}\,\mathbf{a}_3$ | = | $\frac{1}{2}a\,\hat{\mathbf{x}} + \frac{1}{2}a\,\hat{\mathbf{y}} + \frac{1}{2}c\,\hat{\mathbf{z}}$ | $(1d)$ | $NH_4$ |
| $\mathbf{B}_4$ | = | $z_4\,\mathbf{a}_3$ | = | $z_4 c\,\hat{\mathbf{z}}$ | $(2g)$ | Cl II |
| $\mathbf{B}_5$ | = | $-z_4\,\mathbf{a}_3$ | = | $-z_4 c\,\hat{\mathbf{z}}$ | (2$g$) | Cl II |

**Geometry files:**

- CIF: pp. 1690
- POSCAR: pp. 1690

# $K_2PtCl_4$ ($H1_5$) Structure: A4B2C_tP7_123_j_e_a

http://aflow.org/prototype-encyclopedia/A4B2C_tP7_123_j_e_a

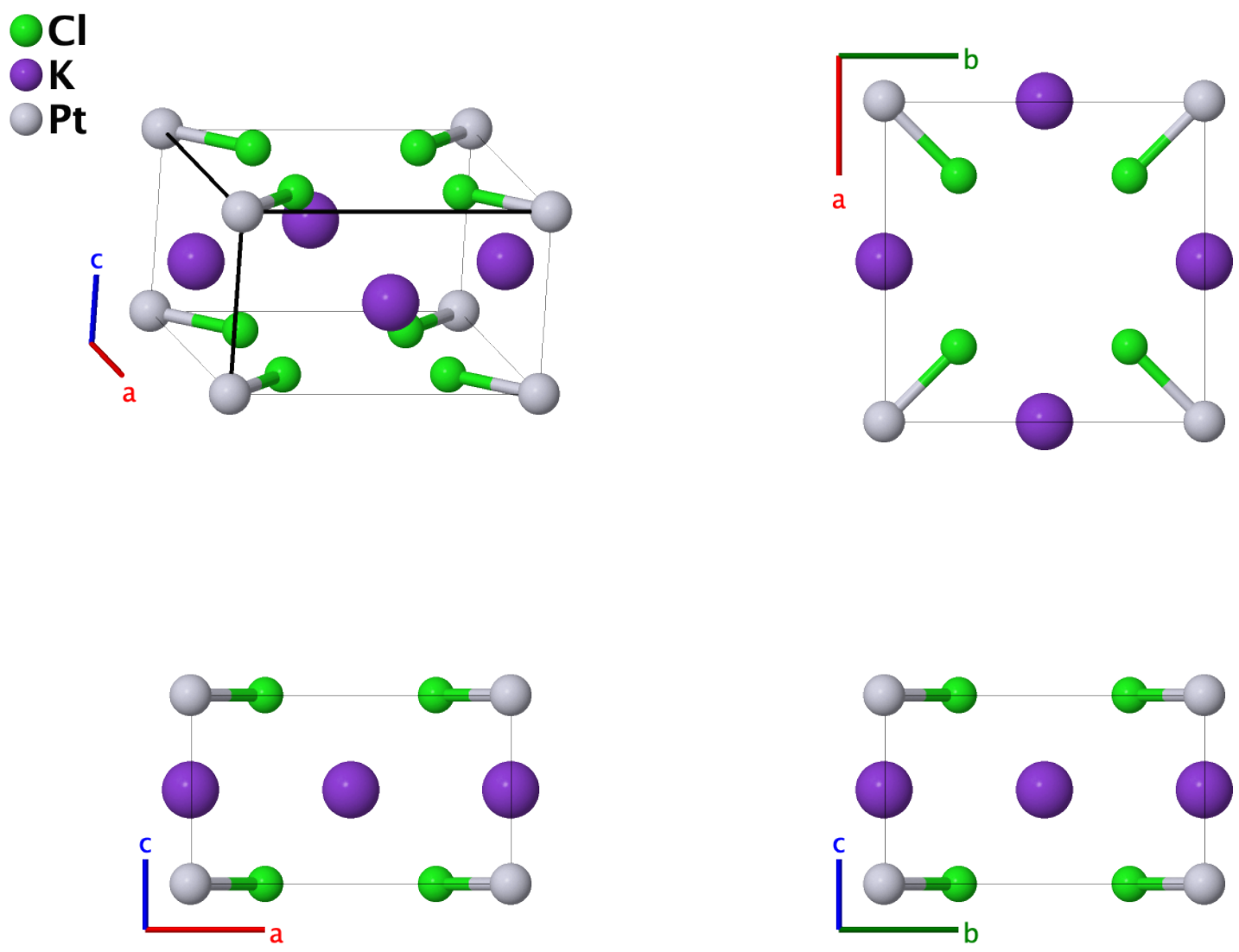

| | | |
|---|---|---|
| **Prototype** | : | $Cl_4K_2Pt$ |
| **AFLOW prototype label** | : | A4B2C_tP7_123_j_e_a |
| ***Strukturbericht* designation** | : | $H1_5$ |
| **Pearson symbol** | : | tP7 |
| **Space group number** | : | 123 |
| **Space group symbol** | : | $P4/mmm$ |
| **AFLOW prototype command** | : | `aflow --proto=A4B2C_tP7_123_j_e_a`<br>`--params=`$a, c/a, x_3$ |

**Other compounds with this structure**

- $K_2PdCl_4$

**Simple Tetragonal primitive vectors:**

$$\mathbf{a}_1 = a\,\hat{\mathbf{x}}$$
$$\mathbf{a}_2 = a\,\hat{\mathbf{y}}$$
$$\mathbf{a}_3 = c\,\hat{\mathbf{z}}$$

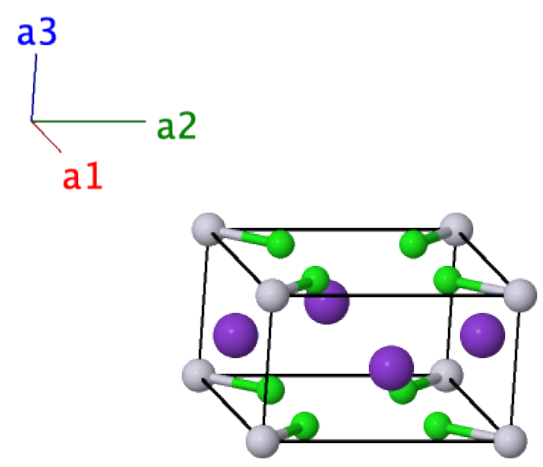

**Basis vectors:**

| | | Lattice Coordinates | | Cartesian Coordinates | Wyckoff Position | Atom Type |
|---|---|---|---|---|---|---|
| $\mathbf{B}_1$ | = | $0\,\mathbf{a}_1 + 0\,\mathbf{a}_2 + 0\,\mathbf{a}_3$ | = | $0\,\hat{\mathbf{x}} + 0\,\hat{\mathbf{y}} + 0\,\hat{\mathbf{z}}$ | (1$a$) | Pt |
| $\mathbf{B}_2$ | = | $\frac{1}{2}\,\mathbf{a}_2 + \frac{1}{2}\,\mathbf{a}_3$ | = | $\frac{1}{2}a\,\hat{\mathbf{y}} + \frac{1}{2}c\,\hat{\mathbf{z}}$ | (2$e$) | K |
| $\mathbf{B}_3$ | = | $\frac{1}{2}\,\mathbf{a}_1 + \frac{1}{2}\,\mathbf{a}_3$ | = | $\frac{1}{2}a\,\hat{\mathbf{x}} + \frac{1}{2}c\,\hat{\mathbf{z}}$ | (2$e$) | K |
| $\mathbf{B}_4$ | = | $x_3\,\mathbf{a}_1 + x_3\,\mathbf{a}_2$ | = | $x_3a\,\hat{\mathbf{x}} + x_3a\,\hat{\mathbf{y}}$ | (4$j$) | Cl |
| $\mathbf{B}_5$ | = | $-x_3\,\mathbf{a}_1 - x_3\,\mathbf{a}_2$ | = | $-x_3a\,\hat{\mathbf{x}} - x_3a\,\hat{\mathbf{y}}$ | (4$j$) | Cl |
| $\mathbf{B}_6$ | = | $-x_3\,\mathbf{a}_1 + x_3\,\mathbf{a}_2$ | = | $-x_3a\,\hat{\mathbf{x}} + x_3a\,\hat{\mathbf{y}}$ | (4$j$) | Cl |
| $\mathbf{B}_7$ | = | $x_3\,\mathbf{a}_1 - x_3\,\mathbf{a}_2$ | = | $x_3a\,\hat{\mathbf{x}} - x_3a\,\hat{\mathbf{y}}$ | (4$j$) | Cl |

**Geometry files:**

- CIF: pp. 1690
- POSCAR: pp. 1691

# $E6_1$ ($Sr(OH)_2(H_2O)_8$) (*Obsolete*) Structure: A8B2C_tP11_123_r_f_a

http://aflow.org/prototype-encyclopedia/A8B2C_tP11_123_r_f_a

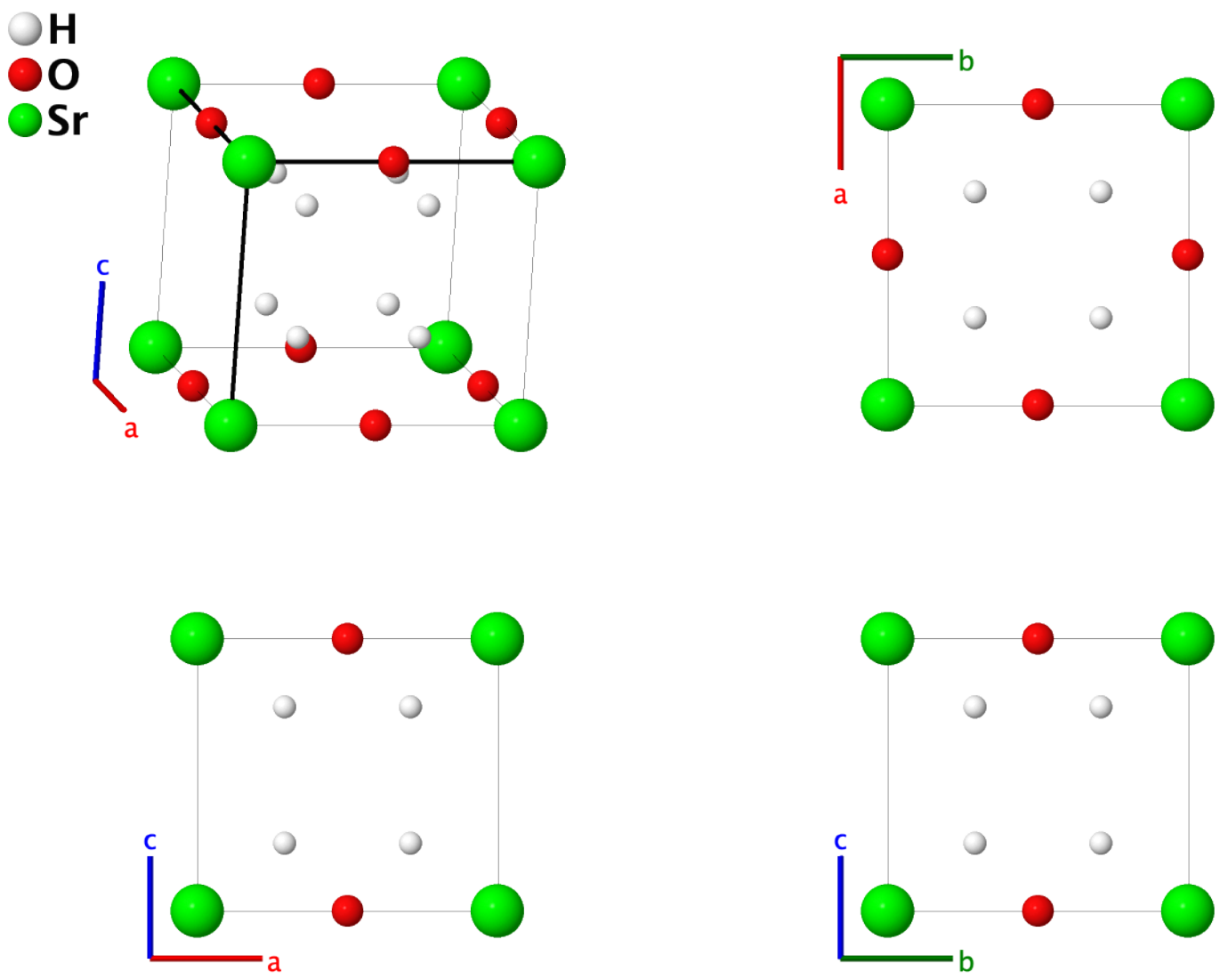

| | | |
|---|---|---|
| **Prototype** | : | $Sr(OH)_2(H_2O)_8$ |
| **AFLOW prototype label** | : | A8B2C_tP11_123_r_f_a |
| ***Strukturbericht* designation** | : | $E6_1$ |
| **Pearson symbol** | : | tP11 |
| **Space group number** | : | 123 |
| **Space group symbol** | : | $P4/mmm$ |
| **AFLOW prototype command** | : | `aflow --proto=A8B2C_tP11_123_r_f_a`<br>`--params=`$a, c/a, x_3, z_3$ |

- This structure was designated $E6_1$ by (Hermann, 1937). It has been superseded by the structure determined by (Ricci, 2005). We present it for historical interest.

**Simple Tetragonal primitive vectors:**

$$\mathbf{a}_1 = a\,\hat{\mathbf{x}}$$
$$\mathbf{a}_2 = a\,\hat{\mathbf{y}}$$
$$\mathbf{a}_3 = c\,\hat{\mathbf{z}}$$

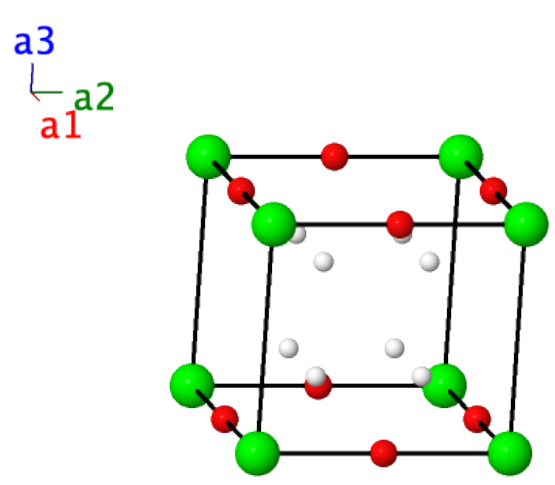

**Basis vectors:**

| | | Lattice Coordinates | | Cartesian Coordinates | Wyckoff Position | Atom Type |
|---|---|---|---|---|---|---|
| $\mathbf{B}_1$ | = | $0\,\mathbf{a}_1 + 0\,\mathbf{a}_2 + 0\,\mathbf{a}_3$ | = | $0\,\hat{\mathbf{x}} + 0\,\hat{\mathbf{y}} + 0\,\hat{\mathbf{z}}$ | $(1a)$ | Sr |
| $\mathbf{B}_2$ | = | $\frac{1}{2}\,\mathbf{a}_2$ | = | $\frac{1}{2}a\,\hat{\mathbf{y}}$ | $(2f)$ | OH |
| $\mathbf{B}_3$ | = | $\frac{1}{2}\,\mathbf{a}_1$ | = | $\frac{1}{2}a\,\hat{\mathbf{x}}$ | (2$f$) | OH |
| $\mathbf{B}_4$ | = | $x_3\,\mathbf{a}_1 + x_3\,\mathbf{a}_2 + z_3\,\mathbf{a}_3$ | = | $x_3a\,\hat{\mathbf{x}} + x_3a\,\hat{\mathbf{y}} + z_3c\,\hat{\mathbf{z}}$ | (8$r$) | $H_2O$ |
| $\mathbf{B}_5$ | = | $-x_3\,\mathbf{a}_1 - x_3\,\mathbf{a}_2 + z_3\,\mathbf{a}_3$ | = | $-x_3a\,\hat{\mathbf{x}} - x_3a\,\hat{\mathbf{y}} + z_3c\,\hat{\mathbf{z}}$ | (8$r$) | $H_2O$ |
| $\mathbf{B}_6$ | = | $-x_3\,\mathbf{a}_1 + x_3\,\mathbf{a}_2 + z_3\,\mathbf{a}_3$ | = | $-x_3a\,\hat{\mathbf{x}} + x_3a\,\hat{\mathbf{y}} + z_3c\,\hat{\mathbf{z}}$ | (8$r$) | $H_2O$ |
| $\mathbf{B}_7$ | = | $x_3\,\mathbf{a}_1 - x_3\,\mathbf{a}_2 + z_3\,\mathbf{a}_3$ | = | $x_3a\,\hat{\mathbf{x}} - x_3a\,\hat{\mathbf{y}} + z_3c\,\hat{\mathbf{z}}$ | (8$r$) | $H_2O$ |
| $\mathbf{B}_8$ | = | $-x_3\,\mathbf{a}_1 + x_3\,\mathbf{a}_2 - z_3\,\mathbf{a}_3$ | = | $-x_3a\,\hat{\mathbf{x}} + x_3a\,\hat{\mathbf{y}} - z_3c\,\hat{\mathbf{z}}$ | (8$r$) | $H_2O$ |
| $\mathbf{B}_9$ | = | $x_3\,\mathbf{a}_1 - x_3\,\mathbf{a}_2 - z_3\,\mathbf{a}_3$ | = | $x_3a\,\hat{\mathbf{x}} - x_3a\,\hat{\mathbf{y}} - z_3c\,\hat{\mathbf{z}}$ | (8$r$) | $H_2O$ |
| $\mathbf{B}_{10}$ | = | $x_3\,\mathbf{a}_1 + x_3\,\mathbf{a}_2 - z_3\,\mathbf{a}_3$ | = | $x_3a\,\hat{\mathbf{x}} + x_3a\,\hat{\mathbf{y}} - z_3c\,\hat{\mathbf{z}}$ | (8$r$) | $H_2O$ |
| $\mathbf{B}_{11}$ | = | $-x_3\,\mathbf{a}_1 - x_3\,\mathbf{a}_2 - z_3\,\mathbf{a}_3$ | = | $-x_3a\,\hat{\mathbf{x}} - x_3a\,\hat{\mathbf{y}} - z_3c\,\hat{\mathbf{z}}$ | (8$r$) | $H_2O$ |

**Geometry files:**

- CIF: pp. 1691
- POSCAR: pp. 1691

# $E6_2$ [$SrO_2(H_2O)_8$] (*possibly obsolete*) Structure: A8B2C_tP11_123_r_h_a

http://aflow.org/prototype-encyclopedia/A8B2C_tP11_123_r_h_a

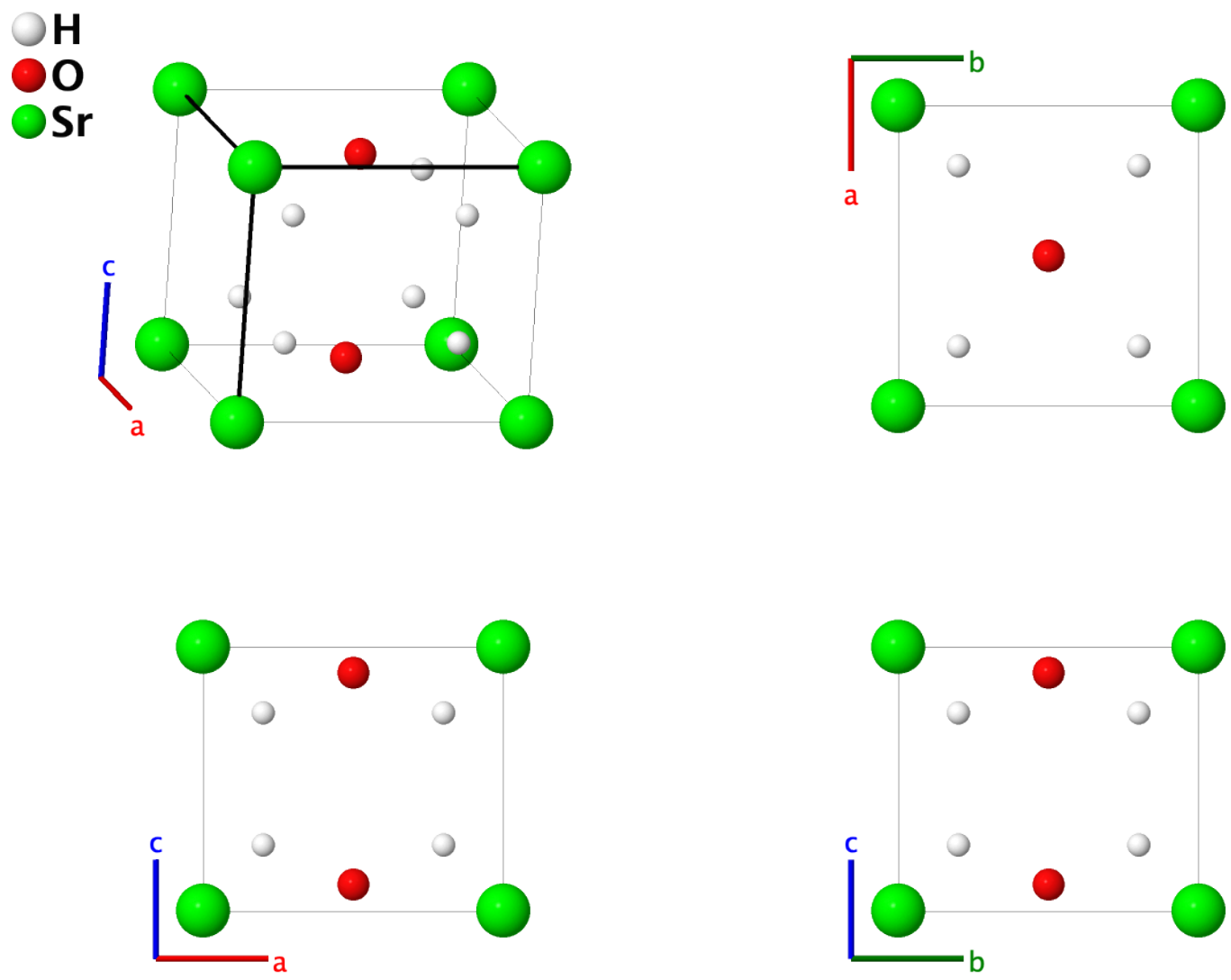

| | | |
|---|---|---|
| **Prototype** | : | $(H_2O)_8SrO_2$ |
| **AFLOW prototype label** | : | A8B2C_tP11_123_r_h_a |
| ***Strukturbericht* designation** | : | $E6_2$ |
| **Pearson symbol** | : | tP11 |
| **Space group number** | : | 123 |
| **Space group symbol** | : | $P4/mmm$ |
| **AFLOW prototype command** | : | `aflow --proto=A8B2C_tP11_123_r_h_a` `--params=`$a, c/a, z_2, x_3, z_3$ |

**Other compounds with this structure**

- $CaO_2(H_2O)_8$ and $BrO_2(H_2O)_8$

- This structure was designated $E6_2$ by (Hermann, 1937). (Shineman, 1951) suggest that this should be replaced by their $CaO_2(H_2O)_8$ structure.
- The positions of the hydrogen atoms in the water molecule were not determined, so we only provide the positions of the oyxgen atoms (labeled as $H_2O$).

**Simple Tetragonal primitive vectors:**

$$\mathbf{a}_1 = a\,\hat{\mathbf{x}}$$
$$\mathbf{a}_2 = a\,\hat{\mathbf{y}}$$
$$\mathbf{a}_3 = c\,\hat{\mathbf{z}}$$

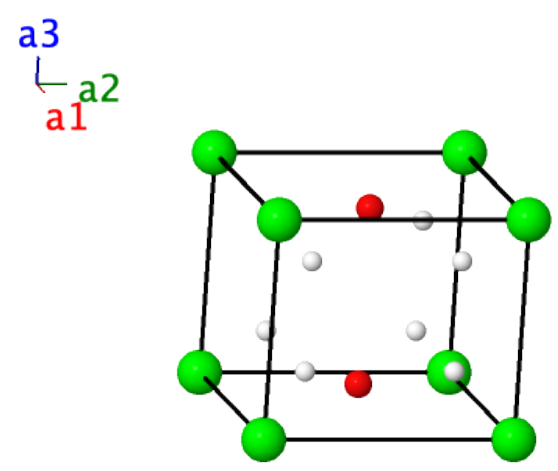

**Basis vectors:**

| | | Lattice Coordinates | | Cartesian Coordinates | Wyckoff Position | Atom Type |
|---|---|---|---|---|---|---|
| $\mathbf{B}_1$ | = | $0\,\mathbf{a}_1 + 0\,\mathbf{a}_2 + 0\,\mathbf{a}_3$ | = | $0\,\hat{\mathbf{x}} + 0\,\hat{\mathbf{y}} + 0\,\hat{\mathbf{z}}$ | $(1a)$ | Sr |
| $\mathbf{B}_2$ | = | $\frac{1}{2}\,\mathbf{a}_1 + \frac{1}{2}\,\mathbf{a}_2 + z_2\,\mathbf{a}_3$ | = | $\frac{1}{2}a\,\hat{\mathbf{x}} + \frac{1}{2}a\,\hat{\mathbf{y}} + z_2c\,\hat{\mathbf{z}}$ | $(2h)$ | O |
| $\mathbf{B}_3$ | = | $\frac{1}{2}\,\mathbf{a}_1 + \frac{1}{2}\,\mathbf{a}_2 - z_2\,\mathbf{a}_3$ | = | $\frac{1}{2}a\,\hat{\mathbf{x}} + \frac{1}{2}a\,\hat{\mathbf{y}} - z_2c\,\hat{\mathbf{z}}$ | $(2h)$ | O |
| $\mathbf{B}_4$ | = | $x_3\,\mathbf{a}_1 + x_3\,\mathbf{a}_2 + z_3\,\mathbf{a}_3$ | = | $x_3a\,\hat{\mathbf{x}} + x_3a\,\hat{\mathbf{y}} + z_3c\,\hat{\mathbf{z}}$ | $(8r)$ | $H_2O$ |
| $\mathbf{B}_5$ | = | $-x_3\,\mathbf{a}_1 - x_3\,\mathbf{a}_2 + z_3\,\mathbf{a}_3$ | = | $-x_3a\,\hat{\mathbf{x}} - x_3a\,\hat{\mathbf{y}} + z_3c\,\hat{\mathbf{z}}$ | $(8r)$ | $H_2O$ |
| $\mathbf{B}_6$ | = | $-x_3\,\mathbf{a}_1 + x_3\,\mathbf{a}_2 + z_3\,\mathbf{a}_3$ | = | $-x_3a\,\hat{\mathbf{x}} + x_3a\,\hat{\mathbf{y}} + z_3c\,\hat{\mathbf{z}}$ | $(8r)$ | $H_2O$ |
| $\mathbf{B}_7$ | = | $x_3\,\mathbf{a}_1 - x_3\,\mathbf{a}_2 + z_3\,\mathbf{a}_3$ | = | $x_3a\,\hat{\mathbf{x}} - x_3a\,\hat{\mathbf{y}} + z_3c\,\hat{\mathbf{z}}$ | $(8r)$ | $H_2O$ |
| $\mathbf{B}_8$ | = | $-x_3\,\mathbf{a}_1 + x_3\,\mathbf{a}_2 - z_3\,\mathbf{a}_3$ | = | $-x_3a\,\hat{\mathbf{x}} + x_3a\,\hat{\mathbf{y}} - z_3c\,\hat{\mathbf{z}}$ | $(8r)$ | $H_2O$ |
| $\mathbf{B}_9$ | = | $x_3\,\mathbf{a}_1 - x_3\,\mathbf{a}_2 - z_3\,\mathbf{a}_3$ | = | $x_3a\,\hat{\mathbf{x}} - x_3a\,\hat{\mathbf{y}} - z_3c\,\hat{\mathbf{z}}$ | $(8r)$ | $H_2O$ |
| $\mathbf{B}_{10}$ | = | $x_3\,\mathbf{a}_1 + x_3\,\mathbf{a}_2 - z_3\,\mathbf{a}_3$ | = | $x_3a\,\hat{\mathbf{x}} + x_3a\,\hat{\mathbf{y}} - z_3c\,\hat{\mathbf{z}}$ | $(8r)$ | $H_2O$ |
| $\mathbf{B}_{11}$ | = | $-x_3\,\mathbf{a}_1 - x_3\,\mathbf{a}_2 - z_3\,\mathbf{a}_3$ | = | $-x_3a\,\hat{\mathbf{x}} - x_3a\,\hat{\mathbf{y}} - z_3c\,\hat{\mathbf{z}}$ | $(8r)$ | $H_2O$ |

**Found in:**

**Geometry files:**

- CIF: pp. 1691
- POSCAR: pp. 1692

# $TlAlF_4$ ($H0_8$) Structure: AB4C_tP6_123_d_eh_a

http://aflow.org/prototype-encyclopedia/AB4C_tP6_123_d_eh_a

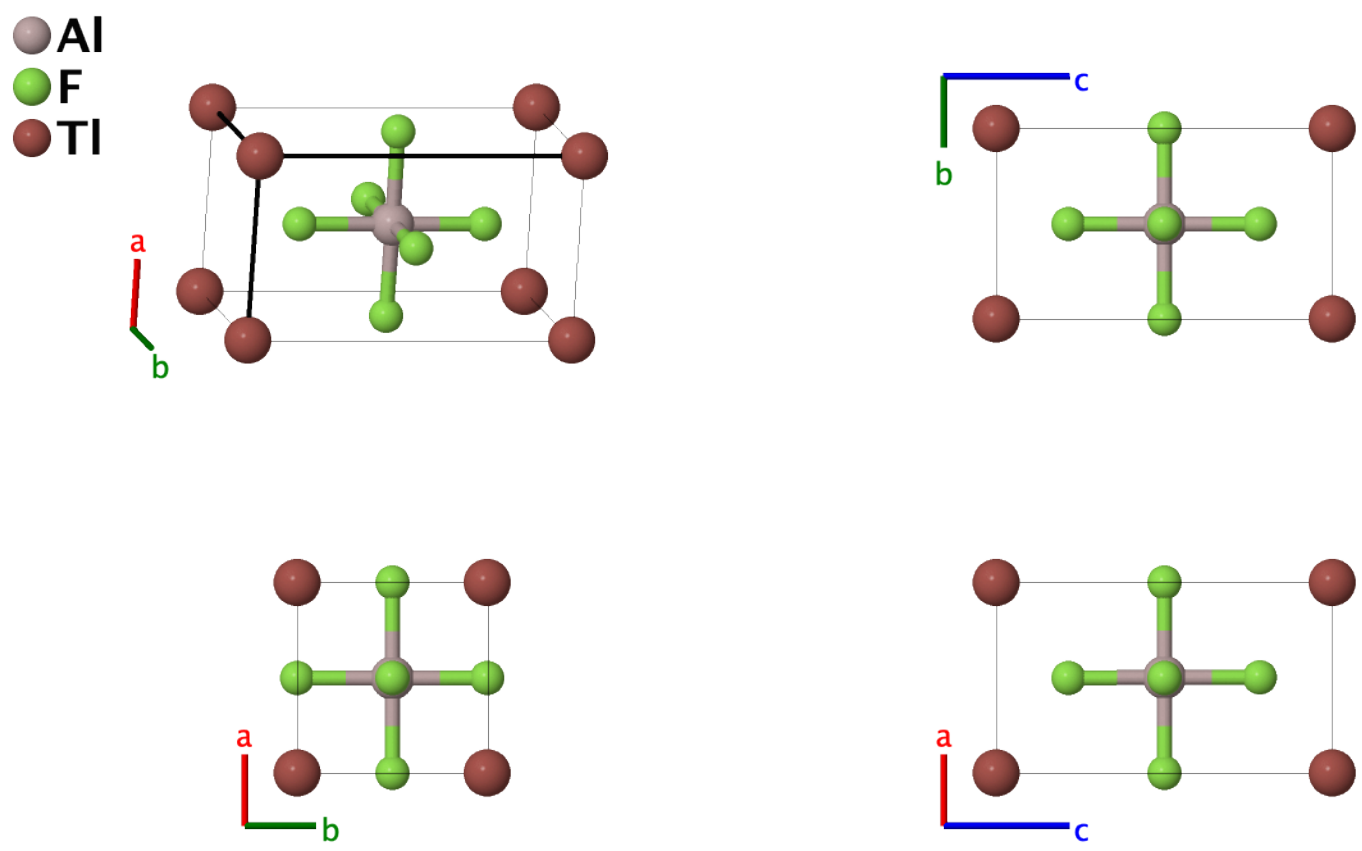

| | | |
|---|---|---|
| **Prototype** | : | $AlF_4Tl$ |
| **AFLOW prototype label** | : | AB4C_tP6_123_d_eh_a |
| ***Strukturbericht* designation** | : | $H0_8$ |
| **Pearson symbol** | : | tP6 |
| **Space group number** | : | 123 |
| **Space group symbol** | : | $P4/mmm$ |
| **AFLOW prototype command** | : | `aflow --proto=AB4C_tP6_123_d_eh_a --params=`$a, c/a, z_4$ |

### Other compounds with this structure

- $\beta$-$RbFeF_4$, $CsFeF_4$, $KAlF_4$, and $RbAlF_4$

### Simple Tetragonal primitive vectors:

$$\mathbf{a}_1 = a\,\hat{\mathbf{x}}$$
$$\mathbf{a}_2 = a\,\hat{\mathbf{y}}$$
$$\mathbf{a}_3 = c\,\hat{\mathbf{z}}$$

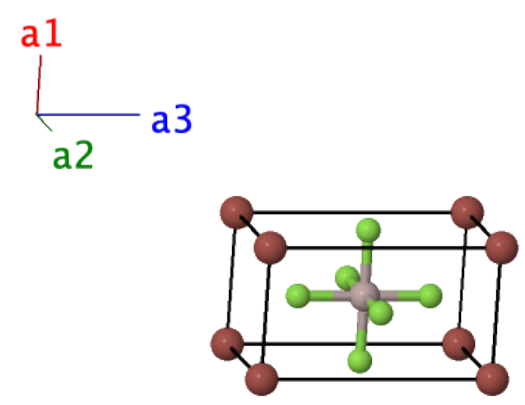

### Basis vectors:

| | | Lattice Coordinates | | Cartesian Coordinates | Wyckoff Position | Atom Type |
|---|---|---|---|---|---|---|
| $\mathbf{B}_1$ | = | $0\,\mathbf{a}_1 + 0\,\mathbf{a}_2 + 0\,\mathbf{a}_3$ | = | $0\,\hat{\mathbf{x}} + 0\,\hat{\mathbf{y}} + 0\,\hat{\mathbf{z}}$ | $(1a)$ | Tl |
| $\mathbf{B}_2$ | = | $\frac{1}{2}\,\mathbf{a}_1 + \frac{1}{2}\,\mathbf{a}_2 + \frac{1}{2}\,\mathbf{a}_3$ | = | $\frac{1}{2}a\,\hat{\mathbf{x}} + \frac{1}{2}a\,\hat{\mathbf{y}} + \frac{1}{2}c\,\hat{\mathbf{z}}$ | $(1d)$ | Al |
| $\mathbf{B}_3$ | = | $\frac{1}{2}\,\mathbf{a}_2 + \frac{1}{2}\,\mathbf{a}_3$ | = | $\frac{1}{2}a\,\hat{\mathbf{y}} + \frac{1}{2}c\,\hat{\mathbf{z}}$ | $(2e)$ | F I |
| $\mathbf{B}_4$ | = | $\frac{1}{2}\,\mathbf{a}_1 + \frac{1}{2}\,\mathbf{a}_3$ | = | $\frac{1}{2}a\,\hat{\mathbf{x}} + \frac{1}{2}c\,\hat{\mathbf{z}}$ | $(2e)$ | F I |
| $\mathbf{B}_5$ | = | $\frac{1}{2}\,\mathbf{a}_1 + \frac{1}{2}\,\mathbf{a}_2 + z_4\,\mathbf{a}_3$ | = | $\frac{1}{2}a\,\hat{\mathbf{x}} + \frac{1}{2}a\,\hat{\mathbf{y}} + z_4c\,\hat{\mathbf{z}}$ | $(2h)$ | F II |
| $\mathbf{B}_6$ | = | $\frac{1}{2}\,\mathbf{a}_1 + \frac{1}{2}\,\mathbf{a}_2 - z_4\,\mathbf{a}_3$ | = | $\frac{1}{2}a\,\hat{\mathbf{x}} + \frac{1}{2}a\,\hat{\mathbf{y}} - z_4c\,\hat{\mathbf{z}}$ | $(2h)$ | F II |

**Geometry files:**
- CIF: pp. 1692
- POSCAR: pp. 1692

# $\delta$-CuTi ($L2_a$) Structure: AB_tP2_123_a_d

http://aflow.org/prototype-encyclopedia/AB_tP2_123_a_d.CuTi

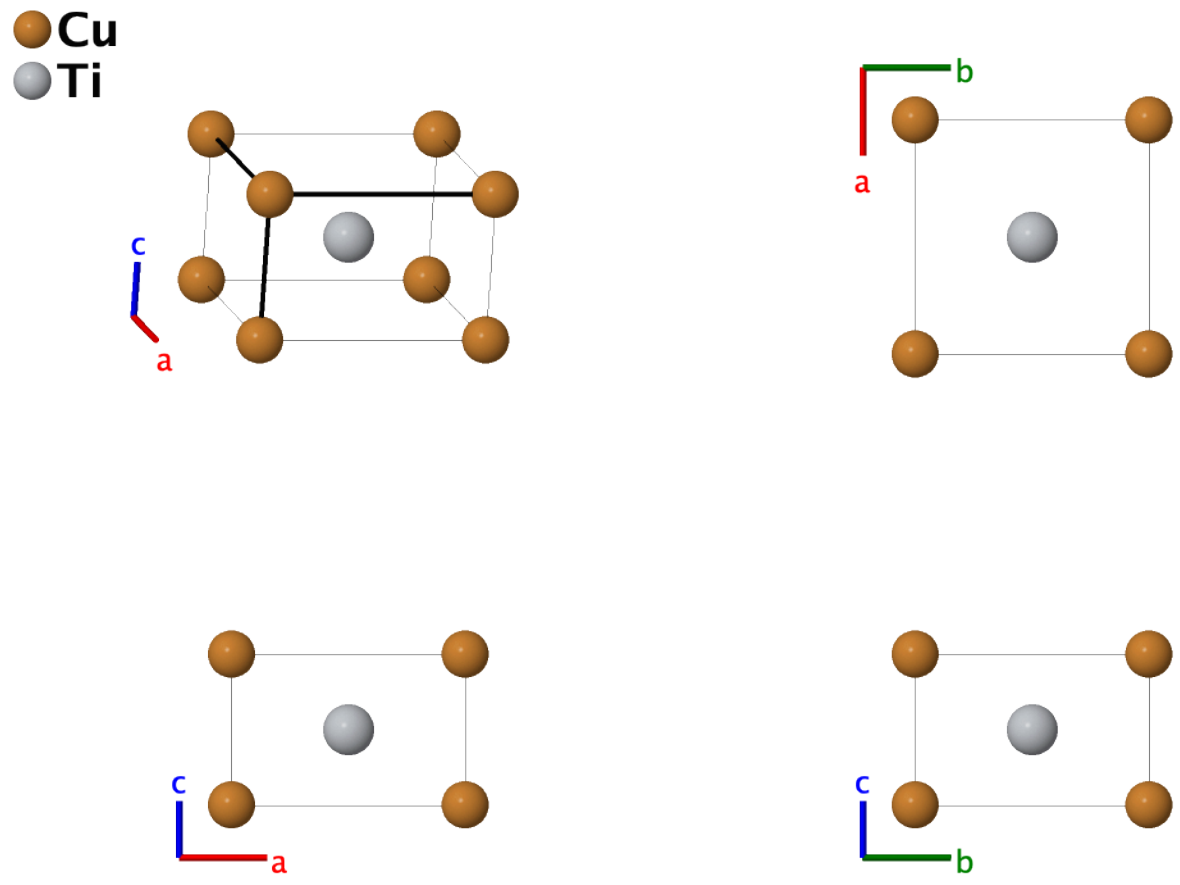

| | | |
|---|---|---|
| **Prototype** | : | CuTi |
| **AFLOW prototype label** | : | AB_tP2_123_a_d |
| ***Strukturbericht* designation** | : | $L2_a$ |
| **Pearson symbol** | : | tP2 |
| **Space group number** | : | 123 |
| **Space group symbol** | : | $P4/mmm$ |
| **AFLOW prototype command** | : | `aflow --proto=AB_tP2_123_a_d --params=`$a, c/a$ |

### Other compounds with this structure

- HgMn

---

- This structure has the same AFLOW designation, AB_tP2_123_a_d, as CuAu ($L1_0$). The only difference in the structures is the c/a ratio. $L1_0$ has $c/a \approx \sqrt{2}$, characteristitic of face-centered cubic ordering, while $L2_a$ has $c/a \approx 1$, a body-centered cubic-like system.

### Simple Tetragonal primitive vectors:

$$\mathbf{a}_1 = a\,\hat{\mathbf{x}}$$
$$\mathbf{a}_2 = a\,\hat{\mathbf{y}}$$
$$\mathbf{a}_3 = c\,\hat{\mathbf{z}}$$

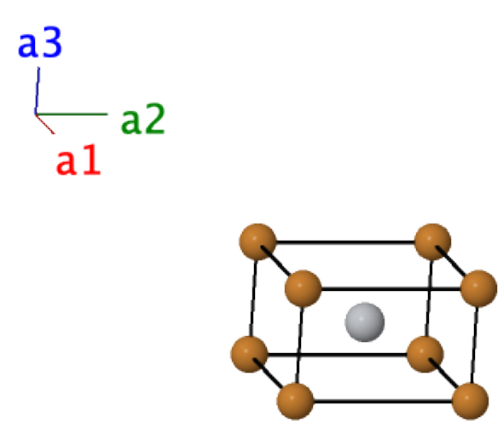

### Basis vectors:

| | | Lattice Coordinates | | Cartesian Coordinates | Wyckoff Position | Atom Type |
|---|---|---|---|---|---|---|
| $\mathbf{B}_1$ | = | $0\,\mathbf{a}_1 + 0\,\mathbf{a}_2 + 0\,\mathbf{a}_3$ | = | $0\,\hat{\mathbf{x}} + 0\,\hat{\mathbf{y}} + 0\,\hat{\mathbf{z}}$ | $(1a)$ | Cu |
| $\mathbf{B}_2$ | = | $\frac{1}{2}\,\mathbf{a}_1 + \frac{1}{2}\,\mathbf{a}_2 + \frac{1}{2}\,\mathbf{a}_3$ | = | $\frac{1}{2}a\,\hat{\mathbf{x}} + \frac{1}{2}a\,\hat{\mathbf{y}} + \frac{1}{2}c\,\hat{\mathbf{z}}$ | $(1d)$ | Ti |

**Found in:**

**Geometry files:**

- CIF: pp. 1692
- POSCAR: pp. 1693

# $CaO_2(H_2O)_8$ Structure: AB8C2_tP22_124_a_n_h

http://aflow.org/prototype-encyclopedia/AB8C2_tP22_124_a_n_h

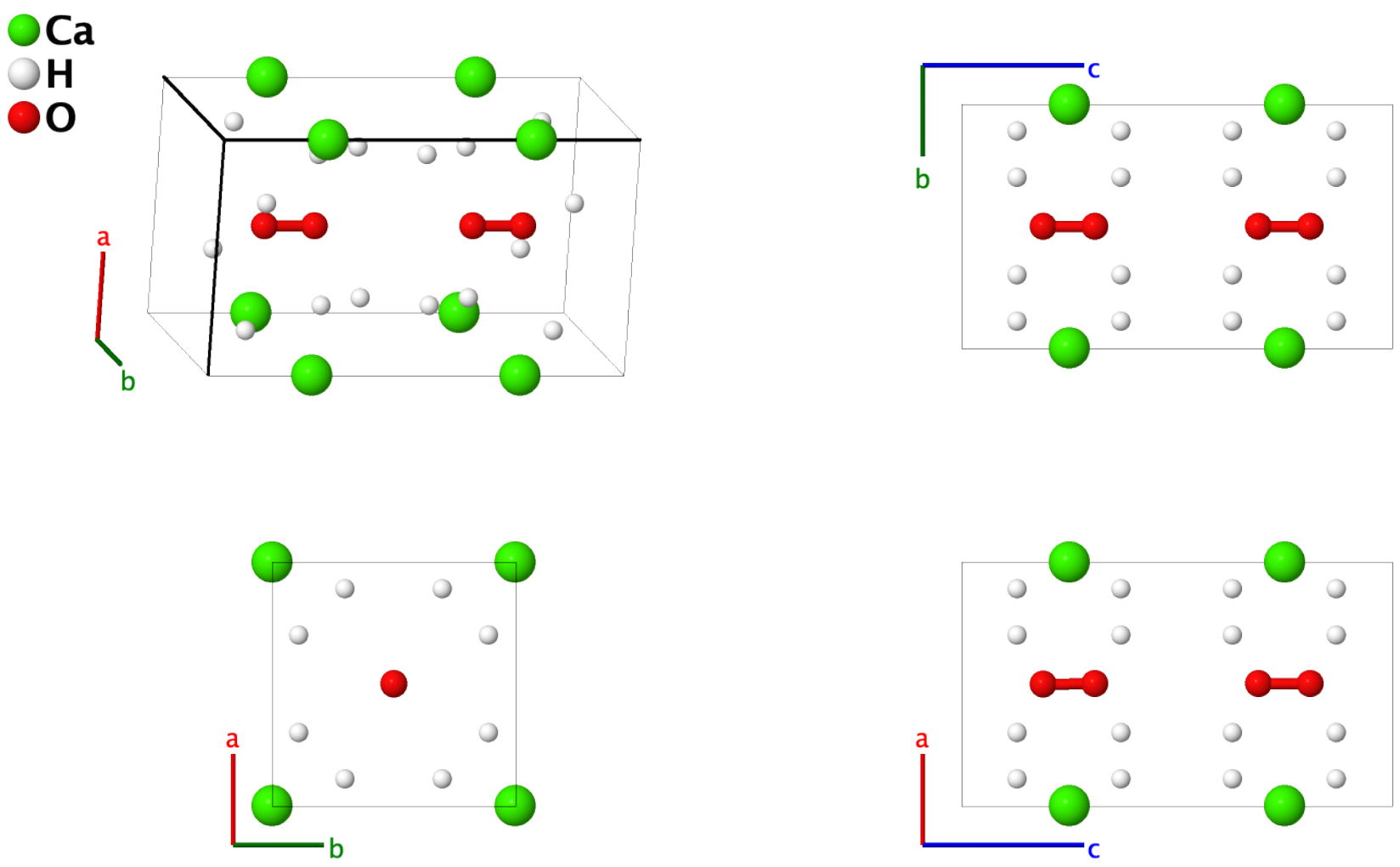

| | | |
|---|---|---|
| **Prototype** | : | $Ca(H_2O)_8O_2$ |
| **AFLOW prototype label** | : | AB8C2_tP22_124_a_n_h |
| ***Strukturbericht* designation** | : | None |
| **Pearson symbol** | : | tP22 |
| **Space group number** | : | 124 |
| **Space group symbol** | : | $P4/mcc$ |
| **AFLOW prototype command** | : | `aflow --proto=AB8C2_tP22_124_a_n_h --params=`$a, c/a, z_2, x_3, y_3, z_3$ |

**Other compounds with this structure**

- $BrO_2(H_2O)_8$ and $SaO_2(H_2O)_8$

- This structure was proposed by (Shineman, 1951) to replace the $E6_2$ $SrO_2(H_2O)_8$ structure. The unit cell is doubled in the (001) direction, and the oxygens are now molecular.
- The positions of the hydrogen atom in the water molecule were not determined, so we only provide the oxygen position (labeled as $H_2O$).

**Simple Tetragonal primitive vectors:**

$$\mathbf{a}_1 = a\,\hat{\mathbf{x}}$$
$$\mathbf{a}_2 = a\,\hat{\mathbf{y}}$$
$$\mathbf{a}_3 = c\,\hat{\mathbf{z}}$$

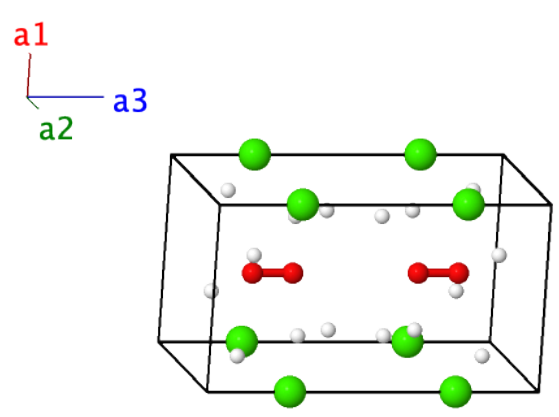

**Basis vectors:**

| | | Lattice Coordinates | | Cartesian Coordinates | Wyckoff Position | Atom Type |
|---|---|---|---|---|---|---|
| $\mathbf{B}_1$ | = | $\frac{1}{4}\,\mathbf{a}_3$ | = | $\frac{1}{4}c\,\hat{\mathbf{z}}$ | (2$a$) | Ca |
| $\mathbf{B}_2$ | = | $\frac{3}{4}\,\mathbf{a}_3$ | = | $\frac{3}{4}c\,\hat{\mathbf{z}}$ | (2$a$) | Ca |
| $\mathbf{B}_3$ | = | $\frac{1}{2}\,\mathbf{a}_1 + \frac{1}{2}\,\mathbf{a}_2 + z_2\,\mathbf{a}_3$ | = | $\frac{1}{2}a\,\hat{\mathbf{x}} + \frac{1}{2}a\,\hat{\mathbf{y}} + z_2c\,\hat{\mathbf{z}}$ | (4$h$) | O |
| $\mathbf{B}_4$ | = | $\frac{1}{2}\,\mathbf{a}_1 + \frac{1}{2}\,\mathbf{a}_2 + \left(\frac{1}{2} - z_2\right)\,\mathbf{a}_3$ | = | $\frac{1}{2}a\,\hat{\mathbf{x}} + \frac{1}{2}a\,\hat{\mathbf{y}} + \left(\frac{1}{2} - z_2\right)c\,\hat{\mathbf{z}}$ | (4$h$) | O |
| $\mathbf{B}_5$ | = | $\frac{1}{2}\,\mathbf{a}_1 + \frac{1}{2}\,\mathbf{a}_2 - z_2\,\mathbf{a}_3$ | = | $\frac{1}{2}a\,\hat{\mathbf{x}} + \frac{1}{2}a\,\hat{\mathbf{y}} - z_2c\,\hat{\mathbf{z}}$ | (4$h$) | O |
| $\mathbf{B}_6$ | = | $\frac{1}{2}\,\mathbf{a}_1 + \frac{1}{2}\,\mathbf{a}_2 + \left(\frac{1}{2} + z_2\right)\,\mathbf{a}_3$ | = | $\frac{1}{2}a\,\hat{\mathbf{x}} + \frac{1}{2}a\,\hat{\mathbf{y}} + \left(\frac{1}{2} + z_2\right)c\,\hat{\mathbf{z}}$ | (4$h$) | O |
| $\mathbf{B}_7$ | = | $x_3\,\mathbf{a}_1 + y_3\,\mathbf{a}_2 + z_3\,\mathbf{a}_3$ | = | $x_3a\,\hat{\mathbf{x}} + y_3a\,\hat{\mathbf{y}} + z_3c\,\hat{\mathbf{z}}$ | (16$n$) | $H_2O$ |
| $\mathbf{B}_8$ | = | $-x_3\,\mathbf{a}_1 - y_3\,\mathbf{a}_2 + z_3\,\mathbf{a}_3$ | = | $-x_3a\,\hat{\mathbf{x}} - y_3a\,\hat{\mathbf{y}} + z_3c\,\hat{\mathbf{z}}$ | (16$n$) | $H_2O$ |
| $\mathbf{B}_9$ | = | $-y_3\,\mathbf{a}_1 + x_3\,\mathbf{a}_2 + z_3\,\mathbf{a}_3$ | = | $-y_3a\,\hat{\mathbf{x}} + x_3a\,\hat{\mathbf{y}} + z_3c\,\hat{\mathbf{z}}$ | (16$n$) | $H_2O$ |
| $\mathbf{B}_{10}$ | = | $y_3\,\mathbf{a}_1 - x_3\,\mathbf{a}_2 + z_3\,\mathbf{a}_3$ | = | $y_3a\,\hat{\mathbf{x}} - x_3a\,\hat{\mathbf{y}} + z_3c\,\hat{\mathbf{z}}$ | (16$n$) | $H_2O$ |
| $\mathbf{B}_{11}$ | = | $-x_3\,\mathbf{a}_1 + y_3\,\mathbf{a}_2 + \left(\frac{1}{2} - z_3\right)\,\mathbf{a}_3$ | = | $-x_3a\,\hat{\mathbf{x}} + y_3a\,\hat{\mathbf{y}} + \left(\frac{1}{2} - z_3\right)c\,\hat{\mathbf{z}}$ | (16$n$) | $H_2O$ |
| $\mathbf{B}_{12}$ | = | $x_3\,\mathbf{a}_1 - y_3\,\mathbf{a}_2 + \left(\frac{1}{2} - z_3\right)\,\mathbf{a}_3$ | = | $x_3a\,\hat{\mathbf{x}} - y_3a\,\hat{\mathbf{y}} + \left(\frac{1}{2} - z_3\right)c\,\hat{\mathbf{z}}$ | (16$n$) | $H_2O$ |
| $\mathbf{B}_{13}$ | = | $y_3\,\mathbf{a}_1 + x_3\,\mathbf{a}_2 + \left(\frac{1}{2} - z_3\right)\,\mathbf{a}_3$ | = | $y_3a\,\hat{\mathbf{x}} + x_3a\,\hat{\mathbf{y}} + \left(\frac{1}{2} - z_3\right)c\,\hat{\mathbf{z}}$ | (16$n$) | $H_2O$ |
| $\mathbf{B}_{14}$ | = | $-y_3\,\mathbf{a}_1 - x_3\,\mathbf{a}_2 + \left(\frac{1}{2} - z_3\right)\,\mathbf{a}_3$ | = | $-y_3a\,\hat{\mathbf{x}} - x_3a\,\hat{\mathbf{y}} + \left(\frac{1}{2} - z_3\right)c\,\hat{\mathbf{z}}$ | (16$n$) | $H_2O$ |
| $\mathbf{B}_{15}$ | = | $-x_3\,\mathbf{a}_1 - y_3\,\mathbf{a}_2 - z_3\,\mathbf{a}_3$ | = | $-x_3a\,\hat{\mathbf{x}} - y_3a\,\hat{\mathbf{y}} - z_3c\,\hat{\mathbf{z}}$ | (16$n$) | $H_2O$ |
| $\mathbf{B}_{16}$ | = | $x_3\,\mathbf{a}_1 + y_3\,\mathbf{a}_2 - z_3\,\mathbf{a}_3$ | = | $x_3a\,\hat{\mathbf{x}} + y_3a\,\hat{\mathbf{y}} - z_3c\,\hat{\mathbf{z}}$ | (16$n$) | $H_2O$ |
| $\mathbf{B}_{17}$ | = | $y_3\,\mathbf{a}_1 - x_3\,\mathbf{a}_2 - z_3\,\mathbf{a}_3$ | = | $y_3a\,\hat{\mathbf{x}} - x_3a\,\hat{\mathbf{y}} - z_3c\,\hat{\mathbf{z}}$ | (16$n$) | $H_2O$ |
| $\mathbf{B}_{18}$ | = | $-y_3\,\mathbf{a}_1 + x_3\,\mathbf{a}_2 - z_3\,\mathbf{a}_3$ | = | $-y_3a\,\hat{\mathbf{x}} + x_3a\,\hat{\mathbf{y}} - z_3c\,\hat{\mathbf{z}}$ | (16$n$) | $H_2O$ |
| $\mathbf{B}_{19}$ | = | $x_3\,\mathbf{a}_1 - y_3\,\mathbf{a}_2 + \left(\frac{1}{2} + z_3\right)\,\mathbf{a}_3$ | = | $x_3a\,\hat{\mathbf{x}} - y_3a\,\hat{\mathbf{y}} + \left(\frac{1}{2} + z_3\right)c\,\hat{\mathbf{z}}$ | (16$n$) | $H_2O$ |
| $\mathbf{B}_{20}$ | = | $-x_3\,\mathbf{a}_1 + y_3\,\mathbf{a}_2 + \left(\frac{1}{2} + z_3\right)\,\mathbf{a}_3$ | = | $-x_3a\,\hat{\mathbf{x}} + y_3a\,\hat{\mathbf{y}} + \left(\frac{1}{2} + z_3\right)c\,\hat{\mathbf{z}}$ | (16$n$) | $H_2O$ |
| $\mathbf{B}_{21}$ | = | $-y_3\,\mathbf{a}_1 - x_3\,\mathbf{a}_2 + \left(\frac{1}{2} + z_3\right)\,\mathbf{a}_3$ | = | $-y_3a\,\hat{\mathbf{x}} - x_3a\,\hat{\mathbf{y}} + \left(\frac{1}{2} + z_3\right)c\,\hat{\mathbf{z}}$ | (16$n$) | $H_2O$ |
| $\mathbf{B}_{22}$ | = | $y_3\,\mathbf{a}_1 + x_3\,\mathbf{a}_2 + \left(\frac{1}{2} + z_3\right)\,\mathbf{a}_3$ | = | $y_3a\,\hat{\mathbf{x}} + x_3a\,\hat{\mathbf{y}} + \left(\frac{1}{2} + z_3\right)c\,\hat{\mathbf{z}}$ | (16$n$) | $H_2O$ |

**Geometry files:**

- CIF: pp. 1693
- POSCAR: pp. 1693

# Vesuvianite ($Ca_{10}Al_4(Mg,Fe)_2Si_9O_{34}(OH)_4$, $S2_3$) Structure: A4B10C2D34E4F9_tP252_126_k_ce2k_f_h8k_k_d2k

http://aflow.org/prototype-encyclopedia/A4B10C2D34E4F9_tP252_126_k_ce2k_f_h8k_k_d2k

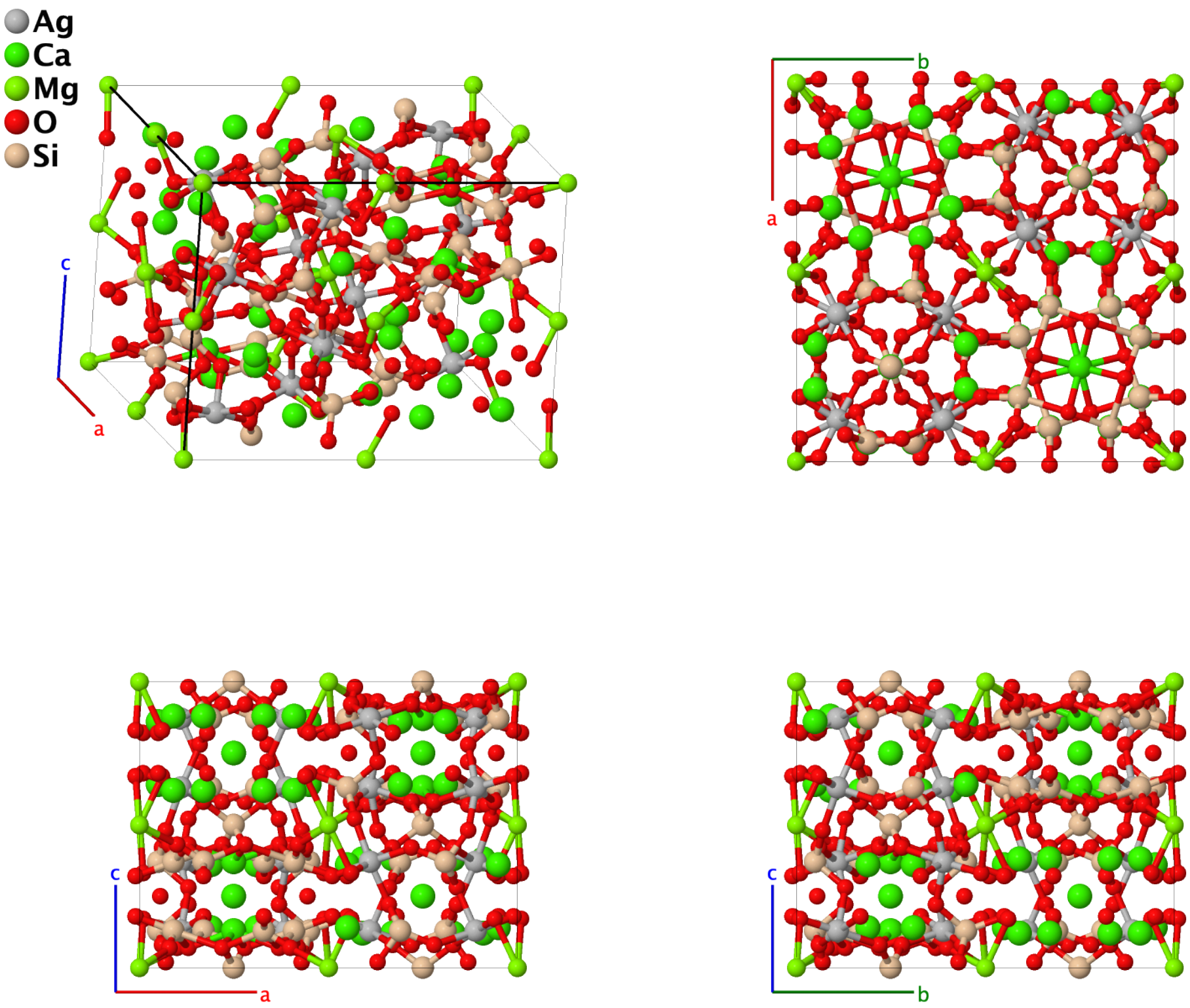

| | | |
|---|---|---|
| **Prototype** | : | $Al_4Ca_{10}(Mg,Fe)_2O_{34}(OH)_4Si_9$ |
| **AFLOW prototype label** | : | A4B10C2D34E4F9_tP252_126_k_ce2k_f_h8k_k_d2k |
| ***Strukturbericht* designation** | : | $S2_3$ |
| **Pearson symbol** | : | tP252 |
| **Space group number** | : | 126 |
| **Space group symbol** | : | $P4/nnc$ |
| **AFLOW prototype command** | : | `aflow --proto=A4B10C2D34E4F9_tP252_126_k_ce2k_f_h8k_k_d2k --params=`$a, c/a, z_3, x_5, x_6, y_6, z_6, x_7, y_7, z_7, x_8, y_8, z_8, x_9, y_9, z_9, x_{10}, y_{10}, z_{10}, x_{11}, y_{11}, z_{11}, x_{12}, y_{12}, z_{12}, x_{13}, y_{13}, z_{13}, x_{14}, y_{14}, z_{14}, x_{15}, y_{15}, z_{15}, x_{16}, y_{16}, z_{16}, x_{17}, y_{17}, z_{17}, x_{18}, y_{18}, z_{18}, x_{19}, y_{19}, z_{19}$ |

- Vesuvianite, also known as idocrase, comes in a variety of compositions and structures, see *e.g.*, (Allen, 1992) and (Rucklidge, 1975) and references therein. For our example we use the original structure of (Warren, 1931), where the magnesium ($8f$) site is filled by a random (Mg,Fe) alloy. The positions of the hydrogen atoms in the OH ions were not determined, so we only give the positions of the oyxgen atoms (labeled as OH).

## Simple Tetragonal primitive vectors:

$$\mathbf{a}_1 = a\,\hat{\mathbf{x}}$$
$$\mathbf{a}_2 = a\,\hat{\mathbf{y}}$$
$$\mathbf{a}_3 = c\,\hat{\mathbf{z}}$$

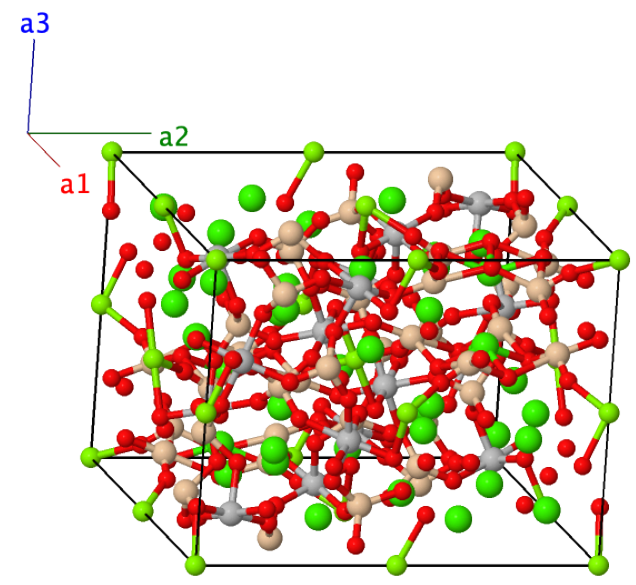

## Basis vectors:

| | | Lattice Coordinates | | Cartesian Coordinates | Wyckoff Position | Atom Type |
|---|---|---|---|---|---|---|
| $\mathbf{B}_1$ | = | $\frac{1}{4}\,\mathbf{a}_1 + \frac{3}{4}\,\mathbf{a}_2 + \frac{3}{4}\,\mathbf{a}_3$ | = | $\frac{1}{4}a\,\hat{\mathbf{x}} + \frac{3}{4}a\,\hat{\mathbf{y}} + \frac{3}{4}c\,\hat{\mathbf{z}}$ | (4$c$) | Ca I |
| $\mathbf{B}_2$ | = | $\frac{3}{4}\,\mathbf{a}_1 + \frac{1}{4}\,\mathbf{a}_2 + \frac{3}{4}\,\mathbf{a}_3$ | = | $\frac{3}{4}a\,\hat{\mathbf{x}} + \frac{1}{4}a\,\hat{\mathbf{y}} + \frac{3}{4}c\,\hat{\mathbf{z}}$ | (4$c$) | Ca I |
| $\mathbf{B}_3$ | = | $\frac{3}{4}\,\mathbf{a}_1 + \frac{1}{4}\,\mathbf{a}_2 + \frac{1}{4}\,\mathbf{a}_3$ | = | $\frac{3}{4}a\,\hat{\mathbf{x}} + \frac{1}{4}a\,\hat{\mathbf{y}} + \frac{1}{4}c\,\hat{\mathbf{z}}$ | (4$c$) | Ca I |
| $\mathbf{B}_4$ | = | $\frac{1}{4}\,\mathbf{a}_1 + \frac{3}{4}\,\mathbf{a}_2 + \frac{1}{4}\,\mathbf{a}_3$ | = | $\frac{1}{4}a\,\hat{\mathbf{x}} + \frac{3}{4}a\,\hat{\mathbf{y}} + \frac{1}{4}c\,\hat{\mathbf{z}}$ | (4$c$) | Ca I |
| $\mathbf{B}_5$ | = | $\frac{1}{4}\,\mathbf{a}_1 + \frac{3}{4}\,\mathbf{a}_2$ | = | $\frac{1}{4}a\,\hat{\mathbf{x}} + \frac{3}{4}a\,\hat{\mathbf{y}}$ | (4$d$) | Si I |
| $\mathbf{B}_6$ | = | $\frac{3}{4}\,\mathbf{a}_1 + \frac{1}{4}\,\mathbf{a}_2$ | = | $\frac{3}{4}a\,\hat{\mathbf{x}} + \frac{1}{4}a\,\hat{\mathbf{y}}$ | (4$d$) | Si I |
| $\mathbf{B}_7$ | = | $\frac{1}{4}\,\mathbf{a}_1 + \frac{3}{4}\,\mathbf{a}_2 + \frac{1}{2}\,\mathbf{a}_3$ | = | $\frac{1}{4}a\,\hat{\mathbf{x}} + \frac{3}{4}a\,\hat{\mathbf{y}} + \frac{1}{2}c\,\hat{\mathbf{z}}$ | (4$d$) | Si I |
| $\mathbf{B}_8$ | = | $\frac{3}{4}\,\mathbf{a}_1 + \frac{1}{4}\,\mathbf{a}_2 + \frac{1}{2}\,\mathbf{a}_3$ | = | $\frac{3}{4}a\,\hat{\mathbf{x}} + \frac{1}{4}a\,\hat{\mathbf{y}} + \frac{1}{2}c\,\hat{\mathbf{z}}$ | (4$d$) | Si I |
| $\mathbf{B}_9$ | = | $\frac{1}{4}\,\mathbf{a}_1 + \frac{1}{4}\,\mathbf{a}_2 + z_3\,\mathbf{a}_3$ | = | $\frac{1}{4}a\,\hat{\mathbf{x}} + \frac{1}{4}a\,\hat{\mathbf{y}} + z_3c\,\hat{\mathbf{z}}$ | (4$e$) | Ca II |
| $\mathbf{B}_{10}$ | = | $\frac{1}{4}\,\mathbf{a}_1 + \frac{1}{4}\,\mathbf{a}_2 + \left(\frac{1}{2} - z_3\right)\,\mathbf{a}_3$ | = | $\frac{1}{4}a\,\hat{\mathbf{x}} + \frac{1}{4}a\,\hat{\mathbf{y}} + \left(\frac{1}{2} - z_3\right)c\,\hat{\mathbf{z}}$ | (4$e$) | Ca II |
| $\mathbf{B}_{11}$ | = | $\frac{3}{4}\,\mathbf{a}_1 + \frac{3}{4}\,\mathbf{a}_2 - z_3\,\mathbf{a}_3$ | = | $\frac{3}{4}a\,\hat{\mathbf{x}} + \frac{3}{4}a\,\hat{\mathbf{y}} - z_3c\,\hat{\mathbf{z}}$ | (4$e$) | Ca II |
| $\mathbf{B}_{12}$ | = | $\frac{3}{4}\,\mathbf{a}_1 + \frac{3}{4}\,\mathbf{a}_2 + \left(\frac{1}{2} + z_3\right)\,\mathbf{a}_3$ | = | $\frac{3}{4}a\,\hat{\mathbf{x}} + \frac{3}{4}a\,\hat{\mathbf{y}} + \left(\frac{1}{2} + z_3\right)c\,\hat{\mathbf{z}}$ | (4$e$) | Ca II |
| $\mathbf{B}_{13}$ | = | $0\,\mathbf{a}_1 + 0\,\mathbf{a}_2 + 0\,\mathbf{a}_3$ | = | $0\,\hat{\mathbf{x}} + 0\,\hat{\mathbf{y}} + 0\,\hat{\mathbf{z}}$ | (8$f$) | Mg |
| $\mathbf{B}_{14}$ | = | $\frac{1}{2}\,\mathbf{a}_1 + \frac{1}{2}\,\mathbf{a}_2$ | = | $\frac{1}{2}a\,\hat{\mathbf{x}} + \frac{1}{2}a\,\hat{\mathbf{y}}$ | (8$f$) | Mg |
| $\mathbf{B}_{15}$ | = | $\frac{1}{2}\,\mathbf{a}_1$ | = | $\frac{1}{2}a\,\hat{\mathbf{x}}$ | (8$f$) | Mg |
| $\mathbf{B}_{16}$ | = | $\frac{1}{2}\,\mathbf{a}_2$ | = | $\frac{1}{2}a\,\hat{\mathbf{y}}$ | (8$f$) | Mg |
| $\mathbf{B}_{17}$ | = | $\frac{1}{2}\,\mathbf{a}_1 + \frac{1}{2}\,\mathbf{a}_3$ | = | $\frac{1}{2}a\,\hat{\mathbf{x}} + \frac{1}{2}c\,\hat{\mathbf{z}}$ | (8$f$) | Mg |
| $\mathbf{B}_{18}$ | = | $\frac{1}{2}\,\mathbf{a}_2 + \frac{1}{2}\,\mathbf{a}_3$ | = | $\frac{1}{2}a\,\hat{\mathbf{y}} + \frac{1}{2}c\,\hat{\mathbf{z}}$ | (8$f$) | Mg |
| $\mathbf{B}_{19}$ | = | $\frac{1}{2}\,\mathbf{a}_3$ | = | $\frac{1}{2}c\,\hat{\mathbf{z}}$ | (8$f$) | Mg |
| $\mathbf{B}_{20}$ | = | $\frac{1}{2}\,\mathbf{a}_1 + \frac{1}{2}\,\mathbf{a}_2 + \frac{1}{2}\,\mathbf{a}_3$ | = | $\frac{1}{2}a\,\hat{\mathbf{x}} + \frac{1}{2}a\,\hat{\mathbf{y}} + \frac{1}{2}c\,\hat{\mathbf{z}}$ | (8$f$) | Mg |
| $\mathbf{B}_{21}$ | = | $x_5\,\mathbf{a}_1 + x_5\,\mathbf{a}_2 + \frac{1}{4}\,\mathbf{a}_3$ | = | $x_5a\,\hat{\mathbf{x}} + x_5a\,\hat{\mathbf{y}} + \frac{1}{4}c\,\hat{\mathbf{z}}$ | (8$h$) | O I |
| $\mathbf{B}_{22}$ | = | $\left(\frac{1}{2} - x_5\right)\,\mathbf{a}_1 + \left(\frac{1}{2} - x_5\right)\,\mathbf{a}_2 + \frac{1}{4}\,\mathbf{a}_3$ | = | $\left(\frac{1}{2} - x_5\right)a\,\hat{\mathbf{x}} + \left(\frac{1}{2} - x_5\right)a\,\hat{\mathbf{y}} + \frac{1}{4}c\,\hat{\mathbf{z}}$ | (8$h$) | O I |
| $\mathbf{B}_{23}$ | = | $\left(\frac{1}{2} - x_5\right)\,\mathbf{a}_1 + x_5\,\mathbf{a}_2 + \frac{1}{4}\,\mathbf{a}_3$ | = | $\left(\frac{1}{2} - x_5\right)a\,\hat{\mathbf{x}} + x_5a\,\hat{\mathbf{y}} + \frac{1}{4}c\,\hat{\mathbf{z}}$ | (8$h$) | O I |
| $\mathbf{B}_{24}$ | = | $x_5\,\mathbf{a}_1 + \left(\frac{1}{2} - x_5\right)\,\mathbf{a}_2 + \frac{1}{4}\,\mathbf{a}_3$ | = | $x_5a\,\hat{\mathbf{x}} + \left(\frac{1}{2} - x_5\right)a\,\hat{\mathbf{y}} + \frac{1}{4}c\,\hat{\mathbf{z}}$ | (8$h$) | O I |
| $\mathbf{B}_{25}$ | = | $-x_5\,\mathbf{a}_1 - x_5\,\mathbf{a}_2 + \frac{3}{4}\,\mathbf{a}_3$ | = | $-x_5a\,\hat{\mathbf{x}} - x_5a\,\hat{\mathbf{y}} + \frac{3}{4}c\,\hat{\mathbf{z}}$ | (8$h$) | O I |
| $\mathbf{B}_{26}$ | = | $\left(\frac{1}{2} + x_5\right)\,\mathbf{a}_1 + \left(\frac{1}{2} + x_5\right)\,\mathbf{a}_2 + \frac{3}{4}\,\mathbf{a}_3$ | = | $\left(\frac{1}{2} + x_5\right)a\,\hat{\mathbf{x}} + \left(\frac{1}{2} + x_5\right)a\,\hat{\mathbf{y}} + \frac{3}{4}c\,\hat{\mathbf{z}}$ | (8$h$) | O I |
| $\mathbf{B}_{27}$ | = | $\left(\frac{1}{2} + x_5\right)\,\mathbf{a}_1 - x_5\,\mathbf{a}_2 + \frac{3}{4}\,\mathbf{a}_3$ | = | $\left(\frac{1}{2} + x_5\right)a\,\hat{\mathbf{x}} - x_5a\,\hat{\mathbf{y}} + \frac{3}{4}c\,\hat{\mathbf{z}}$ | (8$h$) | O I |
| $\mathbf{B}_{28}$ | = | $-x_{5}\,\mathbf{a}_{1}+\left(\frac{1}{2}+x_{5}\right)\,\mathbf{a}_{2}+\frac{3}{4}\,\mathbf{a}_{3}$ | = | $-x_{5}a\,\hat{\mathbf{x}}+\left(\frac{1}{2}+x_{5}\right)a\,\hat{\mathbf{y}}+\frac{3}{4}c\,\hat{\mathbf{z}}$ | (8*h*) | O I |
| $\mathbf{B}_{29}$ | = | $x_{6}\,\mathbf{a}_{1}+y_{6}\,\mathbf{a}_{2}+z_{6}\,\mathbf{a}_{3}$ | = | $x_{6}a\,\hat{\mathbf{x}}+y_{6}a\,\hat{\mathbf{y}}+z_{6}c\,\hat{\mathbf{z}}$ | (16*k*) | Ag |
| $\mathbf{B}_{30}$ | = | $\left(\frac{1}{2}-x_{6}\right)\,\mathbf{a}_{1}+\left(\frac{1}{2}-y_{6}\right)\,\mathbf{a}_{2}+z_{6}\,\mathbf{a}_{3}$ | = | $\left(\frac{1}{2}-x_{6}\right)a\,\hat{\mathbf{x}}+\left(\frac{1}{2}-y_{6}\right)a\,\hat{\mathbf{y}}+z_{6}c\,\hat{\mathbf{z}}$ | (16*k*) | Ag |
| $\mathbf{B}_{31}$ | = | $\left(\frac{1}{2}-y_{6}\right)\,\mathbf{a}_{1}+x_{6}\,\mathbf{a}_{2}+z_{6}\,\mathbf{a}_{3}$ | = | $\left(\frac{1}{2}-y_{6}\right)a\,\hat{\mathbf{x}}+x_{6}a\,\hat{\mathbf{y}}+z_{6}c\,\hat{\mathbf{z}}$ | (16*k*) | Ag |
| $\mathbf{B}_{32}$ | = | $y_{6}\,\mathbf{a}_{1}+\left(\frac{1}{2}-x_{6}\right)\,\mathbf{a}_{2}+z_{6}\,\mathbf{a}_{3}$ | = | $y_{6}a\,\hat{\mathbf{x}}+\left(\frac{1}{2}-x_{6}\right)a\,\hat{\mathbf{y}}+z_{6}c\,\hat{\mathbf{z}}$ | (16*k*) | Ag |
| $\mathbf{B}_{33}$ | = | $\left(\frac{1}{2}-x_{6}\right)\,\mathbf{a}_{1}+y_{6}\,\mathbf{a}_{2}+\left(\frac{1}{2}-z_{6}\right)\,\mathbf{a}_{3}$ | = | $\left(\frac{1}{2}-x_{6}\right)a\,\hat{\mathbf{x}}+y_{6}a\,\hat{\mathbf{y}}+\left(\frac{1}{2}-z_{6}\right)c\,\hat{\mathbf{z}}$ | (16*k*) | Ag |
| $\mathbf{B}_{34}$ | = | $x_{6}\,\mathbf{a}_{1}+\left(\frac{1}{2}-y_{6}\right)\,\mathbf{a}_{2}+\left(\frac{1}{2}-z_{6}\right)\,\mathbf{a}_{3}$ | = | $x_{6}a\,\hat{\mathbf{x}}+\left(\frac{1}{2}-y_{6}\right)a\,\hat{\mathbf{y}}+\left(\frac{1}{2}-z_{6}\right)c\,\hat{\mathbf{z}}$ | (16*k*) | Ag |
| $\mathbf{B}_{35}$ | = | $y_{6}\,\mathbf{a}_{1}+x_{6}\,\mathbf{a}_{2}+\left(\frac{1}{2}-z_{6}\right)\,\mathbf{a}_{3}$ | = | $y_{6}a\,\hat{\mathbf{x}}+x_{6}a\,\hat{\mathbf{y}}+\left(\frac{1}{2}-z_{6}\right)c\,\hat{\mathbf{z}}$ | (16*k*) | Ag |
| $\mathbf{B}_{36}$ | = | $\left(\frac{1}{2}-y_{6}\right)\,\mathbf{a}_{1}+\left(\frac{1}{2}-x_{6}\right)\,\mathbf{a}_{2}+\left(\frac{1}{2}-z_{6}\right)\,\mathbf{a}_{3}$ | = | $\left(\frac{1}{2}-y_{6}\right)a\,\hat{\mathbf{x}}+\left(\frac{1}{2}-x_{6}\right)a\,\hat{\mathbf{y}}+\left(\frac{1}{2}-z_{6}\right)c\,\hat{\mathbf{z}}$ | (16*k*) | Ag |
| $\mathbf{B}_{37}$ | = | $-x_{6}\,\mathbf{a}_{1}-y_{6}\,\mathbf{a}_{2}-z_{6}\,\mathbf{a}_{3}$ | = | $-x_{6}a\,\hat{\mathbf{x}}-y_{6}a\,\hat{\mathbf{y}}-z_{6}c\,\hat{\mathbf{z}}$ | (16*k*) | Ag |
| $\mathbf{B}_{38}$ | = | $\left(\frac{1}{2}+x_{6}\right)\,\mathbf{a}_{1}+\left(\frac{1}{2}+y_{6}\right)\,\mathbf{a}_{2}-z_{6}\,\mathbf{a}_{3}$ | = | $\left(\frac{1}{2}+x_{6}\right)a\,\hat{\mathbf{x}}+\left(\frac{1}{2}+y_{6}\right)a\,\hat{\mathbf{y}}-z_{6}c\,\hat{\mathbf{z}}$ | (16*k*) | Ag |
| $\mathbf{B}_{39}$ | = | $\left(\frac{1}{2}+y_{6}\right)\,\mathbf{a}_{1}-x_{6}\,\mathbf{a}_{2}-z_{6}\,\mathbf{a}_{3}$ | = | $\left(\frac{1}{2}+y_{6}\right)a\,\hat{\mathbf{x}}-x_{6}a\,\hat{\mathbf{y}}-z_{6}c\,\hat{\mathbf{z}}$ | (16*k*) | Ag |
| $\mathbf{B}_{40}$ | = | $-y_{6}\,\mathbf{a}_{1}+\left(\frac{1}{2}+x_{6}\right)\,\mathbf{a}_{2}-z_{6}\,\mathbf{a}_{3}$ | = | $-y_{6}a\,\hat{\mathbf{x}}+\left(\frac{1}{2}+x_{6}\right)a\,\hat{\mathbf{y}}-z_{6}c\,\hat{\mathbf{z}}$ | (16*k*) | Ag |
| $\mathbf{B}_{41}$ | = | $\left(\frac{1}{2}+x_{6}\right)\,\mathbf{a}_{1}-y_{6}\,\mathbf{a}_{2}+\left(\frac{1}{2}+z_{6}\right)\,\mathbf{a}_{3}$ | = | $\left(\frac{1}{2}+x_{6}\right)a\,\hat{\mathbf{x}}-y_{6}a\,\hat{\mathbf{y}}+\left(\frac{1}{2}+z_{6}\right)c\,\hat{\mathbf{z}}$ | (16*k*) | Ag |
| $\mathbf{B}_{42}$ | = | $-x_{6}\,\mathbf{a}_{1}+\left(\frac{1}{2}+y_{6}\right)\,\mathbf{a}_{2}+\left(\frac{1}{2}+z_{6}\right)\,\mathbf{a}_{3}$ | = | $-x_{6}a\,\hat{\mathbf{x}}+\left(\frac{1}{2}+y_{6}\right)a\,\hat{\mathbf{y}}+\left(\frac{1}{2}+z_{6}\right)c\,\hat{\mathbf{z}}$ | (16*k*) | Ag |
| $\mathbf{B}_{43}$ | = | $-y_{6}\,\mathbf{a}_{1}-x_{6}\,\mathbf{a}_{2}+\left(\frac{1}{2}+z_{6}\right)\,\mathbf{a}_{3}$ | = | $-y_{6}a\,\hat{\mathbf{x}}-x_{6}a\,\hat{\mathbf{y}}+\left(\frac{1}{2}+z_{6}\right)c\,\hat{\mathbf{z}}$ | (16*k*) | Ag |
| $\mathbf{B}_{44}$ | = | $\left(\frac{1}{2}+y_{6}\right)\,\mathbf{a}_{1}+\left(\frac{1}{2}+x_{6}\right)\,\mathbf{a}_{2}+\left(\frac{1}{2}+z_{6}\right)\,\mathbf{a}_{3}$ | = | $\left(\frac{1}{2}+y_{6}\right)a\,\hat{\mathbf{x}}+\left(\frac{1}{2}+x_{6}\right)a\,\hat{\mathbf{y}}+\left(\frac{1}{2}+z_{6}\right)c\,\hat{\mathbf{z}}$ | (16*k*) | Ag |
| $\mathbf{B}_{45}$ | = | $x_{7}\,\mathbf{a}_{1}+y_{7}\,\mathbf{a}_{2}+z_{7}\,\mathbf{a}_{3}$ | = | $x_{7}a\,\hat{\mathbf{x}}+y_{7}a\,\hat{\mathbf{y}}+z_{7}c\,\hat{\mathbf{z}}$ | (16*k*) | Ca III |
| $\mathbf{B}_{46}$ | = | $\left(\frac{1}{2}-x_{7}\right)\,\mathbf{a}_{1}+\left(\frac{1}{2}-y_{7}\right)\,\mathbf{a}_{2}+z_{7}\,\mathbf{a}_{3}$ | = | $\left(\frac{1}{2}-x_{7}\right)a\,\hat{\mathbf{x}}+\left(\frac{1}{2}-y_{7}\right)a\,\hat{\mathbf{y}}+z_{7}c\,\hat{\mathbf{z}}$ | (16*k*) | Ca III |
| $\mathbf{B}_{47}$ | = | $\left(\frac{1}{2}-y_{7}\right)\,\mathbf{a}_{1}+x_{7}\,\mathbf{a}_{2}+z_{7}\,\mathbf{a}_{3}$ | = | $\left(\frac{1}{2}-y_{7}\right)a\,\hat{\mathbf{x}}+x_{7}a\,\hat{\mathbf{y}}+z_{7}c\,\hat{\mathbf{z}}$ | (16*k*) | Ca III |
| $\mathbf{B}_{48}$ | = | $y_{7}\,\mathbf{a}_{1}+\left(\frac{1}{2}-x_{7}\right)\,\mathbf{a}_{2}+z_{7}\,\mathbf{a}_{3}$ | = | $y_{7}a\,\hat{\mathbf{x}}+\left(\frac{1}{2}-x_{7}\right)a\,\hat{\mathbf{y}}+z_{7}c\,\hat{\mathbf{z}}$ | (16*k*) | Ca III |
| $\mathbf{B}_{49}$ | = | $\left(\frac{1}{2}-x_{7}\right)\,\mathbf{a}_{1}+y_{7}\,\mathbf{a}_{2}+\left(\frac{1}{2}-z_{7}\right)\,\mathbf{a}_{3}$ | = | $\left(\frac{1}{2}-x_{7}\right)a\,\hat{\mathbf{x}}+y_{7}a\,\hat{\mathbf{y}}+\left(\frac{1}{2}-z_{7}\right)c\,\hat{\mathbf{z}}$ | (16*k*) | Ca III |
| $\mathbf{B}_{50}$ | = | $x_{7}\,\mathbf{a}_{1}+\left(\frac{1}{2}-y_{7}\right)\,\mathbf{a}_{2}+\left(\frac{1}{2}-z_{7}\right)\,\mathbf{a}_{3}$ | = | $x_{7}a\,\hat{\mathbf{x}}+\left(\frac{1}{2}-y_{7}\right)a\,\hat{\mathbf{y}}+\left(\frac{1}{2}-z_{7}\right)c\,\hat{\mathbf{z}}$ | (16*k*) | Ca III |
| $\mathbf{B}_{51}$ | = | $y_{7}\,\mathbf{a}_{1}+x_{7}\,\mathbf{a}_{2}+\left(\frac{1}{2}-z_{7}\right)\,\mathbf{a}_{3}$ | = | $y_{7}a\,\hat{\mathbf{x}}+x_{7}a\,\hat{\mathbf{y}}+\left(\frac{1}{2}-z_{7}\right)c\,\hat{\mathbf{z}}$ | (16*k*) | Ca III |
| $\mathbf{B}_{52}$ | = | $\left(\frac{1}{2}-y_{7}\right)\,\mathbf{a}_{1}+\left(\frac{1}{2}-x_{7}\right)\,\mathbf{a}_{2}+\left(\frac{1}{2}-z_{7}\right)\,\mathbf{a}_{3}$ | = | $\left(\frac{1}{2}-y_{7}\right)a\,\hat{\mathbf{x}}+\left(\frac{1}{2}-x_{7}\right)a\,\hat{\mathbf{y}}+\left(\frac{1}{2}-z_{7}\right)c\,\hat{\mathbf{z}}$ | (16*k*) | Ca III |
| $\mathbf{B}_{53}$ | = | $-x_{7}\,\mathbf{a}_{1}-y_{7}\,\mathbf{a}_{2}-z_{7}\,\mathbf{a}_{3}$ | = | $-x_{7}a\,\hat{\mathbf{x}}-y_{7}a\,\hat{\mathbf{y}}-z_{7}c\,\hat{\mathbf{z}}$ | (16*k*) | Ca III |
| $\mathbf{B}_{54}$ | = | $\left(\frac{1}{2}+x_{7}\right)\,\mathbf{a}_{1}+\left(\frac{1}{2}+y_{7}\right)\,\mathbf{a}_{2}-z_{7}\,\mathbf{a}_{3}$ | = | $\left(\frac{1}{2}+x_{7}\right)a\,\hat{\mathbf{x}}+\left(\frac{1}{2}+y_{7}\right)a\,\hat{\mathbf{y}}-z_{7}c\,\hat{\mathbf{z}}$ | (16*k*) | Ca III |
| $\mathbf{B}_{55}$ | = | $\left(\frac{1}{2}+y_{7}\right)\,\mathbf{a}_{1}-x_{7}\,\mathbf{a}_{2}-z_{7}\,\mathbf{a}_{3}$ | = | $\left(\frac{1}{2}+y_{7}\right)a\,\hat{\mathbf{x}}-x_{7}a\,\hat{\mathbf{y}}-z_{7}c\,\hat{\mathbf{z}}$ | (16*k*) | Ca III |
| $\mathbf{B}_{56}$ | = | $-y_{7}\,\mathbf{a}_{1}+\left(\frac{1}{2}+x_{7}\right)\,\mathbf{a}_{2}-z_{7}\,\mathbf{a}_{3}$ | = | $-y_{7}a\,\hat{\mathbf{x}}+\left(\frac{1}{2}+x_{7}\right)a\,\hat{\mathbf{y}}-z_{7}c\,\hat{\mathbf{z}}$ | (16*k*) | Ca III |
| $\mathbf{B}_{57}$ | = | $\left(\frac{1}{2}+x_{7}\right)\,\mathbf{a}_{1}-y_{7}\,\mathbf{a}_{2}+\left(\frac{1}{2}+z_{7}\right)\,\mathbf{a}_{3}$ | = | $\left(\frac{1}{2}+x_{7}\right)a\,\hat{\mathbf{x}}-y_{7}a\,\hat{\mathbf{y}}+\left(\frac{1}{2}+z_{7}\right)c\,\hat{\mathbf{z}}$ | (16*k*) | Ca III |
| $\mathbf{B}_{58}$ | = | $-x_{7}\,\mathbf{a}_{1}+\left(\frac{1}{2}+y_{7}\right)\,\mathbf{a}_{2}+\left(\frac{1}{2}+z_{7}\right)\,\mathbf{a}_{3}$ | = | $-x_{7}a\,\hat{\mathbf{x}}+\left(\frac{1}{2}+y_{7}\right)a\,\hat{\mathbf{y}}+\left(\frac{1}{2}+z_{7}\right)c\,\hat{\mathbf{z}}$ | (16*k*) | Ca III |
| $\mathbf{B}_{59}$ | = | $-y_{7}\,\mathbf{a}_{1}-x_{7}\,\mathbf{a}_{2}+\left(\frac{1}{2}+z_{7}\right)\,\mathbf{a}_{3}$ | = | $-y_{7}a\,\hat{\mathbf{x}}-x_{7}a\,\hat{\mathbf{y}}+\left(\frac{1}{2}+z_{7}\right)c\,\hat{\mathbf{z}}$ | (16*k*) | Ca III |
| $\mathbf{B}_{60}$ | = | $\left(\frac{1}{2}+y_{7}\right)\,\mathbf{a}_{1}+\left(\frac{1}{2}+x_{7}\right)\,\mathbf{a}_{2}+\left(\frac{1}{2}+z_{7}\right)\,\mathbf{a}_{3}$ | = | $\left(\frac{1}{2}+y_{7}\right)a\,\hat{\mathbf{x}}+\left(\frac{1}{2}+x_{7}\right)a\,\hat{\mathbf{y}}+\left(\frac{1}{2}+z_{7}\right)c\,\hat{\mathbf{z}}$ | (16*k*) | Ca III |
| $\mathbf{B}_{61}$ | = | $x_8 \, \mathbf{a}_1 + y_8 \, \mathbf{a}_2 + z_8 \, \mathbf{a}_3$ | = | $x_8 a \, \mathbf{\hat{x}} + y_8 a \, \mathbf{\hat{y}} + z_8 c \, \mathbf{\hat{z}}$ | (16$k$) | Ca IV |
| $\mathbf{B}_{62}$ | = | $\left(\frac{1}{2} - x_8\right) \mathbf{a}_1 + \left(\frac{1}{2} - y_8\right) \mathbf{a}_2 + z_8 \, \mathbf{a}_3$ | = | $\left(\frac{1}{2} - x_8\right) a \, \mathbf{\hat{x}} + \left(\frac{1}{2} - y_8\right) a \, \mathbf{\hat{y}} + z_8 c \, \mathbf{\hat{z}}$ | (16$k$) | Ca IV |
| $\mathbf{B}_{63}$ | = | $\left(\frac{1}{2} - y_8\right) \mathbf{a}_1 + x_8 \, \mathbf{a}_2 + z_8 \, \mathbf{a}_3$ | = | $\left(\frac{1}{2} - y_8\right) a \, \mathbf{\hat{x}} + x_8 a \, \mathbf{\hat{y}} + z_8 c \, \mathbf{\hat{z}}$ | (16$k$) | Ca IV |
| $\mathbf{B}_{64}$ | = | $y_8 \, \mathbf{a}_1 + \left(\frac{1}{2} - x_8\right) \mathbf{a}_2 + z_8 \, \mathbf{a}_3$ | = | $y_8 a \, \mathbf{\hat{x}} + \left(\frac{1}{2} - x_8\right) a \, \mathbf{\hat{y}} + z_8 c \, \mathbf{\hat{z}}$ | (16$k$) | Ca IV |
| $\mathbf{B}_{65}$ | = | $\left(\frac{1}{2} - x_8\right) \mathbf{a}_1 + y_8 \, \mathbf{a}_2 + \left(\frac{1}{2} - z_8\right) \mathbf{a}_3$ | = | $\left(\frac{1}{2} - x_8\right) a \, \mathbf{\hat{x}} + y_8 a \, \mathbf{\hat{y}} + \left(\frac{1}{2} - z_8\right) c \, \mathbf{\hat{z}}$ | (16$k$) | Ca IV |
| $\mathbf{B}_{66}$ | = | $x_8 \, \mathbf{a}_1 + \left(\frac{1}{2} - y_8\right) \mathbf{a}_2 + \left(\frac{1}{2} - z_8\right) \mathbf{a}_3$ | = | $x_8 a \, \mathbf{\hat{x}} + \left(\frac{1}{2} - y_8\right) a \, \mathbf{\hat{y}} + \left(\frac{1}{2} - z_8\right) c \, \mathbf{\hat{z}}$ | (16$k$) | Ca IV |
| $\mathbf{B}_{67}$ | = | $y_8 \, \mathbf{a}_1 + x_8 \, \mathbf{a}_2 + \left(\frac{1}{2} - z_8\right) \mathbf{a}_3$ | = | $y_8 a \, \mathbf{\hat{x}} + x_8 a \, \mathbf{\hat{y}} + \left(\frac{1}{2} - z_8\right) c \, \mathbf{\hat{z}}$ | (16$k$) | Ca IV |
| $\mathbf{B}_{68}$ | = | $\left(\frac{1}{2} - y_8\right) \mathbf{a}_1 + \left(\frac{1}{2} - x_8\right) \mathbf{a}_2 + \left(\frac{1}{2} - z_8\right) \mathbf{a}_3$ | = | $\left(\frac{1}{2} - y_8\right) a \, \mathbf{\hat{x}} + \left(\frac{1}{2} - x_8\right) a \, \mathbf{\hat{y}} + \left(\frac{1}{2} - z_8\right) c \, \mathbf{\hat{z}}$ | (16$k$) | Ca IV |
| $\mathbf{B}_{69}$ | = | $-x_8 \, \mathbf{a}_1 - y_8 \, \mathbf{a}_2 - z_8 \, \mathbf{a}_3$ | = | $-x_8 a \, \mathbf{\hat{x}} - y_8 a \, \mathbf{\hat{y}} - z_8 c \, \mathbf{\hat{z}}$ | (16$k$) | Ca IV |
| $\mathbf{B}_{70}$ | = | $\left(\frac{1}{2} + x_8\right) \mathbf{a}_1 + \left(\frac{1}{2} + y_8\right) \mathbf{a}_2 - z_8 \, \mathbf{a}_3$ | = | $\left(\frac{1}{2} + x_8\right) a \, \mathbf{\hat{x}} + \left(\frac{1}{2} + y_8\right) a \, \mathbf{\hat{y}} - z_8 c \, \mathbf{\hat{z}}$ | (16$k$) | Ca IV |
| $\mathbf{B}_{71}$ | = | $\left(\frac{1}{2} + y_8\right) \mathbf{a}_1 - x_8 \, \mathbf{a}_2 - z_8 \, \mathbf{a}_3$ | = | $\left(\frac{1}{2} + y_8\right) a \, \mathbf{\hat{x}} - x_8 a \, \mathbf{\hat{y}} - z_8 c \, \mathbf{\hat{z}}$ | (16$k$) | Ca IV |
| $\mathbf{B}_{72}$ | = | $-y_8 \, \mathbf{a}_1 + \left(\frac{1}{2} + x_8\right) \mathbf{a}_2 - z_8 \, \mathbf{a}_3$ | = | $-y_8 a \, \mathbf{\hat{x}} + \left(\frac{1}{2} + x_8\right) a \, \mathbf{\hat{y}} - z_8 c \, \mathbf{\hat{z}}$ | (16$k$) | Ca IV |
| $\mathbf{B}_{73}$ | = | $\left(\frac{1}{2} + x_8\right) \mathbf{a}_1 - y_8 \, \mathbf{a}_2 + \left(\frac{1}{2} + z_8\right) \mathbf{a}_3$ | = | $\left(\frac{1}{2} + x_8\right) a \, \mathbf{\hat{x}} - y_8 a \, \mathbf{\hat{y}} + \left(\frac{1}{2} + z_8\right) c \, \mathbf{\hat{z}}$ | (16$k$) | Ca IV |
| $\mathbf{B}_{74}$ | = | $-x_8 \, \mathbf{a}_1 + \left(\frac{1}{2} + y_8\right) \mathbf{a}_2 + \left(\frac{1}{2} + z_8\right) \mathbf{a}_3$ | = | $-x_8 a \, \mathbf{\hat{x}} + \left(\frac{1}{2} + y_8\right) a \, \mathbf{\hat{y}} + \left(\frac{1}{2} + z_8\right) c \, \mathbf{\hat{z}}$ | (16$k$) | Ca IV |
| $\mathbf{B}_{75}$ | = | $-y_8 \, \mathbf{a}_1 - x_8 \, \mathbf{a}_2 + \left(\frac{1}{2} + z_8\right) \mathbf{a}_3$ | = | $-y_8 a \, \mathbf{\hat{x}} - x_8 a \, \mathbf{\hat{y}} + \left(\frac{1}{2} + z_8\right) c \, \mathbf{\hat{z}}$ | (16$k$) | Ca IV |
| $\mathbf{B}_{76}$ | = | $\left(\frac{1}{2} + y_8\right) \mathbf{a}_1 + \left(\frac{1}{2} + x_8\right) \mathbf{a}_2 + \left(\frac{1}{2} + z_8\right) \mathbf{a}_3$ | = | $\left(\frac{1}{2} + y_8\right) a \, \mathbf{\hat{x}} + \left(\frac{1}{2} + x_8\right) a \, \mathbf{\hat{y}} + \left(\frac{1}{2} + z_8\right) c \, \mathbf{\hat{z}}$ | (16$k$) | Ca IV |
| $\mathbf{B}_{77}$ | = | $x_9 \, \mathbf{a}_1 + y_9 \, \mathbf{a}_2 + z_9 \, \mathbf{a}_3$ | = | $x_9 a \, \mathbf{\hat{x}} + y_9 a \, \mathbf{\hat{y}} + z_9 c \, \mathbf{\hat{z}}$ | (16$k$) | O II |
| $\mathbf{B}_{78}$ | = | $\left(\frac{1}{2} - x_9\right) \mathbf{a}_1 + \left(\frac{1}{2} - y_9\right) \mathbf{a}_2 + z_9 \, \mathbf{a}_3$ | = | $\left(\frac{1}{2} - x_9\right) a \, \mathbf{\hat{x}} + \left(\frac{1}{2} - y_9\right) a \, \mathbf{\hat{y}} + z_9 c \, \mathbf{\hat{z}}$ | (16$k$) | O II |
| $\mathbf{B}_{79}$ | = | $\left(\frac{1}{2} - y_9\right) \mathbf{a}_1 + x_9 \, \mathbf{a}_2 + z_9 \, \mathbf{a}_3$ | = | $\left(\frac{1}{2} - y_9\right) a \, \mathbf{\hat{x}} + x_9 a \, \mathbf{\hat{y}} + z_9 c \, \mathbf{\hat{z}}$ | (16$k$) | O II |
| $\mathbf{B}_{80}$ | = | $y_9 \, \mathbf{a}_1 + \left(\frac{1}{2} - x_9\right) \mathbf{a}_2 + z_9 \, \mathbf{a}_3$ | = | $y_9 a \, \mathbf{\hat{x}} + \left(\frac{1}{2} - x_9\right) a \, \mathbf{\hat{y}} + z_9 c \, \mathbf{\hat{z}}$ | (16$k$) | O II |
| $\mathbf{B}_{81}$ | = | $\left(\frac{1}{2} - x_9\right) \mathbf{a}_1 + y_9 \, \mathbf{a}_2 + \left(\frac{1}{2} - z_9\right) \mathbf{a}_3$ | = | $\left(\frac{1}{2} - x_9\right) a \, \mathbf{\hat{x}} + y_9 a \, \mathbf{\hat{y}} + \left(\frac{1}{2} - z_9\right) c \, \mathbf{\hat{z}}$ | (16$k$) | O II |
| $\mathbf{B}_{82}$ | = | $x_9 \, \mathbf{a}_1 + \left(\frac{1}{2} - y_9\right) \mathbf{a}_2 + \left(\frac{1}{2} - z_9\right) \mathbf{a}_3$ | = | $x_9 a \, \mathbf{\hat{x}} + \left(\frac{1}{2} - y_9\right) a \, \mathbf{\hat{y}} + \left(\frac{1}{2} - z_9\right) c \, \mathbf{\hat{z}}$ | (16$k$) | O II |
| $\mathbf{B}_{83}$ | = | $y_9 \, \mathbf{a}_1 + x_9 \, \mathbf{a}_2 + \left(\frac{1}{2} - z_9\right) \mathbf{a}_3$ | = | $y_9 a \, \mathbf{\hat{x}} + x_9 a \, \mathbf{\hat{y}} + \left(\frac{1}{2} - z_9\right) c \, \mathbf{\hat{z}}$ | (16$k$) | O II |
| $\mathbf{B}_{84}$ | = | $\left(\frac{1}{2} - y_9\right) \mathbf{a}_1 + \left(\frac{1}{2} - x_9\right) \mathbf{a}_2 + \left(\frac{1}{2} - z_9\right) \mathbf{a}_3$ | = | $\left(\frac{1}{2} - y_9\right) a \, \mathbf{\hat{x}} + \left(\frac{1}{2} - x_9\right) a \, \mathbf{\hat{y}} + \left(\frac{1}{2} - z_9\right) c \, \mathbf{\hat{z}}$ | (16$k$) | O II |
| $\mathbf{B}_{85}$ | = | $-x_9 \, \mathbf{a}_1 - y_9 \, \mathbf{a}_2 - z_9 \, \mathbf{a}_3$ | = | $-x_9 a \, \mathbf{\hat{x}} - y_9 a \, \mathbf{\hat{y}} - z_9 c \, \mathbf{\hat{z}}$ | (16$k$) | O II |
| $\mathbf{B}_{86}$ | = | $\left(\frac{1}{2} + x_9\right) \mathbf{a}_1 + \left(\frac{1}{2} + y_9\right) \mathbf{a}_2 - z_9 \, \mathbf{a}_3$ | = | $\left(\frac{1}{2} + x_9\right) a \, \mathbf{\hat{x}} + \left(\frac{1}{2} + y_9\right) a \, \mathbf{\hat{y}} - z_9 c \, \mathbf{\hat{z}}$ | (16$k$) | O II |
| $\mathbf{B}_{87}$ | = | $\left(\frac{1}{2} + y_9\right) \mathbf{a}_1 - x_9 \, \mathbf{a}_2 - z_9 \, \mathbf{a}_3$ | = | $\left(\frac{1}{2} + y_9\right) a \, \mathbf{\hat{x}} - x_9 a \, \mathbf{\hat{y}} - z_9 c \, \mathbf{\hat{z}}$ | (16$k$) | O II |
| $\mathbf{B}_{88}$ | = | $-y_9 \, \mathbf{a}_1 + \left(\frac{1}{2} + x_9\right) \mathbf{a}_2 - z_9 \, \mathbf{a}_3$ | = | $-y_9 a \, \mathbf{\hat{x}} + \left(\frac{1}{2} + x_9\right) a \, \mathbf{\hat{y}} - z_9 c \, \mathbf{\hat{z}}$ | (16$k$) | O II |
| $\mathbf{B}_{89}$ | = | $\left(\frac{1}{2} + x_9\right) \mathbf{a}_1 - y_9 \, \mathbf{a}_2 + \left(\frac{1}{2} + z_9\right) \mathbf{a}_3$ | = | $\left(\frac{1}{2} + x_9\right) a \, \mathbf{\hat{x}} - y_9 a \, \mathbf{\hat{y}} + \left(\frac{1}{2} + z_9\right) c \, \mathbf{\hat{z}}$ | (16$k$) | O II |
| $\mathbf{B}_{90}$ | = | $-x_9 \, \mathbf{a}_1 + \left(\frac{1}{2} + y_9\right) \mathbf{a}_2 + \left(\frac{1}{2} + z_9\right) \mathbf{a}_3$ | = | $-x_9 a \, \mathbf{\hat{x}} + \left(\frac{1}{2} + y_9\right) a \, \mathbf{\hat{y}} + \left(\frac{1}{2} + z_9\right) c \, \mathbf{\hat{z}}$ | (16$k$) | O II |
| $\mathbf{B}_{91}$ | = | $-y_9 \, \mathbf{a}_1 - x_9 \, \mathbf{a}_2 + \left(\frac{1}{2} + z_9\right) \mathbf{a}_3$ | = | $-y_9 a \, \mathbf{\hat{x}} - x_9 a \, \mathbf{\hat{y}} + \left(\frac{1}{2} + z_9\right) c \, \mathbf{\hat{z}}$ | (16$k$) | O II |
| $\mathbf{B}_{92}$ | = | $\left(\frac{1}{2} + y_9\right) \mathbf{a}_1 + \left(\frac{1}{2} + x_9\right) \mathbf{a}_2 + \left(\frac{1}{2} + z_9\right) \mathbf{a}_3$ | = | $\left(\frac{1}{2} + y_9\right) a \, \mathbf{\hat{x}} + \left(\frac{1}{2} + x_9\right) a \, \mathbf{\hat{y}} + \left(\frac{1}{2} + z_9\right) c \, \mathbf{\hat{z}}$ | (16$k$) | O II |
| $\mathbf{B}_{93}$ | = | $x_{10} \, \mathbf{a}_1 + y_{10} \, \mathbf{a}_2 + z_{10} \, \mathbf{a}_3$ | = | $x_{10} a \, \mathbf{\hat{x}} + y_{10} a \, \mathbf{\hat{y}} + z_{10} c \, \mathbf{\hat{z}}$ | (16$k$) | O III |
| $\mathbf{B}_{94}$ | = | $\left(\frac{1}{2} - x_{10}\right) \mathbf{a}_1 + \left(\frac{1}{2} - y_{10}\right) \mathbf{a}_2 + z_{10} \mathbf{a}_3$ | = | $\left(\frac{1}{2} - x_{10}\right) a \hat{\mathbf{x}} + \left(\frac{1}{2} - y_{10}\right) a \hat{\mathbf{y}} + z_{10} c \hat{\mathbf{z}}$ | (16$k$) | O III |
| $\mathbf{B}_{95}$ | = | $\left(\frac{1}{2} - y_{10}\right) \mathbf{a}_1 + x_{10} \mathbf{a}_2 + z_{10} \mathbf{a}_3$ | = | $\left(\frac{1}{2} - y_{10}\right) a \hat{\mathbf{x}} + x_{10} a \hat{\mathbf{y}} + z_{10} c \hat{\mathbf{z}}$ | (16$k$) | O III |
| $\mathbf{B}_{96}$ | = | $y_{10} \mathbf{a}_1 + \left(\frac{1}{2} - x_{10}\right) \mathbf{a}_2 + z_{10} \mathbf{a}_3$ | = | $y_{10} a \hat{\mathbf{x}} + \left(\frac{1}{2} - x_{10}\right) a \hat{\mathbf{y}} + z_{10} c \hat{\mathbf{z}}$ | (16$k$) | O III |
| $\mathbf{B}_{97}$ | = | $\left(\frac{1}{2} - x_{10}\right) \mathbf{a}_1 + y_{10} \mathbf{a}_2 + \left(\frac{1}{2} - z_{10}\right) \mathbf{a}_3$ | = | $\left(\frac{1}{2} - x_{10}\right) a \hat{\mathbf{x}} + y_{10} a \hat{\mathbf{y}} + \left(\frac{1}{2} - z_{10}\right) c \hat{\mathbf{z}}$ | (16$k$) | O III |
| $\mathbf{B}_{98}$ | = | $x_{10} \mathbf{a}_1 + \left(\frac{1}{2} - y_{10}\right) \mathbf{a}_2 + \left(\frac{1}{2} - z_{10}\right) \mathbf{a}_3$ | = | $x_{10} a \hat{\mathbf{x}} + \left(\frac{1}{2} - y_{10}\right) a \hat{\mathbf{y}} + \left(\frac{1}{2} - z_{10}\right) c \hat{\mathbf{z}}$ | (16$k$) | O III |
| $\mathbf{B}_{99}$ | = | $y_{10} \mathbf{a}_1 + x_{10} \mathbf{a}_2 + \left(\frac{1}{2} - z_{10}\right) \mathbf{a}_3$ | = | $y_{10} a \hat{\mathbf{x}} + x_{10} a \hat{\mathbf{y}} + \left(\frac{1}{2} - z_{10}\right) c \hat{\mathbf{z}}$ | (16$k$) | O III |
| $\mathbf{B}_{100}$ | = | $\left(\frac{1}{2} - y_{10}\right) \mathbf{a}_1 + \left(\frac{1}{2} - x_{10}\right) \mathbf{a}_2 + \left(\frac{1}{2} - z_{10}\right) \mathbf{a}_3$ | = | $\left(\frac{1}{2} - y_{10}\right) a \hat{\mathbf{x}} + \left(\frac{1}{2} - x_{10}\right) a \hat{\mathbf{y}} + \left(\frac{1}{2} - z_{10}\right) c \hat{\mathbf{z}}$ | (16$k$) | O III |
| $\mathbf{B}_{101}$ | = | $-x_{10} \mathbf{a}_1 - y_{10} \mathbf{a}_2 - z_{10} \mathbf{a}_3$ | = | $-x_{10} a \hat{\mathbf{x}} - y_{10} a \hat{\mathbf{y}} - z_{10} c \hat{\mathbf{z}}$ | (16$k$) | O III |
| $\mathbf{B}_{102}$ | = | $\left(\frac{1}{2} + x_{10}\right) \mathbf{a}_1 + \left(\frac{1}{2} + y_{10}\right) \mathbf{a}_2 - z_{10} \mathbf{a}_3$ | = | $\left(\frac{1}{2} + x_{10}\right) a \hat{\mathbf{x}} + \left(\frac{1}{2} + y_{10}\right) a \hat{\mathbf{y}} - z_{10} c \hat{\mathbf{z}}$ | (16$k$) | O III |
| $\mathbf{B}_{103}$ | = | $\left(\frac{1}{2} + y_{10}\right) \mathbf{a}_1 - x_{10} \mathbf{a}_2 - z_{10} \mathbf{a}_3$ | = | $\left(\frac{1}{2} + y_{10}\right) a \hat{\mathbf{x}} - x_{10} a \hat{\mathbf{y}} - z_{10} c \hat{\mathbf{z}}$ | (16$k$) | O III |
| $\mathbf{B}_{104}$ | = | $-y_{10} \mathbf{a}_1 + \left(\frac{1}{2} + x_{10}\right) \mathbf{a}_2 - z_{10} \mathbf{a}_3$ | = | $-y_{10} a \hat{\mathbf{x}} + \left(\frac{1}{2} + x_{10}\right) a \hat{\mathbf{y}} - z_{10} c \hat{\mathbf{z}}$ | (16$k$) | O III |
| $\mathbf{B}_{105}$ | = | $\left(\frac{1}{2} + x_{10}\right) \mathbf{a}_1 - y_{10} \mathbf{a}_2 + \left(\frac{1}{2} + z_{10}\right) \mathbf{a}_3$ | = | $\left(\frac{1}{2} + x_{10}\right) a \hat{\mathbf{x}} - y_{10} a \hat{\mathbf{y}} + \left(\frac{1}{2} + z_{10}\right) c \hat{\mathbf{z}}$ | (16$k$) | O III |
| $\mathbf{B}_{106}$ | = | $-x_{10} \mathbf{a}_1 + \left(\frac{1}{2} + y_{10}\right) \mathbf{a}_2 + \left(\frac{1}{2} + z_{10}\right) \mathbf{a}_3$ | = | $-x_{10} a \hat{\mathbf{x}} + \left(\frac{1}{2} + y_{10}\right) a \hat{\mathbf{y}} + \left(\frac{1}{2} + z_{10}\right) c \hat{\mathbf{z}}$ | (16$k$) | O III |
| $\mathbf{B}_{107}$ | = | $-y_{10} \mathbf{a}_1 - x_{10} \mathbf{a}_2 + \left(\frac{1}{2} + z_{10}\right) \mathbf{a}_3$ | = | $-y_{10} a \hat{\mathbf{x}} - x_{10} a \hat{\mathbf{y}} + \left(\frac{1}{2} + z_{10}\right) c \hat{\mathbf{z}}$ | (16$k$) | O III |
| $\mathbf{B}_{108}$ | = | $\left(\frac{1}{2} + y_{10}\right) \mathbf{a}_1 + \left(\frac{1}{2} + x_{10}\right) \mathbf{a}_2 + \left(\frac{1}{2} + z_{10}\right) \mathbf{a}_3$ | = | $\left(\frac{1}{2} + y_{10}\right) a \hat{\mathbf{x}} + \left(\frac{1}{2} + x_{10}\right) a \hat{\mathbf{y}} + \left(\frac{1}{2} + z_{10}\right) c \hat{\mathbf{z}}$ | (16$k$) | O III |
| $\mathbf{B}_{109}$ | = | $x_{11} \mathbf{a}_1 + y_{11} \mathbf{a}_2 + z_{11} \mathbf{a}_3$ | = | $x_{11} a \hat{\mathbf{x}} + y_{11} a \hat{\mathbf{y}} + z_{11} c \hat{\mathbf{z}}$ | (16$k$) | O IV |
| $\mathbf{B}_{110}$ | = | $\left(\frac{1}{2} - x_{11}\right) \mathbf{a}_1 + \left(\frac{1}{2} - y_{11}\right) \mathbf{a}_2 + z_{11} \mathbf{a}_3$ | = | $\left(\frac{1}{2} - x_{11}\right) a \hat{\mathbf{x}} + \left(\frac{1}{2} - y_{11}\right) a \hat{\mathbf{y}} + z_{11} c \hat{\mathbf{z}}$ | (16$k$) | O IV |
| $\mathbf{B}_{111}$ | = | $\left(\frac{1}{2} - y_{11}\right) \mathbf{a}_1 + x_{11} \mathbf{a}_2 + z_{11} \mathbf{a}_3$ | = | $\left(\frac{1}{2} - y_{11}\right) a \hat{\mathbf{x}} + x_{11} a \hat{\mathbf{y}} + z_{11} c \hat{\mathbf{z}}$ | (16$k$) | O IV |
| $\mathbf{B}_{112}$ | = | $y_{11} \mathbf{a}_1 + \left(\frac{1}{2} - x_{11}\right) \mathbf{a}_2 + z_{11} \mathbf{a}_3$ | = | $y_{11} a \hat{\mathbf{x}} + \left(\frac{1}{2} - x_{11}\right) a \hat{\mathbf{y}} + z_{11} c \hat{\mathbf{z}}$ | (16$k$) | O IV |
| $\mathbf{B}_{113}$ | = | $\left(\frac{1}{2} - x_{11}\right) \mathbf{a}_1 + y_{11} \mathbf{a}_2 + \left(\frac{1}{2} - z_{11}\right) \mathbf{a}_3$ | = | $\left(\frac{1}{2} - x_{11}\right) a \hat{\mathbf{x}} + y_{11} a \hat{\mathbf{y}} + \left(\frac{1}{2} - z_{11}\right) c \hat{\mathbf{z}}$ | (16$k$) | O IV |
| $\mathbf{B}_{114}$ | = | $x_{11} \mathbf{a}_1 + \left(\frac{1}{2} - y_{11}\right) \mathbf{a}_2 + \left(\frac{1}{2} - z_{11}\right) \mathbf{a}_3$ | = | $x_{11} a \hat{\mathbf{x}} + \left(\frac{1}{2} - y_{11}\right) a \hat{\mathbf{y}} + \left(\frac{1}{2} - z_{11}\right) c \hat{\mathbf{z}}$ | (16$k$) | O IV |
| $\mathbf{B}_{115}$ | = | $y_{11} \mathbf{a}_1 + x_{11} \mathbf{a}_2 + \left(\frac{1}{2} - z_{11}\right) \mathbf{a}_3$ | = | $y_{11} a \hat{\mathbf{x}} + x_{11} a \hat{\mathbf{y}} + \left(\frac{1}{2} - z_{11}\right) c \hat{\mathbf{z}}$ | (16$k$) | O IV |
| $\mathbf{B}_{116}$ | = | $\left(\frac{1}{2} - y_{11}\right) \mathbf{a}_1 + \left(\frac{1}{2} - x_{11}\right) \mathbf{a}_2 + \left(\frac{1}{2} - z_{11}\right) \mathbf{a}_3$ | = | $\left(\frac{1}{2} - y_{11}\right) a \hat{\mathbf{x}} + \left(\frac{1}{2} - x_{11}\right) a \hat{\mathbf{y}} + \left(\frac{1}{2} - z_{11}\right) c \hat{\mathbf{z}}$ | (16$k$) | O IV |
| $\mathbf{B}_{117}$ | = | $-x_{11} \mathbf{a}_1 - y_{11} \mathbf{a}_2 - z_{11} \mathbf{a}_3$ | = | $-x_{11} a \hat{\mathbf{x}} - y_{11} a \hat{\mathbf{y}} - z_{11} c \hat{\mathbf{z}}$ | (16$k$) | O IV |
| $\mathbf{B}_{118}$ | = | $\left(\frac{1}{2} + x_{11}\right) \mathbf{a}_1 + \left(\frac{1}{2} + y_{11}\right) \mathbf{a}_2 - z_{11} \mathbf{a}_3$ | = | $\left(\frac{1}{2} + x_{11}\right) a \hat{\mathbf{x}} + \left(\frac{1}{2} + y_{11}\right) a \hat{\mathbf{y}} - z_{11} c \hat{\mathbf{z}}$ | (16$k$) | O IV |
| $\mathbf{B}_{119}$ | = | $\left(\frac{1}{2} + y_{11}\right) \mathbf{a}_1 - x_{11} \mathbf{a}_2 - z_{11} \mathbf{a}_3$ | = | $\left(\frac{1}{2} + y_{11}\right) a \hat{\mathbf{x}} - x_{11} a \hat{\mathbf{y}} - z_{11} c \hat{\mathbf{z}}$ | (16$k$) | O IV |
| $\mathbf{B}_{120}$ | = | $-y_{11} \mathbf{a}_1 + \left(\frac{1}{2} + x_{11}\right) \mathbf{a}_2 - z_{11} \mathbf{a}_3$ | = | $-y_{11} a \hat{\mathbf{x}} + \left(\frac{1}{2} + x_{11}\right) a \hat{\mathbf{y}} - z_{11} c \hat{\mathbf{z}}$ | (16$k$) | O IV |
| $\mathbf{B}_{121}$ | = | $\left(\frac{1}{2} + x_{11}\right) \mathbf{a}_1 - y_{11} \mathbf{a}_2 + \left(\frac{1}{2} + z_{11}\right) \mathbf{a}_3$ | = | $\left(\frac{1}{2} + x_{11}\right) a \hat{\mathbf{x}} - y_{11} a \hat{\mathbf{y}} + \left(\frac{1}{2} + z_{11}\right) c \hat{\mathbf{z}}$ | (16$k$) | O IV |
| $\mathbf{B}_{122}$ | = | $-x_{11} \mathbf{a}_1 + \left(\frac{1}{2} + y_{11}\right) \mathbf{a}_2 + \left(\frac{1}{2} + z_{11}\right) \mathbf{a}_3$ | = | $-x_{11} a \hat{\mathbf{x}} + \left(\frac{1}{2} + y_{11}\right) a \hat{\mathbf{y}} + \left(\frac{1}{2} + z_{11}\right) c \hat{\mathbf{z}}$ | (16$k$) | O IV |
| $\mathbf{B}_{123}$ | = | $-y_{11} \mathbf{a}_1 - x_{11} \mathbf{a}_2 + \left(\frac{1}{2} + z_{11}\right) \mathbf{a}_3$ | = | $-y_{11} a \hat{\mathbf{x}} - x_{11} a \hat{\mathbf{y}} + \left(\frac{1}{2} + z_{11}\right) c \hat{\mathbf{z}}$ | (16$k$) | O IV |
| $\mathbf{B}_{124}$ | = | $\left(\frac{1}{2} + y_{11}\right) \mathbf{a}_1 + \left(\frac{1}{2} + x_{11}\right) \mathbf{a}_2 + \left(\frac{1}{2} + z_{11}\right) \mathbf{a}_3$ | = | $\left(\frac{1}{2} + y_{11}\right) a \hat{\mathbf{x}} + \left(\frac{1}{2} + x_{11}\right) a \hat{\mathbf{y}} + \left(\frac{1}{2} + z_{11}\right) c \hat{\mathbf{z}}$ | (16$k$) | O IV |
| $\mathbf{B}_{125}$ | = | $x_{12} \mathbf{a}_1 + y_{12} \mathbf{a}_2 + z_{12} \mathbf{a}_3$ | = | $x_{12} a \hat{\mathbf{x}} + y_{12} a \hat{\mathbf{y}} + z_{12} c \hat{\mathbf{z}}$ | (16$k$) | O V |
| $\mathbf{B}_{126}$ | = | $\left(\frac{1}{2} - x_{12}\right) \mathbf{a}_1 + \left(\frac{1}{2} - y_{12}\right) \mathbf{a}_2 + z_{12} \mathbf{a}_3$ | = | $\left(\frac{1}{2} - x_{12}\right) a \hat{\mathbf{x}} + \left(\frac{1}{2} - y_{12}\right) a \hat{\mathbf{y}} + z_{12} c \hat{\mathbf{z}}$ | (16$k$) | O V |
| $\mathbf{B}_{127}$ | = | $\left(\frac{1}{2} - y_{12}\right) \mathbf{a}_1 + x_{12} \mathbf{a}_2 + z_{12} \mathbf{a}_3$ | = | $\left(\frac{1}{2} - y_{12}\right) a \hat{\mathbf{x}} + x_{12} a \hat{\mathbf{y}} + z_{12} c \hat{\mathbf{z}}$ | (16$k$) | O V |
| $\mathbf{B}_{128}$ | = | $y_{12} \mathbf{a}_1 + \left(\frac{1}{2} - x_{12}\right) \mathbf{a}_2 + z_{12} \mathbf{a}_3$ | = | $y_{12} a \hat{\mathbf{x}} + \left(\frac{1}{2} - x_{12}\right) a \hat{\mathbf{y}} + z_{12} c \hat{\mathbf{z}}$ | (16$k$) | O V |
| $\mathbf{B}_{129}$ | = | $\left(\frac{1}{2} - x_{12}\right) \mathbf{a}_1 + y_{12} \mathbf{a}_2 + \left(\frac{1}{2} - z_{12}\right) \mathbf{a}_3$ | = | $\left(\frac{1}{2} - x_{12}\right) a \hat{\mathbf{x}} + y_{12} a \hat{\mathbf{y}} + \left(\frac{1}{2} - z_{12}\right) c \hat{\mathbf{z}}$ | (16$k$) | O V |
| $\mathbf{B}_{130}$ | = | $x_{12} \mathbf{a}_1 + \left(\frac{1}{2} - y_{12}\right) \mathbf{a}_2 + \left(\frac{1}{2} - z_{12}\right) \mathbf{a}_3$ | = | $x_{12} a \hat{\mathbf{x}} + \left(\frac{1}{2} - y_{12}\right) a \hat{\mathbf{y}} + \left(\frac{1}{2} - z_{12}\right) c \hat{\mathbf{z}}$ | (16$k$) | O V |
| $\mathbf{B}_{131}$ | = | $y_{12} \mathbf{a}_1 + x_{12} \mathbf{a}_2 + \left(\frac{1}{2} - z_{12}\right) \mathbf{a}_3$ | = | $y_{12} a \hat{\mathbf{x}} + x_{12} a \hat{\mathbf{y}} + \left(\frac{1}{2} - z_{12}\right) c \hat{\mathbf{z}}$ | (16$k$) | O V |
| $\mathbf{B}_{132}$ | = | $\left(\frac{1}{2} - y_{12}\right) \mathbf{a}_1 + \left(\frac{1}{2} - x_{12}\right) \mathbf{a}_2 + \left(\frac{1}{2} - z_{12}\right) \mathbf{a}_3$ | = | $\left(\frac{1}{2} - y_{12}\right) a \hat{\mathbf{x}} + \left(\frac{1}{2} - x_{12}\right) a \hat{\mathbf{y}} + \left(\frac{1}{2} - z_{12}\right) c \hat{\mathbf{z}}$ | (16$k$) | O V |
| $\mathbf{B}_{133}$ | = | $-x_{12} \mathbf{a}_1 - y_{12} \mathbf{a}_2 - z_{12} \mathbf{a}_3$ | = | $-x_{12} a \hat{\mathbf{x}} - y_{12} a \hat{\mathbf{y}} - z_{12} c \hat{\mathbf{z}}$ | (16$k$) | O V |
| $\mathbf{B}_{134}$ | = | $\left(\frac{1}{2} + x_{12}\right) \mathbf{a}_1 + \left(\frac{1}{2} + y_{12}\right) \mathbf{a}_2 - z_{12} \mathbf{a}_3$ | = | $\left(\frac{1}{2} + x_{12}\right) a \hat{\mathbf{x}} + \left(\frac{1}{2} + y_{12}\right) a \hat{\mathbf{y}} - z_{12} c \hat{\mathbf{z}}$ | (16$k$) | O V |
| $\mathbf{B}_{135}$ | = | $\left(\frac{1}{2} + y_{12}\right) \mathbf{a}_1 - x_{12} \mathbf{a}_2 - z_{12} \mathbf{a}_3$ | = | $\left(\frac{1}{2} + y_{12}\right) a \hat{\mathbf{x}} - x_{12} a \hat{\mathbf{y}} - z_{12} c \hat{\mathbf{z}}$ | (16$k$) | O V |
| $\mathbf{B}_{136}$ | = | $-y_{12} \mathbf{a}_1 + \left(\frac{1}{2} + x_{12}\right) \mathbf{a}_2 - z_{12} \mathbf{a}_3$ | = | $-y_{12} a \hat{\mathbf{x}} + \left(\frac{1}{2} + x_{12}\right) a \hat{\mathbf{y}} - z_{12} c \hat{\mathbf{z}}$ | (16$k$) | O V |
| $\mathbf{B}_{137}$ | = | $\left(\frac{1}{2} + x_{12}\right) \mathbf{a}_1 - y_{12} \mathbf{a}_2 + \left(\frac{1}{2} + z_{12}\right) \mathbf{a}_3$ | = | $\left(\frac{1}{2} + x_{12}\right) a \hat{\mathbf{x}} - y_{12} a \hat{\mathbf{y}} + \left(\frac{1}{2} + z_{12}\right) c \hat{\mathbf{z}}$ | (16$k$) | O V |
| $\mathbf{B}_{138}$ | = | $-x_{12} \mathbf{a}_1 + \left(\frac{1}{2} + y_{12}\right) \mathbf{a}_2 + \left(\frac{1}{2} + z_{12}\right) \mathbf{a}_3$ | = | $-x_{12} a \hat{\mathbf{x}} + \left(\frac{1}{2} + y_{12}\right) a \hat{\mathbf{y}} + \left(\frac{1}{2} + z_{12}\right) c \hat{\mathbf{z}}$ | (16$k$) | O V |
| $\mathbf{B}_{139}$ | = | $-y_{12} \mathbf{a}_1 - x_{12} \mathbf{a}_2 + \left(\frac{1}{2} + z_{12}\right) \mathbf{a}_3$ | = | $-y_{12} a \hat{\mathbf{x}} - x_{12} a \hat{\mathbf{y}} + \left(\frac{1}{2} + z_{12}\right) c \hat{\mathbf{z}}$ | (16$k$) | O V |
| $\mathbf{B}_{140}$ | = | $\left(\frac{1}{2} + y_{12}\right) \mathbf{a}_1 + \left(\frac{1}{2} + x_{12}\right) \mathbf{a}_2 + \left(\frac{1}{2} + z_{12}\right) \mathbf{a}_3$ | = | $\left(\frac{1}{2} + y_{12}\right) a \hat{\mathbf{x}} + \left(\frac{1}{2} + x_{12}\right) a \hat{\mathbf{y}} + \left(\frac{1}{2} + z_{12}\right) c \hat{\mathbf{z}}$ | (16$k$) | O V |
| $\mathbf{B}_{141}$ | = | $x_{13} \mathbf{a}_1 + y_{13} \mathbf{a}_2 + z_{13} \mathbf{a}_3$ | = | $x_{13} a \hat{\mathbf{x}} + y_{13} a \hat{\mathbf{y}} + z_{13} c \hat{\mathbf{z}}$ | (16$k$) | O VI |
| $\mathbf{B}_{142}$ | = | $\left(\frac{1}{2} - x_{13}\right) \mathbf{a}_1 + \left(\frac{1}{2} - y_{13}\right) \mathbf{a}_2 + z_{13} \mathbf{a}_3$ | = | $\left(\frac{1}{2} - x_{13}\right) a \hat{\mathbf{x}} + \left(\frac{1}{2} - y_{13}\right) a \hat{\mathbf{y}} + z_{13} c \hat{\mathbf{z}}$ | (16$k$) | O VI |
| $\mathbf{B}_{143}$ | = | $\left(\frac{1}{2} - y_{13}\right) \mathbf{a}_1 + x_{13} \mathbf{a}_2 + z_{13} \mathbf{a}_3$ | = | $\left(\frac{1}{2} - y_{13}\right) a \hat{\mathbf{x}} + x_{13} a \hat{\mathbf{y}} + z_{13} c \hat{\mathbf{z}}$ | (16$k$) | O VI |
| $\mathbf{B}_{144}$ | = | $y_{13} \mathbf{a}_1 + \left(\frac{1}{2} - x_{13}\right) \mathbf{a}_2 + z_{13} \mathbf{a}_3$ | = | $y_{13} a \hat{\mathbf{x}} + \left(\frac{1}{2} - x_{13}\right) a \hat{\mathbf{y}} + z_{13} c \hat{\mathbf{z}}$ | (16$k$) | O VI |
| $\mathbf{B}_{145}$ | = | $\left(\frac{1}{2} - x_{13}\right) \mathbf{a}_1 + y_{13} \mathbf{a}_2 + \left(\frac{1}{2} - z_{13}\right) \mathbf{a}_3$ | = | $\left(\frac{1}{2} - x_{13}\right) a \hat{\mathbf{x}} + y_{13} a \hat{\mathbf{y}} + \left(\frac{1}{2} - z_{13}\right) c \hat{\mathbf{z}}$ | (16$k$) | O VI |
| $\mathbf{B}_{146}$ | = | $x_{13} \mathbf{a}_1 + \left(\frac{1}{2} - y_{13}\right) \mathbf{a}_2 + \left(\frac{1}{2} - z_{13}\right) \mathbf{a}_3$ | = | $x_{13} a \hat{\mathbf{x}} + \left(\frac{1}{2} - y_{13}\right) a \hat{\mathbf{y}} + \left(\frac{1}{2} - z_{13}\right) c \hat{\mathbf{z}}$ | (16$k$) | O VI |
| $\mathbf{B}_{147}$ | = | $y_{13} \mathbf{a}_1 + x_{13} \mathbf{a}_2 + \left(\frac{1}{2} - z_{13}\right) \mathbf{a}_3$ | = | $y_{13} a \hat{\mathbf{x}} + x_{13} a \hat{\mathbf{y}} + \left(\frac{1}{2} - z_{13}\right) c \hat{\mathbf{z}}$ | (16$k$) | O VI |
| $\mathbf{B}_{148}$ | = | $\left(\frac{1}{2} - y_{13}\right) \mathbf{a}_1 + \left(\frac{1}{2} - x_{13}\right) \mathbf{a}_2 + \left(\frac{1}{2} - z_{13}\right) \mathbf{a}_3$ | = | $\left(\frac{1}{2} - y_{13}\right) a \hat{\mathbf{x}} + \left(\frac{1}{2} - x_{13}\right) a \hat{\mathbf{y}} + \left(\frac{1}{2} - z_{13}\right) c \hat{\mathbf{z}}$ | (16$k$) | O VI |
| $\mathbf{B}_{149}$ | = | $-x_{13} \mathbf{a}_1 - y_{13} \mathbf{a}_2 - z_{13} \mathbf{a}_3$ | = | $-x_{13} a \hat{\mathbf{x}} - y_{13} a \hat{\mathbf{y}} - z_{13} c \hat{\mathbf{z}}$ | (16$k$) | O VI |
| $\mathbf{B}_{150}$ | = | $\left(\frac{1}{2} + x_{13}\right) \mathbf{a}_1 + \left(\frac{1}{2} + y_{13}\right) \mathbf{a}_2 - z_{13} \mathbf{a}_3$ | = | $\left(\frac{1}{2} + x_{13}\right) a \hat{\mathbf{x}} + \left(\frac{1}{2} + y_{13}\right) a \hat{\mathbf{y}} - z_{13} c \hat{\mathbf{z}}$ | (16$k$) | O VI |
| $\mathbf{B}_{151}$ | = | $\left(\frac{1}{2} + y_{13}\right) \mathbf{a}_1 - x_{13} \mathbf{a}_2 - z_{13} \mathbf{a}_3$ | = | $\left(\frac{1}{2} + y_{13}\right) a \hat{\mathbf{x}} - x_{13} a \hat{\mathbf{y}} - z_{13} c \hat{\mathbf{z}}$ | (16$k$) | O VI |
| $\mathbf{B}_{152}$ | = | $-y_{13} \mathbf{a}_1 + \left(\frac{1}{2} + x_{13}\right) \mathbf{a}_2 - z_{13} \mathbf{a}_3$ | = | $-y_{13} a \hat{\mathbf{x}} + \left(\frac{1}{2} + x_{13}\right) a \hat{\mathbf{y}} - z_{13} c \hat{\mathbf{z}}$ | (16$k$) | O VI |
| $\mathbf{B}_{153}$ | = | $\left(\frac{1}{2} + x_{13}\right) \mathbf{a}_1 - y_{13} \mathbf{a}_2 + \left(\frac{1}{2} + z_{13}\right) \mathbf{a}_3$ | = | $\left(\frac{1}{2} + x_{13}\right) a \hat{\mathbf{x}} - y_{13} a \hat{\mathbf{y}} + \left(\frac{1}{2} + z_{13}\right) c \hat{\mathbf{z}}$ | (16$k$) | O VI |
| $\mathbf{B}_{154}$ | = | $-x_{13} \mathbf{a}_1 + \left(\frac{1}{2} + y_{13}\right) \mathbf{a}_2 + \left(\frac{1}{2} + z_{13}\right) \mathbf{a}_3$ | = | $-x_{13} a \hat{\mathbf{x}} + \left(\frac{1}{2} + y_{13}\right) a \hat{\mathbf{y}} + \left(\frac{1}{2} + z_{13}\right) c \hat{\mathbf{z}}$ | (16$k$) | O VI |
| $\mathbf{B}_{155}$ | = | $-y_{13} \mathbf{a}_1 - x_{13} \mathbf{a}_2 + \left(\frac{1}{2} + z_{13}\right) \mathbf{a}_3$ | = | $-y_{13} a \hat{\mathbf{x}} - x_{13} a \hat{\mathbf{y}} + \left(\frac{1}{2} + z_{13}\right) c \hat{\mathbf{z}}$ | (16$k$) | O VI |
| $\mathbf{B}_{156}$ | = | $\left(\frac{1}{2} + y_{13}\right) \mathbf{a}_1 + \left(\frac{1}{2} + x_{13}\right) \mathbf{a}_2 + \left(\frac{1}{2} + z_{13}\right) \mathbf{a}_3$ | = | $\left(\frac{1}{2} + y_{13}\right) a \hat{\mathbf{x}} + \left(\frac{1}{2} + x_{13}\right) a \hat{\mathbf{y}} + \left(\frac{1}{2} + z_{13}\right) c \hat{\mathbf{z}}$ | (16$k$) | O VI |
| $\mathbf{B}_{157}$ | = | $x_{14} \mathbf{a}_1 + y_{14} \mathbf{a}_2 + z_{14} \mathbf{a}_3$ | = | $x_{14} a \hat{\mathbf{x}} + y_{14} a \hat{\mathbf{y}} + z_{14} c \hat{\mathbf{z}}$ | (16$k$) | O VII |
| $\mathbf{B}_{158}$ | = | $\left(\frac{1}{2}-x_{14}\right)\mathbf{a}_1+\left(\frac{1}{2}-y_{14}\right)\mathbf{a}_2+z_{14}\,\mathbf{a}_3$ | = | $\left(\frac{1}{2}-x_{14}\right)a\,\hat{\mathbf{x}}+\left(\frac{1}{2}-y_{14}\right)a\,\hat{\mathbf{y}}+z_{14}c\,\hat{\mathbf{z}}$ | (16*k*) | O VII |
| $\mathbf{B}_{159}$ | = | $\left(\frac{1}{2}-y_{14}\right)\mathbf{a}_1+x_{14}\,\mathbf{a}_2+z_{14}\,\mathbf{a}_3$ | = | $\left(\frac{1}{2}-y_{14}\right)a\,\hat{\mathbf{x}}+x_{14}a\,\hat{\mathbf{y}}+z_{14}c\,\hat{\mathbf{z}}$ | (16*k*) | O VII |
| $\mathbf{B}_{160}$ | = | $y_{14}\,\mathbf{a}_1+\left(\frac{1}{2}-x_{14}\right)\mathbf{a}_2+z_{14}\,\mathbf{a}_3$ | = | $y_{14}a\,\hat{\mathbf{x}}+\left(\frac{1}{2}-x_{14}\right)a\,\hat{\mathbf{y}}+z_{14}c\,\hat{\mathbf{z}}$ | (16*k*) | O VII |
| $\mathbf{B}_{161}$ | = | $\left(\frac{1}{2}-x_{14}\right)\mathbf{a}_1+y_{14}\,\mathbf{a}_2+\left(\frac{1}{2}-z_{14}\right)\mathbf{a}_3$ | = | $\left(\frac{1}{2}-x_{14}\right)a\,\hat{\mathbf{x}}+y_{14}a\,\hat{\mathbf{y}}+\left(\frac{1}{2}-z_{14}\right)c\,\hat{\mathbf{z}}$ | (16*k*) | O VII |
| $\mathbf{B}_{162}$ | = | $x_{14}\,\mathbf{a}_1+\left(\frac{1}{2}-y_{14}\right)\mathbf{a}_2+\left(\frac{1}{2}-z_{14}\right)\mathbf{a}_3$ | = | $x_{14}a\,\hat{\mathbf{x}}+\left(\frac{1}{2}-y_{14}\right)a\,\hat{\mathbf{y}}+\left(\frac{1}{2}-z_{14}\right)c\,\hat{\mathbf{z}}$ | (16*k*) | O VII |
| $\mathbf{B}_{163}$ | = | $y_{14}\,\mathbf{a}_1+x_{14}\,\mathbf{a}_2+\left(\frac{1}{2}-z_{14}\right)\mathbf{a}_3$ | = | $y_{14}a\,\hat{\mathbf{x}}+x_{14}a\,\hat{\mathbf{y}}+\left(\frac{1}{2}-z_{14}\right)c\,\hat{\mathbf{z}}$ | (16*k*) | O VII |
| $\mathbf{B}_{164}$ | = | $\left(\frac{1}{2}-y_{14}\right)\mathbf{a}_1+\left(\frac{1}{2}-x_{14}\right)\mathbf{a}_2+\left(\frac{1}{2}-z_{14}\right)\mathbf{a}_3$ | = | $\left(\frac{1}{2}-y_{14}\right)a\,\hat{\mathbf{x}}+\left(\frac{1}{2}-x_{14}\right)a\,\hat{\mathbf{y}}+\left(\frac{1}{2}-z_{14}\right)c\,\hat{\mathbf{z}}$ | (16*k*) | O VII |
| $\mathbf{B}_{165}$ | = | $-x_{14}\,\mathbf{a}_1-y_{14}\,\mathbf{a}_2-z_{14}\,\mathbf{a}_3$ | = | $-x_{14}a\,\hat{\mathbf{x}}-y_{14}a\,\hat{\mathbf{y}}-z_{14}c\,\hat{\mathbf{z}}$ | (16*k*) | O VII |
| $\mathbf{B}_{166}$ | = | $\left(\frac{1}{2}+x_{14}\right)\mathbf{a}_1+\left(\frac{1}{2}+y_{14}\right)\mathbf{a}_2-z_{14}\,\mathbf{a}_3$ | = | $\left(\frac{1}{2}+x_{14}\right)a\,\hat{\mathbf{x}}+\left(\frac{1}{2}+y_{14}\right)a\,\hat{\mathbf{y}}-z_{14}c\,\hat{\mathbf{z}}$ | (16*k*) | O VII |
| $\mathbf{B}_{167}$ | = | $\left(\frac{1}{2}+y_{14}\right)\mathbf{a}_1-x_{14}\,\mathbf{a}_2-z_{14}\,\mathbf{a}_3$ | = | $\left(\frac{1}{2}+y_{14}\right)a\,\hat{\mathbf{x}}-x_{14}a\,\hat{\mathbf{y}}-z_{14}c\,\hat{\mathbf{z}}$ | (16*k*) | O VII |
| $\mathbf{B}_{168}$ | = | $-y_{14}\,\mathbf{a}_1+\left(\frac{1}{2}+x_{14}\right)\mathbf{a}_2-z_{14}\,\mathbf{a}_3$ | = | $-y_{14}a\,\hat{\mathbf{x}}+\left(\frac{1}{2}+x_{14}\right)a\,\hat{\mathbf{y}}-z_{14}c\,\hat{\mathbf{z}}$ | (16*k*) | O VII |
| $\mathbf{B}_{169}$ | = | $\left(\frac{1}{2}+x_{14}\right)\mathbf{a}_1-y_{14}\,\mathbf{a}_2+\left(\frac{1}{2}+z_{14}\right)\mathbf{a}_3$ | = | $\left(\frac{1}{2}+x_{14}\right)a\,\hat{\mathbf{x}}-y_{14}a\,\hat{\mathbf{y}}+\left(\frac{1}{2}+z_{14}\right)c\,\hat{\mathbf{z}}$ | (16*k*) | O VII |
| $\mathbf{B}_{170}$ | = | $-x_{14}\,\mathbf{a}_1+\left(\frac{1}{2}+y_{14}\right)\mathbf{a}_2+\left(\frac{1}{2}+z_{14}\right)\mathbf{a}_3$ | = | $-x_{14}a\,\hat{\mathbf{x}}+\left(\frac{1}{2}+y_{14}\right)a\,\hat{\mathbf{y}}+\left(\frac{1}{2}+z_{14}\right)c\,\hat{\mathbf{z}}$ | (16*k*) | O VII |
| $\mathbf{B}_{171}$ | = | $-y_{14}\,\mathbf{a}_1-x_{14}\,\mathbf{a}_2+\left(\frac{1}{2}+z_{14}\right)\mathbf{a}_3$ | = | $-y_{14}a\,\hat{\mathbf{x}}-x_{14}a\,\hat{\mathbf{y}}+\left(\frac{1}{2}+z_{14}\right)c\,\hat{\mathbf{z}}$ | (16*k*) | O VII |
| $\mathbf{B}_{172}$ | = | $\left(\frac{1}{2}+y_{14}\right)\mathbf{a}_1+\left(\frac{1}{2}+x_{14}\right)\mathbf{a}_2+\left(\frac{1}{2}+z_{14}\right)\mathbf{a}_3$ | = | $\left(\frac{1}{2}+y_{14}\right)a\,\hat{\mathbf{x}}+\left(\frac{1}{2}+x_{14}\right)a\,\hat{\mathbf{y}}+\left(\frac{1}{2}+z_{14}\right)c\,\hat{\mathbf{z}}$ | (16*k*) | O VII |
| $\mathbf{B}_{173}$ | = | $x_{15}\,\mathbf{a}_1+y_{15}\,\mathbf{a}_2+z_{15}\,\mathbf{a}_3$ | = | $x_{15}a\,\hat{\mathbf{x}}+y_{15}a\,\hat{\mathbf{y}}+z_{15}c\,\hat{\mathbf{z}}$ | (16*k*) | O VIII |
| $\mathbf{B}_{174}$ | = | $\left(\frac{1}{2}-x_{15}\right)\mathbf{a}_1+\left(\frac{1}{2}-y_{15}\right)\mathbf{a}_2+z_{15}\,\mathbf{a}_3$ | = | $\left(\frac{1}{2}-x_{15}\right)a\,\hat{\mathbf{x}}+\left(\frac{1}{2}-y_{15}\right)a\,\hat{\mathbf{y}}+z_{15}c\,\hat{\mathbf{z}}$ | (16*k*) | O VIII |
| $\mathbf{B}_{175}$ | = | $\left(\frac{1}{2}-y_{15}\right)\mathbf{a}_1+x_{15}\,\mathbf{a}_2+z_{15}\,\mathbf{a}_3$ | = | $\left(\frac{1}{2}-y_{15}\right)a\,\hat{\mathbf{x}}+x_{15}a\,\hat{\mathbf{y}}+z_{15}c\,\hat{\mathbf{z}}$ | (16*k*) | O VIII |
| $\mathbf{B}_{176}$ | = | $y_{15}\,\mathbf{a}_1+\left(\frac{1}{2}-x_{15}\right)\mathbf{a}_2+z_{15}\,\mathbf{a}_3$ | = | $y_{15}a\,\hat{\mathbf{x}}+\left(\frac{1}{2}-x_{15}\right)a\,\hat{\mathbf{y}}+z_{15}c\,\hat{\mathbf{z}}$ | (16*k*) | O VIII |
| $\mathbf{B}_{177}$ | = | $\left(\frac{1}{2}-x_{15}\right)\mathbf{a}_1+y_{15}\,\mathbf{a}_2+\left(\frac{1}{2}-z_{15}\right)\mathbf{a}_3$ | = | $\left(\frac{1}{2}-x_{15}\right)a\,\hat{\mathbf{x}}+y_{15}a\,\hat{\mathbf{y}}+\left(\frac{1}{2}-z_{15}\right)c\,\hat{\mathbf{z}}$ | (16*k*) | O VIII |
| $\mathbf{B}_{178}$ | = | $x_{15}\,\mathbf{a}_1+\left(\frac{1}{2}-y_{15}\right)\mathbf{a}_2+\left(\frac{1}{2}-z_{15}\right)\mathbf{a}_3$ | = | $x_{15}a\,\hat{\mathbf{x}}+\left(\frac{1}{2}-y_{15}\right)a\,\hat{\mathbf{y}}+\left(\frac{1}{2}-z_{15}\right)c\,\hat{\mathbf{z}}$ | (16*k*) | O VIII |
| $\mathbf{B}_{179}$ | = | $y_{15}\,\mathbf{a}_1+x_{15}\,\mathbf{a}_2+\left(\frac{1}{2}-z_{15}\right)\mathbf{a}_3$ | = | $y_{15}a\,\hat{\mathbf{x}}+x_{15}a\,\hat{\mathbf{y}}+\left(\frac{1}{2}-z_{15}\right)c\,\hat{\mathbf{z}}$ | (16*k*) | O VIII |
| $\mathbf{B}_{180}$ | = | $\left(\frac{1}{2}-y_{15}\right)\mathbf{a}_1+\left(\frac{1}{2}-x_{15}\right)\mathbf{a}_2+\left(\frac{1}{2}-z_{15}\right)\mathbf{a}_3$ | = | $\left(\frac{1}{2}-y_{15}\right)a\,\hat{\mathbf{x}}+\left(\frac{1}{2}-x_{15}\right)a\,\hat{\mathbf{y}}+\left(\frac{1}{2}-z_{15}\right)c\,\hat{\mathbf{z}}$ | (16*k*) | O VIII |
| $\mathbf{B}_{181}$ | = | $-x_{15}\,\mathbf{a}_1-y_{15}\,\mathbf{a}_2-z_{15}\,\mathbf{a}_3$ | = | $-x_{15}a\,\hat{\mathbf{x}}-y_{15}a\,\hat{\mathbf{y}}-z_{15}c\,\hat{\mathbf{z}}$ | (16*k*) | O VIII |
| $\mathbf{B}_{182}$ | = | $\left(\frac{1}{2}+x_{15}\right)\mathbf{a}_1+\left(\frac{1}{2}+y_{15}\right)\mathbf{a}_2-z_{15}\,\mathbf{a}_3$ | = | $\left(\frac{1}{2}+x_{15}\right)a\,\hat{\mathbf{x}}+\left(\frac{1}{2}+y_{15}\right)a\,\hat{\mathbf{y}}-z_{15}c\,\hat{\mathbf{z}}$ | (16*k*) | O VIII |
| $\mathbf{B}_{183}$ | = | $\left(\frac{1}{2}+y_{15}\right)\mathbf{a}_1-x_{15}\,\mathbf{a}_2-z_{15}\,\mathbf{a}_3$ | = | $\left(\frac{1}{2}+y_{15}\right)a\,\hat{\mathbf{x}}-x_{15}a\,\hat{\mathbf{y}}-z_{15}c\,\hat{\mathbf{z}}$ | (16*k*) | O VIII |
| $\mathbf{B}_{184}$ | = | $-y_{15}\,\mathbf{a}_1+\left(\frac{1}{2}+x_{15}\right)\mathbf{a}_2-z_{15}\,\mathbf{a}_3$ | = | $-y_{15}a\,\hat{\mathbf{x}}+\left(\frac{1}{2}+x_{15}\right)a\,\hat{\mathbf{y}}-z_{15}c\,\hat{\mathbf{z}}$ | (16*k*) | O VIII |
| $\mathbf{B}_{185}$ | = | $\left(\frac{1}{2}+x_{15}\right)\mathbf{a}_1-y_{15}\,\mathbf{a}_2+\left(\frac{1}{2}+z_{15}\right)\mathbf{a}_3$ | = | $\left(\frac{1}{2}+x_{15}\right)a\,\hat{\mathbf{x}}-y_{15}a\,\hat{\mathbf{y}}+\left(\frac{1}{2}+z_{15}\right)c\,\hat{\mathbf{z}}$ | (16*k*) | O VIII |
| $\mathbf{B}_{186}$ | = | $-x_{15}\,\mathbf{a}_1+\left(\frac{1}{2}+y_{15}\right)\mathbf{a}_2+\left(\frac{1}{2}+z_{15}\right)\mathbf{a}_3$ | = | $-x_{15}a\,\hat{\mathbf{x}}+\left(\frac{1}{2}+y_{15}\right)a\,\hat{\mathbf{y}}+\left(\frac{1}{2}+z_{15}\right)c\,\hat{\mathbf{z}}$ | (16*k*) | O VIII |
| $\mathbf{B}_{187}$ | = | $-y_{15}\,\mathbf{a}_1-x_{15}\,\mathbf{a}_2+\left(\frac{1}{2}+z_{15}\right)\mathbf{a}_3$ | = | $-y_{15}a\,\hat{\mathbf{x}}-x_{15}a\,\hat{\mathbf{y}}+\left(\frac{1}{2}+z_{15}\right)c\,\hat{\mathbf{z}}$ | (16*k*) | O VIII |
| $\mathbf{B}_{188}$ | = | $\left(\frac{1}{2}+y_{15}\right)\mathbf{a}_1+\left(\frac{1}{2}+x_{15}\right)\mathbf{a}_2+\left(\frac{1}{2}+z_{15}\right)\mathbf{a}_3$ | = | $\left(\frac{1}{2}+y_{15}\right)a\,\hat{\mathbf{x}}+\left(\frac{1}{2}+x_{15}\right)a\,\hat{\mathbf{y}}+\left(\frac{1}{2}+z_{15}\right)c\,\hat{\mathbf{z}}$ | (16*k*) | O VIII |
| $\mathbf{B}_{189}$ | = | $x_{16}\,\mathbf{a}_1+y_{16}\,\mathbf{a}_2+z_{16}\,\mathbf{a}_3$ | = | $x_{16}a\,\hat{\mathbf{x}}+y_{16}a\,\hat{\mathbf{y}}+z_{16}c\,\hat{\mathbf{z}}$ | (16*k*) | O IX |

$$\mathbf{B}_{190} = \left(\frac{1}{2} - x_{16}\right)\mathbf{a}_1 + \left(\frac{1}{2} - y_{16}\right)\mathbf{a}_2 + z_{16}\,\mathbf{a}_3 = \left(\frac{1}{2} - x_{16}\right)a\,\hat{\mathbf{x}} + \left(\frac{1}{2} - y_{16}\right)a\,\hat{\mathbf{y}} + z_{16}c\,\hat{\mathbf{z}} \quad (16k) \quad \text{O IX}$$

$$\mathbf{B}_{191} = \left(\frac{1}{2} - y_{16}\right)\mathbf{a}_1 + x_{16}\,\mathbf{a}_2 + z_{16}\,\mathbf{a}_3 = \left(\frac{1}{2} - y_{16}\right)a\,\hat{\mathbf{x}} + x_{16}a\,\hat{\mathbf{y}} + z_{16}c\,\hat{\mathbf{z}} \quad (16k) \quad \text{O IX}$$

$$\mathbf{B}_{192} = y_{16}\,\mathbf{a}_1 + \left(\frac{1}{2} - x_{16}\right)\mathbf{a}_2 + z_{16}\,\mathbf{a}_3 = y_{16}a\,\hat{\mathbf{x}} + \left(\frac{1}{2} - x_{16}\right)a\,\hat{\mathbf{y}} + z_{16}c\,\hat{\mathbf{z}} \quad (16k) \quad \text{O IX}$$

$$\mathbf{B}_{193} = \left(\frac{1}{2} - x_{16}\right)\mathbf{a}_1 + y_{16}\,\mathbf{a}_2 + \left(\frac{1}{2} - z_{16}\right)\mathbf{a}_3 = \left(\frac{1}{2} - x_{16}\right)a\,\hat{\mathbf{x}} + y_{16}a\,\hat{\mathbf{y}} + \left(\frac{1}{2} - z_{16}\right)c\,\hat{\mathbf{z}} \quad (16k) \quad \text{O IX}$$

$$\mathbf{B}_{194} = x_{16}\,\mathbf{a}_1 + \left(\frac{1}{2} - y_{16}\right)\mathbf{a}_2 + \left(\frac{1}{2} - z_{16}\right)\mathbf{a}_3 = x_{16}a\,\hat{\mathbf{x}} + \left(\frac{1}{2} - y_{16}\right)a\,\hat{\mathbf{y}} + \left(\frac{1}{2} - z_{16}\right)c\,\hat{\mathbf{z}} \quad (16k) \quad \text{O IX}$$

$$\mathbf{B}_{195} = y_{16}\,\mathbf{a}_1 + x_{16}\,\mathbf{a}_2 + \left(\frac{1}{2} - z_{16}\right)\mathbf{a}_3 = y_{16}a\,\hat{\mathbf{x}} + x_{16}a\,\hat{\mathbf{y}} + \left(\frac{1}{2} - z_{16}\right)c\,\hat{\mathbf{z}} \quad (16k) \quad \text{O IX}$$

$$\mathbf{B}_{196} = \left(\frac{1}{2} - y_{16}\right)\mathbf{a}_1 + \left(\frac{1}{2} - x_{16}\right)\mathbf{a}_2 + \left(\frac{1}{2} - z_{16}\right)\mathbf{a}_3 = \left(\frac{1}{2} - y_{16}\right)a\,\hat{\mathbf{x}} + \left(\frac{1}{2} - x_{16}\right)a\,\hat{\mathbf{y}} + \left(\frac{1}{2} - z_{16}\right)c\,\hat{\mathbf{z}} \quad (16k) \quad \text{O IX}$$

$$\mathbf{B}_{197} = -x_{16}\,\mathbf{a}_1 - y_{16}\,\mathbf{a}_2 - z_{16}\,\mathbf{a}_3 = -x_{16}a\,\hat{\mathbf{x}} - y_{16}a\,\hat{\mathbf{y}} - z_{16}c\,\hat{\mathbf{z}} \quad (16k) \quad \text{O IX}$$

$$\mathbf{B}_{198} = \left(\frac{1}{2} + x_{16}\right)\mathbf{a}_1 + \left(\frac{1}{2} + y_{16}\right)\mathbf{a}_2 - z_{16}\,\mathbf{a}_3 = \left(\frac{1}{2} + x_{16}\right)a\,\hat{\mathbf{x}} + \left(\frac{1}{2} + y_{16}\right)a\,\hat{\mathbf{y}} - z_{16}c\,\hat{\mathbf{z}} \quad (16k) \quad \text{O IX}$$

$$\mathbf{B}_{199} = \left(\frac{1}{2} + y_{16}\right)\mathbf{a}_1 - x_{16}\,\mathbf{a}_2 - z_{16}\,\mathbf{a}_3 = \left(\frac{1}{2} + y_{16}\right)a\,\hat{\mathbf{x}} - x_{16}a\,\hat{\mathbf{y}} - z_{16}c\,\hat{\mathbf{z}} \quad (16k) \quad \text{O IX}$$

$$\mathbf{B}_{200} = -y_{16}\,\mathbf{a}_1 + \left(\frac{1}{2} + x_{16}\right)\mathbf{a}_2 - z_{16}\,\mathbf{a}_3 = -y_{16}a\,\hat{\mathbf{x}} + \left(\frac{1}{2} + x_{16}\right)a\,\hat{\mathbf{y}} - z_{16}c\,\hat{\mathbf{z}} \quad (16k) \quad \text{O IX}$$

$$\mathbf{B}_{201} = \left(\frac{1}{2} + x_{16}\right)\mathbf{a}_1 - y_{16}\,\mathbf{a}_2 + \left(\frac{1}{2} + z_{16}\right)\mathbf{a}_3 = \left(\frac{1}{2} + x_{16}\right)a\,\hat{\mathbf{x}} - y_{16}a\,\hat{\mathbf{y}} + \left(\frac{1}{2} + z_{16}\right)c\,\hat{\mathbf{z}} \quad (16k) \quad \text{O IX}$$

$$\mathbf{B}_{202} = -x_{16}\,\mathbf{a}_1 + \left(\frac{1}{2} + y_{16}\right)\mathbf{a}_2 + \left(\frac{1}{2} + z_{16}\right)\mathbf{a}_3 = -x_{16}a\,\hat{\mathbf{x}} + \left(\frac{1}{2} + y_{16}\right)a\,\hat{\mathbf{y}} + \left(\frac{1}{2} + z_{16}\right)c\,\hat{\mathbf{z}} \quad (16k) \quad \text{O IX}$$

$$\mathbf{B}_{203} = -y_{16}\,\mathbf{a}_1 - x_{16}\,\mathbf{a}_2 + \left(\frac{1}{2} + z_{16}\right)\mathbf{a}_3 = -y_{16}a\,\hat{\mathbf{x}} - x_{16}a\,\hat{\mathbf{y}} + \left(\frac{1}{2} + z_{16}\right)c\,\hat{\mathbf{z}} \quad (16k) \quad \text{O IX}$$

$$\mathbf{B}_{204} = \left(\frac{1}{2} + y_{16}\right)\mathbf{a}_1 + \left(\frac{1}{2} + x_{16}\right)\mathbf{a}_2 + \left(\frac{1}{2} + z_{16}\right)\mathbf{a}_3 = \left(\frac{1}{2} + y_{16}\right)a\,\hat{\mathbf{x}} + \left(\frac{1}{2} + x_{16}\right)a\,\hat{\mathbf{y}} + \left(\frac{1}{2} + z_{16}\right)c\,\hat{\mathbf{z}} \quad (16k) \quad \text{O IX}$$

$$\mathbf{B}_{205} = x_{17}\,\mathbf{a}_1 + y_{17}\,\mathbf{a}_2 + z_{17}\,\mathbf{a}_3 = x_{17}a\,\hat{\mathbf{x}} + y_{17}a\,\hat{\mathbf{y}} + z_{17}c\,\hat{\mathbf{z}} \quad (16k) \quad \text{OH}$$

$$\mathbf{B}_{206} = \left(\frac{1}{2} - x_{17}\right)\mathbf{a}_1 + \left(\frac{1}{2} - y_{17}\right)\mathbf{a}_2 + z_{17}\,\mathbf{a}_3 = \left(\frac{1}{2} - x_{17}\right)a\,\hat{\mathbf{x}} + \left(\frac{1}{2} - y_{17}\right)a\,\hat{\mathbf{y}} + z_{17}c\,\hat{\mathbf{z}} \quad (16k) \quad \text{OH}$$

$$\mathbf{B}_{207} = \left(\frac{1}{2} - y_{17}\right)\mathbf{a}_1 + x_{17}\,\mathbf{a}_2 + z_{17}\,\mathbf{a}_3 = \left(\frac{1}{2} - y_{17}\right)a\,\hat{\mathbf{x}} + x_{17}a\,\hat{\mathbf{y}} + z_{17}c\,\hat{\mathbf{z}} \quad (16k) \quad \text{OH}$$

$$\mathbf{B}_{208} = y_{17}\,\mathbf{a}_1 + \left(\frac{1}{2} - x_{17}\right)\mathbf{a}_2 + z_{17}\,\mathbf{a}_3 = y_{17}a\,\hat{\mathbf{x}} + \left(\frac{1}{2} - x_{17}\right)a\,\hat{\mathbf{y}} + z_{17}c\,\hat{\mathbf{z}} \quad (16k) \quad \text{OH}$$

$$\mathbf{B}_{209} = \left(\frac{1}{2} - x_{17}\right)\mathbf{a}_1 + y_{17}\,\mathbf{a}_2 + \left(\frac{1}{2} - z_{17}\right)\mathbf{a}_3 = \left(\frac{1}{2} - x_{17}\right)a\,\hat{\mathbf{x}} + y_{17}a\,\hat{\mathbf{y}} + \left(\frac{1}{2} - z_{17}\right)c\,\hat{\mathbf{z}} \quad (16k) \quad \text{OH}$$

$$\mathbf{B}_{210} = x_{17}\,\mathbf{a}_1 + \left(\frac{1}{2} - y_{17}\right)\mathbf{a}_2 + \left(\frac{1}{2} - z_{17}\right)\mathbf{a}_3 = x_{17}a\,\hat{\mathbf{x}} + \left(\frac{1}{2} - y_{17}\right)a\,\hat{\mathbf{y}} + \left(\frac{1}{2} - z_{17}\right)c\,\hat{\mathbf{z}} \quad (16k) \quad \text{OH}$$

$$\mathbf{B}_{211} = y_{17}\,\mathbf{a}_1 + x_{17}\,\mathbf{a}_2 + \left(\frac{1}{2} - z_{17}\right)\mathbf{a}_3 = y_{17}a\,\hat{\mathbf{x}} + x_{17}a\,\hat{\mathbf{y}} + \left(\frac{1}{2} - z_{17}\right)c\,\hat{\mathbf{z}} \quad (16k) \quad \text{OH}$$

$$\mathbf{B}_{212} = \left(\frac{1}{2} - y_{17}\right)\mathbf{a}_1 + \left(\frac{1}{2} - x_{17}\right)\mathbf{a}_2 + \left(\frac{1}{2} - z_{17}\right)\mathbf{a}_3 = \left(\frac{1}{2} - y_{17}\right)a\,\hat{\mathbf{x}} + \left(\frac{1}{2} - x_{17}\right)a\,\hat{\mathbf{y}} + \left(\frac{1}{2} - z_{17}\right)c\,\hat{\mathbf{z}} \quad (16k) \quad \text{OH}$$

$$\mathbf{B}_{213} = -x_{17}\,\mathbf{a}_1 - y_{17}\,\mathbf{a}_2 - z_{17}\,\mathbf{a}_3 = -x_{17}a\,\hat{\mathbf{x}} - y_{17}a\,\hat{\mathbf{y}} - z_{17}c\,\hat{\mathbf{z}} \quad (16k) \quad \text{OH}$$

$$\mathbf{B}_{214} = \left(\frac{1}{2} + x_{17}\right)\mathbf{a}_1 + \left(\frac{1}{2} + y_{17}\right)\mathbf{a}_2 - z_{17}\,\mathbf{a}_3 = \left(\frac{1}{2} + x_{17}\right)a\,\hat{\mathbf{x}} + \left(\frac{1}{2} + y_{17}\right)a\,\hat{\mathbf{y}} - z_{17}c\,\hat{\mathbf{z}} \quad (16k) \quad \text{OH}$$

$$\mathbf{B}_{215} = \left(\frac{1}{2} + y_{17}\right)\mathbf{a}_1 - x_{17}\,\mathbf{a}_2 - z_{17}\,\mathbf{a}_3 = \left(\frac{1}{2} + y_{17}\right)a\,\hat{\mathbf{x}} - x_{17}a\,\hat{\mathbf{y}} - z_{17}c\,\hat{\mathbf{z}} \quad (16k) \quad \text{OH}$$

$$\mathbf{B}_{216} = -y_{17}\,\mathbf{a}_1 + \left(\frac{1}{2} + x_{17}\right)\mathbf{a}_2 - z_{17}\,\mathbf{a}_3 = -y_{17}a\,\hat{\mathbf{x}} + \left(\frac{1}{2} + x_{17}\right)a\,\hat{\mathbf{y}} - z_{17}c\,\hat{\mathbf{z}} \quad (16k) \quad \text{OH}$$

$$\mathbf{B}_{217} = \left(\frac{1}{2} + x_{17}\right)\mathbf{a}_1 - y_{17}\,\mathbf{a}_2 + \left(\frac{1}{2} + z_{17}\right)\mathbf{a}_3 = \left(\frac{1}{2} + x_{17}\right)a\,\hat{\mathbf{x}} - y_{17}a\,\hat{\mathbf{y}} + \left(\frac{1}{2} + z_{17}\right)c\,\hat{\mathbf{z}} \quad (16k) \quad \text{OH}$$

$$\mathbf{B}_{218} = -x_{17}\,\mathbf{a}_1 + \left(\frac{1}{2} + y_{17}\right)\mathbf{a}_2 + \left(\frac{1}{2} + z_{17}\right)\mathbf{a}_3 = -x_{17}a\,\hat{\mathbf{x}} + \left(\frac{1}{2} + y_{17}\right)a\,\hat{\mathbf{y}} + \left(\frac{1}{2} + z_{17}\right)c\,\hat{\mathbf{z}} \quad (16k) \quad \text{OH}$$

$$\mathbf{B}_{219} = -y_{17}\,\mathbf{a}_1 - x_{17}\,\mathbf{a}_2 + \left(\frac{1}{2} + z_{17}\right)\mathbf{a}_3 = -y_{17}a\,\hat{\mathbf{x}} - x_{17}a\,\hat{\mathbf{y}} + \left(\frac{1}{2} + z_{17}\right)c\,\hat{\mathbf{z}} \quad (16k) \quad \text{OH}$$

$$\mathbf{B}_{220} = \left(\frac{1}{2} + y_{17}\right)\mathbf{a}_1 + \left(\frac{1}{2} + x_{17}\right)\mathbf{a}_2 + \left(\frac{1}{2} + z_{17}\right)\mathbf{a}_3 = \left(\frac{1}{2} + y_{17}\right)a\,\hat{\mathbf{x}} + \left(\frac{1}{2} + x_{17}\right)a\,\hat{\mathbf{y}} + \left(\frac{1}{2} + z_{17}\right)c\,\hat{\mathbf{z}} \quad (16k) \quad \text{OH}$$

$$\mathbf{B}_{221} = x_{18}\,\mathbf{a}_1 + y_{18}\,\mathbf{a}_2 + z_{18}\,\mathbf{a}_3 = x_{18}a\,\hat{\mathbf{x}} + y_{18}a\,\hat{\mathbf{y}} + z_{18}c\,\hat{\mathbf{z}} \quad (16k) \quad \text{Si II}$$

| | | | | | | |
|---|---|---|---|---|---|---|
| $\mathbf{B}_{222}$ | = | $\left(\frac{1}{2} - x_{18}\right) \mathbf{a}_1 + \left(\frac{1}{2} - y_{18}\right) \mathbf{a}_2 + z_{18} \mathbf{a}_3$ | = | $\left(\frac{1}{2} - x_{18}\right) a \hat{\mathbf{x}} + \left(\frac{1}{2} - y_{18}\right) a \hat{\mathbf{y}} + z_{18} c \hat{\mathbf{z}}$ | (16$k$) | Si II |
| $\mathbf{B}_{223}$ | = | $\left(\frac{1}{2} - y_{18}\right) \mathbf{a}_1 + x_{18} \mathbf{a}_2 + z_{18} \mathbf{a}_3$ | = | $\left(\frac{1}{2} - y_{18}\right) a \hat{\mathbf{x}} + x_{18} a \hat{\mathbf{y}} + z_{18} c \hat{\mathbf{z}}$ | (16$k$) | Si II |
| $\mathbf{B}_{224}$ | = | $y_{18} \mathbf{a}_1 + \left(\frac{1}{2} - x_{18}\right) \mathbf{a}_2 + z_{18} \mathbf{a}_3$ | = | $y_{18} a \hat{\mathbf{x}} + \left(\frac{1}{2} - x_{18}\right) a \hat{\mathbf{y}} + z_{18} c \hat{\mathbf{z}}$ | (16$k$) | Si II |
| $\mathbf{B}_{225}$ | = | $\left(\frac{1}{2} - x_{18}\right) \mathbf{a}_1 + y_{18} \mathbf{a}_2 + \left(\frac{1}{2} - z_{18}\right) \mathbf{a}_3$ | = | $\left(\frac{1}{2} - x_{18}\right) a \hat{\mathbf{x}} + y_{18} a \hat{\mathbf{y}} + \left(\frac{1}{2} - z_{18}\right) c \hat{\mathbf{z}}$ | (16$k$) | Si II |
| $\mathbf{B}_{226}$ | = | $x_{18} \mathbf{a}_1 + \left(\frac{1}{2} - y_{18}\right) \mathbf{a}_2 + \left(\frac{1}{2} - z_{18}\right) \mathbf{a}_3$ | = | $x_{18} a \hat{\mathbf{x}} + \left(\frac{1}{2} - y_{18}\right) a \hat{\mathbf{y}} + \left(\frac{1}{2} - z_{18}\right) c \hat{\mathbf{z}}$ | (16$k$) | Si II |
| $\mathbf{B}_{227}$ | = | $y_{18} \mathbf{a}_1 + x_{18} \mathbf{a}_2 + \left(\frac{1}{2} - z_{18}\right) \mathbf{a}_3$ | = | $y_{18} a \hat{\mathbf{x}} + x_{18} a \hat{\mathbf{y}} + \left(\frac{1}{2} - z_{18}\right) c \hat{\mathbf{z}}$ | (16$k$) | Si II |
| $\mathbf{B}_{228}$ | = | $\left(\frac{1}{2} - y_{18}\right) \mathbf{a}_1 + \left(\frac{1}{2} - x_{18}\right) \mathbf{a}_2 + \left(\frac{1}{2} - z_{18}\right) \mathbf{a}_3$ | = | $\left(\frac{1}{2} - y_{18}\right) a \hat{\mathbf{x}} + \left(\frac{1}{2} - x_{18}\right) a \hat{\mathbf{y}} + \left(\frac{1}{2} - z_{18}\right) c \hat{\mathbf{z}}$ | (16$k$) | Si II |
| $\mathbf{B}_{229}$ | = | $-x_{18} \mathbf{a}_1 - y_{18} \mathbf{a}_2 - z_{18} \mathbf{a}_3$ | = | $-x_{18} a \hat{\mathbf{x}} - y_{18} a \hat{\mathbf{y}} - z_{18} c \hat{\mathbf{z}}$ | (16$k$) | Si II |
| $\mathbf{B}_{230}$ | = | $\left(\frac{1}{2} + x_{18}\right) \mathbf{a}_1 + \left(\frac{1}{2} + y_{18}\right) \mathbf{a}_2 - z_{18} \mathbf{a}_3$ | = | $\left(\frac{1}{2} + x_{18}\right) a \hat{\mathbf{x}} + \left(\frac{1}{2} + y_{18}\right) a \hat{\mathbf{y}} - z_{18} c \hat{\mathbf{z}}$ | (16$k$) | Si II |
| $\mathbf{B}_{231}$ | = | $\left(\frac{1}{2} + y_{18}\right) \mathbf{a}_1 - x_{18} \mathbf{a}_2 - z_{18} \mathbf{a}_3$ | = | $\left(\frac{1}{2} + y_{18}\right) a \hat{\mathbf{x}} - x_{18} a \hat{\mathbf{y}} - z_{18} c \hat{\mathbf{z}}$ | (16$k$) | Si II |
| $\mathbf{B}_{232}$ | = | $-y_{18} \mathbf{a}_1 + \left(\frac{1}{2} + x_{18}\right) \mathbf{a}_2 - z_{18} \mathbf{a}_3$ | = | $-y_{18} a \hat{\mathbf{x}} + \left(\frac{1}{2} + x_{18}\right) a \hat{\mathbf{y}} - z_{18} c \hat{\mathbf{z}}$ | (16$k$) | Si II |
| $\mathbf{B}_{233}$ | = | $\left(\frac{1}{2} + x_{18}\right) \mathbf{a}_1 - y_{18} \mathbf{a}_2 + \left(\frac{1}{2} + z_{18}\right) \mathbf{a}_3$ | = | $\left(\frac{1}{2} + x_{18}\right) a \hat{\mathbf{x}} - y_{18} a \hat{\mathbf{y}} + \left(\frac{1}{2} + z_{18}\right) c \hat{\mathbf{z}}$ | (16$k$) | Si II |
| $\mathbf{B}_{234}$ | = | $-x_{18} \mathbf{a}_1 + \left(\frac{1}{2} + y_{18}\right) \mathbf{a}_2 + \left(\frac{1}{2} + z_{18}\right) \mathbf{a}_3$ | = | $-x_{18} a \hat{\mathbf{x}} + \left(\frac{1}{2} + y_{18}\right) a \hat{\mathbf{y}} + \left(\frac{1}{2} + z_{18}\right) c \hat{\mathbf{z}}$ | (16$k$) | Si II |
| $\mathbf{B}_{235}$ | = | $-y_{18} \mathbf{a}_1 - x_{18} \mathbf{a}_2 + \left(\frac{1}{2} + z_{18}\right) \mathbf{a}_3$ | = | $-y_{18} a \hat{\mathbf{x}} - x_{18} a \hat{\mathbf{y}} + \left(\frac{1}{2} + z_{18}\right) c \hat{\mathbf{z}}$ | (16$k$) | Si II |
| $\mathbf{B}_{236}$ | = | $\left(\frac{1}{2} + y_{18}\right) \mathbf{a}_1 + \left(\frac{1}{2} + x_{18}\right) \mathbf{a}_2 + \left(\frac{1}{2} + z_{18}\right) \mathbf{a}_3$ | = | $\left(\frac{1}{2} + y_{18}\right) a \hat{\mathbf{x}} + \left(\frac{1}{2} + x_{18}\right) a \hat{\mathbf{y}} + \left(\frac{1}{2} + z_{18}\right) c \hat{\mathbf{z}}$ | (16$k$) | Si II |
| $\mathbf{B}_{237}$ | = | $x_{19} \mathbf{a}_1 + y_{19} \mathbf{a}_2 + z_{19} \mathbf{a}_3$ | = | $x_{19} a \hat{\mathbf{x}} + y_{19} a \hat{\mathbf{y}} + z_{19} c \hat{\mathbf{z}}$ | (16$k$) | Si III |
| $\mathbf{B}_{238}$ | = | $\left(\frac{1}{2} - x_{19}\right) \mathbf{a}_1 + \left(\frac{1}{2} - y_{19}\right) \mathbf{a}_2 + z_{19} \mathbf{a}_3$ | = | $\left(\frac{1}{2} - x_{19}\right) a \hat{\mathbf{x}} + \left(\frac{1}{2} - y_{19}\right) a \hat{\mathbf{y}} + z_{19} c \hat{\mathbf{z}}$ | (16$k$) | Si III |
| $\mathbf{B}_{239}$ | = | $\left(\frac{1}{2} - y_{19}\right) \mathbf{a}_1 + x_{19} \mathbf{a}_2 + z_{19} \mathbf{a}_3$ | = | $\left(\frac{1}{2} - y_{19}\right) a \hat{\mathbf{x}} + x_{19} a \hat{\mathbf{y}} + z_{19} c \hat{\mathbf{z}}$ | (16$k$) | Si III |
| $\mathbf{B}_{240}$ | = | $y_{19} \mathbf{a}_1 + \left(\frac{1}{2} - x_{19}\right) \mathbf{a}_2 + z_{19} \mathbf{a}_3$ | = | $y_{19} a \hat{\mathbf{x}} + \left(\frac{1}{2} - x_{19}\right) a \hat{\mathbf{y}} + z_{19} c \hat{\mathbf{z}}$ | (16$k$) | Si III |
| $\mathbf{B}_{241}$ | = | $\left(\frac{1}{2} - x_{19}\right) \mathbf{a}_1 + y_{19} \mathbf{a}_2 + \left(\frac{1}{2} - z_{19}\right) \mathbf{a}_3$ | = | $\left(\frac{1}{2} - x_{19}\right) a \hat{\mathbf{x}} + y_{19} a \hat{\mathbf{y}} + \left(\frac{1}{2} - z_{19}\right) c \hat{\mathbf{z}}$ | (16$k$) | Si III |
| $\mathbf{B}_{242}$ | = | $x_{19} \mathbf{a}_1 + \left(\frac{1}{2} - y_{19}\right) \mathbf{a}_2 + \left(\frac{1}{2} - z_{19}\right) \mathbf{a}_3$ | = | $x_{19} a \hat{\mathbf{x}} + \left(\frac{1}{2} - y_{19}\right) a \hat{\mathbf{y}} + \left(\frac{1}{2} - z_{19}\right) c \hat{\mathbf{z}}$ | (16$k$) | Si III |
| $\mathbf{B}_{243}$ | = | $y_{19} \mathbf{a}_1 + x_{19} \mathbf{a}_2 + \left(\frac{1}{2} - z_{19}\right) \mathbf{a}_3$ | = | $y_{19} a \hat{\mathbf{x}} + x_{19} a \hat{\mathbf{y}} + \left(\frac{1}{2} - z_{19}\right) c \hat{\mathbf{z}}$ | (16$k$) | Si III |
| $\mathbf{B}_{244}$ | = | $\left(\frac{1}{2} - y_{19}\right) \mathbf{a}_1 + \left(\frac{1}{2} - x_{19}\right) \mathbf{a}_2 + \left(\frac{1}{2} - z_{19}\right) \mathbf{a}_3$ | = | $\left(\frac{1}{2} - y_{19}\right) a \hat{\mathbf{x}} + \left(\frac{1}{2} - x_{19}\right) a \hat{\mathbf{y}} + \left(\frac{1}{2} - z_{19}\right) c \hat{\mathbf{z}}$ | (16$k$) | Si III |
| $\mathbf{B}_{245}$ | = | $-x_{19} \mathbf{a}_1 - y_{19} \mathbf{a}_2 - z_{19} \mathbf{a}_3$ | = | $-x_{19} a \hat{\mathbf{x}} - y_{19} a \hat{\mathbf{y}} - z_{19} c \hat{\mathbf{z}}$ | (16$k$) | Si III |
| $\mathbf{B}_{246}$ | = | $\left(\frac{1}{2} + x_{19}\right) \mathbf{a}_1 + \left(\frac{1}{2} + y_{19}\right) \mathbf{a}_2 - z_{19} \mathbf{a}_3$ | = | $\left(\frac{1}{2} + x_{19}\right) a \hat{\mathbf{x}} + \left(\frac{1}{2} + y_{19}\right) a \hat{\mathbf{y}} - z_{19} c \hat{\mathbf{z}}$ | (16$k$) | Si III |
| $\mathbf{B}_{247}$ | = | $\left(\frac{1}{2} + y_{19}\right) \mathbf{a}_1 - x_{19} \mathbf{a}_2 - z_{19} \mathbf{a}_3$ | = | $\left(\frac{1}{2} + y_{19}\right) a \hat{\mathbf{x}} - x_{19} a \hat{\mathbf{y}} - z_{19} c \hat{\mathbf{z}}$ | (16$k$) | Si III |
| $\mathbf{B}_{248}$ | = | $-y_{19} \mathbf{a}_1 + \left(\frac{1}{2} + x_{19}\right) \mathbf{a}_2 - z_{19} \mathbf{a}_3$ | = | $-y_{19} a \hat{\mathbf{x}} + \left(\frac{1}{2} + x_{19}\right) a \hat{\mathbf{y}} - z_{19} c \hat{\mathbf{z}}$ | (16$k$) | Si III |
| $\mathbf{B}_{249}$ | = | $\left(\frac{1}{2} + x_{19}\right) \mathbf{a}_1 - y_{19} \mathbf{a}_2 + \left(\frac{1}{2} + z_{19}\right) \mathbf{a}_3$ | = | $\left(\frac{1}{2} + x_{19}\right) a \hat{\mathbf{x}} - y_{19} a \hat{\mathbf{y}} + \left(\frac{1}{2} + z_{19}\right) c \hat{\mathbf{z}}$ | (16$k$) | Si III |
| $\mathbf{B}_{250}$ | = | $-x_{19} \mathbf{a}_1 + \left(\frac{1}{2} + y_{19}\right) \mathbf{a}_2 + \left(\frac{1}{2} + z_{19}\right) \mathbf{a}_3$ | = | $-x_{19} a \hat{\mathbf{x}} + \left(\frac{1}{2} + y_{19}\right) a \hat{\mathbf{y}} + \left(\frac{1}{2} + z_{19}\right) c \hat{\mathbf{z}}$ | (16$k$) | Si III |
| $\mathbf{B}_{251}$ | = | $-y_{19} \mathbf{a}_1 - x_{19} \mathbf{a}_2 + \left(\frac{1}{2} + z_{19}\right) \mathbf{a}_3$ | = | $-y_{19} a \hat{\mathbf{x}} - x_{19} a \hat{\mathbf{y}} + \left(\frac{1}{2} + z_{19}\right) c \hat{\mathbf{z}}$ | (16$k$) | Si III |
| $\mathbf{B}_{252}$ | = | $\left(\frac{1}{2} + y_{19}\right) \mathbf{a}_1 + \left(\frac{1}{2} + x_{19}\right) \mathbf{a}_2 + \left(\frac{1}{2} + z_{19}\right) \mathbf{a}_3$ | = | $\left(\frac{1}{2} + y_{19}\right) a \hat{\mathbf{x}} + \left(\frac{1}{2} + x_{19}\right) a \hat{\mathbf{y}} + \left(\frac{1}{2} + z_{19}\right) c \hat{\mathbf{z}}$ | (16$k$) | Si III |

**Geometry files:**

- CIF: pp. 1693
- POSCAR: pp. 1694

# $Ag[Co(NH_3)_2(NO_2)_4]$ ($J1_9$) Structure: ABC4D2E8_tP32_126_a_b_h_e_k

http://aflow.org/prototype-encyclopedia/ABC4D2E8_tP32_126_a_b_h_e_k

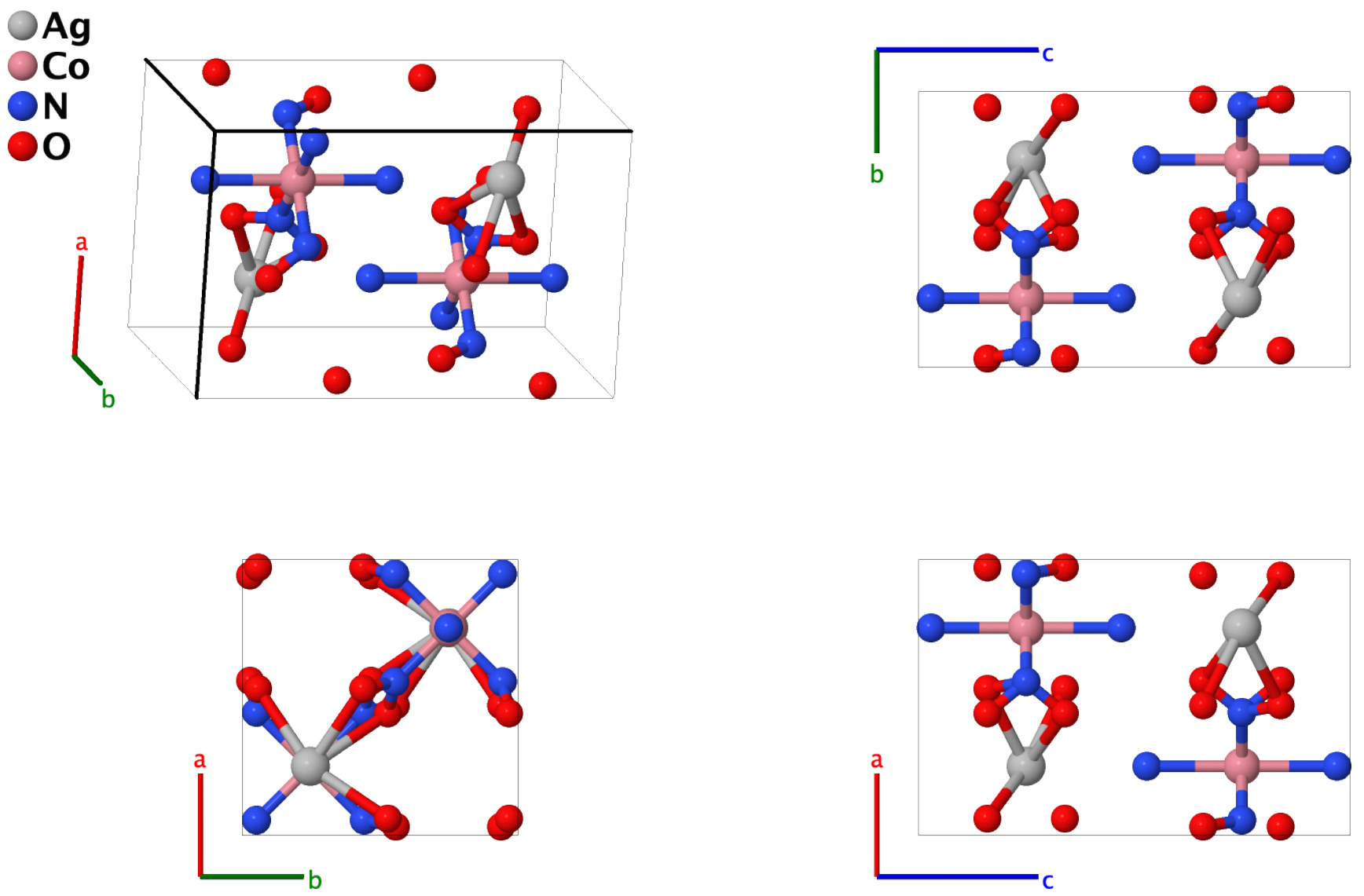

| | | |
|---|---|---|
| **Prototype** | : | $AgCoN_4(NH_3)_2O_8$ |
| **AFLOW prototype label** | : | ABC4D2E8_tP32_126_a_b_h_e_k |
| ***Strukturbericht* designation** | : | $J1_9$ |
| **Pearson symbol** | : | tP32 |
| **Space group number** | : | 126 |
| **Space group symbol** | : | $P4/nnc$ |
| **AFLOW prototype command** | : | `aflow --proto=ABC4D2E8_tP32_126_a_b_h_e_k --params=`$a, c/a, z_3, x_4, x_5, y_5, z_5$ |

- The positions of the hydrogen atoms in the ammonia molecule are not determined, so we only provide the positions of the nitrogen atoms (labeled as $NH_3$).
- (Wells, 1936) gives the lattice coordinates in setting 1 of space group #126. We used FINDSYM to change this to the standard setting 2.

**Simple Tetragonal primitive vectors:**

$$\mathbf{a}_1 = a\,\hat{\mathbf{x}}$$
$$\mathbf{a}_2 = a\,\hat{\mathbf{y}}$$
$$\mathbf{a}_3 = c\,\hat{\mathbf{z}}$$

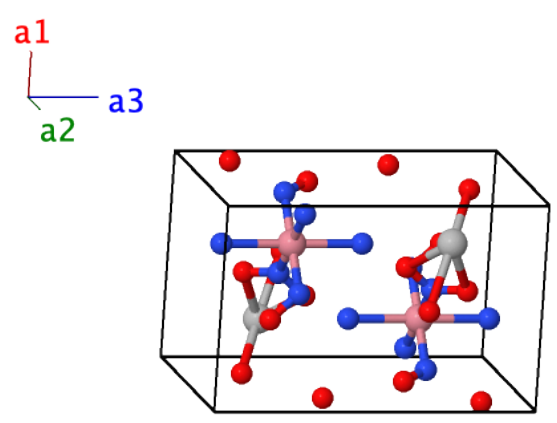

**Basis vectors:**

| | | Lattice Coordinates | | Cartesian Coordinates | Wyckoff Position | Atom Type |
|---|---|---|---|---|---|---|
| $\mathbf{B}_1$ | = | $\frac{1}{4}\,\mathbf{a}_1 + \frac{1}{4}\,\mathbf{a}_2 + \frac{1}{4}\,\mathbf{a}_3$ | = | $\frac{1}{4}a\,\hat{\mathbf{x}} + \frac{1}{4}a\,\hat{\mathbf{y}} + \frac{1}{4}c\,\hat{\mathbf{z}}$ | (2a) | Ag |
| $\mathbf{B}_{2}$ | = | $\frac{3}{4}\,\mathbf{a}_{1} + \frac{3}{4}\,\mathbf{a}_{2} + \frac{3}{4}\,\mathbf{a}_{3}$ | = | $\frac{3}{4}a\,\hat{\mathbf{x}} + \frac{3}{4}a\,\hat{\mathbf{y}} + \frac{3}{4}c\,\hat{\mathbf{z}}$ | (2$a$) | Ag |
| $\mathbf{B}_{3}$ | = | $\frac{1}{4}\,\mathbf{a}_{1} + \frac{1}{4}\,\mathbf{a}_{2} + \frac{3}{4}\,\mathbf{a}_{3}$ | = | $\frac{1}{4}a\,\hat{\mathbf{x}} + \frac{1}{4}a\,\hat{\mathbf{y}} + \frac{3}{4}c\,\hat{\mathbf{z}}$ | (2$b$) | Co |
| $\mathbf{B}_{4}$ | = | $\frac{3}{4}\,\mathbf{a}_{1} + \frac{3}{4}\,\mathbf{a}_{2} + \frac{1}{4}\,\mathbf{a}_{3}$ | = | $\frac{3}{4}a\,\hat{\mathbf{x}} + \frac{3}{4}a\,\hat{\mathbf{y}} + \frac{1}{4}c\,\hat{\mathbf{z}}$ | (2$b$) | Co |
| $\mathbf{B}_{5}$ | = | $\frac{1}{4}\,\mathbf{a}_{1} + \frac{1}{4}\,\mathbf{a}_{2} + z_{3}\,\mathbf{a}_{3}$ | = | $\frac{1}{4}a\,\hat{\mathbf{x}} + \frac{1}{4}a\,\hat{\mathbf{y}} + z_{3}c\,\hat{\mathbf{z}}$ | (4$e$) | $NH_3$ |
| $\mathbf{B}_{6}$ | = | $\frac{1}{4}\,\mathbf{a}_{1} + \frac{1}{4}\,\mathbf{a}_{2} + \left(\frac{1}{2} - z_{3}\right)\,\mathbf{a}_{3}$ | = | $\frac{1}{4}a\,\hat{\mathbf{x}} + \frac{1}{4}a\,\hat{\mathbf{y}} + \left(\frac{1}{2} - z_{3}\right)c\,\hat{\mathbf{z}}$ | (4$e$) | $NH_3$ |
| $\mathbf{B}_{7}$ | = | $\frac{3}{4}\,\mathbf{a}_{1} + \frac{3}{4}\,\mathbf{a}_{2} - z_{3}\,\mathbf{a}_{3}$ | = | $\frac{3}{4}a\,\hat{\mathbf{x}} + \frac{3}{4}a\,\hat{\mathbf{y}} - z_{3}c\,\hat{\mathbf{z}}$ | (4$e$) | $NH_3$ |
| $\mathbf{B}_{8}$ | = | $\frac{3}{4}\,\mathbf{a}_{1} + \frac{3}{4}\,\mathbf{a}_{2} + \left(\frac{1}{2} + z_{3}\right)\,\mathbf{a}_{3}$ | = | $\frac{3}{4}a\,\hat{\mathbf{x}} + \frac{3}{4}a\,\hat{\mathbf{y}} + \left(\frac{1}{2} + z_{3}\right)c\,\hat{\mathbf{z}}$ | (4$e$) | $NH_3$ |
| $\mathbf{B}_{9}$ | = | $x_{4}\,\mathbf{a}_{1} + x_{4}\,\mathbf{a}_{2} + \frac{1}{4}\,\mathbf{a}_{3}$ | = | $x_{4}a\,\hat{\mathbf{x}} + x_{4}a\,\hat{\mathbf{y}} + \frac{1}{4}c\,\hat{\mathbf{z}}$ | (8$h$) | N |
| $\mathbf{B}_{10}$ | = | $\left(\frac{1}{2} - x_{4}\right)\,\mathbf{a}_{1} + \left(\frac{1}{2} - x_{4}\right)\,\mathbf{a}_{2} + \frac{1}{4}\,\mathbf{a}_{3}$ | = | $\left(\frac{1}{2} - x_{4}\right)a\,\hat{\mathbf{x}} + \left(\frac{1}{2} - x_{4}\right)a\,\hat{\mathbf{y}} + \frac{1}{4}c\,\hat{\mathbf{z}}$ | (8$h$) | N |
| $\mathbf{B}_{11}$ | = | $\left(\frac{1}{2} - x_{4}\right)\,\mathbf{a}_{1} + x_{4}\,\mathbf{a}_{2} + \frac{1}{4}\,\mathbf{a}_{3}$ | = | $\left(\frac{1}{2} - x_{4}\right)a\,\hat{\mathbf{x}} + x_{4}a\,\hat{\mathbf{y}} + \frac{1}{4}c\,\hat{\mathbf{z}}$ | (8$h$) | N |
| $\mathbf{B}_{12}$ | = | $x_{4}\,\mathbf{a}_{1} + \left(\frac{1}{2} - x_{4}\right)\,\mathbf{a}_{2} + \frac{1}{4}\,\mathbf{a}_{3}$ | = | $x_{4}a\,\hat{\mathbf{x}} + \left(\frac{1}{2} - x_{4}\right)a\,\hat{\mathbf{y}} + \frac{1}{4}c\,\hat{\mathbf{z}}$ | (8$h$) | N |
| $\mathbf{B}_{13}$ | = | $-x_{4}\,\mathbf{a}_{1} - x_{4}\,\mathbf{a}_{2} + \frac{3}{4}\,\mathbf{a}_{3}$ | = | $-x_{4}a\,\hat{\mathbf{x}} - x_{4}a\,\hat{\mathbf{y}} + \frac{3}{4}c\,\hat{\mathbf{z}}$ | (8$h$) | N |
| $\mathbf{B}_{14}$ | = | $\left(\frac{1}{2} + x_{4}\right)\,\mathbf{a}_{1} + \left(\frac{1}{2} + x_{4}\right)\,\mathbf{a}_{2} + \frac{3}{4}\,\mathbf{a}_{3}$ | = | $\left(\frac{1}{2} + x_{4}\right)a\,\hat{\mathbf{x}} + \left(\frac{1}{2} + x_{4}\right)a\,\hat{\mathbf{y}} + \frac{3}{4}c\,\hat{\mathbf{z}}$ | (8$h$) | N |
| $\mathbf{B}_{15}$ | = | $\left(\frac{1}{2} + x_{4}\right)\,\mathbf{a}_{1} - x_{4}\,\mathbf{a}_{2} + \frac{3}{4}\,\mathbf{a}_{3}$ | = | $\left(\frac{1}{2} + x_{4}\right)a\,\hat{\mathbf{x}} - x_{4}a\,\hat{\mathbf{y}} + \frac{3}{4}c\,\hat{\mathbf{z}}$ | (8$h$) | N |
| $\mathbf{B}_{16}$ | = | $-x_{4}\,\mathbf{a}_{1} + \left(\frac{1}{2} + x_{4}\right)\,\mathbf{a}_{2} + \frac{3}{4}\,\mathbf{a}_{3}$ | = | $-x_{4}a\,\hat{\mathbf{x}} + \left(\frac{1}{2} + x_{4}\right)a\,\hat{\mathbf{y}} + \frac{3}{4}c\,\hat{\mathbf{z}}$ | (8$h$) | N |
| $\mathbf{B}_{17}$ | = | $x_{5}\,\mathbf{a}_{1} + y_{5}\,\mathbf{a}_{2} + z_{5}\,\mathbf{a}_{3}$ | = | $x_{5}a\,\hat{\mathbf{x}} + y_{5}a\,\hat{\mathbf{y}} + z_{5}c\,\hat{\mathbf{z}}$ | (16$k$) | O |
| $\mathbf{B}_{18}$ | = | $\left(\frac{1}{2} - x_{5}\right)\,\mathbf{a}_{1} + \left(\frac{1}{2} - y_{5}\right)\,\mathbf{a}_{2} + z_{5}\,\mathbf{a}_{3}$ | = | $\left(\frac{1}{2} - x_{5}\right)a\,\hat{\mathbf{x}} + \left(\frac{1}{2} - y_{5}\right)a\,\hat{\mathbf{y}} + z_{5}c\,\hat{\mathbf{z}}$ | (16$k$) | O |
| $\mathbf{B}_{19}$ | = | $\left(\frac{1}{2} - y_{5}\right)\,\mathbf{a}_{1} + x_{5}\,\mathbf{a}_{2} + z_{5}\,\mathbf{a}_{3}$ | = | $\left(\frac{1}{2} - y_{5}\right)a\,\hat{\mathbf{x}} + x_{5}a\,\hat{\mathbf{y}} + z_{5}c\,\hat{\mathbf{z}}$ | (16$k$) | O |
| $\mathbf{B}_{20}$ | = | $y_{5}\,\mathbf{a}_{1} + \left(\frac{1}{2} - x_{5}\right)\,\mathbf{a}_{2} + z_{5}\,\mathbf{a}_{3}$ | = | $y_{5}a\,\hat{\mathbf{x}} + \left(\frac{1}{2} - x_{5}\right)a\,\hat{\mathbf{y}} + z_{5}c\,\hat{\mathbf{z}}$ | (16$k$) | O |
| $\mathbf{B}_{21}$ | = | $\left(\frac{1}{2} - x_{5}\right)\,\mathbf{a}_{1} + y_{5}\,\mathbf{a}_{2} + \left(\frac{1}{2} - z_{5}\right)\,\mathbf{a}_{3}$ | = | $\left(\frac{1}{2} - x_{5}\right)a\,\hat{\mathbf{x}} + y_{5}a\,\hat{\mathbf{y}} + \left(\frac{1}{2} - z_{5}\right)c\,\hat{\mathbf{z}}$ | (16$k$) | O |
| $\mathbf{B}_{22}$ | = | $x_{5}\,\mathbf{a}_{1} + \left(\frac{1}{2} - y_{5}\right)\,\mathbf{a}_{2} + \left(\frac{1}{2} - z_{5}\right)\,\mathbf{a}_{3}$ | = | $x_{5}a\,\hat{\mathbf{x}} + \left(\frac{1}{2} - y_{5}\right)a\,\hat{\mathbf{y}} + \left(\frac{1}{2} - z_{5}\right)c\,\hat{\mathbf{z}}$ | (16$k$) | O |
| $\mathbf{B}_{23}$ | = | $y_{5}\,\mathbf{a}_{1} + x_{5}\,\mathbf{a}_{2} + \left(\frac{1}{2} - z_{5}\right)\,\mathbf{a}_{3}$ | = | $y_{5}a\,\hat{\mathbf{x}} + x_{5}a\,\hat{\mathbf{y}} + \left(\frac{1}{2} - z_{5}\right)c\,\hat{\mathbf{z}}$ | (16$k$) | O |
| $\mathbf{B}_{24}$ | = | $\left(\frac{1}{2} - y_{5}\right)\,\mathbf{a}_{1} + \left(\frac{1}{2} - x_{5}\right)\,\mathbf{a}_{2} + \left(\frac{1}{2} - z_{5}\right)\,\mathbf{a}_{3}$ | = | $\left(\frac{1}{2} - y_{5}\right)a\,\hat{\mathbf{x}} + \left(\frac{1}{2} - x_{5}\right)a\,\hat{\mathbf{y}} + \left(\frac{1}{2} - z_{5}\right)c\,\hat{\mathbf{z}}$ | (16$k$) | O |
| $\mathbf{B}_{25}$ | = | $-x_{5}\,\mathbf{a}_{1} - y_{5}\,\mathbf{a}_{2} - z_{5}\,\mathbf{a}_{3}$ | = | $-x_{5}a\,\hat{\mathbf{x}} - y_{5}a\,\hat{\mathbf{y}} - z_{5}c\,\hat{\mathbf{z}}$ | (16$k$) | O |
| $\mathbf{B}_{26}$ | = | $\left(\frac{1}{2} + x_{5}\right)\,\mathbf{a}_{1} + \left(\frac{1}{2} + y_{5}\right)\,\mathbf{a}_{2} - z_{5}\,\mathbf{a}_{3}$ | = | $\left(\frac{1}{2} + x_{5}\right)a\,\hat{\mathbf{x}} + \left(\frac{1}{2} + y_{5}\right)a\,\hat{\mathbf{y}} - z_{5}c\,\hat{\mathbf{z}}$ | (16$k$) | O |
| $\mathbf{B}_{27}$ | = | $\left(\frac{1}{2} + y_{5}\right)\,\mathbf{a}_{1} - x_{5}\,\mathbf{a}_{2} - z_{5}\,\mathbf{a}_{3}$ | = | $\left(\frac{1}{2} + y_{5}\right)a\,\hat{\mathbf{x}} - x_{5}a\,\hat{\mathbf{y}} - z_{5}c\,\hat{\mathbf{z}}$ | (16$k$) | O |
| $\mathbf{B}_{28}$ | = | $-y_{5}\,\mathbf{a}_{1} + \left(\frac{1}{2} + x_{5}\right)\,\mathbf{a}_{2} - z_{5}\,\mathbf{a}_{3}$ | = | $-y_{5}a\,\hat{\mathbf{x}} + \left(\frac{1}{2} + x_{5}\right)a\,\hat{\mathbf{y}} - z_{5}c\,\hat{\mathbf{z}}$ | (16$k$) | O |
| $\mathbf{B}_{29}$ | = | $\left(\frac{1}{2} + x_{5}\right)\,\mathbf{a}_{1} - y_{5}\,\mathbf{a}_{2} + \left(\frac{1}{2} + z_{5}\right)\,\mathbf{a}_{3}$ | = | $\left(\frac{1}{2} + x_{5}\right)a\,\hat{\mathbf{x}} - y_{5}a\,\hat{\mathbf{y}} + \left(\frac{1}{2} + z_{5}\right)c\,\hat{\mathbf{z}}$ | (16$k$) | O |
| $\mathbf{B}_{30}$ | = | $-x_{5}\,\mathbf{a}_{1} + \left(\frac{1}{2} + y_{5}\right)\,\mathbf{a}_{2} + \left(\frac{1}{2} + z_{5}\right)\,\mathbf{a}_{3}$ | = | $-x_{5}a\,\hat{\mathbf{x}} + \left(\frac{1}{2} + y_{5}\right)a\,\hat{\mathbf{y}} + \left(\frac{1}{2} + z_{5}\right)c\,\hat{\mathbf{z}}$ | (16$k$) | O |
| $\mathbf{B}_{31}$ | = | $-y_{5}\,\mathbf{a}_{1} - x_{5}\,\mathbf{a}_{2} + \left(\frac{1}{2} + z_{5}\right)\,\mathbf{a}_{3}$ | = | $-y_{5}a\,\hat{\mathbf{x}} - x_{5}a\,\hat{\mathbf{y}} + \left(\frac{1}{2} + z_{5}\right)c\,\hat{\mathbf{z}}$ | (16$k$) | O |
| $\mathbf{B}_{32}$ | = | $\left(\frac{1}{2} + y_{5}\right)\,\mathbf{a}_{1} + \left(\frac{1}{2} + x_{5}\right)\,\mathbf{a}_{2} + \left(\frac{1}{2} + z_{5}\right)\,\mathbf{a}_{3}$ | = | $\left(\frac{1}{2} + y_{5}\right)a\,\hat{\mathbf{x}} + \left(\frac{1}{2} + x_{5}\right)a\,\hat{\mathbf{y}} + \left(\frac{1}{2} + z_{5}\right)c\,\hat{\mathbf{z}}$ | (16$k$) | O |

**Found in:**

---

**Geometry files:**

- CIF: pp. 1695
- POSCAR: pp. 1696

# $Pd(NH_3)_4Cl_2{\cdot}H_2O$ ($H4_9$) Structure: A2BC4D_tP16_127_h_d_i_a

http://aflow.org/prototype-encyclopedia/A2BC4D_tP16_127_h_d_i_a

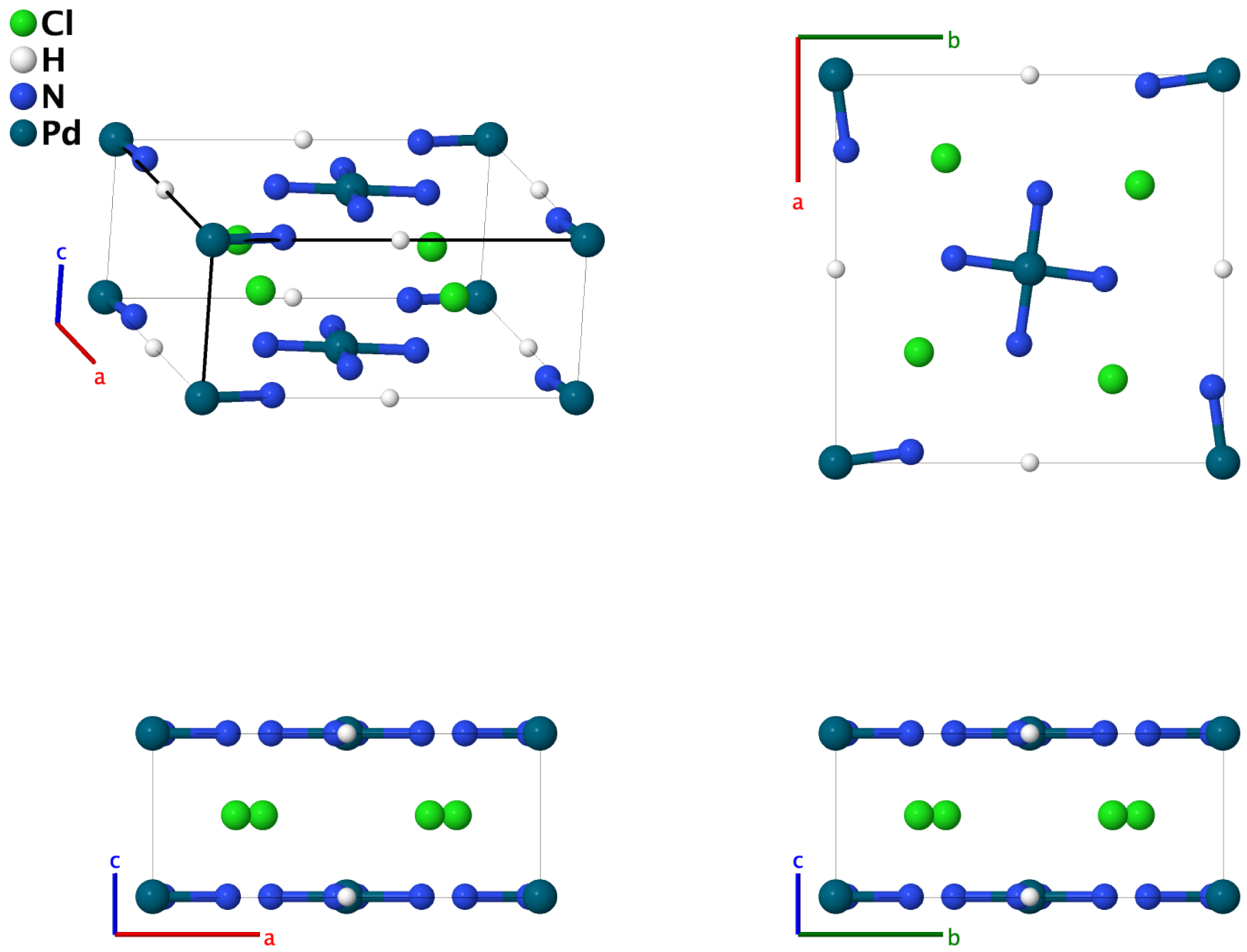

| | | |
|---|---|---|
| **Prototype** | : | $Cl_2(H_2O)(NH_3)_4Pd$ |
| **AFLOW prototype label** | : | A2BC4D_tP16_127_h_d_i_a |
| ***Strukturbericht* designation** | : | $H4_9$ |
| **Pearson symbol** | : | tP16 |
| **Space group number** | : | 127 |
| **Space group symbol** | : | $P4/mbm$ |
| **AFLOW prototype command** | : | `aflow --proto=A2BC4D_tP16_127_h_d_i_a --params=`$a, c/a, x_3, x_4, y_4$ |

- The positions of the hydrogen atoms have not been measured, assuming they are actually fixed. We group them together with the central atoms of their molecules.

**Simple Tetragonal primitive vectors:**

$$\mathbf{a}_1 = a\,\hat{\mathbf{x}}$$
$$\mathbf{a}_2 = a\,\hat{\mathbf{y}}$$
$$\mathbf{a}_3 = c\,\hat{\mathbf{z}}$$

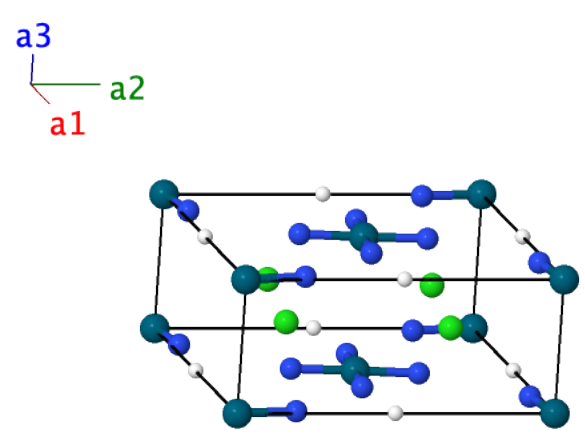

**Basis vectors:**

| | | Lattice Coordinates | | Cartesian Coordinates | Wyckoff Position | Atom Type |
|---|---|---|---|---|---|---|
| $\mathbf{B}_1$ | = | $0\,\mathbf{a}_1 + 0\,\mathbf{a}_2 + 0\,\mathbf{a}_3$ | = | $0\,\hat{\mathbf{x}} + 0\,\hat{\mathbf{y}} + 0\,\hat{\mathbf{z}}$ | (2$a$) | Pd |
| $\mathbf{B}_2$ | = | $\frac{1}{2}\,\mathbf{a}_1 + \frac{1}{2}\,\mathbf{a}_2$ | = | $\frac{1}{2}a\,\hat{\mathbf{x}} + \frac{1}{2}a\,\hat{\mathbf{y}}$ | (2$a$) | Pd |
| $\mathbf{B}_3$ | = | $\frac{1}{2}\,\mathbf{a}_2$ | = | $\frac{1}{2}a\,\hat{\mathbf{y}}$ | (2$d$) | $H_2O$ |
| $\mathbf{B}_4$ | = | $\frac{1}{2}\,\mathbf{a}_1$ | = | $\frac{1}{2}a\,\hat{\mathbf{x}}$ | (2$d$) | $H_2O$ |
| $\mathbf{B}_5$ | = | $x_3\,\mathbf{a}_1 + \left(\frac{1}{2} + x_3\right)\,\mathbf{a}_2 + \frac{1}{2}\,\mathbf{a}_3$ | = | $x_3a\,\hat{\mathbf{x}} + \left(\frac{1}{2} + x_3\right)a\,\hat{\mathbf{y}} + \frac{1}{2}c\,\hat{\mathbf{z}}$ | (4$h$) | Cl |
| $\mathbf{B}_6$ | = | $-x_3\,\mathbf{a}_1 + \left(\frac{1}{2} - x_3\right)\,\mathbf{a}_2 + \frac{1}{2}\,\mathbf{a}_3$ | = | $-x_3a\,\hat{\mathbf{x}} + \left(\frac{1}{2} - x_3\right)a\,\hat{\mathbf{y}} + \frac{1}{2}c\,\hat{\mathbf{z}}$ | (4$h$) | Cl |
| $\mathbf{B}_7$ | = | $\left(\frac{1}{2} - x_3\right)\,\mathbf{a}_1 + x_3\,\mathbf{a}_2 + \frac{1}{2}\,\mathbf{a}_3$ | = | $\left(\frac{1}{2} - x_3\right)a\,\hat{\mathbf{x}} + x_3a\,\hat{\mathbf{y}} + \frac{1}{2}c\,\hat{\mathbf{z}}$ | (4$h$) | Cl |
| $\mathbf{B}_8$ | = | $\left(\frac{1}{2} + x_3\right)\,\mathbf{a}_1 - x_3\,\mathbf{a}_2 + \frac{1}{2}\,\mathbf{a}_3$ | = | $\left(\frac{1}{2} + x_3\right)a\,\hat{\mathbf{x}} - x_3a\,\hat{\mathbf{y}} + \frac{1}{2}c\,\hat{\mathbf{z}}$ | (4$h$) | Cl |
| $\mathbf{B}_9$ | = | $x_4\,\mathbf{a}_1 + y_4\,\mathbf{a}_2$ | = | $x_4a\,\hat{\mathbf{x}} + y_4a\,\hat{\mathbf{y}}$ | (8$i$) | $NH_3$ |
| $\mathbf{B}_{10}$ | = | $-x_4\,\mathbf{a}_1 - y_4\,\mathbf{a}_2$ | = | $-x_4a\,\hat{\mathbf{x}} - y_4a\,\hat{\mathbf{y}}$ | (8$i$) | $NH_3$ |
| $\mathbf{B}_{11}$ | = | $-y_4\,\mathbf{a}_1 + x_4\,\mathbf{a}_2$ | = | $-y_4a\,\hat{\mathbf{x}} + x_4a\,\hat{\mathbf{y}}$ | (8$i$) | $NH_3$ |
| $\mathbf{B}_{12}$ | = | $y_4\,\mathbf{a}_1 - x_4\,\mathbf{a}_2$ | = | $y_4a\,\hat{\mathbf{x}} - x_4a\,\hat{\mathbf{y}}$ | (8$i$) | $NH_3$ |
| $\mathbf{B}_{13}$ | = | $\left(\frac{1}{2} - x_4\right)\,\mathbf{a}_1 + \left(\frac{1}{2} + y_4\right)\,\mathbf{a}_2$ | = | $\left(\frac{1}{2} - x_4\right)a\,\hat{\mathbf{x}} + \left(\frac{1}{2} + y_4\right)a\,\hat{\mathbf{y}}$ | (8$i$) | $NH_3$ |
| $\mathbf{B}_{14}$ | = | $\left(\frac{1}{2} + x_4\right)\,\mathbf{a}_1 + \left(\frac{1}{2} - y_4\right)\,\mathbf{a}_2$ | = | $\left(\frac{1}{2} + x_4\right)a\,\hat{\mathbf{x}} + \left(\frac{1}{2} - y_4\right)a\,\hat{\mathbf{y}}$ | (8$i$) | $NH_3$ |
| $\mathbf{B}_{15}$ | = | $\left(\frac{1}{2} + y_4\right)\,\mathbf{a}_1 + \left(\frac{1}{2} + x_4\right)\,\mathbf{a}_2$ | = | $\left(\frac{1}{2} + y_4\right)a\,\hat{\mathbf{x}} + \left(\frac{1}{2} + x_4\right)a\,\hat{\mathbf{y}}$ | (8$i$) | $NH_3$ |
| $\mathbf{B}_{16}$ | = | $\left(\frac{1}{2} - y_4\right)\,\mathbf{a}_1 + \left(\frac{1}{2} - x_4\right)\,\mathbf{a}_2$ | = | $\left(\frac{1}{2} - y_4\right)a\,\hat{\mathbf{x}} + \left(\frac{1}{2} - x_4\right)a\,\hat{\mathbf{y}}$ | (8$i$) | $NH_3$ |

**Geometry files:**

- CIF: pp. 1696
- POSCAR: pp. 1696

# Phosgenite [$Pb_2Cl_2(CO_3)$] Structure: AB2C3D2_tP32_127_g_eh_gk_k

http://aflow.org/prototype-encyclopedia/AB2C3D2_tP32_127_g_eh_gk_k

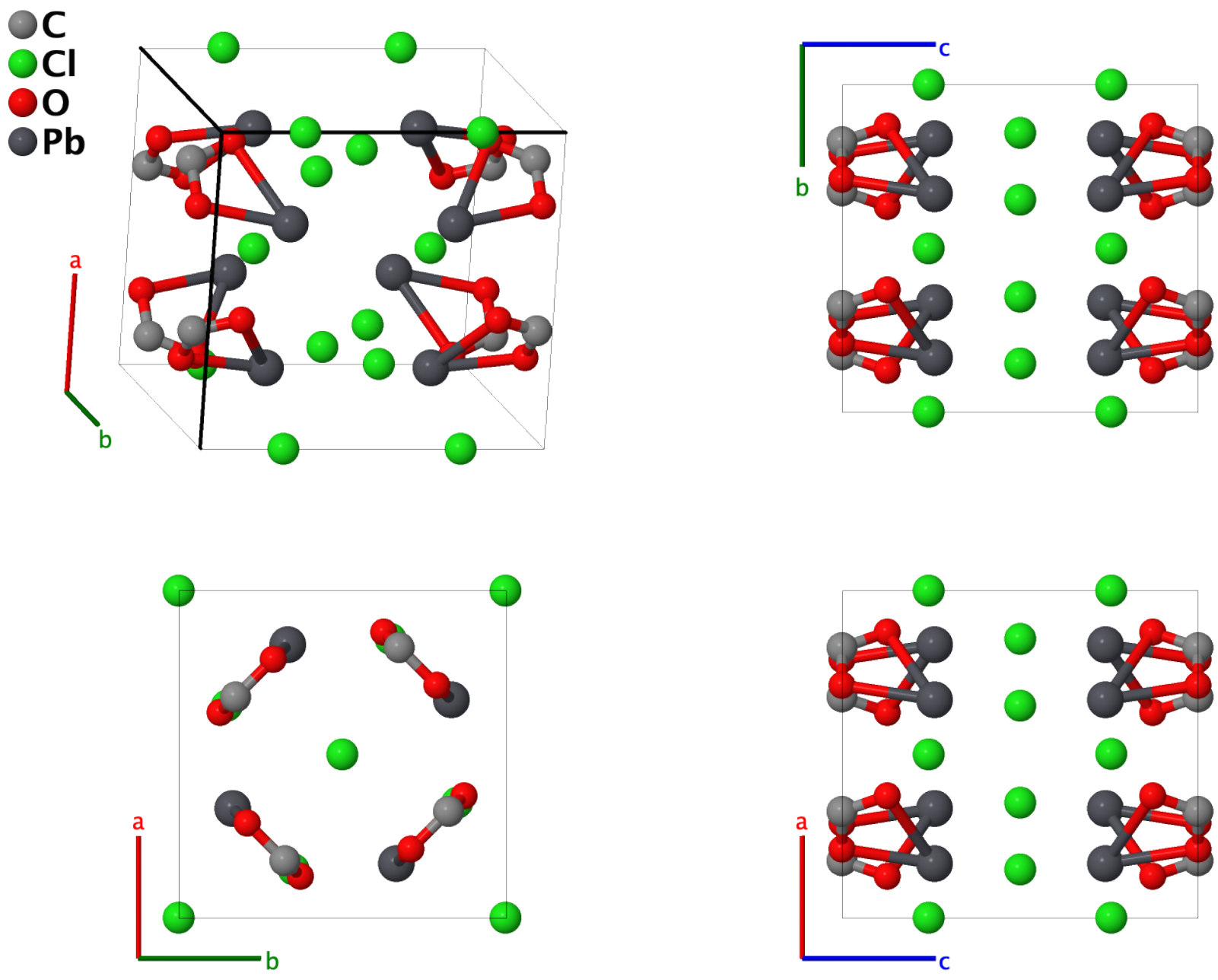

| | | |
|---|---|---|
| **Prototype** | : | $CCl_2O_3Pb_2$ |
| **AFLOW prototype label** | : | AB2C3D2_tP32_127_g_eh_gk_k |
| ***Strukturbericht* designation** | : | None |
| **Pearson symbol** | : | tP32 |
| **Space group number** | : | 127 |
| **Space group symbol** | : | $P4/mbm$ |
| **AFLOW prototype command** | : | `aflow --proto=AB2C3D2_tP32_127_g_eh_gk_k`<br>`--params=`$a, c/a, z_1, x_2, x_3, x_4, x_5, z_5, x_6, z_6$ |

- (Onotaro, 1934) made the first determination of the structure of phosgenite, which was given *Strukturbericht* designation $G7_5$ by (Gottfried). (Giuseppetti, 1974) and others found that the actual structure has a unit cell twice that of $G7_5$. This newer structure is presented here.

**Simple Tetragonal primitive vectors:**

$$\mathbf{a}_1 = a\,\hat{\mathbf{x}}$$
$$\mathbf{a}_2 = a\,\hat{\mathbf{y}}$$
$$\mathbf{a}_3 = c\,\hat{\mathbf{z}}$$

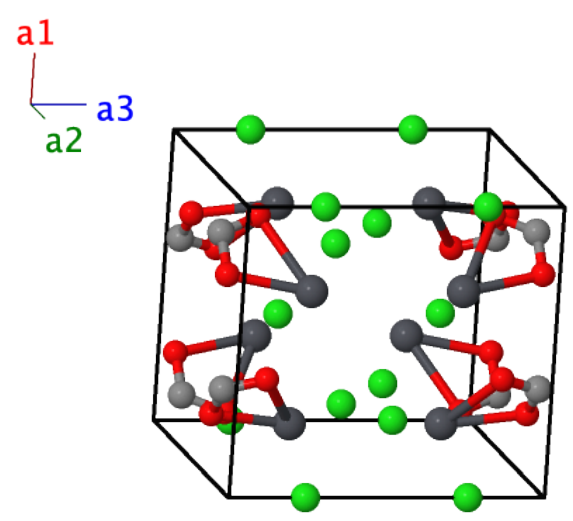

**Basis vectors:**

| | | Lattice Coordinates | | Cartesian Coordinates | Wyckoff Position | Atom Type |
|---|---|---|---|---|---|---|
| $\mathbf{B}_1$ | = | $z_1\,\mathbf{a}_3$ | = | $z_1 c\,\hat{\mathbf{z}}$ | (4e) | Cl I |
| $\mathbf{B}_2$ | = | $\frac{1}{2}\,\mathbf{a}_1 + \frac{1}{2}\,\mathbf{a}_2 - z_1\,\mathbf{a}_3$ | = | $\frac{1}{2}a\,\hat{\mathbf{x}} + \frac{1}{2}a\,\hat{\mathbf{y}} - z_1 c\,\hat{\mathbf{z}}$ | (4e) | Cl I |
| $\mathbf{B}_3$ | = | $-z_1\,\mathbf{a}_3$ | = | $-z_1 c\,\hat{\mathbf{z}}$ | (4e) | Cl I |
| $\mathbf{B}_4$ | = | $\frac{1}{2}\,\mathbf{a}_1 + \frac{1}{2}\,\mathbf{a}_2 + z_1\,\mathbf{a}_3$ | = | $\frac{1}{2}a\,\hat{\mathbf{x}} + \frac{1}{2}a\,\hat{\mathbf{y}} + z_1 c\,\hat{\mathbf{z}}$ | (4e) | Cl I |
| $\mathbf{B}_5$ | = | $x_2\,\mathbf{a}_1 + \left(\frac{1}{2} + x_2\right)\,\mathbf{a}_2$ | = | $x_2 a\,\hat{\mathbf{x}} + \left(\frac{1}{2} + x_2\right)a\,\hat{\mathbf{y}}$ | (4g) | C |
| $\mathbf{B}_6$ | = | $-x_2\,\mathbf{a}_1 + \left(\frac{1}{2} - x_2\right)\,\mathbf{a}_2$ | = | $-x_2 a\,\hat{\mathbf{x}} + \left(\frac{1}{2} - x_2\right)a\,\hat{\mathbf{y}}$ | (4g) | C |
| $\mathbf{B}_7$ | = | $\left(\frac{1}{2} - x_2\right)\,\mathbf{a}_1 + x_2\,\mathbf{a}_2$ | = | $\left(\frac{1}{2} - x_2\right)a\,\hat{\mathbf{x}} + x_2 a\,\hat{\mathbf{y}}$ | (4g) | C |
| $\mathbf{B}_8$ | = | $\left(\frac{1}{2} + x_2\right)\,\mathbf{a}_1 - x_2\,\mathbf{a}_2$ | = | $\left(\frac{1}{2} + x_2\right)a\,\hat{\mathbf{x}} - x_2 a\,\hat{\mathbf{y}}$ | (4g) | C |
| $\mathbf{B}_9$ | = | $x_3\,\mathbf{a}_1 + \left(\frac{1}{2} + x_3\right)\,\mathbf{a}_2$ | = | $x_3 a\,\hat{\mathbf{x}} + \left(\frac{1}{2} + x_3\right)a\,\hat{\mathbf{y}}$ | (4g) | O I |
| $\mathbf{B}_{10}$ | = | $-x_3\,\mathbf{a}_1 + \left(\frac{1}{2} - x_3\right)\,\mathbf{a}_2$ | = | $-x_3 a\,\hat{\mathbf{x}} + \left(\frac{1}{2} - x_3\right)a\,\hat{\mathbf{y}}$ | (4g) | O I |
| $\mathbf{B}_{11}$ | = | $\left(\frac{1}{2} - x_3\right)\,\mathbf{a}_1 + x_3\,\mathbf{a}_2$ | = | $\left(\frac{1}{2} - x_3\right)a\,\hat{\mathbf{x}} + x_3 a\,\hat{\mathbf{y}}$ | (4g) | O I |
| $\mathbf{B}_{12}$ | = | $\left(\frac{1}{2} + x_3\right)\,\mathbf{a}_1 - x_3\,\mathbf{a}_2$ | = | $\left(\frac{1}{2} + x_3\right)a\,\hat{\mathbf{x}} - x_3 a\,\hat{\mathbf{y}}$ | (4g) | O I |
| $\mathbf{B}_{13}$ | = | $x_4\,\mathbf{a}_1 + \left(\frac{1}{2} + x_4\right)\,\mathbf{a}_2 + \frac{1}{2}\,\mathbf{a}_3$ | = | $x_4 a\,\hat{\mathbf{x}} + \left(\frac{1}{2} + x_4\right)a\,\hat{\mathbf{y}} + \frac{1}{2}c\,\hat{\mathbf{z}}$ | (4h) | Cl II |
| $\mathbf{B}_{14}$ | = | $-x_4\,\mathbf{a}_1 + \left(\frac{1}{2} - x_4\right)\,\mathbf{a}_2 + \frac{1}{2}\,\mathbf{a}_3$ | = | $-x_4 a\,\hat{\mathbf{x}} + \left(\frac{1}{2} - x_4\right)a\,\hat{\mathbf{y}} + \frac{1}{2}c\,\hat{\mathbf{z}}$ | (4h) | Cl II |
| $\mathbf{B}_{15}$ | = | $\left(\frac{1}{2} - x_4\right)\,\mathbf{a}_1 + x_4\,\mathbf{a}_2 + \frac{1}{2}\,\mathbf{a}_3$ | = | $\left(\frac{1}{2} - x_4\right)a\,\hat{\mathbf{x}} + x_4 a\,\hat{\mathbf{y}} + \frac{1}{2}c\,\hat{\mathbf{z}}$ | (4h) | Cl II |
| $\mathbf{B}_{16}$ | = | $\left(\frac{1}{2} + x_4\right)\,\mathbf{a}_1 - x_4\,\mathbf{a}_2 + \frac{1}{2}\,\mathbf{a}_3$ | = | $\left(\frac{1}{2} + x_4\right)a\,\hat{\mathbf{x}} - x_4 a\,\hat{\mathbf{y}} + \frac{1}{2}c\,\hat{\mathbf{z}}$ | (4h) | Cl II |
| $\mathbf{B}_{17}$ | = | $x_5\,\mathbf{a}_1 + \left(\frac{1}{2} + x_5\right)\,\mathbf{a}_2 + z_5\,\mathbf{a}_3$ | = | $x_5 a\,\hat{\mathbf{x}} + \left(\frac{1}{2} + x_5\right)a\,\hat{\mathbf{y}} + z_5 c\,\hat{\mathbf{z}}$ | (8k) | O II |
| $\mathbf{B}_{18}$ | = | $-x_5\,\mathbf{a}_1 + \left(\frac{1}{2} - x_5\right)\,\mathbf{a}_2 + z_5\,\mathbf{a}_3$ | = | $-x_5 a\,\hat{\mathbf{x}} + \left(\frac{1}{2} - x_5\right)a\,\hat{\mathbf{y}} + z_5 c\,\hat{\mathbf{z}}$ | (8k) | O II |
| $\mathbf{B}_{19}$ | = | $\left(\frac{1}{2} - x_5\right)\,\mathbf{a}_1 + x_5\,\mathbf{a}_2 + z_5\,\mathbf{a}_3$ | = | $\left(\frac{1}{2} - x_5\right)a\,\hat{\mathbf{x}} + x_5 a\,\hat{\mathbf{y}} + z_5 c\,\hat{\mathbf{z}}$ | (8k) | O II |
| $\mathbf{B}_{20}$ | = | $\left(\frac{1}{2} + x_5\right)\,\mathbf{a}_1 - x_5\,\mathbf{a}_2 + z_5\,\mathbf{a}_3$ | = | $\left(\frac{1}{2} + x_5\right)a\,\hat{\mathbf{x}} - x_5 a\,\hat{\mathbf{y}} + z_5 c\,\hat{\mathbf{z}}$ | (8k) | O II |
| $\mathbf{B}_{21}$ | = | $\left(\frac{1}{2} - x_5\right)\,\mathbf{a}_1 + x_5\,\mathbf{a}_2 - z_5\,\mathbf{a}_3$ | = | $\left(\frac{1}{2} - x_5\right)a\,\hat{\mathbf{x}} + x_5 a\,\hat{\mathbf{y}} - z_5 c\,\hat{\mathbf{z}}$ | (8k) | O II |
| $\mathbf{B}_{22}$ | = | $\left(\frac{1}{2} + x_5\right)\,\mathbf{a}_1 - x_5\,\mathbf{a}_2 - z_5\,\mathbf{a}_3$ | = | $\left(\frac{1}{2} + x_5\right)a\,\hat{\mathbf{x}} - x_5 a\,\hat{\mathbf{y}} - z_5 c\,\hat{\mathbf{z}}$ | (8k) | O II |
| $\mathbf{B}_{23}$ | = | $x_5\,\mathbf{a}_1 + \left(\frac{1}{2} + x_5\right)\,\mathbf{a}_2 - z_5\,\mathbf{a}_3$ | = | $x_5 a\,\hat{\mathbf{x}} + \left(\frac{1}{2} + x_5\right)a\,\hat{\mathbf{y}} - z_5 c\,\hat{\mathbf{z}}$ | (8k) | O II |
| $\mathbf{B}_{24}$ | = | $-x_5\,\mathbf{a}_1 + \left(\frac{1}{2} - x_5\right)\,\mathbf{a}_2 - z_5\,\mathbf{a}_3$ | = | $-x_5 a\,\hat{\mathbf{x}} + \left(\frac{1}{2} - x_5\right)a\,\hat{\mathbf{y}} - z_5 c\,\hat{\mathbf{z}}$ | (8k) | O II |
| $\mathbf{B}_{25}$ | = | $x_6\,\mathbf{a}_1 + \left(\frac{1}{2} + x_6\right)\,\mathbf{a}_2 + z_6\,\mathbf{a}_3$ | = | $x_6 a\,\hat{\mathbf{x}} + \left(\frac{1}{2} + x_6\right)a\,\hat{\mathbf{y}} + z_6 c\,\hat{\mathbf{z}}$ | (8k) | Pb |
| $\mathbf{B}_{26}$ | = | $-x_6\,\mathbf{a}_1 + \left(\frac{1}{2} - x_6\right)\,\mathbf{a}_2 + z_6\,\mathbf{a}_3$ | = | $-x_6 a\,\hat{\mathbf{x}} + \left(\frac{1}{2} - x_6\right)a\,\hat{\mathbf{y}} + z_6 c\,\hat{\mathbf{z}}$ | (8k) | Pb |
| $\mathbf{B}_{27}$ | = | $\left(\frac{1}{2} - x_6\right)\,\mathbf{a}_1 + x_6\,\mathbf{a}_2 + z_6\,\mathbf{a}_3$ | = | $\left(\frac{1}{2} - x_6\right)a\,\hat{\mathbf{x}} + x_6 a\,\hat{\mathbf{y}} + z_6 c\,\hat{\mathbf{z}}$ | (8k) | Pb |
| $\mathbf{B}_{28}$ | = | $\left(\frac{1}{2} + x_6\right)\,\mathbf{a}_1 - x_6\,\mathbf{a}_2 + z_6\,\mathbf{a}_3$ | = | $\left(\frac{1}{2} + x_6\right)a\,\hat{\mathbf{x}} - x_6 a\,\hat{\mathbf{y}} + z_6 c\,\hat{\mathbf{z}}$ | (8k) | Pb |
| $\mathbf{B}_{29}$ | = | $\left(\frac{1}{2} - x_6\right)\,\mathbf{a}_1 + x_6\,\mathbf{a}_2 - z_6\,\mathbf{a}_3$ | = | $\left(\frac{1}{2} - x_6\right)a\,\hat{\mathbf{x}} + x_6 a\,\hat{\mathbf{y}} - z_6 c\,\hat{\mathbf{z}}$ | (8k) | Pb |
| $\mathbf{B}_{30}$ | = | $\left(\frac{1}{2} + x_6\right)\,\mathbf{a}_1 - x_6\,\mathbf{a}_2 - z_6\,\mathbf{a}_3$ | = | $\left(\frac{1}{2} + x_6\right)a\,\hat{\mathbf{x}} - x_6 a\,\hat{\mathbf{y}} - z_6 c\,\hat{\mathbf{z}}$ | (8k) | Pb |
| $\mathbf{B}_{31}$ | = | $x_6\,\mathbf{a}_1 + \left(\frac{1}{2} + x_6\right)\,\mathbf{a}_2 - z_6\,\mathbf{a}_3$ | = | $x_6 a\,\hat{\mathbf{x}} + \left(\frac{1}{2} + x_6\right)a\,\hat{\mathbf{y}} - z_6 c\,\hat{\mathbf{z}}$ | (8k) | Pb |
| $\mathbf{B}_{32}$ | = | $-x_6\,\mathbf{a}_1 + \left(\frac{1}{2} - x_6\right)\,\mathbf{a}_2 - z_6\,\mathbf{a}_3$ | = | $-x_6 a\,\hat{\mathbf{x}} + \left(\frac{1}{2} - x_6\right)a\,\hat{\mathbf{y}} - z_6 c\,\hat{\mathbf{z}}$ | (8k) | Pb |

---

**Geometry files:**
- CIF: pp. 1696
- POSCAR: pp. 1697

# Chiolite ($Na_5Al_3F_{14}$, $K7_5$) Structure: A3B14C5_tP44_128_ac_ehi_bg

http://aflow.org/prototype-encyclopedia/A3B14C5_tP44_128_ac_ehi_bg

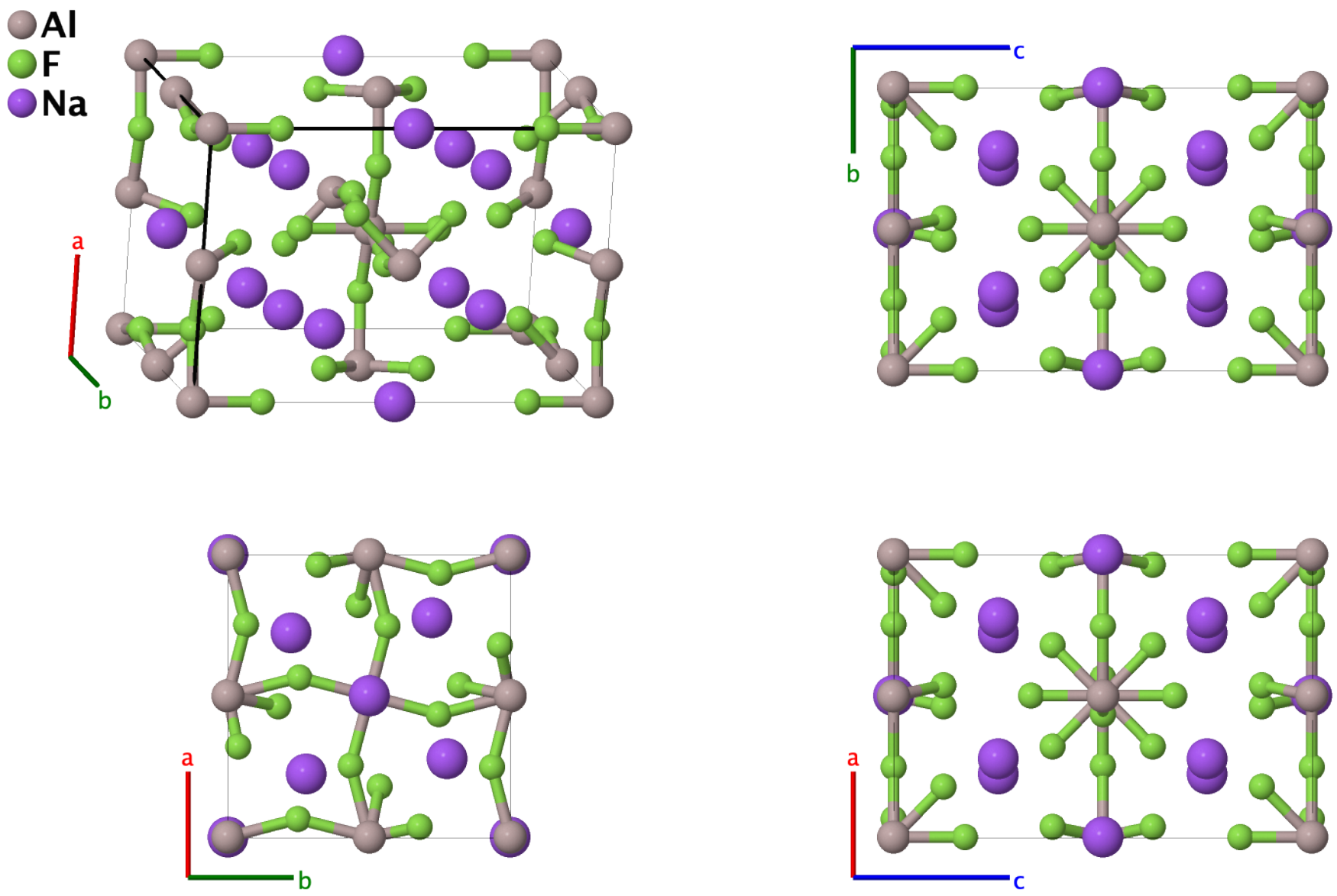

| | | |
|---|---|---|
| **Prototype** | : | $Al_3F_{14}Na_5$ |
| **AFLOW prototype label** | : | A3B14C5_tP44_128_ac_ehi_bg |
| ***Strukturbericht* designation** | : | $K7_5$ |
| **Pearson symbol** | : | tP44 |
| **Space group number** | : | 128 |
| **Space group symbol** | : | $P4/mnc$ |
| **AFLOW prototype command** | : | `aflow --proto=A3B14C5_tP44_128_ac_ehi_bg` `--params=`$a, c/a, z_4, x_5, x_6, y_6, x_7, y_7, z_7$ |

**Simple Tetragonal primitive vectors:**

$$\mathbf{a}_1 = a\,\hat{\mathbf{x}}$$
$$\mathbf{a}_2 = a\,\hat{\mathbf{y}}$$
$$\mathbf{a}_3 = c\,\hat{\mathbf{z}}$$

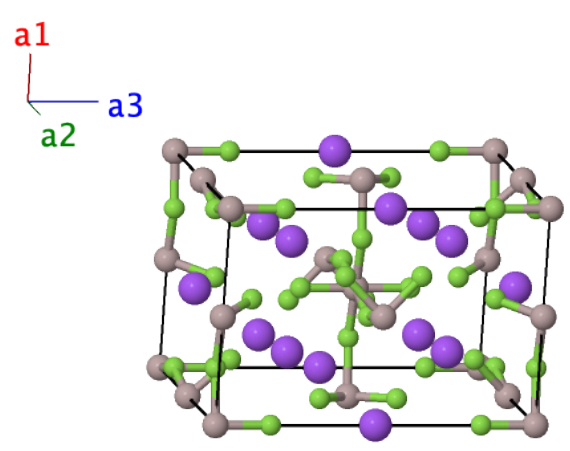

**Basis vectors:**

| | | Lattice Coordinates | | Cartesian Coordinates | Wyckoff Position | Atom Type |
|---|---|---|---|---|---|---|
| $\mathbf{B}_1$ | = | $0\,\mathbf{a}_1 + 0\,\mathbf{a}_2 + 0\,\mathbf{a}_3$ | = | $0\,\hat{\mathbf{x}} + 0\,\hat{\mathbf{y}} + 0\,\hat{\mathbf{z}}$ | (2$a$) | Al I |
| $\mathbf{B}_2$ | = | $\frac{1}{2}\,\mathbf{a}_1 + \frac{1}{2}\,\mathbf{a}_2 + \frac{1}{2}\,\mathbf{a}_3$ | = | $\frac{1}{2}a\,\hat{\mathbf{x}} + \frac{1}{2}a\,\hat{\mathbf{y}} + \frac{1}{2}c\,\hat{\mathbf{z}}$ | (2$a$) | Al I |
| $\mathbf{B}_3$ | = | $\frac{1}{2}\,\mathbf{a}_3$ | = | $\frac{1}{2}c\,\hat{\mathbf{z}}$ | (2$b$) | Na I |
| $\mathbf{B}_4$ | = | $\frac{1}{2}\,\mathbf{a}_1 + \frac{1}{2}\,\mathbf{a}_2$ | = | $\frac{1}{2}a\,\hat{\mathbf{x}} + \frac{1}{2}a\,\hat{\mathbf{y}}$ | (2$b$) | Na I |
| $\mathbf{B}_5$ | = | $\frac{1}{2}\,\mathbf{a}_2$ | = | $\frac{1}{2}a\,\hat{\mathbf{y}}$ | (4$c$) | Al II |
| $\mathbf{B}_6$ | = | $\frac{1}{2}\,\mathbf{a}_1$ | = | $\frac{1}{2}a\,\hat{\mathbf{x}}$ | (4$c$) | Al II |
| $\mathbf{B}_7$ | = | $\frac{1}{2}\,\mathbf{a}_1 + \frac{1}{2}\,\mathbf{a}_3$ | = | $\frac{1}{2}a\,\hat{\mathbf{x}} + \frac{1}{2}c\,\hat{\mathbf{z}}$ | (4$c$) | Al II |
| $\mathbf{B}_8$ | = | $\frac{1}{2}\,\mathbf{a}_2 + \frac{1}{2}\,\mathbf{a}_3$ | = | $\frac{1}{2}a\,\hat{\mathbf{y}} + \frac{1}{2}c\,\hat{\mathbf{z}}$ | (4$c$) | Al II |
| $\mathbf{B}_9$ | = | $z_4\,\mathbf{a}_3$ | = | $z_4c\,\hat{\mathbf{z}}$ | (4$e$) | F I |
| $\mathbf{B}_{10}$ | = | $\frac{1}{2}\,\mathbf{a}_1 + \frac{1}{2}\,\mathbf{a}_2 + \left(\frac{1}{2} - z_4\right)\,\mathbf{a}_3$ | = | $\frac{1}{2}a\,\hat{\mathbf{x}} + \frac{1}{2}a\,\hat{\mathbf{y}} + \left(\frac{1}{2} - z_4\right)c\,\hat{\mathbf{z}}$ | (4$e$) | F I |
| $\mathbf{B}_{11}$ | = | $-z_4\,\mathbf{a}_3$ | = | $-z_4c\,\hat{\mathbf{z}}$ | (4$e$) | F I |
| $\mathbf{B}_{12}$ | = | $\frac{1}{2}\,\mathbf{a}_1 + \frac{1}{2}\,\mathbf{a}_2 + \left(\frac{1}{2} + z_4\right)\,\mathbf{a}_3$ | = | $\frac{1}{2}a\,\hat{\mathbf{x}} + \frac{1}{2}a\,\hat{\mathbf{y}} + \left(\frac{1}{2} + z_4\right)c\,\hat{\mathbf{z}}$ | (4$e$) | F I |
| $\mathbf{B}_{13}$ | = | $x_5\,\mathbf{a}_1 + \left(\frac{1}{2} + x_5\right)\,\mathbf{a}_2 + \frac{1}{4}\,\mathbf{a}_3$ | = | $x_5a\,\hat{\mathbf{x}} + \left(\frac{1}{2} + x_5\right)a\,\hat{\mathbf{y}} + \frac{1}{4}c\,\hat{\mathbf{z}}$ | (8$g$) | Na II |
| $\mathbf{B}_{14}$ | = | $-x_5\,\mathbf{a}_1 + \left(\frac{1}{2} - x_5\right)\,\mathbf{a}_2 + \frac{1}{4}\,\mathbf{a}_3$ | = | $-x_5a\,\hat{\mathbf{x}} + \left(\frac{1}{2} - x_5\right)a\,\hat{\mathbf{y}} + \frac{1}{4}c\,\hat{\mathbf{z}}$ | (8$g$) | Na II |
| $\mathbf{B}_{15}$ | = | $\left(\frac{1}{2} - x_5\right)\,\mathbf{a}_1 + x_5\,\mathbf{a}_2 + \frac{1}{4}\,\mathbf{a}_3$ | = | $\left(\frac{1}{2} - x_5\right)a\,\hat{\mathbf{x}} + x_5a\,\hat{\mathbf{y}} + \frac{1}{4}c\,\hat{\mathbf{z}}$ | (8$g$) | Na II |
| $\mathbf{B}_{16}$ | = | $\left(\frac{1}{2} + x_5\right)\,\mathbf{a}_1 - x_5\,\mathbf{a}_2 + \frac{1}{4}\,\mathbf{a}_3$ | = | $\left(\frac{1}{2} + x_5\right)a\,\hat{\mathbf{x}} - x_5a\,\hat{\mathbf{y}} + \frac{1}{4}c\,\hat{\mathbf{z}}$ | (8$g$) | Na II |
| $\mathbf{B}_{17}$ | = | $-x_5\,\mathbf{a}_1 + \left(\frac{1}{2} - x_5\right)\,\mathbf{a}_2 + \frac{3}{4}\,\mathbf{a}_3$ | = | $-x_5a\,\hat{\mathbf{x}} + \left(\frac{1}{2} - x_5\right)a\,\hat{\mathbf{y}} + \frac{3}{4}c\,\hat{\mathbf{z}}$ | (8$g$) | Na II |
| $\mathbf{B}_{18}$ | = | $x_5\,\mathbf{a}_1 + \left(\frac{1}{2} + x_5\right)\,\mathbf{a}_2 + \frac{3}{4}\,\mathbf{a}_3$ | = | $x_5a\,\hat{\mathbf{x}} + \left(\frac{1}{2} + x_5\right)a\,\hat{\mathbf{y}} + \frac{3}{4}c\,\hat{\mathbf{z}}$ | (8$g$) | Na II |
| $\mathbf{B}_{19}$ | = | $\left(\frac{1}{2} + x_5\right)\,\mathbf{a}_1 - x_5\,\mathbf{a}_2 + \frac{3}{4}\,\mathbf{a}_3$ | = | $\left(\frac{1}{2} + x_5\right)a\,\hat{\mathbf{x}} - x_5a\,\hat{\mathbf{y}} + \frac{3}{4}c\,\hat{\mathbf{z}}$ | (8$g$) | Na II |
| $\mathbf{B}_{20}$ | = | $\left(\frac{1}{2} - x_5\right)\,\mathbf{a}_1 + x_5\,\mathbf{a}_2 + \frac{3}{4}\,\mathbf{a}_3$ | = | $\left(\frac{1}{2} - x_5\right)a\,\hat{\mathbf{x}} + x_5a\,\hat{\mathbf{y}} + \frac{3}{4}c\,\hat{\mathbf{z}}$ | (8$g$) | Na II |
| $\mathbf{B}_{21}$ | = | $x_6\,\mathbf{a}_1 + y_6\,\mathbf{a}_2$ | = | $x_6a\,\hat{\mathbf{x}} + y_6a\,\hat{\mathbf{y}}$ | (8$h$) | F II |
| $\mathbf{B}_{22}$ | = | $-x_6\,\mathbf{a}_1 - y_6\,\mathbf{a}_2$ | = | $-x_6a\,\hat{\mathbf{x}} - y_6a\,\hat{\mathbf{y}}$ | (8$h$) | F II |
| $\mathbf{B}_{23}$ | = | $-y_6\,\mathbf{a}_1 + x_6\,\mathbf{a}_2$ | = | $-y_6a\,\hat{\mathbf{x}} + x_6a\,\hat{\mathbf{y}}$ | (8$h$) | F II |
| $\mathbf{B}_{24}$ | = | $y_6\,\mathbf{a}_1 - x_6\,\mathbf{a}_2$ | = | $y_6a\,\hat{\mathbf{x}} - x_6a\,\hat{\mathbf{y}}$ | (8$h$) | F II |
| $\mathbf{B}_{25}$ | = | $\left(\frac{1}{2} - x_6\right)\,\mathbf{a}_1 + \left(\frac{1}{2} + y_6\right)\,\mathbf{a}_2 + \frac{1}{2}\,\mathbf{a}_3$ | = | $\left(\frac{1}{2} - x_6\right)a\,\hat{\mathbf{x}} + \left(\frac{1}{2} + y_6\right)a\,\hat{\mathbf{y}} + \frac{1}{2}c\,\hat{\mathbf{z}}$ | (8$h$) | F II |
| $\mathbf{B}_{26}$ | = | $\left(\frac{1}{2} + x_6\right)\,\mathbf{a}_1 + \left(\frac{1}{2} - y_6\right)\,\mathbf{a}_2 + \frac{1}{2}\,\mathbf{a}_3$ | = | $\left(\frac{1}{2} + x_6\right)a\,\hat{\mathbf{x}} + \left(\frac{1}{2} - y_6\right)a\,\hat{\mathbf{y}} + \frac{1}{2}c\,\hat{\mathbf{z}}$ | (8$h$) | F II |
| $\mathbf{B}_{27}$ | = | $\left(\frac{1}{2} + y_6\right)\,\mathbf{a}_1 + \left(\frac{1}{2} + x_6\right)\,\mathbf{a}_2 + \frac{1}{2}\,\mathbf{a}_3$ | = | $\left(\frac{1}{2} + y_6\right)a\,\hat{\mathbf{x}} + \left(\frac{1}{2} + x_6\right)a\,\hat{\mathbf{y}} + \frac{1}{2}c\,\hat{\mathbf{z}}$ | (8$h$) | F II |
| $\mathbf{B}_{28}$ | = | $\left(\frac{1}{2} - y_6\right)\,\mathbf{a}_1 + \left(\frac{1}{2} - x_6\right)\,\mathbf{a}_2 + \frac{1}{2}\,\mathbf{a}_3$ | = | $\left(\frac{1}{2} - y_6\right)a\,\hat{\mathbf{x}} + \left(\frac{1}{2} - x_6\right)a\,\hat{\mathbf{y}} + \frac{1}{2}c\,\hat{\mathbf{z}}$ | (8$h$) | F II |
| $\mathbf{B}_{29}$ | = | $x_7\,\mathbf{a}_1 + y_7\,\mathbf{a}_2 + z_7\,\mathbf{a}_3$ | = | $x_7a\,\hat{\mathbf{x}} + y_7a\,\hat{\mathbf{y}} + z_7c\,\hat{\mathbf{z}}$ | (16$i$) | F III |
| $\mathbf{B}_{30}$ | = | $-x_7\,\mathbf{a}_1 - y_7\,\mathbf{a}_2 + z_7\,\mathbf{a}_3$ | = | $-x_7a\,\hat{\mathbf{x}} - y_7a\,\hat{\mathbf{y}} + z_7c\,\hat{\mathbf{z}}$ | (16$i$) | F III |
| $\mathbf{B}_{31}$ | = | $-y_7\,\mathbf{a}_1 + x_7\,\mathbf{a}_2 + z_7\,\mathbf{a}_3$ | = | $-y_7a\,\hat{\mathbf{x}} + x_7a\,\hat{\mathbf{y}} + z_7c\,\hat{\mathbf{z}}$ | (16$i$) | F III |
| $\mathbf{B}_{32}$ | = | $y_7\,\mathbf{a}_1 - x_7\,\mathbf{a}_2 + z_7\,\mathbf{a}_3$ | = | $y_7a\,\hat{\mathbf{x}} - x_7a\,\hat{\mathbf{y}} + z_7c\,\hat{\mathbf{z}}$ | (16$i$) | F III |
| $\mathbf{B}_{33}$ | = | $\left(\frac{1}{2} - x_7\right)\,\mathbf{a}_1 + \left(\frac{1}{2} + y_7\right)\,\mathbf{a}_2 + \left(\frac{1}{2} - z_7\right)\,\mathbf{a}_3$ | = | $\left(\frac{1}{2} - x_7\right)a\,\hat{\mathbf{x}} + \left(\frac{1}{2} + y_7\right)a\,\hat{\mathbf{y}} + \left(\frac{1}{2} - z_7\right)c\,\hat{\mathbf{z}}$ | (16$i$) | F III |
| $\mathbf{B}_{34}$ | = | $\left(\frac{1}{2} + x_7\right)\,\mathbf{a}_1 + \left(\frac{1}{2} - y_7\right)\,\mathbf{a}_2 + \left(\frac{1}{2} - z_7\right)\,\mathbf{a}_3$ | = | $\left(\frac{1}{2} + x_7\right)a\,\hat{\mathbf{x}} + \left(\frac{1}{2} - y_7\right)a\,\hat{\mathbf{y}} + \left(\frac{1}{2} - z_7\right)c\,\hat{\mathbf{z}}$ | (16$i$) | F III |
| $\mathbf{B}_{35}$ | = | $\left(\frac{1}{2} + y_7\right)\,\mathbf{a}_1 + \left(\frac{1}{2} + x_7\right)\,\mathbf{a}_2 + \left(\frac{1}{2} - z_7\right)\,\mathbf{a}_3$ | = | $\left(\frac{1}{2} + y_7\right)a\,\hat{\mathbf{x}} + \left(\frac{1}{2} + x_7\right)a\,\hat{\mathbf{y}} + \left(\frac{1}{2} - z_7\right)c\,\hat{\mathbf{z}}$ | (16$i$) | F III |
| $\mathbf{B}_{36}$ | = | $\left(\frac{1}{2} - y_7\right)\,\mathbf{a}_1 + \left(\frac{1}{2} - x_7\right)\,\mathbf{a}_2 + \left(\frac{1}{2} - z_7\right)\,\mathbf{a}_3$ | = | $\left(\frac{1}{2} - y_7\right)a\,\hat{\mathbf{x}} + \left(\frac{1}{2} - x_7\right)a\,\hat{\mathbf{y}} + \left(\frac{1}{2} - z_7\right)c\,\hat{\mathbf{z}}$ | (16$i$) | F III |
| $\mathbf{B}_{37}$ | = | $-x_7\,\mathbf{a}_1 - y_7\,\mathbf{a}_2 - z_7\,\mathbf{a}_3$ | = | $-x_7a\,\hat{\mathbf{x}} - y_7a\,\hat{\mathbf{y}} - z_7c\,\hat{\mathbf{z}}$ | (16$i$) | F III |
| $\mathbf{B}_{38}$ | = | $x_7\,\mathbf{a}_1 + y_7\,\mathbf{a}_2 - z_7\,\mathbf{a}_3$ | = | $x_7 a\,\hat{\mathbf{x}} + y_7 a\,\hat{\mathbf{y}} - z_7 c\,\hat{\mathbf{z}}$ | (16*i*) | F III |
| $\mathbf{B}_{39}$ | = | $y_7\,\mathbf{a}_1 - x_7\,\mathbf{a}_2 - z_7\,\mathbf{a}_3$ | = | $y_7 a\,\hat{\mathbf{x}} - x_7 a\,\hat{\mathbf{y}} - z_7 c\,\hat{\mathbf{z}}$ | (16*i*) | F III |
| $\mathbf{B}_{40}$ | = | $-y_7\,\mathbf{a}_1 + x_7\,\mathbf{a}_2 - z_7\,\mathbf{a}_3$ | = | $-y_7 a\,\hat{\mathbf{x}} + x_7 a\,\hat{\mathbf{y}} - z_7 c\,\hat{\mathbf{z}}$ | (16*i*) | F III |
| $\mathbf{B}_{41}$ | = | $\left(\frac{1}{2} + x_7\right)\mathbf{a}_1 + \left(\frac{1}{2} - y_7\right)\mathbf{a}_2 + \left(\frac{1}{2} + z_7\right)\mathbf{a}_3$ | = | $\left(\frac{1}{2} + x_7\right)a\,\hat{\mathbf{x}} + \left(\frac{1}{2} - y_7\right)a\,\hat{\mathbf{y}} + \left(\frac{1}{2} + z_7\right)c\,\hat{\mathbf{z}}$ | (16*i*) | F III |
| $\mathbf{B}_{42}$ | = | $\left(\frac{1}{2} - x_7\right)\mathbf{a}_1 + \left(\frac{1}{2} + y_7\right)\mathbf{a}_2 + \left(\frac{1}{2} + z_7\right)\mathbf{a}_3$ | = | $\left(\frac{1}{2} - x_7\right)a\,\hat{\mathbf{x}} + \left(\frac{1}{2} + y_7\right)a\,\hat{\mathbf{y}} + \left(\frac{1}{2} + z_7\right)c\,\hat{\mathbf{z}}$ | (16*i*) | F III |
| $\mathbf{B}_{43}$ | = | $\left(\frac{1}{2} - y_7\right)\mathbf{a}_1 + \left(\frac{1}{2} - x_7\right)\mathbf{a}_2 + \left(\frac{1}{2} + z_7\right)\mathbf{a}_3$ | = | $\left(\frac{1}{2} - y_7\right)a\,\hat{\mathbf{x}} + \left(\frac{1}{2} - x_7\right)a\,\hat{\mathbf{y}} + \left(\frac{1}{2} + z_7\right)c\,\hat{\mathbf{z}}$ | (16*i*) | F III |
| $\mathbf{B}_{44}$ | = | $\left(\frac{1}{2} + y_7\right)\mathbf{a}_1 + \left(\frac{1}{2} + x_7\right)\mathbf{a}_2 + \left(\frac{1}{2} + z_7\right)\mathbf{a}_3$ | = | $\left(\frac{1}{2} + y_7\right)a\,\hat{\mathbf{x}} + \left(\frac{1}{2} + x_7\right)a\,\hat{\mathbf{y}} + \left(\frac{1}{2} + z_7\right)c\,\hat{\mathbf{z}}$ | (16*i*) | F III |

---

**Found in:**

---

**Geometry files:**

- CIF: pp. 1697
- POSCAR: pp. 1697

# Apophyllite ($KCa_4Si_8O_{20}F{\cdot}8H_2O$, $S5_2$) Structure: A4BC16DE28F8_tP116_128_h_a_2i_b_g3i_i

http://aflow.org/prototype-encyclopedia/A4BC16DE28F8_tP116_128_h_a_2i_b_g3i_i

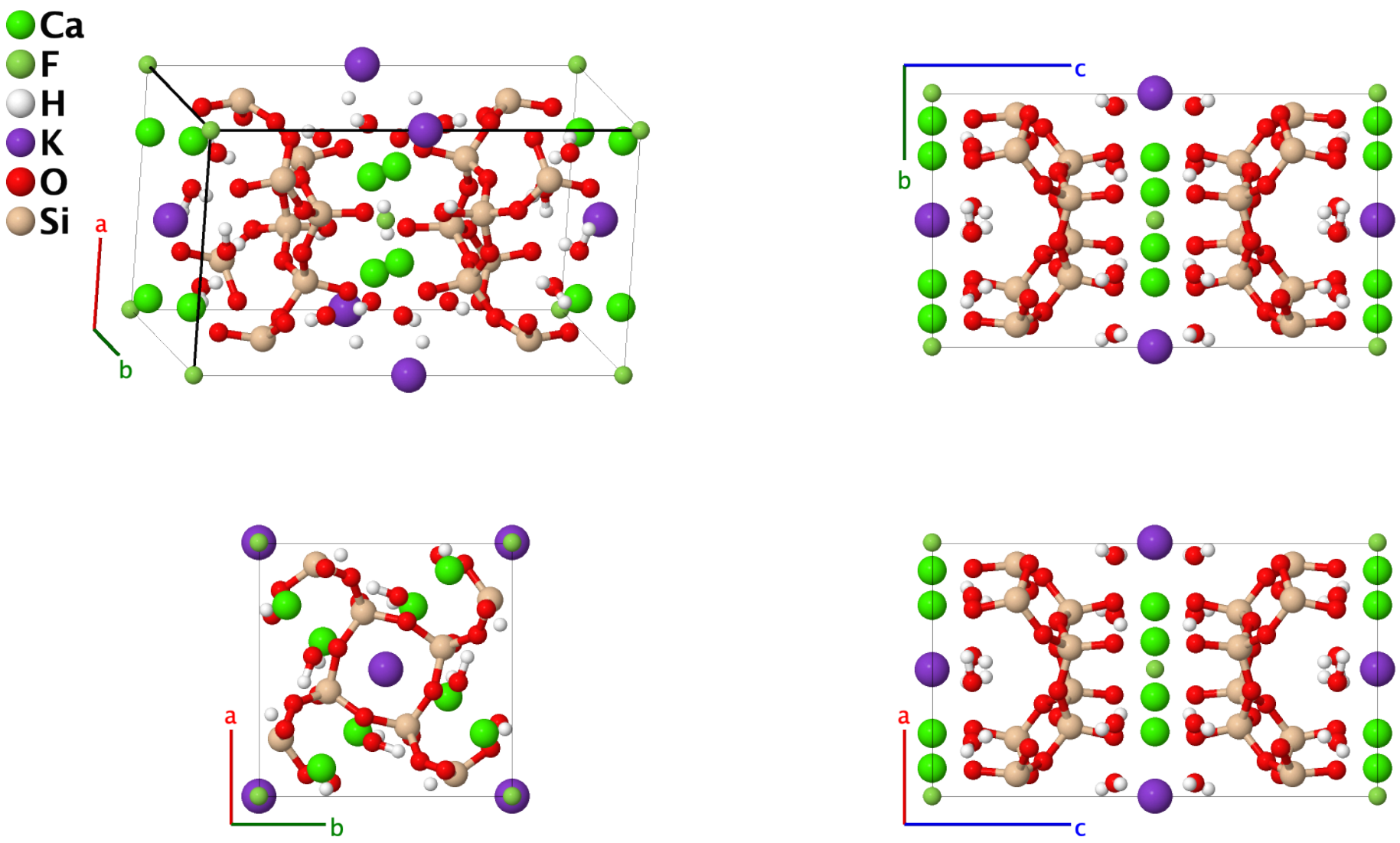

**Prototype** : $Ca_4FH_{16}KO_{28}Si_8$

**AFLOW prototype label** : A4BC16DE28F8_tP116_128_h_a_2i_b_g3i_i

***Strukturbericht* designation** : $S5_2$

**Pearson symbol** : tP116

**Space group number** : 128

**Space group symbol** : $P4/mnc$

**AFLOW prototype command** : `aflow --proto=A4BC16DE28F8_tP116_128_h_a_2i_b_g3i_i`
`--params=`$a, c/a, x_3, x_4, y_4, x_5, y_5, z_5, x_6, y_6, z_6, x_7, y_7, z_7, x_8, y_8, z_8, x_9, y_9, z_9, x_{10}, y_{10}, z_{10}$

- Although we use the structure found by (Chao, 1971), we should note that there is some disagreement between the Chao's X-ray diffraction data and the neutron diffraction data taken by (Prince, 1971): while both agree on the positions of the heavy atoms, Prince's work suggests that some of the hydrogens may form OH radicals rather than water molecules.
- In any case, the fluorine atoms are usually partially replaced by OH radicals. This sample, which is predominantly fluorine, is technically labeled apophyllite-(KF).

**Simple Tetragonal primitive vectors:**

$$\mathbf{a}_1 = a\,\hat{\mathbf{x}}$$
$$\mathbf{a}_2 = a\,\hat{\mathbf{y}}$$
$$\mathbf{a}_3 = c\,\hat{\mathbf{z}}$$

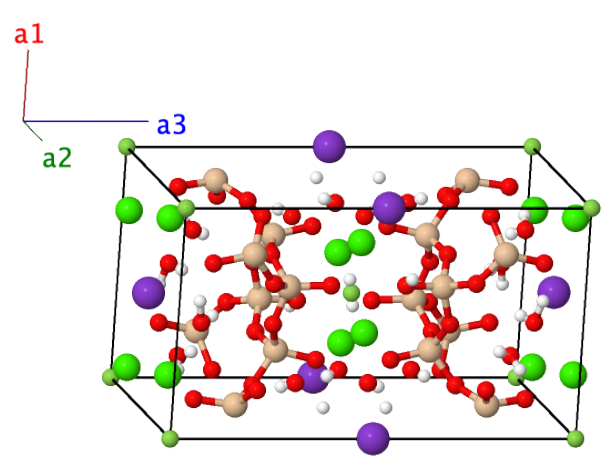

**Basis vectors:**

| | | Lattice Coordinates | | Cartesian Coordinates | Wyckoff Position | Atom Type |
|---|---|---|---|---|---|---|
| $\mathbf{B}_{1}$ | = | $0\,\mathbf{a}_{1}+0\,\mathbf{a}_{2}+0\,\mathbf{a}_{3}$ | = | $0\,\hat{\mathbf{x}}+0\,\hat{\mathbf{y}}+0\,\hat{\mathbf{z}}$ | (2$a$) | F |
| $\mathbf{B}_{2}$ | = | $\frac{1}{2}\,\mathbf{a}_{1}+\frac{1}{2}\,\mathbf{a}_{2}+\frac{1}{2}\,\mathbf{a}_{3}$ | = | $\frac{1}{2}a\,\hat{\mathbf{x}}+\frac{1}{2}a\,\hat{\mathbf{y}}+\frac{1}{2}c\,\hat{\mathbf{z}}$ | (2$a$) | F |
| $\mathbf{B}_{3}$ | = | $\frac{1}{2}\,\mathbf{a}_{3}$ | = | $\frac{1}{2}c\,\hat{\mathbf{z}}$ | (2$b$) | K |
| $\mathbf{B}_{4}$ | = | $\frac{1}{2}\,\mathbf{a}_{1}+\frac{1}{2}\,\mathbf{a}_{2}$ | = | $\frac{1}{2}a\,\hat{\mathbf{x}}+\frac{1}{2}a\,\hat{\mathbf{y}}$ | (2$b$) | K |
| $\mathbf{B}_{5}$ | = | $x_{3}\,\mathbf{a}_{1}+\left(\frac{1}{2}+x_{3}\right)\,\mathbf{a}_{2}+\frac{1}{4}\,\mathbf{a}_{3}$ | = | $x_{3}a\,\hat{\mathbf{x}}+\left(\frac{1}{2}+x_{3}\right)a\,\hat{\mathbf{y}}+\frac{1}{4}c\,\hat{\mathbf{z}}$ | (8$g$) | O I |
| $\mathbf{B}_{6}$ | = | $-x_{3}\,\mathbf{a}_{1}+\left(\frac{1}{2}-x_{3}\right)\,\mathbf{a}_{2}+\frac{1}{4}\,\mathbf{a}_{3}$ | = | $-x_{3}a\,\hat{\mathbf{x}}+\left(\frac{1}{2}-x_{3}\right)a\,\hat{\mathbf{y}}+\frac{1}{4}c\,\hat{\mathbf{z}}$ | (8$g$) | O I |
| $\mathbf{B}_{7}$ | = | $\left(\frac{1}{2}-x_{3}\right)\,\mathbf{a}_{1}+x_{3}\,\mathbf{a}_{2}+\frac{1}{4}\,\mathbf{a}_{3}$ | = | $\left(\frac{1}{2}-x_{3}\right)a\,\hat{\mathbf{x}}+x_{3}a\,\hat{\mathbf{y}}+\frac{1}{4}c\,\hat{\mathbf{z}}$ | (8$g$) | O I |
| $\mathbf{B}_{8}$ | = | $\left(\frac{1}{2}+x_{3}\right)\,\mathbf{a}_{1}-x_{3}\,\mathbf{a}_{2}+\frac{1}{4}\,\mathbf{a}_{3}$ | = | $\left(\frac{1}{2}+x_{3}\right)a\,\hat{\mathbf{x}}-x_{3}a\,\hat{\mathbf{y}}+\frac{1}{4}c\,\hat{\mathbf{z}}$ | (8$g$) | O I |
| $\mathbf{B}_{9}$ | = | $-x_{3}\,\mathbf{a}_{1}+\left(\frac{1}{2}-x_{3}\right)\,\mathbf{a}_{2}+\frac{3}{4}\,\mathbf{a}_{3}$ | = | $-x_{3}a\,\hat{\mathbf{x}}+\left(\frac{1}{2}-x_{3}\right)a\,\hat{\mathbf{y}}+\frac{3}{4}c\,\hat{\mathbf{z}}$ | (8$g$) | O I |
| $\mathbf{B}_{10}$ | = | $x_{3}\,\mathbf{a}_{1}+\left(\frac{1}{2}+x_{3}\right)\,\mathbf{a}_{2}+\frac{3}{4}\,\mathbf{a}_{3}$ | = | $x_{3}a\,\hat{\mathbf{x}}+\left(\frac{1}{2}+x_{3}\right)a\,\hat{\mathbf{y}}+\frac{3}{4}c\,\hat{\mathbf{z}}$ | (8$g$) | O I |
| $\mathbf{B}_{11}$ | = | $\left(\frac{1}{2}+x_{3}\right)\,\mathbf{a}_{1}-x_{3}\,\mathbf{a}_{2}+\frac{3}{4}\,\mathbf{a}_{3}$ | = | $\left(\frac{1}{2}+x_{3}\right)a\,\hat{\mathbf{x}}-x_{3}a\,\hat{\mathbf{y}}+\frac{3}{4}c\,\hat{\mathbf{z}}$ | (8$g$) | O I |
| $\mathbf{B}_{12}$ | = | $\left(\frac{1}{2}-x_{3}\right)\,\mathbf{a}_{1}+x_{3}\,\mathbf{a}_{2}+\frac{3}{4}\,\mathbf{a}_{3}$ | = | $\left(\frac{1}{2}-x_{3}\right)a\,\hat{\mathbf{x}}+x_{3}a\,\hat{\mathbf{y}}+\frac{3}{4}c\,\hat{\mathbf{z}}$ | (8$g$) | O I |
| $\mathbf{B}_{13}$ | = | $x_{4}\,\mathbf{a}_{1}+y_{4}\,\mathbf{a}_{2}$ | = | $x_{4}a\,\hat{\mathbf{x}}+y_{4}a\,\hat{\mathbf{y}}$ | (8$h$) | Ca |
| $\mathbf{B}_{14}$ | = | $-x_{4}\,\mathbf{a}_{1}-y_{4}\,\mathbf{a}_{2}$ | = | $-x_{4}a\,\hat{\mathbf{x}}-y_{4}a\,\hat{\mathbf{y}}$ | (8$h$) | Ca |
| $\mathbf{B}_{15}$ | = | $-y_{4}\,\mathbf{a}_{1}+x_{4}\,\mathbf{a}_{2}$ | = | $-y_{4}a\,\hat{\mathbf{x}}+x_{4}a\,\hat{\mathbf{y}}$ | (8$h$) | Ca |
| $\mathbf{B}_{16}$ | = | $y_{4}\,\mathbf{a}_{1}-x_{4}\,\mathbf{a}_{2}$ | = | $y_{4}a\,\hat{\mathbf{x}}-x_{4}a\,\hat{\mathbf{y}}$ | (8$h$) | Ca |
| $\mathbf{B}_{17}$ | = | $\left(\frac{1}{2}-x_{4}\right)\,\mathbf{a}_{1}+\left(\frac{1}{2}+y_{4}\right)\,\mathbf{a}_{2}+\frac{1}{2}\,\mathbf{a}_{3}$ | = | $\left(\frac{1}{2}-x_{4}\right)a\,\hat{\mathbf{x}}+\left(\frac{1}{2}+y_{4}\right)a\,\hat{\mathbf{y}}+\frac{1}{2}c\,\hat{\mathbf{z}}$ | (8$h$) | Ca |
| $\mathbf{B}_{18}$ | = | $\left(\frac{1}{2}+x_{4}\right)\,\mathbf{a}_{1}+\left(\frac{1}{2}-y_{4}\right)\,\mathbf{a}_{2}+\frac{1}{2}\,\mathbf{a}_{3}$ | = | $\left(\frac{1}{2}+x_{4}\right)a\,\hat{\mathbf{x}}+\left(\frac{1}{2}-y_{4}\right)a\,\hat{\mathbf{y}}+\frac{1}{2}c\,\hat{\mathbf{z}}$ | (8$h$) | Ca |
| $\mathbf{B}_{19}$ | = | $\left(\frac{1}{2}+y_{4}\right)\,\mathbf{a}_{1}+\left(\frac{1}{2}+x_{4}\right)\,\mathbf{a}_{2}+\frac{1}{2}\,\mathbf{a}_{3}$ | = | $\left(\frac{1}{2}+y_{4}\right)a\,\hat{\mathbf{x}}+\left(\frac{1}{2}+x_{4}\right)a\,\hat{\mathbf{y}}+\frac{1}{2}c\,\hat{\mathbf{z}}$ | (8$h$) | Ca |
| $\mathbf{B}_{20}$ | = | $\left(\frac{1}{2}-y_{4}\right)\,\mathbf{a}_{1}+\left(\frac{1}{2}-x_{4}\right)\,\mathbf{a}_{2}+\frac{1}{2}\,\mathbf{a}_{3}$ | = | $\left(\frac{1}{2}-y_{4}\right)a\,\hat{\mathbf{x}}+\left(\frac{1}{2}-x_{4}\right)a\,\hat{\mathbf{y}}+\frac{1}{2}c\,\hat{\mathbf{z}}$ | (8$h$) | Ca |
| $\mathbf{B}_{21}$ | = | $x_{5}\,\mathbf{a}_{1}+y_{5}\,\mathbf{a}_{2}+z_{5}\,\mathbf{a}_{3}$ | = | $x_{5}a\,\hat{\mathbf{x}}+y_{5}a\,\hat{\mathbf{y}}+z_{5}c\,\hat{\mathbf{z}}$ | (16$i$) | H I |
| $\mathbf{B}_{22}$ | = | $-x_{5}\,\mathbf{a}_{1}-y_{5}\,\mathbf{a}_{2}+z_{5}\,\mathbf{a}_{3}$ | = | $-x_{5}a\,\hat{\mathbf{x}}-y_{5}a\,\hat{\mathbf{y}}+z_{5}c\,\hat{\mathbf{z}}$ | (16$i$) | H I |
| $\mathbf{B}_{23}$ | = | $-y_{5}\,\mathbf{a}_{1}+x_{5}\,\mathbf{a}_{2}+z_{5}\,\mathbf{a}_{3}$ | = | $-y_{5}a\,\hat{\mathbf{x}}+x_{5}a\,\hat{\mathbf{y}}+z_{5}c\,\hat{\mathbf{z}}$ | (16$i$) | H I |
| $\mathbf{B}_{24}$ | = | $y_{5}\,\mathbf{a}_{1}-x_{5}\,\mathbf{a}_{2}+z_{5}\,\mathbf{a}_{3}$ | = | $y_{5}a\,\hat{\mathbf{x}}-x_{5}a\,\hat{\mathbf{y}}+z_{5}c\,\hat{\mathbf{z}}$ | (16$i$) | H I |
| $\mathbf{B}_{25}$ | = | $\left(\frac{1}{2}-x_{5}\right)\,\mathbf{a}_{1}+\left(\frac{1}{2}+y_{5}\right)\,\mathbf{a}_{2}+\left(\frac{1}{2}-z_{5}\right)\,\mathbf{a}_{3}$ | = | $\left(\frac{1}{2}-x_{5}\right)a\,\hat{\mathbf{x}}+\left(\frac{1}{2}+y_{5}\right)a\,\hat{\mathbf{y}}+\left(\frac{1}{2}-z_{5}\right)c\,\hat{\mathbf{z}}$ | (16$i$) | H I |
| $\mathbf{B}_{26}$ | = | $\left(\frac{1}{2}+x_{5}\right)\,\mathbf{a}_{1}+\left(\frac{1}{2}-y_{5}\right)\,\mathbf{a}_{2}+\left(\frac{1}{2}-z_{5}\right)\,\mathbf{a}_{3}$ | = | $\left(\frac{1}{2}+x_{5}\right)a\,\hat{\mathbf{x}}+\left(\frac{1}{2}-y_{5}\right)a\,\hat{\mathbf{y}}+\left(\frac{1}{2}-z_{5}\right)c\,\hat{\mathbf{z}}$ | (16$i$) | H I |
| $\mathbf{B}_{27}$ | = | $\left(\frac{1}{2}+y_{5}\right)\,\mathbf{a}_{1}+\left(\frac{1}{2}+x_{5}\right)\,\mathbf{a}_{2}+\left(\frac{1}{2}-z_{5}\right)\,\mathbf{a}_{3}$ | = | $\left(\frac{1}{2}+y_{5}\right)a\,\hat{\mathbf{x}}+\left(\frac{1}{2}+x_{5}\right)a\,\hat{\mathbf{y}}+\left(\frac{1}{2}-z_{5}\right)c\,\hat{\mathbf{z}}$ | (16$i$) | H I |
| $\mathbf{B}_{28}$ | = | $\left(\frac{1}{2}-y_{5}\right)\,\mathbf{a}_{1}+\left(\frac{1}{2}-x_{5}\right)\,\mathbf{a}_{2}+\left(\frac{1}{2}-z_{5}\right)\,\mathbf{a}_{3}$ | = | $\left(\frac{1}{2}-y_{5}\right)a\,\hat{\mathbf{x}}+\left(\frac{1}{2}-x_{5}\right)a\,\hat{\mathbf{y}}+\left(\frac{1}{2}-z_{5}\right)c\,\hat{\mathbf{z}}$ | (16$i$) | H I |
| $\mathbf{B}_{29}$ | = | $-x_{5}\,\mathbf{a}_{1}-y_{5}\,\mathbf{a}_{2}-z_{5}\,\mathbf{a}_{3}$ | = | $-x_{5}a\,\hat{\mathbf{x}}-y_{5}a\,\hat{\mathbf{y}}-z_{5}c\,\hat{\mathbf{z}}$ | (16$i$) | H I |
| $\mathbf{B}_{30}$ | = | $x_{5}\,\mathbf{a}_{1}+y_{5}\,\mathbf{a}_{2}-z_{5}\,\mathbf{a}_{3}$ | = | $x_{5}a\,\hat{\mathbf{x}}+y_{5}a\,\hat{\mathbf{y}}-z_{5}c\,\hat{\mathbf{z}}$ | (16$i$) | H I |
| $\mathbf{B}_{31}$ | = | $y_{5}\,\mathbf{a}_{1}-x_{5}\,\mathbf{a}_{2}-z_{5}\,\mathbf{a}_{3}$ | = | $y_{5}a\,\hat{\mathbf{x}}-x_{5}a\,\hat{\mathbf{y}}-z_{5}c\,\hat{\mathbf{z}}$ | (16$i$) | H I |
| $\mathbf{B}_{32}$ | = | $-y_{5}\,\mathbf{a}_{1}+x_{5}\,\mathbf{a}_{2}-z_{5}\,\mathbf{a}_{3}$ | = | $-y_{5}a\,\hat{\mathbf{x}}+x_{5}a\,\hat{\mathbf{y}}-z_{5}c\,\hat{\mathbf{z}}$ | (16$i$) | H I |
| $\mathbf{B}_{33}$ | = | $\left(\frac{1}{2}+x_5\right)\mathbf{a}_1+\left(\frac{1}{2}-y_5\right)\mathbf{a}_2+\left(\frac{1}{2}+z_5\right)\mathbf{a}_3$ | = | $\left(\frac{1}{2}+x_5\right)a\,\hat{\mathbf{x}}+\left(\frac{1}{2}-y_5\right)a\,\hat{\mathbf{y}}+\left(\frac{1}{2}+z_5\right)c\,\hat{\mathbf{z}}$ | (16$i$) | H I |
| $\mathbf{B}_{34}$ | = | $\left(\frac{1}{2}-x_5\right)\mathbf{a}_1+\left(\frac{1}{2}+y_5\right)\mathbf{a}_2+\left(\frac{1}{2}+z_5\right)\mathbf{a}_3$ | = | $\left(\frac{1}{2}-x_5\right)a\,\hat{\mathbf{x}}+\left(\frac{1}{2}+y_5\right)a\,\hat{\mathbf{y}}+\left(\frac{1}{2}+z_5\right)c\,\hat{\mathbf{z}}$ | (16$i$) | H I |
| $\mathbf{B}_{35}$ | = | $\left(\frac{1}{2}-y_5\right)\mathbf{a}_1+\left(\frac{1}{2}-x_5\right)\mathbf{a}_2+\left(\frac{1}{2}+z_5\right)\mathbf{a}_3$ | = | $\left(\frac{1}{2}-y_5\right)a\,\hat{\mathbf{x}}+\left(\frac{1}{2}-x_5\right)a\,\hat{\mathbf{y}}+\left(\frac{1}{2}+z_5\right)c\,\hat{\mathbf{z}}$ | (16$i$) | H I |
| $\mathbf{B}_{36}$ | = | $\left(\frac{1}{2}+y_5\right)\mathbf{a}_1+\left(\frac{1}{2}+x_5\right)\mathbf{a}_2+\left(\frac{1}{2}+z_5\right)\mathbf{a}_3$ | = | $\left(\frac{1}{2}+y_5\right)a\,\hat{\mathbf{x}}+\left(\frac{1}{2}+x_5\right)a\,\hat{\mathbf{y}}+\left(\frac{1}{2}+z_5\right)c\,\hat{\mathbf{z}}$ | (16$i$) | H I |
| $\mathbf{B}_{37}$ | = | $x_6\,\mathbf{a}_1+y_6\,\mathbf{a}_2+z_6\,\mathbf{a}_3$ | = | $x_6a\,\hat{\mathbf{x}}+y_6a\,\hat{\mathbf{y}}+z_6c\,\hat{\mathbf{z}}$ | (16$i$) | H II |
| $\mathbf{B}_{38}$ | = | $-x_6\,\mathbf{a}_1-y_6\,\mathbf{a}_2+z_6\,\mathbf{a}_3$ | = | $-x_6a\,\hat{\mathbf{x}}-y_6a\,\hat{\mathbf{y}}+z_6c\,\hat{\mathbf{z}}$ | (16$i$) | H II |
| $\mathbf{B}_{39}$ | = | $-y_6\,\mathbf{a}_1+x_6\,\mathbf{a}_2+z_6\,\mathbf{a}_3$ | = | $-y_6a\,\hat{\mathbf{x}}+x_6a\,\hat{\mathbf{y}}+z_6c\,\hat{\mathbf{z}}$ | (16$i$) | H II |
| $\mathbf{B}_{40}$ | = | $y_6\,\mathbf{a}_1-x_6\,\mathbf{a}_2+z_6\,\mathbf{a}_3$ | = | $y_6a\,\hat{\mathbf{x}}-x_6a\,\hat{\mathbf{y}}+z_6c\,\hat{\mathbf{z}}$ | (16$i$) | H II |
| $\mathbf{B}_{41}$ | = | $\left(\frac{1}{2}-x_6\right)\mathbf{a}_1+\left(\frac{1}{2}+y_6\right)\mathbf{a}_2+\left(\frac{1}{2}-z_6\right)\mathbf{a}_3$ | = | $\left(\frac{1}{2}-x_6\right)a\,\hat{\mathbf{x}}+\left(\frac{1}{2}+y_6\right)a\,\hat{\mathbf{y}}+\left(\frac{1}{2}-z_6\right)c\,\hat{\mathbf{z}}$ | (16$i$) | H II |
| $\mathbf{B}_{42}$ | = | $\left(\frac{1}{2}+x_6\right)\mathbf{a}_1+\left(\frac{1}{2}-y_6\right)\mathbf{a}_2+\left(\frac{1}{2}-z_6\right)\mathbf{a}_3$ | = | $\left(\frac{1}{2}+x_6\right)a\,\hat{\mathbf{x}}+\left(\frac{1}{2}-y_6\right)a\,\hat{\mathbf{y}}+\left(\frac{1}{2}-z_6\right)c\,\hat{\mathbf{z}}$ | (16$i$) | H II |
| $\mathbf{B}_{43}$ | = | $\left(\frac{1}{2}+y_6\right)\mathbf{a}_1+\left(\frac{1}{2}+x_6\right)\mathbf{a}_2+\left(\frac{1}{2}-z_6\right)\mathbf{a}_3$ | = | $\left(\frac{1}{2}+y_6\right)a\,\hat{\mathbf{x}}+\left(\frac{1}{2}+x_6\right)a\,\hat{\mathbf{y}}+\left(\frac{1}{2}-z_6\right)c\,\hat{\mathbf{z}}$ | (16$i$) | H II |
| $\mathbf{B}_{44}$ | = | $\left(\frac{1}{2}-y_6\right)\mathbf{a}_1+\left(\frac{1}{2}-x_6\right)\mathbf{a}_2+\left(\frac{1}{2}-z_6\right)\mathbf{a}_3$ | = | $\left(\frac{1}{2}-y_6\right)a\,\hat{\mathbf{x}}+\left(\frac{1}{2}-x_6\right)a\,\hat{\mathbf{y}}+\left(\frac{1}{2}-z_6\right)c\,\hat{\mathbf{z}}$ | (16$i$) | H II |
| $\mathbf{B}_{45}$ | = | $-x_6\,\mathbf{a}_1-y_6\,\mathbf{a}_2-z_6\,\mathbf{a}_3$ | = | $-x_6a\,\hat{\mathbf{x}}-y_6a\,\hat{\mathbf{y}}-z_6c\,\hat{\mathbf{z}}$ | (16$i$) | H II |
| $\mathbf{B}_{46}$ | = | $x_6\,\mathbf{a}_1+y_6\,\mathbf{a}_2-z_6\,\mathbf{a}_3$ | = | $x_6a\,\hat{\mathbf{x}}+y_6a\,\hat{\mathbf{y}}-z_6c\,\hat{\mathbf{z}}$ | (16$i$) | H II |
| $\mathbf{B}_{47}$ | = | $y_6\,\mathbf{a}_1-x_6\,\mathbf{a}_2-z_6\,\mathbf{a}_3$ | = | $y_6a\,\hat{\mathbf{x}}-x_6a\,\hat{\mathbf{y}}-z_6c\,\hat{\mathbf{z}}$ | (16$i$) | H II |
| $\mathbf{B}_{48}$ | = | $-y_6\,\mathbf{a}_1+x_6\,\mathbf{a}_2-z_6\,\mathbf{a}_3$ | = | $-y_6a\,\hat{\mathbf{x}}+x_6a\,\hat{\mathbf{y}}-z_6c\,\hat{\mathbf{z}}$ | (16$i$) | H II |
| $\mathbf{B}_{49}$ | = | $\left(\frac{1}{2}+x_6\right)\mathbf{a}_1+\left(\frac{1}{2}-y_6\right)\mathbf{a}_2+\left(\frac{1}{2}+z_6\right)\mathbf{a}_3$ | = | $\left(\frac{1}{2}+x_6\right)a\,\hat{\mathbf{x}}+\left(\frac{1}{2}-y_6\right)a\,\hat{\mathbf{y}}+\left(\frac{1}{2}+z_6\right)c\,\hat{\mathbf{z}}$ | (16$i$) | H II |
| $\mathbf{B}_{50}$ | = | $\left(\frac{1}{2}-x_6\right)\mathbf{a}_1+\left(\frac{1}{2}+y_6\right)\mathbf{a}_2+\left(\frac{1}{2}+z_6\right)\mathbf{a}_3$ | = | $\left(\frac{1}{2}-x_6\right)a\,\hat{\mathbf{x}}+\left(\frac{1}{2}+y_6\right)a\,\hat{\mathbf{y}}+\left(\frac{1}{2}+z_6\right)c\,\hat{\mathbf{z}}$ | (16$i$) | H II |
| $\mathbf{B}_{51}$ | = | $\left(\frac{1}{2}-y_6\right)\mathbf{a}_1+\left(\frac{1}{2}-x_6\right)\mathbf{a}_2+\left(\frac{1}{2}+z_6\right)\mathbf{a}_3$ | = | $\left(\frac{1}{2}-y_6\right)a\,\hat{\mathbf{x}}+\left(\frac{1}{2}-x_6\right)a\,\hat{\mathbf{y}}+\left(\frac{1}{2}+z_6\right)c\,\hat{\mathbf{z}}$ | (16$i$) | H II |
| $\mathbf{B}_{52}$ | = | $\left(\frac{1}{2}+y_6\right)\mathbf{a}_1+\left(\frac{1}{2}+x_6\right)\mathbf{a}_2+\left(\frac{1}{2}+z_6\right)\mathbf{a}_3$ | = | $\left(\frac{1}{2}+y_6\right)a\,\hat{\mathbf{x}}+\left(\frac{1}{2}+x_6\right)a\,\hat{\mathbf{y}}+\left(\frac{1}{2}+z_6\right)c\,\hat{\mathbf{z}}$ | (16$i$) | H II |
| $\mathbf{B}_{53}$ | = | $x_7\,\mathbf{a}_1+y_7\,\mathbf{a}_2+z_7\,\mathbf{a}_3$ | = | $x_7a\,\hat{\mathbf{x}}+y_7a\,\hat{\mathbf{y}}+z_7c\,\hat{\mathbf{z}}$ | (16$i$) | O II |
| $\mathbf{B}_{54}$ | = | $-x_7\,\mathbf{a}_1-y_7\,\mathbf{a}_2+z_7\,\mathbf{a}_3$ | = | $-x_7a\,\hat{\mathbf{x}}-y_7a\,\hat{\mathbf{y}}+z_7c\,\hat{\mathbf{z}}$ | (16$i$) | O II |
| $\mathbf{B}_{55}$ | = | $-y_7\,\mathbf{a}_1+x_7\,\mathbf{a}_2+z_7\,\mathbf{a}_3$ | = | $-y_7a\,\hat{\mathbf{x}}+x_7a\,\hat{\mathbf{y}}+z_7c\,\hat{\mathbf{z}}$ | (16$i$) | O II |
| $\mathbf{B}_{56}$ | = | $y_7\,\mathbf{a}_1-x_7\,\mathbf{a}_2+z_7\,\mathbf{a}_3$ | = | $y_7a\,\hat{\mathbf{x}}-x_7a\,\hat{\mathbf{y}}+z_7c\,\hat{\mathbf{z}}$ | (16$i$) | O II |
| $\mathbf{B}_{57}$ | = | $\left(\frac{1}{2}-x_7\right)\mathbf{a}_1+\left(\frac{1}{2}+y_7\right)\mathbf{a}_2+\left(\frac{1}{2}-z_7\right)\mathbf{a}_3$ | = | $\left(\frac{1}{2}-x_7\right)a\,\hat{\mathbf{x}}+\left(\frac{1}{2}+y_7\right)a\,\hat{\mathbf{y}}+\left(\frac{1}{2}-z_7\right)c\,\hat{\mathbf{z}}$ | (16$i$) | O II |
| $\mathbf{B}_{58}$ | = | $\left(\frac{1}{2}+x_7\right)\mathbf{a}_1+\left(\frac{1}{2}-y_7\right)\mathbf{a}_2+\left(\frac{1}{2}-z_7\right)\mathbf{a}_3$ | = | $\left(\frac{1}{2}+x_7\right)a\,\hat{\mathbf{x}}+\left(\frac{1}{2}-y_7\right)a\,\hat{\mathbf{y}}+\left(\frac{1}{2}-z_7\right)c\,\hat{\mathbf{z}}$ | (16$i$) | O II |
| $\mathbf{B}_{59}$ | = | $\left(\frac{1}{2}+y_7\right)\mathbf{a}_1+\left(\frac{1}{2}+x_7\right)\mathbf{a}_2+\left(\frac{1}{2}-z_7\right)\mathbf{a}_3$ | = | $\left(\frac{1}{2}+y_7\right)a\,\hat{\mathbf{x}}+\left(\frac{1}{2}+x_7\right)a\,\hat{\mathbf{y}}+\left(\frac{1}{2}-z_7\right)c\,\hat{\mathbf{z}}$ | (16$i$) | O II |
| $\mathbf{B}_{60}$ | = | $\left(\frac{1}{2}-y_7\right)\mathbf{a}_1+\left(\frac{1}{2}-x_7\right)\mathbf{a}_2+\left(\frac{1}{2}-z_7\right)\mathbf{a}_3$ | = | $\left(\frac{1}{2}-y_7\right)a\,\hat{\mathbf{x}}+\left(\frac{1}{2}-x_7\right)a\,\hat{\mathbf{y}}+\left(\frac{1}{2}-z_7\right)c\,\hat{\mathbf{z}}$ | (16$i$) | O II |
| $\mathbf{B}_{61}$ | = | $-x_7\,\mathbf{a}_1-y_7\,\mathbf{a}_2-z_7\,\mathbf{a}_3$ | = | $-x_7a\,\hat{\mathbf{x}}-y_7a\,\hat{\mathbf{y}}-z_7c\,\hat{\mathbf{z}}$ | (16$i$) | O II |
| $\mathbf{B}_{62}$ | = | $x_7\,\mathbf{a}_1+y_7\,\mathbf{a}_2-z_7\,\mathbf{a}_3$ | = | $x_7a\,\hat{\mathbf{x}}+y_7a\,\hat{\mathbf{y}}-z_7c\,\hat{\mathbf{z}}$ | (16$i$) | O II |
| $\mathbf{B}_{63}$ | = | $y_7\,\mathbf{a}_1-x_7\,\mathbf{a}_2-z_7\,\mathbf{a}_3$ | = | $y_7a\,\hat{\mathbf{x}}-x_7a\,\hat{\mathbf{y}}-z_7c\,\hat{\mathbf{z}}$ | (16$i$) | O II |
| $\mathbf{B}_{64}$ | = | $-y_7\,\mathbf{a}_1+x_7\,\mathbf{a}_2-z_7\,\mathbf{a}_3$ | = | $-y_7a\,\hat{\mathbf{x}}+x_7a\,\hat{\mathbf{y}}-z_7c\,\hat{\mathbf{z}}$ | (16$i$) | O II |
| $\mathbf{B}_{65}$ | = | $\left(\frac{1}{2}+x_7\right)\mathbf{a}_1+\left(\frac{1}{2}-y_7\right)\mathbf{a}_2+\left(\frac{1}{2}+z_7\right)\mathbf{a}_3$ | = | $\left(\frac{1}{2}+x_7\right)a\,\hat{\mathbf{x}}+\left(\frac{1}{2}-y_7\right)a\,\hat{\mathbf{y}}+\left(\frac{1}{2}+z_7\right)c\,\hat{\mathbf{z}}$ | (16$i$) | O II |
| $\mathbf{B}_{66}$ | = | $\left(\frac{1}{2}-x_7\right)\mathbf{a}_1+\left(\frac{1}{2}+y_7\right)\mathbf{a}_2+\left(\frac{1}{2}+z_7\right)\mathbf{a}_3$ | = | $\left(\frac{1}{2}-x_7\right)a\,\hat{\mathbf{x}}+\left(\frac{1}{2}+y_7\right)a\,\hat{\mathbf{y}}+\left(\frac{1}{2}+z_7\right)c\,\hat{\mathbf{z}}$ | (16$i$) | O II |
| $\mathbf{B}_{67}$ | = | $\left(\frac{1}{2}-y_7\right)\mathbf{a}_1+\left(\frac{1}{2}-x_7\right)\mathbf{a}_2+\left(\frac{1}{2}+z_7\right)\mathbf{a}_3$ | = | $\left(\frac{1}{2}-y_7\right)a\,\hat{\mathbf{x}}+\left(\frac{1}{2}-x_7\right)a\,\hat{\mathbf{y}}+\left(\frac{1}{2}+z_7\right)c\,\hat{\mathbf{z}}$ | (16$i$) | O II |
| $\mathbf{B}_{68}$ | = | $\left(\frac{1}{2}+y_7\right)\mathbf{a}_1+\left(\frac{1}{2}+x_7\right)\mathbf{a}_2+\left(\frac{1}{2}+z_7\right)\mathbf{a}_3$ | = | $\left(\frac{1}{2}+y_7\right)a\,\hat{\mathbf{x}}+\left(\frac{1}{2}+x_7\right)a\,\hat{\mathbf{y}}+\left(\frac{1}{2}+z_7\right)c\,\hat{\mathbf{z}}$ | (16$i$) | O II |
| $\mathbf{B}_{69}$ | = | $x_8\,\mathbf{a}_1+y_8\,\mathbf{a}_2+z_8\,\mathbf{a}_3$ | = | $x_8a\,\hat{\mathbf{x}}+y_8a\,\hat{\mathbf{y}}+z_8c\,\hat{\mathbf{z}}$ | (16$i$) | O III |
| $\mathbf{B}_{70}$ | = | $-x_8\,\mathbf{a}_1-y_8\,\mathbf{a}_2+z_8\,\mathbf{a}_3$ | = | $-x_8a\,\hat{\mathbf{x}}-y_8a\,\hat{\mathbf{y}}+z_8c\,\hat{\mathbf{z}}$ | (16$i$) | O III |
| $\mathbf{B}_{71}$ | = | $-y_8\,\mathbf{a}_1+x_8\,\mathbf{a}_2+z_8\,\mathbf{a}_3$ | = | $-y_8a\,\hat{\mathbf{x}}+x_8a\,\hat{\mathbf{y}}+z_8c\,\hat{\mathbf{z}}$ | (16$i$) | O III |
| $\mathbf{B}_{72}$ | = | $y_8\,\mathbf{a}_1-x_8\,\mathbf{a}_2+z_8\,\mathbf{a}_3$ | = | $y_8a\,\hat{\mathbf{x}}-x_8a\,\hat{\mathbf{y}}+z_8c\,\hat{\mathbf{z}}$ | (16$i$) | O III |
| $\mathbf{B}_{73}$ | = | $\left(\frac{1}{2}-x_8\right)\mathbf{a}_1+\left(\frac{1}{2}+y_8\right)\mathbf{a}_2+\left(\frac{1}{2}-z_8\right)\mathbf{a}_3$ | = | $\left(\frac{1}{2}-x_8\right)a\,\hat{\mathbf{x}}+\left(\frac{1}{2}+y_8\right)a\,\hat{\mathbf{y}}+\left(\frac{1}{2}-z_8\right)c\,\hat{\mathbf{z}}$ | (16$i$) | O III |
| $\mathbf{B}_{74}$ | = | $\left(\frac{1}{2}+x_8\right)\mathbf{a}_1+\left(\frac{1}{2}-y_8\right)\mathbf{a}_2+\left(\frac{1}{2}-z_8\right)\mathbf{a}_3$ | = | $\left(\frac{1}{2}+x_8\right)a\,\hat{\mathbf{x}}+\left(\frac{1}{2}-y_8\right)a\,\hat{\mathbf{y}}+\left(\frac{1}{2}-z_8\right)c\,\hat{\mathbf{z}}$ | (16$i$) | O III |
| $\mathbf{B}_{75}$ | = | $\left(\frac{1}{2}+y_8\right)\mathbf{a}_1+\left(\frac{1}{2}+x_8\right)\mathbf{a}_2+\left(\frac{1}{2}-z_8\right)\mathbf{a}_3$ | = | $\left(\frac{1}{2}+y_8\right)a\,\hat{\mathbf{x}}+\left(\frac{1}{2}+x_8\right)a\,\hat{\mathbf{y}}+\left(\frac{1}{2}-z_8\right)c\,\hat{\mathbf{z}}$ | (16$i$) | O III |
| $\mathbf{B}_{76}$ | = | $\left(\frac{1}{2}-y_8\right)\mathbf{a}_1+\left(\frac{1}{2}-x_8\right)\mathbf{a}_2+\left(\frac{1}{2}-z_8\right)\mathbf{a}_3$ | = | $\left(\frac{1}{2}-y_8\right)a\,\hat{\mathbf{x}}+\left(\frac{1}{2}-x_8\right)a\,\hat{\mathbf{y}}+\left(\frac{1}{2}-z_8\right)c\,\hat{\mathbf{z}}$ | (16$i$) | O III |
| $\mathbf{B}_{77}$ | = | $-x_8\,\mathbf{a}_1-y_8\,\mathbf{a}_2-z_8\,\mathbf{a}_3$ | = | $-x_8a\,\hat{\mathbf{x}}-y_8a\,\hat{\mathbf{y}}-z_8c\,\hat{\mathbf{z}}$ | (16$i$) | O III |
| $\mathbf{B}_{78}$ | = | $x_8\,\mathbf{a}_1+y_8\,\mathbf{a}_2-z_8\,\mathbf{a}_3$ | = | $x_8a\,\hat{\mathbf{x}}+y_8a\,\hat{\mathbf{y}}-z_8c\,\hat{\mathbf{z}}$ | (16$i$) | O III |
| $\mathbf{B}_{79}$ | = | $y_8\,\mathbf{a}_1-x_8\,\mathbf{a}_2-z_8\,\mathbf{a}_3$ | = | $y_8a\,\hat{\mathbf{x}}-x_8a\,\hat{\mathbf{y}}-z_8c\,\hat{\mathbf{z}}$ | (16$i$) | O III |
| $\mathbf{B}_{80}$ | = | $-y_8\,\mathbf{a}_1+x_8\,\mathbf{a}_2-z_8\,\mathbf{a}_3$ | = | $-y_8a\,\hat{\mathbf{x}}+x_8a\,\hat{\mathbf{y}}-z_8c\,\hat{\mathbf{z}}$ | (16$i$) | O III |
| $\mathbf{B}_{81}$ | = | $\left(\frac{1}{2}+x_8\right)\mathbf{a}_1+\left(\frac{1}{2}-y_8\right)\mathbf{a}_2+\left(\frac{1}{2}+z_8\right)\mathbf{a}_3$ | = | $\left(\frac{1}{2}+x_8\right)a\,\hat{\mathbf{x}}+\left(\frac{1}{2}-y_8\right)a\,\hat{\mathbf{y}}+\left(\frac{1}{2}+z_8\right)c\,\hat{\mathbf{z}}$ | (16$i$) | O III |
| $\mathbf{B}_{82}$ | = | $\left(\frac{1}{2}-x_8\right)\mathbf{a}_1+\left(\frac{1}{2}+y_8\right)\mathbf{a}_2+\left(\frac{1}{2}+z_8\right)\mathbf{a}_3$ | = | $\left(\frac{1}{2}-x_8\right)a\,\hat{\mathbf{x}}+\left(\frac{1}{2}+y_8\right)a\,\hat{\mathbf{y}}+\left(\frac{1}{2}+z_8\right)c\,\hat{\mathbf{z}}$ | (16$i$) | O III |
| $\mathbf{B}_{83}$ | = | $\left(\frac{1}{2} - y_8\right) \mathbf{a}_1 + \left(\frac{1}{2} - x_8\right) \mathbf{a}_2 + \left(\frac{1}{2} + z_8\right) \mathbf{a}_3$ | = | $\left(\frac{1}{2} - y_8\right) a \, \mathbf{\hat{x}} + \left(\frac{1}{2} - x_8\right) a \, \mathbf{\hat{y}} + \left(\frac{1}{2} + z_8\right) c \, \mathbf{\hat{z}}$ | (16*i*) | O III |
| $\mathbf{B}_{84}$ | = | $\left(\frac{1}{2} + y_8\right) \mathbf{a}_1 + \left(\frac{1}{2} + x_8\right) \mathbf{a}_2 + \left(\frac{1}{2} + z_8\right) \mathbf{a}_3$ | = | $\left(\frac{1}{2} + y_8\right) a \, \mathbf{\hat{x}} + \left(\frac{1}{2} + x_8\right) a \, \mathbf{\hat{y}} + \left(\frac{1}{2} + z_8\right) c \, \mathbf{\hat{z}}$ | (16*i*) | O III |
| $\mathbf{B}_{85}$ | = | $x_9 \, \mathbf{a}_1 + y_9 \, \mathbf{a}_2 + z_9 \, \mathbf{a}_3$ | = | $x_9 a \, \mathbf{\hat{x}} + y_9 a \, \mathbf{\hat{y}} + z_9 c \, \mathbf{\hat{z}}$ | (16*i*) | O IV |
| $\mathbf{B}_{86}$ | = | $-x_9 \, \mathbf{a}_1 - y_9 \, \mathbf{a}_2 + z_9 \, \mathbf{a}_3$ | = | $-x_9 a \, \mathbf{\hat{x}} - y_9 a \, \mathbf{\hat{y}} + z_9 c \, \mathbf{\hat{z}}$ | (16*i*) | O IV |
| $\mathbf{B}_{87}$ | = | $-y_9 \, \mathbf{a}_1 + x_9 \, \mathbf{a}_2 + z_9 \, \mathbf{a}_3$ | = | $-y_9 a \, \mathbf{\hat{x}} + x_9 a \, \mathbf{\hat{y}} + z_9 c \, \mathbf{\hat{z}}$ | (16*i*) | O IV |
| $\mathbf{B}_{88}$ | = | $y_9 \, \mathbf{a}_1 - x_9 \, \mathbf{a}_2 + z_9 \, \mathbf{a}_3$ | = | $y_9 a \, \mathbf{\hat{x}} - x_9 a \, \mathbf{\hat{y}} + z_9 c \, \mathbf{\hat{z}}$ | (16*i*) | O IV |
| $\mathbf{B}_{89}$ | = | $\left(\frac{1}{2} - x_9\right) \mathbf{a}_1 + \left(\frac{1}{2} + y_9\right) \mathbf{a}_2 + \left(\frac{1}{2} - z_9\right) \mathbf{a}_3$ | = | $\left(\frac{1}{2} - x_9\right) a \, \mathbf{\hat{x}} + \left(\frac{1}{2} + y_9\right) a \, \mathbf{\hat{y}} + \left(\frac{1}{2} - z_9\right) c \, \mathbf{\hat{z}}$ | (16*i*) | O IV |
| $\mathbf{B}_{90}$ | = | $\left(\frac{1}{2} + x_9\right) \mathbf{a}_1 + \left(\frac{1}{2} - y_9\right) \mathbf{a}_2 + \left(\frac{1}{2} - z_9\right) \mathbf{a}_3$ | = | $\left(\frac{1}{2} + x_9\right) a \, \mathbf{\hat{x}} + \left(\frac{1}{2} - y_9\right) a \, \mathbf{\hat{y}} + \left(\frac{1}{2} - z_9\right) c \, \mathbf{\hat{z}}$ | (16*i*) | O IV |
| $\mathbf{B}_{91}$ | = | $\left(\frac{1}{2} + y_9\right) \mathbf{a}_1 + \left(\frac{1}{2} + x_9\right) \mathbf{a}_2 + \left(\frac{1}{2} - z_9\right) \mathbf{a}_3$ | = | $\left(\frac{1}{2} + y_9\right) a \, \mathbf{\hat{x}} + \left(\frac{1}{2} + x_9\right) a \, \mathbf{\hat{y}} + \left(\frac{1}{2} - z_9\right) c \, \mathbf{\hat{z}}$ | (16*i*) | O IV |
| $\mathbf{B}_{92}$ | = | $\left(\frac{1}{2} - y_9\right) \mathbf{a}_1 + \left(\frac{1}{2} - x_9\right) \mathbf{a}_2 + \left(\frac{1}{2} - z_9\right) \mathbf{a}_3$ | = | $\left(\frac{1}{2} - y_9\right) a \, \mathbf{\hat{x}} + \left(\frac{1}{2} - x_9\right) a \, \mathbf{\hat{y}} + \left(\frac{1}{2} - z_9\right) c \, \mathbf{\hat{z}}$ | (16*i*) | O IV |
| $\mathbf{B}_{93}$ | = | $-x_9 \, \mathbf{a}_1 - y_9 \, \mathbf{a}_2 - z_9 \, \mathbf{a}_3$ | = | $-x_9 a \, \mathbf{\hat{x}} - y_9 a \, \mathbf{\hat{y}} - z_9 c \, \mathbf{\hat{z}}$ | (16*i*) | O IV |
| $\mathbf{B}_{94}$ | = | $x_9 \, \mathbf{a}_1 + y_9 \, \mathbf{a}_2 - z_9 \, \mathbf{a}_3$ | = | $x_9 a \, \mathbf{\hat{x}} + y_9 a \, \mathbf{\hat{y}} - z_9 c \, \mathbf{\hat{z}}$ | (16*i*) | O IV |
| $\mathbf{B}_{95}$ | = | $y_9 \, \mathbf{a}_1 - x_9 \, \mathbf{a}_2 - z_9 \, \mathbf{a}_3$ | = | $y_9 a \, \mathbf{\hat{x}} - x_9 a \, \mathbf{\hat{y}} - z_9 c \, \mathbf{\hat{z}}$ | (16*i*) | O IV |
| $\mathbf{B}_{96}$ | = | $-y_9 \, \mathbf{a}_1 + x_9 \, \mathbf{a}_2 - z_9 \, \mathbf{a}_3$ | = | $-y_9 a \, \mathbf{\hat{x}} + x_9 a \, \mathbf{\hat{y}} - z_9 c \, \mathbf{\hat{z}}$ | (16*i*) | O IV |
| $\mathbf{B}_{97}$ | = | $\left(\frac{1}{2} + x_9\right) \mathbf{a}_1 + \left(\frac{1}{2} - y_9\right) \mathbf{a}_2 + \left(\frac{1}{2} + z_9\right) \mathbf{a}_3$ | = | $\left(\frac{1}{2} + x_9\right) a \, \mathbf{\hat{x}} + \left(\frac{1}{2} - y_9\right) a \, \mathbf{\hat{y}} + \left(\frac{1}{2} + z_9\right) c \, \mathbf{\hat{z}}$ | (16*i*) | O IV |
| $\mathbf{B}_{98}$ | = | $\left(\frac{1}{2} - x_9\right) \mathbf{a}_1 + \left(\frac{1}{2} + y_9\right) \mathbf{a}_2 + \left(\frac{1}{2} + z_9\right) \mathbf{a}_3$ | = | $\left(\frac{1}{2} - x_9\right) a \, \mathbf{\hat{x}} + \left(\frac{1}{2} + y_9\right) a \, \mathbf{\hat{y}} + \left(\frac{1}{2} + z_9\right) c \, \mathbf{\hat{z}}$ | (16*i*) | O IV |
| $\mathbf{B}_{99}$ | = | $\left(\frac{1}{2} - y_9\right) \mathbf{a}_1 + \left(\frac{1}{2} - x_9\right) \mathbf{a}_2 + \left(\frac{1}{2} + z_9\right) \mathbf{a}_3$ | = | $\left(\frac{1}{2} - y_9\right) a \, \mathbf{\hat{x}} + \left(\frac{1}{2} - x_9\right) a \, \mathbf{\hat{y}} + \left(\frac{1}{2} + z_9\right) c \, \mathbf{\hat{z}}$ | (16*i*) | O IV |
| $\mathbf{B}_{100}$ | = | $\left(\frac{1}{2} + y_9\right) \mathbf{a}_1 + \left(\frac{1}{2} + x_9\right) \mathbf{a}_2 + \left(\frac{1}{2} + z_9\right) \mathbf{a}_3$ | = | $\left(\frac{1}{2} + y_9\right) a \, \mathbf{\hat{x}} + \left(\frac{1}{2} + x_9\right) a \, \mathbf{\hat{y}} + \left(\frac{1}{2} + z_9\right) c \, \mathbf{\hat{z}}$ | (16*i*) | O IV |
| $\mathbf{B}_{101}$ | = | $x_{10} \, \mathbf{a}_1 + y_{10} \, \mathbf{a}_2 + z_{10} \, \mathbf{a}_3$ | = | $x_{10} a \, \mathbf{\hat{x}} + y_{10} a \, \mathbf{\hat{y}} + z_{10} c \, \mathbf{\hat{z}}$ | (16*i*) | Si |
| $\mathbf{B}_{102}$ | = | $-x_{10} \, \mathbf{a}_1 - y_{10} \, \mathbf{a}_2 + z_{10} \, \mathbf{a}_3$ | = | $-x_{10} a \, \mathbf{\hat{x}} - y_{10} a \, \mathbf{\hat{y}} + z_{10} c \, \mathbf{\hat{z}}$ | (16*i*) | Si |
| $\mathbf{B}_{103}$ | = | $-y_{10} \, \mathbf{a}_1 + x_{10} \, \mathbf{a}_2 + z_{10} \, \mathbf{a}_3$ | = | $-y_{10} a \, \mathbf{\hat{x}} + x_{10} a \, \mathbf{\hat{y}} + z_{10} c \, \mathbf{\hat{z}}$ | (16*i*) | Si |
| $\mathbf{B}_{104}$ | = | $y_{10} \, \mathbf{a}_1 - x_{10} \, \mathbf{a}_2 + z_{10} \, \mathbf{a}_3$ | = | $y_{10} a \, \mathbf{\hat{x}} - x_{10} a \, \mathbf{\hat{y}} + z_{10} c \, \mathbf{\hat{z}}$ | (16*i*) | Si |
| $\mathbf{B}_{105}$ | = | $\left(\frac{1}{2} - x_{10}\right) \mathbf{a}_1 + \left(\frac{1}{2} + y_{10}\right) \mathbf{a}_2 + \left(\frac{1}{2} - z_{10}\right) \mathbf{a}_3$ | = | $\left(\frac{1}{2} - x_{10}\right) a \, \mathbf{\hat{x}} + \left(\frac{1}{2} + y_{10}\right) a \, \mathbf{\hat{y}} + \left(\frac{1}{2} - z_{10}\right) c \, \mathbf{\hat{z}}$ | (16*i*) | Si |
| $\mathbf{B}_{106}$ | = | $\left(\frac{1}{2} + x_{10}\right) \mathbf{a}_1 + \left(\frac{1}{2} - y_{10}\right) \mathbf{a}_2 + \left(\frac{1}{2} - z_{10}\right) \mathbf{a}_3$ | = | $\left(\frac{1}{2} + x_{10}\right) a \, \mathbf{\hat{x}} + \left(\frac{1}{2} - y_{10}\right) a \, \mathbf{\hat{y}} + \left(\frac{1}{2} - z_{10}\right) c \, \mathbf{\hat{z}}$ | (16*i*) | Si |
| $\mathbf{B}_{107}$ | = | $\left(\frac{1}{2} + y_{10}\right) \mathbf{a}_1 + \left(\frac{1}{2} + x_{10}\right) \mathbf{a}_2 + \left(\frac{1}{2} - z_{10}\right) \mathbf{a}_3$ | = | $\left(\frac{1}{2} + y_{10}\right) a \, \mathbf{\hat{x}} + \left(\frac{1}{2} + x_{10}\right) a \, \mathbf{\hat{y}} + \left(\frac{1}{2} - z_{10}\right) c \, \mathbf{\hat{z}}$ | (16*i*) | Si |
| $\mathbf{B}_{108}$ | = | $\left(\frac{1}{2} - y_{10}\right) \mathbf{a}_1 + \left(\frac{1}{2} - x_{10}\right) \mathbf{a}_2 + \left(\frac{1}{2} - z_{10}\right) \mathbf{a}_3$ | = | $\left(\frac{1}{2} - y_{10}\right) a \, \hat{\mathbf{x}} + \left(\frac{1}{2} - x_{10}\right) a \, \hat{\mathbf{y}} + \left(\frac{1}{2} - z_{10}\right) c \, \hat{\mathbf{z}}$ | (16*i*) | Si |
| $\mathbf{B}_{109}$ | = | $-x_{10} \, \mathbf{a}_1 - y_{10} \, \mathbf{a}_2 - z_{10} \, \mathbf{a}_3$ | = | $-x_{10} a \, \hat{\mathbf{x}} - y_{10} a \, \hat{\mathbf{y}} - z_{10} c \, \hat{\mathbf{z}}$ | (16*i*) | Si |
| $\mathbf{B}_{110}$ | = | $x_{10} \, \mathbf{a}_1 + y_{10} \, \mathbf{a}_2 - z_{10} \, \mathbf{a}_3$ | = | $x_{10} a \, \hat{\mathbf{x}} + y_{10} a \, \hat{\mathbf{y}} - z_{10} c \, \hat{\mathbf{z}}$ | (16*i*) | Si |
| $\mathbf{B}_{111}$ | = | $y_{10} \, \mathbf{a}_1 - x_{10} \, \mathbf{a}_2 - z_{10} \, \mathbf{a}_3$ | = | $y_{10} a \, \hat{\mathbf{x}} - x_{10} a \, \hat{\mathbf{y}} - z_{10} c \, \hat{\mathbf{z}}$ | (16*i*) | Si |
| $\mathbf{B}_{112}$ | = | $-y_{10} \, \mathbf{a}_1 + x_{10} \, \mathbf{a}_2 - z_{10} \, \mathbf{a}_3$ | = | $-y_{10} a \, \hat{\mathbf{x}} + x_{10} a \, \hat{\mathbf{y}} - z_{10} c \, \hat{\mathbf{z}}$ | (16*i*) | Si |
| $\mathbf{B}_{113}$ | = | $\left(\frac{1}{2} + x_{10}\right) \mathbf{a}_1 + \left(\frac{1}{2} - y_{10}\right) \mathbf{a}_2 + \left(\frac{1}{2} + z_{10}\right) \mathbf{a}_3$ | = | $\left(\frac{1}{2} + x_{10}\right) a \, \hat{\mathbf{x}} + \left(\frac{1}{2} - y_{10}\right) a \, \hat{\mathbf{y}} + \left(\frac{1}{2} + z_{10}\right) c \, \hat{\mathbf{z}}$ | (16*i*) | Si |
| $\mathbf{B}_{114}$ | = | $\left(\frac{1}{2} - x_{10}\right) \mathbf{a}_1 + \left(\frac{1}{2} + y_{10}\right) \mathbf{a}_2 + \left(\frac{1}{2} + z_{10}\right) \mathbf{a}_3$ | = | $\left(\frac{1}{2} - x_{10}\right) a \, \hat{\mathbf{x}} + \left(\frac{1}{2} + y_{10}\right) a \, \hat{\mathbf{y}} + \left(\frac{1}{2} + z_{10}\right) c \, \hat{\mathbf{z}}$ | (16*i*) | Si |
| $\mathbf{B}_{115}$ | = | $\left(\frac{1}{2} - y_{10}\right) \mathbf{a}_1 + \left(\frac{1}{2} - x_{10}\right) \mathbf{a}_2 + \left(\frac{1}{2} + z_{10}\right) \mathbf{a}_3$ | = | $\left(\frac{1}{2} - y_{10}\right) a \, \hat{\mathbf{x}} + \left(\frac{1}{2} - x_{10}\right) a \, \hat{\mathbf{y}} + \left(\frac{1}{2} + z_{10}\right) c \, \hat{\mathbf{z}}$ | (16*i*) | Si |
| $\mathbf{B}_{116}$ | = | $\left(\frac{1}{2} + y_{10}\right) \mathbf{a}_1 + \left(\frac{1}{2} + x_{10}\right) \mathbf{a}_2 + \left(\frac{1}{2} + z_{10}\right) \mathbf{a}_3$ | = | $\left(\frac{1}{2} + y_{10}\right) a \, \hat{\mathbf{x}} + \left(\frac{1}{2} + x_{10}\right) a \, \hat{\mathbf{y}} + \left(\frac{1}{2} + z_{10}\right) c \, \hat{\mathbf{z}}$ | (16*i*) | Si |

---

---

**Geometry files:**

- CIF: pp. 1698
- POSCAR: pp. 1698

# $CaBe_2Ge_2$ Structure: A2BC2_tP10_129_ac_c_bc

http://aflow.org/prototype-encyclopedia/A2BC2_tP10_129_ac_c_bc

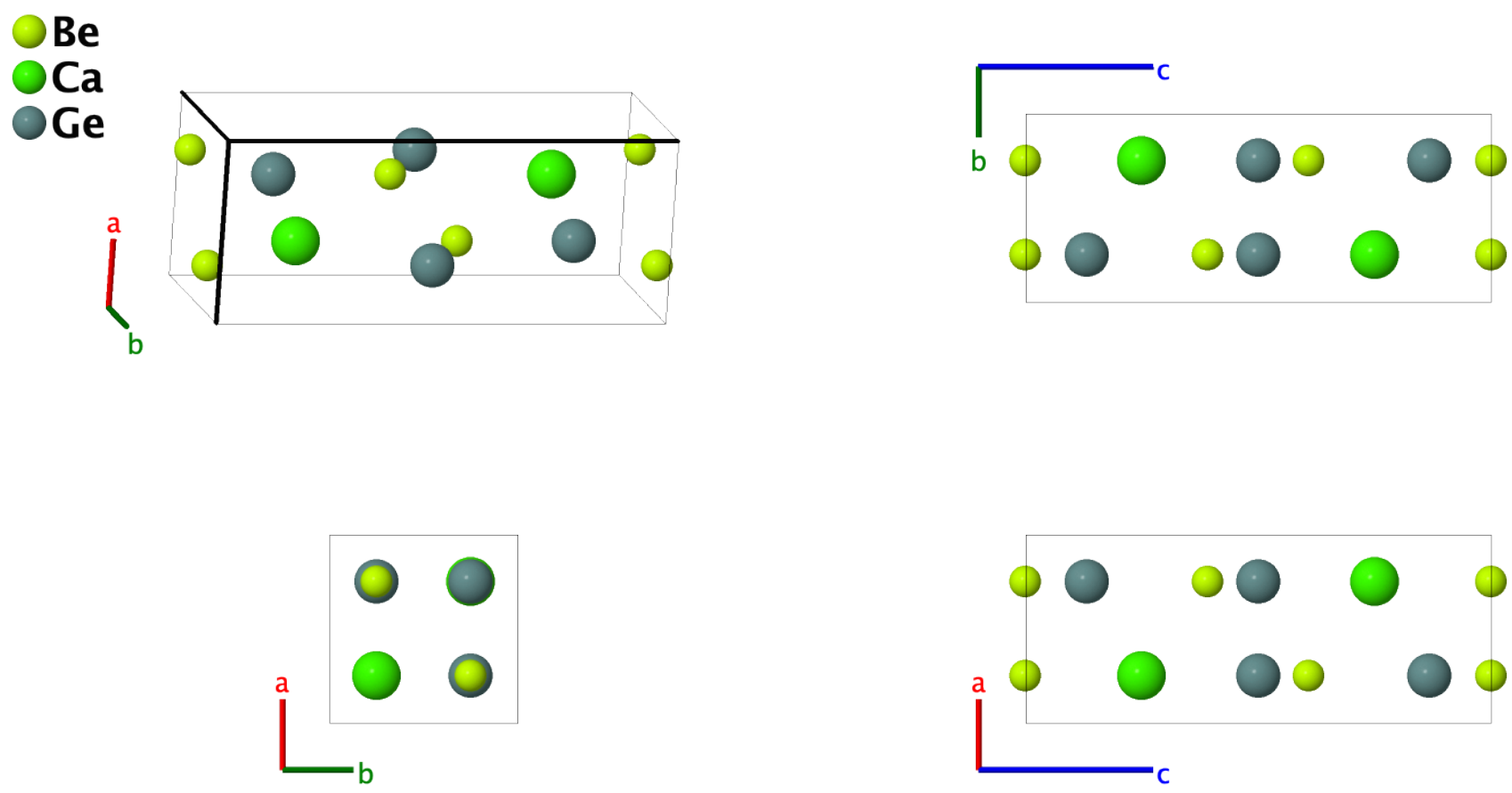

| | | |
|---|---|---|
| **Prototype** | : | $Be_2CaGe_2$ |
| **AFLOW prototype label** | : | A2BC2_tP10_129_ac_c_bc |
| ***Strukturbericht* designation** | : | None |
| **Pearson symbol** | : | tP10 |
| **Space group number** | : | 129 |
| **Space group symbol** | : | $P4/nmm$ |
| **AFLOW prototype command** | : | `aflow --proto=A2BC2_tP10_129_ac_c_bc --params=`$a, c/a, z_3, z_4, z_5$ |

### Other compounds with this structure

- $BaMg_2Pb_2$, $BaPd_2Sb_2$, $BaZn_2Sn_2$, $EuAu_2Al_2$, $EuPd_2Sb_2$, $EuPt_2Ge_2$, $LaPt_2Bi_2$, $LaPt_2Ge_2$, $LiPd_2Bi_2$, $SrPd_2Sb_2$, $SrPt_2As_2$, and $ThIr_2Si_2$

- This is a ternary form of the $D1_3$ ($BaAl_4$) structure. The atomic positions are approximately the same as in the conventional cell of $BaAl_4$, but the distribution of the atoms on those sites and the resulting relaxation leads to a different structure.
- Space group $P4/nmm$ #129 has two settings, but both have the same origin of the $z$-axis, so either setting will do here. We chose our standard setting 2 here.

### Simple Tetragonal primitive vectors:

$$\mathbf{a}_1 = a\,\hat{\mathbf{x}}$$
$$\mathbf{a}_2 = a\,\hat{\mathbf{y}}$$
$$\mathbf{a}_3 = c\,\hat{\mathbf{z}}$$

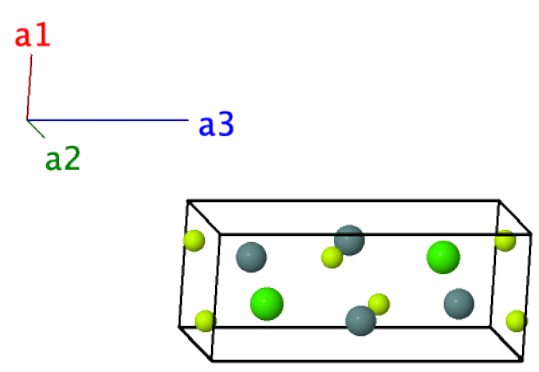

### Basis vectors:

| | | Lattice Coordinates | | Cartesian Coordinates | Wyckoff Position | Atom Type |
|---|---|---|---|---|---|---|
| $\mathbf{B}_1$ | = | $\frac{3}{4}\mathbf{a}_1 + \frac{1}{4}\mathbf{a}_2$ | = | $\frac{3}{4}a\,\hat{\mathbf{x}} + \frac{1}{4}a\,\hat{\mathbf{y}}$ | (2$a$) | Be I |
| $\mathbf{B}_2$ | = | $\frac{1}{4}\,\mathbf{a}_1 + \frac{3}{4}\,\mathbf{a}_2$ | = | $\frac{1}{4}a\,\hat{\mathbf{x}} + \frac{3}{4}a\,\hat{\mathbf{y}}$ | $(2a)$ | Be I |
| $\mathbf{B}_3$ | = | $\frac{3}{4}\,\mathbf{a}_1 + \frac{1}{4}\,\mathbf{a}_2 + \frac{1}{2}\,\mathbf{a}_3$ | = | $\frac{3}{4}a\,\hat{\mathbf{x}} + \frac{1}{4}a\,\hat{\mathbf{y}} + \frac{1}{2}c\,\hat{\mathbf{z}}$ | $(2b)$ | Ge I |
| $\mathbf{B}_4$ | = | $\frac{1}{4}\,\mathbf{a}_1 + \frac{3}{4}\,\mathbf{a}_2 + \frac{1}{2}\,\mathbf{a}_3$ | = | $\frac{1}{4}a\,\hat{\mathbf{x}} + \frac{3}{4}a\,\hat{\mathbf{y}} + \frac{1}{2}c\,\hat{\mathbf{z}}$ | $(2b)$ | Ge I |
| $\mathbf{B}_5$ | = | $\frac{1}{4}\,\mathbf{a}_1 + \frac{1}{4}\,\mathbf{a}_2 + z_3\,\mathbf{a}_3$ | = | $\frac{1}{4}a\,\hat{\mathbf{x}} + \frac{1}{4}a\,\hat{\mathbf{y}} + z_3c\,\hat{\mathbf{z}}$ | $(2c)$ | Be II |
| $\mathbf{B}_6$ | = | $\frac{3}{4}\,\mathbf{a}_1 + \frac{3}{4}\,\mathbf{a}_2 - z_3\,\mathbf{a}_3$ | = | $\frac{3}{4}a\,\hat{\mathbf{x}} + \frac{3}{4}a\,\hat{\mathbf{y}} - z_3c\,\hat{\mathbf{z}}$ | $(2c)$ | Be II |
| $\mathbf{B}_7$ | = | $\frac{1}{4}\,\mathbf{a}_1 + \frac{1}{4}\,\mathbf{a}_2 + z_4\,\mathbf{a}_3$ | = | $\frac{1}{4}a\,\hat{\mathbf{x}} + \frac{1}{4}a\,\hat{\mathbf{y}} + z_4c\,\hat{\mathbf{z}}$ | $(2c)$ | Ca |
| $\mathbf{B}_8$ | = | $\frac{3}{4}\,\mathbf{a}_1 + \frac{3}{4}\,\mathbf{a}_2 - z_4\,\mathbf{a}_3$ | = | $\frac{3}{4}a\,\hat{\mathbf{x}} + \frac{3}{4}a\,\hat{\mathbf{y}} - z_4c\,\hat{\mathbf{z}}$ | $(2c)$ | Ca |
| $\mathbf{B}_9$ | = | $\frac{1}{4}\,\mathbf{a}_1 + \frac{1}{4}\,\mathbf{a}_2 + z_5\,\mathbf{a}_3$ | = | $\frac{1}{4}a\,\hat{\mathbf{x}} + \frac{1}{4}a\,\hat{\mathbf{y}} + z_5c\,\hat{\mathbf{z}}$ | $(2c)$ | Ge II |
| $\mathbf{B}_{10}$ | = | $\frac{3}{4}\,\mathbf{a}_1 + \frac{3}{4}\,\mathbf{a}_2 - z_5\,\mathbf{a}_3$ | = | $\frac{3}{4}a\,\hat{\mathbf{x}} + \frac{3}{4}a\,\hat{\mathbf{y}} - z_5c\,\hat{\mathbf{z}}$ | $(2c)$ | Ge II |

**Geometry files:**

- CIF: pp. 1699
- POSCAR: pp. 1699

# Meta-autunite (I) [$Ca(UO_2)_2(PO_4)_2 \cdot 6H_2O$, $H5_{10}$] Structure: AB4C6DE_tP26_129_c_j_2ci_a_c

http://aflow.org/prototype-encyclopedia/AB4C6DE_tP26_129_c_j_2ci_a_c

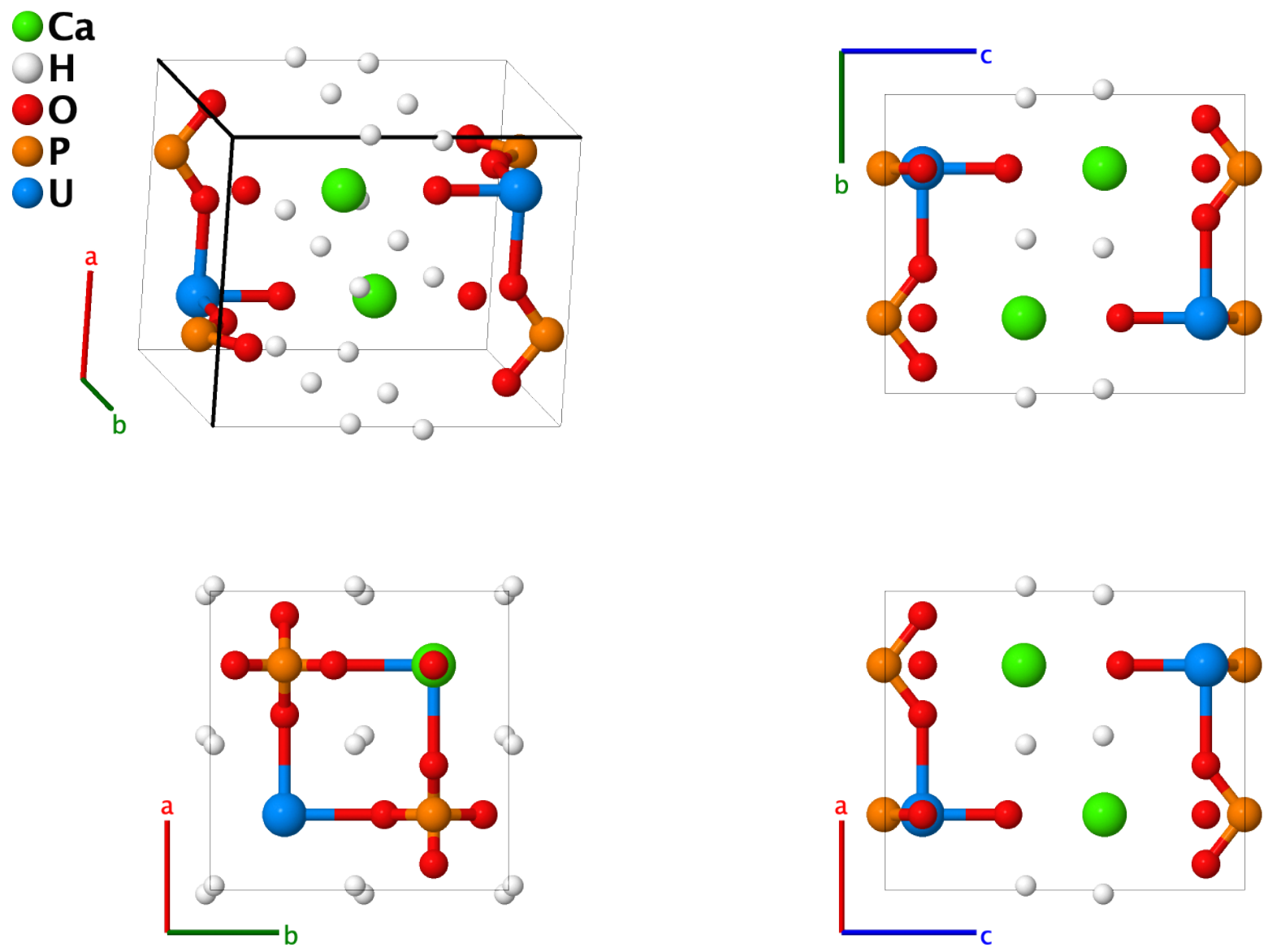

| | | |
|---|---|---|
| **Prototype** | : | $Ca(H_2O)_6O_{12}P_2U_2$ |
| **AFLOW prototype label** | : | AB4C6DE_tP26_129_c_j_2ci_a_c |
| ***Strukturbericht* designation** | : | $H5_{10}$ |
| **Pearson symbol** | : | tP26 |
| **Space group number** | : | 129 |
| **Space group symbol** | : | $P4/nmm$ |
| **AFLOW prototype command** | : | `aflow --proto=AB4C6DE_tP26_129_c_j_2ci_a_c` `--params=`$a, c/a, z_2, z_3, z_4, z_5, y_6, z_6, x_7, z_7$ |

- Autunite $Ca(UO_2)_2(PO_4)_2 \cdot nH_2O$, is found in three varieties: naturally occurring autunite, with $n \gtrsim 10$, and meta-autunite (I), which is partially dehydrated, $6 \gtrsim n \gtrsim 10$. Further dehydration in the laboratory produces meta-autunite (II).
- (Beintema, 1938) proposed a structure for meta-autunite (I), which (Herrmann, 1941) designated $H5_{10}$. He did not locate the calcium and oxygen atoms nor the water molecules. This structure was improved by (Makarov, 1960), and we include it here as our prototype for $H5_{10}$. The Ca-I site is 50% occupied, while the $H_2O$ site is 75% occupied.
- The AFLOW label models the structure as if the sites were fully occupied.

**Simple Tetragonal primitive vectors:**

$$\mathbf{a}_1 = a\,\hat{\mathbf{x}}$$
$$\mathbf{a}_2 = a\,\hat{\mathbf{y}}$$
$$\mathbf{a}_3 = c\,\hat{\mathbf{z}}$$

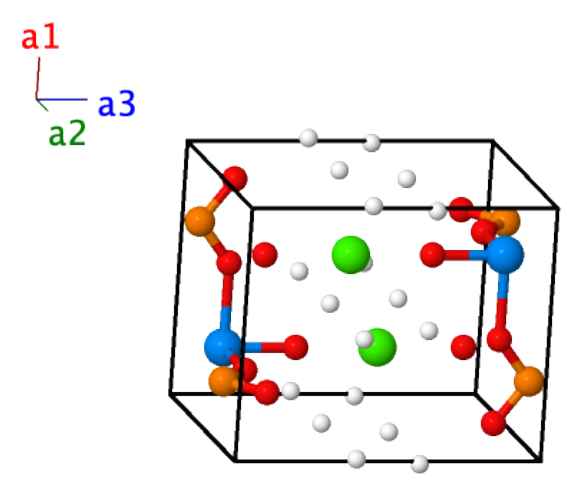

**Basis vectors:**

| | | Lattice Coordinates | | Cartesian Coordinates | Wyckoff Position | Atom Type |
|---|---|---|---|---|---|---|
| $\mathbf{B}_1$ | = | $\frac{3}{4}\,\mathbf{a}_1 + \frac{1}{4}\,\mathbf{a}_2$ | = | $\frac{3}{4}a\,\hat{\mathbf{x}} + \frac{1}{4}a\,\hat{\mathbf{y}}$ | (2$a$) | P |
| $\mathbf{B}_2$ | = | $\frac{1}{4}\,\mathbf{a}_1 + \frac{3}{4}\,\mathbf{a}_2$ | = | $\frac{1}{4}a\,\hat{\mathbf{x}} + \frac{3}{4}a\,\hat{\mathbf{y}}$ | (2$a$) | P |
| $\mathbf{B}_3$ | = | $\frac{1}{4}\,\mathbf{a}_1 + \frac{1}{4}\,\mathbf{a}_2 + z_2\,\mathbf{a}_3$ | = | $\frac{1}{4}a\,\hat{\mathbf{x}} + \frac{1}{4}a\,\hat{\mathbf{y}} + z_2c\,\hat{\mathbf{z}}$ | (2$c$) | Ca |
| $\mathbf{B}_4$ | = | $\frac{3}{4}\,\mathbf{a}_1 + \frac{3}{4}\,\mathbf{a}_2 - z_2\,\mathbf{a}_3$ | = | $\frac{3}{4}a\,\hat{\mathbf{x}} + \frac{3}{4}a\,\hat{\mathbf{y}} - z_2c\,\hat{\mathbf{z}}$ | (2$c$) | Ca |
| $\mathbf{B}_5$ | = | $\frac{1}{4}\,\mathbf{a}_1 + \frac{1}{4}\,\mathbf{a}_2 + z_3\,\mathbf{a}_3$ | = | $\frac{1}{4}a\,\hat{\mathbf{x}} + \frac{1}{4}a\,\hat{\mathbf{y}} + z_3c\,\hat{\mathbf{z}}$ | (2$c$) | O I |
| $\mathbf{B}_6$ | = | $\frac{3}{4}\,\mathbf{a}_1 + \frac{3}{4}\,\mathbf{a}_2 - z_3\,\mathbf{a}_3$ | = | $\frac{3}{4}a\,\hat{\mathbf{x}} + \frac{3}{4}a\,\hat{\mathbf{y}} - z_3c\,\hat{\mathbf{z}}$ | (2$c$) | O I |
| $\mathbf{B}_7$ | = | $\frac{1}{4}\,\mathbf{a}_1 + \frac{1}{4}\,\mathbf{a}_2 + z_4\,\mathbf{a}_3$ | = | $\frac{1}{4}a\,\hat{\mathbf{x}} + \frac{1}{4}a\,\hat{\mathbf{y}} + z_4c\,\hat{\mathbf{z}}$ | (2$c$) | O II |
| $\mathbf{B}_8$ | = | $\frac{3}{4}\,\mathbf{a}_1 + \frac{3}{4}\,\mathbf{a}_2 - z_4\,\mathbf{a}_3$ | = | $\frac{3}{4}a\,\hat{\mathbf{x}} + \frac{3}{4}a\,\hat{\mathbf{y}} - z_4c\,\hat{\mathbf{z}}$ | (2$c$) | O II |
| $\mathbf{B}_9$ | = | $\frac{1}{4}\,\mathbf{a}_1 + \frac{1}{4}\,\mathbf{a}_2 + z_5\,\mathbf{a}_3$ | = | $\frac{1}{4}a\,\hat{\mathbf{x}} + \frac{1}{4}a\,\hat{\mathbf{y}} + z_5c\,\hat{\mathbf{z}}$ | (2$c$) | U |
| $\mathbf{B}_{10}$ | = | $\frac{3}{4}\,\mathbf{a}_1 + \frac{3}{4}\,\mathbf{a}_2 - z_5\,\mathbf{a}_3$ | = | $\frac{3}{4}a\,\hat{\mathbf{x}} + \frac{3}{4}a\,\hat{\mathbf{y}} - z_5c\,\hat{\mathbf{z}}$ | (2$c$) | U |
| $\mathbf{B}_{11}$ | = | $\frac{1}{4}\,\mathbf{a}_1 + y_6\,\mathbf{a}_2 + z_6\,\mathbf{a}_3$ | = | $\frac{1}{4}a\,\hat{\mathbf{x}} + y_6a\,\hat{\mathbf{y}} + z_6c\,\hat{\mathbf{z}}$ | (8$i$) | O III |
| $\mathbf{B}_{12}$ | = | $\frac{1}{4}\,\mathbf{a}_1 + \left(\frac{1}{2} - y_6\right)\,\mathbf{a}_2 + z_6\,\mathbf{a}_3$ | = | $\frac{1}{4}a\,\hat{\mathbf{x}} + \left(\frac{1}{2} - y_6\right)a\,\hat{\mathbf{y}} + z_6c\,\hat{\mathbf{z}}$ | (8$i$) | O III |
| $\mathbf{B}_{13}$ | = | $\left(\frac{1}{2} - y_6\right)\,\mathbf{a}_1 + \frac{1}{4}\,\mathbf{a}_2 + z_6\,\mathbf{a}_3$ | = | $\left(\frac{1}{2} - y_6\right)a\,\hat{\mathbf{x}} + \frac{1}{4}a\,\hat{\mathbf{y}} + z_6c\,\hat{\mathbf{z}}$ | (8$i$) | O III |
| $\mathbf{B}_{14}$ | = | $y_6\,\mathbf{a}_1 + \frac{1}{4}\,\mathbf{a}_2 + z_6\,\mathbf{a}_3$ | = | $y_6a\,\hat{\mathbf{x}} + \frac{1}{4}a\,\hat{\mathbf{y}} + z_6c\,\hat{\mathbf{z}}$ | (8$i$) | O III |
| $\mathbf{B}_{15}$ | = | $\frac{3}{4}\,\mathbf{a}_1 + \left(\frac{1}{2} + y_6\right)\,\mathbf{a}_2 - z_6\,\mathbf{a}_3$ | = | $\frac{3}{4}a\,\hat{\mathbf{x}} + \left(\frac{1}{2} + y_6\right)a\,\hat{\mathbf{y}} - z_6c\,\hat{\mathbf{z}}$ | (8$i$) | O III |
| $\mathbf{B}_{16}$ | = | $\frac{3}{4}\,\mathbf{a}_1 - y_6\,\mathbf{a}_2 - z_6\,\mathbf{a}_3$ | = | $\frac{3}{4}a\,\hat{\mathbf{x}} - y_6a\,\hat{\mathbf{y}} - z_6c\,\hat{\mathbf{z}}$ | (8$i$) | O III |
| $\mathbf{B}_{17}$ | = | $\left(\frac{1}{2} + y_6\right)\,\mathbf{a}_1 + \frac{3}{4}\,\mathbf{a}_2 - z_6\,\mathbf{a}_3$ | = | $\left(\frac{1}{2} + y_6\right)a\,\hat{\mathbf{x}} + \frac{3}{4}a\,\hat{\mathbf{y}} - z_6c\,\hat{\mathbf{z}}$ | (8$i$) | O III |
| $\mathbf{B}_{18}$ | = | $-y_6\,\mathbf{a}_1 + \frac{3}{4}\,\mathbf{a}_2 - z_6\,\mathbf{a}_3$ | = | $-y_6a\,\hat{\mathbf{x}} + \frac{3}{4}a\,\hat{\mathbf{y}} - z_6c\,\hat{\mathbf{z}}$ | (8$i$) | O III |
| $\mathbf{B}_{19}$ | = | $x_7\,\mathbf{a}_1 + x_7\,\mathbf{a}_2 + z_7\,\mathbf{a}_3$ | = | $x_7a\,\hat{\mathbf{x}} + x_7a\,\hat{\mathbf{y}} + z_7c\,\hat{\mathbf{z}}$ | (8$j$) | $H_2O$ |
| $\mathbf{B}_{20}$ | = | $\left(\frac{1}{2} - x_7\right)\,\mathbf{a}_1 + \left(\frac{1}{2} - x_7\right)\,\mathbf{a}_2 + z_7\,\mathbf{a}_3$ | = | $\left(\frac{1}{2} - x_7\right)a\,\hat{\mathbf{x}} + \left(\frac{1}{2} - x_7\right)a\,\hat{\mathbf{y}} + z_7c\,\hat{\mathbf{z}}$ | (8$j$) | $H_2O$ |
| $\mathbf{B}_{21}$ | = | $\left(\frac{1}{2} - x_7\right)\,\mathbf{a}_1 + x_7\,\mathbf{a}_2 + z_7\,\mathbf{a}_3$ | = | $\left(\frac{1}{2} - x_7\right)a\,\hat{\mathbf{x}} + x_7a\,\hat{\mathbf{y}} + z_7c\,\hat{\mathbf{z}}$ | (8$j$) | $H_2O$ |
| $\mathbf{B}_{22}$ | = | $x_7\,\mathbf{a}_1 + \left(\frac{1}{2} - x_7\right)\,\mathbf{a}_2 + z_7\,\mathbf{a}_3$ | = | $x_7a\,\hat{\mathbf{x}} + \left(\frac{1}{2} - x_7\right)a\,\hat{\mathbf{y}} + z_7c\,\hat{\mathbf{z}}$ | (8$j$) | $H_2O$ |
| $\mathbf{B}_{23}$ | = | $-x_7\,\mathbf{a}_1 + \left(\frac{1}{2} + x_7\right)\,\mathbf{a}_2 - z_7\,\mathbf{a}_3$ | = | $-x_7a\,\hat{\mathbf{x}} + \left(\frac{1}{2} + x_7\right)a\,\hat{\mathbf{y}} - z_7c\,\hat{\mathbf{z}}$ | (8$j$) | $H_2O$ |
| $\mathbf{B}_{24}$ | = | $\left(\frac{1}{2} + x_7\right)\,\mathbf{a}_1 - x_7\,\mathbf{a}_2 - z_7\,\mathbf{a}_3$ | = | $\left(\frac{1}{2} + x_7\right)a\,\hat{\mathbf{x}} - x_7a\,\hat{\mathbf{y}} - z_7c\,\hat{\mathbf{z}}$ | (8$j$) | $H_2O$ |
| $\mathbf{B}_{25}$ | = | $\left(\frac{1}{2} + x_7\right)\,\mathbf{a}_1 + \left(\frac{1}{2} + x_7\right)\,\mathbf{a}_2 - z_7\,\mathbf{a}_3$ | = | $\left(\frac{1}{2} + x_7\right)a\,\hat{\mathbf{x}} + \left(\frac{1}{2} + x_7\right)a\,\hat{\mathbf{y}} - z_7c\,\hat{\mathbf{z}}$ | (8$j$) | $H_2O$ |
| $\mathbf{B}_{26}$ | = | $-x_7\,\mathbf{a}_1 - x_7\,\mathbf{a}_2 - z_7\,\mathbf{a}_3$ | = | $-x_7a\,\hat{\mathbf{x}} - x_7a\,\hat{\mathbf{y}} - z_7c\,\hat{\mathbf{z}}$ | (8$j$) | $H_2O$ |

**Found in:**

**Geometry files:**

- CIF: pp. 1699
- POSCAR: pp. 1700

# $NH_4Br$ ($B25$) Structure: AB4C_tP12_129_c_i_a

http://aflow.org/prototype-encyclopedia/AB4C_tP12_129_c_i_a

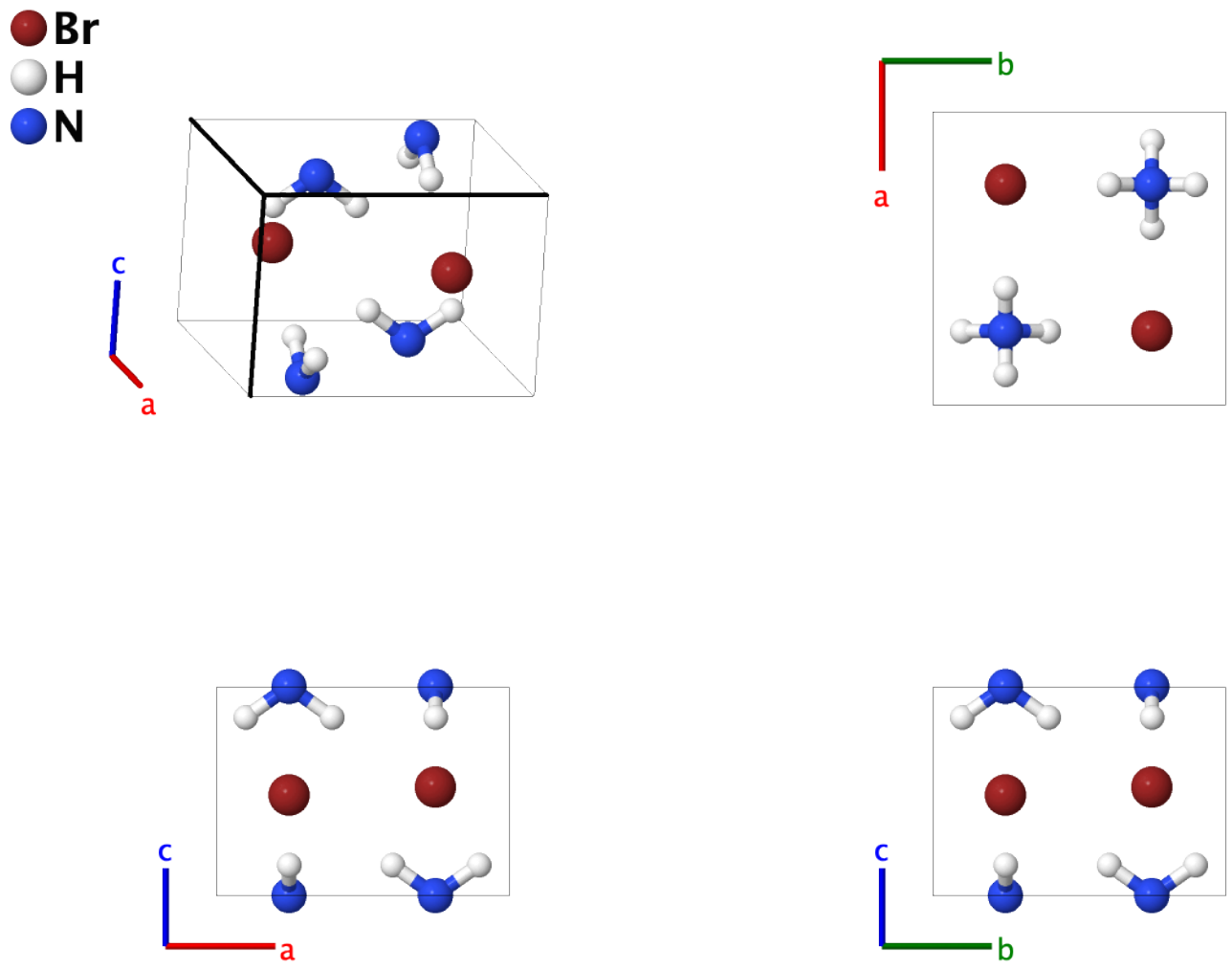

| | | |
|---|---|---|
| **Prototype** | : | $BrH_4N$ |
| **AFLOW prototype label** | : | AB4C_tP12_129_c_i_a |
| ***Strukturbericht* designation** | : | $B25$ |
| **Pearson symbol** | : | tP12 |
| **Space group number** | : | 129 |
| **Space group symbol** | : | $P4/nmm$ |
| **AFLOW prototype command** | : | `aflow --proto=AB4C_tP12_129_c_i_a --params=`$a, c/a, z_2, y_3, z_3$ |

---

**Other compounds with this structure**

- $NH_4I$

---

- Data was taken at 100 °C using deuterium.
- The atomic positions were given for setting 1 of space group #129. We have shifted this to setting 2, placing the origin at the inversion site of the structure.

---

**Simple Tetragonal primitive vectors:**

$$\mathbf{a}_1 = a\,\hat{\mathbf{x}}$$
$$\mathbf{a}_2 = a\,\hat{\mathbf{y}}$$
$$\mathbf{a}_3 = c\,\hat{\mathbf{z}}$$

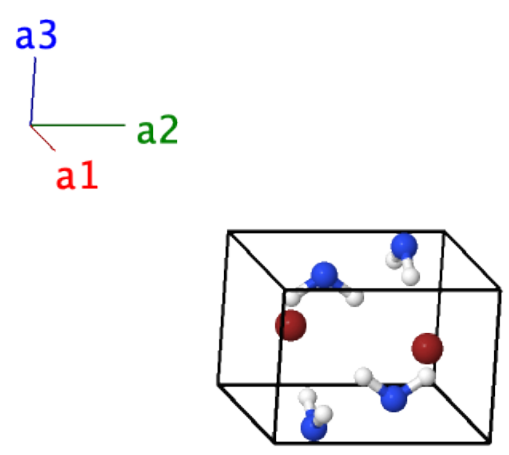

---

**Basis vectors:**

| | | Lattice Coordinates | | Cartesian Coordinates | Wyckoff Position | Atom Type |
|---|---|---|---|---|---|---|
| $\mathbf{B}_1$ | = | $\frac{3}{4}\,\mathbf{a}_1 + \frac{1}{4}\,\mathbf{a}_2$ | = | $\frac{3}{4}a\,\hat{\mathbf{x}} + \frac{1}{4}a\,\hat{\mathbf{y}}$ | $(2a)$ | N |
| $\mathbf{B}_2$ | = | $\frac{1}{4}\,\mathbf{a}_1 + \frac{3}{4}\,\mathbf{a}_2$ | = | $\frac{1}{4}a\,\hat{\mathbf{x}} + \frac{3}{4}a\,\hat{\mathbf{y}}$ | $(2a)$ | N |
| $\mathbf{B}_3$ | = | $\frac{1}{4}\,\mathbf{a}_1 + \frac{1}{4}\,\mathbf{a}_2 + z_2\,\mathbf{a}_3$ | = | $\frac{1}{4}a\,\hat{\mathbf{x}} + \frac{1}{4}a\,\hat{\mathbf{y}} + z_2c\,\hat{\mathbf{z}}$ | $(2c)$ | Br |
| $\mathbf{B}_4$ | = | $\frac{3}{4}\,\mathbf{a}_1 + \frac{3}{4}\,\mathbf{a}_2 - z_2\,\mathbf{a}_3$ | = | $\frac{3}{4}a\,\hat{\mathbf{x}} + \frac{3}{4}a\,\hat{\mathbf{y}} - z_2c\,\hat{\mathbf{z}}$ | $(2c)$ | Br |
| $\mathbf{B}_5$ | = | $\frac{1}{4}\,\mathbf{a}_1 + y_3\,\mathbf{a}_2 + z_3\,\mathbf{a}_3$ | = | $\frac{1}{4}a\,\hat{\mathbf{x}} + y_3a\,\hat{\mathbf{y}} + z_3c\,\hat{\mathbf{z}}$ | $(8i)$ | H |
| $\mathbf{B}_6$ | = | $\frac{1}{4}\,\mathbf{a}_1 + \left(\frac{1}{2} - y_3\right)\,\mathbf{a}_2 + z_3\,\mathbf{a}_3$ | = | $\frac{1}{4}a\,\hat{\mathbf{x}} + \left(\frac{1}{2} - y_3\right)a\,\hat{\mathbf{y}} + z_3c\,\hat{\mathbf{z}}$ | $(8i)$ | H |
| $\mathbf{B}_7$ | = | $\left(\frac{1}{2} - y_3\right)\,\mathbf{a}_1 + \frac{1}{4}\,\mathbf{a}_2 + z_3\,\mathbf{a}_3$ | = | $\left(\frac{1}{2} - y_3\right)a\,\hat{\mathbf{x}} + \frac{1}{4}a\,\hat{\mathbf{y}} + z_3c\,\hat{\mathbf{z}}$ | $(8i)$ | H |
| $\mathbf{B}_8$ | = | $y_3\,\mathbf{a}_1 + \frac{1}{4}\,\mathbf{a}_2 + z_3\,\mathbf{a}_3$ | = | $y_3a\,\hat{\mathbf{x}} + \frac{1}{4}a\,\hat{\mathbf{y}} + z_3c\,\hat{\mathbf{z}}$ | $(8i)$ | H |
| $\mathbf{B}_9$ | = | $\frac{3}{4}\,\mathbf{a}_1 + \left(\frac{1}{2} + y_3\right)\,\mathbf{a}_2 - z_3\,\mathbf{a}_3$ | = | $\frac{3}{4}a\,\hat{\mathbf{x}} + \left(\frac{1}{2} + y_3\right)a\,\hat{\mathbf{y}} - z_3c\,\hat{\mathbf{z}}$ | $(8i)$ | H |
| $\mathbf{B}_{10}$ | = | $\frac{3}{4}\,\mathbf{a}_1 - y_3\,\mathbf{a}_2 - z_3\,\mathbf{a}_3$ | = | $\frac{3}{4}a\,\hat{\mathbf{x}} - y_3a\,\hat{\mathbf{y}} - z_3c\,\hat{\mathbf{z}}$ | $(8i)$ | H |
| $\mathbf{B}_{11}$ | = | $\left(\frac{1}{2} + y_3\right)\,\mathbf{a}_1 + \frac{3}{4}\,\mathbf{a}_2 - z_3\,\mathbf{a}_3$ | = | $\left(\frac{1}{2} + y_3\right)a\,\hat{\mathbf{x}} + \frac{3}{4}a\,\hat{\mathbf{y}} - z_3c\,\hat{\mathbf{z}}$ | $(8i)$ | H |
| $\mathbf{B}_{12}$ | = | $-y_3\,\mathbf{a}_1 + \frac{3}{4}\,\mathbf{a}_2 - z_3\,\mathbf{a}_3$ | = | $-y_3a\,\hat{\mathbf{x}} + \frac{3}{4}a\,\hat{\mathbf{y}} - z_3c\,\hat{\mathbf{z}}$ | $(8i)$ | H |

**Geometry files:**

- CIF: pp. 1700
- POSCAR: pp. 1700

# LaOAgS Structure: ABCD_tP8_129_b_c_a_c

http://aflow.org/prototype-encyclopedia/ABCD_tP8_129_b_c_a_c

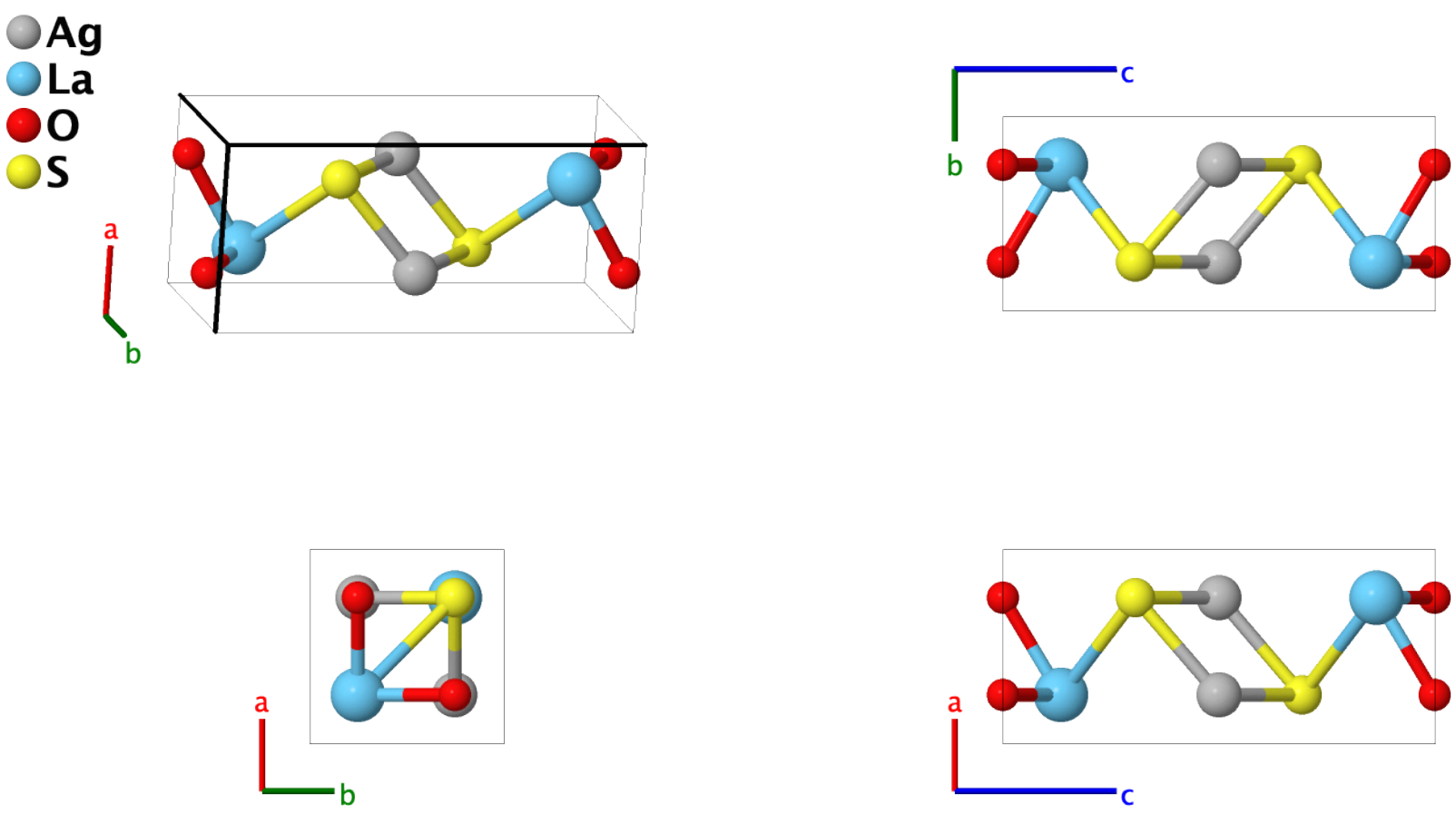

| | | |
|---|---|---|
| **Prototype** | : | AgLaOS |
| **AFLOW prototype label** | : | ABCD_tP8_129_b_c_a_c |
| ***Strukturbericht* designation** | : | None |
| **Pearson symbol** | : | tP8 |
| **Space group number** | : | 129 |
| **Space group symbol** | : | $P4/nmm$ |
| **AFLOW prototype command** | : | `aflow --proto=ABCD_tP8_129_b_c_a_c --params=`$a, c/a, z_3, z_4$ |

**Other compounds with this structure**

- LaAgSeO and LaAgTeO

- This structure has the same space group and Wyckoff positions as AsCuSiZr, but there is a substantial difference in the $c/a$ ratio leading to different bonding between the layers.

**Simple Tetragonal primitive vectors:**

$$\mathbf{a}_1 = a\,\hat{\mathbf{x}}$$
$$\mathbf{a}_2 = a\,\hat{\mathbf{y}}$$
$$\mathbf{a}_3 = c\,\hat{\mathbf{z}}$$

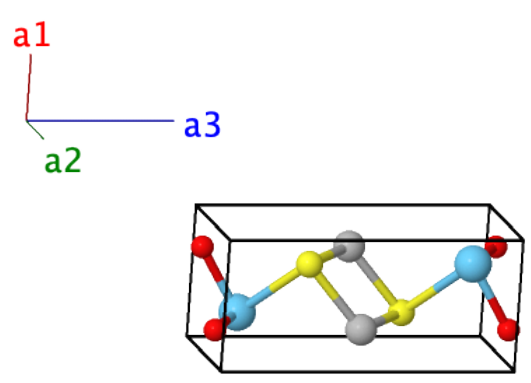

**Basis vectors:**

| | | Lattice Coordinates | | Cartesian Coordinates | Wyckoff Position | Atom Type |
|---|---|---|---|---|---|---|
| $\mathbf{B}_1$ | = | $\frac{3}{4}\,\mathbf{a}_1 + \frac{1}{4}\,\mathbf{a}_2$ | = | $\frac{3}{4}a\,\hat{\mathbf{x}} + \frac{1}{4}a\,\hat{\mathbf{y}}$ | (2$a$) | O |
| $\mathbf{B}_2$ | = | $\frac{1}{4}\,\mathbf{a}_1 + \frac{3}{4}\,\mathbf{a}_2$ | = | $\frac{1}{4}a\,\hat{\mathbf{x}} + \frac{3}{4}a\,\hat{\mathbf{y}}$ | (2$a$) | O |
| $\mathbf{B}_3$ | = | $\frac{3}{4}\,\mathbf{a}_1 + \frac{1}{4}\,\mathbf{a}_2 + \frac{1}{2}\,\mathbf{a}_3$ | = | $\frac{3}{4}a\,\hat{\mathbf{x}} + \frac{1}{4}a\,\hat{\mathbf{y}} + \frac{1}{2}c\,\hat{\mathbf{z}}$ | (2$b$) | Ag |
| $\mathbf{B}_4$ | = | $\frac{1}{4}\,\mathbf{a}_1 + \frac{3}{4}\,\mathbf{a}_2 + \frac{1}{2}\,\mathbf{a}_3$ | = | $\frac{1}{4}a\,\hat{\mathbf{x}} + \frac{3}{4}a\,\hat{\mathbf{y}} + \frac{1}{2}c\,\hat{\mathbf{z}}$ | (2$b$) | Ag |
| $\mathbf{B}_5$ | = | $\frac{1}{4}\,\mathbf{a}_1 + \frac{1}{4}\,\mathbf{a}_2 + z_3\,\mathbf{a}_3$ | = | $\frac{1}{4}a\,\hat{\mathbf{x}} + \frac{1}{4}a\,\hat{\mathbf{y}} + z_3c\,\hat{\mathbf{z}}$ | (2$c$) | La |
| $\mathbf{B}_6$ | = | $\frac{3}{4}\,\mathbf{a}_1 + \frac{3}{4}\,\mathbf{a}_2 - z_3\,\mathbf{a}_3$ | = | $\frac{3}{4}a\,\hat{\mathbf{x}} + \frac{3}{4}a\,\hat{\mathbf{y}} - z_3c\,\hat{\mathbf{z}}$ | (2$c$) | La |
| $\mathbf{B}_7$ | = | $\frac{1}{4}\,\mathbf{a}_1 + \frac{1}{4}\,\mathbf{a}_2 + z_4\,\mathbf{a}_3$ | = | $\frac{1}{4}a\,\hat{\mathbf{x}} + \frac{1}{4}a\,\hat{\mathbf{y}} + z_4c\,\hat{\mathbf{z}}$ | (2$c$) | S |
| $\mathbf{B}_8$ | = | $\frac{3}{4}\,\mathbf{a}_1 + \frac{3}{4}\,\mathbf{a}_2 - z_4\,\mathbf{a}_3$ | = | $\frac{3}{4}a\,\hat{\mathbf{x}} + \frac{3}{4}a\,\hat{\mathbf{y}} - z_4c\,\hat{\mathbf{z}}$ | (2$c$) | S |

---

**Found in:**

---

**Geometry files:**

- CIF: pp. 1700
- POSCAR: pp. 1701

# $Sr(OH)_2(H_2O)_8$ Structure: A18B10C_tP116_130_2c4g_2c2g_a

http://aflow.org/prototype-encyclopedia/A18B10C_tP116_130_2c4g_2c2g_a

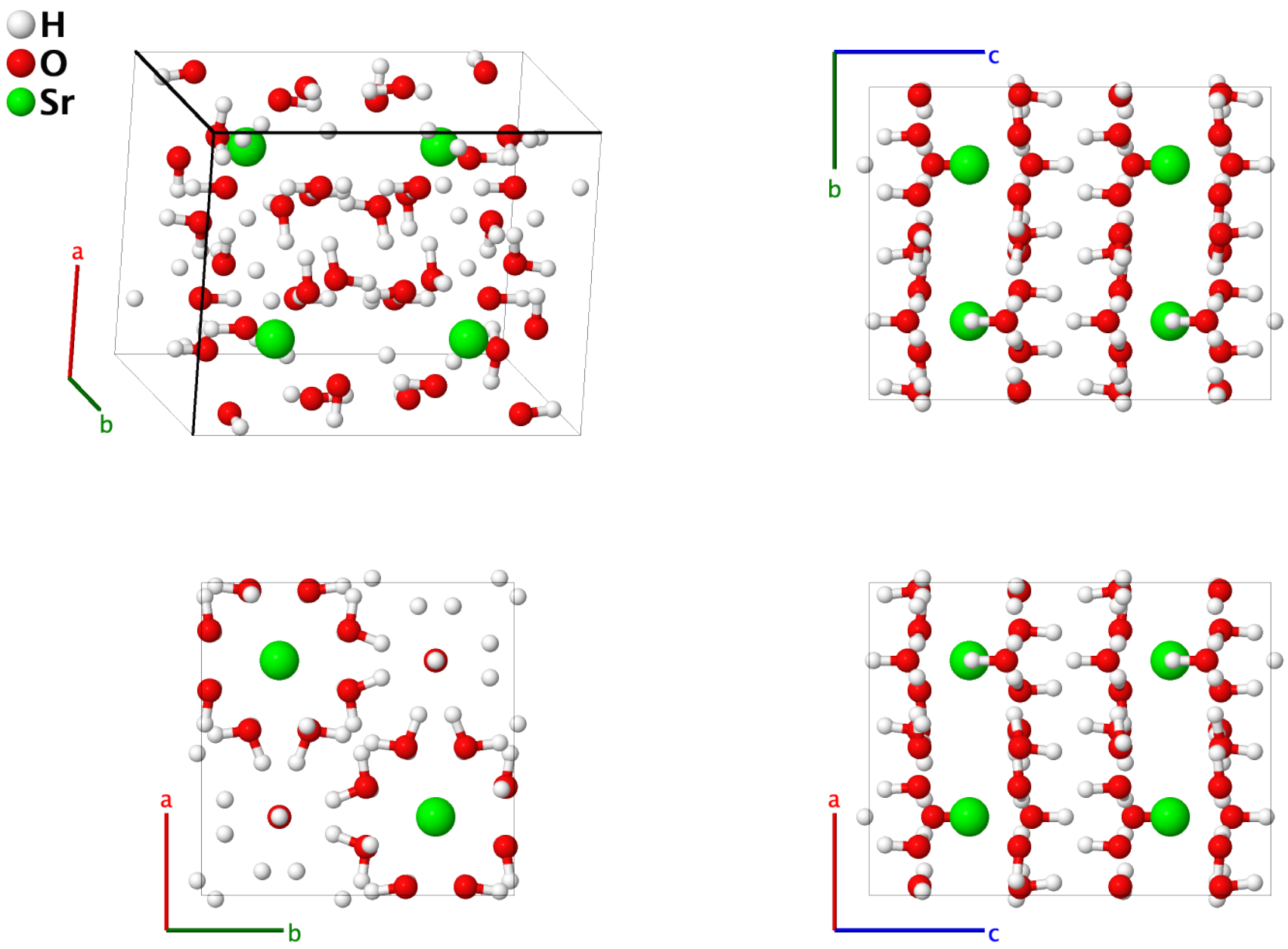

| | | |
|---|---|---|
| **Prototype** | : | $H_{18}O_{10}Sr$ |
| **AFLOW prototype label** | : | A18B10C_tP116_130_2c4g_2c2g_a |
| ***Strukturbericht* designation** | : | None |
| **Pearson symbol** | : | tP116 |
| **Space group number** | : | 130 |
| **Space group symbol** | : | $P4/ncc$ |
| **AFLOW prototype command** | : | `aflow --proto=A18B10C_tP116_130_2c4g_2c2g_a --params=`$a, c/a, z_2, z_3, z_4, z_5, x_6, y_6, z_6, x_7, y_7, z_7, x_8, y_8, z_8, x_9, y_9, z_9, x_{10}, y_{10}, z_{10}, x_{11}, y_{11}, z_{11}$ |

- This determination of the crystal structure of $Sr(OH)_2(H_2O)_8$ improves upon the structure found by (Natta, 1928), which was given the *Strukturbericht* designation $E6_1$ by (Hermann, 1937). The new structure quadruples the size of the unit cell, locates the hydrogen atoms, and changes the space group.
- We use the data taken by (Ricci, 2005) at 20 K.

**Simple Tetragonal primitive vectors:**

$$\mathbf{a}_1 = a\,\hat{\mathbf{x}}$$
$$\mathbf{a}_2 = a\,\hat{\mathbf{y}}$$
$$\mathbf{a}_3 = c\,\hat{\mathbf{z}}$$

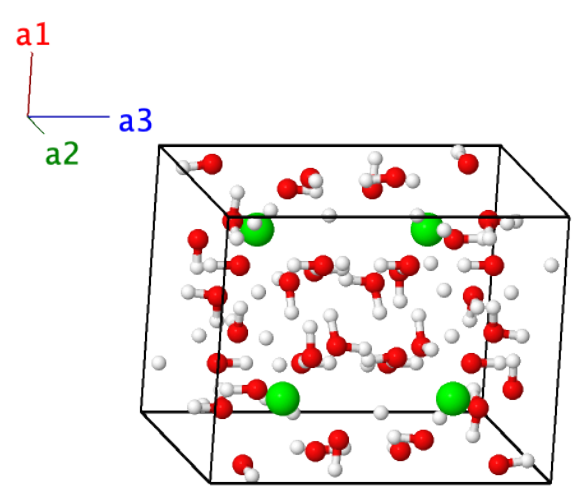

**Basis vectors:**

| | | Lattice Coordinates | | Cartesian Coordinates | Wyckoff Position | Atom Type |
|---|---|---|---|---|---|---|
| $\mathbf{B}_{1}$ | = | $\frac{3}{4}\,\mathbf{a}_{1}+\frac{1}{4}\,\mathbf{a}_{2}+\frac{1}{4}\,\mathbf{a}_{3}$ | = | $\frac{3}{4}a\,\mathbf{\hat{x}}+\frac{1}{4}a\,\mathbf{\hat{y}}+\frac{1}{4}c\,\mathbf{\hat{z}}$ | (4$a$) | Sr |
| $\mathbf{B}_{2}$ | = | $\frac{1}{4}\,\mathbf{a}_{1}+\frac{3}{4}\,\mathbf{a}_{2}+\frac{1}{4}\,\mathbf{a}_{3}$ | = | $\frac{1}{4}a\,\mathbf{\hat{x}}+\frac{3}{4}a\,\mathbf{\hat{y}}+\frac{1}{4}c\,\mathbf{\hat{z}}$ | (4$a$) | Sr |
| $\mathbf{B}_{3}$ | = | $\frac{1}{4}\,\mathbf{a}_{1}+\frac{3}{4}\,\mathbf{a}_{2}+\frac{3}{4}\,\mathbf{a}_{3}$ | = | $\frac{1}{4}a\,\mathbf{\hat{x}}+\frac{3}{4}a\,\mathbf{\hat{y}}+\frac{3}{4}c\,\mathbf{\hat{z}}$ | (4$a$) | Sr |
| $\mathbf{B}_{4}$ | = | $\frac{3}{4}\,\mathbf{a}_{1}+\frac{1}{4}\,\mathbf{a}_{2}+\frac{3}{4}\,\mathbf{a}_{3}$ | = | $\frac{3}{4}a\,\mathbf{\hat{x}}+\frac{1}{4}a\,\mathbf{\hat{y}}+\frac{3}{4}c\,\mathbf{\hat{z}}$ | (4$a$) | Sr |
| $\mathbf{B}_{5}$ | = | $\frac{1}{4}\,\mathbf{a}_{1}+\frac{1}{4}\,\mathbf{a}_{2}+z_{2}\,\mathbf{a}_{3}$ | = | $\frac{1}{4}a\,\mathbf{\hat{x}}+\frac{1}{4}a\,\mathbf{\hat{y}}+z_{2}c\,\mathbf{\hat{z}}$ | (4$c$) | H I |
| $\mathbf{B}_{6}$ | = | $\frac{3}{4}\,\mathbf{a}_{1}+\frac{3}{4}\,\mathbf{a}_{2}+\left(\frac{1}{2}-z_{2}\right)\,\mathbf{a}_{3}$ | = | $\frac{3}{4}a\,\mathbf{\hat{x}}+\frac{3}{4}a\,\mathbf{\hat{y}}+\left(\frac{1}{2}-z_{2}\right)c\,\mathbf{\hat{z}}$ | (4$c$) | H I |
| $\mathbf{B}_{7}$ | = | $\frac{3}{4}\,\mathbf{a}_{1}+\frac{3}{4}\,\mathbf{a}_{2}-z_{2}\,\mathbf{a}_{3}$ | = | $\frac{3}{4}a\,\mathbf{\hat{x}}+\frac{3}{4}a\,\mathbf{\hat{y}}-z_{2}c\,\mathbf{\hat{z}}$ | (4$c$) | H I |
| $\mathbf{B}_{8}$ | = | $\frac{1}{4}\,\mathbf{a}_{1}+\frac{1}{4}\,\mathbf{a}_{2}+\left(\frac{1}{2}+z_{2}\right)\,\mathbf{a}_{3}$ | = | $\frac{1}{4}a\,\mathbf{\hat{x}}+\frac{1}{4}a\,\mathbf{\hat{y}}+\left(\frac{1}{2}+z_{2}\right)c\,\mathbf{\hat{z}}$ | (4$c$) | H I |
| $\mathbf{B}_{9}$ | = | $\frac{1}{4}\,\mathbf{a}_{1}+\frac{1}{4}\,\mathbf{a}_{2}+z_{3}\,\mathbf{a}_{3}$ | = | $\frac{1}{4}a\,\mathbf{\hat{x}}+\frac{1}{4}a\,\mathbf{\hat{y}}+z_{3}c\,\mathbf{\hat{z}}$ | (4$c$) | H II |
| $\mathbf{B}_{10}$ | = | $\frac{3}{4}\,\mathbf{a}_{1}+\frac{3}{4}\,\mathbf{a}_{2}+\left(\frac{1}{2}-z_{3}\right)\,\mathbf{a}_{3}$ | = | $\frac{3}{4}a\,\mathbf{\hat{x}}+\frac{3}{4}a\,\mathbf{\hat{y}}+\left(\frac{1}{2}-z_{3}\right)c\,\mathbf{\hat{z}}$ | (4$c$) | H II |
| $\mathbf{B}_{11}$ | = | $\frac{3}{4}\,\mathbf{a}_{1}+\frac{3}{4}\,\mathbf{a}_{2}-z_{3}\,\mathbf{a}_{3}$ | = | $\frac{3}{4}a\,\mathbf{\hat{x}}+\frac{3}{4}a\,\mathbf{\hat{y}}-z_{3}c\,\mathbf{\hat{z}}$ | (4$c$) | H II |
| $\mathbf{B}_{12}$ | = | $\frac{1}{4}\,\mathbf{a}_{1}+\frac{1}{4}\,\mathbf{a}_{2}+\left(\frac{1}{2}+z_{3}\right)\,\mathbf{a}_{3}$ | = | $\frac{1}{4}a\,\mathbf{\hat{x}}+\frac{1}{4}a\,\mathbf{\hat{y}}+\left(\frac{1}{2}+z_{3}\right)c\,\mathbf{\hat{z}}$ | (4$c$) | H II |
| $\mathbf{B}_{13}$ | = | $\frac{1}{4}\,\mathbf{a}_{1}+\frac{1}{4}\,\mathbf{a}_{2}+z_{4}\,\mathbf{a}_{3}$ | = | $\frac{1}{4}a\,\mathbf{\hat{x}}+\frac{1}{4}a\,\mathbf{\hat{y}}+z_{4}c\,\mathbf{\hat{z}}$ | (4$c$) | O I |
| $\mathbf{B}_{14}$ | = | $\frac{3}{4}\,\mathbf{a}_{1}+\frac{3}{4}\,\mathbf{a}_{2}+\left(\frac{1}{2}-z_{4}\right)\,\mathbf{a}_{3}$ | = | $\frac{3}{4}a\,\mathbf{\hat{x}}+\frac{3}{4}a\,\mathbf{\hat{y}}+\left(\frac{1}{2}-z_{4}\right)c\,\mathbf{\hat{z}}$ | (4$c$) | O I |
| $\mathbf{B}_{15}$ | = | $\frac{3}{4}\,\mathbf{a}_{1}+\frac{3}{4}\,\mathbf{a}_{2}-z_{4}\,\mathbf{a}_{3}$ | = | $\frac{3}{4}a\,\mathbf{\hat{x}}+\frac{3}{4}a\,\mathbf{\hat{y}}-z_{4}c\,\mathbf{\hat{z}}$ | (4$c$) | O I |
| $\mathbf{B}_{16}$ | = | $\frac{1}{4}\,\mathbf{a}_{1}+\frac{1}{4}\,\mathbf{a}_{2}+\left(\frac{1}{2}+z_{4}\right)\,\mathbf{a}_{3}$ | = | $\frac{1}{4}a\,\mathbf{\hat{x}}+\frac{1}{4}a\,\mathbf{\hat{y}}+\left(\frac{1}{2}+z_{4}\right)c\,\mathbf{\hat{z}}$ | (4$c$) | O I |
| $\mathbf{B}_{17}$ | = | $\frac{1}{4}\,\mathbf{a}_{1}+\frac{1}{4}\,\mathbf{a}_{2}+z_{5}\,\mathbf{a}_{3}$ | = | $\frac{1}{4}a\,\mathbf{\hat{x}}+\frac{1}{4}a\,\mathbf{\hat{y}}+z_{5}c\,\mathbf{\hat{z}}$ | (4$c$) | O II |
| $\mathbf{B}_{18}$ | = | $\frac{3}{4}\,\mathbf{a}_{1}+\frac{3}{4}\,\mathbf{a}_{2}+\left(\frac{1}{2}-z_{5}\right)\,\mathbf{a}_{3}$ | = | $\frac{3}{4}a\,\mathbf{\hat{x}}+\frac{3}{4}a\,\mathbf{\hat{y}}+\left(\frac{1}{2}-z_{5}\right)c\,\mathbf{\hat{z}}$ | (4$c$) | O II |
| $\mathbf{B}_{19}$ | = | $\frac{3}{4}\,\mathbf{a}_{1}+\frac{3}{4}\,\mathbf{a}_{2}-z_{5}\,\mathbf{a}_{3}$ | = | $\frac{3}{4}a\,\mathbf{\hat{x}}+\frac{3}{4}a\,\mathbf{\hat{y}}-z_{5}c\,\mathbf{\hat{z}}$ | (4$c$) | O II |
| $\mathbf{B}_{20}$ | = | $\frac{1}{4}\,\mathbf{a}_{1}+\frac{1}{4}\,\mathbf{a}_{2}+\left(\frac{1}{2}+z_{5}\right)\,\mathbf{a}_{3}$ | = | $\frac{1}{4}a\,\mathbf{\hat{x}}+\frac{1}{4}a\,\mathbf{\hat{y}}+\left(\frac{1}{2}+z_{5}\right)c\,\mathbf{\hat{z}}$ | (4$c$) | O II |
| $\mathbf{B}_{21}$ | = | $x_{6}\,\mathbf{a}_{1}+y_{6}\,\mathbf{a}_{2}+z_{6}\,\mathbf{a}_{3}$ | = | $x_{6}a\,\mathbf{\hat{x}}+y_{6}a\,\mathbf{\hat{y}}+z_{6}c\,\mathbf{\hat{z}}$ | (16$g$) | H III |
| $\mathbf{B}_{22}$ | = | $\left(\frac{1}{2}-x_{6}\right)\,\mathbf{a}_{1}+\left(\frac{1}{2}-y_{6}\right)\,\mathbf{a}_{2}+z_{6}\,\mathbf{a}_{3}$ | = | $\left(\frac{1}{2}-x_{6}\right)a\,\mathbf{\hat{x}}+\left(\frac{1}{2}-y_{6}\right)a\,\mathbf{\hat{y}}+z_{6}c\,\mathbf{\hat{z}}$ | (16$g$) | H III |
| $\mathbf{B}_{23}$ | = | $\left(\frac{1}{2}-y_{6}\right)\,\mathbf{a}_{1}+x_{6}\,\mathbf{a}_{2}+z_{6}\,\mathbf{a}_{3}$ | = | $\left(\frac{1}{2}-y_{6}\right)a\,\mathbf{\hat{x}}+x_{6}a\,\mathbf{\hat{y}}+z_{6}c\,\mathbf{\hat{z}}$ | (16$g$) | H III |
| $\mathbf{B}_{24}$ | = | $y_{6}\,\mathbf{a}_{1}+\left(\frac{1}{2}-x_{6}\right)\,\mathbf{a}_{2}+z_{6}\,\mathbf{a}_{3}$ | = | $y_{6}a\,\mathbf{\hat{x}}+\left(\frac{1}{2}-x_{6}\right)a\,\mathbf{\hat{y}}+z_{6}c\,\mathbf{\hat{z}}$ | (16$g$) | H III |
| $\mathbf{B}_{25}$ | = | $-x_{6}\,\mathbf{a}_{1}+\left(\frac{1}{2}+y_{6}\right)\,\mathbf{a}_{2}+\left(\frac{1}{2}-z_{6}\right)\,\mathbf{a}_{3}$ | = | $-x_{6}a\,\mathbf{\hat{x}}+\left(\frac{1}{2}+y_{6}\right)a\,\mathbf{\hat{y}}+\left(\frac{1}{2}-z_{6}\right)c\,\mathbf{\hat{z}}$ | (16$g$) | H III |
| $\mathbf{B}_{26}$ | = | $\left(\frac{1}{2}+x_{6}\right)\,\mathbf{a}_{1}-y_{6}\,\mathbf{a}_{2}+\left(\frac{1}{2}-z_{6}\right)\,\mathbf{a}_{3}$ | = | $\left(\frac{1}{2}+x_{6}\right)a\,\mathbf{\hat{x}}-y_{6}a\,\mathbf{\hat{y}}+\left(\frac{1}{2}-z_{6}\right)c\,\mathbf{\hat{z}}$ | (16$g$) | H III |
| $\mathbf{B}_{27}$ | = | $\left(\frac{1}{2}+y_{6}\right)\,\mathbf{a}_{1}+\left(\frac{1}{2}+x_{6}\right)\,\mathbf{a}_{2}+\left(\frac{1}{2}-z_{6}\right)\,\mathbf{a}_{3}$ | = | $\left(\frac{1}{2}+y_{6}\right)a\,\mathbf{\hat{x}}+\left(\frac{1}{2}+x_{6}\right)a\,\mathbf{\hat{y}}+\left(\frac{1}{2}-z_{6}\right)c\,\mathbf{\hat{z}}$ | (16$g$) | H III |
| $\mathbf{B}_{28}$ | = | $-y_{6}\,\mathbf{a}_{1}-x_{6}\,\mathbf{a}_{2}+\left(\frac{1}{2}-z_{6}\right)\,\mathbf{a}_{3}$ | = | $-y_{6}a\,\mathbf{\hat{x}}-x_{6}a\,\mathbf{\hat{y}}+\left(\frac{1}{2}-z_{6}\right)c\,\mathbf{\hat{z}}$ | (16$g$) | H III |
| $\mathbf{B}_{29}$ | = | $-x_{6}\,\mathbf{a}_{1}-y_{6}\,\mathbf{a}_{2}-z_{6}\,\mathbf{a}_{3}$ | = | $-x_{6}a\,\mathbf{\hat{x}}-y_{6}a\,\mathbf{\hat{y}}-z_{6}c\,\mathbf{\hat{z}}$ | (16$g$) | H III |
| $\mathbf{B}_{30}$ | = | $\left(\frac{1}{2}+x_{6}\right)\,\mathbf{a}_{1}+\left(\frac{1}{2}+y_{6}\right)\,\mathbf{a}_{2}-z_{6}\,\mathbf{a}_{3}$ | = | $\left(\frac{1}{2}+x_{6}\right)a\,\mathbf{\hat{x}}+\left(\frac{1}{2}+y_{6}\right)a\,\mathbf{\hat{y}}-z_{6}c\,\mathbf{\hat{z}}$ | (16$g$) | H III |
| $\mathbf{B}_{31}$ | = | $\left(\frac{1}{2}+y_{6}\right)\,\mathbf{a}_{1}-x_{6}\,\mathbf{a}_{2}-z_{6}\,\mathbf{a}_{3}$ | = | $\left(\frac{1}{2}+y_{6}\right)a\,\mathbf{\hat{x}}-x_{6}a\,\mathbf{\hat{y}}-z_{6}c\,\mathbf{\hat{z}}$ | (16$g$) | H III |
| $\mathbf{B}_{32}$ | = | $-y_{6}\,\mathbf{a}_{1}+\left(\frac{1}{2}+x_{6}\right)\,\mathbf{a}_{2}-z_{6}\,\mathbf{a}_{3}$ | = | $-y_{6}a\,\mathbf{\hat{x}}+\left(\frac{1}{2}+x_{6}\right)a\,\mathbf{\hat{y}}-z_{6}c\,\mathbf{\hat{z}}$ | (16$g$) | H III |
| $\mathbf{B}_{33}$ | = | $x_{6}\,\mathbf{a}_{1}+\left(\frac{1}{2}-y_{6}\right)\,\mathbf{a}_{2}+\left(\frac{1}{2}+z_{6}\right)\,\mathbf{a}_{3}$ | = | $x_{6}a\,\mathbf{\hat{x}}+\left(\frac{1}{2}-y_{6}\right)a\,\mathbf{\hat{y}}+\left(\frac{1}{2}+z_{6}\right)c\,\mathbf{\hat{z}}$ | (16$g$) | H III |
| $\mathbf{B}_{34}$ | = | $\left(\frac{1}{2}-x_{6}\right)\,\mathbf{a}_{1}+y_{6}\,\mathbf{a}_{2}+\left(\frac{1}{2}+z_{6}\right)\,\mathbf{a}_{3}$ | = | $\left(\frac{1}{2}-x_{6}\right)a\,\mathbf{\hat{x}}+y_{6}a\,\mathbf{\hat{y}}+\left(\frac{1}{2}+z_{6}\right)c\,\mathbf{\hat{z}}$ | (16$g$) | H III |
| $\mathbf{B}_{35}$ | = | $\left(\frac{1}{2} - y_6\right)\mathbf{a}_1 + \left(\frac{1}{2} - x_6\right)\mathbf{a}_2 + \left(\frac{1}{2} + z_6\right)\mathbf{a}_3$ | = | $\left(\frac{1}{2} - y_6\right)a\,\hat{\mathbf{x}} + \left(\frac{1}{2} - x_6\right)a\,\hat{\mathbf{y}} + \left(\frac{1}{2} + z_6\right)c\,\hat{\mathbf{z}}$ | (16$g$) | H III |
| $\mathbf{B}_{36}$ | = | $y_6\,\mathbf{a}_1 + x_6\,\mathbf{a}_2 + \left(\frac{1}{2} + z_6\right)\mathbf{a}_3$ | = | $y_6a\,\hat{\mathbf{x}} + x_6a\,\hat{\mathbf{y}} + \left(\frac{1}{2} + z_6\right)c\,\hat{\mathbf{z}}$ | (16$g$) | H III |
| $\mathbf{B}_{37}$ | = | $x_7\,\mathbf{a}_1 + y_7\,\mathbf{a}_2 + z_7\,\mathbf{a}_3$ | = | $x_7a\,\hat{\mathbf{x}} + y_7a\,\hat{\mathbf{y}} + z_7c\,\hat{\mathbf{z}}$ | (16$g$) | H IV |
| $\mathbf{B}_{38}$ | = | $\left(\frac{1}{2} - x_7\right)\mathbf{a}_1 + \left(\frac{1}{2} - y_7\right)\mathbf{a}_2 + z_7\,\mathbf{a}_3$ | = | $\left(\frac{1}{2} - x_7\right)a\,\hat{\mathbf{x}} + \left(\frac{1}{2} - y_7\right)a\,\hat{\mathbf{y}} + z_7c\,\hat{\mathbf{z}}$ | (16$g$) | H IV |
| $\mathbf{B}_{39}$ | = | $\left(\frac{1}{2} - y_7\right)\mathbf{a}_1 + x_7\,\mathbf{a}_2 + z_7\,\mathbf{a}_3$ | = | $\left(\frac{1}{2} - y_7\right)a\,\hat{\mathbf{x}} + x_7a\,\hat{\mathbf{y}} + z_7c\,\hat{\mathbf{z}}$ | (16$g$) | H IV |
| $\mathbf{B}_{40}$ | = | $y_7\,\mathbf{a}_1 + \left(\frac{1}{2} - x_7\right)\mathbf{a}_2 + z_7\,\mathbf{a}_3$ | = | $y_7a\,\hat{\mathbf{x}} + \left(\frac{1}{2} - x_7\right)a\,\hat{\mathbf{y}} + z_7c\,\hat{\mathbf{z}}$ | (16$g$) | H IV |
| $\mathbf{B}_{41}$ | = | $-x_7\,\mathbf{a}_1 + \left(\frac{1}{2} + y_7\right)\mathbf{a}_2 + \left(\frac{1}{2} - z_7\right)\mathbf{a}_3$ | = | $-x_7a\,\hat{\mathbf{x}} + \left(\frac{1}{2} + y_7\right)a\,\hat{\mathbf{y}} + \left(\frac{1}{2} - z_7\right)c\,\hat{\mathbf{z}}$ | (16$g$) | H IV |
| $\mathbf{B}_{42}$ | = | $\left(\frac{1}{2} + x_7\right)\mathbf{a}_1 - y_7\,\mathbf{a}_2 + \left(\frac{1}{2} - z_7\right)\mathbf{a}_3$ | = | $\left(\frac{1}{2} + x_7\right)a\,\hat{\mathbf{x}} - y_7a\,\hat{\mathbf{y}} + \left(\frac{1}{2} - z_7\right)c\,\hat{\mathbf{z}}$ | (16$g$) | H IV |
| $\mathbf{B}_{43}$ | = | $\left(\frac{1}{2} + y_7\right)\mathbf{a}_1 + \left(\frac{1}{2} + x_7\right)\mathbf{a}_2 + \left(\frac{1}{2} - z_7\right)\mathbf{a}_3$ | = | $\left(\frac{1}{2} + y_7\right)a\,\hat{\mathbf{x}} + \left(\frac{1}{2} + x_7\right)a\,\hat{\mathbf{y}} + \left(\frac{1}{2} - z_7\right)c\,\hat{\mathbf{z}}$ | (16$g$) | H IV |
| $\mathbf{B}_{44}$ | = | $-y_7\,\mathbf{a}_1 - x_7\,\mathbf{a}_2 + \left(\frac{1}{2} - z_7\right)\mathbf{a}_3$ | = | $-y_7a\,\hat{\mathbf{x}} - x_7a\,\hat{\mathbf{y}} + \left(\frac{1}{2} - z_7\right)c\,\hat{\mathbf{z}}$ | (16$g$) | H IV |
| $\mathbf{B}_{45}$ | = | $-x_7\,\mathbf{a}_1 - y_7\,\mathbf{a}_2 - z_7\,\mathbf{a}_3$ | = | $-x_7a\,\hat{\mathbf{x}} - y_7a\,\hat{\mathbf{y}} - z_7c\,\hat{\mathbf{z}}$ | (16$g$) | H IV |
| $\mathbf{B}_{46}$ | = | $\left(\frac{1}{2} + x_7\right)\mathbf{a}_1 + \left(\frac{1}{2} + y_7\right)\mathbf{a}_2 - z_7\,\mathbf{a}_3$ | = | $\left(\frac{1}{2} + x_7\right)a\,\hat{\mathbf{x}} + \left(\frac{1}{2} + y_7\right)a\,\hat{\mathbf{y}} - z_7c\,\hat{\mathbf{z}}$ | (16$g$) | H IV |
| $\mathbf{B}_{47}$ | = | $\left(\frac{1}{2} + y_7\right)\mathbf{a}_1 - x_7\,\mathbf{a}_2 - z_7\,\mathbf{a}_3$ | = | $\left(\frac{1}{2} + y_7\right)a\,\hat{\mathbf{x}} - x_7a\,\hat{\mathbf{y}} - z_7c\,\hat{\mathbf{z}}$ | (16$g$) | H IV |
| $\mathbf{B}_{48}$ | = | $-y_7\,\mathbf{a}_1 + \left(\frac{1}{2} + x_7\right)\mathbf{a}_2 - z_7\,\mathbf{a}_3$ | = | $-y_7a\,\hat{\mathbf{x}} + \left(\frac{1}{2} + x_7\right)a\,\hat{\mathbf{y}} - z_7c\,\hat{\mathbf{z}}$ | (16$g$) | H IV |
| $\mathbf{B}_{49}$ | = | $x_7\,\mathbf{a}_1 + \left(\frac{1}{2} - y_7\right)\mathbf{a}_2 + \left(\frac{1}{2} + z_7\right)\mathbf{a}_3$ | = | $x_7a\,\hat{\mathbf{x}} + \left(\frac{1}{2} - y_7\right)a\,\hat{\mathbf{y}} + \left(\frac{1}{2} + z_7\right)c\,\hat{\mathbf{z}}$ | (16$g$) | H IV |
| $\mathbf{B}_{50}$ | = | $\left(\frac{1}{2} - x_7\right)\mathbf{a}_1 + y_7\,\mathbf{a}_2 + \left(\frac{1}{2} + z_7\right)\mathbf{a}_3$ | = | $\left(\frac{1}{2} - x_7\right)a\,\hat{\mathbf{x}} + y_7a\,\hat{\mathbf{y}} + \left(\frac{1}{2} + z_7\right)c\,\hat{\mathbf{z}}$ | (16$g$) | H IV |
| $\mathbf{B}_{51}$ | = | $\left(\frac{1}{2} - y_7\right)\mathbf{a}_1 + \left(\frac{1}{2} - x_7\right)\mathbf{a}_2 + \left(\frac{1}{2} + z_7\right)\mathbf{a}_3$ | = | $\left(\frac{1}{2} - y_7\right)a\,\hat{\mathbf{x}} + \left(\frac{1}{2} - x_7\right)a\,\hat{\mathbf{y}} + \left(\frac{1}{2} + z_7\right)c\,\hat{\mathbf{z}}$ | (16$g$) | H IV |
| $\mathbf{B}_{52}$ | = | $y_7\,\mathbf{a}_1 + x_7\,\mathbf{a}_2 + \left(\frac{1}{2} + z_7\right)\mathbf{a}_3$ | = | $y_7a\,\hat{\mathbf{x}} + x_7a\,\hat{\mathbf{y}} + \left(\frac{1}{2} + z_7\right)c\,\hat{\mathbf{z}}$ | (16$g$) | H IV |
| $\mathbf{B}_{53}$ | = | $x_8\,\mathbf{a}_1 + y_8\,\mathbf{a}_2 + z_8\,\mathbf{a}_3$ | = | $x_8a\,\hat{\mathbf{x}} + y_8a\,\hat{\mathbf{y}} + z_8c\,\hat{\mathbf{z}}$ | (16$g$) | H V |
| $\mathbf{B}_{54}$ | = | $\left(\frac{1}{2} - x_8\right)\mathbf{a}_1 + \left(\frac{1}{2} - y_8\right)\mathbf{a}_2 + z_8\,\mathbf{a}_3$ | = | $\left(\frac{1}{2} - x_8\right)a\,\hat{\mathbf{x}} + \left(\frac{1}{2} - y_8\right)a\,\hat{\mathbf{y}} + z_8c\,\hat{\mathbf{z}}$ | (16$g$) | H V |
| $\mathbf{B}_{55}$ | = | $\left(\frac{1}{2} - y_8\right)\mathbf{a}_1 + x_8\,\mathbf{a}_2 + z_8\,\mathbf{a}_3$ | = | $\left(\frac{1}{2} - y_8\right)a\,\hat{\mathbf{x}} + x_8a\,\hat{\mathbf{y}} + z_8c\,\hat{\mathbf{z}}$ | (16$g$) | H V |
| $\mathbf{B}_{56}$ | = | $y_8\,\mathbf{a}_1 + \left(\frac{1}{2} - x_8\right)\mathbf{a}_2 + z_8\,\mathbf{a}_3$ | = | $y_8a\,\hat{\mathbf{x}} + \left(\frac{1}{2} - x_8\right)a\,\hat{\mathbf{y}} + z_8c\,\hat{\mathbf{z}}$ | (16$g$) | H V |
| $\mathbf{B}_{57}$ | = | $-x_8\,\mathbf{a}_1 + \left(\frac{1}{2} + y_8\right)\mathbf{a}_2 + \left(\frac{1}{2} - z_8\right)\mathbf{a}_3$ | = | $-x_8a\,\hat{\mathbf{x}} + \left(\frac{1}{2} + y_8\right)a\,\hat{\mathbf{y}} + \left(\frac{1}{2} - z_8\right)c\,\hat{\mathbf{z}}$ | (16$g$) | H V |
| $\mathbf{B}_{58}$ | = | $\left(\frac{1}{2} + x_8\right)\mathbf{a}_1 - y_8\,\mathbf{a}_2 + \left(\frac{1}{2} - z_8\right)\mathbf{a}_3$ | = | $\left(\frac{1}{2} + x_8\right)a\,\hat{\mathbf{x}} - y_8a\,\hat{\mathbf{y}} + \left(\frac{1}{2} - z_8\right)c\,\hat{\mathbf{z}}$ | (16$g$) | H V |
| $\mathbf{B}_{59}$ | = | $\left(\frac{1}{2} + y_8\right)\mathbf{a}_1 + \left(\frac{1}{2} + x_8\right)\mathbf{a}_2 + \left(\frac{1}{2} - z_8\right)\mathbf{a}_3$ | = | $\left(\frac{1}{2} + y_8\right)a\,\hat{\mathbf{x}} + \left(\frac{1}{2} + x_8\right)a\,\hat{\mathbf{y}} + \left(\frac{1}{2} - z_8\right)c\,\hat{\mathbf{z}}$ | (16$g$) | H V |
| $\mathbf{B}_{60}$ | = | $-y_8\,\mathbf{a}_1 - x_8\,\mathbf{a}_2 + \left(\frac{1}{2} - z_8\right)\mathbf{a}_3$ | = | $-y_8a\,\hat{\mathbf{x}} - x_8a\,\hat{\mathbf{y}} + \left(\frac{1}{2} - z_8\right)c\,\hat{\mathbf{z}}$ | (16$g$) | H V |
| $\mathbf{B}_{61}$ | = | $-x_8\,\mathbf{a}_1 - y_8\,\mathbf{a}_2 - z_8\,\mathbf{a}_3$ | = | $-x_8a\,\hat{\mathbf{x}} - y_8a\,\hat{\mathbf{y}} - z_8c\,\hat{\mathbf{z}}$ | (16$g$) | H V |
| $\mathbf{B}_{62}$ | = | $\left(\frac{1}{2} + x_8\right)\mathbf{a}_1 + \left(\frac{1}{2} + y_8\right)\mathbf{a}_2 - z_8\,\mathbf{a}_3$ | = | $\left(\frac{1}{2} + x_8\right)a\,\hat{\mathbf{x}} + \left(\frac{1}{2} + y_8\right)a\,\hat{\mathbf{y}} - z_8c\,\hat{\mathbf{z}}$ | (16$g$) | H V |
| $\mathbf{B}_{63}$ | = | $\left(\frac{1}{2} + y_8\right)\mathbf{a}_1 - x_8\,\mathbf{a}_2 - z_8\,\mathbf{a}_3$ | = | $\left(\frac{1}{2} + y_8\right)a\,\hat{\mathbf{x}} - x_8a\,\hat{\mathbf{y}} - z_8c\,\hat{\mathbf{z}}$ | (16$g$) | H V |
| $\mathbf{B}_{64}$ | = | $-y_8\,\mathbf{a}_1 + \left(\frac{1}{2} + x_8\right)\mathbf{a}_2 - z_8\,\mathbf{a}_3$ | = | $-y_8a\,\hat{\mathbf{x}} + \left(\frac{1}{2} + x_8\right)a\,\hat{\mathbf{y}} - z_8c\,\hat{\mathbf{z}}$ | (16$g$) | H V |
| $\mathbf{B}_{65}$ | = | $x_8\,\mathbf{a}_1 + \left(\frac{1}{2} - y_8\right)\mathbf{a}_2 + \left(\frac{1}{2} + z_8\right)\mathbf{a}_3$ | = | $x_8a\,\hat{\mathbf{x}} + \left(\frac{1}{2} - y_8\right)a\,\hat{\mathbf{y}} + \left(\frac{1}{2} + z_8\right)c\,\hat{\mathbf{z}}$ | (16$g$) | H V |
| $\mathbf{B}_{66}$ | = | $\left(\frac{1}{2} - x_8\right)\mathbf{a}_1 + y_8\,\mathbf{a}_2 + \left(\frac{1}{2} + z_8\right)\mathbf{a}_3$ | = | $\left(\frac{1}{2} - x_8\right)a\,\hat{\mathbf{x}} + y_8a\,\hat{\mathbf{y}} + \left(\frac{1}{2} + z_8\right)c\,\hat{\mathbf{z}}$ | (16$g$) | H V |
| $\mathbf{B}_{67}$ | = | $\left(\frac{1}{2}-y_8\right)\mathbf{a}_1+\left(\frac{1}{2}-x_8\right)\mathbf{a}_2+\left(\frac{1}{2}+z_8\right)\mathbf{a}_3$ | = | $\left(\frac{1}{2}-y_8\right)a\,\hat{\mathbf{x}}+\left(\frac{1}{2}-x_8\right)a\,\hat{\mathbf{y}}+\left(\frac{1}{2}+z_8\right)c\,\hat{\mathbf{z}}$ | (16$g$) | H V |
| $\mathbf{B}_{68}$ | = | $y_8\,\mathbf{a}_1+x_8\,\mathbf{a}_2+\left(\frac{1}{2}+z_8\right)\mathbf{a}_3$ | = | $y_8a\,\hat{\mathbf{x}}+x_8a\,\hat{\mathbf{y}}+\left(\frac{1}{2}+z_8\right)c\,\hat{\mathbf{z}}$ | (16$g$) | H V |
| $\mathbf{B}_{69}$ | = | $x_9\,\mathbf{a}_1+y_9\,\mathbf{a}_2+z_9\,\mathbf{a}_3$ | = | $x_9a\,\hat{\mathbf{x}}+y_9a\,\hat{\mathbf{y}}+z_9c\,\hat{\mathbf{z}}$ | (16$g$) | H VI |
| $\mathbf{B}_{70}$ | = | $\left(\frac{1}{2}-x_9\right)\mathbf{a}_1+\left(\frac{1}{2}-y_9\right)\mathbf{a}_2+z_9\,\mathbf{a}_3$ | = | $\left(\frac{1}{2}-x_9\right)a\,\hat{\mathbf{x}}+\left(\frac{1}{2}-y_9\right)a\,\hat{\mathbf{y}}+z_9c\,\hat{\mathbf{z}}$ | (16$g$) | H VI |
| $\mathbf{B}_{71}$ | = | $\left(\frac{1}{2}-y_9\right)\mathbf{a}_1+x_9\,\mathbf{a}_2+z_9\,\mathbf{a}_3$ | = | $\left(\frac{1}{2}-y_9\right)a\,\hat{\mathbf{x}}+x_9a\,\hat{\mathbf{y}}+z_9c\,\hat{\mathbf{z}}$ | (16$g$) | H VI |
| $\mathbf{B}_{72}$ | = | $y_9\,\mathbf{a}_1+\left(\frac{1}{2}-x_9\right)\mathbf{a}_2+z_9\,\mathbf{a}_3$ | = | $y_9a\,\hat{\mathbf{x}}+\left(\frac{1}{2}-x_9\right)a\,\hat{\mathbf{y}}+z_9c\,\hat{\mathbf{z}}$ | (16$g$) | H VI |
| $\mathbf{B}_{73}$ | = | $-x_9\,\mathbf{a}_1+\left(\frac{1}{2}+y_9\right)\mathbf{a}_2+\left(\frac{1}{2}-z_9\right)\mathbf{a}_3$ | = | $-x_9a\,\hat{\mathbf{x}}+\left(\frac{1}{2}+y_9\right)a\,\hat{\mathbf{y}}+\left(\frac{1}{2}-z_9\right)c\,\hat{\mathbf{z}}$ | (16$g$) | H VI |
| $\mathbf{B}_{74}$ | = | $\left(\frac{1}{2}+x_9\right)\mathbf{a}_1-y_9\,\mathbf{a}_2+\left(\frac{1}{2}-z_9\right)\mathbf{a}_3$ | = | $\left(\frac{1}{2}+x_9\right)a\,\hat{\mathbf{x}}-y_9a\,\hat{\mathbf{y}}+\left(\frac{1}{2}-z_9\right)c\,\hat{\mathbf{z}}$ | (16$g$) | H VI |
| $\mathbf{B}_{75}$ | = | $\left(\frac{1}{2}+y_9\right)\mathbf{a}_1+\left(\frac{1}{2}+x_9\right)\mathbf{a}_2+\left(\frac{1}{2}-z_9\right)\mathbf{a}_3$ | = | $\left(\frac{1}{2}+y_9\right)a\,\hat{\mathbf{x}}+\left(\frac{1}{2}+x_9\right)a\,\hat{\mathbf{y}}+\left(\frac{1}{2}-z_9\right)c\,\hat{\mathbf{z}}$ | (16$g$) | H VI |
| $\mathbf{B}_{76}$ | = | $-y_9\,\mathbf{a}_1-x_9\,\mathbf{a}_2+\left(\frac{1}{2}-z_9\right)\mathbf{a}_3$ | = | $-y_9a\,\hat{\mathbf{x}}-x_9a\,\hat{\mathbf{y}}+\left(\frac{1}{2}-z_9\right)c\,\hat{\mathbf{z}}$ | (16$g$) | H VI |
| $\mathbf{B}_{77}$ | = | $-x_9\,\mathbf{a}_1-y_9\,\mathbf{a}_2-z_9\,\mathbf{a}_3$ | = | $-x_9a\,\hat{\mathbf{x}}-y_9a\,\hat{\mathbf{y}}-z_9c\,\hat{\mathbf{z}}$ | (16$g$) | H VI |
| $\mathbf{B}_{78}$ | = | $\left(\frac{1}{2}+x_9\right)\mathbf{a}_1+\left(\frac{1}{2}+y_9\right)\mathbf{a}_2-z_9\,\mathbf{a}_3$ | = | $\left(\frac{1}{2}+x_9\right)a\,\hat{\mathbf{x}}+\left(\frac{1}{2}+y_9\right)a\,\hat{\mathbf{y}}-z_9c\,\hat{\mathbf{z}}$ | (16$g$) | H VI |
| $\mathbf{B}_{79}$ | = | $\left(\frac{1}{2}+y_9\right)\mathbf{a}_1-x_9\,\mathbf{a}_2-z_9\,\mathbf{a}_3$ | = | $\left(\frac{1}{2}+y_9\right)a\,\hat{\mathbf{x}}-x_9a\,\hat{\mathbf{y}}-z_9c\,\hat{\mathbf{z}}$ | (16$g$) | H VI |
| $\mathbf{B}_{80}$ | = | $-y_9\,\mathbf{a}_1+\left(\frac{1}{2}+x_9\right)\mathbf{a}_2-z_9\,\mathbf{a}_3$ | = | $-y_9a\,\hat{\mathbf{x}}+\left(\frac{1}{2}+x_9\right)a\,\hat{\mathbf{y}}-z_9c\,\hat{\mathbf{z}}$ | (16$g$) | H VI |
| $\mathbf{B}_{81}$ | = | $x_9\,\mathbf{a}_1+\left(\frac{1}{2}-y_9\right)\mathbf{a}_2+\left(\frac{1}{2}+z_9\right)\mathbf{a}_3$ | = | $x_9a\,\hat{\mathbf{x}}+\left(\frac{1}{2}-y_9\right)a\,\hat{\mathbf{y}}+\left(\frac{1}{2}+z_9\right)c\,\hat{\mathbf{z}}$ | (16$g$) | H VI |
| $\mathbf{B}_{82}$ | = | $\left(\frac{1}{2}-x_9\right)\mathbf{a}_1+y_9\,\mathbf{a}_2+\left(\frac{1}{2}+z_9\right)\mathbf{a}_3$ | = | $\left(\frac{1}{2}-x_9\right)a\,\hat{\mathbf{x}}+y_9a\,\hat{\mathbf{y}}+\left(\frac{1}{2}+z_9\right)c\,\hat{\mathbf{z}}$ | (16$g$) | H VI |
| $\mathbf{B}_{83}$ | = | $\left(\frac{1}{2}-y_9\right)\mathbf{a}_1+\left(\frac{1}{2}-x_9\right)\mathbf{a}_2+\left(\frac{1}{2}+z_9\right)\mathbf{a}_3$ | = | $\left(\frac{1}{2}-y_9\right)a\,\hat{\mathbf{x}}+\left(\frac{1}{2}-x_9\right)a\,\hat{\mathbf{y}}+\left(\frac{1}{2}+z_9\right)c\,\hat{\mathbf{z}}$ | (16$g$) | H VI |
| $\mathbf{B}_{84}$ | = | $y_9\,\mathbf{a}_1+x_9\,\mathbf{a}_2+\left(\frac{1}{2}+z_9\right)\mathbf{a}_3$ | = | $y_9a\,\hat{\mathbf{x}}+x_9a\,\hat{\mathbf{y}}+\left(\frac{1}{2}+z_9\right)c\,\hat{\mathbf{z}}$ | (16$g$) | H VI |
| $\mathbf{B}_{85}$ | = | $x_{10}\,\mathbf{a}_1+y_{10}\,\mathbf{a}_2+z_{10}\,\mathbf{a}_3$ | = | $x_{10}a\,\hat{\mathbf{x}}+y_{10}a\,\hat{\mathbf{y}}+z_{10}c\,\hat{\mathbf{z}}$ | (16$g$) | O III |
| $\mathbf{B}_{86}$ | = | $\left(\frac{1}{2}-x_{10}\right)\mathbf{a}_1+\left(\frac{1}{2}-y_{10}\right)\mathbf{a}_2+z_{10}\,\mathbf{a}_3$ | = | $\left(\frac{1}{2}-x_{10}\right)a\,\hat{\mathbf{x}}+\left(\frac{1}{2}-y_{10}\right)a\,\hat{\mathbf{y}}+z_{10}c\,\hat{\mathbf{z}}$ | (16$g$) | O III |
| $\mathbf{B}_{87}$ | = | $\left(\frac{1}{2}-y_{10}\right)\mathbf{a}_1+x_{10}\,\mathbf{a}_2+z_{10}\,\mathbf{a}_3$ | = | $\left(\frac{1}{2}-y_{10}\right)a\,\hat{\mathbf{x}}+x_{10}a\,\hat{\mathbf{y}}+z_{10}c\,\hat{\mathbf{z}}$ | (16$g$) | O III |
| $\mathbf{B}_{88}$ | = | $y_{10}\,\mathbf{a}_1+\left(\frac{1}{2}-x_{10}\right)\mathbf{a}_2+z_{10}\,\mathbf{a}_3$ | = | $y_{10}a\,\hat{\mathbf{x}}+\left(\frac{1}{2}-x_{10}\right)a\,\hat{\mathbf{y}}+z_{10}c\,\hat{\mathbf{z}}$ | (16$g$) | O III |
| $\mathbf{B}_{89}$ | = | $-x_{10}\,\mathbf{a}_1+\left(\frac{1}{2}+y_{10}\right)\mathbf{a}_2+\left(\frac{1}{2}-z_{10}\right)\mathbf{a}_3$ | = | $-x_{10}a\,\hat{\mathbf{x}}+\left(\frac{1}{2}+y_{10}\right)a\,\hat{\mathbf{y}}+\left(\frac{1}{2}-z_{10}\right)c\,\hat{\mathbf{z}}$ | (16$g$) | O III |
| $\mathbf{B}_{90}$ | = | $\left(\frac{1}{2}+x_{10}\right)\mathbf{a}_1-y_{10}\,\mathbf{a}_2+\left(\frac{1}{2}-z_{10}\right)\mathbf{a}_3$ | = | $\left(\frac{1}{2}+x_{10}\right)a\,\hat{\mathbf{x}}-y_{10}a\,\hat{\mathbf{y}}+\left(\frac{1}{2}-z_{10}\right)c\,\hat{\mathbf{z}}$ | (16$g$) | O III |
| $\mathbf{B}_{91}$ | = | $\left(\frac{1}{2}+y_{10}\right)\mathbf{a}_1+\left(\frac{1}{2}+x_{10}\right)\mathbf{a}_2+\left(\frac{1}{2}-z_{10}\right)\mathbf{a}_3$ | = | $\left(\frac{1}{2}+y_{10}\right)a\,\hat{\mathbf{x}}+\left(\frac{1}{2}+x_{10}\right)a\,\hat{\mathbf{y}}+\left(\frac{1}{2}-z_{10}\right)c\,\hat{\mathbf{z}}$ | (16$g$) | O III |
| $\mathbf{B}_{92}$ | = | $-y_{10}\,\mathbf{a}_1-x_{10}\,\mathbf{a}_2+\left(\frac{1}{2}-z_{10}\right)\mathbf{a}_3$ | = | $-y_{10}a\,\hat{\mathbf{x}}-x_{10}a\,\hat{\mathbf{y}}+\left(\frac{1}{2}-z_{10}\right)c\,\hat{\mathbf{z}}$ | (16$g$) | O III |
| $\mathbf{B}_{93}$ | = | $-x_{10}\,\mathbf{a}_1-y_{10}\,\mathbf{a}_2-z_{10}\,\mathbf{a}_3$ | = | $-x_{10}a\,\hat{\mathbf{x}}-y_{10}a\,\hat{\mathbf{y}}-z_{10}c\,\hat{\mathbf{z}}$ | (16$g$) | O III |
| $\mathbf{B}_{94}$ | = | $\left(\frac{1}{2}+x_{10}\right)\mathbf{a}_1+\left(\frac{1}{2}+y_{10}\right)\mathbf{a}_2-z_{10}\,\mathbf{a}_3$ | = | $\left(\frac{1}{2}+x_{10}\right)a\,\hat{\mathbf{x}}+\left(\frac{1}{2}+y_{10}\right)a\,\hat{\mathbf{y}}-z_{10}c\,\hat{\mathbf{z}}$ | (16$g$) | O III |
| $\mathbf{B}_{95}$ | = | $\left(\frac{1}{2}+y_{10}\right)\mathbf{a}_1-x_{10}\,\mathbf{a}_2-z_{10}\,\mathbf{a}_3$ | = | $\left(\frac{1}{2}+y_{10}\right)a\,\hat{\mathbf{x}}-x_{10}a\,\hat{\mathbf{y}}-z_{10}c\,\hat{\mathbf{z}}$ | (16$g$) | O III |
| $\mathbf{B}_{96}$ | = | $-y_{10}\,\mathbf{a}_1+\left(\frac{1}{2}+x_{10}\right)\mathbf{a}_2-z_{10}\,\mathbf{a}_3$ | = | $-y_{10}a\,\hat{\mathbf{x}}+\left(\frac{1}{2}+x_{10}\right)a\,\hat{\mathbf{y}}-z_{10}c\,\hat{\mathbf{z}}$ | (16$g$) | O III |
| $\mathbf{B}_{97}$ | = | $x_{10}\,\mathbf{a}_1+\left(\frac{1}{2}-y_{10}\right)\mathbf{a}_2+\left(\frac{1}{2}+z_{10}\right)\mathbf{a}_3$ | = | $x_{10}a\,\hat{\mathbf{x}}+\left(\frac{1}{2}-y_{10}\right)a\,\hat{\mathbf{y}}+\left(\frac{1}{2}+z_{10}\right)c\,\hat{\mathbf{z}}$ | (16$g$) | O III |
| $\mathbf{B}_{98}$ | = | $\left(\frac{1}{2}-x_{10}\right)\mathbf{a}_1+y_{10}\,\mathbf{a}_2+\left(\frac{1}{2}+z_{10}\right)\mathbf{a}_3$ | = | $\left(\frac{1}{2}-x_{10}\right)a\,\hat{\mathbf{x}}+y_{10}a\,\hat{\mathbf{y}}+\left(\frac{1}{2}+z_{10}\right)c\,\hat{\mathbf{z}}$ | (16$g$) | O III |
| $\mathbf{B}_{99}$ | = | $\left(\frac{1}{2}-y_{10}\right)\mathbf{a}_1+\left(\frac{1}{2}-x_{10}\right)\mathbf{a}_2+\left(\frac{1}{2}+z_{10}\right)\mathbf{a}_3$ | = | $\left(\frac{1}{2}-y_{10}\right)a\,\hat{\mathbf{x}}+\left(\frac{1}{2}-x_{10}\right)a\,\hat{\mathbf{y}}+\left(\frac{1}{2}+z_{10}\right)c\,\hat{\mathbf{z}}$ | (16$g$) | O III |
| $\mathbf{B}_{100}$ | = | $y_{10}\,\mathbf{a}_1+x_{10}\,\mathbf{a}_2+\left(\frac{1}{2}+z_{10}\right)\mathbf{a}_3$ | = | $y_{10}a\,\hat{\mathbf{x}}+x_{10}a\,\hat{\mathbf{y}}+\left(\frac{1}{2}+z_{10}\right)c\,\hat{\mathbf{z}}$ | (16$g$) | O III |
| $\mathbf{B}_{101}$ | = | $x_{11}\,\mathbf{a}_1+y_{11}\,\mathbf{a}_2+z_{11}\,\mathbf{a}_3$ | = | $x_{11}a\,\hat{\mathbf{x}}+y_{11}a\,\hat{\mathbf{y}}+z_{11}c\,\hat{\mathbf{z}}$ | (16$g$) | O IV |
| $\mathbf{B}_{102}$ | = | $\left(\frac{1}{2}-x_{11}\right)\mathbf{a}_1+\left(\frac{1}{2}-y_{11}\right)\mathbf{a}_2+z_{11}\,\mathbf{a}_3$ | = | $\left(\frac{1}{2}-x_{11}\right)a\,\hat{\mathbf{x}}+\left(\frac{1}{2}-y_{11}\right)a\,\hat{\mathbf{y}}+z_{11}c\,\hat{\mathbf{z}}$ | (16$g$) | O IV |
| $\mathbf{B}_{103}$ | = | $\left(\frac{1}{2}-y_{11}\right)\mathbf{a}_1+x_{11}\,\mathbf{a}_2+z_{11}\,\mathbf{a}_3$ | = | $\left(\frac{1}{2}-y_{11}\right)a\,\hat{\mathbf{x}}+x_{11}a\,\hat{\mathbf{y}}+z_{11}c\,\hat{\mathbf{z}}$ | (16$g$) | O IV |
| $\mathbf{B}_{104}$ | = | $y_{11}\,\mathbf{a}_1+\left(\frac{1}{2}-x_{11}\right)\mathbf{a}_2+z_{11}\,\mathbf{a}_3$ | = | $y_{11}a\,\hat{\mathbf{x}}+\left(\frac{1}{2}-x_{11}\right)a\,\hat{\mathbf{y}}+z_{11}c\,\hat{\mathbf{z}}$ | (16$g$) | O IV |
| $\mathbf{B}_{105}$ | = | $-x_{11}\,\mathbf{a}_1+\left(\frac{1}{2}+y_{11}\right)\mathbf{a}_2+\left(\frac{1}{2}-z_{11}\right)\mathbf{a}_3$ | = | $-x_{11}a\,\hat{\mathbf{x}}+\left(\frac{1}{2}+y_{11}\right)a\,\hat{\mathbf{y}}+\left(\frac{1}{2}-z_{11}\right)c\,\hat{\mathbf{z}}$ | (16$g$) | O IV |
| $\mathbf{B}_{106}$ | = | $\left(\frac{1}{2}+x_{11}\right)\mathbf{a}_1-y_{11}\,\mathbf{a}_2+\left(\frac{1}{2}-z_{11}\right)\mathbf{a}_3$ | = | $\left(\frac{1}{2}+x_{11}\right)a\,\hat{\mathbf{x}}-y_{11}a\,\hat{\mathbf{y}}+\left(\frac{1}{2}-z_{11}\right)c\,\hat{\mathbf{z}}$ | (16$g$) | O IV |
| $\mathbf{B}_{107}$ | = | $\left(\frac{1}{2}+y_{11}\right)\mathbf{a}_1+\left(\frac{1}{2}+x_{11}\right)\mathbf{a}_2+\left(\frac{1}{2}-z_{11}\right)\mathbf{a}_3$ | = | $\left(\frac{1}{2}+y_{11}\right)a\,\hat{\mathbf{x}}+\left(\frac{1}{2}+x_{11}\right)a\,\hat{\mathbf{y}}+\left(\frac{1}{2}-z_{11}\right)c\,\hat{\mathbf{z}}$ | (16$g$) | O IV |
| $\mathbf{B}_{108}$ | = | $-y_{11}\,\mathbf{a}_1-x_{11}\,\mathbf{a}_2+\left(\frac{1}{2}-z_{11}\right)\mathbf{a}_3$ | = | $-y_{11}a\,\hat{\mathbf{x}}-x_{11}a\,\hat{\mathbf{y}}+\left(\frac{1}{2}-z_{11}\right)c\,\hat{\mathbf{z}}$ | (16$g$) | O IV |
| $\mathbf{B}_{109}$ | = | $-x_{11}\,\mathbf{a}_1-y_{11}\,\mathbf{a}_2-z_{11}\,\mathbf{a}_3$ | = | $-x_{11}a\,\hat{\mathbf{x}}-y_{11}a\,\hat{\mathbf{y}}-z_{11}c\,\hat{\mathbf{z}}$ | (16$g$) | O IV |
| $\mathbf{B}_{110}$ | = | $\left(\frac{1}{2}+x_{11}\right)\mathbf{a}_1+\left(\frac{1}{2}+y_{11}\right)\mathbf{a}_2-z_{11}\,\mathbf{a}_3$ | = | $\left(\frac{1}{2}+x_{11}\right)a\,\hat{\mathbf{x}}+\left(\frac{1}{2}+y_{11}\right)a\,\hat{\mathbf{y}}-z_{11}c\,\hat{\mathbf{z}}$ | (16$g$) | O IV |
| $\mathbf{B}_{111}$ | = | $\left(\frac{1}{2}+y_{11}\right)\mathbf{a}_1-x_{11}\,\mathbf{a}_2-z_{11}\,\mathbf{a}_3$ | = | $\left(\frac{1}{2}+y_{11}\right)a\,\hat{\mathbf{x}}-x_{11}a\,\hat{\mathbf{y}}-z_{11}c\,\hat{\mathbf{z}}$ | (16$g$) | O IV |
| $\mathbf{B}_{112}$ | = | $-y_{11}\,\mathbf{a}_1+\left(\frac{1}{2}+x_{11}\right)\mathbf{a}_2-z_{11}\,\mathbf{a}_3$ | = | $-y_{11}a\,\hat{\mathbf{x}}+\left(\frac{1}{2}+x_{11}\right)a\,\hat{\mathbf{y}}-z_{11}c\,\hat{\mathbf{z}}$ | (16$g$) | O IV |
| $\mathbf{B}_{113}$ | = | $x_{11}\,\mathbf{a}_1+\left(\frac{1}{2}-y_{11}\right)\mathbf{a}_2+\left(\frac{1}{2}+z_{11}\right)\mathbf{a}_3$ | = | $x_{11}a\,\hat{\mathbf{x}}+\left(\frac{1}{2}-y_{11}\right)a\,\hat{\mathbf{y}}+\left(\frac{1}{2}+z_{11}\right)c\,\hat{\mathbf{z}}$ | (16$g$) | O IV |
| $\mathbf{B}_{114}$ | = | $\left(\frac{1}{2}-x_{11}\right)\mathbf{a}_1+y_{11}\,\mathbf{a}_2+\left(\frac{1}{2}+z_{11}\right)\mathbf{a}_3$ | = | $\left(\frac{1}{2}-x_{11}\right)a\,\hat{\mathbf{x}}+y_{11}a\,\hat{\mathbf{y}}+\left(\frac{1}{2}+z_{11}\right)c\,\hat{\mathbf{z}}$ | (16$g$) | O IV |
| $\mathbf{B}_{115}$ | = | $\left(\frac{1}{2}-y_{11}\right)\mathbf{a}_1+\left(\frac{1}{2}-x_{11}\right)\mathbf{a}_2+\left(\frac{1}{2}+z_{11}\right)\mathbf{a}_3$ | = | $\left(\frac{1}{2}-y_{11}\right)a\,\hat{\mathbf{x}}+\left(\frac{1}{2}-x_{11}\right)a\,\hat{\mathbf{y}}+\left(\frac{1}{2}+z_{11}\right)c\,\hat{\mathbf{z}}$ | (16$g$) | O IV |
| $\mathbf{B}_{116}$ | = | $y_{11}\,\mathbf{a}_1+x_{11}\,\mathbf{a}_2+\left(\frac{1}{2}+z_{11}\right)\mathbf{a}_3$ | = | $y_{11}a\,\hat{\mathbf{x}}+x_{11}a\,\hat{\mathbf{y}}+\left(\frac{1}{2}+z_{11}\right)c\,\hat{\mathbf{z}}$ | (16$g$) | O IV |

**Geometry files:**

- CIF: pp. 1701
- POSCAR: pp. 1701

# $\alpha$-$WO_3$ Structure: A3B_tP16_130_cf_c

http://aflow.org/prototype-encyclopedia/A3B_tP16_130_cf_c

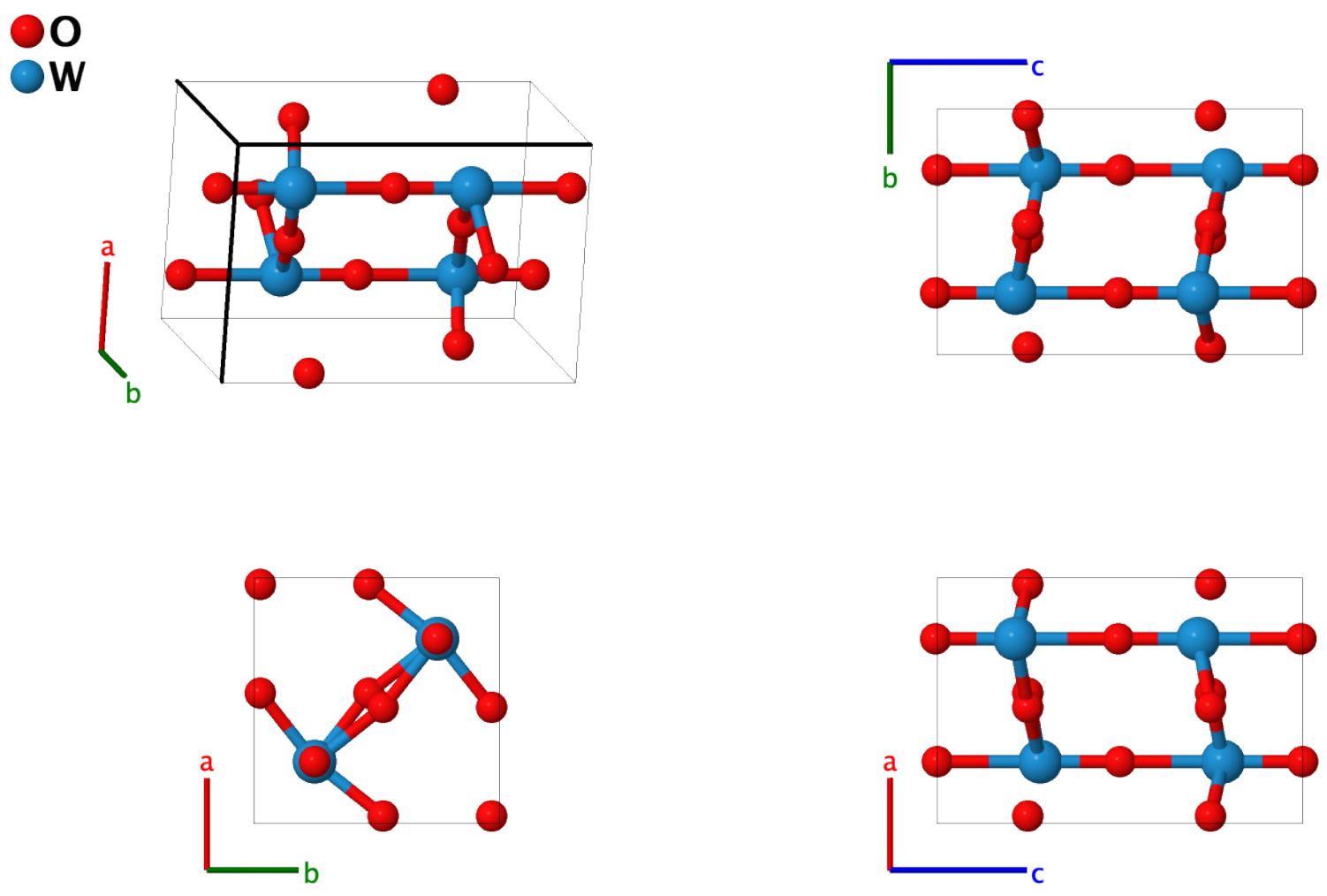

| | | |
|---|---|---|
| **Prototype** | : | $O_3W$ |
| **AFLOW prototype label** | : | A3B_tP16_130_cf_c |
| ***Strukturbericht* designation** | : | None |
| **Pearson symbol** | : | tP16 |
| **Space group number** | : | 130 |
| **Space group symbol** | : | $P4/ncc$ |
| **AFLOW prototype command** | : | `aflow --proto=A3B_tP16_130_cf_c --params=`$a, c/a, z_1, z_2, x_3$ |

- All stable phases of $WO_3$ are distortions of the cubic $\alpha$-$ReO_3$ ($D0_9$) phase. (Woodward, 1997 and Vogt, 1999) The known stable phases and their approximate temperature ranges are:
  - $\alpha$-$WO_3$ (1010-1170 K) (Vogt, 1999), this structure
  - $\beta$-$WO_3$ (600-1170 K) (Vogt, 1999)
  - $\gamma$-$WO_3$ (290-600 K) (Vogt, 1999)
  - $\delta$-$WO_3$ (230-290 K) (Diehl, 1978)
  - $\epsilon$-$WO_3$ (below 23 K) (Woodward, 1997)
- In addition, several other structures have been proposed and/or found:
  - The original $D0_{10}$ structure (Bräkken, 1931), (Hermann, 1937) superseded by $\delta$-$WO_3$
  - Original $\beta$-$WO_3$ (Salje, 1977)
  - Hexagonal $WO_3$ (Gerand, 1979) (metastable)

## Simple Tetragonal primitive vectors:

$$\mathbf{a}_1 = a\,\hat{\mathbf{x}}$$
$$\mathbf{a}_2 = a\,\hat{\mathbf{y}}$$
$$\mathbf{a}_3 = c\,\hat{\mathbf{z}}$$

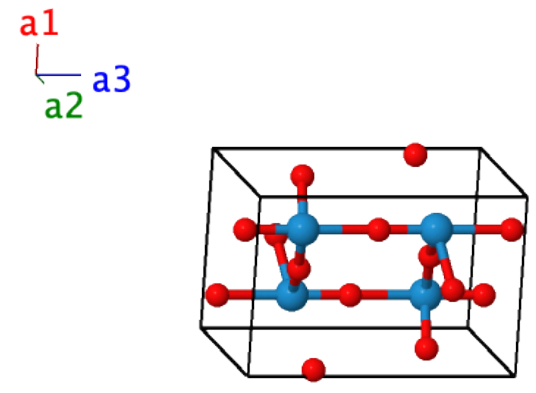

## Basis vectors:

| | | Lattice Coordinates | | Cartesian Coordinates | Wyckoff Position | Atom Type |
|---|---|---|---|---|---|---|
| $\mathbf{B}_1$ | = | $\frac{1}{4}\,\mathbf{a}_1 + \frac{1}{4}\,\mathbf{a}_2 + z_1\,\mathbf{a}_3$ | = | $\frac{1}{4}a\,\hat{\mathbf{x}} + \frac{1}{4}a\,\hat{\mathbf{y}} + z_1 c\,\hat{\mathbf{z}}$ | (4$c$) | O I |
| $\mathbf{B}_2$ | = | $\frac{3}{4}\,\mathbf{a}_1 + \frac{3}{4}\,\mathbf{a}_2 + \left(\frac{1}{2} - z_1\right)\,\mathbf{a}_3$ | = | $\frac{3}{4}a\,\hat{\mathbf{x}} + \frac{3}{4}a\,\hat{\mathbf{y}} + \left(\frac{1}{2} - z_1\right)c\,\hat{\mathbf{z}}$ | (4$c$) | O I |
| $\mathbf{B}_3$ | = | $\frac{3}{4}\,\mathbf{a}_1 + \frac{3}{4}\,\mathbf{a}_2 - z_1\,\mathbf{a}_3$ | = | $\frac{3}{4}a\,\hat{\mathbf{x}} + \frac{3}{4}a\,\hat{\mathbf{y}} - z_1 c\,\hat{\mathbf{z}}$ | (4$c$) | O I |
| $\mathbf{B}_4$ | = | $\frac{1}{4}\,\mathbf{a}_1 + \frac{1}{4}\,\mathbf{a}_2 + \left(\frac{1}{2} + z_1\right)\,\mathbf{a}_3$ | = | $\frac{1}{4}a\,\hat{\mathbf{x}} + \frac{1}{4}a\,\hat{\mathbf{y}} + \left(\frac{1}{2} + z_1\right)c\,\hat{\mathbf{z}}$ | (4$c$) | O I |
| $\mathbf{B}_5$ | = | $\frac{1}{4}\,\mathbf{a}_1 + \frac{1}{4}\,\mathbf{a}_2 + z_2\,\mathbf{a}_3$ | = | $\frac{1}{4}a\,\hat{\mathbf{x}} + \frac{1}{4}a\,\hat{\mathbf{y}} + z_2 c\,\hat{\mathbf{z}}$ | (4$c$) | W |
| $\mathbf{B}_6$ | = | $\frac{3}{4}\,\mathbf{a}_1 + \frac{3}{4}\,\mathbf{a}_2 + \left(\frac{1}{2} - z_2\right)\,\mathbf{a}_3$ | = | $\frac{3}{4}a\,\hat{\mathbf{x}} + \frac{3}{4}a\,\hat{\mathbf{y}} + \left(\frac{1}{2} - z_2\right)c\,\hat{\mathbf{z}}$ | (4$c$) | W |
| $\mathbf{B}_7$ | = | $\frac{3}{4}\,\mathbf{a}_1 + \frac{3}{4}\,\mathbf{a}_2 - z_2\,\mathbf{a}_3$ | = | $\frac{3}{4}a\,\hat{\mathbf{x}} + \frac{3}{4}a\,\hat{\mathbf{y}} - z_2 c\,\hat{\mathbf{z}}$ | (4$c$) | W |
| $\mathbf{B}_8$ | = | $\frac{1}{4}\,\mathbf{a}_1 + \frac{1}{4}\,\mathbf{a}_2 + \left(\frac{1}{2} + z_2\right)\,\mathbf{a}_3$ | = | $\frac{1}{4}a\,\hat{\mathbf{x}} + \frac{1}{4}a\,\hat{\mathbf{y}} + \left(\frac{1}{2} + z_2\right)c\,\hat{\mathbf{z}}$ | (4$c$) | W |
| $\mathbf{B}_9$ | = | $x_3\,\mathbf{a}_1 - x_3\,\mathbf{a}_2 + \frac{1}{4}\,\mathbf{a}_3$ | = | $x_3 a\,\hat{\mathbf{x}} - x_3 a\,\hat{\mathbf{y}} + \frac{1}{4}c\,\hat{\mathbf{z}}$ | (8$f$) | O II |
| $\mathbf{B}_{10}$ | = | $\left(\frac{1}{2} - x_3\right)\,\mathbf{a}_1 + \left(\frac{1}{2} + x_3\right)\,\mathbf{a}_2 + \frac{1}{4}\,\mathbf{a}_3$ | = | $\left(\frac{1}{2} - x_3\right)a\,\hat{\mathbf{x}} + \left(\frac{1}{2} + x_3\right)a\,\hat{\mathbf{y}} + \frac{1}{4}c\,\hat{\mathbf{z}}$ | (8$f$) | O II |
| $\mathbf{B}_{11}$ | = | $\left(\frac{1}{2} + x_3\right)\,\mathbf{a}_1 + x_3\,\mathbf{a}_2 + \frac{1}{4}\,\mathbf{a}_3$ | = | $\left(\frac{1}{2} + x_3\right)a\,\hat{\mathbf{x}} + x_3 a\,\hat{\mathbf{y}} + \frac{1}{4}c\,\hat{\mathbf{z}}$ | (8$f$) | O II |
| $\mathbf{B}_{12}$ | = | $-x_3\,\mathbf{a}_1 + \left(\frac{1}{2} - x_3\right)\,\mathbf{a}_2 + \frac{1}{4}\,\mathbf{a}_3$ | = | $-x_3 a\,\hat{\mathbf{x}} + \left(\frac{1}{2} - x_3\right)a\,\hat{\mathbf{y}} + \frac{1}{4}c\,\hat{\mathbf{z}}$ | (8$f$) | O II |
| $\mathbf{B}_{13}$ | = | $-x_3\,\mathbf{a}_1 + x_3\,\mathbf{a}_2 + \frac{3}{4}\,\mathbf{a}_3$ | = | $-x_3 a\,\hat{\mathbf{x}} + x_3 a\,\hat{\mathbf{y}} + \frac{3}{4}c\,\hat{\mathbf{z}}$ | (8$f$) | O II |
| $\mathbf{B}_{14}$ | = | $\left(\frac{1}{2} + x_3\right)\,\mathbf{a}_1 + \left(\frac{1}{2} - x_3\right)\,\mathbf{a}_2 + \frac{3}{4}\,\mathbf{a}_3$ | = | $\left(\frac{1}{2} + x_3\right)a\,\hat{\mathbf{x}} + \left(\frac{1}{2} - x_3\right)a\,\hat{\mathbf{y}} + \frac{3}{4}c\,\hat{\mathbf{z}}$ | (8$f$) | O II |
| $\mathbf{B}_{15}$ | = | $\left(\frac{1}{2} - x_3\right)\,\mathbf{a}_1 - x_3\,\mathbf{a}_2 + \frac{3}{4}\,\mathbf{a}_3$ | = | $\left(\frac{1}{2} - x_3\right)a\,\hat{\mathbf{x}} - x_3 a\,\hat{\mathbf{y}} + \frac{3}{4}c\,\hat{\mathbf{z}}$ | (8$f$) | O II |
| $\mathbf{B}_{16}$ | = | $x_3\,\mathbf{a}_1 + \left(\frac{1}{2} + x_3\right)\,\mathbf{a}_2 + \frac{3}{4}\,\mathbf{a}_3$ | = | $x_3 a\,\hat{\mathbf{x}} + \left(\frac{1}{2} + x_3\right)a\,\hat{\mathbf{y}} + \frac{3}{4}c\,\hat{\mathbf{z}}$ | (8$f$) | O II |

**Geometry files:**
- CIF: pp. 1702
- POSCAR: pp. 1702

# $Zr_3Al_2$ Structure: A2B3_tP20_136_j_dfg

http://aflow.org/prototype-encyclopedia/A2B3_tP20_136_j_dfg

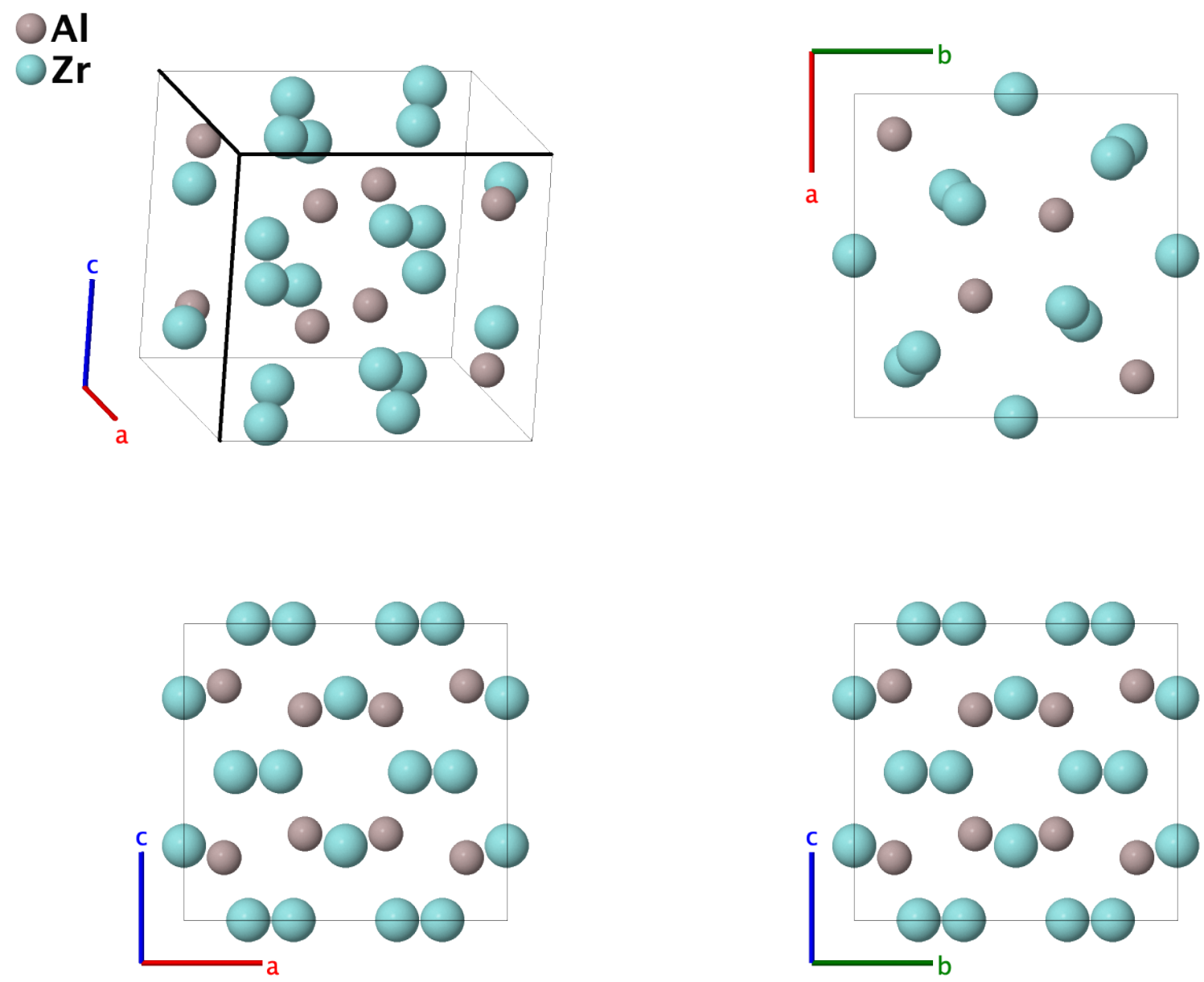

| | | |
|---|---|---|
| **Prototype** | : | $Al_2Zr_3$ |
| **AFLOW prototype label** | : | A2B3_tP20_136_j_dfg |
| ***Strukturbericht* designation** | : | None |
| **Pearson symbol** | : | tP20 |
| **Space group number** | : | 136 |
| **Space group symbol** | : | $P4_2/mnm$ |
| **AFLOW prototype command** | : | `aflow --proto=A2B3_tP20_136_j_dfg` `--params=`$a, c/a, x_2, x_3, x_4, z_4$ |

**Other compounds with this structure**

- $Hf_3Al_2$ and $Y_3Al_2$

**Simple Tetragonal primitive vectors:**

$$\mathbf{a}_1 = a\,\hat{\mathbf{x}}$$
$$\mathbf{a}_2 = a\,\hat{\mathbf{y}}$$
$$\mathbf{a}_3 = c\,\hat{\mathbf{z}}$$

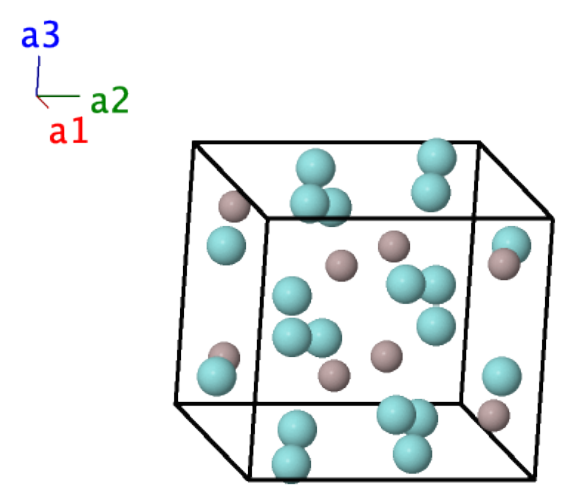

**Basis vectors:**

| | | Lattice Coordinates | | Cartesian Coordinates | Wyckoff Position | Atom Type |
|---|---|---|---|---|---|---|
| $\mathbf{B}_1$ | = | $\frac{1}{2}\,\mathbf{a}_2 + \frac{1}{4}\,\mathbf{a}_3$ | = | $\frac{1}{2}a\,\hat{\mathbf{y}} + \frac{1}{4}c\,\hat{\mathbf{z}}$ | (4$d$) | Zr I |
| $\mathbf{B}_2$ | = | $\frac{1}{2}\,\mathbf{a}_2 + \frac{3}{4}\,\mathbf{a}_3$ | = | $\frac{1}{2}a\,\hat{\mathbf{y}} + \frac{3}{4}c\,\hat{\mathbf{z}}$ | (4$d$) | Zr I |
| $\mathbf{B}_{3}$ | = | $\frac{1}{2}\,\mathbf{a}_{1} + \frac{1}{4}\,\mathbf{a}_{3}$ | = | $\frac{1}{2}a\,\mathbf{\hat{x}} + \frac{1}{4}c\,\mathbf{\hat{z}}$ | (4$d$) | Zr I |
| $\mathbf{B}_{4}$ | = | $\frac{1}{2}\,\mathbf{a}_{1} + \frac{3}{4}\,\mathbf{a}_{3}$ | = | $\frac{1}{2}a\,\mathbf{\hat{x}} + \frac{3}{4}c\,\mathbf{\hat{z}}$ | (4$d$) | Zr I |
| $\mathbf{B}_{5}$ | = | $x_{2}\,\mathbf{a}_{1} + x_{2}\,\mathbf{a}_{2}$ | = | $x_{2}a\,\mathbf{\hat{x}} + x_{2}a\,\mathbf{\hat{y}}$ | (4$f$) | Zr II |
| $\mathbf{B}_{6}$ | = | $-x_{2}\,\mathbf{a}_{1} - x_{2}\,\mathbf{a}_{2}$ | = | $-x_{2}a\,\mathbf{\hat{x}} - x_{2}a\,\mathbf{\hat{y}}$ | (4$f$) | Zr II |
| $\mathbf{B}_{7}$ | = | $\left(\frac{1}{2} - x_{2}\right)\,\mathbf{a}_{1} + \left(\frac{1}{2} + x_{2}\right)\,\mathbf{a}_{2} + \frac{1}{2}\,\mathbf{a}_{3}$ | = | $\left(\frac{1}{2} - x_{2}\right)a\,\mathbf{\hat{x}} + \left(\frac{1}{2} + x_{2}\right)a\,\mathbf{\hat{y}} + \frac{1}{2}c\,\mathbf{\hat{z}}$ | (4$f$) | Zr II |
| $\mathbf{B}_{8}$ | = | $\left(\frac{1}{2} + x_{2}\right)\,\mathbf{a}_{1} + \left(\frac{1}{2} - x_{2}\right)\,\mathbf{a}_{2} + \frac{1}{2}\,\mathbf{a}_{3}$ | = | $\left(\frac{1}{2} + x_{2}\right)a\,\mathbf{\hat{x}} + \left(\frac{1}{2} - x_{2}\right)a\,\mathbf{\hat{y}} + \frac{1}{2}c\,\mathbf{\hat{z}}$ | (4$f$) | Zr II |
| $\mathbf{B}_{9}$ | = | $x_{3}\,\mathbf{a}_{1} - x_{3}\,\mathbf{a}_{2}$ | = | $x_{3}a\,\mathbf{\hat{x}} - x_{3}a\,\mathbf{\hat{y}}$ | (4$g$) | Zr III |
| $\mathbf{B}_{10}$ | = | $-x_{3}\,\mathbf{a}_{1} + x_{3}\,\mathbf{a}_{2}$ | = | $-x_{3}a\,\mathbf{\hat{x}} + x_{3}a\,\mathbf{\hat{y}}$ | (4$g$) | Zr III |
| $\mathbf{B}_{11}$ | = | $\left(\frac{1}{2} + x_{3}\right)\,\mathbf{a}_{1} + \left(\frac{1}{2} + x_{3}\right)\,\mathbf{a}_{2} + \frac{1}{2}\,\mathbf{a}_{3}$ | = | $\left(\frac{1}{2} + x_{3}\right)a\,\mathbf{\hat{x}} + \left(\frac{1}{2} + x_{3}\right)a\,\mathbf{\hat{y}} + \frac{1}{2}c\,\mathbf{\hat{z}}$ | (4$g$) | Zr III |
| $\mathbf{B}_{12}$ | = | $\left(\frac{1}{2} - x_{3}\right)\,\mathbf{a}_{1} + \left(\frac{1}{2} - x_{3}\right)\,\mathbf{a}_{2} + \frac{1}{2}\,\mathbf{a}_{3}$ | = | $\left(\frac{1}{2} - x_{3}\right)a\,\mathbf{\hat{x}} + \left(\frac{1}{2} - x_{3}\right)a\,\mathbf{\hat{y}} + \frac{1}{2}c\,\mathbf{\hat{z}}$ | (4$g$) | Zr III |
| $\mathbf{B}_{13}$ | = | $x_{4}\,\mathbf{a}_{1} + x_{4}\,\mathbf{a}_{2} + z_{4}\,\mathbf{a}_{3}$ | = | $x_{4}a\,\mathbf{\hat{x}} + x_{4}a\,\mathbf{\hat{y}} + z_{4}c\,\mathbf{\hat{z}}$ | (8$j$) | Al |
| $\mathbf{B}_{14}$ | = | $-x_{4}\,\mathbf{a}_{1} - x_{4}\,\mathbf{a}_{2} + z_{4}\,\mathbf{a}_{3}$ | = | $-x_{4}a\,\mathbf{\hat{x}} - x_{4}a\,\mathbf{\hat{y}} + z_{4}c\,\mathbf{\hat{z}}$ | (8$j$) | Al |
| $\mathbf{B}_{15}$ | = | $\left(\frac{1}{2} - x_{4}\right)\,\mathbf{a}_{1} + \left(\frac{1}{2} + x_{4}\right)\,\mathbf{a}_{2} + \left(\frac{1}{2} + z_{4}\right)\,\mathbf{a}_{3}$ | = | $\left(\frac{1}{2} - x_{4}\right)a\,\mathbf{\hat{x}} + \left(\frac{1}{2} + x_{4}\right)a\,\mathbf{\hat{y}} + \left(\frac{1}{2} + z_{4}\right)c\,\mathbf{\hat{z}}$ | (8$j$) | Al |
| $\mathbf{B}_{16}$ | = | $\left(\frac{1}{2} + x_{4}\right)\,\mathbf{a}_{1} + \left(\frac{1}{2} - x_{4}\right)\,\mathbf{a}_{2} + \left(\frac{1}{2} + z_{4}\right)\,\mathbf{a}_{3}$ | = | $\left(\frac{1}{2} + x_{4}\right)a\,\mathbf{\hat{x}} + \left(\frac{1}{2} - x_{4}\right)a\,\mathbf{\hat{y}} + \left(\frac{1}{2} + z_{4}\right)c\,\mathbf{\hat{z}}$ | (8$j$) | Al |
| $\mathbf{B}_{17}$ | = | $\left(\frac{1}{2} - x_{4}\right)\,\mathbf{a}_{1} + \left(\frac{1}{2} + x_{4}\right)\,\mathbf{a}_{2} + \left(\frac{1}{2} - z_{4}\right)\,\mathbf{a}_{3}$ | = | $\left(\frac{1}{2} - x_{4}\right)a\,\mathbf{\hat{x}} + \left(\frac{1}{2} + x_{4}\right)a\,\mathbf{\hat{y}} + \left(\frac{1}{2} - z_{4}\right)c\,\mathbf{\hat{z}}$ | (8$j$) | Al |
| $\mathbf{B}_{18}$ | = | $\left(\frac{1}{2} + x_{4}\right)\,\mathbf{a}_{1} + \left(\frac{1}{2} - x_{4}\right)\,\mathbf{a}_{2} + \left(\frac{1}{2} - z_{4}\right)\,\mathbf{a}_{3}$ | = | $\left(\frac{1}{2} + x_{4}\right)a\,\mathbf{\hat{x}} + \left(\frac{1}{2} - x_{4}\right)a\,\mathbf{\hat{y}} + \left(\frac{1}{2} - z_{4}\right)c\,\mathbf{\hat{z}}$ | (8$j$) | Al |
| $\mathbf{B}_{19}$ | = | $x_{4}\,\mathbf{a}_{1} + x_{4}\,\mathbf{a}_{2} - z_{4}\,\mathbf{a}_{3}$ | = | $x_{4}a\,\mathbf{\hat{x}} + x_{4}a\,\mathbf{\hat{y}} - z_{4}c\,\mathbf{\hat{z}}$ | (8$j$) | Al |
| $\mathbf{B}_{20}$ | = | $-x_{4}\,\mathbf{a}_{1} - x_{4}\,\mathbf{a}_{2} - z_{4}\,\mathbf{a}_{3}$ | = | $-x_{4}a\,\mathbf{\hat{x}} - x_{4}a\,\mathbf{\hat{y}} - z_{4}c\,\mathbf{\hat{z}}$ | (8$j$) | Al |

**Found in:**

**Geometry files:**

- CIF: pp. 1702
- POSCAR: pp. 1703

# $ZrFe_4Si_2$ Structure: A4B2C_tP14_136_i_g_b

http://aflow.org/prototype-encyclopedia/A4B2C_tP14_136_i_g_b

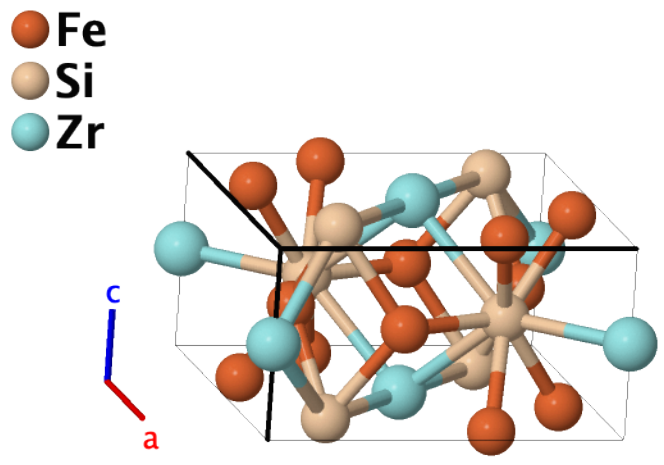

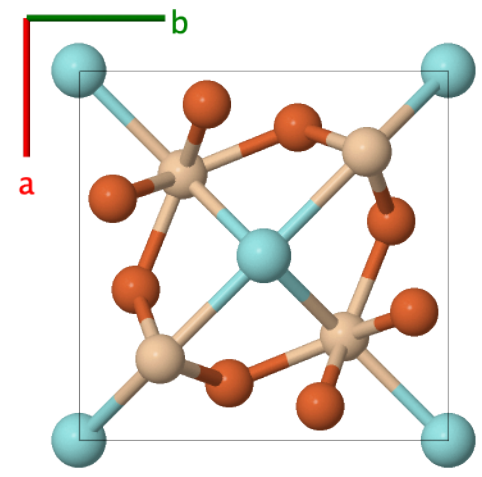

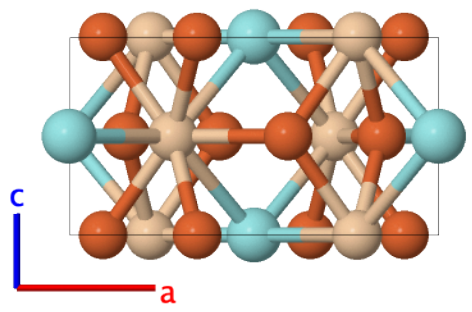

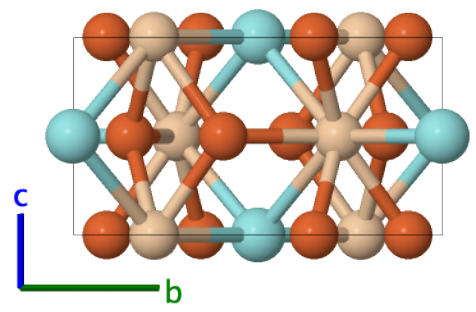

| | | |
|---|---|---|
| **Prototype** | : | $Fe_4Si_2Zr$ |
| **AFLOW prototype label** | : | A4B2C_tP14_136_i_g_b |
| ***Strukturbericht* designation** | : | None |
| **Pearson symbol** | : | tP14 |
| **Space group number** | : | 136 |
| **Space group symbol** | : | $P4_2/mnm$ |
| **AFLOW prototype command** | : | `aflow --proto=A4B2C_tP14_136_i_g_b` <br> `--params=`$a, c/a, x_2, x_3, y_3$ |

### Other compounds with this structure

- $BaCd_4Pt_2$, $DyFe_4Ge_2$, $DyNi_4As_2$, $ErFe_4Ge_2$, $ErNi_4P_2$, $GdNi_4As_2$, $GdRe_4Si_2$, $HoFe_4Ge_2$, $LuFe_4Ge_2$, $LuNi_4As_2$, $LuRe_4Si_2$, $ScFe_4P_2$, $ScFe_4Si_2$, $ScNi_4As_2$, $SmRe_4Si_2$, $SrCd_4Pt_2$, $TbRe_4Si_2$, $TmFe_4Ge_2$, $TmRe_4Si_2$, $UMn_4P_2$, $YFe_4Ge_2$, $YNi_4As_2$, $YNi_4P_2$, $YRe_4Si_2$, $YbNi_4P_2$, $ZrFe_4P_2$, and $ZrNi_4As_2$

### Simple Tetragonal primitive vectors:

$$\mathbf{a}_1 = a\,\hat{\mathbf{x}}$$
$$\mathbf{a}_2 = a\,\hat{\mathbf{y}}$$
$$\mathbf{a}_3 = c\,\hat{\mathbf{z}}$$

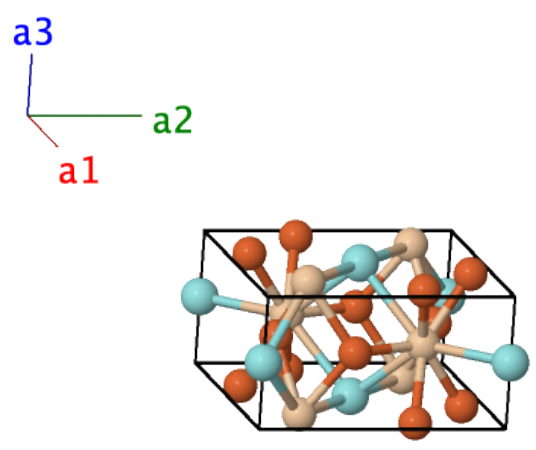

### Basis vectors:

| | | Lattice Coordinates | | Cartesian Coordinates | Wyckoff Position | Atom Type |
|---|---|---|---|---|---|---|
| $\mathbf{B}_1$ | = | $\frac{1}{2}\,\mathbf{a}_3$ | = | $\frac{1}{2}c\,\hat{\mathbf{z}}$ | (2b) | Zr |
| $\mathbf{B}_2$ | = | $\frac{1}{2}\,\mathbf{a}_1 + \frac{1}{2}\,\mathbf{a}_2$ | = | $\frac{1}{2}a\,\hat{\mathbf{x}} + \frac{1}{2}a\,\hat{\mathbf{y}}$ | (2$b$) | Zr |
| $\mathbf{B}_3$ | = | $x_2\,\mathbf{a}_1 - x_2\,\mathbf{a}_2$ | = | $x_2a\,\hat{\mathbf{x}} - x_2a\,\hat{\mathbf{y}}$ | (4$g$) | Si |
| $\mathbf{B}_4$ | = | $-x_2\,\mathbf{a}_1 + x_2\,\mathbf{a}_2$ | = | $-x_2a\,\hat{\mathbf{x}} + x_2a\,\hat{\mathbf{y}}$ | (4$g$) | Si |
| $\mathbf{B}_5$ | = | $\left(\frac{1}{2} + x_2\right)\,\mathbf{a}_1 + \left(\frac{1}{2} + x_2\right)\,\mathbf{a}_2 + \frac{1}{2}\,\mathbf{a}_3$ | = | $\left(\frac{1}{2} + x_2\right)a\,\hat{\mathbf{x}} + \left(\frac{1}{2} + x_2\right)a\,\hat{\mathbf{y}} + \frac{1}{2}c\,\hat{\mathbf{z}}$ | (4$g$) | Si |
| $\mathbf{B}_6$ | = | $\left(\frac{1}{2} - x_2\right)\,\mathbf{a}_1 + \left(\frac{1}{2} - x_2\right)\,\mathbf{a}_2 + \frac{1}{2}\,\mathbf{a}_3$ | = | $\left(\frac{1}{2} - x_2\right)a\,\hat{\mathbf{x}} + \left(\frac{1}{2} - x_2\right)a\,\hat{\mathbf{y}} + \frac{1}{2}c\,\hat{\mathbf{z}}$ | (4$g$) | Si |
| $\mathbf{B}_7$ | = | $x_3\,\mathbf{a}_1 + y_3\,\mathbf{a}_2$ | = | $x_3a\,\hat{\mathbf{x}} + y_3a\,\hat{\mathbf{y}}$ | (8$i$) | Fe |
| $\mathbf{B}_8$ | = | $-x_3\,\mathbf{a}_1 - y_3\,\mathbf{a}_2$ | = | $-x_3a\,\hat{\mathbf{x}} - y_3a\,\hat{\mathbf{y}}$ | (8$i$) | Fe |
| $\mathbf{B}_9$ | = | $\left(\frac{1}{2} - y_3\right)\,\mathbf{a}_1 + \left(\frac{1}{2} + x_3\right)\,\mathbf{a}_2 + \frac{1}{2}\,\mathbf{a}_3$ | = | $\left(\frac{1}{2} - y_3\right)a\,\hat{\mathbf{x}} + \left(\frac{1}{2} + x_3\right)a\,\hat{\mathbf{y}} + \frac{1}{2}c\,\hat{\mathbf{z}}$ | (8$i$) | Fe |
| $\mathbf{B}_{10}$ | = | $\left(\frac{1}{2} + y_3\right)\,\mathbf{a}_1 + \left(\frac{1}{2} - x_3\right)\,\mathbf{a}_2 + \frac{1}{2}\,\mathbf{a}_3$ | = | $\left(\frac{1}{2} + y_3\right)a\,\hat{\mathbf{x}} + \left(\frac{1}{2} - x_3\right)a\,\hat{\mathbf{y}} + \frac{1}{2}c\,\hat{\mathbf{z}}$ | (8$i$) | Fe |
| $\mathbf{B}_{11}$ | = | $\left(\frac{1}{2} - x_3\right)\,\mathbf{a}_1 + \left(\frac{1}{2} + y_3\right)\,\mathbf{a}_2 + \frac{1}{2}\,\mathbf{a}_3$ | = | $\left(\frac{1}{2} - x_3\right)a\,\hat{\mathbf{x}} + \left(\frac{1}{2} + y_3\right)a\,\hat{\mathbf{y}} + \frac{1}{2}c\,\hat{\mathbf{z}}$ | (8$i$) | Fe |
| $\mathbf{B}_{12}$ | = | $\left(\frac{1}{2} + x_3\right)\,\mathbf{a}_1 + \left(\frac{1}{2} - y_3\right)\,\mathbf{a}_2 + \frac{1}{2}\,\mathbf{a}_3$ | = | $\left(\frac{1}{2} + x_3\right)a\,\hat{\mathbf{x}} + \left(\frac{1}{2} - y_3\right)a\,\hat{\mathbf{y}} + \frac{1}{2}c\,\hat{\mathbf{z}}$ | (8$i$) | Fe |
| $\mathbf{B}_{13}$ | = | $y_3\,\mathbf{a}_1 + x_3\,\mathbf{a}_2$ | = | $y_3a\,\hat{\mathbf{x}} + x_3a\,\hat{\mathbf{y}}$ | (8$i$) | Fe |
| $\mathbf{B}_{14}$ | = | $-y_3\,\mathbf{a}_1 - x_3\,\mathbf{a}_2$ | = | $-y_3a\,\hat{\mathbf{x}} - x_3a\,\hat{\mathbf{y}}$ | (8$i$) | Fe |

**Geometry files:**

- CIF: pp. 1703
- POSCAR: pp. 1703

# $K_2CuCl_4{\cdot}2H_2O$ ($H4_1$) Structure: A4BC4D2E2_tP26_136_fg_a_j_d_e

http://aflow.org/prototype-encyclopedia/A4BC4D2E2_tP26_136_fg_a_j_d_e

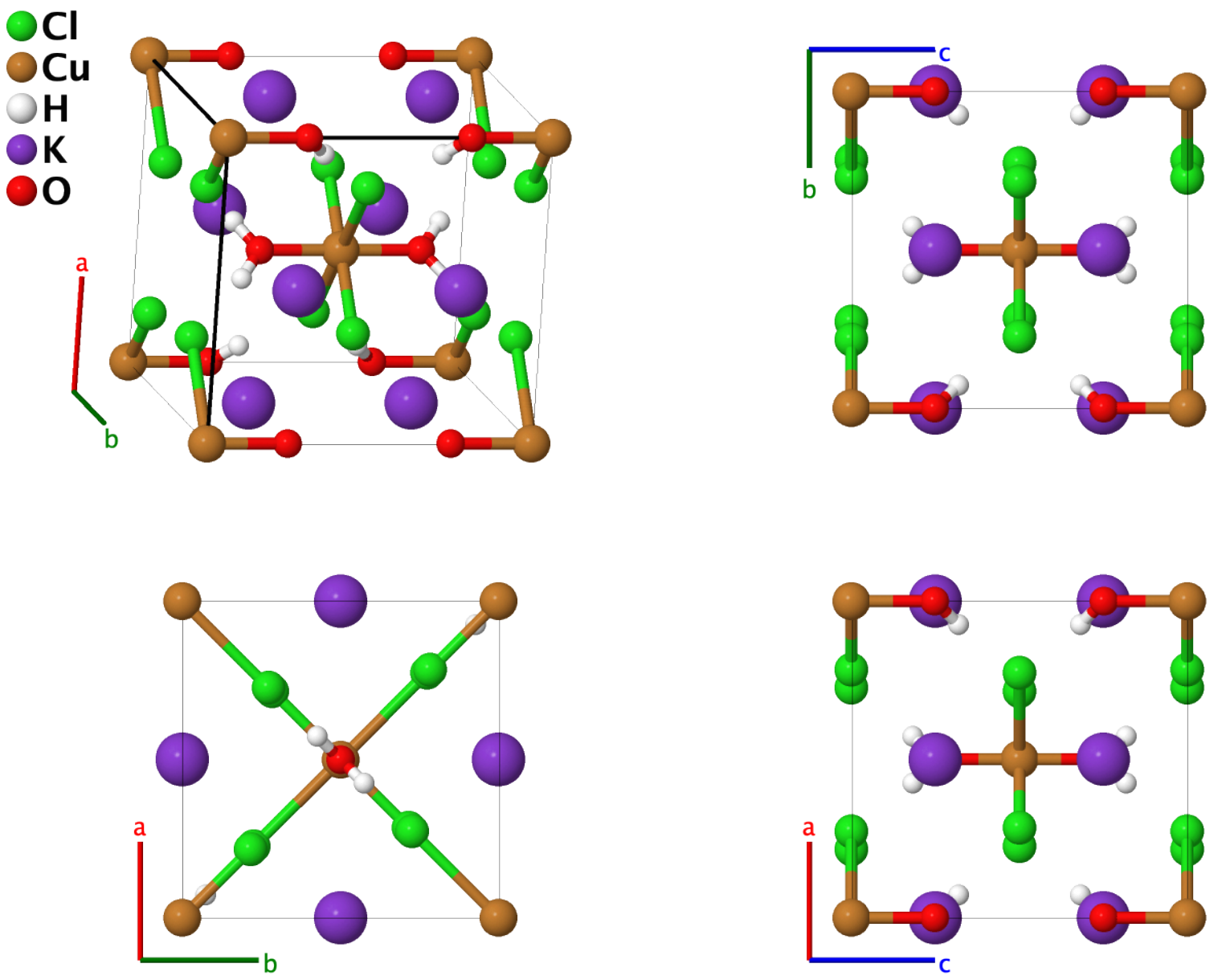

| | | |
|---|---|---|
| **Prototype** | : | $Cl_4CuH_4K_2O_2$ |
| **AFLOW prototype label** | : | A4BC4D2E2_tP26_136_fg_a_j_d_e |
| ***Strukturbericht* designation** | : | $H4_1$ |
| **Pearson symbol** | : | tP26 |
| **Space group number** | : | 136 |
| **Space group symbol** | : | $P4_2/mnm$ |
| **AFLOW prototype command** | : | `aflow --proto=A4BC4D2E2_tP26_136_fg_a_j_d_e` `--params=`$a, c/a, z_3, x_4, x_5, x_6, z_6$ |

**Other compounds with this structure**

- $Rb_2CuCl_4 \cdot 2H_2O$, $(NH_4)_2CuCl_4 \cdot 2H_2O$, and $(NH_4)_2CuBr_4 \cdot 2H_2O$

**Simple Tetragonal primitive vectors:**

$$\mathbf{a}_1 = a\,\hat{\mathbf{x}}$$
$$\mathbf{a}_2 = a\,\hat{\mathbf{y}}$$
$$\mathbf{a}_3 = c\,\hat{\mathbf{z}}$$

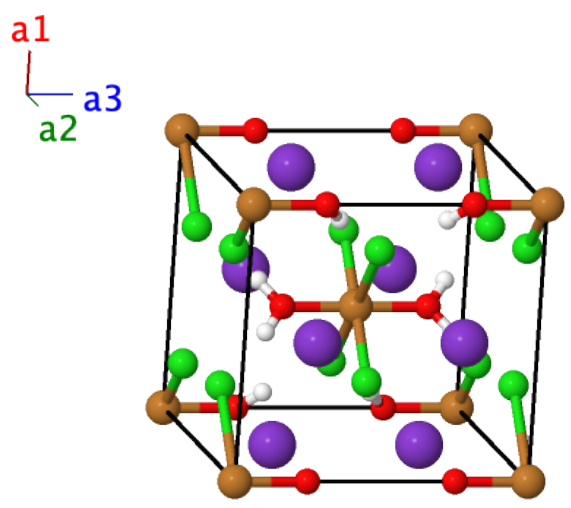

**Basis vectors:**

| | | Lattice Coordinates | | Cartesian Coordinates | Wyckoff Position | Atom Type |
|---|---|---|---|---|---|---|
| $\mathbf{B}_1$ | = | $0\,\mathbf{a}_1 + 0\,\mathbf{a}_2 + 0\,\mathbf{a}_3$ | = | $0\,\hat{\mathbf{x}} + 0\,\hat{\mathbf{y}} + 0\,\hat{\mathbf{z}}$ | (2a) | Cu |
| $\mathbf{B}_2$ | = | $\frac{1}{2}\,\mathbf{a}_1 + \frac{1}{2}\,\mathbf{a}_2 + \frac{1}{2}\,\mathbf{a}_3$ | = | $\frac{1}{2}a\,\hat{\mathbf{x}} + \frac{1}{2}a\,\hat{\mathbf{y}} + \frac{1}{2}c\,\hat{\mathbf{z}}$ | $(2a)$ | Cu |
| $\mathbf{B}_3$ | = | $\frac{1}{2}\,\mathbf{a}_2 + \frac{1}{4}\,\mathbf{a}_3$ | = | $\frac{1}{2}a\,\hat{\mathbf{y}} + \frac{1}{4}c\,\hat{\mathbf{z}}$ | $(4d)$ | K |
| $\mathbf{B}_4$ | = | $\frac{1}{2}\,\mathbf{a}_2 + \frac{3}{4}\,\mathbf{a}_3$ | = | $\frac{1}{2}a\,\hat{\mathbf{y}} + \frac{3}{4}c\,\hat{\mathbf{z}}$ | $(4d)$ | K |
| $\mathbf{B}_5$ | = | $\frac{1}{2}\,\mathbf{a}_1 + \frac{1}{4}\,\mathbf{a}_3$ | = | $\frac{1}{2}a\,\hat{\mathbf{x}} + \frac{1}{4}c\,\hat{\mathbf{z}}$ | $(4d)$ | K |
| $\mathbf{B}_6$ | = | $\frac{1}{2}\,\mathbf{a}_1 + \frac{3}{4}\,\mathbf{a}_3$ | = | $\frac{1}{2}a\,\hat{\mathbf{x}} + \frac{3}{4}c\,\hat{\mathbf{z}}$ | $(4d)$ | K |
| $\mathbf{B}_7$ | = | $z_3\,\mathbf{a}_3$ | = | $z_3c\,\hat{\mathbf{z}}$ | $(4e)$ | O |
| $\mathbf{B}_8$ | = | $\frac{1}{2}\,\mathbf{a}_1 + \frac{1}{2}\,\mathbf{a}_2 + \left(\frac{1}{2} + z_3\right)\,\mathbf{a}_3$ | = | $\frac{1}{2}a\,\hat{\mathbf{x}} + \frac{1}{2}a\,\hat{\mathbf{y}} + \left(\frac{1}{2} + z_3\right)c\,\hat{\mathbf{z}}$ | $(4e)$ | O |
| $\mathbf{B}_9$ | = | $\frac{1}{2}\,\mathbf{a}_1 + \frac{1}{2}\,\mathbf{a}_2 + \left(\frac{1}{2} - z_3\right)\,\mathbf{a}_3$ | = | $\frac{1}{2}a\,\hat{\mathbf{x}} + \frac{1}{2}a\,\hat{\mathbf{y}} + \left(\frac{1}{2} - z_3\right)c\,\hat{\mathbf{z}}$ | $(4e)$ | O |
| $\mathbf{B}_{10}$ | = | $-z_3\,\mathbf{a}_3$ | = | $-z_3c\,\hat{\mathbf{z}}$ | $(4e)$ | O |
| $\mathbf{B}_{11}$ | = | $x_4\,\mathbf{a}_1 + x_4\,\mathbf{a}_2$ | = | $x_4a\,\hat{\mathbf{x}} + x_4a\,\hat{\mathbf{y}}$ | $(4f)$ | Cl I |
| $\mathbf{B}_{12}$ | = | $-x_4\,\mathbf{a}_1 - x_4\,\mathbf{a}_2$ | = | $-x_4a\,\hat{\mathbf{x}} - x_4a\,\hat{\mathbf{y}}$ | $(4f)$ | Cl I |
| $\mathbf{B}_{13}$ | = | $\left(\frac{1}{2} - x_4\right)\,\mathbf{a}_1 + \left(\frac{1}{2} + x_4\right)\,\mathbf{a}_2 + \frac{1}{2}\,\mathbf{a}_3$ | = | $\left(\frac{1}{2} - x_4\right)a\,\hat{\mathbf{x}} + \left(\frac{1}{2} + x_4\right)a\,\hat{\mathbf{y}} + \frac{1}{2}c\,\hat{\mathbf{z}}$ | $(4f)$ | Cl I |
| $\mathbf{B}_{14}$ | = | $\left(\frac{1}{2} + x_4\right)\,\mathbf{a}_1 + \left(\frac{1}{2} - x_4\right)\,\mathbf{a}_2 + \frac{1}{2}\,\mathbf{a}_3$ | = | $\left(\frac{1}{2} + x_4\right)a\,\hat{\mathbf{x}} + \left(\frac{1}{2} - x_4\right)a\,\hat{\mathbf{y}} + \frac{1}{2}c\,\hat{\mathbf{z}}$ | $(4f)$ | Cl I |
| $\mathbf{B}_{15}$ | = | $x_5\,\mathbf{a}_1 - x_5\,\mathbf{a}_2$ | = | $x_5a\,\hat{\mathbf{x}} - x_5a\,\hat{\mathbf{y}}$ | $(4g)$ | Cl II |
| $\mathbf{B}_{16}$ | = | $-x_5\,\mathbf{a}_1 + x_5\,\mathbf{a}_2$ | = | $-x_5a\,\hat{\mathbf{x}} + x_5a\,\hat{\mathbf{y}}$ | $(4g)$ | Cl II |
| $\mathbf{B}_{17}$ | = | $\left(\frac{1}{2} + x_5\right)\,\mathbf{a}_1 + \left(\frac{1}{2} + x_5\right)\,\mathbf{a}_2 + \frac{1}{2}\,\mathbf{a}_3$ | = | $\left(\frac{1}{2} + x_5\right)a\,\hat{\mathbf{x}} + \left(\frac{1}{2} + x_5\right)a\,\hat{\mathbf{y}} + \frac{1}{2}c\,\hat{\mathbf{z}}$ | $(4g)$ | Cl II |
| $\mathbf{B}_{18}$ | = | $\left(\frac{1}{2} - x_5\right)\,\mathbf{a}_1 + \left(\frac{1}{2} - x_5\right)\,\mathbf{a}_2 + \frac{1}{2}\,\mathbf{a}_3$ | = | $\left(\frac{1}{2} - x_5\right)a\,\hat{\mathbf{x}} + \left(\frac{1}{2} - x_5\right)a\,\hat{\mathbf{y}} + \frac{1}{2}c\,\hat{\mathbf{z}}$ | $(4g)$ | Cl II |
| $\mathbf{B}_{19}$ | = | $x_6\,\mathbf{a}_1 + x_6\,\mathbf{a}_2 + z_6\,\mathbf{a}_3$ | = | $x_6a\,\hat{\mathbf{x}} + x_6a\,\hat{\mathbf{y}} + z_6c\,\hat{\mathbf{z}}$ | $(8j)$ | H |
| $\mathbf{B}_{20}$ | = | $-x_6\,\mathbf{a}_1 - x_6\,\mathbf{a}_2 + z_6\,\mathbf{a}_3$ | = | $-x_6a\,\hat{\mathbf{x}} - x_6a\,\hat{\mathbf{y}} + z_6c\,\hat{\mathbf{z}}$ | $(8j)$ | H |
| $\mathbf{B}_{21}$ | = | $\left(\frac{1}{2} - x_6\right)\,\mathbf{a}_1 + \left(\frac{1}{2} + x_6\right)\,\mathbf{a}_2 + \left(\frac{1}{2} + z_6\right)\,\mathbf{a}_3$ | = | $\left(\frac{1}{2} - x_6\right)a\,\hat{\mathbf{x}} + \left(\frac{1}{2} + x_6\right)a\,\hat{\mathbf{y}} + \left(\frac{1}{2} + z_6\right)c\,\hat{\mathbf{z}}$ | $(8j)$ | H |
| $\mathbf{B}_{22}$ | = | $\left(\frac{1}{2} + x_6\right)\,\mathbf{a}_1 + \left(\frac{1}{2} - x_6\right)\,\mathbf{a}_2 + \left(\frac{1}{2} + z_6\right)\,\mathbf{a}_3$ | = | $\left(\frac{1}{2} + x_6\right)a\,\hat{\mathbf{x}} + \left(\frac{1}{2} - x_6\right)a\,\hat{\mathbf{y}} + \left(\frac{1}{2} + z_6\right)c\,\hat{\mathbf{z}}$ | $(8j)$ | H |
| $\mathbf{B}_{23}$ | = | $\left(\frac{1}{2} - x_6\right)\,\mathbf{a}_1 + \left(\frac{1}{2} + x_6\right)\,\mathbf{a}_2 + \left(\frac{1}{2} - z_6\right)\,\mathbf{a}_3$ | = | $\left(\frac{1}{2} - x_6\right)a\,\hat{\mathbf{x}} + \left(\frac{1}{2} + x_6\right)a\,\hat{\mathbf{y}} + \left(\frac{1}{2} - z_6\right)c\,\hat{\mathbf{z}}$ | $(8j)$ | H |
| $\mathbf{B}_{24}$ | = | $\left(\frac{1}{2} + x_6\right)\,\mathbf{a}_1 + \left(\frac{1}{2} - x_6\right)\,\mathbf{a}_2 + \left(\frac{1}{2} - z_6\right)\,\mathbf{a}_3$ | = | $\left(\frac{1}{2} + x_6\right)a\,\hat{\mathbf{x}} + \left(\frac{1}{2} - x_6\right)a\,\hat{\mathbf{y}} + \left(\frac{1}{2} - z_6\right)c\,\hat{\mathbf{z}}$ | $(8j)$ | H |
| $\mathbf{B}_{25}$ | = | $x_6\,\mathbf{a}_1 + x_6\,\mathbf{a}_2 - z_6\,\mathbf{a}_3$ | = | $x_6a\,\hat{\mathbf{x}} + x_6a\,\hat{\mathbf{y}} - z_6c\,\hat{\mathbf{z}}$ | $(8j)$ | H |
| $\mathbf{B}_{26}$ | = | $-x_6\,\mathbf{a}_1 - x_6\,\mathbf{a}_2 - z_6\,\mathbf{a}_3$ | = | $-x_6a\,\hat{\mathbf{x}} - x_6a\,\hat{\mathbf{y}} - z_6c\,\hat{\mathbf{z}}$ | $(8j)$ | H |

**Geometry files:**

- CIF: pp. 1703
- POSCAR: pp. 1704

# $Nd_2Fe_{14}B$ Structure: AB14C2_tP68_136_f_ce2j2k_fg

http://aflow.org/prototype-encyclopedia/AB14C2_tP68_136_f_ce2j2k_fg

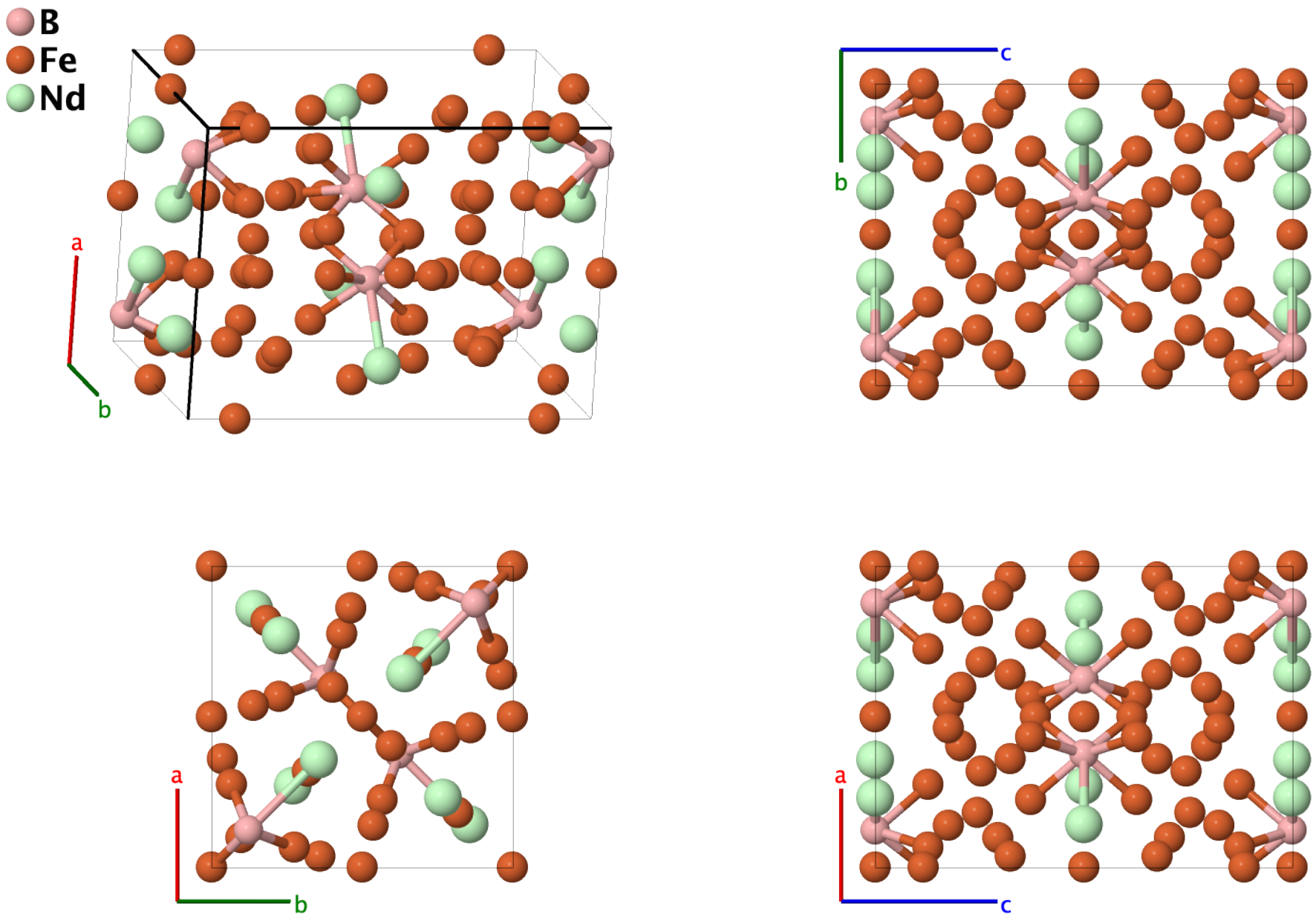

| | | |
|---|---|---|
| **Prototype** | : | $BFe_{14}Nd_2$ |
| **AFLOW prototype label** | : | AB14C2_tP68_136_f_ce2j2k_fg |
| ***Strukturbericht* designation** | : | None |
| **Pearson symbol** | : | tP68 |
| **Space group number** | : | 136 |
| **Space group symbol** | : | $P4_2/mnm$ |
| **AFLOW prototype command** | : | `aflow --proto=AB14C2_tP68_136_f_ce2j2k_fg` `--params=`$a, c/a, z_2, x_3, x_4, x_5, x_6, z_6, x_7, z_7, x_8, y_8, z_8, x_9, y_9, z_9$ |

**Simple Tetragonal primitive vectors:**

$$\mathbf{a}_1 = a\,\hat{\mathbf{x}}$$
$$\mathbf{a}_2 = a\,\hat{\mathbf{y}}$$
$$\mathbf{a}_3 = c\,\hat{\mathbf{z}}$$

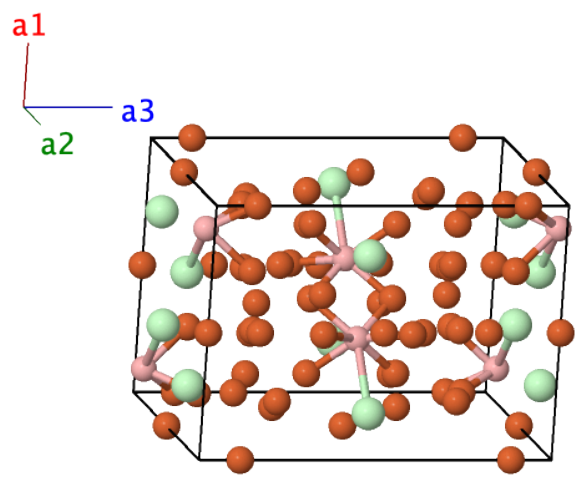

**Basis vectors:**

| | | Lattice Coordinates | | Cartesian Coordinates | Wyckoff Position | Atom Type |
|---|---|---|---|---|---|---|
| $\mathbf{B}_1$ | = | $\frac{1}{2}\,\mathbf{a}_2$ | = | $\frac{1}{2}a\,\hat{\mathbf{y}}$ | (4$c$) | Fe I |
| $\mathbf{B}_2$ | = | $\frac{1}{2}\,\mathbf{a}_2 + \frac{1}{2}\,\mathbf{a}_3$ | = | $\frac{1}{2}a\,\hat{\mathbf{y}} + \frac{1}{2}c\,\hat{\mathbf{z}}$ | (4$c$) | Fe I |
| $\mathbf{B}_3$ | = | $\frac{1}{2}\,\mathbf{a}_1 + \frac{1}{2}\,\mathbf{a}_3$ | = | $\frac{1}{2}a\,\hat{\mathbf{x}} + \frac{1}{2}c\,\hat{\mathbf{z}}$ | (4$c$) | Fe I |
| $\mathbf{B}_4$ | = | $\frac{1}{2}\,\mathbf{a}_1$ | = | $\frac{1}{2}a\,\hat{\mathbf{x}}$ | (4$c$) | Fe I |
| $\mathbf{B}_5$ | = | $z_2\,\mathbf{a}_3$ | = | $z_2c\,\hat{\mathbf{z}}$ | (4$e$) | Fe II |
| $\mathbf{B}_{6}$ | = | $\frac{1}{2}\,\mathbf{a}_{1} + \frac{1}{2}\,\mathbf{a}_{2} + \left(\frac{1}{2} + z_{2}\right)\,\mathbf{a}_{3}$ | = | $\frac{1}{2}a\,\mathbf{\hat{x}} + \frac{1}{2}a\,\mathbf{\hat{y}} + \left(\frac{1}{2} + z_{2}\right)c\,\mathbf{\hat{z}}$ | (4$e$) | Fe II |
| $\mathbf{B}_{7}$ | = | $\frac{1}{2}\,\mathbf{a}_{1} + \frac{1}{2}\,\mathbf{a}_{2} + \left(\frac{1}{2} - z_{2}\right)\,\mathbf{a}_{3}$ | = | $\frac{1}{2}a\,\mathbf{\hat{x}} + \frac{1}{2}a\,\mathbf{\hat{y}} + \left(\frac{1}{2} - z_{2}\right)c\,\mathbf{\hat{z}}$ | (4$e$) | Fe II |
| $\mathbf{B}_{8}$ | = | $-z_{2}\,\mathbf{a}_{3}$ | = | $-z_{2}c\,\mathbf{\hat{z}}$ | (4$e$) | Fe II |
| $\mathbf{B}_{9}$ | = | $x_{3}\,\mathbf{a}_{1} + x_{3}\,\mathbf{a}_{2}$ | = | $x_{3}a\,\mathbf{\hat{x}} + x_{3}a\,\mathbf{\hat{y}}$ | (4$f$) | B |
| $\mathbf{B}_{10}$ | = | $-x_{3}\,\mathbf{a}_{1} - x_{3}\,\mathbf{a}_{2}$ | = | $-x_{3}a\,\mathbf{\hat{x}} - x_{3}a\,\mathbf{\hat{y}}$ | (4$f$) | B |
| $\mathbf{B}_{11}$ | = | $\left(\frac{1}{2} - x_{3}\right)\,\mathbf{a}_{1} + \left(\frac{1}{2} + x_{3}\right)\,\mathbf{a}_{2} + \frac{1}{2}\,\mathbf{a}_{3}$ | = | $\left(\frac{1}{2} - x_{3}\right)a\,\mathbf{\hat{x}} + \left(\frac{1}{2} + x_{3}\right)a\,\mathbf{\hat{y}} + \frac{1}{2}c\,\mathbf{\hat{z}}$ | (4$f$) | B |
| $\mathbf{B}_{12}$ | = | $\left(\frac{1}{2} + x_{3}\right)\,\mathbf{a}_{1} + \left(\frac{1}{2} - x_{3}\right)\,\mathbf{a}_{2} + \frac{1}{2}\,\mathbf{a}_{3}$ | = | $\left(\frac{1}{2} + x_{3}\right)a\,\mathbf{\hat{x}} + \left(\frac{1}{2} - x_{3}\right)a\,\mathbf{\hat{y}} + \frac{1}{2}c\,\mathbf{\hat{z}}$ | (4$f$) | B |
| $\mathbf{B}_{13}$ | = | $x_{4}\,\mathbf{a}_{1} + x_{4}\,\mathbf{a}_{2}$ | = | $x_{4}a\,\mathbf{\hat{x}} + x_{4}a\,\mathbf{\hat{y}}$ | (4$f$) | Nd I |
| $\mathbf{B}_{14}$ | = | $-x_{4}\,\mathbf{a}_{1} - x_{4}\,\mathbf{a}_{2}$ | = | $-x_{4}a\,\mathbf{\hat{x}} - x_{4}a\,\mathbf{\hat{y}}$ | (4$f$) | Nd I |
| $\mathbf{B}_{15}$ | = | $\left(\frac{1}{2} - x_{4}\right)\,\mathbf{a}_{1} + \left(\frac{1}{2} + x_{4}\right)\,\mathbf{a}_{2} + \frac{1}{2}\,\mathbf{a}_{3}$ | = | $\left(\frac{1}{2} - x_{4}\right)a\,\mathbf{\hat{x}} + \left(\frac{1}{2} + x_{4}\right)a\,\mathbf{\hat{y}} + \frac{1}{2}c\,\mathbf{\hat{z}}$ | (4$f$) | Nd I |
| $\mathbf{B}_{16}$ | = | $\left(\frac{1}{2} + x_{4}\right)\,\mathbf{a}_{1} + \left(\frac{1}{2} - x_{4}\right)\,\mathbf{a}_{2} + \frac{1}{2}\,\mathbf{a}_{3}$ | = | $\left(\frac{1}{2} + x_{4}\right)a\,\mathbf{\hat{x}} + \left(\frac{1}{2} - x_{4}\right)a\,\mathbf{\hat{y}} + \frac{1}{2}c\,\mathbf{\hat{z}}$ | (4$f$) | Nd I |
| $\mathbf{B}_{17}$ | = | $x_{5}\,\mathbf{a}_{1} - x_{5}\,\mathbf{a}_{2}$ | = | $x_{5}a\,\mathbf{\hat{x}} - x_{5}a\,\mathbf{\hat{y}}$ | (4$g$) | Nd II |
| $\mathbf{B}_{18}$ | = | $-x_{5}\,\mathbf{a}_{1} + x_{5}\,\mathbf{a}_{2}$ | = | $-x_{5}a\,\mathbf{\hat{x}} + x_{5}a\,\mathbf{\hat{y}}$ | (4$g$) | Nd II |
| $\mathbf{B}_{19}$ | = | $\left(\frac{1}{2} + x_{5}\right)\,\mathbf{a}_{1} + \left(\frac{1}{2} + x_{5}\right)\,\mathbf{a}_{2} + \frac{1}{2}\,\mathbf{a}_{3}$ | = | $\left(\frac{1}{2} + x_{5}\right)a\,\mathbf{\hat{x}} + \left(\frac{1}{2} + x_{5}\right)a\,\mathbf{\hat{y}} + \frac{1}{2}c\,\mathbf{\hat{z}}$ | (4$g$) | Nd II |
| $\mathbf{B}_{20}$ | = | $\left(\frac{1}{2} - x_{5}\right)\,\mathbf{a}_{1} + \left(\frac{1}{2} - x_{5}\right)\,\mathbf{a}_{2} + \frac{1}{2}\,\mathbf{a}_{3}$ | = | $\left(\frac{1}{2} - x_{5}\right)a\,\mathbf{\hat{x}} + \left(\frac{1}{2} - x_{5}\right)a\,\mathbf{\hat{y}} + \frac{1}{2}c\,\mathbf{\hat{z}}$ | (4$g$) | Nd II |
| $\mathbf{B}_{21}$ | = | $x_{6}\,\mathbf{a}_{1} + x_{6}\,\mathbf{a}_{2} + z_{6}\,\mathbf{a}_{3}$ | = | $x_{6}a\,\mathbf{\hat{x}} + x_{6}a\,\mathbf{\hat{y}} + z_{6}c\,\mathbf{\hat{z}}$ | (8$j$) | Fe III |
| $\mathbf{B}_{22}$ | = | $-x_{6}\,\mathbf{a}_{1} - x_{6}\,\mathbf{a}_{2} + z_{6}\,\mathbf{a}_{3}$ | = | $-x_{6}a\,\mathbf{\hat{x}} - x_{6}a\,\mathbf{\hat{y}} + z_{6}c\,\mathbf{\hat{z}}$ | (8$j$) | Fe III |
| $\mathbf{B}_{23}$ | = | $\left(\frac{1}{2} - x_{6}\right)\,\mathbf{a}_{1} + \left(\frac{1}{2} + x_{6}\right)\,\mathbf{a}_{2} + \left(\frac{1}{2} + z_{6}\right)\,\mathbf{a}_{3}$ | = | $\left(\frac{1}{2} - x_{6}\right)a\,\mathbf{\hat{x}} + \left(\frac{1}{2} + x_{6}\right)a\,\mathbf{\hat{y}} + \left(\frac{1}{2} + z_{6}\right)c\,\mathbf{\hat{z}}$ | (8$j$) | Fe III |
| $\mathbf{B}_{24}$ | = | $\left(\frac{1}{2} + x_{6}\right)\,\mathbf{a}_{1} + \left(\frac{1}{2} - x_{6}\right)\,\mathbf{a}_{2} + \left(\frac{1}{2} + z_{6}\right)\,\mathbf{a}_{3}$ | = | $\left(\frac{1}{2} + x_{6}\right)a\,\mathbf{\hat{x}} + \left(\frac{1}{2} - x_{6}\right)a\,\mathbf{\hat{y}} + \left(\frac{1}{2} + z_{6}\right)c\,\mathbf{\hat{z}}$ | (8$j$) | Fe III |
| $\mathbf{B}_{25}$ | = | $\left(\frac{1}{2} - x_{6}\right)\,\mathbf{a}_{1} + \left(\frac{1}{2} + x_{6}\right)\,\mathbf{a}_{2} + \left(\frac{1}{2} - z_{6}\right)\,\mathbf{a}_{3}$ | = | $\left(\frac{1}{2} - x_{6}\right)a\,\mathbf{\hat{x}} + \left(\frac{1}{2} + x_{6}\right)a\,\mathbf{\hat{y}} + \left(\frac{1}{2} - z_{6}\right)c\,\mathbf{\hat{z}}$ | (8$j$) | Fe III |
| $\mathbf{B}_{26}$ | = | $\left(\frac{1}{2} + x_{6}\right)\,\mathbf{a}_{1} + \left(\frac{1}{2} - x_{6}\right)\,\mathbf{a}_{2} + \left(\frac{1}{2} - z_{6}\right)\,\mathbf{a}_{3}$ | = | $\left(\frac{1}{2} + x_{6}\right)a\,\mathbf{\hat{x}} + \left(\frac{1}{2} - x_{6}\right)a\,\mathbf{\hat{y}} + \left(\frac{1}{2} - z_{6}\right)c\,\mathbf{\hat{z}}$ | (8$j$) | Fe III |
| $\mathbf{B}_{27}$ | = | $x_{6}\,\mathbf{a}_{1} + x_{6}\,\mathbf{a}_{2} - z_{6}\,\mathbf{a}_{3}$ | = | $x_{6}a\,\mathbf{\hat{x}} + x_{6}a\,\mathbf{\hat{y}} - z_{6}c\,\mathbf{\hat{z}}$ | (8$j$) | Fe III |
| $\mathbf{B}_{28}$ | = | $-x_{6}\,\mathbf{a}_{1} - x_{6}\,\mathbf{a}_{2} - z_{6}\,\mathbf{a}_{3}$ | = | $-x_{6}a\,\mathbf{\hat{x}} - x_{6}a\,\mathbf{\hat{y}} - z_{6}c\,\mathbf{\hat{z}}$ | (8$j$) | Fe III |
| $\mathbf{B}_{29}$ | = | $x_{7}\,\mathbf{a}_{1} + x_{7}\,\mathbf{a}_{2} + z_{7}\,\mathbf{a}_{3}$ | = | $x_{7}a\,\mathbf{\hat{x}} + x_{7}a\,\mathbf{\hat{y}} + z_{7}c\,\mathbf{\hat{z}}$ | (8$j$) | Fe IV |
| $\mathbf{B}_{30}$ | = | $-x_{7}\,\mathbf{a}_{1} - x_{7}\,\mathbf{a}_{2} + z_{7}\,\mathbf{a}_{3}$ | = | $-x_{7}a\,\mathbf{\hat{x}} - x_{7}a\,\mathbf{\hat{y}} + z_{7}c\,\mathbf{\hat{z}}$ | (8$j$) | Fe IV |
| $\mathbf{B}_{31}$ | = | $\left(\frac{1}{2} - x_{7}\right)\,\mathbf{a}_{1} + \left(\frac{1}{2} + x_{7}\right)\,\mathbf{a}_{2} + \left(\frac{1}{2} + z_{7}\right)\,\mathbf{a}_{3}$ | = | $\left(\frac{1}{2} - x_{7}\right)a\,\mathbf{\hat{x}} + \left(\frac{1}{2} + x_{7}\right)a\,\mathbf{\hat{y}} + \left(\frac{1}{2} + z_{7}\right)c\,\mathbf{\hat{z}}$ | (8$j$) | Fe IV |
| $\mathbf{B}_{32}$ | = | $\left(\frac{1}{2} + x_{7}\right)\,\mathbf{a}_{1} + \left(\frac{1}{2} - x_{7}\right)\,\mathbf{a}_{2} + \left(\frac{1}{2} + z_{7}\right)\,\mathbf{a}_{3}$ | = | $\left(\frac{1}{2} + x_{7}\right)a\,\mathbf{\hat{x}} + \left(\frac{1}{2} - x_{7}\right)a\,\mathbf{\hat{y}} + \left(\frac{1}{2} + z_{7}\right)c\,\mathbf{\hat{z}}$ | (8$j$) | Fe IV |
| $\mathbf{B}_{33}$ | = | $\left(\frac{1}{2} - x_{7}\right)\,\mathbf{a}_{1} + \left(\frac{1}{2} + x_{7}\right)\,\mathbf{a}_{2} + \left(\frac{1}{2} - z_{7}\right)\,\mathbf{a}_{3}$ | = | $\left(\frac{1}{2} - x_{7}\right)a\,\mathbf{\hat{x}} + \left(\frac{1}{2} + x_{7}\right)a\,\mathbf{\hat{y}} + \left(\frac{1}{2} - z_{7}\right)c\,\mathbf{\hat{z}}$ | (8$j$) | Fe IV |
| $\mathbf{B}_{34}$ | = | $\left(\frac{1}{2} + x_{7}\right)\,\mathbf{a}_{1} + \left(\frac{1}{2} - x_{7}\right)\,\mathbf{a}_{2} + \left(\frac{1}{2} - z_{7}\right)\,\mathbf{a}_{3}$ | = | $\left(\frac{1}{2} + x_{7}\right)a\,\mathbf{\hat{x}} + \left(\frac{1}{2} - x_{7}\right)a\,\mathbf{\hat{y}} + \left(\frac{1}{2} - z_{7}\right)c\,\mathbf{\hat{z}}$ | (8$j$) | Fe IV |
| $\mathbf{B}_{35}$ | = | $x_{7}\,\mathbf{a}_{1} + x_{7}\,\mathbf{a}_{2} - z_{7}\,\mathbf{a}_{3}$ | = | $x_{7}a\,\mathbf{\hat{x}} + x_{7}a\,\mathbf{\hat{y}} - z_{7}c\,\mathbf{\hat{z}}$ | (8$j$) | Fe IV |
| $\mathbf{B}_{36}$ | = | $-x_7\,\mathbf{a}_1 - x_7\,\mathbf{a}_2 - z_7\,\mathbf{a}_3$ | = | $-x_7 a\,\hat{\mathbf{x}} - x_7 a\,\hat{\mathbf{y}} - z_7 c\,\hat{\mathbf{z}}$ | $(8j)$ | Fe IV |
| $\mathbf{B}_{37}$ | = | $x_8\,\mathbf{a}_1 + y_8\,\mathbf{a}_2 + z_8\,\mathbf{a}_3$ | = | $x_8 a\,\hat{\mathbf{x}} + y_8 a\,\hat{\mathbf{y}} + z_8 c\,\hat{\mathbf{z}}$ | $(16k)$ | Fe V |
| $\mathbf{B}_{38}$ | = | $-x_8\,\mathbf{a}_1 - y_8\,\mathbf{a}_2 + z_8\,\mathbf{a}_3$ | = | $-x_8 a\,\hat{\mathbf{x}} - y_8 a\,\hat{\mathbf{y}} + z_8 c\,\hat{\mathbf{z}}$ | $(16k)$ | Fe V |
| $\mathbf{B}_{39}$ | = | $\left(\frac{1}{2} - y_8\right)\mathbf{a}_1 + \left(\frac{1}{2} + x_8\right)\mathbf{a}_2 + \left(\frac{1}{2} + z_8\right)\mathbf{a}_3$ | = | $\left(\frac{1}{2} - y_8\right)a\,\hat{\mathbf{x}} + \left(\frac{1}{2} + x_8\right)a\,\hat{\mathbf{y}} + \left(\frac{1}{2} + z_8\right)c\,\hat{\mathbf{z}}$ | $(16k)$ | Fe V |
| $\mathbf{B}_{40}$ | = | $\left(\frac{1}{2} + y_8\right)\mathbf{a}_1 + \left(\frac{1}{2} - x_8\right)\mathbf{a}_2 + \left(\frac{1}{2} + z_8\right)\mathbf{a}_3$ | = | $\left(\frac{1}{2} + y_8\right)a\,\hat{\mathbf{x}} + \left(\frac{1}{2} - x_8\right)a\,\hat{\mathbf{y}} + \left(\frac{1}{2} + z_8\right)c\,\hat{\mathbf{z}}$ | $(16k)$ | Fe V |
| $\mathbf{B}_{41}$ | = | $\left(\frac{1}{2} - x_8\right)\mathbf{a}_1 + \left(\frac{1}{2} + y_8\right)\mathbf{a}_2 + \left(\frac{1}{2} - z_8\right)\mathbf{a}_3$ | = | $\left(\frac{1}{2} - x_8\right)a\,\hat{\mathbf{x}} + \left(\frac{1}{2} + y_8\right)a\,\hat{\mathbf{y}} + \left(\frac{1}{2} - z_8\right)c\,\hat{\mathbf{z}}$ | $(16k)$ | Fe V |
| $\mathbf{B}_{42}$ | = | $\left(\frac{1}{2} + x_8\right)\mathbf{a}_1 + \left(\frac{1}{2} - y_8\right)\mathbf{a}_2 + \left(\frac{1}{2} - z_8\right)\mathbf{a}_3$ | = | $\left(\frac{1}{2} + x_8\right)a\,\hat{\mathbf{x}} + \left(\frac{1}{2} - y_8\right)a\,\hat{\mathbf{y}} + \left(\frac{1}{2} - z_8\right)c\,\hat{\mathbf{z}}$ | $(16k)$ | Fe V |
| $\mathbf{B}_{43}$ | = | $y_8\,\mathbf{a}_1 + x_8\,\mathbf{a}_2 - z_8\,\mathbf{a}_3$ | = | $y_8 a\,\hat{\mathbf{x}} + x_8 a\,\hat{\mathbf{y}} - z_8 c\,\hat{\mathbf{z}}$ | $(16k)$ | Fe V |
| $\mathbf{B}_{44}$ | = | $-y_8\,\mathbf{a}_1 - x_8\,\mathbf{a}_2 - z_8\,\mathbf{a}_3$ | = | $-y_8 a\,\hat{\mathbf{x}} - x_8 a\,\hat{\mathbf{y}} - z_8 c\,\hat{\mathbf{z}}$ | $(16k)$ | Fe V |
| $\mathbf{B}_{45}$ | = | $-x_8\,\mathbf{a}_1 - y_8\,\mathbf{a}_2 - z_8\,\mathbf{a}_3$ | = | $-x_8 a\,\hat{\mathbf{x}} - y_8 a\,\hat{\mathbf{y}} - z_8 c\,\hat{\mathbf{z}}$ | $(16k)$ | Fe V |
| $\mathbf{B}_{46}$ | = | $x_8\,\mathbf{a}_1 + y_8\,\mathbf{a}_2 - z_8\,\mathbf{a}_3$ | = | $x_8 a\,\hat{\mathbf{x}} + y_8 a\,\hat{\mathbf{y}} - z_8 c\,\hat{\mathbf{z}}$ | $(16k)$ | Fe V |
| $\mathbf{B}_{47}$ | = | $\left(\frac{1}{2} + y_8\right)\mathbf{a}_1 + \left(\frac{1}{2} - x_8\right)\mathbf{a}_2 + \left(\frac{1}{2} - z_8\right)\mathbf{a}_3$ | = | $\left(\frac{1}{2} + y_8\right)a\,\hat{\mathbf{x}} + \left(\frac{1}{2} - x_8\right)a\,\hat{\mathbf{y}} + \left(\frac{1}{2} - z_8\right)c\,\hat{\mathbf{z}}$ | $(16k)$ | Fe V |
| $\mathbf{B}_{48}$ | = | $\left(\frac{1}{2} - y_8\right)\mathbf{a}_1 + \left(\frac{1}{2} + x_8\right)\mathbf{a}_2 + \left(\frac{1}{2} - z_8\right)\mathbf{a}_3$ | = | $\left(\frac{1}{2} - y_8\right)a\,\hat{\mathbf{x}} + \left(\frac{1}{2} + x_8\right)a\,\hat{\mathbf{y}} + \left(\frac{1}{2} - z_8\right)c\,\hat{\mathbf{z}}$ | $(16k)$ | Fe V |
| $\mathbf{B}_{49}$ | = | $\left(\frac{1}{2} + x_8\right)\mathbf{a}_1 + \left(\frac{1}{2} - y_8\right)\mathbf{a}_2 + \left(\frac{1}{2} + z_8\right)\mathbf{a}_3$ | = | $\left(\frac{1}{2} + x_8\right)a\,\hat{\mathbf{x}} + \left(\frac{1}{2} - y_8\right)a\,\hat{\mathbf{y}} + \left(\frac{1}{2} + z_8\right)c\,\hat{\mathbf{z}}$ | $(16k)$ | Fe V |
| $\mathbf{B}_{50}$ | = | $\left(\frac{1}{2} - x_8\right)\mathbf{a}_1 + \left(\frac{1}{2} + y_8\right)\mathbf{a}_2 + \left(\frac{1}{2} + z_8\right)\mathbf{a}_3$ | = | $\left(\frac{1}{2} - x_8\right)a\,\hat{\mathbf{x}} + \left(\frac{1}{2} + y_8\right)a\,\hat{\mathbf{y}} + \left(\frac{1}{2} + z_8\right)c\,\hat{\mathbf{z}}$ | $(16k)$ | Fe V |
| $\mathbf{B}_{51}$ | = | $-y_8\,\mathbf{a}_1 - x_8\,\mathbf{a}_2 + z_8\,\mathbf{a}_3$ | = | $-y_8 a\,\hat{\mathbf{x}} - x_8 a\,\hat{\mathbf{y}} + z_8 c\,\hat{\mathbf{z}}$ | $(16k)$ | Fe V |
| $\mathbf{B}_{52}$ | = | $y_8\,\mathbf{a}_1 + x_8\,\mathbf{a}_2 + z_8\,\mathbf{a}_3$ | = | $y_8 a\,\hat{\mathbf{x}} + x_8 a\,\hat{\mathbf{y}} + z_8 c\,\hat{\mathbf{z}}$ | $(16k)$ | Fe V |
| $\mathbf{B}_{53}$ | = | $x_9\,\mathbf{a}_1 + y_9\,\mathbf{a}_2 + z_9\,\mathbf{a}_3$ | = | $x_9 a\,\hat{\mathbf{x}} + y_9 a\,\hat{\mathbf{y}} + z_9 c\,\hat{\mathbf{z}}$ | $(16k)$ | Fe VI |
| $\mathbf{B}_{54}$ | = | $-x_9\,\mathbf{a}_1 - y_9\,\mathbf{a}_2 + z_9\,\mathbf{a}_3$ | = | $-x_9 a\,\hat{\mathbf{x}} - y_9 a\,\hat{\mathbf{y}} + z_9 c\,\hat{\mathbf{z}}$ | $(16k)$ | Fe VI |
| $\mathbf{B}_{55}$ | = | $\left(\frac{1}{2} - y_9\right)\mathbf{a}_1 + \left(\frac{1}{2} + x_9\right)\mathbf{a}_2 + \left(\frac{1}{2} + z_9\right)\mathbf{a}_3$ | = | $\left(\frac{1}{2} - y_9\right)a\,\hat{\mathbf{x}} + \left(\frac{1}{2} + x_9\right)a\,\hat{\mathbf{y}} + \left(\frac{1}{2} + z_9\right)c\,\hat{\mathbf{z}}$ | $(16k)$ | Fe VI |
| $\mathbf{B}_{56}$ | = | $\left(\frac{1}{2} + y_9\right)\mathbf{a}_1 + \left(\frac{1}{2} - x_9\right)\mathbf{a}_2 + \left(\frac{1}{2} + z_9\right)\mathbf{a}_3$ | = | $\left(\frac{1}{2} + y_9\right)a\,\hat{\mathbf{x}} + \left(\frac{1}{2} - x_9\right)a\,\hat{\mathbf{y}} + \left(\frac{1}{2} + z_9\right)c\,\hat{\mathbf{z}}$ | $(16k)$ | Fe VI |
| $\mathbf{B}_{57}$ | = | $\left(\frac{1}{2} - x_9\right)\mathbf{a}_1 + \left(\frac{1}{2} + y_9\right)\mathbf{a}_2 + \left(\frac{1}{2} - z_9\right)\mathbf{a}_3$ | = | $\left(\frac{1}{2} - x_9\right)a\,\hat{\mathbf{x}} + \left(\frac{1}{2} + y_9\right)a\,\hat{\mathbf{y}} + \left(\frac{1}{2} - z_9\right)c\,\hat{\mathbf{z}}$ | $(16k)$ | Fe VI |
| $\mathbf{B}_{58}$ | = | $\left(\frac{1}{2} + x_9\right)\mathbf{a}_1 + \left(\frac{1}{2} - y_9\right)\mathbf{a}_2 + \left(\frac{1}{2} - z_9\right)\mathbf{a}_3$ | = | $\left(\frac{1}{2} + x_9\right)a\,\hat{\mathbf{x}} + \left(\frac{1}{2} - y_9\right)a\,\hat{\mathbf{y}} + \left(\frac{1}{2} - z_9\right)c\,\hat{\mathbf{z}}$ | $(16k)$ | Fe VI |
| $\mathbf{B}_{59}$ | = | $y_9\,\mathbf{a}_1 + x_9\,\mathbf{a}_2 - z_9\,\mathbf{a}_3$ | = | $y_9 a\,\hat{\mathbf{x}} + x_9 a\,\hat{\mathbf{y}} - z_9 c\,\hat{\mathbf{z}}$ | $(16k)$ | Fe VI |
| $\mathbf{B}_{60}$ | = | $-y_9\,\mathbf{a}_1 - x_9\,\mathbf{a}_2 - z_9\,\mathbf{a}_3$ | = | $-y_9 a\,\hat{\mathbf{x}} - x_9 a\,\hat{\mathbf{y}} - z_9 c\,\hat{\mathbf{z}}$ | $(16k)$ | Fe VI |
| $\mathbf{B}_{61}$ | = | $-x_9\,\mathbf{a}_1 - y_9\,\mathbf{a}_2 - z_9\,\mathbf{a}_3$ | = | $-x_9 a\,\hat{\mathbf{x}} - y_9 a\,\hat{\mathbf{y}} - z_9 c\,\hat{\mathbf{z}}$ | $(16k)$ | Fe VI |
| $\mathbf{B}_{62}$ | = | $x_9\,\mathbf{a}_1 + y_9\,\mathbf{a}_2 - z_9\,\mathbf{a}_3$ | = | $x_9 a\,\hat{\mathbf{x}} + y_9 a\,\hat{\mathbf{y}} - z_9 c\,\hat{\mathbf{z}}$ | $(16k)$ | Fe VI |
| $\mathbf{B}_{63}$ | = | $\left(\frac{1}{2}+y_9\right)\mathbf{a}_1+\left(\frac{1}{2}-x_9\right)\mathbf{a}_2+\left(\frac{1}{2}-z_9\right)\mathbf{a}_3$ | = | $\left(\frac{1}{2}+y_9\right)a\,\hat{\mathbf{x}}+\left(\frac{1}{2}-x_9\right)a\,\hat{\mathbf{y}}+\left(\frac{1}{2}-z_9\right)c\,\hat{\mathbf{z}}$ | (16$k$) | Fe VI |
| $\mathbf{B}_{64}$ | = | $\left(\frac{1}{2}-y_9\right)\mathbf{a}_1+\left(\frac{1}{2}+x_9\right)\mathbf{a}_2+\left(\frac{1}{2}-z_9\right)\mathbf{a}_3$ | = | $\left(\frac{1}{2}-y_9\right)a\,\hat{\mathbf{x}}+\left(\frac{1}{2}+x_9\right)a\,\hat{\mathbf{y}}+\left(\frac{1}{2}-z_9\right)c\,\hat{\mathbf{z}}$ | (16$k$) | Fe VI |
| $\mathbf{B}_{65}$ | = | $\left(\frac{1}{2}+x_9\right)\mathbf{a}_1+\left(\frac{1}{2}-y_9\right)\mathbf{a}_2+\left(\frac{1}{2}+z_9\right)\mathbf{a}_3$ | = | $\left(\frac{1}{2}+x_9\right)a\,\hat{\mathbf{x}}+\left(\frac{1}{2}-y_9\right)a\,\hat{\mathbf{y}}+\left(\frac{1}{2}+z_9\right)c\,\hat{\mathbf{z}}$ | (16$k$) | Fe VI |
| $\mathbf{B}_{66}$ | = | $\left(\frac{1}{2}-x_9\right)\mathbf{a}_1+\left(\frac{1}{2}+y_9\right)\mathbf{a}_2+\left(\frac{1}{2}+z_9\right)\mathbf{a}_3$ | = | $\left(\frac{1}{2}-x_9\right)a\,\hat{\mathbf{x}}+\left(\frac{1}{2}+y_9\right)a\,\hat{\mathbf{y}}+\left(\frac{1}{2}+z_9\right)c\,\hat{\mathbf{z}}$ | (16$k$) | Fe VI |
| $\mathbf{B}_{67}$ | = | $-y_9\,\mathbf{a}_1-x_9\,\mathbf{a}_2+z_9\,\mathbf{a}_3$ | = | $-y_9a\,\hat{\mathbf{x}}-x_9a\,\hat{\mathbf{y}}+z_9c\,\hat{\mathbf{z}}$ | (16$k$) | Fe VI |
| $\mathbf{B}_{68}$ | = | $y_9\,\mathbf{a}_1+x_9\,\mathbf{a}_2+z_9\,\mathbf{a}_3$ | = | $y_9a\,\hat{\mathbf{x}}+x_9a\,\hat{\mathbf{y}}+z_9c\,\hat{\mathbf{z}}$ | (16$k$) | Fe VI |

**Geometry files:**

- CIF: pp. 1704
- POSCAR: pp. 1704

# $Sr_4Ti_3O_{10}$ Structure: A10B4C3_tI34_139_c2eg_2e_ae

http://aflow.org/prototype-encyclopedia/A10B4C3_tI34_139_c2eg_2e_ae

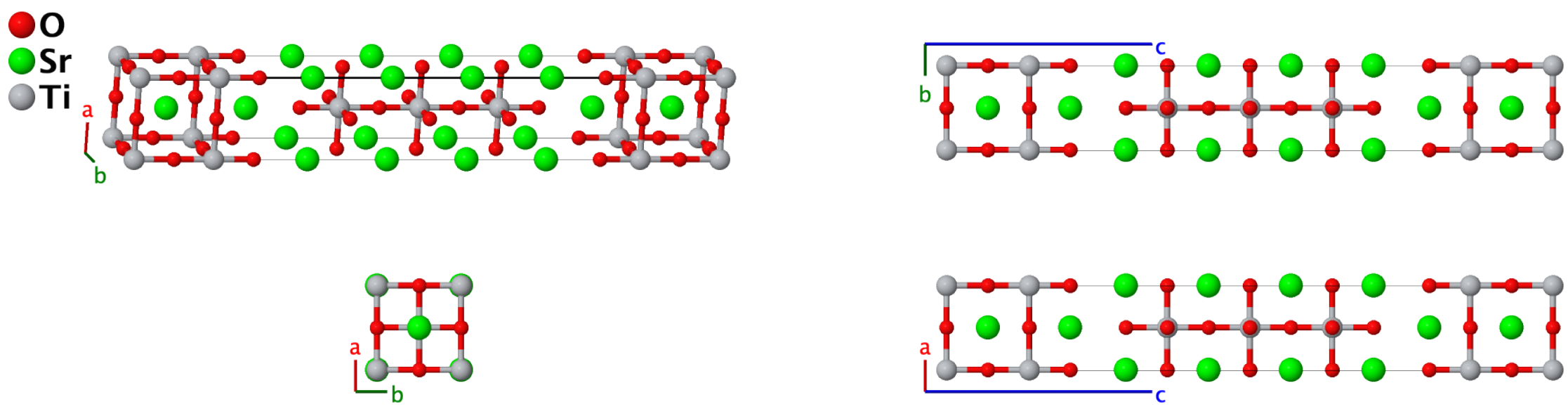

| | | |
|---|---|---|
| **Prototype** | : | $O_{10}Sr_4Ti_3$ |
| **AFLOW prototype label** | : | A10B4C3_tI34_139_c2eg_2e_ae |
| ***Strukturbericht* designation** | : | None |
| **Pearson symbol** | : | tI34 |
| **Space group number** | : | 139 |
| **Space group symbol** | : | $I4/mmm$ |
| **AFLOW prototype command** | : | `aflow --proto=A10B4C3_tI34_139_c2eg_2e_ae --params=`$a, c/a, z_3, z_4, z_5, z_6, z_7, z_8$ |

**Other compounds with this structure**

- $K_2La_2Ti_3O_{10}$, $Li_2Eu_2Ti_3O_{10}$, $Na_2Eu_2Ti_3O_{10}$, and $Na_2Sr_2Nb_2MnO_{10}$

**Body-centered Tetragonal primitive vectors:**

$$\mathbf{a}_1 = -\frac{1}{2}\,a\,\hat{\mathbf{x}} + \frac{1}{2}\,a\,\hat{\mathbf{y}} + \frac{1}{2}\,c\,\hat{\mathbf{z}}$$
$$\mathbf{a}_2 = \frac{1}{2}\,a\,\hat{\mathbf{x}} - \frac{1}{2}\,a\,\hat{\mathbf{y}} + \frac{1}{2}\,c\,\hat{\mathbf{z}}$$
$$\mathbf{a}_3 = \frac{1}{2}\,a\,\hat{\mathbf{x}} + \frac{1}{2}\,a\,\hat{\mathbf{y}} - \frac{1}{2}\,c\,\hat{\mathbf{z}}$$

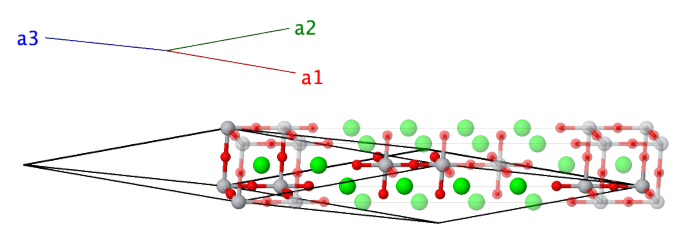

**Basis vectors:**

| | | Lattice Coordinates | | Cartesian Coordinates | Wyckoff Position | Atom Type |
|---|---|---|---|---|---|---|
| $\mathbf{B}_1$ | = | $0\,\mathbf{a}_1 + 0\,\mathbf{a}_2 + 0\,\mathbf{a}_3$ | = | $0\,\hat{\mathbf{x}} + 0\,\hat{\mathbf{y}} + 0\,\hat{\mathbf{z}}$ | $(2a)$ | Ti I |
| $\mathbf{B}_2$ | = | $\frac{1}{2}\,\mathbf{a}_1 + \frac{1}{2}\,\mathbf{a}_3$ | = | $\frac{1}{2}a\,\hat{\mathbf{y}}$ | $(4c)$ | O I |
| $\mathbf{B}_3$ | = | $\frac{1}{2}\,\mathbf{a}_2 + \frac{1}{2}\,\mathbf{a}_3$ | = | $\frac{1}{2}a\,\hat{\mathbf{x}}$ | $(4c)$ | O I |
| $\mathbf{B}_4$ | = | $z_3\,\mathbf{a}_1 + z_3\,\mathbf{a}_2$ | = | $z_3c\,\hat{\mathbf{z}}$ | $(4e)$ | O II |
| $\mathbf{B}_5$ | = | $-z_3\,\mathbf{a}_1 - z_3\,\mathbf{a}_2$ | = | $-z_3c\,\hat{\mathbf{z}}$ | $(4e)$ | O II |
| $\mathbf{B}_6$ | = | $z_4\,\mathbf{a}_1 + z_4\,\mathbf{a}_2$ | = | $z_4c\,\hat{\mathbf{z}}$ | $(4e)$ | O III |
| $\mathbf{B}_7$ | = | $-z_4\,\mathbf{a}_1 - z_4\,\mathbf{a}_2$ | = | $-z_4c\,\hat{\mathbf{z}}$ | $(4e)$ | O III |
| $\mathbf{B}_8$ | = | $z_5\,\mathbf{a}_1 + z_5\,\mathbf{a}_2$ | = | $z_5c\,\hat{\mathbf{z}}$ | $(4e)$ | Sr I |
| $\mathbf{B}_9$ | = | $-z_5\,\mathbf{a}_1 - z_5\,\mathbf{a}_2$ | = | $-z_5c\,\hat{\mathbf{z}}$ | $(4e)$ | Sr I |
| $\mathbf{B}_{10}$ | = | $z_6\,\mathbf{a}_1 + z_6\,\mathbf{a}_2$ | = | $z_6c\,\hat{\mathbf{z}}$ | $(4e)$ | Sr II |
| $\mathbf{B}_{11}$ | = | $-z_6\,\mathbf{a}_1 - z_6\,\mathbf{a}_2$ | = | $-z_6c\,\hat{\mathbf{z}}$ | $(4e)$ | Sr II |

$$\mathbf{B}_{12} = z_7\,\mathbf{a}_1 + z_7\,\mathbf{a}_2 = z_7 c\,\hat{\mathbf{z}} \quad (4e) \quad \text{Ti II}$$

$$\mathbf{B}_{13} = -z_7\,\mathbf{a}_1 - z_7\,\mathbf{a}_2 = -z_7 c\,\hat{\mathbf{z}} \quad (4e) \quad \text{Ti II}$$

$$\mathbf{B}_{14} = \left(\frac{1}{2} + z_8\right)\mathbf{a}_1 + z_8\,\mathbf{a}_2 + \frac{1}{2}\,\mathbf{a}_3 = \frac{1}{2}a\,\hat{\mathbf{y}} + z_8 c\,\hat{\mathbf{z}} \quad (8g) \quad \text{O IV}$$

$$\mathbf{B}_{15} = z_8\,\mathbf{a}_1 + \left(\frac{1}{2} + z_8\right)\mathbf{a}_2 + \frac{1}{2}\,\mathbf{a}_3 = \frac{1}{2}a\,\hat{\mathbf{x}} + z_8 c\,\hat{\mathbf{z}} \quad (8g) \quad \text{O IV}$$

$$\mathbf{B}_{16} = \left(\frac{1}{2} - z_8\right)\mathbf{a}_1 - z_8\,\mathbf{a}_2 + \frac{1}{2}\,\mathbf{a}_3 = \frac{1}{2}a\,\hat{\mathbf{y}} - z_8 c\,\hat{\mathbf{z}} \quad (8g) \quad \text{O IV}$$

$$\mathbf{B}_{17} = -z_8\,\mathbf{a}_1 + \left(\frac{1}{2} - z_8\right)\mathbf{a}_2 + \frac{1}{2}\,\mathbf{a}_3 = \frac{1}{2}a\,\hat{\mathbf{x}} - z_8 c\,\hat{\mathbf{z}} \quad (8g) \quad \text{O IV}$$

---

**Found in:**

- Wikipedia, *Ruddlesden-Popper phase*, `https://en.wikipedia.org/wiki/Ruddlesden-Popper_phase`. $A_3B_2X_7$ series.

---

**Geometry files:**

- CIF: pp. 1705
- POSCAR: pp. 1705

# $TlCo_2S_2$ Structure: A2B2C_tI10_139_d_e_a

http://aflow.org/prototype-encyclopedia/A2B2C_tI10_139_d_e_a.TlCo2S2

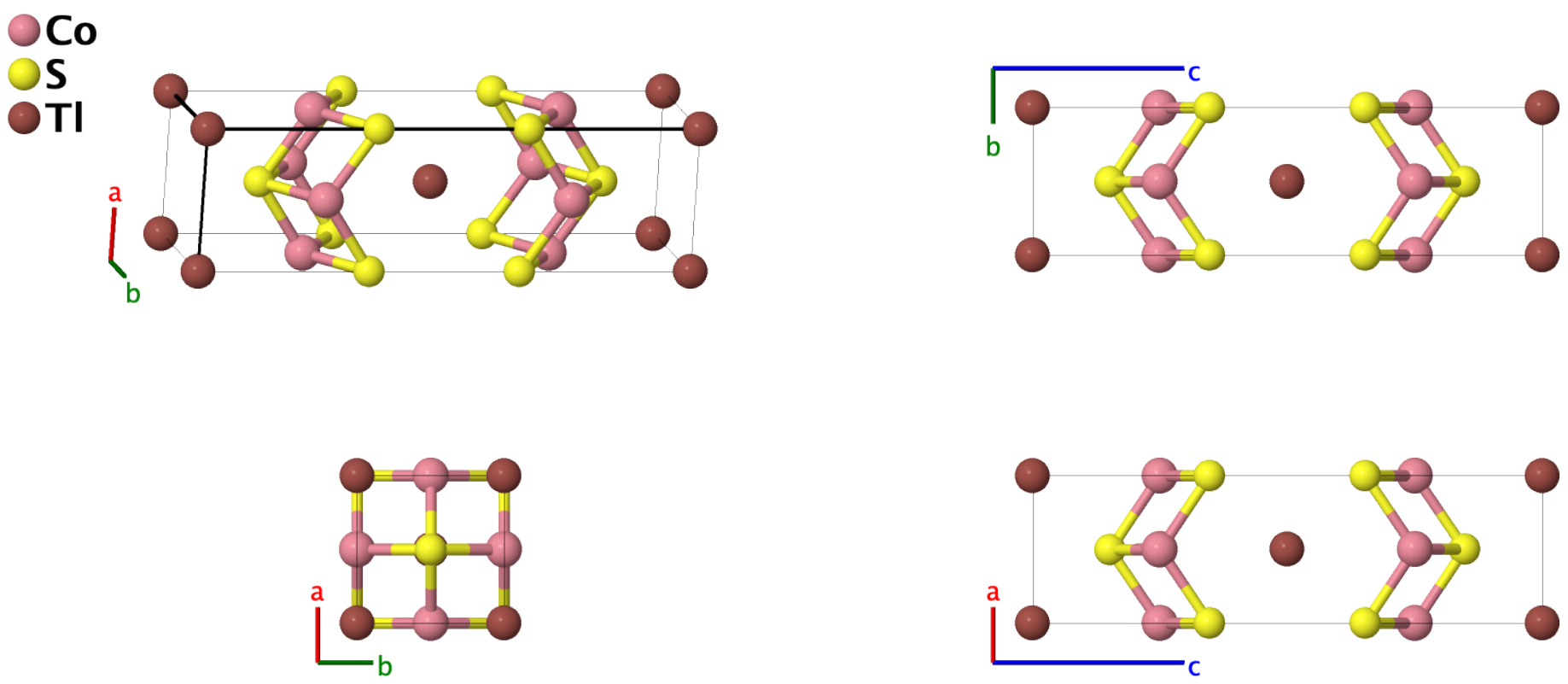

| | | |
|---|---|---|
| **Prototype** | : | $Co_2S_2Tl$ |
| **AFLOW prototype label** | : | A2B2C_tI10_139_d_e_a |
| ***Strukturbericht* designation** | : | None |
| **Pearson symbol** | : | tI10 |
| **Space group number** | : | 139 |
| **Space group symbol** | : | $I4/mmm$ |
| **AFLOW prototype command** | : | `aflow --proto=A2B2C_tI10_139_d_e_a`<br>`--params=`$a, c/a, z_3$ |

### Other compounds with this structure

- $BaCr_2As_2$, $BaFe_2As_2$, $BaP_2Zn_2$, $BaSr_2As_2$, $BiCe_2O_2$, $KCo_2Se_2$, $TeCe_2O_2$, $TlCo_2Se_2$, $TlCu_2S_2$, $TlCu_2Se_2$, $TlCu_2Te_2$, $TlFe_2S_2$, $TlFe_2Se_2$, $TlNi_2S_2$, $TlNi_2Se_2$, and $Cl(Li_{0.25}Be_{0.75})_2O_2$

- This is a ternary form of the $D1_3$ ($BaAl_4$) structure. It differs from the $ThCr_2Si_2$ structure in that here $c/a > 3$.
- Other authors designate $TlCu_2Se_2$ or $Cl(Li_{0.25}Be_{0.75})_2O_2$ as the prototype for this structure, but (Klepp, 1978) give complete structural information for $TlCo_2S_2$, so we use that as the prototype.

### Body-centered Tetragonal primitive vectors:

$$\mathbf{a}_1 = -\frac{1}{2}\,a\,\mathbf{\hat{x}} + \frac{1}{2}\,a\,\mathbf{\hat{y}} + \frac{1}{2}\,c\,\mathbf{\hat{z}}$$
$$\mathbf{a}_2 = \frac{1}{2}\,a\,\mathbf{\hat{x}} - \frac{1}{2}\,a\,\mathbf{\hat{y}} + \frac{1}{2}\,c\,\mathbf{\hat{z}}$$
$$\mathbf{a}_3 = \frac{1}{2}\,a\,\mathbf{\hat{x}} + \frac{1}{2}\,a\,\mathbf{\hat{y}} - \frac{1}{2}\,c\,\mathbf{\hat{z}}$$

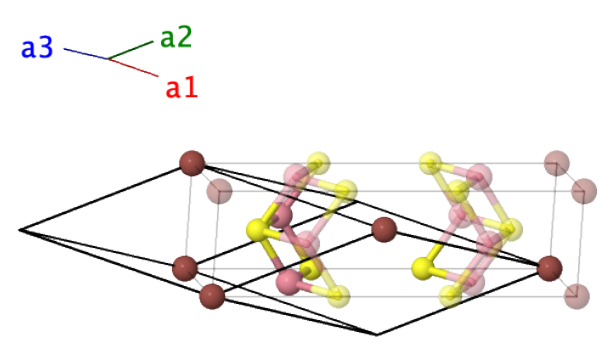

### Basis vectors:

| | | Lattice Coordinates | | Cartesian Coordinates | Wyckoff Position | Atom Type |
|---|---|---|---|---|---|---|
| $\mathbf{B}_1$ | = | $0\,\mathbf{a}_1 + 0\,\mathbf{a}_2 + 0\,\mathbf{a}_3$ | = | $0\,\mathbf{\hat{x}} + 0\,\mathbf{\hat{y}} + 0\,\mathbf{\hat{z}}$ | (2$a$) | Tl |
| $\mathbf{B}_2$ | = | $\frac{3}{4}\,\mathbf{a}_1 + \frac{1}{4}\,\mathbf{a}_2 + \frac{1}{2}\,\mathbf{a}_3$ | = | $\frac{1}{2}a\,\mathbf{\hat{y}} + \frac{1}{4}c\,\mathbf{\hat{z}}$ | (4$d$) | Co |
| $\mathbf{B}_3$ | = | $\frac{1}{4}\,\mathbf{a}_1 + \frac{3}{4}\,\mathbf{a}_2 + \frac{1}{2}\,\mathbf{a}_3$ | = | $\frac{1}{2}a\,\mathbf{\hat{x}} + \frac{1}{4}c\,\mathbf{\hat{z}}$ | (4$d$) | Co |
| $\mathbf{B}_4$ | = | $z_3\,\mathbf{a}_1 + z_3\,\mathbf{a}_2$ | = | $z_3c\,\mathbf{\hat{z}}$ | (4$e$) | S |
| $\mathbf{B}_5$ | = | $-z_3\,\mathbf{a}_1 - z_3\,\mathbf{a}_2$ | = | $-z_3c\,\mathbf{\hat{z}}$ | (4$e$) | S |

**Found in:**

---

**Geometry files:**

- CIF: pp. 1705
- POSCAR: pp. 1706

# $ThCr_2Si_2$ Structure: A2B2C_tI10_139_d_e_a

http://aflow.org/prototype-encyclopedia/A2B2C_tI10_139_d_e_a

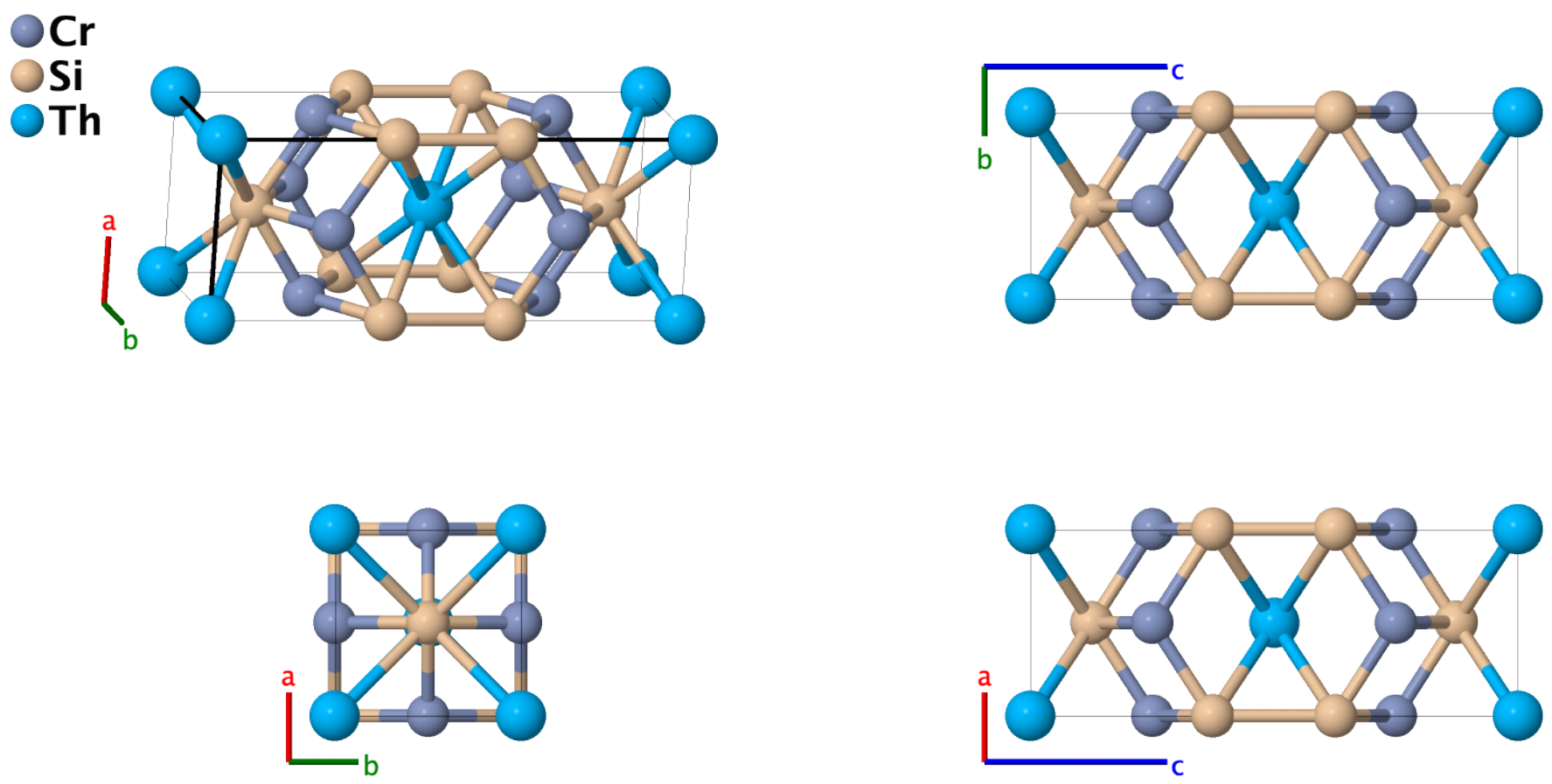

| | | |
|---|---|---|
| **Prototype** | : | $Cr_2Si_2Th$ |
| **AFLOW prototype label** | : | A2B2C_tI10_139_d_e_a |
| ***Strukturbericht* designation** | : | None |
| **Pearson symbol** | : | tI10 |
| **Space group number** | : | 139 |
| **Space group symbol** | : | $I4/mmm$ |
| **AFLOW prototype command** | : | `aflow --proto=A2B2C_tI10_139_d_e_a --params=`$a, c/a, z_3$ |

### Other compounds with this structure

- $BaAl_2Ge_2$, $BaFe_2As_2$, $BaFe_2P_2$, $BaMn_2Ge_2$, $BaNi_2As_2$, $BaNi_2Ge_2$, $BaRh_2B_2$, $CaAl_2Zn_2$, $CaAu_2Si_2$, $CaFe_2As_2$, $CaMn_2Ge_2$, $CaNi_2Ge_2$, $CeAl_2Ga_2$, $CsCo_2As_2$, $CsCo_2P_2$, $CsFe_2As_2$, $CsFe_2P_2$, $CsIr_2As_2$, $CsIr_2P_2$, $CsRh_2As_2$, $CsRh_2P_2$, $CsRu_2As_2$, $CsRu_2P_2$, $DyCr_2Si_2$, $EuCo_2As_2$, $EuCo_2P_2$, $EuFe_2P_2$, $EuRh_2P_2$, $KNi_2S_2$, $KNi_2Se_2$, $LaCo_2P_2$, $LaRu_2P_2$, $LuFe_2B_2$, $LuRu_2Si_2$, $PuCr_2Si_2$, $SrCo_2P_2$, $SrFe_2As_2$, $SrFe_2P_2$, $SrNi_2P_2$, $SrRh_2P_2$, $SrRu_2P_2$, $ThCu_2Si_2$, $ThMn_2Ge_2$, $ThMn_2Si_2$, $ThNi_2Si_2$, $UCr_2Si_2$, and $YNi_2Ge_2$

---

- (Shatruk, 2019) refers to this as "the perovskite of intermetallics." The list of compounds above is by no means complete.
- This is a ternary form of the $D1_3$ ($BaAl_4$) structure.
- The structures generally identified as having prototype $ThCr_2Si_2$ actually divide into two distinct regions, based on $c/a$ ratio. We assign structures with $c/a \leq 3$ to the $ThCr_2Si_2$ structure, and those with $c/a > 3$ to the $TlCo_2S_2$ structure.

### Body-centered Tetragonal primitive vectors:

$$
\begin{aligned}
\mathbf{a}_1 &= -\tfrac{1}{2}\,a\,\mathbf{\hat{x}} + \tfrac{1}{2}\,a\,\mathbf{\hat{y}} + \tfrac{1}{2}\,c\,\mathbf{\hat{z}} \\
\mathbf{a}_2 &= \tfrac{1}{2}\,a\,\mathbf{\hat{x}} - \tfrac{1}{2}\,a\,\mathbf{\hat{y}} + \tfrac{1}{2}\,c\,\mathbf{\hat{z}} \\
\mathbf{a}_3 &= \tfrac{1}{2}\,a\,\mathbf{\hat{x}} + \tfrac{1}{2}\,a\,\mathbf{\hat{y}} - \tfrac{1}{2}\,c\,\mathbf{\hat{z}}
\end{aligned}
$$

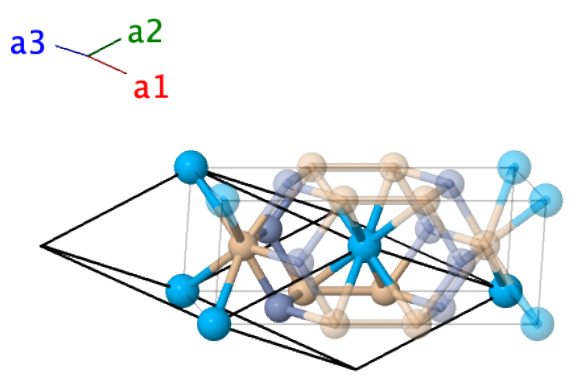

### Basis vectors:

| | | Lattice Coordinates | | Cartesian Coordinates | Wyckoff Position | Atom Type |
|---|---|---|---|---|---|---|
| $\mathbf{B}_1$ | = | $0\,\mathbf{a}_1 + 0\,\mathbf{a}_2 + 0\,\mathbf{a}_3$ | = | $0\,\hat{\mathbf{x}} + 0\,\hat{\mathbf{y}} + 0\,\hat{\mathbf{z}}$ | $(2a)$ | Th |
| $\mathbf{B}_2$ | = | $\frac{3}{4}\,\mathbf{a}_1 + \frac{1}{4}\,\mathbf{a}_2 + \frac{1}{2}\,\mathbf{a}_3$ | = | $\frac{1}{2}a\,\hat{\mathbf{y}} + \frac{1}{4}c\,\hat{\mathbf{z}}$ | $(4d)$ | Cr |
| $\mathbf{B}_3$ | = | $\frac{1}{4}\,\mathbf{a}_1 + \frac{3}{4}\,\mathbf{a}_2 + \frac{1}{2}\,\mathbf{a}_3$ | = | $\frac{1}{2}a\,\hat{\mathbf{x}} + \frac{1}{4}c\,\hat{\mathbf{z}}$ | $(4d)$ | Cr |
| $\mathbf{B}_4$ | = | $z_3\,\mathbf{a}_1 + z_3\,\mathbf{a}_2$ | = | $z_3c\,\hat{\mathbf{z}}$ | $(4e)$ | Si |
| $\mathbf{B}_5$ | = | $-z_3\,\mathbf{a}_1 - z_3\,\mathbf{a}_2$ | = | $-z_3c\,\hat{\mathbf{z}}$ | $(4e)$ | Si |

**Geometry files:**

- CIF: pp. 1706
- POSCAR: pp. 1706

# $Au_2Nb_3$ Structure: A2B3_tI10_139_e_ae

http://aflow.org/prototype-encyclopedia/A2B3_tI10_139_e_ae

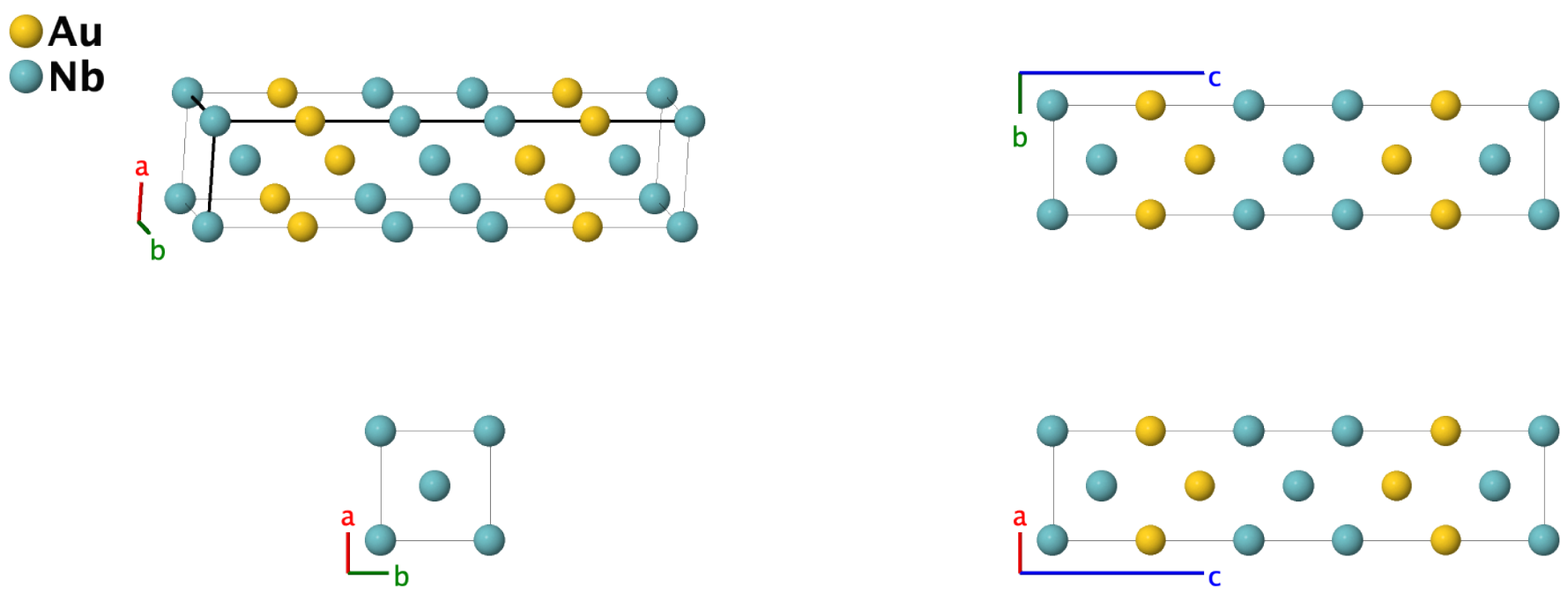

| | | |
|---|---|---|
| **Prototype** | : | $Au_2Nb_3$ |
| **AFLOW prototype label** | : | A2B3_tI10_139_e_ae |
| ***Strukturbericht* designation** | : | None |
| **Pearson symbol** | : | tI10 |
| **Space group number** | : | 139 |
| **Space group symbol** | : | $I4/mmm$ |
| **AFLOW prototype command** | : | `aflow --proto=A2B3_tI10_139_e_ae --params=`$a, c/a, z_2, z_3$ |

**Other compounds with this structure**

- $Os_2Al_3$

- (Schubert, 1960) listed the lattice constants in "kX" units. We used the conversion factor 1/1.00207789 to convert their units into Ångstroms (Arblaster, 1997).
- When $c/a = 5$, $z_2 = 1/5$ and $z_3 = 2/5$, the atoms are on the sites of a body-centered cubic lattice.

**Body-centered Tetragonal primitive vectors:**

$$\mathbf{a}_1 = -\frac{1}{2}\,a\,\hat{\mathbf{x}} + \frac{1}{2}\,a\,\hat{\mathbf{y}} + \frac{1}{2}\,c\,\hat{\mathbf{z}}$$
$$\mathbf{a}_2 = \frac{1}{2}\,a\,\hat{\mathbf{x}} - \frac{1}{2}\,a\,\hat{\mathbf{y}} + \frac{1}{2}\,c\,\hat{\mathbf{z}}$$
$$\mathbf{a}_3 = \frac{1}{2}\,a\,\hat{\mathbf{x}} + \frac{1}{2}\,a\,\hat{\mathbf{y}} - \frac{1}{2}\,c\,\hat{\mathbf{z}}$$

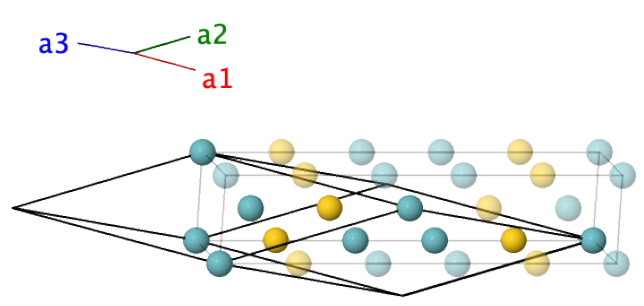

**Basis vectors:**

| | | Lattice Coordinates | | Cartesian Coordinates | Wyckoff Position | Atom Type |
|---|---|---|---|---|---|---|
| $\mathbf{B}_1$ | = | $0\,\mathbf{a}_1 + 0\,\mathbf{a}_2 + 0\,\mathbf{a}_3$ | = | $0\,\hat{\mathbf{x}} + 0\,\hat{\mathbf{y}} + 0\,\hat{\mathbf{z}}$ | (2$a$) | Nb I |
| $\mathbf{B}_2$ | = | $z_2\,\mathbf{a}_1 + z_2\,\mathbf{a}_2$ | = | $z_2 c\,\hat{\mathbf{z}}$ | (4$e$) | Au |
| $\mathbf{B}_3$ | = | $-z_2\,\mathbf{a}_1 - z_2\,\mathbf{a}_2$ | = | $-z_2 c\,\hat{\mathbf{z}}$ | (4$e$) | Au |
| $\mathbf{B}_4$ | = | $z_3\,\mathbf{a}_1 + z_3\,\mathbf{a}_2$ | = | $z_3 c\,\hat{\mathbf{z}}$ | (4$e$) | Nb II |
| $\mathbf{B}_5$ | = | $-z_3\,\mathbf{a}_1 - z_3\,\mathbf{a}_2$ | = | $-z_3 c\,\hat{\mathbf{z}}$ | (4$e$) | Nb II |

**Found in:**

---

**Geometry files:**

- CIF: pp. 1706
- POSCAR: pp. 1707

# $CaC_2$-I ($C11_a$) Structure: A2B_tI6_139_e_a

http://aflow.org/prototype-encyclopedia/A2B_tI6_139_e_a

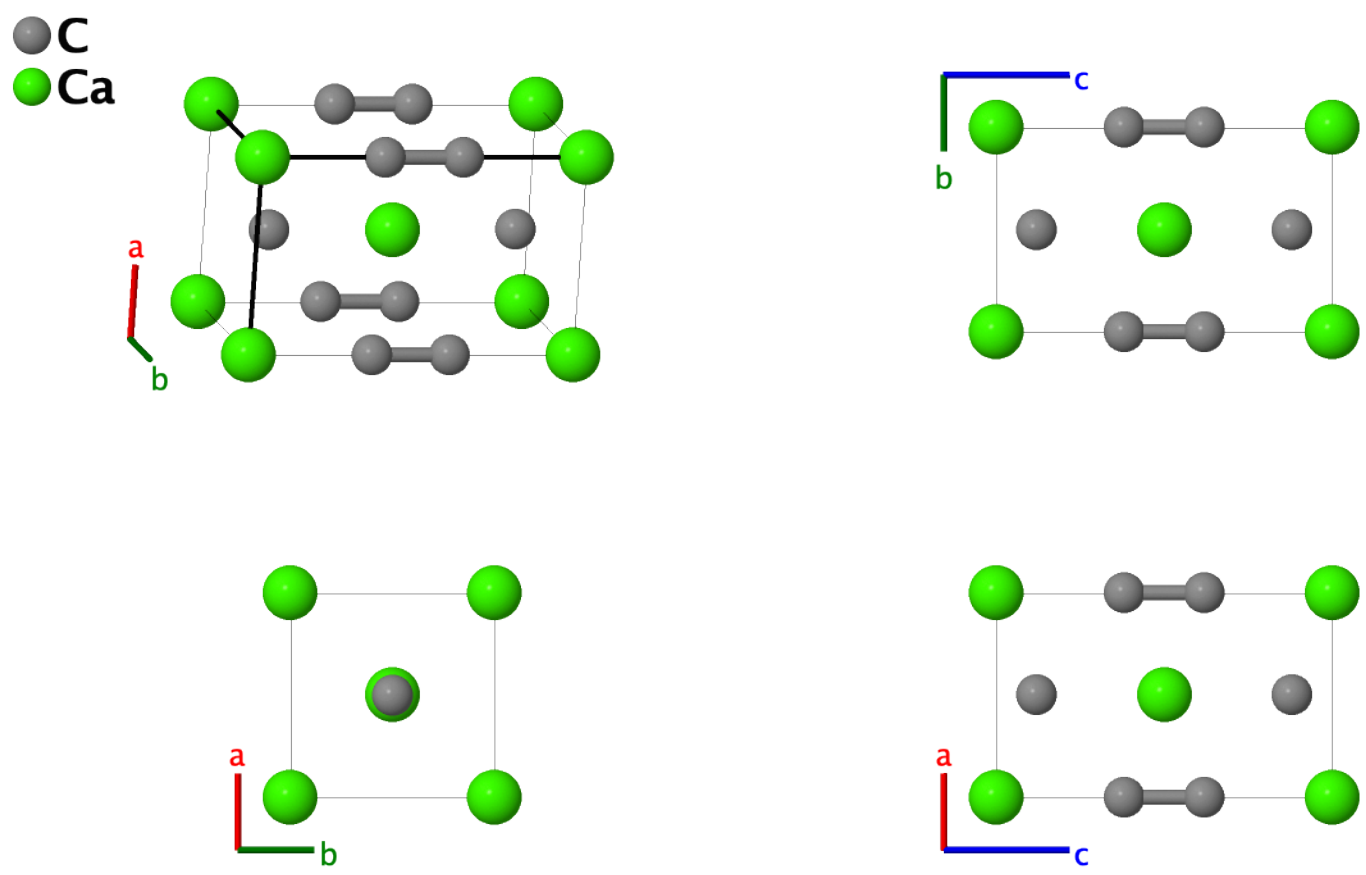

| | | |
|---|---|---|
| **Prototype** | : | $C_2Ca$ |
| **AFLOW prototype label** | : | A2B_tI6_139_e_a |
| ***Strukturbericht* designation** | : | $C11_a$ |
| **Pearson symbol** | : | tI6 |
| **Space group number** | : | 139 |
| **Space group symbol** | : | $I4/mmm$ |
| **AFLOW prototype command** | : | `aflow --proto=A2B_tI6_139_e_a`<br>`--params=`$a, c/a, z_2$ |

- (Ewald, 1931) designated both $CaC_2$ and $MoSi_2$ as *Strukturbericht* $C11$. (Smithells,1955) separated them into $C11_a$ ($CaC_2$) and $C11_b$ ($MoSi_2$, AB2_tI6_139_a_e).
- This is the stable room-temperature structure for $CaC_2$. There is also a meta-stable room temperature structure $CaCl_2$-III (A2B_mC12_12_2i_i), and a low-temperature structure $CaCl_2$, which has the $ThC_2$ ($C_g$) structure.
- (v. Stackelberg, 1930) describes this structure in a face-centered-tetragonal setting. We follow (Smithells, 1955) and place it in the equivalent body-centered tetragonal setting.

**Body-centered Tetragonal primitive vectors:**

$$\mathbf{a}_1 = -\frac{1}{2}\,a\,\hat{\mathbf{x}} + \frac{1}{2}\,a\,\hat{\mathbf{y}} + \frac{1}{2}\,c\,\hat{\mathbf{z}}$$
$$\mathbf{a}_2 = \frac{1}{2}\,a\,\hat{\mathbf{x}} - \frac{1}{2}\,a\,\hat{\mathbf{y}} + \frac{1}{2}\,c\,\hat{\mathbf{z}}$$
$$\mathbf{a}_3 = \frac{1}{2}\,a\,\hat{\mathbf{x}} + \frac{1}{2}\,a\,\hat{\mathbf{y}} - \frac{1}{2}\,c\,\hat{\mathbf{z}}$$

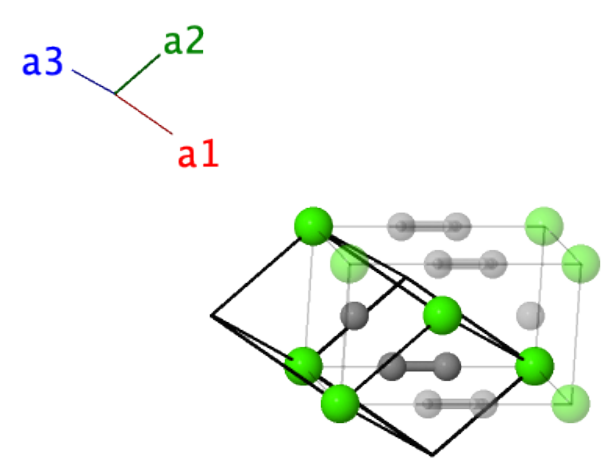

**Basis vectors:**

| | | Lattice Coordinates | | Cartesian Coordinates | Wyckoff Position | Atom Type |
|---|---|---|---|---|---|---|
| $\mathbf{B}_1$ | = | $0\,\mathbf{a}_1 + 0\,\mathbf{a}_2 + 0\,\mathbf{a}_3$ | = | $0\,\hat{\mathbf{x}} + 0\,\hat{\mathbf{y}} + 0\,\hat{\mathbf{z}}$ | (2$a$) | Ca |
| $\mathbf{B}_2$ | = | $z_2\,\mathbf{a}_1 + z_2\,\mathbf{a}_2$ | = | $z_2 c\,\hat{\mathbf{z}}$ | (4$e$) | C |
| $\mathbf{B}_3$ | = | $-z_2\,\mathbf{a}_1 - z_2\,\mathbf{a}_2$ | = | $-z_2 c\,\hat{\mathbf{z}}$ | (4$e$) | C |

**Geometry files:**

- CIF: pp. 1707
- POSCAR: pp. 1707

# $K_2OsO_2Cl_4$ ($J1_5$) Structure: A4B2C2D_tI18_139_h_d_e_a

http://aflow.org/prototype-encyclopedia/A4B2C2D_tI18_139_h_d_e_a

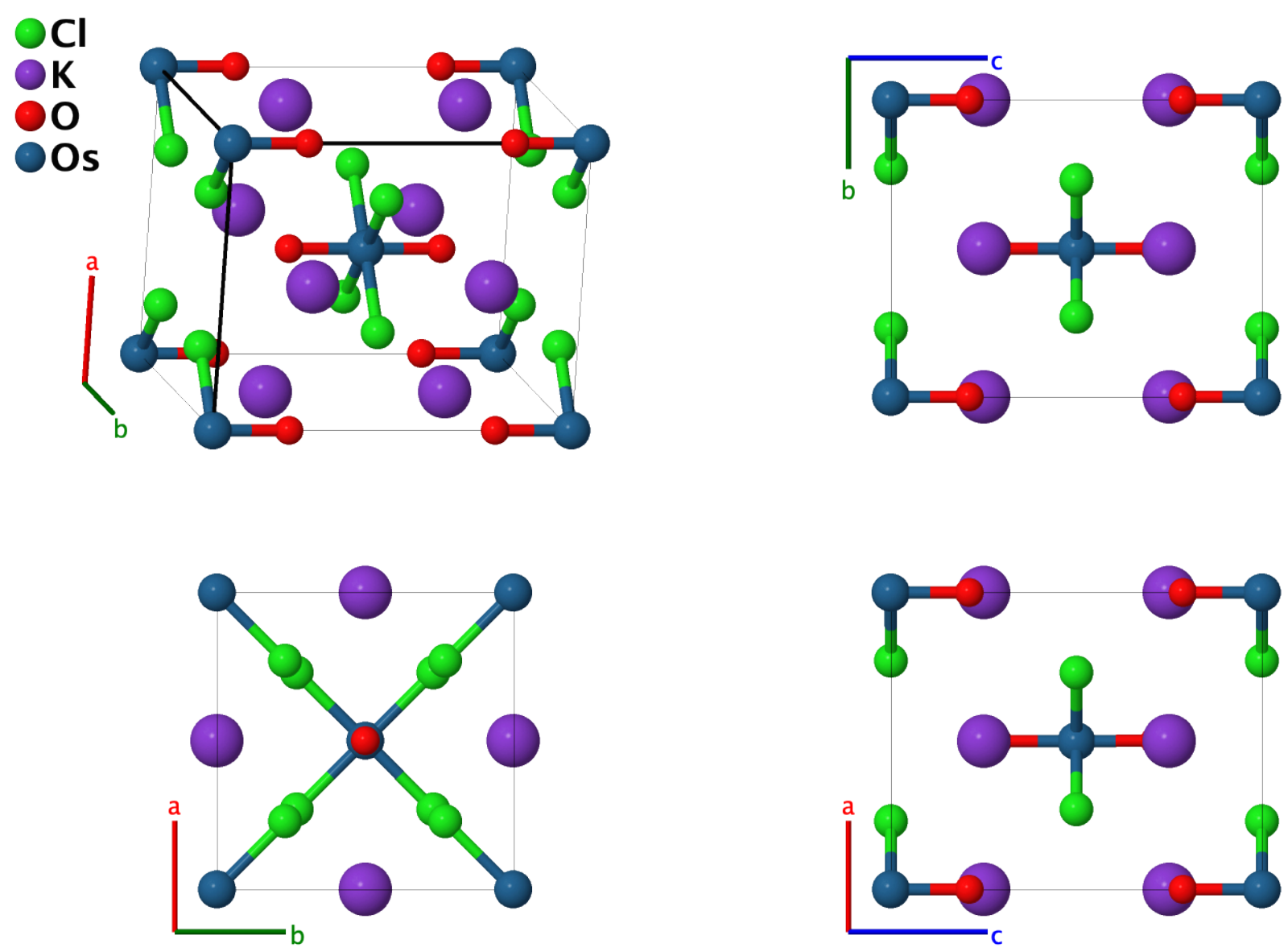

| | | |
|---|---|---|
| **Prototype** | : | $Cl_4K_2O_2Os$ |
| **AFLOW prototype label** | : | A4B2C2D_tI18_139_h_d_e_a |
| ***Strukturbericht* designation** | : | $J1_5$ |
| **Pearson symbol** | : | tI18 |
| **Space group number** | : | 139 |
| **Space group symbol** | : | $I4/mmm$ |
| **AFLOW prototype command** | : | `aflow --proto=A4B2C2D_tI18_139_h_d_e_a --params=`$a, c/a, z_3, x_4$ |

**Body-centered Tetragonal primitive vectors:**

$$\mathbf{a}_1 = -\frac{1}{2}\,a\,\hat{\mathbf{x}} + \frac{1}{2}\,a\,\hat{\mathbf{y}} + \frac{1}{2}\,c\,\hat{\mathbf{z}}$$

$$\mathbf{a}_2 = \frac{1}{2}\,a\,\hat{\mathbf{x}} - \frac{1}{2}\,a\,\hat{\mathbf{y}} + \frac{1}{2}\,c\,\hat{\mathbf{z}}$$

$$\mathbf{a}_3 = \frac{1}{2}\,a\,\hat{\mathbf{x}} + \frac{1}{2}\,a\,\hat{\mathbf{y}} - \frac{1}{2}\,c\,\hat{\mathbf{z}}$$

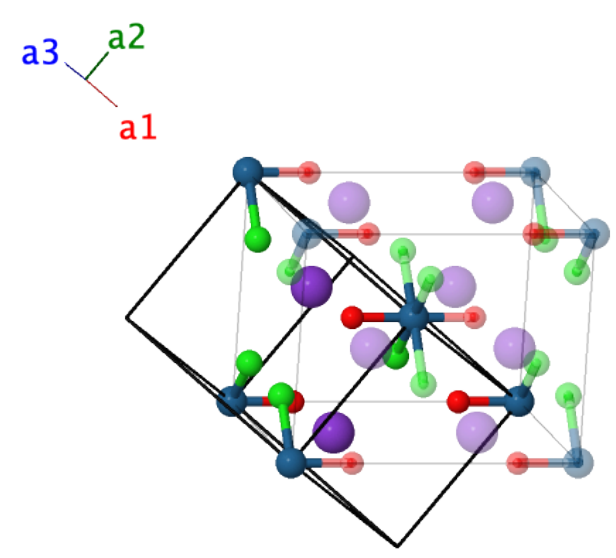

**Basis vectors:**

| | | Lattice Coordinates | | Cartesian Coordinates | Wyckoff Position | Atom Type |
|---|---|---|---|---|---|---|
| $\mathbf{B}_1$ | = | $0\,\mathbf{a}_1 + 0\,\mathbf{a}_2 + 0\,\mathbf{a}_3$ | = | $0\,\hat{\mathbf{x}} + 0\,\hat{\mathbf{y}} + 0\,\hat{\mathbf{z}}$ | (2$a$) | Os |
| $\mathbf{B}_2$ | = | $\frac{3}{4}\,\mathbf{a}_1 + \frac{1}{4}\,\mathbf{a}_2 + \frac{1}{2}\,\mathbf{a}_3$ | = | $\frac{1}{2}a\,\hat{\mathbf{y}} + \frac{1}{4}c\,\hat{\mathbf{z}}$ | (4$d$) | K |
| $\mathbf{B}_3$ | = | $\frac{1}{4}\,\mathbf{a}_1 + \frac{3}{4}\,\mathbf{a}_2 + \frac{1}{2}\,\mathbf{a}_3$ | = | $\frac{1}{2}a\,\hat{\mathbf{x}} + \frac{1}{4}c\,\hat{\mathbf{z}}$ | (4$d$) | K |
| $\mathbf{B}_4$ | = | $z_3\,\mathbf{a}_1 + z_3\,\mathbf{a}_2$ | = | $z_3 c\,\hat{\mathbf{z}}$ | (4$e$) | O |
| $\mathbf{B}_5$ | = | $-z_3\,\mathbf{a}_1 - z_3\,\mathbf{a}_2$ | = | $-z_3 c\,\hat{\mathbf{z}}$ | (4$e$) | O |
| $\mathbf{B}_6$ | = | $x_4\,\mathbf{a}_1 + x_4\,\mathbf{a}_2 + 2x_4\,\mathbf{a}_3$ | = | $x_4 a\,\hat{\mathbf{x}} + x_4 a\,\hat{\mathbf{y}}$ | (8$h$) | Cl |
| $\mathbf{B}_7$ | = | $-x_4\,\mathbf{a}_1 - x_4\,\mathbf{a}_2 - 2x_4\,\mathbf{a}_3$ | = | $-x_4 a\,\hat{\mathbf{x}} - x_4 a\,\hat{\mathbf{y}}$ | (8$h$) | Cl |
| $\mathbf{B}_8$ | = | $x_4\,\mathbf{a}_1 - x_4\,\mathbf{a}_2$ | = | $-x_4 a\,\hat{\mathbf{x}} + x_4 a\,\hat{\mathbf{y}}$ | (8$h$) | Cl |
| $\mathbf{B}_9$ | = | $-x_4\,\mathbf{a}_1 + x_4\,\mathbf{a}_2$ | = | $x_4 a\,\hat{\mathbf{x}} - x_4 a\,\hat{\mathbf{y}}$ | (8$h$) | Cl |

**Geometry files:**

- CIF: pp. 1708
- POSCAR: pp. 1708

# $K_2NiF_4$ Structure: A4B2C_tI14_139_ce_e_a

http://aflow.org/prototype-encyclopedia/A4B2C_tI14_139_ce_e_a

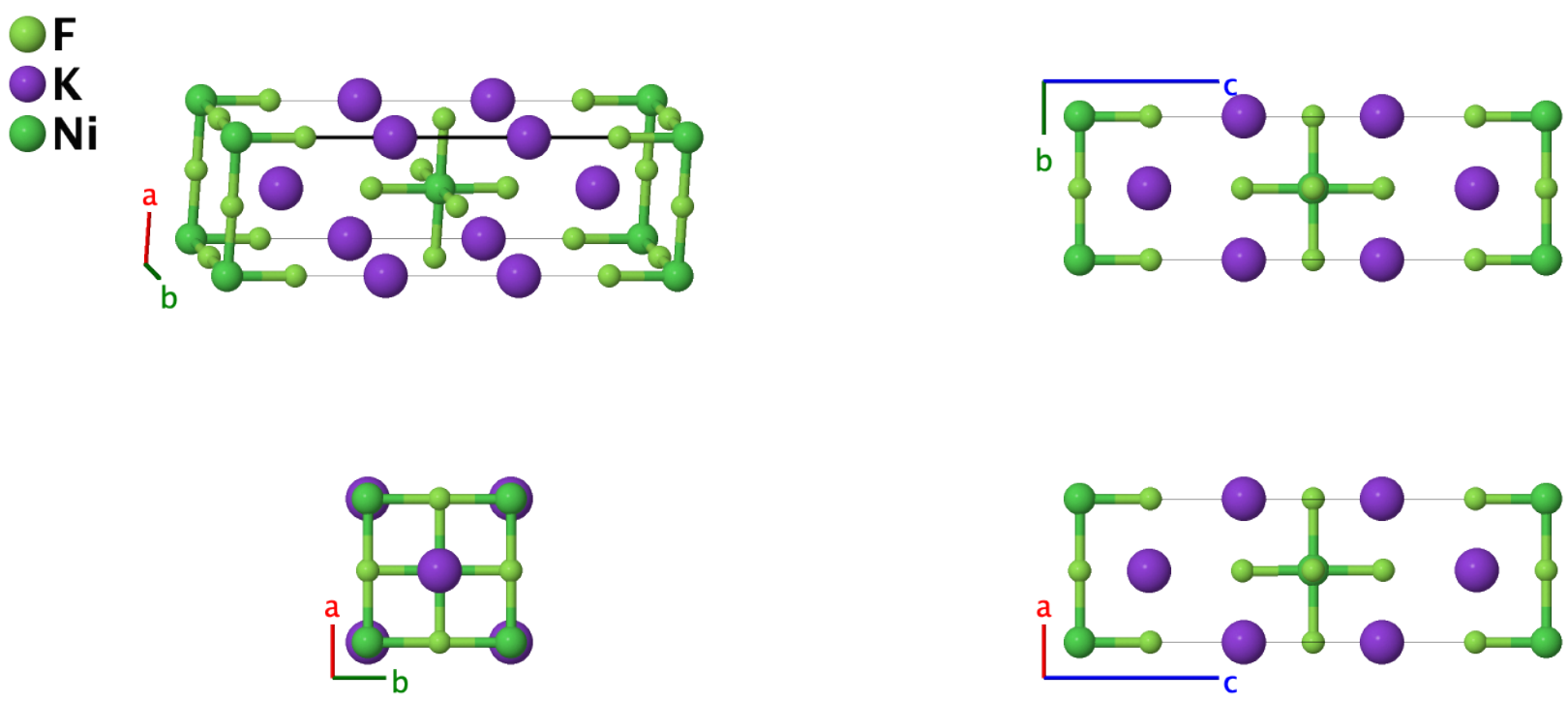

| | | |
|---|---|---|
| **Prototype** | : | $F_4K_2Ni$ |
| **AFLOW prototype label** | : | A4B2C_tI14_139_ce_e_a |
| ***Strukturbericht* designation** | : | None |
| **Pearson symbol** | : | tI14 |
| **Space group number** | : | 139 |
| **Space group symbol** | : | $I4/mmm$ |
| **AFLOW prototype command** | : | `aflow --proto=A4B2C_tI14_139_ce_e_a --params=`$a, c/a, z_3, z_4$ |

### Other compounds with this structure

- $Ca_2MnO_4$, $La_2PdO_4$, $Nd_2CuO_4$, $Sr_2MnO_4$, $Sr_2RuO_4$, $Sr_2TiO_4$, $Sr_2VO_4$, $(Fe,La)_2SrO_4$, $(Sr,La)_2AlO_4$, and $(Sr,La)_2CoO_4$

---

- This is the parent compound of the simplest layered-perovskite Ruddlesden-Popper series (Wikipedia). The series also includes the parent of the high-$T_c$ cuprates, $(La,Ba)_2CuO_4$, but we keep that separate as it represents a new class of materials.

### Body-centered Tetragonal primitive vectors:

$$\mathbf{a}_1 = -\frac{1}{2}\,a\,\hat{\mathbf{x}} + \frac{1}{2}\,a\,\hat{\mathbf{y}} + \frac{1}{2}\,c\,\hat{\mathbf{z}}$$
$$\mathbf{a}_2 = \frac{1}{2}\,a\,\hat{\mathbf{x}} - \frac{1}{2}\,a\,\hat{\mathbf{y}} + \frac{1}{2}\,c\,\hat{\mathbf{z}}$$
$$\mathbf{a}_3 = \frac{1}{2}\,a\,\hat{\mathbf{x}} + \frac{1}{2}\,a\,\hat{\mathbf{y}} - \frac{1}{2}\,c\,\hat{\mathbf{z}}$$

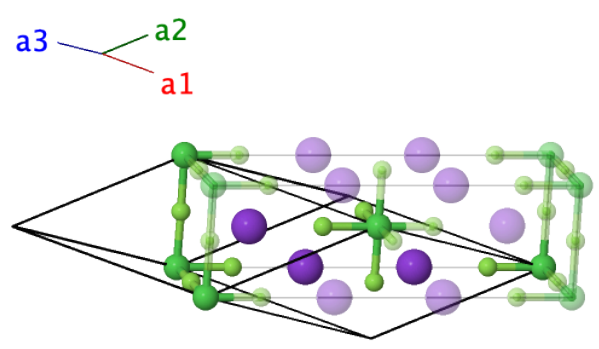

### Basis vectors:

| | | Lattice Coordinates | | Cartesian Coordinates | Wyckoff Position | Atom Type |
|---|---|---|---|---|---|---|
| $\mathbf{B}_1$ | = | $0\,\mathbf{a}_1 + 0\,\mathbf{a}_2 + 0\,\mathbf{a}_3$ | = | $0\,\hat{\mathbf{x}} + 0\,\hat{\mathbf{y}} + 0\,\hat{\mathbf{z}}$ | $(2a)$ | Ni |
| $\mathbf{B}_2$ | = | $\frac{1}{2}\,\mathbf{a}_1 + \frac{1}{2}\,\mathbf{a}_3$ | = | $\frac{1}{2}a\,\hat{\mathbf{y}}$ | $(4c)$ | F I |
| $\mathbf{B}_3$ | = | $\frac{1}{2}\,\mathbf{a}_2 + \frac{1}{2}\,\mathbf{a}_3$ | = | $\frac{1}{2}a\,\hat{\mathbf{x}}$ | $(4c)$ | F I |
| $\mathbf{B}_4$ | = | $z_3\,\mathbf{a}_1 + z_3\,\mathbf{a}_2$ | = | $z_3 c\,\hat{\mathbf{z}}$ | $(4e)$ | F II |
| $\mathbf{B}_5$ | = | $-z_3\,\mathbf{a}_1 - z_3\,\mathbf{a}_2$ | = | $-z_3 c\,\hat{\mathbf{z}}$ | $(4e)$ | F II |
| $\mathbf{B}_6$ | = | $z_4\,\mathbf{a}_1 + z_4\,\mathbf{a}_2$ | = | $z_4 c\,\hat{\mathbf{z}}$ | (4$e$) | K |
| $\mathbf{B}_7$ | = | $-z_4\,\mathbf{a}_1 - z_4\,\mathbf{a}_2$ | = | $-z_4 c\,\hat{\mathbf{z}}$ | (4$e$) | K |

---

**Found in:**

- Wikipedia, *Ruddlesden-Popper phase*, `https://en.wikipedia.org/wiki/Ruddlesden-Popper_phase`. $A_3B_2X_7$ series.

---

**Geometry files:**

- CIF: pp. 1708
- POSCAR: pp. 1709

# $K_3TlCl_6{\cdot}2H_2O$ ($J3_1$) Structure: A6B2C3D_tI168_139_egikl2m_ejn_bh2n_acf

http://aflow.org/prototype-encyclopedia/A6B2C3D_tI168_139_egikl2m_ejn_bh2n_acf

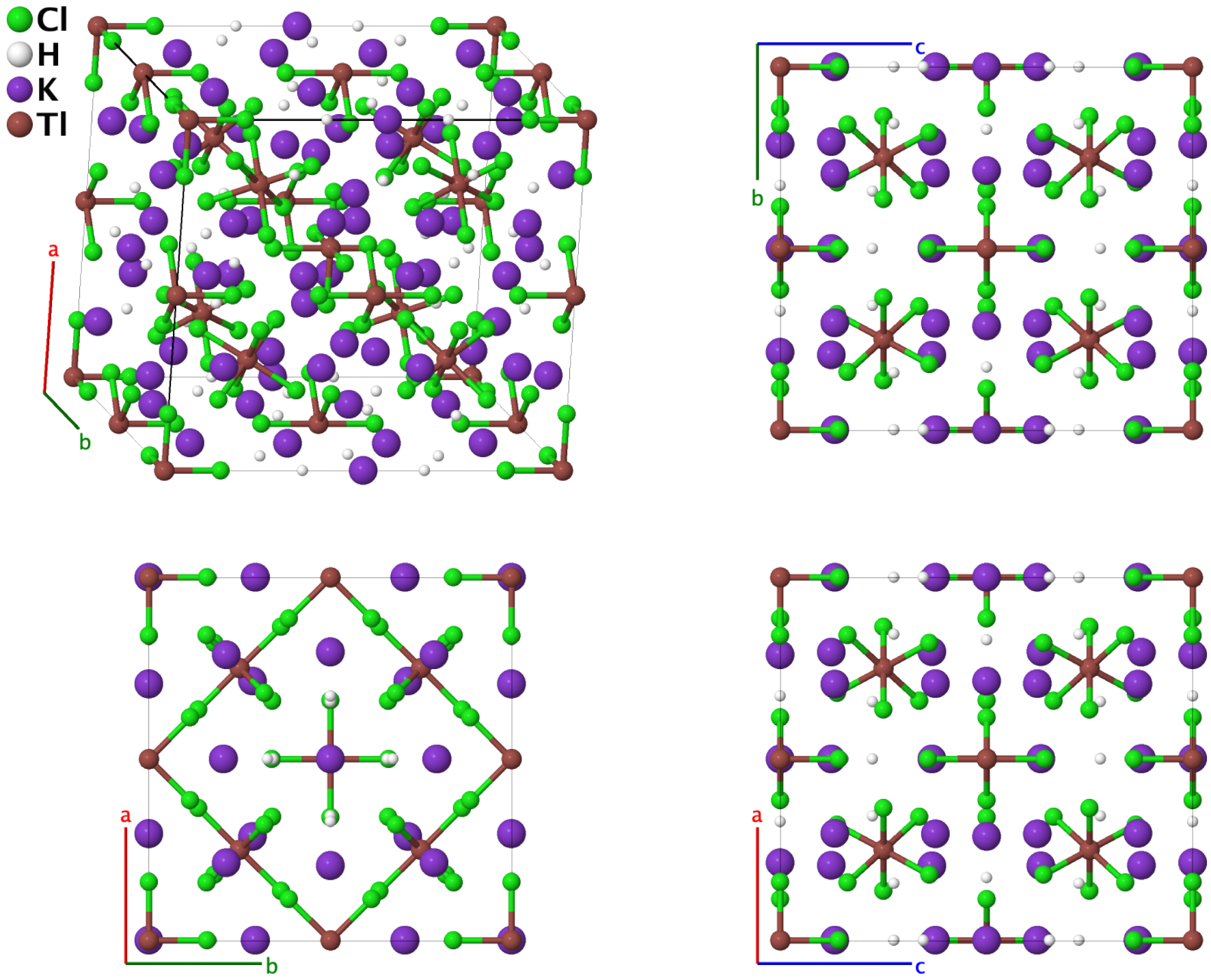

| | | |
|---|---|---|
| **Prototype** | : | $Cl_6(H_2O)_2K_3Tl$ |
| **AFLOW prototype label** | : | A6B2C3D_tI168_139_egikl2m_ejn_bh2n_acf |
| ***Strukturbericht* designation** | : | $J3_1$ |
| **Pearson symbol** | : | tI168 |
| **Space group number** | : | 139 |
| **Space group symbol** | : | $I4/mmm$ |
| **AFLOW prototype command** | : | `aflow --proto=A6B2C3D_tI168_139_egikl2m_ejn_bh2n_acf` `--params=`$a, c/a, z_4, z_5, z_7, x_8, x_9, x_{10}, x_{11}, x_{12}, y_{12}, x_{13}, z_{13}, x_{14}, z_{14}, y_{15}, z_{15}, y_{16}, z_{16}, y_{17}, z_{17}$ |

- The positions of the hydrogen atoms in the water molecules were not determined, so we only provide the positions of the oxygen atoms (labeled as $H_2O$).

**Body-centered Tetragonal primitive vectors:**

$$\mathbf{a}_1 = -\frac{1}{2}\,a\,\hat{\mathbf{x}} + \frac{1}{2}\,a\,\hat{\mathbf{y}} + \frac{1}{2}\,c\,\hat{\mathbf{z}}$$
$$\mathbf{a}_2 = \frac{1}{2}\,a\,\hat{\mathbf{x}} - \frac{1}{2}\,a\,\hat{\mathbf{y}} + \frac{1}{2}\,c\,\hat{\mathbf{z}}$$
$$\mathbf{a}_3 = \frac{1}{2}\,a\,\hat{\mathbf{x}} + \frac{1}{2}\,a\,\hat{\mathbf{y}} - \frac{1}{2}\,c\,\hat{\mathbf{z}}$$

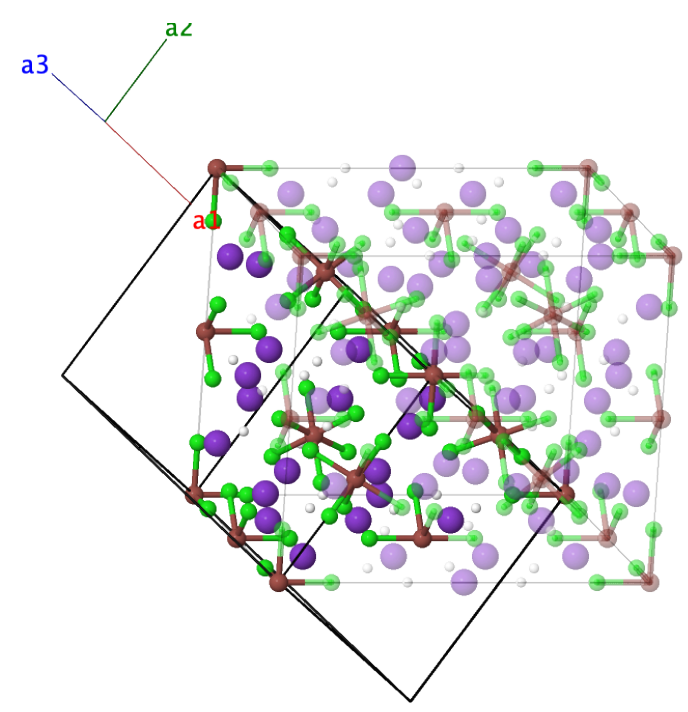

**Basis vectors:**

| | | Lattice Coordinates | | Cartesian Coordinates | Wyckoff Position | Atom Type |
|---|---|---|---|---|---|---|
| $\mathbf{B}_1$ | = | $0\,\mathbf{a}_1 + 0\,\mathbf{a}_2 + 0\,\mathbf{a}_3$ | = | $0\,\hat{\mathbf{x}} + 0\,\hat{\mathbf{y}} + 0\,\hat{\mathbf{z}}$ | (2$a$) | Tl I |
| $\mathbf{B}_2$ | = | $\frac{1}{2}\,\mathbf{a}_1 + \frac{1}{2}\,\mathbf{a}_2$ | = | $\frac{1}{2}c\,\hat{\mathbf{z}}$ | (2$b$) | K I |
| $\mathbf{B}_3$ | = | $\frac{1}{2}\,\mathbf{a}_1 + \frac{1}{2}\,\mathbf{a}_3$ | = | $\frac{1}{2}a\,\hat{\mathbf{y}}$ | (4$c$) | Tl II |
| $\mathbf{B}_4$ | = | $\frac{1}{2}\,\mathbf{a}_2 + \frac{1}{2}\,\mathbf{a}_3$ | = | $\frac{1}{2}a\,\hat{\mathbf{x}}$ | (4$c$) | Tl II |
| $\mathbf{B}_5$ | = | $z_4\,\mathbf{a}_1 + z_4\,\mathbf{a}_2$ | = | $z_4c\,\hat{\mathbf{z}}$ | (4$e$) | Cl I |
| $\mathbf{B}_6$ | = | $-z_4\,\mathbf{a}_1 - z_4\,\mathbf{a}_2$ | = | $-z_4c\,\hat{\mathbf{z}}$ | (4$e$) | Cl I |
| $\mathbf{B}_7$ | = | $z_5\,\mathbf{a}_1 + z_5\,\mathbf{a}_2$ | = | $z_5c\,\hat{\mathbf{z}}$ | (4$e$) | $H_2O$ I |
| $\mathbf{B}_8$ | = | $-z_5\,\mathbf{a}_1 - z_5\,\mathbf{a}_2$ | = | $-z_5c\,\hat{\mathbf{z}}$ | (4$e$) | $H_2O$ I |
| $\mathbf{B}_9$ | = | $\frac{1}{2}\,\mathbf{a}_1 + \frac{1}{2}\,\mathbf{a}_2 + \frac{1}{2}\,\mathbf{a}_3$ | = | $\frac{1}{4}a\,\hat{\mathbf{x}} + \frac{1}{4}a\,\hat{\mathbf{y}} + \frac{1}{4}c\,\hat{\mathbf{z}}$ | (8$f$) | Tl III |
| $\mathbf{B}_{10}$ | = | $\frac{1}{2}\,\mathbf{a}_3$ | = | $\frac{1}{4}a\,\hat{\mathbf{x}} + \frac{1}{4}a\,\hat{\mathbf{y}} - \frac{1}{4}c\,\hat{\mathbf{z}}$ | (8$f$) | Tl III |
| $\mathbf{B}_{11}$ | = | $\frac{1}{2}\,\mathbf{a}_1$ | = | $\frac{3}{4}a\,\hat{\mathbf{x}} + \frac{1}{4}a\,\hat{\mathbf{y}} + \frac{1}{4}c\,\hat{\mathbf{z}}$ | (8$f$) | Tl III |
| $\mathbf{B}_{12}$ | = | $\frac{1}{2}\,\mathbf{a}_2$ | = | $\frac{1}{4}a\,\hat{\mathbf{x}} + \frac{3}{4}a\,\hat{\mathbf{y}} + \frac{1}{4}c\,\hat{\mathbf{z}}$ | (8$f$) | Tl III |
| $\mathbf{B}_{13}$ | = | $\left(\frac{1}{2} + z_7\right)\,\mathbf{a}_1 + z_7\,\mathbf{a}_2 + \frac{1}{2}\,\mathbf{a}_3$ | = | $\frac{1}{2}a\,\hat{\mathbf{y}} + z_7c\,\hat{\mathbf{z}}$ | (8$g$) | Cl II |
| $\mathbf{B}_{14}$ | = | $z_7\,\mathbf{a}_1 + \left(\frac{1}{2} + z_7\right)\,\mathbf{a}_2 + \frac{1}{2}\,\mathbf{a}_3$ | = | $\frac{1}{2}a\,\hat{\mathbf{x}} + z_7c\,\hat{\mathbf{z}}$ | (8$g$) | Cl II |
| $\mathbf{B}_{15}$ | = | $\left(\frac{1}{2} - z_7\right)\,\mathbf{a}_1 - z_7\,\mathbf{a}_2 + \frac{1}{2}\,\mathbf{a}_3$ | = | $\frac{1}{2}a\,\hat{\mathbf{y}} - z_7c\,\hat{\mathbf{z}}$ | (8$g$) | Cl II |
| $\mathbf{B}_{16}$ | = | $-z_7\,\mathbf{a}_1 + \left(\frac{1}{2} - z_7\right)\,\mathbf{a}_2 + \frac{1}{2}\,\mathbf{a}_3$ | = | $\frac{1}{2}a\,\hat{\mathbf{x}} - z_7c\,\hat{\mathbf{z}}$ | (8$g$) | Cl II |
| $\mathbf{B}_{17}$ | = | $x_8\,\mathbf{a}_1 + x_8\,\mathbf{a}_2 + 2x_8\,\mathbf{a}_3$ | = | $x_8a\,\hat{\mathbf{x}} + x_8a\,\hat{\mathbf{y}}$ | (8$h$) | K II |
| $\mathbf{B}_{18}$ | = | $-x_8\,\mathbf{a}_1 - x_8\,\mathbf{a}_2 - 2x_8\,\mathbf{a}_3$ | = | $-x_8a\,\hat{\mathbf{x}} - x_8a\,\hat{\mathbf{y}}$ | (8$h$) | K II |
| $\mathbf{B}_{19}$ | = | $x_8\,\mathbf{a}_1 - x_8\,\mathbf{a}_2$ | = | $-x_8a\,\hat{\mathbf{x}} + x_8a\,\hat{\mathbf{y}}$ | (8$h$) | K II |
| $\mathbf{B}_{20}$ | = | $-x_8\,\mathbf{a}_1 + x_8\,\mathbf{a}_2$ | = | $x_8a\,\hat{\mathbf{x}} - x_8a\,\hat{\mathbf{y}}$ | (8$h$) | K II |
| $\mathbf{B}_{21}$ | = | $x_9\,\mathbf{a}_2 + x_9\,\mathbf{a}_3$ | = | $x_9a\,\hat{\mathbf{x}}$ | (8$i$) | Cl III |
| $\mathbf{B}_{22}$ | = | $-x_9\,\mathbf{a}_2 - x_9\,\mathbf{a}_3$ | = | $-x_9a\,\hat{\mathbf{x}}$ | (8$i$) | Cl III |
| $\mathbf{B}_{23}$ | = | $x_9\,\mathbf{a}_1 + x_9\,\mathbf{a}_3$ | = | $x_9a\,\hat{\mathbf{y}}$ | (8$i$) | Cl III |
| $\mathbf{B}_{24}$ | = | $-x_9\,\mathbf{a}_1 - x_9\,\mathbf{a}_3$ | = | $-x_9a\,\hat{\mathbf{y}}$ | (8$i$) | Cl III |
| $\mathbf{B}_{25}$ | = | $\frac{1}{2}\,\mathbf{a}_1 + x_{10}\,\mathbf{a}_2 + \left(\frac{1}{2} + x_{10}\right)\,\mathbf{a}_3$ | = | $x_{10}a\,\hat{\mathbf{x}} + \frac{1}{2}a\,\hat{\mathbf{y}}$ | (8$j$) | $H_2O$ II |
| $\mathbf{B}_{26}$ | = | $\frac{1}{2}\,\mathbf{a}_1 - x_{10}\,\mathbf{a}_2 + \left(\frac{1}{2} - x_{10}\right)\,\mathbf{a}_3$ | = | $-x_{10}a\,\hat{\mathbf{x}} + \frac{1}{2}a\,\hat{\mathbf{y}}$ | (8$j$) | $H_2O$ II |
| $\mathbf{B}_{27}$ | = | $x_{10}\,\mathbf{a}_1 + \frac{1}{2}\,\mathbf{a}_2 + \left(\frac{1}{2} + x_{10}\right)\,\mathbf{a}_3$ | = | $\frac{1}{2}a\,\hat{\mathbf{x}} + x_{10}a\,\hat{\mathbf{y}}$ | (8$j$) | $H_2O$ II |
| $\mathbf{B}_{28}$ | = | $-x_{10}\,\mathbf{a}_1 + \frac{1}{2}\,\mathbf{a}_2 + \left(\frac{1}{2} - x_{10}\right)\,\mathbf{a}_3$ | = | $\frac{1}{2}a\,\hat{\mathbf{x}} - x_{10}a\,\hat{\mathbf{y}}$ | (8$j$) | $H_2O$ II |
| $\mathbf{B}_{29}$ | = | $\left(\frac{3}{4} + x_{11}\right)\,\mathbf{a}_1 + \left(\frac{1}{4} + x_{11}\right)\,\mathbf{a}_2 + \left(\frac{1}{2} + 2x_{11}\right)\,\mathbf{a}_3$ | = | $x_{11}a\,\hat{\mathbf{x}} + \left(\frac{1}{2} + x_{11}\right)a\,\hat{\mathbf{y}} + \frac{1}{4}c\,\hat{\mathbf{z}}$ | (16$k$) | Cl IV |
| $\mathbf{B}_{30}$ | = | $\left(\frac{3}{4} - x_{11}\right)\,\mathbf{a}_1 + \left(\frac{1}{4} - x_{11}\right)\,\mathbf{a}_2 + \left(\frac{1}{2} - 2x_{11}\right)\,\mathbf{a}_3$ | = | $-x_{11}a\,\hat{\mathbf{x}} + \left(\frac{1}{2} - x_{11}\right)a\,\hat{\mathbf{y}} + \frac{1}{4}c\,\hat{\mathbf{z}}$ | (16$k$) | Cl IV |
| $\mathbf{B}_{31}$ | = | $\left(\frac{1}{4} + x_{11}\right)\,\mathbf{a}_1 + \left(\frac{3}{4} - x_{11}\right)\,\mathbf{a}_2 + \frac{1}{2}\,\mathbf{a}_3$ | = | $\left(\frac{1}{2} - x_{11}\right)a\,\hat{\mathbf{x}} + x_{11}a\,\hat{\mathbf{y}} + \frac{1}{4}c\,\hat{\mathbf{z}}$ | (16$k$) | Cl IV |
| $\mathbf{B}_{32}$ | = | $\left(\frac{1}{4} - x_{11}\right)\,\mathbf{a}_1 + \left(\frac{3}{4} + x_{11}\right)\,\mathbf{a}_2 + \frac{1}{2}\,\mathbf{a}_3$ | = | $\left(\frac{1}{2} + x_{11}\right)a\,\hat{\mathbf{x}} - x_{11}a\,\hat{\mathbf{y}} + \frac{1}{4}c\,\hat{\mathbf{z}}$ | (16$k$) | Cl IV |
| $\mathbf{B}_{33}$ | = | $\left(\frac{1}{4} - x_{11}\right)\,\mathbf{a}_1 + \left(\frac{3}{4} - x_{11}\right)\,\mathbf{a}_2 + \left(\frac{1}{2} - 2x_{11}\right)\,\mathbf{a}_3$ | = | $\left(\frac{1}{2} - x_{11}\right)a\,\hat{\mathbf{x}} - x_{11}a\,\hat{\mathbf{y}} + \frac{1}{4}c\,\hat{\mathbf{z}}$ | (16$k$) | Cl IV |
| $\mathbf{B}_{34}$ | = | $\left(\frac{1}{4} + x_{11}\right) \mathbf{a}_1 + \left(\frac{3}{4} + x_{11}\right) \mathbf{a}_2 + \left(\frac{1}{2} + 2x_{11}\right) \mathbf{a}_3$ | = | $\left(\frac{1}{2} + x_{11}\right) a \, \hat{\mathbf{x}} + x_{11} a \, \hat{\mathbf{y}} + \frac{1}{4} c \, \hat{\mathbf{z}}$ | (16*k*) | Cl IV |
| $\mathbf{B}_{35}$ | = | $\left(\frac{3}{4} - x_{11}\right) \mathbf{a}_1 + \left(\frac{1}{4} + x_{11}\right) \mathbf{a}_2 + \frac{1}{2} \mathbf{a}_3$ | = | $x_{11} a \, \hat{\mathbf{x}} + \left(\frac{1}{2} - x_{11}\right) a \, \hat{\mathbf{y}} + \frac{1}{4} c \, \hat{\mathbf{z}}$ | (16*k*) | Cl IV |
| $\mathbf{B}_{36}$ | = | $\left(\frac{3}{4} + x_{11}\right) \mathbf{a}_1 + \left(\frac{1}{4} - x_{11}\right) \mathbf{a}_2 + \frac{1}{2} \mathbf{a}_3$ | = | $-x_{11} a \, \hat{\mathbf{x}} + \left(\frac{1}{2} + x_{11}\right) a \, \hat{\mathbf{y}} + \frac{1}{4} c \, \hat{\mathbf{z}}$ | (16*k*) | Cl IV |
| $\mathbf{B}_{37}$ | = | $y_{12} \, \mathbf{a}_1 + x_{12} \, \mathbf{a}_2 + (x_{12} + y_{12}) \, \mathbf{a}_3$ | = | $x_{12} a \, \hat{\mathbf{x}} + y_{12} a \, \hat{\mathbf{y}}$ | (16*l*) | Cl V |
| $\mathbf{B}_{38}$ | = | $-y_{12} \, \mathbf{a}_1 - x_{12} \, \mathbf{a}_2 + (-x_{12} - y_{12}) \, \mathbf{a}_3$ | = | $-x_{12} a \, \hat{\mathbf{x}} - y_{12} a \, \hat{\mathbf{y}}$ | (16*l*) | Cl V |
| $\mathbf{B}_{39}$ | = | $x_{12} \, \mathbf{a}_1 - y_{12} \, \mathbf{a}_2 + (x_{12} - y_{12}) \, \mathbf{a}_3$ | = | $-y_{12} a \, \hat{\mathbf{x}} + x_{12} a \, \hat{\mathbf{y}}$ | (16*l*) | Cl V |
| $\mathbf{B}_{40}$ | = | $-x_{12} \, \mathbf{a}_1 + y_{12} \, \mathbf{a}_2 + (-x_{12} + y_{12}) \, \mathbf{a}_3$ | = | $y_{12} a \, \hat{\mathbf{x}} - x_{12} a \, \hat{\mathbf{y}}$ | (16*l*) | Cl V |
| $\mathbf{B}_{41}$ | = | $y_{12} \, \mathbf{a}_1 - x_{12} \, \mathbf{a}_2 + (-x_{12} + y_{12}) \, \mathbf{a}_3$ | = | $-x_{12} a \, \hat{\mathbf{x}} + y_{12} a \, \hat{\mathbf{y}}$ | (16*l*) | Cl V |
| $\mathbf{B}_{42}$ | = | $-y_{12} \, \mathbf{a}_1 + x_{12} \, \mathbf{a}_2 + (x_{12} - y_{12}) \, \mathbf{a}_3$ | = | $x_{12} a \, \hat{\mathbf{x}} - y_{12} a \, \hat{\mathbf{y}}$ | (16*l*) | Cl V |
| $\mathbf{B}_{43}$ | = | $x_{12} \, \mathbf{a}_1 + y_{12} \, \mathbf{a}_2 + (x_{12} + y_{12}) \, \mathbf{a}_3$ | = | $y_{12} a \, \hat{\mathbf{x}} + x_{12} a \, \hat{\mathbf{y}}$ | (16*l*) | Cl V |
| $\mathbf{B}_{44}$ | = | $-x_{12} \, \mathbf{a}_1 - y_{12} \, \mathbf{a}_2 + (-x_{12} - y_{12}) \, \mathbf{a}_3$ | = | $-y_{12} a \, \hat{\mathbf{x}} - x_{12} a \, \hat{\mathbf{y}}$ | (16*l*) | Cl V |
| $\mathbf{B}_{45}$ | = | $(x_{13} + z_{13}) \, \mathbf{a}_1 + (x_{13} + z_{13}) \, \mathbf{a}_2 + 2x_{13} \, \mathbf{a}_3$ | = | $x_{13} a \, \hat{\mathbf{x}} + x_{13} a \, \hat{\mathbf{y}} + z_{13} c \, \hat{\mathbf{z}}$ | (16*m*) | Cl VI |
| $\mathbf{B}_{46}$ | = | $(-x_{13} + z_{13}) \, \mathbf{a}_1 + (-x_{13} + z_{13}) \, \mathbf{a}_2 - 2x_{13} \, \mathbf{a}_3$ | = | $-x_{13} a \, \hat{\mathbf{x}} - x_{13} a \, \hat{\mathbf{y}} + z_{13} c \, \hat{\mathbf{z}}$ | (16*m*) | Cl VI |
| $\mathbf{B}_{47}$ | = | $(x_{13} + z_{13}) \, \mathbf{a}_1 + (-x_{13} + z_{13}) \, \mathbf{a}_2$ | = | $-x_{13} a \, \hat{\mathbf{x}} + x_{13} a \, \hat{\mathbf{y}} + z_{13} c \, \hat{\mathbf{z}}$ | (16*m*) | Cl VI |
| $\mathbf{B}_{48}$ | = | $(-x_{13} + z_{13}) \, \mathbf{a}_1 + (x_{13} + z_{13}) \, \mathbf{a}_2$ | = | $x_{13} a \, \hat{\mathbf{x}} - x_{13} a \, \hat{\mathbf{y}} + z_{13} c \, \hat{\mathbf{z}}$ | (16*m*) | Cl VI |
| $\mathbf{B}_{49}$ | = | $(x_{13} - z_{13}) \, \mathbf{a}_1 + (-x_{13} - z_{13}) \, \mathbf{a}_2$ | = | $-x_{13} a \, \hat{\mathbf{x}} + x_{13} a \, \hat{\mathbf{y}} - z_{13} c \, \hat{\mathbf{z}}$ | (16*m*) | Cl VI |
| $\mathbf{B}_{50}$ | = | $(-x_{13} - z_{13}) \, \mathbf{a}_1 + (x_{13} - z_{13}) \, \mathbf{a}_2$ | = | $x_{13} a \, \hat{\mathbf{x}} - x_{13} a \, \hat{\mathbf{y}} - z_{13} c \, \hat{\mathbf{z}}$ | (16*m*) | Cl VI |
| $\mathbf{B}_{51}$ | = | $(x_{13} - z_{13}) \, \mathbf{a}_1 + (x_{13} - z_{13}) \, \mathbf{a}_2 + 2x_{13} \, \mathbf{a}_3$ | = | $x_{13} a \, \hat{\mathbf{x}} + x_{13} a \, \hat{\mathbf{y}} - z_{13} c \, \hat{\mathbf{z}}$ | (16*m*) | Cl VI |
| $\mathbf{B}_{52}$ | = | $(-x_{13} - z_{13}) \, \mathbf{a}_1 + (-x_{13} - z_{13}) \, \mathbf{a}_2 - 2x_{13} \, \mathbf{a}_3$ | = | $-x_{13} a \, \hat{\mathbf{x}} - x_{13} a \, \hat{\mathbf{y}} - z_{13} c \, \hat{\mathbf{z}}$ | (16*m*) | Cl VI |
| $\mathbf{B}_{53}$ | = | $(x_{14} + z_{14}) \, \mathbf{a}_1 + (x_{14} + z_{14}) \, \mathbf{a}_2 + 2x_{14} \, \mathbf{a}_3$ | = | $x_{14} a \, \hat{\mathbf{x}} + x_{14} a \, \hat{\mathbf{y}} + z_{14} c \, \hat{\mathbf{z}}$ | (16*m*) | Cl VII |
| $\mathbf{B}_{54}$ | = | $(-x_{14} + z_{14}) \, \mathbf{a}_1 + (-x_{14} + z_{14}) \, \mathbf{a}_2 - 2x_{14} \, \mathbf{a}_3$ | = | $-x_{14} a \, \hat{\mathbf{x}} - x_{14} a \, \hat{\mathbf{y}} + z_{14} c \, \hat{\mathbf{z}}$ | (16*m*) | Cl VII |
| $\mathbf{B}_{55}$ | = | $(x_{14} + z_{14}) \, \mathbf{a}_1 + (-x_{14} + z_{14}) \, \mathbf{a}_2$ | = | $-x_{14} a \, \hat{\mathbf{x}} + x_{14} a \, \hat{\mathbf{y}} + z_{14} c \, \hat{\mathbf{z}}$ | (16*m*) | Cl VII |
| $\mathbf{B}_{56}$ | = | $(-x_{14} + z_{14}) \, \mathbf{a}_1 + (x_{14} + z_{14}) \, \mathbf{a}_2$ | = | $x_{14} a \, \hat{\mathbf{x}} - x_{14} a \, \hat{\mathbf{y}} + z_{14} c \, \hat{\mathbf{z}}$ | (16*m*) | Cl VII |
| $\mathbf{B}_{57}$ | = | $(x_{14} - z_{14}) \, \mathbf{a}_1 + (-x_{14} - z_{14}) \, \mathbf{a}_2$ | = | $-x_{14} a \, \hat{\mathbf{x}} + x_{14} a \, \hat{\mathbf{y}} - z_{14} c \, \hat{\mathbf{z}}$ | (16*m*) | Cl VII |
| $\mathbf{B}_{58}$ | = | $(-x_{14} - z_{14}) \, \mathbf{a}_1 + (x_{14} - z_{14}) \, \mathbf{a}_2$ | = | $x_{14} a \, \hat{\mathbf{x}} - x_{14} a \, \hat{\mathbf{y}} - z_{14} c \, \hat{\mathbf{z}}$ | (16*m*) | Cl VII |
| $\mathbf{B}_{59}$ | = | $(x_{14} - z_{14}) \, \mathbf{a}_1 + (x_{14} - z_{14}) \, \mathbf{a}_2 + 2x_{14} \, \mathbf{a}_3$ | = | $x_{14} a \, \hat{\mathbf{x}} + x_{14} a \, \hat{\mathbf{y}} - z_{14} c \, \hat{\mathbf{z}}$ | (16*m*) | Cl VII |
| $\mathbf{B}_{60}$ | = | $(-x_{14} - z_{14}) \, \mathbf{a}_1 + (-x_{14} - z_{14}) \, \mathbf{a}_2 - 2x_{14} \, \mathbf{a}_3$ | = | $-x_{14} a \, \hat{\mathbf{x}} - x_{14} a \, \hat{\mathbf{y}} - z_{14} c \, \hat{\mathbf{z}}$ | (16*m*) | Cl VII |
| $\mathbf{B}_{61}$ | = | $(y_{15} + z_{15}) \, \mathbf{a}_1 + z_{15} \, \mathbf{a}_2 + y_{15} \, \mathbf{a}_3$ | = | $y_{15} a \, \hat{\mathbf{y}} + z_{15} c \, \hat{\mathbf{z}}$ | (16*n*) | $H_2O$ III |
| $\mathbf{B}_{62}$ | = | $(-y_{15} + z_{15}) \, \mathbf{a}_1 + z_{15} \, \mathbf{a}_2 - y_{15} \, \mathbf{a}_3$ | = | $-y_{15} a \, \hat{\mathbf{y}} + z_{15} c \, \hat{\mathbf{z}}$ | (16*n*) | $H_2O$ III |
| $\mathbf{B}_{63}$ | = | $z_{15} \, \mathbf{a}_1 + (-y_{15} + z_{15}) \, \mathbf{a}_2 - y_{15} \, \mathbf{a}_3$ | = | $-y_{15} a \, \hat{\mathbf{x}} + z_{15} c \, \hat{\mathbf{z}}$ | (16*n*) | $H_2O$ III |
| $\mathbf{B}_{64}$ | = | $z_{15} \, \mathbf{a}_1 + (y_{15} + z_{15}) \, \mathbf{a}_2 + y_{15} \, \mathbf{a}_3$ | = | $y_{15} a \, \hat{\mathbf{x}} + z_{15} c \, \hat{\mathbf{z}}$ | (16*n*) | $H_2O$ III |
| $\mathbf{B}_{65}$ | = | $(y_{15} - z_{15}) \, \mathbf{a}_1 - z_{15} \, \mathbf{a}_2 + y_{15} \, \mathbf{a}_3$ | = | $y_{15} a \, \hat{\mathbf{y}} - z_{15} c \, \hat{\mathbf{z}}$ | (16*n*) | $H_2O$ III |
| $\mathbf{B}_{66}$ | = | $(-y_{15} - z_{15}) \, \mathbf{a}_1 - z_{15} \, \mathbf{a}_2 - y_{15} \, \mathbf{a}_3$ | = | $-y_{15} a \, \hat{\mathbf{y}} - z_{15} c \, \hat{\mathbf{z}}$ | (16*n*) | $H_2O$ III |
| $\mathbf{B}_{67}$ | = | $-z_{15} \, \mathbf{a}_1 + (y_{15} - z_{15}) \, \mathbf{a}_2 + y_{15} \, \mathbf{a}_3$ | = | $y_{15} a \, \hat{\mathbf{x}} - z_{15} c \, \hat{\mathbf{z}}$ | (16*n*) | $H_2O$ III |
| $\mathbf{B}_{68}$ | = | $-z_{15}\,\mathbf{a}_1 + \left(-y_{15} - z_{15}\right)\,\mathbf{a}_2 - y_{15}\,\mathbf{a}_3$ | = | $-y_{15}a\,\mathbf{\hat{x}} - z_{15}c\,\mathbf{\hat{z}}$ | $(16n)$ | $H_2O$ III |
| $\mathbf{B}_{69}$ | = | $\left(y_{16} + z_{16}\right)\,\mathbf{a}_1 + z_{16}\,\mathbf{a}_2 + y_{16}\,\mathbf{a}_3$ | = | $y_{16}a\,\mathbf{\hat{y}} + z_{16}c\,\mathbf{\hat{z}}$ | $(16n)$ | K III |
| $\mathbf{B}_{70}$ | = | $\left(-y_{16} + z_{16}\right)\,\mathbf{a}_1 + z_{16}\,\mathbf{a}_2 - y_{16}\,\mathbf{a}_3$ | = | $-y_{16}a\,\mathbf{\hat{y}} + z_{16}c\,\mathbf{\hat{z}}$ | $(16n)$ | K III |
| $\mathbf{B}_{71}$ | = | $z_{16}\,\mathbf{a}_1 + \left(-y_{16} + z_{16}\right)\,\mathbf{a}_2 - y_{16}\,\mathbf{a}_3$ | = | $-y_{16}a\,\mathbf{\hat{x}} + z_{16}c\,\mathbf{\hat{z}}$ | $(16n)$ | K III |
| $\mathbf{B}_{72}$ | = | $z_{16}\,\mathbf{a}_1 + \left(y_{16} + z_{16}\right)\,\mathbf{a}_2 + y_{16}\,\mathbf{a}_3$ | = | $y_{16}a\,\mathbf{\hat{x}} + z_{16}c\,\mathbf{\hat{z}}$ | $(16n)$ | K III |
| $\mathbf{B}_{73}$ | = | $\left(y_{16} - z_{16}\right)\,\mathbf{a}_1 - z_{16}\,\mathbf{a}_2 + y_{16}\,\mathbf{a}_3$ | = | $y_{16}a\,\mathbf{\hat{y}} - z_{16}c\,\mathbf{\hat{z}}$ | $(16n)$ | K III |
| $\mathbf{B}_{74}$ | = | $\left(-y_{16} - z_{16}\right)\,\mathbf{a}_1 - z_{16}\,\mathbf{a}_2 - y_{16}\,\mathbf{a}_3$ | = | $-y_{16}a\,\mathbf{\hat{y}} - z_{16}c\,\mathbf{\hat{z}}$ | $(16n)$ | K III |
| $\mathbf{B}_{75}$ | = | $-z_{16}\,\mathbf{a}_1 + \left(y_{16} - z_{16}\right)\,\mathbf{a}_2 + y_{16}\,\mathbf{a}_3$ | = | $y_{16}a\,\mathbf{\hat{x}} - z_{16}c\,\mathbf{\hat{z}}$ | $(16n)$ | K III |
| $\mathbf{B}_{76}$ | = | $-z_{16}\,\mathbf{a}_1 + \left(-y_{16} - z_{16}\right)\,\mathbf{a}_2 - y_{16}\,\mathbf{a}_3$ | = | $-y_{16}a\,\mathbf{\hat{x}} - z_{16}c\,\mathbf{\hat{z}}$ | $(16n)$ | K III |
| $\mathbf{B}_{77}$ | = | $\left(y_{17} + z_{17}\right)\,\mathbf{a}_1 + z_{17}\,\mathbf{a}_2 + y_{17}\,\mathbf{a}_3$ | = | $y_{17}a\,\mathbf{\hat{y}} + z_{17}c\,\mathbf{\hat{z}}$ | $(16n)$ | K IV |
| $\mathbf{B}_{78}$ | = | $\left(-y_{17} + z_{17}\right)\,\mathbf{a}_1 + z_{17}\,\mathbf{a}_2 - y_{17}\,\mathbf{a}_3$ | = | $-y_{17}a\,\mathbf{\hat{y}} + z_{17}c\,\mathbf{\hat{z}}$ | $(16n)$ | K IV |
| $\mathbf{B}_{79}$ | = | $z_{17}\,\mathbf{a}_1 + \left(-y_{17} + z_{17}\right)\,\mathbf{a}_2 - y_{17}\,\mathbf{a}_3$ | = | $-y_{17}a\,\mathbf{\hat{x}} + z_{17}c\,\mathbf{\hat{z}}$ | $(16n)$ | K IV |
| $\mathbf{B}_{80}$ | = | $z_{17}\,\mathbf{a}_1 + \left(y_{17} + z_{17}\right)\,\mathbf{a}_2 + y_{17}\,\mathbf{a}_3$ | = | $y_{17}a\,\mathbf{\hat{x}} + z_{17}c\,\mathbf{\hat{z}}$ | $(16n)$ | K IV |
| $\mathbf{B}_{81}$ | = | $\left(y_{17} - z_{17}\right)\,\mathbf{a}_1 - z_{17}\,\mathbf{a}_2 + y_{17}\,\mathbf{a}_3$ | = | $y_{17}a\,\mathbf{\hat{y}} - z_{17}c\,\mathbf{\hat{z}}$ | $(16n)$ | K IV |
| $\mathbf{B}_{82}$ | = | $\left(-y_{17} - z_{17}\right)\,\mathbf{a}_1 - z_{17}\,\mathbf{a}_2 - y_{17}\,\mathbf{a}_3$ | = | $-y_{17}a\,\mathbf{\hat{y}} - z_{17}c\,\mathbf{\hat{z}}$ | $(16n)$ | K IV |
| $\mathbf{B}_{83}$ | = | $-z_{17}\,\mathbf{a}_1 + \left(y_{17} - z_{17}\right)\,\mathbf{a}_2 + y_{17}\,\mathbf{a}_3$ | = | $y_{17}a\,\mathbf{\hat{x}} - z_{17}c\,\mathbf{\hat{z}}$ | $(16n)$ | K IV |
| $\mathbf{B}_{84}$ | = | $-z_{17}\,\mathbf{a}_1 + \left(-y_{17} - z_{17}\right)\,\mathbf{a}_2 - y_{17}\,\mathbf{a}_3$ | = | $-y_{17}a\,\mathbf{\hat{x}} - z_{17}c\,\mathbf{\hat{z}}$ | $(16n)$ | K IV |

---

---

**Geometry files:**

- CIF: pp. 1709
- POSCAR: pp. 1709

# $Sr_3Ti_2O_7$ Structure: A7B3C2_tI24_139_aeg_be_e

http://aflow.org/prototype-encyclopedia/A7B3C2_tI24_139_aeg_be_e

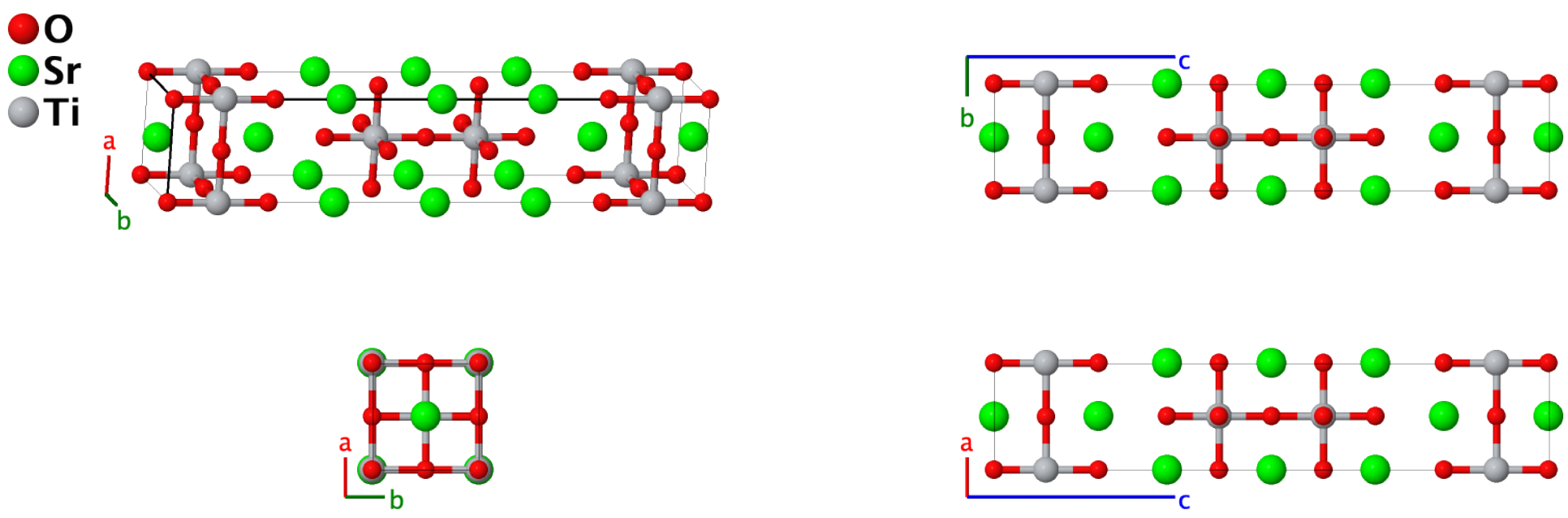

| | | |
|---|---|---|
| **Prototype** | : | $O_7Sr_3Ti_2$ |
| **AFLOW prototype label** | : | A7B3C2_tI24_139_aeg_be_e |
| ***Strukturbericht* designation** | : | None |
| **Pearson symbol** | : | tI24 |
| **Space group number** | : | 139 |
| **Space group symbol** | : | $I4/mmm$ |
| **AFLOW prototype command** | : | `aflow --proto=A7B3C2_tI24_139_aeg_be_e`<br>`--params=`$a, c/a, z_3, z_4, z_5, z_6$ |

### Other compounds with this structure

- $Ca_3Ti_2O_7$, $Sr_3Ru_2O_7$, $BaLa_2Fe_2O_7$, and $SrTb_2Fe_2O_7$

### Body-centered Tetragonal primitive vectors:

$$\mathbf{a}_1 = -\frac{1}{2}\,a\,\hat{\mathbf{x}} + \frac{1}{2}\,a\,\hat{\mathbf{y}} + \frac{1}{2}\,c\,\hat{\mathbf{z}}$$
$$\mathbf{a}_2 = \frac{1}{2}\,a\,\hat{\mathbf{x}} - \frac{1}{2}\,a\,\hat{\mathbf{y}} + \frac{1}{2}\,c\,\hat{\mathbf{z}}$$
$$\mathbf{a}_3 = \frac{1}{2}\,a\,\hat{\mathbf{x}} + \frac{1}{2}\,a\,\hat{\mathbf{y}} - \frac{1}{2}\,c\,\hat{\mathbf{z}}$$

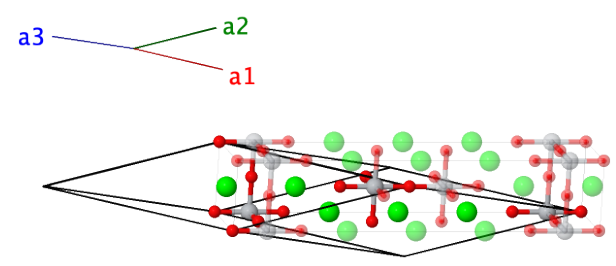

### Basis vectors:

| | | Lattice Coordinates | | Cartesian Coordinates | Wyckoff Position | Atom Type |
|---|---|---|---|---|---|---|
| $\mathbf{B}_1$ | = | $0\,\mathbf{a}_1 + 0\,\mathbf{a}_2 + 0\,\mathbf{a}_3$ | = | $0\,\hat{\mathbf{x}} + 0\,\hat{\mathbf{y}} + 0\,\hat{\mathbf{z}}$ | (2$a$) | O I |
| $\mathbf{B}_2$ | = | $\frac{1}{2}\,\mathbf{a}_1 + \frac{1}{2}\,\mathbf{a}_2$ | = | $\frac{1}{2}c\,\hat{\mathbf{z}}$ | (2$b$) | Sr I |
| $\mathbf{B}_3$ | = | $z_3\,\mathbf{a}_1 + z_3\,\mathbf{a}_2$ | = | $z_3c\,\hat{\mathbf{z}}$ | (4$e$) | O II |
| $\mathbf{B}_4$ | = | $-z_3\,\mathbf{a}_1 - z_3\,\mathbf{a}_2$ | = | $-z_3c\,\hat{\mathbf{z}}$ | (4$e$) | O II |
| $\mathbf{B}_5$ | = | $z_4\,\mathbf{a}_1 + z_4\,\mathbf{a}_2$ | = | $z_4c\,\hat{\mathbf{z}}$ | (4$e$) | Sr II |
| $\mathbf{B}_6$ | = | $-z_4\,\mathbf{a}_1 - z_4\,\mathbf{a}_2$ | = | $-z_4c\,\hat{\mathbf{z}}$ | (4$e$) | Sr II |
| $\mathbf{B}_7$ | = | $z_5\,\mathbf{a}_1 + z_5\,\mathbf{a}_2$ | = | $z_5c\,\hat{\mathbf{z}}$ | (4$e$) | Ti |
| $\mathbf{B}_8$ | = | $-z_5\,\mathbf{a}_1 - z_5\,\mathbf{a}_2$ | = | $-z_5c\,\hat{\mathbf{z}}$ | (4$e$) | Ti |
| $\mathbf{B}_9$ | = | $\left(\frac{1}{2} + z_6\right)\,\mathbf{a}_1 + z_6\,\mathbf{a}_2 + \frac{1}{2}\,\mathbf{a}_3$ | = | $\frac{1}{2}a\,\hat{\mathbf{y}} + z_6c\,\hat{\mathbf{z}}$ | (8$g$) | O III |
| $\mathbf{B}_{10}$ | = | $z_6\,\mathbf{a}_1 + \left(\frac{1}{2} + z_6\right)\,\mathbf{a}_2 + \frac{1}{2}\,\mathbf{a}_3$ | = | $\frac{1}{2}a\,\hat{\mathbf{x}} + z_6c\,\hat{\mathbf{z}}$ | (8$g$) | O III |
| $\mathbf{B}_{11}$ | = | $\left(\frac{1}{2} - z_6\right)\mathbf{a}_1 - z_6\,\mathbf{a}_2 + \frac{1}{2}\,\mathbf{a}_3$ | = | $\frac{1}{2}a\,\hat{\mathbf{y}} - z_6 c\,\hat{\mathbf{z}}$ | (8$g$) | O III |
| $\mathbf{B}_{12}$ | = | $-z_6\,\mathbf{a}_1 + \left(\frac{1}{2} - z_6\right)\mathbf{a}_2 + \frac{1}{2}\,\mathbf{a}_3$ | = | $\frac{1}{2}a\,\hat{\mathbf{x}} - z_6 c\,\hat{\mathbf{z}}$ | (8$g$) | O III |

**Found in:**

- Wikipedia, *Ruddlesden-Popper phase*, `https://en.wikipedia.org/wiki/Ruddlesden-Popper_phase`. $A_3B_2X_7$ series.

**Geometry files:**

- CIF: pp. 1710
- POSCAR: pp. 1710

# $Fe_8N$ ($D2_g$) Structure: A8B_tI18_139_deh_a

http://aflow.org/prototype-encyclopedia/A8B_tI18_139_deh_a

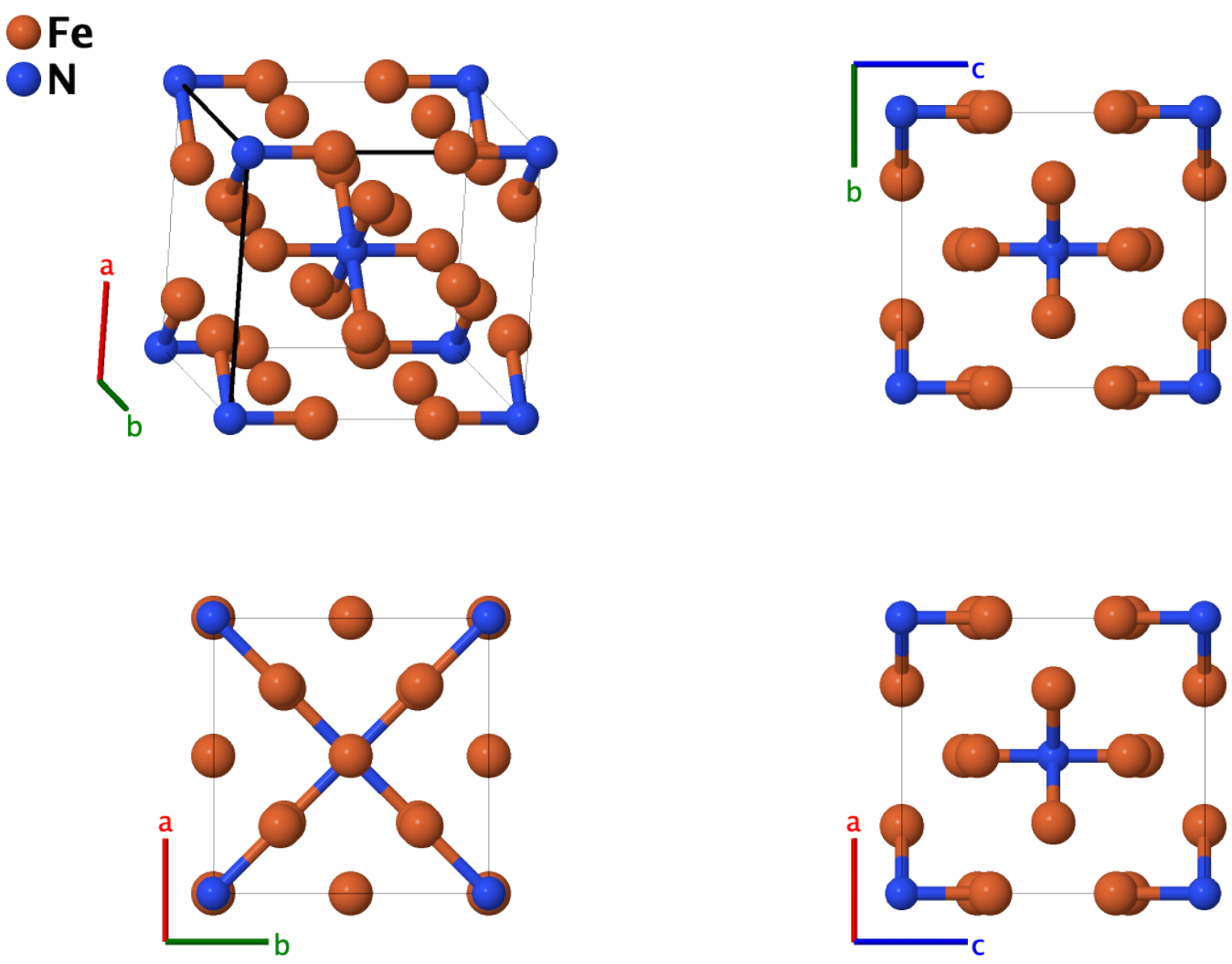

| | | |
|---|---|---|
| **Prototype** | : | $Fe_8N$ |
| **AFLOW prototype label** | : | A8B_tI18_139_deh_a |
| ***Strukturbericht* designation** | : | $D2_g$ |
| **Pearson symbol** | : | tI18 |
| **Space group number** | : | 139 |
| **Space group symbol** | : | $I4/mmm$ |
| **AFLOW prototype command** | : | `aflow --proto=A8B_tI18_139_deh_a --params=`$a, c/a, z_3, x_4$ |

- (Jack, 1951) and others refer to this structure as $\alpha''$-$Fe_{16}N_2$.
- (Yamashita, 2012) determined the structure by examining crystals containing various amounts of $\alpha''$-$Fe_{16}N_2$ and $\alpha$-Fe (bcc iron). We use their data from the sample that was 90% $\alpha''$-$Fe_{16}N_2$. Their measurements were confirmed by first principles calculations, and are in agreement with the first principles calculations of (Sims, 2012).

**Body-centered Tetragonal primitive vectors:**

$$\mathbf{a}_1 = -\frac{1}{2}\,a\,\hat{\mathbf{x}} + \frac{1}{2}\,a\,\hat{\mathbf{y}} + \frac{1}{2}\,c\,\hat{\mathbf{z}}$$
$$\mathbf{a}_2 = \frac{1}{2}\,a\,\hat{\mathbf{x}} - \frac{1}{2}\,a\,\hat{\mathbf{y}} + \frac{1}{2}\,c\,\hat{\mathbf{z}}$$
$$\mathbf{a}_3 = \frac{1}{2}\,a\,\hat{\mathbf{x}} + \frac{1}{2}\,a\,\hat{\mathbf{y}} - \frac{1}{2}\,c\,\hat{\mathbf{z}}$$

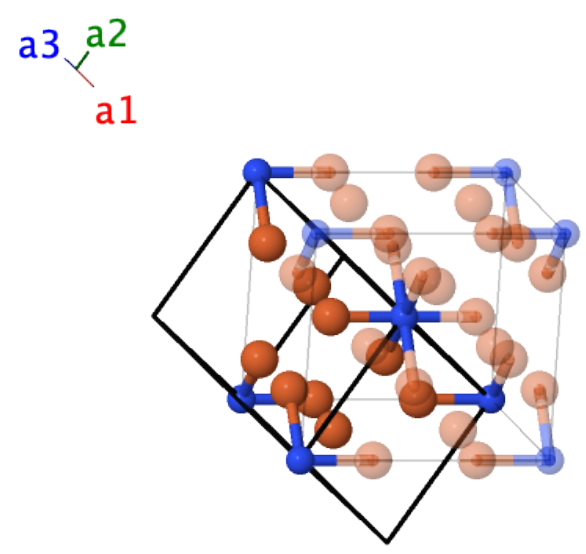

**Basis vectors:**

| | | Lattice Coordinates | | Cartesian Coordinates | Wyckoff Position | Atom Type |
|---|---|---|---|---|---|---|
| $\mathbf{B}_1$ | = | $0\,\mathbf{a}_1 + 0\,\mathbf{a}_2 + 0\,\mathbf{a}_3$ | = | $0\,\hat{\mathbf{x}} + 0\,\hat{\mathbf{y}} + 0\,\hat{\mathbf{z}}$ | (2$a$) | N |
| $\mathbf{B}_2$ | = | $\frac{3}{4}\,\mathbf{a}_1 + \frac{1}{4}\,\mathbf{a}_2 + \frac{1}{2}\,\mathbf{a}_3$ | = | $\frac{1}{2}a\,\hat{\mathbf{y}} + \frac{1}{4}c\,\hat{\mathbf{z}}$ | (4$d$) | Fe I |
| $\mathbf{B}_3$ | = | $\frac{1}{4}\,\mathbf{a}_1 + \frac{3}{4}\,\mathbf{a}_2 + \frac{1}{2}\,\mathbf{a}_3$ | = | $\frac{1}{2}a\,\hat{\mathbf{x}} + \frac{1}{4}c\,\hat{\mathbf{z}}$ | $(4d)$ | Fe I |
| $\mathbf{B}_4$ | = | $z_3\,\mathbf{a}_1 + z_3\,\mathbf{a}_2$ | = | $z_3c\,\hat{\mathbf{z}}$ | $(4e)$ | Fe II |
| $\mathbf{B}_5$ | = | $-z_3\,\mathbf{a}_1 - z_3\,\mathbf{a}_2$ | = | $-z_3c\,\hat{\mathbf{z}}$ | $(4e)$ | Fe II |
| $\mathbf{B}_6$ | = | $x_4\,\mathbf{a}_1 + x_4\,\mathbf{a}_2 + 2x_4\,\mathbf{a}_3$ | = | $x_4a\,\hat{\mathbf{x}} + x_4a\,\hat{\mathbf{y}}$ | $(8h)$ | Fe III |
| $\mathbf{B}_7$ | = | $-x_4\,\mathbf{a}_1 - x_4\,\mathbf{a}_2 - 2x_4\,\mathbf{a}_3$ | = | $-x_4a\,\hat{\mathbf{x}} - x_4a\,\hat{\mathbf{y}}$ | $(8h)$ | Fe III |
| $\mathbf{B}_8$ | = | $x_4\,\mathbf{a}_1 - x_4\,\mathbf{a}_2$ | = | $-x_4a\,\hat{\mathbf{x}} + x_4a\,\hat{\mathbf{y}}$ | $(8h)$ | Fe III |
| $\mathbf{B}_9$ | = | $-x_4\,\mathbf{a}_1 + x_4\,\mathbf{a}_2$ | = | $x_4a\,\hat{\mathbf{x}} - x_4a\,\hat{\mathbf{y}}$ | $(8h)$ | Fe III |

---

---

**Geometry files:**

- CIF: pp. 1710
- POSCAR: pp. 1711

# $Li_2CN_2$ Structure: AB2C2_tI10_139_a_d_e

http://aflow.org/prototype-encyclopedia/AB2C2_tI10_139_a_d_e.Li2CN2

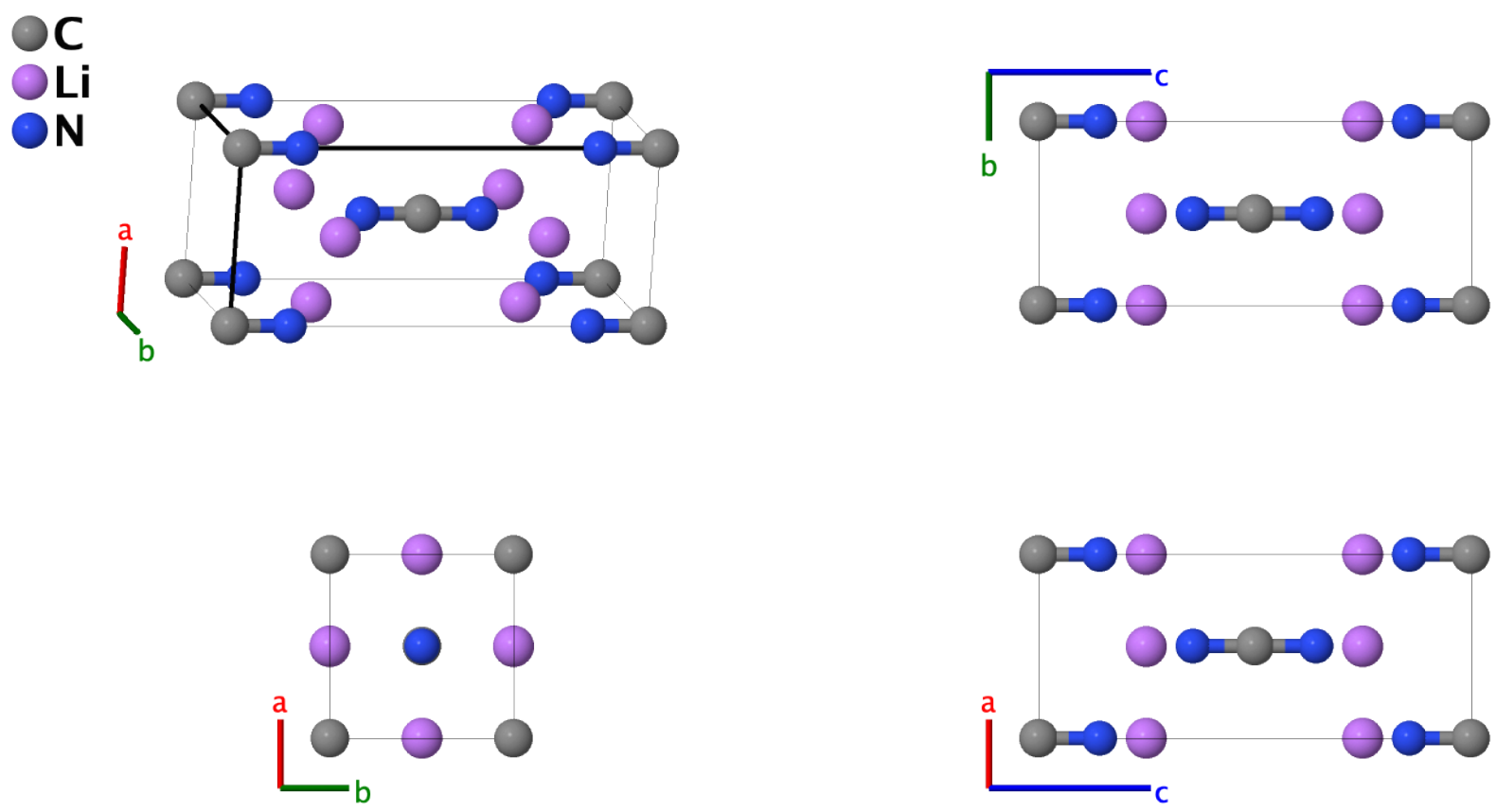

| | | |
|---|---|---|
| **Prototype** | : | $CLi_2N_2$ |
| **AFLOW prototype label** | : | AB2C2_tI10_139_a_d_e |
| ***Strukturbericht* designation** | : | None |
| **Pearson symbol** | : | tI10 |
| **Space group number** | : | 139 |
| **Space group symbol** | : | $I4/mmm$ |
| **AFLOW prototype command** | : | `aflow --proto=AB2C2_tI10_139_a_d_e`<br>`--params=`$a, c/a, z_3$ |

- This structure has the same AFLOW prototype label, AB2C2_tI10_139_a_d_e, as autunite, $Ca(UO_2)_2(PO_4)_2{\cdot}10.5H_2O$ ($H5_9$). They are generated by the same symmetry operations with different sets of parameters (`--params`) specified in their corresponding CIF files.

**Body-centered Tetragonal primitive vectors:**

$$\mathbf{a}_1 = -\frac{1}{2}\,a\,\hat{\mathbf{x}} + \frac{1}{2}\,a\,\hat{\mathbf{y}} + \frac{1}{2}\,c\,\hat{\mathbf{z}}$$
$$\mathbf{a}_2 = \frac{1}{2}\,a\,\hat{\mathbf{x}} - \frac{1}{2}\,a\,\hat{\mathbf{y}} + \frac{1}{2}\,c\,\hat{\mathbf{z}}$$
$$\mathbf{a}_3 = \frac{1}{2}\,a\,\hat{\mathbf{x}} + \frac{1}{2}\,a\,\hat{\mathbf{y}} - \frac{1}{2}\,c\,\hat{\mathbf{z}}$$

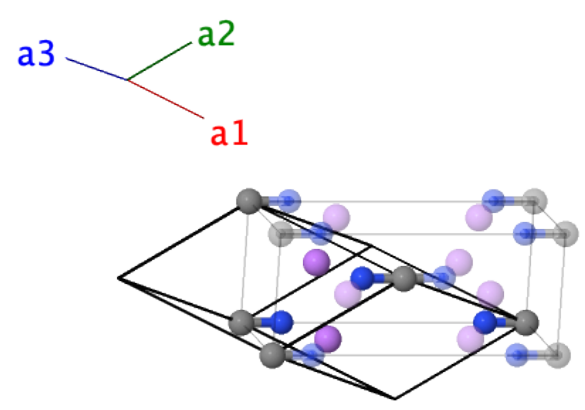

**Basis vectors:**

| | | Lattice Coordinates | | Cartesian Coordinates | Wyckoff Position | Atom Type |
|---|---|---|---|---|---|---|
| $\mathbf{B}_1$ | = | $0\,\mathbf{a}_1 + 0\,\mathbf{a}_2 + 0\,\mathbf{a}_3$ | = | $0\,\hat{\mathbf{x}} + 0\,\hat{\mathbf{y}} + 0\,\hat{\mathbf{z}}$ | (2*a*) | C |
| $\mathbf{B}_2$ | = | $\frac{3}{4}\,\mathbf{a}_1 + \frac{1}{4}\,\mathbf{a}_2 + \frac{1}{2}\,\mathbf{a}_3$ | = | $\frac{1}{2}a\,\hat{\mathbf{y}} + \frac{1}{4}c\,\hat{\mathbf{z}}$ | (4*d*) | Li |
| $\mathbf{B}_3$ | = | $\frac{1}{4}\,\mathbf{a}_1 + \frac{3}{4}\,\mathbf{a}_2 + \frac{1}{2}\,\mathbf{a}_3$ | = | $\frac{1}{2}a\,\hat{\mathbf{x}} + \frac{1}{4}c\,\hat{\mathbf{z}}$ | (4*d*) | Li |
| $\mathbf{B}_4$ | = | $z_3\,\mathbf{a}_1 + z_3\,\mathbf{a}_2$ | = | $z_3 c\,\hat{\mathbf{z}}$ | (4*e*) | N |
| $\mathbf{B}_5$ | = | $-z_3\,\mathbf{a}_1 - z_3\,\mathbf{a}_2$ | = | $-z_3 c\,\hat{\mathbf{z}}$ | (4*e*) | N |

---

**Geometry files:**

- CIF: pp. 1711
- POSCAR: pp. 1711

# $H5_9$ [Autunite, $Ca(UO_2)_2(PO_4)_2 \cdot 10\frac{1}{2}H_2O$] (*obsolete*) Structure: AB2C2_tI10_139_a_d_e

http://aflow.org/prototype-encyclopedia/AB2C2_tI10_139_a_d_e

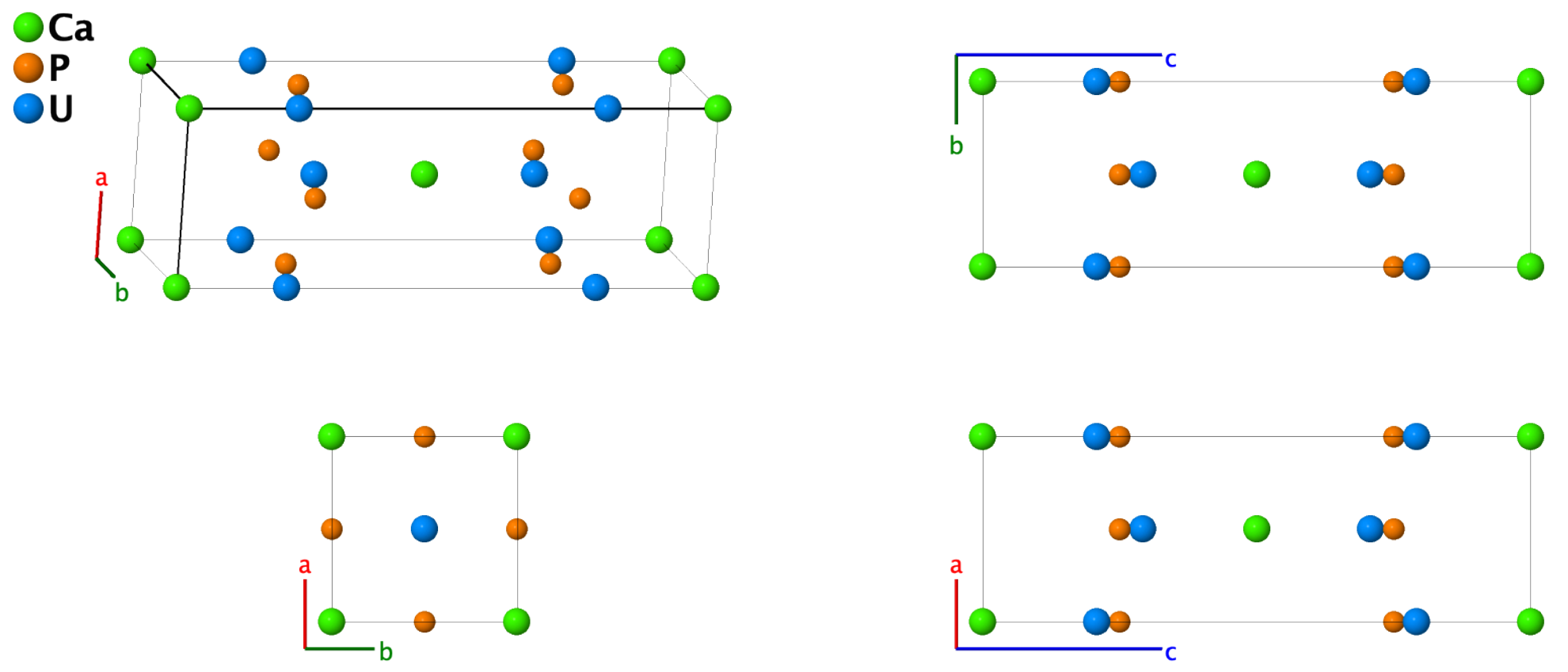

| | | |
|---|---|---|
| **Prototype** | : | $Ca(UO_2)_2(PO_4)_2 \cdot 10\frac{1}{2}H_2O$ |
| **AFLOW prototype label** | : | AB2C2_tI10_139_a_d_e |
| ***Strukturbericht* designation** | : | $H5_9$ |
| **Pearson symbol** | : | tI10 |
| **Space group number** | : | 139 |
| **Space group symbol** | : | $I4/mmm$ |
| **AFLOW prototype command** | : | `aflow --proto=AB2C2_tI10_139_a_d_e` `--params=`$a, c/a, z_3$ |

- Autunite $Ca(UO_2)_2(PO_4)_2 \cdot nH_2O$, is found in three varieties: naturally occurring Autunite, with $n \gtrsim 10$, and meta-autunite (I), which is partially dehydrated, $6 \gtrsim n \gtrsim 10$. Further dehydration in the laboratory produces meta-autunite (II).
- (Beintema, 1938) proposed this structure for autunite, which (Herrmann, 1941) designated $H5_9$. This original tetragonal structure did not locate the oxygen atoms or the water molecules. Later determinations of this structure (Locock, 2003) located all the atoms in the crystal for $n = 11$, and found the structure to be orthorhombic.

**Body-centered Tetragonal primitive vectors:**

$$\mathbf{a}_1 = -\frac{1}{2}\,a\,\hat{\mathbf{x}} + \frac{1}{2}\,a\,\hat{\mathbf{y}} + \frac{1}{2}\,c\,\hat{\mathbf{z}}$$
$$\mathbf{a}_2 = \frac{1}{2}\,a\,\hat{\mathbf{x}} - \frac{1}{2}\,a\,\hat{\mathbf{y}} + \frac{1}{2}\,c\,\hat{\mathbf{z}}$$
$$\mathbf{a}_3 = \frac{1}{2}\,a\,\hat{\mathbf{x}} + \frac{1}{2}\,a\,\hat{\mathbf{y}} - \frac{1}{2}\,c\,\hat{\mathbf{z}}$$

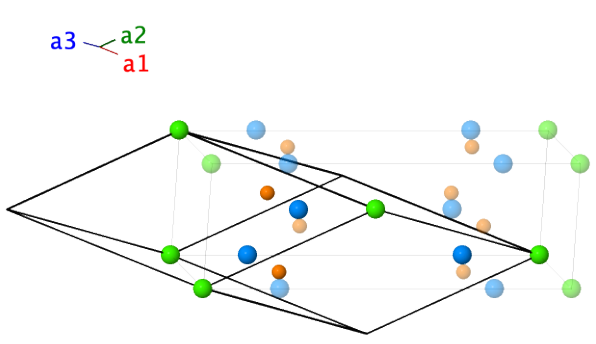

**Basis vectors:**

| | | Lattice Coordinates | | Cartesian Coordinates | Wyckoff Position | Atom Type |
|---|---|---|---|---|---|---|
| $\mathbf{B}_1$ | = | $0\,\mathbf{a}_1 + 0\,\mathbf{a}_2 + 0\,\mathbf{a}_3$ | = | $0\,\hat{\mathbf{x}} + 0\,\hat{\mathbf{y}} + 0\,\hat{\mathbf{z}}$ | (2*a*) | Ca |
| $\mathbf{B}_2$ | = | $\frac{3}{4}\,\mathbf{a}_1 + \frac{1}{4}\,\mathbf{a}_2 + \frac{1}{2}\,\mathbf{a}_3$ | = | $\frac{1}{2}a\,\hat{\mathbf{y}} + \frac{1}{4}c\,\hat{\mathbf{z}}$ | (4*d*) | P |
| $\mathbf{B}_3$ | = | $\frac{1}{4}\,\mathbf{a}_1 + \frac{3}{4}\,\mathbf{a}_2 + \frac{1}{2}\,\mathbf{a}_3$ | = | $\frac{1}{2}a\,\hat{\mathbf{x}} + \frac{1}{4}c\,\hat{\mathbf{z}}$ | (4*d*) | P |
| $\mathbf{B}_4$ | = | $z_3\,\mathbf{a}_1 + z_3\,\mathbf{a}_2$ | = | $z_3 c\,\hat{\mathbf{z}}$ | (4$e$) | U |
| $\mathbf{B}_5$ | = | $-z_3\,\mathbf{a}_1 - z_3\,\mathbf{a}_2$ | = | $-z_3 c\,\hat{\mathbf{z}}$ | (4$e$) | U |

**Geometry files:**

- CIF: pp. 1711
- POSCAR: pp. 1712

# $AuCsCl_3$ ($K7_6$) Structure: AB3C_tI20_139_ab_eh_d

http://aflow.org/prototype-encyclopedia/AB3C_tI20_139_ab_eh_d

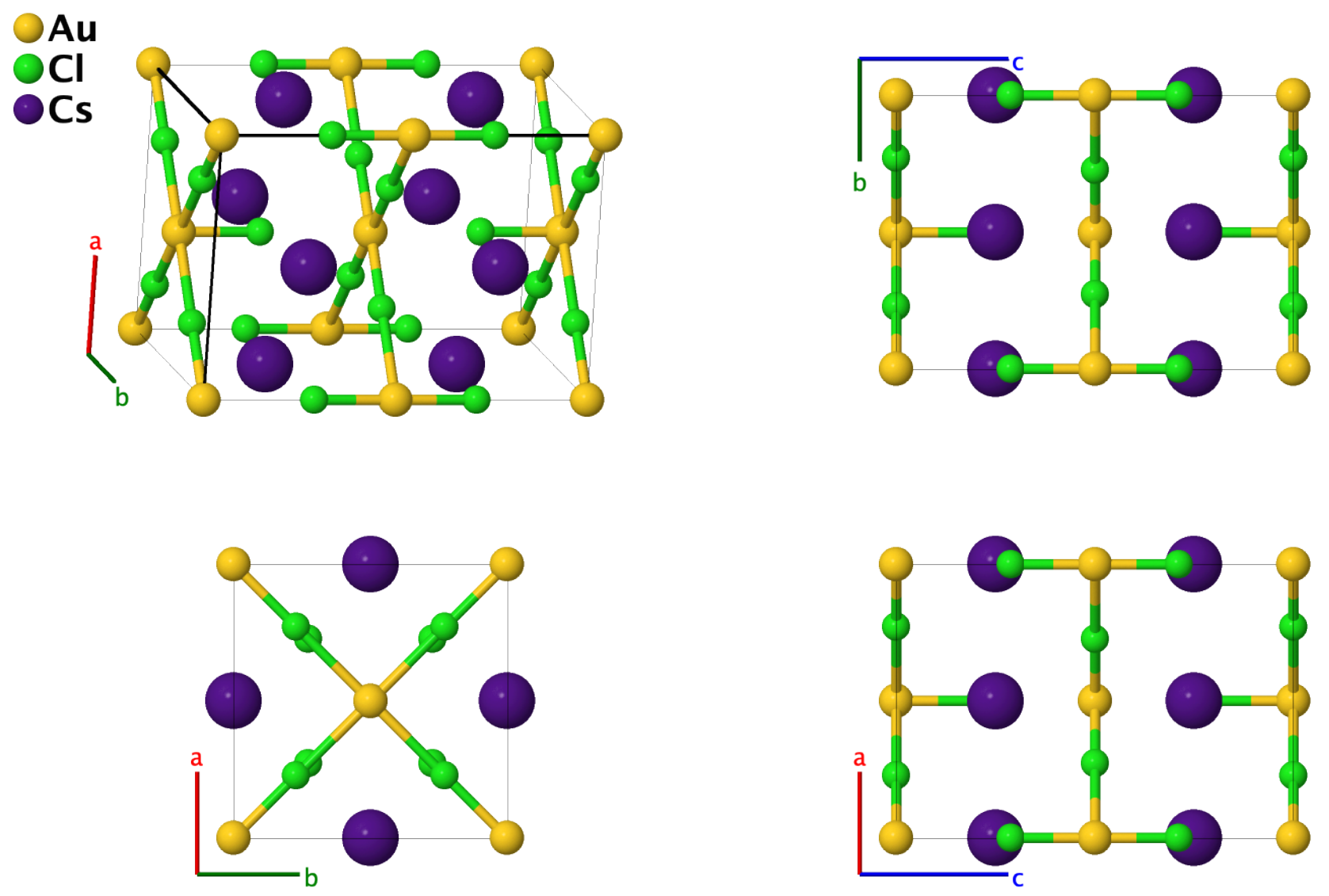

| | | |
|---|---|---|
| **Prototype** | : | $AuCl_3Cs$ |
| **AFLOW prototype label** | : | AB3C_tI20_139_ab_eh_d |
| ***Strukturbericht* designation** | : | $K7_6$ |
| **Pearson symbol** | : | tI20 |
| **Space group number** | : | 139 |
| **Space group symbol** | : | $I4/mmm$ |
| **AFLOW prototype command** | : | `aflow --proto=AB3C_tI20_139_ab_eh_d` <br> `--params=`$a, c/a, z_4, x_5$ |

### Other compounds with this structure

- $AgAuCs_2Cl_6$

### Body-centered Tetragonal primitive vectors:

$$\mathbf{a}_1 = -\frac{1}{2}\,a\,\hat{\mathbf{x}} + \frac{1}{2}\,a\,\hat{\mathbf{y}} + \frac{1}{2}\,c\,\hat{\mathbf{z}}$$
$$\mathbf{a}_2 = \frac{1}{2}\,a\,\hat{\mathbf{x}} - \frac{1}{2}\,a\,\hat{\mathbf{y}} + \frac{1}{2}\,c\,\hat{\mathbf{z}}$$
$$\mathbf{a}_3 = \frac{1}{2}\,a\,\hat{\mathbf{x}} + \frac{1}{2}\,a\,\hat{\mathbf{y}} - \frac{1}{2}\,c\,\hat{\mathbf{z}}$$

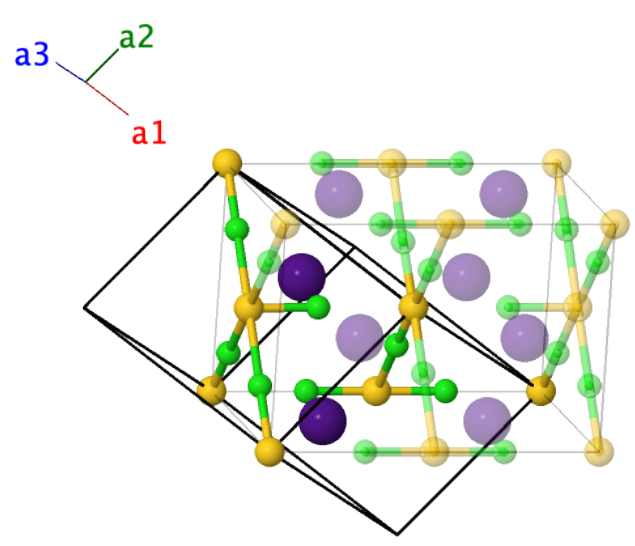

### Basis vectors:

| | | Lattice Coordinates | | Cartesian Coordinates | Wyckoff Position | Atom Type |
|---|---|---|---|---|---|---|
| $\mathbf{B}_1$ | = | $0\,\mathbf{a}_1 + 0\,\mathbf{a}_2 + 0\,\mathbf{a}_3$ | = | $0\,\hat{\mathbf{x}} + 0\,\hat{\mathbf{y}} + 0\,\hat{\mathbf{z}}$ | (2*a*) | Au I |
| $\mathbf{B}_2$ | = | $\frac{1}{2}\,\mathbf{a}_1 + \frac{1}{2}\,\mathbf{a}_2$ | = | $\frac{1}{2}c\,\hat{\mathbf{z}}$ | (2*b*) | Au II |
| $\mathbf{B}_3$ | = | $\frac{3}{4}\,\mathbf{a}_1 + \frac{1}{4}\,\mathbf{a}_2 + \frac{1}{2}\,\mathbf{a}_3$ | = | $\frac{1}{2}a\,\hat{\mathbf{y}} + \frac{1}{4}c\,\hat{\mathbf{z}}$ | (4*d*) | Cs |
| $\mathbf{B}_4$ | = | $\frac{1}{4}\,\mathbf{a}_1 + \frac{3}{4}\,\mathbf{a}_2 + \frac{1}{2}\,\mathbf{a}_3$ | = | $\frac{1}{2}a\,\hat{\mathbf{x}} + \frac{1}{4}c\,\hat{\mathbf{z}}$ | (4*d*) | Cs |
| $\mathbf{B}_5$ | = | $z_4\,\mathbf{a}_1 + z_4\,\mathbf{a}_2$ | = | $z_4 c\,\hat{\mathbf{z}}$ | $(4e)$ | Cl I |
| $\mathbf{B}_6$ | = | $-z_4\,\mathbf{a}_1 - z_4\,\mathbf{a}_2$ | = | $-z_4 c\,\hat{\mathbf{z}}$ | $(4e)$ | Cl I |
| $\mathbf{B}_7$ | = | $x_5\,\mathbf{a}_1 + x_5\,\mathbf{a}_2 + 2x_5\,\mathbf{a}_3$ | = | $x_5 a\,\hat{\mathbf{x}} + x_5 a\,\hat{\mathbf{y}}$ | $(8h)$ | Cl II |
| $\mathbf{B}_8$ | = | $-x_5\,\mathbf{a}_1 - x_5\,\mathbf{a}_2 - 2x_5\,\mathbf{a}_3$ | = | $-x_5 a\,\hat{\mathbf{x}} - x_5 a\,\hat{\mathbf{y}}$ | $(8h)$ | Cl II |
| $\mathbf{B}_9$ | = | $x_5\,\mathbf{a}_1 - x_5\,\mathbf{a}_2$ | = | $-x_5 a\,\hat{\mathbf{x}} + x_5 a\,\hat{\mathbf{y}}$ | $(8h)$ | Cl II |
| $\mathbf{B}_{10}$ | = | $-x_5\,\mathbf{a}_1 + x_5\,\mathbf{a}_2$ | = | $x_5 a\,\hat{\mathbf{x}} - x_5 a\,\hat{\mathbf{y}}$ | $(8h)$ | Cl II |

**Geometry files:**

- CIF: pp. 1712
- POSCAR: pp. 1712

# “Martensite Type” $FeC_x$ ($x \leq 0.06$) ($L2_0$) Structure: AB_tI4_139_b_a

http://aflow.org/prototype-encyclopedia/AB_tI4_139_b_a

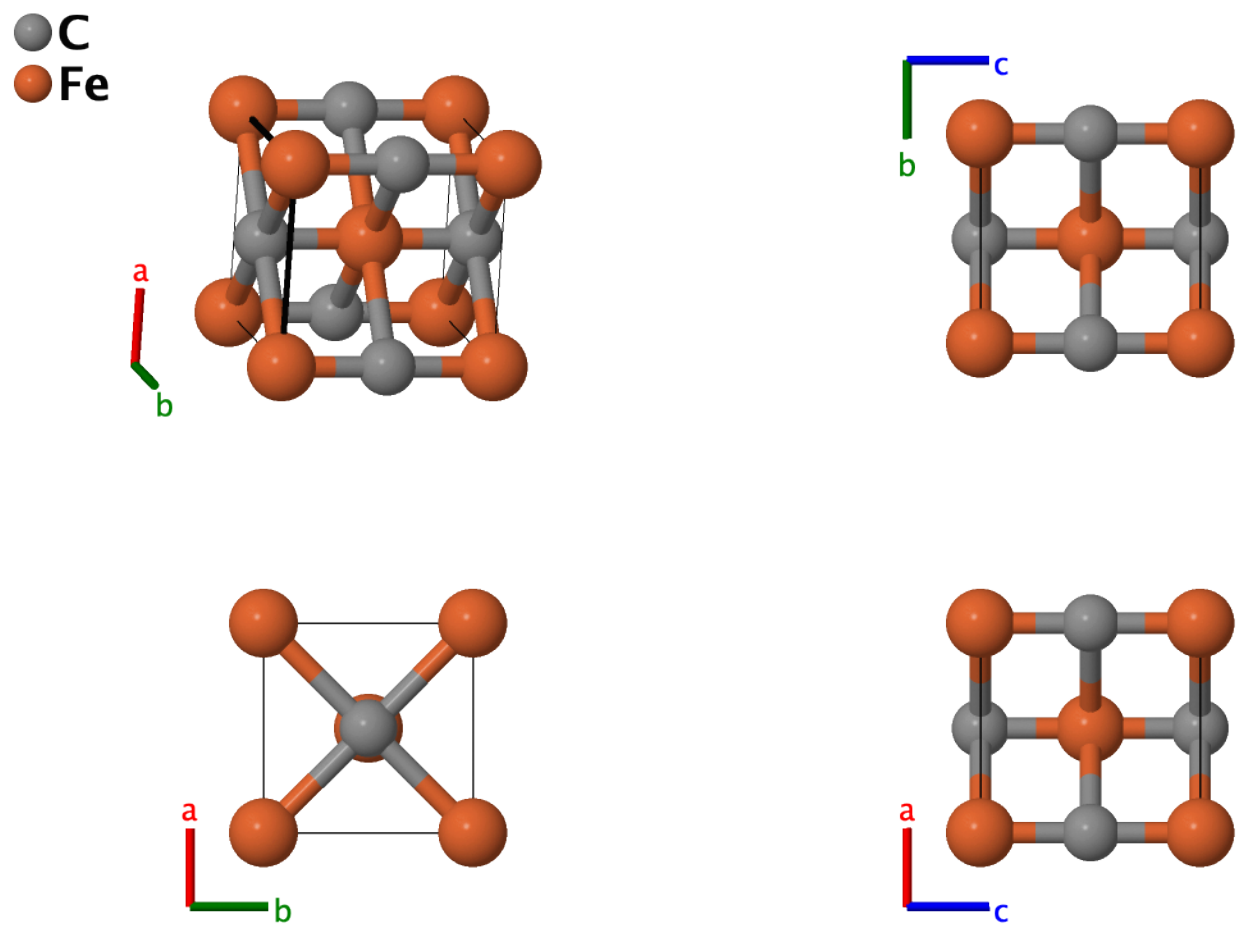

| | | |
|---|---|---|
| **Prototype** | : | $C_xFe$ |
| **AFLOW prototype label** | : | AB_tI4_139_b_a |
| ***Strukturbericht* designation** | : | $L2_0$ |
| **Pearson symbol** | : | tI4 |
| **Space group number** | : | 139 |
| **Space group symbol** | : | $I4/mmm$ |
| **AFLOW prototype command** | : | `aflow --proto=AB_tI4_139_b_a --params=`$a, c/a$ |

- This should *not* be considered a true model of Martensite, a very complex material whose structure is sensitive to composition (Blackwell, 1996; Sherby, 2008). However, (Ewald, 1931) assign this the label $L'2_0$, saying that this structure is “$\alpha$-Fe with 6 atomic percent C.” Although Ewald and Hermann present a long discussion of martensite (pp. 489, 576-583), they do not give this exact crystal structure, which we take from (Brandes, 1992) and (Pearson, 1967).
- (Villars, 1995) and (Westbrook, 1995) ignore this structure and give the $L'2$ label to the $ThH_2$ structure. We previously followed this notation, but in the interest of historical accuracy will relabel that structure with the name given to it by (Brandes, 1992) and (Pearson, 1967), $L'2_b$.
- We should note that these last two references list the current structure simply as $L'2$.

**Body-centered Tetragonal primitive vectors:**

$$\mathbf{a}_1 = -\frac{1}{2}\,a\,\hat{\mathbf{x}} + \frac{1}{2}\,a\,\hat{\mathbf{y}} + \frac{1}{2}\,c\,\hat{\mathbf{z}}$$
$$\mathbf{a}_2 = \frac{1}{2}\,a\,\hat{\mathbf{x}} - \frac{1}{2}\,a\,\hat{\mathbf{y}} + \frac{1}{2}\,c\,\hat{\mathbf{z}}$$
$$\mathbf{a}_3 = \frac{1}{2}\,a\,\hat{\mathbf{x}} + \frac{1}{2}\,a\,\hat{\mathbf{y}} - \frac{1}{2}\,c\,\hat{\mathbf{z}}$$

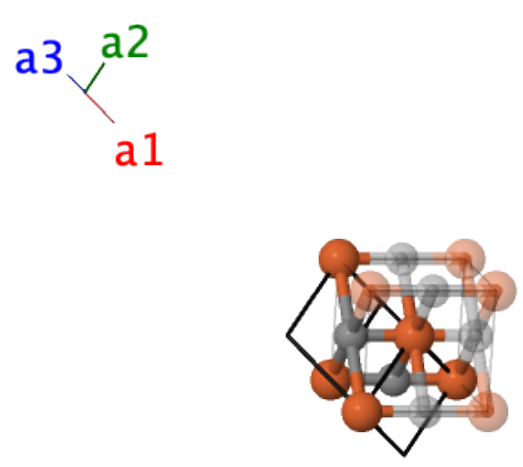

**Basis vectors:**

| | | Lattice Coordinates | | Cartesian Coordinates | Wyckoff Position | Atom Type |
|---|---|---|---|---|---|---|
| $\mathbf{B}_1$ | = | $0\,\mathbf{a}_1 + 0\,\mathbf{a}_2 + 0\,\mathbf{a}_3$ | = | $0\,\hat{\mathbf{x}} + 0\,\hat{\mathbf{y}} + 0\,\hat{\mathbf{z}}$ | $(2a)$ | Fe |
| $\mathbf{B}_2$ | = | $\frac{1}{2}\,\mathbf{a}_1 + \frac{1}{2}\,\mathbf{a}_2$ | = | $\frac{1}{2}c\,\hat{\mathbf{z}}$ | $(2b)$ | C |

---

**Found in:**

---

**Geometry files:**

- CIF: pp. 1713
- POSCAR: pp. 1713

# $V_4SiSb_2$ Structure: A2BC4_tI28_140_h_a_k

http://aflow.org/prototype-encyclopedia/A2BC4_tI28_140_h_a_k

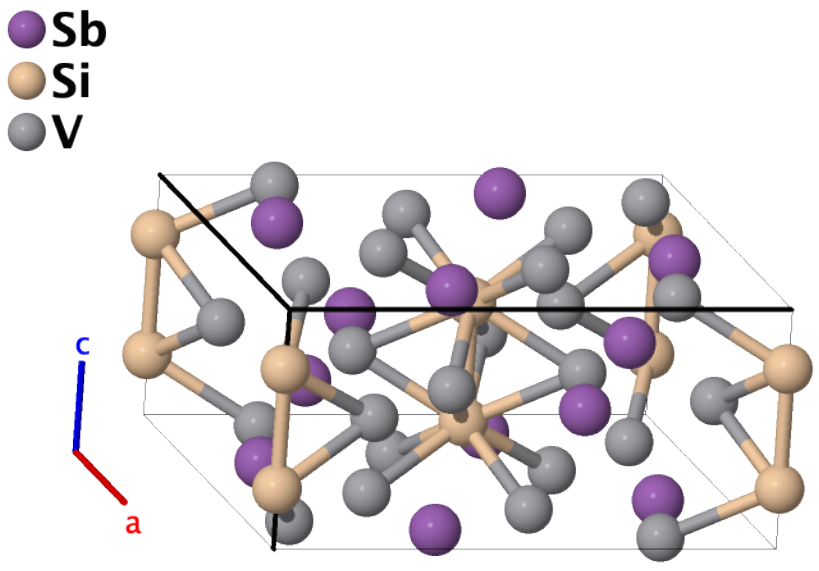

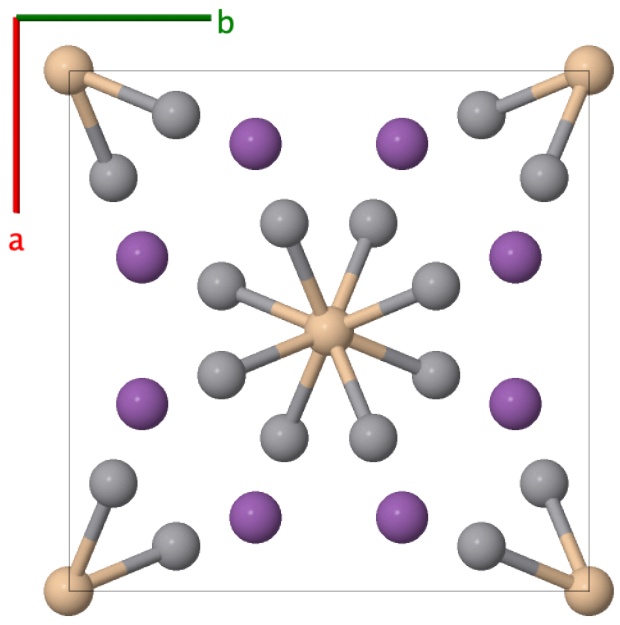

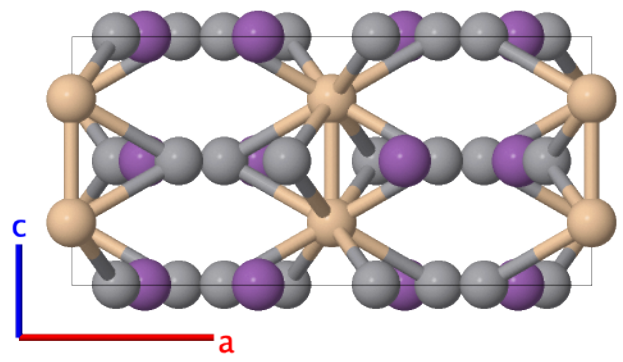

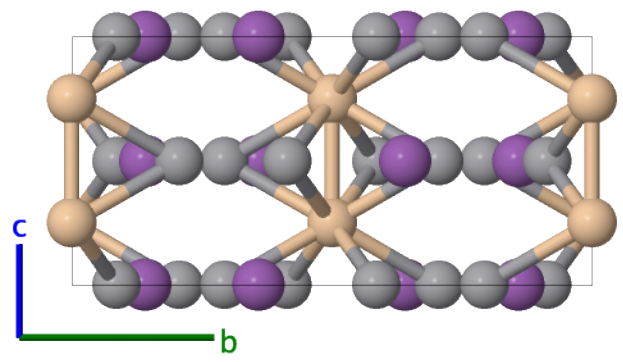

| | | |
|---|---|---|
| **Prototype** | : | $Sb_2SiV_4$ |
| **AFLOW prototype label** | : | A2BC4_tI28_140_h_a_k |
| ***Strukturbericht* designation** | : | None |
| **Pearson symbol** | : | tI28 |
| **Space group number** | : | 140 |
| **Space group symbol** | : | $I4/mcm$ |
| **AFLOW prototype command** | : | `aflow --proto=A2BC4_tI28_140_h_a_k`<br>`--params=`$a, c/a, x_2, x_3, y_3$ |

### Other compounds with this structure

- $Ti_4CoBi_2$, $Ti_4CrBi_2$, $Ti_4FeBi_2$, $Ti_4MnBi_2$, and $Ti_4NiBi_2$

---

- This is the ternary version of the $D2_c$ $U_6Mn$ structure. This can also be identified as a defected version of the $D8_m$ $W_5Si_3$ structure.

### Body-centered Tetragonal primitive vectors:

$$\mathbf{a}_1 = -\frac{1}{2}\,a\,\hat{\mathbf{x}} + \frac{1}{2}\,a\,\hat{\mathbf{y}} + \frac{1}{2}\,c\,\hat{\mathbf{z}}$$
$$\mathbf{a}_2 = \frac{1}{2}\,a\,\hat{\mathbf{x}} - \frac{1}{2}\,a\,\hat{\mathbf{y}} + \frac{1}{2}\,c\,\hat{\mathbf{z}}$$
$$\mathbf{a}_3 = \frac{1}{2}\,a\,\hat{\mathbf{x}} + \frac{1}{2}\,a\,\hat{\mathbf{y}} - \frac{1}{2}\,c\,\hat{\mathbf{z}}$$

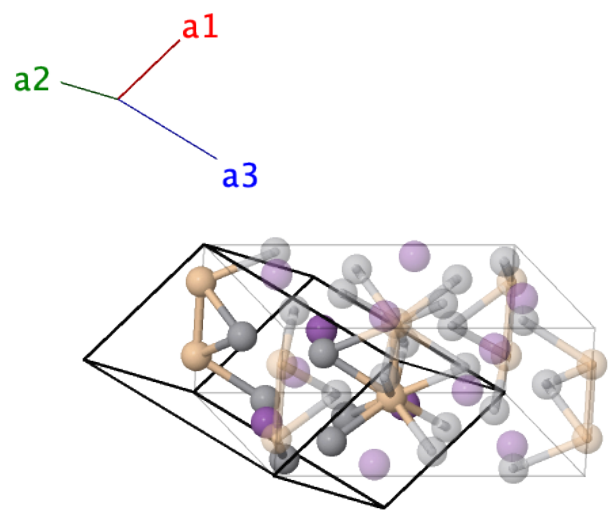

**Basis vectors:**

| | | Lattice Coordinates | | Cartesian Coordinates | Wyckoff Position | Atom Type |
|---|---|---|---|---|---|---|
| $\mathbf{B}_1$ | = | $\frac{1}{4}\,\mathbf{a}_1 + \frac{1}{4}\,\mathbf{a}_2$ | = | $\frac{1}{4}c\,\hat{\mathbf{z}}$ | (4$a$) | Si |
| $\mathbf{B}_2$ | = | $\frac{3}{4}\,\mathbf{a}_1 + \frac{3}{4}\,\mathbf{a}_2$ | = | $\frac{3}{4}c\,\hat{\mathbf{z}}$ | (4$a$) | Si |
| $\mathbf{B}_3$ | = | $\left(\frac{1}{2} + x_2\right)\,\mathbf{a}_1 + x_2\,\mathbf{a}_2 + \left(\frac{1}{2} + 2x_2\right)\,\mathbf{a}_3$ | = | $x_2a\,\hat{\mathbf{x}} + \left(\frac{1}{2} + x_2\right)a\,\hat{\mathbf{y}}$ | (8$h$) | Sb |
| $\mathbf{B}_4$ | = | $\left(\frac{1}{2} - x_2\right)\,\mathbf{a}_1 - x_2\,\mathbf{a}_2 + \left(\frac{1}{2} - 2x_2\right)\,\mathbf{a}_3$ | = | $-x_2a\,\hat{\mathbf{x}} + \left(\frac{1}{2} - x_2\right)a\,\hat{\mathbf{y}}$ | (8$h$) | Sb |
| $\mathbf{B}_5$ | = | $x_2\,\mathbf{a}_1 + \left(\frac{1}{2} - x_2\right)\,\mathbf{a}_2 + \frac{1}{2}\,\mathbf{a}_3$ | = | $\left(\frac{1}{2} - x_2\right)a\,\hat{\mathbf{x}} + x_2a\,\hat{\mathbf{y}}$ | (8$h$) | Sb |
| $\mathbf{B}_6$ | = | $-x_2\,\mathbf{a}_1 + \left(\frac{1}{2} + x_2\right)\,\mathbf{a}_2 + \frac{1}{2}\,\mathbf{a}_3$ | = | $\left(\frac{1}{2} + x_2\right)a\,\hat{\mathbf{x}} - x_2a\,\hat{\mathbf{y}}$ | (8$h$) | Sb |
| $\mathbf{B}_7$ | = | $y_3\,\mathbf{a}_1 + x_3\,\mathbf{a}_2 + (x_3 + y_3)\,\mathbf{a}_3$ | = | $x_3a\,\hat{\mathbf{x}} + y_3a\,\hat{\mathbf{y}}$ | (16$k$) | V |
| $\mathbf{B}_8$ | = | $-y_3\,\mathbf{a}_1 - x_3\,\mathbf{a}_2 + (-x_3 - y_3)\,\mathbf{a}_3$ | = | $-x_3a\,\hat{\mathbf{x}} - y_3a\,\hat{\mathbf{y}}$ | (16$k$) | V |
| $\mathbf{B}_9$ | = | $x_3\,\mathbf{a}_1 - y_3\,\mathbf{a}_2 + (x_3 - y_3)\,\mathbf{a}_3$ | = | $-y_3a\,\hat{\mathbf{x}} + x_3a\,\hat{\mathbf{y}}$ | (16$k$) | V |
| $\mathbf{B}_{10}$ | = | $-x_3\,\mathbf{a}_1 + y_3\,\mathbf{a}_2 + (-x_3 + y_3)\,\mathbf{a}_3$ | = | $y_3a\,\hat{\mathbf{x}} - x_3a\,\hat{\mathbf{y}}$ | (16$k$) | V |
| $\mathbf{B}_{11}$ | = | $\left(\frac{1}{2} + y_3\right)\,\mathbf{a}_1 + \left(\frac{1}{2} - x_3\right)\,\mathbf{a}_2 + (-x_3 + y_3)\,\mathbf{a}_3$ | = | $-x_3a\,\hat{\mathbf{x}} + y_3a\,\hat{\mathbf{y}} + \frac{1}{2}c\,\hat{\mathbf{z}}$ | (16$k$) | V |
| $\mathbf{B}_{12}$ | = | $\left(\frac{1}{2} - y_3\right)\,\mathbf{a}_1 + \left(\frac{1}{2} + x_3\right)\,\mathbf{a}_2 + (x_3 - y_3)\,\mathbf{a}_3$ | = | $x_3a\,\hat{\mathbf{x}} - y_3a\,\hat{\mathbf{y}} + \frac{1}{2}c\,\hat{\mathbf{z}}$ | (16$k$) | V |
| $\mathbf{B}_{13}$ | = | $\left(\frac{1}{2} + x_3\right)\,\mathbf{a}_1 + \left(\frac{1}{2} + y_3\right)\,\mathbf{a}_2 + (x_3 + y_3)\,\mathbf{a}_3$ | = | $y_3a\,\hat{\mathbf{x}} + x_3a\,\hat{\mathbf{y}} + \frac{1}{2}c\,\hat{\mathbf{z}}$ | (16$k$) | V |
| $\mathbf{B}_{14}$ | = | $\left(\frac{1}{2} - x_3\right)\,\mathbf{a}_1 + \left(\frac{1}{2} - y_3\right)\,\mathbf{a}_2 + (-x_3 - y_3)\,\mathbf{a}_3$ | = | $-y_3a\,\hat{\mathbf{x}} - x_3a\,\hat{\mathbf{y}} + \frac{1}{2}c\,\hat{\mathbf{z}}$ | (16$k$) | V |

**Geometry files:**

- CIF: pp. 1713
- POSCAR: pp. 1713

# $KHF_2$ ($F5_2$) Structure: A2BC_tI16_140_h_d_a

http://aflow.org/prototype-encyclopedia/A2BC_tI16_140_h_d_a

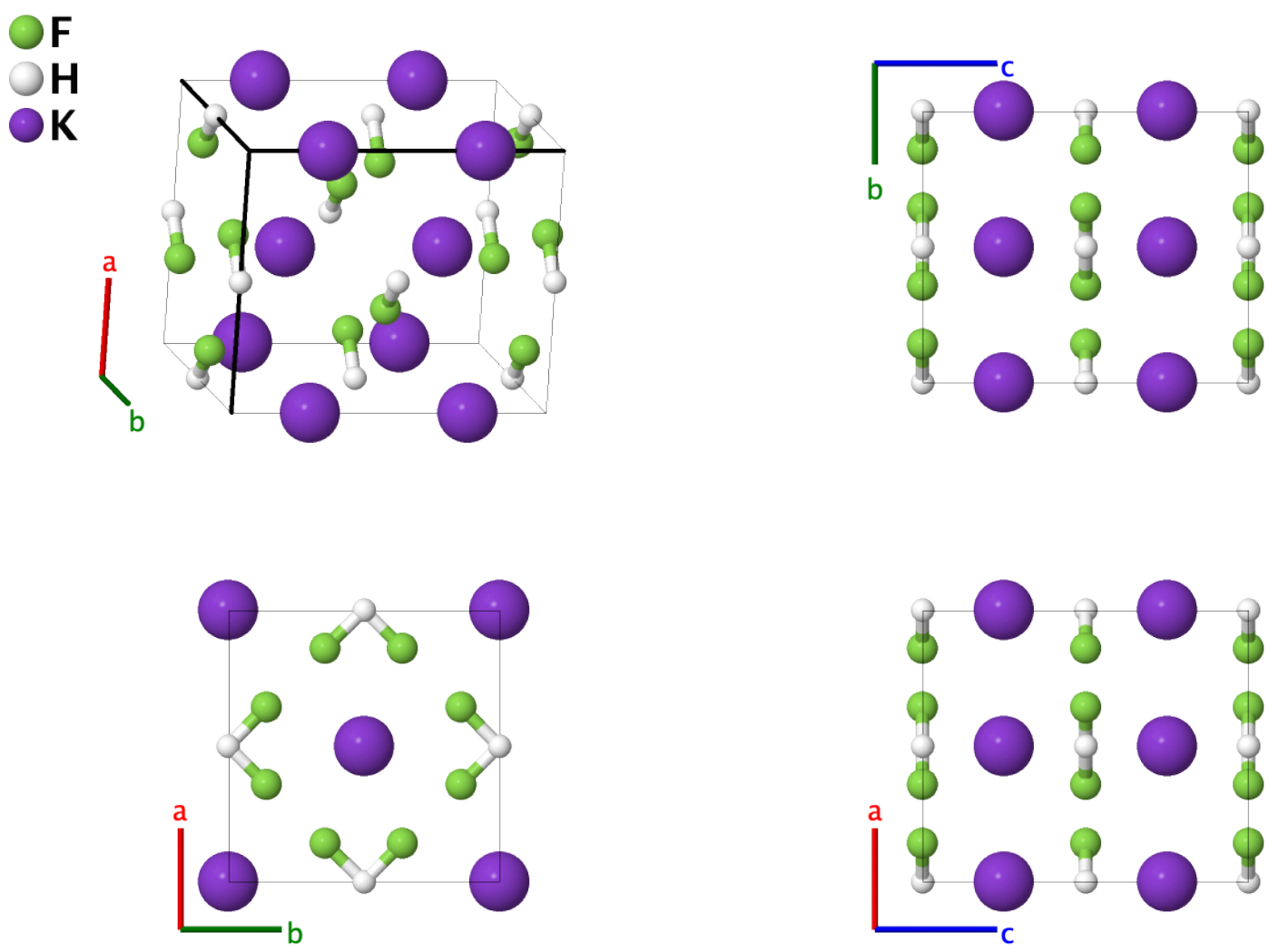

| | | |
|---|---|---|
| **Prototype** | : | $F_2HK$ |
| **AFLOW prototype label** | : | A2BC_tI16_140_h_d_a |
| ***Strukturbericht* designation** | : | $F5_2$ |
| **Pearson symbol** | : | tI16 |
| **Space group number** | : | 140 |
| **Space group symbol** | : | $I4/mcm$ |
| **AFLOW prototype command** | : | `aflow --proto=A2BC_tI16_140_h_d_a --params=`$a, c/a, x_3$ |

**Other compounds with this structure**

- $KN_3$ and $CsN_3$

- (Bozorth, 1923) originally determined the lattice constants of $KHF_2$ along with the positions of the potassium and fluorine atoms. He also assumed that the hydrogen atoms were on the (4*d*) Wyckoff sites. Both (Peterson, 1952) and (Ibers, 1964) confirmed his data. All three papers point out that it is possible that the hydrogen atoms are on half-filled (8*h*) sites, which would reduce to the (4*d*) site as the *x* coordinate approached zero. In $KN_3$ and $CsN_3$ the nitrogen atoms occupy the (4*d*) and (8*h*) Wyckoff sites, while the cation occupies the (4*a*) site.

**Body-centered Tetragonal primitive vectors:**

$$\mathbf{a}_1 = -\frac{1}{2}\,a\,\hat{\mathbf{x}} + \frac{1}{2}\,a\,\hat{\mathbf{y}} + \frac{1}{2}\,c\,\hat{\mathbf{z}}$$
$$\mathbf{a}_2 = \frac{1}{2}\,a\,\hat{\mathbf{x}} - \frac{1}{2}\,a\,\hat{\mathbf{y}} + \frac{1}{2}\,c\,\hat{\mathbf{z}}$$
$$\mathbf{a}_3 = \frac{1}{2}\,a\,\hat{\mathbf{x}} + \frac{1}{2}\,a\,\hat{\mathbf{y}} - \frac{1}{2}\,c\,\hat{\mathbf{z}}$$

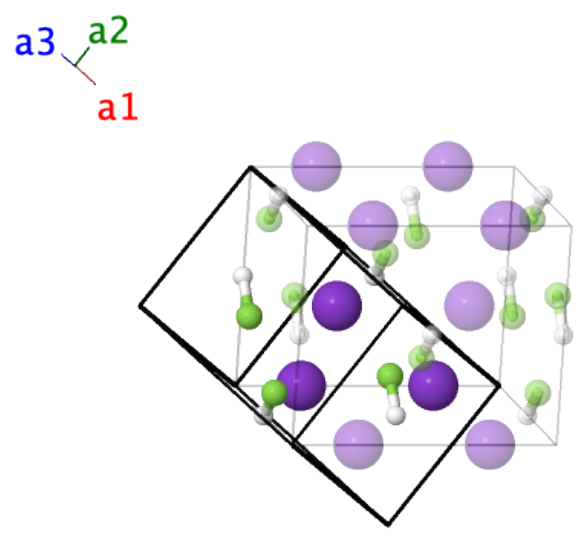

**Basis vectors:**

| | | Lattice Coordinates | | Cartesian Coordinates | Wyckoff Position | Atom Type |
|---|---|---|---|---|---|---|
| $\mathbf{B}_1$ | = | $\frac{1}{4}\,\mathbf{a}_1 + \frac{1}{4}\,\mathbf{a}_2$ | = | $\frac{1}{4}c\,\hat{\mathbf{z}}$ | (4$a$) | K |
| $\mathbf{B}_2$ | = | $\frac{3}{4}\,\mathbf{a}_1 + \frac{3}{4}\,\mathbf{a}_2$ | = | $\frac{3}{4}c\,\hat{\mathbf{z}}$ | (4$a$) | K |
| $\mathbf{B}_3$ | = | $\frac{1}{2}\,\mathbf{a}_1 + \frac{1}{2}\,\mathbf{a}_3$ | = | $\frac{1}{2}a\,\hat{\mathbf{y}}$ | (4$d$) | H |
| $\mathbf{B}_4$ | = | $\frac{1}{2}\,\mathbf{a}_2 + \frac{1}{2}\,\mathbf{a}_3$ | = | $\frac{1}{2}a\,\hat{\mathbf{x}}$ | (4$d$) | H |
| $\mathbf{B}_5$ | = | $\left(\frac{1}{2} + x_3\right)\,\mathbf{a}_1 + x_3\,\mathbf{a}_2 + \left(\frac{1}{2} + 2x_3\right)\,\mathbf{a}_3$ | = | $x_3a\,\hat{\mathbf{x}} + \left(\frac{1}{2} + x_3\right)a\,\hat{\mathbf{y}}$ | (8$h$) | F |
| $\mathbf{B}_6$ | = | $\left(\frac{1}{2} - x_3\right)\,\mathbf{a}_1 - x_3\,\mathbf{a}_2 + \left(\frac{1}{2} - 2x_3\right)\,\mathbf{a}_3$ | = | $-x_3a\,\hat{\mathbf{x}} + \left(\frac{1}{2} - x_3\right)a\,\hat{\mathbf{y}}$ | (8$h$) | F |
| $\mathbf{B}_7$ | = | $x_3\,\mathbf{a}_1 + \left(\frac{1}{2} - x_3\right)\,\mathbf{a}_2 + \frac{1}{2}\,\mathbf{a}_3$ | = | $\left(\frac{1}{2} - x_3\right)a\,\hat{\mathbf{x}} + x_3a\,\hat{\mathbf{y}}$ | (8$h$) | F |
| $\mathbf{B}_8$ | = | $-x_3\,\mathbf{a}_1 + \left(\frac{1}{2} + x_3\right)\,\mathbf{a}_2 + \frac{1}{2}\,\mathbf{a}_3$ | = | $\left(\frac{1}{2} + x_3\right)a\,\hat{\mathbf{x}} - x_3a\,\hat{\mathbf{y}}$ | (8$h$) | F |

---

---

**Geometry files:**

- CIF: pp. 1714
- POSCAR: pp. 1714

# $Pu_{31}Rh_{20}$ Structure: A31B20_tI204_140_b2gh3m_ac2fh3l

http://aflow.org/prototype-encyclopedia/A31B20_tI204_140_b2gh3m_ac2fh3l

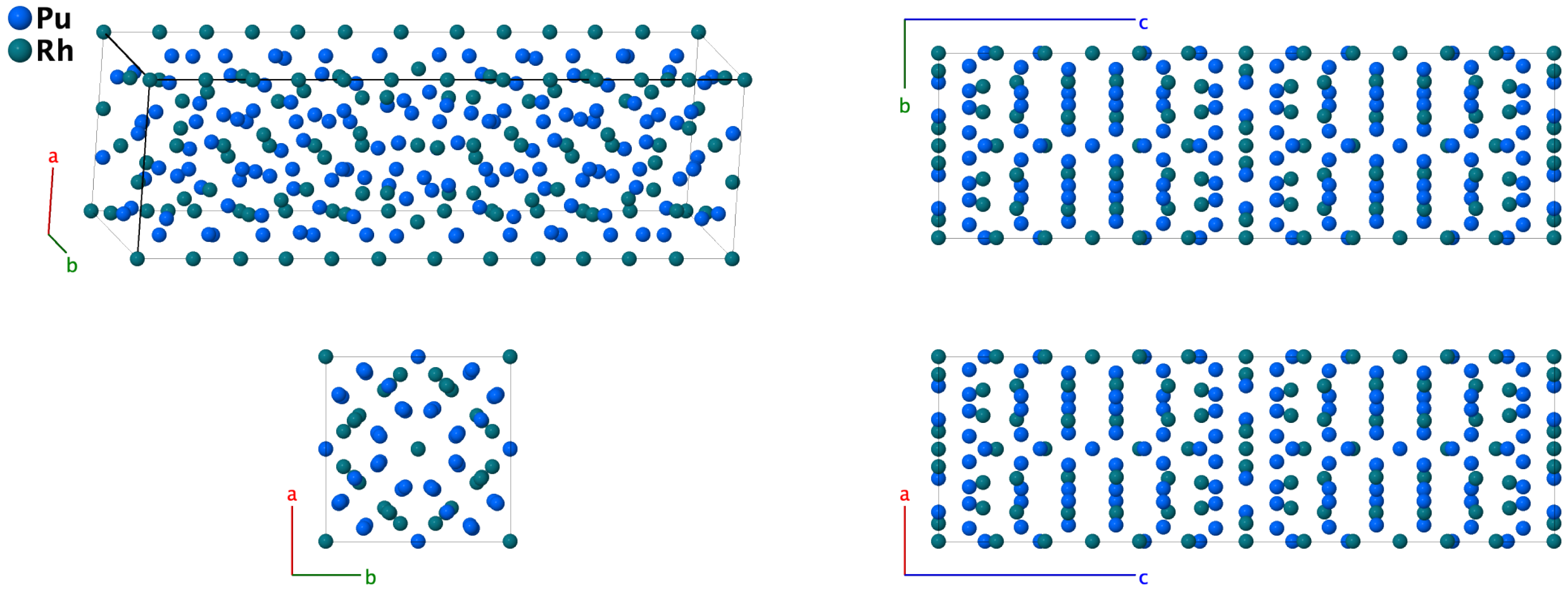

| | | |
|---|---|---|
| **Prototype** | : | $Pu_{31}Rh_{20}$ |
| **AFLOW prototype label** | : | A31B20_tI204_140_b2gh3m_ac2fh3l |
| ***Strukturbericht* designation** | : | None |
| **Pearson symbol** | : | tI204 |
| **Space group number** | : | 140 |
| **Space group symbol** | : | $I4/mcm$ |
| **AFLOW prototype command** | : | `aflow --proto=A31B20_tI204_140_b2gh3m_ac2fh3l` `--params=`$a, c/a, z_4, z_5, z_6, z_7, x_8, x_9, x_{10}, z_{10}, x_{11}, z_{11}, x_{12}, z_{12}, x_{13}, y_{13}, z_{13}, x_{14}, y_{14}, z_{14}, x_{15}, y_{15}, z_{15}$ |

### Other compounds with this structure

- $Pu_{31}Pt_{20}$ and $Ca_{31}Sn_{20}$

### Body-centered Tetragonal primitive vectors:

$$\mathbf{a}_1 = -\frac{1}{2}\,a\,\hat{\mathbf{x}} + \frac{1}{2}\,a\,\hat{\mathbf{y}} + \frac{1}{2}\,c\,\hat{\mathbf{z}}$$
$$\mathbf{a}_2 = \frac{1}{2}\,a\,\hat{\mathbf{x}} - \frac{1}{2}\,a\,\hat{\mathbf{y}} + \frac{1}{2}\,c\,\hat{\mathbf{z}}$$
$$\mathbf{a}_3 = \frac{1}{2}\,a\,\hat{\mathbf{x}} + \frac{1}{2}\,a\,\hat{\mathbf{y}} - \frac{1}{2}\,c\,\hat{\mathbf{z}}$$

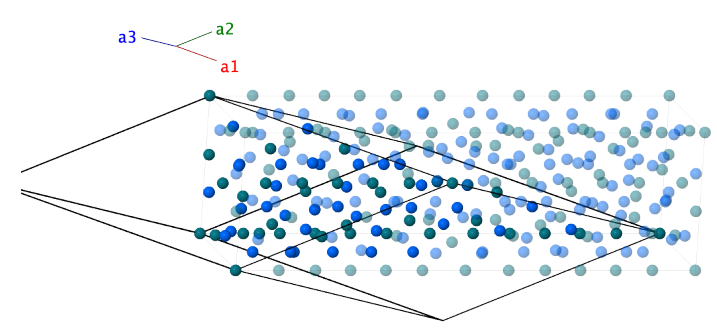

### Basis vectors:

| | | Lattice Coordinates | | Cartesian Coordinates | Wyckoff Position | Atom Type |
|---|---|---|---|---|---|---|
| $\mathbf{B}_1$ | = | $\frac{1}{4}\,\mathbf{a}_1 + \frac{1}{4}\,\mathbf{a}_2$ | = | $\frac{1}{4}c\,\hat{\mathbf{z}}$ | (4$a$) | Rh I |
| $\mathbf{B}_2$ | = | $\frac{3}{4}\,\mathbf{a}_1 + \frac{3}{4}\,\mathbf{a}_2$ | = | $\frac{3}{4}c\,\hat{\mathbf{z}}$ | (4$a$) | Rh I |
| $\mathbf{B}_3$ | = | $\frac{3}{4}\,\mathbf{a}_1 + \frac{1}{4}\,\mathbf{a}_2 + \frac{1}{2}\,\mathbf{a}_3$ | = | $\frac{1}{2}a\,\hat{\mathbf{y}} + \frac{1}{4}c\,\hat{\mathbf{z}}$ | (4$b$) | Pu I |
| $\mathbf{B}_4$ | = | $\frac{1}{4}\,\mathbf{a}_1 + \frac{3}{4}\,\mathbf{a}_2 + \frac{1}{2}\,\mathbf{a}_3$ | = | $\frac{1}{2}a\,\hat{\mathbf{x}} + \frac{1}{4}c\,\hat{\mathbf{z}}$ | (4$b$) | Pu I |
| $\mathbf{B}_5$ | = | $0\,\mathbf{a}_1 + 0\,\mathbf{a}_2 + 0\,\mathbf{a}_3$ | = | $0\,\hat{\mathbf{x}} + 0\,\hat{\mathbf{y}} + 0\,\hat{\mathbf{z}}$ | (4$c$) | Rh II |
| $\mathbf{B}_6$ | = | $\frac{1}{2}\,\mathbf{a}_1 + \frac{1}{2}\,\mathbf{a}_2$ | = | $\frac{1}{2}c\,\hat{\mathbf{z}}$ | (4$c$) | Rh II |
| $\mathbf{B}_7$ | = | $z_4\,\mathbf{a}_1 + z_4\,\mathbf{a}_2$ | = | $z_4 c\,\hat{\mathbf{z}}$ | (8$f$) | Rh III |

| | | | | |
|---|---|---|---|---|
| $\mathbf{B}_8 = \left(\frac{1}{2} - z_4\right)\mathbf{a}_1 + \left(\frac{1}{2} - z_4\right)\mathbf{a}_2$ | $=$ | $\left(\frac{1}{2} - z_4\right)c\,\hat{\mathbf{z}}$ | $(8f)$ | Rh III |
| $\mathbf{B}_9 = -z_4\,\mathbf{a}_1 - z_4\,\mathbf{a}_2$ | $=$ | $-z_4c\,\hat{\mathbf{z}}$ | $(8f)$ | Rh III |
| $\mathbf{B}_{10} = \left(\frac{1}{2} + z_4\right)\mathbf{a}_1 + \left(\frac{1}{2} + z_4\right)\mathbf{a}_2$ | $=$ | $\left(\frac{1}{2} + z_4\right)c\,\hat{\mathbf{z}}$ | $(8f)$ | Rh III |
| $\mathbf{B}_{11} = z_5\,\mathbf{a}_1 + z_5\,\mathbf{a}_2$ | $=$ | $z_5c\,\hat{\mathbf{z}}$ | $(8f)$ | Rh IV |
| $\mathbf{B}_{12} = \left(\frac{1}{2} - z_5\right)\mathbf{a}_1 + \left(\frac{1}{2} - z_5\right)\mathbf{a}_2$ | $=$ | $\left(\frac{1}{2} - z_5\right)c\,\hat{\mathbf{z}}$ | $(8f)$ | Rh IV |
| $\mathbf{B}_{13} = -z_5\,\mathbf{a}_1 - z_5\,\mathbf{a}_2$ | $=$ | $-z_5c\,\hat{\mathbf{z}}$ | $(8f)$ | Rh IV |
| $\mathbf{B}_{14} = \left(\frac{1}{2} + z_5\right)\mathbf{a}_1 + \left(\frac{1}{2} + z_5\right)\mathbf{a}_2$ | $=$ | $\left(\frac{1}{2} + z_5\right)c\,\hat{\mathbf{z}}$ | $(8f)$ | Rh IV |
| $\mathbf{B}_{15} = \left(\frac{1}{2} + z_6\right)\mathbf{a}_1 + z_6\,\mathbf{a}_2 + \frac{1}{2}\,\mathbf{a}_3$ | $=$ | $\frac{1}{2}a\,\hat{\mathbf{y}} + z_6c\,\hat{\mathbf{z}}$ | $(8g)$ | Pu II |
| $\mathbf{B}_{16} = z_6\,\mathbf{a}_1 + \left(\frac{1}{2} + z_6\right)\mathbf{a}_2 + \frac{1}{2}\,\mathbf{a}_3$ | $=$ | $\frac{1}{2}a\,\hat{\mathbf{x}} + z_6c\,\hat{\mathbf{z}}$ | $(8g)$ | Pu II |
| $\mathbf{B}_{17} = -z_6\,\mathbf{a}_1 + \left(\frac{1}{2} - z_6\right)\mathbf{a}_2 + \frac{1}{2}\,\mathbf{a}_3$ | $=$ | $\frac{1}{2}a\,\hat{\mathbf{x}} - z_6c\,\hat{\mathbf{z}}$ | $(8g)$ | Pu II |
| $\mathbf{B}_{18} = \left(\frac{1}{2} - z_6\right)\mathbf{a}_1 - z_6\,\mathbf{a}_2 + \frac{1}{2}\,\mathbf{a}_3$ | $=$ | $\frac{1}{2}a\,\hat{\mathbf{y}} - z_6c\,\hat{\mathbf{z}}$ | $(8g)$ | Pu II |
| $\mathbf{B}_{19} = \left(\frac{1}{2} + z_7\right)\mathbf{a}_1 + z_7\,\mathbf{a}_2 + \frac{1}{2}\,\mathbf{a}_3$ | $=$ | $\frac{1}{2}a\,\hat{\mathbf{y}} + z_7c\,\hat{\mathbf{z}}$ | $(8g)$ | Pu III |
| $\mathbf{B}_{20} = z_7\,\mathbf{a}_1 + \left(\frac{1}{2} + z_7\right)\mathbf{a}_2 + \frac{1}{2}\,\mathbf{a}_3$ | $=$ | $\frac{1}{2}a\,\hat{\mathbf{x}} + z_7c\,\hat{\mathbf{z}}$ | $(8g)$ | Pu III |
| $\mathbf{B}_{21} = -z_7\,\mathbf{a}_1 + \left(\frac{1}{2} - z_7\right)\mathbf{a}_2 + \frac{1}{2}\,\mathbf{a}_3$ | $=$ | $\frac{1}{2}a\,\hat{\mathbf{x}} - z_7c\,\hat{\mathbf{z}}$ | $(8g)$ | Pu III |
| $\mathbf{B}_{22} = \left(\frac{1}{2} - z_7\right)\mathbf{a}_1 - z_7\,\mathbf{a}_2 + \frac{1}{2}\,\mathbf{a}_3$ | $=$ | $\frac{1}{2}a\,\hat{\mathbf{y}} - z_7c\,\hat{\mathbf{z}}$ | $(8g)$ | Pu III |
| $\mathbf{B}_{23} = \left(\frac{1}{2} + x_8\right)\mathbf{a}_1 + x_8\,\mathbf{a}_2 + \left(\frac{1}{2} + 2x_8\right)\mathbf{a}_3$ | $=$ | $x_8a\,\hat{\mathbf{x}} + \left(\frac{1}{2} + x_8\right)a\,\hat{\mathbf{y}}$ | $(8h)$ | Pu IV |
| $\mathbf{B}_{24} = \left(\frac{1}{2} - x_8\right)\mathbf{a}_1 - x_8\,\mathbf{a}_2 + \left(\frac{1}{2} - 2x_8\right)\mathbf{a}_3$ | $=$ | $-x_8a\,\hat{\mathbf{x}} + \left(\frac{1}{2} - x_8\right)a\,\hat{\mathbf{y}}$ | $(8h)$ | Pu IV |
| $\mathbf{B}_{25} = x_8\,\mathbf{a}_1 + \left(\frac{1}{2} - x_8\right)\mathbf{a}_2 + \frac{1}{2}\,\mathbf{a}_3$ | $=$ | $\left(\frac{1}{2} - x_8\right)a\,\hat{\mathbf{x}} + x_8a\,\hat{\mathbf{y}}$ | $(8h)$ | Pu IV |
| $\mathbf{B}_{26} = -x_8\,\mathbf{a}_1 + \left(\frac{1}{2} + x_8\right)\mathbf{a}_2 + \frac{1}{2}\,\mathbf{a}_3$ | $=$ | $\left(\frac{1}{2} + x_8\right)a\,\hat{\mathbf{x}} - x_8a\,\hat{\mathbf{y}}$ | $(8h)$ | Pu IV |
| $\mathbf{B}_{27} = \left(\frac{1}{2} + x_9\right)\mathbf{a}_1 + x_9\,\mathbf{a}_2 + \left(\frac{1}{2} + 2x_9\right)\mathbf{a}_3$ | $=$ | $x_9a\,\hat{\mathbf{x}} + \left(\frac{1}{2} + x_9\right)a\,\hat{\mathbf{y}}$ | $(8h)$ | Rh V |
| $\mathbf{B}_{28} = \left(\frac{1}{2} - x_9\right)\mathbf{a}_1 - x_9\,\mathbf{a}_2 + \left(\frac{1}{2} - 2x_9\right)\mathbf{a}_3$ | $=$ | $-x_9a\,\hat{\mathbf{x}} + \left(\frac{1}{2} - x_9\right)a\,\hat{\mathbf{y}}$ | $(8h)$ | Rh V |
| $\mathbf{B}_{29} = x_9\,\mathbf{a}_1 + \left(\frac{1}{2} - x_9\right)\mathbf{a}_2 + \frac{1}{2}\,\mathbf{a}_3$ | $=$ | $\left(\frac{1}{2} - x_9\right)a\,\hat{\mathbf{x}} + x_9a\,\hat{\mathbf{y}}$ | $(8h)$ | Rh V |
| $\mathbf{B}_{30} = -x_9\,\mathbf{a}_1 + \left(\frac{1}{2} + x_9\right)\mathbf{a}_2 + \frac{1}{2}\,\mathbf{a}_3$ | $=$ | $\left(\frac{1}{2} + x_9\right)a\,\hat{\mathbf{x}} - x_9a\,\hat{\mathbf{y}}$ | $(8h)$ | Rh V |
| $\mathbf{B}_{31} = \left(\frac{1}{2} + x_{10} + z_{10}\right)\mathbf{a}_1 + (x_{10} + z_{10})\,\mathbf{a}_2 + \left(\frac{1}{2} + 2x_{10}\right)\mathbf{a}_3$ | $=$ | $x_{10}a\,\hat{\mathbf{x}} + \left(\frac{1}{2} + x_{10}\right)a\,\hat{\mathbf{y}} + z_{10}c\,\hat{\mathbf{z}}$ | $(16l)$ | Rh VI |
| $\mathbf{B}_{32} = \left(\frac{1}{2} - x_{10} + z_{10}\right)\mathbf{a}_1 + (-x_{10} + z_{10})\,\mathbf{a}_2 + \left(\frac{1}{2} - 2x_{10}\right)\mathbf{a}_3$ | $=$ | $-x_{10}a\,\hat{\mathbf{x}} + \left(\frac{1}{2} - x_{10}\right)a\,\hat{\mathbf{y}} + z_{10}c\,\hat{\mathbf{z}}$ | $(16l)$ | Rh VI |
| $\mathbf{B}_{33} = (x_{10} + z_{10})\,\mathbf{a}_1 + \left(\frac{1}{2} - x_{10} + z_{10}\right)\mathbf{a}_2 + \frac{1}{2}\,\mathbf{a}_3$ | $=$ | $\left(\frac{1}{2} - x_{10}\right)a\,\hat{\mathbf{x}} + x_{10}a\,\hat{\mathbf{y}} + z_{10}c\,\hat{\mathbf{z}}$ | $(16l)$ | Rh VI |
| $\mathbf{B}_{34} = (-x_{10} + z_{10})\,\mathbf{a}_1 + \left(\frac{1}{2} + x_{10} + z_{10}\right)\mathbf{a}_2 + \frac{1}{2}\,\mathbf{a}_3$ | $=$ | $\left(\frac{1}{2} + x_{10}\right)a\,\hat{\mathbf{x}} - x_{10}a\,\hat{\mathbf{y}} + z_{10}c\,\hat{\mathbf{z}}$ | $(16l)$ | Rh VI |
| $\mathbf{B}_{35} = (x_{10} - z_{10})\,\mathbf{a}_1 + \left(\frac{1}{2} - x_{10} - z_{10}\right)\mathbf{a}_2 + \frac{1}{2}\,\mathbf{a}_3$ | $=$ | $\left(\frac{1}{2} - x_{10}\right)a\,\hat{\mathbf{x}} + x_{10}a\,\hat{\mathbf{y}} - z_{10}c\,\hat{\mathbf{z}}$ | $(16l)$ | Rh VI |
| $\mathbf{B}_{36} = (-x_{10} - z_{10})\,\mathbf{a}_1 + \left(\frac{1}{2} + x_{10} - z_{10}\right)\mathbf{a}_2 + \frac{1}{2}\,\mathbf{a}_3$ | $=$ | $\left(\frac{1}{2} + x_{10}\right)a\,\hat{\mathbf{x}} - x_{10}a\,\hat{\mathbf{y}} - z_{10}c\,\hat{\mathbf{z}}$ | $(16l)$ | Rh VI |
| $\mathbf{B}_{37} = \left(\frac{1}{2} + x_{10} - z_{10}\right)\mathbf{a}_1 + (x_{10} - z_{10})\,\mathbf{a}_2 + \left(\frac{1}{2} + 2x_{10}\right)\mathbf{a}_3$ | $=$ | $x_{10}a\,\hat{\mathbf{x}} + \left(\frac{1}{2} + x_{10}\right)a\,\hat{\mathbf{y}} - z_{10}c\,\hat{\mathbf{z}}$ | $(16l)$ | Rh VI |
| $\mathbf{B}_{38} = \left(\frac{1}{2} - x_{10} - z_{10}\right)\mathbf{a}_1 + (-x_{10} - z_{10})\,\mathbf{a}_2 + \left(\frac{1}{2} - 2x_{10}\right)\mathbf{a}_3$ | $=$ | $-x_{10}a\,\hat{\mathbf{x}} + \left(\frac{1}{2} - x_{10}\right)a\,\hat{\mathbf{y}} - z_{10}c\,\hat{\mathbf{z}}$ | $(16l)$ | Rh VI |

| | | | | | | |
|---|---|---|---|---|---|---|
| $\mathbf{B}_{39}$ | = | $\left(\frac{1}{2}+x_{11}+z_{11}\right)\mathbf{a}_1+\left(x_{11}+z_{11}\right)\mathbf{a}_2+\left(\frac{1}{2}+2x_{11}\right)\mathbf{a}_3$ | = | $x_{11}a\,\hat{\mathbf{x}}+\left(\frac{1}{2}+x_{11}\right)a\,\hat{\mathbf{y}}+z_{11}c\,\hat{\mathbf{z}}$ | (16$l$) | Rh VII |
| $\mathbf{B}_{40}$ | = | $\left(\frac{1}{2}-x_{11}+z_{11}\right)\mathbf{a}_1+\left(-x_{11}+z_{11}\right)\mathbf{a}_2+\left(\frac{1}{2}-2x_{11}\right)\mathbf{a}_3$ | = | $-x_{11}a\,\hat{\mathbf{x}}+\left(\frac{1}{2}-x_{11}\right)a\,\hat{\mathbf{y}}+z_{11}c\,\hat{\mathbf{z}}$ | (16$l$) | Rh VII |
| $\mathbf{B}_{41}$ | = | $\left(x_{11}+z_{11}\right)\mathbf{a}_1+\left(\frac{1}{2}-x_{11}+z_{11}\right)\mathbf{a}_2+\frac{1}{2}\,\mathbf{a}_3$ | = | $\left(\frac{1}{2}-x_{11}\right)a\,\hat{\mathbf{x}}+x_{11}a\,\hat{\mathbf{y}}+z_{11}c\,\hat{\mathbf{z}}$ | (16$l$) | Rh VII |
| $\mathbf{B}_{42}$ | = | $\left(-x_{11}+z_{11}\right)\mathbf{a}_1+\left(\frac{1}{2}+x_{11}+z_{11}\right)\mathbf{a}_2+\frac{1}{2}\,\mathbf{a}_3$ | = | $\left(\frac{1}{2}+x_{11}\right)a\,\hat{\mathbf{x}}-x_{11}a\,\hat{\mathbf{y}}+z_{11}c\,\hat{\mathbf{z}}$ | (16$l$) | Rh VII |
| $\mathbf{B}_{43}$ | = | $\left(x_{11}-z_{11}\right)\mathbf{a}_1+\left(\frac{1}{2}-x_{11}-z_{11}\right)\mathbf{a}_2+\frac{1}{2}\,\mathbf{a}_3$ | = | $\left(\frac{1}{2}-x_{11}\right)a\,\hat{\mathbf{x}}+x_{11}a\,\hat{\mathbf{y}}-z_{11}c\,\hat{\mathbf{z}}$ | (16$l$) | Rh VII |
| $\mathbf{B}_{44}$ | = | $\left(-x_{11}-z_{11}\right)\mathbf{a}_1+\left(\frac{1}{2}+x_{11}-z_{11}\right)\mathbf{a}_2+\frac{1}{2}\,\mathbf{a}_3$ | = | $\left(\frac{1}{2}+x_{11}\right)a\,\hat{\mathbf{x}}-x_{11}a\,\hat{\mathbf{y}}-z_{11}c\,\hat{\mathbf{z}}$ | (16$l$) | Rh VII |
| $\mathbf{B}_{45}$ | = | $\left(\frac{1}{2}+x_{11}-z_{11}\right)\mathbf{a}_1+\left(x_{11}-z_{11}\right)\mathbf{a}_2+\left(\frac{1}{2}+2x_{11}\right)\mathbf{a}_3$ | = | $x_{11}a\,\hat{\mathbf{x}}+\left(\frac{1}{2}+x_{11}\right)a\,\hat{\mathbf{y}}-z_{11}c\,\hat{\mathbf{z}}$ | (16$l$) | Rh VII |
| $\mathbf{B}_{46}$ | = | $\left(\frac{1}{2}-x_{11}-z_{11}\right)\mathbf{a}_1+\left(-x_{11}-z_{11}\right)\mathbf{a}_2+\left(\frac{1}{2}-2x_{11}\right)\mathbf{a}_3$ | = | $-x_{11}a\,\hat{\mathbf{x}}+\left(\frac{1}{2}-x_{11}\right)a\,\hat{\mathbf{y}}-z_{11}c\,\hat{\mathbf{z}}$ | (16$l$) | Rh VII |
| $\mathbf{B}_{47}$ | = | $\left(\frac{1}{2}+x_{12}+z_{12}\right)\mathbf{a}_1+\left(x_{12}+z_{12}\right)\mathbf{a}_2+\left(\frac{1}{2}+2x_{12}\right)\mathbf{a}_3$ | = | $x_{12}a\,\hat{\mathbf{x}}+\left(\frac{1}{2}+x_{12}\right)a\,\hat{\mathbf{y}}+z_{12}c\,\hat{\mathbf{z}}$ | (16$l$) | Rh VIII |
| $\mathbf{B}_{48}$ | = | $\left(\frac{1}{2}-x_{12}+z_{12}\right)\mathbf{a}_1+\left(-x_{12}+z_{12}\right)\mathbf{a}_2+\left(\frac{1}{2}-2x_{12}\right)\mathbf{a}_3$ | = | $-x_{12}a\,\hat{\mathbf{x}}+\left(\frac{1}{2}-x_{12}\right)a\,\hat{\mathbf{y}}+z_{12}c\,\hat{\mathbf{z}}$ | (16$l$) | Rh VIII |
| $\mathbf{B}_{49}$ | = | $\left(x_{12}+z_{12}\right)\mathbf{a}_1+\left(\frac{1}{2}-x_{12}+z_{12}\right)\mathbf{a}_2+\frac{1}{2}\,\mathbf{a}_3$ | = | $\left(\frac{1}{2}-x_{12}\right)a\,\hat{\mathbf{x}}+x_{12}a\,\hat{\mathbf{y}}+z_{12}c\,\hat{\mathbf{z}}$ | (16$l$) | Rh VIII |
| $\mathbf{B}_{50}$ | = | $\left(-x_{12}+z_{12}\right)\mathbf{a}_1+\left(\frac{1}{2}+x_{12}+z_{12}\right)\mathbf{a}_2+\frac{1}{2}\,\mathbf{a}_3$ | = | $\left(\frac{1}{2}+x_{12}\right)a\,\hat{\mathbf{x}}-x_{12}a\,\hat{\mathbf{y}}+z_{12}c\,\hat{\mathbf{z}}$ | (16$l$) | Rh VIII |
| $\mathbf{B}_{51}$ | = | $\left(x_{12}-z_{12}\right)\mathbf{a}_1+\left(\frac{1}{2}-x_{12}-z_{12}\right)\mathbf{a}_2+\frac{1}{2}\,\mathbf{a}_3$ | = | $\left(\frac{1}{2}-x_{12}\right)a\,\hat{\mathbf{x}}+x_{12}a\,\hat{\mathbf{y}}-z_{12}c\,\hat{\mathbf{z}}$ | (16$l$) | Rh VIII |
| $\mathbf{B}_{52}$ | = | $\left(-x_{12}-z_{12}\right)\mathbf{a}_1+\left(\frac{1}{2}+x_{12}-z_{12}\right)\mathbf{a}_2+\frac{1}{2}\,\mathbf{a}_3$ | = | $\left(\frac{1}{2}+x_{12}\right)a\,\hat{\mathbf{x}}-x_{12}a\,\hat{\mathbf{y}}-z_{12}c\,\hat{\mathbf{z}}$ | (16$l$) | Rh VIII |
| $\mathbf{B}_{53}$ | = | $\left(\frac{1}{2}+x_{12}-z_{12}\right)\mathbf{a}_1+\left(x_{12}-z_{12}\right)\mathbf{a}_2+\left(\frac{1}{2}+2x_{12}\right)\mathbf{a}_3$ | = | $x_{12}a\,\hat{\mathbf{x}}+\left(\frac{1}{2}+x_{12}\right)a\,\hat{\mathbf{y}}-z_{12}c\,\hat{\mathbf{z}}$ | (16$l$) | Rh VIII |
| $\mathbf{B}_{54}$ | = | $\left(\frac{1}{2}-x_{12}-z_{12}\right)\mathbf{a}_1+\left(-x_{12}-z_{12}\right)\mathbf{a}_2+\left(\frac{1}{2}-2x_{12}\right)\mathbf{a}_3$ | = | $-x_{12}a\,\hat{\mathbf{x}}+\left(\frac{1}{2}-x_{12}\right)a\,\hat{\mathbf{y}}-z_{12}c\,\hat{\mathbf{z}}$ | (16$l$) | Rh VIII |
| $\mathbf{B}_{55}$ | = | $\left(y_{13}+z_{13}\right)\mathbf{a}_1+\left(x_{13}+z_{13}\right)\mathbf{a}_2+\left(x_{13}+y_{13}\right)\mathbf{a}_3$ | = | $x_{13}a\,\hat{\mathbf{x}}+y_{13}a\,\hat{\mathbf{y}}+z_{13}c\,\hat{\mathbf{z}}$ | (32$m$) | Pu V |
| $\mathbf{B}_{56}$ | = | $\left(-y_{13}+z_{13}\right)\mathbf{a}_1+\left(-x_{13}+z_{13}\right)\mathbf{a}_2+\left(-x_{13}-y_{13}\right)\mathbf{a}_3$ | = | $-x_{13}a\,\hat{\mathbf{x}}-y_{13}a\,\hat{\mathbf{y}}+z_{13}c\,\hat{\mathbf{z}}$ | (32$m$) | Pu V |
| $\mathbf{B}_{57}$ | = | $\left(x_{13}+z_{13}\right)\mathbf{a}_1+\left(-y_{13}+z_{13}\right)\mathbf{a}_2+\left(x_{13}-y_{13}\right)\mathbf{a}_3$ | = | $-y_{13}a\,\hat{\mathbf{x}}+x_{13}a\,\hat{\mathbf{y}}+z_{13}c\,\hat{\mathbf{z}}$ | (32$m$) | Pu V |
| $\mathbf{B}_{58}$ | = | $\left(-x_{13}+z_{13}\right)\mathbf{a}_1+\left(y_{13}+z_{13}\right)\mathbf{a}_2+\left(-x_{13}+y_{13}\right)\mathbf{a}_3$ | = | $y_{13}a\,\hat{\mathbf{x}}-x_{13}a\,\hat{\mathbf{y}}+z_{13}c\,\hat{\mathbf{z}}$ | (32$m$) | Pu V |
| $\mathbf{B}_{59}$ | = | $\left(\frac{1}{2}+y_{13}-z_{13}\right)\mathbf{a}_1+\left(\frac{1}{2}-x_{13}-z_{13}\right)\mathbf{a}_2+\left(-x_{13}+y_{13}\right)\mathbf{a}_3$ | = | $-x_{13}a\,\hat{\mathbf{x}}+y_{13}a\,\hat{\mathbf{y}}+\left(\frac{1}{2}-z_{13}\right)c\,\hat{\mathbf{z}}$ | (32$m$) | Pu V |
| $\mathbf{B}_{60}$ | = | $\left(\frac{1}{2} - y_{13} - z_{13}\right) \mathbf{a}_1 + \left(\frac{1}{2} + x_{13} - z_{13}\right) \mathbf{a}_2 + (x_{13} - y_{13}) \mathbf{a}_3$ | = | $x_{13} a \hat{\mathbf{x}} - y_{13} a \hat{\mathbf{y}} + \left(\frac{1}{2} - z_{13}\right) c \hat{\mathbf{z}}$ | (32$m$) | Pu V |
| $\mathbf{B}_{61}$ | = | $\left(\frac{1}{2} + x_{13} - z_{13}\right) \mathbf{a}_1 + \left(\frac{1}{2} + y_{13} - z_{13}\right) \mathbf{a}_2 + (x_{13} + y_{13}) \mathbf{a}_3$ | = | $y_{13} a \hat{\mathbf{x}} + x_{13} a \hat{\mathbf{y}} + \left(\frac{1}{2} - z_{13}\right) c \hat{\mathbf{z}}$ | (32$m$) | Pu V |
| $\mathbf{B}_{62}$ | = | $\left(\frac{1}{2} - x_{13} - z_{13}\right) \mathbf{a}_1 + \left(\frac{1}{2} - y_{13} - z_{13}\right) \mathbf{a}_2 + (-x_{13} - y_{13}) \mathbf{a}_3$ | = | $-y_{13} a \hat{\mathbf{x}} - x_{13} a \hat{\mathbf{y}} + \left(\frac{1}{2} - z_{13}\right) c \hat{\mathbf{z}}$ | (32$m$) | Pu V |
| $\mathbf{B}_{63}$ | = | $(-y_{13} - z_{13}) \mathbf{a}_1 + (-x_{13} - z_{13}) \mathbf{a}_2 + (-x_{13} - y_{13}) \mathbf{a}_3$ | = | $-x_{13} a \hat{\mathbf{x}} - y_{13} a \hat{\mathbf{y}} - z_{13} c \hat{\mathbf{z}}$ | (32$m$) | Pu V |
| $\mathbf{B}_{64}$ | = | $(y_{13} - z_{13}) \mathbf{a}_1 + (x_{13} - z_{13}) \mathbf{a}_2 + (x_{13} + y_{13}) \mathbf{a}_3$ | = | $x_{13} a \hat{\mathbf{x}} + y_{13} a \hat{\mathbf{y}} - z_{13} c \hat{\mathbf{z}}$ | (32$m$) | Pu V |
| $\mathbf{B}_{65}$ | = | $(-x_{13} - z_{13}) \mathbf{a}_1 + (y_{13} - z_{13}) \mathbf{a}_2 + (-x_{13} + y_{13}) \mathbf{a}_3$ | = | $y_{13} a \hat{\mathbf{x}} - x_{13} a \hat{\mathbf{y}} - z_{13} c \hat{\mathbf{z}}$ | (32$m$) | Pu V |
| $\mathbf{B}_{66}$ | = | $(x_{13} - z_{13}) \mathbf{a}_1 + (-y_{13} - z_{13}) \mathbf{a}_2 + (x_{13} - y_{13}) \mathbf{a}_3$ | = | $-y_{13} a \hat{\mathbf{x}} + x_{13} a \hat{\mathbf{y}} - z_{13} c \hat{\mathbf{z}}$ | (32$m$) | Pu V |
| $\mathbf{B}_{67}$ | = | $\left(\frac{1}{2} - y_{13} + z_{13}\right) \mathbf{a}_1 + \left(\frac{1}{2} + x_{13} + z_{13}\right) \mathbf{a}_2 + (x_{13} - y_{13}) \mathbf{a}_3$ | = | $x_{13} a \hat{\mathbf{x}} - y_{13} a \hat{\mathbf{y}} + \left(\frac{1}{2} + z_{13}\right) c \hat{\mathbf{z}}$ | (32$m$) | Pu V |
| $\mathbf{B}_{68}$ | = | $\left(\frac{1}{2} + y_{13} + z_{13}\right) \mathbf{a}_1 + \left(\frac{1}{2} - x_{13} + z_{13}\right) \mathbf{a}_2 + (-x_{13} + y_{13}) \mathbf{a}_3$ | = | $-x_{13} a \hat{\mathbf{x}} + y_{13} a \hat{\mathbf{y}} + \left(\frac{1}{2} + z_{13}\right) c \hat{\mathbf{z}}$ | (32$m$) | Pu V |
| $\mathbf{B}_{69}$ | = | $\left(\frac{1}{2} - x_{13} + z_{13}\right) \mathbf{a}_1 + \left(\frac{1}{2} - y_{13} + z_{13}\right) \mathbf{a}_2 + (-x_{13} - y_{13}) \mathbf{a}_3$ | = | $-y_{13} a \hat{\mathbf{x}} - x_{13} a \hat{\mathbf{y}} + \left(\frac{1}{2} + z_{13}\right) c \hat{\mathbf{z}}$ | (32$m$) | Pu V |
| $\mathbf{B}_{70}$ | = | $\left(\frac{1}{2} + x_{13} + z_{13}\right) \mathbf{a}_1 + \left(\frac{1}{2} + y_{13} + z_{13}\right) \mathbf{a}_2 + (x_{13} + y_{13}) \mathbf{a}_3$ | = | $y_{13} a \hat{\mathbf{x}} + x_{13} a \hat{\mathbf{y}} + \left(\frac{1}{2} + z_{13}\right) c \hat{\mathbf{z}}$ | (32$m$) | Pu V |
| $\mathbf{B}_{71}$ | = | $(y_{14} + z_{14}) \mathbf{a}_1 + (x_{14} + z_{14}) \mathbf{a}_2 + (x_{14} + y_{14}) \mathbf{a}_3$ | = | $x_{14} a \hat{\mathbf{x}} + y_{14} a \hat{\mathbf{y}} + z_{14} c \hat{\mathbf{z}}$ | (32$m$) | Pu VI |
| $\mathbf{B}_{72}$ | = | $(-y_{14} + z_{14}) \mathbf{a}_1 + (-x_{14} + z_{14}) \mathbf{a}_2 + (-x_{14} - y_{14}) \mathbf{a}_3$ | = | $-x_{14} a \hat{\mathbf{x}} - y_{14} a \hat{\mathbf{y}} + z_{14} c \hat{\mathbf{z}}$ | (32$m$) | Pu VI |
| $\mathbf{B}_{73}$ | = | $(x_{14} + z_{14}) \mathbf{a}_1 + (-y_{14} + z_{14}) \mathbf{a}_2 + (x_{14} - y_{14}) \mathbf{a}_3$ | = | $-y_{14} a \hat{\mathbf{x}} + x_{14} a \hat{\mathbf{y}} + z_{14} c \hat{\mathbf{z}}$ | (32$m$) | Pu VI |
| $\mathbf{B}_{74}$ | = | $(-x_{14} + z_{14}) \mathbf{a}_1 + (y_{14} + z_{14}) \mathbf{a}_2 + (-x_{14} + y_{14}) \mathbf{a}_3$ | = | $y_{14} a \hat{\mathbf{x}} - x_{14} a \hat{\mathbf{y}} + z_{14} c \hat{\mathbf{z}}$ | (32$m$) | Pu VI |
| $\mathbf{B}_{75}$ | = | $\left(\frac{1}{2} + y_{14} - z_{14}\right) \mathbf{a}_1 + \left(\frac{1}{2} - x_{14} - z_{14}\right) \mathbf{a}_2 + (-x_{14} + y_{14}) \mathbf{a}_3$ | = | $-x_{14} a \hat{\mathbf{x}} + y_{14} a \hat{\mathbf{y}} + \left(\frac{1}{2} - z_{14}\right) c \hat{\mathbf{z}}$ | (32$m$) | Pu VI |
| $\mathbf{B}_{76}$ | = | $\left(\frac{1}{2} - y_{14} - z_{14}\right) \mathbf{a}_1 + \left(\frac{1}{2} + x_{14} - z_{14}\right) \mathbf{a}_2 + (x_{14} - y_{14}) \mathbf{a}_3$ | = | $x_{14} a \hat{\mathbf{x}} - y_{14} a \hat{\mathbf{y}} + \left(\frac{1}{2} - z_{14}\right) c \hat{\mathbf{z}}$ | (32$m$) | Pu VI |
| $\mathbf{B}_{77}$ | = | $\left(\frac{1}{2} + x_{14} - z_{14}\right) \mathbf{a}_1 + \left(\frac{1}{2} + y_{14} - z_{14}\right) \mathbf{a}_2 + (x_{14} + y_{14}) \mathbf{a}_3$ | = | $y_{14} a \hat{\mathbf{x}} + x_{14} a \hat{\mathbf{y}} + \left(\frac{1}{2} - z_{14}\right) c \hat{\mathbf{z}}$ | (32$m$) | Pu VI |
| $\mathbf{B}_{78}$ | = | $\left(\frac{1}{2} - x_{14} - z_{14}\right) \mathbf{a}_1 + \left(\frac{1}{2} - y_{14} - z_{14}\right) \mathbf{a}_2 + (-x_{14} - y_{14}) \mathbf{a}_3$ | = | $-y_{14} a \hat{\mathbf{x}} - x_{14} a \hat{\mathbf{y}} + \left(\frac{1}{2} - z_{14}\right) c \hat{\mathbf{z}}$ | (32$m$) | Pu VI |
| $\mathbf{B}_{79}$ | = | $(-y_{14} - z_{14}) \mathbf{a}_1 + (-x_{14} - z_{14}) \mathbf{a}_2 + (-x_{14} - y_{14}) \mathbf{a}_3$ | = | $-x_{14} a \hat{\mathbf{x}} - y_{14} a \hat{\mathbf{y}} - z_{14} c \hat{\mathbf{z}}$ | (32$m$) | Pu VI |
| $\mathbf{B}_{80}$ | = | $(y_{14} - z_{14}) \mathbf{a}_1 + (x_{14} - z_{14}) \mathbf{a}_2 + (x_{14} + y_{14}) \mathbf{a}_3$ | = | $x_{14} a \hat{\mathbf{x}} + y_{14} a \hat{\mathbf{y}} - z_{14} c \hat{\mathbf{z}}$ | (32$m$) | Pu VI |
| $\mathbf{B}_{81}$ | = | $(-x_{14} - z_{14}) \mathbf{a}_1 + (y_{14} - z_{14}) \mathbf{a}_2 + (-x_{14} + y_{14}) \mathbf{a}_3$ | = | $y_{14} a \hat{\mathbf{x}} - x_{14} a \hat{\mathbf{y}} - z_{14} c \hat{\mathbf{z}}$ | (32$m$) | Pu VI |
| $\mathbf{B}_{82}$ | = | $(x_{14}-z_{14})\,\mathbf{a}_1+(-y_{14}-z_{14})\,\mathbf{a}_2+(x_{14}-y_{14})\,\mathbf{a}_3$ | = | $-y_{14}a\,\hat{\mathbf{x}}+x_{14}a\,\hat{\mathbf{y}}-z_{14}c\,\hat{\mathbf{z}}$ | (32$m$) | Pu VI |
| $\mathbf{B}_{83}$ | = | $\left(\frac{1}{2}-y_{14}+z_{14}\right)\mathbf{a}_1+\left(\frac{1}{2}+x_{14}+z_{14}\right)\mathbf{a}_2+(x_{14}-y_{14})\,\mathbf{a}_3$ | = | $x_{14}a\,\hat{\mathbf{x}}-y_{14}a\,\hat{\mathbf{y}}+\left(\frac{1}{2}+z_{14}\right)c\,\hat{\mathbf{z}}$ | (32$m$) | Pu VI |
| $\mathbf{B}_{84}$ | = | $\left(\frac{1}{2}+y_{14}+z_{14}\right)\mathbf{a}_1+\left(\frac{1}{2}-x_{14}+z_{14}\right)\mathbf{a}_2+(-x_{14}+y_{14})\,\mathbf{a}_3$ | = | $-x_{14}a\,\hat{\mathbf{x}}+y_{14}a\,\hat{\mathbf{y}}+\left(\frac{1}{2}+z_{14}\right)c\,\hat{\mathbf{z}}$ | (32$m$) | Pu VI |
| $\mathbf{B}_{85}$ | = | $\left(\frac{1}{2}-x_{14}+z_{14}\right)\mathbf{a}_1+\left(\frac{1}{2}-y_{14}+z_{14}\right)\mathbf{a}_2+(-x_{14}-y_{14})\,\mathbf{a}_3$ | = | $-y_{14}a\,\hat{\mathbf{x}}-x_{14}a\,\hat{\mathbf{y}}+\left(\frac{1}{2}+z_{14}\right)c\,\hat{\mathbf{z}}$ | (32$m$) | Pu VI |
| $\mathbf{B}_{86}$ | = | $\left(\frac{1}{2}+x_{14}+z_{14}\right)\mathbf{a}_1+\left(\frac{1}{2}+y_{14}+z_{14}\right)\mathbf{a}_2+(x_{14}+y_{14})\,\mathbf{a}_3$ | = | $y_{14}a\,\hat{\mathbf{x}}+x_{14}a\,\hat{\mathbf{y}}+\left(\frac{1}{2}+z_{14}\right)c\,\hat{\mathbf{z}}$ | (32$m$) | Pu VI |
| $\mathbf{B}_{87}$ | = | $(y_{15}+z_{15})\,\mathbf{a}_1+(x_{15}+z_{15})\,\mathbf{a}_2+(x_{15}+y_{15})\,\mathbf{a}_3$ | = | $x_{15}a\,\hat{\mathbf{x}}+y_{15}a\,\hat{\mathbf{y}}+z_{15}c\,\hat{\mathbf{z}}$ | (32$m$) | Pu VII |
| $\mathbf{B}_{88}$ | = | $(-y_{15}+z_{15})\,\mathbf{a}_1+(-x_{15}+z_{15})\,\mathbf{a}_2+(-x_{15}-y_{15})\,\mathbf{a}_3$ | = | $-x_{15}a\,\hat{\mathbf{x}}-y_{15}a\,\hat{\mathbf{y}}+z_{15}c\,\hat{\mathbf{z}}$ | (32$m$) | Pu VII |
| $\mathbf{B}_{89}$ | = | $(x_{15}+z_{15})\,\mathbf{a}_1+(-y_{15}+z_{15})\,\mathbf{a}_2+(x_{15}-y_{15})\,\mathbf{a}_3$ | = | $-y_{15}a\,\hat{\mathbf{x}}+x_{15}a\,\hat{\mathbf{y}}+z_{15}c\,\hat{\mathbf{z}}$ | (32$m$) | Pu VII |
| $\mathbf{B}_{90}$ | = | $(-x_{15}+z_{15})\,\mathbf{a}_1+(y_{15}+z_{15})\,\mathbf{a}_2+(-x_{15}+y_{15})\,\mathbf{a}_3$ | = | $y_{15}a\,\hat{\mathbf{x}}-x_{15}a\,\hat{\mathbf{y}}+z_{15}c\,\hat{\mathbf{z}}$ | (32$m$) | Pu VII |
| $\mathbf{B}_{91}$ | = | $\left(\frac{1}{2}+y_{15}-z_{15}\right)\mathbf{a}_1+\left(\frac{1}{2}-x_{15}-z_{15}\right)\mathbf{a}_2+(-x_{15}+y_{15})\,\mathbf{a}_3$ | = | $-x_{15}a\,\hat{\mathbf{x}}+y_{15}a\,\hat{\mathbf{y}}+\left(\frac{1}{2}-z_{15}\right)c\,\hat{\mathbf{z}}$ | (32$m$) | Pu VII |
| $\mathbf{B}_{92}$ | = | $\left(\frac{1}{2}-y_{15}-z_{15}\right)\mathbf{a}_1+\left(\frac{1}{2}+x_{15}-z_{15}\right)\mathbf{a}_2+(x_{15}-y_{15})\,\mathbf{a}_3$ | = | $x_{15}a\,\hat{\mathbf{x}}-y_{15}a\,\hat{\mathbf{y}}+\left(\frac{1}{2}-z_{15}\right)c\,\hat{\mathbf{z}}$ | (32$m$) | Pu VII |
| $\mathbf{B}_{93}$ | = | $\left(\frac{1}{2}+x_{15}-z_{15}\right)\mathbf{a}_1+\left(\frac{1}{2}+y_{15}-z_{15}\right)\mathbf{a}_2+(x_{15}+y_{15})\,\mathbf{a}_3$ | = | $y_{15}a\,\hat{\mathbf{x}}+x_{15}a\,\hat{\mathbf{y}}+\left(\frac{1}{2}-z_{15}\right)c\,\hat{\mathbf{z}}$ | (32$m$) | Pu VII |
| $\mathbf{B}_{94}$ | = | $\left(\frac{1}{2}-x_{15}-z_{15}\right)\mathbf{a}_1+\left(\frac{1}{2}-y_{15}-z_{15}\right)\mathbf{a}_2+(-x_{15}-y_{15})\,\mathbf{a}_3$ | = | $-y_{15}a\,\hat{\mathbf{x}}-x_{15}a\,\hat{\mathbf{y}}+\left(\frac{1}{2}-z_{15}\right)c\,\hat{\mathbf{z}}$ | (32$m$) | Pu VII |
| $\mathbf{B}_{95}$ | = | $(-y_{15}-z_{15})\,\mathbf{a}_1+(-x_{15}-z_{15})\,\mathbf{a}_2+(-x_{15}-y_{15})\,\mathbf{a}_3$ | = | $-x_{15}a\,\hat{\mathbf{x}}-y_{15}a\,\hat{\mathbf{y}}-z_{15}c\,\hat{\mathbf{z}}$ | (32$m$) | Pu VII |
| $\mathbf{B}_{96}$ | = | $(y_{15}-z_{15})\,\mathbf{a}_1+(x_{15}-z_{15})\,\mathbf{a}_2+(x_{15}+y_{15})\,\mathbf{a}_3$ | = | $x_{15}a\,\hat{\mathbf{x}}+y_{15}a\,\hat{\mathbf{y}}-z_{15}c\,\hat{\mathbf{z}}$ | (32$m$) | Pu VII |
| $\mathbf{B}_{97}$ | = | $(-x_{15}-z_{15})\,\mathbf{a}_1+(y_{15}-z_{15})\,\mathbf{a}_2+(-x_{15}+y_{15})\,\mathbf{a}_3$ | = | $y_{15}a\,\hat{\mathbf{x}}-x_{15}a\,\hat{\mathbf{y}}-z_{15}c\,\hat{\mathbf{z}}$ | (32$m$) | Pu VII |
| $\mathbf{B}_{98}$ | = | $(x_{15}-z_{15})\,\mathbf{a}_1+(-y_{15}-z_{15})\,\mathbf{a}_2+(x_{15}-y_{15})\,\mathbf{a}_3$ | = | $-y_{15}a\,\hat{\mathbf{x}}+x_{15}a\,\hat{\mathbf{y}}-z_{15}c\,\hat{\mathbf{z}}$ | (32$m$) | Pu VII |
| $\mathbf{B}_{99}$ | = | $\left(\frac{1}{2}-y_{15}+z_{15}\right)\mathbf{a}_1+\left(\frac{1}{2}+x_{15}+z_{15}\right)\mathbf{a}_2+(x_{15}-y_{15})\,\mathbf{a}_3$ | = | $x_{15}a\,\hat{\mathbf{x}}-y_{15}a\,\hat{\mathbf{y}}+\left(\frac{1}{2}+z_{15}\right)c\,\hat{\mathbf{z}}$ | (32$m$) | Pu VII |
| $\mathbf{B}_{100}$ | = | $\left(\frac{1}{2}+y_{15}+z_{15}\right)\mathbf{a}_1+\left(\frac{1}{2}-x_{15}+z_{15}\right)\mathbf{a}_2+(-x_{15}+y_{15})\,\mathbf{a}_3$ | = | $-x_{15}a\,\hat{\mathbf{x}}+y_{15}a\,\hat{\mathbf{y}}+\left(\frac{1}{2}+z_{15}\right)c\,\hat{\mathbf{z}}$ | (32$m$) | Pu VII |
| $\mathbf{B}_{101}$ | = | $\left(\frac{1}{2}-x_{15}+z_{15}\right)\mathbf{a}_1+\left(\frac{1}{2}-y_{15}+z_{15}\right)\mathbf{a}_2+(-x_{15}-y_{15})\,\mathbf{a}_3$ | = | $-y_{15}a\,\hat{\mathbf{x}}-x_{15}a\,\hat{\mathbf{y}}+\left(\frac{1}{2}+z_{15}\right)c\,\hat{\mathbf{z}}$ | (32$m$) | Pu VII |
| $\mathbf{B}_{102}$ | = | $\left(\frac{1}{2}+x_{15}+z_{15}\right)\mathbf{a}_1+\left(\frac{1}{2}+y_{15}+z_{15}\right)\mathbf{a}_2+(x_{15}+y_{15})\,\mathbf{a}_3$ | = | $y_{15}a\,\hat{\mathbf{x}}+x_{15}a\,\hat{\mathbf{y}}+\left(\frac{1}{2}+z_{15}\right)c\,\hat{\mathbf{z}}$ | (32$m$) | Pu VII |

---

---

**Geometry files:**
- CIF: pp. 1714
- POSCAR: pp. 1715

# $NH_4Pb_2Br_5$ ($K3_4$) Structure: A5BC2_tI32_140_bl_a_h

http://aflow.org/prototype-encyclopedia/A5BC2_tI32_140_bl_a_h

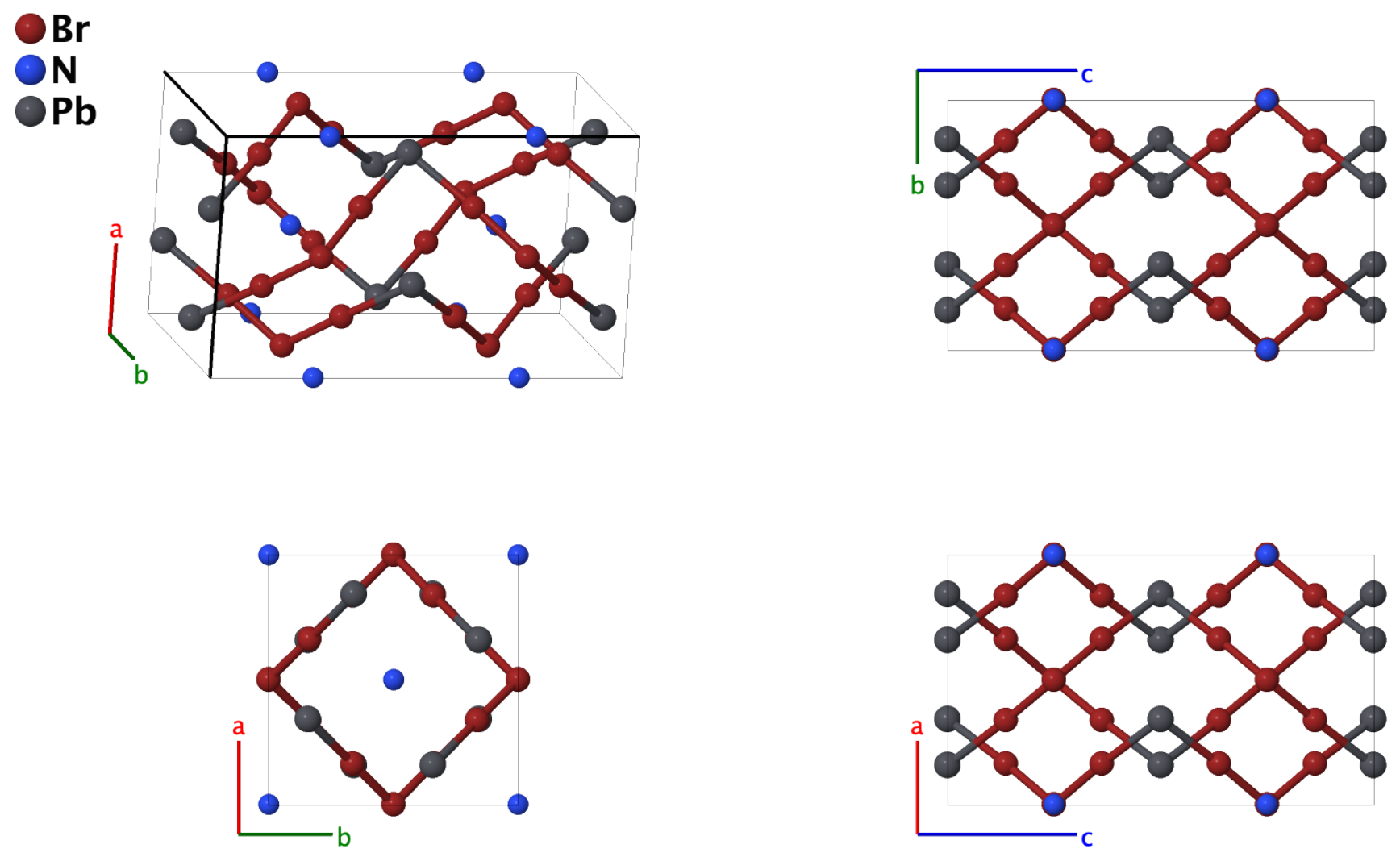

| | | |
|---|---|---|
| **Prototype** | : | $Br_5(NH_4)Pb_2$ |
| **AFLOW prototype label** | : | A5BC2_tI32_140_bl_a_h |
| ***Strukturbericht* designation** | : | $K3_4$ |
| **Pearson symbol** | : | tI32 |
| **Space group number** | : | 140 |
| **Space group symbol** | : | $I4/mcm$ |
| **AFLOW prototype command** | : | `aflow --proto=A5BC2_tI32_140_bl_a_h`<br>`--params=`$a, c/a, x_3, x_4, z_4$ |

**Other compounds with this structure**

- $KPb_2Br_5$, $RbPb_2Br_5$, $KSn_2Br_5$, $KSn_2I_5$, $InSn_2Br_5$, $InSn_2I_5$, $CsSn_2Br_5$, $TlSn_2Br_5$, $RbSn_2Br_5$, $InPb_2I_5$, and $ISe_2Tl_5$

- The positions of the hydrogen atoms in the $NH_4$ ions were not determined, so we only provide the positions of the nitrogen atoms (labeled as $NH_4$).

**Body-centered Tetragonal primitive vectors:**

$$\mathbf{a}_1 = -\frac{1}{2}\,a\,\hat{\mathbf{x}} + \frac{1}{2}\,a\,\hat{\mathbf{y}} + \frac{1}{2}\,c\,\hat{\mathbf{z}}$$
$$\mathbf{a}_2 = \frac{1}{2}\,a\,\hat{\mathbf{x}} - \frac{1}{2}\,a\,\hat{\mathbf{y}} + \frac{1}{2}\,c\,\hat{\mathbf{z}}$$
$$\mathbf{a}_3 = \frac{1}{2}\,a\,\hat{\mathbf{x}} + \frac{1}{2}\,a\,\hat{\mathbf{y}} - \frac{1}{2}\,c\,\hat{\mathbf{z}}$$

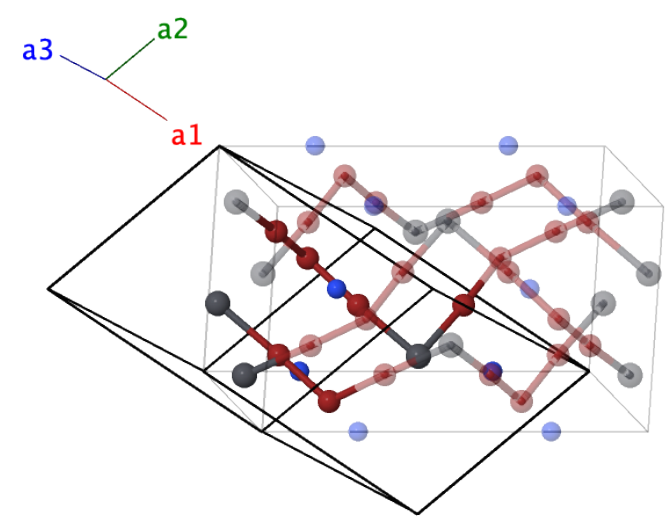

**Basis vectors:**

| | | Lattice Coordinates | | Cartesian Coordinates | Wyckoff Position | Atom Type |
|---|---|---|---|---|---|---|
| $\mathbf{B}_1$ | = | $\frac{1}{4}\,\mathbf{a}_1 + \frac{1}{4}\,\mathbf{a}_2$ | = | $\frac{1}{4}c\,\hat{\mathbf{z}}$ | (4$a$) | $NH_4$ |
| $\mathbf{B}_2$ | = | $\frac{3}{4}\,\mathbf{a}_1 + \frac{3}{4}\,\mathbf{a}_2$ | = | $\frac{3}{4}c\,\hat{\mathbf{z}}$ | (4$a$) | $NH_4$ |
| $\mathbf{B}_3$ | = | $\frac{3}{4}\,\mathbf{a}_1 + \frac{1}{4}\,\mathbf{a}_2 + \frac{1}{2}\,\mathbf{a}_3$ | = | $\frac{1}{2}a\,\hat{\mathbf{y}} + \frac{1}{4}c\,\hat{\mathbf{z}}$ | (4$b$) | Br I |
| $\mathbf{B}_4$ | = | $\frac{1}{4}\,\mathbf{a}_1 + \frac{3}{4}\,\mathbf{a}_2 + \frac{1}{2}\,\mathbf{a}_3$ | = | $\frac{1}{2}a\,\hat{\mathbf{x}} + \frac{1}{4}c\,\hat{\mathbf{z}}$ | (4$b$) | Br I |
| $\mathbf{B}_5$ | = | $\left(\frac{1}{2} + x_3\right)\,\mathbf{a}_1 + x_3\,\mathbf{a}_2 + \left(\frac{1}{2} + 2x_3\right)\,\mathbf{a}_3$ | = | $x_3a\,\hat{\mathbf{x}} + \left(\frac{1}{2} + x_3\right)a\,\hat{\mathbf{y}}$ | (8$h$) | Pb |
| $\mathbf{B}_6$ | = | $\left(\frac{1}{2} - x_3\right)\,\mathbf{a}_1 - x_3\,\mathbf{a}_2 + \left(\frac{1}{2} - 2x_3\right)\,\mathbf{a}_3$ | = | $-x_3a\,\hat{\mathbf{x}} + \left(\frac{1}{2} - x_3\right)a\,\hat{\mathbf{y}}$ | (8$h$) | Pb |
| $\mathbf{B}_7$ | = | $x_3\,\mathbf{a}_1 + \left(\frac{1}{2} - x_3\right)\,\mathbf{a}_2 + \frac{1}{2}\,\mathbf{a}_3$ | = | $\left(\frac{1}{2} - x_3\right)a\,\hat{\mathbf{x}} + x_3a\,\hat{\mathbf{y}}$ | (8$h$) | Pb |
| $\mathbf{B}_8$ | = | $-x_3\,\mathbf{a}_1 + \left(\frac{1}{2} + x_3\right)\,\mathbf{a}_2 + \frac{1}{2}\,\mathbf{a}_3$ | = | $\left(\frac{1}{2} + x_3\right)a\,\hat{\mathbf{x}} - x_3a\,\hat{\mathbf{y}}$ | (8$h$) | Pb |
| $\mathbf{B}_9$ | = | $\left(\frac{1}{2} + x_4 + z_4\right)\,\mathbf{a}_1 + (x_4 + z_4)\,\mathbf{a}_2 + \left(\frac{1}{2} + 2x_4\right)\,\mathbf{a}_3$ | = | $x_4a\,\hat{\mathbf{x}} + \left(\frac{1}{2} + x_4\right)a\,\hat{\mathbf{y}} + z_4c\,\hat{\mathbf{z}}$ | (16$l$) | Br II |
| $\mathbf{B}_{10}$ | = | $\left(\frac{1}{2} - x_4 + z_4\right)\,\mathbf{a}_1 + (-x_4 + z_4)\,\mathbf{a}_2 + \left(\frac{1}{2} - 2x_4\right)\,\mathbf{a}_3$ | = | $-x_4a\,\hat{\mathbf{x}} + \left(\frac{1}{2} - x_4\right)a\,\hat{\mathbf{y}} + z_4c\,\hat{\mathbf{z}}$ | (16$l$) | Br II |
| $\mathbf{B}_{11}$ | = | $(x_4 + z_4)\,\mathbf{a}_1 + \left(\frac{1}{2} - x_4 + z_4\right)\,\mathbf{a}_2 + \frac{1}{2}\,\mathbf{a}_3$ | = | $\left(\frac{1}{2} - x_4\right)a\,\hat{\mathbf{x}} + x_4a\,\hat{\mathbf{y}} + z_4c\,\hat{\mathbf{z}}$ | (16$l$) | Br II |
| $\mathbf{B}_{12}$ | = | $(-x_4 + z_4)\,\mathbf{a}_1 + \left(\frac{1}{2} + x_4 + z_4\right)\,\mathbf{a}_2 + \frac{1}{2}\,\mathbf{a}_3$ | = | $\left(\frac{1}{2} + x_4\right)a\,\hat{\mathbf{x}} - x_4a\,\hat{\mathbf{y}} + z_4c\,\hat{\mathbf{z}}$ | (16$l$) | Br II |
| $\mathbf{B}_{13}$ | = | $(x_4 - z_4)\,\mathbf{a}_1 + \left(\frac{1}{2} - x_4 - z_4\right)\,\mathbf{a}_2 + \frac{1}{2}\,\mathbf{a}_3$ | = | $\left(\frac{1}{2} - x_4\right)a\,\hat{\mathbf{x}} + x_4a\,\hat{\mathbf{y}} - z_4c\,\hat{\mathbf{z}}$ | (16$l$) | Br II |
| $\mathbf{B}_{14}$ | = | $(-x_4 - z_4)\,\mathbf{a}_1 + \left(\frac{1}{2} + x_4 - z_4\right)\,\mathbf{a}_2 + \frac{1}{2}\,\mathbf{a}_3$ | = | $\left(\frac{1}{2} + x_4\right)a\,\hat{\mathbf{x}} - x_4a\,\hat{\mathbf{y}} - z_4c\,\hat{\mathbf{z}}$ | (16$l$) | Br II |
| $\mathbf{B}_{15}$ | = | $\left(\frac{1}{2} + x_4 - z_4\right)\,\mathbf{a}_1 + (x_4 - z_4)\,\mathbf{a}_2 + \left(\frac{1}{2} + 2x_4\right)\,\mathbf{a}_3$ | = | $x_4a\,\hat{\mathbf{x}} + \left(\frac{1}{2} + x_4\right)a\,\hat{\mathbf{y}} - z_4c\,\hat{\mathbf{z}}$ | (16$l$) | Br II |
| $\mathbf{B}_{16}$ | = | $\left(\frac{1}{2} - x_4 - z_4\right)\,\mathbf{a}_1 + (-x_4 - z_4)\,\mathbf{a}_2 + \left(\frac{1}{2} - 2x_4\right)\,\mathbf{a}_3$ | = | $-x_4a\,\hat{\mathbf{x}} + \left(\frac{1}{2} - x_4\right)a\,\hat{\mathbf{y}} - z_4c\,\hat{\mathbf{z}}$ | (16$l$) | Br II |

**Found in:**

**Geometry files:**

- CIF: pp. 1715
- POSCAR: pp. 1716

# $Cs_3CoCl_5$ ($K3_1$) Structure: A5BC3_tI36_140_cl_b_ah

http://aflow.org/prototype-encyclopedia/A5BC3_tI36_140_cl_b_ah

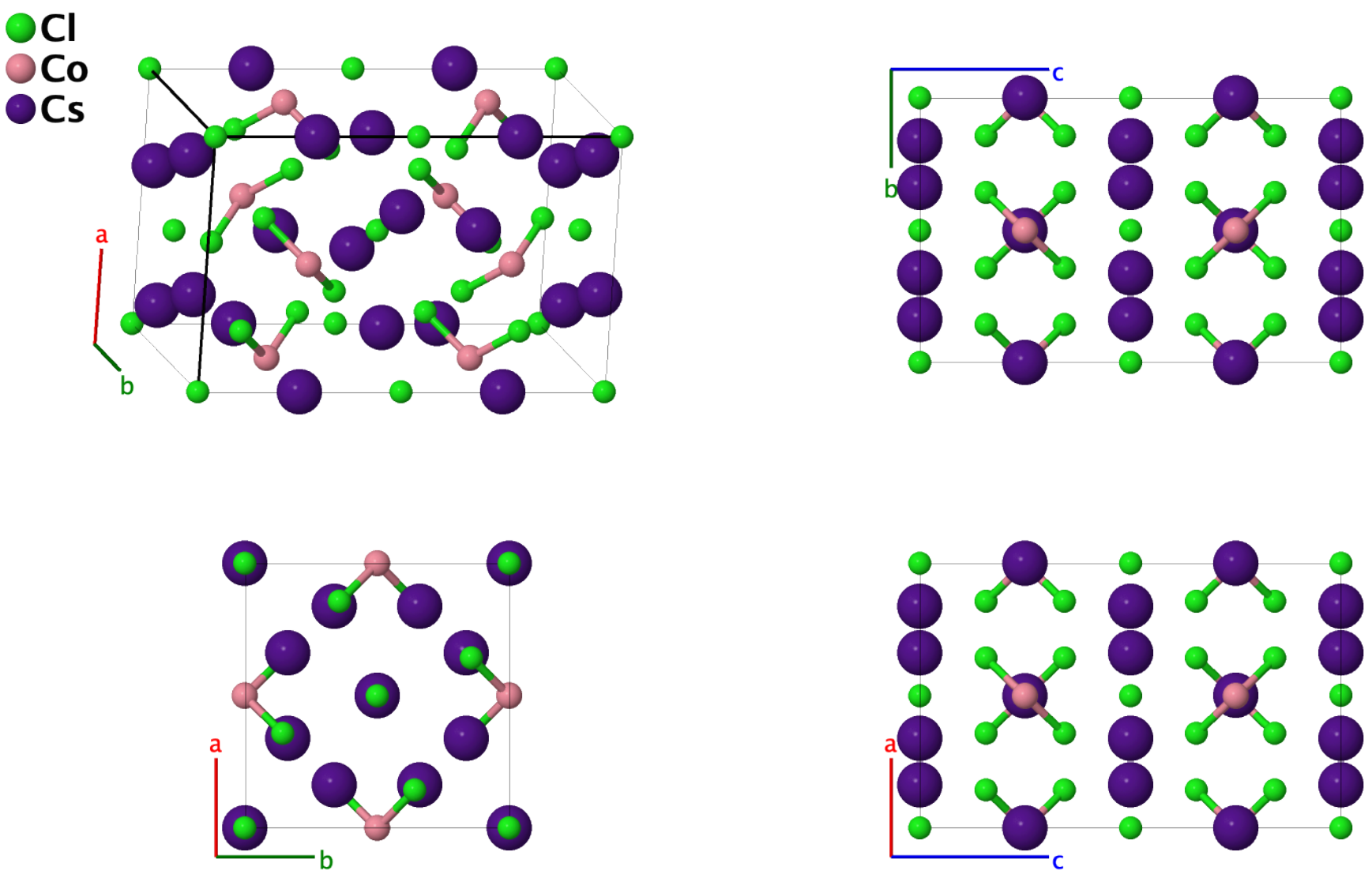

| | | |
|---|---|---|
| **Prototype** | : | $Cl_5CoCs_3$ |
| **AFLOW prototype label** | : | A5BC3_tI36_140_cl_b_ah |
| ***Strukturbericht* designation** | : | $K3_1$ |
| **Pearson symbol** | : | tI36 |
| **Space group number** | : | 140 |
| **Space group symbol** | : | $I4/mcm$ |
| **AFLOW prototype command** | : | `aflow --proto=A5BC3_tI36_140_cl_b_ah`<br>`--params=`$a, c/a, x_4, x_5, z_5$ |

### Other compounds with this structure

- $Cs_3CoBr_5$, $CsBa_5Ti_2Se_9Cl$, $Rb_f3ZnH_5$, $Cs_3ZnH_5$, $(Sr_{3-x}A_x)AlO_4F$, ($A$ = Ba, Ca), $Ba_3MO_5$, ($M$ = tetravalent metal, and $M_3Mg(BH_4)_5$, ($M$ = Rb, Cs)

- We show the structure using the data taken at 4.2 K.

### Body-centered Tetragonal primitive vectors:

$$\mathbf{a}_1 = -\frac{1}{2}\,a\,\hat{\mathbf{x}} + \frac{1}{2}\,a\,\hat{\mathbf{y}} + \frac{1}{2}\,c\,\hat{\mathbf{z}}$$
$$\mathbf{a}_2 = \frac{1}{2}\,a\,\hat{\mathbf{x}} - \frac{1}{2}\,a\,\hat{\mathbf{y}} + \frac{1}{2}\,c\,\hat{\mathbf{z}}$$
$$\mathbf{a}_3 = \frac{1}{2}\,a\,\hat{\mathbf{x}} + \frac{1}{2}\,a\,\hat{\mathbf{y}} - \frac{1}{2}\,c\,\hat{\mathbf{z}}$$

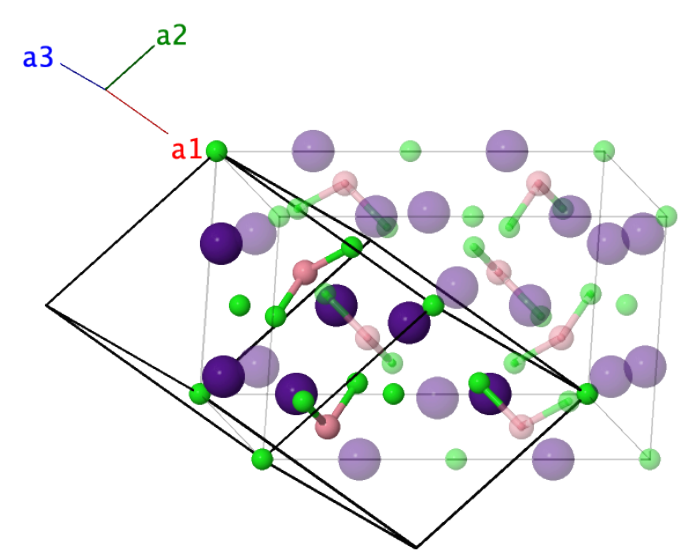

### Basis vectors:

| | | Lattice Coordinates | | Cartesian Coordinates | Wyckoff Position | Atom Type |
|---|---|---|---|---|---|---|
| $\mathbf{B}_1$ | = | $\frac{1}{4}\,\mathbf{a}_1 + \frac{1}{4}\,\mathbf{a}_2$ | = | $\frac{1}{4}c\,\hat{\mathbf{z}}$ | (4$a$) | Cs I |
| $\mathbf{B}_{2}$ | = | $\frac{3}{4}\,\mathbf{a}_{1}+\frac{3}{4}\,\mathbf{a}_{2}$ | = | $\frac{3}{4}c\,\hat{\mathbf{z}}$ | (4$a$) | Cs I |
| $\mathbf{B}_{3}$ | = | $\frac{3}{4}\,\mathbf{a}_{1}+\frac{1}{4}\,\mathbf{a}_{2}+\frac{1}{2}\,\mathbf{a}_{3}$ | = | $\frac{1}{2}a\,\hat{\mathbf{y}}+\frac{1}{4}c\,\hat{\mathbf{z}}$ | (4$b$) | Co |
| $\mathbf{B}_{4}$ | = | $\frac{1}{4}\,\mathbf{a}_{1}+\frac{3}{4}\,\mathbf{a}_{2}+\frac{1}{2}\,\mathbf{a}_{3}$ | = | $\frac{1}{2}a\,\hat{\mathbf{x}}+\frac{1}{4}c\,\hat{\mathbf{z}}$ | (4$b$) | Co |
| $\mathbf{B}_{5}$ | = | $0\,\mathbf{a}_{1}+0\,\mathbf{a}_{2}+0\,\mathbf{a}_{3}$ | = | $0\,\hat{\mathbf{x}}+0\,\hat{\mathbf{y}}+0\,\hat{\mathbf{z}}$ | (4$c$) | Cl I |
| $\mathbf{B}_{6}$ | = | $\frac{1}{2}\,\mathbf{a}_{1}+\frac{1}{2}\,\mathbf{a}_{2}$ | = | $\frac{1}{2}c\,\hat{\mathbf{z}}$ | (4$c$) | Cl I |
| $\mathbf{B}_{7}$ | = | $\left(\frac{1}{2}+x_{4}\right)\,\mathbf{a}_{1}+x_{4}\,\mathbf{a}_{2}+\left(\frac{1}{2}+2x_{4}\right)\,\mathbf{a}_{3}$ | = | $x_{4}a\,\hat{\mathbf{x}}+\left(\frac{1}{2}+x_{4}\right)a\,\hat{\mathbf{y}}$ | (8$h$) | Cs II |
| $\mathbf{B}_{8}$ | = | $\left(\frac{1}{2}-x_{4}\right)\,\mathbf{a}_{1}-x_{4}\,\mathbf{a}_{2}+\left(\frac{1}{2}-2x_{4}\right)\,\mathbf{a}_{3}$ | = | $-x_{4}a\,\hat{\mathbf{x}}+\left(\frac{1}{2}-x_{4}\right)a\,\hat{\mathbf{y}}$ | (8$h$) | Cs II |
| $\mathbf{B}_{9}$ | = | $x_{4}\,\mathbf{a}_{1}+\left(\frac{1}{2}-x_{4}\right)\,\mathbf{a}_{2}+\frac{1}{2}\,\mathbf{a}_{3}$ | = | $\left(\frac{1}{2}-x_{4}\right)a\,\hat{\mathbf{x}}+x_{4}a\,\hat{\mathbf{y}}$ | (8$h$) | Cs II |
| $\mathbf{B}_{10}$ | = | $-x_{4}\,\mathbf{a}_{1}+\left(\frac{1}{2}+x_{4}\right)\,\mathbf{a}_{2}+\frac{1}{2}\,\mathbf{a}_{3}$ | = | $\left(\frac{1}{2}+x_{4}\right)a\,\hat{\mathbf{x}}-x_{4}a\,\hat{\mathbf{y}}$ | (8$h$) | Cs II |
| $\mathbf{B}_{11}$ | = | $\left(\frac{1}{2}+x_{5}+z_{5}\right)\,\mathbf{a}_{1}+\left(x_{5}+z_{5}\right)\,\mathbf{a}_{2}+\left(\frac{1}{2}+2x_{5}\right)\,\mathbf{a}_{3}$ | = | $x_{5}a\,\hat{\mathbf{x}}+\left(\frac{1}{2}+x_{5}\right)a\,\hat{\mathbf{y}}+z_{5}c\,\hat{\mathbf{z}}$ | (16$l$) | Cl II |
| $\mathbf{B}_{12}$ | = | $\left(\frac{1}{2}-x_{5}+z_{5}\right)\,\mathbf{a}_{1}+\left(-x_{5}+z_{5}\right)\,\mathbf{a}_{2}+\left(\frac{1}{2}-2x_{5}\right)\,\mathbf{a}_{3}$ | = | $-x_{5}a\,\hat{\mathbf{x}}+\left(\frac{1}{2}-x_{5}\right)a\,\hat{\mathbf{y}}+z_{5}c\,\hat{\mathbf{z}}$ | (16$l$) | Cl II |
| $\mathbf{B}_{13}$ | = | $\left(x_{5}+z_{5}\right)\,\mathbf{a}_{1}+\left(\frac{1}{2}-x_{5}+z_{5}\right)\,\mathbf{a}_{2}+\frac{1}{2}\,\mathbf{a}_{3}$ | = | $\left(\frac{1}{2}-x_{5}\right)a\,\hat{\mathbf{x}}+x_{5}a\,\hat{\mathbf{y}}+z_{5}c\,\hat{\mathbf{z}}$ | (16$l$) | Cl II |
| $\mathbf{B}_{14}$ | = | $\left(-x_{5}+z_{5}\right)\,\mathbf{a}_{1}+\left(\frac{1}{2}+x_{5}+z_{5}\right)\,\mathbf{a}_{2}+\frac{1}{2}\,\mathbf{a}_{3}$ | = | $\left(\frac{1}{2}+x_{5}\right)a\,\hat{\mathbf{x}}-x_{5}a\,\hat{\mathbf{y}}+z_{5}c\,\hat{\mathbf{z}}$ | (16$l$) | Cl II |
| $\mathbf{B}_{15}$ | = | $\left(x_{5}-z_{5}\right)\,\mathbf{a}_{1}+\left(\frac{1}{2}-x_{5}-z_{5}\right)\,\mathbf{a}_{2}+\frac{1}{2}\,\mathbf{a}_{3}$ | = | $\left(\frac{1}{2}-x_{5}\right)a\,\hat{\mathbf{x}}+x_{5}a\,\hat{\mathbf{y}}-z_{5}c\,\hat{\mathbf{z}}$ | (16$l$) | Cl II |
| $\mathbf{B}_{16}$ | = | $\left(-x_{5}-z_{5}\right)\,\mathbf{a}_{1}+\left(\frac{1}{2}+x_{5}-z_{5}\right)\,\mathbf{a}_{2}+\frac{1}{2}\,\mathbf{a}_{3}$ | = | $\left(\frac{1}{2}+x_{5}\right)a\,\hat{\mathbf{x}}-x_{5}a\,\hat{\mathbf{y}}-z_{5}c\,\hat{\mathbf{z}}$ | (16$l$) | Cl II |
| $\mathbf{B}_{17}$ | = | $\left(\frac{1}{2}+x_{5}-z_{5}\right)\,\mathbf{a}_{1}+\left(x_{5}-z_{5}\right)\,\mathbf{a}_{2}+\left(\frac{1}{2}+2x_{5}\right)\,\mathbf{a}_{3}$ | = | $x_{5}a\,\hat{\mathbf{x}}+\left(\frac{1}{2}+x_{5}\right)a\,\hat{\mathbf{y}}-z_{5}c\,\hat{\mathbf{z}}$ | (16$l$) | Cl II |
| $\mathbf{B}_{18}$ | = | $\left(\frac{1}{2}-x_{5}-z_{5}\right)\,\mathbf{a}_{1}+\left(-x_{5}-z_{5}\right)\,\mathbf{a}_{2}+\left(\frac{1}{2}-2x_{5}\right)\,\mathbf{a}_{3}$ | = | $-x_{5}a\,\hat{\mathbf{x}}+\left(\frac{1}{2}-x_{5}\right)a\,\hat{\mathbf{y}}-z_{5}c\,\hat{\mathbf{z}}$ | (16$l$) | Cl II |

**Geometry files:**

- CIF: pp. 1716
- POSCAR: pp. 1716

# $U_6Mn$ ($D2_c$) Structure: AB6_tI28_140_a_hk

http://aflow.org/prototype-encyclopedia/AB6_tI28_140_a_hk

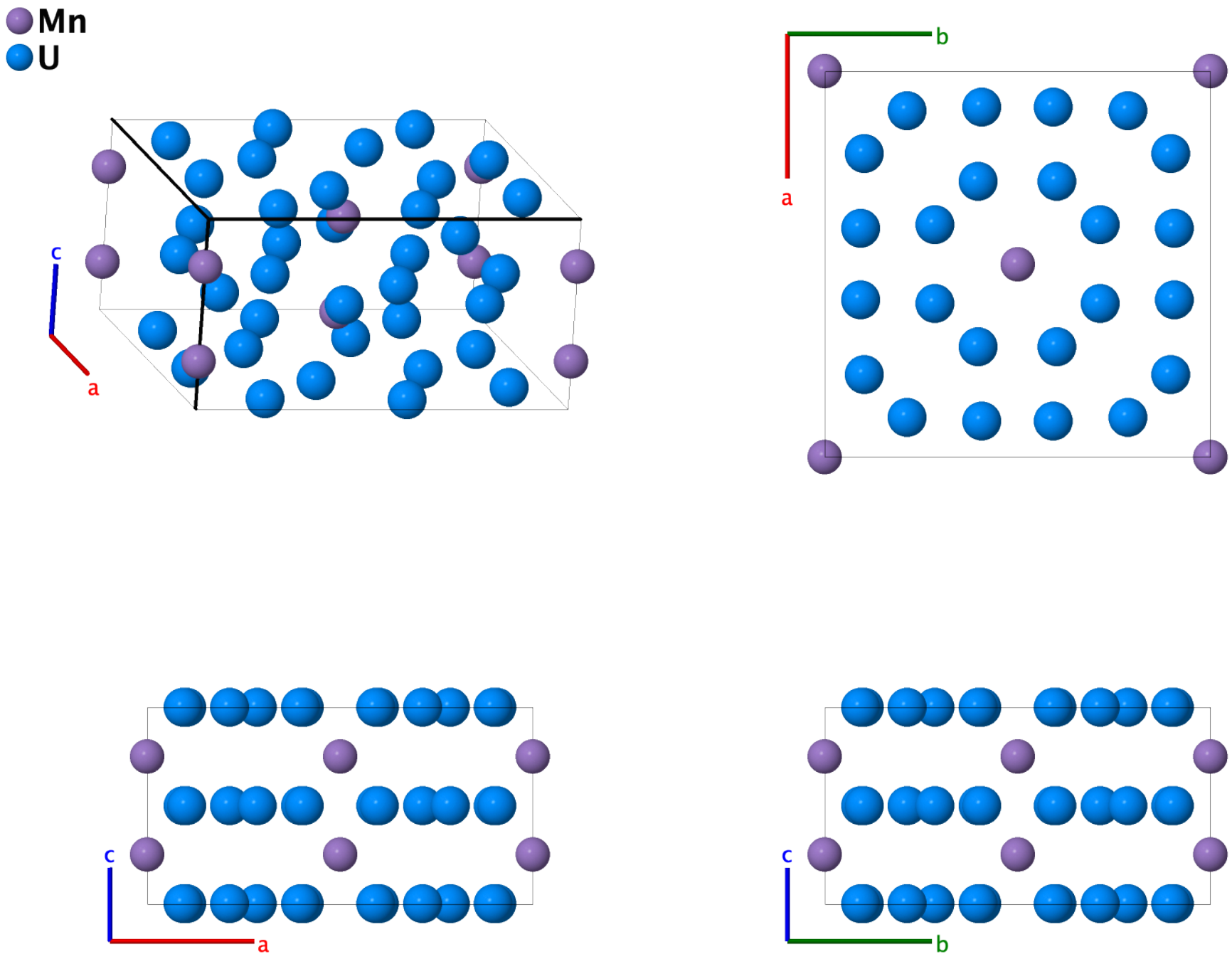

| | | |
|---|---|---|
| **Prototype** | : | $MnU_6$ |
| **AFLOW prototype label** | : | AB6_tI28_140_a_hk |
| ***Strukturbericht* designation** | : | $D2_c$ |
| **Pearson symbol** | : | tI28 |
| **Space group number** | : | 140 |
| **Space group symbol** | : | $I4/mcm$ |
| **AFLOW prototype command** | : | `aflow --proto=AB6_tI28_140_a_hk`<br>`--params=`$a, c/a, x_2, x_3, y_3$ |

**Other compounds with this structure**

- $U_6Co$, $U_6Fe$, and $U_6Ni$

- This structure is closely related to the $V_4SiSb_2$ structure. This can also be identified as a defected version of the $D8_m$ $W_5Si_3$ structure.

**Body-centered Tetragonal primitive vectors:**

$$
\begin{aligned}
\mathbf{a}_1 &= -\tfrac{1}{2}\,a\,\hat{\mathbf{x}} + \tfrac{1}{2}\,a\,\hat{\mathbf{y}} + \tfrac{1}{2}\,c\,\hat{\mathbf{z}} \\
\mathbf{a}_2 &= \tfrac{1}{2}\,a\,\hat{\mathbf{x}} - \tfrac{1}{2}\,a\,\hat{\mathbf{y}} + \tfrac{1}{2}\,c\,\hat{\mathbf{z}} \\
\mathbf{a}_3 &= \tfrac{1}{2}\,a\,\hat{\mathbf{x}} + \tfrac{1}{2}\,a\,\hat{\mathbf{y}} - \tfrac{1}{2}\,c\,\hat{\mathbf{z}}
\end{aligned}
$$

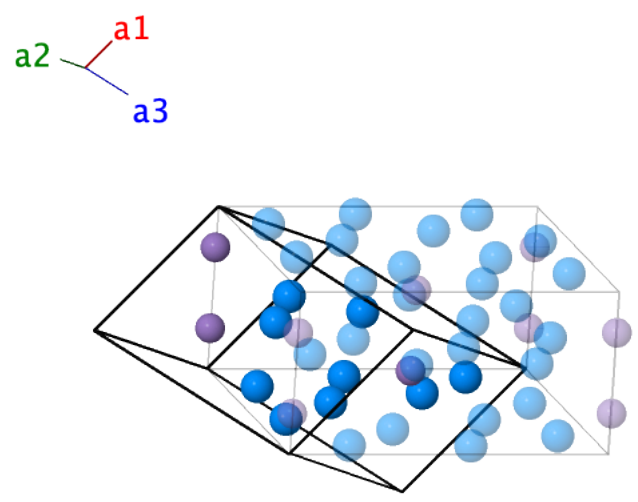

**Basis vectors:**

| | | Lattice Coordinates | | Cartesian Coordinates | Wyckoff Position | Atom Type |
|---|---|---|---|---|---|---|
| $\mathbf{B}_1$ | = | $\frac{1}{4}\,\mathbf{a}_1 + \frac{1}{4}\,\mathbf{a}_2$ | = | $\frac{1}{4}c\,\hat{\mathbf{z}}$ | $(4a)$ | Mn |
| $\mathbf{B}_2$ | = | $\frac{3}{4}\,\mathbf{a}_1 + \frac{3}{4}\,\mathbf{a}_2$ | = | $\frac{3}{4}c\,\hat{\mathbf{z}}$ | $(4a)$ | Mn |
| $\mathbf{B}_3$ | = | $\left(\frac{1}{2} + x_2\right)\,\mathbf{a}_1 + x_2\,\mathbf{a}_2 + \left(\frac{1}{2} + 2x_2\right)\,\mathbf{a}_3$ | = | $x_2a\,\hat{\mathbf{x}} + \left(\frac{1}{2} + x_2\right)a\,\hat{\mathbf{y}}$ | $(8h)$ | U I |
| $\mathbf{B}_4$ | = | $\left(\frac{1}{2} - x_2\right)\,\mathbf{a}_1 - x_2\,\mathbf{a}_2 + \left(\frac{1}{2} - 2x_2\right)\,\mathbf{a}_3$ | = | $-x_2a\,\hat{\mathbf{x}} + \left(\frac{1}{2} - x_2\right)a\,\hat{\mathbf{y}}$ | $(8h)$ | U I |
| $\mathbf{B}_5$ | = | $x_2\,\mathbf{a}_1 + \left(\frac{1}{2} - x_2\right)\,\mathbf{a}_2 + \frac{1}{2}\,\mathbf{a}_3$ | = | $\left(\frac{1}{2} - x_2\right)a\,\hat{\mathbf{x}} + x_2a\,\hat{\mathbf{y}}$ | $(8h)$ | U I |
| $\mathbf{B}_6$ | = | $-x_2\,\mathbf{a}_1 + \left(\frac{1}{2} + x_2\right)\,\mathbf{a}_2 + \frac{1}{2}\,\mathbf{a}_3$ | = | $\left(\frac{1}{2} + x_2\right)a\,\hat{\mathbf{x}} - x_2a\,\hat{\mathbf{y}}$ | $(8h)$ | U I |
| $\mathbf{B}_7$ | = | $y_3\,\mathbf{a}_1 + x_3\,\mathbf{a}_2 + (x_3 + y_3)\,\mathbf{a}_3$ | = | $x_3a\,\hat{\mathbf{x}} + y_3a\,\hat{\mathbf{y}}$ | $(16k)$ | U II |
| $\mathbf{B}_8$ | = | $-y_3\,\mathbf{a}_1 - x_3\,\mathbf{a}_2 + (-x_3 - y_3)\,\mathbf{a}_3$ | = | $-x_3a\,\hat{\mathbf{x}} - y_3a\,\hat{\mathbf{y}}$ | $(16k)$ | U II |
| $\mathbf{B}_9$ | = | $x_3\,\mathbf{a}_1 - y_3\,\mathbf{a}_2 + (x_3 - y_3)\,\mathbf{a}_3$ | = | $-y_3a\,\hat{\mathbf{x}} + x_3a\,\hat{\mathbf{y}}$ | $(16k)$ | U II |
| $\mathbf{B}_{10}$ | = | $-x_3\,\mathbf{a}_1 + y_3\,\mathbf{a}_2 + (-x_3 + y_3)\,\mathbf{a}_3$ | = | $y_3a\,\hat{\mathbf{x}} - x_3a\,\hat{\mathbf{y}}$ | $(16k)$ | U II |
| $\mathbf{B}_{11}$ | = | $\left(\frac{1}{2} + y_3\right)\,\mathbf{a}_1 + \left(\frac{1}{2} - x_3\right)\,\mathbf{a}_2 + (-x_3 + y_3)\,\mathbf{a}_3$ | = | $-x_3a\,\hat{\mathbf{x}} + y_3a\,\hat{\mathbf{y}} + \frac{1}{2}c\,\hat{\mathbf{z}}$ | $(16k)$ | U II |
| $\mathbf{B}_{12}$ | = | $\left(\frac{1}{2} - y_3\right)\,\mathbf{a}_1 + \left(\frac{1}{2} + x_3\right)\,\mathbf{a}_2 + (x_3 - y_3)\,\mathbf{a}_3$ | = | $x_3a\,\hat{\mathbf{x}} - y_3a\,\hat{\mathbf{y}} + \frac{1}{2}c\,\hat{\mathbf{z}}$ | $(16k)$ | U II |
| $\mathbf{B}_{13}$ | = | $\left(\frac{1}{2} + x_3\right)\,\mathbf{a}_1 + \left(\frac{1}{2} + y_3\right)\,\mathbf{a}_2 + (x_3 + y_3)\,\mathbf{a}_3$ | = | $y_3a\,\hat{\mathbf{x}} + x_3a\,\hat{\mathbf{y}} + \frac{1}{2}c\,\hat{\mathbf{z}}$ | $(16k)$ | U II |
| $\mathbf{B}_{14}$ | = | $\left(\frac{1}{2} - x_3\right)\,\mathbf{a}_1 + \left(\frac{1}{2} - y_3\right)\,\mathbf{a}_2 + (-x_3 - y_3)\,\mathbf{a}_3$ | = | $-y_3a\,\hat{\mathbf{x}} - x_3a\,\hat{\mathbf{y}} + \frac{1}{2}c\,\hat{\mathbf{z}}$ | $(16k)$ | U II |

**Found in:**

**Geometry files:**

- CIF: pp. 1716
- POSCAR: pp. 1717

# $BaCd_{11}$ Structure: AB11_tI48_141_a_bdi

http://aflow.org/prototype-encyclopedia/AB11_tI48_141_a_bdi

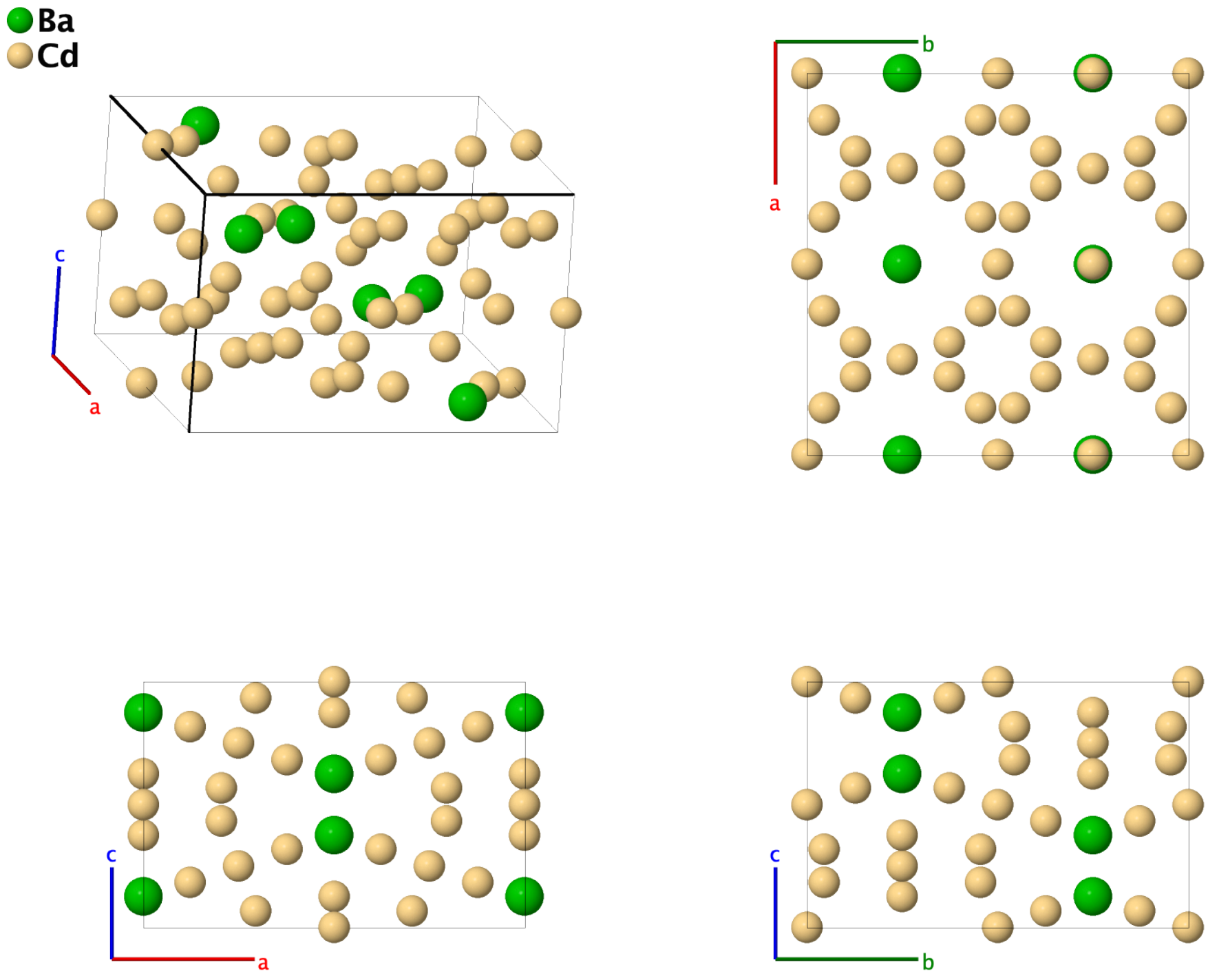

| | | |
|---|---|---|
| **Prototype** | : | $BaCd_{11}$ |
| **AFLOW prototype label** | : | AB11_tI48_141_a_bdi |
| ***Strukturbericht* designation** | : | None |
| **Pearson symbol** | : | tI48 |
| **Space group number** | : | 141 |
| **Space group symbol** | : | $I4_1/amd$ |
| **AFLOW prototype command** | : | `aflow --proto=AB11_tI48_141_a_bdi`<br>`--params=`$a, c/a, x_4, y_4, z_4$ |

### Other compounds with this structure

- $SrCd_{11}$, $LaZn_{11}$, $CeZn_{11}$, and $PrZn_{11}$

- (Sanderson, 1953) gave the atomic positions in setting 1 of space group #141. We used FINDSYM to transform this to the standard setting 2.

## Body-centered Tetragonal primitive vectors:

$$\mathbf{a}_1 = -\frac{1}{2}\,a\,\hat{\mathbf{x}} + \frac{1}{2}\,a\,\hat{\mathbf{y}} + \frac{1}{2}\,c\,\hat{\mathbf{z}}$$
$$\mathbf{a}_2 = \frac{1}{2}\,a\,\hat{\mathbf{x}} - \frac{1}{2}\,a\,\hat{\mathbf{y}} + \frac{1}{2}\,c\,\hat{\mathbf{z}}$$
$$\mathbf{a}_3 = \frac{1}{2}\,a\,\hat{\mathbf{x}} + \frac{1}{2}\,a\,\hat{\mathbf{y}} - \frac{1}{2}\,c\,\hat{\mathbf{z}}$$

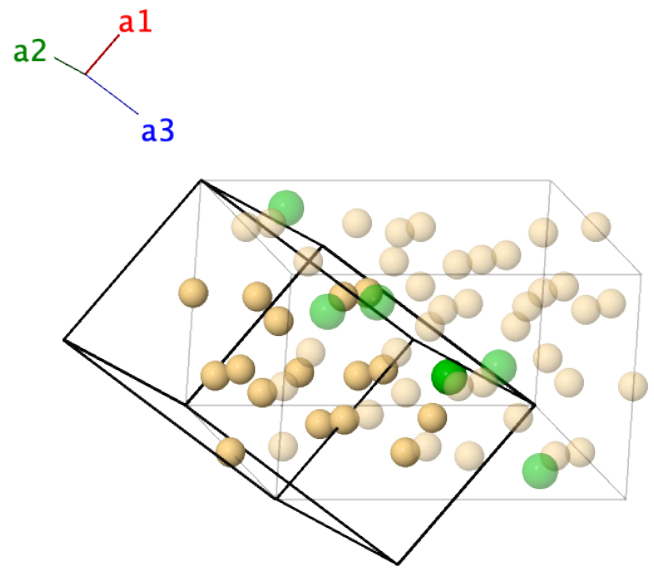

## Basis vectors:

| | | Lattice Coordinates | | Cartesian Coordinates | Wyckoff Position | Atom Type |
|---|---|---|---|---|---|---|
| $\mathbf{B}_1$ | = | $\frac{7}{8}\,\mathbf{a}_1 + \frac{1}{8}\,\mathbf{a}_2 + \frac{3}{4}\,\mathbf{a}_3$ | = | $\frac{3}{4}a\,\hat{\mathbf{y}} + \frac{1}{8}c\,\hat{\mathbf{z}}$ | (4$a$) | Ba |
| $\mathbf{B}_2$ | = | $\frac{1}{8}\,\mathbf{a}_1 + \frac{7}{8}\,\mathbf{a}_2 + \frac{1}{4}\,\mathbf{a}_3$ | = | $\frac{1}{2}a\,\hat{\mathbf{x}} + \frac{3}{4}a\,\hat{\mathbf{y}} + \frac{3}{8}c\,\hat{\mathbf{z}}$ | (4$a$) | Ba |
| $\mathbf{B}_3$ | = | $\frac{5}{8}\,\mathbf{a}_1 + \frac{3}{8}\,\mathbf{a}_2 + \frac{1}{4}\,\mathbf{a}_3$ | = | $\frac{1}{4}a\,\hat{\mathbf{y}} + \frac{3}{8}c\,\hat{\mathbf{z}}$ | (4$b$) | Cd I |
| $\mathbf{B}_4$ | = | $\frac{3}{8}\,\mathbf{a}_1 + \frac{5}{8}\,\mathbf{a}_2 + \frac{3}{4}\,\mathbf{a}_3$ | = | $\frac{1}{2}a\,\hat{\mathbf{x}} + \frac{1}{4}a\,\hat{\mathbf{y}} + \frac{1}{8}c\,\hat{\mathbf{z}}$ | (4$b$) | Cd I |
| $\mathbf{B}_5$ | = | $\frac{1}{2}\,\mathbf{a}_1 + \frac{1}{2}\,\mathbf{a}_2$ | = | $\frac{1}{2}c\,\hat{\mathbf{z}}$ | (8$d$) | Cd II |
| $\mathbf{B}_6$ | = | $\frac{1}{2}\,\mathbf{a}_2 + \frac{1}{2}\,\mathbf{a}_3$ | = | $\frac{1}{2}a\,\hat{\mathbf{x}}$ | (8$d$) | Cd II |
| $\mathbf{B}_7$ | = | $\frac{1}{2}\,\mathbf{a}_1$ | = | $-\frac{1}{4}a\,\hat{\mathbf{x}} + \frac{1}{4}a\,\hat{\mathbf{y}} + \frac{1}{4}c\,\hat{\mathbf{z}}$ | (8$d$) | Cd II |
| $\mathbf{B}_8$ | = | $\frac{1}{2}\,\mathbf{a}_1 + \frac{1}{2}\,\mathbf{a}_2 + \frac{1}{2}\,\mathbf{a}_3$ | = | $\frac{1}{4}a\,\hat{\mathbf{x}} + \frac{1}{4}a\,\hat{\mathbf{y}} + \frac{1}{4}c\,\hat{\mathbf{z}}$ | (8$d$) | Cd II |
| $\mathbf{B}_9$ | = | $(y_4 + z_4)\,\mathbf{a}_1 + (x_4 + z_4)\,\mathbf{a}_2 + (x_4 + y_4)\,\mathbf{a}_3$ | = | $x_4a\,\hat{\mathbf{x}} + y_4a\,\hat{\mathbf{y}} + z_4c\,\hat{\mathbf{z}}$ | (32$i$) | Cd III |
| $\mathbf{B}_{10}$ | = | $\left(\frac{1}{2} - y_4 + z_4\right)\mathbf{a}_1 + (-x_4 + z_4)\,\mathbf{a}_2 + \left(\frac{1}{2} - x_4 - y_4\right)\mathbf{a}_3$ | = | $-x_4a\,\hat{\mathbf{x}} + \left(\frac{1}{2} - y_4\right)a\,\hat{\mathbf{y}} + z_4c\,\hat{\mathbf{z}}$ | (32$i$) | Cd III |
| $\mathbf{B}_{11}$ | = | $(x_4 + z_4)\,\mathbf{a}_1 + \left(\frac{1}{2} - y_4 + z_4\right)\mathbf{a}_2 + (x_4 - y_4)\,\mathbf{a}_3$ | = | $\left(\frac{1}{4} - y_4\right)a\,\hat{\mathbf{x}} + \left(\frac{3}{4} + x_4\right)a\,\hat{\mathbf{y}} + \left(\frac{1}{4} + z_4\right)c\,\hat{\mathbf{z}}$ | (32$i$) | Cd III |
| $\mathbf{B}_{12}$ | = | $(-x_4 + z_4)\,\mathbf{a}_1 + (y_4 + z_4)\,\mathbf{a}_2 + \left(\frac{1}{2} - x_4 + y_4\right)\mathbf{a}_3$ | = | $\left(\frac{1}{4} + y_4\right)a\,\hat{\mathbf{x}} + \left(\frac{1}{4} - x_4\right)a\,\hat{\mathbf{y}} + \left(\frac{3}{4} + z_4\right)c\,\hat{\mathbf{z}}$ | (32$i$) | Cd III |
| $\mathbf{B}_{13}$ | = | $\left(\frac{1}{2} + y_4 - z_4\right)\mathbf{a}_1 + (-x_4 - z_4)\,\mathbf{a}_2 + \left(\frac{1}{2} - x_4 + y_4\right)\mathbf{a}_3$ | = | $-x_4a\,\hat{\mathbf{x}} + \left(\frac{1}{2} + y_4\right)a\,\hat{\mathbf{y}} - z_4c\,\hat{\mathbf{z}}$ | (32$i$) | Cd III |
| $\mathbf{B}_{14}$ | = | $(-y_4 - z_4)\,\mathbf{a}_1 + (x_4 - z_4)\,\mathbf{a}_2 + (x_4 - y_4)\,\mathbf{a}_3$ | = | $x_4a\,\hat{\mathbf{x}} - y_4a\,\hat{\mathbf{y}} - z_4c\,\hat{\mathbf{z}}$ | (32$i$) | Cd III |
| $\mathbf{B}_{15}$ | = | $(x_4 - z_4)\,\mathbf{a}_1 + \left(\frac{1}{2} + y_4 - z_4\right)\mathbf{a}_2 + (x_4 + y_4)\,\mathbf{a}_3$ | = | $\left(\frac{1}{4} + y_4\right)a\,\hat{\mathbf{x}} + \left(\frac{3}{4} + x_4\right)a\,\hat{\mathbf{y}} + \left(\frac{1}{4} - z_4\right)c\,\hat{\mathbf{z}}$ | (32$i$) | Cd III |
| $\mathbf{B}_{16}$ | = | $(-x_4 - z_4)\,\mathbf{a}_1 + (-y_4 - z_4)\,\mathbf{a}_2 + \left(\frac{1}{2} - x_4 - y_4\right)\mathbf{a}_3$ | = | $\left(\frac{1}{4} - y_4\right)a\,\hat{\mathbf{x}} + \left(\frac{1}{4} - x_4\right)a\,\hat{\mathbf{y}} + \left(\frac{3}{4} - z_4\right)c\,\hat{\mathbf{z}}$ | (32$i$) | Cd III |
| $\mathbf{B}_{17}$ | = | $(-y_4 - z_4)\,\mathbf{a}_1 + (-x_4 - z_4)\,\mathbf{a}_2 + (-x_4 - y_4)\,\mathbf{a}_3$ | = | $-x_4a\,\hat{\mathbf{x}} - y_4a\,\hat{\mathbf{y}} - z_4c\,\hat{\mathbf{z}}$ | (32$i$) | Cd III |
| $\mathbf{B}_{18}$ | = | $\left(\frac{1}{2} + y_4 - z_4\right)\mathbf{a}_1 + (x_4 - z_4)\,\mathbf{a}_2 + \left(\frac{1}{2} + x_4 + y_4\right)\mathbf{a}_3$ | = | $x_4a\,\hat{\mathbf{x}} + \left(\frac{1}{2} + y_4\right)a\,\hat{\mathbf{y}} - z_4c\,\hat{\mathbf{z}}$ | (32$i$) | Cd III |
| $\mathbf{B}_{19}$ | = | $(-x_4 - z_4)\,\mathbf{a}_1 + \left(\frac{1}{2} + y_4 - z_4\right)\mathbf{a}_2 + (-x_4 + y_4)\,\mathbf{a}_3$ | = | $\left(\frac{1}{4} + y_4\right)a\,\hat{\mathbf{x}} - a\left(x_4 + \frac{1}{4}\right)\hat{\mathbf{y}} + \left(\frac{1}{4} - z_4\right)c\,\hat{\mathbf{z}}$ | (32$i$) | Cd III |
| $\mathbf{B}_{20}$ | = | $(x_4 - z_4)\,\mathbf{a}_1 + (-y_4 - z_4)\,\mathbf{a}_2 + \left(\frac{1}{2} + x_4 - y_4\right)\,\mathbf{a}_3$ | = | $\left(\frac{1}{4} - y_4\right) a\,\mathbf{\hat{x}} + \left(\frac{1}{4} + x_4\right) a\,\mathbf{\hat{y}} - c\left(z_4 + \frac{1}{4}\right)\,\mathbf{\hat{z}}$ | (32*i*) | Cd III |
| $\mathbf{B}_{21}$ | = | $\left(\frac{1}{2} - y_4 + z_4\right)\,\mathbf{a}_1 + (x_4 + z_4)\,\mathbf{a}_2 + \left(\frac{1}{2} + x_4 - y_4\right)\,\mathbf{a}_3$ | = | $x_4 a\,\mathbf{\hat{x}} + \left(\frac{1}{2} - y_4\right) a\,\mathbf{\hat{y}} + z_4 c\,\mathbf{\hat{z}}$ | (32*i*) | Cd III |
| $\mathbf{B}_{22}$ | = | $(y_4 + z_4)\,\mathbf{a}_1 + (-x_4 + z_4)\,\mathbf{a}_2 + (-x_4 + y_4)\,\mathbf{a}_3$ | = | $-x_4 a\,\mathbf{\hat{x}} + y_4 a\,\mathbf{\hat{y}} + z_4 c\,\mathbf{\hat{z}}$ | (32*i*) | Cd III |
| $\mathbf{B}_{23}$ | = | $(-x_4 + z_4)\,\mathbf{a}_1 + \left(\frac{1}{2} - y_4 + z_4\right)\,\mathbf{a}_2 + (-x_4 - y_4)\,\mathbf{a}_3$ | = | $\left(\frac{1}{4} - y_4\right) a\,\mathbf{\hat{x}} - a\left(x_4 + \frac{1}{4}\right)\,\mathbf{\hat{y}} + \left(\frac{1}{4} + z_4\right) c\,\mathbf{\hat{z}}$ | (32*i*) | Cd III |
| $\mathbf{B}_{24}$ | = | $(x_4 + z_4)\,\mathbf{a}_1 + (y_4 + z_4)\,\mathbf{a}_2 + \left(\frac{1}{2} + x_4 + y_4\right)\,\mathbf{a}_3$ | = | $\left(\frac{1}{4} + y_4\right) a\,\mathbf{\hat{x}} + \left(\frac{1}{4} + x_4\right) a\,\mathbf{\hat{y}} + \left(-\frac{1}{4} + z_4\right) c\,\mathbf{\hat{z}}$ | (32*i*) | Cd III |

---

---

**Geometry files:**

- CIF: pp. 1717
- POSCAR: pp. 1717

# Analcime ($NaAlSi_2O_6 \cdot H_2O$, $S6_1$) Structure: A2B2C3D12E4_tI184_142_f_f_be_3g_g

http://aflow.org/prototype-encyclopedia/A2B2C3D12E4_tI184_142_f_f_be_3g_g

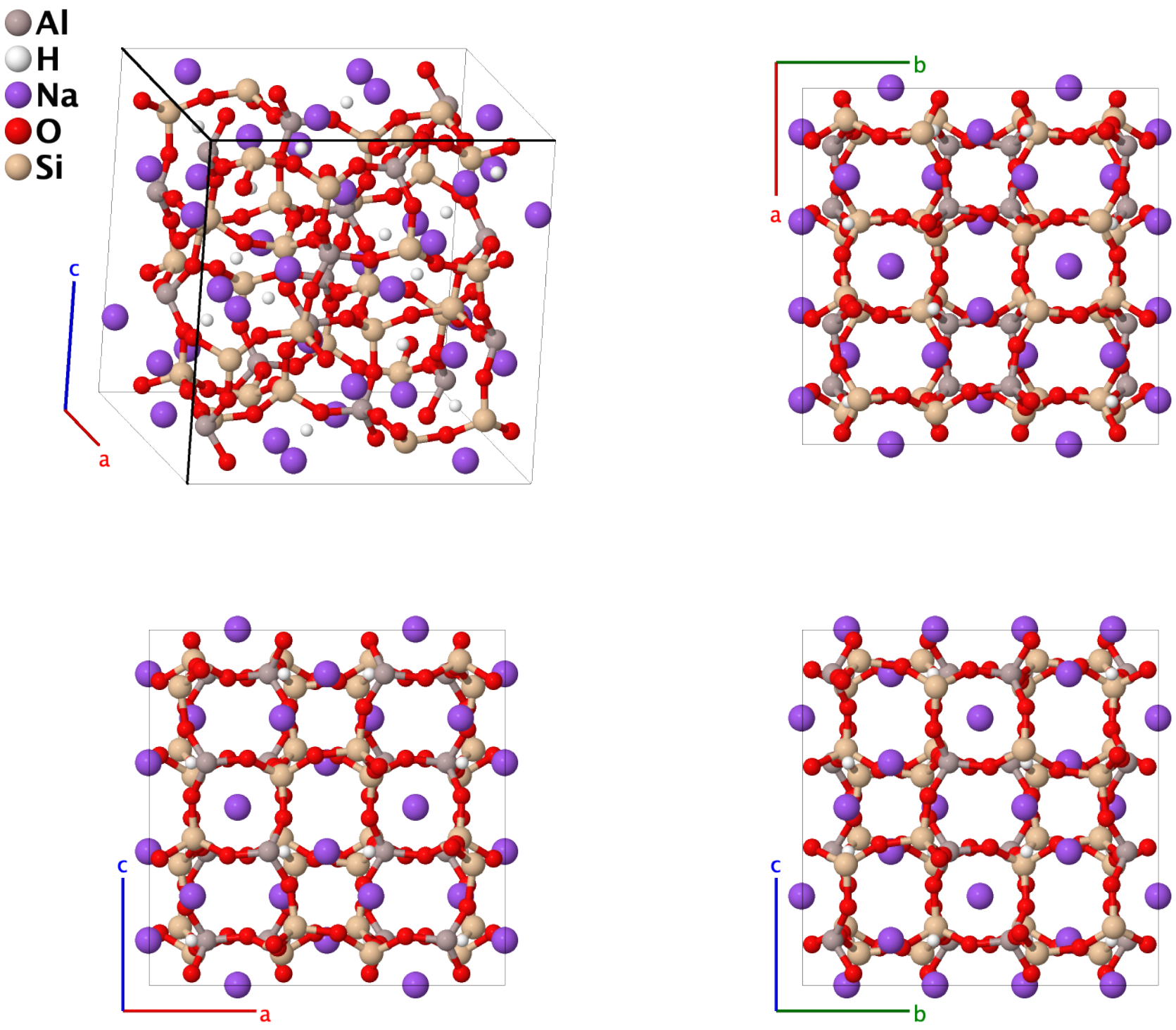

| | | |
|---|---|---|
| **Prototype** | : | $Al(H_2O)NaO_6Si_2$ |
| **AFLOW prototype label** | : | A2B2C3D12E4_tI184_142_f_f_be_3g_g |
| ***Strukturbericht* designation** | : | $S6_1$ |
| **Pearson symbol** | : | tI184 |
| **Space group number** | : | 142 |
| **Space group symbol** | : | $I4_1/acd$ |
| **AFLOW prototype command** | : | `aflow --proto=A2B2C3D12E4_tI184_142_f_f_be_3g_g` `--params=`$a, c/a, x_2, y_2, x_3, x_4, x_5, y_5, z_5, x_6, y_6, z_6, x_7, y_7, z_7, x_8, y_8, z_8$ |

- Analcime is pseudo-cubic, space group $Ia\overline{3}d$ #230, but (Hartwig, 1931) showed that the structure actually is tetragonal, space group $I4_1/acd$ #142. (Hermann, 1937) listed both structures in defining *Strukturbericht* type $S6_1$.
- Analcime can have many crystal structures, as found in *e.g.,* (Mazzi, 1978) and (Pechar, 1988). We take our prototype from (Mazzi, 1978), using the data from their sample ANA 1, which is in space group $I4_1/acd$ #142. The major difference between this structure and that found by (Hartwig, 1931) is that sodium atoms were found at the (8*b*) Wyckoff position (Na-I). (Mazzi, 1978) recover the stoichiometry by noting that site Na-I (8*b*) is only occupied 23% of the time, while site Na-II (16*e*) has 82% occupancy. In addition, our Si site (32*e*) is actually 47% aluminum and 53% silicon, and the site which we, following (Hartwig, 1931), label Al (16*f*) is actually 98% silicon and only 2% aluminum. These fractions are highly dependent upon the choice of sample.
- Removing all sodium from the (8*b*) site results in the original structure found by (Hartwig, 1931) and defined as $S6_1$ by (Hermann, 1937).
- The positions of the hydrogen atoms in the water molecules were not determined, so we only provide the positions of the oxygen atoms (labeled as $H_2O$).

**Body-centered Tetragonal primitive vectors:**

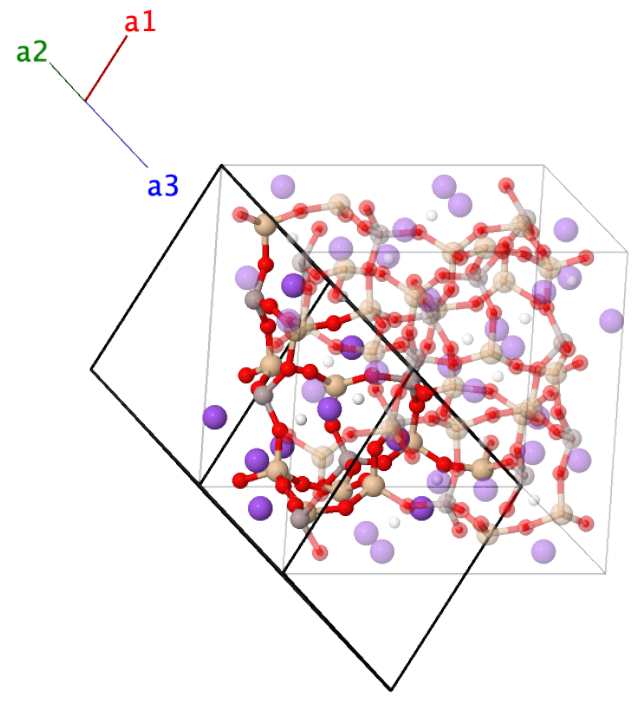

$$
\begin{aligned}
\mathbf{a}_1 &= -\frac{1}{2}\,a\,\hat{\mathbf{x}} + \frac{1}{2}\,a\,\hat{\mathbf{y}} + \frac{1}{2}\,c\,\hat{\mathbf{z}} \\
\mathbf{a}_2 &= \frac{1}{2}\,a\,\hat{\mathbf{x}} - \frac{1}{2}\,a\,\hat{\mathbf{y}} + \frac{1}{2}\,c\,\hat{\mathbf{z}} \\
\mathbf{a}_3 &= \frac{1}{2}\,a\,\hat{\mathbf{x}} + \frac{1}{2}\,a\,\hat{\mathbf{y}} - \frac{1}{2}\,c\,\hat{\mathbf{z}}
\end{aligned}
$$

**Basis vectors:**

| | | Lattice Coordinates | | Cartesian Coordinates | Wyckoff Position | Atom Type |
|---|---|---|---|---|---|---|
| $\mathbf{B}_1$ | = | $\frac{3}{8}\,\mathbf{a}_1 + \frac{1}{8}\,\mathbf{a}_2 + \frac{1}{4}\,\mathbf{a}_3$ | = | $\frac{1}{4}a\,\hat{\mathbf{y}} + \frac{1}{8}c\,\hat{\mathbf{z}}$ | (8$b$) | Na I |
| $\mathbf{B}_2$ | = | $\frac{1}{8}\,\mathbf{a}_1 + \frac{3}{8}\,\mathbf{a}_2 + \frac{3}{4}\,\mathbf{a}_3$ | = | $\frac{1}{2}a\,\hat{\mathbf{x}} + \frac{1}{4}a\,\hat{\mathbf{y}} - \frac{1}{8}c\,\hat{\mathbf{z}}$ | (8$b$) | Na I |
| $\mathbf{B}_3$ | = | $\frac{5}{8}\,\mathbf{a}_1 + \frac{7}{8}\,\mathbf{a}_2 + \frac{3}{4}\,\mathbf{a}_3$ | = | $\frac{1}{2}a\,\hat{\mathbf{x}} + \frac{1}{4}a\,\hat{\mathbf{y}} + \frac{3}{8}c\,\hat{\mathbf{z}}$ | (8$b$) | Na I |
| $\mathbf{B}_4$ | = | $\frac{7}{8}\,\mathbf{a}_1 + \frac{5}{8}\,\mathbf{a}_2 + \frac{1}{4}\,\mathbf{a}_3$ | = | $\frac{1}{4}a\,\hat{\mathbf{y}} + \frac{5}{8}c\,\hat{\mathbf{z}}$ | (8$b$) | Na I |
| $\mathbf{B}_5$ | = | $\frac{1}{4}\,\mathbf{a}_1 + \left(\frac{1}{4} + x_2\right)\,\mathbf{a}_2 + x_2\,\mathbf{a}_3$ | = | $x_2 a\,\hat{\mathbf{x}} + \frac{1}{4}c\,\hat{\mathbf{z}}$ | (16$e$) | Na II |
| $\mathbf{B}_6$ | = | $\frac{3}{4}\,\mathbf{a}_1 + \left(\frac{1}{4} - x_2\right)\,\mathbf{a}_2 + \left(\frac{1}{2} - x_2\right)\,\mathbf{a}_3$ | = | $-x_2 a\,\hat{\mathbf{x}} + \frac{1}{2}a\,\hat{\mathbf{y}} + \frac{1}{4}c\,\hat{\mathbf{z}}$ | (16$e$) | Na II |
| $\mathbf{B}_7$ | = | $\left(\frac{1}{4} + x_2\right)\,\mathbf{a}_1 + \frac{3}{4}\,\mathbf{a}_2 + x_2\,\mathbf{a}_3$ | = | $\frac{1}{4}a\,\hat{\mathbf{x}} + \left(\frac{3}{4} + x_2\right)a\,\hat{\mathbf{y}} + \frac{1}{2}c\,\hat{\mathbf{z}}$ | (16$e$) | Na II |
| $\mathbf{B}_8$ | = | $\left(\frac{1}{4} - x_2\right)\,\mathbf{a}_1 + \frac{1}{4}\,\mathbf{a}_2 + \left(\frac{1}{2} - x_2\right)\,\mathbf{a}_3$ | = | $\frac{1}{4}a\,\hat{\mathbf{x}} + \left(\frac{1}{4} - x_2\right)a\,\hat{\mathbf{y}}$ | (16$e$) | Na II |
| $\mathbf{B}_9$ | = | $\frac{3}{4}\,\mathbf{a}_1 + \left(\frac{3}{4} - x_2\right)\,\mathbf{a}_2 - x_2\,\mathbf{a}_3$ | = | $-x_2 a\,\hat{\mathbf{x}} + \frac{3}{4}c\,\hat{\mathbf{z}}$ | (16$e$) | Na II |
| $\mathbf{B}_{10}$ | = | $\frac{1}{4}\,\mathbf{a}_1 + \left(\frac{3}{4} + x_2\right)\,\mathbf{a}_2 + \left(\frac{1}{2} + x_2\right)\,\mathbf{a}_3$ | = | $\left(\frac{1}{2} + x_2\right)a\,\hat{\mathbf{x}} + \frac{1}{4}c\,\hat{\mathbf{z}}$ | (16$e$) | Na II |
| $\mathbf{B}_{11}$ | = | $\left(\frac{1}{4} - x_2 + y_2\right)\,\mathbf{a}_1 + \left(\frac{3}{4} + y_2\right)\,\mathbf{a}_2 - x_2\,\mathbf{a}_3$ | = | $\frac{1}{4}a\,\hat{\mathbf{x}} - a\left(x_2 + \frac{1}{4}\right)\,\hat{\mathbf{y}} + \left(\frac{1}{2} + y_2\right)c\,\hat{\mathbf{z}}$ | (16$e$) | Na II |
| $\mathbf{B}_{12}$ | = | $\left(\frac{3}{4} + x_2\right)\,\mathbf{a}_1 + \frac{3}{4}\,\mathbf{a}_2 + \left(\frac{1}{2} + x_2\right)\,\mathbf{a}_3$ | = | $\frac{1}{4}a\,\hat{\mathbf{x}} + \left(\frac{1}{4} + x_2\right)a\,\hat{\mathbf{y}} + \frac{1}{2}c\,\hat{\mathbf{z}}$ | (16$e$) | Na II |
| $\mathbf{B}_{13}$ | = | $\left(\frac{3}{8} + x_3\right)\,\mathbf{a}_1 + \left(\frac{1}{8} + x_3\right)\,\mathbf{a}_2 + \left(\frac{1}{4} + 2x_3\right)\,\mathbf{a}_3$ | = | $x_3 a\,\hat{\mathbf{x}} + \left(\frac{1}{4} + x_3\right)a\,\hat{\mathbf{y}} + \frac{1}{8}c\,\hat{\mathbf{z}}$ | (16$f$) | Al |
| $\mathbf{B}_{14}$ | = | $\left(\frac{3}{8} - x_3\right)\,\mathbf{a}_1 + \left(\frac{1}{8} - x_3\right)\,\mathbf{a}_2 + \left(\frac{1}{4} - 2x_3\right)\,\mathbf{a}_3$ | = | $-x_3 a\,\hat{\mathbf{x}} + \left(\frac{1}{4} - x_3\right)a\,\hat{\mathbf{y}} + \frac{1}{8}c\,\hat{\mathbf{z}}$ | (16$f$) | Al |
| $\mathbf{B}_{15}$ | = | $\left(\frac{1}{8} + x_3\right)\,\mathbf{a}_1 + \left(\frac{3}{8} - x_3\right)\,\mathbf{a}_2 + \frac{3}{4}\,\mathbf{a}_3$ | = | $\left(\frac{1}{2} - x_3\right)a\,\hat{\mathbf{x}} + \left(\frac{1}{4} + x_3\right)a\,\hat{\mathbf{y}} - \frac{1}{8}c\,\hat{\mathbf{z}}$ | (16$f$) | Al |
| $\mathbf{B}_{16}$ | = | $\left(\frac{1}{8} - x_3\right)\,\mathbf{a}_1 + \left(\frac{3}{8} + x_3\right)\,\mathbf{a}_2 + \frac{3}{4}\,\mathbf{a}_3$ | = | $\left(\frac{1}{2} + x_3\right)a\,\hat{\mathbf{x}} + \left(\frac{1}{4} - x_3\right)a\,\hat{\mathbf{y}} + \frac{7}{8}c\,\hat{\mathbf{z}}$ | (16$f$) | Al |
| $\mathbf{B}_{17}$ | = | $\left(\frac{5}{8} - x_3\right)\,\mathbf{a}_1 + \left(\frac{7}{8} - x_3\right)\,\mathbf{a}_2 + \left(\frac{3}{4} - 2x_3\right)\,\mathbf{a}_3$ | = | $\left(\frac{1}{2} - x_3\right)a\,\hat{\mathbf{x}} + \left(\frac{1}{4} - x_3\right)a\,\hat{\mathbf{y}} + \frac{3}{8}c\,\hat{\mathbf{z}}$ | (16$f$) | Al |
| $\mathbf{B}_{18}$ | = | $\left(\frac{5}{8} + x_3\right)\,\mathbf{a}_1 + \left(\frac{7}{8} + x_3\right)\,\mathbf{a}_2 + \left(\frac{3}{4} + 2x_3\right)\,\mathbf{a}_3$ | = | $\left(\frac{1}{2} + x_3\right)a\,\hat{\mathbf{x}} + \left(\frac{1}{4} + x_3\right)a\,\hat{\mathbf{y}} + \frac{3}{8}c\,\hat{\mathbf{z}}$ | (16$f$) | Al |
| $\mathbf{B}_{19}$ | = | $\left(\frac{7}{8} - x_3\right)\,\mathbf{a}_1 + \left(\frac{5}{8} + x_3\right)\,\mathbf{a}_2 + \frac{1}{4}\,\mathbf{a}_3$ | = | $x_3 a\,\hat{\mathbf{x}} + \left(\frac{1}{4} - x_3\right)a\,\hat{\mathbf{y}} + \frac{5}{8}c\,\hat{\mathbf{z}}$ | (16$f$) | Al |
| $\mathbf{B}_{20}$ | = | $\left(\frac{7}{8} + x_3\right)\,\mathbf{a}_1 + \left(\frac{5}{8} - x_3\right)\,\mathbf{a}_2 + \frac{1}{4}\,\mathbf{a}_3$ | = | $-x_3 a\,\hat{\mathbf{x}} + \left(\frac{1}{4} + x_3\right)a\,\hat{\mathbf{y}} + \frac{5}{8}c\,\hat{\mathbf{z}}$ | (16$f$) | Al |
| $\mathbf{B}_{21}$ | = | $\left(\frac{3}{8} + x_4\right)\,\mathbf{a}_1 + \left(\frac{1}{8} + x_4\right)\,\mathbf{a}_2 + \left(\frac{1}{4} + 2x_4\right)\,\mathbf{a}_3$ | = | $x_4 a\,\hat{\mathbf{x}} + \left(\frac{1}{4} + x_4\right)a\,\hat{\mathbf{y}} + \frac{1}{8}c\,\hat{\mathbf{z}}$ | (16$f$) | $H_2O$ |
| $\mathbf{B}_{22}$ | = | $\left(\frac{3}{8} - x_4\right)\,\mathbf{a}_1 + \left(\frac{1}{8} - x_4\right)\,\mathbf{a}_2 + \left(\frac{1}{4} - 2x_4\right)\,\mathbf{a}_3$ | = | $-x_4 a\,\hat{\mathbf{x}} + \left(\frac{1}{4} - x_4\right)a\,\hat{\mathbf{y}} + \frac{1}{8}c\,\hat{\mathbf{z}}$ | (16$f$) | $H_2O$ |
| $\mathbf{B}_{23}$ | = | $\left(\frac{1}{8}+x_4\right)\mathbf{a}_1+\left(\frac{3}{8}-x_4\right)\mathbf{a}_2+\frac{3}{4}\mathbf{a}_3$ | = | $\left(\frac{1}{2}-x_4\right)a\,\hat{\mathbf{x}}+\left(\frac{1}{4}+x_4\right)a\,\hat{\mathbf{y}}-\frac{1}{8}c\,\hat{\mathbf{z}}$ | (16$f$) | $H_2O$ |
| $\mathbf{B}_{24}$ | = | $\left(\frac{1}{8}-x_4\right)\mathbf{a}_1+\left(\frac{3}{8}+x_4\right)\mathbf{a}_2+\frac{3}{4}\mathbf{a}_3$ | = | $\left(\frac{1}{2}+x_4\right)a\,\hat{\mathbf{x}}+\left(\frac{1}{4}-x_4\right)a\,\hat{\mathbf{y}}+\frac{7}{8}c\,\hat{\mathbf{z}}$ | (16$f$) | $H_2O$ |
| $\mathbf{B}_{25}$ | = | $\left(\frac{5}{8}-x_4\right)\mathbf{a}_1+\left(\frac{7}{8}-x_4\right)\mathbf{a}_2+\left(\frac{3}{4}-2x_4\right)\mathbf{a}_3$ | = | $\left(\frac{1}{2}-x_4\right)a\,\hat{\mathbf{x}}+\left(\frac{1}{4}-x_4\right)a\,\hat{\mathbf{y}}+\frac{3}{8}c\,\hat{\mathbf{z}}$ | (16$f$) | $H_2O$ |
| $\mathbf{B}_{26}$ | = | $\left(\frac{5}{8}+x_4\right)\mathbf{a}_1+\left(\frac{7}{8}+x_4\right)\mathbf{a}_2+\left(\frac{3}{4}+2x_4\right)\mathbf{a}_3$ | = | $\left(\frac{1}{2}+x_4\right)a\,\hat{\mathbf{x}}+\left(\frac{1}{4}+x_4\right)a\,\hat{\mathbf{y}}+\frac{3}{8}c\,\hat{\mathbf{z}}$ | (16$f$) | $H_2O$ |
| $\mathbf{B}_{27}$ | = | $\left(\frac{7}{8}-x_4\right)\mathbf{a}_1+\left(\frac{5}{8}+x_4\right)\mathbf{a}_2+\frac{1}{4}\mathbf{a}_3$ | = | $x_4a\,\hat{\mathbf{x}}+\left(\frac{1}{4}-x_4\right)a\,\hat{\mathbf{y}}+\frac{5}{8}c\,\hat{\mathbf{z}}$ | (16$f$) | $H_2O$ |
| $\mathbf{B}_{28}$ | = | $\left(\frac{7}{8}+x_4\right)\mathbf{a}_1+\left(\frac{5}{8}-x_4\right)\mathbf{a}_2+\frac{1}{4}\mathbf{a}_3$ | = | $-x_4a\,\hat{\mathbf{x}}+\left(\frac{1}{4}+x_4\right)a\,\hat{\mathbf{y}}+\frac{5}{8}c\,\hat{\mathbf{z}}$ | (16$f$) | $H_2O$ |
| $\mathbf{B}_{29}$ | = | $(y_5+z_5)\,\mathbf{a}_1+(x_5+z_5)\,\mathbf{a}_2+(x_5+y_5)\,\mathbf{a}_3$ | = | $x_5a\,\hat{\mathbf{x}}+y_5a\,\hat{\mathbf{y}}+z_5c\,\hat{\mathbf{z}}$ | (32$g$) | O I |
| $\mathbf{B}_{30}$ | = | $\left(\frac{1}{2}-y_5+z_5\right)\mathbf{a}_1+(-x_5+z_5)\,\mathbf{a}_2+\left(\frac{1}{2}-x_5-y_5\right)\mathbf{a}_3$ | = | $-x_5a\,\hat{\mathbf{x}}+\left(\frac{1}{2}-y_5\right)a\,\hat{\mathbf{y}}+z_5c\,\hat{\mathbf{z}}$ | (32$g$) | O I |
| $\mathbf{B}_{31}$ | = | $(x_5+z_5)\,\mathbf{a}_1+\left(\frac{1}{2}-y_5+z_5\right)\mathbf{a}_2+(x_5-y_5)\,\mathbf{a}_3$ | = | $\left(\frac{1}{4}-y_5\right)a\,\hat{\mathbf{x}}+\left(\frac{3}{4}+x_5\right)a\,\hat{\mathbf{y}}+\left(\frac{1}{4}+z_5\right)c\,\hat{\mathbf{z}}$ | (32$g$) | O I |
| $\mathbf{B}_{32}$ | = | $(-x_5+z_5)\,\mathbf{a}_1+(y_5+z_5)\,\mathbf{a}_2+\left(\frac{1}{2}-x_5+y_5\right)\mathbf{a}_3$ | = | $\left(\frac{1}{4}+y_5\right)a\,\hat{\mathbf{x}}+\left(\frac{1}{4}-x_5\right)a\,\hat{\mathbf{y}}+\left(\frac{3}{4}+z_5\right)c\,\hat{\mathbf{z}}$ | (32$g$) | O I |
| $\mathbf{B}_{33}$ | = | $(y_5-z_5)\,\mathbf{a}_1+\left(\frac{1}{2}-x_5-z_5\right)\mathbf{a}_2+\left(\frac{1}{2}-x_5+y_5\right)\mathbf{a}_3$ | = | $\left(\frac{1}{2}-x_5\right)a\,\hat{\mathbf{x}}+y_5a\,\hat{\mathbf{y}}-z_5c\,\hat{\mathbf{z}}$ | (32$g$) | O I |
| $\mathbf{B}_{34}$ | = | $\left(\frac{1}{2}-y_5-z_5\right)\mathbf{a}_1+\left(\frac{1}{2}+x_5-z_5\right)\mathbf{a}_2+(x_5-y_5)\,\mathbf{a}_3$ | = | $x_5a\,\hat{\mathbf{x}}-y_5a\,\hat{\mathbf{y}}+\left(\frac{1}{2}-z_5\right)c\,\hat{\mathbf{z}}$ | (32$g$) | O I |
| $\mathbf{B}_{35}$ | = | $\left(\frac{1}{2}+x_5-z_5\right)\mathbf{a}_1+(y_5-z_5)\,\mathbf{a}_2+(x_5+y_5)\,\mathbf{a}_3$ | = | $\left(-\frac{1}{4}+y_5\right)a\,\hat{\mathbf{x}}+\left(\frac{1}{4}+x_5\right)a\,\hat{\mathbf{y}}+\left(\frac{1}{4}-z_5\right)c\,\hat{\mathbf{z}}$ | (32$g$) | O I |
| $\mathbf{B}_{36}$ | = | $\left(\frac{1}{2}-x_5-z_5\right)\mathbf{a}_1+\left(\frac{1}{2}-y_5-z_5\right)\mathbf{a}_2+\left(\frac{1}{2}-x_5-y_5\right)\mathbf{a}_3$ | = | $\left(\frac{1}{4}-y_5\right)a\,\hat{\mathbf{x}}+\left(\frac{1}{4}-x_5\right)a\,\hat{\mathbf{y}}+\left(\frac{1}{4}-z_5\right)c\,\hat{\mathbf{z}}$ | (32$g$) | O I |
| $\mathbf{B}_{37}$ | = | $(-y_5-z_5)\,\mathbf{a}_1+(-x_5-z_5)\,\mathbf{a}_2+(-x_5-y_5)\,\mathbf{a}_3$ | = | $-x_5a\,\hat{\mathbf{x}}-y_5a\,\hat{\mathbf{y}}-z_5c\,\hat{\mathbf{z}}$ | (32$g$) | O I |
| $\mathbf{B}_{38}$ | = | $\left(\frac{1}{2}+y_5-z_5\right)\mathbf{a}_1+(x_5-z_5)\,\mathbf{a}_2+\left(\frac{1}{2}+x_5+y_5\right)\mathbf{a}_3$ | = | $x_5a\,\hat{\mathbf{x}}+\left(\frac{1}{2}+y_5\right)a\,\hat{\mathbf{y}}-z_5c\,\hat{\mathbf{z}}$ | (32$g$) | O I |
| $\mathbf{B}_{39}$ | = | $(-x_5-z_5)\,\mathbf{a}_1+\left(\frac{1}{2}+y_5-z_5\right)\mathbf{a}_2+(-x_5+y_5)\,\mathbf{a}_3$ | = | $\left(\frac{1}{4}+y_5\right)a\,\hat{\mathbf{x}}-a\left(x_5+\frac{1}{4}\right)\hat{\mathbf{y}}+\left(\frac{1}{4}-z_5\right)c\,\hat{\mathbf{z}}$ | (32$g$) | O I |
| $\mathbf{B}_{40}$ | = | $(x_5-z_5)\,\mathbf{a}_1+(-y_5-z_5)\,\mathbf{a}_2+\left(\frac{1}{2}+x_5-y_5\right)\mathbf{a}_3$ | = | $\left(\frac{1}{4}-y_5\right)a\,\hat{\mathbf{x}}+\left(\frac{1}{4}+x_5\right)a\,\hat{\mathbf{y}}-c\left(z_5+\frac{1}{4}\right)\hat{\mathbf{z}}$ | (32$g$) | O I |
| $\mathbf{B}_{41}$ | = | $(-y_5+z_5)\,\mathbf{a}_1+\left(\frac{1}{2}+x_5+z_5\right)\mathbf{a}_2+\left(\frac{1}{2}+x_5-y_5\right)\mathbf{a}_3$ | = | $\left(\frac{1}{2}+x_5\right)a\,\hat{\mathbf{x}}-y_5a\,\hat{\mathbf{y}}+z_5c\,\hat{\mathbf{z}}$ | (32$g$) | O I |
| $\mathbf{B}_{42}$ | = | $\left(\frac{1}{2}+y_5+z_5\right)\mathbf{a}_1+\left(\frac{1}{2}-x_5+z_5\right)\mathbf{a}_2+(-x_5+y_5)\,\mathbf{a}_3$ | = | $-x_5a\,\hat{\mathbf{x}}+y_5a\,\hat{\mathbf{y}}+\left(\frac{1}{2}+z_5\right)c\,\hat{\mathbf{z}}$ | (32$g$) | O I |
| $\mathbf{B}_{43}$ | = | $\left(\frac{1}{2}-x_5+z_5\right)\mathbf{a}_1+(-y_5+z_5)\,\mathbf{a}_2+(-x_5-y_5)\,\mathbf{a}_3$ | = | $\left(\frac{3}{4}-y_5\right)a\,\hat{\mathbf{x}}+\left(\frac{1}{4}-x_5\right)a\,\hat{\mathbf{y}}+\left(\frac{1}{4}+z_5\right)c\,\hat{\mathbf{z}}$ | (32$g$) | O I |
| $\mathbf{B}_{44}$ | = | $\left(\frac{1}{2}+x_5+z_5\right)\mathbf{a}_1+\left(\frac{1}{2}+y_5+z_5\right)\mathbf{a}_2+\left(\frac{1}{2}+x_5+y_5\right)\mathbf{a}_3$ | = | $\left(\frac{1}{4}+y_5\right)a\,\hat{\mathbf{x}}+\left(\frac{1}{4}+x_5\right)a\,\hat{\mathbf{y}}+\left(\frac{1}{4}+z_5\right)c\,\hat{\mathbf{z}}$ | (32$g$) | O I |

$$\mathbf{B}_{45} = (y_6+z_6)\,\mathbf{a}_1 + (x_6+z_6)\,\mathbf{a}_2 + (x_6+y_6)\,\mathbf{a}_3 = x_6a\,\hat{\mathbf{x}} + y_6a\,\hat{\mathbf{y}} + z_6c\,\hat{\mathbf{z}} \quad (32g) \quad \text{O II}$$

$$\mathbf{B}_{46} = \left(\frac{1}{2}-y_6+z_6\right)\mathbf{a}_1 + (-x_6+z_6)\,\mathbf{a}_2 + \left(\frac{1}{2}-x_6-y_6\right)\mathbf{a}_3 = -x_6a\,\hat{\mathbf{x}} + \left(\frac{1}{2}-y_6\right)a\,\hat{\mathbf{y}} + z_6c\,\hat{\mathbf{z}} \quad (32g) \quad \text{O II}$$

$$\mathbf{B}_{47} = (x_6+z_6)\,\mathbf{a}_1 + \left(\frac{1}{2}-y_6+z_6\right)\mathbf{a}_2 + (x_6-y_6)\,\mathbf{a}_3 = \left(\frac{1}{4}-y_6\right)a\,\hat{\mathbf{x}} + \left(\frac{3}{4}+x_6\right)a\,\hat{\mathbf{y}} + \left(\frac{1}{4}+z_6\right)c\,\hat{\mathbf{z}} \quad (32g) \quad \text{O II}$$

$$\mathbf{B}_{48} = (-x_6+z_6)\,\mathbf{a}_1 + (y_6+z_6)\,\mathbf{a}_2 + \left(\frac{1}{2}-x_6+y_6\right)\mathbf{a}_3 = \left(\frac{1}{4}+y_6\right)a\,\hat{\mathbf{x}} + \left(\frac{1}{4}-x_6\right)a\,\hat{\mathbf{y}} + \left(\frac{3}{4}+z_6\right)c\,\hat{\mathbf{z}} \quad (32g) \quad \text{O II}$$

$$\mathbf{B}_{49} = (y_6-z_6)\,\mathbf{a}_1 + \left(\frac{1}{2}-x_6-z_6\right)\mathbf{a}_2 + \left(\frac{1}{2}-x_6+y_6\right)\mathbf{a}_3 = \left(\frac{1}{2}-x_6\right)a\,\hat{\mathbf{x}} + y_6a\,\hat{\mathbf{y}} - z_6c\,\hat{\mathbf{z}} \quad (32g) \quad \text{O II}$$

$$\mathbf{B}_{50} = \left(\frac{1}{2}-y_6-z_6\right)\mathbf{a}_1 + \left(\frac{1}{2}+x_6-z_6\right)\mathbf{a}_2 + (x_6-y_6)\,\mathbf{a}_3 = x_6a\,\hat{\mathbf{x}} - y_6a\,\hat{\mathbf{y}} + \left(\frac{1}{2}-z_6\right)c\,\hat{\mathbf{z}} \quad (32g) \quad \text{O II}$$

$$\mathbf{B}_{51} = \left(\frac{1}{2}+x_6-z_6\right)\mathbf{a}_1 + (y_6-z_6)\,\mathbf{a}_2 + (x_6+y_6)\,\mathbf{a}_3 = \left(-\frac{1}{4}+y_6\right)a\,\hat{\mathbf{x}} + \left(\frac{1}{4}+x_6\right)a\,\hat{\mathbf{y}} + \left(\frac{1}{4}-z_6\right)c\,\hat{\mathbf{z}} \quad (32g) \quad \text{O II}$$

$$\mathbf{B}_{52} = \left(\frac{1}{2}-x_6-z_6\right)\mathbf{a}_1 + \left(\frac{1}{2}-y_6-z_6\right)\mathbf{a}_2 + \left(\frac{1}{2}-x_6-y_6\right)\mathbf{a}_3 = \left(\frac{1}{4}-y_6\right)a\,\hat{\mathbf{x}} + \left(\frac{1}{4}-x_6\right)a\,\hat{\mathbf{y}} + \left(\frac{1}{4}-z_6\right)c\,\hat{\mathbf{z}} \quad (32g) \quad \text{O II}$$

$$\mathbf{B}_{53} = (-y_6-z_6)\,\mathbf{a}_1 + (-x_6-z_6)\,\mathbf{a}_2 + (-x_6-y_6)\,\mathbf{a}_3 = -x_6a\,\hat{\mathbf{x}} - y_6a\,\hat{\mathbf{y}} - z_6c\,\hat{\mathbf{z}} \quad (32g) \quad \text{O II}$$

$$\mathbf{B}_{54} = \left(\frac{1}{2}+y_6-z_6\right)\mathbf{a}_1 + (x_6-z_6)\,\mathbf{a}_2 + \left(\frac{1}{2}+x_6+y_6\right)\mathbf{a}_3 = x_6a\,\hat{\mathbf{x}} + \left(\frac{1}{2}+y_6\right)a\,\hat{\mathbf{y}} - z_6c\,\hat{\mathbf{z}} \quad (32g) \quad \text{O II}$$

$$\mathbf{B}_{55} = (-x_6-z_6)\,\mathbf{a}_1 + \left(\frac{1}{2}+y_6-z_6\right)\mathbf{a}_2 + (-x_6+y_6)\,\mathbf{a}_3 = \left(\frac{1}{4}+y_6\right)a\,\hat{\mathbf{x}} - a\left(x_6+\frac{1}{4}\right)\hat{\mathbf{y}} + \left(\frac{1}{4}-z_6\right)c\,\hat{\mathbf{z}} \quad (32g) \quad \text{O II}$$

$$\mathbf{B}_{56} = (x_6-z_6)\,\mathbf{a}_1 + (-y_6-z_6)\,\mathbf{a}_2 + \left(\frac{1}{2}+x_6-y_6\right)\mathbf{a}_3 = \left(\frac{1}{4}-y_6\right)a\,\hat{\mathbf{x}} + \left(\frac{1}{4}+x_6\right)a\,\hat{\mathbf{y}} - c\left(z_6+\frac{1}{4}\right)\hat{\mathbf{z}} \quad (32g) \quad \text{O II}$$

$$\mathbf{B}_{57} = (-y_6+z_6)\,\mathbf{a}_1 + \left(\frac{1}{2}+x_6+z_6\right)\mathbf{a}_2 + \left(\frac{1}{2}+x_6-y_6\right)\mathbf{a}_3 = \left(\frac{1}{2}+x_6\right)a\,\hat{\mathbf{x}} - y_6a\,\hat{\mathbf{y}} + z_6c\,\hat{\mathbf{z}} \quad (32g) \quad \text{O II}$$

$$\mathbf{B}_{58} = \left(\frac{1}{2}+y_6+z_6\right)\mathbf{a}_1 + \left(\frac{1}{2}-x_6+z_6\right)\mathbf{a}_2 + (-x_6+y_6)\,\mathbf{a}_3 = -x_6a\,\hat{\mathbf{x}} + y_6a\,\hat{\mathbf{y}} + \left(\frac{1}{2}+z_6\right)c\,\hat{\mathbf{z}} \quad (32g) \quad \text{O II}$$

$$\mathbf{B}_{59} = \left(\frac{1}{2}-x_6+z_6\right)\mathbf{a}_1 + (-y_6+z_6)\,\mathbf{a}_2 + (-x_6-y_6)\,\mathbf{a}_3 = \left(\frac{3}{4}-y_6\right)a\,\hat{\mathbf{x}} + \left(\frac{1}{4}-x_6\right)a\,\hat{\mathbf{y}} + \left(\frac{1}{4}+z_6\right)c\,\hat{\mathbf{z}} \quad (32g) \quad \text{O II}$$

$$\mathbf{B}_{60} = \left(\frac{1}{2}+x_6+z_6\right)\mathbf{a}_1 + \left(\frac{1}{2}+y_6+z_6\right)\mathbf{a}_2 + \left(\frac{1}{2}+x_6+y_6\right)\mathbf{a}_3 = \left(\frac{1}{4}+y_6\right)a\,\hat{\mathbf{x}} + \left(\frac{1}{4}+x_6\right)a\,\hat{\mathbf{y}} + \left(\frac{1}{4}+z_6\right)c\,\hat{\mathbf{z}} \quad (32g) \quad \text{O II}$$

$$\mathbf{B}_{61} = (y_7+z_7)\,\mathbf{a}_1 + (x_7+z_7)\,\mathbf{a}_2 + (x_7+y_7)\,\mathbf{a}_3 = x_7a\,\hat{\mathbf{x}} + y_7a\,\hat{\mathbf{y}} + z_7c\,\hat{\mathbf{z}} \quad (32g) \quad \text{O III}$$

$$\mathbf{B}_{62} = \left(\frac{1}{2}-y_7+z_7\right)\mathbf{a}_1 + (-x_7+z_7)\,\mathbf{a}_2 + \left(\frac{1}{2}-x_7-y_7\right)\mathbf{a}_3 = -x_7a\,\hat{\mathbf{x}} + \left(\frac{1}{2}-y_7\right)a\,\hat{\mathbf{y}} + z_7c\,\hat{\mathbf{z}} \quad (32g) \quad \text{O III}$$

$$\mathbf{B}_{63} = (x_7+z_7)\,\mathbf{a}_1 + \left(\frac{1}{2}-y_7+z_7\right)\mathbf{a}_2 + (x_7-y_7)\,\mathbf{a}_3 = \left(\frac{1}{4}-y_7\right)a\,\hat{\mathbf{x}} + \left(\frac{3}{4}+x_7\right)a\,\hat{\mathbf{y}} + \left(\frac{1}{4}+z_7\right)c\,\hat{\mathbf{z}} \quad (32g) \quad \text{O III}$$

$$\mathbf{B}_{64} = (-x_7+z_7)\,\mathbf{a}_1 + (y_7+z_7)\,\mathbf{a}_2 + \left(\frac{1}{2}-x_7+y_7\right)\mathbf{a}_3 = \left(\frac{1}{4}+y_7\right)a\,\hat{\mathbf{x}} + \left(\frac{1}{4}-x_7\right)a\,\hat{\mathbf{y}} + \left(\frac{3}{4}+z_7\right)c\,\hat{\mathbf{z}} \quad (32g) \quad \text{O III}$$

$$\mathbf{B}_{65} = (y_7-z_7)\,\mathbf{a}_1 + \left(\frac{1}{2}-x_7-z_7\right)\mathbf{a}_2 + \left(\frac{1}{2}-x_7+y_7\right)\mathbf{a}_3 = \left(\frac{1}{2}-x_7\right)a\,\hat{\mathbf{x}} + y_7a\,\hat{\mathbf{y}} - z_7c\,\hat{\mathbf{z}} \quad (32g) \quad \text{O III}$$

| | | | | | | |
|---|---|---|---|---|---|---|
| $\mathbf{B}_{66}$ | = | $\left(\frac{1}{2} - y_7 - z_7\right) \mathbf{a}_1 + \left(\frac{1}{2} + x_7 - z_7\right) \mathbf{a}_2 + (x_7 - y_7) \mathbf{a}_3$ | = | $x_7 a \hat{\mathbf{x}} - y_7 a \hat{\mathbf{y}} + \left(\frac{1}{2} - z_7\right) c \hat{\mathbf{z}}$ | (32$g$) | O III |
| $\mathbf{B}_{67}$ | = | $\left(\frac{1}{2} + x_7 - z_7\right) \mathbf{a}_1 + (y_7 - z_7) \mathbf{a}_2 + (x_7 + y_7) \mathbf{a}_3$ | = | $\left(-\frac{1}{4} + y_7\right) a \hat{\mathbf{x}} + \left(\frac{1}{4} + x_7\right) a \hat{\mathbf{y}} + \left(\frac{1}{4} - z_7\right) c \hat{\mathbf{z}}$ | (32$g$) | O III |
| $\mathbf{B}_{68}$ | = | $\left(\frac{1}{2} - x_7 - z_7\right) \mathbf{a}_1 + \left(\frac{1}{2} - y_7 - z_7\right) \mathbf{a}_2 + \left(\frac{1}{2} - x_7 - y_7\right) \mathbf{a}_3$ | = | $\left(\frac{1}{4} - y_7\right) a \hat{\mathbf{x}} + \left(\frac{1}{4} - x_7\right) a \hat{\mathbf{y}} + \left(\frac{1}{4} - z_7\right) c \hat{\mathbf{z}}$ | (32$g$) | O III |
| $\mathbf{B}_{69}$ | = | $(-y_7 - z_7) \mathbf{a}_1 + (-x_7 - z_7) \mathbf{a}_2 + (-x_7 - y_7) \mathbf{a}_3$ | = | $-x_7 a \hat{\mathbf{x}} - y_7 a \hat{\mathbf{y}} - z_7 c \hat{\mathbf{z}}$ | (32$g$) | O III |
| $\mathbf{B}_{70}$ | = | $\left(\frac{1}{2} + y_7 - z_7\right) \mathbf{a}_1 + (x_7 - z_7) \mathbf{a}_2 + \left(\frac{1}{2} + x_7 + y_7\right) \mathbf{a}_3$ | = | $x_7 a \hat{\mathbf{x}} + \left(\frac{1}{2} + y_7\right) a \hat{\mathbf{y}} - z_7 c \hat{\mathbf{z}}$ | (32$g$) | O III |
| $\mathbf{B}_{71}$ | = | $(-x_7 - z_7) \mathbf{a}_1 + \left(\frac{1}{2} + y_7 - z_7\right) \mathbf{a}_2 + (-x_7 + y_7) \mathbf{a}_3$ | = | $\left(\frac{1}{4} + y_7\right) a \hat{\mathbf{x}} - a \left(x_7 + \frac{1}{4}\right) \hat{\mathbf{y}} + \left(\frac{1}{4} - z_7\right) c \hat{\mathbf{z}}$ | (32$g$) | O III |
| $\mathbf{B}_{72}$ | = | $(x_7 - z_7) \mathbf{a}_1 + (-y_7 - z_7) \mathbf{a}_2 + \left(\frac{1}{2} + x_7 - y_7\right) \mathbf{a}_3$ | = | $\left(\frac{1}{4} - y_7\right) a \hat{\mathbf{x}} + \left(\frac{1}{4} + x_7\right) a \hat{\mathbf{y}} - c \left(z_7 + \frac{1}{4}\right) \hat{\mathbf{z}}$ | (32$g$) | O III |
| $\mathbf{B}_{73}$ | = | $(-y_7 + z_7) \mathbf{a}_1 + \left(\frac{1}{2} + x_7 + z_7\right) \mathbf{a}_2 + \left(\frac{1}{2} + x_7 - y_7\right) \mathbf{a}_3$ | = | $\left(\frac{1}{2} + x_7\right) a \hat{\mathbf{x}} - y_7 a \hat{\mathbf{y}} + z_7 c \hat{\mathbf{z}}$ | (32$g$) | O III |
| $\mathbf{B}_{74}$ | = | $\left(\frac{1}{2} + y_7 + z_7\right) \mathbf{a}_1 + \left(\frac{1}{2} - x_7 + z_7\right) \mathbf{a}_2 + (-x_7 + y_7) \mathbf{a}_3$ | = | $-x_7 a \hat{\mathbf{x}} + y_7 a \hat{\mathbf{y}} + \left(\frac{1}{2} + z_7\right) c \hat{\mathbf{z}}$ | (32$g$) | O III |
| $\mathbf{B}_{75}$ | = | $\left(\frac{1}{2} - x_7 + z_7\right) \mathbf{a}_1 + (-y_7 + z_7) \mathbf{a}_2 + (-x_7 - y_7) \mathbf{a}_3$ | = | $\left(\frac{3}{4} - y_7\right) a \hat{\mathbf{x}} + \left(\frac{1}{4} - x_7\right) a \hat{\mathbf{y}} + \left(\frac{1}{4} + z_7\right) c \hat{\mathbf{z}}$ | (32$g$) | O III |
| $\mathbf{B}_{76}$ | = | $\left(\frac{1}{2} + x_7 + z_7\right) \mathbf{a}_1 + \left(\frac{1}{2} + y_7 + z_7\right) \mathbf{a}_2 + \left(\frac{1}{2} + x_7 + y_7\right) \mathbf{a}_3$ | = | $\left(\frac{1}{4} + y_7\right) a \hat{\mathbf{x}} + \left(\frac{1}{4} + x_7\right) a \hat{\mathbf{y}} + \left(\frac{1}{4} + z_7\right) c \hat{\mathbf{z}}$ | (32$g$) | O III |
| $\mathbf{B}_{77}$ | = | $(y_8 + z_8) \mathbf{a}_1 + (x_8 + z_8) \mathbf{a}_2 + (x_8 + y_8) \mathbf{a}_3$ | = | $x_8 a \hat{\mathbf{x}} + y_8 a \hat{\mathbf{y}} + z_8 c \hat{\mathbf{z}}$ | (32$g$) | Si |
| $\mathbf{B}_{78}$ | = | $\left(\frac{1}{2} - y_8 + z_8\right) \mathbf{a}_1 + (-x_8 + z_8) \mathbf{a}_2 + \left(\frac{1}{2} - x_8 - y_8\right) \mathbf{a}_3$ | = | $-x_8 a \hat{\mathbf{x}} + \left(\frac{1}{2} - y_8\right) a \hat{\mathbf{y}} + z_8 c \hat{\mathbf{z}}$ | (32$g$) | Si |
| $\mathbf{B}_{79}$ | = | $(x_8 + z_8) \mathbf{a}_1 + \left(\frac{1}{2} - y_8 + z_8\right) \mathbf{a}_2 + (x_8 - y_8) \mathbf{a}_3$ | = | $\left(\frac{1}{4} - y_8\right) a \hat{\mathbf{x}} + \left(\frac{3}{4} + x_8\right) a \hat{\mathbf{y}} + \left(\frac{1}{4} + z_8\right) c \hat{\mathbf{z}}$ | (32$g$) | Si |
| $\mathbf{B}_{80}$ | = | $(-x_8 + z_8) \mathbf{a}_1 + (y_8 + z_8) \mathbf{a}_2 + \left(\frac{1}{2} - x_8 + y_8\right) \mathbf{a}_3$ | = | $\left(\frac{1}{4} + y_8\right) a \hat{\mathbf{x}} + \left(\frac{1}{4} - x_8\right) a \hat{\mathbf{y}} + \left(\frac{3}{4} + z_8\right) c \hat{\mathbf{z}}$ | (32$g$) | Si |
| $\mathbf{B}_{81}$ | = | $(y_8 - z_8) \mathbf{a}_1 + \left(\frac{1}{2} - x_8 - z_8\right) \mathbf{a}_2 + \left(\frac{1}{2} - x_8 + y_8\right) \mathbf{a}_3$ | = | $\left(\frac{1}{2} - x_8\right) a \hat{\mathbf{x}} + y_8 a \hat{\mathbf{y}} - z_8 c \hat{\mathbf{z}}$ | (32$g$) | Si |
| $\mathbf{B}_{82}$ | = | $\left(\frac{1}{2} - y_8 - z_8\right) \mathbf{a}_1 + \left(\frac{1}{2} + x_8 - z_8\right) \mathbf{a}_2 + (x_8 - y_8) \mathbf{a}_3$ | = | $x_8 a \hat{\mathbf{x}} - y_8 a \hat{\mathbf{y}} + \left(\frac{1}{2} - z_8\right) c \hat{\mathbf{z}}$ | (32$g$) | Si |
| $\mathbf{B}_{83}$ | = | $\left(\frac{1}{2} + x_8 - z_8\right) \mathbf{a}_1 + (y_8 - z_8) \mathbf{a}_2 + (x_8 + y_8) \mathbf{a}_3$ | = | $\left(-\frac{1}{4} + y_8\right) a \hat{\mathbf{x}} + \left(\frac{1}{4} + x_8\right) a \hat{\mathbf{y}} + \left(\frac{1}{4} - z_8\right) c \hat{\mathbf{z}}$ | (32$g$) | Si |
| $\mathbf{B}_{84}$ | = | $\left(\frac{1}{2} - x_8 - z_8\right) \mathbf{a}_1 + \left(\frac{1}{2} - y_8 - z_8\right) \mathbf{a}_2 + \left(\frac{1}{2} - x_8 - y_8\right) \mathbf{a}_3$ | = | $\left(\frac{1}{4} - y_8\right) a \hat{\mathbf{x}} + \left(\frac{1}{4} - x_8\right) a \hat{\mathbf{y}} + \left(\frac{1}{4} - z_8\right) c \hat{\mathbf{z}}$ | (32$g$) | Si |
| $\mathbf{B}_{85}$ | = | $(-y_8 - z_8) \mathbf{a}_1 + (-x_8 - z_8) \mathbf{a}_2 + (-x_8 - y_8) \mathbf{a}_3$ | = | $-x_8 a \hat{\mathbf{x}} - y_8 a \hat{\mathbf{y}} - z_8 c \hat{\mathbf{z}}$ | (32$g$) | Si |
| $\mathbf{B}_{86}$ | = | $\left(\frac{1}{2} + y_8 - z_8\right) \mathbf{a}_1 + (x_8 - z_8) \mathbf{a}_2 + \left(\frac{1}{2} + x_8 + y_8\right) \mathbf{a}_3$ | = | $x_8 a \hat{\mathbf{x}} + \left(\frac{1}{2} + y_8\right) a \hat{\mathbf{y}} - z_8 c \hat{\mathbf{z}}$ | (32$g$) | Si |
| $\mathbf{B}_{87}$ | = | $(-x_8 - z_8)\,\mathbf{a}_1 + \left(\frac{1}{2} + y_8 - z_8\right)\mathbf{a}_2 + (-x_8 + y_8)\,\mathbf{a}_3$ | = | $\left(\frac{1}{4} + y_8\right) a\,\hat{\mathbf{x}} - a\left(x_8 + \frac{1}{4}\right)\hat{\mathbf{y}} + \left(\frac{1}{4} - z_8\right) c\,\hat{\mathbf{z}}$ | (32$g$) | Si |
| $\mathbf{B}_{88}$ | = | $(x_8 - z_8)\,\mathbf{a}_1 + (-y_8 - z_8)\,\mathbf{a}_2 + \left(\frac{1}{2} + x_8 - y_8\right)\mathbf{a}_3$ | = | $\left(\frac{1}{4} - y_8\right) a\,\hat{\mathbf{x}} + \left(\frac{1}{4} + x_8\right) a\,\hat{\mathbf{y}} - c\left(z_8 + \frac{1}{4}\right)\hat{\mathbf{z}}$ | (32$g$) | Si |
| $\mathbf{B}_{89}$ | = | $(-y_8 + z_8)\,\mathbf{a}_1 + \left(\frac{1}{2} + x_8 + z_8\right)\mathbf{a}_2 + \left(\frac{1}{2} + x_8 - y_8\right)\mathbf{a}_3$ | = | $\left(\frac{1}{2} + x_8\right) a\,\hat{\mathbf{x}} - y_8 a\,\hat{\mathbf{y}} + z_8 c\,\hat{\mathbf{z}}$ | (32$g$) | Si |
| $\mathbf{B}_{90}$ | = | $\left(\frac{1}{2} + y_8 + z_8\right)\mathbf{a}_1 + \left(\frac{1}{2} - x_8 + z_8\right)\mathbf{a}_2 + (-x_8 + y_8)\,\mathbf{a}_3$ | = | $-x_8 a\,\hat{\mathbf{x}} + y_8 a\,\hat{\mathbf{y}} + \left(\frac{1}{2} + z_8\right) c\,\hat{\mathbf{z}}$ | (32$g$) | Si |
| $\mathbf{B}_{91}$ | = | $\left(\frac{1}{2} - x_8 + z_8\right)\mathbf{a}_1 + (-y_8 + z_8)\,\mathbf{a}_2 + (-x_8 - y_8)\,\mathbf{a}_3$ | = | $\left(\frac{3}{4} - y_8\right) a\,\hat{\mathbf{x}} + \left(\frac{1}{4} - x_8\right) a\,\hat{\mathbf{y}} + \left(\frac{1}{4} + z_8\right) c\,\hat{\mathbf{z}}$ | (32$g$) | Si |
| $\mathbf{B}_{92}$ | = | $\left(\frac{1}{2} + x_8 + z_8\right)\mathbf{a}_1 + \left(\frac{1}{2} + y_8 + z_8\right)\mathbf{a}_2 + \left(\frac{1}{2} + x_8 + y_8\right)\mathbf{a}_3$ | = | $\left(\frac{1}{4} + y_8\right) a\,\hat{\mathbf{x}} + \left(\frac{1}{4} + x_8\right) a\,\hat{\mathbf{y}} + \left(\frac{1}{4} + z_8\right) c\,\hat{\mathbf{z}}$ | (32$g$) | Si |

---

**Found in:**

---

**Geometry files:**

- CIF: pp. 1718
- POSCAR: pp. 1718

# $Cd_3As_2$ Structure: A2B3_tI160_142_deg_3g

http://aflow.org/prototype-encyclopedia/A2B3_tI160_142_deg_3g

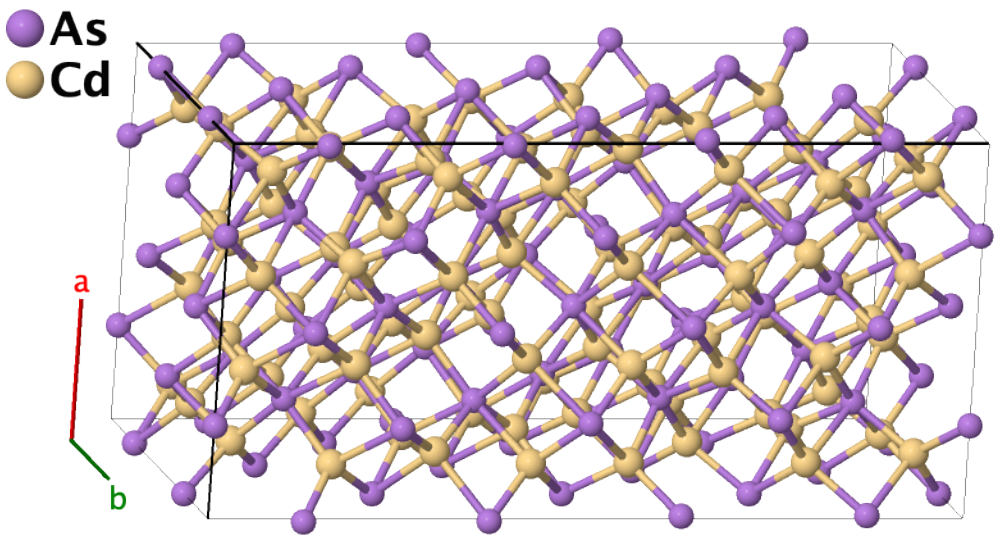

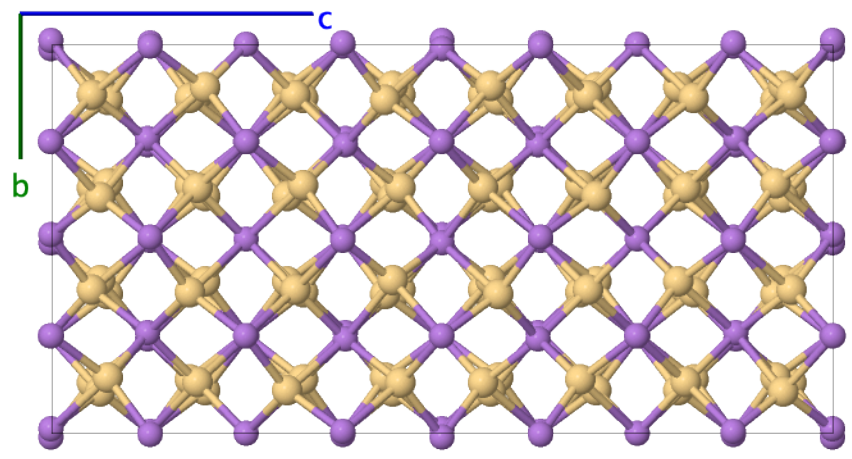

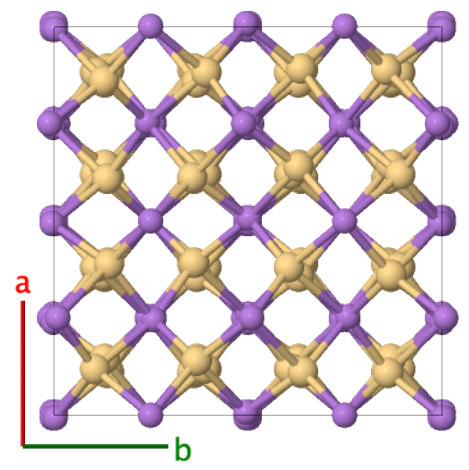

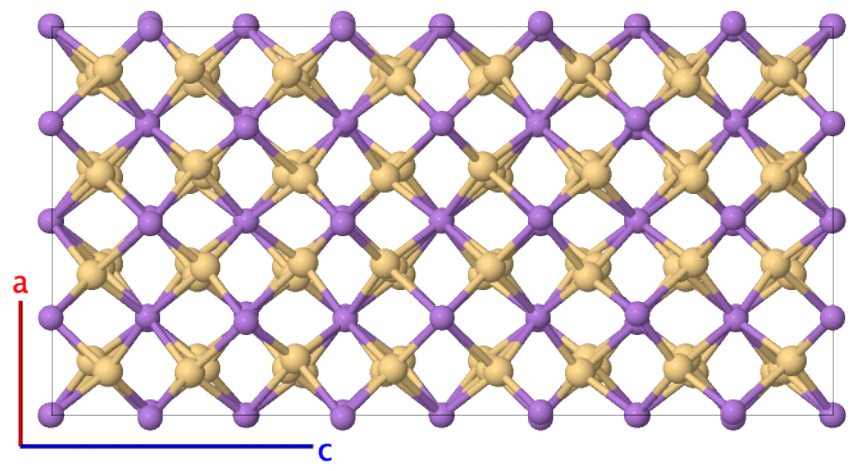

| | | |
|---|---|---|
| **Prototype** | : | $As_2Cd_3$ |
| **AFLOW prototype label** | : | A2B3_tI160_142_deg_3g |
| ***Strukturbericht* designation** | : | None |
| **Pearson symbol** | : | tI160 |
| **Space group number** | : | 142 |
| **Space group symbol** | : | $I4_1/acd$ |
| **AFLOW prototype command** | : | `aflow --proto=A2B3_tI160_142_deg_3g` `--params=`$a, c/a, z_1, x_2, y_2, x_3, y_3, z_3, x_4, y_4, z_4, x_5, y_5, z_5, x_6, y_6, z_6$ |

---

**Body-centered Tetragonal primitive vectors:**

$$\mathbf{a}_1 = -\frac{1}{2}\,a\,\hat{\mathbf{x}} + \frac{1}{2}\,a\,\hat{\mathbf{y}} + \frac{1}{2}\,c\,\hat{\mathbf{z}}$$
$$\mathbf{a}_2 = \frac{1}{2}\,a\,\hat{\mathbf{x}} - \frac{1}{2}\,a\,\hat{\mathbf{y}} + \frac{1}{2}\,c\,\hat{\mathbf{z}}$$
$$\mathbf{a}_3 = \frac{1}{2}\,a\,\hat{\mathbf{x}} + \frac{1}{2}\,a\,\hat{\mathbf{y}} - \frac{1}{2}\,c\,\hat{\mathbf{z}}$$

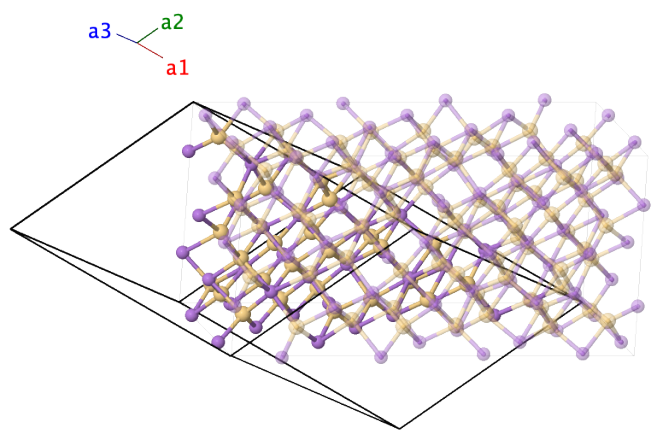

---

**Basis vectors:**

| | | Lattice Coordinates | | Cartesian Coordinates | Wyckoff Position | Atom Type |
|---|---|---|---|---|---|---|
| $\mathbf{B}_1$ | = | $\left(\frac{1}{4}+z_1\right)\mathbf{a}_1 + z_1\,\mathbf{a}_2 + \frac{1}{4}\,\mathbf{a}_3$ | = | $\frac{1}{4}a\,\hat{\mathbf{y}} + z_1 c\,\hat{\mathbf{z}}$ | (16$d$) | As I |
| $\mathbf{B}_2$ | = | $z_1\,\mathbf{a}_1 + \left(\frac{1}{4}+z_1\right)\mathbf{a}_2 + \frac{3}{4}\,\mathbf{a}_3$ | = | $\frac{1}{2}a\,\hat{\mathbf{x}} + \frac{1}{4}a\,\hat{\mathbf{y}} + \left(-\frac{1}{4}+z_1\right)c\,\hat{\mathbf{z}}$ | (16$d$) | As I |
| $\mathbf{B}_3$ | = | $\left(\frac{1}{4}-z_1\right)\mathbf{a}_1 + \left(\frac{1}{2}-z_1\right)\mathbf{a}_2 + \frac{3}{4}\,\mathbf{a}_3$ | = | $\frac{1}{2}a\,\hat{\mathbf{x}} + \frac{1}{4}a\,\hat{\mathbf{y}} - z_1 c\,\hat{\mathbf{z}}$ | (16$d$) | As I |
| $\mathbf{B}_4$ | = | $\left(\frac{1}{2}-z_1\right)\mathbf{a}_1 + \left(\frac{1}{4}-z_1\right)\mathbf{a}_2 + \frac{1}{4}\,\mathbf{a}_3$ | = | $\frac{1}{4}a\,\hat{\mathbf{y}} + \left(\frac{1}{4}-z_1\right)c\,\hat{\mathbf{z}}$ | (16$d$) | As I |
| $\mathbf{B}_5$ | = | $\left(\frac{3}{4}-z_1\right)\mathbf{a}_1 - z_1\,\mathbf{a}_2 + \frac{3}{4}\,\mathbf{a}_3$ | = | $\frac{3}{4}a\,\hat{\mathbf{y}} - z_1 c\,\hat{\mathbf{z}}$ | (16$d$) | As I |
| $\mathbf{B}_6$ | = | $-z_1\,\mathbf{a}_1 + \left(\frac{3}{4}-z_1\right)\mathbf{a}_2 + \frac{1}{4}\,\mathbf{a}_3$ | = | $\frac{1}{2}a\,\hat{\mathbf{x}} - \frac{1}{4}a\,\hat{\mathbf{y}} + \left(\frac{1}{4}-z_1\right)c\,\hat{\mathbf{z}}$ | (16$d$) | As I |
| $\mathbf{B}_7$ | = | $\left(\frac{3}{4}+z_1\right)\mathbf{a}_1 + \left(\frac{1}{2}+z_1\right)\mathbf{a}_2 + \frac{1}{4}\,\mathbf{a}_3$ | = | $\frac{1}{4}a\,\hat{\mathbf{y}} + \left(\frac{1}{2}+z_1\right)c\,\hat{\mathbf{z}}$ | (16$d$) | As I |
| $\mathbf{B}_8$ | = | $\left(\frac{1}{2}+z_1\right)\mathbf{a}_1+\left(\frac{3}{4}+z_1\right)\mathbf{a}_2+\frac{3}{4}\mathbf{a}_3$ | = | $\frac{1}{2}a\,\hat{\mathbf{x}}+\frac{1}{4}a\,\hat{\mathbf{y}}+\left(\frac{1}{4}+z_1\right)c\,\hat{\mathbf{z}}$ | (16*d*) | As I |
| $\mathbf{B}_9$ | = | $\frac{1}{4}\mathbf{a}_1+\left(\frac{1}{4}+x_2\right)\mathbf{a}_2+x_2\,\mathbf{a}_3$ | = | $x_2a\,\hat{\mathbf{x}}+\frac{1}{4}c\,\hat{\mathbf{z}}$ | (16*e*) | As II |
| $\mathbf{B}_{10}$ | = | $\frac{3}{4}\mathbf{a}_1+\left(\frac{1}{4}-x_2\right)\mathbf{a}_2+\left(\frac{1}{2}-x_2\right)\mathbf{a}_3$ | = | $-x_2a\,\hat{\mathbf{x}}+\frac{1}{2}a\,\hat{\mathbf{y}}+\frac{1}{4}c\,\hat{\mathbf{z}}$ | (16*e*) | As II |
| $\mathbf{B}_{11}$ | = | $\left(\frac{1}{4}+x_2\right)\mathbf{a}_1+\frac{3}{4}\mathbf{a}_2+x_2\,\mathbf{a}_3$ | = | $\frac{1}{4}a\,\hat{\mathbf{x}}+\left(\frac{3}{4}+x_2\right)a\,\hat{\mathbf{y}}+\frac{1}{2}c\,\hat{\mathbf{z}}$ | (16*e*) | As II |
| $\mathbf{B}_{12}$ | = | $\left(\frac{1}{4}-x_2\right)\mathbf{a}_1+\frac{1}{4}\mathbf{a}_2+\left(\frac{1}{2}-x_2\right)\mathbf{a}_3$ | = | $\frac{1}{4}a\,\hat{\mathbf{x}}+\left(\frac{1}{4}-x_2\right)a\,\hat{\mathbf{y}}$ | (16*e*) | As II |
| $\mathbf{B}_{13}$ | = | $\frac{3}{4}\mathbf{a}_1+\left(\frac{3}{4}-x_2\right)\mathbf{a}_2-x_2\,\mathbf{a}_3$ | = | $-x_2a\,\hat{\mathbf{x}}+\frac{3}{4}c\,\hat{\mathbf{z}}$ | (16*e*) | As II |
| $\mathbf{B}_{14}$ | = | $\frac{1}{4}\mathbf{a}_1+\left(\frac{3}{4}+x_2\right)\mathbf{a}_2+\left(\frac{1}{2}+x_2\right)\mathbf{a}_3$ | = | $\left(\frac{1}{2}+x_2\right)a\,\hat{\mathbf{x}}+\frac{1}{4}c\,\hat{\mathbf{z}}$ | (16*e*) | As II |
| $\mathbf{B}_{15}$ | = | $\left(\frac{1}{4}-x_2+y_2\right)\mathbf{a}_1+\left(\frac{3}{4}+y_2\right)\mathbf{a}_2-x_2\,\mathbf{a}_3$ | = | $\frac{1}{4}a\,\hat{\mathbf{x}}-a\left(x_2+\frac{1}{4}\right)\hat{\mathbf{y}}+\left(\frac{1}{2}+y_2\right)c\,\hat{\mathbf{z}}$ | (16*e*) | As II |
| $\mathbf{B}_{16}$ | = | $\left(\frac{3}{4}+x_2\right)\mathbf{a}_1+\frac{3}{4}\mathbf{a}_2+\left(\frac{1}{2}+x_2\right)\mathbf{a}_3$ | = | $\frac{1}{4}a\,\hat{\mathbf{x}}+\left(\frac{1}{4}+x_2\right)a\,\hat{\mathbf{y}}+\frac{1}{2}c\,\hat{\mathbf{z}}$ | (16*e*) | As II |
| $\mathbf{B}_{17}$ | = | $(y_3+z_3)\,\mathbf{a}_1+(x_3+z_3)\,\mathbf{a}_2+(x_3+y_3)\,\mathbf{a}_3$ | = | $x_3a\,\hat{\mathbf{x}}+y_3a\,\hat{\mathbf{y}}+z_3c\,\hat{\mathbf{z}}$ | (32*g*) | As III |
| $\mathbf{B}_{18}$ | = | $\left(\frac{1}{2}-y_3+z_3\right)\mathbf{a}_1+(-x_3+z_3)\,\mathbf{a}_2+\left(\frac{1}{2}-x_3-y_3\right)\mathbf{a}_3$ | = | $-x_3a\,\hat{\mathbf{x}}+\left(\frac{1}{2}-y_3\right)a\,\hat{\mathbf{y}}+z_3c\,\hat{\mathbf{z}}$ | (32*g*) | As III |
| $\mathbf{B}_{19}$ | = | $(x_3+z_3)\,\mathbf{a}_1+\left(\frac{1}{2}-y_3+z_3\right)\mathbf{a}_2+(x_3-y_3)\,\mathbf{a}_3$ | = | $\left(\frac{1}{4}-y_3\right)a\,\hat{\mathbf{x}}+\left(\frac{3}{4}+x_3\right)a\,\hat{\mathbf{y}}+\left(\frac{1}{4}+z_3\right)c\,\hat{\mathbf{z}}$ | (32*g*) | As III |
| $\mathbf{B}_{20}$ | = | $(-x_3+z_3)\,\mathbf{a}_1+(y_3+z_3)\,\mathbf{a}_2+\left(\frac{1}{2}-x_3+y_3\right)\mathbf{a}_3$ | = | $\left(\frac{1}{4}+y_3\right)a\,\hat{\mathbf{x}}+\left(\frac{1}{4}-x_3\right)a\,\hat{\mathbf{y}}+\left(\frac{3}{4}+z_3\right)c\,\hat{\mathbf{z}}$ | (32*g*) | As III |
| $\mathbf{B}_{21}$ | = | $(y_3-z_3)\,\mathbf{a}_1+\left(\frac{1}{2}-x_3-z_3\right)\mathbf{a}_2+\left(\frac{1}{2}-x_3+y_3\right)\mathbf{a}_3$ | = | $\left(\frac{1}{2}-x_3\right)a\,\hat{\mathbf{x}}+y_3a\,\hat{\mathbf{y}}-z_3c\,\hat{\mathbf{z}}$ | (32*g*) | As III |
| $\mathbf{B}_{22}$ | = | $\left(\frac{1}{2}-y_3-z_3\right)\mathbf{a}_1+\left(\frac{1}{2}+x_3-z_3\right)\mathbf{a}_2+(x_3-y_3)\,\mathbf{a}_3$ | = | $x_3a\,\hat{\mathbf{x}}-y_3a\,\hat{\mathbf{y}}+\left(\frac{1}{2}-z_3\right)c\,\hat{\mathbf{z}}$ | (32*g*) | As III |
| $\mathbf{B}_{23}$ | = | $\left(\frac{1}{2}+x_3-z_3\right)\mathbf{a}_1+(y_3-z_3)\,\mathbf{a}_2+(x_3+y_3)\,\mathbf{a}_3$ | = | $\left(-\frac{1}{4}+y_3\right)a\,\hat{\mathbf{x}}+\left(\frac{1}{4}+x_3\right)a\,\hat{\mathbf{y}}+\left(\frac{1}{4}-z_3\right)c\,\hat{\mathbf{z}}$ | (32*g*) | As III |
| $\mathbf{B}_{24}$ | = | $\left(\frac{1}{2}-x_3-z_3\right)\mathbf{a}_1+\left(\frac{1}{2}-y_3-z_3\right)\mathbf{a}_2+\left(\frac{1}{2}-x_3-y_3\right)\mathbf{a}_3$ | = | $\left(\frac{1}{4}-y_3\right)a\,\hat{\mathbf{x}}+\left(\frac{1}{4}-x_3\right)a\,\hat{\mathbf{y}}+\left(\frac{1}{4}-z_3\right)c\,\hat{\mathbf{z}}$ | (32*g*) | As III |
| $\mathbf{B}_{25}$ | = | $(-y_3-z_3)\,\mathbf{a}_1+(-x_3-z_3)\,\mathbf{a}_2+(-x_3-y_3)\,\mathbf{a}_3$ | = | $-x_3a\,\hat{\mathbf{x}}-y_3a\,\hat{\mathbf{y}}-z_3c\,\hat{\mathbf{z}}$ | (32*g*) | As III |
| $\mathbf{B}_{26}$ | = | $\left(\frac{1}{2}+y_3-z_3\right)\mathbf{a}_1+(x_3-z_3)\,\mathbf{a}_2+\left(\frac{1}{2}+x_3+y_3\right)\mathbf{a}_3$ | = | $x_3a\,\hat{\mathbf{x}}+\left(\frac{1}{2}+y_3\right)a\,\hat{\mathbf{y}}-z_3c\,\hat{\mathbf{z}}$ | (32*g*) | As III |
| $\mathbf{B}_{27}$ | = | $(-x_3-z_3)\,\mathbf{a}_1+\left(\frac{1}{2}+y_3-z_3\right)\mathbf{a}_2+(-x_3+y_3)\,\mathbf{a}_3$ | = | $\left(\frac{1}{4}+y_3\right)a\,\hat{\mathbf{x}}-a\left(x_3+\frac{1}{4}\right)\hat{\mathbf{y}}+\left(\frac{1}{4}-z_3\right)c\,\hat{\mathbf{z}}$ | (32*g*) | As III |
| $\mathbf{B}_{28}$ | = | $(x_3-z_3)\,\mathbf{a}_1+(-y_3-z_3)\,\mathbf{a}_2+\left(\frac{1}{2}+x_3-y_3\right)\mathbf{a}_3$ | = | $\left(\frac{1}{4}-y_3\right)a\,\hat{\mathbf{x}}+\left(\frac{1}{4}+x_3\right)a\,\hat{\mathbf{y}}-c\left(z_3+\frac{1}{4}\right)\hat{\mathbf{z}}$ | (32*g*) | As III |
| $\mathbf{B}_{29}$ | = | $(-y_3+z_3)\,\mathbf{a}_1+\left(\frac{1}{2}+x_3+z_3\right)\mathbf{a}_2+\left(\frac{1}{2}+x_3-y_3\right)\mathbf{a}_3$ | = | $\left(\frac{1}{2}+x_3\right)a\,\hat{\mathbf{x}}-y_3a\,\hat{\mathbf{y}}+z_3c\,\hat{\mathbf{z}}$ | (32*g*) | As III |
| $\mathbf{B}_{30}$ | = | $\left(\frac{1}{2}+y_3+z_3\right)\mathbf{a}_1+\left(\frac{1}{2}-x_3+z_3\right)\mathbf{a}_2+(-x_3+y_3)\,\mathbf{a}_3$ | = | $-x_3a\,\hat{\mathbf{x}}+y_3a\,\hat{\mathbf{y}}+\left(\frac{1}{2}+z_3\right)c\,\hat{\mathbf{z}}$ | (32*g*) | As III |
| $\mathbf{B}_{31}$ | = | $\left(\frac{1}{2}-x_3+z_3\right)\mathbf{a}_1+(-y_3+z_3)\,\mathbf{a}_2+(-x_3-y_3)\,\mathbf{a}_3$ | = | $\left(\frac{3}{4}-y_3\right)a\,\hat{\mathbf{x}}+\left(\frac{1}{4}-x_3\right)a\,\hat{\mathbf{y}}+\left(\frac{1}{4}+z_3\right)c\,\hat{\mathbf{z}}$ | (32*g*) | As III |
| $\mathbf{B}_{32}$ | = | $\left(\frac{1}{2}+x_3+z_3\right)\mathbf{a}_1+\left(\frac{1}{2}+y_3+z_3\right)\mathbf{a}_2+\left(\frac{1}{2}+x_3+y_3\right)\mathbf{a}_3$ | = | $\left(\frac{1}{4}+y_3\right)a\,\hat{\mathbf{x}}+\left(\frac{1}{4}+x_3\right)a\,\hat{\mathbf{y}}+\left(\frac{1}{4}+z_3\right)c\,\hat{\mathbf{z}}$ | (32$g$) | As III |
| $\mathbf{B}_{33}$ | = | $(y_4+z_4)\,\mathbf{a}_1+(x_4+z_4)\,\mathbf{a}_2+(x_4+y_4)\,\mathbf{a}_3$ | = | $x_4a\,\hat{\mathbf{x}}+y_4a\,\hat{\mathbf{y}}+z_4c\,\hat{\mathbf{z}}$ | (32$g$) | Cd I |
| $\mathbf{B}_{34}$ | = | $\left(\frac{1}{2}-y_4+z_4\right)\mathbf{a}_1+(-x_4+z_4)\,\mathbf{a}_2+\left(\frac{1}{2}-x_4-y_4\right)\mathbf{a}_3$ | = | $-x_4a\,\hat{\mathbf{x}}+\left(\frac{1}{2}-y_4\right)a\,\hat{\mathbf{y}}+z_4c\,\hat{\mathbf{z}}$ | (32$g$) | Cd I |
| $\mathbf{B}_{35}$ | = | $(x_4+z_4)\,\mathbf{a}_1+\left(\frac{1}{2}-y_4+z_4\right)\mathbf{a}_2+(x_4-y_4)\,\mathbf{a}_3$ | = | $\left(\frac{1}{4}-y_4\right)a\,\hat{\mathbf{x}}+\left(\frac{3}{4}+x_4\right)a\,\hat{\mathbf{y}}+\left(\frac{1}{4}+z_4\right)c\,\hat{\mathbf{z}}$ | (32$g$) | Cd I |
| $\mathbf{B}_{36}$ | = | $(-x_4+z_4)\,\mathbf{a}_1+(y_4+z_4)\,\mathbf{a}_2+\left(\frac{1}{2}-x_4+y_4\right)\mathbf{a}_3$ | = | $\left(\frac{1}{4}+y_4\right)a\,\hat{\mathbf{x}}+\left(\frac{1}{4}-x_4\right)a\,\hat{\mathbf{y}}+\left(\frac{3}{4}+z_4\right)c\,\hat{\mathbf{z}}$ | (32$g$) | Cd I |
| $\mathbf{B}_{37}$ | = | $(y_4-z_4)\,\mathbf{a}_1+\left(\frac{1}{2}-x_4-z_4\right)\mathbf{a}_2+\left(\frac{1}{2}-x_4+y_4\right)\mathbf{a}_3$ | = | $\left(\frac{1}{2}-x_4\right)a\,\hat{\mathbf{x}}+y_4a\,\hat{\mathbf{y}}-z_4c\,\hat{\mathbf{z}}$ | (32$g$) | Cd I |
| $\mathbf{B}_{38}$ | = | $\left(\frac{1}{2}-y_4-z_4\right)\mathbf{a}_1+\left(\frac{1}{2}+x_4-z_4\right)\mathbf{a}_2+(x_4-y_4)\,\mathbf{a}_3$ | = | $x_4a\,\hat{\mathbf{x}}-y_4a\,\hat{\mathbf{y}}+\left(\frac{1}{2}-z_4\right)c\,\hat{\mathbf{z}}$ | (32$g$) | Cd I |
| $\mathbf{B}_{39}$ | = | $\left(\frac{1}{2}+x_4-z_4\right)\mathbf{a}_1+(y_4-z_4)\,\mathbf{a}_2+(x_4+y_4)\,\mathbf{a}_3$ | = | $\left(-\frac{1}{4}+y_4\right)a\,\hat{\mathbf{x}}+\left(\frac{1}{4}+x_4\right)a\,\hat{\mathbf{y}}+\left(\frac{1}{4}-z_4\right)c\,\hat{\mathbf{z}}$ | (32$g$) | Cd I |
| $\mathbf{B}_{40}$ | = | $\left(\frac{1}{2}-x_4-z_4\right)\mathbf{a}_1+\left(\frac{1}{2}-y_4-z_4\right)\mathbf{a}_2+\left(\frac{1}{2}-x_4-y_4\right)\mathbf{a}_3$ | = | $\left(\frac{1}{4}-y_4\right)a\,\hat{\mathbf{x}}+\left(\frac{1}{4}-x_4\right)a\,\hat{\mathbf{y}}+\left(\frac{1}{4}-z_4\right)c\,\hat{\mathbf{z}}$ | (32$g$) | Cd I |
| $\mathbf{B}_{41}$ | = | $(-y_4-z_4)\,\mathbf{a}_1+(-x_4-z_4)\,\mathbf{a}_2+(-x_4-y_4)\,\mathbf{a}_3$ | = | $-x_4a\,\hat{\mathbf{x}}-y_4a\,\hat{\mathbf{y}}-z_4c\,\hat{\mathbf{z}}$ | (32$g$) | Cd I |
| $\mathbf{B}_{42}$ | = | $\left(\frac{1}{2}+y_4-z_4\right)\mathbf{a}_1+(x_4-z_4)\,\mathbf{a}_2+\left(\frac{1}{2}+x_4+y_4\right)\mathbf{a}_3$ | = | $x_4a\,\hat{\mathbf{x}}+\left(\frac{1}{2}+y_4\right)a\,\hat{\mathbf{y}}-z_4c\,\hat{\mathbf{z}}$ | (32$g$) | Cd I |
| $\mathbf{B}_{43}$ | = | $(-x_4-z_4)\,\mathbf{a}_1+\left(\frac{1}{2}+y_4-z_4\right)\mathbf{a}_2+(-x_4+y_4)\,\mathbf{a}_3$ | = | $\left(\frac{1}{4}+y_4\right)a\,\hat{\mathbf{x}}-a\left(x_4+\frac{1}{4}\right)\hat{\mathbf{y}}+\left(\frac{1}{4}-z_4\right)c\,\hat{\mathbf{z}}$ | (32$g$) | Cd I |
| $\mathbf{B}_{44}$ | = | $(x_4-z_4)\,\mathbf{a}_1+(-y_4-z_4)\,\mathbf{a}_2+\left(\frac{1}{2}+x_4-y_4\right)\mathbf{a}_3$ | = | $\left(\frac{1}{4}-y_4\right)a\,\hat{\mathbf{x}}+\left(\frac{1}{4}+x_4\right)a\,\hat{\mathbf{y}}-c\left(z_4+\frac{1}{4}\right)\hat{\mathbf{z}}$ | (32$g$) | Cd I |
| $\mathbf{B}_{45}$ | = | $(-y_4+z_4)\,\mathbf{a}_1+\left(\frac{1}{2}+x_4+z_4\right)\mathbf{a}_2+\left(\frac{1}{2}+x_4-y_4\right)\mathbf{a}_3$ | = | $\left(\frac{1}{2}+x_4\right)a\,\hat{\mathbf{x}}-y_4a\,\hat{\mathbf{y}}+z_4c\,\hat{\mathbf{z}}$ | (32$g$) | Cd I |
| $\mathbf{B}_{46}$ | = | $\left(\frac{1}{2}+y_4+z_4\right)\mathbf{a}_1+\left(\frac{1}{2}-x_4+z_4\right)\mathbf{a}_2+(-x_4+y_4)\,\mathbf{a}_3$ | = | $-x_4a\,\hat{\mathbf{x}}+y_4a\,\hat{\mathbf{y}}+\left(\frac{1}{2}+z_4\right)c\,\hat{\mathbf{z}}$ | (32$g$) | Cd I |
| $\mathbf{B}_{47}$ | = | $\left(\frac{1}{2}-x_4+z_4\right)\mathbf{a}_1+(-y_4+z_4)\,\mathbf{a}_2+(-x_4-y_4)\,\mathbf{a}_3$ | = | $\left(\frac{3}{4}-y_4\right)a\,\hat{\mathbf{x}}+\left(\frac{1}{4}-x_4\right)a\,\hat{\mathbf{y}}+\left(\frac{1}{4}+z_4\right)c\,\hat{\mathbf{z}}$ | (32$g$) | Cd I |
| $\mathbf{B}_{48}$ | = | $\left(\frac{1}{2}+x_4+z_4\right)\mathbf{a}_1+\left(\frac{1}{2}+y_4+z_4\right)\mathbf{a}_2+\left(\frac{1}{2}+x_4+y_4\right)\mathbf{a}_3$ | = | $\left(\frac{1}{4}+y_4\right)a\,\hat{\mathbf{x}}+\left(\frac{1}{4}+x_4\right)a\,\hat{\mathbf{y}}+\left(\frac{1}{4}+z_4\right)c\,\hat{\mathbf{z}}$ | (32$g$) | Cd I |
| $\mathbf{B}_{49}$ | = | $(y_5+z_5)\,\mathbf{a}_1+(x_5+z_5)\,\mathbf{a}_2+(x_5+y_5)\,\mathbf{a}_3$ | = | $x_5a\,\hat{\mathbf{x}}+y_5a\,\hat{\mathbf{y}}+z_5c\,\hat{\mathbf{z}}$ | (32$g$) | Cd II |
| $\mathbf{B}_{50}$ | = | $\left(\frac{1}{2}-y_5+z_5\right)\mathbf{a}_1+(-x_5+z_5)\,\mathbf{a}_2+\left(\frac{1}{2}-x_5-y_5\right)\mathbf{a}_3$ | = | $-x_5a\,\hat{\mathbf{x}}+\left(\frac{1}{2}-y_5\right)a\,\hat{\mathbf{y}}+z_5c\,\hat{\mathbf{z}}$ | (32$g$) | Cd II |
| $\mathbf{B}_{51}$ | = | $(x_5+z_5)\,\mathbf{a}_1+\left(\frac{1}{2}-y_5+z_5\right)\mathbf{a}_2+(x_5-y_5)\,\mathbf{a}_3$ | = | $\left(\frac{1}{4}-y_5\right)a\,\hat{\mathbf{x}}+\left(\frac{3}{4}+x_5\right)a\,\hat{\mathbf{y}}+\left(\frac{1}{4}+z_5\right)c\,\hat{\mathbf{z}}$ | (32$g$) | Cd II |
| $\mathbf{B}_{52}$ | = | $(-x_5+z_5)\,\mathbf{a}_1+(y_5+z_5)\,\mathbf{a}_2+\left(\frac{1}{2}-x_5+y_5\right)\mathbf{a}_3$ | = | $\left(\frac{1}{4}+y_5\right)a\,\hat{\mathbf{x}}+\left(\frac{1}{4}-x_5\right)a\,\hat{\mathbf{y}}+\left(\frac{3}{4}+z_5\right)c\,\hat{\mathbf{z}}$ | (32$g$) | Cd II |
| $\mathbf{B}_{53}$ | = | $(y_5 - z_5)\,\mathbf{a}_1 + \left(\frac{1}{2} - x_5 - z_5\right)\mathbf{a}_2 + \left(\frac{1}{2} - x_5 + y_5\right)\mathbf{a}_3$ | = | $\left(\frac{1}{2} - x_5\right)a\,\hat{\mathbf{x}} + y_5 a\,\hat{\mathbf{y}} - z_5 c\,\hat{\mathbf{z}}$ | (32$g$) | Cd II |
| $\mathbf{B}_{54}$ | = | $\left(\frac{1}{2} - y_5 - z_5\right)\mathbf{a}_1 + \left(\frac{1}{2} + x_5 - z_5\right)\mathbf{a}_2 + (x_5 - y_5)\,\mathbf{a}_3$ | = | $x_5 a\,\hat{\mathbf{x}} - y_5 a\,\hat{\mathbf{y}} + \left(\frac{1}{2} - z_5\right)c\,\hat{\mathbf{z}}$ | (32$g$) | Cd II |
| $\mathbf{B}_{55}$ | = | $\left(\frac{1}{2} + x_5 - z_5\right)\mathbf{a}_1 + (y_5 - z_5)\,\mathbf{a}_2 + (x_5 + y_5)\,\mathbf{a}_3$ | = | $\left(-\frac{1}{4} + y_5\right)a\,\hat{\mathbf{x}} + \left(\frac{1}{4} + x_5\right)a\,\hat{\mathbf{y}} + \left(\frac{1}{4} - z_5\right)c\,\hat{\mathbf{z}}$ | (32$g$) | Cd II |
| $\mathbf{B}_{56}$ | = | $\left(\frac{1}{2} - x_5 - z_5\right)\mathbf{a}_1 + \left(\frac{1}{2} - y_5 - z_5\right)\mathbf{a}_2 + \left(\frac{1}{2} - x_5 - y_5\right)\mathbf{a}_3$ | = | $\left(\frac{1}{4} - y_5\right)a\,\hat{\mathbf{x}} + \left(\frac{1}{4} - x_5\right)a\,\hat{\mathbf{y}} + \left(\frac{1}{4} - z_5\right)c\,\hat{\mathbf{z}}$ | (32$g$) | Cd II |
| $\mathbf{B}_{57}$ | = | $(-y_5 - z_5)\,\mathbf{a}_1 + (-x_5 - z_5)\,\mathbf{a}_2 + (-x_5 - y_5)\,\mathbf{a}_3$ | = | $-x_5 a\,\hat{\mathbf{x}} - y_5 a\,\hat{\mathbf{y}} - z_5 c\,\hat{\mathbf{z}}$ | (32$g$) | Cd II |
| $\mathbf{B}_{58}$ | = | $\left(\frac{1}{2} + y_5 - z_5\right)\mathbf{a}_1 + (x_5 - z_5)\,\mathbf{a}_2 + \left(\frac{1}{2} + x_5 + y_5\right)\mathbf{a}_3$ | = | $x_5 a\,\hat{\mathbf{x}} + \left(\frac{1}{2} + y_5\right)a\,\hat{\mathbf{y}} - z_5 c\,\hat{\mathbf{z}}$ | (32$g$) | Cd II |
| $\mathbf{B}_{59}$ | = | $(-x_5 - z_5)\,\mathbf{a}_1 + \left(\frac{1}{2} + y_5 - z_5\right)\mathbf{a}_2 + (-x_5 + y_5)\,\mathbf{a}_3$ | = | $\left(\frac{1}{4} + y_5\right)a\,\hat{\mathbf{x}} - a\left(x_5 + \frac{1}{4}\right)\hat{\mathbf{y}} + \left(\frac{1}{4} - z_5\right)c\,\hat{\mathbf{z}}$ | (32$g$) | Cd II |
| $\mathbf{B}_{60}$ | = | $(x_5 - z_5)\,\mathbf{a}_1 + (-y_5 - z_5)\,\mathbf{a}_2 + \left(\frac{1}{2} + x_5 - y_5\right)\mathbf{a}_3$ | = | $\left(\frac{1}{4} - y_5\right)a\,\hat{\mathbf{x}} + \left(\frac{1}{4} + x_5\right)a\,\hat{\mathbf{y}} - c\left(z_5 + \frac{1}{4}\right)\hat{\mathbf{z}}$ | (32$g$) | Cd II |
| $\mathbf{B}_{61}$ | = | $(-y_5 + z_5)\,\mathbf{a}_1 + \left(\frac{1}{2} + x_5 + z_5\right)\mathbf{a}_2 + \left(\frac{1}{2} + x_5 - y_5\right)\mathbf{a}_3$ | = | $\left(\frac{1}{2} + x_5\right)a\,\hat{\mathbf{x}} - y_5 a\,\hat{\mathbf{y}} + z_5 c\,\hat{\mathbf{z}}$ | (32$g$) | Cd II |
| $\mathbf{B}_{62}$ | = | $\left(\frac{1}{2} + y_5 + z_5\right)\mathbf{a}_1 + \left(\frac{1}{2} - x_5 + z_5\right)\mathbf{a}_2 + (-x_5 + y_5)\,\mathbf{a}_3$ | = | $-x_5 a\,\hat{\mathbf{x}} + y_5 a\,\hat{\mathbf{y}} + \left(\frac{1}{2} + z_5\right)c\,\hat{\mathbf{z}}$ | (32$g$) | Cd II |
| $\mathbf{B}_{63}$ | = | $\left(\frac{1}{2} - x_5 + z_5\right)\mathbf{a}_1 + (-y_5 + z_5)\,\mathbf{a}_2 + (-x_5 - y_5)\,\mathbf{a}_3$ | = | $\left(\frac{3}{4} - y_5\right)a\,\hat{\mathbf{x}} + \left(\frac{1}{4} - x_5\right)a\,\hat{\mathbf{y}} + \left(\frac{1}{4} + z_5\right)c\,\hat{\mathbf{z}}$ | (32$g$) | Cd II |
| $\mathbf{B}_{64}$ | = | $\left(\frac{1}{2} + x_5 + z_5\right)\mathbf{a}_1 + \left(\frac{1}{2} + y_5 + z_5\right)\mathbf{a}_2 + \left(\frac{1}{2} + x_5 + y_5\right)\mathbf{a}_3$ | = | $\left(\frac{1}{4} + y_5\right)a\,\hat{\mathbf{x}} + \left(\frac{1}{4} + x_5\right)a\,\hat{\mathbf{y}} + \left(\frac{1}{4} + z_5\right)c\,\hat{\mathbf{z}}$ | (32$g$) | Cd II |
| $\mathbf{B}_{65}$ | = | $(y_6 + z_6)\,\mathbf{a}_1 + (x_6 + z_6)\,\mathbf{a}_2 + (x_6 + y_6)\,\mathbf{a}_3$ | = | $x_6 a\,\hat{\mathbf{x}} + y_6 a\,\hat{\mathbf{y}} + z_6 c\,\hat{\mathbf{z}}$ | (32$g$) | Cd III |
| $\mathbf{B}_{66}$ | = | $\left(\frac{1}{2} - y_6 + z_6\right)\mathbf{a}_1 + (-x_6 + z_6)\,\mathbf{a}_2 + \left(\frac{1}{2} - x_6 - y_6\right)\mathbf{a}_3$ | = | $-x_6 a\,\hat{\mathbf{x}} + \left(\frac{1}{2} - y_6\right)a\,\hat{\mathbf{y}} + z_6 c\,\hat{\mathbf{z}}$ | (32$g$) | Cd III |
| $\mathbf{B}_{67}$ | = | $(x_6 + z_6)\,\mathbf{a}_1 + \left(\frac{1}{2} - y_6 + z_6\right)\mathbf{a}_2 + (x_6 - y_6)\,\mathbf{a}_3$ | = | $\left(\frac{1}{4} - y_6\right)a\,\hat{\mathbf{x}} + \left(\frac{3}{4} + x_6\right)a\,\hat{\mathbf{y}} + \left(\frac{1}{4} + z_6\right)c\,\hat{\mathbf{z}}$ | (32$g$) | Cd III |
| $\mathbf{B}_{68}$ | = | $(-x_6 + z_6)\,\mathbf{a}_1 + (y_6 + z_6)\,\mathbf{a}_2 + \left(\frac{1}{2} - x_6 + y_6\right)\mathbf{a}_3$ | = | $\left(\frac{1}{4} + y_6\right)a\,\hat{\mathbf{x}} + \left(\frac{1}{4} - x_6\right)a\,\hat{\mathbf{y}} + \left(\frac{3}{4} + z_6\right)c\,\hat{\mathbf{z}}$ | (32$g$) | Cd III |
| $\mathbf{B}_{69}$ | = | $(y_6 - z_6)\,\mathbf{a}_1 + \left(\frac{1}{2} - x_6 - z_6\right)\mathbf{a}_2 + \left(\frac{1}{2} - x_6 + y_6\right)\mathbf{a}_3$ | = | $\left(\frac{1}{2} - x_6\right)a\,\hat{\mathbf{x}} + y_6 a\,\hat{\mathbf{y}} - z_6 c\,\hat{\mathbf{z}}$ | (32$g$) | Cd III |
| $\mathbf{B}_{70}$ | = | $\left(\frac{1}{2} - y_6 - z_6\right)\mathbf{a}_1 + \left(\frac{1}{2} + x_6 - z_6\right)\mathbf{a}_2 + (x_6 - y_6)\,\mathbf{a}_3$ | = | $x_6 a\,\hat{\mathbf{x}} - y_6 a\,\hat{\mathbf{y}} + \left(\frac{1}{2} - z_6\right)c\,\hat{\mathbf{z}}$ | (32$g$) | Cd III |
| $\mathbf{B}_{71}$ | = | $\left(\frac{1}{2} + x_6 - z_6\right)\mathbf{a}_1 + (y_6 - z_6)\,\mathbf{a}_2 + (x_6 + y_6)\,\mathbf{a}_3$ | = | $\left(-\frac{1}{4} + y_6\right)a\,\hat{\mathbf{x}} + \left(\frac{1}{4} + x_6\right)a\,\hat{\mathbf{y}} + \left(\frac{1}{4} - z_6\right)c\,\hat{\mathbf{z}}$ | (32$g$) | Cd III |
| $\mathbf{B}_{72}$ | = | $\left(\frac{1}{2} - x_6 - z_6\right)\mathbf{a}_1 + \left(\frac{1}{2} - y_6 - z_6\right)\mathbf{a}_2 + \left(\frac{1}{2} - x_6 - y_6\right)\mathbf{a}_3$ | = | $\left(\frac{1}{4} - y_6\right)a\,\hat{\mathbf{x}} + \left(\frac{1}{4} - x_6\right)a\,\hat{\mathbf{y}} + \left(\frac{1}{4} - z_6\right)c\,\hat{\mathbf{z}}$ | (32$g$) | Cd III |
| $\mathbf{B}_{73}$ | = | $(-y_6 - z_6)\,\mathbf{a}_1 + (-x_6 - z_6)\,\mathbf{a}_2 + (-x_6 - y_6)\,\mathbf{a}_3$ | = | $-x_6 a\,\hat{\mathbf{x}} - y_6 a\,\hat{\mathbf{y}} - z_6 c\,\hat{\mathbf{z}}$ | (32$g$) | Cd III |
| $\mathbf{B}_{74}$ | = | $\left(\frac{1}{2}+y_6-z_6\right)\mathbf{a}_1+(x_6-z_6)\,\mathbf{a}_2+\left(\frac{1}{2}+x_6+y_6\right)\mathbf{a}_3$ | = | $x_6a\,\mathbf{\hat{x}}+\left(\frac{1}{2}+y_6\right)a\,\mathbf{\hat{y}}-z_6c\,\mathbf{\hat{z}}$ | (32$g$) | Cd III |
| $\mathbf{B}_{75}$ | = | $(-x_6-z_6)\,\mathbf{a}_1+\left(\frac{1}{2}+y_6-z_6\right)\mathbf{a}_2+(-x_6+y_6)\,\mathbf{a}_3$ | = | $\left(\frac{1}{4}+y_6\right)a\,\mathbf{\hat{x}}-a\left(x_6+\frac{1}{4}\right)\mathbf{\hat{y}}+\left(\frac{1}{4}-z_6\right)c\,\mathbf{\hat{z}}$ | (32$g$) | Cd III |
| $\mathbf{B}_{76}$ | = | $(x_6-z_6)\,\mathbf{a}_1+(-y_6-z_6)\,\mathbf{a}_2+\left(\frac{1}{2}+x_6-y_6\right)\mathbf{a}_3$ | = | $\left(\frac{1}{4}-y_6\right)a\,\mathbf{\hat{x}}+\left(\frac{1}{4}+x_6\right)a\,\mathbf{\hat{y}}-c\left(z_6+\frac{1}{4}\right)\mathbf{\hat{z}}$ | (32$g$) | Cd III |
| $\mathbf{B}_{77}$ | = | $(-y_6+z_6)\,\mathbf{a}_1+\left(\frac{1}{2}+x_6+z_6\right)\mathbf{a}_2+\left(\frac{1}{2}+x_6-y_6\right)\mathbf{a}_3$ | = | $\left(\frac{1}{2}+x_6\right)a\,\mathbf{\hat{x}}-y_6a\,\mathbf{\hat{y}}+z_6c\,\mathbf{\hat{z}}$ | (32$g$) | Cd III |
| $\mathbf{B}_{78}$ | = | $\left(\frac{1}{2}+y_6+z_6\right)\mathbf{a}_1+\left(\frac{1}{2}-x_6+z_6\right)\mathbf{a}_2+(-x_6+y_6)\,\mathbf{a}_3$ | = | $-x_6a\,\mathbf{\hat{x}}+y_6a\,\mathbf{\hat{y}}+\left(\frac{1}{2}+z_6\right)c\,\mathbf{\hat{z}}$ | (32$g$) | Cd III |
| $\mathbf{B}_{79}$ | = | $\left(\frac{1}{2}-x_6+z_6\right)\mathbf{a}_1+(-y_6+z_6)\,\mathbf{a}_2+(-x_6-y_6)\,\mathbf{a}_3$ | = | $\left(\frac{3}{4}-y_6\right)a\,\mathbf{\hat{x}}+\left(\frac{1}{4}-x_6\right)a\,\mathbf{\hat{y}}+\left(\frac{1}{4}+z_6\right)c\,\mathbf{\hat{z}}$ | (32$g$) | Cd III |
| $\mathbf{B}_{80}$ | = | $\left(\frac{1}{2}+x_6+z_6\right)\mathbf{a}_1+\left(\frac{1}{2}+y_6+z_6\right)\mathbf{a}_2+\left(\frac{1}{2}+x_6+y_6\right)\mathbf{a}_3$ | = | $\left(\frac{1}{4}+y_6\right)a\,\mathbf{\hat{x}}+\left(\frac{1}{4}+x_6\right)a\,\mathbf{\hat{y}}+\left(\frac{1}{4}+z_6\right)c\,\mathbf{\hat{z}}$ | (32$g$) | Cd III |

**Geometry files:**

- CIF: pp. 1719
- POSCAR: pp. 1719

# $La_3BWO_9$ ($P3$) Structure: AB3C9D_hP28_143_2a_2d_6d_bc

http://aflow.org/prototype-encyclopedia/AB3C9D_hP28_143_2a_2d_6d_bc

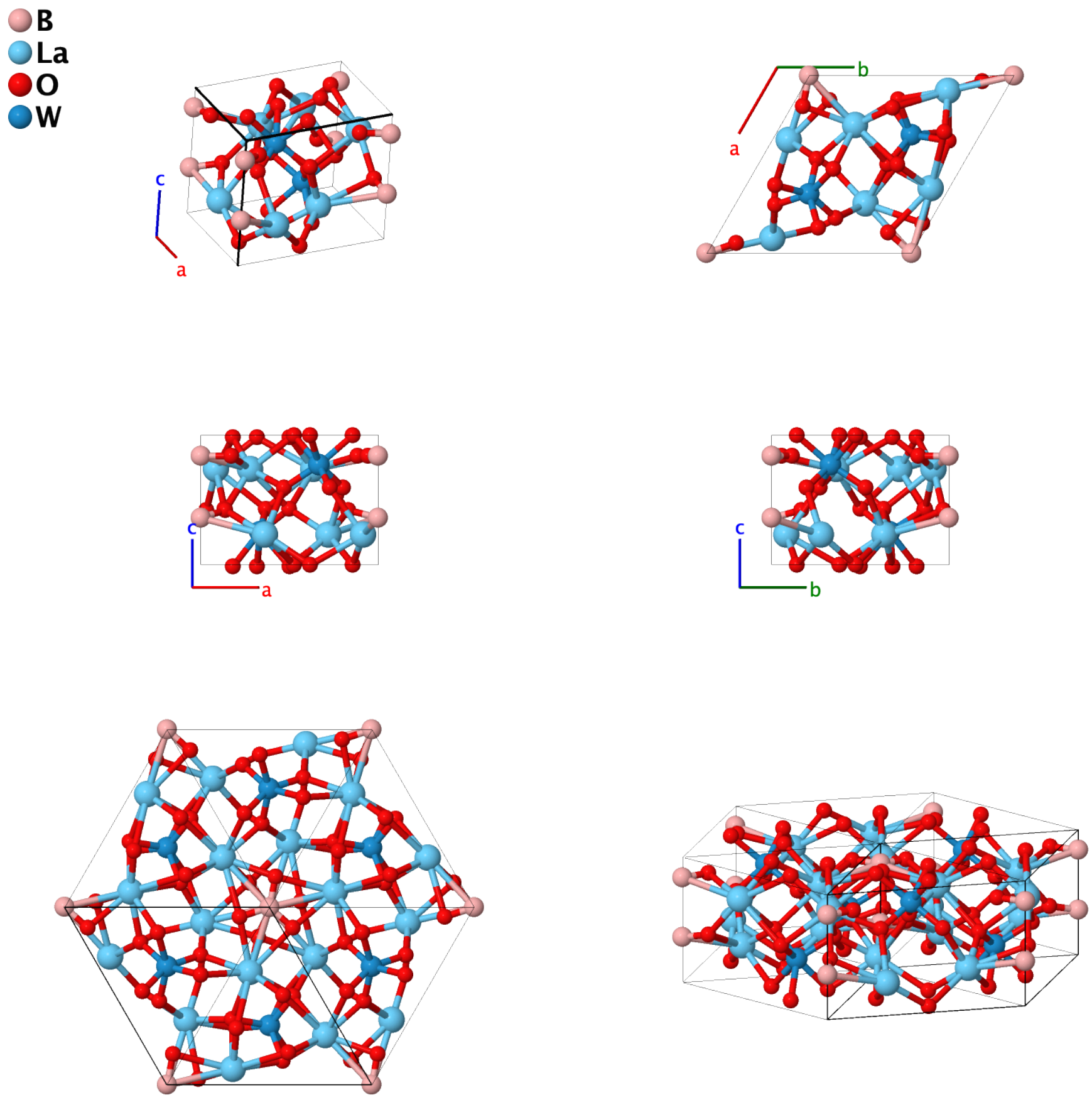

**Prototype** : $BLa_3O_9W$

**AFLOW prototype label** : AB3C9D_hP28_143_2a_2d_6d_bc

***Strukturbericht* designation** : None

**Pearson symbol** : hP28

**Space group number** : 143

**Space group symbol** : $P3$

**AFLOW prototype command** : `aflow --proto=AB3C9D_hP28_143_2a_2d_6d_bc`
`--params=`$a, c/a, z_1, z_2, z_3, z_4, x_5, y_5, z_5, x_6, y_6, z_6, x_7, y_7, z_7, x_8, y_8, z_8, x_9, y_9, z_9, x_{10}, y_{10}, z_{10}, x_{11}, y_{11}, z_{11}, x_{12}, y_{12}, z_{12}$

---

- Most refinements of the $BLa_3O_9W$ structure, including (Ashtar, 2020) place it in hexagonal space group $P6_3$ #173. (Han, 2018) find a better fit to the data by refining it in the trigonal $P3$ #143 space group, which places the lanthanum atoms on two independent crystallographic sites. As this may be due to the presence of bismuth impurities, which (Han, 2018) place on the lanthanum site, while (Ashtar, 2020) claim to have very pure samples, we withhold judgment on which structure is correct and present both.

- Space group $P3$ does not specify the origin of the $z$-axis. Here it is set so that the coordinate of the (1$b$) tungsten atom is $z_3 = 1/4$.

- (Han, 2018) mislabel the Wyckoff positions of the boron and tungsten atoms. We show the correct positions based on

the coordinates in their table, and the correspondence with the $P6_3$ structure.

---

**Trigonal Hexagonal primitive vectors:**

$$\mathbf{a}_1 = \frac{1}{2}\,a\,\hat{\mathbf{x}} - \frac{\sqrt{3}}{2}\,a\,\hat{\mathbf{y}}$$
$$\mathbf{a}_2 = \frac{1}{2}\,a\,\hat{\mathbf{x}} + \frac{\sqrt{3}}{2}\,a\,\hat{\mathbf{y}}$$
$$\mathbf{a}_3 = c\,\hat{\mathbf{z}}$$

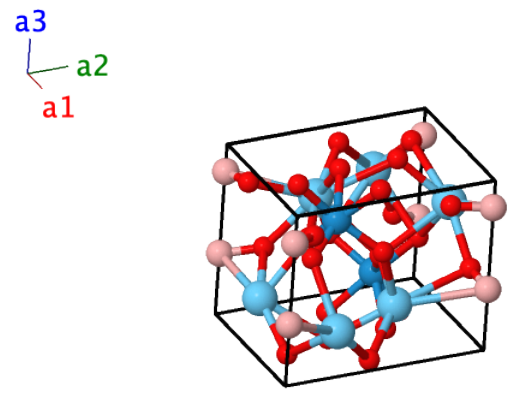

---

**Basis vectors:**

| | | Lattice Coordinates | | Cartesian Coordinates | Wyckoff Position | Atom Type |
|---|---|---|---|---|---|---|
| $\mathbf{B}_1$ | = | $z_1\,\mathbf{a}_3$ | = | $z_1 c\,\hat{\mathbf{z}}$ | (1$a$) | B I |
| $\mathbf{B}_2$ | = | $z_2\,\mathbf{a}_3$ | = | $z_2 c\,\hat{\mathbf{z}}$ | (1$a$) | B II |
| $\mathbf{B}_3$ | = | $\frac{1}{3}\,\mathbf{a}_1 + \frac{2}{3}\,\mathbf{a}_2 + z_3\,\mathbf{a}_3$ | = | $\frac{1}{2}a\,\hat{\mathbf{x}} + \frac{1}{2\sqrt{3}}a\,\hat{\mathbf{y}} + z_3 c\,\hat{\mathbf{z}}$ | (1$b$) | W I |
| $\mathbf{B}_4$ | = | $\frac{2}{3}\,\mathbf{a}_1 + \frac{1}{3}\,\mathbf{a}_2 + z_4\,\mathbf{a}_3$ | = | $\frac{1}{2}a\,\hat{\mathbf{x}} - \frac{1}{2\sqrt{3}}a\,\hat{\mathbf{y}} + z_4 c\,\hat{\mathbf{z}}$ | (1$c$) | W II |
| $\mathbf{B}_5$ | = | $x_5\,\mathbf{a}_1 + y_5\,\mathbf{a}_2 + z_5\,\mathbf{a}_3$ | = | $\frac{1}{2}\left(x_5 + y_5\right)a\,\hat{\mathbf{x}} + \frac{\sqrt{3}}{2}\left(-x_5 + y_5\right)a\,\hat{\mathbf{y}} + z_5 c\,\hat{\mathbf{z}}$ | (3$d$) | La I |
| $\mathbf{B}_6$ | = | $-y_5\,\mathbf{a}_1 + \left(x_5 - y_5\right)\mathbf{a}_2 + z_5\,\mathbf{a}_3$ | = | $\left(\frac{1}{2}x_5 - y_5\right)a\,\hat{\mathbf{x}} + \frac{\sqrt{3}}{2}x_5 a\,\hat{\mathbf{y}} + z_5 c\,\hat{\mathbf{z}}$ | (3$d$) | La I |
| $\mathbf{B}_7$ | = | $\left(-x_5 + y_5\right)\mathbf{a}_1 - x_5\,\mathbf{a}_2 + z_5\,\mathbf{a}_3$ | = | $\left(-x_5 + \frac{1}{2}y_5\right)a\,\hat{\mathbf{x}} - \frac{\sqrt{3}}{2}y_5 a\,\hat{\mathbf{y}} + z_5 c\,\hat{\mathbf{z}}$ | (3$d$) | La I |
| $\mathbf{B}_8$ | = | $x_6\,\mathbf{a}_1 + y_6\,\mathbf{a}_2 + z_6\,\mathbf{a}_3$ | = | $\frac{1}{2}\left(x_6 + y_6\right)a\,\hat{\mathbf{x}} + \frac{\sqrt{3}}{2}\left(-x_6 + y_6\right)a\,\hat{\mathbf{y}} + z_6 c\,\hat{\mathbf{z}}$ | (3$d$) | La II |
| $\mathbf{B}_9$ | = | $-y_6\,\mathbf{a}_1 + \left(x_6 - y_6\right)\mathbf{a}_2 + z_6\,\mathbf{a}_3$ | = | $\left(\frac{1}{2}x_6 - y_6\right)a\,\hat{\mathbf{x}} + \frac{\sqrt{3}}{2}x_6 a\,\hat{\mathbf{y}} + z_6 c\,\hat{\mathbf{z}}$ | (3$d$) | La II |
| $\mathbf{B}_{10}$ | = | $\left(-x_6 + y_6\right)\mathbf{a}_1 - x_6\,\mathbf{a}_2 + z_6\,\mathbf{a}_3$ | = | $\left(-x_6 + \frac{1}{2}y_6\right)a\,\hat{\mathbf{x}} - \frac{\sqrt{3}}{2}y_6 a\,\hat{\mathbf{y}} + z_6 c\,\hat{\mathbf{z}}$ | (3$d$) | La II |
| $\mathbf{B}_{11}$ | = | $x_7\,\mathbf{a}_1 + y_7\,\mathbf{a}_2 + z_7\,\mathbf{a}_3$ | = | $\frac{1}{2}\left(x_7 + y_7\right)a\,\hat{\mathbf{x}} + \frac{\sqrt{3}}{2}\left(-x_7 + y_7\right)a\,\hat{\mathbf{y}} + z_7 c\,\hat{\mathbf{z}}$ | (3$d$) | O I |
| $\mathbf{B}_{12}$ | = | $-y_7\,\mathbf{a}_1 + \left(x_7 - y_7\right)\mathbf{a}_2 + z_7\,\mathbf{a}_3$ | = | $\left(\frac{1}{2}x_7 - y_7\right)a\,\hat{\mathbf{x}} + \frac{\sqrt{3}}{2}x_7 a\,\hat{\mathbf{y}} + z_7 c\,\hat{\mathbf{z}}$ | (3$d$) | O I |
| $\mathbf{B}_{13}$ | = | $\left(-x_7 + y_7\right)\mathbf{a}_1 - x_7\,\mathbf{a}_2 + z_7\,\mathbf{a}_3$ | = | $\left(-x_7 + \frac{1}{2}y_7\right)a\,\hat{\mathbf{x}} - \frac{\sqrt{3}}{2}y_7 a\,\hat{\mathbf{y}} + z_7 c\,\hat{\mathbf{z}}$ | (3$d$) | O I |
| $\mathbf{B}_{14}$ | = | $x_8\,\mathbf{a}_1 + y_8\,\mathbf{a}_2 + z_8\,\mathbf{a}_3$ | = | $\frac{1}{2}\left(x_8 + y_8\right)a\,\hat{\mathbf{x}} + \frac{\sqrt{3}}{2}\left(-x_8 + y_8\right)a\,\hat{\mathbf{y}} + z_8 c\,\hat{\mathbf{z}}$ | (3$d$) | O II |
| $\mathbf{B}_{15}$ | = | $-y_8\,\mathbf{a}_1 + \left(x_8 - y_8\right)\mathbf{a}_2 + z_8\,\mathbf{a}_3$ | = | $\left(\frac{1}{2}x_8 - y_8\right)a\,\hat{\mathbf{x}} + \frac{\sqrt{3}}{2}x_8 a\,\hat{\mathbf{y}} + z_8 c\,\hat{\mathbf{z}}$ | (3$d$) | O II |
| $\mathbf{B}_{16}$ | = | $\left(-x_8 + y_8\right)\mathbf{a}_1 - x_8\,\mathbf{a}_2 + z_8\,\mathbf{a}_3$ | = | $\left(-x_8 + \frac{1}{2}y_8\right)a\,\hat{\mathbf{x}} - \frac{\sqrt{3}}{2}y_8 a\,\hat{\mathbf{y}} + z_8 c\,\hat{\mathbf{z}}$ | (3$d$) | O II |
| $\mathbf{B}_{17}$ | = | $x_9\,\mathbf{a}_1 + y_9\,\mathbf{a}_2 + z_9\,\mathbf{a}_3$ | = | $\frac{1}{2}\left(x_9 + y_9\right)a\,\hat{\mathbf{x}} + \frac{\sqrt{3}}{2}\left(-x_9 + y_9\right)a\,\hat{\mathbf{y}} + z_9 c\,\hat{\mathbf{z}}$ | (3$d$) | O III |
| $\mathbf{B}_{18}$ | = | $-y_9\,\mathbf{a}_1 + \left(x_9 - y_9\right)\mathbf{a}_2 + z_9\,\mathbf{a}_3$ | = | $\left(\frac{1}{2}x_9 - y_9\right)a\,\hat{\mathbf{x}} + \frac{\sqrt{3}}{2}x_9 a\,\hat{\mathbf{y}} + z_9 c\,\hat{\mathbf{z}}$ | (3$d$) | O III |
| $\mathbf{B}_{19}$ | = | $\left(-x_9 + y_9\right)\mathbf{a}_1 - x_9\,\mathbf{a}_2 + z_9\,\mathbf{a}_3$ | = | $\left(-x_9 + \frac{1}{2}y_9\right)a\,\hat{\mathbf{x}} - \frac{\sqrt{3}}{2}y_9 a\,\hat{\mathbf{y}} + z_9 c\,\hat{\mathbf{z}}$ | (3$d$) | O III |
| $\mathbf{B}_{20}$ | = | $x_{10}\,\mathbf{a}_1 + y_{10}\,\mathbf{a}_2 + z_{10}\,\mathbf{a}_3$ | = | $\frac{1}{2}\left(x_{10} + y_{10}\right)a\,\hat{\mathbf{x}} + \frac{\sqrt{3}}{2}\left(-x_{10} + y_{10}\right)a\,\hat{\mathbf{y}} + z_{10} c\,\hat{\mathbf{z}}$ | (3$d$) | O IV |
| $\mathbf{B}_{21}$ | = | $-y_{10}\,\mathbf{a}_1 + \left(x_{10} - y_{10}\right)\mathbf{a}_2 + z_{10}\,\mathbf{a}_3$ | = | $\left(\frac{1}{2}x_{10} - y_{10}\right)a\,\hat{\mathbf{x}} + \frac{\sqrt{3}}{2}x_{10} a\,\hat{\mathbf{y}} + z_{10} c\,\hat{\mathbf{z}}$ | (3$d$) | O IV |
| $\mathbf{B}_{22}$ | = | $\left(-x_{10} + y_{10}\right)\mathbf{a}_1 - x_{10}\,\mathbf{a}_2 + z_{10}\,\mathbf{a}_3$ | = | $\left(-x_{10} + \frac{1}{2}y_{10}\right)a\,\hat{\mathbf{x}} - \frac{\sqrt{3}}{2}y_{10} a\,\hat{\mathbf{y}} + z_{10} c\,\hat{\mathbf{z}}$ | (3$d$) | O IV |
| $\mathbf{B}_{23}$ | = | $x_{11}\,\mathbf{a}_1 + y_{11}\,\mathbf{a}_2 + z_{11}\,\mathbf{a}_3$ | = | $\frac{1}{2}\left(x_{11} + y_{11}\right)a\,\hat{\mathbf{x}} + \frac{\sqrt{3}}{2}\left(-x_{11} + y_{11}\right)a\,\hat{\mathbf{y}} + z_{11} c\,\hat{\mathbf{z}}$ | (3$d$) | O V |
| $\mathbf{B}_{24}$ | = | $-y_{11}\,\mathbf{a}_1 + \left(x_{11} - y_{11}\right)\mathbf{a}_2 + z_{11}\,\mathbf{a}_3$ | = | $\left(\frac{1}{2}x_{11} - y_{11}\right)a\,\hat{\mathbf{x}} + \frac{\sqrt{3}}{2}x_{11} a\,\hat{\mathbf{y}} + z_{11} c\,\hat{\mathbf{z}}$ | (3$d$) | O V |
| $\mathbf{B}_{25}$ | = | $(-x_{11}+y_{11})\,\mathbf{a}_1 - x_{11}\,\mathbf{a}_2 + z_{11}\,\mathbf{a}_3$ | = | $\left(-x_{11}+\frac{1}{2}y_{11}\right)a\,\hat{\mathbf{x}} - \frac{\sqrt{3}}{2}y_{11}a\,\hat{\mathbf{y}} + z_{11}c\,\hat{\mathbf{z}}$ | (3$d$) | O V |
| $\mathbf{B}_{26}$ | = | $x_{12}\,\mathbf{a}_1 + y_{12}\,\mathbf{a}_2 + z_{12}\,\mathbf{a}_3$ | = | $\frac{1}{2}(x_{12}+y_{12})\,a\,\hat{\mathbf{x}} + \frac{\sqrt{3}}{2}(-x_{12}+y_{12})\,a\,\hat{\mathbf{y}} + z_{12}c\,\hat{\mathbf{z}}$ | (3$d$) | O VI |
| $\mathbf{B}_{27}$ | = | $-y_{12}\,\mathbf{a}_1 + (x_{12}-y_{12})\,\mathbf{a}_2 + z_{12}\,\mathbf{a}_3$ | = | $\left(\frac{1}{2}x_{12}-y_{12}\right)a\,\hat{\mathbf{x}} + \frac{\sqrt{3}}{2}x_{12}a\,\hat{\mathbf{y}} + z_{12}c\,\hat{\mathbf{z}}$ | (3$d$) | O VI |
| $\mathbf{B}_{28}$ | = | $(-x_{12}+y_{12})\,\mathbf{a}_1 - x_{12}\,\mathbf{a}_2 + z_{12}\,\mathbf{a}_3$ | = | $\left(-x_{12}+\frac{1}{2}y_{12}\right)a\,\hat{\mathbf{x}} - \frac{\sqrt{3}}{2}y_{12}a\,\hat{\mathbf{y}} + z_{12}c\,\hat{\mathbf{z}}$ | (3$d$) | O VI |

**Geometry files:**

- CIF: pp. 1720
- POSCAR: pp. 1720

# $RbNO_3$ (IV) Structure: AB3C_hP45_144_3a_9a_3a

http://aflow.org/prototype-encyclopedia/AB3C_hP45_144_3a_9a_3a

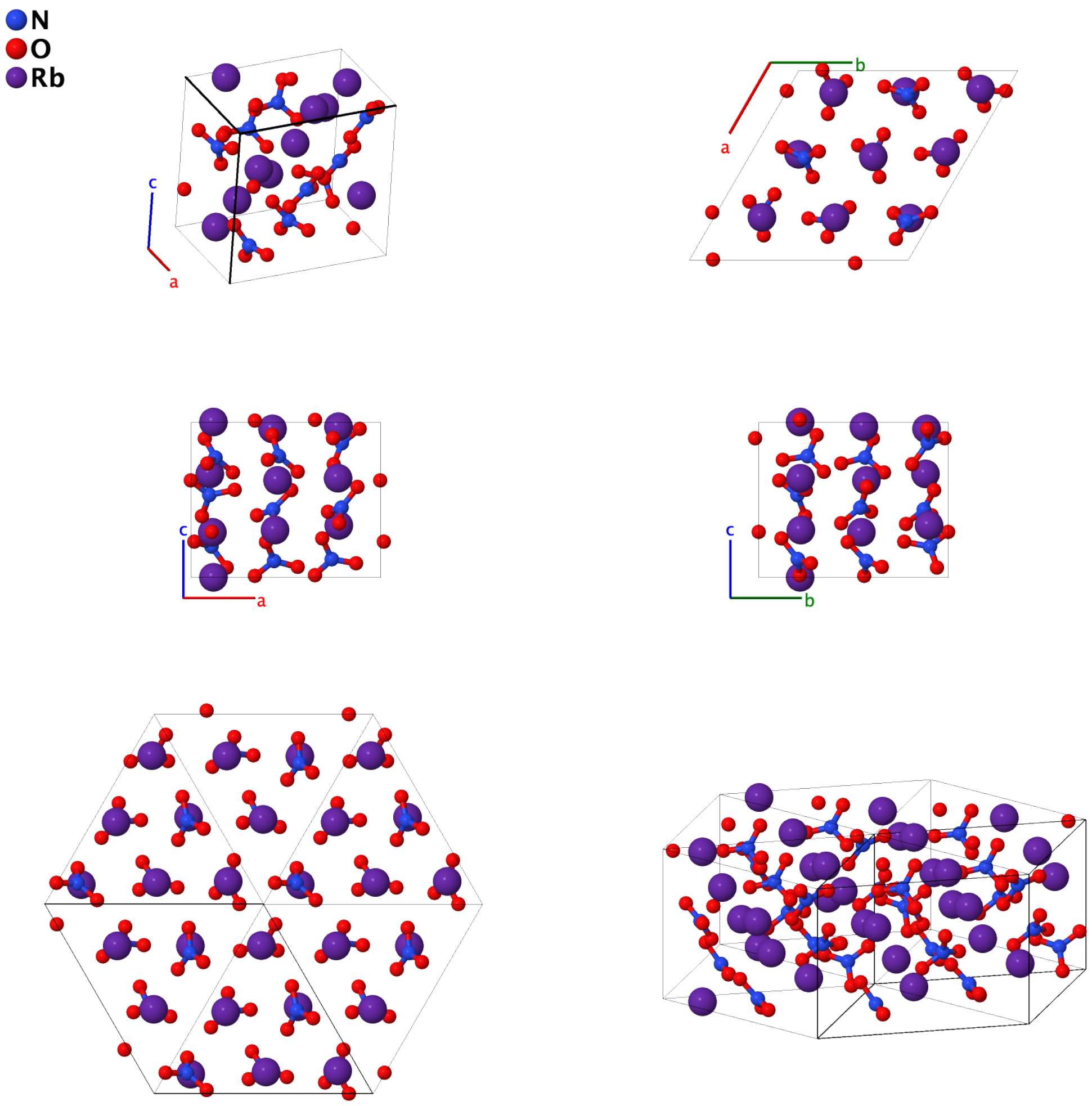

| | | |
|---|---|---|
| **Prototype** | : | $NO_3Rb$ |
| **AFLOW prototype label** | : | AB3C_hP45_144_3a_9a_3a |
| ***Strukturbericht* designation** | : | None |
| **Pearson symbol** | : | hP45 |
| **Space group number** | : | 144 |
| **Space group symbol** | : | $P3_1$ |
| **AFLOW prototype command** | : | `aflow --proto=AB3C_hP45_144_3a_9a_3a`<br>`--params=`$a, c/a, x_1, y_1, z_1, x_2, y_2, z_2, x_3, y_3, z_3, x_4, y_4, z_4, x_5, y_5, z_5, x_6, y_6, z_6, x_7, y_7, z_7, x_8, y_8, z_8, x_9, y_9, z_9, x_{10}, y_{10}, z_{10}, x_{11}, y_{11}, z_{11}, x_{12}, y_{12}, z_{12}, x_{13}, y_{13}, z_{13}, x_{14}, y_{14}, z_{14}, x_{15}, y_{15}, z_{15}$ |

**Other compounds with this structure**

- $CsNO_3$ (II) and $TlNO_3$ (III)

- $RbNO_3$ takes on four distinct phases with increasing temperature (Shamsuzzoha, 1988). Phase IV, presented here, is the ground state, stable up to 437 K. Phase III, stable in the range 437-492 K, is in the cesium chloride ($B2$) structure, with rubidium atoms on the cesium sites and orientationally disordered $NO_3$ ions on the chlorine sites. Phase II, stable up to 564 K, is possibly a body-centered cubic structure with up to eight formula units in the conventional unit cell. Phase I, stable up to the melting point, is thought to be a sodium chloride ($B1$) structure, with $NO_3$ again playing the role of chlorine.

- The phase IV structure also exists in the enantiomorphic space group $P3_2$ #145. (Shamsuzzoha, 1982)
- We used the data taken at room temperature, 298 K.

---

**Trigonal Hexagonal primitive vectors:**

$$
\begin{aligned}
\mathbf{a}_1 &= \tfrac{1}{2}\,a\,\hat{\mathbf{x}} - \tfrac{\sqrt{3}}{2}\,a\,\hat{\mathbf{y}} \\
\mathbf{a}_2 &= \tfrac{1}{2}\,a\,\hat{\mathbf{x}} + \tfrac{\sqrt{3}}{2}\,a\,\hat{\mathbf{y}} \\
\mathbf{a}_3 &= c\,\hat{\mathbf{z}}
\end{aligned}
$$

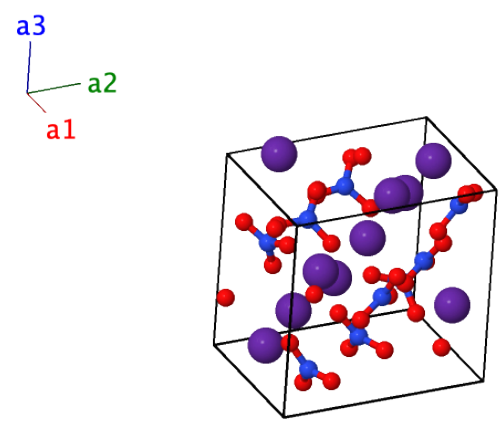

---

**Basis vectors:**

| | | Lattice Coordinates | | Cartesian Coordinates | Wyckoff Position | Atom Type |
|---|---|---|---|---|---|---|
| $\mathbf{B}_1$ | = | $x_1\,\mathbf{a}_1 + y_1\,\mathbf{a}_2 + z_1\,\mathbf{a}_3$ | = | $\frac{1}{2}(x_1 + y_1)\,a\,\hat{\mathbf{x}} + \frac{\sqrt{3}}{2}(-x_1 + y_1)\,a\,\hat{\mathbf{y}} + z_1 c\,\hat{\mathbf{z}}$ | (3$a$) | N I |
| $\mathbf{B}_2$ | = | $-y_1\,\mathbf{a}_1 + (x_1 - y_1)\,\mathbf{a}_2 + \left(\frac{1}{3} + z_1\right)\mathbf{a}_3$ | = | $\left(\frac{1}{2}x_1 - y_1\right)a\,\hat{\mathbf{x}} + \frac{\sqrt{3}}{2}x_1 a\,\hat{\mathbf{y}} + \left(\frac{1}{3} + z_1\right)c\,\hat{\mathbf{z}}$ | (3$a$) | N I |
| $\mathbf{B}_3$ | = | $(-x_1 + y_1)\,\mathbf{a}_1 - x_1\,\mathbf{a}_2 + \left(\frac{2}{3} + z_1\right)\mathbf{a}_3$ | = | $\left(-x_1 + \frac{1}{2}y_1\right)a\,\hat{\mathbf{x}} - \frac{\sqrt{3}}{2}y_1 a\,\hat{\mathbf{y}} + \left(\frac{2}{3} + z_1\right)c\,\hat{\mathbf{z}}$ | (3$a$) | N I |
| $\mathbf{B}_4$ | = | $x_2\,\mathbf{a}_1 + y_2\,\mathbf{a}_2 + z_2\,\mathbf{a}_3$ | = | $\frac{1}{2}(x_2 + y_2)\,a\,\hat{\mathbf{x}} + \frac{\sqrt{3}}{2}(-x_2 + y_2)\,a\,\hat{\mathbf{y}} + z_2 c\,\hat{\mathbf{z}}$ | (3$a$) | N II |
| $\mathbf{B}_5$ | = | $-y_2\,\mathbf{a}_1 + (x_2 - y_2)\,\mathbf{a}_2 + \left(\frac{1}{3} + z_2\right)\mathbf{a}_3$ | = | $\left(\frac{1}{2}x_2 - y_2\right)a\,\hat{\mathbf{x}} + \frac{\sqrt{3}}{2}x_2 a\,\hat{\mathbf{y}} + \left(\frac{1}{3} + z_2\right)c\,\hat{\mathbf{z}}$ | (3$a$) | N II |
| $\mathbf{B}_6$ | = | $(-x_2 + y_2)\,\mathbf{a}_1 - x_2\,\mathbf{a}_2 + \left(\frac{2}{3} + z_2\right)\mathbf{a}_3$ | = | $\left(-x_2 + \frac{1}{2}y_2\right)a\,\hat{\mathbf{x}} - \frac{\sqrt{3}}{2}y_2 a\,\hat{\mathbf{y}} + \left(\frac{2}{3} + z_2\right)c\,\hat{\mathbf{z}}$ | (3$a$) | N II |
| $\mathbf{B}_7$ | = | $x_3\,\mathbf{a}_1 + y_3\,\mathbf{a}_2 + z_3\,\mathbf{a}_3$ | = | $\frac{1}{2}(x_3 + y_3)\,a\,\hat{\mathbf{x}} + \frac{\sqrt{3}}{2}(-x_3 + y_3)\,a\,\hat{\mathbf{y}} + z_3 c\,\hat{\mathbf{z}}$ | (3$a$) | N III |
| $\mathbf{B}_8$ | = | $-y_3\,\mathbf{a}_1 + (x_3 - y_3)\,\mathbf{a}_2 + \left(\frac{1}{3} + z_3\right)\mathbf{a}_3$ | = | $\left(\frac{1}{2}x_3 - y_3\right)a\,\hat{\mathbf{x}} + \frac{\sqrt{3}}{2}x_3 a\,\hat{\mathbf{y}} + \left(\frac{1}{3} + z_3\right)c\,\hat{\mathbf{z}}$ | (3$a$) | N III |
| $\mathbf{B}_9$ | = | $(-x_3 + y_3)\,\mathbf{a}_1 - x_3\,\mathbf{a}_2 + \left(\frac{2}{3} + z_3\right)\mathbf{a}_3$ | = | $\left(-x_3 + \frac{1}{2}y_3\right)a\,\hat{\mathbf{x}} - \frac{\sqrt{3}}{2}y_3 a\,\hat{\mathbf{y}} + \left(\frac{2}{3} + z_3\right)c\,\hat{\mathbf{z}}$ | (3$a$) | N III |
| $\mathbf{B}_{10}$ | = | $x_4\,\mathbf{a}_1 + y_4\,\mathbf{a}_2 + z_4\,\mathbf{a}_3$ | = | $\frac{1}{2}(x_4 + y_4)\,a\,\hat{\mathbf{x}} + \frac{\sqrt{3}}{2}(-x_4 + y_4)\,a\,\hat{\mathbf{y}} + z_4 c\,\hat{\mathbf{z}}$ | (3$a$) | O I |
| $\mathbf{B}_{11}$ | = | $-y_4\,\mathbf{a}_1 + (x_4 - y_4)\,\mathbf{a}_2 + \left(\frac{1}{3} + z_4\right)\mathbf{a}_3$ | = | $\left(\frac{1}{2}x_4 - y_4\right)a\,\hat{\mathbf{x}} + \frac{\sqrt{3}}{2}x_4 a\,\hat{\mathbf{y}} + \left(\frac{1}{3} + z_4\right)c\,\hat{\mathbf{z}}$ | (3$a$) | O I |
| $\mathbf{B}_{12}$ | = | $(-x_4 + y_4)\,\mathbf{a}_1 - x_4\,\mathbf{a}_2 + \left(\frac{2}{3} + z_4\right)\mathbf{a}_3$ | = | $\left(-x_4 + \frac{1}{2}y_4\right)a\,\hat{\mathbf{x}} - \frac{\sqrt{3}}{2}y_4 a\,\hat{\mathbf{y}} + \left(\frac{2}{3} + z_4\right)c\,\hat{\mathbf{z}}$ | (3$a$) | O I |
| $\mathbf{B}_{13}$ | = | $x_5\,\mathbf{a}_1 + y_5\,\mathbf{a}_2 + z_5\,\mathbf{a}_3$ | = | $\frac{1}{2}(x_5 + y_5)\,a\,\hat{\mathbf{x}} + \frac{\sqrt{3}}{2}(-x_5 + y_5)\,a\,\hat{\mathbf{y}} + z_5 c\,\hat{\mathbf{z}}$ | (3$a$) | O II |
| $\mathbf{B}_{14}$ | = | $-y_5\,\mathbf{a}_1 + (x_5 - y_5)\,\mathbf{a}_2 + \left(\frac{1}{3} + z_5\right)\mathbf{a}_3$ | = | $\left(\frac{1}{2}x_5 - y_5\right)a\,\hat{\mathbf{x}} + \frac{\sqrt{3}}{2}x_5a\,\hat{\mathbf{y}} + \left(\frac{1}{3} + z_5\right)c\,\hat{\mathbf{z}}$ | (3*a*) | O II |
| $\mathbf{B}_{15}$ | = | $(-x_5 + y_5)\,\mathbf{a}_1 - x_5\,\mathbf{a}_2 + \left(\frac{2}{3} + z_5\right)\mathbf{a}_3$ | = | $\left(-x_5 + \frac{1}{2}y_5\right)a\,\hat{\mathbf{x}} - \frac{\sqrt{3}}{2}y_5a\,\hat{\mathbf{y}} + \left(\frac{2}{3} + z_5\right)c\,\hat{\mathbf{z}}$ | (3*a*) | O II |
| $\mathbf{B}_{16}$ | = | $x_6\,\mathbf{a}_1 + y_6\,\mathbf{a}_2 + z_6\,\mathbf{a}_3$ | = | $\frac{1}{2}(x_6 + y_6)\,a\,\hat{\mathbf{x}} + \frac{\sqrt{3}}{2}(-x_6 + y_6)\,a\,\hat{\mathbf{y}} + z_6c\,\hat{\mathbf{z}}$ | (3*a*) | O III |
| $\mathbf{B}_{17}$ | = | $-y_6\,\mathbf{a}_1 + (x_6 - y_6)\,\mathbf{a}_2 + \left(\frac{1}{3} + z_6\right)\mathbf{a}_3$ | = | $\left(\frac{1}{2}x_6 - y_6\right)a\,\hat{\mathbf{x}} + \frac{\sqrt{3}}{2}x_6a\,\hat{\mathbf{y}} + \left(\frac{1}{3} + z_6\right)c\,\hat{\mathbf{z}}$ | (3*a*) | O III |
| $\mathbf{B}_{18}$ | = | $(-x_6 + y_6)\,\mathbf{a}_1 - x_6\,\mathbf{a}_2 + \left(\frac{2}{3} + z_6\right)\mathbf{a}_3$ | = | $\left(-x_6 + \frac{1}{2}y_6\right)a\,\hat{\mathbf{x}} - \frac{\sqrt{3}}{2}y_6a\,\hat{\mathbf{y}} + \left(\frac{2}{3} + z_6\right)c\,\hat{\mathbf{z}}$ | (3*a*) | O III |
| $\mathbf{B}_{19}$ | = | $x_7\,\mathbf{a}_1 + y_7\,\mathbf{a}_2 + z_7\,\mathbf{a}_3$ | = | $\frac{1}{2}(x_7 + y_7)\,a\,\hat{\mathbf{x}} + \frac{\sqrt{3}}{2}(-x_7 + y_7)\,a\,\hat{\mathbf{y}} + z_7c\,\hat{\mathbf{z}}$ | (3*a*) | O IV |
| $\mathbf{B}_{20}$ | = | $-y_7\,\mathbf{a}_1 + (x_7 - y_7)\,\mathbf{a}_2 + \left(\frac{1}{3} + z_7\right)\mathbf{a}_3$ | = | $\left(\frac{1}{2}x_7 - y_7\right)a\,\hat{\mathbf{x}} + \frac{\sqrt{3}}{2}x_7a\,\hat{\mathbf{y}} + \left(\frac{1}{3} + z_7\right)c\,\hat{\mathbf{z}}$ | (3*a*) | O IV |
| $\mathbf{B}_{21}$ | = | $(-x_7 + y_7)\,\mathbf{a}_1 - x_7\,\mathbf{a}_2 + \left(\frac{2}{3} + z_7\right)\mathbf{a}_3$ | = | $\left(-x_7 + \frac{1}{2}y_7\right)a\,\hat{\mathbf{x}} - \frac{\sqrt{3}}{2}y_7a\,\hat{\mathbf{y}} + \left(\frac{2}{3} + z_7\right)c\,\hat{\mathbf{z}}$ | (3*a*) | O IV |
| $\mathbf{B}_{22}$ | = | $x_8\,\mathbf{a}_1 + y_8\,\mathbf{a}_2 + z_8\,\mathbf{a}_3$ | = | $\frac{1}{2}(x_8 + y_8)\,a\,\hat{\mathbf{x}} + \frac{\sqrt{3}}{2}(-x_8 + y_8)\,a\,\hat{\mathbf{y}} + z_8c\,\hat{\mathbf{z}}$ | (3*a*) | O V |
| $\mathbf{B}_{23}$ | = | $-y_8\,\mathbf{a}_1 + (x_8 - y_8)\,\mathbf{a}_2 + \left(\frac{1}{3} + z_8\right)\mathbf{a}_3$ | = | $\left(\frac{1}{2}x_8 - y_8\right)a\,\hat{\mathbf{x}} + \frac{\sqrt{3}}{2}x_8a\,\hat{\mathbf{y}} + \left(\frac{1}{3} + z_8\right)c\,\hat{\mathbf{z}}$ | (3*a*) | O V |
| $\mathbf{B}_{24}$ | = | $(-x_8 + y_8)\,\mathbf{a}_1 - x_8\,\mathbf{a}_2 + \left(\frac{2}{3} + z_8\right)\mathbf{a}_3$ | = | $\left(-x_8 + \frac{1}{2}y_8\right)a\,\hat{\mathbf{x}} - \frac{\sqrt{3}}{2}y_8a\,\hat{\mathbf{y}} + \left(\frac{2}{3} + z_8\right)c\,\hat{\mathbf{z}}$ | (3*a*) | O V |
| $\mathbf{B}_{25}$ | = | $x_9\,\mathbf{a}_1 + y_9\,\mathbf{a}_2 + z_9\,\mathbf{a}_3$ | = | $\frac{1}{2}(x_9 + y_9)\,a\,\hat{\mathbf{x}} + \frac{\sqrt{3}}{2}(-x_9 + y_9)\,a\,\hat{\mathbf{y}} + z_9c\,\hat{\mathbf{z}}$ | (3*a*) | O VI |
| $\mathbf{B}_{26}$ | = | $-y_9\,\mathbf{a}_1 + (x_9 - y_9)\,\mathbf{a}_2 + \left(\frac{1}{3} + z_9\right)\mathbf{a}_3$ | = | $\left(\frac{1}{2}x_9 - y_9\right)a\,\hat{\mathbf{x}} + \frac{\sqrt{3}}{2}x_9a\,\hat{\mathbf{y}} + \left(\frac{1}{3} + z_9\right)c\,\hat{\mathbf{z}}$ | (3*a*) | O VI |
| $\mathbf{B}_{27}$ | = | $(-x_9 + y_9)\,\mathbf{a}_1 - x_9\,\mathbf{a}_2 + \left(\frac{2}{3} + z_9\right)\mathbf{a}_3$ | = | $\left(-x_9 + \frac{1}{2}y_9\right)a\,\hat{\mathbf{x}} - \frac{\sqrt{3}}{2}y_9a\,\hat{\mathbf{y}} + \left(\frac{2}{3} + z_9\right)c\,\hat{\mathbf{z}}$ | (3*a*) | O VI |
| $\mathbf{B}_{28}$ | = | $x_{10}\,\mathbf{a}_1 + y_{10}\,\mathbf{a}_2 + z_{10}\,\mathbf{a}_3$ | = | $\frac{1}{2}(x_{10} + y_{10})\,a\,\hat{\mathbf{x}} + \frac{\sqrt{3}}{2}(-x_{10} + y_{10})\,a\,\hat{\mathbf{y}} + z_{10}c\,\hat{\mathbf{z}}$ | (3*a*) | O VII |
| $\mathbf{B}_{29}$ | = | $-y_{10}\,\mathbf{a}_1 + (x_{10} - y_{10})\,\mathbf{a}_2 + \left(\frac{1}{3} + z_{10}\right)\mathbf{a}_3$ | = | $\left(\frac{1}{2}x_{10} - y_{10}\right)a\,\hat{\mathbf{x}} + \frac{\sqrt{3}}{2}x_{10}a\,\hat{\mathbf{y}} + \left(\frac{1}{3} + z_{10}\right)c\,\hat{\mathbf{z}}$ | (3*a*) | O VII |
| $\mathbf{B}_{30}$ | = | $(-x_{10} + y_{10})\,\mathbf{a}_1 - x_{10}\,\mathbf{a}_2 + \left(\frac{2}{3} + z_{10}\right)\mathbf{a}_3$ | = | $\left(-x_{10} + \frac{1}{2}y_{10}\right)a\,\hat{\mathbf{x}} - \frac{\sqrt{3}}{2}y_{10}a\,\hat{\mathbf{y}} + \left(\frac{2}{3} + z_{10}\right)c\,\hat{\mathbf{z}}$ | (3*a*) | O VII |
| $\mathbf{B}_{31}$ | = | $x_{11}\,\mathbf{a}_1 + y_{11}\,\mathbf{a}_2 + z_{11}\,\mathbf{a}_3$ | = | $\frac{1}{2}(x_{11} + y_{11})\,a\,\hat{\mathbf{x}} + \frac{\sqrt{3}}{2}(-x_{11} + y_{11})\,a\,\hat{\mathbf{y}} + z_{11}c\,\hat{\mathbf{z}}$ | (3*a*) | O VIII |
| $\mathbf{B}_{32}$ | = | $-y_{11}\,\mathbf{a}_1 + (x_{11} - y_{11})\,\mathbf{a}_2 + \left(\frac{1}{3} + z_{11}\right)\mathbf{a}_3$ | = | $\left(\frac{1}{2}x_{11} - y_{11}\right)a\,\hat{\mathbf{x}} + \frac{\sqrt{3}}{2}x_{11}a\,\hat{\mathbf{y}} + \left(\frac{1}{3} + z_{11}\right)c\,\hat{\mathbf{z}}$ | (3*a*) | O VIII |
| $\mathbf{B}_{33}$ | = | $(-x_{11} + y_{11})\,\mathbf{a}_1 - x_{11}\,\mathbf{a}_2 + \left(\frac{2}{3} + z_{11}\right)\mathbf{a}_3$ | = | $\left(-x_{11} + \frac{1}{2}y_{11}\right)a\,\hat{\mathbf{x}} - \frac{\sqrt{3}}{2}y_{11}a\,\hat{\mathbf{y}} + \left(\frac{2}{3} + z_{11}\right)c\,\hat{\mathbf{z}}$ | (3*a*) | O VIII |
| $\mathbf{B}_{34}$ | = | $x_{12}\,\mathbf{a}_1 + y_{12}\,\mathbf{a}_2 + z_{12}\,\mathbf{a}_3$ | = | $\frac{1}{2}\left(x_{12} + y_{12}\right) a\,\hat{\mathbf{x}} + \frac{\sqrt{3}}{2}\left(-x_{12} + y_{12}\right) a\,\hat{\mathbf{y}} + z_{12} c\,\hat{\mathbf{z}}$ | (3a) | O IX |
| $\mathbf{B}_{35}$ | = | $-y_{12}\,\mathbf{a}_1 + \left(x_{12} - y_{12}\right)\,\mathbf{a}_2 + \left(\frac{1}{3} + z_{12}\right)\,\mathbf{a}_3$ | = | $\left(\frac{1}{2}x_{12} - y_{12}\right) a\,\hat{\mathbf{x}} + \frac{\sqrt{3}}{2}x_{12} a\,\hat{\mathbf{y}} + \left(\frac{1}{3} + z_{12}\right) c\,\hat{\mathbf{z}}$ | (3a) | O IX |
| $\mathbf{B}_{36}$ | = | $\left(-x_{12} + y_{12}\right)\,\mathbf{a}_1 - x_{12}\,\mathbf{a}_2 + \left(\frac{2}{3} + z_{12}\right)\,\mathbf{a}_3$ | = | $\left(-x_{12} + \frac{1}{2}y_{12}\right) a\,\hat{\mathbf{x}} - \frac{\sqrt{3}}{2}y_{12} a\,\hat{\mathbf{y}} + \left(\frac{2}{3} + z_{12}\right) c\,\hat{\mathbf{z}}$ | (3a) | O IX |
| $\mathbf{B}_{37}$ | = | $x_{13}\,\mathbf{a}_1 + y_{13}\,\mathbf{a}_2 + z_{13}\,\mathbf{a}_3$ | = | $\frac{1}{2}\left(x_{13} + y_{13}\right) a\,\hat{\mathbf{x}} + \frac{\sqrt{3}}{2}\left(-x_{13} + y_{13}\right) a\,\hat{\mathbf{y}} + z_{13} c\,\hat{\mathbf{z}}$ | (3a) | Rb I |
| $\mathbf{B}_{38}$ | = | $-y_{13}\,\mathbf{a}_1 + \left(x_{13} - y_{13}\right)\,\mathbf{a}_2 + \left(\frac{1}{3} + z_{13}\right)\,\mathbf{a}_3$ | = | $\left(\frac{1}{2}x_{13} - y_{13}\right) a\,\hat{\mathbf{x}} + \frac{\sqrt{3}}{2}x_{13} a\,\hat{\mathbf{y}} + \left(\frac{1}{3} + z_{13}\right) c\,\hat{\mathbf{z}}$ | (3a) | Rb I |
| $\mathbf{B}_{39}$ | = | $\left(-x_{13} + y_{13}\right)\,\mathbf{a}_1 - x_{13}\,\mathbf{a}_2 + \left(\frac{2}{3} + z_{13}\right)\,\mathbf{a}_3$ | = | $\left(-x_{13} + \frac{1}{2}y_{13}\right) a\,\hat{\mathbf{x}} - \frac{\sqrt{3}}{2}y_{13} a\,\hat{\mathbf{y}} + \left(\frac{2}{3} + z_{13}\right) c\,\hat{\mathbf{z}}$ | (3a) | Rb I |
| $\mathbf{B}_{40}$ | = | $x_{14}\,\mathbf{a}_1 + y_{14}\,\mathbf{a}_2 + z_{14}\,\mathbf{a}_3$ | = | $\frac{1}{2}\left(x_{14} + y_{14}\right) a\,\hat{\mathbf{x}} + \frac{\sqrt{3}}{2}\left(-x_{14} + y_{14}\right) a\,\hat{\mathbf{y}} + z_{14} c\,\hat{\mathbf{z}}$ | (3a) | Rb II |
| $\mathbf{B}_{41}$ | = | $-y_{14}\,\mathbf{a}_1 + \left(x_{14} - y_{14}\right)\,\mathbf{a}_2 + \left(\frac{1}{3} + z_{14}\right)\,\mathbf{a}_3$ | = | $\left(\frac{1}{2}x_{14} - y_{14}\right) a\,\hat{\mathbf{x}} + \frac{\sqrt{3}}{2}x_{14} a\,\hat{\mathbf{y}} + \left(\frac{1}{3} + z_{14}\right) c\,\hat{\mathbf{z}}$ | (3a) | Rb II |
| $\mathbf{B}_{42}$ | = | $\left(-x_{14} + y_{14}\right)\,\mathbf{a}_1 - x_{14}\,\mathbf{a}_2 + \left(\frac{2}{3} + z_{14}\right)\,\mathbf{a}_3$ | = | $\left(-x_{14} + \frac{1}{2}y_{14}\right) a\,\hat{\mathbf{x}} - \frac{\sqrt{3}}{2}y_{14} a\,\hat{\mathbf{y}} + \left(\frac{2}{3} + z_{14}\right) c\,\hat{\mathbf{z}}$ | (3a) | Rb II |
| $\mathbf{B}_{43}$ | = | $x_{15}\,\mathbf{a}_1 + y_{15}\,\mathbf{a}_2 + z_{15}\,\mathbf{a}_3$ | = | $\frac{1}{2}\left(x_{15} + y_{15}\right) a\,\hat{\mathbf{x}} + \frac{\sqrt{3}}{2}\left(-x_{15} + y_{15}\right) a\,\hat{\mathbf{y}} + z_{15} c\,\hat{\mathbf{z}}$ | (3a) | Rb III |
| $\mathbf{B}_{44}$ | = | $-y_{15}\,\mathbf{a}_1 + \left(x_{15} - y_{15}\right)\,\mathbf{a}_2 + \left(\frac{1}{3} + z_{15}\right)\,\mathbf{a}_3$ | = | $\left(\frac{1}{2}x_{15} - y_{15}\right) a\,\hat{\mathbf{x}} + \frac{\sqrt{3}}{2}x_{15} a\,\hat{\mathbf{y}} + \left(\frac{1}{3} + z_{15}\right) c\,\hat{\mathbf{z}}$ | (3a) | Rb III |
| $\mathbf{B}_{45}$ | = | $\left(-x_{15} + y_{15}\right)\,\mathbf{a}_1 - x_{15}\,\mathbf{a}_2 + \left(\frac{2}{3} + z_{15}\right)\,\mathbf{a}_3$ | = | $\left(-x_{15} + \frac{1}{2}y_{15}\right) a\,\hat{\mathbf{x}} - \frac{\sqrt{3}}{2}y_{15} a\,\hat{\mathbf{y}} + \left(\frac{2}{3} + z_{15}\right) c\,\hat{\mathbf{z}}$ | (3a) | Rb III |

---

---

**Geometry files:**

- CIF: pp. 1720
- POSCAR: pp. 1721

# $Na_2SO_3$ ($G3_2$) Structure: A2B3C_hP12_147_abd_g_d

http://aflow.org/prototype-encyclopedia/A2B3C_hP12_147_abd_g_d

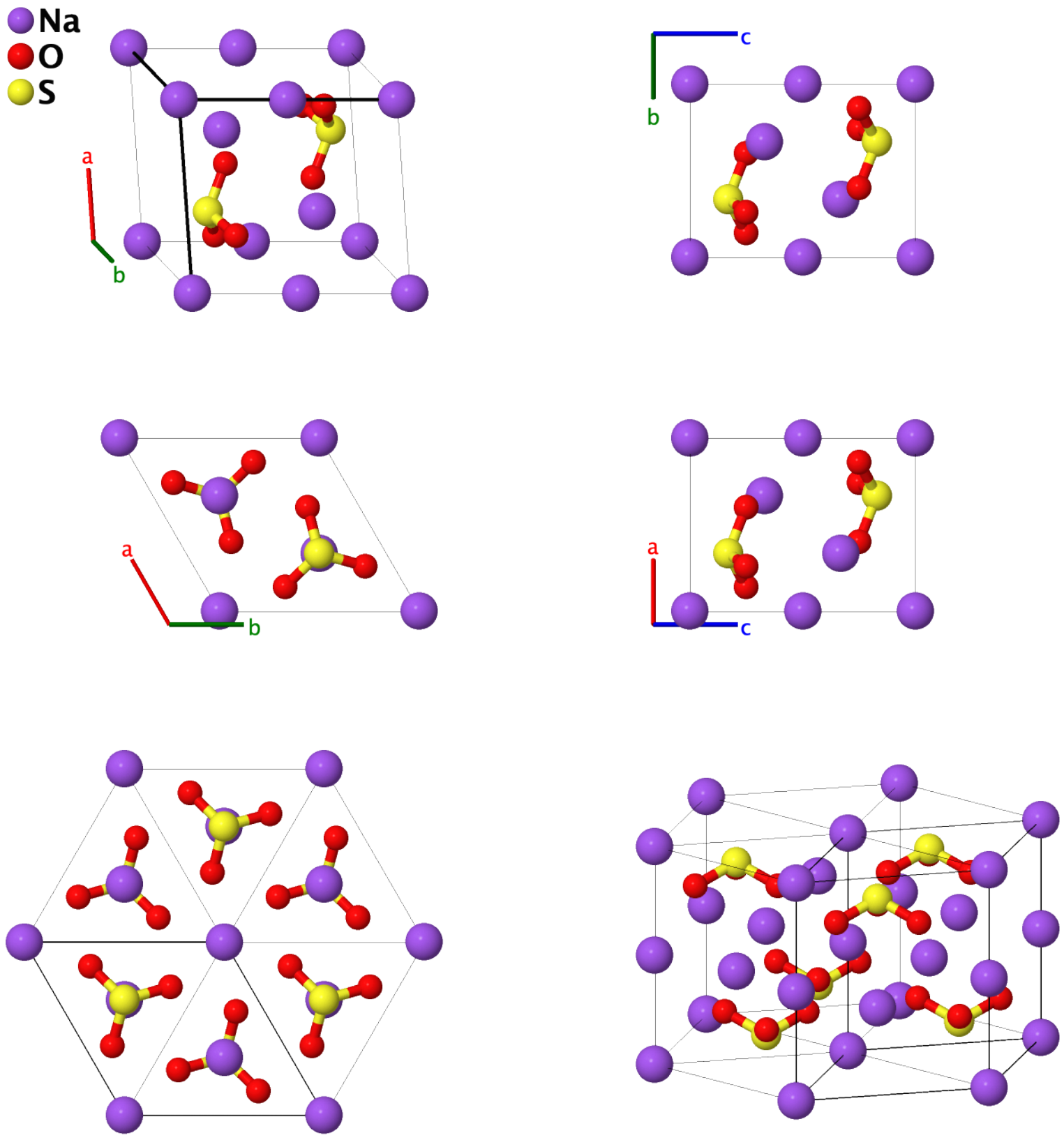

| | | |
|---|---|---|
| **Prototype** | : | $Na_2O_3S$ |
| **AFLOW prototype label** | : | A2B3C_hP12_147_abd_g_d |
| ***Strukturbericht* designation** | : | $G3_2$ |
| **Pearson symbol** | : | hP12 |
| **Space group number** | : | 147 |
| **Space group symbol** | : | $P\bar{3}$ |
| **AFLOW prototype command** | : | `aflow --proto=A2B3C_hP12_147_abd_g_d --params=`$a, c/a, z_3, z_4, x_5, y_5, z_5$ |

**Trigonal Hexagonal primitive vectors:**

$$\mathbf{a}_1 = \frac{1}{2}\,a\,\hat{\mathbf{x}} - \frac{\sqrt{3}}{2}\,a\,\hat{\mathbf{y}}$$
$$\mathbf{a}_2 = \frac{1}{2}\,a\,\hat{\mathbf{x}} + \frac{\sqrt{3}}{2}\,a\,\hat{\mathbf{y}}$$
$$\mathbf{a}_3 = c\,\hat{\mathbf{z}}$$

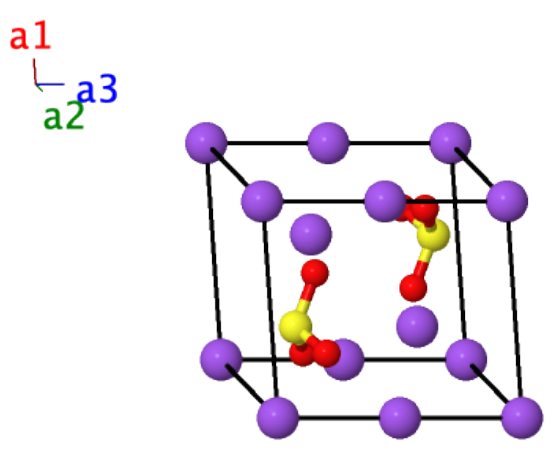

**Basis vectors:**

| | | Lattice Coordinates | | Cartesian Coordinates | Wyckoff Position | Atom Type |
|---|---|---|---|---|---|---|
| $\mathbf{B}_1$ | = | $0\,\mathbf{a}_1 + 0\,\mathbf{a}_2 + 0\,\mathbf{a}_3$ | = | $0\,\hat{\mathbf{x}} + 0\,\hat{\mathbf{y}} + 0\,\hat{\mathbf{z}}$ | $(1a)$ | Na I |
| $\mathbf{B}_2$ | = | $\frac{1}{2}\,\mathbf{a}_3$ | = | $\frac{1}{2}c\,\hat{\mathbf{z}}$ | (1$b$) | Na II |
| $\mathbf{B}_3$ | = | $\frac{1}{3}\,\mathbf{a}_1 + \frac{2}{3}\,\mathbf{a}_2 + z_3\,\mathbf{a}_3$ | = | $\frac{1}{2}a\,\hat{\mathbf{x}} + \frac{1}{2\sqrt{3}}a\,\hat{\mathbf{y}} + z_3c\,\hat{\mathbf{z}}$ | (2$d$) | Na III |
| $\mathbf{B}_4$ | = | $\frac{2}{3}\,\mathbf{a}_1 + \frac{1}{3}\,\mathbf{a}_2 - z_3\,\mathbf{a}_3$ | = | $\frac{1}{2}a\,\hat{\mathbf{x}} - \frac{1}{2\sqrt{3}}a\,\hat{\mathbf{y}} - z_3c\,\hat{\mathbf{z}}$ | (2$d$) | Na III |
| $\mathbf{B}_5$ | = | $\frac{1}{3}\,\mathbf{a}_1 + \frac{2}{3}\,\mathbf{a}_2 + z_4\,\mathbf{a}_3$ | = | $\frac{1}{2}a\,\hat{\mathbf{x}} + \frac{1}{2\sqrt{3}}a\,\hat{\mathbf{y}} + z_4c\,\hat{\mathbf{z}}$ | (2$d$) | S |
| $\mathbf{B}_6$ | = | $\frac{2}{3}\,\mathbf{a}_1 + \frac{1}{3}\,\mathbf{a}_2 - z_4\,\mathbf{a}_3$ | = | $\frac{1}{2}a\,\hat{\mathbf{x}} - \frac{1}{2\sqrt{3}}a\,\hat{\mathbf{y}} - z_4c\,\hat{\mathbf{z}}$ | (2$d$) | S |
| $\mathbf{B}_7$ | = | $x_5\,\mathbf{a}_1 + y_5\,\mathbf{a}_2 + z_5\,\mathbf{a}_3$ | = | $\frac{1}{2}\left(x_5 + y_5\right)a\,\hat{\mathbf{x}} + \frac{\sqrt{3}}{2}\left(-x_5 + y_5\right)a\,\hat{\mathbf{y}} + z_5c\,\hat{\mathbf{z}}$ | (6$g$) | O |
| $\mathbf{B}_8$ | = | $-y_5\,\mathbf{a}_1 + \left(x_5 - y_5\right)\,\mathbf{a}_2 + z_5\,\mathbf{a}_3$ | = | $\left(\frac{1}{2}x_5 - y_5\right)a\,\hat{\mathbf{x}} + \frac{\sqrt{3}}{2}x_5a\,\hat{\mathbf{y}} + z_5c\,\hat{\mathbf{z}}$ | (6$g$) | O |
| $\mathbf{B}_9$ | = | $\left(-x_5 + y_5\right)\,\mathbf{a}_1 - x_5\,\mathbf{a}_2 + z_5\,\mathbf{a}_3$ | = | $\left(-x_5 + \frac{1}{2}y_5\right)a\,\hat{\mathbf{x}} - \frac{\sqrt{3}}{2}y_5a\,\hat{\mathbf{y}} + z_5c\,\hat{\mathbf{z}}$ | (6$g$) | O |
| $\mathbf{B}_{10}$ | = | $-x_5\,\mathbf{a}_1 - y_5\,\mathbf{a}_2 - z_5\,\mathbf{a}_3$ | = | $-\frac{1}{2}\left(x_5 + y_5\right)a\,\hat{\mathbf{x}} + \frac{\sqrt{3}}{2}\left(x_5 - y_5\right)a\,\hat{\mathbf{y}} - z_5c\,\hat{\mathbf{z}}$ | (6$g$) | O |
| $\mathbf{B}_{11}$ | = | $y_5\,\mathbf{a}_1 + \left(-x_5 + y_5\right)\,\mathbf{a}_2 - z_5\,\mathbf{a}_3$ | = | $\left(-\frac{1}{2}x_5 + y_5\right)a\,\hat{\mathbf{x}} - \frac{\sqrt{3}}{2}x_5a\,\hat{\mathbf{y}} - z_5c\,\hat{\mathbf{z}}$ | (6$g$) | O |
| $\mathbf{B}_{12}$ | = | $\left(x_5 - y_5\right)\,\mathbf{a}_1 + x_5\,\mathbf{a}_2 - z_5\,\mathbf{a}_3$ | = | $\left(x_5 - \frac{1}{2}y_5\right)a\,\hat{\mathbf{x}} + \frac{\sqrt{3}}{2}y_5a\,\hat{\mathbf{y}} - z_5c\,\hat{\mathbf{z}}$ | (6$g$) | O |

**Geometry files:**

- CIF: pp. 1721
- POSCAR: pp. 1721

# Dolomite [$MgCa(CO_3)_2$, $G1_1$] Structure: A2BCD6_hR10_148_c_a_b_f

http://aflow.org/prototype-encyclopedia/A2BCD6_hR10_148_c_a_b_f

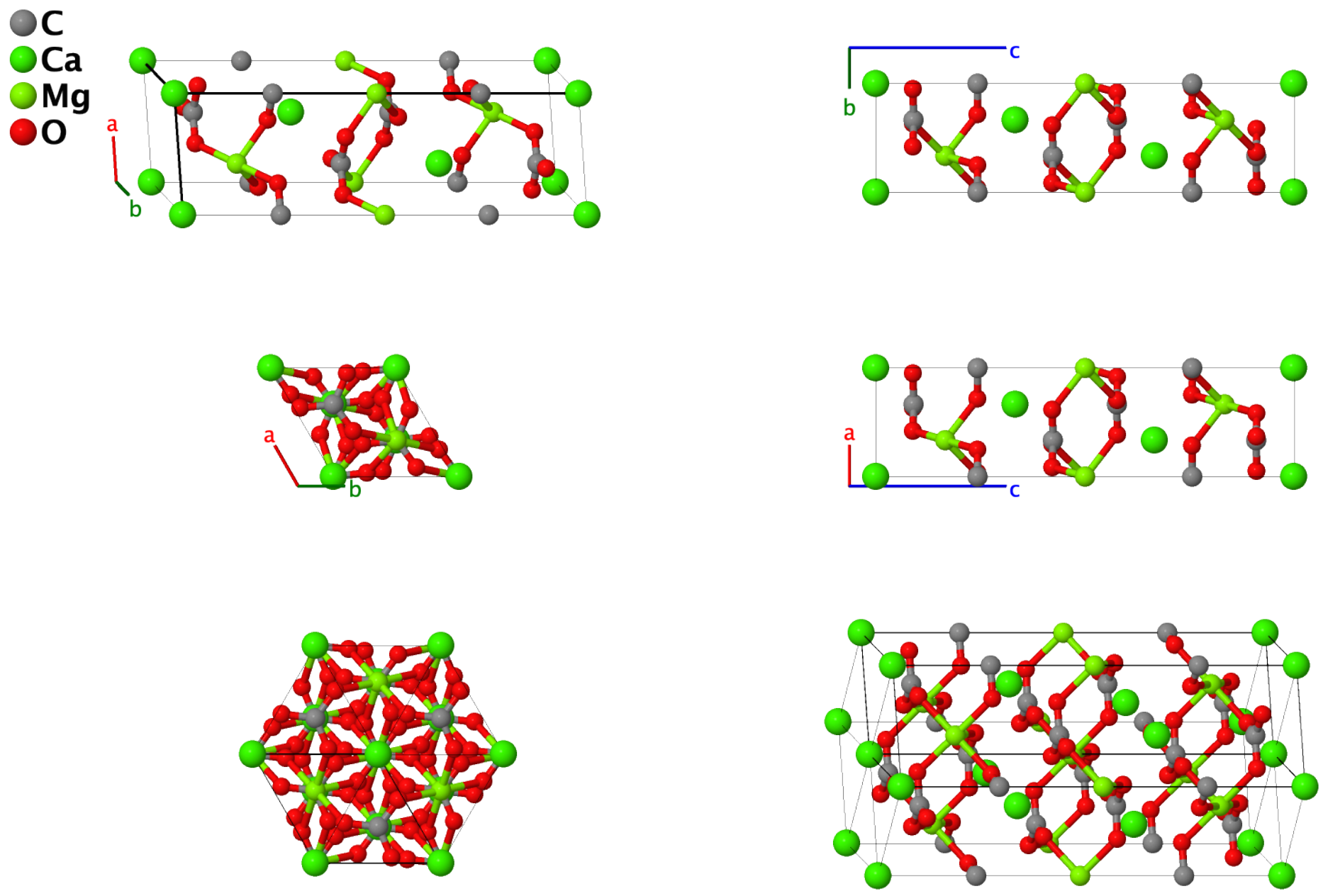

| | | |
|---|---|---|
| **Prototype** | : | $C_2CaMgO_6$ |
| **AFLOW prototype label** | : | A2BCD6_hR10_148_c_a_b_f |
| ***Strukturbericht* designation** | : | $G1_1$ |
| **Pearson symbol** | : | hR10 |
| **Space group number** | : | 148 |
| **Space group symbol** | : | $R\bar{3}$ |
| **AFLOW prototype command** | : | `aflow --proto=A2BCD6_hR10_148_c_a_b_f [--hex]` `--params=`$a, c/a, x_3, x_4, y_4, z_4$ |

- Data was taken at 24 °C.

**Rhombohedral primitive vectors:**

$$\mathbf{a}_1 = \frac{1}{2}\,a\,\hat{\mathbf{x}} - \frac{1}{2\sqrt{3}}\,a\,\hat{\mathbf{y}} + \frac{1}{3}\,c\,\hat{\mathbf{z}}$$
$$\mathbf{a}_2 = \frac{1}{\sqrt{3}}\,a\,\hat{\mathbf{y}} + \frac{1}{3}\,c\,\hat{\mathbf{z}}$$
$$\mathbf{a}_3 = -\frac{1}{2}\,a\,\hat{\mathbf{x}} - \frac{1}{2\sqrt{3}}\,a\,\hat{\mathbf{y}} + \frac{1}{3}\,c\,\hat{\mathbf{z}}$$

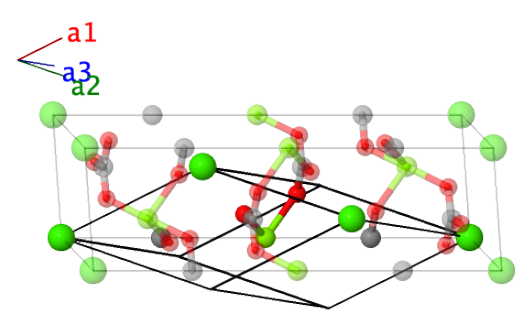

**Basis vectors:**

| | | Lattice Coordinates | | Cartesian Coordinates | Wyckoff Position | Atom Type |
|---|---|---|---|---|---|---|
| $\mathbf{B}_1$ | = | $0\,\mathbf{a}_1 + 0\,\mathbf{a}_2 + 0\,\mathbf{a}_3$ | = | $0\,\hat{\mathbf{x}} + 0\,\hat{\mathbf{y}} + 0\,\hat{\mathbf{z}}$ | (1$a$) | Ca |
| $\mathbf{B}_2$ | = | $\frac{1}{2}\,\mathbf{a}_1 + \frac{1}{2}\,\mathbf{a}_2 + \frac{1}{2}\,\mathbf{a}_3$ | = | $\frac{1}{2}c\,\hat{\mathbf{z}}$ | (1$b$) | Mg |
| $\mathbf{B}_3$ | = | $x_3\,\mathbf{a}_1 + x_3\,\mathbf{a}_2 + x_3\,\mathbf{a}_3$ | = | $x_3 c\,\hat{\mathbf{z}}$ | (2$c$) | C |

$$\mathbf{B}_4 = -x_3\,\mathbf{a}_1 - x_3\,\mathbf{a}_2 - x_3\,\mathbf{a}_3 = -x_3 c\,\hat{\mathbf{z}} \quad (2c) \quad \text{C}$$

$$\mathbf{B}_5 = x_4\,\mathbf{a}_1 + y_4\,\mathbf{a}_2 + z_4\,\mathbf{a}_3 = \frac{1}{2}(x_4 - z_4)\,a\,\hat{\mathbf{x}} + \left(-\frac{1}{2\sqrt{3}}x_4 + \frac{1}{\sqrt{3}}y_4 - \frac{1}{2\sqrt{3}}z_4\right)a\,\hat{\mathbf{y}} + \frac{1}{3}(x_4 + y_4 + z_4)\,c\,\hat{\mathbf{z}} \quad (6f) \quad \text{O}$$

$$\mathbf{B}_6 = z_4\,\mathbf{a}_1 + x_4\,\mathbf{a}_2 + y_4\,\mathbf{a}_3 = \frac{1}{2}(-y_4 + z_4)\,a\,\hat{\mathbf{x}} + \left(\frac{1}{\sqrt{3}}x_4 - \frac{1}{2\sqrt{3}}y_4 - \frac{1}{2\sqrt{3}}z_4\right)a\,\hat{\mathbf{y}} + \frac{1}{3}(x_4 + y_4 + z_4)\,c\,\hat{\mathbf{z}} \quad (6f) \quad \text{O}$$

$$\mathbf{B}_7 = y_4\,\mathbf{a}_1 + z_4\,\mathbf{a}_2 + x_4\,\mathbf{a}_3 = \frac{1}{2}(-x_4 + y_4)\,a\,\hat{\mathbf{x}} + \left(-\frac{1}{2\sqrt{3}}x_4 - \frac{1}{2\sqrt{3}}y_4 + \frac{1}{\sqrt{3}}z_4\right)a\,\hat{\mathbf{y}} + \frac{1}{3}(x_4 + y_4 + z_4)\,c\,\hat{\mathbf{z}} \quad (6f) \quad \text{O}$$

$$\mathbf{B}_8 = -x_4\,\mathbf{a}_1 - y_4\,\mathbf{a}_2 - z_4\,\mathbf{a}_3 = \frac{1}{2}(-x_4 + z_4)\,a\,\hat{\mathbf{x}} + \left(\frac{1}{2\sqrt{3}}x_4 - \frac{1}{\sqrt{3}}y_4 + \frac{1}{2\sqrt{3}}z_4\right)a\,\hat{\mathbf{y}} - \frac{1}{3}(x_4 + y_4 + z_4)\,c\,\hat{\mathbf{z}} \quad (6f) \quad \text{O}$$

$$\mathbf{B}_9 = -z_4\,\mathbf{a}_1 - x_4\,\mathbf{a}_2 - y_4\,\mathbf{a}_3 = \frac{1}{2}(y_4 - z_4)\,a\,\hat{\mathbf{x}} + \left(-\frac{1}{\sqrt{3}}x_4 + \frac{1}{2\sqrt{3}}y_4 + \frac{1}{2\sqrt{3}}z_4\right)a\,\hat{\mathbf{y}} - \frac{1}{3}(x_4 + y_4 + z_4)\,c\,\hat{\mathbf{z}} \quad (6f) \quad \text{O}$$

$$\mathbf{B}_{10} = -y_4\,\mathbf{a}_1 - z_4\,\mathbf{a}_2 - x_4\,\mathbf{a}_3 = \frac{1}{2}(x_4 - y_4)\,a\,\hat{\mathbf{x}} + \left(\frac{1}{2\sqrt{3}}x_4 + \frac{1}{2\sqrt{3}}y_4 - \frac{1}{\sqrt{3}}z_4\right)a\,\hat{\mathbf{y}} - \frac{1}{3}(x_4 + y_4 + z_4)\,c\,\hat{\mathbf{z}} \quad (6f) \quad \text{O}$$

---

---

**Geometry files:**

- CIF: pp. 1721
- POSCAR: pp. 1722

# $K_2Sn(OH)_6$ ($H6_2$) Structure: A6B2C6D_hR15_148_f_c_f_a

http://aflow.org/prototype-encyclopedia/A6B2C6D_hR15_148_f_c_f_a

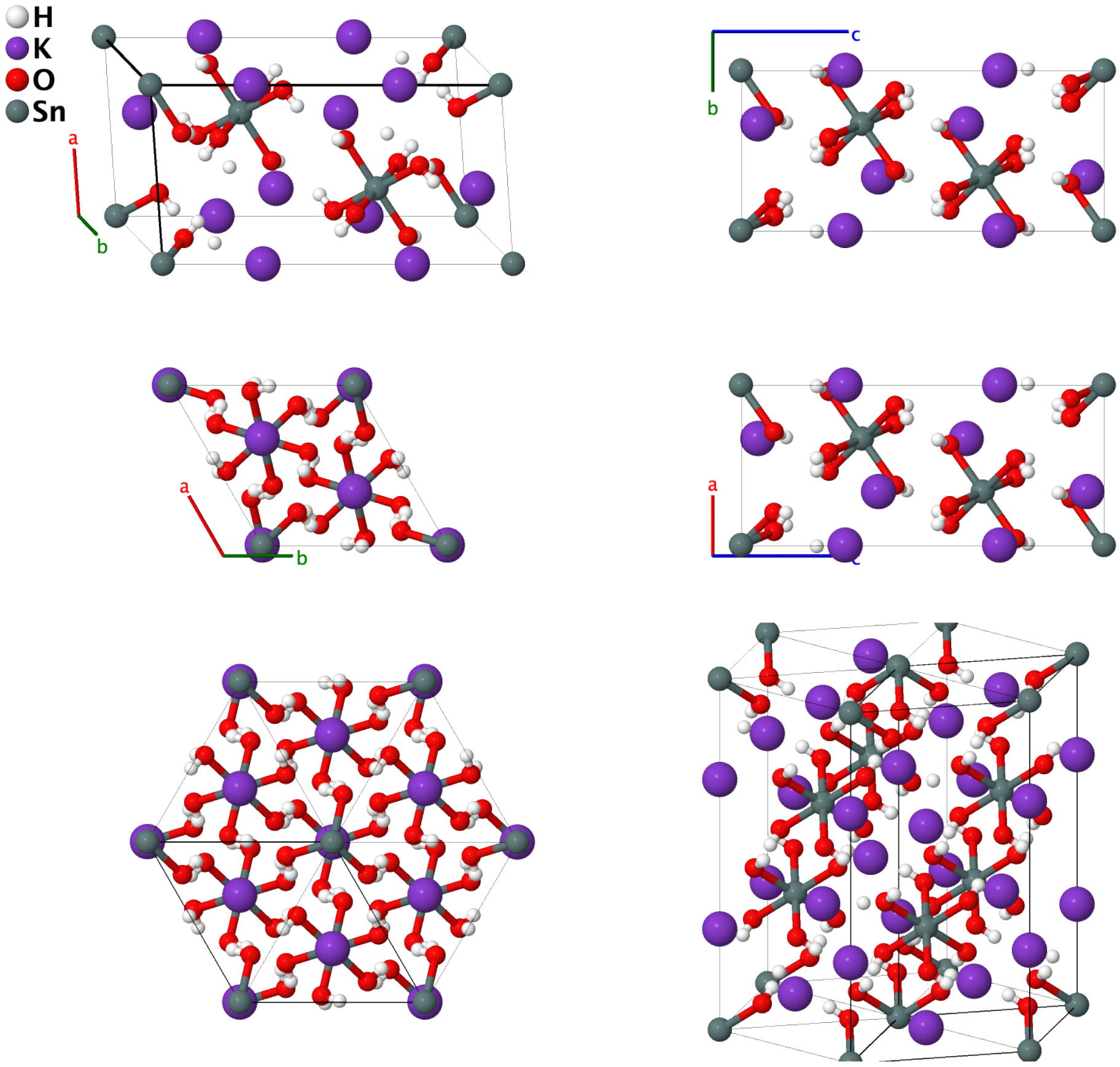

| | | |
|---|---|---|
| **Prototype** | : | $H_6K_2O_6Sn$ |
| **AFLOW prototype label** | : | A6B2C6D_hR15_148_f_c_f_a |
| ***Strukturbericht* designation** | : | $H6_2$ |
| **Pearson symbol** | : | hR15 |
| **Space group number** | : | 148 |
| **Space group symbol** | : | $R\bar{3}$ |
| **AFLOW prototype command** | : | `aflow --proto=A6B2C6D_hR15_148_f_c_f_a [--hex]` `--params=`$a, c/a, x_2, x_3, y_3, z_3, x_4, y_4, z_4$ |

### Other compounds with this structure

- $K_2Pt(OH)_6$ and $Na_2Sn(OH)_6$

- (Wyckoff, 1928) found the positions of the potassium and tin atoms in this structure. He could not locate the oxygen atoms, and so was unable to give an unequivocal determination of the space group, beyond noting that it was rhombohedral. We use the modern structure of (Jacobs, 2000), which agrees with Wyckoff in the positioning of the potassium and tin atoms, and finds both the hydrogen and oxygen locations.

- (Ewald, 1931) gave this the *Strukturbericht* designation $H6_2$. (Hermann, 1937) began relabeling the $H6$ structures as $I1$, and (Gottfried, 1937) relabeled them as $J1$, but neither they nor successor volumes revisited this structure, so it was never formally relabeled $J1_2$. We will therefore leave it as $H6_2$.

**Rhombohedral primitive vectors:**

$$\mathbf{a}_1 = \frac{1}{2}\,a\,\hat{\mathbf{x}} - \frac{1}{2\sqrt{3}}\,a\,\hat{\mathbf{y}} + \frac{1}{3}\,c\,\hat{\mathbf{z}}$$
$$\mathbf{a}_2 = \frac{1}{\sqrt{3}}\,a\,\hat{\mathbf{y}} + \frac{1}{3}\,c\,\hat{\mathbf{z}}$$
$$\mathbf{a}_3 = -\frac{1}{2}\,a\,\hat{\mathbf{x}} - \frac{1}{2\sqrt{3}}\,a\,\hat{\mathbf{y}} + \frac{1}{3}\,c\,\hat{\mathbf{z}}$$

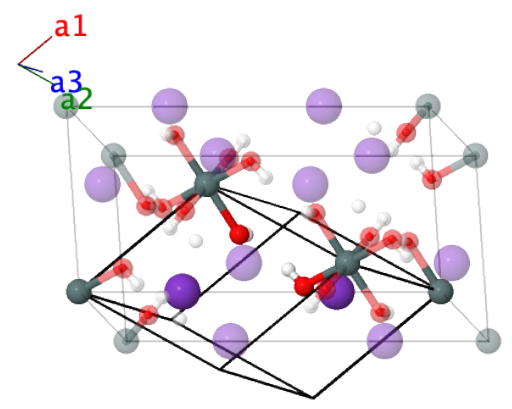

**Basis vectors:**

| | | Lattice Coordinates | | Cartesian Coordinates | Wyckoff Position | Atom Type |
|---|---|---|---|---|---|---|
| $\mathbf{B}_1$ | = | $0\,\mathbf{a}_1 + 0\,\mathbf{a}_2 + 0\,\mathbf{a}_3$ | = | $0\,\hat{\mathbf{x}} + 0\,\hat{\mathbf{y}} + 0\,\hat{\mathbf{z}}$ | (1$a$) | Sn |
| $\mathbf{B}_2$ | = | $x_2\,\mathbf{a}_1 + x_2\,\mathbf{a}_2 + x_2\,\mathbf{a}_3$ | = | $x_2 c\,\hat{\mathbf{z}}$ | (2$c$) | K |
| $\mathbf{B}_3$ | = | $-x_2\,\mathbf{a}_1 - x_2\,\mathbf{a}_2 - x_2\,\mathbf{a}_3$ | = | $-x_2 c\,\hat{\mathbf{z}}$ | (2$c$) | K |
| $\mathbf{B}_4$ | = | $x_3\,\mathbf{a}_1 + y_3\,\mathbf{a}_2 + z_3\,\mathbf{a}_3$ | = | $\frac{1}{2}(x_3 - z_3)\,a\,\hat{\mathbf{x}} + \left(-\frac{1}{2\sqrt{3}}x_3 + \frac{1}{\sqrt{3}}y_3 - \frac{1}{2\sqrt{3}}z_3\right)a\,\hat{\mathbf{y}} + \frac{1}{3}(x_3 + y_3 + z_3)\,c\,\hat{\mathbf{z}}$ | (6$f$) | H |
| $\mathbf{B}_5$ | = | $z_3\,\mathbf{a}_1 + x_3\,\mathbf{a}_2 + y_3\,\mathbf{a}_3$ | = | $\frac{1}{2}(-y_3 + z_3)\,a\,\hat{\mathbf{x}} + \left(\frac{1}{\sqrt{3}}x_3 - \frac{1}{2\sqrt{3}}y_3 - \frac{1}{2\sqrt{3}}z_3\right)a\,\hat{\mathbf{y}} + \frac{1}{3}(x_3 + y_3 + z_3)\,c\,\hat{\mathbf{z}}$ | (6$f$) | H |
| $\mathbf{B}_6$ | = | $y_3\,\mathbf{a}_1 + z_3\,\mathbf{a}_2 + x_3\,\mathbf{a}_3$ | = | $\frac{1}{2}(-x_3 + y_3)\,a\,\hat{\mathbf{x}} + \left(-\frac{1}{2\sqrt{3}}x_3 - \frac{1}{2\sqrt{3}}y_3 + \frac{1}{\sqrt{3}}z_3\right)a\,\hat{\mathbf{y}} + \frac{1}{3}(x_3 + y_3 + z_3)\,c\,\hat{\mathbf{z}}$ | (6$f$) | H |
| $\mathbf{B}_7$ | = | $-x_3\,\mathbf{a}_1 - y_3\,\mathbf{a}_2 - z_3\,\mathbf{a}_3$ | = | $\frac{1}{2}(-x_3 + z_3)\,a\,\hat{\mathbf{x}} + \left(\frac{1}{2\sqrt{3}}x_3 - \frac{1}{\sqrt{3}}y_3 + \frac{1}{2\sqrt{3}}z_3\right)a\,\hat{\mathbf{y}} - \frac{1}{3}(x_3 + y_3 + z_3)\,c\,\hat{\mathbf{z}}$ | (6$f$) | H |
| $\mathbf{B}_8$ | = | $-z_3\,\mathbf{a}_1 - x_3\,\mathbf{a}_2 - y_3\,\mathbf{a}_3$ | = | $\frac{1}{2}(y_3 - z_3)\,a\,\hat{\mathbf{x}} + \left(-\frac{1}{\sqrt{3}}x_3 + \frac{1}{2\sqrt{3}}y_3 + \frac{1}{2\sqrt{3}}z_3\right)a\,\hat{\mathbf{y}} - \frac{1}{3}(x_3 + y_3 + z_3)\,c\,\hat{\mathbf{z}}$ | (6$f$) | H |
| $\mathbf{B}_9$ | = | $-y_3\,\mathbf{a}_1 - z_3\,\mathbf{a}_2 - x_3\,\mathbf{a}_3$ | = | $\frac{1}{2}(x_3 - y_3)\,a\,\hat{\mathbf{x}} + \left(\frac{1}{2\sqrt{3}}x_3 + \frac{1}{2\sqrt{3}}y_3 - \frac{1}{\sqrt{3}}z_3\right)a\,\hat{\mathbf{y}} - \frac{1}{3}(x_3 + y_3 + z_3)\,c\,\hat{\mathbf{z}}$ | (6$f$) | H |
| $\mathbf{B}_{10}$ | = | $x_4\,\mathbf{a}_1 + y_4\,\mathbf{a}_2 + z_4\,\mathbf{a}_3$ | = | $\frac{1}{2}(x_4 - z_4)\,a\,\hat{\mathbf{x}} + \left(-\frac{1}{2\sqrt{3}}x_4 + \frac{1}{\sqrt{3}}y_4 - \frac{1}{2\sqrt{3}}z_4\right)a\,\hat{\mathbf{y}} + \frac{1}{3}(x_4 + y_4 + z_4)\,c\,\hat{\mathbf{z}}$ | (6$f$) | O |
| $\mathbf{B}_{11}$ | = | $z_4\,\mathbf{a}_1 + x_4\,\mathbf{a}_2 + y_4\,\mathbf{a}_3$ | = | $\frac{1}{2}(-y_4 + z_4)\,a\,\hat{\mathbf{x}} + \left(\frac{1}{\sqrt{3}}x_4 - \frac{1}{2\sqrt{3}}y_4 - \frac{1}{2\sqrt{3}}z_4\right)a\,\hat{\mathbf{y}} + \frac{1}{3}(x_4 + y_4 + z_4)\,c\,\hat{\mathbf{z}}$ | (6$f$) | O |
| $\mathbf{B}_{12}$ | = | $y_4\,\mathbf{a}_1 + z_4\,\mathbf{a}_2 + x_4\,\mathbf{a}_3$ | = | $\frac{1}{2}(-x_4 + y_4)\,a\,\hat{\mathbf{x}} + \left(-\frac{1}{2\sqrt{3}}x_4 - \frac{1}{2\sqrt{3}}y_4 + \frac{1}{\sqrt{3}}z_4\right)a\,\hat{\mathbf{y}} + \frac{1}{3}(x_4 + y_4 + z_4)\,c\,\hat{\mathbf{z}}$ | (6$f$) | O |
| $\mathbf{B}_{13}$ | = | $-x_4\,\mathbf{a}_1 - y_4\,\mathbf{a}_2 - z_4\,\mathbf{a}_3$ | = | $\frac{1}{2}(-x_4 + z_4)\,a\,\hat{\mathbf{x}} + \left(\frac{1}{2\sqrt{3}}x_4 - \frac{1}{\sqrt{3}}y_4 + \frac{1}{2\sqrt{3}}z_4\right)a\,\hat{\mathbf{y}} - \frac{1}{3}(x_4 + y_4 + z_4)\,c\,\hat{\mathbf{z}}$ | (6$f$) | O |
| $\mathbf{B}_{14}$ | = | $-z_4\,\mathbf{a}_1 - x_4\,\mathbf{a}_2 - y_4\,\mathbf{a}_3$ | = | $\frac{1}{2}(y_4 - z_4)\,a\,\hat{\mathbf{x}} + \left(-\frac{1}{\sqrt{3}}x_4 + \frac{1}{2\sqrt{3}}y_4 + \frac{1}{2\sqrt{3}}z_4\right)a\,\hat{\mathbf{y}} - \frac{1}{3}(x_4 + y_4 + z_4)\,c\,\hat{\mathbf{z}}$ | (6$f$) | O |
| $\mathbf{B}_{15}$ | = | $-y_4\,\mathbf{a}_1 - z_4\,\mathbf{a}_2 - x_4\,\mathbf{a}_3$ | = | $\frac{1}{2}(x_4 - y_4)\,a\,\hat{\mathbf{x}} + \left(\frac{1}{2\sqrt{3}}x_4 + \frac{1}{2\sqrt{3}}y_4 - \frac{1}{\sqrt{3}}z_4\right)a\,\hat{\mathbf{y}} - \frac{1}{3}(x_4 + y_4 + z_4)\,c\,\hat{\mathbf{z}}$ | (6$f$) | O |

**Found in:**

---

**Geometry files:**

- CIF: pp. 1722
- POSCAR: pp. 1722

# $Ni(H_2O)_6SnCl_6$ ($I6_1$) Structure: A6B6CD_hR14_148_f_f_b_a

http://aflow.org/prototype-encyclopedia/A6B6CD_hR14_148_f_f_b_a

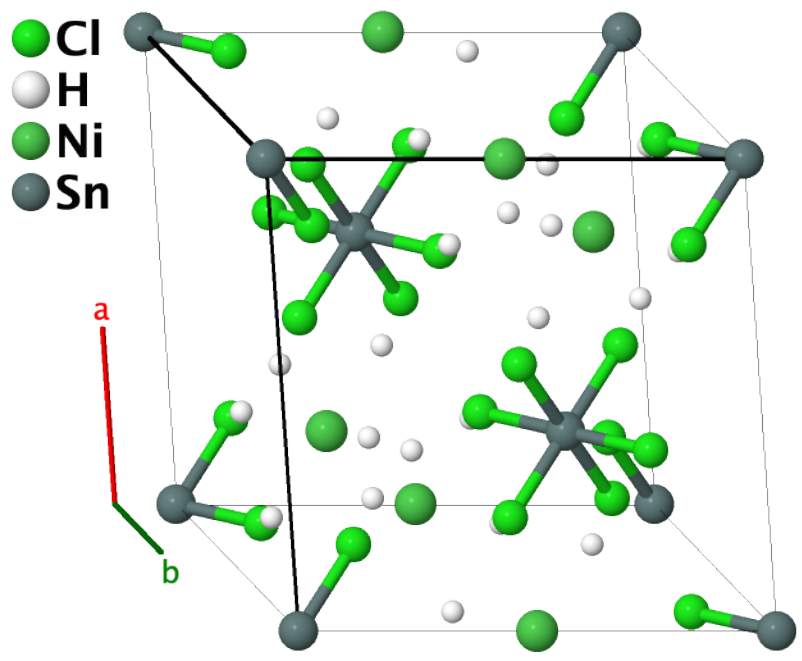

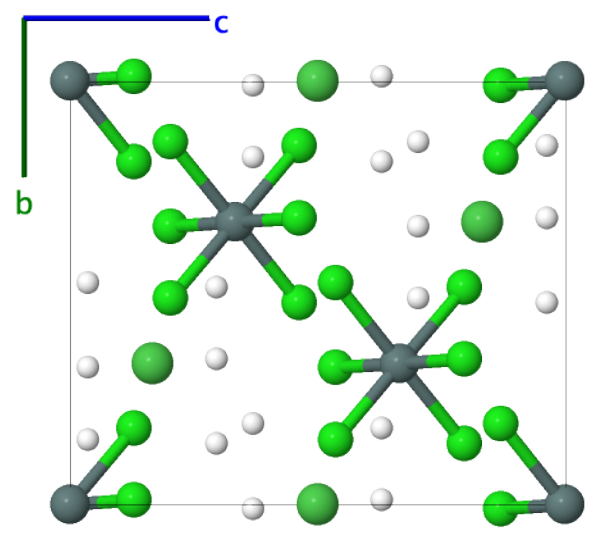

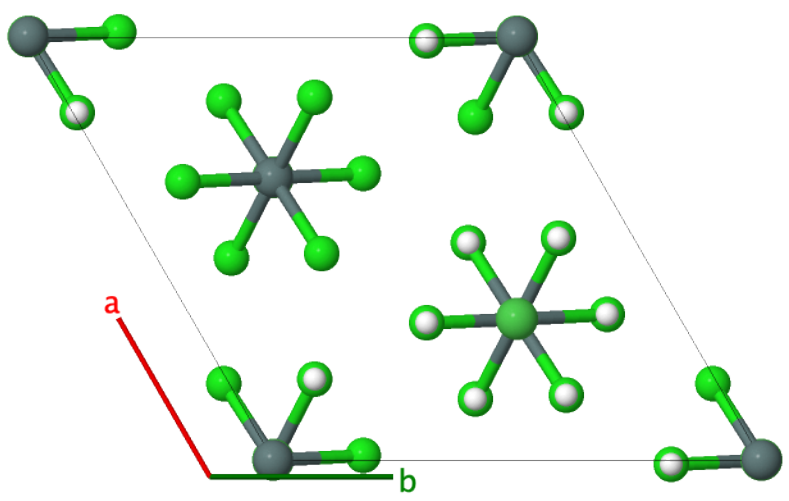

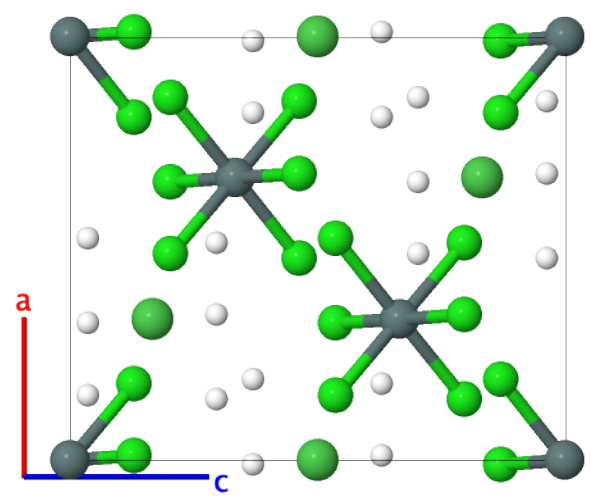

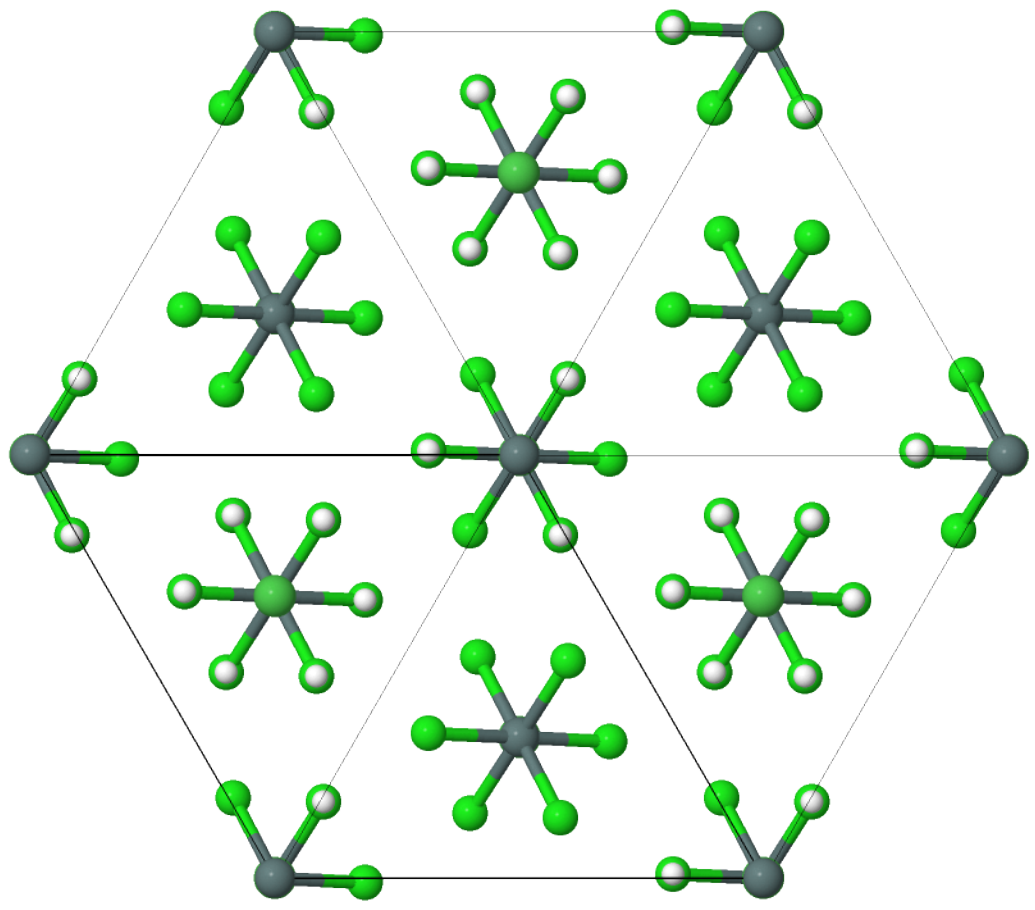

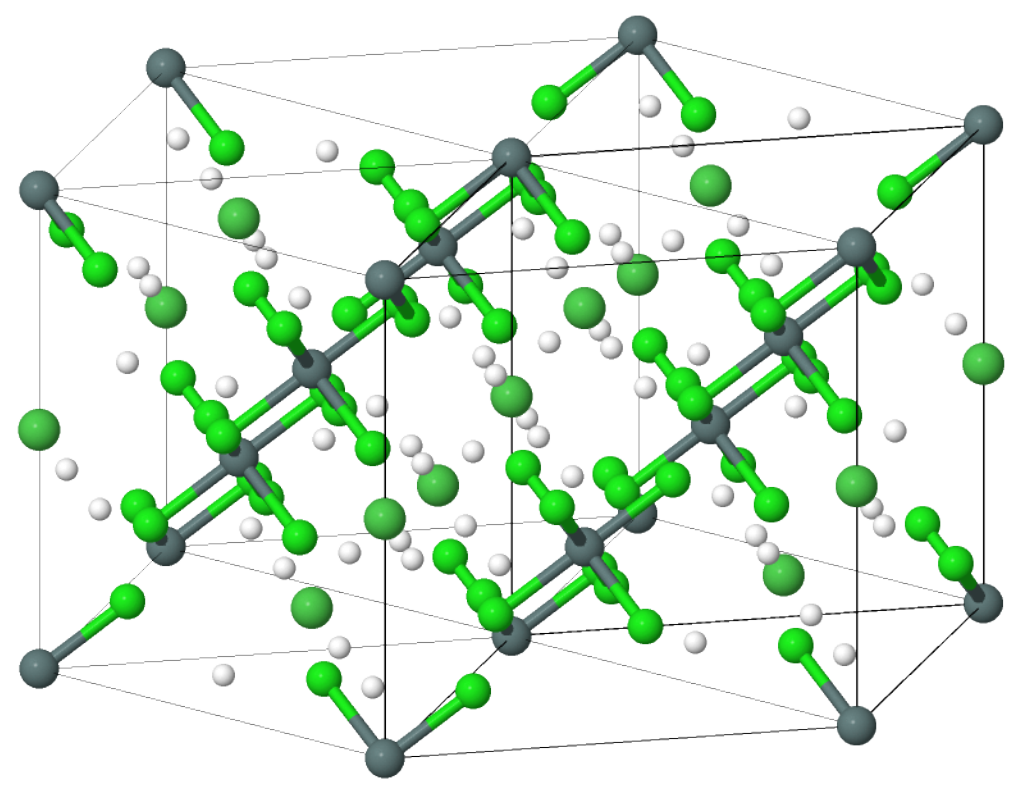

| | | |
|---|---|---|
| **Prototype** | : | $Cl_6(H_2O)_6NiSn$ |
| **AFLOW prototype label** | : | A6B6CD_hR14_148_f_f_b_a |
| ***Strukturbericht* designation** | : | $I6_1$ |
| **Pearson symbol** | : | hR14 |
| **Space group number** | : | 148 |
| **Space group symbol** | : | $R\bar{3}$ |
| **AFLOW prototype command** | : | `aflow --proto=A6B6CD_hR14_148_f_f_b_a [--hex]` `--params=`$a, c/a, x_3, y_3, z_3, x_4, y_4, z_4$ |

**Other compounds with this structure**

- $Ca(H_2O)_6SnF_6$, $Co(H_2O)_6PtF_6$, $Co(H_2O)_6SiF_6$, $Co(NH_3)_6Co(CN)_6$, $Fe(H_2O)_6SiF_6$, $Mg(H_2O)_6SiF_6$, $Mg(H_2O)_6SnF_6$, $Mg(H_2O)_6TiF_6$, $Mn(H_2O)_6SiF_6$, $Ni(H_2O)_6SiF_6$, $Zn(H_2O)_6SiF_6$, $Zn(H_2O)_6SnF_6$, $Zn(H_2O)_6TiF_6$, and $Zn(H_2O)_6ZrF_6$

- $Ni(H_2O)_6SnCl_6$ is the prototype for a large class of molecular crystals with the form $MG_6LR_6$, where $MG_6$ is a cation and $LR_6$ is an anion. (Hermann, 1937) gave this the *Strukturbericht* designation $I6_1$. (Gottfried, 1937) changed the $I$ designations to $J$, so this should have become $J6_1$, but it was never referenced in any form in later volumes of *Strukturbericht*. Since $I6_1$ is the only designation for this structure in the literature, we use it rather than $J6_1$.

- The positions of the hydrogen atoms in the water molecules were not determined, so we only provide the positions of the oxygen atoms (labeled as $H_2O$).

**Rhombohedral primitive vectors:**

$$\mathbf{a}_1 = \frac{1}{2}\,a\,\hat{\mathbf{x}} - \frac{1}{2\sqrt{3}}\,a\,\hat{\mathbf{y}} + \frac{1}{3}\,c\,\hat{\mathbf{z}}$$

$$\mathbf{a}_2 = \frac{1}{\sqrt{3}}\,a\,\hat{\mathbf{y}} + \frac{1}{3}\,c\,\hat{\mathbf{z}}$$

$$\mathbf{a}_3 = -\frac{1}{2}\,a\,\hat{\mathbf{x}} - \frac{1}{2\sqrt{3}}\,a\,\hat{\mathbf{y}} + \frac{1}{3}\,c\,\hat{\mathbf{z}}$$

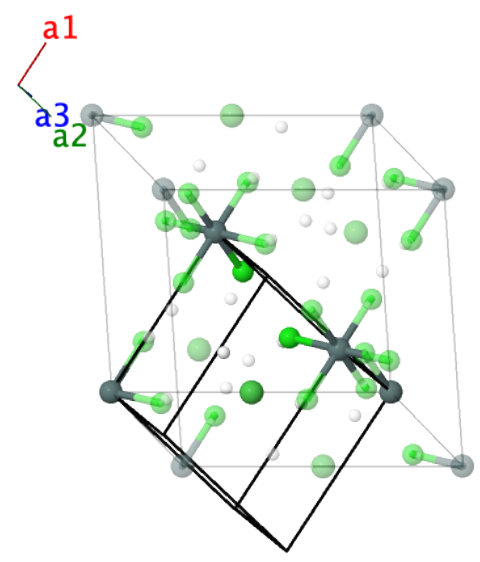

**Basis vectors:**

| | | Lattice Coordinates | | Cartesian Coordinates | Wyckoff Position | Atom Type |
|---|---|---|---|---|---|---|
| $\mathbf{B}_1$ | = | $0\,\mathbf{a}_1 + 0\,\mathbf{a}_2 + 0\,\mathbf{a}_3$ | = | $0\,\hat{\mathbf{x}} + 0\,\hat{\mathbf{y}} + 0\,\hat{\mathbf{z}}$ | (1$a$) | Sn |
| $\mathbf{B}_2$ | = | $\frac{1}{2}\,\mathbf{a}_1 + \frac{1}{2}\,\mathbf{a}_2 + \frac{1}{2}\,\mathbf{a}_3$ | = | $\frac{1}{2}c\,\hat{\mathbf{z}}$ | (1$b$) | Ni |
| $\mathbf{B}_3$ | = | $x_3\,\mathbf{a}_1 + y_3\,\mathbf{a}_2 + z_3\,\mathbf{a}_3$ | = | $\frac{1}{2}\left(x_3 - z_3\right)a\,\hat{\mathbf{x}} + \left(-\frac{1}{2\sqrt{3}}x_3 + \frac{1}{\sqrt{3}}y_3 - \frac{1}{2\sqrt{3}}z_3\right)a\,\hat{\mathbf{y}} + \frac{1}{3}\left(x_3 + y_3 + z_3\right)c\,\hat{\mathbf{z}}$ | (6$f$) | Cl |
| $\mathbf{B}_4$ | = | $z_3\,\mathbf{a}_1 + x_3\,\mathbf{a}_2 + y_3\,\mathbf{a}_3$ | = | $\frac{1}{2}\left(-y_3 + z_3\right)a\,\hat{\mathbf{x}} + \left(\frac{1}{\sqrt{3}}x_3 - \frac{1}{2\sqrt{3}}y_3 - \frac{1}{2\sqrt{3}}z_3\right)a\,\hat{\mathbf{y}} + \frac{1}{3}\left(x_3 + y_3 + z_3\right)c\,\hat{\mathbf{z}}$ | (6$f$) | Cl |
| $\mathbf{B}_5$ | = | $y_3\,\mathbf{a}_1 + z_3\,\mathbf{a}_2 + x_3\,\mathbf{a}_3$ | = | $\frac{1}{2}\left(-x_3 + y_3\right)a\,\hat{\mathbf{x}} + \left(-\frac{1}{2\sqrt{3}}x_3 - \frac{1}{2\sqrt{3}}y_3 + \frac{1}{\sqrt{3}}z_3\right)a\,\hat{\mathbf{y}} + \frac{1}{3}\left(x_3 + y_3 + z_3\right)c\,\hat{\mathbf{z}}$ | (6$f$) | Cl |
| $\mathbf{B}_6$ | = | $-x_3\,\mathbf{a}_1 - y_3\,\mathbf{a}_2 - z_3\,\mathbf{a}_3$ | = | $\frac{1}{2}\left(-x_3 + z_3\right)a\,\hat{\mathbf{x}} + \left(\frac{1}{2\sqrt{3}}x_3 - \frac{1}{\sqrt{3}}y_3 + \frac{1}{2\sqrt{3}}z_3\right)a\,\hat{\mathbf{y}} - \frac{1}{3}\left(x_3 + y_3 + z_3\right)c\,\hat{\mathbf{z}}$ | (6$f$) | Cl |
| $\mathbf{B}_7$ | = | $-z_3\,\mathbf{a}_1 - x_3\,\mathbf{a}_2 - y_3\,\mathbf{a}_3$ | = | $\frac{1}{2}\left(y_3 - z_3\right)a\,\hat{\mathbf{x}} + \left(-\frac{1}{\sqrt{3}}x_3 + \frac{1}{2\sqrt{3}}y_3 + \frac{1}{2\sqrt{3}}z_3\right)a\,\hat{\mathbf{y}} - \frac{1}{3}\left(x_3 + y_3 + z_3\right)c\,\hat{\mathbf{z}}$ | (6$f$) | Cl |
| $\mathbf{B}_8$ | = | $-y_3\,\mathbf{a}_1 - z_3\,\mathbf{a}_2 - x_3\,\mathbf{a}_3$ | = | $\frac{1}{2}\left(x_3 - y_3\right)a\,\hat{\mathbf{x}} + \left(\frac{1}{2\sqrt{3}}x_3 + \frac{1}{2\sqrt{3}}y_3 - \frac{1}{\sqrt{3}}z_3\right)a\,\hat{\mathbf{y}} - \frac{1}{3}\left(x_3 + y_3 + z_3\right)c\,\hat{\mathbf{z}}$ | (6$f$) | Cl |
| $\mathbf{B}_9$ | = | $x_4\,\mathbf{a}_1 + y_4\,\mathbf{a}_2 + z_4\,\mathbf{a}_3$ | = | $\frac{1}{2}\left(x_4 - z_4\right)a\,\hat{\mathbf{x}} + \left(-\frac{1}{2\sqrt{3}}x_4 + \frac{1}{\sqrt{3}}y_4 - \frac{1}{2\sqrt{3}}z_4\right)a\,\hat{\mathbf{y}} + \frac{1}{3}\left(x_4 + y_4 + z_4\right)c\,\hat{\mathbf{z}}$ | (6$f$) | $H_2O$ |
| $\mathbf{B}_{10}$ | = | $z_4\,\mathbf{a}_1 + x_4\,\mathbf{a}_2 + y_4\,\mathbf{a}_3$ | = | $\frac{1}{2}\left(-y_4 + z_4\right)a\,\hat{\mathbf{x}} + \left(\frac{1}{\sqrt{3}}x_4 - \frac{1}{2\sqrt{3}}y_4 - \frac{1}{2\sqrt{3}}z_4\right)a\,\hat{\mathbf{y}} + \frac{1}{3}\left(x_4 + y_4 + z_4\right)c\,\hat{\mathbf{z}}$ | (6$f$) | $H_2O$ |
| $\mathbf{B}_{11}$ | = | $y_4\,\mathbf{a}_1 + z_4\,\mathbf{a}_2 + x_4\,\mathbf{a}_3$ | = | $\frac{1}{2}\left(-x_4 + y_4\right)a\,\hat{\mathbf{x}} + \left(-\frac{1}{2\sqrt{3}}x_4 - \frac{1}{2\sqrt{3}}y_4 + \frac{1}{\sqrt{3}}z_4\right)a\,\hat{\mathbf{y}} + \frac{1}{3}\left(x_4 + y_4 + z_4\right)c\,\hat{\mathbf{z}}$ | (6$f$) | $H_2O$ |
| $\mathbf{B}_{12}$ | = | $-x_4\,\mathbf{a}_1 - y_4\,\mathbf{a}_2 - z_4\,\mathbf{a}_3$ | = | $\frac{1}{2}\left(-x_4 + z_4\right)a\,\hat{\mathbf{x}} + \left(\frac{1}{2\sqrt{3}}x_4 - \frac{1}{\sqrt{3}}y_4 + \frac{1}{2\sqrt{3}}z_4\right)a\,\hat{\mathbf{y}} - \frac{1}{3}\left(x_4 + y_4 + z_4\right)c\,\hat{\mathbf{z}}$ | (6$f$) | $H_2O$ |
| $\mathbf{B}_{13}$ | = | $-z_4\,\mathbf{a}_1 - x_4\,\mathbf{a}_2 - y_4\,\mathbf{a}_3$ | = | $\frac{1}{2}\,(y_4 - z_4)\,a\,\hat{\mathbf{x}} + \left(-\frac{1}{\sqrt{3}}x_4 + \frac{1}{2\sqrt{3}}y_4 + \frac{1}{2\sqrt{3}}z_4\right)a\,\hat{\mathbf{y}} - \frac{1}{3}\,(x_4 + y_4 + z_4)\,c\,\hat{\mathbf{z}}$ | (6$f$) | $H_2O$ |
| $\mathbf{B}_{14}$ | = | $-y_4\,\mathbf{a}_1 - z_4\,\mathbf{a}_2 - x_4\,\mathbf{a}_3$ | = | $\frac{1}{2}\,(x_4 - y_4)\,a\,\hat{\mathbf{x}} + \left(\frac{1}{2\sqrt{3}}x_4 + \frac{1}{2\sqrt{3}}y_4 - \frac{1}{\sqrt{3}}z_4\right)a\,\hat{\mathbf{y}} - \frac{1}{3}\,(x_4 + y_4 + z_4)\,c\,\hat{\mathbf{z}}$ | (6$f$) | $H_2O$ |

**Found in:**

**Geometry files:**

- CIF: pp. 1722
- POSCAR: pp. 1723

# $Li_7TaO_6$ Structure: A8B6C_hR15_148_cf_f_a

http://aflow.org/prototype-encyclopedia/A8B6C_hR15_148_cf_f_a

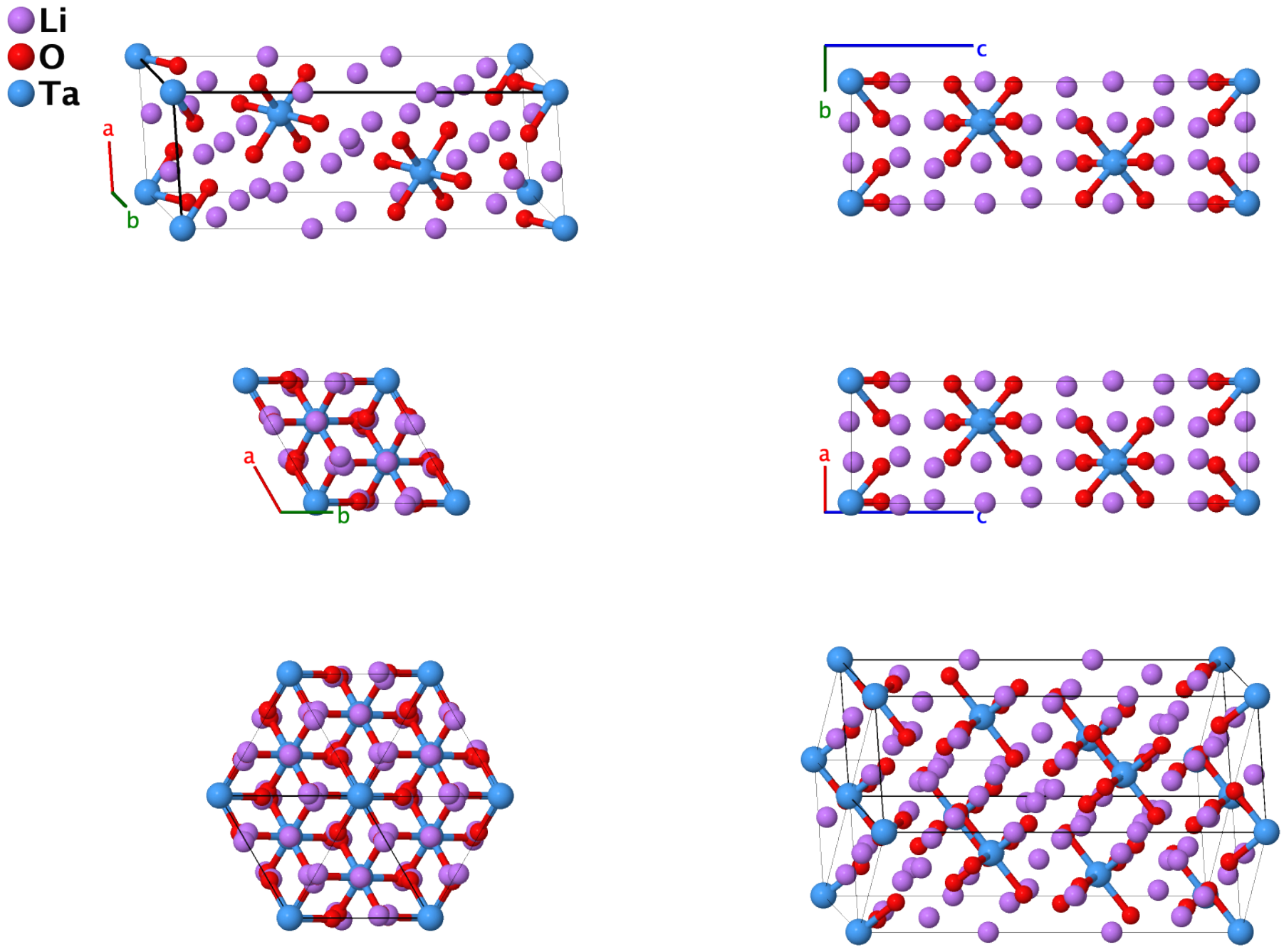

| | | |
|---|---|---|
| **Prototype** | : | $Li_7O_6Ta$ |
| **AFLOW prototype label** | : | A8B6C_hR15_148_cf_f_a |
| ***Strukturbericht* designation** | : | None |
| **Pearson symbol** | : | hR15 |
| **Space group number** | : | 148 |
| **Space group symbol** | : | $R\bar{3}$ |
| **AFLOW prototype command** | : | `aflow --proto=A8B6C_hR15_148_cf_f_a [--hex]` `--params=`$a, c/a, x_2, x_3, y_3, z_3, x_4, y_4, z_4$ |

- The Li-I (2c) site is half-occupied.

**Rhombohedral primitive vectors:**

$$\mathbf{a}_1 = \frac{1}{2}\,a\,\hat{\mathbf{x}} - \frac{1}{2\sqrt{3}}\,a\,\hat{\mathbf{y}} + \frac{1}{3}\,c\,\hat{\mathbf{z}}$$
$$\mathbf{a}_2 = \frac{1}{\sqrt{3}}\,a\,\hat{\mathbf{y}} + \frac{1}{3}\,c\,\hat{\mathbf{z}}$$
$$\mathbf{a}_3 = -\frac{1}{2}\,a\,\hat{\mathbf{x}} - \frac{1}{2\sqrt{3}}\,a\,\hat{\mathbf{y}} + \frac{1}{3}\,c\,\hat{\mathbf{z}}$$

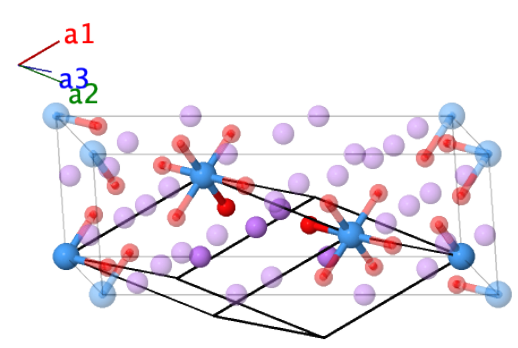

**Basis vectors:**

| | | Lattice Coordinates | | Cartesian Coordinates | Wyckoff Position | Atom Type |
|---|---|---|---|---|---|---|
| $\mathbf{B}_1$ | = | $0\,\mathbf{a}_1 + 0\,\mathbf{a}_2 + 0\,\mathbf{a}_3$ | = | $0\,\hat{\mathbf{x}} + 0\,\hat{\mathbf{y}} + 0\,\hat{\mathbf{z}}$ | (1$a$) | Ta |
| $\mathbf{B}_2$ | = | $x_2\,\mathbf{a}_1 + x_2\,\mathbf{a}_2 + x_2\,\mathbf{a}_3$ | = | $x_2 c\,\hat{\mathbf{z}}$ | (2$c$) | Li I |
| $\mathbf{B}_3$ | = | $-x_2\,\mathbf{a}_1 - x_2\,\mathbf{a}_2 - x_2\,\mathbf{a}_3$ | = | $-x_2 c\,\hat{\mathbf{z}}$ | (2$c$) | Li I |
| $\mathbf{B}_4$ | = | $x_3\,\mathbf{a}_1 + y_3\,\mathbf{a}_2 + z_3\,\mathbf{a}_3$ | = | $\frac{1}{2}(x_3 - z_3)\,a\,\mathbf{\hat{x}} + \left(-\frac{1}{2\sqrt{3}}x_3 + \frac{1}{\sqrt{3}}y_3 - \frac{1}{2\sqrt{3}}z_3\right)a\,\mathbf{\hat{y}} + \frac{1}{3}(x_3 + y_3 + z_3)\,c\,\mathbf{\hat{z}}$ | (6$f$) | Li II |
| $\mathbf{B}_5$ | = | $z_3\,\mathbf{a}_1 + x_3\,\mathbf{a}_2 + y_3\,\mathbf{a}_3$ | = | $\frac{1}{2}(-y_3 + z_3)\,a\,\mathbf{\hat{x}} + \left(\frac{1}{\sqrt{3}}x_3 - \frac{1}{2\sqrt{3}}y_3 - \frac{1}{2\sqrt{3}}z_3\right)a\,\mathbf{\hat{y}} + \frac{1}{3}(x_3 + y_3 + z_3)\,c\,\mathbf{\hat{z}}$ | (6$f$) | Li II |
| $\mathbf{B}_6$ | = | $y_3\,\mathbf{a}_1 + z_3\,\mathbf{a}_2 + x_3\,\mathbf{a}_3$ | = | $\frac{1}{2}(-x_3 + y_3)\,a\,\mathbf{\hat{x}} + \left(-\frac{1}{2\sqrt{3}}x_3 - \frac{1}{2\sqrt{3}}y_3 + \frac{1}{\sqrt{3}}z_3\right)a\,\mathbf{\hat{y}} + \frac{1}{3}(x_3 + y_3 + z_3)\,c\,\mathbf{\hat{z}}$ | (6$f$) | Li II |
| $\mathbf{B}_7$ | = | $-x_3\,\mathbf{a}_1 - y_3\,\mathbf{a}_2 - z_3\,\mathbf{a}_3$ | = | $\frac{1}{2}(-x_3 + z_3)\,a\,\mathbf{\hat{x}} + \left(\frac{1}{2\sqrt{3}}x_3 - \frac{1}{\sqrt{3}}y_3 + \frac{1}{2\sqrt{3}}z_3\right)a\,\mathbf{\hat{y}} - \frac{1}{3}(x_3 + y_3 + z_3)\,c\,\mathbf{\hat{z}}$ | (6$f$) | Li II |
| $\mathbf{B}_8$ | = | $-z_3\,\mathbf{a}_1 - x_3\,\mathbf{a}_2 - y_3\,\mathbf{a}_3$ | = | $\frac{1}{2}(y_3 - z_3)\,a\,\mathbf{\hat{x}} + \left(-\frac{1}{\sqrt{3}}x_3 + \frac{1}{2\sqrt{3}}y_3 + \frac{1}{2\sqrt{3}}z_3\right)a\,\mathbf{\hat{y}} - \frac{1}{3}(x_3 + y_3 + z_3)\,c\,\mathbf{\hat{z}}$ | (6$f$) | Li II |
| $\mathbf{B}_9$ | = | $-y_3\,\mathbf{a}_1 - z_3\,\mathbf{a}_2 - x_3\,\mathbf{a}_3$ | = | $\frac{1}{2}(x_3 - y_3)\,a\,\mathbf{\hat{x}} + \left(\frac{1}{2\sqrt{3}}x_3 + \frac{1}{2\sqrt{3}}y_3 - \frac{1}{\sqrt{3}}z_3\right)a\,\mathbf{\hat{y}} - \frac{1}{3}(x_3 + y_3 + z_3)\,c\,\mathbf{\hat{z}}$ | (6$f$) | Li II |
| $\mathbf{B}_{10}$ | = | $x_4\,\mathbf{a}_1 + y_4\,\mathbf{a}_2 + z_4\,\mathbf{a}_3$ | = | $\frac{1}{2}(x_4 - z_4)\,a\,\mathbf{\hat{x}} + \left(-\frac{1}{2\sqrt{3}}x_4 + \frac{1}{\sqrt{3}}y_4 - \frac{1}{2\sqrt{3}}z_4\right)a\,\mathbf{\hat{y}} + \frac{1}{3}(x_4 + y_4 + z_4)\,c\,\mathbf{\hat{z}}$ | (6$f$) | O |
| $\mathbf{B}_{11}$ | = | $z_4\,\mathbf{a}_1 + x_4\,\mathbf{a}_2 + y_4\,\mathbf{a}_3$ | = | $\frac{1}{2}(-y_4 + z_4)\,a\,\mathbf{\hat{x}} + \left(\frac{1}{\sqrt{3}}x_4 - \frac{1}{2\sqrt{3}}y_4 - \frac{1}{2\sqrt{3}}z_4\right)a\,\mathbf{\hat{y}} + \frac{1}{3}(x_4 + y_4 + z_4)\,c\,\mathbf{\hat{z}}$ | (6$f$) | O |
| $\mathbf{B}_{12}$ | = | $y_4\,\mathbf{a}_1 + z_4\,\mathbf{a}_2 + x_4\,\mathbf{a}_3$ | = | $\frac{1}{2}(-x_4 + y_4)\,a\,\mathbf{\hat{x}} + \left(-\frac{1}{2\sqrt{3}}x_4 - \frac{1}{2\sqrt{3}}y_4 + \frac{1}{\sqrt{3}}z_4\right)a\,\mathbf{\hat{y}} + \frac{1}{3}(x_4 + y_4 + z_4)\,c\,\mathbf{\hat{z}}$ | (6$f$) | O |
| $\mathbf{B}_{13}$ | = | $-x_4\,\mathbf{a}_1 - y_4\,\mathbf{a}_2 - z_4\,\mathbf{a}_3$ | = | $\frac{1}{2}(-x_4 + z_4)\,a\,\mathbf{\hat{x}} + \left(\frac{1}{2\sqrt{3}}x_4 - \frac{1}{\sqrt{3}}y_4 + \frac{1}{2\sqrt{3}}z_4\right)a\,\mathbf{\hat{y}} - \frac{1}{3}(x_4 + y_4 + z_4)\,c\,\mathbf{\hat{z}}$ | (6$f$) | O |
| $\mathbf{B}_{14}$ | = | $-z_4\,\mathbf{a}_1 - x_4\,\mathbf{a}_2 - y_4\,\mathbf{a}_3$ | = | $\frac{1}{2}(y_4 - z_4)\,a\,\mathbf{\hat{x}} + \left(-\frac{1}{\sqrt{3}}x_4 + \frac{1}{2\sqrt{3}}y_4 + \frac{1}{2\sqrt{3}}z_4\right)a\,\mathbf{\hat{y}} - \frac{1}{3}(x_4 + y_4 + z_4)\,c\,\mathbf{\hat{z}}$ | (6$f$) | O |
| $\mathbf{B}_{15}$ | = | $-y_4\,\mathbf{a}_1 - z_4\,\mathbf{a}_2 - x_4\,\mathbf{a}_3$ | = | $\frac{1}{2}(x_4 - y_4)\,a\,\mathbf{\hat{x}} + \left(\frac{1}{2\sqrt{3}}x_4 + \frac{1}{2\sqrt{3}}y_4 - \frac{1}{\sqrt{3}}z_4\right)a\,\mathbf{\hat{y}} - \frac{1}{3}(x_4 + y_4 + z_4)\,c\,\mathbf{\hat{z}}$ | (6$f$) | O |

**Found in:**

**Geometry files:**

- CIF: pp. 1723
- POSCAR: pp. 1723

# $SrCl_2 \cdot (H_2O)_6$ Structure: A2B12C6D_hP21_150_d_2g_ef_a

http://aflow.org/prototype-encyclopedia/A2B12C6D_hP21_150_d_2g_ef_a

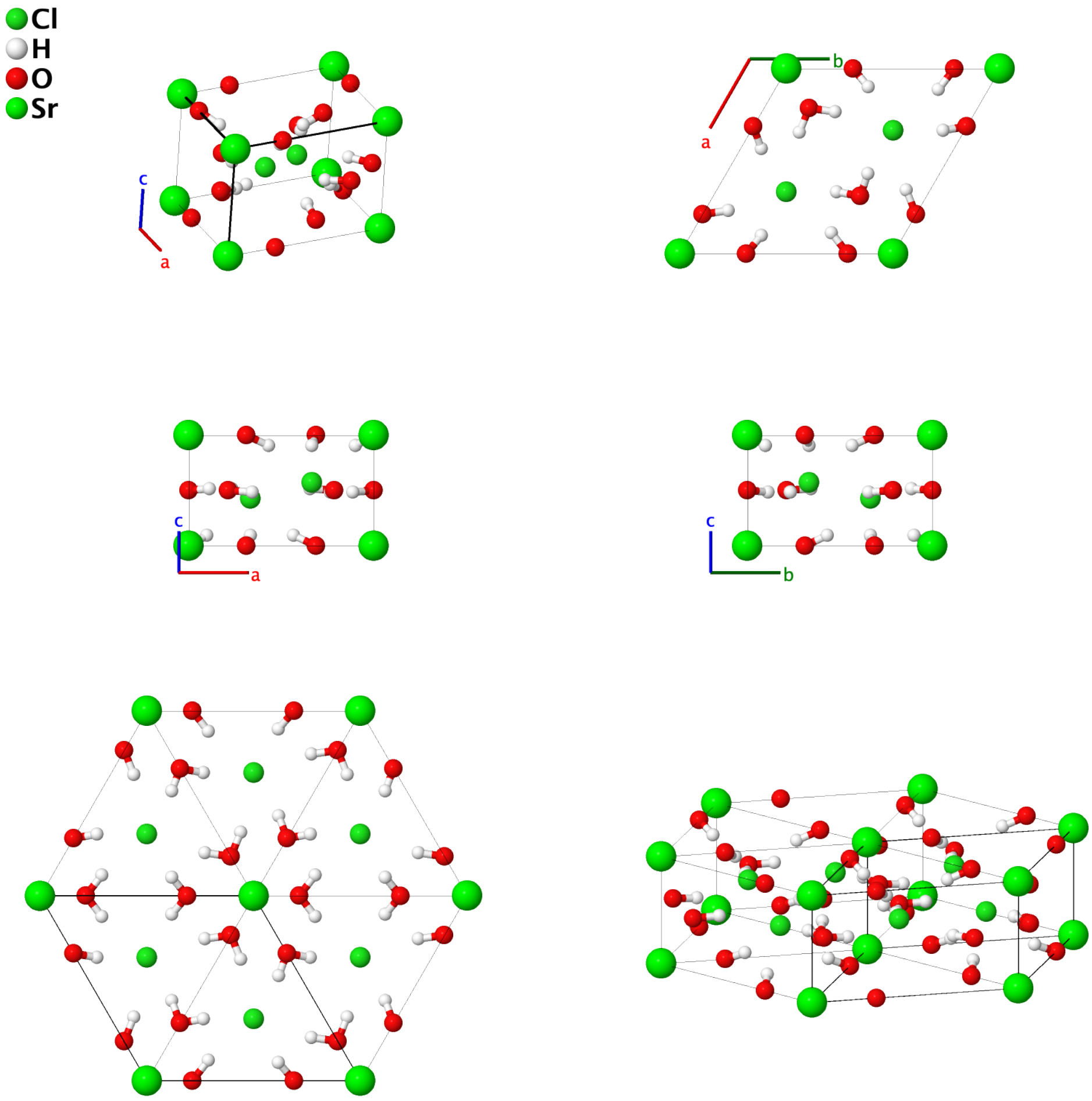

| | | |
|---|---|---|
| **Prototype** | : | $Cl_2H_{12}O_6Sr$ |
| **AFLOW prototype label** | : | A2B12C6D_hP21_150_d_2g_ef_a |
| ***Strukturbericht* designation** | : | None |
| **Pearson symbol** | : | hP21 |
| **Space group number** | : | 150 |
| **Space group symbol** | : | $P321$ |
| **AFLOW prototype command** | : | `aflow --proto=A2B12C6D_hP21_150_d_2g_ef_a --params=`$a, c/a, z_2, x_3, x_4, x_5, y_5, z_5, x_6, y_6, z_6$ |

### Other compounds with this structure

- $CaCl_2 \cdot (H_2O)_6$ and $CaBr_2 \cdot (H_2O)_6$

- This is a redetermination of the $I1_3$ structure. In addition to locating the hydrogen atoms in the water molecule, (Agron, 1986) show that the space group is $P321$ #150, not $P\bar{3}$ #147.

**Trigonal Hexagonal primitive vectors:**

$$\mathbf{a}_1 = \frac{1}{2}\,a\,\hat{\mathbf{x}} - \frac{\sqrt{3}}{2}\,a\,\hat{\mathbf{y}}$$
$$\mathbf{a}_2 = \frac{1}{2}\,a\,\hat{\mathbf{x}} + \frac{\sqrt{3}}{2}\,a\,\hat{\mathbf{y}}$$
$$\mathbf{a}_3 = c\,\hat{\mathbf{z}}$$

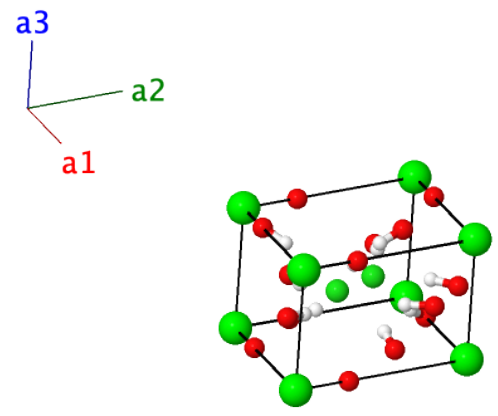

**Basis vectors:**

| | | Lattice Coordinates | | Cartesian Coordinates | Wyckoff Position | Atom Type |
|---|---|---|---|---|---|---|
| $\mathbf{B}_1$ | = | $0\,\mathbf{a}_1 + 0\,\mathbf{a}_2 + 0\,\mathbf{a}_3$ | = | $0\,\hat{\mathbf{x}} + 0\,\hat{\mathbf{y}} + 0\,\hat{\mathbf{z}}$ | (1$a$) | Sr |
| $\mathbf{B}_2$ | = | $\frac{1}{3}\,\mathbf{a}_1 + \frac{2}{3}\,\mathbf{a}_2 + z_2\,\mathbf{a}_3$ | = | $\frac{1}{2}a\,\hat{\mathbf{x}} + \frac{1}{2\sqrt{3}}a\,\hat{\mathbf{y}} + z_2c\,\hat{\mathbf{z}}$ | (2$d$) | Cl |
| $\mathbf{B}_3$ | = | $\frac{2}{3}\,\mathbf{a}_1 + \frac{1}{3}\,\mathbf{a}_2 - z_2\,\mathbf{a}_3$ | = | $\frac{1}{2}a\,\hat{\mathbf{x}} - \frac{1}{2\sqrt{3}}a\,\hat{\mathbf{y}} - z_2c\,\hat{\mathbf{z}}$ | (2$d$) | Cl |
| $\mathbf{B}_4$ | = | $x_3\,\mathbf{a}_1$ | = | $\frac{1}{2}x_3a\,\hat{\mathbf{x}} - \frac{\sqrt{3}}{2}x_3a\,\hat{\mathbf{y}}$ | (3$e$) | O I |
| $\mathbf{B}_5$ | = | $x_3\,\mathbf{a}_2$ | = | $\frac{1}{2}x_3a\,\hat{\mathbf{x}} + \frac{\sqrt{3}}{2}x_3a\,\hat{\mathbf{y}}$ | (3$e$) | O I |
| $\mathbf{B}_6$ | = | $-x_3\,\mathbf{a}_1 - x_3\,\mathbf{a}_2$ | = | $-x_3a\,\hat{\mathbf{x}}$ | (3$e$) | O I |
| $\mathbf{B}_7$ | = | $x_4\,\mathbf{a}_1 + \frac{1}{2}\,\mathbf{a}_3$ | = | $\frac{1}{2}x_4a\,\hat{\mathbf{x}} - \frac{\sqrt{3}}{2}x_4a\,\hat{\mathbf{y}} + \frac{1}{2}c\,\hat{\mathbf{z}}$ | (3$f$) | O II |
| $\mathbf{B}_8$ | = | $x_4\,\mathbf{a}_2 + \frac{1}{2}\,\mathbf{a}_3$ | = | $\frac{1}{2}x_4a\,\hat{\mathbf{x}} + \frac{\sqrt{3}}{2}x_4a\,\hat{\mathbf{y}} + \frac{1}{2}c\,\hat{\mathbf{z}}$ | (3$f$) | O II |
| $\mathbf{B}_9$ | = | $-x_4\,\mathbf{a}_1 - x_4\,\mathbf{a}_2 + \frac{1}{2}\,\mathbf{a}_3$ | = | $-x_4a\,\hat{\mathbf{x}} + \frac{1}{2}c\,\hat{\mathbf{z}}$ | (3$f$) | O II |
| $\mathbf{B}_{10}$ | = | $x_5\,\mathbf{a}_1 + y_5\,\mathbf{a}_2 + z_5\,\mathbf{a}_3$ | = | $\frac{1}{2}\,(x_5 + y_5)\,a\,\hat{\mathbf{x}} + \frac{\sqrt{3}}{2}\,(-x_5 + y_5)\,a\,\hat{\mathbf{y}} + z_5c\,\hat{\mathbf{z}}$ | (6$g$) | H I |
| $\mathbf{B}_{11}$ | = | $-y_5\,\mathbf{a}_1 + (x_5 - y_5)\,\mathbf{a}_2 + z_5\,\mathbf{a}_3$ | = | $\left(\frac{1}{2}x_5 - y_5\right)a\,\hat{\mathbf{x}} + \frac{\sqrt{3}}{2}x_5a\,\hat{\mathbf{y}} + z_5c\,\hat{\mathbf{z}}$ | (6$g$) | H I |
| $\mathbf{B}_{12}$ | = | $(-x_5 + y_5)\,\mathbf{a}_1 - x_5\,\mathbf{a}_2 + z_5\,\mathbf{a}_3$ | = | $\left(-x_5 + \frac{1}{2}y_5\right)a\,\hat{\mathbf{x}} - \frac{\sqrt{3}}{2}y_5a\,\hat{\mathbf{y}} + z_5c\,\hat{\mathbf{z}}$ | (6$g$) | H I |
| $\mathbf{B}_{13}$ | = | $y_5\,\mathbf{a}_1 + x_5\,\mathbf{a}_2 - z_5\,\mathbf{a}_3$ | = | $\frac{1}{2}\,(x_5 + y_5)\,a\,\hat{\mathbf{x}} + \frac{\sqrt{3}}{2}\,(x_5 - y_5)\,a\,\hat{\mathbf{y}} - z_5c\,\hat{\mathbf{z}}$ | (6$g$) | H I |
| $\mathbf{B}_{14}$ | = | $(x_5 - y_5)\,\mathbf{a}_1 - y_5\,\mathbf{a}_2 - z_5\,\mathbf{a}_3$ | = | $\left(\frac{1}{2}x_5 - y_5\right)a\,\hat{\mathbf{x}} - \frac{\sqrt{3}}{2}x_5a\,\hat{\mathbf{y}} - z_5c\,\hat{\mathbf{z}}$ | (6$g$) | H I |
| $\mathbf{B}_{15}$ | = | $-x_5\,\mathbf{a}_1 + (-x_5 + y_5)\,\mathbf{a}_2 - z_5\,\mathbf{a}_3$ | = | $\left(-x_5 + \frac{1}{2}y_5\right)a\,\hat{\mathbf{x}} + \frac{\sqrt{3}}{2}y_5a\,\hat{\mathbf{y}} - z_5c\,\hat{\mathbf{z}}$ | (6$g$) | H I |
| $\mathbf{B}_{16}$ | = | $x_6\,\mathbf{a}_1 + y_6\,\mathbf{a}_2 + z_6\,\mathbf{a}_3$ | = | $\frac{1}{2}\,(x_6 + y_6)\,a\,\hat{\mathbf{x}} + \frac{\sqrt{3}}{2}\,(-x_6 + y_6)\,a\,\hat{\mathbf{y}} + z_6c\,\hat{\mathbf{z}}$ | (6$g$) | H II |
| $\mathbf{B}_{17}$ | = | $-y_6\,\mathbf{a}_1 + (x_6 - y_6)\,\mathbf{a}_2 + z_6\,\mathbf{a}_3$ | = | $\left(\frac{1}{2}x_6 - y_6\right)a\,\hat{\mathbf{x}} + \frac{\sqrt{3}}{2}x_6a\,\hat{\mathbf{y}} + z_6c\,\hat{\mathbf{z}}$ | (6$g$) | H II |
| $\mathbf{B}_{18}$ | = | $(-x_6 + y_6)\,\mathbf{a}_1 - x_6\,\mathbf{a}_2 + z_6\,\mathbf{a}_3$ | = | $\left(-x_6 + \frac{1}{2}y_6\right)a\,\hat{\mathbf{x}} - \frac{\sqrt{3}}{2}y_6a\,\hat{\mathbf{y}} + z_6c\,\hat{\mathbf{z}}$ | (6$g$) | H II |
| $\mathbf{B}_{19}$ | = | $y_6\,\mathbf{a}_1 + x_6\,\mathbf{a}_2 - z_6\,\mathbf{a}_3$ | = | $\frac{1}{2}\,(x_6 + y_6)\,a\,\hat{\mathbf{x}} + \frac{\sqrt{3}}{2}\,(x_6 - y_6)\,a\,\hat{\mathbf{y}} - z_6c\,\hat{\mathbf{z}}$ | (6$g$) | H II |
| $\mathbf{B}_{20}$ | = | $(x_6 - y_6)\,\mathbf{a}_1 - y_6\,\mathbf{a}_2 - z_6\,\mathbf{a}_3$ | = | $\left(\frac{1}{2}x_6 - y_6\right)a\,\hat{\mathbf{x}} - \frac{\sqrt{3}}{2}x_6a\,\hat{\mathbf{y}} - z_6c\,\hat{\mathbf{z}}$ | (6$g$) | H II |
| $\mathbf{B}_{21}$ | = | $-x_6\,\mathbf{a}_1 + (-x_6 + y_6)\,\mathbf{a}_2 - z_6\,\mathbf{a}_3$ | = | $\left(-x_6 + \frac{1}{2}y_6\right)a\,\hat{\mathbf{x}} + \frac{\sqrt{3}}{2}y_6a\,\hat{\mathbf{y}} - z_6c\,\hat{\mathbf{z}}$ | (6$g$) | H II |

**Found in:**

**Geometry files:**
- CIF: pp. 1723
- POSCAR: pp. 1724

# $Cs_3As_2Cl_9$ ($K7_3$) Structure: A2B9C3_hP14_150_d_eg_ad

http://aflow.org/prototype-encyclopedia/A2B9C3_hP14_150_d_eg_ad

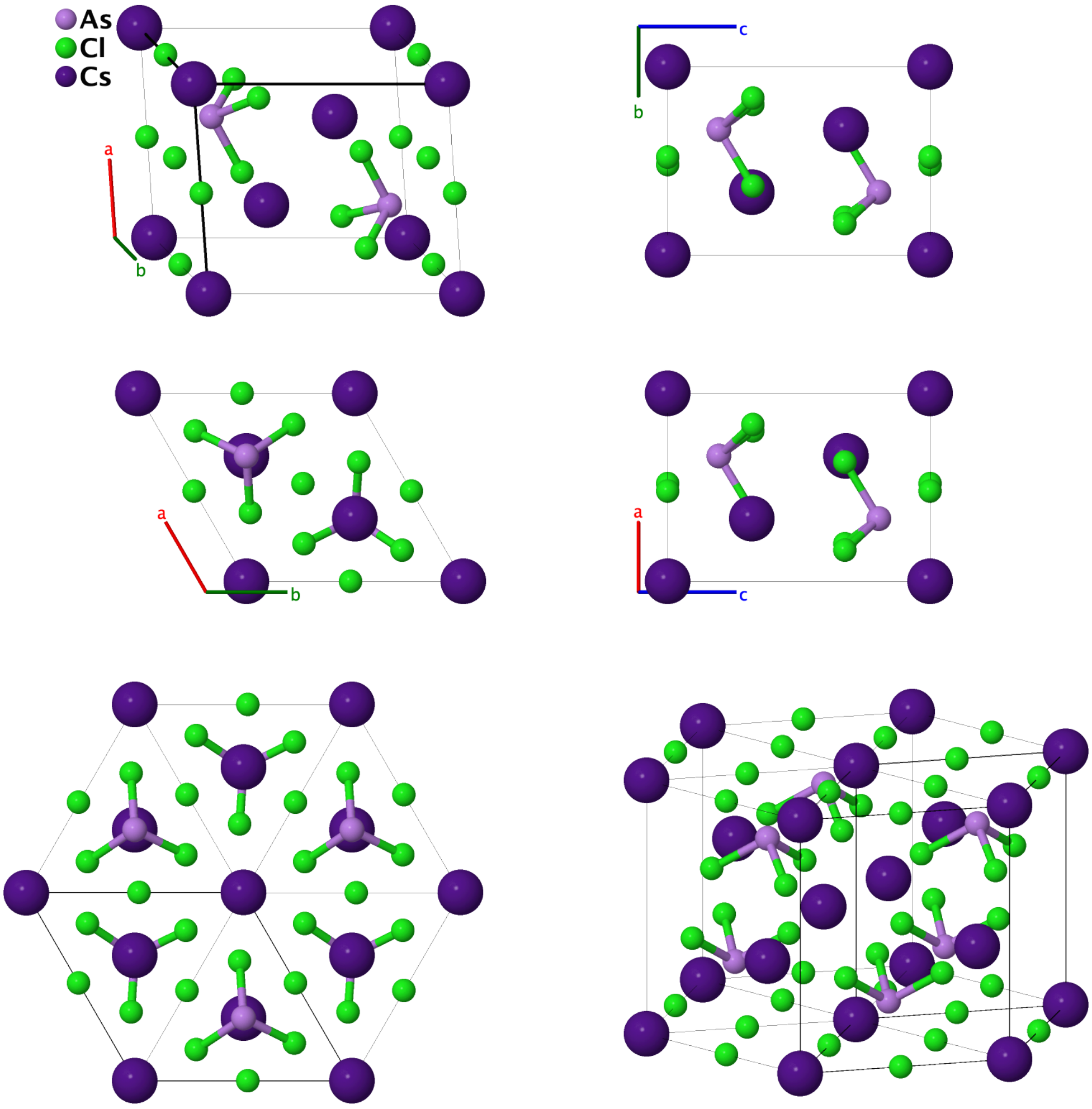

| | | |
|---|---|---|
| **Prototype** | : | $As_2Cl_9Cs_3$ |
| **AFLOW prototype label** | : | A2B9C3_hP14_150_d_eg_ad |
| ***Strukturbericht* designation** | : | $K7_3$ |
| **Pearson symbol** | : | hP14 |
| **Space group number** | : | 150 |
| **Space group symbol** | : | $P321$ |
| **AFLOW prototype command** | : | `aflow --proto=A2B9C3_hP14_150_d_eg_ad --params=`$a, c/a, z_2, z_3, x_4, x_5, y_5, z_5$ |

**Trigonal Hexagonal primitive vectors:**

$$\mathbf{a}_1 = \frac{1}{2}\,a\,\hat{\mathbf{x}} - \frac{\sqrt{3}}{2}\,a\,\hat{\mathbf{y}}$$
$$\mathbf{a}_2 = \frac{1}{2}\,a\,\hat{\mathbf{x}} + \frac{\sqrt{3}}{2}\,a\,\hat{\mathbf{y}}$$
$$\mathbf{a}_3 = c\,\hat{\mathbf{z}}$$

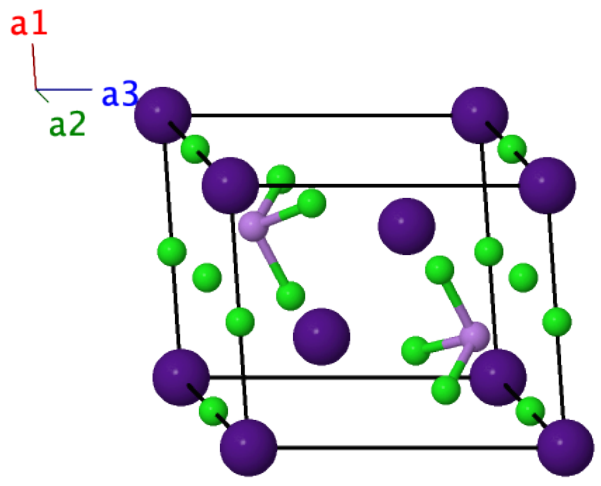

**Basis vectors:**

| | | Lattice Coordinates | | Cartesian Coordinates | Wyckoff Position | Atom Type |
|---|---|---|---|---|---|---|
| $\mathbf{B}_1$ | = | $0\,\mathbf{a}_1 + 0\,\mathbf{a}_2 + 0\,\mathbf{a}_3$ | = | $0\,\hat{\mathbf{x}} + 0\,\hat{\mathbf{y}} + 0\,\hat{\mathbf{z}}$ | (1*a*) | Cs I |
| $\mathbf{B}_2$ | = | $\frac{1}{3}\,\mathbf{a}_1 + \frac{2}{3}\,\mathbf{a}_2 + z_2\,\mathbf{a}_3$ | = | $\frac{1}{2}a\,\hat{\mathbf{x}} + \frac{1}{2\sqrt{3}}a\,\hat{\mathbf{y}} + z_2c\,\hat{\mathbf{z}}$ | (2*d*) | As |
| $\mathbf{B}_3$ | = | $\frac{2}{3}\,\mathbf{a}_1 + \frac{1}{3}\,\mathbf{a}_2 - z_2\,\mathbf{a}_3$ | = | $\frac{1}{2}a\,\hat{\mathbf{x}} - \frac{1}{2\sqrt{3}}a\,\hat{\mathbf{y}} - z_2c\,\hat{\mathbf{z}}$ | (2*d*) | As |
| $\mathbf{B}_4$ | = | $\frac{1}{3}\,\mathbf{a}_1 + \frac{2}{3}\,\mathbf{a}_2 + z_3\,\mathbf{a}_3$ | = | $\frac{1}{2}a\,\hat{\mathbf{x}} + \frac{1}{2\sqrt{3}}a\,\hat{\mathbf{y}} + z_3c\,\hat{\mathbf{z}}$ | (2*d*) | Cs II |
| $\mathbf{B}_5$ | = | $\frac{2}{3}\,\mathbf{a}_1 + \frac{1}{3}\,\mathbf{a}_2 - z_3\,\mathbf{a}_3$ | = | $\frac{1}{2}a\,\hat{\mathbf{x}} - \frac{1}{2\sqrt{3}}a\,\hat{\mathbf{y}} - z_3c\,\hat{\mathbf{z}}$ | (2*d*) | Cs II |
| $\mathbf{B}_6$ | = | $x_4\,\mathbf{a}_1$ | = | $\frac{1}{2}x_4a\,\hat{\mathbf{x}} - \frac{\sqrt{3}}{2}x_4a\,\hat{\mathbf{y}}$ | (3*e*) | Cl I |
| $\mathbf{B}_7$ | = | $x_4\,\mathbf{a}_2$ | = | $\frac{1}{2}x_4a\,\hat{\mathbf{x}} + \frac{\sqrt{3}}{2}x_4a\,\hat{\mathbf{y}}$ | (3*e*) | Cl I |
| $\mathbf{B}_8$ | = | $-x_4\,\mathbf{a}_1 - x_4\,\mathbf{a}_2$ | = | $-x_4a\,\hat{\mathbf{x}}$ | (3*e*) | Cl I |
| $\mathbf{B}_9$ | = | $x_5\,\mathbf{a}_1 + y_5\,\mathbf{a}_2 + z_5\,\mathbf{a}_3$ | = | $\frac{1}{2}\left(x_5 + y_5\right)a\,\hat{\mathbf{x}} + \frac{\sqrt{3}}{2}\left(-x_5 + y_5\right)a\,\hat{\mathbf{y}} + z_5c\,\hat{\mathbf{z}}$ | (6*g*) | Cl II |
| $\mathbf{B}_{10}$ | = | $-y_5\,\mathbf{a}_1 + (x_5 - y_5)\,\mathbf{a}_2 + z_5\,\mathbf{a}_3$ | = | $\left(\frac{1}{2}x_5 - y_5\right)a\,\hat{\mathbf{x}} + \frac{\sqrt{3}}{2}x_5a\,\hat{\mathbf{y}} + z_5c\,\hat{\mathbf{z}}$ | (6*g*) | Cl II |
| $\mathbf{B}_{11}$ | = | $(-x_5 + y_5)\,\mathbf{a}_1 - x_5\,\mathbf{a}_2 + z_5\,\mathbf{a}_3$ | = | $\left(-x_5 + \frac{1}{2}y_5\right)a\,\hat{\mathbf{x}} - \frac{\sqrt{3}}{2}y_5a\,\hat{\mathbf{y}} + z_5c\,\hat{\mathbf{z}}$ | (6*g*) | Cl II |
| $\mathbf{B}_{12}$ | = | $y_5\,\mathbf{a}_1 + x_5\,\mathbf{a}_2 - z_5\,\mathbf{a}_3$ | = | $\frac{1}{2}\left(x_5 + y_5\right)a\,\hat{\mathbf{x}} + \frac{\sqrt{3}}{2}\left(x_5 - y_5\right)a\,\hat{\mathbf{y}} - z_5c\,\hat{\mathbf{z}}$ | (6*g*) | Cl II |
| $\mathbf{B}_{13}$ | = | $(x_5 - y_5)\,\mathbf{a}_1 - y_5\,\mathbf{a}_2 - z_5\,\mathbf{a}_3$ | = | $\left(\frac{1}{2}x_5 - y_5\right)a\,\hat{\mathbf{x}} - \frac{\sqrt{3}}{2}x_5a\,\hat{\mathbf{y}} - z_5c\,\hat{\mathbf{z}}$ | (6*g*) | Cl II |
| $\mathbf{B}_{14}$ | = | $-x_5\,\mathbf{a}_1 + (-x_5 + y_5)\,\mathbf{a}_2 - z_5\,\mathbf{a}_3$ | = | $\left(-x_5 + \frac{1}{2}y_5\right)a\,\hat{\mathbf{x}} + \frac{\sqrt{3}}{2}y_5a\,\hat{\mathbf{y}} - z_5c\,\hat{\mathbf{z}}$ | (6*g*) | Cl II |

**Geometry files:**

- CIF: pp. 1724
- POSCAR: pp. 1724

# Paralstonite ($BaCa(CO_3)_2$) Structure: AB2CD6_hP30_150_e_c2d_f_3g

http://aflow.org/prototype-encyclopedia/AB2CD6_hP30_150_e_c2d_f_3g

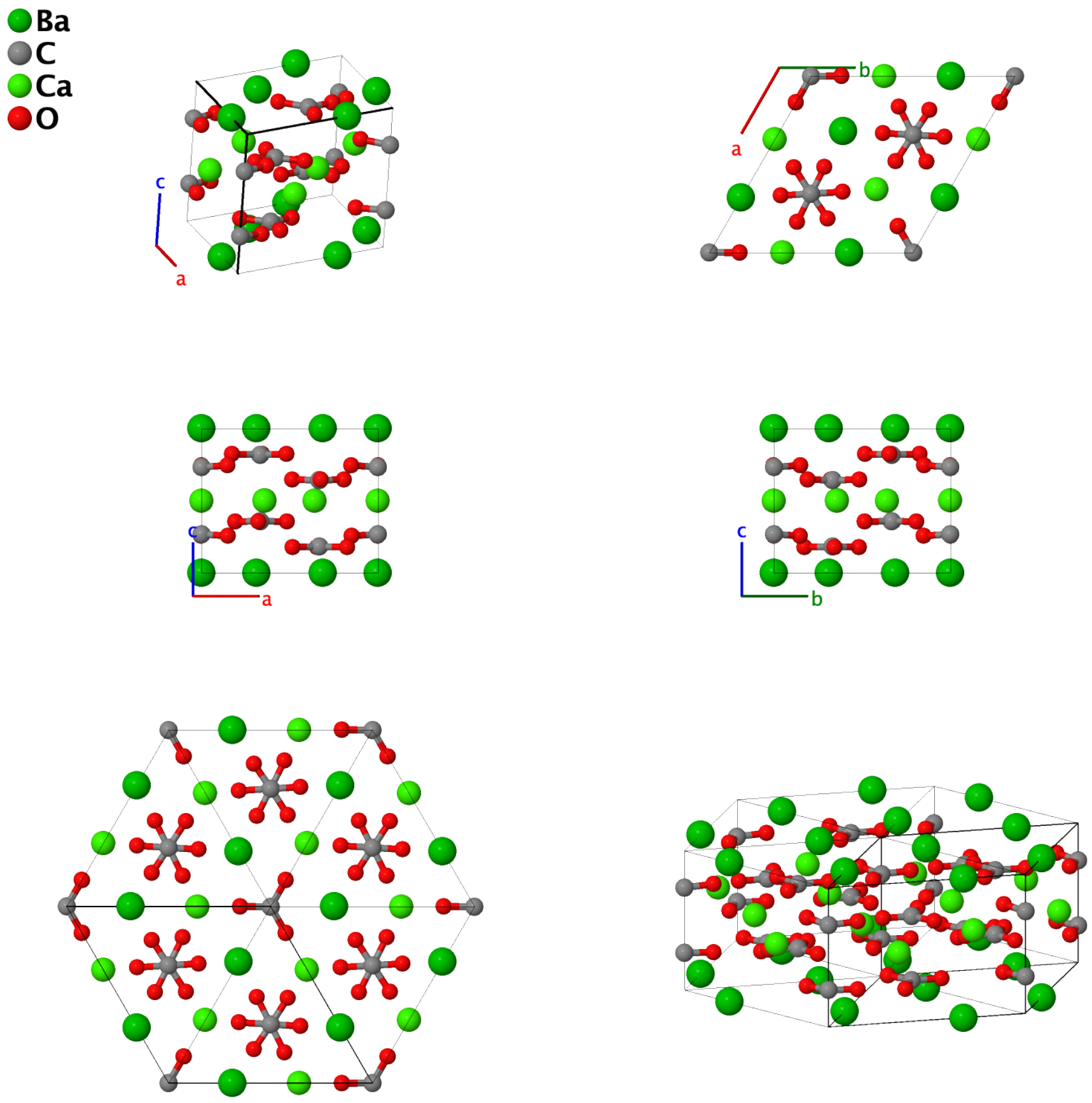

| | | |
|---|---|---|
| **Prototype** | : | $BaC_2CaO_6$ |
| **AFLOW prototype label** | : | AB2CD6_hP30_150_e_c2d_f_3g |
| ***Strukturbericht* designation** | : | None |
| **Pearson symbol** | : | hP30 |
| **Space group number** | : | 150 |
| **Space group symbol** | : | $P321$ |
| **AFLOW prototype command** | : | `aflow --proto=AB2CD6_hP30_150_e_c2d_f_3g` `--params=`$a, c/a, z_1, z_2, z_3, x_4, x_5, x_6, y_6, z_6, x_7, y_7, z_7, x_8, y_8, z_8$ |

- $BaCa(CO_3)_2$ comes in a variety of crystal structures (Spahr, 2019):
  - monoclinic barytocalcite, space group $P2_1/m$ #11,
  - trigonal paralstonite, space group $P321$ #150 (the current structure),
  - triclinic alstonite, space group $P1$ #1 or $P\bar{1}$ #2 (Sartori, 1975), and
  - a new monoclinic structure, space group $C2$ #5, synthesized by (Spahr, 2019), and lacking the centrosymmetric character of barytocalcite.
- We were unable to obtain a copy of (Effenberger, 1980), so we use the data provided by (Downs, 2003).

**Trigonal Hexagonal primitive vectors:**

$$
\begin{aligned}
\mathbf{a}_1 &= \tfrac{1}{2}\, a\, \hat{\mathbf{x}} - \tfrac{\sqrt{3}}{2}\, a\, \hat{\mathbf{y}} \\
\mathbf{a}_2 &= \tfrac{1}{2}\, a\, \hat{\mathbf{x}} + \tfrac{\sqrt{3}}{2}\, a\, \hat{\mathbf{y}} \\
\mathbf{a}_3 &= c\, \hat{\mathbf{z}}
\end{aligned}
$$

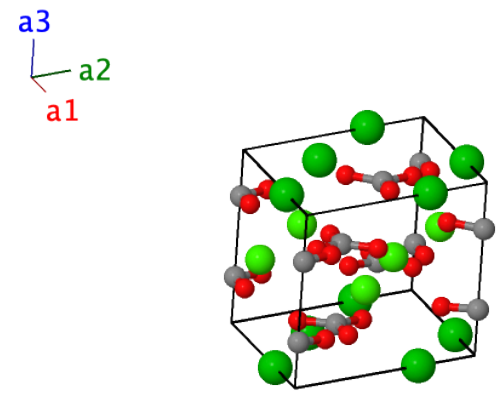

**Basis vectors:**

| | | Lattice Coordinates | | Cartesian Coordinates | Wyckoff Position | Atom Type |
|---|---|---|---|---|---|---|
| $\mathbf{B}_1$ | = | $z_1\,\mathbf{a}_3$ | = | $z_1 c\,\hat{\mathbf{z}}$ | (2c) | C I |
| $\mathbf{B}_2$ | = | $-z_1\,\mathbf{a}_3$ | = | $-z_1 c\,\hat{\mathbf{z}}$ | (2c) | C I |
| $\mathbf{B}_3$ | = | $\frac{1}{3}\,\mathbf{a}_1 + \frac{2}{3}\,\mathbf{a}_2 + z_2\,\mathbf{a}_3$ | = | $\frac{1}{2}a\,\hat{\mathbf{x}} + \frac{1}{2\sqrt{3}}a\,\hat{\mathbf{y}} + z_2 c\,\hat{\mathbf{z}}$ | (2d) | C II |
| $\mathbf{B}_4$ | = | $\frac{2}{3}\,\mathbf{a}_1 + \frac{1}{3}\,\mathbf{a}_2 - z_2\,\mathbf{a}_3$ | = | $\frac{1}{2}a\,\hat{\mathbf{x}} - \frac{1}{2\sqrt{3}}a\,\hat{\mathbf{y}} - z_2 c\,\hat{\mathbf{z}}$ | (2d) | C II |
| $\mathbf{B}_5$ | = | $\frac{1}{3}\,\mathbf{a}_1 + \frac{2}{3}\,\mathbf{a}_2 + z_3\,\mathbf{a}_3$ | = | $\frac{1}{2}a\,\hat{\mathbf{x}} + \frac{1}{2\sqrt{3}}a\,\hat{\mathbf{y}} + z_3 c\,\hat{\mathbf{z}}$ | (2d) | C III |
| $\mathbf{B}_6$ | = | $\frac{2}{3}\,\mathbf{a}_1 + \frac{1}{3}\,\mathbf{a}_2 - z_3\,\mathbf{a}_3$ | = | $\frac{1}{2}a\,\hat{\mathbf{x}} - \frac{1}{2\sqrt{3}}a\,\hat{\mathbf{y}} - z_3 c\,\hat{\mathbf{z}}$ | (2d) | C III |
| $\mathbf{B}_7$ | = | $x_4\,\mathbf{a}_1$ | = | $\frac{1}{2}x_4 a\,\hat{\mathbf{x}} - \frac{\sqrt{3}}{2}x_4 a\,\hat{\mathbf{y}}$ | (3e) | Ba |
| $\mathbf{B}_8$ | = | $x_4\,\mathbf{a}_2$ | = | $\frac{1}{2}x_4 a\,\hat{\mathbf{x}} + \frac{\sqrt{3}}{2}x_4 a\,\hat{\mathbf{y}}$ | (3e) | Ba |
| $\mathbf{B}_9$ | = | $-x_4\,\mathbf{a}_1 - x_4\,\mathbf{a}_2$ | = | $-x_4 a\,\hat{\mathbf{x}}$ | (3e) | Ba |
| $\mathbf{B}_{10}$ | = | $x_5\,\mathbf{a}_1 + \frac{1}{2}\,\mathbf{a}_3$ | = | $\frac{1}{2}x_5 a\,\hat{\mathbf{x}} - \frac{\sqrt{3}}{2}x_5 a\,\hat{\mathbf{y}} + \frac{1}{2}c\,\hat{\mathbf{z}}$ | (3f) | Ca |
| $\mathbf{B}_{11}$ | = | $x_5\,\mathbf{a}_2 + \frac{1}{2}\,\mathbf{a}_3$ | = | $\frac{1}{2}x_5 a\,\hat{\mathbf{x}} + \frac{\sqrt{3}}{2}x_5 a\,\hat{\mathbf{y}} + \frac{1}{2}c\,\hat{\mathbf{z}}$ | (3f) | Ca |
| $\mathbf{B}_{12}$ | = | $-x_5\,\mathbf{a}_1 - x_5\,\mathbf{a}_2 + \frac{1}{2}\,\mathbf{a}_3$ | = | $-x_5 a\,\hat{\mathbf{x}} + \frac{1}{2}c\,\hat{\mathbf{z}}$ | (3f) | Ca |
| $\mathbf{B}_{13}$ | = | $x_6\,\mathbf{a}_1 + y_6\,\mathbf{a}_2 + z_6\,\mathbf{a}_3$ | = | $\frac{1}{2}(x_6 + y_6)\,a\,\hat{\mathbf{x}} + \frac{\sqrt{3}}{2}(-x_6 + y_6)\,a\,\hat{\mathbf{y}} + z_6 c\,\hat{\mathbf{z}}$ | (6g) | O I |
| $\mathbf{B}_{14}$ | = | $-y_6\,\mathbf{a}_1 + (x_6 - y_6)\,\mathbf{a}_2 + z_6\,\mathbf{a}_3$ | = | $\left(\frac{1}{2}x_6 - y_6\right)a\,\hat{\mathbf{x}} + \frac{\sqrt{3}}{2}x_6 a\,\hat{\mathbf{y}} + z_6 c\,\hat{\mathbf{z}}$ | (6g) | O I |
| $\mathbf{B}_{15}$ | = | $(-x_6 + y_6)\,\mathbf{a}_1 - x_6\,\mathbf{a}_2 + z_6\,\mathbf{a}_3$ | = | $\left(-x_6 + \frac{1}{2}y_6\right)a\,\hat{\mathbf{x}} - \frac{\sqrt{3}}{2}y_6 a\,\hat{\mathbf{y}} + z_6 c\,\hat{\mathbf{z}}$ | (6g) | O I |
| $\mathbf{B}_{16}$ | = | $y_6\,\mathbf{a}_1 + x_6\,\mathbf{a}_2 - z_6\,\mathbf{a}_3$ | = | $\frac{1}{2}(x_6 + y_6)\,a\,\hat{\mathbf{x}} + \frac{\sqrt{3}}{2}(x_6 - y_6)\,a\,\hat{\mathbf{y}} - z_6 c\,\hat{\mathbf{z}}$ | (6g) | O I |
| $\mathbf{B}_{17}$ | = | $(x_6 - y_6)\,\mathbf{a}_1 - y_6\,\mathbf{a}_2 - z_6\,\mathbf{a}_3$ | = | $\left(\frac{1}{2}x_6 - y_6\right)a\,\hat{\mathbf{x}} - \frac{\sqrt{3}}{2}x_6 a\,\hat{\mathbf{y}} - z_6 c\,\hat{\mathbf{z}}$ | (6g) | O I |
| $\mathbf{B}_{18}$ | = | $-x_6\,\mathbf{a}_1 + (-x_6 + y_6)\,\mathbf{a}_2 - z_6\,\mathbf{a}_3$ | = | $\left(-x_6 + \frac{1}{2}y_6\right)a\,\hat{\mathbf{x}} + \frac{\sqrt{3}}{2}y_6 a\,\hat{\mathbf{y}} - z_6 c\,\hat{\mathbf{z}}$ | (6g) | O I |
| $\mathbf{B}_{19}$ | = | $x_7\,\mathbf{a}_1 + y_7\,\mathbf{a}_2 + z_7\,\mathbf{a}_3$ | = | $\frac{1}{2}(x_7 + y_7)\,a\,\hat{\mathbf{x}} + \frac{\sqrt{3}}{2}(-x_7 + y_7)\,a\,\hat{\mathbf{y}} + z_7 c\,\hat{\mathbf{z}}$ | (6g) | O II |
| $\mathbf{B}_{20}$ | = | $-y_7\,\mathbf{a}_1 + (x_7 - y_7)\,\mathbf{a}_2 + z_7\,\mathbf{a}_3$ | = | $\left(\frac{1}{2}x_7 - y_7\right)a\,\hat{\mathbf{x}} + \frac{\sqrt{3}}{2}x_7 a\,\hat{\mathbf{y}} + z_7 c\,\hat{\mathbf{z}}$ | (6g) | O II |
| $\mathbf{B}_{21}$ | = | $(-x_7 + y_7)\,\mathbf{a}_1 - x_7\,\mathbf{a}_2 + z_7\,\mathbf{a}_3$ | = | $\left(-x_7 + \frac{1}{2}y_7\right)a\,\hat{\mathbf{x}} - \frac{\sqrt{3}}{2}y_7 a\,\hat{\mathbf{y}} + z_7 c\,\hat{\mathbf{z}}$ | (6g) | O II |
| $\mathbf{B}_{22}$ | = | $y_7\,\mathbf{a}_1 + x_7\,\mathbf{a}_2 - z_7\,\mathbf{a}_3$ | = | $\frac{1}{2}(x_7 + y_7)\,a\,\hat{\mathbf{x}} + \frac{\sqrt{3}}{2}(x_7 - y_7)\,a\,\hat{\mathbf{y}} - z_7 c\,\hat{\mathbf{z}}$ | (6g) | O II |
| $\mathbf{B}_{23}$ | = | $(x_7 - y_7)\,\mathbf{a}_1 - y_7\,\mathbf{a}_2 - z_7\,\mathbf{a}_3$ | = | $\left(\frac{1}{2}x_7 - y_7\right)a\,\hat{\mathbf{x}} - \frac{\sqrt{3}}{2}x_7 a\,\hat{\mathbf{y}} - z_7 c\,\hat{\mathbf{z}}$ | (6g) | O II |
| $\mathbf{B}_{24}$ | = | $-x_7\,\mathbf{a}_1 + (-x_7 + y_7)\,\mathbf{a}_2 - z_7\,\mathbf{a}_3$ | = | $\left(-x_7 + \frac{1}{2}y_7\right)a\,\hat{\mathbf{x}} + \frac{\sqrt{3}}{2}y_7 a\,\hat{\mathbf{y}} - z_7 c\,\hat{\mathbf{z}}$ | (6g) | O II |
| $\mathbf{B}_{25}$ | = | $x_8\,\mathbf{a}_1 + y_8\,\mathbf{a}_2 + z_8\,\mathbf{a}_3$ | = | $\frac{1}{2}(x_8 + y_8)\,a\,\hat{\mathbf{x}} + \frac{\sqrt{3}}{2}(-x_8 + y_8)\,a\,\hat{\mathbf{y}} + z_8 c\,\hat{\mathbf{z}}$ | (6g) | O III |
| $\mathbf{B}_{26}$ | = | $-y_8\,\mathbf{a}_1 + (x_8 - y_8)\,\mathbf{a}_2 + z_8\,\mathbf{a}_3$ | = | $\left(\frac{1}{2}x_8 - y_8\right)a\,\hat{\mathbf{x}} + \frac{\sqrt{3}}{2}x_8 a\,\hat{\mathbf{y}} + z_8 c\,\hat{\mathbf{z}}$ | (6g) | O III |
| $\mathbf{B}_{27}$ | = | $(-x_8 + y_8)\,\mathbf{a}_1 - x_8\,\mathbf{a}_2 + z_8\,\mathbf{a}_3$ | = | $\left(-x_8 + \frac{1}{2}y_8\right)a\,\hat{\mathbf{x}} - \frac{\sqrt{3}}{2}y_8 a\,\hat{\mathbf{y}} + z_8 c\,\hat{\mathbf{z}}$ | (6g) | O III |
| $\mathbf{B}_{28}$ | = | $y_8\,\mathbf{a}_1 + x_8\,\mathbf{a}_2 - z_8\,\mathbf{a}_3$ | = | $\frac{1}{2}\left(x_8 + y_8\right) a\,\hat{\mathbf{x}} + \frac{\sqrt{3}}{2}\left(x_8 - y_8\right) a\,\hat{\mathbf{y}} - z_8 c\,\hat{\mathbf{z}}$ | (6$g$) | O III |
| $\mathbf{B}_{29}$ | = | $\left(x_8 - y_8\right)\,\mathbf{a}_1 - y_8\,\mathbf{a}_2 - z_8\,\mathbf{a}_3$ | = | $\left(\frac{1}{2}x_8 - y_8\right) a\,\hat{\mathbf{x}} - \frac{\sqrt{3}}{2}x_8 a\,\hat{\mathbf{y}} - z_8 c\,\hat{\mathbf{z}}$ | (6$g$) | O III |
| $\mathbf{B}_{30}$ | = | $-x_8\,\mathbf{a}_1 + \left(-x_8 + y_8\right)\,\mathbf{a}_2 - z_8\,\mathbf{a}_3$ | = | $\left(-x_8 + \frac{1}{2}y_8\right) a\,\hat{\mathbf{x}} + \frac{\sqrt{3}}{2}y_8 a\,\hat{\mathbf{y}} - z_8 c\,\hat{\mathbf{z}}$ | (6$g$) | O III |

**Found in:**

**Geometry files:**

- CIF: pp. 1724
- POSCAR: pp. 1725

# $KSO_3$ ($K1_1$) Structure: AB3C_hP30_150_ef_3g_c2d

http://aflow.org/prototype-encyclopedia/AB3C_hP30_150_ef_3g_c2d

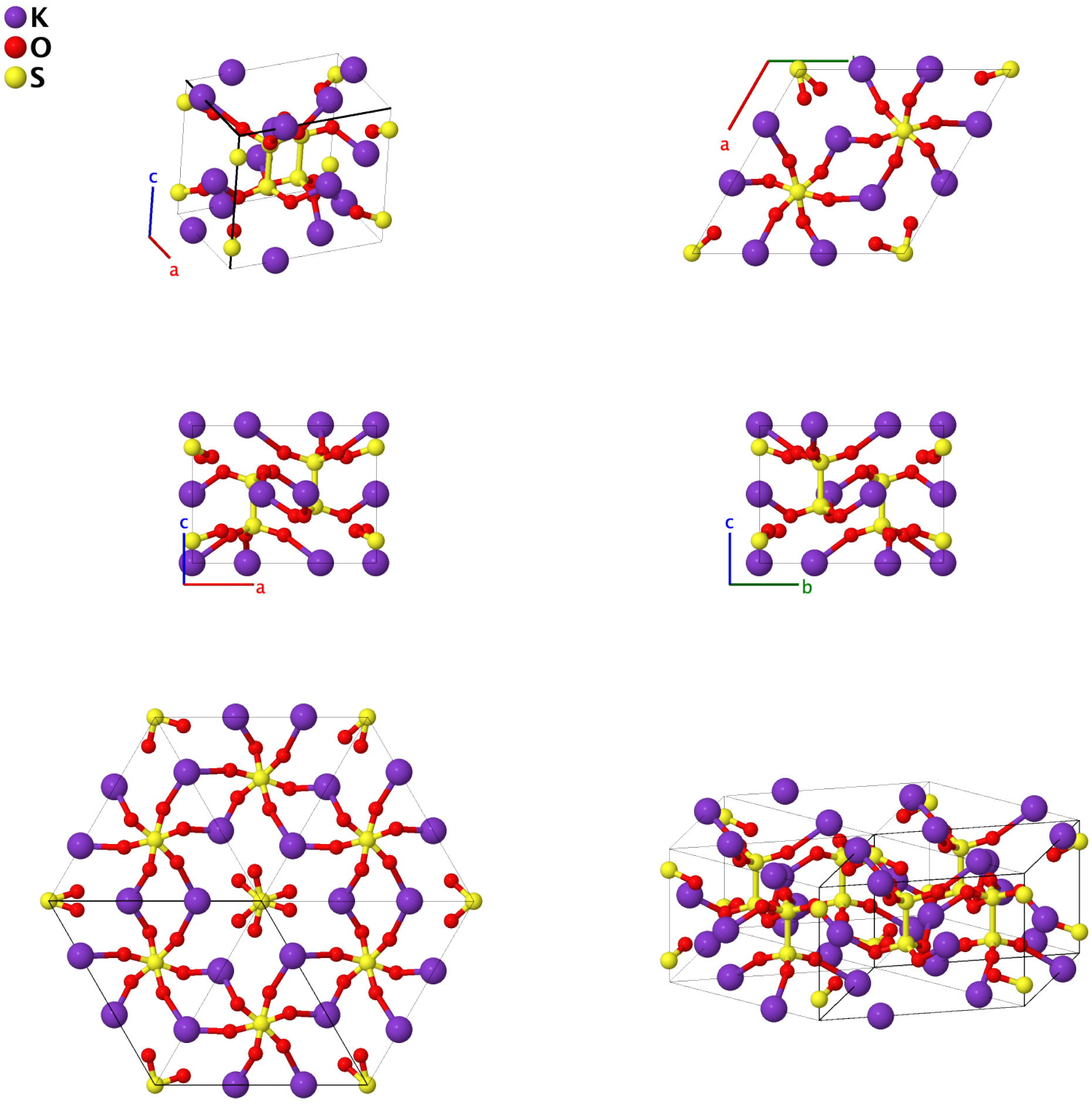

| | | |
|---|---|---|
| **Prototype** | : | $KO_3S$ |
| **AFLOW prototype label** | : | AB3C_hP30_150_ef_3g_c2d |
| ***Strukturbericht* designation** | : | $K1_1$ |
| **Pearson symbol** | : | hP30 |
| **Space group number** | : | 150 |
| **Space group symbol** | : | $P321$ |
| **AFLOW prototype command** | : | `aflow --proto=AB3C_hP30_150_ef_3g_c2d`<br>`--params=`$a, c/a, z_1, z_2, z_3, x_4, x_5, x_6, y_6, z_6, x_7, y_7, z_7, x_8, y_8, z_8$ |

### Other compounds with this structure

- $RbSO_3$

**Trigonal Hexagonal primitive vectors:**

$$\mathbf{a}_1 = \frac{1}{2}\,a\,\hat{\mathbf{x}} - \frac{\sqrt{3}}{2}\,a\,\hat{\mathbf{y}}$$
$$\mathbf{a}_2 = \frac{1}{2}\,a\,\hat{\mathbf{x}} + \frac{\sqrt{3}}{2}\,a\,\hat{\mathbf{y}}$$
$$\mathbf{a}_3 = c\,\hat{\mathbf{z}}$$

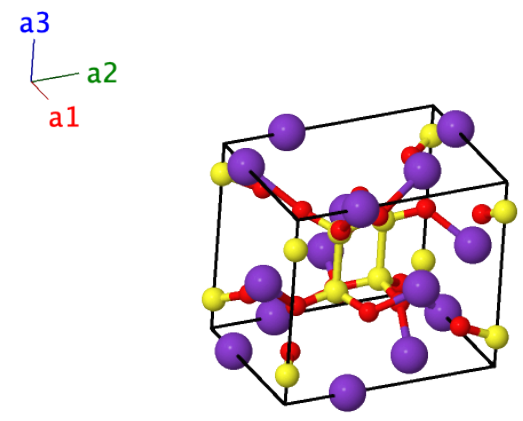

**Basis vectors:**

| | | Lattice Coordinates | | Cartesian Coordinates | Wyckoff Position | Atom Type |
|---|---|---|---|---|---|---|
| $\mathbf{B}_1$ | = | $z_1\,\mathbf{a}_3$ | = | $z_1 c\,\hat{\mathbf{z}}$ | (2$c$) | S I |
| $\mathbf{B}_2$ | = | $-z_1\,\mathbf{a}_3$ | = | $-z_1 c\,\hat{\mathbf{z}}$ | (2$c$) | S I |
| $\mathbf{B}_3$ | = | $\frac{1}{3}\,\mathbf{a}_1 + \frac{2}{3}\,\mathbf{a}_2 + z_2\,\mathbf{a}_3$ | = | $\frac{1}{2}a\,\hat{\mathbf{x}} + \frac{1}{2\sqrt{3}}a\,\hat{\mathbf{y}} + z_2 c\,\hat{\mathbf{z}}$ | (2$d$) | S II |
| $\mathbf{B}_4$ | = | $\frac{2}{3}\,\mathbf{a}_1 + \frac{1}{3}\,\mathbf{a}_2 - z_2\,\mathbf{a}_3$ | = | $\frac{1}{2}a\,\hat{\mathbf{x}} - \frac{1}{2\sqrt{3}}a\,\hat{\mathbf{y}} - z_2 c\,\hat{\mathbf{z}}$ | (2$d$) | S II |
| $\mathbf{B}_5$ | = | $\frac{1}{3}\,\mathbf{a}_1 + \frac{2}{3}\,\mathbf{a}_2 + z_3\,\mathbf{a}_3$ | = | $\frac{1}{2}a\,\hat{\mathbf{x}} + \frac{1}{2\sqrt{3}}a\,\hat{\mathbf{y}} + z_3 c\,\hat{\mathbf{z}}$ | (2$d$) | S III |
| $\mathbf{B}_6$ | = | $\frac{2}{3}\,\mathbf{a}_1 + \frac{1}{3}\,\mathbf{a}_2 - z_3\,\mathbf{a}_3$ | = | $\frac{1}{2}a\,\hat{\mathbf{x}} - \frac{1}{2\sqrt{3}}a\,\hat{\mathbf{y}} - z_3 c\,\hat{\mathbf{z}}$ | (2$d$) | S III |
| $\mathbf{B}_7$ | = | $x_4\,\mathbf{a}_1$ | = | $\frac{1}{2}x_4 a\,\hat{\mathbf{x}} - \frac{\sqrt{3}}{2}x_4 a\,\hat{\mathbf{y}}$ | (3$e$) | K I |
| $\mathbf{B}_8$ | = | $x_4\,\mathbf{a}_2$ | = | $\frac{1}{2}x_4 a\,\hat{\mathbf{x}} + \frac{\sqrt{3}}{2}x_4 a\,\hat{\mathbf{y}}$ | (3$e$) | K I |
| $\mathbf{B}_9$ | = | $-x_4\,\mathbf{a}_1 - x_4\,\mathbf{a}_2$ | = | $-x_4 a\,\hat{\mathbf{x}}$ | (3$e$) | K I |
| $\mathbf{B}_{10}$ | = | $x_5\,\mathbf{a}_1 + \frac{1}{2}\,\mathbf{a}_3$ | = | $\frac{1}{2}x_5 a\,\hat{\mathbf{x}} - \frac{\sqrt{3}}{2}x_5 a\,\hat{\mathbf{y}} + \frac{1}{2}c\,\hat{\mathbf{z}}$ | (3$f$) | K II |
| $\mathbf{B}_{11}$ | = | $x_5\,\mathbf{a}_2 + \frac{1}{2}\,\mathbf{a}_3$ | = | $\frac{1}{2}x_5 a\,\hat{\mathbf{x}} + \frac{\sqrt{3}}{2}x_5 a\,\hat{\mathbf{y}} + \frac{1}{2}c\,\hat{\mathbf{z}}$ | (3$f$) | K II |
| $\mathbf{B}_{12}$ | = | $-x_5\,\mathbf{a}_1 - x_5\,\mathbf{a}_2 + \frac{1}{2}\,\mathbf{a}_3$ | = | $-x_5 a\,\hat{\mathbf{x}} + \frac{1}{2}c\,\hat{\mathbf{z}}$ | (3$f$) | K II |
| $\mathbf{B}_{13}$ | = | $x_6\,\mathbf{a}_1 + y_6\,\mathbf{a}_2 + z_6\,\mathbf{a}_3$ | = | $\frac{1}{2}\left(x_6 + y_6\right)a\,\hat{\mathbf{x}} + \frac{\sqrt{3}}{2}\left(-x_6 + y_6\right)a\,\hat{\mathbf{y}} + z_6 c\,\hat{\mathbf{z}}$ | (6$g$) | O I |
| $\mathbf{B}_{14}$ | = | $-y_6\,\mathbf{a}_1 + \left(x_6 - y_6\right)\mathbf{a}_2 + z_6\,\mathbf{a}_3$ | = | $\left(\frac{1}{2}x_6 - y_6\right)a\,\hat{\mathbf{x}} + \frac{\sqrt{3}}{2}x_6 a\,\hat{\mathbf{y}} + z_6 c\,\hat{\mathbf{z}}$ | (6$g$) | O I |
| $\mathbf{B}_{15}$ | = | $\left(-x_6 + y_6\right)\mathbf{a}_1 - x_6\,\mathbf{a}_2 + z_6\,\mathbf{a}_3$ | = | $\left(-x_6 + \frac{1}{2}y_6\right)a\,\hat{\mathbf{x}} - \frac{\sqrt{3}}{2}y_6 a\,\hat{\mathbf{y}} + z_6 c\,\hat{\mathbf{z}}$ | (6$g$) | O I |
| $\mathbf{B}_{16}$ | = | $y_6\,\mathbf{a}_1 + x_6\,\mathbf{a}_2 - z_6\,\mathbf{a}_3$ | = | $\frac{1}{2}\left(x_6 + y_6\right)a\,\hat{\mathbf{x}} + \frac{\sqrt{3}}{2}\left(x_6 - y_6\right)a\,\hat{\mathbf{y}} - z_6 c\,\hat{\mathbf{z}}$ | (6$g$) | O I |
| $\mathbf{B}_{17}$ | = | $\left(x_6 - y_6\right)\mathbf{a}_1 - y_6\,\mathbf{a}_2 - z_6\,\mathbf{a}_3$ | = | $\left(\frac{1}{2}x_6 - y_6\right)a\,\hat{\mathbf{x}} - \frac{\sqrt{3}}{2}x_6 a\,\hat{\mathbf{y}} - z_6 c\,\hat{\mathbf{z}}$ | (6$g$) | O I |
| $\mathbf{B}_{18}$ | = | $-x_6\,\mathbf{a}_1 + \left(-x_6 + y_6\right)\mathbf{a}_2 - z_6\,\mathbf{a}_3$ | = | $\left(-x_6 + \frac{1}{2}y_6\right)a\,\hat{\mathbf{x}} + \frac{\sqrt{3}}{2}y_6 a\,\hat{\mathbf{y}} - z_6 c\,\hat{\mathbf{z}}$ | (6$g$) | O I |
| $\mathbf{B}_{19}$ | = | $x_7\,\mathbf{a}_1 + y_7\,\mathbf{a}_2 + z_7\,\mathbf{a}_3$ | = | $\frac{1}{2}\left(x_7 + y_7\right)a\,\hat{\mathbf{x}} + \frac{\sqrt{3}}{2}\left(-x_7 + y_7\right)a\,\hat{\mathbf{y}} + z_7 c\,\hat{\mathbf{z}}$ | (6$g$) | O II |
| $\mathbf{B}_{20}$ | = | $-y_7\,\mathbf{a}_1 + \left(x_7 - y_7\right)\mathbf{a}_2 + z_7\,\mathbf{a}_3$ | = | $\left(\frac{1}{2}x_7 - y_7\right)a\,\hat{\mathbf{x}} + \frac{\sqrt{3}}{2}x_7 a\,\hat{\mathbf{y}} + z_7 c\,\hat{\mathbf{z}}$ | (6$g$) | O II |
| $\mathbf{B}_{21}$ | = | $\left(-x_7 + y_7\right)\mathbf{a}_1 - x_7\,\mathbf{a}_2 + z_7\,\mathbf{a}_3$ | = | $\left(-x_7 + \frac{1}{2}y_7\right)a\,\hat{\mathbf{x}} - \frac{\sqrt{3}}{2}y_7 a\,\hat{\mathbf{y}} + z_7 c\,\hat{\mathbf{z}}$ | (6$g$) | O II |
| $\mathbf{B}_{22}$ | = | $y_7\,\mathbf{a}_1 + x_7\,\mathbf{a}_2 - z_7\,\mathbf{a}_3$ | = | $\frac{1}{2}\left(x_7 + y_7\right)a\,\hat{\mathbf{x}} + \frac{\sqrt{3}}{2}\left(x_7 - y_7\right)a\,\hat{\mathbf{y}} - z_7 c\,\hat{\mathbf{z}}$ | (6$g$) | O II |
| $\mathbf{B}_{23}$ | = | $\left(x_7 - y_7\right)\mathbf{a}_1 - y_7\,\mathbf{a}_2 - z_7\,\mathbf{a}_3$ | = | $\left(\frac{1}{2}x_7 - y_7\right)a\,\hat{\mathbf{x}} - \frac{\sqrt{3}}{2}x_7 a\,\hat{\mathbf{y}} - z_7 c\,\hat{\mathbf{z}}$ | (6$g$) | O II |
| $\mathbf{B}_{24}$ | = | $-x_7\,\mathbf{a}_1 + \left(-x_7 + y_7\right)\mathbf{a}_2 - z_7\,\mathbf{a}_3$ | = | $\left(-x_7 + \frac{1}{2}y_7\right)a\,\hat{\mathbf{x}} + \frac{\sqrt{3}}{2}y_7 a\,\hat{\mathbf{y}} - z_7 c\,\hat{\mathbf{z}}$ | (6$g$) | O II |
| $\mathbf{B}_{25}$ | = | $x_8\,\mathbf{a}_1 + y_8\,\mathbf{a}_2 + z_8\,\mathbf{a}_3$ | = | $\frac{1}{2}\left(x_8 + y_8\right)a\,\hat{\mathbf{x}} + \frac{\sqrt{3}}{2}\left(-x_8 + y_8\right)a\,\hat{\mathbf{y}} + z_8 c\,\hat{\mathbf{z}}$ | (6$g$) | O III |
| $\mathbf{B}_{26}$ | = | $-y_8\,\mathbf{a}_1 + \left(x_8 - y_8\right)\mathbf{a}_2 + z_8\,\mathbf{a}_3$ | = | $\left(\frac{1}{2}x_8 - y_8\right)a\,\hat{\mathbf{x}} + \frac{\sqrt{3}}{2}x_8 a\,\hat{\mathbf{y}} + z_8 c\,\hat{\mathbf{z}}$ | (6$g$) | O III |
| $\mathbf{B}_{27}$ | = | $\left(-x_8 + y_8\right)\mathbf{a}_1 - x_8\,\mathbf{a}_2 + z_8\,\mathbf{a}_3$ | = | $\left(-x_8 + \frac{1}{2}y_8\right)a\,\hat{\mathbf{x}} - \frac{\sqrt{3}}{2}y_8 a\,\hat{\mathbf{y}} + z_8 c\,\hat{\mathbf{z}}$ | (6$g$) | O III |
| $\mathbf{B}_{28}$ | = | $y_8\,\mathbf{a}_1 + x_8\,\mathbf{a}_2 - z_8\,\mathbf{a}_3$ | = | $\frac{1}{2}\left(x_8 + y_8\right) a\,\hat{\mathbf{x}} + \frac{\sqrt{3}}{2}\left(x_8 - y_8\right) a\,\hat{\mathbf{y}} - z_8 c\,\hat{\mathbf{z}}$ | (6g) | O III |
| $\mathbf{B}_{29}$ | = | $\left(x_8 - y_8\right)\,\mathbf{a}_1 - y_8\,\mathbf{a}_2 - z_8\,\mathbf{a}_3$ | = | $\left(\frac{1}{2}x_8 - y_8\right) a\,\hat{\mathbf{x}} - \frac{\sqrt{3}}{2}x_8 a\,\hat{\mathbf{y}} - z_8 c\,\hat{\mathbf{z}}$ | (6g) | O III |
| $\mathbf{B}_{30}$ | = | $-x_8\,\mathbf{a}_1 + \left(-x_8 + y_8\right)\,\mathbf{a}_2 - z_8\,\mathbf{a}_3$ | = | $\left(-x_8 + \frac{1}{2}y_8\right) a\,\hat{\mathbf{x}} + \frac{\sqrt{3}}{2}y_8 a\,\hat{\mathbf{y}} - z_8 c\,\hat{\mathbf{z}}$ | (6g) | O III |

---

**Found in:**

---

**Geometry files:**

- CIF: pp. 1725
- POSCAR: pp. 1725

# Steklite [$KAl(SO_4)_2$, $H3_2$] Structure: ABC8D2_hP12_150_b_a_dg_d

http://aflow.org/prototype-encyclopedia/ABC8D2_hP12_150_b_a_dg_d

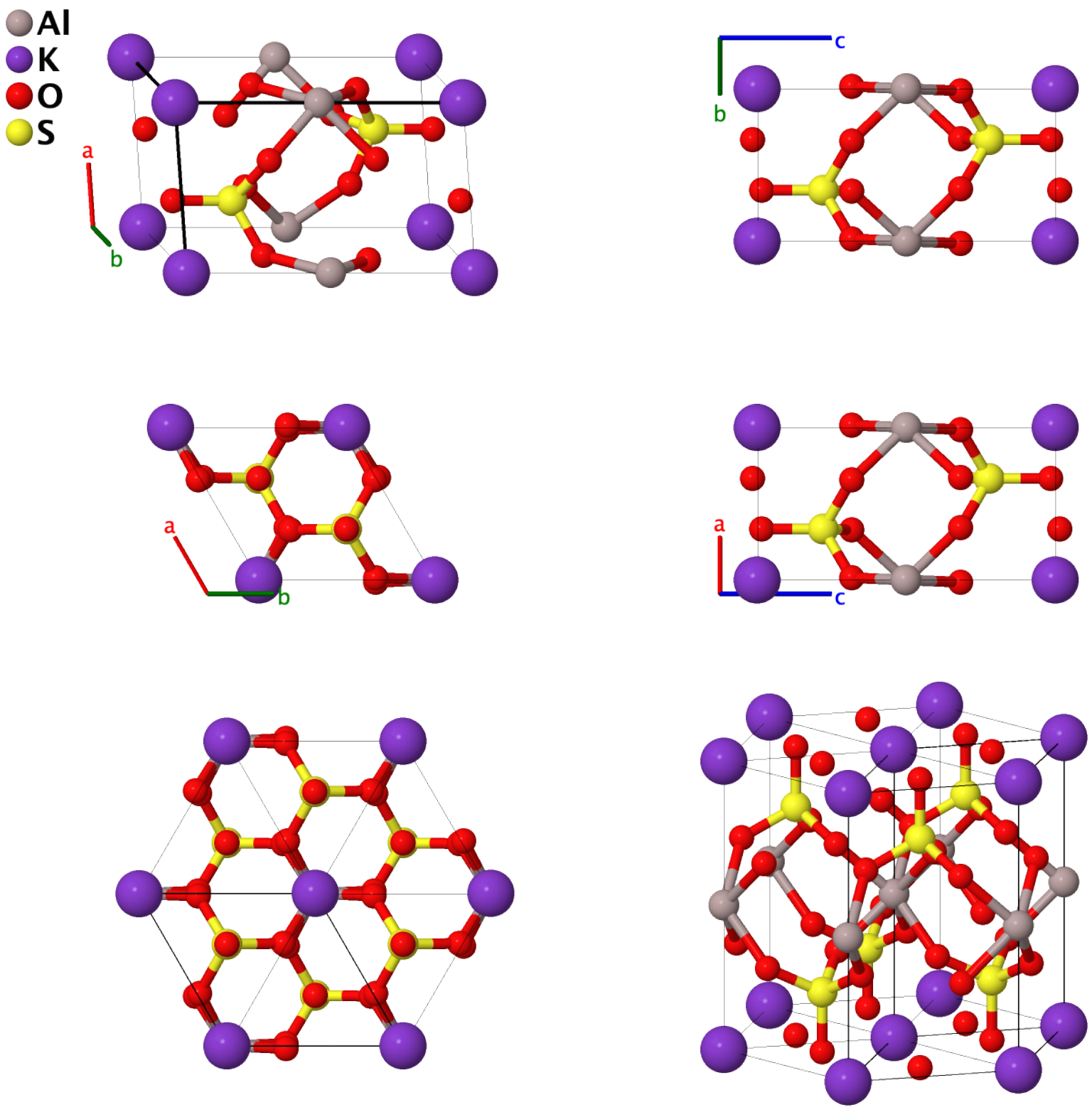

| | | |
|---|---|---|
| **Prototype** | : | $AlKO_8S_2$ |
| **AFLOW prototype label** | : | ABC8D2_hP12_150_b_a_dg_d |
| ***Strukturbericht* designation** | : | $H3_2$ |
| **Pearson symbol** | : | hP12 |
| **Space group number** | : | 150 |
| **Space group symbol** | : | $P321$ |
| **AFLOW prototype command** | : | `aflow --proto=ABC8D2_hP12_150_b_a_dg_d --params=`$a, c/a, z_3, z_4, x_5, y_5, z_5$ |

**Other compounds with this structure**

- $NH_4(Al,Fe)(SO_4)_2$ (godovikovite), $KCr(SO_4)_2$, and $RbCr(SO_4)_2$

- This has been a rather difficult structure to follow through the literature. (Villars, 2016) quotes the structure given by (Manoli, 1970), but gives the space group as $P\bar{3}$ #147. (Murashko, 2013) lists the space group as both $P312$ #149 and $P321$ #150, as well as listing obviously incorrect Wyckoff positions.

- (West, 2008) states that the simple structure is in $P\bar{3}$ but that it may be doubled along the $c$ axis and be in space group $P321$.

- After correcting Murashko's results, we find that all of these interpretations yield essentially the same structure in a given layer, and only differ as the structure is reflected through the $z = 0$ plane. As it is not clear which structure is correct, we will use the original $H3_2$ structure given by (Hermann, 1937).

- Steklite is the name of the mineral form of this compound (Murashko, 2013). (Hermann, 1937) simply calls it *Wasserfreier Alaun* (anhydrous alum). For hydrated alum, $KAl(SO_4)_2 \cdot 12H_2O$, see the $H4_{13}$ structure.

---

**Trigonal Hexagonal primitive vectors:**

$$
\begin{aligned}
\mathbf{a}_1 &= \tfrac{1}{2}\,a\,\hat{\mathbf{x}} - \tfrac{\sqrt{3}}{2}\,a\,\hat{\mathbf{y}} \\
\mathbf{a}_2 &= \tfrac{1}{2}\,a\,\hat{\mathbf{x}} + \tfrac{\sqrt{3}}{2}\,a\,\hat{\mathbf{y}} \\
\mathbf{a}_3 &= c\,\hat{\mathbf{z}}
\end{aligned}
$$

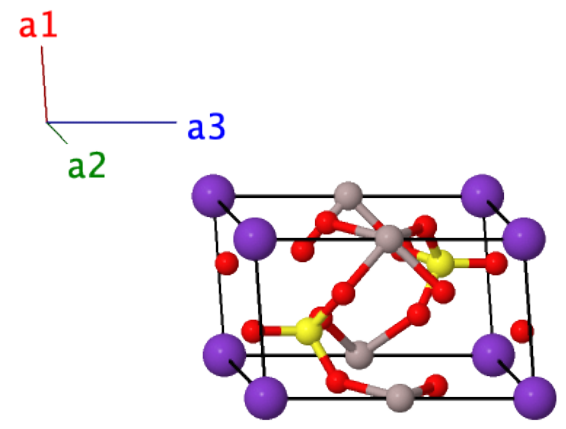

---

**Basis vectors:**

| | | Lattice Coordinates | | Cartesian Coordinates | Wyckoff Position | Atom Type |
|---|---|---|---|---|---|---|
| $\mathbf{B}_1$ | = | $0\,\mathbf{a}_1 + 0\,\mathbf{a}_2 + 0\,\mathbf{a}_3$ | = | $0\,\hat{\mathbf{x}} + 0\,\hat{\mathbf{y}} + 0\,\hat{\mathbf{z}}$ | (1a) | K |
| $\mathbf{B}_2$ | = | $\frac{1}{2}\,\mathbf{a}_3$ | = | $\frac{1}{2}c\,\hat{\mathbf{z}}$ | (1b) | Al |
| $\mathbf{B}_3$ | = | $\frac{1}{3}\,\mathbf{a}_1 + \frac{2}{3}\,\mathbf{a}_2 + z_3\,\mathbf{a}_3$ | = | $\frac{1}{2}a\,\hat{\mathbf{x}} + \frac{1}{2\sqrt{3}}a\,\hat{\mathbf{y}} + z_3c\,\hat{\mathbf{z}}$ | (2d) | O I |
| $\mathbf{B}_4$ | = | $\frac{2}{3}\,\mathbf{a}_1 + \frac{1}{3}\,\mathbf{a}_2 - z_3\,\mathbf{a}_3$ | = | $\frac{1}{2}a\,\hat{\mathbf{x}} - \frac{1}{2\sqrt{3}}a\,\hat{\mathbf{y}} - z_3c\,\hat{\mathbf{z}}$ | (2d) | O I |
| $\mathbf{B}_5$ | = | $\frac{1}{3}\,\mathbf{a}_1 + \frac{2}{3}\,\mathbf{a}_2 + z_4\,\mathbf{a}_3$ | = | $\frac{1}{2}a\,\hat{\mathbf{x}} + \frac{1}{2\sqrt{3}}a\,\hat{\mathbf{y}} + z_4c\,\hat{\mathbf{z}}$ | (2d) | S |
| $\mathbf{B}_6$ | = | $\frac{2}{3}\,\mathbf{a}_1 + \frac{1}{3}\,\mathbf{a}_2 - z_4\,\mathbf{a}_3$ | = | $\frac{1}{2}a\,\hat{\mathbf{x}} - \frac{1}{2\sqrt{3}}a\,\hat{\mathbf{y}} - z_4c\,\hat{\mathbf{z}}$ | (2d) | S |
| $\mathbf{B}_7$ | = | $x_5\,\mathbf{a}_1 + y_5\,\mathbf{a}_2 + z_5\,\mathbf{a}_3$ | = | $\frac{1}{2}(x_5+y_5)\,a\,\hat{\mathbf{x}} + \frac{\sqrt{3}}{2}(-x_5+y_5)\,a\,\hat{\mathbf{y}} + z_5c\,\hat{\mathbf{z}}$ | (6g) | O II |
| $\mathbf{B}_8$ | = | $-y_5\,\mathbf{a}_1 + (x_5-y_5)\,\mathbf{a}_2 + z_5\,\mathbf{a}_3$ | = | $\left(\frac{1}{2}x_5 - y_5\right)a\,\hat{\mathbf{x}} + \frac{\sqrt{3}}{2}x_5a\,\hat{\mathbf{y}} + z_5c\,\hat{\mathbf{z}}$ | (6g) | O II |
| $\mathbf{B}_9$ | = | $(-x_5+y_5)\,\mathbf{a}_1 - x_5\,\mathbf{a}_2 + z_5\,\mathbf{a}_3$ | = | $\left(-x_5 + \frac{1}{2}y_5\right)a\,\hat{\mathbf{x}} - \frac{\sqrt{3}}{2}y_5a\,\hat{\mathbf{y}} + z_5c\,\hat{\mathbf{z}}$ | (6g) | O II |
| $\mathbf{B}_{10}$ | = | $y_5\,\mathbf{a}_1 + x_5\,\mathbf{a}_2 - z_5\,\mathbf{a}_3$ | = | $\frac{1}{2}(x_5+y_5)\,a\,\hat{\mathbf{x}} + \frac{\sqrt{3}}{2}(x_5-y_5)\,a\,\hat{\mathbf{y}} - z_5c\,\hat{\mathbf{z}}$ | (6g) | O II |
| $\mathbf{B}_{11}$ | = | $(x_5-y_5)\,\mathbf{a}_1 - y_5\,\mathbf{a}_2 - z_5\,\mathbf{a}_3$ | = | $\left(\frac{1}{2}x_5 - y_5\right)a\,\hat{\mathbf{x}} - \frac{\sqrt{3}}{2}x_5a\,\hat{\mathbf{y}} - z_5c\,\hat{\mathbf{z}}$ | (6g) | O II |
| $\mathbf{B}_{12}$ | = | $-x_5\,\mathbf{a}_1 + (-x_5+y_5)\,\mathbf{a}_2 - z_5\,\mathbf{a}_3$ | = | $\left(-x_5 + \frac{1}{2}y_5\right)a\,\hat{\mathbf{x}} + \frac{\sqrt{3}}{2}y_5a\,\hat{\mathbf{y}} - z_5c\,\hat{\mathbf{z}}$ | (6g) | O II |

---

**Found in:**

---

**Geometry files:**
- CIF: pp. 1726
- POSCAR: pp. 1726

# $KBe_2BO_3F_2$ Structure: AB2C2DE3_hR9_155_b_c_c_a_e

http://aflow.org/prototype-encyclopedia/AB2C2DE3_hR9_155_b_c_c_a_e

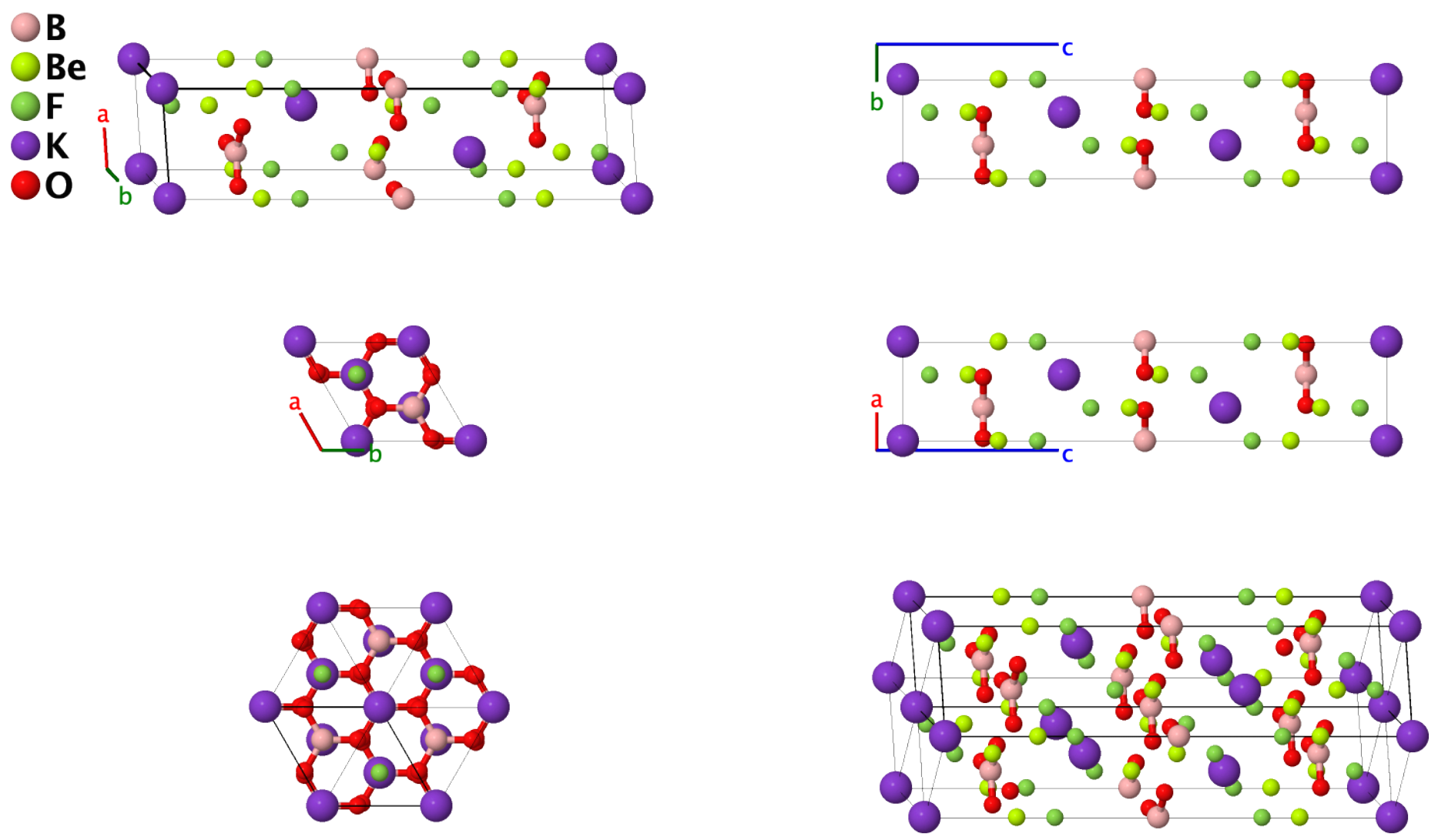

| | | |
|---|---|---|
| **Prototype** | : | $BBe_2F_2KO_3$ |
| **AFLOW prototype label** | : | AB2C2DE3_hR9_155_b_c_c_a_e |
| ***Strukturbericht* designation** | : | None |
| **Pearson symbol** | : | hR9 |
| **Space group number** | : | 155 |
| **Space group symbol** | : | $R32$ |
| **AFLOW prototype command** | : | `aflow --proto=AB2C2DE3_hR9_155_b_c_c_a_e [--hex]`<br>`--params=`$a, c/a, x_3, x_4, y_5$ |

**Rhombohedral primitive vectors:**

$$\mathbf{a}_1 = \frac{1}{2}\,a\,\hat{\mathbf{x}} - \frac{1}{2\sqrt{3}}\,a\,\hat{\mathbf{y}} + \frac{1}{3}\,c\,\hat{\mathbf{z}}$$
$$\mathbf{a}_2 = \frac{1}{\sqrt{3}}\,a\,\hat{\mathbf{y}} + \frac{1}{3}\,c\,\hat{\mathbf{z}}$$
$$\mathbf{a}_3 = -\frac{1}{2}\,a\,\hat{\mathbf{x}} - \frac{1}{2\sqrt{3}}\,a\,\hat{\mathbf{y}} + \frac{1}{3}\,c\,\hat{\mathbf{z}}$$

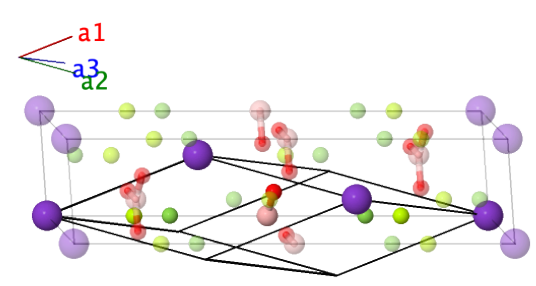

**Basis vectors:**

| | | Lattice Coordinates | | Cartesian Coordinates | Wyckoff Position | Atom Type |
|---|---|---|---|---|---|---|
| $\mathbf{B}_1$ | = | $0\,\mathbf{a}_1 + 0\,\mathbf{a}_2 + 0\,\mathbf{a}_3$ | = | $0\,\hat{\mathbf{x}} + 0\,\hat{\mathbf{y}} + 0\,\hat{\mathbf{z}}$ | (1$a$) | K |
| $\mathbf{B}_2$ | = | $\frac{1}{2}\,\mathbf{a}_1 + \frac{1}{2}\,\mathbf{a}_2 + \frac{1}{2}\,\mathbf{a}_3$ | = | $\frac{1}{2}c\,\hat{\mathbf{z}}$ | (1$b$) | B |
| $\mathbf{B}_3$ | = | $x_3\,\mathbf{a}_1 + x_3\,\mathbf{a}_2 + x_3\,\mathbf{a}_3$ | = | $x_3 c\,\hat{\mathbf{z}}$ | (2$c$) | Be |
| $\mathbf{B}_4$ | = | $-x_3\,\mathbf{a}_1 - x_3\,\mathbf{a}_2 - x_3\,\mathbf{a}_3$ | = | $-x_3 c\,\hat{\mathbf{z}}$ | (2$c$) | Be |
| $\mathbf{B}_5$ | = | $x_4\,\mathbf{a}_1 + x_4\,\mathbf{a}_2 + x_4\,\mathbf{a}_3$ | = | $x_4 c\,\hat{\mathbf{z}}$ | (2$c$) | F |
| $\mathbf{B}_6$ | = | $-x_4\,\mathbf{a}_1 - x_4\,\mathbf{a}_2 - x_4\,\mathbf{a}_3$ | = | $-x_4 c\,\hat{\mathbf{z}}$ | (2$c$) | F |
| $\mathbf{B}_7$ | = | $\frac{1}{2}\,\mathbf{a}_1 + y_5\,\mathbf{a}_2 - y_5\,\mathbf{a}_3$ | = | $\left(\frac{1}{4} + \frac{1}{2}y_5\right)a\,\hat{\mathbf{x}} + \left(-\frac{1}{4\sqrt{3}} + \frac{\sqrt{3}}{2}y_5\right)a\,\hat{\mathbf{y}} + \frac{1}{6}c\,\hat{\mathbf{z}}$ | (3$e$) | O |
| $\mathbf{B}_8$ | = | $-y_5\,\mathbf{a}_1 + \frac{1}{2}\,\mathbf{a}_2 + y_5\,\mathbf{a}_3$ | = | $-y_5 a\,\hat{\mathbf{x}} + \frac{1}{2\sqrt{3}}a\,\hat{\mathbf{y}} + \frac{1}{6}c\,\hat{\mathbf{z}}$ | (3$e$) | O |
| $\mathbf{B}_9$ | = | $y_5\,\mathbf{a}_1 - y_5\,\mathbf{a}_2 + \frac{1}{2}\,\mathbf{a}_3$ | = | $\left(-\frac{1}{4} + \frac{1}{2}y_5\right)a\,\hat{\mathbf{x}} - a\left(\frac{\sqrt{3}}{2}y_5 + \frac{1}{4\sqrt{3}}\right)\hat{\mathbf{y}} + \frac{1}{6}c\,\hat{\mathbf{z}}$ | $(3e)$ | O |

**Geometry files:**

- CIF: pp. 1726
- POSCAR: pp. 1726

# $SbI_3S_{24}$ Structure: A3B24C_hR28_160_b_2b3c_a

http://aflow.org/prototype-encyclopedia/A3B24C_hR28_160_b_2b3c_a

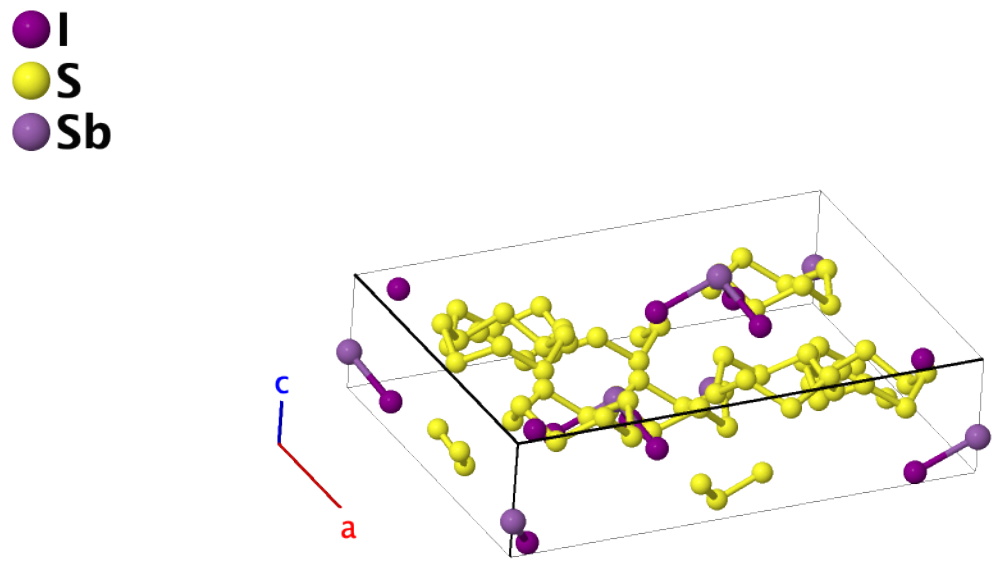

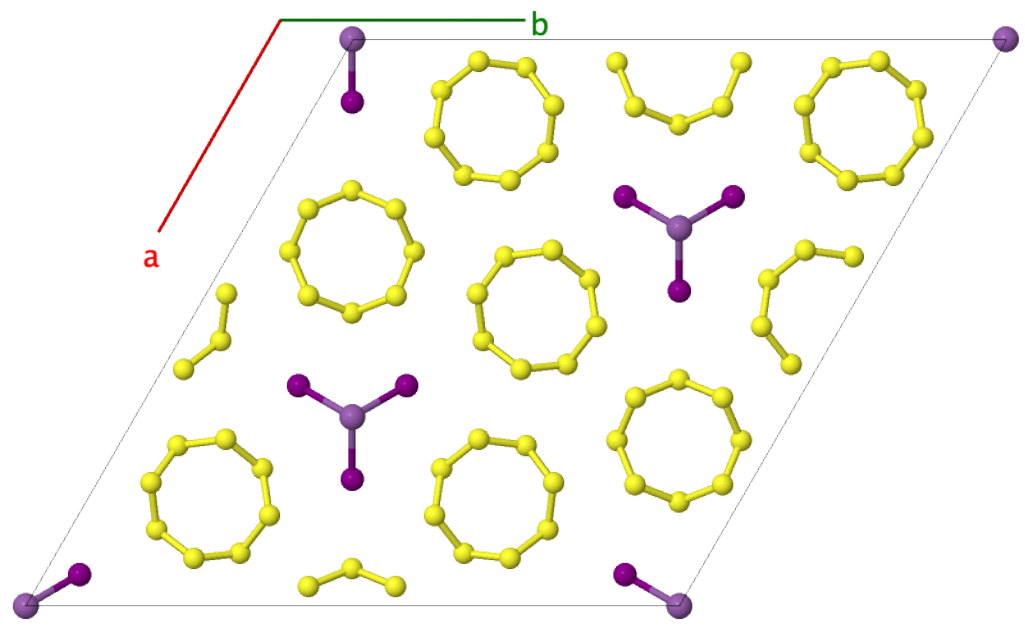

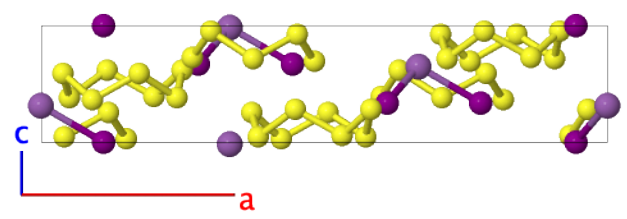

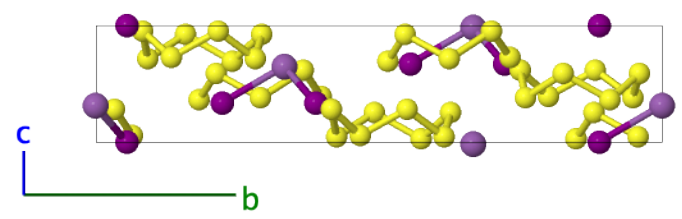

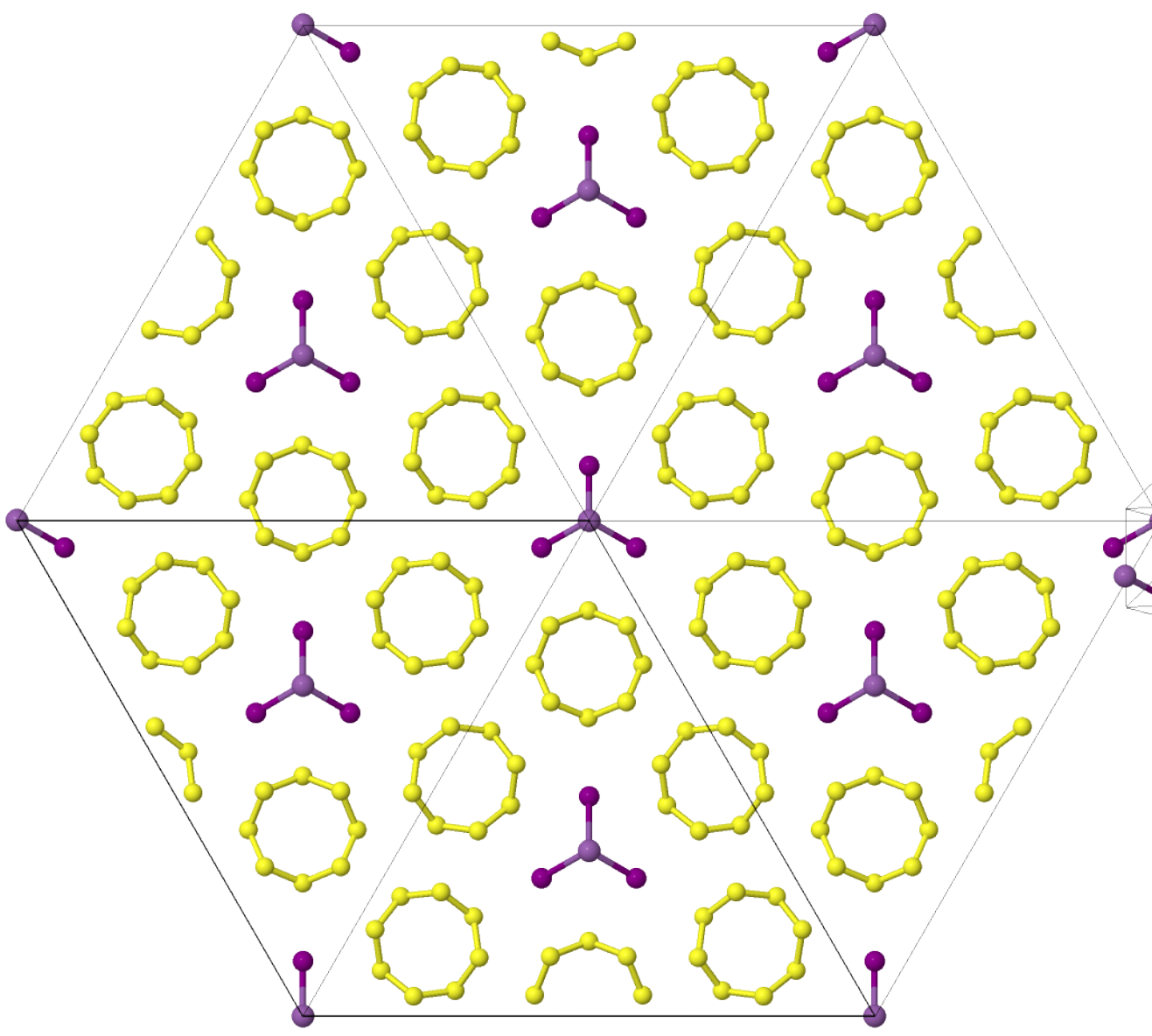

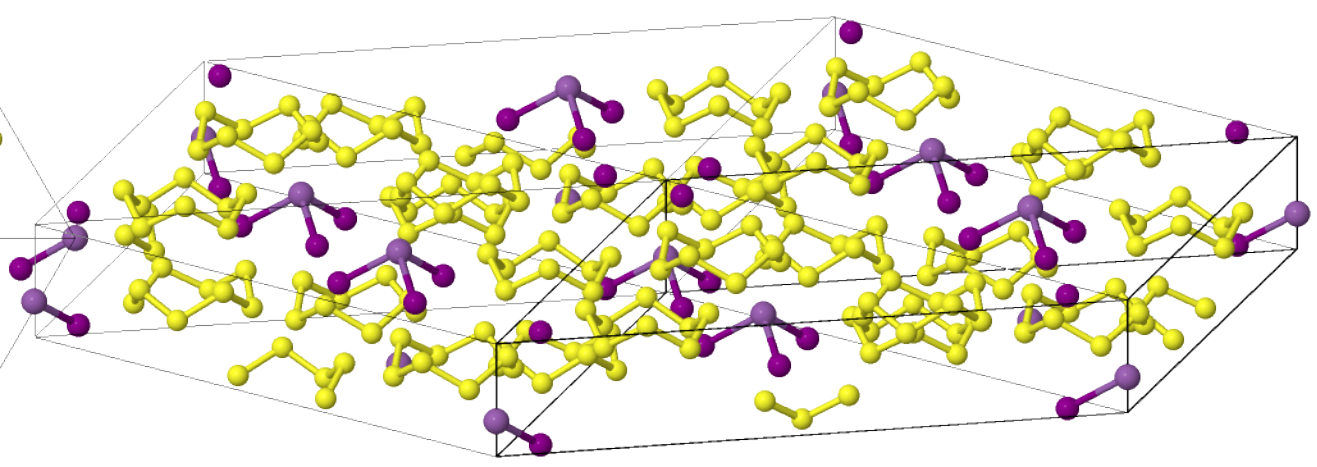

| | | |
|---|---|---|
| **Prototype** | : | $I_3S_{24}Sb$ |
| **AFLOW prototype label** | : | A3B24C_hR28_160_b_2b3c_a |
| ***Strukturbericht* designation** | : | None |
| **Pearson symbol** | : | hR28 |
| **Space group number** | : | 160 |
| **Space group symbol** | : | $R3m$ |
| **AFLOW prototype command** | : | `aflow --proto=A3B24C_hR28_160_b_2b3c_a [--hex]` `--params=`$a, c/a, x_1, x_2, z_2, x_3, z_3, x_4, z_4, x_5, y_5, z_5, x_6, y_6, z_6, x_7, y_7, z_7$ |

- Since there are three $S_8$ molecules in this structure, (Bjorvatten, 1963) refer to it as $SbI_3$:3$S_8$.
- Space group $R3m$ #160 does not fix the zero of the $z$-axis. Here it is set to coincide with the plane of the iodine atoms.

**Rhombohedral primitive vectors:**

$$\mathbf{a}_1 = \frac{1}{2}\,a\,\hat{\mathbf{x}} - \frac{1}{2\sqrt{3}}\,a\,\hat{\mathbf{y}} + \frac{1}{3}\,c\,\hat{\mathbf{z}}$$
$$\mathbf{a}_2 = \frac{1}{\sqrt{3}}\,a\,\hat{\mathbf{y}} + \frac{1}{3}\,c\,\hat{\mathbf{z}}$$
$$\mathbf{a}_3 = -\frac{1}{2}\,a\,\hat{\mathbf{x}} - \frac{1}{2\sqrt{3}}\,a\,\hat{\mathbf{y}} + \frac{1}{3}\,c\,\hat{\mathbf{z}}$$

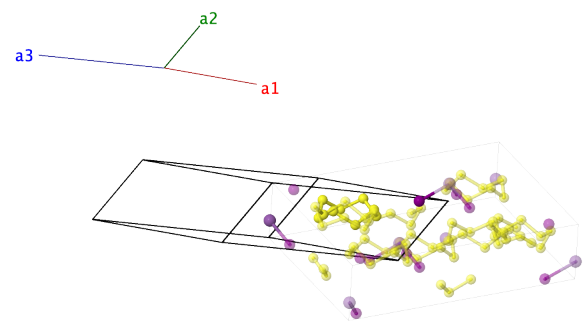

**Basis vectors:**

| | | Lattice Coordinates | | Cartesian Coordinates | Wyckoff Position | Atom Type |
|---|---|---|---|---|---|---|
| $\mathbf{B}_1$ | = | $x_1\,\mathbf{a}_1 + x_1\,\mathbf{a}_2 + x_1\,\mathbf{a}_3$ | = | $x_1 c\,\hat{\mathbf{z}}$ | (1$a$) | Sb |
| $\mathbf{B}_2$ | = | $x_2\,\mathbf{a}_1 + x_2\,\mathbf{a}_2 + z_2\,\mathbf{a}_3$ | = | $\frac{1}{2}(x_2 - z_2)\,a\,\hat{\mathbf{x}} + \frac{1}{2\sqrt{3}}(x_2 - z_2)\,a\,\hat{\mathbf{y}} + \left(\frac{2}{3}x_2 + \frac{1}{3}z_2\right)c\,\hat{\mathbf{z}}$ | (3$b$) | I |
| $\mathbf{B}_3$ | = | $z_2\,\mathbf{a}_1 + x_2\,\mathbf{a}_2 + x_2\,\mathbf{a}_3$ | = | $\frac{1}{2}(-x_2 + z_2)\,a\,\hat{\mathbf{x}} + \frac{1}{2\sqrt{3}}(x_2 - z_2)\,a\,\hat{\mathbf{y}} + \left(\frac{2}{3}x_2 + \frac{1}{3}z_2\right)c\,\hat{\mathbf{z}}$ | (3$b$) | I |
| $\mathbf{B}_4$ | = | $x_2\,\mathbf{a}_1 + z_2\,\mathbf{a}_2 + x_2\,\mathbf{a}_3$ | = | $\frac{1}{\sqrt{3}}(-x_2 + z_2)\,a\,\hat{\mathbf{y}} + \left(\frac{2}{3}x_2 + \frac{1}{3}z_2\right)c\,\hat{\mathbf{z}}$ | (3$b$) | I |
| $\mathbf{B}_5$ | = | $x_3\,\mathbf{a}_1 + x_3\,\mathbf{a}_2 + z_3\,\mathbf{a}_3$ | = | $\frac{1}{2}(x_3 - z_3)\,a\,\hat{\mathbf{x}} + \frac{1}{2\sqrt{3}}(x_3 - z_3)\,a\,\hat{\mathbf{y}} + \left(\frac{2}{3}x_3 + \frac{1}{3}z_3\right)c\,\hat{\mathbf{z}}$ | (3$b$) | S I |
| $\mathbf{B}_6$ | = | $z_3\,\mathbf{a}_1 + x_3\,\mathbf{a}_2 + x_3\,\mathbf{a}_3$ | = | $\frac{1}{2}(-x_3 + z_3)\,a\,\hat{\mathbf{x}} + \frac{1}{2\sqrt{3}}(x_3 - z_3)\,a\,\hat{\mathbf{y}} + \left(\frac{2}{3}x_3 + \frac{1}{3}z_3\right)c\,\hat{\mathbf{z}}$ | (3$b$) | S I |
| $\mathbf{B}_7$ | = | $x_3\,\mathbf{a}_1 + z_3\,\mathbf{a}_2 + x_3\,\mathbf{a}_3$ | = | $\frac{1}{\sqrt{3}}(-x_3 + z_3)\,a\,\hat{\mathbf{y}} + \left(\frac{2}{3}x_3 + \frac{1}{3}z_3\right)c\,\hat{\mathbf{z}}$ | (3$b$) | S I |
| $\mathbf{B}_8$ | = | $x_4\,\mathbf{a}_1 + x_4\,\mathbf{a}_2 + z_4\,\mathbf{a}_3$ | = | $\frac{1}{2}(x_4 - z_4)\,a\,\hat{\mathbf{x}} + \frac{1}{2\sqrt{3}}(x_4 - z_4)\,a\,\hat{\mathbf{y}} + \left(\frac{2}{3}x_4 + \frac{1}{3}z_4\right)c\,\hat{\mathbf{z}}$ | (3$b$) | S II |
| $\mathbf{B}_9$ | = | $z_4\,\mathbf{a}_1 + x_4\,\mathbf{a}_2 + x_4\,\mathbf{a}_3$ | = | $\frac{1}{2}(-x_4 + z_4)\,a\,\hat{\mathbf{x}} + \frac{1}{2\sqrt{3}}(x_4 - z_4)\,a\,\hat{\mathbf{y}} + \left(\frac{2}{3}x_4 + \frac{1}{3}z_4\right)c\,\hat{\mathbf{z}}$ | (3$b$) | S II |
| $\mathbf{B}_{10}$ | = | $x_4\,\mathbf{a}_1 + z_4\,\mathbf{a}_2 + x_4\,\mathbf{a}_3$ | = | $\frac{1}{\sqrt{3}}(-x_4 + z_4)\,a\,\hat{\mathbf{y}} + \left(\frac{2}{3}x_4 + \frac{1}{3}z_4\right)c\,\hat{\mathbf{z}}$ | (3$b$) | S II |
| $\mathbf{B}_{11}$ | = | $x_5\,\mathbf{a}_1 + y_5\,\mathbf{a}_2 + z_5\,\mathbf{a}_3$ | = | $\frac{1}{2}(x_5 - z_5)\,a\,\hat{\mathbf{x}} + \left(-\frac{1}{2\sqrt{3}}x_5 + \frac{1}{\sqrt{3}}y_5 - \frac{1}{2\sqrt{3}}z_5\right)a\,\hat{\mathbf{y}} + \frac{1}{3}(x_5 + y_5 + z_5)\,c\,\hat{\mathbf{z}}$ | (6$c$) | S III |
| $\mathbf{B}_{12}$ | = | $z_5\,\mathbf{a}_1 + x_5\,\mathbf{a}_2 + y_5\,\mathbf{a}_3$ | = | $\frac{1}{2}(-y_5 + z_5)\,a\,\hat{\mathbf{x}} + \left(\frac{1}{\sqrt{3}}x_5 - \frac{1}{2\sqrt{3}}y_5 - \frac{1}{2\sqrt{3}}z_5\right)a\,\hat{\mathbf{y}} + \frac{1}{3}(x_5 + y_5 + z_5)\,c\,\hat{\mathbf{z}}$ | (6$c$) | S III |
| $\mathbf{B}_{13}$ | = | $y_5\,\mathbf{a}_1 + z_5\,\mathbf{a}_2 + x_5\,\mathbf{a}_3$ | = | $\frac{1}{2}(-x_5 + y_5)\,a\,\hat{\mathbf{x}} + \left(-\frac{1}{2\sqrt{3}}x_5 - \frac{1}{2\sqrt{3}}y_5 + \frac{1}{\sqrt{3}}z_5\right)a\,\hat{\mathbf{y}} + \frac{1}{3}(x_5 + y_5 + z_5)\,c\,\hat{\mathbf{z}}$ | (6$c$) | S III |
| $\mathbf{B}_{14}$ | = | $z_5\,\mathbf{a}_1 + y_5\,\mathbf{a}_2 + x_5\,\mathbf{a}_3$ | = | $\frac{1}{2}(-x_5 + z_5)\,a\,\hat{\mathbf{x}} + \left(-\frac{1}{2\sqrt{3}}x_5 + \frac{1}{\sqrt{3}}y_5 - \frac{1}{2\sqrt{3}}z_5\right)a\,\hat{\mathbf{y}} + \frac{1}{3}(x_5 + y_5 + z_5)\,c\,\hat{\mathbf{z}}$ | (6$c$) | S III |
| $\mathbf{B}_{15}$ | = | $y_5\,\mathbf{a}_1 + x_5\,\mathbf{a}_2 + z_5\,\mathbf{a}_3$ | = | $\frac{1}{2}(y_5 - z_5)\,a\,\hat{\mathbf{x}} + \left(\frac{1}{\sqrt{3}}x_5 - \frac{1}{2\sqrt{3}}y_5 - \frac{1}{2\sqrt{3}}z_5\right)a\,\hat{\mathbf{y}} + \frac{1}{3}(x_5 + y_5 + z_5)\,c\,\hat{\mathbf{z}}$ | (6$c$) | S III |
| $\mathbf{B}_{16}$ | = | $x_5\,\mathbf{a}_1 + z_5\,\mathbf{a}_2 + y_5\,\mathbf{a}_3$ | = | $\frac{1}{2}(x_5 - y_5)\,a\,\hat{\mathbf{x}} + \left(-\frac{1}{2\sqrt{3}}x_5 - \frac{1}{2\sqrt{3}}y_5 + \frac{1}{\sqrt{3}}z_5\right)a\,\hat{\mathbf{y}} + \frac{1}{3}(x_5 + y_5 + z_5)\,c\,\hat{\mathbf{z}}$ | (6$c$) | S III |
| $\mathbf{B}_{17}$ | = | $x_6\,\mathbf{a}_1 + y_6\,\mathbf{a}_2 + z_6\,\mathbf{a}_3$ | = | $\frac{1}{2}(x_6 - z_6)\,a\,\hat{\mathbf{x}} + \left(-\frac{1}{2\sqrt{3}}x_6 + \frac{1}{\sqrt{3}}y_6 - \frac{1}{2\sqrt{3}}z_6\right)a\,\hat{\mathbf{y}} + \frac{1}{3}(x_6 + y_6 + z_6)\,c\,\hat{\mathbf{z}}$ | (6$c$) | S IV |
| $\mathbf{B}_{18}$ | = | $z_6\,\mathbf{a}_1 + x_6\,\mathbf{a}_2 + y_6\,\mathbf{a}_3$ | = | $\frac{1}{2}\left(-y_6 + z_6\right) a\,\mathbf{\hat{x}} + \left(\frac{1}{\sqrt{3}}x_6 - \frac{1}{2\sqrt{3}}y_6 - \frac{1}{2\sqrt{3}}z_6\right) a\,\mathbf{\hat{y}} + \frac{1}{3}\left(x_6 + y_6 + z_6\right) c\,\mathbf{\hat{z}}$ | (6$c$) | S IV |
| $\mathbf{B}_{19}$ | = | $y_6\,\mathbf{a}_1 + z_6\,\mathbf{a}_2 + x_6\,\mathbf{a}_3$ | = | $\frac{1}{2}\left(-x_6 + y_6\right) a\,\mathbf{\hat{x}} + \left(-\frac{1}{2\sqrt{3}}x_6 - \frac{1}{2\sqrt{3}}y_6 + \frac{1}{\sqrt{3}}z_6\right) a\,\mathbf{\hat{y}} + \frac{1}{3}\left(x_6 + y_6 + z_6\right) c\,\mathbf{\hat{z}}$ | (6$c$) | S IV |
| $\mathbf{B}_{20}$ | = | $z_6\,\mathbf{a}_1 + y_6\,\mathbf{a}_2 + x_6\,\mathbf{a}_3$ | = | $\frac{1}{2}\left(-x_6 + z_6\right) a\,\mathbf{\hat{x}} + \left(-\frac{1}{2\sqrt{3}}x_6 + \frac{1}{\sqrt{3}}y_6 - \frac{1}{2\sqrt{3}}z_6\right) a\,\mathbf{\hat{y}} + \frac{1}{3}\left(x_6 + y_6 + z_6\right) c\,\mathbf{\hat{z}}$ | (6$c$) | S IV |
| $\mathbf{B}_{21}$ | = | $y_6\,\mathbf{a}_1 + x_6\,\mathbf{a}_2 + z_6\,\mathbf{a}_3$ | = | $\frac{1}{2}\left(y_6 - z_6\right) a\,\mathbf{\hat{x}} + \left(\frac{1}{\sqrt{3}}x_6 - \frac{1}{2\sqrt{3}}y_6 - \frac{1}{2\sqrt{3}}z_6\right) a\,\mathbf{\hat{y}} + \frac{1}{3}\left(x_6 + y_6 + z_6\right) c\,\mathbf{\hat{z}}$ | (6$c$) | S IV |
| $\mathbf{B}_{22}$ | = | $x_6\,\mathbf{a}_1 + z_6\,\mathbf{a}_2 + y_6\,\mathbf{a}_3$ | = | $\frac{1}{2}\left(x_6 - y_6\right) a\,\mathbf{\hat{x}} + \left(-\frac{1}{2\sqrt{3}}x_6 - \frac{1}{2\sqrt{3}}y_6 + \frac{1}{\sqrt{3}}z_6\right) a\,\mathbf{\hat{y}} + \frac{1}{3}\left(x_6 + y_6 + z_6\right) c\,\mathbf{\hat{z}}$ | (6$c$) | S IV |
| $\mathbf{B}_{23}$ | = | $x_7\,\mathbf{a}_1 + y_7\,\mathbf{a}_2 + z_7\,\mathbf{a}_3$ | = | $\frac{1}{2}\left(x_7 - z_7\right) a\,\mathbf{\hat{x}} + \left(-\frac{1}{2\sqrt{3}}x_7 + \frac{1}{\sqrt{3}}y_7 - \frac{1}{2\sqrt{3}}z_7\right) a\,\mathbf{\hat{y}} + \frac{1}{3}\left(x_7 + y_7 + z_7\right) c\,\mathbf{\hat{z}}$ | (6$c$) | S V |
| $\mathbf{B}_{24}$ | = | $z_7\,\mathbf{a}_1 + x_7\,\mathbf{a}_2 + y_7\,\mathbf{a}_3$ | = | $\frac{1}{2}\left(-y_7 + z_7\right) a\,\mathbf{\hat{x}} + \left(\frac{1}{\sqrt{3}}x_7 - \frac{1}{2\sqrt{3}}y_7 - \frac{1}{2\sqrt{3}}z_7\right) a\,\mathbf{\hat{y}} + \frac{1}{3}\left(x_7 + y_7 + z_7\right) c\,\mathbf{\hat{z}}$ | (6$c$) | S V |
| $\mathbf{B}_{25}$ | = | $y_7\,\mathbf{a}_1 + z_7\,\mathbf{a}_2 + x_7\,\mathbf{a}_3$ | = | $\frac{1}{2}\left(-x_7 + y_7\right) a\,\mathbf{\hat{x}} + \left(-\frac{1}{2\sqrt{3}}x_7 - \frac{1}{2\sqrt{3}}y_7 + \frac{1}{\sqrt{3}}z_7\right) a\,\mathbf{\hat{y}} + \frac{1}{3}\left(x_7 + y_7 + z_7\right) c\,\mathbf{\hat{z}}$ | (6$c$) | S V |
| $\mathbf{B}_{26}$ | = | $z_7\,\mathbf{a}_1 + y_7\,\mathbf{a}_2 + x_7\,\mathbf{a}_3$ | = | $\frac{1}{2}\left(-x_7 + z_7\right) a\,\mathbf{\hat{x}} + \left(-\frac{1}{2\sqrt{3}}x_7 + \frac{1}{\sqrt{3}}y_7 - \frac{1}{2\sqrt{3}}z_7\right) a\,\mathbf{\hat{y}} + \frac{1}{3}\left(x_7 + y_7 + z_7\right) c\,\mathbf{\hat{z}}$ | (6$c$) | S V |
| $\mathbf{B}_{27}$ | = | $y_7\,\mathbf{a}_1 + x_7\,\mathbf{a}_2 + z_7\,\mathbf{a}_3$ | = | $\frac{1}{2}\left(y_7 - z_7\right) a\,\mathbf{\hat{x}} + \left(\frac{1}{\sqrt{3}}x_7 - \frac{1}{2\sqrt{3}}y_7 - \frac{1}{2\sqrt{3}}z_7\right) a\,\mathbf{\hat{y}} + \frac{1}{3}\left(x_7 + y_7 + z_7\right) c\,\mathbf{\hat{z}}$ | (6$c$) | S V |
| $\mathbf{B}_{28}$ | = | $x_7\,\mathbf{a}_1 + z_7\,\mathbf{a}_2 + y_7\,\mathbf{a}_3$ | = | $\frac{1}{2}\left(x_7 - y_7\right) a\,\mathbf{\hat{x}} + \left(-\frac{1}{2\sqrt{3}}x_7 - \frac{1}{2\sqrt{3}}y_7 + \frac{1}{\sqrt{3}}z_7\right) a\,\mathbf{\hat{y}} + \frac{1}{3}\left(x_7 + y_7 + z_7\right) c\,\mathbf{\hat{z}}$ | (6$c$) | S V |

**Geometry files:**

- CIF: pp. 1727
- POSCAR: pp. 1727

# $Fe_3PO_7$ Structure: A3B7C_hR11_160_b_a2b_a

http://aflow.org/prototype-encyclopedia/A3B7C_hR11_160_b_a2b_a

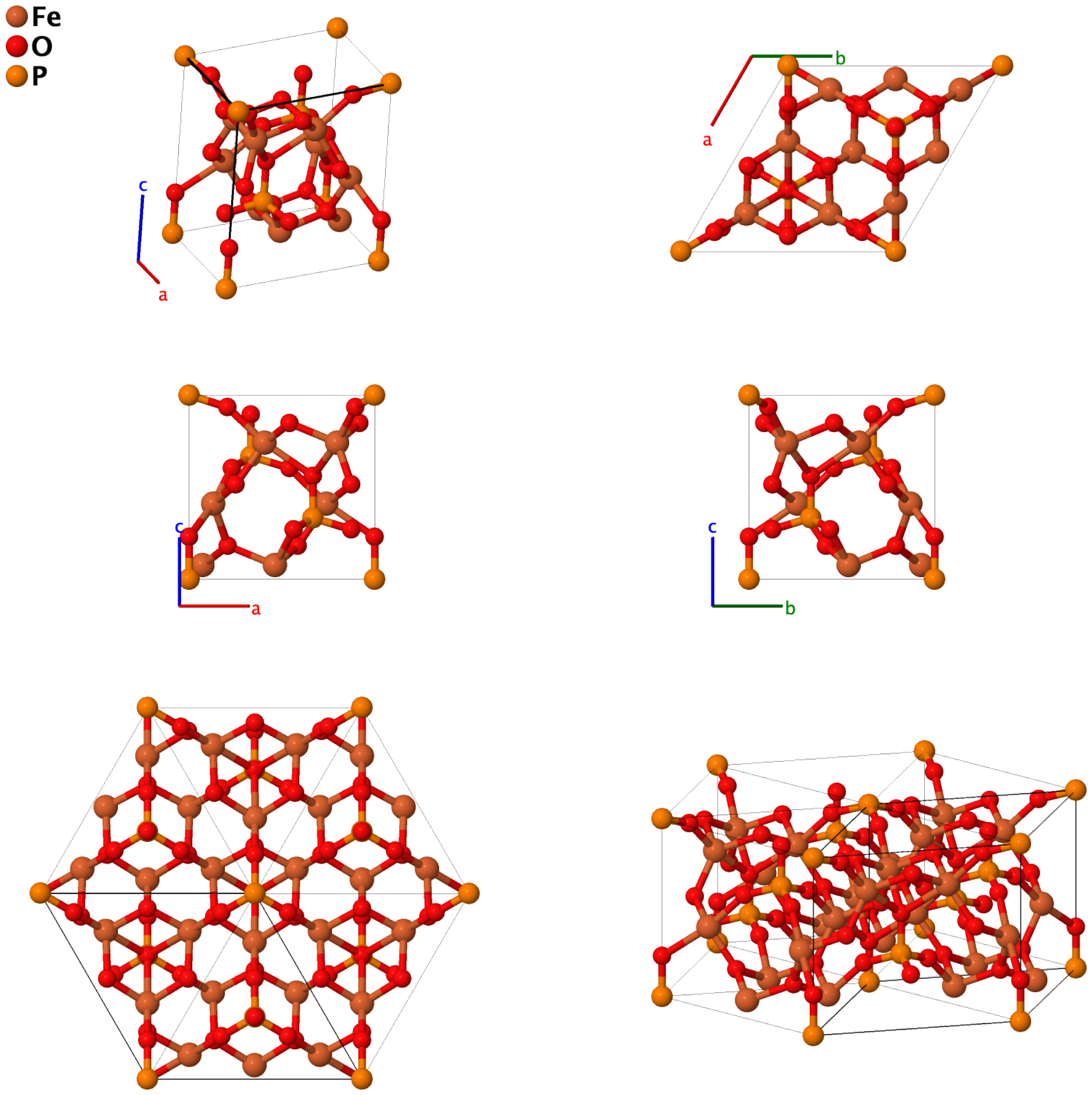

| | | |
|---|---|---|
| **Prototype** | : | $Fe_3O_7P$ |
| **AFLOW prototype label** | : | A3B7C_hR11_160_b_a2b_a |
| ***Strukturbericht* designation** | : | None |
| **Pearson symbol** | : | hR11 |
| **Space group number** | : | 160 |
| **Space group symbol** | : | $R3m$ |
| **AFLOW prototype command** | : | `aflow --proto=A3B7C_hR11_160_b_a2b_a [--hex]` <br> `--params=`$a, c/a, x_1, x_2, x_3, z_3, x_4, z_4, x_5, z_5$ |

- Space group $R3m$ #160 allows an arbitary definition of the zero of the $z$-axis. Here we select $z_1 = 0$, putting the phosphorous atom at the origin.

**Rhombohedral primitive vectors:**

$$\mathbf{a}_1 = \frac{1}{2}\,a\,\hat{\mathbf{x}} - \frac{1}{2\sqrt{3}}\,a\,\hat{\mathbf{y}} + \frac{1}{3}\,c\,\hat{\mathbf{z}}$$
$$\mathbf{a}_2 = \frac{1}{\sqrt{3}}\,a\,\hat{\mathbf{y}} + \frac{1}{3}\,c\,\hat{\mathbf{z}}$$
$$\mathbf{a}_3 = -\frac{1}{2}\,a\,\hat{\mathbf{x}} - \frac{1}{2\sqrt{3}}\,a\,\hat{\mathbf{y}} + \frac{1}{3}\,c\,\hat{\mathbf{z}}$$

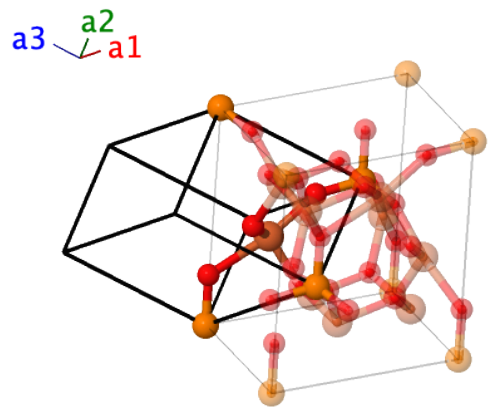

**Basis vectors:**

| | | Lattice Coordinates | | Cartesian Coordinates | Wyckoff Position | Atom Type |
|---|---|---|---|---|---|---|
| $\mathbf{B}_1$ | = | $x_1\,\mathbf{a}_1 + x_1\,\mathbf{a}_2 + x_1\,\mathbf{a}_3$ | = | $x_1 c\,\hat{\mathbf{z}}$ | (1$a$) | O I |
| $\mathbf{B}_2$ | = | $x_2\,\mathbf{a}_1 + x_2\,\mathbf{a}_2 + x_2\,\mathbf{a}_3$ | = | $x_2 c\,\hat{\mathbf{z}}$ | (1$a$) | P |
| $\mathbf{B}_3$ | = | $x_3\,\mathbf{a}_1 + x_3\,\mathbf{a}_2 + z_3\,\mathbf{a}_3$ | = | $\frac{1}{2}\left(x_3 - z_3\right)a\,\hat{\mathbf{x}} + \frac{1}{2\sqrt{3}}\left(x_3 - z_3\right)a\,\hat{\mathbf{y}} + \left(\frac{2}{3}x_3 + \frac{1}{3}z_3\right)c\,\hat{\mathbf{z}}$ | (3$b$) | Fe |
| $\mathbf{B}_4$ | = | $z_3\,\mathbf{a}_1 + x_3\,\mathbf{a}_2 + x_3\,\mathbf{a}_3$ | = | $\frac{1}{2}\left(-x_3 + z_3\right)a\,\hat{\mathbf{x}} + \frac{1}{2\sqrt{3}}\left(x_3 - z_3\right)a\,\hat{\mathbf{y}} + \left(\frac{2}{3}x_3 + \frac{1}{3}z_3\right)c\,\hat{\mathbf{z}}$ | (3$b$) | Fe |
| $\mathbf{B}_5$ | = | $x_3\,\mathbf{a}_1 + z_3\,\mathbf{a}_2 + x_3\,\mathbf{a}_3$ | = | $\frac{1}{\sqrt{3}}\left(-x_3 + z_3\right)a\,\hat{\mathbf{y}} + \left(\frac{2}{3}x_3 + \frac{1}{3}z_3\right)c\,\hat{\mathbf{z}}$ | (3$b$) | Fe |
| $\mathbf{B}_6$ | = | $x_4\,\mathbf{a}_1 + x_4\,\mathbf{a}_2 + z_4\,\mathbf{a}_3$ | = | $\frac{1}{2}\left(x_4 - z_4\right)a\,\hat{\mathbf{x}} + \frac{1}{2\sqrt{3}}\left(x_4 - z_4\right)a\,\hat{\mathbf{y}} + \left(\frac{2}{3}x_4 + \frac{1}{3}z_4\right)c\,\hat{\mathbf{z}}$ | (3$b$) | O II |
| $\mathbf{B}_7$ | = | $z_4\,\mathbf{a}_1 + x_4\,\mathbf{a}_2 + x_4\,\mathbf{a}_3$ | = | $\frac{1}{2}\left(-x_4 + z_4\right)a\,\hat{\mathbf{x}} + \frac{1}{2\sqrt{3}}\left(x_4 - z_4\right)a\,\hat{\mathbf{y}} + \left(\frac{2}{3}x_4 + \frac{1}{3}z_4\right)c\,\hat{\mathbf{z}}$ | (3$b$) | O II |
| $\mathbf{B}_8$ | = | $x_4\,\mathbf{a}_1 + z_4\,\mathbf{a}_2 + x_4\,\mathbf{a}_3$ | = | $\frac{1}{\sqrt{3}}\left(-x_4 + z_4\right)a\,\hat{\mathbf{y}} + \left(\frac{2}{3}x_4 + \frac{1}{3}z_4\right)c\,\hat{\mathbf{z}}$ | (3$b$) | O II |
| $\mathbf{B}_9$ | = | $x_5\,\mathbf{a}_1 + x_5\,\mathbf{a}_2 + z_5\,\mathbf{a}_3$ | = | $\frac{1}{2}\left(x_5 - z_5\right)a\,\hat{\mathbf{x}} + \frac{1}{2\sqrt{3}}\left(x_5 - z_5\right)a\,\hat{\mathbf{y}} + \left(\frac{2}{3}x_5 + \frac{1}{3}z_5\right)c\,\hat{\mathbf{z}}$ | (3$b$) | O III |
| $\mathbf{B}_{10}$ | = | $z_5\,\mathbf{a}_1 + x_5\,\mathbf{a}_2 + x_5\,\mathbf{a}_3$ | = | $\frac{1}{2}\left(-x_5 + z_5\right)a\,\hat{\mathbf{x}} + \frac{1}{2\sqrt{3}}\left(x_5 - z_5\right)a\,\hat{\mathbf{y}} + \left(\frac{2}{3}x_5 + \frac{1}{3}z_5\right)c\,\hat{\mathbf{z}}$ | (3$b$) | O III |
| $\mathbf{B}_{11}$ | = | $x_5\,\mathbf{a}_1 + z_5\,\mathbf{a}_2 + x_5\,\mathbf{a}_3$ | = | $\frac{1}{\sqrt{3}}\left(-x_5 + z_5\right)a\,\hat{\mathbf{y}} + \left(\frac{2}{3}x_5 + \frac{1}{3}z_5\right)c\,\hat{\mathbf{z}}$ | (3$b$) | O III |

**Found in:**

**Geometry files:**

- CIF: pp. 1727
- POSCAR: pp. 1728

# Cronstedtite {Fe(Fe,Si)[(OH)$_2$,O]O$_3$, $S5_7$} Structure: AB3C2D_hR7_160_a_b_2a_a

http://aflow.org/prototype-encyclopedia/AB3C2D_hR7_160_a_b_2a_a

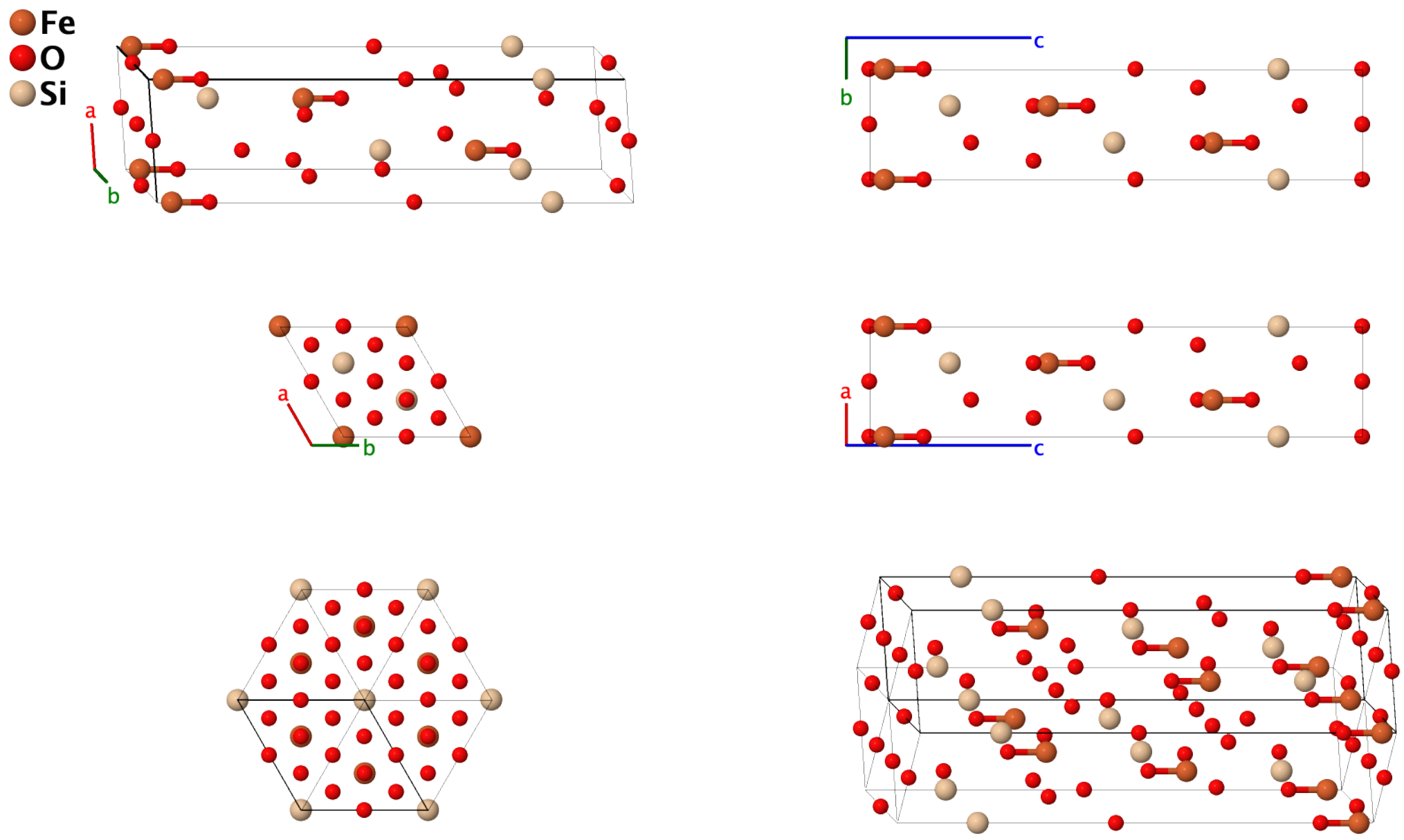

| | | |
|---|---|---|
| **Prototype** | : | FeO$_3$(OH)$_2$Si |
| **AFLOW prototype label** | : | AB3C2D_hR7_160_a_b_2a_a |
| ***Strukturbericht* designation** | : | $S5_7$ |
| **Pearson symbol** | : | hR7 |
| **Space group number** | : | 160 |
| **Space group symbol** | : | $R3m$ |
| **AFLOW prototype command** | : | `aflow --proto=AB3C2D_hR7_160_a_b_2a_a [--hex]` `--params=`$a, c/a, x_1, x_2, x_3, x_4, x_5, z_5$ |

- This particular structure of cronstedtite (also spelled cronstedite) is technically cronstedtite-3R, as there are three layers in the hexagonal cell, repeated in rhombohedral order. It is characterized by a high degree of disorder. The site we have labeled Si is better written as (Si,Fe), and the site labeled OH-I is actually [(OH)$_2$,O]. In addition the pure iron site, Fe (1a), is only occupied 90% of the time.
- As noted by (Frondel, 1962), there are a large number of varieties of cronstedtite, which each type having a different stacking sequence in the hexagonal cell.
- (Sterling, 1939) and (Hermann, 1943) put this structure in space group $R3$ #146, but the coordinates given by Sterling resolve to space group $R3m$ #160, and so we list this structure there.
- This is the final structure given a *Strukturbericht* designation.

**Rhombohedral primitive vectors:**

$$\begin{aligned}
\mathbf{a}_1 &= \frac{1}{2}\,a\,\hat{\mathbf{x}} - \frac{1}{2\sqrt{3}}\,a\,\hat{\mathbf{y}} + \frac{1}{3}\,c\,\hat{\mathbf{z}} \\
\mathbf{a}_2 &= \frac{1}{\sqrt{3}}\,a\,\hat{\mathbf{y}} + \frac{1}{3}\,c\,\hat{\mathbf{z}} \\
\mathbf{a}_3 &= -\frac{1}{2}\,a\,\hat{\mathbf{x}} - \frac{1}{2\sqrt{3}}\,a\,\hat{\mathbf{y}} + \frac{1}{3}\,c\,\hat{\mathbf{z}}
\end{aligned}$$

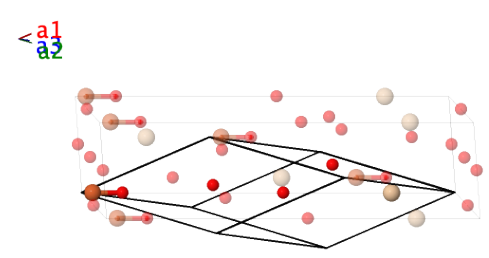

**Basis vectors:**

| | | Lattice Coordinates | | Cartesian Coordinates | Wyckoff Position | Atom Type |
|---|---|---|---|---|---|---|
| $\mathbf{B}_1$ | = | $x_1\,\mathbf{a}_1 + x_1\,\mathbf{a}_2 + x_1\,\mathbf{a}_3$ | = | $x_1 c\,\hat{\mathbf{z}}$ | (1$a$) | Fe |
| $\mathbf{B}_2$ | = | $x_2\,\mathbf{a}_1 + x_2\,\mathbf{a}_2 + x_2\,\mathbf{a}_3$ | = | $x_2 c\,\hat{\mathbf{z}}$ | (1$a$) | OH I |
| $\mathbf{B}_3$ | = | $x_3\,\mathbf{a}_1 + x_3\,\mathbf{a}_2 + x_3\,\mathbf{a}_3$ | = | $x_3 c\,\hat{\mathbf{z}}$ | (1$a$) | OH II |
| $\mathbf{B}_4$ | = | $x_4\,\mathbf{a}_1 + x_4\,\mathbf{a}_2 + x_4\,\mathbf{a}_3$ | = | $x_4 c\,\hat{\mathbf{z}}$ | (1$a$) | Si |
| $\mathbf{B}_5$ | = | $x_5\,\mathbf{a}_1 + x_5\,\mathbf{a}_2 + z_5\,\mathbf{a}_3$ | = | $\frac{1}{2}\,(x_5 - z_5)\,a\,\hat{\mathbf{x}} + \frac{1}{2\sqrt{3}}\,(x_5 - z_5)\,a\,\hat{\mathbf{y}} + \left(\frac{2}{3}x_5 + \frac{1}{3}z_5\right)c\,\hat{\mathbf{z}}$ | (3$b$) | O |
| $\mathbf{B}_6$ | = | $z_5\,\mathbf{a}_1 + x_5\,\mathbf{a}_2 + x_5\,\mathbf{a}_3$ | = | $\frac{1}{2}\,(-x_5 + z_5)\,a\,\hat{\mathbf{x}} + \frac{1}{2\sqrt{3}}\,(x_5 - z_5)\,a\,\hat{\mathbf{y}} + \left(\frac{2}{3}x_5 + \frac{1}{3}z_5\right)c\,\hat{\mathbf{z}}$ | (3$b$) | O |
| $\mathbf{B}_7$ | = | $x_5\,\mathbf{a}_1 + z_5\,\mathbf{a}_2 + x_5\,\mathbf{a}_3$ | = | $\frac{1}{\sqrt{3}}\,(-x_5 + z_5)\,a\,\hat{\mathbf{y}} + \left(\frac{2}{3}x_5 + \frac{1}{3}z_5\right)c\,\hat{\mathbf{z}}$ | (3$b$) | O |

**Found in:**

**Geometry files:**

- CIF: pp. 1728
- POSCAR: pp. 1728

# Low-Temperature $GaMo_4S_8$ Structure: AB4C8_hR13_160_a_ab_2a2b

http://aflow.org/prototype-encyclopedia/AB4C8_hR13_160_a_ab_2a2b

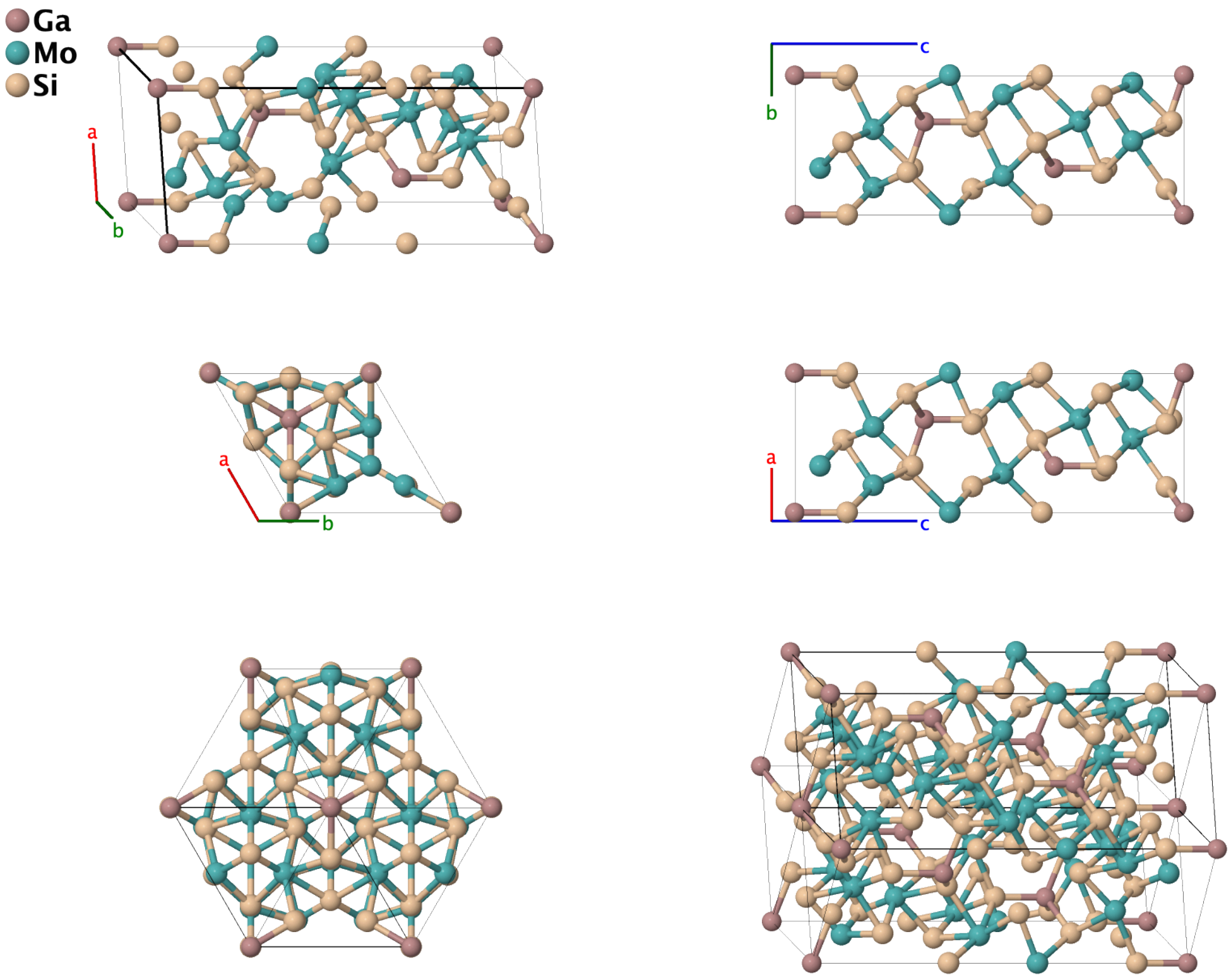

| | | |
|---|---|---|
| **Prototype** | : | $GaMo_4S_8$ |
| **AFLOW prototype label** | : | AB4C8_hR13_160_a_ab_2a2b |
| ***Strukturbericht* designation** | : | None |
| **Pearson symbol** | : | hR13 |
| **Space group number** | : | 160 |
| **Space group symbol** | : | $R3m$ |
| **AFLOW prototype command** | : | `aflow --proto=AB4C8_hR13_160_a_ab_2a2b [--hex]` `--params=`$a, c/a, x_1, x_2, x_3, x_4, x_5, z_5, x_6, z_6, x_7, z_7$ |

- At temperatures below 45 K $GaMo_4S_8$ transforms from its high-temperature cubic structure to this rhombohedral structure.
- We use the data from (François, 1991) at 8 K.
- Space group $R3m$ #160 allows an arbitrary choice of the zero of the $z$-axis. Here this use used to place the gallium atom at the origin ($z_1 = 0$).

**Rhombohedral primitive vectors:**

$$\mathbf{a}_1 = \frac{1}{2}\,a\,\hat{\mathbf{x}} - \frac{1}{2\sqrt{3}}\,a\,\hat{\mathbf{y}} + \frac{1}{3}\,c\,\hat{\mathbf{z}}$$
$$\mathbf{a}_2 = \frac{1}{\sqrt{3}}\,a\,\hat{\mathbf{y}} + \frac{1}{3}\,c\,\hat{\mathbf{z}}$$
$$\mathbf{a}_3 = -\frac{1}{2}\,a\,\hat{\mathbf{x}} - \frac{1}{2\sqrt{3}}\,a\,\hat{\mathbf{y}} + \frac{1}{3}\,c\,\hat{\mathbf{z}}$$

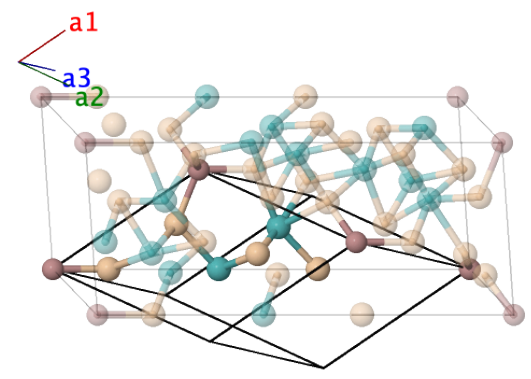

**Basis vectors:**

| | | Lattice Coordinates | | Cartesian Coordinates | Wyckoff Position | Atom Type |
|---|---|---|---|---|---|---|
| $\mathbf{B}_1$ | = | $x_1\,\mathbf{a}_1 + x_1\,\mathbf{a}_2 + x_1\,\mathbf{a}_3$ | = | $x_1 c\,\hat{\mathbf{z}}$ | (1$a$) | Ga |
| $\mathbf{B}_2$ | = | $x_2\,\mathbf{a}_1 + x_2\,\mathbf{a}_2 + x_2\,\mathbf{a}_3$ | = | $x_2 c\,\hat{\mathbf{z}}$ | (1$a$) | Mo I |
| $\mathbf{B}_3$ | = | $x_3\,\mathbf{a}_1 + x_3\,\mathbf{a}_2 + x_3\,\mathbf{a}_3$ | = | $x_3 c\,\hat{\mathbf{z}}$ | (1$a$) | Si I |
| $\mathbf{B}_4$ | = | $x_4\,\mathbf{a}_1 + x_4\,\mathbf{a}_2 + x_4\,\mathbf{a}_3$ | = | $x_4 c\,\hat{\mathbf{z}}$ | (1$a$) | Si II |
| $\mathbf{B}_5$ | = | $x_5\,\mathbf{a}_1 + x_5\,\mathbf{a}_2 + z_5\,\mathbf{a}_3$ | = | $\frac{1}{2}(x_5 - z_5)\,a\,\hat{\mathbf{x}} + \frac{1}{2\sqrt{3}}(x_5 - z_5)\,a\,\hat{\mathbf{y}} + \left(\frac{2}{3}x_5 + \frac{1}{3}z_5\right)c\,\hat{\mathbf{z}}$ | (3$b$) | Mo II |
| $\mathbf{B}_6$ | = | $z_5\,\mathbf{a}_1 + x_5\,\mathbf{a}_2 + x_5\,\mathbf{a}_3$ | = | $\frac{1}{2}(-x_5 + z_5)\,a\,\hat{\mathbf{x}} + \frac{1}{2\sqrt{3}}(x_5 - z_5)\,a\,\hat{\mathbf{y}} + \left(\frac{2}{3}x_5 + \frac{1}{3}z_5\right)c\,\hat{\mathbf{z}}$ | (3$b$) | Mo II |
| $\mathbf{B}_7$ | = | $x_5\,\mathbf{a}_1 + z_5\,\mathbf{a}_2 + x_5\,\mathbf{a}_3$ | = | $\frac{1}{\sqrt{3}}(-x_5 + z_5)\,a\,\hat{\mathbf{y}} + \left(\frac{2}{3}x_5 + \frac{1}{3}z_5\right)c\,\hat{\mathbf{z}}$ | (3$b$) | Mo II |
| $\mathbf{B}_8$ | = | $x_6\,\mathbf{a}_1 + x_6\,\mathbf{a}_2 + z_6\,\mathbf{a}_3$ | = | $\frac{1}{2}(x_6 - z_6)\,a\,\hat{\mathbf{x}} + \frac{1}{2\sqrt{3}}(x_6 - z_6)\,a\,\hat{\mathbf{y}} + \left(\frac{2}{3}x_6 + \frac{1}{3}z_6\right)c\,\hat{\mathbf{z}}$ | (3$b$) | Si III |
| $\mathbf{B}_9$ | = | $z_6\,\mathbf{a}_1 + x_6\,\mathbf{a}_2 + x_6\,\mathbf{a}_3$ | = | $\frac{1}{2}(-x_6 + z_6)\,a\,\hat{\mathbf{x}} + \frac{1}{2\sqrt{3}}(x_6 - z_6)\,a\,\hat{\mathbf{y}} + \left(\frac{2}{3}x_6 + \frac{1}{3}z_6\right)c\,\hat{\mathbf{z}}$ | (3$b$) | Si III |
| $\mathbf{B}_{10}$ | = | $x_6\,\mathbf{a}_1 + z_6\,\mathbf{a}_2 + x_6\,\mathbf{a}_3$ | = | $\frac{1}{\sqrt{3}}(-x_6 + z_6)\,a\,\hat{\mathbf{y}} + \left(\frac{2}{3}x_6 + \frac{1}{3}z_6\right)c\,\hat{\mathbf{z}}$ | (3$b$) | Si III |
| $\mathbf{B}_{11}$ | = | $x_7\,\mathbf{a}_1 + x_7\,\mathbf{a}_2 + z_7\,\mathbf{a}_3$ | = | $\frac{1}{2}(x_7 - z_7)\,a\,\hat{\mathbf{x}} + \frac{1}{2\sqrt{3}}(x_7 - z_7)\,a\,\hat{\mathbf{y}} + \left(\frac{2}{3}x_7 + \frac{1}{3}z_7\right)c\,\hat{\mathbf{z}}$ | (3$b$) | Si IV |
| $\mathbf{B}_{12}$ | = | $z_7\,\mathbf{a}_1 + x_7\,\mathbf{a}_2 + x_7\,\mathbf{a}_3$ | = | $\frac{1}{2}(-x_7 + z_7)\,a\,\hat{\mathbf{x}} + \frac{1}{2\sqrt{3}}(x_7 - z_7)\,a\,\hat{\mathbf{y}} + \left(\frac{2}{3}x_7 + \frac{1}{3}z_7\right)c\,\hat{\mathbf{z}}$ | (3$b$) | Si IV |
| $\mathbf{B}_{13}$ | = | $x_7\,\mathbf{a}_1 + z_7\,\mathbf{a}_2 + x_7\,\mathbf{a}_3$ | = | $\frac{1}{\sqrt{3}}(-x_7 + z_7)\,a\,\hat{\mathbf{y}} + \left(\frac{2}{3}x_7 + \frac{1}{3}z_7\right)c\,\hat{\mathbf{z}}$ | (3$b$) | Si IV |

**Geometry files:**

- CIF: pp. 1728
- POSCAR: pp. 1729

# $KBrO_3$ ($G0_7$) Structure: ABC3_hR5_160_a_a_b

http://aflow.org/prototype-encyclopedia/ABC3_hR5_160_a_a_b.KBrO3

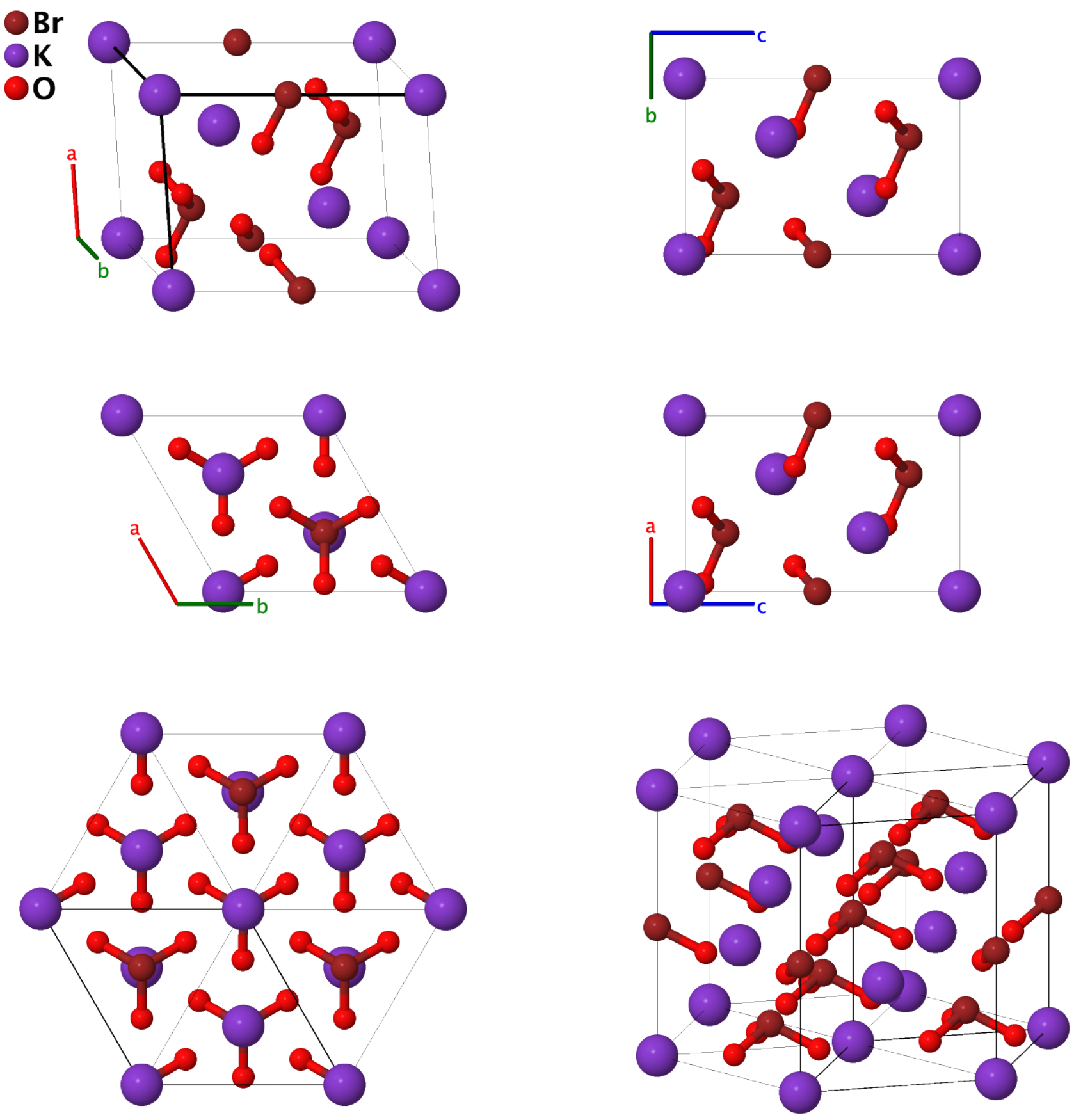

| | | |
|---|---|---|
| **Prototype** | : | $BrKO_3$ |
| **AFLOW prototype label** | : | ABC3_hR5_160_a_a_b |
| ***Strukturbericht* designation** | : | $G0_7$ |
| **Pearson symbol** | : | hR5 |
| **Space group number** | : | 160 |
| **Space group symbol** | : | $R3m$ |
| **AFLOW prototype command** | : | `aflow --proto=ABC3_hR5_160_a_a_b [--hex]` `--params=`$a, c/a, x_1, x_2, x_3, z_3$ |

### Other compounds with this structure

- $KNO_3$, $\gamma$-$KNO_3$, and $RbNO_3$

- $\gamma$-$KNO_3$ and $KBrO_3$ ($G0_7$) have the same AFLOW prototype label, ABC3_hR5_160_a_a_b. They are generated by the same symmetry operations with different sets of parameters (`--params`) specified in their corresponding CIF files.

**Rhombohedral primitive vectors:**

$$\mathbf{a}_1 = \frac{1}{2}\,a\,\hat{\mathbf{x}} - \frac{1}{2\sqrt{3}}\,a\,\hat{\mathbf{y}} + \frac{1}{3}\,c\,\hat{\mathbf{z}}$$
$$\mathbf{a}_2 = \frac{1}{\sqrt{3}}\,a\,\hat{\mathbf{y}} + \frac{1}{3}\,c\,\hat{\mathbf{z}}$$
$$\mathbf{a}_3 = -\frac{1}{2}\,a\,\hat{\mathbf{x}} - \frac{1}{2\sqrt{3}}\,a\,\hat{\mathbf{y}} + \frac{1}{3}\,c\,\hat{\mathbf{z}}$$

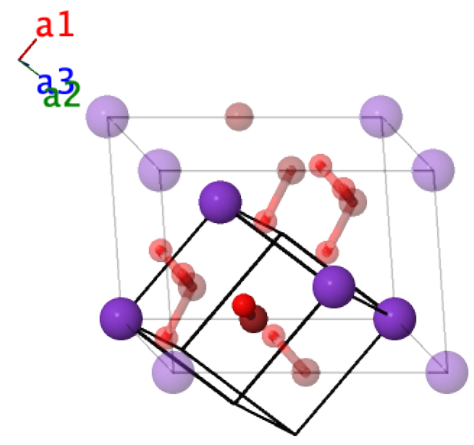

**Basis vectors:**

| | | Lattice Coordinates | | Cartesian Coordinates | Wyckoff Position | Atom Type |
|---|---|---|---|---|---|---|
| $\mathbf{B}_1$ | = | $x_1\,\mathbf{a}_1 + x_1\,\mathbf{a}_2 + x_1\,\mathbf{a}_3$ | = | $x_1 c\,\hat{\mathbf{z}}$ | (1$a$) | Br |
| $\mathbf{B}_2$ | = | $x_2\,\mathbf{a}_1 + x_2\,\mathbf{a}_2 + x_2\,\mathbf{a}_3$ | = | $x_2 c\,\hat{\mathbf{z}}$ | (1$a$) | K |
| $\mathbf{B}_3$ | = | $x_3\,\mathbf{a}_1 + x_3\,\mathbf{a}_2 + z_3\,\mathbf{a}_3$ | = | $\frac{1}{2}\left(x_3 - z_3\right)a\,\hat{\mathbf{x}} + \frac{1}{2\sqrt{3}}\left(x_3 - z_3\right)a\,\hat{\mathbf{y}} + \left(\frac{2}{3}x_3 + \frac{1}{3}z_3\right)c\,\hat{\mathbf{z}}$ | (3$b$) | O |
| $\mathbf{B}_4$ | = | $z_3\,\mathbf{a}_1 + x_3\,\mathbf{a}_2 + x_3\,\mathbf{a}_3$ | = | $\frac{1}{2}\left(-x_3 + z_3\right)a\,\hat{\mathbf{x}} + \frac{1}{2\sqrt{3}}\left(x_3 - z_3\right)a\,\hat{\mathbf{y}} + \left(\frac{2}{3}x_3 + \frac{1}{3}z_3\right)c\,\hat{\mathbf{z}}$ | (3$b$) | O |
| $\mathbf{B}_5$ | = | $x_3\,\mathbf{a}_1 + z_3\,\mathbf{a}_2 + x_3\,\mathbf{a}_3$ | = | $\frac{1}{\sqrt{3}}\left(-x_3 + z_3\right)a\,\hat{\mathbf{y}} + \left(\frac{2}{3}x_3 + \frac{1}{3}z_3\right)c\,\hat{\mathbf{z}}$ | (3$b$) | O |

**Geometry files:**

- CIF: pp. 1729
- POSCAR: pp. 1729

# $\gamma$-Potassium Nitrate ($KNO_3$) Structure: ABC3_hR5_160_a_a_b

http://aflow.org/prototype-encyclopedia/ABC3_hR5_160_a_a_b

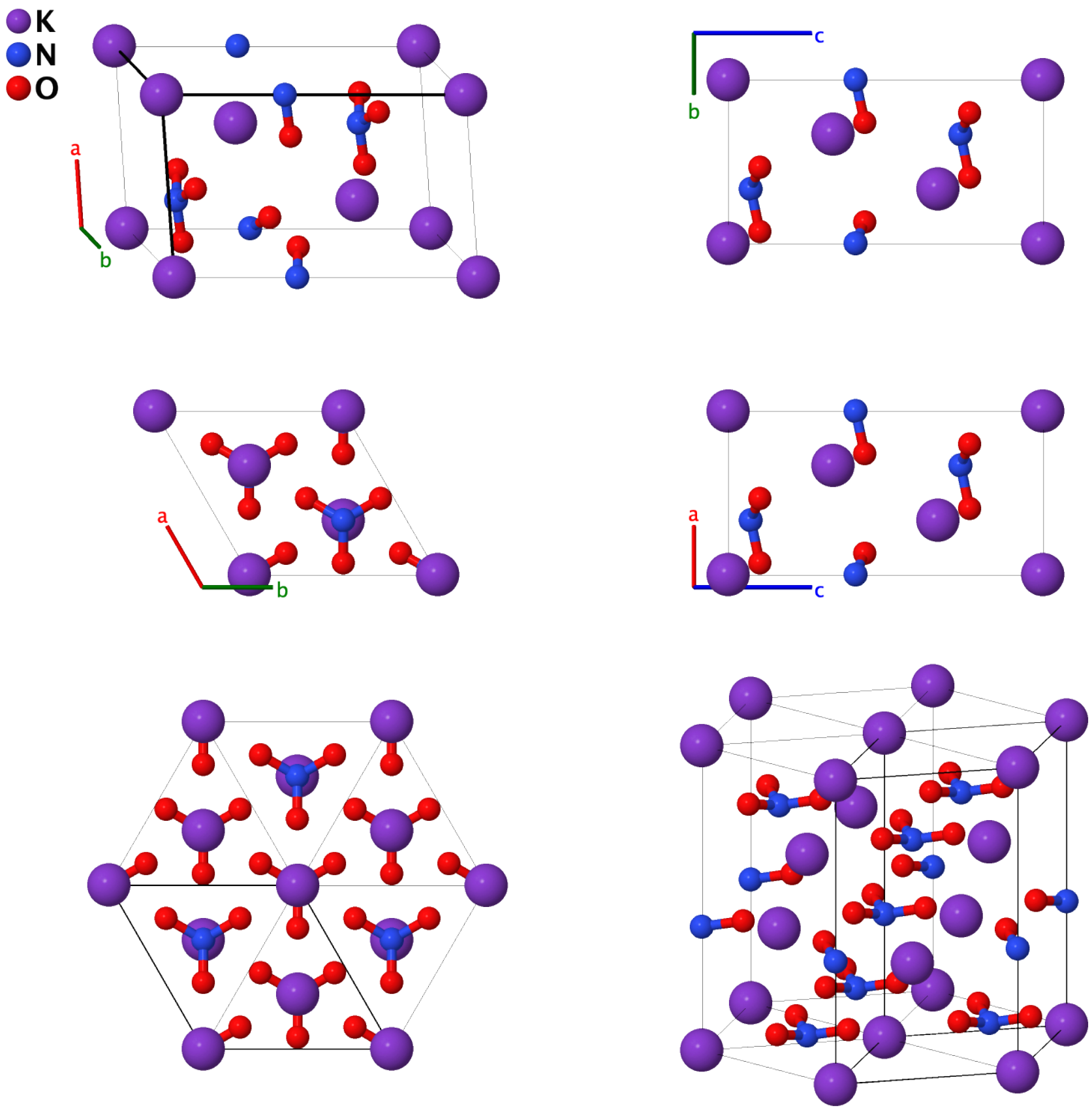

| | | |
|---|---|---|
| **Prototype** | : | $KNO_3$ |
| **AFLOW prototype label** | : | ABC3_hR5_160_a_a_b |
| ***Strukturbericht* designation** | : | None |
| **Pearson symbol** | : | hR5 |
| **Space group number** | : | 160 |
| **Space group symbol** | : | $R3m$ |
| **AFLOW prototype command** | : | `aflow --proto=ABC3_hR5_160_a_a_b [--hex]` `--params=`$a, c/a, x_1, x_2, x_3, z_3$ |

### Other compounds with this structure

- $NH_4ClO_4$

- On heating, $\alpha$-$KNO_3$ (either Structure I or Structure II) transforms into $\beta$-$KNO_3$ at 128 °C. When heated above 200 °C and then cooled, the $\beta$-phase transforms into the metastable ferroelectric $\gamma$-$KNO_3$ phase, which can remain down to room temperature.
- (Nimmo, 1976) give the data for $\gamma$-$KNO_3$ taken at 91 °C.
- Although this is isostructural with the $KBrO_3$ ($G0_7$) structure, we have included it here to facilitate the comparison of the various $KNO_3$ phases. $\gamma$-$KNO_3$ and $KBrO_3$ ($G0_7$) have the same AFLOW prototype label, ABC3_hR5_160_a_a_b. They are generated by the same symmetry operations with different sets of parameters (`--params`) specified in their corresponding CIF files.

**Rhombohedral primitive vectors:**

$$\mathbf{a}_1 = \frac{1}{2}\,a\,\hat{\mathbf{x}} - \frac{1}{2\sqrt{3}}\,a\,\hat{\mathbf{y}} + \frac{1}{3}\,c\,\hat{\mathbf{z}}$$
$$\mathbf{a}_2 = \frac{1}{\sqrt{3}}\,a\,\hat{\mathbf{y}} + \frac{1}{3}\,c\,\hat{\mathbf{z}}$$
$$\mathbf{a}_3 = -\frac{1}{2}\,a\,\hat{\mathbf{x}} - \frac{1}{2\sqrt{3}}\,a\,\hat{\mathbf{y}} + \frac{1}{3}\,c\,\hat{\mathbf{z}}$$

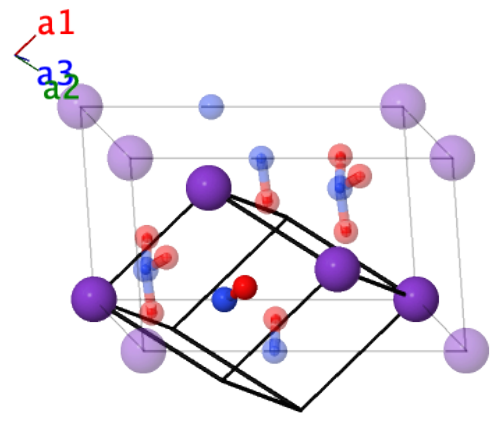

**Basis vectors:**

| | | Lattice Coordinates | | Cartesian Coordinates | Wyckoff Position | Atom Type |
|---|---|---|---|---|---|---|
| $\mathbf{B}_1$ | = | $x_1\,\mathbf{a}_1 + x_1\,\mathbf{a}_2 + x_1\,\mathbf{a}_3$ | = | $x_1 c\,\hat{\mathbf{z}}$ | (1$a$) | K |
| $\mathbf{B}_2$ | = | $x_2\,\mathbf{a}_1 + x_2\,\mathbf{a}_2 + x_2\,\mathbf{a}_3$ | = | $x_2 c\,\hat{\mathbf{z}}$ | (1$a$) | N |
| $\mathbf{B}_3$ | = | $x_3\,\mathbf{a}_1 + x_3\,\mathbf{a}_2 + z_3\,\mathbf{a}_3$ | = | $\frac{1}{2}\left(x_3 - z_3\right)a\,\hat{\mathbf{x}} + \frac{1}{2\sqrt{3}}\left(x_3 - z_3\right)a\,\hat{\mathbf{y}} + \left(\frac{2}{3}x_3 + \frac{1}{3}z_3\right)c\,\hat{\mathbf{z}}$ | (3$b$) | O |
| $\mathbf{B}_4$ | = | $z_3\,\mathbf{a}_1 + x_3\,\mathbf{a}_2 + x_3\,\mathbf{a}_3$ | = | $\frac{1}{2}\left(-x_3 + z_3\right)a\,\hat{\mathbf{x}} + \frac{1}{2\sqrt{3}}\left(x_3 - z_3\right)a\,\hat{\mathbf{y}} + \left(\frac{2}{3}x_3 + \frac{1}{3}z_3\right)c\,\hat{\mathbf{z}}$ | (3$b$) | O |
| $\mathbf{B}_5$ | = | $x_3\,\mathbf{a}_1 + z_3\,\mathbf{a}_2 + x_3\,\mathbf{a}_3$ | = | $\frac{1}{\sqrt{3}}\left(-x_3 + z_3\right)a\,\hat{\mathbf{y}} + \left(\frac{2}{3}x_3 + \frac{1}{3}z_3\right)c\,\hat{\mathbf{z}}$ | (3$b$) | O |

**Found in:**

**Geometry files:**

- CIF: pp. 1729
- POSCAR: pp. 1730

# $\alpha$-$BaB_2O_4$ (Low-Temperature) Structure: A2BC4_hR42_161_2b_b_4b

http://aflow.org/prototype-encyclopedia/A2BC4_hR42_161_2b_b_4b

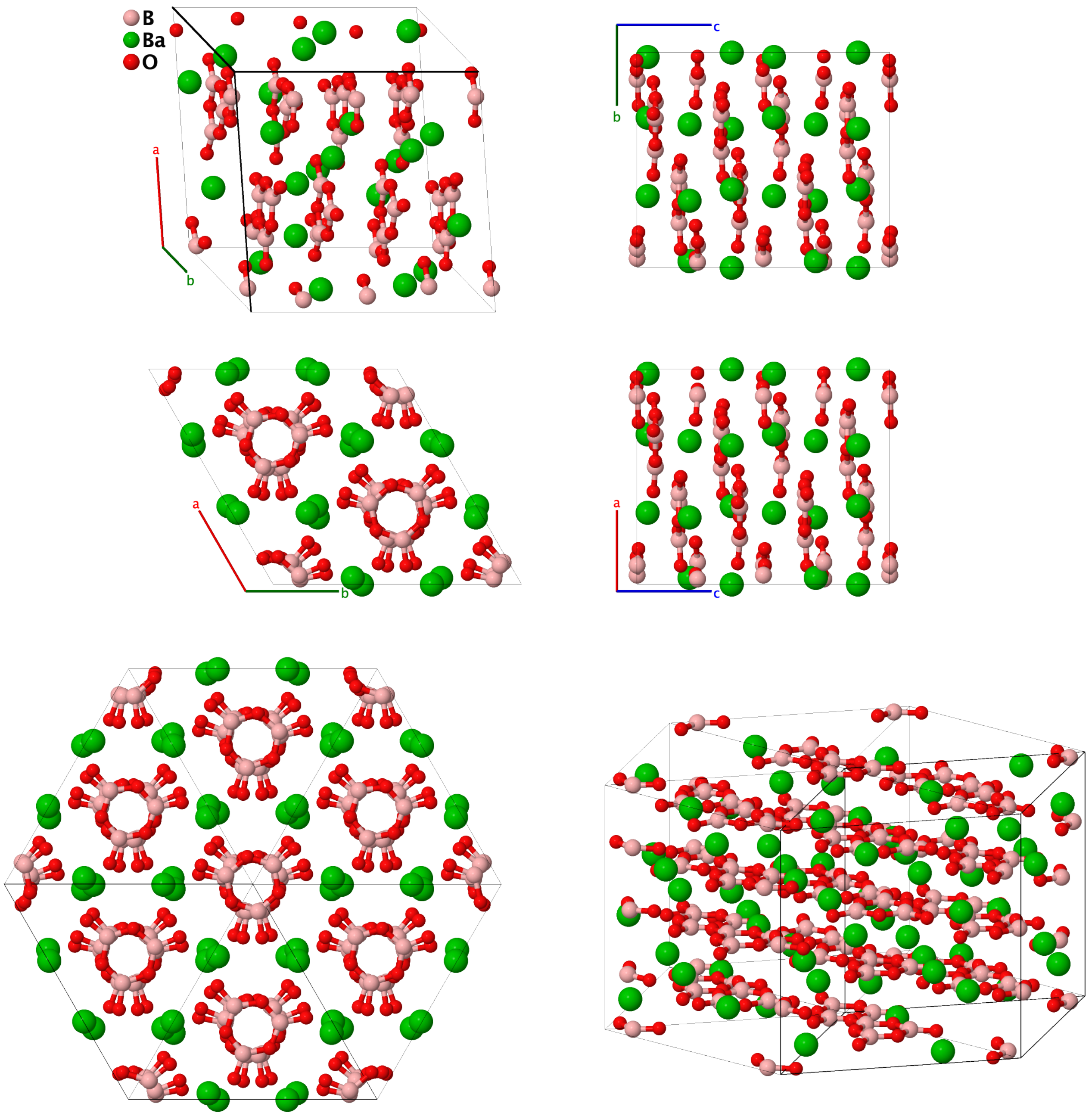

| | | |
|---|---|---|
| **Prototype** | : | $B_2BaO_4$ |
| **AFLOW prototype label** | : | A2BC4_hR42_161_2b_b_4b |
| ***Strukturbericht* designation** | : | None |
| **Pearson symbol** | : | hR42 |
| **Space group number** | : | 161 |
| **Space group symbol** | : | $R3c$ |
| **AFLOW prototype command** | : | `aflow --proto=A2BC4_hR42_161_2b_b_4b [--hex]` `--params=`$a, c/a, x_1, y_1, z_1, x_2, y_2, z_2, x_3, y_3, z_3, x_4, y_4, z_4, x_5, y_5, z_5, x_6, y_6, z_6, x_7, y_7, z_7$ |

- This is the low-temperature structure. Heating to temperatures between 100-400 °C transforms it into $\beta$-$BaB_2O_4$. The principle difference between the two forms is the lack of inversion symmetry in the low-temperature structure.

**Rhombohedral primitive vectors:**

$$\mathbf{a}_1 = \frac{1}{2}\,a\,\hat{\mathbf{x}} - \frac{1}{2\sqrt{3}}\,a\,\hat{\mathbf{y}} + \frac{1}{3}\,c\,\hat{\mathbf{z}}$$
$$\mathbf{a}_2 = \frac{1}{\sqrt{3}}\,a\,\hat{\mathbf{y}} + \frac{1}{3}\,c\,\hat{\mathbf{z}}$$
$$\mathbf{a}_3 = -\frac{1}{2}\,a\,\hat{\mathbf{x}} - \frac{1}{2\sqrt{3}}\,a\,\hat{\mathbf{y}} + \frac{1}{3}\,c\,\hat{\mathbf{z}}$$

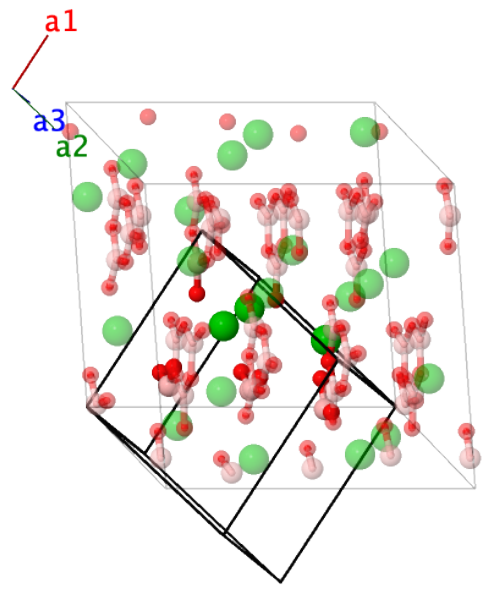

**Basis vectors:**

| | | Lattice Coordinates | | Cartesian Coordinates | Wyckoff Position | Atom Type |
|---|---|---|---|---|---|---|
| $\mathbf{B}_1$ | = | $x_1\,\mathbf{a}_1 + y_1\,\mathbf{a}_2 + z_1\,\mathbf{a}_3$ | = | $\frac{1}{2}(x_1 - z_1)\,a\,\hat{\mathbf{x}} + \left(-\frac{1}{2\sqrt{3}}x_1 + \frac{1}{\sqrt{3}}y_1 - \frac{1}{2\sqrt{3}}z_1\right)a\,\hat{\mathbf{y}} + \frac{1}{3}(x_1 + y_1 + z_1)\,c\,\hat{\mathbf{z}}$ | (6*b*) | B I |
| $\mathbf{B}_2$ | = | $z_1\,\mathbf{a}_1 + x_1\,\mathbf{a}_2 + y_1\,\mathbf{a}_3$ | = | $\frac{1}{2}(-y_1 + z_1)\,a\,\hat{\mathbf{x}} + \left(\frac{1}{\sqrt{3}}x_1 - \frac{1}{2\sqrt{3}}y_1 - \frac{1}{2\sqrt{3}}z_1\right)a\,\hat{\mathbf{y}} + \frac{1}{3}(x_1 + y_1 + z_1)\,c\,\hat{\mathbf{z}}$ | (6*b*) | B I |
| $\mathbf{B}_3$ | = | $y_1\,\mathbf{a}_1 + z_1\,\mathbf{a}_2 + x_1\,\mathbf{a}_3$ | = | $\frac{1}{2}(-x_1 + y_1)\,a\,\hat{\mathbf{x}} + \left(-\frac{1}{2\sqrt{3}}x_1 - \frac{1}{2\sqrt{3}}y_1 + \frac{1}{\sqrt{3}}z_1\right)a\,\hat{\mathbf{y}} + \frac{1}{3}(x_1 + y_1 + z_1)\,c\,\hat{\mathbf{z}}$ | (6*b*) | B I |
| $\mathbf{B}_4$ | = | $\left(\frac{1}{2} + z_1\right)\mathbf{a}_1 + \left(\frac{1}{2} + y_1\right)\mathbf{a}_2 + \left(\frac{1}{2} + x_1\right)\mathbf{a}_3$ | = | $\frac{1}{2}(-x_1 + z_1)\,a\,\hat{\mathbf{x}} + \left(-\frac{1}{2\sqrt{3}}x_1 + \frac{1}{\sqrt{3}}y_1 - \frac{1}{2\sqrt{3}}z_1\right)a\,\hat{\mathbf{y}} + \left(\frac{1}{2} + \frac{1}{3}x_1 + \frac{1}{3}y_1 + \frac{1}{3}z_1\right)c\,\hat{\mathbf{z}}$ | (6*b*) | B I |
| $\mathbf{B}_5$ | = | $\left(\frac{1}{2} + y_1\right)\mathbf{a}_1 + \left(\frac{1}{2} + x_1\right)\mathbf{a}_2 + \left(\frac{1}{2} + z_1\right)\mathbf{a}_3$ | = | $\frac{1}{2}(y_1 - z_1)\,a\,\hat{\mathbf{x}} + \left(\frac{1}{\sqrt{3}}x_1 - \frac{1}{2\sqrt{3}}y_1 - \frac{1}{2\sqrt{3}}z_1\right)a\,\hat{\mathbf{y}} + \left(\frac{1}{2} + \frac{1}{3}x_1 + \frac{1}{3}y_1 + \frac{1}{3}z_1\right)c\,\hat{\mathbf{z}}$ | (6*b*) | B I |
| $\mathbf{B}_6$ | = | $\left(\frac{1}{2} + x_1\right)\mathbf{a}_1 + \left(\frac{1}{2} + z_1\right)\mathbf{a}_2 + \left(\frac{1}{2} + y_1\right)\mathbf{a}_3$ | = | $\frac{1}{2}(x_1 - y_1)\,a\,\hat{\mathbf{x}} + \left(-\frac{1}{2\sqrt{3}}x_1 - \frac{1}{2\sqrt{3}}y_1 + \frac{1}{\sqrt{3}}z_1\right)a\,\hat{\mathbf{y}} + \left(\frac{1}{2} + \frac{1}{3}x_1 + \frac{1}{3}y_1 + \frac{1}{3}z_1\right)c\,\hat{\mathbf{z}}$ | (6*b*) | B I |
| $\mathbf{B}_7$ | = | $x_2\,\mathbf{a}_1 + y_2\,\mathbf{a}_2 + z_2\,\mathbf{a}_3$ | = | $\frac{1}{2}(x_2 - z_2)\,a\,\hat{\mathbf{x}} + \left(-\frac{1}{2\sqrt{3}}x_2 + \frac{1}{\sqrt{3}}y_2 - \frac{1}{2\sqrt{3}}z_2\right)a\,\hat{\mathbf{y}} + \frac{1}{3}(x_2 + y_2 + z_2)\,c\,\hat{\mathbf{z}}$ | (6*b*) | B II |
| $\mathbf{B}_8$ | = | $z_2\,\mathbf{a}_1 + x_2\,\mathbf{a}_2 + y_2\,\mathbf{a}_3$ | = | $\frac{1}{2}(-y_2 + z_2)\,a\,\hat{\mathbf{x}} + \left(\frac{1}{\sqrt{3}}x_2 - \frac{1}{2\sqrt{3}}y_2 - \frac{1}{2\sqrt{3}}z_2\right)a\,\hat{\mathbf{y}} + \frac{1}{3}(x_2 + y_2 + z_2)\,c\,\hat{\mathbf{z}}$ | (6*b*) | B II |
| $\mathbf{B}_9$ | = | $y_2\,\mathbf{a}_1 + z_2\,\mathbf{a}_2 + x_2\,\mathbf{a}_3$ | = | $\frac{1}{2}(-x_2 + y_2)\,a\,\hat{\mathbf{x}} + \left(-\frac{1}{2\sqrt{3}}x_2 - \frac{1}{2\sqrt{3}}y_2 + \frac{1}{\sqrt{3}}z_2\right)a\,\hat{\mathbf{y}} + \frac{1}{3}(x_2 + y_2 + z_2)\,c\,\hat{\mathbf{z}}$ | (6*b*) | B II |
| $\mathbf{B}_{10}$ | = | $\left(\frac{1}{2}+z_2\right)\mathbf{a}_1+\left(\frac{1}{2}+y_2\right)\mathbf{a}_2+\left(\frac{1}{2}+x_2\right)\mathbf{a}_3$ | = | $\frac{1}{2}\left(-x_2+z_2\right)a\,\mathbf{\hat{x}}+\left(-\frac{1}{2\sqrt{3}}x_2+\frac{1}{\sqrt{3}}y_2-\frac{1}{2\sqrt{3}}z_2\right)a\,\mathbf{\hat{y}}+\left(\frac{1}{2}+\frac{1}{3}x_2+\frac{1}{3}y_2+\frac{1}{3}z_2\right)c\,\mathbf{\hat{z}}$ | (6$b$) | B II |
| $\mathbf{B}_{11}$ | = | $\left(\frac{1}{2}+y_2\right)\mathbf{a}_1+\left(\frac{1}{2}+x_2\right)\mathbf{a}_2+\left(\frac{1}{2}+z_2\right)\mathbf{a}_3$ | = | $\frac{1}{2}\left(y_2-z_2\right)a\,\mathbf{\hat{x}}+\left(\frac{1}{\sqrt{3}}x_2-\frac{1}{2\sqrt{3}}y_2-\frac{1}{2\sqrt{3}}z_2\right)a\,\mathbf{\hat{y}}+\left(\frac{1}{2}+\frac{1}{3}x_2+\frac{1}{3}y_2+\frac{1}{3}z_2\right)c\,\mathbf{\hat{z}}$ | (6$b$) | B II |
| $\mathbf{B}_{12}$ | = | $\left(\frac{1}{2}+x_2\right)\mathbf{a}_1+\left(\frac{1}{2}+z_2\right)\mathbf{a}_2+\left(\frac{1}{2}+y_2\right)\mathbf{a}_3$ | = | $\frac{1}{2}\left(x_2-y_2\right)a\,\mathbf{\hat{x}}+\left(-\frac{1}{2\sqrt{3}}x_2-\frac{1}{2\sqrt{3}}y_2+\frac{1}{\sqrt{3}}z_2\right)a\,\mathbf{\hat{y}}+\left(\frac{1}{2}+\frac{1}{3}x_2+\frac{1}{3}y_2+\frac{1}{3}z_2\right)c\,\mathbf{\hat{z}}$ | (6$b$) | B II |
| $\mathbf{B}_{13}$ | = | $x_3\,\mathbf{a}_1+y_3\,\mathbf{a}_2+z_3\,\mathbf{a}_3$ | = | $\frac{1}{2}\left(x_3-z_3\right)a\,\mathbf{\hat{x}}+\left(-\frac{1}{2\sqrt{3}}x_3+\frac{1}{\sqrt{3}}y_3-\frac{1}{2\sqrt{3}}z_3\right)a\,\mathbf{\hat{y}}+\frac{1}{3}\left(x_3+y_3+z_3\right)c\,\mathbf{\hat{z}}$ | (6$b$) | Ba |
| $\mathbf{B}_{14}$ | = | $z_3\,\mathbf{a}_1+x_3\,\mathbf{a}_2+y_3\,\mathbf{a}_3$ | = | $\frac{1}{2}\left(-y_3+z_3\right)a\,\mathbf{\hat{x}}+\left(\frac{1}{\sqrt{3}}x_3-\frac{1}{2\sqrt{3}}y_3-\frac{1}{2\sqrt{3}}z_3\right)a\,\mathbf{\hat{y}}+\frac{1}{3}\left(x_3+y_3+z_3\right)c\,\mathbf{\hat{z}}$ | (6$b$) | Ba |
| $\mathbf{B}_{15}$ | = | $y_3\,\mathbf{a}_1+z_3\,\mathbf{a}_2+x_3\,\mathbf{a}_3$ | = | $\frac{1}{2}\left(-x_3+y_3\right)a\,\mathbf{\hat{x}}+\left(-\frac{1}{2\sqrt{3}}x_3-\frac{1}{2\sqrt{3}}y_3+\frac{1}{\sqrt{3}}z_3\right)a\,\mathbf{\hat{y}}+\frac{1}{3}\left(x_3+y_3+z_3\right)c\,\mathbf{\hat{z}}$ | (6$b$) | Ba |
| $\mathbf{B}_{16}$ | = | $\left(\frac{1}{2}+z_3\right)\mathbf{a}_1+\left(\frac{1}{2}+y_3\right)\mathbf{a}_2+\left(\frac{1}{2}+x_3\right)\mathbf{a}_3$ | = | $\frac{1}{2}\left(-x_3+z_3\right)a\,\mathbf{\hat{x}}+\left(-\frac{1}{2\sqrt{3}}x_3+\frac{1}{\sqrt{3}}y_3-\frac{1}{2\sqrt{3}}z_3\right)a\,\mathbf{\hat{y}}+\left(\frac{1}{2}+\frac{1}{3}x_3+\frac{1}{3}y_3+\frac{1}{3}z_3\right)c\,\mathbf{\hat{z}}$ | (6$b$) | Ba |
| $\mathbf{B}_{17}$ | = | $\left(\frac{1}{2}+y_3\right)\mathbf{a}_1+\left(\frac{1}{2}+x_3\right)\mathbf{a}_2+\left(\frac{1}{2}+z_3\right)\mathbf{a}_3$ | = | $\frac{1}{2}\left(y_3-z_3\right)a\,\mathbf{\hat{x}}+\left(\frac{1}{\sqrt{3}}x_3-\frac{1}{2\sqrt{3}}y_3-\frac{1}{2\sqrt{3}}z_3\right)a\,\mathbf{\hat{y}}+\left(\frac{1}{2}+\frac{1}{3}x_3+\frac{1}{3}y_3+\frac{1}{3}z_3\right)c\,\mathbf{\hat{z}}$ | (6$b$) | Ba |
| $\mathbf{B}_{18}$ | = | $\left(\frac{1}{2}+x_3\right)\mathbf{a}_1+\left(\frac{1}{2}+z_3\right)\mathbf{a}_2+\left(\frac{1}{2}+y_3\right)\mathbf{a}_3$ | = | $\frac{1}{2}\left(x_3-y_3\right)a\,\mathbf{\hat{x}}+\left(-\frac{1}{2\sqrt{3}}x_3-\frac{1}{2\sqrt{3}}y_3+\frac{1}{\sqrt{3}}z_3\right)a\,\mathbf{\hat{y}}+\left(\frac{1}{2}+\frac{1}{3}x_3+\frac{1}{3}y_3+\frac{1}{3}z_3\right)c\,\mathbf{\hat{z}}$ | (6$b$) | Ba |
| $\mathbf{B}_{19}$ | = | $x_4\,\mathbf{a}_1+y_4\,\mathbf{a}_2+z_4\,\mathbf{a}_3$ | = | $\frac{1}{2}\left(x_4-z_4\right)a\,\mathbf{\hat{x}}+\left(-\frac{1}{2\sqrt{3}}x_4+\frac{1}{\sqrt{3}}y_4-\frac{1}{2\sqrt{3}}z_4\right)a\,\mathbf{\hat{y}}+\frac{1}{3}\left(x_4+y_4+z_4\right)c\,\mathbf{\hat{z}}$ | (6$b$) | O I |
| $\mathbf{B}_{20}$ | = | $z_4\,\mathbf{a}_1+x_4\,\mathbf{a}_2+y_4\,\mathbf{a}_3$ | = | $\frac{1}{2}\left(-y_4+z_4\right)a\,\mathbf{\hat{x}}+\left(\frac{1}{\sqrt{3}}x_4-\frac{1}{2\sqrt{3}}y_4-\frac{1}{2\sqrt{3}}z_4\right)a\,\mathbf{\hat{y}}+\frac{1}{3}\left(x_4+y_4+z_4\right)c\,\mathbf{\hat{z}}$ | (6$b$) | O I |
| $\mathbf{B}_{21}$ | = | $y_4\,\mathbf{a}_1+z_4\,\mathbf{a}_2+x_4\,\mathbf{a}_3$ | = | $\frac{1}{2}\left(-x_4+y_4\right)a\,\mathbf{\hat{x}}+\left(-\frac{1}{2\sqrt{3}}x_4-\frac{1}{2\sqrt{3}}y_4+\frac{1}{\sqrt{3}}z_4\right)a\,\mathbf{\hat{y}}+\frac{1}{3}\left(x_4+y_4+z_4\right)c\,\mathbf{\hat{z}}$ | (6$b$) | O I |
| $\mathbf{B}_{22}$ | = | $\left(\frac{1}{2}+z_4\right)\mathbf{a}_1+\left(\frac{1}{2}+y_4\right)\mathbf{a}_2+\left(\frac{1}{2}+x_4\right)\mathbf{a}_3$ | = | $\frac{1}{2}\left(-x_4+z_4\right)a\,\mathbf{\hat{x}}+\left(-\frac{1}{2\sqrt{3}}x_4+\frac{1}{\sqrt{3}}y_4-\frac{1}{2\sqrt{3}}z_4\right)a\,\mathbf{\hat{y}}+\left(\frac{1}{2}+\frac{1}{3}x_4+\frac{1}{3}y_4+\frac{1}{3}z_4\right)c\,\mathbf{\hat{z}}$ | (6$b$) | O I |

$$\mathbf{B}_{23} = \left(\frac{1}{2}+y_4\right)\mathbf{a}_1 + \left(\frac{1}{2}+x_4\right)\mathbf{a}_2 + \left(\frac{1}{2}+z_4\right)\mathbf{a}_3 = \frac{1}{2}\left(y_4-z_4\right)a\,\hat{\mathbf{x}} + \left(\frac{1}{\sqrt{3}}x_4 - \frac{1}{2\sqrt{3}}y_4 - \frac{1}{2\sqrt{3}}z_4\right)a\,\hat{\mathbf{y}} + \left(\frac{1}{2}+\frac{1}{3}x_4+\frac{1}{3}y_4+\frac{1}{3}z_4\right)c\,\hat{\mathbf{z}} \quad (6b) \quad \text{O I}$$

$$\mathbf{B}_{24} = \left(\frac{1}{2}+x_4\right)\mathbf{a}_1 + \left(\frac{1}{2}+z_4\right)\mathbf{a}_2 + \left(\frac{1}{2}+y_4\right)\mathbf{a}_3 = \frac{1}{2}\left(x_4-y_4\right)a\,\hat{\mathbf{x}} + \left(-\frac{1}{2\sqrt{3}}x_4 - \frac{1}{2\sqrt{3}}y_4 + \frac{1}{\sqrt{3}}z_4\right)a\,\hat{\mathbf{y}} + \left(\frac{1}{2}+\frac{1}{3}x_4+\frac{1}{3}y_4+\frac{1}{3}z_4\right)c\,\hat{\mathbf{z}} \quad (6b) \quad \text{O I}$$

$$\mathbf{B}_{25} = x_5\,\mathbf{a}_1 + y_5\,\mathbf{a}_2 + z_5\,\mathbf{a}_3 = \frac{1}{2}\left(x_5-z_5\right)a\,\hat{\mathbf{x}} + \left(-\frac{1}{2\sqrt{3}}x_5 + \frac{1}{\sqrt{3}}y_5 - \frac{1}{2\sqrt{3}}z_5\right)a\,\hat{\mathbf{y}} + \frac{1}{3}\left(x_5+y_5+z_5\right)c\,\hat{\mathbf{z}} \quad (6b) \quad \text{O II}$$

$$\mathbf{B}_{26} = z_5\,\mathbf{a}_1 + x_5\,\mathbf{a}_2 + y_5\,\mathbf{a}_3 = \frac{1}{2}\left(-y_5+z_5\right)a\,\hat{\mathbf{x}} + \left(\frac{1}{\sqrt{3}}x_5 - \frac{1}{2\sqrt{3}}y_5 - \frac{1}{2\sqrt{3}}z_5\right)a\,\hat{\mathbf{y}} + \frac{1}{3}\left(x_5+y_5+z_5\right)c\,\hat{\mathbf{z}} \quad (6b) \quad \text{O II}$$

$$\mathbf{B}_{27} = y_5\,\mathbf{a}_1 + z_5\,\mathbf{a}_2 + x_5\,\mathbf{a}_3 = \frac{1}{2}\left(-x_5+y_5\right)a\,\hat{\mathbf{x}} + \left(-\frac{1}{2\sqrt{3}}x_5 - \frac{1}{2\sqrt{3}}y_5 + \frac{1}{\sqrt{3}}z_5\right)a\,\hat{\mathbf{y}} + \frac{1}{3}\left(x_5+y_5+z_5\right)c\,\hat{\mathbf{z}} \quad (6b) \quad \text{O II}$$

$$\mathbf{B}_{28} = \left(\frac{1}{2}+z_5\right)\mathbf{a}_1 + \left(\frac{1}{2}+y_5\right)\mathbf{a}_2 + \left(\frac{1}{2}+x_5\right)\mathbf{a}_3 = \frac{1}{2}\left(-x_5+z_5\right)a\,\hat{\mathbf{x}} + \left(-\frac{1}{2\sqrt{3}}x_5 + \frac{1}{\sqrt{3}}y_5 - \frac{1}{2\sqrt{3}}z_5\right)a\,\hat{\mathbf{y}} + \left(\frac{1}{2}+\frac{1}{3}x_5+\frac{1}{3}y_5+\frac{1}{3}z_5\right)c\,\hat{\mathbf{z}} \quad (6b) \quad \text{O II}$$

$$\mathbf{B}_{29} = \left(\frac{1}{2}+y_5\right)\mathbf{a}_1 + \left(\frac{1}{2}+x_5\right)\mathbf{a}_2 + \left(\frac{1}{2}+z_5\right)\mathbf{a}_3 = \frac{1}{2}\left(y_5-z_5\right)a\,\hat{\mathbf{x}} + \left(\frac{1}{\sqrt{3}}x_5 - \frac{1}{2\sqrt{3}}y_5 - \frac{1}{2\sqrt{3}}z_5\right)a\,\hat{\mathbf{y}} + \left(\frac{1}{2}+\frac{1}{3}x_5+\frac{1}{3}y_5+\frac{1}{3}z_5\right)c\,\hat{\mathbf{z}} \quad (6b) \quad \text{O II}$$

$$\mathbf{B}_{30} = \left(\frac{1}{2}+x_5\right)\mathbf{a}_1 + \left(\frac{1}{2}+z_5\right)\mathbf{a}_2 + \left(\frac{1}{2}+y_5\right)\mathbf{a}_3 = \frac{1}{2}\left(x_5-y_5\right)a\,\hat{\mathbf{x}} + \left(-\frac{1}{2\sqrt{3}}x_5 - \frac{1}{2\sqrt{3}}y_5 + \frac{1}{\sqrt{3}}z_5\right)a\,\hat{\mathbf{y}} + \left(\frac{1}{2}+\frac{1}{3}x_5+\frac{1}{3}y_5+\frac{1}{3}z_5\right)c\,\hat{\mathbf{z}} \quad (6b) \quad \text{O II}$$

$$\mathbf{B}_{31} = x_6\,\mathbf{a}_1 + y_6\,\mathbf{a}_2 + z_6\,\mathbf{a}_3 = \frac{1}{2}\left(x_6-z_6\right)a\,\hat{\mathbf{x}} + \left(-\frac{1}{2\sqrt{3}}x_6 + \frac{1}{\sqrt{3}}y_6 - \frac{1}{2\sqrt{3}}z_6\right)a\,\hat{\mathbf{y}} + \frac{1}{3}\left(x_6+y_6+z_6\right)c\,\hat{\mathbf{z}} \quad (6b) \quad \text{O III}$$

$$\mathbf{B}_{32} = z_6\,\mathbf{a}_1 + x_6\,\mathbf{a}_2 + y_6\,\mathbf{a}_3 = \frac{1}{2}\left(-y_6+z_6\right)a\,\hat{\mathbf{x}} + \left(\frac{1}{\sqrt{3}}x_6 - \frac{1}{2\sqrt{3}}y_6 - \frac{1}{2\sqrt{3}}z_6\right)a\,\hat{\mathbf{y}} + \frac{1}{3}\left(x_6+y_6+z_6\right)c\,\hat{\mathbf{z}} \quad (6b) \quad \text{O III}$$

$$\mathbf{B}_{33} = y_6\,\mathbf{a}_1 + z_6\,\mathbf{a}_2 + x_6\,\mathbf{a}_3 = \frac{1}{2}\left(-x_6+y_6\right)a\,\hat{\mathbf{x}} + \left(-\frac{1}{2\sqrt{3}}x_6 - \frac{1}{2\sqrt{3}}y_6 + \frac{1}{\sqrt{3}}z_6\right)a\,\hat{\mathbf{y}} + \frac{1}{3}\left(x_6+y_6+z_6\right)c\,\hat{\mathbf{z}} \quad (6b) \quad \text{O III}$$

$$\mathbf{B}_{34} = \left(\frac{1}{2}+z_6\right)\mathbf{a}_1 + \left(\frac{1}{2}+y_6\right)\mathbf{a}_2 + \left(\frac{1}{2}+x_6\right)\mathbf{a}_3 = \frac{1}{2}\left(-x_6+z_6\right)a\,\hat{\mathbf{x}} + \left(-\frac{1}{2\sqrt{3}}x_6 + \frac{1}{\sqrt{3}}y_6 - \frac{1}{2\sqrt{3}}z_6\right)a\,\hat{\mathbf{y}} + \left(\frac{1}{2}+\frac{1}{3}x_6+\frac{1}{3}y_6+\frac{1}{3}z_6\right)c\,\hat{\mathbf{z}} \quad (6b) \quad \text{O III}$$

$$\mathbf{B}_{35} = \left(\frac{1}{2}+y_6\right)\mathbf{a}_1 + \left(\frac{1}{2}+x_6\right)\mathbf{a}_2 + \left(\frac{1}{2}+z_6\right)\mathbf{a}_3 = \frac{1}{2}\left(y_6-z_6\right)a\,\hat{\mathbf{x}} + \left(\frac{1}{\sqrt{3}}x_6 - \frac{1}{2\sqrt{3}}y_6 - \frac{1}{2\sqrt{3}}z_6\right)a\,\hat{\mathbf{y}} + \left(\frac{1}{2}+\frac{1}{3}x_6+\frac{1}{3}y_6+\frac{1}{3}z_6\right)c\,\hat{\mathbf{z}} \quad (6b) \quad \text{O III}$$

| | | | | | | |
|---|---|---|---|---|---|---|
| $\mathbf{B}_{36}$ | = | $\left(\frac{1}{2}+x_6\right)\mathbf{a}_1+\left(\frac{1}{2}+z_6\right)\mathbf{a}_2+\left(\frac{1}{2}+y_6\right)\mathbf{a}_3$ | = | $\frac{1}{2}(x_6-y_6)\,a\,\mathbf{\hat{x}}+\left(-\frac{1}{2\sqrt{3}}x_6-\frac{1}{2\sqrt{3}}y_6+\frac{1}{\sqrt{3}}z_6\right)a\,\mathbf{\hat{y}}+\left(\frac{1}{2}+\frac{1}{3}x_6+\frac{1}{3}y_6+\frac{1}{3}z_6\right)c\,\mathbf{\hat{z}}$ | (6$b$) | O III |
| $\mathbf{B}_{37}$ | = | $x_7\,\mathbf{a}_1+y_7\,\mathbf{a}_2+z_7\,\mathbf{a}_3$ | = | $\frac{1}{2}(x_7-z_7)\,a\,\mathbf{\hat{x}}+\left(-\frac{1}{2\sqrt{3}}x_7+\frac{1}{\sqrt{3}}y_7-\frac{1}{2\sqrt{3}}z_7\right)a\,\mathbf{\hat{y}}+\frac{1}{3}(x_7+y_7+z_7)\,c\,\mathbf{\hat{z}}$ | (6$b$) | O IV |
| $\mathbf{B}_{38}$ | = | $z_7\,\mathbf{a}_1+x_7\,\mathbf{a}_2+y_7\,\mathbf{a}_3$ | = | $\frac{1}{2}(-y_7+z_7)\,a\,\mathbf{\hat{x}}+\left(\frac{1}{\sqrt{3}}x_7-\frac{1}{2\sqrt{3}}y_7-\frac{1}{2\sqrt{3}}z_7\right)a\,\mathbf{\hat{y}}+\frac{1}{3}(x_7+y_7+z_7)\,c\,\mathbf{\hat{z}}$ | (6$b$) | O IV |
| $\mathbf{B}_{39}$ | = | $y_7\,\mathbf{a}_1+z_7\,\mathbf{a}_2+x_7\,\mathbf{a}_3$ | = | $\frac{1}{2}(-x_7+y_7)\,a\,\mathbf{\hat{x}}+\left(-\frac{1}{2\sqrt{3}}x_7-\frac{1}{2\sqrt{3}}y_7+\frac{1}{\sqrt{3}}z_7\right)a\,\mathbf{\hat{y}}+\frac{1}{3}(x_7+y_7+z_7)\,c\,\mathbf{\hat{z}}$ | (6$b$) | O IV |
| $\mathbf{B}_{40}$ | = | $\left(\frac{1}{2}+z_7\right)\mathbf{a}_1+\left(\frac{1}{2}+y_7\right)\mathbf{a}_2+\left(\frac{1}{2}+x_7\right)\mathbf{a}_3$ | = | $\frac{1}{2}(-x_7+z_7)\,a\,\mathbf{\hat{x}}+\left(-\frac{1}{2\sqrt{3}}x_7+\frac{1}{\sqrt{3}}y_7-\frac{1}{2\sqrt{3}}z_7\right)a\,\mathbf{\hat{y}}+\left(\frac{1}{2}+\frac{1}{3}x_7+\frac{1}{3}y_7+\frac{1}{3}z_7\right)c\,\mathbf{\hat{z}}$ | (6$b$) | O IV |
| $\mathbf{B}_{41}$ | = | $\left(\frac{1}{2}+y_7\right)\mathbf{a}_1+\left(\frac{1}{2}+x_7\right)\mathbf{a}_2+\left(\frac{1}{2}+z_7\right)\mathbf{a}_3$ | = | $\frac{1}{2}(y_7-z_7)\,a\,\mathbf{\hat{x}}+\left(\frac{1}{\sqrt{3}}x_7-\frac{1}{2\sqrt{3}}y_7-\frac{1}{2\sqrt{3}}z_7\right)a\,\mathbf{\hat{y}}+\left(\frac{1}{2}+\frac{1}{3}x_7+\frac{1}{3}y_7+\frac{1}{3}z_7\right)c\,\mathbf{\hat{z}}$ | (6$b$) | O IV |
| $\mathbf{B}_{42}$ | = | $\left(\frac{1}{2}+x_7\right)\mathbf{a}_1+\left(\frac{1}{2}+z_7\right)\mathbf{a}_2+\left(\frac{1}{2}+y_7\right)\mathbf{a}_3$ | = | $\frac{1}{2}(x_7-y_7)\,a\,\mathbf{\hat{x}}+\left(-\frac{1}{2\sqrt{3}}x_7-\frac{1}{2\sqrt{3}}y_7+\frac{1}{\sqrt{3}}z_7\right)a\,\mathbf{\hat{y}}+\left(\frac{1}{2}+\frac{1}{3}x_7+\frac{1}{3}y_7+\frac{1}{3}z_7\right)c\,\mathbf{\hat{z}}$ | (6$b$) | O IV |

**Geometry files:**

- CIF: pp. 1730
- POSCAR: pp. 1730

# $I1_3$ ($SrCl_2 \cdot (H_2O)_6$) (*obsolete*) Structure: A2B6C_hP9_162_d_k_a

http://aflow.org/prototype-encyclopedia/A2B6C_hP9_162_d_k_a

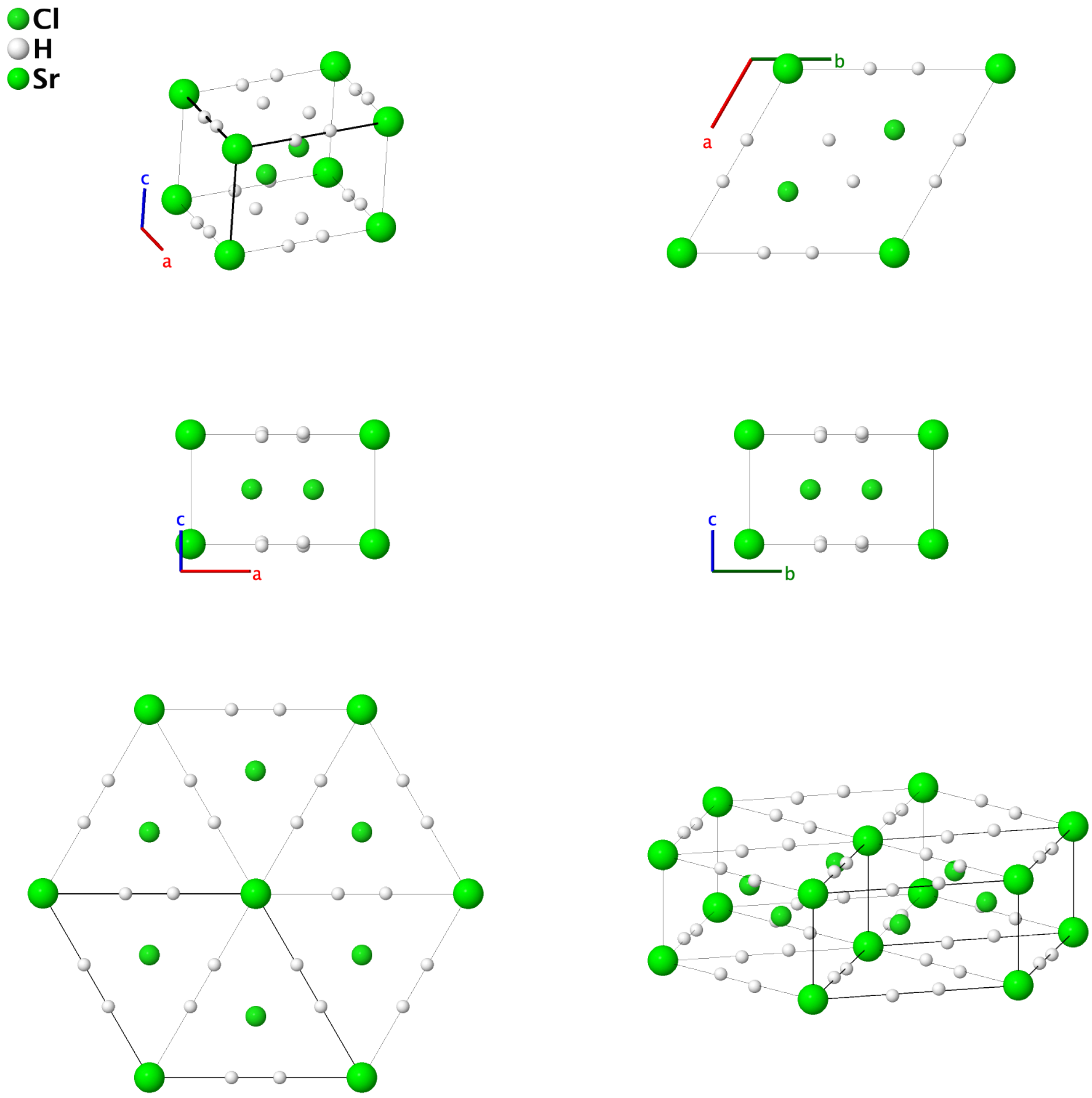

| | | |
|---|---|---|
| **Prototype** | : | $Cl_2(H_2O)_6Sr$ |
| **AFLOW prototype label** | : | A2B6C_hP9_162_d_k_a |
| ***Strukturbericht* designation** | : | $I1_3$ |
| **Pearson symbol** | : | hP9 |
| **Space group number** | : | 162 |
| **Space group symbol** | : | $P\bar{3}1m$ |
| **AFLOW prototype command** | : | `aflow --proto=A2B6C_hP9_162_d_k_a --params=`$a, c/a, x_3, z_3$ |

**Other compounds with this structure**

- $CaCl_2 \cdot (H_2O)_6$, $CaBr_2 \cdot (H_2O)_6$, $SrBr_2 \cdot (H_2O)_6$, $CaI_2 \cdot (H_2O)_6$, $SrI_2 \cdot (H_2O)_6$, and $BaI_2 \cdot (H_2O)_6$

- (Hermann, 1937) Gives this the *Strukturbericht* designation $I1_3$, but gives the prototype as $K_2Pt(SCN)_6$. As we discussed on the $K_2Pt(SCN)_6$ ($H6_3$) page, the difference between these two structures is significant, so we will use the original $H6_3$ designation for $K_2Pt(SCN)_6$, and $I1_3$ for $SrCl_2 \cdot (H_2O)_6$.
- In any case, (Aagon, 1986) and others have shown that the correct space group of this structure is $P321$ #150. We discuss the corrected structure on the $SrCl_2 \cdot (H_2O)_6$ page.

- Using the notation of (Gottfried, 1937) this could also be designated the $J1_3$ structure. That designation was never used in any *Strukturbericht* volume, so we will use $I1_3$ here.

- The positions of the hydrogen atoms in the water molecules were not determined, so we only provide the positions of the oxygen atoms (labeled as $H_2O$).

---

**Trigonal Hexagonal primitive vectors:**

$$\mathbf{a}_1 = \frac{1}{2}\,a\,\hat{\mathbf{x}} - \frac{\sqrt{3}}{2}\,a\,\hat{\mathbf{y}}$$
$$\mathbf{a}_2 = \frac{1}{2}\,a\,\hat{\mathbf{x}} + \frac{\sqrt{3}}{2}\,a\,\hat{\mathbf{y}}$$
$$\mathbf{a}_3 = c\,\hat{\mathbf{z}}$$

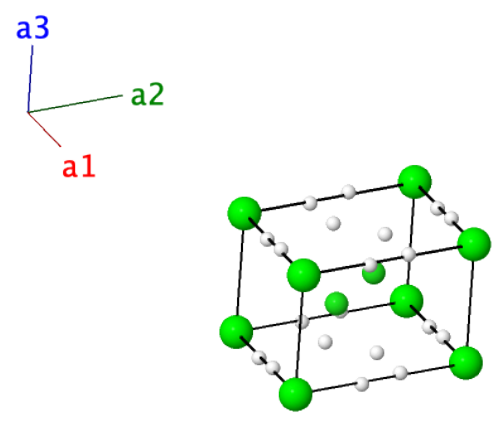

---

**Basis vectors:**

| | | Lattice Coordinates | | Cartesian Coordinates | Wyckoff Position | Atom Type |
|---|---|---|---|---|---|---|
| $\mathbf{B}_1$ | = | $0\,\mathbf{a}_1 + 0\,\mathbf{a}_2 + 0\,\mathbf{a}_3$ | = | $0\,\hat{\mathbf{x}} + 0\,\hat{\mathbf{y}} + 0\,\hat{\mathbf{z}}$ | (1$a$) | Sr |
| $\mathbf{B}_2$ | = | $\frac{1}{3}\,\mathbf{a}_1 + \frac{2}{3}\,\mathbf{a}_2 + \frac{1}{2}\,\mathbf{a}_3$ | = | $\frac{1}{2}a\,\hat{\mathbf{x}} + \frac{1}{2\sqrt{3}}a\,\hat{\mathbf{y}} + \frac{1}{2}c\,\hat{\mathbf{z}}$ | (2$d$) | Cl |
| $\mathbf{B}_3$ | = | $\frac{2}{3}\,\mathbf{a}_1 + \frac{1}{3}\,\mathbf{a}_2 + \frac{1}{2}\,\mathbf{a}_3$ | = | $\frac{1}{2}a\,\hat{\mathbf{x}} - \frac{1}{2\sqrt{3}}a\,\hat{\mathbf{y}} + \frac{1}{2}c\,\hat{\mathbf{z}}$ | (2$d$) | Cl |
| $\mathbf{B}_4$ | = | $x_3\,\mathbf{a}_1 + z_3\,\mathbf{a}_3$ | = | $\frac{1}{2}x_3a\,\hat{\mathbf{x}} - \frac{\sqrt{3}}{2}x_3a\,\hat{\mathbf{y}} + z_3c\,\hat{\mathbf{z}}$ | (6$k$) | $H_2O$ |
| $\mathbf{B}_5$ | = | $x_3\,\mathbf{a}_2 + z_3\,\mathbf{a}_3$ | = | $\frac{1}{2}x_3a\,\hat{\mathbf{x}} + \frac{\sqrt{3}}{2}x_3a\,\hat{\mathbf{y}} + z_3c\,\hat{\mathbf{z}}$ | (6$k$) | $H_2O$ |
| $\mathbf{B}_6$ | = | $-x_3\,\mathbf{a}_1 - x_3\,\mathbf{a}_2 + z_3\,\mathbf{a}_3$ | = | $-x_3a\,\hat{\mathbf{x}} + z_3c\,\hat{\mathbf{z}}$ | (6$k$) | $H_2O$ |
| $\mathbf{B}_7$ | = | $-x_3\,\mathbf{a}_2 - z_3\,\mathbf{a}_3$ | = | $-\frac{1}{2}x_3a\,\hat{\mathbf{x}} - \frac{\sqrt{3}}{2}x_3a\,\hat{\mathbf{y}} - z_3c\,\hat{\mathbf{z}}$ | (6$k$) | $H_2O$ |
| $\mathbf{B}_8$ | = | $-x_3\,\mathbf{a}_1 - z_3\,\mathbf{a}_3$ | = | $-\frac{1}{2}x_3a\,\hat{\mathbf{x}} + \frac{\sqrt{3}}{2}x_3a\,\hat{\mathbf{y}} - z_3c\,\hat{\mathbf{z}}$ | (6$k$) | $H_2O$ |
| $\mathbf{B}_9$ | = | $x_3\,\mathbf{a}_1 + x_3\,\mathbf{a}_2 - z_3\,\mathbf{a}_3$ | = | $x_3a\,\hat{\mathbf{x}} - z_3c\,\hat{\mathbf{z}}$ | (6$k$) | $H_2O$ |

---

**Found in:**

---

**Geometry files:**

- CIF: pp. 1730
- POSCAR: pp. 1731

# Rosiaite ($PbSb_2O_6$) Structure: A6BC2_hP9_162_k_a_d

http://aflow.org/prototype-encyclopedia/A6BC2_hP9_162_k_a_d

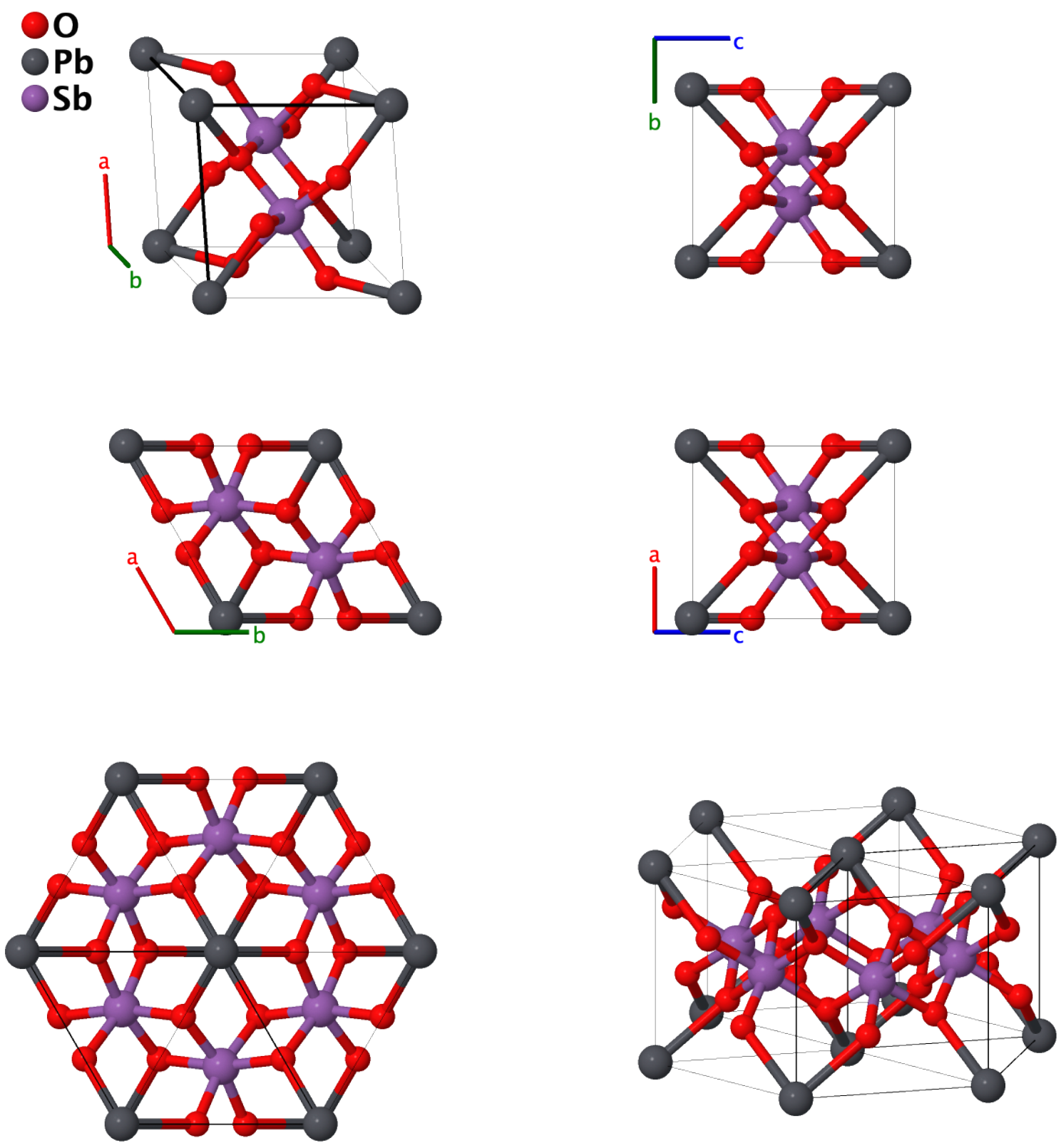

| | | |
|---|---|---|
| **Prototype** | : | $O_6PbSb_2$ |
| **AFLOW prototype label** | : | A6BC2_hP9_162_k_a_d |
| ***Strukturbericht* designation** | : | None |
| **Pearson symbol** | : | hP9 |
| **Space group number** | : | 162 |
| **Space group symbol** | : | $P\bar{3}1m$ |
| **AFLOW prototype command** | : | `aflow --proto=A6BC2_hP9_162_k_a_d` <br> `--params=`$a, c/a, x_3, z_3$ |

**Other compounds with this structure**

- $MnSeTeO_6$, $MnSnTeO_6$, and $PbTeGeO_6$

- This is the ternary form of the $L'3_2$ ($\beta$-$V_2N$) structure.

**Trigonal Hexagonal primitive vectors:**

$$\mathbf{a}_1 = \frac{1}{2}\,a\,\hat{\mathbf{x}} - \frac{\sqrt{3}}{2}\,a\,\hat{\mathbf{y}}$$
$$\mathbf{a}_2 = \frac{1}{2}\,a\,\hat{\mathbf{x}} + \frac{\sqrt{3}}{2}\,a\,\hat{\mathbf{y}}$$
$$\mathbf{a}_3 = c\,\hat{\mathbf{z}}$$

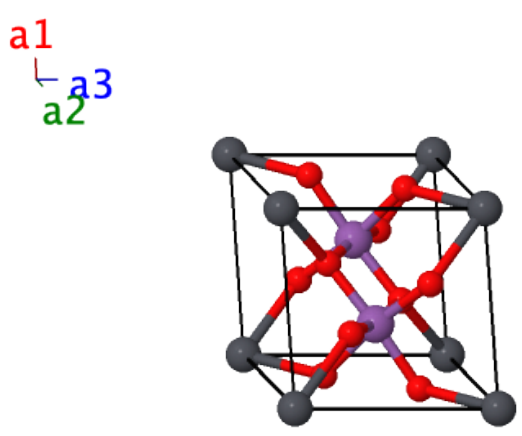

**Basis vectors:**

| | | Lattice Coordinates | | Cartesian Coordinates | Wyckoff Position | Atom Type |
|---|---|---|---|---|---|---|
| $\mathbf{B}_1$ | = | $0\,\mathbf{a}_1 + 0\,\mathbf{a}_2 + 0\,\mathbf{a}_3$ | = | $0\,\hat{\mathbf{x}} + 0\,\hat{\mathbf{y}} + 0\,\hat{\mathbf{z}}$ | (1$a$) | Pb |
| $\mathbf{B}_2$ | = | $\frac{1}{3}\,\mathbf{a}_1 + \frac{2}{3}\,\mathbf{a}_2 + \frac{1}{2}\,\mathbf{a}_3$ | = | $\frac{1}{2}a\,\hat{\mathbf{x}} + \frac{1}{2\sqrt{3}}a\,\hat{\mathbf{y}} + \frac{1}{2}c\,\hat{\mathbf{z}}$ | (2$d$) | Sb |
| $\mathbf{B}_3$ | = | $\frac{2}{3}\,\mathbf{a}_1 + \frac{1}{3}\,\mathbf{a}_2 + \frac{1}{2}\,\mathbf{a}_3$ | = | $\frac{1}{2}a\,\hat{\mathbf{x}} - \frac{1}{2\sqrt{3}}a\,\hat{\mathbf{y}} + \frac{1}{2}c\,\hat{\mathbf{z}}$ | (2$d$) | Sb |
| $\mathbf{B}_4$ | = | $x_3\,\mathbf{a}_1 + z_3\,\mathbf{a}_3$ | = | $\frac{1}{2}x_3a\,\hat{\mathbf{x}} - \frac{\sqrt{3}}{2}x_3a\,\hat{\mathbf{y}} + z_3c\,\hat{\mathbf{z}}$ | (6$k$) | O |
| $\mathbf{B}_5$ | = | $x_3\,\mathbf{a}_2 + z_3\,\mathbf{a}_3$ | = | $\frac{1}{2}x_3a\,\hat{\mathbf{x}} + \frac{\sqrt{3}}{2}x_3a\,\hat{\mathbf{y}} + z_3c\,\hat{\mathbf{z}}$ | (6$k$) | O |
| $\mathbf{B}_6$ | = | $-x_3\,\mathbf{a}_1 - x_3\,\mathbf{a}_2 + z_3\,\mathbf{a}_3$ | = | $-x_3a\,\hat{\mathbf{x}} + z_3c\,\hat{\mathbf{z}}$ | (6$k$) | O |
| $\mathbf{B}_7$ | = | $-x_3\,\mathbf{a}_2 - z_3\,\mathbf{a}_3$ | = | $-\frac{1}{2}x_3a\,\hat{\mathbf{x}} - \frac{\sqrt{3}}{2}x_3a\,\hat{\mathbf{y}} - z_3c\,\hat{\mathbf{z}}$ | (6$k$) | O |
| $\mathbf{B}_8$ | = | $-x_3\,\mathbf{a}_1 - z_3\,\mathbf{a}_3$ | = | $-\frac{1}{2}x_3a\,\hat{\mathbf{x}} + \frac{\sqrt{3}}{2}x_3a\,\hat{\mathbf{y}} - z_3c\,\hat{\mathbf{z}}$ | (6$k$) | O |
| $\mathbf{B}_9$ | = | $x_3\,\mathbf{a}_1 + x_3\,\mathbf{a}_2 - z_3\,\mathbf{a}_3$ | = | $x_3a\,\hat{\mathbf{x}} - z_3c\,\hat{\mathbf{z}}$ | (6$k$) | O |

**Geometry files:**

- CIF: pp. 1731
- POSCAR: pp. 1731

# $NaSbF_4(OH)_2$ ($J1_{12}$) Structure: A6BC_hP16_163_i_b_c

http://aflow.org/prototype-encyclopedia/A6BC_hP16_163_i_b_c

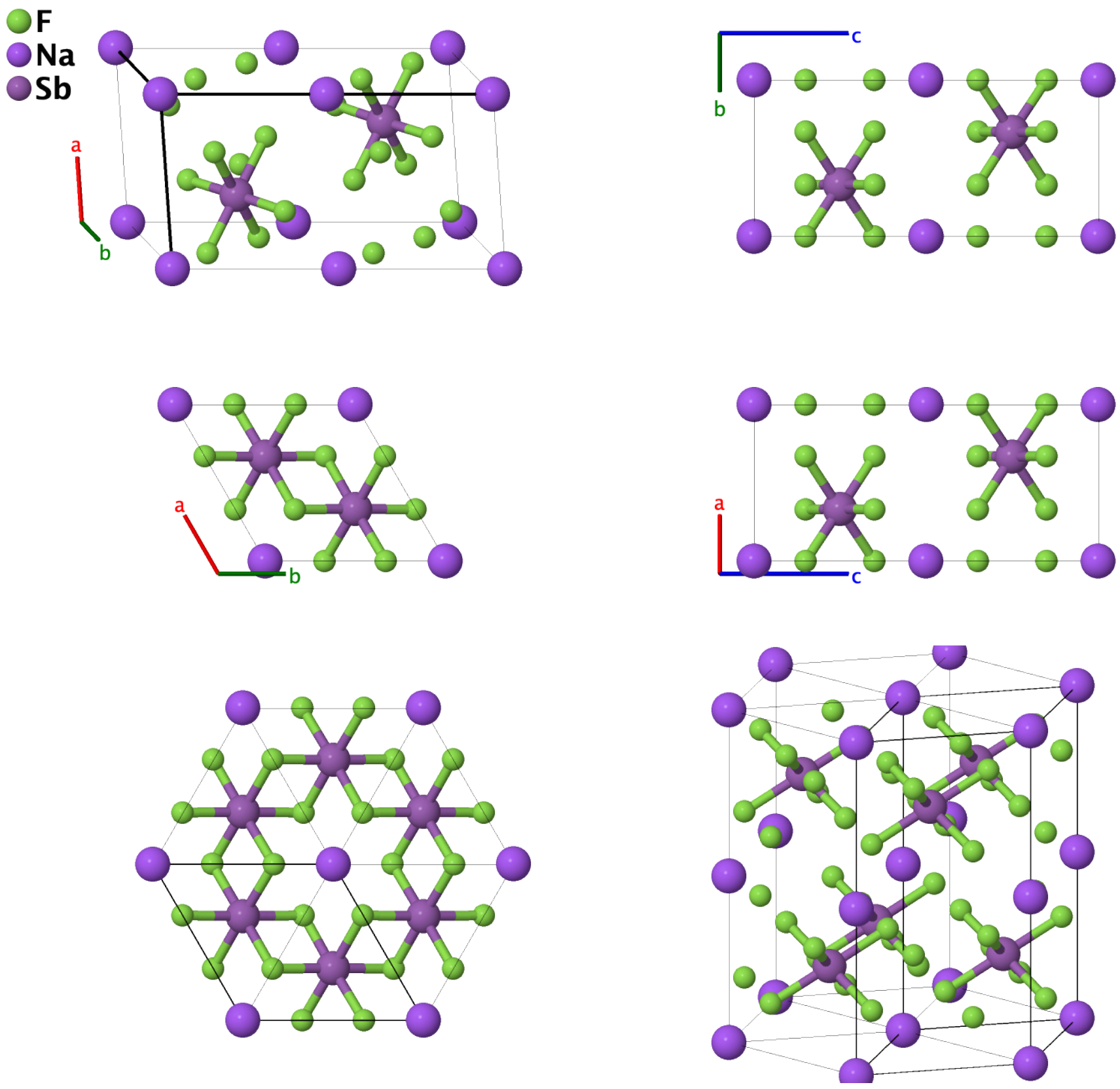

| | | |
|---|---|---|
| **Prototype** | : | $F_4NaSb$ |
| **AFLOW prototype label** | : | A6BC_hP16_163_i_b_c |
| ***Strukturbericht* designation** | : | $J1_{12}$ |
| **Pearson symbol** | : | hP16 |
| **Space group number** | : | 163 |
| **Space group symbol** | : | $P\bar{3}1c$ |
| **AFLOW prototype command** | : | `aflow --proto=A6BC_hP16_163_i_b_c`<br>`--params=`$a, c/a, x_3, y_3, z_3$ |

- The ($2i$) site, which we show as occupied by fluorine, is actually occupied by a random mixture of fluorine atoms (67%) and OH radicals (33%).
- Although the replacement of fluorine by OH does not affect the shape of the Sb-(F,OH)$_6$ ions, it has a profound effect on the structure, as can be seen by looking at $NaSbF_6$ and $NaSb(OH)_6$ ($J1_{11}$).

**Trigonal Hexagonal primitive vectors:**

$$
\begin{aligned}
\mathbf{a}_1 &= \tfrac{1}{2}\,a\,\hat{\mathbf{x}} - \tfrac{\sqrt{3}}{2}\,a\,\hat{\mathbf{y}} \\
\mathbf{a}_2 &= \tfrac{1}{2}\,a\,\hat{\mathbf{x}} + \tfrac{\sqrt{3}}{2}\,a\,\hat{\mathbf{y}} \\
\mathbf{a}_3 &= c\,\hat{\mathbf{z}}
\end{aligned}
$$

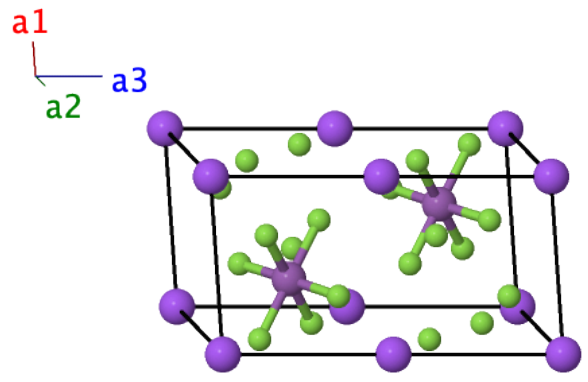

**Basis vectors:**

| | | Lattice Coordinates | | Cartesian Coordinates | Wyckoff Position | Atom Type |
|---|---|---|---|---|---|---|
| $\mathbf{B}_1$ | = | $0\,\mathbf{a}_1 + 0\,\mathbf{a}_2 + 0\,\mathbf{a}_3$ | = | $0\,\hat{\mathbf{x}} + 0\,\hat{\mathbf{y}} + 0\,\hat{\mathbf{z}}$ | (2*b*) | Na |
| $\mathbf{B}_2$ | = | $\frac{1}{2}\,\mathbf{a}_3$ | = | $\frac{1}{2}c\,\hat{\mathbf{z}}$ | (2*b*) | Na |
| $\mathbf{B}_3$ | = | $\frac{1}{3}\,\mathbf{a}_1 + \frac{2}{3}\,\mathbf{a}_2 + \frac{1}{4}\,\mathbf{a}_3$ | = | $\frac{1}{2}a\,\hat{\mathbf{x}} + \frac{1}{2\sqrt{3}}a\,\hat{\mathbf{y}} + \frac{1}{4}c\,\hat{\mathbf{z}}$ | (2*c*) | Sb |
| $\mathbf{B}_4$ | = | $\frac{2}{3}\,\mathbf{a}_1 + \frac{1}{3}\,\mathbf{a}_2 + \frac{3}{4}\,\mathbf{a}_3$ | = | $\frac{1}{2}a\,\hat{\mathbf{x}} - \frac{1}{2\sqrt{3}}a\,\hat{\mathbf{y}} + \frac{3}{4}c\,\hat{\mathbf{z}}$ | (2*c*) | Sb |
| $\mathbf{B}_5$ | = | $x_3\,\mathbf{a}_1 + y_3\,\mathbf{a}_2 + z_3\,\mathbf{a}_3$ | = | $\frac{1}{2}\,(x_3+y_3)\,a\,\hat{\mathbf{x}} + \frac{\sqrt{3}}{2}\,(-x_3+y_3)\,a\,\hat{\mathbf{y}} + z_3c\,\hat{\mathbf{z}}$ | (12*i*) | F |
| $\mathbf{B}_6$ | = | $-y_3\,\mathbf{a}_1 + (x_3-y_3)\,\mathbf{a}_2 + z_3\,\mathbf{a}_3$ | = | $\left(\frac{1}{2}x_3 - y_3\right)a\,\hat{\mathbf{x}} + \frac{\sqrt{3}}{2}x_3a\,\hat{\mathbf{y}} + z_3c\,\hat{\mathbf{z}}$ | (12*i*) | F |
| $\mathbf{B}_7$ | = | $(-x_3+y_3)\,\mathbf{a}_1 - x_3\,\mathbf{a}_2 + z_3\,\mathbf{a}_3$ | = | $\left(-x_3 + \frac{1}{2}y_3\right)a\,\hat{\mathbf{x}} - \frac{\sqrt{3}}{2}y_3a\,\hat{\mathbf{y}} + z_3c\,\hat{\mathbf{z}}$ | (12*i*) | F |
| $\mathbf{B}_8$ | = | $-y_3\,\mathbf{a}_1 - x_3\,\mathbf{a}_2 + \left(\frac{1}{2} - z_3\right)\mathbf{a}_3$ | = | $-\frac{1}{2}\,(x_3+y_3)\,a\,\hat{\mathbf{x}} + \frac{\sqrt{3}}{2}\,(-x_3+y_3)\,a\,\hat{\mathbf{y}} + \left(\frac{1}{2} - z_3\right)c\,\hat{\mathbf{z}}$ | (12*i*) | F |
| $\mathbf{B}_9$ | = | $(-x_3+y_3)\,\mathbf{a}_1 + y_3\,\mathbf{a}_2 + \left(\frac{1}{2} - z_3\right)\mathbf{a}_3$ | = | $\left(-\frac{1}{2}x_3 + y_3\right)a\,\hat{\mathbf{x}} + \frac{\sqrt{3}}{2}x_3a\,\hat{\mathbf{y}} + \left(\frac{1}{2} - z_3\right)c\,\hat{\mathbf{z}}$ | (12*i*) | F |
| $\mathbf{B}_{10}$ | = | $x_3\,\mathbf{a}_1 + (x_3-y_3)\,\mathbf{a}_2 + \left(\frac{1}{2} - z_3\right)\mathbf{a}_3$ | = | $\left(x_3 - \frac{1}{2}y_3\right)a\,\hat{\mathbf{x}} - \frac{\sqrt{3}}{2}y_3a\,\hat{\mathbf{y}} + \left(\frac{1}{2} - z_3\right)c\,\hat{\mathbf{z}}$ | (12*i*) | F |
| $\mathbf{B}_{11}$ | = | $-x_3\,\mathbf{a}_1 - y_3\,\mathbf{a}_2 - z_3\,\mathbf{a}_3$ | = | $-\frac{1}{2}\,(x_3+y_3)\,a\,\hat{\mathbf{x}} + \frac{\sqrt{3}}{2}\,(x_3-y_3)\,a\,\hat{\mathbf{y}} - z_3c\,\hat{\mathbf{z}}$ | (12*i*) | F |
| $\mathbf{B}_{12}$ | = | $y_3\,\mathbf{a}_1 + (-x_3+y_3)\,\mathbf{a}_2 - z_3\,\mathbf{a}_3$ | = | $\left(-\frac{1}{2}x_3 + y_3\right)a\,\hat{\mathbf{x}} - \frac{\sqrt{3}}{2}x_3a\,\hat{\mathbf{y}} - z_3c\,\hat{\mathbf{z}}$ | (12*i*) | F |
| $\mathbf{B}_{13}$ | = | $(x_3-y_3)\,\mathbf{a}_1 + x_3\,\mathbf{a}_2 - z_3\,\mathbf{a}_3$ | = | $\left(x_3 - \frac{1}{2}y_3\right)a\,\hat{\mathbf{x}} + \frac{\sqrt{3}}{2}y_3a\,\hat{\mathbf{y}} - z_3c\,\hat{\mathbf{z}}$ | (12*i*) | F |
| $\mathbf{B}_{14}$ | = | $y_3\,\mathbf{a}_1 + x_3\,\mathbf{a}_2 + \left(\frac{1}{2} + z_3\right)\mathbf{a}_3$ | = | $\frac{1}{2}\,(x_3+y_3)\,a\,\hat{\mathbf{x}} + \frac{\sqrt{3}}{2}\,(x_3-y_3)\,a\,\hat{\mathbf{y}} + \left(\frac{1}{2} + z_3\right)c\,\hat{\mathbf{z}}$ | (12*i*) | F |
| $\mathbf{B}_{15}$ | = | $(x_3-y_3)\,\mathbf{a}_1 - y_3\,\mathbf{a}_2 + \left(\frac{1}{2} + z_3\right)\mathbf{a}_3$ | = | $\left(\frac{1}{2}x_3 - y_3\right)a\,\hat{\mathbf{x}} - \frac{\sqrt{3}}{2}x_3a\,\hat{\mathbf{y}} + \left(\frac{1}{2} + z_3\right)c\,\hat{\mathbf{z}}$ | (12*i*) | F |
| $\mathbf{B}_{16}$ | = | $-x_3\,\mathbf{a}_1 + (-x_3+y_3)\,\mathbf{a}_2 + \left(\frac{1}{2} + z_3\right)\mathbf{a}_3$ | = | $\left(-x_3 + \frac{1}{2}y_3\right)a\,\hat{\mathbf{x}} + \frac{\sqrt{3}}{2}y_3a\,\hat{\mathbf{y}} + \left(\frac{1}{2} + z_3\right)c\,\hat{\mathbf{z}}$ | (12*i*) | F |

**Found in:**

**Geometry files:**

- CIF: pp. 1731
- POSCAR: pp. 1732

# Colquiriite ($LiCaAlF_6$) Structure: ABC6D_hP18_163_d_b_i_c

http://aflow.org/prototype-encyclopedia/ABC6D_hP18_163_d_b_i_c

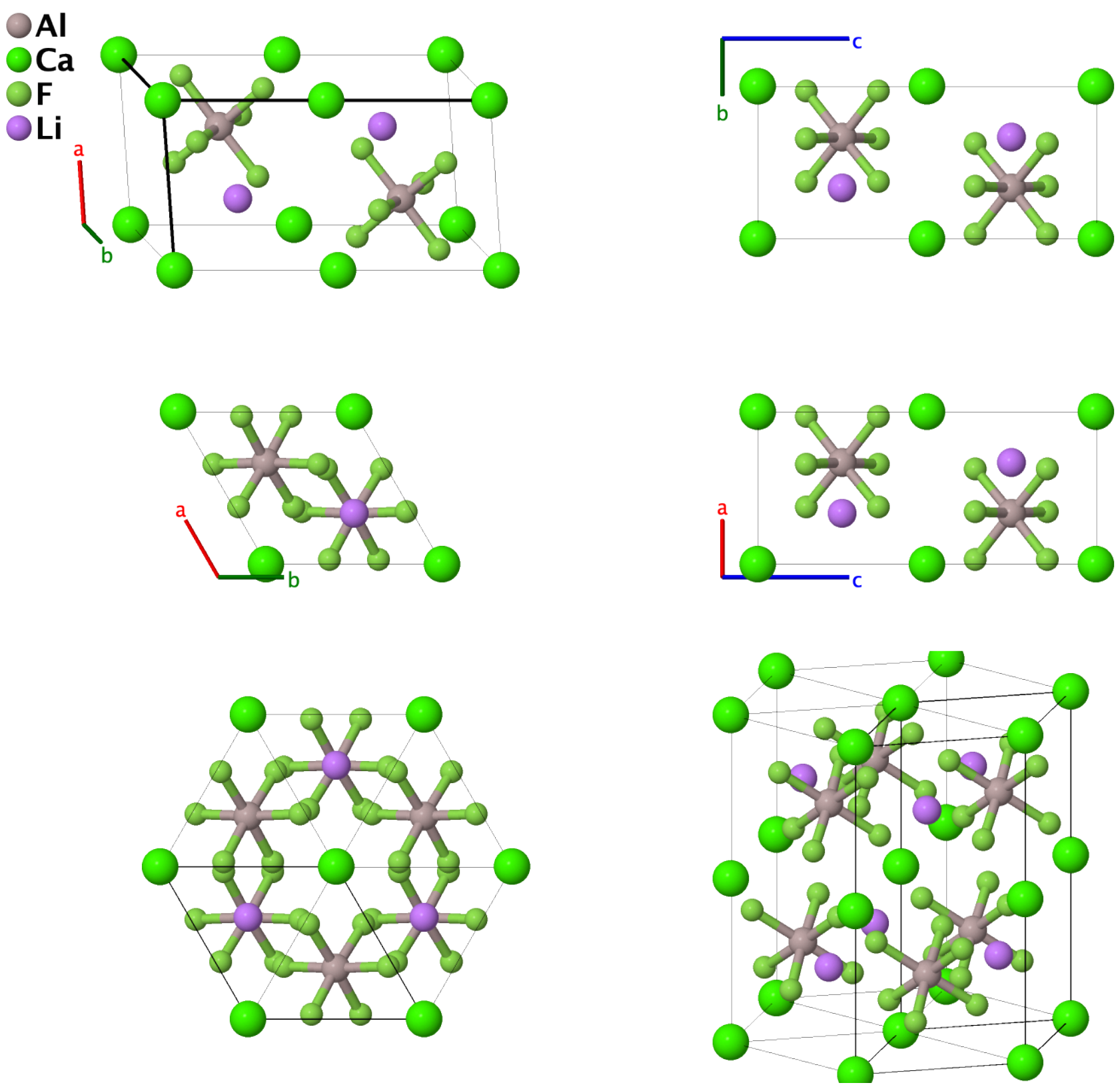

| | | |
|---|---|---|
| **Prototype** | : | $AlCaF_6Li$ |
| **AFLOW prototype label** | : | ABC6D_hP18_163_d_b_i_c |
| ***Strukturbericht* designation** | : | None |
| **Pearson symbol** | : | hP18 |
| **Space group number** | : | 163 |
| **Space group symbol** | : | $P\bar{3}1c$ |
| **AFLOW prototype command** | : | `aflow --proto=ABC6D_hP18_163_d_b_i_c` `--params=`$a, c/a, x_4, y_4, z_4$ |

**Other compounds with this structure**

- $LiSrAlF_6$

- This is the low-temperature standard-pressure structure of $LiCaAlF_6$. We used the data taken at 300K and ambient pressure.

### Trigonal Hexagonal primitive vectors:

$$\mathbf{a}_1 = \frac{1}{2}\,a\,\hat{\mathbf{x}} - \frac{\sqrt{3}}{2}\,a\,\hat{\mathbf{y}}$$
$$\mathbf{a}_2 = \frac{1}{2}\,a\,\hat{\mathbf{x}} + \frac{\sqrt{3}}{2}\,a\,\hat{\mathbf{y}}$$
$$\mathbf{a}_3 = c\,\hat{\mathbf{z}}$$

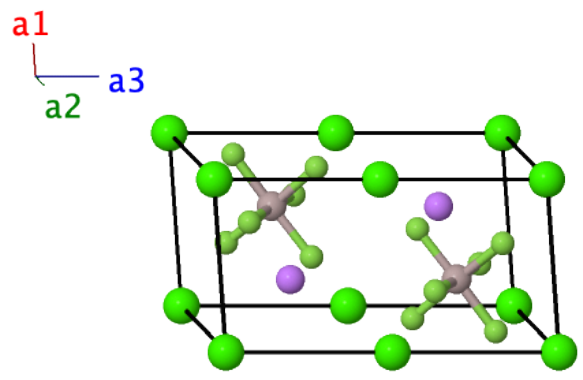

### Basis vectors:

| | | Lattice Coordinates | | Cartesian Coordinates | Wyckoff Position | Atom Type |
|---|---|---|---|---|---|---|
| $\mathbf{B}_1$ | = | $0\,\mathbf{a}_1 + 0\,\mathbf{a}_2 + 0\,\mathbf{a}_3$ | = | $0\,\hat{\mathbf{x}} + 0\,\hat{\mathbf{y}} + 0\,\hat{\mathbf{z}}$ | (2*b*) | Ca |
| $\mathbf{B}_2$ | = | $\frac{1}{2}\,\mathbf{a}_3$ | = | $\frac{1}{2}c\,\hat{\mathbf{z}}$ | (2*b*) | Ca |
| $\mathbf{B}_3$ | = | $\frac{1}{3}\,\mathbf{a}_1 + \frac{2}{3}\,\mathbf{a}_2 + \frac{1}{4}\,\mathbf{a}_3$ | = | $\frac{1}{2}a\,\hat{\mathbf{x}} + \frac{1}{2\sqrt{3}}a\,\hat{\mathbf{y}} + \frac{1}{4}c\,\hat{\mathbf{z}}$ | (2*c*) | Li |
| $\mathbf{B}_4$ | = | $\frac{2}{3}\,\mathbf{a}_1 + \frac{1}{3}\,\mathbf{a}_2 + \frac{3}{4}\,\mathbf{a}_3$ | = | $\frac{1}{2}a\,\hat{\mathbf{x}} - \frac{1}{2\sqrt{3}}a\,\hat{\mathbf{y}} + \frac{3}{4}c\,\hat{\mathbf{z}}$ | (2*c*) | Li |
| $\mathbf{B}_5$ | = | $\frac{2}{3}\,\mathbf{a}_1 + \frac{1}{3}\,\mathbf{a}_2 + \frac{1}{4}\,\mathbf{a}_3$ | = | $\frac{1}{2}a\,\hat{\mathbf{x}} - \frac{1}{2\sqrt{3}}a\,\hat{\mathbf{y}} + \frac{1}{4}c\,\hat{\mathbf{z}}$ | (2*d*) | Al |
| $\mathbf{B}_6$ | = | $\frac{1}{3}\,\mathbf{a}_1 + \frac{2}{3}\,\mathbf{a}_2 + \frac{3}{4}\,\mathbf{a}_3$ | = | $\frac{1}{2}a\,\hat{\mathbf{x}} + \frac{1}{2\sqrt{3}}a\,\hat{\mathbf{y}} + \frac{3}{4}c\,\hat{\mathbf{z}}$ | (2*d*) | Al |
| $\mathbf{B}_7$ | = | $x_4\,\mathbf{a}_1 + y_4\,\mathbf{a}_2 + z_4\,\mathbf{a}_3$ | = | $\frac{1}{2}(x_4 + y_4)\,a\,\hat{\mathbf{x}} + \frac{\sqrt{3}}{2}(-x_4 + y_4)\,a\,\hat{\mathbf{y}} + z_4 c\,\hat{\mathbf{z}}$ | (12*i*) | F |
| $\mathbf{B}_8$ | = | $-y_4\,\mathbf{a}_1 + (x_4 - y_4)\,\mathbf{a}_2 + z_4\,\mathbf{a}_3$ | = | $\left(\frac{1}{2}x_4 - y_4\right)a\,\hat{\mathbf{x}} + \frac{\sqrt{3}}{2}x_4 a\,\hat{\mathbf{y}} + z_4 c\,\hat{\mathbf{z}}$ | (12*i*) | F |
| $\mathbf{B}_9$ | = | $(-x_4 + y_4)\,\mathbf{a}_1 - x_4\,\mathbf{a}_2 + z_4\,\mathbf{a}_3$ | = | $\left(-x_4 + \frac{1}{2}y_4\right)a\,\hat{\mathbf{x}} - \frac{\sqrt{3}}{2}y_4 a\,\hat{\mathbf{y}} + z_4 c\,\hat{\mathbf{z}}$ | (12*i*) | F |
| $\mathbf{B}_{10}$ | = | $-y_4\,\mathbf{a}_1 - x_4\,\mathbf{a}_2 + \left(\frac{1}{2} - z_4\right)\mathbf{a}_3$ | = | $-\frac{1}{2}(x_4 + y_4)\,a\,\hat{\mathbf{x}} + \frac{\sqrt{3}}{2}(-x_4 + y_4)\,a\,\hat{\mathbf{y}} + \left(\frac{1}{2} - z_4\right)c\,\hat{\mathbf{z}}$ | (12*i*) | F |
| $\mathbf{B}_{11}$ | = | $(-x_4 + y_4)\,\mathbf{a}_1 + y_4\,\mathbf{a}_2 + \left(\frac{1}{2} - z_4\right)\mathbf{a}_3$ | = | $\left(-\frac{1}{2}x_4 + y_4\right)a\,\hat{\mathbf{x}} + \frac{\sqrt{3}}{2}x_4 a\,\hat{\mathbf{y}} + \left(\frac{1}{2} - z_4\right)c\,\hat{\mathbf{z}}$ | (12*i*) | F |
| $\mathbf{B}_{12}$ | = | $x_4\,\mathbf{a}_1 + (x_4 - y_4)\,\mathbf{a}_2 + \left(\frac{1}{2} - z_4\right)\mathbf{a}_3$ | = | $\left(x_4 - \frac{1}{2}y_4\right)a\,\hat{\mathbf{x}} - \frac{\sqrt{3}}{2}y_4 a\,\hat{\mathbf{y}} + \left(\frac{1}{2} - z_4\right)c\,\hat{\mathbf{z}}$ | (12*i*) | F |
| $\mathbf{B}_{13}$ | = | $-x_4\,\mathbf{a}_1 - y_4\,\mathbf{a}_2 - z_4\,\mathbf{a}_3$ | = | $-\frac{1}{2}(x_4 + y_4)\,a\,\hat{\mathbf{x}} + \frac{\sqrt{3}}{2}(x_4 - y_4)\,a\,\hat{\mathbf{y}} - z_4 c\,\hat{\mathbf{z}}$ | (12*i*) | F |
| $\mathbf{B}_{14}$ | = | $y_4\,\mathbf{a}_1 + (-x_4 + y_4)\,\mathbf{a}_2 - z_4\,\mathbf{a}_3$ | = | $\left(-\frac{1}{2}x_4 + y_4\right)a\,\hat{\mathbf{x}} - \frac{\sqrt{3}}{2}x_4 a\,\hat{\mathbf{y}} - z_4 c\,\hat{\mathbf{z}}$ | (12*i*) | F |
| $\mathbf{B}_{15}$ | = | $(x_4 - y_4)\,\mathbf{a}_1 + x_4\,\mathbf{a}_2 - z_4\,\mathbf{a}_3$ | = | $\left(x_4 - \frac{1}{2}y_4\right)a\,\hat{\mathbf{x}} + \frac{\sqrt{3}}{2}y_4 a\,\hat{\mathbf{y}} - z_4 c\,\hat{\mathbf{z}}$ | (12*i*) | F |
| $\mathbf{B}_{16}$ | = | $y_4\,\mathbf{a}_1 + x_4\,\mathbf{a}_2 + \left(\frac{1}{2} + z_4\right)\mathbf{a}_3$ | = | $\frac{1}{2}(x_4 + y_4)\,a\,\hat{\mathbf{x}} + \frac{\sqrt{3}}{2}(x_4 - y_4)\,a\,\hat{\mathbf{y}} + \left(\frac{1}{2} + z_4\right)c\,\hat{\mathbf{z}}$ | (12*i*) | F |
| $\mathbf{B}_{17}$ | = | $(x_4 - y_4)\,\mathbf{a}_1 - y_4\,\mathbf{a}_2 + \left(\frac{1}{2} + z_4\right)\mathbf{a}_3$ | = | $\left(\frac{1}{2}x_4 - y_4\right)a\,\hat{\mathbf{x}} - \frac{\sqrt{3}}{2}x_4 a\,\hat{\mathbf{y}} + \left(\frac{1}{2} + z_4\right)c\,\hat{\mathbf{z}}$ | (12*i*) | F |
| $\mathbf{B}_{18}$ | = | $-x_4\,\mathbf{a}_1 + (-x_4 + y_4)\,\mathbf{a}_2 + \left(\frac{1}{2} + z_4\right)\mathbf{a}_3$ | = | $\left(-x_4 + \frac{1}{2}y_4\right)a\,\hat{\mathbf{x}} + \frac{\sqrt{3}}{2}y_4 a\,\hat{\mathbf{y}} + \left(\frac{1}{2} + z_4\right)c\,\hat{\mathbf{z}}$ | (12*i*) | F |

**Geometry files:**

- CIF: pp. 1732
- POSCAR: pp. 1732

# Predicted $Li_2MgH_{16}$ 300 GPa Structure: A16B2C_hP19_164_2d2i_d_b

http://aflow.org/prototype-encyclopedia/A16B2C_hP19_164_2d2i_d_b

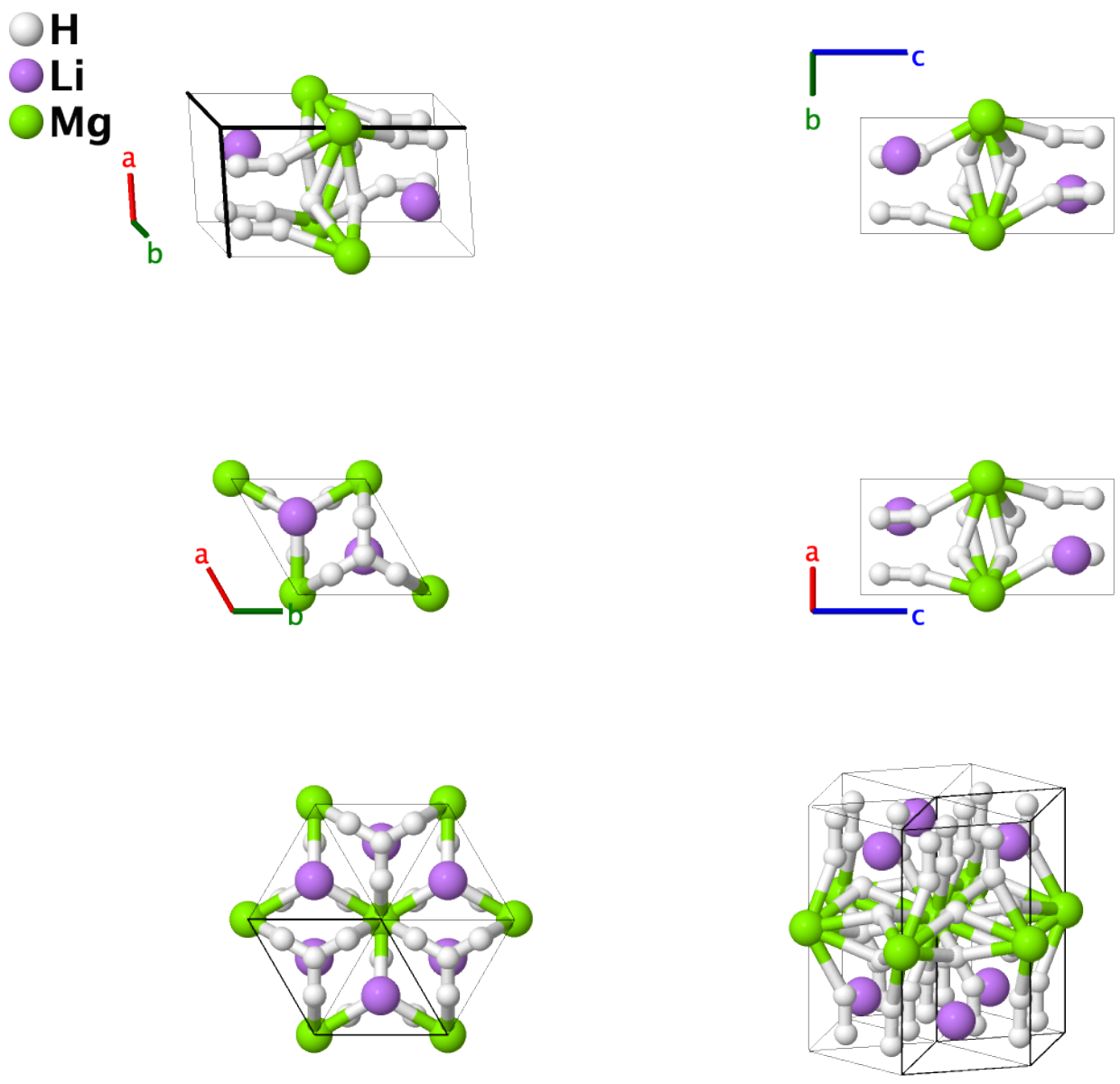

| | | |
|---|---|---|
| **Prototype** | : | $H_{16}Li_2Mg$ |
| **AFLOW prototype label** | : | A16B2C_hP19_164_2d2i_d_b |
| ***Strukturbericht* designation** | : | None |
| **Pearson symbol** | : | hP19 |
| **Space group number** | : | 164 |
| **Space group symbol** | : | $P\bar{3}m1$ |
| **AFLOW prototype command** | : | `aflow --proto=A16B2C_hP19_164_2d2i_d_b`<br>`--params=`$a, c/a, z_2, z_3, z_4, x_5, z_5, x_6, z_6$ |

- This structure is predicted to be the zero-temperature ground state of $Li_2MgH_{16}$ at 300 GPa. It is primarily of interest because a metastable cubic structure with the same composition is predicted to be superconducting at 250 GPa with $T_c = 430 - 473$K.

**Trigonal Hexagonal primitive vectors:**

$$\mathbf{a}_1 = \frac{1}{2}\,a\,\hat{\mathbf{x}} - \frac{\sqrt{3}}{2}\,a\,\hat{\mathbf{y}}$$
$$\mathbf{a}_2 = \frac{1}{2}\,a\,\hat{\mathbf{x}} + \frac{\sqrt{3}}{2}\,a\,\hat{\mathbf{y}}$$
$$\mathbf{a}_3 = c\,\hat{\mathbf{z}}$$

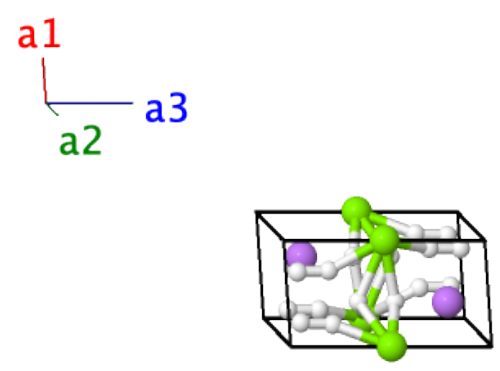

**Basis vectors:**

| | | Lattice Coordinates | | Cartesian Coordinates | Wyckoff Position | Atom Type |
|---|---|---|---|---|---|---|
| $\mathbf{B}_1$ | = | $\frac{1}{2}\,\mathbf{a}_3$ | = | $\frac{1}{2}c\,\hat{\mathbf{z}}$ | (1$b$) | Mg |
| $\mathbf{B}_{2}$ | = | $\frac{1}{3}\,\mathbf{a}_{1} + \frac{2}{3}\,\mathbf{a}_{2} + z_{2}\,\mathbf{a}_{3}$ | = | $\frac{1}{2}a\,\hat{\mathbf{x}} + \frac{1}{2\sqrt{3}}a\,\hat{\mathbf{y}} + z_{2}c\,\hat{\mathbf{z}}$ | (2$d$) | H I |
| $\mathbf{B}_{3}$ | = | $\frac{2}{3}\,\mathbf{a}_{1} + \frac{1}{3}\,\mathbf{a}_{2} - z_{2}\,\mathbf{a}_{3}$ | = | $\frac{1}{2}a\,\hat{\mathbf{x}} - \frac{1}{2\sqrt{3}}a\,\hat{\mathbf{y}} - z_{2}c\,\hat{\mathbf{z}}$ | (2$d$) | H I |
| $\mathbf{B}_{4}$ | = | $\frac{1}{3}\,\mathbf{a}_{1} + \frac{2}{3}\,\mathbf{a}_{2} + z_{3}\,\mathbf{a}_{3}$ | = | $\frac{1}{2}a\,\hat{\mathbf{x}} + \frac{1}{2\sqrt{3}}a\,\hat{\mathbf{y}} + z_{3}c\,\hat{\mathbf{z}}$ | (2$d$) | H II |
| $\mathbf{B}_{5}$ | = | $\frac{2}{3}\,\mathbf{a}_{1} + \frac{1}{3}\,\mathbf{a}_{2} - z_{3}\,\mathbf{a}_{3}$ | = | $\frac{1}{2}a\,\hat{\mathbf{x}} - \frac{1}{2\sqrt{3}}a\,\hat{\mathbf{y}} - z_{3}c\,\hat{\mathbf{z}}$ | (2$d$) | H II |
| $\mathbf{B}_{6}$ | = | $\frac{1}{3}\,\mathbf{a}_{1} + \frac{2}{3}\,\mathbf{a}_{2} + z_{4}\,\mathbf{a}_{3}$ | = | $\frac{1}{2}a\,\hat{\mathbf{x}} + \frac{1}{2\sqrt{3}}a\,\hat{\mathbf{y}} + z_{4}c\,\hat{\mathbf{z}}$ | (2$d$) | Li |
| $\mathbf{B}_{7}$ | = | $\frac{2}{3}\,\mathbf{a}_{1} + \frac{1}{3}\,\mathbf{a}_{2} - z_{4}\,\mathbf{a}_{3}$ | = | $\frac{1}{2}a\,\hat{\mathbf{x}} - \frac{1}{2\sqrt{3}}a\,\hat{\mathbf{y}} - z_{4}c\,\hat{\mathbf{z}}$ | (2$d$) | Li |
| $\mathbf{B}_{8}$ | = | $x_{5}\,\mathbf{a}_{1} - x_{5}\,\mathbf{a}_{2} + z_{5}\,\mathbf{a}_{3}$ | = | $-\sqrt{3}x_{5}a\,\hat{\mathbf{y}} + z_{5}c\,\hat{\mathbf{z}}$ | (6$i$) | H III |
| $\mathbf{B}_{9}$ | = | $x_{5}\,\mathbf{a}_{1} + 2x_{5}\,\mathbf{a}_{2} + z_{5}\,\mathbf{a}_{3}$ | = | $\frac{3}{2}x_{5}a\,\hat{\mathbf{x}} + \frac{\sqrt{3}}{2}x_{5}a\,\hat{\mathbf{y}} + z_{5}c\,\hat{\mathbf{z}}$ | (6$i$) | H III |
| $\mathbf{B}_{10}$ | = | $-2x_{5}\,\mathbf{a}_{1} - x_{5}\,\mathbf{a}_{2} + z_{5}\,\mathbf{a}_{3}$ | = | $-\frac{3}{2}x_{5}a\,\hat{\mathbf{x}} + \frac{\sqrt{3}}{2}x_{5}a\,\hat{\mathbf{y}} + z_{5}c\,\hat{\mathbf{z}}$ | (6$i$) | H III |
| $\mathbf{B}_{11}$ | = | $-x_{5}\,\mathbf{a}_{1} + x_{5}\,\mathbf{a}_{2} - z_{5}\,\mathbf{a}_{3}$ | = | $\sqrt{3}x_{5}a\,\hat{\mathbf{y}} - z_{5}c\,\hat{\mathbf{z}}$ | (6$i$) | H III |
| $\mathbf{B}_{12}$ | = | $2x_{5}\,\mathbf{a}_{1} + x_{5}\,\mathbf{a}_{2} - z_{5}\,\mathbf{a}_{3}$ | = | $\frac{3}{2}x_{5}a\,\hat{\mathbf{x}} - \frac{\sqrt{3}}{2}x_{5}a\,\hat{\mathbf{y}} - z_{5}c\,\hat{\mathbf{z}}$ | (6$i$) | H III |
| $\mathbf{B}_{13}$ | = | $-x_{5}\,\mathbf{a}_{1} - 2x_{5}\,\mathbf{a}_{2} - z_{5}\,\mathbf{a}_{3}$ | = | $-\frac{3}{2}x_{5}a\,\hat{\mathbf{x}} - \frac{\sqrt{3}}{2}x_{5}a\,\hat{\mathbf{y}} - z_{5}c\,\hat{\mathbf{z}}$ | (6$i$) | H III |
| $\mathbf{B}_{14}$ | = | $x_{6}\,\mathbf{a}_{1} - x_{6}\,\mathbf{a}_{2} + z_{6}\,\mathbf{a}_{3}$ | = | $-\sqrt{3}x_{6}a\,\hat{\mathbf{y}} + z_{6}c\,\hat{\mathbf{z}}$ | (6$i$) | H IV |
| $\mathbf{B}_{15}$ | = | $x_{6}\,\mathbf{a}_{1} + 2x_{6}\,\mathbf{a}_{2} + z_{6}\,\mathbf{a}_{3}$ | = | $\frac{3}{2}x_{6}a\,\hat{\mathbf{x}} + \frac{\sqrt{3}}{2}x_{6}a\,\hat{\mathbf{y}} + z_{6}c\,\hat{\mathbf{z}}$ | (6$i$) | H IV |
| $\mathbf{B}_{16}$ | = | $-2x_{6}\,\mathbf{a}_{1} - x_{6}\,\mathbf{a}_{2} + z_{6}\,\mathbf{a}_{3}$ | = | $-\frac{3}{2}x_{6}a\,\hat{\mathbf{x}} + \frac{\sqrt{3}}{2}x_{6}a\,\hat{\mathbf{y}} + z_{6}c\,\hat{\mathbf{z}}$ | (6$i$) | H IV |
| $\mathbf{B}_{17}$ | = | $-x_{6}\,\mathbf{a}_{1} + x_{6}\,\mathbf{a}_{2} - z_{6}\,\mathbf{a}_{3}$ | = | $\sqrt{3}x_{6}a\,\hat{\mathbf{y}} - z_{6}c\,\hat{\mathbf{z}}$ | (6$i$) | H IV |
| $\mathbf{B}_{18}$ | = | $2x_{6}\,\mathbf{a}_{1} + x_{6}\,\mathbf{a}_{2} - z_{6}\,\mathbf{a}_{3}$ | = | $\frac{3}{2}x_{6}a\,\hat{\mathbf{x}} - \frac{\sqrt{3}}{2}x_{6}a\,\hat{\mathbf{y}} - z_{6}c\,\hat{\mathbf{z}}$ | (6$i$) | H IV |
| $\mathbf{B}_{19}$ | = | $-x_{6}\,\mathbf{a}_{1} - 2x_{6}\,\mathbf{a}_{2} - z_{6}\,\mathbf{a}_{3}$ | = | $-\frac{3}{2}x_{6}a\,\hat{\mathbf{x}} - \frac{\sqrt{3}}{2}x_{6}a\,\hat{\mathbf{y}} - z_{6}c\,\hat{\mathbf{z}}$ | (6$i$) | H IV |

**Geometry files:**

- CIF: pp. 1732
- POSCAR: pp. 1733

# $Ce_2O_2S$ Structure: A2B2C_hP5_164_d_d_a

http://aflow.org/prototype-encyclopedia/A2B2C_hP5_164_d_d_a

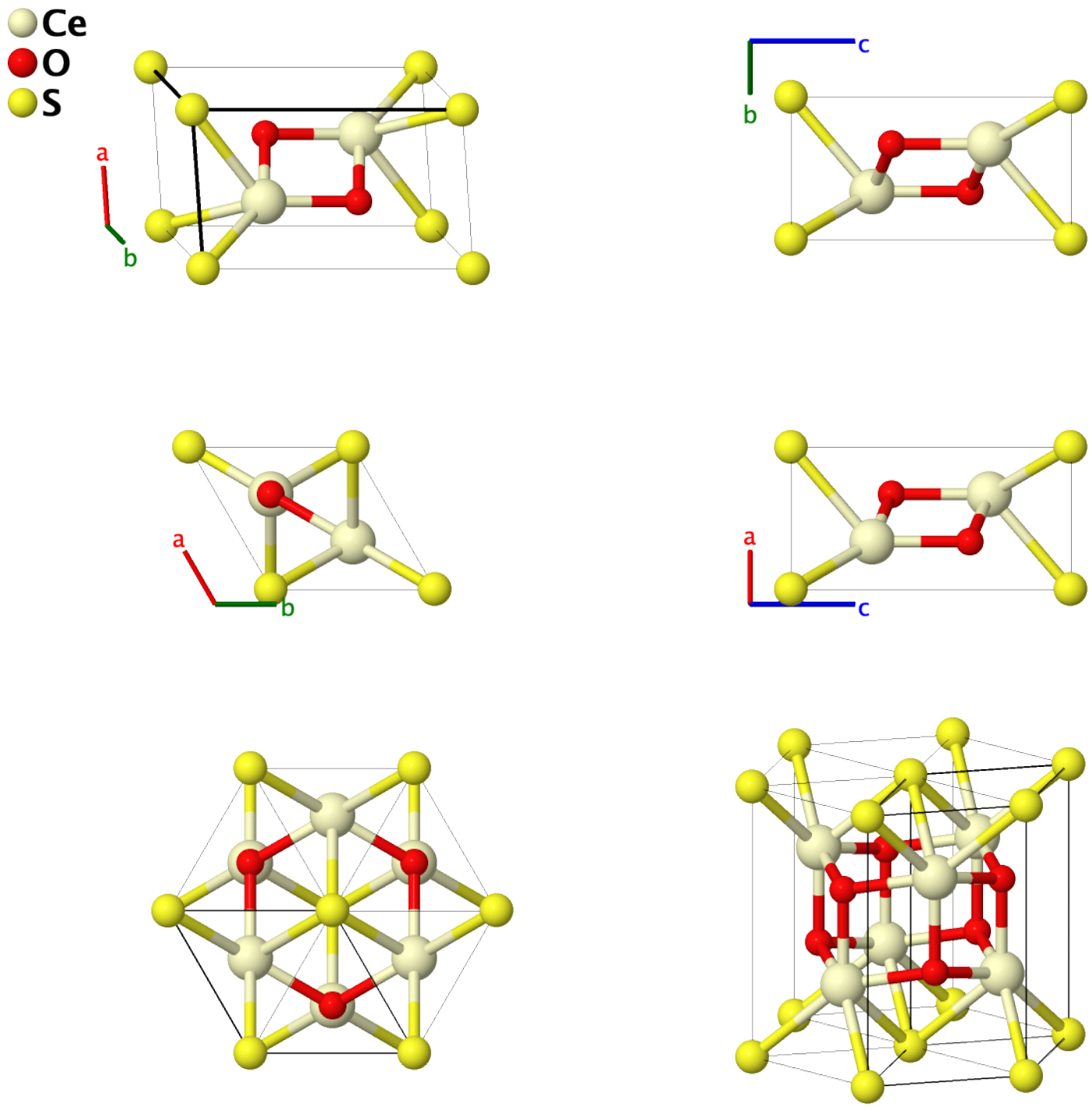

| | | |
|---|---|---|
| **Prototype** | : | $Ce_2O_2S$ |
| **AFLOW prototype label** | : | A2B2C_hP5_164_d_d_a |
| ***Strukturbericht* designation** | : | None |
| **Pearson symbol** | : | hP5 |
| **Space group number** | : | 164 |
| **Space group symbol** | : | $P\bar{3}m1$ |
| **AFLOW prototype command** | : | `aflow --proto=A2B2C_hP5_164_d_d_a --params=`$a, c/a, z_2, z_3$ |

**Other compounds with this structure**

- $Ce_2O_2S$, $Ce_2Se_2S$, $La_2O_2S$, $Pu_2O_2S$, $CeCuZnP_2$, $DyCuZnP_2$, $ErCuZnP_2$, $GdCuZnP_2$, $HoCuZnP_2$, $LaCuZnP_2$, $LuCuZnP_2$, $NdCuZnP_2$, $PrCuZnP_2$, $ScCuZnP_2$, $SmCuZnP_2$, $TbCuZnP_2$, $TmCuZnP_2$, $YCuZnP_2$, $YbCuZnP_2$, $CaAl_2Si_2$, $CaAs_2Bi_2$, $CaAs_2Mn_2$, $CaAs_2P_2$, $CaAs_2Sb_2$, $CaBi_2Mn_2$, $SrAs_2Mn_2$, $SrAs_2P_2$, $SrAs_2Sb_2$, and $YbMnCuP_2$

- This is the ternary form of the $D5_2$ $La_2O_3$ structure and the $D5_{13}$ $Al_3Ni_2$ structure. We have separated it from the parents because of the large number of compounds involved.
- Authors after (Zachariasen, 1949) often use $CaAl_2Si_2$ as the prototype for this structure.

**Trigonal Hexagonal primitive vectors:**

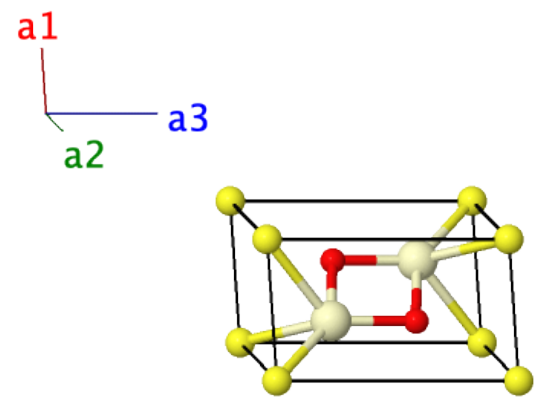

$$\mathbf{a}_1 = \frac{1}{2}\,a\,\hat{\mathbf{x}} - \frac{\sqrt{3}}{2}\,a\,\hat{\mathbf{y}}$$
$$\mathbf{a}_2 = \frac{1}{2}\,a\,\hat{\mathbf{x}} + \frac{\sqrt{3}}{2}\,a\,\hat{\mathbf{y}}$$
$$\mathbf{a}_3 = c\,\hat{\mathbf{z}}$$

**Basis vectors:**

| | | Lattice Coordinates | | Cartesian Coordinates | Wyckoff Position | Atom Type |
|---|---|---|---|---|---|---|
| $\mathbf{B}_1$ | = | $0\,\mathbf{a}_1 + 0\,\mathbf{a}_2 + 0\,\mathbf{a}_3$ | = | $0\,\hat{\mathbf{x}} + 0\,\hat{\mathbf{y}} + 0\,\hat{\mathbf{z}}$ | (1$a$) | S |
| $\mathbf{B}_2$ | = | $\frac{1}{3}\,\mathbf{a}_1 + \frac{2}{3}\,\mathbf{a}_2 + z_2\,\mathbf{a}_3$ | = | $\frac{1}{2}a\,\hat{\mathbf{x}} + \frac{1}{2\sqrt{3}}a\,\hat{\mathbf{y}} + z_2c\,\hat{\mathbf{z}}$ | (2$d$) | Ce |
| $\mathbf{B}_3$ | = | $\frac{2}{3}\,\mathbf{a}_1 + \frac{1}{3}\,\mathbf{a}_2 - z_2\,\mathbf{a}_3$ | = | $\frac{1}{2}a\,\hat{\mathbf{x}} - \frac{1}{2\sqrt{3}}a\,\hat{\mathbf{y}} - z_2c\,\hat{\mathbf{z}}$ | (2$d$) | Ce |
| $\mathbf{B}_4$ | = | $\frac{1}{3}\,\mathbf{a}_1 + \frac{2}{3}\,\mathbf{a}_2 + z_3\,\mathbf{a}_3$ | = | $\frac{1}{2}a\,\hat{\mathbf{x}} + \frac{1}{2\sqrt{3}}a\,\hat{\mathbf{y}} + z_3c\,\hat{\mathbf{z}}$ | (2$d$) | O |
| $\mathbf{B}_5$ | = | $\frac{2}{3}\,\mathbf{a}_1 + \frac{1}{3}\,\mathbf{a}_2 - z_3\,\mathbf{a}_3$ | = | $\frac{1}{2}a\,\hat{\mathbf{x}} - \frac{1}{2\sqrt{3}}a\,\hat{\mathbf{y}} - z_3c\,\hat{\mathbf{z}}$ | (2$d$) | O |

**Geometry files:**

- CIF: pp. 1733
- POSCAR: pp. 1733

# Brucite [$Mg(OH)_2$] Structure: A2BC2_hP5_164_d_a_d

http://aflow.org/prototype-encyclopedia/A2BC2_hP5_164_d_a_d

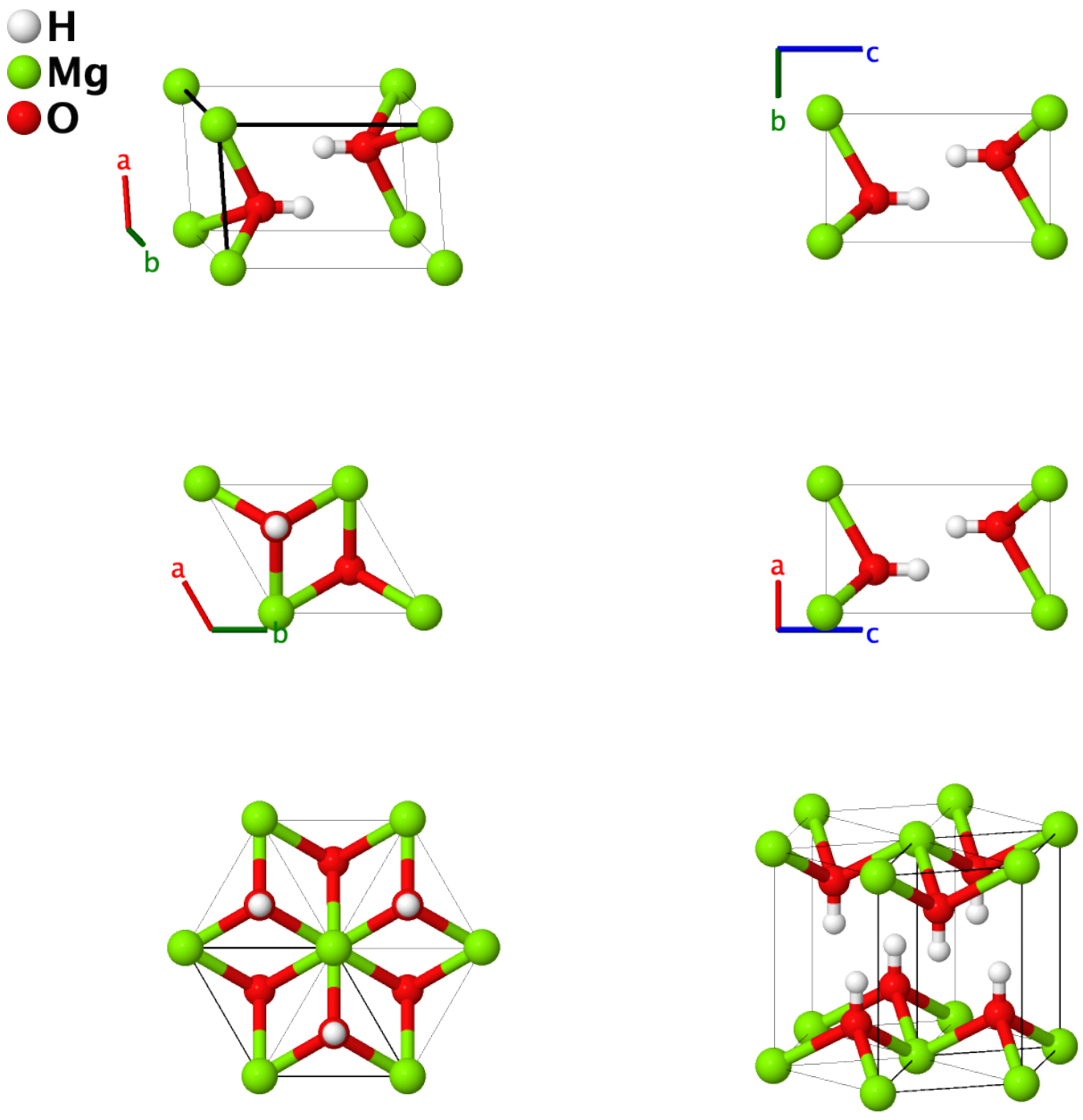

| | | |
|---|---|---|
| **Prototype** | : | $H_2MgO_2$ |
| **AFLOW prototype label** | : | A2BC2_hP5_164_d_a_d |
| ***Strukturbericht* designation** | : | None |
| **Pearson symbol** | : | hP5 |
| **Space group number** | : | 164 |
| **Space group symbol** | : | $P\bar{3}m1$ |
| **AFLOW prototype command** | : | `aflow --proto=A2BC2_hP5_164_d_a_d` `--params=`$a, c/a, z_2, z_3$ |

**Other compounds with this structure**

- $Ca(OH)_2$ (Portlandite), $Fe(OH)_2$, $Mn(OH)_2$ (Pyrochroite), $Ni(OH)_2$ (Theophrastite), and $\beta$-$Co(OH)_2$

- We used the data from (Catti, 1995) at ambient pressure. In some Brucite-like materials the hydrogen atoms are displaced to the (6$i$) Wyckoff positions $(x, -x, z)$ of space group #164, and these sites are 1/3 occupied.

**Trigonal Hexagonal primitive vectors:**

$$\mathbf{a}_1 = \frac{1}{2}\,a\,\hat{\mathbf{x}} - \frac{\sqrt{3}}{2}\,a\,\hat{\mathbf{y}}$$
$$\mathbf{a}_2 = \frac{1}{2}\,a\,\hat{\mathbf{x}} + \frac{\sqrt{3}}{2}\,a\,\hat{\mathbf{y}}$$
$$\mathbf{a}_3 = c\,\hat{\mathbf{z}}$$

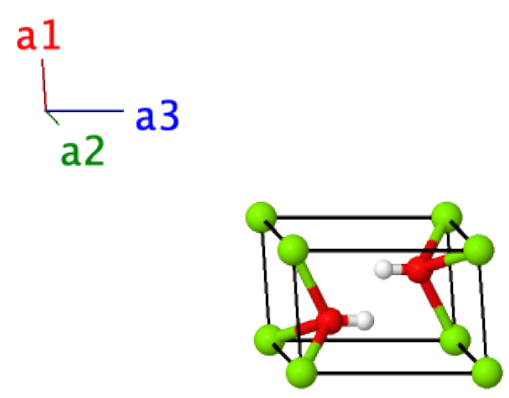

**Basis vectors:**

| | | Lattice Coordinates | | Cartesian Coordinates | Wyckoff Position | Atom Type |
|---|---|---|---|---|---|---|
| $\mathbf{B}_1$ | = | $0\,\mathbf{a}_1 + 0\,\mathbf{a}_2 + 0\,\mathbf{a}_3$ | = | $0\,\hat{\mathbf{x}} + 0\,\hat{\mathbf{y}} + 0\,\hat{\mathbf{z}}$ | $(1a)$ | Mg |
| $\mathbf{B}_2$ | = | $\frac{1}{3}\,\mathbf{a}_1 + \frac{2}{3}\,\mathbf{a}_2 + z_2\,\mathbf{a}_3$ | = | $\frac{1}{2}a\,\hat{\mathbf{x}} + \frac{1}{2\sqrt{3}}a\,\hat{\mathbf{y}} + z_2c\,\hat{\mathbf{z}}$ | $(2d)$ | H |
| $\mathbf{B}_3$ | = | $\frac{2}{3}\,\mathbf{a}_1 + \frac{1}{3}\,\mathbf{a}_2 - z_2\,\mathbf{a}_3$ | = | $\frac{1}{2}a\,\hat{\mathbf{x}} - \frac{1}{2\sqrt{3}}a\,\hat{\mathbf{y}} - z_2c\,\hat{\mathbf{z}}$ | $(2d)$ | H |
| $\mathbf{B}_4$ | = | $\frac{1}{3}\,\mathbf{a}_1 + \frac{2}{3}\,\mathbf{a}_2 + z_3\,\mathbf{a}_3$ | = | $\frac{1}{2}a\,\hat{\mathbf{x}} + \frac{1}{2\sqrt{3}}a\,\hat{\mathbf{y}} + z_3c\,\hat{\mathbf{z}}$ | $(2d)$ | O |
| $\mathbf{B}_5$ | = | $\frac{2}{3}\,\mathbf{a}_1 + \frac{1}{3}\,\mathbf{a}_2 - z_3\,\mathbf{a}_3$ | = | $\frac{1}{2}a\,\hat{\mathbf{x}} - \frac{1}{2\sqrt{3}}a\,\hat{\mathbf{y}} - z_3c\,\hat{\mathbf{z}}$ | $(2d)$ | O |

---

---

**Geometry files:**

- CIF: pp. 1733
- POSCAR: pp. 1734

# $K_2Pt(SCN)_6$ ($H6_3$) Structure: A2BC6_hP9_164_d_a_i

http://aflow.org/prototype-encyclopedia/A2BC6_hP9_164_d_a_i

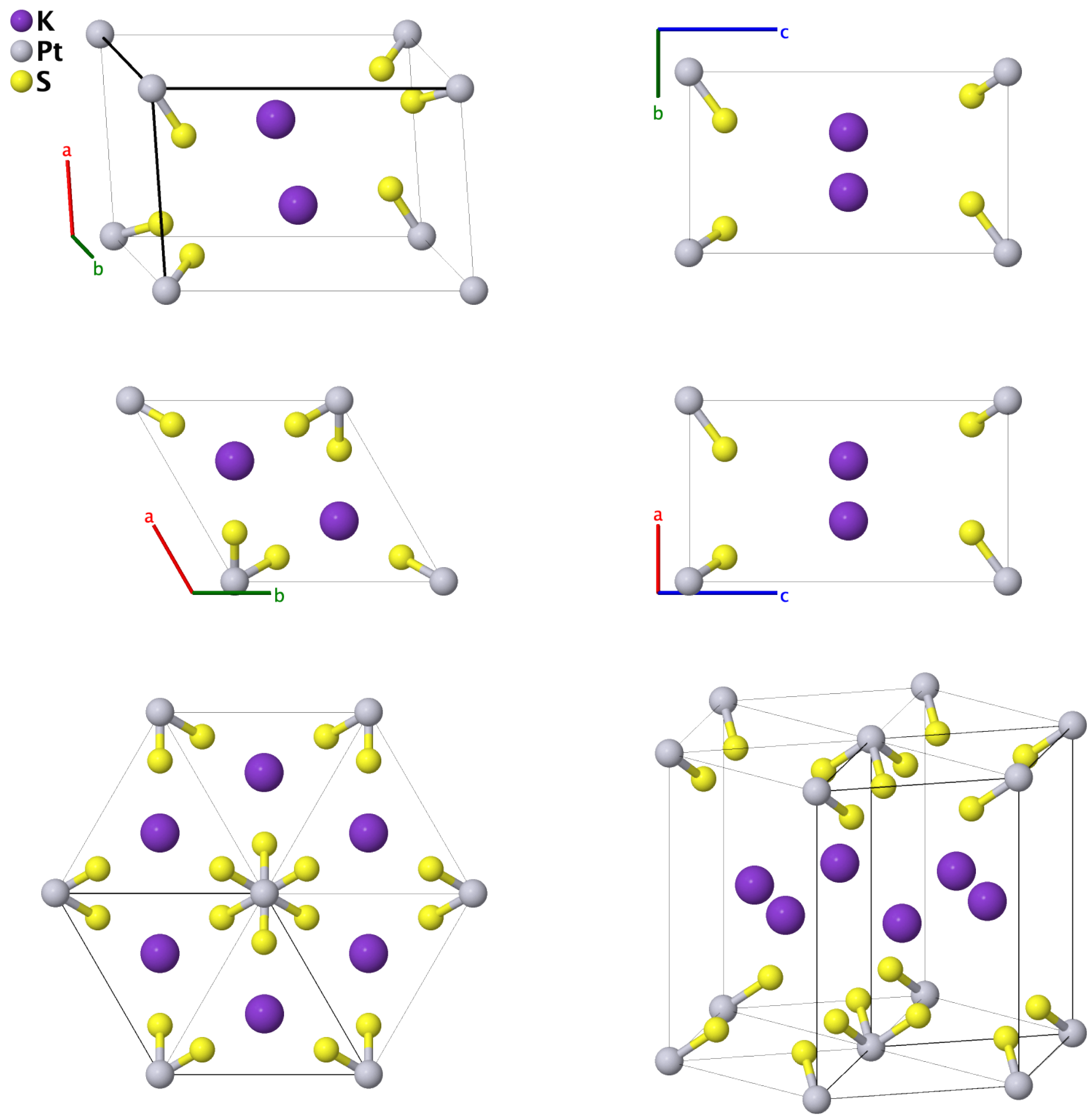

| | | |
|---|---|---|
| **Prototype** | : | $K_2Pt(SCN)_6$ |
| **AFLOW prototype label** | : | A2BC6_hP9_164_d_a_i |
| ***Strukturbericht* designation** | : | $H6_3$ |
| **Pearson symbol** | : | hP9 |
| **Space group number** | : | 164 |
| **Space group symbol** | : | $P\bar{3}m1$ |
| **AFLOW prototype command** | : | `aflow --proto=A2BC6_hP9_164_d_a_i`<br>`--params=`$a, c/a, z_2, x_3, z_3$ |

**Other compounds with this structure**

- $(NH4)_2Pt(SCN)_6$ and $Rb_2Pt(SCN)_6$

- The structure here must be regarded at tentative. Even the "official" labeling of the structure is uncertain:
  - (Hendricks, 1928) made the only structural study of this compound that we have been able to find. Unfortunately, they were not able to determine the positions of the carbon and silicon atoms. Presumably $(SCN)^-$ forms a straight-line ionic system making a roughly 105° angle with the Pt-S line, as it does in the hydrated form of this compound $K_2Pt(SCN)_6 \cdot 2H_2O$.
  - (Hendricks, 1928) were also uncertain about the space group, giving two possibilities: $P\bar{3}1m$ #162 and $P\bar{3}m1$ #164. They prefer the latter, so we present this as a $P\bar{3}m1$ #164 structure.

– The positions of the potassium and sulfur ions are not well determined. (Hendricks, 1928) give $z_2 \approx 0.5$, $x_3 = 0.10 - 0.17$, and $z_3 = 0.09 - 0.135$. We used the average values for the S-III coordinates.
– While (Ewald, 1931) gave this the *Strukturbericht* designation $H6_3$, (Hermann, 1937) lists it as $I1_3$, "previously $H6_3$." However, they then go on to describe structures of the form $SrCl_2(H_2O)_6$, with a very different $c/a$ ratio.
– (Ewald, 1931) originally used the $H6$ *Strukturbericht* category for compounds of the form $B(X)_6$, but (Herman, 1937) and following volumes use both $I$ and $J$ for $B(X)_6$, sometimes changing a previous $H6$ designation to $I1$ or $J1$. In general we will follow this change in notation, but given the striking difference in structures between $K_2Pt(SCN)_6$ and $CaCl_2(H_2O)_6$ we will continue to use $H6_3$ for $K_2Pt(SCN)_6$ and $I1_3$ for $CaCl_2(H_2O)_6$.

**Trigonal Hexagonal primitive vectors:**

$$\mathbf{a}_1 = \frac{1}{2}\,a\,\hat{\mathbf{x}} - \frac{\sqrt{3}}{2}\,a\,\hat{\mathbf{y}}$$
$$\mathbf{a}_2 = \frac{1}{2}\,a\,\hat{\mathbf{x}} + \frac{\sqrt{3}}{2}\,a\,\hat{\mathbf{y}}$$
$$\mathbf{a}_3 = c\,\hat{\mathbf{z}}$$

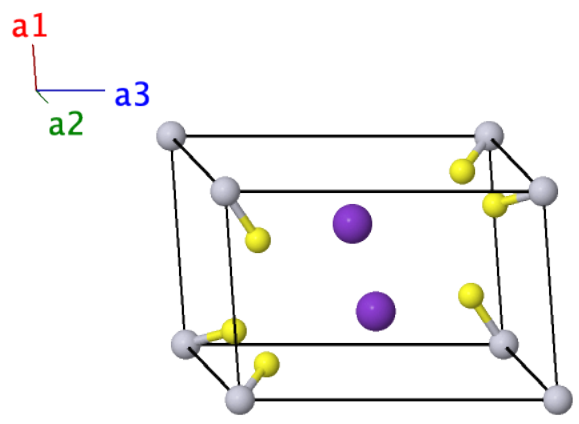

**Basis vectors:**

| | | Lattice Coordinates | | Cartesian Coordinates | Wyckoff Position | Atom Type |
|---|---|---|---|---|---|---|
| $\mathbf{B}_1$ | = | $0\,\mathbf{a}_1 + 0\,\mathbf{a}_2 + 0\,\mathbf{a}_3$ | = | $0\,\hat{\mathbf{x}} + 0\,\hat{\mathbf{y}} + 0\,\hat{\mathbf{z}}$ | (1$a$) | Pt |
| $\mathbf{B}_2$ | = | $\frac{1}{3}\,\mathbf{a}_1 + \frac{2}{3}\,\mathbf{a}_2 + z_2\,\mathbf{a}_3$ | = | $\frac{1}{2}a\,\hat{\mathbf{x}} + \frac{1}{2\sqrt{3}}a\,\hat{\mathbf{y}} + z_2c\,\hat{\mathbf{z}}$ | (2$d$) | K |
| $\mathbf{B}_3$ | = | $\frac{2}{3}\,\mathbf{a}_1 + \frac{1}{3}\,\mathbf{a}_2 - z_2\,\mathbf{a}_3$ | = | $\frac{1}{2}a\,\hat{\mathbf{x}} - \frac{1}{2\sqrt{3}}a\,\hat{\mathbf{y}} - z_2c\,\hat{\mathbf{z}}$ | (2$d$) | K |
| $\mathbf{B}_4$ | = | $x_3\,\mathbf{a}_1 - x_3\,\mathbf{a}_2 + z_3\,\mathbf{a}_3$ | = | $-\sqrt{3}x_3a\,\hat{\mathbf{y}} + z_3c\,\hat{\mathbf{z}}$ | (6$i$) | S |
| $\mathbf{B}_5$ | = | $x_3\,\mathbf{a}_1 + 2x_3\,\mathbf{a}_2 + z_3\,\mathbf{a}_3$ | = | $\frac{3}{2}x_3a\,\hat{\mathbf{x}} + \frac{\sqrt{3}}{2}x_3a\,\hat{\mathbf{y}} + z_3c\,\hat{\mathbf{z}}$ | (6$i$) | S |
| $\mathbf{B}_6$ | = | $-2x_3\,\mathbf{a}_1 - x_3\,\mathbf{a}_2 + z_3\,\mathbf{a}_3$ | = | $-\frac{3}{2}x_3a\,\hat{\mathbf{x}} + \frac{\sqrt{3}}{2}x_3a\,\hat{\mathbf{y}} + z_3c\,\hat{\mathbf{z}}$ | (6$i$) | S |
| $\mathbf{B}_7$ | = | $-x_3\,\mathbf{a}_1 + x_3\,\mathbf{a}_2 - z_3\,\mathbf{a}_3$ | = | $\sqrt{3}x_3a\,\hat{\mathbf{y}} - z_3c\,\hat{\mathbf{z}}$ | (6$i$) | S |
| $\mathbf{B}_8$ | = | $2x_3\,\mathbf{a}_1 + x_3\,\mathbf{a}_2 - z_3\,\mathbf{a}_3$ | = | $\frac{3}{2}x_3a\,\hat{\mathbf{x}} - \frac{\sqrt{3}}{2}x_3a\,\hat{\mathbf{y}} - z_3c\,\hat{\mathbf{z}}$ | (6$i$) | S |
| $\mathbf{B}_9$ | = | $-x_3\,\mathbf{a}_1 - 2x_3\,\mathbf{a}_2 - z_3\,\mathbf{a}_3$ | = | $-\frac{3}{2}x_3a\,\hat{\mathbf{x}} - \frac{\sqrt{3}}{2}x_3a\,\hat{\mathbf{y}} - z_3c\,\hat{\mathbf{z}}$ | (6$i$) | S |

**Found in:**

- P. P. Ewald and C. Hermann, eds., *Strukturbericht 1913-1928* (Akademische Verlagsgesellschaft M. B. H., Leipzig, 1931).

**Geometry files:**

- CIF: pp. 1734
- POSCAR: pp. 1734

# Bararite (Trigonal $(NH_4)_2SiF_6$, $J1_6$) Structure: A6B2C_hP9_164_i_d_a

http://aflow.org/prototype-encyclopedia/A6B2C_hP9_164_i_d_a

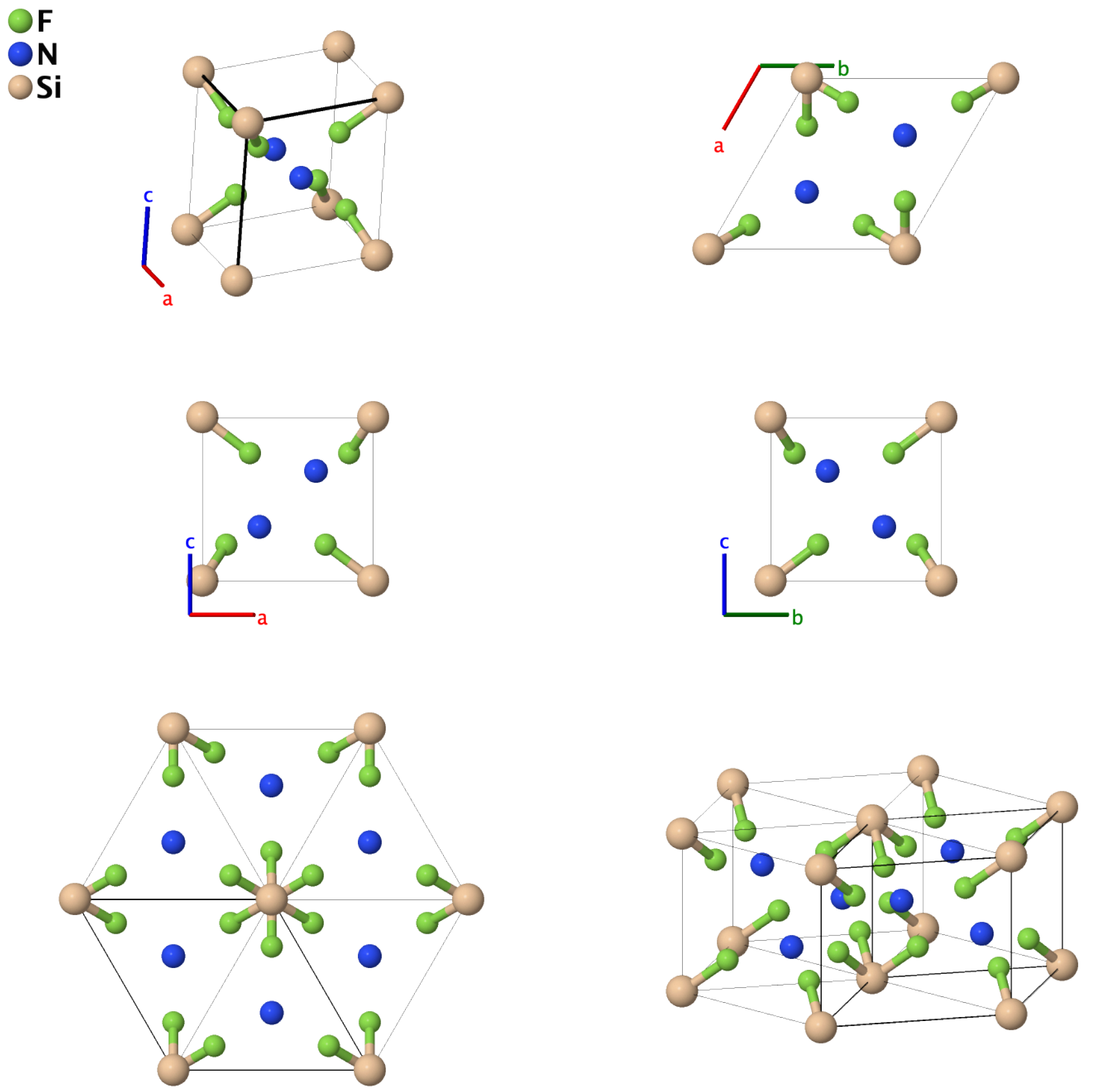

| | | |
|---|---|---|
| **Prototype** | : | $F_6(NH_4)_2Si$ |
| **AFLOW prototype label** | : | A6B2C_hP9_164_i_d_a |
| ***Strukturbericht* designation** | : | $J1_6$ |
| **Pearson symbol** | : | hP9 |
| **Space group number** | : | 164 |
| **Space group symbol** | : | $P\bar{3}m1$ |
| **AFLOW prototype command** | : | `aflow --proto=A6B2C_hP9_164_i_d_a` `--params=`$a, c/a, z_2, x_3, z_3$ |

- Bararite is a trigonal form of $(NH_4)_2SiF_6$, metastable at room temperature. The room temperature stable form is cubic cryptohalite, which takes on the $J1_1$ structure. Except for the hydrogen atoms, this structure is very similar to $J1_{13}$, $K_2GeF_6$. (Schlemper, 1966) state that the hydrogen atoms are on (2d) and (6i) sites, but were not able to determine the coordinates because of large thermal fluctuations. They were to study the system at 77 K, but we have not found any evidence that this work was ever published.
- The positions of the hydrogen atoms in the $NH_4$ ions were not determined, so we only provide the positions of the nitrogen atoms (labeled as $NH_4$).

## Trigonal Hexagonal primitive vectors:

$$
\begin{aligned}
\mathbf{a}_1 &= \tfrac{1}{2}\,a\,\mathbf{\hat{x}} - \tfrac{\sqrt{3}}{2}\,a\,\mathbf{\hat{y}} \\
\mathbf{a}_2 &= \tfrac{1}{2}\,a\,\mathbf{\hat{x}} + \tfrac{\sqrt{3}}{2}\,a\,\mathbf{\hat{y}} \\
\mathbf{a}_3 &= c\,\mathbf{\hat{z}}
\end{aligned}
$$

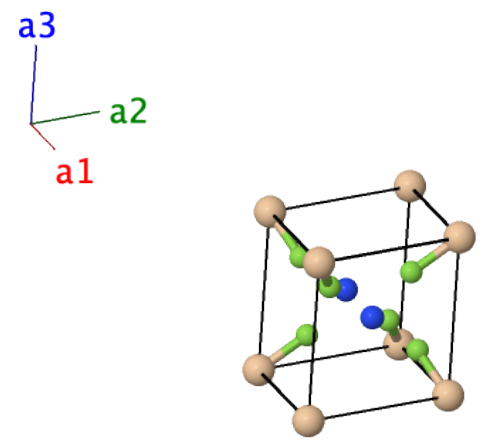

## Basis vectors:

| | | Lattice Coordinates | | Cartesian Coordinates | Wyckoff Position | Atom Type |
|---|---|---|---|---|---|---|
| $\mathbf{B}_1$ | = | $0\,\mathbf{a}_1 + 0\,\mathbf{a}_2 + 0\,\mathbf{a}_3$ | = | $0\,\mathbf{\hat{x}} + 0\,\mathbf{\hat{y}} + 0\,\mathbf{\hat{z}}$ | (1$a$) | Si |
| $\mathbf{B}_2$ | = | $\frac{1}{3}\,\mathbf{a}_1 + \frac{2}{3}\,\mathbf{a}_2 + z_2\,\mathbf{a}_3$ | = | $\frac{1}{2}a\,\mathbf{\hat{x}} + \frac{1}{2\sqrt{3}}a\,\mathbf{\hat{y}} + z_2c\,\mathbf{\hat{z}}$ | (2$d$) | $NH_4$ |
| $\mathbf{B}_3$ | = | $\frac{2}{3}\,\mathbf{a}_1 + \frac{1}{3}\,\mathbf{a}_2 - z_2\,\mathbf{a}_3$ | = | $\frac{1}{2}a\,\mathbf{\hat{x}} - \frac{1}{2\sqrt{3}}a\,\mathbf{\hat{y}} - z_2c\,\mathbf{\hat{z}}$ | (2$d$) | $NH_4$ |
| $\mathbf{B}_4$ | = | $x_3\,\mathbf{a}_1 - x_3\,\mathbf{a}_2 + z_3\,\mathbf{a}_3$ | = | $-\sqrt{3}x_3a\,\mathbf{\hat{y}} + z_3c\,\mathbf{\hat{z}}$ | (6$i$) | F |
| $\mathbf{B}_5$ | = | $x_3\,\mathbf{a}_1 + 2x_3\,\mathbf{a}_2 + z_3\,\mathbf{a}_3$ | = | $\frac{3}{2}x_3a\,\mathbf{\hat{x}} + \frac{\sqrt{3}}{2}x_3a\,\mathbf{\hat{y}} + z_3c\,\mathbf{\hat{z}}$ | (6$i$) | F |
| $\mathbf{B}_6$ | = | $-2x_3\,\mathbf{a}_1 - x_3\,\mathbf{a}_2 + z_3\,\mathbf{a}_3$ | = | $-\frac{3}{2}x_3a\,\mathbf{\hat{x}} + \frac{\sqrt{3}}{2}x_3a\,\mathbf{\hat{y}} + z_3c\,\mathbf{\hat{z}}$ | (6$i$) | F |
| $\mathbf{B}_7$ | = | $-x_3\,\mathbf{a}_1 + x_3\,\mathbf{a}_2 - z_3\,\mathbf{a}_3$ | = | $\sqrt{3}x_3a\,\mathbf{\hat{y}} - z_3c\,\mathbf{\hat{z}}$ | (6$i$) | F |
| $\mathbf{B}_8$ | = | $2x_3\,\mathbf{a}_1 + x_3\,\mathbf{a}_2 - z_3\,\mathbf{a}_3$ | = | $\frac{3}{2}x_3a\,\mathbf{\hat{x}} - \frac{\sqrt{3}}{2}x_3a\,\mathbf{\hat{y}} - z_3c\,\mathbf{\hat{z}}$ | (6$i$) | F |
| $\mathbf{B}_9$ | = | $-x_3\,\mathbf{a}_1 - 2x_3\,\mathbf{a}_2 - z_3\,\mathbf{a}_3$ | = | $-\frac{3}{2}x_3a\,\mathbf{\hat{x}} - \frac{\sqrt{3}}{2}x_3a\,\mathbf{\hat{y}} - z_3c\,\mathbf{\hat{z}}$ | (6$i$) | F |

## Found in:

## Geometry files:

- CIF: pp. 1734
- POSCAR: pp. 1735

# $K_2GeF_6$ ($J1_{13}$) Structure: A6BC2_hP9_164_i_a_d

http://aflow.org/prototype-encyclopedia/A6BC2_hP9_164_i_a_d

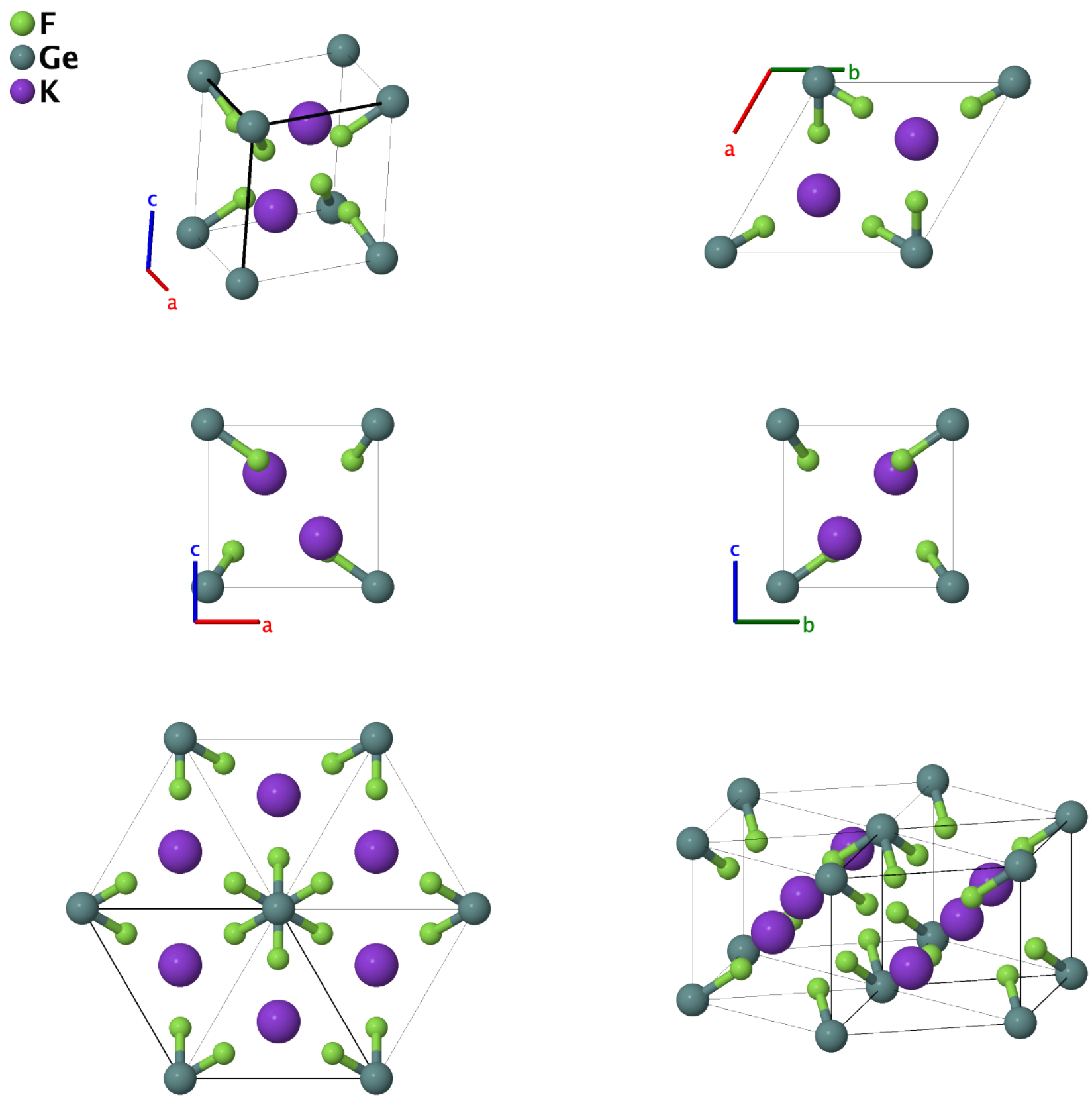

| | | |
|---|---|---|
| **Prototype** | : | $F_6GeK_2$ |
| **AFLOW prototype label** | : | A6BC2_hP9_164_i_a_d |
| ***Strukturbericht* designation** | : | $J1_{13}$ |
| **Pearson symbol** | : | hP9 |
| **Space group number** | : | 164 |
| **Space group symbol** | : | $P\bar{3}m1$ |
| **AFLOW prototype command** | : | `aflow --proto=A6BC2_hP9_164_i_a_d --params=`$a, c/a, z_2, x_3, z_3$ |

**Other compounds with this structure**

- $(NH_4)_2GeF_6$, $Cs_2CeCl_6$, $K_2PtF_6$, and $K_2SiF_6$

**Trigonal Hexagonal primitive vectors:**

$$\mathbf{a}_1 = \frac{1}{2}\,a\,\hat{\mathbf{x}} - \frac{\sqrt{3}}{2}\,a\,\hat{\mathbf{y}}$$
$$\mathbf{a}_2 = \frac{1}{2}\,a\,\hat{\mathbf{x}} + \frac{\sqrt{3}}{2}\,a\,\hat{\mathbf{y}}$$
$$\mathbf{a}_3 = c\,\hat{\mathbf{z}}$$

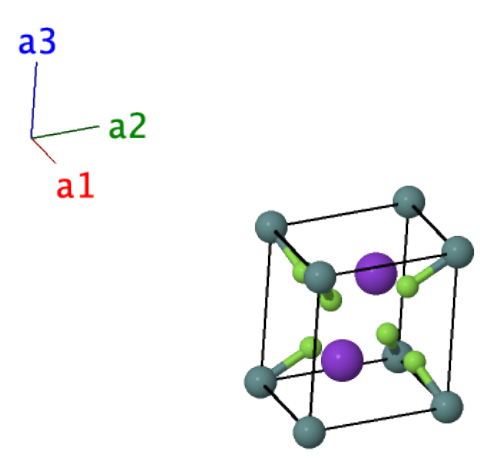

**Basis vectors:**

| | | Lattice Coordinates | | Cartesian Coordinates | Wyckoff Position | Atom Type |
|---|---|---|---|---|---|---|
| $\mathbf{B}_1$ | = | $0\,\mathbf{a}_1 + 0\,\mathbf{a}_2 + 0\,\mathbf{a}_3$ | = | $0\,\hat{\mathbf{x}} + 0\,\hat{\mathbf{y}} + 0\,\hat{\mathbf{z}}$ | $(1a)$ | Ge |
| $\mathbf{B}_2$ | = | $\frac{1}{3}\,\mathbf{a}_1 + \frac{2}{3}\,\mathbf{a}_2 + z_2\,\mathbf{a}_3$ | = | $\frac{1}{2}a\,\hat{\mathbf{x}} + \frac{1}{2\sqrt{3}}a\,\hat{\mathbf{y}} + z_2c\,\hat{\mathbf{z}}$ | $(2d)$ | K |
| $\mathbf{B}_3$ | = | $\frac{2}{3}\,\mathbf{a}_1 + \frac{1}{3}\,\mathbf{a}_2 - z_2\,\mathbf{a}_3$ | = | $\frac{1}{2}a\,\hat{\mathbf{x}} - \frac{1}{2\sqrt{3}}a\,\hat{\mathbf{y}} - z_2c\,\hat{\mathbf{z}}$ | $(2d)$ | K |
| $\mathbf{B}_4$ | = | $x_3\,\mathbf{a}_1 - x_3\,\mathbf{a}_2 + z_3\,\mathbf{a}_3$ | = | $-\sqrt{3}x_3a\,\hat{\mathbf{y}} + z_3c\,\hat{\mathbf{z}}$ | $(6i)$ | F |
| $\mathbf{B}_5$ | = | $x_3\,\mathbf{a}_1 + 2x_3\,\mathbf{a}_2 + z_3\,\mathbf{a}_3$ | = | $\frac{3}{2}x_3a\,\hat{\mathbf{x}} + \frac{\sqrt{3}}{2}x_3a\,\hat{\mathbf{y}} + z_3c\,\hat{\mathbf{z}}$ | $(6i)$ | F |
| $\mathbf{B}_6$ | = | $-2x_3\,\mathbf{a}_1 - x_3\,\mathbf{a}_2 + z_3\,\mathbf{a}_3$ | = | $-\frac{3}{2}x_3a\,\hat{\mathbf{x}} + \frac{\sqrt{3}}{2}x_3a\,\hat{\mathbf{y}} + z_3c\,\hat{\mathbf{z}}$ | $(6i)$ | F |
| $\mathbf{B}_7$ | = | $-x_3\,\mathbf{a}_1 + x_3\,\mathbf{a}_2 - z_3\,\mathbf{a}_3$ | = | $\sqrt{3}x_3a\,\hat{\mathbf{y}} - z_3c\,\hat{\mathbf{z}}$ | $(6i)$ | F |
| $\mathbf{B}_8$ | = | $2x_3\,\mathbf{a}_1 + x_3\,\mathbf{a}_2 - z_3\,\mathbf{a}_3$ | = | $\frac{3}{2}x_3a\,\hat{\mathbf{x}} - \frac{\sqrt{3}}{2}x_3a\,\hat{\mathbf{y}} - z_3c\,\hat{\mathbf{z}}$ | $(6i)$ | F |
| $\mathbf{B}_9$ | = | $-x_3\,\mathbf{a}_1 - 2x_3\,\mathbf{a}_2 - z_3\,\mathbf{a}_3$ | = | $-\frac{3}{2}x_3a\,\hat{\mathbf{x}} - \frac{\sqrt{3}}{2}x_3a\,\hat{\mathbf{y}} - z_3c\,\hat{\mathbf{z}}$ | $(6i)$ | F |

**Geometry files:**

- CIF: pp. 1735
- POSCAR: pp. 1735

# Jacutingaite ($Pt_2HgSe_3$) Structure: AB2C3_hP12_164_d_ae_i

http://aflow.org/prototype-encyclopedia/AB2C3_hP12_164_d_ae_i

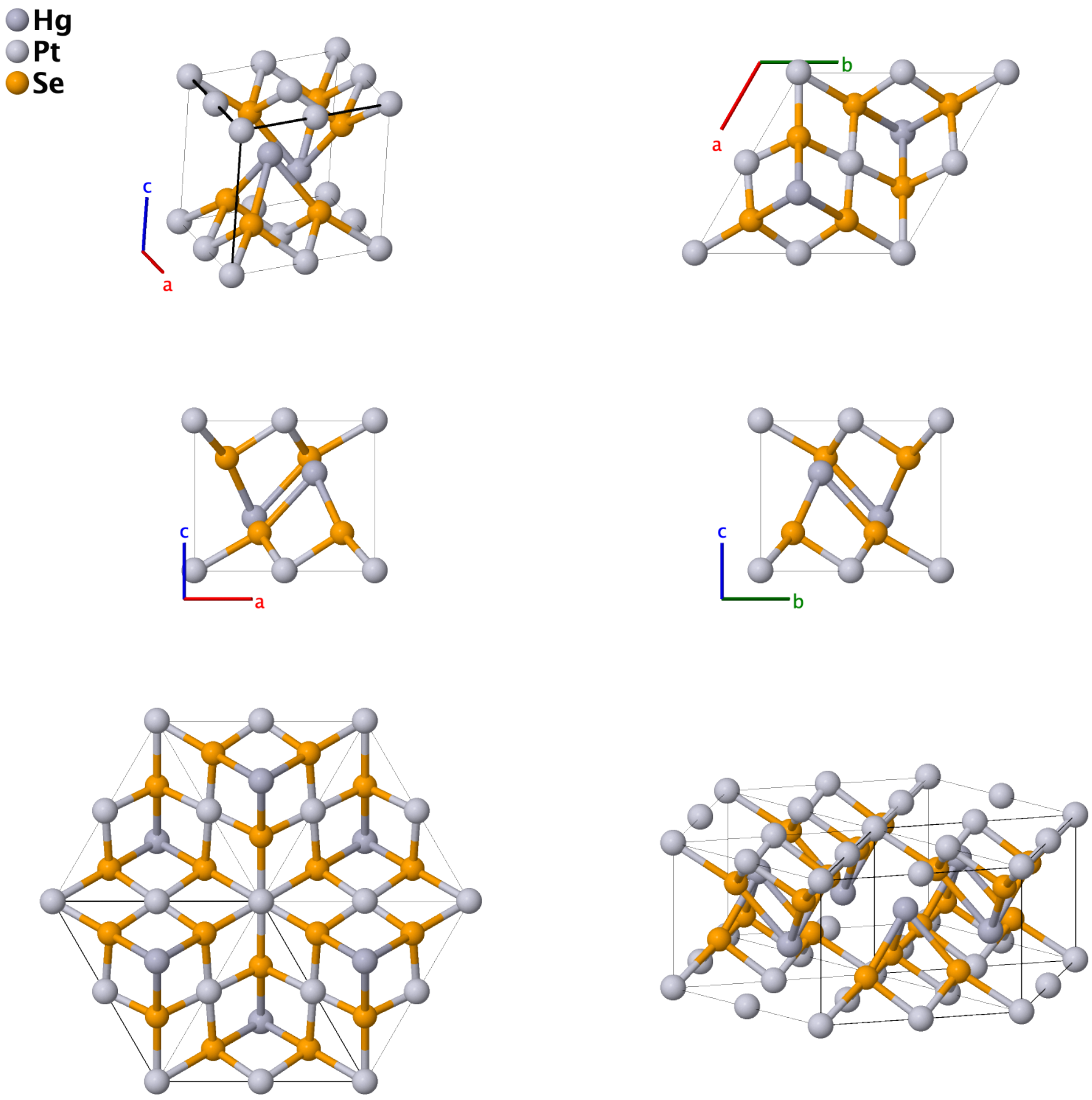

**Prototype** : $HgPt_2Se_3$

**AFLOW prototype label** : AB2C3_hP12_164_d_ae_i

***Strukturbericht* designation** : None

**Pearson symbol** : hP12

**Space group number** : 164

**Space group symbol** : $P\bar{3}m1$

**AFLOW prototype command** : `aflow --proto=AB2C3_hP12_164_d_ae_i`
`--params=`$a, c/a, z_2, x_4, z_4$

- Table 4 of (Vymazalová, 2012) has a small error: the positions of the Hg(1) atoms should be written $(1/3, 2/3, 0.3507)$ rather than $(1/3, 1/3, 0.3507)$.

**Trigonal Hexagonal primitive vectors:**

$$\mathbf{a}_1 = \frac{1}{2}\,a\,\hat{\mathbf{x}} - \frac{\sqrt{3}}{2}\,a\,\hat{\mathbf{y}}$$
$$\mathbf{a}_2 = \frac{1}{2}\,a\,\hat{\mathbf{x}} + \frac{\sqrt{3}}{2}\,a\,\hat{\mathbf{y}}$$
$$\mathbf{a}_3 = c\,\hat{\mathbf{z}}$$

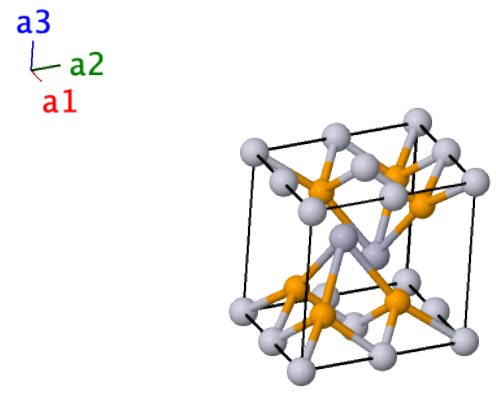

**Basis vectors:**

| | | Lattice Coordinates | | Cartesian Coordinates | Wyckoff Position | Atom Type |
|---|---|---|---|---|---|---|
| $\mathbf{B}_1$ | = | $0\,\mathbf{a}_1 + 0\,\mathbf{a}_2 + 0\,\mathbf{a}_3$ | = | $0\,\hat{\mathbf{x}} + 0\,\hat{\mathbf{y}} + 0\,\hat{\mathbf{z}}$ | $(1a)$ | Pt I |
| $\mathbf{B}_2$ | = | $\frac{1}{3}\,\mathbf{a}_1 + \frac{2}{3}\,\mathbf{a}_2 + z_2\,\mathbf{a}_3$ | = | $\frac{1}{2}a\,\hat{\mathbf{x}} + \frac{1}{2\sqrt{3}}a\,\hat{\mathbf{y}} + z_2c\,\hat{\mathbf{z}}$ | $(2d)$ | Hg |
| $\mathbf{B}_3$ | = | $\frac{2}{3}\,\mathbf{a}_1 + \frac{1}{3}\,\mathbf{a}_2 - z_2\,\mathbf{a}_3$ | = | $\frac{1}{2}a\,\hat{\mathbf{x}} - \frac{1}{2\sqrt{3}}a\,\hat{\mathbf{y}} - z_2c\,\hat{\mathbf{z}}$ | $(2d)$ | Hg |
| $\mathbf{B}_4$ | = | $\frac{1}{2}\,\mathbf{a}_1$ | = | $\frac{1}{4}a\,\hat{\mathbf{x}} - \frac{\sqrt{3}}{4}a\,\hat{\mathbf{y}}$ | $(3e)$ | Pt II |
| $\mathbf{B}_5$ | = | $\frac{1}{2}\,\mathbf{a}_2$ | = | $\frac{1}{4}a\,\hat{\mathbf{x}} + \frac{\sqrt{3}}{4}a\,\hat{\mathbf{y}}$ | $(3e)$ | Pt II |
| $\mathbf{B}_6$ | = | $\frac{1}{2}\,\mathbf{a}_1 + \frac{1}{2}\,\mathbf{a}_2$ | = | $\frac{1}{2}a\,\hat{\mathbf{x}}$ | $(3e)$ | Pt II |
| $\mathbf{B}_7$ | = | $x_4\,\mathbf{a}_1 - x_4\,\mathbf{a}_2 + z_4\,\mathbf{a}_3$ | = | $-\sqrt{3}x_4a\,\hat{\mathbf{y}} + z_4c\,\hat{\mathbf{z}}$ | $(6i)$ | Se |
| $\mathbf{B}_8$ | = | $x_4\,\mathbf{a}_1 + 2x_4\,\mathbf{a}_2 + z_4\,\mathbf{a}_3$ | = | $\frac{3}{2}x_4a\,\hat{\mathbf{x}} + \frac{\sqrt{3}}{2}x_4a\,\hat{\mathbf{y}} + z_4c\,\hat{\mathbf{z}}$ | $(6i)$ | Se |
| $\mathbf{B}_9$ | = | $-2x_4\,\mathbf{a}_1 - x_4\,\mathbf{a}_2 + z_4\,\mathbf{a}_3$ | = | $-\frac{3}{2}x_4a\,\hat{\mathbf{x}} + \frac{\sqrt{3}}{2}x_4a\,\hat{\mathbf{y}} + z_4c\,\hat{\mathbf{z}}$ | $(6i)$ | Se |
| $\mathbf{B}_{10}$ | = | $-x_4\,\mathbf{a}_1 + x_4\,\mathbf{a}_2 - z_4\,\mathbf{a}_3$ | = | $\sqrt{3}x_4a\,\hat{\mathbf{y}} - z_4c\,\hat{\mathbf{z}}$ | $(6i)$ | Se |
| $\mathbf{B}_{11}$ | = | $2x_4\,\mathbf{a}_1 + x_4\,\mathbf{a}_2 - z_4\,\mathbf{a}_3$ | = | $\frac{3}{2}x_4a\,\hat{\mathbf{x}} - \frac{\sqrt{3}}{2}x_4a\,\hat{\mathbf{y}} - z_4c\,\hat{\mathbf{z}}$ | $(6i)$ | Se |
| $\mathbf{B}_{12}$ | = | $-x_4\,\mathbf{a}_1 - 2x_4\,\mathbf{a}_2 - z_4\,\mathbf{a}_3$ | = | $-\frac{3}{2}x_4a\,\hat{\mathbf{x}} - \frac{\sqrt{3}}{2}x_4a\,\hat{\mathbf{y}} - z_4c\,\hat{\mathbf{z}}$ | $(6i)$ | Se |

**Found in:**

**Geometry files:**

- CIF: pp. 1735
- POSCAR: pp. 1736

# $D0_{13}$ ($AlCl_3$) (*obsolete*) Structure: AB3_hP4_164_b_ad

http://aflow.org/prototype-encyclopedia/AB3_hP4_164_b_ad

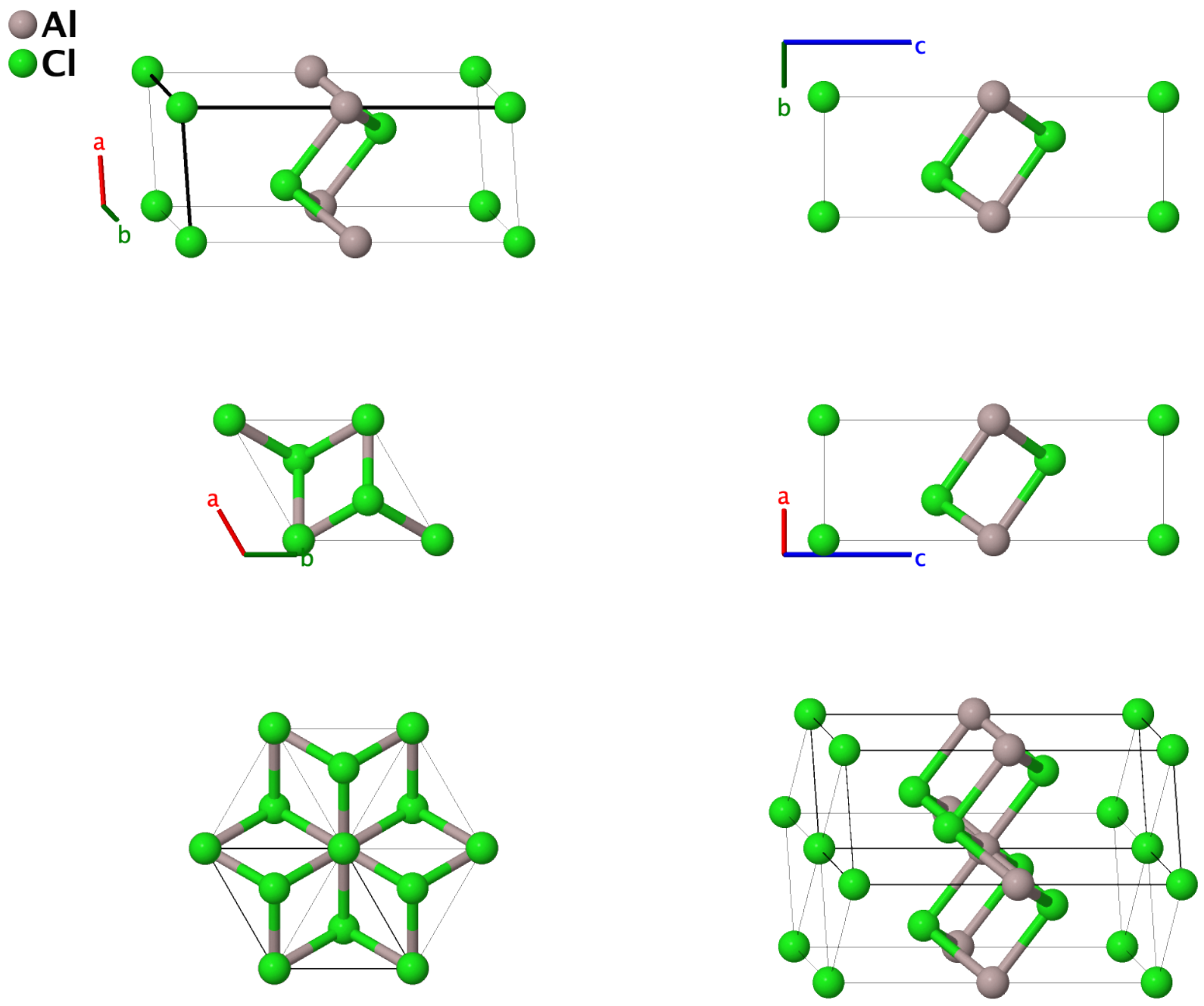

| | | |
|---|---|---|
| **Prototype** | : | $AlCl_3$ |
| **AFLOW prototype label** | : | AB3_hP4_164_b_ad |
| ***Strukturbericht* designation** | : | $D0_{13}$ |
| **Pearson symbol** | : | hP4 |
| **Space group number** | : | 164 |
| **Space group symbol** | : | $P\bar{3}m1$ |
| **AFLOW prototype command** | : | `aflow --proto=AB3_hP4_164_b_ad` `--params=`$a, c/a, z_3$ |

- This structure was suggested by (Laschkarew, 1930) and designated as *Strukturbericht* $D0_{13}$ by (Hermann, 1937). However, in the next edition of *Strukturbericht*, (Gottfried, 1937) designated the $AlCl_3$ structure found by (Ketellar, 1935) as $D0_{15}$. This structure, however, has a lower symmetry than the currently accepted structure, which we previously designated $D0_{15}$ and is body-centered orthorhombic, space group $C2/m$ #12.

**Trigonal Hexagonal primitive vectors:**

$$\mathbf{a}_1 = \frac{1}{2}\,a\,\hat{\mathbf{x}} - \frac{\sqrt{3}}{2}\,a\,\hat{\mathbf{y}}$$
$$\mathbf{a}_2 = \frac{1}{2}\,a\,\hat{\mathbf{x}} + \frac{\sqrt{3}}{2}\,a\,\hat{\mathbf{y}}$$
$$\mathbf{a}_3 = c\,\hat{\mathbf{z}}$$

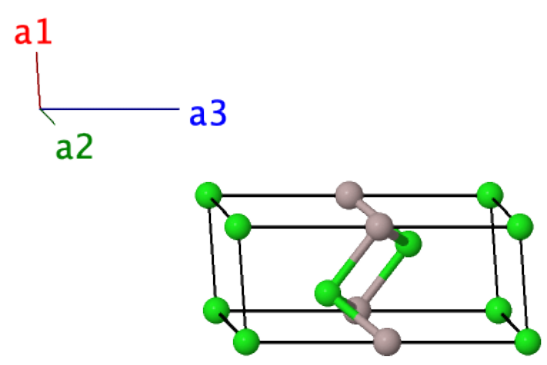

**Basis vectors:**

| | | Lattice Coordinates | | Cartesian Coordinates | Wyckoff Position | Atom Type |
|---|---|---|---|---|---|---|
| $\mathbf{B}_1$ | = | $0\,\mathbf{a}_1 + 0\,\mathbf{a}_2 + 0\,\mathbf{a}_3$ | = | $0\,\hat{\mathbf{x}} + 0\,\hat{\mathbf{y}} + 0\,\hat{\mathbf{z}}$ | $(1a)$ | Cl I |
| $\mathbf{B}_2$ | = | $\frac{1}{2}\,\mathbf{a}_3$ | = | $\frac{1}{2}c\,\hat{\mathbf{z}}$ | $(1b)$ | Al |
| $\mathbf{B}_3$ | = | $\frac{1}{3}\,\mathbf{a}_1 + \frac{2}{3}\,\mathbf{a}_2 + z_3\,\mathbf{a}_3$ | = | $\frac{1}{2}a\,\hat{\mathbf{x}} + \frac{1}{2\sqrt{3}}a\,\hat{\mathbf{y}} + z_3c\,\hat{\mathbf{z}}$ | $(2d)$ | Cl II |
| $\mathbf{B}_4$ | = | $\frac{2}{3}\,\mathbf{a}_1 + \frac{1}{3}\,\mathbf{a}_2 - z_3\,\mathbf{a}_3$ | = | $\frac{1}{2}a\,\hat{\mathbf{x}} - \frac{1}{2\sqrt{3}}a\,\hat{\mathbf{y}} - z_3c\,\hat{\mathbf{z}}$ | $(2d)$ | Cl II |

**Geometry files:**

- CIF: pp. 1736
- POSCAR: pp. 1736

# Nevskite (BiSe) Structure: AB_hP12_164_c2d_c2d

http://aflow.org/prototype-encyclopedia/AB_hP12_164_c2d_c2d

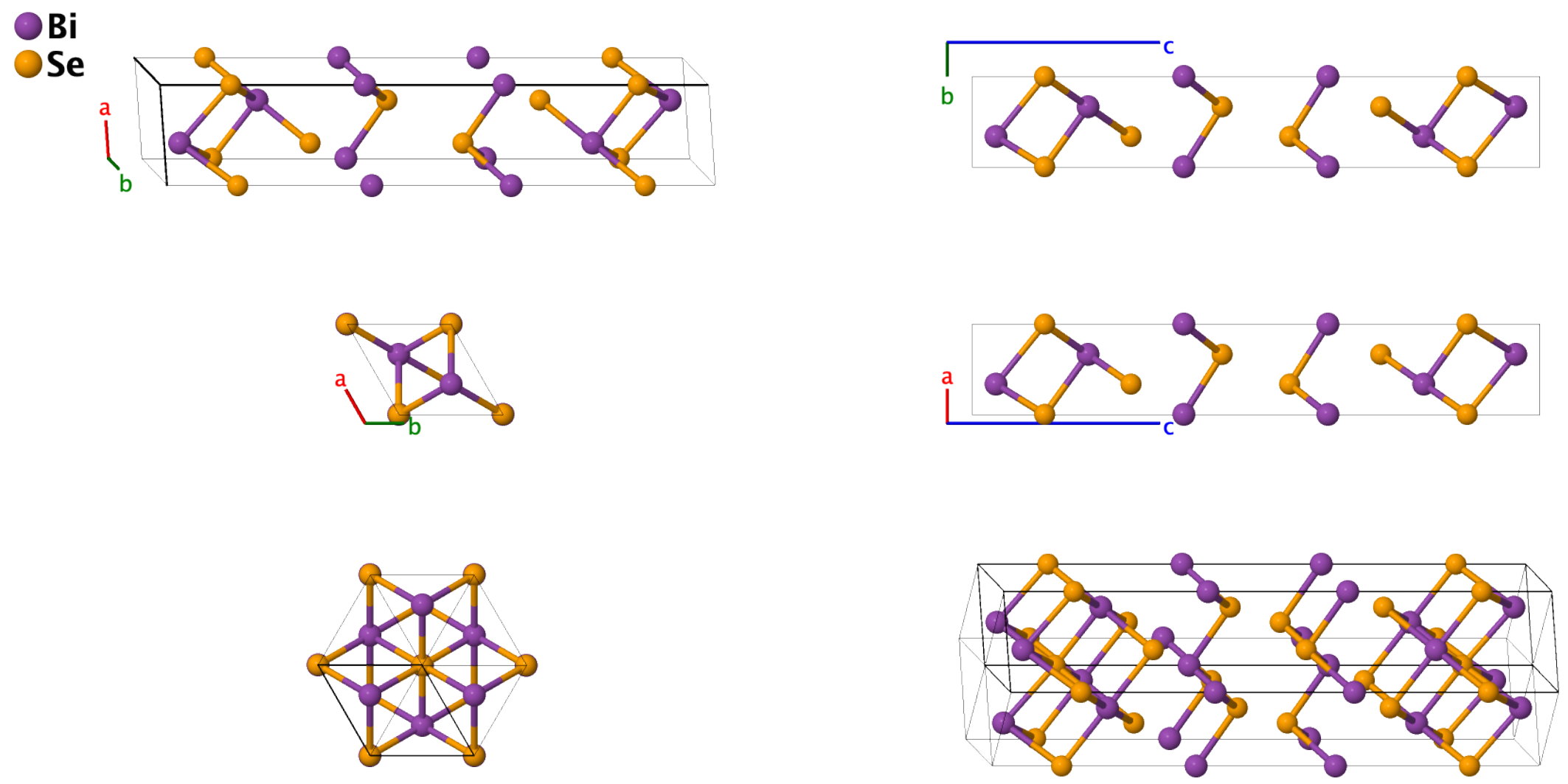

| | | |
|---|---|---|
| **Prototype** | : | BiSe |
| **AFLOW prototype label** | : | AB_hP12_164_c2d_c2d |
| ***Strukturbericht* designation** | : | None |
| **Pearson symbol** | : | hP12 |
| **Space group number** | : | 164 |
| **Space group symbol** | : | $P\bar{3}m1$ |
| **AFLOW prototype command** | : | `aflow --proto=AB_hP12_164_c2d_c2d` `--params=`$a, c/a, z_1, z_2, z_3, z_4, z_5, z_6$ |

### Other compounds with this structure

- BiTe (Tsumoite) and $Bi(S_{0.56}Te_{0.44})$, (Ingodite)

### Trigonal Hexagonal primitive vectors:

$$\mathbf{a}_1 = \frac{1}{2}\,a\,\hat{\mathbf{x}} - \frac{\sqrt{3}}{2}\,a\,\hat{\mathbf{y}}$$
$$\mathbf{a}_2 = \frac{1}{2}\,a\,\hat{\mathbf{x}} + \frac{\sqrt{3}}{2}\,a\,\hat{\mathbf{y}}$$
$$\mathbf{a}_3 = c\,\hat{\mathbf{z}}$$

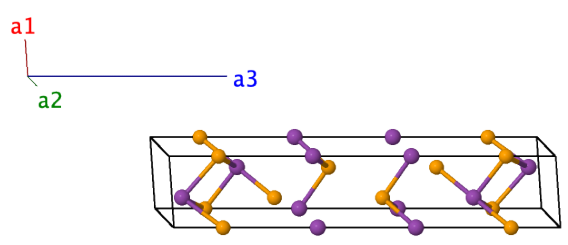

### Basis vectors:

| | | Lattice Coordinates | | Cartesian Coordinates | Wyckoff Position | Atom Type |
|---|---|---|---|---|---|---|
| $\mathbf{B}_1$ | = | $z_1\,\mathbf{a}_3$ | = | $z_1 c\,\hat{\mathbf{z}}$ | (2*c*) | Bi I |
| $\mathbf{B}_2$ | = | $-z_1\,\mathbf{a}_3$ | = | $-z_1 c\,\hat{\mathbf{z}}$ | (2*c*) | Bi I |
| $\mathbf{B}_3$ | = | $z_2\,\mathbf{a}_3$ | = | $z_2 c\,\hat{\mathbf{z}}$ | (2*c*) | Se I |
| $\mathbf{B}_4$ | = | $-z_2\,\mathbf{a}_3$ | = | $-z_2 c\,\hat{\mathbf{z}}$ | (2*c*) | Se I |
| $\mathbf{B}_5$ | = | $\frac{1}{3}\,\mathbf{a}_1 + \frac{2}{3}\,\mathbf{a}_2 + z_3\,\mathbf{a}_3$ | = | $\frac{1}{2}a\,\hat{\mathbf{x}} + \frac{1}{2\sqrt{3}}a\,\hat{\mathbf{y}} + z_3 c\,\hat{\mathbf{z}}$ | (2*d*) | Bi II |
| $\mathbf{B}_6$ | = | $\frac{2}{3}\,\mathbf{a}_1 + \frac{1}{3}\,\mathbf{a}_2 - z_3\,\mathbf{a}_3$ | = | $\frac{1}{2}a\,\hat{\mathbf{x}} - \frac{1}{2\sqrt{3}}a\,\hat{\mathbf{y}} - z_3 c\,\hat{\mathbf{z}}$ | (2*d*) | Bi II |
| $\mathbf{B}_7$ | = | $\frac{1}{3}\,\mathbf{a}_1 + \frac{2}{3}\,\mathbf{a}_2 + z_4\,\mathbf{a}_3$ | = | $\frac{1}{2}a\,\mathbf{\hat{x}} + \frac{1}{2\sqrt{3}}a\,\mathbf{\hat{y}} + z_4c\,\mathbf{\hat{z}}$ | $(2d)$ | Bi III |
| $\mathbf{B}_8$ | = | $\frac{2}{3}\,\mathbf{a}_1 + \frac{1}{3}\,\mathbf{a}_2 - z_4\,\mathbf{a}_3$ | = | $\frac{1}{2}a\,\mathbf{\hat{x}} - \frac{1}{2\sqrt{3}}a\,\mathbf{\hat{y}} - z_4c\,\mathbf{\hat{z}}$ | $(2d)$ | Bi III |
| $\mathbf{B}_9$ | = | $\frac{1}{3}\,\mathbf{a}_1 + \frac{2}{3}\,\mathbf{a}_2 + z_5\,\mathbf{a}_3$ | = | $\frac{1}{2}a\,\mathbf{\hat{x}} + \frac{1}{2\sqrt{3}}a\,\mathbf{\hat{y}} + z_5c\,\mathbf{\hat{z}}$ | $(2d)$ | Se II |
| $\mathbf{B}_{10}$ | = | $\frac{2}{3}\,\mathbf{a}_1 + \frac{1}{3}\,\mathbf{a}_2 - z_5\,\mathbf{a}_3$ | = | $\frac{1}{2}a\,\mathbf{\hat{x}} - \frac{1}{2\sqrt{3}}a\,\mathbf{\hat{y}} - z_5c\,\mathbf{\hat{z}}$ | $(2d)$ | Se II |
| $\mathbf{B}_{11}$ | = | $\frac{1}{3}\,\mathbf{a}_1 + \frac{2}{3}\,\mathbf{a}_2 + z_6\,\mathbf{a}_3$ | = | $\frac{1}{2}a\,\mathbf{\hat{x}} + \frac{1}{2\sqrt{3}}a\,\mathbf{\hat{y}} + z_6c\,\mathbf{\hat{z}}$ | $(2d)$ | Se III |
| $\mathbf{B}_{12}$ | = | $\frac{2}{3}\,\mathbf{a}_1 + \frac{1}{3}\,\mathbf{a}_2 - z_6\,\mathbf{a}_3$ | = | $\frac{1}{2}a\,\mathbf{\hat{x}} - \frac{1}{2\sqrt{3}}a\,\mathbf{\hat{y}} - z_6c\,\mathbf{\hat{z}}$ | $(2d)$ | Se III |

**Geometry files:**

- CIF: pp. 1736
- POSCAR: pp. 1737

# $B_{13}C_2$ "$B_4C$" ($D1_g$) Structure: A13B2_hR15_166_b2h_c

http://aflow.org/prototype-encyclopedia/A13B2_hR15_166_b2h_c

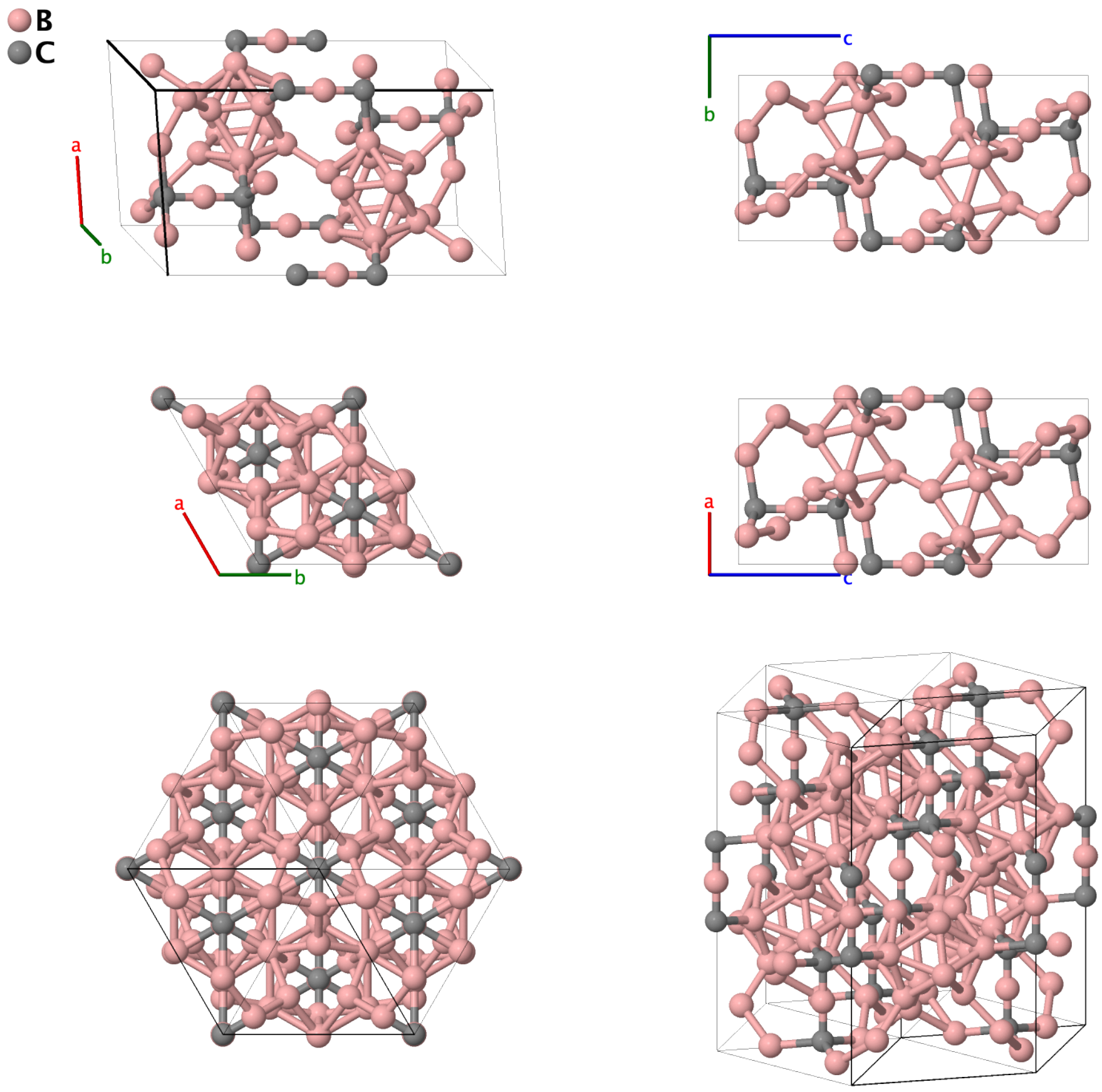

| | | |
|---|---|---|
| **Prototype** | : | $B_{13}C_2$ |
| **AFLOW prototype label** | : | A13B2_hR15_166_b2h_c |
| ***Strukturbericht* designation** | : | $D1_g$ |
| **Pearson symbol** | : | hR15 |
| **Space group number** | : | 166 |
| **Space group symbol** | : | $R\bar{3}m$ |
| **AFLOW prototype command** | : | `aflow --proto=A13B2_hR15_166_b2h_c [--hex]` `--params=`$a, c/a, x_2, x_3, z_3, x_4, z_4$ |

### Other compounds with this structure

- $B_{1-x}C_x$ ($0.08 < x < 0.20$) and $B_{13}P_2$, $B_4Si$, $B_6O$

- This structure has a rather complicated history:
  - It is difficult to determine the species of atoms at a given site because of the similar electronic and nuclear cross sections of $^{11}B$ and $^{12}C$ (Domnich, 2011).
  - Early investigations (Clark, 1943) assumed the structure was $B_4C$, with the extra carbon atom replacing the boron on the (1*b*) site. [Note that Clark has an error in the coordinates of one set of boron atoms, giving a boron-boron distance of less than 1 Å. This error is repeated in (Brandes, 1992) and (Wykcoff, 1964).]

- In reality, concentrations can range from 8-20% carbon (Domnich, 2011).
- (Larson, 1986) states that the (1$b$) site in $B_{13}C_2$ is boron, and as the structure becomes more carbon-rich the carbon atoms replace boron in the icosahedra. We follow this and use the structure determined by (Will, 1976) as our reference.
- (Lazzari, 1999) states that excess electrons go on the "polar" sites of the icosahedron, *i.e.* the sites closest to the carbon atoms on the chains (the B-III atoms in our notation).

---

**Rhombohedral primitive vectors:**

$$
\begin{aligned}
\mathbf{a}_1 &= \tfrac{1}{2}\,a\,\hat{\mathbf{x}} - \tfrac{1}{2\sqrt{3}}\,a\,\hat{\mathbf{y}} + \tfrac{1}{3}\,c\,\hat{\mathbf{z}} \\
\mathbf{a}_2 &= \tfrac{1}{\sqrt{3}}\,a\,\hat{\mathbf{y}} + \tfrac{1}{3}\,c\,\hat{\mathbf{z}} \\
\mathbf{a}_3 &= -\tfrac{1}{2}\,a\,\hat{\mathbf{x}} - \tfrac{1}{2\sqrt{3}}\,a\,\hat{\mathbf{y}} + \tfrac{1}{3}\,c\,\hat{\mathbf{z}}
\end{aligned}
$$

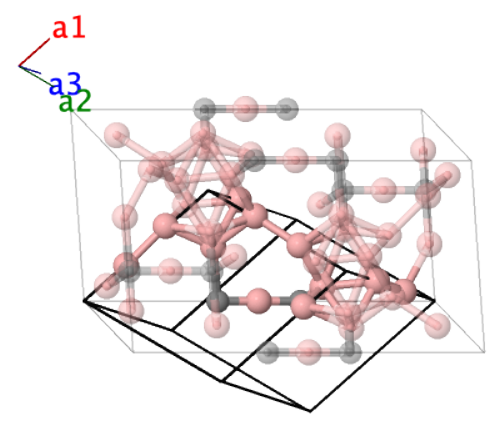

---

**Basis vectors:**

| | Lattice Coordinates | | Cartesian Coordinates | Wyckoff Position | Atom Type |
|---|---|---|---|---|---|
| $\mathbf{B}_1$ = | $\frac{1}{2}\,\mathbf{a}_1 + \frac{1}{2}\,\mathbf{a}_2 + \frac{1}{2}\,\mathbf{a}_3$ | = | $\frac{1}{2}c\,\hat{\mathbf{z}}$ | (1$b$) | B I |
| $\mathbf{B}_2$ = | $x_2\,\mathbf{a}_1 + x_2\,\mathbf{a}_2 + x_2\,\mathbf{a}_3$ | = | $x_2 c\,\hat{\mathbf{z}}$ | (2$c$) | C |
| $\mathbf{B}_3$ = | $-x_2\,\mathbf{a}_1 - x_2\,\mathbf{a}_2 - x_2\,\mathbf{a}_3$ | = | $-x_2 c\,\hat{\mathbf{z}}$ | (2$c$) | C |
| $\mathbf{B}_4$ = | $x_3\,\mathbf{a}_1 + x_3\,\mathbf{a}_2 + z_3\,\mathbf{a}_3$ | = | $\frac{1}{2}(x_3 - z_3)\,a\,\hat{\mathbf{x}} + \frac{1}{2\sqrt{3}}(x_3 - z_3)\,a\,\hat{\mathbf{y}} + \left(\frac{2}{3}x_3 + \frac{1}{3}z_3\right)c\,\hat{\mathbf{z}}$ | (6$h$) | B II |
| $\mathbf{B}_5$ = | $z_3\,\mathbf{a}_1 + x_3\,\mathbf{a}_2 + x_3\,\mathbf{a}_3$ | = | $\frac{1}{2}(-x_3 + z_3)\,a\,\hat{\mathbf{x}} + \frac{1}{2\sqrt{3}}(x_3 - z_3)\,a\,\hat{\mathbf{y}} + \left(\frac{2}{3}x_3 + \frac{1}{3}z_3\right)c\,\hat{\mathbf{z}}$ | (6$h$) | B II |
| $\mathbf{B}_6$ = | $x_3\,\mathbf{a}_1 + z_3\,\mathbf{a}_2 + x_3\,\mathbf{a}_3$ | = | $\frac{1}{\sqrt{3}}(-x_3 + z_3)\,a\,\hat{\mathbf{y}} + \left(\frac{2}{3}x_3 + \frac{1}{3}z_3\right)c\,\hat{\mathbf{z}}$ | (6$h$) | B II |
| $\mathbf{B}_7$ = | $-z_3\,\mathbf{a}_1 - x_3\,\mathbf{a}_2 - x_3\,\mathbf{a}_3$ | = | $\frac{1}{2}(x_3 - z_3)\,a\,\hat{\mathbf{x}} + \frac{1}{2\sqrt{3}}(-x_3 + z_3)\,a\,\hat{\mathbf{y}} - c\left(\frac{2}{3}x_3 + \frac{1}{3}z_3\right)\hat{\mathbf{z}}$ | (6$h$) | B II |
| $\mathbf{B}_8$ = | $-x_3\,\mathbf{a}_1 - x_3\,\mathbf{a}_2 - z_3\,\mathbf{a}_3$ | = | $\frac{1}{2}(-x_3 + z_3)\,a\,\hat{\mathbf{x}} + \frac{1}{2\sqrt{3}}(-x_3 + z_3)\,a\,\hat{\mathbf{y}} - c\left(\frac{2}{3}x_3 + \frac{1}{3}z_3\right)\hat{\mathbf{z}}$ | (6$h$) | B II |
| $\mathbf{B}_9$ = | $-x_3\,\mathbf{a}_1 - z_3\,\mathbf{a}_2 - x_3\,\mathbf{a}_3$ | = | $\frac{1}{\sqrt{3}}(x_3 - z_3)\,a\,\hat{\mathbf{y}} - c\left(\frac{2}{3}x_3 + \frac{1}{3}z_3\right)\hat{\mathbf{z}}$ | (6$h$) | B II |
| $\mathbf{B}_{10}$ = | $x_4\,\mathbf{a}_1 + x_4\,\mathbf{a}_2 + z_4\,\mathbf{a}_3$ | = | $\frac{1}{2}(x_4 - z_4)\,a\,\hat{\mathbf{x}} + \frac{1}{2\sqrt{3}}(x_4 - z_4)\,a\,\hat{\mathbf{y}} + \left(\frac{2}{3}x_4 + \frac{1}{3}z_4\right)c\,\hat{\mathbf{z}}$ | (6$h$) | B III |
| $\mathbf{B}_{11}$ = | $z_4\,\mathbf{a}_1 + x_4\,\mathbf{a}_2 + x_4\,\mathbf{a}_3$ | = | $\frac{1}{2}(-x_4 + z_4)\,a\,\hat{\mathbf{x}} + \frac{1}{2\sqrt{3}}(x_4 - z_4)\,a\,\hat{\mathbf{y}} + \left(\frac{2}{3}x_4 + \frac{1}{3}z_4\right)c\,\hat{\mathbf{z}}$ | (6$h$) | B III |
| $\mathbf{B}_{12}$ = | $x_4\,\mathbf{a}_1 + z_4\,\mathbf{a}_2 + x_4\,\mathbf{a}_3$ | = | $\frac{1}{\sqrt{3}}(-x_4 + z_4)\,a\,\hat{\mathbf{y}} + \left(\frac{2}{3}x_4 + \frac{1}{3}z_4\right)c\,\hat{\mathbf{z}}$ | (6$h$) | B III |
| $\mathbf{B}_{13}$ = | $-z_4\,\mathbf{a}_1 - x_4\,\mathbf{a}_2 - x_4\,\mathbf{a}_3$ | = | $\frac{1}{2}(x_4 - z_4)\,a\,\hat{\mathbf{x}} + \frac{1}{2\sqrt{3}}(-x_4 + z_4)\,a\,\hat{\mathbf{y}} - c\left(\frac{2}{3}x_4 + \frac{1}{3}z_4\right)\hat{\mathbf{z}}$ | (6$h$) | B III |
| $\mathbf{B}_{14}$ = | $-x_4\,\mathbf{a}_1 - x_4\,\mathbf{a}_2 - z_4\,\mathbf{a}_3$ | = | $\frac{1}{2}(-x_4 + z_4)\,a\,\hat{\mathbf{x}} + \frac{1}{2\sqrt{3}}(-x_4 + z_4)\,a\,\hat{\mathbf{y}} - c\left(\frac{2}{3}x_4 + \frac{1}{3}z_4\right)\hat{\mathbf{z}}$ | (6$h$) | B III |
| $\mathbf{B}_{15}$ = | $-x_4\,\mathbf{a}_1 - z_4\,\mathbf{a}_2 - x_4\,\mathbf{a}_3$ | = | $\frac{1}{\sqrt{3}}(x_4 - z_4)\,a\,\hat{\mathbf{y}} - c\left(\frac{2}{3}x_4 + \frac{1}{3}z_4\right)\hat{\mathbf{z}}$ | (6$h$) | B III |

---

---

**Geometry files:**
- CIF: pp. 1737
- POSCAR: pp. 1737

# $MnBi_2Te_4$ Structure: A2BC4_hR7_166_c_a_2c

http://aflow.org/prototype-encyclopedia/A2BC4_hR7_166_c_a_2c

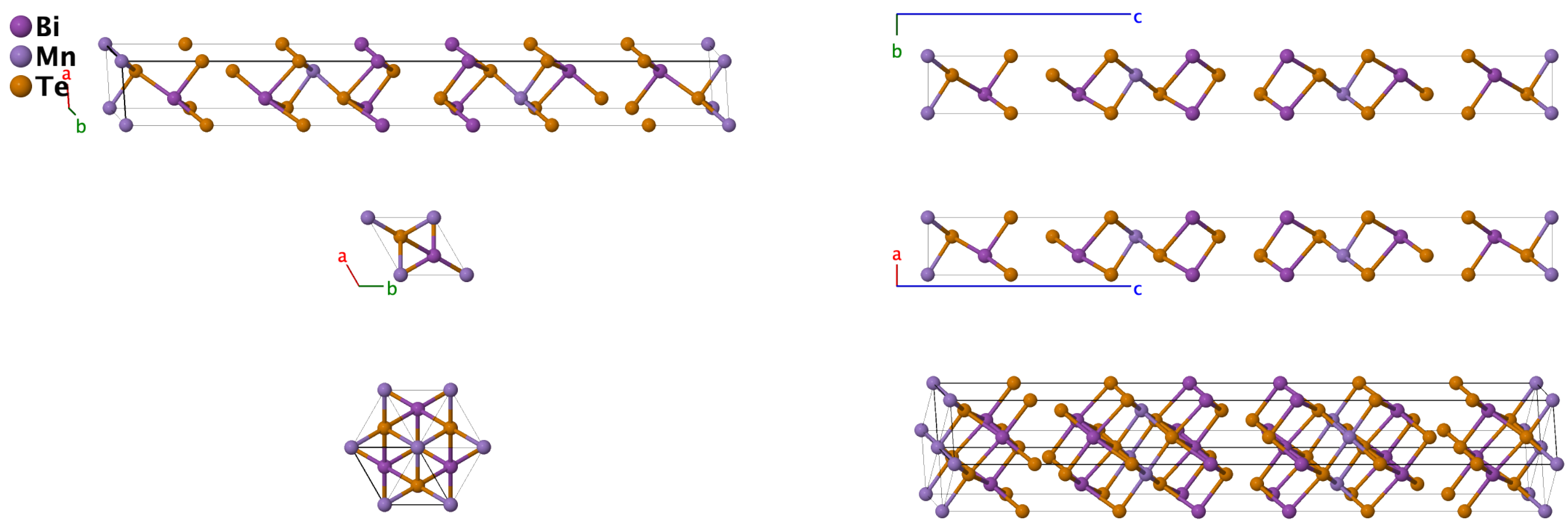

| | | |
|---|---|---|
| **Prototype** | : | $Bi_2MnTe_4$ |
| **AFLOW prototype label** | : | A2BC4_hR7_166_c_a_2c |
| ***Strukturbericht* designation** | : | None |
| **Pearson symbol** | : | hR7 |
| **Space group number** | : | 166 |
| **Space group symbol** | : | $R\bar{3}m$ |
| **AFLOW prototype command** | : | `aflow --proto=A2BC4_hR7_166_c_a_2c [--hex]` `--params=`$a, c/a, x_2, x_3, x_4$ |

- We use the data taken at 10 K. Except for the magnetic ordering there is no substantial change in the structure up to room temperature.

**Rhombohedral primitive vectors:**

$$\mathbf{a}_1 = \frac{1}{2}\,a\,\hat{\mathbf{x}} - \frac{1}{2\sqrt{3}}\,a\,\hat{\mathbf{y}} + \frac{1}{3}\,c\,\hat{\mathbf{z}}$$
$$\mathbf{a}_2 = \frac{1}{\sqrt{3}}\,a\,\hat{\mathbf{y}} + \frac{1}{3}\,c\,\hat{\mathbf{z}}$$
$$\mathbf{a}_3 = -\frac{1}{2}\,a\,\hat{\mathbf{x}} - \frac{1}{2\sqrt{3}}\,a\,\hat{\mathbf{y}} + \frac{1}{3}\,c\,\hat{\mathbf{z}}$$

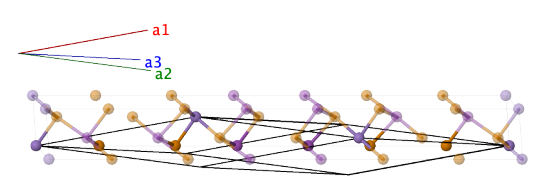

**Basis vectors:**

| | | Lattice Coordinates | | Cartesian Coordinates | Wyckoff Position | Atom Type |
|---|---|---|---|---|---|---|
| $\mathbf{B}_1$ | = | $0\,\mathbf{a}_1 + 0\,\mathbf{a}_2 + 0\,\mathbf{a}_3$ | = | $0\,\hat{\mathbf{x}} + 0\,\hat{\mathbf{y}} + 0\,\hat{\mathbf{z}}$ | (1$a$) | Mn |
| $\mathbf{B}_2$ | = | $x_2\,\mathbf{a}_1 + x_2\,\mathbf{a}_2 + x_2\,\mathbf{a}_3$ | = | $x_2 c\,\hat{\mathbf{z}}$ | (2$c$) | Bi |
| $\mathbf{B}_3$ | = | $-x_2\,\mathbf{a}_1 - x_2\,\mathbf{a}_2 - x_2\,\mathbf{a}_3$ | = | $-x_2 c\,\hat{\mathbf{z}}$ | (2$c$) | Bi |
| $\mathbf{B}_4$ | = | $x_3\,\mathbf{a}_1 + x_3\,\mathbf{a}_2 + x_3\,\mathbf{a}_3$ | = | $x_3 c\,\hat{\mathbf{z}}$ | (2$c$) | Te I |
| $\mathbf{B}_5$ | = | $-x_3\,\mathbf{a}_1 - x_3\,\mathbf{a}_2 - x_3\,\mathbf{a}_3$ | = | $-x_3 c\,\hat{\mathbf{z}}$ | (2$c$) | Te I |
| $\mathbf{B}_6$ | = | $x_4\,\mathbf{a}_1 + x_4\,\mathbf{a}_2 + x_4\,\mathbf{a}_3$ | = | $x_4 c\,\hat{\mathbf{z}}$ | (2$c$) | Te II |
| $\mathbf{B}_7$ | = | $-x_4\,\mathbf{a}_1 - x_4\,\mathbf{a}_2 - x_4\,\mathbf{a}_3$ | = | $-x_4 c\,\hat{\mathbf{z}}$ | (2$c$) | Te II |

**Geometry files:**

- CIF: pp. 1737
- POSCAR: pp. 1738

# Shandite ($Ni_3Pb_2S_2$) Structure: A3B2C2_hR7_166_d_ab_c

http://aflow.org/prototype-encyclopedia/A3B2C2_hR7_166_d_ab_c

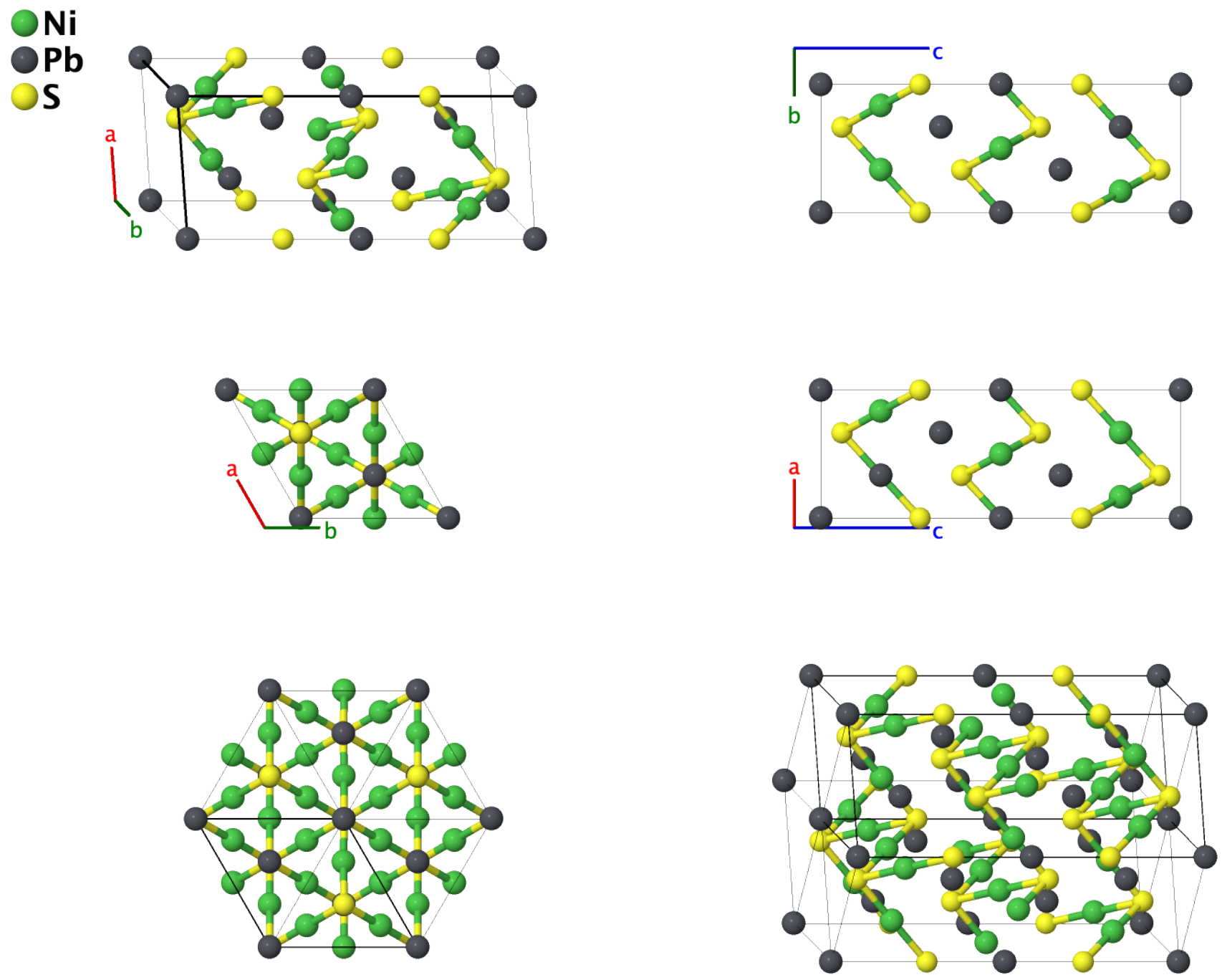

| | | |
|---|---|---|
| **Prototype** | : | $Ni_3Pb_2S_2$ |
| **AFLOW prototype label** | : | A3B2C2_hR7_166_d_ab_c |
| ***Strukturbericht* designation** | : | None |
| **Pearson symbol** | : | hR7 |
| **Space group number** | : | 166 |
| **Space group symbol** | : | $R\bar{3}m$ |
| **AFLOW prototype command** | : | `aflow --proto=A3B2C2_hR7_166_d_ab_c [--hex]` `--params=`$a, c/a, x_3$ |

**Other compounds with this structure**

- $Co_3In_2S_2$, $Co_3Sn_2S_2$, $Ni_3In_2S_2$, $Ni_3Sn_2S_2$, and $O_3K_2Sn_2$

**Rhombohedral primitive vectors:**

$$\mathbf{a}_1 = \frac{1}{2}\,a\,\hat{\mathbf{x}} - \frac{1}{2\sqrt{3}}\,a\,\hat{\mathbf{y}} + \frac{1}{3}\,c\,\hat{\mathbf{z}}$$
$$\mathbf{a}_2 = \frac{1}{\sqrt{3}}\,a\,\hat{\mathbf{y}} + \frac{1}{3}\,c\,\hat{\mathbf{z}}$$
$$\mathbf{a}_3 = -\frac{1}{2}\,a\,\hat{\mathbf{x}} - \frac{1}{2\sqrt{3}}\,a\,\hat{\mathbf{y}} + \frac{1}{3}\,c\,\hat{\mathbf{z}}$$

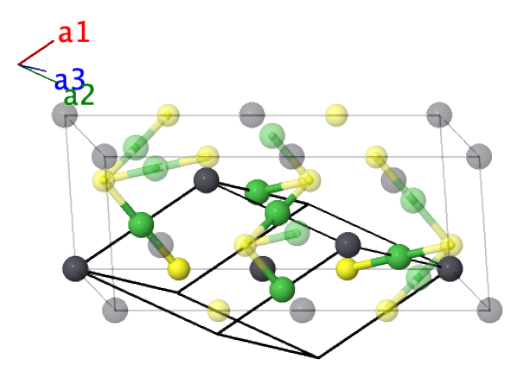

**Basis vectors:**

| | | Lattice Coordinates | | Cartesian Coordinates | Wyckoff Position | Atom Type |
|---|---|---|---|---|---|---|
| $\mathbf{B}_1$ | = | $0\,\mathbf{a}_1 + 0\,\mathbf{a}_2 + 0\,\mathbf{a}_3$ | = | $0\,\hat{\mathbf{x}} + 0\,\hat{\mathbf{y}} + 0\,\hat{\mathbf{z}}$ | $(1a)$ | Pb I |
| $\mathbf{B}_2$ | = | $\frac{1}{2}\,\mathbf{a}_1 + \frac{1}{2}\,\mathbf{a}_2 + \frac{1}{2}\,\mathbf{a}_3$ | = | $\frac{1}{2}c\,\hat{\mathbf{z}}$ | $(1b)$ | Pb II |
| $\mathbf{B}_3$ | = | $x_3\,\mathbf{a}_1 + x_3\,\mathbf{a}_2 + x_3\,\mathbf{a}_3$ | = | $x_3 c\,\hat{\mathbf{z}}$ | $(2c)$ | S |
| $\mathbf{B}_4$ | = | $-x_3\,\mathbf{a}_1 - x_3\,\mathbf{a}_2 - x_3\,\mathbf{a}_3$ | = | $-x_3 c\,\hat{\mathbf{z}}$ | $(2c)$ | S |
| $\mathbf{B}_5$ | = | $\frac{1}{2}\,\mathbf{a}_1$ | = | $\frac{1}{4}a\,\hat{\mathbf{x}} - \frac{1}{4\sqrt{3}}a\,\hat{\mathbf{y}} + \frac{1}{6}c\,\hat{\mathbf{z}}$ | $(3d)$ | Ni |
| $\mathbf{B}_6$ | = | $\frac{1}{2}\,\mathbf{a}_2$ | = | $\frac{1}{2\sqrt{3}}a\,\hat{\mathbf{y}} + \frac{1}{6}c\,\hat{\mathbf{z}}$ | $(3d)$ | Ni |
| $\mathbf{B}_7$ | = | $\frac{1}{2}\,\mathbf{a}_3$ | = | $-\frac{1}{4}a\,\hat{\mathbf{x}} - \frac{1}{4\sqrt{3}}a\,\hat{\mathbf{y}} + \frac{1}{6}c\,\hat{\mathbf{z}}$ | $(3d)$ | Ni |

**Geometry files:**

- CIF: pp. 1738
- POSCAR: pp. 1738

# Chabazite ($Ca_{1.4}Sr_{0.3}Al_{3.8}Si_{8.3}O_{24}\cdot 13H_2O$, $S3_4$ (I)) Structure: A5B21C24D12_hR62_166_a2c_ehi_fg2h_i

http://aflow.org/prototype-encyclopedia/A5B21C24D12_hR62_166_a2c_ehi_fg2h_i

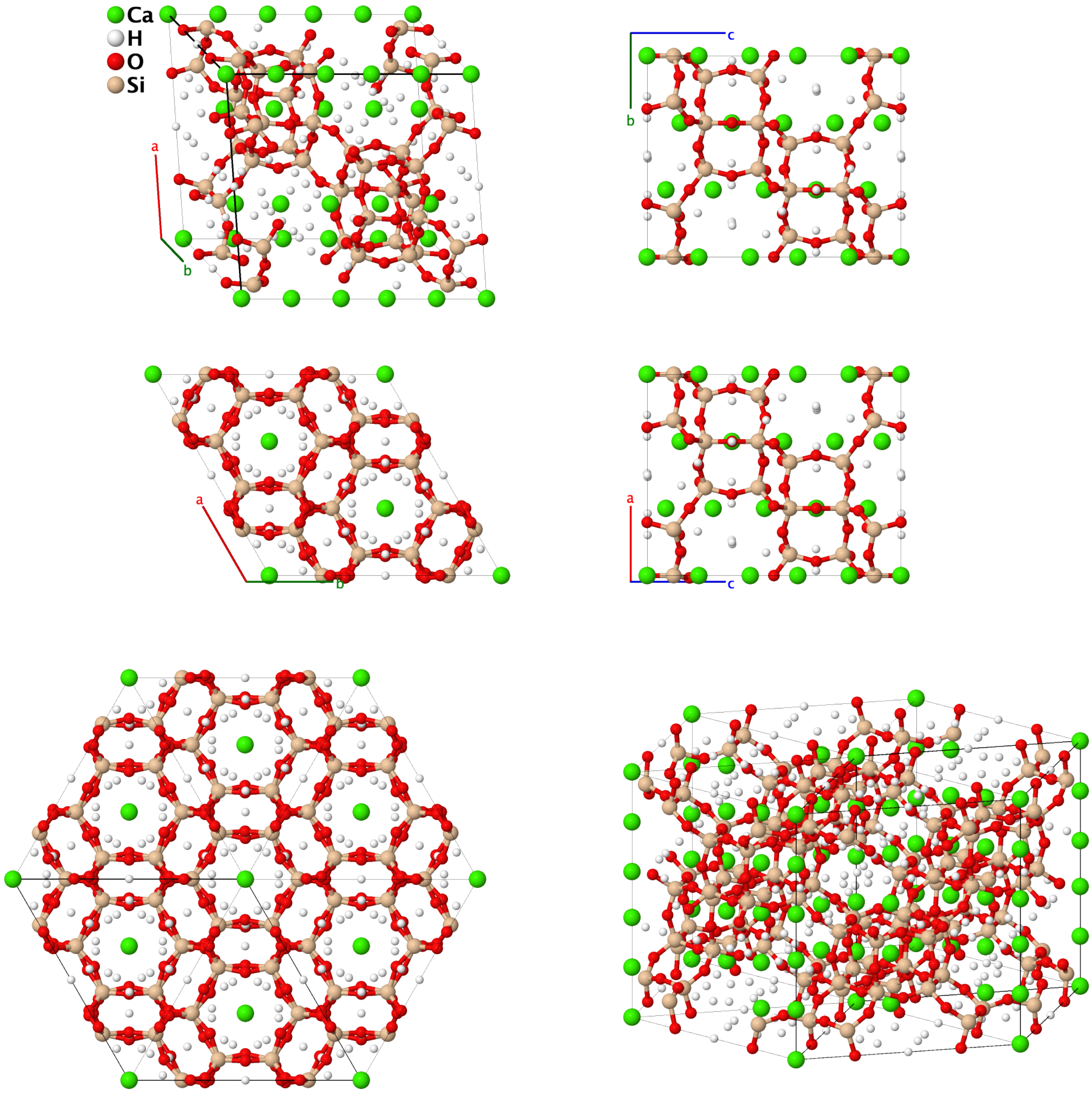

| | | |
|---|---|---|
| **Prototype** | : | $(Ca_{1.4},Sr_{0.3})(H_2O)_{13}O_{24}(Si_{8.3},Al_{3.8})$ |
| **AFLOW prototype label** | : | A5B21C24D12_hR62_166_a2c_ehi_fg2h_i |
| ***Strukturbericht* designation** | : | $S3_4$ (I) |
| **Pearson symbol** | : | hR62 |
| **Space group number** | : | 166 |
| **Space group symbol** | : | $R\bar{3}m$ |
| **AFLOW prototype command** | : | `aflow --proto=A5B21C24D12_hR62_166_a2c_ehi_fg2h_i [--hex]` `--params=`$a, c/a, x_2, x_3, x_5, x_6, x_7, z_7, x_8, z_8, x_9, z_9, x_{10}, y_{10}, z_{10}, x_{11}, y_{11}, z_{11}$ |

- The structure given here, from (Calligaris, 1982), has substantially the same Wyckoff positions as the structure desig-

nated $S3_4$ by (Gottfried, 1937). The largest change is the splitting of the calcium site from one stoichiometric $(2c)$ site to three partially filled sites, one $(1a)$ and two $(2c)$.

- (Calligaris, 1982) did not give the exact positions of the water molecules, so we use the positions given by (Downs, 2003). However, this leaves us with only 7.7 water molecules per formula unit, rather than the 13 claimed by (Calligaris, 1982).
- Only the oxygen sites are fully occupied. For the remaining sites, according to (Downs, 2003),
  - The $(1a)$ site is 9.06% Ca and 1.94% Sr, with the remaining sites vacant.
  - The first $(2c)$ site is 43.65% Ca and 9.35% Sr.
  - The second $(2c)$ site is 19.76% Ca and 4.24% Sr.
  - The $(3e)$ is only occupied by a water molecule 50% of the time.
  - The $(6h)$ $H_2O$ site is only occupied 57% of the time.
  - The $(12i)$ $H_2O$ site is only occupied 23% of the time.
  - The remaining $(12i)$ is 69% Si and 31% Al.
- (Gottfried, 1937) gave chabazite the *Strukturbericht* designation $S3_4$. However, (Gottfried, 1940) ignored this and designated catapleiite, $Na_2ZrSi_3O_9{\cdot}2H_2O$ as *Strukturbericht* $S3_4$. We resolve this by using $S3_4$ (I) to designate chabazite and $S3_4$ (II) to designate catapleiite.

---

**Rhombohedral primitive vectors:**

$$\mathbf{a}_1 = \frac{1}{2}\,a\,\hat{\mathbf{x}} - \frac{1}{2\sqrt{3}}\,a\,\hat{\mathbf{y}} + \frac{1}{3}\,c\,\hat{\mathbf{z}}$$
$$\mathbf{a}_2 = \frac{1}{\sqrt{3}}\,a\,\hat{\mathbf{y}} + \frac{1}{3}\,c\,\hat{\mathbf{z}}$$
$$\mathbf{a}_3 = -\frac{1}{2}\,a\,\hat{\mathbf{x}} - \frac{1}{2\sqrt{3}}\,a\,\hat{\mathbf{y}} + \frac{1}{3}\,c\,\hat{\mathbf{z}}$$

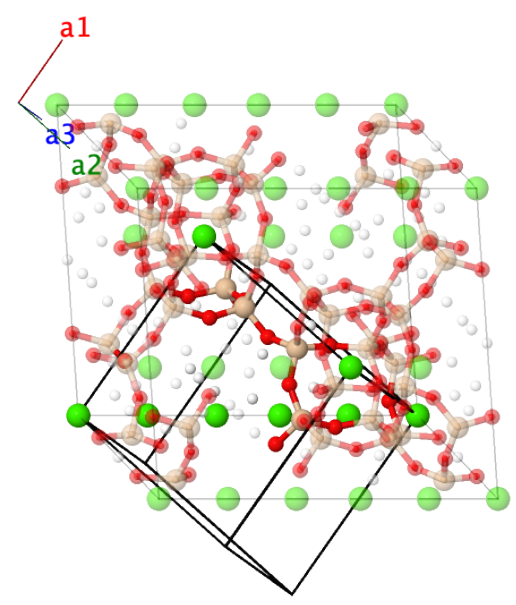

---

**Basis vectors:**

| | | Lattice Coordinates | | Cartesian Coordinates | Wyckoff Position | Atom Type |
|---|---|---|---|---|---|---|
| $\mathbf{B}_1$ | = | $0\,\mathbf{a}_1 + 0\,\mathbf{a}_2 + 0\,\mathbf{a}_3$ | = | $0\,\hat{\mathbf{x}} + 0\,\hat{\mathbf{y}} + 0\,\hat{\mathbf{z}}$ | $(1a)$ | Ca I |
| $\mathbf{B}_2$ | = | $x_2\,\mathbf{a}_1 + x_2\,\mathbf{a}_2 + x_2\,\mathbf{a}_3$ | = | $x_2 c\,\hat{\mathbf{z}}$ | $(2c)$ | Ca II |
| $\mathbf{B}_3$ | = | $-x_2\,\mathbf{a}_1 - x_2\,\mathbf{a}_2 - x_2\,\mathbf{a}_3$ | = | $-x_2 c\,\hat{\mathbf{z}}$ | $(2c)$ | Ca II |
| $\mathbf{B}_4$ | = | $x_3\,\mathbf{a}_1 + x_3\,\mathbf{a}_2 + x_3\,\mathbf{a}_3$ | = | $x_3 c\,\hat{\mathbf{z}}$ | $(2c)$ | Ca III |
| $\mathbf{B}_5$ | = | $-x_3\,\mathbf{a}_1 - x_3\,\mathbf{a}_2 - x_3\,\mathbf{a}_3$ | = | $-x_3 c\,\hat{\mathbf{z}}$ | $(2c)$ | Ca III |
| $\mathbf{B}_6$ | = | $\frac{1}{2}\,\mathbf{a}_2 + \frac{1}{2}\,\mathbf{a}_3$ | = | $-\frac{1}{4}a\,\hat{\mathbf{x}} + \frac{1}{4\sqrt{3}}a\,\hat{\mathbf{y}} + \frac{1}{3}c\,\hat{\mathbf{z}}$ | $(3e)$ | $H_2O$ I |
| $\mathbf{B}_7$ | = | $\frac{1}{2}\,\mathbf{a}_1 + \frac{1}{2}\,\mathbf{a}_3$ | = | $-\frac{1}{2\sqrt{3}}a\,\hat{\mathbf{y}} + \frac{1}{3}c\,\hat{\mathbf{z}}$ | $(3e)$ | $H_2O$ I |
| $\mathbf{B}_8$ | = | $\frac{1}{2}\,\mathbf{a}_1 + \frac{1}{2}\,\mathbf{a}_2$ | = | $\frac{1}{4}a\,\hat{\mathbf{x}} + \frac{1}{4\sqrt{3}}a\,\hat{\mathbf{y}} + \frac{1}{3}c\,\hat{\mathbf{z}}$ | $(3e)$ | $H_2O$ I |
| $\mathbf{B}_9$ | = | $x_5\,\mathbf{a}_1 - x_5\,\mathbf{a}_2$ | = | $\frac{1}{2}x_5 a\,\hat{\mathbf{x}} - \frac{\sqrt{3}}{2}x_5 a\,\hat{\mathbf{y}}$ | $(6f)$ | O I |
| $\mathbf{B}_{10}$ | = | $x_5\,\mathbf{a}_2 - x_5\,\mathbf{a}_3$ | = | $\frac{1}{2}x_5 a\,\hat{\mathbf{x}} + \frac{\sqrt{3}}{2}x_5 a\,\hat{\mathbf{y}}$ | $(6f)$ | O I |
| $\mathbf{B}_{11}$ | = | $-x_5\,\mathbf{a}_1 + x_5\,\mathbf{a}_3$ | = | $-x_5 a\,\hat{\mathbf{x}}$ | $(6f)$ | O I |
| $\mathbf{B}_{12}$ | = | $-x_5\,\mathbf{a}_1 + x_5\,\mathbf{a}_2$ | = | $-\frac{1}{2}x_5 a\,\hat{\mathbf{x}} + \frac{\sqrt{3}}{2}x_5 a\,\hat{\mathbf{y}}$ | $(6f)$ | O I |
| $\mathbf{B}_{13}$ | = | $-x_5\,\mathbf{a}_2 + x_5\,\mathbf{a}_3$ | = | $-\frac{1}{2}x_5 a\,\hat{\mathbf{x}} - \frac{\sqrt{3}}{2}x_5 a\,\hat{\mathbf{y}}$ | $(6f)$ | O I |
| $\mathbf{B}_{14}$ | = | $x_5\,\mathbf{a}_1 - x_5\,\mathbf{a}_3$ | = | $x_5 a\,\hat{\mathbf{x}}$ | $(6f)$ | O I |
| $\mathbf{B}_{15}$ | = | $x_6\,\mathbf{a}_1 - x_6\,\mathbf{a}_2 + \frac{1}{2}\,\mathbf{a}_3$ | = | $\left(-\frac{1}{4} + \frac{1}{2}x_6\right)a\,\hat{\mathbf{x}} - a\left(\frac{\sqrt{3}}{2}x_6 + \frac{1}{4\sqrt{3}}\right)\hat{\mathbf{y}} + \frac{1}{6}c\,\hat{\mathbf{z}}$ | (6$g$) | O II |
| $\mathbf{B}_{16}$ | = | $\frac{1}{2}\,\mathbf{a}_1 + x_6\,\mathbf{a}_2 - x_6\,\mathbf{a}_3$ | = | $\left(\frac{1}{4} + \frac{1}{2}x_6\right)a\,\hat{\mathbf{x}} + \left(-\frac{1}{4\sqrt{3}} + \frac{\sqrt{3}}{2}x_6\right)a\,\hat{\mathbf{y}} + \frac{1}{6}c\,\hat{\mathbf{z}}$ | (6$g$) | O II |
| $\mathbf{B}_{17}$ | = | $-x_6\,\mathbf{a}_1 + \frac{1}{2}\,\mathbf{a}_2 + x_6\,\mathbf{a}_3$ | = | $-x_6 a\,\hat{\mathbf{x}} + \frac{1}{2\sqrt{3}}a\,\hat{\mathbf{y}} + \frac{1}{6}c\,\hat{\mathbf{z}}$ | (6$g$) | O II |
| $\mathbf{B}_{18}$ | = | $-x_6\,\mathbf{a}_1 + x_6\,\mathbf{a}_2 + \frac{1}{2}\,\mathbf{a}_3$ | = | $-a\left(\frac{1}{2}x_6 + \frac{1}{4}\right)\hat{\mathbf{x}} + \left(-\frac{1}{4\sqrt{3}} + \frac{\sqrt{3}}{2}x_6\right)a\,\hat{\mathbf{y}} + \frac{1}{6}c\,\hat{\mathbf{z}}$ | (6$g$) | O II |
| $\mathbf{B}_{19}$ | = | $\frac{1}{2}\,\mathbf{a}_1 - x_6\,\mathbf{a}_2 + x_6\,\mathbf{a}_3$ | = | $\left(\frac{1}{4} - \frac{1}{2}x_6\right)a\,\hat{\mathbf{x}} - a\left(\frac{\sqrt{3}}{2}x_6 + \frac{1}{4\sqrt{3}}\right)\hat{\mathbf{y}} + \frac{1}{6}c\,\hat{\mathbf{z}}$ | (6$g$) | O II |
| $\mathbf{B}_{20}$ | = | $x_6\,\mathbf{a}_1 + \frac{1}{2}\,\mathbf{a}_2 - x_6\,\mathbf{a}_3$ | = | $x_6 a\,\hat{\mathbf{x}} + \frac{1}{2\sqrt{3}}a\,\hat{\mathbf{y}} + \frac{1}{6}c\,\hat{\mathbf{z}}$ | (6$g$) | O II |
| $\mathbf{B}_{21}$ | = | $x_7\,\mathbf{a}_1 + x_7\,\mathbf{a}_2 + z_7\,\mathbf{a}_3$ | = | $\frac{1}{2}\left(x_7 - z_7\right)a\,\hat{\mathbf{x}} + \frac{1}{2\sqrt{3}}\left(x_7 - z_7\right)a\,\hat{\mathbf{y}} + \left(\frac{2}{3}x_7 + \frac{1}{3}z_7\right)c\,\hat{\mathbf{z}}$ | (6$h$) | $H_2O$ II |
| $\mathbf{B}_{22}$ | = | $z_7\,\mathbf{a}_1 + x_7\,\mathbf{a}_2 + x_7\,\mathbf{a}_3$ | = | $\frac{1}{2}\left(-x_7 + z_7\right)a\,\hat{\mathbf{x}} + \frac{1}{2\sqrt{3}}\left(x_7 - z_7\right)a\,\hat{\mathbf{y}} + \left(\frac{2}{3}x_7 + \frac{1}{3}z_7\right)c\,\hat{\mathbf{z}}$ | (6$h$) | $H_2O$ II |
| $\mathbf{B}_{23}$ | = | $x_7\,\mathbf{a}_1 + z_7\,\mathbf{a}_2 + x_7\,\mathbf{a}_3$ | = | $\frac{1}{\sqrt{3}}\left(-x_7 + z_7\right)a\,\hat{\mathbf{y}} + \left(\frac{2}{3}x_7 + \frac{1}{3}z_7\right)c\,\hat{\mathbf{z}}$ | (6$h$) | $H_2O$ II |
| $\mathbf{B}_{24}$ | = | $-z_7\,\mathbf{a}_1 - x_7\,\mathbf{a}_2 - x_7\,\mathbf{a}_3$ | = | $\frac{1}{2}\left(x_7 - z_7\right)a\,\hat{\mathbf{x}} + \frac{1}{2\sqrt{3}}\left(-x_7 + z_7\right)a\,\hat{\mathbf{y}} - c\left(\frac{2}{3}x_7 + \frac{1}{3}z_7\right)\hat{\mathbf{z}}$ | (6$h$) | $H_2O$ II |
| $\mathbf{B}_{25}$ | = | $-x_7\,\mathbf{a}_1 - x_7\,\mathbf{a}_2 - z_7\,\mathbf{a}_3$ | = | $\frac{1}{2}\left(-x_7 + z_7\right)a\,\hat{\mathbf{x}} + \frac{1}{2\sqrt{3}}\left(-x_7 + z_7\right)a\,\hat{\mathbf{y}} - c\left(\frac{2}{3}x_7 + \frac{1}{3}z_7\right)\hat{\mathbf{z}}$ | (6$h$) | $H_2O$ II |
| $\mathbf{B}_{26}$ | = | $-x_7\,\mathbf{a}_1 - z_7\,\mathbf{a}_2 - x_7\,\mathbf{a}_3$ | = | $\frac{1}{\sqrt{3}}\left(x_7 - z_7\right)a\,\hat{\mathbf{y}} - c\left(\frac{2}{3}x_7 + \frac{1}{3}z_7\right)\hat{\mathbf{z}}$ | (6$h$) | $H_2O$ II |
| $\mathbf{B}_{27}$ | = | $x_8\,\mathbf{a}_1 + x_8\,\mathbf{a}_2 + z_8\,\mathbf{a}_3$ | = | $\frac{1}{2}\left(x_8 - z_8\right)a\,\hat{\mathbf{x}} + \frac{1}{2\sqrt{3}}\left(x_8 - z_8\right)a\,\hat{\mathbf{y}} + \left(\frac{2}{3}x_8 + \frac{1}{3}z_8\right)c\,\hat{\mathbf{z}}$ | (6$h$) | O III |
| $\mathbf{B}_{28}$ | = | $z_8\,\mathbf{a}_1 + x_8\,\mathbf{a}_2 + x_8\,\mathbf{a}_3$ | = | $\frac{1}{2}\left(-x_8 + z_8\right)a\,\hat{\mathbf{x}} + \frac{1}{2\sqrt{3}}\left(x_8 - z_8\right)a\,\hat{\mathbf{y}} + \left(\frac{2}{3}x_8 + \frac{1}{3}z_8\right)c\,\hat{\mathbf{z}}$ | (6$h$) | O III |
| $\mathbf{B}_{29}$ | = | $x_8\,\mathbf{a}_1 + z_8\,\mathbf{a}_2 + x_8\,\mathbf{a}_3$ | = | $\frac{1}{\sqrt{3}}\left(-x_8 + z_8\right)a\,\hat{\mathbf{y}} + \left(\frac{2}{3}x_8 + \frac{1}{3}z_8\right)c\,\hat{\mathbf{z}}$ | (6$h$) | O III |
| $\mathbf{B}_{30}$ | = | $-z_8\,\mathbf{a}_1 - x_8\,\mathbf{a}_2 - x_8\,\mathbf{a}_3$ | = | $\frac{1}{2}\left(x_8 - z_8\right)a\,\hat{\mathbf{x}} + \frac{1}{2\sqrt{3}}\left(-x_8 + z_8\right)a\,\hat{\mathbf{y}} - c\left(\frac{2}{3}x_8 + \frac{1}{3}z_8\right)\hat{\mathbf{z}}$ | (6$h$) | O III |
| $\mathbf{B}_{31}$ | = | $-x_8\,\mathbf{a}_1 - x_8\,\mathbf{a}_2 - z_8\,\mathbf{a}_3$ | = | $\frac{1}{2}\left(-x_8 + z_8\right)a\,\hat{\mathbf{x}} + \frac{1}{2\sqrt{3}}\left(-x_8 + z_8\right)a\,\hat{\mathbf{y}} - c\left(\frac{2}{3}x_8 + \frac{1}{3}z_8\right)\hat{\mathbf{z}}$ | (6$h$) | O III |
| $\mathbf{B}_{32}$ | = | $-x_8\,\mathbf{a}_1 - z_8\,\mathbf{a}_2 - x_8\,\mathbf{a}_3$ | = | $\frac{1}{\sqrt{3}}\left(x_8 - z_8\right)a\,\hat{\mathbf{y}} - c\left(\frac{2}{3}x_8 + \frac{1}{3}z_8\right)\hat{\mathbf{z}}$ | (6$h$) | O III |
| $\mathbf{B}_{33}$ | = | $x_9\,\mathbf{a}_1 + x_9\,\mathbf{a}_2 + z_9\,\mathbf{a}_3$ | = | $\frac{1}{2}\left(x_9 - z_9\right)a\,\hat{\mathbf{x}} + \frac{1}{2\sqrt{3}}\left(x_9 - z_9\right)a\,\hat{\mathbf{y}} + \left(\frac{2}{3}x_9 + \frac{1}{3}z_9\right)c\,\hat{\mathbf{z}}$ | (6$h$) | O IV |
| $\mathbf{B}_{34}$ | = | $z_9\,\mathbf{a}_1 + x_9\,\mathbf{a}_2 + x_9\,\mathbf{a}_3$ | = | $\frac{1}{2}\left(-x_9 + z_9\right)a\,\hat{\mathbf{x}} + \frac{1}{2\sqrt{3}}\left(x_9 - z_9\right)a\,\hat{\mathbf{y}} + \left(\frac{2}{3}x_9 + \frac{1}{3}z_9\right)c\,\hat{\mathbf{z}}$ | (6$h$) | O IV |
| $\mathbf{B}_{35}$ | = | $x_9\,\mathbf{a}_1 + z_9\,\mathbf{a}_2 + x_9\,\mathbf{a}_3$ | = | $\frac{1}{\sqrt{3}}\left(-x_9 + z_9\right)a\,\hat{\mathbf{y}} + \left(\frac{2}{3}x_9 + \frac{1}{3}z_9\right)c\,\hat{\mathbf{z}}$ | (6$h$) | O IV |
| $\mathbf{B}_{36}$ | = | $-z_9\,\mathbf{a}_1 - x_9\,\mathbf{a}_2 - x_9\,\mathbf{a}_3$ | = | $\frac{1}{2}\left(x_9 - z_9\right)a\,\hat{\mathbf{x}} + \frac{1}{2\sqrt{3}}\left(-x_9 + z_9\right)a\,\hat{\mathbf{y}} - c\left(\frac{2}{3}x_9 + \frac{1}{3}z_9\right)\hat{\mathbf{z}}$ | (6$h$) | O IV |
| $\mathbf{B}_{37}$ | = | $-x_9\,\mathbf{a}_1 - x_9\,\mathbf{a}_2 - z_9\,\mathbf{a}_3$ | = | $\frac{1}{2}\left(-x_9 + z_9\right)a\,\hat{\mathbf{x}} + \frac{1}{2\sqrt{3}}\left(-x_9 + z_9\right)a\,\hat{\mathbf{y}} - c\left(\frac{2}{3}x_9 + \frac{1}{3}z_9\right)\hat{\mathbf{z}}$ | (6$h$) | O IV |
| $\mathbf{B}_{38}$ | = | $-x_9\,\mathbf{a}_1 - z_9\,\mathbf{a}_2 - x_9\,\mathbf{a}_3$ | = | $\frac{1}{\sqrt{3}}\left(x_9 - z_9\right)a\,\hat{\mathbf{y}} - c\left(\frac{2}{3}x_9 + \frac{1}{3}z_9\right)\hat{\mathbf{z}}$ | (6$h$) | O IV |
| $\mathbf{B}_{39}$ | = | $x_{10}\,\mathbf{a}_1 + y_{10}\,\mathbf{a}_2 + z_{10}\,\mathbf{a}_3$ | = | $\frac{1}{2}\left(x_{10} - z_{10}\right) a\,\hat{\mathbf{x}} + \left(-\frac{1}{2\sqrt{3}}x_{10} + \frac{1}{\sqrt{3}}y_{10} - \frac{1}{2\sqrt{3}}z_{10}\right) a\,\hat{\mathbf{y}} + \frac{1}{3}\left(x_{10} + y_{10} + z_{10}\right) c\,\hat{\mathbf{z}}$ | (12*i*) | $H_2O$ III |
| $\mathbf{B}_{40}$ | = | $z_{10}\,\mathbf{a}_1 + x_{10}\,\mathbf{a}_2 + y_{10}\,\mathbf{a}_3$ | = | $\frac{1}{2}\left(-y_{10} + z_{10}\right) a\,\hat{\mathbf{x}} + \left(\frac{1}{\sqrt{3}}x_{10} - \frac{1}{2\sqrt{3}}y_{10} - \frac{1}{2\sqrt{3}}z_{10}\right) a\,\hat{\mathbf{y}} + \frac{1}{3}\left(x_{10} + y_{10} + z_{10}\right) c\,\hat{\mathbf{z}}$ | (12*i*) | $H_2O$ III |
| $\mathbf{B}_{41}$ | = | $y_{10}\,\mathbf{a}_1 + z_{10}\,\mathbf{a}_2 + x_{10}\,\mathbf{a}_3$ | = | $\frac{1}{2}\left(-x_{10} + y_{10}\right) a\,\hat{\mathbf{x}} + \left(-\frac{1}{2\sqrt{3}}x_{10} - \frac{1}{2\sqrt{3}}y_{10} + \frac{1}{\sqrt{3}}z_{10}\right) a\,\hat{\mathbf{y}} + \frac{1}{3}\left(x_{10} + y_{10} + z_{10}\right) c\,\hat{\mathbf{z}}$ | (12*i*) | $H_2O$ III |
| $\mathbf{B}_{42}$ | = | $-z_{10}\,\mathbf{a}_1 - y_{10}\,\mathbf{a}_2 - x_{10}\,\mathbf{a}_3$ | = | $\frac{1}{2}\left(x_{10} - z_{10}\right) a\,\hat{\mathbf{x}} + \left(\frac{1}{2\sqrt{3}}x_{10} - \frac{1}{\sqrt{3}}y_{10} + \frac{1}{2\sqrt{3}}z_{10}\right) a\,\hat{\mathbf{y}} - \frac{1}{3}\left(x_{10} + y_{10} + z_{10}\right) c\,\hat{\mathbf{z}}$ | (12*i*) | $H_2O$ III |
| $\mathbf{B}_{43}$ | = | $-y_{10}\,\mathbf{a}_1 - x_{10}\,\mathbf{a}_2 - z_{10}\,\mathbf{a}_3$ | = | $\frac{1}{2}\left(-y_{10} + z_{10}\right) a\,\hat{\mathbf{x}} + \left(-\frac{1}{\sqrt{3}}x_{10} + \frac{1}{2\sqrt{3}}y_{10} + \frac{1}{2\sqrt{3}}z_{10}\right) a\,\hat{\mathbf{y}} - \frac{1}{3}\left(x_{10} + y_{10} + z_{10}\right) c\,\hat{\mathbf{z}}$ | (12*i*) | $H_2O$ III |
| $\mathbf{B}_{44}$ | = | $-x_{10}\,\mathbf{a}_1 - z_{10}\,\mathbf{a}_2 - y_{10}\,\mathbf{a}_3$ | = | $\frac{1}{2}\left(-x_{10} + y_{10}\right) a\,\hat{\mathbf{x}} + \left(\frac{1}{2\sqrt{3}}x_{10} + \frac{1}{2\sqrt{3}}y_{10} - \frac{1}{\sqrt{3}}z_{10}\right) a\,\hat{\mathbf{y}} - \frac{1}{3}\left(x_{10} + y_{10} + z_{10}\right) c\,\hat{\mathbf{z}}$ | (12*i*) | $H_2O$ III |
| $\mathbf{B}_{45}$ | = | $-x_{10}\,\mathbf{a}_1 - y_{10}\,\mathbf{a}_2 - z_{10}\,\mathbf{a}_3$ | = | $\frac{1}{2}\left(-x_{10} + z_{10}\right) a\,\hat{\mathbf{x}} + \left(\frac{1}{2\sqrt{3}}x_{10} - \frac{1}{\sqrt{3}}y_{10} + \frac{1}{2\sqrt{3}}z_{10}\right) a\,\hat{\mathbf{y}} - \frac{1}{3}\left(x_{10} + y_{10} + z_{10}\right) c\,\hat{\mathbf{z}}$ | (12*i*) | $H_2O$ III |
| $\mathbf{B}_{46}$ | = | $-z_{10}\,\mathbf{a}_1 - x_{10}\,\mathbf{a}_2 - y_{10}\,\mathbf{a}_3$ | = | $\frac{1}{2}\left(y_{10} - z_{10}\right) a\,\hat{\mathbf{x}} + \left(-\frac{1}{\sqrt{3}}x_{10} + \frac{1}{2\sqrt{3}}y_{10} + \frac{1}{2\sqrt{3}}z_{10}\right) a\,\hat{\mathbf{y}} - \frac{1}{3}\left(x_{10} + y_{10} + z_{10}\right) c\,\hat{\mathbf{z}}$ | (12*i*) | $H_2O$ III |
| $\mathbf{B}_{47}$ | = | $-y_{10}\,\mathbf{a}_1 - z_{10}\,\mathbf{a}_2 - x_{10}\,\mathbf{a}_3$ | = | $\frac{1}{2}\left(x_{10} - y_{10}\right) a\,\hat{\mathbf{x}} + \left(\frac{1}{2\sqrt{3}}x_{10} + \frac{1}{2\sqrt{3}}y_{10} - \frac{1}{\sqrt{3}}z_{10}\right) a\,\hat{\mathbf{y}} - \frac{1}{3}\left(x_{10} + y_{10} + z_{10}\right) c\,\hat{\mathbf{z}}$ | (12*i*) | $H_2O$ III |
| $\mathbf{B}_{48}$ | = | $z_{10}\,\mathbf{a}_1 + y_{10}\,\mathbf{a}_2 + x_{10}\,\mathbf{a}_3$ | = | $\frac{1}{2}\left(-x_{10} + z_{10}\right) a\,\hat{\mathbf{x}} + \left(-\frac{1}{2\sqrt{3}}x_{10} + \frac{1}{\sqrt{3}}y_{10} - \frac{1}{2\sqrt{3}}z_{10}\right) a\,\hat{\mathbf{y}} + \frac{1}{3}\left(x_{10} + y_{10} + z_{10}\right) c\,\hat{\mathbf{z}}$ | (12*i*) | $H_2O$ III |
| $\mathbf{B}_{49}$ | = | $y_{10}\,\mathbf{a}_1 + x_{10}\,\mathbf{a}_2 + z_{10}\,\mathbf{a}_3$ | = | $\frac{1}{2}\left(y_{10} - z_{10}\right) a\,\hat{\mathbf{x}} + \left(\frac{1}{\sqrt{3}}x_{10} - \frac{1}{2\sqrt{3}}y_{10} - \frac{1}{2\sqrt{3}}z_{10}\right) a\,\hat{\mathbf{y}} + \frac{1}{3}\left(x_{10} + y_{10} + z_{10}\right) c\,\hat{\mathbf{z}}$ | (12*i*) | $H_2O$ III |
| $\mathbf{B}_{50}$ | = | $x_{10}\,\mathbf{a}_1 + z_{10}\,\mathbf{a}_2 + y_{10}\,\mathbf{a}_3$ | = | $\frac{1}{2}\left(x_{10} - y_{10}\right) a\,\hat{\mathbf{x}} + \left(-\frac{1}{2\sqrt{3}}x_{10} - \frac{1}{2\sqrt{3}}y_{10} + \frac{1}{\sqrt{3}}z_{10}\right) a\,\hat{\mathbf{y}} + \frac{1}{3}\left(x_{10} + y_{10} + z_{10}\right) c\,\hat{\mathbf{z}}$ | (12*i*) | $H_2O$ III |
| $\mathbf{B}_{51}$ | = | $x_{11}\,\mathbf{a}_1 + y_{11}\,\mathbf{a}_2 + z_{11}\,\mathbf{a}_3$ | = | $\frac{1}{2}\left(x_{11} - z_{11}\right) a\,\hat{\mathbf{x}} + \left(-\frac{1}{2\sqrt{3}}x_{11} + \frac{1}{\sqrt{3}}y_{11} - \frac{1}{2\sqrt{3}}z_{11}\right) a\,\hat{\mathbf{y}} + \frac{1}{3}\left(x_{11} + y_{11} + z_{11}\right) c\,\hat{\mathbf{z}}$ | (12*i*) | Si |

$$\mathbf{B}_{52} = z_{11}\,\mathbf{a}_1 + x_{11}\,\mathbf{a}_2 + y_{11}\,\mathbf{a}_3 = \frac{1}{2}\left(-y_{11} + z_{11}\right)a\,\hat{\mathbf{x}} + \left(\frac{1}{\sqrt{3}}x_{11} - \frac{1}{2\sqrt{3}}y_{11} - \frac{1}{2\sqrt{3}}z_{11}\right)a\,\hat{\mathbf{y}} + \frac{1}{3}\left(x_{11} + y_{11} + z_{11}\right)c\,\hat{\mathbf{z}} \qquad (12i) \qquad \text{Si}$$

$$\mathbf{B}_{53} = y_{11}\,\mathbf{a}_1 + z_{11}\,\mathbf{a}_2 + x_{11}\,\mathbf{a}_3 = \frac{1}{2}\left(-x_{11} + y_{11}\right)a\,\hat{\mathbf{x}} + \left(-\frac{1}{2\sqrt{3}}x_{11} - \frac{1}{2\sqrt{3}}y_{11} + \frac{1}{\sqrt{3}}z_{11}\right)a\,\hat{\mathbf{y}} + \frac{1}{3}\left(x_{11} + y_{11} + z_{11}\right)c\,\hat{\mathbf{z}} \qquad (12i) \qquad \text{Si}$$

$$\mathbf{B}_{54} = -z_{11}\,\mathbf{a}_1 - y_{11}\,\mathbf{a}_2 - x_{11}\,\mathbf{a}_3 = \frac{1}{2}\left(x_{11} - z_{11}\right)a\,\hat{\mathbf{x}} + \left(\frac{1}{2\sqrt{3}}x_{11} - \frac{1}{\sqrt{3}}y_{11} + \frac{1}{2\sqrt{3}}z_{11}\right)a\,\hat{\mathbf{y}} - \frac{1}{3}\left(x_{11} + y_{11} + z_{11}\right)c\,\hat{\mathbf{z}} \qquad (12i) \qquad \text{Si}$$

$$\mathbf{B}_{55} = -y_{11}\,\mathbf{a}_1 - x_{11}\,\mathbf{a}_2 - z_{11}\,\mathbf{a}_3 = \frac{1}{2}\left(-y_{11} + z_{11}\right)a\,\hat{\mathbf{x}} + \left(-\frac{1}{\sqrt{3}}x_{11} + \frac{1}{2\sqrt{3}}y_{11} + \frac{1}{2\sqrt{3}}z_{11}\right)a\,\hat{\mathbf{y}} - \frac{1}{3}\left(x_{11} + y_{11} + z_{11}\right)c\,\hat{\mathbf{z}} \qquad (12i) \qquad \text{Si}$$

$$\mathbf{B}_{56} = -x_{11}\,\mathbf{a}_1 - z_{11}\,\mathbf{a}_2 - y_{11}\,\mathbf{a}_3 = \frac{1}{2}\left(-x_{11} + y_{11}\right)a\,\hat{\mathbf{x}} + \left(\frac{1}{2\sqrt{3}}x_{11} + \frac{1}{2\sqrt{3}}y_{11} - \frac{1}{\sqrt{3}}z_{11}\right)a\,\hat{\mathbf{y}} - \frac{1}{3}\left(x_{11} + y_{11} + z_{11}\right)c\,\hat{\mathbf{z}} \qquad (12i) \qquad \text{Si}$$

$$\mathbf{B}_{57} = -x_{11}\,\mathbf{a}_1 - y_{11}\,\mathbf{a}_2 - z_{11}\,\mathbf{a}_3 = \frac{1}{2}\left(-x_{11} + z_{11}\right)a\,\hat{\mathbf{x}} + \left(\frac{1}{2\sqrt{3}}x_{11} - \frac{1}{\sqrt{3}}y_{11} + \frac{1}{2\sqrt{3}}z_{11}\right)a\,\hat{\mathbf{y}} - \frac{1}{3}\left(x_{11} + y_{11} + z_{11}\right)c\,\hat{\mathbf{z}} \qquad (12i) \qquad \text{Si}$$

$$\mathbf{B}_{58} = -z_{11}\,\mathbf{a}_1 - x_{11}\,\mathbf{a}_2 - y_{11}\,\mathbf{a}_3 = \frac{1}{2}\left(y_{11} - z_{11}\right)a\,\hat{\mathbf{x}} + \left(-\frac{1}{\sqrt{3}}x_{11} + \frac{1}{2\sqrt{3}}y_{11} + \frac{1}{2\sqrt{3}}z_{11}\right)a\,\hat{\mathbf{y}} - \frac{1}{3}\left(x_{11} + y_{11} + z_{11}\right)c\,\hat{\mathbf{z}} \qquad (12i) \qquad \text{Si}$$

$$\mathbf{B}_{59} = -y_{11}\,\mathbf{a}_1 - z_{11}\,\mathbf{a}_2 - x_{11}\,\mathbf{a}_3 = \frac{1}{2}\left(x_{11} - y_{11}\right)a\,\hat{\mathbf{x}} + \left(\frac{1}{2\sqrt{3}}x_{11} + \frac{1}{2\sqrt{3}}y_{11} - \frac{1}{\sqrt{3}}z_{11}\right)a\,\hat{\mathbf{y}} - \frac{1}{3}\left(x_{11} + y_{11} + z_{11}\right)c\,\hat{\mathbf{z}} \qquad (12i) \qquad \text{Si}$$

$$\mathbf{B}_{60} = z_{11}\,\mathbf{a}_1 + y_{11}\,\mathbf{a}_2 + x_{11}\,\mathbf{a}_3 = \frac{1}{2}\left(-x_{11} + z_{11}\right)a\,\hat{\mathbf{x}} + \left(-\frac{1}{2\sqrt{3}}x_{11} + \frac{1}{\sqrt{3}}y_{11} - \frac{1}{2\sqrt{3}}z_{11}\right)a\,\hat{\mathbf{y}} + \frac{1}{3}\left(x_{11} + y_{11} + z_{11}\right)c\,\hat{\mathbf{z}} \qquad (12i) \qquad \text{Si}$$

$$\mathbf{B}_{61} = y_{11}\,\mathbf{a}_1 + x_{11}\,\mathbf{a}_2 + z_{11}\,\mathbf{a}_3 = \frac{1}{2}\left(y_{11} - z_{11}\right)a\,\hat{\mathbf{x}} + \left(\frac{1}{\sqrt{3}}x_{11} - \frac{1}{2\sqrt{3}}y_{11} - \frac{1}{2\sqrt{3}}z_{11}\right)a\,\hat{\mathbf{y}} + \frac{1}{3}\left(x_{11} + y_{11} + z_{11}\right)c\,\hat{\mathbf{z}} \qquad (12i) \qquad \text{Si}$$

$$\mathbf{B}_{62} = x_{11}\,\mathbf{a}_1 + z_{11}\,\mathbf{a}_2 + y_{11}\,\mathbf{a}_3 = \frac{1}{2}\left(x_{11} - y_{11}\right)a\,\hat{\mathbf{x}} + \left(-\frac{1}{2\sqrt{3}}x_{11} - \frac{1}{2\sqrt{3}}y_{11} + \frac{1}{\sqrt{3}}z_{11}\right)a\,\hat{\mathbf{y}} + \frac{1}{3}\left(x_{11} + y_{11} + z_{11}\right)c\,\hat{\mathbf{z}} \qquad (12i) \qquad \text{Si}$$

---

**Geometry files:**

- CIF: pp. 1738
- POSCAR: pp. 1739

# $CaSi_2$ ($C12$) Structure: AB2_hR6_166_c_2c

http://aflow.org/prototype-encyclopedia/AB2_hR6_166_c_2c

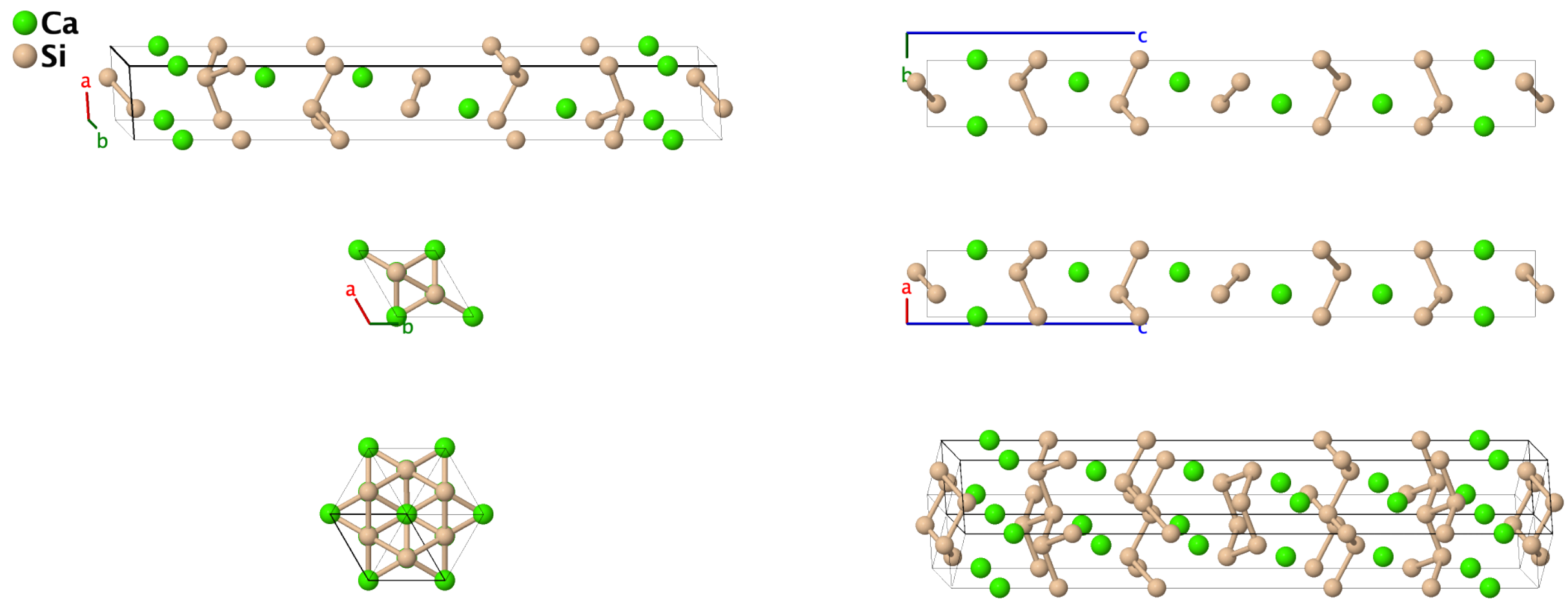

| | | |
|---|---|---|
| **Prototype** | : | $CaSi_2$ |
| **AFLOW prototype label** | : | AB2_hR6_166_c_2c |
| ***Strukturbericht* designation** | : | $C12$ |
| **Pearson symbol** | : | hR6 |
| **Space group number** | : | 166 |
| **Space group symbol** | : | $R\bar{3}m$ |
| **AFLOW prototype command** | : | `aflow --proto=AB2_hR6_166_c_2c [--hex]` `--params=`$a, c/a, x_1, x_2, x_3$ |

### Other compounds with this structure

- $CaGe_2$

### Rhombohedral primitive vectors:

$$\begin{aligned}
\mathbf{a}_1 &= \tfrac{1}{2}\,a\,\hat{\mathbf{x}} - \tfrac{1}{2\sqrt{3}}\,a\,\hat{\mathbf{y}} + \tfrac{1}{3}\,c\,\hat{\mathbf{z}} \\
\mathbf{a}_2 &= \tfrac{1}{\sqrt{3}}\,a\,\hat{\mathbf{y}} + \tfrac{1}{3}\,c\,\hat{\mathbf{z}} \\
\mathbf{a}_3 &= -\tfrac{1}{2}\,a\,\hat{\mathbf{x}} - \tfrac{1}{2\sqrt{3}}\,a\,\hat{\mathbf{y}} + \tfrac{1}{3}\,c\,\hat{\mathbf{z}}
\end{aligned}$$

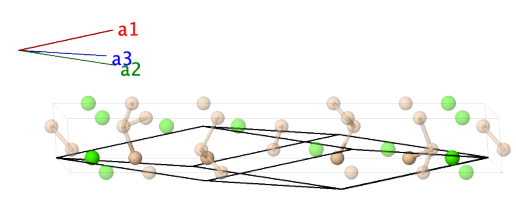

### Basis vectors:

| | | Lattice Coordinates | | Cartesian Coordinates | Wyckoff Position | Atom Type |
|---|---|---|---|---|---|---|
| $\mathbf{B}_1$ | = | $x_1\,\mathbf{a}_1 + x_1\,\mathbf{a}_2 + x_1\,\mathbf{a}_3$ | = | $x_1 c\,\hat{\mathbf{z}}$ | (2$c$) | Ca |
| $\mathbf{B}_2$ | = | $-x_1\,\mathbf{a}_1 - x_1\,\mathbf{a}_2 - x_1\,\mathbf{a}_3$ | = | $-x_1 c\,\hat{\mathbf{z}}$ | (2$c$) | Ca |
| $\mathbf{B}_3$ | = | $x_2\,\mathbf{a}_1 + x_2\,\mathbf{a}_2 + x_2\,\mathbf{a}_3$ | = | $x_2 c\,\hat{\mathbf{z}}$ | (2$c$) | Si I |
| $\mathbf{B}_4$ | = | $-x_2\,\mathbf{a}_1 - x_2\,\mathbf{a}_2 - x_2\,\mathbf{a}_3$ | = | $-x_2 c\,\hat{\mathbf{z}}$ | (2$c$) | Si I |
| $\mathbf{B}_5$ | = | $x_3\,\mathbf{a}_1 + x_3\,\mathbf{a}_2 + x_3\,\mathbf{a}_3$ | = | $x_3 c\,\hat{\mathbf{z}}$ | (2$c$) | Si II |
| $\mathbf{B}_6$ | = | $-x_3\,\mathbf{a}_1 - x_3\,\mathbf{a}_2 - x_3\,\mathbf{a}_3$ | = | $-x_3 c\,\hat{\mathbf{z}}$ | (2$c$) | Si II |

---

**Geometry files:**
- CIF: pp. 1739
- POSCAR: pp. 1740

# Rhombohedral $CuTi_2S_4$ Structure: AB4C2_hR28_166_2c_2c2h_abh

http://aflow.org/prototype-encyclopedia/AB4C2_hR28_166_2c_2c2h_abh

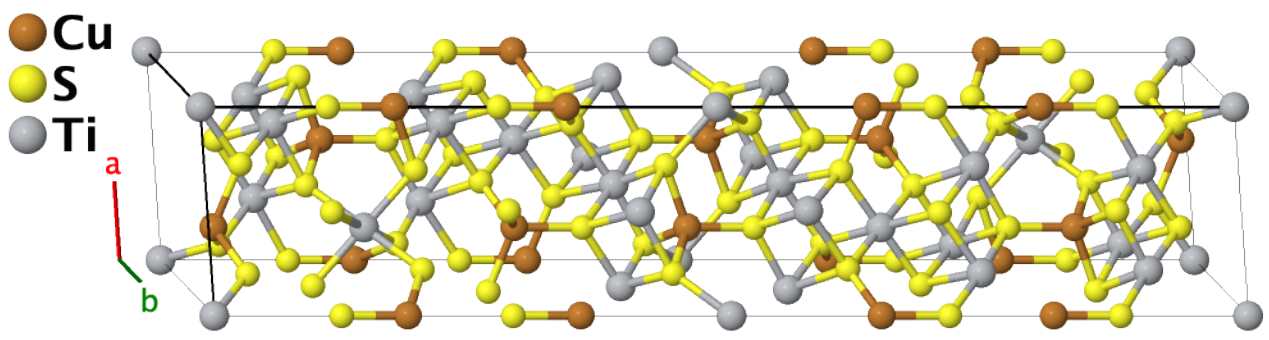

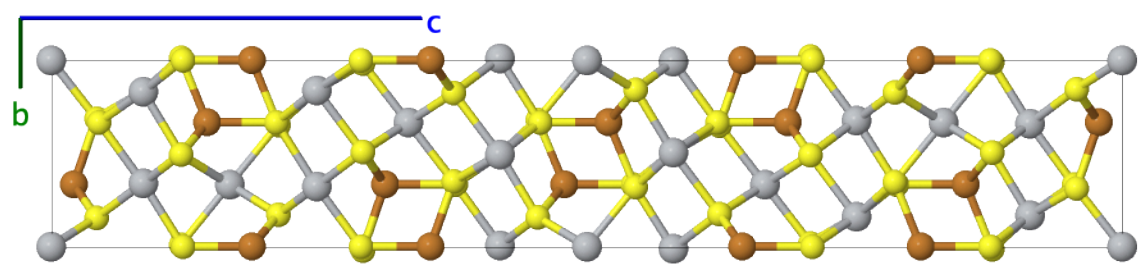

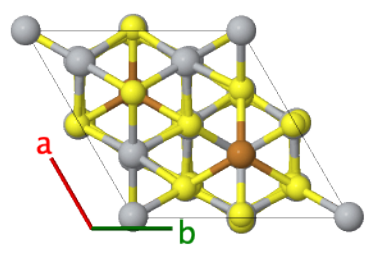

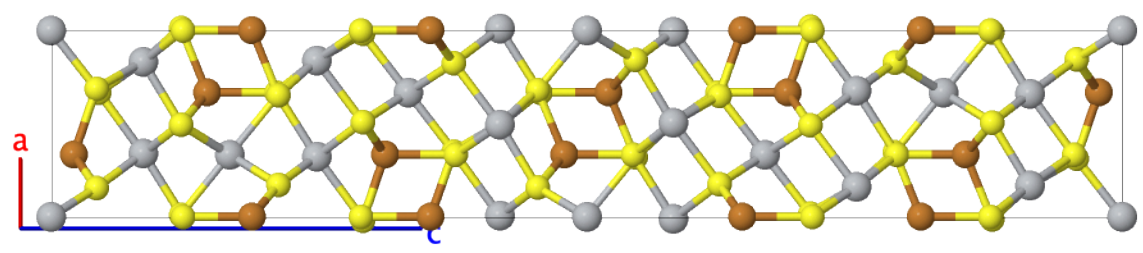

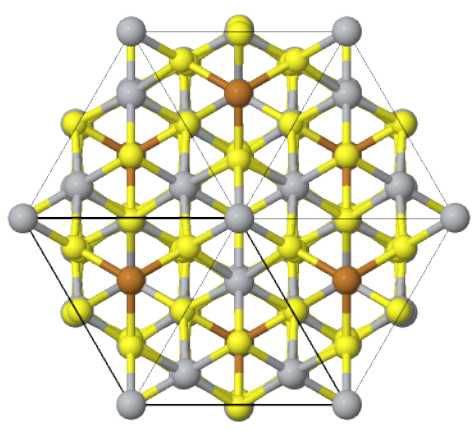

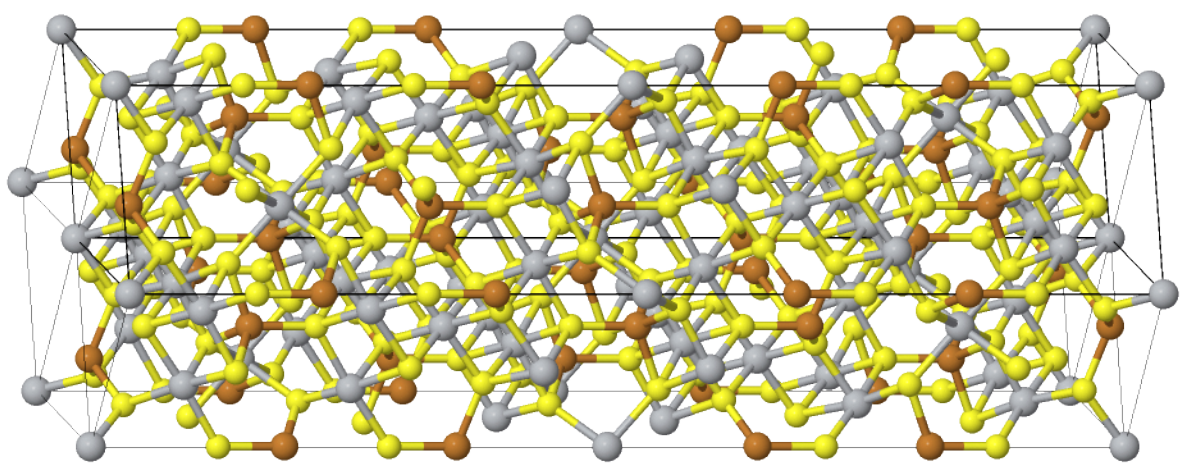

| | | |
|---|---|---|
| **Prototype** | : | $CuS_4Ti_2$ |
| **AFLOW prototype label** | : | AB4C2_hR28_166_2c_2c2h_abh |
| ***Strukturbericht* designation** | : | None |
| **Pearson symbol** | : | hR28 |
| **Space group number** | : | 166 |
| **Space group symbol** | : | $R\bar{3}m$ |
| **AFLOW prototype command** | : | `aflow --proto=AB4C2_hR28_166_2c_2c2h_abh [--hex]` `--params=`$a, c/a, x_3, x_4, x_5, x_6, x_7, z_7, x_8, z_8, x_9, z_9$ |

**Rhombohedral primitive vectors:**

$$\mathbf{a}_1 = \frac{1}{2}\,a\,\hat{\mathbf{x}} - \frac{1}{2\sqrt{3}}\,a\,\hat{\mathbf{y}} + \frac{1}{3}\,c\,\hat{\mathbf{z}}$$
$$\mathbf{a}_2 = \frac{1}{\sqrt{3}}\,a\,\hat{\mathbf{y}} + \frac{1}{3}\,c\,\hat{\mathbf{z}}$$
$$\mathbf{a}_3 = -\frac{1}{2}\,a\,\hat{\mathbf{x}} - \frac{1}{2\sqrt{3}}\,a\,\hat{\mathbf{y}} + \frac{1}{3}\,c\,\hat{\mathbf{z}}$$

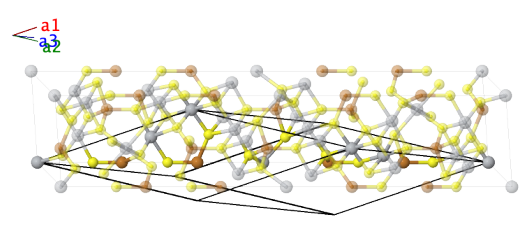

**Basis vectors:**

| | | Lattice Coordinates | | Cartesian Coordinates | Wyckoff Position | Atom Type |
|---|---|---|---|---|---|---|
| $\mathbf{B}_1$ | = | $0\,\mathbf{a}_1 + 0\,\mathbf{a}_2 + 0\,\mathbf{a}_3$ | = | $0\,\hat{\mathbf{x}} + 0\,\hat{\mathbf{y}} + 0\,\hat{\mathbf{z}}$ | (1$a$) | Ti I |
| $\mathbf{B}_2$ | = | $\frac{1}{2}\,\mathbf{a}_1 + \frac{1}{2}\,\mathbf{a}_2 + \frac{1}{2}\,\mathbf{a}_3$ | = | $\frac{1}{2}c\,\hat{\mathbf{z}}$ | (1$b$) | Ti II |
| $\mathbf{B}_3$ | = | $x_3\,\mathbf{a}_1 + x_3\,\mathbf{a}_2 + x_3\,\mathbf{a}_3$ | = | $x_3c\,\hat{\mathbf{z}}$ | (2$c$) | Cu I |
| $\mathbf{B}_4$ | = | $-x_3\,\mathbf{a}_1 - x_3\,\mathbf{a}_2 - x_3\,\mathbf{a}_3$ | = | $-x_3c\,\hat{\mathbf{z}}$ | (2$c$) | Cu I |
| $\mathbf{B}_5$ | = | $x_4\,\mathbf{a}_1 + x_4\,\mathbf{a}_2 + x_4\,\mathbf{a}_3$ | = | $x_4c\,\hat{\mathbf{z}}$ | (2$c$) | Cu II |
| $\mathbf{B}_6$ | = | $-x_4\,\mathbf{a}_1 - x_4\,\mathbf{a}_2 - x_4\,\mathbf{a}_3$ | = | $-x_4c\,\hat{\mathbf{z}}$ | (2$c$) | Cu II |
| $\mathbf{B}_{7}$ | = | $x_{5}\,\mathbf{a}_{1}+x_{5}\,\mathbf{a}_{2}+x_{5}\,\mathbf{a}_{3}$ | = | $x_{5}c\,\hat{\mathbf{z}}$ | (2$c$) | S I |
| $\mathbf{B}_{8}$ | = | $-x_{5}\,\mathbf{a}_{1}-x_{5}\,\mathbf{a}_{2}-x_{5}\,\mathbf{a}_{3}$ | = | $-x_{5}c\,\hat{\mathbf{z}}$ | (2$c$) | S I |
| $\mathbf{B}_{9}$ | = | $x_{6}\,\mathbf{a}_{1}+x_{6}\,\mathbf{a}_{2}+x_{6}\,\mathbf{a}_{3}$ | = | $x_{6}c\,\hat{\mathbf{z}}$ | (2$c$) | S II |
| $\mathbf{B}_{10}$ | = | $-x_{6}\,\mathbf{a}_{1}-x_{6}\,\mathbf{a}_{2}-x_{6}\,\mathbf{a}_{3}$ | = | $-x_{6}c\,\hat{\mathbf{z}}$ | (2$c$) | S II |
| $\mathbf{B}_{11}$ | = | $x_{7}\,\mathbf{a}_{1}+x_{7}\,\mathbf{a}_{2}+z_{7}\,\mathbf{a}_{3}$ | = | $\frac{1}{2}\left(x_{7}-z_{7}\right)a\,\hat{\mathbf{x}}+\frac{1}{2\sqrt{3}}\left(x_{7}-z_{7}\right)a\,\hat{\mathbf{y}}+\left(\frac{2}{3}x_{7}+\frac{1}{3}z_{7}\right)c\,\hat{\mathbf{z}}$ | (6$h$) | S III |
| $\mathbf{B}_{12}$ | = | $z_{7}\,\mathbf{a}_{1}+x_{7}\,\mathbf{a}_{2}+x_{7}\,\mathbf{a}_{3}$ | = | $\frac{1}{2}\left(-x_{7}+z_{7}\right)a\,\hat{\mathbf{x}}+\frac{1}{2\sqrt{3}}\left(x_{7}-z_{7}\right)a\,\hat{\mathbf{y}}+\left(\frac{2}{3}x_{7}+\frac{1}{3}z_{7}\right)c\,\hat{\mathbf{z}}$ | (6$h$) | S III |
| $\mathbf{B}_{13}$ | = | $x_{7}\,\mathbf{a}_{1}+z_{7}\,\mathbf{a}_{2}+x_{7}\,\mathbf{a}_{3}$ | = | $\frac{1}{\sqrt{3}}\left(-x_{7}+z_{7}\right)a\,\hat{\mathbf{y}}+\left(\frac{2}{3}x_{7}+\frac{1}{3}z_{7}\right)c\,\hat{\mathbf{z}}$ | (6$h$) | S III |
| $\mathbf{B}_{14}$ | = | $-z_{7}\,\mathbf{a}_{1}-x_{7}\,\mathbf{a}_{2}-x_{7}\,\mathbf{a}_{3}$ | = | $\frac{1}{2}\left(x_{7}-z_{7}\right)a\,\hat{\mathbf{x}}+\frac{1}{2\sqrt{3}}\left(-x_{7}+z_{7}\right)a\,\hat{\mathbf{y}}-c\left(\frac{2}{3}x_{7}+\frac{1}{3}z_{7}\right)\hat{\mathbf{z}}$ | (6$h$) | S III |
| $\mathbf{B}_{15}$ | = | $-x_{7}\,\mathbf{a}_{1}-x_{7}\,\mathbf{a}_{2}-z_{7}\,\mathbf{a}_{3}$ | = | $\frac{1}{2}\left(-x_{7}+z_{7}\right)a\,\hat{\mathbf{x}}+\frac{1}{2\sqrt{3}}\left(-x_{7}+z_{7}\right)a\,\hat{\mathbf{y}}-c\left(\frac{2}{3}x_{7}+\frac{1}{3}z_{7}\right)\hat{\mathbf{z}}$ | (6$h$) | S III |
| $\mathbf{B}_{16}$ | = | $-x_{7}\,\mathbf{a}_{1}-z_{7}\,\mathbf{a}_{2}-x_{7}\,\mathbf{a}_{3}$ | = | $\frac{1}{\sqrt{3}}\left(x_{7}-z_{7}\right)a\,\hat{\mathbf{y}}-c\left(\frac{2}{3}x_{7}+\frac{1}{3}z_{7}\right)\hat{\mathbf{z}}$ | (6$h$) | S III |
| $\mathbf{B}_{17}$ | = | $x_{8}\,\mathbf{a}_{1}+x_{8}\,\mathbf{a}_{2}+z_{8}\,\mathbf{a}_{3}$ | = | $\frac{1}{2}\left(x_{8}-z_{8}\right)a\,\hat{\mathbf{x}}+\frac{1}{2\sqrt{3}}\left(x_{8}-z_{8}\right)a\,\hat{\mathbf{y}}+\left(\frac{2}{3}x_{8}+\frac{1}{3}z_{8}\right)c\,\hat{\mathbf{z}}$ | (6$h$) | S IV |
| $\mathbf{B}_{18}$ | = | $z_{8}\,\mathbf{a}_{1}+x_{8}\,\mathbf{a}_{2}+x_{8}\,\mathbf{a}_{3}$ | = | $\frac{1}{2}\left(-x_{8}+z_{8}\right)a\,\hat{\mathbf{x}}+\frac{1}{2\sqrt{3}}\left(x_{8}-z_{8}\right)a\,\hat{\mathbf{y}}+\left(\frac{2}{3}x_{8}+\frac{1}{3}z_{8}\right)c\,\hat{\mathbf{z}}$ | (6$h$) | S IV |
| $\mathbf{B}_{19}$ | = | $x_{8}\,\mathbf{a}_{1}+z_{8}\,\mathbf{a}_{2}+x_{8}\,\mathbf{a}_{3}$ | = | $\frac{1}{\sqrt{3}}\left(-x_{8}+z_{8}\right)a\,\hat{\mathbf{y}}+\left(\frac{2}{3}x_{8}+\frac{1}{3}z_{8}\right)c\,\hat{\mathbf{z}}$ | (6$h$) | S IV |
| $\mathbf{B}_{20}$ | = | $-z_{8}\,\mathbf{a}_{1}-x_{8}\,\mathbf{a}_{2}-x_{8}\,\mathbf{a}_{3}$ | = | $\frac{1}{2}\left(x_{8}-z_{8}\right)a\,\hat{\mathbf{x}}+\frac{1}{2\sqrt{3}}\left(-x_{8}+z_{8}\right)a\,\hat{\mathbf{y}}-c\left(\frac{2}{3}x_{8}+\frac{1}{3}z_{8}\right)\hat{\mathbf{z}}$ | (6$h$) | S IV |
| $\mathbf{B}_{21}$ | = | $-x_{8}\,\mathbf{a}_{1}-x_{8}\,\mathbf{a}_{2}-z_{8}\,\mathbf{a}_{3}$ | = | $\frac{1}{2}\left(-x_{8}+z_{8}\right)a\,\hat{\mathbf{x}}+\frac{1}{2\sqrt{3}}\left(-x_{8}+z_{8}\right)a\,\hat{\mathbf{y}}-c\left(\frac{2}{3}x_{8}+\frac{1}{3}z_{8}\right)\hat{\mathbf{z}}$ | (6$h$) | S IV |
| $\mathbf{B}_{22}$ | = | $-x_{8}\,\mathbf{a}_{1}-z_{8}\,\mathbf{a}_{2}-x_{8}\,\mathbf{a}_{3}$ | = | $\frac{1}{\sqrt{3}}\left(x_{8}-z_{8}\right)a\,\hat{\mathbf{y}}-c\left(\frac{2}{3}x_{8}+\frac{1}{3}z_{8}\right)\hat{\mathbf{z}}$ | (6$h$) | S IV |
| $\mathbf{B}_{23}$ | = | $x_{9}\,\mathbf{a}_{1}+x_{9}\,\mathbf{a}_{2}+z_{9}\,\mathbf{a}_{3}$ | = | $\frac{1}{2}\left(x_{9}-z_{9}\right)a\,\hat{\mathbf{x}}+\frac{1}{2\sqrt{3}}\left(x_{9}-z_{9}\right)a\,\hat{\mathbf{y}}+\left(\frac{2}{3}x_{9}+\frac{1}{3}z_{9}\right)c\,\hat{\mathbf{z}}$ | (6$h$) | Ti III |
| $\mathbf{B}_{24}$ | = | $z_{9}\,\mathbf{a}_{1}+x_{9}\,\mathbf{a}_{2}+x_{9}\,\mathbf{a}_{3}$ | = | $\frac{1}{2}\left(-x_{9}+z_{9}\right)a\,\hat{\mathbf{x}}+\frac{1}{2\sqrt{3}}\left(x_{9}-z_{9}\right)a\,\hat{\mathbf{y}}+\left(\frac{2}{3}x_{9}+\frac{1}{3}z_{9}\right)c\,\hat{\mathbf{z}}$ | (6$h$) | Ti III |
| $\mathbf{B}_{25}$ | = | $x_{9}\,\mathbf{a}_{1}+z_{9}\,\mathbf{a}_{2}+x_{9}\,\mathbf{a}_{3}$ | = | $\frac{1}{\sqrt{3}}\left(-x_{9}+z_{9}\right)a\,\hat{\mathbf{y}}+\left(\frac{2}{3}x_{9}+\frac{1}{3}z_{9}\right)c\,\hat{\mathbf{z}}$ | (6$h$) | Ti III |
| $\mathbf{B}_{26}$ | = | $-z_{9}\,\mathbf{a}_{1}-x_{9}\,\mathbf{a}_{2}-x_{9}\,\mathbf{a}_{3}$ | = | $\frac{1}{2}\left(x_{9}-z_{9}\right)a\,\hat{\mathbf{x}}+\frac{1}{2\sqrt{3}}\left(-x_{9}+z_{9}\right)a\,\hat{\mathbf{y}}-c\left(\frac{2}{3}x_{9}+\frac{1}{3}z_{9}\right)\hat{\mathbf{z}}$ | (6$h$) | Ti III |
| $\mathbf{B}_{27}$ | = | $-x_{9}\,\mathbf{a}_{1}-x_{9}\,\mathbf{a}_{2}-z_{9}\,\mathbf{a}_{3}$ | = | $\frac{1}{2}\left(-x_{9}+z_{9}\right)a\,\hat{\mathbf{x}}+\frac{1}{2\sqrt{3}}\left(-x_{9}+z_{9}\right)a\,\hat{\mathbf{y}}-c\left(\frac{2}{3}x_{9}+\frac{1}{3}z_{9}\right)\hat{\mathbf{z}}$ | (6$h$) | Ti III |
| $\mathbf{B}_{28}$ | = | $-x_{9}\,\mathbf{a}_{1}-z_{9}\,\mathbf{a}_{2}-x_{9}\,\mathbf{a}_{3}$ | = | $\frac{1}{\sqrt{3}}\left(x_{9}-z_{9}\right)a\,\hat{\mathbf{y}}-c\left(\frac{2}{3}x_{9}+\frac{1}{3}z_{9}\right)\hat{\mathbf{z}}$ | (6$h$) | Ti III |

**Geometry files:**

- CIF: pp. 1740
- POSCAR: pp. 1740

# $CaCu_4P_2$ Structure: AB4C2_hR7_166_a_2c_c

http://aflow.org/prototype-encyclopedia/AB4C2_hR7_166_a_2c_c

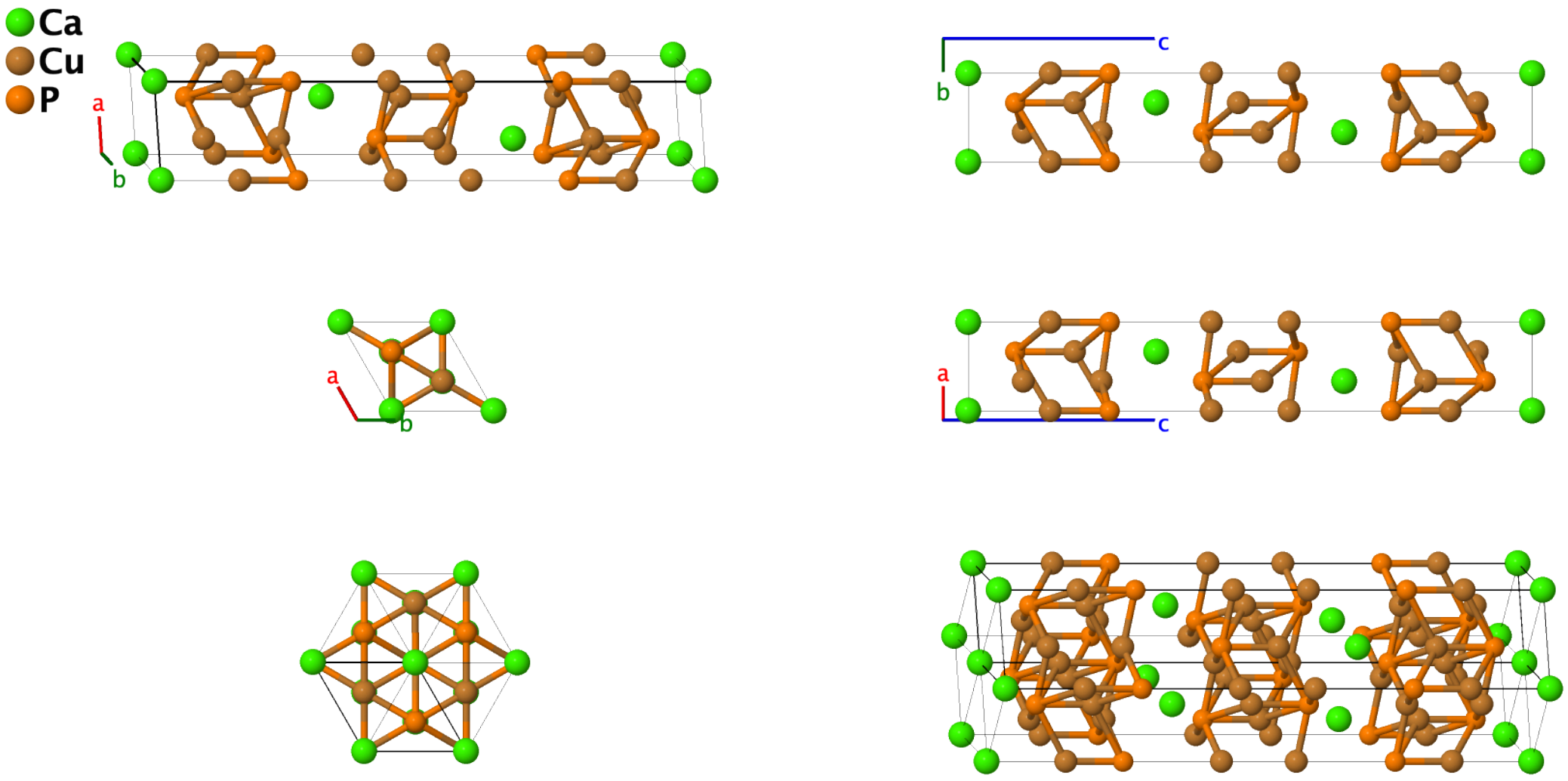

| | | |
|---|---|---|
| **Prototype** | : | $CaCu_4P_2$ |
| **AFLOW prototype label** | : | AB4C2_hR7_166_a_2c_c |
| ***Strukturbericht* designation** | : | None |
| **Pearson symbol** | : | hR7 |
| **Space group number** | : | 166 |
| **Space group symbol** | : | $R\bar{3}m$ |
| **AFLOW prototype command** | : | `aflow --proto=AB4C2_hR7_166_a_2c_c [--hex]` `--params=`$a, c/a, x_2, x_3, x_4$ |

**Other compounds with this structure**

- $BaAg_4As_2$, $CaAg_4As_2$, $EuAg_4As_2$, $EuAg_4Sb_2$, $SrAg_4As_2$, $SrAg_4Sb_2$, and $ZrLi_4Ge_2$

**Rhombohedral primitive vectors:**

$$\mathbf{a}_1 = \frac{1}{2}\,a\,\mathbf{\hat{x}} - \frac{1}{2\sqrt{3}}\,a\,\mathbf{\hat{y}} + \frac{1}{3}\,c\,\mathbf{\hat{z}}$$
$$\mathbf{a}_2 = \frac{1}{\sqrt{3}}\,a\,\mathbf{\hat{y}} + \frac{1}{3}\,c\,\mathbf{\hat{z}}$$
$$\mathbf{a}_3 = -\frac{1}{2}\,a\,\mathbf{\hat{x}} - \frac{1}{2\sqrt{3}}\,a\,\mathbf{\hat{y}} + \frac{1}{3}\,c\,\mathbf{\hat{z}}$$

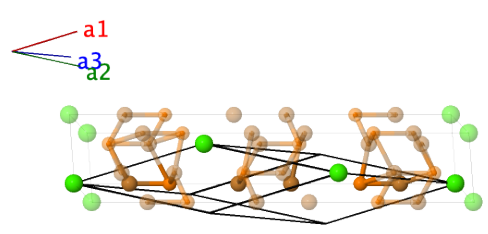

**Basis vectors:**

| | | Lattice Coordinates | | Cartesian Coordinates | Wyckoff Position | Atom Type |
|---|---|---|---|---|---|---|
| $\mathbf{B}_1$ | = | $0\,\mathbf{a}_1 + 0\,\mathbf{a}_2 + 0\,\mathbf{a}_3$ | = | $0\,\mathbf{\hat{x}} + 0\,\mathbf{\hat{y}} + 0\,\mathbf{\hat{z}}$ | $(1a)$ | Ca |
| $\mathbf{B}_2$ | = | $x_2\,\mathbf{a}_1 + x_2\,\mathbf{a}_2 + x_2\,\mathbf{a}_3$ | = | $x_2 c\,\mathbf{\hat{z}}$ | $(2c)$ | Cu I |
| $\mathbf{B}_3$ | = | $-x_2\,\mathbf{a}_1 - x_2\,\mathbf{a}_2 - x_2\,\mathbf{a}_3$ | = | $-x_2 c\,\mathbf{\hat{z}}$ | $(2c)$ | Cu I |
| $\mathbf{B}_4$ | = | $x_3\,\mathbf{a}_1 + x_3\,\mathbf{a}_2 + x_3\,\mathbf{a}_3$ | = | $x_3 c\,\mathbf{\hat{z}}$ | $(2c)$ | Cu II |
| $\mathbf{B}_5$ | = | $-x_3\,\mathbf{a}_1 - x_3\,\mathbf{a}_2 - x_3\,\mathbf{a}_3$ | = | $-x_3 c\,\mathbf{\hat{z}}$ | $(2c)$ | Cu II |
| $\mathbf{B}_6$ | = | $x_4\,\mathbf{a}_1 + x_4\,\mathbf{a}_2 + x_4\,\mathbf{a}_3$ | = | $x_4 c\,\mathbf{\hat{z}}$ | $(2c)$ | P |
| $\mathbf{B}_7$ | = | $-x_4\,\mathbf{a}_1 - x_4\,\mathbf{a}_2 - x_4\,\mathbf{a}_3$ | = | $-x_4 c\,\mathbf{\hat{z}}$ | $(2c)$ | P |

**Geometry files:**
- CIF: pp. 1741
- POSCAR: pp. 1741

# $CaUO_4$ Structure: AB4C_hR6_166_b_2c_a

http://aflow.org/prototype-encyclopedia/AB4C_hR6_166_b_2c_a

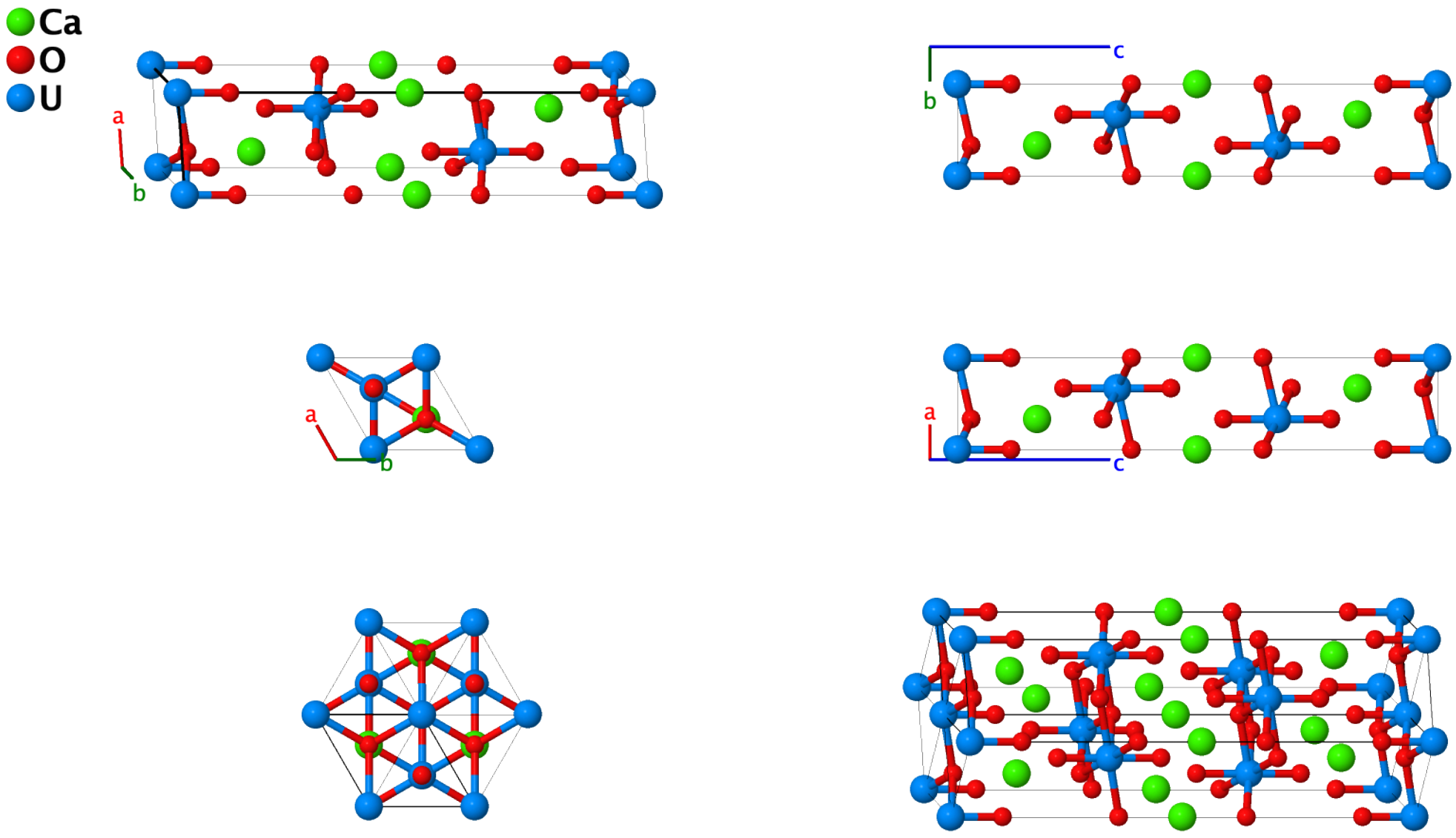

| | | |
|---|---|---|
| **Prototype** | : | $CaO_4U$ |
| **AFLOW prototype label** | : | AB4C_hR6_166_b_2c_a |
| ***Strukturbericht* designation** | : | None |
| **Pearson symbol** | : | hR6 |
| **Space group number** | : | 166 |
| **Space group symbol** | : | $R\bar{3}m$ |
| **AFLOW prototype command** | : | `aflow --proto=AB4C_hR6_166_b_2c_a [--hex]` `--params=`$a, c/a, x_3, x_4$ |

**Other compounds with this structure**

- $NaUO_4$ (Clarkeite)

- Hexagonal settings of this structure can be obtained with the option `--hex`.

**Rhombohedral primitive vectors:**

$$\mathbf{a}_1 = \frac{1}{2}\,a\,\hat{\mathbf{x}} - \frac{1}{2\sqrt{3}}\,a\,\hat{\mathbf{y}} + \frac{1}{3}\,c\,\hat{\mathbf{z}}$$
$$\mathbf{a}_2 = \frac{1}{\sqrt{3}}\,a\,\hat{\mathbf{y}} + \frac{1}{3}\,c\,\hat{\mathbf{z}}$$
$$\mathbf{a}_3 = -\frac{1}{2}\,a\,\hat{\mathbf{x}} - \frac{1}{2\sqrt{3}}\,a\,\hat{\mathbf{y}} + \frac{1}{3}\,c\,\hat{\mathbf{z}}$$

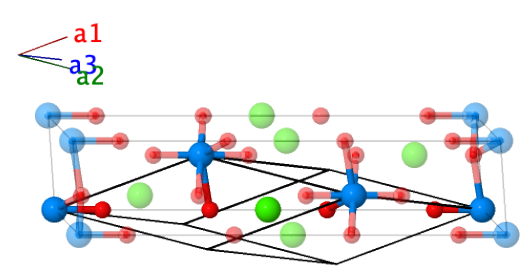

**Basis vectors:**

| | | Lattice Coordinates | | Cartesian Coordinates | Wyckoff Position | Atom Type |
|---|---|---|---|---|---|---|
| $\mathbf{B}_1$ | = | $0\,\mathbf{a}_1 + 0\,\mathbf{a}_2 + 0\,\mathbf{a}_3$ | = | $0\,\hat{\mathbf{x}} + 0\,\hat{\mathbf{y}} + 0\,\hat{\mathbf{z}}$ | (1$a$) | U |
| $\mathbf{B}_2$ | = | $\frac{1}{2}\,\mathbf{a}_1 + \frac{1}{2}\,\mathbf{a}_2 + \frac{1}{2}\,\mathbf{a}_3$ | = | $\frac{1}{2}c\,\hat{\mathbf{z}}$ | (1$b$) | Ca |
| $\mathbf{B}_3$ | = | $x_3\,\mathbf{a}_1 + x_3\,\mathbf{a}_2 + x_3\,\mathbf{a}_3$ | = | $x_3 c\,\hat{\mathbf{z}}$ | (2$c$) | O I |
| $\mathbf{B}_4$ | = | $-x_3\,\mathbf{a}_1 - x_3\,\mathbf{a}_2 - x_3\,\mathbf{a}_3$ | = | $-x_3 c\,\hat{\mathbf{z}}$ | (2$c$) | O I |
| $\mathbf{B}_5$ | = | $x_4\,\mathbf{a}_1 + x_4\,\mathbf{a}_2 + x_4\,\mathbf{a}_3$ | = | $x_4 c\,\hat{\mathbf{z}}$ | (2$c$) | O II |
| $\mathbf{B}_6$ | = | $-x_4\,\mathbf{a}_1 - x_4\,\mathbf{a}_2 - x_4\,\mathbf{a}_3$ | = | $-x_4 c\,\hat{\mathbf{z}}$ | (2$c$) | O II |

---

---

**Geometry files:**

- CIF: pp. 1741
- POSCAR: pp. 1742

# $TaTi_7$ (BCC SQS-16) Structure: AB7_hR16_166_c_c2h

http://aflow.org/prototype-encyclopedia/AB7_hR16_166_c_c2h

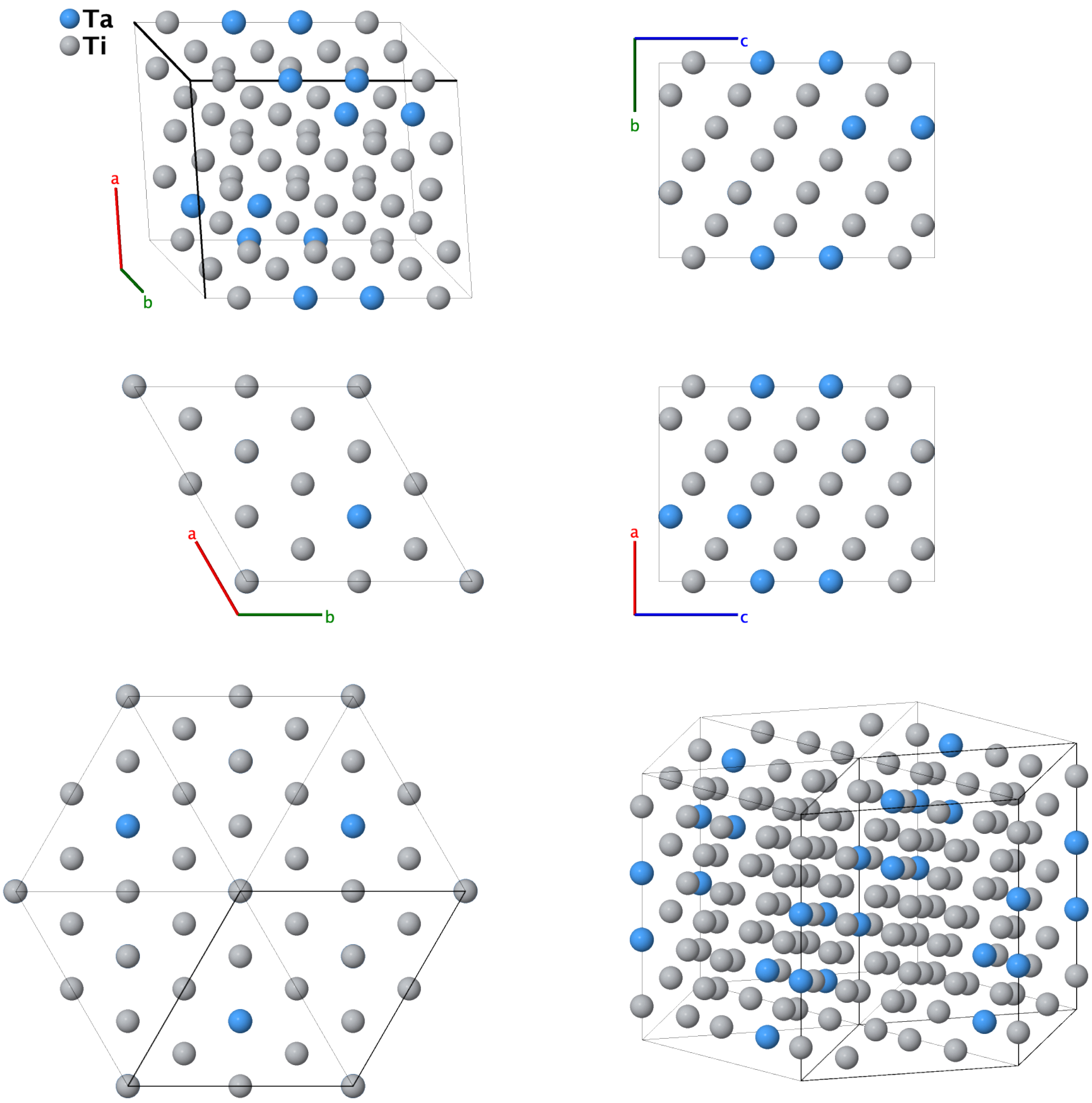

| | | |
|---|---|---|
| **Prototype** | : | $TaTi_7$ |
| **AFLOW prototype label** | : | AB7_hR16_166_c_c2h |
| ***Strukturbericht* designation** | : | None |
| **Pearson symbol** | : | hR16 |
| **Space group number** | : | 166 |
| **Space group symbol** | : | $R\bar{3}m$ |
| **AFLOW prototype command** | : | `aflow --proto=AB7_hR16_166_c_c2h [--hex]` `--params=`$a, c/a, x_1, x_2, x_3, z_3, x_4, z_4$ |

- This is a special quasirandom structure with 16 atoms per unit cell (SQS-16) for the $\beta$-phase (high-temperature austenite) bcc substitutional Ti-Ta alloy (Chakraborty, 2016). This protoype contains 12.5% Ta. Prototypes are listed for other Ta-Ti concentrations: 18.75% Ta (A3B13_oC32_38_ac_a2bcdef), 25% Ta (AB3_mC32_8_4a_4a4b), 31.25% Ta (A5B11_mP16_6_2abc_2a3b3c), and 37.5% Ta (A3B5_oC32_38_abce_abcdf).

**Rhombohedral primitive vectors:**

$$\mathbf{a}_1 = \frac{1}{2}\,a\,\hat{\mathbf{x}} - \frac{1}{2\sqrt{3}}\,a\,\hat{\mathbf{y}} + \frac{1}{3}\,c\,\hat{\mathbf{z}}$$
$$\mathbf{a}_2 = \frac{1}{\sqrt{3}}\,a\,\hat{\mathbf{y}} + \frac{1}{3}\,c\,\hat{\mathbf{z}}$$
$$\mathbf{a}_3 = -\frac{1}{2}\,a\,\hat{\mathbf{x}} - \frac{1}{2\sqrt{3}}\,a\,\hat{\mathbf{y}} + \frac{1}{3}\,c\,\hat{\mathbf{z}}$$

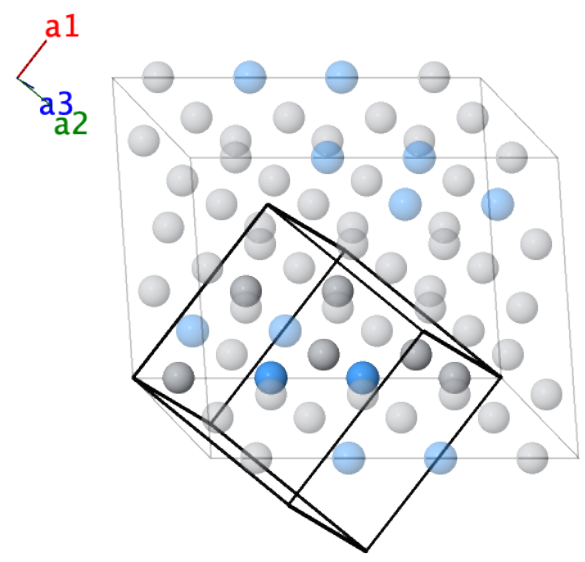

**Basis vectors:**

| | | Lattice Coordinates | | Cartesian Coordinates | Wyckoff Position | Atom Type |
|---|---|---|---|---|---|---|
| $\mathbf{B}_1$ | = | $x_1\,\mathbf{a}_1 + x_1\,\mathbf{a}_2 + x_1\,\mathbf{a}_3$ | = | $x_1 c\,\hat{\mathbf{z}}$ | (2$c$) | Ta |
| $\mathbf{B}_2$ | = | $-x_1\,\mathbf{a}_1 - x_1\,\mathbf{a}_2 - x_1\,\mathbf{a}_3$ | = | $-x_1 c\,\hat{\mathbf{z}}$ | (2$c$) | Ta |
| $\mathbf{B}_3$ | = | $x_2\,\mathbf{a}_1 + x_2\,\mathbf{a}_2 + x_2\,\mathbf{a}_3$ | = | $x_2 c\,\hat{\mathbf{z}}$ | (2$c$) | Ti I |
| $\mathbf{B}_4$ | = | $-x_2\,\mathbf{a}_1 - x_2\,\mathbf{a}_2 - x_2\,\mathbf{a}_3$ | = | $-x_2 c\,\hat{\mathbf{z}}$ | (2$c$) | Ti I |
| $\mathbf{B}_5$ | = | $x_3\,\mathbf{a}_1 + x_3\,\mathbf{a}_2 + z_3\,\mathbf{a}_3$ | = | $\frac{1}{2}(x_3 - z_3)\,a\,\hat{\mathbf{x}} + \frac{1}{2\sqrt{3}}(x_3 - z_3)\,a\,\hat{\mathbf{y}} + \left(\frac{2}{3}x_3 + \frac{1}{3}z_3\right)c\,\hat{\mathbf{z}}$ | (6$h$) | Ti II |
| $\mathbf{B}_6$ | = | $z_3\,\mathbf{a}_1 + x_3\,\mathbf{a}_2 + x_3\,\mathbf{a}_3$ | = | $\frac{1}{2}(-x_3 + z_3)\,a\,\hat{\mathbf{x}} + \frac{1}{2\sqrt{3}}(x_3 - z_3)\,a\,\hat{\mathbf{y}} + \left(\frac{2}{3}x_3 + \frac{1}{3}z_3\right)c\,\hat{\mathbf{z}}$ | (6$h$) | Ti II |
| $\mathbf{B}_7$ | = | $x_3\,\mathbf{a}_1 + z_3\,\mathbf{a}_2 + x_3\,\mathbf{a}_3$ | = | $\frac{1}{\sqrt{3}}(-x_3 + z_3)\,a\,\hat{\mathbf{y}} + \left(\frac{2}{3}x_3 + \frac{1}{3}z_3\right)c\,\hat{\mathbf{z}}$ | (6$h$) | Ti II |
| $\mathbf{B}_8$ | = | $-z_3\,\mathbf{a}_1 - x_3\,\mathbf{a}_2 - x_3\,\mathbf{a}_3$ | = | $\frac{1}{2}(x_3 - z_3)\,a\,\hat{\mathbf{x}} + \frac{1}{2\sqrt{3}}(-x_3 + z_3)\,a\,\hat{\mathbf{y}} - c\left(\frac{2}{3}x_3 + \frac{1}{3}z_3\right)\hat{\mathbf{z}}$ | (6$h$) | Ti II |
| $\mathbf{B}_9$ | = | $-x_3\,\mathbf{a}_1 - x_3\,\mathbf{a}_2 - z_3\,\mathbf{a}_3$ | = | $\frac{1}{2}(-x_3 + z_3)\,a\,\hat{\mathbf{x}} + \frac{1}{2\sqrt{3}}(-x_3 + z_3)\,a\,\hat{\mathbf{y}} - c\left(\frac{2}{3}x_3 + \frac{1}{3}z_3\right)\hat{\mathbf{z}}$ | (6$h$) | Ti II |
| $\mathbf{B}_{10}$ | = | $-x_3\,\mathbf{a}_1 - z_3\,\mathbf{a}_2 - x_3\,\mathbf{a}_3$ | = | $\frac{1}{\sqrt{3}}(x_3 - z_3)\,a\,\hat{\mathbf{y}} - c\left(\frac{2}{3}x_3 + \frac{1}{3}z_3\right)\hat{\mathbf{z}}$ | (6$h$) | Ti II |
| $\mathbf{B}_{11}$ | = | $x_4\,\mathbf{a}_1 + x_4\,\mathbf{a}_2 + z_4\,\mathbf{a}_3$ | = | $\frac{1}{2}(x_4 - z_4)\,a\,\hat{\mathbf{x}} + \frac{1}{2\sqrt{3}}(x_4 - z_4)\,a\,\hat{\mathbf{y}} + \left(\frac{2}{3}x_4 + \frac{1}{3}z_4\right)c\,\hat{\mathbf{z}}$ | (6$h$) | Ti III |
| $\mathbf{B}_{12}$ | = | $z_4\,\mathbf{a}_1 + x_4\,\mathbf{a}_2 + x_4\,\mathbf{a}_3$ | = | $\frac{1}{2}(-x_4 + z_4)\,a\,\hat{\mathbf{x}} + \frac{1}{2\sqrt{3}}(x_4 - z_4)\,a\,\hat{\mathbf{y}} + \left(\frac{2}{3}x_4 + \frac{1}{3}z_4\right)c\,\hat{\mathbf{z}}$ | (6$h$) | Ti III |
| $\mathbf{B}_{13}$ | = | $x_4\,\mathbf{a}_1 + z_4\,\mathbf{a}_2 + x_4\,\mathbf{a}_3$ | = | $\frac{1}{\sqrt{3}}(-x_4 + z_4)\,a\,\hat{\mathbf{y}} + \left(\frac{2}{3}x_4 + \frac{1}{3}z_4\right)c\,\hat{\mathbf{z}}$ | (6$h$) | Ti III |
| $\mathbf{B}_{14}$ | = | $-z_4\,\mathbf{a}_1 - x_4\,\mathbf{a}_2 - x_4\,\mathbf{a}_3$ | = | $\frac{1}{2}(x_4 - z_4)\,a\,\hat{\mathbf{x}} + \frac{1}{2\sqrt{3}}(-x_4 + z_4)\,a\,\hat{\mathbf{y}} - c\left(\frac{2}{3}x_4 + \frac{1}{3}z_4\right)\hat{\mathbf{z}}$ | (6$h$) | Ti III |
| $\mathbf{B}_{15}$ | = | $-x_4\,\mathbf{a}_1 - x_4\,\mathbf{a}_2 - z_4\,\mathbf{a}_3$ | = | $\frac{1}{2}(-x_4 + z_4)\,a\,\hat{\mathbf{x}} + \frac{1}{2\sqrt{3}}(-x_4 + z_4)\,a\,\hat{\mathbf{y}} - c\left(\frac{2}{3}x_4 + \frac{1}{3}z_4\right)\hat{\mathbf{z}}$ | (6$h$) | Ti III |
| $\mathbf{B}_{16}$ | = | $-x_4\,\mathbf{a}_1 - z_4\,\mathbf{a}_2 - x_4\,\mathbf{a}_3$ | = | $\frac{1}{\sqrt{3}}(x_4 - z_4)\,a\,\hat{\mathbf{y}} - c\left(\frac{2}{3}x_4 + \frac{1}{3}z_4\right)\hat{\mathbf{z}}$ | (6$h$) | Ti III |

**Geometry files:**

- CIF: pp. 1742
- POSCAR: pp. 1742

# Rhombohedral Delafossite ($CuFeO_2$) Structure: ABC2_hR4_166_a_b_c

http://aflow.org/prototype-encyclopedia/ABC2_hR4_166_a_b_c.CuFeO2

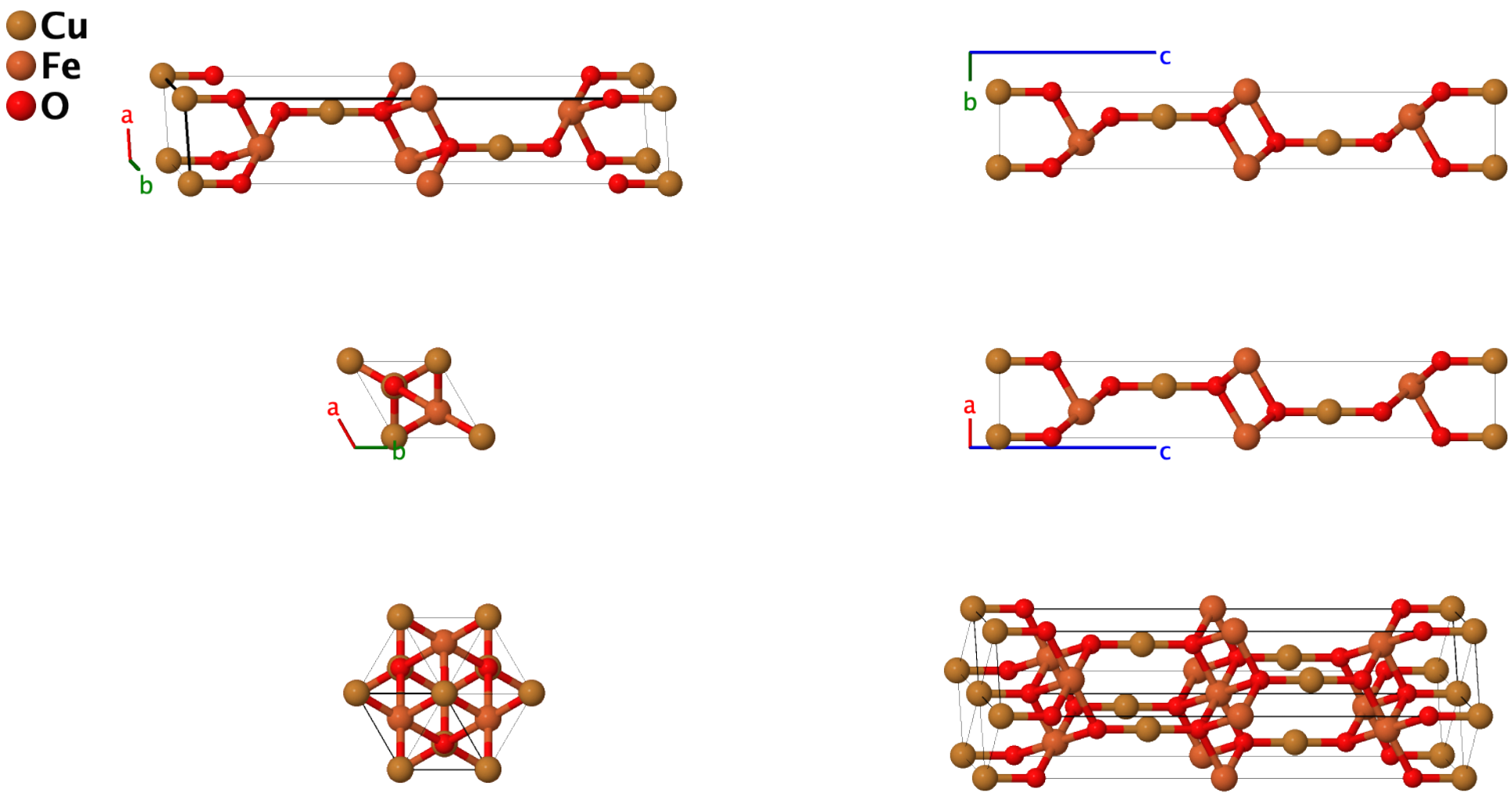

| | | |
|---|---|---|
| **Prototype** | : | $CuFeO_2$ |
| **AFLOW prototype label** | : | ABC2_hR4_166_a_b_c |
| ***Strukturbericht* designation** | : | None |
| **Pearson symbol** | : | hR4 |
| **Space group number** | : | 166 |
| **Space group symbol** | : | $R\bar{3}m$ |
| **AFLOW prototype command** | : | `aflow --proto=ABC2_hR4_166_a_b_c [--hex]` `--params=`$a, c/a, x_3$ |

**Other compounds with this structure**

- $AgAlO_2$, $AgCoO_2$, $AgCrO_2$, $AgFeO_2$, $AgGaO_2$, $AgInO_2$, $AgNiO_2$, $AgRhO_2$, $AgScO_2$, $AgTlO_2$, $CuAlO_2$, $CuCoO_2$, $CuCrO_2$, $CuEuO_2$, $CuGaO_2$, $CuInO_2$, $CuLaO_2$, $CuRhO_2$, $CuScO_2$, $CuYO_2$, $PdCoO_2$, $PdCrO_2$, $PdRhO_2$, and $PtCoO_2$

- Delafossite appears in two forms which differ in the stacking of the layers: rhombohedral, shown here, and hexagonal, prototype $CuAlO_2$. Most of the compounds found in the hexagonal phase can also be found in the rhombohedral structure (Marquardt, 2006).
- Rhombohedral delafossite has the same AFLOW label, ABC2_hR4_166_a_b_c, as caswellsilverite $F5_1$. The difference in the internal parameter $z_3$ causes a large change in the bonding of the two crystals, so we present them as different structures.

**Rhombohedral primitive vectors:**

$$\begin{aligned}
\mathbf{a}_1 &= \tfrac{1}{2}\,a\,\hat{\mathbf{x}} - \tfrac{1}{2\sqrt{3}}\,a\,\hat{\mathbf{y}} + \tfrac{1}{3}\,c\,\hat{\mathbf{z}} \\
\mathbf{a}_2 &= \tfrac{1}{\sqrt{3}}\,a\,\hat{\mathbf{y}} + \tfrac{1}{3}\,c\,\hat{\mathbf{z}} \\
\mathbf{a}_3 &= -\tfrac{1}{2}\,a\,\hat{\mathbf{x}} - \tfrac{1}{2\sqrt{3}}\,a\,\hat{\mathbf{y}} + \tfrac{1}{3}\,c\,\hat{\mathbf{z}}
\end{aligned}$$

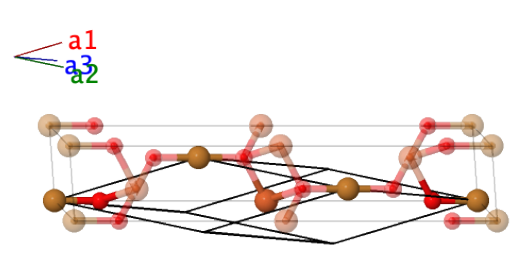

**Basis vectors:**

| | | Lattice Coordinates | | Cartesian Coordinates | Wyckoff Position | Atom Type |
|---|---|---|---|---|---|---|
| $\mathbf{B}_1$ | = | $0\,\mathbf{a}_1 + 0\,\mathbf{a}_2 + 0\,\mathbf{a}_3$ | = | $0\,\hat{\mathbf{x}} + 0\,\hat{\mathbf{y}} + 0\,\hat{\mathbf{z}}$ | (1$a$) | Cu |
| $\mathbf{B}_2$ | = | $\frac{1}{2}\,\mathbf{a}_1 + \frac{1}{2}\,\mathbf{a}_2 + \frac{1}{2}\,\mathbf{a}_3$ | = | $\frac{1}{2}c\,\hat{\mathbf{z}}$ | (1$b$) | Fe |
| $\mathbf{B}_3$ | = | $x_3\,\mathbf{a}_1 + x_3\,\mathbf{a}_2 + x_3\,\mathbf{a}_3$ | = | $x_3c\,\hat{\mathbf{z}}$ | (2$c$) | O |
| $\mathbf{B}_4$ | = | $-x_3\,\mathbf{a}_1 - x_3\,\mathbf{a}_2 - x_3\,\mathbf{a}_3$ | = | $-x_3c\,\hat{\mathbf{z}}$ | (2$c$) | O |

**Geometry files:**

- CIF: pp. 1742
- POSCAR: pp. 1743

# $\beta$-Potassium Nitrate ($KNO_3$) Structure: ABC6_hR8_166_a_b_h

http://aflow.org/prototype-encyclopedia/ABC6_hR8_166_a_b_h

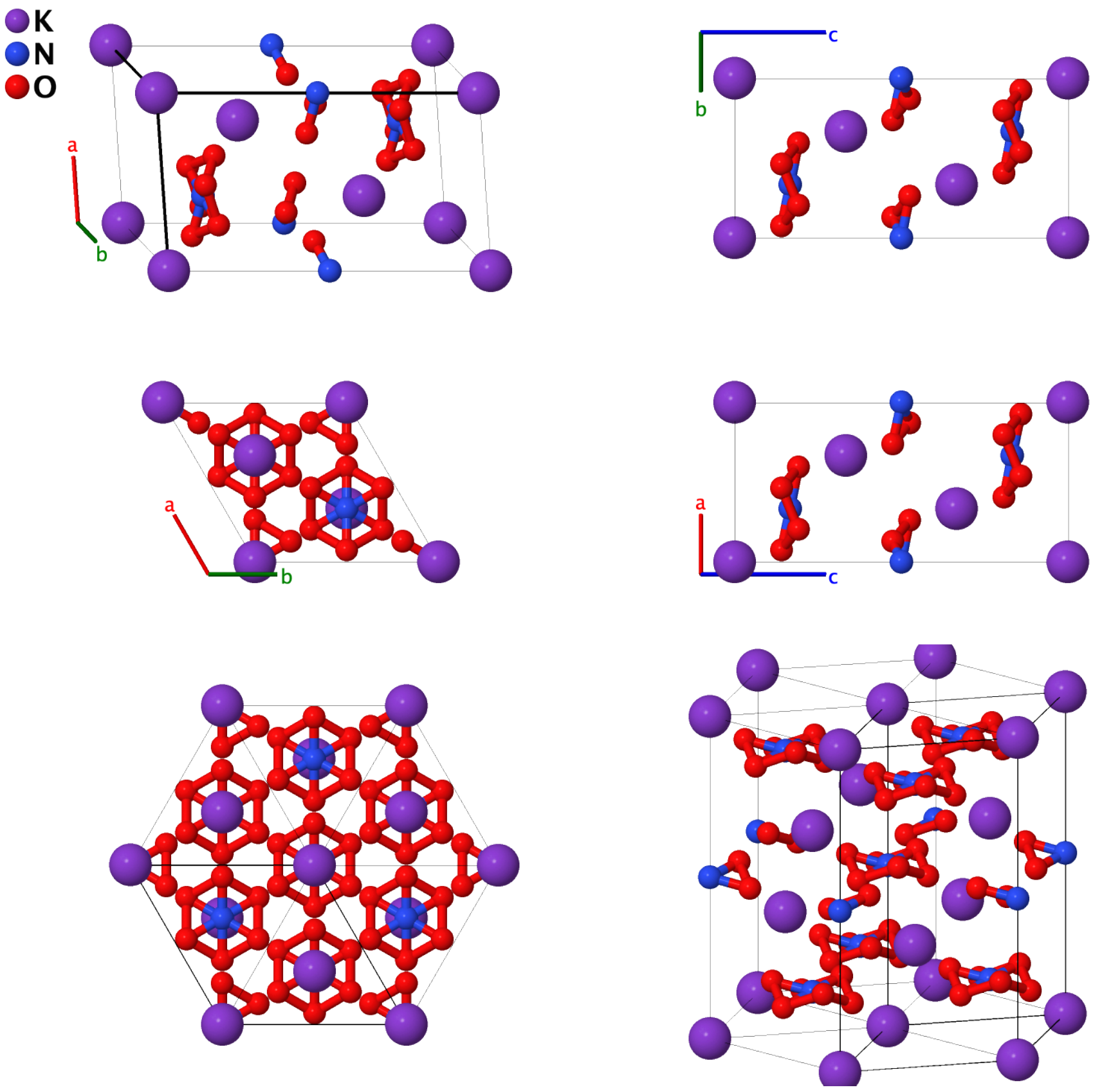

| | | |
|---|---|---|
| **Prototype** | : | $KNO_3$ |
| **AFLOW prototype label** | : | ABC6_hR8_166_a_b_h |
| ***Strukturbericht* designation** | : | None |
| **Pearson symbol** | : | hR8 |
| **Space group number** | : | 166 |
| **Space group symbol** | : | $R\bar{3}m$ |
| **AFLOW prototype command** | : | `aflow --proto=ABC6_hR8_166_a_b_h [--hex]` `--params=`$a, c/a, x_3, z_3$ |

- On heating, $\alpha$-$KNO_3$ (either Structure I or Structure II) transforms into $\beta$-$KNO_3$ at 128 °C. When heated above 200 °C and then cooled, the $\beta$ phase transforms into the metastable ferroelectric $\gamma$-$KNO_3$ phase, which can remain down to room temperature.
- In the $\beta$-phase the oxygen (6h) sites are only 50% filled. Presumably only the $+z$ or $-z$ sites will be occupied around a given nitrogen atom.
- (Nimmo, 1976) give the data for $\beta$-$KNO_3$ taken at 151 °C.

## Rhombohedral primitive vectors:

$$
\begin{aligned}
\mathbf{a}_1 &= \tfrac{1}{2}\,a\,\hat{\mathbf{x}} - \tfrac{1}{2\sqrt{3}}\,a\,\hat{\mathbf{y}} + \tfrac{1}{3}\,c\,\hat{\mathbf{z}} \\
\mathbf{a}_2 &= \tfrac{1}{\sqrt{3}}\,a\,\hat{\mathbf{y}} + \tfrac{1}{3}\,c\,\hat{\mathbf{z}} \\
\mathbf{a}_3 &= -\tfrac{1}{2}\,a\,\hat{\mathbf{x}} - \tfrac{1}{2\sqrt{3}}\,a\,\hat{\mathbf{y}} + \tfrac{1}{3}\,c\,\hat{\mathbf{z}}
\end{aligned}
$$

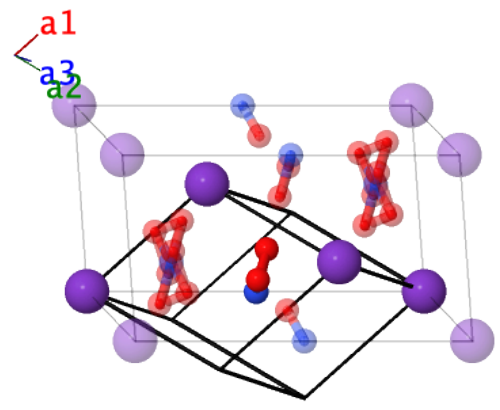

## Basis vectors:

| | | Lattice Coordinates | | Cartesian Coordinates | Wyckoff Position | Atom Type |
|---|---|---|---|---|---|---|
| $\mathbf{B}_1$ | = | $0\,\mathbf{a}_1 + 0\,\mathbf{a}_2 + 0\,\mathbf{a}_3$ | = | $0\,\hat{\mathbf{x}} + 0\,\hat{\mathbf{y}} + 0\,\hat{\mathbf{z}}$ | (1*a*) | K |
| $\mathbf{B}_2$ | = | $\frac{1}{2}\,\mathbf{a}_1 + \frac{1}{2}\,\mathbf{a}_2 + \frac{1}{2}\,\mathbf{a}_3$ | = | $\frac{1}{2}c\,\hat{\mathbf{z}}$ | (1*b*) | N |
| $\mathbf{B}_3$ | = | $x_3\,\mathbf{a}_1 + x_3\,\mathbf{a}_2 + z_3\,\mathbf{a}_3$ | = | $\frac{1}{2}\,(x_3 - z_3)\,a\,\hat{\mathbf{x}} + \frac{1}{2\sqrt{3}}\,(x_3 - z_3)\,a\,\hat{\mathbf{y}} + \left(\frac{2}{3}x_3 + \frac{1}{3}z_3\right)\,c\,\hat{\mathbf{z}}$ | (6*h*) | O |
| $\mathbf{B}_4$ | = | $z_3\,\mathbf{a}_1 + x_3\,\mathbf{a}_2 + x_3\,\mathbf{a}_3$ | = | $\frac{1}{2}\,(-x_3 + z_3)\,a\,\hat{\mathbf{x}} + \frac{1}{2\sqrt{3}}\,(x_3 - z_3)\,a\,\hat{\mathbf{y}} + \left(\frac{2}{3}x_3 + \frac{1}{3}z_3\right)\,c\,\hat{\mathbf{z}}$ | (6*h*) | O |
| $\mathbf{B}_5$ | = | $x_3\,\mathbf{a}_1 + z_3\,\mathbf{a}_2 + x_3\,\mathbf{a}_3$ | = | $\frac{1}{\sqrt{3}}\,(-x_3 + z_3)\,a\,\hat{\mathbf{y}} + \left(\frac{2}{3}x_3 + \frac{1}{3}z_3\right)\,c\,\hat{\mathbf{z}}$ | (6*h*) | O |
| $\mathbf{B}_6$ | = | $-z_3\,\mathbf{a}_1 - x_3\,\mathbf{a}_2 - x_3\,\mathbf{a}_3$ | = | $\frac{1}{2}\,(x_3 - z_3)\,a\,\hat{\mathbf{x}} + \frac{1}{2\sqrt{3}}\,(-x_3 + z_3)\,a\,\hat{\mathbf{y}} - c\left(\frac{2}{3}x_3 + \frac{1}{3}z_3\right)\,\hat{\mathbf{z}}$ | (6*h*) | O |
| $\mathbf{B}_7$ | = | $-x_3\,\mathbf{a}_1 - x_3\,\mathbf{a}_2 - z_3\,\mathbf{a}_3$ | = | $\frac{1}{2}\,(-x_3 + z_3)\,a\,\hat{\mathbf{x}} + \frac{1}{2\sqrt{3}}\,(-x_3 + z_3)\,a\,\hat{\mathbf{y}} - c\left(\frac{2}{3}x_3 + \frac{1}{3}z_3\right)\,\hat{\mathbf{z}}$ | (6*h*) | O |
| $\mathbf{B}_8$ | = | $-x_3\,\mathbf{a}_1 - z_3\,\mathbf{a}_2 - x_3\,\mathbf{a}_3$ | = | $\frac{1}{\sqrt{3}}\,(x_3 - z_3)\,a\,\hat{\mathbf{y}} - c\left(\frac{2}{3}x_3 + \frac{1}{3}z_3\right)\,\hat{\mathbf{z}}$ | (6*h*) | O |

## Found in:

## Geometry files:

- CIF: pp. 1743
- POSCAR: pp. 1743

# K(SH) ($B22$) Structure: AB_hR2_166_a_b

http://aflow.org/prototype-encyclopedia/AB_hR2_166_a_b.KSH

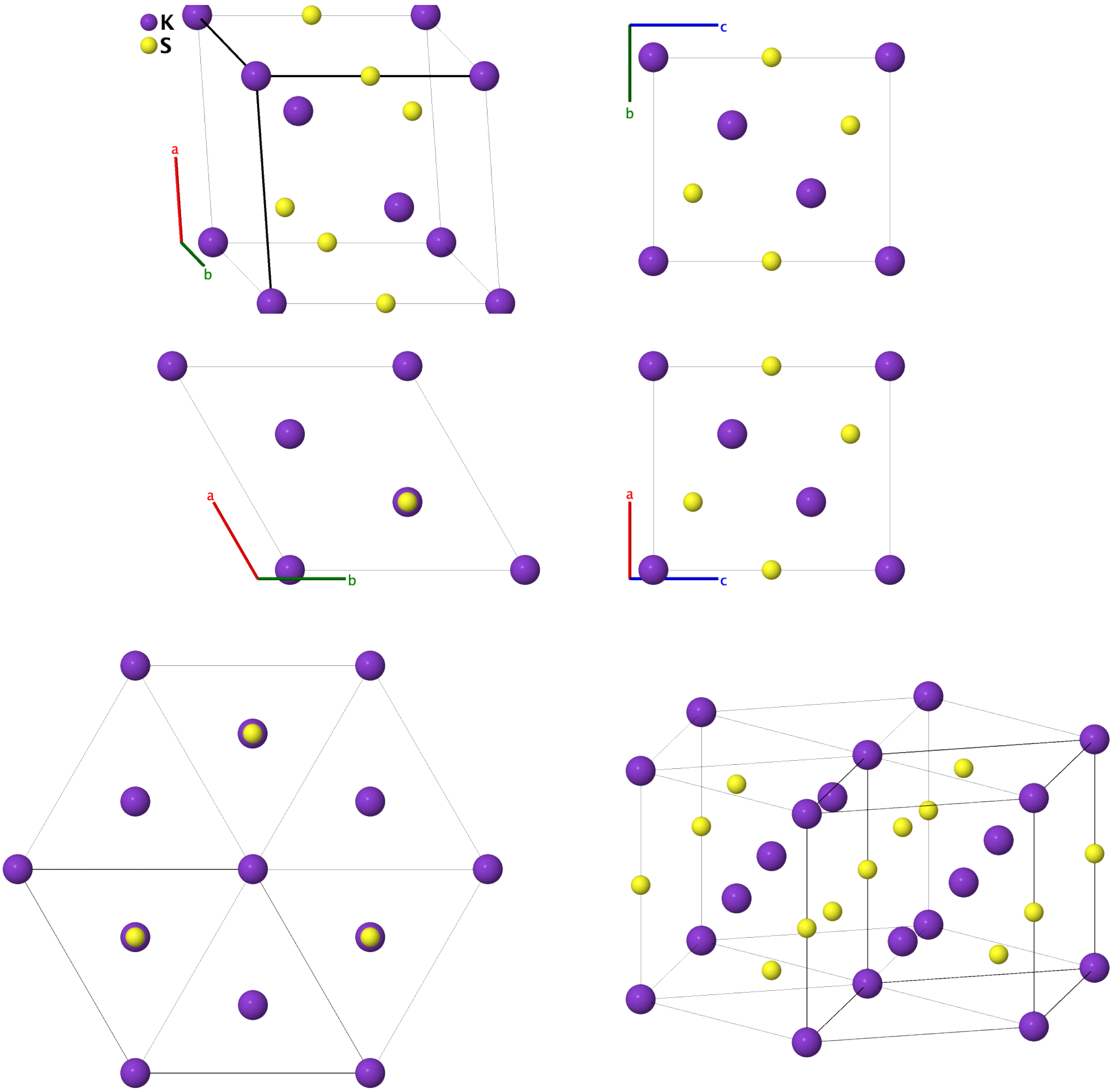

| | | |
|---|---|---|
| **Prototype** | : | KS |
| **AFLOW prototype label** | : | AB_hR2_166_a_b |
| ***Strukturbericht* designation** | : | $B22$ |
| **Pearson symbol** | : | hR2 |
| **Space group number** | : | 166 |
| **Space group symbol** | : | $R\bar{3}m$ |
| **AFLOW prototype command** | : | `aflow --proto=AB_hR2_166_a_b [--hex]`<br>`--params=`$a, c/a$ |

**Other compounds with this structure**

- Na(SH), Rb(SH), and Cs(SH)

- This is the room-temperature structure of KSH. The S and H atoms form an $SH^-$ ion, and so are listed together at the $(1b)$ Wyckoff postition, leading to the listing of this compound in the *B Strukturbericht* category.

---

**Rhombohedral primitive vectors:**

$$\mathbf{a}_1 = \frac{1}{2}\,a\,\hat{\mathbf{x}} - \frac{1}{2\sqrt{3}}\,a\,\hat{\mathbf{y}} + \frac{1}{3}\,c\,\hat{\mathbf{z}}$$
$$\mathbf{a}_2 = \frac{1}{\sqrt{3}}\,a\,\hat{\mathbf{y}} + \frac{1}{3}\,c\,\hat{\mathbf{z}}$$
$$\mathbf{a}_3 = -\frac{1}{2}\,a\,\hat{\mathbf{x}} - \frac{1}{2\sqrt{3}}\,a\,\hat{\mathbf{y}} + \frac{1}{3}\,c\,\hat{\mathbf{z}}$$

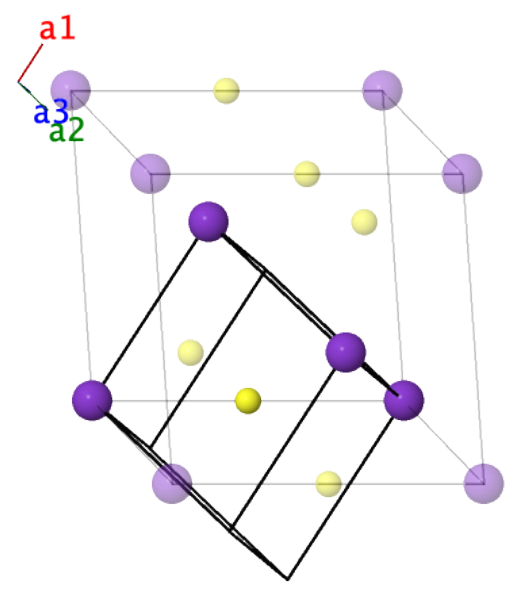

---

**Basis vectors:**

| | | Lattice Coordinates | | Cartesian Coordinates | Wyckoff Position | Atom Type |
|---|---|---|---|---|---|---|
| $\mathbf{B}_1$ | = | $0\,\mathbf{a}_1 + 0\,\mathbf{a}_2 + 0\,\mathbf{a}_3$ | = | $0\,\hat{\mathbf{x}} + 0\,\hat{\mathbf{y}} + 0\,\hat{\mathbf{z}}$ | $(1a)$ | K |
| $\mathbf{B}_2$ | = | $\frac{1}{2}\,\mathbf{a}_1 + \frac{1}{2}\,\mathbf{a}_2 + \frac{1}{2}\,\mathbf{a}_3$ | = | $\frac{1}{2}c\,\hat{\mathbf{z}}$ | $(1b)$ | S |

---

**Found in:**

---

**Geometry files:**

- CIF: pp. 1744
- POSCAR: pp. 1744

# $Zr_{21}Re_{25}$ Structure: A25B21_hR92_167_b2e3f_e3f

http://aflow.org/prototype-encyclopedia/A25B21_hR92_167_b2e3f_e3f

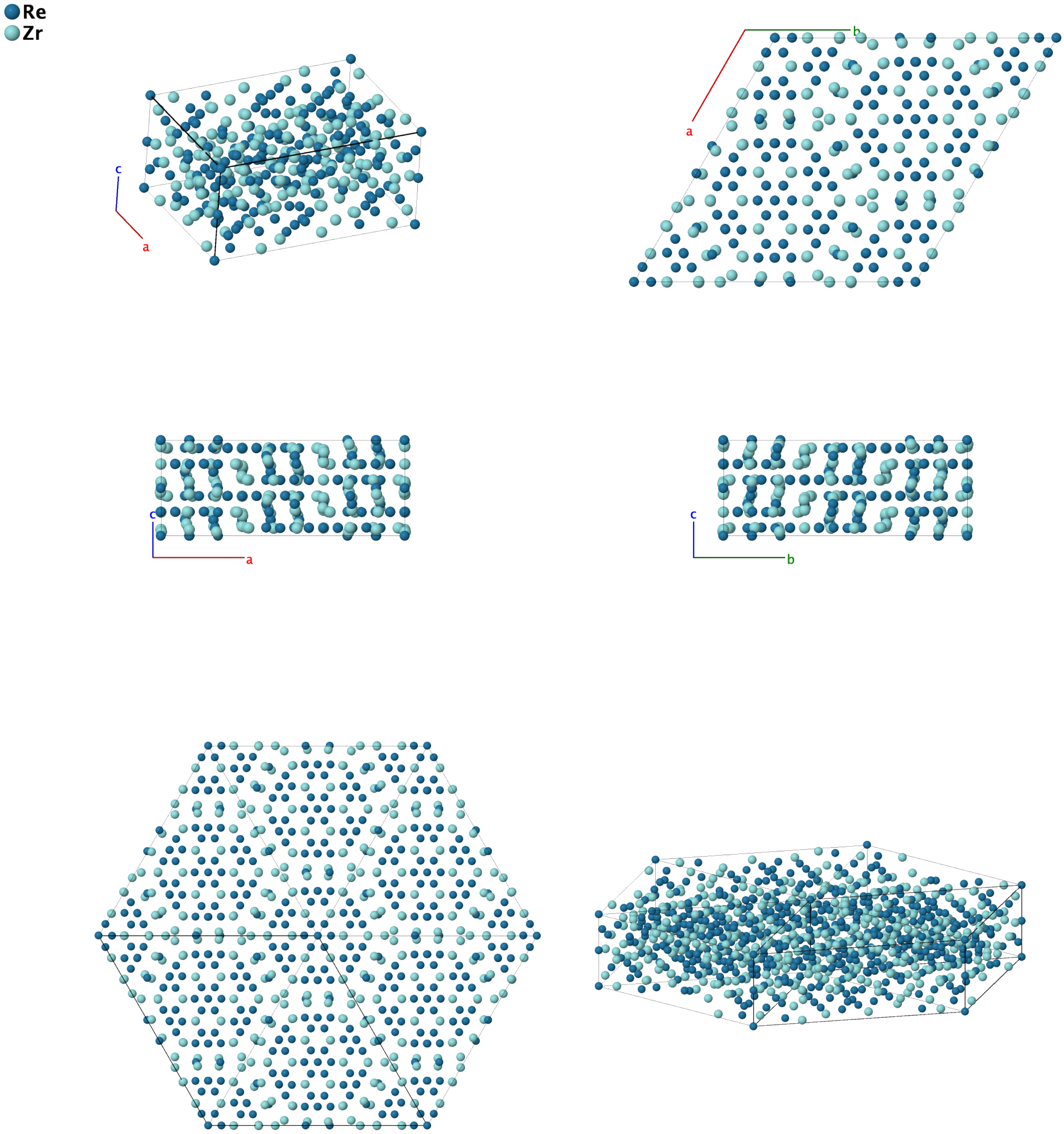

| | | |
|---|---|---|
| **Prototype** | : | $Re_{25}Zr_{21}$ |
| **AFLOW prototype label** | : | A25B21_hR92_167_b2e3f_e3f |
| ***Strukturbericht* designation** | : | None |
| **Pearson symbol** | : | hR92 |
| **Space group number** | : | 167 |
| **Space group symbol** | : | $R\bar{3}c$ |

**AFLOW prototype command** : `aflow --proto=A25B21_hR92_167_b2e3f_e3f [--hex]`
`--params=`$a, c/a, x_{2}, x_{3}, x_{4}, x_{5}, y_{5}, z_{5}, x_{6}, y_{6}, z_{6}, x_{7}, y_{7}, z_{7}, x_{8}, y_{8}, z_{8}, x_{9}, y_{9}, z_{9}, x_{10}, y_{10}, z_{10}$

---

### Other compounds with this structure

- $Mg_{21}Zn_{25}$

---

### Rhombohedral primitive vectors:

$$\mathbf{a}_1 = \tfrac{1}{2}\,a\,\mathbf{\hat{x}} - \tfrac{1}{2\sqrt{3}}\,a\,\mathbf{\hat{y}} + \tfrac{1}{3}\,c\,\mathbf{\hat{z}}$$
$$\mathbf{a}_2 = \tfrac{1}{\sqrt{3}}\,a\,\mathbf{\hat{y}} + \tfrac{1}{3}\,c\,\mathbf{\hat{z}}$$
$$\mathbf{a}_3 = -\tfrac{1}{2}\,a\,\mathbf{\hat{x}} - \tfrac{1}{2\sqrt{3}}\,a\,\mathbf{\hat{y}} + \tfrac{1}{3}\,c\,\mathbf{\hat{z}}$$

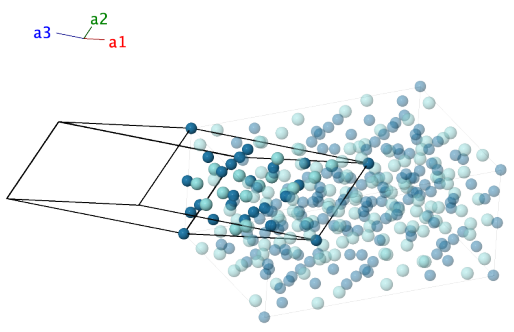

---

### Basis vectors:

| | | Lattice Coordinates | | Cartesian Coordinates | Wyckoff Position | Atom Type |
|---|---|---|---|---|---|---|
| $\mathbf{B}_1$ | = | $0\,\mathbf{a}_1 + 0\,\mathbf{a}_2 + 0\,\mathbf{a}_3$ | = | $0\,\mathbf{\hat{x}} + 0\,\mathbf{\hat{y}} + 0\,\mathbf{\hat{z}}$ | (2*b*) | Re I |
| $\mathbf{B}_2$ | = | $\tfrac{1}{2}\,\mathbf{a}_1 + \tfrac{1}{2}\,\mathbf{a}_2 + \tfrac{1}{2}\,\mathbf{a}_3$ | = | $\tfrac{1}{2}c\,\mathbf{\hat{z}}$ | (2*b*) | Re I |
| $\mathbf{B}_3$ | = | $x_2\,\mathbf{a}_1 + \left(\tfrac{1}{2} - x_2\right)\,\mathbf{a}_2 + \tfrac{1}{4}\,\mathbf{a}_3$ | = | $\left(-\tfrac{1}{8} + \tfrac{1}{2}x_2\right)a\,\mathbf{\hat{x}} + \left(\tfrac{\sqrt{3}}{8} - \tfrac{\sqrt{3}}{2}x_2\right)a\,\mathbf{\hat{y}} + \tfrac{1}{4}c\,\mathbf{\hat{z}}$ | (6*e*) | Re II |
| $\mathbf{B}_4$ | = | $\tfrac{1}{4}\,\mathbf{a}_1 + x_2\,\mathbf{a}_2 + \left(\tfrac{1}{2} - x_2\right)\,\mathbf{a}_3$ | = | $\left(-\tfrac{1}{8} + \tfrac{1}{2}x_2\right)a\,\mathbf{\hat{x}} + \left(-\tfrac{\sqrt{3}}{8} + \tfrac{\sqrt{3}}{2}x_2\right)a\,\mathbf{\hat{y}} + \tfrac{1}{4}c\,\mathbf{\hat{z}}$ | (6*e*) | Re II |
| $\mathbf{B}_5$ | = | $\left(\tfrac{1}{2} - x_2\right)\,\mathbf{a}_1 + \tfrac{1}{4}\,\mathbf{a}_2 + x_2\,\mathbf{a}_3$ | = | $\left(\tfrac{1}{4} - x_2\right)a\,\mathbf{\hat{x}} + \tfrac{1}{4}c\,\mathbf{\hat{z}}$ | (6*e*) | Re II |
| $\mathbf{B}_6$ | = | $-x_2\,\mathbf{a}_1 + \left(\tfrac{1}{2} + x_2\right)\,\mathbf{a}_2 + \tfrac{3}{4}\,\mathbf{a}_3$ | = | $-a\left(\tfrac{1}{2}x_2 + \tfrac{3}{8}\right)\,\mathbf{\hat{x}} + \left(\tfrac{1}{8\sqrt{3}} + \tfrac{\sqrt{3}}{2}x_2\right)a\,\mathbf{\hat{y}} + \tfrac{5}{12}c\,\mathbf{\hat{z}}$ | (6*e*) | Re II |
| $\mathbf{B}_7$ | = | $\tfrac{3}{4}\,\mathbf{a}_1 - x_2\,\mathbf{a}_2 + \left(\tfrac{1}{2} + x_2\right)\,\mathbf{a}_3$ | = | $\left(\tfrac{1}{8} - \tfrac{1}{2}x_2\right)a\,\mathbf{\hat{x}} - a\left(\tfrac{\sqrt{3}}{2}x_2 + \tfrac{5}{8\sqrt{3}}\right)\,\mathbf{\hat{y}} + \tfrac{5}{12}c\,\mathbf{\hat{z}}$ | (6*e*) | Re II |
| $\mathbf{B}_8$ | = | $\left(\tfrac{1}{2} + x_2\right)\,\mathbf{a}_1 + \tfrac{3}{4}\,\mathbf{a}_2 - x_2\,\mathbf{a}_3$ | = | $\left(\tfrac{1}{4} + x_2\right)a\,\mathbf{\hat{x}} + \tfrac{1}{2\sqrt{3}}a\,\mathbf{\hat{y}} + \tfrac{5}{12}c\,\mathbf{\hat{z}}$ | (6*e*) | Re II |
| $\mathbf{B}_9$ | = | $x_3\,\mathbf{a}_1 + \left(\tfrac{1}{2} - x_3\right)\,\mathbf{a}_2 + \tfrac{1}{4}\,\mathbf{a}_3$ | = | $\left(-\tfrac{1}{8} + \tfrac{1}{2}x_3\right)a\,\mathbf{\hat{x}} + \left(\tfrac{\sqrt{3}}{8} - \tfrac{\sqrt{3}}{2}x_3\right)a\,\mathbf{\hat{y}} + \tfrac{1}{4}c\,\mathbf{\hat{z}}$ | (6*e*) | Re III |
| $\mathbf{B}_{10}$ | = | $\tfrac{1}{4}\,\mathbf{a}_1 + x_3\,\mathbf{a}_2 + \left(\tfrac{1}{2} - x_3\right)\,\mathbf{a}_3$ | = | $\left(-\tfrac{1}{8} + \tfrac{1}{2}x_3\right)a\,\mathbf{\hat{x}} + \left(-\tfrac{\sqrt{3}}{8} + \tfrac{\sqrt{3}}{2}x_3\right)a\,\mathbf{\hat{y}} + \tfrac{1}{4}c\,\mathbf{\hat{z}}$ | (6*e*) | Re III |
| $\mathbf{B}_{11}$ | = | $\left(\tfrac{1}{2} - x_3\right)\,\mathbf{a}_1 + \tfrac{1}{4}\,\mathbf{a}_2 + x_3\,\mathbf{a}_3$ | = | $\left(\tfrac{1}{4} - x_3\right)a\,\mathbf{\hat{x}} + \tfrac{1}{4}c\,\mathbf{\hat{z}}$ | (6*e*) | Re III |
| $\mathbf{B}_{12}$ | = | $-x_3\,\mathbf{a}_1 + \left(\tfrac{1}{2} + x_3\right)\,\mathbf{a}_2 + \tfrac{3}{4}\,\mathbf{a}_3$ | = | $-a\left(\tfrac{1}{2}x_3 + \tfrac{3}{8}\right)\,\mathbf{\hat{x}} + \left(\tfrac{1}{8\sqrt{3}} + \tfrac{\sqrt{3}}{2}x_3\right)a\,\mathbf{\hat{y}} + \tfrac{5}{12}c\,\mathbf{\hat{z}}$ | (6*e*) | Re III |
| $\mathbf{B}_{13}$ | = | $\tfrac{3}{4}\,\mathbf{a}_1 - x_3\,\mathbf{a}_2 + \left(\tfrac{1}{2} + x_3\right)\,\mathbf{a}_3$ | = | $\left(\tfrac{1}{8} - \tfrac{1}{2}x_3\right)a\,\mathbf{\hat{x}} - a\left(\tfrac{\sqrt{3}}{2}x_3 + \tfrac{5}{8\sqrt{3}}\right)\,\mathbf{\hat{y}} + \tfrac{5}{12}c\,\mathbf{\hat{z}}$ | (6*e*) | Re III |
| $\mathbf{B}_{14}$ | = | $\left(\tfrac{1}{2} + x_3\right)\,\mathbf{a}_1 + \tfrac{3}{4}\,\mathbf{a}_2 - x_3\,\mathbf{a}_3$ | = | $\left(\tfrac{1}{4} + x_3\right)a\,\mathbf{\hat{x}} + \tfrac{1}{2\sqrt{3}}a\,\mathbf{\hat{y}} + \tfrac{5}{12}c\,\mathbf{\hat{z}}$ | (6*e*) | Re III |
| $\mathbf{B}_{15}$ | = | $x_4\,\mathbf{a}_1 + \left(\frac{1}{2} - x_4\right)\,\mathbf{a}_2 + \frac{1}{4}\,\mathbf{a}_3$ | = | $\left(-\frac{1}{8} + \frac{1}{2}x_4\right)a\,\hat{\mathbf{x}} + \left(\frac{\sqrt{3}}{8} - \frac{\sqrt{3}}{2}x_4\right)a\,\hat{\mathbf{y}} + \frac{1}{4}c\,\hat{\mathbf{z}}$ | (6$e$) | Zr I |
| $\mathbf{B}_{16}$ | = | $\frac{1}{4}\,\mathbf{a}_1 + x_4\,\mathbf{a}_2 + \left(\frac{1}{2} - x_4\right)\,\mathbf{a}_3$ | = | $\left(-\frac{1}{8} + \frac{1}{2}x_4\right)a\,\hat{\mathbf{x}} + \left(-\frac{\sqrt{3}}{8} + \frac{\sqrt{3}}{2}x_4\right)a\,\hat{\mathbf{y}} + \frac{1}{4}c\,\hat{\mathbf{z}}$ | (6$e$) | Zr I |
| $\mathbf{B}_{17}$ | = | $\left(\frac{1}{2} - x_4\right)\,\mathbf{a}_1 + \frac{1}{4}\,\mathbf{a}_2 + x_4\,\mathbf{a}_3$ | = | $\left(\frac{1}{4} - x_4\right)a\,\hat{\mathbf{x}} + \frac{1}{4}c\,\hat{\mathbf{z}}$ | (6$e$) | Zr I |
| $\mathbf{B}_{18}$ | = | $-x_4\,\mathbf{a}_1 + \left(\frac{1}{2} + x_4\right)\,\mathbf{a}_2 + \frac{3}{4}\,\mathbf{a}_3$ | = | $-a\left(\frac{1}{2}x_4 + \frac{3}{8}\right)\,\hat{\mathbf{x}} + \left(\frac{1}{8\sqrt{3}} + \frac{\sqrt{3}}{2}x_4\right)a\,\hat{\mathbf{y}} + \frac{5}{12}c\,\hat{\mathbf{z}}$ | (6$e$) | Zr I |
| $\mathbf{B}_{19}$ | = | $\frac{3}{4}\,\mathbf{a}_1 - x_4\,\mathbf{a}_2 + \left(\frac{1}{2} + x_4\right)\,\mathbf{a}_3$ | = | $\left(\frac{1}{8} - \frac{1}{2}x_4\right)a\,\hat{\mathbf{x}} - a\left(\frac{\sqrt{3}}{2}x_4 + \frac{5}{8\sqrt{3}}\right)\,\hat{\mathbf{y}} + \frac{5}{12}c\,\hat{\mathbf{z}}$ | (6$e$) | Zr I |
| $\mathbf{B}_{20}$ | = | $\left(\frac{1}{2} + x_4\right)\,\mathbf{a}_1 + \frac{3}{4}\,\mathbf{a}_2 - x_4\,\mathbf{a}_3$ | = | $\left(\frac{1}{4} + x_4\right)a\,\hat{\mathbf{x}} + \frac{1}{2\sqrt{3}}a\,\hat{\mathbf{y}} + \frac{5}{12}c\,\hat{\mathbf{z}}$ | (6$e$) | Zr I |
| $\mathbf{B}_{21}$ | = | $x_5\,\mathbf{a}_1 + y_5\,\mathbf{a}_2 + z_5\,\mathbf{a}_3$ | = | $\frac{1}{2}\left(x_5 - z_5\right)a\,\hat{\mathbf{x}} + \left(-\frac{1}{2\sqrt{3}}x_5 + \frac{1}{\sqrt{3}}y_5 - \frac{1}{2\sqrt{3}}z_5\right)a\,\hat{\mathbf{y}} + \frac{1}{3}\left(x_5 + y_5 + z_5\right)c\,\hat{\mathbf{z}}$ | (12$f$) | Re IV |
| $\mathbf{B}_{22}$ | = | $z_5\,\mathbf{a}_1 + x_5\,\mathbf{a}_2 + y_5\,\mathbf{a}_3$ | = | $\frac{1}{2}\left(-y_5 + z_5\right)a\,\hat{\mathbf{x}} + \left(\frac{1}{\sqrt{3}}x_5 - \frac{1}{2\sqrt{3}}y_5 - \frac{1}{2\sqrt{3}}z_5\right)a\,\hat{\mathbf{y}} + \frac{1}{3}\left(x_5 + y_5 + z_5\right)c\,\hat{\mathbf{z}}$ | (12$f$) | Re IV |
| $\mathbf{B}_{23}$ | = | $y_5\,\mathbf{a}_1 + z_5\,\mathbf{a}_2 + x_5\,\mathbf{a}_3$ | = | $\frac{1}{2}\left(-x_5 + y_5\right)a\,\hat{\mathbf{x}} + \left(-\frac{1}{2\sqrt{3}}x_5 - \frac{1}{2\sqrt{3}}y_5 + \frac{1}{\sqrt{3}}z_5\right)a\,\hat{\mathbf{y}} + \frac{1}{3}\left(x_5 + y_5 + z_5\right)c\,\hat{\mathbf{z}}$ | (12$f$) | Re IV |
| $\mathbf{B}_{24}$ | = | $\left(\frac{1}{2} - z_5\right)\,\mathbf{a}_1 + \left(\frac{1}{2} - y_5\right)\,\mathbf{a}_2 + \left(\frac{1}{2} - x_5\right)\,\mathbf{a}_3$ | = | $\frac{1}{2}\left(x_5 - z_5\right)a\,\hat{\mathbf{x}} + \left(\frac{1}{2\sqrt{3}}x_5 - \frac{1}{\sqrt{3}}y_5 + \frac{1}{2\sqrt{3}}z_5\right)a\,\hat{\mathbf{y}} + \left(\frac{1}{2} - \frac{1}{3}x_5 - \frac{1}{3}y_5 - \frac{1}{3}z_5\right)c\,\hat{\mathbf{z}}$ | (12$f$) | Re IV |
| $\mathbf{B}_{25}$ | = | $\left(\frac{1}{2} - y_5\right)\,\mathbf{a}_1 + \left(\frac{1}{2} - x_5\right)\,\mathbf{a}_2 + \left(\frac{1}{2} - z_5\right)\,\mathbf{a}_3$ | = | $\frac{1}{2}\left(-y_5 + z_5\right)a\,\hat{\mathbf{x}} + \left(-\frac{1}{\sqrt{3}}x_5 + \frac{1}{2\sqrt{3}}y_5 + \frac{1}{2\sqrt{3}}z_5\right)a\,\hat{\mathbf{y}} + \left(\frac{1}{2} - \frac{1}{3}x_5 - \frac{1}{3}y_5 - \frac{1}{3}z_5\right)c\,\hat{\mathbf{z}}$ | (12$f$) | Re IV |
| $\mathbf{B}_{26}$ | = | $\left(\frac{1}{2} - x_5\right)\,\mathbf{a}_1 + \left(\frac{1}{2} - z_5\right)\,\mathbf{a}_2 + \left(\frac{1}{2} - y_5\right)\,\mathbf{a}_3$ | = | $\frac{1}{2}\left(-x_5 + y_5\right)a\,\hat{\mathbf{x}} + \left(\frac{1}{2\sqrt{3}}x_5 + \frac{1}{2\sqrt{3}}y_5 - \frac{1}{\sqrt{3}}z_5\right)a\,\hat{\mathbf{y}} + \left(\frac{1}{2} - \frac{1}{3}x_5 - \frac{1}{3}y_5 - \frac{1}{3}z_5\right)c\,\hat{\mathbf{z}}$ | (12$f$) | Re IV |
| $\mathbf{B}_{27}$ | = | $-x_5\,\mathbf{a}_1 - y_5\,\mathbf{a}_2 - z_5\,\mathbf{a}_3$ | = | $\frac{1}{2}\left(-x_5 + z_5\right)a\,\hat{\mathbf{x}} + \left(\frac{1}{2\sqrt{3}}x_5 - \frac{1}{\sqrt{3}}y_5 + \frac{1}{2\sqrt{3}}z_5\right)a\,\hat{\mathbf{y}} - \frac{1}{3}\left(x_5 + y_5 + z_5\right)c\,\hat{\mathbf{z}}$ | (12$f$) | Re IV |
| $\mathbf{B}_{28}$ | = | $-z_5\,\mathbf{a}_1 - x_5\,\mathbf{a}_2 - y_5\,\mathbf{a}_3$ | = | $\frac{1}{2}\left(y_5 - z_5\right)a\,\hat{\mathbf{x}} + \left(-\frac{1}{\sqrt{3}}x_5 + \frac{1}{2\sqrt{3}}y_5 + \frac{1}{2\sqrt{3}}z_5\right)a\,\hat{\mathbf{y}} - \frac{1}{3}\left(x_5 + y_5 + z_5\right)c\,\hat{\mathbf{z}}$ | (12$f$) | Re IV |
| $\mathbf{B}_{29}$ | = | $-y_5\,\mathbf{a}_1 - z_5\,\mathbf{a}_2 - x_5\,\mathbf{a}_3$ | = | $\frac{1}{2}\left(x_5 - y_5\right)a\,\hat{\mathbf{x}} + \left(\frac{1}{2\sqrt{3}}x_5 + \frac{1}{2\sqrt{3}}y_5 - \frac{1}{\sqrt{3}}z_5\right)a\,\hat{\mathbf{y}} - \frac{1}{3}\left(x_5 + y_5 + z_5\right)c\,\hat{\mathbf{z}}$ | (12$f$) | Re IV |
| $\mathbf{B}_{30}$ | = | $\left(\frac{1}{2}+z_5\right)\mathbf{a}_1+\left(\frac{1}{2}+y_5\right)\mathbf{a}_2+\left(\frac{1}{2}+x_5\right)\mathbf{a}_3$ | = | $\frac{1}{2}\left(-x_5+z_5\right)a\,\hat{\mathbf{x}}+\left(-\frac{1}{2\sqrt{3}}x_5+\frac{1}{\sqrt{3}}y_5-\frac{1}{2\sqrt{3}}z_5\right)a\,\hat{\mathbf{y}}+\left(\frac{1}{2}+\frac{1}{3}x_5+\frac{1}{3}y_5+\frac{1}{3}z_5\right)c\,\hat{\mathbf{z}}$ | (12$f$) | Re IV |
| $\mathbf{B}_{31}$ | = | $\left(\frac{1}{2}+y_5\right)\mathbf{a}_1+\left(\frac{1}{2}+x_5\right)\mathbf{a}_2+\left(\frac{1}{2}+z_5\right)\mathbf{a}_3$ | = | $\frac{1}{2}\left(y_5-z_5\right)a\,\hat{\mathbf{x}}+\left(\frac{1}{\sqrt{3}}x_5-\frac{1}{2\sqrt{3}}y_5-\frac{1}{2\sqrt{3}}z_5\right)a\,\hat{\mathbf{y}}+\left(\frac{1}{2}+\frac{1}{3}x_5+\frac{1}{3}y_5+\frac{1}{3}z_5\right)c\,\hat{\mathbf{z}}$ | (12$f$) | Re IV |
| $\mathbf{B}_{32}$ | = | $\left(\frac{1}{2}+x_5\right)\mathbf{a}_1+\left(\frac{1}{2}+z_5\right)\mathbf{a}_2+\left(\frac{1}{2}+y_5\right)\mathbf{a}_3$ | = | $\frac{1}{2}\left(x_5-y_5\right)a\,\hat{\mathbf{x}}+\left(-\frac{1}{2\sqrt{3}}x_5-\frac{1}{2\sqrt{3}}y_5+\frac{1}{\sqrt{3}}z_5\right)a\,\hat{\mathbf{y}}+\left(\frac{1}{2}+\frac{1}{3}x_5+\frac{1}{3}y_5+\frac{1}{3}z_5\right)c\,\hat{\mathbf{z}}$ | (12$f$) | Re IV |
| $\mathbf{B}_{33}$ | = | $x_6\,\mathbf{a}_1+y_6\,\mathbf{a}_2+z_6\,\mathbf{a}_3$ | = | $\frac{1}{2}\left(x_6-z_6\right)a\,\hat{\mathbf{x}}+\left(-\frac{1}{2\sqrt{3}}x_6+\frac{1}{\sqrt{3}}y_6-\frac{1}{2\sqrt{3}}z_6\right)a\,\hat{\mathbf{y}}+\frac{1}{3}\left(x_6+y_6+z_6\right)c\,\hat{\mathbf{z}}$ | (12$f$) | Re V |
| $\mathbf{B}_{34}$ | = | $z_6\,\mathbf{a}_1+x_6\,\mathbf{a}_2+y_6\,\mathbf{a}_3$ | = | $\frac{1}{2}\left(-y_6+z_6\right)a\,\hat{\mathbf{x}}+\left(\frac{1}{\sqrt{3}}x_6-\frac{1}{2\sqrt{3}}y_6-\frac{1}{2\sqrt{3}}z_6\right)a\,\hat{\mathbf{y}}+\frac{1}{3}\left(x_6+y_6+z_6\right)c\,\hat{\mathbf{z}}$ | (12$f$) | Re V |
| $\mathbf{B}_{35}$ | = | $y_6\,\mathbf{a}_1+z_6\,\mathbf{a}_2+x_6\,\mathbf{a}_3$ | = | $\frac{1}{2}\left(-x_6+y_6\right)a\,\hat{\mathbf{x}}+\left(-\frac{1}{2\sqrt{3}}x_6-\frac{1}{2\sqrt{3}}y_6+\frac{1}{\sqrt{3}}z_6\right)a\,\hat{\mathbf{y}}+\frac{1}{3}\left(x_6+y_6+z_6\right)c\,\hat{\mathbf{z}}$ | (12$f$) | Re V |
| $\mathbf{B}_{36}$ | = | $\left(\frac{1}{2}-z_6\right)\mathbf{a}_1+\left(\frac{1}{2}-y_6\right)\mathbf{a}_2+\left(\frac{1}{2}-x_6\right)\mathbf{a}_3$ | = | $\frac{1}{2}\left(x_6-z_6\right)a\,\hat{\mathbf{x}}+\left(\frac{1}{2\sqrt{3}}x_6-\frac{1}{\sqrt{3}}y_6+\frac{1}{2\sqrt{3}}z_6\right)a\,\hat{\mathbf{y}}+\left(\frac{1}{2}-\frac{1}{3}x_6-\frac{1}{3}y_6-\frac{1}{3}z_6\right)c\,\hat{\mathbf{z}}$ | (12$f$) | Re V |
| $\mathbf{B}_{37}$ | = | $\left(\frac{1}{2}-y_6\right)\mathbf{a}_1+\left(\frac{1}{2}-x_6\right)\mathbf{a}_2+\left(\frac{1}{2}-z_6\right)\mathbf{a}_3$ | = | $\frac{1}{2}\left(-y_6+z_6\right)a\,\hat{\mathbf{x}}+\left(-\frac{1}{\sqrt{3}}x_6+\frac{1}{2\sqrt{3}}y_6+\frac{1}{2\sqrt{3}}z_6\right)a\,\hat{\mathbf{y}}+\left(\frac{1}{2}-\frac{1}{3}x_6-\frac{1}{3}y_6-\frac{1}{3}z_6\right)c\,\hat{\mathbf{z}}$ | (12$f$) | Re V |
| $\mathbf{B}_{38}$ | = | $\left(\frac{1}{2}-x_6\right)\mathbf{a}_1+\left(\frac{1}{2}-z_6\right)\mathbf{a}_2+\left(\frac{1}{2}-y_6\right)\mathbf{a}_3$ | = | $\frac{1}{2}\left(-x_6+y_6\right)a\,\hat{\mathbf{x}}+\left(\frac{1}{2\sqrt{3}}x_6+\frac{1}{2\sqrt{3}}y_6-\frac{1}{\sqrt{3}}z_6\right)a\,\hat{\mathbf{y}}+\left(\frac{1}{2}-\frac{1}{3}x_6-\frac{1}{3}y_6-\frac{1}{3}z_6\right)c\,\hat{\mathbf{z}}$ | (12$f$) | Re V |
| $\mathbf{B}_{39}$ | = | $-x_6\,\mathbf{a}_1-y_6\,\mathbf{a}_2-z_6\,\mathbf{a}_3$ | = | $\frac{1}{2}\left(-x_6+z_6\right)a\,\hat{\mathbf{x}}+\left(\frac{1}{2\sqrt{3}}x_6-\frac{1}{\sqrt{3}}y_6+\frac{1}{2\sqrt{3}}z_6\right)a\,\hat{\mathbf{y}}-\frac{1}{3}\left(x_6+y_6+z_6\right)c\,\hat{\mathbf{z}}$ | (12$f$) | Re V |
| $\mathbf{B}_{40}$ | = | $-z_6\,\mathbf{a}_1-x_6\,\mathbf{a}_2-y_6\,\mathbf{a}_3$ | = | $\frac{1}{2}\left(y_6-z_6\right)a\,\hat{\mathbf{x}}+\left(-\frac{1}{\sqrt{3}}x_6+\frac{1}{2\sqrt{3}}y_6+\frac{1}{2\sqrt{3}}z_6\right)a\,\hat{\mathbf{y}}-\frac{1}{3}\left(x_6+y_6+z_6\right)c\,\hat{\mathbf{z}}$ | (12$f$) | Re V |
| $\mathbf{B}_{41}$ | = | $-y_6\,\mathbf{a}_1-z_6\,\mathbf{a}_2-x_6\,\mathbf{a}_3$ | = | $\frac{1}{2}\left(x_6-y_6\right)a\,\hat{\mathbf{x}}+\left(\frac{1}{2\sqrt{3}}x_6+\frac{1}{2\sqrt{3}}y_6-\frac{1}{\sqrt{3}}z_6\right)a\,\hat{\mathbf{y}}-\frac{1}{3}\left(x_6+y_6+z_6\right)c\,\hat{\mathbf{z}}$ | (12$f$) | Re V |
| $\mathbf{B}_{42}$ | = | $\left(\frac{1}{2}+z_6\right)\mathbf{a}_1+\left(\frac{1}{2}+y_6\right)\mathbf{a}_2+\left(\frac{1}{2}+x_6\right)\mathbf{a}_3$ | = | $\frac{1}{2}\left(-x_6+z_6\right)a\,\hat{\mathbf{x}}+\left(-\frac{1}{2\sqrt{3}}x_6+\frac{1}{\sqrt{3}}y_6-\frac{1}{2\sqrt{3}}z_6\right)a\,\hat{\mathbf{y}}+\left(\frac{1}{2}+\frac{1}{3}x_6+\frac{1}{3}y_6+\frac{1}{3}z_6\right)c\,\hat{\mathbf{z}}$ | (12$f$) | Re V |

$$\mathbf{B}_{43} = \left(\frac{1}{2} + y_6\right)\mathbf{a}_1 + \left(\frac{1}{2} + x_6\right)\mathbf{a}_2 + \left(\frac{1}{2} + z_6\right)\mathbf{a}_3 = \frac{1}{2}(y_6 - z_6)\,a\,\hat{\mathbf{x}} + \left(\frac{1}{\sqrt{3}}x_6 - \frac{1}{2\sqrt{3}}y_6 - \frac{1}{2\sqrt{3}}z_6\right)a\,\hat{\mathbf{y}} + \left(\frac{1}{2} + \frac{1}{3}x_6 + \frac{1}{3}y_6 + \frac{1}{3}z_6\right)c\,\hat{\mathbf{z}} \quad (12f) \quad \text{Re V}$$

$$\mathbf{B}_{44} = \left(\frac{1}{2} + x_6\right)\mathbf{a}_1 + \left(\frac{1}{2} + z_6\right)\mathbf{a}_2 + \left(\frac{1}{2} + y_6\right)\mathbf{a}_3 = \frac{1}{2}(x_6 - y_6)\,a\,\hat{\mathbf{x}} + \left(-\frac{1}{2\sqrt{3}}x_6 - \frac{1}{2\sqrt{3}}y_6 + \frac{1}{\sqrt{3}}z_6\right)a\,\hat{\mathbf{y}} + \left(\frac{1}{2} + \frac{1}{3}x_6 + \frac{1}{3}y_6 + \frac{1}{3}z_6\right)c\,\hat{\mathbf{z}} \quad (12f) \quad \text{Re V}$$

$$\mathbf{B}_{45} = x_7\,\mathbf{a}_1 + y_7\,\mathbf{a}_2 + z_7\,\mathbf{a}_3 = \frac{1}{2}(x_7 - z_7)\,a\,\hat{\mathbf{x}} + \left(-\frac{1}{2\sqrt{3}}x_7 + \frac{1}{\sqrt{3}}y_7 - \frac{1}{2\sqrt{3}}z_7\right)a\,\hat{\mathbf{y}} + \frac{1}{3}(x_7 + y_7 + z_7)\,c\,\hat{\mathbf{z}} \quad (12f) \quad \text{Re VI}$$

$$\mathbf{B}_{46} = z_7\,\mathbf{a}_1 + x_7\,\mathbf{a}_2 + y_7\,\mathbf{a}_3 = \frac{1}{2}(-y_7 + z_7)\,a\,\hat{\mathbf{x}} + \left(\frac{1}{\sqrt{3}}x_7 - \frac{1}{2\sqrt{3}}y_7 - \frac{1}{2\sqrt{3}}z_7\right)a\,\hat{\mathbf{y}} + \frac{1}{3}(x_7 + y_7 + z_7)\,c\,\hat{\mathbf{z}} \quad (12f) \quad \text{Re VI}$$

$$\mathbf{B}_{47} = y_7\,\mathbf{a}_1 + z_7\,\mathbf{a}_2 + x_7\,\mathbf{a}_3 = \frac{1}{2}(-x_7 + y_7)\,a\,\hat{\mathbf{x}} + \left(-\frac{1}{2\sqrt{3}}x_7 - \frac{1}{2\sqrt{3}}y_7 + \frac{1}{\sqrt{3}}z_7\right)a\,\hat{\mathbf{y}} + \frac{1}{3}(x_7 + y_7 + z_7)\,c\,\hat{\mathbf{z}} \quad (12f) \quad \text{Re VI}$$

$$\mathbf{B}_{48} = \left(\frac{1}{2} - z_7\right)\mathbf{a}_1 + \left(\frac{1}{2} - y_7\right)\mathbf{a}_2 + \left(\frac{1}{2} - x_7\right)\mathbf{a}_3 = \frac{1}{2}(x_7 - z_7)\,a\,\hat{\mathbf{x}} + \left(\frac{1}{2\sqrt{3}}x_7 - \frac{1}{\sqrt{3}}y_7 + \frac{1}{2\sqrt{3}}z_7\right)a\,\hat{\mathbf{y}} + \left(\frac{1}{2} - \frac{1}{3}x_7 - \frac{1}{3}y_7 - \frac{1}{3}z_7\right)c\,\hat{\mathbf{z}} \quad (12f) \quad \text{Re VI}$$

$$\mathbf{B}_{49} = \left(\frac{1}{2} - y_7\right)\mathbf{a}_1 + \left(\frac{1}{2} - x_7\right)\mathbf{a}_2 + \left(\frac{1}{2} - z_7\right)\mathbf{a}_3 = \frac{1}{2}(-y_7 + z_7)\,a\,\hat{\mathbf{x}} + \left(-\frac{1}{\sqrt{3}}x_7 + \frac{1}{2\sqrt{3}}y_7 + \frac{1}{2\sqrt{3}}z_7\right)a\,\hat{\mathbf{y}} + \left(\frac{1}{2} - \frac{1}{3}x_7 - \frac{1}{3}y_7 - \frac{1}{3}z_7\right)c\,\hat{\mathbf{z}} \quad (12f) \quad \text{Re VI}$$

$$\mathbf{B}_{50} = \left(\frac{1}{2} - x_7\right)\mathbf{a}_1 + \left(\frac{1}{2} - z_7\right)\mathbf{a}_2 + \left(\frac{1}{2} - y_7\right)\mathbf{a}_3 = \frac{1}{2}(-x_7 + y_7)\,a\,\hat{\mathbf{x}} + \left(\frac{1}{2\sqrt{3}}x_7 + \frac{1}{2\sqrt{3}}y_7 - \frac{1}{\sqrt{3}}z_7\right)a\,\hat{\mathbf{y}} + \left(\frac{1}{2} - \frac{1}{3}x_7 - \frac{1}{3}y_7 - \frac{1}{3}z_7\right)c\,\hat{\mathbf{z}} \quad (12f) \quad \text{Re VI}$$

$$\mathbf{B}_{51} = -x_7\,\mathbf{a}_1 - y_7\,\mathbf{a}_2 - z_7\,\mathbf{a}_3 = \frac{1}{2}(-x_7 + z_7)\,a\,\hat{\mathbf{x}} + \left(\frac{1}{2\sqrt{3}}x_7 - \frac{1}{\sqrt{3}}y_7 + \frac{1}{2\sqrt{3}}z_7\right)a\,\hat{\mathbf{y}} - \frac{1}{3}(x_7 + y_7 + z_7)\,c\,\hat{\mathbf{z}} \quad (12f) \quad \text{Re VI}$$

$$\mathbf{B}_{52} = -z_7\,\mathbf{a}_1 - x_7\,\mathbf{a}_2 - y_7\,\mathbf{a}_3 = \frac{1}{2}(y_7 - z_7)\,a\,\hat{\mathbf{x}} + \left(-\frac{1}{\sqrt{3}}x_7 + \frac{1}{2\sqrt{3}}y_7 + \frac{1}{2\sqrt{3}}z_7\right)a\,\hat{\mathbf{y}} - \frac{1}{3}(x_7 + y_7 + z_7)\,c\,\hat{\mathbf{z}} \quad (12f) \quad \text{Re VI}$$

$$\mathbf{B}_{53} = -y_7\,\mathbf{a}_1 - z_7\,\mathbf{a}_2 - x_7\,\mathbf{a}_3 = \frac{1}{2}(x_7 - y_7)\,a\,\hat{\mathbf{x}} + \left(\frac{1}{2\sqrt{3}}x_7 + \frac{1}{2\sqrt{3}}y_7 - \frac{1}{\sqrt{3}}z_7\right)a\,\hat{\mathbf{y}} - \frac{1}{3}(x_7 + y_7 + z_7)\,c\,\hat{\mathbf{z}} \quad (12f) \quad \text{Re VI}$$

$$\mathbf{B}_{54} = \left(\frac{1}{2} + z_7\right)\mathbf{a}_1 + \left(\frac{1}{2} + y_7\right)\mathbf{a}_2 + \left(\frac{1}{2} + x_7\right)\mathbf{a}_3 = \frac{1}{2}(-x_7 + z_7)\,a\,\hat{\mathbf{x}} + \left(-\frac{1}{2\sqrt{3}}x_7 + \frac{1}{\sqrt{3}}y_7 - \frac{1}{2\sqrt{3}}z_7\right)a\,\hat{\mathbf{y}} + \left(\frac{1}{2} + \frac{1}{3}x_7 + \frac{1}{3}y_7 + \frac{1}{3}z_7\right)c\,\hat{\mathbf{z}} \quad (12f) \quad \text{Re VI}$$

$$\mathbf{B}_{55} = \left(\frac{1}{2} + y_7\right)\mathbf{a}_1 + \left(\frac{1}{2} + x_7\right)\mathbf{a}_2 + \left(\frac{1}{2} + z_7\right)\mathbf{a}_3 = \frac{1}{2}(y_7 - z_7)\,a\,\hat{\mathbf{x}} + \left(\frac{1}{\sqrt{3}}x_7 - \frac{1}{2\sqrt{3}}y_7 - \frac{1}{2\sqrt{3}}z_7\right)a\,\hat{\mathbf{y}} + \left(\frac{1}{2} + \frac{1}{3}x_7 + \frac{1}{3}y_7 + \frac{1}{3}z_7\right)c\,\hat{\mathbf{z}} \quad (12f) \quad \text{Re VI}$$

$$\mathbf{B}_{56} = \left(\frac{1}{2}+x_7\right)\mathbf{a}_1 + \left(\frac{1}{2}+z_7\right)\mathbf{a}_2 + \left(\frac{1}{2}+y_7\right)\mathbf{a}_3 = \frac{1}{2}\left(x_7-y_7\right)a\,\hat{\mathbf{x}} + \left(-\frac{1}{2\sqrt{3}}x_7 - \frac{1}{2\sqrt{3}}y_7 + \frac{1}{\sqrt{3}}z_7\right)a\,\hat{\mathbf{y}} + \left(\frac{1}{2}+\frac{1}{3}x_7+\frac{1}{3}y_7+\frac{1}{3}z_7\right)c\,\hat{\mathbf{z}} \quad (12f) \quad \text{Re VI}$$

$$\mathbf{B}_{57} = x_8\,\mathbf{a}_1 + y_8\,\mathbf{a}_2 + z_8\,\mathbf{a}_3 = \frac{1}{2}\left(x_8-z_8\right)a\,\hat{\mathbf{x}} + \left(-\frac{1}{2\sqrt{3}}x_8 + \frac{1}{\sqrt{3}}y_8 - \frac{1}{2\sqrt{3}}z_8\right)a\,\hat{\mathbf{y}} + \frac{1}{3}\left(x_8+y_8+z_8\right)c\,\hat{\mathbf{z}} \quad (12f) \quad \text{Zr II}$$

$$\mathbf{B}_{58} = z_8\,\mathbf{a}_1 + x_8\,\mathbf{a}_2 + y_8\,\mathbf{a}_3 = \frac{1}{2}\left(-y_8+z_8\right)a\,\hat{\mathbf{x}} + \left(\frac{1}{\sqrt{3}}x_8 - \frac{1}{2\sqrt{3}}y_8 - \frac{1}{2\sqrt{3}}z_8\right)a\,\hat{\mathbf{y}} + \frac{1}{3}\left(x_8+y_8+z_8\right)c\,\hat{\mathbf{z}} \quad (12f) \quad \text{Zr II}$$

$$\mathbf{B}_{59} = y_8\,\mathbf{a}_1 + z_8\,\mathbf{a}_2 + x_8\,\mathbf{a}_3 = \frac{1}{2}\left(-x_8+y_8\right)a\,\hat{\mathbf{x}} + \left(-\frac{1}{2\sqrt{3}}x_8 - \frac{1}{2\sqrt{3}}y_8 + \frac{1}{\sqrt{3}}z_8\right)a\,\hat{\mathbf{y}} + \frac{1}{3}\left(x_8+y_8+z_8\right)c\,\hat{\mathbf{z}} \quad (12f) \quad \text{Zr II}$$

$$\mathbf{B}_{60} = \left(\frac{1}{2}-z_8\right)\mathbf{a}_1 + \left(\frac{1}{2}-y_8\right)\mathbf{a}_2 + \left(\frac{1}{2}-x_8\right)\mathbf{a}_3 = \frac{1}{2}\left(x_8-z_8\right)a\,\hat{\mathbf{x}} + \left(\frac{1}{2\sqrt{3}}x_8 - \frac{1}{\sqrt{3}}y_8 + \frac{1}{2\sqrt{3}}z_8\right)a\,\hat{\mathbf{y}} + \left(\frac{1}{2}-\frac{1}{3}x_8-\frac{1}{3}y_8-\frac{1}{3}z_8\right)c\,\hat{\mathbf{z}} \quad (12f) \quad \text{Zr II}$$

$$\mathbf{B}_{61} = \left(\frac{1}{2}-y_8\right)\mathbf{a}_1 + \left(\frac{1}{2}-x_8\right)\mathbf{a}_2 + \left(\frac{1}{2}-z_8\right)\mathbf{a}_3 = \frac{1}{2}\left(-y_8+z_8\right)a\,\hat{\mathbf{x}} + \left(-\frac{1}{\sqrt{3}}x_8 + \frac{1}{2\sqrt{3}}y_8 + \frac{1}{2\sqrt{3}}z_8\right)a\,\hat{\mathbf{y}} + \left(\frac{1}{2}-\frac{1}{3}x_8-\frac{1}{3}y_8-\frac{1}{3}z_8\right)c\,\hat{\mathbf{z}} \quad (12f) \quad \text{Zr II}$$

$$\mathbf{B}_{62} = \left(\frac{1}{2}-x_8\right)\mathbf{a}_1 + \left(\frac{1}{2}-z_8\right)\mathbf{a}_2 + \left(\frac{1}{2}-y_8\right)\mathbf{a}_3 = \frac{1}{2}\left(-x_8+y_8\right)a\,\hat{\mathbf{x}} + \left(\frac{1}{2\sqrt{3}}x_8 + \frac{1}{2\sqrt{3}}y_8 - \frac{1}{\sqrt{3}}z_8\right)a\,\hat{\mathbf{y}} + \left(\frac{1}{2}-\frac{1}{3}x_8-\frac{1}{3}y_8-\frac{1}{3}z_8\right)c\,\hat{\mathbf{z}} \quad (12f) \quad \text{Zr II}$$

$$\mathbf{B}_{63} = -x_8\,\mathbf{a}_1 - y_8\,\mathbf{a}_2 - z_8\,\mathbf{a}_3 = \frac{1}{2}\left(-x_8+z_8\right)a\,\hat{\mathbf{x}} + \left(\frac{1}{2\sqrt{3}}x_8 - \frac{1}{\sqrt{3}}y_8 + \frac{1}{2\sqrt{3}}z_8\right)a\,\hat{\mathbf{y}} - \frac{1}{3}\left(x_8+y_8+z_8\right)c\,\hat{\mathbf{z}} \quad (12f) \quad \text{Zr II}$$

$$\mathbf{B}_{64} = -z_8\,\mathbf{a}_1 - x_8\,\mathbf{a}_2 - y_8\,\mathbf{a}_3 = \frac{1}{2}\left(y_8-z_8\right)a\,\hat{\mathbf{x}} + \left(-\frac{1}{\sqrt{3}}x_8 + \frac{1}{2\sqrt{3}}y_8 + \frac{1}{2\sqrt{3}}z_8\right)a\,\hat{\mathbf{y}} - \frac{1}{3}\left(x_8+y_8+z_8\right)c\,\hat{\mathbf{z}} \quad (12f) \quad \text{Zr II}$$

$$\mathbf{B}_{65} = -y_8\,\mathbf{a}_1 - z_8\,\mathbf{a}_2 - x_8\,\mathbf{a}_3 = \frac{1}{2}\left(x_8-y_8\right)a\,\hat{\mathbf{x}} + \left(\frac{1}{2\sqrt{3}}x_8 + \frac{1}{2\sqrt{3}}y_8 - \frac{1}{\sqrt{3}}z_8\right)a\,\hat{\mathbf{y}} - \frac{1}{3}\left(x_8+y_8+z_8\right)c\,\hat{\mathbf{z}} \quad (12f) \quad \text{Zr II}$$

$$\mathbf{B}_{66} = \left(\frac{1}{2}+z_8\right)\mathbf{a}_1 + \left(\frac{1}{2}+y_8\right)\mathbf{a}_2 + \left(\frac{1}{2}+x_8\right)\mathbf{a}_3 = \frac{1}{2}\left(-x_8+z_8\right)a\,\hat{\mathbf{x}} + \left(-\frac{1}{2\sqrt{3}}x_8 + \frac{1}{\sqrt{3}}y_8 - \frac{1}{2\sqrt{3}}z_8\right)a\,\hat{\mathbf{y}} + \left(\frac{1}{2}+\frac{1}{3}x_8+\frac{1}{3}y_8+\frac{1}{3}z_8\right)c\,\hat{\mathbf{z}} \quad (12f) \quad \text{Zr II}$$

$$\mathbf{B}_{67} = \left(\frac{1}{2}+y_8\right)\mathbf{a}_1 + \left(\frac{1}{2}+x_8\right)\mathbf{a}_2 + \left(\frac{1}{2}+z_8\right)\mathbf{a}_3 = \frac{1}{2}\left(y_8-z_8\right)a\,\hat{\mathbf{x}} + \left(\frac{1}{\sqrt{3}}x_8 - \frac{1}{2\sqrt{3}}y_8 - \frac{1}{2\sqrt{3}}z_8\right)a\,\hat{\mathbf{y}} + \left(\frac{1}{2}+\frac{1}{3}x_8+\frac{1}{3}y_8+\frac{1}{3}z_8\right)c\,\hat{\mathbf{z}} \quad (12f) \quad \text{Zr II}$$

$$\mathbf{B}_{68} = \left(\frac{1}{2}+x_8\right)\mathbf{a}_1 + \left(\frac{1}{2}+z_8\right)\mathbf{a}_2 + \left(\frac{1}{2}+y_8\right)\mathbf{a}_3 = \frac{1}{2}\left(x_8-y_8\right)a\,\hat{\mathbf{x}} + \left(-\frac{1}{2\sqrt{3}}x_8 - \frac{1}{2\sqrt{3}}y_8 + \frac{1}{\sqrt{3}}z_8\right)a\,\hat{\mathbf{y}} + \left(\frac{1}{2}+\frac{1}{3}x_8+\frac{1}{3}y_8+\frac{1}{3}z_8\right)c\,\hat{\mathbf{z}} \quad (12f) \quad \text{Zr II}$$

| | | | | | | |
|---|---|---|---|---|---|---|
| $\mathbf{B}_{69}$ | = | $x_9\,\mathbf{a}_1 + y_9\,\mathbf{a}_2 + z_9\,\mathbf{a}_3$ | = | $\frac{1}{2}\left(x_9 - z_9\right) a\,\mathbf{\hat{x}} + \left(-\frac{1}{2\sqrt{3}}x_9 + \frac{1}{\sqrt{3}}y_9 - \frac{1}{2\sqrt{3}}z_9\right) a\,\mathbf{\hat{y}} + \frac{1}{3}\left(x_9 + y_9 + z_9\right) c\,\mathbf{\hat{z}}$ | $(12f)$ | Zr III |
| $\mathbf{B}_{70}$ | = | $z_9\,\mathbf{a}_1 + x_9\,\mathbf{a}_2 + y_9\,\mathbf{a}_3$ | = | $\frac{1}{2}\left(-y_9 + z_9\right) a\,\mathbf{\hat{x}} + \left(\frac{1}{\sqrt{3}}x_9 - \frac{1}{2\sqrt{3}}y_9 - \frac{1}{2\sqrt{3}}z_9\right) a\,\mathbf{\hat{y}} + \frac{1}{3}\left(x_9 + y_9 + z_9\right) c\,\mathbf{\hat{z}}$ | $(12f)$ | Zr III |
| $\mathbf{B}_{71}$ | = | $y_9\,\mathbf{a}_1 + z_9\,\mathbf{a}_2 + x_9\,\mathbf{a}_3$ | = | $\frac{1}{2}\left(-x_9 + y_9\right) a\,\mathbf{\hat{x}} + \left(-\frac{1}{2\sqrt{3}}x_9 - \frac{1}{2\sqrt{3}}y_9 + \frac{1}{\sqrt{3}}z_9\right) a\,\mathbf{\hat{y}} + \frac{1}{3}\left(x_9 + y_9 + z_9\right) c\,\mathbf{\hat{z}}$ | $(12f)$ | Zr III |
| $\mathbf{B}_{72}$ | = | $\left(\frac{1}{2} - z_9\right)\mathbf{a}_1 + \left(\frac{1}{2} - y_9\right)\mathbf{a}_2 + \left(\frac{1}{2} - x_9\right)\mathbf{a}_3$ | = | $\frac{1}{2}\left(x_9 - z_9\right) a\,\mathbf{\hat{x}} + \left(\frac{1}{2\sqrt{3}}x_9 - \frac{1}{\sqrt{3}}y_9 + \frac{1}{2\sqrt{3}}z_9\right) a\,\mathbf{\hat{y}} + \left(\frac{1}{2} - \frac{1}{3}x_9 - \frac{1}{3}y_9 - \frac{1}{3}z_9\right) c\,\mathbf{\hat{z}}$ | $(12f)$ | Zr III |
| $\mathbf{B}_{73}$ | = | $\left(\frac{1}{2} - y_9\right)\mathbf{a}_1 + \left(\frac{1}{2} - x_9\right)\mathbf{a}_2 + \left(\frac{1}{2} - z_9\right)\mathbf{a}_3$ | = | $\frac{1}{2}\left(-y_9 + z_9\right) a\,\mathbf{\hat{x}} + \left(-\frac{1}{\sqrt{3}}x_9 + \frac{1}{2\sqrt{3}}y_9 + \frac{1}{2\sqrt{3}}z_9\right) a\,\mathbf{\hat{y}} + \left(\frac{1}{2} - \frac{1}{3}x_9 - \frac{1}{3}y_9 - \frac{1}{3}z_9\right) c\,\mathbf{\hat{z}}$ | $(12f)$ | Zr III |
| $\mathbf{B}_{74}$ | = | $\left(\frac{1}{2} - x_9\right)\mathbf{a}_1 + \left(\frac{1}{2} - z_9\right)\mathbf{a}_2 + \left(\frac{1}{2} - y_9\right)\mathbf{a}_3$ | = | $\frac{1}{2}\left(-x_9 + y_9\right) a\,\mathbf{\hat{x}} + \left(\frac{1}{2\sqrt{3}}x_9 + \frac{1}{2\sqrt{3}}y_9 - \frac{1}{\sqrt{3}}z_9\right) a\,\mathbf{\hat{y}} + \left(\frac{1}{2} - \frac{1}{3}x_9 - \frac{1}{3}y_9 - \frac{1}{3}z_9\right) c\,\mathbf{\hat{z}}$ | $(12f)$ | Zr III |
| $\mathbf{B}_{75}$ | = | $-x_9\,\mathbf{a}_1 - y_9\,\mathbf{a}_2 - z_9\,\mathbf{a}_3$ | = | $\frac{1}{2}\left(-x_9 + z_9\right) a\,\mathbf{\hat{x}} + \left(\frac{1}{2\sqrt{3}}x_9 - \frac{1}{\sqrt{3}}y_9 + \frac{1}{2\sqrt{3}}z_9\right) a\,\mathbf{\hat{y}} - \frac{1}{3}\left(x_9 + y_9 + z_9\right) c\,\mathbf{\hat{z}}$ | $(12f)$ | Zr III |
| $\mathbf{B}_{76}$ | = | $-z_9\,\mathbf{a}_1 - x_9\,\mathbf{a}_2 - y_9\,\mathbf{a}_3$ | = | $\frac{1}{2}\left(y_9 - z_9\right) a\,\mathbf{\hat{x}} + \left(-\frac{1}{\sqrt{3}}x_9 + \frac{1}{2\sqrt{3}}y_9 + \frac{1}{2\sqrt{3}}z_9\right) a\,\mathbf{\hat{y}} - \frac{1}{3}\left(x_9 + y_9 + z_9\right) c\,\mathbf{\hat{z}}$ | $(12f)$ | Zr III |
| $\mathbf{B}_{77}$ | = | $-y_9\,\mathbf{a}_1 - z_9\,\mathbf{a}_2 - x_9\,\mathbf{a}_3$ | = | $\frac{1}{2}\left(x_9 - y_9\right) a\,\mathbf{\hat{x}} + \left(\frac{1}{2\sqrt{3}}x_9 + \frac{1}{2\sqrt{3}}y_9 - \frac{1}{\sqrt{3}}z_9\right) a\,\mathbf{\hat{y}} - \frac{1}{3}\left(x_9 + y_9 + z_9\right) c\,\mathbf{\hat{z}}$ | $(12f)$ | Zr III |
| $\mathbf{B}_{78}$ | = | $\left(\frac{1}{2} + z_9\right)\mathbf{a}_1 + \left(\frac{1}{2} + y_9\right)\mathbf{a}_2 + \left(\frac{1}{2} + x_9\right)\mathbf{a}_3$ | = | $\frac{1}{2}\left(-x_9 + z_9\right) a\,\mathbf{\hat{x}} + \left(-\frac{1}{2\sqrt{3}}x_9 + \frac{1}{\sqrt{3}}y_9 - \frac{1}{2\sqrt{3}}z_9\right) a\,\mathbf{\hat{y}} + \left(\frac{1}{2} + \frac{1}{3}x_9 + \frac{1}{3}y_9 + \frac{1}{3}z_9\right) c\,\mathbf{\hat{z}}$ | $(12f)$ | Zr III |
| $\mathbf{B}_{79}$ | = | $\left(\frac{1}{2} + y_9\right)\mathbf{a}_1 + \left(\frac{1}{2} + x_9\right)\mathbf{a}_2 + \left(\frac{1}{2} + z_9\right)\mathbf{a}_3$ | = | $\frac{1}{2}\left(y_9 - z_9\right) a\,\mathbf{\hat{x}} + \left(\frac{1}{\sqrt{3}}x_9 - \frac{1}{2\sqrt{3}}y_9 - \frac{1}{2\sqrt{3}}z_9\right) a\,\mathbf{\hat{y}} + \left(\frac{1}{2} + \frac{1}{3}x_9 + \frac{1}{3}y_9 + \frac{1}{3}z_9\right) c\,\mathbf{\hat{z}}$ | $(12f)$ | Zr III |
| $\mathbf{B}_{80}$ | = | $\left(\frac{1}{2} + x_9\right)\mathbf{a}_1 + \left(\frac{1}{2} + z_9\right)\mathbf{a}_2 + \left(\frac{1}{2} + y_9\right)\mathbf{a}_3$ | = | $\frac{1}{2}\left(x_9 - y_9\right) a\,\mathbf{\hat{x}} + \left(-\frac{1}{2\sqrt{3}}x_9 - \frac{1}{2\sqrt{3}}y_9 + \frac{1}{\sqrt{3}}z_9\right) a\,\mathbf{\hat{y}} + \left(\frac{1}{2} + \frac{1}{3}x_9 + \frac{1}{3}y_9 + \frac{1}{3}z_9\right) c\,\mathbf{\hat{z}}$ | $(12f)$ | Zr III |
| $\mathbf{B}_{81}$ | = | $x_{10}\,\mathbf{a}_1 + y_{10}\,\mathbf{a}_2 + z_{10}\,\mathbf{a}_3$ | = | $\frac{1}{2}\left(x_{10} - z_{10}\right) a\,\mathbf{\hat{x}} + \left(-\frac{1}{2\sqrt{3}}x_{10} + \frac{1}{\sqrt{3}}y_{10} - \frac{1}{2\sqrt{3}}z_{10}\right) a\,\mathbf{\hat{y}} + \frac{1}{3}\left(x_{10} + y_{10} + z_{10}\right) c\,\mathbf{\hat{z}}$ | $(12f)$ | Zr IV |
| $\mathbf{B}_{82}$ | = | $z_{10}\,\mathbf{a}_1 + x_{10}\,\mathbf{a}_2 + y_{10}\,\mathbf{a}_3$ | = | $\frac{1}{2}\left(-y_{10}+z_{10}\right)a\,\hat{\mathbf{x}} + \left(\frac{1}{\sqrt{3}}x_{10} - \frac{1}{2\sqrt{3}}y_{10} - \frac{1}{2\sqrt{3}}z_{10}\right)a\,\hat{\mathbf{y}} + \frac{1}{3}\left(x_{10}+y_{10}+z_{10}\right)c\,\hat{\mathbf{z}}$ | (12$f$) | Zr IV |
| $\mathbf{B}_{83}$ | = | $y_{10}\,\mathbf{a}_1 + z_{10}\,\mathbf{a}_2 + x_{10}\,\mathbf{a}_3$ | = | $\frac{1}{2}\left(-x_{10}+y_{10}\right)a\,\hat{\mathbf{x}} + \left(-\frac{1}{2\sqrt{3}}x_{10} - \frac{1}{2\sqrt{3}}y_{10} + \frac{1}{\sqrt{3}}z_{10}\right)a\,\hat{\mathbf{y}} + \frac{1}{3}\left(x_{10}+y_{10}+z_{10}\right)c\,\hat{\mathbf{z}}$ | (12$f$) | Zr IV |
| $\mathbf{B}_{84}$ | = | $\left(\frac{1}{2}-z_{10}\right)\mathbf{a}_1 + \left(\frac{1}{2}-y_{10}\right)\mathbf{a}_2 + \left(\frac{1}{2}-x_{10}\right)\mathbf{a}_3$ | = | $\frac{1}{2}\left(x_{10}-z_{10}\right)a\,\hat{\mathbf{x}} + \left(\frac{1}{2\sqrt{3}}x_{10} - \frac{1}{\sqrt{3}}y_{10} + \frac{1}{2\sqrt{3}}z_{10}\right)a\,\hat{\mathbf{y}} + \left(\frac{1}{2} - \frac{1}{3}x_{10} - \frac{1}{3}y_{10} - \frac{1}{3}z_{10}\right)c\,\hat{\mathbf{z}}$ | (12$f$) | Zr IV |
| $\mathbf{B}_{85}$ | = | $\left(\frac{1}{2}-y_{10}\right)\mathbf{a}_1 + \left(\frac{1}{2}-x_{10}\right)\mathbf{a}_2 + \left(\frac{1}{2}-z_{10}\right)\mathbf{a}_3$ | = | $\frac{1}{2}\left(-y_{10}+z_{10}\right)a\,\hat{\mathbf{x}} + \left(-\frac{1}{\sqrt{3}}x_{10} + \frac{1}{2\sqrt{3}}y_{10} + \frac{1}{2\sqrt{3}}z_{10}\right)a\,\hat{\mathbf{y}} + \left(\frac{1}{2} - \frac{1}{3}x_{10} - \frac{1}{3}y_{10} - \frac{1}{3}z_{10}\right)c\,\hat{\mathbf{z}}$ | (12$f$) | Zr IV |
| $\mathbf{B}_{86}$ | = | $\left(\frac{1}{2}-x_{10}\right)\mathbf{a}_1 + \left(\frac{1}{2}-z_{10}\right)\mathbf{a}_2 + \left(\frac{1}{2}-y_{10}\right)\mathbf{a}_3$ | = | $\frac{1}{2}\left(-x_{10}+y_{10}\right)a\,\hat{\mathbf{x}} + \left(\frac{1}{2\sqrt{3}}x_{10} + \frac{1}{2\sqrt{3}}y_{10} - \frac{1}{\sqrt{3}}z_{10}\right)a\,\hat{\mathbf{y}} + \left(\frac{1}{2} - \frac{1}{3}x_{10} - \frac{1}{3}y_{10} - \frac{1}{3}z_{10}\right)c\,\hat{\mathbf{z}}$ | (12$f$) | Zr IV |
| $\mathbf{B}_{87}$ | = | $-x_{10}\,\mathbf{a}_1 - y_{10}\,\mathbf{a}_2 - z_{10}\,\mathbf{a}_3$ | = | $\frac{1}{2}\left(-x_{10}+z_{10}\right)a\,\hat{\mathbf{x}} + \left(\frac{1}{2\sqrt{3}}x_{10} - \frac{1}{\sqrt{3}}y_{10} + \frac{1}{2\sqrt{3}}z_{10}\right)a\,\hat{\mathbf{y}} - \frac{1}{3}\left(x_{10}+y_{10}+z_{10}\right)c\,\hat{\mathbf{z}}$ | (12$f$) | Zr IV |
| $\mathbf{B}_{88}$ | = | $-z_{10}\,\mathbf{a}_1 - x_{10}\,\mathbf{a}_2 - y_{10}\,\mathbf{a}_3$ | = | $\frac{1}{2}\left(y_{10}-z_{10}\right)a\,\hat{\mathbf{x}} + \left(-\frac{1}{\sqrt{3}}x_{10} + \frac{1}{2\sqrt{3}}y_{10} + \frac{1}{2\sqrt{3}}z_{10}\right)a\,\hat{\mathbf{y}} - \frac{1}{3}\left(x_{10}+y_{10}+z_{10}\right)c\,\hat{\mathbf{z}}$ | (12$f$) | Zr IV |
| $\mathbf{B}_{89}$ | = | $-y_{10}\,\mathbf{a}_1 - z_{10}\,\mathbf{a}_2 - x_{10}\,\mathbf{a}_3$ | = | $\frac{1}{2}\left(x_{10}-y_{10}\right)a\,\hat{\mathbf{x}} + \left(\frac{1}{2\sqrt{3}}x_{10} + \frac{1}{2\sqrt{3}}y_{10} - \frac{1}{\sqrt{3}}z_{10}\right)a\,\hat{\mathbf{y}} - \frac{1}{3}\left(x_{10}+y_{10}+z_{10}\right)c\,\hat{\mathbf{z}}$ | (12$f$) | Zr IV |
| $\mathbf{B}_{90}$ | = | $\left(\frac{1}{2}+z_{10}\right)\mathbf{a}_1 + \left(\frac{1}{2}+y_{10}\right)\mathbf{a}_2 + \left(\frac{1}{2}+x_{10}\right)\mathbf{a}_3$ | = | $\frac{1}{2}\left(-x_{10}+z_{10}\right)a\,\hat{\mathbf{x}} + \left(-\frac{1}{2\sqrt{3}}x_{10} + \frac{1}{\sqrt{3}}y_{10} - \frac{1}{2\sqrt{3}}z_{10}\right)a\,\hat{\mathbf{y}} + \left(\frac{1}{2} + \frac{1}{3}x_{10} + \frac{1}{3}y_{10} + \frac{1}{3}z_{10}\right)c\,\hat{\mathbf{z}}$ | (12$f$) | Zr IV |
| $\mathbf{B}_{91}$ | = | $\left(\frac{1}{2}+y_{10}\right)\mathbf{a}_1 + \left(\frac{1}{2}+x_{10}\right)\mathbf{a}_2 + \left(\frac{1}{2}+z_{10}\right)\mathbf{a}_3$ | = | $\frac{1}{2}\left(y_{10}-z_{10}\right)a\,\hat{\mathbf{x}} + \left(\frac{1}{\sqrt{3}}x_{10} - \frac{1}{2\sqrt{3}}y_{10} - \frac{1}{2\sqrt{3}}z_{10}\right)a\,\hat{\mathbf{y}} + \left(\frac{1}{2} + \frac{1}{3}x_{10} + \frac{1}{3}y_{10} + \frac{1}{3}z_{10}\right)c\,\hat{\mathbf{z}}$ | (12$f$) | Zr IV |
| $\mathbf{B}_{92}$ | = | $\left(\frac{1}{2}+x_{10}\right)\mathbf{a}_1 + \left(\frac{1}{2}+z_{10}\right)\mathbf{a}_2 + \left(\frac{1}{2}+y_{10}\right)\mathbf{a}_3$ | = | $\frac{1}{2}\left(x_{10}-y_{10}\right)a\,\hat{\mathbf{x}} + \left(-\frac{1}{2\sqrt{3}}x_{10} - \frac{1}{2\sqrt{3}}y_{10} + \frac{1}{\sqrt{3}}z_{10}\right)a\,\hat{\mathbf{y}} + \left(\frac{1}{2} + \frac{1}{3}x_{10} + \frac{1}{3}y_{10} + \frac{1}{3}z_{10}\right)c\,\hat{\mathbf{z}}$ | (12$f$) | Zr IV |

**Geometry files:**

- CIF: pp. 1744
- POSCAR: pp. 1745

# $\beta$-$BaB_2O_4$ (High-Temperature) Structure: A2BC4_hR42_167_f_ac_2f

http://aflow.org/prototype-encyclopedia/A2BC4_hR42_167_f_ac_2f

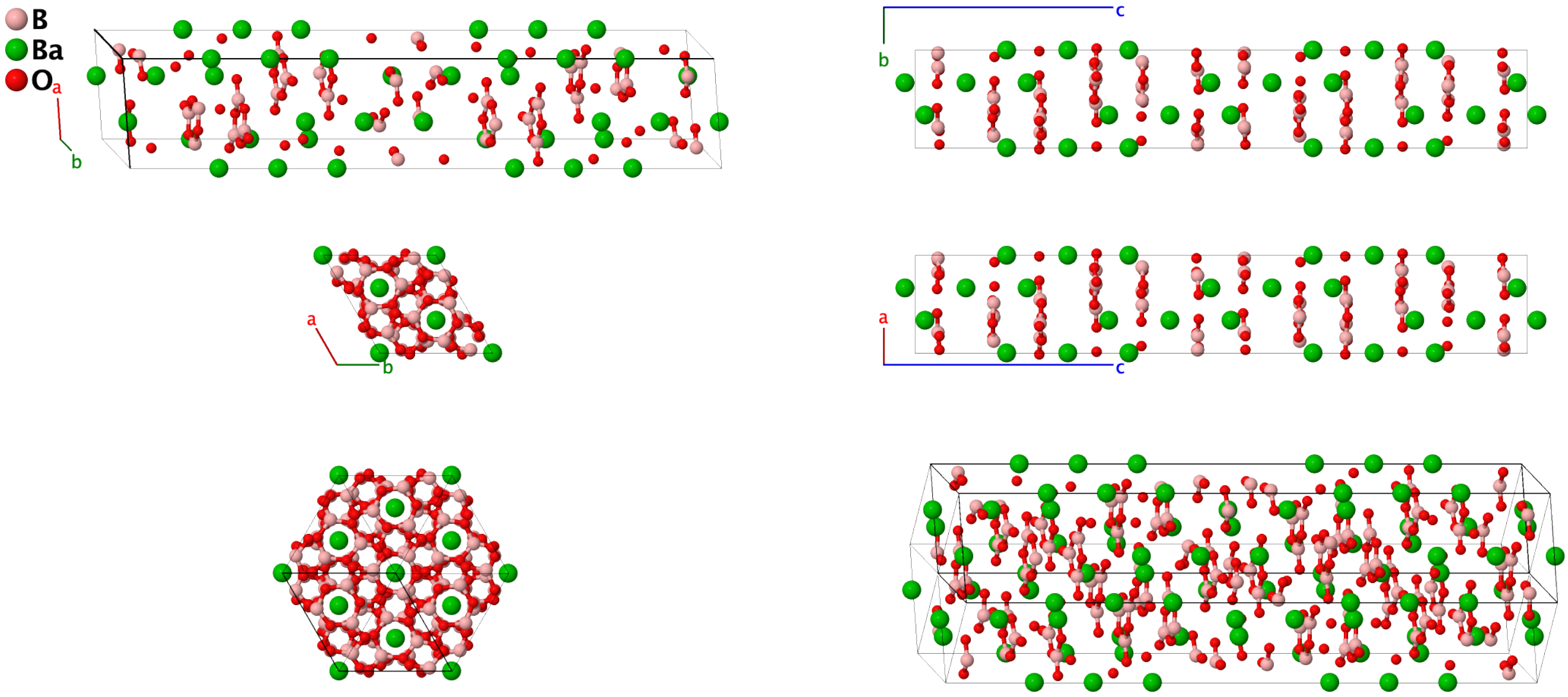

| | | |
|---|---|---|
| **Prototype** | : | $B_2BaO_4$ |
| **AFLOW prototype label** | : | A2BC4_hR42_167_f_ac_2f |
| ***Strukturbericht* designation** | : | None |
| **Pearson symbol** | : | hR42 |
| **Space group number** | : | 167 |
| **Space group symbol** | : | $R\bar{3}c$ |
| **AFLOW prototype command** | : | `aflow --proto=A2BC4_hR42_167_f_ac_2f [--hex]` `--params=`$a, c/a, x_2, x_3, y_3, z_3, x_4, y_4, z_4, x_5, y_5, z_5$ |

- This is the high-temperature structure of $BaB_2O_4$. Cooling to temperatures between 100-400 °C, it transforms into $\alpha$-$BaB_2O_4$. The principle difference between the two forms is the lack of inversion symmetry in the low-temperature structure.
- This structure was determined by cooling liquid $BaB_2O_4$ to room temperature.

**Rhombohedral primitive vectors:**

$$\mathbf{a}_1 = \frac{1}{2}\,a\,\mathbf{\hat{x}} - \frac{1}{2\sqrt{3}}\,a\,\mathbf{\hat{y}} + \frac{1}{3}\,c\,\mathbf{\hat{z}}$$
$$\mathbf{a}_2 = \frac{1}{\sqrt{3}}\,a\,\mathbf{\hat{y}} + \frac{1}{3}\,c\,\mathbf{\hat{z}}$$
$$\mathbf{a}_3 = -\frac{1}{2}\,a\,\mathbf{\hat{x}} - \frac{1}{2\sqrt{3}}\,a\,\mathbf{\hat{y}} + \frac{1}{3}\,c\,\mathbf{\hat{z}}$$

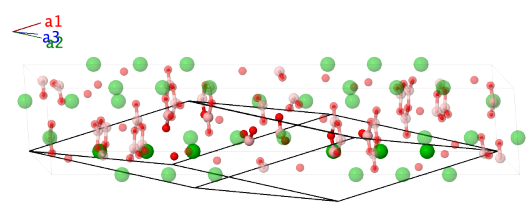

**Basis vectors:**

| | | Lattice Coordinates | | Cartesian Coordinates | Wyckoff Position | Atom Type |
|---|---|---|---|---|---|---|
| $\mathbf{B}_1$ | = | $\frac{1}{4}\,\mathbf{a}_1 + \frac{1}{4}\,\mathbf{a}_2 + \frac{1}{4}\,\mathbf{a}_3$ | = | $\frac{1}{4}c\,\mathbf{\hat{z}}$ | (2$a$) | Ba I |
| $\mathbf{B}_2$ | = | $\frac{3}{4}\,\mathbf{a}_1 + \frac{3}{4}\,\mathbf{a}_2 + \frac{3}{4}\,\mathbf{a}_3$ | = | $\frac{3}{4}c\,\mathbf{\hat{z}}$ | (2$a$) | Ba I |
| $\mathbf{B}_{3}$ | = | $x_{2}\,\mathbf{a}_{1}+x_{2}\,\mathbf{a}_{2}+x_{2}\,\mathbf{a}_{3}$ | = | $x_{2}c\,\hat{\mathbf{z}}$ | (4$c$) | Ba II |
| $\mathbf{B}_{4}$ | = | $\left(\frac{1}{2}-x_{2}\right)\mathbf{a}_{1}+\left(\frac{1}{2}-x_{2}\right)\mathbf{a}_{2}+\left(\frac{1}{2}-x_{2}\right)\mathbf{a}_{3}$ | = | $\left(\frac{1}{2}-x_{2}\right)c\,\hat{\mathbf{z}}$ | (4$c$) | Ba II |
| $\mathbf{B}_{5}$ | = | $-x_{2}\,\mathbf{a}_{1}-x_{2}\,\mathbf{a}_{2}-x_{2}\,\mathbf{a}_{3}$ | = | $-x_{2}c\,\hat{\mathbf{z}}$ | (4$c$) | Ba II |
| $\mathbf{B}_{6}$ | = | $\left(\frac{1}{2}+x_{2}\right)\mathbf{a}_{1}+\left(\frac{1}{2}+x_{2}\right)\mathbf{a}_{2}+\left(\frac{1}{2}+x_{2}\right)\mathbf{a}_{3}$ | = | $\left(\frac{1}{2}+x_{2}\right)c\,\hat{\mathbf{z}}$ | (4$c$) | Ba II |
| $\mathbf{B}_{7}$ | = | $x_{3}\,\mathbf{a}_{1}+y_{3}\,\mathbf{a}_{2}+z_{3}\,\mathbf{a}_{3}$ | = | $\frac{1}{2}\left(x_{3}-z_{3}\right)a\,\hat{\mathbf{x}}+\left(-\frac{1}{2\sqrt{3}}x_{3}+\frac{1}{\sqrt{3}}y_{3}-\frac{1}{2\sqrt{3}}z_{3}\right)a\,\hat{\mathbf{y}}+\frac{1}{3}\left(x_{3}+y_{3}+z_{3}\right)c\,\hat{\mathbf{z}}$ | (12$f$) | B |
| $\mathbf{B}_{8}$ | = | $z_{3}\,\mathbf{a}_{1}+x_{3}\,\mathbf{a}_{2}+y_{3}\,\mathbf{a}_{3}$ | = | $\frac{1}{2}\left(-y_{3}+z_{3}\right)a\,\hat{\mathbf{x}}+\left(\frac{1}{\sqrt{3}}x_{3}-\frac{1}{2\sqrt{3}}y_{3}-\frac{1}{2\sqrt{3}}z_{3}\right)a\,\hat{\mathbf{y}}+\frac{1}{3}\left(x_{3}+y_{3}+z_{3}\right)c\,\hat{\mathbf{z}}$ | (12$f$) | B |
| $\mathbf{B}_{9}$ | = | $y_{3}\,\mathbf{a}_{1}+z_{3}\,\mathbf{a}_{2}+x_{3}\,\mathbf{a}_{3}$ | = | $\frac{1}{2}\left(-x_{3}+y_{3}\right)a\,\hat{\mathbf{x}}+\left(-\frac{1}{2\sqrt{3}}x_{3}-\frac{1}{2\sqrt{3}}y_{3}+\frac{1}{\sqrt{3}}z_{3}\right)a\,\hat{\mathbf{y}}+\frac{1}{3}\left(x_{3}+y_{3}+z_{3}\right)c\,\hat{\mathbf{z}}$ | (12$f$) | B |
| $\mathbf{B}_{10}$ | = | $\left(\frac{1}{2}-z_{3}\right)\mathbf{a}_{1}+\left(\frac{1}{2}-y_{3}\right)\mathbf{a}_{2}+\left(\frac{1}{2}-x_{3}\right)\mathbf{a}_{3}$ | = | $\frac{1}{2}\left(x_{3}-z_{3}\right)a\,\hat{\mathbf{x}}+\left(\frac{1}{2\sqrt{3}}x_{3}-\frac{1}{\sqrt{3}}y_{3}+\frac{1}{2\sqrt{3}}z_{3}\right)a\,\hat{\mathbf{y}}+\left(\frac{1}{2}-\frac{1}{3}x_{3}-\frac{1}{3}y_{3}-\frac{1}{3}z_{3}\right)c\,\hat{\mathbf{z}}$ | (12$f$) | B |
| $\mathbf{B}_{11}$ | = | $\left(\frac{1}{2}-y_{3}\right)\mathbf{a}_{1}+\left(\frac{1}{2}-x_{3}\right)\mathbf{a}_{2}+\left(\frac{1}{2}-z_{3}\right)\mathbf{a}_{3}$ | = | $\frac{1}{2}\left(-y_{3}+z_{3}\right)a\,\hat{\mathbf{x}}+\left(-\frac{1}{\sqrt{3}}x_{3}+\frac{1}{2\sqrt{3}}y_{3}+\frac{1}{2\sqrt{3}}z_{3}\right)a\,\hat{\mathbf{y}}+\left(\frac{1}{2}-\frac{1}{3}x_{3}-\frac{1}{3}y_{3}-\frac{1}{3}z_{3}\right)c\,\hat{\mathbf{z}}$ | (12$f$) | B |
| $\mathbf{B}_{12}$ | = | $\left(\frac{1}{2}-x_{3}\right)\mathbf{a}_{1}+\left(\frac{1}{2}-z_{3}\right)\mathbf{a}_{2}+\left(\frac{1}{2}-y_{3}\right)\mathbf{a}_{3}$ | = | $\frac{1}{2}\left(-x_{3}+y_{3}\right)a\,\hat{\mathbf{x}}+\left(\frac{1}{2\sqrt{3}}x_{3}+\frac{1}{2\sqrt{3}}y_{3}-\frac{1}{\sqrt{3}}z_{3}\right)a\,\hat{\mathbf{y}}+\left(\frac{1}{2}-\frac{1}{3}x_{3}-\frac{1}{3}y_{3}-\frac{1}{3}z_{3}\right)c\,\hat{\mathbf{z}}$ | (12$f$) | B |
| $\mathbf{B}_{13}$ | = | $-x_{3}\,\mathbf{a}_{1}-y_{3}\,\mathbf{a}_{2}-z_{3}\,\mathbf{a}_{3}$ | = | $\frac{1}{2}\left(-x_{3}+z_{3}\right)a\,\hat{\mathbf{x}}+\left(\frac{1}{2\sqrt{3}}x_{3}-\frac{1}{\sqrt{3}}y_{3}+\frac{1}{2\sqrt{3}}z_{3}\right)a\,\hat{\mathbf{y}}-\frac{1}{3}\left(x_{3}+y_{3}+z_{3}\right)c\,\hat{\mathbf{z}}$ | (12$f$) | B |
| $\mathbf{B}_{14}$ | = | $-z_{3}\,\mathbf{a}_{1}-x_{3}\,\mathbf{a}_{2}-y_{3}\,\mathbf{a}_{3}$ | = | $\frac{1}{2}\left(y_{3}-z_{3}\right)a\,\hat{\mathbf{x}}+\left(-\frac{1}{\sqrt{3}}x_{3}+\frac{1}{2\sqrt{3}}y_{3}+\frac{1}{2\sqrt{3}}z_{3}\right)a\,\hat{\mathbf{y}}-\frac{1}{3}\left(x_{3}+y_{3}+z_{3}\right)c\,\hat{\mathbf{z}}$ | (12$f$) | B |
| $\mathbf{B}_{15}$ | = | $-y_{3}\,\mathbf{a}_{1}-z_{3}\,\mathbf{a}_{2}-x_{3}\,\mathbf{a}_{3}$ | = | $\frac{1}{2}\left(x_{3}-y_{3}\right)a\,\hat{\mathbf{x}}+\left(\frac{1}{2\sqrt{3}}x_{3}+\frac{1}{2\sqrt{3}}y_{3}-\frac{1}{\sqrt{3}}z_{3}\right)a\,\hat{\mathbf{y}}-\frac{1}{3}\left(x_{3}+y_{3}+z_{3}\right)c\,\hat{\mathbf{z}}$ | (12$f$) | B |
| $\mathbf{B}_{16}$ | = | $\left(\frac{1}{2}+z_{3}\right)\mathbf{a}_{1}+\left(\frac{1}{2}+y_{3}\right)\mathbf{a}_{2}+\left(\frac{1}{2}+x_{3}\right)\mathbf{a}_{3}$ | = | $\frac{1}{2}\left(-x_{3}+z_{3}\right)a\,\hat{\mathbf{x}}+\left(-\frac{1}{2\sqrt{3}}x_{3}+\frac{1}{\sqrt{3}}y_{3}-\frac{1}{2\sqrt{3}}z_{3}\right)a\,\hat{\mathbf{y}}+\left(\frac{1}{2}+\frac{1}{3}x_{3}+\frac{1}{3}y_{3}+\frac{1}{3}z_{3}\right)c\,\hat{\mathbf{z}}$ | (12$f$) | B |
| $\mathbf{B}_{17}$ | = | $\left(\frac{1}{2}+y_{3}\right)\mathbf{a}_{1}+\left(\frac{1}{2}+x_{3}\right)\mathbf{a}_{2}+\left(\frac{1}{2}+z_{3}\right)\mathbf{a}_{3}$ | = | $\frac{1}{2}\left(y_{3}-z_{3}\right)a\,\hat{\mathbf{x}}+\left(\frac{1}{\sqrt{3}}x_{3}-\frac{1}{2\sqrt{3}}y_{3}-\frac{1}{2\sqrt{3}}z_{3}\right)a\,\hat{\mathbf{y}}+\left(\frac{1}{2}+\frac{1}{3}x_{3}+\frac{1}{3}y_{3}+\frac{1}{3}z_{3}\right)c\,\hat{\mathbf{z}}$ | (12$f$) | B |

$$\begin{array}{llllll}
\mathbf{B}_{18} & = & \left(\frac{1}{2}+x_3\right)\mathbf{a}_1+\left(\frac{1}{2}+z_3\right)\mathbf{a}_2+\left(\frac{1}{2}+y_3\right)\mathbf{a}_3 & = & \frac{1}{2}\left(x_3-y_3\right)a\,\mathbf{\hat{x}}+\left(-\frac{1}{2\sqrt{3}}x_3-\frac{1}{2\sqrt{3}}y_3+\frac{1}{\sqrt{3}}z_3\right)a\,\mathbf{\hat{y}}+\left(\frac{1}{2}+\frac{1}{3}x_3+\frac{1}{3}y_3+\frac{1}{3}z_3\right)c\,\mathbf{\hat{z}} & (12f) & \mathrm{B} \\
\mathbf{B}_{19} & = & x_4\,\mathbf{a}_1+y_4\,\mathbf{a}_2+z_4\,\mathbf{a}_3 & = & \frac{1}{2}\left(x_4-z_4\right)a\,\mathbf{\hat{x}}+\left(-\frac{1}{2\sqrt{3}}x_4+\frac{1}{\sqrt{3}}y_4-\frac{1}{2\sqrt{3}}z_4\right)a\,\mathbf{\hat{y}}+\frac{1}{3}\left(x_4+y_4+z_4\right)c\,\mathbf{\hat{z}} & (12f) & \mathrm{O\,I} \\
\mathbf{B}_{20} & = & z_4\,\mathbf{a}_1+x_4\,\mathbf{a}_2+y_4\,\mathbf{a}_3 & = & \frac{1}{2}\left(-y_4+z_4\right)a\,\mathbf{\hat{x}}+\left(\frac{1}{\sqrt{3}}x_4-\frac{1}{2\sqrt{3}}y_4-\frac{1}{2\sqrt{3}}z_4\right)a\,\mathbf{\hat{y}}+\frac{1}{3}\left(x_4+y_4+z_4\right)c\,\mathbf{\hat{z}} & (12f) & \mathrm{O\,I} \\
\mathbf{B}_{21} & = & y_4\,\mathbf{a}_1+z_4\,\mathbf{a}_2+x_4\,\mathbf{a}_3 & = & \frac{1}{2}\left(-x_4+y_4\right)a\,\mathbf{\hat{x}}+\left(-\frac{1}{2\sqrt{3}}x_4-\frac{1}{2\sqrt{3}}y_4+\frac{1}{\sqrt{3}}z_4\right)a\,\mathbf{\hat{y}}+\frac{1}{3}\left(x_4+y_4+z_4\right)c\,\mathbf{\hat{z}} & (12f) & \mathrm{O\,I} \\
\mathbf{B}_{22} & = & \left(\frac{1}{2}-z_4\right)\mathbf{a}_1+\left(\frac{1}{2}-y_4\right)\mathbf{a}_2+\left(\frac{1}{2}-x_4\right)\mathbf{a}_3 & = & \frac{1}{2}\left(x_4-z_4\right)a\,\mathbf{\hat{x}}+\left(\frac{1}{2\sqrt{3}}x_4-\frac{1}{\sqrt{3}}y_4+\frac{1}{2\sqrt{3}}z_4\right)a\,\mathbf{\hat{y}}+\left(\frac{1}{2}-\frac{1}{3}x_4-\frac{1}{3}y_4-\frac{1}{3}z_4\right)c\,\mathbf{\hat{z}} & (12f) & \mathrm{O\,I} \\
\mathbf{B}_{23} & = & \left(\frac{1}{2}-y_4\right)\mathbf{a}_1+\left(\frac{1}{2}-x_4\right)\mathbf{a}_2+\left(\frac{1}{2}-z_4\right)\mathbf{a}_3 & = & \frac{1}{2}\left(-y_4+z_4\right)a\,\mathbf{\hat{x}}+\left(-\frac{1}{\sqrt{3}}x_4+\frac{1}{2\sqrt{3}}y_4+\frac{1}{2\sqrt{3}}z_4\right)a\,\mathbf{\hat{y}}+\left(\frac{1}{2}-\frac{1}{3}x_4-\frac{1}{3}y_4-\frac{1}{3}z_4\right)c\,\mathbf{\hat{z}} & (12f) & \mathrm{O\,I} \\
\mathbf{B}_{24} & = & \left(\frac{1}{2}-x_4\right)\mathbf{a}_1+\left(\frac{1}{2}-z_4\right)\mathbf{a}_2+\left(\frac{1}{2}-y_4\right)\mathbf{a}_3 & = & \frac{1}{2}\left(-x_4+y_4\right)a\,\mathbf{\hat{x}}+\left(\frac{1}{2\sqrt{3}}x_4+\frac{1}{2\sqrt{3}}y_4-\frac{1}{\sqrt{3}}z_4\right)a\,\mathbf{\hat{y}}+\left(\frac{1}{2}-\frac{1}{3}x_4-\frac{1}{3}y_4-\frac{1}{3}z_4\right)c\,\mathbf{\hat{z}} & (12f) & \mathrm{O\,I} \\
\mathbf{B}_{25} & = & -x_4\,\mathbf{a}_1-y_4\,\mathbf{a}_2-z_4\,\mathbf{a}_3 & = & \frac{1}{2}\left(-x_4+z_4\right)a\,\mathbf{\hat{x}}+\left(\frac{1}{2\sqrt{3}}x_4-\frac{1}{\sqrt{3}}y_4+\frac{1}{2\sqrt{3}}z_4\right)a\,\mathbf{\hat{y}}-\frac{1}{3}\left(x_4+y_4+z_4\right)c\,\mathbf{\hat{z}} & (12f) & \mathrm{O\,I} \\
\mathbf{B}_{26} & = & -z_4\,\mathbf{a}_1-x_4\,\mathbf{a}_2-y_4\,\mathbf{a}_3 & = & \frac{1}{2}\left(y_4-z_4\right)a\,\mathbf{\hat{x}}+\left(-\frac{1}{\sqrt{3}}x_4+\frac{1}{2\sqrt{3}}y_4+\frac{1}{2\sqrt{3}}z_4\right)a\,\mathbf{\hat{y}}-\frac{1}{3}\left(x_4+y_4+z_4\right)c\,\mathbf{\hat{z}} & (12f) & \mathrm{O\,I} \\
\mathbf{B}_{27} & = & -y_4\,\mathbf{a}_1-z_4\,\mathbf{a}_2-x_4\,\mathbf{a}_3 & = & \frac{1}{2}\left(x_4-y_4\right)a\,\mathbf{\hat{x}}+\left(\frac{1}{2\sqrt{3}}x_4+\frac{1}{2\sqrt{3}}y_4-\frac{1}{\sqrt{3}}z_4\right)a\,\mathbf{\hat{y}}-\frac{1}{3}\left(x_4+y_4+z_4\right)c\,\mathbf{\hat{z}} & (12f) & \mathrm{O\,I} \\
\mathbf{B}_{28} & = & \left(\frac{1}{2}+z_4\right)\mathbf{a}_1+\left(\frac{1}{2}+y_4\right)\mathbf{a}_2+\left(\frac{1}{2}+x_4\right)\mathbf{a}_3 & = & \frac{1}{2}\left(-x_4+z_4\right)a\,\mathbf{\hat{x}}+\left(-\frac{1}{2\sqrt{3}}x_4+\frac{1}{\sqrt{3}}y_4-\frac{1}{2\sqrt{3}}z_4\right)a\,\mathbf{\hat{y}}+\left(\frac{1}{2}+\frac{1}{3}x_4+\frac{1}{3}y_4+\frac{1}{3}z_4\right)c\,\mathbf{\hat{z}} & (12f) & \mathrm{O\,I} \\
\mathbf{B}_{29} & = & \left(\frac{1}{2}+y_4\right)\mathbf{a}_1+\left(\frac{1}{2}+x_4\right)\mathbf{a}_2+\left(\frac{1}{2}+z_4\right)\mathbf{a}_3 & = & \frac{1}{2}\left(y_4-z_4\right)a\,\mathbf{\hat{x}}+\left(\frac{1}{\sqrt{3}}x_4-\frac{1}{2\sqrt{3}}y_4-\frac{1}{2\sqrt{3}}z_4\right)a\,\mathbf{\hat{y}}+\left(\frac{1}{2}+\frac{1}{3}x_4+\frac{1}{3}y_4+\frac{1}{3}z_4\right)c\,\mathbf{\hat{z}} & (12f) & \mathrm{O\,I} \\
\mathbf{B}_{30} & = & \left(\frac{1}{2}+x_4\right)\mathbf{a}_1+\left(\frac{1}{2}+z_4\right)\mathbf{a}_2+\left(\frac{1}{2}+y_4\right)\mathbf{a}_3 & = & \frac{1}{2}\left(x_4-y_4\right)a\,\mathbf{\hat{x}}+\left(-\frac{1}{2\sqrt{3}}x_4-\frac{1}{2\sqrt{3}}y_4+\frac{1}{\sqrt{3}}z_4\right)a\,\mathbf{\hat{y}}+\left(\frac{1}{2}+\frac{1}{3}x_4+\frac{1}{3}y_4+\frac{1}{3}z_4\right)c\,\mathbf{\hat{z}} & (12f) & \mathrm{O\,I}
\end{array}$$

$$\mathbf{B}_{31} = x_5\,\mathbf{a}_1 + y_5\,\mathbf{a}_2 + z_5\,\mathbf{a}_3 = \frac{1}{2}(x_5 - z_5)\,a\,\hat{\mathbf{x}} + \left(-\frac{1}{2\sqrt{3}}x_5 + \frac{1}{\sqrt{3}}y_5 - \frac{1}{2\sqrt{3}}z_5\right)a\,\hat{\mathbf{y}} + \frac{1}{3}(x_5 + y_5 + z_5)\,c\,\hat{\mathbf{z}} \quad (12f) \quad \text{O II}$$

$$\mathbf{B}_{32} = z_5\,\mathbf{a}_1 + x_5\,\mathbf{a}_2 + y_5\,\mathbf{a}_3 = \frac{1}{2}(-y_5 + z_5)\,a\,\hat{\mathbf{x}} + \left(\frac{1}{\sqrt{3}}x_5 - \frac{1}{2\sqrt{3}}y_5 - \frac{1}{2\sqrt{3}}z_5\right)a\,\hat{\mathbf{y}} + \frac{1}{3}(x_5 + y_5 + z_5)\,c\,\hat{\mathbf{z}} \quad (12f) \quad \text{O II}$$

$$\mathbf{B}_{33} = y_5\,\mathbf{a}_1 + z_5\,\mathbf{a}_2 + x_5\,\mathbf{a}_3 = \frac{1}{2}(-x_5 + y_5)\,a\,\hat{\mathbf{x}} + \left(-\frac{1}{2\sqrt{3}}x_5 - \frac{1}{2\sqrt{3}}y_5 + \frac{1}{\sqrt{3}}z_5\right)a\,\hat{\mathbf{y}} + \frac{1}{3}(x_5 + y_5 + z_5)\,c\,\hat{\mathbf{z}} \quad (12f) \quad \text{O II}$$

$$\mathbf{B}_{34} = \left(\frac{1}{2} - z_5\right)\mathbf{a}_1 + \left(\frac{1}{2} - y_5\right)\mathbf{a}_2 + \left(\frac{1}{2} - x_5\right)\mathbf{a}_3 = \frac{1}{2}(x_5 - z_5)\,a\,\hat{\mathbf{x}} + \left(\frac{1}{2\sqrt{3}}x_5 - \frac{1}{\sqrt{3}}y_5 + \frac{1}{2\sqrt{3}}z_5\right)a\,\hat{\mathbf{y}} + \left(\frac{1}{2} - \frac{1}{3}x_5 - \frac{1}{3}y_5 - \frac{1}{3}z_5\right)c\,\hat{\mathbf{z}} \quad (12f) \quad \text{O II}$$

$$\mathbf{B}_{35} = \left(\frac{1}{2} - y_5\right)\mathbf{a}_1 + \left(\frac{1}{2} - x_5\right)\mathbf{a}_2 + \left(\frac{1}{2} - z_5\right)\mathbf{a}_3 = \frac{1}{2}(-y_5 + z_5)\,a\,\hat{\mathbf{x}} + \left(-\frac{1}{\sqrt{3}}x_5 + \frac{1}{2\sqrt{3}}y_5 + \frac{1}{2\sqrt{3}}z_5\right)a\,\hat{\mathbf{y}} + \left(\frac{1}{2} - \frac{1}{3}x_5 - \frac{1}{3}y_5 - \frac{1}{3}z_5\right)c\,\hat{\mathbf{z}} \quad (12f) \quad \text{O II}$$

$$\mathbf{B}_{36} = \left(\frac{1}{2} - x_5\right)\mathbf{a}_1 + \left(\frac{1}{2} - z_5\right)\mathbf{a}_2 + \left(\frac{1}{2} - y_5\right)\mathbf{a}_3 = \frac{1}{2}(-x_5 + y_5)\,a\,\hat{\mathbf{x}} + \left(\frac{1}{2\sqrt{3}}x_5 + \frac{1}{2\sqrt{3}}y_5 - \frac{1}{\sqrt{3}}z_5\right)a\,\hat{\mathbf{y}} + \left(\frac{1}{2} - \frac{1}{3}x_5 - \frac{1}{3}y_5 - \frac{1}{3}z_5\right)c\,\hat{\mathbf{z}} \quad (12f) \quad \text{O II}$$

$$\mathbf{B}_{37} = -x_5\,\mathbf{a}_1 - y_5\,\mathbf{a}_2 - z_5\,\mathbf{a}_3 = \frac{1}{2}(-x_5 + z_5)\,a\,\hat{\mathbf{x}} + \left(\frac{1}{2\sqrt{3}}x_5 - \frac{1}{\sqrt{3}}y_5 + \frac{1}{2\sqrt{3}}z_5\right)a\,\hat{\mathbf{y}} - \frac{1}{3}(x_5 + y_5 + z_5)\,c\,\hat{\mathbf{z}} \quad (12f) \quad \text{O II}$$

$$\mathbf{B}_{38} = -z_5\,\mathbf{a}_1 - x_5\,\mathbf{a}_2 - y_5\,\mathbf{a}_3 = \frac{1}{2}(y_5 - z_5)\,a\,\hat{\mathbf{x}} + \left(-\frac{1}{\sqrt{3}}x_5 + \frac{1}{2\sqrt{3}}y_5 + \frac{1}{2\sqrt{3}}z_5\right)a\,\hat{\mathbf{y}} - \frac{1}{3}(x_5 + y_5 + z_5)\,c\,\hat{\mathbf{z}} \quad (12f) \quad \text{O II}$$

$$\mathbf{B}_{39} = -y_5\,\mathbf{a}_1 - z_5\,\mathbf{a}_2 - x_5\,\mathbf{a}_3 = \frac{1}{2}(x_5 - y_5)\,a\,\hat{\mathbf{x}} + \left(\frac{1}{2\sqrt{3}}x_5 + \frac{1}{2\sqrt{3}}y_5 - \frac{1}{\sqrt{3}}z_5\right)a\,\hat{\mathbf{y}} - \frac{1}{3}(x_5 + y_5 + z_5)\,c\,\hat{\mathbf{z}} \quad (12f) \quad \text{O II}$$

$$\mathbf{B}_{40} = \left(\frac{1}{2} + z_5\right)\mathbf{a}_1 + \left(\frac{1}{2} + y_5\right)\mathbf{a}_2 + \left(\frac{1}{2} + x_5\right)\mathbf{a}_3 = \frac{1}{2}(-x_5 + z_5)\,a\,\hat{\mathbf{x}} + \left(-\frac{1}{2\sqrt{3}}x_5 + \frac{1}{\sqrt{3}}y_5 - \frac{1}{2\sqrt{3}}z_5\right)a\,\hat{\mathbf{y}} + \left(\frac{1}{2} + \frac{1}{3}x_5 + \frac{1}{3}y_5 + \frac{1}{3}z_5\right)c\,\hat{\mathbf{z}} \quad (12f) \quad \text{O II}$$

$$\mathbf{B}_{41} = \left(\frac{1}{2} + y_5\right)\mathbf{a}_1 + \left(\frac{1}{2} + x_5\right)\mathbf{a}_2 + \left(\frac{1}{2} + z_5\right)\mathbf{a}_3 = \frac{1}{2}(y_5 - z_5)\,a\,\hat{\mathbf{x}} + \left(\frac{1}{\sqrt{3}}x_5 - \frac{1}{2\sqrt{3}}y_5 - \frac{1}{2\sqrt{3}}z_5\right)a\,\hat{\mathbf{y}} + \left(\frac{1}{2} + \frac{1}{3}x_5 + \frac{1}{3}y_5 + \frac{1}{3}z_5\right)c\,\hat{\mathbf{z}} \quad (12f) \quad \text{O II}$$

$$\mathbf{B}_{42} = \left(\frac{1}{2} + x_5\right)\mathbf{a}_1 + \left(\frac{1}{2} + z_5\right)\mathbf{a}_2 + \left(\frac{1}{2} + y_5\right)\mathbf{a}_3 = \frac{1}{2}(x_5 - y_5)\,a\,\hat{\mathbf{x}} + \left(-\frac{1}{2\sqrt{3}}x_5 - \frac{1}{2\sqrt{3}}y_5 + \frac{1}{\sqrt{3}}z_5\right)a\,\hat{\mathbf{y}} + \left(\frac{1}{2} + \frac{1}{3}x_5 + \frac{1}{3}y_5 + \frac{1}{3}z_5\right)c\,\hat{\mathbf{z}} \quad (12f) \quad \text{O II}$$

---

**Found in:**

- R. Fröhlich, *Crystal Structure of the low-temperature form of* $BaB_2O_4$, Zeitschrift für Kristallographie - Crystalline Materials **168**, 109–112 (1984), doi:10.1524/zkri.1984.168.14.109.

---

**Geometry files:**

- CIF: pp. 1745
- POSCAR: pp. 1746

# $CrCl_3(H_2O)_6$ ($J2_2$) Structure: A3BC6_hR20_167_e_b_f

http://aflow.org/prototype-encyclopedia/A3BC6_hR20_167_e_b_f

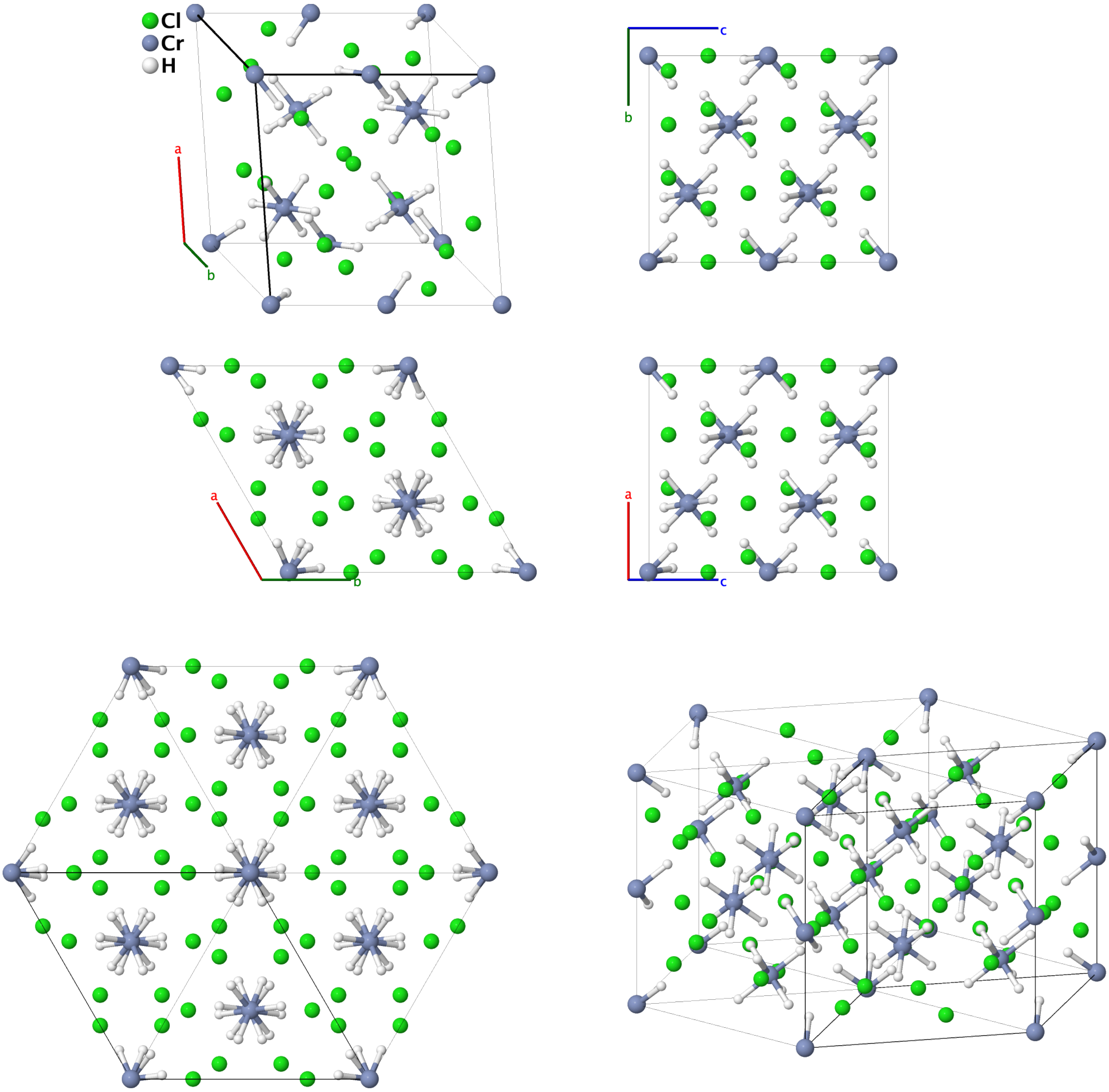

| | | |
|---|---|---|
| **Prototype** | : | $Cl_3Cr(H_2O)_6$ |
| **AFLOW prototype label** | : | A3BC6_hR20_167_e_b_f |
| ***Strukturbericht* designation** | : | $J2_2$ |
| **Pearson symbol** | : | hR20 |
| **Space group number** | : | 167 |
| **Space group symbol** | : | $R\bar{3}c$ |
| **AFLOW prototype command** | : | `aflow --proto=A3BC6_hR20_167_e_b_f [--hex]` `--params=`$a, c/a, x_2, x_3, y_3, z_3$ |

**Other compounds with this structure**

- $AlCl_3(H_2O)_6$

---

- The positions of the hydrogen atoms in the water molecules were not determined, so we only provide the positions of the oxygen atoms (labeled as $H_2O$).

---

**Rhombohedral primitive vectors:**

$$\mathbf{a}_1 = \tfrac{1}{2}\,a\,\hat{\mathbf{x}} - \tfrac{1}{2\sqrt{3}}\,a\,\hat{\mathbf{y}} + \tfrac{1}{3}\,c\,\hat{\mathbf{z}}$$
$$\mathbf{a}_2 = \tfrac{1}{\sqrt{3}}\,a\,\hat{\mathbf{y}} + \tfrac{1}{3}\,c\,\hat{\mathbf{z}}$$
$$\mathbf{a}_3 = -\tfrac{1}{2}\,a\,\hat{\mathbf{x}} - \tfrac{1}{2\sqrt{3}}\,a\,\hat{\mathbf{y}} + \tfrac{1}{3}\,c\,\hat{\mathbf{z}}$$

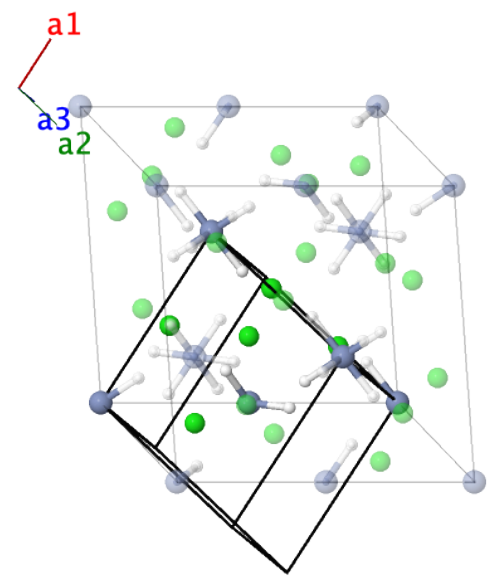

---

**Basis vectors:**

| | | Lattice Coordinates | | Cartesian Coordinates | Wyckoff Position | Atom Type |
|---|---|---|---|---|---|---|
| $\mathbf{B}_1$ | = | $0\,\mathbf{a}_1 + 0\,\mathbf{a}_2 + 0\,\mathbf{a}_3$ | = | $0\,\hat{\mathbf{x}} + 0\,\hat{\mathbf{y}} + 0\,\hat{\mathbf{z}}$ | (2$b$) | Cr |
| $\mathbf{B}_2$ | = | $\frac{1}{2}\,\mathbf{a}_1 + \frac{1}{2}\,\mathbf{a}_2 + \frac{1}{2}\,\mathbf{a}_3$ | = | $\frac{1}{2}c\,\hat{\mathbf{z}}$ | (2$b$) | Cr |
| $\mathbf{B}_3$ | = | $x_2\,\mathbf{a}_1 + \left(\frac{1}{2} - x_2\right)\,\mathbf{a}_2 + \frac{1}{4}\,\mathbf{a}_3$ | = | $\left(-\frac{1}{8} + \frac{1}{2}x_2\right)a\,\hat{\mathbf{x}} + \left(\frac{\sqrt{3}}{8} - \frac{\sqrt{3}}{2}x_2\right)a\,\hat{\mathbf{y}} + \frac{1}{4}c\,\hat{\mathbf{z}}$ | (6$e$) | Cl |
| $\mathbf{B}_4$ | = | $\frac{1}{4}\,\mathbf{a}_1 + x_2\,\mathbf{a}_2 + \left(\frac{1}{2} - x_2\right)\,\mathbf{a}_3$ | = | $\left(-\frac{1}{8} + \frac{1}{2}x_2\right)a\,\hat{\mathbf{x}} + \left(-\frac{\sqrt{3}}{8} + \frac{\sqrt{3}}{2}x_2\right)a\,\hat{\mathbf{y}} + \frac{1}{4}c\,\hat{\mathbf{z}}$ | (6$e$) | Cl |
| $\mathbf{B}_5$ | = | $\left(\frac{1}{2} - x_2\right)\,\mathbf{a}_1 + \frac{1}{4}\,\mathbf{a}_2 + x_2\,\mathbf{a}_3$ | = | $\left(\frac{1}{4} - x_2\right)a\,\hat{\mathbf{x}} + \frac{1}{4}c\,\hat{\mathbf{z}}$ | (6$e$) | Cl |
| $\mathbf{B}_6$ | = | $-x_2\,\mathbf{a}_1 + \left(\frac{1}{2} + x_2\right)\,\mathbf{a}_2 + \frac{3}{4}\,\mathbf{a}_3$ | = | $-a\left(\frac{1}{2}x_2 + \frac{3}{8}\right)\,\hat{\mathbf{x}} + \left(\frac{1}{8\sqrt{3}} + \frac{\sqrt{3}}{2}x_2\right)a\,\hat{\mathbf{y}} + \frac{5}{12}c\,\hat{\mathbf{z}}$ | (6$e$) | Cl |
| $\mathbf{B}_7$ | = | $\frac{3}{4}\,\mathbf{a}_1 - x_2\,\mathbf{a}_2 + \left(\frac{1}{2} + x_2\right)\,\mathbf{a}_3$ | = | $\left(\frac{1}{8} - \frac{1}{2}x_2\right)a\,\hat{\mathbf{x}} - a\left(\frac{\sqrt{3}}{2}x_2 + \frac{5}{8\sqrt{3}}\right)\,\hat{\mathbf{y}} + \frac{5}{12}c\,\hat{\mathbf{z}}$ | (6$e$) | Cl |
| $\mathbf{B}_8$ | = | $\left(\frac{1}{2} + x_2\right)\,\mathbf{a}_1 + \frac{3}{4}\,\mathbf{a}_2 - x_2\,\mathbf{a}_3$ | = | $\left(\frac{1}{4} + x_2\right)a\,\hat{\mathbf{x}} + \frac{1}{2\sqrt{3}}a\,\hat{\mathbf{y}} + \frac{5}{12}c\,\hat{\mathbf{z}}$ | (6$e$) | Cl |
| $\mathbf{B}_9$ | = | $x_3\,\mathbf{a}_1 + y_3\,\mathbf{a}_2 + z_3\,\mathbf{a}_3$ | = | $\frac{1}{2}\left(x_3 - z_3\right)a\,\hat{\mathbf{x}} + \left(-\frac{1}{2\sqrt{3}}x_3 + \frac{1}{\sqrt{3}}y_3 - \frac{1}{2\sqrt{3}}z_3\right)a\,\hat{\mathbf{y}} + \frac{1}{3}\left(x_3 + y_3 + z_3\right)c\,\hat{\mathbf{z}}$ | (12$f$) | $H_2O$ |
| $\mathbf{B}_{10}$ | = | $z_3\,\mathbf{a}_1 + x_3\,\mathbf{a}_2 + y_3\,\mathbf{a}_3$ | = | $\frac{1}{2}\left(-y_3 + z_3\right)a\,\hat{\mathbf{x}} + \left(\frac{1}{\sqrt{3}}x_3 - \frac{1}{2\sqrt{3}}y_3 - \frac{1}{2\sqrt{3}}z_3\right)a\,\hat{\mathbf{y}} + \frac{1}{3}\left(x_3 + y_3 + z_3\right)c\,\hat{\mathbf{z}}$ | (12$f$) | $H_2O$ |
| $\mathbf{B}_{11}$ | = | $y_3\,\mathbf{a}_1 + z_3\,\mathbf{a}_2 + x_3\,\mathbf{a}_3$ | = | $\frac{1}{2}\left(-x_3 + y_3\right)a\,\hat{\mathbf{x}} + \left(-\frac{1}{2\sqrt{3}}x_3 - \frac{1}{2\sqrt{3}}y_3 + \frac{1}{\sqrt{3}}z_3\right)a\,\hat{\mathbf{y}} + \frac{1}{3}\left(x_3 + y_3 + z_3\right)c\,\hat{\mathbf{z}}$ | (12$f$) | $H_2O$ |
| $\mathbf{B}_{12}$ | = | $\left(\frac{1}{2} - z_3\right)\,\mathbf{a}_1 + \left(\frac{1}{2} - y_3\right)\,\mathbf{a}_2 + \left(\frac{1}{2} - x_3\right)\,\mathbf{a}_3$ | = | $\frac{1}{2}\left(x_3 - z_3\right)a\,\hat{\mathbf{x}} + \left(\frac{1}{2\sqrt{3}}x_3 - \frac{1}{\sqrt{3}}y_3 + \frac{1}{2\sqrt{3}}z_3\right)a\,\hat{\mathbf{y}} + \left(\frac{1}{2} - \frac{1}{3}x_3 - \frac{1}{3}y_3 - \frac{1}{3}z_3\right)c\,\hat{\mathbf{z}}$ | (12$f$) | $H_2O$ |
| $\mathbf{B}_{13}$ | = | $\left(\frac{1}{2} - y_3\right)\mathbf{a}_1 + \left(\frac{1}{2} - x_3\right)\mathbf{a}_2 + \left(\frac{1}{2} - z_3\right)\mathbf{a}_3$ | = | $\frac{1}{2}\left(-y_3 + z_3\right) a\,\mathbf{\hat{x}} + \left(-\frac{1}{\sqrt{3}}x_3 + \frac{1}{2\sqrt{3}}y_3 + \frac{1}{2\sqrt{3}}z_3\right) a\,\mathbf{\hat{y}} + \left(\frac{1}{2} - \frac{1}{3}x_3 - \frac{1}{3}y_3 - \frac{1}{3}z_3\right) c\,\mathbf{\hat{z}}$ | (12$f$) | $H_2O$ |
| $\mathbf{B}_{14}$ | = | $\left(\frac{1}{2} - x_3\right)\mathbf{a}_1 + \left(\frac{1}{2} - z_3\right)\mathbf{a}_2 + \left(\frac{1}{2} - y_3\right)\mathbf{a}_3$ | = | $\frac{1}{2}\left(-x_3 + y_3\right) a\,\mathbf{\hat{x}} + \left(\frac{1}{2\sqrt{3}}x_3 + \frac{1}{2\sqrt{3}}y_3 - \frac{1}{\sqrt{3}}z_3\right) a\,\mathbf{\hat{y}} + \left(\frac{1}{2} - \frac{1}{3}x_3 - \frac{1}{3}y_3 - \frac{1}{3}z_3\right) c\,\mathbf{\hat{z}}$ | (12$f$) | $H_2O$ |
| $\mathbf{B}_{15}$ | = | $-x_3\,\mathbf{a}_1 - y_3\,\mathbf{a}_2 - z_3\,\mathbf{a}_3$ | = | $\frac{1}{2}\left(-x_3 + z_3\right) a\,\mathbf{\hat{x}} + \left(\frac{1}{2\sqrt{3}}x_3 - \frac{1}{\sqrt{3}}y_3 + \frac{1}{2\sqrt{3}}z_3\right) a\,\mathbf{\hat{y}} - \frac{1}{3}\left(x_3 + y_3 + z_3\right) c\,\mathbf{\hat{z}}$ | (12$f$) | $H_2O$ |
| $\mathbf{B}_{16}$ | = | $-z_3\,\mathbf{a}_1 - x_3\,\mathbf{a}_2 - y_3\,\mathbf{a}_3$ | = | $\frac{1}{2}\left(y_3 - z_3\right) a\,\mathbf{\hat{x}} + \left(-\frac{1}{\sqrt{3}}x_3 + \frac{1}{2\sqrt{3}}y_3 + \frac{1}{2\sqrt{3}}z_3\right) a\,\mathbf{\hat{y}} - \frac{1}{3}\left(x_3 + y_3 + z_3\right) c\,\mathbf{\hat{z}}$ | (12$f$) | $H_2O$ |
| $\mathbf{B}_{17}$ | = | $-y_3\,\mathbf{a}_1 - z_3\,\mathbf{a}_2 - x_3\,\mathbf{a}_3$ | = | $\frac{1}{2}\left(x_3 - y_3\right) a\,\mathbf{\hat{x}} + \left(\frac{1}{2\sqrt{3}}x_3 + \frac{1}{2\sqrt{3}}y_3 - \frac{1}{\sqrt{3}}z_3\right) a\,\mathbf{\hat{y}} - \frac{1}{3}\left(x_3 + y_3 + z_3\right) c\,\mathbf{\hat{z}}$ | (12$f$) | $H_2O$ |
| $\mathbf{B}_{18}$ | = | $\left(\frac{1}{2} + z_3\right)\mathbf{a}_1 + \left(\frac{1}{2} + y_3\right)\mathbf{a}_2 + \left(\frac{1}{2} + x_3\right)\mathbf{a}_3$ | = | $\frac{1}{2}\left(-x_3 + z_3\right) a\,\mathbf{\hat{x}} + \left(-\frac{1}{2\sqrt{3}}x_3 + \frac{1}{\sqrt{3}}y_3 - \frac{1}{2\sqrt{3}}z_3\right) a\,\mathbf{\hat{y}} + \left(\frac{1}{2} + \frac{1}{3}x_3 + \frac{1}{3}y_3 + \frac{1}{3}z_3\right) c\,\mathbf{\hat{z}}$ | (12$f$) | $H_2O$ |
| $\mathbf{B}_{19}$ | = | $\left(\frac{1}{2} + y_3\right)\mathbf{a}_1 + \left(\frac{1}{2} + x_3\right)\mathbf{a}_2 + \left(\frac{1}{2} + z_3\right)\mathbf{a}_3$ | = | $\frac{1}{2}\left(y_3 - z_3\right) a\,\mathbf{\hat{x}} + \left(\frac{1}{\sqrt{3}}x_3 - \frac{1}{2\sqrt{3}}y_3 - \frac{1}{2\sqrt{3}}z_3\right) a\,\mathbf{\hat{y}} + \left(\frac{1}{2} + \frac{1}{3}x_3 + \frac{1}{3}y_3 + \frac{1}{3}z_3\right) c\,\mathbf{\hat{z}}$ | (12$f$) | $H_2O$ |
| $\mathbf{B}_{20}$ | = | $\left(\frac{1}{2} + x_3\right)\mathbf{a}_1 + \left(\frac{1}{2} + z_3\right)\mathbf{a}_2 + \left(\frac{1}{2} + y_3\right)\mathbf{a}_3$ | = | $\frac{1}{2}\left(x_3 - y_3\right) a\,\mathbf{\hat{x}} + \left(-\frac{1}{2\sqrt{3}}x_3 - \frac{1}{2\sqrt{3}}y_3 + \frac{1}{\sqrt{3}}z_3\right) a\,\mathbf{\hat{y}} + \left(\frac{1}{2} + \frac{1}{3}x_3 + \frac{1}{3}y_3 + \frac{1}{3}z_3\right) c\,\mathbf{\hat{z}}$ | (12$f$) | $H_2O$ |

**Geometry files:**

- CIF: pp. 1746
- POSCAR: pp. 1746

# $FeF_3$ ($D0_{12}$) Structure: A3B_hR8_167_e_b

http://aflow.org/prototype-encyclopedia/A3B_hR8_167_e_b

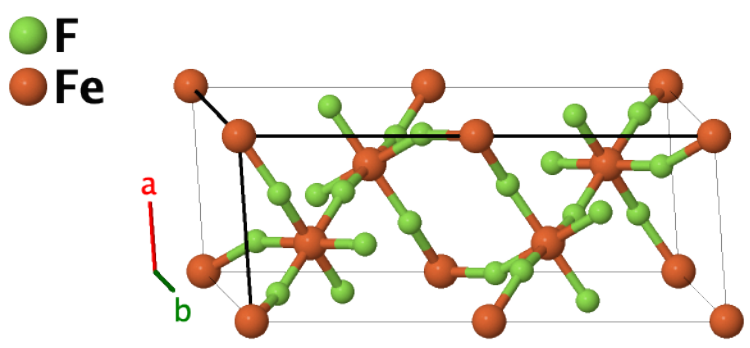

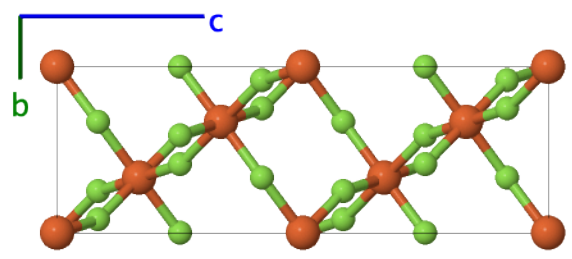

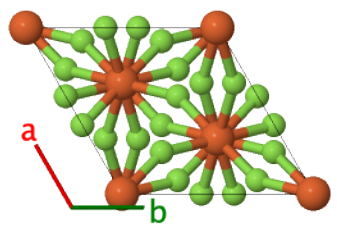

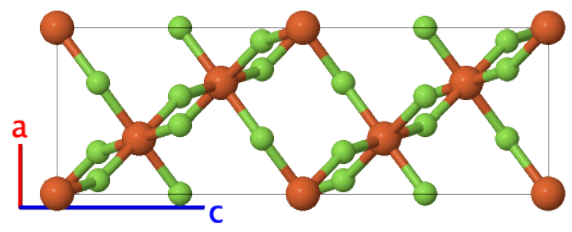

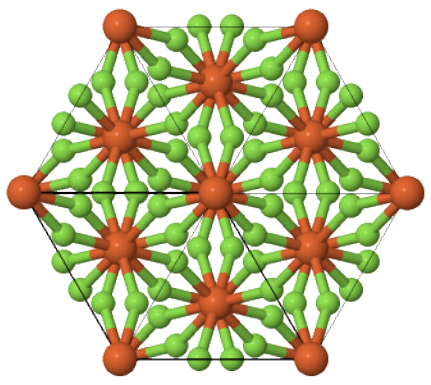

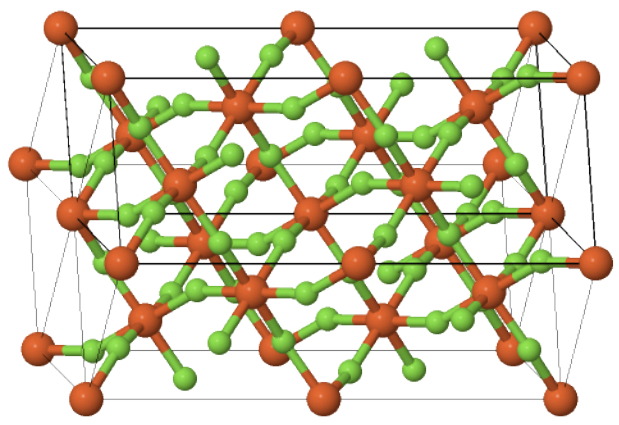

| | | |
|---|---|---|
| **Prototype** | : | $F_3Fe$ |
| **AFLOW prototype label** | : | A3B_hR8_167_e_b |
| ***Strukturbericht* designation** | : | $D0_{12}$ |
| **Pearson symbol** | : | hR8 |
| **Space group number** | : | 167 |
| **Space group symbol** | : | $R\bar{3}c$ |
| **AFLOW prototype command** | : | `aflow --proto=A3B_hR8_167_e_b [--hex]` `--params=`$a, c/a, x_2$ |

### Other compounds with this structure

- $CoF_3$, $RuF_3$, $RhF_3$, $PdF_3$, $IrF_3$, $\alpha$-$AlF_3$, and $AlH_3$

### Rhombohedral primitive vectors:

$$\mathbf{a}_1 = \frac{1}{2}\,a\,\hat{\mathbf{x}} - \frac{1}{2\sqrt{3}}\,a\,\hat{\mathbf{y}} + \frac{1}{3}\,c\,\hat{\mathbf{z}}$$
$$\mathbf{a}_2 = \frac{1}{\sqrt{3}}\,a\,\hat{\mathbf{y}} + \frac{1}{3}\,c\,\hat{\mathbf{z}}$$
$$\mathbf{a}_3 = -\frac{1}{2}\,a\,\hat{\mathbf{x}} - \frac{1}{2\sqrt{3}}\,a\,\hat{\mathbf{y}} + \frac{1}{3}\,c\,\hat{\mathbf{z}}$$

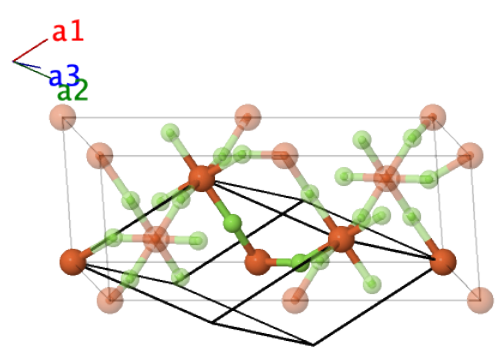

### Basis vectors:

| | | Lattice Coordinates | | Cartesian Coordinates | Wyckoff Position | Atom Type |
|---|---|---|---|---|---|---|
| $\mathbf{B}_1$ | = | $0\,\mathbf{a}_1 + 0\,\mathbf{a}_2 + 0\,\mathbf{a}_3$ | = | $0\,\hat{\mathbf{x}} + 0\,\hat{\mathbf{y}} + 0\,\hat{\mathbf{z}}$ | (2$b$) | Fe |
| $\mathbf{B}_2$ | = | $\frac{1}{2}\,\mathbf{a}_1 + \frac{1}{2}\,\mathbf{a}_2 + \frac{1}{2}\,\mathbf{a}_3$ | = | $\frac{1}{2}c\,\hat{\mathbf{z}}$ | (2$b$) | Fe |
| $\mathbf{B}_3$ | = | $x_2\,\mathbf{a}_1 + \left(\frac{1}{2} - x_2\right)\,\mathbf{a}_2 + \frac{1}{4}\,\mathbf{a}_3$ | = | $\left(-\frac{1}{8} + \frac{1}{2}x_2\right)a\,\hat{\mathbf{x}} + \left(\frac{\sqrt{3}}{8} - \frac{\sqrt{3}}{2}x_2\right)a\,\hat{\mathbf{y}} + \frac{1}{4}c\,\hat{\mathbf{z}}$ | (6$e$) | F |

$$\mathbf{B}_4 = \frac{1}{4}\,\mathbf{a}_1 + x_2\,\mathbf{a}_2 + \left(\frac{1}{2} - x_2\right)\,\mathbf{a}_3 = \left(-\frac{1}{8} + \frac{1}{2}x_2\right)a\,\hat{\mathbf{x}} + \left(-\frac{\sqrt{3}}{8} + \frac{\sqrt{3}}{2}x_2\right)a\,\hat{\mathbf{y}} + \frac{1}{4}c\,\hat{\mathbf{z}} \quad (6e) \quad \text{F}$$

$$\mathbf{B}_5 = \left(\frac{1}{2} - x_2\right)\,\mathbf{a}_1 + \frac{1}{4}\,\mathbf{a}_2 + x_2\,\mathbf{a}_3 = \left(\frac{1}{4} - x_2\right)a\,\hat{\mathbf{x}} + \frac{1}{4}c\,\hat{\mathbf{z}} \quad (6e) \quad \text{F}$$

$$\mathbf{B}_6 = -x_2\,\mathbf{a}_1 + \left(\frac{1}{2} + x_2\right)\,\mathbf{a}_2 + \frac{3}{4}\,\mathbf{a}_3 = -a\left(\frac{1}{2}x_2 + \frac{3}{8}\right)\,\hat{\mathbf{x}} + \left(\frac{1}{8\sqrt{3}} + \frac{\sqrt{3}}{2}x_2\right)a\,\hat{\mathbf{y}} + \frac{5}{12}c\,\hat{\mathbf{z}} \quad (6e) \quad \text{F}$$

$$\mathbf{B}_7 = \frac{3}{4}\,\mathbf{a}_1 - x_2\,\mathbf{a}_2 + \left(\frac{1}{2} + x_2\right)\,\mathbf{a}_3 = \left(\frac{1}{8} - \frac{1}{2}x_2\right)a\,\hat{\mathbf{x}} - a\left(\frac{\sqrt{3}}{2}x_2 + \frac{5}{8\sqrt{3}}\right)\,\hat{\mathbf{y}} + \frac{5}{12}c\,\hat{\mathbf{z}} \quad (6e) \quad \text{F}$$

$$\mathbf{B}_8 = \left(\frac{1}{2} + x_2\right)\,\mathbf{a}_1 + \frac{3}{4}\,\mathbf{a}_2 - x_2\,\mathbf{a}_3 = \left(\frac{1}{4} + x_2\right)a\,\hat{\mathbf{x}} + \frac{1}{2\sqrt{3}}a\,\hat{\mathbf{y}} + \frac{5}{12}c\,\hat{\mathbf{z}} \quad (6e) \quad \text{F}$$

---

---

**Geometry files:**

- CIF: pp. 1746
- POSCAR: pp. 1747

# Rinneite ($K_3NaFeCl_6$) Structure: A6BC3D_hR22_167_f_b_e_a

http://aflow.org/prototype-encyclopedia/A6BC3D_hR22_167_f_b_e_a

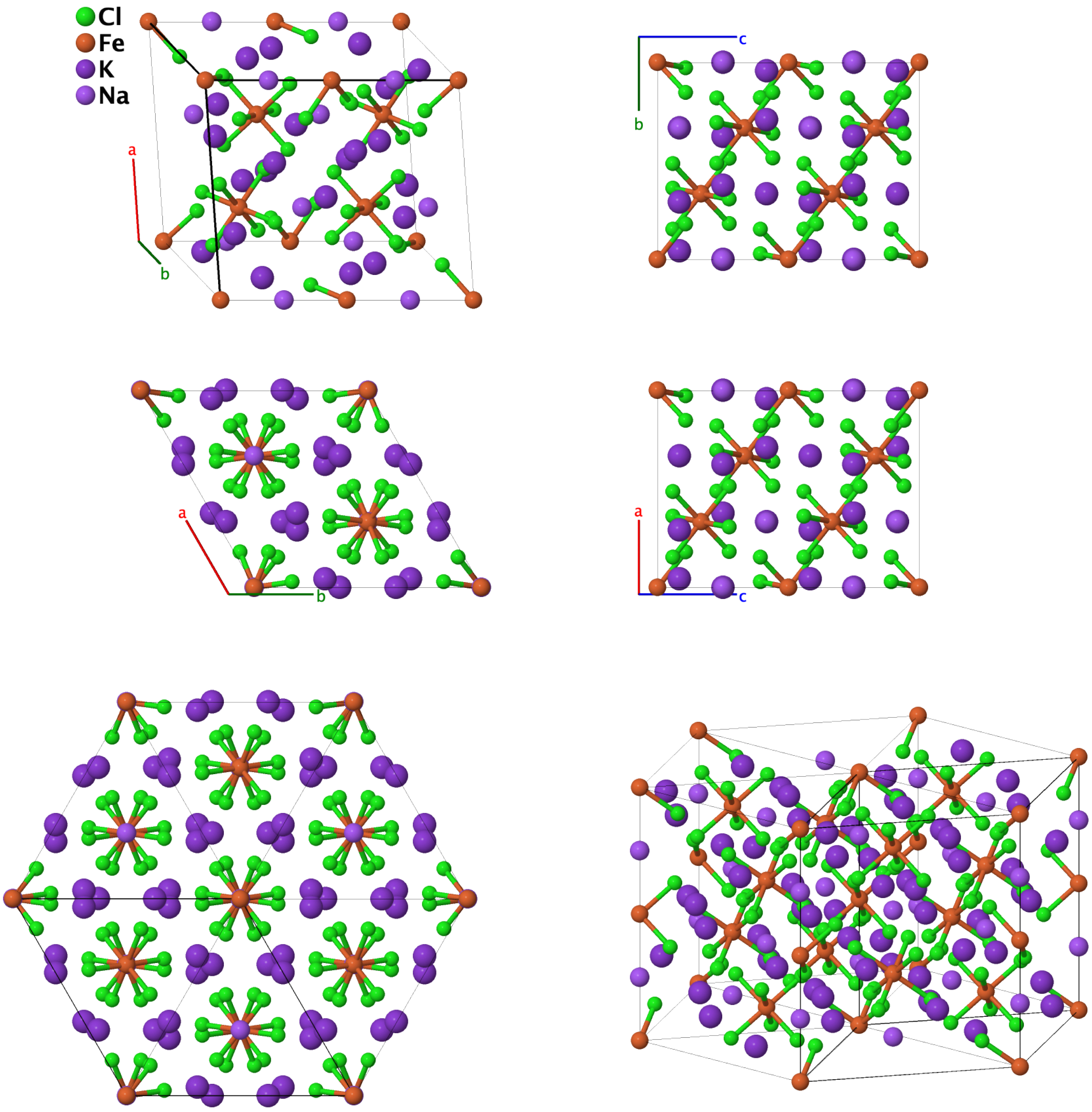

| | | |
|---|---|---|
| **Prototype** | : | $Cl_6FeK_3Na$ |
| **AFLOW prototype label** | : | A6BC3D_hR22_167_f_b_e_a |
| ***Strukturbericht* designation** | : | None |
| **Pearson symbol** | : | hR22 |
| **Space group number** | : | 167 |
| **Space group symbol** | : | $R\bar{3}c$ |
| **AFLOW prototype command** | : | `aflow --proto=A6BC3D_hR22_167_f_b_e_a [--hex]` `--params=`$a, c/a, x_3, x_4, y_4, z_4$ |

**Other compounds with this structure**

- $Ca_3LiRuO_6$, $Sr_3NiIrO_6$, and $K_4CdCl_6$

- We use the data taken at 293 K.

**Rhombohedral primitive vectors:**

$$
\begin{aligned}
\mathbf{a}_1 &= \tfrac{1}{2}\,a\,\hat{\mathbf{x}} - \tfrac{1}{2\sqrt{3}}\,a\,\hat{\mathbf{y}} + \tfrac{1}{3}\,c\,\hat{\mathbf{z}} \\
\mathbf{a}_2 &= \tfrac{1}{\sqrt{3}}\,a\,\hat{\mathbf{y}} + \tfrac{1}{3}\,c\,\hat{\mathbf{z}} \\
\mathbf{a}_3 &= -\tfrac{1}{2}\,a\,\hat{\mathbf{x}} - \tfrac{1}{2\sqrt{3}}\,a\,\hat{\mathbf{y}} + \tfrac{1}{3}\,c\,\hat{\mathbf{z}}
\end{aligned}
$$

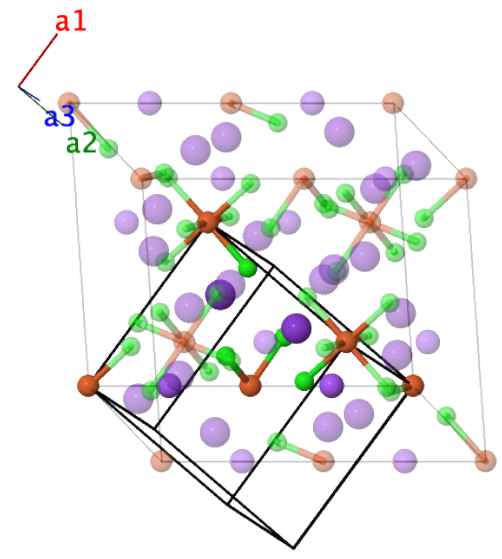

**Basis vectors:**

| | | Lattice Coordinates | | Cartesian Coordinates | Wyckoff Position | Atom Type |
|---|---|---|---|---|---|---|
| $\mathbf{B}_1$ | = | $\frac{1}{4}\,\mathbf{a}_1 + \frac{1}{4}\,\mathbf{a}_2 + \frac{1}{4}\,\mathbf{a}_3$ | = | $\frac{1}{4}c\,\hat{\mathbf{z}}$ | (2*a*) | Na |
| $\mathbf{B}_2$ | = | $\frac{3}{4}\,\mathbf{a}_1 + \frac{3}{4}\,\mathbf{a}_2 + \frac{3}{4}\,\mathbf{a}_3$ | = | $\frac{3}{4}c\,\hat{\mathbf{z}}$ | (2*a*) | Na |
| $\mathbf{B}_3$ | = | $0\,\mathbf{a}_1 + 0\,\mathbf{a}_2 + 0\,\mathbf{a}_3$ | = | $0\,\hat{\mathbf{x}} + 0\,\hat{\mathbf{y}} + 0\,\hat{\mathbf{z}}$ | (2*b*) | Fe |
| $\mathbf{B}_4$ | = | $\frac{1}{2}\,\mathbf{a}_1 + \frac{1}{2}\,\mathbf{a}_2 + \frac{1}{2}\,\mathbf{a}_3$ | = | $\frac{1}{2}c\,\hat{\mathbf{z}}$ | (2*b*) | Fe |
| $\mathbf{B}_5$ | = | $x_3\,\mathbf{a}_1 + \left(\frac{1}{2} - x_3\right)\,\mathbf{a}_2 + \frac{1}{4}\,\mathbf{a}_3$ | = | $\left(-\frac{1}{8} + \frac{1}{2}x_3\right)a\,\hat{\mathbf{x}} + \left(\frac{\sqrt{3}}{8} - \frac{\sqrt{3}}{2}x_3\right)a\,\hat{\mathbf{y}} + \frac{1}{4}c\,\hat{\mathbf{z}}$ | (6*e*) | K |
| $\mathbf{B}_6$ | = | $\frac{1}{4}\,\mathbf{a}_1 + x_3\,\mathbf{a}_2 + \left(\frac{1}{2} - x_3\right)\,\mathbf{a}_3$ | = | $\left(-\frac{1}{8} + \frac{1}{2}x_3\right)a\,\hat{\mathbf{x}} + \left(-\frac{\sqrt{3}}{8} + \frac{\sqrt{3}}{2}x_3\right)a\,\hat{\mathbf{y}} + \frac{1}{4}c\,\hat{\mathbf{z}}$ | (6*e*) | K |
| $\mathbf{B}_7$ | = | $\left(\frac{1}{2} - x_3\right)\,\mathbf{a}_1 + \frac{1}{4}\,\mathbf{a}_2 + x_3\,\mathbf{a}_3$ | = | $\left(\frac{1}{4} - x_3\right)a\,\hat{\mathbf{x}} + \frac{1}{4}c\,\hat{\mathbf{z}}$ | (6*e*) | K |
| $\mathbf{B}_8$ | = | $-x_3\,\mathbf{a}_1 + \left(\frac{1}{2} + x_3\right)\,\mathbf{a}_2 + \frac{3}{4}\,\mathbf{a}_3$ | = | $-a\left(\frac{1}{2}x_3 + \frac{3}{8}\right)\,\hat{\mathbf{x}} + \left(\frac{1}{8\sqrt{3}} + \frac{\sqrt{3}}{2}x_3\right)a\,\hat{\mathbf{y}} + \frac{5}{12}c\,\hat{\mathbf{z}}$ | (6*e*) | K |
| $\mathbf{B}_9$ | = | $\frac{3}{4}\,\mathbf{a}_1 - x_3\,\mathbf{a}_2 + \left(\frac{1}{2} + x_3\right)\,\mathbf{a}_3$ | = | $\left(\frac{1}{8} - \frac{1}{2}x_3\right)a\,\hat{\mathbf{x}} - a\left(\frac{\sqrt{3}}{2}x_3 + \frac{5}{8\sqrt{3}}\right)\,\hat{\mathbf{y}} + \frac{5}{12}c\,\hat{\mathbf{z}}$ | (6*e*) | K |
| $\mathbf{B}_{10}$ | = | $\left(\frac{1}{2} + x_3\right)\,\mathbf{a}_1 + \frac{3}{4}\,\mathbf{a}_2 - x_3\,\mathbf{a}_3$ | = | $\left(\frac{1}{4} + x_3\right)a\,\hat{\mathbf{x}} + \frac{1}{2\sqrt{3}}a\,\hat{\mathbf{y}} + \frac{5}{12}c\,\hat{\mathbf{z}}$ | (6*e*) | K |
| $\mathbf{B}_{11}$ | = | $x_4\,\mathbf{a}_1 + y_4\,\mathbf{a}_2 + z_4\,\mathbf{a}_3$ | = | $\frac{1}{2}\left(x_4 - z_4\right)a\,\hat{\mathbf{x}} + \left(-\frac{1}{2\sqrt{3}}x_4 + \frac{1}{\sqrt{3}}y_4 - \frac{1}{2\sqrt{3}}z_4\right)a\,\hat{\mathbf{y}} + \frac{1}{3}\left(x_4 + y_4 + z_4\right)c\,\hat{\mathbf{z}}$ | (12*f*) | Cl |
| $\mathbf{B}_{12}$ | = | $z_4\,\mathbf{a}_1 + x_4\,\mathbf{a}_2 + y_4\,\mathbf{a}_3$ | = | $\frac{1}{2}\left(-y_4 + z_4\right)a\,\hat{\mathbf{x}} + \left(\frac{1}{\sqrt{3}}x_4 - \frac{1}{2\sqrt{3}}y_4 - \frac{1}{2\sqrt{3}}z_4\right)a\,\hat{\mathbf{y}} + \frac{1}{3}\left(x_4 + y_4 + z_4\right)c\,\hat{\mathbf{z}}$ | (12*f*) | Cl |
| $\mathbf{B}_{13}$ | = | $y_4\,\mathbf{a}_1 + z_4\,\mathbf{a}_2 + x_4\,\mathbf{a}_3$ | = | $\frac{1}{2}\left(-x_4 + y_4\right)a\,\hat{\mathbf{x}} + \left(-\frac{1}{2\sqrt{3}}x_4 - \frac{1}{2\sqrt{3}}y_4 + \frac{1}{\sqrt{3}}z_4\right)a\,\hat{\mathbf{y}} + \frac{1}{3}\left(x_4 + y_4 + z_4\right)c\,\hat{\mathbf{z}}$ | (12*f*) | Cl |
| $\mathbf{B}_{14}$ | = | $\left(\frac{1}{2} - z_4\right)\,\mathbf{a}_1 + \left(\frac{1}{2} - y_4\right)\,\mathbf{a}_2 + \left(\frac{1}{2} - x_4\right)\,\mathbf{a}_3$ | = | $\frac{1}{2}\left(x_4 - z_4\right)a\,\hat{\mathbf{x}} + \left(\frac{1}{2\sqrt{3}}x_4 - \frac{1}{\sqrt{3}}y_4 + \frac{1}{2\sqrt{3}}z_4\right)a\,\hat{\mathbf{y}} + \left(\frac{1}{2} - \frac{1}{3}x_4 - \frac{1}{3}y_4 - \frac{1}{3}z_4\right)c\,\hat{\mathbf{z}}$ | (12*f*) | Cl |
| $\mathbf{B}_{15}$ | = | $\left(\frac{1}{2} - y_4\right)\,\mathbf{a}_1 + \left(\frac{1}{2} - x_4\right)\,\mathbf{a}_2 + \left(\frac{1}{2} - z_4\right)\,\mathbf{a}_3$ | = | $\frac{1}{2}\left(-y_4 + z_4\right)a\,\hat{\mathbf{x}} + \left(-\frac{1}{\sqrt{3}}x_4 + \frac{1}{2\sqrt{3}}y_4 + \frac{1}{2\sqrt{3}}z_4\right)a\,\hat{\mathbf{y}} + \left(\frac{1}{2} - \frac{1}{3}x_4 - \frac{1}{3}y_4 - \frac{1}{3}z_4\right)c\,\hat{\mathbf{z}}$ | (12*f*) | Cl |
| $\mathbf{B}_{16}$ | = | $\left(\frac{1}{2} - x_4\right)\mathbf{a}_1 + \left(\frac{1}{2} - z_4\right)\mathbf{a}_2 + \left(\frac{1}{2} - y_4\right)\mathbf{a}_3$ | = | $\frac{1}{2}\left(-x_4 + y_4\right)a\,\mathbf{\hat{x}} + \left(\frac{1}{2\sqrt{3}}x_4 + \frac{1}{2\sqrt{3}}y_4 - \frac{1}{\sqrt{3}}z_4\right)a\,\mathbf{\hat{y}} + \left(\frac{1}{2} - \frac{1}{3}x_4 - \frac{1}{3}y_4 - \frac{1}{3}z_4\right)c\,\mathbf{\hat{z}}$ | (12$f$) | Cl |
| $\mathbf{B}_{17}$ | = | $-x_4\,\mathbf{a}_1 - y_4\,\mathbf{a}_2 - z_4\,\mathbf{a}_3$ | = | $\frac{1}{2}\left(-x_4 + z_4\right)a\,\mathbf{\hat{x}} + \left(\frac{1}{2\sqrt{3}}x_4 - \frac{1}{\sqrt{3}}y_4 + \frac{1}{2\sqrt{3}}z_4\right)a\,\mathbf{\hat{y}} - \frac{1}{3}\left(x_4 + y_4 + z_4\right)c\,\mathbf{\hat{z}}$ | (12$f$) | Cl |
| $\mathbf{B}_{18}$ | = | $-z_4\,\mathbf{a}_1 - x_4\,\mathbf{a}_2 - y_4\,\mathbf{a}_3$ | = | $\frac{1}{2}\left(y_4 - z_4\right)a\,\mathbf{\hat{x}} + \left(-\frac{1}{\sqrt{3}}x_4 + \frac{1}{2\sqrt{3}}y_4 + \frac{1}{2\sqrt{3}}z_4\right)a\,\mathbf{\hat{y}} - \frac{1}{3}\left(x_4 + y_4 + z_4\right)c\,\mathbf{\hat{z}}$ | (12$f$) | Cl |
| $\mathbf{B}_{19}$ | = | $-y_4\,\mathbf{a}_1 - z_4\,\mathbf{a}_2 - x_4\,\mathbf{a}_3$ | = | $\frac{1}{2}\left(x_4 - y_4\right)a\,\mathbf{\hat{x}} + \left(\frac{1}{2\sqrt{3}}x_4 + \frac{1}{2\sqrt{3}}y_4 - \frac{1}{\sqrt{3}}z_4\right)a\,\mathbf{\hat{y}} - \frac{1}{3}\left(x_4 + y_4 + z_4\right)c\,\mathbf{\hat{z}}$ | (12$f$) | Cl |
| $\mathbf{B}_{20}$ | = | $\left(\frac{1}{2} + z_4\right)\mathbf{a}_1 + \left(\frac{1}{2} + y_4\right)\mathbf{a}_2 + \left(\frac{1}{2} + x_4\right)\mathbf{a}_3$ | = | $\frac{1}{2}\left(-x_4 + z_4\right)a\,\mathbf{\hat{x}} + \left(-\frac{1}{2\sqrt{3}}x_4 + \frac{1}{\sqrt{3}}y_4 - \frac{1}{2\sqrt{3}}z_4\right)a\,\mathbf{\hat{y}} + \left(\frac{1}{2} + \frac{1}{3}x_4 + \frac{1}{3}y_4 + \frac{1}{3}z_4\right)c\,\mathbf{\hat{z}}$ | (12$f$) | Cl |
| $\mathbf{B}_{21}$ | = | $\left(\frac{1}{2} + y_4\right)\mathbf{a}_1 + \left(\frac{1}{2} + x_4\right)\mathbf{a}_2 + \left(\frac{1}{2} + z_4\right)\mathbf{a}_3$ | = | $\frac{1}{2}\left(y_4 - z_4\right)a\,\mathbf{\hat{x}} + \left(\frac{1}{\sqrt{3}}x_4 - \frac{1}{2\sqrt{3}}y_4 - \frac{1}{2\sqrt{3}}z_4\right)a\,\mathbf{\hat{y}} + \left(\frac{1}{2} + \frac{1}{3}x_4 + \frac{1}{3}y_4 + \frac{1}{3}z_4\right)c\,\mathbf{\hat{z}}$ | (12$f$) | Cl |
| $\mathbf{B}_{22}$ | = | $\left(\frac{1}{2} + x_4\right)\mathbf{a}_1 + \left(\frac{1}{2} + z_4\right)\mathbf{a}_2 + \left(\frac{1}{2} + y_4\right)\mathbf{a}_3$ | = | $\frac{1}{2}\left(x_4 - y_4\right)a\,\mathbf{\hat{x}} + \left(-\frac{1}{2\sqrt{3}}x_4 - \frac{1}{2\sqrt{3}}y_4 + \frac{1}{\sqrt{3}}z_4\right)a\,\mathbf{\hat{y}} + \left(\frac{1}{2} + \frac{1}{3}x_4 + \frac{1}{3}y_4 + \frac{1}{3}z_4\right)c\,\mathbf{\hat{z}}$ | (12$f$) | Cl |

**Geometry files:**

- CIF: pp. 1747
- POSCAR: pp. 1748

# $Cs_3Tl_2Cl_9$ ($K7_2$) Structure: A9B3C2_hR28_167_ef_e_c

http://aflow.org/prototype-encyclopedia/A9B3C2_hR28_167_ef_e_c

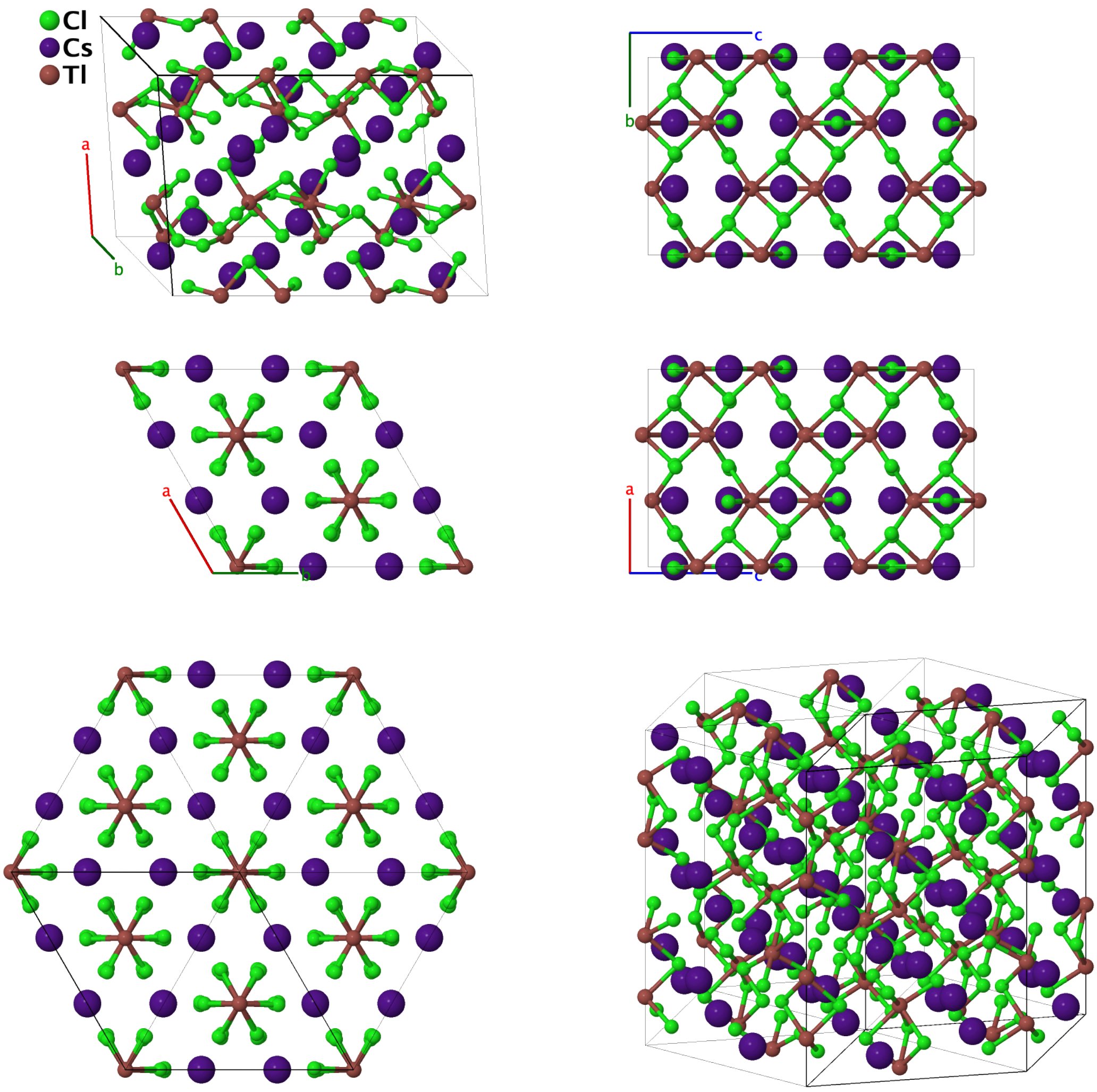

| | | |
|---|---|---|
| **Prototype** | : | $Cl_9Cs_3Tl_2$ |
| **AFLOW prototype label** | : | A9B3C2_hR28_167_ef_e_c |
| ***Strukturbericht* designation** | : | $K7_2$ |
| **Pearson symbol** | : | hR28 |
| **Space group number** | : | 167 |
| **Space group symbol** | : | $R\bar{3}c$ |
| **AFLOW prototype command** | : | `aflow --proto=A9B3C2_hR28_167_ef_e_c [--hex]` `--params=`$a, c/a, x_1, x_2, x_3, x_4, y_4, z_4$ |

**Other compounds with this structure**

- $Cs_3Dy_2Br_9$, $Cs_3Er_2Br_9$, $Cs_3Ho_2Br_9$, $Cs_3Lu_2Cl_9$, $Cs_3Tb_2Br_9$, $Cs_3Yb_2Br_9$, $Ba_3Os_2O_9$, and $Ba_3W_2O_9$

- (Hoard, 1935) followed by (Downs, 2003), give the atomic coordinates in the style of (Wyckoff, 1922), who lists space group $D_{3d}^6$ (the Schönflies notation for space group $R\bar{3}c$) as space group #203, instead of #167, and uses an origin which corresponds to $(1/4, 1/4, 1/4)$ in our lattice coordinates. We used FINDSYM to convert this into our standard setting for space group #167.

---

**Rhombohedral primitive vectors:**

$$\begin{aligned}
\mathbf{a}_1 &= \tfrac{1}{2}\,a\,\hat{\mathbf{x}} - \tfrac{1}{2\sqrt{3}}\,a\,\hat{\mathbf{y}} + \tfrac{1}{3}\,c\,\hat{\mathbf{z}} \\
\mathbf{a}_2 &= \tfrac{1}{\sqrt{3}}\,a\,\hat{\mathbf{y}} + \tfrac{1}{3}\,c\,\hat{\mathbf{z}} \\
\mathbf{a}_3 &= -\tfrac{1}{2}\,a\,\hat{\mathbf{x}} - \tfrac{1}{2\sqrt{3}}\,a\,\hat{\mathbf{y}} + \tfrac{1}{3}\,c\,\hat{\mathbf{z}}
\end{aligned}$$

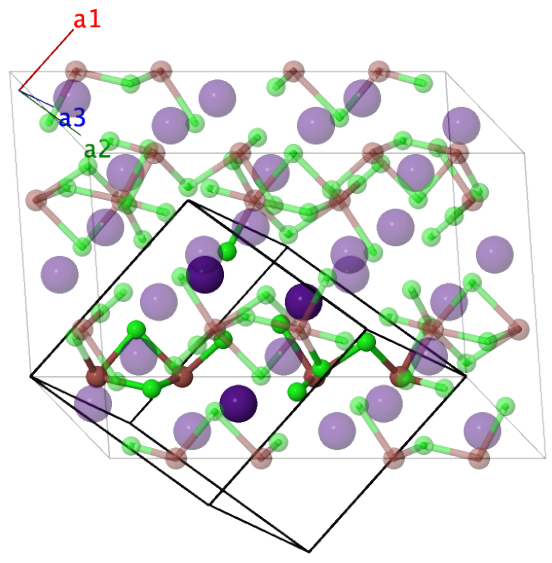

---

**Basis vectors:**

| | | Lattice Coordinates | | Cartesian Coordinates | Wyckoff Position | Atom Type |
|---|---|---|---|---|---|---|
| $\mathbf{B}_1$ | = | $x_1\,\mathbf{a}_1 + x_1\,\mathbf{a}_2 + x_1\,\mathbf{a}_3$ | = | $x_1 c\,\hat{\mathbf{z}}$ | (4$c$) | Tl |
| $\mathbf{B}_2$ | = | $\left(\frac{1}{2} - x_1\right)\mathbf{a}_1 + \left(\frac{1}{2} - x_1\right)\mathbf{a}_2 + \left(\frac{1}{2} - x_1\right)\mathbf{a}_3$ | = | $\left(\frac{1}{2} - x_1\right)c\,\hat{\mathbf{z}}$ | (4$c$) | Tl |
| $\mathbf{B}_3$ | = | $-x_1\,\mathbf{a}_1 - x_1\,\mathbf{a}_2 - x_1\,\mathbf{a}_3$ | = | $-x_1 c\,\hat{\mathbf{z}}$ | (4$c$) | Tl |
| $\mathbf{B}_4$ | = | $\left(\frac{1}{2} + x_1\right)\mathbf{a}_1 + \left(\frac{1}{2} + x_1\right)\mathbf{a}_2 + \left(\frac{1}{2} + x_1\right)\mathbf{a}_3$ | = | $\left(\frac{1}{2} + x_1\right)c\,\hat{\mathbf{z}}$ | (4$c$) | Tl |
| $\mathbf{B}_5$ | = | $x_2\,\mathbf{a}_1 + \left(\frac{1}{2} - x_2\right)\mathbf{a}_2 + \frac{1}{4}\,\mathbf{a}_3$ | = | $\left(-\frac{1}{8} + \frac{1}{2}x_2\right)a\,\hat{\mathbf{x}} + \left(\frac{\sqrt{3}}{8} - \frac{\sqrt{3}}{2}x_2\right)a\,\hat{\mathbf{y}} + \frac{1}{4}c\,\hat{\mathbf{z}}$ | (6$e$) | Cl I |
| $\mathbf{B}_6$ | = | $\frac{1}{4}\,\mathbf{a}_1 + x_2\,\mathbf{a}_2 + \left(\frac{1}{2} - x_2\right)\mathbf{a}_3$ | = | $\left(-\frac{1}{8} + \frac{1}{2}x_2\right)a\,\hat{\mathbf{x}} + \left(-\frac{\sqrt{3}}{8} + \frac{\sqrt{3}}{2}x_2\right)a\,\hat{\mathbf{y}} + \frac{1}{4}c\,\hat{\mathbf{z}}$ | (6$e$) | Cl I |
| $\mathbf{B}_7$ | = | $\left(\frac{1}{2} - x_2\right)\mathbf{a}_1 + \frac{1}{4}\,\mathbf{a}_2 + x_2\,\mathbf{a}_3$ | = | $\left(\frac{1}{4} - x_2\right)a\,\hat{\mathbf{x}} + \frac{1}{4}c\,\hat{\mathbf{z}}$ | (6$e$) | Cl I |
| $\mathbf{B}_8$ | = | $-x_2\,\mathbf{a}_1 + \left(\frac{1}{2} + x_2\right)\mathbf{a}_2 + \frac{3}{4}\,\mathbf{a}_3$ | = | $-a\left(\frac{1}{2}x_2 + \frac{3}{8}\right)\hat{\mathbf{x}} + \left(\frac{1}{8\sqrt{3}} + \frac{\sqrt{3}}{2}x_2\right)a\,\hat{\mathbf{y}} + \frac{5}{12}c\,\hat{\mathbf{z}}$ | (6$e$) | Cl I |
| $\mathbf{B}_9$ | = | $\frac{3}{4}\,\mathbf{a}_1 - x_2\,\mathbf{a}_2 + \left(\frac{1}{2} + x_2\right)\mathbf{a}_3$ | = | $\left(\frac{1}{8} - \frac{1}{2}x_2\right)a\,\hat{\mathbf{x}} - a\left(\frac{\sqrt{3}}{2}x_2 + \frac{5}{8\sqrt{3}}\right)\hat{\mathbf{y}} + \frac{5}{12}c\,\hat{\mathbf{z}}$ | (6$e$) | Cl I |
| $\mathbf{B}_{10}$ | = | $\left(\frac{1}{2} + x_2\right)\mathbf{a}_1 + \frac{3}{4}\,\mathbf{a}_2 - x_2\,\mathbf{a}_3$ | = | $\left(\frac{1}{4} + x_2\right)a\,\hat{\mathbf{x}} + \frac{1}{2\sqrt{3}}a\,\hat{\mathbf{y}} + \frac{5}{12}c\,\hat{\mathbf{z}}$ | (6$e$) | Cl I |
| $\mathbf{B}_{11}$ | = | $x_3\,\mathbf{a}_1 + \left(\frac{1}{2} - x_3\right)\mathbf{a}_2 + \frac{1}{4}\,\mathbf{a}_3$ | = | $\left(-\frac{1}{8} + \frac{1}{2}x_3\right)a\,\hat{\mathbf{x}} + \left(\frac{\sqrt{3}}{8} - \frac{\sqrt{3}}{2}x_3\right)a\,\hat{\mathbf{y}} + \frac{1}{4}c\,\hat{\mathbf{z}}$ | (6$e$) | Cs |
| $\mathbf{B}_{12}$ | = | $\frac{1}{4}\,\mathbf{a}_1 + x_3\,\mathbf{a}_2 + \left(\frac{1}{2} - x_3\right)\mathbf{a}_3$ | = | $\left(-\frac{1}{8} + \frac{1}{2}x_3\right)a\,\hat{\mathbf{x}} + \left(-\frac{\sqrt{3}}{8} + \frac{\sqrt{3}}{2}x_3\right)a\,\hat{\mathbf{y}} + \frac{1}{4}c\,\hat{\mathbf{z}}$ | (6$e$) | Cs |
| $\mathbf{B}_{13}$ | = | $\left(\frac{1}{2} - x_3\right)\mathbf{a}_1 + \frac{1}{4}\,\mathbf{a}_2 + x_3\,\mathbf{a}_3$ | = | $\left(\frac{1}{4} - x_3\right)a\,\hat{\mathbf{x}} + \frac{1}{4}c\,\hat{\mathbf{z}}$ | (6$e$) | Cs |
| $\mathbf{B}_{14}$ | = | $-x_3\,\mathbf{a}_1 + \left(\frac{1}{2} + x_3\right)\mathbf{a}_2 + \frac{3}{4}\,\mathbf{a}_3$ | = | $-a\left(\frac{1}{2}x_3 + \frac{3}{8}\right)\hat{\mathbf{x}} + \left(\frac{1}{8\sqrt{3}} + \frac{\sqrt{3}}{2}x_3\right)a\,\hat{\mathbf{y}} + \frac{5}{12}c\,\hat{\mathbf{z}}$ | (6$e$) | Cs |
| $\mathbf{B}_{15}$ | = | $\frac{3}{4}\,\mathbf{a}_1 - x_3\,\mathbf{a}_2 + \left(\frac{1}{2} + x_3\right)\mathbf{a}_3$ | = | $\left(\frac{1}{8} - \frac{1}{2}x_3\right)a\,\hat{\mathbf{x}} - a\left(\frac{\sqrt{3}}{2}x_3 + \frac{5}{8\sqrt{3}}\right)\hat{\mathbf{y}} + \frac{5}{12}c\,\hat{\mathbf{z}}$ | (6$e$) | Cs |
| $\mathbf{B}_{16}$ | = | $\left(\frac{1}{2} + x_3\right)\mathbf{a}_1 + \frac{3}{4}\,\mathbf{a}_2 - x_3\,\mathbf{a}_3$ | = | $\left(\frac{1}{4} + x_3\right)a\,\hat{\mathbf{x}} + \frac{1}{2\sqrt{3}}a\,\hat{\mathbf{y}} + \frac{5}{12}c\,\hat{\mathbf{z}}$ | (6$e$) | Cs |

$$\mathbf{B}_{17} = x_4\,\mathbf{a}_1 + y_4\,\mathbf{a}_2 + z_4\,\mathbf{a}_3 = \frac{1}{2}(x_4 - z_4)\,a\,\hat{\mathbf{x}} + \left(-\frac{1}{2\sqrt{3}}x_4 + \frac{1}{\sqrt{3}}y_4 - \frac{1}{2\sqrt{3}}z_4\right)a\,\hat{\mathbf{y}} + \frac{1}{3}(x_4 + y_4 + z_4)\,c\,\hat{\mathbf{z}} \quad (12f) \quad \text{Cl II}$$

$$\mathbf{B}_{18} = z_4\,\mathbf{a}_1 + x_4\,\mathbf{a}_2 + y_4\,\mathbf{a}_3 = \frac{1}{2}(-y_4 + z_4)\,a\,\hat{\mathbf{x}} + \left(\frac{1}{\sqrt{3}}x_4 - \frac{1}{2\sqrt{3}}y_4 - \frac{1}{2\sqrt{3}}z_4\right)a\,\hat{\mathbf{y}} + \frac{1}{3}(x_4 + y_4 + z_4)\,c\,\hat{\mathbf{z}} \quad (12f) \quad \text{Cl II}$$

$$\mathbf{B}_{19} = y_4\,\mathbf{a}_1 + z_4\,\mathbf{a}_2 + x_4\,\mathbf{a}_3 = \frac{1}{2}(-x_4 + y_4)\,a\,\hat{\mathbf{x}} + \left(-\frac{1}{2\sqrt{3}}x_4 - \frac{1}{2\sqrt{3}}y_4 + \frac{1}{\sqrt{3}}z_4\right)a\,\hat{\mathbf{y}} + \frac{1}{3}(x_4 + y_4 + z_4)\,c\,\hat{\mathbf{z}} \quad (12f) \quad \text{Cl II}$$

$$\mathbf{B}_{20} = \left(\frac{1}{2} - z_4\right)\mathbf{a}_1 + \left(\frac{1}{2} - y_4\right)\mathbf{a}_2 + \left(\frac{1}{2} - x_4\right)\mathbf{a}_3 = \frac{1}{2}(x_4 - z_4)\,a\,\hat{\mathbf{x}} + \left(\frac{1}{2\sqrt{3}}x_4 - \frac{1}{\sqrt{3}}y_4 + \frac{1}{2\sqrt{3}}z_4\right)a\,\hat{\mathbf{y}} + \left(\frac{1}{2} - \frac{1}{3}x_4 - \frac{1}{3}y_4 - \frac{1}{3}z_4\right)c\,\hat{\mathbf{z}} \quad (12f) \quad \text{Cl II}$$

$$\mathbf{B}_{21} = \left(\frac{1}{2} - y_4\right)\mathbf{a}_1 + \left(\frac{1}{2} - x_4\right)\mathbf{a}_2 + \left(\frac{1}{2} - z_4\right)\mathbf{a}_3 = \frac{1}{2}(-y_4 + z_4)\,a\,\hat{\mathbf{x}} + \left(-\frac{1}{\sqrt{3}}x_4 + \frac{1}{2\sqrt{3}}y_4 + \frac{1}{2\sqrt{3}}z_4\right)a\,\hat{\mathbf{y}} + \left(\frac{1}{2} - \frac{1}{3}x_4 - \frac{1}{3}y_4 - \frac{1}{3}z_4\right)c\,\hat{\mathbf{z}} \quad (12f) \quad \text{Cl II}$$

$$\mathbf{B}_{22} = \left(\frac{1}{2} - x_4\right)\mathbf{a}_1 + \left(\frac{1}{2} - z_4\right)\mathbf{a}_2 + \left(\frac{1}{2} - y_4\right)\mathbf{a}_3 = \frac{1}{2}(-x_4 + y_4)\,a\,\hat{\mathbf{x}} + \left(\frac{1}{2\sqrt{3}}x_4 + \frac{1}{2\sqrt{3}}y_4 - \frac{1}{\sqrt{3}}z_4\right)a\,\hat{\mathbf{y}} + \left(\frac{1}{2} - \frac{1}{3}x_4 - \frac{1}{3}y_4 - \frac{1}{3}z_4\right)c\,\hat{\mathbf{z}} \quad (12f) \quad \text{Cl II}$$

$$\mathbf{B}_{23} = -x_4\,\mathbf{a}_1 - y_4\,\mathbf{a}_2 - z_4\,\mathbf{a}_3 = \frac{1}{2}(-x_4 + z_4)\,a\,\hat{\mathbf{x}} + \left(\frac{1}{2\sqrt{3}}x_4 - \frac{1}{\sqrt{3}}y_4 + \frac{1}{2\sqrt{3}}z_4\right)a\,\hat{\mathbf{y}} - \frac{1}{3}(x_4 + y_4 + z_4)\,c\,\hat{\mathbf{z}} \quad (12f) \quad \text{Cl II}$$

$$\mathbf{B}_{24} = -z_4\,\mathbf{a}_1 - x_4\,\mathbf{a}_2 - y_4\,\mathbf{a}_3 = \frac{1}{2}(y_4 - z_4)\,a\,\hat{\mathbf{x}} + \left(-\frac{1}{\sqrt{3}}x_4 + \frac{1}{2\sqrt{3}}y_4 + \frac{1}{2\sqrt{3}}z_4\right)a\,\hat{\mathbf{y}} - \frac{1}{3}(x_4 + y_4 + z_4)\,c\,\hat{\mathbf{z}} \quad (12f) \quad \text{Cl II}$$

$$\mathbf{B}_{25} = -y_4\,\mathbf{a}_1 - z_4\,\mathbf{a}_2 - x_4\,\mathbf{a}_3 = \frac{1}{2}(x_4 - y_4)\,a\,\hat{\mathbf{x}} + \left(\frac{1}{2\sqrt{3}}x_4 + \frac{1}{2\sqrt{3}}y_4 - \frac{1}{\sqrt{3}}z_4\right)a\,\hat{\mathbf{y}} - \frac{1}{3}(x_4 + y_4 + z_4)\,c\,\hat{\mathbf{z}} \quad (12f) \quad \text{Cl II}$$

$$\mathbf{B}_{26} = \left(\frac{1}{2} + z_4\right)\mathbf{a}_1 + \left(\frac{1}{2} + y_4\right)\mathbf{a}_2 + \left(\frac{1}{2} + x_4\right)\mathbf{a}_3 = \frac{1}{2}(-x_4 + z_4)\,a\,\hat{\mathbf{x}} + \left(-\frac{1}{2\sqrt{3}}x_4 + \frac{1}{\sqrt{3}}y_4 - \frac{1}{2\sqrt{3}}z_4\right)a\,\hat{\mathbf{y}} + \left(\frac{1}{2} + \frac{1}{3}x_4 + \frac{1}{3}y_4 + \frac{1}{3}z_4\right)c\,\hat{\mathbf{z}} \quad (12f) \quad \text{Cl II}$$

$$\mathbf{B}_{27} = \left(\frac{1}{2} + y_4\right)\mathbf{a}_1 + \left(\frac{1}{2} + x_4\right)\mathbf{a}_2 + \left(\frac{1}{2} + z_4\right)\mathbf{a}_3 = \frac{1}{2}(y_4 - z_4)\,a\,\hat{\mathbf{x}} + \left(\frac{1}{\sqrt{3}}x_4 - \frac{1}{2\sqrt{3}}y_4 - \frac{1}{2\sqrt{3}}z_4\right)a\,\hat{\mathbf{y}} + \left(\frac{1}{2} + \frac{1}{3}x_4 + \frac{1}{3}y_4 + \frac{1}{3}z_4\right)c\,\hat{\mathbf{z}} \quad (12f) \quad \text{Cl II}$$

$$\mathbf{B}_{28} = \left(\frac{1}{2} + x_4\right)\mathbf{a}_1 + \left(\frac{1}{2} + z_4\right)\mathbf{a}_2 + \left(\frac{1}{2} + y_4\right)\mathbf{a}_3 = \frac{1}{2}(x_4 - y_4)\,a\,\hat{\mathbf{x}} + \left(-\frac{1}{2\sqrt{3}}x_4 - \frac{1}{2\sqrt{3}}y_4 + \frac{1}{\sqrt{3}}z_4\right)a\,\hat{\mathbf{y}} + \left(\frac{1}{2} + \frac{1}{3}x_4 + \frac{1}{3}y_4 + \frac{1}{3}z_4\right)c\,\hat{\mathbf{z}} \quad (12f) \quad \text{Cl II}$$

---

**Found in:**

---

**Geometry files:**

- CIF: pp. 1748
- POSCAR: pp. 1748

# Crancrinite ($Na_6Ca_2Al_6Si_6O_{24}(CO_3)_2$, $S3_3$ (I)) Structure: A3BCD3E15F3_hP52_173_c_b_b_c_5c_c

http://aflow.org/prototype-encyclopedia/A3BCD3E15F3_hP52_173_c_b_b_c_5c_c

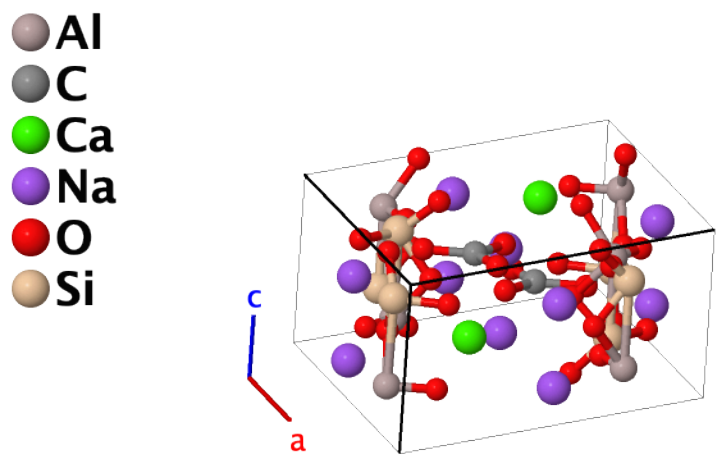

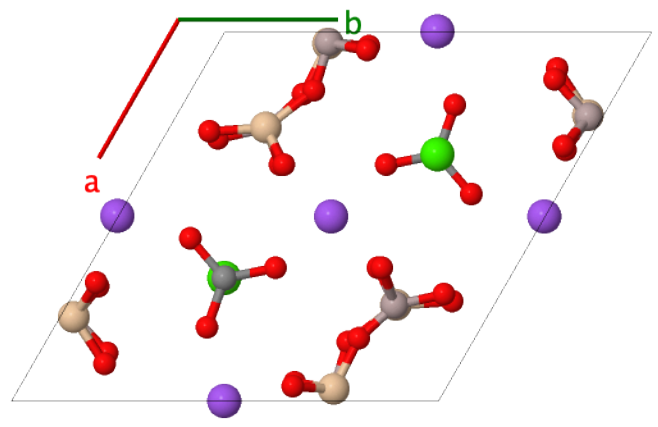

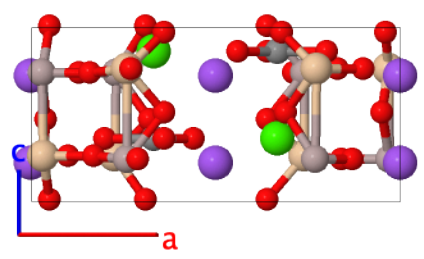

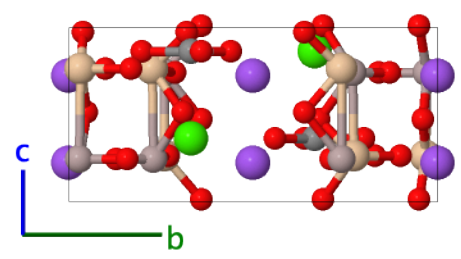

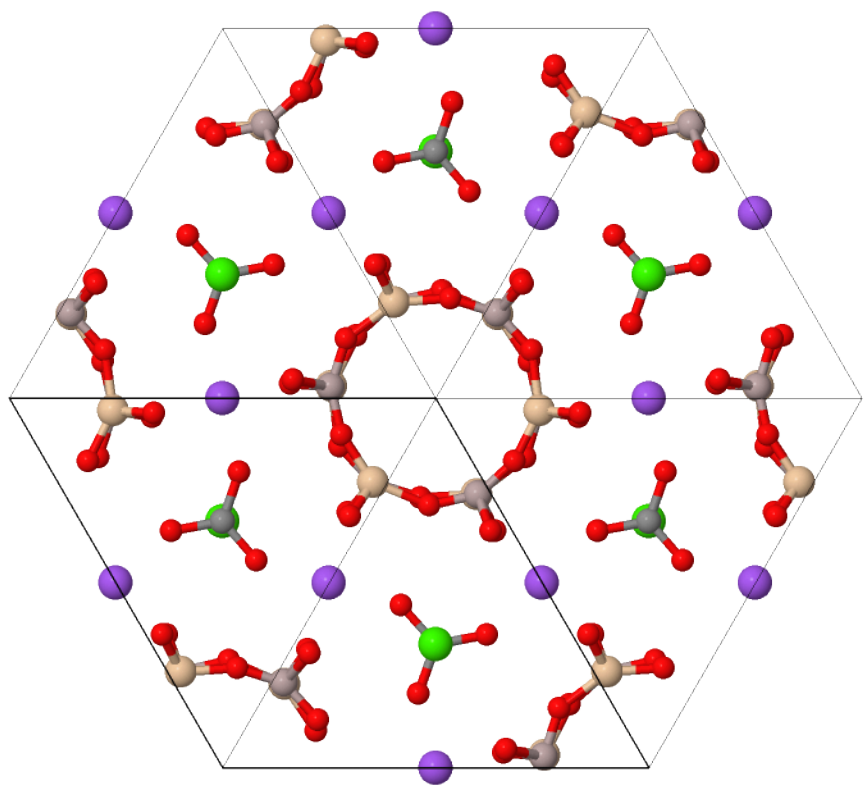

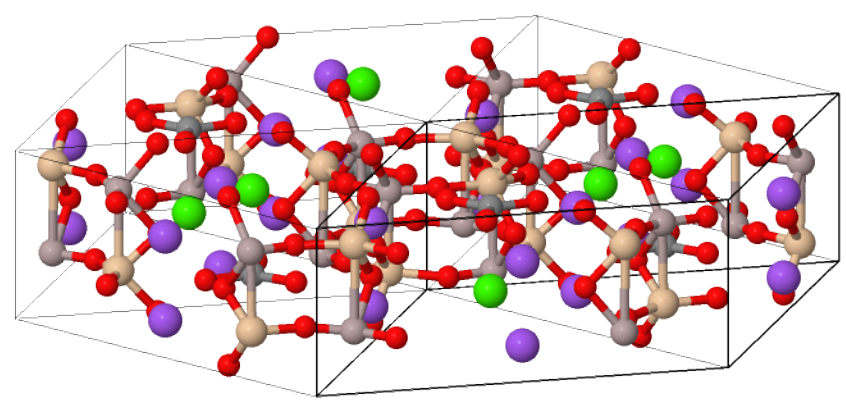

| | | |
|---|---|---|
| **Prototype** | : | $Al_3CCaNa_3O_{15}Si_3$ |
| **AFLOW prototype label** | : | A3BCD3E15F3_hP52_173_c_b_b_c_5c_c |
| ***Strukturbericht* designation** | : | $S3_3$ (I) |
| **Pearson symbol** | : | hP52 |
| **Space group number** | : | 173 |
| **Space group symbol** | : | $P6_3$ |
| **AFLOW prototype command** | : | `aflow --proto=A3BCD3E15F3_hP52_173_c_b_b_c_5c_c --params=`$a, c/a, z_1, z_2, x_3, y_3, z_3, x_4, y_4, z_4, x_5, y_5, z_5, x_6, y_6, z_6, x_7, y_7, z_7, x_8, y_8, z_8, x_9, y_9, z_9, x_{10}, y_{10}, z_{10}$ |

- The term "crancrinite" covers a wide range of mineral compositions, some of which may include OH radicals and $H_2O$ molecules. We use this early determination of the structure as the base.
- There are two major inconsistencies between (Kozu, 1933) and (Gottfried, 1937):
  - (Gottfried, 1937) interchanges the $z$ coordinates of the carbon and calcium ions compared to (Kozu, 1933). Kozu's coordinates are consistent with $CO_3$ radicals, so we use their version.

– (Kozu, 1933) gives $x_{10} = 0.33$ for the $x$ coordinate of silicon. This puts the silicon atoms very close to carbon. We follow (Gottfried, 1937) and take $x_{10} = 0.033$, which gives a reasonable distance.

- (Gottfried, 1937) gave this structure the label $S3_3$, but later (Gottfried, 1938) gave the same label to parawollastonite, $CaSiO_3$. We will refer to crancrinite as $S3_3$ $(I)$, and parawollastonite as $S3_3$ $(II)$.

---

**Hexagonal primitive vectors:**

$$
\begin{aligned}
\mathbf{a}_1 &= \tfrac{1}{2}\,a\,\hat{\mathbf{x}} - \tfrac{\sqrt{3}}{2}\,a\,\hat{\mathbf{y}} \\
\mathbf{a}_2 &= \tfrac{1}{2}\,a\,\hat{\mathbf{x}} + \tfrac{\sqrt{3}}{2}\,a\,\hat{\mathbf{y}} \\
\mathbf{a}_3 &= c\,\hat{\mathbf{z}}
\end{aligned}
$$

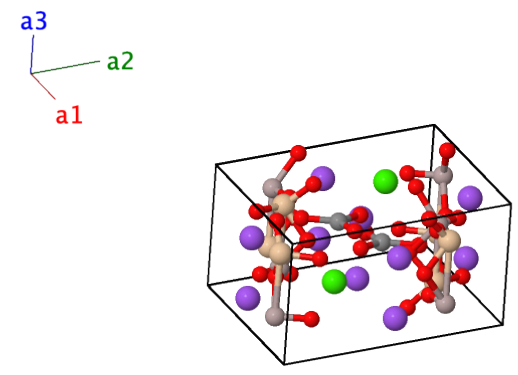

---

**Basis vectors:**

| | | Lattice Coordinates | | Cartesian Coordinates | Wyckoff Position | Atom Type |
|---|---|---|---|---|---|---|
| $\mathbf{B}_1$ | = | $\frac{1}{3}\,\mathbf{a}_1 + \frac{2}{3}\,\mathbf{a}_2 + z_1\,\mathbf{a}_3$ | = | $\frac{1}{2}a\,\hat{\mathbf{x}} + \frac{1}{2\sqrt{3}}a\,\hat{\mathbf{y}} + z_1c\,\hat{\mathbf{z}}$ | (2$b$) | C |
| $\mathbf{B}_2$ | = | $\frac{2}{3}\,\mathbf{a}_1 + \frac{1}{3}\,\mathbf{a}_2 + \left(\frac{1}{2} + z_1\right)\mathbf{a}_3$ | = | $\frac{1}{2}a\,\hat{\mathbf{x}} - \frac{1}{2\sqrt{3}}a\,\hat{\mathbf{y}} + \left(\frac{1}{2} + z_1\right)c\,\hat{\mathbf{z}}$ | (2$b$) | C |
| $\mathbf{B}_3$ | = | $\frac{1}{3}\,\mathbf{a}_1 + \frac{2}{3}\,\mathbf{a}_2 + z_2\,\mathbf{a}_3$ | = | $\frac{1}{2}a\,\hat{\mathbf{x}} + \frac{1}{2\sqrt{3}}a\,\hat{\mathbf{y}} + z_2c\,\hat{\mathbf{z}}$ | (2$b$) | Ca |
| $\mathbf{B}_4$ | = | $\frac{2}{3}\,\mathbf{a}_1 + \frac{1}{3}\,\mathbf{a}_2 + \left(\frac{1}{2} + z_2\right)\mathbf{a}_3$ | = | $\frac{1}{2}a\,\hat{\mathbf{x}} - \frac{1}{2\sqrt{3}}a\,\hat{\mathbf{y}} + \left(\frac{1}{2} + z_2\right)c\,\hat{\mathbf{z}}$ | (2$b$) | Ca |
| $\mathbf{B}_5$ | = | $x_3\,\mathbf{a}_1 + y_3\,\mathbf{a}_2 + z_3\,\mathbf{a}_3$ | = | $\frac{1}{2}\left(x_3 + y_3\right)a\,\hat{\mathbf{x}} + \frac{\sqrt{3}}{2}\left(-x_3 + y_3\right)a\,\hat{\mathbf{y}} + z_3c\,\hat{\mathbf{z}}$ | (6$c$) | Al |
| $\mathbf{B}_6$ | = | $-y_3\,\mathbf{a}_1 + \left(x_3 - y_3\right)\mathbf{a}_2 + z_3\,\mathbf{a}_3$ | = | $\left(\frac{1}{2}x_3 - y_3\right)a\,\hat{\mathbf{x}} + \frac{\sqrt{3}}{2}x_3a\,\hat{\mathbf{y}} + z_3c\,\hat{\mathbf{z}}$ | (6$c$) | Al |
| $\mathbf{B}_7$ | = | $\left(-x_3 + y_3\right)\mathbf{a}_1 - x_3\,\mathbf{a}_2 + z_3\,\mathbf{a}_3$ | = | $\left(-x_3 + \frac{1}{2}y_3\right)a\,\hat{\mathbf{x}} - \frac{\sqrt{3}}{2}y_3a\,\hat{\mathbf{y}} + z_3c\,\hat{\mathbf{z}}$ | (6$c$) | Al |
| $\mathbf{B}_8$ | = | $-x_3\,\mathbf{a}_1 - y_3\,\mathbf{a}_2 + \left(\frac{1}{2} + z_3\right)\mathbf{a}_3$ | = | $-\frac{1}{2}\left(x_3 + y_3\right)a\,\hat{\mathbf{x}} + \frac{\sqrt{3}}{2}\left(x_3 - y_3\right)a\,\hat{\mathbf{y}} + \left(\frac{1}{2} + z_3\right)c\,\hat{\mathbf{z}}$ | (6$c$) | Al |
| $\mathbf{B}_9$ | = | $y_3\,\mathbf{a}_1 + \left(-x_3 + y_3\right)\mathbf{a}_2 + \left(\frac{1}{2} + z_3\right)\mathbf{a}_3$ | = | $\left(-\frac{1}{2}x_3 + y_3\right)a\,\hat{\mathbf{x}} - \frac{\sqrt{3}}{2}x_3a\,\hat{\mathbf{y}} + \left(\frac{1}{2} + z_3\right)c\,\hat{\mathbf{z}}$ | (6$c$) | Al |
| $\mathbf{B}_{10}$ | = | $\left(x_3 - y_3\right)\mathbf{a}_1 + x_3\,\mathbf{a}_2 + \left(\frac{1}{2} + z_3\right)\mathbf{a}_3$ | = | $\left(x_3 - \frac{1}{2}y_3\right)a\,\hat{\mathbf{x}} + \frac{\sqrt{3}}{2}y_3a\,\hat{\mathbf{y}} + \left(\frac{1}{2} + z_3\right)c\,\hat{\mathbf{z}}$ | (6$c$) | Al |
| $\mathbf{B}_{11}$ | = | $x_4\,\mathbf{a}_1 + y_4\,\mathbf{a}_2 + z_4\,\mathbf{a}_3$ | = | $\frac{1}{2}\left(x_4 + y_4\right)a\,\hat{\mathbf{x}} + \frac{\sqrt{3}}{2}\left(-x_4 + y_4\right)a\,\hat{\mathbf{y}} + z_4c\,\hat{\mathbf{z}}$ | (6$c$) | Na |
| $\mathbf{B}_{12}$ | = | $-y_4\,\mathbf{a}_1 + \left(x_4 - y_4\right)\mathbf{a}_2 + z_4\,\mathbf{a}_3$ | = | $\left(\frac{1}{2}x_4 - y_4\right)a\,\hat{\mathbf{x}} + \frac{\sqrt{3}}{2}x_4a\,\hat{\mathbf{y}} + z_4c\,\hat{\mathbf{z}}$ | (6$c$) | Na |
| $\mathbf{B}_{13}$ | = | $\left(-x_4 + y_4\right)\mathbf{a}_1 - x_4\,\mathbf{a}_2 + z_4\,\mathbf{a}_3$ | = | $\left(-x_4 + \frac{1}{2}y_4\right)a\,\hat{\mathbf{x}} - \frac{\sqrt{3}}{2}y_4a\,\hat{\mathbf{y}} + z_4c\,\hat{\mathbf{z}}$ | (6$c$) | Na |
| $\mathbf{B}_{14}$ | = | $-x_4\,\mathbf{a}_1 - y_4\,\mathbf{a}_2 + \left(\frac{1}{2} + z_4\right)\mathbf{a}_3$ | = | $-\frac{1}{2}\left(x_4 + y_4\right)a\,\hat{\mathbf{x}} + \frac{\sqrt{3}}{2}\left(x_4 - y_4\right)a\,\hat{\mathbf{y}} + \left(\frac{1}{2} + z_4\right)c\,\hat{\mathbf{z}}$ | (6$c$) | Na |
| $\mathbf{B}_{15}$ | = | $y_4\,\mathbf{a}_1 + \left(-x_4 + y_4\right)\mathbf{a}_2 + \left(\frac{1}{2} + z_4\right)\mathbf{a}_3$ | = | $\left(-\frac{1}{2}x_4 + y_4\right)a\,\hat{\mathbf{x}} - \frac{\sqrt{3}}{2}x_4a\,\hat{\mathbf{y}} + \left(\frac{1}{2} + z_4\right)c\,\hat{\mathbf{z}}$ | (6$c$) | Na |
| $\mathbf{B}_{16}$ | = | $\left(x_4 - y_4\right)\mathbf{a}_1 + x_4\,\mathbf{a}_2 + \left(\frac{1}{2} + z_4\right)\mathbf{a}_3$ | = | $\left(x_4 - \frac{1}{2}y_4\right)a\,\hat{\mathbf{x}} + \frac{\sqrt{3}}{2}y_4a\,\hat{\mathbf{y}} + \left(\frac{1}{2} + z_4\right)c\,\hat{\mathbf{z}}$ | (6$c$) | Na |
| $\mathbf{B}_{17}$ | = | $x_5\,\mathbf{a}_1 + y_5\,\mathbf{a}_2 + z_5\,\mathbf{a}_3$ | = | $\frac{1}{2}\left(x_5 + y_5\right)a\,\hat{\mathbf{x}} + \frac{\sqrt{3}}{2}\left(-x_5 + y_5\right)a\,\hat{\mathbf{y}} + z_5c\,\hat{\mathbf{z}}$ | (6$c$) | O I |
| $\mathbf{B}_{18}$ | = | $-y_5\,\mathbf{a}_1 + \left(x_5 - y_5\right)\mathbf{a}_2 + z_5\,\mathbf{a}_3$ | = | $\left(\frac{1}{2}x_5 - y_5\right)a\,\hat{\mathbf{x}} + \frac{\sqrt{3}}{2}x_5a\,\hat{\mathbf{y}} + z_5c\,\hat{\mathbf{z}}$ | (6$c$) | O I |
| $\mathbf{B}_{19}$ | = | $\left(-x_5+y_5\right)\,\mathbf{a}_1 - x_5\,\mathbf{a}_2 + z_5\,\mathbf{a}_3$ | = | $\left(-x_5+\frac{1}{2}y_5\right)a\,\mathbf{\hat{x}} - \frac{\sqrt{3}}{2}y_5a\,\mathbf{\hat{y}} + z_5c\,\mathbf{\hat{z}}$ | (6c) | O I |
| $\mathbf{B}_{20}$ | = | $-x_5\,\mathbf{a}_1 - y_5\,\mathbf{a}_2 + \left(\frac{1}{2}+z_5\right)\,\mathbf{a}_3$ | = | $-\frac{1}{2}\left(x_5+y_5\right)a\,\mathbf{\hat{x}} + \frac{\sqrt{3}}{2}\left(x_5-y_5\right)a\,\mathbf{\hat{y}} + \left(\frac{1}{2}+z_5\right)c\,\mathbf{\hat{z}}$ | (6c) | O I |
| $\mathbf{B}_{21}$ | = | $y_5\,\mathbf{a}_1 + \left(-x_5+y_5\right)\,\mathbf{a}_2 + \left(\frac{1}{2}+z_5\right)\,\mathbf{a}_3$ | = | $\left(-\frac{1}{2}x_5+y_5\right)a\,\mathbf{\hat{x}} - \frac{\sqrt{3}}{2}x_5a\,\mathbf{\hat{y}} + \left(\frac{1}{2}+z_5\right)c\,\mathbf{\hat{z}}$ | (6c) | O I |
| $\mathbf{B}_{22}$ | = | $\left(x_5-y_5\right)\,\mathbf{a}_1 + x_5\,\mathbf{a}_2 + \left(\frac{1}{2}+z_5\right)\,\mathbf{a}_3$ | = | $\left(x_5-\frac{1}{2}y_5\right)a\,\mathbf{\hat{x}} + \frac{\sqrt{3}}{2}y_5a\,\mathbf{\hat{y}} + \left(\frac{1}{2}+z_5\right)c\,\mathbf{\hat{z}}$ | (6c) | O I |
| $\mathbf{B}_{23}$ | = | $x_6\,\mathbf{a}_1 + y_6\,\mathbf{a}_2 + z_6\,\mathbf{a}_3$ | = | $\frac{1}{2}\left(x_6+y_6\right)a\,\mathbf{\hat{x}} + \frac{\sqrt{3}}{2}\left(-x_6+y_6\right)a\,\mathbf{\hat{y}} + z_6c\,\mathbf{\hat{z}}$ | (6c) | O II |
| $\mathbf{B}_{24}$ | = | $-y_6\,\mathbf{a}_1 + \left(x_6-y_6\right)\,\mathbf{a}_2 + z_6\,\mathbf{a}_3$ | = | $\left(\frac{1}{2}x_6-y_6\right)a\,\mathbf{\hat{x}} + \frac{\sqrt{3}}{2}x_6a\,\mathbf{\hat{y}} + z_6c\,\mathbf{\hat{z}}$ | (6c) | O II |
| $\mathbf{B}_{25}$ | = | $\left(-x_6+y_6\right)\,\mathbf{a}_1 - x_6\,\mathbf{a}_2 + z_6\,\mathbf{a}_3$ | = | $\left(-x_6+\frac{1}{2}y_6\right)a\,\mathbf{\hat{x}} - \frac{\sqrt{3}}{2}y_6a\,\mathbf{\hat{y}} + z_6c\,\mathbf{\hat{z}}$ | (6c) | O II |
| $\mathbf{B}_{26}$ | = | $-x_6\,\mathbf{a}_1 - y_6\,\mathbf{a}_2 + \left(\frac{1}{2}+z_6\right)\,\mathbf{a}_3$ | = | $-\frac{1}{2}\left(x_6+y_6\right)a\,\mathbf{\hat{x}} + \frac{\sqrt{3}}{2}\left(x_6-y_6\right)a\,\mathbf{\hat{y}} + \left(\frac{1}{2}+z_6\right)c\,\mathbf{\hat{z}}$ | (6c) | O II |
| $\mathbf{B}_{27}$ | = | $y_6\,\mathbf{a}_1 + \left(-x_6+y_6\right)\,\mathbf{a}_2 + \left(\frac{1}{2}+z_6\right)\,\mathbf{a}_3$ | = | $\left(-\frac{1}{2}x_6+y_6\right)a\,\mathbf{\hat{x}} - \frac{\sqrt{3}}{2}x_6a\,\mathbf{\hat{y}} + \left(\frac{1}{2}+z_6\right)c\,\mathbf{\hat{z}}$ | (6c) | O II |
| $\mathbf{B}_{28}$ | = | $\left(x_6-y_6\right)\,\mathbf{a}_1 + x_6\,\mathbf{a}_2 + \left(\frac{1}{2}+z_6\right)\,\mathbf{a}_3$ | = | $\left(x_6-\frac{1}{2}y_6\right)a\,\mathbf{\hat{x}} + \frac{\sqrt{3}}{2}y_6a\,\mathbf{\hat{y}} + \left(\frac{1}{2}+z_6\right)c\,\mathbf{\hat{z}}$ | (6c) | O II |
| $\mathbf{B}_{29}$ | = | $x_7\,\mathbf{a}_1 + y_7\,\mathbf{a}_2 + z_7\,\mathbf{a}_3$ | = | $\frac{1}{2}\left(x_7+y_7\right)a\,\mathbf{\hat{x}} + \frac{\sqrt{3}}{2}\left(-x_7+y_7\right)a\,\mathbf{\hat{y}} + z_7c\,\mathbf{\hat{z}}$ | (6c) | O III |
| $\mathbf{B}_{30}$ | = | $-y_7\,\mathbf{a}_1 + \left(x_7-y_7\right)\,\mathbf{a}_2 + z_7\,\mathbf{a}_3$ | = | $\left(\frac{1}{2}x_7-y_7\right)a\,\mathbf{\hat{x}} + \frac{\sqrt{3}}{2}x_7a\,\mathbf{\hat{y}} + z_7c\,\mathbf{\hat{z}}$ | (6c) | O III |
| $\mathbf{B}_{31}$ | = | $\left(-x_7+y_7\right)\,\mathbf{a}_1 - x_7\,\mathbf{a}_2 + z_7\,\mathbf{a}_3$ | = | $\left(-x_7+\frac{1}{2}y_7\right)a\,\mathbf{\hat{x}} - \frac{\sqrt{3}}{2}y_7a\,\mathbf{\hat{y}} + z_7c\,\mathbf{\hat{z}}$ | (6c) | O III |
| $\mathbf{B}_{32}$ | = | $-x_7\,\mathbf{a}_1 - y_7\,\mathbf{a}_2 + \left(\frac{1}{2}+z_7\right)\,\mathbf{a}_3$ | = | $-\frac{1}{2}\left(x_7+y_7\right)a\,\mathbf{\hat{x}} + \frac{\sqrt{3}}{2}\left(x_7-y_7\right)a\,\mathbf{\hat{y}} + \left(\frac{1}{2}+z_7\right)c\,\mathbf{\hat{z}}$ | (6c) | O III |
| $\mathbf{B}_{33}$ | = | $y_7\,\mathbf{a}_1 + \left(-x_7+y_7\right)\,\mathbf{a}_2 + \left(\frac{1}{2}+z_7\right)\,\mathbf{a}_3$ | = | $\left(-\frac{1}{2}x_7+y_7\right)a\,\mathbf{\hat{x}} - \frac{\sqrt{3}}{2}x_7a\,\mathbf{\hat{y}} + \left(\frac{1}{2}+z_7\right)c\,\mathbf{\hat{z}}$ | (6c) | O III |
| $\mathbf{B}_{34}$ | = | $\left(x_7-y_7\right)\,\mathbf{a}_1 + x_7\,\mathbf{a}_2 + \left(\frac{1}{2}+z_7\right)\,\mathbf{a}_3$ | = | $\left(x_7-\frac{1}{2}y_7\right)a\,\mathbf{\hat{x}} + \frac{\sqrt{3}}{2}y_7a\,\mathbf{\hat{y}} + \left(\frac{1}{2}+z_7\right)c\,\mathbf{\hat{z}}$ | (6c) | O III |
| $\mathbf{B}_{35}$ | = | $x_8\,\mathbf{a}_1 + y_8\,\mathbf{a}_2 + z_8\,\mathbf{a}_3$ | = | $\frac{1}{2}\left(x_8+y_8\right)a\,\mathbf{\hat{x}} + \frac{\sqrt{3}}{2}\left(-x_8+y_8\right)a\,\mathbf{\hat{y}} + z_8c\,\mathbf{\hat{z}}$ | (6c) | O IV |
| $\mathbf{B}_{36}$ | = | $-y_8\,\mathbf{a}_1 + \left(x_8-y_8\right)\,\mathbf{a}_2 + z_8\,\mathbf{a}_3$ | = | $\left(\frac{1}{2}x_8-y_8\right)a\,\mathbf{\hat{x}} + \frac{\sqrt{3}}{2}x_8a\,\mathbf{\hat{y}} + z_8c\,\mathbf{\hat{z}}$ | (6c) | O IV |
| $\mathbf{B}_{37}$ | = | $\left(-x_8+y_8\right)\,\mathbf{a}_1 - x_8\,\mathbf{a}_2 + z_8\,\mathbf{a}_3$ | = | $\left(-x_8+\frac{1}{2}y_8\right)a\,\mathbf{\hat{x}} - \frac{\sqrt{3}}{2}y_8a\,\mathbf{\hat{y}} + z_8c\,\mathbf{\hat{z}}$ | (6c) | O IV |
| $\mathbf{B}_{38}$ | = | $-x_8\,\mathbf{a}_1 - y_8\,\mathbf{a}_2 + \left(\frac{1}{2}+z_8\right)\,\mathbf{a}_3$ | = | $-\frac{1}{2}\left(x_8+y_8\right)a\,\mathbf{\hat{x}} + \frac{\sqrt{3}}{2}\left(x_8-y_8\right)a\,\mathbf{\hat{y}} + \left(\frac{1}{2}+z_8\right)c\,\mathbf{\hat{z}}$ | (6c) | O IV |
| $\mathbf{B}_{39}$ | = | $y_8\,\mathbf{a}_1 + \left(-x_8+y_8\right)\,\mathbf{a}_2 + \left(\frac{1}{2}+z_8\right)\,\mathbf{a}_3$ | = | $\left(-\frac{1}{2}x_8+y_8\right)a\,\mathbf{\hat{x}} - \frac{\sqrt{3}}{2}x_8a\,\mathbf{\hat{y}} + \left(\frac{1}{2}+z_8\right)c\,\mathbf{\hat{z}}$ | (6c) | O IV |
| $\mathbf{B}_{40}$ | = | $\left(x_8-y_8\right)\,\mathbf{a}_1 + x_8\,\mathbf{a}_2 + \left(\frac{1}{2}+z_8\right)\,\mathbf{a}_3$ | = | $\left(x_8-\frac{1}{2}y_8\right)a\,\mathbf{\hat{x}} + \frac{\sqrt{3}}{2}y_8a\,\mathbf{\hat{y}} + \left(\frac{1}{2}+z_8\right)c\,\mathbf{\hat{z}}$ | (6c) | O IV |
| $\mathbf{B}_{41}$ | = | $x_9\,\mathbf{a}_1 + y_9\,\mathbf{a}_2 + z_9\,\mathbf{a}_3$ | = | $\frac{1}{2}\left(x_9+y_9\right)a\,\mathbf{\hat{x}} + \frac{\sqrt{3}}{2}\left(-x_9+y_9\right)a\,\mathbf{\hat{y}} + z_9c\,\mathbf{\hat{z}}$ | (6c) | O V |
| $\mathbf{B}_{42}$ | = | $-y_9\,\mathbf{a}_1 + \left(x_9-y_9\right)\,\mathbf{a}_2 + z_9\,\mathbf{a}_3$ | = | $\left(\frac{1}{2}x_9-y_9\right)a\,\mathbf{\hat{x}} + \frac{\sqrt{3}}{2}x_9a\,\mathbf{\hat{y}} + z_9c\,\mathbf{\hat{z}}$ | (6c) | O V |
| $\mathbf{B}_{43}$ | = | $(-x_9 + y_9)\,\mathbf{a}_1 - x_9\,\mathbf{a}_2 + z_9\,\mathbf{a}_3$ | = | $\left(-x_9 + \frac{1}{2}y_9\right) a\,\hat{\mathbf{x}} - \frac{\sqrt{3}}{2}y_9 a\,\hat{\mathbf{y}} + z_9 c\,\hat{\mathbf{z}}$ | (6$c$) | O V |
| $\mathbf{B}_{44}$ | = | $-x_9\,\mathbf{a}_1 - y_9\,\mathbf{a}_2 + \left(\frac{1}{2} + z_9\right)\mathbf{a}_3$ | = | $-\frac{1}{2}(x_9 + y_9)\,a\,\hat{\mathbf{x}} + \frac{\sqrt{3}}{2}(x_9 - y_9)\,a\,\hat{\mathbf{y}} + \left(\frac{1}{2} + z_9\right) c\,\hat{\mathbf{z}}$ | (6$c$) | O V |
| $\mathbf{B}_{45}$ | = | $y_9\,\mathbf{a}_1 + (-x_9 + y_9)\,\mathbf{a}_2 + \left(\frac{1}{2} + z_9\right)\mathbf{a}_3$ | = | $\left(-\frac{1}{2}x_9 + y_9\right) a\,\hat{\mathbf{x}} - \frac{\sqrt{3}}{2}x_9 a\,\hat{\mathbf{y}} + \left(\frac{1}{2} + z_9\right) c\,\hat{\mathbf{z}}$ | (6$c$) | O V |
| $\mathbf{B}_{46}$ | = | $(x_9 - y_9)\,\mathbf{a}_1 + x_9\,\mathbf{a}_2 + \left(\frac{1}{2} + z_9\right)\mathbf{a}_3$ | = | $\left(x_9 - \frac{1}{2}y_9\right) a\,\hat{\mathbf{x}} + \frac{\sqrt{3}}{2}y_9 a\,\hat{\mathbf{y}} + \left(\frac{1}{2} + z_9\right) c\,\hat{\mathbf{z}}$ | (6$c$) | O V |
| $\mathbf{B}_{47}$ | = | $x_{10}\,\mathbf{a}_1 + y_{10}\,\mathbf{a}_2 + z_{10}\,\mathbf{a}_3$ | = | $\frac{1}{2}(x_{10} + y_{10})\,a\,\hat{\mathbf{x}} + \frac{\sqrt{3}}{2}(-x_{10} + y_{10})\,a\,\hat{\mathbf{y}} + z_{10} c\,\hat{\mathbf{z}}$ | (6$c$) | Si |
| $\mathbf{B}_{48}$ | = | $-y_{10}\,\mathbf{a}_1 + (x_{10} - y_{10})\,\mathbf{a}_2 + z_{10}\,\mathbf{a}_3$ | = | $\left(\frac{1}{2}x_{10} - y_{10}\right) a\,\hat{\mathbf{x}} + \frac{\sqrt{3}}{2}x_{10} a\,\hat{\mathbf{y}} + z_{10} c\,\hat{\mathbf{z}}$ | (6$c$) | Si |
| $\mathbf{B}_{49}$ | = | $(-x_{10} + y_{10})\,\mathbf{a}_1 - x_{10}\,\mathbf{a}_2 + z_{10}\,\mathbf{a}_3$ | = | $\left(-x_{10} + \frac{1}{2}y_{10}\right) a\,\hat{\mathbf{x}} - \frac{\sqrt{3}}{2}y_{10} a\,\hat{\mathbf{y}} + z_{10} c\,\hat{\mathbf{z}}$ | (6$c$) | Si |
| $\mathbf{B}_{50}$ | = | $-x_{10}\,\mathbf{a}_1 - y_{10}\,\mathbf{a}_2 + \left(\frac{1}{2} + z_{10}\right)\mathbf{a}_3$ | = | $-\frac{1}{2}(x_{10} + y_{10})\,a\,\hat{\mathbf{x}} + \frac{\sqrt{3}}{2}(x_{10} - y_{10})\,a\,\hat{\mathbf{y}} + \left(\frac{1}{2} + z_{10}\right) c\,\hat{\mathbf{z}}$ | (6$c$) | Si |
| $\mathbf{B}_{51}$ | = | $y_{10}\,\mathbf{a}_1 + (-x_{10} + y_{10})\,\mathbf{a}_2 + \left(\frac{1}{2} + z_{10}\right)\mathbf{a}_3$ | = | $\left(-\frac{1}{2}x_{10} + y_{10}\right) a\,\hat{\mathbf{x}} - \frac{\sqrt{3}}{2}x_{10} a\,\hat{\mathbf{y}} + \left(\frac{1}{2} + z_{10}\right) c\,\hat{\mathbf{z}}$ | (6$c$) | Si |
| $\mathbf{B}_{52}$ | = | $(x_{10} - y_{10})\,\mathbf{a}_1 + x_{10}\,\mathbf{a}_2 + \left(\frac{1}{2} + z_{10}\right)\mathbf{a}_3$ | = | $\left(x_{10} - \frac{1}{2}y_{10}\right) a\,\hat{\mathbf{x}} + \frac{\sqrt{3}}{2}y_{10} a\,\hat{\mathbf{y}} + \left(\frac{1}{2} + z_{10}\right) c\,\hat{\mathbf{z}}$ | (6$c$) | Si |

---

**Found in:**

---

**Geometry files:**

- CIF: pp. 1748
- POSCAR: pp. 1749

# $La_3CuSiS_7$ Structure: AB3C7D_hP24_173_a_c_b2c_b

http://aflow.org/prototype-encyclopedia/AB3C7D_hP24_173_a_c_b2c_b

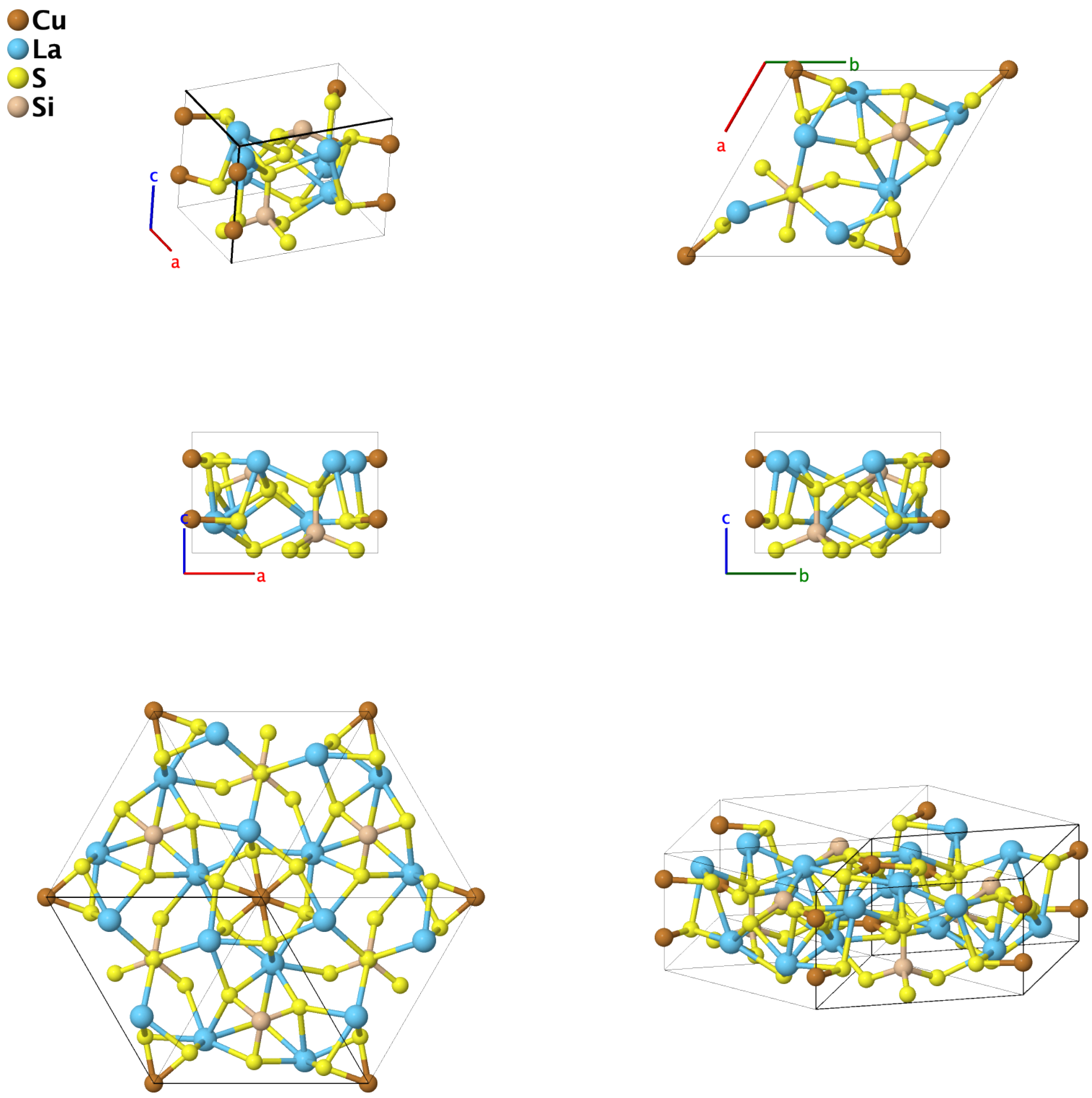

| | | |
|---|---|---|
| **Prototype** | : | $CuLa_3S_7Si$ |
| **AFLOW prototype label** | : | AB3C7D_hP24_173_a_c_b2c_b |
| ***Strukturbericht* designation** | : | None |
| **Pearson symbol** | : | hP24 |
| **Space group number** | : | 173 |
| **Space group symbol** | : | $P6_3$ |
| **AFLOW prototype command** | : | `aflow --proto=AB3C7D_hP24_173_a_c_b2c_b` `--params=`$a, c/a, z_1, z_2, z_3, x_4, y_4, z_4, x_5, y_5, z_5, x_6, y_6, z_6$ |

### Other compounds with this structure

- $Ce_3CuGeS_7$, $Ce_3CuGeSe_7$, $Ce_3CuSnSe_7$, $Dy_3CuGeS_7$, $Gd_3CuGeS_7$, $Gd_3CuGeSe_7$, $Ho_3CuGeSe_7$, $La_3MnFeS_7$, $Nd_3CuGeS_7$, $Nd_3CuGeSe_7$, $Pr_3CuGeS_7$, $Pr_3CuGeSe_7$, $Sm_3CuGeS_7$, $Sm_3CuGeSe_7$, $Tb_3CuGeS_7$, $Tb_3CuGeSe_7$, $Y_3CuSiS_7$, and $Y_3CuSiSe_7$

**Hexagonal primitive vectors:**

$$\mathbf{a}_1 = \frac{1}{2}\,a\,\mathbf{\hat{x}} - \frac{\sqrt{3}}{2}\,a\,\mathbf{\hat{y}}$$
$$\mathbf{a}_2 = \frac{1}{2}\,a\,\mathbf{\hat{x}} + \frac{\sqrt{3}}{2}\,a\,\mathbf{\hat{y}}$$
$$\mathbf{a}_3 = c\,\mathbf{\hat{z}}$$

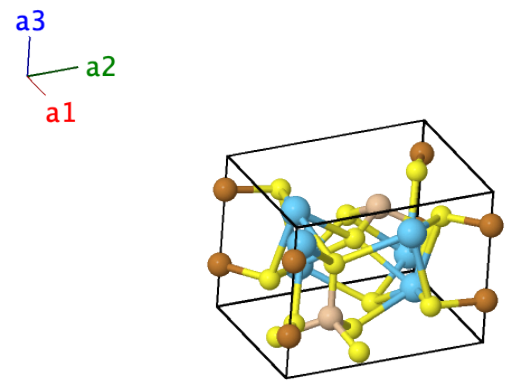

**Basis vectors:**

| | | Lattice Coordinates | | Cartesian Coordinates | Wyckoff Position | Atom Type |
|---|---|---|---|---|---|---|
| $\mathbf{B}_1$ | = | $z_1\,\mathbf{a}_3$ | = | $z_1 c\,\mathbf{\hat{z}}$ | (2$a$) | Cu |
| $\mathbf{B}_2$ | = | $\left(\frac{1}{2}+z_1\right)\mathbf{a}_3$ | = | $\left(\frac{1}{2}+z_1\right)c\,\mathbf{\hat{z}}$ | (2$a$) | Cu |
| $\mathbf{B}_3$ | = | $\frac{1}{3}\,\mathbf{a}_1+\frac{2}{3}\,\mathbf{a}_2+z_2\,\mathbf{a}_3$ | = | $\frac{1}{2}a\,\mathbf{\hat{x}}+\frac{1}{2\sqrt{3}}a\,\mathbf{\hat{y}}+z_2 c\,\mathbf{\hat{z}}$ | (2$b$) | S I |
| $\mathbf{B}_4$ | = | $\frac{2}{3}\,\mathbf{a}_1+\frac{1}{3}\,\mathbf{a}_2+\left(\frac{1}{2}+z_2\right)\mathbf{a}_3$ | = | $\frac{1}{2}a\,\mathbf{\hat{x}}-\frac{1}{2\sqrt{3}}a\,\mathbf{\hat{y}}+\left(\frac{1}{2}+z_2\right)c\,\mathbf{\hat{z}}$ | (2$b$) | S I |
| $\mathbf{B}_5$ | = | $\frac{1}{3}\,\mathbf{a}_1+\frac{2}{3}\,\mathbf{a}_2+z_3\,\mathbf{a}_3$ | = | $\frac{1}{2}a\,\mathbf{\hat{x}}+\frac{1}{2\sqrt{3}}a\,\mathbf{\hat{y}}+z_3 c\,\mathbf{\hat{z}}$ | (2$b$) | Si |
| $\mathbf{B}_6$ | = | $\frac{2}{3}\,\mathbf{a}_1+\frac{1}{3}\,\mathbf{a}_2+\left(\frac{1}{2}+z_3\right)\mathbf{a}_3$ | = | $\frac{1}{2}a\,\mathbf{\hat{x}}-\frac{1}{2\sqrt{3}}a\,\mathbf{\hat{y}}+\left(\frac{1}{2}+z_3\right)c\,\mathbf{\hat{z}}$ | (2$b$) | Si |
| $\mathbf{B}_7$ | = | $x_4\,\mathbf{a}_1+y_4\,\mathbf{a}_2+z_4\,\mathbf{a}_3$ | = | $\frac{1}{2}(x_4+y_4)\,a\,\mathbf{\hat{x}}+\frac{\sqrt{3}}{2}(-x_4+y_4)\,a\,\mathbf{\hat{y}}+z_4 c\,\mathbf{\hat{z}}$ | (6$c$) | La |
| $\mathbf{B}_8$ | = | $-y_4\,\mathbf{a}_1+(x_4-y_4)\,\mathbf{a}_2+z_4\,\mathbf{a}_3$ | = | $\left(\frac{1}{2}x_4-y_4\right)a\,\mathbf{\hat{x}}+\frac{\sqrt{3}}{2}x_4 a\,\mathbf{\hat{y}}+z_4 c\,\mathbf{\hat{z}}$ | (6$c$) | La |
| $\mathbf{B}_9$ | = | $(-x_4+y_4)\,\mathbf{a}_1-x_4\,\mathbf{a}_2+z_4\,\mathbf{a}_3$ | = | $\left(-x_4+\frac{1}{2}y_4\right)a\,\mathbf{\hat{x}}-\frac{\sqrt{3}}{2}y_4 a\,\mathbf{\hat{y}}+z_4 c\,\mathbf{\hat{z}}$ | (6$c$) | La |
| $\mathbf{B}_{10}$ | = | $-x_4\,\mathbf{a}_1-y_4\,\mathbf{a}_2+\left(\frac{1}{2}+z_4\right)\mathbf{a}_3$ | = | $-\frac{1}{2}(x_4+y_4)\,a\,\mathbf{\hat{x}}+\frac{\sqrt{3}}{2}(x_4-y_4)\,a\,\mathbf{\hat{y}}+\left(\frac{1}{2}+z_4\right)c\,\mathbf{\hat{z}}$ | (6$c$) | La |
| $\mathbf{B}_{11}$ | = | $y_4\,\mathbf{a}_1+(-x_4+y_4)\,\mathbf{a}_2+\left(\frac{1}{2}+z_4\right)\mathbf{a}_3$ | = | $\left(-\frac{1}{2}x_4+y_4\right)a\,\mathbf{\hat{x}}-\frac{\sqrt{3}}{2}x_4 a\,\mathbf{\hat{y}}+\left(\frac{1}{2}+z_4\right)c\,\mathbf{\hat{z}}$ | (6$c$) | La |
| $\mathbf{B}_{12}$ | = | $(x_4-y_4)\,\mathbf{a}_1+x_4\,\mathbf{a}_2+\left(\frac{1}{2}+z_4\right)\mathbf{a}_3$ | = | $\left(x_4-\frac{1}{2}y_4\right)a\,\mathbf{\hat{x}}+\frac{\sqrt{3}}{2}y_4 a\,\mathbf{\hat{y}}+\left(\frac{1}{2}+z_4\right)c\,\mathbf{\hat{z}}$ | (6$c$) | La |
| $\mathbf{B}_{13}$ | = | $x_5\,\mathbf{a}_1+y_5\,\mathbf{a}_2+z_5\,\mathbf{a}_3$ | = | $\frac{1}{2}(x_5+y_5)\,a\,\mathbf{\hat{x}}+\frac{\sqrt{3}}{2}(-x_5+y_5)\,a\,\mathbf{\hat{y}}+z_5 c\,\mathbf{\hat{z}}$ | (6$c$) | S II |
| $\mathbf{B}_{14}$ | = | $-y_5\,\mathbf{a}_1+(x_5-y_5)\,\mathbf{a}_2+z_5\,\mathbf{a}_3$ | = | $\left(\frac{1}{2}x_5-y_5\right)a\,\mathbf{\hat{x}}+\frac{\sqrt{3}}{2}x_5 a\,\mathbf{\hat{y}}+z_5 c\,\mathbf{\hat{z}}$ | (6$c$) | S II |
| $\mathbf{B}_{15}$ | = | $(-x_5+y_5)\,\mathbf{a}_1-x_5\,\mathbf{a}_2+z_5\,\mathbf{a}_3$ | = | $\left(-x_5+\frac{1}{2}y_5\right)a\,\mathbf{\hat{x}}-\frac{\sqrt{3}}{2}y_5 a\,\mathbf{\hat{y}}+z_5 c\,\mathbf{\hat{z}}$ | (6$c$) | S II |
| $\mathbf{B}_{16}$ | = | $-x_5\,\mathbf{a}_1-y_5\,\mathbf{a}_2+\left(\frac{1}{2}+z_5\right)\mathbf{a}_3$ | = | $-\frac{1}{2}(x_5+y_5)\,a\,\mathbf{\hat{x}}+\frac{\sqrt{3}}{2}(x_5-y_5)\,a\,\mathbf{\hat{y}}+\left(\frac{1}{2}+z_5\right)c\,\mathbf{\hat{z}}$ | (6$c$) | S II |
| $\mathbf{B}_{17}$ | = | $y_5\,\mathbf{a}_1+(-x_5+y_5)\,\mathbf{a}_2+\left(\frac{1}{2}+z_5\right)\mathbf{a}_3$ | = | $\left(-\frac{1}{2}x_5+y_5\right)a\,\mathbf{\hat{x}}-\frac{\sqrt{3}}{2}x_5 a\,\mathbf{\hat{y}}+\left(\frac{1}{2}+z_5\right)c\,\mathbf{\hat{z}}$ | (6$c$) | S II |
| $\mathbf{B}_{18}$ | = | $(x_5-y_5)\,\mathbf{a}_1+x_5\,\mathbf{a}_2+\left(\frac{1}{2}+z_5\right)\mathbf{a}_3$ | = | $\left(x_5-\frac{1}{2}y_5\right)a\,\mathbf{\hat{x}}+\frac{\sqrt{3}}{2}y_5 a\,\mathbf{\hat{y}}+\left(\frac{1}{2}+z_5\right)c\,\mathbf{\hat{z}}$ | (6$c$) | S II |
| $\mathbf{B}_{19}$ | = | $x_6\,\mathbf{a}_1+y_6\,\mathbf{a}_2+z_6\,\mathbf{a}_3$ | = | $\frac{1}{2}(x_6+y_6)\,a\,\mathbf{\hat{x}}+\frac{\sqrt{3}}{2}(-x_6+y_6)\,a\,\mathbf{\hat{y}}+z_6 c\,\mathbf{\hat{z}}$ | (6$c$) | S III |
| $\mathbf{B}_{20}$ | = | $-y_6\,\mathbf{a}_1+(x_6-y_6)\,\mathbf{a}_2+z_6\,\mathbf{a}_3$ | = | $\left(\frac{1}{2}x_6-y_6\right)a\,\mathbf{\hat{x}}+\frac{\sqrt{3}}{2}x_6 a\,\mathbf{\hat{y}}+z_6 c\,\mathbf{\hat{z}}$ | (6$c$) | S III |
| $\mathbf{B}_{21}$ | = | $(-x_6+y_6)\,\mathbf{a}_1-x_6\,\mathbf{a}_2+z_6\,\mathbf{a}_3$ | = | $\left(-x_6+\frac{1}{2}y_6\right)a\,\mathbf{\hat{x}}-\frac{\sqrt{3}}{2}y_6 a\,\mathbf{\hat{y}}+z_6 c\,\mathbf{\hat{z}}$ | (6$c$) | S III |
| $\mathbf{B}_{22}$ | = | $-x_6\,\mathbf{a}_1 - y_6\,\mathbf{a}_2 + \left(\frac{1}{2} + z_6\right)\mathbf{a}_3$ | = | $-\frac{1}{2}(x_6 + y_6)\,a\,\hat{\mathbf{x}} + \frac{\sqrt{3}}{2}(x_6 - y_6)\,a\,\hat{\mathbf{y}} + \left(\frac{1}{2} + z_6\right)c\,\hat{\mathbf{z}}$ | (6$c$) | S III |
| $\mathbf{B}_{23}$ | = | $y_6\,\mathbf{a}_1 + (-x_6 + y_6)\,\mathbf{a}_2 + \left(\frac{1}{2} + z_6\right)\mathbf{a}_3$ | = | $\left(-\frac{1}{2}x_6 + y_6\right)a\,\hat{\mathbf{x}} - \frac{\sqrt{3}}{2}x_6 a\,\hat{\mathbf{y}} + \left(\frac{1}{2} + z_6\right)c\,\hat{\mathbf{z}}$ | (6$c$) | S III |
| $\mathbf{B}_{24}$ | = | $(x_6 - y_6)\,\mathbf{a}_1 + x_6\,\mathbf{a}_2 + \left(\frac{1}{2} + z_6\right)\mathbf{a}_3$ | = | $\left(x_6 - \frac{1}{2}y_6\right)a\,\hat{\mathbf{x}} + \frac{\sqrt{3}}{2}y_6 a\,\hat{\mathbf{y}} + \left(\frac{1}{2} + z_6\right)c\,\hat{\mathbf{z}}$ | (6$c$) | S III |

---

---

**Geometry files:**

- CIF: pp. 1749
- POSCAR: pp. 1749

# $La_3BWO_9$ ($P6_3$) Structure: AB3C9D_hP28_173_a_c_3c_b

http://aflow.org/prototype-encyclopedia/AB3C9D_hP28_173_a_c_3c_b

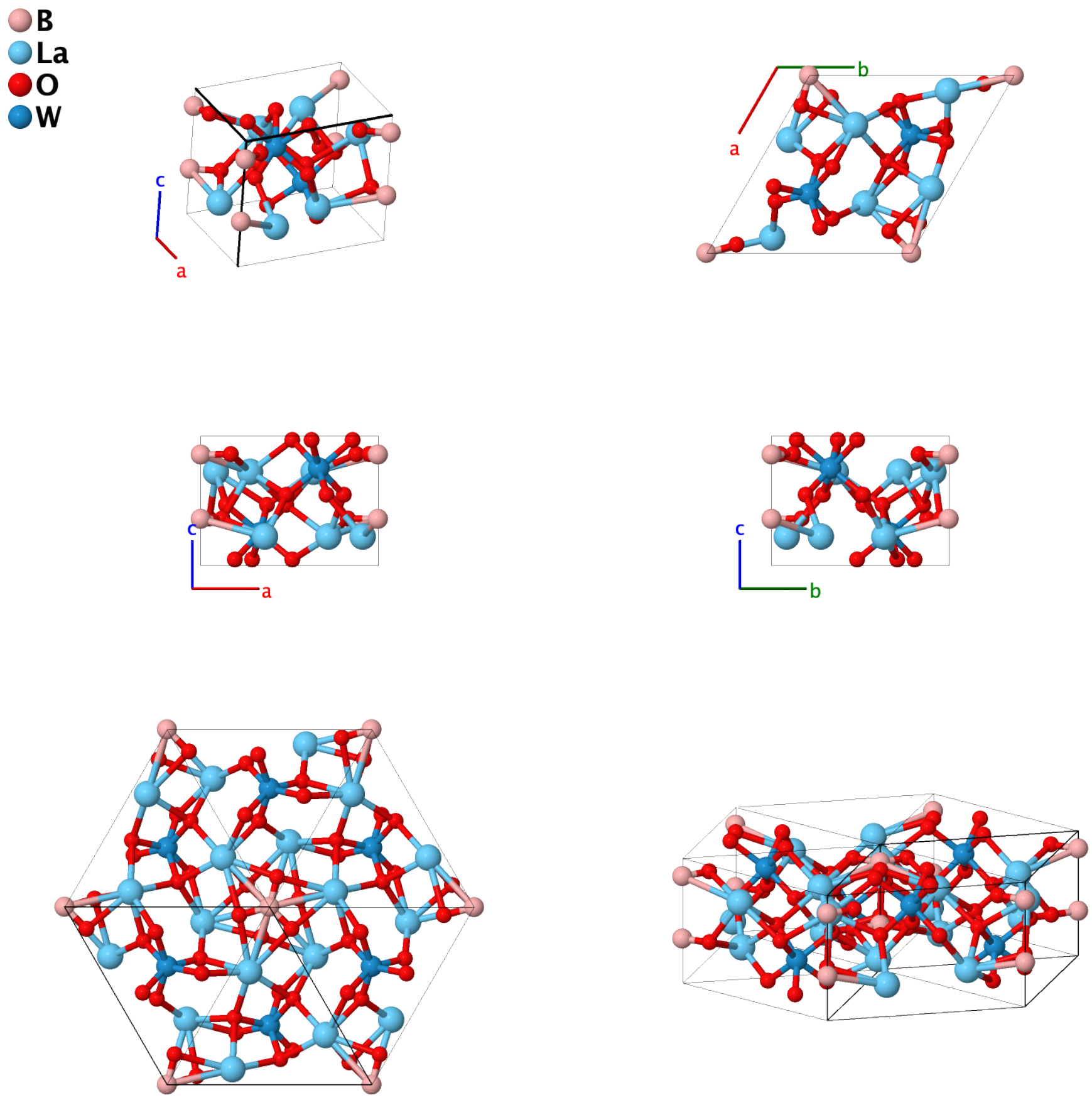

| | | |
|---|---|---|
| **Prototype** | : | $BaLi_3O_9W$ |
| **AFLOW prototype label** | : | AB3C9D_hP28_173_a_c_3c_b |
| ***Strukturbericht* designation** | : | None |
| **Pearson symbol** | : | hP28 |
| **Space group number** | : | 173 |
| **Space group symbol** | : | $P6_3$ |
| **AFLOW prototype command** | : | `aflow --proto=AB3C9D_hP28_173_a_c_3c_b` <br> `--params=`$a, c/a, z_1, z_2, x_3, y_3, z_3, x_4, y_4, z_4, x_5, y_5, z_5, x_6, y_6, z_6$ |

### Other compounds with this structure

- $La_3BWO_9$, $Ce_3BWO_9$, $Nd_3BWO_9$, $Sm_3BWO_9$, $Gd_3BWO_9$, $Tb_3BWO_9$, $Dy_3BWO_9$, and $Ho_3BWO_9$

- Most refinements of the $BaLi_3O_9W$ structure, including (Ashtar, 2020) place it in hexagonal space group $P6_3$ #173. (Han, 2018) find a better fit to the data by refining it in the trigonal $P3$ #143 space group, which places the lanthanum atoms on two independent crystallographic sites. As this may be due to the presence of bismuth impurities, which (Han, 2018) place on the lanthanum site, while (Ashtar, 2020) claim to have very pure samples, we withhold judgment on which structure is correct and present both.

- Space group $P6_3$ does not specify the origin of the $z$-axis. Here it is set so that the coordinate of the tungsten atom is $z_2 = 1/4$.

## Hexagonal primitive vectors:

$$\mathbf{a}_1 = \frac{1}{2}\,a\,\hat{\mathbf{x}} - \frac{\sqrt{3}}{2}\,a\,\hat{\mathbf{y}}$$
$$\mathbf{a}_2 = \frac{1}{2}\,a\,\hat{\mathbf{x}} + \frac{\sqrt{3}}{2}\,a\,\hat{\mathbf{y}}$$
$$\mathbf{a}_3 = c\,\hat{\mathbf{z}}$$

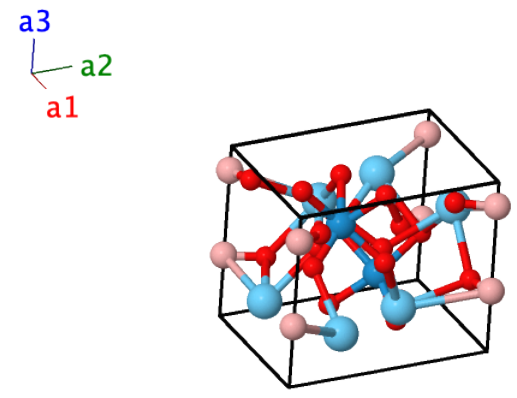

## Basis vectors:

| | Lattice Coordinates | | Cartesian Coordinates | Wyckoff Position | Atom Type |
|---|---|---|---|---|---|
| $\mathbf{B}_1 =$ | $z_1\,\mathbf{a}_3$ | = | $z_1 c\,\hat{\mathbf{z}}$ | (2$a$) | B |
| $\mathbf{B}_2 =$ | $\left(\frac{1}{2}+z_1\right)\mathbf{a}_3$ | = | $\left(\frac{1}{2}+z_1\right)c\,\hat{\mathbf{z}}$ | (2$a$) | B |
| $\mathbf{B}_3 =$ | $\frac{1}{3}\,\mathbf{a}_1 + \frac{2}{3}\,\mathbf{a}_2 + z_2\,\mathbf{a}_3$ | = | $\frac{1}{2}a\,\hat{\mathbf{x}} + \frac{1}{2\sqrt{3}}a\,\hat{\mathbf{y}} + z_2 c\,\hat{\mathbf{z}}$ | (2$b$) | W |
| $\mathbf{B}_4 =$ | $\frac{2}{3}\,\mathbf{a}_1 + \frac{1}{3}\,\mathbf{a}_2 + \left(\frac{1}{2}+z_2\right)\mathbf{a}_3$ | = | $\frac{1}{2}a\,\hat{\mathbf{x}} - \frac{1}{2\sqrt{3}}a\,\hat{\mathbf{y}} + \left(\frac{1}{2}+z_2\right)c\,\hat{\mathbf{z}}$ | (2$b$) | W |
| $\mathbf{B}_5 =$ | $x_3\,\mathbf{a}_1 + y_3\,\mathbf{a}_2 + z_3\,\mathbf{a}_3$ | = | $\frac{1}{2}\left(x_3+y_3\right)a\,\hat{\mathbf{x}} + \frac{\sqrt{3}}{2}\left(-x_3+y_3\right)a\,\hat{\mathbf{y}} + z_3 c\,\hat{\mathbf{z}}$ | (6$c$) | La |
| $\mathbf{B}_6 =$ | $-y_3\,\mathbf{a}_1 + \left(x_3-y_3\right)\mathbf{a}_2 + z_3\,\mathbf{a}_3$ | = | $\left(\frac{1}{2}x_3-y_3\right)a\,\hat{\mathbf{x}} + \frac{\sqrt{3}}{2}x_3 a\,\hat{\mathbf{y}} + z_3 c\,\hat{\mathbf{z}}$ | (6$c$) | La |
| $\mathbf{B}_7 =$ | $\left(-x_3+y_3\right)\mathbf{a}_1 - x_3\,\mathbf{a}_2 + z_3\,\mathbf{a}_3$ | = | $\left(-x_3+\frac{1}{2}y_3\right)a\,\hat{\mathbf{x}} - \frac{\sqrt{3}}{2}y_3 a\,\hat{\mathbf{y}} + z_3 c\,\hat{\mathbf{z}}$ | (6$c$) | La |
| $\mathbf{B}_8 =$ | $-x_3\,\mathbf{a}_1 - y_3\,\mathbf{a}_2 + \left(\frac{1}{2}+z_3\right)\mathbf{a}_3$ | = | $-\frac{1}{2}\left(x_3+y_3\right)a\,\hat{\mathbf{x}} + \frac{\sqrt{3}}{2}\left(x_3-y_3\right)a\,\hat{\mathbf{y}} + \left(\frac{1}{2}+z_3\right)c\,\hat{\mathbf{z}}$ | (6$c$) | La |
| $\mathbf{B}_9 =$ | $y_3\,\mathbf{a}_1 + \left(-x_3+y_3\right)\mathbf{a}_2 + \left(\frac{1}{2}+z_3\right)\mathbf{a}_3$ | = | $\left(-\frac{1}{2}x_3+y_3\right)a\,\hat{\mathbf{x}} - \frac{\sqrt{3}}{2}x_3 a\,\hat{\mathbf{y}} + \left(\frac{1}{2}+z_3\right)c\,\hat{\mathbf{z}}$ | (6$c$) | La |
| $\mathbf{B}_{10} =$ | $\left(x_3-y_3\right)\mathbf{a}_1 + x_3\,\mathbf{a}_2 + \left(\frac{1}{2}+z_3\right)\mathbf{a}_3$ | = | $\left(x_3-\frac{1}{2}y_3\right)a\,\hat{\mathbf{x}} + \frac{\sqrt{3}}{2}y_3 a\,\hat{\mathbf{y}} + \left(\frac{1}{2}+z_3\right)c\,\hat{\mathbf{z}}$ | (6$c$) | La |
| $\mathbf{B}_{11} =$ | $x_4\,\mathbf{a}_1 + y_4\,\mathbf{a}_2 + z_4\,\mathbf{a}_3$ | = | $\frac{1}{2}\left(x_4+y_4\right)a\,\hat{\mathbf{x}} + \frac{\sqrt{3}}{2}\left(-x_4+y_4\right)a\,\hat{\mathbf{y}} + z_4 c\,\hat{\mathbf{z}}$ | (6$c$) | O I |
| $\mathbf{B}_{12} =$ | $-y_4\,\mathbf{a}_1 + \left(x_4-y_4\right)\mathbf{a}_2 + z_4\,\mathbf{a}_3$ | = | $\left(\frac{1}{2}x_4-y_4\right)a\,\hat{\mathbf{x}} + \frac{\sqrt{3}}{2}x_4 a\,\hat{\mathbf{y}} + z_4 c\,\hat{\mathbf{z}}$ | (6$c$) | O I |
| $\mathbf{B}_{13} =$ | $\left(-x_4+y_4\right)\mathbf{a}_1 - x_4\,\mathbf{a}_2 + z_4\,\mathbf{a}_3$ | = | $\left(-x_4+\frac{1}{2}y_4\right)a\,\hat{\mathbf{x}} - \frac{\sqrt{3}}{2}y_4 a\,\hat{\mathbf{y}} + z_4 c\,\hat{\mathbf{z}}$ | (6$c$) | O I |
| $\mathbf{B}_{14} =$ | $-x_4\,\mathbf{a}_1 - y_4\,\mathbf{a}_2 + \left(\frac{1}{2}+z_4\right)\mathbf{a}_3$ | = | $-\frac{1}{2}\left(x_4+y_4\right)a\,\hat{\mathbf{x}} + \frac{\sqrt{3}}{2}\left(x_4-y_4\right)a\,\hat{\mathbf{y}} + \left(\frac{1}{2}+z_4\right)c\,\hat{\mathbf{z}}$ | (6$c$) | O I |
| $\mathbf{B}_{15} =$ | $y_4\,\mathbf{a}_1 + \left(-x_4+y_4\right)\mathbf{a}_2 + \left(\frac{1}{2}+z_4\right)\mathbf{a}_3$ | = | $\left(-\frac{1}{2}x_4+y_4\right)a\,\hat{\mathbf{x}} - \frac{\sqrt{3}}{2}x_4 a\,\hat{\mathbf{y}} + \left(\frac{1}{2}+z_4\right)c\,\hat{\mathbf{z}}$ | (6$c$) | O I |
| $\mathbf{B}_{16} =$ | $\left(x_4-y_4\right)\mathbf{a}_1 + x_4\,\mathbf{a}_2 + \left(\frac{1}{2}+z_4\right)\mathbf{a}_3$ | = | $\left(x_4-\frac{1}{2}y_4\right)a\,\hat{\mathbf{x}} + \frac{\sqrt{3}}{2}y_4 a\,\hat{\mathbf{y}} + \left(\frac{1}{2}+z_4\right)c\,\hat{\mathbf{z}}$ | (6$c$) | O I |
| $\mathbf{B}_{17} =$ | $x_5\,\mathbf{a}_1 + y_5\,\mathbf{a}_2 + z_5\,\mathbf{a}_3$ | = | $\frac{1}{2}\left(x_5+y_5\right)a\,\hat{\mathbf{x}} + \frac{\sqrt{3}}{2}\left(-x_5+y_5\right)a\,\hat{\mathbf{y}} + z_5 c\,\hat{\mathbf{z}}$ | (6$c$) | O II |
| $\mathbf{B}_{18} =$ | $-y_5\,\mathbf{a}_1 + \left(x_5-y_5\right)\mathbf{a}_2 + z_5\,\mathbf{a}_3$ | = | $\left(\frac{1}{2}x_5-y_5\right)a\,\hat{\mathbf{x}} + \frac{\sqrt{3}}{2}x_5 a\,\hat{\mathbf{y}} + z_5 c\,\hat{\mathbf{z}}$ | (6$c$) | O II |
| $\mathbf{B}_{19} =$ | $\left(-x_5+y_5\right)\mathbf{a}_1 - x_5\,\mathbf{a}_2 + z_5\,\mathbf{a}_3$ | = | $\left(-x_5+\frac{1}{2}y_5\right)a\,\hat{\mathbf{x}} - \frac{\sqrt{3}}{2}y_5 a\,\hat{\mathbf{y}} + z_5 c\,\hat{\mathbf{z}}$ | (6$c$) | O II |
| $\mathbf{B}_{20} =$ | $-x_5\,\mathbf{a}_1 - y_5\,\mathbf{a}_2 + \left(\frac{1}{2}+z_5\right)\mathbf{a}_3$ | = | $-\frac{1}{2}\left(x_5+y_5\right)a\,\hat{\mathbf{x}} + \frac{\sqrt{3}}{2}\left(x_5-y_5\right)a\,\hat{\mathbf{y}} + \left(\frac{1}{2}+z_5\right)c\,\hat{\mathbf{z}}$ | (6$c$) | O II |

| | | | | | | |
|---|---|---|---|---|---|---|
| $\mathbf{B}_{21}$ | = | $y_5\,\mathbf{a}_1 + (-x_5 + y_5)\,\mathbf{a}_2 + \left(\frac{1}{2} + z_5\right)\,\mathbf{a}_3$ | = | $\left(-\frac{1}{2}x_5 + y_5\right)a\,\mathbf{\hat{x}} - \frac{\sqrt{3}}{2}x_5 a\,\mathbf{\hat{y}} + \left(\frac{1}{2} + z_5\right)c\,\mathbf{\hat{z}}$ | (6$c$) | O II |
| $\mathbf{B}_{22}$ | = | $(x_5 - y_5)\,\mathbf{a}_1 + x_5\,\mathbf{a}_2 + \left(\frac{1}{2} + z_5\right)\,\mathbf{a}_3$ | = | $\left(x_5 - \frac{1}{2}y_5\right)a\,\mathbf{\hat{x}} + \frac{\sqrt{3}}{2}y_5 a\,\mathbf{\hat{y}} + \left(\frac{1}{2} + z_5\right)c\,\mathbf{\hat{z}}$ | (6$c$) | O II |
| $\mathbf{B}_{23}$ | = | $x_6\,\mathbf{a}_1 + y_6\,\mathbf{a}_2 + z_6\,\mathbf{a}_3$ | = | $\frac{1}{2}(x_6 + y_6)\,a\,\mathbf{\hat{x}} + \frac{\sqrt{3}}{2}(-x_6 + y_6)\,a\,\mathbf{\hat{y}} + z_6 c\,\mathbf{\hat{z}}$ | (6$c$) | O III |
| $\mathbf{B}_{24}$ | = | $-y_6\,\mathbf{a}_1 + (x_6 - y_6)\,\mathbf{a}_2 + z_6\,\mathbf{a}_3$ | = | $\left(\frac{1}{2}x_6 - y_6\right)a\,\mathbf{\hat{x}} + \frac{\sqrt{3}}{2}x_6 a\,\mathbf{\hat{y}} + z_6 c\,\mathbf{\hat{z}}$ | (6$c$) | O III |
| $\mathbf{B}_{25}$ | = | $(-x_6 + y_6)\,\mathbf{a}_1 - x_6\,\mathbf{a}_2 + z_6\,\mathbf{a}_3$ | = | $\left(-x_6 + \frac{1}{2}y_6\right)a\,\mathbf{\hat{x}} - \frac{\sqrt{3}}{2}y_6 a\,\mathbf{\hat{y}} + z_6 c\,\mathbf{\hat{z}}$ | (6$c$) | O III |
| $\mathbf{B}_{26}$ | = | $-x_6\,\mathbf{a}_1 - y_6\,\mathbf{a}_2 + \left(\frac{1}{2} + z_6\right)\,\mathbf{a}_3$ | = | $-\frac{1}{2}(x_6 + y_6)\,a\,\mathbf{\hat{x}} + \frac{\sqrt{3}}{2}(x_6 - y_6)\,a\,\mathbf{\hat{y}} + \left(\frac{1}{2} + z_6\right)c\,\mathbf{\hat{z}}$ | (6$c$) | O III |
| $\mathbf{B}_{27}$ | = | $y_6\,\mathbf{a}_1 + (-x_6 + y_6)\,\mathbf{a}_2 + \left(\frac{1}{2} + z_6\right)\,\mathbf{a}_3$ | = | $\left(-\frac{1}{2}x_6 + y_6\right)a\,\mathbf{\hat{x}} - \frac{\sqrt{3}}{2}x_6 a\,\mathbf{\hat{y}} + \left(\frac{1}{2} + z_6\right)c\,\mathbf{\hat{z}}$ | (6$c$) | O III |
| $\mathbf{B}_{28}$ | = | $(x_6 - y_6)\,\mathbf{a}_1 + x_6\,\mathbf{a}_2 + \left(\frac{1}{2} + z_6\right)\,\mathbf{a}_3$ | = | $\left(x_6 - \frac{1}{2}y_6\right)a\,\mathbf{\hat{x}} + \frac{\sqrt{3}}{2}y_6 a\,\mathbf{\hat{y}} + \left(\frac{1}{2} + z_6\right)c\,\mathbf{\hat{z}}$ | (6$c$) | O III |

---

---

**Geometry files:**

- CIF: pp. 1750
- POSCAR: pp. 1750

# $\alpha$-$LiIO_3$ Structure: ABC3_hP10_173_b_a_c

http://aflow.org/prototype-encyclopedia/ABC3_hP10_173_b_a_c

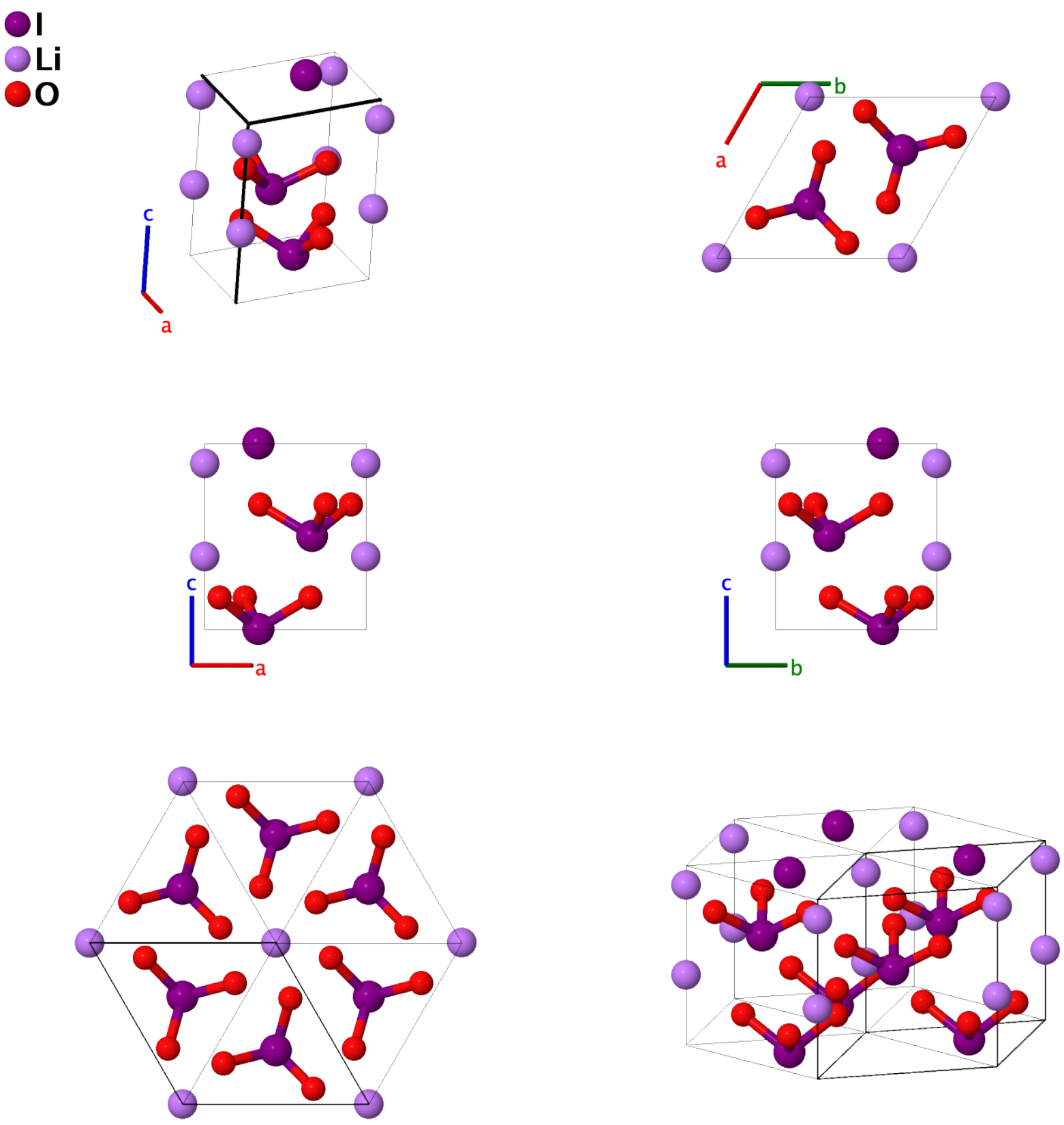

| | | |
|---|---|---|
| **Prototype** | : | $ILiO_3$ |
| **AFLOW prototype label** | : | ABC3_hP10_173_b_a_c |
| ***Strukturbericht* designation** | : | None |
| **Pearson symbol** | : | hP10 |
| **Space group number** | : | 173 |
| **Space group symbol** | : | $P6_3$ |
| **AFLOW prototype command** | : | `aflow --proto=ABC3_hP10_173_b_a_c --params=`$a, c/a, z_1, z_2, x_3, y_3, z_3$ |

- $LiIO_3$ is known to exist in three forms:
  - $\alpha$-$LiIO_3$, stable below 470 K: (Zachariasen, 1931) originally determined that the structure of $\alpha$-$LiIO_3$ was in space group $P6_322$ #182, which (Hermann, 1937) designated *Strukturbericht* $E2_3$. (Rosenzweig, 1966) subsequently determined that this structure was incorrect because of the small sample size, and determined that the true structure was in space group $P6_3$ #173.
  - $\beta$-$LiIO_3$, stable from 573 up to the melting point at 708 K.
  - $\gamma$-$LiIO_3$, stable between the $\alpha$- and $\beta$-phases, with an orthorhombic structure in space group $Pna2_1$ #33.

**Hexagonal primitive vectors:**

$$\mathbf{a}_1 = \tfrac{1}{2}\,a\,\hat{\mathbf{x}} - \tfrac{\sqrt{3}}{2}\,a\,\hat{\mathbf{y}}$$
$$\mathbf{a}_2 = \tfrac{1}{2}\,a\,\hat{\mathbf{x}} + \tfrac{\sqrt{3}}{2}\,a\,\hat{\mathbf{y}}$$
$$\mathbf{a}_3 = c\,\hat{\mathbf{z}}$$

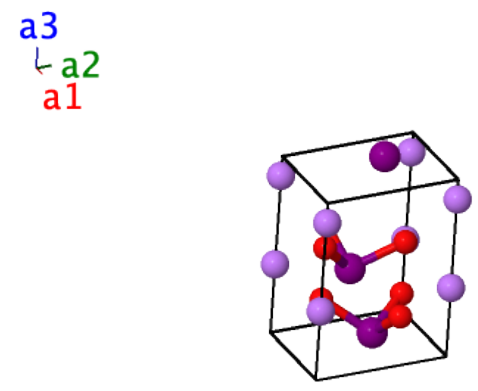

**Basis vectors:**

| | | Lattice Coordinates | | Cartesian Coordinates | Wyckoff Position | Atom Type |
|---|---|---|---|---|---|---|
| $\mathbf{B}_1$ | = | $z_1\,\mathbf{a}_3$ | = | $z_1 c\,\hat{\mathbf{z}}$ | (2$a$) | Li |
| $\mathbf{B}_2$ | = | $\left(\frac{1}{2} + z_1\right)\mathbf{a}_3$ | = | $\left(\frac{1}{2} + z_1\right)c\,\hat{\mathbf{z}}$ | (2$a$) | Li |
| $\mathbf{B}_3$ | = | $\frac{1}{3}\,\mathbf{a}_1 + \frac{2}{3}\,\mathbf{a}_2 + z_2\,\mathbf{a}_3$ | = | $\frac{1}{2}a\,\hat{\mathbf{x}} + \frac{1}{2\sqrt{3}}a\,\hat{\mathbf{y}} + z_2 c\,\hat{\mathbf{z}}$ | (2$b$) | I |
| $\mathbf{B}_4$ | = | $\frac{2}{3}\,\mathbf{a}_1 + \frac{1}{3}\,\mathbf{a}_2 + \left(\frac{1}{2} + z_2\right)\mathbf{a}_3$ | = | $\frac{1}{2}a\,\hat{\mathbf{x}} - \frac{1}{2\sqrt{3}}a\,\hat{\mathbf{y}} + \left(\frac{1}{2} + z_2\right)c\,\hat{\mathbf{z}}$ | (2$b$) | I |
| $\mathbf{B}_5$ | = | $x_3\,\mathbf{a}_1 + y_3\,\mathbf{a}_2 + z_3\,\mathbf{a}_3$ | = | $\frac{1}{2}\left(x_3 + y_3\right)a\,\hat{\mathbf{x}} + \frac{\sqrt{3}}{2}\left(-x_3 + y_3\right)a\,\hat{\mathbf{y}} + z_3 c\,\hat{\mathbf{z}}$ | (6$c$) | O |
| $\mathbf{B}_6$ | = | $-y_3\,\mathbf{a}_1 + \left(x_3 - y_3\right)\mathbf{a}_2 + z_3\,\mathbf{a}_3$ | = | $\left(\frac{1}{2}x_3 - y_3\right)a\,\hat{\mathbf{x}} + \frac{\sqrt{3}}{2}x_3 a\,\hat{\mathbf{y}} + z_3 c\,\hat{\mathbf{z}}$ | (6$c$) | O |
| $\mathbf{B}_7$ | = | $\left(-x_3 + y_3\right)\mathbf{a}_1 - x_3\,\mathbf{a}_2 + z_3\,\mathbf{a}_3$ | = | $\left(-x_3 + \frac{1}{2}y_3\right)a\,\hat{\mathbf{x}} - \frac{\sqrt{3}}{2}y_3 a\,\hat{\mathbf{y}} + z_3 c\,\hat{\mathbf{z}}$ | (6$c$) | O |
| $\mathbf{B}_8$ | = | $-x_3\,\mathbf{a}_1 - y_3\,\mathbf{a}_2 + \left(\frac{1}{2} + z_3\right)\mathbf{a}_3$ | = | $-\frac{1}{2}\left(x_3 + y_3\right)a\,\hat{\mathbf{x}} + \frac{\sqrt{3}}{2}\left(x_3 - y_3\right)a\,\hat{\mathbf{y}} + \left(\frac{1}{2} + z_3\right)c\,\hat{\mathbf{z}}$ | (6$c$) | O |
| $\mathbf{B}_9$ | = | $y_3\,\mathbf{a}_1 + \left(-x_3 + y_3\right)\mathbf{a}_2 + \left(\frac{1}{2} + z_3\right)\mathbf{a}_3$ | = | $\left(-\frac{1}{2}x_3 + y_3\right)a\,\hat{\mathbf{x}} - \frac{\sqrt{3}}{2}x_3 a\,\hat{\mathbf{y}} + \left(\frac{1}{2} + z_3\right)c\,\hat{\mathbf{z}}$ | (6$c$) | O |
| $\mathbf{B}_{10}$ | = | $\left(x_3 - y_3\right)\mathbf{a}_1 + x_3\,\mathbf{a}_2 + \left(\frac{1}{2} + z_3\right)\mathbf{a}_3$ | = | $\left(x_3 - \frac{1}{2}y_3\right)a\,\hat{\mathbf{x}} + \frac{\sqrt{3}}{2}y_3 a\,\hat{\mathbf{y}} + \left(\frac{1}{2} + z_3\right)c\,\hat{\mathbf{z}}$ | (6$c$) | O |

**Geometry files:**

- CIF: pp. 1750
- POSCAR: pp. 1751

# $LiKSO_4$ ($H1_4$) Structure: ABC4D_hP14_173_a_b_bc_b

http://aflow.org/prototype-encyclopedia/ABC4D_hP14_173_a_b_bc_b

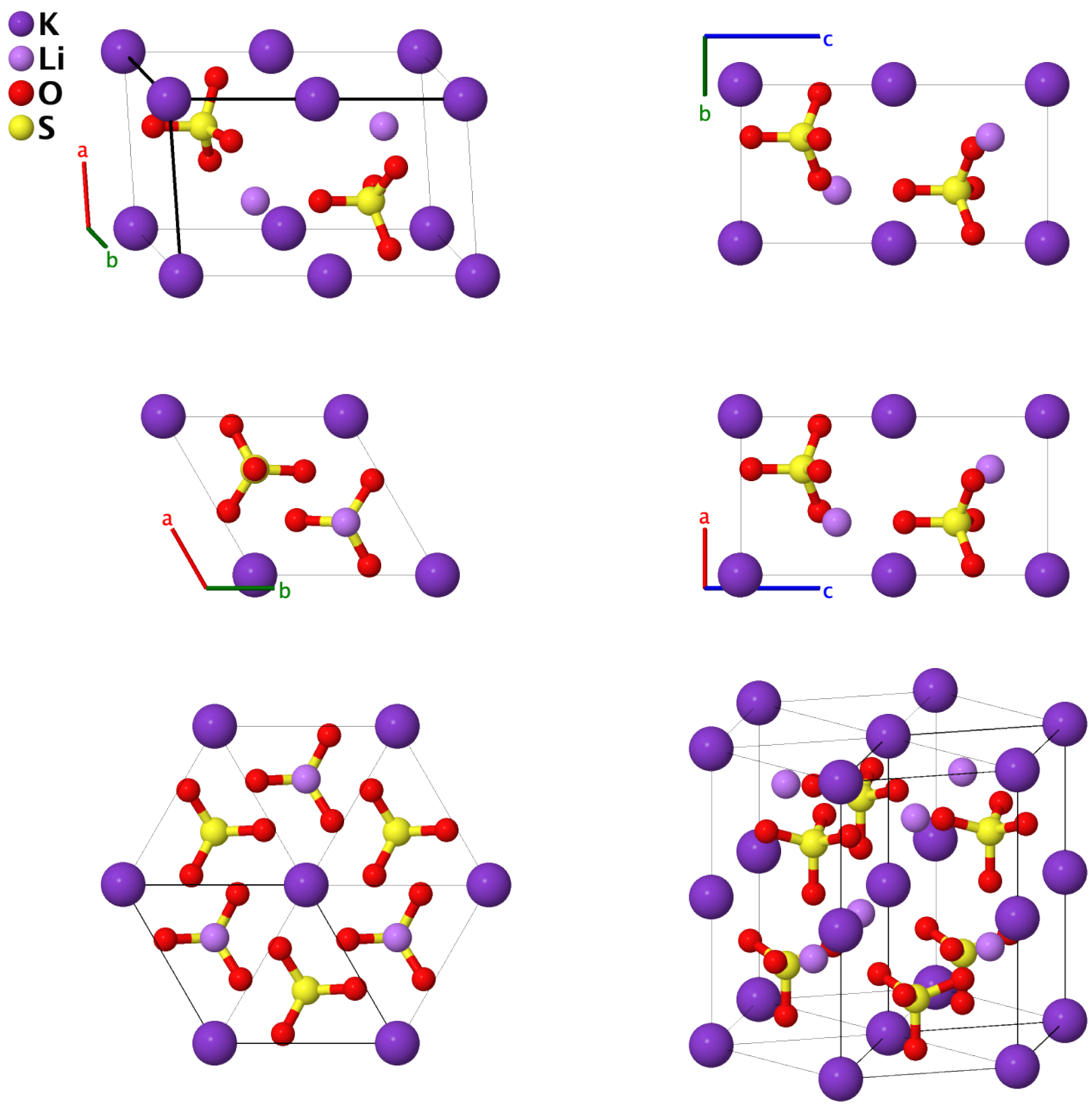

| | | |
|---|---|---|
| **Prototype** | : | $KLiO_4S$ |
| **AFLOW prototype label** | : | ABC4D_hP14_173_a_b_bc_b |
| ***Strukturbericht* designation** | : | $H1_4$ |
| **Pearson symbol** | : | hP14 |
| **Space group number** | : | 173 |
| **Space group symbol** | : | $P6_3$ |
| **AFLOW prototype command** | : | `aflow --proto=ABC4D_hP14_173_a_b_bc_b`<br>`--params=`$a, c/a, z_1, z_2, z_3, z_4, x_5, y_5, z_5$ |

- As there is no center of symmetry in the structure, the point $z = 0$ can be set arbitrarily. Following (Bhakay-Tamhane, 1984) we set $z_1 = 0$.
- While *Struckturbericht* symbols $H1_1$, $H1_2$ and $H1_5$ are identical with $S1_1$, $S1_2$ and $S1_5$, respectively, there is no connection between this structure and $S1_4$, garnet (A2B3C12D3_cI160_230_a_c_h_d).

**Hexagonal primitive vectors:**

$$
\begin{aligned}
\mathbf{a}_1 &= \tfrac{1}{2}\,a\,\hat{\mathbf{x}} - \tfrac{\sqrt{3}}{2}\,a\,\hat{\mathbf{y}} \\
\mathbf{a}_2 &= \tfrac{1}{2}\,a\,\hat{\mathbf{x}} + \tfrac{\sqrt{3}}{2}\,a\,\hat{\mathbf{y}} \\
\mathbf{a}_3 &= c\,\hat{\mathbf{z}}
\end{aligned}
$$

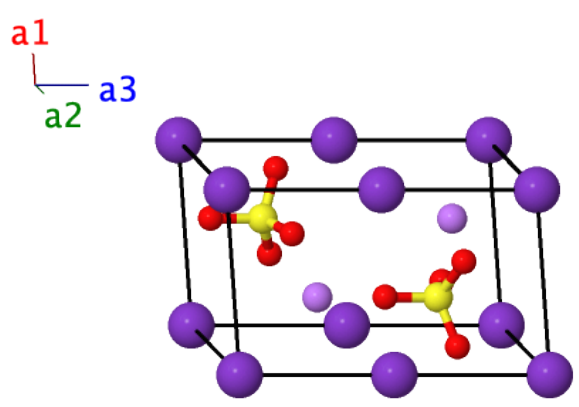

**Basis vectors:**

| | | Lattice Coordinates | | Cartesian Coordinates | Wyckoff Position | Atom Type |
|---|---|---|---|---|---|---|
| $\mathbf{B}_1$ | = | $z_1\,\mathbf{a}_3$ | = | $z_1 c\,\hat{\mathbf{z}}$ | (2$a$) | K |
| $\mathbf{B}_2$ | = | $\left(\frac{1}{2}+z_1\right)\mathbf{a}_3$ | = | $\left(\frac{1}{2}+z_1\right)c\,\hat{\mathbf{z}}$ | (2$a$) | K |
| $\mathbf{B}_3$ | = | $\frac{1}{3}\,\mathbf{a}_1+\frac{2}{3}\,\mathbf{a}_2+z_2\,\mathbf{a}_3$ | = | $\frac{1}{2}a\,\hat{\mathbf{x}}+\frac{1}{2\sqrt{3}}a\,\hat{\mathbf{y}}+z_2c\,\hat{\mathbf{z}}$ | (2$b$) | Li |
| $\mathbf{B}_4$ | = | $\frac{2}{3}\,\mathbf{a}_1+\frac{1}{3}\,\mathbf{a}_2+\left(\frac{1}{2}+z_2\right)\mathbf{a}_3$ | = | $\frac{1}{2}a\,\hat{\mathbf{x}}-\frac{1}{2\sqrt{3}}a\,\hat{\mathbf{y}}+\left(\frac{1}{2}+z_2\right)c\,\hat{\mathbf{z}}$ | (2$b$) | Li |
| $\mathbf{B}_5$ | = | $\frac{1}{3}\,\mathbf{a}_1+\frac{2}{3}\,\mathbf{a}_2+z_3\,\mathbf{a}_3$ | = | $\frac{1}{2}a\,\hat{\mathbf{x}}+\frac{1}{2\sqrt{3}}a\,\hat{\mathbf{y}}+z_3c\,\hat{\mathbf{z}}$ | (2$b$) | O I |
| $\mathbf{B}_6$ | = | $\frac{2}{3}\,\mathbf{a}_1+\frac{1}{3}\,\mathbf{a}_2+\left(\frac{1}{2}+z_3\right)\mathbf{a}_3$ | = | $\frac{1}{2}a\,\hat{\mathbf{x}}-\frac{1}{2\sqrt{3}}a\,\hat{\mathbf{y}}+\left(\frac{1}{2}+z_3\right)c\,\hat{\mathbf{z}}$ | (2$b$) | O I |
| $\mathbf{B}_7$ | = | $\frac{1}{3}\,\mathbf{a}_1+\frac{2}{3}\,\mathbf{a}_2+z_4\,\mathbf{a}_3$ | = | $\frac{1}{2}a\,\hat{\mathbf{x}}+\frac{1}{2\sqrt{3}}a\,\hat{\mathbf{y}}+z_4c\,\hat{\mathbf{z}}$ | (2$b$) | S |
| $\mathbf{B}_8$ | = | $\frac{2}{3}\,\mathbf{a}_1+\frac{1}{3}\,\mathbf{a}_2+\left(\frac{1}{2}+z_4\right)\mathbf{a}_3$ | = | $\frac{1}{2}a\,\hat{\mathbf{x}}-\frac{1}{2\sqrt{3}}a\,\hat{\mathbf{y}}+\left(\frac{1}{2}+z_4\right)c\,\hat{\mathbf{z}}$ | (2$b$) | S |
| $\mathbf{B}_9$ | = | $x_5\,\mathbf{a}_1+y_5\,\mathbf{a}_2+z_5\,\mathbf{a}_3$ | = | $\frac{1}{2}\left(x_5+y_5\right)a\,\hat{\mathbf{x}}+\frac{\sqrt{3}}{2}\left(-x_5+y_5\right)a\,\hat{\mathbf{y}}+z_5c\,\hat{\mathbf{z}}$ | (6$c$) | O II |
| $\mathbf{B}_{10}$ | = | $-y_5\,\mathbf{a}_1+\left(x_5-y_5\right)\mathbf{a}_2+z_5\,\mathbf{a}_3$ | = | $\left(\frac{1}{2}x_5-y_5\right)a\,\hat{\mathbf{x}}+\frac{\sqrt{3}}{2}x_5a\,\hat{\mathbf{y}}+z_5c\,\hat{\mathbf{z}}$ | (6$c$) | O II |
| $\mathbf{B}_{11}$ | = | $\left(-x_5+y_5\right)\mathbf{a}_1-x_5\,\mathbf{a}_2+z_5\,\mathbf{a}_3$ | = | $\left(-x_5+\frac{1}{2}y_5\right)a\,\hat{\mathbf{x}}-\frac{\sqrt{3}}{2}y_5a\,\hat{\mathbf{y}}+z_5c\,\hat{\mathbf{z}}$ | (6$c$) | O II |
| $\mathbf{B}_{12}$ | = | $-x_5\,\mathbf{a}_1-y_5\,\mathbf{a}_2+\left(\frac{1}{2}+z_5\right)\mathbf{a}_3$ | = | $-\frac{1}{2}\left(x_5+y_5\right)a\,\hat{\mathbf{x}}+\frac{\sqrt{3}}{2}\left(x_5-y_5\right)a\,\hat{\mathbf{y}}+\left(\frac{1}{2}+z_5\right)c\,\hat{\mathbf{z}}$ | (6$c$) | O II |
| $\mathbf{B}_{13}$ | = | $y_5\,\mathbf{a}_1+\left(-x_5+y_5\right)\mathbf{a}_2+\left(\frac{1}{2}+z_5\right)\mathbf{a}_3$ | = | $\left(-\frac{1}{2}x_5+y_5\right)a\,\hat{\mathbf{x}}-\frac{\sqrt{3}}{2}x_5a\,\hat{\mathbf{y}}+\left(\frac{1}{2}+z_5\right)c\,\hat{\mathbf{z}}$ | (6$c$) | O II |
| $\mathbf{B}_{14}$ | = | $\left(x_5-y_5\right)\mathbf{a}_1+x_5\,\mathbf{a}_2+\left(\frac{1}{2}+z_5\right)\mathbf{a}_3$ | = | $\left(x_5-\frac{1}{2}y_5\right)a\,\hat{\mathbf{x}}+\frac{\sqrt{3}}{2}y_5a\,\hat{\mathbf{y}}+\left(\frac{1}{2}+z_5\right)c\,\hat{\mathbf{z}}$ | (6$c$) | O II |

**Geometry files:**

- CIF: pp. 1751
- POSCAR: pp. 1751

# $Rh_{20}Si_{13}$ Structure: A10B7_hP34_176_c3h_b2h

http://aflow.org/prototype-encyclopedia/A10B7_hP34_176_c3h_b2h

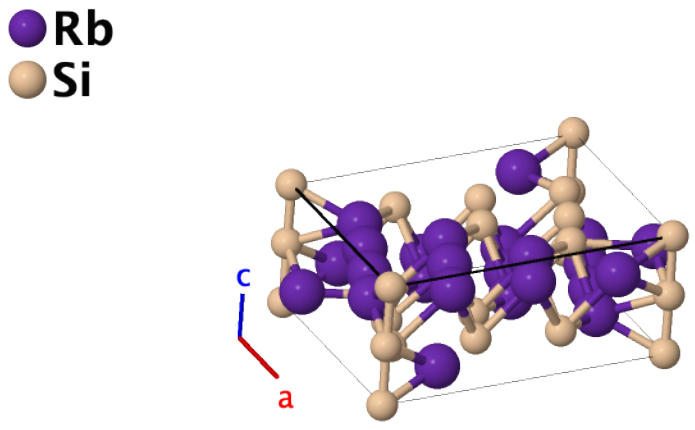

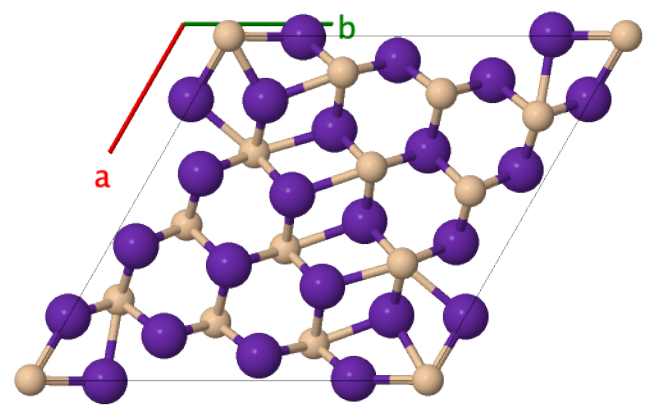

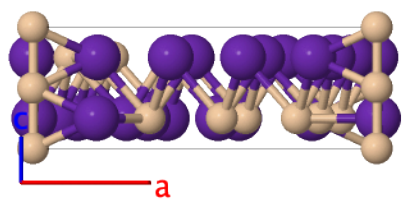

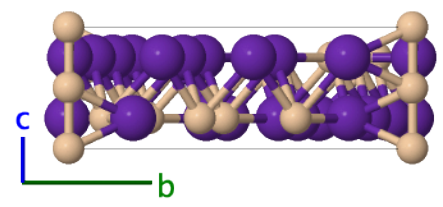

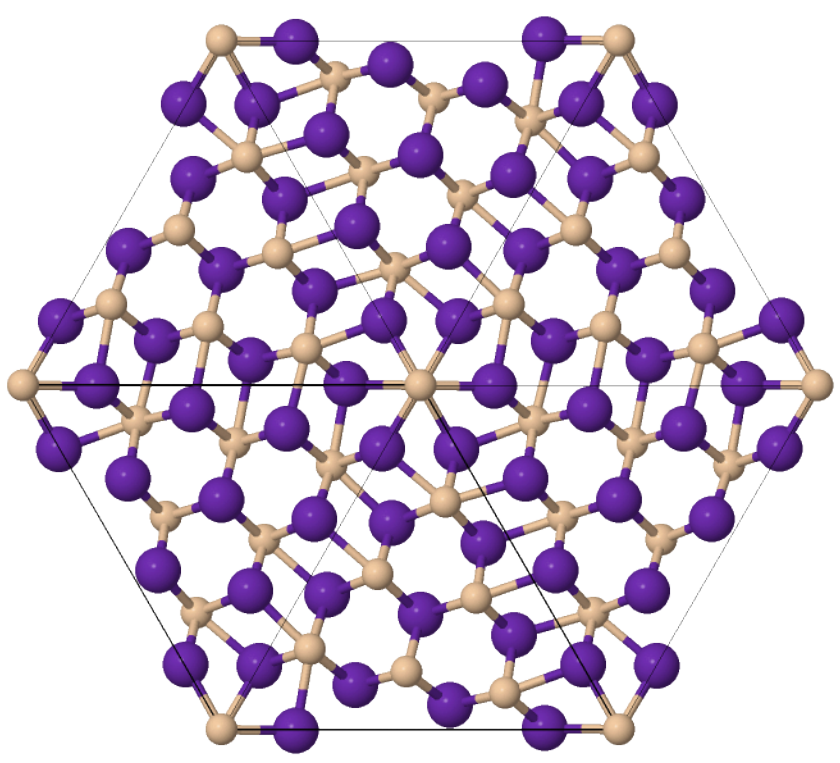

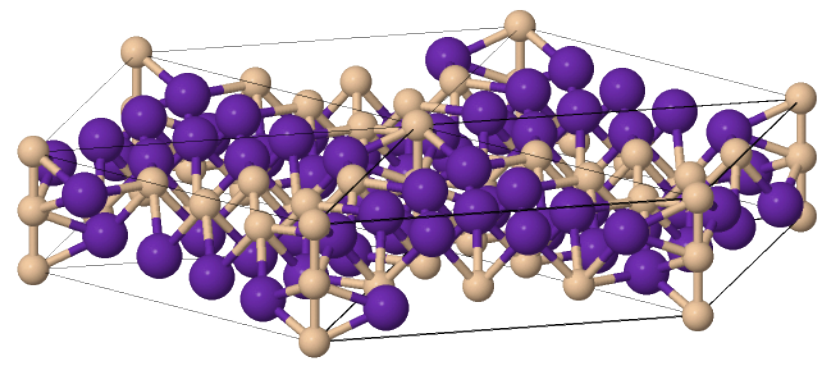

| | | |
|---|---|---|
| **Prototype** | : | $Rh_{20}Si_{13}$ |
| **AFLOW prototype label** | : | A10B7_hP34_176_c3h_b2h |
| ***Strukturbericht* designation** | : | None |
| **Pearson symbol** | : | hP34 |
| **Space group number** | : | 176 |
| **Space group symbol** | : | $P6_3/m$ |
| **AFLOW prototype command** | : | `aflow --proto=A10B7_hP34_176_c3h_b2h` `--params=`$a, c/a, x_3, y_3, x_4, y_4, x_5, y_5, x_6, y_6, x_7, y_7$ |

- The Si-I ($2b$) site is only occupied 50% of the time, which gives the observed stoichiometry.

**Hexagonal primitive vectors:**

$$\mathbf{a}_1 = \frac{1}{2}\,a\,\hat{\mathbf{x}} - \frac{\sqrt{3}}{2}\,a\,\hat{\mathbf{y}}$$
$$\mathbf{a}_2 = \frac{1}{2}\,a\,\hat{\mathbf{x}} + \frac{\sqrt{3}}{2}\,a\,\hat{\mathbf{y}}$$
$$\mathbf{a}_3 = c\,\hat{\mathbf{z}}$$

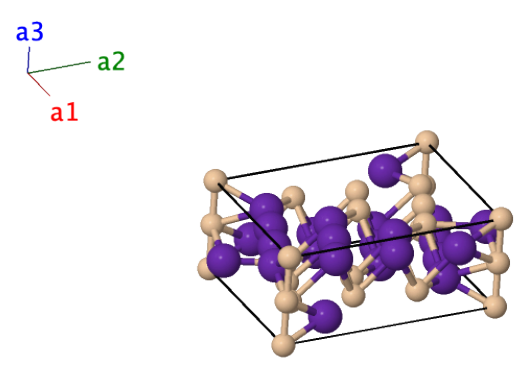

**Basis vectors:**

| | | Lattice Coordinates | | Cartesian Coordinates | Wyckoff Position | Atom Type |
|---|---|---|---|---|---|---|
| $\mathbf{B}_1$ | = | $0\,\mathbf{a}_1 + 0\,\mathbf{a}_2 + 0\,\mathbf{a}_3$ | = | $0\,\hat{\mathbf{x}} + 0\,\hat{\mathbf{y}} + 0\,\hat{\mathbf{z}}$ | (2*b*) | Si I |
| $\mathbf{B}_2$ | = | $\frac{1}{2}\,\mathbf{a}_3$ | = | $\frac{1}{2}c\,\hat{\mathbf{z}}$ | (2*b*) | Si I |
| $\mathbf{B}_3$ | = | $\frac{1}{3}\,\mathbf{a}_1 + \frac{2}{3}\,\mathbf{a}_2 + \frac{1}{4}\,\mathbf{a}_3$ | = | $\frac{1}{2}a\,\hat{\mathbf{x}} + \frac{1}{2\sqrt{3}}a\,\hat{\mathbf{y}} + \frac{1}{4}c\,\hat{\mathbf{z}}$ | (2*c*) | Rb I |
| $\mathbf{B}_4$ | = | $\frac{2}{3}\,\mathbf{a}_1 + \frac{1}{3}\,\mathbf{a}_2 + \frac{3}{4}\,\mathbf{a}_3$ | = | $\frac{1}{2}a\,\hat{\mathbf{x}} - \frac{1}{2\sqrt{3}}a\,\hat{\mathbf{y}} + \frac{3}{4}c\,\hat{\mathbf{z}}$ | (2*c*) | Rb I |
| $\mathbf{B}_5$ | = | $x_3\,\mathbf{a}_1 + y_3\,\mathbf{a}_2 + \frac{1}{4}\,\mathbf{a}_3$ | = | $\frac{1}{2}\left(x_3 + y_3\right)a\,\hat{\mathbf{x}} + \frac{\sqrt{3}}{2}\left(-x_3 + y_3\right)a\,\hat{\mathbf{y}} + \frac{1}{4}c\,\hat{\mathbf{z}}$ | (6*h*) | Rb II |
| $\mathbf{B}_6$ | = | $-y_3\,\mathbf{a}_1 + \left(x_3 - y_3\right)\,\mathbf{a}_2 + \frac{1}{4}\,\mathbf{a}_3$ | = | $\left(\frac{1}{2}x_3 - y_3\right)a\,\hat{\mathbf{x}} + \frac{\sqrt{3}}{2}x_3a\,\hat{\mathbf{y}} + \frac{1}{4}c\,\hat{\mathbf{z}}$ | (6*h*) | Rb II |
| $\mathbf{B}_7$ | = | $\left(-x_3 + y_3\right)\,\mathbf{a}_1 - x_3\,\mathbf{a}_2 + \frac{1}{4}\,\mathbf{a}_3$ | = | $\left(-x_3 + \frac{1}{2}y_3\right)a\,\hat{\mathbf{x}} - \frac{\sqrt{3}}{2}y_3a\,\hat{\mathbf{y}} + \frac{1}{4}c\,\hat{\mathbf{z}}$ | (6*h*) | Rb II |
| $\mathbf{B}_8$ | = | $-x_3\,\mathbf{a}_1 - y_3\,\mathbf{a}_2 + \frac{3}{4}\,\mathbf{a}_3$ | = | $-\frac{1}{2}\left(x_3 + y_3\right)a\,\hat{\mathbf{x}} + \frac{\sqrt{3}}{2}\left(x_3 - y_3\right)a\,\hat{\mathbf{y}} + \frac{3}{4}c\,\hat{\mathbf{z}}$ | (6*h*) | Rb II |
| $\mathbf{B}_9$ | = | $y_3\,\mathbf{a}_1 + \left(-x_3 + y_3\right)\,\mathbf{a}_2 + \frac{3}{4}\,\mathbf{a}_3$ | = | $\left(-\frac{1}{2}x_3 + y_3\right)a\,\hat{\mathbf{x}} - \frac{\sqrt{3}}{2}x_3a\,\hat{\mathbf{y}} + \frac{3}{4}c\,\hat{\mathbf{z}}$ | (6*h*) | Rb II |
| $\mathbf{B}_{10}$ | = | $\left(x_3 - y_3\right)\,\mathbf{a}_1 + x_3\,\mathbf{a}_2 + \frac{3}{4}\,\mathbf{a}_3$ | = | $\left(x_3 - \frac{1}{2}y_3\right)a\,\hat{\mathbf{x}} + \frac{\sqrt{3}}{2}y_3a\,\hat{\mathbf{y}} + \frac{3}{4}c\,\hat{\mathbf{z}}$ | (6*h*) | Rb II |
| $\mathbf{B}_{11}$ | = | $x_4\,\mathbf{a}_1 + y_4\,\mathbf{a}_2 + \frac{1}{4}\,\mathbf{a}_3$ | = | $\frac{1}{2}\left(x_4 + y_4\right)a\,\hat{\mathbf{x}} + \frac{\sqrt{3}}{2}\left(-x_4 + y_4\right)a\,\hat{\mathbf{y}} + \frac{1}{4}c\,\hat{\mathbf{z}}$ | (6*h*) | Rb III |
| $\mathbf{B}_{12}$ | = | $-y_4\,\mathbf{a}_1 + \left(x_4 - y_4\right)\,\mathbf{a}_2 + \frac{1}{4}\,\mathbf{a}_3$ | = | $\left(\frac{1}{2}x_4 - y_4\right)a\,\hat{\mathbf{x}} + \frac{\sqrt{3}}{2}x_4a\,\hat{\mathbf{y}} + \frac{1}{4}c\,\hat{\mathbf{z}}$ | (6*h*) | Rb III |
| $\mathbf{B}_{13}$ | = | $\left(-x_4 + y_4\right)\,\mathbf{a}_1 - x_4\,\mathbf{a}_2 + \frac{1}{4}\,\mathbf{a}_3$ | = | $\left(-x_4 + \frac{1}{2}y_4\right)a\,\hat{\mathbf{x}} - \frac{\sqrt{3}}{2}y_4a\,\hat{\mathbf{y}} + \frac{1}{4}c\,\hat{\mathbf{z}}$ | (6*h*) | Rb III |
| $\mathbf{B}_{14}$ | = | $-x_4\,\mathbf{a}_1 - y_4\,\mathbf{a}_2 + \frac{3}{4}\,\mathbf{a}_3$ | = | $-\frac{1}{2}\left(x_4 + y_4\right)a\,\hat{\mathbf{x}} + \frac{\sqrt{3}}{2}\left(x_4 - y_4\right)a\,\hat{\mathbf{y}} + \frac{3}{4}c\,\hat{\mathbf{z}}$ | (6*h*) | Rb III |
| $\mathbf{B}_{15}$ | = | $y_4\,\mathbf{a}_1 + \left(-x_4 + y_4\right)\,\mathbf{a}_2 + \frac{3}{4}\,\mathbf{a}_3$ | = | $\left(-\frac{1}{2}x_4 + y_4\right)a\,\hat{\mathbf{x}} - \frac{\sqrt{3}}{2}x_4a\,\hat{\mathbf{y}} + \frac{3}{4}c\,\hat{\mathbf{z}}$ | (6*h*) | Rb III |
| $\mathbf{B}_{16}$ | = | $\left(x_4 - y_4\right)\,\mathbf{a}_1 + x_4\,\mathbf{a}_2 + \frac{3}{4}\,\mathbf{a}_3$ | = | $\left(x_4 - \frac{1}{2}y_4\right)a\,\hat{\mathbf{x}} + \frac{\sqrt{3}}{2}y_4a\,\hat{\mathbf{y}} + \frac{3}{4}c\,\hat{\mathbf{z}}$ | (6*h*) | Rb III |
| $\mathbf{B}_{17}$ | = | $x_5\,\mathbf{a}_1 + y_5\,\mathbf{a}_2 + \frac{1}{4}\,\mathbf{a}_3$ | = | $\frac{1}{2}\left(x_5 + y_5\right)a\,\hat{\mathbf{x}} + \frac{\sqrt{3}}{2}\left(-x_5 + y_5\right)a\,\hat{\mathbf{y}} + \frac{1}{4}c\,\hat{\mathbf{z}}$ | (6*h*) | Rb IV |
| $\mathbf{B}_{18}$ | = | $-y_5\,\mathbf{a}_1 + \left(x_5 - y_5\right)\,\mathbf{a}_2 + \frac{1}{4}\,\mathbf{a}_3$ | = | $\left(\frac{1}{2}x_5 - y_5\right)a\,\hat{\mathbf{x}} + \frac{\sqrt{3}}{2}x_5a\,\hat{\mathbf{y}} + \frac{1}{4}c\,\hat{\mathbf{z}}$ | (6*h*) | Rb IV |
| $\mathbf{B}_{19}$ | = | $\left(-x_5 + y_5\right)\,\mathbf{a}_1 - x_5\,\mathbf{a}_2 + \frac{1}{4}\,\mathbf{a}_3$ | = | $\left(-x_5 + \frac{1}{2}y_5\right)a\,\hat{\mathbf{x}} - \frac{\sqrt{3}}{2}y_5a\,\hat{\mathbf{y}} + \frac{1}{4}c\,\hat{\mathbf{z}}$ | (6*h*) | Rb IV |
| $\mathbf{B}_{20}$ | = | $-x_5\,\mathbf{a}_1 - y_5\,\mathbf{a}_2 + \frac{3}{4}\,\mathbf{a}_3$ | = | $-\frac{1}{2}\left(x_5 + y_5\right)a\,\hat{\mathbf{x}} + \frac{\sqrt{3}}{2}\left(x_5 - y_5\right)a\,\hat{\mathbf{y}} + \frac{3}{4}c\,\hat{\mathbf{z}}$ | (6*h*) | Rb IV |
| $\mathbf{B}_{21}$ | = | $y_5\,\mathbf{a}_1 + \left(-x_5 + y_5\right)\,\mathbf{a}_2 + \frac{3}{4}\,\mathbf{a}_3$ | = | $\left(-\frac{1}{2}x_5 + y_5\right)a\,\hat{\mathbf{x}} - \frac{\sqrt{3}}{2}x_5a\,\hat{\mathbf{y}} + \frac{3}{4}c\,\hat{\mathbf{z}}$ | (6*h*) | Rb IV |
| $\mathbf{B}_{22}$ | = | $\left(x_5 - y_5\right)\,\mathbf{a}_1 + x_5\,\mathbf{a}_2 + \frac{3}{4}\,\mathbf{a}_3$ | = | $\left(x_5 - \frac{1}{2}y_5\right)a\,\hat{\mathbf{x}} + \frac{\sqrt{3}}{2}y_5a\,\hat{\mathbf{y}} + \frac{3}{4}c\,\hat{\mathbf{z}}$ | (6*h*) | Rb IV |
| $\mathbf{B}_{23}$ | = | $x_6\,\mathbf{a}_1 + y_6\,\mathbf{a}_2 + \frac{1}{4}\,\mathbf{a}_3$ | = | $\frac{1}{2}\left(x_6 + y_6\right)a\,\hat{\mathbf{x}} + \frac{\sqrt{3}}{2}\left(-x_6 + y_6\right)a\,\hat{\mathbf{y}} + \frac{1}{4}c\,\hat{\mathbf{z}}$ | (6*h*) | Si II |
| $\mathbf{B}_{24}$ | = | $-y_6\,\mathbf{a}_1 + \left(x_6 - y_6\right)\,\mathbf{a}_2 + \frac{1}{4}\,\mathbf{a}_3$ | = | $\left(\frac{1}{2}x_6 - y_6\right)a\,\hat{\mathbf{x}} + \frac{\sqrt{3}}{2}x_6a\,\hat{\mathbf{y}} + \frac{1}{4}c\,\hat{\mathbf{z}}$ | (6*h*) | Si II |
| $\mathbf{B}_{25}$ | = | $\left(-x_6 + y_6\right)\,\mathbf{a}_1 - x_6\,\mathbf{a}_2 + \frac{1}{4}\,\mathbf{a}_3$ | = | $\left(-x_6 + \frac{1}{2}y_6\right)a\,\hat{\mathbf{x}} - \frac{\sqrt{3}}{2}y_6a\,\hat{\mathbf{y}} + \frac{1}{4}c\,\hat{\mathbf{z}}$ | (6*h*) | Si II |
| $\mathbf{B}_{26}$ | = | $-x_6\,\mathbf{a}_1 - y_6\,\mathbf{a}_2 + \frac{3}{4}\,\mathbf{a}_3$ | = | $-\frac{1}{2}\left(x_6 + y_6\right)a\,\hat{\mathbf{x}} + \frac{\sqrt{3}}{2}\left(x_6 - y_6\right)a\,\hat{\mathbf{y}} + \frac{3}{4}c\,\hat{\mathbf{z}}$ | (6*h*) | Si II |
| $\mathbf{B}_{27}$ | = | $y_6\,\mathbf{a}_1 + \left(-x_6 + y_6\right)\,\mathbf{a}_2 + \frac{3}{4}\,\mathbf{a}_3$ | = | $\left(-\frac{1}{2}x_6 + y_6\right)a\,\hat{\mathbf{x}} - \frac{\sqrt{3}}{2}x_6a\,\hat{\mathbf{y}} + \frac{3}{4}c\,\hat{\mathbf{z}}$ | (6*h*) | Si II |
| $\mathbf{B}_{28}$ | = | $\left(x_6 - y_6\right)\,\mathbf{a}_1 + x_6\,\mathbf{a}_2 + \frac{3}{4}\,\mathbf{a}_3$ | = | $\left(x_6 - \frac{1}{2}y_6\right)a\,\hat{\mathbf{x}} + \frac{\sqrt{3}}{2}y_6a\,\hat{\mathbf{y}} + \frac{3}{4}c\,\hat{\mathbf{z}}$ | (6*h*) | Si II |
| $\mathbf{B}_{29}$ | = | $x_7\,\mathbf{a}_1 + y_7\,\mathbf{a}_2 + \frac{1}{4}\,\mathbf{a}_3$ | = | $\frac{1}{2}\left(x_7 + y_7\right)a\,\hat{\mathbf{x}} + \frac{\sqrt{3}}{2}\left(-x_7 + y_7\right)a\,\hat{\mathbf{y}} + \frac{1}{4}c\,\hat{\mathbf{z}}$ | (6*h*) | Si III |
| $\mathbf{B}_{30}$ | = | $-y_7\,\mathbf{a}_1 + \left(x_7 - y_7\right)\,\mathbf{a}_2 + \frac{1}{4}\,\mathbf{a}_3$ | = | $\left(\frac{1}{2}x_7 - y_7\right)a\,\hat{\mathbf{x}} + \frac{\sqrt{3}}{2}x_7a\,\hat{\mathbf{y}} + \frac{1}{4}c\,\hat{\mathbf{z}}$ | (6*h*) | Si III |
| $\mathbf{B}_{31}$ | = | $\left(-x_7 + y_7\right)\,\mathbf{a}_1 - x_7\,\mathbf{a}_2 + \frac{1}{4}\,\mathbf{a}_3$ | = | $\left(-x_7 + \frac{1}{2}y_7\right)a\,\hat{\mathbf{x}} - \frac{\sqrt{3}}{2}y_7a\,\hat{\mathbf{y}} + \frac{1}{4}c\,\hat{\mathbf{z}}$ | (6*h*) | Si III |
| $\mathbf{B}_{32}$ | = | $-x_7\,\mathbf{a}_1 - y_7\,\mathbf{a}_2 + \frac{3}{4}\,\mathbf{a}_3$ | = | $-\frac{1}{2}\left(x_7 + y_7\right)a\,\hat{\mathbf{x}} + \frac{\sqrt{3}}{2}\left(x_7 - y_7\right)a\,\hat{\mathbf{y}} + \frac{3}{4}c\,\hat{\mathbf{z}}$ | (6*h*) | Si III |
| $\mathbf{B}_{33}$ | = | $y_7\,\mathbf{a}_1 + \left(-x_7 + y_7\right)\,\mathbf{a}_2 + \frac{3}{4}\,\mathbf{a}_3$ | = | $\left(-\frac{1}{2}x_7 + y_7\right)a\,\hat{\mathbf{x}} - \frac{\sqrt{3}}{2}x_7a\,\hat{\mathbf{y}} + \frac{3}{4}c\,\hat{\mathbf{z}}$ | (6*h*) | Si III |
| $\mathbf{B}_{34}$ | = | $\left(x_7 - y_7\right)\,\mathbf{a}_1 + x_7\,\mathbf{a}_2 + \frac{3}{4}\,\mathbf{a}_3$ | = | $\left(x_7 - \frac{1}{2}y_7\right)a\,\hat{\mathbf{x}} + \frac{\sqrt{3}}{2}y_7a\,\hat{\mathbf{y}} + \frac{3}{4}c\,\hat{\mathbf{z}}$ | (6*h*) | Si III |

**Geometry files:**

- CIF: pp. 1751
- POSCAR: pp. 1752

# $Th_7S_{12}$ ($D8_k$) Structure: A3B2_hP20_176_2h_ah

http://aflow.org/prototype-encyclopedia/A3B2_hP20_176_2h_ah

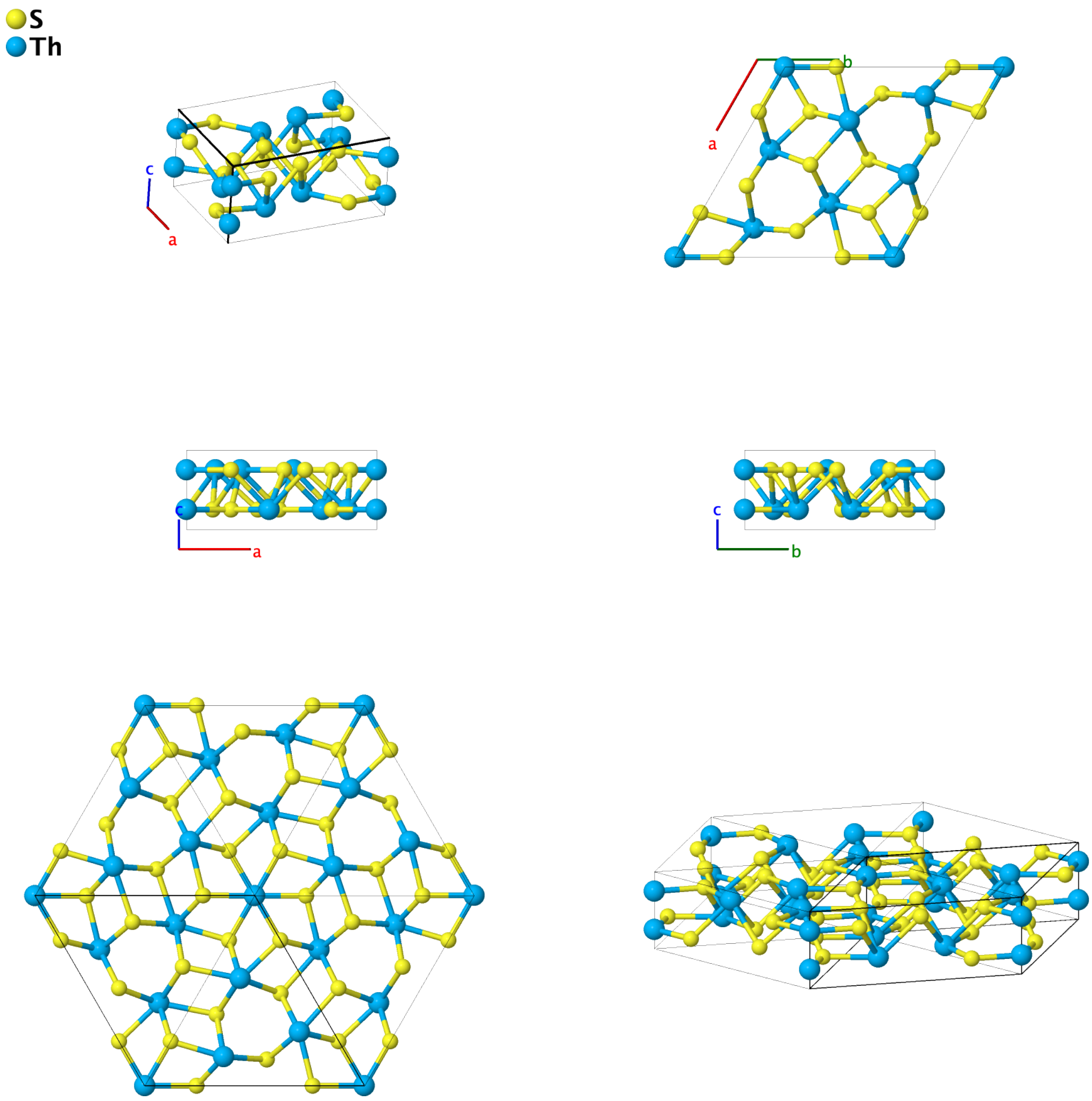

| | | |
|---|---|---|
| **Prototype** | : | $S_{12}Th_7$ |
| **AFLOW prototype label** | : | A3B2_hP20_176_2h_ah |
| ***Strukturbericht* designation** | : | $D8_k$ |
| **Pearson symbol** | : | hP20 |
| **Space group number** | : | 176 |
| **Space group symbol** | : | $P6_3/m$ |
| **AFLOW prototype command** | : | `aflow --proto=A3B2_hP20_176_2h_ah`<br>`--params=`$a, c/a, x_2, y_2, x_3, y_3, x_4, y_4$ |

**Other compounds with this structure**

- $Th_7Se_{12}$ and $(Ga,As)_7Pd_{12}$

- The Th (2*a*) site is half-filled. The stoichiometry for the AFLOW label treats the Th site as fully occupied.

## Hexagonal primitive vectors:

$$\mathbf{a}_1 = \frac{1}{2}\,a\,\hat{\mathbf{x}} - \frac{\sqrt{3}}{2}\,a\,\hat{\mathbf{y}}$$
$$\mathbf{a}_2 = \frac{1}{2}\,a\,\hat{\mathbf{x}} + \frac{\sqrt{3}}{2}\,a\,\hat{\mathbf{y}}$$
$$\mathbf{a}_3 = c\,\hat{\mathbf{z}}$$

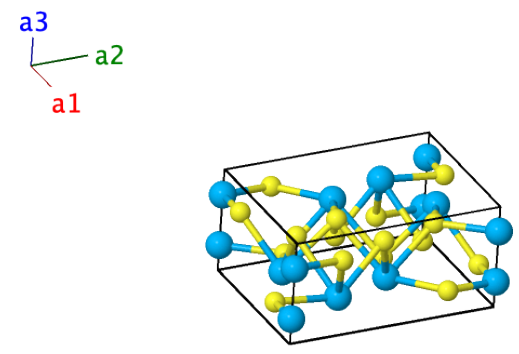

## Basis vectors:

| | | Lattice Coordinates | | Cartesian Coordinates | Wyckoff Position | Atom Type |
|---|---|---|---|---|---|---|
| $\mathbf{B}_1$ | = | $\frac{1}{4}\,\mathbf{a}_3$ | = | $\frac{1}{4}c\,\hat{\mathbf{z}}$ | (2$a$) | Th I |
| $\mathbf{B}_2$ | = | $\frac{3}{4}\,\mathbf{a}_3$ | = | $\frac{3}{4}c\,\hat{\mathbf{z}}$ | (2$a$) | Th I |
| $\mathbf{B}_3$ | = | $x_2\,\mathbf{a}_1 + y_2\,\mathbf{a}_2 + \frac{1}{4}\,\mathbf{a}_3$ | = | $\frac{1}{2}\,(x_2+y_2)\,a\,\hat{\mathbf{x}} + \frac{\sqrt{3}}{2}\,(-x_2+y_2)\,a\,\hat{\mathbf{y}} + \frac{1}{4}c\,\hat{\mathbf{z}}$ | (6$h$) | S I |
| $\mathbf{B}_4$ | = | $-y_2\,\mathbf{a}_1 + (x_2-y_2)\,\mathbf{a}_2 + \frac{1}{4}\,\mathbf{a}_3$ | = | $\left(\frac{1}{2}x_2 - y_2\right)a\,\hat{\mathbf{x}} + \frac{\sqrt{3}}{2}x_2 a\,\hat{\mathbf{y}} + \frac{1}{4}c\,\hat{\mathbf{z}}$ | (6$h$) | S I |
| $\mathbf{B}_5$ | = | $(-x_2+y_2)\,\mathbf{a}_1 - x_2\,\mathbf{a}_2 + \frac{1}{4}\,\mathbf{a}_3$ | = | $\left(-x_2 + \frac{1}{2}y_2\right)a\,\hat{\mathbf{x}} - \frac{\sqrt{3}}{2}y_2 a\,\hat{\mathbf{y}} + \frac{1}{4}c\,\hat{\mathbf{z}}$ | (6$h$) | S I |
| $\mathbf{B}_6$ | = | $-x_2\,\mathbf{a}_1 - y_2\,\mathbf{a}_2 + \frac{3}{4}\,\mathbf{a}_3$ | = | $-\frac{1}{2}\,(x_2+y_2)\,a\,\hat{\mathbf{x}} + \frac{\sqrt{3}}{2}\,(x_2-y_2)\,a\,\hat{\mathbf{y}} + \frac{3}{4}c\,\hat{\mathbf{z}}$ | (6$h$) | S I |
| $\mathbf{B}_7$ | = | $y_2\,\mathbf{a}_1 + (-x_2+y_2)\,\mathbf{a}_2 + \frac{3}{4}\,\mathbf{a}_3$ | = | $\left(-\frac{1}{2}x_2 + y_2\right)a\,\hat{\mathbf{x}} - \frac{\sqrt{3}}{2}x_2 a\,\hat{\mathbf{y}} + \frac{3}{4}c\,\hat{\mathbf{z}}$ | (6$h$) | S I |
| $\mathbf{B}_8$ | = | $(x_2-y_2)\,\mathbf{a}_1 + x_2\,\mathbf{a}_2 + \frac{3}{4}\,\mathbf{a}_3$ | = | $\left(x_2 - \frac{1}{2}y_2\right)a\,\hat{\mathbf{x}} + \frac{\sqrt{3}}{2}y_2 a\,\hat{\mathbf{y}} + \frac{3}{4}c\,\hat{\mathbf{z}}$ | (6$h$) | S I |
| $\mathbf{B}_9$ | = | $x_3\,\mathbf{a}_1 + y_3\,\mathbf{a}_2 + \frac{1}{4}\,\mathbf{a}_3$ | = | $\frac{1}{2}\,(x_3+y_3)\,a\,\hat{\mathbf{x}} + \frac{\sqrt{3}}{2}\,(-x_3+y_3)\,a\,\hat{\mathbf{y}} + \frac{1}{4}c\,\hat{\mathbf{z}}$ | (6$h$) | S II |
| $\mathbf{B}_{10}$ | = | $-y_3\,\mathbf{a}_1 + (x_3-y_3)\,\mathbf{a}_2 + \frac{1}{4}\,\mathbf{a}_3$ | = | $\left(\frac{1}{2}x_3 - y_3\right)a\,\hat{\mathbf{x}} + \frac{\sqrt{3}}{2}x_3 a\,\hat{\mathbf{y}} + \frac{1}{4}c\,\hat{\mathbf{z}}$ | (6$h$) | S II |
| $\mathbf{B}_{11}$ | = | $(-x_3+y_3)\,\mathbf{a}_1 - x_3\,\mathbf{a}_2 + \frac{1}{4}\,\mathbf{a}_3$ | = | $\left(-x_3 + \frac{1}{2}y_3\right)a\,\hat{\mathbf{x}} - \frac{\sqrt{3}}{2}y_3 a\,\hat{\mathbf{y}} + \frac{1}{4}c\,\hat{\mathbf{z}}$ | (6$h$) | S II |
| $\mathbf{B}_{12}$ | = | $-x_3\,\mathbf{a}_1 - y_3\,\mathbf{a}_2 + \frac{3}{4}\,\mathbf{a}_3$ | = | $-\frac{1}{2}\,(x_3+y_3)\,a\,\hat{\mathbf{x}} + \frac{\sqrt{3}}{2}\,(x_3-y_3)\,a\,\hat{\mathbf{y}} + \frac{3}{4}c\,\hat{\mathbf{z}}$ | (6$h$) | S II |
| $\mathbf{B}_{13}$ | = | $y_3\,\mathbf{a}_1 + (-x_3+y_3)\,\mathbf{a}_2 + \frac{3}{4}\,\mathbf{a}_3$ | = | $\left(-\frac{1}{2}x_3 + y_3\right)a\,\hat{\mathbf{x}} - \frac{\sqrt{3}}{2}x_3 a\,\hat{\mathbf{y}} + \frac{3}{4}c\,\hat{\mathbf{z}}$ | (6$h$) | S II |
| $\mathbf{B}_{14}$ | = | $(x_3-y_3)\,\mathbf{a}_1 + x_3\,\mathbf{a}_2 + \frac{3}{4}\,\mathbf{a}_3$ | = | $\left(x_3 - \frac{1}{2}y_3\right)a\,\hat{\mathbf{x}} + \frac{\sqrt{3}}{2}y_3 a\,\hat{\mathbf{y}} + \frac{3}{4}c\,\hat{\mathbf{z}}$ | (6$h$) | S II |
| $\mathbf{B}_{15}$ | = | $x_4\,\mathbf{a}_1 + y_4\,\mathbf{a}_2 + \frac{1}{4}\,\mathbf{a}_3$ | = | $\frac{1}{2}\,(x_4+y_4)\,a\,\hat{\mathbf{x}} + \frac{\sqrt{3}}{2}\,(-x_4+y_4)\,a\,\hat{\mathbf{y}} + \frac{1}{4}c\,\hat{\mathbf{z}}$ | (6$h$) | Th II |
| $\mathbf{B}_{16}$ | = | $-y_4\,\mathbf{a}_1 + (x_4-y_4)\,\mathbf{a}_2 + \frac{1}{4}\,\mathbf{a}_3$ | = | $\left(\frac{1}{2}x_4 - y_4\right)a\,\hat{\mathbf{x}} + \frac{\sqrt{3}}{2}x_4 a\,\hat{\mathbf{y}} + \frac{1}{4}c\,\hat{\mathbf{z}}$ | (6$h$) | Th II |
| $\mathbf{B}_{17}$ | = | $(-x_4+y_4)\,\mathbf{a}_1 - x_4\,\mathbf{a}_2 + \frac{1}{4}\,\mathbf{a}_3$ | = | $\left(-x_4 + \frac{1}{2}y_4\right)a\,\hat{\mathbf{x}} - \frac{\sqrt{3}}{2}y_4 a\,\hat{\mathbf{y}} + \frac{1}{4}c\,\hat{\mathbf{z}}$ | (6$h$) | Th II |
| $\mathbf{B}_{18}$ | = | $-x_4\,\mathbf{a}_1 - y_4\,\mathbf{a}_2 + \frac{3}{4}\,\mathbf{a}_3$ | = | $-\frac{1}{2}\,(x_4+y_4)\,a\,\hat{\mathbf{x}} + \frac{\sqrt{3}}{2}\,(x_4-y_4)\,a\,\hat{\mathbf{y}} + \frac{3}{4}c\,\hat{\mathbf{z}}$ | (6$h$) | Th II |
| $\mathbf{B}_{19}$ | = | $y_4\,\mathbf{a}_1 + (-x_4+y_4)\,\mathbf{a}_2 + \frac{3}{4}\,\mathbf{a}_3$ | = | $\left(-\frac{1}{2}x_4 + y_4\right)a\,\hat{\mathbf{x}} - \frac{\sqrt{3}}{2}x_4 a\,\hat{\mathbf{y}} + \frac{3}{4}c\,\hat{\mathbf{z}}$ | (6$h$) | Th II |
| $\mathbf{B}_{20}$ | = | $(x_4-y_4)\,\mathbf{a}_1 + x_4\,\mathbf{a}_2 + \frac{3}{4}\,\mathbf{a}_3$ | = | $\left(x_4 - \frac{1}{2}y_4\right)a\,\hat{\mathbf{x}} + \frac{\sqrt{3}}{2}y_4 a\,\hat{\mathbf{y}} + \frac{3}{4}c\,\hat{\mathbf{z}}$ | (6$h$) | Th II |

## Geometry files:

- CIF: pp. 1752
- POSCAR: pp. 1752

# $\beta$-$Si_3N_4$ Structure: A4B3_hP14_176_ch_h

http://aflow.org/prototype-encyclopedia/A4B3_hP14_176_ch_h

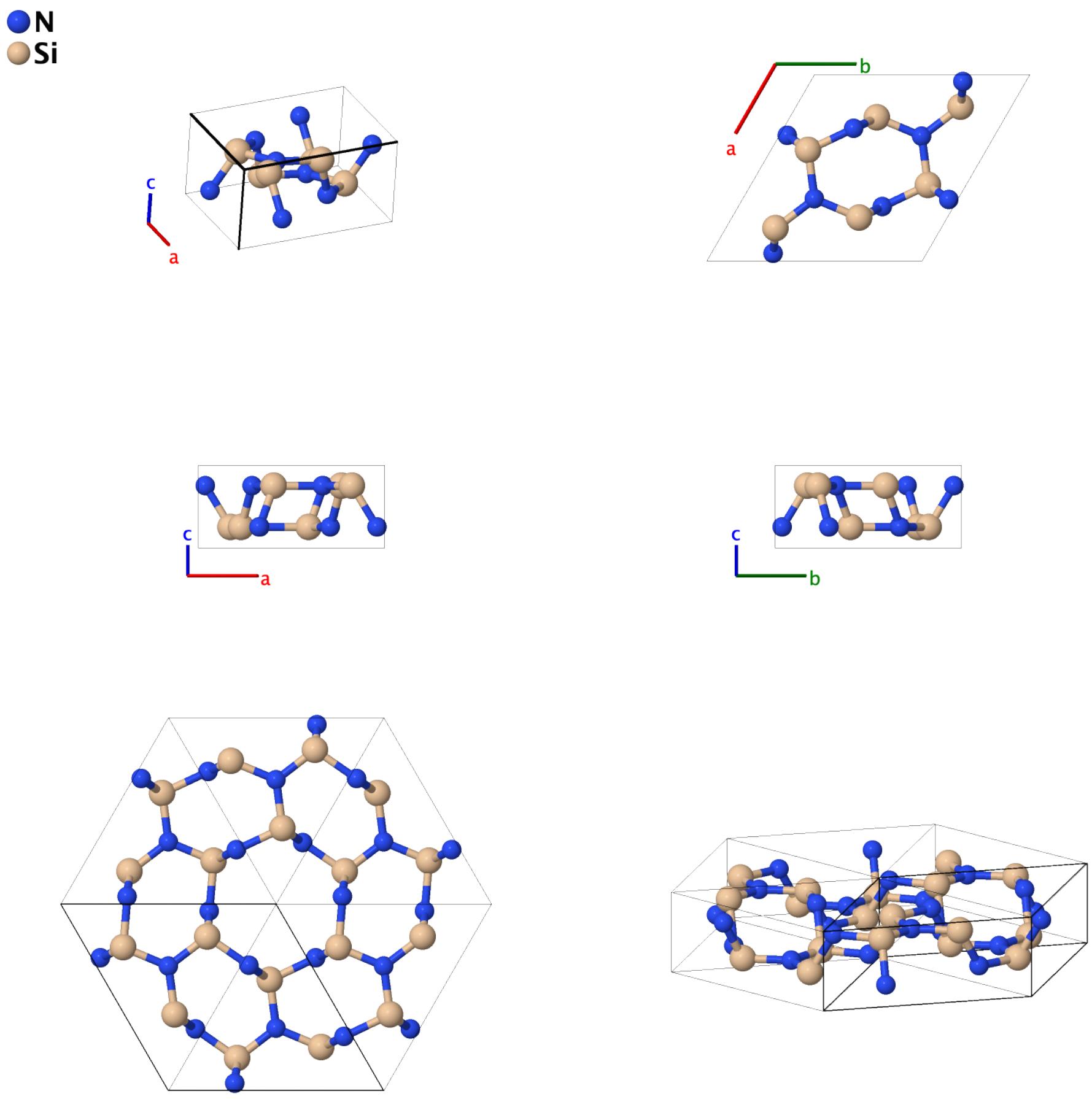

| | | |
|---|---|---|
| **Prototype** | : | $N_4Si_3$ |
| **AFLOW prototype label** | : | A4B3_hP14_176_ch_h |
| ***Strukturbericht* designation** | : | None |
| **Pearson symbol** | : | hP14 |
| **Space group number** | : | 176 |
| **Space group symbol** | : | $P6_3/m$ |
| **AFLOW prototype command** | : | `aflow --proto=A4B3_hP14_176_ch_h` `--params=`$a, c/a, x_2, y_2, x_3, y_3$ |

- (Grun, 1979) places this structure in space group $P6_3$ #173. His structure is nearly indistinguishable from this one.

**Hexagonal primitive vectors:**

$$\mathbf{a}_1 = \frac{1}{2}\,a\,\hat{\mathbf{x}} - \frac{\sqrt{3}}{2}\,a\,\hat{\mathbf{y}}$$
$$\mathbf{a}_2 = \frac{1}{2}\,a\,\hat{\mathbf{x}} + \frac{\sqrt{3}}{2}\,a\,\hat{\mathbf{y}}$$
$$\mathbf{a}_3 = c\,\hat{\mathbf{z}}$$

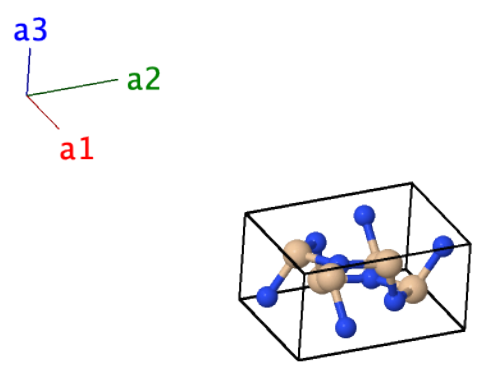

**Basis vectors:**

| | | Lattice Coordinates | | Cartesian Coordinates | Wyckoff Position | Atom Type |
|---|---|---|---|---|---|---|
| $\mathbf{B}_1$ | = | $\frac{1}{3}\,\mathbf{a}_1 + \frac{2}{3}\,\mathbf{a}_2 + \frac{1}{4}\,\mathbf{a}_3$ | = | $\frac{1}{2}a\,\hat{\mathbf{x}} + \frac{1}{2\sqrt{3}}a\,\hat{\mathbf{y}} + \frac{1}{4}c\,\hat{\mathbf{z}}$ | (2$c$) | N I |
| $\mathbf{B}_2$ | = | $\frac{2}{3}\,\mathbf{a}_1 + \frac{1}{3}\,\mathbf{a}_2 + \frac{3}{4}\,\mathbf{a}_3$ | = | $\frac{1}{2}a\,\hat{\mathbf{x}} - \frac{1}{2\sqrt{3}}a\,\hat{\mathbf{y}} + \frac{3}{4}c\,\hat{\mathbf{z}}$ | (2$c$) | N I |
| $\mathbf{B}_3$ | = | $x_2\,\mathbf{a}_1 + y_2\,\mathbf{a}_2 + \frac{1}{4}\,\mathbf{a}_3$ | = | $\frac{1}{2}\left(x_2 + y_2\right)a\,\hat{\mathbf{x}} + \frac{\sqrt{3}}{2}\left(-x_2 + y_2\right)a\,\hat{\mathbf{y}} + \frac{1}{4}c\,\hat{\mathbf{z}}$ | (6$h$) | N II |
| $\mathbf{B}_4$ | = | $-y_2\,\mathbf{a}_1 + \left(x_2 - y_2\right)\mathbf{a}_2 + \frac{1}{4}\,\mathbf{a}_3$ | = | $\left(\frac{1}{2}x_2 - y_2\right)a\,\hat{\mathbf{x}} + \frac{\sqrt{3}}{2}x_2a\,\hat{\mathbf{y}} + \frac{1}{4}c\,\hat{\mathbf{z}}$ | (6$h$) | N II |
| $\mathbf{B}_5$ | = | $\left(-x_2 + y_2\right)\mathbf{a}_1 - x_2\,\mathbf{a}_2 + \frac{1}{4}\,\mathbf{a}_3$ | = | $\left(-x_2 + \frac{1}{2}y_2\right)a\,\hat{\mathbf{x}} - \frac{\sqrt{3}}{2}y_2a\,\hat{\mathbf{y}} + \frac{1}{4}c\,\hat{\mathbf{z}}$ | (6$h$) | N II |
| $\mathbf{B}_6$ | = | $-x_2\,\mathbf{a}_1 - y_2\,\mathbf{a}_2 + \frac{3}{4}\,\mathbf{a}_3$ | = | $-\frac{1}{2}\left(x_2 + y_2\right)a\,\hat{\mathbf{x}} + \frac{\sqrt{3}}{2}\left(x_2 - y_2\right)a\,\hat{\mathbf{y}} + \frac{3}{4}c\,\hat{\mathbf{z}}$ | (6$h$) | N II |
| $\mathbf{B}_7$ | = | $y_2\,\mathbf{a}_1 + \left(-x_2 + y_2\right)\mathbf{a}_2 + \frac{3}{4}\,\mathbf{a}_3$ | = | $\left(-\frac{1}{2}x_2 + y_2\right)a\,\hat{\mathbf{x}} - \frac{\sqrt{3}}{2}x_2a\,\hat{\mathbf{y}} + \frac{3}{4}c\,\hat{\mathbf{z}}$ | (6$h$) | N II |
| $\mathbf{B}_8$ | = | $\left(x_2 - y_2\right)\mathbf{a}_1 + x_2\,\mathbf{a}_2 + \frac{3}{4}\,\mathbf{a}_3$ | = | $\left(x_2 - \frac{1}{2}y_2\right)a\,\hat{\mathbf{x}} + \frac{\sqrt{3}}{2}y_2a\,\hat{\mathbf{y}} + \frac{3}{4}c\,\hat{\mathbf{z}}$ | (6$h$) | N II |
| $\mathbf{B}_9$ | = | $x_3\,\mathbf{a}_1 + y_3\,\mathbf{a}_2 + \frac{1}{4}\,\mathbf{a}_3$ | = | $\frac{1}{2}\left(x_3 + y_3\right)a\,\hat{\mathbf{x}} + \frac{\sqrt{3}}{2}\left(-x_3 + y_3\right)a\,\hat{\mathbf{y}} + \frac{1}{4}c\,\hat{\mathbf{z}}$ | (6$h$) | Si |
| $\mathbf{B}_{10}$ | = | $-y_3\,\mathbf{a}_1 + \left(x_3 - y_3\right)\mathbf{a}_2 + \frac{1}{4}\,\mathbf{a}_3$ | = | $\left(\frac{1}{2}x_3 - y_3\right)a\,\hat{\mathbf{x}} + \frac{\sqrt{3}}{2}x_3a\,\hat{\mathbf{y}} + \frac{1}{4}c\,\hat{\mathbf{z}}$ | (6$h$) | Si |
| $\mathbf{B}_{11}$ | = | $\left(-x_3 + y_3\right)\mathbf{a}_1 - x_3\,\mathbf{a}_2 + \frac{1}{4}\,\mathbf{a}_3$ | = | $\left(-x_3 + \frac{1}{2}y_3\right)a\,\hat{\mathbf{x}} - \frac{\sqrt{3}}{2}y_3a\,\hat{\mathbf{y}} + \frac{1}{4}c\,\hat{\mathbf{z}}$ | (6$h$) | Si |
| $\mathbf{B}_{12}$ | = | $-x_3\,\mathbf{a}_1 - y_3\,\mathbf{a}_2 + \frac{3}{4}\,\mathbf{a}_3$ | = | $-\frac{1}{2}\left(x_3 + y_3\right)a\,\hat{\mathbf{x}} + \frac{\sqrt{3}}{2}\left(x_3 - y_3\right)a\,\hat{\mathbf{y}} + \frac{3}{4}c\,\hat{\mathbf{z}}$ | (6$h$) | Si |
| $\mathbf{B}_{13}$ | = | $y_3\,\mathbf{a}_1 + \left(-x_3 + y_3\right)\mathbf{a}_2 + \frac{3}{4}\,\mathbf{a}_3$ | = | $\left(-\frac{1}{2}x_3 + y_3\right)a\,\hat{\mathbf{x}} - \frac{\sqrt{3}}{2}x_3a\,\hat{\mathbf{y}} + \frac{3}{4}c\,\hat{\mathbf{z}}$ | (6$h$) | Si |
| $\mathbf{B}_{14}$ | = | $\left(x_3 - y_3\right)\mathbf{a}_1 + x_3\,\mathbf{a}_2 + \frac{3}{4}\,\mathbf{a}_3$ | = | $\left(x_3 - \frac{1}{2}y_3\right)a\,\hat{\mathbf{x}} + \frac{\sqrt{3}}{2}y_3a\,\hat{\mathbf{y}} + \frac{3}{4}c\,\hat{\mathbf{z}}$ | (6$h$) | Si |

**Found in:**

**Geometry files:**

- CIF: pp. 1752
- POSCAR: pp. 1753

# Fluorapatite [$Ca_5F(PO_4)_3$, $H5_7$] Structure: A5BC12D3_hP42_176_fh_a_2hi_h

http://aflow.org/prototype-encyclopedia/A5BC12D3_hP42_176_fh_a_2hi_h

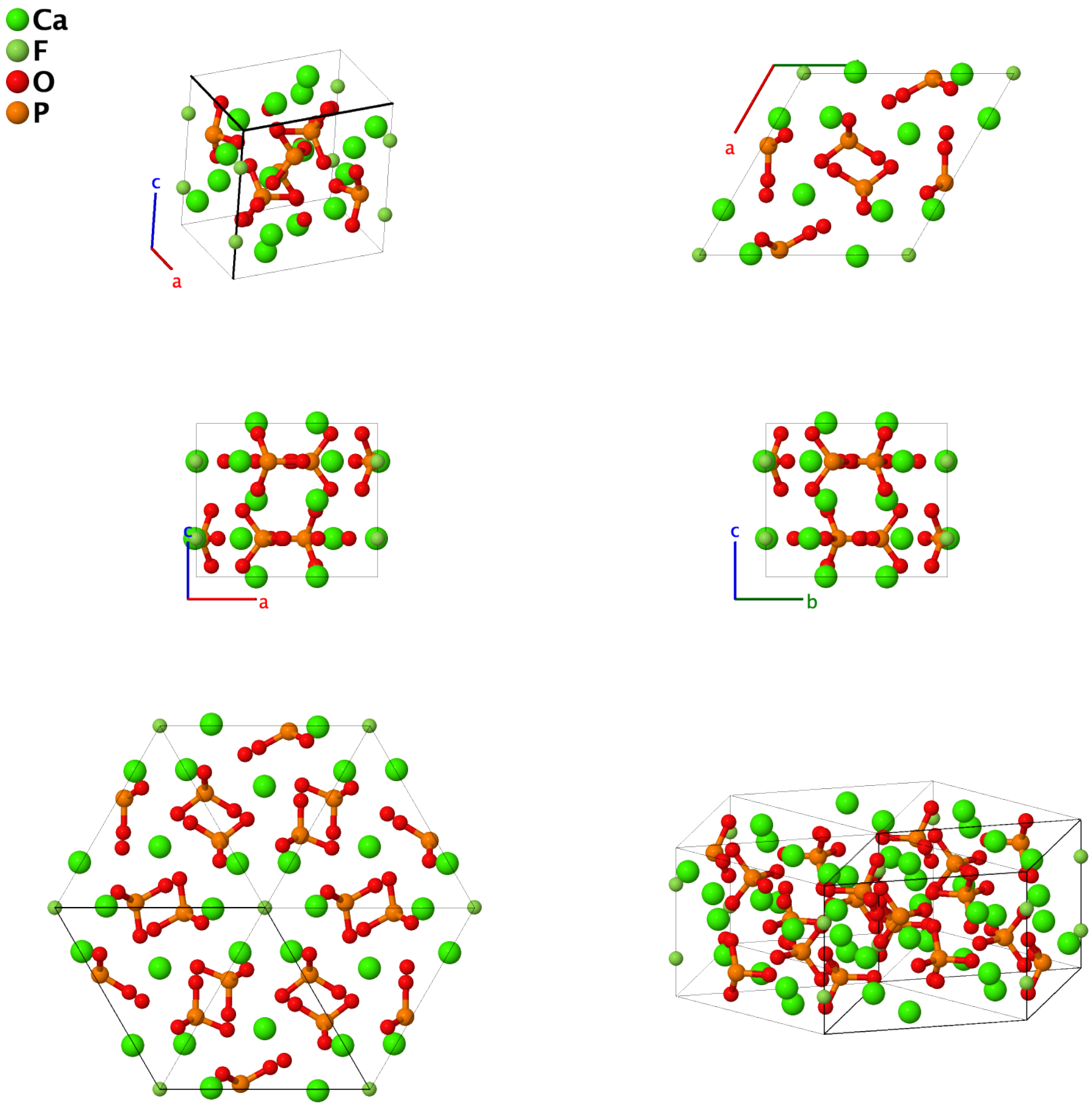

| | | |
|---|---|---|
| **Prototype** | : | $Ca_5FO_{12}P_3$ |
| **AFLOW prototype label** | : | A5BC12D3_hP42_176_fh_a_2hi_h |
| ***Strukturbericht* designation** | : | $H5_7$ |
| **Pearson symbol** | : | hP42 |
| **Space group number** | : | 176 |
| **Space group symbol** | : | $P6_3/m$ |
| **AFLOW prototype command** | : | `aflow --proto=A5BC12D3_hP42_176_fh_a_2hi_h` `--params=`$a, c/a, z_2, x_3, y_3, x_4, y_4, x_5, y_5, x_6, y_6, x_7, y_7, z_7$ |

**Other compounds with this structure**

- $Ca_5OH(PO_4)_3$ (hydroxylapatite) and $Ca_5Cl(PO_4)_3$ (chlorapatite)

- Apatite can be formed with most $M^{2+}$ metallic ions replacing the calcium, and many ions ($AsO_3$, $CO_3$, $Si_3$, *etc.*) replacing the phosphate. While these structures are related to the prototype, they may have slight changes in crystal structure. The phosphate apatites are the main source of phosphorus on Earth. (Hughes, 2002)
- When OH or Cl replaces F, that ion is displaced from the (1*a*) position to the (4*e*) position, with $z = 0.1979$ for OH and 0.4323 for Cl. The substitute ion fills half of the (4*f*) sites.

## Hexagonal primitive vectors:

$$
\begin{aligned}
\mathbf{a}_1 &= \tfrac{1}{2}\,a\,\hat{\mathbf{x}} - \tfrac{\sqrt{3}}{2}\,a\,\hat{\mathbf{y}} \\
\mathbf{a}_2 &= \tfrac{1}{2}\,a\,\hat{\mathbf{x}} + \tfrac{\sqrt{3}}{2}\,a\,\hat{\mathbf{y}} \\
\mathbf{a}_3 &= c\,\hat{\mathbf{z}}
\end{aligned}
$$

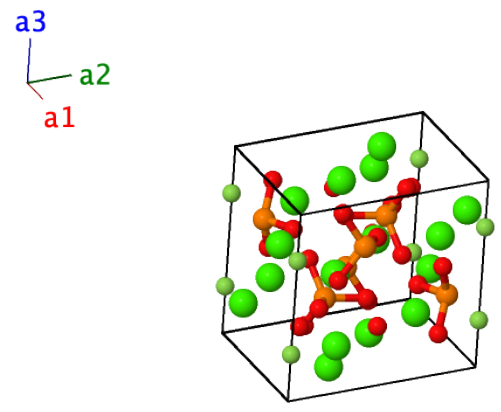

## Basis vectors:

| | | Lattice Coordinates | | Cartesian Coordinates | Wyckoff Position | Atom Type |
|---|---|---|---|---|---|---|
| $\mathbf{B}_1$ | = | $\frac{1}{4}\,\mathbf{a}_3$ | = | $\frac{1}{4}c\,\hat{\mathbf{z}}$ | $(2a)$ | F |
| $\mathbf{B}_2$ | = | $\frac{3}{4}\,\mathbf{a}_3$ | = | $\frac{3}{4}c\,\hat{\mathbf{z}}$ | $(2a)$ | F |
| $\mathbf{B}_3$ | = | $\frac{1}{3}\,\mathbf{a}_1 + \frac{2}{3}\,\mathbf{a}_2 + z_2\,\mathbf{a}_3$ | = | $\frac{1}{2}a\,\hat{\mathbf{x}} + \frac{1}{2\sqrt{3}}a\,\hat{\mathbf{y}} + z_2c\,\hat{\mathbf{z}}$ | $(4f)$ | Ca I |
| $\mathbf{B}_4$ | = | $\frac{2}{3}\,\mathbf{a}_1 + \frac{1}{3}\,\mathbf{a}_2 + \left(\frac{1}{2} + z_2\right)\,\mathbf{a}_3$ | = | $\frac{1}{2}a\,\hat{\mathbf{x}} - \frac{1}{2\sqrt{3}}a\,\hat{\mathbf{y}} + \left(\frac{1}{2} + z_2\right)c\,\hat{\mathbf{z}}$ | $(4f)$ | Ca I |
| $\mathbf{B}_5$ | = | $\frac{2}{3}\,\mathbf{a}_1 + \frac{1}{3}\,\mathbf{a}_2 - z_2\,\mathbf{a}_3$ | = | $\frac{1}{2}a\,\hat{\mathbf{x}} - \frac{1}{2\sqrt{3}}a\,\hat{\mathbf{y}} - z_2c\,\hat{\mathbf{z}}$ | $(4f)$ | Ca I |
| $\mathbf{B}_6$ | = | $\frac{1}{3}\,\mathbf{a}_1 + \frac{2}{3}\,\mathbf{a}_2 + \left(\frac{1}{2} - z_2\right)\,\mathbf{a}_3$ | = | $\frac{1}{2}a\,\hat{\mathbf{x}} + \frac{1}{2\sqrt{3}}a\,\hat{\mathbf{y}} + \left(\frac{1}{2} - z_2\right)c\,\hat{\mathbf{z}}$ | $(4f)$ | Ca I |
| $\mathbf{B}_7$ | = | $x_3\,\mathbf{a}_1 + y_3\,\mathbf{a}_2 + \frac{1}{4}\,\mathbf{a}_3$ | = | $\frac{1}{2}\left(x_3 + y_3\right)a\,\hat{\mathbf{x}} + \frac{\sqrt{3}}{2}\left(-x_3 + y_3\right)a\,\hat{\mathbf{y}} + \frac{1}{4}c\,\hat{\mathbf{z}}$ | $(6h)$ | Ca II |
| $\mathbf{B}_8$ | = | $-y_3\,\mathbf{a}_1 + \left(x_3 - y_3\right)\,\mathbf{a}_2 + \frac{1}{4}\,\mathbf{a}_3$ | = | $\left(\frac{1}{2}x_3 - y_3\right)a\,\hat{\mathbf{x}} + \frac{\sqrt{3}}{2}x_3a\,\hat{\mathbf{y}} + \frac{1}{4}c\,\hat{\mathbf{z}}$ | $(6h)$ | Ca II |
| $\mathbf{B}_9$ | = | $\left(-x_3 + y_3\right)\,\mathbf{a}_1 - x_3\,\mathbf{a}_2 + \frac{1}{4}\,\mathbf{a}_3$ | = | $\left(-x_3 + \frac{1}{2}y_3\right)a\,\hat{\mathbf{x}} - \frac{\sqrt{3}}{2}y_3a\,\hat{\mathbf{y}} + \frac{1}{4}c\,\hat{\mathbf{z}}$ | $(6h)$ | Ca II |
| $\mathbf{B}_{10}$ | = | $-x_3\,\mathbf{a}_1 - y_3\,\mathbf{a}_2 + \frac{3}{4}\,\mathbf{a}_3$ | = | $-\frac{1}{2}\left(x_3 + y_3\right)a\,\hat{\mathbf{x}} + \frac{\sqrt{3}}{2}\left(x_3 - y_3\right)a\,\hat{\mathbf{y}} + \frac{3}{4}c\,\hat{\mathbf{z}}$ | $(6h)$ | Ca II |
| $\mathbf{B}_{11}$ | = | $y_3\,\mathbf{a}_1 + \left(-x_3 + y_3\right)\,\mathbf{a}_2 + \frac{3}{4}\,\mathbf{a}_3$ | = | $\left(-\frac{1}{2}x_3 + y_3\right)a\,\hat{\mathbf{x}} - \frac{\sqrt{3}}{2}x_3a\,\hat{\mathbf{y}} + \frac{3}{4}c\,\hat{\mathbf{z}}$ | $(6h)$ | Ca II |
| $\mathbf{B}_{12}$ | = | $\left(x_3 - y_3\right)\,\mathbf{a}_1 + x_3\,\mathbf{a}_2 + \frac{3}{4}\,\mathbf{a}_3$ | = | $\left(x_3 - \frac{1}{2}y_3\right)a\,\hat{\mathbf{x}} + \frac{\sqrt{3}}{2}y_3a\,\hat{\mathbf{y}} + \frac{3}{4}c\,\hat{\mathbf{z}}$ | $(6h)$ | Ca II |
| $\mathbf{B}_{13}$ | = | $x_4\,\mathbf{a}_1 + y_4\,\mathbf{a}_2 + \frac{1}{4}\,\mathbf{a}_3$ | = | $\frac{1}{2}\left(x_4 + y_4\right)a\,\hat{\mathbf{x}} + \frac{\sqrt{3}}{2}\left(-x_4 + y_4\right)a\,\hat{\mathbf{y}} + \frac{1}{4}c\,\hat{\mathbf{z}}$ | $(6h)$ | O I |
| $\mathbf{B}_{14}$ | = | $-y_4\,\mathbf{a}_1 + \left(x_4 - y_4\right)\,\mathbf{a}_2 + \frac{1}{4}\,\mathbf{a}_3$ | = | $\left(\frac{1}{2}x_4 - y_4\right)a\,\hat{\mathbf{x}} + \frac{\sqrt{3}}{2}x_4a\,\hat{\mathbf{y}} + \frac{1}{4}c\,\hat{\mathbf{z}}$ | $(6h)$ | O I |
| $\mathbf{B}_{15}$ | = | $\left(-x_4 + y_4\right)\,\mathbf{a}_1 - x_4\,\mathbf{a}_2 + \frac{1}{4}\,\mathbf{a}_3$ | = | $\left(-x_4 + \frac{1}{2}y_4\right)a\,\hat{\mathbf{x}} - \frac{\sqrt{3}}{2}y_4a\,\hat{\mathbf{y}} + \frac{1}{4}c\,\hat{\mathbf{z}}$ | $(6h)$ | O I |
| $\mathbf{B}_{16}$ | = | $-x_4\,\mathbf{a}_1 - y_4\,\mathbf{a}_2 + \frac{3}{4}\,\mathbf{a}_3$ | = | $-\frac{1}{2}\left(x_4 + y_4\right)a\,\hat{\mathbf{x}} + \frac{\sqrt{3}}{2}\left(x_4 - y_4\right)a\,\hat{\mathbf{y}} + \frac{3}{4}c\,\hat{\mathbf{z}}$ | $(6h)$ | O I |
| $\mathbf{B}_{17}$ | = | $y_4\,\mathbf{a}_1 + \left(-x_4 + y_4\right)\,\mathbf{a}_2 + \frac{3}{4}\,\mathbf{a}_3$ | = | $\left(-\frac{1}{2}x_4 + y_4\right)a\,\hat{\mathbf{x}} - \frac{\sqrt{3}}{2}x_4a\,\hat{\mathbf{y}} + \frac{3}{4}c\,\hat{\mathbf{z}}$ | $(6h)$ | O I |
| $\mathbf{B}_{18}$ | = | $\left(x_4 - y_4\right)\,\mathbf{a}_1 + x_4\,\mathbf{a}_2 + \frac{3}{4}\,\mathbf{a}_3$ | = | $\left(x_4 - \frac{1}{2}y_4\right)a\,\hat{\mathbf{x}} + \frac{\sqrt{3}}{2}y_4a\,\hat{\mathbf{y}} + \frac{3}{4}c\,\hat{\mathbf{z}}$ | $(6h)$ | O I |
| $\mathbf{B}_{19}$ | = | $x_5\,\mathbf{a}_1 + y_5\,\mathbf{a}_2 + \frac{1}{4}\,\mathbf{a}_3$ | = | $\frac{1}{2}\left(x_5 + y_5\right)a\,\hat{\mathbf{x}} + \frac{\sqrt{3}}{2}\left(-x_5 + y_5\right)a\,\hat{\mathbf{y}} + \frac{1}{4}c\,\hat{\mathbf{z}}$ | $(6h)$ | O II |
| $\mathbf{B}_{20}$ | = | $-y_5\,\mathbf{a}_1 + \left(x_5 - y_5\right)\,\mathbf{a}_2 + \frac{1}{4}\,\mathbf{a}_3$ | = | $\left(\frac{1}{2}x_5 - y_5\right)a\,\hat{\mathbf{x}} + \frac{\sqrt{3}}{2}x_5a\,\hat{\mathbf{y}} + \frac{1}{4}c\,\hat{\mathbf{z}}$ | $(6h)$ | O II |
| $\mathbf{B}_{21}$ | = | $\left(-x_5 + y_5\right)\,\mathbf{a}_1 - x_5\,\mathbf{a}_2 + \frac{1}{4}\,\mathbf{a}_3$ | = | $\left(-x_5 + \frac{1}{2}y_5\right)a\,\hat{\mathbf{x}} - \frac{\sqrt{3}}{2}y_5a\,\hat{\mathbf{y}} + \frac{1}{4}c\,\hat{\mathbf{z}}$ | $(6h)$ | O II |
| $\mathbf{B}_{22}$ | = | $-x_5\,\mathbf{a}_1 - y_5\,\mathbf{a}_2 + \frac{3}{4}\,\mathbf{a}_3$ | = | $-\frac{1}{2}\left(x_5 + y_5\right)a\,\hat{\mathbf{x}} + \frac{\sqrt{3}}{2}\left(x_5 - y_5\right)a\,\hat{\mathbf{y}} + \frac{3}{4}c\,\hat{\mathbf{z}}$ | $(6h)$ | O II |
| $\mathbf{B}_{23}$ | = | $y_5\,\mathbf{a}_1 + \left(-x_5 + y_5\right)\,\mathbf{a}_2 + \frac{3}{4}\,\mathbf{a}_3$ | = | $\left(-\frac{1}{2}x_5 + y_5\right)a\,\hat{\mathbf{x}} - \frac{\sqrt{3}}{2}x_5a\,\hat{\mathbf{y}} + \frac{3}{4}c\,\hat{\mathbf{z}}$ | $(6h)$ | O II |
| $\mathbf{B}_{24}$ | = | $\left(x_5 - y_5\right)\,\mathbf{a}_1 + x_5\,\mathbf{a}_2 + \frac{3}{4}\,\mathbf{a}_3$ | = | $\left(x_5 - \frac{1}{2}y_5\right)a\,\hat{\mathbf{x}} + \frac{\sqrt{3}}{2}y_5a\,\hat{\mathbf{y}} + \frac{3}{4}c\,\hat{\mathbf{z}}$ | $(6h)$ | O II |
| $\mathbf{B}_{25}$ | = | $x_6\,\mathbf{a}_1 + y_6\,\mathbf{a}_2 + \frac{1}{4}\,\mathbf{a}_3$ | = | $\frac{1}{2}\left(x_6 + y_6\right)a\,\mathbf{\hat{x}} + \frac{\sqrt{3}}{2}\left(-x_6 + y_6\right)a\,\mathbf{\hat{y}} + \frac{1}{4}c\,\mathbf{\hat{z}}$ | (6$h$) | P |
| $\mathbf{B}_{26}$ | = | $-y_6\,\mathbf{a}_1 + \left(x_6 - y_6\right)\,\mathbf{a}_2 + \frac{1}{4}\,\mathbf{a}_3$ | = | $\left(\frac{1}{2}x_6 - y_6\right)a\,\mathbf{\hat{x}} + \frac{\sqrt{3}}{2}x_6a\,\mathbf{\hat{y}} + \frac{1}{4}c\,\mathbf{\hat{z}}$ | (6$h$) | P |
| $\mathbf{B}_{27}$ | = | $\left(-x_6 + y_6\right)\,\mathbf{a}_1 - x_6\,\mathbf{a}_2 + \frac{1}{4}\,\mathbf{a}_3$ | = | $\left(-x_6 + \frac{1}{2}y_6\right)a\,\mathbf{\hat{x}} - \frac{\sqrt{3}}{2}y_6a\,\mathbf{\hat{y}} + \frac{1}{4}c\,\mathbf{\hat{z}}$ | (6$h$) | P |
| $\mathbf{B}_{28}$ | = | $-x_6\,\mathbf{a}_1 - y_6\,\mathbf{a}_2 + \frac{3}{4}\,\mathbf{a}_3$ | = | $-\frac{1}{2}\left(x_6 + y_6\right)a\,\mathbf{\hat{x}} + \frac{\sqrt{3}}{2}\left(x_6 - y_6\right)a\,\mathbf{\hat{y}} + \frac{3}{4}c\,\mathbf{\hat{z}}$ | (6$h$) | P |
| $\mathbf{B}_{29}$ | = | $y_6\,\mathbf{a}_1 + \left(-x_6 + y_6\right)\,\mathbf{a}_2 + \frac{3}{4}\,\mathbf{a}_3$ | = | $\left(-\frac{1}{2}x_6 + y_6\right)a\,\mathbf{\hat{x}} - \frac{\sqrt{3}}{2}x_6a\,\mathbf{\hat{y}} + \frac{3}{4}c\,\mathbf{\hat{z}}$ | (6$h$) | P |
| $\mathbf{B}_{30}$ | = | $\left(x_6 - y_6\right)\,\mathbf{a}_1 + x_6\,\mathbf{a}_2 + \frac{3}{4}\,\mathbf{a}_3$ | = | $\left(x_6 - \frac{1}{2}y_6\right)a\,\mathbf{\hat{x}} + \frac{\sqrt{3}}{2}y_6a\,\mathbf{\hat{y}} + \frac{3}{4}c\,\mathbf{\hat{z}}$ | (6$h$) | P |
| $\mathbf{B}_{31}$ | = | $x_7\,\mathbf{a}_1 + y_7\,\mathbf{a}_2 + z_7\,\mathbf{a}_3$ | = | $\frac{1}{2}\left(x_7 + y_7\right)a\,\mathbf{\hat{x}} + \frac{\sqrt{3}}{2}\left(-x_7 + y_7\right)a\,\mathbf{\hat{y}} + z_7c\,\mathbf{\hat{z}}$ | (12$i$) | O III |
| $\mathbf{B}_{32}$ | = | $-y_7\,\mathbf{a}_1 + \left(x_7 - y_7\right)\,\mathbf{a}_2 + z_7\,\mathbf{a}_3$ | = | $\left(\frac{1}{2}x_7 - y_7\right)a\,\mathbf{\hat{x}} + \frac{\sqrt{3}}{2}x_7a\,\mathbf{\hat{y}} + z_7c\,\mathbf{\hat{z}}$ | (12$i$) | O III |
| $\mathbf{B}_{33}$ | = | $\left(-x_7 + y_7\right)\,\mathbf{a}_1 - x_7\,\mathbf{a}_2 + z_7\,\mathbf{a}_3$ | = | $\left(-x_7 + \frac{1}{2}y_7\right)a\,\mathbf{\hat{x}} - \frac{\sqrt{3}}{2}y_7a\,\mathbf{\hat{y}} + z_7c\,\mathbf{\hat{z}}$ | (12$i$) | O III |
| $\mathbf{B}_{34}$ | = | $-x_7\,\mathbf{a}_1 - y_7\,\mathbf{a}_2 + \left(\frac{1}{2} + z_7\right)\,\mathbf{a}_3$ | = | $-\frac{1}{2}\left(x_7 + y_7\right)a\,\mathbf{\hat{x}} + \frac{\sqrt{3}}{2}\left(x_7 - y_7\right)a\,\mathbf{\hat{y}} + \left(\frac{1}{2} + z_7\right)c\,\mathbf{\hat{z}}$ | (12$i$) | O III |
| $\mathbf{B}_{35}$ | = | $y_7\,\mathbf{a}_1 + \left(-x_7 + y_7\right)\,\mathbf{a}_2 + \left(\frac{1}{2} + z_7\right)\,\mathbf{a}_3$ | = | $\left(-\frac{1}{2}x_7 + y_7\right)a\,\mathbf{\hat{x}} - \frac{\sqrt{3}}{2}x_7a\,\mathbf{\hat{y}} + \left(\frac{1}{2} + z_7\right)c\,\mathbf{\hat{z}}$ | (12$i$) | O III |
| $\mathbf{B}_{36}$ | = | $\left(x_7 - y_7\right)\,\mathbf{a}_1 + x_7\,\mathbf{a}_2 + \left(\frac{1}{2} + z_7\right)\,\mathbf{a}_3$ | = | $\left(x_7 - \frac{1}{2}y_7\right)a\,\mathbf{\hat{x}} + \frac{\sqrt{3}}{2}y_7a\,\mathbf{\hat{y}} + \left(\frac{1}{2} + z_7\right)c\,\mathbf{\hat{z}}$ | (12$i$) | O III |
| $\mathbf{B}_{37}$ | = | $-x_7\,\mathbf{a}_1 - y_7\,\mathbf{a}_2 - z_7\,\mathbf{a}_3$ | = | $-\frac{1}{2}\left(x_7 + y_7\right)a\,\mathbf{\hat{x}} + \frac{\sqrt{3}}{2}\left(x_7 - y_7\right)a\,\mathbf{\hat{y}} - z_7c\,\mathbf{\hat{z}}$ | (12$i$) | O III |
| $\mathbf{B}_{38}$ | = | $y_7\,\mathbf{a}_1 + \left(-x_7 + y_7\right)\,\mathbf{a}_2 - z_7\,\mathbf{a}_3$ | = | $\left(-\frac{1}{2}x_7 + y_7\right)a\,\mathbf{\hat{x}} - \frac{\sqrt{3}}{2}x_7a\,\mathbf{\hat{y}} - z_7c\,\mathbf{\hat{z}}$ | (12$i$) | O III |
| $\mathbf{B}_{39}$ | = | $\left(x_7 - y_7\right)\,\mathbf{a}_1 + x_7\,\mathbf{a}_2 - z_7\,\mathbf{a}_3$ | = | $\left(x_7 - \frac{1}{2}y_7\right)a\,\mathbf{\hat{x}} + \frac{\sqrt{3}}{2}y_7a\,\mathbf{\hat{y}} - z_7c\,\mathbf{\hat{z}}$ | (12$i$) | O III |
| $\mathbf{B}_{40}$ | = | $x_7\,\mathbf{a}_1 + y_7\,\mathbf{a}_2 + \left(\frac{1}{2} - z_7\right)\,\mathbf{a}_3$ | = | $\frac{1}{2}\left(x_7 + y_7\right)a\,\mathbf{\hat{x}} + \frac{\sqrt{3}}{2}\left(-x_7 + y_7\right)a\,\mathbf{\hat{y}} + \left(\frac{1}{2} - z_7\right)c\,\mathbf{\hat{z}}$ | (12$i$) | O III |
| $\mathbf{B}_{41}$ | = | $-y_7\,\mathbf{a}_1 + \left(x_7 - y_7\right)\,\mathbf{a}_2 + \left(\frac{1}{2} - z_7\right)\,\mathbf{a}_3$ | = | $\left(\frac{1}{2}x_7 - y_7\right)a\,\mathbf{\hat{x}} + \frac{\sqrt{3}}{2}x_7a\,\mathbf{\hat{y}} + \left(\frac{1}{2} - z_7\right)c\,\mathbf{\hat{z}}$ | (12$i$) | O III |
| $\mathbf{B}_{42}$ | = | $\left(-x_7 + y_7\right)\,\mathbf{a}_1 - x_7\,\mathbf{a}_2 + \left(\frac{1}{2} - z_7\right)\,\mathbf{a}_3$ | = | $\left(-x_7 + \frac{1}{2}y_7\right)a\,\mathbf{\hat{x}} - \frac{\sqrt{3}}{2}y_7a\,\mathbf{\hat{y}} + \left(\frac{1}{2} - z_7\right)c\,\mathbf{\hat{z}}$ | (12$i$) | O III |

**Geometry files:**

- CIF: pp. 1753
- POSCAR: pp. 1753

# $Fe_2(CO)_9$ ($F4_1$) Structure: A9B2C9_hP40_176_hi_f_hi

http://aflow.org/prototype-encyclopedia/A9B2C9_hP40_176_hi_f_hi

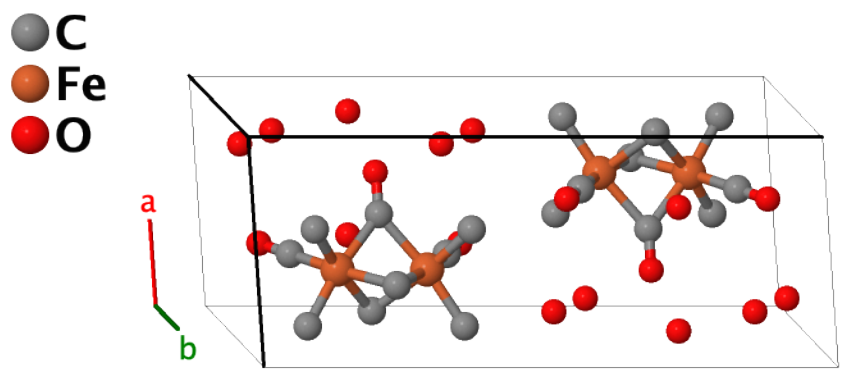

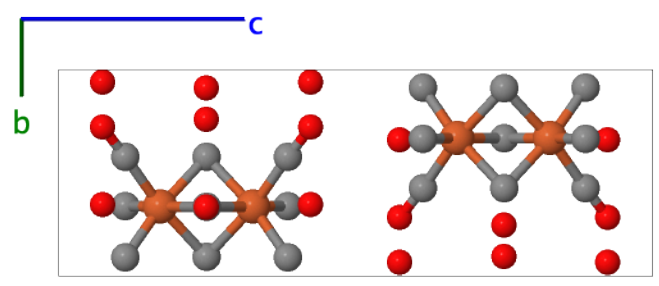

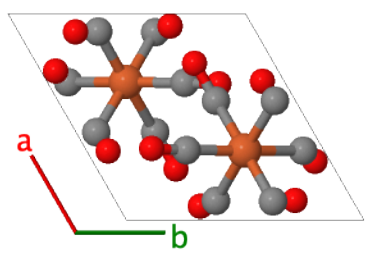

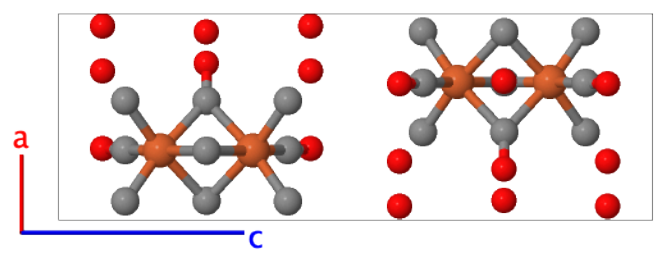

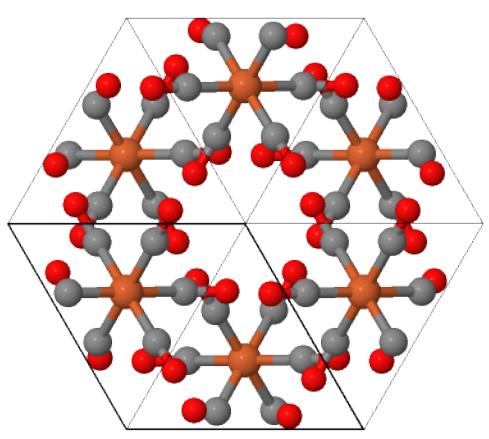

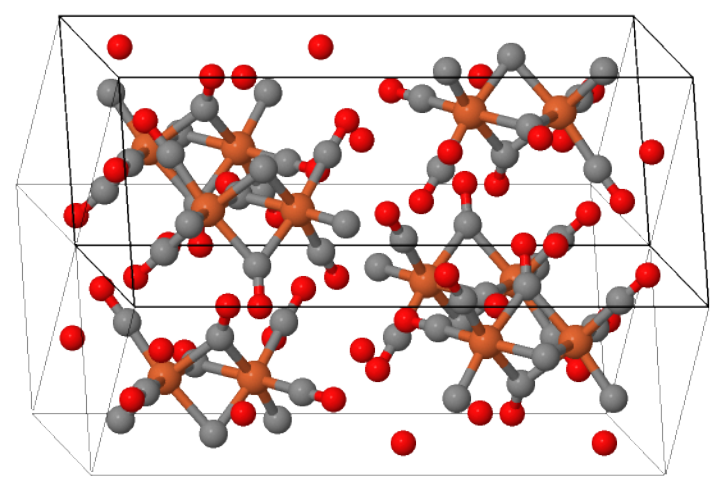

| | | |
|---|---|---|
| **Prototype** | : | $C_9Fe_2O_9$ |
| **AFLOW prototype label** | : | A9B2C9_hP40_176_hi_f_hi |
| ***Strukturbericht* designation** | : | $F4_1$ |
| **Pearson symbol** | : | hP40 |
| **Space group number** | : | 176 |
| **Space group symbol** | : | $P6_3/m$ |
| **AFLOW prototype command** | : | `aflow --proto=A9B2C9_hP40_176_hi_f_hi --params=`$a, c/a, z_1, x_2, y_2, x_3, y_3, x_4, y_4, z_4, x_5, y_5, z_5$ |

---

**Hexagonal primitive vectors:**

$$\mathbf{a}_1 = \frac{1}{2}\,a\,\hat{\mathbf{x}} - \frac{\sqrt{3}}{2}\,a\,\hat{\mathbf{y}}$$
$$\mathbf{a}_2 = \frac{1}{2}\,a\,\hat{\mathbf{x}} + \frac{\sqrt{3}}{2}\,a\,\hat{\mathbf{y}}$$
$$\mathbf{a}_3 = c\,\hat{\mathbf{z}}$$

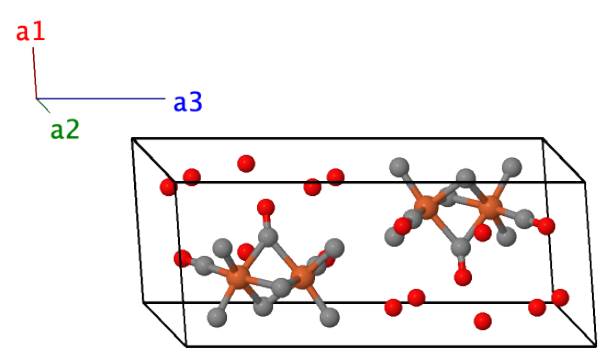

---

**Basis vectors:**

| | | Lattice Coordinates | | Cartesian Coordinates | Wyckoff Position | Atom Type |
|---|---|---|---|---|---|---|
| $\mathbf{B}_1$ | = | $\frac{1}{3}\,\mathbf{a}_1 + \frac{2}{3}\,\mathbf{a}_2 + z_1\,\mathbf{a}_3$ | = | $\frac{1}{2}a\,\hat{\mathbf{x}} + \frac{1}{2\sqrt{3}}a\,\hat{\mathbf{y}} + z_1 c\,\hat{\mathbf{z}}$ | $(4f)$ | Fe |
| $\mathbf{B}_2$ | = | $\frac{2}{3}\,\mathbf{a}_1 + \frac{1}{3}\,\mathbf{a}_2 + \left(\frac{1}{2} + z_1\right)\mathbf{a}_3$ | = | $\frac{1}{2}a\,\hat{\mathbf{x}} - \frac{1}{2\sqrt{3}}a\,\hat{\mathbf{y}} + \left(\frac{1}{2} + z_1\right)c\,\hat{\mathbf{z}}$ | $(4f)$ | Fe |
| $\mathbf{B}_3$ | = | $\frac{2}{3}\,\mathbf{a}_1 + \frac{1}{3}\,\mathbf{a}_2 - z_1\,\mathbf{a}_3$ | = | $\frac{1}{2}a\,\hat{\mathbf{x}} - \frac{1}{2\sqrt{3}}a\,\hat{\mathbf{y}} - z_1 c\,\hat{\mathbf{z}}$ | $(4f)$ | Fe |
| $\mathbf{B}_{4}$ | = | $\frac{1}{3}\,\mathbf{a}_{1} + \frac{2}{3}\,\mathbf{a}_{2} + \left(\frac{1}{2} - z_{1}\right)\,\mathbf{a}_{3}$ | = | $\frac{1}{2}a\,\hat{\mathbf{x}} + \frac{1}{2\sqrt{3}}a\,\hat{\mathbf{y}} + \left(\frac{1}{2} - z_{1}\right)c\,\hat{\mathbf{z}}$ | $(4f)$ | Fe |
| $\mathbf{B}_{5}$ | = | $x_{2}\,\mathbf{a}_{1} + y_{2}\,\mathbf{a}_{2} + \frac{1}{4}\,\mathbf{a}_{3}$ | = | $\frac{1}{2}\left(x_{2} + y_{2}\right)a\,\hat{\mathbf{x}} + \frac{\sqrt{3}}{2}\left(-x_{2} + y_{2}\right)a\,\hat{\mathbf{y}} + \frac{1}{4}c\,\hat{\mathbf{z}}$ | $(6h)$ | C I |
| $\mathbf{B}_{6}$ | = | $-y_{2}\,\mathbf{a}_{1} + \left(x_{2} - y_{2}\right)\,\mathbf{a}_{2} + \frac{1}{4}\,\mathbf{a}_{3}$ | = | $\left(\frac{1}{2}x_{2} - y_{2}\right)a\,\hat{\mathbf{x}} + \frac{\sqrt{3}}{2}x_{2}a\,\hat{\mathbf{y}} + \frac{1}{4}c\,\hat{\mathbf{z}}$ | $(6h)$ | C I |
| $\mathbf{B}_{7}$ | = | $\left(-x_{2} + y_{2}\right)\,\mathbf{a}_{1} - x_{2}\,\mathbf{a}_{2} + \frac{1}{4}\,\mathbf{a}_{3}$ | = | $\left(-x_{2} + \frac{1}{2}y_{2}\right)a\,\hat{\mathbf{x}} - \frac{\sqrt{3}}{2}y_{2}a\,\hat{\mathbf{y}} + \frac{1}{4}c\,\hat{\mathbf{z}}$ | $(6h)$ | C I |
| $\mathbf{B}_{8}$ | = | $-x_{2}\,\mathbf{a}_{1} - y_{2}\,\mathbf{a}_{2} + \frac{3}{4}\,\mathbf{a}_{3}$ | = | $-\frac{1}{2}\left(x_{2} + y_{2}\right)a\,\hat{\mathbf{x}} + \frac{\sqrt{3}}{2}\left(x_{2} - y_{2}\right)a\,\hat{\mathbf{y}} + \frac{3}{4}c\,\hat{\mathbf{z}}$ | $(6h)$ | C I |
| $\mathbf{B}_{9}$ | = | $y_{2}\,\mathbf{a}_{1} + \left(-x_{2} + y_{2}\right)\,\mathbf{a}_{2} + \frac{3}{4}\,\mathbf{a}_{3}$ | = | $\left(-\frac{1}{2}x_{2} + y_{2}\right)a\,\hat{\mathbf{x}} - \frac{\sqrt{3}}{2}x_{2}a\,\hat{\mathbf{y}} + \frac{3}{4}c\,\hat{\mathbf{z}}$ | $(6h)$ | C I |
| $\mathbf{B}_{10}$ | = | $\left(x_{2} - y_{2}\right)\,\mathbf{a}_{1} + x_{2}\,\mathbf{a}_{2} + \frac{3}{4}\,\mathbf{a}_{3}$ | = | $\left(x_{2} - \frac{1}{2}y_{2}\right)a\,\hat{\mathbf{x}} + \frac{\sqrt{3}}{2}y_{2}a\,\hat{\mathbf{y}} + \frac{3}{4}c\,\hat{\mathbf{z}}$ | $(6h)$ | C I |
| $\mathbf{B}_{11}$ | = | $x_{3}\,\mathbf{a}_{1} + y_{3}\,\mathbf{a}_{2} + \frac{1}{4}\,\mathbf{a}_{3}$ | = | $\frac{1}{2}\left(x_{3} + y_{3}\right)a\,\hat{\mathbf{x}} + \frac{\sqrt{3}}{2}\left(-x_{3} + y_{3}\right)a\,\hat{\mathbf{y}} + \frac{1}{4}c\,\hat{\mathbf{z}}$ | $(6h)$ | O I |
| $\mathbf{B}_{12}$ | = | $-y_{3}\,\mathbf{a}_{1} + \left(x_{3} - y_{3}\right)\,\mathbf{a}_{2} + \frac{1}{4}\,\mathbf{a}_{3}$ | = | $\left(\frac{1}{2}x_{3} - y_{3}\right)a\,\hat{\mathbf{x}} + \frac{\sqrt{3}}{2}x_{3}a\,\hat{\mathbf{y}} + \frac{1}{4}c\,\hat{\mathbf{z}}$ | $(6h)$ | O I |
| $\mathbf{B}_{13}$ | = | $\left(-x_{3} + y_{3}\right)\,\mathbf{a}_{1} - x_{3}\,\mathbf{a}_{2} + \frac{1}{4}\,\mathbf{a}_{3}$ | = | $\left(-x_{3} + \frac{1}{2}y_{3}\right)a\,\hat{\mathbf{x}} - \frac{\sqrt{3}}{2}y_{3}a\,\hat{\mathbf{y}} + \frac{1}{4}c\,\hat{\mathbf{z}}$ | $(6h)$ | O I |
| $\mathbf{B}_{14}$ | = | $-x_{3}\,\mathbf{a}_{1} - y_{3}\,\mathbf{a}_{2} + \frac{3}{4}\,\mathbf{a}_{3}$ | = | $-\frac{1}{2}\left(x_{3} + y_{3}\right)a\,\hat{\mathbf{x}} + \frac{\sqrt{3}}{2}\left(x_{3} - y_{3}\right)a\,\hat{\mathbf{y}} + \frac{3}{4}c\,\hat{\mathbf{z}}$ | $(6h)$ | O I |
| $\mathbf{B}_{15}$ | = | $y_{3}\,\mathbf{a}_{1} + \left(-x_{3} + y_{3}\right)\,\mathbf{a}_{2} + \frac{3}{4}\,\mathbf{a}_{3}$ | = | $\left(-\frac{1}{2}x_{3} + y_{3}\right)a\,\hat{\mathbf{x}} - \frac{\sqrt{3}}{2}x_{3}a\,\hat{\mathbf{y}} + \frac{3}{4}c\,\hat{\mathbf{z}}$ | $(6h)$ | O I |
| $\mathbf{B}_{16}$ | = | $\left(x_{3} - y_{3}\right)\,\mathbf{a}_{1} + x_{3}\,\mathbf{a}_{2} + \frac{3}{4}\,\mathbf{a}_{3}$ | = | $\left(x_{3} - \frac{1}{2}y_{3}\right)a\,\hat{\mathbf{x}} + \frac{\sqrt{3}}{2}y_{3}a\,\hat{\mathbf{y}} + \frac{3}{4}c\,\hat{\mathbf{z}}$ | $(6h)$ | O I |
| $\mathbf{B}_{17}$ | = | $x_{4}\,\mathbf{a}_{1} + y_{4}\,\mathbf{a}_{2} + z_{4}\,\mathbf{a}_{3}$ | = | $\frac{1}{2}\left(x_{4} + y_{4}\right)a\,\hat{\mathbf{x}} + \frac{\sqrt{3}}{2}\left(-x_{4} + y_{4}\right)a\,\hat{\mathbf{y}} + z_{4}c\,\hat{\mathbf{z}}$ | $(12i)$ | C II |
| $\mathbf{B}_{18}$ | = | $-y_{4}\,\mathbf{a}_{1} + \left(x_{4} - y_{4}\right)\,\mathbf{a}_{2} + z_{4}\,\mathbf{a}_{3}$ | = | $\left(\frac{1}{2}x_{4} - y_{4}\right)a\,\hat{\mathbf{x}} + \frac{\sqrt{3}}{2}x_{4}a\,\hat{\mathbf{y}} + z_{4}c\,\hat{\mathbf{z}}$ | $(12i)$ | C II |
| $\mathbf{B}_{19}$ | = | $\left(-x_{4} + y_{4}\right)\,\mathbf{a}_{1} - x_{4}\,\mathbf{a}_{2} + z_{4}\,\mathbf{a}_{3}$ | = | $\left(-x_{4} + \frac{1}{2}y_{4}\right)a\,\hat{\mathbf{x}} - \frac{\sqrt{3}}{2}y_{4}a\,\hat{\mathbf{y}} + z_{4}c\,\hat{\mathbf{z}}$ | $(12i)$ | C II |
| $\mathbf{B}_{20}$ | = | $-x_{4}\,\mathbf{a}_{1} - y_{4}\,\mathbf{a}_{2} + \left(\frac{1}{2} + z_{4}\right)\,\mathbf{a}_{3}$ | = | $-\frac{1}{2}\left(x_{4} + y_{4}\right)a\,\hat{\mathbf{x}} + \frac{\sqrt{3}}{2}\left(x_{4} - y_{4}\right)a\,\hat{\mathbf{y}} + \left(\frac{1}{2} + z_{4}\right)c\,\hat{\mathbf{z}}$ | $(12i)$ | C II |
| $\mathbf{B}_{21}$ | = | $y_{4}\,\mathbf{a}_{1} + \left(-x_{4} + y_{4}\right)\,\mathbf{a}_{2} + \left(\frac{1}{2} + z_{4}\right)\,\mathbf{a}_{3}$ | = | $\left(-\frac{1}{2}x_{4} + y_{4}\right)a\,\hat{\mathbf{x}} - \frac{\sqrt{3}}{2}x_{4}a\,\hat{\mathbf{y}} + \left(\frac{1}{2} + z_{4}\right)c\,\hat{\mathbf{z}}$ | $(12i)$ | C II |
| $\mathbf{B}_{22}$ | = | $\left(x_{4} - y_{4}\right)\,\mathbf{a}_{1} + x_{4}\,\mathbf{a}_{2} + \left(\frac{1}{2} + z_{4}\right)\,\mathbf{a}_{3}$ | = | $\left(x_{4} - \frac{1}{2}y_{4}\right)a\,\hat{\mathbf{x}} + \frac{\sqrt{3}}{2}y_{4}a\,\hat{\mathbf{y}} + \left(\frac{1}{2} + z_{4}\right)c\,\hat{\mathbf{z}}$ | $(12i)$ | C II |
| $\mathbf{B}_{23}$ | = | $-x_{4}\,\mathbf{a}_{1} - y_{4}\,\mathbf{a}_{2} - z_{4}\,\mathbf{a}_{3}$ | = | $-\frac{1}{2}\left(x_{4} + y_{4}\right)a\,\hat{\mathbf{x}} + \frac{\sqrt{3}}{2}\left(x_{4} - y_{4}\right)a\,\hat{\mathbf{y}} - z_{4}c\,\hat{\mathbf{z}}$ | $(12i)$ | C II |
| $\mathbf{B}_{24}$ | = | $y_{4}\,\mathbf{a}_{1} + \left(-x_{4} + y_{4}\right)\,\mathbf{a}_{2} - z_{4}\,\mathbf{a}_{3}$ | = | $\left(-\frac{1}{2}x_{4} + y_{4}\right)a\,\hat{\mathbf{x}} - \frac{\sqrt{3}}{2}x_{4}a\,\hat{\mathbf{y}} - z_{4}c\,\hat{\mathbf{z}}$ | $(12i)$ | C II |
| $\mathbf{B}_{25}$ | = | $\left(x_{4} - y_{4}\right)\,\mathbf{a}_{1} + x_{4}\,\mathbf{a}_{2} - z_{4}\,\mathbf{a}_{3}$ | = | $\left(x_{4} - \frac{1}{2}y_{4}\right)a\,\hat{\mathbf{x}} + \frac{\sqrt{3}}{2}y_{4}a\,\hat{\mathbf{y}} - z_{4}c\,\hat{\mathbf{z}}$ | $(12i)$ | C II |
| $\mathbf{B}_{26}$ | = | $x_{4}\,\mathbf{a}_{1} + y_{4}\,\mathbf{a}_{2} + \left(\frac{1}{2} - z_{4}\right)\,\mathbf{a}_{3}$ | = | $\frac{1}{2}\left(x_{4} + y_{4}\right)a\,\hat{\mathbf{x}} + \frac{\sqrt{3}}{2}\left(-x_{4} + y_{4}\right)a\,\hat{\mathbf{y}} + \left(\frac{1}{2} - z_{4}\right)c\,\hat{\mathbf{z}}$ | $(12i)$ | C II |
| $\mathbf{B}_{27}$ | = | $-y_{4}\,\mathbf{a}_{1} + \left(x_{4} - y_{4}\right)\,\mathbf{a}_{2} + \left(\frac{1}{2} - z_{4}\right)\,\mathbf{a}_{3}$ | = | $\left(\frac{1}{2}x_{4} - y_{4}\right)a\,\hat{\mathbf{x}} + \frac{\sqrt{3}}{2}x_{4}a\,\hat{\mathbf{y}} + \left(\frac{1}{2} - z_{4}\right)c\,\hat{\mathbf{z}}$ | $(12i)$ | C II |
| $\mathbf{B}_{28}$ | = | $\left(-x_{4} + y_{4}\right)\,\mathbf{a}_{1} - x_{4}\,\mathbf{a}_{2} + \left(\frac{1}{2} - z_{4}\right)\,\mathbf{a}_{3}$ | = | $\left(-x_{4} + \frac{1}{2}y_{4}\right)a\,\hat{\mathbf{x}} - \frac{\sqrt{3}}{2}y_{4}a\,\hat{\mathbf{y}} + \left(\frac{1}{2} - z_{4}\right)c\,\hat{\mathbf{z}}$ | $(12i)$ | C II |
| $\mathbf{B}_{29}$ | = | $x_{5}\,\mathbf{a}_{1} + y_{5}\,\mathbf{a}_{2} + z_{5}\,\mathbf{a}_{3}$ | = | $\frac{1}{2}\left(x_{5} + y_{5}\right)a\,\hat{\mathbf{x}} + \frac{\sqrt{3}}{2}\left(-x_{5} + y_{5}\right)a\,\hat{\mathbf{y}} + z_{5}c\,\hat{\mathbf{z}}$ | $(12i)$ | O II |
| $\mathbf{B}_{30}$ | = | $-y_{5}\,\mathbf{a}_{1} + \left(x_{5} - y_{5}\right)\,\mathbf{a}_{2} + z_{5}\,\mathbf{a}_{3}$ | = | $\left(\frac{1}{2}x_{5} - y_{5}\right)a\,\hat{\mathbf{x}} + \frac{\sqrt{3}}{2}x_{5}a\,\hat{\mathbf{y}} + z_{5}c\,\hat{\mathbf{z}}$ | $(12i)$ | O II |
| $\mathbf{B}_{31}$ | = | $(-x_5 + y_5)\,\mathbf{a}_1 - x_5\,\mathbf{a}_2 + z_5\,\mathbf{a}_3$ | = | $\left(-x_5 + \frac{1}{2}y_5\right)a\,\hat{\mathbf{x}} - \frac{\sqrt{3}}{2}y_5 a\,\hat{\mathbf{y}} + z_5 c\,\hat{\mathbf{z}}$ | (12*i*) | O II |
| $\mathbf{B}_{32}$ | = | $-x_5\,\mathbf{a}_1 - y_5\,\mathbf{a}_2 + \left(\frac{1}{2} + z_5\right)\mathbf{a}_3$ | = | $-\frac{1}{2}(x_5 + y_5)\,a\,\hat{\mathbf{x}} + \frac{\sqrt{3}}{2}(x_5 - y_5)\,a\,\hat{\mathbf{y}} + \left(\frac{1}{2} + z_5\right)c\,\hat{\mathbf{z}}$ | (12*i*) | O II |
| $\mathbf{B}_{33}$ | = | $y_5\,\mathbf{a}_1 + (-x_5 + y_5)\,\mathbf{a}_2 + \left(\frac{1}{2} + z_5\right)\mathbf{a}_3$ | = | $\left(-\frac{1}{2}x_5 + y_5\right)a\,\hat{\mathbf{x}} - \frac{\sqrt{3}}{2}x_5 a\,\hat{\mathbf{y}} + \left(\frac{1}{2} + z_5\right)c\,\hat{\mathbf{z}}$ | (12*i*) | O II |
| $\mathbf{B}_{34}$ | = | $(x_5 - y_5)\,\mathbf{a}_1 + x_5\,\mathbf{a}_2 + \left(\frac{1}{2} + z_5\right)\mathbf{a}_3$ | = | $\left(x_5 - \frac{1}{2}y_5\right)a\,\hat{\mathbf{x}} + \frac{\sqrt{3}}{2}y_5 a\,\hat{\mathbf{y}} + \left(\frac{1}{2} + z_5\right)c\,\hat{\mathbf{z}}$ | (12*i*) | O II |
| $\mathbf{B}_{35}$ | = | $-x_5\,\mathbf{a}_1 - y_5\,\mathbf{a}_2 - z_5\,\mathbf{a}_3$ | = | $-\frac{1}{2}(x_5 + y_5)\,a\,\hat{\mathbf{x}} + \frac{\sqrt{3}}{2}(x_5 - y_5)\,a\,\hat{\mathbf{y}} - z_5 c\,\hat{\mathbf{z}}$ | (12*i*) | O II |
| $\mathbf{B}_{36}$ | = | $y_5\,\mathbf{a}_1 + (-x_5 + y_5)\,\mathbf{a}_2 - z_5\,\mathbf{a}_3$ | = | $\left(-\frac{1}{2}x_5 + y_5\right)a\,\hat{\mathbf{x}} - \frac{\sqrt{3}}{2}x_5 a\,\hat{\mathbf{y}} - z_5 c\,\hat{\mathbf{z}}$ | (12*i*) | O II |
| $\mathbf{B}_{37}$ | = | $(x_5 - y_5)\,\mathbf{a}_1 + x_5\,\mathbf{a}_2 - z_5\,\mathbf{a}_3$ | = | $\left(x_5 - \frac{1}{2}y_5\right)a\,\hat{\mathbf{x}} + \frac{\sqrt{3}}{2}y_5 a\,\hat{\mathbf{y}} - z_5 c\,\hat{\mathbf{z}}$ | (12*i*) | O II |
| $\mathbf{B}_{38}$ | = | $x_5\,\mathbf{a}_1 + y_5\,\mathbf{a}_2 + \left(\frac{1}{2} - z_5\right)\mathbf{a}_3$ | = | $\frac{1}{2}(x_5 + y_5)\,a\,\hat{\mathbf{x}} + \frac{\sqrt{3}}{2}(-x_5 + y_5)\,a\,\hat{\mathbf{y}} + \left(\frac{1}{2} - z_5\right)c\,\hat{\mathbf{z}}$ | (12*i*) | O II |
| $\mathbf{B}_{39}$ | = | $-y_5\,\mathbf{a}_1 + (x_5 - y_5)\,\mathbf{a}_2 + \left(\frac{1}{2} - z_5\right)\mathbf{a}_3$ | = | $\left(\frac{1}{2}x_5 - y_5\right)a\,\hat{\mathbf{x}} + \frac{\sqrt{3}}{2}x_5 a\,\hat{\mathbf{y}} + \left(\frac{1}{2} - z_5\right)c\,\hat{\mathbf{z}}$ | (12*i*) | O II |
| $\mathbf{B}_{40}$ | = | $(-x_5 + y_5)\,\mathbf{a}_1 - x_5\,\mathbf{a}_2 + \left(\frac{1}{2} - z_5\right)\mathbf{a}_3$ | = | $\left(-x_5 + \frac{1}{2}y_5\right)a\,\hat{\mathbf{x}} - \frac{\sqrt{3}}{2}y_5 a\,\hat{\mathbf{y}} + \left(\frac{1}{2} - z_5\right)c\,\hat{\mathbf{z}}$ | (12*i*) | O II |

---

**Found in:**

---

**Geometry files:**

- CIF: pp. 1753
- POSCAR: pp. 1754

# $K_3W_2Cl_9$ ($K7_1$) Structure: A9B3C2_hP28_176_hi_af_f

http://aflow.org/prototype-encyclopedia/A9B3C2_hP28_176_hi_af_f

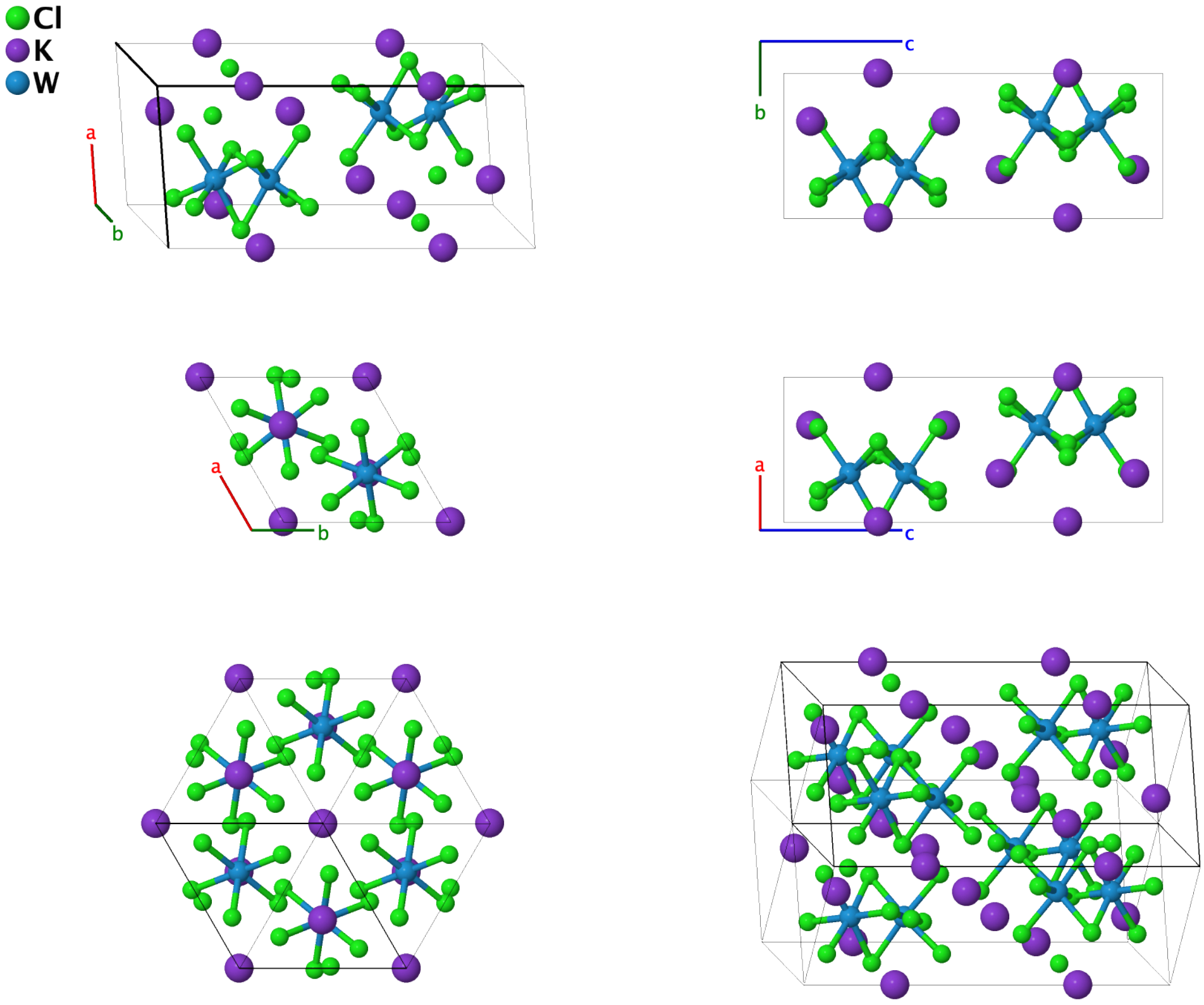

**Prototype** : $Cl_9K_3W_2$

**AFLOW prototype label** : A9B3C2_hP28_176_hi_af_f

***Strukturbericht* designation** : $K7_1$

**Pearson symbol** : hP28

**Space group number** : 176

**Space group symbol** : $P6_3/m$

**AFLOW prototype command** : `aflow --proto=A9B3C2_hP28_176_hi_af_f`
`--params=`$a, c/a, z_2, z_3, x_4, y_4, x_5, y_5, z_5$

---

### Other compounds with this structure

- $K_3Ti_2Cl_9$, $Rb_3Ti_2Cl_9$, $K_3Mo_2Br_9$, $K_3Ti_2Br_9$, and $Rb_3Os_2Br_9$

---

### Hexagonal primitive vectors:

$$
\begin{aligned}
\mathbf{a}_1 &= \tfrac{1}{2}\, a\,\hat{\mathbf{x}} - \tfrac{\sqrt{3}}{2}\, a\,\hat{\mathbf{y}} \\
\mathbf{a}_2 &= \tfrac{1}{2}\, a\,\hat{\mathbf{x}} + \tfrac{\sqrt{3}}{2}\, a\,\hat{\mathbf{y}} \\
\mathbf{a}_3 &= c\,\hat{\mathbf{z}}
\end{aligned}
$$

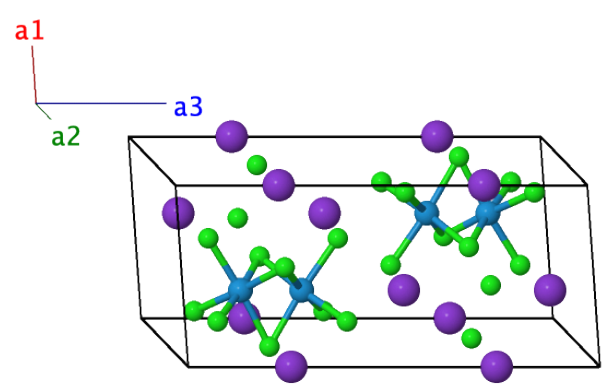

---

### Basis vectors:

| Lattice Coordinates | Cartesian Coordinates | Wyckoff Position | Atom Type |
|---|---|---|---|

| | | | | | | |
|---|---|---|---|---|---|---|
| $\mathbf{B}_1$ | = | $\frac{1}{4}\,\mathbf{a}_3$ | = | $\frac{1}{4}c\,\hat{\mathbf{z}}$ | $(2a)$ | K I |
| $\mathbf{B}_2$ | = | $\frac{3}{4}\,\mathbf{a}_3$ | = | $\frac{3}{4}c\,\hat{\mathbf{z}}$ | $(2a)$ | K I |
| $\mathbf{B}_3$ | = | $\frac{1}{3}\,\mathbf{a}_1 + \frac{2}{3}\,\mathbf{a}_2 + z_2\,\mathbf{a}_3$ | = | $\frac{1}{2}a\,\hat{\mathbf{x}} + \frac{1}{2\sqrt{3}}a\,\hat{\mathbf{y}} + z_2c\,\hat{\mathbf{z}}$ | $(4f)$ | K II |
| $\mathbf{B}_4$ | = | $\frac{2}{3}\,\mathbf{a}_1 + \frac{1}{3}\,\mathbf{a}_2 + \left(\frac{1}{2} + z_2\right)\,\mathbf{a}_3$ | = | $\frac{1}{2}a\,\hat{\mathbf{x}} - \frac{1}{2\sqrt{3}}a\,\hat{\mathbf{y}} + \left(\frac{1}{2} + z_2\right)c\,\hat{\mathbf{z}}$ | $(4f)$ | K II |
| $\mathbf{B}_5$ | = | $\frac{2}{3}\,\mathbf{a}_1 + \frac{1}{3}\,\mathbf{a}_2 - z_2\,\mathbf{a}_3$ | = | $\frac{1}{2}a\,\hat{\mathbf{x}} - \frac{1}{2\sqrt{3}}a\,\hat{\mathbf{y}} - z_2c\,\hat{\mathbf{z}}$ | $(4f)$ | K II |
| $\mathbf{B}_6$ | = | $\frac{1}{3}\,\mathbf{a}_1 + \frac{2}{3}\,\mathbf{a}_2 + \left(\frac{1}{2} - z_2\right)\,\mathbf{a}_3$ | = | $\frac{1}{2}a\,\hat{\mathbf{x}} + \frac{1}{2\sqrt{3}}a\,\hat{\mathbf{y}} + \left(\frac{1}{2} - z_2\right)c\,\hat{\mathbf{z}}$ | $(4f)$ | K II |
| $\mathbf{B}_7$ | = | $\frac{1}{3}\,\mathbf{a}_1 + \frac{2}{3}\,\mathbf{a}_2 + z_3\,\mathbf{a}_3$ | = | $\frac{1}{2}a\,\hat{\mathbf{x}} + \frac{1}{2\sqrt{3}}a\,\hat{\mathbf{y}} + z_3c\,\hat{\mathbf{z}}$ | $(4f)$ | W |
| $\mathbf{B}_8$ | = | $\frac{2}{3}\,\mathbf{a}_1 + \frac{1}{3}\,\mathbf{a}_2 + \left(\frac{1}{2} + z_3\right)\,\mathbf{a}_3$ | = | $\frac{1}{2}a\,\hat{\mathbf{x}} - \frac{1}{2\sqrt{3}}a\,\hat{\mathbf{y}} + \left(\frac{1}{2} + z_3\right)c\,\hat{\mathbf{z}}$ | $(4f)$ | W |
| $\mathbf{B}_9$ | = | $\frac{2}{3}\,\mathbf{a}_1 + \frac{1}{3}\,\mathbf{a}_2 - z_3\,\mathbf{a}_3$ | = | $\frac{1}{2}a\,\hat{\mathbf{x}} - \frac{1}{2\sqrt{3}}a\,\hat{\mathbf{y}} - z_3c\,\hat{\mathbf{z}}$ | $(4f)$ | W |
| $\mathbf{B}_{10}$ | = | $\frac{1}{3}\,\mathbf{a}_1 + \frac{2}{3}\,\mathbf{a}_2 + \left(\frac{1}{2} - z_3\right)\,\mathbf{a}_3$ | = | $\frac{1}{2}a\,\hat{\mathbf{x}} + \frac{1}{2\sqrt{3}}a\,\hat{\mathbf{y}} + \left(\frac{1}{2} - z_3\right)c\,\hat{\mathbf{z}}$ | $(4f)$ | W |
| $\mathbf{B}_{11}$ | = | $x_4\,\mathbf{a}_1 + y_4\,\mathbf{a}_2 + \frac{1}{4}\,\mathbf{a}_3$ | = | $\frac{1}{2}\,(x_4 + y_4)\,a\,\hat{\mathbf{x}} + \frac{\sqrt{3}}{2}\,(-x_4 + y_4)\,a\,\hat{\mathbf{y}} + \frac{1}{4}c\,\hat{\mathbf{z}}$ | $(6h)$ | Cl I |
| $\mathbf{B}_{12}$ | = | $-y_4\,\mathbf{a}_1 + (x_4 - y_4)\,\mathbf{a}_2 + \frac{1}{4}\,\mathbf{a}_3$ | = | $\left(\frac{1}{2}x_4 - y_4\right)a\,\hat{\mathbf{x}} + \frac{\sqrt{3}}{2}x_4a\,\hat{\mathbf{y}} + \frac{1}{4}c\,\hat{\mathbf{z}}$ | $(6h)$ | Cl I |
| $\mathbf{B}_{13}$ | = | $(-x_4 + y_4)\,\mathbf{a}_1 - x_4\,\mathbf{a}_2 + \frac{1}{4}\,\mathbf{a}_3$ | = | $\left(-x_4 + \frac{1}{2}y_4\right)a\,\hat{\mathbf{x}} - \frac{\sqrt{3}}{2}y_4a\,\hat{\mathbf{y}} + \frac{1}{4}c\,\hat{\mathbf{z}}$ | $(6h)$ | Cl I |
| $\mathbf{B}_{14}$ | = | $-x_4\,\mathbf{a}_1 - y_4\,\mathbf{a}_2 + \frac{3}{4}\,\mathbf{a}_3$ | = | $-\frac{1}{2}\,(x_4 + y_4)\,a\,\hat{\mathbf{x}} + \frac{\sqrt{3}}{2}\,(x_4 - y_4)\,a\,\hat{\mathbf{y}} + \frac{3}{4}c\,\hat{\mathbf{z}}$ | $(6h)$ | Cl I |
| $\mathbf{B}_{15}$ | = | $y_4\,\mathbf{a}_1 + (-x_4 + y_4)\,\mathbf{a}_2 + \frac{3}{4}\,\mathbf{a}_3$ | = | $\left(-\frac{1}{2}x_4 + y_4\right)a\,\hat{\mathbf{x}} - \frac{\sqrt{3}}{2}x_4a\,\hat{\mathbf{y}} + \frac{3}{4}c\,\hat{\mathbf{z}}$ | $(6h)$ | Cl I |
| $\mathbf{B}_{16}$ | = | $(x_4 - y_4)\,\mathbf{a}_1 + x_4\,\mathbf{a}_2 + \frac{3}{4}\,\mathbf{a}_3$ | = | $\left(x_4 - \frac{1}{2}y_4\right)a\,\hat{\mathbf{x}} + \frac{\sqrt{3}}{2}y_4a\,\hat{\mathbf{y}} + \frac{3}{4}c\,\hat{\mathbf{z}}$ | $(6h)$ | Cl I |
| $\mathbf{B}_{17}$ | = | $x_5\,\mathbf{a}_1 + y_5\,\mathbf{a}_2 + z_5\,\mathbf{a}_3$ | = | $\frac{1}{2}\,(x_5 + y_5)\,a\,\hat{\mathbf{x}} + \frac{\sqrt{3}}{2}\,(-x_5 + y_5)\,a\,\hat{\mathbf{y}} + z_5c\,\hat{\mathbf{z}}$ | $(12i)$ | Cl II |
| $\mathbf{B}_{18}$ | = | $-y_5\,\mathbf{a}_1 + (x_5 - y_5)\,\mathbf{a}_2 + z_5\,\mathbf{a}_3$ | = | $\left(\frac{1}{2}x_5 - y_5\right)a\,\hat{\mathbf{x}} + \frac{\sqrt{3}}{2}x_5a\,\hat{\mathbf{y}} + z_5c\,\hat{\mathbf{z}}$ | $(12i)$ | Cl II |
| $\mathbf{B}_{19}$ | = | $(-x_5 + y_5)\,\mathbf{a}_1 - x_5\,\mathbf{a}_2 + z_5\,\mathbf{a}_3$ | = | $\left(-x_5 + \frac{1}{2}y_5\right)a\,\hat{\mathbf{x}} - \frac{\sqrt{3}}{2}y_5a\,\hat{\mathbf{y}} + z_5c\,\hat{\mathbf{z}}$ | $(12i)$ | Cl II |
| $\mathbf{B}_{20}$ | = | $-x_5\,\mathbf{a}_1 - y_5\,\mathbf{a}_2 + \left(\frac{1}{2} + z_5\right)\,\mathbf{a}_3$ | = | $-\frac{1}{2}\,(x_5 + y_5)\,a\,\hat{\mathbf{x}} + \frac{\sqrt{3}}{2}\,(x_5 - y_5)\,a\,\hat{\mathbf{y}} + \left(\frac{1}{2} + z_5\right)c\,\hat{\mathbf{z}}$ | $(12i)$ | Cl II |
| $\mathbf{B}_{21}$ | = | $y_5\,\mathbf{a}_1 + (-x_5 + y_5)\,\mathbf{a}_2 + \left(\frac{1}{2} + z_5\right)\,\mathbf{a}_3$ | = | $\left(-\frac{1}{2}x_5 + y_5\right)a\,\hat{\mathbf{x}} - \frac{\sqrt{3}}{2}x_5a\,\hat{\mathbf{y}} + \left(\frac{1}{2} + z_5\right)c\,\hat{\mathbf{z}}$ | $(12i)$ | Cl II |
| $\mathbf{B}_{22}$ | = | $(x_5 - y_5)\,\mathbf{a}_1 + x_5\,\mathbf{a}_2 + \left(\frac{1}{2} + z_5\right)\,\mathbf{a}_3$ | = | $\left(x_5 - \frac{1}{2}y_5\right)a\,\hat{\mathbf{x}} + \frac{\sqrt{3}}{2}y_5a\,\hat{\mathbf{y}} + \left(\frac{1}{2} + z_5\right)c\,\hat{\mathbf{z}}$ | $(12i)$ | Cl II |
| $\mathbf{B}_{23}$ | = | $-x_5\,\mathbf{a}_1 - y_5\,\mathbf{a}_2 - z_5\,\mathbf{a}_3$ | = | $-\frac{1}{2}\,(x_5 + y_5)\,a\,\hat{\mathbf{x}} + \frac{\sqrt{3}}{2}\,(x_5 - y_5)\,a\,\hat{\mathbf{y}} - z_5c\,\hat{\mathbf{z}}$ | $(12i)$ | Cl II |
| $\mathbf{B}_{24}$ | = | $y_5\,\mathbf{a}_1 + (-x_5 + y_5)\,\mathbf{a}_2 - z_5\,\mathbf{a}_3$ | = | $\left(-\frac{1}{2}x_5 + y_5\right)a\,\hat{\mathbf{x}} - \frac{\sqrt{3}}{2}x_5a\,\hat{\mathbf{y}} - z_5c\,\hat{\mathbf{z}}$ | $(12i)$ | Cl II |
| $\mathbf{B}_{25}$ | = | $(x_5 - y_5)\,\mathbf{a}_1 + x_5\,\mathbf{a}_2 - z_5\,\mathbf{a}_3$ | = | $\left(x_5 - \frac{1}{2}y_5\right)a\,\hat{\mathbf{x}} + \frac{\sqrt{3}}{2}y_5a\,\hat{\mathbf{y}} - z_5c\,\hat{\mathbf{z}}$ | $(12i)$ | Cl II |
| $\mathbf{B}_{26}$ | = | $x_5\,\mathbf{a}_1 + y_5\,\mathbf{a}_2 + \left(\frac{1}{2} - z_5\right)\,\mathbf{a}_3$ | = | $\frac{1}{2}\,(x_5 + y_5)\,a\,\hat{\mathbf{x}} + \frac{\sqrt{3}}{2}\,(-x_5 + y_5)\,a\,\hat{\mathbf{y}} + \left(\frac{1}{2} - z_5\right)c\,\hat{\mathbf{z}}$ | $(12i)$ | Cl II |
| $\mathbf{B}_{27}$ | = | $-y_5\,\mathbf{a}_1 + (x_5 - y_5)\,\mathbf{a}_2 + \left(\frac{1}{2} - z_5\right)\,\mathbf{a}_3$ | = | $\left(\frac{1}{2}x_5 - y_5\right)a\,\hat{\mathbf{x}} + \frac{\sqrt{3}}{2}x_5a\,\hat{\mathbf{y}} + \left(\frac{1}{2} - z_5\right)c\,\hat{\mathbf{z}}$ | $(12i)$ | Cl II |
| $\mathbf{B}_{28}$ | = | $(-x_5 + y_5)\,\mathbf{a}_1 - x_5\,\mathbf{a}_2 + \left(\frac{1}{2} - z_5\right)\,\mathbf{a}_3$ | = | $\left(-x_5 + \frac{1}{2}y_5\right)a\,\hat{\mathbf{x}} - \frac{\sqrt{3}}{2}y_5a\,\hat{\mathbf{y}} + \left(\frac{1}{2} - z_5\right)c\,\hat{\mathbf{z}}$ | $(12i)$ | Cl II |

**Geometry files:**

- CIF: pp. 1754
- POSCAR: pp. 1754

# $Hg_2O_2NaI$ Structure: A2BCD2_hP18_180_f_c_b_i

http://aflow.org/prototype-encyclopedia/A2BCD2_hP18_180_f_c_b_i

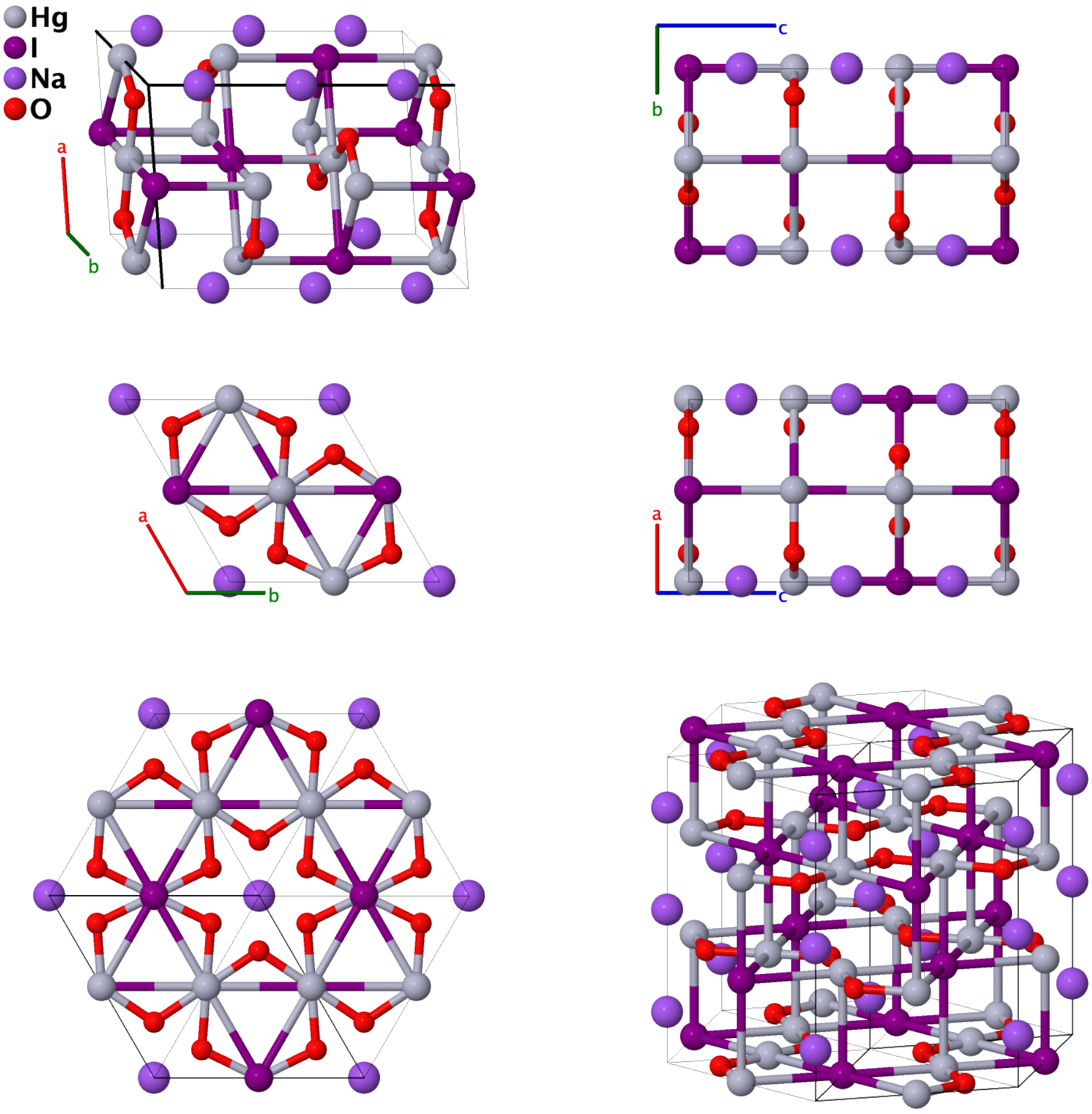

| | | |
|---|---|---|
| **Prototype** | : | $Hg_2INaO_2$ |
| **AFLOW prototype label** | : | A2BCD2_hP18_180_f_c_b_i |
| ***Strukturbericht* designation** | : | None |
| **Pearson symbol** | : | hP18 |
| **Space group number** | : | 180 |
| **Space group symbol** | : | $P6_222$ |
| **AFLOW prototype command** | : | `aflow --proto=A2BCD2_hP18_180_f_c_b_i --params=`$a, c/a, z_3, x_4$ |

**Hexagonal primitive vectors:**

$$\mathbf{a}_1 = \frac{1}{2}\,a\,\hat{\mathbf{x}} - \frac{\sqrt{3}}{2}\,a\,\hat{\mathbf{y}}$$

$$\mathbf{a}_2 = \frac{1}{2}\,a\,\hat{\mathbf{x}} + \frac{\sqrt{3}}{2}\,a\,\hat{\mathbf{y}}$$

$$\mathbf{a}_3 = c\,\hat{\mathbf{z}}$$

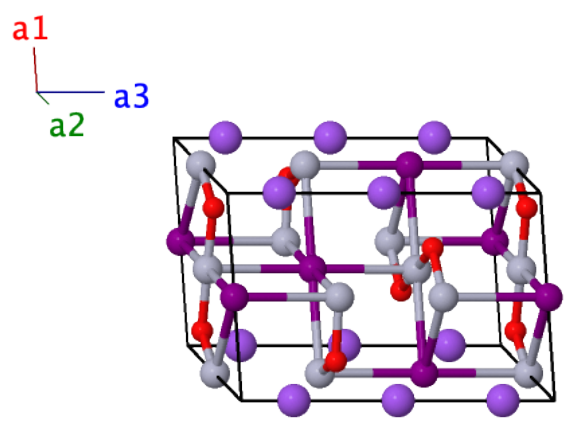

**Basis vectors:**

| | | Lattice Coordinates | | Cartesian Coordinates | Wyckoff Position | Atom Type |
|---|---|---|---|---|---|---|
| $\mathbf{B}_{1}$ | = | $\frac{1}{2}\,\mathbf{a}_{3}$ | = | $\frac{1}{2}c\,\mathbf{\hat{z}}$ | (3$b$) | Na |
| $\mathbf{B}_{2}$ | = | $\frac{1}{6}\,\mathbf{a}_{3}$ | = | $\frac{1}{6}c\,\mathbf{\hat{z}}$ | (3$b$) | Na |
| $\mathbf{B}_{3}$ | = | $\frac{5}{6}\,\mathbf{a}_{3}$ | = | $\frac{5}{6}c\,\mathbf{\hat{z}}$ | (3$b$) | Na |
| $\mathbf{B}_{4}$ | = | $\frac{1}{2}\,\mathbf{a}_{1}$ | = | $\frac{1}{4}a\,\mathbf{\hat{x}} - \frac{\sqrt{3}}{4}a\,\mathbf{\hat{y}}$ | (3$c$) | I |
| $\mathbf{B}_{5}$ | = | $\frac{1}{2}\,\mathbf{a}_{2} + \frac{2}{3}\,\mathbf{a}_{3}$ | = | $\frac{1}{4}a\,\mathbf{\hat{x}} + \frac{\sqrt{3}}{4}a\,\mathbf{\hat{y}} + \frac{2}{3}c\,\mathbf{\hat{z}}$ | (3$c$) | I |
| $\mathbf{B}_{6}$ | = | $\frac{1}{2}\,\mathbf{a}_{1} + \frac{1}{2}\,\mathbf{a}_{2} + \frac{1}{3}\,\mathbf{a}_{3}$ | = | $\frac{1}{2}a\,\mathbf{\hat{x}} + \frac{1}{3}c\,\mathbf{\hat{z}}$ | (3$c$) | I |
| $\mathbf{B}_{7}$ | = | $\frac{1}{2}\,\mathbf{a}_{1} + z_{3}\,\mathbf{a}_{3}$ | = | $\frac{1}{4}a\,\mathbf{\hat{x}} - \frac{\sqrt{3}}{4}a\,\mathbf{\hat{y}} + z_{3}c\,\mathbf{\hat{z}}$ | (6$f$) | Hg |
| $\mathbf{B}_{8}$ | = | $\frac{1}{2}\,\mathbf{a}_{2} + \left(\frac{2}{3} + z_{3}\right)\,\mathbf{a}_{3}$ | = | $\frac{1}{4}a\,\mathbf{\hat{x}} + \frac{\sqrt{3}}{4}a\,\mathbf{\hat{y}} + \left(\frac{2}{3} + z_{3}\right)c\,\mathbf{\hat{z}}$ | (6$f$) | Hg |
| $\mathbf{B}_{9}$ | = | $\frac{1}{2}\,\mathbf{a}_{1} + \frac{1}{2}\,\mathbf{a}_{2} + \left(\frac{1}{3} + z_{3}\right)\,\mathbf{a}_{3}$ | = | $\frac{1}{2}a\,\mathbf{\hat{x}} + \left(\frac{1}{3} + z_{3}\right)c\,\mathbf{\hat{z}}$ | (6$f$) | Hg |
| $\mathbf{B}_{10}$ | = | $\frac{1}{2}\,\mathbf{a}_{2} + \left(\frac{2}{3} - z_{3}\right)\,\mathbf{a}_{3}$ | = | $\frac{1}{4}a\,\mathbf{\hat{x}} + \frac{\sqrt{3}}{4}a\,\mathbf{\hat{y}} + \left(\frac{2}{3} - z_{3}\right)c\,\mathbf{\hat{z}}$ | (6$f$) | Hg |
| $\mathbf{B}_{11}$ | = | $\frac{1}{2}\,\mathbf{a}_{1} - z_{3}\,\mathbf{a}_{3}$ | = | $\frac{1}{4}a\,\mathbf{\hat{x}} - \frac{\sqrt{3}}{4}a\,\mathbf{\hat{y}} - z_{3}c\,\mathbf{\hat{z}}$ | (6$f$) | Hg |
| $\mathbf{B}_{12}$ | = | $\frac{1}{2}\,\mathbf{a}_{1} + \frac{1}{2}\,\mathbf{a}_{2} + \left(\frac{1}{3} - z_{3}\right)\,\mathbf{a}_{3}$ | = | $\frac{1}{2}a\,\mathbf{\hat{x}} + \left(\frac{1}{3} - z_{3}\right)c\,\mathbf{\hat{z}}$ | (6$f$) | Hg |
| $\mathbf{B}_{13}$ | = | $x_{4}\,\mathbf{a}_{1} + 2x_{4}\,\mathbf{a}_{2}$ | = | $\frac{3}{2}x_{4}a\,\mathbf{\hat{x}} + \frac{\sqrt{3}}{2}x_{4}a\,\mathbf{\hat{y}}$ | (6$i$) | O |
| $\mathbf{B}_{14}$ | = | $-2x_{4}\,\mathbf{a}_{1} - x_{4}\,\mathbf{a}_{2} + \frac{2}{3}\,\mathbf{a}_{3}$ | = | $-\frac{3}{2}x_{4}a\,\mathbf{\hat{x}} + \frac{\sqrt{3}}{2}x_{4}a\,\mathbf{\hat{y}} + \frac{2}{3}c\,\mathbf{\hat{z}}$ | (6$i$) | O |
| $\mathbf{B}_{15}$ | = | $x_{4}\,\mathbf{a}_{1} - x_{4}\,\mathbf{a}_{2} + \frac{1}{3}\,\mathbf{a}_{3}$ | = | $-\sqrt{3}x_{4}a\,\mathbf{\hat{y}} + \frac{1}{3}c\,\mathbf{\hat{z}}$ | (6$i$) | O |
| $\mathbf{B}_{16}$ | = | $-x_{4}\,\mathbf{a}_{1} - 2x_{4}\,\mathbf{a}_{2}$ | = | $-\frac{3}{2}x_{4}a\,\mathbf{\hat{x}} - \frac{\sqrt{3}}{2}x_{4}a\,\mathbf{\hat{y}}$ | (6$i$) | O |
| $\mathbf{B}_{17}$ | = | $2x_{4}\,\mathbf{a}_{1} + x_{4}\,\mathbf{a}_{2} + \frac{2}{3}\,\mathbf{a}_{3}$ | = | $\frac{3}{2}x_{4}a\,\mathbf{\hat{x}} - \frac{\sqrt{3}}{2}x_{4}a\,\mathbf{\hat{y}} + \frac{2}{3}c\,\mathbf{\hat{z}}$ | (6$i$) | O |
| $\mathbf{B}_{18}$ | = | $-x_{4}\,\mathbf{a}_{1} + x_{4}\,\mathbf{a}_{2} + \frac{1}{3}\,\mathbf{a}_{3}$ | = | $\sqrt{3}x_{4}a\,\mathbf{\hat{y}} + \frac{1}{3}c\,\mathbf{\hat{z}}$ | (6$i$) | O |

**Geometry files:**

- CIF: pp. 1755
- POSCAR: pp. 1755

# $BaAl_2O_4$ ($H2_8$) Structure: A2BC6_hP18_182_f_b_gh

http://aflow.org/prototype-encyclopedia/A2BC6_hP18_182_f_b_gh

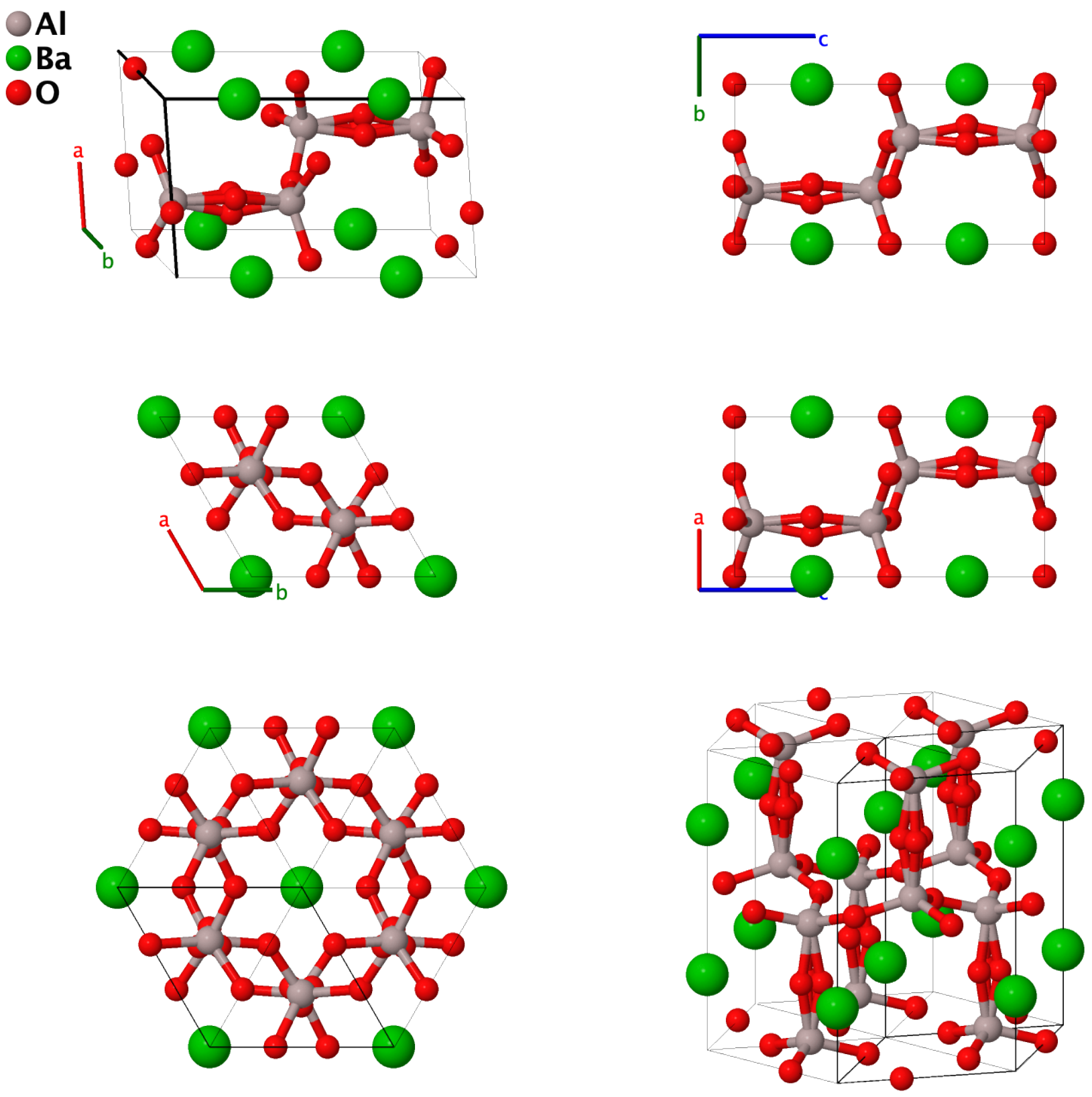

| | | |
|---|---|---|
| **Prototype** | : | $Al_2BaO_4$ |
| **AFLOW prototype label** | : | A2BC6_hP18_182_f_b_gh |
| ***Strukturbericht* designation** | : | $H2_8$ |
| **Pearson symbol** | : | hP18 |
| **Space group number** | : | 182 |
| **Space group symbol** | : | $P6_322$ |
| **AFLOW prototype command** | : | `aflow --proto=A2BC6_hP18_182_f_b_gh`<br>`--params=`$a, c/a, z_2, x_3, x_4$ |

- The structure of $BaAl_2O_4$ was originally determined by (Wallmark, 1937) and designated $H2_8$ by (Gottfried, 1940). This had two oxygen atoms at the (2*c*) Wyckoff position, $\pm(1/3, 2/3, z)$. (Perrotta, 1968) redetermined the structure and found that the best fit to external data required the (2*c*) oxygen atoms to be distributed among the (6*h*) sites. Thus each of the (6*h*) sites in this structure is 1/3 occupied.

**Hexagonal primitive vectors:**

$$\mathbf{a}_1 = \frac{1}{2}\,a\,\hat{\mathbf{x}} - \frac{\sqrt{3}}{2}\,a\,\hat{\mathbf{y}}$$
$$\mathbf{a}_2 = \frac{1}{2}\,a\,\hat{\mathbf{x}} + \frac{\sqrt{3}}{2}\,a\,\hat{\mathbf{y}}$$
$$\mathbf{a}_3 = c\,\hat{\mathbf{z}}$$

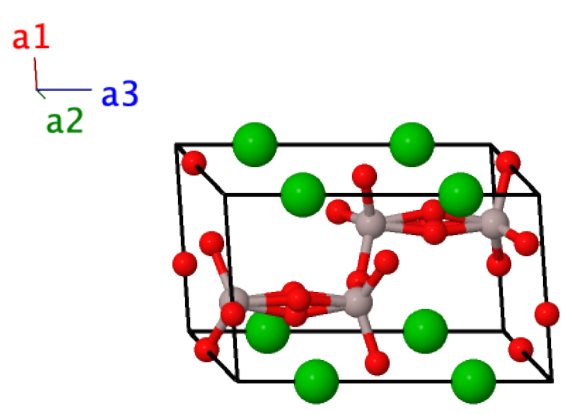

**Basis vectors:**

| | | Lattice Coordinates | | Cartesian Coordinates | Wyckoff Position | Atom Type |
|---|---|---|---|---|---|---|
| $\mathbf{B}_1$ | = | $\frac{1}{4}\,\mathbf{a}_3$ | = | $\frac{1}{4}c\,\hat{\mathbf{z}}$ | $(2b)$ | Ba |
| $\mathbf{B}_2$ | = | $\frac{3}{4}\,\mathbf{a}_3$ | = | $\frac{3}{4}c\,\hat{\mathbf{z}}$ | $(2b)$ | Ba |
| $\mathbf{B}_3$ | = | $\frac{1}{3}\,\mathbf{a}_1 + \frac{2}{3}\,\mathbf{a}_2 + z_2\,\mathbf{a}_3$ | = | $\frac{1}{2}a\,\hat{\mathbf{x}} + \frac{1}{2\sqrt{3}}a\,\hat{\mathbf{y}} + z_2c\,\hat{\mathbf{z}}$ | $(4f)$ | Al |
| $\mathbf{B}_4$ | = | $\frac{2}{3}\,\mathbf{a}_1 + \frac{1}{3}\,\mathbf{a}_2 + \left(\frac{1}{2} + z_2\right)\,\mathbf{a}_3$ | = | $\frac{1}{2}a\,\hat{\mathbf{x}} - \frac{1}{2\sqrt{3}}a\,\hat{\mathbf{y}} + \left(\frac{1}{2} + z_2\right)c\,\hat{\mathbf{z}}$ | $(4f)$ | Al |
| $\mathbf{B}_5$ | = | $\frac{2}{3}\,\mathbf{a}_1 + \frac{1}{3}\,\mathbf{a}_2 - z_2\,\mathbf{a}_3$ | = | $\frac{1}{2}a\,\hat{\mathbf{x}} - \frac{1}{2\sqrt{3}}a\,\hat{\mathbf{y}} - z_2c\,\hat{\mathbf{z}}$ | $(4f)$ | Al |
| $\mathbf{B}_6$ | = | $\frac{1}{3}\,\mathbf{a}_1 + \frac{2}{3}\,\mathbf{a}_2 + \left(\frac{1}{2} - z_2\right)\,\mathbf{a}_3$ | = | $\frac{1}{2}a\,\hat{\mathbf{x}} + \frac{1}{2\sqrt{3}}a\,\hat{\mathbf{y}} + \left(\frac{1}{2} - z_2\right)c\,\hat{\mathbf{z}}$ | $(4f)$ | Al |
| $\mathbf{B}_7$ | = | $x_3\,\mathbf{a}_1$ | = | $\frac{1}{2}x_3a\,\hat{\mathbf{x}} - \frac{\sqrt{3}}{2}x_3a\,\hat{\mathbf{y}}$ | $(6g)$ | O I |
| $\mathbf{B}_8$ | = | $x_3\,\mathbf{a}_2$ | = | $\frac{1}{2}x_3a\,\hat{\mathbf{x}} + \frac{\sqrt{3}}{2}x_3a\,\hat{\mathbf{y}}$ | $(6g)$ | O I |
| $\mathbf{B}_9$ | = | $-x_3\,\mathbf{a}_1 - x_3\,\mathbf{a}_2$ | = | $-x_3a\,\hat{\mathbf{x}}$ | $(6g)$ | O I |
| $\mathbf{B}_{10}$ | = | $-x_3\,\mathbf{a}_1 + \frac{1}{2}\,\mathbf{a}_3$ | = | $-\frac{1}{2}x_3a\,\hat{\mathbf{x}} + \frac{\sqrt{3}}{2}x_3a\,\hat{\mathbf{y}} + \frac{1}{2}c\,\hat{\mathbf{z}}$ | $(6g)$ | O I |
| $\mathbf{B}_{11}$ | = | $-x_3\,\mathbf{a}_2 + \frac{1}{2}\,\mathbf{a}_3$ | = | $-\frac{1}{2}x_3a\,\hat{\mathbf{x}} - \frac{\sqrt{3}}{2}x_3a\,\hat{\mathbf{y}} + \frac{1}{2}c\,\hat{\mathbf{z}}$ | $(6g)$ | O I |
| $\mathbf{B}_{12}$ | = | $x_3\,\mathbf{a}_1 + x_3\,\mathbf{a}_2 + \frac{1}{2}\,\mathbf{a}_3$ | = | $x_3a\,\hat{\mathbf{x}} + \frac{1}{2}c\,\hat{\mathbf{z}}$ | $(6g)$ | O I |
| $\mathbf{B}_{13}$ | = | $x_4\,\mathbf{a}_1 + 2x_4\,\mathbf{a}_2 + \frac{1}{4}\,\mathbf{a}_3$ | = | $\frac{3}{2}x_4a\,\hat{\mathbf{x}} + \frac{\sqrt{3}}{2}x_4a\,\hat{\mathbf{y}} + \frac{1}{4}c\,\hat{\mathbf{z}}$ | $(6h)$ | O II |
| $\mathbf{B}_{14}$ | = | $-2x_4\,\mathbf{a}_1 - x_4\,\mathbf{a}_2 + \frac{1}{4}\,\mathbf{a}_3$ | = | $-\frac{3}{2}x_4a\,\hat{\mathbf{x}} + \frac{\sqrt{3}}{2}x_4a\,\hat{\mathbf{y}} + \frac{1}{4}c\,\hat{\mathbf{z}}$ | $(6h)$ | O II |
| $\mathbf{B}_{15}$ | = | $x_4\,\mathbf{a}_1 - x_4\,\mathbf{a}_2 + \frac{1}{4}\,\mathbf{a}_3$ | = | $-\sqrt{3}x_4a\,\hat{\mathbf{y}} + \frac{1}{4}c\,\hat{\mathbf{z}}$ | $(6h)$ | O II |
| $\mathbf{B}_{16}$ | = | $-x_4\,\mathbf{a}_1 - 2x_4\,\mathbf{a}_2 + \frac{3}{4}\,\mathbf{a}_3$ | = | $-\frac{3}{2}x_4a\,\hat{\mathbf{x}} - \frac{\sqrt{3}}{2}x_4a\,\hat{\mathbf{y}} + \frac{3}{4}c\,\hat{\mathbf{z}}$ | $(6h)$ | O II |
| $\mathbf{B}_{17}$ | = | $2x_4\,\mathbf{a}_1 + x_4\,\mathbf{a}_2 + \frac{3}{4}\,\mathbf{a}_3$ | = | $\frac{3}{2}x_4a\,\hat{\mathbf{x}} - \frac{\sqrt{3}}{2}x_4a\,\hat{\mathbf{y}} + \frac{3}{4}c\,\hat{\mathbf{z}}$ | $(6h)$ | O II |
| $\mathbf{B}_{18}$ | = | $-x_4\,\mathbf{a}_1 + x_4\,\mathbf{a}_2 + \frac{3}{4}\,\mathbf{a}_3$ | = | $\sqrt{3}x_4a\,\hat{\mathbf{y}} + \frac{3}{4}c\,\hat{\mathbf{z}}$ | $(6h)$ | O II |

**Geometry files:**

- CIF: pp. 1755
- POSCAR: pp. 1755

# $E2_3$ ($LiIO_3$) (*obsolete*) Structure: ABC3_hP10_182_c_b_g

http://aflow.org/prototype-encyclopedia/ABC3_hP10_182_c_b_g

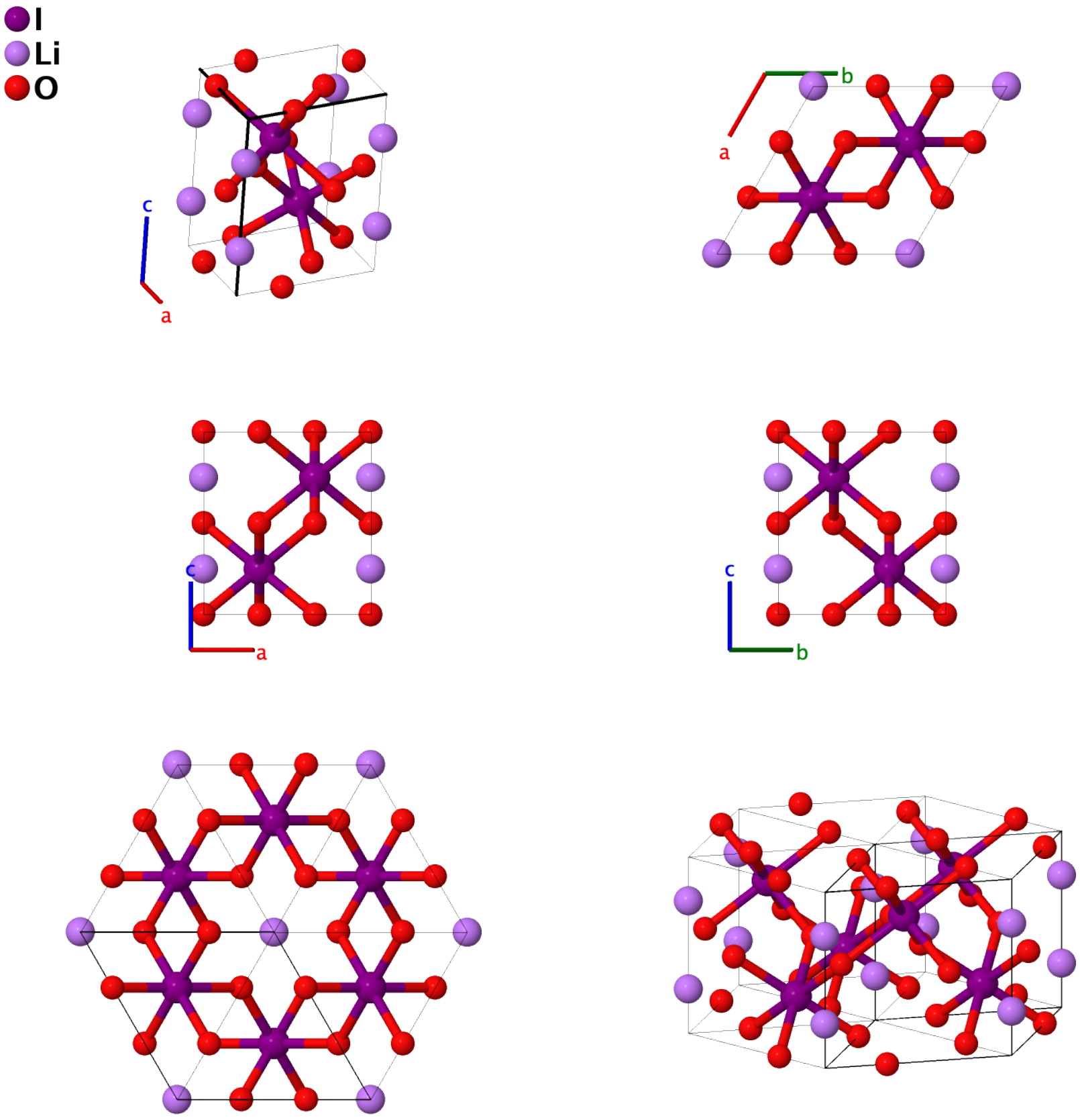

| | | |
|---|---|---|
| **Prototype** | : | $ILiO_3$ |
| **AFLOW prototype label** | : | ABC3_hP10_182_c_b_g |
| ***Strukturbericht* designation** | : | $E2_3$ |
| **Pearson symbol** | : | hP10 |
| **Space group number** | : | 182 |
| **Space group symbol** | : | $P6_322$ |
| **AFLOW prototype command** | : | `aflow --proto=ABC3_hP10_182_c_b_g --params=`$a, c/a, x_3$ |

- $LiIO_3$ is known to exist in three forms:
  - $\alpha$-$LiIO_3$, stable below 273 K: (Zachariasen, 1931) originally determined that this was the structure of $\alpha$-$LiIO_3$, and (Hermann, 1937) designated it as *Strukturbericht* $E2_3$. (Rosenzweig, 1966) subsequently determined that this structure was incorrect because of the small sample size, and determined that the true structure was in space group $P6_3$ #173.
  - $\beta$-$LiIO_3$, stable from 573 K up to the melting point at 708 K.
  - $\gamma$-$LiIO_3$, stable between the $\alpha$- and $\beta$-phases, with an orthorhombic structure.

**Hexagonal primitive vectors:**

$$
\begin{aligned}
\mathbf{a}_1 &= \tfrac{1}{2}\,a\,\hat{\mathbf{x}} - \tfrac{\sqrt{3}}{2}\,a\,\hat{\mathbf{y}} \\
\mathbf{a}_2 &= \tfrac{1}{2}\,a\,\hat{\mathbf{x}} + \tfrac{\sqrt{3}}{2}\,a\,\hat{\mathbf{y}} \\
\mathbf{a}_3 &= c\,\hat{\mathbf{z}}
\end{aligned}
$$

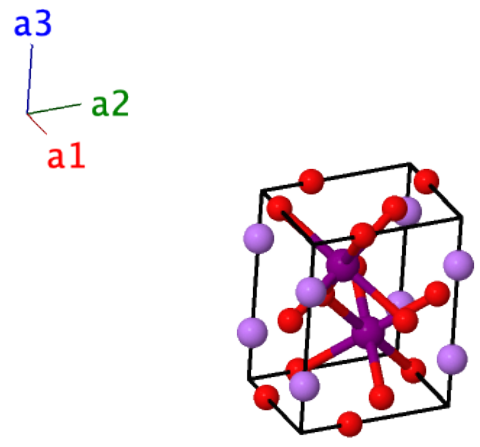

**Basis vectors:**

| | | Lattice Coordinates | | Cartesian Coordinates | Wyckoff Position | Atom Type |
|---|---|---|---|---|---|---|
| $\mathbf{B}_1$ | = | $\frac{1}{4}\,\mathbf{a}_3$ | = | $\frac{1}{4}c\,\hat{\mathbf{z}}$ | $(2b)$ | Li |
| $\mathbf{B}_2$ | = | $\frac{3}{4}\,\mathbf{a}_3$ | = | $\frac{3}{4}c\,\hat{\mathbf{z}}$ | $(2b)$ | Li |
| $\mathbf{B}_3$ | = | $\frac{1}{3}\,\mathbf{a}_1 + \frac{2}{3}\,\mathbf{a}_2 + \frac{1}{4}\,\mathbf{a}_3$ | = | $\frac{1}{2}a\,\hat{\mathbf{x}} + \frac{1}{2\sqrt{3}}a\,\hat{\mathbf{y}} + \frac{1}{4}c\,\hat{\mathbf{z}}$ | $(2c)$ | I |
| $\mathbf{B}_4$ | = | $\frac{2}{3}\,\mathbf{a}_1 + \frac{1}{3}\,\mathbf{a}_2 + \frac{3}{4}\,\mathbf{a}_3$ | = | $\frac{1}{2}a\,\hat{\mathbf{x}} - \frac{1}{2\sqrt{3}}a\,\hat{\mathbf{y}} + \frac{3}{4}c\,\hat{\mathbf{z}}$ | $(2c)$ | I |
| $\mathbf{B}_5$ | = | $x_3\,\mathbf{a}_1$ | = | $\frac{1}{2}x_3a\,\hat{\mathbf{x}} - \frac{\sqrt{3}}{2}x_3a\,\hat{\mathbf{y}}$ | $(6g)$ | O |
| $\mathbf{B}_6$ | = | $x_3\,\mathbf{a}_2$ | = | $\frac{1}{2}x_3a\,\hat{\mathbf{x}} + \frac{\sqrt{3}}{2}x_3a\,\hat{\mathbf{y}}$ | $(6g)$ | O |
| $\mathbf{B}_7$ | = | $-x_3\,\mathbf{a}_1 - x_3\,\mathbf{a}_2$ | = | $-x_3a\,\hat{\mathbf{x}}$ | $(6g)$ | O |
| $\mathbf{B}_8$ | = | $-x_3\,\mathbf{a}_1 + \frac{1}{2}\,\mathbf{a}_3$ | = | $-\frac{1}{2}x_3a\,\hat{\mathbf{x}} + \frac{\sqrt{3}}{2}x_3a\,\hat{\mathbf{y}} + \frac{1}{2}c\,\hat{\mathbf{z}}$ | $(6g)$ | O |
| $\mathbf{B}_9$ | = | $-x_3\,\mathbf{a}_2 + \frac{1}{2}\,\mathbf{a}_3$ | = | $-\frac{1}{2}x_3a\,\hat{\mathbf{x}} - \frac{\sqrt{3}}{2}x_3a\,\hat{\mathbf{y}} + \frac{1}{2}c\,\hat{\mathbf{z}}$ | $(6g)$ | O |
| $\mathbf{B}_{10}$ | = | $x_3\,\mathbf{a}_1 + x_3\,\mathbf{a}_2 + \frac{1}{2}\,\mathbf{a}_3$ | = | $x_3a\,\hat{\mathbf{x}} + \frac{1}{2}c\,\hat{\mathbf{z}}$ | $(6g)$ | O |

**Found in:**

**Geometry files:**

- CIF: pp. 1756
- POSCAR: pp. 1756

# $Zn_2Mo_3O_8$ Structure: A3B8C2_hP26_186_c_ab2c_2b

http://aflow.org/prototype-encyclopedia/A3B8C2_hP26_186_c_ab2c_2b

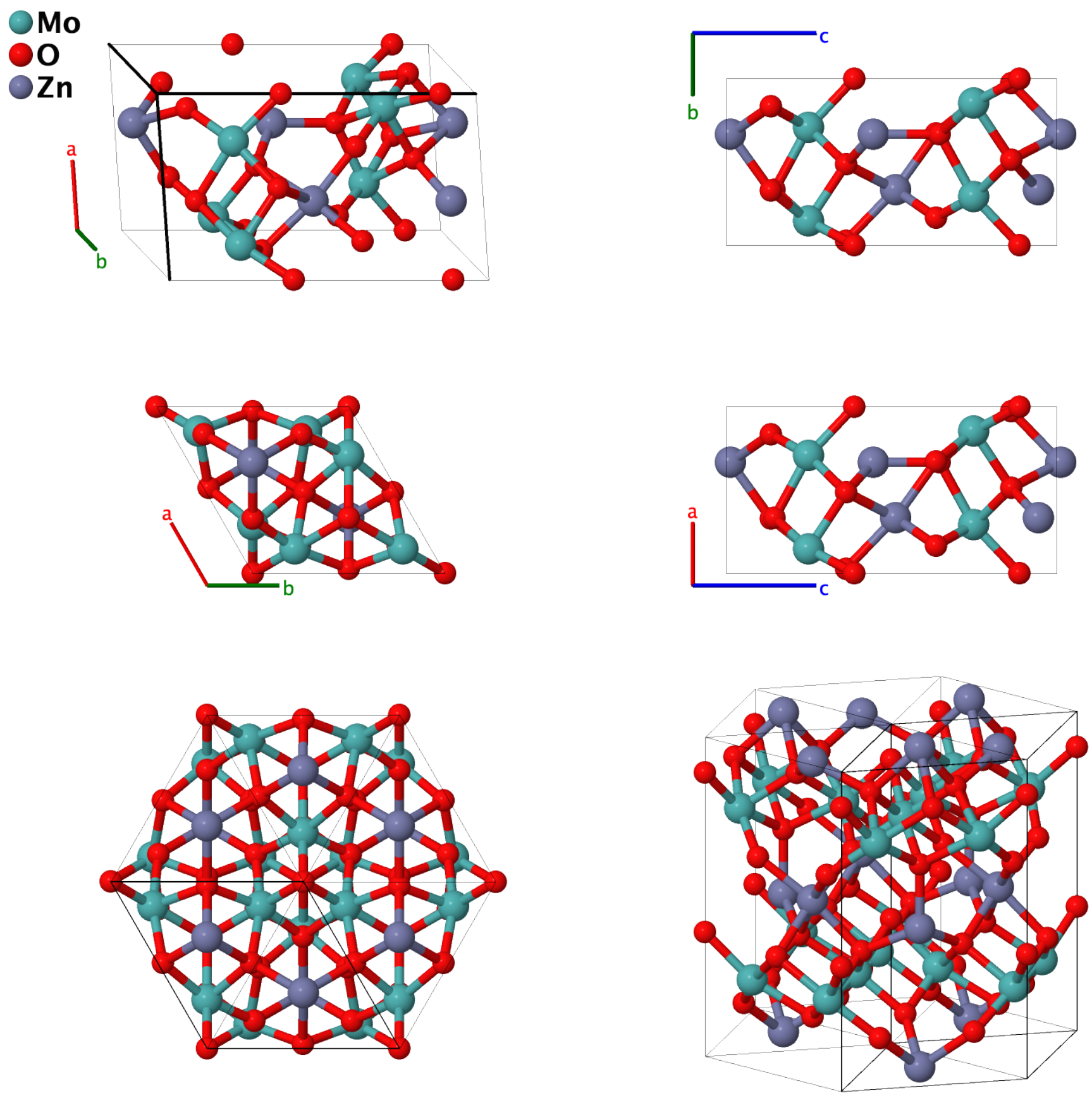

| | | |
|---|---|---|
| **Prototype** | : | $Mo_3O_8Zn_2$ |
| **AFLOW prototype label** | : | A3B8C2_hP26_186_c_ab2c_2b |
| ***Strukturbericht* designation** | : | None |
| **Pearson symbol** | : | hP26 |
| **Space group number** | : | 186 |
| **Space group symbol** | : | $P6_3mc$ |
| **AFLOW prototype command** | : | `aflow --proto=A3B8C2_hP26_186_c_ab2c_2b` `--params=`$a, c/a, z_1, z_2, z_3, z_4, x_5, z_5, x_6, z_6, x_7, z_7$ |

### Other compounds with this structure

- $Cd_2Mo_3O_8$, $Co_2Mo_3O_8$, $Fe_2Mo_3O_8$ (kamiokite), $Mg_2Mo_3O_8$, $Mn_2Mo_3O_8$, and $Ni_2Mo_3O_8$

---

- Space group $P6_3mc$ #186 does not specify the origin of the $z$-coordinate. Here we fix it by setting the $z$-coordinate of the molybdenum atom at $z_5 = 1/4$.

**Hexagonal primitive vectors:**

$$\mathbf{a}_1 = \frac{1}{2}\,a\,\hat{\mathbf{x}} - \frac{\sqrt{3}}{2}\,a\,\hat{\mathbf{y}}$$
$$\mathbf{a}_2 = \frac{1}{2}\,a\,\hat{\mathbf{x}} + \frac{\sqrt{3}}{2}\,a\,\hat{\mathbf{y}}$$
$$\mathbf{a}_3 = c\,\hat{\mathbf{z}}$$

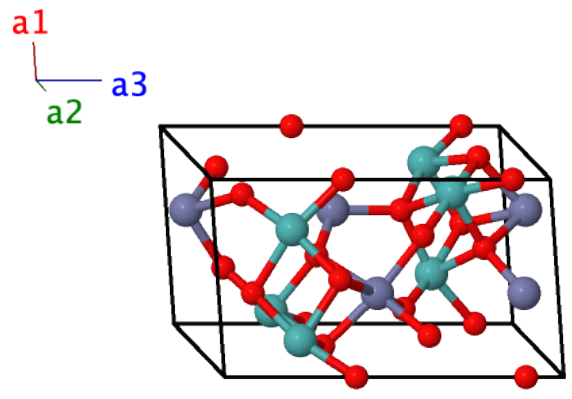

**Basis vectors:**

| | | Lattice Coordinates | | Cartesian Coordinates | Wyckoff Position | Atom Type |
|---|---|---|---|---|---|---|
| $\mathbf{B}_1$ | = | $z_1\,\mathbf{a}_3$ | = | $z_1 c\,\hat{\mathbf{z}}$ | (2$a$) | O I |
| $\mathbf{B}_2$ | = | $\left(\frac{1}{2}+z_1\right)\mathbf{a}_3$ | = | $\left(\frac{1}{2}+z_1\right)c\,\hat{\mathbf{z}}$ | (2$a$) | O I |
| $\mathbf{B}_3$ | = | $\frac{1}{3}\,\mathbf{a}_1 + \frac{2}{3}\,\mathbf{a}_2 + z_2\,\mathbf{a}_3$ | = | $\frac{1}{2}a\,\hat{\mathbf{x}} + \frac{1}{2\sqrt{3}}a\,\hat{\mathbf{y}} + z_2 c\,\hat{\mathbf{z}}$ | (2$b$) | O II |
| $\mathbf{B}_4$ | = | $\frac{2}{3}\,\mathbf{a}_1 + \frac{1}{3}\,\mathbf{a}_2 + \left(\frac{1}{2}+z_2\right)\mathbf{a}_3$ | = | $\frac{1}{2}a\,\hat{\mathbf{x}} - \frac{1}{2\sqrt{3}}a\,\hat{\mathbf{y}} + \left(\frac{1}{2}+z_2\right)c\,\hat{\mathbf{z}}$ | (2$b$) | O II |
| $\mathbf{B}_5$ | = | $\frac{1}{3}\,\mathbf{a}_1 + \frac{2}{3}\,\mathbf{a}_2 + z_3\,\mathbf{a}_3$ | = | $\frac{1}{2}a\,\hat{\mathbf{x}} + \frac{1}{2\sqrt{3}}a\,\hat{\mathbf{y}} + z_3 c\,\hat{\mathbf{z}}$ | (2$b$) | Zn I |
| $\mathbf{B}_6$ | = | $\frac{2}{3}\,\mathbf{a}_1 + \frac{1}{3}\,\mathbf{a}_2 + \left(\frac{1}{2}+z_3\right)\mathbf{a}_3$ | = | $\frac{1}{2}a\,\hat{\mathbf{x}} - \frac{1}{2\sqrt{3}}a\,\hat{\mathbf{y}} + \left(\frac{1}{2}+z_3\right)c\,\hat{\mathbf{z}}$ | (2$b$) | Zn I |
| $\mathbf{B}_7$ | = | $\frac{1}{3}\,\mathbf{a}_1 + \frac{2}{3}\,\mathbf{a}_2 + z_4\,\mathbf{a}_3$ | = | $\frac{1}{2}a\,\hat{\mathbf{x}} + \frac{1}{2\sqrt{3}}a\,\hat{\mathbf{y}} + z_4 c\,\hat{\mathbf{z}}$ | (2$b$) | Zn II |
| $\mathbf{B}_8$ | = | $\frac{2}{3}\,\mathbf{a}_1 + \frac{1}{3}\,\mathbf{a}_2 + \left(\frac{1}{2}+z_4\right)\mathbf{a}_3$ | = | $\frac{1}{2}a\,\hat{\mathbf{x}} - \frac{1}{2\sqrt{3}}a\,\hat{\mathbf{y}} + \left(\frac{1}{2}+z_4\right)c\,\hat{\mathbf{z}}$ | (2$b$) | Zn II |
| $\mathbf{B}_9$ | = | $x_5\,\mathbf{a}_1 - x_5\,\mathbf{a}_2 + z_5\,\mathbf{a}_3$ | = | $-\sqrt{3}x_5 a\,\hat{\mathbf{y}} + z_5 c\,\hat{\mathbf{z}}$ | (6$c$) | Mo |
| $\mathbf{B}_{10}$ | = | $x_5\,\mathbf{a}_1 + 2x_5\,\mathbf{a}_2 + z_5\,\mathbf{a}_3$ | = | $\frac{3}{2}x_5 a\,\hat{\mathbf{x}} + \frac{\sqrt{3}}{2}x_5 a\,\hat{\mathbf{y}} + z_5 c\,\hat{\mathbf{z}}$ | (6$c$) | Mo |
| $\mathbf{B}_{11}$ | = | $-2x_5\,\mathbf{a}_1 - x_5\,\mathbf{a}_2 + z_5\,\mathbf{a}_3$ | = | $-\frac{3}{2}x_5 a\,\hat{\mathbf{x}} + \frac{\sqrt{3}}{2}x_5 a\,\hat{\mathbf{y}} + z_5 c\,\hat{\mathbf{z}}$ | (6$c$) | Mo |
| $\mathbf{B}_{12}$ | = | $-x_5\,\mathbf{a}_1 + x_5\,\mathbf{a}_2 + \left(\frac{1}{2}+z_5\right)\mathbf{a}_3$ | = | $\sqrt{3}x_5 a\,\hat{\mathbf{y}} + \left(\frac{1}{2}+z_5\right)c\,\hat{\mathbf{z}}$ | (6$c$) | Mo |
| $\mathbf{B}_{13}$ | = | $-x_5\,\mathbf{a}_1 - 2x_5\,\mathbf{a}_2 + \left(\frac{1}{2}+z_5\right)\mathbf{a}_3$ | = | $-\frac{3}{2}x_5 a\,\hat{\mathbf{x}} - \frac{\sqrt{3}}{2}x_5 a\,\hat{\mathbf{y}} + \left(\frac{1}{2}+z_5\right)c\,\hat{\mathbf{z}}$ | (6$c$) | Mo |
| $\mathbf{B}_{14}$ | = | $2x_5\,\mathbf{a}_1 + x_5\,\mathbf{a}_2 + \left(\frac{1}{2}+z_5\right)\mathbf{a}_3$ | = | $\frac{3}{2}x_5 a\,\hat{\mathbf{x}} - \frac{\sqrt{3}}{2}x_5 a\,\hat{\mathbf{y}} + \left(\frac{1}{2}+z_5\right)c\,\hat{\mathbf{z}}$ | (6$c$) | Mo |
| $\mathbf{B}_{15}$ | = | $x_6\,\mathbf{a}_1 - x_6\,\mathbf{a}_2 + z_6\,\mathbf{a}_3$ | = | $-\sqrt{3}x_6 a\,\hat{\mathbf{y}} + z_6 c\,\hat{\mathbf{z}}$ | (6$c$) | O III |
| $\mathbf{B}_{16}$ | = | $x_6\,\mathbf{a}_1 + 2x_6\,\mathbf{a}_2 + z_6\,\mathbf{a}_3$ | = | $\frac{3}{2}x_6 a\,\hat{\mathbf{x}} + \frac{\sqrt{3}}{2}x_6 a\,\hat{\mathbf{y}} + z_6 c\,\hat{\mathbf{z}}$ | (6$c$) | O III |
| $\mathbf{B}_{17}$ | = | $-2x_6\,\mathbf{a}_1 - x_6\,\mathbf{a}_2 + z_6\,\mathbf{a}_3$ | = | $-\frac{3}{2}x_6 a\,\hat{\mathbf{x}} + \frac{\sqrt{3}}{2}x_6 a\,\hat{\mathbf{y}} + z_6 c\,\hat{\mathbf{z}}$ | (6$c$) | O III |
| $\mathbf{B}_{18}$ | = | $-x_6\,\mathbf{a}_1 + x_6\,\mathbf{a}_2 + \left(\frac{1}{2}+z_6\right)\mathbf{a}_3$ | = | $\sqrt{3}x_6 a\,\hat{\mathbf{y}} + \left(\frac{1}{2}+z_6\right)c\,\hat{\mathbf{z}}$ | (6$c$) | O III |
| $\mathbf{B}_{19}$ | = | $-x_6\,\mathbf{a}_1 - 2x_6\,\mathbf{a}_2 + \left(\frac{1}{2}+z_6\right)\mathbf{a}_3$ | = | $-\frac{3}{2}x_6 a\,\hat{\mathbf{x}} - \frac{\sqrt{3}}{2}x_6 a\,\hat{\mathbf{y}} + \left(\frac{1}{2}+z_6\right)c\,\hat{\mathbf{z}}$ | (6$c$) | O III |
| $\mathbf{B}_{20}$ | = | $2x_6\,\mathbf{a}_1 + x_6\,\mathbf{a}_2 + \left(\frac{1}{2}+z_6\right)\mathbf{a}_3$ | = | $\frac{3}{2}x_6 a\,\hat{\mathbf{x}} - \frac{\sqrt{3}}{2}x_6 a\,\hat{\mathbf{y}} + \left(\frac{1}{2}+z_6\right)c\,\hat{\mathbf{z}}$ | (6$c$) | O III |
| $\mathbf{B}_{21}$ | = | $x_7\,\mathbf{a}_1 - x_7\,\mathbf{a}_2 + z_7\,\mathbf{a}_3$ | = | $-\sqrt{3}x_7 a\,\hat{\mathbf{y}} + z_7 c\,\hat{\mathbf{z}}$ | (6$c$) | O IV |
| $\mathbf{B}_{22}$ | = | $x_7\,\mathbf{a}_1 + 2x_7\,\mathbf{a}_2 + z_7\,\mathbf{a}_3$ | = | $\frac{3}{2}x_7 a\,\hat{\mathbf{x}} + \frac{\sqrt{3}}{2}x_7 a\,\hat{\mathbf{y}} + z_7 c\,\hat{\mathbf{z}}$ | (6$c$) | O IV |
| $\mathbf{B}_{23}$ | = | $-2x_7\,\mathbf{a}_1 - x_7\,\mathbf{a}_2 + z_7\,\mathbf{a}_3$ | = | $-\frac{3}{2}x_7 a\,\hat{\mathbf{x}} + \frac{\sqrt{3}}{2}x_7 a\,\hat{\mathbf{y}} + z_7 c\,\hat{\mathbf{z}}$ | (6$c$) | O IV |
| $\mathbf{B}_{24}$ | = | $-x_7\,\mathbf{a}_1 + x_7\,\mathbf{a}_2 + \left(\frac{1}{2}+z_7\right)\mathbf{a}_3$ | = | $\sqrt{3}x_7 a\,\hat{\mathbf{y}} + \left(\frac{1}{2}+z_7\right)c\,\hat{\mathbf{z}}$ | (6$c$) | O IV |
| $\mathbf{B}_{25}$ | = | $-x_7\,\mathbf{a}_1 - 2x_7\,\mathbf{a}_2 + \left(\frac{1}{2}+z_7\right)\mathbf{a}_3$ | = | $-\frac{3}{2}x_7 a\,\hat{\mathbf{x}} - \frac{\sqrt{3}}{2}x_7 a\,\hat{\mathbf{y}} + \left(\frac{1}{2}+z_7\right)c\,\hat{\mathbf{z}}$ | (6$c$) | O IV |
| $\mathbf{B}_{26}$ | = | $2x_7\,\mathbf{a}_1 + x_7\,\mathbf{a}_2 + \left(\frac{1}{2}+z_7\right)\mathbf{a}_3$ | = | $\frac{3}{2}x_7 a\,\hat{\mathbf{x}} - \frac{\sqrt{3}}{2}x_7 a\,\hat{\mathbf{y}} + \left(\frac{1}{2}+z_7\right)c\,\hat{\mathbf{z}}$ | (6$c$) | O IV |

**Found in:**

---

**Geometry files:**

- CIF: pp. 1756
- POSCAR: pp. 1756

# $Nd(BrO_3)_3 \cdot 9H_2O$ ($G2_2$) Structure: A3B9CD9_hP44_186_c_3c_b_cd

http://aflow.org/prototype-encyclopedia/A3B9CD9_hP44_186_c_3c_b_cd

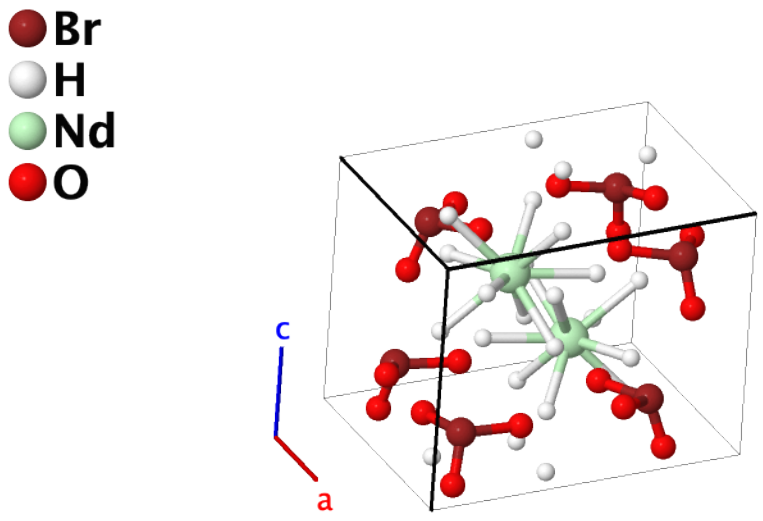

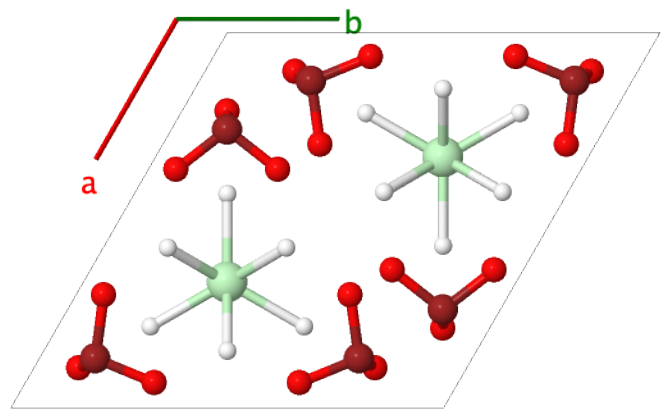

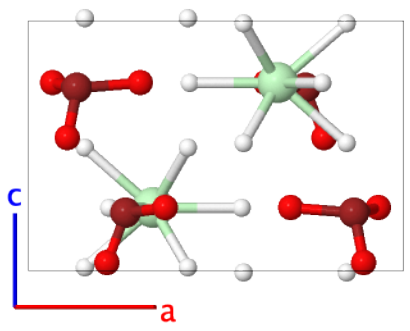

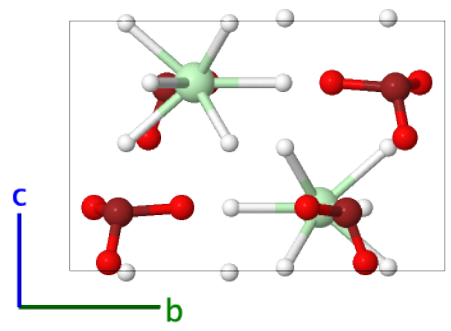

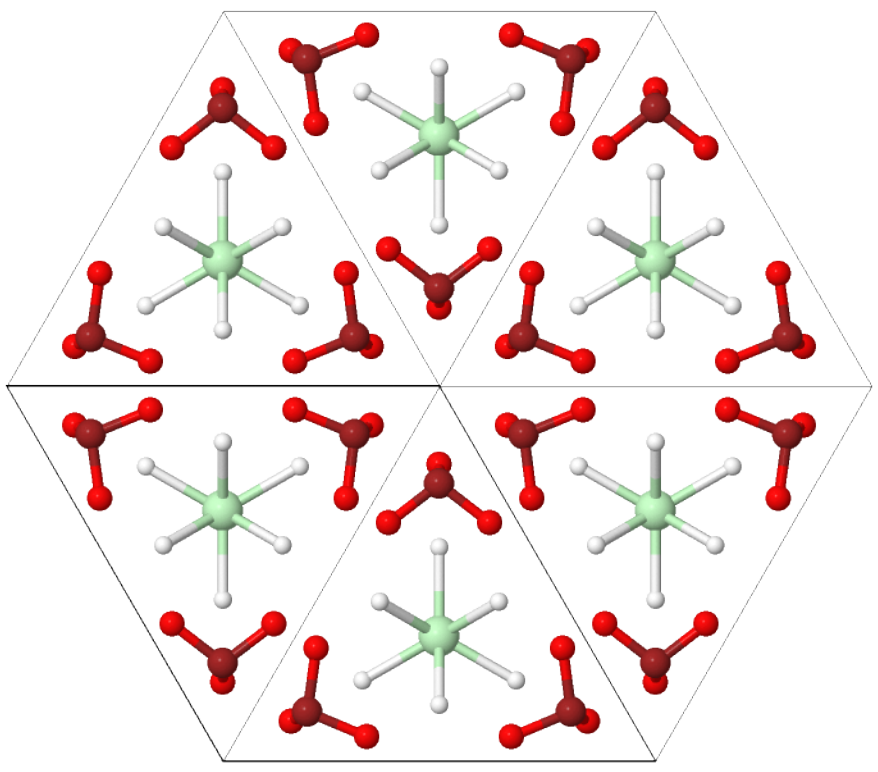

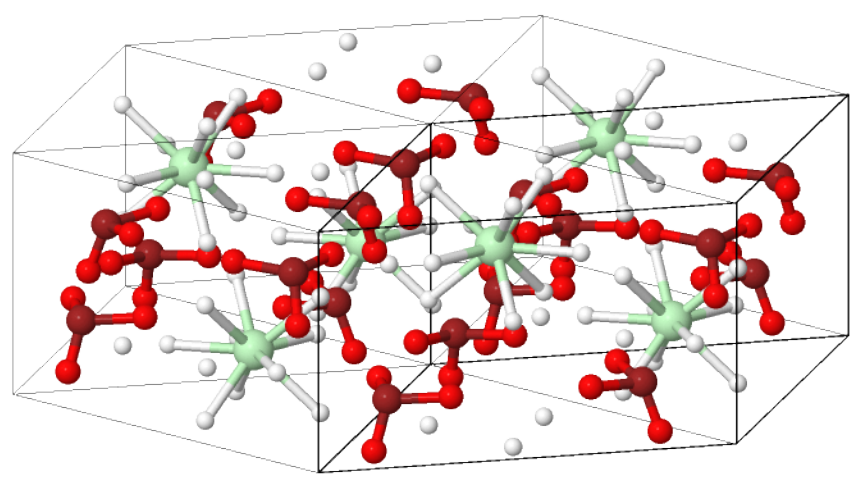

| | | |
|---|---|---|
| **Prototype** | : | $Br_3(H_2O)_9NdO_9$ |
| **AFLOW prototype label** | : | A3B9CD9_hP44_186_c_3c_b_cd |
| ***Strukturbericht* designation** | : | $G2_2$ |
| **Pearson symbol** | : | hP44 |
| **Space group number** | : | 186 |
| **Space group symbol** | : | $P6_3mc$ |
| **AFLOW prototype command** | : | `aflow --proto=A3B9CD9_hP44_186_c_3c_b_cd` `--params=`$a, c/a, z_1, x_2, z_2, x_3, z_3, x_4, z_4, x_5, z_5, x_6, z_6, x_7, y_7, z_7$ |

- The positions of the hydrogen atoms in the water molecules were not determined, so we only provide the oxygen atom positions (labeled as $H_2O$).

**Hexagonal primitive vectors:**

$$\mathbf{a}_1 = \frac{1}{2}\,a\,\mathbf{\hat{x}} - \frac{\sqrt{3}}{2}\,a\,\mathbf{\hat{y}}$$

$$\mathbf{a}_2 = \frac{1}{2}\,a\,\mathbf{\hat{x}} + \frac{\sqrt{3}}{2}\,a\,\mathbf{\hat{y}}$$

$$\mathbf{a}_3 = c\,\mathbf{\hat{z}}$$

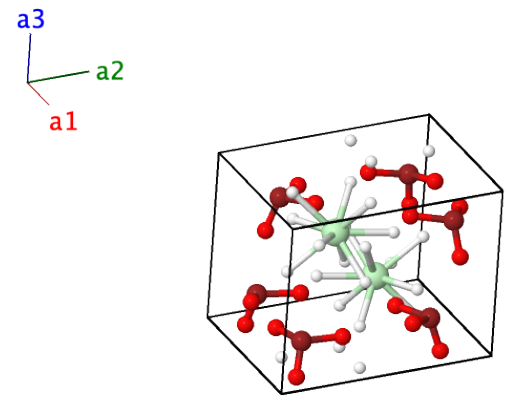

**Basis vectors:**

| | | Lattice Coordinates | | Cartesian Coordinates | Wyckoff Position | Atom Type |
|---|---|---|---|---|---|---|
| $\mathbf{B}_1$ | = | $\frac{1}{3}\,\mathbf{a}_1 + \frac{2}{3}\,\mathbf{a}_2 + z_1\,\mathbf{a}_3$ | = | $\frac{1}{2}a\,\mathbf{\hat{x}} + \frac{1}{2\sqrt{3}}a\,\mathbf{\hat{y}} + z_1 c\,\mathbf{\hat{z}}$ | (2$b$) | Nd |
| $\mathbf{B}_2$ | = | $\frac{2}{3}\,\mathbf{a}_1 + \frac{1}{3}\,\mathbf{a}_2 + \left(\frac{1}{2} + z_1\right)\,\mathbf{a}_3$ | = | $\frac{1}{2}a\,\mathbf{\hat{x}} - \frac{1}{2\sqrt{3}}a\,\mathbf{\hat{y}} + \left(\frac{1}{2} + z_1\right)c\,\mathbf{\hat{z}}$ | (2$b$) | Nd |
| $\mathbf{B}_3$ | = | $x_2\,\mathbf{a}_1 - x_2\,\mathbf{a}_2 + z_2\,\mathbf{a}_3$ | = | $-\sqrt{3}x_2 a\,\mathbf{\hat{y}} + z_2 c\,\mathbf{\hat{z}}$ | (6$c$) | Br |
| $\mathbf{B}_4$ | = | $x_2\,\mathbf{a}_1 + 2x_2\,\mathbf{a}_2 + z_2\,\mathbf{a}_3$ | = | $\frac{3}{2}x_2 a\,\mathbf{\hat{x}} + \frac{\sqrt{3}}{2}x_2 a\,\mathbf{\hat{y}} + z_2 c\,\mathbf{\hat{z}}$ | (6$c$) | Br |
| $\mathbf{B}_5$ | = | $-2x_2\,\mathbf{a}_1 - x_2\,\mathbf{a}_2 + z_2\,\mathbf{a}_3$ | = | $-\frac{3}{2}x_2 a\,\mathbf{\hat{x}} + \frac{\sqrt{3}}{2}x_2 a\,\mathbf{\hat{y}} + z_2 c\,\mathbf{\hat{z}}$ | (6$c$) | Br |
| $\mathbf{B}_6$ | = | $-x_2\,\mathbf{a}_1 + x_2\,\mathbf{a}_2 + \left(\frac{1}{2} + z_2\right)\,\mathbf{a}_3$ | = | $\sqrt{3}x_2 a\,\mathbf{\hat{y}} + \left(\frac{1}{2} + z_2\right)c\,\mathbf{\hat{z}}$ | (6$c$) | Br |
| $\mathbf{B}_7$ | = | $-x_2\,\mathbf{a}_1 - 2x_2\,\mathbf{a}_2 + \left(\frac{1}{2} + z_2\right)\,\mathbf{a}_3$ | = | $-\frac{3}{2}x_2 a\,\mathbf{\hat{x}} - \frac{\sqrt{3}}{2}x_2 a\,\mathbf{\hat{y}} + \left(\frac{1}{2} + z_2\right)c\,\mathbf{\hat{z}}$ | (6$c$) | Br |
| $\mathbf{B}_8$ | = | $2x_2\,\mathbf{a}_1 + x_2\,\mathbf{a}_2 + \left(\frac{1}{2} + z_2\right)\,\mathbf{a}_3$ | = | $\frac{3}{2}x_2 a\,\mathbf{\hat{x}} - \frac{\sqrt{3}}{2}x_2 a\,\mathbf{\hat{y}} + \left(\frac{1}{2} + z_2\right)c\,\mathbf{\hat{z}}$ | (6$c$) | Br |
| $\mathbf{B}_9$ | = | $x_3\,\mathbf{a}_1 - x_3\,\mathbf{a}_2 + z_3\,\mathbf{a}_3$ | = | $-\sqrt{3}x_3 a\,\mathbf{\hat{y}} + z_3 c\,\mathbf{\hat{z}}$ | (6$c$) | $H_2O$ I |
| $\mathbf{B}_{10}$ | = | $x_3\,\mathbf{a}_1 + 2x_3\,\mathbf{a}_2 + z_3\,\mathbf{a}_3$ | = | $\frac{3}{2}x_3 a\,\mathbf{\hat{x}} + \frac{\sqrt{3}}{2}x_3 a\,\mathbf{\hat{y}} + z_3 c\,\mathbf{\hat{z}}$ | (6$c$) | $H_2O$ I |
| $\mathbf{B}_{11}$ | = | $-2x_3\,\mathbf{a}_1 - x_3\,\mathbf{a}_2 + z_3\,\mathbf{a}_3$ | = | $-\frac{3}{2}x_3 a\,\mathbf{\hat{x}} + \frac{\sqrt{3}}{2}x_3 a\,\mathbf{\hat{y}} + z_3 c\,\mathbf{\hat{z}}$ | (6$c$) | $H_2O$ I |
| $\mathbf{B}_{12}$ | = | $-x_3\,\mathbf{a}_1 + x_3\,\mathbf{a}_2 + \left(\frac{1}{2} + z_3\right)\,\mathbf{a}_3$ | = | $\sqrt{3}x_3 a\,\mathbf{\hat{y}} + \left(\frac{1}{2} + z_3\right)c\,\mathbf{\hat{z}}$ | (6$c$) | $H_2O$ I |
| $\mathbf{B}_{13}$ | = | $-x_3\,\mathbf{a}_1 - 2x_3\,\mathbf{a}_2 + \left(\frac{1}{2} + z_3\right)\,\mathbf{a}_3$ | = | $-\frac{3}{2}x_3 a\,\mathbf{\hat{x}} - \frac{\sqrt{3}}{2}x_3 a\,\mathbf{\hat{y}} + \left(\frac{1}{2} + z_3\right)c\,\mathbf{\hat{z}}$ | (6$c$) | $H_2O$ I |
| $\mathbf{B}_{14}$ | = | $2x_3\,\mathbf{a}_1 + x_3\,\mathbf{a}_2 + \left(\frac{1}{2} + z_3\right)\,\mathbf{a}_3$ | = | $\frac{3}{2}x_3 a\,\mathbf{\hat{x}} - \frac{\sqrt{3}}{2}x_3 a\,\mathbf{\hat{y}} + \left(\frac{1}{2} + z_3\right)c\,\mathbf{\hat{z}}$ | (6$c$) | $H_2O$ I |
| $\mathbf{B}_{15}$ | = | $x_4\,\mathbf{a}_1 - x_4\,\mathbf{a}_2 + z_4\,\mathbf{a}_3$ | = | $-\sqrt{3}x_4 a\,\mathbf{\hat{y}} + z_4 c\,\mathbf{\hat{z}}$ | (6$c$) | $H_2O$ II |
| $\mathbf{B}_{16}$ | = | $x_4\,\mathbf{a}_1 + 2x_4\,\mathbf{a}_2 + z_4\,\mathbf{a}_3$ | = | $\frac{3}{2}x_4 a\,\mathbf{\hat{x}} + \frac{\sqrt{3}}{2}x_4 a\,\mathbf{\hat{y}} + z_4 c\,\mathbf{\hat{z}}$ | (6$c$) | $H_2O$ II |
| $\mathbf{B}_{17}$ | = | $-2x_4\,\mathbf{a}_1 - x_4\,\mathbf{a}_2 + z_4\,\mathbf{a}_3$ | = | $-\frac{3}{2}x_4 a\,\mathbf{\hat{x}} + \frac{\sqrt{3}}{2}x_4 a\,\mathbf{\hat{y}} + z_4 c\,\mathbf{\hat{z}}$ | (6$c$) | $H_2O$ II |
| $\mathbf{B}_{18}$ | = | $-x_4\,\mathbf{a}_1 + x_4\,\mathbf{a}_2 + \left(\frac{1}{2} + z_4\right)\,\mathbf{a}_3$ | = | $\sqrt{3}x_4 a\,\mathbf{\hat{y}} + \left(\frac{1}{2} + z_4\right)c\,\mathbf{\hat{z}}$ | (6$c$) | $H_2O$ II |
| $\mathbf{B}_{19}$ | = | $-x_4\,\mathbf{a}_1 - 2x_4\,\mathbf{a}_2 + \left(\frac{1}{2} + z_4\right)\,\mathbf{a}_3$ | = | $-\frac{3}{2}x_4 a\,\mathbf{\hat{x}} - \frac{\sqrt{3}}{2}x_4 a\,\mathbf{\hat{y}} + \left(\frac{1}{2} + z_4\right)c\,\mathbf{\hat{z}}$ | (6$c$) | $H_2O$ II |
| $\mathbf{B}_{20}$ | = | $2x_4\,\mathbf{a}_1 + x_4\,\mathbf{a}_2 + \left(\frac{1}{2} + z_4\right)\,\mathbf{a}_3$ | = | $\frac{3}{2}x_4 a\,\mathbf{\hat{x}} - \frac{\sqrt{3}}{2}x_4 a\,\mathbf{\hat{y}} + \left(\frac{1}{2} + z_4\right)c\,\mathbf{\hat{z}}$ | (6$c$) | $H_2O$ II |
| $\mathbf{B}_{21}$ | = | $x_5\,\mathbf{a}_1 - x_5\,\mathbf{a}_2 + z_5\,\mathbf{a}_3$ | = | $-\sqrt{3}x_5 a\,\mathbf{\hat{y}} + z_5 c\,\mathbf{\hat{z}}$ | (6$c$) | $H_2O$ III |
| $\mathbf{B}_{22}$ | = | $x_5\,\mathbf{a}_1 + 2x_5\,\mathbf{a}_2 + z_5\,\mathbf{a}_3$ | = | $\frac{3}{2}x_5 a\,\mathbf{\hat{x}} + \frac{\sqrt{3}}{2}x_5 a\,\mathbf{\hat{y}} + z_5 c\,\mathbf{\hat{z}}$ | (6$c$) | $H_2O$ III |
| $\mathbf{B}_{23}$ | = | $-2x_5\,\mathbf{a}_1 - x_5\,\mathbf{a}_2 + z_5\,\mathbf{a}_3$ | = | $-\frac{3}{2}x_5 a\,\mathbf{\hat{x}} + \frac{\sqrt{3}}{2}x_5 a\,\mathbf{\hat{y}} + z_5 c\,\mathbf{\hat{z}}$ | (6$c$) | $H_2O$ III |
| $\mathbf{B}_{24}$ | = | $-x_5\,\mathbf{a}_1 + x_5\,\mathbf{a}_2 + \left(\frac{1}{2} + z_5\right)\,\mathbf{a}_3$ | = | $\sqrt{3}x_5 a\,\mathbf{\hat{y}} + \left(\frac{1}{2} + z_5\right)c\,\mathbf{\hat{z}}$ | (6$c$) | $H_2O$ III |
| $\mathbf{B}_{25}$ | = | $-x_5\,\mathbf{a}_1 - 2x_5\,\mathbf{a}_2 + \left(\frac{1}{2} + z_5\right)\,\mathbf{a}_3$ | = | $-\frac{3}{2}x_5 a\,\mathbf{\hat{x}} - \frac{\sqrt{3}}{2}x_5 a\,\mathbf{\hat{y}} + \left(\frac{1}{2} + z_5\right)c\,\mathbf{\hat{z}}$ | (6$c$) | $H_2O$ III |
| $\mathbf{B}_{26}$ | = | $2x_5\,\mathbf{a}_1 + x_5\,\mathbf{a}_2 + \left(\frac{1}{2} + z_5\right)\,\mathbf{a}_3$ | = | $\frac{3}{2}x_5 a\,\mathbf{\hat{x}} - \frac{\sqrt{3}}{2}x_5 a\,\mathbf{\hat{y}} + \left(\frac{1}{2} + z_5\right)c\,\mathbf{\hat{z}}$ | (6$c$) | $H_2O$ III |
| $\mathbf{B}_{27}$ | = | $x_6\,\mathbf{a}_1 - x_6\,\mathbf{a}_2 + z_6\,\mathbf{a}_3$ | = | $-\sqrt{3}x_6 a\,\mathbf{\hat{y}} + z_6 c\,\mathbf{\hat{z}}$ | (6$c$) | O I |
| $\mathbf{B}_{28}$ | = | $x_6\,\mathbf{a}_1 + 2x_6\,\mathbf{a}_2 + z_6\,\mathbf{a}_3$ | = | $\frac{3}{2}x_6 a\,\mathbf{\hat{x}} + \frac{\sqrt{3}}{2}x_6 a\,\mathbf{\hat{y}} + z_6 c\,\mathbf{\hat{z}}$ | (6$c$) | O I |
| $\mathbf{B}_{29}$ | = | $-2x_6\,\mathbf{a}_1 - x_6\,\mathbf{a}_2 + z_6\,\mathbf{a}_3$ | = | $-\frac{3}{2}x_6 a\,\hat{\mathbf{x}} + \frac{\sqrt{3}}{2}x_6 a\,\hat{\mathbf{y}} + z_6 c\,\hat{\mathbf{z}}$ | (6c) | O I |
| $\mathbf{B}_{30}$ | = | $-x_6\,\mathbf{a}_1 + x_6\,\mathbf{a}_2 + \left(\frac{1}{2} + z_6\right)\mathbf{a}_3$ | = | $\sqrt{3}x_6 a\,\hat{\mathbf{y}} + \left(\frac{1}{2} + z_6\right)c\,\hat{\mathbf{z}}$ | (6c) | O I |
| $\mathbf{B}_{31}$ | = | $-x_6\,\mathbf{a}_1 - 2x_6\,\mathbf{a}_2 + \left(\frac{1}{2} + z_6\right)\mathbf{a}_3$ | = | $-\frac{3}{2}x_6 a\,\hat{\mathbf{x}} - \frac{\sqrt{3}}{2}x_6 a\,\hat{\mathbf{y}} + \left(\frac{1}{2} + z_6\right)c\,\hat{\mathbf{z}}$ | (6c) | O I |
| $\mathbf{B}_{32}$ | = | $2x_6\,\mathbf{a}_1 + x_6\,\mathbf{a}_2 + \left(\frac{1}{2} + z_6\right)\mathbf{a}_3$ | = | $\frac{3}{2}x_6 a\,\hat{\mathbf{x}} - \frac{\sqrt{3}}{2}x_6 a\,\hat{\mathbf{y}} + \left(\frac{1}{2} + z_6\right)c\,\hat{\mathbf{z}}$ | (6c) | O I |
| $\mathbf{B}_{33}$ | = | $x_7\,\mathbf{a}_1 + y_7\,\mathbf{a}_2 + z_7\,\mathbf{a}_3$ | = | $\frac{1}{2}(x_7 + y_7)\,a\,\hat{\mathbf{x}} + \frac{\sqrt{3}}{2}(-x_7 + y_7)\,a\,\hat{\mathbf{y}} + z_7 c\,\hat{\mathbf{z}}$ | (12d) | O II |
| $\mathbf{B}_{34}$ | = | $-y_7\,\mathbf{a}_1 + (x_7 - y_7)\,\mathbf{a}_2 + z_7\,\mathbf{a}_3$ | = | $\left(\frac{1}{2}x_7 - y_7\right)a\,\hat{\mathbf{x}} + \frac{\sqrt{3}}{2}x_7 a\,\hat{\mathbf{y}} + z_7 c\,\hat{\mathbf{z}}$ | (12d) | O II |
| $\mathbf{B}_{35}$ | = | $(-x_7 + y_7)\,\mathbf{a}_1 - x_7\,\mathbf{a}_2 + z_7\,\mathbf{a}_3$ | = | $\left(-x_7 + \frac{1}{2}y_7\right)a\,\hat{\mathbf{x}} - \frac{\sqrt{3}}{2}y_7 a\,\hat{\mathbf{y}} + z_7 c\,\hat{\mathbf{z}}$ | (12d) | O II |
| $\mathbf{B}_{36}$ | = | $-x_7\,\mathbf{a}_1 - y_7\,\mathbf{a}_2 + \left(\frac{1}{2} + z_7\right)\mathbf{a}_3$ | = | $-\frac{1}{2}(x_7 + y_7)\,a\,\hat{\mathbf{x}} + \frac{\sqrt{3}}{2}(x_7 - y_7)\,a\,\hat{\mathbf{y}} + \left(\frac{1}{2} + z_7\right)c\,\hat{\mathbf{z}}$ | (12d) | O II |
| $\mathbf{B}_{37}$ | = | $y_7\,\mathbf{a}_1 + (-x_7 + y_7)\,\mathbf{a}_2 + \left(\frac{1}{2} + z_7\right)\mathbf{a}_3$ | = | $\left(-\frac{1}{2}x_7 + y_7\right)a\,\hat{\mathbf{x}} - \frac{\sqrt{3}}{2}x_7 a\,\hat{\mathbf{y}} + \left(\frac{1}{2} + z_7\right)c\,\hat{\mathbf{z}}$ | (12d) | O II |
| $\mathbf{B}_{38}$ | = | $(x_7 - y_7)\,\mathbf{a}_1 + x_7\,\mathbf{a}_2 + \left(\frac{1}{2} + z_7\right)\mathbf{a}_3$ | = | $\left(x_7 - \frac{1}{2}y_7\right)a\,\hat{\mathbf{x}} + \frac{\sqrt{3}}{2}y_7 a\,\hat{\mathbf{y}} + \left(\frac{1}{2} + z_7\right)c\,\hat{\mathbf{z}}$ | (12d) | O II |
| $\mathbf{B}_{39}$ | = | $-y_7\,\mathbf{a}_1 - x_7\,\mathbf{a}_2 + z_7\,\mathbf{a}_3$ | = | $-\frac{1}{2}(x_7 + y_7)\,a\,\hat{\mathbf{x}} + \frac{\sqrt{3}}{2}(-x_7 + y_7)\,a\,\hat{\mathbf{y}} + z_7 c\,\hat{\mathbf{z}}$ | (12d) | O II |
| $\mathbf{B}_{40}$ | = | $(-x_7 + y_7)\,\mathbf{a}_1 + y_7\,\mathbf{a}_2 + z_7\,\mathbf{a}_3$ | = | $\left(-\frac{1}{2}x_7 + y_7\right)a\,\hat{\mathbf{x}} + \frac{\sqrt{3}}{2}x_7 a\,\hat{\mathbf{y}} + z_7 c\,\hat{\mathbf{z}}$ | (12d) | O II |
| $\mathbf{B}_{41}$ | = | $x_7\,\mathbf{a}_1 + (x_7 - y_7)\,\mathbf{a}_2 + z_7\,\mathbf{a}_3$ | = | $\left(x_7 - \frac{1}{2}y_7\right)a\,\hat{\mathbf{x}} - \frac{\sqrt{3}}{2}y_7 a\,\hat{\mathbf{y}} + z_7 c\,\hat{\mathbf{z}}$ | (12d) | O II |
| $\mathbf{B}_{42}$ | = | $y_7\,\mathbf{a}_1 + x_7\,\mathbf{a}_2 + \left(\frac{1}{2} + z_7\right)\mathbf{a}_3$ | = | $\frac{1}{2}(x_7 + y_7)\,a\,\hat{\mathbf{x}} + \frac{\sqrt{3}}{2}(x_7 - y_7)\,a\,\hat{\mathbf{y}} + \left(\frac{1}{2} + z_7\right)c\,\hat{\mathbf{z}}$ | (12d) | O II |
| $\mathbf{B}_{43}$ | = | $(x_7 - y_7)\,\mathbf{a}_1 - y_7\,\mathbf{a}_2 + \left(\frac{1}{2} + z_7\right)\mathbf{a}_3$ | = | $\left(\frac{1}{2}x_7 - y_7\right)a\,\hat{\mathbf{x}} - \frac{\sqrt{3}}{2}x_7 a\,\hat{\mathbf{y}} + \left(\frac{1}{2} + z_7\right)c\,\hat{\mathbf{z}}$ | (12d) | O II |
| $\mathbf{B}_{44}$ | = | $-x_7\,\mathbf{a}_1 + (-x_7 + y_7)\,\mathbf{a}_2 + \left(\frac{1}{2} + z_7\right)\mathbf{a}_3$ | = | $\left(-x_7 + \frac{1}{2}y_7\right)a\,\hat{\mathbf{x}} + \frac{\sqrt{3}}{2}y_7 a\,\hat{\mathbf{y}} + \left(\frac{1}{2} + z_7\right)c\,\hat{\mathbf{z}}$ | (12d) | O II |

**Geometry files:**

- CIF: pp. 1757
- POSCAR: pp. 1757

# Swedenborgite ($NaBe_4SbO_7$, $E9_2$) Structure: A4BC7D_hP26_186_ac_b_a2c_b

http://aflow.org/prototype-encyclopedia/A4BC7D_hP26_186_ac_b_a2c_b

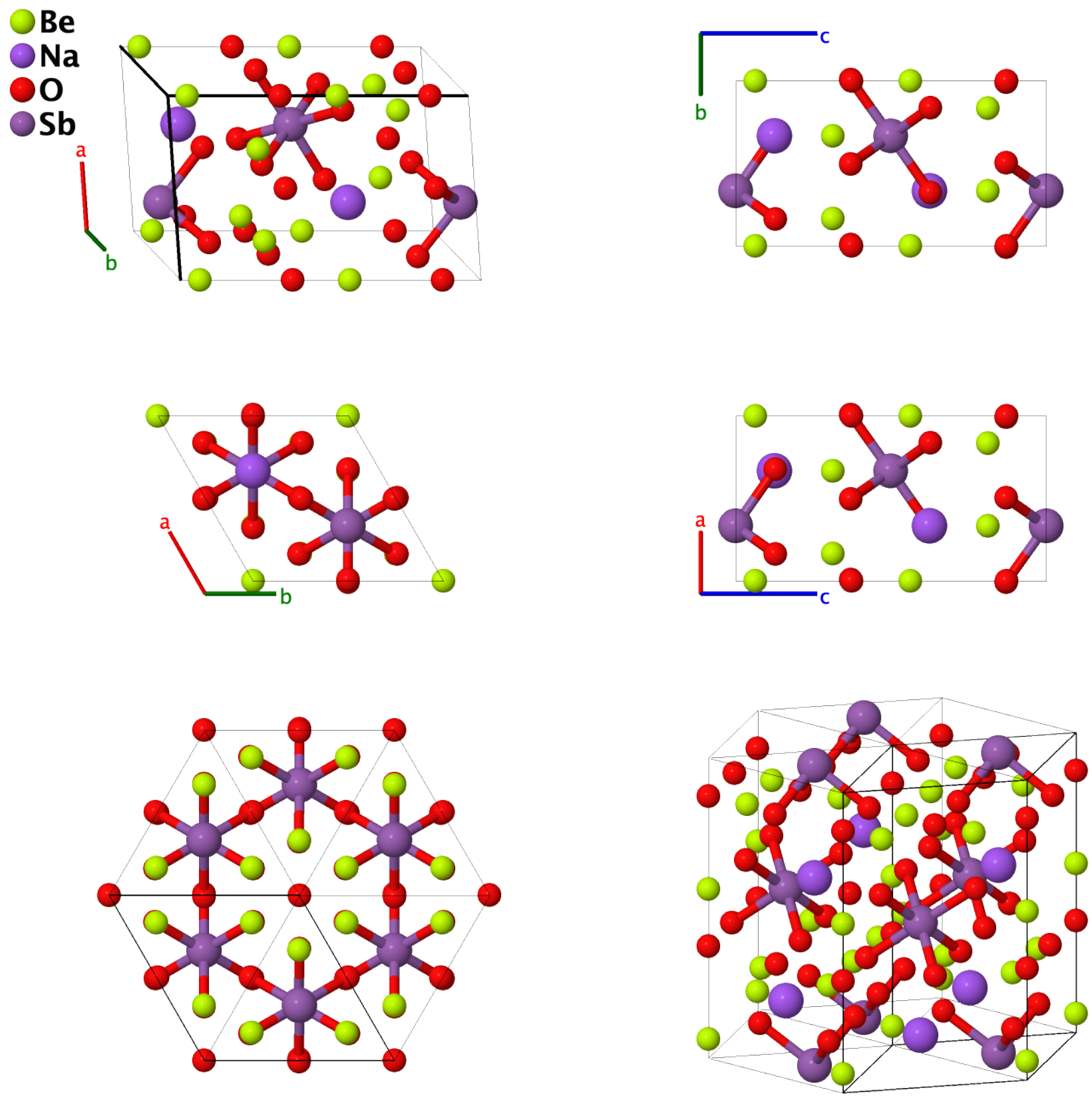

| | | |
|---|---|---|
| **Prototype** | : | $Be_4NaO_7Sb$ |
| **AFLOW prototype label** | : | A4BC7D_hP26_186_ac_b_a2c_b |
| ***Strukturbericht* designation** | : | $E9_2$ |
| **Pearson symbol** | : | hP26 |
| **Space group number** | : | 186 |
| **Space group symbol** | : | $P6_3mc$ |
| **AFLOW prototype command** | : | `aflow --proto=A4BC7D_hP26_186_ac_b_a2c_b --params=`$a, c/a, z_1, z_2, z_3, z_4, x_5, z_5, x_6, z_6, x_7, z_7$ |

### Other compounds with this structure

- $CaBa(Fe_2Mn_2)O_7$, $CaBaCo_3AlO_7$, $CaBaCo_3FeO_7$, $CaBaCo_3ZnO_7$, $CaBaCo_4O_7$, $CaBaFe_4O_7$, $Er_2Si_4N_6C$, $Ho_2Si_4N_6C$, $Tb_2Si_4N_6C$, $Y_2Si_4N_6C$, $YBaCo_3AlO_7$, $YBaCo_3FeO_7$, $YBaCo_4O_7$, $YBaMn_3AlO_7$, $Yb_2Si_4N_6C$, and $YbBaSi_4N_7$

---

- The actual composition of the studied sample is ($Na_{0.89}Ca_{0.04}Be_4SbO_7$), with the sodium (Na-I) site containing 89% sodium, 4% calcium, and 7% vacancies.
- Space group $P6_3mc$ #186 does not fix the origin of the $z$-coordinate. Here the position of the antimony atom (Sb-I) is set so that $z_7 = 0$.

## Hexagonal primitive vectors:

$$\mathbf{a}_1 = \frac{1}{2}\,a\,\hat{\mathbf{x}} - \frac{\sqrt{3}}{2}\,a\,\hat{\mathbf{y}}$$
$$\mathbf{a}_2 = \frac{1}{2}\,a\,\hat{\mathbf{x}} + \frac{\sqrt{3}}{2}\,a\,\hat{\mathbf{y}}$$
$$\mathbf{a}_3 = c\,\hat{\mathbf{z}}$$

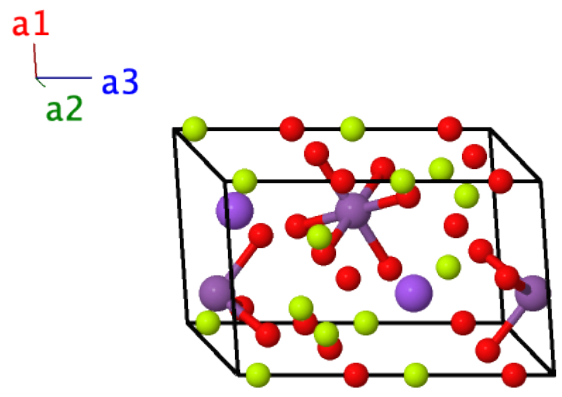

## Basis vectors:

| | | Lattice Coordinates | | Cartesian Coordinates | Wyckoff Position | Atom Type |
|---|---|---|---|---|---|---|
| $\mathbf{B}_1$ | = | $z_1\,\mathbf{a}_3$ | = | $z_1 c\,\hat{\mathbf{z}}$ | (2$a$) | Be I |
| $\mathbf{B}_2$ | = | $\left(\frac{1}{2}+z_1\right)\mathbf{a}_3$ | = | $\left(\frac{1}{2}+z_1\right)c\,\hat{\mathbf{z}}$ | (2$a$) | Be I |
| $\mathbf{B}_3$ | = | $z_2\,\mathbf{a}_3$ | = | $z_2 c\,\hat{\mathbf{z}}$ | (2$a$) | O I |
| $\mathbf{B}_4$ | = | $\left(\frac{1}{2}+z_2\right)\mathbf{a}_3$ | = | $\left(\frac{1}{2}+z_2\right)c\,\hat{\mathbf{z}}$ | (2$a$) | O I |
| $\mathbf{B}_5$ | = | $\frac{1}{3}\,\mathbf{a}_1+\frac{2}{3}\,\mathbf{a}_2+z_3\,\mathbf{a}_3$ | = | $\frac{1}{2}a\,\hat{\mathbf{x}}+\frac{1}{2\sqrt{3}}a\,\hat{\mathbf{y}}+z_3 c\,\hat{\mathbf{z}}$ | (2$b$) | Na |
| $\mathbf{B}_6$ | = | $\frac{2}{3}\,\mathbf{a}_1+\frac{1}{3}\,\mathbf{a}_2+\left(\frac{1}{2}+z_3\right)\mathbf{a}_3$ | = | $\frac{1}{2}a\,\hat{\mathbf{x}}-\frac{1}{2\sqrt{3}}a\,\hat{\mathbf{y}}+\left(\frac{1}{2}+z_3\right)c\,\hat{\mathbf{z}}$ | (2$b$) | Na |
| $\mathbf{B}_7$ | = | $\frac{1}{3}\,\mathbf{a}_1+\frac{2}{3}\,\mathbf{a}_2+z_4\,\mathbf{a}_3$ | = | $\frac{1}{2}a\,\hat{\mathbf{x}}+\frac{1}{2\sqrt{3}}a\,\hat{\mathbf{y}}+z_4 c\,\hat{\mathbf{z}}$ | (2$b$) | Sb |
| $\mathbf{B}_8$ | = | $\frac{2}{3}\,\mathbf{a}_1+\frac{1}{3}\,\mathbf{a}_2+\left(\frac{1}{2}+z_4\right)\mathbf{a}_3$ | = | $\frac{1}{2}a\,\hat{\mathbf{x}}-\frac{1}{2\sqrt{3}}a\,\hat{\mathbf{y}}+\left(\frac{1}{2}+z_4\right)c\,\hat{\mathbf{z}}$ | (2$b$) | Sb |
| $\mathbf{B}_9$ | = | $x_5\,\mathbf{a}_1-x_5\,\mathbf{a}_2+z_5\,\mathbf{a}_3$ | = | $-\sqrt{3}x_5 a\,\hat{\mathbf{y}}+z_5 c\,\hat{\mathbf{z}}$ | (6$c$) | Be II |
| $\mathbf{B}_{10}$ | = | $x_5\,\mathbf{a}_1+2x_5\,\mathbf{a}_2+z_5\,\mathbf{a}_3$ | = | $\frac{3}{2}x_5 a\,\hat{\mathbf{x}}+\frac{\sqrt{3}}{2}x_5 a\,\hat{\mathbf{y}}+z_5 c\,\hat{\mathbf{z}}$ | (6$c$) | Be II |
| $\mathbf{B}_{11}$ | = | $-2x_5\,\mathbf{a}_1-x_5\,\mathbf{a}_2+z_5\,\mathbf{a}_3$ | = | $-\frac{3}{2}x_5 a\,\hat{\mathbf{x}}+\frac{\sqrt{3}}{2}x_5 a\,\hat{\mathbf{y}}+z_5 c\,\hat{\mathbf{z}}$ | (6$c$) | Be II |
| $\mathbf{B}_{12}$ | = | $-x_5\,\mathbf{a}_1+x_5\,\mathbf{a}_2+\left(\frac{1}{2}+z_5\right)\mathbf{a}_3$ | = | $\sqrt{3}x_5 a\,\hat{\mathbf{y}}+\left(\frac{1}{2}+z_5\right)c\,\hat{\mathbf{z}}$ | (6$c$) | Be II |
| $\mathbf{B}_{13}$ | = | $-x_5\,\mathbf{a}_1-2x_5\,\mathbf{a}_2+\left(\frac{1}{2}+z_5\right)\mathbf{a}_3$ | = | $-\frac{3}{2}x_5 a\,\hat{\mathbf{x}}-\frac{\sqrt{3}}{2}x_5 a\,\hat{\mathbf{y}}+\left(\frac{1}{2}+z_5\right)c\,\hat{\mathbf{z}}$ | (6$c$) | Be II |
| $\mathbf{B}_{14}$ | = | $2x_5\,\mathbf{a}_1+x_5\,\mathbf{a}_2+\left(\frac{1}{2}+z_5\right)\mathbf{a}_3$ | = | $\frac{3}{2}x_5 a\,\hat{\mathbf{x}}-\frac{\sqrt{3}}{2}x_5 a\,\hat{\mathbf{y}}+\left(\frac{1}{2}+z_5\right)c\,\hat{\mathbf{z}}$ | (6$c$) | Be II |
| $\mathbf{B}_{15}$ | = | $x_6\,\mathbf{a}_1-x_6\,\mathbf{a}_2+z_6\,\mathbf{a}_3$ | = | $-\sqrt{3}x_6 a\,\hat{\mathbf{y}}+z_6 c\,\hat{\mathbf{z}}$ | (6$c$) | O II |
| $\mathbf{B}_{16}$ | = | $x_6\,\mathbf{a}_1+2x_6\,\mathbf{a}_2+z_6\,\mathbf{a}_3$ | = | $\frac{3}{2}x_6 a\,\hat{\mathbf{x}}+\frac{\sqrt{3}}{2}x_6 a\,\hat{\mathbf{y}}+z_6 c\,\hat{\mathbf{z}}$ | (6$c$) | O II |
| $\mathbf{B}_{17}$ | = | $-2x_6\,\mathbf{a}_1-x_6\,\mathbf{a}_2+z_6\,\mathbf{a}_3$ | = | $-\frac{3}{2}x_6 a\,\hat{\mathbf{x}}+\frac{\sqrt{3}}{2}x_6 a\,\hat{\mathbf{y}}+z_6 c\,\hat{\mathbf{z}}$ | (6$c$) | O II |
| $\mathbf{B}_{18}$ | = | $-x_6\,\mathbf{a}_1+x_6\,\mathbf{a}_2+\left(\frac{1}{2}+z_6\right)\mathbf{a}_3$ | = | $\sqrt{3}x_6 a\,\hat{\mathbf{y}}+\left(\frac{1}{2}+z_6\right)c\,\hat{\mathbf{z}}$ | (6$c$) | O II |
| $\mathbf{B}_{19}$ | = | $-x_6\,\mathbf{a}_1-2x_6\,\mathbf{a}_2+\left(\frac{1}{2}+z_6\right)\mathbf{a}_3$ | = | $-\frac{3}{2}x_6 a\,\hat{\mathbf{x}}-\frac{\sqrt{3}}{2}x_6 a\,\hat{\mathbf{y}}+\left(\frac{1}{2}+z_6\right)c\,\hat{\mathbf{z}}$ | (6$c$) | O II |
| $\mathbf{B}_{20}$ | = | $2x_6\,\mathbf{a}_1+x_6\,\mathbf{a}_2+\left(\frac{1}{2}+z_6\right)\mathbf{a}_3$ | = | $\frac{3}{2}x_6 a\,\hat{\mathbf{x}}-\frac{\sqrt{3}}{2}x_6 a\,\hat{\mathbf{y}}+\left(\frac{1}{2}+z_6\right)c\,\hat{\mathbf{z}}$ | (6$c$) | O II |
| $\mathbf{B}_{21}$ | = | $x_7\,\mathbf{a}_1-x_7\,\mathbf{a}_2+z_7\,\mathbf{a}_3$ | = | $-\sqrt{3}x_7 a\,\hat{\mathbf{y}}+z_7 c\,\hat{\mathbf{z}}$ | (6$c$) | O III |
| $\mathbf{B}_{22}$ | = | $x_7\,\mathbf{a}_1+2x_7\,\mathbf{a}_2+z_7\,\mathbf{a}_3$ | = | $\frac{3}{2}x_7 a\,\hat{\mathbf{x}}+\frac{\sqrt{3}}{2}x_7 a\,\hat{\mathbf{y}}+z_7 c\,\hat{\mathbf{z}}$ | (6$c$) | O III |
| $\mathbf{B}_{23}$ | = | $-2x_7\,\mathbf{a}_1-x_7\,\mathbf{a}_2+z_7\,\mathbf{a}_3$ | = | $-\frac{3}{2}x_7 a\,\hat{\mathbf{x}}+\frac{\sqrt{3}}{2}x_7 a\,\hat{\mathbf{y}}+z_7 c\,\hat{\mathbf{z}}$ | (6$c$) | O III |
| $\mathbf{B}_{24}$ | = | $-x_7\,\mathbf{a}_1+x_7\,\mathbf{a}_2+\left(\frac{1}{2}+z_7\right)\mathbf{a}_3$ | = | $\sqrt{3}x_7 a\,\hat{\mathbf{y}}+\left(\frac{1}{2}+z_7\right)c\,\hat{\mathbf{z}}$ | (6$c$) | O III |
| $\mathbf{B}_{25}$ | = | $-x_7\,\mathbf{a}_1-2x_7\,\mathbf{a}_2+\left(\frac{1}{2}+z_7\right)\mathbf{a}_3$ | = | $-\frac{3}{2}x_7 a\,\hat{\mathbf{x}}-\frac{\sqrt{3}}{2}x_7 a\,\hat{\mathbf{y}}+\left(\frac{1}{2}+z_7\right)c\,\hat{\mathbf{z}}$ | (6$c$) | O III |
| $\mathbf{B}_{26}$ | = | $2x_7\,\mathbf{a}_1+x_7\,\mathbf{a}_2+\left(\frac{1}{2}+z_7\right)\mathbf{a}_3$ | = | $\frac{3}{2}x_7 a\,\hat{\mathbf{x}}-\frac{\sqrt{3}}{2}x_7 a\,\hat{\mathbf{y}}+\left(\frac{1}{2}+z_7\right)c\,\hat{\mathbf{z}}$ | (6$c$) | O III |

---

**Geometry files:**
- CIF: pp. 1757
- POSCAR: pp. 1758

# $C27$ ($CdI_2$) (*questionable*) Structure: AB2_hP6_186_b_ab

http://aflow.org/prototype-encyclopedia/AB2_hP6_186_b_ab

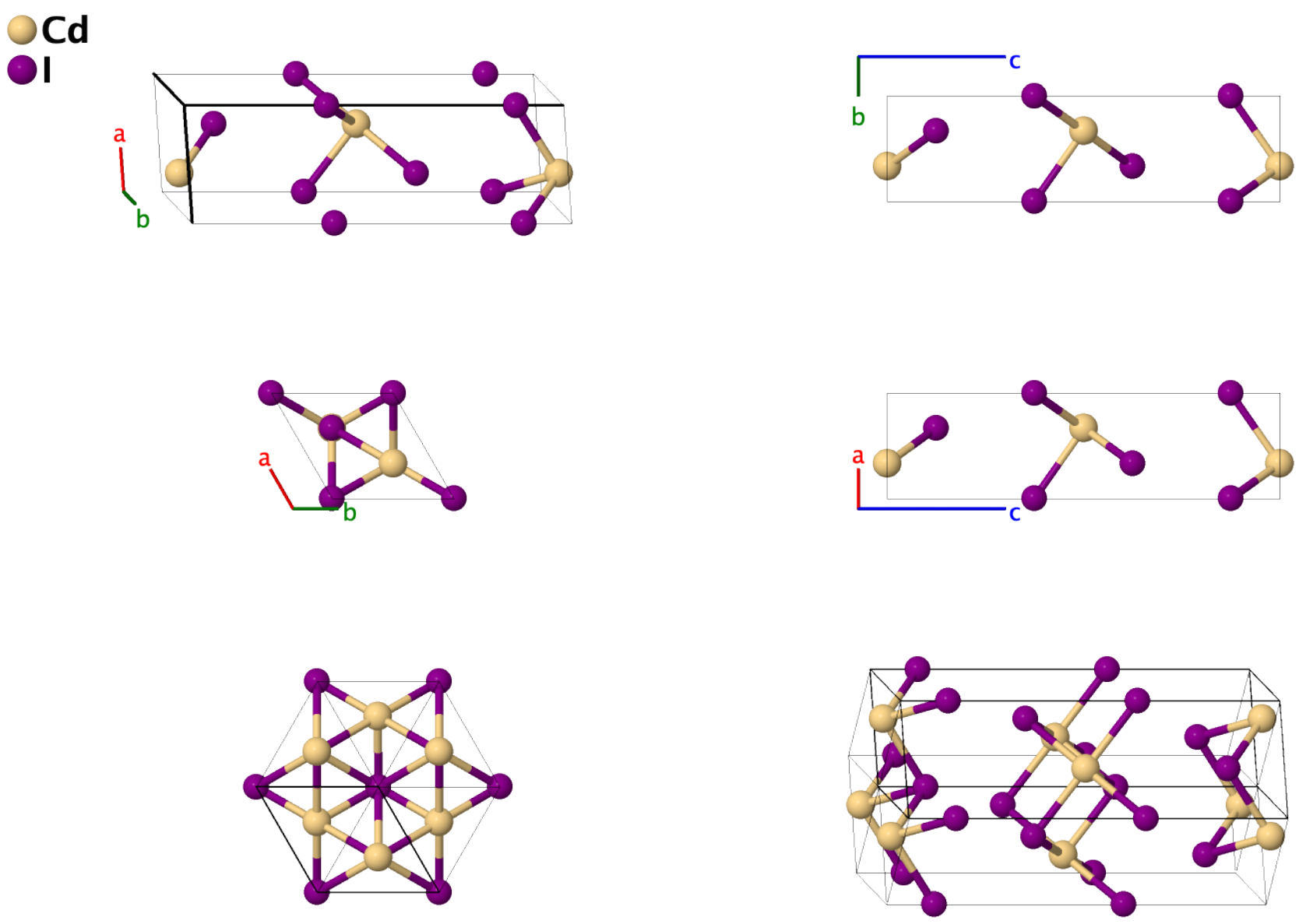

| | | |
|---|---|---|
| **Prototype** | : | $CdI_2$ |
| **AFLOW prototype label** | : | AB2_hP6_186_b_ab |
| ***Strukturbericht* designation** | : | $C27$ |
| **Pearson symbol** | : | hP6 |
| **Space group number** | : | 186 |
| **Space group symbol** | : | $P6_3mc$ |
| **AFLOW prototype command** | : | `aflow --proto=AB2_hP6_186_b_ab --params=`$a, c/a, z_1, z_2, z_3$ |

- This is a modification of the $C6$ ($\omega$ phase) $CdI_2$ structure proposed by (Hassel, 1933) to explain extra lines found in the x-ray spectra. This structure never was accepted by most researchers and does not appear in modern lists of *Strukturbericht* symbols, although (Parthé, 1993) lists it as a stacking varient of $CdI_2$. We include it here as part of the historical record.
- Hassel originally placed a Cd atom at the origin, however both (Gottfried, 1937) and FINDSYM shift the origin as shown.

**Hexagonal primitive vectors:**

$$\mathbf{a}_1 = \frac{1}{2}\,a\,\hat{\mathbf{x}} - \frac{\sqrt{3}}{2}\,a\,\hat{\mathbf{y}}$$
$$\mathbf{a}_2 = \frac{1}{2}\,a\,\hat{\mathbf{x}} + \frac{\sqrt{3}}{2}\,a\,\hat{\mathbf{y}}$$
$$\mathbf{a}_3 = c\,\hat{\mathbf{z}}$$

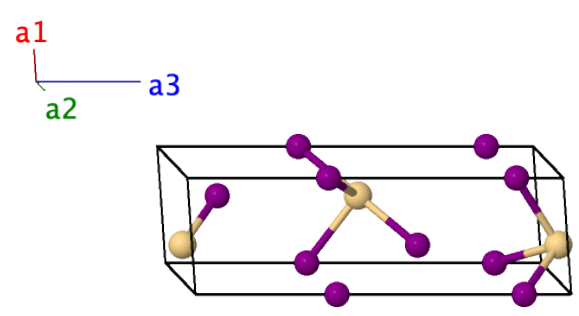

**Basis vectors:**

| | | Lattice Coordinates | | Cartesian Coordinates | Wyckoff Position | Atom Type |
|---|---|---|---|---|---|---|
| $\mathbf{B}_1$ | = | $z_1\,\mathbf{a}_3$ | = | $z_1 c\,\mathbf{\hat{z}}$ | (2$a$) | I I |
| $\mathbf{B}_2$ | = | $\left(\frac{1}{2}+z_1\right)\mathbf{a}_3$ | = | $\left(\frac{1}{2}+z_1\right)c\,\mathbf{\hat{z}}$ | (2$a$) | I I |
| $\mathbf{B}_3$ | = | $\frac{1}{3}\,\mathbf{a}_1+\frac{2}{3}\,\mathbf{a}_2+z_2\,\mathbf{a}_3$ | = | $\frac{1}{2}a\,\mathbf{\hat{x}}+\frac{1}{2\sqrt{3}}a\,\mathbf{\hat{y}}+z_2 c\,\mathbf{\hat{z}}$ | (2$b$) | Cd |
| $\mathbf{B}_4$ | = | $\frac{2}{3}\,\mathbf{a}_1+\frac{1}{3}\,\mathbf{a}_2+\left(\frac{1}{2}+z_2\right)\mathbf{a}_3$ | = | $\frac{1}{2}a\,\mathbf{\hat{x}}-\frac{1}{2\sqrt{3}}a\,\mathbf{\hat{y}}+\left(\frac{1}{2}+z_2\right)c\,\mathbf{\hat{z}}$ | (2$b$) | Cd |
| $\mathbf{B}_5$ | = | $\frac{1}{3}\,\mathbf{a}_1+\frac{2}{3}\,\mathbf{a}_2+z_3\,\mathbf{a}_3$ | = | $\frac{1}{2}a\,\mathbf{\hat{x}}+\frac{1}{2\sqrt{3}}a\,\mathbf{\hat{y}}+z_3 c\,\mathbf{\hat{z}}$ | (2$b$) | I II |
| $\mathbf{B}_6$ | = | $\frac{2}{3}\,\mathbf{a}_1+\frac{1}{3}\,\mathbf{a}_2+\left(\frac{1}{2}+z_3\right)\mathbf{a}_3$ | = | $\frac{1}{2}a\,\mathbf{\hat{x}}-\frac{1}{2\sqrt{3}}a\,\mathbf{\hat{y}}+\left(\frac{1}{2}+z_3\right)c\,\mathbf{\hat{z}}$ | (2$b$) | I II |

---

---

**Geometry files:**

- CIF: pp. 1758
- POSCAR: pp. 1758

# $LiClO_4 \cdot 3H_2O$ ($H4_{18}$) Structure: AB6CD7_hP30_186_b_d_a_b2c

http://aflow.org/prototype-encyclopedia/AB6CD7_hP30_186_b_d_a_b2c

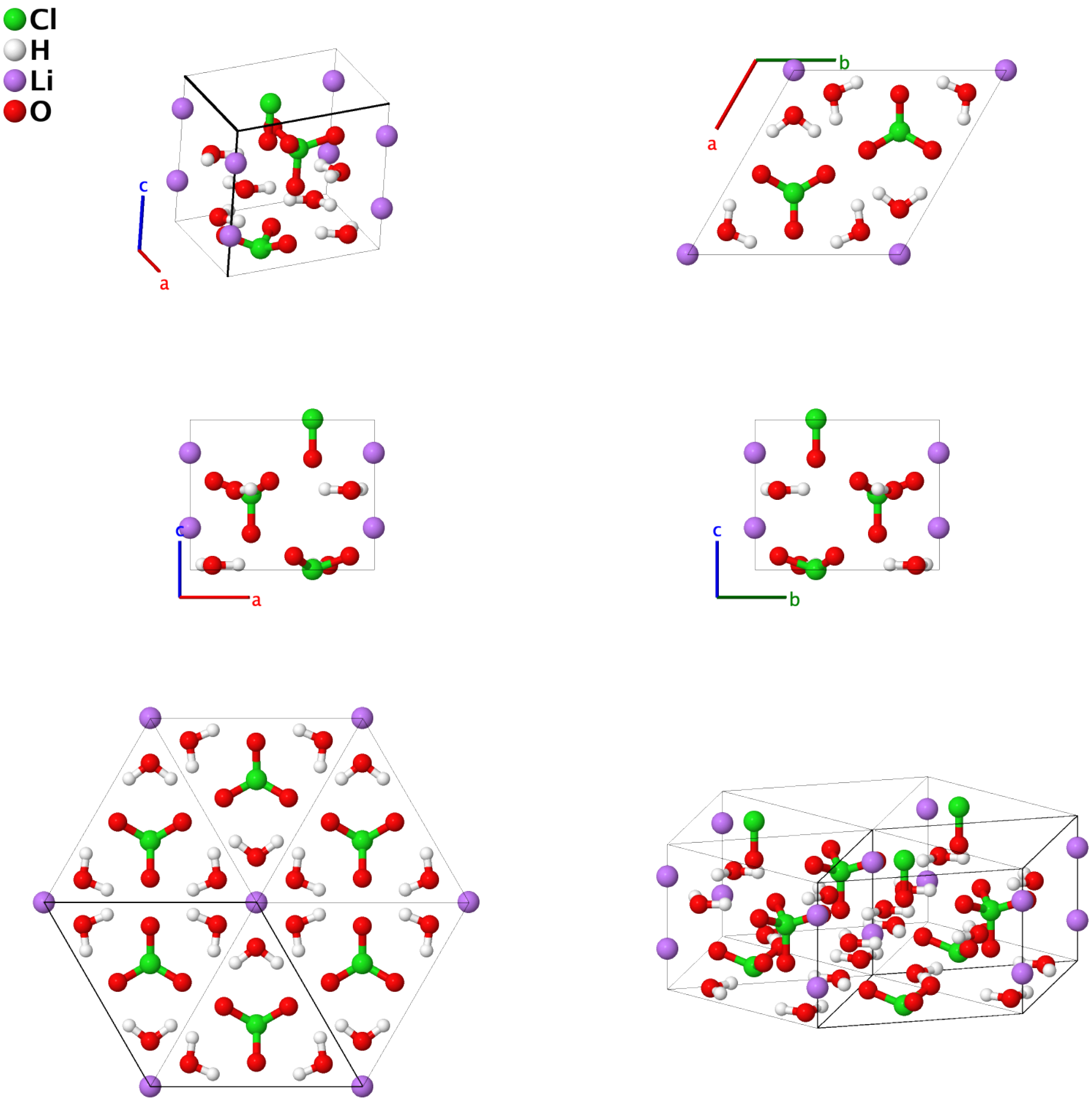

| | | |
|---|---|---|
| **Prototype** | : | $ClH_6LiO_7$ |
| **AFLOW prototype label** | : | AB6CD7_hP30_186_b_d_a_b2c |
| ***Strukturbericht* designation** | : | $H4_{18}$ |
| **Pearson symbol** | : | hP30 |
| **Space group number** | : | 186 |
| **Space group symbol** | : | $P6_3mc$ |
| **AFLOW prototype command** | : | `aflow --proto=AB6CD7_hP30_186_b_d_a_b2c --params=`$a, c/a, z_1, z_2, z_3, x_4, z_4, x_5, z_5, x_6, y_6, z_6$ |

### Other compounds with this structure

- $LiTcO_4 \cdot 3H_2O$

## Hexagonal primitive vectors:

$$\mathbf{a}_1 = \frac{1}{2}\,a\,\hat{\mathbf{x}} - \frac{\sqrt{3}}{2}\,a\,\hat{\mathbf{y}}$$
$$\mathbf{a}_2 = \frac{1}{2}\,a\,\hat{\mathbf{x}} + \frac{\sqrt{3}}{2}\,a\,\hat{\mathbf{y}}$$
$$\mathbf{a}_3 = c\,\hat{\mathbf{z}}$$

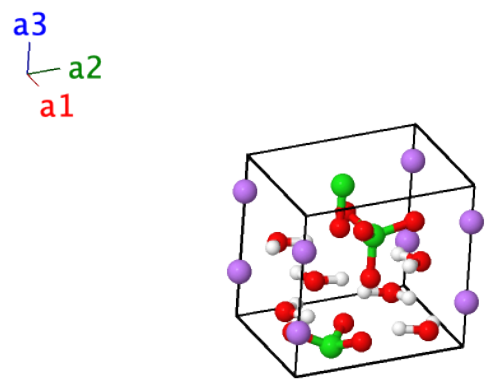

## Basis vectors:

| | | Lattice Coordinates | | Cartesian Coordinates | Wyckoff Position | Atom Type |
|---|---|---|---|---|---|---|
| $\mathbf{B}_1$ | = | $z_1\,\mathbf{a}_3$ | = | $z_1 c\,\hat{\mathbf{z}}$ | (2$a$) | Li |
| $\mathbf{B}_2$ | = | $\left(\frac{1}{2}+z_1\right)\mathbf{a}_3$ | = | $\left(\frac{1}{2}+z_1\right)c\,\hat{\mathbf{z}}$ | (2$a$) | Li |
| $\mathbf{B}_3$ | = | $\frac{1}{3}\,\mathbf{a}_1 + \frac{2}{3}\,\mathbf{a}_2 + z_2\,\mathbf{a}_3$ | = | $\frac{1}{2}a\,\hat{\mathbf{x}} + \frac{1}{2\sqrt{3}}a\,\hat{\mathbf{y}} + z_2 c\,\hat{\mathbf{z}}$ | (2$b$) | Cl |
| $\mathbf{B}_4$ | = | $\frac{2}{3}\,\mathbf{a}_1 + \frac{1}{3}\,\mathbf{a}_2 + \left(\frac{1}{2}+z_2\right)\mathbf{a}_3$ | = | $\frac{1}{2}a\,\hat{\mathbf{x}} - \frac{1}{2\sqrt{3}}a\,\hat{\mathbf{y}} + \left(\frac{1}{2}+z_2\right)c\,\hat{\mathbf{z}}$ | (2$b$) | Cl |
| $\mathbf{B}_5$ | = | $\frac{1}{3}\,\mathbf{a}_1 + \frac{2}{3}\,\mathbf{a}_2 + z_3\,\mathbf{a}_3$ | = | $\frac{1}{2}a\,\hat{\mathbf{x}} + \frac{1}{2\sqrt{3}}a\,\hat{\mathbf{y}} + z_3 c\,\hat{\mathbf{z}}$ | (2$b$) | O I |
| $\mathbf{B}_6$ | = | $\frac{2}{3}\,\mathbf{a}_1 + \frac{1}{3}\,\mathbf{a}_2 + \left(\frac{1}{2}+z_3\right)\mathbf{a}_3$ | = | $\frac{1}{2}a\,\hat{\mathbf{x}} - \frac{1}{2\sqrt{3}}a\,\hat{\mathbf{y}} + \left(\frac{1}{2}+z_3\right)c\,\hat{\mathbf{z}}$ | (2$b$) | O I |
| $\mathbf{B}_7$ | = | $x_4\,\mathbf{a}_1 - x_4\,\mathbf{a}_2 + z_4\,\mathbf{a}_3$ | = | $-\sqrt{3}x_4 a\,\hat{\mathbf{y}} + z_4 c\,\hat{\mathbf{z}}$ | (6$c$) | O II |
| $\mathbf{B}_8$ | = | $x_4\,\mathbf{a}_1 + 2x_4\,\mathbf{a}_2 + z_4\,\mathbf{a}_3$ | = | $\frac{3}{2}x_4 a\,\hat{\mathbf{x}} + \frac{\sqrt{3}}{2}x_4 a\,\hat{\mathbf{y}} + z_4 c\,\hat{\mathbf{z}}$ | (6$c$) | O II |
| $\mathbf{B}_9$ | = | $-2x_4\,\mathbf{a}_1 - x_4\,\mathbf{a}_2 + z_4\,\mathbf{a}_3$ | = | $-\frac{3}{2}x_4 a\,\hat{\mathbf{x}} + \frac{\sqrt{3}}{2}x_4 a\,\hat{\mathbf{y}} + z_4 c\,\hat{\mathbf{z}}$ | (6$c$) | O II |
| $\mathbf{B}_{10}$ | = | $-x_4\,\mathbf{a}_1 + x_4\,\mathbf{a}_2 + \left(\frac{1}{2}+z_4\right)\mathbf{a}_3$ | = | $\sqrt{3}x_4 a\,\hat{\mathbf{y}} + \left(\frac{1}{2}+z_4\right)c\,\hat{\mathbf{z}}$ | (6$c$) | O II |
| $\mathbf{B}_{11}$ | = | $-x_4\,\mathbf{a}_1 - 2x_4\,\mathbf{a}_2 + \left(\frac{1}{2}+z_4\right)\mathbf{a}_3$ | = | $-\frac{3}{2}x_4 a\,\hat{\mathbf{x}} - \frac{\sqrt{3}}{2}x_4 a\,\hat{\mathbf{y}} + \left(\frac{1}{2}+z_4\right)c\,\hat{\mathbf{z}}$ | (6$c$) | O II |
| $\mathbf{B}_{12}$ | = | $2x_4\,\mathbf{a}_1 + x_4\,\mathbf{a}_2 + \left(\frac{1}{2}+z_4\right)\mathbf{a}_3$ | = | $\frac{3}{2}x_4 a\,\hat{\mathbf{x}} - \frac{\sqrt{3}}{2}x_4 a\,\hat{\mathbf{y}} + \left(\frac{1}{2}+z_4\right)c\,\hat{\mathbf{z}}$ | (6$c$) | O II |
| $\mathbf{B}_{13}$ | = | $x_5\,\mathbf{a}_1 - x_5\,\mathbf{a}_2 + z_5\,\mathbf{a}_3$ | = | $-\sqrt{3}x_5 a\,\hat{\mathbf{y}} + z_5 c\,\hat{\mathbf{z}}$ | (6$c$) | O III |
| $\mathbf{B}_{14}$ | = | $x_5\,\mathbf{a}_1 + 2x_5\,\mathbf{a}_2 + z_5\,\mathbf{a}_3$ | = | $\frac{3}{2}x_5 a\,\hat{\mathbf{x}} + \frac{\sqrt{3}}{2}x_5 a\,\hat{\mathbf{y}} + z_5 c\,\hat{\mathbf{z}}$ | (6$c$) | O III |
| $\mathbf{B}_{15}$ | = | $-2x_5\,\mathbf{a}_1 - x_5\,\mathbf{a}_2 + z_5\,\mathbf{a}_3$ | = | $-\frac{3}{2}x_5 a\,\hat{\mathbf{x}} + \frac{\sqrt{3}}{2}x_5 a\,\hat{\mathbf{y}} + z_5 c\,\hat{\mathbf{z}}$ | (6$c$) | O III |
| $\mathbf{B}_{16}$ | = | $-x_5\,\mathbf{a}_1 + x_5\,\mathbf{a}_2 + \left(\frac{1}{2}+z_5\right)\mathbf{a}_3$ | = | $\sqrt{3}x_5 a\,\hat{\mathbf{y}} + \left(\frac{1}{2}+z_5\right)c\,\hat{\mathbf{z}}$ | (6$c$) | O III |
| $\mathbf{B}_{17}$ | = | $-x_5\,\mathbf{a}_1 - 2x_5\,\mathbf{a}_2 + \left(\frac{1}{2}+z_5\right)\mathbf{a}_3$ | = | $-\frac{3}{2}x_5 a\,\hat{\mathbf{x}} - \frac{\sqrt{3}}{2}x_5 a\,\hat{\mathbf{y}} + \left(\frac{1}{2}+z_5\right)c\,\hat{\mathbf{z}}$ | (6$c$) | O III |
| $\mathbf{B}_{18}$ | = | $2x_5\,\mathbf{a}_1 + x_5\,\mathbf{a}_2 + \left(\frac{1}{2}+z_5\right)\mathbf{a}_3$ | = | $\frac{3}{2}x_5 a\,\hat{\mathbf{x}} - \frac{\sqrt{3}}{2}x_5 a\,\hat{\mathbf{y}} + \left(\frac{1}{2}+z_5\right)c\,\hat{\mathbf{z}}$ | (6$c$) | O III |
| $\mathbf{B}_{19}$ | = | $x_6\,\mathbf{a}_1 + y_6\,\mathbf{a}_2 + z_6\,\mathbf{a}_3$ | = | $\frac{1}{2}\left(x_6+y_6\right)a\,\hat{\mathbf{x}} + \frac{\sqrt{3}}{2}\left(-x_6+y_6\right)a\,\hat{\mathbf{y}} + z_6 c\,\hat{\mathbf{z}}$ | (12$d$) | H |
| $\mathbf{B}_{20}$ | = | $-y_6\,\mathbf{a}_1 + \left(x_6-y_6\right)\mathbf{a}_2 + z_6\,\mathbf{a}_3$ | = | $\left(\frac{1}{2}x_6-y_6\right)a\,\hat{\mathbf{x}} + \frac{\sqrt{3}}{2}x_6 a\,\hat{\mathbf{y}} + z_6 c\,\hat{\mathbf{z}}$ | (12$d$) | H |
| $\mathbf{B}_{21}$ | = | $\left(-x_6+y_6\right)\mathbf{a}_1 - x_6\,\mathbf{a}_2 + z_6\,\mathbf{a}_3$ | = | $\left(-x_6+\frac{1}{2}y_6\right)a\,\hat{\mathbf{x}} - \frac{\sqrt{3}}{2}y_6 a\,\hat{\mathbf{y}} + z_6 c\,\hat{\mathbf{z}}$ | (12$d$) | H |
| $\mathbf{B}_{22}$ | = | $-x_6\,\mathbf{a}_1 - y_6\,\mathbf{a}_2 + \left(\frac{1}{2}+z_6\right)\mathbf{a}_3$ | = | $-\frac{1}{2}\left(x_6+y_6\right)a\,\hat{\mathbf{x}} + \frac{\sqrt{3}}{2}\left(x_6-y_6\right)a\,\hat{\mathbf{y}} + \left(\frac{1}{2}+z_6\right)c\,\hat{\mathbf{z}}$ | (12$d$) | H |
| $\mathbf{B}_{23}$ | = | $y_6\,\mathbf{a}_1 + \left(-x_6+y_6\right)\mathbf{a}_2 + \left(\frac{1}{2}+z_6\right)\mathbf{a}_3$ | = | $\left(-\frac{1}{2}x_6+y_6\right)a\,\hat{\mathbf{x}} - \frac{\sqrt{3}}{2}x_6 a\,\hat{\mathbf{y}} + \left(\frac{1}{2}+z_6\right)c\,\hat{\mathbf{z}}$ | (12$d$) | H |
| $\mathbf{B}_{24}$ | = | $\left(x_6-y_6\right)\mathbf{a}_1 + x_6\,\mathbf{a}_2 + \left(\frac{1}{2}+z_6\right)\mathbf{a}_3$ | = | $\left(x_6-\frac{1}{2}y_6\right)a\,\hat{\mathbf{x}} + \frac{\sqrt{3}}{2}y_6 a\,\hat{\mathbf{y}} + \left(\frac{1}{2}+z_6\right)c\,\hat{\mathbf{z}}$ | (12$d$) | H |
| $\mathbf{B}_{25}$ | = | $-y_6\,\mathbf{a}_1 - x_6\,\mathbf{a}_2 + z_6\,\mathbf{a}_3$ | = | $-\frac{1}{2}(x_6+y_6)\,a\,\hat{\mathbf{x}} + \frac{\sqrt{3}}{2}(-x_6+y_6)\,a\,\hat{\mathbf{y}} + z_6c\,\hat{\mathbf{z}}$ | (12$d$) | H |
| $\mathbf{B}_{26}$ | = | $(-x_6+y_6)\,\mathbf{a}_1 + y_6\,\mathbf{a}_2 + z_6\,\mathbf{a}_3$ | = | $\left(-\frac{1}{2}x_6+y_6\right)a\,\hat{\mathbf{x}} + \frac{\sqrt{3}}{2}x_6a\,\hat{\mathbf{y}} + z_6c\,\hat{\mathbf{z}}$ | (12$d$) | H |
| $\mathbf{B}_{27}$ | = | $x_6\,\mathbf{a}_1 + (x_6-y_6)\,\mathbf{a}_2 + z_6\,\mathbf{a}_3$ | = | $\left(x_6-\frac{1}{2}y_6\right)a\,\hat{\mathbf{x}} - \frac{\sqrt{3}}{2}y_6a\,\hat{\mathbf{y}} + z_6c\,\hat{\mathbf{z}}$ | (12$d$) | H |
| $\mathbf{B}_{28}$ | = | $y_6\,\mathbf{a}_1 + x_6\,\mathbf{a}_2 + \left(\frac{1}{2}+z_6\right)\mathbf{a}_3$ | = | $\frac{1}{2}(x_6+y_6)\,a\,\hat{\mathbf{x}} + \frac{\sqrt{3}}{2}(x_6-y_6)\,a\,\hat{\mathbf{y}} + \left(\frac{1}{2}+z_6\right)c\,\hat{\mathbf{z}}$ | (12$d$) | H |
| $\mathbf{B}_{29}$ | = | $(x_6-y_6)\,\mathbf{a}_1 - y_6\,\mathbf{a}_2 + \left(\frac{1}{2}+z_6\right)\mathbf{a}_3$ | = | $\left(\frac{1}{2}x_6-y_6\right)a\,\hat{\mathbf{x}} - \frac{\sqrt{3}}{2}x_6a\,\hat{\mathbf{y}} + \left(\frac{1}{2}+z_6\right)c\,\hat{\mathbf{z}}$ | (12$d$) | H |
| $\mathbf{B}_{30}$ | = | $-x_6\,\mathbf{a}_1+(-x_6+y_6)\,\mathbf{a}_2+\left(\frac{1}{2}+z_6\right)\mathbf{a}_3$ | = | $\left(-x_6+\frac{1}{2}y_6\right)a\,\hat{\mathbf{x}} + \frac{\sqrt{3}}{2}y_6a\,\hat{\mathbf{y}} + \left(\frac{1}{2}+z_6\right)c\,\hat{\mathbf{z}}$ | (12$d$) | H |

**Geometry files:**

- CIF: pp. 1758
- POSCAR: pp. 1759

# Cd(OH)Cl ($E0_3$) Structure: ABCD_hP8_186_b_b_a_a

http://aflow.org/prototype-encyclopedia/ABCD_hP8_186_b_b_a_a

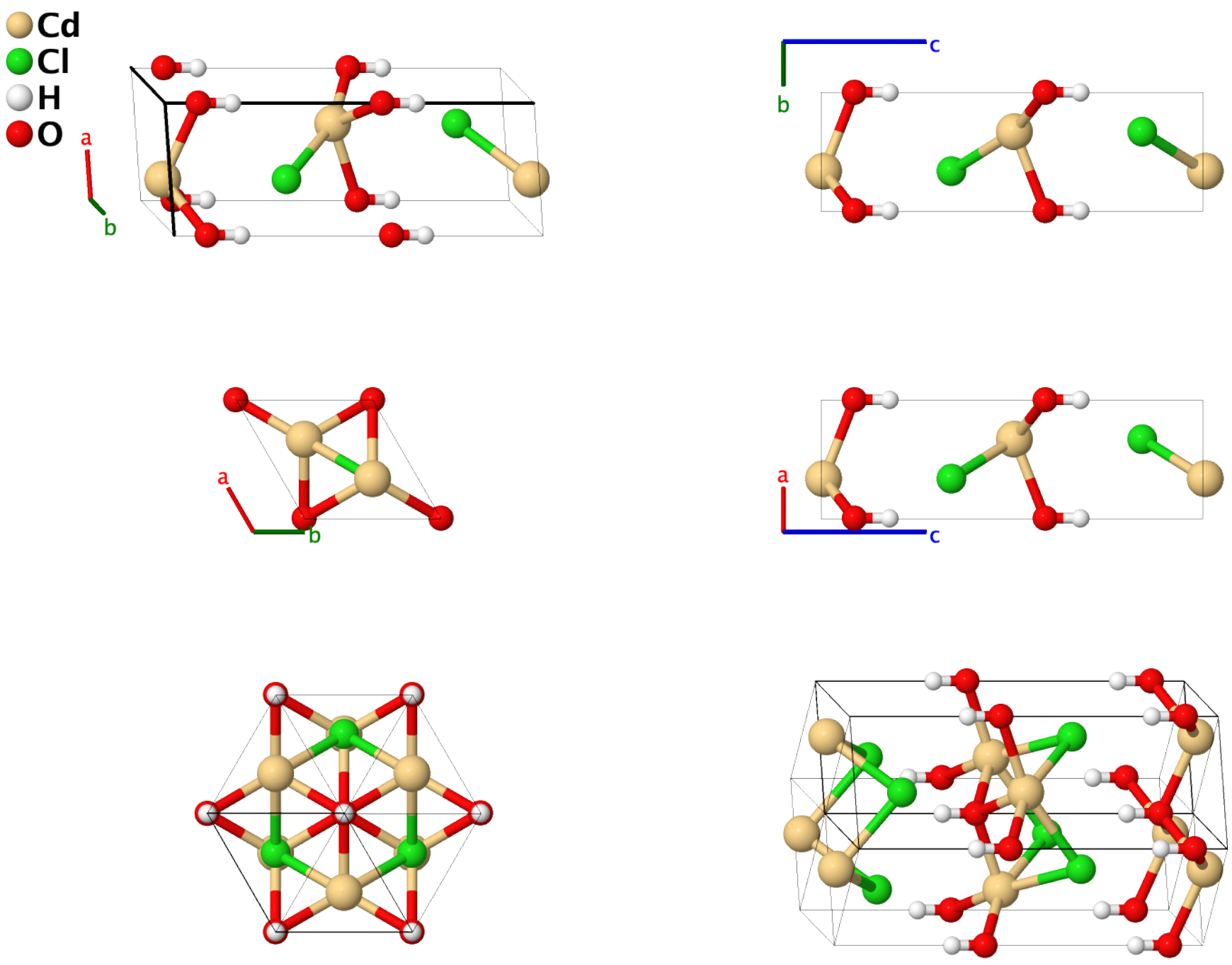

| | | |
|---|---|---|
| **Prototype** | : | CdClHO |
| **AFLOW prototype label** | : | ABCD_hP8_186_b_b_a_a |
| ***Strukturbericht* designation** | : | $E0_3$ |
| **Pearson symbol** | : | hP8 |
| **Space group number** | : | 186 |
| **Space group symbol** | : | $P6_3mc$ |
| **AFLOW prototype command** | : | `aflow --proto=ABCD_hP8_186_b_b_a_a` `--params=`$a, c/a, z_1, z_2, z_3, z_4$ |

- Early determinations of this structure did not locate the hydrogen atoms. (Cudennec, 2000) state that "H atom coordinates were calculated for Cd(OH)Cl."

**Hexagonal primitive vectors:**

$$\mathbf{a}_1 = \frac{1}{2}\,a\,\hat{\mathbf{x}} - \frac{\sqrt{3}}{2}\,a\,\hat{\mathbf{y}}$$
$$\mathbf{a}_2 = \frac{1}{2}\,a\,\hat{\mathbf{x}} + \frac{\sqrt{3}}{2}\,a\,\hat{\mathbf{y}}$$
$$\mathbf{a}_3 = c\,\hat{\mathbf{z}}$$

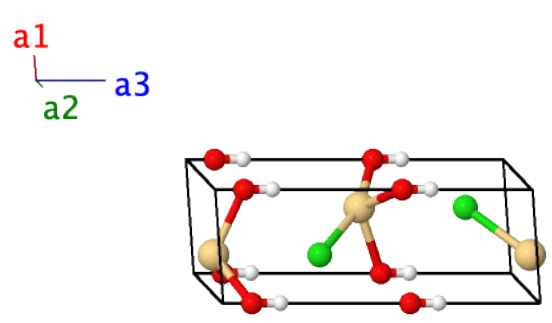

**Basis vectors:**

| | | Lattice Coordinates | | Cartesian Coordinates | Wyckoff Position | Atom Type |
|---|---|---|---|---|---|---|
| $\mathbf{B}_1$ | = | $z_1\,\mathbf{a}_3$ | = | $z_1 c\,\hat{\mathbf{z}}$ | (2$a$) | H |
| $\mathbf{B}_2$ | = | $\left(\frac{1}{2} + z_1\right)\mathbf{a}_3$ | = | $\left(\frac{1}{2} + z_1\right)c\,\hat{\mathbf{z}}$ | (2$a$) | H |
| $\mathbf{B}_3$ | = | $z_2\,\mathbf{a}_3$ | = | $z_2 c\,\hat{\mathbf{z}}$ | (2$a$) | O |
| $\mathbf{B}_4$ | = | $\left(\frac{1}{2}+z_2\right)\mathbf{a}_3$ | = | $\left(\frac{1}{2}+z_2\right)c\,\hat{\mathbf{z}}$ | (2$a$) | O |
| $\mathbf{B}_5$ | = | $\frac{1}{3}\,\mathbf{a}_1+\frac{2}{3}\,\mathbf{a}_2+z_3\,\mathbf{a}_3$ | = | $\frac{1}{2}a\,\hat{\mathbf{x}}+\frac{1}{2\sqrt{3}}a\,\hat{\mathbf{y}}+z_3 c\,\hat{\mathbf{z}}$ | (2$b$) | Cd |
| $\mathbf{B}_6$ | = | $\frac{2}{3}\,\mathbf{a}_1+\frac{1}{3}\,\mathbf{a}_2+\left(\frac{1}{2}+z_3\right)\mathbf{a}_3$ | = | $\frac{1}{2}a\,\hat{\mathbf{x}}-\frac{1}{2\sqrt{3}}a\,\hat{\mathbf{y}}+\left(\frac{1}{2}+z_3\right)c\,\hat{\mathbf{z}}$ | (2$b$) | Cd |
| $\mathbf{B}_7$ | = | $\frac{1}{3}\,\mathbf{a}_1+\frac{2}{3}\,\mathbf{a}_2+z_4\,\mathbf{a}_3$ | = | $\frac{1}{2}a\,\hat{\mathbf{x}}+\frac{1}{2\sqrt{3}}a\,\hat{\mathbf{y}}+z_4 c\,\hat{\mathbf{z}}$ | (2$b$) | Cl |
| $\mathbf{B}_8$ | = | $\frac{2}{3}\,\mathbf{a}_1+\frac{1}{3}\,\mathbf{a}_2+\left(\frac{1}{2}+z_4\right)\mathbf{a}_3$ | = | $\frac{1}{2}a\,\hat{\mathbf{x}}-\frac{1}{2\sqrt{3}}a\,\hat{\mathbf{y}}+\left(\frac{1}{2}+z_4\right)c\,\hat{\mathbf{z}}$ | (2$b$) | Cl |

**Geometry files:**

- CIF: pp. 1759
- POSCAR: pp. 1759

# Cr-233 Quasi-One-Dimensional Superconductor ($K_2Cr_3As_3$) Structure: A3B3C2_hP16_187_jk_jk_ck

http://aflow.org/prototype-encyclopedia/A3B3C2_hP16_187_jk_jk_ck

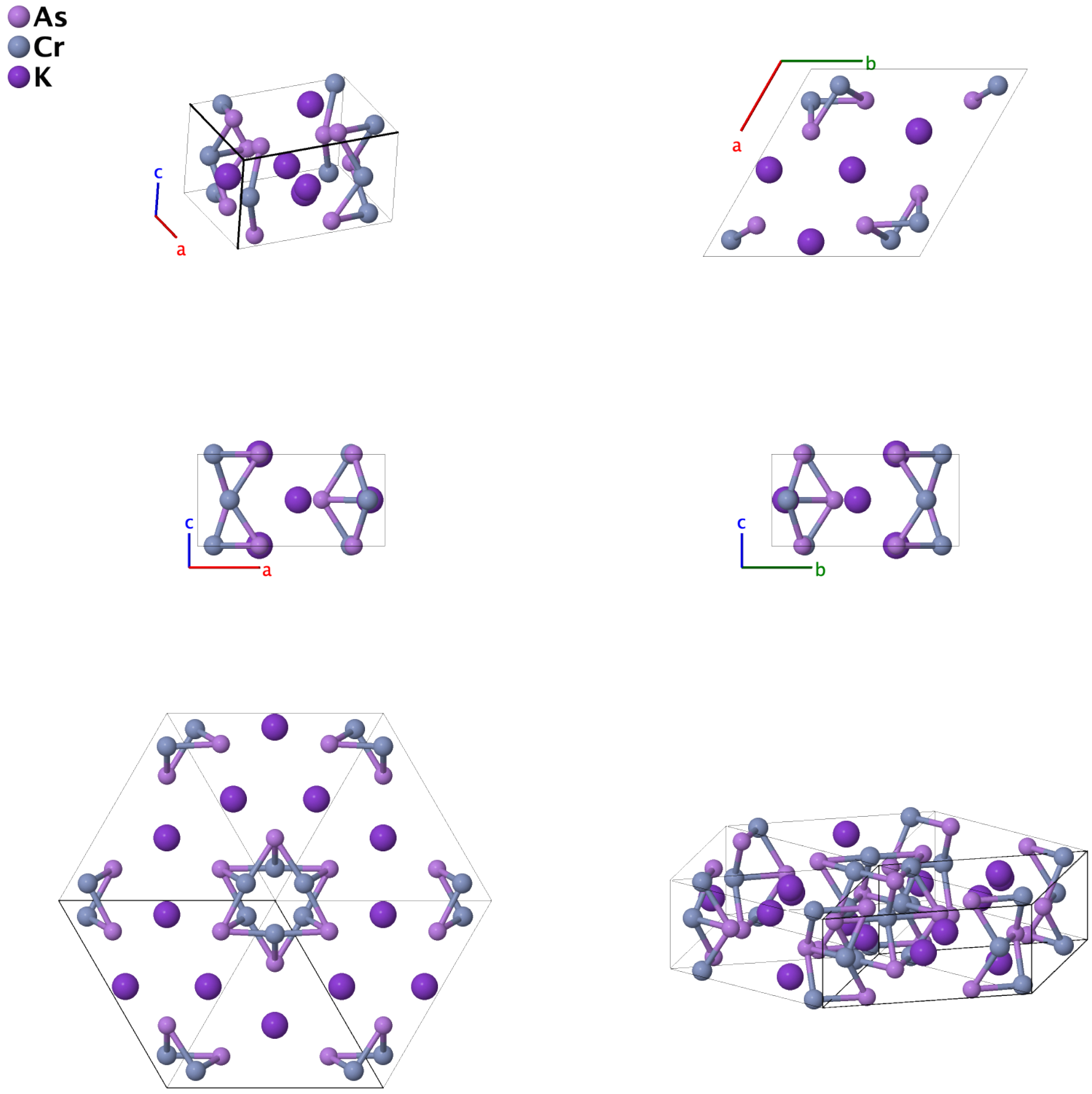

| | | |
|---|---|---|
| **Prototype** | : | $As_3Cr_3K_2$ |
| **AFLOW prototype label** | : | A3B3C2_hP16_187_jk_jk_ck |
| ***Strukturbericht* designation** | : | None |
| **Pearson symbol** | : | hP16 |
| **Space group number** | : | 187 |
| **Space group symbol** | : | $P\bar{6}m2$ |
| **AFLOW prototype command** | : | `aflow --proto=A3B3C2_hP16_187_jk_jk_ck --params=`$a, c/a, x_2, x_3, x_4, x_5, x_6$ |

**Other compounds with this structure**

- $Cs_2Cr_3As_3$, $Rb_2Cr_3As_3$, and $K_2Mo_3As_3$

- Cr-233 designates a class of structures of the form $A_2B_2As_3$, where the 'A' atoms form one-dimensional chains. Several of these compounds have been found to superconduct at temperatures on the order of 5-10 K.

**Hexagonal primitive vectors:**

$$\mathbf{a}_1 = \frac{1}{2}\,a\,\hat{\mathbf{x}} - \frac{\sqrt{3}}{2}\,a\,\hat{\mathbf{y}}$$
$$\mathbf{a}_2 = \frac{1}{2}\,a\,\hat{\mathbf{x}} + \frac{\sqrt{3}}{2}\,a\,\hat{\mathbf{y}}$$
$$\mathbf{a}_3 = c\,\hat{\mathbf{z}}$$

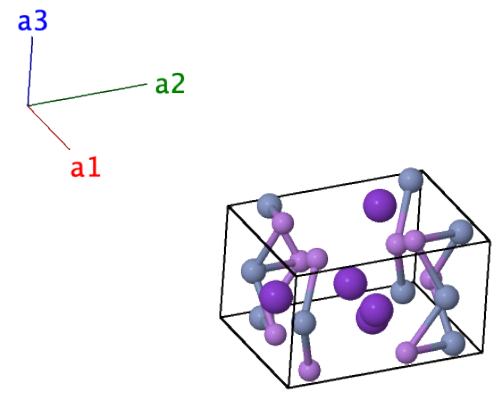

**Basis vectors:**

| | | Lattice Coordinates | | Cartesian Coordinates | Wyckoff Position | Atom Type |
|---|---|---|---|---|---|---|
| $\mathbf{B}_1$ | = | $\frac{1}{3}\,\mathbf{a}_1 + \frac{2}{3}\,\mathbf{a}_2$ | = | $\frac{1}{2}a\,\hat{\mathbf{x}} + \frac{1}{2\sqrt{3}}a\,\hat{\mathbf{y}}$ | $(1c)$ | K I |
| $\mathbf{B}_2$ | = | $x_2\,\mathbf{a}_1 - x_2\,\mathbf{a}_2$ | = | $-\sqrt{3}x_2a\,\hat{\mathbf{y}}$ | $(3j)$ | As I |
| $\mathbf{B}_3$ | = | $x_2\,\mathbf{a}_1 + 2x_2\,\mathbf{a}_2$ | = | $\frac{3}{2}x_2a\,\hat{\mathbf{x}} + \frac{\sqrt{3}}{2}x_2a\,\hat{\mathbf{y}}$ | $(3j)$ | As I |
| $\mathbf{B}_4$ | = | $-2x_2\,\mathbf{a}_1 - x_2\,\mathbf{a}_2$ | = | $-\frac{3}{2}x_2a\,\hat{\mathbf{x}} + \frac{\sqrt{3}}{2}x_2a\,\hat{\mathbf{y}}$ | $(3j)$ | As I |
| $\mathbf{B}_5$ | = | $x_3\,\mathbf{a}_1 - x_3\,\mathbf{a}_2$ | = | $-\sqrt{3}x_3a\,\hat{\mathbf{y}}$ | $(3j)$ | Cr I |
| $\mathbf{B}_6$ | = | $x_3\,\mathbf{a}_1 + 2x_3\,\mathbf{a}_2$ | = | $\frac{3}{2}x_3a\,\hat{\mathbf{x}} + \frac{\sqrt{3}}{2}x_3a\,\hat{\mathbf{y}}$ | $(3j)$ | Cr I |
| $\mathbf{B}_7$ | = | $-2x_3\,\mathbf{a}_1 - x_3\,\mathbf{a}_2$ | = | $-\frac{3}{2}x_3a\,\hat{\mathbf{x}} + \frac{\sqrt{3}}{2}x_3a\,\hat{\mathbf{y}}$ | $(3j)$ | Cr I |
| $\mathbf{B}_8$ | = | $x_4\,\mathbf{a}_1 - x_4\,\mathbf{a}_2 + \frac{1}{2}\,\mathbf{a}_3$ | = | $-\sqrt{3}x_4a\,\hat{\mathbf{y}} + \frac{1}{2}c\,\hat{\mathbf{z}}$ | $(3k)$ | As II |
| $\mathbf{B}_9$ | = | $x_4\,\mathbf{a}_1 + 2x_4\,\mathbf{a}_2 + \frac{1}{2}\,\mathbf{a}_3$ | = | $\frac{3}{2}x_4a\,\hat{\mathbf{x}} + \frac{\sqrt{3}}{2}x_4a\,\hat{\mathbf{y}} + \frac{1}{2}c\,\hat{\mathbf{z}}$ | $(3k)$ | As II |
| $\mathbf{B}_{10}$ | = | $-2x_4\,\mathbf{a}_1 - x_4\,\mathbf{a}_2 + \frac{1}{2}\,\mathbf{a}_3$ | = | $-\frac{3}{2}x_4a\,\hat{\mathbf{x}} + \frac{\sqrt{3}}{2}x_4a\,\hat{\mathbf{y}} + \frac{1}{2}c\,\hat{\mathbf{z}}$ | $(3k)$ | As II |
| $\mathbf{B}_{11}$ | = | $x_5\,\mathbf{a}_1 - x_5\,\mathbf{a}_2 + \frac{1}{2}\,\mathbf{a}_3$ | = | $-\sqrt{3}x_5a\,\hat{\mathbf{y}} + \frac{1}{2}c\,\hat{\mathbf{z}}$ | $(3k)$ | Cr II |
| $\mathbf{B}_{12}$ | = | $x_5\,\mathbf{a}_1 + 2x_5\,\mathbf{a}_2 + \frac{1}{2}\,\mathbf{a}_3$ | = | $\frac{3}{2}x_5a\,\hat{\mathbf{x}} + \frac{\sqrt{3}}{2}x_5a\,\hat{\mathbf{y}} + \frac{1}{2}c\,\hat{\mathbf{z}}$ | $(3k)$ | Cr II |
| $\mathbf{B}_{13}$ | = | $-2x_5\,\mathbf{a}_1 - x_5\,\mathbf{a}_2 + \frac{1}{2}\,\mathbf{a}_3$ | = | $-\frac{3}{2}x_5a\,\hat{\mathbf{x}} + \frac{\sqrt{3}}{2}x_5a\,\hat{\mathbf{y}} + \frac{1}{2}c\,\hat{\mathbf{z}}$ | $(3k)$ | Cr II |
| $\mathbf{B}_{14}$ | = | $x_6\,\mathbf{a}_1 - x_6\,\mathbf{a}_2 + \frac{1}{2}\,\mathbf{a}_3$ | = | $-\sqrt{3}x_6a\,\hat{\mathbf{y}} + \frac{1}{2}c\,\hat{\mathbf{z}}$ | $(3k)$ | K II |
| $\mathbf{B}_{15}$ | = | $x_6\,\mathbf{a}_1 + 2x_6\,\mathbf{a}_2 + \frac{1}{2}\,\mathbf{a}_3$ | = | $\frac{3}{2}x_6a\,\hat{\mathbf{x}} + \frac{\sqrt{3}}{2}x_6a\,\hat{\mathbf{y}} + \frac{1}{2}c\,\hat{\mathbf{z}}$ | $(3k)$ | K II |
| $\mathbf{B}_{16}$ | = | $-2x_6\,\mathbf{a}_1 - x_6\,\mathbf{a}_2 + \frac{1}{2}\,\mathbf{a}_3$ | = | $-\frac{3}{2}x_6a\,\hat{\mathbf{x}} + \frac{\sqrt{3}}{2}x_6a\,\hat{\mathbf{y}} + \frac{1}{2}c\,\hat{\mathbf{z}}$ | $(3k)$ | K II |

**Geometry files:**

- CIF: pp. 1759
- POSCAR: pp. 1760

# $Cs_7O$ Structure: A7B_hP24_187_ai2j2kn_j

http://aflow.org/prototype-encyclopedia/A7B_hP24_187_ai2j2kn_j

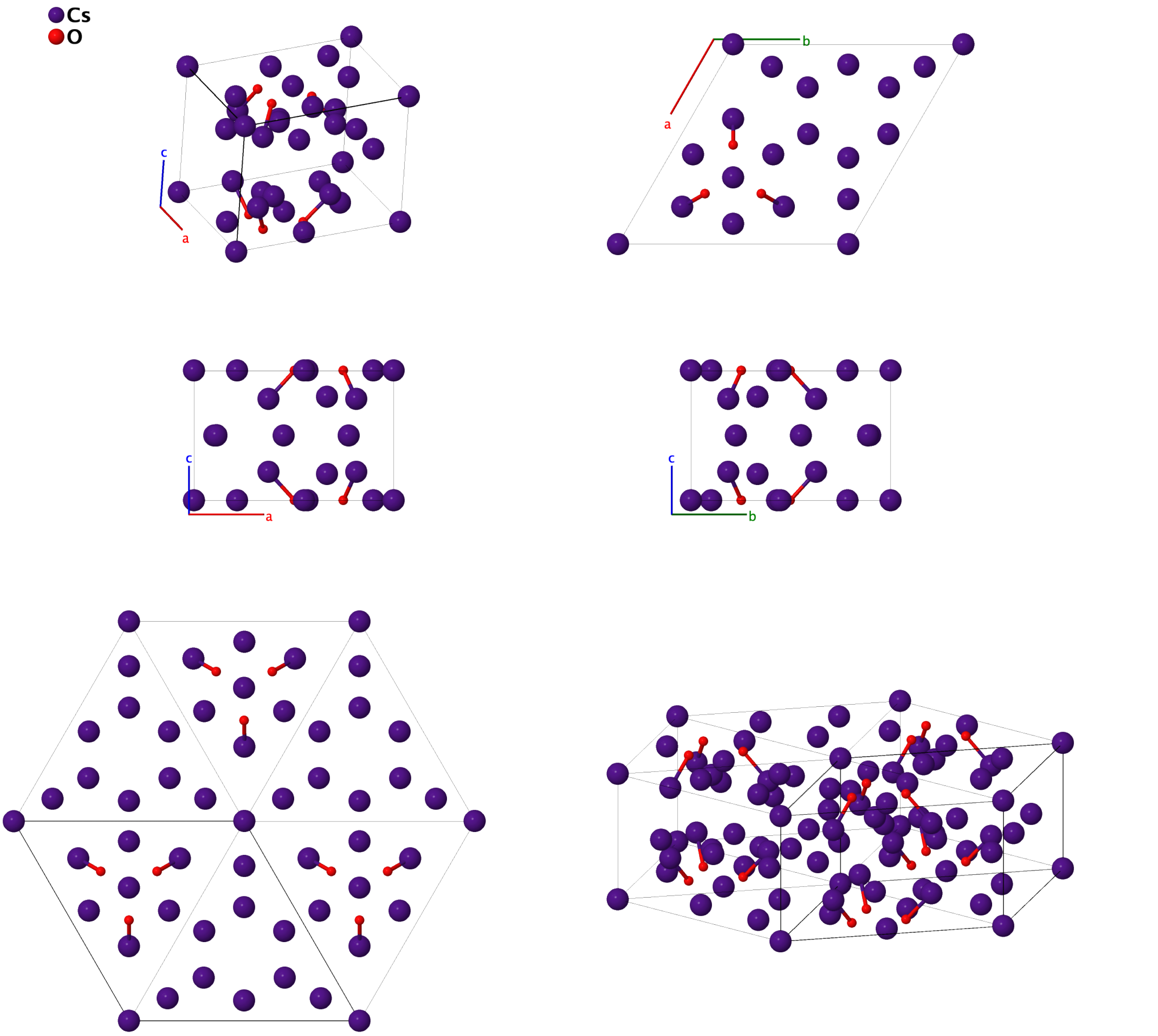

| | | |
|---|---|---|
| **Prototype** | : | $Cs_7O$ |
| **AFLOW prototype label** | : | A7B_hP24_187_ai2j2kn_j |
| ***Strukturbericht* designation** | : | None |
| **Pearson symbol** | : | hP24 |
| **Space group number** | : | 187 |
| **Space group symbol** | : | $P\bar{6}m2$ |
| **AFLOW prototype command** | : | `aflow --proto=A7B_hP24_187_ai2j2kn_j --params=`$a, c/a, z_2, x_3, x_4, x_5, x_6, x_7, x_8, z_8$ |

- This structure is composed of $Cs_{11}O_3$ clusters, similar to the building blocks of the $Cs_{11}O_3$ structure, interlaced with cesium atoms which have approximately the same spacing as in bcc-Cs.

- Lattice constant data was given at -150 °C, while the atomic positions were given at -175 °C.

---

**Hexagonal primitive vectors:**

$$\mathbf{a}_1 = \frac{1}{2}\,a\,\hat{\mathbf{x}} - \frac{\sqrt{3}}{2}\,a\,\hat{\mathbf{y}}$$
$$\mathbf{a}_2 = \frac{1}{2}\,a\,\hat{\mathbf{x}} + \frac{\sqrt{3}}{2}\,a\,\hat{\mathbf{y}}$$
$$\mathbf{a}_3 = c\,\hat{\mathbf{z}}$$

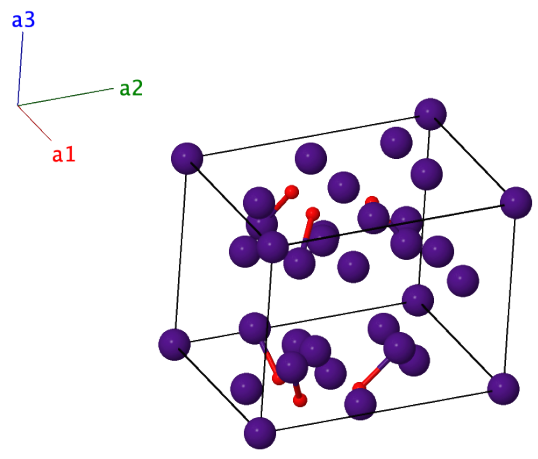

---

**Basis vectors:**

| | | Lattice Coordinates | | Cartesian Coordinates | Wyckoff Position | Atom Type |
|---|---|---|---|---|---|---|
| $\mathbf{B}_1$ | = | $0\,\mathbf{a}_1 + 0\,\mathbf{a}_2 + 0\,\mathbf{a}_3$ | = | $0\,\hat{\mathbf{x}} + 0\,\hat{\mathbf{y}} + 0\,\hat{\mathbf{z}}$ | (1$a$) | Cs I |
| $\mathbf{B}_2$ | = | $\frac{2}{3}\,\mathbf{a}_1 + \frac{1}{3}\,\mathbf{a}_2 + z_2\,\mathbf{a}_3$ | = | $\frac{1}{2}a\,\hat{\mathbf{x}} - \frac{1}{2\sqrt{3}}a\,\hat{\mathbf{y}} + z_2c\,\hat{\mathbf{z}}$ | (2$i$) | Cs II |
| $\mathbf{B}_3$ | = | $\frac{2}{3}\,\mathbf{a}_1 + \frac{1}{3}\,\mathbf{a}_2 - z_2\,\mathbf{a}_3$ | = | $\frac{1}{2}a\,\hat{\mathbf{x}} - \frac{1}{2\sqrt{3}}a\,\hat{\mathbf{y}} - z_2c\,\hat{\mathbf{z}}$ | (2$i$) | Cs II |
| $\mathbf{B}_4$ | = | $x_3\,\mathbf{a}_1 - x_3\,\mathbf{a}_2$ | = | $-\sqrt{3}x_3a\,\hat{\mathbf{y}}$ | (3$j$) | Cs III |
| $\mathbf{B}_5$ | = | $x_3\,\mathbf{a}_1 + 2x_3\,\mathbf{a}_2$ | = | $\frac{3}{2}x_3a\,\hat{\mathbf{x}} + \frac{\sqrt{3}}{2}x_3a\,\hat{\mathbf{y}}$ | (3$j$) | Cs III |
| $\mathbf{B}_6$ | = | $-2x_3\,\mathbf{a}_1 - x_3\,\mathbf{a}_2$ | = | $-\frac{3}{2}x_3a\,\hat{\mathbf{x}} + \frac{\sqrt{3}}{2}x_3a\,\hat{\mathbf{y}}$ | (3$j$) | Cs III |
| $\mathbf{B}_7$ | = | $x_4\,\mathbf{a}_1 - x_4\,\mathbf{a}_2$ | = | $-\sqrt{3}x_4a\,\hat{\mathbf{y}}$ | (3$j$) | Cs IV |
| $\mathbf{B}_8$ | = | $x_4\,\mathbf{a}_1 + 2x_4\,\mathbf{a}_2$ | = | $\frac{3}{2}x_4a\,\hat{\mathbf{x}} + \frac{\sqrt{3}}{2}x_4a\,\hat{\mathbf{y}}$ | (3$j$) | Cs IV |
| $\mathbf{B}_9$ | = | $-2x_4\,\mathbf{a}_1 - x_4\,\mathbf{a}_2$ | = | $-\frac{3}{2}x_4a\,\hat{\mathbf{x}} + \frac{\sqrt{3}}{2}x_4a\,\hat{\mathbf{y}}$ | (3$j$) | Cs IV |
| $\mathbf{B}_{10}$ | = | $x_5\,\mathbf{a}_1 - x_5\,\mathbf{a}_2$ | = | $-\sqrt{3}x_5a\,\hat{\mathbf{y}}$ | (3$j$) | O |
| $\mathbf{B}_{11}$ | = | $x_5\,\mathbf{a}_1 + 2x_5\,\mathbf{a}_2$ | = | $\frac{3}{2}x_5a\,\hat{\mathbf{x}} + \frac{\sqrt{3}}{2}x_5a\,\hat{\mathbf{y}}$ | (3$j$) | O |
| $\mathbf{B}_{12}$ | = | $-2x_5\,\mathbf{a}_1 - x_5\,\mathbf{a}_2$ | = | $-\frac{3}{2}x_5a\,\hat{\mathbf{x}} + \frac{\sqrt{3}}{2}x_5a\,\hat{\mathbf{y}}$ | (3$j$) | O |
| $\mathbf{B}_{13}$ | = | $x_6\,\mathbf{a}_1 - x_6\,\mathbf{a}_2 + \frac{1}{2}\,\mathbf{a}_3$ | = | $-\sqrt{3}x_6a\,\hat{\mathbf{y}} + \frac{1}{2}c\,\hat{\mathbf{z}}$ | (3$k$) | Cs V |
| $\mathbf{B}_{14}$ | = | $x_6\,\mathbf{a}_1 + 2x_6\,\mathbf{a}_2 + \frac{1}{2}\,\mathbf{a}_3$ | = | $\frac{3}{2}x_6a\,\hat{\mathbf{x}} + \frac{\sqrt{3}}{2}x_6a\,\hat{\mathbf{y}} + \frac{1}{2}c\,\hat{\mathbf{z}}$ | (3$k$) | Cs V |
| $\mathbf{B}_{15}$ | = | $-2x_6\,\mathbf{a}_1 - x_6\,\mathbf{a}_2 + \frac{1}{2}\,\mathbf{a}_3$ | = | $-\frac{3}{2}x_6a\,\hat{\mathbf{x}} + \frac{\sqrt{3}}{2}x_6a\,\hat{\mathbf{y}} + \frac{1}{2}c\,\hat{\mathbf{z}}$ | (3$k$) | Cs V |
| $\mathbf{B}_{16}$ | = | $x_7\,\mathbf{a}_1 - x_7\,\mathbf{a}_2 + \frac{1}{2}\,\mathbf{a}_3$ | = | $-\sqrt{3}x_7a\,\hat{\mathbf{y}} + \frac{1}{2}c\,\hat{\mathbf{z}}$ | (3$k$) | Cs VI |
| $\mathbf{B}_{17}$ | = | $x_7\,\mathbf{a}_1 + 2x_7\,\mathbf{a}_2 + \frac{1}{2}\,\mathbf{a}_3$ | = | $\frac{3}{2}x_7a\,\hat{\mathbf{x}} + \frac{\sqrt{3}}{2}x_7a\,\hat{\mathbf{y}} + \frac{1}{2}c\,\hat{\mathbf{z}}$ | (3$k$) | Cs VI |
| $\mathbf{B}_{18}$ | = | $-2x_7\,\mathbf{a}_1 - x_7\,\mathbf{a}_2 + \frac{1}{2}\,\mathbf{a}_3$ | = | $-\frac{3}{2}x_7a\,\hat{\mathbf{x}} + \frac{\sqrt{3}}{2}x_7a\,\hat{\mathbf{y}} + \frac{1}{2}c\,\hat{\mathbf{z}}$ | (3$k$) | Cs VI |
| $\mathbf{B}_{19}$ | = | $x_8\,\mathbf{a}_1 - x_8\,\mathbf{a}_2 + z_8\,\mathbf{a}_3$ | = | $-\sqrt{3}x_8a\,\hat{\mathbf{y}} + z_8c\,\hat{\mathbf{z}}$ | (6$n$) | Cs VII |
| $\mathbf{B}_{20}$ | = | $x_8\,\mathbf{a}_1 + 2x_8\,\mathbf{a}_2 + z_8\,\mathbf{a}_3$ | = | $\frac{3}{2}x_8a\,\hat{\mathbf{x}} + \frac{\sqrt{3}}{2}x_8a\,\hat{\mathbf{y}} + z_8c\,\hat{\mathbf{z}}$ | (6$n$) | Cs VII |
| $\mathbf{B}_{21}$ | = | $-2x_8\,\mathbf{a}_1 - x_8\,\mathbf{a}_2 + z_8\,\mathbf{a}_3$ | = | $-\frac{3}{2}x_8a\,\hat{\mathbf{x}} + \frac{\sqrt{3}}{2}x_8a\,\hat{\mathbf{y}} + z_8c\,\hat{\mathbf{z}}$ | (6$n$) | Cs VII |
| $\mathbf{B}_{22}$ | = | $x_8\,\mathbf{a}_1 - x_8\,\mathbf{a}_2 - z_8\,\mathbf{a}_3$ | = | $-\sqrt{3}x_8a\,\hat{\mathbf{y}} - z_8c\,\hat{\mathbf{z}}$ | (6$n$) | Cs VII |
| $\mathbf{B}_{23}$ | = | $x_8\,\mathbf{a}_1 + 2x_8\,\mathbf{a}_2 - z_8\,\mathbf{a}_3$ | = | $\frac{3}{2}x_8a\,\hat{\mathbf{x}} + \frac{\sqrt{3}}{2}x_8a\,\hat{\mathbf{y}} - z_8c\,\hat{\mathbf{z}}$ | (6$n$) | Cs VII |
| $\mathbf{B}_{24}$ | = | $-2x_8\,\mathbf{a}_1 - x_8\,\mathbf{a}_2 - z_8\,\mathbf{a}_3$ | = | $-\frac{3}{2}x_8a\,\hat{\mathbf{x}} + \frac{\sqrt{3}}{2}x_8a\,\hat{\mathbf{y}} - z_8c\,\hat{\mathbf{z}}$ | (6$n$) | Cs VII |

---

**Found in:**

- T. B. Massalski, H. Okamoto, P. R. Subramanian, and L. Kacprzak, eds., *Binary Alloy Phase Diagrams*, vol. 2 (ASM International, Materials Park, Ohio, USA, 1990), 2$^{nd}$ edn. Cd-Ce to Hf-Rb.

---

**Geometry files:**

- CIF: pp. 1760
- POSCAR: pp. 1760

# ZrNiAl Structure: ABC_hP9_189_g_ad_f

http://aflow.org/prototype-encyclopedia/ABC_hP9_189_g_ad_f

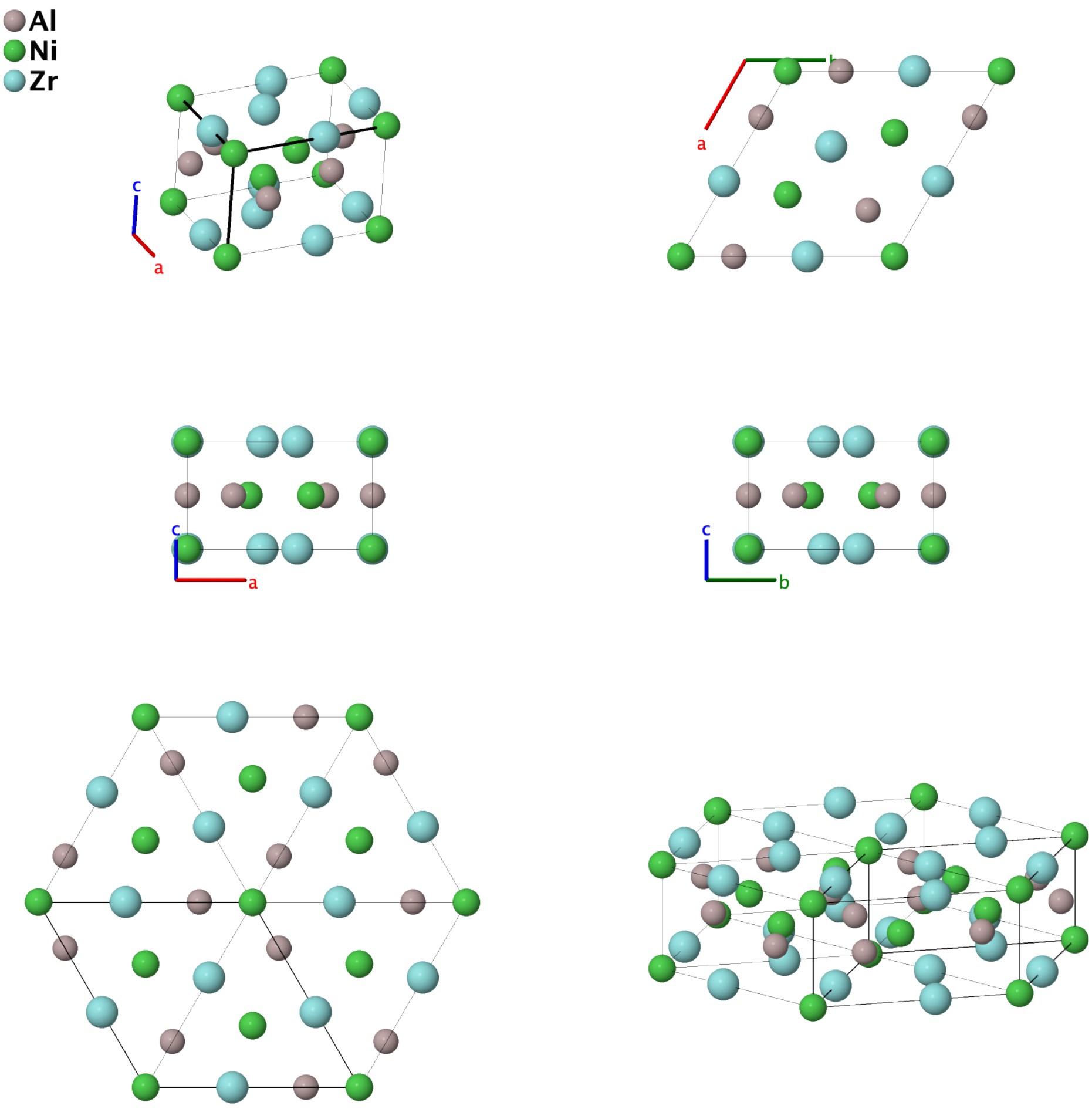

| | | |
|---|---|---|
| **Prototype** | : | AlNiZr |
| **AFLOW prototype label** | : | ABC_hP9_189_g_ad_f |
| ***Strukturbericht* designation** | : | None |
| **Pearson symbol** | : | hP9 |
| **Space group number** | : | 189 |
| **Space group symbol** | : | $P\bar{6}2m$ |
| **AFLOW prototype command** | : | `aflow --proto=ABC_hP9_189_g_ad_f`<br>`--params=`$a, c/a, x_3, x_4$ |

### Other compounds with this structure

- AgAsCa, AgSiYb, AlCoPu, AlCuTm, AlNiTb, DyNiIn, DyNiSn, ErNiAl, FeGaU, FeNiP, GdNiIn, GdNiSn, HoNiIn, RhSnZr, RuSiZr, ScIrP, TbNiIn, and $BSi_2Ni_6$

---

- This is the ternary form of the $Fe_2P$ structure. In the former case the origin was placed so that the potssium atoms were on the (1$b$) and (2$c$) Wyckoff positions, while here we follow (Shved, 2019) and place the nickel atoms on the (1$a$) and (2$d$) sites. This is merely a shift in the origin by $1/2\, c\, \hat{z}$. If the structures have a common origin then they are nearly identical.
- (Shved, 2019) found evidence of 6-10% mixing between the Ni-II (2$d$) and Al (3$g$) sites.

## Hexagonal primitive vectors:

$$
\begin{aligned}
\mathbf{a}_1 &= \tfrac{1}{2}\,a\,\hat{\mathbf{x}} - \tfrac{\sqrt{3}}{2}\,a\,\hat{\mathbf{y}} \\
\mathbf{a}_2 &= \tfrac{1}{2}\,a\,\hat{\mathbf{x}} + \tfrac{\sqrt{3}}{2}\,a\,\hat{\mathbf{y}} \\
\mathbf{a}_3 &= c\,\hat{\mathbf{z}}
\end{aligned}
$$

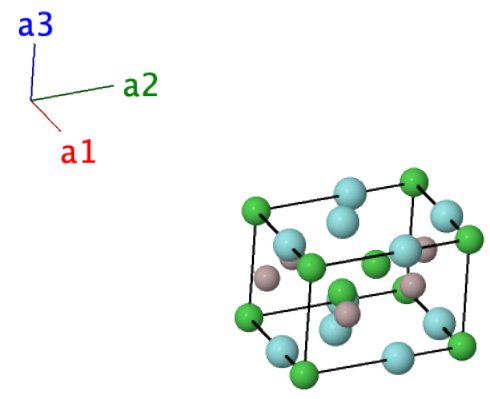

## Basis vectors:

| | | Lattice Coordinates | | Cartesian Coordinates | Wyckoff Position | Atom Type |
|---|---|---|---|---|---|---|
| $\mathbf{B}_1$ | = | $0\,\mathbf{a}_1 + 0\,\mathbf{a}_2 + 0\,\mathbf{a}_3$ | = | $0\,\hat{\mathbf{x}} + 0\,\hat{\mathbf{y}} + 0\,\hat{\mathbf{z}}$ | $(1a)$ | Ni I |
| $\mathbf{B}_2$ | = | $\frac{1}{3}\,\mathbf{a}_1 + \frac{2}{3}\,\mathbf{a}_2 + \frac{1}{2}\,\mathbf{a}_3$ | = | $\frac{1}{2}a\,\hat{\mathbf{x}} + \frac{1}{2\sqrt{3}}a\,\hat{\mathbf{y}} + \frac{1}{2}c\,\hat{\mathbf{z}}$ | $(2d)$ | Ni II |
| $\mathbf{B}_3$ | = | $\frac{2}{3}\,\mathbf{a}_1 + \frac{1}{3}\,\mathbf{a}_2 + \frac{1}{2}\,\mathbf{a}_3$ | = | $\frac{1}{2}a\,\hat{\mathbf{x}} - \frac{1}{2\sqrt{3}}a\,\hat{\mathbf{y}} + \frac{1}{2}c\,\hat{\mathbf{z}}$ | $(2d)$ | Ni II |
| $\mathbf{B}_4$ | = | $x_3\,\mathbf{a}_1$ | = | $\frac{1}{2}x_3a\,\hat{\mathbf{x}} - \frac{\sqrt{3}}{2}x_3a\,\hat{\mathbf{y}}$ | $(3f)$ | Zr |
| $\mathbf{B}_5$ | = | $x_3\,\mathbf{a}_2$ | = | $\frac{1}{2}x_3a\,\hat{\mathbf{x}} + \frac{\sqrt{3}}{2}x_3a\,\hat{\mathbf{y}}$ | $(3f)$ | Zr |
| $\mathbf{B}_6$ | = | $-x_3\,\mathbf{a}_1 - x_3\,\mathbf{a}_2$ | = | $-x_3a\,\hat{\mathbf{x}}$ | $(3f)$ | Zr |
| $\mathbf{B}_7$ | = | $x_4\,\mathbf{a}_1 + \frac{1}{2}\,\mathbf{a}_3$ | = | $\frac{1}{2}x_4a\,\hat{\mathbf{x}} - \frac{\sqrt{3}}{2}x_4a\,\hat{\mathbf{y}} + \frac{1}{2}c\,\hat{\mathbf{z}}$ | $(3g)$ | Al |
| $\mathbf{B}_8$ | = | $x_4\,\mathbf{a}_2 + \frac{1}{2}\,\mathbf{a}_3$ | = | $\frac{1}{2}x_4a\,\hat{\mathbf{x}} + \frac{\sqrt{3}}{2}x_4a\,\hat{\mathbf{y}} + \frac{1}{2}c\,\hat{\mathbf{z}}$ | $(3g)$ | Al |
| $\mathbf{B}_9$ | = | $-x_4\,\mathbf{a}_1 - x_4\,\mathbf{a}_2 + \frac{1}{2}\,\mathbf{a}_3$ | = | $-x_4a\,\hat{\mathbf{x}} + \frac{1}{2}c\,\hat{\mathbf{z}}$ | $(3g)$ | Al |

## Geometry files:

- CIF: pp. 1761
- POSCAR: pp. 1761

# $CsSO_3$ ($K1_2$) Structure: AB3C_hP20_190_ac_i_f

http://aflow.org/prototype-encyclopedia/AB3C_hP20_190_ac_i_f

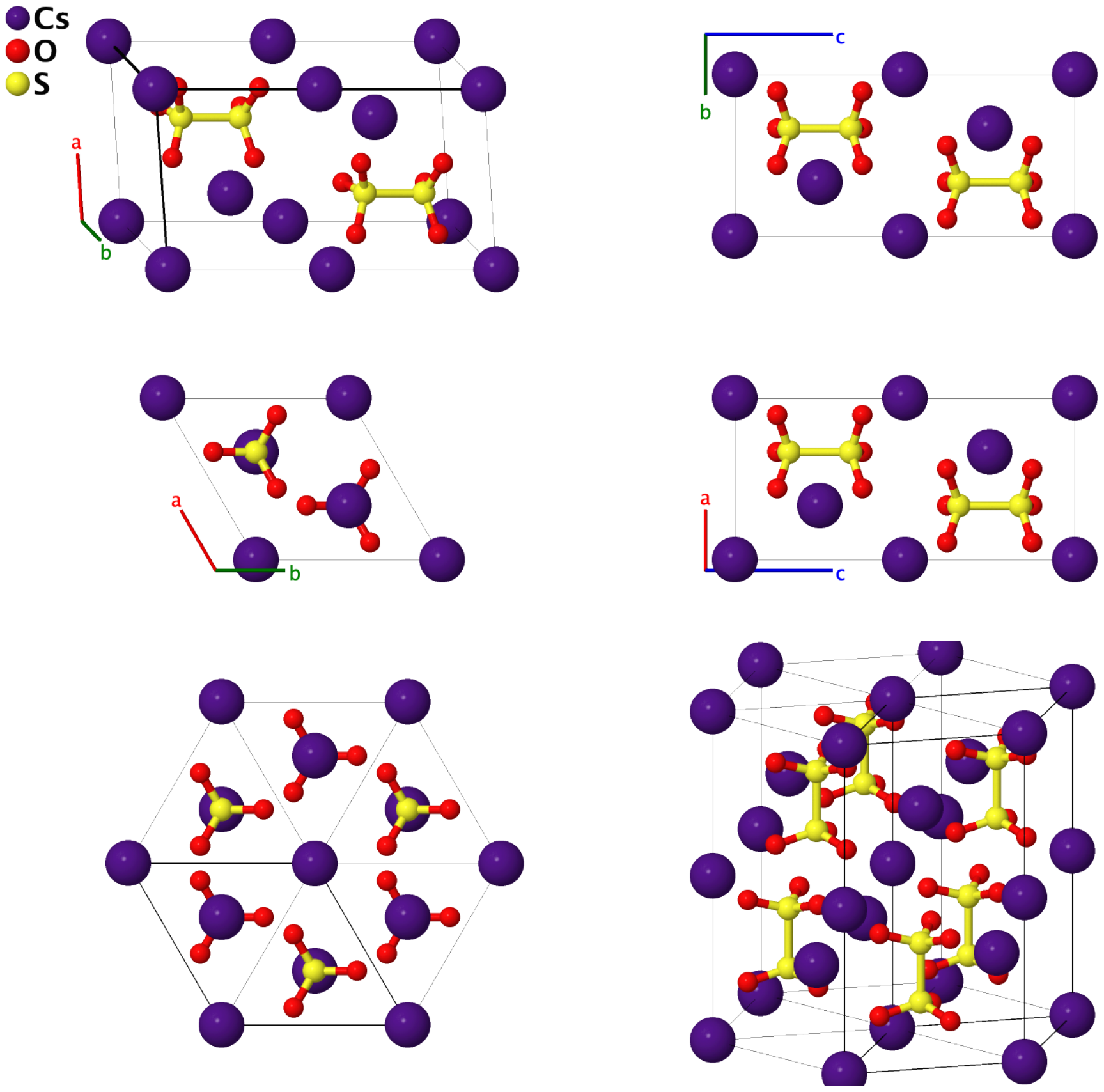

| | | |
|---|---|---|
| **Prototype** | : | $CsO_3S$ |
| **AFLOW prototype label** | : | AB3C_hP20_190_ac_i_f |
| ***Strukturbericht* designation** | : | $K1_2$ |
| **Pearson symbol** | : | hP20 |
| **Space group number** | : | 190 |
| **Space group symbol** | : | $P\bar{6}2c$ |
| **AFLOW prototype command** | : | `aflow --proto=AB3C_hP20_190_ac_i_f`<br>`--params=`$a, c/a, z_3, x_4, y_4, z_4$ |

- (Hägg, 1932) states that the space group might be $P6_322$ #182 or $P\bar{6}c2$ #190. We follow (Downs, 2003) and use the latter. (Hägg, 1932) and (Downs, 2003) set the sulfur coordinate $z_3 = 0.73$. As noted by (Hermann, 1937), this gives a very short S-S bond distance, on the order of 0.5Å. We use the value suggested by (Hermann, 1937), $z_3 = 0.63$, "which gives the $S_2O_6$ group exactly the same shape it has in $K_2S_2O_6$ [*Strukturbericht* $K1_1$]."

**Hexagonal primitive vectors:**

$$
\begin{aligned}
\mathbf{a}_1 &= \tfrac{1}{2}\, a\,\hat{\mathbf{x}} - \tfrac{\sqrt{3}}{2}\, a\,\hat{\mathbf{y}} \\
\mathbf{a}_2 &= \tfrac{1}{2}\, a\,\hat{\mathbf{x}} + \tfrac{\sqrt{3}}{2}\, a\,\hat{\mathbf{y}} \\
\mathbf{a}_3 &= c\,\hat{\mathbf{z}}
\end{aligned}
$$

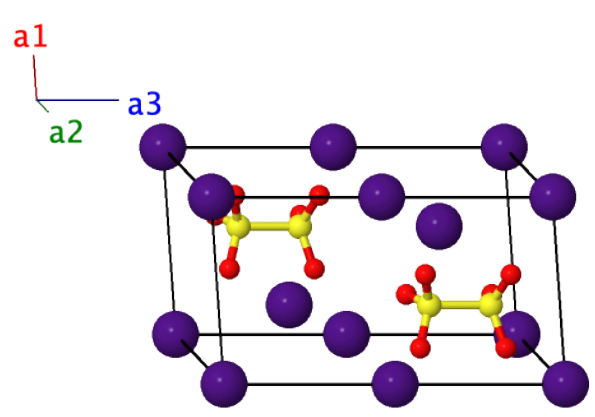

**Basis vectors:**

| | | Lattice Coordinates | | Cartesian Coordinates | Wyckoff Position | Atom Type |
|---|---|---|---|---|---|---|
| $\mathbf{B}_1$ | = | $0\,\mathbf{a}_1 + 0\,\mathbf{a}_2 + 0\,\mathbf{a}_3$ | = | $0\,\hat{\mathbf{x}} + 0\,\hat{\mathbf{y}} + 0\,\hat{\mathbf{z}}$ | (2$a$) | Cs I |
| $\mathbf{B}_2$ | = | $\frac{1}{2}\,\mathbf{a}_3$ | = | $\frac{1}{2}c\,\hat{\mathbf{z}}$ | (2$a$) | Cs I |
| $\mathbf{B}_3$ | = | $\frac{1}{3}\,\mathbf{a}_1 + \frac{2}{3}\,\mathbf{a}_2 + \frac{1}{4}\,\mathbf{a}_3$ | = | $\frac{1}{2}a\,\hat{\mathbf{x}} + \frac{1}{2\sqrt{3}}a\,\hat{\mathbf{y}} + \frac{1}{4}c\,\hat{\mathbf{z}}$ | (2$c$) | Cs II |
| $\mathbf{B}_4$ | = | $\frac{2}{3}\,\mathbf{a}_1 + \frac{1}{3}\,\mathbf{a}_2 + \frac{3}{4}\,\mathbf{a}_3$ | = | $\frac{1}{2}a\,\hat{\mathbf{x}} - \frac{1}{2\sqrt{3}}a\,\hat{\mathbf{y}} + \frac{3}{4}c\,\hat{\mathbf{z}}$ | (2$c$) | Cs II |
| $\mathbf{B}_5$ | = | $\frac{1}{3}\,\mathbf{a}_1 + \frac{2}{3}\,\mathbf{a}_2 + z_3\,\mathbf{a}_3$ | = | $\frac{1}{2}a\,\hat{\mathbf{x}} + \frac{1}{2\sqrt{3}}a\,\hat{\mathbf{y}} + z_3c\,\hat{\mathbf{z}}$ | (4$f$) | S |
| $\mathbf{B}_6$ | = | $\frac{1}{3}\,\mathbf{a}_1 + \frac{2}{3}\,\mathbf{a}_2 + \left(\frac{1}{2} - z_3\right)\,\mathbf{a}_3$ | = | $\frac{1}{2}a\,\hat{\mathbf{x}} + \frac{1}{2\sqrt{3}}a\,\hat{\mathbf{y}} + \left(\frac{1}{2} - z_3\right)c\,\hat{\mathbf{z}}$ | (4$f$) | S |
| $\mathbf{B}_7$ | = | $\frac{2}{3}\,\mathbf{a}_1 + \frac{1}{3}\,\mathbf{a}_2 - z_3\,\mathbf{a}_3$ | = | $\frac{1}{2}a\,\hat{\mathbf{x}} - \frac{1}{2\sqrt{3}}a\,\hat{\mathbf{y}} - z_3c\,\hat{\mathbf{z}}$ | (4$f$) | S |
| $\mathbf{B}_8$ | = | $\frac{2}{3}\,\mathbf{a}_1 + \frac{1}{3}\,\mathbf{a}_2 + \left(\frac{1}{2} + z_3\right)\,\mathbf{a}_3$ | = | $\frac{1}{2}a\,\hat{\mathbf{x}} - \frac{1}{2\sqrt{3}}a\,\hat{\mathbf{y}} + \left(\frac{1}{2} + z_3\right)c\,\hat{\mathbf{z}}$ | (4$f$) | S |
| $\mathbf{B}_9$ | = | $x_4\,\mathbf{a}_1 + y_4\,\mathbf{a}_2 + z_4\,\mathbf{a}_3$ | = | $\frac{1}{2}\left(x_4 + y_4\right)a\,\hat{\mathbf{x}} + \frac{\sqrt{3}}{2}\left(-x_4 + y_4\right)a\,\hat{\mathbf{y}} + z_4c\,\hat{\mathbf{z}}$ | (12$i$) | O |
| $\mathbf{B}_{10}$ | = | $-y_4\,\mathbf{a}_1 + \left(x_4 - y_4\right)\,\mathbf{a}_2 + z_4\,\mathbf{a}_3$ | = | $\left(\frac{1}{2}x_4 - y_4\right)a\,\hat{\mathbf{x}} + \frac{\sqrt{3}}{2}x_4a\,\hat{\mathbf{y}} + z_4c\,\hat{\mathbf{z}}$ | (12$i$) | O |
| $\mathbf{B}_{11}$ | = | $\left(-x_4 + y_4\right)\,\mathbf{a}_1 - x_4\,\mathbf{a}_2 + z_4\,\mathbf{a}_3$ | = | $\left(-x_4 + \frac{1}{2}y_4\right)a\,\hat{\mathbf{x}} - \frac{\sqrt{3}}{2}y_4a\,\hat{\mathbf{y}} + z_4c\,\hat{\mathbf{z}}$ | (12$i$) | O |
| $\mathbf{B}_{12}$ | = | $x_4\,\mathbf{a}_1 + y_4\,\mathbf{a}_2 + \left(\frac{1}{2} - z_4\right)\,\mathbf{a}_3$ | = | $\frac{1}{2}\left(x_4 + y_4\right)a\,\hat{\mathbf{x}} + \frac{\sqrt{3}}{2}\left(-x_4 + y_4\right)a\,\hat{\mathbf{y}} + \left(\frac{1}{2} - z_4\right)c\,\hat{\mathbf{z}}$ | (12$i$) | O |
| $\mathbf{B}_{13}$ | = | $-y_4\,\mathbf{a}_1 + \left(x_4 - y_4\right)\,\mathbf{a}_2 + \left(\frac{1}{2} - z_4\right)\,\mathbf{a}_3$ | = | $\left(\frac{1}{2}x_4 - y_4\right)a\,\hat{\mathbf{x}} + \frac{\sqrt{3}}{2}x_4a\,\hat{\mathbf{y}} + \left(\frac{1}{2} - z_4\right)c\,\hat{\mathbf{z}}$ | (12$i$) | O |
| $\mathbf{B}_{14}$ | = | $\left(-x_4 + y_4\right)\,\mathbf{a}_1 - x_4\,\mathbf{a}_2 + \left(\frac{1}{2} - z_4\right)\,\mathbf{a}_3$ | = | $\left(-x_4 + \frac{1}{2}y_4\right)a\,\hat{\mathbf{x}} - \frac{\sqrt{3}}{2}y_4a\,\hat{\mathbf{y}} + \left(\frac{1}{2} - z_4\right)c\,\hat{\mathbf{z}}$ | (12$i$) | O |
| $\mathbf{B}_{15}$ | = | $y_4\,\mathbf{a}_1 + x_4\,\mathbf{a}_2 - z_4\,\mathbf{a}_3$ | = | $\frac{1}{2}\left(x_4 + y_4\right)a\,\hat{\mathbf{x}} + \frac{\sqrt{3}}{2}\left(x_4 - y_4\right)a\,\hat{\mathbf{y}} - z_4c\,\hat{\mathbf{z}}$ | (12$i$) | O |
| $\mathbf{B}_{16}$ | = | $\left(x_4 - y_4\right)\,\mathbf{a}_1 - y_4\,\mathbf{a}_2 - z_4\,\mathbf{a}_3$ | = | $\left(\frac{1}{2}x_4 - y_4\right)a\,\hat{\mathbf{x}} - \frac{\sqrt{3}}{2}x_4a\,\hat{\mathbf{y}} - z_4c\,\hat{\mathbf{z}}$ | (12$i$) | O |
| $\mathbf{B}_{17}$ | = | $-x_4\,\mathbf{a}_1 + \left(-x_4 + y_4\right)\,\mathbf{a}_2 - z_4\,\mathbf{a}_3$ | = | $\left(-x_4 + \frac{1}{2}y_4\right)a\,\hat{\mathbf{x}} + \frac{\sqrt{3}}{2}y_4a\,\hat{\mathbf{y}} - z_4c\,\hat{\mathbf{z}}$ | (12$i$) | O |
| $\mathbf{B}_{18}$ | = | $y_4\,\mathbf{a}_1 + x_4\,\mathbf{a}_2 + \left(\frac{1}{2} + z_4\right)\,\mathbf{a}_3$ | = | $\frac{1}{2}\left(x_4 + y_4\right)a\,\hat{\mathbf{x}} + \frac{\sqrt{3}}{2}\left(x_4 - y_4\right)a\,\hat{\mathbf{y}} + \left(\frac{1}{2} + z_4\right)c\,\hat{\mathbf{z}}$ | (12$i$) | O |
| $\mathbf{B}_{19}$ | = | $\left(x_4 - y_4\right)\,\mathbf{a}_1 - y_4\,\mathbf{a}_2 + \left(\frac{1}{2} + z_4\right)\,\mathbf{a}_3$ | = | $\left(\frac{1}{2}x_4 - y_4\right)a\,\hat{\mathbf{x}} - \frac{\sqrt{3}}{2}x_4a\,\hat{\mathbf{y}} + \left(\frac{1}{2} + z_4\right)c\,\hat{\mathbf{z}}$ | (12$i$) | O |
| $\mathbf{B}_{20}$ | = | $-x_4\,\mathbf{a}_1 + \left(-x_4 + y_4\right)\,\mathbf{a}_2 + \left(\frac{1}{2} + z_4\right)\,\mathbf{a}_3$ | = | $\left(-x_4 + \frac{1}{2}y_4\right)a\,\hat{\mathbf{x}} + \frac{\sqrt{3}}{2}y_4a\,\hat{\mathbf{y}} + \left(\frac{1}{2} + z_4\right)c\,\hat{\mathbf{z}}$ | (12$i$) | O |

**Geometry files:**

- CIF: pp. 1761

- POSCAR: pp. 1761

# Bastnäsite [$CeF(CO_3)$] Structure: ABCD3_hP36_190_h_g_af_hi

http://aflow.org/prototype-encyclopedia/ABCD3_hP36_190_h_g_af_hi

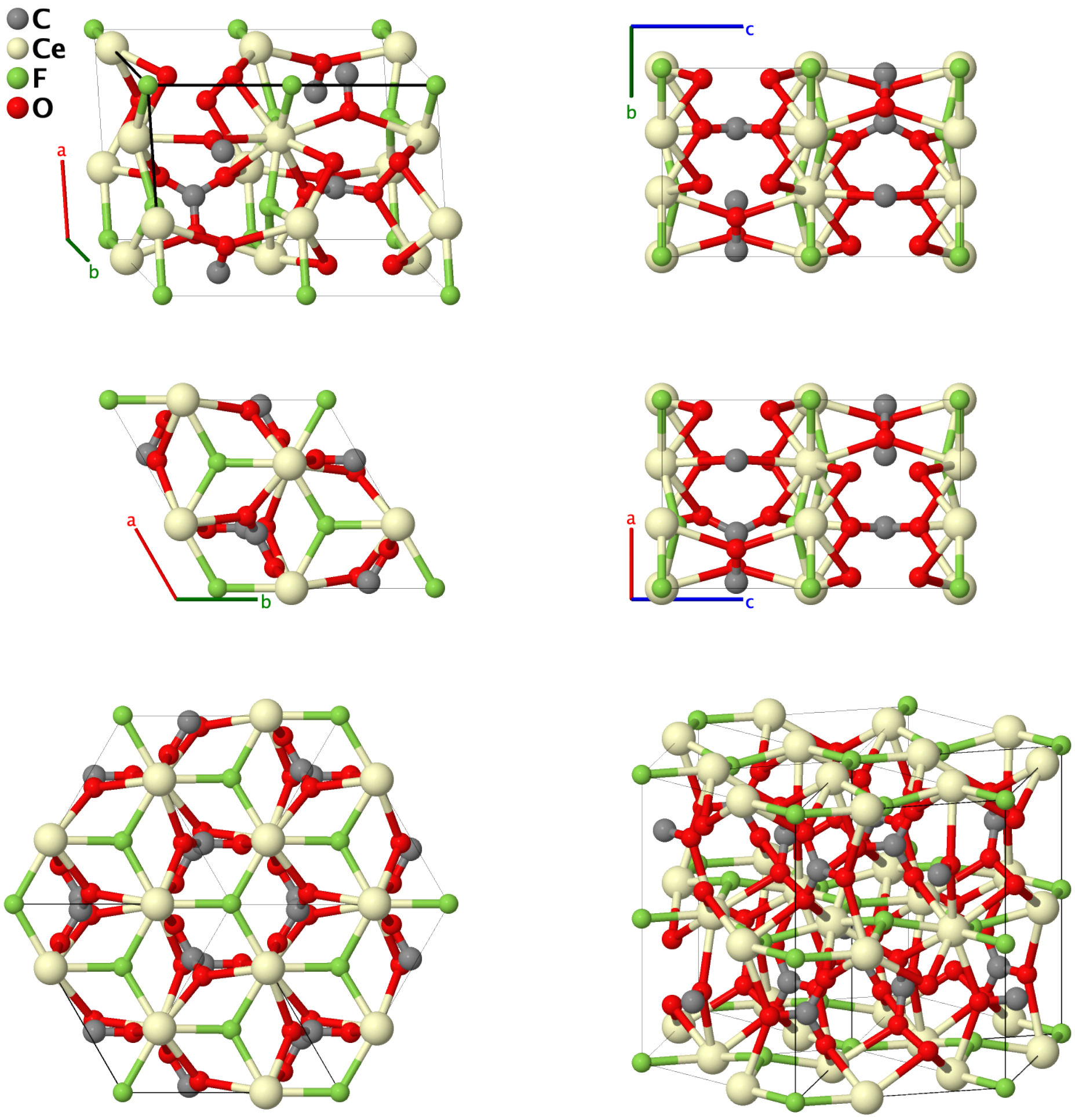

| | | |
|---|---|---|
| **Prototype** | : | $CCeFO_3$ |
| **AFLOW prototype label** | : | ABCD3_hP36_190_h_g_af_hi |
| ***Strukturbericht* designation** | : | $G7_1$ |
| **Pearson symbol** | : | hP36 |
| **Space group number** | : | 190 |
| **Space group symbol** | : | $P\bar{6}2c$ |
| **AFLOW prototype command** | : | `aflow --proto=ABCD3_hP36_190_h_g_af_hi` <br> `--params=`$a, c/a, z_2, x_3, x_4, y_4, x_5, y_5, x_6, y_6, z_6$ |

**Other compounds with this structure**

- $LaF(CO_3)$

- Technically the name bastnäsite belongs to both cerium and lanthanum compounds, and the are usually designated bastnäsite-(Ce) and bastnäsite-(La), respectively.

**Hexagonal primitive vectors:**

$$\mathbf{a}_1 = \frac{1}{2}\,a\,\hat{\mathbf{x}} - \frac{\sqrt{3}}{2}\,a\,\hat{\mathbf{y}}$$
$$\mathbf{a}_2 = \frac{1}{2}\,a\,\hat{\mathbf{x}} + \frac{\sqrt{3}}{2}\,a\,\hat{\mathbf{y}}$$
$$\mathbf{a}_3 = c\,\hat{\mathbf{z}}$$

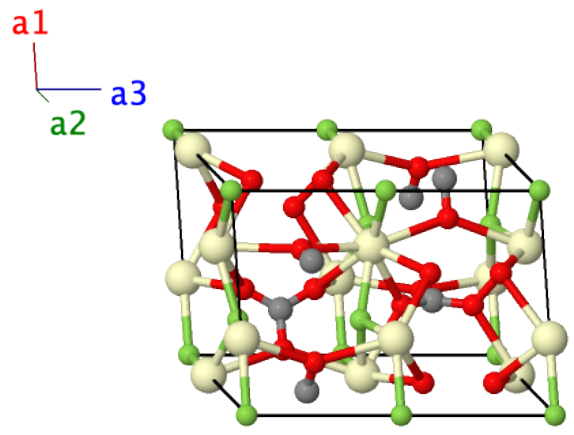

**Basis vectors:**

| | | Lattice Coordinates | | Cartesian Coordinates | Wyckoff Position | Atom Type |
|---|---|---|---|---|---|---|
| $\mathbf{B}_1$ | = | $0\,\mathbf{a}_1 + 0\,\mathbf{a}_2 + 0\,\mathbf{a}_3$ | = | $0\,\hat{\mathbf{x}} + 0\,\hat{\mathbf{y}} + 0\,\hat{\mathbf{z}}$ | (2$a$) | F I |
| $\mathbf{B}_2$ | = | $\frac{1}{2}\,\mathbf{a}_3$ | = | $\frac{1}{2}c\,\hat{\mathbf{z}}$ | (2$a$) | F I |
| $\mathbf{B}_3$ | = | $\frac{1}{3}\,\mathbf{a}_1 + \frac{2}{3}\,\mathbf{a}_2 + z_2\,\mathbf{a}_3$ | = | $\frac{1}{2}a\,\hat{\mathbf{x}} + \frac{1}{2\sqrt{3}}a\,\hat{\mathbf{y}} + z_2c\,\hat{\mathbf{z}}$ | (4$f$) | F II |
| $\mathbf{B}_4$ | = | $\frac{1}{3}\,\mathbf{a}_1 + \frac{2}{3}\,\mathbf{a}_2 + \left(\frac{1}{2} - z_2\right)\,\mathbf{a}_3$ | = | $\frac{1}{2}a\,\hat{\mathbf{x}} + \frac{1}{2\sqrt{3}}a\,\hat{\mathbf{y}} + \left(\frac{1}{2} - z_2\right)c\,\hat{\mathbf{z}}$ | (4$f$) | F II |
| $\mathbf{B}_5$ | = | $\frac{2}{3}\,\mathbf{a}_1 + \frac{1}{3}\,\mathbf{a}_2 - z_2\,\mathbf{a}_3$ | = | $\frac{1}{2}a\,\hat{\mathbf{x}} - \frac{1}{2\sqrt{3}}a\,\hat{\mathbf{y}} - z_2c\,\hat{\mathbf{z}}$ | (4$f$) | F II |
| $\mathbf{B}_6$ | = | $\frac{2}{3}\,\mathbf{a}_1 + \frac{1}{3}\,\mathbf{a}_2 + \left(\frac{1}{2} + z_2\right)\,\mathbf{a}_3$ | = | $\frac{1}{2}a\,\hat{\mathbf{x}} - \frac{1}{2\sqrt{3}}a\,\hat{\mathbf{y}} + \left(\frac{1}{2} + z_2\right)c\,\hat{\mathbf{z}}$ | (4$f$) | F II |
| $\mathbf{B}_7$ | = | $x_3\,\mathbf{a}_1$ | = | $\frac{1}{2}x_3a\,\hat{\mathbf{x}} - \frac{\sqrt{3}}{2}x_3a\,\hat{\mathbf{y}}$ | (6$g$) | Ce |
| $\mathbf{B}_8$ | = | $x_3\,\mathbf{a}_2$ | = | $\frac{1}{2}x_3a\,\hat{\mathbf{x}} + \frac{\sqrt{3}}{2}x_3a\,\hat{\mathbf{y}}$ | (6$g$) | Ce |
| $\mathbf{B}_9$ | = | $-x_3\,\mathbf{a}_1 - x_3\,\mathbf{a}_2$ | = | $-x_3a\,\hat{\mathbf{x}}$ | (6$g$) | Ce |
| $\mathbf{B}_{10}$ | = | $x_3\,\mathbf{a}_1 + \frac{1}{2}\,\mathbf{a}_3$ | = | $\frac{1}{2}x_3a\,\hat{\mathbf{x}} - \frac{\sqrt{3}}{2}x_3a\,\hat{\mathbf{y}} + \frac{1}{2}c\,\hat{\mathbf{z}}$ | (6$g$) | Ce |
| $\mathbf{B}_{11}$ | = | $x_3\,\mathbf{a}_2 + \frac{1}{2}\,\mathbf{a}_3$ | = | $\frac{1}{2}x_3a\,\hat{\mathbf{x}} + \frac{\sqrt{3}}{2}x_3a\,\hat{\mathbf{y}} + \frac{1}{2}c\,\hat{\mathbf{z}}$ | (6$g$) | Ce |
| $\mathbf{B}_{12}$ | = | $-x_3\,\mathbf{a}_1 - x_3\,\mathbf{a}_2 + \frac{1}{2}\,\mathbf{a}_3$ | = | $-x_3a\,\hat{\mathbf{x}} + \frac{1}{2}c\,\hat{\mathbf{z}}$ | (6$g$) | Ce |
| $\mathbf{B}_{13}$ | = | $x_4\,\mathbf{a}_1 + y_4\,\mathbf{a}_2 + \frac{1}{4}\,\mathbf{a}_3$ | = | $\frac{1}{2}\left(x_4 + y_4\right)a\,\hat{\mathbf{x}} + \frac{\sqrt{3}}{2}\left(-x_4 + y_4\right)a\,\hat{\mathbf{y}} + \frac{1}{4}c\,\hat{\mathbf{z}}$ | (6$h$) | C |
| $\mathbf{B}_{14}$ | = | $-y_4\,\mathbf{a}_1 + \left(x_4 - y_4\right)\,\mathbf{a}_2 + \frac{1}{4}\,\mathbf{a}_3$ | = | $\left(\frac{1}{2}x_4 - y_4\right)a\,\hat{\mathbf{x}} + \frac{\sqrt{3}}{2}x_4a\,\hat{\mathbf{y}} + \frac{1}{4}c\,\hat{\mathbf{z}}$ | (6$h$) | C |
| $\mathbf{B}_{15}$ | = | $\left(-x_4 + y_4\right)\,\mathbf{a}_1 - x_4\,\mathbf{a}_2 + \frac{1}{4}\,\mathbf{a}_3$ | = | $\left(-x_4 + \frac{1}{2}y_4\right)a\,\hat{\mathbf{x}} - \frac{\sqrt{3}}{2}y_4a\,\hat{\mathbf{y}} + \frac{1}{4}c\,\hat{\mathbf{z}}$ | (6$h$) | C |
| $\mathbf{B}_{16}$ | = | $y_4\,\mathbf{a}_1 + x_4\,\mathbf{a}_2 + \frac{3}{4}\,\mathbf{a}_3$ | = | $\frac{1}{2}\left(x_4 + y_4\right)a\,\hat{\mathbf{x}} + \frac{\sqrt{3}}{2}\left(x_4 - y_4\right)a\,\hat{\mathbf{y}} + \frac{3}{4}c\,\hat{\mathbf{z}}$ | (6$h$) | C |
| $\mathbf{B}_{17}$ | = | $\left(x_4 - y_4\right)\,\mathbf{a}_1 - y_4\,\mathbf{a}_2 + \frac{3}{4}\,\mathbf{a}_3$ | = | $\left(\frac{1}{2}x_4 - y_4\right)a\,\hat{\mathbf{x}} - \frac{\sqrt{3}}{2}x_4a\,\hat{\mathbf{y}} + \frac{3}{4}c\,\hat{\mathbf{z}}$ | (6$h$) | C |
| $\mathbf{B}_{18}$ | = | $-x_4\,\mathbf{a}_1 + \left(-x_4 + y_4\right)\,\mathbf{a}_2 + \frac{3}{4}\,\mathbf{a}_3$ | = | $\left(-x_4 + \frac{1}{2}y_4\right)a\,\hat{\mathbf{x}} + \frac{\sqrt{3}}{2}y_4a\,\hat{\mathbf{y}} + \frac{3}{4}c\,\hat{\mathbf{z}}$ | (6$h$) | C |
| $\mathbf{B}_{19}$ | = | $x_5\,\mathbf{a}_1 + y_5\,\mathbf{a}_2 + \frac{1}{4}\,\mathbf{a}_3$ | = | $\frac{1}{2}\left(x_5 + y_5\right)a\,\hat{\mathbf{x}} + \frac{\sqrt{3}}{2}\left(-x_5 + y_5\right)a\,\hat{\mathbf{y}} + \frac{1}{4}c\,\hat{\mathbf{z}}$ | (6$h$) | O I |
| $\mathbf{B}_{20}$ | = | $-y_5\,\mathbf{a}_1 + \left(x_5 - y_5\right)\,\mathbf{a}_2 + \frac{1}{4}\,\mathbf{a}_3$ | = | $\left(\frac{1}{2}x_5 - y_5\right)a\,\hat{\mathbf{x}} + \frac{\sqrt{3}}{2}x_5a\,\hat{\mathbf{y}} + \frac{1}{4}c\,\hat{\mathbf{z}}$ | (6$h$) | O I |
| $\mathbf{B}_{21}$ | = | $\left(-x_5 + y_5\right)\,\mathbf{a}_1 - x_5\,\mathbf{a}_2 + \frac{1}{4}\,\mathbf{a}_3$ | = | $\left(-x_5 + \frac{1}{2}y_5\right)a\,\hat{\mathbf{x}} - \frac{\sqrt{3}}{2}y_5a\,\hat{\mathbf{y}} + \frac{1}{4}c\,\hat{\mathbf{z}}$ | (6$h$) | O I |
| $\mathbf{B}_{22}$ | = | $y_5\,\mathbf{a}_1 + x_5\,\mathbf{a}_2 + \frac{3}{4}\,\mathbf{a}_3$ | = | $\frac{1}{2}\left(x_5 + y_5\right)a\,\hat{\mathbf{x}} + \frac{\sqrt{3}}{2}\left(x_5 - y_5\right)a\,\hat{\mathbf{y}} + \frac{3}{4}c\,\hat{\mathbf{z}}$ | (6$h$) | O I |
| $\mathbf{B}_{23}$ | = | $\left(x_5 - y_5\right)\,\mathbf{a}_1 - y_5\,\mathbf{a}_2 + \frac{3}{4}\,\mathbf{a}_3$ | = | $\left(\frac{1}{2}x_5 - y_5\right)a\,\hat{\mathbf{x}} - \frac{\sqrt{3}}{2}x_5a\,\hat{\mathbf{y}} + \frac{3}{4}c\,\hat{\mathbf{z}}$ | (6$h$) | O I |
| $\mathbf{B}_{24}$ | = | $-x_5\,\mathbf{a}_1 + \left(-x_5 + y_5\right)\,\mathbf{a}_2 + \frac{3}{4}\,\mathbf{a}_3$ | = | $\left(-x_5 + \frac{1}{2}y_5\right)a\,\hat{\mathbf{x}} + \frac{\sqrt{3}}{2}y_5a\,\hat{\mathbf{y}} + \frac{3}{4}c\,\hat{\mathbf{z}}$ | (6$h$) | O I |
| $\mathbf{B}_{25}$ | = | $x_6\,\mathbf{a}_1 + y_6\,\mathbf{a}_2 + z_6\,\mathbf{a}_3$ | = | $\frac{1}{2}\left(x_6 + y_6\right)a\,\hat{\mathbf{x}} + \frac{\sqrt{3}}{2}\left(-x_6 + y_6\right)a\,\hat{\mathbf{y}} + z_6c\,\hat{\mathbf{z}}$ | (12$i$) | O II |
| $\mathbf{B}_{26}$ | = | $-y_6\,\mathbf{a}_1 + (x_6 - y_6)\,\mathbf{a}_2 + z_6\,\mathbf{a}_3$ | = | $\left(\frac{1}{2}x_6 - y_6\right) a\,\hat{\mathbf{x}} + \frac{\sqrt{3}}{2}x_6 a\,\hat{\mathbf{y}} + z_6 c\,\hat{\mathbf{z}}$ | (12$i$) | O II |
| $\mathbf{B}_{27}$ | = | $(-x_6 + y_6)\,\mathbf{a}_1 - x_6\,\mathbf{a}_2 + z_6\,\mathbf{a}_3$ | = | $\left(-x_6 + \frac{1}{2}y_6\right) a\,\hat{\mathbf{x}} - \frac{\sqrt{3}}{2}y_6 a\,\hat{\mathbf{y}} + z_6 c\,\hat{\mathbf{z}}$ | (12$i$) | O II |
| $\mathbf{B}_{28}$ | = | $x_6\,\mathbf{a}_1 + y_6\,\mathbf{a}_2 + \left(\frac{1}{2} - z_6\right)\mathbf{a}_3$ | = | $\frac{1}{2}(x_6 + y_6)\,a\,\hat{\mathbf{x}} + \frac{\sqrt{3}}{2}(-x_6 + y_6)\,a\,\hat{\mathbf{y}} + \left(\frac{1}{2} - z_6\right) c\,\hat{\mathbf{z}}$ | (12$i$) | O II |
| $\mathbf{B}_{29}$ | = | $-y_6\,\mathbf{a}_1 + (x_6 - y_6)\,\mathbf{a}_2 + \left(\frac{1}{2} - z_6\right)\mathbf{a}_3$ | = | $\left(\frac{1}{2}x_6 - y_6\right) a\,\hat{\mathbf{x}} + \frac{\sqrt{3}}{2}x_6 a\,\hat{\mathbf{y}} + \left(\frac{1}{2} - z_6\right) c\,\hat{\mathbf{z}}$ | (12$i$) | O II |
| $\mathbf{B}_{30}$ | = | $(-x_6 + y_6)\,\mathbf{a}_1 - x_6\,\mathbf{a}_2 + \left(\frac{1}{2} - z_6\right)\mathbf{a}_3$ | = | $\left(-x_6 + \frac{1}{2}y_6\right) a\,\hat{\mathbf{x}} - \frac{\sqrt{3}}{2}y_6 a\,\hat{\mathbf{y}} + \left(\frac{1}{2} - z_6\right) c\,\hat{\mathbf{z}}$ | (12$i$) | O II |
| $\mathbf{B}_{31}$ | = | $y_6\,\mathbf{a}_1 + x_6\,\mathbf{a}_2 - z_6\,\mathbf{a}_3$ | = | $\frac{1}{2}(x_6 + y_6)\,a\,\hat{\mathbf{x}} + \frac{\sqrt{3}}{2}(x_6 - y_6)\,a\,\hat{\mathbf{y}} - z_6 c\,\hat{\mathbf{z}}$ | (12$i$) | O II |
| $\mathbf{B}_{32}$ | = | $(x_6 - y_6)\,\mathbf{a}_1 - y_6\,\mathbf{a}_2 - z_6\,\mathbf{a}_3$ | = | $\left(\frac{1}{2}x_6 - y_6\right) a\,\hat{\mathbf{x}} - \frac{\sqrt{3}}{2}x_6 a\,\hat{\mathbf{y}} - z_6 c\,\hat{\mathbf{z}}$ | (12$i$) | O II |
| $\mathbf{B}_{33}$ | = | $-x_6\,\mathbf{a}_1 + (-x_6 + y_6)\,\mathbf{a}_2 - z_6\,\mathbf{a}_3$ | = | $\left(-x_6 + \frac{1}{2}y_6\right) a\,\hat{\mathbf{x}} + \frac{\sqrt{3}}{2}y_6 a\,\hat{\mathbf{y}} - z_6 c\,\hat{\mathbf{z}}$ | (12$i$) | O II |
| $\mathbf{B}_{34}$ | = | $y_6\,\mathbf{a}_1 + x_6\,\mathbf{a}_2 + \left(\frac{1}{2} + z_6\right)\mathbf{a}_3$ | = | $\frac{1}{2}(x_6 + y_6)\,a\,\hat{\mathbf{x}} + \frac{\sqrt{3}}{2}(x_6 - y_6)\,a\,\hat{\mathbf{y}} + \left(\frac{1}{2} + z_6\right) c\,\hat{\mathbf{z}}$ | (12$i$) | O II |
| $\mathbf{B}_{35}$ | = | $(x_6 - y_6)\,\mathbf{a}_1 - y_6\,\mathbf{a}_2 + \left(\frac{1}{2} + z_6\right)\mathbf{a}_3$ | = | $\left(\frac{1}{2}x_6 - y_6\right) a\,\hat{\mathbf{x}} - \frac{\sqrt{3}}{2}x_6 a\,\hat{\mathbf{y}} + \left(\frac{1}{2} + z_6\right) c\,\hat{\mathbf{z}}$ | (12$i$) | O II |
| $\mathbf{B}_{36}$ | = | $-x_6\,\mathbf{a}_1 + (-x_6 + y_6)\,\mathbf{a}_2 + \left(\frac{1}{2} + z_6\right)\mathbf{a}_3$ | = | $\left(-x_6 + \frac{1}{2}y_6\right) a\,\hat{\mathbf{x}} + \frac{\sqrt{3}}{2}y_6 a\,\hat{\mathbf{y}} + \left(\frac{1}{2} + z_6\right) c\,\hat{\mathbf{z}}$ | (12$i$) | O II |

---

---

**Geometry files:**

- CIF: pp. 1762
- POSCAR: pp. 1762

# $TiBe_{12}$ (approximate, $D2_a$) Structure: A12B_hP13_191_cdei_a

http://aflow.org/prototype-encyclopedia/A12B_hP13_191_cdei_a

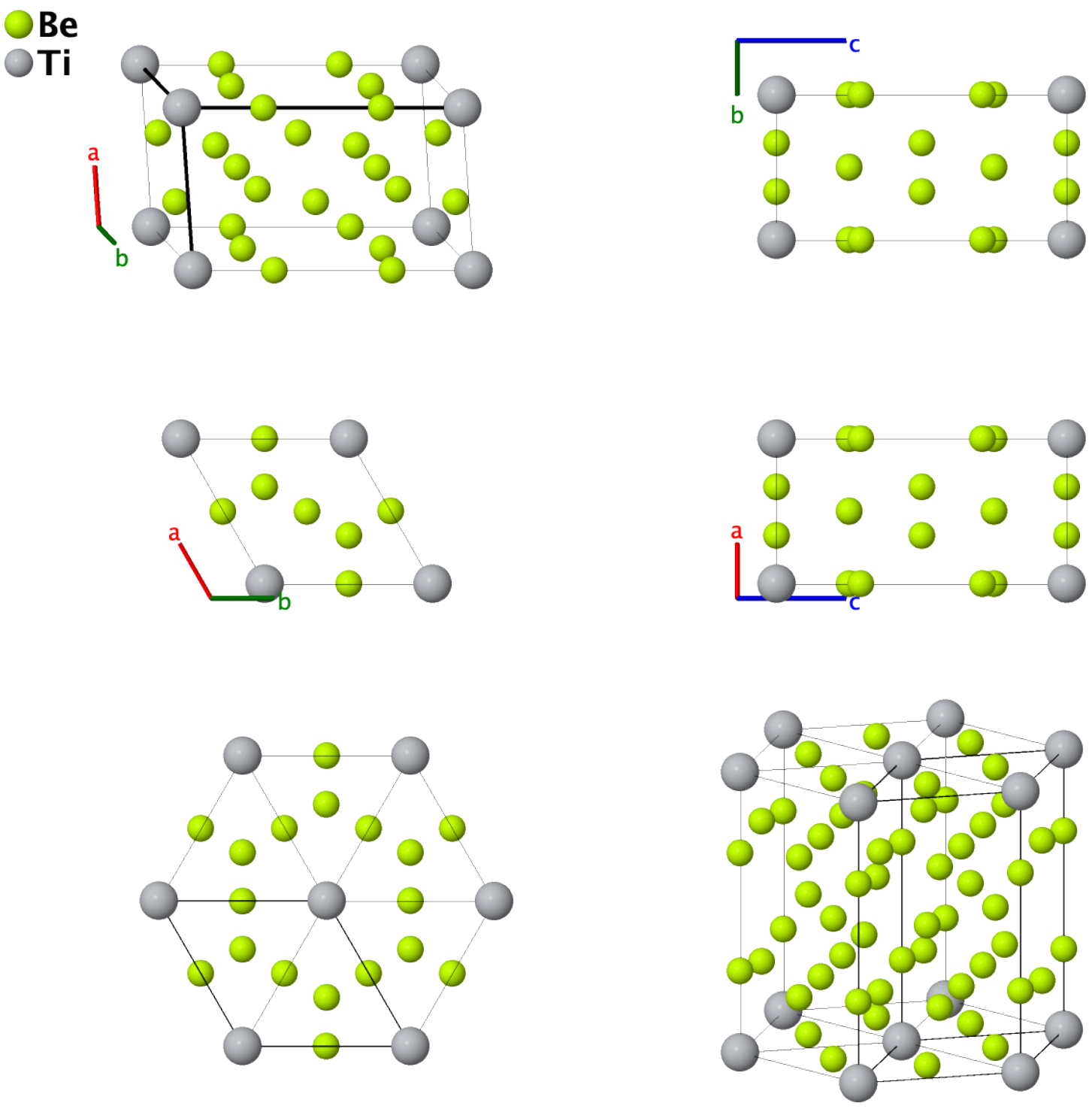

| | | |
|---|---|---|
| **Prototype** | : | $Be_{12}Ti$ |
| **AFLOW prototype label** | : | A12B_hP13_191_cdei_a |
| ***Strukturbericht* designation** | : | $D2_a$ |
| **Pearson symbol** | : | hP13 |
| **Space group number** | : | 191 |
| **Space group symbol** | : | $P6/mmm$ |
| **AFLOW prototype command** | : | `aflow --proto=A12B_hP13_191_cdei_a --params=`$a, c/a, z_4, z_5$ |

- The structure of $TiBe_{12}$ is not settled. (Raeuchle, 1952) described the structure as a somewhat disordered supercell containing 48 atoms with lattice constants $a$ = 29.44 Å and $c$ = 7.33 Å, but they stated that a 'pseudo-cell' existed with dimensions $a$ = 4.23 Å and $c$ = 7.33 Å. This pseudo-cell is described here, and was assigned the $D2_a$ *Strukturbericht* type by Smithells (Brandes, 1992). Rauchle and Rundle suggested that the larger primitive cell was constructed from the multiple pseudo-cells, with the titanium atom alternating between the (1$a$) and (1$b$) $(0, 0, 1/2)$ Wyckoff positions.

- Other experimenters have suggested that the actual structure of $TiBe_{12}$ is tetragonal. (Jackson, 2016) presents first-principles calculations which suggest that the actual structure is of the tetragonal $ThMn_{12}$ ($D2_b$) type.

**Hexagonal primitive vectors:**

$$\mathbf{a}_1 = \frac{1}{2}\,a\,\hat{\mathbf{x}} - \frac{\sqrt{3}}{2}\,a\,\hat{\mathbf{y}}$$
$$\mathbf{a}_2 = \frac{1}{2}\,a\,\hat{\mathbf{x}} + \frac{\sqrt{3}}{2}\,a\,\hat{\mathbf{y}}$$
$$\mathbf{a}_3 = c\,\hat{\mathbf{z}}$$

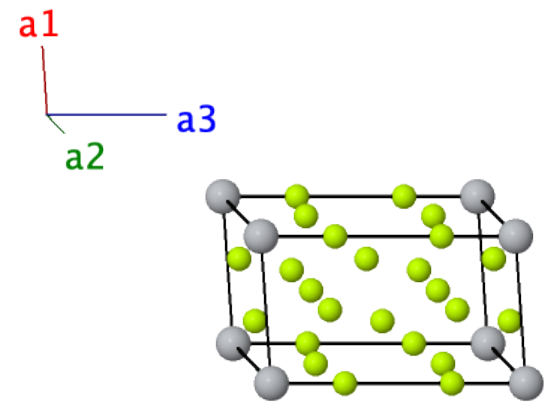

**Basis vectors:**

| | | Lattice Coordinates | | Cartesian Coordinates | Wyckoff Position | Atom Type |
|---|---|---|---|---|---|---|
| $\mathbf{B}_1$ | = | $0\,\mathbf{a}_1 + 0\,\mathbf{a}_2 + 0\,\mathbf{a}_3$ | = | $0\,\hat{\mathbf{x}} + 0\,\hat{\mathbf{y}} + 0\,\hat{\mathbf{z}}$ | $(1a)$ | Ti |
| $\mathbf{B}_2$ | = | $\frac{1}{3}\,\mathbf{a}_1 + \frac{2}{3}\,\mathbf{a}_2$ | = | $\frac{1}{2}a\,\hat{\mathbf{x}} + \frac{1}{2\sqrt{3}}a\,\hat{\mathbf{y}}$ | $(2c)$ | Be I |
| $\mathbf{B}_3$ | = | $\frac{2}{3}\,\mathbf{a}_1 + \frac{1}{3}\,\mathbf{a}_2$ | = | $\frac{1}{2}a\,\hat{\mathbf{x}} - \frac{1}{2\sqrt{3}}a\,\hat{\mathbf{y}}$ | $(2c)$ | Be I |
| $\mathbf{B}_4$ | = | $\frac{1}{3}\,\mathbf{a}_1 + \frac{2}{3}\,\mathbf{a}_2 + \frac{1}{2}\,\mathbf{a}_3$ | = | $\frac{1}{2}a\,\hat{\mathbf{x}} + \frac{1}{2\sqrt{3}}a\,\hat{\mathbf{y}} + \frac{1}{2}c\,\hat{\mathbf{z}}$ | $(2d)$ | Be II |
| $\mathbf{B}_5$ | = | $\frac{2}{3}\,\mathbf{a}_1 + \frac{1}{3}\,\mathbf{a}_2 + \frac{1}{2}\,\mathbf{a}_3$ | = | $\frac{1}{2}a\,\hat{\mathbf{x}} - \frac{1}{2\sqrt{3}}a\,\hat{\mathbf{y}} + \frac{1}{2}c\,\hat{\mathbf{z}}$ | $(2d)$ | Be II |
| $\mathbf{B}_6$ | = | $z_4\,\mathbf{a}_3$ | = | $z_4 c\,\hat{\mathbf{z}}$ | $(2e)$ | Be III |
| $\mathbf{B}_7$ | = | $-z_4\,\mathbf{a}_3$ | = | $-z_4 c\,\hat{\mathbf{z}}$ | $(2e)$ | Be III |
| $\mathbf{B}_8$ | = | $\frac{1}{2}\,\mathbf{a}_1 + z_5\,\mathbf{a}_3$ | = | $\frac{1}{4}a\,\hat{\mathbf{x}} - \frac{\sqrt{3}}{4}a\,\hat{\mathbf{y}} + z_5 c\,\hat{\mathbf{z}}$ | $(6i)$ | Be IV |
| $\mathbf{B}_9$ | = | $\frac{1}{2}\,\mathbf{a}_2 + z_5\,\mathbf{a}_3$ | = | $\frac{1}{4}a\,\hat{\mathbf{x}} + \frac{\sqrt{3}}{4}a\,\hat{\mathbf{y}} + z_5 c\,\hat{\mathbf{z}}$ | $(6i)$ | Be IV |
| $\mathbf{B}_{10}$ | = | $\frac{1}{2}\,\mathbf{a}_1 + \frac{1}{2}\,\mathbf{a}_2 + z_5\,\mathbf{a}_3$ | = | $\frac{1}{2}a\,\hat{\mathbf{x}} + z_5 c\,\hat{\mathbf{z}}$ | $(6i)$ | Be IV |
| $\mathbf{B}_{11}$ | = | $\frac{1}{2}\,\mathbf{a}_2 - z_5\,\mathbf{a}_3$ | = | $\frac{1}{4}a\,\hat{\mathbf{x}} + \frac{\sqrt{3}}{4}a\,\hat{\mathbf{y}} - z_5 c\,\hat{\mathbf{z}}$ | $(6i)$ | Be IV |
| $\mathbf{B}_{12}$ | = | $\frac{1}{2}\,\mathbf{a}_1 - z_5\,\mathbf{a}_3$ | = | $\frac{1}{4}a\,\hat{\mathbf{x}} - \frac{\sqrt{3}}{4}a\,\hat{\mathbf{y}} - z_5 c\,\hat{\mathbf{z}}$ | $(6i)$ | Be IV |
| $\mathbf{B}_{13}$ | = | $\frac{1}{2}\,\mathbf{a}_1 + \frac{1}{2}\,\mathbf{a}_2 - z_5\,\mathbf{a}_3$ | = | $\frac{1}{2}a\,\hat{\mathbf{x}} - z_5 c\,\hat{\mathbf{z}}$ | $(6i)$ | Be IV |

**Found in:**

**Geometry files:**

- CIF: pp. 1762
- POSCAR: pp. 1763

# Hexagonal $WO_3$ Structure: A3B_hP12_191_gl_f

http://aflow.org/prototype-encyclopedia/A3B_hP12_191_gl_f

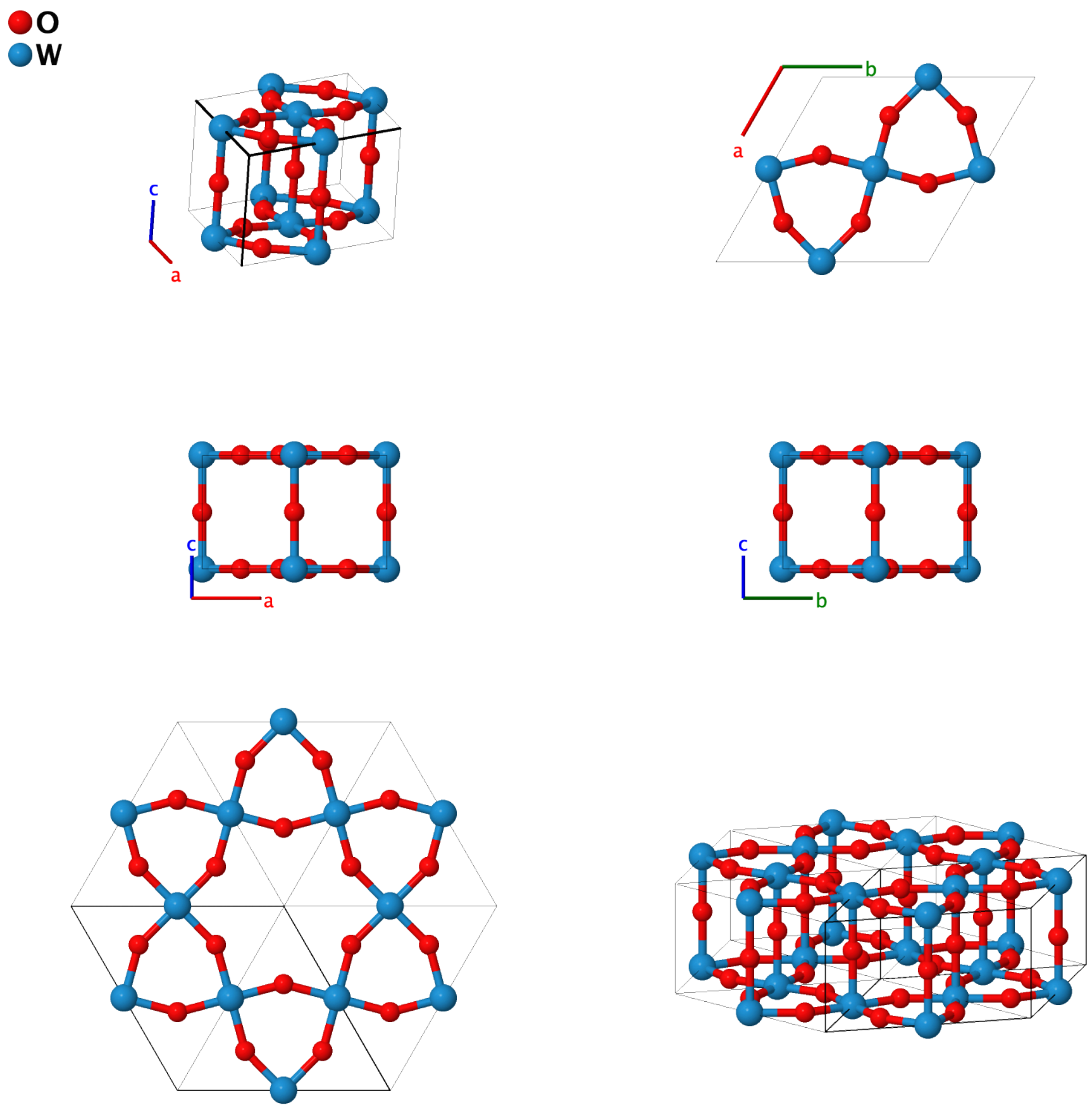

| | | |
|---|---|---|
| **Prototype** | : | $O_3W$ |
| **AFLOW prototype label** | : | A3B_hP12_191_gl_f |
| ***Strukturbericht* designation** | : | None |
| **Pearson symbol** | : | hP12 |
| **Space group number** | : | 191 |
| **Space group symbol** | : | $P6/mmm$ |
| **AFLOW prototype command** | : | `aflow --proto=A3B_hP12_191_gl_f --params=`$a, c/a, x_3$ |

- All stable phases of $WO_3$ are distortions of the cubic $\alpha$-$ReO_3$ ($D0_9$) phase. (Woodward, 1997 and Vogt, 1999) The known stable phases and their approximate temperature ranges are:
  - $\alpha$-$WO_3$ (1010-1170 K) (Vogt, 1999)
  - $\beta$-$WO_3$ (600-1170 K) (Vogt, 1999)
  - $\gamma$-$WO_3$ (290-600 K) (Vogt, 1999)
  - $\delta$-$WO_3$ (230-290 K) (Diehl, 1978)
  - $\epsilon$-$WO_3$ (below 23 K) (Woodward, 1997)
- In addition, several other structures have been proposed and/or found:
  - The original $D0_{10}$ structure (Bräkken, 1931), (Hermann, 1937) superseded by $\delta$-$WO_3$

– Original $\beta$-$WO_3$ (Salje, 1977)
– Hexagonal $WO_3$ (Gerand, 1979) (metastable), this structure

---

**Hexagonal primitive vectors:**

$$\mathbf{a}_1 = \frac{1}{2}\,a\,\hat{\mathbf{x}} - \frac{\sqrt{3}}{2}\,a\,\hat{\mathbf{y}}$$
$$\mathbf{a}_2 = \frac{1}{2}\,a\,\hat{\mathbf{x}} + \frac{\sqrt{3}}{2}\,a\,\hat{\mathbf{y}}$$
$$\mathbf{a}_3 = c\,\hat{\mathbf{z}}$$

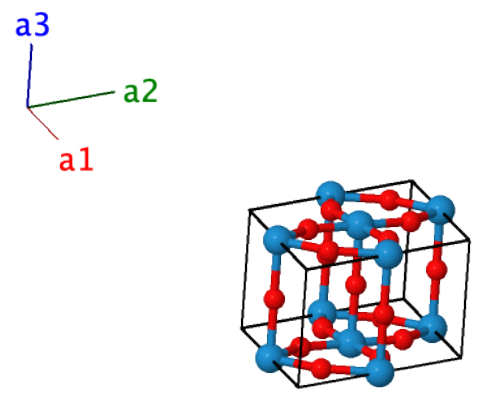

---

**Basis vectors:**

| | | Lattice Coordinates | | Cartesian Coordinates | Wyckoff Position | Atom Type |
|---|---|---|---|---|---|---|
| $\mathbf{B}_1$ | = | $\frac{1}{2}\,\mathbf{a}_1$ | = | $\frac{1}{4}a\,\hat{\mathbf{x}} - \frac{\sqrt{3}}{4}a\,\hat{\mathbf{y}}$ | $(3f)$ | W |
| $\mathbf{B}_2$ | = | $\frac{1}{2}\,\mathbf{a}_2$ | = | $\frac{1}{4}a\,\hat{\mathbf{x}} + \frac{\sqrt{3}}{4}a\,\hat{\mathbf{y}}$ | $(3f)$ | W |
| $\mathbf{B}_3$ | = | $\frac{1}{2}\,\mathbf{a}_1 + \frac{1}{2}\,\mathbf{a}_2$ | = | $\frac{1}{2}a\,\hat{\mathbf{x}}$ | $(3f)$ | W |
| $\mathbf{B}_4$ | = | $\frac{1}{2}\,\mathbf{a}_1 + \frac{1}{2}\,\mathbf{a}_3$ | = | $\frac{1}{4}a\,\hat{\mathbf{x}} - \frac{\sqrt{3}}{4}a\,\hat{\mathbf{y}} + \frac{1}{2}c\,\hat{\mathbf{z}}$ | $(3g)$ | O I |
| $\mathbf{B}_5$ | = | $\frac{1}{2}\,\mathbf{a}_2 + \frac{1}{2}\,\mathbf{a}_3$ | = | $\frac{1}{4}a\,\hat{\mathbf{x}} + \frac{\sqrt{3}}{4}a\,\hat{\mathbf{y}} + \frac{1}{2}c\,\hat{\mathbf{z}}$ | $(3g)$ | O I |
| $\mathbf{B}_6$ | = | $\frac{1}{2}\,\mathbf{a}_1 + \frac{1}{2}\,\mathbf{a}_2 + \frac{1}{2}\,\mathbf{a}_3$ | = | $\frac{1}{2}a\,\hat{\mathbf{x}} + \frac{1}{2}c\,\hat{\mathbf{z}}$ | $(3g)$ | O I |
| $\mathbf{B}_7$ | = | $x_3\,\mathbf{a}_1 + 2x_3\,\mathbf{a}_2$ | = | $\frac{3}{2}x_3a\,\hat{\mathbf{x}} + \frac{\sqrt{3}}{2}x_3a\,\hat{\mathbf{y}}$ | $(6l)$ | O II |
| $\mathbf{B}_8$ | = | $-2x_3\,\mathbf{a}_1 - x_3\,\mathbf{a}_2$ | = | $-\frac{3}{2}x_3a\,\hat{\mathbf{x}} + \frac{\sqrt{3}}{2}x_3a\,\hat{\mathbf{y}}$ | $(6l)$ | O II |
| $\mathbf{B}_9$ | = | $x_3\,\mathbf{a}_1 - x_3\,\mathbf{a}_2$ | = | $-\sqrt{3}x_3a\,\hat{\mathbf{y}}$ | $(6l)$ | O II |
| $\mathbf{B}_{10}$ | = | $-x_3\,\mathbf{a}_1 - 2x_3\,\mathbf{a}_2$ | = | $-\frac{3}{2}x_3a\,\hat{\mathbf{x}} - \frac{\sqrt{3}}{2}x_3a\,\hat{\mathbf{y}}$ | $(6l)$ | O II |
| $\mathbf{B}_{11}$ | = | $2x_3\,\mathbf{a}_1 + x_3\,\mathbf{a}_2$ | = | $\frac{3}{2}x_3a\,\hat{\mathbf{x}} - \frac{\sqrt{3}}{2}x_3a\,\hat{\mathbf{y}}$ | $(6l)$ | O II |
| $\mathbf{B}_{12}$ | = | $-x_3\,\mathbf{a}_1 + x_3\,\mathbf{a}_2$ | = | $\sqrt{3}x_3a\,\hat{\mathbf{y}}$ | $(6l)$ | O II |

---

---

**Geometry files:**

- CIF: pp. 1763

- POSCAR: pp. 1763

# $D0_6$ (Tysonite, $LaF_3$) (*obsolete*) Structure: A3B_hP24_193_ack_g

http://aflow.org/prototype-encyclopedia/A3B_hP24_193_ack_g

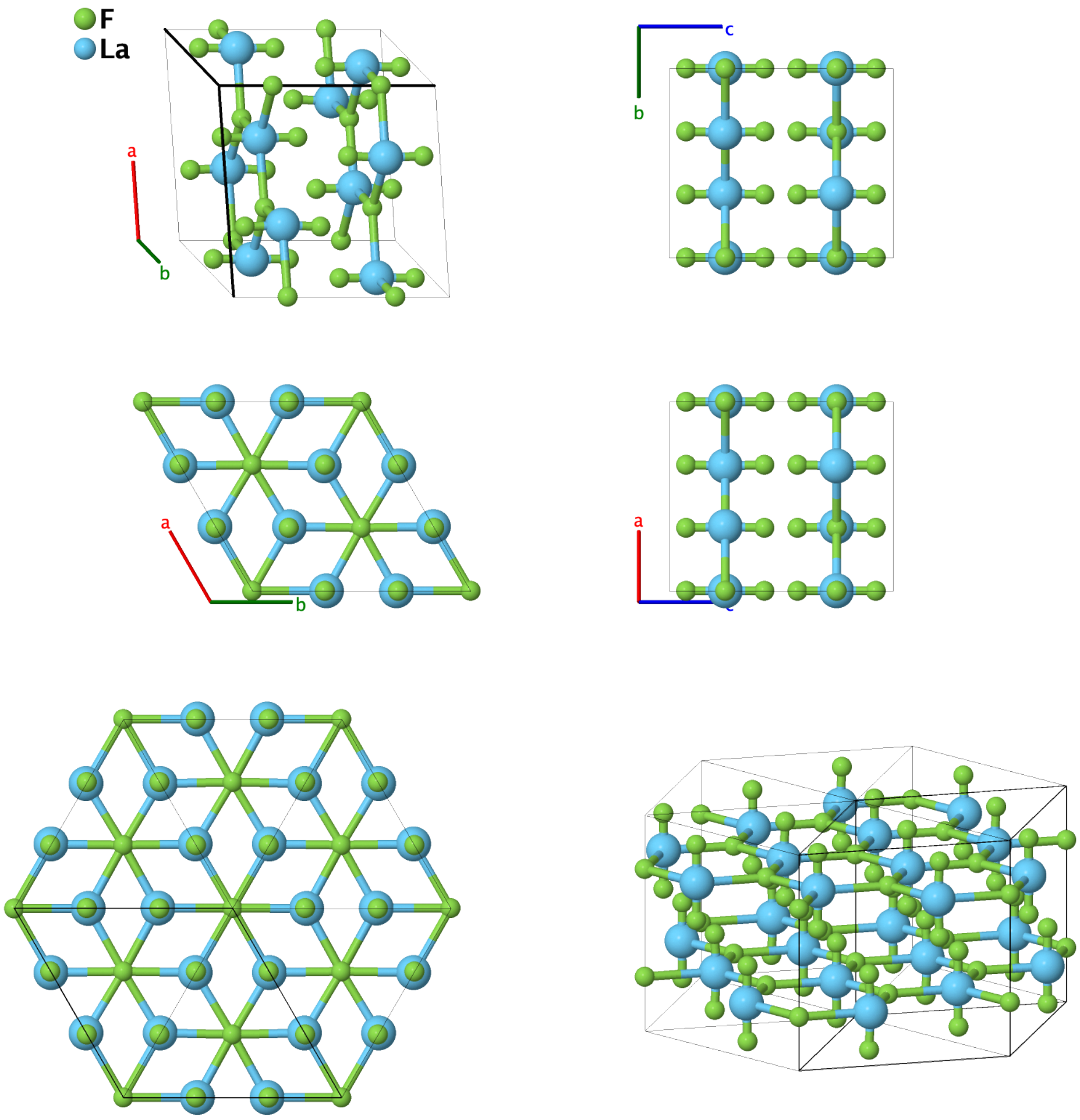

| | | |
|---|---|---|
| **Prototype** | : | $F_3La$ |
| **AFLOW prototype label** | : | A3B_hP24_193_ack_g |
| ***Strukturbericht* designation** | : | $D0_6$ |
| **Pearson symbol** | : | hP24 |
| **Space group number** | : | 193 |
| **Space group symbol** | : | $P6_3/mcm$ |
| **AFLOW prototype command** | : | `aflow --proto=A3B_hP24_193_ack_g` `--params=`$a, c/a, x_3, x_4, z_4$ |

- This stucture was one of several considered by (Hermann, 1937) as a candidate structure for tysonite, and was chosen because it gave the best positioning of the fluorine atoms. Later, (Zalkin, 1985) showed that the structure was trigonal rather than hexagonal, and isostructural with $Cu_3P$ ($D0_{21}$). We keep the original structure as part of the historical record.

**Hexagonal primitive vectors:**

$$\begin{aligned}
\mathbf{a}_1 &= \tfrac{1}{2}\,a\,\hat{\mathbf{x}} - \tfrac{\sqrt{3}}{2}\,a\,\hat{\mathbf{y}} \\
\mathbf{a}_2 &= \tfrac{1}{2}\,a\,\hat{\mathbf{x}} + \tfrac{\sqrt{3}}{2}\,a\,\hat{\mathbf{y}} \\
\mathbf{a}_3 &= c\,\hat{\mathbf{z}}
\end{aligned}$$

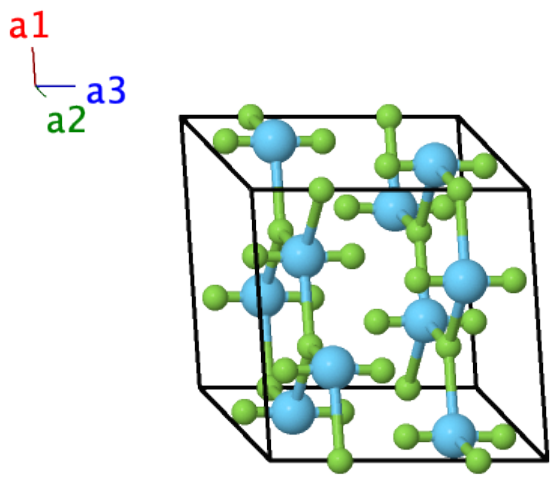

**Basis vectors:**

| | | Lattice Coordinates | | Cartesian Coordinates | Wyckoff Position | Atom Type |
|---|---|---|---|---|---|---|
| $\mathbf{B}_1$ | = | $\frac{1}{4}\,\mathbf{a}_3$ | = | $\frac{1}{4}c\,\hat{\mathbf{z}}$ | (2$a$) | F I |
| $\mathbf{B}_2$ | = | $\frac{3}{4}\,\mathbf{a}_3$ | = | $\frac{3}{4}c\,\hat{\mathbf{z}}$ | (2$a$) | F I |
| $\mathbf{B}_3$ | = | $\frac{1}{3}\,\mathbf{a}_1 + \frac{2}{3}\,\mathbf{a}_2 + \frac{1}{4}\,\mathbf{a}_3$ | = | $\frac{1}{2}a\,\hat{\mathbf{x}} + \frac{1}{2\sqrt{3}}a\,\hat{\mathbf{y}} + \frac{1}{4}c\,\hat{\mathbf{z}}$ | (4$c$) | F II |
| $\mathbf{B}_4$ | = | $\frac{2}{3}\,\mathbf{a}_1 + \frac{1}{3}\,\mathbf{a}_2 + \frac{3}{4}\,\mathbf{a}_3$ | = | $\frac{1}{2}a\,\hat{\mathbf{x}} - \frac{1}{2\sqrt{3}}a\,\hat{\mathbf{y}} + \frac{3}{4}c\,\hat{\mathbf{z}}$ | (4$c$) | F II |
| $\mathbf{B}_5$ | = | $\frac{2}{3}\,\mathbf{a}_1 + \frac{1}{3}\,\mathbf{a}_2 + \frac{1}{4}\,\mathbf{a}_3$ | = | $\frac{1}{2}a\,\hat{\mathbf{x}} - \frac{1}{2\sqrt{3}}a\,\hat{\mathbf{y}} + \frac{1}{4}c\,\hat{\mathbf{z}}$ | (4$c$) | F II |
| $\mathbf{B}_6$ | = | $\frac{1}{3}\,\mathbf{a}_1 + \frac{2}{3}\,\mathbf{a}_2 + \frac{3}{4}\,\mathbf{a}_3$ | = | $\frac{1}{2}a\,\hat{\mathbf{x}} + \frac{1}{2\sqrt{3}}a\,\hat{\mathbf{y}} + \frac{3}{4}c\,\hat{\mathbf{z}}$ | (4$c$) | F II |
| $\mathbf{B}_7$ | = | $x_3\,\mathbf{a}_1 + \frac{1}{4}\,\mathbf{a}_3$ | = | $\frac{1}{2}x_3a\,\hat{\mathbf{x}} - \frac{\sqrt{3}}{2}x_3a\,\hat{\mathbf{y}} + \frac{1}{4}c\,\hat{\mathbf{z}}$ | (6$g$) | La |
| $\mathbf{B}_8$ | = | $x_3\,\mathbf{a}_2 + \frac{1}{4}\,\mathbf{a}_3$ | = | $\frac{1}{2}x_3a\,\hat{\mathbf{x}} + \frac{\sqrt{3}}{2}x_3a\,\hat{\mathbf{y}} + \frac{1}{4}c\,\hat{\mathbf{z}}$ | (6$g$) | La |
| $\mathbf{B}_9$ | = | $-x_3\,\mathbf{a}_1 - x_3\,\mathbf{a}_2 + \frac{1}{4}\,\mathbf{a}_3$ | = | $-x_3a\,\hat{\mathbf{x}} + \frac{1}{4}c\,\hat{\mathbf{z}}$ | (6$g$) | La |
| $\mathbf{B}_{10}$ | = | $-x_3\,\mathbf{a}_1 + \frac{3}{4}\,\mathbf{a}_3$ | = | $-\frac{1}{2}x_3a\,\hat{\mathbf{x}} + \frac{\sqrt{3}}{2}x_3a\,\hat{\mathbf{y}} + \frac{3}{4}c\,\hat{\mathbf{z}}$ | (6$g$) | La |
| $\mathbf{B}_{11}$ | = | $-x_3\,\mathbf{a}_2 + \frac{3}{4}\,\mathbf{a}_3$ | = | $-\frac{1}{2}x_3a\,\hat{\mathbf{x}} - \frac{\sqrt{3}}{2}x_3a\,\hat{\mathbf{y}} + \frac{3}{4}c\,\hat{\mathbf{z}}$ | (6$g$) | La |
| $\mathbf{B}_{12}$ | = | $x_3\,\mathbf{a}_1 + x_3\,\mathbf{a}_2 + \frac{3}{4}\,\mathbf{a}_3$ | = | $x_3a\,\hat{\mathbf{x}} + \frac{3}{4}c\,\hat{\mathbf{z}}$ | (6$g$) | La |
| $\mathbf{B}_{13}$ | = | $x_4\,\mathbf{a}_1 + z_4\,\mathbf{a}_3$ | = | $\frac{1}{2}x_4a\,\hat{\mathbf{x}} - \frac{\sqrt{3}}{2}x_4a\,\hat{\mathbf{y}} + z_4c\,\hat{\mathbf{z}}$ | (12$k$) | F III |
| $\mathbf{B}_{14}$ | = | $x_4\,\mathbf{a}_2 + z_4\,\mathbf{a}_3$ | = | $\frac{1}{2}x_4a\,\hat{\mathbf{x}} + \frac{\sqrt{3}}{2}x_4a\,\hat{\mathbf{y}} + z_4c\,\hat{\mathbf{z}}$ | (12$k$) | F III |
| $\mathbf{B}_{15}$ | = | $-x_4\,\mathbf{a}_1 - x_4\,\mathbf{a}_2 + z_4\,\mathbf{a}_3$ | = | $-x_4a\,\hat{\mathbf{x}} + z_4c\,\hat{\mathbf{z}}$ | (12$k$) | F III |
| $\mathbf{B}_{16}$ | = | $-x_4\,\mathbf{a}_1 + \left(\frac{1}{2} + z_4\right)\,\mathbf{a}_3$ | = | $-\frac{1}{2}x_4a\,\hat{\mathbf{x}} + \frac{\sqrt{3}}{2}x_4a\,\hat{\mathbf{y}} + \left(\frac{1}{2} + z_4\right)c\,\hat{\mathbf{z}}$ | (12$k$) | F III |
| $\mathbf{B}_{17}$ | = | $-x_4\,\mathbf{a}_2 + \left(\frac{1}{2} + z_4\right)\,\mathbf{a}_3$ | = | $-\frac{1}{2}x_4a\,\hat{\mathbf{x}} - \frac{\sqrt{3}}{2}x_4a\,\hat{\mathbf{y}} + \left(\frac{1}{2} + z_4\right)c\,\hat{\mathbf{z}}$ | (12$k$) | F III |
| $\mathbf{B}_{18}$ | = | $x_4\,\mathbf{a}_1 + x_4\,\mathbf{a}_2 + \left(\frac{1}{2} + z_4\right)\,\mathbf{a}_3$ | = | $x_4a\,\hat{\mathbf{x}} + \left(\frac{1}{2} + z_4\right)c\,\hat{\mathbf{z}}$ | (12$k$) | F III |
| $\mathbf{B}_{19}$ | = | $x_4\,\mathbf{a}_2 + \left(\frac{1}{2} - z_4\right)\,\mathbf{a}_3$ | = | $\frac{1}{2}x_4a\,\hat{\mathbf{x}} + \frac{\sqrt{3}}{2}x_4a\,\hat{\mathbf{y}} + \left(\frac{1}{2} - z_4\right)c\,\hat{\mathbf{z}}$ | (12$k$) | F III |
| $\mathbf{B}_{20}$ | = | $x_4\,\mathbf{a}_1 + \left(\frac{1}{2} - z_4\right)\,\mathbf{a}_3$ | = | $\frac{1}{2}x_4a\,\hat{\mathbf{x}} - \frac{\sqrt{3}}{2}x_4a\,\hat{\mathbf{y}} + \left(\frac{1}{2} - z_4\right)c\,\hat{\mathbf{z}}$ | (12$k$) | F III |
| $\mathbf{B}_{21}$ | = | $-x_4\,\mathbf{a}_1 - x_4\,\mathbf{a}_2 + \left(\frac{1}{2} - z_4\right)\,\mathbf{a}_3$ | = | $-x_4a\,\hat{\mathbf{x}} + \left(\frac{1}{2} - z_4\right)c\,\hat{\mathbf{z}}$ | (12$k$) | F III |
| $\mathbf{B}_{22}$ | = | $-x_4\,\mathbf{a}_2 - z_4\,\mathbf{a}_3$ | = | $-\frac{1}{2}x_4a\,\hat{\mathbf{x}} - \frac{\sqrt{3}}{2}x_4a\,\hat{\mathbf{y}} - z_4c\,\hat{\mathbf{z}}$ | (12$k$) | F III |
| $\mathbf{B}_{23}$ | = | $-x_4\,\mathbf{a}_1 - z_4\,\mathbf{a}_3$ | = | $-\frac{1}{2}x_4a\,\hat{\mathbf{x}} + \frac{\sqrt{3}}{2}x_4a\,\hat{\mathbf{y}} - z_4c\,\hat{\mathbf{z}}$ | (12$k$) | F III |
| $\mathbf{B}_{24}$ | = | $x_4\,\mathbf{a}_1 + x_4\,\mathbf{a}_2 - z_4\,\mathbf{a}_3$ | = | $x_4a\,\hat{\mathbf{x}} - z_4c\,\hat{\mathbf{z}}$ | (12$k$) | F III |

**Found in:**

---

**Geometry files:**

- CIF: pp. 1763
- POSCAR: pp. 1764

# $Ti_5Ga_4$ Structure: A4B5_hP18_193_bg_dg

http://aflow.org/prototype-encyclopedia/A4B5_hP18_193_bg_dg

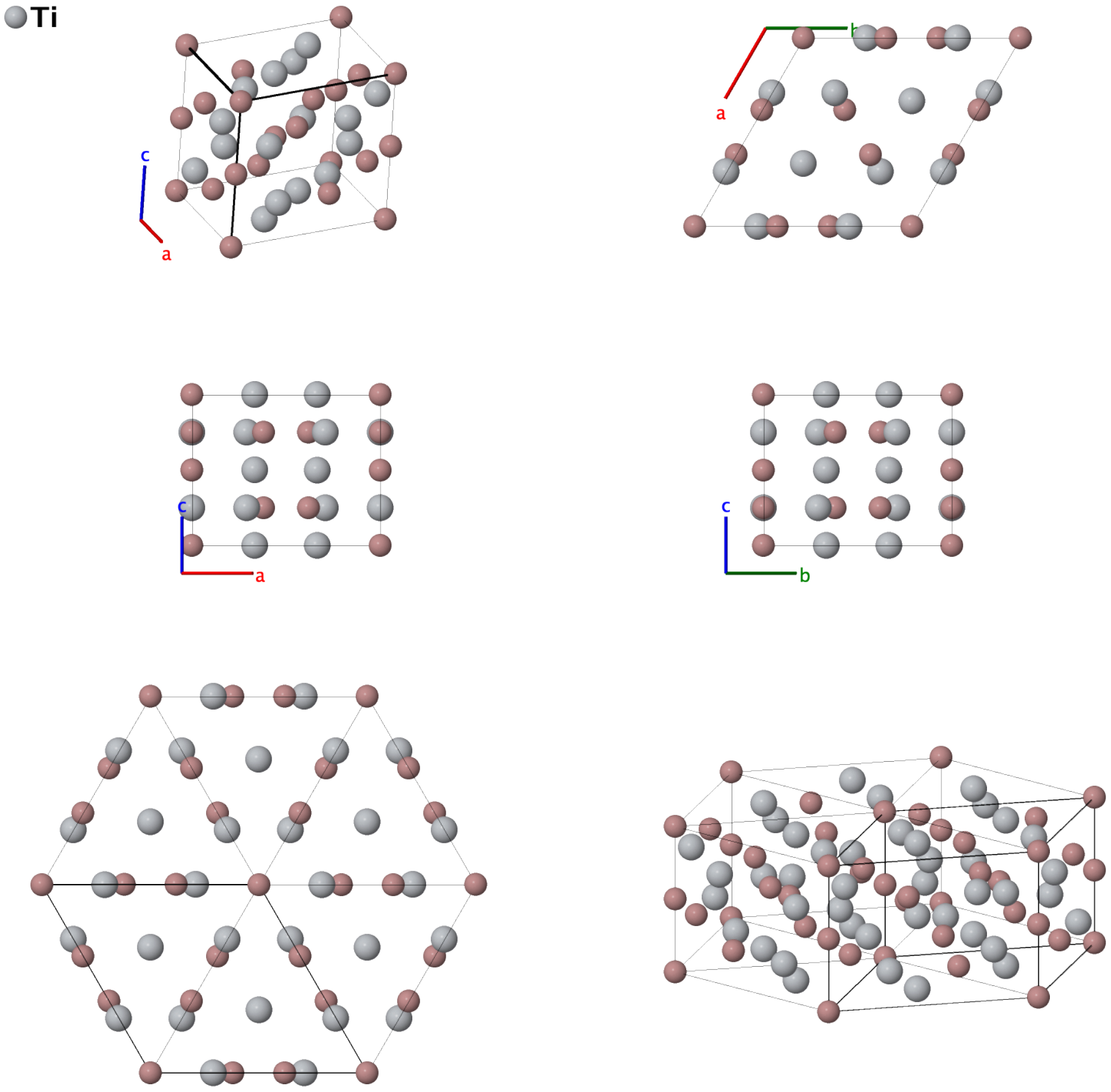

| | | |
|---|---|---|
| **Prototype** | : | $Ga_4Ti_5$ |
| **AFLOW prototype label** | : | A4B5_hP18_193_bg_dg |
| ***Strukturbericht* designation** | : | None |
| **Pearson symbol** | : | hP18 |
| **Space group number** | : | 193 |
| **Space group symbol** | : | $P6_3/mcm$ |
| **AFLOW prototype command** | : | `aflow --proto=A4B5_hP18_193_bg_dg --params=`$a, c/a, x_3, x_4$ |

### Other compounds with this structure

- $Zr_5Ga_4$, $Hf_5Sn_4$, $Ba_3TlTe_5$, and NbIrO

- This is the $D8_8$ structure with the addition of two atoms on the (2$b$) site.

**Hexagonal primitive vectors:**

$$
\begin{aligned}
\mathbf{a}_1 &= \tfrac{1}{2}\,a\,\mathbf{\hat{x}} - \tfrac{\sqrt{3}}{2}\,a\,\mathbf{\hat{y}} \\
\mathbf{a}_2 &= \tfrac{1}{2}\,a\,\mathbf{\hat{x}} + \tfrac{\sqrt{3}}{2}\,a\,\mathbf{\hat{y}} \\
\mathbf{a}_3 &= c\,\mathbf{\hat{z}}
\end{aligned}
$$

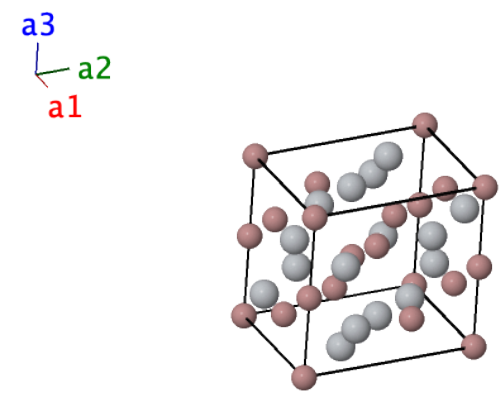

**Basis vectors:**

| | | Lattice Coordinates | | Cartesian Coordinates | Wyckoff Position | Atom Type |
|---|---|---|---|---|---|---|
| $\mathbf{B}_1$ | = | $0\,\mathbf{a}_1 + 0\,\mathbf{a}_2 + 0\,\mathbf{a}_3$ | = | $0\,\mathbf{\hat{x}} + 0\,\mathbf{\hat{y}} + 0\,\mathbf{\hat{z}}$ | $(2b)$ | Ga I |
| $\mathbf{B}_2$ | = | $\frac{1}{2}\,\mathbf{a}_3$ | = | $\frac{1}{2}c\,\mathbf{\hat{z}}$ | $(2b)$ | Ga I |
| $\mathbf{B}_3$ | = | $\frac{1}{3}\,\mathbf{a}_1 + \frac{2}{3}\,\mathbf{a}_2$ | = | $\frac{1}{2}a\,\mathbf{\hat{x}} + \frac{1}{2\sqrt{3}}a\,\mathbf{\hat{y}}$ | $(4d)$ | Ti I |
| $\mathbf{B}_4$ | = | $\frac{2}{3}\,\mathbf{a}_1 + \frac{1}{3}\,\mathbf{a}_2 + \frac{1}{2}\,\mathbf{a}_3$ | = | $\frac{1}{2}a\,\mathbf{\hat{x}} - \frac{1}{2\sqrt{3}}a\,\mathbf{\hat{y}} + \frac{1}{2}c\,\mathbf{\hat{z}}$ | $(4d)$ | Ti I |
| $\mathbf{B}_5$ | = | $\frac{2}{3}\,\mathbf{a}_1 + \frac{1}{3}\,\mathbf{a}_2$ | = | $\frac{1}{2}a\,\mathbf{\hat{x}} - \frac{1}{2\sqrt{3}}a\,\mathbf{\hat{y}}$ | $(4d)$ | Ti I |
| $\mathbf{B}_6$ | = | $\frac{1}{3}\,\mathbf{a}_1 + \frac{2}{3}\,\mathbf{a}_2 + \frac{1}{2}\,\mathbf{a}_3$ | = | $\frac{1}{2}a\,\mathbf{\hat{x}} + \frac{1}{2\sqrt{3}}a\,\mathbf{\hat{y}} + \frac{1}{2}c\,\mathbf{\hat{z}}$ | $(4d)$ | Ti I |
| $\mathbf{B}_7$ | = | $x_3\,\mathbf{a}_1 + \frac{1}{4}\,\mathbf{a}_3$ | = | $\frac{1}{2}x_3a\,\mathbf{\hat{x}} - \frac{\sqrt{3}}{2}x_3a\,\mathbf{\hat{y}} + \frac{1}{4}c\,\mathbf{\hat{z}}$ | $(6g)$ | Ga II |
| $\mathbf{B}_8$ | = | $x_3\,\mathbf{a}_2 + \frac{1}{4}\,\mathbf{a}_3$ | = | $\frac{1}{2}x_3a\,\mathbf{\hat{x}} + \frac{\sqrt{3}}{2}x_3a\,\mathbf{\hat{y}} + \frac{1}{4}c\,\mathbf{\hat{z}}$ | $(6g)$ | Ga II |
| $\mathbf{B}_9$ | = | $-x_3\,\mathbf{a}_1 - x_3\,\mathbf{a}_2 + \frac{1}{4}\,\mathbf{a}_3$ | = | $-x_3a\,\mathbf{\hat{x}} + \frac{1}{4}c\,\mathbf{\hat{z}}$ | $(6g)$ | Ga II |
| $\mathbf{B}_{10}$ | = | $-x_3\,\mathbf{a}_1 + \frac{3}{4}\,\mathbf{a}_3$ | = | $-\frac{1}{2}x_3a\,\mathbf{\hat{x}} + \frac{\sqrt{3}}{2}x_3a\,\mathbf{\hat{y}} + \frac{3}{4}c\,\mathbf{\hat{z}}$ | $(6g)$ | Ga II |
| $\mathbf{B}_{11}$ | = | $-x_3\,\mathbf{a}_2 + \frac{3}{4}\,\mathbf{a}_3$ | = | $-\frac{1}{2}x_3a\,\mathbf{\hat{x}} - \frac{\sqrt{3}}{2}x_3a\,\mathbf{\hat{y}} + \frac{3}{4}c\,\mathbf{\hat{z}}$ | $(6g)$ | Ga II |
| $\mathbf{B}_{12}$ | = | $x_3\,\mathbf{a}_1 + x_3\,\mathbf{a}_2 + \frac{3}{4}\,\mathbf{a}_3$ | = | $x_3a\,\mathbf{\hat{x}} + \frac{3}{4}c\,\mathbf{\hat{z}}$ | $(6g)$ | Ga II |
| $\mathbf{B}_{13}$ | = | $x_4\,\mathbf{a}_1 + \frac{1}{4}\,\mathbf{a}_3$ | = | $\frac{1}{2}x_4a\,\mathbf{\hat{x}} - \frac{\sqrt{3}}{2}x_4a\,\mathbf{\hat{y}} + \frac{1}{4}c\,\mathbf{\hat{z}}$ | $(6g)$ | Ti II |
| $\mathbf{B}_{14}$ | = | $x_4\,\mathbf{a}_2 + \frac{1}{4}\,\mathbf{a}_3$ | = | $\frac{1}{2}x_4a\,\mathbf{\hat{x}} + \frac{\sqrt{3}}{2}x_4a\,\mathbf{\hat{y}} + \frac{1}{4}c\,\mathbf{\hat{z}}$ | $(6g)$ | Ti II |
| $\mathbf{B}_{15}$ | = | $-x_4\,\mathbf{a}_1 - x_4\,\mathbf{a}_2 + \frac{1}{4}\,\mathbf{a}_3$ | = | $-x_4a\,\mathbf{\hat{x}} + \frac{1}{4}c\,\mathbf{\hat{z}}$ | $(6g)$ | Ti II |
| $\mathbf{B}_{16}$ | = | $-x_4\,\mathbf{a}_1 + \frac{3}{4}\,\mathbf{a}_3$ | = | $-\frac{1}{2}x_4a\,\mathbf{\hat{x}} + \frac{\sqrt{3}}{2}x_4a\,\mathbf{\hat{y}} + \frac{3}{4}c\,\mathbf{\hat{z}}$ | $(6g)$ | Ti II |
| $\mathbf{B}_{17}$ | = | $-x_4\,\mathbf{a}_2 + \frac{3}{4}\,\mathbf{a}_3$ | = | $-\frac{1}{2}x_4a\,\mathbf{\hat{x}} - \frac{\sqrt{3}}{2}x_4a\,\mathbf{\hat{y}} + \frac{3}{4}c\,\mathbf{\hat{z}}$ | $(6g)$ | Ti II |
| $\mathbf{B}_{18}$ | = | $x_4\,\mathbf{a}_1 + x_4\,\mathbf{a}_2 + \frac{3}{4}\,\mathbf{a}_3$ | = | $x_4a\,\mathbf{\hat{x}} + \frac{3}{4}c\,\mathbf{\hat{z}}$ | $(6g)$ | Ti II |

**Geometry files:**
- CIF: pp. 1764
- POSCAR: pp. 1764

# Proposed 300 GPa $HfH_{10}$ Structure: A10B_hP22_194_bhj_c

http://aflow.org/prototype-encyclopedia/A10B_hP22_194_bhj_c

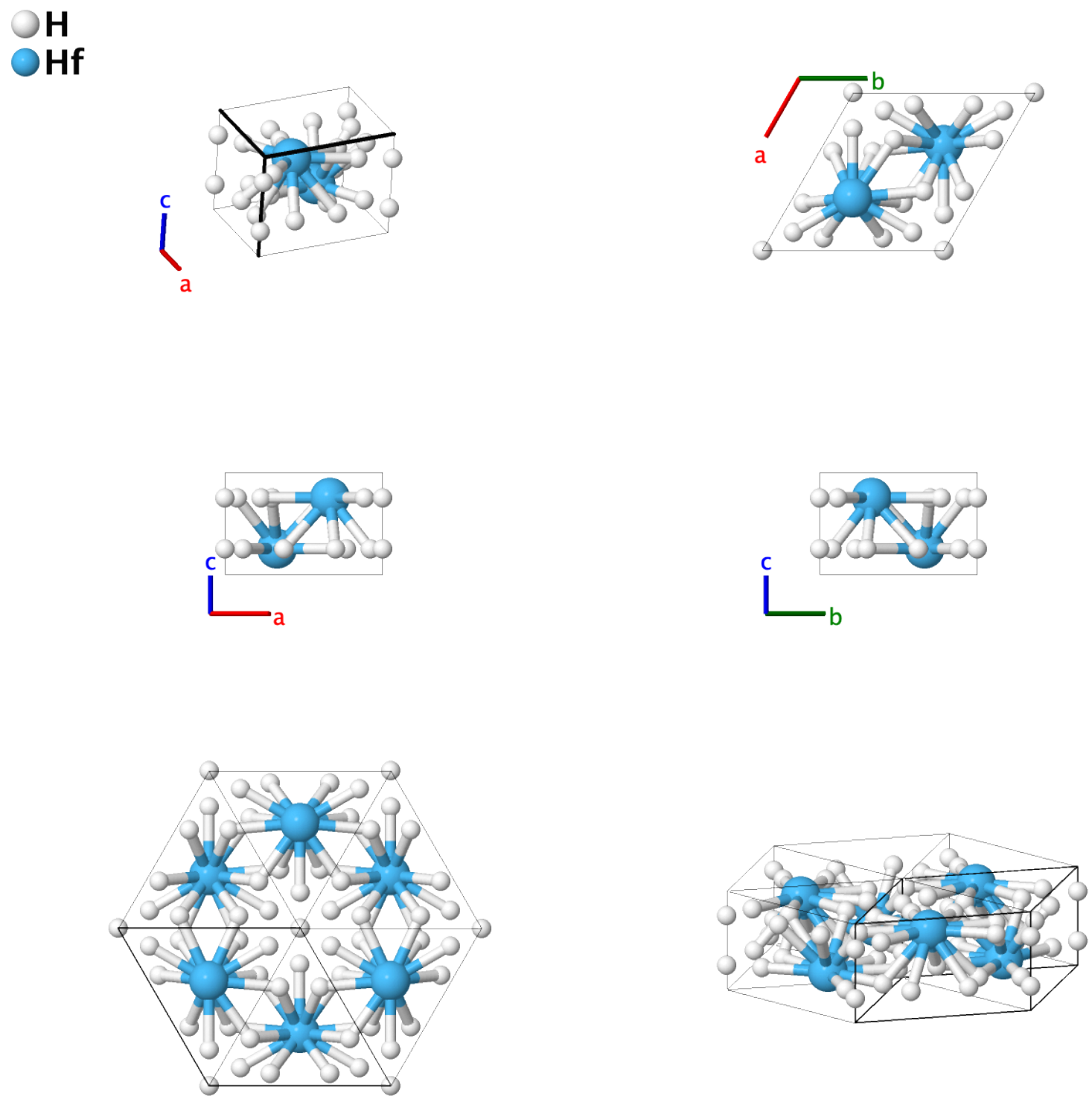

| | | |
|---|---|---|
| **Prototype** | : | $H_{10}Hf$ |
| **AFLOW prototype label** | : | A10B_hP22_194_bhj_c |
| ***Strukturbericht* designation** | : | None |
| **Pearson symbol** | : | hP22 |
| **Space group number** | : | 194 |
| **Space group symbol** | : | $P6_3/mmc$ |
| **AFLOW prototype command** | : | `aflow --proto=A10B_hP22_194_bhj_c --params=`$a, c/a, x_3, x_4, y_4$ |

### Other compounds with this structure

- $LuH_{10}$, $ScH_{10}$, and $ZrH_{10}$

---

- This structure has only been determined computationally. At 300 GPa it is predicted to have a superconducting transitions temperature $T_c$ above 200 K.

**Hexagonal primitive vectors:**

$$
\begin{aligned}
\mathbf{a}_1 &= \tfrac{1}{2}\,a\,\hat{\mathbf{x}} - \tfrac{\sqrt{3}}{2}\,a\,\hat{\mathbf{y}} \\
\mathbf{a}_2 &= \tfrac{1}{2}\,a\,\hat{\mathbf{x}} + \tfrac{\sqrt{3}}{2}\,a\,\hat{\mathbf{y}} \\
\mathbf{a}_3 &= c\,\hat{\mathbf{z}}
\end{aligned}
$$

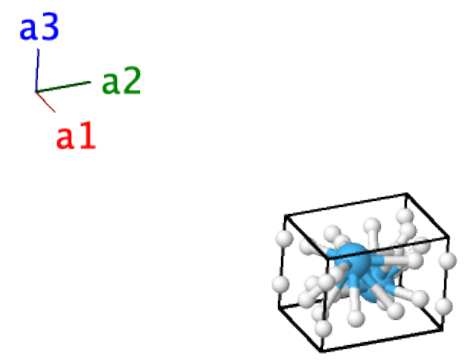

**Basis vectors:**

| | | Lattice Coordinates | | Cartesian Coordinates | Wyckoff Position | Atom Type |
|---|---|---|---|---|---|---|
| $\mathbf{B}_1$ | = | $\frac{1}{4}\,\mathbf{a}_3$ | = | $\frac{1}{4}c\,\hat{\mathbf{z}}$ | (2$b$) | H I |
| $\mathbf{B}_2$ | = | $\frac{3}{4}\,\mathbf{a}_3$ | = | $\frac{3}{4}c\,\hat{\mathbf{z}}$ | (2$b$) | H I |
| $\mathbf{B}_3$ | = | $\frac{1}{3}\,\mathbf{a}_1 + \frac{2}{3}\,\mathbf{a}_2 + \frac{1}{4}\,\mathbf{a}_3$ | = | $\frac{1}{2}a\,\hat{\mathbf{x}} + \frac{1}{2\sqrt{3}}a\,\hat{\mathbf{y}} + \frac{1}{4}c\,\hat{\mathbf{z}}$ | (2$c$) | Hf |
| $\mathbf{B}_4$ | = | $\frac{2}{3}\,\mathbf{a}_1 + \frac{1}{3}\,\mathbf{a}_2 + \frac{3}{4}\,\mathbf{a}_3$ | = | $\frac{1}{2}a\,\hat{\mathbf{x}} - \frac{1}{2\sqrt{3}}a\,\hat{\mathbf{y}} + \frac{3}{4}c\,\hat{\mathbf{z}}$ | (2$c$) | Hf |
| $\mathbf{B}_5$ | = | $x_3\,\mathbf{a}_1 + 2x_3\,\mathbf{a}_2 + \frac{1}{4}\,\mathbf{a}_3$ | = | $\frac{3}{2}x_3a\,\hat{\mathbf{x}} + \frac{\sqrt{3}}{2}x_3a\,\hat{\mathbf{y}} + \frac{1}{4}c\,\hat{\mathbf{z}}$ | (6$h$) | H II |
| $\mathbf{B}_6$ | = | $-2x_3\,\mathbf{a}_1 - x_3\,\mathbf{a}_2 + \frac{1}{4}\,\mathbf{a}_3$ | = | $-\frac{3}{2}x_3a\,\hat{\mathbf{x}} + \frac{\sqrt{3}}{2}x_3a\,\hat{\mathbf{y}} + \frac{1}{4}c\,\hat{\mathbf{z}}$ | (6$h$) | H II |
| $\mathbf{B}_7$ | = | $x_3\,\mathbf{a}_1 - x_3\,\mathbf{a}_2 + \frac{1}{4}\,\mathbf{a}_3$ | = | $-\sqrt{3}x_3a\,\hat{\mathbf{y}} + \frac{1}{4}c\,\hat{\mathbf{z}}$ | (6$h$) | H II |
| $\mathbf{B}_8$ | = | $-x_3\,\mathbf{a}_1 - 2x_3\,\mathbf{a}_2 + \frac{3}{4}\,\mathbf{a}_3$ | = | $-\frac{3}{2}x_3a\,\hat{\mathbf{x}} - \frac{\sqrt{3}}{2}x_3a\,\hat{\mathbf{y}} + \frac{3}{4}c\,\hat{\mathbf{z}}$ | (6$h$) | H II |
| $\mathbf{B}_9$ | = | $2x_3\,\mathbf{a}_1 + x_3\,\mathbf{a}_2 + \frac{3}{4}\,\mathbf{a}_3$ | = | $\frac{3}{2}x_3a\,\hat{\mathbf{x}} - \frac{\sqrt{3}}{2}x_3a\,\hat{\mathbf{y}} + \frac{3}{4}c\,\hat{\mathbf{z}}$ | (6$h$) | H II |
| $\mathbf{B}_{10}$ | = | $-x_3\,\mathbf{a}_1 + x_3\,\mathbf{a}_2 + \frac{3}{4}\,\mathbf{a}_3$ | = | $\sqrt{3}x_3a\,\hat{\mathbf{y}} + \frac{3}{4}c\,\hat{\mathbf{z}}$ | (6$h$) | H II |
| $\mathbf{B}_{11}$ | = | $x_4\,\mathbf{a}_1 + y_4\,\mathbf{a}_2 + \frac{1}{4}\,\mathbf{a}_3$ | = | $\frac{1}{2}(x_4 + y_4)\,a\,\hat{\mathbf{x}} + \frac{\sqrt{3}}{2}(-x_4 + y_4)\,a\,\hat{\mathbf{y}} + \frac{1}{4}c\,\hat{\mathbf{z}}$ | (12$j$) | H III |
| $\mathbf{B}_{12}$ | = | $-y_4\,\mathbf{a}_1 + (x_4 - y_4)\,\mathbf{a}_2 + \frac{1}{4}\,\mathbf{a}_3$ | = | $\left(\frac{1}{2}x_4 - y_4\right)a\,\hat{\mathbf{x}} + \frac{\sqrt{3}}{2}x_4a\,\hat{\mathbf{y}} + \frac{1}{4}c\,\hat{\mathbf{z}}$ | (12$j$) | H III |
| $\mathbf{B}_{13}$ | = | $(-x_4 + y_4)\,\mathbf{a}_1 - x_4\,\mathbf{a}_2 + \frac{1}{4}\,\mathbf{a}_3$ | = | $\left(-x_4 + \frac{1}{2}y_4\right)a\,\hat{\mathbf{x}} - \frac{\sqrt{3}}{2}y_4a\,\hat{\mathbf{y}} + \frac{1}{4}c\,\hat{\mathbf{z}}$ | (12$j$) | H III |
| $\mathbf{B}_{14}$ | = | $-x_4\,\mathbf{a}_1 - y_4\,\mathbf{a}_2 + \frac{3}{4}\,\mathbf{a}_3$ | = | $-\frac{1}{2}(x_4 + y_4)\,a\,\hat{\mathbf{x}} + \frac{\sqrt{3}}{2}(x_4 - y_4)\,a\,\hat{\mathbf{y}} + \frac{3}{4}c\,\hat{\mathbf{z}}$ | (12$j$) | H III |
| $\mathbf{B}_{15}$ | = | $y_4\,\mathbf{a}_1 + (-x_4 + y_4)\,\mathbf{a}_2 + \frac{3}{4}\,\mathbf{a}_3$ | = | $\left(-\frac{1}{2}x_4 + y_4\right)a\,\hat{\mathbf{x}} - \frac{\sqrt{3}}{2}x_4a\,\hat{\mathbf{y}} + \frac{3}{4}c\,\hat{\mathbf{z}}$ | (12$j$) | H III |
| $\mathbf{B}_{16}$ | = | $(x_4 - y_4)\,\mathbf{a}_1 + x_4\,\mathbf{a}_2 + \frac{3}{4}\,\mathbf{a}_3$ | = | $\left(x_4 - \frac{1}{2}y_4\right)a\,\hat{\mathbf{x}} + \frac{\sqrt{3}}{2}y_4a\,\hat{\mathbf{y}} + \frac{3}{4}c\,\hat{\mathbf{z}}$ | (12$j$) | H III |
| $\mathbf{B}_{17}$ | = | $y_4\,\mathbf{a}_1 + x_4\,\mathbf{a}_2 + \frac{3}{4}\,\mathbf{a}_3$ | = | $\frac{1}{2}(x_4 + y_4)\,a\,\hat{\mathbf{x}} + \frac{\sqrt{3}}{2}(x_4 - y_4)\,a\,\hat{\mathbf{y}} + \frac{3}{4}c\,\hat{\mathbf{z}}$ | (12$j$) | H III |
| $\mathbf{B}_{18}$ | = | $(x_4 - y_4)\,\mathbf{a}_1 - y_4\,\mathbf{a}_2 + \frac{3}{4}\,\mathbf{a}_3$ | = | $\left(\frac{1}{2}x_4 - y_4\right)a\,\hat{\mathbf{x}} - \frac{\sqrt{3}}{2}x_4a\,\hat{\mathbf{y}} + \frac{3}{4}c\,\hat{\mathbf{z}}$ | (12$j$) | H III |
| $\mathbf{B}_{19}$ | = | $-x_4\,\mathbf{a}_1 + (-x_4 + y_4)\,\mathbf{a}_2 + \frac{3}{4}\,\mathbf{a}_3$ | = | $\left(-x_4 + \frac{1}{2}y_4\right)a\,\hat{\mathbf{x}} + \frac{\sqrt{3}}{2}y_4a\,\hat{\mathbf{y}} + \frac{3}{4}c\,\hat{\mathbf{z}}$ | (12$j$) | H III |
| $\mathbf{B}_{20}$ | = | $-y_4\,\mathbf{a}_1 - x_4\,\mathbf{a}_2 + \frac{1}{4}\,\mathbf{a}_3$ | = | $-\frac{1}{2}(x_4 + y_4)\,a\,\hat{\mathbf{x}} + \frac{\sqrt{3}}{2}(-x_4 + y_4)\,a\,\hat{\mathbf{y}} + \frac{1}{4}c\,\hat{\mathbf{z}}$ | (12$j$) | H III |
| $\mathbf{B}_{21}$ | = | $(-x_4 + y_4)\,\mathbf{a}_1 + y_4\,\mathbf{a}_2 + \frac{1}{4}\,\mathbf{a}_3$ | = | $\left(-\frac{1}{2}x_4 + y_4\right)a\,\hat{\mathbf{x}} + \frac{\sqrt{3}}{2}x_4a\,\hat{\mathbf{y}} + \frac{1}{4}c\,\hat{\mathbf{z}}$ | (12$j$) | H III |
| $\mathbf{B}_{22}$ | = | $x_4\,\mathbf{a}_1 + (x_4 - y_4)\,\mathbf{a}_2 + \frac{1}{4}\,\mathbf{a}_3$ | = | $\left(x_4 - \frac{1}{2}y_4\right)a\,\hat{\mathbf{x}} - \frac{\sqrt{3}}{2}y_4a\,\hat{\mathbf{y}} + \frac{1}{4}c\,\hat{\mathbf{z}}$ | (12$j$) | H III |

**Geometry files:**

- CIF: pp. 1764
- POSCAR: pp. 1765

# Magnetoplumbite ($PbFe_{12}O_{19}$) Structure: A12B19C_hP64_194_ab2fk_efh2k_d

http://aflow.org/prototype-encyclopedia/A12B19C_hP64_194_ab2fk_efh2k_d

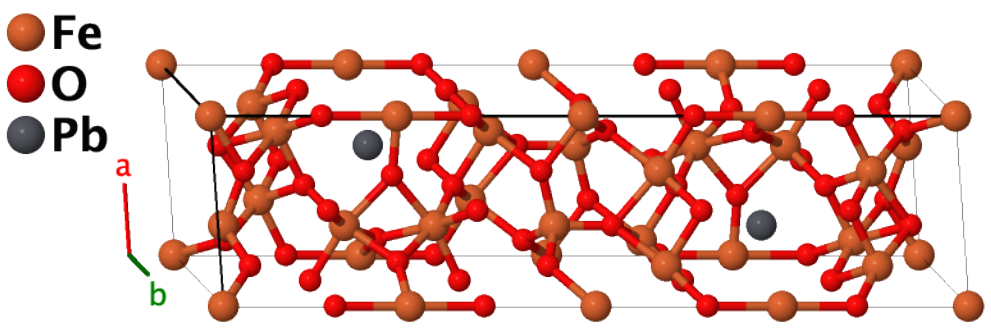

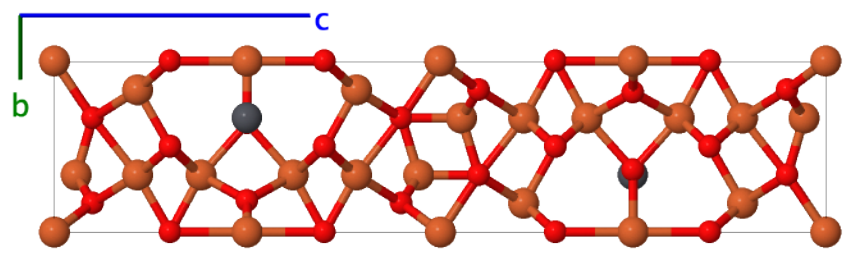

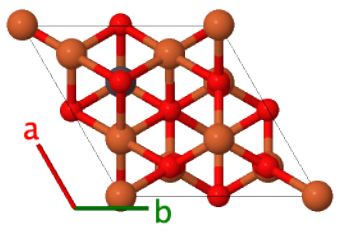

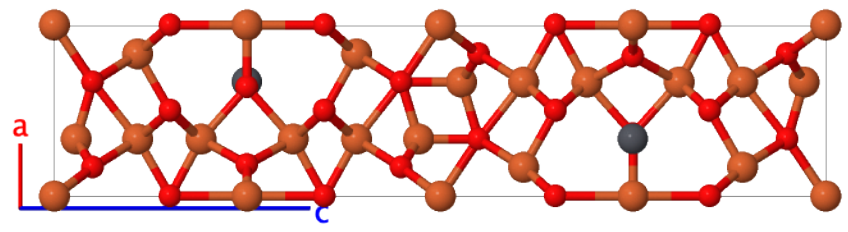

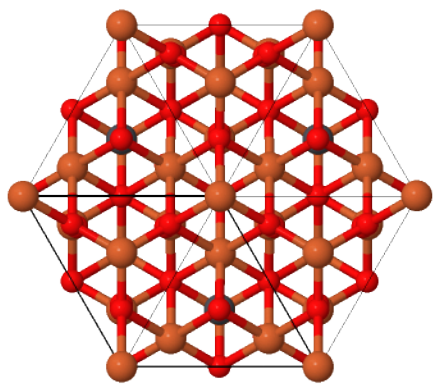

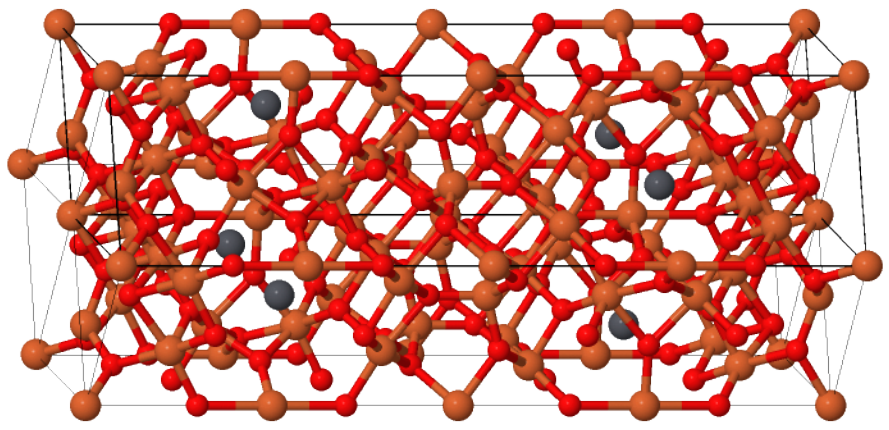

| | | |
|---|---|---|
| **Prototype** | : | $Fe_{12}O_{19}Pb$ |
| **AFLOW prototype label** | : | A12B19C_hP64_194_ab2fk_efh2k_d |
| ***Strukturbericht* designation** | : | None |
| **Pearson symbol** | : | hP64 |
| **Space group number** | : | 194 |
| **Space group symbol** | : | $P6_3/mmc$ |
| **AFLOW prototype command** | : | `aflow --proto=A12B19C_hP64_194_ab2fk_efh2k_d` `--params=`$a, c/a, z_4, z_5, z_6, z_7, x_8, x_9, z_9, x_{10}, z_{10}, x_{11}, z_{11}$ |

### Other compounds with this structure

- $PbAl_{12}O_{19}$, $PbGa_{12}O_{19}$, $PbMn_{12}O_{19}$, $Pb(Co,Ti)_{12}O_{19}$, $BaAl_{12}O_{19}$, $BaGa_{12}O_{19}$, $BaMn_{12}O_{19}$, $Ba(Co,Ti)_{12}O_{19}$, $CaAl_{12}O_{19}$, $CaGa_{12}O_{19}$, $CaMn_{12}O_{19}$, $Ca(Co,Ti)_{12}O_{19}$, $SrAl_{12}O_{19}$, $SrGa_{12}O_{19}$, $SrMn_{12}O_{19}$, and $Sr(Co,Ti)_{12}O_{19}$

---

- In addition to the listed compounds, the lead and iron sites may be alloyed with a wide variety of metals and semi-metals resulting in high-entropy phases (Vinnik, 2019).

### Hexagonal primitive vectors:

$$\mathbf{a}_1 = \frac{1}{2}\,a\,\hat{\mathbf{x}} - \frac{\sqrt{3}}{2}\,a\,\hat{\mathbf{y}}$$
$$\mathbf{a}_2 = \frac{1}{2}\,a\,\hat{\mathbf{x}} + \frac{\sqrt{3}}{2}\,a\,\hat{\mathbf{y}}$$
$$\mathbf{a}_3 = c\,\hat{\mathbf{z}}$$

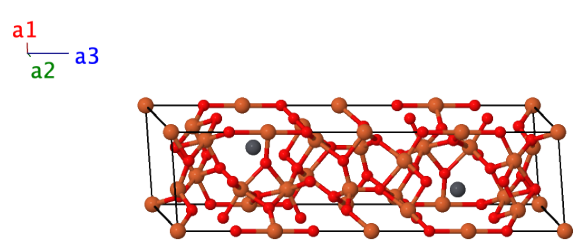

### Basis vectors:

| | | Lattice Coordinates | | Cartesian Coordinates | Wyckoff Position | Atom Type |
|---|---|---|---|---|---|---|
| $\mathbf{B}_1$ | = | $0\,\mathbf{a}_1 + 0\,\mathbf{a}_2 + 0\,\mathbf{a}_3$ | = | $0\,\hat{\mathbf{x}} + 0\,\hat{\mathbf{y}} + 0\,\hat{\mathbf{z}}$ | $(2a)$ | Fe I |
| $\mathbf{B}_2$ | = | $\frac{1}{2}\,\mathbf{a}_3$ | = | $\frac{1}{2}c\,\hat{\mathbf{z}}$ | $(2a)$ | Fe I |
| $\mathbf{B}_3$ | = | $\frac{1}{4}\,\mathbf{a}_3$ | = | $\frac{1}{4}c\,\hat{\mathbf{z}}$ | $(2b)$ | Fe II |
| $\mathbf{B}_4$ | = | $\frac{3}{4}\,\mathbf{a}_3$ | = | $\frac{3}{4}c\,\hat{\mathbf{z}}$ | $(2b)$ | Fe II |
| $\mathbf{B}_5$ | = | $\frac{1}{3}\,\mathbf{a}_1 + \frac{2}{3}\,\mathbf{a}_2 + \frac{3}{4}\,\mathbf{a}_3$ | = | $\frac{1}{2}a\,\hat{\mathbf{x}} + \frac{1}{2\sqrt{3}}a\,\hat{\mathbf{y}} + \frac{3}{4}c\,\hat{\mathbf{z}}$ | $(2d)$ | Pb |
| $\mathbf{B}_6$ | = | $\frac{2}{3}\,\mathbf{a}_1 + \frac{1}{3}\,\mathbf{a}_2 + \frac{1}{4}\,\mathbf{a}_3$ | = | $\frac{1}{2}a\,\hat{\mathbf{x}} - \frac{1}{2\sqrt{3}}a\,\hat{\mathbf{y}} + \frac{1}{4}c\,\hat{\mathbf{z}}$ | $(2d)$ | Pb |
| $\mathbf{B}_7$ | = | $z_4\,\mathbf{a}_3$ | = | $z_4c\,\hat{\mathbf{z}}$ | $(4e)$ | O I |
| $\mathbf{B}_8$ | = | $\left(\frac{1}{2} + z_4\right)\,\mathbf{a}_3$ | = | $\left(\frac{1}{2} + z_4\right)c\,\hat{\mathbf{z}}$ | $(4e)$ | O I |
| $\mathbf{B}_9$ | = | $-z_4\,\mathbf{a}_3$ | = | $-z_4c\,\hat{\mathbf{z}}$ | $(4e)$ | O I |
| $\mathbf{B}_{10}$ | = | $\left(\frac{1}{2} - z_4\right)\,\mathbf{a}_3$ | = | $\left(\frac{1}{2} - z_4\right)c\,\hat{\mathbf{z}}$ | $(4e)$ | O I |
| $\mathbf{B}_{11}$ | = | $\frac{1}{3}\,\mathbf{a}_1 + \frac{2}{3}\,\mathbf{a}_2 + z_5\,\mathbf{a}_3$ | = | $\frac{1}{2}a\,\hat{\mathbf{x}} + \frac{1}{2\sqrt{3}}a\,\hat{\mathbf{y}} + z_5c\,\hat{\mathbf{z}}$ | $(4f)$ | Fe III |
| $\mathbf{B}_{12}$ | = | $\frac{2}{3}\,\mathbf{a}_1 + \frac{1}{3}\,\mathbf{a}_2 + \left(\frac{1}{2} + z_5\right)\,\mathbf{a}_3$ | = | $\frac{1}{2}a\,\hat{\mathbf{x}} - \frac{1}{2\sqrt{3}}a\,\hat{\mathbf{y}} + \left(\frac{1}{2} + z_5\right)c\,\hat{\mathbf{z}}$ | $(4f)$ | Fe III |
| $\mathbf{B}_{13}$ | = | $\frac{2}{3}\,\mathbf{a}_1 + \frac{1}{3}\,\mathbf{a}_2 - z_5\,\mathbf{a}_3$ | = | $\frac{1}{2}a\,\hat{\mathbf{x}} - \frac{1}{2\sqrt{3}}a\,\hat{\mathbf{y}} - z_5c\,\hat{\mathbf{z}}$ | $(4f)$ | Fe III |
| $\mathbf{B}_{14}$ | = | $\frac{1}{3}\,\mathbf{a}_1 + \frac{2}{3}\,\mathbf{a}_2 + \left(\frac{1}{2} - z_5\right)\,\mathbf{a}_3$ | = | $\frac{1}{2}a\,\hat{\mathbf{x}} + \frac{1}{2\sqrt{3}}a\,\hat{\mathbf{y}} + \left(\frac{1}{2} - z_5\right)c\,\hat{\mathbf{z}}$ | $(4f)$ | Fe III |
| $\mathbf{B}_{15}$ | = | $\frac{1}{3}\,\mathbf{a}_1 + \frac{2}{3}\,\mathbf{a}_2 + z_6\,\mathbf{a}_3$ | = | $\frac{1}{2}a\,\hat{\mathbf{x}} + \frac{1}{2\sqrt{3}}a\,\hat{\mathbf{y}} + z_6c\,\hat{\mathbf{z}}$ | $(4f)$ | Fe IV |
| $\mathbf{B}_{16}$ | = | $\frac{2}{3}\,\mathbf{a}_1 + \frac{1}{3}\,\mathbf{a}_2 + \left(\frac{1}{2} + z_6\right)\,\mathbf{a}_3$ | = | $\frac{1}{2}a\,\hat{\mathbf{x}} - \frac{1}{2\sqrt{3}}a\,\hat{\mathbf{y}} + \left(\frac{1}{2} + z_6\right)c\,\hat{\mathbf{z}}$ | $(4f)$ | Fe IV |
| $\mathbf{B}_{17}$ | = | $\frac{2}{3}\,\mathbf{a}_1 + \frac{1}{3}\,\mathbf{a}_2 - z_6\,\mathbf{a}_3$ | = | $\frac{1}{2}a\,\hat{\mathbf{x}} - \frac{1}{2\sqrt{3}}a\,\hat{\mathbf{y}} - z_6c\,\hat{\mathbf{z}}$ | $(4f)$ | Fe IV |
| $\mathbf{B}_{18}$ | = | $\frac{1}{3}\,\mathbf{a}_1 + \frac{2}{3}\,\mathbf{a}_2 + \left(\frac{1}{2} - z_6\right)\,\mathbf{a}_3$ | = | $\frac{1}{2}a\,\hat{\mathbf{x}} + \frac{1}{2\sqrt{3}}a\,\hat{\mathbf{y}} + \left(\frac{1}{2} - z_6\right)c\,\hat{\mathbf{z}}$ | $(4f)$ | Fe IV |
| $\mathbf{B}_{19}$ | = | $\frac{1}{3}\,\mathbf{a}_1 + \frac{2}{3}\,\mathbf{a}_2 + z_7\,\mathbf{a}_3$ | = | $\frac{1}{2}a\,\hat{\mathbf{x}} + \frac{1}{2\sqrt{3}}a\,\hat{\mathbf{y}} + z_7c\,\hat{\mathbf{z}}$ | $(4f)$ | O II |
| $\mathbf{B}_{20}$ | = | $\frac{2}{3}\,\mathbf{a}_1 + \frac{1}{3}\,\mathbf{a}_2 + \left(\frac{1}{2} + z_7\right)\,\mathbf{a}_3$ | = | $\frac{1}{2}a\,\hat{\mathbf{x}} - \frac{1}{2\sqrt{3}}a\,\hat{\mathbf{y}} + \left(\frac{1}{2} + z_7\right)c\,\hat{\mathbf{z}}$ | $(4f)$ | O II |
| $\mathbf{B}_{21}$ | = | $\frac{2}{3}\,\mathbf{a}_1 + \frac{1}{3}\,\mathbf{a}_2 - z_7\,\mathbf{a}_3$ | = | $\frac{1}{2}a\,\hat{\mathbf{x}} - \frac{1}{2\sqrt{3}}a\,\hat{\mathbf{y}} - z_7c\,\hat{\mathbf{z}}$ | $(4f)$ | O II |
| $\mathbf{B}_{22}$ | = | $\frac{1}{3}\,\mathbf{a}_1 + \frac{2}{3}\,\mathbf{a}_2 + \left(\frac{1}{2} - z_7\right)\,\mathbf{a}_3$ | = | $\frac{1}{2}a\,\hat{\mathbf{x}} + \frac{1}{2\sqrt{3}}a\,\hat{\mathbf{y}} + \left(\frac{1}{2} - z_7\right)c\,\hat{\mathbf{z}}$ | $(4f)$ | O II |
| $\mathbf{B}_{23}$ | = | $x_8\,\mathbf{a}_1 + 2x_8\,\mathbf{a}_2 + \frac{1}{4}\,\mathbf{a}_3$ | = | $\frac{3}{2}x_8a\,\hat{\mathbf{x}} + \frac{\sqrt{3}}{2}x_8a\,\hat{\mathbf{y}} + \frac{1}{4}c\,\hat{\mathbf{z}}$ | $(6h)$ | O III |
| $\mathbf{B}_{24}$ | = | $-2x_8\,\mathbf{a}_1 - x_8\,\mathbf{a}_2 + \frac{1}{4}\,\mathbf{a}_3$ | = | $-\frac{3}{2}x_8a\,\hat{\mathbf{x}} + \frac{\sqrt{3}}{2}x_8a\,\hat{\mathbf{y}} + \frac{1}{4}c\,\hat{\mathbf{z}}$ | $(6h)$ | O III |
| $\mathbf{B}_{25}$ | = | $x_8\,\mathbf{a}_1 - x_8\,\mathbf{a}_2 + \frac{1}{4}\,\mathbf{a}_3$ | = | $-\sqrt{3}x_8a\,\hat{\mathbf{y}} + \frac{1}{4}c\,\hat{\mathbf{z}}$ | $(6h)$ | O III |
| $\mathbf{B}_{26}$ | = | $-x_8\,\mathbf{a}_1 - 2x_8\,\mathbf{a}_2 + \frac{3}{4}\,\mathbf{a}_3$ | = | $-\frac{3}{2}x_8a\,\hat{\mathbf{x}} - \frac{\sqrt{3}}{2}x_8a\,\hat{\mathbf{y}} + \frac{3}{4}c\,\hat{\mathbf{z}}$ | $(6h)$ | O III |
| $\mathbf{B}_{27}$ | = | $2x_8\,\mathbf{a}_1 + x_8\,\mathbf{a}_2 + \frac{3}{4}\,\mathbf{a}_3$ | = | $\frac{3}{2}x_8a\,\hat{\mathbf{x}} - \frac{\sqrt{3}}{2}x_8a\,\hat{\mathbf{y}} + \frac{3}{4}c\,\hat{\mathbf{z}}$ | $(6h)$ | O III |
| $\mathbf{B}_{28}$ | = | $-x_8\,\mathbf{a}_1 + x_8\,\mathbf{a}_2 + \frac{3}{4}\,\mathbf{a}_3$ | = | $\sqrt{3}x_8a\,\hat{\mathbf{y}} + \frac{3}{4}c\,\hat{\mathbf{z}}$ | $(6h)$ | O III |
| $\mathbf{B}_{29}$ | = | $x_9\,\mathbf{a}_1 + 2x_9\,\mathbf{a}_2 + z_9\,\mathbf{a}_3$ | = | $\frac{3}{2}x_9a\,\hat{\mathbf{x}} + \frac{\sqrt{3}}{2}x_9a\,\hat{\mathbf{y}} + z_9c\,\hat{\mathbf{z}}$ | $(12k)$ | Fe V |
| $\mathbf{B}_{30}$ | = | $-2x_9\,\mathbf{a}_1 - x_9\,\mathbf{a}_2 + z_9\,\mathbf{a}_3$ | = | $-\frac{3}{2}x_9a\,\hat{\mathbf{x}} + \frac{\sqrt{3}}{2}x_9a\,\hat{\mathbf{y}} + z_9c\,\hat{\mathbf{z}}$ | $(12k)$ | Fe V |
| $\mathbf{B}_{31}$ | = | $x_9\,\mathbf{a}_1 - x_9\,\mathbf{a}_2 + z_9\,\mathbf{a}_3$ | = | $-\sqrt{3}x_9a\,\hat{\mathbf{y}} + z_9c\,\hat{\mathbf{z}}$ | $(12k)$ | Fe V |
| $\mathbf{B}_{32}$ | = | $-x_9\,\mathbf{a}_1 - 2x_9\,\mathbf{a}_2 + \left(\frac{1}{2} + z_9\right)\,\mathbf{a}_3$ | = | $-\frac{3}{2}x_9a\,\hat{\mathbf{x}} - \frac{\sqrt{3}}{2}x_9a\,\hat{\mathbf{y}} + \left(\frac{1}{2} + z_9\right)c\,\hat{\mathbf{z}}$ | $(12k)$ | Fe V |
| $\mathbf{B}_{33}$ | = | $2x_9\,\mathbf{a}_1 + x_9\,\mathbf{a}_2 + \left(\frac{1}{2} + z_9\right)\,\mathbf{a}_3$ | = | $\frac{3}{2}x_9a\,\hat{\mathbf{x}} - \frac{\sqrt{3}}{2}x_9a\,\hat{\mathbf{y}} + \left(\frac{1}{2} + z_9\right)c\,\hat{\mathbf{z}}$ | $(12k)$ | Fe V |
| $\mathbf{B}_{34}$ | = | $-x_9\,\mathbf{a}_1 + x_9\,\mathbf{a}_2 + \left(\frac{1}{2} + z_9\right)\,\mathbf{a}_3$ | = | $\sqrt{3}x_9a\,\hat{\mathbf{y}} + \left(\frac{1}{2} + z_9\right)c\,\hat{\mathbf{z}}$ | $(12k)$ | Fe V |
| $\mathbf{B}_{35}$ | = | $2x_9\,\mathbf{a}_1 + x_9\,\mathbf{a}_2 - z_9\,\mathbf{a}_3$ | = | $\frac{3}{2}x_9a\,\hat{\mathbf{x}} - \frac{\sqrt{3}}{2}x_9a\,\hat{\mathbf{y}} - z_9c\,\hat{\mathbf{z}}$ | $(12k)$ | Fe V |
| $\mathbf{B}_{36}$ | = | $-x_9\,\mathbf{a}_1 - 2x_9\,\mathbf{a}_2 - z_9\,\mathbf{a}_3$ | = | $-\frac{3}{2}x_9a\,\hat{\mathbf{x}} - \frac{\sqrt{3}}{2}x_9a\,\hat{\mathbf{y}} - z_9c\,\hat{\mathbf{z}}$ | $(12k)$ | Fe V |
| $\mathbf{B}_{37}$ | = | $-x_9\,\mathbf{a}_1 + x_9\,\mathbf{a}_2 - z_9\,\mathbf{a}_3$ | = | $\sqrt{3}x_9a\,\hat{\mathbf{y}} - z_9c\,\hat{\mathbf{z}}$ | $(12k)$ | Fe V |
| $\mathbf{B}_{38}$ | = | $-2x_9\,\mathbf{a}_1 - x_9\,\mathbf{a}_2 + \left(\frac{1}{2} - z_9\right)\,\mathbf{a}_3$ | = | $-\frac{3}{2}x_9a\,\hat{\mathbf{x}} + \frac{\sqrt{3}}{2}x_9a\,\hat{\mathbf{y}} + \left(\frac{1}{2} - z_9\right)c\,\hat{\mathbf{z}}$ | $(12k)$ | Fe V |
| $\mathbf{B}_{39}$ | = | $x_9\,\mathbf{a}_1 + 2x_9\,\mathbf{a}_2 + \left(\frac{1}{2} - z_9\right)\,\mathbf{a}_3$ | = | $\frac{3}{2}x_9a\,\hat{\mathbf{x}} + \frac{\sqrt{3}}{2}x_9a\,\hat{\mathbf{y}} + \left(\frac{1}{2} - z_9\right)c\,\hat{\mathbf{z}}$ | $(12k)$ | Fe V |
| $\mathbf{B}_{40}$ | = | $x_9\,\mathbf{a}_1 - x_9\,\mathbf{a}_2 + \left(\frac{1}{2} - z_9\right)\,\mathbf{a}_3$ | = | $-\sqrt{3}x_9a\,\hat{\mathbf{y}} + \left(\frac{1}{2} - z_9\right)c\,\hat{\mathbf{z}}$ | $(12k)$ | Fe V |
| $\mathbf{B}_{41}$ | = | $x_{10}\,\mathbf{a}_1 + 2x_{10}\,\mathbf{a}_2 + z_{10}\,\mathbf{a}_3$ | = | $\frac{3}{2}x_{10}a\,\hat{\mathbf{x}} + \frac{\sqrt{3}}{2}x_{10}a\,\hat{\mathbf{y}} + z_{10}c\,\hat{\mathbf{z}}$ | $(12k)$ | O IV |
| $\mathbf{B}_{42}$ | = | $-2x_{10}\,\mathbf{a}_1 - x_{10}\,\mathbf{a}_2 + z_{10}\,\mathbf{a}_3$ | = | $-\frac{3}{2}x_{10}a\,\hat{\mathbf{x}} + \frac{\sqrt{3}}{2}x_{10}a\,\hat{\mathbf{y}} + z_{10}c\,\hat{\mathbf{z}}$ | $(12k)$ | O IV |
| $\mathbf{B}_{43}$ | = | $x_{10}\,\mathbf{a}_1 - x_{10}\,\mathbf{a}_2 + z_{10}\,\mathbf{a}_3$ | = | $-\sqrt{3}x_{10}a\,\hat{\mathbf{y}} + z_{10}c\,\hat{\mathbf{z}}$ | $(12k)$ | O IV |
| $\mathbf{B}_{44}$ | = | $-x_{10}\,\mathbf{a}_1 - 2x_{10}\,\mathbf{a}_2 + \left(\frac{1}{2} + z_{10}\right)\,\mathbf{a}_3$ | = | $-\frac{3}{2}x_{10}a\,\hat{\mathbf{x}} - \frac{\sqrt{3}}{2}x_{10}a\,\hat{\mathbf{y}} + \left(\frac{1}{2} + z_{10}\right)c\,\hat{\mathbf{z}}$ | $(12k)$ | O IV |
| $\mathbf{B}_{45}$ | = | $2x_{10}\,\mathbf{a}_1 + x_{10}\,\mathbf{a}_2 + \left(\frac{1}{2} + z_{10}\right)\,\mathbf{a}_3$ | = | $\frac{3}{2}x_{10}a\,\hat{\mathbf{x}} - \frac{\sqrt{3}}{2}x_{10}a\,\hat{\mathbf{y}} + \left(\frac{1}{2} + z_{10}\right)c\,\hat{\mathbf{z}}$ | $(12k)$ | O IV |
| $\mathbf{B}_{46}$ | = | $-x_{10}\,\mathbf{a}_1 + x_{10}\,\mathbf{a}_2 + \left(\frac{1}{2} + z_{10}\right)\,\mathbf{a}_3$ | = | $\sqrt{3}x_{10}a\,\hat{\mathbf{y}} + \left(\frac{1}{2} + z_{10}\right)c\,\hat{\mathbf{z}}$ | $(12k)$ | O IV |
| $\mathbf{B}_{47}$ | = | $2x_{10}\,\mathbf{a}_1 + x_{10}\,\mathbf{a}_2 - z_{10}\,\mathbf{a}_3$ | = | $\frac{3}{2}x_{10}a\,\hat{\mathbf{x}} - \frac{\sqrt{3}}{2}x_{10}a\,\hat{\mathbf{y}} - z_{10}c\,\hat{\mathbf{z}}$ | $(12k)$ | O IV |
| $\mathbf{B}_{48}$ | = | $-x_{10}\,\mathbf{a}_1 - 2x_{10}\,\mathbf{a}_2 - z_{10}\,\mathbf{a}_3$ | = | $-\frac{3}{2}x_{10}a\,\hat{\mathbf{x}} - \frac{\sqrt{3}}{2}x_{10}a\,\hat{\mathbf{y}} - z_{10}c\,\hat{\mathbf{z}}$ | $(12k)$ | O IV |
| $\mathbf{B}_{49}$ | = | $-x_{10}\,\mathbf{a}_1 + x_{10}\,\mathbf{a}_2 - z_{10}\,\mathbf{a}_3$ | = | $\sqrt{3}x_{10}a\,\hat{\mathbf{y}} - z_{10}c\,\hat{\mathbf{z}}$ | $(12k)$ | O IV |
| $\mathbf{B}_{50}$ | = | $-2x_{10}\,\mathbf{a}_1 - x_{10}\,\mathbf{a}_2 + \left(\frac{1}{2} - z_{10}\right)\,\mathbf{a}_3$ | = | $-\frac{3}{2}x_{10}a\,\hat{\mathbf{x}} + \frac{\sqrt{3}}{2}x_{10}a\,\hat{\mathbf{y}} + \left(\frac{1}{2} - z_{10}\right)c\,\hat{\mathbf{z}}$ | $(12k)$ | O IV |
| $\mathbf{B}_{51}$ | = | $x_{10}\,\mathbf{a}_1 + 2x_{10}\,\mathbf{a}_2 + \left(\frac{1}{2} - z_{10}\right)\,\mathbf{a}_3$ | = | $\frac{3}{2}x_{10}a\,\hat{\mathbf{x}} + \frac{\sqrt{3}}{2}x_{10}a\,\hat{\mathbf{y}} + \left(\frac{1}{2} - z_{10}\right)c\,\hat{\mathbf{z}}$ | $(12k)$ | O IV |
| $\mathbf{B}_{52}$ | = | $x_{10}\,\mathbf{a}_1 - x_{10}\,\mathbf{a}_2 + \left(\frac{1}{2} - z_{10}\right)\,\mathbf{a}_3$ | = | $-\sqrt{3}x_{10}a\,\hat{\mathbf{y}} + \left(\frac{1}{2} - z_{10}\right)c\,\hat{\mathbf{z}}$ | $(12k)$ | O IV |
| $\mathbf{B}_{53}$ | = | $x_{11}\,\mathbf{a}_1 + 2x_{11}\,\mathbf{a}_2 + z_{11}\,\mathbf{a}_3$ | = | $\frac{3}{2}x_{11}a\,\hat{\mathbf{x}} + \frac{\sqrt{3}}{2}x_{11}a\,\hat{\mathbf{y}} + z_{11}c\,\hat{\mathbf{z}}$ | $(12k)$ | O V |
| $\mathbf{B}_{54}$ | = | $-2x_{11}\,\mathbf{a}_1 - x_{11}\,\mathbf{a}_2 + z_{11}\,\mathbf{a}_3$ | = | $-\frac{3}{2}x_{11}a\,\hat{\mathbf{x}} + \frac{\sqrt{3}}{2}x_{11}a\,\hat{\mathbf{y}} + z_{11}c\,\hat{\mathbf{z}}$ | $(12k)$ | O V |
| $\mathbf{B}_{55}$ | = | $x_{11}\,\mathbf{a}_1 - x_{11}\,\mathbf{a}_2 + z_{11}\,\mathbf{a}_3$ | = | $-\sqrt{3}x_{11}a\,\hat{\mathbf{y}} + z_{11}c\,\hat{\mathbf{z}}$ | $(12k)$ | O V |
| $\mathbf{B}_{56}$ | = | $-x_{11}\,\mathbf{a}_1 - 2x_{11}\,\mathbf{a}_2 + \left(\frac{1}{2} + z_{11}\right)\,\mathbf{a}_3$ | = | $-\frac{3}{2}x_{11}a\,\hat{\mathbf{x}} - \frac{\sqrt{3}}{2}x_{11}a\,\hat{\mathbf{y}} + \left(\frac{1}{2} + z_{11}\right)c\,\hat{\mathbf{z}}$ | $(12k)$ | O V |
| $\mathbf{B}_{57}$ | = | $2x_{11}\,\mathbf{a}_1 + x_{11}\,\mathbf{a}_2 + \left(\frac{1}{2} + z_{11}\right)\,\mathbf{a}_3$ | = | $\frac{3}{2}x_{11}a\,\hat{\mathbf{x}} - \frac{\sqrt{3}}{2}x_{11}a\,\hat{\mathbf{y}} + \left(\frac{1}{2} + z_{11}\right)c\,\hat{\mathbf{z}}$ | $(12k)$ | O V |
| $\mathbf{B}_{58}$ | = | $-x_{11}\,\mathbf{a}_1 + x_{11}\,\mathbf{a}_2 + \left(\frac{1}{2} + z_{11}\right)\,\mathbf{a}_3$ | = | $\sqrt{3}x_{11}a\,\hat{\mathbf{y}} + \left(\frac{1}{2} + z_{11}\right)c\,\hat{\mathbf{z}}$ | $(12k)$ | O V |
| $\mathbf{B}_{59}$ | = | $2x_{11}\,\mathbf{a}_1 + x_{11}\,\mathbf{a}_2 - z_{11}\,\mathbf{a}_3$ | = | $\frac{3}{2}x_{11}a\,\hat{\mathbf{x}} - \frac{\sqrt{3}}{2}x_{11}a\,\hat{\mathbf{y}} - z_{11}c\,\hat{\mathbf{z}}$ | $(12k)$ | O V |
| $\mathbf{B}_{60}$ | = | $-x_{11}\,\mathbf{a}_1 - 2x_{11}\,\mathbf{a}_2 - z_{11}\,\mathbf{a}_3$ | = | $-\frac{3}{2}x_{11}a\,\hat{\mathbf{x}} - \frac{\sqrt{3}}{2}x_{11}a\,\hat{\mathbf{y}} - z_{11}c\,\hat{\mathbf{z}}$ | $(12k)$ | O V |
| $\mathbf{B}_{61}$ | = | $-x_{11}\,\mathbf{a}_1 + x_{11}\,\mathbf{a}_2 - z_{11}\,\mathbf{a}_3$ | = | $\sqrt{3}x_{11}a\,\hat{\mathbf{y}} - z_{11}c\,\hat{\mathbf{z}}$ | $(12k)$ | O V |
| $\mathbf{B}_{62}$ | = | $-2x_{11}\,\mathbf{a}_1 - x_{11}\,\mathbf{a}_2 + \left(\frac{1}{2} - z_{11}\right)\,\mathbf{a}_3$ | = | $-\frac{3}{2}x_{11}a\,\hat{\mathbf{x}} + \frac{\sqrt{3}}{2}x_{11}a\,\hat{\mathbf{y}} + \left(\frac{1}{2} - z_{11}\right)c\,\hat{\mathbf{z}}$ | $(12k)$ | O V |
| $\mathbf{B}_{63}$ | = | $x_{11}\,\mathbf{a}_1 + 2x_{11}\,\mathbf{a}_2 + \left(\frac{1}{2} - z_{11}\right)\,\mathbf{a}_3$ | = | $\frac{3}{2}x_{11}a\,\hat{\mathbf{x}} + \frac{\sqrt{3}}{2}x_{11}a\,\hat{\mathbf{y}} + \left(\frac{1}{2} - z_{11}\right)c\,\hat{\mathbf{z}}$ | $(12k)$ | O V |
| $\mathbf{B}_{64}$ | = | $x_{11}\,\mathbf{a}_1 - x_{11}\,\mathbf{a}_2 + \left(\frac{1}{2} - z_{11}\right)\,\mathbf{a}_3$ | = | $-\sqrt{3}x_{11}a\,\hat{\mathbf{y}} + \left(\frac{1}{2} - z_{11}\right)c\,\hat{\mathbf{z}}$ | $(12k)$ | O V |

**Geometry files:**

- CIF: pp. 1765

- POSCAR: pp. 1766

# $Pt_2Sn_3$ ($D5_b$) Structure: A2B3_hP10_194_f_bf

http://aflow.org/prototype-encyclopedia/A2B3_hP10_194_f_bf

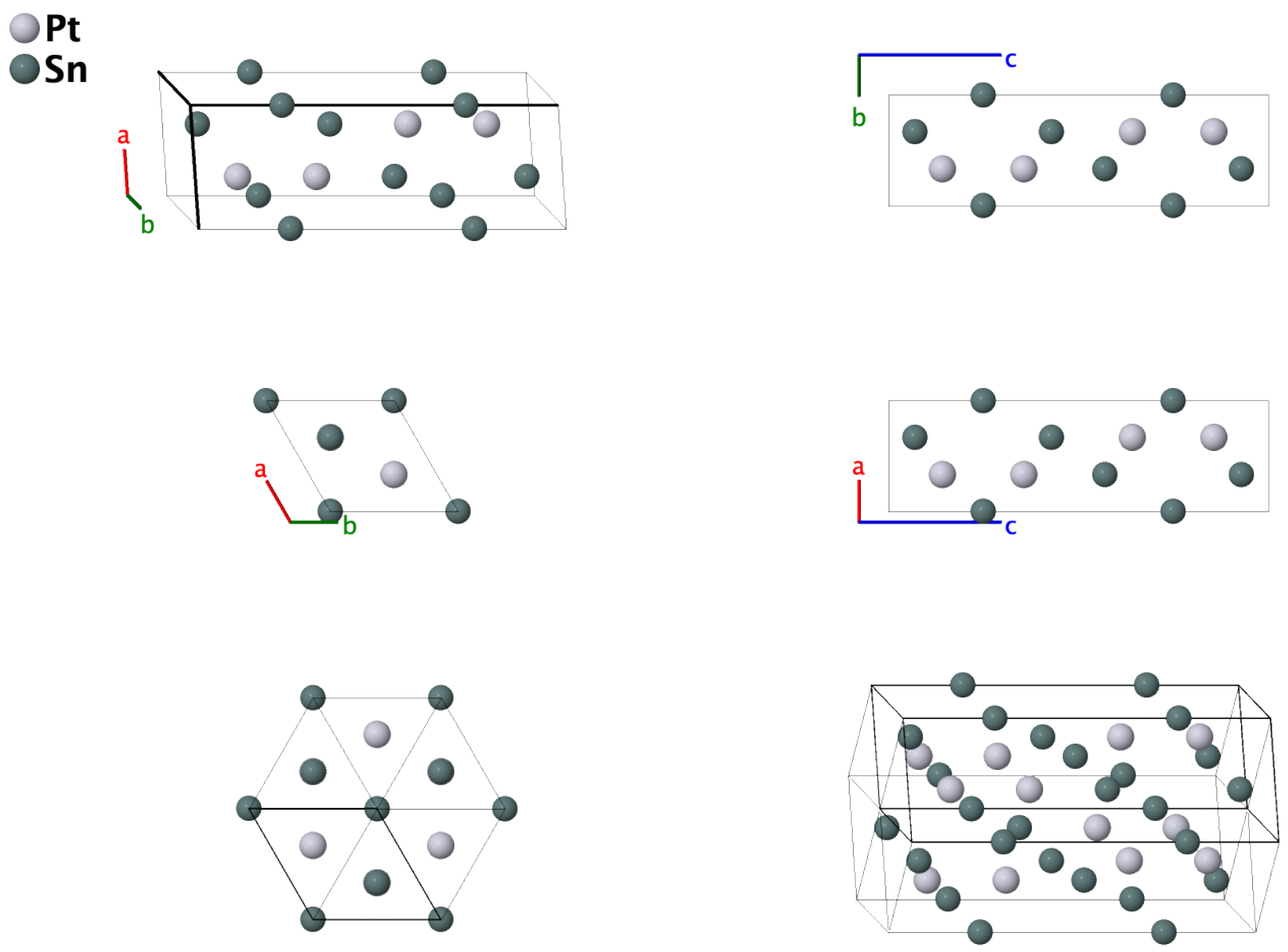

| | | |
|---|---|---|
| **Prototype** | : | $Pt_2Sn_3$ |
| **AFLOW prototype label** | : | A2B3_hP10_194_f_bf |
| ***Strukturbericht* designation** | : | $D5_b$ |
| **Pearson symbol** | : | hP10 |
| **Space group number** | : | 194 |
| **Space group symbol** | : | $P6_3/mmc$ |
| **AFLOW prototype command** | : | `aflow --proto=A2B3_hP10_194_f_bf`<br>`--params=`$a, c/a, z_2, z_3$ |

**Other compounds with this structure**

- $Pt_2Si_3$, $InAu_2Ga_2$, and $Au_4In_3Sn_3$

- We were unable to obtain a copy of (Schubert, 1949), and so we use the structural parameters given by (Pearson, 1958), using (Wood, 1947) to convert the $kX$ units used by Pearson into Ångström units.

**Hexagonal primitive vectors:**

$$\mathbf{a}_1 = \frac{1}{2}\,a\,\hat{\mathbf{x}} - \frac{\sqrt{3}}{2}\,a\,\hat{\mathbf{y}}$$
$$\mathbf{a}_2 = \frac{1}{2}\,a\,\hat{\mathbf{x}} + \frac{\sqrt{3}}{2}\,a\,\hat{\mathbf{y}}$$
$$\mathbf{a}_3 = c\,\hat{\mathbf{z}}$$

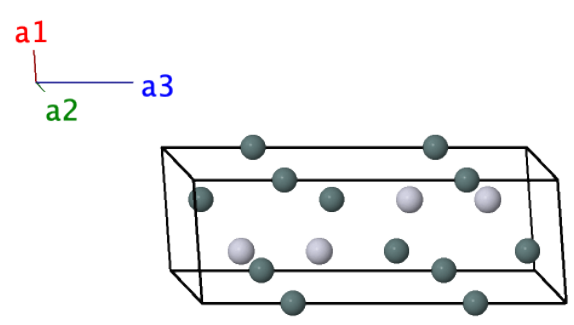

**Basis vectors:**

| | | Lattice Coordinates | | Cartesian Coordinates | Wyckoff Position | Atom Type |
|---|---|---|---|---|---|---|
| $\mathbf{B}_1$ | = | $\frac{1}{4}\,\mathbf{a}_3$ | = | $\frac{1}{4}c\,\hat{\mathbf{z}}$ | (2$b$) | Sn I |
| $\mathbf{B}_2$ | = | $\frac{3}{4}\,\mathbf{a}_3$ | = | $\frac{3}{4}c\,\hat{\mathbf{z}}$ | (2$b$) | Sn I |
| $\mathbf{B}_3$ | = | $\frac{1}{3}\,\mathbf{a}_1 + \frac{2}{3}\,\mathbf{a}_2 + z_2\,\mathbf{a}_3$ | = | $\frac{1}{2}a\,\hat{\mathbf{x}} + \frac{1}{2\sqrt{3}}a\,\hat{\mathbf{y}} + z_2c\,\hat{\mathbf{z}}$ | (4$f$) | Pt |
| $\mathbf{B}_4$ | = | $\frac{2}{3}\,\mathbf{a}_1 + \frac{1}{3}\,\mathbf{a}_2 + \left(\frac{1}{2} + z_2\right)\,\mathbf{a}_3$ | = | $\frac{1}{2}a\,\hat{\mathbf{x}} - \frac{1}{2\sqrt{3}}a\,\hat{\mathbf{y}} + \left(\frac{1}{2} + z_2\right)c\,\hat{\mathbf{z}}$ | (4$f$) | Pt |
| $\mathbf{B}_5$ | = | $\frac{2}{3}\,\mathbf{a}_1 + \frac{1}{3}\,\mathbf{a}_2 - z_2\,\mathbf{a}_3$ | = | $\frac{1}{2}a\,\hat{\mathbf{x}} - \frac{1}{2\sqrt{3}}a\,\hat{\mathbf{y}} - z_2c\,\hat{\mathbf{z}}$ | (4$f$) | Pt |
| $\mathbf{B}_6$ | = | $\frac{1}{3}\,\mathbf{a}_1 + \frac{2}{3}\,\mathbf{a}_2 + \left(\frac{1}{2} - z_2\right)\,\mathbf{a}_3$ | = | $\frac{1}{2}a\,\hat{\mathbf{x}} + \frac{1}{2\sqrt{3}}a\,\hat{\mathbf{y}} + \left(\frac{1}{2} - z_2\right)c\,\hat{\mathbf{z}}$ | (4$f$) | Pt |
| $\mathbf{B}_7$ | = | $\frac{1}{3}\,\mathbf{a}_1 + \frac{2}{3}\,\mathbf{a}_2 + z_3\,\mathbf{a}_3$ | = | $\frac{1}{2}a\,\hat{\mathbf{x}} + \frac{1}{2\sqrt{3}}a\,\hat{\mathbf{y}} + z_3c\,\hat{\mathbf{z}}$ | (4$f$) | Sn II |
| $\mathbf{B}_8$ | = | $\frac{2}{3}\,\mathbf{a}_1 + \frac{1}{3}\,\mathbf{a}_2 + \left(\frac{1}{2} + z_3\right)\,\mathbf{a}_3$ | = | $\frac{1}{2}a\,\hat{\mathbf{x}} - \frac{1}{2\sqrt{3}}a\,\hat{\mathbf{y}} + \left(\frac{1}{2} + z_3\right)c\,\hat{\mathbf{z}}$ | (4$f$) | Sn II |
| $\mathbf{B}_9$ | = | $\frac{2}{3}\,\mathbf{a}_1 + \frac{1}{3}\,\mathbf{a}_2 - z_3\,\mathbf{a}_3$ | = | $\frac{1}{2}a\,\hat{\mathbf{x}} - \frac{1}{2\sqrt{3}}a\,\hat{\mathbf{y}} - z_3c\,\hat{\mathbf{z}}$ | (4$f$) | Sn II |
| $\mathbf{B}_{10}$ | = | $\frac{1}{3}\,\mathbf{a}_1 + \frac{2}{3}\,\mathbf{a}_2 + \left(\frac{1}{2} - z_3\right)\,\mathbf{a}_3$ | = | $\frac{1}{2}a\,\hat{\mathbf{x}} + \frac{1}{2\sqrt{3}}a\,\hat{\mathbf{y}} + \left(\frac{1}{2} - z_3\right)c\,\hat{\mathbf{z}}$ | (4$f$) | Sn II |

**Found in:**

**Geometry files:**

- CIF: pp. 1766
- POSCAR: pp. 1766

# $\beta$-Alumina ($Al_2O_3$, $D5_6$) Structure: A2B3_hP60_194_3fk_cdef2k

http://aflow.org/prototype-encyclopedia/A2B3_hP60_194_3fk_cdef2k

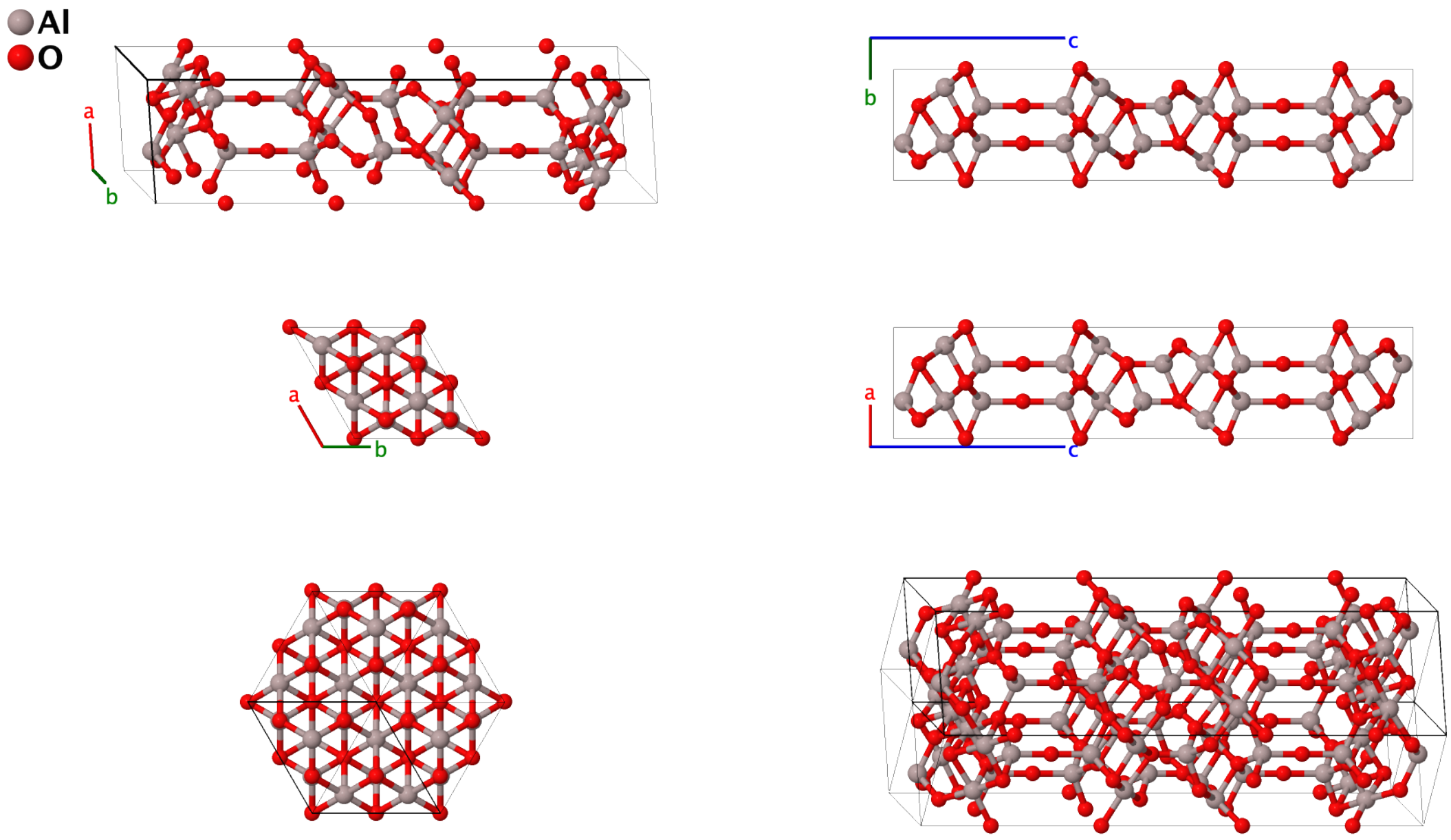

| | | |
|---|---|---|
| **Prototype** | : | $Al_2O_3$ |
| **AFLOW prototype label** | : | A2B3_hP60_194_3fk_cdef2k |
| ***Strukturbericht* designation** | : | $D5_6$ |
| **Pearson symbol** | : | hP60 |
| **Space group number** | : | 194 |
| **Space group symbol** | : | $P6_3/mmc$ |
| **AFLOW prototype command** | : | `aflow --proto=A2B3_hP60_194_3fk_cdef2k`<br>`--params=`$a, c/a, z_3, z_4, z_5, z_6, z_7, x_8, z_8, x_9, z_9, x_{10}, z_{10}$ |

- (Hermann, 1937) assigned this the *Strukturbericht* designation $D5_6$, calling it $\beta$-corundum, and subtitles the section "with small $Na_2O$ impurities." As noted by (Gottfried, 1937) and (Le Cars, 1975), the Na impurities replace Al atoms on the ($4f$) Wyckoff positions, along with some unspecified O sites. Charge neutrality requires that some of the displaced atoms remain in the lattice, and they are moved to ($2a$) Wyckoff positions at (0 0 0) and (0 0 1/2). Here we only list the "ideal" structure.

**Hexagonal primitive vectors:**

$$\mathbf{a}_1 = \frac{1}{2}\,a\,\hat{\mathbf{x}} - \frac{\sqrt{3}}{2}\,a\,\hat{\mathbf{y}}$$
$$\mathbf{a}_2 = \frac{1}{2}\,a\,\hat{\mathbf{x}} + \frac{\sqrt{3}}{2}\,a\,\hat{\mathbf{y}}$$
$$\mathbf{a}_3 = c\,\hat{\mathbf{z}}$$

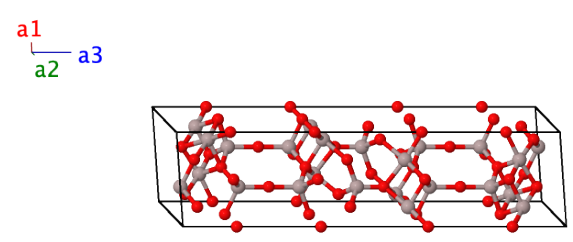

**Basis vectors:**

| | | Lattice Coordinates | | Cartesian Coordinates | Wyckoff Position | Atom Type |
|---|---|---|---|---|---|---|
| $\mathbf{B}_1$ | = | $\frac{1}{3}\,\mathbf{a}_1 + \frac{2}{3}\,\mathbf{a}_2 + \frac{1}{4}\,\mathbf{a}_3$ | = | $\frac{1}{2}a\,\hat{\mathbf{x}} + \frac{1}{2\sqrt{3}}a\,\hat{\mathbf{y}} + \frac{1}{4}c\,\hat{\mathbf{z}}$ | ($2c$) | O I |
| $\mathbf{B}_{2}$ | = | $\frac{2}{3}\,\mathbf{a}_{1}+\frac{1}{3}\,\mathbf{a}_{2}+\frac{3}{4}\,\mathbf{a}_{3}$ | = | $\frac{1}{2}a\,\mathbf{\hat{x}}-\frac{1}{2\sqrt{3}}a\,\mathbf{\hat{y}}+\frac{3}{4}c\,\mathbf{\hat{z}}$ | (2$c$) | O I |
| $\mathbf{B}_{3}$ | = | $\frac{1}{3}\,\mathbf{a}_{1}+\frac{2}{3}\,\mathbf{a}_{2}+\frac{3}{4}\,\mathbf{a}_{3}$ | = | $\frac{1}{2}a\,\mathbf{\hat{x}}+\frac{1}{2\sqrt{3}}a\,\mathbf{\hat{y}}+\frac{3}{4}c\,\mathbf{\hat{z}}$ | (2$d$) | O II |
| $\mathbf{B}_{4}$ | = | $\frac{2}{3}\,\mathbf{a}_{1}+\frac{1}{3}\,\mathbf{a}_{2}+\frac{1}{4}\,\mathbf{a}_{3}$ | = | $\frac{1}{2}a\,\mathbf{\hat{x}}-\frac{1}{2\sqrt{3}}a\,\mathbf{\hat{y}}+\frac{1}{4}c\,\mathbf{\hat{z}}$ | (2$d$) | O II |
| $\mathbf{B}_{5}$ | = | $z_{3}\,\mathbf{a}_{3}$ | = | $z_{3}c\,\mathbf{\hat{z}}$ | (4$e$) | O III |
| $\mathbf{B}_{6}$ | = | $\left(\frac{1}{2}+z_{3}\right)\,\mathbf{a}_{3}$ | = | $\left(\frac{1}{2}+z_{3}\right)c\,\mathbf{\hat{z}}$ | (4$e$) | O III |
| $\mathbf{B}_{7}$ | = | $-z_{3}\,\mathbf{a}_{3}$ | = | $-z_{3}c\,\mathbf{\hat{z}}$ | (4$e$) | O III |
| $\mathbf{B}_{8}$ | = | $\left(\frac{1}{2}-z_{3}\right)\,\mathbf{a}_{3}$ | = | $\left(\frac{1}{2}-z_{3}\right)c\,\mathbf{\hat{z}}$ | (4$e$) | O III |
| $\mathbf{B}_{9}$ | = | $\frac{1}{3}\,\mathbf{a}_{1}+\frac{2}{3}\,\mathbf{a}_{2}+z_{4}\,\mathbf{a}_{3}$ | = | $\frac{1}{2}a\,\mathbf{\hat{x}}+\frac{1}{2\sqrt{3}}a\,\mathbf{\hat{y}}+z_{4}c\,\mathbf{\hat{z}}$ | (4$f$) | Al I |
| $\mathbf{B}_{10}$ | = | $\frac{2}{3}\,\mathbf{a}_{1}+\frac{1}{3}\,\mathbf{a}_{2}+\left(\frac{1}{2}+z_{4}\right)\,\mathbf{a}_{3}$ | = | $\frac{1}{2}a\,\mathbf{\hat{x}}-\frac{1}{2\sqrt{3}}a\,\mathbf{\hat{y}}+\left(\frac{1}{2}+z_{4}\right)c\,\mathbf{\hat{z}}$ | (4$f$) | Al I |
| $\mathbf{B}_{11}$ | = | $\frac{2}{3}\,\mathbf{a}_{1}+\frac{1}{3}\,\mathbf{a}_{2}-z_{4}\,\mathbf{a}_{3}$ | = | $\frac{1}{2}a\,\mathbf{\hat{x}}-\frac{1}{2\sqrt{3}}a\,\mathbf{\hat{y}}-z_{4}c\,\mathbf{\hat{z}}$ | (4$f$) | Al I |
| $\mathbf{B}_{12}$ | = | $\frac{1}{3}\,\mathbf{a}_{1}+\frac{2}{3}\,\mathbf{a}_{2}+\left(\frac{1}{2}-z_{4}\right)\,\mathbf{a}_{3}$ | = | $\frac{1}{2}a\,\mathbf{\hat{x}}+\frac{1}{2\sqrt{3}}a\,\mathbf{\hat{y}}+\left(\frac{1}{2}-z_{4}\right)c\,\mathbf{\hat{z}}$ | (4$f$) | Al I |
| $\mathbf{B}_{13}$ | = | $\frac{1}{3}\,\mathbf{a}_{1}+\frac{2}{3}\,\mathbf{a}_{2}+z_{5}\,\mathbf{a}_{3}$ | = | $\frac{1}{2}a\,\mathbf{\hat{x}}+\frac{1}{2\sqrt{3}}a\,\mathbf{\hat{y}}+z_{5}c\,\mathbf{\hat{z}}$ | (4$f$) | Al II |
| $\mathbf{B}_{14}$ | = | $\frac{2}{3}\,\mathbf{a}_{1}+\frac{1}{3}\,\mathbf{a}_{2}+\left(\frac{1}{2}+z_{5}\right)\,\mathbf{a}_{3}$ | = | $\frac{1}{2}a\,\mathbf{\hat{x}}-\frac{1}{2\sqrt{3}}a\,\mathbf{\hat{y}}+\left(\frac{1}{2}+z_{5}\right)c\,\mathbf{\hat{z}}$ | (4$f$) | Al II |
| $\mathbf{B}_{15}$ | = | $\frac{2}{3}\,\mathbf{a}_{1}+\frac{1}{3}\,\mathbf{a}_{2}-z_{5}\,\mathbf{a}_{3}$ | = | $\frac{1}{2}a\,\mathbf{\hat{x}}-\frac{1}{2\sqrt{3}}a\,\mathbf{\hat{y}}-z_{5}c\,\mathbf{\hat{z}}$ | (4$f$) | Al II |
| $\mathbf{B}_{16}$ | = | $\frac{1}{3}\,\mathbf{a}_{1}+\frac{2}{3}\,\mathbf{a}_{2}+\left(\frac{1}{2}-z_{5}\right)\,\mathbf{a}_{3}$ | = | $\frac{1}{2}a\,\mathbf{\hat{x}}+\frac{1}{2\sqrt{3}}a\,\mathbf{\hat{y}}+\left(\frac{1}{2}-z_{5}\right)c\,\mathbf{\hat{z}}$ | (4$f$) | Al II |
| $\mathbf{B}_{17}$ | = | $\frac{1}{3}\,\mathbf{a}_{1}+\frac{2}{3}\,\mathbf{a}_{2}+z_{6}\,\mathbf{a}_{3}$ | = | $\frac{1}{2}a\,\mathbf{\hat{x}}+\frac{1}{2\sqrt{3}}a\,\mathbf{\hat{y}}+z_{6}c\,\mathbf{\hat{z}}$ | (4$f$) | Al III |
| $\mathbf{B}_{18}$ | = | $\frac{2}{3}\,\mathbf{a}_{1}+\frac{1}{3}\,\mathbf{a}_{2}+\left(\frac{1}{2}+z_{6}\right)\,\mathbf{a}_{3}$ | = | $\frac{1}{2}a\,\mathbf{\hat{x}}-\frac{1}{2\sqrt{3}}a\,\mathbf{\hat{y}}+\left(\frac{1}{2}+z_{6}\right)c\,\mathbf{\hat{z}}$ | (4$f$) | Al III |
| $\mathbf{B}_{19}$ | = | $\frac{2}{3}\,\mathbf{a}_{1}+\frac{1}{3}\,\mathbf{a}_{2}-z_{6}\,\mathbf{a}_{3}$ | = | $\frac{1}{2}a\,\mathbf{\hat{x}}-\frac{1}{2\sqrt{3}}a\,\mathbf{\hat{y}}-z_{6}c\,\mathbf{\hat{z}}$ | (4$f$) | Al III |
| $\mathbf{B}_{20}$ | = | $\frac{1}{3}\,\mathbf{a}_{1}+\frac{2}{3}\,\mathbf{a}_{2}+\left(\frac{1}{2}-z_{6}\right)\,\mathbf{a}_{3}$ | = | $\frac{1}{2}a\,\mathbf{\hat{x}}+\frac{1}{2\sqrt{3}}a\,\mathbf{\hat{y}}+\left(\frac{1}{2}-z_{6}\right)c\,\mathbf{\hat{z}}$ | (4$f$) | Al III |
| $\mathbf{B}_{21}$ | = | $\frac{1}{3}\,\mathbf{a}_{1}+\frac{2}{3}\,\mathbf{a}_{2}+z_{7}\,\mathbf{a}_{3}$ | = | $\frac{1}{2}a\,\mathbf{\hat{x}}+\frac{1}{2\sqrt{3}}a\,\mathbf{\hat{y}}+z_{7}c\,\mathbf{\hat{z}}$ | (4$f$) | O IV |
| $\mathbf{B}_{22}$ | = | $\frac{2}{3}\,\mathbf{a}_{1}+\frac{1}{3}\,\mathbf{a}_{2}+\left(\frac{1}{2}+z_{7}\right)\,\mathbf{a}_{3}$ | = | $\frac{1}{2}a\,\mathbf{\hat{x}}-\frac{1}{2\sqrt{3}}a\,\mathbf{\hat{y}}+\left(\frac{1}{2}+z_{7}\right)c\,\mathbf{\hat{z}}$ | (4$f$) | O IV |
| $\mathbf{B}_{23}$ | = | $\frac{2}{3}\,\mathbf{a}_{1}+\frac{1}{3}\,\mathbf{a}_{2}-z_{7}\,\mathbf{a}_{3}$ | = | $\frac{1}{2}a\,\mathbf{\hat{x}}-\frac{1}{2\sqrt{3}}a\,\mathbf{\hat{y}}-z_{7}c\,\mathbf{\hat{z}}$ | (4$f$) | O IV |
| $\mathbf{B}_{24}$ | = | $\frac{1}{3}\,\mathbf{a}_{1}+\frac{2}{3}\,\mathbf{a}_{2}+\left(\frac{1}{2}-z_{7}\right)\,\mathbf{a}_{3}$ | = | $\frac{1}{2}a\,\mathbf{\hat{x}}+\frac{1}{2\sqrt{3}}a\,\mathbf{\hat{y}}+\left(\frac{1}{2}-z_{7}\right)c\,\mathbf{\hat{z}}$ | (4$f$) | O IV |
| $\mathbf{B}_{25}$ | = | $x_{8}\,\mathbf{a}_{1}+2x_{8}\,\mathbf{a}_{2}+z_{8}\,\mathbf{a}_{3}$ | = | $\frac{3}{2}x_{8}a\,\mathbf{\hat{x}}+\frac{\sqrt{3}}{2}x_{8}a\,\mathbf{\hat{y}}+z_{8}c\,\mathbf{\hat{z}}$ | (12$k$) | Al IV |
| $\mathbf{B}_{26}$ | = | $-2x_{8}\,\mathbf{a}_{1}-x_{8}\,\mathbf{a}_{2}+z_{8}\,\mathbf{a}_{3}$ | = | $-\frac{3}{2}x_{8}a\,\mathbf{\hat{x}}+\frac{\sqrt{3}}{2}x_{8}a\,\mathbf{\hat{y}}+z_{8}c\,\mathbf{\hat{z}}$ | (12$k$) | Al IV |
| $\mathbf{B}_{27}$ | = | $x_{8}\,\mathbf{a}_{1}-x_{8}\,\mathbf{a}_{2}+z_{8}\,\mathbf{a}_{3}$ | = | $-\sqrt{3}x_{8}a\,\mathbf{\hat{y}}+z_{8}c\,\mathbf{\hat{z}}$ | (12$k$) | Al IV |
| $\mathbf{B}_{28}$ | = | $-x_{8}\,\mathbf{a}_{1}-2x_{8}\,\mathbf{a}_{2}+\left(\frac{1}{2}+z_{8}\right)\,\mathbf{a}_{3}$ | = | $-\frac{3}{2}x_{8}a\,\mathbf{\hat{x}}-\frac{\sqrt{3}}{2}x_{8}a\,\mathbf{\hat{y}}+\left(\frac{1}{2}+z_{8}\right)c\,\mathbf{\hat{z}}$ | (12$k$) | Al IV |
| $\mathbf{B}_{29}$ | = | $2x_{8}\,\mathbf{a}_{1}+x_{8}\,\mathbf{a}_{2}+\left(\frac{1}{2}+z_{8}\right)\,\mathbf{a}_{3}$ | = | $\frac{3}{2}x_{8}a\,\mathbf{\hat{x}}-\frac{\sqrt{3}}{2}x_{8}a\,\mathbf{\hat{y}}+\left(\frac{1}{2}+z_{8}\right)c\,\mathbf{\hat{z}}$ | (12$k$) | Al IV |
| $\mathbf{B}_{30}$ | = | $-x_{8}\,\mathbf{a}_{1}+x_{8}\,\mathbf{a}_{2}+\left(\frac{1}{2}+z_{8}\right)\,\mathbf{a}_{3}$ | = | $\sqrt{3}x_{8}a\,\mathbf{\hat{y}}+\left(\frac{1}{2}+z_{8}\right)c\,\mathbf{\hat{z}}$ | (12$k$) | Al IV |
| $\mathbf{B}_{31}$ | = | $2x_{8}\,\mathbf{a}_{1}+x_{8}\,\mathbf{a}_{2}-z_{8}\,\mathbf{a}_{3}$ | = | $\frac{3}{2}x_{8}a\,\mathbf{\hat{x}}-\frac{\sqrt{3}}{2}x_{8}a\,\mathbf{\hat{y}}-z_{8}c\,\mathbf{\hat{z}}$ | (12$k$) | Al IV |
| $\mathbf{B}_{32}$ | = | $-x_{8}\,\mathbf{a}_{1}-2x_{8}\,\mathbf{a}_{2}-z_{8}\,\mathbf{a}_{3}$ | = | $-\frac{3}{2}x_{8}a\,\mathbf{\hat{x}}-\frac{\sqrt{3}}{2}x_{8}a\,\mathbf{\hat{y}}-z_{8}c\,\mathbf{\hat{z}}$ | (12$k$) | Al IV |
| $\mathbf{B}_{33}$ | = | $-x_{8}\,\mathbf{a}_{1}+x_{8}\,\mathbf{a}_{2}-z_{8}\,\mathbf{a}_{3}$ | = | $\sqrt{3}x_{8}a\,\mathbf{\hat{y}}-z_{8}c\,\mathbf{\hat{z}}$ | (12$k$) | Al IV |
| $\mathbf{B}_{34}$ | = | $-2x_{8}\,\mathbf{a}_{1}-x_{8}\,\mathbf{a}_{2}+\left(\frac{1}{2}-z_{8}\right)\,\mathbf{a}_{3}$ | = | $-\frac{3}{2}x_{8}a\,\mathbf{\hat{x}}+\frac{\sqrt{3}}{2}x_{8}a\,\mathbf{\hat{y}}+\left(\frac{1}{2}-z_{8}\right)c\,\mathbf{\hat{z}}$ | (12$k$) | Al IV |
| $\mathbf{B}_{35}$ | = | $x_{8}\,\mathbf{a}_{1}+2x_{8}\,\mathbf{a}_{2}+\left(\frac{1}{2}-z_{8}\right)\,\mathbf{a}_{3}$ | = | $\frac{3}{2}x_{8}a\,\mathbf{\hat{x}}+\frac{\sqrt{3}}{2}x_{8}a\,\mathbf{\hat{y}}+\left(\frac{1}{2}-z_{8}\right)c\,\mathbf{\hat{z}}$ | (12$k$) | Al IV |
| $\mathbf{B}_{36}$ | = | $x_{8}\,\mathbf{a}_{1}-x_{8}\,\mathbf{a}_{2}+\left(\frac{1}{2}-z_{8}\right)\,\mathbf{a}_{3}$ | = | $-\sqrt{3}x_{8}a\,\mathbf{\hat{y}}+\left(\frac{1}{2}-z_{8}\right)c\,\mathbf{\hat{z}}$ | (12$k$) | Al IV |
| $\mathbf{B}_{37}$ | = | $x_{9}\,\mathbf{a}_{1}+2x_{9}\,\mathbf{a}_{2}+z_{9}\,\mathbf{a}_{3}$ | = | $\frac{3}{2}x_{9}a\,\mathbf{\hat{x}}+\frac{\sqrt{3}}{2}x_{9}a\,\mathbf{\hat{y}}+z_{9}c\,\mathbf{\hat{z}}$ | (12$k$) | O V |
| $\mathbf{B}_{38}$ | = | $-2x_9\,\mathbf{a}_1 - x_9\,\mathbf{a}_2 + z_9\,\mathbf{a}_3$ | = | $-\frac{3}{2}x_9a\,\hat{\mathbf{x}} + \frac{\sqrt{3}}{2}x_9a\,\hat{\mathbf{y}} + z_9c\,\hat{\mathbf{z}}$ | $(12k)$ | O V |
| $\mathbf{B}_{39}$ | = | $x_9\,\mathbf{a}_1 - x_9\,\mathbf{a}_2 + z_9\,\mathbf{a}_3$ | = | $-\sqrt{3}x_9a\,\hat{\mathbf{y}} + z_9c\,\hat{\mathbf{z}}$ | $(12k)$ | O V |
| $\mathbf{B}_{40}$ | = | $-x_9\,\mathbf{a}_1 - 2x_9\,\mathbf{a}_2 + \left(\frac{1}{2} + z_9\right)\,\mathbf{a}_3$ | = | $-\frac{3}{2}x_9a\,\hat{\mathbf{x}} - \frac{\sqrt{3}}{2}x_9a\,\hat{\mathbf{y}} + \left(\frac{1}{2} + z_9\right)c\,\hat{\mathbf{z}}$ | $(12k)$ | O V |
| $\mathbf{B}_{41}$ | = | $2x_9\,\mathbf{a}_1 + x_9\,\mathbf{a}_2 + \left(\frac{1}{2} + z_9\right)\,\mathbf{a}_3$ | = | $\frac{3}{2}x_9a\,\hat{\mathbf{x}} - \frac{\sqrt{3}}{2}x_9a\,\hat{\mathbf{y}} + \left(\frac{1}{2} + z_9\right)c\,\hat{\mathbf{z}}$ | $(12k)$ | O V |
| $\mathbf{B}_{42}$ | = | $-x_9\,\mathbf{a}_1 + x_9\,\mathbf{a}_2 + \left(\frac{1}{2} + z_9\right)\,\mathbf{a}_3$ | = | $\sqrt{3}x_9a\,\hat{\mathbf{y}} + \left(\frac{1}{2} + z_9\right)c\,\hat{\mathbf{z}}$ | $(12k)$ | O V |
| $\mathbf{B}_{43}$ | = | $2x_9\,\mathbf{a}_1 + x_9\,\mathbf{a}_2 - z_9\,\mathbf{a}_3$ | = | $\frac{3}{2}x_9a\,\hat{\mathbf{x}} - \frac{\sqrt{3}}{2}x_9a\,\hat{\mathbf{y}} - z_9c\,\hat{\mathbf{z}}$ | $(12k)$ | O V |
| $\mathbf{B}_{44}$ | = | $-x_9\,\mathbf{a}_1 - 2x_9\,\mathbf{a}_2 - z_9\,\mathbf{a}_3$ | = | $-\frac{3}{2}x_9a\,\hat{\mathbf{x}} - \frac{\sqrt{3}}{2}x_9a\,\hat{\mathbf{y}} - z_9c\,\hat{\mathbf{z}}$ | $(12k)$ | O V |
| $\mathbf{B}_{45}$ | = | $-x_9\,\mathbf{a}_1 + x_9\,\mathbf{a}_2 - z_9\,\mathbf{a}_3$ | = | $\sqrt{3}x_9a\,\hat{\mathbf{y}} - z_9c\,\hat{\mathbf{z}}$ | $(12k)$ | O V |
| $\mathbf{B}_{46}$ | = | $-2x_9\,\mathbf{a}_1 - x_9\,\mathbf{a}_2 + \left(\frac{1}{2} - z_9\right)\,\mathbf{a}_3$ | = | $-\frac{3}{2}x_9a\,\hat{\mathbf{x}} + \frac{\sqrt{3}}{2}x_9a\,\hat{\mathbf{y}} + \left(\frac{1}{2} - z_9\right)c\,\hat{\mathbf{z}}$ | $(12k)$ | O V |
| $\mathbf{B}_{47}$ | = | $x_9\,\mathbf{a}_1 + 2x_9\,\mathbf{a}_2 + \left(\frac{1}{2} - z_9\right)\,\mathbf{a}_3$ | = | $\frac{3}{2}x_9a\,\hat{\mathbf{x}} + \frac{\sqrt{3}}{2}x_9a\,\hat{\mathbf{y}} + \left(\frac{1}{2} - z_9\right)c\,\hat{\mathbf{z}}$ | $(12k)$ | O V |
| $\mathbf{B}_{48}$ | = | $x_9\,\mathbf{a}_1 - x_9\,\mathbf{a}_2 + \left(\frac{1}{2} - z_9\right)\,\mathbf{a}_3$ | = | $-\sqrt{3}x_9a\,\hat{\mathbf{y}} + \left(\frac{1}{2} - z_9\right)c\,\hat{\mathbf{z}}$ | $(12k)$ | O V |
| $\mathbf{B}_{49}$ | = | $x_{10}\,\mathbf{a}_1 + 2x_{10}\,\mathbf{a}_2 + z_{10}\,\mathbf{a}_3$ | = | $\frac{3}{2}x_{10}a\,\hat{\mathbf{x}} + \frac{\sqrt{3}}{2}x_{10}a\,\hat{\mathbf{y}} + z_{10}c\,\hat{\mathbf{z}}$ | $(12k)$ | O VI |
| $\mathbf{B}_{50}$ | = | $-2x_{10}\,\mathbf{a}_1 - x_{10}\,\mathbf{a}_2 + z_{10}\,\mathbf{a}_3$ | = | $-\frac{3}{2}x_{10}a\,\hat{\mathbf{x}} + \frac{\sqrt{3}}{2}x_{10}a\,\hat{\mathbf{y}} + z_{10}c\,\hat{\mathbf{z}}$ | $(12k)$ | O VI |
| $\mathbf{B}_{51}$ | = | $x_{10}\,\mathbf{a}_1 - x_{10}\,\mathbf{a}_2 + z_{10}\,\mathbf{a}_3$ | = | $-\sqrt{3}x_{10}a\,\hat{\mathbf{y}} + z_{10}c\,\hat{\mathbf{z}}$ | $(12k)$ | O VI |
| $\mathbf{B}_{52}$ | = | $-x_{10}\,\mathbf{a}_1 - 2x_{10}\,\mathbf{a}_2 + \left(\frac{1}{2} + z_{10}\right)\,\mathbf{a}_3$ | = | $-\frac{3}{2}x_{10}a\,\hat{\mathbf{x}} - \frac{\sqrt{3}}{2}x_{10}a\,\hat{\mathbf{y}} + \left(\frac{1}{2} + z_{10}\right)c\,\hat{\mathbf{z}}$ | $(12k)$ | O VI |
| $\mathbf{B}_{53}$ | = | $2x_{10}\,\mathbf{a}_1 + x_{10}\,\mathbf{a}_2 + \left(\frac{1}{2} + z_{10}\right)\,\mathbf{a}_3$ | = | $\frac{3}{2}x_{10}a\,\hat{\mathbf{x}} - \frac{\sqrt{3}}{2}x_{10}a\,\hat{\mathbf{y}} + \left(\frac{1}{2} + z_{10}\right)c\,\hat{\mathbf{z}}$ | $(12k)$ | O VI |
| $\mathbf{B}_{54}$ | = | $-x_{10}\,\mathbf{a}_1 + x_{10}\,\mathbf{a}_2 + \left(\frac{1}{2} + z_{10}\right)\,\mathbf{a}_3$ | = | $\sqrt{3}x_{10}a\,\hat{\mathbf{y}} + \left(\frac{1}{2} + z_{10}\right)c\,\hat{\mathbf{z}}$ | $(12k)$ | O VI |
| $\mathbf{B}_{55}$ | = | $2x_{10}\,\mathbf{a}_1 + x_{10}\,\mathbf{a}_2 - z_{10}\,\mathbf{a}_3$ | = | $\frac{3}{2}x_{10}a\,\hat{\mathbf{x}} - \frac{\sqrt{3}}{2}x_{10}a\,\hat{\mathbf{y}} - z_{10}c\,\hat{\mathbf{z}}$ | $(12k)$ | O VI |
| $\mathbf{B}_{56}$ | = | $-x_{10}\,\mathbf{a}_1 - 2x_{10}\,\mathbf{a}_2 - z_{10}\,\mathbf{a}_3$ | = | $-\frac{3}{2}x_{10}a\,\hat{\mathbf{x}} - \frac{\sqrt{3}}{2}x_{10}a\,\hat{\mathbf{y}} - z_{10}c\,\hat{\mathbf{z}}$ | $(12k)$ | O VI |
| $\mathbf{B}_{57}$ | = | $-x_{10}\,\mathbf{a}_1 + x_{10}\,\mathbf{a}_2 - z_{10}\,\mathbf{a}_3$ | = | $\sqrt{3}x_{10}a\,\hat{\mathbf{y}} - z_{10}c\,\hat{\mathbf{z}}$ | $(12k)$ | O VI |
| $\mathbf{B}_{58}$ | = | $-2x_{10}\,\mathbf{a}_1 - x_{10}\,\mathbf{a}_2 + \left(\frac{1}{2} - z_{10}\right)\,\mathbf{a}_3$ | = | $-\frac{3}{2}x_{10}a\,\hat{\mathbf{x}} + \frac{\sqrt{3}}{2}x_{10}a\,\hat{\mathbf{y}} + \left(\frac{1}{2} - z_{10}\right)c\,\hat{\mathbf{z}}$ | $(12k)$ | O VI |
| $\mathbf{B}_{59}$ | = | $x_{10}\,\mathbf{a}_1 + 2x_{10}\,\mathbf{a}_2 + \left(\frac{1}{2} - z_{10}\right)\,\mathbf{a}_3$ | = | $\frac{3}{2}x_{10}a\,\hat{\mathbf{x}} + \frac{\sqrt{3}}{2}x_{10}a\,\hat{\mathbf{y}} + \left(\frac{1}{2} - z_{10}\right)c\,\hat{\mathbf{z}}$ | $(12k)$ | O VI |
| $\mathbf{B}_{60}$ | = | $x_{10}\,\mathbf{a}_1 - x_{10}\,\mathbf{a}_2 + \left(\frac{1}{2} - z_{10}\right)\,\mathbf{a}_3$ | = | $-\sqrt{3}x_{10}a\,\hat{\mathbf{y}} + \left(\frac{1}{2} - z_{10}\right)c\,\hat{\mathbf{z}}$ | $(12k)$ | O VI |

**Geometry files:**

- CIF: pp. 1766
- POSCAR: pp. 1767

# $S3_4$ (II) (Catapleiite, $Na_2Zr(SiO_3)_3 \cdot H_2O$) (*obsolete*) Structure: A3B2C9D3E_hP36_194_g_f_hk_h_a

http://aflow.org/prototype-encyclopedia/A3B2C9D3E_hP36_194_g_f_hk_h_a

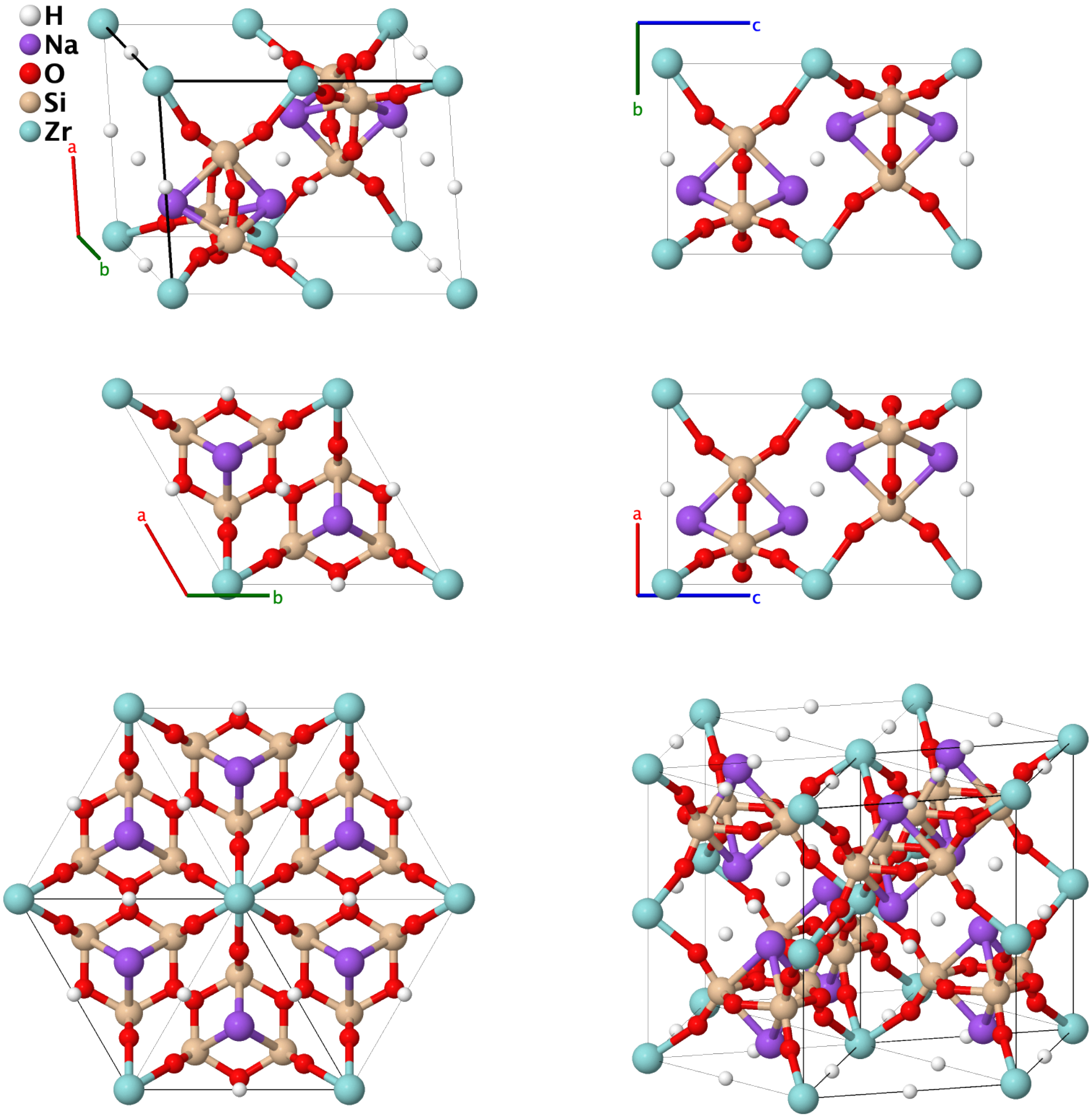

| | | |
|---|---|---|
| **Prototype** | : | $(H_2O)Na_2O_9Si_3Zr$ |
| **AFLOW prototype label** | : | A3B2C9D3E_hP36_194_g_f_hk_h_a |
| ***Strukturbericht* designation** | : | $S3_4$ (II) |
| **Pearson symbol** | : | hP36 |
| **Space group number** | : | 194 |
| **Space group symbol** | : | $P6_3/mmc$ |
| **AFLOW prototype command** | : | `aflow --proto=A3B2C9D3E_hP36_194_g_f_hk_h_a --params=`$a, c/a, z_2, x_4, x_5, x_6, z_6$ |

- This hexagonal structure has been superseded by the monoclinic structure of (Ilyushin, 1981). We present it here for historical interest.
- We were unable to procure the original reference, so we use the data provided by (Gottfried, 1937).
- Four of the (6*g*) sites are randomly occupied by water molecules.
- (Gottfried, 1937) originally gave the *Strukturbericht* designation $S3_4$ to chabazite, but (Gottfried, 1940) gave it to this structure of catapleiite. We resolve this dilemma by using $S3_4$(I) for chabazite and $S3_4$(II) for catapleiite.

**Hexagonal primitive vectors:**

$$
\begin{aligned}
\mathbf{a}_1 &= \tfrac{1}{2}\,a\,\hat{\mathbf{x}} - \tfrac{\sqrt{3}}{2}\,a\,\hat{\mathbf{y}} \\
\mathbf{a}_2 &= \tfrac{1}{2}\,a\,\hat{\mathbf{x}} + \tfrac{\sqrt{3}}{2}\,a\,\hat{\mathbf{y}} \\
\mathbf{a}_3 &= c\,\hat{\mathbf{z}}
\end{aligned}
$$

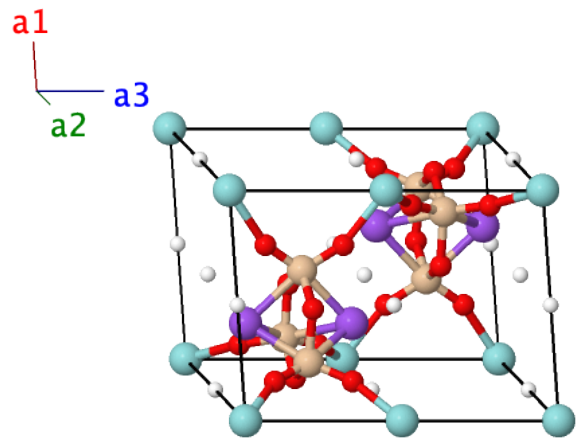

**Basis vectors:**

| | | Lattice Coordinates | | Cartesian Coordinates | Wyckoff Position | Atom Type |
|---|---|---|---|---|---|---|
| $\mathbf{B}_1$ | = | $0\,\mathbf{a}_1 + 0\,\mathbf{a}_2 + 0\,\mathbf{a}_3$ | = | $0\,\hat{\mathbf{x}} + 0\,\hat{\mathbf{y}} + 0\,\hat{\mathbf{z}}$ | (2$a$) | Zr |
| $\mathbf{B}_2$ | = | $\frac{1}{2}\,\mathbf{a}_3$ | = | $\frac{1}{2}c\,\hat{\mathbf{z}}$ | (2$a$) | Zr |
| $\mathbf{B}_3$ | = | $\frac{1}{3}\,\mathbf{a}_1 + \frac{2}{3}\,\mathbf{a}_2 + z_2\,\mathbf{a}_3$ | = | $\frac{1}{2}a\,\hat{\mathbf{x}} + \frac{1}{2\sqrt{3}}a\,\hat{\mathbf{y}} + z_2c\,\hat{\mathbf{z}}$ | (4$f$) | Na |
| $\mathbf{B}_4$ | = | $\frac{2}{3}\,\mathbf{a}_1 + \frac{1}{3}\,\mathbf{a}_2 + \left(\frac{1}{2} + z_2\right)\,\mathbf{a}_3$ | = | $\frac{1}{2}a\,\hat{\mathbf{x}} - \frac{1}{2\sqrt{3}}a\,\hat{\mathbf{y}} + \left(\frac{1}{2} + z_2\right)c\,\hat{\mathbf{z}}$ | (4$f$) | Na |
| $\mathbf{B}_5$ | = | $\frac{2}{3}\,\mathbf{a}_1 + \frac{1}{3}\,\mathbf{a}_2 - z_2\,\mathbf{a}_3$ | = | $\frac{1}{2}a\,\hat{\mathbf{x}} - \frac{1}{2\sqrt{3}}a\,\hat{\mathbf{y}} - z_2c\,\hat{\mathbf{z}}$ | (4$f$) | Na |
| $\mathbf{B}_6$ | = | $\frac{1}{3}\,\mathbf{a}_1 + \frac{2}{3}\,\mathbf{a}_2 + \left(\frac{1}{2} - z_2\right)\,\mathbf{a}_3$ | = | $\frac{1}{2}a\,\hat{\mathbf{x}} + \frac{1}{2\sqrt{3}}a\,\hat{\mathbf{y}} + \left(\frac{1}{2} - z_2\right)c\,\hat{\mathbf{z}}$ | (4$f$) | Na |
| $\mathbf{B}_7$ | = | $\frac{1}{2}\,\mathbf{a}_1$ | = | $\frac{1}{4}a\,\hat{\mathbf{x}} - \frac{\sqrt{3}}{4}a\,\hat{\mathbf{y}}$ | (6$g$) | $H_2O$ |
| $\mathbf{B}_8$ | = | $\frac{1}{2}\,\mathbf{a}_2$ | = | $\frac{1}{4}a\,\hat{\mathbf{x}} + \frac{\sqrt{3}}{4}a\,\hat{\mathbf{y}}$ | (6$g$) | $H_2O$ |
| $\mathbf{B}_9$ | = | $\frac{1}{2}\,\mathbf{a}_1 + \frac{1}{2}\,\mathbf{a}_2$ | = | $\frac{1}{2}a\,\hat{\mathbf{x}}$ | (6$g$) | $H_2O$ |
| $\mathbf{B}_{10}$ | = | $\frac{1}{2}\,\mathbf{a}_1 + \frac{1}{2}\,\mathbf{a}_3$ | = | $\frac{1}{4}a\,\hat{\mathbf{x}} - \frac{\sqrt{3}}{4}a\,\hat{\mathbf{y}} + \frac{1}{2}c\,\hat{\mathbf{z}}$ | (6$g$) | $H_2O$ |
| $\mathbf{B}_{11}$ | = | $\frac{1}{2}\,\mathbf{a}_2 + \frac{1}{2}\,\mathbf{a}_3$ | = | $\frac{1}{4}a\,\hat{\mathbf{x}} + \frac{\sqrt{3}}{4}a\,\hat{\mathbf{y}} + \frac{1}{2}c\,\hat{\mathbf{z}}$ | (6$g$) | $H_2O$ |
| $\mathbf{B}_{12}$ | = | $\frac{1}{2}\,\mathbf{a}_1 + \frac{1}{2}\,\mathbf{a}_2 + \frac{1}{2}\,\mathbf{a}_3$ | = | $\frac{1}{2}a\,\hat{\mathbf{x}} + \frac{1}{2}c\,\hat{\mathbf{z}}$ | (6$g$) | $H_2O$ |
| $\mathbf{B}_{13}$ | = | $x_4\,\mathbf{a}_1 + 2x_4\,\mathbf{a}_2 + \frac{1}{4}\,\mathbf{a}_3$ | = | $\frac{3}{2}x_4a\,\hat{\mathbf{x}} + \frac{\sqrt{3}}{2}x_4a\,\hat{\mathbf{y}} + \frac{1}{4}c\,\hat{\mathbf{z}}$ | (6$h$) | O I |
| $\mathbf{B}_{14}$ | = | $-2x_4\,\mathbf{a}_1 - x_4\,\mathbf{a}_2 + \frac{1}{4}\,\mathbf{a}_3$ | = | $-\frac{3}{2}x_4a\,\hat{\mathbf{x}} + \frac{\sqrt{3}}{2}x_4a\,\hat{\mathbf{y}} + \frac{1}{4}c\,\hat{\mathbf{z}}$ | (6$h$) | O I |
| $\mathbf{B}_{15}$ | = | $x_4\,\mathbf{a}_1 - x_4\,\mathbf{a}_2 + \frac{1}{4}\,\mathbf{a}_3$ | = | $-\sqrt{3}x_4a\,\hat{\mathbf{y}} + \frac{1}{4}c\,\hat{\mathbf{z}}$ | (6$h$) | O I |
| $\mathbf{B}_{16}$ | = | $-x_4\,\mathbf{a}_1 - 2x_4\,\mathbf{a}_2 + \frac{3}{4}\,\mathbf{a}_3$ | = | $-\frac{3}{2}x_4a\,\hat{\mathbf{x}} - \frac{\sqrt{3}}{2}x_4a\,\hat{\mathbf{y}} + \frac{3}{4}c\,\hat{\mathbf{z}}$ | (6$h$) | O I |
| $\mathbf{B}_{17}$ | = | $2x_4\,\mathbf{a}_1 + x_4\,\mathbf{a}_2 + \frac{3}{4}\,\mathbf{a}_3$ | = | $\frac{3}{2}x_4a\,\hat{\mathbf{x}} - \frac{\sqrt{3}}{2}x_4a\,\hat{\mathbf{y}} + \frac{3}{4}c\,\hat{\mathbf{z}}$ | (6$h$) | O I |
| $\mathbf{B}_{18}$ | = | $-x_4\,\mathbf{a}_1 + x_4\,\mathbf{a}_2 + \frac{3}{4}\,\mathbf{a}_3$ | = | $\sqrt{3}x_4a\,\hat{\mathbf{y}} + \frac{3}{4}c\,\hat{\mathbf{z}}$ | (6$h$) | O I |
| $\mathbf{B}_{19}$ | = | $x_5\,\mathbf{a}_1 + 2x_5\,\mathbf{a}_2 + \frac{1}{4}\,\mathbf{a}_3$ | = | $\frac{3}{2}x_5a\,\hat{\mathbf{x}} + \frac{\sqrt{3}}{2}x_5a\,\hat{\mathbf{y}} + \frac{1}{4}c\,\hat{\mathbf{z}}$ | (6$h$) | Si |
| $\mathbf{B}_{20}$ | = | $-2x_5\,\mathbf{a}_1 - x_5\,\mathbf{a}_2 + \frac{1}{4}\,\mathbf{a}_3$ | = | $-\frac{3}{2}x_5a\,\hat{\mathbf{x}} + \frac{\sqrt{3}}{2}x_5a\,\hat{\mathbf{y}} + \frac{1}{4}c\,\hat{\mathbf{z}}$ | (6$h$) | Si |
| $\mathbf{B}_{21}$ | = | $x_5\,\mathbf{a}_1 - x_5\,\mathbf{a}_2 + \frac{1}{4}\,\mathbf{a}_3$ | = | $-\sqrt{3}x_5a\,\hat{\mathbf{y}} + \frac{1}{4}c\,\hat{\mathbf{z}}$ | (6$h$) | Si |
| $\mathbf{B}_{22}$ | = | $-x_5\,\mathbf{a}_1 - 2x_5\,\mathbf{a}_2 + \frac{3}{4}\,\mathbf{a}_3$ | = | $-\frac{3}{2}x_5a\,\hat{\mathbf{x}} - \frac{\sqrt{3}}{2}x_5a\,\hat{\mathbf{y}} + \frac{3}{4}c\,\hat{\mathbf{z}}$ | (6$h$) | Si |
| $\mathbf{B}_{23}$ | = | $2x_5\,\mathbf{a}_1 + x_5\,\mathbf{a}_2 + \frac{3}{4}\,\mathbf{a}_3$ | = | $\frac{3}{2}x_5a\,\hat{\mathbf{x}} - \frac{\sqrt{3}}{2}x_5a\,\hat{\mathbf{y}} + \frac{3}{4}c\,\hat{\mathbf{z}}$ | (6$h$) | Si |
| $\mathbf{B}_{24}$ | = | $-x_5\,\mathbf{a}_1 + x_5\,\mathbf{a}_2 + \frac{3}{4}\,\mathbf{a}_3$ | = | $\sqrt{3}x_5a\,\hat{\mathbf{y}} + \frac{3}{4}c\,\hat{\mathbf{z}}$ | (6$h$) | Si |
| $\mathbf{B}_{25}$ | = | $x_6\,\mathbf{a}_1 + 2x_6\,\mathbf{a}_2 + z_6\,\mathbf{a}_3$ | = | $\frac{3}{2}x_6a\,\hat{\mathbf{x}} + \frac{\sqrt{3}}{2}x_6a\,\hat{\mathbf{y}} + z_6c\,\hat{\mathbf{z}}$ | (12$k$) | O II |
| $\mathbf{B}_{26}$ | = | $-2x_6\,\mathbf{a}_1 - x_6\,\mathbf{a}_2 + z_6\,\mathbf{a}_3$ | = | $-\frac{3}{2}x_6a\,\hat{\mathbf{x}} + \frac{\sqrt{3}}{2}x_6a\,\hat{\mathbf{y}} + z_6c\,\hat{\mathbf{z}}$ | (12$k$) | O II |
| $\mathbf{B}_{27}$ | = | $x_6\,\mathbf{a}_1 - x_6\,\mathbf{a}_2 + z_6\,\mathbf{a}_3$ | = | $-\sqrt{3}x_6a\,\hat{\mathbf{y}} + z_6c\,\hat{\mathbf{z}}$ | (12$k$) | O II |
| $\mathbf{B}_{28}$ | = | $-x_6\,\mathbf{a}_1 - 2x_6\,\mathbf{a}_2 + \left(\frac{1}{2} + z_6\right)\,\mathbf{a}_3$ | = | $-\frac{3}{2}x_6a\,\hat{\mathbf{x}} - \frac{\sqrt{3}}{2}x_6a\,\hat{\mathbf{y}} + \left(\frac{1}{2} + z_6\right)c\,\hat{\mathbf{z}}$ | (12$k$) | O II |
| $\mathbf{B}_{29}$ | = | $2x_6\,\mathbf{a}_1 + x_6\,\mathbf{a}_2 + \left(\frac{1}{2} + z_6\right)\,\mathbf{a}_3$ | = | $\frac{3}{2}x_6a\,\hat{\mathbf{x}} - \frac{\sqrt{3}}{2}x_6a\,\hat{\mathbf{y}} + \left(\frac{1}{2} + z_6\right)c\,\hat{\mathbf{z}}$ | (12$k$) | O II |
| $\mathbf{B}_{30}$ | = | $-x_6\,\mathbf{a}_1 + x_6\,\mathbf{a}_2 + \left(\frac{1}{2} + z_6\right)\,\mathbf{a}_3$ | = | $\sqrt{3}x_6a\,\hat{\mathbf{y}} + \left(\frac{1}{2} + z_6\right)c\,\hat{\mathbf{z}}$ | (12$k$) | O II |
| $\mathbf{B}_{31}$ | = | $2x_6\,\mathbf{a}_1 + x_6\,\mathbf{a}_2 - z_6\,\mathbf{a}_3$ | = | $\frac{3}{2}x_6a\,\hat{\mathbf{x}} - \frac{\sqrt{3}}{2}x_6a\,\hat{\mathbf{y}} - z_6c\,\hat{\mathbf{z}}$ | (12$k$) | O II |
| $\mathbf{B}_{32}$ | = | $-x_6\,\mathbf{a}_1 - 2x_6\,\mathbf{a}_2 - z_6\,\mathbf{a}_3$ | = | $-\frac{3}{2}x_6a\,\hat{\mathbf{x}} - \frac{\sqrt{3}}{2}x_6a\,\hat{\mathbf{y}} - z_6c\,\hat{\mathbf{z}}$ | (12$k$) | O II |
| $\mathbf{B}_{33}$ | = | $-x_6\,\mathbf{a}_1 + x_6\,\mathbf{a}_2 - z_6\,\mathbf{a}_3$ | = | $\sqrt{3}x_6a\,\hat{\mathbf{y}} - z_6c\,\hat{\mathbf{z}}$ | (12$k$) | O II |
| $\mathbf{B}_{34}$ | = | $-2x_6\,\mathbf{a}_1 - x_6\,\mathbf{a}_2 + \left(\frac{1}{2} - z_6\right)\,\mathbf{a}_3$ | = | $-\frac{3}{2}x_6a\,\hat{\mathbf{x}} + \frac{\sqrt{3}}{2}x_6a\,\hat{\mathbf{y}} + \left(\frac{1}{2} - z_6\right)c\,\hat{\mathbf{z}}$ | (12$k$) | O II |
| $\mathbf{B}_{35}$ | = | $x_6\,\mathbf{a}_1 + 2x_6\,\mathbf{a}_2 + \left(\frac{1}{2} - z_6\right)\,\mathbf{a}_3$ | = | $\frac{3}{2}x_6a\,\hat{\mathbf{x}} + \frac{\sqrt{3}}{2}x_6a\,\hat{\mathbf{y}} + \left(\frac{1}{2} - z_6\right)c\,\hat{\mathbf{z}}$ | (12$k$) | O II |
| $\mathbf{B}_{36}$ | = | $x_6\,\mathbf{a}_1 - x_6\,\mathbf{a}_2 + \left(\frac{1}{2} - z_6\right)\,\mathbf{a}_3$ | = | $-\sqrt{3}x_6a\,\hat{\mathbf{y}} + \left(\frac{1}{2} - z_6\right)c\,\hat{\mathbf{z}}$ | (12$k$) | O II |

**Found in:**

**Geometry files:**

- CIF: pp. 1767
- POSCAR: pp. 1768

# $ReB_3$ Structure: A3B_hP8_194_af_c

http://aflow.org/prototype-encyclopedia/A3B_hP8_194_af_c

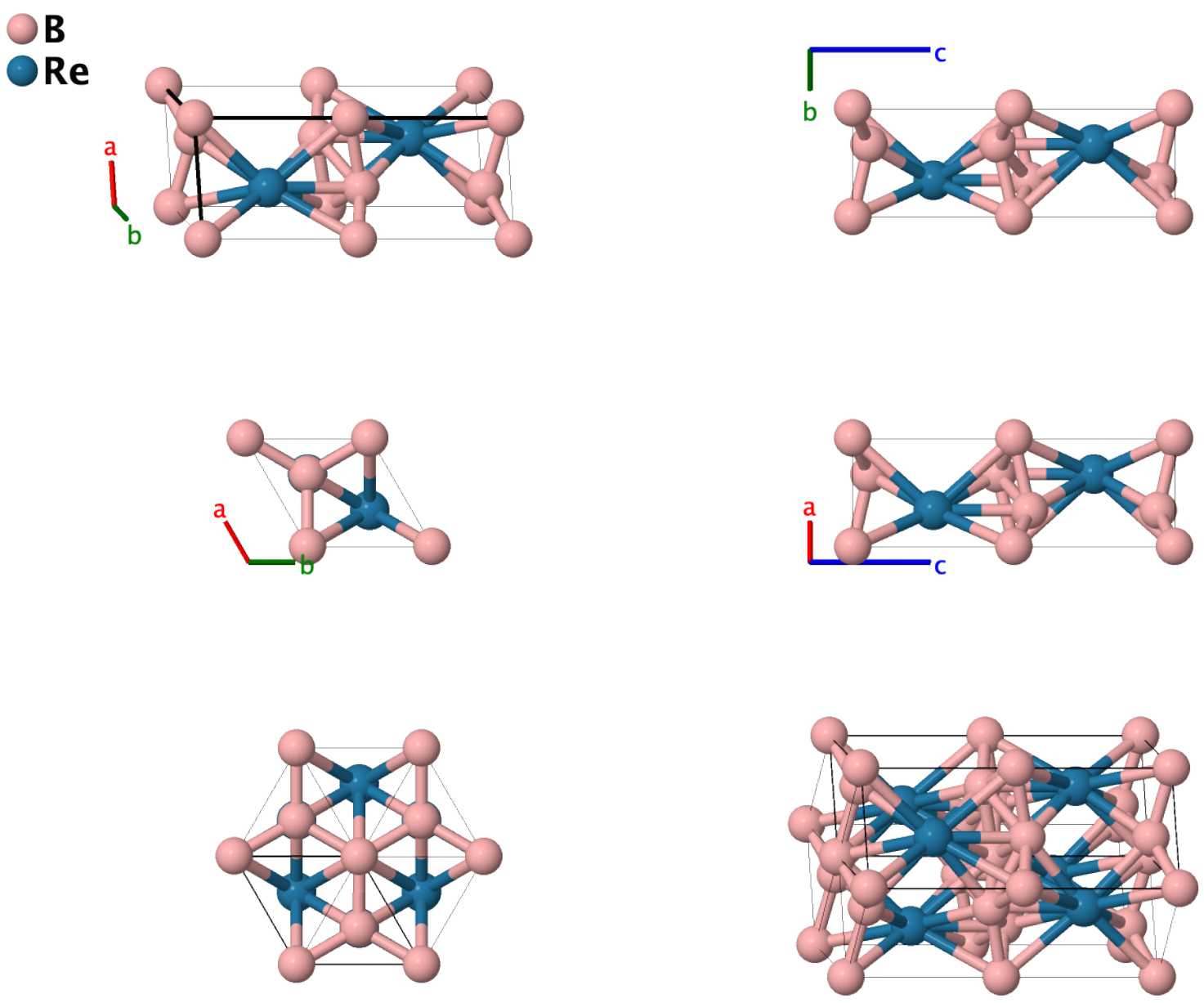

| | | |
|---|---|---|
| **Prototype** | : | $B_3Re$ |
| **AFLOW prototype label** | : | A3B_hP8_194_af_c |
| ***Strukturbericht* designation** | : | None |
| **Pearson symbol** | : | hP8 |
| **Space group number** | : | 194 |
| **Space group symbol** | : | $P6_3/mmc$ |
| **AFLOW prototype command** | : | `aflow --proto=A3B_hP8_194_af_c`<br>`--params=`$a, c/a, z_3$ |

### Other compounds with this structure

- TcB3, $CaNi_2Si$, and $GdPt_2Sn$

- The lattice constants $a$ and $c$ were inferred from the nearest-neighbor distances in (Aronsson, 1960).

### Hexagonal primitive vectors:

$$\mathbf{a}_1 = \frac{1}{2}\,a\,\hat{\mathbf{x}} - \frac{\sqrt{3}}{2}\,a\,\hat{\mathbf{y}}$$
$$\mathbf{a}_2 = \frac{1}{2}\,a\,\hat{\mathbf{x}} + \frac{\sqrt{3}}{2}\,a\,\hat{\mathbf{y}}$$
$$\mathbf{a}_3 = c\,\hat{\mathbf{z}}$$

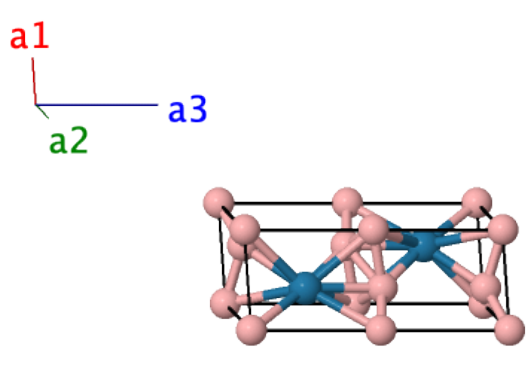

### Basis vectors:

| | | Lattice Coordinates | | Cartesian Coordinates | Wyckoff Position | Atom Type |
|---|---|---|---|---|---|---|
| $\mathbf{B}_1$ | = | $0\,\mathbf{a}_1 + 0\,\mathbf{a}_2 + 0\,\mathbf{a}_3$ | = | $0\,\hat{\mathbf{x}} + 0\,\hat{\mathbf{y}} + 0\,\hat{\mathbf{z}}$ | $(2a)$ | B I |
| $\mathbf{B}_2$ | = | $\frac{1}{2}\,\mathbf{a}_3$ | = | $\frac{1}{2}c\,\hat{\mathbf{z}}$ | $(2a)$ | B I |
| $\mathbf{B}_3$ | = | $\frac{1}{3}\,\mathbf{a}_1 + \frac{2}{3}\,\mathbf{a}_2 + \frac{1}{4}\,\mathbf{a}_3$ | = | $\frac{1}{2}a\,\mathbf{\hat{x}} + \frac{1}{2\sqrt{3}}a\,\mathbf{\hat{y}} + \frac{1}{4}c\,\mathbf{\hat{z}}$ | (2$c$) | Re |
| $\mathbf{B}_4$ | = | $\frac{2}{3}\,\mathbf{a}_1 + \frac{1}{3}\,\mathbf{a}_2 + \frac{3}{4}\,\mathbf{a}_3$ | = | $\frac{1}{2}a\,\mathbf{\hat{x}} - \frac{1}{2\sqrt{3}}a\,\mathbf{\hat{y}} + \frac{3}{4}c\,\mathbf{\hat{z}}$ | (2$c$) | Re |
| $\mathbf{B}_5$ | = | $\frac{1}{3}\,\mathbf{a}_1 + \frac{2}{3}\,\mathbf{a}_2 + z_3\,\mathbf{a}_3$ | = | $\frac{1}{2}a\,\mathbf{\hat{x}} + \frac{1}{2\sqrt{3}}a\,\mathbf{\hat{y}} + z_3c\,\mathbf{\hat{z}}$ | (4$f$) | B II |
| $\mathbf{B}_6$ | = | $\frac{2}{3}\,\mathbf{a}_1 + \frac{1}{3}\,\mathbf{a}_2 + \left(\frac{1}{2} + z_3\right)\,\mathbf{a}_3$ | = | $\frac{1}{2}a\,\mathbf{\hat{x}} - \frac{1}{2\sqrt{3}}a\,\mathbf{\hat{y}} + \left(\frac{1}{2} + z_3\right)c\,\mathbf{\hat{z}}$ | (4$f$) | B II |
| $\mathbf{B}_7$ | = | $\frac{2}{3}\,\mathbf{a}_1 + \frac{1}{3}\,\mathbf{a}_2 - z_3\,\mathbf{a}_3$ | = | $\frac{1}{2}a\,\mathbf{\hat{x}} - \frac{1}{2\sqrt{3}}a\,\mathbf{\hat{y}} - z_3c\,\mathbf{\hat{z}}$ | (4$f$) | B II |
| $\mathbf{B}_8$ | = | $\frac{1}{3}\,\mathbf{a}_1 + \frac{2}{3}\,\mathbf{a}_2 + \left(\frac{1}{2} - z_3\right)\,\mathbf{a}_3$ | = | $\frac{1}{2}a\,\mathbf{\hat{x}} + \frac{1}{2\sqrt{3}}a\,\mathbf{\hat{y}} + \left(\frac{1}{2} - z_3\right)c\,\mathbf{\hat{z}}$ | (4$f$) | B II |

---

**Found in:**

---

**Geometry files:**

- CIF: pp. 1768
- POSCAR: pp. 1768

# $Cs_3Cr_2Cl_9$ Structure: A9B2C3_hP28_194_hk_f_bf

http://aflow.org/prototype-encyclopedia/A9B2C3_hP28_194_hk_f_bf

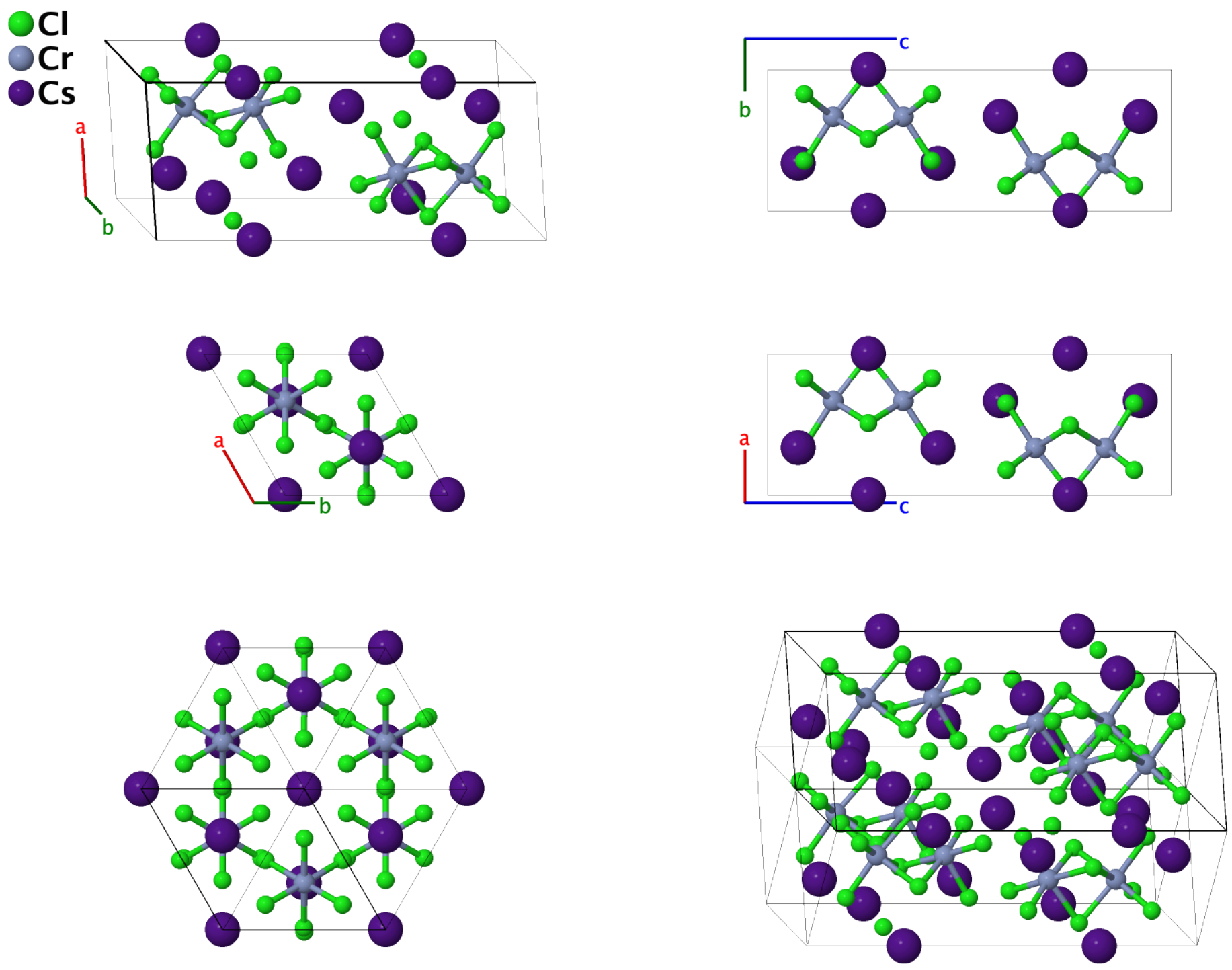

| | | |
|---|---|---|
| **Prototype** | : | $Cl_9Cr_2Cs_3$ |
| **AFLOW prototype label** | : | A9B2C3_hP28_194_hk_f_bf |
| ***Strukturbericht* designation** | : | None |
| **Pearson symbol** | : | hP28 |
| **Space group number** | : | 194 |
| **Space group symbol** | : | $P6_3/mmc$ |
| **AFLOW prototype command** | : | `aflow --proto=A9B2C3_hP28_194_hk_f_bf` `--params=`$a, c/a, z_2, z_3, x_4, x_5, z_5$ |

**Hexagonal primitive vectors:**

$$\mathbf{a}_1 = \frac{1}{2}\,a\,\hat{\mathbf{x}} - \frac{\sqrt{3}}{2}\,a\,\hat{\mathbf{y}}$$
$$\mathbf{a}_2 = \frac{1}{2}\,a\,\hat{\mathbf{x}} + \frac{\sqrt{3}}{2}\,a\,\hat{\mathbf{y}}$$
$$\mathbf{a}_3 = c\,\hat{\mathbf{z}}$$

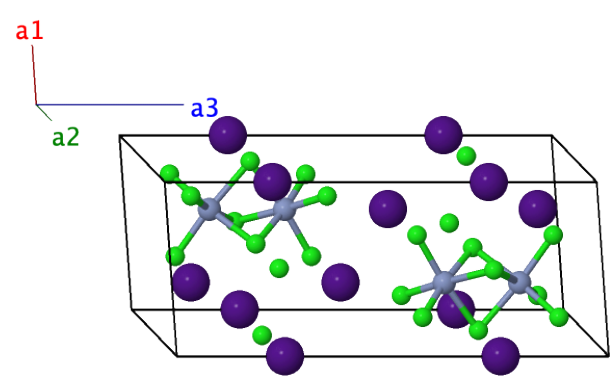

**Basis vectors:**

| | | Lattice Coordinates | | Cartesian Coordinates | Wyckoff Position | Atom Type |
|---|---|---|---|---|---|---|
| $\mathbf{B}_1$ | = | $\frac{1}{4}\,\mathbf{a}_3$ | = | $\frac{1}{4}c\,\hat{\mathbf{z}}$ | (2$b$) | Cs I |
| $\mathbf{B}_2$ | = | $\frac{3}{4}\,\mathbf{a}_3$ | = | $\frac{3}{4}c\,\hat{\mathbf{z}}$ | (2$b$) | Cs I |
| $\mathbf{B}_3$ | = | $\frac{1}{3}\,\mathbf{a}_1 + \frac{2}{3}\,\mathbf{a}_2 + z_2\,\mathbf{a}_3$ | = | $\frac{1}{2}a\,\hat{\mathbf{x}} + \frac{1}{2\sqrt{3}}a\,\hat{\mathbf{y}} + z_2 c\,\hat{\mathbf{z}}$ | (4$f$) | Cr |
| $\mathbf{B}_4$ | = | $\frac{2}{3}\,\mathbf{a}_1 + \frac{1}{3}\,\mathbf{a}_2 + \left(\frac{1}{2} + z_2\right)\,\mathbf{a}_3$ | = | $\frac{1}{2}a\,\hat{\mathbf{x}} - \frac{1}{2\sqrt{3}}a\,\hat{\mathbf{y}} + \left(\frac{1}{2} + z_2\right)c\,\hat{\mathbf{z}}$ | (4$f$) | Cr |
| $\mathbf{B}_{5}$ | = | $\frac{2}{3}\,\mathbf{a}_{1} + \frac{1}{3}\,\mathbf{a}_{2} - z_{2}\,\mathbf{a}_{3}$ | = | $\frac{1}{2}a\,\mathbf{\hat{x}} - \frac{1}{2\sqrt{3}}a\,\mathbf{\hat{y}} - z_{2}c\,\mathbf{\hat{z}}$ | $(4f)$ | Cr |
| $\mathbf{B}_{6}$ | = | $\frac{1}{3}\,\mathbf{a}_{1} + \frac{2}{3}\,\mathbf{a}_{2} + \left(\frac{1}{2} - z_{2}\right)\,\mathbf{a}_{3}$ | = | $\frac{1}{2}a\,\mathbf{\hat{x}} + \frac{1}{2\sqrt{3}}a\,\mathbf{\hat{y}} + \left(\frac{1}{2} - z_{2}\right)c\,\mathbf{\hat{z}}$ | $(4f)$ | Cr |
| $\mathbf{B}_{7}$ | = | $\frac{1}{3}\,\mathbf{a}_{1} + \frac{2}{3}\,\mathbf{a}_{2} + z_{3}\,\mathbf{a}_{3}$ | = | $\frac{1}{2}a\,\mathbf{\hat{x}} + \frac{1}{2\sqrt{3}}a\,\mathbf{\hat{y}} + z_{3}c\,\mathbf{\hat{z}}$ | $(4f)$ | Cs II |
| $\mathbf{B}_{8}$ | = | $\frac{2}{3}\,\mathbf{a}_{1} + \frac{1}{3}\,\mathbf{a}_{2} + \left(\frac{1}{2} + z_{3}\right)\,\mathbf{a}_{3}$ | = | $\frac{1}{2}a\,\mathbf{\hat{x}} - \frac{1}{2\sqrt{3}}a\,\mathbf{\hat{y}} + \left(\frac{1}{2} + z_{3}\right)c\,\mathbf{\hat{z}}$ | $(4f)$ | Cs II |
| $\mathbf{B}_{9}$ | = | $\frac{2}{3}\,\mathbf{a}_{1} + \frac{1}{3}\,\mathbf{a}_{2} - z_{3}\,\mathbf{a}_{3}$ | = | $\frac{1}{2}a\,\mathbf{\hat{x}} - \frac{1}{2\sqrt{3}}a\,\mathbf{\hat{y}} - z_{3}c\,\mathbf{\hat{z}}$ | $(4f)$ | Cs II |
| $\mathbf{B}_{10}$ | = | $\frac{1}{3}\,\mathbf{a}_{1} + \frac{2}{3}\,\mathbf{a}_{2} + \left(\frac{1}{2} - z_{3}\right)\,\mathbf{a}_{3}$ | = | $\frac{1}{2}a\,\mathbf{\hat{x}} + \frac{1}{2\sqrt{3}}a\,\mathbf{\hat{y}} + \left(\frac{1}{2} - z_{3}\right)c\,\mathbf{\hat{z}}$ | $(4f)$ | Cs II |
| $\mathbf{B}_{11}$ | = | $x_{4}\,\mathbf{a}_{1} + 2x_{4}\,\mathbf{a}_{2} + \frac{1}{4}\,\mathbf{a}_{3}$ | = | $\frac{3}{2}x_{4}a\,\mathbf{\hat{x}} + \frac{\sqrt{3}}{2}x_{4}a\,\mathbf{\hat{y}} + \frac{1}{4}c\,\mathbf{\hat{z}}$ | $(6h)$ | Cl I |
| $\mathbf{B}_{12}$ | = | $-2x_{4}\,\mathbf{a}_{1} - x_{4}\,\mathbf{a}_{2} + \frac{1}{4}\,\mathbf{a}_{3}$ | = | $-\frac{3}{2}x_{4}a\,\mathbf{\hat{x}} + \frac{\sqrt{3}}{2}x_{4}a\,\mathbf{\hat{y}} + \frac{1}{4}c\,\mathbf{\hat{z}}$ | $(6h)$ | Cl I |
| $\mathbf{B}_{13}$ | = | $x_{4}\,\mathbf{a}_{1} - x_{4}\,\mathbf{a}_{2} + \frac{1}{4}\,\mathbf{a}_{3}$ | = | $-\sqrt{3}x_{4}a\,\mathbf{\hat{y}} + \frac{1}{4}c\,\mathbf{\hat{z}}$ | $(6h)$ | Cl I |
| $\mathbf{B}_{14}$ | = | $-x_{4}\,\mathbf{a}_{1} - 2x_{4}\,\mathbf{a}_{2} + \frac{3}{4}\,\mathbf{a}_{3}$ | = | $-\frac{3}{2}x_{4}a\,\mathbf{\hat{x}} - \frac{\sqrt{3}}{2}x_{4}a\,\mathbf{\hat{y}} + \frac{3}{4}c\,\mathbf{\hat{z}}$ | $(6h)$ | Cl I |
| $\mathbf{B}_{15}$ | = | $2x_{4}\,\mathbf{a}_{1} + x_{4}\,\mathbf{a}_{2} + \frac{3}{4}\,\mathbf{a}_{3}$ | = | $\frac{3}{2}x_{4}a\,\mathbf{\hat{x}} - \frac{\sqrt{3}}{2}x_{4}a\,\mathbf{\hat{y}} + \frac{3}{4}c\,\mathbf{\hat{z}}$ | $(6h)$ | Cl I |
| $\mathbf{B}_{16}$ | = | $-x_{4}\,\mathbf{a}_{1} + x_{4}\,\mathbf{a}_{2} + \frac{3}{4}\,\mathbf{a}_{3}$ | = | $\sqrt{3}x_{4}a\,\mathbf{\hat{y}} + \frac{3}{4}c\,\mathbf{\hat{z}}$ | $(6h)$ | Cl I |
| $\mathbf{B}_{17}$ | = | $x_{5}\,\mathbf{a}_{1} + 2x_{5}\,\mathbf{a}_{2} + z_{5}\,\mathbf{a}_{3}$ | = | $\frac{3}{2}x_{5}a\,\mathbf{\hat{x}} + \frac{\sqrt{3}}{2}x_{5}a\,\mathbf{\hat{y}} + z_{5}c\,\mathbf{\hat{z}}$ | $(12k)$ | Cl II |
| $\mathbf{B}_{18}$ | = | $-2x_{5}\,\mathbf{a}_{1} - x_{5}\,\mathbf{a}_{2} + z_{5}\,\mathbf{a}_{3}$ | = | $-\frac{3}{2}x_{5}a\,\mathbf{\hat{x}} + \frac{\sqrt{3}}{2}x_{5}a\,\mathbf{\hat{y}} + z_{5}c\,\mathbf{\hat{z}}$ | $(12k)$ | Cl II |
| $\mathbf{B}_{19}$ | = | $x_{5}\,\mathbf{a}_{1} - x_{5}\,\mathbf{a}_{2} + z_{5}\,\mathbf{a}_{3}$ | = | $-\sqrt{3}x_{5}a\,\mathbf{\hat{y}} + z_{5}c\,\mathbf{\hat{z}}$ | $(12k)$ | Cl II |
| $\mathbf{B}_{20}$ | = | $-x_{5}\,\mathbf{a}_{1} - 2x_{5}\,\mathbf{a}_{2} + \left(\frac{1}{2} + z_{5}\right)\,\mathbf{a}_{3}$ | = | $-\frac{3}{2}x_{5}a\,\mathbf{\hat{x}} - \frac{\sqrt{3}}{2}x_{5}a\,\mathbf{\hat{y}} + \left(\frac{1}{2} + z_{5}\right)c\,\mathbf{\hat{z}}$ | $(12k)$ | Cl II |
| $\mathbf{B}_{21}$ | = | $2x_{5}\,\mathbf{a}_{1} + x_{5}\,\mathbf{a}_{2} + \left(\frac{1}{2} + z_{5}\right)\,\mathbf{a}_{3}$ | = | $\frac{3}{2}x_{5}a\,\mathbf{\hat{x}} - \frac{\sqrt{3}}{2}x_{5}a\,\mathbf{\hat{y}} + \left(\frac{1}{2} + z_{5}\right)c\,\mathbf{\hat{z}}$ | $(12k)$ | Cl II |
| $\mathbf{B}_{22}$ | = | $-x_{5}\,\mathbf{a}_{1} + x_{5}\,\mathbf{a}_{2} + \left(\frac{1}{2} + z_{5}\right)\,\mathbf{a}_{3}$ | = | $\sqrt{3}x_{5}a\,\mathbf{\hat{y}} + \left(\frac{1}{2} + z_{5}\right)c\,\mathbf{\hat{z}}$ | $(12k)$ | Cl II |
| $\mathbf{B}_{23}$ | = | $2x_{5}\,\mathbf{a}_{1} + x_{5}\,\mathbf{a}_{2} - z_{5}\,\mathbf{a}_{3}$ | = | $\frac{3}{2}x_{5}a\,\mathbf{\hat{x}} - \frac{\sqrt{3}}{2}x_{5}a\,\mathbf{\hat{y}} - z_{5}c\,\mathbf{\hat{z}}$ | $(12k)$ | Cl II |
| $\mathbf{B}_{24}$ | = | $-x_{5}\,\mathbf{a}_{1} - 2x_{5}\,\mathbf{a}_{2} - z_{5}\,\mathbf{a}_{3}$ | = | $-\frac{3}{2}x_{5}a\,\mathbf{\hat{x}} - \frac{\sqrt{3}}{2}x_{5}a\,\mathbf{\hat{y}} - z_{5}c\,\mathbf{\hat{z}}$ | $(12k)$ | Cl II |
| $\mathbf{B}_{25}$ | = | $-x_{5}\,\mathbf{a}_{1} + x_{5}\,\mathbf{a}_{2} - z_{5}\,\mathbf{a}_{3}$ | = | $\sqrt{3}x_{5}a\,\mathbf{\hat{y}} - z_{5}c\,\mathbf{\hat{z}}$ | $(12k)$ | Cl II |
| $\mathbf{B}_{26}$ | = | $-2x_{5}\,\mathbf{a}_{1} - x_{5}\,\mathbf{a}_{2} + \left(\frac{1}{2} - z_{5}\right)\,\mathbf{a}_{3}$ | = | $-\frac{3}{2}x_{5}a\,\mathbf{\hat{x}} + \frac{\sqrt{3}}{2}x_{5}a\,\mathbf{\hat{y}} + \left(\frac{1}{2} - z_{5}\right)c\,\mathbf{\hat{z}}$ | $(12k)$ | Cl II |
| $\mathbf{B}_{27}$ | = | $x_{5}\,\mathbf{a}_{1} + 2x_{5}\,\mathbf{a}_{2} + \left(\frac{1}{2} - z_{5}\right)\,\mathbf{a}_{3}$ | = | $\frac{3}{2}x_{5}a\,\mathbf{\hat{x}} + \frac{\sqrt{3}}{2}x_{5}a\,\mathbf{\hat{y}} + \left(\frac{1}{2} - z_{5}\right)c\,\mathbf{\hat{z}}$ | $(12k)$ | Cl II |
| $\mathbf{B}_{28}$ | = | $x_{5}\,\mathbf{a}_{1} - x_{5}\,\mathbf{a}_{2} + \left(\frac{1}{2} - z_{5}\right)\,\mathbf{a}_{3}$ | = | $-\sqrt{3}x_{5}a\,\mathbf{\hat{y}} + \left(\frac{1}{2} - z_{5}\right)c\,\mathbf{\hat{z}}$ | $(12k)$ | Cl II |

**Geometry files:**

- CIF: pp. 1768
- POSCAR: pp. 1769

# $Na_{0.74}CoO_2$ Structure: AB2C2_hP10_194_a_bc_f

http://aflow.org/prototype-encyclopedia/AB2C2_hP10_194_a_bc_f

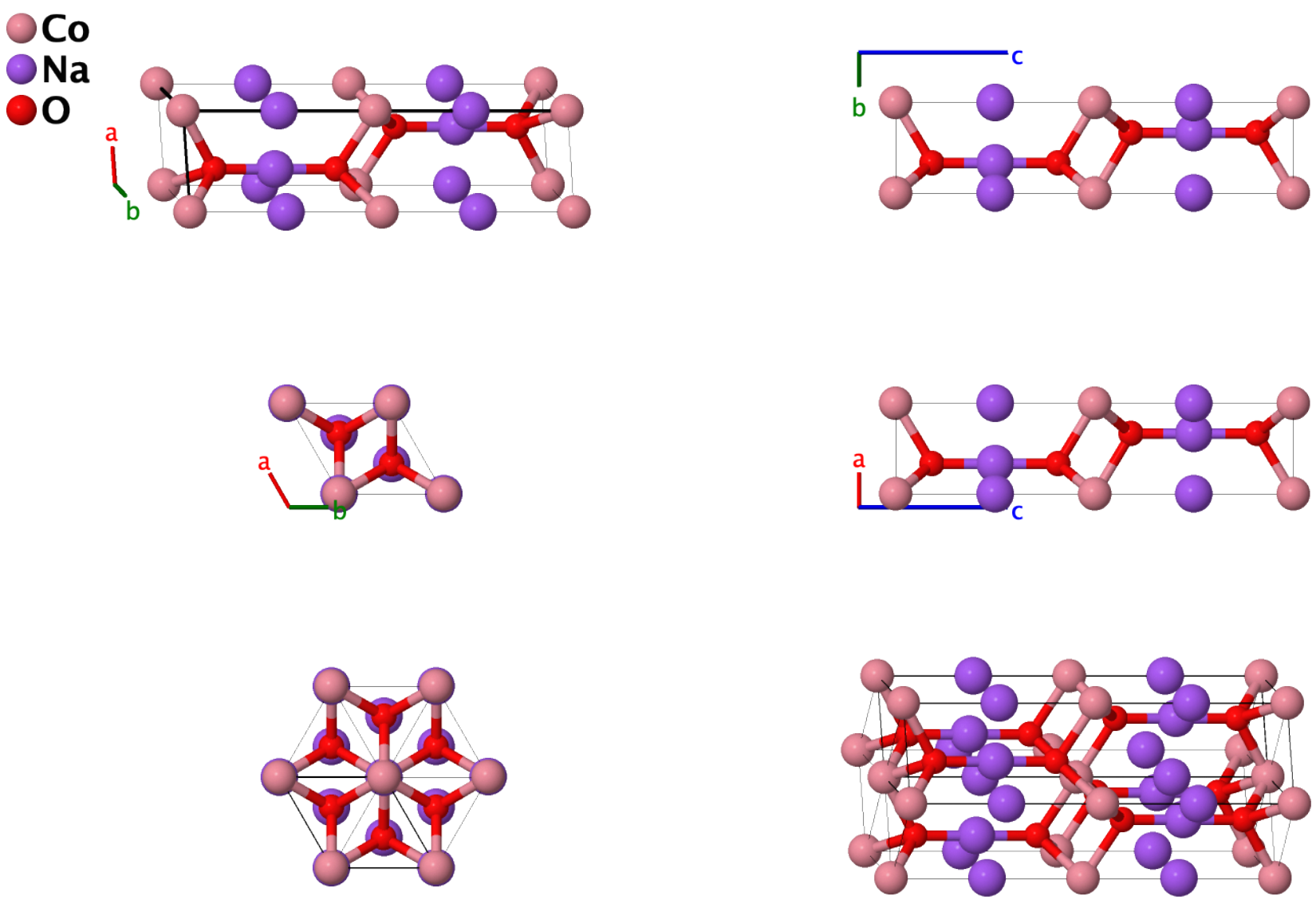

| | | |
|---|---|---|
| **Prototype** | : | $CoNa_{0.74}O_2$ |
| **AFLOW prototype label** | : | AB2C2_hP10_194_a_bc_f |
| ***Strukturbericht* designation** | : | None |
| **Pearson symbol** | : | hP10 |
| **Space group number** | : | 194 |
| **Space group symbol** | : | $P6_3/mmc$ |
| **AFLOW prototype command** | : | `aflow --proto=AB2C2_hP10_194_a_bc_f --params=`$a, c/a, z_4$ |

- $Na_{0.74}CoO_2$ is a high figure-of-merit thermoelectric (Sk, 2019). The sodium sites are only partial filled, with the Na-I (2b) site having 21% occupancy while the Na-II (2c) site is at 51%.

**Hexagonal primitive vectors:**

$$\mathbf{a}_1 = \frac{1}{2}\,a\,\hat{\mathbf{x}} - \frac{\sqrt{3}}{2}\,a\,\hat{\mathbf{y}}$$
$$\mathbf{a}_2 = \frac{1}{2}\,a\,\hat{\mathbf{x}} + \frac{\sqrt{3}}{2}\,a\,\hat{\mathbf{y}}$$
$$\mathbf{a}_3 = c\,\hat{\mathbf{z}}$$

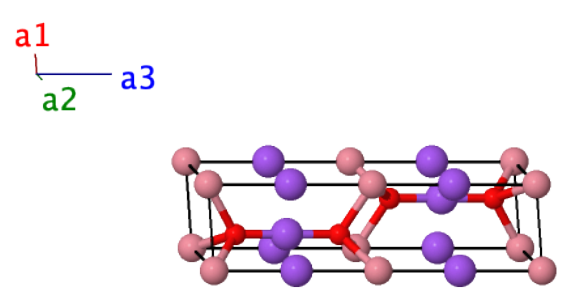

**Basis vectors:**

| | | Lattice Coordinates | | Cartesian Coordinates | Wyckoff Position | Atom Type |
|---|---|---|---|---|---|---|
| $\mathbf{B}_1$ | = | $0\,\mathbf{a}_1 + 0\,\mathbf{a}_2 + 0\,\mathbf{a}_3$ | = | $0\,\hat{\mathbf{x}} + 0\,\hat{\mathbf{y}} + 0\,\hat{\mathbf{z}}$ | (2$a$) | Co |
| $\mathbf{B}_2$ | = | $\frac{1}{2}\,\mathbf{a}_3$ | = | $\frac{1}{2}c\,\hat{\mathbf{z}}$ | (2$a$) | Co |
| $\mathbf{B}_3$ | = | $\frac{1}{4}\,\mathbf{a}_3$ | = | $\frac{1}{4}c\,\hat{\mathbf{z}}$ | (2$b$) | Na I |
| $\mathbf{B}_4$ | = | $\frac{3}{4}\,\mathbf{a}_3$ | = | $\frac{3}{4}c\,\hat{\mathbf{z}}$ | (2$b$) | Na I |
| $\mathbf{B}_5$ | = | $\frac{1}{3}\,\mathbf{a}_1 + \frac{2}{3}\,\mathbf{a}_2 + \frac{1}{4}\,\mathbf{a}_3$ | = | $\frac{1}{2}a\,\hat{\mathbf{x}} + \frac{1}{2\sqrt{3}}a\,\hat{\mathbf{y}} + \frac{1}{4}c\,\hat{\mathbf{z}}$ | (2$c$) | Na II |
| $\mathbf{B}_6$ | = | $\frac{2}{3}\,\mathbf{a}_1 + \frac{1}{3}\,\mathbf{a}_2 + \frac{3}{4}\,\mathbf{a}_3$ | = | $\frac{1}{2}a\,\hat{\mathbf{x}} - \frac{1}{2\sqrt{3}}a\,\hat{\mathbf{y}} + \frac{3}{4}c\,\hat{\mathbf{z}}$ | (2$c$) | Na II |
| $\mathbf{B}_7$ | = | $\frac{1}{3}\,\mathbf{a}_1 + \frac{2}{3}\,\mathbf{a}_2 + z_4\,\mathbf{a}_3$ | = | $\frac{1}{2}a\,\hat{\mathbf{x}} + \frac{1}{2\sqrt{3}}a\,\hat{\mathbf{y}} + z_4c\,\hat{\mathbf{z}}$ | (4$f$) | O |
| $\mathbf{B}_8$ | = | $\frac{2}{3}\,\mathbf{a}_1 + \frac{1}{3}\,\mathbf{a}_2 + \left(\frac{1}{2} + z_4\right)\,\mathbf{a}_3$ | = | $\frac{1}{2}a\,\hat{\mathbf{x}} - \frac{1}{2\sqrt{3}}a\,\hat{\mathbf{y}} + \left(\frac{1}{2} + z_4\right)c\,\hat{\mathbf{z}}$ | (4$f$) | O |
| $\mathbf{B}_9$ | = | $\frac{2}{3}\,\mathbf{a}_1 + \frac{1}{3}\,\mathbf{a}_2 - z_4\,\mathbf{a}_3$ | = | $\frac{1}{2}a\,\hat{\mathbf{x}} - \frac{1}{2\sqrt{3}}a\,\hat{\mathbf{y}} - z_4c\,\hat{\mathbf{z}}$ | (4$f$) | O |
| $\mathbf{B}_{10}$ | = | $\frac{1}{3}\,\mathbf{a}_1 + \frac{2}{3}\,\mathbf{a}_2 + \left(\frac{1}{2} - z_4\right)\,\mathbf{a}_3$ | = | $\frac{1}{2}a\,\hat{\mathbf{x}} + \frac{1}{2\sqrt{3}}a\,\hat{\mathbf{y}} + \left(\frac{1}{2} - z_4\right)c\,\hat{\mathbf{z}}$ | (4$f$) | O |

---

**Found in:**

---

**Geometry files:**

- CIF: pp. 1769
- POSCAR: pp. 1770

# $EuIn_2P_2$ Structure: AB2C2_hP10_194_a_f_f

http://aflow.org/prototype-encyclopedia/AB2C2_hP10_194_a_f_f

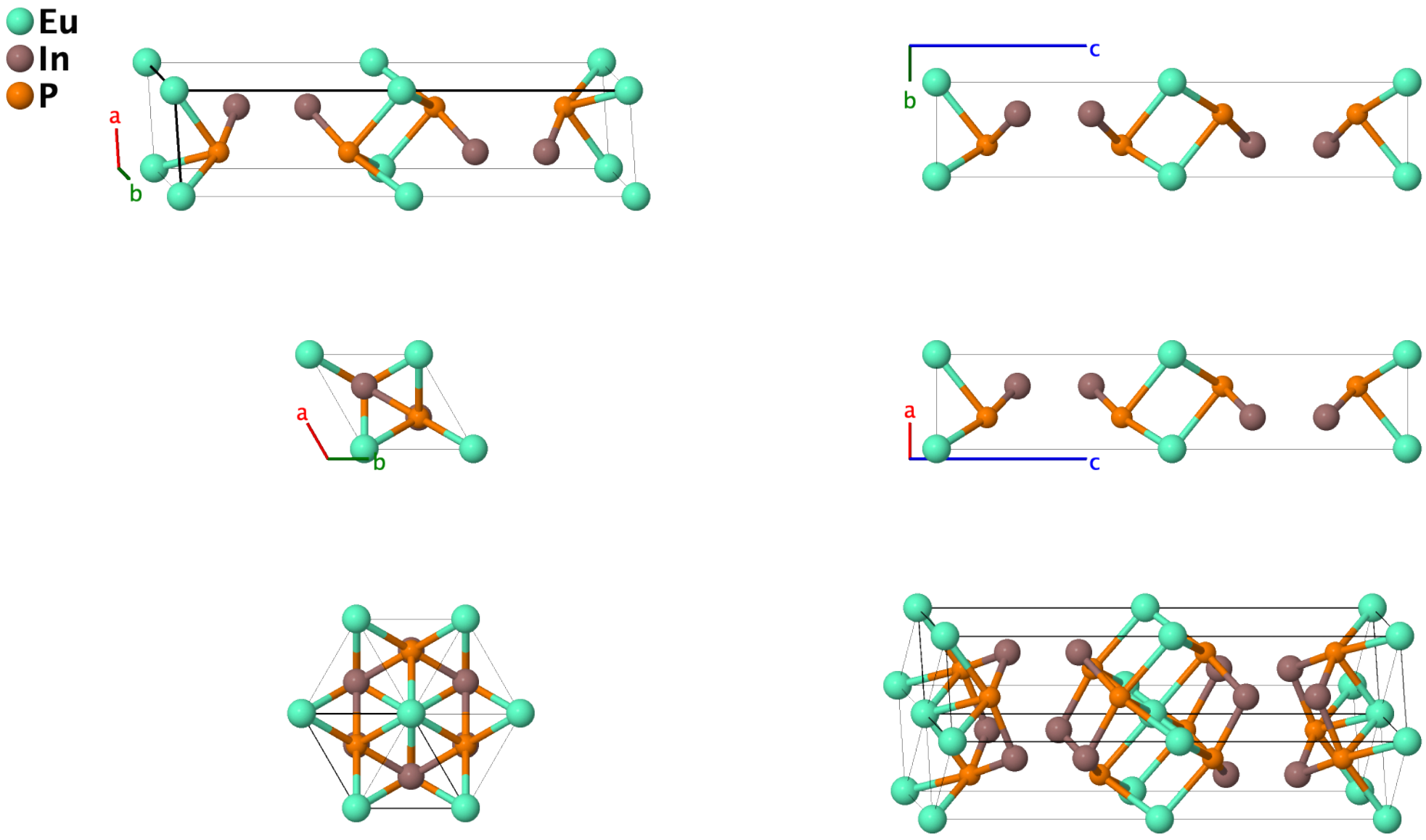

| | | |
|---|---|---|
| **Prototype** | : | $EuIn_2P_2$ |
| **AFLOW prototype label** | : | AB2C2_hP10_194_a_f_f |
| ***Strukturbericht* designation** | : | None |
| **Pearson symbol** | : | hP10 |
| **Space group number** | : | 194 |
| **Space group symbol** | : | $P6_3/mmc$ |
| **AFLOW prototype command** | : | `aflow --proto=AB2C2_hP10_194_a_f_f` <br> `--params=`$a, c/a, z_2, z_3$ |

### Other compounds with this structure

- $EuIn_2As_2$

### Hexagonal primitive vectors:

$$
\begin{aligned}
\mathbf{a}_1 &= \tfrac{1}{2}\,a\,\hat{\mathbf{x}} - \tfrac{\sqrt{3}}{2}\,a\,\hat{\mathbf{y}} \\
\mathbf{a}_2 &= \tfrac{1}{2}\,a\,\hat{\mathbf{x}} + \tfrac{\sqrt{3}}{2}\,a\,\hat{\mathbf{y}} \\
\mathbf{a}_3 &= c\,\hat{\mathbf{z}}
\end{aligned}
$$

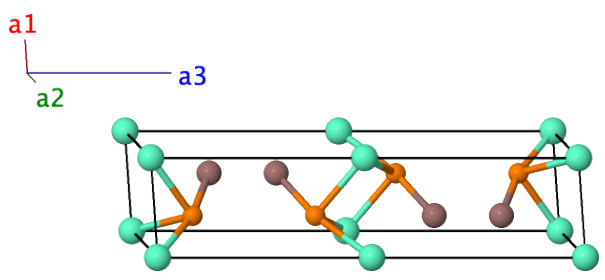

### Basis vectors:

| | | Lattice Coordinates | | Cartesian Coordinates | Wyckoff Position | Atom Type |
|---|---|---|---|---|---|---|
| $\mathbf{B}_1$ | = | $0\,\mathbf{a}_1 + 0\,\mathbf{a}_2 + 0\,\mathbf{a}_3$ | = | $0\,\hat{\mathbf{x}} + 0\,\hat{\mathbf{y}} + 0\,\hat{\mathbf{z}}$ | (2$a$) | Eu |
| $\mathbf{B}_2$ | = | $\frac{1}{2}\,\mathbf{a}_3$ | = | $\frac{1}{2}c\,\hat{\mathbf{z}}$ | (2$a$) | Eu |
| $\mathbf{B}_3$ | = | $\frac{1}{3}\,\mathbf{a}_1 + \frac{2}{3}\,\mathbf{a}_2 + z_2\,\mathbf{a}_3$ | = | $\frac{1}{2}a\,\hat{\mathbf{x}} + \frac{1}{2\sqrt{3}}a\,\hat{\mathbf{y}} + z_2c\,\hat{\mathbf{z}}$ | (4$f$) | In |
| $\mathbf{B}_4$ | = | $\frac{2}{3}\,\mathbf{a}_1 + \frac{1}{3}\,\mathbf{a}_2 + \left(\frac{1}{2} + z_2\right)\,\mathbf{a}_3$ | = | $\frac{1}{2}a\,\hat{\mathbf{x}} - \frac{1}{2\sqrt{3}}a\,\hat{\mathbf{y}} + \left(\frac{1}{2} + z_2\right)c\,\hat{\mathbf{z}}$ | (4$f$) | In |
| $\mathbf{B}_5$ | = | $\frac{2}{3}\,\mathbf{a}_1 + \frac{1}{3}\,\mathbf{a}_2 - z_2\,\mathbf{a}_3$ | = | $\frac{1}{2}a\,\hat{\mathbf{x}} - \frac{1}{2\sqrt{3}}a\,\hat{\mathbf{y}} - z_2c\,\hat{\mathbf{z}}$ | (4$f$) | In |
| $\mathbf{B}_6$ | = | $\frac{1}{3}\,\mathbf{a}_1 + \frac{2}{3}\,\mathbf{a}_2 + \left(\frac{1}{2} - z_2\right)\,\mathbf{a}_3$ | = | $\frac{1}{2}a\,\hat{\mathbf{x}} + \frac{1}{2\sqrt{3}}a\,\hat{\mathbf{y}} + \left(\frac{1}{2} - z_2\right)c\,\hat{\mathbf{z}}$ | (4$f$) | In |
| $\mathbf{B}_7$ | = | $\frac{1}{3}\,\mathbf{a}_1 + \frac{2}{3}\,\mathbf{a}_2 + z_3\,\mathbf{a}_3$ | = | $\frac{1}{2}a\,\hat{\mathbf{x}} + \frac{1}{2\sqrt{3}}a\,\hat{\mathbf{y}} + z_3c\,\hat{\mathbf{z}}$ | (4$f$) | P |
| $\mathbf{B}_8$ | = | $\frac{2}{3}\,\mathbf{a}_1 + \frac{1}{3}\,\mathbf{a}_2 + \left(\frac{1}{2} + z_3\right)\,\mathbf{a}_3$ | = | $\frac{1}{2}a\,\hat{\mathbf{x}} - \frac{1}{2\sqrt{3}}a\,\hat{\mathbf{y}} + \left(\frac{1}{2} + z_3\right)c\,\hat{\mathbf{z}}$ | (4$f$) | P |
| $\mathbf{B}_9$ | = | $\frac{2}{3}\,\mathbf{a}_1 + \frac{1}{3}\,\mathbf{a}_2 - z_3\,\mathbf{a}_3$ | = | $\frac{1}{2}a\,\hat{\mathbf{x}} - \frac{1}{2\sqrt{3}}a\,\hat{\mathbf{y}} - z_3c\,\hat{\mathbf{z}}$ | (4$f$) | P |
| $\mathbf{B}_{10}$ | = | $\frac{1}{3}\,\mathbf{a}_1 + \frac{2}{3}\,\mathbf{a}_2 + \left(\frac{1}{2} - z_3\right)\,\mathbf{a}_3$ | = | $\frac{1}{2}a\,\hat{\mathbf{x}} + \frac{1}{2\sqrt{3}}a\,\hat{\mathbf{y}} + \left(\frac{1}{2} - z_3\right)c\,\hat{\mathbf{z}}$ | (4$f$) | P |

**Geometry files:**

- CIF: pp. 1770
- POSCAR: pp. 1770

# $Lu_2CoGa_3$ Structure: AB3C2_hP24_194_f_k_bh

http://aflow.org/prototype-encyclopedia/AB3C2_hP24_194_f_k_bh

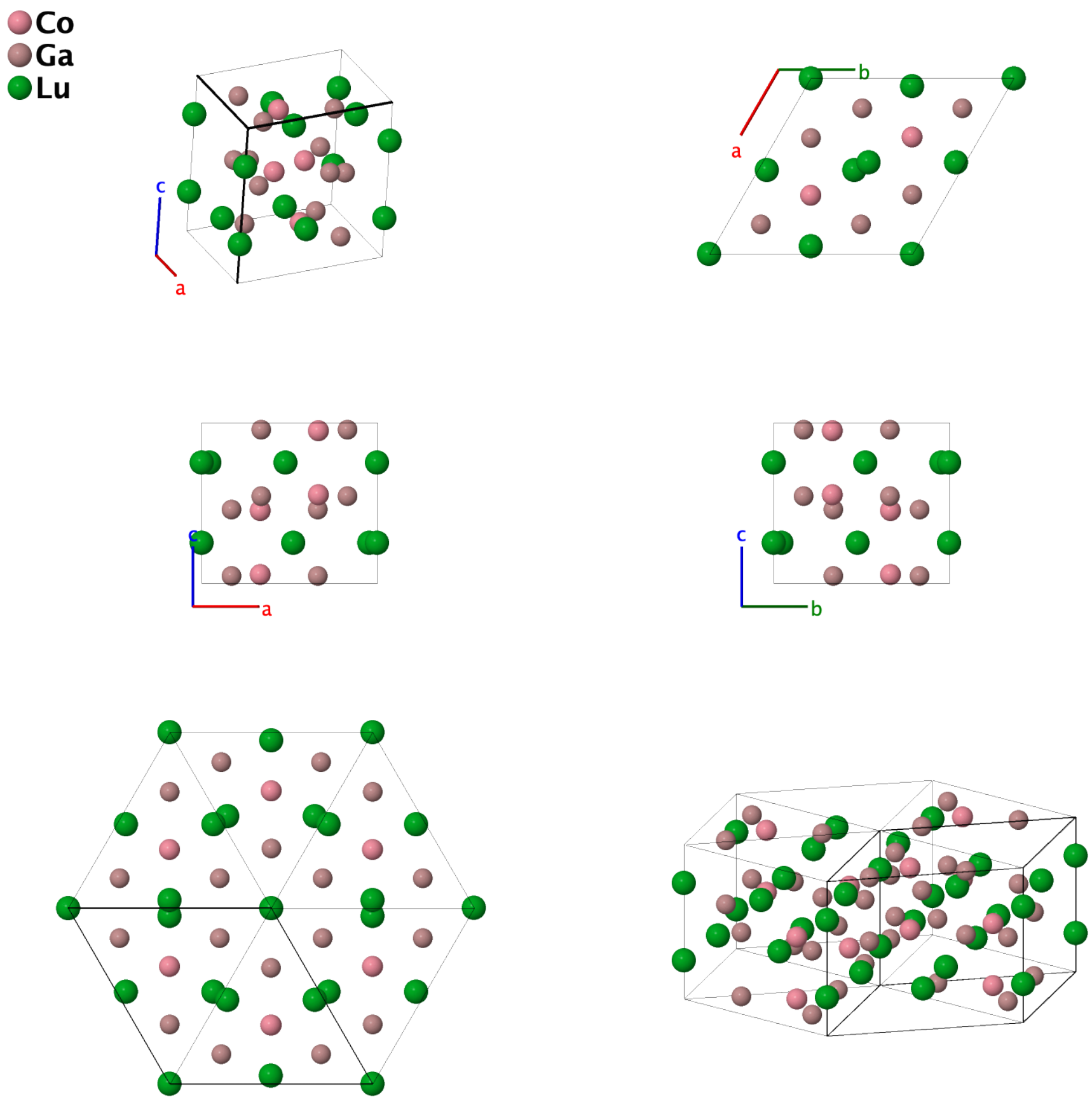

| | | |
|---|---|---|
| **Prototype** | : | $CoGa_3Lu_2$ |
| **AFLOW prototype label** | : | AB3C2_hP24_194_f_k_bh |
| ***Strukturbericht* designation** | : | None |
| **Pearson symbol** | : | hP24 |
| **Space group number** | : | 194 |
| **Space group symbol** | : | $P6_3/mmc$ |
| **AFLOW prototype command** | : | `aflow --proto=AB3C2_hP24_194_f_k_bh` `--params=`$a, c/a, z_2, x_3, x_4, z_4$ |

---

**Other compounds with this structure**

- $Dy_2CoGa_3$, $Er_2CoGa_3$, $Er_2RhSi_3$, $Ho_2CoGa_3$, $Nd_2PdSi_3$, $Tb_2CoGa_3$, $Tm_2CoGa_3$, $Y_2CoGa_3$, and $Yb_2CoGa_3$

---

- Although (Gladyshevskii, 1992) refers to these compounds as "new members of the $AlB_2$ structure family" and this structure is related to $AlB_2$, the two are not isostructural.

**Hexagonal primitive vectors:**

$$\mathbf{a}_1 = \frac{1}{2}\,a\,\hat{\mathbf{x}} - \frac{\sqrt{3}}{2}\,a\,\hat{\mathbf{y}}$$
$$\mathbf{a}_2 = \frac{1}{2}\,a\,\hat{\mathbf{x}} + \frac{\sqrt{3}}{2}\,a\,\hat{\mathbf{y}}$$
$$\mathbf{a}_3 = c\,\hat{\mathbf{z}}$$

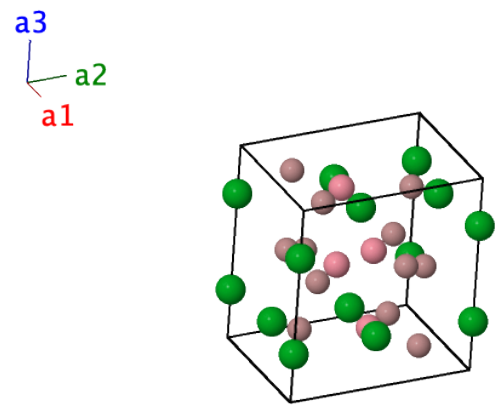

**Basis vectors:**

| | | Lattice Coordinates | | Cartesian Coordinates | Wyckoff Position | Atom Type |
|---|---|---|---|---|---|---|
| $\mathbf{B}_1$ | = | $\frac{1}{4}\,\mathbf{a}_3$ | = | $\frac{1}{4}c\,\hat{\mathbf{z}}$ | (2$b$) | Lu I |
| $\mathbf{B}_2$ | = | $\frac{3}{4}\,\mathbf{a}_3$ | = | $\frac{3}{4}c\,\hat{\mathbf{z}}$ | (2$b$) | Lu I |
| $\mathbf{B}_3$ | = | $\frac{1}{3}\,\mathbf{a}_1 + \frac{2}{3}\,\mathbf{a}_2 + z_2\,\mathbf{a}_3$ | = | $\frac{1}{2}a\,\hat{\mathbf{x}} + \frac{1}{2\sqrt{3}}a\,\hat{\mathbf{y}} + z_2c\,\hat{\mathbf{z}}$ | (4$f$) | Co |
| $\mathbf{B}_4$ | = | $\frac{2}{3}\,\mathbf{a}_1 + \frac{1}{3}\,\mathbf{a}_2 + \left(\frac{1}{2} + z_2\right)\,\mathbf{a}_3$ | = | $\frac{1}{2}a\,\hat{\mathbf{x}} - \frac{1}{2\sqrt{3}}a\,\hat{\mathbf{y}} + \left(\frac{1}{2} + z_2\right)c\,\hat{\mathbf{z}}$ | (4$f$) | Co |
| $\mathbf{B}_5$ | = | $\frac{2}{3}\,\mathbf{a}_1 + \frac{1}{3}\,\mathbf{a}_2 - z_2\,\mathbf{a}_3$ | = | $\frac{1}{2}a\,\hat{\mathbf{x}} - \frac{1}{2\sqrt{3}}a\,\hat{\mathbf{y}} - z_2c\,\hat{\mathbf{z}}$ | (4$f$) | Co |
| $\mathbf{B}_6$ | = | $\frac{1}{3}\,\mathbf{a}_1 + \frac{2}{3}\,\mathbf{a}_2 + \left(\frac{1}{2} - z_2\right)\,\mathbf{a}_3$ | = | $\frac{1}{2}a\,\hat{\mathbf{x}} + \frac{1}{2\sqrt{3}}a\,\hat{\mathbf{y}} + \left(\frac{1}{2} - z_2\right)c\,\hat{\mathbf{z}}$ | (4$f$) | Co |
| $\mathbf{B}_7$ | = | $x_3\,\mathbf{a}_1 + 2x_3\,\mathbf{a}_2 + \frac{1}{4}\,\mathbf{a}_3$ | = | $\frac{3}{2}x_3a\,\hat{\mathbf{x}} + \frac{\sqrt{3}}{2}x_3a\,\hat{\mathbf{y}} + \frac{1}{4}c\,\hat{\mathbf{z}}$ | (6$h$) | Lu II |
| $\mathbf{B}_8$ | = | $-2x_3\,\mathbf{a}_1 - x_3\,\mathbf{a}_2 + \frac{1}{4}\,\mathbf{a}_3$ | = | $-\frac{3}{2}x_3a\,\hat{\mathbf{x}} + \frac{\sqrt{3}}{2}x_3a\,\hat{\mathbf{y}} + \frac{1}{4}c\,\hat{\mathbf{z}}$ | (6$h$) | Lu II |
| $\mathbf{B}_9$ | = | $x_3\,\mathbf{a}_1 - x_3\,\mathbf{a}_2 + \frac{1}{4}\,\mathbf{a}_3$ | = | $-\sqrt{3}x_3a\,\hat{\mathbf{y}} + \frac{1}{4}c\,\hat{\mathbf{z}}$ | (6$h$) | Lu II |
| $\mathbf{B}_{10}$ | = | $-x_3\,\mathbf{a}_1 - 2x_3\,\mathbf{a}_2 + \frac{3}{4}\,\mathbf{a}_3$ | = | $-\frac{3}{2}x_3a\,\hat{\mathbf{x}} - \frac{\sqrt{3}}{2}x_3a\,\hat{\mathbf{y}} + \frac{3}{4}c\,\hat{\mathbf{z}}$ | (6$h$) | Lu II |
| $\mathbf{B}_{11}$ | = | $2x_3\,\mathbf{a}_1 + x_3\,\mathbf{a}_2 + \frac{3}{4}\,\mathbf{a}_3$ | = | $\frac{3}{2}x_3a\,\hat{\mathbf{x}} - \frac{\sqrt{3}}{2}x_3a\,\hat{\mathbf{y}} + \frac{3}{4}c\,\hat{\mathbf{z}}$ | (6$h$) | Lu II |
| $\mathbf{B}_{12}$ | = | $-x_3\,\mathbf{a}_1 + x_3\,\mathbf{a}_2 + \frac{3}{4}\,\mathbf{a}_3$ | = | $\sqrt{3}x_3a\,\hat{\mathbf{y}} + \frac{3}{4}c\,\hat{\mathbf{z}}$ | (6$h$) | Lu II |
| $\mathbf{B}_{13}$ | = | $x_4\,\mathbf{a}_1 + 2x_4\,\mathbf{a}_2 + z_4\,\mathbf{a}_3$ | = | $\frac{3}{2}x_4a\,\hat{\mathbf{x}} + \frac{\sqrt{3}}{2}x_4a\,\hat{\mathbf{y}} + z_4c\,\hat{\mathbf{z}}$ | (12$k$) | Ga |
| $\mathbf{B}_{14}$ | = | $-2x_4\,\mathbf{a}_1 - x_4\,\mathbf{a}_2 + z_4\,\mathbf{a}_3$ | = | $-\frac{3}{2}x_4a\,\hat{\mathbf{x}} + \frac{\sqrt{3}}{2}x_4a\,\hat{\mathbf{y}} + z_4c\,\hat{\mathbf{z}}$ | (12$k$) | Ga |
| $\mathbf{B}_{15}$ | = | $x_4\,\mathbf{a}_1 - x_4\,\mathbf{a}_2 + z_4\,\mathbf{a}_3$ | = | $-\sqrt{3}x_4a\,\hat{\mathbf{y}} + z_4c\,\hat{\mathbf{z}}$ | (12$k$) | Ga |
| $\mathbf{B}_{16}$ | = | $-x_4\,\mathbf{a}_1 - 2x_4\,\mathbf{a}_2 + \left(\frac{1}{2} + z_4\right)\,\mathbf{a}_3$ | = | $-\frac{3}{2}x_4a\,\hat{\mathbf{x}} - \frac{\sqrt{3}}{2}x_4a\,\hat{\mathbf{y}} + \left(\frac{1}{2} + z_4\right)c\,\hat{\mathbf{z}}$ | (12$k$) | Ga |
| $\mathbf{B}_{17}$ | = | $2x_4\,\mathbf{a}_1 + x_4\,\mathbf{a}_2 + \left(\frac{1}{2} + z_4\right)\,\mathbf{a}_3$ | = | $\frac{3}{2}x_4a\,\hat{\mathbf{x}} - \frac{\sqrt{3}}{2}x_4a\,\hat{\mathbf{y}} + \left(\frac{1}{2} + z_4\right)c\,\hat{\mathbf{z}}$ | (12$k$) | Ga |
| $\mathbf{B}_{18}$ | = | $-x_4\,\mathbf{a}_1 + x_4\,\mathbf{a}_2 + \left(\frac{1}{2} + z_4\right)\,\mathbf{a}_3$ | = | $\sqrt{3}x_4a\,\hat{\mathbf{y}} + \left(\frac{1}{2} + z_4\right)c\,\hat{\mathbf{z}}$ | (12$k$) | Ga |
| $\mathbf{B}_{19}$ | = | $2x_4\,\mathbf{a}_1 + x_4\,\mathbf{a}_2 - z_4\,\mathbf{a}_3$ | = | $\frac{3}{2}x_4a\,\hat{\mathbf{x}} - \frac{\sqrt{3}}{2}x_4a\,\hat{\mathbf{y}} - z_4c\,\hat{\mathbf{z}}$ | (12$k$) | Ga |
| $\mathbf{B}_{20}$ | = | $-x_4\,\mathbf{a}_1 - 2x_4\,\mathbf{a}_2 - z_4\,\mathbf{a}_3$ | = | $-\frac{3}{2}x_4a\,\hat{\mathbf{x}} - \frac{\sqrt{3}}{2}x_4a\,\hat{\mathbf{y}} - z_4c\,\hat{\mathbf{z}}$ | (12$k$) | Ga |
| $\mathbf{B}_{21}$ | = | $-x_4\,\mathbf{a}_1 + x_4\,\mathbf{a}_2 - z_4\,\mathbf{a}_3$ | = | $\sqrt{3}x_4a\,\hat{\mathbf{y}} - z_4c\,\hat{\mathbf{z}}$ | (12$k$) | Ga |
| $\mathbf{B}_{22}$ | = | $-2x_4\,\mathbf{a}_1 - x_4\,\mathbf{a}_2 + \left(\frac{1}{2} - z_4\right)\,\mathbf{a}_3$ | = | $-\frac{3}{2}x_4a\,\hat{\mathbf{x}} + \frac{\sqrt{3}}{2}x_4a\,\hat{\mathbf{y}} + \left(\frac{1}{2} - z_4\right)c\,\hat{\mathbf{z}}$ | (12$k$) | Ga |
| $\mathbf{B}_{23}$ | = | $x_4\,\mathbf{a}_1 + 2x_4\,\mathbf{a}_2 + \left(\frac{1}{2} - z_4\right)\,\mathbf{a}_3$ | = | $\frac{3}{2}x_4a\,\hat{\mathbf{x}} + \frac{\sqrt{3}}{2}x_4a\,\hat{\mathbf{y}} + \left(\frac{1}{2} - z_4\right)c\,\hat{\mathbf{z}}$ | (12$k$) | Ga |
| $\mathbf{B}_{24}$ | = | $x_4\,\mathbf{a}_1 - x_4\,\mathbf{a}_2 + \left(\frac{1}{2} - z_4\right)\,\mathbf{a}_3$ | = | $-\sqrt{3}x_4a\,\hat{\mathbf{y}} + \left(\frac{1}{2} - z_4\right)c\,\hat{\mathbf{z}}$ | (12$k$) | Ga |

**Geometry files:**

- CIF: pp. 1770
- POSCAR: pp. 1771

# Hexagonal Delafossite ($CuAlO_2$) Structure: ABC2_hP8_194_a_c_f

http://aflow.org/prototype-encyclopedia/ABC2_hP8_194_a_c_f

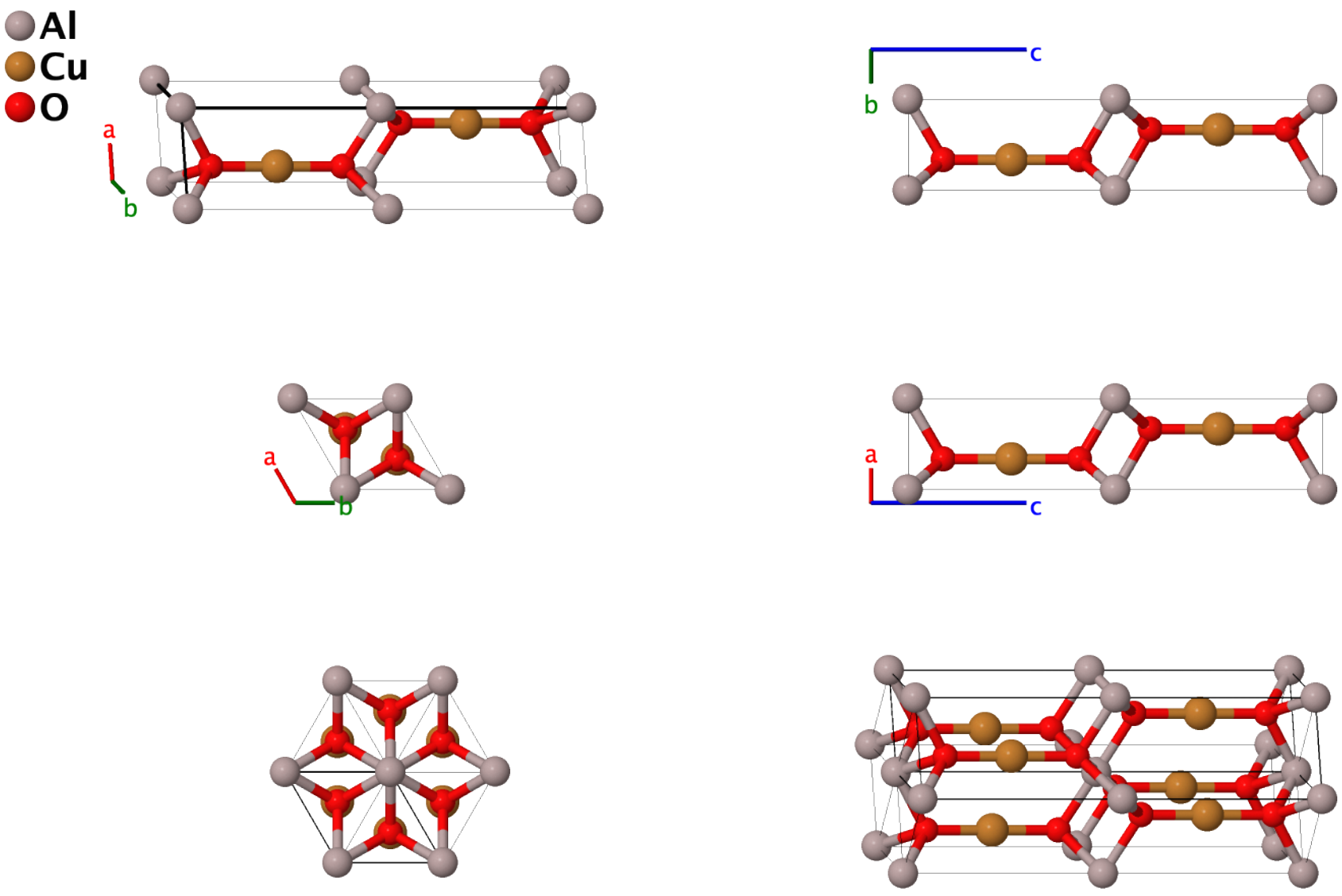

| | | |
|---|---|---|
| **Prototype** | : | $AlCuO_2$ |
| **AFLOW prototype label** | : | ABC2_hP8_194_a_c_f |
| ***Strukturbericht* designation** | : | None |
| **Pearson symbol** | : | hP8 |
| **Space group number** | : | 194 |
| **Space group symbol** | : | $P6_3/mmc$ |
| **AFLOW prototype command** | : | `aflow --proto=ABC2_hP8_194_a_c_f --params=`$a, c/a, z_3$ |

**Other compounds with this structure**

- $CuGaO_2$, $CuScO_2$, and $CuYO_2$

- Delafossite appears in two forms which differ in the stacking of the layers: rhombohedral, prototype $CuFeO_2$, and hexagonal, shown here. Most of the structures found in the hexagonal phase can also be found in the rhombohedral structure (Marquardt, 2006).

**Hexagonal primitive vectors:**

$$\mathbf{a}_1 = \frac{1}{2}\,a\,\hat{\mathbf{x}} - \frac{\sqrt{3}}{2}\,a\,\hat{\mathbf{y}}$$
$$\mathbf{a}_2 = \frac{1}{2}\,a\,\hat{\mathbf{x}} + \frac{\sqrt{3}}{2}\,a\,\hat{\mathbf{y}}$$
$$\mathbf{a}_3 = c\,\hat{\mathbf{z}}$$

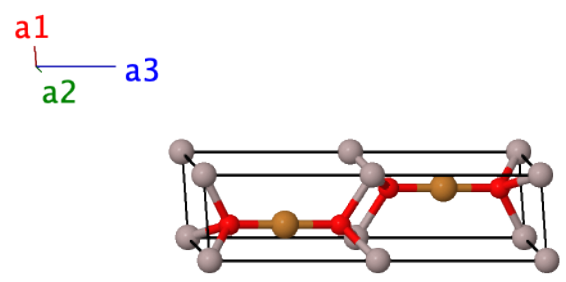

**Basis vectors:**

Lattice Coordinates | Cartesian Coordinates | Wyckoff Position | Atom Type

| | | | | | | |
|---|---|---|---|---|---|---|
| $\mathbf{B}_1$ | = | $0\,\mathbf{a}_1 + 0\,\mathbf{a}_2 + 0\,\mathbf{a}_3$ | = | $0\,\hat{\mathbf{x}} + 0\,\hat{\mathbf{y}} + 0\,\hat{\mathbf{z}}$ | (2$a$) | Al |
| $\mathbf{B}_2$ | = | $\frac{1}{2}\,\mathbf{a}_3$ | = | $\frac{1}{2}c\,\hat{\mathbf{z}}$ | (2$a$) | Al |
| $\mathbf{B}_3$ | = | $\frac{1}{3}\,\mathbf{a}_1 + \frac{2}{3}\,\mathbf{a}_2 + \frac{1}{4}\,\mathbf{a}_3$ | = | $\frac{1}{2}a\,\hat{\mathbf{x}} + \frac{1}{2\sqrt{3}}a\,\hat{\mathbf{y}} + \frac{1}{4}c\,\hat{\mathbf{z}}$ | (2$c$) | Cu |
| $\mathbf{B}_4$ | = | $\frac{2}{3}\,\mathbf{a}_1 + \frac{1}{3}\,\mathbf{a}_2 + \frac{3}{4}\,\mathbf{a}_3$ | = | $\frac{1}{2}a\,\hat{\mathbf{x}} - \frac{1}{2\sqrt{3}}a\,\hat{\mathbf{y}} + \frac{3}{4}c\,\hat{\mathbf{z}}$ | (2$c$) | Cu |
| $\mathbf{B}_5$ | = | $\frac{1}{3}\,\mathbf{a}_1 + \frac{2}{3}\,\mathbf{a}_2 + z_3\,\mathbf{a}_3$ | = | $\frac{1}{2}a\,\hat{\mathbf{x}} + \frac{1}{2\sqrt{3}}a\,\hat{\mathbf{y}} + z_3c\,\hat{\mathbf{z}}$ | (4$f$) | O |
| $\mathbf{B}_6$ | = | $\frac{2}{3}\,\mathbf{a}_1 + \frac{1}{3}\,\mathbf{a}_2 + \left(\frac{1}{2} + z_3\right)\,\mathbf{a}_3$ | = | $\frac{1}{2}a\,\hat{\mathbf{x}} - \frac{1}{2\sqrt{3}}a\,\hat{\mathbf{y}} + \left(\frac{1}{2} + z_3\right)c\,\hat{\mathbf{z}}$ | (4$f$) | O |
| $\mathbf{B}_7$ | = | $\frac{2}{3}\,\mathbf{a}_1 + \frac{1}{3}\,\mathbf{a}_2 - z_3\,\mathbf{a}_3$ | = | $\frac{1}{2}a\,\hat{\mathbf{x}} - \frac{1}{2\sqrt{3}}a\,\hat{\mathbf{y}} - z_3c\,\hat{\mathbf{z}}$ | (4$f$) | O |
| $\mathbf{B}_8$ | = | $\frac{1}{3}\,\mathbf{a}_1 + \frac{2}{3}\,\mathbf{a}_2 + \left(\frac{1}{2} - z_3\right)\,\mathbf{a}_3$ | = | $\frac{1}{2}a\,\hat{\mathbf{x}} + \frac{1}{2\sqrt{3}}a\,\hat{\mathbf{y}} + \left(\frac{1}{2} - z_3\right)c\,\hat{\mathbf{z}}$ | (4$f$) | O |

**Geometry files:**

- CIF: pp. 1771
- POSCAR: pp. 1771

# $LiZn_2$ ($C_k$) Structure: AB_hP4_194_a_c

http://aflow.org/prototype-encyclopedia/AB_hP4_194_a_c

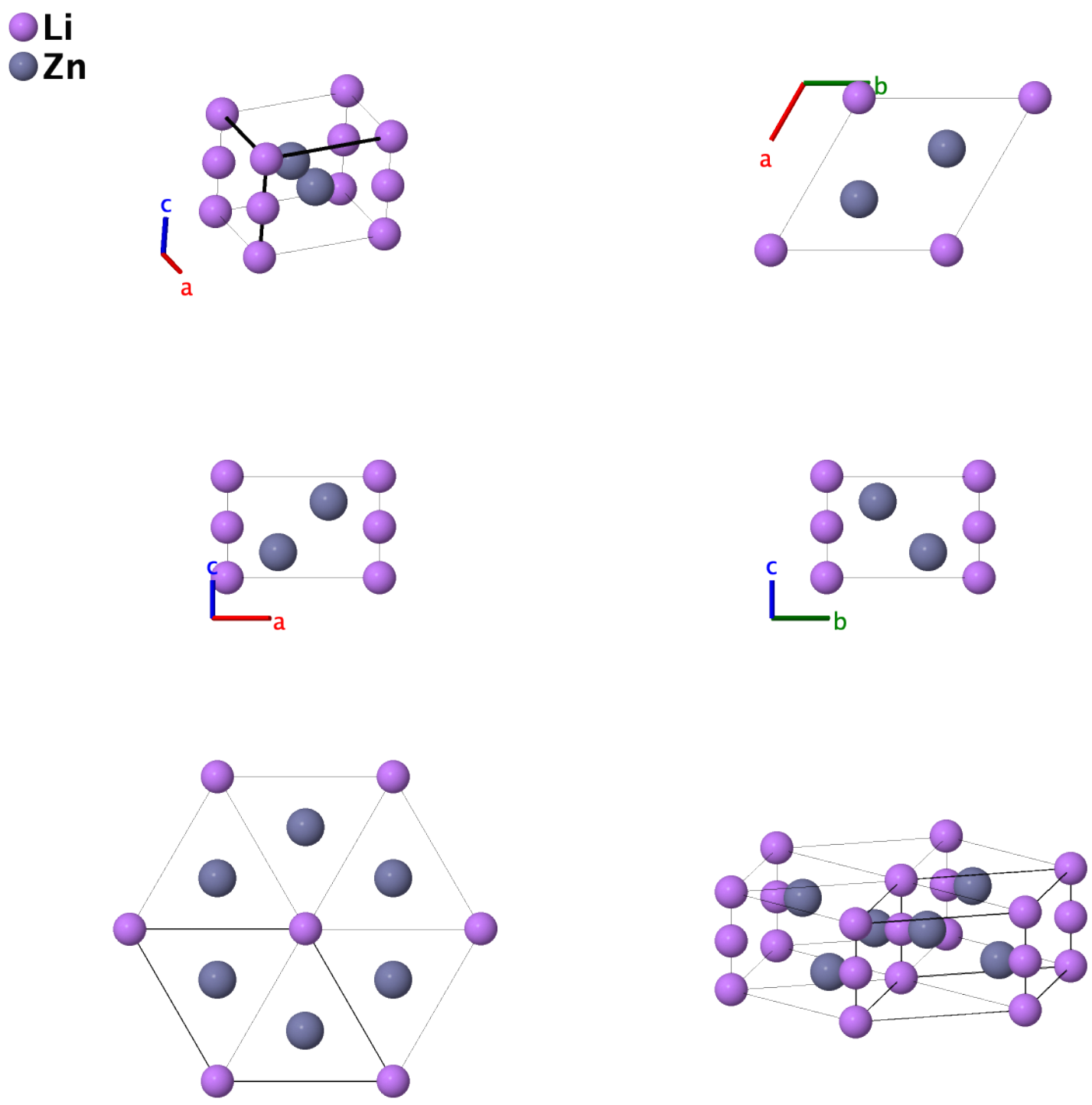

| | | |
|---|---|---|
| **Prototype** | : | $LiZn_2$ |
| **AFLOW prototype label** | : | AB_hP4_194_a_c |
| ***Strukturbericht* designation** | : | $C_k$ |
| **Pearson symbol** | : | hP4 |
| **Space group number** | : | 194 |
| **Space group symbol** | : | $P6_3/mmc$ |
| **AFLOW prototype command** | : | `aflow --proto=AB_hP4_194_a_c`<br>`--params=`$a, c/a$ |

- The actual stoichiometry of this compounds is $Li_{0.8}Zn_2$, so each (2$a$) site is only filled 40% of the time. This means that the short (1.26 Å) distance between the lithium atoms shown here is illusionary.

**Hexagonal primitive vectors:**

$$\mathbf{a}_1 = \frac{1}{2}\,a\,\hat{\mathbf{x}} - \frac{\sqrt{3}}{2}\,a\,\hat{\mathbf{y}}$$
$$\mathbf{a}_2 = \frac{1}{2}\,a\,\hat{\mathbf{x}} + \frac{\sqrt{3}}{2}\,a\,\hat{\mathbf{y}}$$
$$\mathbf{a}_3 = c\,\hat{\mathbf{z}}$$

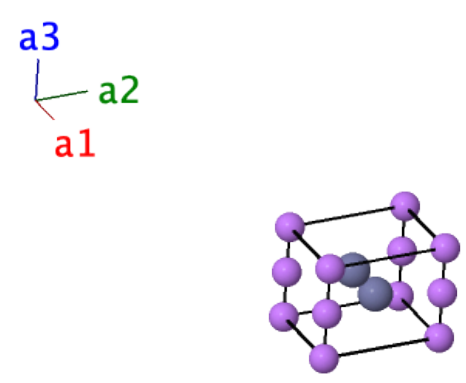

**Basis vectors:**

| | | Lattice Coordinates | | Cartesian Coordinates | Wyckoff Position | Atom Type |
|---|---|---|---|---|---|---|
| $\mathbf{B}_1$ | = | $0\,\mathbf{a}_1 + 0\,\mathbf{a}_2 + 0\,\mathbf{a}_3$ | = | $0\,\hat{\mathbf{x}} + 0\,\hat{\mathbf{y}} + 0\,\hat{\mathbf{z}}$ | (2$a$) | Li |
| $\mathbf{B}_2$ | = | $\frac{1}{2}\,\mathbf{a}_3$ | = | $\frac{1}{2}c\,\hat{\mathbf{z}}$ | (2*a*) | Li |
| $\mathbf{B}_3$ | = | $\frac{1}{3}\,\mathbf{a}_1 + \frac{2}{3}\,\mathbf{a}_2 + \frac{1}{4}\,\mathbf{a}_3$ | = | $\frac{1}{2}a\,\hat{\mathbf{x}} + \frac{1}{2\sqrt{3}}a\,\hat{\mathbf{y}} + \frac{1}{4}c\,\hat{\mathbf{z}}$ | (2*c*) | Zn |
| $\mathbf{B}_4$ | = | $\frac{2}{3}\,\mathbf{a}_1 + \frac{1}{3}\,\mathbf{a}_2 + \frac{3}{4}\,\mathbf{a}_3$ | = | $\frac{1}{2}a\,\hat{\mathbf{x}} - \frac{1}{2\sqrt{3}}a\,\hat{\mathbf{y}} + \frac{3}{4}c\,\hat{\mathbf{z}}$ | (2*c*) | Zn |

---

**Found in:**

---

**Geometry files:**

- CIF: pp. 1771
- POSCAR: pp. 1772

# $Fe_2N$ (approximate, $L'3_0$) Structure: AB_hP4_194_c_a

http://aflow.org/prototype-encyclopedia/AB_hP4_194_c_a.Fe2N

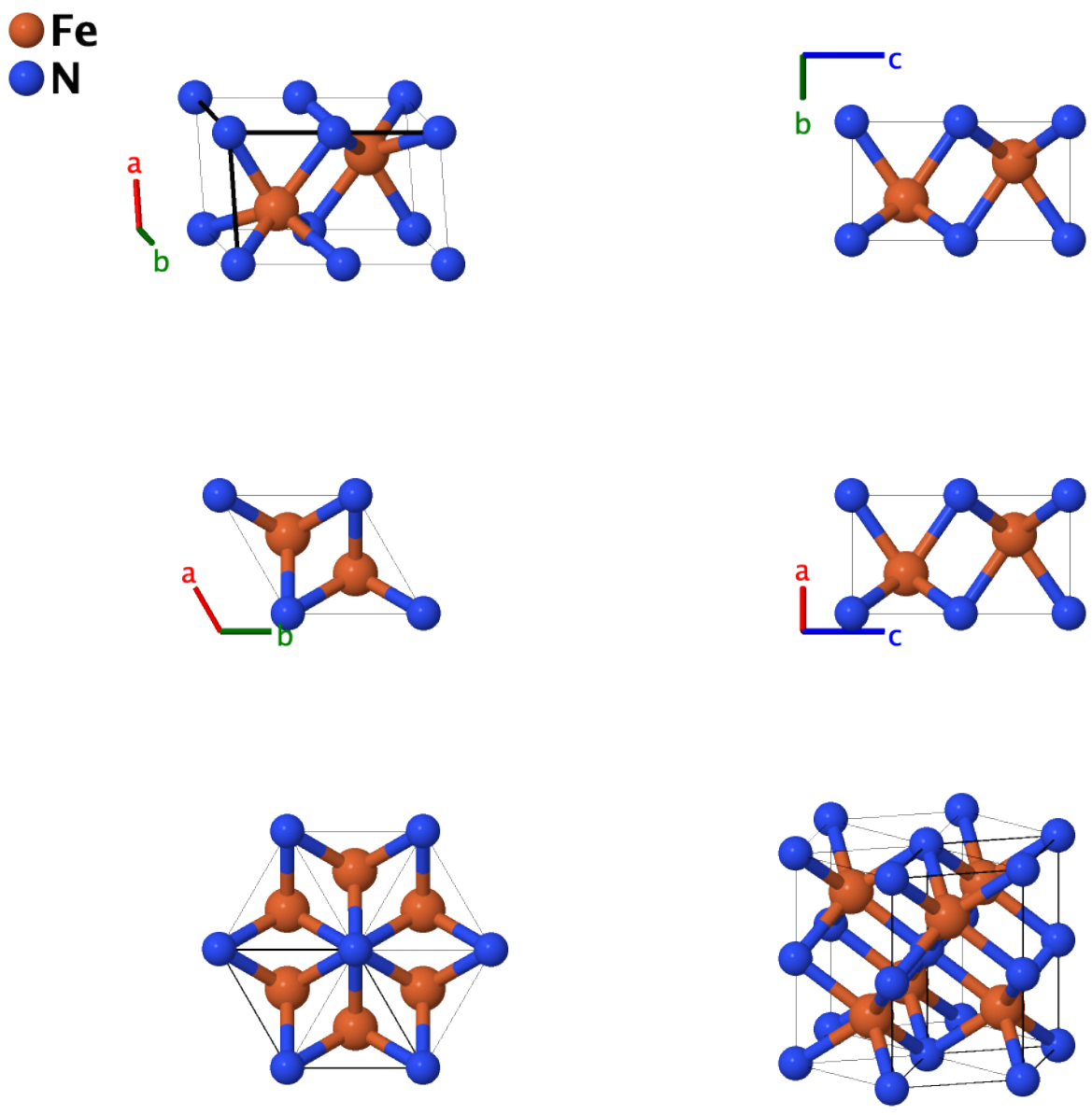

| | | |
|---|---|---|
| **Prototype** | : | $Fe_2N$ |
| **AFLOW prototype label** | : | AB_hP4_194_c_a |
| ***Strukturbericht* designation** | : | $L'3_0$ |
| **Pearson symbol** | : | hP4 |
| **Space group number** | : | 194 |
| **Space group symbol** | : | $P6_3/mmc$ |
| **AFLOW prototype command** | : | `aflow --proto=AB_hP4_194_c_a --params=`$a, c/a$ |

### Other compounds with this structure

- $\beta$-$CTa_2$, $CV_2$, $CW_2$, $Mn_2N$, $Nb_2N$, and $Ta_{\approx 2}N$

- The $L'3_0$ designation is only found in (Smithells, 1955) and (Pearson, 1958), as well as their following volumes. Smithells gives two possible positions of the nitrogen atoms. We use the one that matches Pearson, which (Parthé, 1993) notes is equivalent to the NiAs $B8_1$ structure, but here the ($2a$) nitrogen site is only half-filled to maintain stoichiometry.
- Neither of the defining references gives a source for the structure. It is likely an approximation to $\epsilon$-$Fe_2N$, which has the $\beta$-$V_2N$ $L'3_2$ structure. We take the lattice constants for this structure from the Fe-Fe distances in $\epsilon$-$Fe_2N$ provided by (Hendricks, 1930).
- The similar compounds list is taken from (Pearson, 1967). As it is quite similar to the list for $\beta$-$V_2N$, it is likely that this structure is only an approximation to the correct structure for these compounds as well.
- This structure has the same AFLOW designation, AB_hP4_194_c_a.Fe2N, as $B8_1$. The two structures differ by their lattice constants and the half-occupancy of the ($2a$) site in $L'3_0$.

**Hexagonal primitive vectors:**

$$
\begin{aligned}
\mathbf{a}_1 &= \tfrac{1}{2}\,a\,\hat{\mathbf{x}} - \tfrac{\sqrt{3}}{2}\,a\,\hat{\mathbf{y}} \\
\mathbf{a}_2 &= \tfrac{1}{2}\,a\,\hat{\mathbf{x}} + \tfrac{\sqrt{3}}{2}\,a\,\hat{\mathbf{y}} \\
\mathbf{a}_3 &= c\,\hat{\mathbf{z}}
\end{aligned}
$$

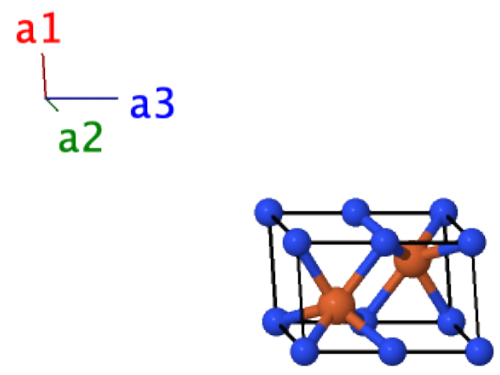

**Basis vectors:**

| | | Lattice Coordinates | | Cartesian Coordinates | Wyckoff Position | Atom Type |
|---|---|---|---|---|---|---|
| $\mathbf{B}_1$ | = | $0\,\mathbf{a}_1 + 0\,\mathbf{a}_2 + 0\,\mathbf{a}_3$ | = | $0\,\hat{\mathbf{x}} + 0\,\hat{\mathbf{y}} + 0\,\hat{\mathbf{z}}$ | (2*a*) | N |
| $\mathbf{B}_2$ | = | $\frac{1}{2}\,\mathbf{a}_3$ | = | $\frac{1}{2}c\,\hat{\mathbf{z}}$ | (2*a*) | N |
| $\mathbf{B}_3$ | = | $\frac{1}{3}\,\mathbf{a}_1 + \frac{2}{3}\,\mathbf{a}_2 + \frac{1}{4}\,\mathbf{a}_3$ | = | $\frac{1}{2}a\,\hat{\mathbf{x}} + \frac{1}{2\sqrt{3}}a\,\hat{\mathbf{y}} + \frac{1}{4}c\,\hat{\mathbf{z}}$ | (2*c*) | Fe |
| $\mathbf{B}_4$ | = | $\frac{2}{3}\,\mathbf{a}_1 + \frac{1}{3}\,\mathbf{a}_2 + \frac{3}{4}\,\mathbf{a}_3$ | = | $\frac{1}{2}a\,\hat{\mathbf{x}} - \frac{1}{2\sqrt{3}}a\,\hat{\mathbf{y}} + \frac{3}{4}c\,\hat{\mathbf{z}}$ | (2*c*) | Fe |

**Geometry files:**

- CIF: pp. 1772
- POSCAR: pp. 1772

# Cubic $Cu_2OSeO_3$ Structure: A2B4C_cP56_198_ab_2a2b_2a

http://aflow.org/prototype-encyclopedia/A2B4C_cP56_198_ab_2a2b_2a

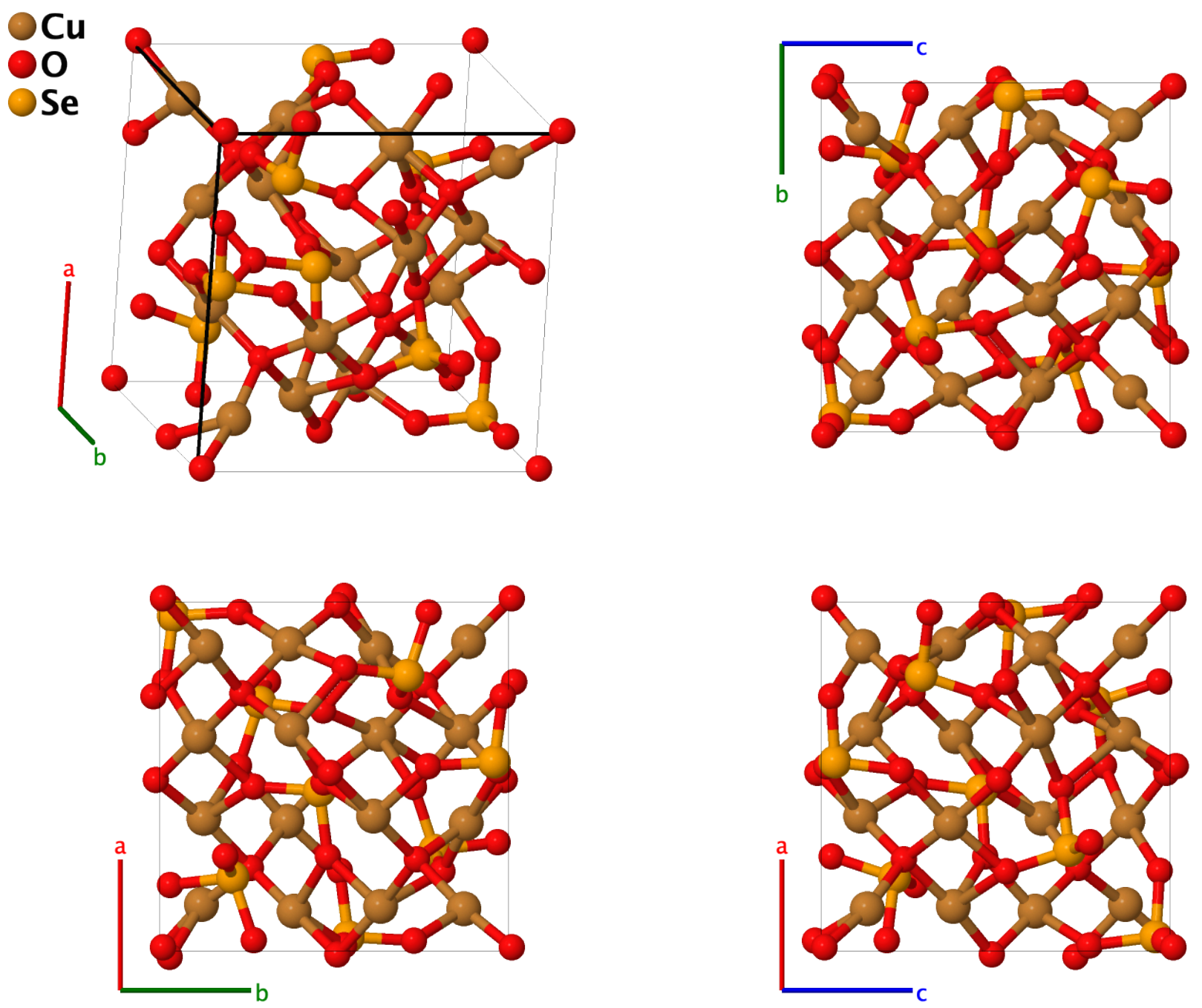

| | | |
|---|---|---|
| **Prototype** | : | $Cu_2O_4Se$ |
| **AFLOW prototype label** | : | A2B4C_cP56_198_ab_2a2b_2a |
| ***Strukturbericht* designation** | : | None |
| **Pearson symbol** | : | cP56 |
| **Space group number** | : | 198 |
| **Space group symbol** | : | $P2_13$ |
| **AFLOW prototype command** | : | `aflow --proto=A2B4C_cP56_198_ab_2a2b_2a` <br> `--params=`$a, x_1, x_2, x_3, x_4, x_5, x_6, y_6, z_6, x_7, y_7, z_7, x_8, y_8, z_8$ |

- This is the cubic phase of $Cu_2OSeO_3$. There is also a monoclinic phase.

**Simple Cubic primitive vectors:**

$$\mathbf{a}_1 = a\,\hat{\mathbf{x}}$$
$$\mathbf{a}_2 = a\,\hat{\mathbf{y}}$$
$$\mathbf{a}_3 = a\,\hat{\mathbf{z}}$$

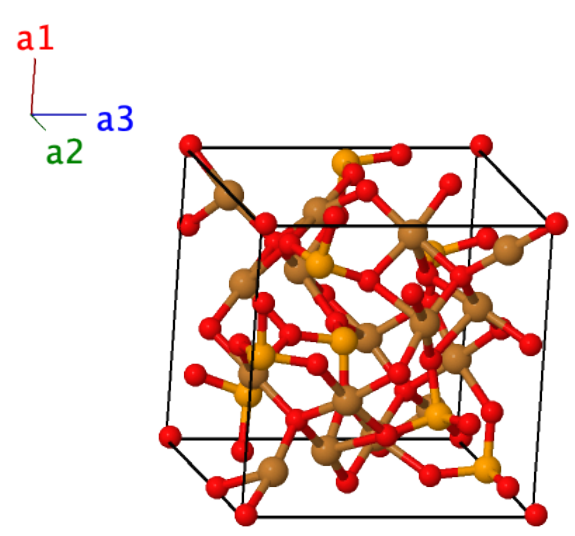

**Basis vectors:**

| | | Lattice Coordinates | | Cartesian Coordinates | Wyckoff Position | Atom Type |
|---|---|---|---|---|---|---|
| $\mathbf{B}_1$ | = | $x_1\,\mathbf{a}_1 + x_1\,\mathbf{a}_2 + x_1\,\mathbf{a}_3$ | = | $x_1 a\,\hat{\mathbf{x}} + x_1 a\,\hat{\mathbf{y}} + x_1 a\,\hat{\mathbf{z}}$ | (4$a$) | Cu I |
| $\mathbf{B}_2$ | = | $\left(\frac{1}{2} - x_1\right)\mathbf{a}_1 - x_1\,\mathbf{a}_2 + \left(\frac{1}{2} + x_1\right)\mathbf{a}_3$ | = | $\left(\frac{1}{2} - x_1\right)a\,\hat{\mathbf{x}} - x_1 a\,\hat{\mathbf{y}} + \left(\frac{1}{2} + x_1\right)a\,\hat{\mathbf{z}}$ | (4$a$) | Cu I |
| $\mathbf{B}_3$ | = | $-x_1\,\mathbf{a}_1 + \left(\frac{1}{2} + x_1\right)\mathbf{a}_2 + \left(\frac{1}{2} - x_1\right)\mathbf{a}_3$ | = | $-x_1 a\,\hat{\mathbf{x}} + \left(\frac{1}{2} + x_1\right)a\,\hat{\mathbf{y}} + \left(\frac{1}{2} - x_1\right)a\,\hat{\mathbf{z}}$ | (4$a$) | Cu I |
| $\mathbf{B}_4$ | = | $\left(\frac{1}{2} + x_1\right)\mathbf{a}_1 + \left(\frac{1}{2} - x_1\right)\mathbf{a}_2 - x_1\,\mathbf{a}_3$ | = | $\left(\frac{1}{2} + x_1\right)a\,\hat{\mathbf{x}} + \left(\frac{1}{2} - x_1\right)a\,\hat{\mathbf{y}} - x_1 a\,\hat{\mathbf{z}}$ | (4$a$) | Cu I |
| $\mathbf{B}_5$ | = | $x_2\,\mathbf{a}_1 + x_2\,\mathbf{a}_2 + x_2\,\mathbf{a}_3$ | = | $x_2 a\,\hat{\mathbf{x}} + x_2 a\,\hat{\mathbf{y}} + x_2 a\,\hat{\mathbf{z}}$ | (4$a$) | O I |
| $\mathbf{B}_6$ | = | $\left(\frac{1}{2} - x_2\right)\mathbf{a}_1 - x_2\,\mathbf{a}_2 + \left(\frac{1}{2} + x_2\right)\mathbf{a}_3$ | = | $\left(\frac{1}{2} - x_2\right)a\,\hat{\mathbf{x}} - x_2 a\,\hat{\mathbf{y}} + \left(\frac{1}{2} + x_2\right)a\,\hat{\mathbf{z}}$ | (4$a$) | O I |
| $\mathbf{B}_7$ | = | $-x_2\,\mathbf{a}_1 + \left(\frac{1}{2} + x_2\right)\mathbf{a}_2 + \left(\frac{1}{2} - x_2\right)\mathbf{a}_3$ | = | $-x_2 a\,\hat{\mathbf{x}} + \left(\frac{1}{2} + x_2\right)a\,\hat{\mathbf{y}} + \left(\frac{1}{2} - x_2\right)a\,\hat{\mathbf{z}}$ | (4$a$) | O I |
| $\mathbf{B}_8$ | = | $\left(\frac{1}{2} + x_2\right)\mathbf{a}_1 + \left(\frac{1}{2} - x_2\right)\mathbf{a}_2 - x_2\,\mathbf{a}_3$ | = | $\left(\frac{1}{2} + x_2\right)a\,\hat{\mathbf{x}} + \left(\frac{1}{2} - x_2\right)a\,\hat{\mathbf{y}} - x_2 a\,\hat{\mathbf{z}}$ | (4$a$) | O I |
| $\mathbf{B}_9$ | = | $x_3\,\mathbf{a}_1 + x_3\,\mathbf{a}_2 + x_3\,\mathbf{a}_3$ | = | $x_3 a\,\hat{\mathbf{x}} + x_3 a\,\hat{\mathbf{y}} + x_3 a\,\hat{\mathbf{z}}$ | (4$a$) | O II |
| $\mathbf{B}_{10}$ | = | $\left(\frac{1}{2} - x_3\right)\mathbf{a}_1 - x_3\,\mathbf{a}_2 + \left(\frac{1}{2} + x_3\right)\mathbf{a}_3$ | = | $\left(\frac{1}{2} - x_3\right)a\,\hat{\mathbf{x}} - x_3 a\,\hat{\mathbf{y}} + \left(\frac{1}{2} + x_3\right)a\,\hat{\mathbf{z}}$ | (4$a$) | O II |
| $\mathbf{B}_{11}$ | = | $-x_3\,\mathbf{a}_1 + \left(\frac{1}{2} + x_3\right)\mathbf{a}_2 + \left(\frac{1}{2} - x_3\right)\mathbf{a}_3$ | = | $-x_3 a\,\hat{\mathbf{x}} + \left(\frac{1}{2} + x_3\right)a\,\hat{\mathbf{y}} + \left(\frac{1}{2} - x_3\right)a\,\hat{\mathbf{z}}$ | (4$a$) | O II |
| $\mathbf{B}_{12}$ | = | $\left(\frac{1}{2} + x_3\right)\mathbf{a}_1 + \left(\frac{1}{2} - x_3\right)\mathbf{a}_2 - x_3\,\mathbf{a}_3$ | = | $\left(\frac{1}{2} + x_3\right)a\,\hat{\mathbf{x}} + \left(\frac{1}{2} - x_3\right)a\,\hat{\mathbf{y}} - x_3 a\,\hat{\mathbf{z}}$ | (4$a$) | O II |
| $\mathbf{B}_{13}$ | = | $x_4\,\mathbf{a}_1 + x_4\,\mathbf{a}_2 + x_4\,\mathbf{a}_3$ | = | $x_4 a\,\hat{\mathbf{x}} + x_4 a\,\hat{\mathbf{y}} + x_4 a\,\hat{\mathbf{z}}$ | (4$a$) | Se I |
| $\mathbf{B}_{14}$ | = | $\left(\frac{1}{2} - x_4\right)\mathbf{a}_1 - x_4\,\mathbf{a}_2 + \left(\frac{1}{2} + x_4\right)\mathbf{a}_3$ | = | $\left(\frac{1}{2} - x_4\right)a\,\hat{\mathbf{x}} - x_4 a\,\hat{\mathbf{y}} + \left(\frac{1}{2} + x_4\right)a\,\hat{\mathbf{z}}$ | (4$a$) | Se I |
| $\mathbf{B}_{15}$ | = | $-x_4\,\mathbf{a}_1 + \left(\frac{1}{2} + x_4\right)\mathbf{a}_2 + \left(\frac{1}{2} - x_4\right)\mathbf{a}_3$ | = | $-x_4 a\,\hat{\mathbf{x}} + \left(\frac{1}{2} + x_4\right)a\,\hat{\mathbf{y}} + \left(\frac{1}{2} - x_4\right)a\,\hat{\mathbf{z}}$ | (4$a$) | Se I |
| $\mathbf{B}_{16}$ | = | $\left(\frac{1}{2} + x_4\right)\mathbf{a}_1 + \left(\frac{1}{2} - x_4\right)\mathbf{a}_2 - x_4\,\mathbf{a}_3$ | = | $\left(\frac{1}{2} + x_4\right)a\,\hat{\mathbf{x}} + \left(\frac{1}{2} - x_4\right)a\,\hat{\mathbf{y}} - x_4 a\,\hat{\mathbf{z}}$ | (4$a$) | Se I |
| $\mathbf{B}_{17}$ | = | $x_5\,\mathbf{a}_1 + x_5\,\mathbf{a}_2 + x_5\,\mathbf{a}_3$ | = | $x_5 a\,\hat{\mathbf{x}} + x_5 a\,\hat{\mathbf{y}} + x_5 a\,\hat{\mathbf{z}}$ | (4$a$) | Se II |
| $\mathbf{B}_{18}$ | = | $\left(\frac{1}{2} - x_5\right)\mathbf{a}_1 - x_5\,\mathbf{a}_2 + \left(\frac{1}{2} + x_5\right)\mathbf{a}_3$ | = | $\left(\frac{1}{2} - x_5\right)a\,\hat{\mathbf{x}} - x_5 a\,\hat{\mathbf{y}} + \left(\frac{1}{2} + x_5\right)a\,\hat{\mathbf{z}}$ | (4$a$) | Se II |
| $\mathbf{B}_{19}$ | = | $-x_5\,\mathbf{a}_1 + \left(\frac{1}{2} + x_5\right)\mathbf{a}_2 + \left(\frac{1}{2} - x_5\right)\mathbf{a}_3$ | = | $-x_5 a\,\hat{\mathbf{x}} + \left(\frac{1}{2} + x_5\right)a\,\hat{\mathbf{y}} + \left(\frac{1}{2} - x_5\right)a\,\hat{\mathbf{z}}$ | (4$a$) | Se II |
| $\mathbf{B}_{20}$ | = | $\left(\frac{1}{2} + x_5\right)\mathbf{a}_1 + \left(\frac{1}{2} - x_5\right)\mathbf{a}_2 - x_5\,\mathbf{a}_3$ | = | $\left(\frac{1}{2} + x_5\right)a\,\hat{\mathbf{x}} + \left(\frac{1}{2} - x_5\right)a\,\hat{\mathbf{y}} - x_5 a\,\hat{\mathbf{z}}$ | (4$a$) | Se II |
| $\mathbf{B}_{21}$ | = | $x_6\,\mathbf{a}_1 + y_6\,\mathbf{a}_2 + z_6\,\mathbf{a}_3$ | = | $x_6 a\,\hat{\mathbf{x}} + y_6 a\,\hat{\mathbf{y}} + z_6 a\,\hat{\mathbf{z}}$ | (12$b$) | Cu II |
| $\mathbf{B}_{22}$ | = | $\left(\frac{1}{2} - x_6\right)\mathbf{a}_1 - y_6\,\mathbf{a}_2 + \left(\frac{1}{2} + z_6\right)\mathbf{a}_3$ | = | $\left(\frac{1}{2} - x_6\right)a\,\hat{\mathbf{x}} - y_6 a\,\hat{\mathbf{y}} + \left(\frac{1}{2} + z_6\right)a\,\hat{\mathbf{z}}$ | (12$b$) | Cu II |
| $\mathbf{B}_{23}$ | = | $-x_6\,\mathbf{a}_1 + \left(\frac{1}{2} + y_6\right)\mathbf{a}_2 + \left(\frac{1}{2} - z_6\right)\mathbf{a}_3$ | = | $-x_6 a\,\hat{\mathbf{x}} + \left(\frac{1}{2} + y_6\right)a\,\hat{\mathbf{y}} + \left(\frac{1}{2} - z_6\right)a\,\hat{\mathbf{z}}$ | (12$b$) | Cu II |
| $\mathbf{B}_{24}$ | = | $\left(\frac{1}{2} + x_6\right)\mathbf{a}_1 + \left(\frac{1}{2} - y_6\right)\mathbf{a}_2 - z_6\,\mathbf{a}_3$ | = | $\left(\frac{1}{2} + x_6\right)a\,\hat{\mathbf{x}} + \left(\frac{1}{2} - y_6\right)a\,\hat{\mathbf{y}} - z_6 a\,\hat{\mathbf{z}}$ | (12$b$) | Cu II |
| $\mathbf{B}_{25}$ | = | $z_6\,\mathbf{a}_1 + x_6\,\mathbf{a}_2 + y_6\,\mathbf{a}_3$ | = | $z_6 a\,\hat{\mathbf{x}} + x_6 a\,\hat{\mathbf{y}} + y_6 a\,\hat{\mathbf{z}}$ | (12$b$) | Cu II |
| $\mathbf{B}_{26}$ | = | $\left(\frac{1}{2} + z_6\right)\mathbf{a}_1 + \left(\frac{1}{2} - x_6\right)\mathbf{a}_2 - y_6\,\mathbf{a}_3$ | = | $\left(\frac{1}{2} + z_6\right)a\,\hat{\mathbf{x}} + \left(\frac{1}{2} - x_6\right)a\,\hat{\mathbf{y}} - y_6 a\,\hat{\mathbf{z}}$ | (12$b$) | Cu II |
| $\mathbf{B}_{27}$ | = | $\left(\frac{1}{2} - z_6\right)\mathbf{a}_1 - x_6\,\mathbf{a}_2 + \left(\frac{1}{2} + y_6\right)\mathbf{a}_3$ | = | $\left(\frac{1}{2} - z_6\right)a\,\hat{\mathbf{x}} - x_6 a\,\hat{\mathbf{y}} + \left(\frac{1}{2} + y_6\right)a\,\hat{\mathbf{z}}$ | (12$b$) | Cu II |
| $\mathbf{B}_{28}$ | = | $-z_6\,\mathbf{a}_1 + \left(\frac{1}{2} + x_6\right)\mathbf{a}_2 + \left(\frac{1}{2} - y_6\right)\mathbf{a}_3$ | = | $-z_6 a\,\hat{\mathbf{x}} + \left(\frac{1}{2} + x_6\right)a\,\hat{\mathbf{y}} + \left(\frac{1}{2} - y_6\right)a\,\hat{\mathbf{z}}$ | (12$b$) | Cu II |
| $\mathbf{B}_{29}$ | = | $y_6\,\mathbf{a}_1 + z_6\,\mathbf{a}_2 + x_6\,\mathbf{a}_3$ | = | $y_6 a\,\hat{\mathbf{x}} + z_6 a\,\hat{\mathbf{y}} + x_6 a\,\hat{\mathbf{z}}$ | (12$b$) | Cu II |
| $\mathbf{B}_{30}$ | = | $-y_6\,\mathbf{a}_1 + \left(\frac{1}{2} + z_6\right)\mathbf{a}_2 + \left(\frac{1}{2} - x_6\right)\mathbf{a}_3$ | = | $-y_6 a\,\hat{\mathbf{x}} + \left(\frac{1}{2} + z_6\right)a\,\hat{\mathbf{y}} + \left(\frac{1}{2} - x_6\right)a\,\hat{\mathbf{z}}$ | (12$b$) | Cu II |
| $\mathbf{B}_{31}$ | = | $\left(\frac{1}{2} + y_6\right)\mathbf{a}_1 + \left(\frac{1}{2} - z_6\right)\mathbf{a}_2 - x_6\,\mathbf{a}_3$ | = | $\left(\frac{1}{2} + y_6\right)a\,\hat{\mathbf{x}} + \left(\frac{1}{2} - z_6\right)a\,\hat{\mathbf{y}} - x_6 a\,\hat{\mathbf{z}}$ | (12$b$) | Cu II |
| $\mathbf{B}_{32}$ | = | $\left(\frac{1}{2} - y_6\right)\mathbf{a}_1 - z_6\,\mathbf{a}_2 + \left(\frac{1}{2} + x_6\right)\mathbf{a}_3$ | = | $\left(\frac{1}{2} - y_6\right)a\,\hat{\mathbf{x}} - z_6 a\,\hat{\mathbf{y}} + \left(\frac{1}{2} + x_6\right)a\,\hat{\mathbf{z}}$ | (12$b$) | Cu II |
| $\mathbf{B}_{33}$ | = | $x_7\,\mathbf{a}_1 + y_7\,\mathbf{a}_2 + z_7\,\mathbf{a}_3$ | = | $x_7 a\,\hat{\mathbf{x}} + y_7 a\,\hat{\mathbf{y}} + z_7 a\,\hat{\mathbf{z}}$ | (12$b$) | O III |
| $\mathbf{B}_{34}$ | = | $\left(\frac{1}{2} - x_7\right)\mathbf{a}_1 - y_7\,\mathbf{a}_2 + \left(\frac{1}{2} + z_7\right)\mathbf{a}_3$ | = | $\left(\frac{1}{2} - x_7\right)a\,\hat{\mathbf{x}} - y_7 a\,\hat{\mathbf{y}} + \left(\frac{1}{2} + z_7\right)a\,\hat{\mathbf{z}}$ | (12$b$) | O III |
| $\mathbf{B}_{35}$ | = | $-x_7\,\mathbf{a}_1 + \left(\frac{1}{2} + y_7\right)\mathbf{a}_2 + \left(\frac{1}{2} - z_7\right)\mathbf{a}_3$ | = | $-x_7 a\,\hat{\mathbf{x}} + \left(\frac{1}{2} + y_7\right)a\,\hat{\mathbf{y}} + \left(\frac{1}{2} - z_7\right)a\,\hat{\mathbf{z}}$ | (12$b$) | O III |
| $\mathbf{B}_{36}$ | = | $\left(\frac{1}{2} + x_7\right)\mathbf{a}_1 + \left(\frac{1}{2} - y_7\right)\mathbf{a}_2 - z_7\,\mathbf{a}_3$ | = | $\left(\frac{1}{2} + x_7\right)a\,\hat{\mathbf{x}} + \left(\frac{1}{2} - y_7\right)a\,\hat{\mathbf{y}} - z_7 a\,\hat{\mathbf{z}}$ | (12$b$) | O III |
| $\mathbf{B}_{37}$ | = | $z_7\,\mathbf{a}_1 + x_7\,\mathbf{a}_2 + y_7\,\mathbf{a}_3$ | = | $z_7 a\,\hat{\mathbf{x}} + x_7 a\,\hat{\mathbf{y}} + y_7 a\,\hat{\mathbf{z}}$ | (12$b$) | O III |
| $\mathbf{B}_{38}$ | = | $\left(\frac{1}{2}+z_7\right)\mathbf{a}_1+\left(\frac{1}{2}-x_7\right)\mathbf{a}_2-y_7\,\mathbf{a}_3$ | = | $\left(\frac{1}{2}+z_7\right)a\,\hat{\mathbf{x}}+\left(\frac{1}{2}-x_7\right)a\,\hat{\mathbf{y}}-y_7a\,\hat{\mathbf{z}}$ | (12$b$) | O III |
| $\mathbf{B}_{39}$ | = | $\left(\frac{1}{2}-z_7\right)\mathbf{a}_1-x_7\,\mathbf{a}_2+\left(\frac{1}{2}+y_7\right)\mathbf{a}_3$ | = | $\left(\frac{1}{2}-z_7\right)a\,\hat{\mathbf{x}}-x_7a\,\hat{\mathbf{y}}+\left(\frac{1}{2}+y_7\right)a\,\hat{\mathbf{z}}$ | (12$b$) | O III |
| $\mathbf{B}_{40}$ | = | $-z_7\,\mathbf{a}_1+\left(\frac{1}{2}+x_7\right)\mathbf{a}_2+\left(\frac{1}{2}-y_7\right)\mathbf{a}_3$ | = | $-z_7a\,\hat{\mathbf{x}}+\left(\frac{1}{2}+x_7\right)a\,\hat{\mathbf{y}}+\left(\frac{1}{2}-y_7\right)a\,\hat{\mathbf{z}}$ | (12$b$) | O III |
| $\mathbf{B}_{41}$ | = | $y_7\,\mathbf{a}_1+z_7\,\mathbf{a}_2+x_7\,\mathbf{a}_3$ | = | $y_7a\,\hat{\mathbf{x}}+z_7a\,\hat{\mathbf{y}}+x_7a\,\hat{\mathbf{z}}$ | (12$b$) | O III |
| $\mathbf{B}_{42}$ | = | $-y_7\,\mathbf{a}_1+\left(\frac{1}{2}+z_7\right)\mathbf{a}_2+\left(\frac{1}{2}-x_7\right)\mathbf{a}_3$ | = | $-y_7a\,\hat{\mathbf{x}}+\left(\frac{1}{2}+z_7\right)a\,\hat{\mathbf{y}}+\left(\frac{1}{2}-x_7\right)a\,\hat{\mathbf{z}}$ | (12$b$) | O III |
| $\mathbf{B}_{43}$ | = | $\left(\frac{1}{2}+y_7\right)\mathbf{a}_1+\left(\frac{1}{2}-z_7\right)\mathbf{a}_2-x_7\,\mathbf{a}_3$ | = | $\left(\frac{1}{2}+y_7\right)a\,\hat{\mathbf{x}}+\left(\frac{1}{2}-z_7\right)a\,\hat{\mathbf{y}}-x_7a\,\hat{\mathbf{z}}$ | (12$b$) | O III |
| $\mathbf{B}_{44}$ | = | $\left(\frac{1}{2}-y_7\right)\mathbf{a}_1-z_7\,\mathbf{a}_2+\left(\frac{1}{2}+x_7\right)\mathbf{a}_3$ | = | $\left(\frac{1}{2}-y_7\right)a\,\hat{\mathbf{x}}-z_7a\,\hat{\mathbf{y}}+\left(\frac{1}{2}+x_7\right)a\,\hat{\mathbf{z}}$ | (12$b$) | O III |
| $\mathbf{B}_{45}$ | = | $x_8\,\mathbf{a}_1+y_8\,\mathbf{a}_2+z_8\,\mathbf{a}_3$ | = | $x_8a\,\hat{\mathbf{x}}+y_8a\,\hat{\mathbf{y}}+z_8a\,\hat{\mathbf{z}}$ | (12$b$) | O IV |
| $\mathbf{B}_{46}$ | = | $\left(\frac{1}{2}-x_8\right)\mathbf{a}_1-y_8\,\mathbf{a}_2+\left(\frac{1}{2}+z_8\right)\mathbf{a}_3$ | = | $\left(\frac{1}{2}-x_8\right)a\,\hat{\mathbf{x}}-y_8a\,\hat{\mathbf{y}}+\left(\frac{1}{2}+z_8\right)a\,\hat{\mathbf{z}}$ | (12$b$) | O IV |
| $\mathbf{B}_{47}$ | = | $-x_8\,\mathbf{a}_1+\left(\frac{1}{2}+y_8\right)\mathbf{a}_2+\left(\frac{1}{2}-z_8\right)\mathbf{a}_3$ | = | $-x_8a\,\hat{\mathbf{x}}+\left(\frac{1}{2}+y_8\right)a\,\hat{\mathbf{y}}+\left(\frac{1}{2}-z_8\right)a\,\hat{\mathbf{z}}$ | (12$b$) | O IV |
| $\mathbf{B}_{48}$ | = | $\left(\frac{1}{2}+x_8\right)\mathbf{a}_1+\left(\frac{1}{2}-y_8\right)\mathbf{a}_2-z_8\,\mathbf{a}_3$ | = | $\left(\frac{1}{2}+x_8\right)a\,\hat{\mathbf{x}}+\left(\frac{1}{2}-y_8\right)a\,\hat{\mathbf{y}}-z_8a\,\hat{\mathbf{z}}$ | (12$b$) | O IV |
| $\mathbf{B}_{49}$ | = | $z_8\,\mathbf{a}_1+x_8\,\mathbf{a}_2+y_8\,\mathbf{a}_3$ | = | $z_8a\,\hat{\mathbf{x}}+x_8a\,\hat{\mathbf{y}}+y_8a\,\hat{\mathbf{z}}$ | (12$b$) | O IV |
| $\mathbf{B}_{50}$ | = | $\left(\frac{1}{2}+z_8\right)\mathbf{a}_1+\left(\frac{1}{2}-x_8\right)\mathbf{a}_2-y_8\,\mathbf{a}_3$ | = | $\left(\frac{1}{2}+z_8\right)a\,\hat{\mathbf{x}}+\left(\frac{1}{2}-x_8\right)a\,\hat{\mathbf{y}}-y_8a\,\hat{\mathbf{z}}$ | (12$b$) | O IV |
| $\mathbf{B}_{51}$ | = | $\left(\frac{1}{2}-z_8\right)\mathbf{a}_1-x_8\,\mathbf{a}_2+\left(\frac{1}{2}+y_8\right)\mathbf{a}_3$ | = | $\left(\frac{1}{2}-z_8\right)a\,\hat{\mathbf{x}}-x_8a\,\hat{\mathbf{y}}+\left(\frac{1}{2}+y_8\right)a\,\hat{\mathbf{z}}$ | (12$b$) | O IV |
| $\mathbf{B}_{52}$ | = | $-z_8\,\mathbf{a}_1+\left(\frac{1}{2}+x_8\right)\mathbf{a}_2+\left(\frac{1}{2}-y_8\right)\mathbf{a}_3$ | = | $-z_8a\,\hat{\mathbf{x}}+\left(\frac{1}{2}+x_8\right)a\,\hat{\mathbf{y}}+\left(\frac{1}{2}-y_8\right)a\,\hat{\mathbf{z}}$ | (12$b$) | O IV |
| $\mathbf{B}_{53}$ | = | $y_8\,\mathbf{a}_1+z_8\,\mathbf{a}_2+x_8\,\mathbf{a}_3$ | = | $y_8a\,\hat{\mathbf{x}}+z_8a\,\hat{\mathbf{y}}+x_8a\,\hat{\mathbf{z}}$ | (12$b$) | O IV |
| $\mathbf{B}_{54}$ | = | $-y_8\,\mathbf{a}_1+\left(\frac{1}{2}+z_8\right)\mathbf{a}_2+\left(\frac{1}{2}-x_8\right)\mathbf{a}_3$ | = | $-y_8a\,\hat{\mathbf{x}}+\left(\frac{1}{2}+z_8\right)a\,\hat{\mathbf{y}}+\left(\frac{1}{2}-x_8\right)a\,\hat{\mathbf{z}}$ | (12$b$) | O IV |
| $\mathbf{B}_{55}$ | = | $\left(\frac{1}{2}+y_8\right)\mathbf{a}_1+\left(\frac{1}{2}-z_8\right)\mathbf{a}_2-x_8\,\mathbf{a}_3$ | = | $\left(\frac{1}{2}+y_8\right)a\,\hat{\mathbf{x}}+\left(\frac{1}{2}-z_8\right)a\,\hat{\mathbf{y}}-x_8a\,\hat{\mathbf{z}}$ | (12$b$) | O IV |
| $\mathbf{B}_{56}$ | = | $\left(\frac{1}{2}-y_8\right)\mathbf{a}_1-z_8\,\mathbf{a}_2+\left(\frac{1}{2}+x_8\right)\mathbf{a}_3$ | = | $\left(\frac{1}{2}-y_8\right)a\,\hat{\mathbf{x}}-z_8a\,\hat{\mathbf{y}}+\left(\frac{1}{2}+x_8\right)a\,\hat{\mathbf{z}}$ | (12$b$) | O IV |

**Geometry files:**

- CIF: pp. 1772
- POSCAR: pp. 1773

# $Na_2CaSiO_4$ ($S6_6$) Structure: AB2C4D_cP32_198_a_2a_ab_a

http://aflow.org/prototype-encyclopedia/AB2C4D_cP32_198_a_2a_ab_a

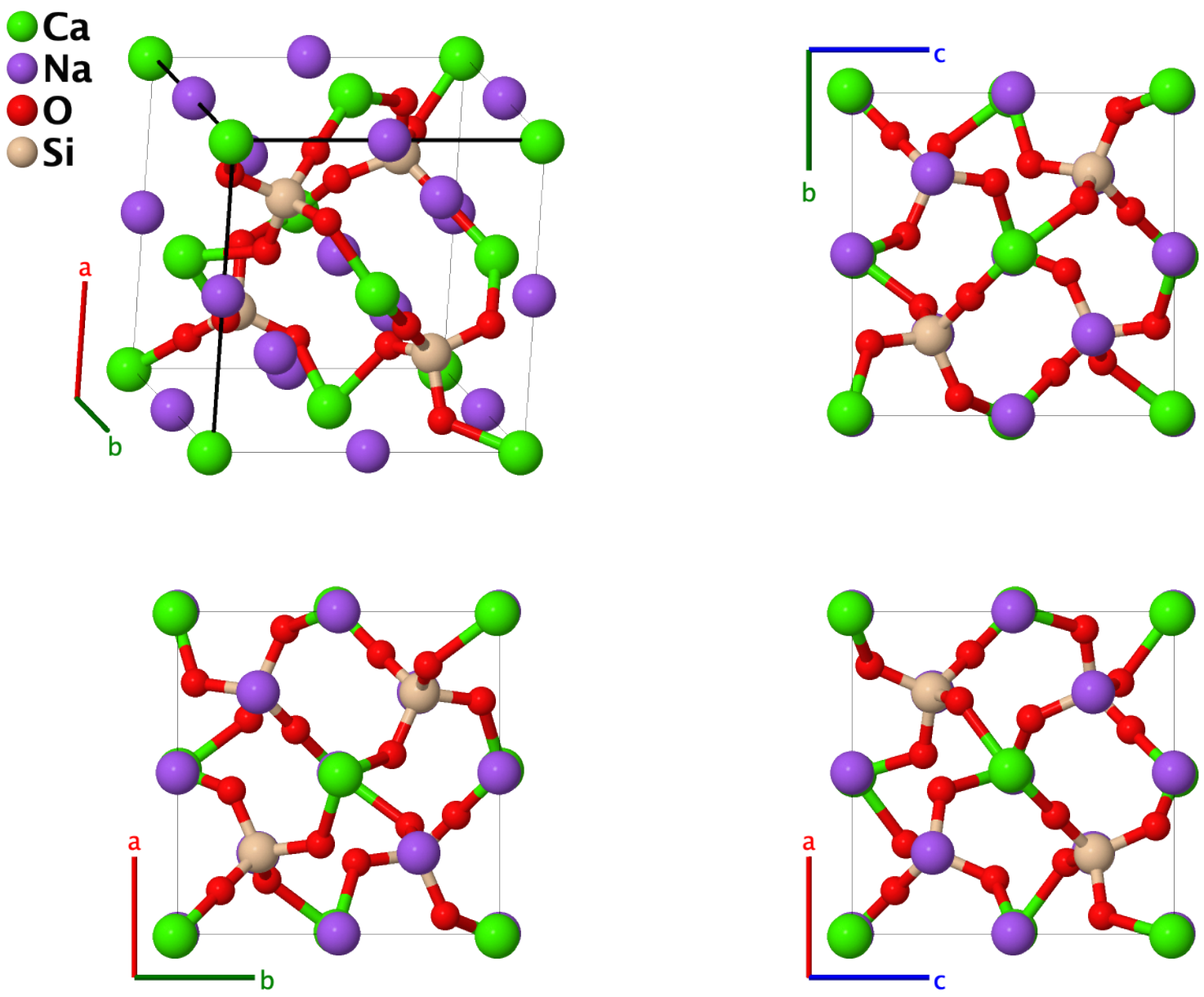

| | | |
|---|---|---|
| **Prototype** | : | $CaNa_2O_4Si$ |
| **AFLOW prototype label** | : | AB2C4D_cP32_198_a_2a_ab_a |
| ***Strukturbericht* designation** | : | $S6_6$ |
| **Pearson symbol** | : | cP32 |
| **Space group number** | : | 198 |
| **Space group symbol** | : | $P2_13$ |
| **AFLOW prototype command** | : | `aflow --proto=AB2C4D_cP32_198_a_2a_ab_a` `--params=`$a, x_1, x_2, x_3, x_4, x_5, x_6, y_6, z_6$ |

### Other compounds with this structure

- $Li_2SrSi_4$

- Substitution of small fractions of rare earth elements onto the Ca site can produce photoluminescence.
- This is the crystal structure found by (Barth, 1932). The coordinates on pp. 159 of (Hermann, 1937) are incorrect, as they do not correctly transform from Barth and Posnjak's coordinates in degrees to fractional coordinates. See (Hermann, 1937) pp. 557-8 for the correct coordinates.
- (Dollase, 1991) found that the $SiO_4$ tetrahedra were orientationally disordered, and that the resulting $Na_2Ca(SiO_4)$ structure took on the Heusler ($L1_2$) structure.

## Simple Cubic primitive vectors:

$$\mathbf{a}_1 = a\,\hat{\mathbf{x}}$$
$$\mathbf{a}_2 = a\,\hat{\mathbf{y}}$$
$$\mathbf{a}_3 = a\,\hat{\mathbf{z}}$$

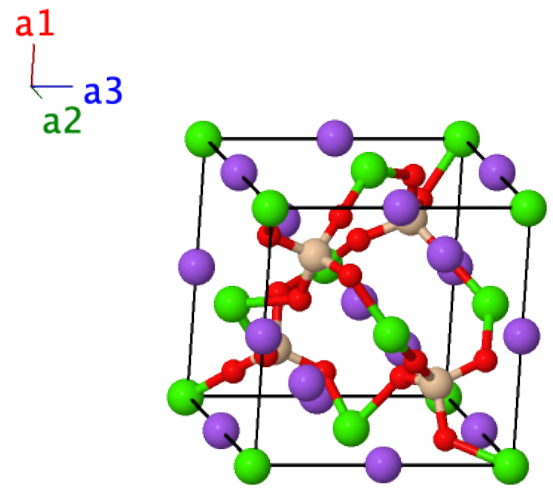

## Basis vectors:

| | | Lattice Coordinates | | Cartesian Coordinates | Wyckoff Position | Atom Type |
|---|---|---|---|---|---|---|
| $\mathbf{B}_1$ | = | $x_1\,\mathbf{a}_1 + x_1\,\mathbf{a}_2 + x_1\,\mathbf{a}_3$ | = | $x_1 a\,\hat{\mathbf{x}} + x_1 a\,\hat{\mathbf{y}} + x_1 a\,\hat{\mathbf{z}}$ | (4$a$) | Ca |
| $\mathbf{B}_2$ | = | $\left(\frac{1}{2} - x_1\right)\mathbf{a}_1 - x_1\,\mathbf{a}_2 + \left(\frac{1}{2} + x_1\right)\mathbf{a}_3$ | = | $\left(\frac{1}{2} - x_1\right)a\,\hat{\mathbf{x}} - x_1 a\,\hat{\mathbf{y}} + \left(\frac{1}{2} + x_1\right)a\,\hat{\mathbf{z}}$ | (4$a$) | Ca |
| $\mathbf{B}_3$ | = | $-x_1\,\mathbf{a}_1 + \left(\frac{1}{2} + x_1\right)\mathbf{a}_2 + \left(\frac{1}{2} - x_1\right)\mathbf{a}_3$ | = | $-x_1 a\,\hat{\mathbf{x}} + \left(\frac{1}{2} + x_1\right)a\,\hat{\mathbf{y}} + \left(\frac{1}{2} - x_1\right)a\,\hat{\mathbf{z}}$ | (4$a$) | Ca |
| $\mathbf{B}_4$ | = | $\left(\frac{1}{2} + x_1\right)\mathbf{a}_1 + \left(\frac{1}{2} - x_1\right)\mathbf{a}_2 - x_1\,\mathbf{a}_3$ | = | $\left(\frac{1}{2} + x_1\right)a\,\hat{\mathbf{x}} + \left(\frac{1}{2} - x_1\right)a\,\hat{\mathbf{y}} - x_1 a\,\hat{\mathbf{z}}$ | (4$a$) | Ca |
| $\mathbf{B}_5$ | = | $x_2\,\mathbf{a}_1 + x_2\,\mathbf{a}_2 + x_2\,\mathbf{a}_3$ | = | $x_2 a\,\hat{\mathbf{x}} + x_2 a\,\hat{\mathbf{y}} + x_2 a\,\hat{\mathbf{z}}$ | (4$a$) | Na I |
| $\mathbf{B}_6$ | = | $\left(\frac{1}{2} - x_2\right)\mathbf{a}_1 - x_2\,\mathbf{a}_2 + \left(\frac{1}{2} + x_2\right)\mathbf{a}_3$ | = | $\left(\frac{1}{2} - x_2\right)a\,\hat{\mathbf{x}} - x_2 a\,\hat{\mathbf{y}} + \left(\frac{1}{2} + x_2\right)a\,\hat{\mathbf{z}}$ | (4$a$) | Na I |
| $\mathbf{B}_7$ | = | $-x_2\,\mathbf{a}_1 + \left(\frac{1}{2} + x_2\right)\mathbf{a}_2 + \left(\frac{1}{2} - x_2\right)\mathbf{a}_3$ | = | $-x_2 a\,\hat{\mathbf{x}} + \left(\frac{1}{2} + x_2\right)a\,\hat{\mathbf{y}} + \left(\frac{1}{2} - x_2\right)a\,\hat{\mathbf{z}}$ | (4$a$) | Na I |
| $\mathbf{B}_8$ | = | $\left(\frac{1}{2} + x_2\right)\mathbf{a}_1 + \left(\frac{1}{2} - x_2\right)\mathbf{a}_2 - x_2\,\mathbf{a}_3$ | = | $\left(\frac{1}{2} + x_2\right)a\,\hat{\mathbf{x}} + \left(\frac{1}{2} - x_2\right)a\,\hat{\mathbf{y}} - x_2 a\,\hat{\mathbf{z}}$ | (4$a$) | Na I |
| $\mathbf{B}_9$ | = | $x_3\,\mathbf{a}_1 + x_3\,\mathbf{a}_2 + x_3\,\mathbf{a}_3$ | = | $x_3 a\,\hat{\mathbf{x}} + x_3 a\,\hat{\mathbf{y}} + x_3 a\,\hat{\mathbf{z}}$ | (4$a$) | Na II |
| $\mathbf{B}_{10}$ | = | $\left(\frac{1}{2} - x_3\right)\mathbf{a}_1 - x_3\,\mathbf{a}_2 + \left(\frac{1}{2} + x_3\right)\mathbf{a}_3$ | = | $\left(\frac{1}{2} - x_3\right)a\,\hat{\mathbf{x}} - x_3 a\,\hat{\mathbf{y}} + \left(\frac{1}{2} + x_3\right)a\,\hat{\mathbf{z}}$ | (4$a$) | Na II |
| $\mathbf{B}_{11}$ | = | $-x_3\,\mathbf{a}_1 + \left(\frac{1}{2} + x_3\right)\mathbf{a}_2 + \left(\frac{1}{2} - x_3\right)\mathbf{a}_3$ | = | $-x_3 a\,\hat{\mathbf{x}} + \left(\frac{1}{2} + x_3\right)a\,\hat{\mathbf{y}} + \left(\frac{1}{2} - x_3\right)a\,\hat{\mathbf{z}}$ | (4$a$) | Na II |
| $\mathbf{B}_{12}$ | = | $\left(\frac{1}{2} + x_3\right)\mathbf{a}_1 + \left(\frac{1}{2} - x_3\right)\mathbf{a}_2 - x_3\,\mathbf{a}_3$ | = | $\left(\frac{1}{2} + x_3\right)a\,\hat{\mathbf{x}} + \left(\frac{1}{2} - x_3\right)a\,\hat{\mathbf{y}} - x_3 a\,\hat{\mathbf{z}}$ | (4$a$) | Na II |
| $\mathbf{B}_{13}$ | = | $x_4\,\mathbf{a}_1 + x_4\,\mathbf{a}_2 + x_4\,\mathbf{a}_3$ | = | $x_4 a\,\hat{\mathbf{x}} + x_4 a\,\hat{\mathbf{y}} + x_4 a\,\hat{\mathbf{z}}$ | (4$a$) | O I |
| $\mathbf{B}_{14}$ | = | $\left(\frac{1}{2} - x_4\right)\mathbf{a}_1 - x_4\,\mathbf{a}_2 + \left(\frac{1}{2} + x_4\right)\mathbf{a}_3$ | = | $\left(\frac{1}{2} - x_4\right)a\,\hat{\mathbf{x}} - x_4 a\,\hat{\mathbf{y}} + \left(\frac{1}{2} + x_4\right)a\,\hat{\mathbf{z}}$ | (4$a$) | O I |
| $\mathbf{B}_{15}$ | = | $-x_4\,\mathbf{a}_1 + \left(\frac{1}{2} + x_4\right)\mathbf{a}_2 + \left(\frac{1}{2} - x_4\right)\mathbf{a}_3$ | = | $-x_4 a\,\hat{\mathbf{x}} + \left(\frac{1}{2} + x_4\right)a\,\hat{\mathbf{y}} + \left(\frac{1}{2} - x_4\right)a\,\hat{\mathbf{z}}$ | (4$a$) | O I |
| $\mathbf{B}_{16}$ | = | $\left(\frac{1}{2} + x_4\right)\mathbf{a}_1 + \left(\frac{1}{2} - x_4\right)\mathbf{a}_2 - x_4\,\mathbf{a}_3$ | = | $\left(\frac{1}{2} + x_4\right)a\,\hat{\mathbf{x}} + \left(\frac{1}{2} - x_4\right)a\,\hat{\mathbf{y}} - x_4 a\,\hat{\mathbf{z}}$ | (4$a$) | O I |
| $\mathbf{B}_{17}$ | = | $x_5\,\mathbf{a}_1 + x_5\,\mathbf{a}_2 + x_5\,\mathbf{a}_3$ | = | $x_5 a\,\hat{\mathbf{x}} + x_5 a\,\hat{\mathbf{y}} + x_5 a\,\hat{\mathbf{z}}$ | (4$a$) | Si |
| $\mathbf{B}_{18}$ | = | $\left(\frac{1}{2} - x_5\right)\mathbf{a}_1 - x_5\,\mathbf{a}_2 + \left(\frac{1}{2} + x_5\right)\mathbf{a}_3$ | = | $\left(\frac{1}{2} - x_5\right)a\,\hat{\mathbf{x}} - x_5 a\,\hat{\mathbf{y}} + \left(\frac{1}{2} + x_5\right)a\,\hat{\mathbf{z}}$ | (4$a$) | Si |
| $\mathbf{B}_{19}$ | = | $-x_5\,\mathbf{a}_1 + \left(\frac{1}{2} + x_5\right)\mathbf{a}_2 + \left(\frac{1}{2} - x_5\right)\mathbf{a}_3$ | = | $-x_5 a\,\hat{\mathbf{x}} + \left(\frac{1}{2} + x_5\right)a\,\hat{\mathbf{y}} + \left(\frac{1}{2} - x_5\right)a\,\hat{\mathbf{z}}$ | (4$a$) | Si |
| $\mathbf{B}_{20}$ | = | $\left(\frac{1}{2} + x_5\right)\mathbf{a}_1 + \left(\frac{1}{2} - x_5\right)\mathbf{a}_2 - x_5\,\mathbf{a}_3$ | = | $\left(\frac{1}{2} + x_5\right)a\,\hat{\mathbf{x}} + \left(\frac{1}{2} - x_5\right)a\,\hat{\mathbf{y}} - x_5 a\,\hat{\mathbf{z}}$ | (4$a$) | Si |
| $\mathbf{B}_{21}$ | = | $x_6\,\mathbf{a}_1 + y_6\,\mathbf{a}_2 + z_6\,\mathbf{a}_3$ | = | $x_6 a\,\hat{\mathbf{x}} + y_6 a\,\hat{\mathbf{y}} + z_6 a\,\hat{\mathbf{z}}$ | (12$b$) | O II |
| $\mathbf{B}_{22}$ | = | $\left(\frac{1}{2} - x_6\right)\mathbf{a}_1 - y_6\,\mathbf{a}_2 + \left(\frac{1}{2} + z_6\right)\mathbf{a}_3$ | = | $\left(\frac{1}{2} - x_6\right)a\,\hat{\mathbf{x}} - y_6 a\,\hat{\mathbf{y}} + \left(\frac{1}{2} + z_6\right)a\,\hat{\mathbf{z}}$ | (12$b$) | O II |
| $\mathbf{B}_{23}$ | = | $-x_6\,\mathbf{a}_1 + \left(\frac{1}{2} + y_6\right)\mathbf{a}_2 + \left(\frac{1}{2} - z_6\right)\mathbf{a}_3$ | = | $-x_6 a\,\hat{\mathbf{x}} + \left(\frac{1}{2} + y_6\right)a\,\hat{\mathbf{y}} + \left(\frac{1}{2} - z_6\right)a\,\hat{\mathbf{z}}$ | (12$b$) | O II |
| $\mathbf{B}_{24}$ | = | $\left(\frac{1}{2} + x_6\right)\mathbf{a}_1 + \left(\frac{1}{2} - y_6\right)\mathbf{a}_2 - z_6\,\mathbf{a}_3$ | = | $\left(\frac{1}{2} + x_6\right)a\,\hat{\mathbf{x}} + \left(\frac{1}{2} - y_6\right)a\,\hat{\mathbf{y}} - z_6 a\,\hat{\mathbf{z}}$ | (12$b$) | O II |
| $\mathbf{B}_{25}$ | = | $z_6\,\mathbf{a}_1 + x_6\,\mathbf{a}_2 + y_6\,\mathbf{a}_3$ | = | $z_6 a\,\hat{\mathbf{x}} + x_6 a\,\hat{\mathbf{y}} + y_6 a\,\hat{\mathbf{z}}$ | (12$b$) | O II |
| $\mathbf{B}_{26}$ | = | $\left(\frac{1}{2} + z_6\right)\mathbf{a}_1 + \left(\frac{1}{2} - x_6\right)\mathbf{a}_2 - y_6\,\mathbf{a}_3$ | = | $\left(\frac{1}{2} + z_6\right)a\,\hat{\mathbf{x}} + \left(\frac{1}{2} - x_6\right)a\,\hat{\mathbf{y}} - y_6 a\,\hat{\mathbf{z}}$ | (12$b$) | O II |
| $\mathbf{B}_{27}$ | = | $\left(\frac{1}{2} - z_6\right)\mathbf{a}_1 - x_6\,\mathbf{a}_2 + \left(\frac{1}{2} + y_6\right)\mathbf{a}_3$ | = | $\left(\frac{1}{2} - z_6\right)a\,\hat{\mathbf{x}} - x_6 a\,\hat{\mathbf{y}} + \left(\frac{1}{2} + y_6\right)a\,\hat{\mathbf{z}}$ | (12$b$) | O II |

$$
\begin{array}{lllll}
\mathbf{B}_{28} & = & -z_6\,\mathbf{a}_1 + \left(\frac{1}{2} + x_6\right)\,\mathbf{a}_2 + \left(\frac{1}{2} - y_6\right)\,\mathbf{a}_3 & = & -z_6 a\,\hat{\mathbf{x}} + \left(\frac{1}{2} + x_6\right) a\,\hat{\mathbf{y}} + \left(\frac{1}{2} - y_6\right) a\,\hat{\mathbf{z}} & (12b) & \text{O II} \\
\mathbf{B}_{29} & = & y_6\,\mathbf{a}_1 + z_6\,\mathbf{a}_2 + x_6\,\mathbf{a}_3 & = & y_6 a\,\hat{\mathbf{x}} + z_6 a\,\hat{\mathbf{y}} + x_6 a\,\hat{\mathbf{z}} & (12b) & \text{O II} \\
\mathbf{B}_{30} & = & -y_6\,\mathbf{a}_1 + \left(\frac{1}{2} + z_6\right)\,\mathbf{a}_2 + \left(\frac{1}{2} - x_6\right)\,\mathbf{a}_3 & = & -y_6 a\,\hat{\mathbf{x}} + \left(\frac{1}{2} + z_6\right) a\,\hat{\mathbf{y}} + \left(\frac{1}{2} - x_6\right) a\,\hat{\mathbf{z}} & (12b) & \text{O II} \\
\mathbf{B}_{31} & = & \left(\frac{1}{2} + y_6\right)\,\mathbf{a}_1 + \left(\frac{1}{2} - z_6\right)\,\mathbf{a}_2 - x_6\,\mathbf{a}_3 & = & \left(\frac{1}{2} + y_6\right) a\,\hat{\mathbf{x}} + \left(\frac{1}{2} - z_6\right) a\,\hat{\mathbf{y}} - x_6 a\,\hat{\mathbf{z}} & (12b) & \text{O II} \\
\mathbf{B}_{32} & = & \left(\frac{1}{2} - y_6\right)\,\mathbf{a}_1 - z_6\,\mathbf{a}_2 + \left(\frac{1}{2} + x_6\right)\,\mathbf{a}_3 & = & \left(\frac{1}{2} - y_6\right) a\,\hat{\mathbf{x}} - z_6 a\,\hat{\mathbf{y}} + \left(\frac{1}{2} + x_6\right) a\,\hat{\mathbf{z}} & (12b) & \text{O II}
\end{array}
$$

**Geometry files:**

- CIF: pp. 1773
- POSCAR: pp. 1773

# $\alpha$-Carnegieite ($NaAlSiO_4$, $S6_5$) Structure: ABC4D_cP28_198_a_a_ab_a

http://aflow.org/prototype-encyclopedia/ABC4D_cP28_198_a_a_ab_a

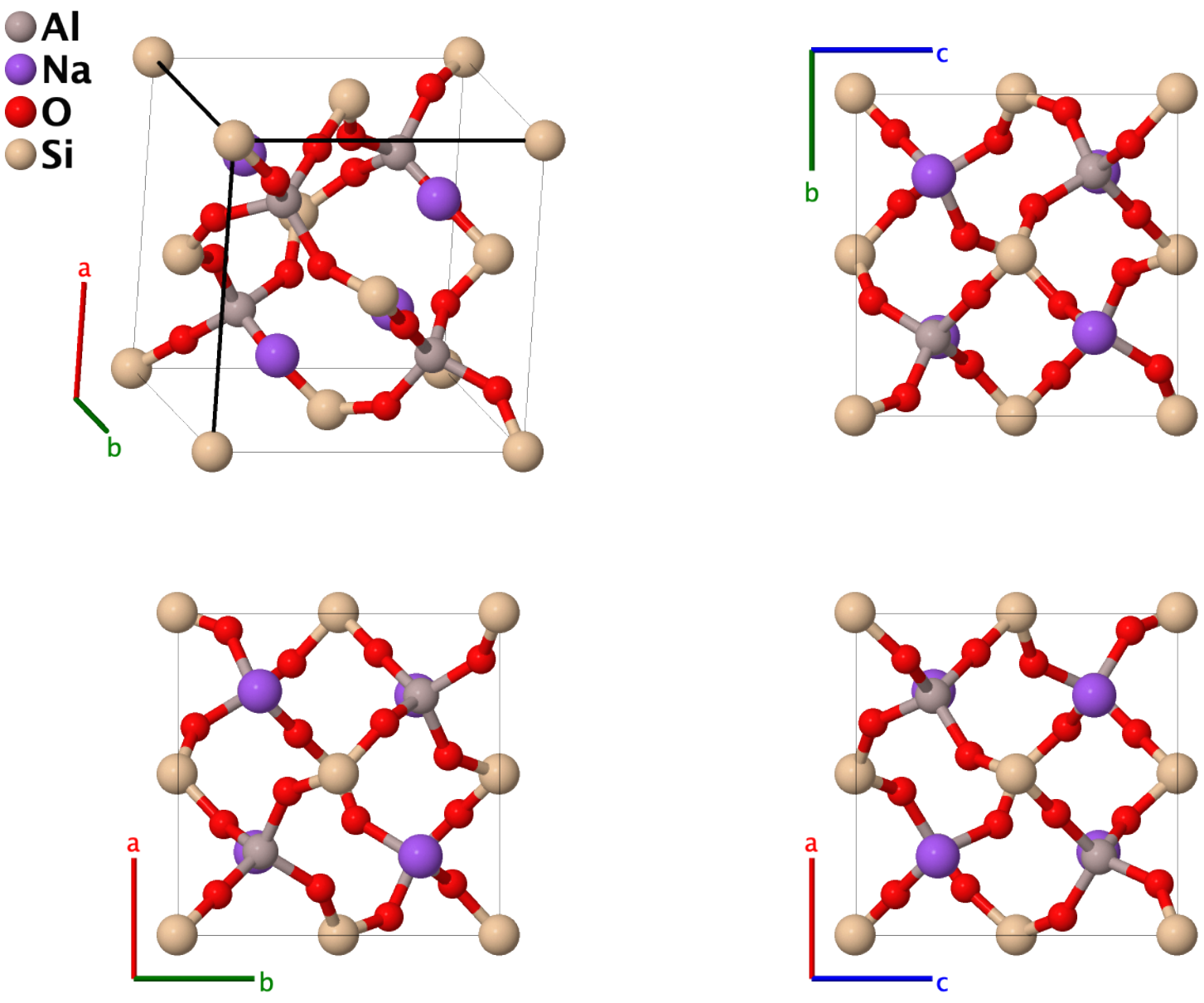

| | | |
|---|---|---|
| **Prototype** | : | $AlNaO_4Si$ |
| **AFLOW prototype label** | : | ABC4D_cP28_198_a_a_ab_a |
| ***Strukturbericht* designation** | : | $S6_5$ |
| **Pearson symbol** | : | cP28 |
| **Space group number** | : | 198 |
| **Space group symbol** | : | $P2_13$ |
| **AFLOW prototype command** | : | `aflow --proto=ABC4D_cP28_198_a_a_ab_a --params=`$a, x_1, x_2, x_3, x_4, x_5, y_5, z_5$ |

- This high-temperature form of carnegieite is stable above 970 K. To our knowledge, the atomic positions of the low temperature $\beta$-Carnegieite structure have not been determined.

**Simple Cubic primitive vectors:**

$$\mathbf{a}_1 = a\,\hat{\mathbf{x}}$$
$$\mathbf{a}_2 = a\,\hat{\mathbf{y}}$$
$$\mathbf{a}_3 = a\,\hat{\mathbf{z}}$$

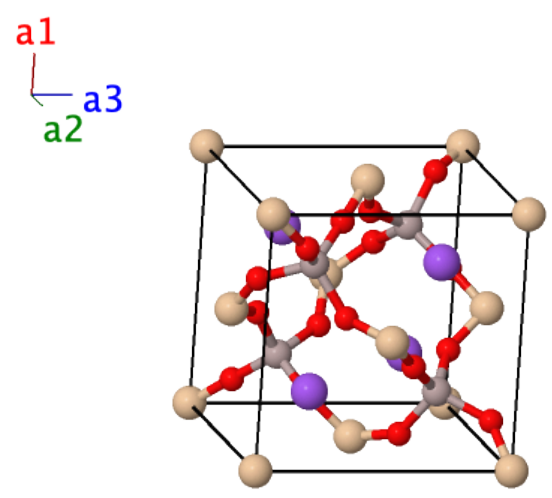

**Basis vectors:**

| | | Lattice Coordinates | | Cartesian Coordinates | Wyckoff Position | Atom Type |
|---|---|---|---|---|---|---|
| $\mathbf{B}_1$ | = | $x_1\,\mathbf{a}_1 + x_1\,\mathbf{a}_2 + x_1\,\mathbf{a}_3$ | = | $x_1 a\,\hat{\mathbf{x}} + x_1 a\,\hat{\mathbf{y}} + x_1 a\,\hat{\mathbf{z}}$ | (4$a$) | Al |
| $\mathbf{B}_2$ | = | $\left(\frac{1}{2} - x_1\right) \mathbf{a}_1 - x_1 \mathbf{a}_2 + \left(\frac{1}{2} + x_1\right) \mathbf{a}_3$ | = | $\left(\frac{1}{2} - x_1\right) a \hat{\mathbf{x}} - x_1 a \hat{\mathbf{y}} + \left(\frac{1}{2} + x_1\right) a \hat{\mathbf{z}}$ | (4$a$) | Al |
| $\mathbf{B}_3$ | = | $-x_1 \mathbf{a}_1 + \left(\frac{1}{2} + x_1\right) \mathbf{a}_2 + \left(\frac{1}{2} - x_1\right) \mathbf{a}_3$ | = | $-x_1 a \hat{\mathbf{x}} + \left(\frac{1}{2} + x_1\right) a \hat{\mathbf{y}} + \left(\frac{1}{2} - x_1\right) a \hat{\mathbf{z}}$ | (4$a$) | Al |
| $\mathbf{B}_4$ | = | $\left(\frac{1}{2} + x_1\right) \mathbf{a}_1 + \left(\frac{1}{2} - x_1\right) \mathbf{a}_2 - x_1 \mathbf{a}_3$ | = | $\left(\frac{1}{2} + x_1\right) a \hat{\mathbf{x}} + \left(\frac{1}{2} - x_1\right) a \hat{\mathbf{y}} - x_1 a \hat{\mathbf{z}}$ | (4$a$) | Al |
| $\mathbf{B}_5$ | = | $x_2 \mathbf{a}_1 + x_2 \mathbf{a}_2 + x_2 \mathbf{a}_3$ | = | $x_2 a \hat{\mathbf{x}} + x_2 a \hat{\mathbf{y}} + x_2 a \hat{\mathbf{z}}$ | (4$a$) | Na |
| $\mathbf{B}_6$ | = | $\left(\frac{1}{2} - x_2\right) \mathbf{a}_1 - x_2 \mathbf{a}_2 + \left(\frac{1}{2} + x_2\right) \mathbf{a}_3$ | = | $\left(\frac{1}{2} - x_2\right) a \hat{\mathbf{x}} - x_2 a \hat{\mathbf{y}} + \left(\frac{1}{2} + x_2\right) a \hat{\mathbf{z}}$ | (4$a$) | Na |
| $\mathbf{B}_7$ | = | $-x_2 \mathbf{a}_1 + \left(\frac{1}{2} + x_2\right) \mathbf{a}_2 + \left(\frac{1}{2} - x_2\right) \mathbf{a}_3$ | = | $-x_2 a \hat{\mathbf{x}} + \left(\frac{1}{2} + x_2\right) a \hat{\mathbf{y}} + \left(\frac{1}{2} - x_2\right) a \hat{\mathbf{z}}$ | (4$a$) | Na |
| $\mathbf{B}_8$ | = | $\left(\frac{1}{2} + x_2\right) \mathbf{a}_1 + \left(\frac{1}{2} - x_2\right) \mathbf{a}_2 - x_2 \mathbf{a}_3$ | = | $\left(\frac{1}{2} + x_2\right) a \hat{\mathbf{x}} + \left(\frac{1}{2} - x_2\right) a \hat{\mathbf{y}} - x_2 a \hat{\mathbf{z}}$ | (4$a$) | Na |
| $\mathbf{B}_9$ | = | $x_3 \mathbf{a}_1 + x_3 \mathbf{a}_2 + x_3 \mathbf{a}_3$ | = | $x_3 a \hat{\mathbf{x}} + x_3 a \hat{\mathbf{y}} + x_3 a \hat{\mathbf{z}}$ | (4$a$) | O I |
| $\mathbf{B}_{10}$ | = | $\left(\frac{1}{2} - x_3\right) \mathbf{a}_1 - x_3 \mathbf{a}_2 + \left(\frac{1}{2} + x_3\right) \mathbf{a}_3$ | = | $\left(\frac{1}{2} - x_3\right) a \hat{\mathbf{x}} - x_3 a \hat{\mathbf{y}} + \left(\frac{1}{2} + x_3\right) a \hat{\mathbf{z}}$ | (4$a$) | O I |
| $\mathbf{B}_{11}$ | = | $-x_3 \mathbf{a}_1 + \left(\frac{1}{2} + x_3\right) \mathbf{a}_2 + \left(\frac{1}{2} - x_3\right) \mathbf{a}_3$ | = | $-x_3 a \hat{\mathbf{x}} + \left(\frac{1}{2} + x_3\right) a \hat{\mathbf{y}} + \left(\frac{1}{2} - x_3\right) a \hat{\mathbf{z}}$ | (4$a$) | O I |
| $\mathbf{B}_{12}$ | = | $\left(\frac{1}{2} + x_3\right) \mathbf{a}_1 + \left(\frac{1}{2} - x_3\right) \mathbf{a}_2 - x_3 \mathbf{a}_3$ | = | $\left(\frac{1}{2} + x_3\right) a \hat{\mathbf{x}} + \left(\frac{1}{2} - x_3\right) a \hat{\mathbf{y}} - x_3 a \hat{\mathbf{z}}$ | (4$a$) | O I |
| $\mathbf{B}_{13}$ | = | $x_4 \mathbf{a}_1 + x_4 \mathbf{a}_2 + x_4 \mathbf{a}_3$ | = | $x_4 a \hat{\mathbf{x}} + x_4 a \hat{\mathbf{y}} + x_4 a \hat{\mathbf{z}}$ | (4$a$) | Si |
| $\mathbf{B}_{14}$ | = | $\left(\frac{1}{2} - x_4\right) \mathbf{a}_1 - x_4 \mathbf{a}_2 + \left(\frac{1}{2} + x_4\right) \mathbf{a}_3$ | = | $\left(\frac{1}{2} - x_4\right) a \hat{\mathbf{x}} - x_4 a \hat{\mathbf{y}} + \left(\frac{1}{2} + x_4\right) a \hat{\mathbf{z}}$ | (4$a$) | Si |
| $\mathbf{B}_{15}$ | = | $-x_4 \mathbf{a}_1 + \left(\frac{1}{2} + x_4\right) \mathbf{a}_2 + \left(\frac{1}{2} - x_4\right) \mathbf{a}_3$ | = | $-x_4 a \hat{\mathbf{x}} + \left(\frac{1}{2} + x_4\right) a \hat{\mathbf{y}} + \left(\frac{1}{2} - x_4\right) a \hat{\mathbf{z}}$ | (4$a$) | Si |
| $\mathbf{B}_{16}$ | = | $\left(\frac{1}{2} + x_4\right) \mathbf{a}_1 + \left(\frac{1}{2} - x_4\right) \mathbf{a}_2 - x_4 \mathbf{a}_3$ | = | $\left(\frac{1}{2} + x_4\right) a \hat{\mathbf{x}} + \left(\frac{1}{2} - x_4\right) a \hat{\mathbf{y}} - x_4 a \hat{\mathbf{z}}$ | (4$a$) | Si |
| $\mathbf{B}_{17}$ | = | $x_5 \mathbf{a}_1 + y_5 \mathbf{a}_2 + z_5 \mathbf{a}_3$ | = | $x_5 a \hat{\mathbf{x}} + y_5 a \hat{\mathbf{y}} + z_5 a \hat{\mathbf{z}}$ | (12$b$) | O II |
| $\mathbf{B}_{18}$ | = | $\left(\frac{1}{2} - x_5\right) \mathbf{a}_1 - y_5 \mathbf{a}_2 + \left(\frac{1}{2} + z_5\right) \mathbf{a}_3$ | = | $\left(\frac{1}{2} - x_5\right) a \hat{\mathbf{x}} - y_5 a \hat{\mathbf{y}} + \left(\frac{1}{2} + z_5\right) a \hat{\mathbf{z}}$ | (12$b$) | O II |
| $\mathbf{B}_{19}$ | = | $-x_5 \mathbf{a}_1 + \left(\frac{1}{2} + y_5\right) \mathbf{a}_2 + \left(\frac{1}{2} - z_5\right) \mathbf{a}_3$ | = | $-x_5 a \hat{\mathbf{x}} + \left(\frac{1}{2} + y_5\right) a \hat{\mathbf{y}} + \left(\frac{1}{2} - z_5\right) a \hat{\mathbf{z}}$ | (12$b$) | O II |
| $\mathbf{B}_{20}$ | = | $\left(\frac{1}{2} + x_5\right) \mathbf{a}_1 + \left(\frac{1}{2} - y_5\right) \mathbf{a}_2 - z_5 \mathbf{a}_3$ | = | $\left(\frac{1}{2} + x_5\right) a \hat{\mathbf{x}} + \left(\frac{1}{2} - y_5\right) a \hat{\mathbf{y}} - z_5 a \hat{\mathbf{z}}$ | (12$b$) | O II |
| $\mathbf{B}_{21}$ | = | $z_5 \mathbf{a}_1 + x_5 \mathbf{a}_2 + y_5 \mathbf{a}_3$ | = | $z_5 a \hat{\mathbf{x}} + x_5 a \hat{\mathbf{y}} + y_5 a \hat{\mathbf{z}}$ | (12$b$) | O II |
| $\mathbf{B}_{22}$ | = | $\left(\frac{1}{2} + z_5\right) \mathbf{a}_1 + \left(\frac{1}{2} - x_5\right) \mathbf{a}_2 - y_5 \mathbf{a}_3$ | = | $\left(\frac{1}{2} + z_5\right) a \hat{\mathbf{x}} + \left(\frac{1}{2} - x_5\right) a \hat{\mathbf{y}} - y_5 a \hat{\mathbf{z}}$ | (12$b$) | O II |
| $\mathbf{B}_{23}$ | = | $\left(\frac{1}{2} - z_5\right) \mathbf{a}_1 - x_5 \mathbf{a}_2 + \left(\frac{1}{2} + y_5\right) \mathbf{a}_3$ | = | $\left(\frac{1}{2} - z_5\right) a \hat{\mathbf{x}} - x_5 a \hat{\mathbf{y}} + \left(\frac{1}{2} + y_5\right) a \hat{\mathbf{z}}$ | (12$b$) | O II |
| $\mathbf{B}_{24}$ | = | $-z_5 \mathbf{a}_1 + \left(\frac{1}{2} + x_5\right) \mathbf{a}_2 + \left(\frac{1}{2} - y_5\right) \mathbf{a}_3$ | = | $-z_5 a \hat{\mathbf{x}} + \left(\frac{1}{2} + x_5\right) a \hat{\mathbf{y}} + \left(\frac{1}{2} - y_5\right) a \hat{\mathbf{z}}$ | (12$b$) | O II |
| $\mathbf{B}_{25}$ | = | $y_5 \mathbf{a}_1 + z_5 \mathbf{a}_2 + x_5 \mathbf{a}_3$ | = | $y_5 a \hat{\mathbf{x}} + z_5 a \hat{\mathbf{y}} + x_5 a \hat{\mathbf{z}}$ | (12$b$) | O II |
| $\mathbf{B}_{26}$ | = | $-y_5 \mathbf{a}_1 + \left(\frac{1}{2} + z_5\right) \mathbf{a}_2 + \left(\frac{1}{2} - x_5\right) \mathbf{a}_3$ | = | $-y_5 a \hat{\mathbf{x}} + \left(\frac{1}{2} + z_5\right) a \hat{\mathbf{y}} + \left(\frac{1}{2} - x_5\right) a \hat{\mathbf{z}}$ | (12$b$) | O II |
| $\mathbf{B}_{27}$ | = | $\left(\frac{1}{2} + y_5\right) \mathbf{a}_1 + \left(\frac{1}{2} - z_5\right) \mathbf{a}_2 - x_5 \mathbf{a}_3$ | = | $\left(\frac{1}{2} + y_5\right) a \hat{\mathbf{x}} + \left(\frac{1}{2} - z_5\right) a \hat{\mathbf{y}} - x_5 a \hat{\mathbf{z}}$ | (12$b$) | O II |
| $\mathbf{B}_{28}$ | = | $\left(\frac{1}{2} - y_5\right) \mathbf{a}_1 - z_5 \mathbf{a}_2 + \left(\frac{1}{2} + x_5\right) \mathbf{a}_3$ | = | $\left(\frac{1}{2} - y_5\right) a \hat{\mathbf{x}} - z_5 a \hat{\mathbf{y}} + \left(\frac{1}{2} + x_5\right) a \hat{\mathbf{z}}$ | (12$b$) | O II |

**Geometry files:**

- CIF: pp. 1774
- POSCAR: pp. 1774

# $C26_a$ ($NO_2$) (*obsolete*) Structure: AB2_cI36_199_b_c

http://aflow.org/prototype-encyclopedia/AB2_cI36_199_b_c

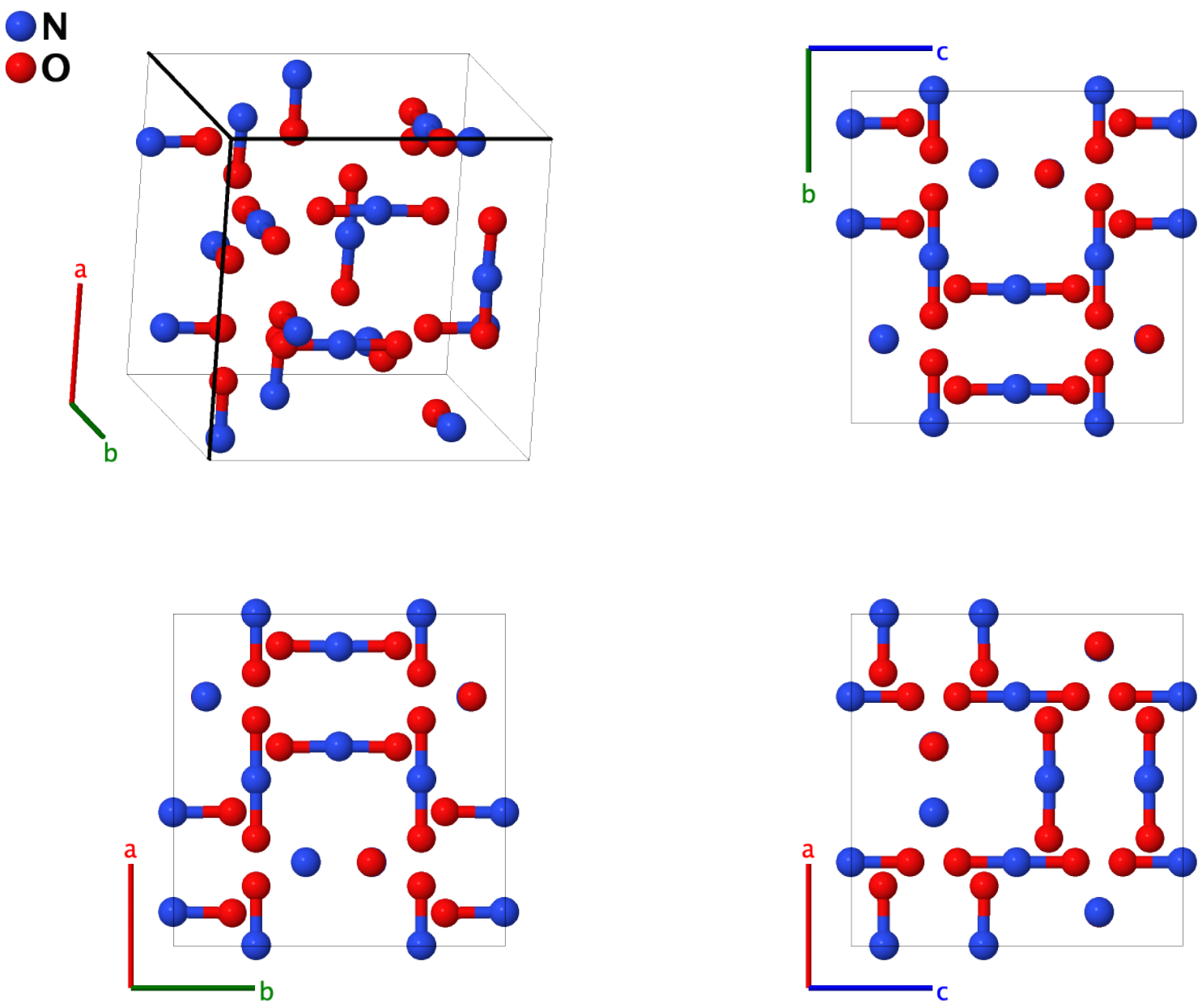

| | | |
|---|---|---|
| **Prototype** | : | $NO_2$ |
| **AFLOW prototype label** | : | AB2_cI36_199_b_c |
| ***Strukturbericht* designation** | : | $C26_a$ |
| **Pearson symbol** | : | cI36 |
| **Space group number** | : | 199 |
| **Space group symbol** | : | $I2_13$ |
| **AFLOW prototype command** | : | `aflow --proto=AB2_cI36_199_b_c --params=`$a, x_1, x_2, y_2, z_2$ |

- (Hermann, 1937) listed two possible structures for the low temperature solid cubic phase of $NO_2$, which were given *Strukturbericht* designations $C26_a$ and $C26_b$, the only structures with Roman subscripts in the original series.
- $C26_a$ (AB2_cI36_199_b_c) was set in space group $I2_13$ #199. Hermann noted that this structure has a very short distance (1.88 Å) between oxygen atoms on different $NO_2$ molecules, and that this structure does not form the $(NO_2)_2$ aggregate molecule found in the $C26_b$ structure, making "making this proposed structure very unlikely."
- Recognizing this, (Hendricks, 1931) suggested that $NO_2$ was actually in space group $I23$ #197. (Hermann, 1997) gave this structure the $C26_b$ designation, but noted that based on Hendricks's atomic positions the space group was actually $Im\overline{3}$ #204.
- The modern accepted structure for $NO_2$ (AB2_cI36_204_d_g) is set in space group $Im\overline{3}$ #204, confirming Hendricks. We follow (Villars, 2005) and give this the $C26$ designation.

**Body-centered Cubic primitive vectors:**

$$\mathbf{a}_1 = -\frac{1}{2}\,a\,\hat{\mathbf{x}} + \frac{1}{2}\,a\,\hat{\mathbf{y}} + \frac{1}{2}\,a\,\hat{\mathbf{z}}$$
$$\mathbf{a}_2 = \frac{1}{2}\,a\,\hat{\mathbf{x}} - \frac{1}{2}\,a\,\hat{\mathbf{y}} + \frac{1}{2}\,a\,\hat{\mathbf{z}}$$
$$\mathbf{a}_3 = \frac{1}{2}\,a\,\hat{\mathbf{x}} + \frac{1}{2}\,a\,\hat{\mathbf{y}} - \frac{1}{2}\,a\,\hat{\mathbf{z}}$$

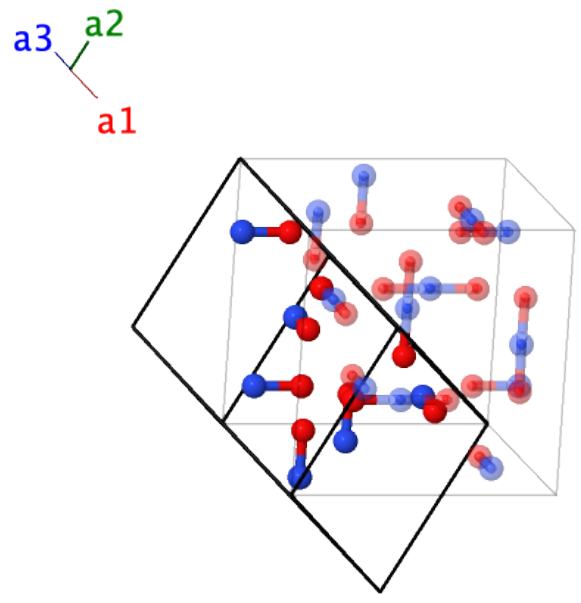

**Basis vectors:**

| | | Lattice Coordinates | | Cartesian Coordinates | Wyckoff Position | Atom Type |
|---|---|---|---|---|---|---|
| $\mathbf{B}_1$ | = | $\frac{1}{4}\,\mathbf{a}_1 + \left(\frac{1}{4} + x_1\right)\,\mathbf{a}_2 + x_1\,\mathbf{a}_3$ | = | $x_1 a\,\hat{\mathbf{x}} + \frac{1}{4} a\,\hat{\mathbf{z}}$ | (12$b$) | N |
| $\mathbf{B}_2$ | = | $\frac{3}{4}\,\mathbf{a}_1 + \left(\frac{1}{4} - x_1\right)\,\mathbf{a}_2 + \left(\frac{1}{2} - x_1\right)\,\mathbf{a}_3$ | = | $-x_1 a\,\hat{\mathbf{x}} + \frac{1}{2} a\,\hat{\mathbf{y}} + \frac{1}{4} a\,\hat{\mathbf{z}}$ | (12$b$) | N |
| $\mathbf{B}_3$ | = | $x_1\,\mathbf{a}_1 + \frac{1}{4}\,\mathbf{a}_2 + \left(\frac{1}{4} + x_1\right)\,\mathbf{a}_3$ | = | $\frac{1}{4} a\,\hat{\mathbf{x}} + x_1 a\,\hat{\mathbf{y}}$ | (12$b$) | N |
| $\mathbf{B}_4$ | = | $\left(\frac{1}{2} - x_1\right)\,\mathbf{a}_1 + \frac{3}{4}\,\mathbf{a}_2 + \left(\frac{1}{4} - x_1\right)\,\mathbf{a}_3$ | = | $\frac{1}{4} a\,\hat{\mathbf{x}} - x_1 a\,\hat{\mathbf{y}} + \frac{1}{2} a\,\hat{\mathbf{z}}$ | (12$b$) | N |
| $\mathbf{B}_5$ | = | $\left(\frac{1}{4} + x_1\right)\,\mathbf{a}_1 + x_1\,\mathbf{a}_2 + \frac{1}{4}\,\mathbf{a}_3$ | = | $\frac{1}{4} a\,\hat{\mathbf{y}} + x_1 a\,\hat{\mathbf{z}}$ | (12$b$) | N |
| $\mathbf{B}_6$ | = | $\left(\frac{1}{4} - x_1\right)\,\mathbf{a}_1 + \left(\frac{1}{2} - x_1\right)\,\mathbf{a}_2 + \frac{3}{4}\,\mathbf{a}_3$ | = | $\frac{1}{2} a\,\hat{\mathbf{x}} + \frac{1}{4} a\,\hat{\mathbf{y}} - x_1 a\,\hat{\mathbf{z}}$ | (12$b$) | N |
| $\mathbf{B}_7$ | = | $(y_2 + z_2)\,\mathbf{a}_1 + (x_2 + z_2)\,\mathbf{a}_2 + (x_2 + y_2)\,\mathbf{a}_3$ | = | $x_2 a\,\hat{\mathbf{x}} + y_2 a\,\hat{\mathbf{y}} + z_2 a\,\hat{\mathbf{z}}$ | (24$c$) | O |
| $\mathbf{B}_8$ | = | $\left(\frac{1}{2} - y_2 + z_2\right)\,\mathbf{a}_1 + (-x_2 + z_2)\,\mathbf{a}_2 + \left(\frac{1}{2} - x_2 - y_2\right)\,\mathbf{a}_3$ | = | $-x_2 a\,\hat{\mathbf{x}} + \left(\frac{1}{2} - y_2\right) a\,\hat{\mathbf{y}} + z_2 a\,\hat{\mathbf{z}}$ | (24$c$) | O |
| $\mathbf{B}_9$ | = | $(y_2 - z_2)\,\mathbf{a}_1 + \left(\frac{1}{2} - x_2 - z_2\right)\,\mathbf{a}_2 + \left(\frac{1}{2} - x_2 + y_2\right)\,\mathbf{a}_3$ | = | $\left(\frac{1}{2} - x_2\right) a\,\hat{\mathbf{x}} + y_2 a\,\hat{\mathbf{y}} - z_2 a\,\hat{\mathbf{z}}$ | (24$c$) | O |
| $\mathbf{B}_{10}$ | = | $\left(\frac{1}{2} - y_2 - z_2\right)\,\mathbf{a}_1 + \left(\frac{1}{2} + x_2 - z_2\right)\,\mathbf{a}_2 + (x_2 - y_2)\,\mathbf{a}_3$ | = | $x_2 a\,\hat{\mathbf{x}} - y_2 a\,\hat{\mathbf{y}} + \left(\frac{1}{2} - z_2\right) a\,\hat{\mathbf{z}}$ | (24$c$) | O |
| $\mathbf{B}_{11}$ | = | $(x_2 + y_2)\,\mathbf{a}_1 + (y_2 + z_2)\,\mathbf{a}_2 + (x_2 + z_2)\,\mathbf{a}_3$ | = | $z_2 a\,\hat{\mathbf{x}} + x_2 a\,\hat{\mathbf{y}} + y_2 a\,\hat{\mathbf{z}}$ | (24$c$) | O |
| $\mathbf{B}_{12}$ | = | $\left(\frac{1}{2} - x_2 - y_2\right)\,\mathbf{a}_1 + \left(\frac{1}{2} - y_2 + z_2\right)\,\mathbf{a}_2 + (-x_2 + z_2)\,\mathbf{a}_3$ | = | $z_2 a\,\hat{\mathbf{x}} - x_2 a\,\hat{\mathbf{y}} + \left(\frac{1}{2} - y_2\right) a\,\hat{\mathbf{z}}$ | (24$c$) | O |
| $\mathbf{B}_{13}$ | = | $\left(\frac{1}{2} - x_2 + y_2\right)\,\mathbf{a}_1 + (y_2 - z_2)\,\mathbf{a}_2 + \left(\frac{1}{2} - x_2 - z_2\right)\,\mathbf{a}_3$ | = | $-z_2 a\,\hat{\mathbf{x}} + \left(\frac{1}{2} - x_2\right) a\,\hat{\mathbf{y}} + y_2 a\,\hat{\mathbf{z}}$ | (24$c$) | O |
| $\mathbf{B}_{14}$ | = | $(x_2 - y_2)\,\mathbf{a}_1 + \left(\frac{1}{2} - y_2 - z_2\right)\,\mathbf{a}_2 + \left(\frac{1}{2} + x_2 - z_2\right)\,\mathbf{a}_3$ | = | $\left(\frac{1}{2} - z_2\right) a\,\hat{\mathbf{x}} + x_2 a\,\hat{\mathbf{y}} - y_2 a\,\hat{\mathbf{z}}$ | (24$c$) | O |
| $\mathbf{B}_{15}$ | = | $(x_2 + z_2)\,\mathbf{a}_1 + (x_2 + y_2)\,\mathbf{a}_2 + (y_2 + z_2)\,\mathbf{a}_3$ | = | $y_2 a\,\hat{\mathbf{x}} + z_2 a\,\hat{\mathbf{y}} + x_2 a\,\hat{\mathbf{z}}$ | (24$c$) | O |
| $\mathbf{B}_{16}$ | = | $(-x_2 + z_2)\,\mathbf{a}_1 + \left(\frac{1}{2} - x_2 - y_2\right)\,\mathbf{a}_2 + \left(\frac{1}{2} - y_2 + z_2\right)\,\mathbf{a}_3$ | = | $\left(\frac{1}{2} - y_2\right) a\,\hat{\mathbf{x}} + z_2 a\,\hat{\mathbf{y}} - x_2 a\,\hat{\mathbf{z}}$ | (24$c$) | O |
| $\mathbf{B}_{17}$ | = | $\left(\frac{1}{2} - x_2 - z_2\right)\,\mathbf{a}_1 + \left(\frac{1}{2} - x_2 + y_2\right)\,\mathbf{a}_2 + (y_2 - z_2)\,\mathbf{a}_3$ | = | $y_2 a\,\hat{\mathbf{x}} - z_2 a\,\hat{\mathbf{y}} + \left(\frac{1}{2} - x_2\right) a\,\hat{\mathbf{z}}$ | (24$c$) | O |
| $\mathbf{B}_{18}$ | = | $\left(\frac{1}{2} + x_2 - z_2\right)\,\mathbf{a}_1 + (x_2 - y_2)\,\mathbf{a}_2 + \left(\frac{1}{2} - y_2 - z_2\right)\,\mathbf{a}_3$ | = | $-y_2 a\,\hat{\mathbf{x}} + \left(\frac{1}{2} - z_2\right) a\,\hat{\mathbf{y}} + x_2 a\,\hat{\mathbf{z}}$ | (24$c$) | O |

**Found in:**

---

**Geometry files:**

- CIF: pp. 1774
- POSCAR: pp. 1775

# $Bi_3Ru_3O_{11}$ Structure: A3B11C3_cP68_201_be_efh_g

http://aflow.org/prototype-encyclopedia/A3B11C3_cP68_201_be_efh_g

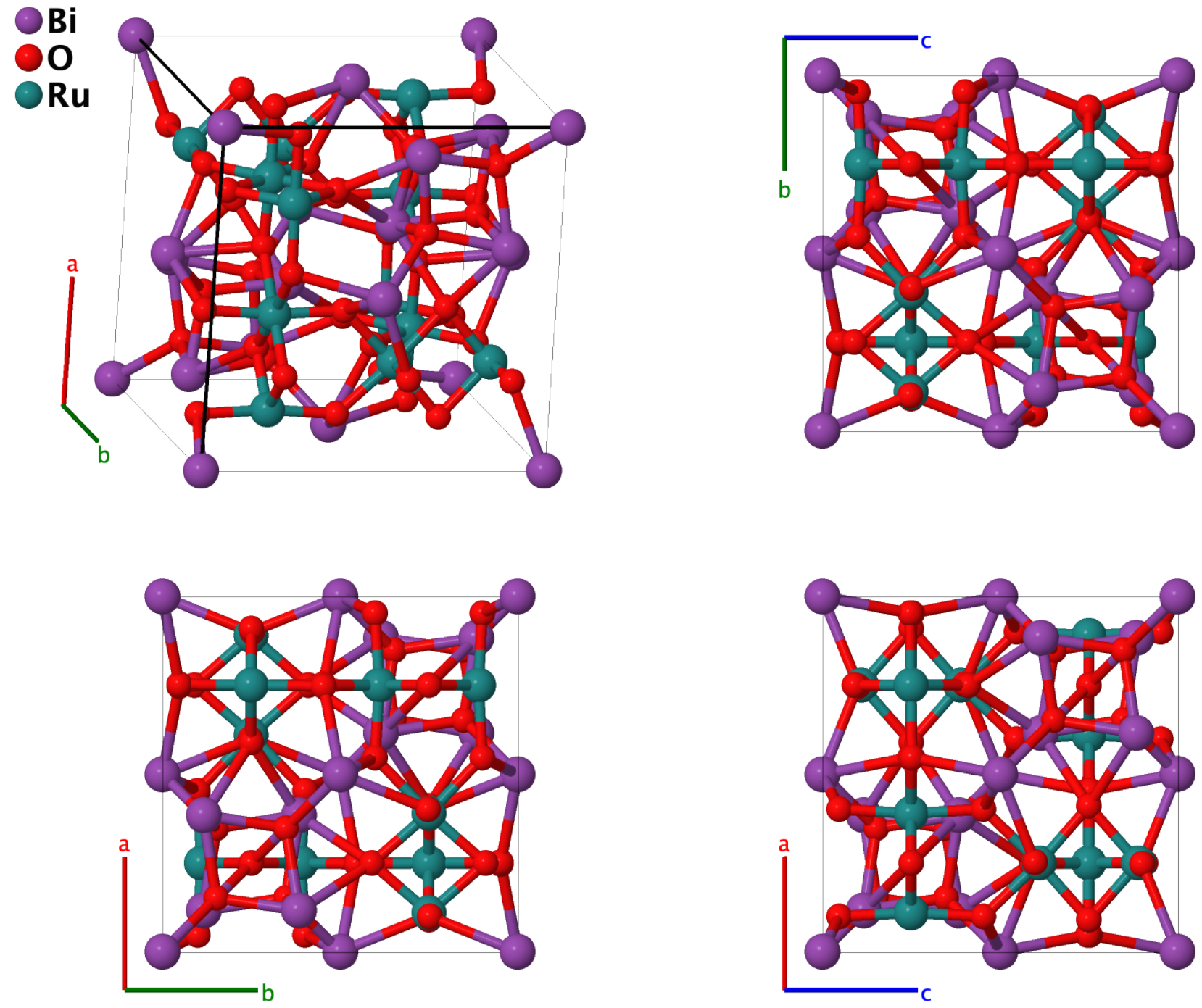

| | | |
|---|---|---|
| **Prototype** | : | $Bi_3O_{11}Ru_3$ |
| **AFLOW prototype label** | : | A3B11C3_cP68_201_be_efh_g |
| ***Strukturbericht* designation** | : | None |
| **Pearson symbol** | : | cP68 |
| **Space group number** | : | 201 |
| **Space group symbol** | : | $Pn\bar{3}$ |
| **AFLOW prototype command** | : | `aflow --proto=A3B11C3_cP68_201_be_efh_g` `--params=`$a, x_2, x_3, x_4, x_5, x_6, y_6, z_6$ |

**Simple Cubic primitive vectors:**

$$\mathbf{a}_1 = a\,\hat{\mathbf{x}}$$
$$\mathbf{a}_2 = a\,\hat{\mathbf{y}}$$
$$\mathbf{a}_3 = a\,\hat{\mathbf{z}}$$

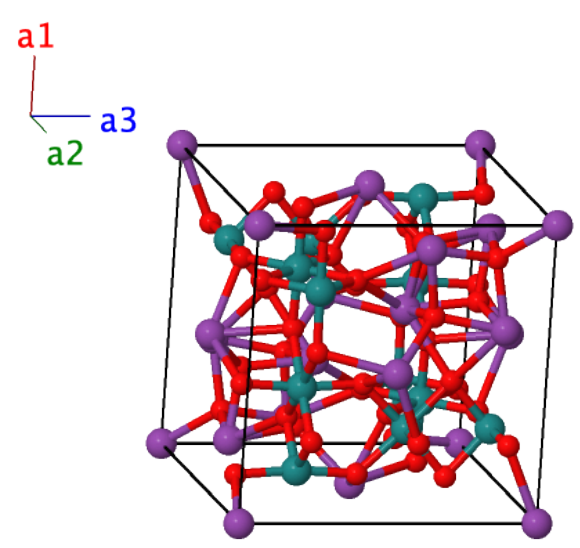

**Basis vectors:**

| | | Lattice Coordinates | | Cartesian Coordinates | Wyckoff Position | Atom Type |
|---|---|---|---|---|---|---|
| $\mathbf{B}_1$ | = | $0\,\mathbf{a}_1 + 0\,\mathbf{a}_2 + 0\,\mathbf{a}_3$ | = | $0\,\hat{\mathbf{x}} + 0\,\hat{\mathbf{y}} + 0\,\hat{\mathbf{z}}$ | (4$b$) | Bi I |
| $\mathbf{B}_2$ | = | $\frac{1}{2}\,\mathbf{a}_1 + \frac{1}{2}\,\mathbf{a}_2$ | = | $\frac{1}{2}a\,\hat{\mathbf{x}} + \frac{1}{2}a\,\hat{\mathbf{y}}$ | (4$b$) | Bi I |
| $\mathbf{B}_3$ | = | $\frac{1}{2}\,\mathbf{a}_1 + \frac{1}{2}\,\mathbf{a}_3$ | = | $\frac{1}{2}a\,\hat{\mathbf{x}} + \frac{1}{2}a\,\hat{\mathbf{z}}$ | (4$b$) | Bi I |
| $\mathbf{B}_4$ | = | $\frac{1}{2}\,\mathbf{a}_2 + \frac{1}{2}\,\mathbf{a}_3$ | = | $\frac{1}{2}a\,\hat{\mathbf{y}} + \frac{1}{2}a\,\hat{\mathbf{z}}$ | $(4b)$ | Bi I |
| $\mathbf{B}_5$ | = | $x_2\,\mathbf{a}_1 + x_2\,\mathbf{a}_2 + x_2\,\mathbf{a}_3$ | = | $x_2a\,\hat{\mathbf{x}} + x_2a\,\hat{\mathbf{y}} + x_2a\,\hat{\mathbf{z}}$ | $(8e)$ | Bi II |
| $\mathbf{B}_6$ | = | $\left(\frac{1}{2} - x_2\right)\mathbf{a}_1 + \left(\frac{1}{2} - x_2\right)\mathbf{a}_2 + x_2\,\mathbf{a}_3$ | = | $\left(\frac{1}{2} - x_2\right)a\,\hat{\mathbf{x}} + \left(\frac{1}{2} - x_2\right)a\,\hat{\mathbf{y}} + x_2a\,\hat{\mathbf{z}}$ | $(8e)$ | Bi II |
| $\mathbf{B}_7$ | = | $\left(\frac{1}{2} - x_2\right)\mathbf{a}_1 + x_2\,\mathbf{a}_2 + \left(\frac{1}{2} - x_2\right)\mathbf{a}_3$ | = | $\left(\frac{1}{2} - x_2\right)a\,\hat{\mathbf{x}} + x_2a\,\hat{\mathbf{y}} + \left(\frac{1}{2} - x_2\right)a\,\hat{\mathbf{z}}$ | $(8e)$ | Bi II |
| $\mathbf{B}_8$ | = | $x_2\,\mathbf{a}_1 + \left(\frac{1}{2} - x_2\right)\mathbf{a}_2 + \left(\frac{1}{2} - x_2\right)\mathbf{a}_3$ | = | $x_2a\,\hat{\mathbf{x}} + \left(\frac{1}{2} - x_2\right)a\,\hat{\mathbf{y}} + \left(\frac{1}{2} - x_2\right)a\,\hat{\mathbf{z}}$ | $(8e)$ | Bi II |
| $\mathbf{B}_9$ | = | $-x_2\,\mathbf{a}_1 - x_2\,\mathbf{a}_2 - x_2\,\mathbf{a}_3$ | = | $-x_2a\,\hat{\mathbf{x}} - x_2a\,\hat{\mathbf{y}} - x_2a\,\hat{\mathbf{z}}$ | $(8e)$ | Bi II |
| $\mathbf{B}_{10}$ | = | $\left(\frac{1}{2} + x_2\right)\mathbf{a}_1 + \left(\frac{1}{2} + x_2\right)\mathbf{a}_2 - x_2\,\mathbf{a}_3$ | = | $\left(\frac{1}{2} + x_2\right)a\,\hat{\mathbf{x}} + \left(\frac{1}{2} + x_2\right)a\,\hat{\mathbf{y}} - x_2a\,\hat{\mathbf{z}}$ | $(8e)$ | Bi II |
| $\mathbf{B}_{11}$ | = | $\left(\frac{1}{2} + x_2\right)\mathbf{a}_1 - x_2\,\mathbf{a}_2 + \left(\frac{1}{2} + x_2\right)\mathbf{a}_3$ | = | $\left(\frac{1}{2} + x_2\right)a\,\hat{\mathbf{x}} - x_2a\,\hat{\mathbf{y}} + \left(\frac{1}{2} + x_2\right)a\,\hat{\mathbf{z}}$ | $(8e)$ | Bi II |
| $\mathbf{B}_{12}$ | = | $-x_2\,\mathbf{a}_1 + \left(\frac{1}{2} + x_2\right)\mathbf{a}_2 + \left(\frac{1}{2} + x_2\right)\mathbf{a}_3$ | = | $-x_2a\,\hat{\mathbf{x}} + \left(\frac{1}{2} + x_2\right)a\,\hat{\mathbf{y}} + \left(\frac{1}{2} + x_2\right)a\,\hat{\mathbf{z}}$ | $(8e)$ | Bi II |
| $\mathbf{B}_{13}$ | = | $x_3\,\mathbf{a}_1 + x_3\,\mathbf{a}_2 + x_3\,\mathbf{a}_3$ | = | $x_3a\,\hat{\mathbf{x}} + x_3a\,\hat{\mathbf{y}} + x_3a\,\hat{\mathbf{z}}$ | $(8e)$ | O I |
| $\mathbf{B}_{14}$ | = | $\left(\frac{1}{2} - x_3\right)\mathbf{a}_1 + \left(\frac{1}{2} - x_3\right)\mathbf{a}_2 + x_3\,\mathbf{a}_3$ | = | $\left(\frac{1}{2} - x_3\right)a\,\hat{\mathbf{x}} + \left(\frac{1}{2} - x_3\right)a\,\hat{\mathbf{y}} + x_3a\,\hat{\mathbf{z}}$ | $(8e)$ | O I |
| $\mathbf{B}_{15}$ | = | $\left(\frac{1}{2} - x_3\right)\mathbf{a}_1 + x_3\,\mathbf{a}_2 + \left(\frac{1}{2} - x_3\right)\mathbf{a}_3$ | = | $\left(\frac{1}{2} - x_3\right)a\,\hat{\mathbf{x}} + x_3a\,\hat{\mathbf{y}} + \left(\frac{1}{2} - x_3\right)a\,\hat{\mathbf{z}}$ | $(8e)$ | O I |
| $\mathbf{B}_{16}$ | = | $x_3\,\mathbf{a}_1 + \left(\frac{1}{2} - x_3\right)\mathbf{a}_2 + \left(\frac{1}{2} - x_3\right)\mathbf{a}_3$ | = | $x_3a\,\hat{\mathbf{x}} + \left(\frac{1}{2} - x_3\right)a\,\hat{\mathbf{y}} + \left(\frac{1}{2} - x_3\right)a\,\hat{\mathbf{z}}$ | $(8e)$ | O I |
| $\mathbf{B}_{17}$ | = | $-x_3\,\mathbf{a}_1 - x_3\,\mathbf{a}_2 - x_3\,\mathbf{a}_3$ | = | $-x_3a\,\hat{\mathbf{x}} - x_3a\,\hat{\mathbf{y}} - x_3a\,\hat{\mathbf{z}}$ | $(8e)$ | O I |
| $\mathbf{B}_{18}$ | = | $\left(\frac{1}{2} + x_3\right)\mathbf{a}_1 + \left(\frac{1}{2} + x_3\right)\mathbf{a}_2 - x_3\,\mathbf{a}_3$ | = | $\left(\frac{1}{2} + x_3\right)a\,\hat{\mathbf{x}} + \left(\frac{1}{2} + x_3\right)a\,\hat{\mathbf{y}} - x_3a\,\hat{\mathbf{z}}$ | $(8e)$ | O I |
| $\mathbf{B}_{19}$ | = | $\left(\frac{1}{2} + x_3\right)\mathbf{a}_1 - x_3\,\mathbf{a}_2 + \left(\frac{1}{2} + x_3\right)\mathbf{a}_3$ | = | $\left(\frac{1}{2} + x_3\right)a\,\hat{\mathbf{x}} - x_3a\,\hat{\mathbf{y}} + \left(\frac{1}{2} + x_3\right)a\,\hat{\mathbf{z}}$ | $(8e)$ | O I |
| $\mathbf{B}_{20}$ | = | $-x_3\,\mathbf{a}_1 + \left(\frac{1}{2} + x_3\right)\mathbf{a}_2 + \left(\frac{1}{2} + x_3\right)\mathbf{a}_3$ | = | $-x_3a\,\hat{\mathbf{x}} + \left(\frac{1}{2} + x_3\right)a\,\hat{\mathbf{y}} + \left(\frac{1}{2} + x_3\right)a\,\hat{\mathbf{z}}$ | $(8e)$ | O I |
| $\mathbf{B}_{21}$ | = | $x_4\,\mathbf{a}_1 + \frac{1}{4}\,\mathbf{a}_2 + \frac{1}{4}\,\mathbf{a}_3$ | = | $x_4a\,\hat{\mathbf{x}} + \frac{1}{4}a\,\hat{\mathbf{y}} + \frac{1}{4}a\,\hat{\mathbf{z}}$ | $(12f)$ | O II |
| $\mathbf{B}_{22}$ | = | $\left(\frac{1}{2} - x_4\right)\mathbf{a}_1 + \frac{1}{4}\,\mathbf{a}_2 + \frac{1}{4}\,\mathbf{a}_3$ | = | $\left(\frac{1}{2} - x_4\right)a\,\hat{\mathbf{x}} + \frac{1}{4}a\,\hat{\mathbf{y}} + \frac{1}{4}a\,\hat{\mathbf{z}}$ | $(12f)$ | O II |
| $\mathbf{B}_{23}$ | = | $\frac{1}{4}\,\mathbf{a}_1 + x_4\,\mathbf{a}_2 + \frac{1}{4}\,\mathbf{a}_3$ | = | $\frac{1}{4}a\,\hat{\mathbf{x}} + x_4a\,\hat{\mathbf{y}} + \frac{1}{4}a\,\hat{\mathbf{z}}$ | $(12f)$ | O II |
| $\mathbf{B}_{24}$ | = | $\frac{1}{4}\,\mathbf{a}_1 + \left(\frac{1}{2} - x_4\right)\mathbf{a}_2 + \frac{1}{4}\,\mathbf{a}_3$ | = | $\frac{1}{4}a\,\hat{\mathbf{x}} + \left(\frac{1}{2} - x_4\right)a\,\hat{\mathbf{y}} + \frac{1}{4}a\,\hat{\mathbf{z}}$ | $(12f)$ | O II |
| $\mathbf{B}_{25}$ | = | $\frac{1}{4}\,\mathbf{a}_1 + \frac{1}{4}\,\mathbf{a}_2 + x_4\,\mathbf{a}_3$ | = | $\frac{1}{4}a\,\hat{\mathbf{x}} + \frac{1}{4}a\,\hat{\mathbf{y}} + x_4a\,\hat{\mathbf{z}}$ | $(12f)$ | O II |
| $\mathbf{B}_{26}$ | = | $\frac{1}{4}\,\mathbf{a}_1 + \frac{1}{4}\,\mathbf{a}_2 + \left(\frac{1}{2} - x_4\right)\mathbf{a}_3$ | = | $\frac{1}{4}a\,\hat{\mathbf{x}} + \frac{1}{4}a\,\hat{\mathbf{y}} + \left(\frac{1}{2} - x_4\right)a\,\hat{\mathbf{z}}$ | $(12f)$ | O II |
| $\mathbf{B}_{27}$ | = | $-x_4\,\mathbf{a}_1 + \frac{3}{4}\,\mathbf{a}_2 + \frac{3}{4}\,\mathbf{a}_3$ | = | $-x_4a\,\hat{\mathbf{x}} + \frac{3}{4}a\,\hat{\mathbf{y}} + \frac{3}{4}a\,\hat{\mathbf{z}}$ | $(12f)$ | O II |
| $\mathbf{B}_{28}$ | = | $\left(\frac{1}{2} + x_4\right)\mathbf{a}_1 + \frac{3}{4}\,\mathbf{a}_2 + \frac{3}{4}\,\mathbf{a}_3$ | = | $\left(\frac{1}{2} + x_4\right)a\,\hat{\mathbf{x}} + \frac{3}{4}a\,\hat{\mathbf{y}} + \frac{3}{4}a\,\hat{\mathbf{z}}$ | $(12f)$ | O II |
| $\mathbf{B}_{29}$ | = | $\frac{3}{4}\,\mathbf{a}_1 - x_4\,\mathbf{a}_2 + \frac{3}{4}\,\mathbf{a}_3$ | = | $\frac{3}{4}a\,\hat{\mathbf{x}} - x_4a\,\hat{\mathbf{y}} + \frac{3}{4}a\,\hat{\mathbf{z}}$ | $(12f)$ | O II |
| $\mathbf{B}_{30}$ | = | $\frac{3}{4}\,\mathbf{a}_1 + \left(\frac{1}{2} + x_4\right)\mathbf{a}_2 + \frac{3}{4}\,\mathbf{a}_3$ | = | $\frac{3}{4}a\,\hat{\mathbf{x}} + \left(\frac{1}{2} + x_4\right)a\,\hat{\mathbf{y}} + \frac{3}{4}a\,\hat{\mathbf{z}}$ | $(12f)$ | O II |
| $\mathbf{B}_{31}$ | = | $\frac{3}{4}\,\mathbf{a}_1 + \frac{3}{4}\,\mathbf{a}_2 - x_4\,\mathbf{a}_3$ | = | $\frac{3}{4}a\,\hat{\mathbf{x}} + \frac{3}{4}a\,\hat{\mathbf{y}} - x_4a\,\hat{\mathbf{z}}$ | $(12f)$ | O II |
| $\mathbf{B}_{32}$ | = | $\frac{3}{4}\,\mathbf{a}_1 + \frac{3}{4}\,\mathbf{a}_2 + \left(\frac{1}{2} + x_4\right)\mathbf{a}_3$ | = | $\frac{3}{4}a\,\hat{\mathbf{x}} + \frac{3}{4}a\,\hat{\mathbf{y}} + \left(\frac{1}{2} + x_4\right)a\,\hat{\mathbf{z}}$ | $(12f)$ | O II |
| $\mathbf{B}_{33}$ | = | $x_5\,\mathbf{a}_1 + \frac{3}{4}\,\mathbf{a}_2 + \frac{1}{4}\,\mathbf{a}_3$ | = | $x_5a\,\hat{\mathbf{x}} + \frac{3}{4}a\,\hat{\mathbf{y}} + \frac{1}{4}a\,\hat{\mathbf{z}}$ | $(12g)$ | Ru |
| $\mathbf{B}_{34}$ | = | $\left(\frac{1}{2} - x_5\right)\mathbf{a}_1 + \frac{3}{4}\,\mathbf{a}_2 + \frac{1}{4}\,\mathbf{a}_3$ | = | $\left(\frac{1}{2} - x_5\right)a\,\hat{\mathbf{x}} + \frac{3}{4}a\,\hat{\mathbf{y}} + \frac{1}{4}a\,\hat{\mathbf{z}}$ | $(12g)$ | Ru |
| $\mathbf{B}_{35}$ | = | $\frac{1}{4}\,\mathbf{a}_1 + x_5\,\mathbf{a}_2 + \frac{3}{4}\,\mathbf{a}_3$ | = | $\frac{1}{4}a\,\hat{\mathbf{x}} + x_5a\,\hat{\mathbf{y}} + \frac{3}{4}a\,\hat{\mathbf{z}}$ | $(12g)$ | Ru |
| $\mathbf{B}_{36}$ | = | $\frac{1}{4}\,\mathbf{a}_1 + \left(\frac{1}{2} - x_5\right)\mathbf{a}_2 + \frac{3}{4}\,\mathbf{a}_3$ | = | $\frac{1}{4}a\,\hat{\mathbf{x}} + \left(\frac{1}{2} - x_5\right)a\,\hat{\mathbf{y}} + \frac{3}{4}a\,\hat{\mathbf{z}}$ | $(12g)$ | Ru |
| $\mathbf{B}_{37}$ | = | $\frac{3}{4}\,\mathbf{a}_1 + \frac{1}{4}\,\mathbf{a}_2 + x_5\,\mathbf{a}_3$ | = | $\frac{3}{4}a\,\hat{\mathbf{x}} + \frac{1}{4}a\,\hat{\mathbf{y}} + x_5a\,\hat{\mathbf{z}}$ | $(12g)$ | Ru |
| $\mathbf{B}_{38}$ | = | $\frac{3}{4}\,\mathbf{a}_1 + \frac{1}{4}\,\mathbf{a}_2 + \left(\frac{1}{2} - x_5\right)\mathbf{a}_3$ | = | $\frac{3}{4}a\,\hat{\mathbf{x}} + \frac{1}{4}a\,\hat{\mathbf{y}} + \left(\frac{1}{2} - x_5\right)a\,\hat{\mathbf{z}}$ | $(12g)$ | Ru |
| $\mathbf{B}_{39}$ | = | $-x_5\,\mathbf{a}_1 + \frac{1}{4}\,\mathbf{a}_2 + \frac{3}{4}\,\mathbf{a}_3$ | = | $-x_5a\,\hat{\mathbf{x}} + \frac{1}{4}a\,\hat{\mathbf{y}} + \frac{3}{4}a\,\hat{\mathbf{z}}$ | $(12g)$ | Ru |
| $\mathbf{B}_{40}$ | = | $\left(\frac{1}{2} + x_{5}\right) \mathbf{a}_{1} + \frac{1}{4} \mathbf{a}_{2} + \frac{3}{4} \mathbf{a}_{3}$ | = | $\left(\frac{1}{2} + x_{5}\right) a \mathbf{\hat{x}} + \frac{1}{4} a \mathbf{\hat{y}} + \frac{3}{4} a \mathbf{\hat{z}}$ | (12$g$) | Ru |
| $\mathbf{B}_{41}$ | = | $\frac{3}{4} \mathbf{a}_{1} - x_{5} \mathbf{a}_{2} + \frac{1}{4} \mathbf{a}_{3}$ | = | $\frac{3}{4} a \mathbf{\hat{x}} - x_{5} a \mathbf{\hat{y}} + \frac{1}{4} a \mathbf{\hat{z}}$ | (12$g$) | Ru |
| $\mathbf{B}_{42}$ | = | $\frac{3}{4} \mathbf{a}_{1} + \left(\frac{1}{2} + x_{5}\right) \mathbf{a}_{2} + \frac{1}{4} \mathbf{a}_{3}$ | = | $\frac{3}{4} a \mathbf{\hat{x}} + \left(\frac{1}{2} + x_{5}\right) a \mathbf{\hat{y}} + \frac{1}{4} a \mathbf{\hat{z}}$ | (12$g$) | Ru |
| $\mathbf{B}_{43}$ | = | $\frac{1}{4} \mathbf{a}_{1} + \frac{3}{4} \mathbf{a}_{2} - x_{5} \mathbf{a}_{3}$ | = | $\frac{1}{4} a \mathbf{\hat{x}} + \frac{3}{4} a \mathbf{\hat{y}} - x_{5} a \mathbf{\hat{z}}$ | (12$g$) | Ru |
| $\mathbf{B}_{44}$ | = | $\frac{1}{4} \mathbf{a}_{1} + \frac{3}{4} \mathbf{a}_{2} + \left(\frac{1}{2} + x_{5}\right) \mathbf{a}_{3}$ | = | $\frac{1}{4} a \mathbf{\hat{x}} + \frac{3}{4} a \mathbf{\hat{y}} + \left(\frac{1}{2} + x_{5}\right) a \mathbf{\hat{z}}$ | (12$g$) | Ru |
| $\mathbf{B}_{45}$ | = | $x_{6} \mathbf{a}_{1} + y_{6} \mathbf{a}_{2} + z_{6} \mathbf{a}_{3}$ | = | $x_{6} a \mathbf{\hat{x}} + y_{6} a \mathbf{\hat{y}} + z_{6} a \mathbf{\hat{z}}$ | (24$h$) | O III |
| $\mathbf{B}_{46}$ | = | $\left(\frac{1}{2} - x_{6}\right) \mathbf{a}_{1} + \left(\frac{1}{2} - y_{6}\right) \mathbf{a}_{2} + z_{6} \mathbf{a}_{3}$ | = | $\left(\frac{1}{2} - x_{6}\right) a \mathbf{\hat{x}} + \left(\frac{1}{2} - y_{6}\right) a \mathbf{\hat{y}} + z_{6} a \mathbf{\hat{z}}$ | (24$h$) | O III |
| $\mathbf{B}_{47}$ | = | $\left(\frac{1}{2} - x_{6}\right) \mathbf{a}_{1} + y_{6} \mathbf{a}_{2} + \left(\frac{1}{2} - z_{6}\right) \mathbf{a}_{3}$ | = | $\left(\frac{1}{2} - x_{6}\right) a \mathbf{\hat{x}} + y_{6} a \mathbf{\hat{y}} + \left(\frac{1}{2} - z_{6}\right) a \mathbf{\hat{z}}$ | (24$h$) | O III |
| $\mathbf{B}_{48}$ | = | $x_{6} \mathbf{a}_{1} + \left(\frac{1}{2} - y_{6}\right) \mathbf{a}_{2} + \left(\frac{1}{2} - z_{6}\right) \mathbf{a}_{3}$ | = | $x_{6} a \mathbf{\hat{x}} + \left(\frac{1}{2} - y_{6}\right) a \mathbf{\hat{y}} + \left(\frac{1}{2} - z_{6}\right) a \mathbf{\hat{z}}$ | (24$h$) | O III |
| $\mathbf{B}_{49}$ | = | $z_{6} \mathbf{a}_{1} + x_{6} \mathbf{a}_{2} + y_{6} \mathbf{a}_{3}$ | = | $z_{6} a \mathbf{\hat{x}} + x_{6} a \mathbf{\hat{y}} + y_{6} a \mathbf{\hat{z}}$ | (24$h$) | O III |
| $\mathbf{B}_{50}$ | = | $z_{6} \mathbf{a}_{1} + \left(\frac{1}{2} - x_{6}\right) \mathbf{a}_{2} + \left(\frac{1}{2} - y_{6}\right) \mathbf{a}_{3}$ | = | $z_{6} a \mathbf{\hat{x}} + \left(\frac{1}{2} - x_{6}\right) a \mathbf{\hat{y}} + \left(\frac{1}{2} - y_{6}\right) a \mathbf{\hat{z}}$ | (24$h$) | O III |
| $\mathbf{B}_{51}$ | = | $\left(\frac{1}{2} - z_{6}\right) \mathbf{a}_{1} + \left(\frac{1}{2} - x_{6}\right) \mathbf{a}_{2} + y_{6} \mathbf{a}_{3}$ | = | $\left(\frac{1}{2} - z_{6}\right) a \mathbf{\hat{x}} + \left(\frac{1}{2} - x_{6}\right) a \mathbf{\hat{y}} + y_{6} a \mathbf{\hat{z}}$ | (24$h$) | O III |
| $\mathbf{B}_{52}$ | = | $\left(\frac{1}{2} - z_{6}\right) \mathbf{a}_{1} + x_{6} \mathbf{a}_{2} + \left(\frac{1}{2} - y_{6}\right) \mathbf{a}_{3}$ | = | $\left(\frac{1}{2} - z_{6}\right) a \mathbf{\hat{x}} + x_{6} a \mathbf{\hat{y}} + \left(\frac{1}{2} - y_{6}\right) a \mathbf{\hat{z}}$ | (24$h$) | O III |
| $\mathbf{B}_{53}$ | = | $y_{6} \mathbf{a}_{1} + z_{6} \mathbf{a}_{2} + x_{6} \mathbf{a}_{3}$ | = | $y_{6} a \mathbf{\hat{x}} + z_{6} a \mathbf{\hat{y}} + x_{6} a \mathbf{\hat{z}}$ | (24$h$) | O III |
| $\mathbf{B}_{54}$ | = | $\left(\frac{1}{2} - y_{6}\right) \mathbf{a}_{1} + z_{6} \mathbf{a}_{2} + \left(\frac{1}{2} - x_{6}\right) \mathbf{a}_{3}$ | = | $\left(\frac{1}{2} - y_{6}\right) a \mathbf{\hat{x}} + z_{6} a \mathbf{\hat{y}} + \left(\frac{1}{2} - x_{6}\right) a \mathbf{\hat{z}}$ | (24$h$) | O III |
| $\mathbf{B}_{55}$ | = | $y_{6} \mathbf{a}_{1} + \left(\frac{1}{2} - z_{6}\right) \mathbf{a}_{2} + \left(\frac{1}{2} - x_{6}\right) \mathbf{a}_{3}$ | = | $y_{6} a \mathbf{\hat{x}} + \left(\frac{1}{2} - z_{6}\right) a \mathbf{\hat{y}} + \left(\frac{1}{2} - x_{6}\right) a \mathbf{\hat{z}}$ | (24$h$) | O III |
| $\mathbf{B}_{56}$ | = | $\left(\frac{1}{2} - y_{6}\right) \mathbf{a}_{1} + \left(\frac{1}{2} - z_{6}\right) \mathbf{a}_{2} + x_{6} \mathbf{a}_{3}$ | = | $\left(\frac{1}{2} - y_{6}\right) a \mathbf{\hat{x}} + \left(\frac{1}{2} - z_{6}\right) a \mathbf{\hat{y}} + x_{6} a \mathbf{\hat{z}}$ | (24$h$) | O III |
| $\mathbf{B}_{57}$ | = | $-x_{6} \mathbf{a}_{1} - y_{6} \mathbf{a}_{2} - z_{6} \mathbf{a}_{3}$ | = | $-x_{6} a \mathbf{\hat{x}} - y_{6} a \mathbf{\hat{y}} - z_{6} a \mathbf{\hat{z}}$ | (24$h$) | O III |
| $\mathbf{B}_{58}$ | = | $\left(\frac{1}{2} + x_{6}\right) \mathbf{a}_{1} + \left(\frac{1}{2} + y_{6}\right) \mathbf{a}_{2} - z_{6} \mathbf{a}_{3}$ | = | $\left(\frac{1}{2} + x_{6}\right) a \mathbf{\hat{x}} + \left(\frac{1}{2} + y_{6}\right) a \mathbf{\hat{y}} - z_{6} a \mathbf{\hat{z}}$ | (24$h$) | O III |
| $\mathbf{B}_{59}$ | = | $\left(\frac{1}{2} + x_{6}\right) \mathbf{a}_{1} - y_{6} \mathbf{a}_{2} + \left(\frac{1}{2} + z_{6}\right) \mathbf{a}_{3}$ | = | $\left(\frac{1}{2} + x_{6}\right) a \mathbf{\hat{x}} - y_{6} a \mathbf{\hat{y}} + \left(\frac{1}{2} + z_{6}\right) a \mathbf{\hat{z}}$ | (24$h$) | O III |
| $\mathbf{B}_{60}$ | = | $-x_{6} \mathbf{a}_{1} + \left(\frac{1}{2} + y_{6}\right) \mathbf{a}_{2} + \left(\frac{1}{2} + z_{6}\right) \mathbf{a}_{3}$ | = | $-x_{6} a \mathbf{\hat{x}} + \left(\frac{1}{2} + y_{6}\right) a \mathbf{\hat{y}} + \left(\frac{1}{2} + z_{6}\right) a \mathbf{\hat{z}}$ | (24$h$) | O III |
| $\mathbf{B}_{61}$ | = | $-z_{6} \mathbf{a}_{1} - x_{6} \mathbf{a}_{2} - y_{6} \mathbf{a}_{3}$ | = | $-z_{6} a \mathbf{\hat{x}} - x_{6} a \mathbf{\hat{y}} - y_{6} a \mathbf{\hat{z}}$ | (24$h$) | O III |
| $\mathbf{B}_{62}$ | = | $-z_{6} \mathbf{a}_{1} + \left(\frac{1}{2} + x_{6}\right) \mathbf{a}_{2} + \left(\frac{1}{2} + y_{6}\right) \mathbf{a}_{3}$ | = | $-z_{6} a \mathbf{\hat{x}} + \left(\frac{1}{2} + x_{6}\right) a \mathbf{\hat{y}} + \left(\frac{1}{2} + y_{6}\right) a \mathbf{\hat{z}}$ | (24$h$) | O III |
| $\mathbf{B}_{63}$ | = | $\left(\frac{1}{2} + z_{6}\right) \mathbf{a}_{1} + \left(\frac{1}{2} + x_{6}\right) \mathbf{a}_{2} - y_{6} \mathbf{a}_{3}$ | = | $\left(\frac{1}{2} + z_{6}\right) a \mathbf{\hat{x}} + \left(\frac{1}{2} + x_{6}\right) a \mathbf{\hat{y}} - y_{6} a \mathbf{\hat{z}}$ | (24$h$) | O III |
| $\mathbf{B}_{64}$ | = | $\left(\frac{1}{2} + z_{6}\right) \mathbf{a}_{1} - x_{6} \mathbf{a}_{2} + \left(\frac{1}{2} + y_{6}\right) \mathbf{a}_{3}$ | = | $\left(\frac{1}{2} + z_{6}\right) a \mathbf{\hat{x}} - x_{6} a \mathbf{\hat{y}} + \left(\frac{1}{2} + y_{6}\right) a \mathbf{\hat{z}}$ | (24$h$) | O III |
| $\mathbf{B}_{65}$ | = | $-y_{6} \mathbf{a}_{1} - z_{6} \mathbf{a}_{2} - x_{6} \mathbf{a}_{3}$ | = | $-y_{6} a \mathbf{\hat{x}} - z_{6} a \mathbf{\hat{y}} - x_{6} a \mathbf{\hat{z}}$ | (24$h$) | O III |
| $\mathbf{B}_{66}$ | = | $\left(\frac{1}{2} + y_{6}\right) \mathbf{a}_{1} - z_{6} \mathbf{a}_{2} + \left(\frac{1}{2} + x_{6}\right) \mathbf{a}_{3}$ | = | $\left(\frac{1}{2} + y_{6}\right) a \mathbf{\hat{x}} - z_{6} a \mathbf{\hat{y}} + \left(\frac{1}{2} + x_{6}\right) a \mathbf{\hat{z}}$ | (24$h$) | O III |
| $\mathbf{B}_{67}$ | = | $-y_{6} \mathbf{a}_{1} + \left(\frac{1}{2} + z_{6}\right) \mathbf{a}_{2} + \left(\frac{1}{2} + x_{6}\right) \mathbf{a}_{3}$ | = | $-y_{6} a \mathbf{\hat{x}} + \left(\frac{1}{2} + z_{6}\right) a \mathbf{\hat{y}} + \left(\frac{1}{2} + x_{6}\right) a \mathbf{\hat{z}}$ | (24$h$) | O III |
| $\mathbf{B}_{68}$ | = | $\left(\frac{1}{2} + y_{6}\right) \mathbf{a}_{1} + \left(\frac{1}{2} + z_{6}\right) \mathbf{a}_{2} - x_{6} \mathbf{a}_{3}$ | = | $\left(\frac{1}{2} + y_{6}\right) a \mathbf{\hat{x}} + \left(\frac{1}{2} + z_{6}\right) a \mathbf{\hat{y}} - x_{6} a \mathbf{\hat{z}}$ | (24$h$) | O III |

**Found in:**

**Geometry files:**

- CIF: pp. 1775
- POSCAR: pp. 1775

# $K_3Co(NO_2)_6$ ($J2_4$) Structure: AB3C6D12_cF88_202_a_bc_e_h

http://aflow.org/prototype-encyclopedia/AB3C6D12_cF88_202_a_bc_e_h

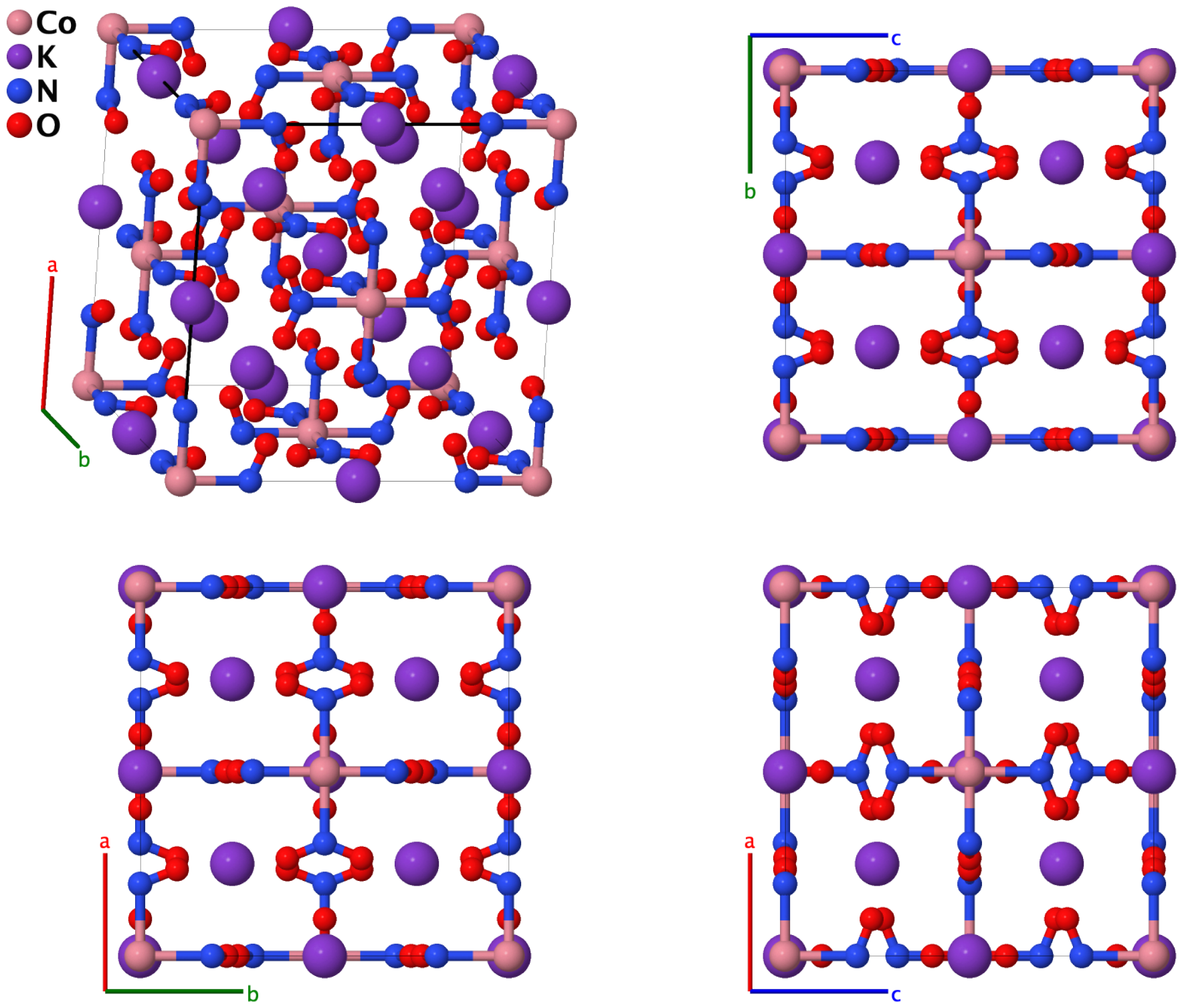

| | | |
|---|---|---|
| **Prototype** | : | $CoK_3N_6O_{12}$ |
| **AFLOW prototype label** | : | AB3C6D12_cF88_202_a_bc_e_h |
| ***Strukturbericht* designation** | : | $J2_4$ |
| **Pearson symbol** | : | cF88 |
| **Space group number** | : | 202 |
| **Space group symbol** | : | $Fm\bar{3}$ |
| **AFLOW prototype command** | : | `aflow --proto=AB3C6D12_cF88_202_a_bc_e_h --params=`$a, x_4, y_5, z_5$ |

### Other compounds with this structure

- $(NH_4)_2AgBi(NO_2)_6$, $(NH_4)_2LiBi(NO_2)_6$, $(NH_4)_2NaBi(NO_2)_6$, $(NH_4)_2NaCo(NO_2)_6$, $(NH_4)_2NaRh(NO_2)_6$, $(NH_4)_3Co(NO_2)_6$, $Cs_2AgBi(NO_2)_6$, $Cs_2LiBi(NO_2)_6$, $Cs_2NaBi(NO_2)_6$, $Cs_3Bi(NO_2)_6$, $K_2LiBi(NO_2)_6$, $K_2NaBi(NO_2)_6$, $K_2NaCo(NO_2)_6$, $K_2PbCu(NO_2)_6$, $K_3Ca(NO_2)_6$, $Rb_2AgBi(NO_2)_6$, $Rb_2NaBi(NO_2)_6$, $Tl_2AgBi(NO_2)_6$, $Tl_2LiBi(NO_2)_6$, and $Tl_2NaBi(NO_2)_6$

## Face-centered Cubic primitive vectors:

$$\mathbf{a}_1 = \frac{1}{2}\,a\,\hat{\mathbf{y}} + \frac{1}{2}\,a\,\hat{\mathbf{z}}$$
$$\mathbf{a}_2 = \frac{1}{2}\,a\,\hat{\mathbf{x}} + \frac{1}{2}\,a\,\hat{\mathbf{z}}$$
$$\mathbf{a}_3 = \frac{1}{2}\,a\,\hat{\mathbf{x}} + \frac{1}{2}\,a\,\hat{\mathbf{y}}$$

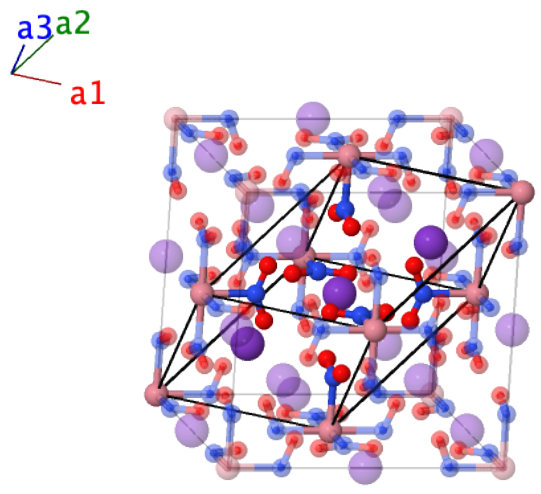

## Basis vectors:

| | | Lattice Coordinates | | Cartesian Coordinates | Wyckoff Position | Atom Type |
|---|---|---|---|---|---|---|
| $\mathbf{B}_1$ | = | $0\,\mathbf{a}_1 + 0\,\mathbf{a}_2 + 0\,\mathbf{a}_3$ | = | $0\,\hat{\mathbf{x}} + 0\,\hat{\mathbf{y}} + 0\,\hat{\mathbf{z}}$ | (4$a$) | Co |
| $\mathbf{B}_2$ | = | $\frac{1}{2}\,\mathbf{a}_1 + \frac{1}{2}\,\mathbf{a}_2 + \frac{1}{2}\,\mathbf{a}_3$ | = | $\frac{1}{2}a\,\hat{\mathbf{x}} + \frac{1}{2}a\,\hat{\mathbf{y}} + \frac{1}{2}a\,\hat{\mathbf{z}}$ | (4$b$) | K I |
| $\mathbf{B}_3$ | = | $\frac{1}{4}\,\mathbf{a}_1 + \frac{1}{4}\,\mathbf{a}_2 + \frac{1}{4}\,\mathbf{a}_3$ | = | $\frac{1}{4}a\,\hat{\mathbf{x}} + \frac{1}{4}a\,\hat{\mathbf{y}} + \frac{1}{4}a\,\hat{\mathbf{z}}$ | (8$c$) | K II |
| $\mathbf{B}_4$ | = | $\frac{3}{4}\,\mathbf{a}_1 + \frac{3}{4}\,\mathbf{a}_2 + \frac{3}{4}\,\mathbf{a}_3$ | = | $\frac{3}{4}a\,\hat{\mathbf{x}} + \frac{3}{4}a\,\hat{\mathbf{y}} + \frac{3}{4}a\,\hat{\mathbf{z}}$ | (8$c$) | K II |
| $\mathbf{B}_5$ | = | $-x_4\,\mathbf{a}_1 + x_4\,\mathbf{a}_2 + x_4\,\mathbf{a}_3$ | = | $x_4a\,\hat{\mathbf{x}}$ | (24$e$) | N |
| $\mathbf{B}_6$ | = | $x_4\,\mathbf{a}_1 - x_4\,\mathbf{a}_2 - x_4\,\mathbf{a}_3$ | = | $-x_4a\,\hat{\mathbf{x}}$ | (24$e$) | N |
| $\mathbf{B}_7$ | = | $x_4\,\mathbf{a}_1 - x_4\,\mathbf{a}_2 + x_4\,\mathbf{a}_3$ | = | $x_4a\,\hat{\mathbf{y}}$ | (24$e$) | N |
| $\mathbf{B}_8$ | = | $-x_4\,\mathbf{a}_1 + x_4\,\mathbf{a}_2 - x_4\,\mathbf{a}_3$ | = | $-x_4a\,\hat{\mathbf{y}}$ | (24$e$) | N |
| $\mathbf{B}_9$ | = | $x_4\,\mathbf{a}_1 + x_4\,\mathbf{a}_2 - x_4\,\mathbf{a}_3$ | = | $x_4a\,\hat{\mathbf{z}}$ | (24$e$) | N |
| $\mathbf{B}_{10}$ | = | $-x_4\,\mathbf{a}_1 - x_4\,\mathbf{a}_2 + x_4\,\mathbf{a}_3$ | = | $-x_4a\,\hat{\mathbf{z}}$ | (24$e$) | N |
| $\mathbf{B}_{11}$ | = | $(y_5 + z_5)\,\mathbf{a}_1 + (-y_5 + z_5)\,\mathbf{a}_2 + (y_5 - z_5)\,\mathbf{a}_3$ | = | $y_5a\,\hat{\mathbf{y}} + z_5a\,\hat{\mathbf{z}}$ | (48$h$) | O |
| $\mathbf{B}_{12}$ | = | $(-y_5 + z_5)\,\mathbf{a}_1 + (y_5 + z_5)\,\mathbf{a}_2 + (-y_5 - z_5)\,\mathbf{a}_3$ | = | $-y_5a\,\hat{\mathbf{y}} + z_5a\,\hat{\mathbf{z}}$ | (48$h$) | O |
| $\mathbf{B}_{13}$ | = | $(y_5 - z_5)\,\mathbf{a}_1 + (-y_5 - z_5)\,\mathbf{a}_2 + (y_5 + z_5)\,\mathbf{a}_3$ | = | $y_5a\,\hat{\mathbf{y}} - z_5a\,\hat{\mathbf{z}}$ | (48$h$) | O |
| $\mathbf{B}_{14}$ | = | $(-y_5 - z_5)\,\mathbf{a}_1 + (y_5 - z_5)\,\mathbf{a}_2 + (-y_5 + z_5)\,\mathbf{a}_3$ | = | $-y_5a\,\hat{\mathbf{y}} - z_5a\,\hat{\mathbf{z}}$ | (48$h$) | O |
| $\mathbf{B}_{15}$ | = | $(y_5 - z_5)\,\mathbf{a}_1 + (y_5 + z_5)\,\mathbf{a}_2 + (-y_5 + z_5)\,\mathbf{a}_3$ | = | $z_5a\,\hat{\mathbf{x}} + y_5a\,\hat{\mathbf{z}}$ | (48$h$) | O |
| $\mathbf{B}_{16}$ | = | $(-y_5 - z_5)\,\mathbf{a}_1 + (-y_5 + z_5)\,\mathbf{a}_2 + (y_5 + z_5)\,\mathbf{a}_3$ | = | $z_5a\,\hat{\mathbf{x}} - y_5a\,\hat{\mathbf{z}}$ | (48$h$) | O |
| $\mathbf{B}_{17}$ | = | $(y_5 + z_5)\,\mathbf{a}_1 + (y_5 - z_5)\,\mathbf{a}_2 + (-y_5 - z_5)\,\mathbf{a}_3$ | = | $-z_5a\,\hat{\mathbf{x}} + y_5a\,\hat{\mathbf{z}}$ | (48$h$) | O |
| $\mathbf{B}_{18}$ | = | $(-y_5 + z_5)\,\mathbf{a}_1 + (-y_5 - z_5)\,\mathbf{a}_2 + (y_5 - z_5)\,\mathbf{a}_3$ | = | $-z_5a\,\hat{\mathbf{x}} - y_5a\,\hat{\mathbf{z}}$ | (48$h$) | O |
| $\mathbf{B}_{19}$ | = | $(-y_5 + z_5)\,\mathbf{a}_1 + (y_5 - z_5)\,\mathbf{a}_2 + (y_5 + z_5)\,\mathbf{a}_3$ | = | $y_5a\,\hat{\mathbf{x}} + z_5a\,\hat{\mathbf{y}}$ | (48$h$) | O |
| $\mathbf{B}_{20}$ | = | $(y_5 + z_5)\,\mathbf{a}_1 + (-y_5 - z_5)\,\mathbf{a}_2 + (-y_5 + z_5)\,\mathbf{a}_3$ | = | $-y_5a\,\hat{\mathbf{x}} + z_5a\,\hat{\mathbf{y}}$ | (48$h$) | O |
| $\mathbf{B}_{21}$ | = | $(-y_5 - z_5)\,\mathbf{a}_1 + (y_5 + z_5)\,\mathbf{a}_2 + (y_5 - z_5)\,\mathbf{a}_3$ | = | $y_5a\,\hat{\mathbf{x}} - z_5a\,\hat{\mathbf{y}}$ | (48$h$) | O |
| $\mathbf{B}_{22}$ | = | $(y_5 - z_5)\,\mathbf{a}_1 + (-y_5 + z_5)\,\mathbf{a}_2 + (-y_5 - z_5)\,\mathbf{a}_3$ | = | $-y_5a\,\hat{\mathbf{x}} - z_5a\,\hat{\mathbf{y}}$ | (48$h$) | O |

**Geometry files:**
- CIF: pp. 1776
- POSCAR: pp. 1776

# $LaFe_4P_{12}$ Structure: A4BC12_cI34_204_c_a_g

http://aflow.org/prototype-encyclopedia/A4BC12_cI34_204_c_a_g

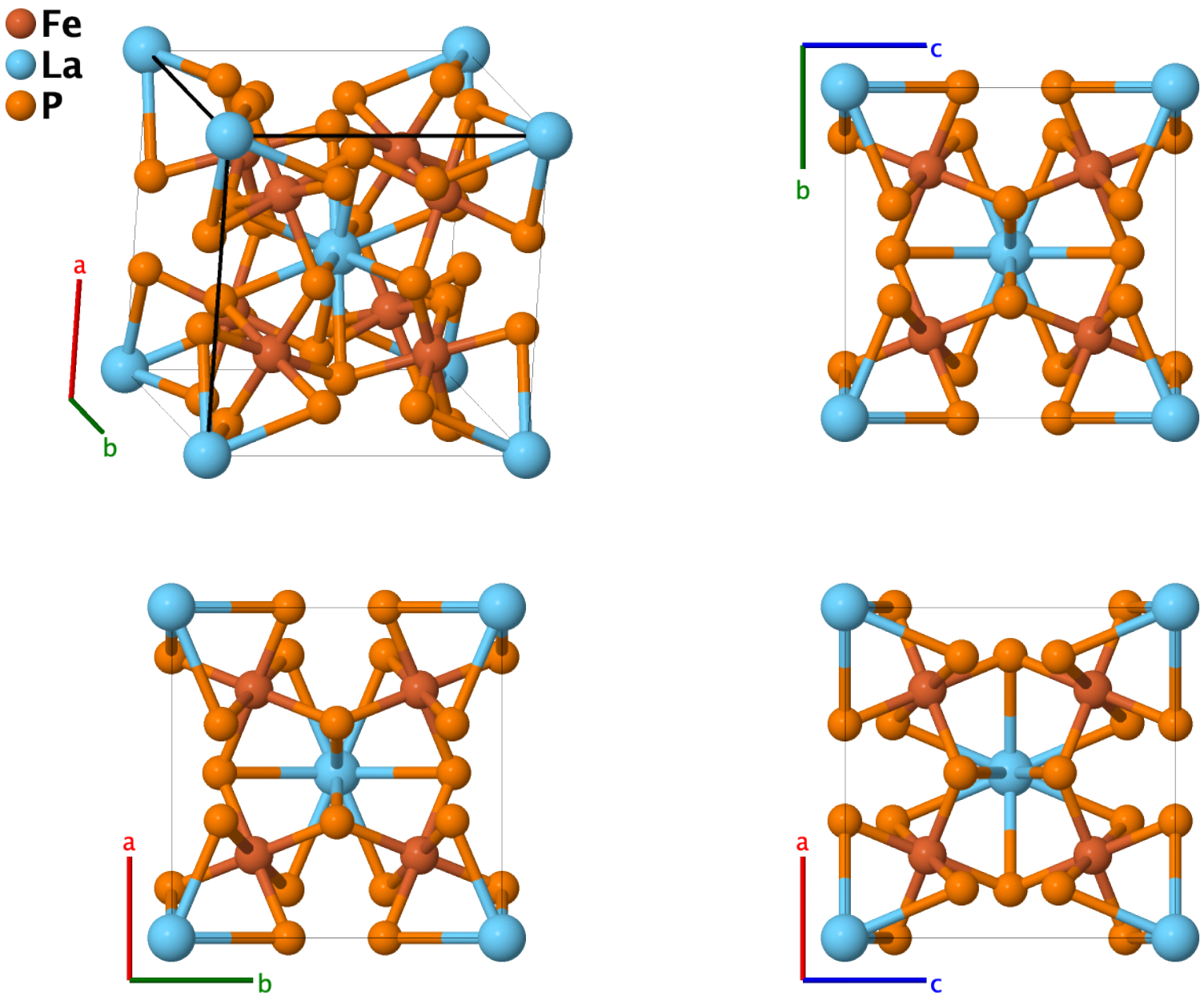

| | | |
|---|---|---|
| **Prototype** | : | $Fe_4LaP_{12}$ |
| **AFLOW prototype label** | : | A4BC12_cI34_204_c_a_g |
| ***Strukturbericht* designation** | : | None |
| **Pearson symbol** | : | cI34 |
| **Space group number** | : | 204 |
| **Space group symbol** | : | $Im\bar{3}$ |
| **AFLOW prototype command** | : | `aflow --proto=A4BC12_cI34_204_c_a_g --params=`$a, y_3, z_3$ |

### Other compounds with this structure

- $CeFe_4P_{12}$, $EuFe_4P_{12}$, $NdFe_4P_{12}$, $PrFe_4P_{12}$, $SmFe_4P_{12}$, $CeRu_4P_{12}$, $EuRu_4P_{12}$, $LaRu_4P_{12}$, $NdRu_4P_{12}$, $PrRu_4P_{12}$, $SmRu_4P_{12}$, $CeOs_4P_{12}$, $LaOs_4P_{12}$, $NdOs_4P_{12}$, $PrOs_4P_{12}$, $SmOs_4P_{12}$, $CeFe_4Sb_{12}$, $LaFe_4Sb_{12}$, $PrFe_4Sb_{12}$, $SmFe_4Sb_{12}$, $CeRu_4Sb_{12}$, $EuRu_4Sb_{12}$, $LaRu_4Sb_{12}$, $NdRu_4Sb_{12}$, $PrRu_4Sb_{12}$, $SmRu_4Sb_{12}$, $CeOs_4Sb_{12}$, $EuOs_4Sb_{12}$, $LaOs_4Sb_{12}$, $NdOs_4Sb_{12}$, $PrOs_4Sb_{12}$, and $SmOs_4Sb_{12}$

- This is a filled skutterudite ($CoAs_3$, $D0_2$) structure. Removing the rare-earth atom from the ($2a$) position leaves the skutterudite structure.

### Body-centered Cubic primitive vectors:

$$\mathbf{a}_1 = -\frac{1}{2}\,a\,\hat{\mathbf{x}} + \frac{1}{2}\,a\,\hat{\mathbf{y}} + \frac{1}{2}\,a\,\hat{\mathbf{z}}$$
$$\mathbf{a}_2 = \frac{1}{2}\,a\,\hat{\mathbf{x}} - \frac{1}{2}\,a\,\hat{\mathbf{y}} + \frac{1}{2}\,a\,\hat{\mathbf{z}}$$
$$\mathbf{a}_3 = \frac{1}{2}\,a\,\hat{\mathbf{x}} + \frac{1}{2}\,a\,\hat{\mathbf{y}} - \frac{1}{2}\,a\,\hat{\mathbf{z}}$$

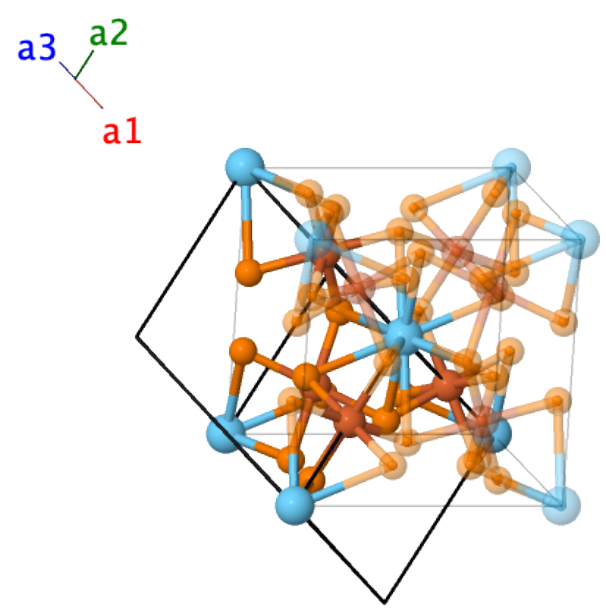

**Basis vectors:**

| | | Lattice Coordinates | | Cartesian Coordinates | Wyckoff Position | Atom Type |
|---|---|---|---|---|---|---|
| $\mathbf{B}_1$ | = | $0\,\mathbf{a}_1 + 0\,\mathbf{a}_2 + 0\,\mathbf{a}_3$ | = | $0\,\hat{\mathbf{x}} + 0\,\hat{\mathbf{y}} + 0\,\hat{\mathbf{z}}$ | (2$a$) | La |
| $\mathbf{B}_2$ | = | $\frac{1}{2}\,\mathbf{a}_1 + \frac{1}{2}\,\mathbf{a}_2 + \frac{1}{2}\,\mathbf{a}_3$ | = | $\frac{1}{4}a\,\hat{\mathbf{x}} + \frac{1}{4}a\,\hat{\mathbf{y}} + \frac{1}{4}a\,\hat{\mathbf{z}}$ | (8$c$) | Fe |
| $\mathbf{B}_3$ | = | $\frac{1}{2}\,\mathbf{a}_3$ | = | $\frac{1}{4}a\,\hat{\mathbf{x}} + \frac{1}{4}a\,\hat{\mathbf{y}} - \frac{1}{4}a\,\hat{\mathbf{z}}$ | (8$c$) | Fe |
| $\mathbf{B}_4$ | = | $\frac{1}{2}\,\mathbf{a}_2$ | = | $\frac{1}{4}a\,\hat{\mathbf{x}} - \frac{1}{4}a\,\hat{\mathbf{y}} + \frac{1}{4}a\,\hat{\mathbf{z}}$ | (8$c$) | Fe |
| $\mathbf{B}_5$ | = | $\frac{1}{2}\,\mathbf{a}_1$ | = | $-\frac{1}{4}a\,\hat{\mathbf{x}} + \frac{1}{4}a\,\hat{\mathbf{y}} + \frac{1}{4}a\,\hat{\mathbf{z}}$ | (8$c$) | Fe |
| $\mathbf{B}_6$ | = | $(y_3 + z_3)\,\mathbf{a}_1 + z_3\,\mathbf{a}_2 + y_3\,\mathbf{a}_3$ | = | $y_3a\,\hat{\mathbf{y}} + z_3a\,\hat{\mathbf{z}}$ | (24$g$) | P |
| $\mathbf{B}_7$ | = | $(-y_3 + z_3)\,\mathbf{a}_1 + z_3\,\mathbf{a}_2 - y_3\,\mathbf{a}_3$ | = | $-y_3a\,\hat{\mathbf{y}} + z_3a\,\hat{\mathbf{z}}$ | (24$g$) | P |
| $\mathbf{B}_8$ | = | $(y_3 - z_3)\,\mathbf{a}_1 - z_3\,\mathbf{a}_2 + y_3\,\mathbf{a}_3$ | = | $y_3a\,\hat{\mathbf{y}} - z_3a\,\hat{\mathbf{z}}$ | (24$g$) | P |
| $\mathbf{B}_9$ | = | $(-y_3 - z_3)\,\mathbf{a}_1 - z_3\,\mathbf{a}_2 - y_3\,\mathbf{a}_3$ | = | $-y_3a\,\hat{\mathbf{y}} - z_3a\,\hat{\mathbf{z}}$ | (24$g$) | P |
| $\mathbf{B}_{10}$ | = | $y_3\,\mathbf{a}_1 + (y_3 + z_3)\,\mathbf{a}_2 + z_3\,\mathbf{a}_3$ | = | $z_3a\,\hat{\mathbf{x}} + y_3a\,\hat{\mathbf{z}}$ | (24$g$) | P |
| $\mathbf{B}_{11}$ | = | $-y_3\,\mathbf{a}_1 + (-y_3 + z_3)\,\mathbf{a}_2 + z_3\,\mathbf{a}_3$ | = | $z_3a\,\hat{\mathbf{x}} - y_3a\,\hat{\mathbf{z}}$ | (24$g$) | P |
| $\mathbf{B}_{12}$ | = | $y_3\,\mathbf{a}_1 + (y_3 - z_3)\,\mathbf{a}_2 - z_3\,\mathbf{a}_3$ | = | $-z_3a\,\hat{\mathbf{x}} + y_3a\,\hat{\mathbf{z}}$ | (24$g$) | P |
| $\mathbf{B}_{13}$ | = | $-y_3\,\mathbf{a}_1 + (-y_3 - z_3)\,\mathbf{a}_2 - z_3\,\mathbf{a}_3$ | = | $-z_3a\,\hat{\mathbf{x}} - y_3a\,\hat{\mathbf{z}}$ | (24$g$) | P |
| $\mathbf{B}_{14}$ | = | $z_3\,\mathbf{a}_1 + y_3\,\mathbf{a}_2 + (y_3 + z_3)\,\mathbf{a}_3$ | = | $y_3a\,\hat{\mathbf{x}} + z_3a\,\hat{\mathbf{y}}$ | (24$g$) | P |
| $\mathbf{B}_{15}$ | = | $z_3\,\mathbf{a}_1 - y_3\,\mathbf{a}_2 + (-y_3 + z_3)\,\mathbf{a}_3$ | = | $-y_3a\,\hat{\mathbf{x}} + z_3a\,\hat{\mathbf{y}}$ | (24$g$) | P |
| $\mathbf{B}_{16}$ | = | $-z_3\,\mathbf{a}_1 + y_3\,\mathbf{a}_2 + (y_3 - z_3)\,\mathbf{a}_3$ | = | $y_3a\,\hat{\mathbf{x}} - z_3a\,\hat{\mathbf{y}}$ | (24$g$) | P |
| $\mathbf{B}_{17}$ | = | $-z_3\,\mathbf{a}_1 - y_3\,\mathbf{a}_2 + (-y_3 - z_3)\,\mathbf{a}_3$ | = | $-y_3a\,\hat{\mathbf{x}} - z_3a\,\hat{\mathbf{y}}$ | (24$g$) | P |

**Found in:**

**Geometry files:**

- CIF: pp. 1776
- POSCAR: pp. 1777

# $NaMn_7O_{12}$ Structure: A7BC12_cI40_204_bc_a_g

http://aflow.org/prototype-encyclopedia/A7BC12_cI40_204_bc_a_g

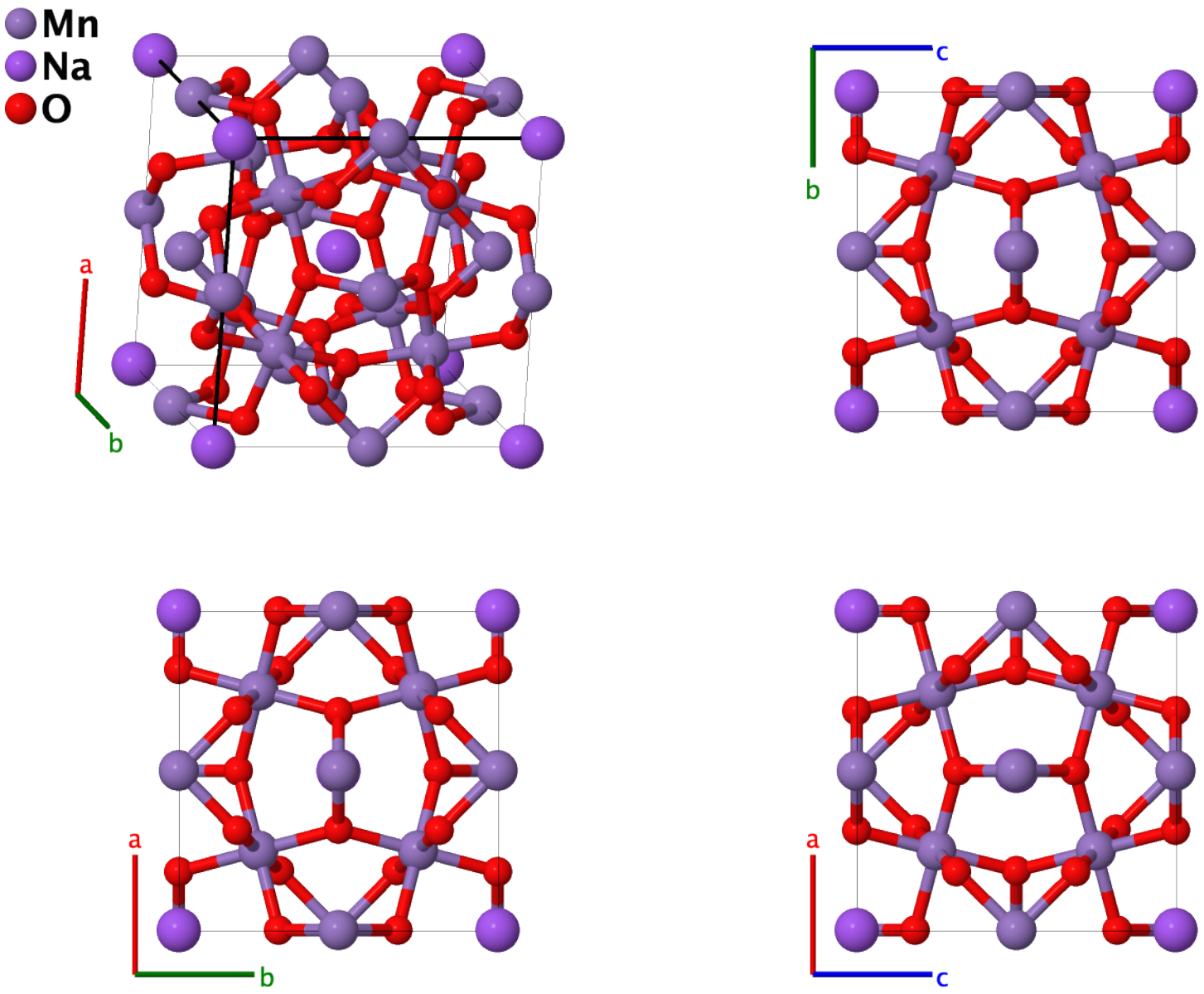

| | | |
|---|---|---|
| **Prototype** | : | $Mn_7NaO_{12}$ |
| **AFLOW prototype label** | : | A7BC12_cI40_204_bc_a_g |
| ***Strukturbericht* designation** | : | None |
| **Pearson symbol** | : | cI40 |
| **Space group number** | : | 204 |
| **Space group symbol** | : | $Im\bar{3}$ |
| **AFLOW prototype command** | : | `aflow --proto=A7BC12_cI40_204_bc_a_g --params=`$a, y_4, z_4$ |

**Other compounds with this structure**

- $CaCu_3Ge_4O_{12}$ and $CaCu_3Mn_4O_{12}$

- This is a double perovskite structure, and is only stable above 3 Gbar and above room temperature, but it is metastable under ambient conditions (Gilioli 2005ab). The actual composition of the measured sample is $Na_{0.95}Mn_{7.05}O_{12}$, with the excess manganese displacing some of the sodium atoms on the (2$a$) site.

**Body-centered Cubic primitive vectors:**

$$\mathbf{a}_1 = -\frac{1}{2}\,a\,\hat{\mathbf{x}} + \frac{1}{2}\,a\,\hat{\mathbf{y}} + \frac{1}{2}\,a\,\hat{\mathbf{z}}$$
$$\mathbf{a}_2 = \frac{1}{2}\,a\,\hat{\mathbf{x}} - \frac{1}{2}\,a\,\hat{\mathbf{y}} + \frac{1}{2}\,a\,\hat{\mathbf{z}}$$
$$\mathbf{a}_3 = \frac{1}{2}\,a\,\hat{\mathbf{x}} + \frac{1}{2}\,a\,\hat{\mathbf{y}} - \frac{1}{2}\,a\,\hat{\mathbf{z}}$$

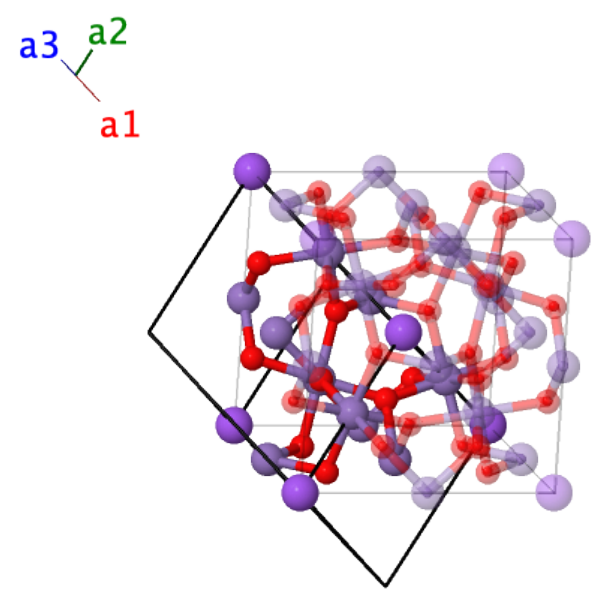

**Basis vectors:**

| | | Lattice Coordinates | | Cartesian Coordinates | Wyckoff Position | Atom Type |
|---|---|---|---|---|---|---|
| $\mathbf{B}_1$ | = | $0\,\mathbf{a}_1 + 0\,\mathbf{a}_2 + 0\,\mathbf{a}_3$ | = | $0\,\hat{\mathbf{x}} + 0\,\hat{\mathbf{y}} + 0\,\hat{\mathbf{z}}$ | $(2a)$ | Na |
| $\mathbf{B}_2$ | = | $\frac{1}{2}\,\mathbf{a}_2 + \frac{1}{2}\,\mathbf{a}_3$ | = | $\frac{1}{2}a\,\hat{\mathbf{x}}$ | $(6b)$ | Mn I |
| $\mathbf{B}_3$ | = | $\frac{1}{2}\,\mathbf{a}_1 + \frac{1}{2}\,\mathbf{a}_3$ | = | $\frac{1}{2}a\,\hat{\mathbf{y}}$ | $(6b)$ | Mn I |
| $\mathbf{B}_4$ | = | $\frac{1}{2}\,\mathbf{a}_1 + \frac{1}{2}\,\mathbf{a}_2$ | = | $\frac{1}{2}a\,\hat{\mathbf{z}}$ | $(6b)$ | Mn I |
| $\mathbf{B}_5$ | = | $\frac{1}{2}\,\mathbf{a}_1 + \frac{1}{2}\,\mathbf{a}_2 + \frac{1}{2}\,\mathbf{a}_3$ | = | $\frac{1}{4}a\,\hat{\mathbf{x}} + \frac{1}{4}a\,\hat{\mathbf{y}} + \frac{1}{4}a\,\hat{\mathbf{z}}$ | $(8c)$ | Mn II |
| $\mathbf{B}_6$ | = | $\frac{1}{2}\,\mathbf{a}_3$ | = | $\frac{1}{4}a\,\hat{\mathbf{x}} + \frac{1}{4}a\,\hat{\mathbf{y}} - \frac{1}{4}a\,\hat{\mathbf{z}}$ | $(8c)$ | Mn II |
| $\mathbf{B}_7$ | = | $\frac{1}{2}\,\mathbf{a}_2$ | = | $\frac{1}{4}a\,\hat{\mathbf{x}} - \frac{1}{4}a\,\hat{\mathbf{y}} + \frac{1}{4}a\,\hat{\mathbf{z}}$ | $(8c)$ | Mn II |
| $\mathbf{B}_8$ | = | $\frac{1}{2}\,\mathbf{a}_1$ | = | $-\frac{1}{4}a\,\hat{\mathbf{x}} + \frac{1}{4}a\,\hat{\mathbf{y}} + \frac{1}{4}a\,\hat{\mathbf{z}}$ | $(8c)$ | Mn II |
| $\mathbf{B}_9$ | = | $(y_4 + z_4)\,\mathbf{a}_1 + z_4\,\mathbf{a}_2 + y_4\,\mathbf{a}_3$ | = | $y_4a\,\hat{\mathbf{y}} + z_4a\,\hat{\mathbf{z}}$ | $(24g)$ | O |
| $\mathbf{B}_{10}$ | = | $(-y_4 + z_4)\,\mathbf{a}_1 + z_4\,\mathbf{a}_2 - y_4\,\mathbf{a}_3$ | = | $-y_4a\,\hat{\mathbf{y}} + z_4a\,\hat{\mathbf{z}}$ | $(24g)$ | O |
| $\mathbf{B}_{11}$ | = | $(y_4 - z_4)\,\mathbf{a}_1 - z_4\,\mathbf{a}_2 + y_4\,\mathbf{a}_3$ | = | $y_4a\,\hat{\mathbf{y}} - z_4a\,\hat{\mathbf{z}}$ | $(24g)$ | O |
| $\mathbf{B}_{12}$ | = | $(-y_4 - z_4)\,\mathbf{a}_1 - z_4\,\mathbf{a}_2 - y_4\,\mathbf{a}_3$ | = | $-y_4a\,\hat{\mathbf{y}} - z_4a\,\hat{\mathbf{z}}$ | $(24g)$ | O |
| $\mathbf{B}_{13}$ | = | $y_4\,\mathbf{a}_1 + (y_4 + z_4)\,\mathbf{a}_2 + z_4\,\mathbf{a}_3$ | = | $z_4a\,\hat{\mathbf{x}} + y_4a\,\hat{\mathbf{z}}$ | $(24g)$ | O |
| $\mathbf{B}_{14}$ | = | $-y_4\,\mathbf{a}_1 + (-y_4 + z_4)\,\mathbf{a}_2 + z_4\,\mathbf{a}_3$ | = | $z_4a\,\hat{\mathbf{x}} - y_4a\,\hat{\mathbf{z}}$ | $(24g)$ | O |
| $\mathbf{B}_{15}$ | = | $y_4\,\mathbf{a}_1 + (y_4 - z_4)\,\mathbf{a}_2 - z_4\,\mathbf{a}_3$ | = | $-z_4a\,\hat{\mathbf{x}} + y_4a\,\hat{\mathbf{z}}$ | $(24g)$ | O |
| $\mathbf{B}_{16}$ | = | $-y_4\,\mathbf{a}_1 + (-y_4 - z_4)\,\mathbf{a}_2 - z_4\,\mathbf{a}_3$ | = | $-z_4a\,\hat{\mathbf{x}} - y_4a\,\hat{\mathbf{z}}$ | $(24g)$ | O |
| $\mathbf{B}_{17}$ | = | $z_4\,\mathbf{a}_1 + y_4\,\mathbf{a}_2 + (y_4 + z_4)\,\mathbf{a}_3$ | = | $y_4a\,\hat{\mathbf{x}} + z_4a\,\hat{\mathbf{y}}$ | $(24g)$ | O |
| $\mathbf{B}_{18}$ | = | $z_4\,\mathbf{a}_1 - y_4\,\mathbf{a}_2 + (-y_4 + z_4)\,\mathbf{a}_3$ | = | $-y_4a\,\hat{\mathbf{x}} + z_4a\,\hat{\mathbf{y}}$ | $(24g)$ | O |
| $\mathbf{B}_{19}$ | = | $-z_4\,\mathbf{a}_1 + y_4\,\mathbf{a}_2 + (y_4 - z_4)\,\mathbf{a}_3$ | = | $y_4a\,\hat{\mathbf{x}} - z_4a\,\hat{\mathbf{y}}$ | $(24g)$ | O |
| $\mathbf{B}_{20}$ | = | $-z_4\,\mathbf{a}_1 - y_4\,\mathbf{a}_2 + (-y_4 - z_4)\,\mathbf{a}_3$ | = | $-y_4a\,\hat{\mathbf{x}} - z_4a\,\hat{\mathbf{y}}$ | $(24g)$ | O |

**Geometry files:**

- CIF: pp. 1777
- POSCAR: pp. 1778

# $NO_2$ (Modern, $C26$) Structure: AB2_cI36_204_d_g

http://aflow.org/prototype-encyclopedia/AB2_cI36_204_d_g

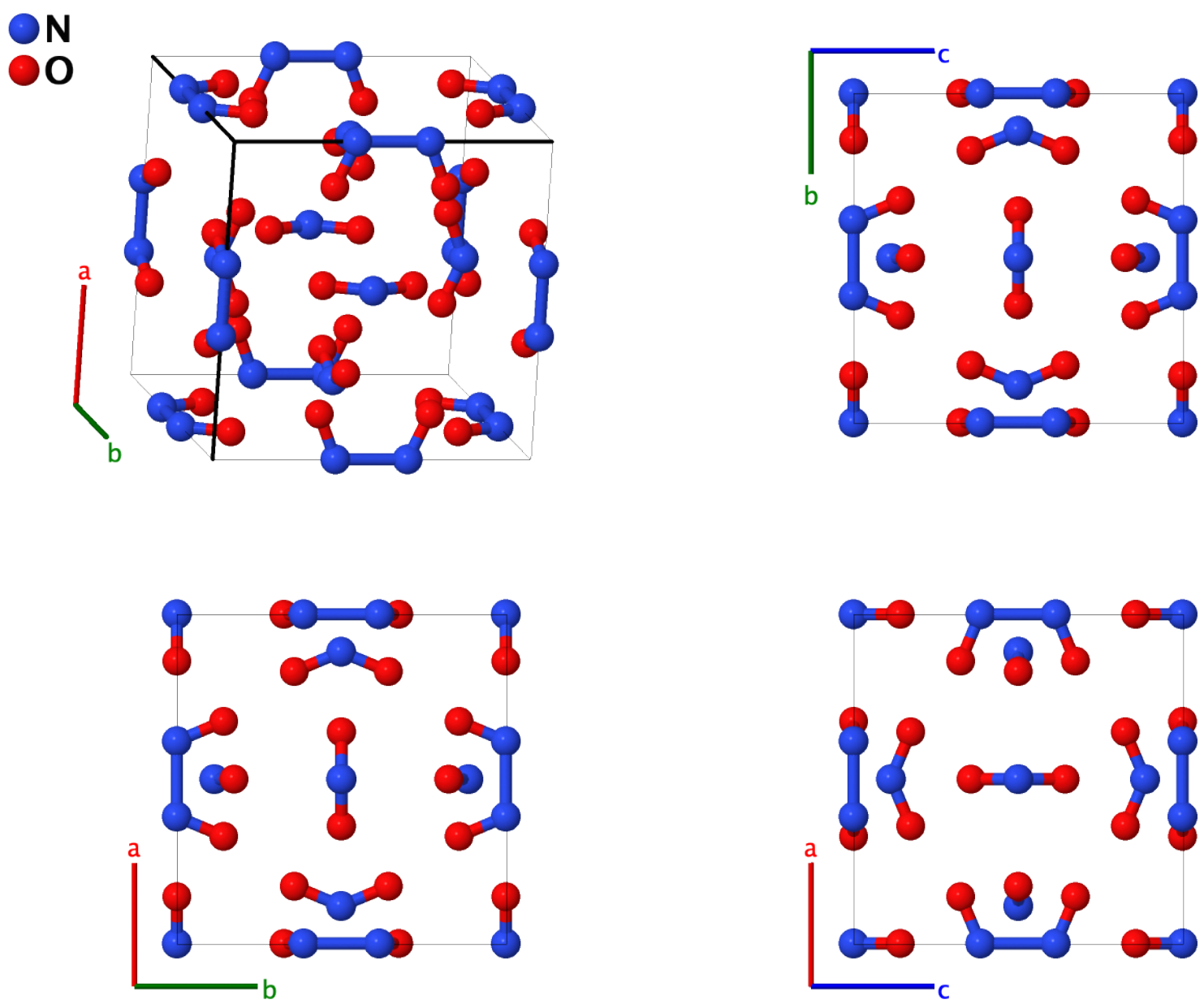

| | | |
|---|---|---|
| **Prototype** | : | $NO_2$ |
| **AFLOW prototype label** | : | AB2_cI36_204_d_g |
| ***Strukturbericht* designation** | : | $C26$ |
| **Pearson symbol** | : | cI36 |
| **Space group number** | : | 204 |
| **Space group symbol** | : | $Im\bar{3}$ |
| **AFLOW prototype command** | : | `aflow --proto=AB2_cI36_204_d_g --params=`$a, x_1, y_2, z_2$ |

- (Hermann, 1937) listed two possible structures for the low temperature solid cubic phase of $NO_2$, which were given *Strukturbericht* designations $C26_a$ and $C26_b$, the only structures with Roman subscripts in the original series.
- $C26_a$ (AB2_cI36_199_b_c) was set in space group $I2_13$ #199. Hermann noted that this structure has a very short distance (1.88 Å) between oxygen atoms on different $NO_2$ molecules, and that this structure does not form the $(NO_2)_2$ aggregate molecule found in the $C26_b$ structure, making "making this proposed structure very unlikely."
- Recognizing this, (Hendricks, 1931) suggested that $NO_2$ was actually in space group $I23$ #197. (Hermann, 1997) gave this structure the $C26_b$ designation, but noted that based on Hendricks's atomic positions the space group was actually $Im\bar{3}$ #204.
- The modern accepted structure for $NO_2$ (AB2_cI36_204_d_g) is set in space group $Im\bar{3}$ #204, confirming Hendricks. We follow (Villars, 2005) and give this the $C26$ designation.
- We used the experimental data for this phase collected at 20 K.

### Body-centered Cubic primitive vectors:

$$\mathbf{a}_1 = -\frac{1}{2}\,a\,\hat{\mathbf{x}} + \frac{1}{2}\,a\,\hat{\mathbf{y}} + \frac{1}{2}\,a\,\hat{\mathbf{z}}$$
$$\mathbf{a}_2 = \frac{1}{2}\,a\,\hat{\mathbf{x}} - \frac{1}{2}\,a\,\hat{\mathbf{y}} + \frac{1}{2}\,a\,\hat{\mathbf{z}}$$
$$\mathbf{a}_3 = \frac{1}{2}\,a\,\hat{\mathbf{x}} + \frac{1}{2}\,a\,\hat{\mathbf{y}} - \frac{1}{2}\,a\,\hat{\mathbf{z}}$$

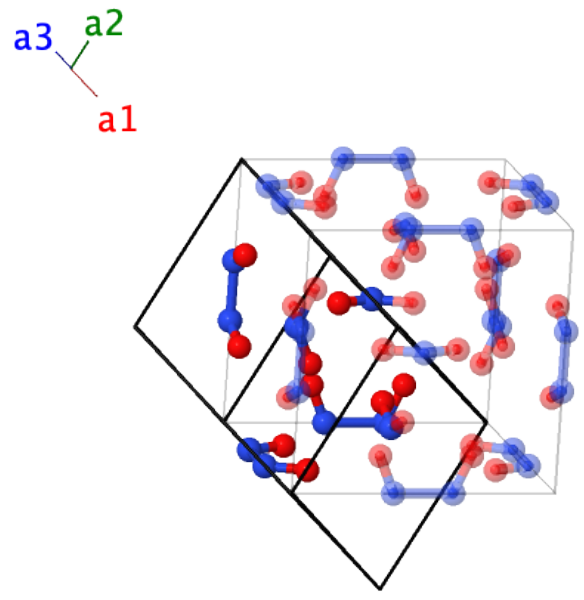

### Basis vectors:

| | | Lattice Coordinates | | Cartesian Coordinates | Wyckoff Position | Atom Type |
|---|---|---|---|---|---|---|
| $\mathbf{B}_1$ | = | $x_1\,\mathbf{a}_2 + x_1\,\mathbf{a}_3$ | = | $x_1 a\,\hat{\mathbf{x}}$ | (12$d$) | N |
| $\mathbf{B}_2$ | = | $-x_1\,\mathbf{a}_2 - x_1\,\mathbf{a}_3$ | = | $-x_1 a\,\hat{\mathbf{x}}$ | (12$d$) | N |
| $\mathbf{B}_3$ | = | $x_1\,\mathbf{a}_1 + x_1\,\mathbf{a}_3$ | = | $x_1 a\,\hat{\mathbf{y}}$ | (12$d$) | N |
| $\mathbf{B}_4$ | = | $-x_1\,\mathbf{a}_1 - x_1\,\mathbf{a}_3$ | = | $-x_1 a\,\hat{\mathbf{y}}$ | (12$d$) | N |
| $\mathbf{B}_5$ | = | $x_1\,\mathbf{a}_1 + x_1\,\mathbf{a}_2$ | = | $x_1 a\,\hat{\mathbf{z}}$ | (12$d$) | N |
| $\mathbf{B}_6$ | = | $-x_1\,\mathbf{a}_1 - x_1\,\mathbf{a}_2$ | = | $-x_1 a\,\hat{\mathbf{z}}$ | (12$d$) | N |
| $\mathbf{B}_7$ | = | $(y_2 + z_2)\,\mathbf{a}_1 + z_2\,\mathbf{a}_2 + y_2\,\mathbf{a}_3$ | = | $y_2 a\,\hat{\mathbf{y}} + z_2 a\,\hat{\mathbf{z}}$ | (24$g$) | O |
| $\mathbf{B}_8$ | = | $(-y_2 + z_2)\,\mathbf{a}_1 + z_2\,\mathbf{a}_2 - y_2\,\mathbf{a}_3$ | = | $-y_2 a\,\hat{\mathbf{y}} + z_2 a\,\hat{\mathbf{z}}$ | (24$g$) | O |
| $\mathbf{B}_9$ | = | $(y_2 - z_2)\,\mathbf{a}_1 - z_2\,\mathbf{a}_2 + y_2\,\mathbf{a}_3$ | = | $y_2 a\,\hat{\mathbf{y}} - z_2 a\,\hat{\mathbf{z}}$ | (24$g$) | O |
| $\mathbf{B}_{10}$ | = | $(-y_2 - z_2)\,\mathbf{a}_1 - z_2\,\mathbf{a}_2 - y_2\,\mathbf{a}_3$ | = | $-y_2 a\,\hat{\mathbf{y}} - z_2 a\,\hat{\mathbf{z}}$ | (24$g$) | O |
| $\mathbf{B}_{11}$ | = | $y_2\,\mathbf{a}_1 + (y_2 + z_2)\,\mathbf{a}_2 + z_2\,\mathbf{a}_3$ | = | $z_2 a\,\hat{\mathbf{x}} + y_2 a\,\hat{\mathbf{z}}$ | (24$g$) | O |
| $\mathbf{B}_{12}$ | = | $-y_2\,\mathbf{a}_1 + (-y_2 + z_2)\,\mathbf{a}_2 + z_2\,\mathbf{a}_3$ | = | $z_2 a\,\hat{\mathbf{x}} - y_2 a\,\hat{\mathbf{z}}$ | (24$g$) | O |
| $\mathbf{B}_{13}$ | = | $y_2\,\mathbf{a}_1 + (y_2 - z_2)\,\mathbf{a}_2 - z_2\,\mathbf{a}_3$ | = | $-z_2 a\,\hat{\mathbf{x}} + y_2 a\,\hat{\mathbf{z}}$ | (24$g$) | O |
| $\mathbf{B}_{14}$ | = | $-y_2\,\mathbf{a}_1 + (-y_2 - z_2)\,\mathbf{a}_2 - z_2\,\mathbf{a}_3$ | = | $-z_2 a\,\hat{\mathbf{x}} - y_2 a\,\hat{\mathbf{z}}$ | (24$g$) | O |
| $\mathbf{B}_{15}$ | = | $z_2\,\mathbf{a}_1 + y_2\,\mathbf{a}_2 + (y_2 + z_2)\,\mathbf{a}_3$ | = | $y_2 a\,\hat{\mathbf{x}} + z_2 a\,\hat{\mathbf{y}}$ | (24$g$) | O |
| $\mathbf{B}_{16}$ | = | $z_2\,\mathbf{a}_1 - y_2\,\mathbf{a}_2 + (-y_2 + z_2)\,\mathbf{a}_3$ | = | $-y_2 a\,\hat{\mathbf{x}} + z_2 a\,\hat{\mathbf{y}}$ | (24$g$) | O |
| $\mathbf{B}_{17}$ | = | $-z_2\,\mathbf{a}_1 + y_2\,\mathbf{a}_2 + (y_2 - z_2)\,\mathbf{a}_3$ | = | $y_2 a\,\hat{\mathbf{x}} - z_2 a\,\hat{\mathbf{y}}$ | (24$g$) | O |
| $\mathbf{B}_{18}$ | = | $-z_2\,\mathbf{a}_1 - y_2\,\mathbf{a}_2 + (-y_2 - z_2)\,\mathbf{a}_3$ | = | $-y_2 a\,\hat{\mathbf{x}} - z_2 a\,\hat{\mathbf{y}}$ | (24$g$) | O |

### Found in:

### Geometry files:

- CIF: pp. 1778
- POSCAR: pp. 1778

# $Zn(BrO_3)_2 \cdot 6H_2O$ ($J1_{10}$) Structure: A2B6C6D_cP60_205_c_d_d_a

http://aflow.org/prototype-encyclopedia/A2B6C6D_cP60_205_c_d_d_a

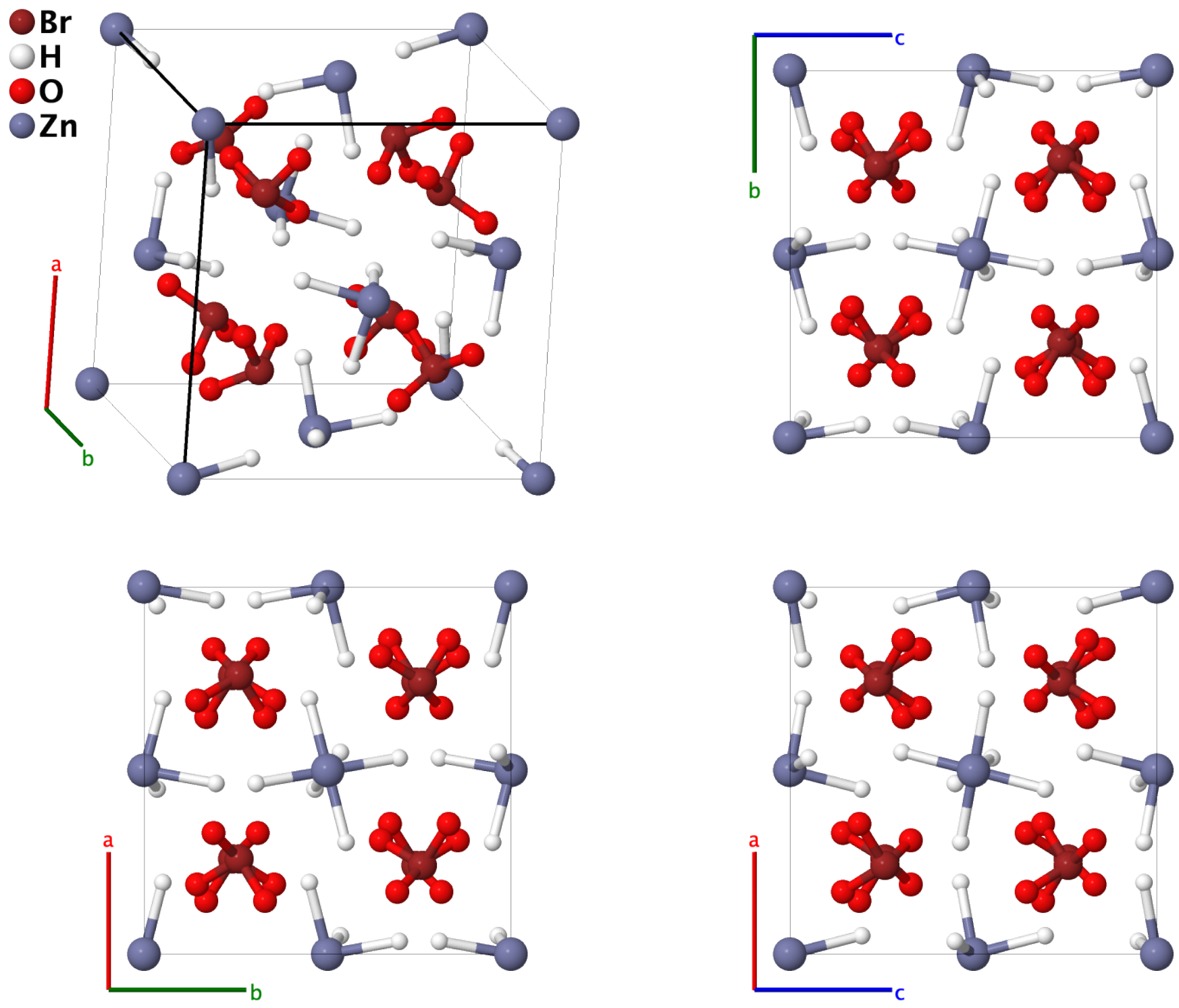

| | | |
|---|---|---|
| **Prototype** | : | $Br_2(H_2O)_6O_6Zn$ |
| **AFLOW prototype label** | : | A2B6C6D_cP60_205_c_d_d_a |
| ***Strukturbericht* designation** | : | $J1_{10}$ |
| **Pearson symbol** | : | cP60 |
| **Space group number** | : | 205 |
| **Space group symbol** | : | $Pa\bar{3}$ |
| **AFLOW prototype command** | : | `aflow --proto=A2B6C6D_cP60_205_c_d_d_a --params=`$a, x_2, x_3, y_3, z_3, x_4, y_4, z_4$ |

- The positions of the hydrogen atoms in the water molecules were not determined, so we only provide the positions of the oxygen atom (labeled as $H_2O$).

**Simple Cubic primitive vectors:**

$$\mathbf{a}_1 = a\,\hat{\mathbf{x}}$$
$$\mathbf{a}_2 = a\,\hat{\mathbf{y}}$$
$$\mathbf{a}_3 = a\,\hat{\mathbf{z}}$$

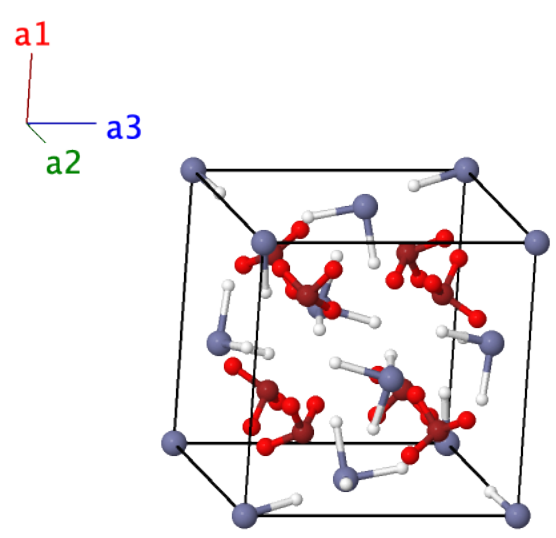

**Basis vectors:**

| | | Lattice Coordinates | | Cartesian Coordinates | Wyckoff Position | Atom Type |
|---|---|---|---|---|---|---|
| $\mathbf{B}_{1}$ | = | $0\,\mathbf{a}_{1}+0\,\mathbf{a}_{2}+0\,\mathbf{a}_{3}$ | = | $0\,\hat{\mathbf{x}}+0\,\hat{\mathbf{y}}+0\,\hat{\mathbf{z}}$ | (4$a$) | Zn |
| $\mathbf{B}_{2}$ | = | $\frac{1}{2}\,\mathbf{a}_{1}+\frac{1}{2}\,\mathbf{a}_{3}$ | = | $\frac{1}{2}a\,\hat{\mathbf{x}}+\frac{1}{2}a\,\hat{\mathbf{z}}$ | (4$a$) | Zn |
| $\mathbf{B}_{3}$ | = | $\frac{1}{2}\,\mathbf{a}_{2}+\frac{1}{2}\,\mathbf{a}_{3}$ | = | $\frac{1}{2}a\,\hat{\mathbf{y}}+\frac{1}{2}a\,\hat{\mathbf{z}}$ | (4$a$) | Zn |
| $\mathbf{B}_{4}$ | = | $\frac{1}{2}\,\mathbf{a}_{1}+\frac{1}{2}\,\mathbf{a}_{2}$ | = | $\frac{1}{2}a\,\hat{\mathbf{x}}+\frac{1}{2}a\,\hat{\mathbf{y}}$ | (4$a$) | Zn |
| $\mathbf{B}_{5}$ | = | $x_{2}\,\mathbf{a}_{1}+x_{2}\,\mathbf{a}_{2}+x_{2}\,\mathbf{a}_{3}$ | = | $x_{2}a\,\hat{\mathbf{x}}+x_{2}a\,\hat{\mathbf{y}}+x_{2}a\,\hat{\mathbf{z}}$ | (8$c$) | Br |
| $\mathbf{B}_{6}$ | = | $\left(\frac{1}{2}-x_{2}\right)\,\mathbf{a}_{1}-x_{2}\,\mathbf{a}_{2}+\left(\frac{1}{2}+x_{2}\right)\,\mathbf{a}_{3}$ | = | $\left(\frac{1}{2}-x_{2}\right)a\,\hat{\mathbf{x}}-x_{2}a\,\hat{\mathbf{y}}+\left(\frac{1}{2}+x_{2}\right)a\,\hat{\mathbf{z}}$ | (8$c$) | Br |
| $\mathbf{B}_{7}$ | = | $-x_{2}\,\mathbf{a}_{1}+\left(\frac{1}{2}+x_{2}\right)\,\mathbf{a}_{2}+\left(\frac{1}{2}-x_{2}\right)\,\mathbf{a}_{3}$ | = | $-x_{2}a\,\hat{\mathbf{x}}+\left(\frac{1}{2}+x_{2}\right)a\,\hat{\mathbf{y}}+\left(\frac{1}{2}-x_{2}\right)a\,\hat{\mathbf{z}}$ | (8$c$) | Br |
| $\mathbf{B}_{8}$ | = | $\left(\frac{1}{2}+x_{2}\right)\,\mathbf{a}_{1}+\left(\frac{1}{2}-x_{2}\right)\,\mathbf{a}_{2}-x_{2}\,\mathbf{a}_{3}$ | = | $\left(\frac{1}{2}+x_{2}\right)a\,\hat{\mathbf{x}}+\left(\frac{1}{2}-x_{2}\right)a\,\hat{\mathbf{y}}-x_{2}a\,\hat{\mathbf{z}}$ | (8$c$) | Br |
| $\mathbf{B}_{9}$ | = | $-x_{2}\,\mathbf{a}_{1}-x_{2}\,\mathbf{a}_{2}-x_{2}\,\mathbf{a}_{3}$ | = | $-x_{2}a\,\hat{\mathbf{x}}-x_{2}a\,\hat{\mathbf{y}}-x_{2}a\,\hat{\mathbf{z}}$ | (8$c$) | Br |
| $\mathbf{B}_{10}$ | = | $\left(\frac{1}{2}+x_{2}\right)\,\mathbf{a}_{1}+x_{2}\,\mathbf{a}_{2}+\left(\frac{1}{2}-x_{2}\right)\,\mathbf{a}_{3}$ | = | $\left(\frac{1}{2}+x_{2}\right)a\,\hat{\mathbf{x}}+x_{2}a\,\hat{\mathbf{y}}+\left(\frac{1}{2}-x_{2}\right)a\,\hat{\mathbf{z}}$ | (8$c$) | Br |
| $\mathbf{B}_{11}$ | = | $x_{2}\,\mathbf{a}_{1}+\left(\frac{1}{2}-x_{2}\right)\,\mathbf{a}_{2}+\left(\frac{1}{2}+x_{2}\right)\,\mathbf{a}_{3}$ | = | $x_{2}a\,\hat{\mathbf{x}}+\left(\frac{1}{2}-x_{2}\right)a\,\hat{\mathbf{y}}+\left(\frac{1}{2}+x_{2}\right)a\,\hat{\mathbf{z}}$ | (8$c$) | Br |
| $\mathbf{B}_{12}$ | = | $\left(\frac{1}{2}-x_{2}\right)\,\mathbf{a}_{1}+\left(\frac{1}{2}+x_{2}\right)\,\mathbf{a}_{2}+x_{2}\,\mathbf{a}_{3}$ | = | $\left(\frac{1}{2}-x_{2}\right)a\,\hat{\mathbf{x}}+\left(\frac{1}{2}+x_{2}\right)a\,\hat{\mathbf{y}}+x_{2}a\,\hat{\mathbf{z}}$ | (8$c$) | Br |
| $\mathbf{B}_{13}$ | = | $x_{3}\,\mathbf{a}_{1}+y_{3}\,\mathbf{a}_{2}+z_{3}\,\mathbf{a}_{3}$ | = | $x_{3}a\,\hat{\mathbf{x}}+y_{3}a\,\hat{\mathbf{y}}+z_{3}a\,\hat{\mathbf{z}}$ | (24$d$) | $H_2O$ |
| $\mathbf{B}_{14}$ | = | $\left(\frac{1}{2}-x_{3}\right)\,\mathbf{a}_{1}-y_{3}\,\mathbf{a}_{2}+\left(\frac{1}{2}+z_{3}\right)\,\mathbf{a}_{3}$ | = | $\left(\frac{1}{2}-x_{3}\right)a\,\hat{\mathbf{x}}-y_{3}a\,\hat{\mathbf{y}}+\left(\frac{1}{2}+z_{3}\right)a\,\hat{\mathbf{z}}$ | (24$d$) | $H_2O$ |
| $\mathbf{B}_{15}$ | = | $-x_{3}\,\mathbf{a}_{1}+\left(\frac{1}{2}+y_{3}\right)\,\mathbf{a}_{2}+\left(\frac{1}{2}-z_{3}\right)\,\mathbf{a}_{3}$ | = | $-x_{3}a\,\hat{\mathbf{x}}+\left(\frac{1}{2}+y_{3}\right)a\,\hat{\mathbf{y}}+\left(\frac{1}{2}-z_{3}\right)a\,\hat{\mathbf{z}}$ | (24$d$) | $H_2O$ |
| $\mathbf{B}_{16}$ | = | $\left(\frac{1}{2}+x_{3}\right)\,\mathbf{a}_{1}+\left(\frac{1}{2}-y_{3}\right)\,\mathbf{a}_{2}-z_{3}\,\mathbf{a}_{3}$ | = | $\left(\frac{1}{2}+x_{3}\right)a\,\hat{\mathbf{x}}+\left(\frac{1}{2}-y_{3}\right)a\,\hat{\mathbf{y}}-z_{3}a\,\hat{\mathbf{z}}$ | (24$d$) | $H_2O$ |
| $\mathbf{B}_{17}$ | = | $z_{3}\,\mathbf{a}_{1}+x_{3}\,\mathbf{a}_{2}+y_{3}\,\mathbf{a}_{3}$ | = | $z_{3}a\,\hat{\mathbf{x}}+x_{3}a\,\hat{\mathbf{y}}+y_{3}a\,\hat{\mathbf{z}}$ | (24$d$) | $H_2O$ |
| $\mathbf{B}_{18}$ | = | $\left(\frac{1}{2}+z_{3}\right)\,\mathbf{a}_{1}+\left(\frac{1}{2}-x_{3}\right)\,\mathbf{a}_{2}-y_{3}\,\mathbf{a}_{3}$ | = | $\left(\frac{1}{2}+z_{3}\right)a\,\hat{\mathbf{x}}+\left(\frac{1}{2}-x_{3}\right)a\,\hat{\mathbf{y}}-y_{3}a\,\hat{\mathbf{z}}$ | (24$d$) | $H_2O$ |
| $\mathbf{B}_{19}$ | = | $\left(\frac{1}{2}-z_{3}\right)\,\mathbf{a}_{1}-x_{3}\,\mathbf{a}_{2}+\left(\frac{1}{2}+y_{3}\right)\,\mathbf{a}_{3}$ | = | $\left(\frac{1}{2}-z_{3}\right)a\,\hat{\mathbf{x}}-x_{3}a\,\hat{\mathbf{y}}+\left(\frac{1}{2}+y_{3}\right)a\,\hat{\mathbf{z}}$ | (24$d$) | $H_2O$ |
| $\mathbf{B}_{20}$ | = | $-z_{3}\,\mathbf{a}_{1}+\left(\frac{1}{2}+x_{3}\right)\,\mathbf{a}_{2}+\left(\frac{1}{2}-y_{3}\right)\,\mathbf{a}_{3}$ | = | $-z_{3}a\,\hat{\mathbf{x}}+\left(\frac{1}{2}+x_{3}\right)a\,\hat{\mathbf{y}}+\left(\frac{1}{2}-y_{3}\right)a\,\hat{\mathbf{z}}$ | (24$d$) | $H_2O$ |
| $\mathbf{B}_{21}$ | = | $y_{3}\,\mathbf{a}_{1}+z_{3}\,\mathbf{a}_{2}+x_{3}\,\mathbf{a}_{3}$ | = | $y_{3}a\,\hat{\mathbf{x}}+z_{3}a\,\hat{\mathbf{y}}+x_{3}a\,\hat{\mathbf{z}}$ | (24$d$) | $H_2O$ |
| $\mathbf{B}_{22}$ | = | $-y_{3}\,\mathbf{a}_{1}+\left(\frac{1}{2}+z_{3}\right)\,\mathbf{a}_{2}+\left(\frac{1}{2}-x_{3}\right)\,\mathbf{a}_{3}$ | = | $-y_{3}a\,\hat{\mathbf{x}}+\left(\frac{1}{2}+z_{3}\right)a\,\hat{\mathbf{y}}+\left(\frac{1}{2}-x_{3}\right)a\,\hat{\mathbf{z}}$ | (24$d$) | $H_2O$ |
| $\mathbf{B}_{23}$ | = | $\left(\frac{1}{2}+y_{3}\right)\,\mathbf{a}_{1}+\left(\frac{1}{2}-z_{3}\right)\,\mathbf{a}_{2}-x_{3}\,\mathbf{a}_{3}$ | = | $\left(\frac{1}{2}+y_{3}\right)a\,\hat{\mathbf{x}}+\left(\frac{1}{2}-z_{3}\right)a\,\hat{\mathbf{y}}-x_{3}a\,\hat{\mathbf{z}}$ | (24$d$) | $H_2O$ |
| $\mathbf{B}_{24}$ | = | $\left(\frac{1}{2}-y_{3}\right)\,\mathbf{a}_{1}-z_{3}\,\mathbf{a}_{2}+\left(\frac{1}{2}+x_{3}\right)\,\mathbf{a}_{3}$ | = | $\left(\frac{1}{2}-y_{3}\right)a\,\hat{\mathbf{x}}-z_{3}a\,\hat{\mathbf{y}}+\left(\frac{1}{2}+x_{3}\right)a\,\hat{\mathbf{z}}$ | (24$d$) | $H_2O$ |
| $\mathbf{B}_{25}$ | = | $-x_{3}\,\mathbf{a}_{1}-y_{3}\,\mathbf{a}_{2}-z_{3}\,\mathbf{a}_{3}$ | = | $-x_{3}a\,\hat{\mathbf{x}}-y_{3}a\,\hat{\mathbf{y}}-z_{3}a\,\hat{\mathbf{z}}$ | (24$d$) | $H_2O$ |
| $\mathbf{B}_{26}$ | = | $\left(\frac{1}{2}+x_{3}\right)\,\mathbf{a}_{1}+y_{3}\,\mathbf{a}_{2}+\left(\frac{1}{2}-z_{3}\right)\,\mathbf{a}_{3}$ | = | $\left(\frac{1}{2}+x_{3}\right)a\,\hat{\mathbf{x}}+y_{3}a\,\hat{\mathbf{y}}+\left(\frac{1}{2}-z_{3}\right)a\,\hat{\mathbf{z}}$ | (24$d$) | $H_2O$ |
| $\mathbf{B}_{27}$ | = | $x_{3}\,\mathbf{a}_{1}+\left(\frac{1}{2}-y_{3}\right)\,\mathbf{a}_{2}+\left(\frac{1}{2}+z_{3}\right)\,\mathbf{a}_{3}$ | = | $x_{3}a\,\hat{\mathbf{x}}+\left(\frac{1}{2}-y_{3}\right)a\,\hat{\mathbf{y}}+\left(\frac{1}{2}+z_{3}\right)a\,\hat{\mathbf{z}}$ | (24$d$) | $H_2O$ |
| $\mathbf{B}_{28}$ | = | $\left(\frac{1}{2}-x_{3}\right)\,\mathbf{a}_{1}+\left(\frac{1}{2}+y_{3}\right)\,\mathbf{a}_{2}+z_{3}\,\mathbf{a}_{3}$ | = | $\left(\frac{1}{2}-x_{3}\right)a\,\hat{\mathbf{x}}+\left(\frac{1}{2}+y_{3}\right)a\,\hat{\mathbf{y}}+z_{3}a\,\hat{\mathbf{z}}$ | (24$d$) | $H_2O$ |
| $\mathbf{B}_{29}$ | = | $-z_{3}\,\mathbf{a}_{1}-x_{3}\,\mathbf{a}_{2}-y_{3}\,\mathbf{a}_{3}$ | = | $-z_{3}a\,\hat{\mathbf{x}}-x_{3}a\,\hat{\mathbf{y}}-y_{3}a\,\hat{\mathbf{z}}$ | (24$d$) | $H_2O$ |
| $\mathbf{B}_{30}$ | = | $\left(\frac{1}{2}-z_{3}\right)\,\mathbf{a}_{1}+\left(\frac{1}{2}+x_{3}\right)\,\mathbf{a}_{2}+y_{3}\,\mathbf{a}_{3}$ | = | $\left(\frac{1}{2}-z_{3}\right)a\,\hat{\mathbf{x}}+\left(\frac{1}{2}+x_{3}\right)a\,\hat{\mathbf{y}}+y_{3}a\,\hat{\mathbf{z}}$ | (24$d$) | $H_2O$ |
| $\mathbf{B}_{31}$ | = | $\left(\frac{1}{2}+z_{3}\right)\,\mathbf{a}_{1}+x_{3}\,\mathbf{a}_{2}+\left(\frac{1}{2}-y_{3}\right)\,\mathbf{a}_{3}$ | = | $\left(\frac{1}{2}+z_{3}\right)a\,\hat{\mathbf{x}}+x_{3}a\,\hat{\mathbf{y}}+\left(\frac{1}{2}-y_{3}\right)a\,\hat{\mathbf{z}}$ | (24$d$) | $H_2O$ |
| $\mathbf{B}_{32}$ | = | $z_{3}\,\mathbf{a}_{1}+\left(\frac{1}{2}-x_{3}\right)\,\mathbf{a}_{2}+\left(\frac{1}{2}+y_{3}\right)\,\mathbf{a}_{3}$ | = | $z_{3}a\,\hat{\mathbf{x}}+\left(\frac{1}{2}-x_{3}\right)a\,\hat{\mathbf{y}}+\left(\frac{1}{2}+y_{3}\right)a\,\hat{\mathbf{z}}$ | (24$d$) | $H_2O$ |
| $\mathbf{B}_{33}$ | = | $-y_{3}\,\mathbf{a}_{1}-z_{3}\,\mathbf{a}_{2}-x_{3}\,\mathbf{a}_{3}$ | = | $-y_{3}a\,\hat{\mathbf{x}}-z_{3}a\,\hat{\mathbf{y}}-x_{3}a\,\hat{\mathbf{z}}$ | (24$d$) | $H_2O$ |
| $\mathbf{B}_{34}$ | = | $y_{3}\,\mathbf{a}_{1}+\left(\frac{1}{2}-z_{3}\right)\,\mathbf{a}_{2}+\left(\frac{1}{2}+x_{3}\right)\,\mathbf{a}_{3}$ | = | $y_{3}a\,\hat{\mathbf{x}}+\left(\frac{1}{2}-z_{3}\right)a\,\hat{\mathbf{y}}+\left(\frac{1}{2}+x_{3}\right)a\,\hat{\mathbf{z}}$ | (24$d$) | $H_2O$ |
| $\mathbf{B}_{35}$ | = | $\left(\frac{1}{2}-y_{3}\right)\,\mathbf{a}_{1}+\left(\frac{1}{2}+z_{3}\right)\,\mathbf{a}_{2}+x_{3}\,\mathbf{a}_{3}$ | = | $\left(\frac{1}{2}-y_{3}\right)a\,\hat{\mathbf{x}}+\left(\frac{1}{2}+z_{3}\right)a\,\hat{\mathbf{y}}+x_{3}a\,\hat{\mathbf{z}}$ | (24$d$) | $H_2O$ |
| $\mathbf{B}_{36}$ | = | $\left(\frac{1}{2}+y_{3}\right)\,\mathbf{a}_{1}+z_{3}\,\mathbf{a}_{2}+\left(\frac{1}{2}-x_{3}\right)\,\mathbf{a}_{3}$ | = | $\left(\frac{1}{2}+y_{3}\right)a\,\mathbf{\hat{x}}+z_{3}a\,\mathbf{\hat{y}}+\left(\frac{1}{2}-x_{3}\right)a\,\mathbf{\hat{z}}$ | (24$d$) | $H_2O$ |
| $\mathbf{B}_{37}$ | = | $x_{4}\,\mathbf{a}_{1}+y_{4}\,\mathbf{a}_{2}+z_{4}\,\mathbf{a}_{3}$ | = | $x_{4}a\,\mathbf{\hat{x}}+y_{4}a\,\mathbf{\hat{y}}+z_{4}a\,\mathbf{\hat{z}}$ | (24$d$) | O |
| $\mathbf{B}_{38}$ | = | $\left(\frac{1}{2}-x_{4}\right)\,\mathbf{a}_{1}-y_{4}\,\mathbf{a}_{2}+\left(\frac{1}{2}+z_{4}\right)\,\mathbf{a}_{3}$ | = | $\left(\frac{1}{2}-x_{4}\right)a\,\mathbf{\hat{x}}-y_{4}a\,\mathbf{\hat{y}}+\left(\frac{1}{2}+z_{4}\right)a\,\mathbf{\hat{z}}$ | (24$d$) | O |
| $\mathbf{B}_{39}$ | = | $-x_{4}\,\mathbf{a}_{1}+\left(\frac{1}{2}+y_{4}\right)\,\mathbf{a}_{2}+\left(\frac{1}{2}-z_{4}\right)\,\mathbf{a}_{3}$ | = | $-x_{4}a\,\mathbf{\hat{x}}+\left(\frac{1}{2}+y_{4}\right)a\,\mathbf{\hat{y}}+\left(\frac{1}{2}-z_{4}\right)a\,\mathbf{\hat{z}}$ | (24$d$) | O |
| $\mathbf{B}_{40}$ | = | $\left(\frac{1}{2}+x_{4}\right)\,\mathbf{a}_{1}+\left(\frac{1}{2}-y_{4}\right)\,\mathbf{a}_{2}-z_{4}\,\mathbf{a}_{3}$ | = | $\left(\frac{1}{2}+x_{4}\right)a\,\mathbf{\hat{x}}+\left(\frac{1}{2}-y_{4}\right)a\,\mathbf{\hat{y}}-z_{4}a\,\mathbf{\hat{z}}$ | (24$d$) | O |
| $\mathbf{B}_{41}$ | = | $z_{4}\,\mathbf{a}_{1}+x_{4}\,\mathbf{a}_{2}+y_{4}\,\mathbf{a}_{3}$ | = | $z_{4}a\,\mathbf{\hat{x}}+x_{4}a\,\mathbf{\hat{y}}+y_{4}a\,\mathbf{\hat{z}}$ | (24$d$) | O |
| $\mathbf{B}_{42}$ | = | $\left(\frac{1}{2}+z_{4}\right)\,\mathbf{a}_{1}+\left(\frac{1}{2}-x_{4}\right)\,\mathbf{a}_{2}-y_{4}\,\mathbf{a}_{3}$ | = | $\left(\frac{1}{2}+z_{4}\right)a\,\mathbf{\hat{x}}+\left(\frac{1}{2}-x_{4}\right)a\,\mathbf{\hat{y}}-y_{4}a\,\mathbf{\hat{z}}$ | (24$d$) | O |
| $\mathbf{B}_{43}$ | = | $\left(\frac{1}{2}-z_{4}\right)\,\mathbf{a}_{1}-x_{4}\,\mathbf{a}_{2}+\left(\frac{1}{2}+y_{4}\right)\,\mathbf{a}_{3}$ | = | $\left(\frac{1}{2}-z_{4}\right)a\,\mathbf{\hat{x}}-x_{4}a\,\mathbf{\hat{y}}+\left(\frac{1}{2}+y_{4}\right)a\,\mathbf{\hat{z}}$ | (24$d$) | O |
| $\mathbf{B}_{44}$ | = | $-z_{4}\,\mathbf{a}_{1}+\left(\frac{1}{2}+x_{4}\right)\,\mathbf{a}_{2}+\left(\frac{1}{2}-y_{4}\right)\,\mathbf{a}_{3}$ | = | $-z_{4}a\,\mathbf{\hat{x}}+\left(\frac{1}{2}+x_{4}\right)a\,\mathbf{\hat{y}}+\left(\frac{1}{2}-y_{4}\right)a\,\mathbf{\hat{z}}$ | (24$d$) | O |
| $\mathbf{B}_{45}$ | = | $y_{4}\,\mathbf{a}_{1}+z_{4}\,\mathbf{a}_{2}+x_{4}\,\mathbf{a}_{3}$ | = | $y_{4}a\,\mathbf{\hat{x}}+z_{4}a\,\mathbf{\hat{y}}+x_{4}a\,\mathbf{\hat{z}}$ | (24$d$) | O |
| $\mathbf{B}_{46}$ | = | $-y_{4}\,\mathbf{a}_{1}+\left(\frac{1}{2}+z_{4}\right)\,\mathbf{a}_{2}+\left(\frac{1}{2}-x_{4}\right)\,\mathbf{a}_{3}$ | = | $-y_{4}a\,\mathbf{\hat{x}}+\left(\frac{1}{2}+z_{4}\right)a\,\mathbf{\hat{y}}+\left(\frac{1}{2}-x_{4}\right)a\,\mathbf{\hat{z}}$ | (24$d$) | O |
| $\mathbf{B}_{47}$ | = | $\left(\frac{1}{2}+y_{4}\right)\,\mathbf{a}_{1}+\left(\frac{1}{2}-z_{4}\right)\,\mathbf{a}_{2}-x_{4}\,\mathbf{a}_{3}$ | = | $\left(\frac{1}{2}+y_{4}\right)a\,\mathbf{\hat{x}}+\left(\frac{1}{2}-z_{4}\right)a\,\mathbf{\hat{y}}-x_{4}a\,\mathbf{\hat{z}}$ | (24$d$) | O |
| $\mathbf{B}_{48}$ | = | $\left(\frac{1}{2}-y_{4}\right)\,\mathbf{a}_{1}-z_{4}\,\mathbf{a}_{2}+\left(\frac{1}{2}+x_{4}\right)\,\mathbf{a}_{3}$ | = | $\left(\frac{1}{2}-y_{4}\right)a\,\mathbf{\hat{x}}-z_{4}a\,\mathbf{\hat{y}}+\left(\frac{1}{2}+x_{4}\right)a\,\mathbf{\hat{z}}$ | (24$d$) | O |
| $\mathbf{B}_{49}$ | = | $-x_{4}\,\mathbf{a}_{1}-y_{4}\,\mathbf{a}_{2}-z_{4}\,\mathbf{a}_{3}$ | = | $-x_{4}a\,\mathbf{\hat{x}}-y_{4}a\,\mathbf{\hat{y}}-z_{4}a\,\mathbf{\hat{z}}$ | (24$d$) | O |
| $\mathbf{B}_{50}$ | = | $\left(\frac{1}{2}+x_{4}\right)\,\mathbf{a}_{1}+y_{4}\,\mathbf{a}_{2}+\left(\frac{1}{2}-z_{4}\right)\,\mathbf{a}_{3}$ | = | $\left(\frac{1}{2}+x_{4}\right)a\,\mathbf{\hat{x}}+y_{4}a\,\mathbf{\hat{y}}+\left(\frac{1}{2}-z_{4}\right)a\,\mathbf{\hat{z}}$ | (24$d$) | O |
| $\mathbf{B}_{51}$ | = | $x_{4}\,\mathbf{a}_{1}+\left(\frac{1}{2}-y_{4}\right)\,\mathbf{a}_{2}+\left(\frac{1}{2}+z_{4}\right)\,\mathbf{a}_{3}$ | = | $x_{4}a\,\mathbf{\hat{x}}+\left(\frac{1}{2}-y_{4}\right)a\,\mathbf{\hat{y}}+\left(\frac{1}{2}+z_{4}\right)a\,\mathbf{\hat{z}}$ | (24$d$) | O |
| $\mathbf{B}_{52}$ | = | $\left(\frac{1}{2}-x_{4}\right)\,\mathbf{a}_{1}+\left(\frac{1}{2}+y_{4}\right)\,\mathbf{a}_{2}+z_{4}\,\mathbf{a}_{3}$ | = | $\left(\frac{1}{2}-x_{4}\right)a\,\mathbf{\hat{x}}+\left(\frac{1}{2}+y_{4}\right)a\,\mathbf{\hat{y}}+z_{4}a\,\mathbf{\hat{z}}$ | (24$d$) | O |
| $\mathbf{B}_{53}$ | = | $-z_{4}\,\mathbf{a}_{1}-x_{4}\,\mathbf{a}_{2}-y_{4}\,\mathbf{a}_{3}$ | = | $-z_{4}a\,\mathbf{\hat{x}}-x_{4}a\,\mathbf{\hat{y}}-y_{4}a\,\mathbf{\hat{z}}$ | (24$d$) | O |
| $\mathbf{B}_{54}$ | = | $\left(\frac{1}{2}-z_{4}\right)\,\mathbf{a}_{1}+\left(\frac{1}{2}+x_{4}\right)\,\mathbf{a}_{2}+y_{4}\,\mathbf{a}_{3}$ | = | $\left(\frac{1}{2}-z_{4}\right)a\,\mathbf{\hat{x}}+\left(\frac{1}{2}+x_{4}\right)a\,\mathbf{\hat{y}}+y_{4}a\,\mathbf{\hat{z}}$ | (24$d$) | O |
| $\mathbf{B}_{55}$ | = | $\left(\frac{1}{2}+z_{4}\right)\,\mathbf{a}_{1}+x_{4}\,\mathbf{a}_{2}+\left(\frac{1}{2}-y_{4}\right)\,\mathbf{a}_{3}$ | = | $\left(\frac{1}{2}+z_{4}\right)a\,\mathbf{\hat{x}}+x_{4}a\,\mathbf{\hat{y}}+\left(\frac{1}{2}-y_{4}\right)a\,\mathbf{\hat{z}}$ | (24$d$) | O |
| $\mathbf{B}_{56}$ | = | $z_{4}\,\mathbf{a}_{1}+\left(\frac{1}{2}-x_{4}\right)\,\mathbf{a}_{2}+\left(\frac{1}{2}+y_{4}\right)\,\mathbf{a}_{3}$ | = | $z_{4}a\,\mathbf{\hat{x}}+\left(\frac{1}{2}-x_{4}\right)a\,\mathbf{\hat{y}}+\left(\frac{1}{2}+y_{4}\right)a\,\mathbf{\hat{z}}$ | (24$d$) | O |
| $\mathbf{B}_{57}$ | = | $-y_{4}\,\mathbf{a}_{1}-z_{4}\,\mathbf{a}_{2}-x_{4}\,\mathbf{a}_{3}$ | = | $-y_{4}a\,\mathbf{\hat{x}}-z_{4}a\,\mathbf{\hat{y}}-x_{4}a\,\mathbf{\hat{z}}$ | (24$d$) | O |
| $\mathbf{B}_{58}$ | = | $y_{4}\,\mathbf{a}_{1}+\left(\frac{1}{2}-z_{4}\right)\,\mathbf{a}_{2}+\left(\frac{1}{2}+x_{4}\right)\,\mathbf{a}_{3}$ | = | $y_{4}a\,\mathbf{\hat{x}}+\left(\frac{1}{2}-z_{4}\right)a\,\mathbf{\hat{y}}+\left(\frac{1}{2}+x_{4}\right)a\,\mathbf{\hat{z}}$ | (24$d$) | O |
| $\mathbf{B}_{59}$ | = | $\left(\frac{1}{2}-y_{4}\right)\,\mathbf{a}_{1}+\left(\frac{1}{2}+z_{4}\right)\,\mathbf{a}_{2}+x_{4}\,\mathbf{a}_{3}$ | = | $\left(\frac{1}{2}-y_{4}\right)a\,\mathbf{\hat{x}}+\left(\frac{1}{2}+z_{4}\right)a\,\mathbf{\hat{y}}+x_{4}a\,\mathbf{\hat{z}}$ | (24$d$) | O |
| $\mathbf{B}_{60}$ | = | $\left(\frac{1}{2}+y_{4}\right)\,\mathbf{a}_{1}+z_{4}\,\mathbf{a}_{2}+\left(\frac{1}{2}-x_{4}\right)\,\mathbf{a}_{3}$ | = | $\left(\frac{1}{2}+y_{4}\right)a\,\mathbf{\hat{x}}+z_{4}a\,\mathbf{\hat{y}}+\left(\frac{1}{2}-x_{4}\right)a\,\mathbf{\hat{z}}$ | (24$d$) | O |

**Found in:**

**Geometry files:**

- CIF: pp. 1779
- POSCAR: pp. 1779

# $H6_4$ [$Ni(NO_3)_2(NH_3)_6$] (*obsolete*) Structure: A2B6CD6_cP60_205_c_d_a_d

http://aflow.org/prototype-encyclopedia/A2B6CD6_cP60_205_c_d_a_d

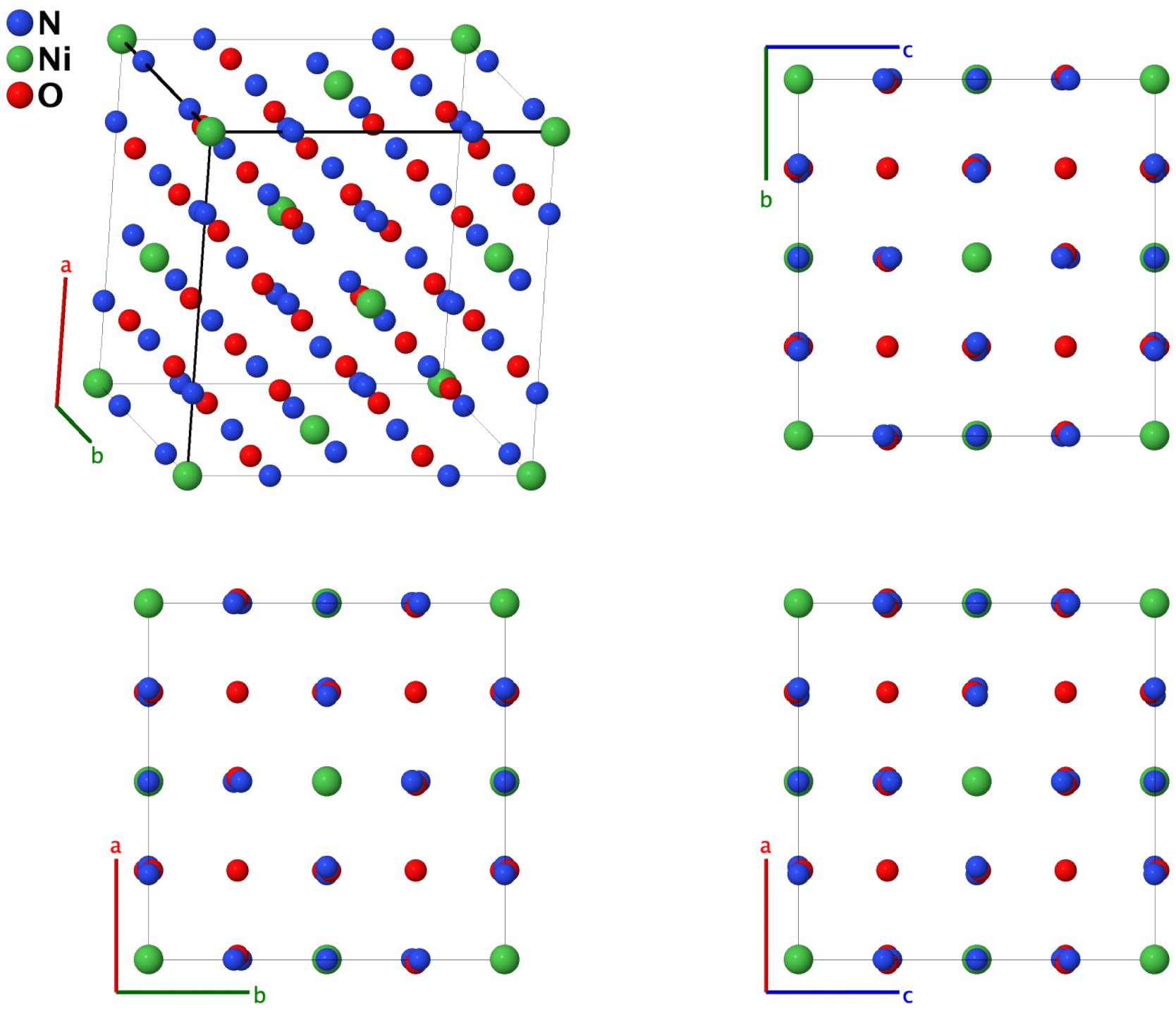

| | | |
|---|---|---|
| **Prototype** | : | $N_2(NH_3)_6NiO_6$ |
| **AFLOW prototype label** | : | A2B6CD6_cP60_205_c_d_a_d |
| ***Strukturbericht* designation** | : | None |
| **Pearson symbol** | : | cP60 |
| **Space group number** | : | 205 |
| **Space group symbol** | : | $Pa\bar{3}$ |
| **AFLOW prototype command** | : | `aflow --proto=A2B6CD6_cP60_205_c_d_a_d` `--params=`$a, x_2, x_3, y_3, z_3, x_4, y_4, z_4$ |

- (Wyckoff, 1922) determined this approximate structure. In his paper, non-zero coordinates were $x_2 = 1/4$ for the nitrogen (8*c*) atoms, $x_3 = v$, "where $v$ is somewhat less than 0.25," for the $NH_3$ (24*d*) molecules, and $x_4 = y_4 = 1/4$, $z_4 = v'$, where v' "should not deviate far from 0." If we take these coordinates as written the space group becomes $Fm\bar{3}m$ #225 rather than Wyckoff's $Pa\bar{3}$ #205, so we adjusted $v$ and $v'$ slightly to put the system in his space group.

- (Ewald, 1931) gave this the *Strukturbericht* designation *H*61, or $H6_1$ in later notation. (Hermann, 1937) moved it to $I1_4$ in their "list of type descriptions," but no other volume of *Strukturbericht* refers to it at all. Accordingly, we will designate this structure by its original label.

- This structure is an idealized appoximation to the true structure of $Ni(NO_3)_2(NH_3)_6$. (Bigoli, 1971) showed that the correct structure is triclinic, with space group $P\bar{1}$.

- The positions of the hydrogen atoms in the ammonia molecules were not determined, so we only provide the positions of the nitrogen atoms (labeled as $NH_3$).

## Simple Cubic primitive vectors:

$$\mathbf{a}_1 = a\,\hat{\mathbf{x}}$$
$$\mathbf{a}_2 = a\,\hat{\mathbf{y}}$$
$$\mathbf{a}_3 = a\,\hat{\mathbf{z}}$$

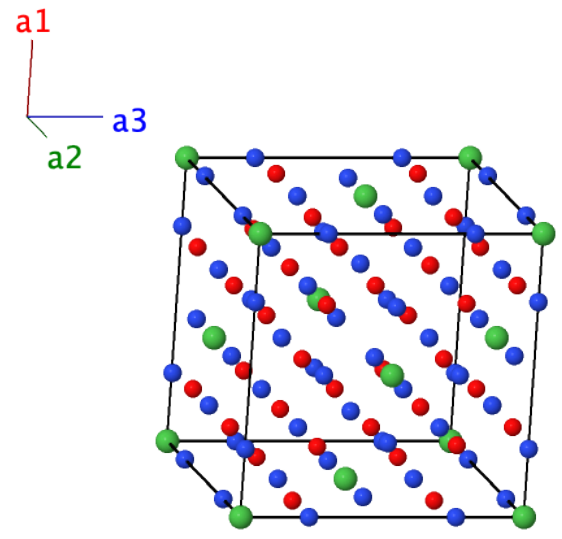

## Basis vectors:

| | | Lattice Coordinates | | Cartesian Coordinates | Wyckoff Position | Atom Type |
|---|---|---|---|---|---|---|
| $\mathbf{B}_1$ | = | $0\,\mathbf{a}_1 + 0\,\mathbf{a}_2 + 0\,\mathbf{a}_3$ | = | $0\,\hat{\mathbf{x}} + 0\,\hat{\mathbf{y}} + 0\,\hat{\mathbf{z}}$ | (4$a$) | Ni |
| $\mathbf{B}_2$ | = | $\frac{1}{2}\,\mathbf{a}_1 + \frac{1}{2}\,\mathbf{a}_3$ | = | $\frac{1}{2}a\,\hat{\mathbf{x}} + \frac{1}{2}a\,\hat{\mathbf{z}}$ | (4$a$) | Ni |
| $\mathbf{B}_3$ | = | $\frac{1}{2}\,\mathbf{a}_2 + \frac{1}{2}\,\mathbf{a}_3$ | = | $\frac{1}{2}a\,\hat{\mathbf{y}} + \frac{1}{2}a\,\hat{\mathbf{z}}$ | (4$a$) | Ni |
| $\mathbf{B}_4$ | = | $\frac{1}{2}\,\mathbf{a}_1 + \frac{1}{2}\,\mathbf{a}_2$ | = | $\frac{1}{2}a\,\hat{\mathbf{x}} + \frac{1}{2}a\,\hat{\mathbf{y}}$ | (4$a$) | Ni |
| $\mathbf{B}_5$ | = | $x_2\,\mathbf{a}_1 + x_2\,\mathbf{a}_2 + x_2\,\mathbf{a}_3$ | = | $x_2a\,\hat{\mathbf{x}} + x_2a\,\hat{\mathbf{y}} + x_2a\,\hat{\mathbf{z}}$ | (8$c$) | N |
| $\mathbf{B}_6$ | = | $\left(\frac{1}{2} - x_2\right)\,\mathbf{a}_1 - x_2\,\mathbf{a}_2 + \left(\frac{1}{2} + x_2\right)\,\mathbf{a}_3$ | = | $\left(\frac{1}{2} - x_2\right)a\,\hat{\mathbf{x}} - x_2a\,\hat{\mathbf{y}} + \left(\frac{1}{2} + x_2\right)a\,\hat{\mathbf{z}}$ | (8$c$) | N |
| $\mathbf{B}_7$ | = | $-x_2\,\mathbf{a}_1 + \left(\frac{1}{2} + x_2\right)\,\mathbf{a}_2 + \left(\frac{1}{2} - x_2\right)\,\mathbf{a}_3$ | = | $-x_2a\,\hat{\mathbf{x}} + \left(\frac{1}{2} + x_2\right)a\,\hat{\mathbf{y}} + \left(\frac{1}{2} - x_2\right)a\,\hat{\mathbf{z}}$ | (8$c$) | N |
| $\mathbf{B}_8$ | = | $\left(\frac{1}{2} + x_2\right)\,\mathbf{a}_1 + \left(\frac{1}{2} - x_2\right)\,\mathbf{a}_2 - x_2\,\mathbf{a}_3$ | = | $\left(\frac{1}{2} + x_2\right)a\,\hat{\mathbf{x}} + \left(\frac{1}{2} - x_2\right)a\,\hat{\mathbf{y}} - x_2a\,\hat{\mathbf{z}}$ | (8$c$) | N |
| $\mathbf{B}_9$ | = | $-x_2\,\mathbf{a}_1 - x_2\,\mathbf{a}_2 - x_2\,\mathbf{a}_3$ | = | $-x_2a\,\hat{\mathbf{x}} - x_2a\,\hat{\mathbf{y}} - x_2a\,\hat{\mathbf{z}}$ | (8$c$) | N |
| $\mathbf{B}_{10}$ | = | $\left(\frac{1}{2} + x_2\right)\,\mathbf{a}_1 + x_2\,\mathbf{a}_2 + \left(\frac{1}{2} - x_2\right)\,\mathbf{a}_3$ | = | $\left(\frac{1}{2} + x_2\right)a\,\hat{\mathbf{x}} + x_2a\,\hat{\mathbf{y}} + \left(\frac{1}{2} - x_2\right)a\,\hat{\mathbf{z}}$ | (8$c$) | N |
| $\mathbf{B}_{11}$ | = | $x_2\,\mathbf{a}_1 + \left(\frac{1}{2} - x_2\right)\,\mathbf{a}_2 + \left(\frac{1}{2} + x_2\right)\,\mathbf{a}_3$ | = | $x_2a\,\hat{\mathbf{x}} + \left(\frac{1}{2} - x_2\right)a\,\hat{\mathbf{y}} + \left(\frac{1}{2} + x_2\right)a\,\hat{\mathbf{z}}$ | (8$c$) | N |
| $\mathbf{B}_{12}$ | = | $\left(\frac{1}{2} - x_2\right)\,\mathbf{a}_1 + \left(\frac{1}{2} + x_2\right)\,\mathbf{a}_2 + x_2\,\mathbf{a}_3$ | = | $\left(\frac{1}{2} - x_2\right)a\,\hat{\mathbf{x}} + \left(\frac{1}{2} + x_2\right)a\,\hat{\mathbf{y}} + x_2a\,\hat{\mathbf{z}}$ | (8$c$) | N |
| $\mathbf{B}_{13}$ | = | $x_3\,\mathbf{a}_1 + y_3\,\mathbf{a}_2 + z_3\,\mathbf{a}_3$ | = | $x_3a\,\hat{\mathbf{x}} + y_3a\,\hat{\mathbf{y}} + z_3a\,\hat{\mathbf{z}}$ | (24$d$) | $NH_3$ |
| $\mathbf{B}_{14}$ | = | $\left(\frac{1}{2} - x_3\right)\,\mathbf{a}_1 - y_3\,\mathbf{a}_2 + \left(\frac{1}{2} + z_3\right)\,\mathbf{a}_3$ | = | $\left(\frac{1}{2} - x_3\right)a\,\hat{\mathbf{x}} - y_3a\,\hat{\mathbf{y}} + \left(\frac{1}{2} + z_3\right)a\,\hat{\mathbf{z}}$ | (24$d$) | $NH_3$ |
| $\mathbf{B}_{15}$ | = | $-x_3\,\mathbf{a}_1 + \left(\frac{1}{2} + y_3\right)\,\mathbf{a}_2 + \left(\frac{1}{2} - z_3\right)\,\mathbf{a}_3$ | = | $-x_3a\,\hat{\mathbf{x}} + \left(\frac{1}{2} + y_3\right)a\,\hat{\mathbf{y}} + \left(\frac{1}{2} - z_3\right)a\,\hat{\mathbf{z}}$ | (24$d$) | $NH_3$ |
| $\mathbf{B}_{16}$ | = | $\left(\frac{1}{2} + x_3\right)\,\mathbf{a}_1 + \left(\frac{1}{2} - y_3\right)\,\mathbf{a}_2 - z_3\,\mathbf{a}_3$ | = | $\left(\frac{1}{2} + x_3\right)a\,\hat{\mathbf{x}} + \left(\frac{1}{2} - y_3\right)a\,\hat{\mathbf{y}} - z_3a\,\hat{\mathbf{z}}$ | (24$d$) | $NH_3$ |
| $\mathbf{B}_{17}$ | = | $z_3\,\mathbf{a}_1 + x_3\,\mathbf{a}_2 + y_3\,\mathbf{a}_3$ | = | $z_3a\,\hat{\mathbf{x}} + x_3a\,\hat{\mathbf{y}} + y_3a\,\hat{\mathbf{z}}$ | (24$d$) | $NH_3$ |
| $\mathbf{B}_{18}$ | = | $\left(\frac{1}{2} + z_3\right)\,\mathbf{a}_1 + \left(\frac{1}{2} - x_3\right)\,\mathbf{a}_2 - y_3\,\mathbf{a}_3$ | = | $\left(\frac{1}{2} + z_3\right)a\,\hat{\mathbf{x}} + \left(\frac{1}{2} - x_3\right)a\,\hat{\mathbf{y}} - y_3a\,\hat{\mathbf{z}}$ | (24$d$) | $NH_3$ |
| $\mathbf{B}_{19}$ | = | $\left(\frac{1}{2} - z_3\right)\,\mathbf{a}_1 - x_3\,\mathbf{a}_2 + \left(\frac{1}{2} + y_3\right)\,\mathbf{a}_3$ | = | $\left(\frac{1}{2} - z_3\right)a\,\hat{\mathbf{x}} - x_3a\,\hat{\mathbf{y}} + \left(\frac{1}{2} + y_3\right)a\,\hat{\mathbf{z}}$ | (24$d$) | $NH_3$ |
| $\mathbf{B}_{20}$ | = | $-z_3\,\mathbf{a}_1 + \left(\frac{1}{2} + x_3\right)\,\mathbf{a}_2 + \left(\frac{1}{2} - y_3\right)\,\mathbf{a}_3$ | = | $-z_3a\,\hat{\mathbf{x}} + \left(\frac{1}{2} + x_3\right)a\,\hat{\mathbf{y}} + \left(\frac{1}{2} - y_3\right)a\,\hat{\mathbf{z}}$ | (24$d$) | $NH_3$ |
| $\mathbf{B}_{21}$ | = | $y_3\,\mathbf{a}_1 + z_3\,\mathbf{a}_2 + x_3\,\mathbf{a}_3$ | = | $y_3a\,\hat{\mathbf{x}} + z_3a\,\hat{\mathbf{y}} + x_3a\,\hat{\mathbf{z}}$ | (24$d$) | $NH_3$ |
| $\mathbf{B}_{22}$ | = | $-y_3\,\mathbf{a}_1 + \left(\frac{1}{2} + z_3\right)\,\mathbf{a}_2 + \left(\frac{1}{2} - x_3\right)\,\mathbf{a}_3$ | = | $-y_3a\,\hat{\mathbf{x}} + \left(\frac{1}{2} + z_3\right)a\,\hat{\mathbf{y}} + \left(\frac{1}{2} - x_3\right)a\,\hat{\mathbf{z}}$ | (24$d$) | $NH_3$ |
| $\mathbf{B}_{23}$ | = | $\left(\frac{1}{2} + y_3\right)\,\mathbf{a}_1 + \left(\frac{1}{2} - z_3\right)\,\mathbf{a}_2 - x_3\,\mathbf{a}_3$ | = | $\left(\frac{1}{2} + y_3\right)a\,\hat{\mathbf{x}} + \left(\frac{1}{2} - z_3\right)a\,\hat{\mathbf{y}} - x_3a\,\hat{\mathbf{z}}$ | (24$d$) | $NH_3$ |
| $\mathbf{B}_{24}$ | = | $\left(\frac{1}{2} - y_3\right)\,\mathbf{a}_1 - z_3\,\mathbf{a}_2 + \left(\frac{1}{2} + x_3\right)\,\mathbf{a}_3$ | = | $\left(\frac{1}{2} - y_3\right)a\,\hat{\mathbf{x}} - z_3a\,\hat{\mathbf{y}} + \left(\frac{1}{2} + x_3\right)a\,\hat{\mathbf{z}}$ | (24$d$) | $NH_3$ |
| $\mathbf{B}_{25}$ | = | $-x_3\,\mathbf{a}_1 - y_3\,\mathbf{a}_2 - z_3\,\mathbf{a}_3$ | = | $-x_3a\,\hat{\mathbf{x}} - y_3a\,\hat{\mathbf{y}} - z_3a\,\hat{\mathbf{z}}$ | (24$d$) | $NH_3$ |
| $\mathbf{B}_{26}$ | = | $\left(\frac{1}{2} + x_3\right)\,\mathbf{a}_1 + y_3\,\mathbf{a}_2 + \left(\frac{1}{2} - z_3\right)\,\mathbf{a}_3$ | = | $\left(\frac{1}{2} + x_3\right)a\,\hat{\mathbf{x}} + y_3a\,\hat{\mathbf{y}} + \left(\frac{1}{2} - z_3\right)a\,\hat{\mathbf{z}}$ | (24$d$) | $NH_3$ |
| $\mathbf{B}_{27}$ | = | $x_3\,\mathbf{a}_1 + \left(\frac{1}{2} - y_3\right)\,\mathbf{a}_2 + \left(\frac{1}{2} + z_3\right)\,\mathbf{a}_3$ | = | $x_3a\,\hat{\mathbf{x}} + \left(\frac{1}{2} - y_3\right)a\,\hat{\mathbf{y}} + \left(\frac{1}{2} + z_3\right)a\,\hat{\mathbf{z}}$ | (24$d$) | $NH_3$ |
| $\mathbf{B}_{28}$ | = | $\left(\frac{1}{2} - x_3\right) \mathbf{a}_1 + \left(\frac{1}{2} + y_3\right) \mathbf{a}_2 + z_3 \mathbf{a}_3$ | = | $\left(\frac{1}{2} - x_3\right) a \hat{\mathbf{x}} + \left(\frac{1}{2} + y_3\right) a \hat{\mathbf{y}} + z_3 a \hat{\mathbf{z}}$ | (24$d$) | $NH_3$ |
| $\mathbf{B}_{29}$ | = | $-z_3 \mathbf{a}_1 - x_3 \mathbf{a}_2 - y_3 \mathbf{a}_3$ | = | $-z_3 a \hat{\mathbf{x}} - x_3 a \hat{\mathbf{y}} - y_3 a \hat{\mathbf{z}}$ | (24$d$) | $NH_3$ |
| $\mathbf{B}_{30}$ | = | $\left(\frac{1}{2} - z_3\right) \mathbf{a}_1 + \left(\frac{1}{2} + x_3\right) \mathbf{a}_2 + y_3 \mathbf{a}_3$ | = | $\left(\frac{1}{2} - z_3\right) a \hat{\mathbf{x}} + \left(\frac{1}{2} + x_3\right) a \hat{\mathbf{y}} + y_3 a \hat{\mathbf{z}}$ | (24$d$) | $NH_3$ |
| $\mathbf{B}_{31}$ | = | $\left(\frac{1}{2} + z_3\right) \mathbf{a}_1 + x_3 \mathbf{a}_2 + \left(\frac{1}{2} - y_3\right) \mathbf{a}_3$ | = | $\left(\frac{1}{2} + z_3\right) a \hat{\mathbf{x}} + x_3 a \hat{\mathbf{y}} + \left(\frac{1}{2} - y_3\right) a \hat{\mathbf{z}}$ | (24$d$) | $NH_3$ |
| $\mathbf{B}_{32}$ | = | $z_3 \mathbf{a}_1 + \left(\frac{1}{2} - x_3\right) \mathbf{a}_2 + \left(\frac{1}{2} + y_3\right) \mathbf{a}_3$ | = | $z_3 a \hat{\mathbf{x}} + \left(\frac{1}{2} - x_3\right) a \hat{\mathbf{y}} + \left(\frac{1}{2} + y_3\right) a \hat{\mathbf{z}}$ | (24$d$) | $NH_3$ |
| $\mathbf{B}_{33}$ | = | $-y_3 \mathbf{a}_1 - z_3 \mathbf{a}_2 - x_3 \mathbf{a}_3$ | = | $-y_3 a \hat{\mathbf{x}} - z_3 a \hat{\mathbf{y}} - x_3 a \hat{\mathbf{z}}$ | (24$d$) | $NH_3$ |
| $\mathbf{B}_{34}$ | = | $y_3 \mathbf{a}_1 + \left(\frac{1}{2} - z_3\right) \mathbf{a}_2 + \left(\frac{1}{2} + x_3\right) \mathbf{a}_3$ | = | $y_3 a \hat{\mathbf{x}} + \left(\frac{1}{2} - z_3\right) a \hat{\mathbf{y}} + \left(\frac{1}{2} + x_3\right) a \hat{\mathbf{z}}$ | (24$d$) | $NH_3$ |
| $\mathbf{B}_{35}$ | = | $\left(\frac{1}{2} - y_3\right) \mathbf{a}_1 + \left(\frac{1}{2} + z_3\right) \mathbf{a}_2 + x_3 \mathbf{a}_3$ | = | $\left(\frac{1}{2} - y_3\right) a \hat{\mathbf{x}} + \left(\frac{1}{2} + z_3\right) a \hat{\mathbf{y}} + x_3 a \hat{\mathbf{z}}$ | (24$d$) | $NH_3$ |
| $\mathbf{B}_{36}$ | = | $\left(\frac{1}{2} + y_3\right) \mathbf{a}_1 + z_3 \mathbf{a}_2 + \left(\frac{1}{2} - x_3\right) \mathbf{a}_3$ | = | $\left(\frac{1}{2} + y_3\right) a \hat{\mathbf{x}} + z_3 a \hat{\mathbf{y}} + \left(\frac{1}{2} - x_3\right) a \hat{\mathbf{z}}$ | (24$d$) | $NH_3$ |
| $\mathbf{B}_{37}$ | = | $x_4 \mathbf{a}_1 + y_4 \mathbf{a}_2 + z_4 \mathbf{a}_3$ | = | $x_4 a \hat{\mathbf{x}} + y_4 a \hat{\mathbf{y}} + z_4 a \hat{\mathbf{z}}$ | (24$d$) | O |
| $\mathbf{B}_{38}$ | = | $\left(\frac{1}{2} - x_4\right) \mathbf{a}_1 - y_4 \mathbf{a}_2 + \left(\frac{1}{2} + z_4\right) \mathbf{a}_3$ | = | $\left(\frac{1}{2} - x_4\right) a \hat{\mathbf{x}} - y_4 a \hat{\mathbf{y}} + \left(\frac{1}{2} + z_4\right) a \hat{\mathbf{z}}$ | (24$d$) | O |
| $\mathbf{B}_{39}$ | = | $-x_4 \mathbf{a}_1 + \left(\frac{1}{2} + y_4\right) \mathbf{a}_2 + \left(\frac{1}{2} - z_4\right) \mathbf{a}_3$ | = | $-x_4 a \hat{\mathbf{x}} + \left(\frac{1}{2} + y_4\right) a \hat{\mathbf{y}} + \left(\frac{1}{2} - z_4\right) a \hat{\mathbf{z}}$ | (24$d$) | O |
| $\mathbf{B}_{40}$ | = | $\left(\frac{1}{2} + x_4\right) \mathbf{a}_1 + \left(\frac{1}{2} - y_4\right) \mathbf{a}_2 - z_4 \mathbf{a}_3$ | = | $\left(\frac{1}{2} + x_4\right) a \hat{\mathbf{x}} + \left(\frac{1}{2} - y_4\right) a \hat{\mathbf{y}} - z_4 a \hat{\mathbf{z}}$ | (24$d$) | O |
| $\mathbf{B}_{41}$ | = | $z_4 \mathbf{a}_1 + x_4 \mathbf{a}_2 + y_4 \mathbf{a}_3$ | = | $z_4 a \hat{\mathbf{x}} + x_4 a \hat{\mathbf{y}} + y_4 a \hat{\mathbf{z}}$ | (24$d$) | O |
| $\mathbf{B}_{42}$ | = | $\left(\frac{1}{2} + z_4\right) \mathbf{a}_1 + \left(\frac{1}{2} - x_4\right) \mathbf{a}_2 - y_4 \mathbf{a}_3$ | = | $\left(\frac{1}{2} + z_4\right) a \hat{\mathbf{x}} + \left(\frac{1}{2} - x_4\right) a \hat{\mathbf{y}} - y_4 a \hat{\mathbf{z}}$ | (24$d$) | O |
| $\mathbf{B}_{43}$ | = | $\left(\frac{1}{2} - z_4\right) \mathbf{a}_1 - x_4 \mathbf{a}_2 + \left(\frac{1}{2} + y_4\right) \mathbf{a}_3$ | = | $\left(\frac{1}{2} - z_4\right) a \hat{\mathbf{x}} - x_4 a \hat{\mathbf{y}} + \left(\frac{1}{2} + y_4\right) a \hat{\mathbf{z}}$ | (24$d$) | O |
| $\mathbf{B}_{44}$ | = | $-z_4 \mathbf{a}_1 + \left(\frac{1}{2} + x_4\right) \mathbf{a}_2 + \left(\frac{1}{2} - y_4\right) \mathbf{a}_3$ | = | $-z_4 a \hat{\mathbf{x}} + \left(\frac{1}{2} + x_4\right) a \hat{\mathbf{y}} + \left(\frac{1}{2} - y_4\right) a \hat{\mathbf{z}}$ | (24$d$) | O |
| $\mathbf{B}_{45}$ | = | $y_4 \mathbf{a}_1 + z_4 \mathbf{a}_2 + x_4 \mathbf{a}_3$ | = | $y_4 a \hat{\mathbf{x}} + z_4 a \hat{\mathbf{y}} + x_4 a \hat{\mathbf{z}}$ | (24$d$) | O |
| $\mathbf{B}_{46}$ | = | $-y_4 \mathbf{a}_1 + \left(\frac{1}{2} + z_4\right) \mathbf{a}_2 + \left(\frac{1}{2} - x_4\right) \mathbf{a}_3$ | = | $-y_4 a \hat{\mathbf{x}} + \left(\frac{1}{2} + z_4\right) a \hat{\mathbf{y}} + \left(\frac{1}{2} - x_4\right) a \hat{\mathbf{z}}$ | (24$d$) | O |
| $\mathbf{B}_{47}$ | = | $\left(\frac{1}{2} + y_4\right) \mathbf{a}_1 + \left(\frac{1}{2} - z_4\right) \mathbf{a}_2 - x_4 \mathbf{a}_3$ | = | $\left(\frac{1}{2} + y_4\right) a \hat{\mathbf{x}} + \left(\frac{1}{2} - z_4\right) a \hat{\mathbf{y}} - x_4 a \hat{\mathbf{z}}$ | (24$d$) | O |
| $\mathbf{B}_{48}$ | = | $\left(\frac{1}{2} - y_4\right) \mathbf{a}_1 - z_4 \mathbf{a}_2 + \left(\frac{1}{2} + x_4\right) \mathbf{a}_3$ | = | $\left(\frac{1}{2} - y_4\right) a \hat{\mathbf{x}} - z_4 a \hat{\mathbf{y}} + \left(\frac{1}{2} + x_4\right) a \hat{\mathbf{z}}$ | (24$d$) | O |
| $\mathbf{B}_{49}$ | = | $-x_4 \mathbf{a}_1 - y_4 \mathbf{a}_2 - z_4 \mathbf{a}_3$ | = | $-x_4 a \hat{\mathbf{x}} - y_4 a \hat{\mathbf{y}} - z_4 a \hat{\mathbf{z}}$ | (24$d$) | O |
| $\mathbf{B}_{50}$ | = | $\left(\frac{1}{2} + x_4\right) \mathbf{a}_1 + y_4 \mathbf{a}_2 + \left(\frac{1}{2} - z_4\right) \mathbf{a}_3$ | = | $\left(\frac{1}{2} + x_4\right) a \hat{\mathbf{x}} + y_4 a \hat{\mathbf{y}} + \left(\frac{1}{2} - z_4\right) a \hat{\mathbf{z}}$ | (24$d$) | O |
| $\mathbf{B}_{51}$ | = | $x_4 \mathbf{a}_1 + \left(\frac{1}{2} - y_4\right) \mathbf{a}_2 + \left(\frac{1}{2} + z_4\right) \mathbf{a}_3$ | = | $x_4 a \hat{\mathbf{x}} + \left(\frac{1}{2} - y_4\right) a \hat{\mathbf{y}} + \left(\frac{1}{2} + z_4\right) a \hat{\mathbf{z}}$ | (24$d$) | O |
| $\mathbf{B}_{52}$ | = | $\left(\frac{1}{2} - x_4\right) \mathbf{a}_1 + \left(\frac{1}{2} + y_4\right) \mathbf{a}_2 + z_4 \mathbf{a}_3$ | = | $\left(\frac{1}{2} - x_4\right) a \hat{\mathbf{x}} + \left(\frac{1}{2} + y_4\right) a \hat{\mathbf{y}} + z_4 a \hat{\mathbf{z}}$ | (24$d$) | O |
| $\mathbf{B}_{53}$ | = | $-z_4 \mathbf{a}_1 - x_4 \mathbf{a}_2 - y_4 \mathbf{a}_3$ | = | $-z_4 a \hat{\mathbf{x}} - x_4 a \hat{\mathbf{y}} - y_4 a \hat{\mathbf{z}}$ | (24$d$) | O |
| $\mathbf{B}_{54}$ | = | $\left(\frac{1}{2} - z_4\right) \mathbf{a}_1 + \left(\frac{1}{2} + x_4\right) \mathbf{a}_2 + y_4 \mathbf{a}_3$ | = | $\left(\frac{1}{2} - z_4\right) a \hat{\mathbf{x}} + \left(\frac{1}{2} + x_4\right) a \hat{\mathbf{y}} + y_4 a \hat{\mathbf{z}}$ | (24$d$) | O |
| $\mathbf{B}_{55}$ | = | $\left(\frac{1}{2} + z_4\right) \mathbf{a}_1 + x_4 \mathbf{a}_2 + \left(\frac{1}{2} - y_4\right) \mathbf{a}_3$ | = | $\left(\frac{1}{2} + z_4\right) a \hat{\mathbf{x}} + x_4 a \hat{\mathbf{y}} + \left(\frac{1}{2} - y_4\right) a \hat{\mathbf{z}}$ | (24$d$) | O |
| $\mathbf{B}_{56}$ | = | $z_4 \mathbf{a}_1 + \left(\frac{1}{2} - x_4\right) \mathbf{a}_2 + \left(\frac{1}{2} + y_4\right) \mathbf{a}_3$ | = | $z_4 a \hat{\mathbf{x}} + \left(\frac{1}{2} - x_4\right) a \hat{\mathbf{y}} + \left(\frac{1}{2} + y_4\right) a \hat{\mathbf{z}}$ | (24$d$) | O |
| $\mathbf{B}_{57}$ | = | $-y_4 \mathbf{a}_1 - z_4 \mathbf{a}_2 - x_4 \mathbf{a}_3$ | = | $-y_4 a \hat{\mathbf{x}} - z_4 a \hat{\mathbf{y}} - x_4 a \hat{\mathbf{z}}$ | (24$d$) | O |
| $\mathbf{B}_{58}$ | = | $y_4 \mathbf{a}_1 + \left(\frac{1}{2} - z_4\right) \mathbf{a}_2 + \left(\frac{1}{2} + x_4\right) \mathbf{a}_3$ | = | $y_4 a \hat{\mathbf{x}} + \left(\frac{1}{2} - z_4\right) a \hat{\mathbf{y}} + \left(\frac{1}{2} + x_4\right) a \hat{\mathbf{z}}$ | (24$d$) | O |
| $\mathbf{B}_{59}$ | = | $\left(\frac{1}{2} - y_4\right) \mathbf{a}_1 + \left(\frac{1}{2} + z_4\right) \mathbf{a}_2 + x_4 \mathbf{a}_3$ | = | $\left(\frac{1}{2} - y_4\right) a \hat{\mathbf{x}} + \left(\frac{1}{2} + z_4\right) a \hat{\mathbf{y}} + x_4 a \hat{\mathbf{z}}$ | (24$d$) | O |
| $\mathbf{B}_{60}$ | = | $\left(\frac{1}{2} + y_4\right) \mathbf{a}_1 + z_4 \mathbf{a}_2 + \left(\frac{1}{2} - x_4\right) \mathbf{a}_3$ | = | $\left(\frac{1}{2} + y_4\right) a \hat{\mathbf{x}} + z_4 a \hat{\mathbf{y}} + \left(\frac{1}{2} - x_4\right) a \hat{\mathbf{z}}$ | (24$d$) | O |

**Found in:**

---

**Geometry files:**

- CIF: pp. 1779
- POSCAR: pp. 1780

# $Pb(NO_3)_2$ ($G2_1$) Structure: A2B6C_cP36_205_c_d_a

http://aflow.org/prototype-encyclopedia/A2B6C_cP36_205_c_d_a

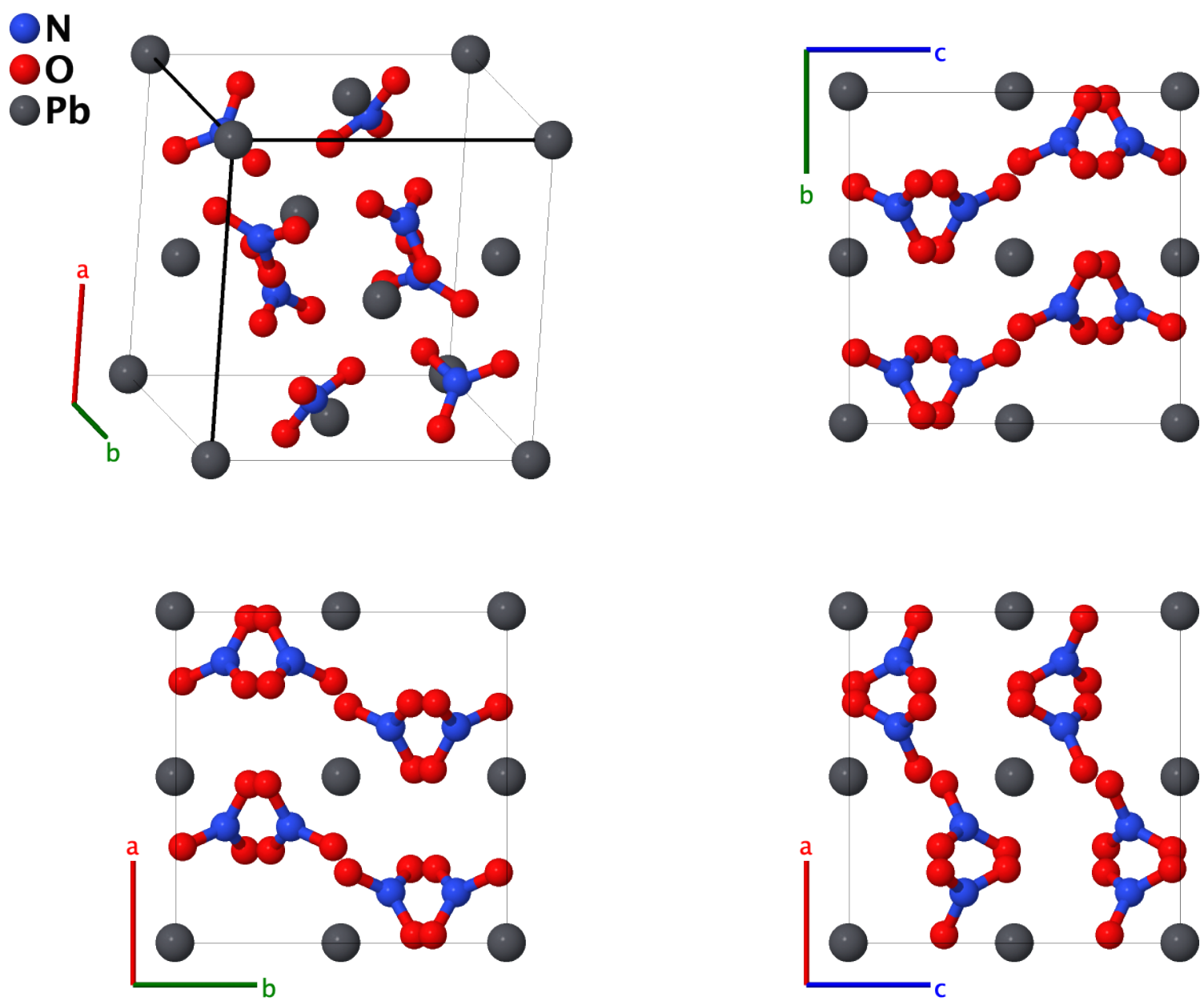

| | | |
|---|---|---|
| **Prototype** | : | $N_2O_6Pb$ |
| **AFLOW prototype label** | : | A2B6C_cP36_205_c_d_a |
| ***Strukturbericht* designation** | : | $G2_1$ |
| **Pearson symbol** | : | cP36 |
| **Space group number** | : | 205 |
| **Space group symbol** | : | $Pa\bar{3}$ |
| **AFLOW prototype command** | : | `aflow --proto=A2B6C_cP36_205_c_d_a --params=`$a, x_2, x_3, y_3, z_3$ |

**Simple Cubic primitive vectors:**

$$\mathbf{a}_1 = a\,\mathbf{\hat{x}}$$
$$\mathbf{a}_2 = a\,\mathbf{\hat{y}}$$
$$\mathbf{a}_3 = a\,\mathbf{\hat{z}}$$

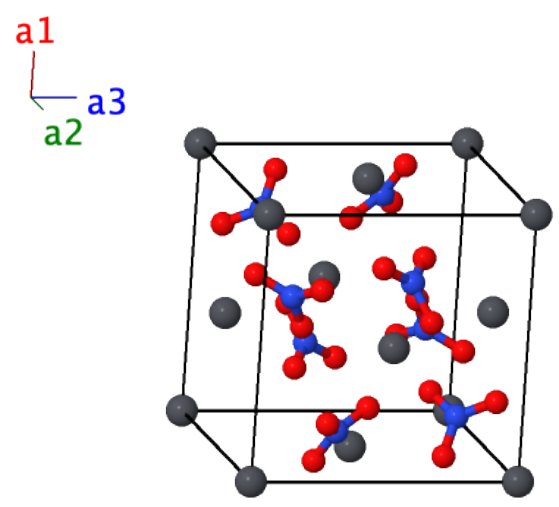

**Basis vectors:**

| | | Lattice Coordinates | | Cartesian Coordinates | Wyckoff Position | Atom Type |
|---|---|---|---|---|---|---|
| $\mathbf{B}_1$ | = | $0\,\mathbf{a}_1 + 0\,\mathbf{a}_2 + 0\,\mathbf{a}_3$ | = | $0\,\mathbf{\hat{x}} + 0\,\mathbf{\hat{y}} + 0\,\mathbf{\hat{z}}$ | (4$a$) | Pb |
| $\mathbf{B}_2$ | = | $\frac{1}{2}\,\mathbf{a}_1 + \frac{1}{2}\,\mathbf{a}_3$ | = | $\frac{1}{2}a\,\mathbf{\hat{x}} + \frac{1}{2}a\,\mathbf{\hat{z}}$ | (4$a$) | Pb |
| $\mathbf{B}_3$ | = | $\frac{1}{2}\,\mathbf{a}_2 + \frac{1}{2}\,\mathbf{a}_3$ | = | $\frac{1}{2}a\,\mathbf{\hat{y}} + \frac{1}{2}a\,\mathbf{\hat{z}}$ | (4$a$) | Pb |
| $\mathbf{B}_4$ | = | $\frac{1}{2}\,\mathbf{a}_1 + \frac{1}{2}\,\mathbf{a}_2$ | = | $\frac{1}{2}a\,\mathbf{\hat{x}} + \frac{1}{2}a\,\mathbf{\hat{y}}$ | (4$a$) | Pb |
| $\mathbf{B}_5$ | = | $x_2\,\mathbf{a}_1 + x_2\,\mathbf{a}_2 + x_2\,\mathbf{a}_3$ | = | $x_2 a\,\hat{\mathbf{x}} + x_2 a\,\hat{\mathbf{y}} + x_2 a\,\hat{\mathbf{z}}$ | $(8c)$ | N |
| $\mathbf{B}_6$ | = | $\left(\frac{1}{2} - x_2\right)\mathbf{a}_1 - x_2\,\mathbf{a}_2 + \left(\frac{1}{2} + x_2\right)\mathbf{a}_3$ | = | $\left(\frac{1}{2} - x_2\right)a\,\hat{\mathbf{x}} - x_2 a\,\hat{\mathbf{y}} + \left(\frac{1}{2} + x_2\right)a\,\hat{\mathbf{z}}$ | $(8c)$ | N |
| $\mathbf{B}_7$ | = | $-x_2\,\mathbf{a}_1 + \left(\frac{1}{2} + x_2\right)\mathbf{a}_2 + \left(\frac{1}{2} - x_2\right)\mathbf{a}_3$ | = | $-x_2 a\,\hat{\mathbf{x}} + \left(\frac{1}{2} + x_2\right)a\,\hat{\mathbf{y}} + \left(\frac{1}{2} - x_2\right)a\,\hat{\mathbf{z}}$ | $(8c)$ | N |
| $\mathbf{B}_8$ | = | $\left(\frac{1}{2} + x_2\right)\mathbf{a}_1 + \left(\frac{1}{2} - x_2\right)\mathbf{a}_2 - x_2\,\mathbf{a}_3$ | = | $\left(\frac{1}{2} + x_2\right)a\,\hat{\mathbf{x}} + \left(\frac{1}{2} - x_2\right)a\,\hat{\mathbf{y}} - x_2 a\,\hat{\mathbf{z}}$ | $(8c)$ | N |
| $\mathbf{B}_9$ | = | $-x_2\,\mathbf{a}_1 - x_2\,\mathbf{a}_2 - x_2\,\mathbf{a}_3$ | = | $-x_2 a\,\hat{\mathbf{x}} - x_2 a\,\hat{\mathbf{y}} - x_2 a\,\hat{\mathbf{z}}$ | $(8c)$ | N |
| $\mathbf{B}_{10}$ | = | $\left(\frac{1}{2} + x_2\right)\mathbf{a}_1 + x_2\,\mathbf{a}_2 + \left(\frac{1}{2} - x_2\right)\mathbf{a}_3$ | = | $\left(\frac{1}{2} + x_2\right)a\,\hat{\mathbf{x}} + x_2 a\,\hat{\mathbf{y}} + \left(\frac{1}{2} - x_2\right)a\,\hat{\mathbf{z}}$ | $(8c)$ | N |
| $\mathbf{B}_{11}$ | = | $x_2\,\mathbf{a}_1 + \left(\frac{1}{2} - x_2\right)\mathbf{a}_2 + \left(\frac{1}{2} + x_2\right)\mathbf{a}_3$ | = | $x_2 a\,\hat{\mathbf{x}} + \left(\frac{1}{2} - x_2\right)a\,\hat{\mathbf{y}} + \left(\frac{1}{2} + x_2\right)a\,\hat{\mathbf{z}}$ | $(8c)$ | N |
| $\mathbf{B}_{12}$ | = | $\left(\frac{1}{2} - x_2\right)\mathbf{a}_1 + \left(\frac{1}{2} + x_2\right)\mathbf{a}_2 + x_2\,\mathbf{a}_3$ | = | $\left(\frac{1}{2} - x_2\right)a\,\hat{\mathbf{x}} + \left(\frac{1}{2} + x_2\right)a\,\hat{\mathbf{y}} + x_2 a\,\hat{\mathbf{z}}$ | $(8c)$ | N |
| $\mathbf{B}_{13}$ | = | $x_3\,\mathbf{a}_1 + y_3\,\mathbf{a}_2 + z_3\,\mathbf{a}_3$ | = | $x_3 a\,\hat{\mathbf{x}} + y_3 a\,\hat{\mathbf{y}} + z_3 a\,\hat{\mathbf{z}}$ | $(24d)$ | O |
| $\mathbf{B}_{14}$ | = | $\left(\frac{1}{2} - x_3\right)\mathbf{a}_1 - y_3\,\mathbf{a}_2 + \left(\frac{1}{2} + z_3\right)\mathbf{a}_3$ | = | $\left(\frac{1}{2} - x_3\right)a\,\hat{\mathbf{x}} - y_3 a\,\hat{\mathbf{y}} + \left(\frac{1}{2} + z_3\right)a\,\hat{\mathbf{z}}$ | $(24d)$ | O |
| $\mathbf{B}_{15}$ | = | $-x_3\,\mathbf{a}_1 + \left(\frac{1}{2} + y_3\right)\mathbf{a}_2 + \left(\frac{1}{2} - z_3\right)\mathbf{a}_3$ | = | $-x_3 a\,\hat{\mathbf{x}} + \left(\frac{1}{2} + y_3\right)a\,\hat{\mathbf{y}} + \left(\frac{1}{2} - z_3\right)a\,\hat{\mathbf{z}}$ | $(24d)$ | O |
| $\mathbf{B}_{16}$ | = | $\left(\frac{1}{2} + x_3\right)\mathbf{a}_1 + \left(\frac{1}{2} - y_3\right)\mathbf{a}_2 - z_3\,\mathbf{a}_3$ | = | $\left(\frac{1}{2} + x_3\right)a\,\hat{\mathbf{x}} + \left(\frac{1}{2} - y_3\right)a\,\hat{\mathbf{y}} - z_3 a\,\hat{\mathbf{z}}$ | $(24d)$ | O |
| $\mathbf{B}_{17}$ | = | $z_3\,\mathbf{a}_1 + x_3\,\mathbf{a}_2 + y_3\,\mathbf{a}_3$ | = | $z_3 a\,\hat{\mathbf{x}} + x_3 a\,\hat{\mathbf{y}} + y_3 a\,\hat{\mathbf{z}}$ | $(24d)$ | O |
| $\mathbf{B}_{18}$ | = | $\left(\frac{1}{2} + z_3\right)\mathbf{a}_1 + \left(\frac{1}{2} - x_3\right)\mathbf{a}_2 - y_3\,\mathbf{a}_3$ | = | $\left(\frac{1}{2} + z_3\right)a\,\hat{\mathbf{x}} + \left(\frac{1}{2} - x_3\right)a\,\hat{\mathbf{y}} - y_3 a\,\hat{\mathbf{z}}$ | $(24d)$ | O |
| $\mathbf{B}_{19}$ | = | $\left(\frac{1}{2} - z_3\right)\mathbf{a}_1 - x_3\,\mathbf{a}_2 + \left(\frac{1}{2} + y_3\right)\mathbf{a}_3$ | = | $\left(\frac{1}{2} - z_3\right)a\,\hat{\mathbf{x}} - x_3 a\,\hat{\mathbf{y}} + \left(\frac{1}{2} + y_3\right)a\,\hat{\mathbf{z}}$ | $(24d)$ | O |
| $\mathbf{B}_{20}$ | = | $-z_3\,\mathbf{a}_1 + \left(\frac{1}{2} + x_3\right)\mathbf{a}_2 + \left(\frac{1}{2} - y_3\right)\mathbf{a}_3$ | = | $-z_3 a\,\hat{\mathbf{x}} + \left(\frac{1}{2} + x_3\right)a\,\hat{\mathbf{y}} + \left(\frac{1}{2} - y_3\right)a\,\hat{\mathbf{z}}$ | $(24d)$ | O |
| $\mathbf{B}_{21}$ | = | $y_3\,\mathbf{a}_1 + z_3\,\mathbf{a}_2 + x_3\,\mathbf{a}_3$ | = | $y_3 a\,\hat{\mathbf{x}} + z_3 a\,\hat{\mathbf{y}} + x_3 a\,\hat{\mathbf{z}}$ | $(24d)$ | O |
| $\mathbf{B}_{22}$ | = | $-y_3\,\mathbf{a}_1 + \left(\frac{1}{2} + z_3\right)\mathbf{a}_2 + \left(\frac{1}{2} - x_3\right)\mathbf{a}_3$ | = | $-y_3 a\,\hat{\mathbf{x}} + \left(\frac{1}{2} + z_3\right)a\,\hat{\mathbf{y}} + \left(\frac{1}{2} - x_3\right)a\,\hat{\mathbf{z}}$ | $(24d)$ | O |
| $\mathbf{B}_{23}$ | = | $\left(\frac{1}{2} + y_3\right)\mathbf{a}_1 + \left(\frac{1}{2} - z_3\right)\mathbf{a}_2 - x_3\,\mathbf{a}_3$ | = | $\left(\frac{1}{2} + y_3\right)a\,\hat{\mathbf{x}} + \left(\frac{1}{2} - z_3\right)a\,\hat{\mathbf{y}} - x_3 a\,\hat{\mathbf{z}}$ | $(24d)$ | O |
| $\mathbf{B}_{24}$ | = | $\left(\frac{1}{2} - y_3\right)\mathbf{a}_1 - z_3\,\mathbf{a}_2 + \left(\frac{1}{2} + x_3\right)\mathbf{a}_3$ | = | $\left(\frac{1}{2} - y_3\right)a\,\hat{\mathbf{x}} - z_3 a\,\hat{\mathbf{y}} + \left(\frac{1}{2} + x_3\right)a\,\hat{\mathbf{z}}$ | $(24d)$ | O |
| $\mathbf{B}_{25}$ | = | $-x_3\,\mathbf{a}_1 - y_3\,\mathbf{a}_2 - z_3\,\mathbf{a}_3$ | = | $-x_3 a\,\hat{\mathbf{x}} - y_3 a\,\hat{\mathbf{y}} - z_3 a\,\hat{\mathbf{z}}$ | $(24d)$ | O |
| $\mathbf{B}_{26}$ | = | $\left(\frac{1}{2} + x_3\right)\mathbf{a}_1 + y_3\,\mathbf{a}_2 + \left(\frac{1}{2} - z_3\right)\mathbf{a}_3$ | = | $\left(\frac{1}{2} + x_3\right)a\,\hat{\mathbf{x}} + y_3 a\,\hat{\mathbf{y}} + \left(\frac{1}{2} - z_3\right)a\,\hat{\mathbf{z}}$ | $(24d)$ | O |
| $\mathbf{B}_{27}$ | = | $x_3\,\mathbf{a}_1 + \left(\frac{1}{2} - y_3\right)\mathbf{a}_2 + \left(\frac{1}{2} + z_3\right)\mathbf{a}_3$ | = | $x_3 a\,\hat{\mathbf{x}} + \left(\frac{1}{2} - y_3\right)a\,\hat{\mathbf{y}} + \left(\frac{1}{2} + z_3\right)a\,\hat{\mathbf{z}}$ | $(24d)$ | O |
| $\mathbf{B}_{28}$ | = | $\left(\frac{1}{2} - x_3\right)\mathbf{a}_1 + \left(\frac{1}{2} + y_3\right)\mathbf{a}_2 + z_3\,\mathbf{a}_3$ | = | $\left(\frac{1}{2} - x_3\right)a\,\hat{\mathbf{x}} + \left(\frac{1}{2} + y_3\right)a\,\hat{\mathbf{y}} + z_3 a\,\hat{\mathbf{z}}$ | $(24d)$ | O |
| $\mathbf{B}_{29}$ | = | $-z_3\,\mathbf{a}_1 - x_3\,\mathbf{a}_2 - y_3\,\mathbf{a}_3$ | = | $-z_3 a\,\hat{\mathbf{x}} - x_3 a\,\hat{\mathbf{y}} - y_3 a\,\hat{\mathbf{z}}$ | $(24d)$ | O |
| $\mathbf{B}_{30}$ | = | $\left(\frac{1}{2} - z_3\right)\mathbf{a}_1 + \left(\frac{1}{2} + x_3\right)\mathbf{a}_2 + y_3\,\mathbf{a}_3$ | = | $\left(\frac{1}{2} - z_3\right)a\,\hat{\mathbf{x}} + \left(\frac{1}{2} + x_3\right)a\,\hat{\mathbf{y}} + y_3 a\,\hat{\mathbf{z}}$ | $(24d)$ | O |
| $\mathbf{B}_{31}$ | = | $\left(\frac{1}{2} + z_3\right)\mathbf{a}_1 + x_3\,\mathbf{a}_2 + \left(\frac{1}{2} - y_3\right)\mathbf{a}_3$ | = | $\left(\frac{1}{2} + z_3\right)a\,\hat{\mathbf{x}} + x_3 a\,\hat{\mathbf{y}} + \left(\frac{1}{2} - y_3\right)a\,\hat{\mathbf{z}}$ | $(24d)$ | O |
| $\mathbf{B}_{32}$ | = | $z_3\,\mathbf{a}_1 + \left(\frac{1}{2} - x_3\right)\mathbf{a}_2 + \left(\frac{1}{2} + y_3\right)\mathbf{a}_3$ | = | $z_3 a\,\hat{\mathbf{x}} + \left(\frac{1}{2} - x_3\right)a\,\hat{\mathbf{y}} + \left(\frac{1}{2} + y_3\right)a\,\hat{\mathbf{z}}$ | $(24d)$ | O |
| $\mathbf{B}_{33}$ | = | $-y_3\,\mathbf{a}_1 - z_3\,\mathbf{a}_2 - x_3\,\mathbf{a}_3$ | = | $-y_3 a\,\hat{\mathbf{x}} - z_3 a\,\hat{\mathbf{y}} - x_3 a\,\hat{\mathbf{z}}$ | $(24d)$ | O |
| $\mathbf{B}_{34}$ | = | $y_3\,\mathbf{a}_1 + \left(\frac{1}{2} - z_3\right)\mathbf{a}_2 + \left(\frac{1}{2} + x_3\right)\mathbf{a}_3$ | = | $y_3 a\,\hat{\mathbf{x}} + \left(\frac{1}{2} - z_3\right)a\,\hat{\mathbf{y}} + \left(\frac{1}{2} + x_3\right)a\,\hat{\mathbf{z}}$ | $(24d)$ | O |
| $\mathbf{B}_{35}$ | = | $\left(\frac{1}{2} - y_3\right)\mathbf{a}_1 + \left(\frac{1}{2} + z_3\right)\mathbf{a}_2 + x_3\,\mathbf{a}_3$ | = | $\left(\frac{1}{2} - y_3\right)a\,\hat{\mathbf{x}} + \left(\frac{1}{2} + z_3\right)a\,\hat{\mathbf{y}} + x_3 a\,\hat{\mathbf{z}}$ | $(24d)$ | O |
| $\mathbf{B}_{36}$ | = | $\left(\frac{1}{2} + y_3\right)\mathbf{a}_1 + z_3\,\mathbf{a}_2 + \left(\frac{1}{2} - x_3\right)\mathbf{a}_3$ | = | $\left(\frac{1}{2} + y_3\right)a\,\hat{\mathbf{x}} + z_3 a\,\hat{\mathbf{y}} + \left(\frac{1}{2} - x_3\right)a\,\hat{\mathbf{z}}$ | $(24d)$ | O |

**Geometry files:**

- CIF: pp. 1780
- POSCAR: pp. 1780

# $CaB_2O_4$ (IV) Structure: A2BC4_cP84_205_d_ac_2d

http://aflow.org/prototype-encyclopedia/A2BC4_cP84_205_d_ac_2d

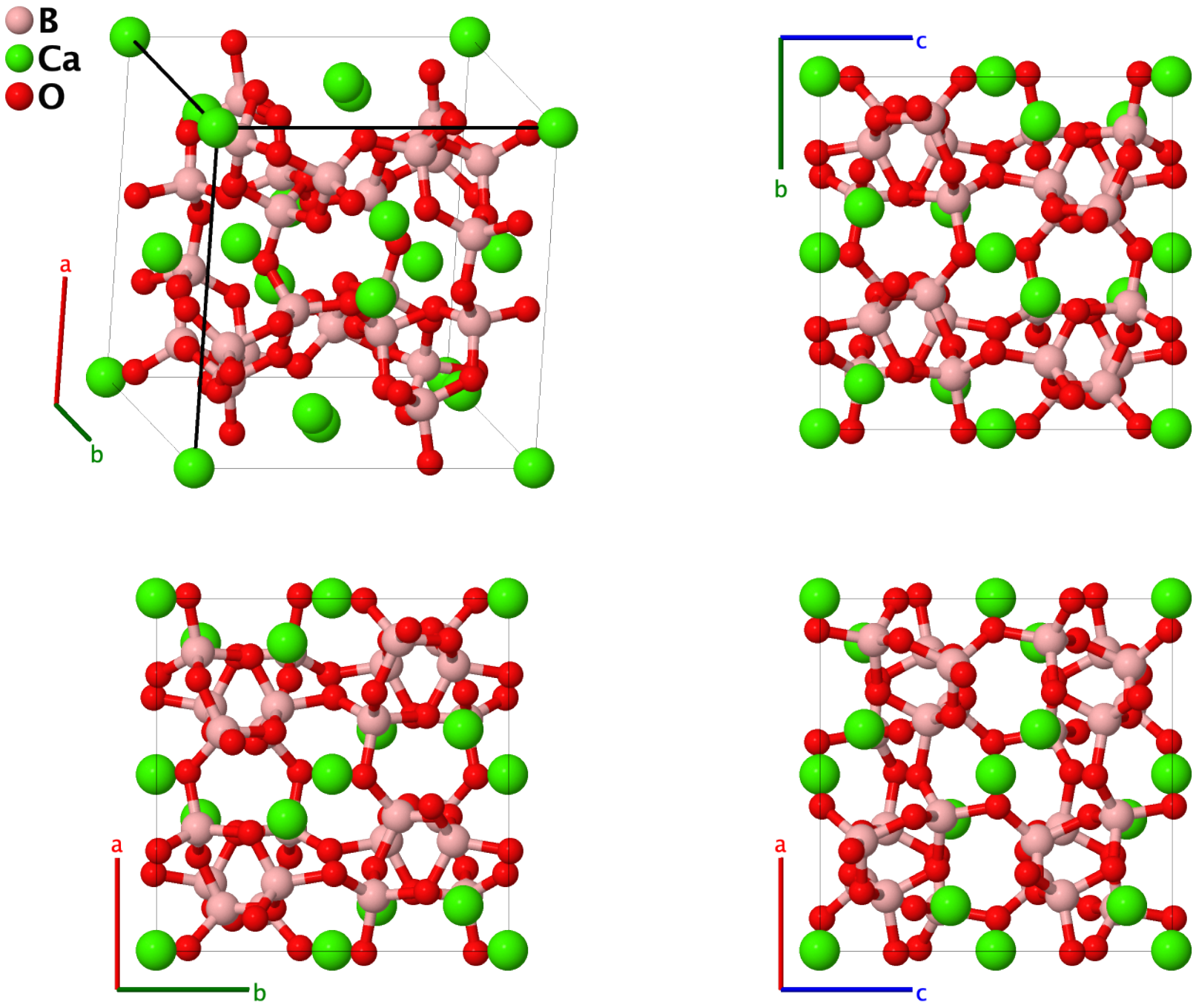

| | | |
|---|---|---|
| **Prototype** | : | $B_2CaO_4$ |
| **AFLOW prototype label** | : | A2BC4_cP84_205_d_ac_2d |
| ***Strukturbericht* designation** | : | None |
| **Pearson symbol** | : | cP84 |
| **Space group number** | : | 205 |
| **Space group symbol** | : | $Pa\bar{3}$ |
| **AFLOW prototype command** | : | `aflow --proto=A2BC4_cP84_205_d_ac_2d` `--params=`$a, x_2, x_3, y_3, z_3, x_4, y_4, z_4, x_5, y_5, z_5$ |

- $CaB_2O_4$ exists in at least four phases (Marezio, 1969):
- I - The ground state, stable below 1.2 GPa, *Strukturbericht* $E3_2$.
- II – Orthorhombic high pressure structure, stable between 1.2 and 1.5 GPa, presumably calciborite.
- III – Orthorhombic high pressure structure, stable between 1.5 and 2.5 GPa.
- IV – Cubic high pressure structure, stable between 2.5 and 4.0 GPa (this structure).

**Simple Cubic primitive vectors:**

$$\mathbf{a}_1 = a\,\hat{\mathbf{x}}$$
$$\mathbf{a}_2 = a\,\hat{\mathbf{y}}$$
$$\mathbf{a}_3 = a\,\hat{\mathbf{z}}$$

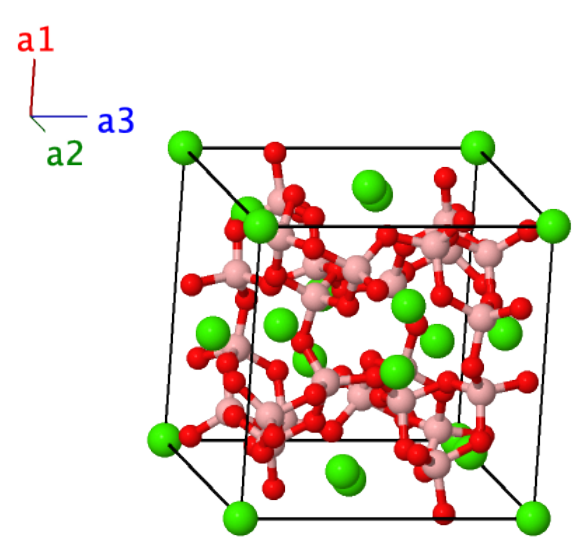

**Basis vectors:**

| | | Lattice Coordinates | | Cartesian Coordinates | Wyckoff Position | Atom Type |
|---|---|---|---|---|---|---|
| $\mathbf{B}_1$ | = | $0\,\mathbf{a}_1 + 0\,\mathbf{a}_2 + 0\,\mathbf{a}_3$ | = | $0\,\hat{\mathbf{x}} + 0\,\hat{\mathbf{y}} + 0\,\hat{\mathbf{z}}$ | (4$a$) | Ca I |
| $\mathbf{B}_2$ | = | $\frac{1}{2}\,\mathbf{a}_1 + \frac{1}{2}\,\mathbf{a}_3$ | = | $\frac{1}{2}a\,\hat{\mathbf{x}} + \frac{1}{2}a\,\hat{\mathbf{z}}$ | (4$a$) | Ca I |
| $\mathbf{B}_3$ | = | $\frac{1}{2}\,\mathbf{a}_2 + \frac{1}{2}\,\mathbf{a}_3$ | = | $\frac{1}{2}a\,\hat{\mathbf{y}} + \frac{1}{2}a\,\hat{\mathbf{z}}$ | (4$a$) | Ca I |
| $\mathbf{B}_4$ | = | $\frac{1}{2}\,\mathbf{a}_1 + \frac{1}{2}\,\mathbf{a}_2$ | = | $\frac{1}{2}a\,\hat{\mathbf{x}} + \frac{1}{2}a\,\hat{\mathbf{y}}$ | (4$a$) | Ca I |
| $\mathbf{B}_5$ | = | $x_2\,\mathbf{a}_1 + x_2\,\mathbf{a}_2 + x_2\,\mathbf{a}_3$ | = | $x_2a\,\hat{\mathbf{x}} + x_2a\,\hat{\mathbf{y}} + x_2a\,\hat{\mathbf{z}}$ | (8$c$) | Ca II |
| $\mathbf{B}_6$ | = | $\left(\frac{1}{2} - x_2\right)\,\mathbf{a}_1 - x_2\,\mathbf{a}_2 + \left(\frac{1}{2} + x_2\right)\,\mathbf{a}_3$ | = | $\left(\frac{1}{2} - x_2\right)a\,\hat{\mathbf{x}} - x_2a\,\hat{\mathbf{y}} + \left(\frac{1}{2} + x_2\right)a\,\hat{\mathbf{z}}$ | (8$c$) | Ca II |
| $\mathbf{B}_7$ | = | $-x_2\,\mathbf{a}_1 + \left(\frac{1}{2} + x_2\right)\,\mathbf{a}_2 + \left(\frac{1}{2} - x_2\right)\,\mathbf{a}_3$ | = | $-x_2a\,\hat{\mathbf{x}} + \left(\frac{1}{2} + x_2\right)a\,\hat{\mathbf{y}} + \left(\frac{1}{2} - x_2\right)a\,\hat{\mathbf{z}}$ | (8$c$) | Ca II |
| $\mathbf{B}_8$ | = | $\left(\frac{1}{2} + x_2\right)\,\mathbf{a}_1 + \left(\frac{1}{2} - x_2\right)\,\mathbf{a}_2 - x_2\,\mathbf{a}_3$ | = | $\left(\frac{1}{2} + x_2\right)a\,\hat{\mathbf{x}} + \left(\frac{1}{2} - x_2\right)a\,\hat{\mathbf{y}} - x_2a\,\hat{\mathbf{z}}$ | (8$c$) | Ca II |
| $\mathbf{B}_9$ | = | $-x_2\,\mathbf{a}_1 - x_2\,\mathbf{a}_2 - x_2\,\mathbf{a}_3$ | = | $-x_2a\,\hat{\mathbf{x}} - x_2a\,\hat{\mathbf{y}} - x_2a\,\hat{\mathbf{z}}$ | (8$c$) | Ca II |
| $\mathbf{B}_{10}$ | = | $\left(\frac{1}{2} + x_2\right)\,\mathbf{a}_1 + x_2\,\mathbf{a}_2 + \left(\frac{1}{2} - x_2\right)\,\mathbf{a}_3$ | = | $\left(\frac{1}{2} + x_2\right)a\,\hat{\mathbf{x}} + x_2a\,\hat{\mathbf{y}} + \left(\frac{1}{2} - x_2\right)a\,\hat{\mathbf{z}}$ | (8$c$) | Ca II |
| $\mathbf{B}_{11}$ | = | $x_2\,\mathbf{a}_1 + \left(\frac{1}{2} - x_2\right)\,\mathbf{a}_2 + \left(\frac{1}{2} + x_2\right)\,\mathbf{a}_3$ | = | $x_2a\,\hat{\mathbf{x}} + \left(\frac{1}{2} - x_2\right)a\,\hat{\mathbf{y}} + \left(\frac{1}{2} + x_2\right)a\,\hat{\mathbf{z}}$ | (8$c$) | Ca II |
| $\mathbf{B}_{12}$ | = | $\left(\frac{1}{2} - x_2\right)\,\mathbf{a}_1 + \left(\frac{1}{2} + x_2\right)\,\mathbf{a}_2 + x_2\,\mathbf{a}_3$ | = | $\left(\frac{1}{2} - x_2\right)a\,\hat{\mathbf{x}} + \left(\frac{1}{2} + x_2\right)a\,\hat{\mathbf{y}} + x_2a\,\hat{\mathbf{z}}$ | (8$c$) | Ca II |
| $\mathbf{B}_{13}$ | = | $x_3\,\mathbf{a}_1 + y_3\,\mathbf{a}_2 + z_3\,\mathbf{a}_3$ | = | $x_3a\,\hat{\mathbf{x}} + y_3a\,\hat{\mathbf{y}} + z_3a\,\hat{\mathbf{z}}$ | (24$d$) | B |
| $\mathbf{B}_{14}$ | = | $\left(\frac{1}{2} - x_3\right)\,\mathbf{a}_1 - y_3\,\mathbf{a}_2 + \left(\frac{1}{2} + z_3\right)\,\mathbf{a}_3$ | = | $\left(\frac{1}{2} - x_3\right)a\,\hat{\mathbf{x}} - y_3a\,\hat{\mathbf{y}} + \left(\frac{1}{2} + z_3\right)a\,\hat{\mathbf{z}}$ | (24$d$) | B |
| $\mathbf{B}_{15}$ | = | $-x_3\,\mathbf{a}_1 + \left(\frac{1}{2} + y_3\right)\,\mathbf{a}_2 + \left(\frac{1}{2} - z_3\right)\,\mathbf{a}_3$ | = | $-x_3a\,\hat{\mathbf{x}} + \left(\frac{1}{2} + y_3\right)a\,\hat{\mathbf{y}} + \left(\frac{1}{2} - z_3\right)a\,\hat{\mathbf{z}}$ | (24$d$) | B |
| $\mathbf{B}_{16}$ | = | $\left(\frac{1}{2} + x_3\right)\,\mathbf{a}_1 + \left(\frac{1}{2} - y_3\right)\,\mathbf{a}_2 - z_3\,\mathbf{a}_3$ | = | $\left(\frac{1}{2} + x_3\right)a\,\hat{\mathbf{x}} + \left(\frac{1}{2} - y_3\right)a\,\hat{\mathbf{y}} - z_3a\,\hat{\mathbf{z}}$ | (24$d$) | B |
| $\mathbf{B}_{17}$ | = | $z_3\,\mathbf{a}_1 + x_3\,\mathbf{a}_2 + y_3\,\mathbf{a}_3$ | = | $z_3a\,\hat{\mathbf{x}} + x_3a\,\hat{\mathbf{y}} + y_3a\,\hat{\mathbf{z}}$ | (24$d$) | B |
| $\mathbf{B}_{18}$ | = | $\left(\frac{1}{2} + z_3\right)\,\mathbf{a}_1 + \left(\frac{1}{2} - x_3\right)\,\mathbf{a}_2 - y_3\,\mathbf{a}_3$ | = | $\left(\frac{1}{2} + z_3\right)a\,\hat{\mathbf{x}} + \left(\frac{1}{2} - x_3\right)a\,\hat{\mathbf{y}} - y_3a\,\hat{\mathbf{z}}$ | (24$d$) | B |
| $\mathbf{B}_{19}$ | = | $\left(\frac{1}{2} - z_3\right)\,\mathbf{a}_1 - x_3\,\mathbf{a}_2 + \left(\frac{1}{2} + y_3\right)\,\mathbf{a}_3$ | = | $\left(\frac{1}{2} - z_3\right)a\,\hat{\mathbf{x}} - x_3a\,\hat{\mathbf{y}} + \left(\frac{1}{2} + y_3\right)a\,\hat{\mathbf{z}}$ | (24$d$) | B |
| $\mathbf{B}_{20}$ | = | $-z_3\,\mathbf{a}_1 + \left(\frac{1}{2} + x_3\right)\,\mathbf{a}_2 + \left(\frac{1}{2} - y_3\right)\,\mathbf{a}_3$ | = | $-z_3a\,\hat{\mathbf{x}} + \left(\frac{1}{2} + x_3\right)a\,\hat{\mathbf{y}} + \left(\frac{1}{2} - y_3\right)a\,\hat{\mathbf{z}}$ | (24$d$) | B |
| $\mathbf{B}_{21}$ | = | $y_3\,\mathbf{a}_1 + z_3\,\mathbf{a}_2 + x_3\,\mathbf{a}_3$ | = | $y_3a\,\hat{\mathbf{x}} + z_3a\,\hat{\mathbf{y}} + x_3a\,\hat{\mathbf{z}}$ | (24$d$) | B |
| $\mathbf{B}_{22}$ | = | $-y_3\,\mathbf{a}_1 + \left(\frac{1}{2} + z_3\right)\,\mathbf{a}_2 + \left(\frac{1}{2} - x_3\right)\,\mathbf{a}_3$ | = | $-y_3a\,\hat{\mathbf{x}} + \left(\frac{1}{2} + z_3\right)a\,\hat{\mathbf{y}} + \left(\frac{1}{2} - x_3\right)a\,\hat{\mathbf{z}}$ | (24$d$) | B |
| $\mathbf{B}_{23}$ | = | $\left(\frac{1}{2} + y_3\right)\,\mathbf{a}_1 + \left(\frac{1}{2} - z_3\right)\,\mathbf{a}_2 - x_3\,\mathbf{a}_3$ | = | $\left(\frac{1}{2} + y_3\right)a\,\hat{\mathbf{x}} + \left(\frac{1}{2} - z_3\right)a\,\hat{\mathbf{y}} - x_3a\,\hat{\mathbf{z}}$ | (24$d$) | B |
| $\mathbf{B}_{24}$ | = | $\left(\frac{1}{2} - y_3\right)\,\mathbf{a}_1 - z_3\,\mathbf{a}_2 + \left(\frac{1}{2} + x_3\right)\,\mathbf{a}_3$ | = | $\left(\frac{1}{2} - y_3\right)a\,\hat{\mathbf{x}} - z_3a\,\hat{\mathbf{y}} + \left(\frac{1}{2} + x_3\right)a\,\hat{\mathbf{z}}$ | (24$d$) | B |
| $\mathbf{B}_{25}$ | = | $-x_3\,\mathbf{a}_1 - y_3\,\mathbf{a}_2 - z_3\,\mathbf{a}_3$ | = | $-x_3a\,\hat{\mathbf{x}} - y_3a\,\hat{\mathbf{y}} - z_3a\,\hat{\mathbf{z}}$ | (24$d$) | B |
| $\mathbf{B}_{26}$ | = | $\left(\frac{1}{2} + x_3\right)\,\mathbf{a}_1 + y_3\,\mathbf{a}_2 + \left(\frac{1}{2} - z_3\right)\,\mathbf{a}_3$ | = | $\left(\frac{1}{2} + x_3\right)a\,\hat{\mathbf{x}} + y_3a\,\hat{\mathbf{y}} + \left(\frac{1}{2} - z_3\right)a\,\hat{\mathbf{z}}$ | (24$d$) | B |
| $\mathbf{B}_{27}$ | = | $x_3\,\mathbf{a}_1 + \left(\frac{1}{2} - y_3\right)\,\mathbf{a}_2 + \left(\frac{1}{2} + z_3\right)\,\mathbf{a}_3$ | = | $x_3a\,\hat{\mathbf{x}} + \left(\frac{1}{2} - y_3\right)a\,\hat{\mathbf{y}} + \left(\frac{1}{2} + z_3\right)a\,\hat{\mathbf{z}}$ | (24$d$) | B |
| $\mathbf{B}_{28}$ | = | $\left(\frac{1}{2} - x_3\right)\,\mathbf{a}_1 + \left(\frac{1}{2} + y_3\right)\,\mathbf{a}_2 + z_3\,\mathbf{a}_3$ | = | $\left(\frac{1}{2} - x_3\right)a\,\hat{\mathbf{x}} + \left(\frac{1}{2} + y_3\right)a\,\hat{\mathbf{y}} + z_3a\,\hat{\mathbf{z}}$ | (24$d$) | B |
| $\mathbf{B}_{29}$ | = | $-z_3\,\mathbf{a}_1 - x_3\,\mathbf{a}_2 - y_3\,\mathbf{a}_3$ | = | $-z_3a\,\hat{\mathbf{x}} - x_3a\,\hat{\mathbf{y}} - y_3a\,\hat{\mathbf{z}}$ | (24$d$) | B |
| $\mathbf{B}_{30}$ | = | $\left(\frac{1}{2} - z_3\right)\,\mathbf{a}_1 + \left(\frac{1}{2} + x_3\right)\,\mathbf{a}_2 + y_3\,\mathbf{a}_3$ | = | $\left(\frac{1}{2} - z_3\right)a\,\hat{\mathbf{x}} + \left(\frac{1}{2} + x_3\right)a\,\hat{\mathbf{y}} + y_3a\,\hat{\mathbf{z}}$ | (24$d$) | B |
| $\mathbf{B}_{31}$ | = | $\left(\frac{1}{2} + z_3\right)\,\mathbf{a}_1 + x_3\,\mathbf{a}_2 + \left(\frac{1}{2} - y_3\right)\,\mathbf{a}_3$ | = | $\left(\frac{1}{2} + z_3\right)a\,\hat{\mathbf{x}} + x_3a\,\hat{\mathbf{y}} + \left(\frac{1}{2} - y_3\right)a\,\hat{\mathbf{z}}$ | (24$d$) | B |
| $\mathbf{B}_{32}$ | = | $z_3\,\mathbf{a}_1 + \left(\frac{1}{2} - x_3\right)\,\mathbf{a}_2 + \left(\frac{1}{2} + y_3\right)\,\mathbf{a}_3$ | = | $z_3a\,\hat{\mathbf{x}} + \left(\frac{1}{2} - x_3\right)a\,\hat{\mathbf{y}} + \left(\frac{1}{2} + y_3\right)a\,\hat{\mathbf{z}}$ | (24$d$) | B |
| $\mathbf{B}_{33}$ | = | $-y_3\,\mathbf{a}_1 - z_3\,\mathbf{a}_2 - x_3\,\mathbf{a}_3$ | = | $-y_3a\,\hat{\mathbf{x}} - z_3a\,\hat{\mathbf{y}} - x_3a\,\hat{\mathbf{z}}$ | (24$d$) | B |
| $\mathbf{B}_{34}$ | = | $y_3\,\mathbf{a}_1 + \left(\frac{1}{2} - z_3\right)\,\mathbf{a}_2 + \left(\frac{1}{2} + x_3\right)\,\mathbf{a}_3$ | = | $y_3a\,\hat{\mathbf{x}} + \left(\frac{1}{2} - z_3\right)a\,\hat{\mathbf{y}} + \left(\frac{1}{2} + x_3\right)a\,\hat{\mathbf{z}}$ | (24$d$) | B |
| $\mathbf{B}_{35}$ | = | $\left(\frac{1}{2} - y_3\right) \mathbf{a}_1 + \left(\frac{1}{2} + z_3\right) \mathbf{a}_2 + x_3 \mathbf{a}_3$ | = | $\left(\frac{1}{2} - y_3\right) a \hat{\mathbf{x}} + \left(\frac{1}{2} + z_3\right) a \hat{\mathbf{y}} + x_3 a \hat{\mathbf{z}}$ | (24$d$) | B |
| $\mathbf{B}_{36}$ | = | $\left(\frac{1}{2} + y_3\right) \mathbf{a}_1 + z_3 \mathbf{a}_2 + \left(\frac{1}{2} - x_3\right) \mathbf{a}_3$ | = | $\left(\frac{1}{2} + y_3\right) a \hat{\mathbf{x}} + z_3 a \hat{\mathbf{y}} + \left(\frac{1}{2} - x_3\right) a \hat{\mathbf{z}}$ | (24$d$) | B |
| $\mathbf{B}_{37}$ | = | $x_4 \mathbf{a}_1 + y_4 \mathbf{a}_2 + z_4 \mathbf{a}_3$ | = | $x_4 a \hat{\mathbf{x}} + y_4 a \hat{\mathbf{y}} + z_4 a \hat{\mathbf{z}}$ | (24$d$) | O I |
| $\mathbf{B}_{38}$ | = | $\left(\frac{1}{2} - x_4\right) \mathbf{a}_1 - y_4 \mathbf{a}_2 + \left(\frac{1}{2} + z_4\right) \mathbf{a}_3$ | = | $\left(\frac{1}{2} - x_4\right) a \hat{\mathbf{x}} - y_4 a \hat{\mathbf{y}} + \left(\frac{1}{2} + z_4\right) a \hat{\mathbf{z}}$ | (24$d$) | O I |
| $\mathbf{B}_{39}$ | = | $-x_4 \mathbf{a}_1 + \left(\frac{1}{2} + y_4\right) \mathbf{a}_2 + \left(\frac{1}{2} - z_4\right) \mathbf{a}_3$ | = | $-x_4 a \hat{\mathbf{x}} + \left(\frac{1}{2} + y_4\right) a \hat{\mathbf{y}} + \left(\frac{1}{2} - z_4\right) a \hat{\mathbf{z}}$ | (24$d$) | O I |
| $\mathbf{B}_{40}$ | = | $\left(\frac{1}{2} + x_4\right) \mathbf{a}_1 + \left(\frac{1}{2} - y_4\right) \mathbf{a}_2 - z_4 \mathbf{a}_3$ | = | $\left(\frac{1}{2} + x_4\right) a \hat{\mathbf{x}} + \left(\frac{1}{2} - y_4\right) a \hat{\mathbf{y}} - z_4 a \hat{\mathbf{z}}$ | (24$d$) | O I |
| $\mathbf{B}_{41}$ | = | $z_4 \mathbf{a}_1 + x_4 \mathbf{a}_2 + y_4 \mathbf{a}_3$ | = | $z_4 a \hat{\mathbf{x}} + x_4 a \hat{\mathbf{y}} + y_4 a \hat{\mathbf{z}}$ | (24$d$) | O I |
| $\mathbf{B}_{42}$ | = | $\left(\frac{1}{2} + z_4\right) \mathbf{a}_1 + \left(\frac{1}{2} - x_4\right) \mathbf{a}_2 - y_4 \mathbf{a}_3$ | = | $\left(\frac{1}{2} + z_4\right) a \hat{\mathbf{x}} + \left(\frac{1}{2} - x_4\right) a \hat{\mathbf{y}} - y_4 a \hat{\mathbf{z}}$ | (24$d$) | O I |
| $\mathbf{B}_{43}$ | = | $\left(\frac{1}{2} - z_4\right) \mathbf{a}_1 - x_4 \mathbf{a}_2 + \left(\frac{1}{2} + y_4\right) \mathbf{a}_3$ | = | $\left(\frac{1}{2} - z_4\right) a \hat{\mathbf{x}} - x_4 a \hat{\mathbf{y}} + \left(\frac{1}{2} + y_4\right) a \hat{\mathbf{z}}$ | (24$d$) | O I |
| $\mathbf{B}_{44}$ | = | $-z_4 \mathbf{a}_1 + \left(\frac{1}{2} + x_4\right) \mathbf{a}_2 + \left(\frac{1}{2} - y_4\right) \mathbf{a}_3$ | = | $-z_4 a \hat{\mathbf{x}} + \left(\frac{1}{2} + x_4\right) a \hat{\mathbf{y}} + \left(\frac{1}{2} - y_4\right) a \hat{\mathbf{z}}$ | (24$d$) | O I |
| $\mathbf{B}_{45}$ | = | $y_4 \mathbf{a}_1 + z_4 \mathbf{a}_2 + x_4 \mathbf{a}_3$ | = | $y_4 a \hat{\mathbf{x}} + z_4 a \hat{\mathbf{y}} + x_4 a \hat{\mathbf{z}}$ | (24$d$) | O I |
| $\mathbf{B}_{46}$ | = | $-y_4 \mathbf{a}_1 + \left(\frac{1}{2} + z_4\right) \mathbf{a}_2 + \left(\frac{1}{2} - x_4\right) \mathbf{a}_3$ | = | $-y_4 a \hat{\mathbf{x}} + \left(\frac{1}{2} + z_4\right) a \hat{\mathbf{y}} + \left(\frac{1}{2} - x_4\right) a \hat{\mathbf{z}}$ | (24$d$) | O I |
| $\mathbf{B}_{47}$ | = | $\left(\frac{1}{2} + y_4\right) \mathbf{a}_1 + \left(\frac{1}{2} - z_4\right) \mathbf{a}_2 - x_4 \mathbf{a}_3$ | = | $\left(\frac{1}{2} + y_4\right) a \hat{\mathbf{x}} + \left(\frac{1}{2} - z_4\right) a \hat{\mathbf{y}} - x_4 a \hat{\mathbf{z}}$ | (24$d$) | O I |
| $\mathbf{B}_{48}$ | = | $\left(\frac{1}{2} - y_4\right) \mathbf{a}_1 - z_4 \mathbf{a}_2 + \left(\frac{1}{2} + x_4\right) \mathbf{a}_3$ | = | $\left(\frac{1}{2} - y_4\right) a \hat{\mathbf{x}} - z_4 a \hat{\mathbf{y}} + \left(\frac{1}{2} + x_4\right) a \hat{\mathbf{z}}$ | (24$d$) | O I |
| $\mathbf{B}_{49}$ | = | $-x_4 \mathbf{a}_1 - y_4 \mathbf{a}_2 - z_4 \mathbf{a}_3$ | = | $-x_4 a \hat{\mathbf{x}} - y_4 a \hat{\mathbf{y}} - z_4 a \hat{\mathbf{z}}$ | (24$d$) | O I |
| $\mathbf{B}_{50}$ | = | $\left(\frac{1}{2} + x_4\right) \mathbf{a}_1 + y_4 \mathbf{a}_2 + \left(\frac{1}{2} - z_4\right) \mathbf{a}_3$ | = | $\left(\frac{1}{2} + x_4\right) a \hat{\mathbf{x}} + y_4 a \hat{\mathbf{y}} + \left(\frac{1}{2} - z_4\right) a \hat{\mathbf{z}}$ | (24$d$) | O I |
| $\mathbf{B}_{51}$ | = | $x_4 \mathbf{a}_1 + \left(\frac{1}{2} - y_4\right) \mathbf{a}_2 + \left(\frac{1}{2} + z_4\right) \mathbf{a}_3$ | = | $x_4 a \hat{\mathbf{x}} + \left(\frac{1}{2} - y_4\right) a \hat{\mathbf{y}} + \left(\frac{1}{2} + z_4\right) a \hat{\mathbf{z}}$ | (24$d$) | O I |
| $\mathbf{B}_{52}$ | = | $\left(\frac{1}{2} - x_4\right) \mathbf{a}_1 + \left(\frac{1}{2} + y_4\right) \mathbf{a}_2 + z_4 \mathbf{a}_3$ | = | $\left(\frac{1}{2} - x_4\right) a \hat{\mathbf{x}} + \left(\frac{1}{2} + y_4\right) a \hat{\mathbf{y}} + z_4 a \hat{\mathbf{z}}$ | (24$d$) | O I |
| $\mathbf{B}_{53}$ | = | $-z_4 \mathbf{a}_1 - x_4 \mathbf{a}_2 - y_4 \mathbf{a}_3$ | = | $-z_4 a \hat{\mathbf{x}} - x_4 a \hat{\mathbf{y}} - y_4 a \hat{\mathbf{z}}$ | (24$d$) | O I |
| $\mathbf{B}_{54}$ | = | $\left(\frac{1}{2} - z_4\right) \mathbf{a}_1 + \left(\frac{1}{2} + x_4\right) \mathbf{a}_2 + y_4 \mathbf{a}_3$ | = | $\left(\frac{1}{2} - z_4\right) a \hat{\mathbf{x}} + \left(\frac{1}{2} + x_4\right) a \hat{\mathbf{y}} + y_4 a \hat{\mathbf{z}}$ | (24$d$) | O I |
| $\mathbf{B}_{55}$ | = | $\left(\frac{1}{2} + z_4\right) \mathbf{a}_1 + x_4 \mathbf{a}_2 + \left(\frac{1}{2} - y_4\right) \mathbf{a}_3$ | = | $\left(\frac{1}{2} + z_4\right) a \hat{\mathbf{x}} + x_4 a \hat{\mathbf{y}} + \left(\frac{1}{2} - y_4\right) a \hat{\mathbf{z}}$ | (24$d$) | O I |
| $\mathbf{B}_{56}$ | = | $z_4 \mathbf{a}_1 + \left(\frac{1}{2} - x_4\right) \mathbf{a}_2 + \left(\frac{1}{2} + y_4\right) \mathbf{a}_3$ | = | $z_4 a \hat{\mathbf{x}} + \left(\frac{1}{2} - x_4\right) a \hat{\mathbf{y}} + \left(\frac{1}{2} + y_4\right) a \hat{\mathbf{z}}$ | (24$d$) | O I |
| $\mathbf{B}_{57}$ | = | $-y_4 \mathbf{a}_1 - z_4 \mathbf{a}_2 - x_4 \mathbf{a}_3$ | = | $-y_4 a \hat{\mathbf{x}} - z_4 a \hat{\mathbf{y}} - x_4 a \hat{\mathbf{z}}$ | (24$d$) | O I |
| $\mathbf{B}_{58}$ | = | $y_4 \mathbf{a}_1 + \left(\frac{1}{2} - z_4\right) \mathbf{a}_2 + \left(\frac{1}{2} + x_4\right) \mathbf{a}_3$ | = | $y_4 a \hat{\mathbf{x}} + \left(\frac{1}{2} - z_4\right) a \hat{\mathbf{y}} + \left(\frac{1}{2} + x_4\right) a \hat{\mathbf{z}}$ | (24$d$) | O I |
| $\mathbf{B}_{59}$ | = | $\left(\frac{1}{2} - y_4\right) \mathbf{a}_1 + \left(\frac{1}{2} + z_4\right) \mathbf{a}_2 + x_4 \mathbf{a}_3$ | = | $\left(\frac{1}{2} - y_4\right) a \hat{\mathbf{x}} + \left(\frac{1}{2} + z_4\right) a \hat{\mathbf{y}} + x_4 a \hat{\mathbf{z}}$ | (24$d$) | O I |
| $\mathbf{B}_{60}$ | = | $\left(\frac{1}{2} + y_4\right) \mathbf{a}_1 + z_4 \mathbf{a}_2 + \left(\frac{1}{2} - x_4\right) \mathbf{a}_3$ | = | $\left(\frac{1}{2} + y_4\right) a \hat{\mathbf{x}} + z_4 a \hat{\mathbf{y}} + \left(\frac{1}{2} - x_4\right) a \hat{\mathbf{z}}$ | (24$d$) | O I |
| $\mathbf{B}_{61}$ | = | $x_5 \mathbf{a}_1 + y_5 \mathbf{a}_2 + z_5 \mathbf{a}_3$ | = | $x_5 a \hat{\mathbf{x}} + y_5 a \hat{\mathbf{y}} + z_5 a \hat{\mathbf{z}}$ | (24$d$) | O II |
| $\mathbf{B}_{62}$ | = | $\left(\frac{1}{2} - x_5\right) \mathbf{a}_1 - y_5 \mathbf{a}_2 + \left(\frac{1}{2} + z_5\right) \mathbf{a}_3$ | = | $\left(\frac{1}{2} - x_5\right) a \hat{\mathbf{x}} - y_5 a \hat{\mathbf{y}} + \left(\frac{1}{2} + z_5\right) a \hat{\mathbf{z}}$ | (24$d$) | O II |
| $\mathbf{B}_{63}$ | = | $-x_5 \mathbf{a}_1 + \left(\frac{1}{2} + y_5\right) \mathbf{a}_2 + \left(\frac{1}{2} - z_5\right) \mathbf{a}_3$ | = | $-x_5 a \hat{\mathbf{x}} + \left(\frac{1}{2} + y_5\right) a \hat{\mathbf{y}} + \left(\frac{1}{2} - z_5\right) a \hat{\mathbf{z}}$ | (24$d$) | O II |
| $\mathbf{B}_{64}$ | = | $\left(\frac{1}{2} + x_5\right) \mathbf{a}_1 + \left(\frac{1}{2} - y_5\right) \mathbf{a}_2 - z_5 \mathbf{a}_3$ | = | $\left(\frac{1}{2} + x_5\right) a \hat{\mathbf{x}} + \left(\frac{1}{2} - y_5\right) a \hat{\mathbf{y}} - z_5 a \hat{\mathbf{z}}$ | (24$d$) | O II |
| $\mathbf{B}_{65}$ | = | $z_5 \mathbf{a}_1 + x_5 \mathbf{a}_2 + y_5 \mathbf{a}_3$ | = | $z_5 a \hat{\mathbf{x}} + x_5 a \hat{\mathbf{y}} + y_5 a \hat{\mathbf{z}}$ | (24$d$) | O II |
| $\mathbf{B}_{66}$ | = | $\left(\frac{1}{2} + z_5\right) \mathbf{a}_1 + \left(\frac{1}{2} - x_5\right) \mathbf{a}_2 - y_5 \mathbf{a}_3$ | = | $\left(\frac{1}{2} + z_5\right) a \hat{\mathbf{x}} + \left(\frac{1}{2} - x_5\right) a \hat{\mathbf{y}} - y_5 a \hat{\mathbf{z}}$ | (24$d$) | O II |
| $\mathbf{B}_{67}$ | = | $\left(\frac{1}{2} - z_5\right) \mathbf{a}_1 - x_5 \mathbf{a}_2 + \left(\frac{1}{2} + y_5\right) \mathbf{a}_3$ | = | $\left(\frac{1}{2} - z_5\right) a \hat{\mathbf{x}} - x_5 a \hat{\mathbf{y}} + \left(\frac{1}{2} + y_5\right) a \hat{\mathbf{z}}$ | (24$d$) | O II |
| $\mathbf{B}_{68}$ | = | $-z_5 \mathbf{a}_1 + \left(\frac{1}{2} + x_5\right) \mathbf{a}_2 + \left(\frac{1}{2} - y_5\right) \mathbf{a}_3$ | = | $-z_5 a \hat{\mathbf{x}} + \left(\frac{1}{2} + x_5\right) a \hat{\mathbf{y}} + \left(\frac{1}{2} - y_5\right) a \hat{\mathbf{z}}$ | (24$d$) | O II |
| $\mathbf{B}_{69}$ | = | $y_5 \mathbf{a}_1 + z_5 \mathbf{a}_2 + x_5 \mathbf{a}_3$ | = | $y_5 a \hat{\mathbf{x}} + z_5 a \hat{\mathbf{y}} + x_5 a \hat{\mathbf{z}}$ | (24$d$) | O II |
| $\mathbf{B}_{70}$ | = | $-y_5 \mathbf{a}_1 + \left(\frac{1}{2} + z_5\right) \mathbf{a}_2 + \left(\frac{1}{2} - x_5\right) \mathbf{a}_3$ | = | $-y_5 a \hat{\mathbf{x}} + \left(\frac{1}{2} + z_5\right) a \hat{\mathbf{y}} + \left(\frac{1}{2} - x_5\right) a \hat{\mathbf{z}}$ | (24$d$) | O II |
| $\mathbf{B}_{71}$ | = | $\left(\frac{1}{2}+y_5\right)\mathbf{a}_1+\left(\frac{1}{2}-z_5\right)\mathbf{a}_2-x_5\,\mathbf{a}_3$ | = | $\left(\frac{1}{2}+y_5\right)a\,\hat{\mathbf{x}}+\left(\frac{1}{2}-z_5\right)a\,\hat{\mathbf{y}}-x_5a\,\hat{\mathbf{z}}$ | (24$d$) | O II |
| $\mathbf{B}_{72}$ | = | $\left(\frac{1}{2}-y_5\right)\mathbf{a}_1-z_5\,\mathbf{a}_2+\left(\frac{1}{2}+x_5\right)\mathbf{a}_3$ | = | $\left(\frac{1}{2}-y_5\right)a\,\hat{\mathbf{x}}-z_5a\,\hat{\mathbf{y}}+\left(\frac{1}{2}+x_5\right)a\,\hat{\mathbf{z}}$ | (24$d$) | O II |
| $\mathbf{B}_{73}$ | = | $-x_5\,\mathbf{a}_1-y_5\,\mathbf{a}_2-z_5\,\mathbf{a}_3$ | = | $-x_5a\,\hat{\mathbf{x}}-y_5a\,\hat{\mathbf{y}}-z_5a\,\hat{\mathbf{z}}$ | (24$d$) | O II |
| $\mathbf{B}_{74}$ | = | $\left(\frac{1}{2}+x_5\right)\mathbf{a}_1+y_5\,\mathbf{a}_2+\left(\frac{1}{2}-z_5\right)\mathbf{a}_3$ | = | $\left(\frac{1}{2}+x_5\right)a\,\hat{\mathbf{x}}+y_5a\,\hat{\mathbf{y}}+\left(\frac{1}{2}-z_5\right)a\,\hat{\mathbf{z}}$ | (24$d$) | O II |
| $\mathbf{B}_{75}$ | = | $x_5\,\mathbf{a}_1+\left(\frac{1}{2}-y_5\right)\mathbf{a}_2+\left(\frac{1}{2}+z_5\right)\mathbf{a}_3$ | = | $x_5a\,\hat{\mathbf{x}}+\left(\frac{1}{2}-y_5\right)a\,\hat{\mathbf{y}}+\left(\frac{1}{2}+z_5\right)a\,\hat{\mathbf{z}}$ | (24$d$) | O II |
| $\mathbf{B}_{76}$ | = | $\left(\frac{1}{2}-x_5\right)\mathbf{a}_1+\left(\frac{1}{2}+y_5\right)\mathbf{a}_2+z_5\,\mathbf{a}_3$ | = | $\left(\frac{1}{2}-x_5\right)a\,\hat{\mathbf{x}}+\left(\frac{1}{2}+y_5\right)a\,\hat{\mathbf{y}}+z_5a\,\hat{\mathbf{z}}$ | (24$d$) | O II |
| $\mathbf{B}_{77}$ | = | $-z_5\,\mathbf{a}_1-x_5\,\mathbf{a}_2-y_5\,\mathbf{a}_3$ | = | $-z_5a\,\hat{\mathbf{x}}-x_5a\,\hat{\mathbf{y}}-y_5a\,\hat{\mathbf{z}}$ | (24$d$) | O II |
| $\mathbf{B}_{78}$ | = | $\left(\frac{1}{2}-z_5\right)\mathbf{a}_1+\left(\frac{1}{2}+x_5\right)\mathbf{a}_2+y_5\,\mathbf{a}_3$ | = | $\left(\frac{1}{2}-z_5\right)a\,\hat{\mathbf{x}}+\left(\frac{1}{2}+x_5\right)a\,\hat{\mathbf{y}}+y_5a\,\hat{\mathbf{z}}$ | (24$d$) | O II |
| $\mathbf{B}_{79}$ | = | $\left(\frac{1}{2}+z_5\right)\mathbf{a}_1+x_5\,\mathbf{a}_2+\left(\frac{1}{2}-y_5\right)\mathbf{a}_3$ | = | $\left(\frac{1}{2}+z_5\right)a\,\hat{\mathbf{x}}+x_5a\,\hat{\mathbf{y}}+\left(\frac{1}{2}-y_5\right)a\,\hat{\mathbf{z}}$ | (24$d$) | O II |
| $\mathbf{B}_{80}$ | = | $z_5\,\mathbf{a}_1+\left(\frac{1}{2}-x_5\right)\mathbf{a}_2+\left(\frac{1}{2}+y_5\right)\mathbf{a}_3$ | = | $z_5a\,\hat{\mathbf{x}}+\left(\frac{1}{2}-x_5\right)a\,\hat{\mathbf{y}}+\left(\frac{1}{2}+y_5\right)a\,\hat{\mathbf{z}}$ | (24$d$) | O II |
| $\mathbf{B}_{81}$ | = | $-y_5\,\mathbf{a}_1-z_5\,\mathbf{a}_2-x_5\,\mathbf{a}_3$ | = | $-y_5a\,\hat{\mathbf{x}}-z_5a\,\hat{\mathbf{y}}-x_5a\,\hat{\mathbf{z}}$ | (24$d$) | O II |
| $\mathbf{B}_{82}$ | = | $y_5\,\mathbf{a}_1+\left(\frac{1}{2}-z_5\right)\mathbf{a}_2+\left(\frac{1}{2}+x_5\right)\mathbf{a}_3$ | = | $y_5a\,\hat{\mathbf{x}}+\left(\frac{1}{2}-z_5\right)a\,\hat{\mathbf{y}}+\left(\frac{1}{2}+x_5\right)a\,\hat{\mathbf{z}}$ | (24$d$) | O II |
| $\mathbf{B}_{83}$ | = | $\left(\frac{1}{2}-y_5\right)\mathbf{a}_1+\left(\frac{1}{2}+z_5\right)\mathbf{a}_2+x_5\,\mathbf{a}_3$ | = | $\left(\frac{1}{2}-y_5\right)a\,\hat{\mathbf{x}}+\left(\frac{1}{2}+z_5\right)a\,\hat{\mathbf{y}}+x_5a\,\hat{\mathbf{z}}$ | (24$d$) | O II |
| $\mathbf{B}_{84}$ | = | $\left(\frac{1}{2}+y_5\right)\mathbf{a}_1+z_5\,\mathbf{a}_2+\left(\frac{1}{2}-x_5\right)\mathbf{a}_3$ | = | $\left(\frac{1}{2}+y_5\right)a\,\hat{\mathbf{x}}+z_5a\,\hat{\mathbf{y}}+\left(\frac{1}{2}-x_5\right)a\,\hat{\mathbf{z}}$ | (24$d$) | O II |

**Geometry files:**

- CIF: pp. 1781
- POSCAR: pp. 1781

# $SnI_4$ ($D1_1$) Structure: A4B_cP40_205_cd_c

http://aflow.org/prototype-encyclopedia/A4B_cP40_205_cd_c

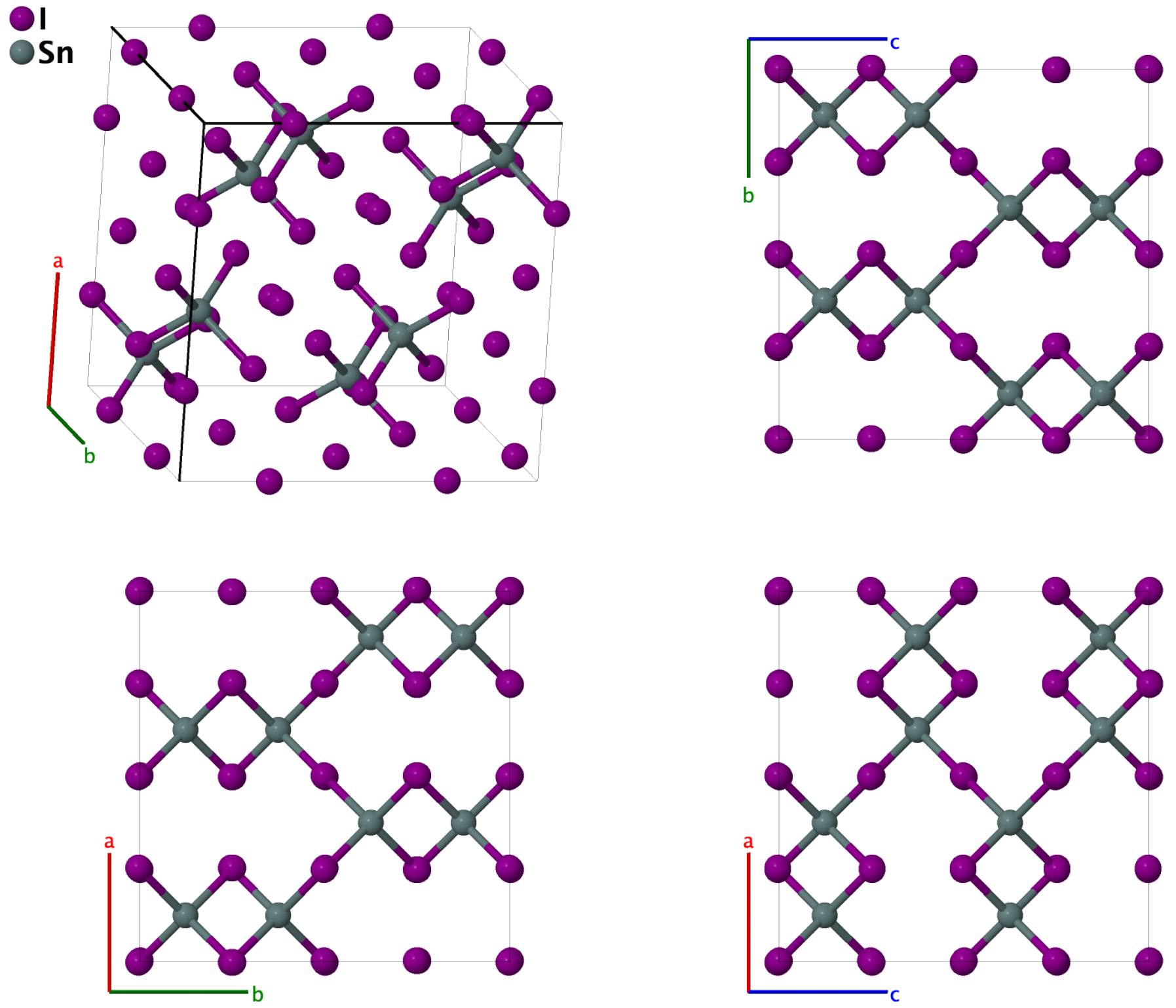

| | | |
|---|---|---|
| **Prototype** | : | $I_4Sn$ |
| **AFLOW prototype label** | : | A4B_cP40_205_cd_c |
| ***Strukturbericht* designation** | : | $D1_1$ |
| **Pearson symbol** | : | cP40 |
| **Space group number** | : | 205 |
| **Space group symbol** | : | $Pa\bar{3}$ |
| **AFLOW prototype command** | : | `aflow --proto=A4B_cP40_205_cd_c --params=`$a, x_1, x_2, x_3, y_3, z_3$ |

**Other compounds with this structure**

- $SiI_4$, $TiBr_4$, $TiI_4$, and $Ni(CO)_4$

**Simple Cubic primitive vectors:**

$$\mathbf{a}_1 = a\,\hat{\mathbf{x}}$$
$$\mathbf{a}_2 = a\,\hat{\mathbf{y}}$$
$$\mathbf{a}_3 = a\,\hat{\mathbf{z}}$$

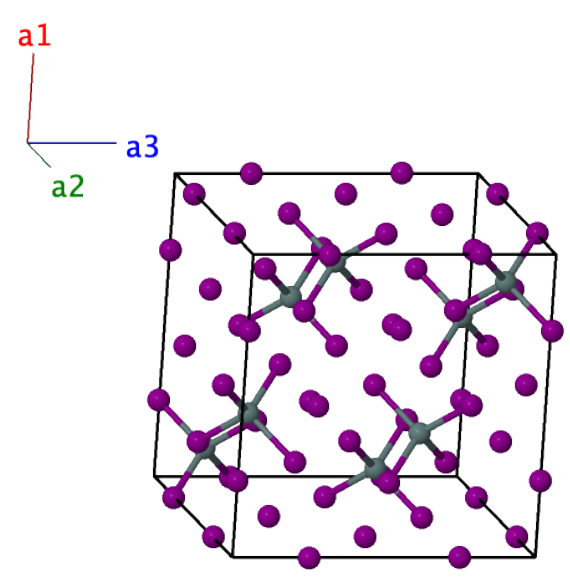

**Basis vectors:**

| | | Lattice Coordinates | | Cartesian Coordinates | Wyckoff Position | Atom Type |
|---|---|---|---|---|---|---|
| $\mathbf{B}_{1}$ | = | $x_1\,\mathbf{a}_1 + x_1\,\mathbf{a}_2 + x_1\,\mathbf{a}_3$ | = | $x_1 a\,\hat{\mathbf{x}} + x_1 a\,\hat{\mathbf{y}} + x_1 a\,\hat{\mathbf{z}}$ | (8$c$) | I I |
| $\mathbf{B}_{2}$ | = | $\left(\frac{1}{2} - x_1\right)\mathbf{a}_1 - x_1\,\mathbf{a}_2 + \left(\frac{1}{2} + x_1\right)\mathbf{a}_3$ | = | $\left(\frac{1}{2} - x_1\right)a\,\hat{\mathbf{x}} - x_1 a\,\hat{\mathbf{y}} + \left(\frac{1}{2} + x_1\right)a\,\hat{\mathbf{z}}$ | (8$c$) | I I |
| $\mathbf{B}_{3}$ | = | $-x_1\,\mathbf{a}_1 + \left(\frac{1}{2} + x_1\right)\mathbf{a}_2 + \left(\frac{1}{2} - x_1\right)\mathbf{a}_3$ | = | $-x_1 a\,\hat{\mathbf{x}} + \left(\frac{1}{2} + x_1\right)a\,\hat{\mathbf{y}} + \left(\frac{1}{2} - x_1\right)a\,\hat{\mathbf{z}}$ | (8$c$) | I I |
| $\mathbf{B}_{4}$ | = | $\left(\frac{1}{2} + x_1\right)\mathbf{a}_1 + \left(\frac{1}{2} - x_1\right)\mathbf{a}_2 - x_1\,\mathbf{a}_3$ | = | $\left(\frac{1}{2} + x_1\right)a\,\hat{\mathbf{x}} + \left(\frac{1}{2} - x_1\right)a\,\hat{\mathbf{y}} - x_1 a\,\hat{\mathbf{z}}$ | (8$c$) | I I |
| $\mathbf{B}_{5}$ | = | $-x_1\,\mathbf{a}_1 - x_1\,\mathbf{a}_2 - x_1\,\mathbf{a}_3$ | = | $-x_1 a\,\hat{\mathbf{x}} - x_1 a\,\hat{\mathbf{y}} - x_1 a\,\hat{\mathbf{z}}$ | (8$c$) | I I |
| $\mathbf{B}_{6}$ | = | $\left(\frac{1}{2} + x_1\right)\mathbf{a}_1 + x_1\,\mathbf{a}_2 + \left(\frac{1}{2} - x_1\right)\mathbf{a}_3$ | = | $\left(\frac{1}{2} + x_1\right)a\,\hat{\mathbf{x}} + x_1 a\,\hat{\mathbf{y}} + \left(\frac{1}{2} - x_1\right)a\,\hat{\mathbf{z}}$ | (8$c$) | I I |
| $\mathbf{B}_{7}$ | = | $x_1\,\mathbf{a}_1 + \left(\frac{1}{2} - x_1\right)\mathbf{a}_2 + \left(\frac{1}{2} + x_1\right)\mathbf{a}_3$ | = | $x_1 a\,\hat{\mathbf{x}} + \left(\frac{1}{2} - x_1\right)a\,\hat{\mathbf{y}} + \left(\frac{1}{2} + x_1\right)a\,\hat{\mathbf{z}}$ | (8$c$) | I I |
| $\mathbf{B}_{8}$ | = | $\left(\frac{1}{2} - x_1\right)\mathbf{a}_1 + \left(\frac{1}{2} + x_1\right)\mathbf{a}_2 + x_1\,\mathbf{a}_3$ | = | $\left(\frac{1}{2} - x_1\right)a\,\hat{\mathbf{x}} + \left(\frac{1}{2} + x_1\right)a\,\hat{\mathbf{y}} + x_1 a\,\hat{\mathbf{z}}$ | (8$c$) | I I |
| $\mathbf{B}_{9}$ | = | $x_2\,\mathbf{a}_1 + x_2\,\mathbf{a}_2 + x_2\,\mathbf{a}_3$ | = | $x_2 a\,\hat{\mathbf{x}} + x_2 a\,\hat{\mathbf{y}} + x_2 a\,\hat{\mathbf{z}}$ | (8$c$) | Sn |
| $\mathbf{B}_{10}$ | = | $\left(\frac{1}{2} - x_2\right)\mathbf{a}_1 - x_2\,\mathbf{a}_2 + \left(\frac{1}{2} + x_2\right)\mathbf{a}_3$ | = | $\left(\frac{1}{2} - x_2\right)a\,\hat{\mathbf{x}} - x_2 a\,\hat{\mathbf{y}} + \left(\frac{1}{2} + x_2\right)a\,\hat{\mathbf{z}}$ | (8$c$) | Sn |
| $\mathbf{B}_{11}$ | = | $-x_2\,\mathbf{a}_1 + \left(\frac{1}{2} + x_2\right)\mathbf{a}_2 + \left(\frac{1}{2} - x_2\right)\mathbf{a}_3$ | = | $-x_2 a\,\hat{\mathbf{x}} + \left(\frac{1}{2} + x_2\right)a\,\hat{\mathbf{y}} + \left(\frac{1}{2} - x_2\right)a\,\hat{\mathbf{z}}$ | (8$c$) | Sn |
| $\mathbf{B}_{12}$ | = | $\left(\frac{1}{2} + x_2\right)\mathbf{a}_1 + \left(\frac{1}{2} - x_2\right)\mathbf{a}_2 - x_2\,\mathbf{a}_3$ | = | $\left(\frac{1}{2} + x_2\right)a\,\hat{\mathbf{x}} + \left(\frac{1}{2} - x_2\right)a\,\hat{\mathbf{y}} - x_2 a\,\hat{\mathbf{z}}$ | (8$c$) | Sn |
| $\mathbf{B}_{13}$ | = | $-x_2\,\mathbf{a}_1 - x_2\,\mathbf{a}_2 - x_2\,\mathbf{a}_3$ | = | $-x_2 a\,\hat{\mathbf{x}} - x_2 a\,\hat{\mathbf{y}} - x_2 a\,\hat{\mathbf{z}}$ | (8$c$) | Sn |
| $\mathbf{B}_{14}$ | = | $\left(\frac{1}{2} + x_2\right)\mathbf{a}_1 + x_2\,\mathbf{a}_2 + \left(\frac{1}{2} - x_2\right)\mathbf{a}_3$ | = | $\left(\frac{1}{2} + x_2\right)a\,\hat{\mathbf{x}} + x_2 a\,\hat{\mathbf{y}} + \left(\frac{1}{2} - x_2\right)a\,\hat{\mathbf{z}}$ | (8$c$) | Sn |
| $\mathbf{B}_{15}$ | = | $x_2\,\mathbf{a}_1 + \left(\frac{1}{2} - x_2\right)\mathbf{a}_2 + \left(\frac{1}{2} + x_2\right)\mathbf{a}_3$ | = | $x_2 a\,\hat{\mathbf{x}} + \left(\frac{1}{2} - x_2\right)a\,\hat{\mathbf{y}} + \left(\frac{1}{2} + x_2\right)a\,\hat{\mathbf{z}}$ | (8$c$) | Sn |
| $\mathbf{B}_{16}$ | = | $\left(\frac{1}{2} - x_2\right)\mathbf{a}_1 + \left(\frac{1}{2} + x_2\right)\mathbf{a}_2 + x_2\,\mathbf{a}_3$ | = | $\left(\frac{1}{2} - x_2\right)a\,\hat{\mathbf{x}} + \left(\frac{1}{2} + x_2\right)a\,\hat{\mathbf{y}} + x_2 a\,\hat{\mathbf{z}}$ | (8$c$) | Sn |
| $\mathbf{B}_{17}$ | = | $x_3\,\mathbf{a}_1 + y_3\,\mathbf{a}_2 + z_3\,\mathbf{a}_3$ | = | $x_3 a\,\hat{\mathbf{x}} + y_3 a\,\hat{\mathbf{y}} + z_3 a\,\hat{\mathbf{z}}$ | (24$d$) | I II |
| $\mathbf{B}_{18}$ | = | $\left(\frac{1}{2} - x_3\right)\mathbf{a}_1 - y_3\,\mathbf{a}_2 + \left(\frac{1}{2} + z_3\right)\mathbf{a}_3$ | = | $\left(\frac{1}{2} - x_3\right)a\,\hat{\mathbf{x}} - y_3 a\,\hat{\mathbf{y}} + \left(\frac{1}{2} + z_3\right)a\,\hat{\mathbf{z}}$ | (24$d$) | I II |
| $\mathbf{B}_{19}$ | = | $-x_3\,\mathbf{a}_1 + \left(\frac{1}{2} + y_3\right)\mathbf{a}_2 + \left(\frac{1}{2} - z_3\right)\mathbf{a}_3$ | = | $-x_3 a\,\hat{\mathbf{x}} + \left(\frac{1}{2} + y_3\right)a\,\hat{\mathbf{y}} + \left(\frac{1}{2} - z_3\right)a\,\hat{\mathbf{z}}$ | (24$d$) | I II |
| $\mathbf{B}_{20}$ | = | $\left(\frac{1}{2} + x_3\right)\mathbf{a}_1 + \left(\frac{1}{2} - y_3\right)\mathbf{a}_2 - z_3\,\mathbf{a}_3$ | = | $\left(\frac{1}{2} + x_3\right)a\,\hat{\mathbf{x}} + \left(\frac{1}{2} - y_3\right)a\,\hat{\mathbf{y}} - z_3 a\,\hat{\mathbf{z}}$ | (24$d$) | I II |
| $\mathbf{B}_{21}$ | = | $z_3\,\mathbf{a}_1 + x_3\,\mathbf{a}_2 + y_3\,\mathbf{a}_3$ | = | $z_3 a\,\hat{\mathbf{x}} + x_3 a\,\hat{\mathbf{y}} + y_3 a\,\hat{\mathbf{z}}$ | (24$d$) | I II |
| $\mathbf{B}_{22}$ | = | $\left(\frac{1}{2} + z_3\right)\mathbf{a}_1 + \left(\frac{1}{2} - x_3\right)\mathbf{a}_2 - y_3\,\mathbf{a}_3$ | = | $\left(\frac{1}{2} + z_3\right)a\,\hat{\mathbf{x}} + \left(\frac{1}{2} - x_3\right)a\,\hat{\mathbf{y}} - y_3 a\,\hat{\mathbf{z}}$ | (24$d$) | I II |
| $\mathbf{B}_{23}$ | = | $\left(\frac{1}{2} - z_3\right)\mathbf{a}_1 - x_3\,\mathbf{a}_2 + \left(\frac{1}{2} + y_3\right)\mathbf{a}_3$ | = | $\left(\frac{1}{2} - z_3\right)a\,\hat{\mathbf{x}} - x_3 a\,\hat{\mathbf{y}} + \left(\frac{1}{2} + y_3\right)a\,\hat{\mathbf{z}}$ | (24$d$) | I II |
| $\mathbf{B}_{24}$ | = | $-z_3\,\mathbf{a}_1 + \left(\frac{1}{2} + x_3\right)\mathbf{a}_2 + \left(\frac{1}{2} - y_3\right)\mathbf{a}_3$ | = | $-z_3 a\,\hat{\mathbf{x}} + \left(\frac{1}{2} + x_3\right)a\,\hat{\mathbf{y}} + \left(\frac{1}{2} - y_3\right)a\,\hat{\mathbf{z}}$ | (24$d$) | I II |
| $\mathbf{B}_{25}$ | = | $y_3\,\mathbf{a}_1 + z_3\,\mathbf{a}_2 + x_3\,\mathbf{a}_3$ | = | $y_3 a\,\hat{\mathbf{x}} + z_3 a\,\hat{\mathbf{y}} + x_3 a\,\hat{\mathbf{z}}$ | (24$d$) | I II |
| $\mathbf{B}_{26}$ | = | $-y_3\,\mathbf{a}_1 + \left(\frac{1}{2} + z_3\right)\mathbf{a}_2 + \left(\frac{1}{2} - x_3\right)\mathbf{a}_3$ | = | $-y_3 a\,\hat{\mathbf{x}} + \left(\frac{1}{2} + z_3\right)a\,\hat{\mathbf{y}} + \left(\frac{1}{2} - x_3\right)a\,\hat{\mathbf{z}}$ | (24$d$) | I II |
| $\mathbf{B}_{27}$ | = | $\left(\frac{1}{2} + y_3\right)\mathbf{a}_1 + \left(\frac{1}{2} - z_3\right)\mathbf{a}_2 - x_3\,\mathbf{a}_3$ | = | $\left(\frac{1}{2} + y_3\right)a\,\hat{\mathbf{x}} + \left(\frac{1}{2} - z_3\right)a\,\hat{\mathbf{y}} - x_3 a\,\hat{\mathbf{z}}$ | (24$d$) | I II |
| $\mathbf{B}_{28}$ | = | $\left(\frac{1}{2} - y_3\right)\mathbf{a}_1 - z_3\,\mathbf{a}_2 + \left(\frac{1}{2} + x_3\right)\mathbf{a}_3$ | = | $\left(\frac{1}{2} - y_3\right)a\,\hat{\mathbf{x}} - z_3 a\,\hat{\mathbf{y}} + \left(\frac{1}{2} + x_3\right)a\,\hat{\mathbf{z}}$ | (24$d$) | I II |
| $\mathbf{B}_{29}$ | = | $-x_3\,\mathbf{a}_1 - y_3\,\mathbf{a}_2 - z_3\,\mathbf{a}_3$ | = | $-x_3 a\,\hat{\mathbf{x}} - y_3 a\,\hat{\mathbf{y}} - z_3 a\,\hat{\mathbf{z}}$ | (24$d$) | I II |
| $\mathbf{B}_{30}$ | = | $\left(\frac{1}{2} + x_3\right)\mathbf{a}_1 + y_3\,\mathbf{a}_2 + \left(\frac{1}{2} - z_3\right)\mathbf{a}_3$ | = | $\left(\frac{1}{2} + x_3\right)a\,\hat{\mathbf{x}} + y_3 a\,\hat{\mathbf{y}} + \left(\frac{1}{2} - z_3\right)a\,\hat{\mathbf{z}}$ | (24$d$) | I II |
| $\mathbf{B}_{31}$ | = | $x_3\,\mathbf{a}_1 + \left(\frac{1}{2} - y_3\right)\mathbf{a}_2 + \left(\frac{1}{2} + z_3\right)\mathbf{a}_3$ | = | $x_3 a\,\hat{\mathbf{x}} + \left(\frac{1}{2} - y_3\right)a\,\hat{\mathbf{y}} + \left(\frac{1}{2} + z_3\right)a\,\hat{\mathbf{z}}$ | (24$d$) | I II |
| $\mathbf{B}_{32}$ | = | $\left(\frac{1}{2} - x_3\right)\mathbf{a}_1 + \left(\frac{1}{2} + y_3\right)\mathbf{a}_2 + z_3\,\mathbf{a}_3$ | = | $\left(\frac{1}{2} - x_3\right)a\,\hat{\mathbf{x}} + \left(\frac{1}{2} + y_3\right)a\,\hat{\mathbf{y}} + z_3 a\,\hat{\mathbf{z}}$ | (24$d$) | I II |
| $\mathbf{B}_{33}$ | = | $-z_3\,\mathbf{a}_1 - x_3\,\mathbf{a}_2 - y_3\,\mathbf{a}_3$ | = | $-z_3 a\,\hat{\mathbf{x}} - x_3 a\,\hat{\mathbf{y}} - y_3 a\,\hat{\mathbf{z}}$ | (24$d$) | I II |
| $\mathbf{B}_{34}$ | = | $\left(\frac{1}{2} - z_3\right)\mathbf{a}_1 + \left(\frac{1}{2} + x_3\right)\mathbf{a}_2 + y_3\,\mathbf{a}_3$ | = | $\left(\frac{1}{2} - z_3\right)a\,\hat{\mathbf{x}} + \left(\frac{1}{2} + x_3\right)a\,\hat{\mathbf{y}} + y_3 a\,\hat{\mathbf{z}}$ | (24$d$) | I II |
| $\mathbf{B}_{35}$ | = | $\left(\frac{1}{2} + z_3\right)\mathbf{a}_1 + x_3\,\mathbf{a}_2 + \left(\frac{1}{2} - y_3\right)\mathbf{a}_3$ | = | $\left(\frac{1}{2} + z_3\right)a\,\hat{\mathbf{x}} + x_3 a\,\hat{\mathbf{y}} + \left(\frac{1}{2} - y_3\right)a\,\hat{\mathbf{z}}$ | (24$d$) | I II |
| $\mathbf{B}_{36}$ | = | $z_3\,\mathbf{a}_1 + \left(\frac{1}{2} - x_3\right)\mathbf{a}_2 + \left(\frac{1}{2} + y_3\right)\mathbf{a}_3$ | = | $z_3 a\,\hat{\mathbf{x}} + \left(\frac{1}{2} - x_3\right) a\,\hat{\mathbf{y}} + \left(\frac{1}{2} + y_3\right) a\,\hat{\mathbf{z}}$ | (24$d$) | I II |
| $\mathbf{B}_{37}$ | = | $-y_3\,\mathbf{a}_1 - z_3\,\mathbf{a}_2 - x_3\,\mathbf{a}_3$ | = | $-y_3 a\,\hat{\mathbf{x}} - z_3 a\,\hat{\mathbf{y}} - x_3 a\,\hat{\mathbf{z}}$ | (24$d$) | I II |
| $\mathbf{B}_{38}$ | = | $y_3\,\mathbf{a}_1 + \left(\frac{1}{2} - z_3\right)\mathbf{a}_2 + \left(\frac{1}{2} + x_3\right)\mathbf{a}_3$ | = | $y_3 a\,\hat{\mathbf{x}} + \left(\frac{1}{2} - z_3\right) a\,\hat{\mathbf{y}} + \left(\frac{1}{2} + x_3\right) a\,\hat{\mathbf{z}}$ | (24$d$) | I II |
| $\mathbf{B}_{39}$ | = | $\left(\frac{1}{2} - y_3\right)\mathbf{a}_1 + \left(\frac{1}{2} + z_3\right)\mathbf{a}_2 + x_3\,\mathbf{a}_3$ | = | $\left(\frac{1}{2} - y_3\right) a\,\hat{\mathbf{x}} + \left(\frac{1}{2} + z_3\right) a\,\hat{\mathbf{y}} + x_3 a\,\hat{\mathbf{z}}$ | (24$d$) | I II |
| $\mathbf{B}_{40}$ | = | $\left(\frac{1}{2} + y_3\right)\mathbf{a}_1 + z_3\,\mathbf{a}_2 + \left(\frac{1}{2} - x_3\right)\mathbf{a}_3$ | = | $\left(\frac{1}{2} + y_3\right) a\,\hat{\mathbf{x}} + z_3 a\,\hat{\mathbf{y}} + \left(\frac{1}{2} - x_3\right) a\,\hat{\mathbf{z}}$ | (24$d$) | I II |

**Found in:**

**Geometry files:**

- CIF: pp. 1782
- POSCAR: pp. 1782

# $NaSbF_6$ Structure: A6BC_cP32_205_d_b_a

http://aflow.org/prototype-encyclopedia/A6BC_cP32_205_d_b_a

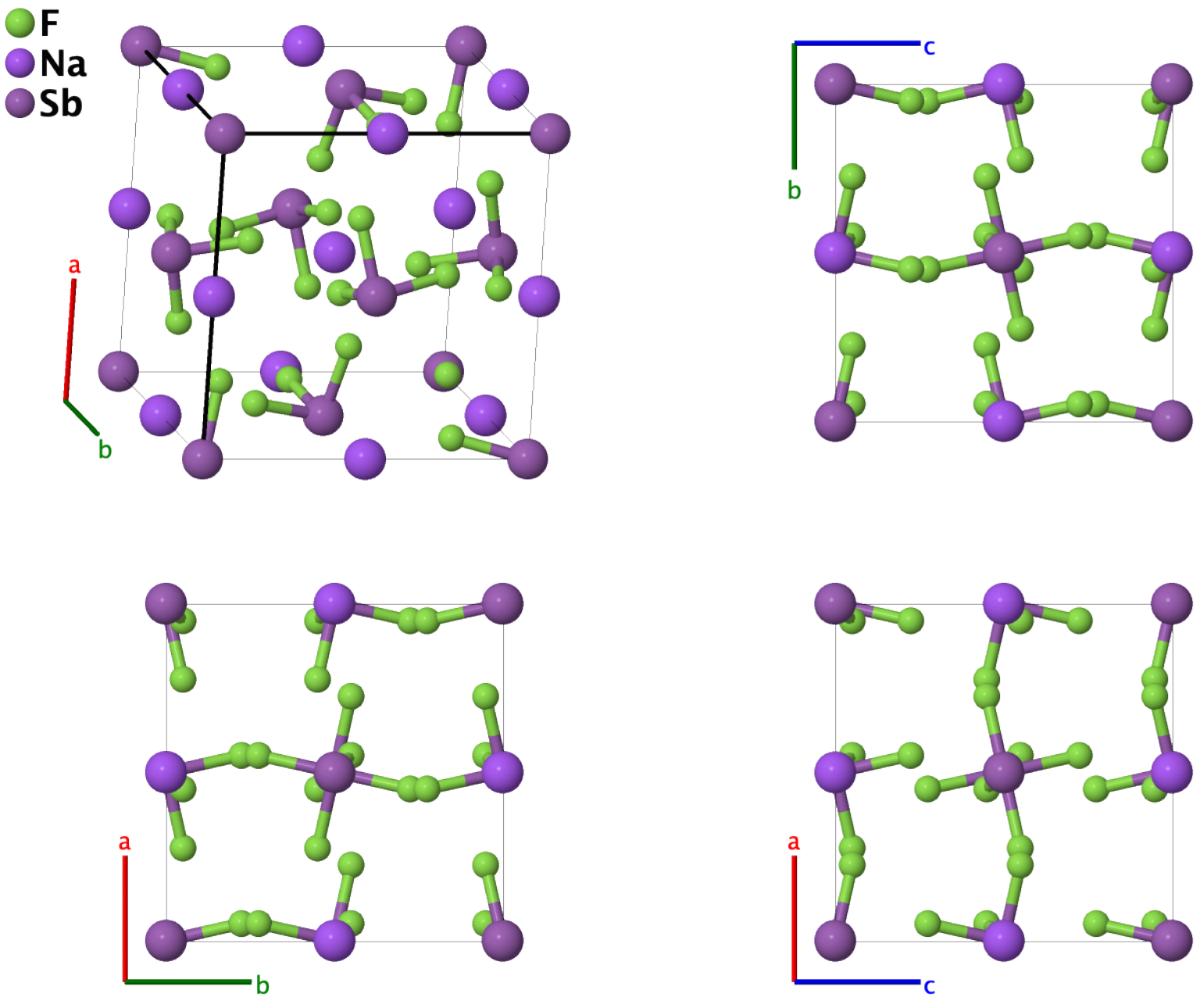

| | | |
|---|---|---|
| **Prototype** | : | $F_6NaSb$ |
| **AFLOW prototype label** | : | A6BC_cP32_205_d_b_a |
| ***Strukturbericht* designation** | : | None |
| **Pearson symbol** | : | cP32 |
| **Space group number** | : | 205 |
| **Space group symbol** | : | $Pa\bar{3}$ |
| **AFLOW prototype command** | : | `aflow --proto=A6BC_cP32_205_d_b_a`<br>`--params=`$a, x_3, y_3, z_3$ |

- This is the pure fluorine end-point of the $NaSbF_{6-x}(OH)_x$ compounds. Although the replacement of fluoride by OH does not affect the shape of the Sb-(F,OH)$_6$ ions, it has a profound effect on the structure, as can be seen by looking at and $NaSb(OH)_6$ ($J1_{11}$) and $NaSbF_4(OH)_2$ ($J1_{12}$).

**Simple Cubic primitive vectors:**

$$\mathbf{a}_1 = a\,\hat{\mathbf{x}}$$
$$\mathbf{a}_2 = a\,\hat{\mathbf{y}}$$
$$\mathbf{a}_3 = a\,\hat{\mathbf{z}}$$

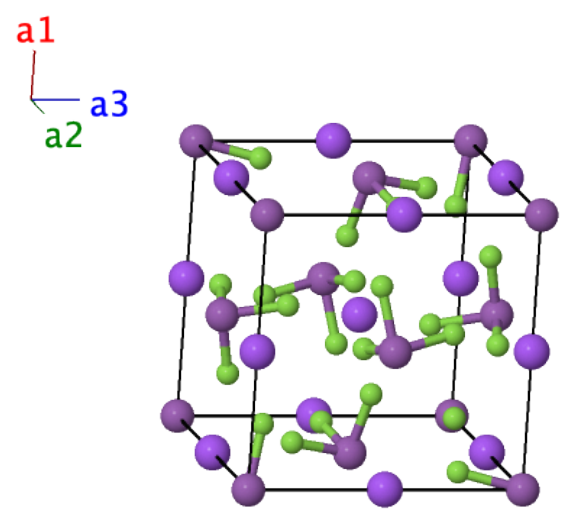

**Basis vectors:**

| | | Lattice Coordinates | | Cartesian Coordinates | Wyckoff Position | Atom Type |
|---|---|---|---|---|---|---|
| $\mathbf{B}_1$ | = | $0\,\mathbf{a}_1 + 0\,\mathbf{a}_2 + 0\,\mathbf{a}_3$ | = | $0\,\hat{\mathbf{x}} + 0\,\hat{\mathbf{y}} + 0\,\hat{\mathbf{z}}$ | (4$a$) | Sb |
| $\mathbf{B}_2$ | = | $\frac{1}{2}\,\mathbf{a}_1 + \frac{1}{2}\,\mathbf{a}_3$ | = | $\frac{1}{2}a\,\hat{\mathbf{x}} + \frac{1}{2}a\,\hat{\mathbf{z}}$ | (4$a$) | Sb |
| $\mathbf{B}_3$ | = | $\frac{1}{2}\,\mathbf{a}_2 + \frac{1}{2}\,\mathbf{a}_3$ | = | $\frac{1}{2}a\,\hat{\mathbf{y}} + \frac{1}{2}a\,\hat{\mathbf{z}}$ | (4$a$) | Sb |
| $\mathbf{B}_4$ | = | $\frac{1}{2}\,\mathbf{a}_1 + \frac{1}{2}\,\mathbf{a}_2$ | = | $\frac{1}{2}a\,\hat{\mathbf{x}} + \frac{1}{2}a\,\hat{\mathbf{y}}$ | (4$a$) | Sb |
| $\mathbf{B}_5$ | = | $\frac{1}{2}\,\mathbf{a}_1 + \frac{1}{2}\,\mathbf{a}_2 + \frac{1}{2}\,\mathbf{a}_3$ | = | $\frac{1}{2}a\,\hat{\mathbf{x}} + \frac{1}{2}a\,\hat{\mathbf{y}} + \frac{1}{2}a\,\hat{\mathbf{z}}$ | (4$b$) | Na |
| $\mathbf{B}_6$ | = | $\frac{1}{2}\,\mathbf{a}_2$ | = | $\frac{1}{2}a\,\hat{\mathbf{y}}$ | (4$b$) | Na |
| $\mathbf{B}_7$ | = | $\frac{1}{2}\,\mathbf{a}_1$ | = | $\frac{1}{2}a\,\hat{\mathbf{x}}$ | (4$b$) | Na |
| $\mathbf{B}_8$ | = | $\frac{1}{2}\,\mathbf{a}_3$ | = | $\frac{1}{2}a\,\hat{\mathbf{z}}$ | (4$b$) | Na |
| $\mathbf{B}_9$ | = | $x_3\,\mathbf{a}_1 + y_3\,\mathbf{a}_2 + z_3\,\mathbf{a}_3$ | = | $x_3a\,\hat{\mathbf{x}} + y_3a\,\hat{\mathbf{y}} + z_3a\,\hat{\mathbf{z}}$ | (24$d$) | F |
| $\mathbf{B}_{10}$ | = | $\left(\frac{1}{2} - x_3\right)\,\mathbf{a}_1 - y_3\,\mathbf{a}_2 + \left(\frac{1}{2} + z_3\right)\,\mathbf{a}_3$ | = | $\left(\frac{1}{2} - x_3\right)a\,\hat{\mathbf{x}} - y_3a\,\hat{\mathbf{y}} + \left(\frac{1}{2} + z_3\right)a\,\hat{\mathbf{z}}$ | (24$d$) | F |
| $\mathbf{B}_{11}$ | = | $-x_3\,\mathbf{a}_1 + \left(\frac{1}{2} + y_3\right)\,\mathbf{a}_2 + \left(\frac{1}{2} - z_3\right)\,\mathbf{a}_3$ | = | $-x_3a\,\hat{\mathbf{x}} + \left(\frac{1}{2} + y_3\right)a\,\hat{\mathbf{y}} + \left(\frac{1}{2} - z_3\right)a\,\hat{\mathbf{z}}$ | (24$d$) | F |
| $\mathbf{B}_{12}$ | = | $\left(\frac{1}{2} + x_3\right)\,\mathbf{a}_1 + \left(\frac{1}{2} - y_3\right)\,\mathbf{a}_2 - z_3\,\mathbf{a}_3$ | = | $\left(\frac{1}{2} + x_3\right)a\,\hat{\mathbf{x}} + \left(\frac{1}{2} - y_3\right)a\,\hat{\mathbf{y}} - z_3a\,\hat{\mathbf{z}}$ | (24$d$) | F |
| $\mathbf{B}_{13}$ | = | $z_3\,\mathbf{a}_1 + x_3\,\mathbf{a}_2 + y_3\,\mathbf{a}_3$ | = | $z_3a\,\hat{\mathbf{x}} + x_3a\,\hat{\mathbf{y}} + y_3a\,\hat{\mathbf{z}}$ | (24$d$) | F |
| $\mathbf{B}_{14}$ | = | $\left(\frac{1}{2} + z_3\right)\,\mathbf{a}_1 + \left(\frac{1}{2} - x_3\right)\,\mathbf{a}_2 - y_3\,\mathbf{a}_3$ | = | $\left(\frac{1}{2} + z_3\right)a\,\hat{\mathbf{x}} + \left(\frac{1}{2} - x_3\right)a\,\hat{\mathbf{y}} - y_3a\,\hat{\mathbf{z}}$ | (24$d$) | F |
| $\mathbf{B}_{15}$ | = | $\left(\frac{1}{2} - z_3\right)\,\mathbf{a}_1 - x_3\,\mathbf{a}_2 + \left(\frac{1}{2} + y_3\right)\,\mathbf{a}_3$ | = | $\left(\frac{1}{2} - z_3\right)a\,\hat{\mathbf{x}} - x_3a\,\hat{\mathbf{y}} + \left(\frac{1}{2} + y_3\right)a\,\hat{\mathbf{z}}$ | (24$d$) | F |
| $\mathbf{B}_{16}$ | = | $-z_3\,\mathbf{a}_1 + \left(\frac{1}{2} + x_3\right)\,\mathbf{a}_2 + \left(\frac{1}{2} - y_3\right)\,\mathbf{a}_3$ | = | $-z_3a\,\hat{\mathbf{x}} + \left(\frac{1}{2} + x_3\right)a\,\hat{\mathbf{y}} + \left(\frac{1}{2} - y_3\right)a\,\hat{\mathbf{z}}$ | (24$d$) | F |
| $\mathbf{B}_{17}$ | = | $y_3\,\mathbf{a}_1 + z_3\,\mathbf{a}_2 + x_3\,\mathbf{a}_3$ | = | $y_3a\,\hat{\mathbf{x}} + z_3a\,\hat{\mathbf{y}} + x_3a\,\hat{\mathbf{z}}$ | (24$d$) | F |
| $\mathbf{B}_{18}$ | = | $-y_3\,\mathbf{a}_1 + \left(\frac{1}{2} + z_3\right)\,\mathbf{a}_2 + \left(\frac{1}{2} - x_3\right)\,\mathbf{a}_3$ | = | $-y_3a\,\hat{\mathbf{x}} + \left(\frac{1}{2} + z_3\right)a\,\hat{\mathbf{y}} + \left(\frac{1}{2} - x_3\right)a\,\hat{\mathbf{z}}$ | (24$d$) | F |
| $\mathbf{B}_{19}$ | = | $\left(\frac{1}{2} + y_3\right)\,\mathbf{a}_1 + \left(\frac{1}{2} - z_3\right)\,\mathbf{a}_2 - x_3\,\mathbf{a}_3$ | = | $\left(\frac{1}{2} + y_3\right)a\,\hat{\mathbf{x}} + \left(\frac{1}{2} - z_3\right)a\,\hat{\mathbf{y}} - x_3a\,\hat{\mathbf{z}}$ | (24$d$) | F |
| $\mathbf{B}_{20}$ | = | $\left(\frac{1}{2} - y_3\right)\,\mathbf{a}_1 - z_3\,\mathbf{a}_2 + \left(\frac{1}{2} + x_3\right)\,\mathbf{a}_3$ | = | $\left(\frac{1}{2} - y_3\right)a\,\hat{\mathbf{x}} - z_3a\,\hat{\mathbf{y}} + \left(\frac{1}{2} + x_3\right)a\,\hat{\mathbf{z}}$ | (24$d$) | F |
| $\mathbf{B}_{21}$ | = | $-x_3\,\mathbf{a}_1 - y_3\,\mathbf{a}_2 - z_3\,\mathbf{a}_3$ | = | $-x_3a\,\hat{\mathbf{x}} - y_3a\,\hat{\mathbf{y}} - z_3a\,\hat{\mathbf{z}}$ | (24$d$) | F |
| $\mathbf{B}_{22}$ | = | $\left(\frac{1}{2} + x_3\right)\,\mathbf{a}_1 + y_3\,\mathbf{a}_2 + \left(\frac{1}{2} - z_3\right)\,\mathbf{a}_3$ | = | $\left(\frac{1}{2} + x_3\right)a\,\hat{\mathbf{x}} + y_3a\,\hat{\mathbf{y}} + \left(\frac{1}{2} - z_3\right)a\,\hat{\mathbf{z}}$ | (24$d$) | F |
| $\mathbf{B}_{23}$ | = | $x_3\,\mathbf{a}_1 + \left(\frac{1}{2} - y_3\right)\,\mathbf{a}_2 + \left(\frac{1}{2} + z_3\right)\,\mathbf{a}_3$ | = | $x_3a\,\hat{\mathbf{x}} + \left(\frac{1}{2} - y_3\right)a\,\hat{\mathbf{y}} + \left(\frac{1}{2} + z_3\right)a\,\hat{\mathbf{z}}$ | (24$d$) | F |
| $\mathbf{B}_{24}$ | = | $\left(\frac{1}{2} - x_3\right)\,\mathbf{a}_1 + \left(\frac{1}{2} + y_3\right)\,\mathbf{a}_2 + z_3\,\mathbf{a}_3$ | = | $\left(\frac{1}{2} - x_3\right)a\,\hat{\mathbf{x}} + \left(\frac{1}{2} + y_3\right)a\,\hat{\mathbf{y}} + z_3a\,\hat{\mathbf{z}}$ | (24$d$) | F |
| $\mathbf{B}_{25}$ | = | $-z_3\,\mathbf{a}_1 - x_3\,\mathbf{a}_2 - y_3\,\mathbf{a}_3$ | = | $-z_3a\,\hat{\mathbf{x}} - x_3a\,\hat{\mathbf{y}} - y_3a\,\hat{\mathbf{z}}$ | (24$d$) | F |
| $\mathbf{B}_{26}$ | = | $\left(\frac{1}{2} - z_3\right)\,\mathbf{a}_1 + \left(\frac{1}{2} + x_3\right)\,\mathbf{a}_2 + y_3\,\mathbf{a}_3$ | = | $\left(\frac{1}{2} - z_3\right)a\,\hat{\mathbf{x}} + \left(\frac{1}{2} + x_3\right)a\,\hat{\mathbf{y}} + y_3a\,\hat{\mathbf{z}}$ | (24$d$) | F |
| $\mathbf{B}_{27}$ | = | $\left(\frac{1}{2} + z_3\right)\,\mathbf{a}_1 + x_3\,\mathbf{a}_2 + \left(\frac{1}{2} - y_3\right)\,\mathbf{a}_3$ | = | $\left(\frac{1}{2} + z_3\right)a\,\hat{\mathbf{x}} + x_3a\,\hat{\mathbf{y}} + \left(\frac{1}{2} - y_3\right)a\,\hat{\mathbf{z}}$ | (24$d$) | F |
| $\mathbf{B}_{28}$ | = | $z_3\,\mathbf{a}_1 + \left(\frac{1}{2} - x_3\right)\,\mathbf{a}_2 + \left(\frac{1}{2} + y_3\right)\,\mathbf{a}_3$ | = | $z_3a\,\hat{\mathbf{x}} + \left(\frac{1}{2} - x_3\right)a\,\hat{\mathbf{y}} + \left(\frac{1}{2} + y_3\right)a\,\hat{\mathbf{z}}$ | (24$d$) | F |
| $\mathbf{B}_{29}$ | = | $-y_3\,\mathbf{a}_1 - z_3\,\mathbf{a}_2 - x_3\,\mathbf{a}_3$ | = | $-y_3a\,\hat{\mathbf{x}} - z_3a\,\hat{\mathbf{y}} - x_3a\,\hat{\mathbf{z}}$ | (24$d$) | F |
| $\mathbf{B}_{30}$ | = | $y_3\,\mathbf{a}_1 + \left(\frac{1}{2} - z_3\right)\,\mathbf{a}_2 + \left(\frac{1}{2} + x_3\right)\,\mathbf{a}_3$ | = | $y_3a\,\hat{\mathbf{x}} + \left(\frac{1}{2} - z_3\right)a\,\hat{\mathbf{y}} + \left(\frac{1}{2} + x_3\right)a\,\hat{\mathbf{z}}$ | (24$d$) | F |
| $\mathbf{B}_{31}$ | = | $\left(\frac{1}{2} - y_3\right)\,\mathbf{a}_1 + \left(\frac{1}{2} + z_3\right)\,\mathbf{a}_2 + x_3\,\mathbf{a}_3$ | = | $\left(\frac{1}{2} - y_3\right)a\,\hat{\mathbf{x}} + \left(\frac{1}{2} + z_3\right)a\,\hat{\mathbf{y}} + x_3a\,\hat{\mathbf{z}}$ | (24$d$) | F |
| $\mathbf{B}_{32}$ | = | $\left(\frac{1}{2} + y_3\right)\,\mathbf{a}_1 + z_3\,\mathbf{a}_2 + \left(\frac{1}{2} - x_3\right)\,\mathbf{a}_3$ | = | $\left(\frac{1}{2} + y_3\right)a\,\hat{\mathbf{x}} + z_3a\,\hat{\mathbf{y}} + \left(\frac{1}{2} - x_3\right)a\,\hat{\mathbf{z}}$ | (24$d$) | F |

**Geometry files:**

- CIF: pp. 1782
- POSCAR: pp. 1783

# $ZrP_2O_7$ High-Temperature ($K6_1$) Structure: A7B2C_cP40_205_bd_c_a

http://aflow.org/prototype-encyclopedia/A7B2C_cP40_205_bd_c_a

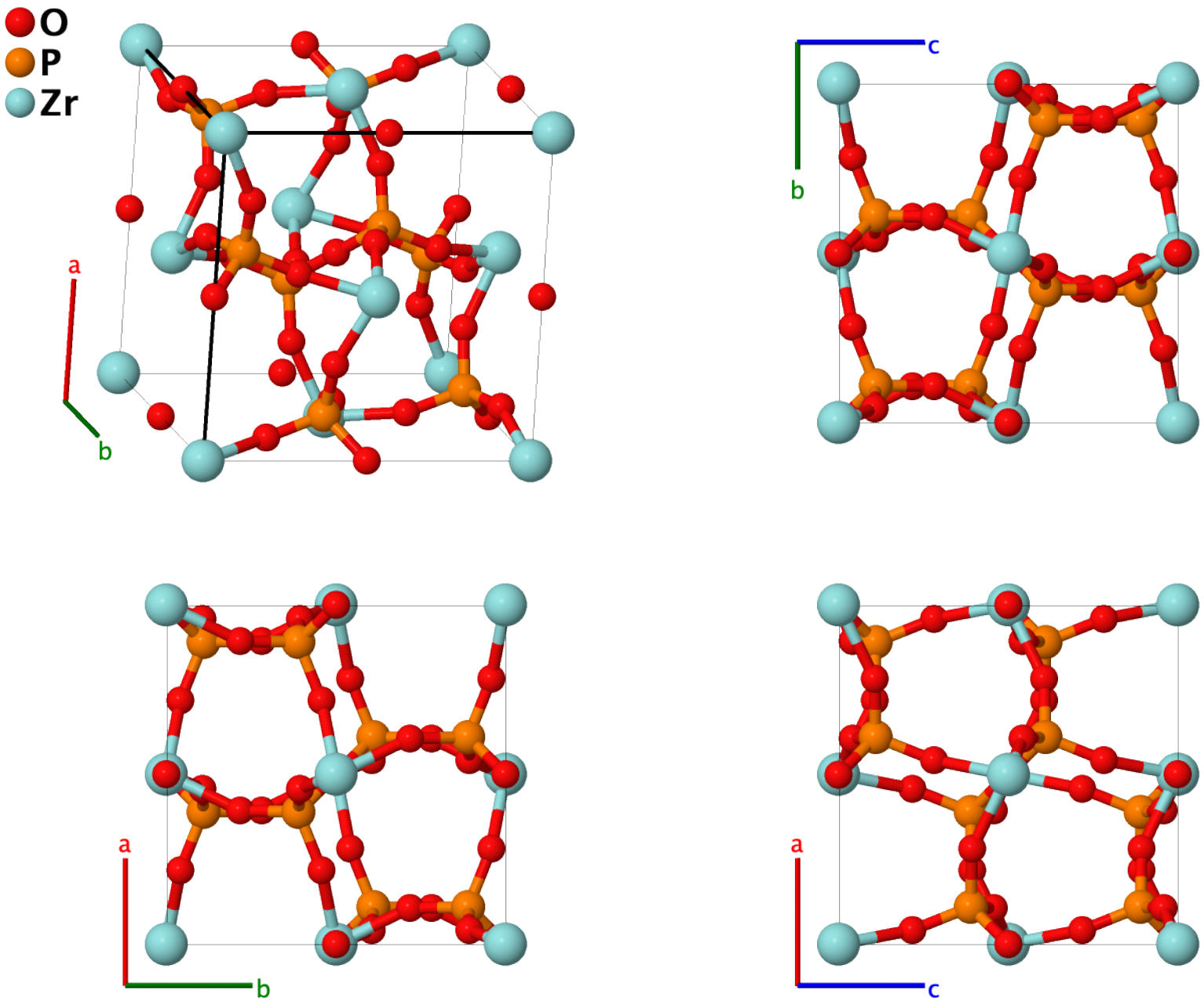

| | | |
|---|---|---|
| **Prototype** | : | $P_2O_7Zr$ |
| **AFLOW prototype label** | : | A7B2C_cP40_205_bd_c_a |
| ***Strukturbericht* designation** | : | $K6_1$ |
| **Pearson symbol** | : | cP40 |
| **Space group number** | : | 205 |
| **Space group symbol** | : | $Pa\bar{3}$ |
| **AFLOW prototype command** | : | `aflow --proto=A7B2C_cP40_205_bd_c_a` `--params=`$a, x_3, x_4, y_4, z_4$ |

**Other compounds with this structure**

- $AB_2O_7$, ($A$ = Si, Ge, Sn, Pb, Ti, Zr, Hf, Mo, W, Re, Ce, U, *etc.*; $B$ = P, V, As)

- This is the high temperature form of all the structures listed. The low temperature structure depends on the composition. Below 290 °C, $ZrP_2O_7$ transforms to an orthorhombic structure, space group *Pbca* #61, with 136 unique crystallographic positions and 1080 atomic sites. See (Birkedal, 2006) and (Stinton, 2006) for more details.

**Simple Cubic primitive vectors:**

$$\mathbf{a}_1 = a\,\hat{\mathbf{x}}$$
$$\mathbf{a}_2 = a\,\hat{\mathbf{y}}$$
$$\mathbf{a}_3 = a\,\hat{\mathbf{z}}$$

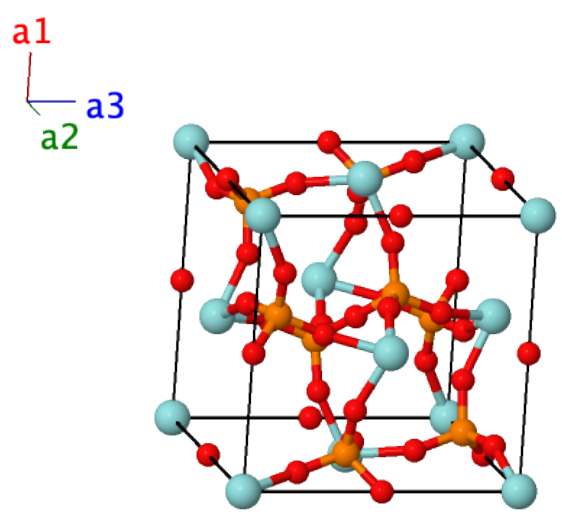

**Basis vectors:**

| | | Lattice Coordinates | | Cartesian Coordinates | Wyckoff Position | Atom Type |
|---|---|---|---|---|---|---|
| $\mathbf{B}_1$ | = | $0\,\mathbf{a}_1 + 0\,\mathbf{a}_2 + 0\,\mathbf{a}_3$ | = | $0\,\hat{\mathbf{x}} + 0\,\hat{\mathbf{y}} + 0\,\hat{\mathbf{z}}$ | (4$a$) | Zr |
| $\mathbf{B}_2$ | = | $\frac{1}{2}\,\mathbf{a}_1 + \frac{1}{2}\,\mathbf{a}_3$ | = | $\frac{1}{2}a\,\hat{\mathbf{x}} + \frac{1}{2}a\,\hat{\mathbf{z}}$ | (4$a$) | Zr |
| $\mathbf{B}_3$ | = | $\frac{1}{2}\,\mathbf{a}_2 + \frac{1}{2}\,\mathbf{a}_3$ | = | $\frac{1}{2}a\,\hat{\mathbf{y}} + \frac{1}{2}a\,\hat{\mathbf{z}}$ | (4$a$) | Zr |
| $\mathbf{B}_4$ | = | $\frac{1}{2}\,\mathbf{a}_1 + \frac{1}{2}\,\mathbf{a}_2$ | = | $\frac{1}{2}a\,\hat{\mathbf{x}} + \frac{1}{2}a\,\hat{\mathbf{y}}$ | (4$a$) | Zr |
| $\mathbf{B}_5$ | = | $\frac{1}{2}\,\mathbf{a}_1 + \frac{1}{2}\,\mathbf{a}_2 + \frac{1}{2}\,\mathbf{a}_3$ | = | $\frac{1}{2}a\,\hat{\mathbf{x}} + \frac{1}{2}a\,\hat{\mathbf{y}} + \frac{1}{2}a\,\hat{\mathbf{z}}$ | (4$b$) | O I |
| $\mathbf{B}_6$ | = | $\frac{1}{2}\,\mathbf{a}_2$ | = | $\frac{1}{2}a\,\hat{\mathbf{y}}$ | (4$b$) | O I |
| $\mathbf{B}_7$ | = | $\frac{1}{2}\,\mathbf{a}_1$ | = | $\frac{1}{2}a\,\hat{\mathbf{x}}$ | (4$b$) | O I |
| $\mathbf{B}_8$ | = | $\frac{1}{2}\,\mathbf{a}_3$ | = | $\frac{1}{2}a\,\hat{\mathbf{z}}$ | (4$b$) | O I |
| $\mathbf{B}_9$ | = | $x_3\,\mathbf{a}_1 + x_3\,\mathbf{a}_2 + x_3\,\mathbf{a}_3$ | = | $x_3a\,\hat{\mathbf{x}} + x_3a\,\hat{\mathbf{y}} + x_3a\,\hat{\mathbf{z}}$ | (8$c$) | P |
| $\mathbf{B}_{10}$ | = | $\left(\frac{1}{2} - x_3\right)\mathbf{a}_1 - x_3\,\mathbf{a}_2 + \left(\frac{1}{2} + x_3\right)\mathbf{a}_3$ | = | $\left(\frac{1}{2} - x_3\right)a\,\hat{\mathbf{x}} - x_3a\,\hat{\mathbf{y}} + \left(\frac{1}{2} + x_3\right)a\,\hat{\mathbf{z}}$ | (8$c$) | P |
| $\mathbf{B}_{11}$ | = | $-x_3\,\mathbf{a}_1 + \left(\frac{1}{2} + x_3\right)\mathbf{a}_2 + \left(\frac{1}{2} - x_3\right)\mathbf{a}_3$ | = | $-x_3a\,\hat{\mathbf{x}} + \left(\frac{1}{2} + x_3\right)a\,\hat{\mathbf{y}} + \left(\frac{1}{2} - x_3\right)a\,\hat{\mathbf{z}}$ | (8$c$) | P |
| $\mathbf{B}_{12}$ | = | $\left(\frac{1}{2} + x_3\right)\mathbf{a}_1 + \left(\frac{1}{2} - x_3\right)\mathbf{a}_2 - x_3\,\mathbf{a}_3$ | = | $\left(\frac{1}{2} + x_3\right)a\,\hat{\mathbf{x}} + \left(\frac{1}{2} - x_3\right)a\,\hat{\mathbf{y}} - x_3a\,\hat{\mathbf{z}}$ | (8$c$) | P |
| $\mathbf{B}_{13}$ | = | $-x_3\,\mathbf{a}_1 - x_3\,\mathbf{a}_2 - x_3\,\mathbf{a}_3$ | = | $-x_3a\,\hat{\mathbf{x}} - x_3a\,\hat{\mathbf{y}} - x_3a\,\hat{\mathbf{z}}$ | (8$c$) | P |
| $\mathbf{B}_{14}$ | = | $\left(\frac{1}{2} + x_3\right)\mathbf{a}_1 + x_3\,\mathbf{a}_2 + \left(\frac{1}{2} - x_3\right)\mathbf{a}_3$ | = | $\left(\frac{1}{2} + x_3\right)a\,\hat{\mathbf{x}} + x_3a\,\hat{\mathbf{y}} + \left(\frac{1}{2} - x_3\right)a\,\hat{\mathbf{z}}$ | (8$c$) | P |
| $\mathbf{B}_{15}$ | = | $x_3\,\mathbf{a}_1 + \left(\frac{1}{2} - x_3\right)\mathbf{a}_2 + \left(\frac{1}{2} + x_3\right)\mathbf{a}_3$ | = | $x_3a\,\hat{\mathbf{x}} + \left(\frac{1}{2} - x_3\right)a\,\hat{\mathbf{y}} + \left(\frac{1}{2} + x_3\right)a\,\hat{\mathbf{z}}$ | (8$c$) | P |
| $\mathbf{B}_{16}$ | = | $\left(\frac{1}{2} - x_3\right)\mathbf{a}_1 + \left(\frac{1}{2} + x_3\right)\mathbf{a}_2 + x_3\,\mathbf{a}_3$ | = | $\left(\frac{1}{2} - x_3\right)a\,\hat{\mathbf{x}} + \left(\frac{1}{2} + x_3\right)a\,\hat{\mathbf{y}} + x_3a\,\hat{\mathbf{z}}$ | (8$c$) | P |
| $\mathbf{B}_{17}$ | = | $x_4\,\mathbf{a}_1 + y_4\,\mathbf{a}_2 + z_4\,\mathbf{a}_3$ | = | $x_4a\,\hat{\mathbf{x}} + y_4a\,\hat{\mathbf{y}} + z_4a\,\hat{\mathbf{z}}$ | (24$d$) | O II |
| $\mathbf{B}_{18}$ | = | $\left(\frac{1}{2} - x_4\right)\mathbf{a}_1 - y_4\,\mathbf{a}_2 + \left(\frac{1}{2} + z_4\right)\mathbf{a}_3$ | = | $\left(\frac{1}{2} - x_4\right)a\,\hat{\mathbf{x}} - y_4a\,\hat{\mathbf{y}} + \left(\frac{1}{2} + z_4\right)a\,\hat{\mathbf{z}}$ | (24$d$) | O II |
| $\mathbf{B}_{19}$ | = | $-x_4\,\mathbf{a}_1 + \left(\frac{1}{2} + y_4\right)\mathbf{a}_2 + \left(\frac{1}{2} - z_4\right)\mathbf{a}_3$ | = | $-x_4a\,\hat{\mathbf{x}} + \left(\frac{1}{2} + y_4\right)a\,\hat{\mathbf{y}} + \left(\frac{1}{2} - z_4\right)a\,\hat{\mathbf{z}}$ | (24$d$) | O II |
| $\mathbf{B}_{20}$ | = | $\left(\frac{1}{2} + x_4\right)\mathbf{a}_1 + \left(\frac{1}{2} - y_4\right)\mathbf{a}_2 - z_4\,\mathbf{a}_3$ | = | $\left(\frac{1}{2} + x_4\right)a\,\hat{\mathbf{x}} + \left(\frac{1}{2} - y_4\right)a\,\hat{\mathbf{y}} - z_4a\,\hat{\mathbf{z}}$ | (24$d$) | O II |
| $\mathbf{B}_{21}$ | = | $z_4\,\mathbf{a}_1 + x_4\,\mathbf{a}_2 + y_4\,\mathbf{a}_3$ | = | $z_4a\,\hat{\mathbf{x}} + x_4a\,\hat{\mathbf{y}} + y_4a\,\hat{\mathbf{z}}$ | (24$d$) | O II |
| $\mathbf{B}_{22}$ | = | $\left(\frac{1}{2} + z_4\right)\mathbf{a}_1 + \left(\frac{1}{2} - x_4\right)\mathbf{a}_2 - y_4\,\mathbf{a}_3$ | = | $\left(\frac{1}{2} + z_4\right)a\,\hat{\mathbf{x}} + \left(\frac{1}{2} - x_4\right)a\,\hat{\mathbf{y}} - y_4a\,\hat{\mathbf{z}}$ | (24$d$) | O II |
| $\mathbf{B}_{23}$ | = | $\left(\frac{1}{2} - z_4\right)\mathbf{a}_1 - x_4\,\mathbf{a}_2 + \left(\frac{1}{2} + y_4\right)\mathbf{a}_3$ | = | $\left(\frac{1}{2} - z_4\right)a\,\hat{\mathbf{x}} - x_4a\,\hat{\mathbf{y}} + \left(\frac{1}{2} + y_4\right)a\,\hat{\mathbf{z}}$ | (24$d$) | O II |
| $\mathbf{B}_{24}$ | = | $-z_4\,\mathbf{a}_1 + \left(\frac{1}{2} + x_4\right)\mathbf{a}_2 + \left(\frac{1}{2} - y_4\right)\mathbf{a}_3$ | = | $-z_4a\,\hat{\mathbf{x}} + \left(\frac{1}{2} + x_4\right)a\,\hat{\mathbf{y}} + \left(\frac{1}{2} - y_4\right)a\,\hat{\mathbf{z}}$ | (24$d$) | O II |
| $\mathbf{B}_{25}$ | = | $y_4\,\mathbf{a}_1 + z_4\,\mathbf{a}_2 + x_4\,\mathbf{a}_3$ | = | $y_4a\,\hat{\mathbf{x}} + z_4a\,\hat{\mathbf{y}} + x_4a\,\hat{\mathbf{z}}$ | (24$d$) | O II |
| $\mathbf{B}_{26}$ | = | $-y_4\,\mathbf{a}_1 + \left(\frac{1}{2} + z_4\right)\mathbf{a}_2 + \left(\frac{1}{2} - x_4\right)\mathbf{a}_3$ | = | $-y_4a\,\hat{\mathbf{x}} + \left(\frac{1}{2} + z_4\right)a\,\hat{\mathbf{y}} + \left(\frac{1}{2} - x_4\right)a\,\hat{\mathbf{z}}$ | (24$d$) | O II |
| $\mathbf{B}_{27}$ | = | $\left(\frac{1}{2} + y_4\right)\mathbf{a}_1 + \left(\frac{1}{2} - z_4\right)\mathbf{a}_2 - x_4\,\mathbf{a}_3$ | = | $\left(\frac{1}{2} + y_4\right)a\,\hat{\mathbf{x}} + \left(\frac{1}{2} - z_4\right)a\,\hat{\mathbf{y}} - x_4a\,\hat{\mathbf{z}}$ | (24$d$) | O II |
| $\mathbf{B}_{28}$ | = | $\left(\frac{1}{2} - y_4\right)\mathbf{a}_1 - z_4\,\mathbf{a}_2 + \left(\frac{1}{2} + x_4\right)\mathbf{a}_3$ | = | $\left(\frac{1}{2} - y_4\right)a\,\hat{\mathbf{x}} - z_4a\,\hat{\mathbf{y}} + \left(\frac{1}{2} + x_4\right)a\,\hat{\mathbf{z}}$ | (24$d$) | O II |
| $\mathbf{B}_{29}$ | = | $-x_4\,\mathbf{a}_1 - y_4\,\mathbf{a}_2 - z_4\,\mathbf{a}_3$ | = | $-x_4a\,\hat{\mathbf{x}} - y_4a\,\hat{\mathbf{y}} - z_4a\,\hat{\mathbf{z}}$ | (24$d$) | O II |
| $\mathbf{B}_{30}$ | = | $\left(\frac{1}{2} + x_4\right)\mathbf{a}_1 + y_4\,\mathbf{a}_2 + \left(\frac{1}{2} - z_4\right)\mathbf{a}_3$ | = | $\left(\frac{1}{2} + x_4\right)a\,\hat{\mathbf{x}} + y_4a\,\hat{\mathbf{y}} + \left(\frac{1}{2} - z_4\right)a\,\hat{\mathbf{z}}$ | (24$d$) | O II |
| $\mathbf{B}_{31}$ | = | $x_4\,\mathbf{a}_1 + \left(\frac{1}{2} - y_4\right)\mathbf{a}_2 + \left(\frac{1}{2} + z_4\right)\mathbf{a}_3$ | = | $x_4a\,\hat{\mathbf{x}} + \left(\frac{1}{2} - y_4\right)a\,\hat{\mathbf{y}} + \left(\frac{1}{2} + z_4\right)a\,\hat{\mathbf{z}}$ | (24$d$) | O II |
| $\mathbf{B}_{32}$ | = | $\left(\frac{1}{2} - x_4\right)\mathbf{a}_1 + \left(\frac{1}{2} + y_4\right)\mathbf{a}_2 + z_4\,\mathbf{a}_3$ | = | $\left(\frac{1}{2} - x_4\right)a\,\hat{\mathbf{x}} + \left(\frac{1}{2} + y_4\right)a\,\hat{\mathbf{y}} + z_4a\,\hat{\mathbf{z}}$ | (24$d$) | O II |
| $\mathbf{B}_{33}$ | = | $-z_4\,\mathbf{a}_1 - x_4\,\mathbf{a}_2 - y_4\,\mathbf{a}_3$ | = | $-z_4a\,\hat{\mathbf{x}} - x_4a\,\hat{\mathbf{y}} - y_4a\,\hat{\mathbf{z}}$ | (24$d$) | O II |
| $\mathbf{B}_{34}$ | = | $\left(\frac{1}{2} - z_4\right)\mathbf{a}_1 + \left(\frac{1}{2} + x_4\right)\mathbf{a}_2 + y_4\,\mathbf{a}_3$ | = | $\left(\frac{1}{2} - z_4\right)a\,\hat{\mathbf{x}} + \left(\frac{1}{2} + x_4\right)a\,\hat{\mathbf{y}} + y_4a\,\hat{\mathbf{z}}$ | (24$d$) | O II |
| $\mathbf{B}_{35}$ | = | $\left(\frac{1}{2}+z_4\right)\mathbf{a}_1 + x_4\,\mathbf{a}_2 + \left(\frac{1}{2}-y_4\right)\mathbf{a}_3$ | = | $\left(\frac{1}{2}+z_4\right)a\,\hat{\mathbf{x}} + x_4 a\,\hat{\mathbf{y}} + \left(\frac{1}{2}-y_4\right)a\,\hat{\mathbf{z}}$ | (24$d$) | O II |
| $\mathbf{B}_{36}$ | = | $z_4\,\mathbf{a}_1 + \left(\frac{1}{2}-x_4\right)\mathbf{a}_2 + \left(\frac{1}{2}+y_4\right)\mathbf{a}_3$ | = | $z_4 a\,\hat{\mathbf{x}} + \left(\frac{1}{2}-x_4\right)a\,\hat{\mathbf{y}} + \left(\frac{1}{2}+y_4\right)a\,\hat{\mathbf{z}}$ | (24$d$) | O II |
| $\mathbf{B}_{37}$ | = | $-y_4\,\mathbf{a}_1 - z_4\,\mathbf{a}_2 - x_4\,\mathbf{a}_3$ | = | $-y_4 a\,\hat{\mathbf{x}} - z_4 a\,\hat{\mathbf{y}} - x_4 a\,\hat{\mathbf{z}}$ | (24$d$) | O II |
| $\mathbf{B}_{38}$ | = | $y_4\,\mathbf{a}_1 + \left(\frac{1}{2}-z_4\right)\mathbf{a}_2 + \left(\frac{1}{2}+x_4\right)\mathbf{a}_3$ | = | $y_4 a\,\hat{\mathbf{x}} + \left(\frac{1}{2}-z_4\right)a\,\hat{\mathbf{y}} + \left(\frac{1}{2}+x_4\right)a\,\hat{\mathbf{z}}$ | (24$d$) | O II |
| $\mathbf{B}_{39}$ | = | $\left(\frac{1}{2}-y_4\right)\mathbf{a}_1 + \left(\frac{1}{2}+z_4\right)\mathbf{a}_2 + x_4\,\mathbf{a}_3$ | = | $\left(\frac{1}{2}-y_4\right)a\,\hat{\mathbf{x}} + \left(\frac{1}{2}+z_4\right)a\,\hat{\mathbf{y}} + x_4 a\,\hat{\mathbf{z}}$ | (24$d$) | O II |
| $\mathbf{B}_{40}$ | = | $\left(\frac{1}{2}+y_4\right)\mathbf{a}_1 + z_4\,\mathbf{a}_2 + \left(\frac{1}{2}-x_4\right)\mathbf{a}_3$ | = | $\left(\frac{1}{2}+y_4\right)a\,\hat{\mathbf{x}} + z_4 a\,\hat{\mathbf{y}} + \left(\frac{1}{2}-x_4\right)a\,\hat{\mathbf{z}}$ | (24$d$) | O II |

---

**Found in:**

---

**Geometry files:**

- CIF: pp. 1783
- POSCAR: pp. 1783

# $NaCr(SO_4)_2 \cdot 12H_2O$ Alum Structure: AB12CD8E2_cP96_205_a_2d_b_cd_c

http://aflow.org/prototype-encyclopedia/AB12CD8E2_cP96_205_a_2d_b_cd_c

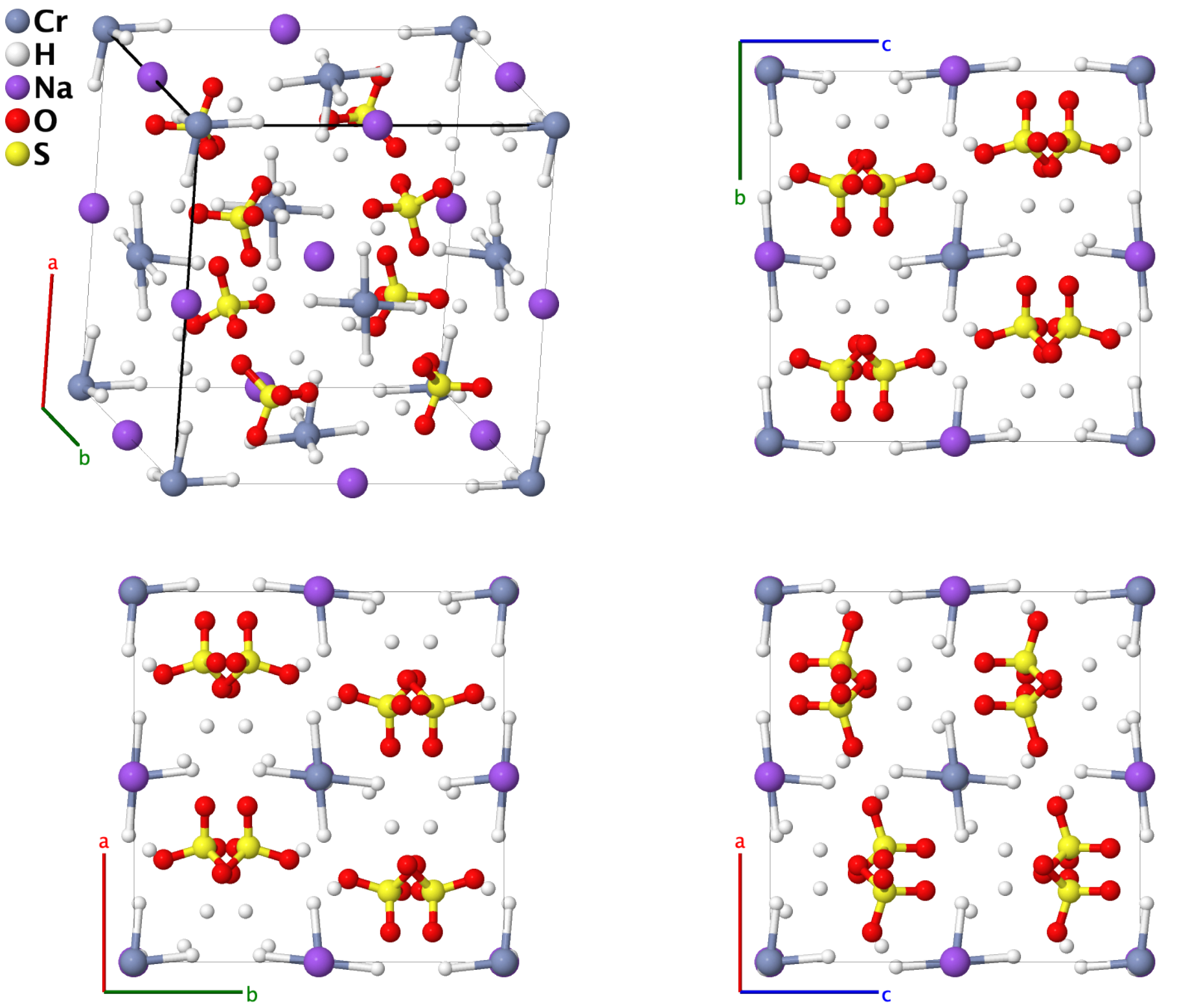

| | | |
|---|---|---|
| **Prototype** | : | $Cr(H_2O)_{12}NaO_8S_2$ |
| **AFLOW prototype label** | : | AB12CD8E2_cP96_205_a_2d_b_cd_c |
| ***Strukturbericht* designation** | : | None |
| **Pearson symbol** | : | cP96 |
| **Space group number** | : | 205 |
| **Space group symbol** | : | $Pa\bar{3}$ |
| **AFLOW prototype command** | : | `aflow --proto=AB12CD8E2_cP96_205_a_2d_b_cd_c` <br> `--params=`$a, x_3, x_4, x_5, y_5, z_5, x_6, y_6, z_6, x_7, y_7, z_7$ |

### Other compounds with this structure

- $CsCr(SO_4)_2 \cdot 12H_2O$

---

- The alums have the general formula $AB(XO_4)_2 \cdot 12H_2O$, where $A$ is a monovalent ion, $B$ is a trivalent ion, and $X$ is a chalcogen. In most cases atom $B$ is aluminum and atom $X$ is sulfur, leading to the name alum.
- All alums have their room-temperature form in space group $Pa\bar{3}$ #205, but the bonding between the $A$ and $B$ ions and the $XO_4$ complex can be quite different.
- (Lipson, 1935ab) described three general forms of alum based on the sizes of the monovalent ions. Each of these forms was given a *Strukturbericht* designation by (Gottfried, 1937):
  - $\alpha$-alum, with intermediate sized ions, prototype $KAl(SO_4)_2 \cdot 12H_2O$, $H4_{13}$,

– $\beta$-alum, with large ions, prototype $(NH_3CH_3)Al(SO_4)_2{\cdot}12H_2O$, $H4_{14}$, and
  – $\gamma$-alum, with small ions, prototype $NaAl(SO_4)_2{\cdot}12H_2O$, $H4_{15}$.

- This classification scheme is not complete, *e.g.*, (Ledsham, 1968) points out that $NaCr(SO_4)_2{\cdot}12H_2O$ (this structure) does not fit into any of these categories, and that the actual structure depends on the combination of monovalent and trivalent ions.

- As noted above, the $Pa\overline{3}$ structures of alum are the room temperature form. As the temperature decreases the alum structure may transform. For example, in the temperature range 150-170 K the $\beta$-alum $(NH_3CH_3)Al(SO_4)_2{\cdot}12H_2O$ transforms into an orthorhombic structure with fully ordered $NH_3CH_3$ ions.

- The positions of the hydrogen atoms in the water molecules were not determined, so we only provide the positions of the oxygen atoms (labeled as $H_2O$).

---

**Simple Cubic primitive vectors:**

$$\mathbf{a}_1 = a\,\hat{\mathbf{x}}$$
$$\mathbf{a}_2 = a\,\hat{\mathbf{y}}$$
$$\mathbf{a}_3 = a\,\hat{\mathbf{z}}$$

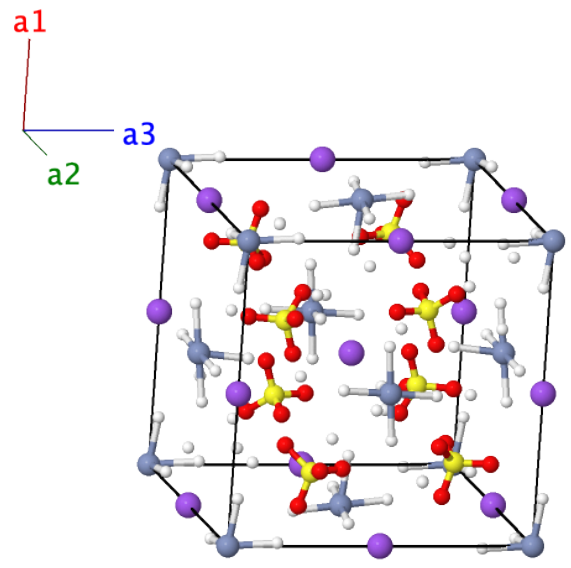

---

**Basis vectors:**

| | | Lattice Coordinates | | Cartesian Coordinates | Wyckoff Position | Atom Type |
|---|---|---|---|---|---|---|
| $\mathbf{B}_1$ | = | $0\,\mathbf{a}_1 + 0\,\mathbf{a}_2 + 0\,\mathbf{a}_3$ | = | $0\,\hat{\mathbf{x}} + 0\,\hat{\mathbf{y}} + 0\,\hat{\mathbf{z}}$ | (4*a*) | Cr |
| $\mathbf{B}_2$ | = | $\frac{1}{2}\,\mathbf{a}_1 + \frac{1}{2}\,\mathbf{a}_3$ | = | $\frac{1}{2}a\,\hat{\mathbf{x}} + \frac{1}{2}a\,\hat{\mathbf{z}}$ | (4*a*) | Cr |
| $\mathbf{B}_3$ | = | $\frac{1}{2}\,\mathbf{a}_2 + \frac{1}{2}\,\mathbf{a}_3$ | = | $\frac{1}{2}a\,\hat{\mathbf{y}} + \frac{1}{2}a\,\hat{\mathbf{z}}$ | (4*a*) | Cr |
| $\mathbf{B}_4$ | = | $\frac{1}{2}\,\mathbf{a}_1 + \frac{1}{2}\,\mathbf{a}_2$ | = | $\frac{1}{2}a\,\hat{\mathbf{x}} + \frac{1}{2}a\,\hat{\mathbf{y}}$ | (4*a*) | Cr |
| $\mathbf{B}_5$ | = | $\frac{1}{2}\,\mathbf{a}_1 + \frac{1}{2}\,\mathbf{a}_2 + \frac{1}{2}\,\mathbf{a}_3$ | = | $\frac{1}{2}a\,\hat{\mathbf{x}} + \frac{1}{2}a\,\hat{\mathbf{y}} + \frac{1}{2}a\,\hat{\mathbf{z}}$ | (4*b*) | Na |
| $\mathbf{B}_6$ | = | $\frac{1}{2}\,\mathbf{a}_2$ | = | $\frac{1}{2}a\,\hat{\mathbf{y}}$ | (4*b*) | Na |
| $\mathbf{B}_7$ | = | $\frac{1}{2}\,\mathbf{a}_1$ | = | $\frac{1}{2}a\,\hat{\mathbf{x}}$ | (4*b*) | Na |
| $\mathbf{B}_8$ | = | $\frac{1}{2}\,\mathbf{a}_3$ | = | $\frac{1}{2}a\,\hat{\mathbf{z}}$ | (4*b*) | Na |
| $\mathbf{B}_9$ | = | $x_3\,\mathbf{a}_1 + x_3\,\mathbf{a}_2 + x_3\,\mathbf{a}_3$ | = | $x_3a\,\hat{\mathbf{x}} + x_3a\,\hat{\mathbf{y}} + x_3a\,\hat{\mathbf{z}}$ | (8*c*) | O I |
| $\mathbf{B}_{10}$ | = | $\left(\frac{1}{2} - x_3\right)\mathbf{a}_1 - x_3\,\mathbf{a}_2 + \left(\frac{1}{2} + x_3\right)\mathbf{a}_3$ | = | $\left(\frac{1}{2} - x_3\right)a\,\hat{\mathbf{x}} - x_3a\,\hat{\mathbf{y}} + \left(\frac{1}{2} + x_3\right)a\,\hat{\mathbf{z}}$ | (8*c*) | O I |
| $\mathbf{B}_{11}$ | = | $-x_3\,\mathbf{a}_1 + \left(\frac{1}{2} + x_3\right)\mathbf{a}_2 + \left(\frac{1}{2} - x_3\right)\mathbf{a}_3$ | = | $-x_3a\,\hat{\mathbf{x}} + \left(\frac{1}{2} + x_3\right)a\,\hat{\mathbf{y}} + \left(\frac{1}{2} - x_3\right)a\,\hat{\mathbf{z}}$ | (8*c*) | O I |
| $\mathbf{B}_{12}$ | = | $\left(\frac{1}{2} + x_3\right)\mathbf{a}_1 + \left(\frac{1}{2} - x_3\right)\mathbf{a}_2 - x_3\,\mathbf{a}_3$ | = | $\left(\frac{1}{2} + x_3\right)a\,\hat{\mathbf{x}} + \left(\frac{1}{2} - x_3\right)a\,\hat{\mathbf{y}} - x_3a\,\hat{\mathbf{z}}$ | (8*c*) | O I |
| $\mathbf{B}_{13}$ | = | $-x_3\,\mathbf{a}_1 - x_3\,\mathbf{a}_2 - x_3\,\mathbf{a}_3$ | = | $-x_3a\,\hat{\mathbf{x}} - x_3a\,\hat{\mathbf{y}} - x_3a\,\hat{\mathbf{z}}$ | (8*c*) | O I |
| $\mathbf{B}_{14}$ | = | $\left(\frac{1}{2} + x_3\right)\mathbf{a}_1 + x_3\,\mathbf{a}_2 + \left(\frac{1}{2} - x_3\right)\mathbf{a}_3$ | = | $\left(\frac{1}{2} + x_3\right)a\,\hat{\mathbf{x}} + x_3a\,\hat{\mathbf{y}} + \left(\frac{1}{2} - x_3\right)a\,\hat{\mathbf{z}}$ | (8*c*) | O I |
| $\mathbf{B}_{15}$ | = | $x_3\,\mathbf{a}_1 + \left(\frac{1}{2} - x_3\right)\mathbf{a}_2 + \left(\frac{1}{2} + x_3\right)\mathbf{a}_3$ | = | $x_3a\,\hat{\mathbf{x}} + \left(\frac{1}{2} - x_3\right)a\,\hat{\mathbf{y}} + \left(\frac{1}{2} + x_3\right)a\,\hat{\mathbf{z}}$ | (8*c*) | O I |
| $\mathbf{B}_{16}$ | = | $\left(\frac{1}{2} - x_3\right)\mathbf{a}_1 + \left(\frac{1}{2} + x_3\right)\mathbf{a}_2 + x_3\,\mathbf{a}_3$ | = | $\left(\frac{1}{2} - x_3\right)a\,\hat{\mathbf{x}} + \left(\frac{1}{2} + x_3\right)a\,\hat{\mathbf{y}} + x_3a\,\hat{\mathbf{z}}$ | (8*c*) | O I |
| $\mathbf{B}_{17}$ | = | $x_4\,\mathbf{a}_1 + x_4\,\mathbf{a}_2 + x_4\,\mathbf{a}_3$ | = | $x_4a\,\hat{\mathbf{x}} + x_4a\,\hat{\mathbf{y}} + x_4a\,\hat{\mathbf{z}}$ | (8*c*) | S |
| $\mathbf{B}_{18}$ | = | $\left(\frac{1}{2} - x_4\right)\mathbf{a}_1 - x_4\,\mathbf{a}_2 + \left(\frac{1}{2} + x_4\right)\mathbf{a}_3$ | = | $\left(\frac{1}{2} - x_4\right)a\,\hat{\mathbf{x}} - x_4a\,\hat{\mathbf{y}} + \left(\frac{1}{2} + x_4\right)a\,\hat{\mathbf{z}}$ | (8*c*) | S |
| $\mathbf{B}_{19}$ | = | $-x_4\,\mathbf{a}_1 + \left(\frac{1}{2}+x_4\right)\mathbf{a}_2 + \left(\frac{1}{2}-x_4\right)\mathbf{a}_3$ | = | $-x_4 a\,\hat{\mathbf{x}} + \left(\frac{1}{2}+x_4\right)a\,\hat{\mathbf{y}} + \left(\frac{1}{2}-x_4\right)a\,\hat{\mathbf{z}}$ | (8$c$) | S |
| $\mathbf{B}_{20}$ | = | $\left(\frac{1}{2}+x_4\right)\mathbf{a}_1 + \left(\frac{1}{2}-x_4\right)\mathbf{a}_2 - x_4\,\mathbf{a}_3$ | = | $\left(\frac{1}{2}+x_4\right)a\,\hat{\mathbf{x}} + \left(\frac{1}{2}-x_4\right)a\,\hat{\mathbf{y}} - x_4 a\,\hat{\mathbf{z}}$ | (8$c$) | S |
| $\mathbf{B}_{21}$ | = | $-x_4\,\mathbf{a}_1 - x_4\,\mathbf{a}_2 - x_4\,\mathbf{a}_3$ | = | $-x_4 a\,\hat{\mathbf{x}} - x_4 a\,\hat{\mathbf{y}} - x_4 a\,\hat{\mathbf{z}}$ | (8$c$) | S |
| $\mathbf{B}_{22}$ | = | $\left(\frac{1}{2}+x_4\right)\mathbf{a}_1 + x_4\,\mathbf{a}_2 + \left(\frac{1}{2}-x_4\right)\mathbf{a}_3$ | = | $\left(\frac{1}{2}+x_4\right)a\,\hat{\mathbf{x}} + x_4 a\,\hat{\mathbf{y}} + \left(\frac{1}{2}-x_4\right)a\,\hat{\mathbf{z}}$ | (8$c$) | S |
| $\mathbf{B}_{23}$ | = | $x_4\,\mathbf{a}_1 + \left(\frac{1}{2}-x_4\right)\mathbf{a}_2 + \left(\frac{1}{2}+x_4\right)\mathbf{a}_3$ | = | $x_4 a\,\hat{\mathbf{x}} + \left(\frac{1}{2}-x_4\right)a\,\hat{\mathbf{y}} + \left(\frac{1}{2}+x_4\right)a\,\hat{\mathbf{z}}$ | (8$c$) | S |
| $\mathbf{B}_{24}$ | = | $\left(\frac{1}{2}-x_4\right)\mathbf{a}_1 + \left(\frac{1}{2}+x_4\right)\mathbf{a}_2 + x_4\,\mathbf{a}_3$ | = | $\left(\frac{1}{2}-x_4\right)a\,\hat{\mathbf{x}} + \left(\frac{1}{2}+x_4\right)a\,\hat{\mathbf{y}} + x_4 a\,\hat{\mathbf{z}}$ | (8$c$) | S |
| $\mathbf{B}_{25}$ | = | $x_5\,\mathbf{a}_1 + y_5\,\mathbf{a}_2 + z_5\,\mathbf{a}_3$ | = | $x_5 a\,\hat{\mathbf{x}} + y_5 a\,\hat{\mathbf{y}} + z_5 a\,\hat{\mathbf{z}}$ | (24$d$) | $H_2O$ I |
| $\mathbf{B}_{26}$ | = | $\left(\frac{1}{2}-x_5\right)\mathbf{a}_1 - y_5\,\mathbf{a}_2 + \left(\frac{1}{2}+z_5\right)\mathbf{a}_3$ | = | $\left(\frac{1}{2}-x_5\right)a\,\hat{\mathbf{x}} - y_5 a\,\hat{\mathbf{y}} + \left(\frac{1}{2}+z_5\right)a\,\hat{\mathbf{z}}$ | (24$d$) | $H_2O$ I |
| $\mathbf{B}_{27}$ | = | $-x_5\,\mathbf{a}_1 + \left(\frac{1}{2}+y_5\right)\mathbf{a}_2 + \left(\frac{1}{2}-z_5\right)\mathbf{a}_3$ | = | $-x_5 a\,\hat{\mathbf{x}} + \left(\frac{1}{2}+y_5\right)a\,\hat{\mathbf{y}} + \left(\frac{1}{2}-z_5\right)a\,\hat{\mathbf{z}}$ | (24$d$) | $H_2O$ I |
| $\mathbf{B}_{28}$ | = | $\left(\frac{1}{2}+x_5\right)\mathbf{a}_1 + \left(\frac{1}{2}-y_5\right)\mathbf{a}_2 - z_5\,\mathbf{a}_3$ | = | $\left(\frac{1}{2}+x_5\right)a\,\hat{\mathbf{x}} + \left(\frac{1}{2}-y_5\right)a\,\hat{\mathbf{y}} - z_5 a\,\hat{\mathbf{z}}$ | (24$d$) | $H_2O$ I |
| $\mathbf{B}_{29}$ | = | $z_5\,\mathbf{a}_1 + x_5\,\mathbf{a}_2 + y_5\,\mathbf{a}_3$ | = | $z_5 a\,\hat{\mathbf{x}} + x_5 a\,\hat{\mathbf{y}} + y_5 a\,\hat{\mathbf{z}}$ | (24$d$) | $H_2O$ I |
| $\mathbf{B}_{30}$ | = | $\left(\frac{1}{2}+z_5\right)\mathbf{a}_1 + \left(\frac{1}{2}-x_5\right)\mathbf{a}_2 - y_5\,\mathbf{a}_3$ | = | $\left(\frac{1}{2}+z_5\right)a\,\hat{\mathbf{x}} + \left(\frac{1}{2}-x_5\right)a\,\hat{\mathbf{y}} - y_5 a\,\hat{\mathbf{z}}$ | (24$d$) | $H_2O$ I |
| $\mathbf{B}_{31}$ | = | $\left(\frac{1}{2}-z_5\right)\mathbf{a}_1 - x_5\,\mathbf{a}_2 + \left(\frac{1}{2}+y_5\right)\mathbf{a}_3$ | = | $\left(\frac{1}{2}-z_5\right)a\,\hat{\mathbf{x}} - x_5 a\,\hat{\mathbf{y}} + \left(\frac{1}{2}+y_5\right)a\,\hat{\mathbf{z}}$ | (24$d$) | $H_2O$ I |
| $\mathbf{B}_{32}$ | = | $-z_5\,\mathbf{a}_1 + \left(\frac{1}{2}+x_5\right)\mathbf{a}_2 + \left(\frac{1}{2}-y_5\right)\mathbf{a}_3$ | = | $-z_5 a\,\hat{\mathbf{x}} + \left(\frac{1}{2}+x_5\right)a\,\hat{\mathbf{y}} + \left(\frac{1}{2}-y_5\right)a\,\hat{\mathbf{z}}$ | (24$d$) | $H_2O$ I |
| $\mathbf{B}_{33}$ | = | $y_5\,\mathbf{a}_1 + z_5\,\mathbf{a}_2 + x_5\,\mathbf{a}_3$ | = | $y_5 a\,\hat{\mathbf{x}} + z_5 a\,\hat{\mathbf{y}} + x_5 a\,\hat{\mathbf{z}}$ | (24$d$) | $H_2O$ I |
| $\mathbf{B}_{34}$ | = | $-y_5\,\mathbf{a}_1 + \left(\frac{1}{2}+z_5\right)\mathbf{a}_2 + \left(\frac{1}{2}-x_5\right)\mathbf{a}_3$ | = | $-y_5 a\,\hat{\mathbf{x}} + \left(\frac{1}{2}+z_5\right)a\,\hat{\mathbf{y}} + \left(\frac{1}{2}-x_5\right)a\,\hat{\mathbf{z}}$ | (24$d$) | $H_2O$ I |
| $\mathbf{B}_{35}$ | = | $\left(\frac{1}{2}+y_5\right)\mathbf{a}_1 + \left(\frac{1}{2}-z_5\right)\mathbf{a}_2 - x_5\,\mathbf{a}_3$ | = | $\left(\frac{1}{2}+y_5\right)a\,\hat{\mathbf{x}} + \left(\frac{1}{2}-z_5\right)a\,\hat{\mathbf{y}} - x_5 a\,\hat{\mathbf{z}}$ | (24$d$) | $H_2O$ I |
| $\mathbf{B}_{36}$ | = | $\left(\frac{1}{2}-y_5\right)\mathbf{a}_1 - z_5\,\mathbf{a}_2 + \left(\frac{1}{2}+x_5\right)\mathbf{a}_3$ | = | $\left(\frac{1}{2}-y_5\right)a\,\hat{\mathbf{x}} - z_5 a\,\hat{\mathbf{y}} + \left(\frac{1}{2}+x_5\right)a\,\hat{\mathbf{z}}$ | (24$d$) | $H_2O$ I |
| $\mathbf{B}_{37}$ | = | $-x_5\,\mathbf{a}_1 - y_5\,\mathbf{a}_2 - z_5\,\mathbf{a}_3$ | = | $-x_5 a\,\hat{\mathbf{x}} - y_5 a\,\hat{\mathbf{y}} - z_5 a\,\hat{\mathbf{z}}$ | (24$d$) | $H_2O$ I |
| $\mathbf{B}_{38}$ | = | $\left(\frac{1}{2}+x_5\right)\mathbf{a}_1 + y_5\,\mathbf{a}_2 + \left(\frac{1}{2}-z_5\right)\mathbf{a}_3$ | = | $\left(\frac{1}{2}+x_5\right)a\,\hat{\mathbf{x}} + y_5 a\,\hat{\mathbf{y}} + \left(\frac{1}{2}-z_5\right)a\,\hat{\mathbf{z}}$ | (24$d$) | $H_2O$ I |
| $\mathbf{B}_{39}$ | = | $x_5\,\mathbf{a}_1 + \left(\frac{1}{2}-y_5\right)\mathbf{a}_2 + \left(\frac{1}{2}+z_5\right)\mathbf{a}_3$ | = | $x_5 a\,\hat{\mathbf{x}} + \left(\frac{1}{2}-y_5\right)a\,\hat{\mathbf{y}} + \left(\frac{1}{2}+z_5\right)a\,\hat{\mathbf{z}}$ | (24$d$) | $H_2O$ I |
| $\mathbf{B}_{40}$ | = | $\left(\frac{1}{2}-x_5\right)\mathbf{a}_1 + \left(\frac{1}{2}+y_5\right)\mathbf{a}_2 + z_5\,\mathbf{a}_3$ | = | $\left(\frac{1}{2}-x_5\right)a\,\hat{\mathbf{x}} + \left(\frac{1}{2}+y_5\right)a\,\hat{\mathbf{y}} + z_5 a\,\hat{\mathbf{z}}$ | (24$d$) | $H_2O$ I |
| $\mathbf{B}_{41}$ | = | $-z_5\,\mathbf{a}_1 - x_5\,\mathbf{a}_2 - y_5\,\mathbf{a}_3$ | = | $-z_5 a\,\hat{\mathbf{x}} - x_5 a\,\hat{\mathbf{y}} - y_5 a\,\hat{\mathbf{z}}$ | (24$d$) | $H_2O$ I |
| $\mathbf{B}_{42}$ | = | $\left(\frac{1}{2}-z_5\right)\mathbf{a}_1 + \left(\frac{1}{2}+x_5\right)\mathbf{a}_2 + y_5\,\mathbf{a}_3$ | = | $\left(\frac{1}{2}-z_5\right)a\,\hat{\mathbf{x}} + \left(\frac{1}{2}+x_5\right)a\,\hat{\mathbf{y}} + y_5 a\,\hat{\mathbf{z}}$ | (24$d$) | $H_2O$ I |
| $\mathbf{B}_{43}$ | = | $\left(\frac{1}{2}+z_5\right)\mathbf{a}_1 + x_5\,\mathbf{a}_2 + \left(\frac{1}{2}-y_5\right)\mathbf{a}_3$ | = | $\left(\frac{1}{2}+z_5\right)a\,\hat{\mathbf{x}} + x_5 a\,\hat{\mathbf{y}} + \left(\frac{1}{2}-y_5\right)a\,\hat{\mathbf{z}}$ | (24$d$) | $H_2O$ I |
| $\mathbf{B}_{44}$ | = | $z_5\,\mathbf{a}_1 + \left(\frac{1}{2}-x_5\right)\mathbf{a}_2 + \left(\frac{1}{2}+y_5\right)\mathbf{a}_3$ | = | $z_5 a\,\hat{\mathbf{x}} + \left(\frac{1}{2}-x_5\right)a\,\hat{\mathbf{y}} + \left(\frac{1}{2}+y_5\right)a\,\hat{\mathbf{z}}$ | (24$d$) | $H_2O$ I |
| $\mathbf{B}_{45}$ | = | $-y_5\,\mathbf{a}_1 - z_5\,\mathbf{a}_2 - x_5\,\mathbf{a}_3$ | = | $-y_5 a\,\hat{\mathbf{x}} - z_5 a\,\hat{\mathbf{y}} - x_5 a\,\hat{\mathbf{z}}$ | (24$d$) | $H_2O$ I |
| $\mathbf{B}_{46}$ | = | $y_5\,\mathbf{a}_1 + \left(\frac{1}{2}-z_5\right)\mathbf{a}_2 + \left(\frac{1}{2}+x_5\right)\mathbf{a}_3$ | = | $y_5 a\,\hat{\mathbf{x}} + \left(\frac{1}{2}-z_5\right)a\,\hat{\mathbf{y}} + \left(\frac{1}{2}+x_5\right)a\,\hat{\mathbf{z}}$ | (24$d$) | $H_2O$ I |
| $\mathbf{B}_{47}$ | = | $\left(\frac{1}{2}-y_5\right)\mathbf{a}_1 + \left(\frac{1}{2}+z_5\right)\mathbf{a}_2 + x_5\,\mathbf{a}_3$ | = | $\left(\frac{1}{2}-y_5\right)a\,\hat{\mathbf{x}} + \left(\frac{1}{2}+z_5\right)a\,\hat{\mathbf{y}} + x_5 a\,\hat{\mathbf{z}}$ | (24$d$) | $H_2O$ I |
| $\mathbf{B}_{48}$ | = | $\left(\frac{1}{2}+y_5\right)\mathbf{a}_1 + z_5\,\mathbf{a}_2 + \left(\frac{1}{2}-x_5\right)\mathbf{a}_3$ | = | $\left(\frac{1}{2}+y_5\right)a\,\hat{\mathbf{x}} + z_5 a\,\hat{\mathbf{y}} + \left(\frac{1}{2}-x_5\right)a\,\hat{\mathbf{z}}$ | (24$d$) | $H_2O$ I |
| $\mathbf{B}_{49}$ | = | $x_6\,\mathbf{a}_1 + y_6\,\mathbf{a}_2 + z_6\,\mathbf{a}_3$ | = | $x_6 a\,\hat{\mathbf{x}} + y_6 a\,\hat{\mathbf{y}} + z_6 a\,\hat{\mathbf{z}}$ | (24$d$) | $H_2O$ II |
| $\mathbf{B}_{50}$ | = | $\left(\frac{1}{2}-x_6\right)\mathbf{a}_1 - y_6\,\mathbf{a}_2 + \left(\frac{1}{2}+z_6\right)\mathbf{a}_3$ | = | $\left(\frac{1}{2}-x_6\right)a\,\hat{\mathbf{x}} - y_6 a\,\hat{\mathbf{y}} + \left(\frac{1}{2}+z_6\right)a\,\hat{\mathbf{z}}$ | (24$d$) | $H_2O$ II |
| $\mathbf{B}_{51}$ | = | $-x_6\,\mathbf{a}_1 + \left(\frac{1}{2}+y_6\right)\mathbf{a}_2 + \left(\frac{1}{2}-z_6\right)\mathbf{a}_3$ | = | $-x_6 a\,\hat{\mathbf{x}} + \left(\frac{1}{2}+y_6\right)a\,\hat{\mathbf{y}} + \left(\frac{1}{2}-z_6\right)a\,\hat{\mathbf{z}}$ | (24$d$) | $H_2O$ II |
| $\mathbf{B}_{52}$ | = | $\left(\frac{1}{2}+x_6\right)\mathbf{a}_1 + \left(\frac{1}{2}-y_6\right)\mathbf{a}_2 - z_6\,\mathbf{a}_3$ | = | $\left(\frac{1}{2}+x_6\right)a\,\hat{\mathbf{x}} + \left(\frac{1}{2}-y_6\right)a\,\hat{\mathbf{y}} - z_6 a\,\hat{\mathbf{z}}$ | (24$d$) | $H_2O$ II |
| $\mathbf{B}_{53}$ | = | $z_6\,\mathbf{a}_1 + x_6\,\mathbf{a}_2 + y_6\,\mathbf{a}_3$ | = | $z_6 a\,\hat{\mathbf{x}} + x_6 a\,\hat{\mathbf{y}} + y_6 a\,\hat{\mathbf{z}}$ | (24$d$) | $H_2O$ II |
| $\mathbf{B}_{54}$ | = | $\left(\frac{1}{2}+z_6\right)\mathbf{a}_1 + \left(\frac{1}{2}-x_6\right)\mathbf{a}_2 - y_6\,\mathbf{a}_3$ | = | $\left(\frac{1}{2}+z_6\right)a\,\hat{\mathbf{x}} + \left(\frac{1}{2}-x_6\right)a\,\hat{\mathbf{y}} - y_6 a\,\hat{\mathbf{z}}$ | (24$d$) | $H_2O$ II |
| $\mathbf{B}_{55}$ | = | $\left(\frac{1}{2} - z_6\right)\mathbf{a}_1 - x_6\,\mathbf{a}_2 + \left(\frac{1}{2} + y_6\right)\mathbf{a}_3$ | = | $\left(\frac{1}{2} - z_6\right)a\,\hat{\mathbf{x}} - x_6 a\,\hat{\mathbf{y}} + \left(\frac{1}{2} + y_6\right)a\,\hat{\mathbf{z}}$ | (24$d$) | $H_2O$ II |
| $\mathbf{B}_{56}$ | = | $-z_6\,\mathbf{a}_1 + \left(\frac{1}{2} + x_6\right)\mathbf{a}_2 + \left(\frac{1}{2} - y_6\right)\mathbf{a}_3$ | = | $-z_6 a\,\hat{\mathbf{x}} + \left(\frac{1}{2} + x_6\right)a\,\hat{\mathbf{y}} + \left(\frac{1}{2} - y_6\right)a\,\hat{\mathbf{z}}$ | (24$d$) | $H_2O$ II |
| $\mathbf{B}_{57}$ | = | $y_6\,\mathbf{a}_1 + z_6\,\mathbf{a}_2 + x_6\,\mathbf{a}_3$ | = | $y_6 a\,\hat{\mathbf{x}} + z_6 a\,\hat{\mathbf{y}} + x_6 a\,\hat{\mathbf{z}}$ | (24$d$) | $H_2O$ II |
| $\mathbf{B}_{58}$ | = | $-y_6\,\mathbf{a}_1 + \left(\frac{1}{2} + z_6\right)\mathbf{a}_2 + \left(\frac{1}{2} - x_6\right)\mathbf{a}_3$ | = | $-y_6 a\,\hat{\mathbf{x}} + \left(\frac{1}{2} + z_6\right)a\,\hat{\mathbf{y}} + \left(\frac{1}{2} - x_6\right)a\,\hat{\mathbf{z}}$ | (24$d$) | $H_2O$ II |
| $\mathbf{B}_{59}$ | = | $\left(\frac{1}{2} + y_6\right)\mathbf{a}_1 + \left(\frac{1}{2} - z_6\right)\mathbf{a}_2 - x_6\,\mathbf{a}_3$ | = | $\left(\frac{1}{2} + y_6\right)a\,\hat{\mathbf{x}} + \left(\frac{1}{2} - z_6\right)a\,\hat{\mathbf{y}} - x_6 a\,\hat{\mathbf{z}}$ | (24$d$) | $H_2O$ II |
| $\mathbf{B}_{60}$ | = | $\left(\frac{1}{2} - y_6\right)\mathbf{a}_1 - z_6\,\mathbf{a}_2 + \left(\frac{1}{2} + x_6\right)\mathbf{a}_3$ | = | $\left(\frac{1}{2} - y_6\right)a\,\hat{\mathbf{x}} - z_6 a\,\hat{\mathbf{y}} + \left(\frac{1}{2} + x_6\right)a\,\hat{\mathbf{z}}$ | (24$d$) | $H_2O$ II |
| $\mathbf{B}_{61}$ | = | $-x_6\,\mathbf{a}_1 - y_6\,\mathbf{a}_2 - z_6\,\mathbf{a}_3$ | = | $-x_6 a\,\hat{\mathbf{x}} - y_6 a\,\hat{\mathbf{y}} - z_6 a\,\hat{\mathbf{z}}$ | (24$d$) | $H_2O$ II |
| $\mathbf{B}_{62}$ | = | $\left(\frac{1}{2} + x_6\right)\mathbf{a}_1 + y_6\,\mathbf{a}_2 + \left(\frac{1}{2} - z_6\right)\mathbf{a}_3$ | = | $\left(\frac{1}{2} + x_6\right)a\,\hat{\mathbf{x}} + y_6 a\,\hat{\mathbf{y}} + \left(\frac{1}{2} - z_6\right)a\,\hat{\mathbf{z}}$ | (24$d$) | $H_2O$ II |
| $\mathbf{B}_{63}$ | = | $x_6\,\mathbf{a}_1 + \left(\frac{1}{2} - y_6\right)\mathbf{a}_2 + \left(\frac{1}{2} + z_6\right)\mathbf{a}_3$ | = | $x_6 a\,\hat{\mathbf{x}} + \left(\frac{1}{2} - y_6\right)a\,\hat{\mathbf{y}} + \left(\frac{1}{2} + z_6\right)a\,\hat{\mathbf{z}}$ | (24$d$) | $H_2O$ II |
| $\mathbf{B}_{64}$ | = | $\left(\frac{1}{2} - x_6\right)\mathbf{a}_1 + \left(\frac{1}{2} + y_6\right)\mathbf{a}_2 + z_6\,\mathbf{a}_3$ | = | $\left(\frac{1}{2} - x_6\right)a\,\hat{\mathbf{x}} + \left(\frac{1}{2} + y_6\right)a\,\hat{\mathbf{y}} + z_6 a\,\hat{\mathbf{z}}$ | (24$d$) | $H_2O$ II |
| $\mathbf{B}_{65}$ | = | $-z_6\,\mathbf{a}_1 - x_6\,\mathbf{a}_2 - y_6\,\mathbf{a}_3$ | = | $-z_6 a\,\hat{\mathbf{x}} - x_6 a\,\hat{\mathbf{y}} - y_6 a\,\hat{\mathbf{z}}$ | (24$d$) | $H_2O$ II |
| $\mathbf{B}_{66}$ | = | $\left(\frac{1}{2} - z_6\right)\mathbf{a}_1 + \left(\frac{1}{2} + x_6\right)\mathbf{a}_2 + y_6\,\mathbf{a}_3$ | = | $\left(\frac{1}{2} - z_6\right)a\,\hat{\mathbf{x}} + \left(\frac{1}{2} + x_6\right)a\,\hat{\mathbf{y}} + y_6 a\,\hat{\mathbf{z}}$ | (24$d$) | $H_2O$ II |
| $\mathbf{B}_{67}$ | = | $\left(\frac{1}{2} + z_6\right)\mathbf{a}_1 + x_6\,\mathbf{a}_2 + \left(\frac{1}{2} - y_6\right)\mathbf{a}_3$ | = | $\left(\frac{1}{2} + z_6\right)a\,\hat{\mathbf{x}} + x_6 a\,\hat{\mathbf{y}} + \left(\frac{1}{2} - y_6\right)a\,\hat{\mathbf{z}}$ | (24$d$) | $H_2O$ II |
| $\mathbf{B}_{68}$ | = | $z_6\,\mathbf{a}_1 + \left(\frac{1}{2} - x_6\right)\mathbf{a}_2 + \left(\frac{1}{2} + y_6\right)\mathbf{a}_3$ | = | $z_6 a\,\hat{\mathbf{x}} + \left(\frac{1}{2} - x_6\right)a\,\hat{\mathbf{y}} + \left(\frac{1}{2} + y_6\right)a\,\hat{\mathbf{z}}$ | (24$d$) | $H_2O$ II |
| $\mathbf{B}_{69}$ | = | $-y_6\,\mathbf{a}_1 - z_6\,\mathbf{a}_2 - x_6\,\mathbf{a}_3$ | = | $-y_6 a\,\hat{\mathbf{x}} - z_6 a\,\hat{\mathbf{y}} - x_6 a\,\hat{\mathbf{z}}$ | (24$d$) | $H_2O$ II |
| $\mathbf{B}_{70}$ | = | $y_6\,\mathbf{a}_1 + \left(\frac{1}{2} - z_6\right)\mathbf{a}_2 + \left(\frac{1}{2} + x_6\right)\mathbf{a}_3$ | = | $y_6 a\,\hat{\mathbf{x}} + \left(\frac{1}{2} - z_6\right)a\,\hat{\mathbf{y}} + \left(\frac{1}{2} + x_6\right)a\,\hat{\mathbf{z}}$ | (24$d$) | $H_2O$ II |
| $\mathbf{B}_{71}$ | = | $\left(\frac{1}{2} - y_6\right)\mathbf{a}_1 + \left(\frac{1}{2} + z_6\right)\mathbf{a}_2 + x_6\,\mathbf{a}_3$ | = | $\left(\frac{1}{2} - y_6\right)a\,\hat{\mathbf{x}} + \left(\frac{1}{2} + z_6\right)a\,\hat{\mathbf{y}} + x_6 a\,\hat{\mathbf{z}}$ | (24$d$) | $H_2O$ II |
| $\mathbf{B}_{72}$ | = | $\left(\frac{1}{2} + y_6\right)\mathbf{a}_1 + z_6\,\mathbf{a}_2 + \left(\frac{1}{2} - x_6\right)\mathbf{a}_3$ | = | $\left(\frac{1}{2} + y_6\right)a\,\hat{\mathbf{x}} + z_6 a\,\hat{\mathbf{y}} + \left(\frac{1}{2} - x_6\right)a\,\hat{\mathbf{z}}$ | (24$d$) | $H_2O$ II |
| $\mathbf{B}_{73}$ | = | $x_7\,\mathbf{a}_1 + y_7\,\mathbf{a}_2 + z_7\,\mathbf{a}_3$ | = | $x_7 a\,\hat{\mathbf{x}} + y_7 a\,\hat{\mathbf{y}} + z_7 a\,\hat{\mathbf{z}}$ | (24$d$) | O II |
| $\mathbf{B}_{74}$ | = | $\left(\frac{1}{2} - x_7\right)\mathbf{a}_1 - y_7\,\mathbf{a}_2 + \left(\frac{1}{2} + z_7\right)\mathbf{a}_3$ | = | $\left(\frac{1}{2} - x_7\right)a\,\hat{\mathbf{x}} - y_7 a\,\hat{\mathbf{y}} + \left(\frac{1}{2} + z_7\right)a\,\hat{\mathbf{z}}$ | (24$d$) | O II |
| $\mathbf{B}_{75}$ | = | $-x_7\,\mathbf{a}_1 + \left(\frac{1}{2} + y_7\right)\mathbf{a}_2 + \left(\frac{1}{2} - z_7\right)\mathbf{a}_3$ | = | $-x_7 a\,\hat{\mathbf{x}} + \left(\frac{1}{2} + y_7\right)a\,\hat{\mathbf{y}} + \left(\frac{1}{2} - z_7\right)a\,\hat{\mathbf{z}}$ | (24$d$) | O II |
| $\mathbf{B}_{76}$ | = | $\left(\frac{1}{2} + x_7\right)\mathbf{a}_1 + \left(\frac{1}{2} - y_7\right)\mathbf{a}_2 - z_7\,\mathbf{a}_3$ | = | $\left(\frac{1}{2} + x_7\right)a\,\hat{\mathbf{x}} + \left(\frac{1}{2} - y_7\right)a\,\hat{\mathbf{y}} - z_7 a\,\hat{\mathbf{z}}$ | (24$d$) | O II |
| $\mathbf{B}_{77}$ | = | $z_7\,\mathbf{a}_1 + x_7\,\mathbf{a}_2 + y_7\,\mathbf{a}_3$ | = | $z_7 a\,\hat{\mathbf{x}} + x_7 a\,\hat{\mathbf{y}} + y_7 a\,\hat{\mathbf{z}}$ | (24$d$) | O II |
| $\mathbf{B}_{78}$ | = | $\left(\frac{1}{2} + z_7\right)\mathbf{a}_1 + \left(\frac{1}{2} - x_7\right)\mathbf{a}_2 - y_7\,\mathbf{a}_3$ | = | $\left(\frac{1}{2} + z_7\right)a\,\hat{\mathbf{x}} + \left(\frac{1}{2} - x_7\right)a\,\hat{\mathbf{y}} - y_7 a\,\hat{\mathbf{z}}$ | (24$d$) | O II |
| $\mathbf{B}_{79}$ | = | $\left(\frac{1}{2} - z_7\right)\mathbf{a}_1 - x_7\,\mathbf{a}_2 + \left(\frac{1}{2} + y_7\right)\mathbf{a}_3$ | = | $\left(\frac{1}{2} - z_7\right)a\,\hat{\mathbf{x}} - x_7 a\,\hat{\mathbf{y}} + \left(\frac{1}{2} + y_7\right)a\,\hat{\mathbf{z}}$ | (24$d$) | O II |
| $\mathbf{B}_{80}$ | = | $-z_7\,\mathbf{a}_1 + \left(\frac{1}{2} + x_7\right)\mathbf{a}_2 + \left(\frac{1}{2} - y_7\right)\mathbf{a}_3$ | = | $-z_7 a\,\hat{\mathbf{x}} + \left(\frac{1}{2} + x_7\right)a\,\hat{\mathbf{y}} + \left(\frac{1}{2} - y_7\right)a\,\hat{\mathbf{z}}$ | (24$d$) | O II |
| $\mathbf{B}_{81}$ | = | $y_7\,\mathbf{a}_1 + z_7\,\mathbf{a}_2 + x_7\,\mathbf{a}_3$ | = | $y_7 a\,\hat{\mathbf{x}} + z_7 a\,\hat{\mathbf{y}} + x_7 a\,\hat{\mathbf{z}}$ | (24$d$) | O II |
| $\mathbf{B}_{82}$ | = | $-y_7\,\mathbf{a}_1 + \left(\frac{1}{2} + z_7\right)\mathbf{a}_2 + \left(\frac{1}{2} - x_7\right)\mathbf{a}_3$ | = | $-y_7 a\,\hat{\mathbf{x}} + \left(\frac{1}{2} + z_7\right)a\,\hat{\mathbf{y}} + \left(\frac{1}{2} - x_7\right)a\,\hat{\mathbf{z}}$ | (24$d$) | O II |
| $\mathbf{B}_{83}$ | = | $\left(\frac{1}{2} + y_7\right)\mathbf{a}_1 + \left(\frac{1}{2} - z_7\right)\mathbf{a}_2 - x_7\,\mathbf{a}_3$ | = | $\left(\frac{1}{2} + y_7\right)a\,\hat{\mathbf{x}} + \left(\frac{1}{2} - z_7\right)a\,\hat{\mathbf{y}} - x_7 a\,\hat{\mathbf{z}}$ | (24$d$) | O II |
| $\mathbf{B}_{84}$ | = | $\left(\frac{1}{2} - y_7\right)\mathbf{a}_1 - z_7\,\mathbf{a}_2 + \left(\frac{1}{2} + x_7\right)\mathbf{a}_3$ | = | $\left(\frac{1}{2} - y_7\right)a\,\hat{\mathbf{x}} - z_7 a\,\hat{\mathbf{y}} + \left(\frac{1}{2} + x_7\right)a\,\hat{\mathbf{z}}$ | (24$d$) | O II |
| $\mathbf{B}_{85}$ | = | $-x_7\,\mathbf{a}_1 - y_7\,\mathbf{a}_2 - z_7\,\mathbf{a}_3$ | = | $-x_7 a\,\hat{\mathbf{x}} - y_7 a\,\hat{\mathbf{y}} - z_7 a\,\hat{\mathbf{z}}$ | (24$d$) | O II |
| $\mathbf{B}_{86}$ | = | $\left(\frac{1}{2} + x_7\right)\mathbf{a}_1 + y_7\,\mathbf{a}_2 + \left(\frac{1}{2} - z_7\right)\mathbf{a}_3$ | = | $\left(\frac{1}{2} + x_7\right)a\,\hat{\mathbf{x}} + y_7 a\,\hat{\mathbf{y}} + \left(\frac{1}{2} - z_7\right)a\,\hat{\mathbf{z}}$ | (24$d$) | O II |
| $\mathbf{B}_{87}$ | = | $x_7\,\mathbf{a}_1 + \left(\frac{1}{2} - y_7\right)\mathbf{a}_2 + \left(\frac{1}{2} + z_7\right)\mathbf{a}_3$ | = | $x_7 a\,\hat{\mathbf{x}} + \left(\frac{1}{2} - y_7\right)a\,\hat{\mathbf{y}} + \left(\frac{1}{2} + z_7\right)a\,\hat{\mathbf{z}}$ | (24$d$) | O II |
| $\mathbf{B}_{88}$ | = | $\left(\frac{1}{2} - x_7\right)\mathbf{a}_1 + \left(\frac{1}{2} + y_7\right)\mathbf{a}_2 + z_7\,\mathbf{a}_3$ | = | $\left(\frac{1}{2} - x_7\right)a\,\hat{\mathbf{x}} + \left(\frac{1}{2} + y_7\right)a\,\hat{\mathbf{y}} + z_7 a\,\hat{\mathbf{z}}$ | (24$d$) | O II |
| $\mathbf{B}_{89}$ | = | $-z_7\,\mathbf{a}_1 - x_7\,\mathbf{a}_2 - y_7\,\mathbf{a}_3$ | = | $-z_7 a\,\hat{\mathbf{x}} - x_7 a\,\hat{\mathbf{y}} - y_7 a\,\hat{\mathbf{z}}$ | (24$d$) | O II |
| $\mathbf{B}_{90}$ | = | $\left(\frac{1}{2} - z_7\right)\mathbf{a}_1 + \left(\frac{1}{2} + x_7\right)\mathbf{a}_2 + y_7\,\mathbf{a}_3$ | = | $\left(\frac{1}{2} - z_7\right)a\,\hat{\mathbf{x}} + \left(\frac{1}{2} + x_7\right)a\,\hat{\mathbf{y}} + y_7 a\,\hat{\mathbf{z}}$ | (24$d$) | O II |
| $\mathbf{B}_{91}$ | = | $\left(\frac{1}{2}+z_7\right)\mathbf{a}_1 + x_7\,\mathbf{a}_2 + \left(\frac{1}{2}-y_7\right)\mathbf{a}_3$ | = | $\left(\frac{1}{2}+z_7\right)a\,\hat{\mathbf{x}} + x_7 a\,\hat{\mathbf{y}} + \left(\frac{1}{2}-y_7\right)a\,\hat{\mathbf{z}}$ | $(24d)$ | O II |
| $\mathbf{B}_{92}$ | = | $z_7\,\mathbf{a}_1 + \left(\frac{1}{2}-x_7\right)\mathbf{a}_2 + \left(\frac{1}{2}+y_7\right)\mathbf{a}_3$ | = | $z_7 a\,\hat{\mathbf{x}} + \left(\frac{1}{2}-x_7\right)a\,\hat{\mathbf{y}} + \left(\frac{1}{2}+y_7\right)a\,\hat{\mathbf{z}}$ | $(24d)$ | O II |
| $\mathbf{B}_{93}$ | = | $-y_7\,\mathbf{a}_1 - z_7\,\mathbf{a}_2 - x_7\,\mathbf{a}_3$ | = | $-y_7 a\,\hat{\mathbf{x}} - z_7 a\,\hat{\mathbf{y}} - x_7 a\,\hat{\mathbf{z}}$ | $(24d)$ | O II |
| $\mathbf{B}_{94}$ | = | $y_7\,\mathbf{a}_1 + \left(\frac{1}{2}-z_7\right)\mathbf{a}_2 + \left(\frac{1}{2}+x_7\right)\mathbf{a}_3$ | = | $y_7 a\,\hat{\mathbf{x}} + \left(\frac{1}{2}-z_7\right)a\,\hat{\mathbf{y}} + \left(\frac{1}{2}+x_7\right)a\,\hat{\mathbf{z}}$ | $(24d)$ | O II |
| $\mathbf{B}_{95}$ | = | $\left(\frac{1}{2}-y_7\right)\mathbf{a}_1 + \left(\frac{1}{2}+z_7\right)\mathbf{a}_2 + x_7\,\mathbf{a}_3$ | = | $\left(\frac{1}{2}-y_7\right)a\,\hat{\mathbf{x}} + \left(\frac{1}{2}+z_7\right)a\,\hat{\mathbf{y}} + x_7 a\,\hat{\mathbf{z}}$ | $(24d)$ | O II |
| $\mathbf{B}_{96}$ | = | $\left(\frac{1}{2}+y_7\right)\mathbf{a}_1 + z_7\,\mathbf{a}_2 + \left(\frac{1}{2}-x_7\right)\mathbf{a}_3$ | = | $\left(\frac{1}{2}+y_7\right)a\,\hat{\mathbf{x}} + z_7 a\,\hat{\mathbf{y}} + \left(\frac{1}{2}-x_7\right)a\,\hat{\mathbf{z}}$ | $(24d)$ | O II |

---

---

**Geometry files:**

- CIF: pp. 1784
- POSCAR: pp. 1784

# $\gamma$-Alum [$AlNa(SO_4)_2\cdot12H_2O$, $H4_{15}$] Structure: AB24CD20E2_cP192_205_a_4d_b_c3d_c

http://aflow.org/prototype-encyclopedia/AB24CD20E2_cP192_205_a_4d_b_c3d_c

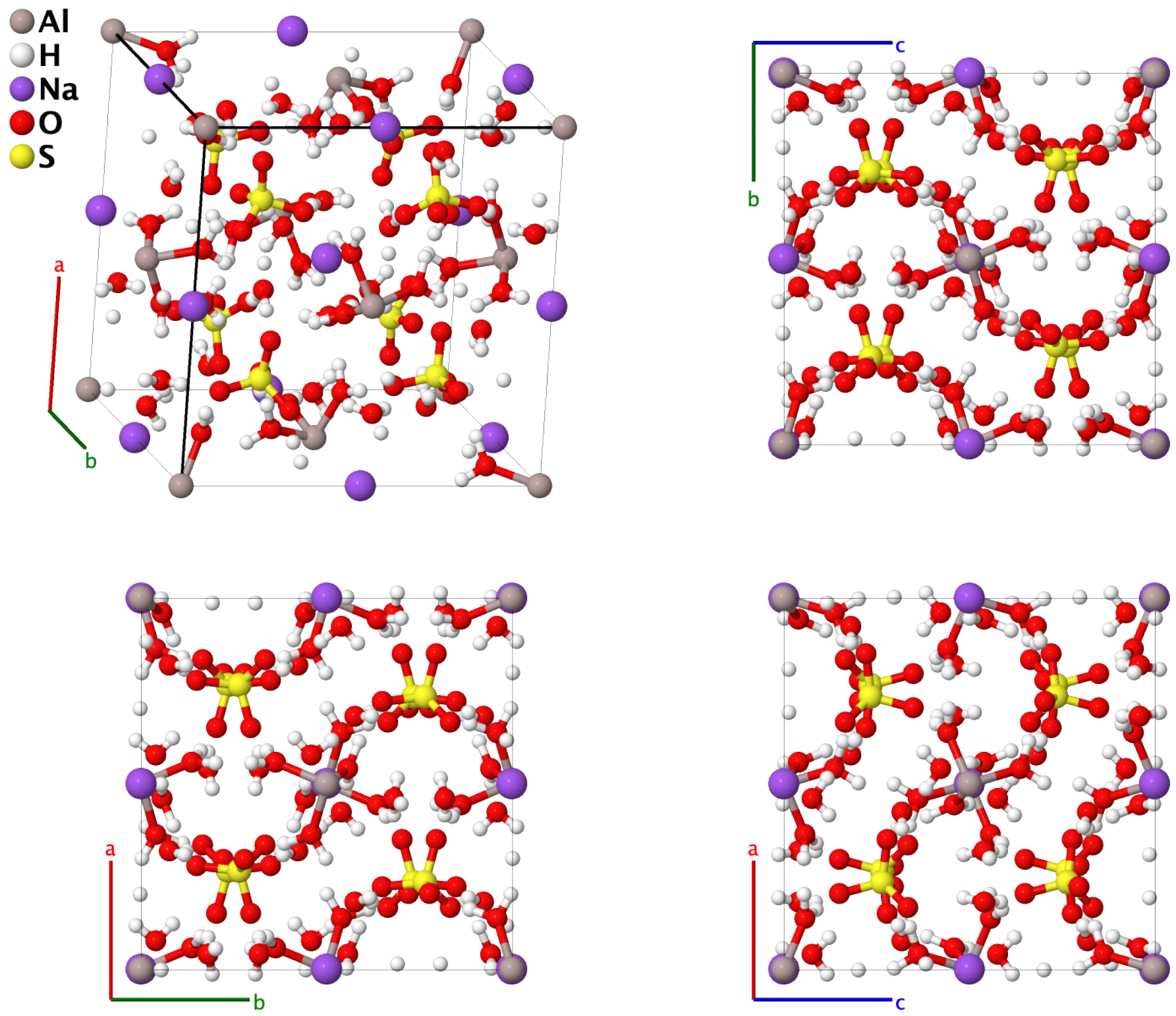

| | | |
|---|---|---|
| **Prototype** | : | $AlH_{24}NaO_{20}S_2$ |
| **AFLOW prototype label** | : | AB24CD20E2_cP192_205_a_4d_b_c3d_c |
| ***Strukturbericht* designation** | : | $H4_{15}$ |
| **Pearson symbol** | : | cP192 |
| **Space group number** | : | 205 |
| **Space group symbol** | : | $Pa\bar{3}$ |
| **AFLOW prototype command** | : | `aflow --proto=AB24CD20E2_cP192_205_a_4d_b_c3d_c` `--params=`$a, x_3, x_4, x_5, y_5, z_5, x_6, y_6, z_6, x_7, y_7, z_7, x_8, y_8, z_8, x_9, y_9, z_9, x_{10}, y_{10}, z_{10}, x_{11}, y_{11}, z_{11}$ |

- The alums have the general formula $AB(XO_4)_2\cdot12H_2O$, where $A$ is a monovalent ion, $B$ is a trivalent ion, and $X$ is a chalcogen. In most cases atom $B$ is aluminum and atom $X$ is sulfur, leading to the name alum.
- All alums have their room-temperature form in space group $Pa\bar{3}$ #205, but the bonding between the $A$ and $B$ ions and the $XO_4$ complex can be quite different.
- (Lipson, 1935ab) described three general forms of alum based on the sizes of the monovalent ions. Each of these forms was given a *Strukturbericht* designation by (Gottfried, 1937):
  - $\alpha$-alum, with intermediate sized ions, prototype $KAl(SO_4)_2\cdot12H_2O$, $H4_{13}$,
  - $\beta$-alum, with large ions, prototype $(NH_3CH_3)Al(SO_4)_2\cdot12H_2O$, $H4_{14}$, and
  - $\gamma$-alum, with small ions, prototype $NaAl(SO_4)_2\cdot12H_2O$, $H4_{15}$ (this structure).

- This classification scheme is not complete, *e.g.*, (Ledsham, 1968) points out that $NaCr(SO_4)_2 \cdot 12H_2O$ does not fit into any of these categories, and that the actual structure depends on the combination of monovalent and trivalent ions.

- As noted above, the $Pa\overline{3}$ structures of alum are the room temperature form. As the temperature decreases the alum structure may transform. For example, in the temperature range 150-170 K the $\beta$-alum $(NH_3CH_3)Al(SO_4)_2 \cdot 12H_2O$ transforms into an orthorhombic structure with fully ordered $NH_3CH_3$ ions.

---

**Simple Cubic primitive vectors:**

$$
\begin{aligned}
\mathbf{a}_1 &= a\,\hat{\mathbf{x}} \\
\mathbf{a}_2 &= a\,\hat{\mathbf{y}} \\
\mathbf{a}_3 &= a\,\hat{\mathbf{z}}
\end{aligned}
$$

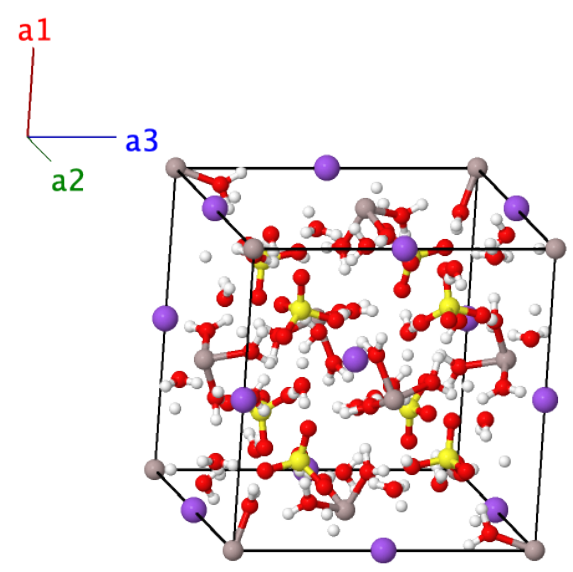

---

**Basis vectors:**

| | | Lattice Coordinates | | Cartesian Coordinates | Wyckoff Position | Atom Type |
|---|---|---|---|---|---|---|
| $\mathbf{B}_1$ | = | $0\,\mathbf{a}_1 + 0\,\mathbf{a}_2 + 0\,\mathbf{a}_3$ | = | $0\,\hat{\mathbf{x}} + 0\,\hat{\mathbf{y}} + 0\,\hat{\mathbf{z}}$ | (4*a*) | Al |
| $\mathbf{B}_2$ | = | $\frac{1}{2}\,\mathbf{a}_1 + \frac{1}{2}\,\mathbf{a}_3$ | = | $\frac{1}{2}a\,\hat{\mathbf{x}} + \frac{1}{2}a\,\hat{\mathbf{z}}$ | (4*a*) | Al |
| $\mathbf{B}_3$ | = | $\frac{1}{2}\,\mathbf{a}_2 + \frac{1}{2}\,\mathbf{a}_3$ | = | $\frac{1}{2}a\,\hat{\mathbf{y}} + \frac{1}{2}a\,\hat{\mathbf{z}}$ | (4*a*) | Al |
| $\mathbf{B}_4$ | = | $\frac{1}{2}\,\mathbf{a}_1 + \frac{1}{2}\,\mathbf{a}_2$ | = | $\frac{1}{2}a\,\hat{\mathbf{x}} + \frac{1}{2}a\,\hat{\mathbf{y}}$ | (4*a*) | Al |
| $\mathbf{B}_5$ | = | $\frac{1}{2}\,\mathbf{a}_1 + \frac{1}{2}\,\mathbf{a}_2 + \frac{1}{2}\,\mathbf{a}_3$ | = | $\frac{1}{2}a\,\hat{\mathbf{x}} + \frac{1}{2}a\,\hat{\mathbf{y}} + \frac{1}{2}a\,\hat{\mathbf{z}}$ | (4*b*) | Na |
| $\mathbf{B}_6$ | = | $\frac{1}{2}\,\mathbf{a}_2$ | = | $\frac{1}{2}a\,\hat{\mathbf{y}}$ | (4*b*) | Na |
| $\mathbf{B}_7$ | = | $\frac{1}{2}\,\mathbf{a}_1$ | = | $\frac{1}{2}a\,\hat{\mathbf{x}}$ | (4*b*) | Na |
| $\mathbf{B}_8$ | = | $\frac{1}{2}\,\mathbf{a}_3$ | = | $\frac{1}{2}a\,\hat{\mathbf{z}}$ | (4*b*) | Na |
| $\mathbf{B}_9$ | = | $x_3\,\mathbf{a}_1 + x_3\,\mathbf{a}_2 + x_3\,\mathbf{a}_3$ | = | $x_3a\,\hat{\mathbf{x}} + x_3a\,\hat{\mathbf{y}} + x_3a\,\hat{\mathbf{z}}$ | (8*c*) | O I |
| $\mathbf{B}_{10}$ | = | $\left(\frac{1}{2} - x_3\right)\,\mathbf{a}_1 - x_3\,\mathbf{a}_2 + \left(\frac{1}{2} + x_3\right)\,\mathbf{a}_3$ | = | $\left(\frac{1}{2} - x_3\right)a\,\hat{\mathbf{x}} - x_3a\,\hat{\mathbf{y}} + \left(\frac{1}{2} + x_3\right)a\,\hat{\mathbf{z}}$ | (8*c*) | O I |
| $\mathbf{B}_{11}$ | = | $-x_3\,\mathbf{a}_1 + \left(\frac{1}{2} + x_3\right)\,\mathbf{a}_2 + \left(\frac{1}{2} - x_3\right)\,\mathbf{a}_3$ | = | $-x_3a\,\hat{\mathbf{x}} + \left(\frac{1}{2} + x_3\right)a\,\hat{\mathbf{y}} + \left(\frac{1}{2} - x_3\right)a\,\hat{\mathbf{z}}$ | (8*c*) | O I |
| $\mathbf{B}_{12}$ | = | $\left(\frac{1}{2} + x_3\right)\,\mathbf{a}_1 + \left(\frac{1}{2} - x_3\right)\,\mathbf{a}_2 - x_3\,\mathbf{a}_3$ | = | $\left(\frac{1}{2} + x_3\right)a\,\hat{\mathbf{x}} + \left(\frac{1}{2} - x_3\right)a\,\hat{\mathbf{y}} - x_3a\,\hat{\mathbf{z}}$ | (8*c*) | O I |
| $\mathbf{B}_{13}$ | = | $-x_3\,\mathbf{a}_1 - x_3\,\mathbf{a}_2 - x_3\,\mathbf{a}_3$ | = | $-x_3a\,\hat{\mathbf{x}} - x_3a\,\hat{\mathbf{y}} - x_3a\,\hat{\mathbf{z}}$ | (8*c*) | O I |
| $\mathbf{B}_{14}$ | = | $\left(\frac{1}{2} + x_3\right)\,\mathbf{a}_1 + x_3\,\mathbf{a}_2 + \left(\frac{1}{2} - x_3\right)\,\mathbf{a}_3$ | = | $\left(\frac{1}{2} + x_3\right)a\,\hat{\mathbf{x}} + x_3a\,\hat{\mathbf{y}} + \left(\frac{1}{2} - x_3\right)a\,\hat{\mathbf{z}}$ | (8*c*) | O I |
| $\mathbf{B}_{15}$ | = | $x_3\,\mathbf{a}_1 + \left(\frac{1}{2} - x_3\right)\,\mathbf{a}_2 + \left(\frac{1}{2} + x_3\right)\,\mathbf{a}_3$ | = | $x_3a\,\hat{\mathbf{x}} + \left(\frac{1}{2} - x_3\right)a\,\hat{\mathbf{y}} + \left(\frac{1}{2} + x_3\right)a\,\hat{\mathbf{z}}$ | (8*c*) | O I |
| $\mathbf{B}_{16}$ | = | $\left(\frac{1}{2} - x_3\right)\,\mathbf{a}_1 + \left(\frac{1}{2} + x_3\right)\,\mathbf{a}_2 + x_3\,\mathbf{a}_3$ | = | $\left(\frac{1}{2} - x_3\right)a\,\hat{\mathbf{x}} + \left(\frac{1}{2} + x_3\right)a\,\hat{\mathbf{y}} + x_3a\,\hat{\mathbf{z}}$ | (8*c*) | O I |
| $\mathbf{B}_{17}$ | = | $x_4\,\mathbf{a}_1 + x_4\,\mathbf{a}_2 + x_4\,\mathbf{a}_3$ | = | $x_4a\,\hat{\mathbf{x}} + x_4a\,\hat{\mathbf{y}} + x_4a\,\hat{\mathbf{z}}$ | (8*c*) | S |
| $\mathbf{B}_{18}$ | = | $\left(\frac{1}{2} - x_4\right)\,\mathbf{a}_1 - x_4\,\mathbf{a}_2 + \left(\frac{1}{2} + x_4\right)\,\mathbf{a}_3$ | = | $\left(\frac{1}{2} - x_4\right)a\,\hat{\mathbf{x}} - x_4a\,\hat{\mathbf{y}} + \left(\frac{1}{2} + x_4\right)a\,\hat{\mathbf{z}}$ | (8*c*) | S |
| $\mathbf{B}_{19}$ | = | $-x_4\,\mathbf{a}_1 + \left(\frac{1}{2} + x_4\right)\,\mathbf{a}_2 + \left(\frac{1}{2} - x_4\right)\,\mathbf{a}_3$ | = | $-x_4a\,\hat{\mathbf{x}} + \left(\frac{1}{2} + x_4\right)a\,\hat{\mathbf{y}} + \left(\frac{1}{2} - x_4\right)a\,\hat{\mathbf{z}}$ | (8*c*) | S |
| $\mathbf{B}_{20}$ | = | $\left(\frac{1}{2} + x_4\right)\,\mathbf{a}_1 + \left(\frac{1}{2} - x_4\right)\,\mathbf{a}_2 - x_4\,\mathbf{a}_3$ | = | $\left(\frac{1}{2} + x_4\right)a\,\hat{\mathbf{x}} + \left(\frac{1}{2} - x_4\right)a\,\hat{\mathbf{y}} - x_4a\,\hat{\mathbf{z}}$ | (8*c*) | S |
| $\mathbf{B}_{21}$ | = | $-x_4\,\mathbf{a}_1 - x_4\,\mathbf{a}_2 - x_4\,\mathbf{a}_3$ | = | $-x_4a\,\hat{\mathbf{x}} - x_4a\,\hat{\mathbf{y}} - x_4a\,\hat{\mathbf{z}}$ | (8*c*) | S |
| $\mathbf{B}_{22}$ | = | $\left(\frac{1}{2} + x_4\right)\,\mathbf{a}_1 + x_4\,\mathbf{a}_2 + \left(\frac{1}{2} - x_4\right)\,\mathbf{a}_3$ | = | $\left(\frac{1}{2} + x_4\right)a\,\hat{\mathbf{x}} + x_4a\,\hat{\mathbf{y}} + \left(\frac{1}{2} - x_4\right)a\,\hat{\mathbf{z}}$ | (8*c*) | S |
| $\mathbf{B}_{23}$ | = | $x_4\,\mathbf{a}_1 + \left(\frac{1}{2} - x_4\right)\,\mathbf{a}_2 + \left(\frac{1}{2} + x_4\right)\,\mathbf{a}_3$ | = | $x_4a\,\hat{\mathbf{x}} + \left(\frac{1}{2} - x_4\right)a\,\hat{\mathbf{y}} + \left(\frac{1}{2} + x_4\right)a\,\hat{\mathbf{z}}$ | (8*c*) | S |
| $\mathbf{B}_{24}$ | = | $\left(\frac{1}{2} - x_4\right)\mathbf{a}_1 + \left(\frac{1}{2} + x_4\right)\mathbf{a}_2 + x_4\,\mathbf{a}_3$ | = | $\left(\frac{1}{2} - x_4\right)a\,\mathbf{\hat{x}} + \left(\frac{1}{2} + x_4\right)a\,\mathbf{\hat{y}} + x_4 a\,\mathbf{\hat{z}}$ | (8$c$) | S |
| $\mathbf{B}_{25}$ | = | $x_5\,\mathbf{a}_1 + y_5\,\mathbf{a}_2 + z_5\,\mathbf{a}_3$ | = | $x_5 a\,\mathbf{\hat{x}} + y_5 a\,\mathbf{\hat{y}} + z_5 a\,\mathbf{\hat{z}}$ | (24$d$) | H I |
| $\mathbf{B}_{26}$ | = | $\left(\frac{1}{2} - x_5\right)\mathbf{a}_1 - y_5\,\mathbf{a}_2 + \left(\frac{1}{2} + z_5\right)\mathbf{a}_3$ | = | $\left(\frac{1}{2} - x_5\right)a\,\mathbf{\hat{x}} - y_5 a\,\mathbf{\hat{y}} + \left(\frac{1}{2} + z_5\right)a\,\mathbf{\hat{z}}$ | (24$d$) | H I |
| $\mathbf{B}_{27}$ | = | $-x_5\,\mathbf{a}_1 + \left(\frac{1}{2} + y_5\right)\mathbf{a}_2 + \left(\frac{1}{2} - z_5\right)\mathbf{a}_3$ | = | $-x_5 a\,\mathbf{\hat{x}} + \left(\frac{1}{2} + y_5\right)a\,\mathbf{\hat{y}} + \left(\frac{1}{2} - z_5\right)a\,\mathbf{\hat{z}}$ | (24$d$) | H I |
| $\mathbf{B}_{28}$ | = | $\left(\frac{1}{2} + x_5\right)\mathbf{a}_1 + \left(\frac{1}{2} - y_5\right)\mathbf{a}_2 - z_5\,\mathbf{a}_3$ | = | $\left(\frac{1}{2} + x_5\right)a\,\mathbf{\hat{x}} + \left(\frac{1}{2} - y_5\right)a\,\mathbf{\hat{y}} - z_5 a\,\mathbf{\hat{z}}$ | (24$d$) | H I |
| $\mathbf{B}_{29}$ | = | $z_5\,\mathbf{a}_1 + x_5\,\mathbf{a}_2 + y_5\,\mathbf{a}_3$ | = | $z_5 a\,\mathbf{\hat{x}} + x_5 a\,\mathbf{\hat{y}} + y_5 a\,\mathbf{\hat{z}}$ | (24$d$) | H I |
| $\mathbf{B}_{30}$ | = | $\left(\frac{1}{2} + z_5\right)\mathbf{a}_1 + \left(\frac{1}{2} - x_5\right)\mathbf{a}_2 - y_5\,\mathbf{a}_3$ | = | $\left(\frac{1}{2} + z_5\right)a\,\mathbf{\hat{x}} + \left(\frac{1}{2} - x_5\right)a\,\mathbf{\hat{y}} - y_5 a\,\mathbf{\hat{z}}$ | (24$d$) | H I |
| $\mathbf{B}_{31}$ | = | $\left(\frac{1}{2} - z_5\right)\mathbf{a}_1 - x_5\,\mathbf{a}_2 + \left(\frac{1}{2} + y_5\right)\mathbf{a}_3$ | = | $\left(\frac{1}{2} - z_5\right)a\,\mathbf{\hat{x}} - x_5 a\,\mathbf{\hat{y}} + \left(\frac{1}{2} + y_5\right)a\,\mathbf{\hat{z}}$ | (24$d$) | H I |
| $\mathbf{B}_{32}$ | = | $-z_5\,\mathbf{a}_1 + \left(\frac{1}{2} + x_5\right)\mathbf{a}_2 + \left(\frac{1}{2} - y_5\right)\mathbf{a}_3$ | = | $-z_5 a\,\mathbf{\hat{x}} + \left(\frac{1}{2} + x_5\right)a\,\mathbf{\hat{y}} + \left(\frac{1}{2} - y_5\right)a\,\mathbf{\hat{z}}$ | (24$d$) | H I |
| $\mathbf{B}_{33}$ | = | $y_5\,\mathbf{a}_1 + z_5\,\mathbf{a}_2 + x_5\,\mathbf{a}_3$ | = | $y_5 a\,\mathbf{\hat{x}} + z_5 a\,\mathbf{\hat{y}} + x_5 a\,\mathbf{\hat{z}}$ | (24$d$) | H I |
| $\mathbf{B}_{34}$ | = | $-y_5\,\mathbf{a}_1 + \left(\frac{1}{2} + z_5\right)\mathbf{a}_2 + \left(\frac{1}{2} - x_5\right)\mathbf{a}_3$ | = | $-y_5 a\,\mathbf{\hat{x}} + \left(\frac{1}{2} + z_5\right)a\,\mathbf{\hat{y}} + \left(\frac{1}{2} - x_5\right)a\,\mathbf{\hat{z}}$ | (24$d$) | H I |
| $\mathbf{B}_{35}$ | = | $\left(\frac{1}{2} + y_5\right)\mathbf{a}_1 + \left(\frac{1}{2} - z_5\right)\mathbf{a}_2 - x_5\,\mathbf{a}_3$ | = | $\left(\frac{1}{2} + y_5\right)a\,\mathbf{\hat{x}} + \left(\frac{1}{2} - z_5\right)a\,\mathbf{\hat{y}} - x_5 a\,\mathbf{\hat{z}}$ | (24$d$) | H I |
| $\mathbf{B}_{36}$ | = | $\left(\frac{1}{2} - y_5\right)\mathbf{a}_1 - z_5\,\mathbf{a}_2 + \left(\frac{1}{2} + x_5\right)\mathbf{a}_3$ | = | $\left(\frac{1}{2} - y_5\right)a\,\mathbf{\hat{x}} - z_5 a\,\mathbf{\hat{y}} + \left(\frac{1}{2} + x_5\right)a\,\mathbf{\hat{z}}$ | (24$d$) | H I |
| $\mathbf{B}_{37}$ | = | $-x_5\,\mathbf{a}_1 - y_5\,\mathbf{a}_2 - z_5\,\mathbf{a}_3$ | = | $-x_5 a\,\mathbf{\hat{x}} - y_5 a\,\mathbf{\hat{y}} - z_5 a\,\mathbf{\hat{z}}$ | (24$d$) | H I |
| $\mathbf{B}_{38}$ | = | $\left(\frac{1}{2} + x_5\right)\mathbf{a}_1 + y_5\,\mathbf{a}_2 + \left(\frac{1}{2} - z_5\right)\mathbf{a}_3$ | = | $\left(\frac{1}{2} + x_5\right)a\,\mathbf{\hat{x}} + y_5 a\,\mathbf{\hat{y}} + \left(\frac{1}{2} - z_5\right)a\,\mathbf{\hat{z}}$ | (24$d$) | H I |
| $\mathbf{B}_{39}$ | = | $x_5\,\mathbf{a}_1 + \left(\frac{1}{2} - y_5\right)\mathbf{a}_2 + \left(\frac{1}{2} + z_5\right)\mathbf{a}_3$ | = | $x_5 a\,\mathbf{\hat{x}} + \left(\frac{1}{2} - y_5\right)a\,\mathbf{\hat{y}} + \left(\frac{1}{2} + z_5\right)a\,\mathbf{\hat{z}}$ | (24$d$) | H I |
| $\mathbf{B}_{40}$ | = | $\left(\frac{1}{2} - x_5\right)\mathbf{a}_1 + \left(\frac{1}{2} + y_5\right)\mathbf{a}_2 + z_5\,\mathbf{a}_3$ | = | $\left(\frac{1}{2} - x_5\right)a\,\mathbf{\hat{x}} + \left(\frac{1}{2} + y_5\right)a\,\mathbf{\hat{y}} + z_5 a\,\mathbf{\hat{z}}$ | (24$d$) | H I |
| $\mathbf{B}_{41}$ | = | $-z_5\,\mathbf{a}_1 - x_5\,\mathbf{a}_2 - y_5\,\mathbf{a}_3$ | = | $-z_5 a\,\mathbf{\hat{x}} - x_5 a\,\mathbf{\hat{y}} - y_5 a\,\mathbf{\hat{z}}$ | (24$d$) | H I |
| $\mathbf{B}_{42}$ | = | $\left(\frac{1}{2} - z_5\right)\mathbf{a}_1 + \left(\frac{1}{2} + x_5\right)\mathbf{a}_2 + y_5\,\mathbf{a}_3$ | = | $\left(\frac{1}{2} - z_5\right)a\,\mathbf{\hat{x}} + \left(\frac{1}{2} + x_5\right)a\,\mathbf{\hat{y}} + y_5 a\,\mathbf{\hat{z}}$ | (24$d$) | H I |
| $\mathbf{B}_{43}$ | = | $\left(\frac{1}{2} + z_5\right)\mathbf{a}_1 + x_5\,\mathbf{a}_2 + \left(\frac{1}{2} - y_5\right)\mathbf{a}_3$ | = | $\left(\frac{1}{2} + z_5\right)a\,\mathbf{\hat{x}} + x_5 a\,\mathbf{\hat{y}} + \left(\frac{1}{2} - y_5\right)a\,\mathbf{\hat{z}}$ | (24$d$) | H I |
| $\mathbf{B}_{44}$ | = | $z_5\,\mathbf{a}_1 + \left(\frac{1}{2} - x_5\right)\mathbf{a}_2 + \left(\frac{1}{2} + y_5\right)\mathbf{a}_3$ | = | $z_5 a\,\mathbf{\hat{x}} + \left(\frac{1}{2} - x_5\right)a\,\mathbf{\hat{y}} + \left(\frac{1}{2} + y_5\right)a\,\mathbf{\hat{z}}$ | (24$d$) | H I |
| $\mathbf{B}_{45}$ | = | $-y_5\,\mathbf{a}_1 - z_5\,\mathbf{a}_2 - x_5\,\mathbf{a}_3$ | = | $-y_5 a\,\mathbf{\hat{x}} - z_5 a\,\mathbf{\hat{y}} - x_5 a\,\mathbf{\hat{z}}$ | (24$d$) | H I |
| $\mathbf{B}_{46}$ | = | $y_5\,\mathbf{a}_1 + \left(\frac{1}{2} - z_5\right)\mathbf{a}_2 + \left(\frac{1}{2} + x_5\right)\mathbf{a}_3$ | = | $y_5 a\,\mathbf{\hat{x}} + \left(\frac{1}{2} - z_5\right)a\,\mathbf{\hat{y}} + \left(\frac{1}{2} + x_5\right)a\,\mathbf{\hat{z}}$ | (24$d$) | H I |
| $\mathbf{B}_{47}$ | = | $\left(\frac{1}{2} - y_5\right)\mathbf{a}_1 + \left(\frac{1}{2} + z_5\right)\mathbf{a}_2 + x_5\,\mathbf{a}_3$ | = | $\left(\frac{1}{2} - y_5\right)a\,\mathbf{\hat{x}} + \left(\frac{1}{2} + z_5\right)a\,\mathbf{\hat{y}} + x_5 a\,\mathbf{\hat{z}}$ | (24$d$) | H I |
| $\mathbf{B}_{48}$ | = | $\left(\frac{1}{2} + y_5\right)\mathbf{a}_1 + z_5\,\mathbf{a}_2 + \left(\frac{1}{2} - x_5\right)\mathbf{a}_3$ | = | $\left(\frac{1}{2} + y_5\right)a\,\mathbf{\hat{x}} + z_5 a\,\mathbf{\hat{y}} + \left(\frac{1}{2} - x_5\right)a\,\mathbf{\hat{z}}$ | (24$d$) | H I |
| $\mathbf{B}_{49}$ | = | $x_6\,\mathbf{a}_1 + y_6\,\mathbf{a}_2 + z_6\,\mathbf{a}_3$ | = | $x_6 a\,\mathbf{\hat{x}} + y_6 a\,\mathbf{\hat{y}} + z_6 a\,\mathbf{\hat{z}}$ | (24$d$) | H II |
| $\mathbf{B}_{50}$ | = | $\left(\frac{1}{2} - x_6\right)\mathbf{a}_1 - y_6\,\mathbf{a}_2 + \left(\frac{1}{2} + z_6\right)\mathbf{a}_3$ | = | $\left(\frac{1}{2} - x_6\right)a\,\mathbf{\hat{x}} - y_6 a\,\mathbf{\hat{y}} + \left(\frac{1}{2} + z_6\right)a\,\mathbf{\hat{z}}$ | (24$d$) | H II |
| $\mathbf{B}_{51}$ | = | $-x_6\,\mathbf{a}_1 + \left(\frac{1}{2} + y_6\right)\mathbf{a}_2 + \left(\frac{1}{2} - z_6\right)\mathbf{a}_3$ | = | $-x_6 a\,\mathbf{\hat{x}} + \left(\frac{1}{2} + y_6\right)a\,\mathbf{\hat{y}} + \left(\frac{1}{2} - z_6\right)a\,\mathbf{\hat{z}}$ | (24$d$) | H II |
| $\mathbf{B}_{52}$ | = | $\left(\frac{1}{2} + x_6\right)\mathbf{a}_1 + \left(\frac{1}{2} - y_6\right)\mathbf{a}_2 - z_6\,\mathbf{a}_3$ | = | $\left(\frac{1}{2} + x_6\right)a\,\mathbf{\hat{x}} + \left(\frac{1}{2} - y_6\right)a\,\mathbf{\hat{y}} - z_6 a\,\mathbf{\hat{z}}$ | (24$d$) | H II |
| $\mathbf{B}_{53}$ | = | $z_6\,\mathbf{a}_1 + x_6\,\mathbf{a}_2 + y_6\,\mathbf{a}_3$ | = | $z_6 a\,\mathbf{\hat{x}} + x_6 a\,\mathbf{\hat{y}} + y_6 a\,\mathbf{\hat{z}}$ | (24$d$) | H II |
| $\mathbf{B}_{54}$ | = | $\left(\frac{1}{2} + z_6\right)\mathbf{a}_1 + \left(\frac{1}{2} - x_6\right)\mathbf{a}_2 - y_6\,\mathbf{a}_3$ | = | $\left(\frac{1}{2} + z_6\right)a\,\mathbf{\hat{x}} + \left(\frac{1}{2} - x_6\right)a\,\mathbf{\hat{y}} - y_6 a\,\mathbf{\hat{z}}$ | (24$d$) | H II |
| $\mathbf{B}_{55}$ | = | $\left(\frac{1}{2} - z_6\right)\mathbf{a}_1 - x_6\,\mathbf{a}_2 + \left(\frac{1}{2} + y_6\right)\mathbf{a}_3$ | = | $\left(\frac{1}{2} - z_6\right)a\,\mathbf{\hat{x}} - x_6 a\,\mathbf{\hat{y}} + \left(\frac{1}{2} + y_6\right)a\,\mathbf{\hat{z}}$ | (24$d$) | H II |
| $\mathbf{B}_{56}$ | = | $-z_6\,\mathbf{a}_1 + \left(\frac{1}{2} + x_6\right)\mathbf{a}_2 + \left(\frac{1}{2} - y_6\right)\mathbf{a}_3$ | = | $-z_6 a\,\mathbf{\hat{x}} + \left(\frac{1}{2} + x_6\right)a\,\mathbf{\hat{y}} + \left(\frac{1}{2} - y_6\right)a\,\mathbf{\hat{z}}$ | (24$d$) | H II |
| $\mathbf{B}_{57}$ | = | $y_6\,\mathbf{a}_1 + z_6\,\mathbf{a}_2 + x_6\,\mathbf{a}_3$ | = | $y_6 a\,\mathbf{\hat{x}} + z_6 a\,\mathbf{\hat{y}} + x_6 a\,\mathbf{\hat{z}}$ | (24$d$) | H II |
| $\mathbf{B}_{58}$ | = | $-y_6\,\mathbf{a}_1 + \left(\frac{1}{2} + z_6\right)\mathbf{a}_2 + \left(\frac{1}{2} - x_6\right)\mathbf{a}_3$ | = | $-y_6 a\,\mathbf{\hat{x}} + \left(\frac{1}{2} + z_6\right)a\,\mathbf{\hat{y}} + \left(\frac{1}{2} - x_6\right)a\,\mathbf{\hat{z}}$ | (24$d$) | H II |
| $\mathbf{B}_{59}$ | = | $\left(\frac{1}{2} + y_6\right)\mathbf{a}_1 + \left(\frac{1}{2} - z_6\right)\mathbf{a}_2 - x_6\,\mathbf{a}_3$ | = | $\left(\frac{1}{2} + y_6\right)a\,\mathbf{\hat{x}} + \left(\frac{1}{2} - z_6\right)a\,\mathbf{\hat{y}} - x_6 a\,\mathbf{\hat{z}}$ | (24$d$) | H II |
| $\mathbf{B}_{60}$ | = | $\left(\frac{1}{2} - y_6\right) \mathbf{a}_1 - z_6 \mathbf{a}_2 + \left(\frac{1}{2} + x_6\right) \mathbf{a}_3$ | = | $\left(\frac{1}{2} - y_6\right) a \hat{\mathbf{x}} - z_6 a \hat{\mathbf{y}} + \left(\frac{1}{2} + x_6\right) a \hat{\mathbf{z}}$ | $(24d)$ | H II |
| $\mathbf{B}_{61}$ | = | $-x_6 \mathbf{a}_1 - y_6 \mathbf{a}_2 - z_6 \mathbf{a}_3$ | = | $-x_6 a \hat{\mathbf{x}} - y_6 a \hat{\mathbf{y}} - z_6 a \hat{\mathbf{z}}$ | $(24d)$ | H II |
| $\mathbf{B}_{62}$ | = | $\left(\frac{1}{2} + x_6\right) \mathbf{a}_1 + y_6 \mathbf{a}_2 + \left(\frac{1}{2} - z_6\right) \mathbf{a}_3$ | = | $\left(\frac{1}{2} + x_6\right) a \hat{\mathbf{x}} + y_6 a \hat{\mathbf{y}} + \left(\frac{1}{2} - z_6\right) a \hat{\mathbf{z}}$ | $(24d)$ | H II |
| $\mathbf{B}_{63}$ | = | $x_6 \mathbf{a}_1 + \left(\frac{1}{2} - y_6\right) \mathbf{a}_2 + \left(\frac{1}{2} + z_6\right) \mathbf{a}_3$ | = | $x_6 a \hat{\mathbf{x}} + \left(\frac{1}{2} - y_6\right) a \hat{\mathbf{y}} + \left(\frac{1}{2} + z_6\right) a \hat{\mathbf{z}}$ | $(24d)$ | H II |
| $\mathbf{B}_{64}$ | = | $\left(\frac{1}{2} - x_6\right) \mathbf{a}_1 + \left(\frac{1}{2} + y_6\right) \mathbf{a}_2 + z_6 \mathbf{a}_3$ | = | $\left(\frac{1}{2} - x_6\right) a \hat{\mathbf{x}} + \left(\frac{1}{2} + y_6\right) a \hat{\mathbf{y}} + z_6 a \hat{\mathbf{z}}$ | $(24d)$ | H II |
| $\mathbf{B}_{65}$ | = | $-z_6 \mathbf{a}_1 - x_6 \mathbf{a}_2 - y_6 \mathbf{a}_3$ | = | $-z_6 a \hat{\mathbf{x}} - x_6 a \hat{\mathbf{y}} - y_6 a \hat{\mathbf{z}}$ | $(24d)$ | H II |
| $\mathbf{B}_{66}$ | = | $\left(\frac{1}{2} - z_6\right) \mathbf{a}_1 + \left(\frac{1}{2} + x_6\right) \mathbf{a}_2 + y_6 \mathbf{a}_3$ | = | $\left(\frac{1}{2} - z_6\right) a \hat{\mathbf{x}} + \left(\frac{1}{2} + x_6\right) a \hat{\mathbf{y}} + y_6 a \hat{\mathbf{z}}$ | $(24d)$ | H II |
| $\mathbf{B}_{67}$ | = | $\left(\frac{1}{2} + z_6\right) \mathbf{a}_1 + x_6 \mathbf{a}_2 + \left(\frac{1}{2} - y_6\right) \mathbf{a}_3$ | = | $\left(\frac{1}{2} + z_6\right) a \hat{\mathbf{x}} + x_6 a \hat{\mathbf{y}} + \left(\frac{1}{2} - y_6\right) a \hat{\mathbf{z}}$ | $(24d)$ | H II |
| $\mathbf{B}_{68}$ | = | $z_6 \mathbf{a}_1 + \left(\frac{1}{2} - x_6\right) \mathbf{a}_2 + \left(\frac{1}{2} + y_6\right) \mathbf{a}_3$ | = | $z_6 a \hat{\mathbf{x}} + \left(\frac{1}{2} - x_6\right) a \hat{\mathbf{y}} + \left(\frac{1}{2} + y_6\right) a \hat{\mathbf{z}}$ | $(24d)$ | H II |
| $\mathbf{B}_{69}$ | = | $-y_6 \mathbf{a}_1 - z_6 \mathbf{a}_2 - x_6 \mathbf{a}_3$ | = | $-y_6 a \hat{\mathbf{x}} - z_6 a \hat{\mathbf{y}} - x_6 a \hat{\mathbf{z}}$ | $(24d)$ | H II |
| $\mathbf{B}_{70}$ | = | $y_6 \mathbf{a}_1 + \left(\frac{1}{2} - z_6\right) \mathbf{a}_2 + \left(\frac{1}{2} + x_6\right) \mathbf{a}_3$ | = | $y_6 a \hat{\mathbf{x}} + \left(\frac{1}{2} - z_6\right) a \hat{\mathbf{y}} + \left(\frac{1}{2} + x_6\right) a \hat{\mathbf{z}}$ | $(24d)$ | H II |
| $\mathbf{B}_{71}$ | = | $\left(\frac{1}{2} - y_6\right) \mathbf{a}_1 + \left(\frac{1}{2} + z_6\right) \mathbf{a}_2 + x_6 \mathbf{a}_3$ | = | $\left(\frac{1}{2} - y_6\right) a \hat{\mathbf{x}} + \left(\frac{1}{2} + z_6\right) a \hat{\mathbf{y}} + x_6 a \hat{\mathbf{z}}$ | $(24d)$ | H II |
| $\mathbf{B}_{72}$ | = | $\left(\frac{1}{2} + y_6\right) \mathbf{a}_1 + z_6 \mathbf{a}_2 + \left(\frac{1}{2} - x_6\right) \mathbf{a}_3$ | = | $\left(\frac{1}{2} + y_6\right) a \hat{\mathbf{x}} + z_6 a \hat{\mathbf{y}} + \left(\frac{1}{2} - x_6\right) a \hat{\mathbf{z}}$ | $(24d)$ | H II |
| $\mathbf{B}_{73}$ | = | $x_7 \mathbf{a}_1 + y_7 \mathbf{a}_2 + z_7 \mathbf{a}_3$ | = | $x_7 a \hat{\mathbf{x}} + y_7 a \hat{\mathbf{y}} + z_7 a \hat{\mathbf{z}}$ | $(24d)$ | H III |
| $\mathbf{B}_{74}$ | = | $\left(\frac{1}{2} - x_7\right) \mathbf{a}_1 - y_7 \mathbf{a}_2 + \left(\frac{1}{2} + z_7\right) \mathbf{a}_3$ | = | $\left(\frac{1}{2} - x_7\right) a \hat{\mathbf{x}} - y_7 a \hat{\mathbf{y}} + \left(\frac{1}{2} + z_7\right) a \hat{\mathbf{z}}$ | $(24d)$ | H III |
| $\mathbf{B}_{75}$ | = | $-x_7 \mathbf{a}_1 + \left(\frac{1}{2} + y_7\right) \mathbf{a}_2 + \left(\frac{1}{2} - z_7\right) \mathbf{a}_3$ | = | $-x_7 a \hat{\mathbf{x}} + \left(\frac{1}{2} + y_7\right) a \hat{\mathbf{y}} + \left(\frac{1}{2} - z_7\right) a \hat{\mathbf{z}}$ | $(24d)$ | H III |
| $\mathbf{B}_{76}$ | = | $\left(\frac{1}{2} + x_7\right) \mathbf{a}_1 + \left(\frac{1}{2} - y_7\right) \mathbf{a}_2 - z_7 \mathbf{a}_3$ | = | $\left(\frac{1}{2} + x_7\right) a \hat{\mathbf{x}} + \left(\frac{1}{2} - y_7\right) a \hat{\mathbf{y}} - z_7 a \hat{\mathbf{z}}$ | $(24d)$ | H III |
| $\mathbf{B}_{77}$ | = | $z_7 \mathbf{a}_1 + x_7 \mathbf{a}_2 + y_7 \mathbf{a}_3$ | = | $z_7 a \hat{\mathbf{x}} + x_7 a \hat{\mathbf{y}} + y_7 a \hat{\mathbf{z}}$ | $(24d)$ | H III |
| $\mathbf{B}_{78}$ | = | $\left(\frac{1}{2} + z_7\right) \mathbf{a}_1 + \left(\frac{1}{2} - x_7\right) \mathbf{a}_2 - y_7 \mathbf{a}_3$ | = | $\left(\frac{1}{2} + z_7\right) a \hat{\mathbf{x}} + \left(\frac{1}{2} - x_7\right) a \hat{\mathbf{y}} - y_7 a \hat{\mathbf{z}}$ | $(24d)$ | H III |
| $\mathbf{B}_{79}$ | = | $\left(\frac{1}{2} - z_7\right) \mathbf{a}_1 - x_7 \mathbf{a}_2 + \left(\frac{1}{2} + y_7\right) \mathbf{a}_3$ | = | $\left(\frac{1}{2} - z_7\right) a \hat{\mathbf{x}} - x_7 a \hat{\mathbf{y}} + \left(\frac{1}{2} + y_7\right) a \hat{\mathbf{z}}$ | $(24d)$ | H III |
| $\mathbf{B}_{80}$ | = | $-z_7 \mathbf{a}_1 + \left(\frac{1}{2} + x_7\right) \mathbf{a}_2 + \left(\frac{1}{2} - y_7\right) \mathbf{a}_3$ | = | $-z_7 a \hat{\mathbf{x}} + \left(\frac{1}{2} + x_7\right) a \hat{\mathbf{y}} + \left(\frac{1}{2} - y_7\right) a \hat{\mathbf{z}}$ | $(24d)$ | H III |
| $\mathbf{B}_{81}$ | = | $y_7 \mathbf{a}_1 + z_7 \mathbf{a}_2 + x_7 \mathbf{a}_3$ | = | $y_7 a \hat{\mathbf{x}} + z_7 a \hat{\mathbf{y}} + x_7 a \hat{\mathbf{z}}$ | $(24d)$ | H III |
| $\mathbf{B}_{82}$ | = | $-y_7 \mathbf{a}_1 + \left(\frac{1}{2} + z_7\right) \mathbf{a}_2 + \left(\frac{1}{2} - x_7\right) \mathbf{a}_3$ | = | $-y_7 a \hat{\mathbf{x}} + \left(\frac{1}{2} + z_7\right) a \hat{\mathbf{y}} + \left(\frac{1}{2} - x_7\right) a \hat{\mathbf{z}}$ | $(24d)$ | H III |
| $\mathbf{B}_{83}$ | = | $\left(\frac{1}{2} + y_7\right) \mathbf{a}_1 + \left(\frac{1}{2} - z_7\right) \mathbf{a}_2 - x_7 \mathbf{a}_3$ | = | $\left(\frac{1}{2} + y_7\right) a \hat{\mathbf{x}} + \left(\frac{1}{2} - z_7\right) a \hat{\mathbf{y}} - x_7 a \hat{\mathbf{z}}$ | $(24d)$ | H III |
| $\mathbf{B}_{84}$ | = | $\left(\frac{1}{2} - y_7\right) \mathbf{a}_1 - z_7 \mathbf{a}_2 + \left(\frac{1}{2} + x_7\right) \mathbf{a}_3$ | = | $\left(\frac{1}{2} - y_7\right) a \hat{\mathbf{x}} - z_7 a \hat{\mathbf{y}} + \left(\frac{1}{2} + x_7\right) a \hat{\mathbf{z}}$ | $(24d)$ | H III |
| $\mathbf{B}_{85}$ | = | $-x_7 \mathbf{a}_1 - y_7 \mathbf{a}_2 - z_7 \mathbf{a}_3$ | = | $-x_7 a \hat{\mathbf{x}} - y_7 a \hat{\mathbf{y}} - z_7 a \hat{\mathbf{z}}$ | $(24d)$ | H III |
| $\mathbf{B}_{86}$ | = | $\left(\frac{1}{2} + x_7\right) \mathbf{a}_1 + y_7 \mathbf{a}_2 + \left(\frac{1}{2} - z_7\right) \mathbf{a}_3$ | = | $\left(\frac{1}{2} + x_7\right) a \hat{\mathbf{x}} + y_7 a \hat{\mathbf{y}} + \left(\frac{1}{2} - z_7\right) a \hat{\mathbf{z}}$ | $(24d)$ | H III |
| $\mathbf{B}_{87}$ | = | $x_7 \mathbf{a}_1 + \left(\frac{1}{2} - y_7\right) \mathbf{a}_2 + \left(\frac{1}{2} + z_7\right) \mathbf{a}_3$ | = | $x_7 a \hat{\mathbf{x}} + \left(\frac{1}{2} - y_7\right) a \hat{\mathbf{y}} + \left(\frac{1}{2} + z_7\right) a \hat{\mathbf{z}}$ | $(24d)$ | H III |
| $\mathbf{B}_{88}$ | = | $\left(\frac{1}{2} - x_7\right) \mathbf{a}_1 + \left(\frac{1}{2} + y_7\right) \mathbf{a}_2 + z_7 \mathbf{a}_3$ | = | $\left(\frac{1}{2} - x_7\right) a \hat{\mathbf{x}} + \left(\frac{1}{2} + y_7\right) a \hat{\mathbf{y}} + z_7 a \hat{\mathbf{z}}$ | $(24d)$ | H III |
| $\mathbf{B}_{89}$ | = | $-z_7 \mathbf{a}_1 - x_7 \mathbf{a}_2 - y_7 \mathbf{a}_3$ | = | $-z_7 a \hat{\mathbf{x}} - x_7 a \hat{\mathbf{y}} - y_7 a \hat{\mathbf{z}}$ | $(24d)$ | H III |
| $\mathbf{B}_{90}$ | = | $\left(\frac{1}{2} - z_7\right) \mathbf{a}_1 + \left(\frac{1}{2} + x_7\right) \mathbf{a}_2 + y_7 \mathbf{a}_3$ | = | $\left(\frac{1}{2} - z_7\right) a \hat{\mathbf{x}} + \left(\frac{1}{2} + x_7\right) a \hat{\mathbf{y}} + y_7 a \hat{\mathbf{z}}$ | $(24d)$ | H III |
| $\mathbf{B}_{91}$ | = | $\left(\frac{1}{2} + z_7\right) \mathbf{a}_1 + x_7 \mathbf{a}_2 + \left(\frac{1}{2} - y_7\right) \mathbf{a}_3$ | = | $\left(\frac{1}{2} + z_7\right) a \hat{\mathbf{x}} + x_7 a \hat{\mathbf{y}} + \left(\frac{1}{2} - y_7\right) a \hat{\mathbf{z}}$ | $(24d)$ | H III |
| $\mathbf{B}_{92}$ | = | $z_7 \mathbf{a}_1 + \left(\frac{1}{2} - x_7\right) \mathbf{a}_2 + \left(\frac{1}{2} + y_7\right) \mathbf{a}_3$ | = | $z_7 a \hat{\mathbf{x}} + \left(\frac{1}{2} - x_7\right) a \hat{\mathbf{y}} + \left(\frac{1}{2} + y_7\right) a \hat{\mathbf{z}}$ | $(24d)$ | H III |
| $\mathbf{B}_{93}$ | = | $-y_7 \mathbf{a}_1 - z_7 \mathbf{a}_2 - x_7 \mathbf{a}_3$ | = | $-y_7 a \hat{\mathbf{x}} - z_7 a \hat{\mathbf{y}} - x_7 a \hat{\mathbf{z}}$ | $(24d)$ | H III |
| $\mathbf{B}_{94}$ | = | $y_7 \mathbf{a}_1 + \left(\frac{1}{2} - z_7\right) \mathbf{a}_2 + \left(\frac{1}{2} + x_7\right) \mathbf{a}_3$ | = | $y_7 a \hat{\mathbf{x}} + \left(\frac{1}{2} - z_7\right) a \hat{\mathbf{y}} + \left(\frac{1}{2} + x_7\right) a \hat{\mathbf{z}}$ | $(24d)$ | H III |
| $\mathbf{B}_{95}$ | = | $\left(\frac{1}{2} - y_7\right) \mathbf{a}_1 + \left(\frac{1}{2} + z_7\right) \mathbf{a}_2 + x_7 \mathbf{a}_3$ | = | $\left(\frac{1}{2} - y_7\right) a \hat{\mathbf{x}} + \left(\frac{1}{2} + z_7\right) a \hat{\mathbf{y}} + x_7 a \hat{\mathbf{z}}$ | $(24d)$ | H III |
| $\mathbf{B}_{96}$ | = | $\left(\frac{1}{2}+y_7\right)\mathbf{a}_1+z_7\,\mathbf{a}_2+\left(\frac{1}{2}-x_7\right)\mathbf{a}_3$ | = | $\left(\frac{1}{2}+y_7\right)a\,\hat{\mathbf{x}}+z_7a\,\hat{\mathbf{y}}+\left(\frac{1}{2}-x_7\right)a\,\hat{\mathbf{z}}$ | (24$d$) | H III |
| $\mathbf{B}_{97}$ | = | $x_8\,\mathbf{a}_1+y_8\,\mathbf{a}_2+z_8\,\mathbf{a}_3$ | = | $x_8a\,\hat{\mathbf{x}}+y_8a\,\hat{\mathbf{y}}+z_8a\,\hat{\mathbf{z}}$ | (24$d$) | H IV |
| $\mathbf{B}_{98}$ | = | $\left(\frac{1}{2}-x_8\right)\mathbf{a}_1-y_8\,\mathbf{a}_2+\left(\frac{1}{2}+z_8\right)\mathbf{a}_3$ | = | $\left(\frac{1}{2}-x_8\right)a\,\hat{\mathbf{x}}-y_8a\,\hat{\mathbf{y}}+\left(\frac{1}{2}+z_8\right)a\,\hat{\mathbf{z}}$ | (24$d$) | H IV |
| $\mathbf{B}_{99}$ | = | $-x_8\,\mathbf{a}_1+\left(\frac{1}{2}+y_8\right)\mathbf{a}_2+\left(\frac{1}{2}-z_8\right)\mathbf{a}_3$ | = | $-x_8a\,\hat{\mathbf{x}}+\left(\frac{1}{2}+y_8\right)a\,\hat{\mathbf{y}}+\left(\frac{1}{2}-z_8\right)a\,\hat{\mathbf{z}}$ | (24$d$) | H IV |
| $\mathbf{B}_{100}$ | = | $\left(\frac{1}{2}+x_8\right)\mathbf{a}_1+\left(\frac{1}{2}-y_8\right)\mathbf{a}_2-z_8\,\mathbf{a}_3$ | = | $\left(\frac{1}{2}+x_8\right)a\,\hat{\mathbf{x}}+\left(\frac{1}{2}-y_8\right)a\,\hat{\mathbf{y}}-z_8a\,\hat{\mathbf{z}}$ | (24$d$) | H IV |
| $\mathbf{B}_{101}$ | = | $z_8\,\mathbf{a}_1+x_8\,\mathbf{a}_2+y_8\,\mathbf{a}_3$ | = | $z_8a\,\hat{\mathbf{x}}+x_8a\,\hat{\mathbf{y}}+y_8a\,\hat{\mathbf{z}}$ | (24$d$) | H IV |
| $\mathbf{B}_{102}$ | = | $\left(\frac{1}{2}+z_8\right)\mathbf{a}_1+\left(\frac{1}{2}-x_8\right)\mathbf{a}_2-y_8\,\mathbf{a}_3$ | = | $\left(\frac{1}{2}+z_8\right)a\,\hat{\mathbf{x}}+\left(\frac{1}{2}-x_8\right)a\,\hat{\mathbf{y}}-y_8a\,\hat{\mathbf{z}}$ | (24$d$) | H IV |
| $\mathbf{B}_{103}$ | = | $\left(\frac{1}{2}-z_8\right)\mathbf{a}_1-x_8\,\mathbf{a}_2+\left(\frac{1}{2}+y_8\right)\mathbf{a}_3$ | = | $\left(\frac{1}{2}-z_8\right)a\,\hat{\mathbf{x}}-x_8a\,\hat{\mathbf{y}}+\left(\frac{1}{2}+y_8\right)a\,\hat{\mathbf{z}}$ | (24$d$) | H IV |
| $\mathbf{B}_{104}$ | = | $-z_8\,\mathbf{a}_1+\left(\frac{1}{2}+x_8\right)\mathbf{a}_2+\left(\frac{1}{2}-y_8\right)\mathbf{a}_3$ | = | $-z_8a\,\hat{\mathbf{x}}+\left(\frac{1}{2}+x_8\right)a\,\hat{\mathbf{y}}+\left(\frac{1}{2}-y_8\right)a\,\hat{\mathbf{z}}$ | (24$d$) | H IV |
| $\mathbf{B}_{105}$ | = | $y_8\,\mathbf{a}_1+z_8\,\mathbf{a}_2+x_8\,\mathbf{a}_3$ | = | $y_8a\,\hat{\mathbf{x}}+z_8a\,\hat{\mathbf{y}}+x_8a\,\hat{\mathbf{z}}$ | (24$d$) | H IV |
| $\mathbf{B}_{106}$ | = | $-y_8\,\mathbf{a}_1+\left(\frac{1}{2}+z_8\right)\mathbf{a}_2+\left(\frac{1}{2}-x_8\right)\mathbf{a}_3$ | = | $-y_8a\,\hat{\mathbf{x}}+\left(\frac{1}{2}+z_8\right)a\,\hat{\mathbf{y}}+\left(\frac{1}{2}-x_8\right)a\,\hat{\mathbf{z}}$ | (24$d$) | H IV |
| $\mathbf{B}_{107}$ | = | $\left(\frac{1}{2}+y_8\right)\mathbf{a}_1+\left(\frac{1}{2}-z_8\right)\mathbf{a}_2-x_8\,\mathbf{a}_3$ | = | $\left(\frac{1}{2}+y_8\right)a\,\hat{\mathbf{x}}+\left(\frac{1}{2}-z_8\right)a\,\hat{\mathbf{y}}-x_8a\,\hat{\mathbf{z}}$ | (24$d$) | H IV |
| $\mathbf{B}_{108}$ | = | $\left(\frac{1}{2}-y_8\right)\mathbf{a}_1-z_8\,\mathbf{a}_2+\left(\frac{1}{2}+x_8\right)\mathbf{a}_3$ | = | $\left(\frac{1}{2}-y_8\right)a\,\hat{\mathbf{x}}-z_8a\,\hat{\mathbf{y}}+\left(\frac{1}{2}+x_8\right)a\,\hat{\mathbf{z}}$ | (24$d$) | H IV |
| $\mathbf{B}_{109}$ | = | $-x_8\,\mathbf{a}_1-y_8\,\mathbf{a}_2-z_8\,\mathbf{a}_3$ | = | $-x_8a\,\hat{\mathbf{x}}-y_8a\,\hat{\mathbf{y}}-z_8a\,\hat{\mathbf{z}}$ | (24$d$) | H IV |
| $\mathbf{B}_{110}$ | = | $\left(\frac{1}{2}+x_8\right)\mathbf{a}_1+y_8\,\mathbf{a}_2+\left(\frac{1}{2}-z_8\right)\mathbf{a}_3$ | = | $\left(\frac{1}{2}+x_8\right)a\,\hat{\mathbf{x}}+y_8a\,\hat{\mathbf{y}}+\left(\frac{1}{2}-z_8\right)a\,\hat{\mathbf{z}}$ | (24$d$) | H IV |
| $\mathbf{B}_{111}$ | = | $x_8\,\mathbf{a}_1+\left(\frac{1}{2}-y_8\right)\mathbf{a}_2+\left(\frac{1}{2}+z_8\right)\mathbf{a}_3$ | = | $x_8a\,\hat{\mathbf{x}}+\left(\frac{1}{2}-y_8\right)a\,\hat{\mathbf{y}}+\left(\frac{1}{2}+z_8\right)a\,\hat{\mathbf{z}}$ | (24$d$) | H IV |
| $\mathbf{B}_{112}$ | = | $\left(\frac{1}{2}-x_8\right)\mathbf{a}_1+\left(\frac{1}{2}+y_8\right)\mathbf{a}_2+z_8\,\mathbf{a}_3$ | = | $\left(\frac{1}{2}-x_8\right)a\,\hat{\mathbf{x}}+\left(\frac{1}{2}+y_8\right)a\,\hat{\mathbf{y}}+z_8a\,\hat{\mathbf{z}}$ | (24$d$) | H IV |
| $\mathbf{B}_{113}$ | = | $-z_8\,\mathbf{a}_1-x_8\,\mathbf{a}_2-y_8\,\mathbf{a}_3$ | = | $-z_8a\,\hat{\mathbf{x}}-x_8a\,\hat{\mathbf{y}}-y_8a\,\hat{\mathbf{z}}$ | (24$d$) | H IV |
| $\mathbf{B}_{114}$ | = | $\left(\frac{1}{2}-z_8\right)\mathbf{a}_1+\left(\frac{1}{2}+x_8\right)\mathbf{a}_2+y_8\,\mathbf{a}_3$ | = | $\left(\frac{1}{2}-z_8\right)a\,\hat{\mathbf{x}}+\left(\frac{1}{2}+x_8\right)a\,\hat{\mathbf{y}}+y_8a\,\hat{\mathbf{z}}$ | (24$d$) | H IV |
| $\mathbf{B}_{115}$ | = | $\left(\frac{1}{2}+z_8\right)\mathbf{a}_1+x_8\,\mathbf{a}_2+\left(\frac{1}{2}-y_8\right)\mathbf{a}_3$ | = | $\left(\frac{1}{2}+z_8\right)a\,\hat{\mathbf{x}}+x_8a\,\hat{\mathbf{y}}+\left(\frac{1}{2}-y_8\right)a\,\hat{\mathbf{z}}$ | (24$d$) | H IV |
| $\mathbf{B}_{116}$ | = | $z_8\,\mathbf{a}_1+\left(\frac{1}{2}-x_8\right)\mathbf{a}_2+\left(\frac{1}{2}+y_8\right)\mathbf{a}_3$ | = | $z_8a\,\hat{\mathbf{x}}+\left(\frac{1}{2}-x_8\right)a\,\hat{\mathbf{y}}+\left(\frac{1}{2}+y_8\right)a\,\hat{\mathbf{z}}$ | (24$d$) | H IV |
| $\mathbf{B}_{117}$ | = | $-y_8\,\mathbf{a}_1-z_8\,\mathbf{a}_2-x_8\,\mathbf{a}_3$ | = | $-y_8a\,\hat{\mathbf{x}}-z_8a\,\hat{\mathbf{y}}-x_8a\,\hat{\mathbf{z}}$ | (24$d$) | H IV |
| $\mathbf{B}_{118}$ | = | $y_8\,\mathbf{a}_1+\left(\frac{1}{2}-z_8\right)\mathbf{a}_2+\left(\frac{1}{2}+x_8\right)\mathbf{a}_3$ | = | $y_8a\,\hat{\mathbf{x}}+\left(\frac{1}{2}-z_8\right)a\,\hat{\mathbf{y}}+\left(\frac{1}{2}+x_8\right)a\,\hat{\mathbf{z}}$ | (24$d$) | H IV |
| $\mathbf{B}_{119}$ | = | $\left(\frac{1}{2}-y_8\right)\mathbf{a}_1+\left(\frac{1}{2}+z_8\right)\mathbf{a}_2+x_8\,\mathbf{a}_3$ | = | $\left(\frac{1}{2}-y_8\right)a\,\hat{\mathbf{x}}+\left(\frac{1}{2}+z_8\right)a\,\hat{\mathbf{y}}+x_8a\,\hat{\mathbf{z}}$ | (24$d$) | H IV |
| $\mathbf{B}_{120}$ | = | $\left(\frac{1}{2}+y_8\right)\mathbf{a}_1+z_8\,\mathbf{a}_2+\left(\frac{1}{2}-x_8\right)\mathbf{a}_3$ | = | $\left(\frac{1}{2}+y_8\right)a\,\hat{\mathbf{x}}+z_8a\,\hat{\mathbf{y}}+\left(\frac{1}{2}-x_8\right)a\,\hat{\mathbf{z}}$ | (24$d$) | H IV |
| $\mathbf{B}_{121}$ | = | $x_9\,\mathbf{a}_1+y_9\,\mathbf{a}_2+z_9\,\mathbf{a}_3$ | = | $x_9a\,\hat{\mathbf{x}}+y_9a\,\hat{\mathbf{y}}+z_9a\,\hat{\mathbf{z}}$ | (24$d$) | O II |
| $\mathbf{B}_{122}$ | = | $\left(\frac{1}{2}-x_9\right)\mathbf{a}_1-y_9\,\mathbf{a}_2+\left(\frac{1}{2}+z_9\right)\mathbf{a}_3$ | = | $\left(\frac{1}{2}-x_9\right)a\,\hat{\mathbf{x}}-y_9a\,\hat{\mathbf{y}}+\left(\frac{1}{2}+z_9\right)a\,\hat{\mathbf{z}}$ | (24$d$) | O II |
| $\mathbf{B}_{123}$ | = | $-x_9\,\mathbf{a}_1+\left(\frac{1}{2}+y_9\right)\mathbf{a}_2+\left(\frac{1}{2}-z_9\right)\mathbf{a}_3$ | = | $-x_9a\,\hat{\mathbf{x}}+\left(\frac{1}{2}+y_9\right)a\,\hat{\mathbf{y}}+\left(\frac{1}{2}-z_9\right)a\,\hat{\mathbf{z}}$ | (24$d$) | O II |
| $\mathbf{B}_{124}$ | = | $\left(\frac{1}{2}+x_9\right)\mathbf{a}_1+\left(\frac{1}{2}-y_9\right)\mathbf{a}_2-z_9\,\mathbf{a}_3$ | = | $\left(\frac{1}{2}+x_9\right)a\,\hat{\mathbf{x}}+\left(\frac{1}{2}-y_9\right)a\,\hat{\mathbf{y}}-z_9a\,\hat{\mathbf{z}}$ | (24$d$) | O II |
| $\mathbf{B}_{125}$ | = | $z_9\,\mathbf{a}_1+x_9\,\mathbf{a}_2+y_9\,\mathbf{a}_3$ | = | $z_9a\,\hat{\mathbf{x}}+x_9a\,\hat{\mathbf{y}}+y_9a\,\hat{\mathbf{z}}$ | (24$d$) | O II |
| $\mathbf{B}_{126}$ | = | $\left(\frac{1}{2}+z_9\right)\mathbf{a}_1+\left(\frac{1}{2}-x_9\right)\mathbf{a}_2-y_9\,\mathbf{a}_3$ | = | $\left(\frac{1}{2}+z_9\right)a\,\hat{\mathbf{x}}+\left(\frac{1}{2}-x_9\right)a\,\hat{\mathbf{y}}-y_9a\,\hat{\mathbf{z}}$ | (24$d$) | O II |
| $\mathbf{B}_{127}$ | = | $\left(\frac{1}{2}-z_9\right)\mathbf{a}_1-x_9\,\mathbf{a}_2+\left(\frac{1}{2}+y_9\right)\mathbf{a}_3$ | = | $\left(\frac{1}{2}-z_9\right)a\,\hat{\mathbf{x}}-x_9a\,\hat{\mathbf{y}}+\left(\frac{1}{2}+y_9\right)a\,\hat{\mathbf{z}}$ | (24$d$) | O II |
| $\mathbf{B}_{128}$ | = | $-z_9\,\mathbf{a}_1+\left(\frac{1}{2}+x_9\right)\mathbf{a}_2+\left(\frac{1}{2}-y_9\right)\mathbf{a}_3$ | = | $-z_9a\,\hat{\mathbf{x}}+\left(\frac{1}{2}+x_9\right)a\,\hat{\mathbf{y}}+\left(\frac{1}{2}-y_9\right)a\,\hat{\mathbf{z}}$ | (24$d$) | O II |
| $\mathbf{B}_{129}$ | = | $y_9\,\mathbf{a}_1+z_9\,\mathbf{a}_2+x_9\,\mathbf{a}_3$ | = | $y_9a\,\hat{\mathbf{x}}+z_9a\,\hat{\mathbf{y}}+x_9a\,\hat{\mathbf{z}}$ | (24$d$) | O II |
| $\mathbf{B}_{130}$ | = | $-y_9\,\mathbf{a}_1+\left(\frac{1}{2}+z_9\right)\mathbf{a}_2+\left(\frac{1}{2}-x_9\right)\mathbf{a}_3$ | = | $-y_9a\,\hat{\mathbf{x}}+\left(\frac{1}{2}+z_9\right)a\,\hat{\mathbf{y}}+\left(\frac{1}{2}-x_9\right)a\,\hat{\mathbf{z}}$ | (24$d$) | O II |
| $\mathbf{B}_{131}$ | = | $\left(\frac{1}{2}+y_9\right)\mathbf{a}_1+\left(\frac{1}{2}-z_9\right)\mathbf{a}_2-x_9\,\mathbf{a}_3$ | = | $\left(\frac{1}{2}+y_9\right)a\,\hat{\mathbf{x}}+\left(\frac{1}{2}-z_9\right)a\,\hat{\mathbf{y}}-x_9a\,\hat{\mathbf{z}}$ | (24$d$) | O II |
| $\mathbf{B}_{132}$ | = | $\left(\frac{1}{2} - y_{9}\right) \mathbf{a}_{1} - z_{9} \mathbf{a}_{2} + \left(\frac{1}{2} + x_{9}\right) \mathbf{a}_{3}$ | = | $\left(\frac{1}{2} - y_{9}\right) a \hat{\mathbf{x}} - z_{9} a \hat{\mathbf{y}} + \left(\frac{1}{2} + x_{9}\right) a \hat{\mathbf{z}}$ | (24*d*) | O II |
| $\mathbf{B}_{133}$ | = | $-x_{9} \mathbf{a}_{1} - y_{9} \mathbf{a}_{2} - z_{9} \mathbf{a}_{3}$ | = | $-x_{9} a \hat{\mathbf{x}} - y_{9} a \hat{\mathbf{y}} - z_{9} a \hat{\mathbf{z}}$ | (24*d*) | O II |
| $\mathbf{B}_{134}$ | = | $\left(\frac{1}{2} + x_{9}\right) \mathbf{a}_{1} + y_{9} \mathbf{a}_{2} + \left(\frac{1}{2} - z_{9}\right) \mathbf{a}_{3}$ | = | $\left(\frac{1}{2} + x_{9}\right) a \hat{\mathbf{x}} + y_{9} a \hat{\mathbf{y}} + \left(\frac{1}{2} - z_{9}\right) a \hat{\mathbf{z}}$ | (24*d*) | O II |
| $\mathbf{B}_{135}$ | = | $x_{9} \mathbf{a}_{1} + \left(\frac{1}{2} - y_{9}\right) \mathbf{a}_{2} + \left(\frac{1}{2} + z_{9}\right) \mathbf{a}_{3}$ | = | $x_{9} a \hat{\mathbf{x}} + \left(\frac{1}{2} - y_{9}\right) a \hat{\mathbf{y}} + \left(\frac{1}{2} + z_{9}\right) a \hat{\mathbf{z}}$ | (24*d*) | O II |
| $\mathbf{B}_{136}$ | = | $\left(\frac{1}{2} - x_{9}\right) \mathbf{a}_{1} + \left(\frac{1}{2} + y_{9}\right) \mathbf{a}_{2} + z_{9} \mathbf{a}_{3}$ | = | $\left(\frac{1}{2} - x_{9}\right) a \hat{\mathbf{x}} + \left(\frac{1}{2} + y_{9}\right) a \hat{\mathbf{y}} + z_{9} a \hat{\mathbf{z}}$ | (24*d*) | O II |
| $\mathbf{B}_{137}$ | = | $-z_{9} \mathbf{a}_{1} - x_{9} \mathbf{a}_{2} - y_{9} \mathbf{a}_{3}$ | = | $-z_{9} a \hat{\mathbf{x}} - x_{9} a \hat{\mathbf{y}} - y_{9} a \hat{\mathbf{z}}$ | (24*d*) | O II |
| $\mathbf{B}_{138}$ | = | $\left(\frac{1}{2} - z_{9}\right) \mathbf{a}_{1} + \left(\frac{1}{2} + x_{9}\right) \mathbf{a}_{2} + y_{9} \mathbf{a}_{3}$ | = | $\left(\frac{1}{2} - z_{9}\right) a \hat{\mathbf{x}} + \left(\frac{1}{2} + x_{9}\right) a \hat{\mathbf{y}} + y_{9} a \hat{\mathbf{z}}$ | (24*d*) | O II |
| $\mathbf{B}_{139}$ | = | $\left(\frac{1}{2} + z_{9}\right) \mathbf{a}_{1} + x_{9} \mathbf{a}_{2} + \left(\frac{1}{2} - y_{9}\right) \mathbf{a}_{3}$ | = | $\left(\frac{1}{2} + z_{9}\right) a \hat{\mathbf{x}} + x_{9} a \hat{\mathbf{y}} + \left(\frac{1}{2} - y_{9}\right) a \hat{\mathbf{z}}$ | (24*d*) | O II |
| $\mathbf{B}_{140}$ | = | $z_{9} \mathbf{a}_{1} + \left(\frac{1}{2} - x_{9}\right) \mathbf{a}_{2} + \left(\frac{1}{2} + y_{9}\right) \mathbf{a}_{3}$ | = | $z_{9} a \hat{\mathbf{x}} + \left(\frac{1}{2} - x_{9}\right) a \hat{\mathbf{y}} + \left(\frac{1}{2} + y_{9}\right) a \hat{\mathbf{z}}$ | (24*d*) | O II |
| $\mathbf{B}_{141}$ | = | $-y_{9} \mathbf{a}_{1} - z_{9} \mathbf{a}_{2} - x_{9} \mathbf{a}_{3}$ | = | $-y_{9} a \hat{\mathbf{x}} - z_{9} a \hat{\mathbf{y}} - x_{9} a \hat{\mathbf{z}}$ | (24*d*) | O II |
| $\mathbf{B}_{142}$ | = | $y_{9} \mathbf{a}_{1} + \left(\frac{1}{2} - z_{9}\right) \mathbf{a}_{2} + \left(\frac{1}{2} + x_{9}\right) \mathbf{a}_{3}$ | = | $y_{9} a \hat{\mathbf{x}} + \left(\frac{1}{2} - z_{9}\right) a \hat{\mathbf{y}} + \left(\frac{1}{2} + x_{9}\right) a \hat{\mathbf{z}}$ | (24*d*) | O II |
| $\mathbf{B}_{143}$ | = | $\left(\frac{1}{2} - y_{9}\right) \mathbf{a}_{1} + \left(\frac{1}{2} + z_{9}\right) \mathbf{a}_{2} + x_{9} \mathbf{a}_{3}$ | = | $\left(\frac{1}{2} - y_{9}\right) a \hat{\mathbf{x}} + \left(\frac{1}{2} + z_{9}\right) a \hat{\mathbf{y}} + x_{9} a \hat{\mathbf{z}}$ | (24*d*) | O II |
| $\mathbf{B}_{144}$ | = | $\left(\frac{1}{2} + y_{9}\right) \mathbf{a}_{1} + z_{9} \mathbf{a}_{2} + \left(\frac{1}{2} - x_{9}\right) \mathbf{a}_{3}$ | = | $\left(\frac{1}{2} + y_{9}\right) a \hat{\mathbf{x}} + z_{9} a \hat{\mathbf{y}} + \left(\frac{1}{2} - x_{9}\right) a \hat{\mathbf{z}}$ | (24*d*) | O II |
| $\mathbf{B}_{145}$ | = | $x_{10} \mathbf{a}_{1} + y_{10} \mathbf{a}_{2} + z_{10} \mathbf{a}_{3}$ | = | $x_{10} a \hat{\mathbf{x}} + y_{10} a \hat{\mathbf{y}} + z_{10} a \hat{\mathbf{z}}$ | (24*d*) | O III |
| $\mathbf{B}_{146}$ | = | $\left(\frac{1}{2} - x_{10}\right) \mathbf{a}_{1} - y_{10} \mathbf{a}_{2} + \left(\frac{1}{2} + z_{10}\right) \mathbf{a}_{3}$ | = | $\left(\frac{1}{2} - x_{10}\right) a \hat{\mathbf{x}} - y_{10} a \hat{\mathbf{y}} + \left(\frac{1}{2} + z_{10}\right) a \hat{\mathbf{z}}$ | (24*d*) | O III |
| $\mathbf{B}_{147}$ | = | $-x_{10} \mathbf{a}_{1} + \left(\frac{1}{2} + y_{10}\right) \mathbf{a}_{2} + \left(\frac{1}{2} - z_{10}\right) \mathbf{a}_{3}$ | = | $-x_{10} a \hat{\mathbf{x}} + \left(\frac{1}{2} + y_{10}\right) a \hat{\mathbf{y}} + \left(\frac{1}{2} - z_{10}\right) a \hat{\mathbf{z}}$ | (24*d*) | O III |
| $\mathbf{B}_{148}$ | = | $\left(\frac{1}{2} + x_{10}\right) \mathbf{a}_{1} + \left(\frac{1}{2} - y_{10}\right) \mathbf{a}_{2} - z_{10} \mathbf{a}_{3}$ | = | $\left(\frac{1}{2} + x_{10}\right) a \hat{\mathbf{x}} + \left(\frac{1}{2} - y_{10}\right) a \hat{\mathbf{y}} - z_{10} a \hat{\mathbf{z}}$ | (24*d*) | O III |
| $\mathbf{B}_{149}$ | = | $z_{10} \mathbf{a}_{1} + x_{10} \mathbf{a}_{2} + y_{10} \mathbf{a}_{3}$ | = | $z_{10} a \hat{\mathbf{x}} + x_{10} a \hat{\mathbf{y}} + y_{10} a \hat{\mathbf{z}}$ | (24*d*) | O III |
| $\mathbf{B}_{150}$ | = | $\left(\frac{1}{2} + z_{10}\right) \mathbf{a}_{1} + \left(\frac{1}{2} - x_{10}\right) \mathbf{a}_{2} - y_{10} \mathbf{a}_{3}$ | = | $\left(\frac{1}{2} + z_{10}\right) a \hat{\mathbf{x}} + \left(\frac{1}{2} - x_{10}\right) a \hat{\mathbf{y}} - y_{10} a \hat{\mathbf{z}}$ | (24*d*) | O III |
| $\mathbf{B}_{151}$ | = | $\left(\frac{1}{2} - z_{10}\right) \mathbf{a}_{1} - x_{10} \mathbf{a}_{2} + \left(\frac{1}{2} + y_{10}\right) \mathbf{a}_{3}$ | = | $\left(\frac{1}{2} - z_{10}\right) a \hat{\mathbf{x}} - x_{10} a \hat{\mathbf{y}} + \left(\frac{1}{2} + y_{10}\right) a \hat{\mathbf{z}}$ | (24*d*) | O III |
| $\mathbf{B}_{152}$ | = | $-z_{10} \mathbf{a}_{1} + \left(\frac{1}{2} + x_{10}\right) \mathbf{a}_{2} + \left(\frac{1}{2} - y_{10}\right) \mathbf{a}_{3}$ | = | $-z_{10} a \hat{\mathbf{x}} + \left(\frac{1}{2} + x_{10}\right) a \hat{\mathbf{y}} + \left(\frac{1}{2} - y_{10}\right) a \hat{\mathbf{z}}$ | (24*d*) | O III |
| $\mathbf{B}_{153}$ | = | $y_{10} \mathbf{a}_{1} + z_{10} \mathbf{a}_{2} + x_{10} \mathbf{a}_{3}$ | = | $y_{10} a \hat{\mathbf{x}} + z_{10} a \hat{\mathbf{y}} + x_{10} a \hat{\mathbf{z}}$ | (24*d*) | O III |
| $\mathbf{B}_{154}$ | = | $-y_{10} \mathbf{a}_{1} + \left(\frac{1}{2} + z_{10}\right) \mathbf{a}_{2} + \left(\frac{1}{2} - x_{10}\right) \mathbf{a}_{3}$ | = | $-y_{10} a \hat{\mathbf{x}} + \left(\frac{1}{2} + z_{10}\right) a \hat{\mathbf{y}} + \left(\frac{1}{2} - x_{10}\right) a \hat{\mathbf{z}}$ | (24*d*) | O III |
| $\mathbf{B}_{155}$ | = | $\left(\frac{1}{2} + y_{10}\right) \mathbf{a}_{1} + \left(\frac{1}{2} - z_{10}\right) \mathbf{a}_{2} - x_{10} \mathbf{a}_{3}$ | = | $\left(\frac{1}{2} + y_{10}\right) a \hat{\mathbf{x}} + \left(\frac{1}{2} - z_{10}\right) a \hat{\mathbf{y}} - x_{10} a \hat{\mathbf{z}}$ | (24*d*) | O III |
| $\mathbf{B}_{156}$ | = | $\left(\frac{1}{2} - y_{10}\right) \mathbf{a}_{1} - z_{10} \mathbf{a}_{2} + \left(\frac{1}{2} + x_{10}\right) \mathbf{a}_{3}$ | = | $\left(\frac{1}{2} - y_{10}\right) a \hat{\mathbf{x}} - z_{10} a \hat{\mathbf{y}} + \left(\frac{1}{2} + x_{10}\right) a \hat{\mathbf{z}}$ | (24*d*) | O III |
| $\mathbf{B}_{157}$ | = | $-x_{10} \mathbf{a}_{1} - y_{10} \mathbf{a}_{2} - z_{10} \mathbf{a}_{3}$ | = | $-x_{10} a \hat{\mathbf{x}} - y_{10} a \hat{\mathbf{y}} - z_{10} a \hat{\mathbf{z}}$ | (24*d*) | O III |
| $\mathbf{B}_{158}$ | = | $\left(\frac{1}{2} + x_{10}\right) \mathbf{a}_{1} + y_{10} \mathbf{a}_{2} + \left(\frac{1}{2} - z_{10}\right) \mathbf{a}_{3}$ | = | $\left(\frac{1}{2} + x_{10}\right) a \hat{\mathbf{x}} + y_{10} a \hat{\mathbf{y}} + \left(\frac{1}{2} - z_{10}\right) a \hat{\mathbf{z}}$ | (24*d*) | O III |
| $\mathbf{B}_{159}$ | = | $x_{10} \mathbf{a}_{1} + \left(\frac{1}{2} - y_{10}\right) \mathbf{a}_{2} + \left(\frac{1}{2} + z_{10}\right) \mathbf{a}_{3}$ | = | $x_{10} a \hat{\mathbf{x}} + \left(\frac{1}{2} - y_{10}\right) a \hat{\mathbf{y}} + \left(\frac{1}{2} + z_{10}\right) a \hat{\mathbf{z}}$ | (24*d*) | O III |
| $\mathbf{B}_{160}$ | = | $\left(\frac{1}{2} - x_{10}\right) \mathbf{a}_{1} + \left(\frac{1}{2} + y_{10}\right) \mathbf{a}_{2} + z_{10} \mathbf{a}_{3}$ | = | $\left(\frac{1}{2} - x_{10}\right) a \hat{\mathbf{x}} + \left(\frac{1}{2} + y_{10}\right) a \hat{\mathbf{y}} + z_{10} a \hat{\mathbf{z}}$ | (24*d*) | O III |
| $\mathbf{B}_{161}$ | = | $-z_{10} \mathbf{a}_{1} - x_{10} \mathbf{a}_{2} - y_{10} \mathbf{a}_{3}$ | = | $-z_{10} a \hat{\mathbf{x}} - x_{10} a \hat{\mathbf{y}} - y_{10} a \hat{\mathbf{z}}$ | (24*d*) | O III |
| $\mathbf{B}_{162}$ | = | $\left(\frac{1}{2} - z_{10}\right) \mathbf{a}_{1} + \left(\frac{1}{2} + x_{10}\right) \mathbf{a}_{2} + y_{10} \mathbf{a}_{3}$ | = | $\left(\frac{1}{2} - z_{10}\right) a \hat{\mathbf{x}} + \left(\frac{1}{2} + x_{10}\right) a \hat{\mathbf{y}} + y_{10} a \hat{\mathbf{z}}$ | (24*d*) | O III |
| $\mathbf{B}_{163}$ | = | $\left(\frac{1}{2} + z_{10}\right) \mathbf{a}_{1} + x_{10} \mathbf{a}_{2} + \left(\frac{1}{2} - y_{10}\right) \mathbf{a}_{3}$ | = | $\left(\frac{1}{2} + z_{10}\right) a \hat{\mathbf{x}} + x_{10} a \hat{\mathbf{y}} + \left(\frac{1}{2} - y_{10}\right) a \hat{\mathbf{z}}$ | (24*d*) | O III |
| $\mathbf{B}_{164}$ | = | $z_{10} \mathbf{a}_{1} + \left(\frac{1}{2} - x_{10}\right) \mathbf{a}_{2} + \left(\frac{1}{2} + y_{10}\right) \mathbf{a}_{3}$ | = | $z_{10} a \hat{\mathbf{x}} + \left(\frac{1}{2} - x_{10}\right) a \hat{\mathbf{y}} + \left(\frac{1}{2} + y_{10}\right) a \hat{\mathbf{z}}$ | (24*d*) | O III |
| $\mathbf{B}_{165}$ | = | $-y_{10} \mathbf{a}_{1} - z_{10} \mathbf{a}_{2} - x_{10} \mathbf{a}_{3}$ | = | $-y_{10} a \hat{\mathbf{x}} - z_{10} a \hat{\mathbf{y}} - x_{10} a \hat{\mathbf{z}}$ | (24*d*) | O III |
| $\mathbf{B}_{166}$ | = | $y_{10}\,\mathbf{a}_1+\left(\frac{1}{2}-z_{10}\right)\mathbf{a}_2+\left(\frac{1}{2}+x_{10}\right)\mathbf{a}_3$ | = | $y_{10}a\,\hat{\mathbf{x}}+\left(\frac{1}{2}-z_{10}\right)a\,\hat{\mathbf{y}}+\left(\frac{1}{2}+x_{10}\right)a\,\hat{\mathbf{z}}$ | (24*d*) | O III |
| $\mathbf{B}_{167}$ | = | $\left(\frac{1}{2}-y_{10}\right)\mathbf{a}_1+\left(\frac{1}{2}+z_{10}\right)\mathbf{a}_2+x_{10}\,\mathbf{a}_3$ | = | $\left(\frac{1}{2}-y_{10}\right)a\,\hat{\mathbf{x}}+\left(\frac{1}{2}+z_{10}\right)a\,\hat{\mathbf{y}}+x_{10}a\,\hat{\mathbf{z}}$ | (24*d*) | O III |
| $\mathbf{B}_{168}$ | = | $\left(\frac{1}{2}+y_{10}\right)\mathbf{a}_1+z_{10}\,\mathbf{a}_2+\left(\frac{1}{2}-x_{10}\right)\mathbf{a}_3$ | = | $\left(\frac{1}{2}+y_{10}\right)a\,\hat{\mathbf{x}}+z_{10}a\,\hat{\mathbf{y}}+\left(\frac{1}{2}-x_{10}\right)a\,\hat{\mathbf{z}}$ | (24*d*) | O III |
| $\mathbf{B}_{169}$ | = | $x_{11}\,\mathbf{a}_1+y_{11}\,\mathbf{a}_2+z_{11}\,\mathbf{a}_3$ | = | $x_{11}a\,\hat{\mathbf{x}}+y_{11}a\,\hat{\mathbf{y}}+z_{11}a\,\hat{\mathbf{z}}$ | (24*d*) | O IV |
| $\mathbf{B}_{170}$ | = | $\left(\frac{1}{2}-x_{11}\right)\mathbf{a}_1-y_{11}\,\mathbf{a}_2+\left(\frac{1}{2}+z_{11}\right)\mathbf{a}_3$ | = | $\left(\frac{1}{2}-x_{11}\right)a\,\hat{\mathbf{x}}-y_{11}a\,\hat{\mathbf{y}}+\left(\frac{1}{2}+z_{11}\right)a\,\hat{\mathbf{z}}$ | (24*d*) | O IV |
| $\mathbf{B}_{171}$ | = | $-x_{11}\,\mathbf{a}_1+\left(\frac{1}{2}+y_{11}\right)\mathbf{a}_2+\left(\frac{1}{2}-z_{11}\right)\mathbf{a}_3$ | = | $-x_{11}a\,\hat{\mathbf{x}}+\left(\frac{1}{2}+y_{11}\right)a\,\hat{\mathbf{y}}+\left(\frac{1}{2}-z_{11}\right)a\,\hat{\mathbf{z}}$ | (24*d*) | O IV |
| $\mathbf{B}_{172}$ | = | $\left(\frac{1}{2}+x_{11}\right)\mathbf{a}_1+\left(\frac{1}{2}-y_{11}\right)\mathbf{a}_2-z_{11}\,\mathbf{a}_3$ | = | $\left(\frac{1}{2}+x_{11}\right)a\,\hat{\mathbf{x}}+\left(\frac{1}{2}-y_{11}\right)a\,\hat{\mathbf{y}}-z_{11}a\,\hat{\mathbf{z}}$ | (24*d*) | O IV |
| $\mathbf{B}_{173}$ | = | $z_{11}\,\mathbf{a}_1+x_{11}\,\mathbf{a}_2+y_{11}\,\mathbf{a}_3$ | = | $z_{11}a\,\hat{\mathbf{x}}+x_{11}a\,\hat{\mathbf{y}}+y_{11}a\,\hat{\mathbf{z}}$ | (24*d*) | O IV |
| $\mathbf{B}_{174}$ | = | $\left(\frac{1}{2}+z_{11}\right)\mathbf{a}_1+\left(\frac{1}{2}-x_{11}\right)\mathbf{a}_2-y_{11}\,\mathbf{a}_3$ | = | $\left(\frac{1}{2}+z_{11}\right)a\,\hat{\mathbf{x}}+\left(\frac{1}{2}-x_{11}\right)a\,\hat{\mathbf{y}}-y_{11}a\,\hat{\mathbf{z}}$ | (24*d*) | O IV |
| $\mathbf{B}_{175}$ | = | $\left(\frac{1}{2}-z_{11}\right)\mathbf{a}_1-x_{11}\,\mathbf{a}_2+\left(\frac{1}{2}+y_{11}\right)\mathbf{a}_3$ | = | $\left(\frac{1}{2}-z_{11}\right)a\,\hat{\mathbf{x}}-x_{11}a\,\hat{\mathbf{y}}+\left(\frac{1}{2}+y_{11}\right)a\,\hat{\mathbf{z}}$ | (24*d*) | O IV |
| $\mathbf{B}_{176}$ | = | $-z_{11}\,\mathbf{a}_1+\left(\frac{1}{2}+x_{11}\right)\mathbf{a}_2+\left(\frac{1}{2}-y_{11}\right)\mathbf{a}_3$ | = | $-z_{11}a\,\hat{\mathbf{x}}+\left(\frac{1}{2}+x_{11}\right)a\,\hat{\mathbf{y}}+\left(\frac{1}{2}-y_{11}\right)a\,\hat{\mathbf{z}}$ | (24*d*) | O IV |
| $\mathbf{B}_{177}$ | = | $y_{11}\,\mathbf{a}_1+z_{11}\,\mathbf{a}_2+x_{11}\,\mathbf{a}_3$ | = | $y_{11}a\,\hat{\mathbf{x}}+z_{11}a\,\hat{\mathbf{y}}+x_{11}a\,\hat{\mathbf{z}}$ | (24*d*) | O IV |
| $\mathbf{B}_{178}$ | = | $-y_{11}\,\mathbf{a}_1+\left(\frac{1}{2}+z_{11}\right)\mathbf{a}_2+\left(\frac{1}{2}-x_{11}\right)\mathbf{a}_3$ | = | $-y_{11}a\,\hat{\mathbf{x}}+\left(\frac{1}{2}+z_{11}\right)a\,\hat{\mathbf{y}}+\left(\frac{1}{2}-x_{11}\right)a\,\hat{\mathbf{z}}$ | (24*d*) | O IV |
| $\mathbf{B}_{179}$ | = | $\left(\frac{1}{2}+y_{11}\right)\mathbf{a}_1+\left(\frac{1}{2}-z_{11}\right)\mathbf{a}_2-x_{11}\,\mathbf{a}_3$ | = | $\left(\frac{1}{2}+y_{11}\right)a\,\hat{\mathbf{x}}+\left(\frac{1}{2}-z_{11}\right)a\,\hat{\mathbf{y}}-x_{11}a\,\hat{\mathbf{z}}$ | (24*d*) | O IV |
| $\mathbf{B}_{180}$ | = | $\left(\frac{1}{2}-y_{11}\right)\mathbf{a}_1-z_{11}\,\mathbf{a}_2+\left(\frac{1}{2}+x_{11}\right)\mathbf{a}_3$ | = | $\left(\frac{1}{2}-y_{11}\right)a\,\hat{\mathbf{x}}-z_{11}a\,\hat{\mathbf{y}}+\left(\frac{1}{2}+x_{11}\right)a\,\hat{\mathbf{z}}$ | (24*d*) | O IV |
| $\mathbf{B}_{181}$ | = | $-x_{11}\,\mathbf{a}_1-y_{11}\,\mathbf{a}_2-z_{11}\,\mathbf{a}_3$ | = | $-x_{11}a\,\hat{\mathbf{x}}-y_{11}a\,\hat{\mathbf{y}}-z_{11}a\,\hat{\mathbf{z}}$ | (24*d*) | O IV |
| $\mathbf{B}_{182}$ | = | $\left(\frac{1}{2}+x_{11}\right)\mathbf{a}_1+y_{11}\,\mathbf{a}_2+\left(\frac{1}{2}-z_{11}\right)\mathbf{a}_3$ | = | $\left(\frac{1}{2}+x_{11}\right)a\,\hat{\mathbf{x}}+y_{11}a\,\hat{\mathbf{y}}+\left(\frac{1}{2}-z_{11}\right)a\,\hat{\mathbf{z}}$ | (24*d*) | O IV |
| $\mathbf{B}_{183}$ | = | $x_{11}\,\mathbf{a}_1+\left(\frac{1}{2}-y_{11}\right)\mathbf{a}_2+\left(\frac{1}{2}+z_{11}\right)\mathbf{a}_3$ | = | $x_{11}a\,\hat{\mathbf{x}}+\left(\frac{1}{2}-y_{11}\right)a\,\hat{\mathbf{y}}+\left(\frac{1}{2}+z_{11}\right)a\,\hat{\mathbf{z}}$ | (24*d*) | O IV |
| $\mathbf{B}_{184}$ | = | $\left(\frac{1}{2}-x_{11}\right)\mathbf{a}_1+\left(\frac{1}{2}+y_{11}\right)\mathbf{a}_2+z_{11}\,\mathbf{a}_3$ | = | $\left(\frac{1}{2}-x_{11}\right)a\,\hat{\mathbf{x}}+\left(\frac{1}{2}+y_{11}\right)a\,\hat{\mathbf{y}}+z_{11}a\,\hat{\mathbf{z}}$ | (24*d*) | O IV |
| $\mathbf{B}_{185}$ | = | $-z_{11}\,\mathbf{a}_1-x_{11}\,\mathbf{a}_2-y_{11}\,\mathbf{a}_3$ | = | $-z_{11}a\,\hat{\mathbf{x}}-x_{11}a\,\hat{\mathbf{y}}-y_{11}a\,\hat{\mathbf{z}}$ | (24*d*) | O IV |
| $\mathbf{B}_{186}$ | = | $\left(\frac{1}{2}-z_{11}\right)\mathbf{a}_1+\left(\frac{1}{2}+x_{11}\right)\mathbf{a}_2+y_{11}\,\mathbf{a}_3$ | = | $\left(\frac{1}{2}-z_{11}\right)a\,\hat{\mathbf{x}}+\left(\frac{1}{2}+x_{11}\right)a\,\hat{\mathbf{y}}+y_{11}a\,\hat{\mathbf{z}}$ | (24*d*) | O IV |
| $\mathbf{B}_{187}$ | = | $\left(\frac{1}{2}+z_{11}\right)\mathbf{a}_1+x_{11}\,\mathbf{a}_2+\left(\frac{1}{2}-y_{11}\right)\mathbf{a}_3$ | = | $\left(\frac{1}{2}+z_{11}\right)a\,\hat{\mathbf{x}}+x_{11}a\,\hat{\mathbf{y}}+\left(\frac{1}{2}-y_{11}\right)a\,\hat{\mathbf{z}}$ | (24*d*) | O IV |
| $\mathbf{B}_{188}$ | = | $z_{11}\,\mathbf{a}_1+\left(\frac{1}{2}-x_{11}\right)\mathbf{a}_2+\left(\frac{1}{2}+y_{11}\right)\mathbf{a}_3$ | = | $z_{11}a\,\hat{\mathbf{x}}+\left(\frac{1}{2}-x_{11}\right)a\,\hat{\mathbf{y}}+\left(\frac{1}{2}+y_{11}\right)a\,\hat{\mathbf{z}}$ | (24*d*) | O IV |
| $\mathbf{B}_{189}$ | = | $-y_{11}\,\mathbf{a}_1-z_{11}\,\mathbf{a}_2-x_{11}\,\mathbf{a}_3$ | = | $-y_{11}a\,\hat{\mathbf{x}}-z_{11}a\,\hat{\mathbf{y}}-x_{11}a\,\hat{\mathbf{z}}$ | (24*d*) | O IV |
| $\mathbf{B}_{190}$ | = | $y_{11}\,\mathbf{a}_1+\left(\frac{1}{2}-z_{11}\right)\mathbf{a}_2+\left(\frac{1}{2}+x_{11}\right)\mathbf{a}_3$ | = | $y_{11}a\,\hat{\mathbf{x}}+\left(\frac{1}{2}-z_{11}\right)a\,\hat{\mathbf{y}}+\left(\frac{1}{2}+x_{11}\right)a\,\hat{\mathbf{z}}$ | (24*d*) | O IV |
| $\mathbf{B}_{191}$ | = | $\left(\frac{1}{2}-y_{11}\right)\mathbf{a}_1+\left(\frac{1}{2}+z_{11}\right)\mathbf{a}_2+x_{11}\,\mathbf{a}_3$ | = | $\left(\frac{1}{2}-y_{11}\right)a\,\hat{\mathbf{x}}+\left(\frac{1}{2}+z_{11}\right)a\,\hat{\mathbf{y}}+x_{11}a\,\hat{\mathbf{z}}$ | (24*d*) | O IV |
| $\mathbf{B}_{192}$ | = | $\left(\frac{1}{2}+y_{11}\right)\mathbf{a}_1+z_{11}\,\mathbf{a}_2+\left(\frac{1}{2}-x_{11}\right)\mathbf{a}_3$ | = | $\left(\frac{1}{2}+y_{11}\right)a\,\hat{\mathbf{x}}+z_{11}a\,\hat{\mathbf{y}}+\left(\frac{1}{2}-x_{11}\right)a\,\hat{\mathbf{z}}$ | (24*d*) | O IV |

**Found in:**

---

**Geometry files:**

- CIF: pp. 1785
- POSCAR: pp. 1785

# $\alpha$-Alum [$KAl(SO_4)_2 \cdot 12H_2O$, $H4_{13}$] Structure: AB24CD28E2_cP224_205_a_4d_b_2c4d_c

http://aflow.org/prototype-encyclopedia/AB24CD28E2_cP224_205_a_4d_b_2c4d_c

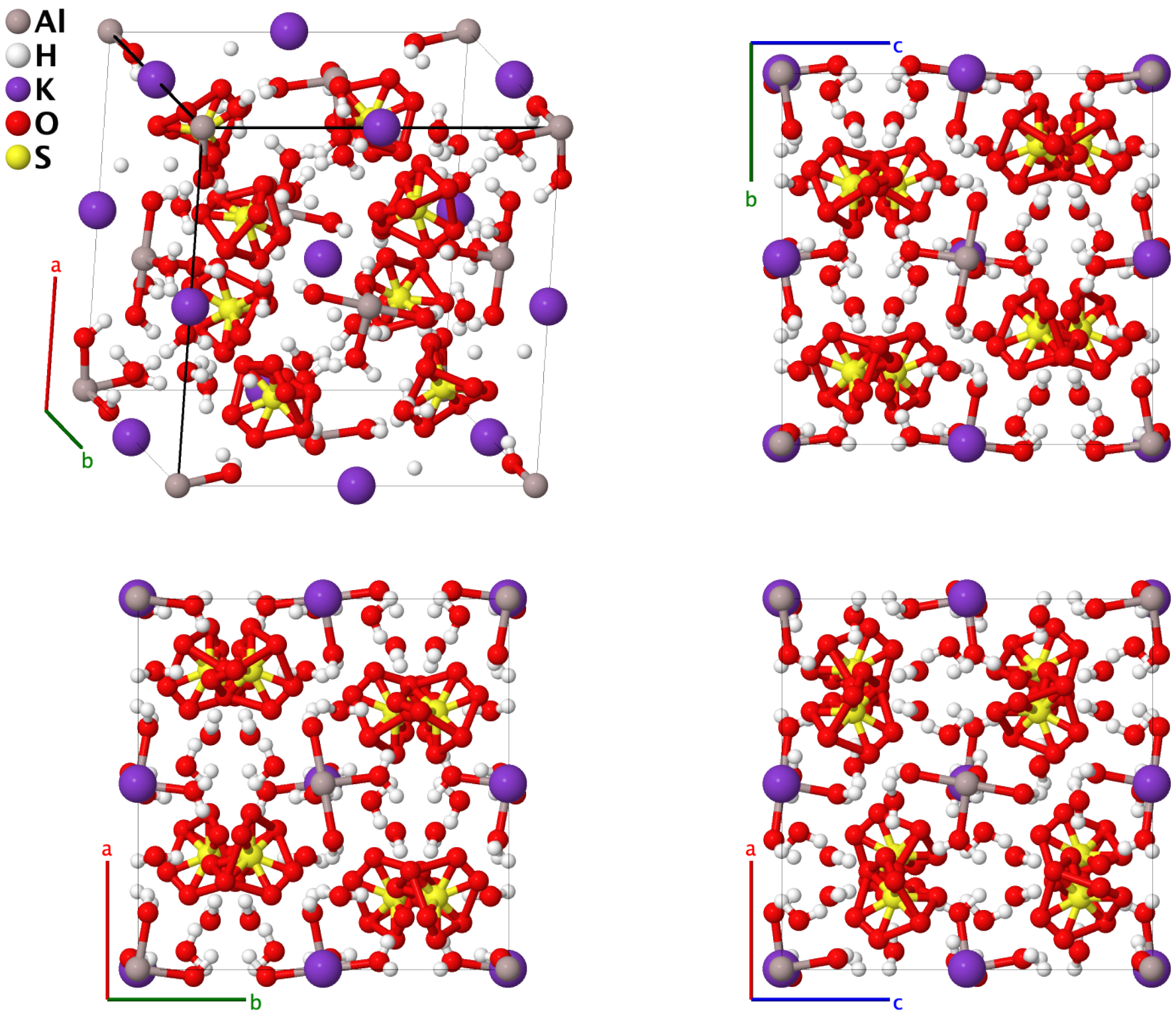

| | | |
|---|---|---|
| **Prototype** | : | $AlH_{24}KO_{20}S_2$ |
| **AFLOW prototype label** | : | AB24CD28E2_cP224_205_a_4d_b_2c4d_c |
| ***Strukturbericht* designation** | : | $H4_{13}$ |
| **Pearson symbol** | : | cP224 |
| **Space group number** | : | 205 |
| **Space group symbol** | : | $Pa\bar{3}$ |
| **AFLOW prototype command** | : | `aflow --proto=AB24CD28E2_cP224_205_a_4d_b_2c4d_c` `--params=`$a, x_3, x_4, x_5, x_6, y_6, z_6, x_7, y_7, z_7, x_8, y_8, z_8, x_9, y_9, z_9, x_{10}, y_{10}, z_{10}, x_{11}, y_{11}, z_{11}, x_{12}, y_{12}, z_{12}, x_{13}, y_{13}, z_{13}$ |

**Other compounds with this structure**

- $NH_4Al(SO_4)_2{\cdot}12H_2O$, $KCr(SO_4)_2{\cdot}12H_2O$, $RbAl(SO_4)_2{\cdot}12H_2O$, and $KAl(SeO_4)_2{\cdot}12H_2O$

- The alums have the general formula $AB(XO_4)_2{\cdot}12H_2O$, where $A$ is a monovalent ion, $B$ is a trivalent ion, and $X$ is a chalcogen. In most cases atom $B$ is aluminum and atom $X$ is sulfur, leading to the name alum.
- All alums have their room-temperature form in space group $Pa\bar{3}$ #205, but the bonding between the $A$ and $B$ ions and the $XO_4$ complex can be quite different.
- (Lipson, 1935ab) described three general forms of alum based on the sizes of the monovalent ions. Each of these forms was given a *Strukturbericht* designation by (Gottfried, 1937):
  - $\alpha$-alum, with intermediate sized ions, prototype $KAl(SO_4)_2{\cdot}12H_2O$, $H4_{13}$ (this structure),

– [$\beta$-alum](), with large ions, prototype $(NH_3CH_3)Al(SO_4)_2{\cdot}12H_2O$, $H4_{14}$, and
– $\gamma$-alum, with small ions, prototype $NaAl(SO_4)_2{\cdot}12H_2O$, $H4_{15}$.

- This classification scheme is not complete, *e.g.*, (Ledsham, 1968) points out that $NaCr(SO_4)_2{\cdot}12H_2O$ does not fit into any of these categories, and that the actual structure depends on the combination of monovalent and trivalent ions.
- As noted above, the $Pa\overline{3}$ structures of alum are the room temperature form. As the temperature decreases the alum structure may transform. For example, in the temperature range 150-170 K the $\beta$-alum $(NH_3CH_3)Al(SO_4)_2{\cdot}12H_2O$ transforms into an orthorhombic structure with fully ordered $NH_3CH_3$ ions.
- (Ewald, 1931) designated the $\alpha$-alum structure determined by (Cork, 1927) as $H4_2$, but when the superior structural determination of (Beevers, 1934) appeared this type was abandoned, and the new structure was given the designation $H4_{13}$. (Nyburg, 2000) were able to determine the positions of the hydrogen atoms, so we use their improved structure as our prototype.
- The oxygen atoms around the sulfur are statistically distributed in two ways: the probability of the O-I and O-III sites being occupied is 78.7%, while the probability of the O-II/O-IV combination being occupied is 21.3%.

---

**Simple Cubic primitive vectors:**

$$\mathbf{a}_1 = a\,\hat{\mathbf{x}}$$
$$\mathbf{a}_2 = a\,\hat{\mathbf{y}}$$
$$\mathbf{a}_3 = a\,\hat{\mathbf{z}}$$

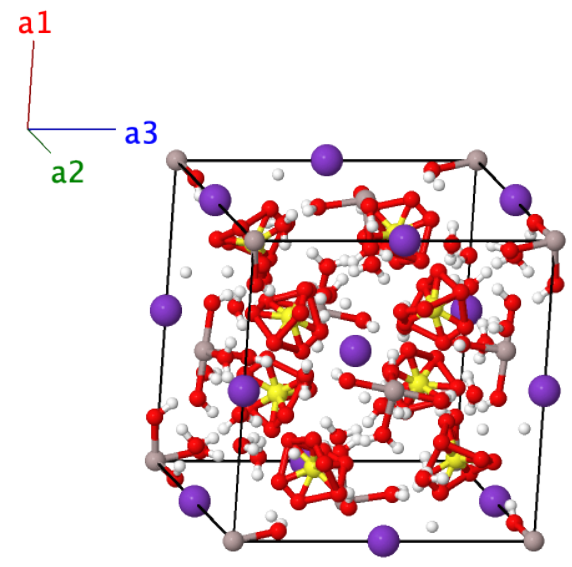

---

**Basis vectors:**

| | | Lattice Coordinates | | Cartesian Coordinates | Wyckoff Position | Atom Type |
|---|---|---|---|---|---|---|
| $\mathbf{B}_1$ | = | $0\,\mathbf{a}_1 + 0\,\mathbf{a}_2 + 0\,\mathbf{a}_3$ | = | $0\,\hat{\mathbf{x}} + 0\,\hat{\mathbf{y}} + 0\,\hat{\mathbf{z}}$ | (4$a$) | Al |
| $\mathbf{B}_2$ | = | $\frac{1}{2}\,\mathbf{a}_1 + \frac{1}{2}\,\mathbf{a}_3$ | = | $\frac{1}{2}a\,\hat{\mathbf{x}} + \frac{1}{2}a\,\hat{\mathbf{z}}$ | (4$a$) | Al |
| $\mathbf{B}_3$ | = | $\frac{1}{2}\,\mathbf{a}_2 + \frac{1}{2}\,\mathbf{a}_3$ | = | $\frac{1}{2}a\,\hat{\mathbf{y}} + \frac{1}{2}a\,\hat{\mathbf{z}}$ | (4$a$) | Al |
| $\mathbf{B}_4$ | = | $\frac{1}{2}\,\mathbf{a}_1 + \frac{1}{2}\,\mathbf{a}_2$ | = | $\frac{1}{2}a\,\hat{\mathbf{x}} + \frac{1}{2}a\,\hat{\mathbf{y}}$ | (4$a$) | Al |
| $\mathbf{B}_5$ | = | $\frac{1}{2}\,\mathbf{a}_1 + \frac{1}{2}\,\mathbf{a}_2 + \frac{1}{2}\,\mathbf{a}_3$ | = | $\frac{1}{2}a\,\hat{\mathbf{x}} + \frac{1}{2}a\,\hat{\mathbf{y}} + \frac{1}{2}a\,\hat{\mathbf{z}}$ | (4$b$) | K |
| $\mathbf{B}_6$ | = | $\frac{1}{2}\,\mathbf{a}_2$ | = | $\frac{1}{2}a\,\hat{\mathbf{y}}$ | (4$b$) | K |
| $\mathbf{B}_7$ | = | $\frac{1}{2}\,\mathbf{a}_1$ | = | $\frac{1}{2}a\,\hat{\mathbf{x}}$ | (4$b$) | K |
| $\mathbf{B}_8$ | = | $\frac{1}{2}\,\mathbf{a}_3$ | = | $\frac{1}{2}a\,\hat{\mathbf{z}}$ | (4$b$) | K |
| $\mathbf{B}_9$ | = | $x_3\,\mathbf{a}_1 + x_3\,\mathbf{a}_2 + x_3\,\mathbf{a}_3$ | = | $x_3a\,\hat{\mathbf{x}} + x_3a\,\hat{\mathbf{y}} + x_3a\,\hat{\mathbf{z}}$ | (8$c$) | O I |
| $\mathbf{B}_{10}$ | = | $\left(\frac{1}{2} - x_3\right)\mathbf{a}_1 - x_3\,\mathbf{a}_2 + \left(\frac{1}{2} + x_3\right)\mathbf{a}_3$ | = | $\left(\frac{1}{2} - x_3\right)a\,\hat{\mathbf{x}} - x_3a\,\hat{\mathbf{y}} + \left(\frac{1}{2} + x_3\right)a\,\hat{\mathbf{z}}$ | (8$c$) | O I |
| $\mathbf{B}_{11}$ | = | $-x_3\,\mathbf{a}_1 + \left(\frac{1}{2} + x_3\right)\mathbf{a}_2 + \left(\frac{1}{2} - x_3\right)\mathbf{a}_3$ | = | $-x_3a\,\hat{\mathbf{x}} + \left(\frac{1}{2} + x_3\right)a\,\hat{\mathbf{y}} + \left(\frac{1}{2} - x_3\right)a\,\hat{\mathbf{z}}$ | (8$c$) | O I |
| $\mathbf{B}_{12}$ | = | $\left(\frac{1}{2} + x_3\right)\mathbf{a}_1 + \left(\frac{1}{2} - x_3\right)\mathbf{a}_2 - x_3\,\mathbf{a}_3$ | = | $\left(\frac{1}{2} + x_3\right)a\,\hat{\mathbf{x}} + \left(\frac{1}{2} - x_3\right)a\,\hat{\mathbf{y}} - x_3a\,\hat{\mathbf{z}}$ | (8$c$) | O I |
| $\mathbf{B}_{13}$ | = | $-x_3\,\mathbf{a}_1 - x_3\,\mathbf{a}_2 - x_3\,\mathbf{a}_3$ | = | $-x_3a\,\hat{\mathbf{x}} - x_3a\,\hat{\mathbf{y}} - x_3a\,\hat{\mathbf{z}}$ | (8$c$) | O I |
| $\mathbf{B}_{14}$ | = | $\left(\frac{1}{2} + x_3\right)\mathbf{a}_1 + x_3\,\mathbf{a}_2 + \left(\frac{1}{2} - x_3\right)\mathbf{a}_3$ | = | $\left(\frac{1}{2} + x_3\right)a\,\hat{\mathbf{x}} + x_3a\,\hat{\mathbf{y}} + \left(\frac{1}{2} - x_3\right)a\,\hat{\mathbf{z}}$ | (8$c$) | O I |
| $\mathbf{B}_{15}$ | = | $x_3\,\mathbf{a}_1 + \left(\frac{1}{2} - x_3\right)\mathbf{a}_2 + \left(\frac{1}{2} + x_3\right)\mathbf{a}_3$ | = | $x_3a\,\hat{\mathbf{x}} + \left(\frac{1}{2} - x_3\right)a\,\hat{\mathbf{y}} + \left(\frac{1}{2} + x_3\right)a\,\hat{\mathbf{z}}$ | (8$c$) | O I |
| $\mathbf{B}_{16}$ | = | $\left(\frac{1}{2} - x_3\right)\mathbf{a}_1 + \left(\frac{1}{2} + x_3\right)\mathbf{a}_2 + x_3\,\mathbf{a}_3$ | = | $\left(\frac{1}{2} - x_3\right)a\,\hat{\mathbf{x}} + \left(\frac{1}{2} + x_3\right)a\,\hat{\mathbf{y}} + x_3a\,\hat{\mathbf{z}}$ | (8$c$) | O I |
| $\mathbf{B}_{17}$ | = | $x_4\,\mathbf{a}_1 + x_4\,\mathbf{a}_2 + x_4\,\mathbf{a}_3$ | = | $x_4 a\,\hat{\mathbf{x}} + x_4 a\,\hat{\mathbf{y}} + x_4 a\,\hat{\mathbf{z}}$ | $(8c)$ | O II |
| $\mathbf{B}_{18}$ | = | $\left(\frac{1}{2} - x_4\right)\mathbf{a}_1 - x_4\,\mathbf{a}_2 + \left(\frac{1}{2} + x_4\right)\mathbf{a}_3$ | = | $\left(\frac{1}{2} - x_4\right)a\,\hat{\mathbf{x}} - x_4 a\,\hat{\mathbf{y}} + \left(\frac{1}{2} + x_4\right)a\,\hat{\mathbf{z}}$ | $(8c)$ | O II |
| $\mathbf{B}_{19}$ | = | $-x_4\,\mathbf{a}_1 + \left(\frac{1}{2} + x_4\right)\mathbf{a}_2 + \left(\frac{1}{2} - x_4\right)\mathbf{a}_3$ | = | $-x_4 a\,\hat{\mathbf{x}} + \left(\frac{1}{2} + x_4\right)a\,\hat{\mathbf{y}} + \left(\frac{1}{2} - x_4\right)a\,\hat{\mathbf{z}}$ | $(8c)$ | O II |
| $\mathbf{B}_{20}$ | = | $\left(\frac{1}{2} + x_4\right)\mathbf{a}_1 + \left(\frac{1}{2} - x_4\right)\mathbf{a}_2 - x_4\,\mathbf{a}_3$ | = | $\left(\frac{1}{2} + x_4\right)a\,\hat{\mathbf{x}} + \left(\frac{1}{2} - x_4\right)a\,\hat{\mathbf{y}} - x_4 a\,\hat{\mathbf{z}}$ | $(8c)$ | O II |
| $\mathbf{B}_{21}$ | = | $-x_4\,\mathbf{a}_1 - x_4\,\mathbf{a}_2 - x_4\,\mathbf{a}_3$ | = | $-x_4 a\,\hat{\mathbf{x}} - x_4 a\,\hat{\mathbf{y}} - x_4 a\,\hat{\mathbf{z}}$ | $(8c)$ | O II |
| $\mathbf{B}_{22}$ | = | $\left(\frac{1}{2} + x_4\right)\mathbf{a}_1 + x_4\,\mathbf{a}_2 + \left(\frac{1}{2} - x_4\right)\mathbf{a}_3$ | = | $\left(\frac{1}{2} + x_4\right)a\,\hat{\mathbf{x}} + x_4 a\,\hat{\mathbf{y}} + \left(\frac{1}{2} - x_4\right)a\,\hat{\mathbf{z}}$ | $(8c)$ | O II |
| $\mathbf{B}_{23}$ | = | $x_4\,\mathbf{a}_1 + \left(\frac{1}{2} - x_4\right)\mathbf{a}_2 + \left(\frac{1}{2} + x_4\right)\mathbf{a}_3$ | = | $x_4 a\,\hat{\mathbf{x}} + \left(\frac{1}{2} - x_4\right)a\,\hat{\mathbf{y}} + \left(\frac{1}{2} + x_4\right)a\,\hat{\mathbf{z}}$ | $(8c)$ | O II |
| $\mathbf{B}_{24}$ | = | $\left(\frac{1}{2} - x_4\right)\mathbf{a}_1 + \left(\frac{1}{2} + x_4\right)\mathbf{a}_2 + x_4\,\mathbf{a}_3$ | = | $\left(\frac{1}{2} - x_4\right)a\,\hat{\mathbf{x}} + \left(\frac{1}{2} + x_4\right)a\,\hat{\mathbf{y}} + x_4 a\,\hat{\mathbf{z}}$ | $(8c)$ | O II |
| $\mathbf{B}_{25}$ | = | $x_5\,\mathbf{a}_1 + x_5\,\mathbf{a}_2 + x_5\,\mathbf{a}_3$ | = | $x_5 a\,\hat{\mathbf{x}} + x_5 a\,\hat{\mathbf{y}} + x_5 a\,\hat{\mathbf{z}}$ | $(8c)$ | S |
| $\mathbf{B}_{26}$ | = | $\left(\frac{1}{2} - x_5\right)\mathbf{a}_1 - x_5\,\mathbf{a}_2 + \left(\frac{1}{2} + x_5\right)\mathbf{a}_3$ | = | $\left(\frac{1}{2} - x_5\right)a\,\hat{\mathbf{x}} - x_5 a\,\hat{\mathbf{y}} + \left(\frac{1}{2} + x_5\right)a\,\hat{\mathbf{z}}$ | $(8c)$ | S |
| $\mathbf{B}_{27}$ | = | $-x_5\,\mathbf{a}_1 + \left(\frac{1}{2} + x_5\right)\mathbf{a}_2 + \left(\frac{1}{2} - x_5\right)\mathbf{a}_3$ | = | $-x_5 a\,\hat{\mathbf{x}} + \left(\frac{1}{2} + x_5\right)a\,\hat{\mathbf{y}} + \left(\frac{1}{2} - x_5\right)a\,\hat{\mathbf{z}}$ | $(8c)$ | S |
| $\mathbf{B}_{28}$ | = | $\left(\frac{1}{2} + x_5\right)\mathbf{a}_1 + \left(\frac{1}{2} - x_5\right)\mathbf{a}_2 - x_5\,\mathbf{a}_3$ | = | $\left(\frac{1}{2} + x_5\right)a\,\hat{\mathbf{x}} + \left(\frac{1}{2} - x_5\right)a\,\hat{\mathbf{y}} - x_5 a\,\hat{\mathbf{z}}$ | $(8c)$ | S |
| $\mathbf{B}_{29}$ | = | $-x_5\,\mathbf{a}_1 - x_5\,\mathbf{a}_2 - x_5\,\mathbf{a}_3$ | = | $-x_5 a\,\hat{\mathbf{x}} - x_5 a\,\hat{\mathbf{y}} - x_5 a\,\hat{\mathbf{z}}$ | $(8c)$ | S |
| $\mathbf{B}_{30}$ | = | $\left(\frac{1}{2} + x_5\right)\mathbf{a}_1 + x_5\,\mathbf{a}_2 + \left(\frac{1}{2} - x_5\right)\mathbf{a}_3$ | = | $\left(\frac{1}{2} + x_5\right)a\,\hat{\mathbf{x}} + x_5 a\,\hat{\mathbf{y}} + \left(\frac{1}{2} - x_5\right)a\,\hat{\mathbf{z}}$ | $(8c)$ | S |
| $\mathbf{B}_{31}$ | = | $x_5\,\mathbf{a}_1 + \left(\frac{1}{2} - x_5\right)\mathbf{a}_2 + \left(\frac{1}{2} + x_5\right)\mathbf{a}_3$ | = | $x_5 a\,\hat{\mathbf{x}} + \left(\frac{1}{2} - x_5\right)a\,\hat{\mathbf{y}} + \left(\frac{1}{2} + x_5\right)a\,\hat{\mathbf{z}}$ | $(8c)$ | S |
| $\mathbf{B}_{32}$ | = | $\left(\frac{1}{2} - x_5\right)\mathbf{a}_1 + \left(\frac{1}{2} + x_5\right)\mathbf{a}_2 + x_5\,\mathbf{a}_3$ | = | $\left(\frac{1}{2} - x_5\right)a\,\hat{\mathbf{x}} + \left(\frac{1}{2} + x_5\right)a\,\hat{\mathbf{y}} + x_5 a\,\hat{\mathbf{z}}$ | $(8c)$ | S |
| $\mathbf{B}_{33}$ | = | $x_6\,\mathbf{a}_1 + y_6\,\mathbf{a}_2 + z_6\,\mathbf{a}_3$ | = | $x_6 a\,\hat{\mathbf{x}} + y_6 a\,\hat{\mathbf{y}} + z_6 a\,\hat{\mathbf{z}}$ | $(24d)$ | H I |
| $\mathbf{B}_{34}$ | = | $\left(\frac{1}{2} - x_6\right)\mathbf{a}_1 - y_6\,\mathbf{a}_2 + \left(\frac{1}{2} + z_6\right)\mathbf{a}_3$ | = | $\left(\frac{1}{2} - x_6\right)a\,\hat{\mathbf{x}} - y_6 a\,\hat{\mathbf{y}} + \left(\frac{1}{2} + z_6\right)a\,\hat{\mathbf{z}}$ | $(24d)$ | H I |
| $\mathbf{B}_{35}$ | = | $-x_6\,\mathbf{a}_1 + \left(\frac{1}{2} + y_6\right)\mathbf{a}_2 + \left(\frac{1}{2} - z_6\right)\mathbf{a}_3$ | = | $-x_6 a\,\hat{\mathbf{x}} + \left(\frac{1}{2} + y_6\right)a\,\hat{\mathbf{y}} + \left(\frac{1}{2} - z_6\right)a\,\hat{\mathbf{z}}$ | $(24d)$ | H I |
| $\mathbf{B}_{36}$ | = | $\left(\frac{1}{2} + x_6\right)\mathbf{a}_1 + \left(\frac{1}{2} - y_6\right)\mathbf{a}_2 - z_6\,\mathbf{a}_3$ | = | $\left(\frac{1}{2} + x_6\right)a\,\hat{\mathbf{x}} + \left(\frac{1}{2} - y_6\right)a\,\hat{\mathbf{y}} - z_6 a\,\hat{\mathbf{z}}$ | $(24d)$ | H I |
| $\mathbf{B}_{37}$ | = | $z_6\,\mathbf{a}_1 + x_6\,\mathbf{a}_2 + y_6\,\mathbf{a}_3$ | = | $z_6 a\,\hat{\mathbf{x}} + x_6 a\,\hat{\mathbf{y}} + y_6 a\,\hat{\mathbf{z}}$ | $(24d)$ | H I |
| $\mathbf{B}_{38}$ | = | $\left(\frac{1}{2} + z_6\right)\mathbf{a}_1 + \left(\frac{1}{2} - x_6\right)\mathbf{a}_2 - y_6\,\mathbf{a}_3$ | = | $\left(\frac{1}{2} + z_6\right)a\,\hat{\mathbf{x}} + \left(\frac{1}{2} - x_6\right)a\,\hat{\mathbf{y}} - y_6 a\,\hat{\mathbf{z}}$ | $(24d)$ | H I |
| $\mathbf{B}_{39}$ | = | $\left(\frac{1}{2} - z_6\right)\mathbf{a}_1 - x_6\,\mathbf{a}_2 + \left(\frac{1}{2} + y_6\right)\mathbf{a}_3$ | = | $\left(\frac{1}{2} - z_6\right)a\,\hat{\mathbf{x}} - x_6 a\,\hat{\mathbf{y}} + \left(\frac{1}{2} + y_6\right)a\,\hat{\mathbf{z}}$ | $(24d)$ | H I |
| $\mathbf{B}_{40}$ | = | $-z_6\,\mathbf{a}_1 + \left(\frac{1}{2} + x_6\right)\mathbf{a}_2 + \left(\frac{1}{2} - y_6\right)\mathbf{a}_3$ | = | $-z_6 a\,\hat{\mathbf{x}} + \left(\frac{1}{2} + x_6\right)a\,\hat{\mathbf{y}} + \left(\frac{1}{2} - y_6\right)a\,\hat{\mathbf{z}}$ | $(24d)$ | H I |
| $\mathbf{B}_{41}$ | = | $y_6\,\mathbf{a}_1 + z_6\,\mathbf{a}_2 + x_6\,\mathbf{a}_3$ | = | $y_6 a\,\hat{\mathbf{x}} + z_6 a\,\hat{\mathbf{y}} + x_6 a\,\hat{\mathbf{z}}$ | $(24d)$ | H I |
| $\mathbf{B}_{42}$ | = | $-y_6\,\mathbf{a}_1 + \left(\frac{1}{2} + z_6\right)\mathbf{a}_2 + \left(\frac{1}{2} - x_6\right)\mathbf{a}_3$ | = | $-y_6 a\,\hat{\mathbf{x}} + \left(\frac{1}{2} + z_6\right)a\,\hat{\mathbf{y}} + \left(\frac{1}{2} - x_6\right)a\,\hat{\mathbf{z}}$ | $(24d)$ | H I |
| $\mathbf{B}_{43}$ | = | $\left(\frac{1}{2} + y_6\right)\mathbf{a}_1 + \left(\frac{1}{2} - z_6\right)\mathbf{a}_2 - x_6\,\mathbf{a}_3$ | = | $\left(\frac{1}{2} + y_6\right)a\,\hat{\mathbf{x}} + \left(\frac{1}{2} - z_6\right)a\,\hat{\mathbf{y}} - x_6 a\,\hat{\mathbf{z}}$ | $(24d)$ | H I |
| $\mathbf{B}_{44}$ | = | $\left(\frac{1}{2} - y_6\right)\mathbf{a}_1 - z_6\,\mathbf{a}_2 + \left(\frac{1}{2} + x_6\right)\mathbf{a}_3$ | = | $\left(\frac{1}{2} - y_6\right)a\,\hat{\mathbf{x}} - z_6 a\,\hat{\mathbf{y}} + \left(\frac{1}{2} + x_6\right)a\,\hat{\mathbf{z}}$ | $(24d)$ | H I |
| $\mathbf{B}_{45}$ | = | $-x_6\,\mathbf{a}_1 - y_6\,\mathbf{a}_2 - z_6\,\mathbf{a}_3$ | = | $-x_6 a\,\hat{\mathbf{x}} - y_6 a\,\hat{\mathbf{y}} - z_6 a\,\hat{\mathbf{z}}$ | $(24d)$ | H I |
| $\mathbf{B}_{46}$ | = | $\left(\frac{1}{2} + x_6\right)\mathbf{a}_1 + y_6\,\mathbf{a}_2 + \left(\frac{1}{2} - z_6\right)\mathbf{a}_3$ | = | $\left(\frac{1}{2} + x_6\right)a\,\hat{\mathbf{x}} + y_6 a\,\hat{\mathbf{y}} + \left(\frac{1}{2} - z_6\right)a\,\hat{\mathbf{z}}$ | $(24d)$ | H I |
| $\mathbf{B}_{47}$ | = | $x_6\,\mathbf{a}_1 + \left(\frac{1}{2} - y_6\right)\mathbf{a}_2 + \left(\frac{1}{2} + z_6\right)\mathbf{a}_3$ | = | $x_6 a\,\hat{\mathbf{x}} + \left(\frac{1}{2} - y_6\right)a\,\hat{\mathbf{y}} + \left(\frac{1}{2} + z_6\right)a\,\hat{\mathbf{z}}$ | $(24d)$ | H I |
| $\mathbf{B}_{48}$ | = | $\left(\frac{1}{2} - x_6\right)\mathbf{a}_1 + \left(\frac{1}{2} + y_6\right)\mathbf{a}_2 + z_6\,\mathbf{a}_3$ | = | $\left(\frac{1}{2} - x_6\right)a\,\hat{\mathbf{x}} + \left(\frac{1}{2} + y_6\right)a\,\hat{\mathbf{y}} + z_6 a\,\hat{\mathbf{z}}$ | $(24d)$ | H I |
| $\mathbf{B}_{49}$ | = | $-z_6\,\mathbf{a}_1 - x_6\,\mathbf{a}_2 - y_6\,\mathbf{a}_3$ | = | $-z_6 a\,\hat{\mathbf{x}} - x_6 a\,\hat{\mathbf{y}} - y_6 a\,\hat{\mathbf{z}}$ | $(24d)$ | H I |
| $\mathbf{B}_{50}$ | = | $\left(\frac{1}{2} - z_6\right)\mathbf{a}_1 + \left(\frac{1}{2} + x_6\right)\mathbf{a}_2 + y_6\,\mathbf{a}_3$ | = | $\left(\frac{1}{2} - z_6\right)a\,\hat{\mathbf{x}} + \left(\frac{1}{2} + x_6\right)a\,\hat{\mathbf{y}} + y_6 a\,\hat{\mathbf{z}}$ | $(24d)$ | H I |
| $\mathbf{B}_{51}$ | = | $\left(\frac{1}{2} + z_6\right)\mathbf{a}_1 + x_6\,\mathbf{a}_2 + \left(\frac{1}{2} - y_6\right)\mathbf{a}_3$ | = | $\left(\frac{1}{2} + z_6\right)a\,\hat{\mathbf{x}} + x_6 a\,\hat{\mathbf{y}} + \left(\frac{1}{2} - y_6\right)a\,\hat{\mathbf{z}}$ | $(24d)$ | H I |
| $\mathbf{B}_{52}$ | = | $z_6\,\mathbf{a}_1 + \left(\frac{1}{2} - x_6\right)\mathbf{a}_2 + \left(\frac{1}{2} + y_6\right)\mathbf{a}_3$ | = | $z_6 a\,\hat{\mathbf{x}} + \left(\frac{1}{2} - x_6\right)a\,\hat{\mathbf{y}} + \left(\frac{1}{2} + y_6\right)a\,\hat{\mathbf{z}}$ | $(24d)$ | H I |

$$\mathbf{B}_{53} = -y_6\,\mathbf{a}_1 - z_6\,\mathbf{a}_2 - x_6\,\mathbf{a}_3 = -y_6 a\,\hat{\mathbf{x}} - z_6 a\,\hat{\mathbf{y}} - x_6 a\,\hat{\mathbf{z}} \qquad (24d) \qquad \text{H I}$$

$$\mathbf{B}_{54} = y_6\,\mathbf{a}_1 + \left(\frac{1}{2} - z_6\right)\mathbf{a}_2 + \left(\frac{1}{2} + x_6\right)\mathbf{a}_3 = y_6 a\,\hat{\mathbf{x}} + \left(\frac{1}{2} - z_6\right)a\,\hat{\mathbf{y}} + \left(\frac{1}{2} + x_6\right)a\,\hat{\mathbf{z}} \qquad (24d) \qquad \text{H I}$$

$$\mathbf{B}_{55} = \left(\frac{1}{2} - y_6\right)\mathbf{a}_1 + \left(\frac{1}{2} + z_6\right)\mathbf{a}_2 + x_6\,\mathbf{a}_3 = \left(\frac{1}{2} - y_6\right)a\,\hat{\mathbf{x}} + \left(\frac{1}{2} + z_6\right)a\,\hat{\mathbf{y}} + x_6 a\,\hat{\mathbf{z}} \qquad (24d) \qquad \text{H I}$$

$$\mathbf{B}_{56} = \left(\frac{1}{2} + y_6\right)\mathbf{a}_1 + z_6\,\mathbf{a}_2 + \left(\frac{1}{2} - x_6\right)\mathbf{a}_3 = \left(\frac{1}{2} + y_6\right)a\,\hat{\mathbf{x}} + z_6 a\,\hat{\mathbf{y}} + \left(\frac{1}{2} - x_6\right)a\,\hat{\mathbf{z}} \qquad (24d) \qquad \text{H I}$$

$$\mathbf{B}_{57} = x_7\,\mathbf{a}_1 + y_7\,\mathbf{a}_2 + z_7\,\mathbf{a}_3 = x_7 a\,\hat{\mathbf{x}} + y_7 a\,\hat{\mathbf{y}} + z_7 a\,\hat{\mathbf{z}} \qquad (24d) \qquad \text{H II}$$

$$\mathbf{B}_{58} = \left(\frac{1}{2} - x_7\right)\mathbf{a}_1 - y_7\,\mathbf{a}_2 + \left(\frac{1}{2} + z_7\right)\mathbf{a}_3 = \left(\frac{1}{2} - x_7\right)a\,\hat{\mathbf{x}} - y_7 a\,\hat{\mathbf{y}} + \left(\frac{1}{2} + z_7\right)a\,\hat{\mathbf{z}} \qquad (24d) \qquad \text{H II}$$

$$\mathbf{B}_{59} = -x_7\,\mathbf{a}_1 + \left(\frac{1}{2} + y_7\right)\mathbf{a}_2 + \left(\frac{1}{2} - z_7\right)\mathbf{a}_3 = -x_7 a\,\hat{\mathbf{x}} + \left(\frac{1}{2} + y_7\right)a\,\hat{\mathbf{y}} + \left(\frac{1}{2} - z_7\right)a\,\hat{\mathbf{z}} \qquad (24d) \qquad \text{H II}$$

$$\mathbf{B}_{60} = \left(\frac{1}{2} + x_7\right)\mathbf{a}_1 + \left(\frac{1}{2} - y_7\right)\mathbf{a}_2 - z_7\,\mathbf{a}_3 = \left(\frac{1}{2} + x_7\right)a\,\hat{\mathbf{x}} + \left(\frac{1}{2} - y_7\right)a\,\hat{\mathbf{y}} - z_7 a\,\hat{\mathbf{z}} \qquad (24d) \qquad \text{H II}$$

$$\mathbf{B}_{61} = z_7\,\mathbf{a}_1 + x_7\,\mathbf{a}_2 + y_7\,\mathbf{a}_3 = z_7 a\,\hat{\mathbf{x}} + x_7 a\,\hat{\mathbf{y}} + y_7 a\,\hat{\mathbf{z}} \qquad (24d) \qquad \text{H II}$$

$$\mathbf{B}_{62} = \left(\frac{1}{2} + z_7\right)\mathbf{a}_1 + \left(\frac{1}{2} - x_7\right)\mathbf{a}_2 - y_7\,\mathbf{a}_3 = \left(\frac{1}{2} + z_7\right)a\,\hat{\mathbf{x}} + \left(\frac{1}{2} - x_7\right)a\,\hat{\mathbf{y}} - y_7 a\,\hat{\mathbf{z}} \qquad (24d) \qquad \text{H II}$$

$$\mathbf{B}_{63} = \left(\frac{1}{2} - z_7\right)\mathbf{a}_1 - x_7\,\mathbf{a}_2 + \left(\frac{1}{2} + y_7\right)\mathbf{a}_3 = \left(\frac{1}{2} - z_7\right)a\,\hat{\mathbf{x}} - x_7 a\,\hat{\mathbf{y}} + \left(\frac{1}{2} + y_7\right)a\,\hat{\mathbf{z}} \qquad (24d) \qquad \text{H II}$$

$$\mathbf{B}_{64} = -z_7\,\mathbf{a}_1 + \left(\frac{1}{2} + x_7\right)\mathbf{a}_2 + \left(\frac{1}{2} - y_7\right)\mathbf{a}_3 = -z_7 a\,\hat{\mathbf{x}} + \left(\frac{1}{2} + x_7\right)a\,\hat{\mathbf{y}} + \left(\frac{1}{2} - y_7\right)a\,\hat{\mathbf{z}} \qquad (24d) \qquad \text{H II}$$

$$\mathbf{B}_{65} = y_7\,\mathbf{a}_1 + z_7\,\mathbf{a}_2 + x_7\,\mathbf{a}_3 = y_7 a\,\hat{\mathbf{x}} + z_7 a\,\hat{\mathbf{y}} + x_7 a\,\hat{\mathbf{z}} \qquad (24d) \qquad \text{H II}$$

$$\mathbf{B}_{66} = -y_7\,\mathbf{a}_1 + \left(\frac{1}{2} + z_7\right)\mathbf{a}_2 + \left(\frac{1}{2} - x_7\right)\mathbf{a}_3 = -y_7 a\,\hat{\mathbf{x}} + \left(\frac{1}{2} + z_7\right)a\,\hat{\mathbf{y}} + \left(\frac{1}{2} - x_7\right)a\,\hat{\mathbf{z}} \qquad (24d) \qquad \text{H II}$$

$$\mathbf{B}_{67} = \left(\frac{1}{2} + y_7\right)\mathbf{a}_1 + \left(\frac{1}{2} - z_7\right)\mathbf{a}_2 - x_7\,\mathbf{a}_3 = \left(\frac{1}{2} + y_7\right)a\,\hat{\mathbf{x}} + \left(\frac{1}{2} - z_7\right)a\,\hat{\mathbf{y}} - x_7 a\,\hat{\mathbf{z}} \qquad (24d) \qquad \text{H II}$$

$$\mathbf{B}_{68} = \left(\frac{1}{2} - y_7\right)\mathbf{a}_1 - z_7\,\mathbf{a}_2 + \left(\frac{1}{2} + x_7\right)\mathbf{a}_3 = \left(\frac{1}{2} - y_7\right)a\,\hat{\mathbf{x}} - z_7 a\,\hat{\mathbf{y}} + \left(\frac{1}{2} + x_7\right)a\,\hat{\mathbf{z}} \qquad (24d) \qquad \text{H II}$$

$$\mathbf{B}_{69} = -x_7\,\mathbf{a}_1 - y_7\,\mathbf{a}_2 - z_7\,\mathbf{a}_3 = -x_7 a\,\hat{\mathbf{x}} - y_7 a\,\hat{\mathbf{y}} - z_7 a\,\hat{\mathbf{z}} \qquad (24d) \qquad \text{H II}$$

$$\mathbf{B}_{70} = \left(\frac{1}{2} + x_7\right)\mathbf{a}_1 + y_7\,\mathbf{a}_2 + \left(\frac{1}{2} - z_7\right)\mathbf{a}_3 = \left(\frac{1}{2} + x_7\right)a\,\hat{\mathbf{x}} + y_7 a\,\hat{\mathbf{y}} + \left(\frac{1}{2} - z_7\right)a\,\hat{\mathbf{z}} \qquad (24d) \qquad \text{H II}$$

$$\mathbf{B}_{71} = x_7\,\mathbf{a}_1 + \left(\frac{1}{2} - y_7\right)\mathbf{a}_2 + \left(\frac{1}{2} + z_7\right)\mathbf{a}_3 = x_7 a\,\hat{\mathbf{x}} + \left(\frac{1}{2} - y_7\right)a\,\hat{\mathbf{y}} + \left(\frac{1}{2} + z_7\right)a\,\hat{\mathbf{z}} \qquad (24d) \qquad \text{H II}$$

$$\mathbf{B}_{72} = \left(\frac{1}{2} - x_7\right)\mathbf{a}_1 + \left(\frac{1}{2} + y_7\right)\mathbf{a}_2 + z_7\,\mathbf{a}_3 = \left(\frac{1}{2} - x_7\right)a\,\hat{\mathbf{x}} + \left(\frac{1}{2} + y_7\right)a\,\hat{\mathbf{y}} + z_7 a\,\hat{\mathbf{z}} \qquad (24d) \qquad \text{H II}$$

$$\mathbf{B}_{73} = -z_7\,\mathbf{a}_1 - x_7\,\mathbf{a}_2 - y_7\,\mathbf{a}_3 = -z_7 a\,\hat{\mathbf{x}} - x_7 a\,\hat{\mathbf{y}} - y_7 a\,\hat{\mathbf{z}} \qquad (24d) \qquad \text{H II}$$

$$\mathbf{B}_{74} = \left(\frac{1}{2} - z_7\right)\mathbf{a}_1 + \left(\frac{1}{2} + x_7\right)\mathbf{a}_2 + y_7\,\mathbf{a}_3 = \left(\frac{1}{2} - z_7\right)a\,\hat{\mathbf{x}} + \left(\frac{1}{2} + x_7\right)a\,\hat{\mathbf{y}} + y_7 a\,\hat{\mathbf{z}} \qquad (24d) \qquad \text{H II}$$

$$\mathbf{B}_{75} = \left(\frac{1}{2} + z_7\right)\mathbf{a}_1 + x_7\,\mathbf{a}_2 + \left(\frac{1}{2} - y_7\right)\mathbf{a}_3 = \left(\frac{1}{2} + z_7\right)a\,\hat{\mathbf{x}} + x_7 a\,\hat{\mathbf{y}} + \left(\frac{1}{2} - y_7\right)a\,\hat{\mathbf{z}} \qquad (24d) \qquad \text{H II}$$

$$\mathbf{B}_{76} = z_7\,\mathbf{a}_1 + \left(\frac{1}{2} - x_7\right)\mathbf{a}_2 + \left(\frac{1}{2} + y_7\right)\mathbf{a}_3 = z_7 a\,\hat{\mathbf{x}} + \left(\frac{1}{2} - x_7\right)a\,\hat{\mathbf{y}} + \left(\frac{1}{2} + y_7\right)a\,\hat{\mathbf{z}} \qquad (24d) \qquad \text{H II}$$

$$\mathbf{B}_{77} = -y_7\,\mathbf{a}_1 - z_7\,\mathbf{a}_2 - x_7\,\mathbf{a}_3 = -y_7 a\,\hat{\mathbf{x}} - z_7 a\,\hat{\mathbf{y}} - x_7 a\,\hat{\mathbf{z}} \qquad (24d) \qquad \text{H II}$$

$$\mathbf{B}_{78} = y_7\,\mathbf{a}_1 + \left(\frac{1}{2} - z_7\right)\mathbf{a}_2 + \left(\frac{1}{2} + x_7\right)\mathbf{a}_3 = y_7 a\,\hat{\mathbf{x}} + \left(\frac{1}{2} - z_7\right)a\,\hat{\mathbf{y}} + \left(\frac{1}{2} + x_7\right)a\,\hat{\mathbf{z}} \qquad (24d) \qquad \text{H II}$$

$$\mathbf{B}_{79} = \left(\frac{1}{2} - y_7\right)\mathbf{a}_1 + \left(\frac{1}{2} + z_7\right)\mathbf{a}_2 + x_7\,\mathbf{a}_3 = \left(\frac{1}{2} - y_7\right)a\,\hat{\mathbf{x}} + \left(\frac{1}{2} + z_7\right)a\,\hat{\mathbf{y}} + x_7 a\,\hat{\mathbf{z}} \qquad (24d) \qquad \text{H II}$$

$$\mathbf{B}_{80} = \left(\frac{1}{2} + y_7\right)\mathbf{a}_1 + z_7\,\mathbf{a}_2 + \left(\frac{1}{2} - x_7\right)\mathbf{a}_3 = \left(\frac{1}{2} + y_7\right)a\,\hat{\mathbf{x}} + z_7 a\,\hat{\mathbf{y}} + \left(\frac{1}{2} - x_7\right)a\,\hat{\mathbf{z}} \qquad (24d) \qquad \text{H II}$$

$$\mathbf{B}_{81} = x_8\,\mathbf{a}_1 + y_8\,\mathbf{a}_2 + z_8\,\mathbf{a}_3 = x_8 a\,\hat{\mathbf{x}} + y_8 a\,\hat{\mathbf{y}} + z_8 a\,\hat{\mathbf{z}} \qquad (24d) \qquad \text{H III}$$

$$\mathbf{B}_{82} = \left(\frac{1}{2} - x_8\right)\mathbf{a}_1 - y_8\,\mathbf{a}_2 + \left(\frac{1}{2} + z_8\right)\mathbf{a}_3 = \left(\frac{1}{2} - x_8\right)a\,\hat{\mathbf{x}} - y_8 a\,\hat{\mathbf{y}} + \left(\frac{1}{2} + z_8\right)a\,\hat{\mathbf{z}} \qquad (24d) \qquad \text{H III}$$

$$\mathbf{B}_{83} = -x_8\,\mathbf{a}_1 + \left(\frac{1}{2} + y_8\right)\mathbf{a}_2 + \left(\frac{1}{2} - z_8\right)\mathbf{a}_3 = -x_8 a\,\hat{\mathbf{x}} + \left(\frac{1}{2} + y_8\right)a\,\hat{\mathbf{y}} + \left(\frac{1}{2} - z_8\right)a\,\hat{\mathbf{z}} \qquad (24d) \qquad \text{H III}$$

$$\mathbf{B}_{84} = \left(\frac{1}{2} + x_8\right)\mathbf{a}_1 + \left(\frac{1}{2} - y_8\right)\mathbf{a}_2 - z_8\,\mathbf{a}_3 = \left(\frac{1}{2} + x_8\right)a\,\hat{\mathbf{x}} + \left(\frac{1}{2} - y_8\right)a\,\hat{\mathbf{y}} - z_8 a\,\hat{\mathbf{z}} \qquad (24d) \qquad \text{H III}$$

$$\mathbf{B}_{85} = z_8\,\mathbf{a}_1 + x_8\,\mathbf{a}_2 + y_8\,\mathbf{a}_3 = z_8 a\,\hat{\mathbf{x}} + x_8 a\,\hat{\mathbf{y}} + y_8 a\,\hat{\mathbf{z}} \qquad (24d) \qquad \text{H III}$$

$$\mathbf{B}_{86} = \left(\frac{1}{2} + z_8\right)\mathbf{a}_1 + \left(\frac{1}{2} - x_8\right)\mathbf{a}_2 - y_8\,\mathbf{a}_3 = \left(\frac{1}{2} + z_8\right)a\,\hat{\mathbf{x}} + \left(\frac{1}{2} - x_8\right)a\,\hat{\mathbf{y}} - y_8 a\,\hat{\mathbf{z}} \qquad (24d) \qquad \text{H III}$$

$$\mathbf{B}_{87} = \left(\frac{1}{2} - z_8\right)\mathbf{a}_1 - x_8\,\mathbf{a}_2 + \left(\frac{1}{2} + y_8\right)\mathbf{a}_3 = \left(\frac{1}{2} - z_8\right)a\,\hat{\mathbf{x}} - x_8 a\,\hat{\mathbf{y}} + \left(\frac{1}{2} + y_8\right)a\,\hat{\mathbf{z}} \qquad (24d) \qquad \text{H III}$$

$$\mathbf{B}_{88} = -z_8\,\mathbf{a}_1 + \left(\frac{1}{2} + x_8\right)\mathbf{a}_2 + \left(\frac{1}{2} - y_8\right)\mathbf{a}_3 = -z_8 a\,\hat{\mathbf{x}} + \left(\frac{1}{2} + x_8\right)a\,\hat{\mathbf{y}} + \left(\frac{1}{2} - y_8\right)a\,\hat{\mathbf{z}} \qquad (24d) \qquad \text{H III}$$

| | | | | | | |
|---|---|---|---|---|---|---|
| $\mathbf{B}_{89}$ | = | $y_8\,\mathbf{a}_1 + z_8\,\mathbf{a}_2 + x_8\,\mathbf{a}_3$ | = | $y_8 a\,\hat{\mathbf{x}} + z_8 a\,\hat{\mathbf{y}} + x_8 a\,\hat{\mathbf{z}}$ | $(24d)$ | H III |
| $\mathbf{B}_{90}$ | = | $-y_8\,\mathbf{a}_1 + \left(\frac{1}{2} + z_8\right)\mathbf{a}_2 + \left(\frac{1}{2} - x_8\right)\mathbf{a}_3$ | = | $-y_8 a\,\hat{\mathbf{x}} + \left(\frac{1}{2} + z_8\right) a\,\hat{\mathbf{y}} + \left(\frac{1}{2} - x_8\right) a\,\hat{\mathbf{z}}$ | $(24d)$ | H III |
| $\mathbf{B}_{91}$ | = | $\left(\frac{1}{2} + y_8\right)\mathbf{a}_1 + \left(\frac{1}{2} - z_8\right)\mathbf{a}_2 - x_8\,\mathbf{a}_3$ | = | $\left(\frac{1}{2} + y_8\right) a\,\hat{\mathbf{x}} + \left(\frac{1}{2} - z_8\right) a\,\hat{\mathbf{y}} - x_8 a\,\hat{\mathbf{z}}$ | $(24d)$ | H III |
| $\mathbf{B}_{92}$ | = | $\left(\frac{1}{2} - y_8\right)\mathbf{a}_1 - z_8\,\mathbf{a}_2 + \left(\frac{1}{2} + x_8\right)\mathbf{a}_3$ | = | $\left(\frac{1}{2} - y_8\right) a\,\hat{\mathbf{x}} - z_8 a\,\hat{\mathbf{y}} + \left(\frac{1}{2} + x_8\right) a\,\hat{\mathbf{z}}$ | $(24d)$ | H III |
| $\mathbf{B}_{93}$ | = | $-x_8\,\mathbf{a}_1 - y_8\,\mathbf{a}_2 - z_8\,\mathbf{a}_3$ | = | $-x_8 a\,\hat{\mathbf{x}} - y_8 a\,\hat{\mathbf{y}} - z_8 a\,\hat{\mathbf{z}}$ | $(24d)$ | H III |
| $\mathbf{B}_{94}$ | = | $\left(\frac{1}{2} + x_8\right)\mathbf{a}_1 + y_8\,\mathbf{a}_2 + \left(\frac{1}{2} - z_8\right)\mathbf{a}_3$ | = | $\left(\frac{1}{2} + x_8\right) a\,\hat{\mathbf{x}} + y_8 a\,\hat{\mathbf{y}} + \left(\frac{1}{2} - z_8\right) a\,\hat{\mathbf{z}}$ | $(24d)$ | H III |
| $\mathbf{B}_{95}$ | = | $x_8\,\mathbf{a}_1 + \left(\frac{1}{2} - y_8\right)\mathbf{a}_2 + \left(\frac{1}{2} + z_8\right)\mathbf{a}_3$ | = | $x_8 a\,\hat{\mathbf{x}} + \left(\frac{1}{2} - y_8\right) a\,\hat{\mathbf{y}} + \left(\frac{1}{2} + z_8\right) a\,\hat{\mathbf{z}}$ | $(24d)$ | H III |
| $\mathbf{B}_{96}$ | = | $\left(\frac{1}{2} - x_8\right)\mathbf{a}_1 + \left(\frac{1}{2} + y_8\right)\mathbf{a}_2 + z_8\,\mathbf{a}_3$ | = | $\left(\frac{1}{2} - x_8\right) a\,\hat{\mathbf{x}} + \left(\frac{1}{2} + y_8\right) a\,\hat{\mathbf{y}} + z_8 a\,\hat{\mathbf{z}}$ | $(24d)$ | H III |
| $\mathbf{B}_{97}$ | = | $-z_8\,\mathbf{a}_1 - x_8\,\mathbf{a}_2 - y_8\,\mathbf{a}_3$ | = | $-z_8 a\,\hat{\mathbf{x}} - x_8 a\,\hat{\mathbf{y}} - y_8 a\,\hat{\mathbf{z}}$ | $(24d)$ | H III |
| $\mathbf{B}_{98}$ | = | $\left(\frac{1}{2} - z_8\right)\mathbf{a}_1 + \left(\frac{1}{2} + x_8\right)\mathbf{a}_2 + y_8\,\mathbf{a}_3$ | = | $\left(\frac{1}{2} - z_8\right) a\,\hat{\mathbf{x}} + \left(\frac{1}{2} + x_8\right) a\,\hat{\mathbf{y}} + y_8 a\,\hat{\mathbf{z}}$ | $(24d)$ | H III |
| $\mathbf{B}_{99}$ | = | $\left(\frac{1}{2} + z_8\right)\mathbf{a}_1 + x_8\,\mathbf{a}_2 + \left(\frac{1}{2} - y_8\right)\mathbf{a}_3$ | = | $\left(\frac{1}{2} + z_8\right) a\,\hat{\mathbf{x}} + x_8 a\,\hat{\mathbf{y}} + \left(\frac{1}{2} - y_8\right) a\,\hat{\mathbf{z}}$ | $(24d)$ | H III |
| $\mathbf{B}_{100}$ | = | $z_8\,\mathbf{a}_1 + \left(\frac{1}{2} - x_8\right)\mathbf{a}_2 + \left(\frac{1}{2} + y_8\right)\mathbf{a}_3$ | = | $z_8 a\,\hat{\mathbf{x}} + \left(\frac{1}{2} - x_8\right) a\,\hat{\mathbf{y}} + \left(\frac{1}{2} + y_8\right) a\,\hat{\mathbf{z}}$ | $(24d)$ | H III |
| $\mathbf{B}_{101}$ | = | $-y_8\,\mathbf{a}_1 - z_8\,\mathbf{a}_2 - x_8\,\mathbf{a}_3$ | = | $-y_8 a\,\hat{\mathbf{x}} - z_8 a\,\hat{\mathbf{y}} - x_8 a\,\hat{\mathbf{z}}$ | $(24d)$ | H III |
| $\mathbf{B}_{102}$ | = | $y_8\,\mathbf{a}_1 + \left(\frac{1}{2} - z_8\right)\mathbf{a}_2 + \left(\frac{1}{2} + x_8\right)\mathbf{a}_3$ | = | $y_8 a\,\hat{\mathbf{x}} + \left(\frac{1}{2} - z_8\right) a\,\hat{\mathbf{y}} + \left(\frac{1}{2} + x_8\right) a\,\hat{\mathbf{z}}$ | $(24d)$ | H III |
| $\mathbf{B}_{103}$ | = | $\left(\frac{1}{2} - y_8\right)\mathbf{a}_1 + \left(\frac{1}{2} + z_8\right)\mathbf{a}_2 + x_8\,\mathbf{a}_3$ | = | $\left(\frac{1}{2} - y_8\right) a\,\hat{\mathbf{x}} + \left(\frac{1}{2} + z_8\right) a\,\hat{\mathbf{y}} + x_8 a\,\hat{\mathbf{z}}$ | $(24d)$ | H III |
| $\mathbf{B}_{104}$ | = | $\left(\frac{1}{2} + y_8\right)\mathbf{a}_1 + z_8\,\mathbf{a}_2 + \left(\frac{1}{2} - x_8\right)\mathbf{a}_3$ | = | $\left(\frac{1}{2} + y_8\right) a\,\hat{\mathbf{x}} + z_8 a\,\hat{\mathbf{y}} + \left(\frac{1}{2} - x_8\right) a\,\hat{\mathbf{z}}$ | $(24d)$ | H III |
| $\mathbf{B}_{105}$ | = | $x_9\,\mathbf{a}_1 + y_9\,\mathbf{a}_2 + z_9\,\mathbf{a}_3$ | = | $x_9 a\,\hat{\mathbf{x}} + y_9 a\,\hat{\mathbf{y}} + z_9 a\,\hat{\mathbf{z}}$ | $(24d)$ | H IV |
| $\mathbf{B}_{106}$ | = | $\left(\frac{1}{2} - x_9\right)\mathbf{a}_1 - y_9\,\mathbf{a}_2 + \left(\frac{1}{2} + z_9\right)\mathbf{a}_3$ | = | $\left(\frac{1}{2} - x_9\right) a\,\hat{\mathbf{x}} - y_9 a\,\hat{\mathbf{y}} + \left(\frac{1}{2} + z_9\right) a\,\hat{\mathbf{z}}$ | $(24d)$ | H IV |
| $\mathbf{B}_{107}$ | = | $-x_9\,\mathbf{a}_1 + \left(\frac{1}{2} + y_9\right)\mathbf{a}_2 + \left(\frac{1}{2} - z_9\right)\mathbf{a}_3$ | = | $-x_9 a\,\hat{\mathbf{x}} + \left(\frac{1}{2} + y_9\right) a\,\hat{\mathbf{y}} + \left(\frac{1}{2} - z_9\right) a\,\hat{\mathbf{z}}$ | $(24d)$ | H IV |
| $\mathbf{B}_{108}$ | = | $\left(\frac{1}{2} + x_9\right)\mathbf{a}_1 + \left(\frac{1}{2} - y_9\right)\mathbf{a}_2 - z_9\,\mathbf{a}_3$ | = | $\left(\frac{1}{2} + x_9\right) a\,\hat{\mathbf{x}} + \left(\frac{1}{2} - y_9\right) a\,\hat{\mathbf{y}} - z_9 a\,\hat{\mathbf{z}}$ | $(24d)$ | H IV |
| $\mathbf{B}_{109}$ | = | $z_9\,\mathbf{a}_1 + x_9\,\mathbf{a}_2 + y_9\,\mathbf{a}_3$ | = | $z_9 a\,\hat{\mathbf{x}} + x_9 a\,\hat{\mathbf{y}} + y_9 a\,\hat{\mathbf{z}}$ | $(24d)$ | H IV |
| $\mathbf{B}_{110}$ | = | $\left(\frac{1}{2} + z_9\right)\mathbf{a}_1 + \left(\frac{1}{2} - x_9\right)\mathbf{a}_2 - y_9\,\mathbf{a}_3$ | = | $\left(\frac{1}{2} + z_9\right) a\,\hat{\mathbf{x}} + \left(\frac{1}{2} - x_9\right) a\,\hat{\mathbf{y}} - y_9 a\,\hat{\mathbf{z}}$ | $(24d)$ | H IV |
| $\mathbf{B}_{111}$ | = | $\left(\frac{1}{2} - z_9\right)\mathbf{a}_1 - x_9\,\mathbf{a}_2 + \left(\frac{1}{2} + y_9\right)\mathbf{a}_3$ | = | $\left(\frac{1}{2} - z_9\right) a\,\hat{\mathbf{x}} - x_9 a\,\hat{\mathbf{y}} + \left(\frac{1}{2} + y_9\right) a\,\hat{\mathbf{z}}$ | $(24d)$ | H IV |
| $\mathbf{B}_{112}$ | = | $-z_9\,\mathbf{a}_1 + \left(\frac{1}{2} + x_9\right)\mathbf{a}_2 + \left(\frac{1}{2} - y_9\right)\mathbf{a}_3$ | = | $-z_9 a\,\hat{\mathbf{x}} + \left(\frac{1}{2} + x_9\right) a\,\hat{\mathbf{y}} + \left(\frac{1}{2} - y_9\right) a\,\hat{\mathbf{z}}$ | $(24d)$ | H IV |
| $\mathbf{B}_{113}$ | = | $y_9\,\mathbf{a}_1 + z_9\,\mathbf{a}_2 + x_9\,\mathbf{a}_3$ | = | $y_9 a\,\hat{\mathbf{x}} + z_9 a\,\hat{\mathbf{y}} + x_9 a\,\hat{\mathbf{z}}$ | $(24d)$ | H IV |
| $\mathbf{B}_{114}$ | = | $-y_9\,\mathbf{a}_1 + \left(\frac{1}{2} + z_9\right)\mathbf{a}_2 + \left(\frac{1}{2} - x_9\right)\mathbf{a}_3$ | = | $-y_9 a\,\hat{\mathbf{x}} + \left(\frac{1}{2} + z_9\right) a\,\hat{\mathbf{y}} + \left(\frac{1}{2} - x_9\right) a\,\hat{\mathbf{z}}$ | $(24d)$ | H IV |
| $\mathbf{B}_{115}$ | = | $\left(\frac{1}{2} + y_9\right)\mathbf{a}_1 + \left(\frac{1}{2} - z_9\right)\mathbf{a}_2 - x_9\,\mathbf{a}_3$ | = | $\left(\frac{1}{2} + y_9\right) a\,\hat{\mathbf{x}} + \left(\frac{1}{2} - z_9\right) a\,\hat{\mathbf{y}} - x_9 a\,\hat{\mathbf{z}}$ | $(24d)$ | H IV |
| $\mathbf{B}_{116}$ | = | $\left(\frac{1}{2} - y_9\right)\mathbf{a}_1 - z_9\,\mathbf{a}_2 + \left(\frac{1}{2} + x_9\right)\mathbf{a}_3$ | = | $\left(\frac{1}{2} - y_9\right) a\,\hat{\mathbf{x}} - z_9 a\,\hat{\mathbf{y}} + \left(\frac{1}{2} + x_9\right) a\,\hat{\mathbf{z}}$ | $(24d)$ | H IV |
| $\mathbf{B}_{117}$ | = | $-x_9\,\mathbf{a}_1 - y_9\,\mathbf{a}_2 - z_9\,\mathbf{a}_3$ | = | $-x_9 a\,\hat{\mathbf{x}} - y_9 a\,\hat{\mathbf{y}} - z_9 a\,\hat{\mathbf{z}}$ | $(24d)$ | H IV |
| $\mathbf{B}_{118}$ | = | $\left(\frac{1}{2} + x_9\right)\mathbf{a}_1 + y_9\,\mathbf{a}_2 + \left(\frac{1}{2} - z_9\right)\mathbf{a}_3$ | = | $\left(\frac{1}{2} + x_9\right) a\,\hat{\mathbf{x}} + y_9 a\,\hat{\mathbf{y}} + \left(\frac{1}{2} - z_9\right) a\,\hat{\mathbf{z}}$ | $(24d)$ | H IV |
| $\mathbf{B}_{119}$ | = | $x_9\,\mathbf{a}_1 + \left(\frac{1}{2} - y_9\right)\mathbf{a}_2 + \left(\frac{1}{2} + z_9\right)\mathbf{a}_3$ | = | $x_9 a\,\hat{\mathbf{x}} + \left(\frac{1}{2} - y_9\right) a\,\hat{\mathbf{y}} + \left(\frac{1}{2} + z_9\right) a\,\hat{\mathbf{z}}$ | $(24d)$ | H IV |
| $\mathbf{B}_{120}$ | = | $\left(\frac{1}{2} - x_9\right)\mathbf{a}_1 + \left(\frac{1}{2} + y_9\right)\mathbf{a}_2 + z_9\,\mathbf{a}_3$ | = | $\left(\frac{1}{2} - x_9\right) a\,\hat{\mathbf{x}} + \left(\frac{1}{2} + y_9\right) a\,\hat{\mathbf{y}} + z_9 a\,\hat{\mathbf{z}}$ | $(24d)$ | H IV |
| $\mathbf{B}_{121}$ | = | $-z_9\,\mathbf{a}_1 - x_9\,\mathbf{a}_2 - y_9\,\mathbf{a}_3$ | = | $-z_9 a\,\hat{\mathbf{x}} - x_9 a\,\hat{\mathbf{y}} - y_9 a\,\hat{\mathbf{z}}$ | $(24d)$ | H IV |
| $\mathbf{B}_{122}$ | = | $\left(\frac{1}{2} - z_9\right)\mathbf{a}_1 + \left(\frac{1}{2} + x_9\right)\mathbf{a}_2 + y_9\,\mathbf{a}_3$ | = | $\left(\frac{1}{2} - z_9\right) a\,\hat{\mathbf{x}} + \left(\frac{1}{2} + x_9\right) a\,\hat{\mathbf{y}} + y_9 a\,\hat{\mathbf{z}}$ | $(24d)$ | H IV |
| $\mathbf{B}_{123}$ | = | $\left(\frac{1}{2} + z_9\right)\mathbf{a}_1 + x_9\,\mathbf{a}_2 + \left(\frac{1}{2} - y_9\right)\mathbf{a}_3$ | = | $\left(\frac{1}{2} + z_9\right) a\,\hat{\mathbf{x}} + x_9 a\,\hat{\mathbf{y}} + \left(\frac{1}{2} - y_9\right) a\,\hat{\mathbf{z}}$ | $(24d)$ | H IV |
| $\mathbf{B}_{124}$ | = | $z_9\,\mathbf{a}_1 + \left(\frac{1}{2} - x_9\right)\mathbf{a}_2 + \left(\frac{1}{2} + y_9\right)\mathbf{a}_3$ | = | $z_9 a\,\hat{\mathbf{x}} + \left(\frac{1}{2} - x_9\right) a\,\hat{\mathbf{y}} + \left(\frac{1}{2} + y_9\right) a\,\hat{\mathbf{z}}$ | $(24d)$ | H IV |
| $\mathbf{B}_{125}$ | = | $-y_9\,\mathbf{a}_1 - z_9\,\mathbf{a}_2 - x_9\,\mathbf{a}_3$ | = | $-y_9 a\,\hat{\mathbf{x}} - z_9 a\,\hat{\mathbf{y}} - x_9 a\,\hat{\mathbf{z}}$ | (24*d*) | H IV |
| $\mathbf{B}_{126}$ | = | $y_9\,\mathbf{a}_1 + \left(\frac{1}{2} - z_9\right)\mathbf{a}_2 + \left(\frac{1}{2} + x_9\right)\mathbf{a}_3$ | = | $y_9 a\,\hat{\mathbf{x}} + \left(\frac{1}{2} - z_9\right)a\,\hat{\mathbf{y}} + \left(\frac{1}{2} + x_9\right)a\,\hat{\mathbf{z}}$ | (24*d*) | H IV |
| $\mathbf{B}_{127}$ | = | $\left(\frac{1}{2} - y_9\right)\mathbf{a}_1 + \left(\frac{1}{2} + z_9\right)\mathbf{a}_2 + x_9\,\mathbf{a}_3$ | = | $\left(\frac{1}{2} - y_9\right)a\,\hat{\mathbf{x}} + \left(\frac{1}{2} + z_9\right)a\,\hat{\mathbf{y}} + x_9 a\,\hat{\mathbf{z}}$ | (24*d*) | H IV |
| $\mathbf{B}_{128}$ | = | $\left(\frac{1}{2} + y_9\right)\mathbf{a}_1 + z_9\,\mathbf{a}_2 + \left(\frac{1}{2} - x_9\right)\mathbf{a}_3$ | = | $\left(\frac{1}{2} + y_9\right)a\,\hat{\mathbf{x}} + z_9 a\,\hat{\mathbf{y}} + \left(\frac{1}{2} - x_9\right)a\,\hat{\mathbf{z}}$ | (24*d*) | H IV |
| $\mathbf{B}_{129}$ | = | $x_{10}\,\mathbf{a}_1 + y_{10}\,\mathbf{a}_2 + z_{10}\,\mathbf{a}_3$ | = | $x_{10} a\,\hat{\mathbf{x}} + y_{10} a\,\hat{\mathbf{y}} + z_{10} a\,\hat{\mathbf{z}}$ | (24*d*) | O III |
| $\mathbf{B}_{130}$ | = | $\left(\frac{1}{2} - x_{10}\right)\mathbf{a}_1 - y_{10}\,\mathbf{a}_2 + \left(\frac{1}{2} + z_{10}\right)\mathbf{a}_3$ | = | $\left(\frac{1}{2} - x_{10}\right)a\,\hat{\mathbf{x}} - y_{10} a\,\hat{\mathbf{y}} + \left(\frac{1}{2} + z_{10}\right)a\,\hat{\mathbf{z}}$ | (24*d*) | O III |
| $\mathbf{B}_{131}$ | = | $-x_{10}\,\mathbf{a}_1 + \left(\frac{1}{2} + y_{10}\right)\mathbf{a}_2 + \left(\frac{1}{2} - z_{10}\right)\mathbf{a}_3$ | = | $-x_{10} a\,\hat{\mathbf{x}} + \left(\frac{1}{2} + y_{10}\right)a\,\hat{\mathbf{y}} + \left(\frac{1}{2} - z_{10}\right)a\,\hat{\mathbf{z}}$ | (24*d*) | O III |
| $\mathbf{B}_{132}$ | = | $\left(\frac{1}{2} + x_{10}\right)\mathbf{a}_1 + \left(\frac{1}{2} - y_{10}\right)\mathbf{a}_2 - z_{10}\,\mathbf{a}_3$ | = | $\left(\frac{1}{2} + x_{10}\right)a\,\hat{\mathbf{x}} + \left(\frac{1}{2} - y_{10}\right)a\,\hat{\mathbf{y}} - z_{10} a\,\hat{\mathbf{z}}$ | (24*d*) | O III |
| $\mathbf{B}_{133}$ | = | $z_{10}\,\mathbf{a}_1 + x_{10}\,\mathbf{a}_2 + y_{10}\,\mathbf{a}_3$ | = | $z_{10} a\,\hat{\mathbf{x}} + x_{10} a\,\hat{\mathbf{y}} + y_{10} a\,\hat{\mathbf{z}}$ | (24*d*) | O III |
| $\mathbf{B}_{134}$ | = | $\left(\frac{1}{2} + z_{10}\right)\mathbf{a}_1 + \left(\frac{1}{2} - x_{10}\right)\mathbf{a}_2 - y_{10}\,\mathbf{a}_3$ | = | $\left(\frac{1}{2} + z_{10}\right)a\,\hat{\mathbf{x}} + \left(\frac{1}{2} - x_{10}\right)a\,\hat{\mathbf{y}} - y_{10} a\,\hat{\mathbf{z}}$ | (24*d*) | O III |
| $\mathbf{B}_{135}$ | = | $\left(\frac{1}{2} - z_{10}\right)\mathbf{a}_1 - x_{10}\,\mathbf{a}_2 + \left(\frac{1}{2} + y_{10}\right)\mathbf{a}_3$ | = | $\left(\frac{1}{2} - z_{10}\right)a\,\hat{\mathbf{x}} - x_{10} a\,\hat{\mathbf{y}} + \left(\frac{1}{2} + y_{10}\right)a\,\hat{\mathbf{z}}$ | (24*d*) | O III |
| $\mathbf{B}_{136}$ | = | $-z_{10}\,\mathbf{a}_1 + \left(\frac{1}{2} + x_{10}\right)\mathbf{a}_2 + \left(\frac{1}{2} - y_{10}\right)\mathbf{a}_3$ | = | $-z_{10} a\,\hat{\mathbf{x}} + \left(\frac{1}{2} + x_{10}\right)a\,\hat{\mathbf{y}} + \left(\frac{1}{2} - y_{10}\right)a\,\hat{\mathbf{z}}$ | (24*d*) | O III |
| $\mathbf{B}_{137}$ | = | $y_{10}\,\mathbf{a}_1 + z_{10}\,\mathbf{a}_2 + x_{10}\,\mathbf{a}_3$ | = | $y_{10} a\,\hat{\mathbf{x}} + z_{10} a\,\hat{\mathbf{y}} + x_{10} a\,\hat{\mathbf{z}}$ | (24*d*) | O III |
| $\mathbf{B}_{138}$ | = | $-y_{10}\,\mathbf{a}_1 + \left(\frac{1}{2} + z_{10}\right)\mathbf{a}_2 + \left(\frac{1}{2} - x_{10}\right)\mathbf{a}_3$ | = | $-y_{10} a\,\hat{\mathbf{x}} + \left(\frac{1}{2} + z_{10}\right)a\,\hat{\mathbf{y}} + \left(\frac{1}{2} - x_{10}\right)a\,\hat{\mathbf{z}}$ | (24*d*) | O III |
| $\mathbf{B}_{139}$ | = | $\left(\frac{1}{2} + y_{10}\right)\mathbf{a}_1 + \left(\frac{1}{2} - z_{10}\right)\mathbf{a}_2 - x_{10}\,\mathbf{a}_3$ | = | $\left(\frac{1}{2} + y_{10}\right)a\,\hat{\mathbf{x}} + \left(\frac{1}{2} - z_{10}\right)a\,\hat{\mathbf{y}} - x_{10} a\,\hat{\mathbf{z}}$ | (24*d*) | O III |
| $\mathbf{B}_{140}$ | = | $\left(\frac{1}{2} - y_{10}\right)\mathbf{a}_1 - z_{10}\,\mathbf{a}_2 + \left(\frac{1}{2} + x_{10}\right)\mathbf{a}_3$ | = | $\left(\frac{1}{2} - y_{10}\right)a\,\hat{\mathbf{x}} - z_{10} a\,\hat{\mathbf{y}} + \left(\frac{1}{2} + x_{10}\right)a\,\hat{\mathbf{z}}$ | (24*d*) | O III |
| $\mathbf{B}_{141}$ | = | $-x_{10}\,\mathbf{a}_1 - y_{10}\,\mathbf{a}_2 - z_{10}\,\mathbf{a}_3$ | = | $-x_{10} a\,\hat{\mathbf{x}} - y_{10} a\,\hat{\mathbf{y}} - z_{10} a\,\hat{\mathbf{z}}$ | (24*d*) | O III |
| $\mathbf{B}_{142}$ | = | $\left(\frac{1}{2} + x_{10}\right)\mathbf{a}_1 + y_{10}\,\mathbf{a}_2 + \left(\frac{1}{2} - z_{10}\right)\mathbf{a}_3$ | = | $\left(\frac{1}{2} + x_{10}\right)a\,\hat{\mathbf{x}} + y_{10} a\,\hat{\mathbf{y}} + \left(\frac{1}{2} - z_{10}\right)a\,\hat{\mathbf{z}}$ | (24*d*) | O III |
| $\mathbf{B}_{143}$ | = | $x_{10}\,\mathbf{a}_1 + \left(\frac{1}{2} - y_{10}\right)\mathbf{a}_2 + \left(\frac{1}{2} + z_{10}\right)\mathbf{a}_3$ | = | $x_{10} a\,\hat{\mathbf{x}} + \left(\frac{1}{2} - y_{10}\right)a\,\hat{\mathbf{y}} + \left(\frac{1}{2} + z_{10}\right)a\,\hat{\mathbf{z}}$ | (24*d*) | O III |
| $\mathbf{B}_{144}$ | = | $\left(\frac{1}{2} - x_{10}\right)\mathbf{a}_1 + \left(\frac{1}{2} + y_{10}\right)\mathbf{a}_2 + z_{10}\,\mathbf{a}_3$ | = | $\left(\frac{1}{2} - x_{10}\right)a\,\hat{\mathbf{x}} + \left(\frac{1}{2} + y_{10}\right)a\,\hat{\mathbf{y}} + z_{10} a\,\hat{\mathbf{z}}$ | (24*d*) | O III |
| $\mathbf{B}_{145}$ | = | $-z_{10}\,\mathbf{a}_1 - x_{10}\,\mathbf{a}_2 - y_{10}\,\mathbf{a}_3$ | = | $-z_{10} a\,\hat{\mathbf{x}} - x_{10} a\,\hat{\mathbf{y}} - y_{10} a\,\hat{\mathbf{z}}$ | (24*d*) | O III |
| $\mathbf{B}_{146}$ | = | $\left(\frac{1}{2} - z_{10}\right)\mathbf{a}_1 + \left(\frac{1}{2} + x_{10}\right)\mathbf{a}_2 + y_{10}\,\mathbf{a}_3$ | = | $\left(\frac{1}{2} - z_{10}\right)a\,\hat{\mathbf{x}} + \left(\frac{1}{2} + x_{10}\right)a\,\hat{\mathbf{y}} + y_{10} a\,\hat{\mathbf{z}}$ | (24*d*) | O III |
| $\mathbf{B}_{147}$ | = | $\left(\frac{1}{2} + z_{10}\right)\mathbf{a}_1 + x_{10}\,\mathbf{a}_2 + \left(\frac{1}{2} - y_{10}\right)\mathbf{a}_3$ | = | $\left(\frac{1}{2} + z_{10}\right)a\,\hat{\mathbf{x}} + x_{10} a\,\hat{\mathbf{y}} + \left(\frac{1}{2} - y_{10}\right)a\,\hat{\mathbf{z}}$ | (24*d*) | O III |
| $\mathbf{B}_{148}$ | = | $z_{10}\,\mathbf{a}_1 + \left(\frac{1}{2} - x_{10}\right)\mathbf{a}_2 + \left(\frac{1}{2} + y_{10}\right)\mathbf{a}_3$ | = | $z_{10} a\,\hat{\mathbf{x}} + \left(\frac{1}{2} - x_{10}\right)a\,\hat{\mathbf{y}} + \left(\frac{1}{2} + y_{10}\right)a\,\hat{\mathbf{z}}$ | (24*d*) | O III |
| $\mathbf{B}_{149}$ | = | $-y_{10}\,\mathbf{a}_1 - z_{10}\,\mathbf{a}_2 - x_{10}\,\mathbf{a}_3$ | = | $-y_{10} a\,\hat{\mathbf{x}} - z_{10} a\,\hat{\mathbf{y}} - x_{10} a\,\hat{\mathbf{z}}$ | (24*d*) | O III |
| $\mathbf{B}_{150}$ | = | $y_{10}\,\mathbf{a}_1 + \left(\frac{1}{2} - z_{10}\right)\mathbf{a}_2 + \left(\frac{1}{2} + x_{10}\right)\mathbf{a}_3$ | = | $y_{10} a\,\hat{\mathbf{x}} + \left(\frac{1}{2} - z_{10}\right)a\,\hat{\mathbf{y}} + \left(\frac{1}{2} + x_{10}\right)a\,\hat{\mathbf{z}}$ | (24*d*) | O III |
| $\mathbf{B}_{151}$ | = | $\left(\frac{1}{2} - y_{10}\right)\mathbf{a}_1 + \left(\frac{1}{2} + z_{10}\right)\mathbf{a}_2 + x_{10}\,\mathbf{a}_3$ | = | $\left(\frac{1}{2} - y_{10}\right)a\,\hat{\mathbf{x}} + \left(\frac{1}{2} + z_{10}\right)a\,\hat{\mathbf{y}} + x_{10} a\,\hat{\mathbf{z}}$ | (24*d*) | O III |
| $\mathbf{B}_{152}$ | = | $\left(\frac{1}{2} + y_{10}\right)\mathbf{a}_1 + z_{10}\,\mathbf{a}_2 + \left(\frac{1}{2} - x_{10}\right)\mathbf{a}_3$ | = | $\left(\frac{1}{2} + y_{10}\right)a\,\hat{\mathbf{x}} + z_{10} a\,\hat{\mathbf{y}} + \left(\frac{1}{2} - x_{10}\right)a\,\hat{\mathbf{z}}$ | (24*d*) | O III |
| $\mathbf{B}_{153}$ | = | $x_{11}\,\mathbf{a}_1 + y_{11}\,\mathbf{a}_2 + z_{11}\,\mathbf{a}_3$ | = | $x_{11} a\,\hat{\mathbf{x}} + y_{11} a\,\hat{\mathbf{y}} + z_{11} a\,\hat{\mathbf{z}}$ | (24*d*) | O IV |
| $\mathbf{B}_{154}$ | = | $\left(\frac{1}{2} - x_{11}\right)\mathbf{a}_1 - y_{11}\,\mathbf{a}_2 + \left(\frac{1}{2} + z_{11}\right)\mathbf{a}_3$ | = | $\left(\frac{1}{2} - x_{11}\right)a\,\hat{\mathbf{x}} - y_{11} a\,\hat{\mathbf{y}} + \left(\frac{1}{2} + z_{11}\right)a\,\hat{\mathbf{z}}$ | (24*d*) | O IV |
| $\mathbf{B}_{155}$ | = | $-x_{11}\,\mathbf{a}_1 + \left(\frac{1}{2} + y_{11}\right)\mathbf{a}_2 + \left(\frac{1}{2} - z_{11}\right)\mathbf{a}_3$ | = | $-x_{11} a\,\hat{\mathbf{x}} + \left(\frac{1}{2} + y_{11}\right)a\,\hat{\mathbf{y}} + \left(\frac{1}{2} - z_{11}\right)a\,\hat{\mathbf{z}}$ | (24*d*) | O IV |
| $\mathbf{B}_{156}$ | = | $\left(\frac{1}{2} + x_{11}\right)\mathbf{a}_1 + \left(\frac{1}{2} - y_{11}\right)\mathbf{a}_2 - z_{11}\,\mathbf{a}_3$ | = | $\left(\frac{1}{2} + x_{11}\right)a\,\hat{\mathbf{x}} + \left(\frac{1}{2} - y_{11}\right)a\,\hat{\mathbf{y}} - z_{11} a\,\hat{\mathbf{z}}$ | (24*d*) | O IV |
| $\mathbf{B}_{157}$ | = | $z_{11}\,\mathbf{a}_1 + x_{11}\,\mathbf{a}_2 + y_{11}\,\mathbf{a}_3$ | = | $z_{11} a\,\hat{\mathbf{x}} + x_{11} a\,\hat{\mathbf{y}} + y_{11} a\,\hat{\mathbf{z}}$ | (24*d*) | O IV |
| $\mathbf{B}_{158}$ | = | $\left(\frac{1}{2}+z_{11}\right)\mathbf{a}_1+\left(\frac{1}{2}-x_{11}\right)\mathbf{a}_2-y_{11}\,\mathbf{a}_3$ | = | $\left(\frac{1}{2}+z_{11}\right)a\,\hat{\mathbf{x}}+\left(\frac{1}{2}-x_{11}\right)a\,\hat{\mathbf{y}}-y_{11}a\,\hat{\mathbf{z}}$ | (24$d$) | O IV |
| $\mathbf{B}_{159}$ | = | $\left(\frac{1}{2}-z_{11}\right)\mathbf{a}_1-x_{11}\,\mathbf{a}_2+\left(\frac{1}{2}+y_{11}\right)\mathbf{a}_3$ | = | $\left(\frac{1}{2}-z_{11}\right)a\,\hat{\mathbf{x}}-x_{11}a\,\hat{\mathbf{y}}+\left(\frac{1}{2}+y_{11}\right)a\,\hat{\mathbf{z}}$ | (24$d$) | O IV |
| $\mathbf{B}_{160}$ | = | $-z_{11}\,\mathbf{a}_1+\left(\frac{1}{2}+x_{11}\right)\mathbf{a}_2+\left(\frac{1}{2}-y_{11}\right)\mathbf{a}_3$ | = | $-z_{11}a\,\hat{\mathbf{x}}+\left(\frac{1}{2}+x_{11}\right)a\,\hat{\mathbf{y}}+\left(\frac{1}{2}-y_{11}\right)a\,\hat{\mathbf{z}}$ | (24$d$) | O IV |
| $\mathbf{B}_{161}$ | = | $y_{11}\,\mathbf{a}_1+z_{11}\,\mathbf{a}_2+x_{11}\,\mathbf{a}_3$ | = | $y_{11}a\,\hat{\mathbf{x}}+z_{11}a\,\hat{\mathbf{y}}+x_{11}a\,\hat{\mathbf{z}}$ | (24$d$) | O IV |
| $\mathbf{B}_{162}$ | = | $-y_{11}\,\mathbf{a}_1+\left(\frac{1}{2}+z_{11}\right)\mathbf{a}_2+\left(\frac{1}{2}-x_{11}\right)\mathbf{a}_3$ | = | $-y_{11}a\,\hat{\mathbf{x}}+\left(\frac{1}{2}+z_{11}\right)a\,\hat{\mathbf{y}}+\left(\frac{1}{2}-x_{11}\right)a\,\hat{\mathbf{z}}$ | (24$d$) | O IV |
| $\mathbf{B}_{163}$ | = | $\left(\frac{1}{2}+y_{11}\right)\mathbf{a}_1+\left(\frac{1}{2}-z_{11}\right)\mathbf{a}_2-x_{11}\,\mathbf{a}_3$ | = | $\left(\frac{1}{2}+y_{11}\right)a\,\hat{\mathbf{x}}+\left(\frac{1}{2}-z_{11}\right)a\,\hat{\mathbf{y}}-x_{11}a\,\hat{\mathbf{z}}$ | (24$d$) | O IV |
| $\mathbf{B}_{164}$ | = | $\left(\frac{1}{2}-y_{11}\right)\mathbf{a}_1-z_{11}\,\mathbf{a}_2+\left(\frac{1}{2}+x_{11}\right)\mathbf{a}_3$ | = | $\left(\frac{1}{2}-y_{11}\right)a\,\hat{\mathbf{x}}-z_{11}a\,\hat{\mathbf{y}}+\left(\frac{1}{2}+x_{11}\right)a\,\hat{\mathbf{z}}$ | (24$d$) | O IV |
| $\mathbf{B}_{165}$ | = | $-x_{11}\,\mathbf{a}_1-y_{11}\,\mathbf{a}_2-z_{11}\,\mathbf{a}_3$ | = | $-x_{11}a\,\hat{\mathbf{x}}-y_{11}a\,\hat{\mathbf{y}}-z_{11}a\,\hat{\mathbf{z}}$ | (24$d$) | O IV |
| $\mathbf{B}_{166}$ | = | $\left(\frac{1}{2}+x_{11}\right)\mathbf{a}_1+y_{11}\,\mathbf{a}_2+\left(\frac{1}{2}-z_{11}\right)\mathbf{a}_3$ | = | $\left(\frac{1}{2}+x_{11}\right)a\,\hat{\mathbf{x}}+y_{11}a\,\hat{\mathbf{y}}+\left(\frac{1}{2}-z_{11}\right)a\,\hat{\mathbf{z}}$ | (24$d$) | O IV |
| $\mathbf{B}_{167}$ | = | $x_{11}\,\mathbf{a}_1+\left(\frac{1}{2}-y_{11}\right)\mathbf{a}_2+\left(\frac{1}{2}+z_{11}\right)\mathbf{a}_3$ | = | $x_{11}a\,\hat{\mathbf{x}}+\left(\frac{1}{2}-y_{11}\right)a\,\hat{\mathbf{y}}+\left(\frac{1}{2}+z_{11}\right)a\,\hat{\mathbf{z}}$ | (24$d$) | O IV |
| $\mathbf{B}_{168}$ | = | $\left(\frac{1}{2}-x_{11}\right)\mathbf{a}_1+\left(\frac{1}{2}+y_{11}\right)\mathbf{a}_2+z_{11}\,\mathbf{a}_3$ | = | $\left(\frac{1}{2}-x_{11}\right)a\,\hat{\mathbf{x}}+\left(\frac{1}{2}+y_{11}\right)a\,\hat{\mathbf{y}}+z_{11}a\,\hat{\mathbf{z}}$ | (24$d$) | O IV |
| $\mathbf{B}_{169}$ | = | $-z_{11}\,\mathbf{a}_1-x_{11}\,\mathbf{a}_2-y_{11}\,\mathbf{a}_3$ | = | $-z_{11}a\,\hat{\mathbf{x}}-x_{11}a\,\hat{\mathbf{y}}-y_{11}a\,\hat{\mathbf{z}}$ | (24$d$) | O IV |
| $\mathbf{B}_{170}$ | = | $\left(\frac{1}{2}-z_{11}\right)\mathbf{a}_1+\left(\frac{1}{2}+x_{11}\right)\mathbf{a}_2+y_{11}\,\mathbf{a}_3$ | = | $\left(\frac{1}{2}-z_{11}\right)a\,\hat{\mathbf{x}}+\left(\frac{1}{2}+x_{11}\right)a\,\hat{\mathbf{y}}+y_{11}a\,\hat{\mathbf{z}}$ | (24$d$) | O IV |
| $\mathbf{B}_{171}$ | = | $\left(\frac{1}{2}+z_{11}\right)\mathbf{a}_1+x_{11}\,\mathbf{a}_2+\left(\frac{1}{2}-y_{11}\right)\mathbf{a}_3$ | = | $\left(\frac{1}{2}+z_{11}\right)a\,\hat{\mathbf{x}}+x_{11}a\,\hat{\mathbf{y}}+\left(\frac{1}{2}-y_{11}\right)a\,\hat{\mathbf{z}}$ | (24$d$) | O IV |
| $\mathbf{B}_{172}$ | = | $z_{11}\,\mathbf{a}_1+\left(\frac{1}{2}-x_{11}\right)\mathbf{a}_2+\left(\frac{1}{2}+y_{11}\right)\mathbf{a}_3$ | = | $z_{11}a\,\hat{\mathbf{x}}+\left(\frac{1}{2}-x_{11}\right)a\,\hat{\mathbf{y}}+\left(\frac{1}{2}+y_{11}\right)a\,\hat{\mathbf{z}}$ | (24$d$) | O IV |
| $\mathbf{B}_{173}$ | = | $-y_{11}\,\mathbf{a}_1-z_{11}\,\mathbf{a}_2-x_{11}\,\mathbf{a}_3$ | = | $-y_{11}a\,\hat{\mathbf{x}}-z_{11}a\,\hat{\mathbf{y}}-x_{11}a\,\hat{\mathbf{z}}$ | (24$d$) | O IV |
| $\mathbf{B}_{174}$ | = | $y_{11}\,\mathbf{a}_1+\left(\frac{1}{2}-z_{11}\right)\mathbf{a}_2+\left(\frac{1}{2}+x_{11}\right)\mathbf{a}_3$ | = | $y_{11}a\,\hat{\mathbf{x}}+\left(\frac{1}{2}-z_{11}\right)a\,\hat{\mathbf{y}}+\left(\frac{1}{2}+x_{11}\right)a\,\hat{\mathbf{z}}$ | (24$d$) | O IV |
| $\mathbf{B}_{175}$ | = | $\left(\frac{1}{2}-y_{11}\right)\mathbf{a}_1+\left(\frac{1}{2}+z_{11}\right)\mathbf{a}_2+x_{11}\,\mathbf{a}_3$ | = | $\left(\frac{1}{2}-y_{11}\right)a\,\hat{\mathbf{x}}+\left(\frac{1}{2}+z_{11}\right)a\,\hat{\mathbf{y}}+x_{11}a\,\hat{\mathbf{z}}$ | (24$d$) | O IV |
| $\mathbf{B}_{176}$ | = | $\left(\frac{1}{2}+y_{11}\right)\mathbf{a}_1+z_{11}\,\mathbf{a}_2+\left(\frac{1}{2}-x_{11}\right)\mathbf{a}_3$ | = | $\left(\frac{1}{2}+y_{11}\right)a\,\hat{\mathbf{x}}+z_{11}a\,\hat{\mathbf{y}}+\left(\frac{1}{2}-x_{11}\right)a\,\hat{\mathbf{z}}$ | (24$d$) | O IV |
| $\mathbf{B}_{177}$ | = | $x_{12}\,\mathbf{a}_1+y_{12}\,\mathbf{a}_2+z_{12}\,\mathbf{a}_3$ | = | $x_{12}a\,\hat{\mathbf{x}}+y_{12}a\,\hat{\mathbf{y}}+z_{12}a\,\hat{\mathbf{z}}$ | (24$d$) | O V |
| $\mathbf{B}_{178}$ | = | $\left(\frac{1}{2}-x_{12}\right)\mathbf{a}_1-y_{12}\,\mathbf{a}_2+\left(\frac{1}{2}+z_{12}\right)\mathbf{a}_3$ | = | $\left(\frac{1}{2}-x_{12}\right)a\,\hat{\mathbf{x}}-y_{12}a\,\hat{\mathbf{y}}+\left(\frac{1}{2}+z_{12}\right)a\,\hat{\mathbf{z}}$ | (24$d$) | O V |
| $\mathbf{B}_{179}$ | = | $-x_{12}\,\mathbf{a}_1+\left(\frac{1}{2}+y_{12}\right)\mathbf{a}_2+\left(\frac{1}{2}-z_{12}\right)\mathbf{a}_3$ | = | $-x_{12}a\,\hat{\mathbf{x}}+\left(\frac{1}{2}+y_{12}\right)a\,\hat{\mathbf{y}}+\left(\frac{1}{2}-z_{12}\right)a\,\hat{\mathbf{z}}$ | (24$d$) | O V |
| $\mathbf{B}_{180}$ | = | $\left(\frac{1}{2}+x_{12}\right)\mathbf{a}_1+\left(\frac{1}{2}-y_{12}\right)\mathbf{a}_2-z_{12}\,\mathbf{a}_3$ | = | $\left(\frac{1}{2}+x_{12}\right)a\,\hat{\mathbf{x}}+\left(\frac{1}{2}-y_{12}\right)a\,\hat{\mathbf{y}}-z_{12}a\,\hat{\mathbf{z}}$ | (24$d$) | O V |
| $\mathbf{B}_{181}$ | = | $z_{12}\,\mathbf{a}_1+x_{12}\,\mathbf{a}_2+y_{12}\,\mathbf{a}_3$ | = | $z_{12}a\,\hat{\mathbf{x}}+x_{12}a\,\hat{\mathbf{y}}+y_{12}a\,\hat{\mathbf{z}}$ | (24$d$) | O V |
| $\mathbf{B}_{182}$ | = | $\left(\frac{1}{2}+z_{12}\right)\mathbf{a}_1+\left(\frac{1}{2}-x_{12}\right)\mathbf{a}_2-y_{12}\,\mathbf{a}_3$ | = | $\left(\frac{1}{2}+z_{12}\right)a\,\hat{\mathbf{x}}+\left(\frac{1}{2}-x_{12}\right)a\,\hat{\mathbf{y}}-y_{12}a\,\hat{\mathbf{z}}$ | (24$d$) | O V |
| $\mathbf{B}_{183}$ | = | $\left(\frac{1}{2}-z_{12}\right)\mathbf{a}_1-x_{12}\,\mathbf{a}_2+\left(\frac{1}{2}+y_{12}\right)\mathbf{a}_3$ | = | $\left(\frac{1}{2}-z_{12}\right)a\,\hat{\mathbf{x}}-x_{12}a\,\hat{\mathbf{y}}+\left(\frac{1}{2}+y_{12}\right)a\,\hat{\mathbf{z}}$ | (24$d$) | O V |
| $\mathbf{B}_{184}$ | = | $-z_{12}\,\mathbf{a}_1+\left(\frac{1}{2}+x_{12}\right)\mathbf{a}_2+\left(\frac{1}{2}-y_{12}\right)\mathbf{a}_3$ | = | $-z_{12}a\,\hat{\mathbf{x}}+\left(\frac{1}{2}+x_{12}\right)a\,\hat{\mathbf{y}}+\left(\frac{1}{2}-y_{12}\right)a\,\hat{\mathbf{z}}$ | (24$d$) | O V |
| $\mathbf{B}_{185}$ | = | $y_{12}\,\mathbf{a}_1+z_{12}\,\mathbf{a}_2+x_{12}\,\mathbf{a}_3$ | = | $y_{12}a\,\hat{\mathbf{x}}+z_{12}a\,\hat{\mathbf{y}}+x_{12}a\,\hat{\mathbf{z}}$ | (24$d$) | O V |
| $\mathbf{B}_{186}$ | = | $-y_{12}\,\mathbf{a}_1+\left(\frac{1}{2}+z_{12}\right)\mathbf{a}_2+\left(\frac{1}{2}-x_{12}\right)\mathbf{a}_3$ | = | $-y_{12}a\,\hat{\mathbf{x}}+\left(\frac{1}{2}+z_{12}\right)a\,\hat{\mathbf{y}}+\left(\frac{1}{2}-x_{12}\right)a\,\hat{\mathbf{z}}$ | (24$d$) | O V |
| $\mathbf{B}_{187}$ | = | $\left(\frac{1}{2}+y_{12}\right)\mathbf{a}_1+\left(\frac{1}{2}-z_{12}\right)\mathbf{a}_2-x_{12}\,\mathbf{a}_3$ | = | $\left(\frac{1}{2}+y_{12}\right)a\,\hat{\mathbf{x}}+\left(\frac{1}{2}-z_{12}\right)a\,\hat{\mathbf{y}}-x_{12}a\,\hat{\mathbf{z}}$ | (24$d$) | O V |
| $\mathbf{B}_{188}$ | = | $\left(\frac{1}{2}-y_{12}\right)\mathbf{a}_1-z_{12}\,\mathbf{a}_2+\left(\frac{1}{2}+x_{12}\right)\mathbf{a}_3$ | = | $\left(\frac{1}{2}-y_{12}\right)a\,\hat{\mathbf{x}}-z_{12}a\,\hat{\mathbf{y}}+\left(\frac{1}{2}+x_{12}\right)a\,\hat{\mathbf{z}}$ | (24$d$) | O V |
| $\mathbf{B}_{189}$ | = | $-x_{12}\,\mathbf{a}_1-y_{12}\,\mathbf{a}_2-z_{12}\,\mathbf{a}_3$ | = | $-x_{12}a\,\hat{\mathbf{x}}-y_{12}a\,\hat{\mathbf{y}}-z_{12}a\,\hat{\mathbf{z}}$ | (24$d$) | O V |
| $\mathbf{B}_{190}$ | = | $\left(\frac{1}{2}+x_{12}\right)\mathbf{a}_1+y_{12}\,\mathbf{a}_2+\left(\frac{1}{2}-z_{12}\right)\mathbf{a}_3$ | = | $\left(\frac{1}{2}+x_{12}\right)a\,\hat{\mathbf{x}}+y_{12}a\,\hat{\mathbf{y}}+\left(\frac{1}{2}-z_{12}\right)a\,\hat{\mathbf{z}}$ | (24$d$) | O V |
| $\mathbf{B}_{191}$ | = | $x_{12}\,\mathbf{a}_1+\left(\frac{1}{2}-y_{12}\right)\mathbf{a}_2+\left(\frac{1}{2}+z_{12}\right)\mathbf{a}_3$ | = | $x_{12}a\,\hat{\mathbf{x}}+\left(\frac{1}{2}-y_{12}\right)a\,\hat{\mathbf{y}}+\left(\frac{1}{2}+z_{12}\right)a\,\hat{\mathbf{z}}$ | (24$d$) | O V |
| $\mathbf{B}_{192}$ | = | $\left(\frac{1}{2}-x_{12}\right)\mathbf{a}_1+\left(\frac{1}{2}+y_{12}\right)\mathbf{a}_2+z_{12}\,\mathbf{a}_3$ | = | $\left(\frac{1}{2}-x_{12}\right)a\,\hat{\mathbf{x}}+\left(\frac{1}{2}+y_{12}\right)a\,\hat{\mathbf{y}}+z_{12}a\,\hat{\mathbf{z}}$ | (24$d$) | O V |
| $\mathbf{B}_{193}$ | = | $-z_{12}\,\mathbf{a}_1-x_{12}\,\mathbf{a}_2-y_{12}\,\mathbf{a}_3$ | = | $-z_{12}a\,\hat{\mathbf{x}}-x_{12}a\,\hat{\mathbf{y}}-y_{12}a\,\hat{\mathbf{z}}$ | (24$d$) | O V |
| $\mathbf{B}_{194}$ | = | $\left(\frac{1}{2}-z_{12}\right)\mathbf{a}_1+\left(\frac{1}{2}+x_{12}\right)\mathbf{a}_2+y_{12}\,\mathbf{a}_3$ | = | $\left(\frac{1}{2}-z_{12}\right)a\,\hat{\mathbf{x}}+\left(\frac{1}{2}+x_{12}\right)a\,\hat{\mathbf{y}}+y_{12}a\,\hat{\mathbf{z}}$ | (24$d$) | O V |
| $\mathbf{B}_{195}$ | = | $\left(\frac{1}{2}+z_{12}\right)\mathbf{a}_1+x_{12}\,\mathbf{a}_2+\left(\frac{1}{2}-y_{12}\right)\mathbf{a}_3$ | = | $\left(\frac{1}{2}+z_{12}\right)a\,\hat{\mathbf{x}}+x_{12}a\,\hat{\mathbf{y}}+\left(\frac{1}{2}-y_{12}\right)a\,\hat{\mathbf{z}}$ | (24$d$) | O V |
| $\mathbf{B}_{196}$ | = | $z_{12}\,\mathbf{a}_1+\left(\frac{1}{2}-x_{12}\right)\mathbf{a}_2+\left(\frac{1}{2}+y_{12}\right)\mathbf{a}_3$ | = | $z_{12}a\,\hat{\mathbf{x}}+\left(\frac{1}{2}-x_{12}\right)a\,\hat{\mathbf{y}}+\left(\frac{1}{2}+y_{12}\right)a\,\hat{\mathbf{z}}$ | (24$d$) | O V |
| $\mathbf{B}_{197}$ | = | $-y_{12}\,\mathbf{a}_1-z_{12}\,\mathbf{a}_2-x_{12}\,\mathbf{a}_3$ | = | $-y_{12}a\,\hat{\mathbf{x}}-z_{12}a\,\hat{\mathbf{y}}-x_{12}a\,\hat{\mathbf{z}}$ | (24$d$) | O V |
| $\mathbf{B}_{198}$ | = | $y_{12}\,\mathbf{a}_1+\left(\frac{1}{2}-z_{12}\right)\mathbf{a}_2+\left(\frac{1}{2}+x_{12}\right)\mathbf{a}_3$ | = | $y_{12}a\,\hat{\mathbf{x}}+\left(\frac{1}{2}-z_{12}\right)a\,\hat{\mathbf{y}}+\left(\frac{1}{2}+x_{12}\right)a\,\hat{\mathbf{z}}$ | (24$d$) | O V |
| $\mathbf{B}_{199}$ | = | $\left(\frac{1}{2}-y_{12}\right)\mathbf{a}_1+\left(\frac{1}{2}+z_{12}\right)\mathbf{a}_2+x_{12}\,\mathbf{a}_3$ | = | $\left(\frac{1}{2}-y_{12}\right)a\,\hat{\mathbf{x}}+\left(\frac{1}{2}+z_{12}\right)a\,\hat{\mathbf{y}}+x_{12}a\,\hat{\mathbf{z}}$ | (24$d$) | O V |
| $\mathbf{B}_{200}$ | = | $\left(\frac{1}{2}+y_{12}\right)\mathbf{a}_1+z_{12}\,\mathbf{a}_2+\left(\frac{1}{2}-x_{12}\right)\mathbf{a}_3$ | = | $\left(\frac{1}{2}+y_{12}\right)a\,\hat{\mathbf{x}}+z_{12}a\,\hat{\mathbf{y}}+\left(\frac{1}{2}-x_{12}\right)a\,\hat{\mathbf{z}}$ | (24$d$) | O V |
| $\mathbf{B}_{201}$ | = | $x_{13}\,\mathbf{a}_1+y_{13}\,\mathbf{a}_2+z_{13}\,\mathbf{a}_3$ | = | $x_{13}a\,\hat{\mathbf{x}}+y_{13}a\,\hat{\mathbf{y}}+z_{13}a\,\hat{\mathbf{z}}$ | (24$d$) | O VI |
| $\mathbf{B}_{202}$ | = | $\left(\frac{1}{2}-x_{13}\right)\mathbf{a}_1-y_{13}\,\mathbf{a}_2+\left(\frac{1}{2}+z_{13}\right)\mathbf{a}_3$ | = | $\left(\frac{1}{2}-x_{13}\right)a\,\hat{\mathbf{x}}-y_{13}a\,\hat{\mathbf{y}}+\left(\frac{1}{2}+z_{13}\right)a\,\hat{\mathbf{z}}$ | (24$d$) | O VI |
| $\mathbf{B}_{203}$ | = | $-x_{13}\,\mathbf{a}_1+\left(\frac{1}{2}+y_{13}\right)\mathbf{a}_2+\left(\frac{1}{2}-z_{13}\right)\mathbf{a}_3$ | = | $-x_{13}a\,\hat{\mathbf{x}}+\left(\frac{1}{2}+y_{13}\right)a\,\hat{\mathbf{y}}+\left(\frac{1}{2}-z_{13}\right)a\,\hat{\mathbf{z}}$ | (24$d$) | O VI |
| $\mathbf{B}_{204}$ | = | $\left(\frac{1}{2}+x_{13}\right)\mathbf{a}_1+\left(\frac{1}{2}-y_{13}\right)\mathbf{a}_2-z_{13}\,\mathbf{a}_3$ | = | $\left(\frac{1}{2}+x_{13}\right)a\,\hat{\mathbf{x}}+\left(\frac{1}{2}-y_{13}\right)a\,\hat{\mathbf{y}}-z_{13}a\,\hat{\mathbf{z}}$ | (24$d$) | O VI |
| $\mathbf{B}_{205}$ | = | $z_{13}\,\mathbf{a}_1+x_{13}\,\mathbf{a}_2+y_{13}\,\mathbf{a}_3$ | = | $z_{13}a\,\hat{\mathbf{x}}+x_{13}a\,\hat{\mathbf{y}}+y_{13}a\,\hat{\mathbf{z}}$ | (24$d$) | O VI |
| $\mathbf{B}_{206}$ | = | $\left(\frac{1}{2}+z_{13}\right)\mathbf{a}_1+\left(\frac{1}{2}-x_{13}\right)\mathbf{a}_2-y_{13}\,\mathbf{a}_3$ | = | $\left(\frac{1}{2}+z_{13}\right)a\,\hat{\mathbf{x}}+\left(\frac{1}{2}-x_{13}\right)a\,\hat{\mathbf{y}}-y_{13}a\,\hat{\mathbf{z}}$ | (24$d$) | O VI |
| $\mathbf{B}_{207}$ | = | $\left(\frac{1}{2}-z_{13}\right)\mathbf{a}_1-x_{13}\,\mathbf{a}_2+\left(\frac{1}{2}+y_{13}\right)\mathbf{a}_3$ | = | $\left(\frac{1}{2}-z_{13}\right)a\,\hat{\mathbf{x}}-x_{13}a\,\hat{\mathbf{y}}+\left(\frac{1}{2}+y_{13}\right)a\,\hat{\mathbf{z}}$ | (24$d$) | O VI |
| $\mathbf{B}_{208}$ | = | $-z_{13}\,\mathbf{a}_1+\left(\frac{1}{2}+x_{13}\right)\mathbf{a}_2+\left(\frac{1}{2}-y_{13}\right)\mathbf{a}_3$ | = | $-z_{13}a\,\hat{\mathbf{x}}+\left(\frac{1}{2}+x_{13}\right)a\,\hat{\mathbf{y}}+\left(\frac{1}{2}-y_{13}\right)a\,\hat{\mathbf{z}}$ | (24$d$) | O VI |
| $\mathbf{B}_{209}$ | = | $y_{13}\,\mathbf{a}_1+z_{13}\,\mathbf{a}_2+x_{13}\,\mathbf{a}_3$ | = | $y_{13}a\,\hat{\mathbf{x}}+z_{13}a\,\hat{\mathbf{y}}+x_{13}a\,\hat{\mathbf{z}}$ | (24$d$) | O VI |
| $\mathbf{B}_{210}$ | = | $-y_{13}\,\mathbf{a}_1+\left(\frac{1}{2}+z_{13}\right)\mathbf{a}_2+\left(\frac{1}{2}-x_{13}\right)\mathbf{a}_3$ | = | $-y_{13}a\,\hat{\mathbf{x}}+\left(\frac{1}{2}+z_{13}\right)a\,\hat{\mathbf{y}}+\left(\frac{1}{2}-x_{13}\right)a\,\hat{\mathbf{z}}$ | (24$d$) | O VI |
| $\mathbf{B}_{211}$ | = | $\left(\frac{1}{2}+y_{13}\right)\mathbf{a}_1+\left(\frac{1}{2}-z_{13}\right)\mathbf{a}_2-x_{13}\,\mathbf{a}_3$ | = | $\left(\frac{1}{2}+y_{13}\right)a\,\hat{\mathbf{x}}+\left(\frac{1}{2}-z_{13}\right)a\,\hat{\mathbf{y}}-x_{13}a\,\hat{\mathbf{z}}$ | (24$d$) | O VI |
| $\mathbf{B}_{212}$ | = | $\left(\frac{1}{2}-y_{13}\right)\mathbf{a}_1-z_{13}\,\mathbf{a}_2+\left(\frac{1}{2}+x_{13}\right)\mathbf{a}_3$ | = | $\left(\frac{1}{2}-y_{13}\right)a\,\hat{\mathbf{x}}-z_{13}a\,\hat{\mathbf{y}}+\left(\frac{1}{2}+x_{13}\right)a\,\hat{\mathbf{z}}$ | (24$d$) | O VI |
| $\mathbf{B}_{213}$ | = | $-x_{13}\,\mathbf{a}_1-y_{13}\,\mathbf{a}_2-z_{13}\,\mathbf{a}_3$ | = | $-x_{13}a\,\hat{\mathbf{x}}-y_{13}a\,\hat{\mathbf{y}}-z_{13}a\,\hat{\mathbf{z}}$ | (24$d$) | O VI |
| $\mathbf{B}_{214}$ | = | $\left(\frac{1}{2}+x_{13}\right)\mathbf{a}_1+y_{13}\,\mathbf{a}_2+\left(\frac{1}{2}-z_{13}\right)\mathbf{a}_3$ | = | $\left(\frac{1}{2}+x_{13}\right)a\,\hat{\mathbf{x}}+y_{13}a\,\hat{\mathbf{y}}+\left(\frac{1}{2}-z_{13}\right)a\,\hat{\mathbf{z}}$ | (24$d$) | O VI |
| $\mathbf{B}_{215}$ | = | $x_{13}\,\mathbf{a}_1+\left(\frac{1}{2}-y_{13}\right)\mathbf{a}_2+\left(\frac{1}{2}+z_{13}\right)\mathbf{a}_3$ | = | $x_{13}a\,\hat{\mathbf{x}}+\left(\frac{1}{2}-y_{13}\right)a\,\hat{\mathbf{y}}+\left(\frac{1}{2}+z_{13}\right)a\,\hat{\mathbf{z}}$ | (24$d$) | O VI |
| $\mathbf{B}_{216}$ | = | $\left(\frac{1}{2}-x_{13}\right)\mathbf{a}_1+\left(\frac{1}{2}+y_{13}\right)\mathbf{a}_2+z_{13}\,\mathbf{a}_3$ | = | $\left(\frac{1}{2}-x_{13}\right)a\,\hat{\mathbf{x}}+\left(\frac{1}{2}+y_{13}\right)a\,\hat{\mathbf{y}}+z_{13}a\,\hat{\mathbf{z}}$ | (24$d$) | O VI |
| $\mathbf{B}_{217}$ | = | $-z_{13}\,\mathbf{a}_1-x_{13}\,\mathbf{a}_2-y_{13}\,\mathbf{a}_3$ | = | $-z_{13}a\,\hat{\mathbf{x}}-x_{13}a\,\hat{\mathbf{y}}-y_{13}a\,\hat{\mathbf{z}}$ | (24$d$) | O VI |
| $\mathbf{B}_{218}$ | = | $\left(\frac{1}{2}-z_{13}\right)\mathbf{a}_1+\left(\frac{1}{2}+x_{13}\right)\mathbf{a}_2+y_{13}\,\mathbf{a}_3$ | = | $\left(\frac{1}{2}-z_{13}\right)a\,\hat{\mathbf{x}}+\left(\frac{1}{2}+x_{13}\right)a\,\hat{\mathbf{y}}+y_{13}a\,\hat{\mathbf{z}}$ | (24$d$) | O VI |
| $\mathbf{B}_{219}$ | = | $\left(\frac{1}{2}+z_{13}\right)\mathbf{a}_1+x_{13}\,\mathbf{a}_2+\left(\frac{1}{2}-y_{13}\right)\mathbf{a}_3$ | = | $\left(\frac{1}{2}+z_{13}\right)a\,\hat{\mathbf{x}}+x_{13}a\,\hat{\mathbf{y}}+\left(\frac{1}{2}-y_{13}\right)a\,\hat{\mathbf{z}}$ | (24$d$) | O VI |
| $\mathbf{B}_{220}$ | = | $z_{13}\,\mathbf{a}_1+\left(\frac{1}{2}-x_{13}\right)\mathbf{a}_2+\left(\frac{1}{2}+y_{13}\right)\mathbf{a}_3$ | = | $z_{13}a\,\hat{\mathbf{x}}+\left(\frac{1}{2}-x_{13}\right)a\,\hat{\mathbf{y}}+\left(\frac{1}{2}+y_{13}\right)a\,\hat{\mathbf{z}}$ | (24$d$) | O VI |
| $\mathbf{B}_{221}$ | = | $-y_{13}\,\mathbf{a}_1-z_{13}\,\mathbf{a}_2-x_{13}\,\mathbf{a}_3$ | = | $-y_{13}a\,\hat{\mathbf{x}}-z_{13}a\,\hat{\mathbf{y}}-x_{13}a\,\hat{\mathbf{z}}$ | (24$d$) | O VI |
| $\mathbf{B}_{222}$ | = | $y_{13}\,\mathbf{a}_1+\left(\frac{1}{2}-z_{13}\right)\mathbf{a}_2+\left(\frac{1}{2}+x_{13}\right)\mathbf{a}_3$ | = | $y_{13}a\,\hat{\mathbf{x}}+\left(\frac{1}{2}-z_{13}\right)a\,\hat{\mathbf{y}}+\left(\frac{1}{2}+x_{13}\right)a\,\hat{\mathbf{z}}$ | (24$d$) | O VI |
| $\mathbf{B}_{223}$ | = | $\left(\frac{1}{2}-y_{13}\right)\mathbf{a}_1+\left(\frac{1}{2}+z_{13}\right)\mathbf{a}_2+x_{13}\,\mathbf{a}_3$ | = | $\left(\frac{1}{2}-y_{13}\right)a\,\hat{\mathbf{x}}+\left(\frac{1}{2}+z_{13}\right)a\,\hat{\mathbf{y}}+x_{13}a\,\hat{\mathbf{z}}$ | (24$d$) | O VI |

$$\mathbf{B}_{224} = \left(\frac{1}{2}+y_{13}\right)\mathbf{a}_1+z_{13}\,\mathbf{a}_2+\left(\frac{1}{2}-x_{13}\right)\mathbf{a}_3 = \left(\frac{1}{2}+y_{13}\right)a\,\hat{\mathbf{x}}+z_{13}a\,\hat{\mathbf{y}}+\left(\frac{1}{2}-x_{13}\right)a\,\hat{\mathbf{z}} \qquad (24d) \qquad \text{O VI}$$

---

---

**Geometry files:**

- CIF: pp. 1786
- POSCAR: pp. 1787

# $\beta$-Alum [$Al(NH_3CH_3)_2(SO_4)_2{\cdot}12H_2O$, $H4_{14}$] Structure: AB2C36D2E20F2_cP252_205_a_c_6d_c_c3d_c

http://aflow.org/prototype-encyclopedia/AB2C36D2E20F2_cP252_205_a_c_6d_c_c3d_c

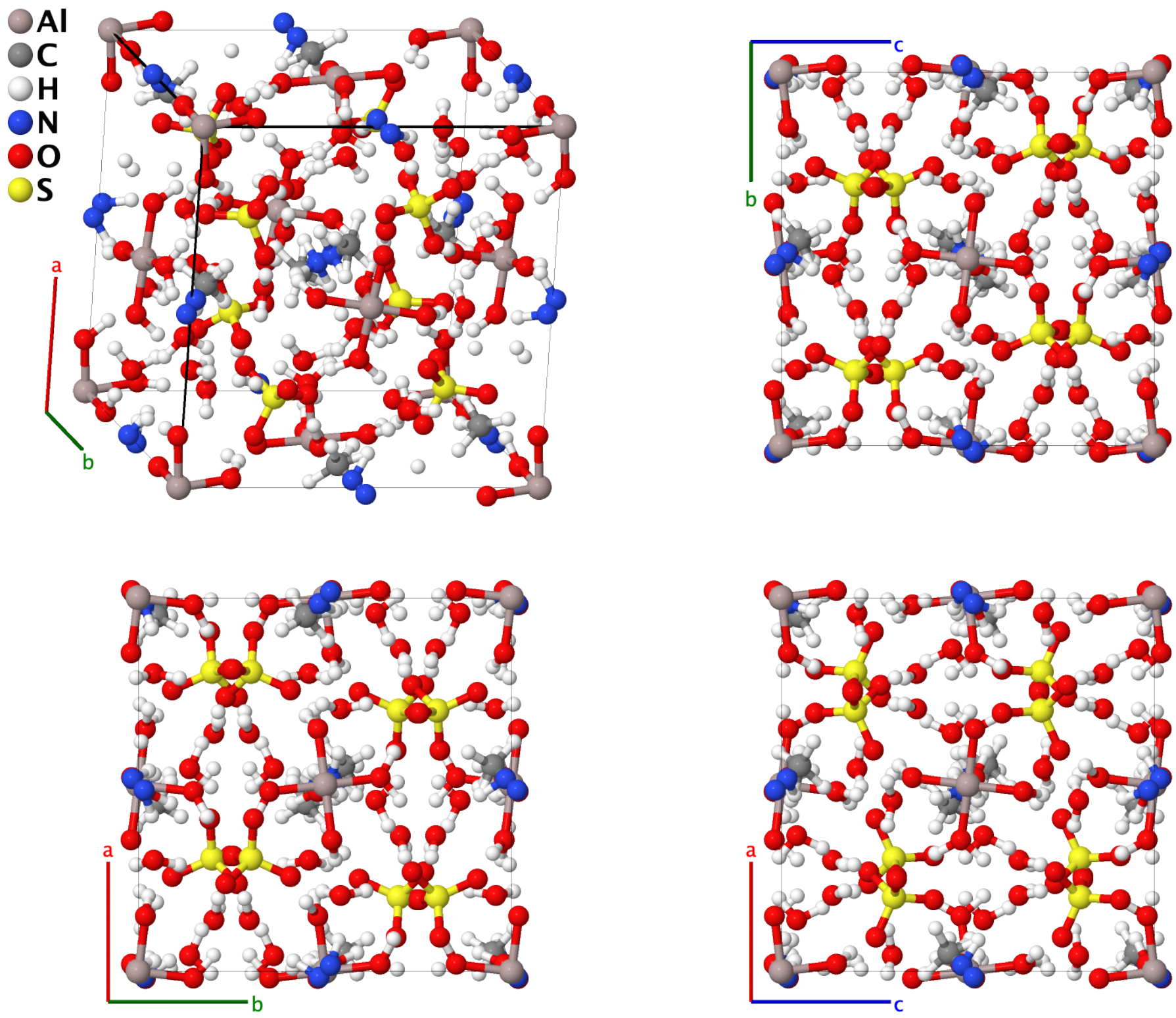

| | | |
|---|---|---|
| **Prototype** | : | $AlC_2H_{36}N_2O_{20}S_2$ |
| **AFLOW prototype label** | : | AB2C36D2E20F2_cP252_205_a_c_6d_c_c3d_c |
| ***Strukturbericht* designation** | : | $H4_{14}$ |
| **Pearson symbol** | : | cP252 |
| **Space group number** | : | 205 |
| **Space group symbol** | : | $Pa\bar{3}$ |
| **AFLOW prototype command** | : | `aflow --proto=AB2C36D2E20F2_cP252_205_a_c_6d_c_c3d_c --params=`$a, x_2, x_3, x_4, x_5, x_6, y_6, z_6, x_7, y_7, z_7, x_8, y_8, z_8, x_9, y_9, z_9, x_{10}, y_{10}, z_{10}, x_{11}, y_{11}, z_{11}, x_{12}, y_{12}, z_{12}, x_{13}, y_{13}, z_{13}, x_{14}, y_{14}, z_{14}$ |

### Other compounds with this structure

- $AlCs(SO_4)_2{\cdot}12H_2O$

---

- The alums have the general formula $AB(XO_4)_2{\cdot}12H_2O$, where $A$ is a monovalent ion, $B$ is a trivalent ion, and $X$ is a chalcogen. In most cases atom $B$ is aluminum and atom $X$ is sulfur, leading to the name alum.
- All alums have their room-temperature form in space group $Pa\bar{3}$ #205, but the bonding between the $A$ and $B$ ions and the $XO_4$ complex can be quite different.
- (Lipson, 1935ab) described three general forms of alum based on the sizes of the monovalent ions. Each of these forms was given a *Strukturbericht* designation by (Gottfried, 1937):
  - $\alpha$-alum, with intermediate sized ions, prototype $KAl(SO_4)_2{\cdot}12H_2O$, $H4_{13}$,

- $\beta$-alum, with large ions, prototype $(NH_3CH_3)Al(SO_4)_2{\cdot}12H_2O$, $H4_{14}$ (this structure), and
- $\gamma$-alum, with small ions, prototype $NaAl(SO_4)_2{\cdot}12H_2O$, $H4_{15}$.

- This classification scheme is not complete, *e.g.*, (Ledsham, 1968) points out that $NaCr(SO_4)_2{\cdot}12H_2O$ does not fit into any of these categories, and that the actual structure depends on the combination of monovalent and trivalent ions.
- As noted above, the $Pa\overline{3}$ structures of alum are the room temperature form. As the temperature decreases the alum structure may transform. For example, in the temperature range 150-170 K the $\beta$-alum $(NH_3CH_3)Al(SO_4)_2{\cdot}12H_2O$ transforms into an orthorhombic structure with fully ordered $NH_3CH_3$ ions.
- This structure was originally determined by (Lipsom, 1935c), could only determine that the $NH_3CH_3$ ion occupied the (4*b*) Wyckoff position. (Abdeen, 1981) showed that the ion was statistically distributed at two possible sites. The C-N bond distance is 1.4 Å, slightly smaller than the 1.51 Å distance observed in the low temperature structure. At any site, one of the two nitrogen positions is occupied, along with the carbon position 1.4 Å away. Six hydrogen positions from the (H-I) and (H-II) sites are then occupied.

---

**Simple Cubic primitive vectors:**

$$\mathbf{a}_1 = a\,\hat{\mathbf{x}}$$
$$\mathbf{a}_2 = a\,\hat{\mathbf{y}}$$
$$\mathbf{a}_3 = a\,\hat{\mathbf{z}}$$

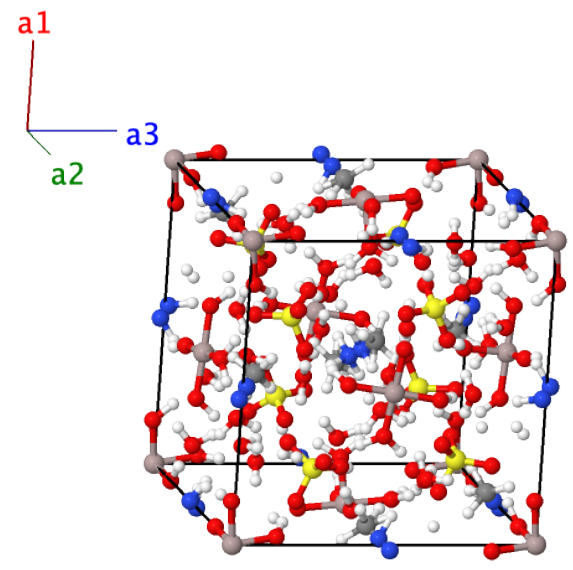

---

**Basis vectors:**

| | | Lattice Coordinates | | Cartesian Coordinates | Wyckoff Position | Atom Type |
|---|---|---|---|---|---|---|
| $\mathbf{B}_1$ | = | $0\,\mathbf{a}_1 + 0\,\mathbf{a}_2 + 0\,\mathbf{a}_3$ | = | $0\,\hat{\mathbf{x}} + 0\,\hat{\mathbf{y}} + 0\,\hat{\mathbf{z}}$ | (4*a*) | Al |
| $\mathbf{B}_2$ | = | $\frac{1}{2}\,\mathbf{a}_1 + \frac{1}{2}\,\mathbf{a}_3$ | = | $\frac{1}{2}a\,\hat{\mathbf{x}} + \frac{1}{2}a\,\hat{\mathbf{z}}$ | (4*a*) | Al |
| $\mathbf{B}_3$ | = | $\frac{1}{2}\,\mathbf{a}_2 + \frac{1}{2}\,\mathbf{a}_3$ | = | $\frac{1}{2}a\,\hat{\mathbf{y}} + \frac{1}{2}a\,\hat{\mathbf{z}}$ | (4*a*) | Al |
| $\mathbf{B}_4$ | = | $\frac{1}{2}\,\mathbf{a}_1 + \frac{1}{2}\,\mathbf{a}_2$ | = | $\frac{1}{2}a\,\hat{\mathbf{x}} + \frac{1}{2}a\,\hat{\mathbf{y}}$ | (4*a*) | Al |
| $\mathbf{B}_5$ | = | $x_2\,\mathbf{a}_1 + x_2\,\mathbf{a}_2 + x_2\,\mathbf{a}_3$ | = | $x_2a\,\hat{\mathbf{x}} + x_2a\,\hat{\mathbf{y}} + x_2a\,\hat{\mathbf{z}}$ | (8*c*) | C |
| $\mathbf{B}_6$ | = | $\left(\frac{1}{2} - x_2\right)\mathbf{a}_1 - x_2\,\mathbf{a}_2 + \left(\frac{1}{2} + x_2\right)\mathbf{a}_3$ | = | $\left(\frac{1}{2} - x_2\right)a\,\hat{\mathbf{x}} - x_2a\,\hat{\mathbf{y}} + \left(\frac{1}{2} + x_2\right)a\,\hat{\mathbf{z}}$ | (8*c*) | C |
| $\mathbf{B}_7$ | = | $-x_2\,\mathbf{a}_1 + \left(\frac{1}{2} + x_2\right)\mathbf{a}_2 + \left(\frac{1}{2} - x_2\right)\mathbf{a}_3$ | = | $-x_2a\,\hat{\mathbf{x}} + \left(\frac{1}{2} + x_2\right)a\,\hat{\mathbf{y}} + \left(\frac{1}{2} - x_2\right)a\,\hat{\mathbf{z}}$ | (8*c*) | C |
| $\mathbf{B}_8$ | = | $\left(\frac{1}{2} + x_2\right)\mathbf{a}_1 + \left(\frac{1}{2} - x_2\right)\mathbf{a}_2 - x_2\,\mathbf{a}_3$ | = | $\left(\frac{1}{2} + x_2\right)a\,\hat{\mathbf{x}} + \left(\frac{1}{2} - x_2\right)a\,\hat{\mathbf{y}} - x_2a\,\hat{\mathbf{z}}$ | (8*c*) | C |
| $\mathbf{B}_9$ | = | $-x_2\,\mathbf{a}_1 - x_2\,\mathbf{a}_2 - x_2\,\mathbf{a}_3$ | = | $-x_2a\,\hat{\mathbf{x}} - x_2a\,\hat{\mathbf{y}} - x_2a\,\hat{\mathbf{z}}$ | (8*c*) | C |
| $\mathbf{B}_{10}$ | = | $\left(\frac{1}{2} + x_2\right)\mathbf{a}_1 + x_2\,\mathbf{a}_2 + \left(\frac{1}{2} - x_2\right)\mathbf{a}_3$ | = | $\left(\frac{1}{2} + x_2\right)a\,\hat{\mathbf{x}} + x_2a\,\hat{\mathbf{y}} + \left(\frac{1}{2} - x_2\right)a\,\hat{\mathbf{z}}$ | (8*c*) | C |
| $\mathbf{B}_{11}$ | = | $x_2\,\mathbf{a}_1 + \left(\frac{1}{2} - x_2\right)\mathbf{a}_2 + \left(\frac{1}{2} + x_2\right)\mathbf{a}_3$ | = | $x_2a\,\hat{\mathbf{x}} + \left(\frac{1}{2} - x_2\right)a\,\hat{\mathbf{y}} + \left(\frac{1}{2} + x_2\right)a\,\hat{\mathbf{z}}$ | (8*c*) | C |
| $\mathbf{B}_{12}$ | = | $\left(\frac{1}{2} - x_2\right)\mathbf{a}_1 + \left(\frac{1}{2} + x_2\right)\mathbf{a}_2 + x_2\,\mathbf{a}_3$ | = | $\left(\frac{1}{2} - x_2\right)a\,\hat{\mathbf{x}} + \left(\frac{1}{2} + x_2\right)a\,\hat{\mathbf{y}} + x_2a\,\hat{\mathbf{z}}$ | (8*c*) | C |
| $\mathbf{B}_{13}$ | = | $x_3\,\mathbf{a}_1 + x_3\,\mathbf{a}_2 + x_3\,\mathbf{a}_3$ | = | $x_3a\,\hat{\mathbf{x}} + x_3a\,\hat{\mathbf{y}} + x_3a\,\hat{\mathbf{z}}$ | (8*c*) | N |
| $\mathbf{B}_{14}$ | = | $\left(\frac{1}{2} - x_3\right)\mathbf{a}_1 - x_3\,\mathbf{a}_2 + \left(\frac{1}{2} + x_3\right)\mathbf{a}_3$ | = | $\left(\frac{1}{2} - x_3\right)a\,\hat{\mathbf{x}} - x_3a\,\hat{\mathbf{y}} + \left(\frac{1}{2} + x_3\right)a\,\hat{\mathbf{z}}$ | (8*c*) | N |
| $\mathbf{B}_{15}$ | = | $-x_3\,\mathbf{a}_1 + \left(\frac{1}{2} + x_3\right)\mathbf{a}_2 + \left(\frac{1}{2} - x_3\right)\mathbf{a}_3$ | = | $-x_3a\,\hat{\mathbf{x}} + \left(\frac{1}{2} + x_3\right)a\,\hat{\mathbf{y}} + \left(\frac{1}{2} - x_3\right)a\,\hat{\mathbf{z}}$ | (8*c*) | N |
| $\mathbf{B}_{16}$ | = | $\left(\frac{1}{2} + x_3\right)\mathbf{a}_1 + \left(\frac{1}{2} - x_3\right)\mathbf{a}_2 - x_3\,\mathbf{a}_3$ | = | $\left(\frac{1}{2} + x_3\right)a\,\hat{\mathbf{x}} + \left(\frac{1}{2} - x_3\right)a\,\hat{\mathbf{y}} - x_3a\,\hat{\mathbf{z}}$ | (8*c*) | N |
| $\mathbf{B}_{17}$ | = | $-x_3\,\mathbf{a}_1 - x_3\,\mathbf{a}_2 - x_3\,\mathbf{a}_3$ | = | $-x_3a\,\hat{\mathbf{x}} - x_3a\,\hat{\mathbf{y}} - x_3a\,\hat{\mathbf{z}}$ | (8*c*) | N |
| $\mathbf{B}_{18}$ | = | $\left(\frac{1}{2}+x_3\right)\mathbf{a}_1+x_3\,\mathbf{a}_2+\left(\frac{1}{2}-x_3\right)\mathbf{a}_3$ | = | $\left(\frac{1}{2}+x_3\right)a\,\hat{\mathbf{x}}+x_3a\,\hat{\mathbf{y}}+\left(\frac{1}{2}-x_3\right)a\,\hat{\mathbf{z}}$ | (8$c$) | N |
| $\mathbf{B}_{19}$ | = | $x_3\,\mathbf{a}_1+\left(\frac{1}{2}-x_3\right)\mathbf{a}_2+\left(\frac{1}{2}+x_3\right)\mathbf{a}_3$ | = | $x_3a\,\hat{\mathbf{x}}+\left(\frac{1}{2}-x_3\right)a\,\hat{\mathbf{y}}+\left(\frac{1}{2}+x_3\right)a\,\hat{\mathbf{z}}$ | (8$c$) | N |
| $\mathbf{B}_{20}$ | = | $\left(\frac{1}{2}-x_3\right)\mathbf{a}_1+\left(\frac{1}{2}+x_3\right)\mathbf{a}_2+x_3\,\mathbf{a}_3$ | = | $\left(\frac{1}{2}-x_3\right)a\,\hat{\mathbf{x}}+\left(\frac{1}{2}+x_3\right)a\,\hat{\mathbf{y}}+x_3a\,\hat{\mathbf{z}}$ | (8$c$) | N |
| $\mathbf{B}_{21}$ | = | $x_4\,\mathbf{a}_1+x_4\,\mathbf{a}_2+x_4\,\mathbf{a}_3$ | = | $x_4a\,\hat{\mathbf{x}}+x_4a\,\hat{\mathbf{y}}+x_4a\,\hat{\mathbf{z}}$ | (8$c$) | O I |
| $\mathbf{B}_{22}$ | = | $\left(\frac{1}{2}-x_4\right)\mathbf{a}_1-x_4\,\mathbf{a}_2+\left(\frac{1}{2}+x_4\right)\mathbf{a}_3$ | = | $\left(\frac{1}{2}-x_4\right)a\,\hat{\mathbf{x}}-x_4a\,\hat{\mathbf{y}}+\left(\frac{1}{2}+x_4\right)a\,\hat{\mathbf{z}}$ | (8$c$) | O I |
| $\mathbf{B}_{23}$ | = | $-x_4\,\mathbf{a}_1+\left(\frac{1}{2}+x_4\right)\mathbf{a}_2+\left(\frac{1}{2}-x_4\right)\mathbf{a}_3$ | = | $-x_4a\,\hat{\mathbf{x}}+\left(\frac{1}{2}+x_4\right)a\,\hat{\mathbf{y}}+\left(\frac{1}{2}-x_4\right)a\,\hat{\mathbf{z}}$ | (8$c$) | O I |
| $\mathbf{B}_{24}$ | = | $\left(\frac{1}{2}+x_4\right)\mathbf{a}_1+\left(\frac{1}{2}-x_4\right)\mathbf{a}_2-x_4\,\mathbf{a}_3$ | = | $\left(\frac{1}{2}+x_4\right)a\,\hat{\mathbf{x}}+\left(\frac{1}{2}-x_4\right)a\,\hat{\mathbf{y}}-x_4a\,\hat{\mathbf{z}}$ | (8$c$) | O I |
| $\mathbf{B}_{25}$ | = | $-x_4\,\mathbf{a}_1-x_4\,\mathbf{a}_2-x_4\,\mathbf{a}_3$ | = | $-x_4a\,\hat{\mathbf{x}}-x_4a\,\hat{\mathbf{y}}-x_4a\,\hat{\mathbf{z}}$ | (8$c$) | O I |
| $\mathbf{B}_{26}$ | = | $\left(\frac{1}{2}+x_4\right)\mathbf{a}_1+x_4\,\mathbf{a}_2+\left(\frac{1}{2}-x_4\right)\mathbf{a}_3$ | = | $\left(\frac{1}{2}+x_4\right)a\,\hat{\mathbf{x}}+x_4a\,\hat{\mathbf{y}}+\left(\frac{1}{2}-x_4\right)a\,\hat{\mathbf{z}}$ | (8$c$) | O I |
| $\mathbf{B}_{27}$ | = | $x_4\,\mathbf{a}_1+\left(\frac{1}{2}-x_4\right)\mathbf{a}_2+\left(\frac{1}{2}+x_4\right)\mathbf{a}_3$ | = | $x_4a\,\hat{\mathbf{x}}+\left(\frac{1}{2}-x_4\right)a\,\hat{\mathbf{y}}+\left(\frac{1}{2}+x_4\right)a\,\hat{\mathbf{z}}$ | (8$c$) | O I |
| $\mathbf{B}_{28}$ | = | $\left(\frac{1}{2}-x_4\right)\mathbf{a}_1+\left(\frac{1}{2}+x_4\right)\mathbf{a}_2+x_4\,\mathbf{a}_3$ | = | $\left(\frac{1}{2}-x_4\right)a\,\hat{\mathbf{x}}+\left(\frac{1}{2}+x_4\right)a\,\hat{\mathbf{y}}+x_4a\,\hat{\mathbf{z}}$ | (8$c$) | O I |
| $\mathbf{B}_{29}$ | = | $x_5\,\mathbf{a}_1+x_5\,\mathbf{a}_2+x_5\,\mathbf{a}_3$ | = | $x_5a\,\hat{\mathbf{x}}+x_5a\,\hat{\mathbf{y}}+x_5a\,\hat{\mathbf{z}}$ | (8$c$) | S |
| $\mathbf{B}_{30}$ | = | $\left(\frac{1}{2}-x_5\right)\mathbf{a}_1-x_5\,\mathbf{a}_2+\left(\frac{1}{2}+x_5\right)\mathbf{a}_3$ | = | $\left(\frac{1}{2}-x_5\right)a\,\hat{\mathbf{x}}-x_5a\,\hat{\mathbf{y}}+\left(\frac{1}{2}+x_5\right)a\,\hat{\mathbf{z}}$ | (8$c$) | S |
| $\mathbf{B}_{31}$ | = | $-x_5\,\mathbf{a}_1+\left(\frac{1}{2}+x_5\right)\mathbf{a}_2+\left(\frac{1}{2}-x_5\right)\mathbf{a}_3$ | = | $-x_5a\,\hat{\mathbf{x}}+\left(\frac{1}{2}+x_5\right)a\,\hat{\mathbf{y}}+\left(\frac{1}{2}-x_5\right)a\,\hat{\mathbf{z}}$ | (8$c$) | S |
| $\mathbf{B}_{32}$ | = | $\left(\frac{1}{2}+x_5\right)\mathbf{a}_1+\left(\frac{1}{2}-x_5\right)\mathbf{a}_2-x_5\,\mathbf{a}_3$ | = | $\left(\frac{1}{2}+x_5\right)a\,\hat{\mathbf{x}}+\left(\frac{1}{2}-x_5\right)a\,\hat{\mathbf{y}}-x_5a\,\hat{\mathbf{z}}$ | (8$c$) | S |
| $\mathbf{B}_{33}$ | = | $-x_5\,\mathbf{a}_1-x_5\,\mathbf{a}_2-x_5\,\mathbf{a}_3$ | = | $-x_5a\,\hat{\mathbf{x}}-x_5a\,\hat{\mathbf{y}}-x_5a\,\hat{\mathbf{z}}$ | (8$c$) | S |
| $\mathbf{B}_{34}$ | = | $\left(\frac{1}{2}+x_5\right)\mathbf{a}_1+x_5\,\mathbf{a}_2+\left(\frac{1}{2}-x_5\right)\mathbf{a}_3$ | = | $\left(\frac{1}{2}+x_5\right)a\,\hat{\mathbf{x}}+x_5a\,\hat{\mathbf{y}}+\left(\frac{1}{2}-x_5\right)a\,\hat{\mathbf{z}}$ | (8$c$) | S |
| $\mathbf{B}_{35}$ | = | $x_5\,\mathbf{a}_1+\left(\frac{1}{2}-x_5\right)\mathbf{a}_2+\left(\frac{1}{2}+x_5\right)\mathbf{a}_3$ | = | $x_5a\,\hat{\mathbf{x}}+\left(\frac{1}{2}-x_5\right)a\,\hat{\mathbf{y}}+\left(\frac{1}{2}+x_5\right)a\,\hat{\mathbf{z}}$ | (8$c$) | S |
| $\mathbf{B}_{36}$ | = | $\left(\frac{1}{2}-x_5\right)\mathbf{a}_1+\left(\frac{1}{2}+x_5\right)\mathbf{a}_2+x_5\,\mathbf{a}_3$ | = | $\left(\frac{1}{2}-x_5\right)a\,\hat{\mathbf{x}}+\left(\frac{1}{2}+x_5\right)a\,\hat{\mathbf{y}}+x_5a\,\hat{\mathbf{z}}$ | (8$c$) | S |
| $\mathbf{B}_{37}$ | = | $x_6\,\mathbf{a}_1+y_6\,\mathbf{a}_2+z_6\,\mathbf{a}_3$ | = | $x_6a\,\hat{\mathbf{x}}+y_6a\,\hat{\mathbf{y}}+z_6a\,\hat{\mathbf{z}}$ | (24$d$) | H I |
| $\mathbf{B}_{38}$ | = | $\left(\frac{1}{2}-x_6\right)\mathbf{a}_1-y_6\,\mathbf{a}_2+\left(\frac{1}{2}+z_6\right)\mathbf{a}_3$ | = | $\left(\frac{1}{2}-x_6\right)a\,\hat{\mathbf{x}}-y_6a\,\hat{\mathbf{y}}+\left(\frac{1}{2}+z_6\right)a\,\hat{\mathbf{z}}$ | (24$d$) | H I |
| $\mathbf{B}_{39}$ | = | $-x_6\,\mathbf{a}_1+\left(\frac{1}{2}+y_6\right)\mathbf{a}_2+\left(\frac{1}{2}-z_6\right)\mathbf{a}_3$ | = | $-x_6a\,\hat{\mathbf{x}}+\left(\frac{1}{2}+y_6\right)a\,\hat{\mathbf{y}}+\left(\frac{1}{2}-z_6\right)a\,\hat{\mathbf{z}}$ | (24$d$) | H I |
| $\mathbf{B}_{40}$ | = | $\left(\frac{1}{2}+x_6\right)\mathbf{a}_1+\left(\frac{1}{2}-y_6\right)\mathbf{a}_2-z_6\,\mathbf{a}_3$ | = | $\left(\frac{1}{2}+x_6\right)a\,\hat{\mathbf{x}}+\left(\frac{1}{2}-y_6\right)a\,\hat{\mathbf{y}}-z_6a\,\hat{\mathbf{z}}$ | (24$d$) | H I |
| $\mathbf{B}_{41}$ | = | $z_6\,\mathbf{a}_1+x_6\,\mathbf{a}_2+y_6\,\mathbf{a}_3$ | = | $z_6a\,\hat{\mathbf{x}}+x_6a\,\hat{\mathbf{y}}+y_6a\,\hat{\mathbf{z}}$ | (24$d$) | H I |
| $\mathbf{B}_{42}$ | = | $\left(\frac{1}{2}+z_6\right)\mathbf{a}_1+\left(\frac{1}{2}-x_6\right)\mathbf{a}_2-y_6\,\mathbf{a}_3$ | = | $\left(\frac{1}{2}+z_6\right)a\,\hat{\mathbf{x}}+\left(\frac{1}{2}-x_6\right)a\,\hat{\mathbf{y}}-y_6a\,\hat{\mathbf{z}}$ | (24$d$) | H I |
| $\mathbf{B}_{43}$ | = | $\left(\frac{1}{2}-z_6\right)\mathbf{a}_1-x_6\,\mathbf{a}_2+\left(\frac{1}{2}+y_6\right)\mathbf{a}_3$ | = | $\left(\frac{1}{2}-z_6\right)a\,\hat{\mathbf{x}}-x_6a\,\hat{\mathbf{y}}+\left(\frac{1}{2}+y_6\right)a\,\hat{\mathbf{z}}$ | (24$d$) | H I |
| $\mathbf{B}_{44}$ | = | $-z_6\,\mathbf{a}_1+\left(\frac{1}{2}+x_6\right)\mathbf{a}_2+\left(\frac{1}{2}-y_6\right)\mathbf{a}_3$ | = | $-z_6a\,\hat{\mathbf{x}}+\left(\frac{1}{2}+x_6\right)a\,\hat{\mathbf{y}}+\left(\frac{1}{2}-y_6\right)a\,\hat{\mathbf{z}}$ | (24$d$) | H I |
| $\mathbf{B}_{45}$ | = | $y_6\,\mathbf{a}_1+z_6\,\mathbf{a}_2+x_6\,\mathbf{a}_3$ | = | $y_6a\,\hat{\mathbf{x}}+z_6a\,\hat{\mathbf{y}}+x_6a\,\hat{\mathbf{z}}$ | (24$d$) | H I |
| $\mathbf{B}_{46}$ | = | $-y_6\,\mathbf{a}_1+\left(\frac{1}{2}+z_6\right)\mathbf{a}_2+\left(\frac{1}{2}-x_6\right)\mathbf{a}_3$ | = | $-y_6a\,\hat{\mathbf{x}}+\left(\frac{1}{2}+z_6\right)a\,\hat{\mathbf{y}}+\left(\frac{1}{2}-x_6\right)a\,\hat{\mathbf{z}}$ | (24$d$) | H I |
| $\mathbf{B}_{47}$ | = | $\left(\frac{1}{2}+y_6\right)\mathbf{a}_1+\left(\frac{1}{2}-z_6\right)\mathbf{a}_2-x_6\,\mathbf{a}_3$ | = | $\left(\frac{1}{2}+y_6\right)a\,\hat{\mathbf{x}}+\left(\frac{1}{2}-z_6\right)a\,\hat{\mathbf{y}}-x_6a\,\hat{\mathbf{z}}$ | (24$d$) | H I |
| $\mathbf{B}_{48}$ | = | $\left(\frac{1}{2}-y_6\right)\mathbf{a}_1-z_6\,\mathbf{a}_2+\left(\frac{1}{2}+x_6\right)\mathbf{a}_3$ | = | $\left(\frac{1}{2}-y_6\right)a\,\hat{\mathbf{x}}-z_6a\,\hat{\mathbf{y}}+\left(\frac{1}{2}+x_6\right)a\,\hat{\mathbf{z}}$ | (24$d$) | H I |
| $\mathbf{B}_{49}$ | = | $-x_6\,\mathbf{a}_1-y_6\,\mathbf{a}_2-z_6\,\mathbf{a}_3$ | = | $-x_6a\,\hat{\mathbf{x}}-y_6a\,\hat{\mathbf{y}}-z_6a\,\hat{\mathbf{z}}$ | (24$d$) | H I |
| $\mathbf{B}_{50}$ | = | $\left(\frac{1}{2}+x_6\right)\mathbf{a}_1+y_6\,\mathbf{a}_2+\left(\frac{1}{2}-z_6\right)\mathbf{a}_3$ | = | $\left(\frac{1}{2}+x_6\right)a\,\hat{\mathbf{x}}+y_6a\,\hat{\mathbf{y}}+\left(\frac{1}{2}-z_6\right)a\,\hat{\mathbf{z}}$ | (24$d$) | H I |
| $\mathbf{B}_{51}$ | = | $x_6\,\mathbf{a}_1+\left(\frac{1}{2}-y_6\right)\mathbf{a}_2+\left(\frac{1}{2}+z_6\right)\mathbf{a}_3$ | = | $x_6a\,\hat{\mathbf{x}}+\left(\frac{1}{2}-y_6\right)a\,\hat{\mathbf{y}}+\left(\frac{1}{2}+z_6\right)a\,\hat{\mathbf{z}}$ | (24$d$) | H I |
| $\mathbf{B}_{52}$ | = | $\left(\frac{1}{2}-x_6\right)\mathbf{a}_1+\left(\frac{1}{2}+y_6\right)\mathbf{a}_2+z_6\,\mathbf{a}_3$ | = | $\left(\frac{1}{2}-x_6\right)a\,\hat{\mathbf{x}}+\left(\frac{1}{2}+y_6\right)a\,\hat{\mathbf{y}}+z_6a\,\hat{\mathbf{z}}$ | (24$d$) | H I |
| $\mathbf{B}_{53}$ | = | $-z_6\,\mathbf{a}_1-x_6\,\mathbf{a}_2-y_6\,\mathbf{a}_3$ | = | $-z_6a\,\hat{\mathbf{x}}-x_6a\,\hat{\mathbf{y}}-y_6a\,\hat{\mathbf{z}}$ | (24$d$) | H I |
| $\mathbf{B}_{54}$ | = | $\left(\frac{1}{2}-z_6\right)\mathbf{a}_1+\left(\frac{1}{2}+x_6\right)\mathbf{a}_2+y_6\,\mathbf{a}_3$ | = | $\left(\frac{1}{2}-z_6\right)a\,\mathbf{\hat{x}}+\left(\frac{1}{2}+x_6\right)a\,\mathbf{\hat{y}}+y_6a\,\mathbf{\hat{z}}$ | (24$d$) | H I |
| $\mathbf{B}_{55}$ | = | $\left(\frac{1}{2}+z_6\right)\mathbf{a}_1+x_6\,\mathbf{a}_2+\left(\frac{1}{2}-y_6\right)\mathbf{a}_3$ | = | $\left(\frac{1}{2}+z_6\right)a\,\mathbf{\hat{x}}+x_6a\,\mathbf{\hat{y}}+\left(\frac{1}{2}-y_6\right)a\,\mathbf{\hat{z}}$ | (24$d$) | H I |
| $\mathbf{B}_{56}$ | = | $z_6\,\mathbf{a}_1+\left(\frac{1}{2}-x_6\right)\mathbf{a}_2+\left(\frac{1}{2}+y_6\right)\mathbf{a}_3$ | = | $z_6a\,\mathbf{\hat{x}}+\left(\frac{1}{2}-x_6\right)a\,\mathbf{\hat{y}}+\left(\frac{1}{2}+y_6\right)a\,\mathbf{\hat{z}}$ | (24$d$) | H I |
| $\mathbf{B}_{57}$ | = | $-y_6\,\mathbf{a}_1-z_6\,\mathbf{a}_2-x_6\,\mathbf{a}_3$ | = | $-y_6a\,\mathbf{\hat{x}}-z_6a\,\mathbf{\hat{y}}-x_6a\,\mathbf{\hat{z}}$ | (24$d$) | H I |
| $\mathbf{B}_{58}$ | = | $y_6\,\mathbf{a}_1+\left(\frac{1}{2}-z_6\right)\mathbf{a}_2+\left(\frac{1}{2}+x_6\right)\mathbf{a}_3$ | = | $y_6a\,\mathbf{\hat{x}}+\left(\frac{1}{2}-z_6\right)a\,\mathbf{\hat{y}}+\left(\frac{1}{2}+x_6\right)a\,\mathbf{\hat{z}}$ | (24$d$) | H I |
| $\mathbf{B}_{59}$ | = | $\left(\frac{1}{2}-y_6\right)\mathbf{a}_1+\left(\frac{1}{2}+z_6\right)\mathbf{a}_2+x_6\,\mathbf{a}_3$ | = | $\left(\frac{1}{2}-y_6\right)a\,\mathbf{\hat{x}}+\left(\frac{1}{2}+z_6\right)a\,\mathbf{\hat{y}}+x_6a\,\mathbf{\hat{z}}$ | (24$d$) | H I |
| $\mathbf{B}_{60}$ | = | $\left(\frac{1}{2}+y_6\right)\mathbf{a}_1+z_6\,\mathbf{a}_2+\left(\frac{1}{2}-x_6\right)\mathbf{a}_3$ | = | $\left(\frac{1}{2}+y_6\right)a\,\mathbf{\hat{x}}+z_6a\,\mathbf{\hat{y}}+\left(\frac{1}{2}-x_6\right)a\,\mathbf{\hat{z}}$ | (24$d$) | H I |
| $\mathbf{B}_{61}$ | = | $x_7\,\mathbf{a}_1+y_7\,\mathbf{a}_2+z_7\,\mathbf{a}_3$ | = | $x_7a\,\mathbf{\hat{x}}+y_7a\,\mathbf{\hat{y}}+z_7a\,\mathbf{\hat{z}}$ | (24$d$) | H II |
| $\mathbf{B}_{62}$ | = | $\left(\frac{1}{2}-x_7\right)\mathbf{a}_1-y_7\,\mathbf{a}_2+\left(\frac{1}{2}+z_7\right)\mathbf{a}_3$ | = | $\left(\frac{1}{2}-x_7\right)a\,\mathbf{\hat{x}}-y_7a\,\mathbf{\hat{y}}+\left(\frac{1}{2}+z_7\right)a\,\mathbf{\hat{z}}$ | (24$d$) | H II |
| $\mathbf{B}_{63}$ | = | $-x_7\,\mathbf{a}_1+\left(\frac{1}{2}+y_7\right)\mathbf{a}_2+\left(\frac{1}{2}-z_7\right)\mathbf{a}_3$ | = | $-x_7a\,\mathbf{\hat{x}}+\left(\frac{1}{2}+y_7\right)a\,\mathbf{\hat{y}}+\left(\frac{1}{2}-z_7\right)a\,\mathbf{\hat{z}}$ | (24$d$) | H II |
| $\mathbf{B}_{64}$ | = | $\left(\frac{1}{2}+x_7\right)\mathbf{a}_1+\left(\frac{1}{2}-y_7\right)\mathbf{a}_2-z_7\,\mathbf{a}_3$ | = | $\left(\frac{1}{2}+x_7\right)a\,\mathbf{\hat{x}}+\left(\frac{1}{2}-y_7\right)a\,\mathbf{\hat{y}}-z_7a\,\mathbf{\hat{z}}$ | (24$d$) | H II |
| $\mathbf{B}_{65}$ | = | $z_7\,\mathbf{a}_1+x_7\,\mathbf{a}_2+y_7\,\mathbf{a}_3$ | = | $z_7a\,\mathbf{\hat{x}}+x_7a\,\mathbf{\hat{y}}+y_7a\,\mathbf{\hat{z}}$ | (24$d$) | H II |
| $\mathbf{B}_{66}$ | = | $\left(\frac{1}{2}+z_7\right)\mathbf{a}_1+\left(\frac{1}{2}-x_7\right)\mathbf{a}_2-y_7\,\mathbf{a}_3$ | = | $\left(\frac{1}{2}+z_7\right)a\,\mathbf{\hat{x}}+\left(\frac{1}{2}-x_7\right)a\,\mathbf{\hat{y}}-y_7a\,\mathbf{\hat{z}}$ | (24$d$) | H II |
| $\mathbf{B}_{67}$ | = | $\left(\frac{1}{2}-z_7\right)\mathbf{a}_1-x_7\,\mathbf{a}_2+\left(\frac{1}{2}+y_7\right)\mathbf{a}_3$ | = | $\left(\frac{1}{2}-z_7\right)a\,\mathbf{\hat{x}}-x_7a\,\mathbf{\hat{y}}+\left(\frac{1}{2}+y_7\right)a\,\mathbf{\hat{z}}$ | (24$d$) | H II |
| $\mathbf{B}_{68}$ | = | $-z_7\,\mathbf{a}_1+\left(\frac{1}{2}+x_7\right)\mathbf{a}_2+\left(\frac{1}{2}-y_7\right)\mathbf{a}_3$ | = | $-z_7a\,\mathbf{\hat{x}}+\left(\frac{1}{2}+x_7\right)a\,\mathbf{\hat{y}}+\left(\frac{1}{2}-y_7\right)a\,\mathbf{\hat{z}}$ | (24$d$) | H II |
| $\mathbf{B}_{69}$ | = | $y_7\,\mathbf{a}_1+z_7\,\mathbf{a}_2+x_7\,\mathbf{a}_3$ | = | $y_7a\,\mathbf{\hat{x}}+z_7a\,\mathbf{\hat{y}}+x_7a\,\mathbf{\hat{z}}$ | (24$d$) | H II |
| $\mathbf{B}_{70}$ | = | $-y_7\,\mathbf{a}_1+\left(\frac{1}{2}+z_7\right)\mathbf{a}_2+\left(\frac{1}{2}-x_7\right)\mathbf{a}_3$ | = | $-y_7a\,\mathbf{\hat{x}}+\left(\frac{1}{2}+z_7\right)a\,\mathbf{\hat{y}}+\left(\frac{1}{2}-x_7\right)a\,\mathbf{\hat{z}}$ | (24$d$) | H II |
| $\mathbf{B}_{71}$ | = | $\left(\frac{1}{2}+y_7\right)\mathbf{a}_1+\left(\frac{1}{2}-z_7\right)\mathbf{a}_2-x_7\,\mathbf{a}_3$ | = | $\left(\frac{1}{2}+y_7\right)a\,\mathbf{\hat{x}}+\left(\frac{1}{2}-z_7\right)a\,\mathbf{\hat{y}}-x_7a\,\mathbf{\hat{z}}$ | (24$d$) | H II |
| $\mathbf{B}_{72}$ | = | $\left(\frac{1}{2}-y_7\right)\mathbf{a}_1-z_7\,\mathbf{a}_2+\left(\frac{1}{2}+x_7\right)\mathbf{a}_3$ | = | $\left(\frac{1}{2}-y_7\right)a\,\mathbf{\hat{x}}-z_7a\,\mathbf{\hat{y}}+\left(\frac{1}{2}+x_7\right)a\,\mathbf{\hat{z}}$ | (24$d$) | H II |
| $\mathbf{B}_{73}$ | = | $-x_7\,\mathbf{a}_1-y_7\,\mathbf{a}_2-z_7\,\mathbf{a}_3$ | = | $-x_7a\,\mathbf{\hat{x}}-y_7a\,\mathbf{\hat{y}}-z_7a\,\mathbf{\hat{z}}$ | (24$d$) | H II |
| $\mathbf{B}_{74}$ | = | $\left(\frac{1}{2}+x_7\right)\mathbf{a}_1+y_7\,\mathbf{a}_2+\left(\frac{1}{2}-z_7\right)\mathbf{a}_3$ | = | $\left(\frac{1}{2}+x_7\right)a\,\mathbf{\hat{x}}+y_7a\,\mathbf{\hat{y}}+\left(\frac{1}{2}-z_7\right)a\,\mathbf{\hat{z}}$ | (24$d$) | H II |
| $\mathbf{B}_{75}$ | = | $x_7\,\mathbf{a}_1+\left(\frac{1}{2}-y_7\right)\mathbf{a}_2+\left(\frac{1}{2}+z_7\right)\mathbf{a}_3$ | = | $x_7a\,\mathbf{\hat{x}}+\left(\frac{1}{2}-y_7\right)a\,\mathbf{\hat{y}}+\left(\frac{1}{2}+z_7\right)a\,\mathbf{\hat{z}}$ | (24$d$) | H II |
| $\mathbf{B}_{76}$ | = | $\left(\frac{1}{2}-x_7\right)\mathbf{a}_1+\left(\frac{1}{2}+y_7\right)\mathbf{a}_2+z_7\,\mathbf{a}_3$ | = | $\left(\frac{1}{2}-x_7\right)a\,\mathbf{\hat{x}}+\left(\frac{1}{2}+y_7\right)a\,\mathbf{\hat{y}}+z_7a\,\mathbf{\hat{z}}$ | (24$d$) | H II |
| $\mathbf{B}_{77}$ | = | $-z_7\,\mathbf{a}_1-x_7\,\mathbf{a}_2-y_7\,\mathbf{a}_3$ | = | $-z_7a\,\mathbf{\hat{x}}-x_7a\,\mathbf{\hat{y}}-y_7a\,\mathbf{\hat{z}}$ | (24$d$) | H II |
| $\mathbf{B}_{78}$ | = | $\left(\frac{1}{2}-z_7\right)\mathbf{a}_1+\left(\frac{1}{2}+x_7\right)\mathbf{a}_2+y_7\,\mathbf{a}_3$ | = | $\left(\frac{1}{2}-z_7\right)a\,\mathbf{\hat{x}}+\left(\frac{1}{2}+x_7\right)a\,\mathbf{\hat{y}}+y_7a\,\mathbf{\hat{z}}$ | (24$d$) | H II |
| $\mathbf{B}_{79}$ | = | $\left(\frac{1}{2}+z_7\right)\mathbf{a}_1+x_7\,\mathbf{a}_2+\left(\frac{1}{2}-y_7\right)\mathbf{a}_3$ | = | $\left(\frac{1}{2}+z_7\right)a\,\mathbf{\hat{x}}+x_7a\,\mathbf{\hat{y}}+\left(\frac{1}{2}-y_7\right)a\,\mathbf{\hat{z}}$ | (24$d$) | H II |
| $\mathbf{B}_{80}$ | = | $z_7\,\mathbf{a}_1+\left(\frac{1}{2}-x_7\right)\mathbf{a}_2+\left(\frac{1}{2}+y_7\right)\mathbf{a}_3$ | = | $z_7a\,\mathbf{\hat{x}}+\left(\frac{1}{2}-x_7\right)a\,\mathbf{\hat{y}}+\left(\frac{1}{2}+y_7\right)a\,\mathbf{\hat{z}}$ | (24$d$) | H II |
| $\mathbf{B}_{81}$ | = | $-y_7\,\mathbf{a}_1-z_7\,\mathbf{a}_2-x_7\,\mathbf{a}_3$ | = | $-y_7a\,\mathbf{\hat{x}}-z_7a\,\mathbf{\hat{y}}-x_7a\,\mathbf{\hat{z}}$ | (24$d$) | H II |
| $\mathbf{B}_{82}$ | = | $y_7\,\mathbf{a}_1+\left(\frac{1}{2}-z_7\right)\mathbf{a}_2+\left(\frac{1}{2}+x_7\right)\mathbf{a}_3$ | = | $y_7a\,\mathbf{\hat{x}}+\left(\frac{1}{2}-z_7\right)a\,\mathbf{\hat{y}}+\left(\frac{1}{2}+x_7\right)a\,\mathbf{\hat{z}}$ | (24$d$) | H II |
| $\mathbf{B}_{83}$ | = | $\left(\frac{1}{2}-y_7\right)\mathbf{a}_1+\left(\frac{1}{2}+z_7\right)\mathbf{a}_2+x_7\,\mathbf{a}_3$ | = | $\left(\frac{1}{2}-y_7\right)a\,\mathbf{\hat{x}}+\left(\frac{1}{2}+z_7\right)a\,\mathbf{\hat{y}}+x_7a\,\mathbf{\hat{z}}$ | (24$d$) | H II |
| $\mathbf{B}_{84}$ | = | $\left(\frac{1}{2}+y_7\right)\mathbf{a}_1+z_7\,\mathbf{a}_2+\left(\frac{1}{2}-x_7\right)\mathbf{a}_3$ | = | $\left(\frac{1}{2}+y_7\right)a\,\mathbf{\hat{x}}+z_7a\,\mathbf{\hat{y}}+\left(\frac{1}{2}-x_7\right)a\,\mathbf{\hat{z}}$ | (24$d$) | H II |
| $\mathbf{B}_{85}$ | = | $x_8\,\mathbf{a}_1+y_8\,\mathbf{a}_2+z_8\,\mathbf{a}_3$ | = | $x_8a\,\mathbf{\hat{x}}+y_8a\,\mathbf{\hat{y}}+z_8a\,\mathbf{\hat{z}}$ | (24$d$) | H III |
| $\mathbf{B}_{86}$ | = | $\left(\frac{1}{2}-x_8\right)\mathbf{a}_1-y_8\,\mathbf{a}_2+\left(\frac{1}{2}+z_8\right)\mathbf{a}_3$ | = | $\left(\frac{1}{2}-x_8\right)a\,\mathbf{\hat{x}}-y_8a\,\mathbf{\hat{y}}+\left(\frac{1}{2}+z_8\right)a\,\mathbf{\hat{z}}$ | (24$d$) | H III |
| $\mathbf{B}_{87}$ | = | $-x_8\,\mathbf{a}_1+\left(\frac{1}{2}+y_8\right)\mathbf{a}_2+\left(\frac{1}{2}-z_8\right)\mathbf{a}_3$ | = | $-x_8a\,\mathbf{\hat{x}}+\left(\frac{1}{2}+y_8\right)a\,\mathbf{\hat{y}}+\left(\frac{1}{2}-z_8\right)a\,\mathbf{\hat{z}}$ | (24$d$) | H III |
| $\mathbf{B}_{88}$ | = | $\left(\frac{1}{2}+x_8\right)\mathbf{a}_1+\left(\frac{1}{2}-y_8\right)\mathbf{a}_2-z_8\,\mathbf{a}_3$ | = | $\left(\frac{1}{2}+x_8\right)a\,\mathbf{\hat{x}}+\left(\frac{1}{2}-y_8\right)a\,\mathbf{\hat{y}}-z_8a\,\mathbf{\hat{z}}$ | (24$d$) | H III |
| $\mathbf{B}_{89}$ | = | $z_8\,\mathbf{a}_1+x_8\,\mathbf{a}_2+y_8\,\mathbf{a}_3$ | = | $z_8a\,\mathbf{\hat{x}}+x_8a\,\mathbf{\hat{y}}+y_8a\,\mathbf{\hat{z}}$ | (24$d$) | H III |
| $\mathbf{B}_{90}$ | = | $\left(\frac{1}{2}+z_8\right)\mathbf{a}_1+\left(\frac{1}{2}-x_8\right)\mathbf{a}_2-y_8\,\mathbf{a}_3$ | = | $\left(\frac{1}{2}+z_8\right)a\,\hat{\mathbf{x}}+\left(\frac{1}{2}-x_8\right)a\,\hat{\mathbf{y}}-y_8a\,\hat{\mathbf{z}}$ | (24$d$) | H III |
| $\mathbf{B}_{91}$ | = | $\left(\frac{1}{2}-z_8\right)\mathbf{a}_1-x_8\,\mathbf{a}_2+\left(\frac{1}{2}+y_8\right)\mathbf{a}_3$ | = | $\left(\frac{1}{2}-z_8\right)a\,\hat{\mathbf{x}}-x_8a\,\hat{\mathbf{y}}+\left(\frac{1}{2}+y_8\right)a\,\hat{\mathbf{z}}$ | (24$d$) | H III |
| $\mathbf{B}_{92}$ | = | $-z_8\,\mathbf{a}_1+\left(\frac{1}{2}+x_8\right)\mathbf{a}_2+\left(\frac{1}{2}-y_8\right)\mathbf{a}_3$ | = | $-z_8a\,\hat{\mathbf{x}}+\left(\frac{1}{2}+x_8\right)a\,\hat{\mathbf{y}}+\left(\frac{1}{2}-y_8\right)a\,\hat{\mathbf{z}}$ | (24$d$) | H III |
| $\mathbf{B}_{93}$ | = | $y_8\,\mathbf{a}_1+z_8\,\mathbf{a}_2+x_8\,\mathbf{a}_3$ | = | $y_8a\,\hat{\mathbf{x}}+z_8a\,\hat{\mathbf{y}}+x_8a\,\hat{\mathbf{z}}$ | (24$d$) | H III |
| $\mathbf{B}_{94}$ | = | $-y_8\,\mathbf{a}_1+\left(\frac{1}{2}+z_8\right)\mathbf{a}_2+\left(\frac{1}{2}-x_8\right)\mathbf{a}_3$ | = | $-y_8a\,\hat{\mathbf{x}}+\left(\frac{1}{2}+z_8\right)a\,\hat{\mathbf{y}}+\left(\frac{1}{2}-x_8\right)a\,\hat{\mathbf{z}}$ | (24$d$) | H III |
| $\mathbf{B}_{95}$ | = | $\left(\frac{1}{2}+y_8\right)\mathbf{a}_1+\left(\frac{1}{2}-z_8\right)\mathbf{a}_2-x_8\,\mathbf{a}_3$ | = | $\left(\frac{1}{2}+y_8\right)a\,\hat{\mathbf{x}}+\left(\frac{1}{2}-z_8\right)a\,\hat{\mathbf{y}}-x_8a\,\hat{\mathbf{z}}$ | (24$d$) | H III |
| $\mathbf{B}_{96}$ | = | $\left(\frac{1}{2}-y_8\right)\mathbf{a}_1-z_8\,\mathbf{a}_2+\left(\frac{1}{2}+x_8\right)\mathbf{a}_3$ | = | $\left(\frac{1}{2}-y_8\right)a\,\hat{\mathbf{x}}-z_8a\,\hat{\mathbf{y}}+\left(\frac{1}{2}+x_8\right)a\,\hat{\mathbf{z}}$ | (24$d$) | H III |
| $\mathbf{B}_{97}$ | = | $-x_8\,\mathbf{a}_1-y_8\,\mathbf{a}_2-z_8\,\mathbf{a}_3$ | = | $-x_8a\,\hat{\mathbf{x}}-y_8a\,\hat{\mathbf{y}}-z_8a\,\hat{\mathbf{z}}$ | (24$d$) | H III |
| $\mathbf{B}_{98}$ | = | $\left(\frac{1}{2}+x_8\right)\mathbf{a}_1+y_8\,\mathbf{a}_2+\left(\frac{1}{2}-z_8\right)\mathbf{a}_3$ | = | $\left(\frac{1}{2}+x_8\right)a\,\hat{\mathbf{x}}+y_8a\,\hat{\mathbf{y}}+\left(\frac{1}{2}-z_8\right)a\,\hat{\mathbf{z}}$ | (24$d$) | H III |
| $\mathbf{B}_{99}$ | = | $x_8\,\mathbf{a}_1+\left(\frac{1}{2}-y_8\right)\mathbf{a}_2+\left(\frac{1}{2}+z_8\right)\mathbf{a}_3$ | = | $x_8a\,\hat{\mathbf{x}}+\left(\frac{1}{2}-y_8\right)a\,\hat{\mathbf{y}}+\left(\frac{1}{2}+z_8\right)a\,\hat{\mathbf{z}}$ | (24$d$) | H III |
| $\mathbf{B}_{100}$ | = | $\left(\frac{1}{2}-x_8\right)\mathbf{a}_1+\left(\frac{1}{2}+y_8\right)\mathbf{a}_2+z_8\,\mathbf{a}_3$ | = | $\left(\frac{1}{2}-x_8\right)a\,\hat{\mathbf{x}}+\left(\frac{1}{2}+y_8\right)a\,\hat{\mathbf{y}}+z_8a\,\hat{\mathbf{z}}$ | (24$d$) | H III |
| $\mathbf{B}_{101}$ | = | $-z_8\,\mathbf{a}_1-x_8\,\mathbf{a}_2-y_8\,\mathbf{a}_3$ | = | $-z_8a\,\hat{\mathbf{x}}-x_8a\,\hat{\mathbf{y}}-y_8a\,\hat{\mathbf{z}}$ | (24$d$) | H III |
| $\mathbf{B}_{102}$ | = | $\left(\frac{1}{2}-z_8\right)\mathbf{a}_1+\left(\frac{1}{2}+x_8\right)\mathbf{a}_2+y_8\,\mathbf{a}_3$ | = | $\left(\frac{1}{2}-z_8\right)a\,\hat{\mathbf{x}}+\left(\frac{1}{2}+x_8\right)a\,\hat{\mathbf{y}}+y_8a\,\hat{\mathbf{z}}$ | (24$d$) | H III |
| $\mathbf{B}_{103}$ | = | $\left(\frac{1}{2}+z_8\right)\mathbf{a}_1+x_8\,\mathbf{a}_2+\left(\frac{1}{2}-y_8\right)\mathbf{a}_3$ | = | $\left(\frac{1}{2}+z_8\right)a\,\hat{\mathbf{x}}+x_8a\,\hat{\mathbf{y}}+\left(\frac{1}{2}-y_8\right)a\,\hat{\mathbf{z}}$ | (24$d$) | H III |
| $\mathbf{B}_{104}$ | = | $z_8\,\mathbf{a}_1+\left(\frac{1}{2}-x_8\right)\mathbf{a}_2+\left(\frac{1}{2}+y_8\right)\mathbf{a}_3$ | = | $z_8a\,\hat{\mathbf{x}}+\left(\frac{1}{2}-x_8\right)a\,\hat{\mathbf{y}}+\left(\frac{1}{2}+y_8\right)a\,\hat{\mathbf{z}}$ | (24$d$) | H III |
| $\mathbf{B}_{105}$ | = | $-y_8\,\mathbf{a}_1-z_8\,\mathbf{a}_2-x_8\,\mathbf{a}_3$ | = | $-y_8a\,\hat{\mathbf{x}}-z_8a\,\hat{\mathbf{y}}-x_8a\,\hat{\mathbf{z}}$ | (24$d$) | H III |
| $\mathbf{B}_{106}$ | = | $y_8\,\mathbf{a}_1+\left(\frac{1}{2}-z_8\right)\mathbf{a}_2+\left(\frac{1}{2}+x_8\right)\mathbf{a}_3$ | = | $y_8a\,\hat{\mathbf{x}}+\left(\frac{1}{2}-z_8\right)a\,\hat{\mathbf{y}}+\left(\frac{1}{2}+x_8\right)a\,\hat{\mathbf{z}}$ | (24$d$) | H III |
| $\mathbf{B}_{107}$ | = | $\left(\frac{1}{2}-y_8\right)\mathbf{a}_1+\left(\frac{1}{2}+z_8\right)\mathbf{a}_2+x_8\,\mathbf{a}_3$ | = | $\left(\frac{1}{2}-y_8\right)a\,\hat{\mathbf{x}}+\left(\frac{1}{2}+z_8\right)a\,\hat{\mathbf{y}}+x_8a\,\hat{\mathbf{z}}$ | (24$d$) | H III |
| $\mathbf{B}_{108}$ | = | $\left(\frac{1}{2}+y_8\right)\mathbf{a}_1+z_8\,\mathbf{a}_2+\left(\frac{1}{2}-x_8\right)\mathbf{a}_3$ | = | $\left(\frac{1}{2}+y_8\right)a\,\hat{\mathbf{x}}+z_8a\,\hat{\mathbf{y}}+\left(\frac{1}{2}-x_8\right)a\,\hat{\mathbf{z}}$ | (24$d$) | H III |
| $\mathbf{B}_{109}$ | = | $x_9\,\mathbf{a}_1+y_9\,\mathbf{a}_2+z_9\,\mathbf{a}_3$ | = | $x_9a\,\hat{\mathbf{x}}+y_9a\,\hat{\mathbf{y}}+z_9a\,\hat{\mathbf{z}}$ | (24$d$) | H IV |
| $\mathbf{B}_{110}$ | = | $\left(\frac{1}{2}-x_9\right)\mathbf{a}_1-y_9\,\mathbf{a}_2+\left(\frac{1}{2}+z_9\right)\mathbf{a}_3$ | = | $\left(\frac{1}{2}-x_9\right)a\,\hat{\mathbf{x}}-y_9a\,\hat{\mathbf{y}}+\left(\frac{1}{2}+z_9\right)a\,\hat{\mathbf{z}}$ | (24$d$) | H IV |
| $\mathbf{B}_{111}$ | = | $-x_9\,\mathbf{a}_1+\left(\frac{1}{2}+y_9\right)\mathbf{a}_2+\left(\frac{1}{2}-z_9\right)\mathbf{a}_3$ | = | $-x_9a\,\hat{\mathbf{x}}+\left(\frac{1}{2}+y_9\right)a\,\hat{\mathbf{y}}+\left(\frac{1}{2}-z_9\right)a\,\hat{\mathbf{z}}$ | (24$d$) | H IV |
| $\mathbf{B}_{112}$ | = | $\left(\frac{1}{2}+x_9\right)\mathbf{a}_1+\left(\frac{1}{2}-y_9\right)\mathbf{a}_2-z_9\,\mathbf{a}_3$ | = | $\left(\frac{1}{2}+x_9\right)a\,\hat{\mathbf{x}}+\left(\frac{1}{2}-y_9\right)a\,\hat{\mathbf{y}}-z_9a\,\hat{\mathbf{z}}$ | (24$d$) | H IV |
| $\mathbf{B}_{113}$ | = | $z_9\,\mathbf{a}_1+x_9\,\mathbf{a}_2+y_9\,\mathbf{a}_3$ | = | $z_9a\,\hat{\mathbf{x}}+x_9a\,\hat{\mathbf{y}}+y_9a\,\hat{\mathbf{z}}$ | (24$d$) | H IV |
| $\mathbf{B}_{114}$ | = | $\left(\frac{1}{2}+z_9\right)\mathbf{a}_1+\left(\frac{1}{2}-x_9\right)\mathbf{a}_2-y_9\,\mathbf{a}_3$ | = | $\left(\frac{1}{2}+z_9\right)a\,\hat{\mathbf{x}}+\left(\frac{1}{2}-x_9\right)a\,\hat{\mathbf{y}}-y_9a\,\hat{\mathbf{z}}$ | (24$d$) | H IV |
| $\mathbf{B}_{115}$ | = | $\left(\frac{1}{2}-z_9\right)\mathbf{a}_1-x_9\,\mathbf{a}_2+\left(\frac{1}{2}+y_9\right)\mathbf{a}_3$ | = | $\left(\frac{1}{2}-z_9\right)a\,\hat{\mathbf{x}}-x_9a\,\hat{\mathbf{y}}+\left(\frac{1}{2}+y_9\right)a\,\hat{\mathbf{z}}$ | (24$d$) | H IV |
| $\mathbf{B}_{116}$ | = | $-z_9\,\mathbf{a}_1+\left(\frac{1}{2}+x_9\right)\mathbf{a}_2+\left(\frac{1}{2}-y_9\right)\mathbf{a}_3$ | = | $-z_9a\,\hat{\mathbf{x}}+\left(\frac{1}{2}+x_9\right)a\,\hat{\mathbf{y}}+\left(\frac{1}{2}-y_9\right)a\,\hat{\mathbf{z}}$ | (24$d$) | H IV |
| $\mathbf{B}_{117}$ | = | $y_9\,\mathbf{a}_1+z_9\,\mathbf{a}_2+x_9\,\mathbf{a}_3$ | = | $y_9a\,\hat{\mathbf{x}}+z_9a\,\hat{\mathbf{y}}+x_9a\,\hat{\mathbf{z}}$ | (24$d$) | H IV |
| $\mathbf{B}_{118}$ | = | $-y_9\,\mathbf{a}_1+\left(\frac{1}{2}+z_9\right)\mathbf{a}_2+\left(\frac{1}{2}-x_9\right)\mathbf{a}_3$ | = | $-y_9a\,\hat{\mathbf{x}}+\left(\frac{1}{2}+z_9\right)a\,\hat{\mathbf{y}}+\left(\frac{1}{2}-x_9\right)a\,\hat{\mathbf{z}}$ | (24$d$) | H IV |
| $\mathbf{B}_{119}$ | = | $\left(\frac{1}{2}+y_9\right)\mathbf{a}_1+\left(\frac{1}{2}-z_9\right)\mathbf{a}_2-x_9\,\mathbf{a}_3$ | = | $\left(\frac{1}{2}+y_9\right)a\,\hat{\mathbf{x}}+\left(\frac{1}{2}-z_9\right)a\,\hat{\mathbf{y}}-x_9a\,\hat{\mathbf{z}}$ | (24$d$) | H IV |
| $\mathbf{B}_{120}$ | = | $\left(\frac{1}{2}-y_9\right)\mathbf{a}_1-z_9\,\mathbf{a}_2+\left(\frac{1}{2}+x_9\right)\mathbf{a}_3$ | = | $\left(\frac{1}{2}-y_9\right)a\,\hat{\mathbf{x}}-z_9a\,\hat{\mathbf{y}}+\left(\frac{1}{2}+x_9\right)a\,\hat{\mathbf{z}}$ | (24$d$) | H IV |
| $\mathbf{B}_{121}$ | = | $-x_9\,\mathbf{a}_1-y_9\,\mathbf{a}_2-z_9\,\mathbf{a}_3$ | = | $-x_9a\,\hat{\mathbf{x}}-y_9a\,\hat{\mathbf{y}}-z_9a\,\hat{\mathbf{z}}$ | (24$d$) | H IV |
| $\mathbf{B}_{122}$ | = | $\left(\frac{1}{2}+x_9\right)\mathbf{a}_1+y_9\,\mathbf{a}_2+\left(\frac{1}{2}-z_9\right)\mathbf{a}_3$ | = | $\left(\frac{1}{2}+x_9\right)a\,\hat{\mathbf{x}}+y_9a\,\hat{\mathbf{y}}+\left(\frac{1}{2}-z_9\right)a\,\hat{\mathbf{z}}$ | (24$d$) | H IV |
| $\mathbf{B}_{123}$ | = | $x_9\,\mathbf{a}_1+\left(\frac{1}{2}-y_9\right)\mathbf{a}_2+\left(\frac{1}{2}+z_9\right)\mathbf{a}_3$ | = | $x_9a\,\hat{\mathbf{x}}+\left(\frac{1}{2}-y_9\right)a\,\hat{\mathbf{y}}+\left(\frac{1}{2}+z_9\right)a\,\hat{\mathbf{z}}$ | (24$d$) | H IV |
| $\mathbf{B}_{124}$ | = | $\left(\frac{1}{2}-x_9\right)\mathbf{a}_1+\left(\frac{1}{2}+y_9\right)\mathbf{a}_2+z_9\,\mathbf{a}_3$ | = | $\left(\frac{1}{2}-x_9\right)a\,\hat{\mathbf{x}}+\left(\frac{1}{2}+y_9\right)a\,\hat{\mathbf{y}}+z_9a\,\hat{\mathbf{z}}$ | (24$d$) | H IV |
| $\mathbf{B}_{125}$ | = | $-z_9\,\mathbf{a}_1-x_9\,\mathbf{a}_2-y_9\,\mathbf{a}_3$ | = | $-z_9a\,\hat{\mathbf{x}}-x_9a\,\hat{\mathbf{y}}-y_9a\,\hat{\mathbf{z}}$ | (24$d$) | H IV |
| $\mathbf{B}_{126}$ | = | $\left(\frac{1}{2} - z_9\right)\mathbf{a}_1 + \left(\frac{1}{2} + x_9\right)\mathbf{a}_2 + y_9\,\mathbf{a}_3$ | = | $\left(\frac{1}{2} - z_9\right)a\,\hat{\mathbf{x}} + \left(\frac{1}{2} + x_9\right)a\,\hat{\mathbf{y}} + y_9 a\,\hat{\mathbf{z}}$ | (24$d$) | H IV |
| $\mathbf{B}_{127}$ | = | $\left(\frac{1}{2} + z_9\right)\mathbf{a}_1 + x_9\,\mathbf{a}_2 + \left(\frac{1}{2} - y_9\right)\mathbf{a}_3$ | = | $\left(\frac{1}{2} + z_9\right)a\,\hat{\mathbf{x}} + x_9 a\,\hat{\mathbf{y}} + \left(\frac{1}{2} - y_9\right)a\,\hat{\mathbf{z}}$ | (24$d$) | H IV |
| $\mathbf{B}_{128}$ | = | $z_9\,\mathbf{a}_1 + \left(\frac{1}{2} - x_9\right)\mathbf{a}_2 + \left(\frac{1}{2} + y_9\right)\mathbf{a}_3$ | = | $z_9 a\,\hat{\mathbf{x}} + \left(\frac{1}{2} - x_9\right)a\,\hat{\mathbf{y}} + \left(\frac{1}{2} + y_9\right)a\,\hat{\mathbf{z}}$ | (24$d$) | H IV |
| $\mathbf{B}_{129}$ | = | $-y_9\,\mathbf{a}_1 - z_9\,\mathbf{a}_2 - x_9\,\mathbf{a}_3$ | = | $-y_9 a\,\hat{\mathbf{x}} - z_9 a\,\hat{\mathbf{y}} - x_9 a\,\hat{\mathbf{z}}$ | (24$d$) | H IV |
| $\mathbf{B}_{130}$ | = | $y_9\,\mathbf{a}_1 + \left(\frac{1}{2} - z_9\right)\mathbf{a}_2 + \left(\frac{1}{2} + x_9\right)\mathbf{a}_3$ | = | $y_9 a\,\hat{\mathbf{x}} + \left(\frac{1}{2} - z_9\right)a\,\hat{\mathbf{y}} + \left(\frac{1}{2} + x_9\right)a\,\hat{\mathbf{z}}$ | (24$d$) | H IV |
| $\mathbf{B}_{131}$ | = | $\left(\frac{1}{2} - y_9\right)\mathbf{a}_1 + \left(\frac{1}{2} + z_9\right)\mathbf{a}_2 + x_9\,\mathbf{a}_3$ | = | $\left(\frac{1}{2} - y_9\right)a\,\hat{\mathbf{x}} + \left(\frac{1}{2} + z_9\right)a\,\hat{\mathbf{y}} + x_9 a\,\hat{\mathbf{z}}$ | (24$d$) | H IV |
| $\mathbf{B}_{132}$ | = | $\left(\frac{1}{2} + y_9\right)\mathbf{a}_1 + z_9\,\mathbf{a}_2 + \left(\frac{1}{2} - x_9\right)\mathbf{a}_3$ | = | $\left(\frac{1}{2} + y_9\right)a\,\hat{\mathbf{x}} + z_9 a\,\hat{\mathbf{y}} + \left(\frac{1}{2} - x_9\right)a\,\hat{\mathbf{z}}$ | (24$d$) | H IV |
| $\mathbf{B}_{133}$ | = | $x_{10}\,\mathbf{a}_1 + y_{10}\,\mathbf{a}_2 + z_{10}\,\mathbf{a}_3$ | = | $x_{10} a\,\hat{\mathbf{x}} + y_{10} a\,\hat{\mathbf{y}} + z_{10} a\,\hat{\mathbf{z}}$ | (24$d$) | H V |
| $\mathbf{B}_{134}$ | = | $\left(\frac{1}{2} - x_{10}\right)\mathbf{a}_1 - y_{10}\,\mathbf{a}_2 + \left(\frac{1}{2} + z_{10}\right)\mathbf{a}_3$ | = | $\left(\frac{1}{2} - x_{10}\right)a\,\hat{\mathbf{x}} - y_{10} a\,\hat{\mathbf{y}} + \left(\frac{1}{2} + z_{10}\right)a\,\hat{\mathbf{z}}$ | (24$d$) | H V |
| $\mathbf{B}_{135}$ | = | $-x_{10}\,\mathbf{a}_1 + \left(\frac{1}{2} + y_{10}\right)\mathbf{a}_2 + \left(\frac{1}{2} - z_{10}\right)\mathbf{a}_3$ | = | $-x_{10} a\,\hat{\mathbf{x}} + \left(\frac{1}{2} + y_{10}\right)a\,\hat{\mathbf{y}} + \left(\frac{1}{2} - z_{10}\right)a\,\hat{\mathbf{z}}$ | (24$d$) | H V |
| $\mathbf{B}_{136}$ | = | $\left(\frac{1}{2} + x_{10}\right)\mathbf{a}_1 + \left(\frac{1}{2} - y_{10}\right)\mathbf{a}_2 - z_{10}\,\mathbf{a}_3$ | = | $\left(\frac{1}{2} + x_{10}\right)a\,\hat{\mathbf{x}} + \left(\frac{1}{2} - y_{10}\right)a\,\hat{\mathbf{y}} - z_{10} a\,\hat{\mathbf{z}}$ | (24$d$) | H V |
| $\mathbf{B}_{137}$ | = | $z_{10}\,\mathbf{a}_1 + x_{10}\,\mathbf{a}_2 + y_{10}\,\mathbf{a}_3$ | = | $z_{10} a\,\hat{\mathbf{x}} + x_{10} a\,\hat{\mathbf{y}} + y_{10} a\,\hat{\mathbf{z}}$ | (24$d$) | H V |
| $\mathbf{B}_{138}$ | = | $\left(\frac{1}{2} + z_{10}\right)\mathbf{a}_1 + \left(\frac{1}{2} - x_{10}\right)\mathbf{a}_2 - y_{10}\,\mathbf{a}_3$ | = | $\left(\frac{1}{2} + z_{10}\right)a\,\hat{\mathbf{x}} + \left(\frac{1}{2} - x_{10}\right)a\,\hat{\mathbf{y}} - y_{10} a\,\hat{\mathbf{z}}$ | (24$d$) | H V |
| $\mathbf{B}_{139}$ | = | $\left(\frac{1}{2} - z_{10}\right)\mathbf{a}_1 - x_{10}\,\mathbf{a}_2 + \left(\frac{1}{2} + y_{10}\right)\mathbf{a}_3$ | = | $\left(\frac{1}{2} - z_{10}\right)a\,\hat{\mathbf{x}} - x_{10} a\,\hat{\mathbf{y}} + \left(\frac{1}{2} + y_{10}\right)a\,\hat{\mathbf{z}}$ | (24$d$) | H V |
| $\mathbf{B}_{140}$ | = | $-z_{10}\,\mathbf{a}_1 + \left(\frac{1}{2} + x_{10}\right)\mathbf{a}_2 + \left(\frac{1}{2} - y_{10}\right)\mathbf{a}_3$ | = | $-z_{10} a\,\hat{\mathbf{x}} + \left(\frac{1}{2} + x_{10}\right)a\,\hat{\mathbf{y}} + \left(\frac{1}{2} - y_{10}\right)a\,\hat{\mathbf{z}}$ | (24$d$) | H V |
| $\mathbf{B}_{141}$ | = | $y_{10}\,\mathbf{a}_1 + z_{10}\,\mathbf{a}_2 + x_{10}\,\mathbf{a}_3$ | = | $y_{10} a\,\hat{\mathbf{x}} + z_{10} a\,\hat{\mathbf{y}} + x_{10} a\,\hat{\mathbf{z}}$ | (24$d$) | H V |
| $\mathbf{B}_{142}$ | = | $-y_{10}\,\mathbf{a}_1 + \left(\frac{1}{2} + z_{10}\right)\mathbf{a}_2 + \left(\frac{1}{2} - x_{10}\right)\mathbf{a}_3$ | = | $-y_{10} a\,\hat{\mathbf{x}} + \left(\frac{1}{2} + z_{10}\right)a\,\hat{\mathbf{y}} + \left(\frac{1}{2} - x_{10}\right)a\,\hat{\mathbf{z}}$ | (24$d$) | H V |
| $\mathbf{B}_{143}$ | = | $\left(\frac{1}{2} + y_{10}\right)\mathbf{a}_1 + \left(\frac{1}{2} - z_{10}\right)\mathbf{a}_2 - x_{10}\,\mathbf{a}_3$ | = | $\left(\frac{1}{2} + y_{10}\right)a\,\hat{\mathbf{x}} + \left(\frac{1}{2} - z_{10}\right)a\,\hat{\mathbf{y}} - x_{10} a\,\hat{\mathbf{z}}$ | (24$d$) | H V |
| $\mathbf{B}_{144}$ | = | $\left(\frac{1}{2} - y_{10}\right)\mathbf{a}_1 - z_{10}\,\mathbf{a}_2 + \left(\frac{1}{2} + x_{10}\right)\mathbf{a}_3$ | = | $\left(\frac{1}{2} - y_{10}\right)a\,\hat{\mathbf{x}} - z_{10} a\,\hat{\mathbf{y}} + \left(\frac{1}{2} + x_{10}\right)a\,\hat{\mathbf{z}}$ | (24$d$) | H V |
| $\mathbf{B}_{145}$ | = | $-x_{10}\,\mathbf{a}_1 - y_{10}\,\mathbf{a}_2 - z_{10}\,\mathbf{a}_3$ | = | $-x_{10} a\,\hat{\mathbf{x}} - y_{10} a\,\hat{\mathbf{y}} - z_{10} a\,\hat{\mathbf{z}}$ | (24$d$) | H V |
| $\mathbf{B}_{146}$ | = | $\left(\frac{1}{2} + x_{10}\right)\mathbf{a}_1 + y_{10}\,\mathbf{a}_2 + \left(\frac{1}{2} - z_{10}\right)\mathbf{a}_3$ | = | $\left(\frac{1}{2} + x_{10}\right)a\,\hat{\mathbf{x}} + y_{10} a\,\hat{\mathbf{y}} + \left(\frac{1}{2} - z_{10}\right)a\,\hat{\mathbf{z}}$ | (24$d$) | H V |
| $\mathbf{B}_{147}$ | = | $x_{10}\,\mathbf{a}_1 + \left(\frac{1}{2} - y_{10}\right)\mathbf{a}_2 + \left(\frac{1}{2} + z_{10}\right)\mathbf{a}_3$ | = | $x_{10} a\,\hat{\mathbf{x}} + \left(\frac{1}{2} - y_{10}\right)a\,\hat{\mathbf{y}} + \left(\frac{1}{2} + z_{10}\right)a\,\hat{\mathbf{z}}$ | (24$d$) | H V |
| $\mathbf{B}_{148}$ | = | $\left(\frac{1}{2} - x_{10}\right)\mathbf{a}_1 + \left(\frac{1}{2} + y_{10}\right)\mathbf{a}_2 + z_{10}\,\mathbf{a}_3$ | = | $\left(\frac{1}{2} - x_{10}\right)a\,\hat{\mathbf{x}} + \left(\frac{1}{2} + y_{10}\right)a\,\hat{\mathbf{y}} + z_{10} a\,\hat{\mathbf{z}}$ | (24$d$) | H V |
| $\mathbf{B}_{149}$ | = | $-z_{10}\,\mathbf{a}_1 - x_{10}\,\mathbf{a}_2 - y_{10}\,\mathbf{a}_3$ | = | $-z_{10} a\,\hat{\mathbf{x}} - x_{10} a\,\hat{\mathbf{y}} - y_{10} a\,\hat{\mathbf{z}}$ | (24$d$) | H V |
| $\mathbf{B}_{150}$ | = | $\left(\frac{1}{2} - z_{10}\right)\mathbf{a}_1 + \left(\frac{1}{2} + x_{10}\right)\mathbf{a}_2 + y_{10}\,\mathbf{a}_3$ | = | $\left(\frac{1}{2} - z_{10}\right)a\,\hat{\mathbf{x}} + \left(\frac{1}{2} + x_{10}\right)a\,\hat{\mathbf{y}} + y_{10} a\,\hat{\mathbf{z}}$ | (24$d$) | H V |
| $\mathbf{B}_{151}$ | = | $\left(\frac{1}{2} + z_{10}\right)\mathbf{a}_1 + x_{10}\,\mathbf{a}_2 + \left(\frac{1}{2} - y_{10}\right)\mathbf{a}_3$ | = | $\left(\frac{1}{2} + z_{10}\right)a\,\hat{\mathbf{x}} + x_{10} a\,\hat{\mathbf{y}} + \left(\frac{1}{2} - y_{10}\right)a\,\hat{\mathbf{z}}$ | (24$d$) | H V |
| $\mathbf{B}_{152}$ | = | $z_{10}\,\mathbf{a}_1 + \left(\frac{1}{2} - x_{10}\right)\mathbf{a}_2 + \left(\frac{1}{2} + y_{10}\right)\mathbf{a}_3$ | = | $z_{10} a\,\hat{\mathbf{x}} + \left(\frac{1}{2} - x_{10}\right)a\,\hat{\mathbf{y}} + \left(\frac{1}{2} + y_{10}\right)a\,\hat{\mathbf{z}}$ | (24$d$) | H V |
| $\mathbf{B}_{153}$ | = | $-y_{10}\,\mathbf{a}_1 - z_{10}\,\mathbf{a}_2 - x_{10}\,\mathbf{a}_3$ | = | $-y_{10} a\,\hat{\mathbf{x}} - z_{10} a\,\hat{\mathbf{y}} - x_{10} a\,\hat{\mathbf{z}}$ | (24$d$) | H V |
| $\mathbf{B}_{154}$ | = | $y_{10}\,\mathbf{a}_1 + \left(\frac{1}{2} - z_{10}\right)\mathbf{a}_2 + \left(\frac{1}{2} + x_{10}\right)\mathbf{a}_3$ | = | $y_{10} a\,\hat{\mathbf{x}} + \left(\frac{1}{2} - z_{10}\right)a\,\hat{\mathbf{y}} + \left(\frac{1}{2} + x_{10}\right)a\,\hat{\mathbf{z}}$ | (24$d$) | H V |
| $\mathbf{B}_{155}$ | = | $\left(\frac{1}{2} - y_{10}\right)\mathbf{a}_1 + \left(\frac{1}{2} + z_{10}\right)\mathbf{a}_2 + x_{10}\,\mathbf{a}_3$ | = | $\left(\frac{1}{2} - y_{10}\right)a\,\hat{\mathbf{x}} + \left(\frac{1}{2} + z_{10}\right)a\,\hat{\mathbf{y}} + x_{10} a\,\hat{\mathbf{z}}$ | (24$d$) | H V |
| $\mathbf{B}_{156}$ | = | $\left(\frac{1}{2} + y_{10}\right)\mathbf{a}_1 + z_{10}\,\mathbf{a}_2 + \left(\frac{1}{2} - x_{10}\right)\mathbf{a}_3$ | = | $\left(\frac{1}{2} + y_{10}\right)a\,\hat{\mathbf{x}} + z_{10} a\,\hat{\mathbf{y}} + \left(\frac{1}{2} - x_{10}\right)a\,\hat{\mathbf{z}}$ | (24$d$) | H V |
| $\mathbf{B}_{157}$ | = | $x_{11}\,\mathbf{a}_1 + y_{11}\,\mathbf{a}_2 + z_{11}\,\mathbf{a}_3$ | = | $x_{11} a\,\hat{\mathbf{x}} + y_{11} a\,\hat{\mathbf{y}} + z_{11} a\,\hat{\mathbf{z}}$ | (24$d$) | H VI |
| $\mathbf{B}_{158}$ | = | $\left(\frac{1}{2} - x_{11}\right)\mathbf{a}_1 - y_{11}\,\mathbf{a}_2 + \left(\frac{1}{2} + z_{11}\right)\mathbf{a}_3$ | = | $\left(\frac{1}{2} - x_{11}\right)a\,\hat{\mathbf{x}} - y_{11} a\,\hat{\mathbf{y}} + \left(\frac{1}{2} + z_{11}\right)a\,\hat{\mathbf{z}}$ | (24$d$) | H VI |
| $\mathbf{B}_{159}$ | = | $-x_{11}\,\mathbf{a}_1 + \left(\frac{1}{2} + y_{11}\right)\,\mathbf{a}_2 + \left(\frac{1}{2} - z_{11}\right)\,\mathbf{a}_3$ | = | $-x_{11}a\,\hat{\mathbf{x}} + \left(\frac{1}{2} + y_{11}\right)a\,\hat{\mathbf{y}} + \left(\frac{1}{2} - z_{11}\right)a\,\hat{\mathbf{z}}$ | (24$d$) | H VI |
| $\mathbf{B}_{160}$ | = | $\left(\frac{1}{2} + x_{11}\right)\,\mathbf{a}_1 + \left(\frac{1}{2} - y_{11}\right)\,\mathbf{a}_2 - z_{11}\,\mathbf{a}_3$ | = | $\left(\frac{1}{2} + x_{11}\right)a\,\hat{\mathbf{x}} + \left(\frac{1}{2} - y_{11}\right)a\,\hat{\mathbf{y}} - z_{11}a\,\hat{\mathbf{z}}$ | (24$d$) | H VI |
| $\mathbf{B}_{161}$ | = | $z_{11}\,\mathbf{a}_1 + x_{11}\,\mathbf{a}_2 + y_{11}\,\mathbf{a}_3$ | = | $z_{11}a\,\hat{\mathbf{x}} + x_{11}a\,\hat{\mathbf{y}} + y_{11}a\,\hat{\mathbf{z}}$ | (24$d$) | H VI |
| $\mathbf{B}_{162}$ | = | $\left(\frac{1}{2} + z_{11}\right)\,\mathbf{a}_1 + \left(\frac{1}{2} - x_{11}\right)\,\mathbf{a}_2 - y_{11}\,\mathbf{a}_3$ | = | $\left(\frac{1}{2} + z_{11}\right)a\,\hat{\mathbf{x}} + \left(\frac{1}{2} - x_{11}\right)a\,\hat{\mathbf{y}} - y_{11}a\,\hat{\mathbf{z}}$ | (24$d$) | H VI |
| $\mathbf{B}_{163}$ | = | $\left(\frac{1}{2} - z_{11}\right)\,\mathbf{a}_1 - x_{11}\,\mathbf{a}_2 + \left(\frac{1}{2} + y_{11}\right)\,\mathbf{a}_3$ | = | $\left(\frac{1}{2} - z_{11}\right)a\,\hat{\mathbf{x}} - x_{11}a\,\hat{\mathbf{y}} + \left(\frac{1}{2} + y_{11}\right)a\,\hat{\mathbf{z}}$ | (24$d$) | H VI |
| $\mathbf{B}_{164}$ | = | $-z_{11}\,\mathbf{a}_1 + \left(\frac{1}{2} + x_{11}\right)\,\mathbf{a}_2 + \left(\frac{1}{2} - y_{11}\right)\,\mathbf{a}_3$ | = | $-z_{11}a\,\hat{\mathbf{x}} + \left(\frac{1}{2} + x_{11}\right)a\,\hat{\mathbf{y}} + \left(\frac{1}{2} - y_{11}\right)a\,\hat{\mathbf{z}}$ | (24$d$) | H VI |
| $\mathbf{B}_{165}$ | = | $y_{11}\,\mathbf{a}_1 + z_{11}\,\mathbf{a}_2 + x_{11}\,\mathbf{a}_3$ | = | $y_{11}a\,\hat{\mathbf{x}} + z_{11}a\,\hat{\mathbf{y}} + x_{11}a\,\hat{\mathbf{z}}$ | (24$d$) | H VI |
| $\mathbf{B}_{166}$ | = | $-y_{11}\,\mathbf{a}_1 + \left(\frac{1}{2} + z_{11}\right)\,\mathbf{a}_2 + \left(\frac{1}{2} - x_{11}\right)\,\mathbf{a}_3$ | = | $-y_{11}a\,\hat{\mathbf{x}} + \left(\frac{1}{2} + z_{11}\right)a\,\hat{\mathbf{y}} + \left(\frac{1}{2} - x_{11}\right)a\,\hat{\mathbf{z}}$ | (24$d$) | H VI |
| $\mathbf{B}_{167}$ | = | $\left(\frac{1}{2} + y_{11}\right)\,\mathbf{a}_1 + \left(\frac{1}{2} - z_{11}\right)\,\mathbf{a}_2 - x_{11}\,\mathbf{a}_3$ | = | $\left(\frac{1}{2} + y_{11}\right)a\,\hat{\mathbf{x}} + \left(\frac{1}{2} - z_{11}\right)a\,\hat{\mathbf{y}} - x_{11}a\,\hat{\mathbf{z}}$ | (24$d$) | H VI |
| $\mathbf{B}_{168}$ | = | $\left(\frac{1}{2} - y_{11}\right)\,\mathbf{a}_1 - z_{11}\,\mathbf{a}_2 + \left(\frac{1}{2} + x_{11}\right)\,\mathbf{a}_3$ | = | $\left(\frac{1}{2} - y_{11}\right)a\,\hat{\mathbf{x}} - z_{11}a\,\hat{\mathbf{y}} + \left(\frac{1}{2} + x_{11}\right)a\,\hat{\mathbf{z}}$ | (24$d$) | H VI |
| $\mathbf{B}_{169}$ | = | $-x_{11}\,\mathbf{a}_1 - y_{11}\,\mathbf{a}_2 - z_{11}\,\mathbf{a}_3$ | = | $-x_{11}a\,\hat{\mathbf{x}} - y_{11}a\,\hat{\mathbf{y}} - z_{11}a\,\hat{\mathbf{z}}$ | (24$d$) | H VI |
| $\mathbf{B}_{170}$ | = | $\left(\frac{1}{2} + x_{11}\right)\,\mathbf{a}_1 + y_{11}\,\mathbf{a}_2 + \left(\frac{1}{2} - z_{11}\right)\,\mathbf{a}_3$ | = | $\left(\frac{1}{2} + x_{11}\right)a\,\hat{\mathbf{x}} + y_{11}a\,\hat{\mathbf{y}} + \left(\frac{1}{2} - z_{11}\right)a\,\hat{\mathbf{z}}$ | (24$d$) | H VI |
| $\mathbf{B}_{171}$ | = | $x_{11}\,\mathbf{a}_1 + \left(\frac{1}{2} - y_{11}\right)\,\mathbf{a}_2 + \left(\frac{1}{2} + z_{11}\right)\,\mathbf{a}_3$ | = | $x_{11}a\,\hat{\mathbf{x}} + \left(\frac{1}{2} - y_{11}\right)a\,\hat{\mathbf{y}} + \left(\frac{1}{2} + z_{11}\right)a\,\hat{\mathbf{z}}$ | (24$d$) | H VI |
| $\mathbf{B}_{172}$ | = | $\left(\frac{1}{2} - x_{11}\right)\,\mathbf{a}_1 + \left(\frac{1}{2} + y_{11}\right)\,\mathbf{a}_2 + z_{11}\,\mathbf{a}_3$ | = | $\left(\frac{1}{2} - x_{11}\right)a\,\hat{\mathbf{x}} + \left(\frac{1}{2} + y_{11}\right)a\,\hat{\mathbf{y}} + z_{11}a\,\hat{\mathbf{z}}$ | (24$d$) | H VI |
| $\mathbf{B}_{173}$ | = | $-z_{11}\,\mathbf{a}_1 - x_{11}\,\mathbf{a}_2 - y_{11}\,\mathbf{a}_3$ | = | $-z_{11}a\,\hat{\mathbf{x}} - x_{11}a\,\hat{\mathbf{y}} - y_{11}a\,\hat{\mathbf{z}}$ | (24$d$) | H VI |
| $\mathbf{B}_{174}$ | = | $\left(\frac{1}{2} - z_{11}\right)\,\mathbf{a}_1 + \left(\frac{1}{2} + x_{11}\right)\,\mathbf{a}_2 + y_{11}\,\mathbf{a}_3$ | = | $\left(\frac{1}{2} - z_{11}\right)a\,\hat{\mathbf{x}} + \left(\frac{1}{2} + x_{11}\right)a\,\hat{\mathbf{y}} + y_{11}a\,\hat{\mathbf{z}}$ | (24$d$) | H VI |
| $\mathbf{B}_{175}$ | = | $\left(\frac{1}{2} + z_{11}\right)\,\mathbf{a}_1 + x_{11}\,\mathbf{a}_2 + \left(\frac{1}{2} - y_{11}\right)\,\mathbf{a}_3$ | = | $\left(\frac{1}{2} + z_{11}\right)a\,\hat{\mathbf{x}} + x_{11}a\,\hat{\mathbf{y}} + \left(\frac{1}{2} - y_{11}\right)a\,\hat{\mathbf{z}}$ | (24$d$) | H VI |
| $\mathbf{B}_{176}$ | = | $z_{11}\,\mathbf{a}_1 + \left(\frac{1}{2} - x_{11}\right)\,\mathbf{a}_2 + \left(\frac{1}{2} + y_{11}\right)\,\mathbf{a}_3$ | = | $z_{11}a\,\hat{\mathbf{x}} + \left(\frac{1}{2} - x_{11}\right)a\,\hat{\mathbf{y}} + \left(\frac{1}{2} + y_{11}\right)a\,\hat{\mathbf{z}}$ | (24$d$) | H VI |
| $\mathbf{B}_{177}$ | = | $-y_{11}\,\mathbf{a}_1 - z_{11}\,\mathbf{a}_2 - x_{11}\,\mathbf{a}_3$ | = | $-y_{11}a\,\hat{\mathbf{x}} - z_{11}a\,\hat{\mathbf{y}} - x_{11}a\,\hat{\mathbf{z}}$ | (24$d$) | H VI |
| $\mathbf{B}_{178}$ | = | $y_{11}\,\mathbf{a}_1 + \left(\frac{1}{2} - z_{11}\right)\,\mathbf{a}_2 + \left(\frac{1}{2} + x_{11}\right)\,\mathbf{a}_3$ | = | $y_{11}a\,\hat{\mathbf{x}} + \left(\frac{1}{2} - z_{11}\right)a\,\hat{\mathbf{y}} + \left(\frac{1}{2} + x_{11}\right)a\,\hat{\mathbf{z}}$ | (24$d$) | H VI |
| $\mathbf{B}_{179}$ | = | $\left(\frac{1}{2} - y_{11}\right)\,\mathbf{a}_1 + \left(\frac{1}{2} + z_{11}\right)\,\mathbf{a}_2 + x_{11}\,\mathbf{a}_3$ | = | $\left(\frac{1}{2} - y_{11}\right)a\,\hat{\mathbf{x}} + \left(\frac{1}{2} + z_{11}\right)a\,\hat{\mathbf{y}} + x_{11}a\,\hat{\mathbf{z}}$ | (24$d$) | H VI |
| $\mathbf{B}_{180}$ | = | $\left(\frac{1}{2} + y_{11}\right)\,\mathbf{a}_1 + z_{11}\,\mathbf{a}_2 + \left(\frac{1}{2} - x_{11}\right)\,\mathbf{a}_3$ | = | $\left(\frac{1}{2} + y_{11}\right)a\,\hat{\mathbf{x}} + z_{11}a\,\hat{\mathbf{y}} + \left(\frac{1}{2} - x_{11}\right)a\,\hat{\mathbf{z}}$ | (24$d$) | H VI |
| $\mathbf{B}_{181}$ | = | $x_{12}\,\mathbf{a}_1 + y_{12}\,\mathbf{a}_2 + z_{12}\,\mathbf{a}_3$ | = | $x_{12}a\,\hat{\mathbf{x}} + y_{12}a\,\hat{\mathbf{y}} + z_{12}a\,\hat{\mathbf{z}}$ | (24$d$) | O II |
| $\mathbf{B}_{182}$ | = | $\left(\frac{1}{2} - x_{12}\right)\,\mathbf{a}_1 - y_{12}\,\mathbf{a}_2 + \left(\frac{1}{2} + z_{12}\right)\,\mathbf{a}_3$ | = | $\left(\frac{1}{2} - x_{12}\right)a\,\hat{\mathbf{x}} - y_{12}a\,\hat{\mathbf{y}} + \left(\frac{1}{2} + z_{12}\right)a\,\hat{\mathbf{z}}$ | (24$d$) | O II |
| $\mathbf{B}_{183}$ | = | $-x_{12}\,\mathbf{a}_1 + \left(\frac{1}{2} + y_{12}\right)\,\mathbf{a}_2 + \left(\frac{1}{2} - z_{12}\right)\,\mathbf{a}_3$ | = | $-x_{12}a\,\hat{\mathbf{x}} + \left(\frac{1}{2} + y_{12}\right)a\,\hat{\mathbf{y}} + \left(\frac{1}{2} - z_{12}\right)a\,\hat{\mathbf{z}}$ | (24$d$) | O II |
| $\mathbf{B}_{184}$ | = | $\left(\frac{1}{2} + x_{12}\right)\,\mathbf{a}_1 + \left(\frac{1}{2} - y_{12}\right)\,\mathbf{a}_2 - z_{12}\,\mathbf{a}_3$ | = | $\left(\frac{1}{2} + x_{12}\right)a\,\hat{\mathbf{x}} + \left(\frac{1}{2} - y_{12}\right)a\,\hat{\mathbf{y}} - z_{12}a\,\hat{\mathbf{z}}$ | (24$d$) | O II |
| $\mathbf{B}_{185}$ | = | $z_{12}\,\mathbf{a}_1 + x_{12}\,\mathbf{a}_2 + y_{12}\,\mathbf{a}_3$ | = | $z_{12}a\,\hat{\mathbf{x}} + x_{12}a\,\hat{\mathbf{y}} + y_{12}a\,\hat{\mathbf{z}}$ | (24$d$) | O II |
| $\mathbf{B}_{186}$ | = | $\left(\frac{1}{2} + z_{12}\right)\,\mathbf{a}_1 + \left(\frac{1}{2} - x_{12}\right)\,\mathbf{a}_2 - y_{12}\,\mathbf{a}_3$ | = | $\left(\frac{1}{2} + z_{12}\right)a\,\hat{\mathbf{x}} + \left(\frac{1}{2} - x_{12}\right)a\,\hat{\mathbf{y}} - y_{12}a\,\hat{\mathbf{z}}$ | (24$d$) | O II |
| $\mathbf{B}_{187}$ | = | $\left(\frac{1}{2} - z_{12}\right)\,\mathbf{a}_1 - x_{12}\,\mathbf{a}_2 + \left(\frac{1}{2} + y_{12}\right)\,\mathbf{a}_3$ | = | $\left(\frac{1}{2} - z_{12}\right)a\,\hat{\mathbf{x}} - x_{12}a\,\hat{\mathbf{y}} + \left(\frac{1}{2} + y_{12}\right)a\,\hat{\mathbf{z}}$ | (24$d$) | O II |
| $\mathbf{B}_{188}$ | = | $-z_{12}\,\mathbf{a}_1 + \left(\frac{1}{2} + x_{12}\right)\,\mathbf{a}_2 + \left(\frac{1}{2} - y_{12}\right)\,\mathbf{a}_3$ | = | $-z_{12}a\,\hat{\mathbf{x}} + \left(\frac{1}{2} + x_{12}\right)a\,\hat{\mathbf{y}} + \left(\frac{1}{2} - y_{12}\right)a\,\hat{\mathbf{z}}$ | (24$d$) | O II |
| $\mathbf{B}_{189}$ | = | $y_{12}\,\mathbf{a}_1 + z_{12}\,\mathbf{a}_2 + x_{12}\,\mathbf{a}_3$ | = | $y_{12}a\,\hat{\mathbf{x}} + z_{12}a\,\hat{\mathbf{y}} + x_{12}a\,\hat{\mathbf{z}}$ | (24$d$) | O II |
| $\mathbf{B}_{190}$ | = | $-y_{12}\,\mathbf{a}_1 + \left(\frac{1}{2} + z_{12}\right)\,\mathbf{a}_2 + \left(\frac{1}{2} - x_{12}\right)\,\mathbf{a}_3$ | = | $-y_{12}a\,\hat{\mathbf{x}} + \left(\frac{1}{2} + z_{12}\right)a\,\hat{\mathbf{y}} + \left(\frac{1}{2} - x_{12}\right)a\,\hat{\mathbf{z}}$ | (24$d$) | O II |
| $\mathbf{B}_{191}$ | = | $\left(\frac{1}{2}+y_{12}\right)\mathbf{a}_1+\left(\frac{1}{2}-z_{12}\right)\mathbf{a}_2-x_{12}\,\mathbf{a}_3$ | = | $\left(\frac{1}{2}+y_{12}\right)a\,\hat{\mathbf{x}}+\left(\frac{1}{2}-z_{12}\right)a\,\hat{\mathbf{y}}-x_{12}a\,\hat{\mathbf{z}}$ | $(24d)$ | O II |
| $\mathbf{B}_{192}$ | = | $\left(\frac{1}{2}-y_{12}\right)\mathbf{a}_1-z_{12}\,\mathbf{a}_2+\left(\frac{1}{2}+x_{12}\right)\mathbf{a}_3$ | = | $\left(\frac{1}{2}-y_{12}\right)a\,\hat{\mathbf{x}}-z_{12}a\,\hat{\mathbf{y}}+\left(\frac{1}{2}+x_{12}\right)a\,\hat{\mathbf{z}}$ | $(24d)$ | O II |
| $\mathbf{B}_{193}$ | = | $-x_{12}\,\mathbf{a}_1-y_{12}\,\mathbf{a}_2-z_{12}\,\mathbf{a}_3$ | = | $-x_{12}a\,\hat{\mathbf{x}}-y_{12}a\,\hat{\mathbf{y}}-z_{12}a\,\hat{\mathbf{z}}$ | $(24d)$ | O II |
| $\mathbf{B}_{194}$ | = | $\left(\frac{1}{2}+x_{12}\right)\mathbf{a}_1+y_{12}\,\mathbf{a}_2+\left(\frac{1}{2}-z_{12}\right)\mathbf{a}_3$ | = | $\left(\frac{1}{2}+x_{12}\right)a\,\hat{\mathbf{x}}+y_{12}a\,\hat{\mathbf{y}}+\left(\frac{1}{2}-z_{12}\right)a\,\hat{\mathbf{z}}$ | $(24d)$ | O II |
| $\mathbf{B}_{195}$ | = | $x_{12}\,\mathbf{a}_1+\left(\frac{1}{2}-y_{12}\right)\mathbf{a}_2+\left(\frac{1}{2}+z_{12}\right)\mathbf{a}_3$ | = | $x_{12}a\,\hat{\mathbf{x}}+\left(\frac{1}{2}-y_{12}\right)a\,\hat{\mathbf{y}}+\left(\frac{1}{2}+z_{12}\right)a\,\hat{\mathbf{z}}$ | $(24d)$ | O II |
| $\mathbf{B}_{196}$ | = | $\left(\frac{1}{2}-x_{12}\right)\mathbf{a}_1+\left(\frac{1}{2}+y_{12}\right)\mathbf{a}_2+z_{12}\,\mathbf{a}_3$ | = | $\left(\frac{1}{2}-x_{12}\right)a\,\hat{\mathbf{x}}+\left(\frac{1}{2}+y_{12}\right)a\,\hat{\mathbf{y}}+z_{12}a\,\hat{\mathbf{z}}$ | $(24d)$ | O II |
| $\mathbf{B}_{197}$ | = | $-z_{12}\,\mathbf{a}_1-x_{12}\,\mathbf{a}_2-y_{12}\,\mathbf{a}_3$ | = | $-z_{12}a\,\hat{\mathbf{x}}-x_{12}a\,\hat{\mathbf{y}}-y_{12}a\,\hat{\mathbf{z}}$ | $(24d)$ | O II |
| $\mathbf{B}_{198}$ | = | $\left(\frac{1}{2}-z_{12}\right)\mathbf{a}_1+\left(\frac{1}{2}+x_{12}\right)\mathbf{a}_2+y_{12}\,\mathbf{a}_3$ | = | $\left(\frac{1}{2}-z_{12}\right)a\,\hat{\mathbf{x}}+\left(\frac{1}{2}+x_{12}\right)a\,\hat{\mathbf{y}}+y_{12}a\,\hat{\mathbf{z}}$ | $(24d)$ | O II |
| $\mathbf{B}_{199}$ | = | $\left(\frac{1}{2}+z_{12}\right)\mathbf{a}_1+x_{12}\,\mathbf{a}_2+\left(\frac{1}{2}-y_{12}\right)\mathbf{a}_3$ | = | $\left(\frac{1}{2}+z_{12}\right)a\,\hat{\mathbf{x}}+x_{12}a\,\hat{\mathbf{y}}+\left(\frac{1}{2}-y_{12}\right)a\,\hat{\mathbf{z}}$ | $(24d)$ | O II |
| $\mathbf{B}_{200}$ | = | $z_{12}\,\mathbf{a}_1+\left(\frac{1}{2}-x_{12}\right)\mathbf{a}_2+\left(\frac{1}{2}+y_{12}\right)\mathbf{a}_3$ | = | $z_{12}a\,\hat{\mathbf{x}}+\left(\frac{1}{2}-x_{12}\right)a\,\hat{\mathbf{y}}+\left(\frac{1}{2}+y_{12}\right)a\,\hat{\mathbf{z}}$ | $(24d)$ | O II |
| $\mathbf{B}_{201}$ | = | $-y_{12}\,\mathbf{a}_1-z_{12}\,\mathbf{a}_2-x_{12}\,\mathbf{a}_3$ | = | $-y_{12}a\,\hat{\mathbf{x}}-z_{12}a\,\hat{\mathbf{y}}-x_{12}a\,\hat{\mathbf{z}}$ | $(24d)$ | O II |
| $\mathbf{B}_{202}$ | = | $y_{12}\,\mathbf{a}_1+\left(\frac{1}{2}-z_{12}\right)\mathbf{a}_2+\left(\frac{1}{2}+x_{12}\right)\mathbf{a}_3$ | = | $y_{12}a\,\hat{\mathbf{x}}+\left(\frac{1}{2}-z_{12}\right)a\,\hat{\mathbf{y}}+\left(\frac{1}{2}+x_{12}\right)a\,\hat{\mathbf{z}}$ | $(24d)$ | O II |
| $\mathbf{B}_{203}$ | = | $\left(\frac{1}{2}-y_{12}\right)\mathbf{a}_1+\left(\frac{1}{2}+z_{12}\right)\mathbf{a}_2+x_{12}\,\mathbf{a}_3$ | = | $\left(\frac{1}{2}-y_{12}\right)a\,\hat{\mathbf{x}}+\left(\frac{1}{2}+z_{12}\right)a\,\hat{\mathbf{y}}+x_{12}a\,\hat{\mathbf{z}}$ | $(24d)$ | O II |
| $\mathbf{B}_{204}$ | = | $\left(\frac{1}{2}+y_{12}\right)\mathbf{a}_1+z_{12}\,\mathbf{a}_2+\left(\frac{1}{2}-x_{12}\right)\mathbf{a}_3$ | = | $\left(\frac{1}{2}+y_{12}\right)a\,\hat{\mathbf{x}}+z_{12}a\,\hat{\mathbf{y}}+\left(\frac{1}{2}-x_{12}\right)a\,\hat{\mathbf{z}}$ | $(24d)$ | O II |
| $\mathbf{B}_{205}$ | = | $x_{13}\,\mathbf{a}_1+y_{13}\,\mathbf{a}_2+z_{13}\,\mathbf{a}_3$ | = | $x_{13}a\,\hat{\mathbf{x}}+y_{13}a\,\hat{\mathbf{y}}+z_{13}a\,\hat{\mathbf{z}}$ | $(24d)$ | O III |
| $\mathbf{B}_{206}$ | = | $\left(\frac{1}{2}-x_{13}\right)\mathbf{a}_1-y_{13}\,\mathbf{a}_2+\left(\frac{1}{2}+z_{13}\right)\mathbf{a}_3$ | = | $\left(\frac{1}{2}-x_{13}\right)a\,\hat{\mathbf{x}}-y_{13}a\,\hat{\mathbf{y}}+\left(\frac{1}{2}+z_{13}\right)a\,\hat{\mathbf{z}}$ | $(24d)$ | O III |
| $\mathbf{B}_{207}$ | = | $-x_{13}\,\mathbf{a}_1+\left(\frac{1}{2}+y_{13}\right)\mathbf{a}_2+\left(\frac{1}{2}-z_{13}\right)\mathbf{a}_3$ | = | $-x_{13}a\,\hat{\mathbf{x}}+\left(\frac{1}{2}+y_{13}\right)a\,\hat{\mathbf{y}}+\left(\frac{1}{2}-z_{13}\right)a\,\hat{\mathbf{z}}$ | $(24d)$ | O III |
| $\mathbf{B}_{208}$ | = | $\left(\frac{1}{2}+x_{13}\right)\mathbf{a}_1+\left(\frac{1}{2}-y_{13}\right)\mathbf{a}_2-z_{13}\,\mathbf{a}_3$ | = | $\left(\frac{1}{2}+x_{13}\right)a\,\hat{\mathbf{x}}+\left(\frac{1}{2}-y_{13}\right)a\,\hat{\mathbf{y}}-z_{13}a\,\hat{\mathbf{z}}$ | $(24d)$ | O III |
| $\mathbf{B}_{209}$ | = | $z_{13}\,\mathbf{a}_1+x_{13}\,\mathbf{a}_2+y_{13}\,\mathbf{a}_3$ | = | $z_{13}a\,\hat{\mathbf{x}}+x_{13}a\,\hat{\mathbf{y}}+y_{13}a\,\hat{\mathbf{z}}$ | $(24d)$ | O III |
| $\mathbf{B}_{210}$ | = | $\left(\frac{1}{2}+z_{13}\right)\mathbf{a}_1+\left(\frac{1}{2}-x_{13}\right)\mathbf{a}_2-y_{13}\,\mathbf{a}_3$ | = | $\left(\frac{1}{2}+z_{13}\right)a\,\hat{\mathbf{x}}+\left(\frac{1}{2}-x_{13}\right)a\,\hat{\mathbf{y}}-y_{13}a\,\hat{\mathbf{z}}$ | $(24d)$ | O III |
| $\mathbf{B}_{211}$ | = | $\left(\frac{1}{2}-z_{13}\right)\mathbf{a}_1-x_{13}\,\mathbf{a}_2+\left(\frac{1}{2}+y_{13}\right)\mathbf{a}_3$ | = | $\left(\frac{1}{2}-z_{13}\right)a\,\hat{\mathbf{x}}-x_{13}a\,\hat{\mathbf{y}}+\left(\frac{1}{2}+y_{13}\right)a\,\hat{\mathbf{z}}$ | $(24d)$ | O III |
| $\mathbf{B}_{212}$ | = | $-z_{13}\,\mathbf{a}_1+\left(\frac{1}{2}+x_{13}\right)\mathbf{a}_2+\left(\frac{1}{2}-y_{13}\right)\mathbf{a}_3$ | = | $-z_{13}a\,\hat{\mathbf{x}}+\left(\frac{1}{2}+x_{13}\right)a\,\hat{\mathbf{y}}+\left(\frac{1}{2}-y_{13}\right)a\,\hat{\mathbf{z}}$ | $(24d)$ | O III |
| $\mathbf{B}_{213}$ | = | $y_{13}\,\mathbf{a}_1+z_{13}\,\mathbf{a}_2+x_{13}\,\mathbf{a}_3$ | = | $y_{13}a\,\hat{\mathbf{x}}+z_{13}a\,\hat{\mathbf{y}}+x_{13}a\,\hat{\mathbf{z}}$ | $(24d)$ | O III |
| $\mathbf{B}_{214}$ | = | $-y_{13}\,\mathbf{a}_1+\left(\frac{1}{2}+z_{13}\right)\mathbf{a}_2+\left(\frac{1}{2}-x_{13}\right)\mathbf{a}_3$ | = | $-y_{13}a\,\hat{\mathbf{x}}+\left(\frac{1}{2}+z_{13}\right)a\,\hat{\mathbf{y}}+\left(\frac{1}{2}-x_{13}\right)a\,\hat{\mathbf{z}}$ | $(24d)$ | O III |
| $\mathbf{B}_{215}$ | = | $\left(\frac{1}{2}+y_{13}\right)\mathbf{a}_1+\left(\frac{1}{2}-z_{13}\right)\mathbf{a}_2-x_{13}\,\mathbf{a}_3$ | = | $\left(\frac{1}{2}+y_{13}\right)a\,\hat{\mathbf{x}}+\left(\frac{1}{2}-z_{13}\right)a\,\hat{\mathbf{y}}-x_{13}a\,\hat{\mathbf{z}}$ | $(24d)$ | O III |
| $\mathbf{B}_{216}$ | = | $\left(\frac{1}{2}-y_{13}\right)\mathbf{a}_1-z_{13}\,\mathbf{a}_2+\left(\frac{1}{2}+x_{13}\right)\mathbf{a}_3$ | = | $\left(\frac{1}{2}-y_{13}\right)a\,\hat{\mathbf{x}}-z_{13}a\,\hat{\mathbf{y}}+\left(\frac{1}{2}+x_{13}\right)a\,\hat{\mathbf{z}}$ | $(24d)$ | O III |
| $\mathbf{B}_{217}$ | = | $-x_{13}\,\mathbf{a}_1-y_{13}\,\mathbf{a}_2-z_{13}\,\mathbf{a}_3$ | = | $-x_{13}a\,\hat{\mathbf{x}}-y_{13}a\,\hat{\mathbf{y}}-z_{13}a\,\hat{\mathbf{z}}$ | $(24d)$ | O III |
| $\mathbf{B}_{218}$ | = | $\left(\frac{1}{2}+x_{13}\right)\mathbf{a}_1+y_{13}\,\mathbf{a}_2+\left(\frac{1}{2}-z_{13}\right)\mathbf{a}_3$ | = | $\left(\frac{1}{2}+x_{13}\right)a\,\hat{\mathbf{x}}+y_{13}a\,\hat{\mathbf{y}}+\left(\frac{1}{2}-z_{13}\right)a\,\hat{\mathbf{z}}$ | $(24d)$ | O III |
| $\mathbf{B}_{219}$ | = | $x_{13}\,\mathbf{a}_1+\left(\frac{1}{2}-y_{13}\right)\mathbf{a}_2+\left(\frac{1}{2}+z_{13}\right)\mathbf{a}_3$ | = | $x_{13}a\,\hat{\mathbf{x}}+\left(\frac{1}{2}-y_{13}\right)a\,\hat{\mathbf{y}}+\left(\frac{1}{2}+z_{13}\right)a\,\hat{\mathbf{z}}$ | $(24d)$ | O III |
| $\mathbf{B}_{220}$ | = | $\left(\frac{1}{2}-x_{13}\right)\mathbf{a}_1+\left(\frac{1}{2}+y_{13}\right)\mathbf{a}_2+z_{13}\,\mathbf{a}_3$ | = | $\left(\frac{1}{2}-x_{13}\right)a\,\hat{\mathbf{x}}+\left(\frac{1}{2}+y_{13}\right)a\,\hat{\mathbf{y}}+z_{13}a\,\hat{\mathbf{z}}$ | $(24d)$ | O III |
| $\mathbf{B}_{221}$ | = | $-z_{13}\,\mathbf{a}_1-x_{13}\,\mathbf{a}_2-y_{13}\,\mathbf{a}_3$ | = | $-z_{13}a\,\hat{\mathbf{x}}-x_{13}a\,\hat{\mathbf{y}}-y_{13}a\,\hat{\mathbf{z}}$ | $(24d)$ | O III |
| $\mathbf{B}_{222}$ | = | $\left(\frac{1}{2}-z_{13}\right)\mathbf{a}_1+\left(\frac{1}{2}+x_{13}\right)\mathbf{a}_2+y_{13}\,\mathbf{a}_3$ | = | $\left(\frac{1}{2}-z_{13}\right)a\,\hat{\mathbf{x}}+\left(\frac{1}{2}+x_{13}\right)a\,\hat{\mathbf{y}}+y_{13}a\,\hat{\mathbf{z}}$ | $(24d)$ | O III |
| $\mathbf{B}_{223}$ | = | $\left(\frac{1}{2}+z_{13}\right)\mathbf{a}_1+x_{13}\,\mathbf{a}_2+\left(\frac{1}{2}-y_{13}\right)\mathbf{a}_3$ | = | $\left(\frac{1}{2}+z_{13}\right)a\,\hat{\mathbf{x}}+x_{13}a\,\hat{\mathbf{y}}+\left(\frac{1}{2}-y_{13}\right)a\,\hat{\mathbf{z}}$ | $(24d)$ | O III |
| $\mathbf{B}_{224}$ | = | $z_{13}\,\mathbf{a}_1+\left(\frac{1}{2}-x_{13}\right)\mathbf{a}_2+\left(\frac{1}{2}+y_{13}\right)\mathbf{a}_3$ | = | $z_{13}a\,\hat{\mathbf{x}}+\left(\frac{1}{2}-x_{13}\right)a\,\hat{\mathbf{y}}+\left(\frac{1}{2}+y_{13}\right)a\,\hat{\mathbf{z}}$ | $(24d)$ | O III |
| $\mathbf{B}_{225}$ | = | $-y_{13}\,\mathbf{a}_1 - z_{13}\,\mathbf{a}_2 - x_{13}\,\mathbf{a}_3$ | = | $-y_{13}a\,\mathbf{\hat{x}} - z_{13}a\,\mathbf{\hat{y}} - x_{13}a\,\mathbf{\hat{z}}$ | (24*d*) | O III |
| $\mathbf{B}_{226}$ | = | $y_{13}\,\mathbf{a}_1 + \left(\frac{1}{2} - z_{13}\right)\,\mathbf{a}_2 + \left(\frac{1}{2} + x_{13}\right)\,\mathbf{a}_3$ | = | $y_{13}a\,\mathbf{\hat{x}} + \left(\frac{1}{2} - z_{13}\right)a\,\mathbf{\hat{y}} + \left(\frac{1}{2} + x_{13}\right)a\,\mathbf{\hat{z}}$ | (24*d*) | O III |
| $\mathbf{B}_{227}$ | = | $\left(\frac{1}{2} - y_{13}\right)\,\mathbf{a}_1 + \left(\frac{1}{2} + z_{13}\right)\,\mathbf{a}_2 + x_{13}\,\mathbf{a}_3$ | = | $\left(\frac{1}{2} - y_{13}\right)a\,\mathbf{\hat{x}} + \left(\frac{1}{2} + z_{13}\right)a\,\mathbf{\hat{y}} + x_{13}a\,\mathbf{\hat{z}}$ | (24*d*) | O III |
| $\mathbf{B}_{228}$ | = | $\left(\frac{1}{2} + y_{13}\right)\,\mathbf{a}_1 + z_{13}\,\mathbf{a}_2 + \left(\frac{1}{2} - x_{13}\right)\,\mathbf{a}_3$ | = | $\left(\frac{1}{2} + y_{13}\right)a\,\mathbf{\hat{x}} + z_{13}a\,\mathbf{\hat{y}} + \left(\frac{1}{2} - x_{13}\right)a\,\mathbf{\hat{z}}$ | (24*d*) | O III |
| $\mathbf{B}_{229}$ | = | $x_{14}\,\mathbf{a}_1 + y_{14}\,\mathbf{a}_2 + z_{14}\,\mathbf{a}_3$ | = | $x_{14}a\,\mathbf{\hat{x}} + y_{14}a\,\mathbf{\hat{y}} + z_{14}a\,\mathbf{\hat{z}}$ | (24*d*) | O IV |
| $\mathbf{B}_{230}$ | = | $\left(\frac{1}{2} - x_{14}\right)\,\mathbf{a}_1 - y_{14}\,\mathbf{a}_2 + \left(\frac{1}{2} + z_{14}\right)\,\mathbf{a}_3$ | = | $\left(\frac{1}{2} - x_{14}\right)a\,\mathbf{\hat{x}} - y_{14}a\,\mathbf{\hat{y}} + \left(\frac{1}{2} + z_{14}\right)a\,\mathbf{\hat{z}}$ | (24*d*) | O IV |
| $\mathbf{B}_{231}$ | = | $-x_{14}\,\mathbf{a}_1 + \left(\frac{1}{2} + y_{14}\right)\,\mathbf{a}_2 + \left(\frac{1}{2} - z_{14}\right)\,\mathbf{a}_3$ | = | $-x_{14}a\,\mathbf{\hat{x}} + \left(\frac{1}{2} + y_{14}\right)a\,\mathbf{\hat{y}} + \left(\frac{1}{2} - z_{14}\right)a\,\mathbf{\hat{z}}$ | (24*d*) | O IV |
| $\mathbf{B}_{232}$ | = | $\left(\frac{1}{2} + x_{14}\right)\,\mathbf{a}_1 + \left(\frac{1}{2} - y_{14}\right)\,\mathbf{a}_2 - z_{14}\,\mathbf{a}_3$ | = | $\left(\frac{1}{2} + x_{14}\right)a\,\mathbf{\hat{x}} + \left(\frac{1}{2} - y_{14}\right)a\,\mathbf{\hat{y}} - z_{14}a\,\mathbf{\hat{z}}$ | (24*d*) | O IV |
| $\mathbf{B}_{233}$ | = | $z_{14}\,\mathbf{a}_1 + x_{14}\,\mathbf{a}_2 + y_{14}\,\mathbf{a}_3$ | = | $z_{14}a\,\mathbf{\hat{x}} + x_{14}a\,\mathbf{\hat{y}} + y_{14}a\,\mathbf{\hat{z}}$ | (24*d*) | O IV |
| $\mathbf{B}_{234}$ | = | $\left(\frac{1}{2} + z_{14}\right)\,\mathbf{a}_1 + \left(\frac{1}{2} - x_{14}\right)\,\mathbf{a}_2 - y_{14}\,\mathbf{a}_3$ | = | $\left(\frac{1}{2} + z_{14}\right)a\,\mathbf{\hat{x}} + \left(\frac{1}{2} - x_{14}\right)a\,\mathbf{\hat{y}} - y_{14}a\,\mathbf{\hat{z}}$ | (24*d*) | O IV |
| $\mathbf{B}_{235}$ | = | $\left(\frac{1}{2} - z_{14}\right)\,\mathbf{a}_1 - x_{14}\,\mathbf{a}_2 + \left(\frac{1}{2} + y_{14}\right)\,\mathbf{a}_3$ | = | $\left(\frac{1}{2} - z_{14}\right)a\,\mathbf{\hat{x}} - x_{14}a\,\mathbf{\hat{y}} + \left(\frac{1}{2} + y_{14}\right)a\,\mathbf{\hat{z}}$ | (24*d*) | O IV |
| $\mathbf{B}_{236}$ | = | $-z_{14}\,\mathbf{a}_1 + \left(\frac{1}{2} + x_{14}\right)\,\mathbf{a}_2 + \left(\frac{1}{2} - y_{14}\right)\,\mathbf{a}_3$ | = | $-z_{14}a\,\mathbf{\hat{x}} + \left(\frac{1}{2} + x_{14}\right)a\,\mathbf{\hat{y}} + \left(\frac{1}{2} - y_{14}\right)a\,\mathbf{\hat{z}}$ | (24*d*) | O IV |
| $\mathbf{B}_{237}$ | = | $y_{14}\,\mathbf{a}_1 + z_{14}\,\mathbf{a}_2 + x_{14}\,\mathbf{a}_3$ | = | $y_{14}a\,\mathbf{\hat{x}} + z_{14}a\,\mathbf{\hat{y}} + x_{14}a\,\mathbf{\hat{z}}$ | (24*d*) | O IV |
| $\mathbf{B}_{238}$ | = | $-y_{14}\,\mathbf{a}_1 + \left(\frac{1}{2} + z_{14}\right)\,\mathbf{a}_2 + \left(\frac{1}{2} - x_{14}\right)\,\mathbf{a}_3$ | = | $-y_{14}a\,\mathbf{\hat{x}} + \left(\frac{1}{2} + z_{14}\right)a\,\mathbf{\hat{y}} + \left(\frac{1}{2} - x_{14}\right)a\,\mathbf{\hat{z}}$ | (24*d*) | O IV |
| $\mathbf{B}_{239}$ | = | $\left(\frac{1}{2} + y_{14}\right)\,\mathbf{a}_1 + \left(\frac{1}{2} - z_{14}\right)\,\mathbf{a}_2 - x_{14}\,\mathbf{a}_3$ | = | $\left(\frac{1}{2} + y_{14}\right)a\,\mathbf{\hat{x}} + \left(\frac{1}{2} - z_{14}\right)a\,\mathbf{\hat{y}} - x_{14}a\,\mathbf{\hat{z}}$ | (24*d*) | O IV |
| $\mathbf{B}_{240}$ | = | $\left(\frac{1}{2} - y_{14}\right)\,\mathbf{a}_1 - z_{14}\,\mathbf{a}_2 + \left(\frac{1}{2} + x_{14}\right)\,\mathbf{a}_3$ | = | $\left(\frac{1}{2} - y_{14}\right)a\,\mathbf{\hat{x}} - z_{14}a\,\mathbf{\hat{y}} + \left(\frac{1}{2} + x_{14}\right)a\,\mathbf{\hat{z}}$ | (24*d*) | O IV |
| $\mathbf{B}_{241}$ | = | $-x_{14}\,\mathbf{a}_1 - y_{14}\,\mathbf{a}_2 - z_{14}\,\mathbf{a}_3$ | = | $-x_{14}a\,\mathbf{\hat{x}} - y_{14}a\,\mathbf{\hat{y}} - z_{14}a\,\mathbf{\hat{z}}$ | (24*d*) | O IV |
| $\mathbf{B}_{242}$ | = | $\left(\frac{1}{2} + x_{14}\right)\,\mathbf{a}_1 + y_{14}\,\mathbf{a}_2 + \left(\frac{1}{2} - z_{14}\right)\,\mathbf{a}_3$ | = | $\left(\frac{1}{2} + x_{14}\right)a\,\mathbf{\hat{x}} + y_{14}a\,\mathbf{\hat{y}} + \left(\frac{1}{2} - z_{14}\right)a\,\mathbf{\hat{z}}$ | (24*d*) | O IV |
| $\mathbf{B}_{243}$ | = | $x_{14}\,\mathbf{a}_1 + \left(\frac{1}{2} - y_{14}\right)\,\mathbf{a}_2 + \left(\frac{1}{2} + z_{14}\right)\,\mathbf{a}_3$ | = | $x_{14}a\,\mathbf{\hat{x}} + \left(\frac{1}{2} - y_{14}\right)a\,\mathbf{\hat{y}} + \left(\frac{1}{2} + z_{14}\right)a\,\mathbf{\hat{z}}$ | (24*d*) | O IV |
| $\mathbf{B}_{244}$ | = | $\left(\frac{1}{2} - x_{14}\right)\,\mathbf{a}_1 + \left(\frac{1}{2} + y_{14}\right)\,\mathbf{a}_2 + z_{14}\,\mathbf{a}_3$ | = | $\left(\frac{1}{2} - x_{14}\right)a\,\mathbf{\hat{x}} + \left(\frac{1}{2} + y_{14}\right)a\,\mathbf{\hat{y}} + z_{14}a\,\mathbf{\hat{z}}$ | (24*d*) | O IV |
| $\mathbf{B}_{245}$ | = | $-z_{14}\,\mathbf{a}_1 - x_{14}\,\mathbf{a}_2 - y_{14}\,\mathbf{a}_3$ | = | $-z_{14}a\,\mathbf{\hat{x}} - x_{14}a\,\mathbf{\hat{y}} - y_{14}a\,\mathbf{\hat{z}}$ | (24*d*) | O IV |
| $\mathbf{B}_{246}$ | = | $\left(\frac{1}{2} - z_{14}\right)\,\mathbf{a}_1 + \left(\frac{1}{2} + x_{14}\right)\,\mathbf{a}_2 + y_{14}\,\mathbf{a}_3$ | = | $\left(\frac{1}{2} - z_{14}\right)a\,\mathbf{\hat{x}} + \left(\frac{1}{2} + x_{14}\right)a\,\mathbf{\hat{y}} + y_{14}a\,\mathbf{\hat{z}}$ | (24*d*) | O IV |
| $\mathbf{B}_{247}$ | = | $\left(\frac{1}{2} + z_{14}\right)\,\mathbf{a}_1 + x_{14}\,\mathbf{a}_2 + \left(\frac{1}{2} - y_{14}\right)\,\mathbf{a}_3$ | = | $\left(\frac{1}{2} + z_{14}\right)a\,\mathbf{\hat{x}} + x_{14}a\,\mathbf{\hat{y}} + \left(\frac{1}{2} - y_{14}\right)a\,\mathbf{\hat{z}}$ | (24*d*) | O IV |
| $\mathbf{B}_{248}$ | = | $z_{14}\,\mathbf{a}_1 + \left(\frac{1}{2} - x_{14}\right)\,\mathbf{a}_2 + \left(\frac{1}{2} + y_{14}\right)\,\mathbf{a}_3$ | = | $z_{14}a\,\mathbf{\hat{x}} + \left(\frac{1}{2} - x_{14}\right)a\,\mathbf{\hat{y}} + \left(\frac{1}{2} + y_{14}\right)a\,\mathbf{\hat{z}}$ | (24*d*) | O IV |
| $\mathbf{B}_{249}$ | = | $-y_{14}\,\mathbf{a}_1 - z_{14}\,\mathbf{a}_2 - x_{14}\,\mathbf{a}_3$ | = | $-y_{14}a\,\mathbf{\hat{x}} - z_{14}a\,\mathbf{\hat{y}} - x_{14}a\,\mathbf{\hat{z}}$ | (24*d*) | O IV |
| $\mathbf{B}_{250}$ | = | $y_{14}\,\mathbf{a}_1 + \left(\frac{1}{2} - z_{14}\right)\,\mathbf{a}_2 + \left(\frac{1}{2} + x_{14}\right)\,\mathbf{a}_3$ | = | $y_{14}a\,\mathbf{\hat{x}} + \left(\frac{1}{2} - z_{14}\right)a\,\mathbf{\hat{y}} + \left(\frac{1}{2} + x_{14}\right)a\,\mathbf{\hat{z}}$ | (24*d*) | O IV |
| $\mathbf{B}_{251}$ | = | $\left(\frac{1}{2} - y_{14}\right)\,\mathbf{a}_1 + \left(\frac{1}{2} + z_{14}\right)\,\mathbf{a}_2 + x_{14}\,\mathbf{a}_3$ | = | $\left(\frac{1}{2} - y_{14}\right)a\,\mathbf{\hat{x}} + \left(\frac{1}{2} + z_{14}\right)a\,\mathbf{\hat{y}} + x_{14}a\,\mathbf{\hat{z}}$ | (24*d*) | O IV |
| $\mathbf{B}_{252}$ | = | $\left(\frac{1}{2} + y_{14}\right)\,\mathbf{a}_1 + z_{14}\,\mathbf{a}_2 + \left(\frac{1}{2} - x_{14}\right)\,\mathbf{a}_3$ | = | $\left(\frac{1}{2} + y_{14}\right)a\,\mathbf{\hat{x}} + z_{14}a\,\mathbf{\hat{y}} + \left(\frac{1}{2} - x_{14}\right)a\,\mathbf{\hat{z}}$ | (24*d*) | O IV |

---

**Geometry files:**
- CIF: pp. 1788
- POSCAR: pp. 1788

# Maghemite ($\gamma$-$Fe_2O_3$, $D5_7$) Structure: A2B3_cP60_212_bcd_ace

http://aflow.org/prototype-encyclopedia/A2B3_cP60_212_bcd_ace

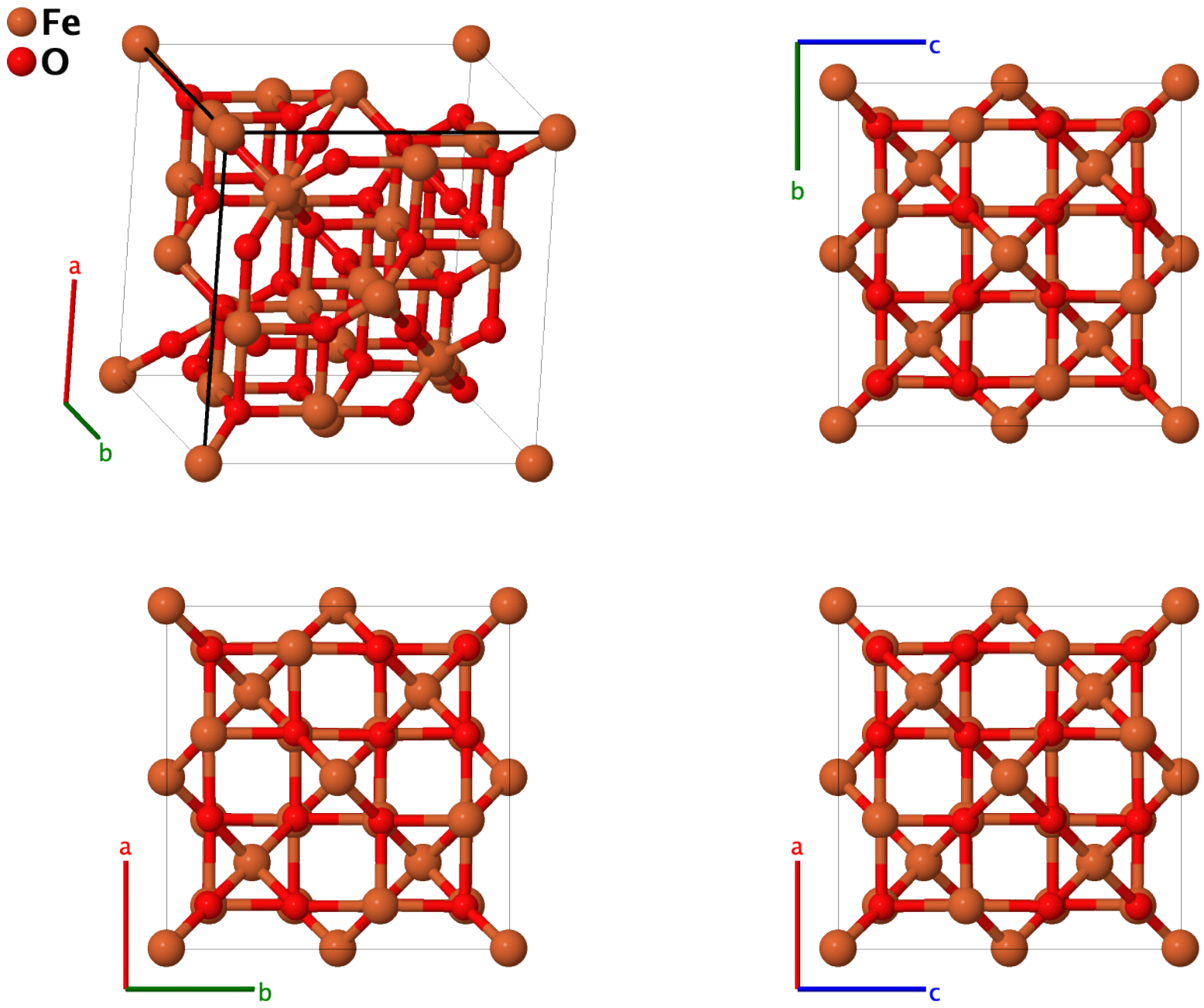

| | | |
|---|---|---|
| **Prototype** | : | $Fe_2O_3$ |
| **AFLOW prototype label** | : | A2B3_cP60_212_bcd_ace |
| ***Strukturbericht* designation** | : | $D5_7$ |
| **Pearson symbol** | : | cP60 |
| **Space group number** | : | 212 |
| **Space group symbol** | : | $P4_332$ |
| **AFLOW prototype command** | : | `aflow --proto=A2B3_cP60_212_bcd_ace --params=`$a, x_3, x_4, y_5, x_6, y_6, z_6$ |

**Other compounds with this structure**

- $\gamma$-$Al_2O_3$ ($\gamma$-corundum)

- (Hermann, 1937) gives $\gamma$-$Al_2O_3$ as the prototype for *Strukturbericht* $D5_7$, but states that the data for $\gamma$-$Fe_2O_3$ is more reliable and presents the data for the later compound, which we use as the prototype. This is a rock-salt ($B1$) structure with defects. This structure can also be expressed in the enantiomorphic space group $P4_132$ #213.

**Simple Cubic primitive vectors:**

$$\mathbf{a}_1 = a\,\hat{\mathbf{x}}$$
$$\mathbf{a}_2 = a\,\hat{\mathbf{y}}$$
$$\mathbf{a}_3 = a\,\hat{\mathbf{z}}$$

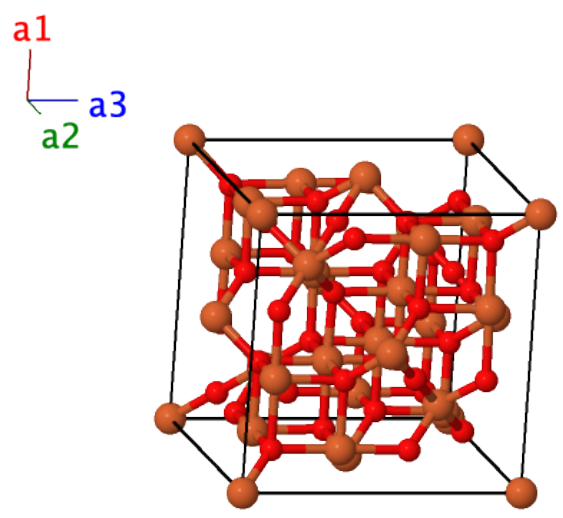

**Basis vectors:**

| | | Lattice Coordinates | | Cartesian Coordinates | Wyckoff Position | Atom Type |
|---|---|---|---|---|---|---|
| $\mathbf{B}_1$ | = | $\frac{1}{8}\,\mathbf{a}_1 + \frac{1}{8}\,\mathbf{a}_2 + \frac{1}{8}\,\mathbf{a}_3$ | = | $\frac{1}{8}a\,\hat{\mathbf{x}} + \frac{1}{8}a\,\hat{\mathbf{y}} + \frac{1}{8}a\,\hat{\mathbf{z}}$ | $(4a)$ | O I |
| $\mathbf{B}_2$ | = | $\frac{3}{8}\,\mathbf{a}_1 + \frac{7}{8}\,\mathbf{a}_2 + \frac{5}{8}\,\mathbf{a}_3$ | = | $\frac{3}{8}a\,\hat{\mathbf{x}} + \frac{7}{8}a\,\hat{\mathbf{y}} + \frac{5}{8}a\,\hat{\mathbf{z}}$ | $(4a)$ | O I |
| $\mathbf{B}_3$ | = | $\frac{7}{8}\,\mathbf{a}_1 + \frac{5}{8}\,\mathbf{a}_2 + \frac{3}{8}\,\mathbf{a}_3$ | = | $\frac{7}{8}a\,\hat{\mathbf{x}} + \frac{5}{8}a\,\hat{\mathbf{y}} + \frac{3}{8}a\,\hat{\mathbf{z}}$ | $(4a)$ | O I |
| $\mathbf{B}_4$ | = | $\frac{5}{8}\,\mathbf{a}_1 + \frac{3}{8}\,\mathbf{a}_2 + \frac{7}{8}\,\mathbf{a}_3$ | = | $\frac{5}{8}a\,\hat{\mathbf{x}} + \frac{3}{8}a\,\hat{\mathbf{y}} + \frac{7}{8}a\,\hat{\mathbf{z}}$ | $(4a)$ | O I |
| $\mathbf{B}_5$ | = | $\frac{5}{8}\,\mathbf{a}_1 + \frac{5}{8}\,\mathbf{a}_2 + \frac{5}{8}\,\mathbf{a}_3$ | = | $\frac{5}{8}a\,\hat{\mathbf{x}} + \frac{5}{8}a\,\hat{\mathbf{y}} + \frac{5}{8}a\,\hat{\mathbf{z}}$ | $(4b)$ | Fe I |
| $\mathbf{B}_6$ | = | $\frac{7}{8}\,\mathbf{a}_1 + \frac{3}{8}\,\mathbf{a}_2 + \frac{1}{8}\,\mathbf{a}_3$ | = | $\frac{7}{8}a\,\hat{\mathbf{x}} + \frac{3}{8}a\,\hat{\mathbf{y}} + \frac{1}{8}a\,\hat{\mathbf{z}}$ | $(4b)$ | Fe I |
| $\mathbf{B}_7$ | = | $\frac{3}{8}\,\mathbf{a}_1 + \frac{1}{8}\,\mathbf{a}_2 + \frac{7}{8}\,\mathbf{a}_3$ | = | $\frac{3}{8}a\,\hat{\mathbf{x}} + \frac{1}{8}a\,\hat{\mathbf{y}} + \frac{7}{8}a\,\hat{\mathbf{z}}$ | $(4b)$ | Fe I |
| $\mathbf{B}_8$ | = | $\frac{1}{8}\,\mathbf{a}_1 + \frac{7}{8}\,\mathbf{a}_2 + \frac{3}{8}\,\mathbf{a}_3$ | = | $\frac{1}{8}a\,\hat{\mathbf{x}} + \frac{7}{8}a\,\hat{\mathbf{y}} + \frac{3}{8}a\,\hat{\mathbf{z}}$ | $(4b)$ | Fe I |
| $\mathbf{B}_9$ | = | $x_3\,\mathbf{a}_1 + x_3\,\mathbf{a}_2 + x_3\,\mathbf{a}_3$ | = | $x_3a\,\hat{\mathbf{x}} + x_3a\,\hat{\mathbf{y}} + x_3a\,\hat{\mathbf{z}}$ | $(8c)$ | Fe II |
| $\mathbf{B}_{10}$ | = | $\left(\frac{1}{2} - x_3\right)\mathbf{a}_1 - x_3\,\mathbf{a}_2 + \left(\frac{1}{2} + x_3\right)\mathbf{a}_3$ | = | $\left(\frac{1}{2} - x_3\right)a\,\hat{\mathbf{x}} - x_3a\,\hat{\mathbf{y}} + \left(\frac{1}{2} + x_3\right)a\,\hat{\mathbf{z}}$ | $(8c)$ | Fe II |
| $\mathbf{B}_{11}$ | = | $-x_3\,\mathbf{a}_1 + \left(\frac{1}{2} + x_3\right)\mathbf{a}_2 + \left(\frac{1}{2} - x_3\right)\mathbf{a}_3$ | = | $-x_3a\,\hat{\mathbf{x}} + \left(\frac{1}{2} + x_3\right)a\,\hat{\mathbf{y}} + \left(\frac{1}{2} - x_3\right)a\,\hat{\mathbf{z}}$ | $(8c)$ | Fe II |
| $\mathbf{B}_{12}$ | = | $\left(\frac{1}{2} + x_3\right)\mathbf{a}_1 + \left(\frac{1}{2} - x_3\right)\mathbf{a}_2 - x_3\,\mathbf{a}_3$ | = | $\left(\frac{1}{2} + x_3\right)a\,\hat{\mathbf{x}} + \left(\frac{1}{2} - x_3\right)a\,\hat{\mathbf{y}} - x_3a\,\hat{\mathbf{z}}$ | $(8c)$ | Fe II |
| $\mathbf{B}_{13}$ | = | $\left(\frac{1}{4} + x_3\right)\mathbf{a}_1 + \left(\frac{3}{4} + x_3\right)\mathbf{a}_2 + \left(\frac{3}{4} - x_3\right)\mathbf{a}_3$ | = | $\left(\frac{1}{4} + x_3\right)a\,\hat{\mathbf{x}} + \left(\frac{3}{4} + x_3\right)a\,\hat{\mathbf{y}} + \left(\frac{3}{4} - x_3\right)a\,\hat{\mathbf{z}}$ | $(8c)$ | Fe II |
| $\mathbf{B}_{14}$ | = | $\left(\frac{1}{4} - x_3\right)\mathbf{a}_1 + \left(\frac{1}{4} - x_3\right)\mathbf{a}_2 + \left(\frac{1}{4} - x_3\right)\mathbf{a}_3$ | = | $\left(\frac{1}{4} - x_3\right)a\,\hat{\mathbf{x}} + \left(\frac{1}{4} - x_3\right)a\,\hat{\mathbf{y}} + \left(\frac{1}{4} - x_3\right)a\,\hat{\mathbf{z}}$ | $(8c)$ | Fe II |
| $\mathbf{B}_{15}$ | = | $\left(\frac{3}{4} + x_3\right)\mathbf{a}_1 + \left(\frac{3}{4} - x_3\right)\mathbf{a}_2 + \left(\frac{1}{4} + x_3\right)\mathbf{a}_3$ | = | $\left(\frac{3}{4} + x_3\right)a\,\hat{\mathbf{x}} + \left(\frac{3}{4} - x_3\right)a\,\hat{\mathbf{y}} + \left(\frac{1}{4} + x_3\right)a\,\hat{\mathbf{z}}$ | $(8c)$ | Fe II |
| $\mathbf{B}_{16}$ | = | $\left(\frac{3}{4} - x_3\right)\mathbf{a}_1 + \left(\frac{1}{4} + x_3\right)\mathbf{a}_2 + \left(\frac{3}{4} + x_3\right)\mathbf{a}_3$ | = | $\left(\frac{3}{4} - x_3\right)a\,\hat{\mathbf{x}} + \left(\frac{1}{4} + x_3\right)a\,\hat{\mathbf{y}} + \left(\frac{3}{4} + x_3\right)a\,\hat{\mathbf{z}}$ | $(8c)$ | Fe II |
| $\mathbf{B}_{17}$ | = | $x_4\,\mathbf{a}_1 + x_4\,\mathbf{a}_2 + x_4\,\mathbf{a}_3$ | = | $x_4a\,\hat{\mathbf{x}} + x_4a\,\hat{\mathbf{y}} + x_4a\,\hat{\mathbf{z}}$ | $(8c)$ | O II |
| $\mathbf{B}_{18}$ | = | $\left(\frac{1}{2} - x_4\right)\mathbf{a}_1 - x_4\,\mathbf{a}_2 + \left(\frac{1}{2} + x_4\right)\mathbf{a}_3$ | = | $\left(\frac{1}{2} - x_4\right)a\,\hat{\mathbf{x}} - x_4a\,\hat{\mathbf{y}} + \left(\frac{1}{2} + x_4\right)a\,\hat{\mathbf{z}}$ | $(8c)$ | O II |
| $\mathbf{B}_{19}$ | = | $-x_4\,\mathbf{a}_1 + \left(\frac{1}{2} + x_4\right)\mathbf{a}_2 + \left(\frac{1}{2} - x_4\right)\mathbf{a}_3$ | = | $-x_4a\,\hat{\mathbf{x}} + \left(\frac{1}{2} + x_4\right)a\,\hat{\mathbf{y}} + \left(\frac{1}{2} - x_4\right)a\,\hat{\mathbf{z}}$ | $(8c)$ | O II |
| $\mathbf{B}_{20}$ | = | $\left(\frac{1}{2} + x_4\right)\mathbf{a}_1 + \left(\frac{1}{2} - x_4\right)\mathbf{a}_2 - x_4\,\mathbf{a}_3$ | = | $\left(\frac{1}{2} + x_4\right)a\,\hat{\mathbf{x}} + \left(\frac{1}{2} - x_4\right)a\,\hat{\mathbf{y}} - x_4a\,\hat{\mathbf{z}}$ | $(8c)$ | O II |
| $\mathbf{B}_{21}$ | = | $\left(\frac{1}{4} + x_4\right)\mathbf{a}_1 + \left(\frac{3}{4} + x_4\right)\mathbf{a}_2 + \left(\frac{3}{4} - x_4\right)\mathbf{a}_3$ | = | $\left(\frac{1}{4} + x_4\right)a\,\hat{\mathbf{x}} + \left(\frac{3}{4} + x_4\right)a\,\hat{\mathbf{y}} + \left(\frac{3}{4} - x_4\right)a\,\hat{\mathbf{z}}$ | $(8c)$ | O II |
| $\mathbf{B}_{22}$ | = | $\left(\frac{1}{4} - x_4\right)\mathbf{a}_1 + \left(\frac{1}{4} - x_4\right)\mathbf{a}_2 + \left(\frac{1}{4} - x_4\right)\mathbf{a}_3$ | = | $\left(\frac{1}{4} - x_4\right)a\,\hat{\mathbf{x}} + \left(\frac{1}{4} - x_4\right)a\,\hat{\mathbf{y}} + \left(\frac{1}{4} - x_4\right)a\,\hat{\mathbf{z}}$ | $(8c)$ | O II |
| $\mathbf{B}_{23}$ | = | $\left(\frac{3}{4} + x_4\right)\mathbf{a}_1 + \left(\frac{3}{4} - x_4\right)\mathbf{a}_2 + \left(\frac{1}{4} + x_4\right)\mathbf{a}_3$ | = | $\left(\frac{3}{4} + x_4\right)a\,\hat{\mathbf{x}} + \left(\frac{3}{4} - x_4\right)a\,\hat{\mathbf{y}} + \left(\frac{1}{4} + x_4\right)a\,\hat{\mathbf{z}}$ | $(8c)$ | O II |
| $\mathbf{B}_{24}$ | = | $\left(\frac{3}{4} - x_4\right)\mathbf{a}_1 + \left(\frac{1}{4} + x_4\right)\mathbf{a}_2 + \left(\frac{3}{4} + x_4\right)\mathbf{a}_3$ | = | $\left(\frac{3}{4} - x_4\right)a\,\hat{\mathbf{x}} + \left(\frac{1}{4} + x_4\right)a\,\hat{\mathbf{y}} + \left(\frac{3}{4} + x_4\right)a\,\hat{\mathbf{z}}$ | $(8c)$ | O II |
| $\mathbf{B}_{25}$ | = | $\frac{1}{8}\,\mathbf{a}_1 + y_5\,\mathbf{a}_2 + \left(\frac{1}{4} - y_5\right)\mathbf{a}_3$ | = | $\frac{1}{8}a\,\hat{\mathbf{x}} + y_5a\,\hat{\mathbf{y}} + \left(\frac{1}{4} - y_5\right)a\,\hat{\mathbf{z}}$ | $(12d)$ | Fe III |
| $\mathbf{B}_{26}$ | = | $\frac{3}{8}\,\mathbf{a}_1 - y_5\,\mathbf{a}_2 + \left(\frac{3}{4} - y_5\right)\mathbf{a}_3$ | = | $\frac{3}{8}a\,\hat{\mathbf{x}} - y_5a\,\hat{\mathbf{y}} + \left(\frac{3}{4} - y_5\right)a\,\hat{\mathbf{z}}$ | $(12d)$ | Fe III |
| $\mathbf{B}_{27}$ | = | $\frac{7}{8}\,\mathbf{a}_1 + \left(\frac{1}{2} + y_5\right)\mathbf{a}_2 + \left(\frac{1}{4} + y_5\right)\mathbf{a}_3$ | = | $\frac{7}{8}a\,\hat{\mathbf{x}} + \left(\frac{1}{2} + y_5\right)a\,\hat{\mathbf{y}} + \left(\frac{1}{4} + y_5\right)a\,\hat{\mathbf{z}}$ | $(12d)$ | Fe III |
| $\mathbf{B}_{28}$ | = | $\frac{5}{8}\,\mathbf{a}_1 + \left(\frac{1}{2} - y_5\right)\,\mathbf{a}_2 + \left(\frac{3}{4} + y_5\right)\,\mathbf{a}_3$ | = | $\frac{5}{8}a\,\hat{\mathbf{x}} + \left(\frac{1}{2} - y_5\right)a\,\hat{\mathbf{y}} + \left(\frac{3}{4} + y_5\right)a\,\hat{\mathbf{z}}$ | (12$d$) | Fe III |
| $\mathbf{B}_{29}$ | = | $\left(\frac{1}{4} - y_5\right)\,\mathbf{a}_1 + \frac{1}{8}\,\mathbf{a}_2 + y_5\,\mathbf{a}_3$ | = | $\left(\frac{1}{4} - y_5\right)a\,\hat{\mathbf{x}} + \frac{1}{8}a\,\hat{\mathbf{y}} + y_5a\,\hat{\mathbf{z}}$ | (12$d$) | Fe III |
| $\mathbf{B}_{30}$ | = | $\left(\frac{3}{4} - y_5\right)\,\mathbf{a}_1 + \frac{3}{8}\,\mathbf{a}_2 - y_5\,\mathbf{a}_3$ | = | $\left(\frac{3}{4} - y_5\right)a\,\hat{\mathbf{x}} + \frac{3}{8}a\,\hat{\mathbf{y}} - y_5a\,\hat{\mathbf{z}}$ | (12$d$) | Fe III |
| $\mathbf{B}_{31}$ | = | $\left(\frac{1}{4} + y_5\right)\,\mathbf{a}_1 + \frac{7}{8}\,\mathbf{a}_2 + \left(\frac{1}{2} + y_5\right)\,\mathbf{a}_3$ | = | $\left(\frac{1}{4} + y_5\right)a\,\hat{\mathbf{x}} + \frac{7}{8}a\,\hat{\mathbf{y}} + \left(\frac{1}{2} + y_5\right)a\,\hat{\mathbf{z}}$ | (12$d$) | Fe III |
| $\mathbf{B}_{32}$ | = | $\left(\frac{3}{4} + y_5\right)\,\mathbf{a}_1 + \frac{5}{8}\,\mathbf{a}_2 + \left(\frac{1}{2} - y_5\right)\,\mathbf{a}_3$ | = | $\left(\frac{3}{4} + y_5\right)a\,\hat{\mathbf{x}} + \frac{5}{8}a\,\hat{\mathbf{y}} + \left(\frac{1}{2} - y_5\right)a\,\hat{\mathbf{z}}$ | (12$d$) | Fe III |
| $\mathbf{B}_{33}$ | = | $y_5\,\mathbf{a}_1 + \left(\frac{1}{4} - y_5\right)\,\mathbf{a}_2 + \frac{1}{8}\,\mathbf{a}_3$ | = | $y_5a\,\hat{\mathbf{x}} + \left(\frac{1}{4} - y_5\right)a\,\hat{\mathbf{y}} + \frac{1}{8}a\,\hat{\mathbf{z}}$ | (12$d$) | Fe III |
| $\mathbf{B}_{34}$ | = | $-y_5\,\mathbf{a}_1 + \left(\frac{3}{4} - y_5\right)\,\mathbf{a}_2 + \frac{3}{8}\,\mathbf{a}_3$ | = | $-y_5a\,\hat{\mathbf{x}} + \left(\frac{3}{4} - y_5\right)a\,\hat{\mathbf{y}} + \frac{3}{8}a\,\hat{\mathbf{z}}$ | (12$d$) | Fe III |
| $\mathbf{B}_{35}$ | = | $\left(\frac{1}{2} + y_5\right)\,\mathbf{a}_1 + \left(\frac{1}{4} + y_5\right)\,\mathbf{a}_2 + \frac{7}{8}\,\mathbf{a}_3$ | = | $\left(\frac{1}{2} + y_5\right)a\,\hat{\mathbf{x}} + \left(\frac{1}{4} + y_5\right)a\,\hat{\mathbf{y}} + \frac{7}{8}a\,\hat{\mathbf{z}}$ | (12$d$) | Fe III |
| $\mathbf{B}_{36}$ | = | $\left(\frac{1}{2} - y_5\right)\,\mathbf{a}_1 + \left(\frac{3}{4} + y_5\right)\,\mathbf{a}_2 + \frac{5}{8}\,\mathbf{a}_3$ | = | $\left(\frac{1}{2} - y_5\right)a\,\hat{\mathbf{x}} + \left(\frac{3}{4} + y_5\right)a\,\hat{\mathbf{y}} + \frac{5}{8}a\,\hat{\mathbf{z}}$ | (12$d$) | Fe III |
| $\mathbf{B}_{37}$ | = | $x_6\,\mathbf{a}_1 + y_6\,\mathbf{a}_2 + z_6\,\mathbf{a}_3$ | = | $x_6a\,\hat{\mathbf{x}} + y_6a\,\hat{\mathbf{y}} + z_6a\,\hat{\mathbf{z}}$ | (24$e$) | O III |
| $\mathbf{B}_{38}$ | = | $\left(\frac{1}{2} - x_6\right)\,\mathbf{a}_1 - y_6\,\mathbf{a}_2 + \left(\frac{1}{2} + z_6\right)\,\mathbf{a}_3$ | = | $\left(\frac{1}{2} - x_6\right)a\,\hat{\mathbf{x}} - y_6a\,\hat{\mathbf{y}} + \left(\frac{1}{2} + z_6\right)a\,\hat{\mathbf{z}}$ | (24$e$) | O III |
| $\mathbf{B}_{39}$ | = | $-x_6\,\mathbf{a}_1 + \left(\frac{1}{2} + y_6\right)\,\mathbf{a}_2 + \left(\frac{1}{2} - z_6\right)\,\mathbf{a}_3$ | = | $-x_6a\,\hat{\mathbf{x}} + \left(\frac{1}{2} + y_6\right)a\,\hat{\mathbf{y}} + \left(\frac{1}{2} - z_6\right)a\,\hat{\mathbf{z}}$ | (24$e$) | O III |
| $\mathbf{B}_{40}$ | = | $\left(\frac{1}{2} + x_6\right)\,\mathbf{a}_1 + \left(\frac{1}{2} - y_6\right)\,\mathbf{a}_2 - z_6\,\mathbf{a}_3$ | = | $\left(\frac{1}{2} + x_6\right)a\,\hat{\mathbf{x}} + \left(\frac{1}{2} - y_6\right)a\,\hat{\mathbf{y}} - z_6a\,\hat{\mathbf{z}}$ | (24$e$) | O III |
| $\mathbf{B}_{41}$ | = | $z_6\,\mathbf{a}_1 + x_6\,\mathbf{a}_2 + y_6\,\mathbf{a}_3$ | = | $z_6a\,\hat{\mathbf{x}} + x_6a\,\hat{\mathbf{y}} + y_6a\,\hat{\mathbf{z}}$ | (24$e$) | O III |
| $\mathbf{B}_{42}$ | = | $\left(\frac{1}{2} + z_6\right)\,\mathbf{a}_1 + \left(\frac{1}{2} - x_6\right)\,\mathbf{a}_2 - y_6\,\mathbf{a}_3$ | = | $\left(\frac{1}{2} + z_6\right)a\,\hat{\mathbf{x}} + \left(\frac{1}{2} - x_6\right)a\,\hat{\mathbf{y}} - y_6a\,\hat{\mathbf{z}}$ | (24$e$) | O III |
| $\mathbf{B}_{43}$ | = | $\left(\frac{1}{2} - z_6\right)\,\mathbf{a}_1 - x_6\,\mathbf{a}_2 + \left(\frac{1}{2} + y_6\right)\,\mathbf{a}_3$ | = | $\left(\frac{1}{2} - z_6\right)a\,\hat{\mathbf{x}} - x_6a\,\hat{\mathbf{y}} + \left(\frac{1}{2} + y_6\right)a\,\hat{\mathbf{z}}$ | (24$e$) | O III |
| $\mathbf{B}_{44}$ | = | $-z_6\,\mathbf{a}_1 + \left(\frac{1}{2} + x_6\right)\,\mathbf{a}_2 + \left(\frac{1}{2} - y_6\right)\,\mathbf{a}_3$ | = | $-z_6a\,\hat{\mathbf{x}} + \left(\frac{1}{2} + x_6\right)a\,\hat{\mathbf{y}} + \left(\frac{1}{2} - y_6\right)a\,\hat{\mathbf{z}}$ | (24$e$) | O III |
| $\mathbf{B}_{45}$ | = | $y_6\,\mathbf{a}_1 + z_6\,\mathbf{a}_2 + x_6\,\mathbf{a}_3$ | = | $y_6a\,\hat{\mathbf{x}} + z_6a\,\hat{\mathbf{y}} + x_6a\,\hat{\mathbf{z}}$ | (24$e$) | O III |
| $\mathbf{B}_{46}$ | = | $-y_6\,\mathbf{a}_1 + \left(\frac{1}{2} + z_6\right)\,\mathbf{a}_2 + \left(\frac{1}{2} - x_6\right)\,\mathbf{a}_3$ | = | $-y_6a\,\hat{\mathbf{x}} + \left(\frac{1}{2} + z_6\right)a\,\hat{\mathbf{y}} + \left(\frac{1}{2} - x_6\right)a\,\hat{\mathbf{z}}$ | (24$e$) | O III |
| $\mathbf{B}_{47}$ | = | $\left(\frac{1}{2} + y_6\right)\,\mathbf{a}_1 + \left(\frac{1}{2} - z_6\right)\,\mathbf{a}_2 - x_6\,\mathbf{a}_3$ | = | $\left(\frac{1}{2} + y_6\right)a\,\hat{\mathbf{x}} + \left(\frac{1}{2} - z_6\right)a\,\hat{\mathbf{y}} - x_6a\,\hat{\mathbf{z}}$ | (24$e$) | O III |
| $\mathbf{B}_{48}$ | = | $\left(\frac{1}{2} - y_6\right)\,\mathbf{a}_1 - z_6\,\mathbf{a}_2 + \left(\frac{1}{2} + x_6\right)\,\mathbf{a}_3$ | = | $\left(\frac{1}{2} - y_6\right)a\,\hat{\mathbf{x}} - z_6a\,\hat{\mathbf{y}} + \left(\frac{1}{2} + x_6\right)a\,\hat{\mathbf{z}}$ | (24$e$) | O III |
| $\mathbf{B}_{49}$ | = | $\left(\frac{1}{4} + y_6\right)\,\mathbf{a}_1 + \left(\frac{3}{4} + x_6\right)\,\mathbf{a}_2 + \left(\frac{3}{4} - z_6\right)\,\mathbf{a}_3$ | = | $\left(\frac{1}{4} + y_6\right)a\,\hat{\mathbf{x}} + \left(\frac{3}{4} + x_6\right)a\,\hat{\mathbf{y}} + \left(\frac{3}{4} - z_6\right)a\,\hat{\mathbf{z}}$ | (24$e$) | O III |
| $\mathbf{B}_{50}$ | = | $\left(\frac{1}{4} - y_6\right)\,\mathbf{a}_1 + \left(\frac{1}{4} - x_6\right)\,\mathbf{a}_2 + \left(\frac{1}{4} - z_6\right)\,\mathbf{a}_3$ | = | $\left(\frac{1}{4} - y_6\right)a\,\hat{\mathbf{x}} + \left(\frac{1}{4} - x_6\right)a\,\hat{\mathbf{y}} + \left(\frac{1}{4} - z_6\right)a\,\hat{\mathbf{z}}$ | (24$e$) | O III |
| $\mathbf{B}_{51}$ | = | $\left(\frac{3}{4} + y_6\right)\,\mathbf{a}_1 + \left(\frac{3}{4} - x_6\right)\,\mathbf{a}_2 + \left(\frac{1}{4} + z_6\right)\,\mathbf{a}_3$ | = | $\left(\frac{3}{4} + y_6\right)a\,\hat{\mathbf{x}} + \left(\frac{3}{4} - x_6\right)a\,\hat{\mathbf{y}} + \left(\frac{1}{4} + z_6\right)a\,\hat{\mathbf{z}}$ | (24$e$) | O III |
| $\mathbf{B}_{52}$ | = | $\left(\frac{3}{4} - y_6\right)\,\mathbf{a}_1 + \left(\frac{1}{4} + x_6\right)\,\mathbf{a}_2 + \left(\frac{3}{4} + z_6\right)\,\mathbf{a}_3$ | = | $\left(\frac{3}{4} - y_6\right)a\,\hat{\mathbf{x}} + \left(\frac{1}{4} + x_6\right)a\,\hat{\mathbf{y}} + \left(\frac{3}{4} + z_6\right)a\,\hat{\mathbf{z}}$ | (24$e$) | O III |
| $\mathbf{B}_{53}$ | = | $\left(\frac{1}{4} + x_6\right)\,\mathbf{a}_1 + \left(\frac{3}{4} + z_6\right)\,\mathbf{a}_2 + \left(\frac{3}{4} - y_6\right)\,\mathbf{a}_3$ | = | $\left(\frac{1}{4} + x_6\right)a\,\hat{\mathbf{x}} + \left(\frac{3}{4} + z_6\right)a\,\hat{\mathbf{y}} + \left(\frac{3}{4} - y_6\right)a\,\hat{\mathbf{z}}$ | (24$e$) | O III |
| $\mathbf{B}_{54}$ | = | $\left(\frac{3}{4} - x_6\right)\,\mathbf{a}_1 + \left(\frac{1}{4} + z_6\right)\,\mathbf{a}_2 + \left(\frac{3}{4} + y_6\right)\,\mathbf{a}_3$ | = | $\left(\frac{3}{4} - x_6\right)a\,\hat{\mathbf{x}} + \left(\frac{1}{4} + z_6\right)a\,\hat{\mathbf{y}} + \left(\frac{3}{4} + y_6\right)a\,\hat{\mathbf{z}}$ | (24$e$) | O III |
| $\mathbf{B}_{55}$ | = | $\left(\frac{1}{4} - x_6\right)\,\mathbf{a}_1 + \left(\frac{1}{4} - z_6\right)\,\mathbf{a}_2 + \left(\frac{1}{4} - y_6\right)\,\mathbf{a}_3$ | = | $\left(\frac{1}{4} - x_6\right)a\,\hat{\mathbf{x}} + \left(\frac{1}{4} - z_6\right)a\,\hat{\mathbf{y}} + \left(\frac{1}{4} - y_6\right)a\,\hat{\mathbf{z}}$ | (24$e$) | O III |
| $\mathbf{B}_{56}$ | = | $\left(\frac{3}{4} + x_6\right)\,\mathbf{a}_1 + \left(\frac{3}{4} - z_6\right)\,\mathbf{a}_2 + \left(\frac{1}{4} + y_6\right)\,\mathbf{a}_3$ | = | $\left(\frac{3}{4} + x_6\right)a\,\hat{\mathbf{x}} + \left(\frac{3}{4} - z_6\right)a\,\hat{\mathbf{y}} + \left(\frac{1}{4} + y_6\right)a\,\hat{\mathbf{z}}$ | (24$e$) | O III |
| $\mathbf{B}_{57}$ | = | $\left(\frac{1}{4}+z_6\right)\mathbf{a}_1+\left(\frac{3}{4}+y_6\right)\mathbf{a}_2+\left(\frac{3}{4}-x_6\right)\mathbf{a}_3$ | = | $\left(\frac{1}{4}+z_6\right)a\,\hat{\mathbf{x}}+\left(\frac{3}{4}+y_6\right)a\,\hat{\mathbf{y}}+\left(\frac{3}{4}-x_6\right)a\,\hat{\mathbf{z}}$ | (24$e$) | O III |
| $\mathbf{B}_{58}$ | = | $\left(\frac{3}{4}+z_6\right)\mathbf{a}_1+\left(\frac{3}{4}-y_6\right)\mathbf{a}_2+\left(\frac{1}{4}+x_6\right)\mathbf{a}_3$ | = | $\left(\frac{3}{4}+z_6\right)a\,\hat{\mathbf{x}}+\left(\frac{3}{4}-y_6\right)a\,\hat{\mathbf{y}}+\left(\frac{1}{4}+x_6\right)a\,\hat{\mathbf{z}}$ | (24$e$) | O III |
| $\mathbf{B}_{59}$ | = | $\left(\frac{3}{4}-z_6\right)\mathbf{a}_1+\left(\frac{1}{4}+y_6\right)\mathbf{a}_2+\left(\frac{3}{4}+x_6\right)\mathbf{a}_3$ | = | $\left(\frac{3}{4}-z_6\right)a\,\hat{\mathbf{x}}+\left(\frac{1}{4}+y_6\right)a\,\hat{\mathbf{y}}+\left(\frac{3}{4}+x_6\right)a\,\hat{\mathbf{z}}$ | (24$e$) | O III |
| $\mathbf{B}_{60}$ | = | $\left(\frac{1}{4}-z_6\right)\mathbf{a}_1+\left(\frac{1}{4}-y_6\right)\mathbf{a}_2+\left(\frac{1}{4}-x_6\right)\mathbf{a}_3$ | = | $\left(\frac{1}{4}-z_6\right)a\,\hat{\mathbf{x}}+\left(\frac{1}{4}-y_6\right)a\,\hat{\mathbf{y}}+\left(\frac{1}{4}-x_6\right)a\,\hat{\mathbf{z}}$ | (24$e$) | O III |

---

**Found in:**

---

**Geometry files:**

- CIF: pp. 1790
- POSCAR: pp. 1790

# $Al_2Mo_3C$ Structure: A2BC3_cP24_213_c_a_d

http://aflow.org/prototype-encyclopedia/A2BC3_cP24_213_c_a_d

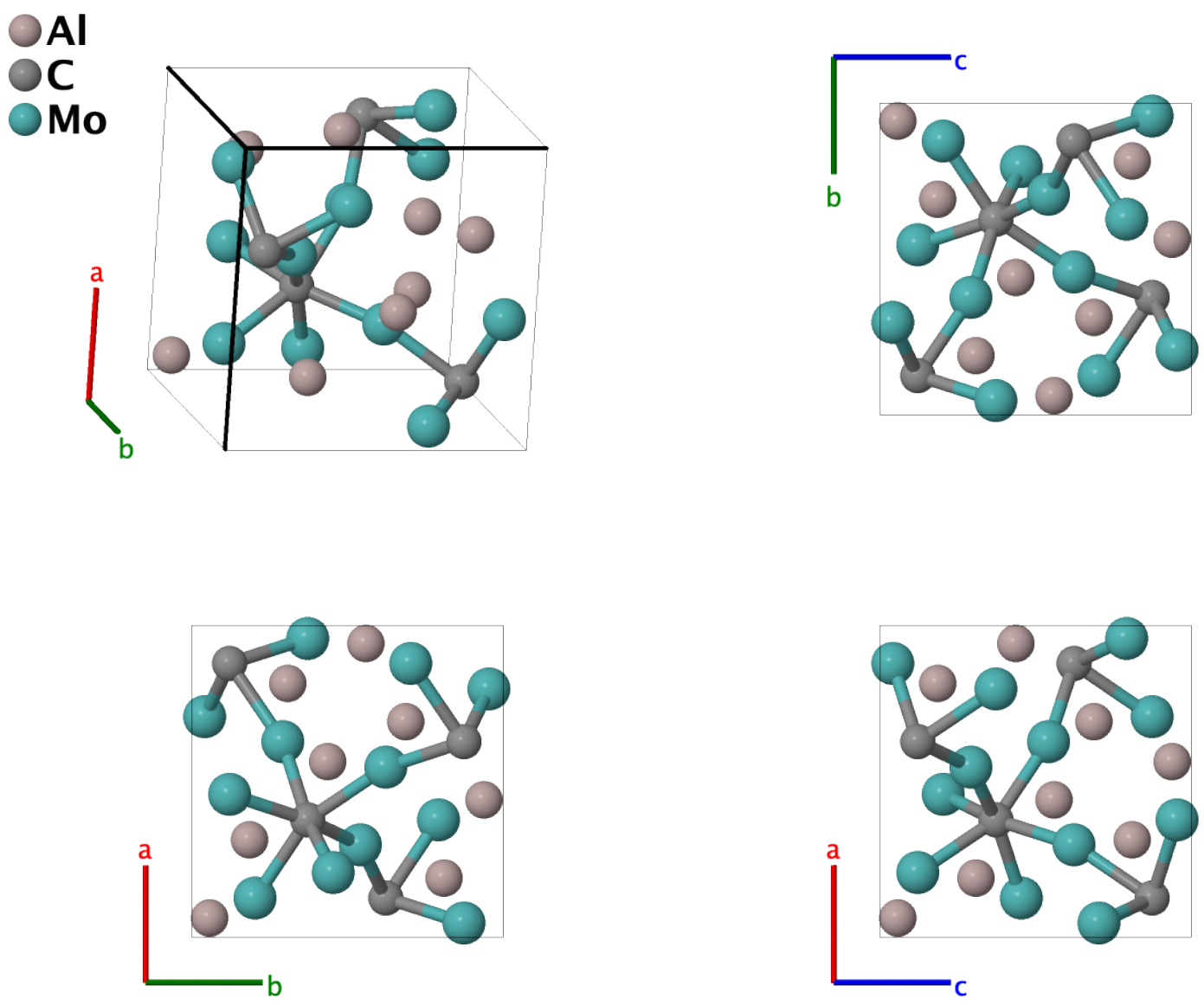

| | | |
|---|---|---|
| **Prototype** | : | $Al_2CMo_3$ |
| **AFLOW prototype label** | : | A2BC3_cP24_213_c_a_d |
| ***Strukturbericht* designation** | : | None |
| **Pearson symbol** | : | cP24 |
| **Space group number** | : | 213 |
| **Space group symbol** | : | $P4_132$ |
| **AFLOW prototype command** | : | `aflow --proto=A2BC3_cP24_213_c_a_d` <br> `--params=`$a, x_2, y_3$ |

**Other compounds with this structure**

- $Ag_2Pd_3Sn$, $Al_2Nb_3C$, $Al_2Nb_3N$, $Al_2Ta_3C$, $Cr_2Re_3B$, $Li_2Pd_3B$, $Li_2Pt_3B$, $Mn_2Rh_3P$, $Ni_2W_3N$, $Rh_2Mo_3N$, and $(Fe_{2-x}Rh_x)Mo_3N$

- This is a "filled" $\beta$-Mn ($A13$) structure, with the aluminum and molybdenum atoms almost exactly on the sites of the manganese atoms in $A13$.

**Simple Cubic primitive vectors:**

$$\mathbf{a}_1 = a\,\hat{\mathbf{x}}$$
$$\mathbf{a}_2 = a\,\hat{\mathbf{y}}$$
$$\mathbf{a}_3 = a\,\hat{\mathbf{z}}$$

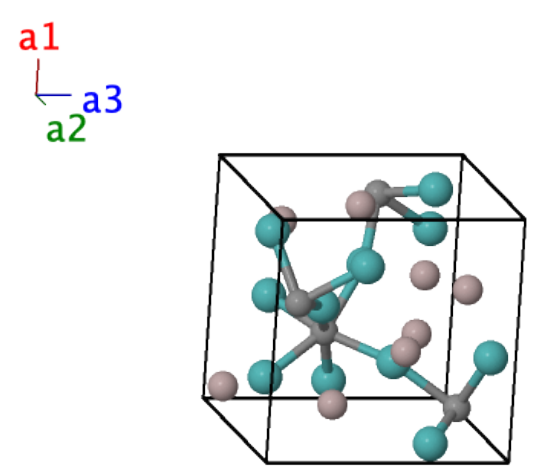

**Basis vectors:**

| | | Lattice Coordinates | | Cartesian Coordinates | Wyckoff Position | Atom Type |
|---|---|---|---|---|---|---|
| $\mathbf{B}_1$ | = | $\frac{3}{8}\,\mathbf{a}_1 + \frac{3}{8}\,\mathbf{a}_2 + \frac{3}{8}\,\mathbf{a}_3$ | = | $\frac{3}{8}a\,\hat{\mathbf{x}} + \frac{3}{8}a\,\hat{\mathbf{y}} + \frac{3}{8}a\,\hat{\mathbf{z}}$ | (4$a$) | C |
| $\mathbf{B}_2$ | = | $\frac{1}{8}\,\mathbf{a}_1 + \frac{5}{8}\,\mathbf{a}_2 + \frac{7}{8}\,\mathbf{a}_3$ | = | $\frac{1}{8}a\,\hat{\mathbf{x}} + \frac{5}{8}a\,\hat{\mathbf{y}} + \frac{7}{8}a\,\hat{\mathbf{z}}$ | (4$a$) | C |
| $\mathbf{B}_3$ | = | $\frac{5}{8}\,\mathbf{a}_1 + \frac{7}{8}\,\mathbf{a}_2 + \frac{1}{8}\,\mathbf{a}_3$ | = | $\frac{5}{8}a\,\hat{\mathbf{x}} + \frac{7}{8}a\,\hat{\mathbf{y}} + \frac{1}{8}a\,\hat{\mathbf{z}}$ | (4$a$) | C |
| $\mathbf{B}_4$ | = | $\frac{7}{8}\,\mathbf{a}_1 + \frac{1}{8}\,\mathbf{a}_2 + \frac{5}{8}\,\mathbf{a}_3$ | = | $\frac{7}{8}a\,\hat{\mathbf{x}} + \frac{1}{8}a\,\hat{\mathbf{y}} + \frac{5}{8}a\,\hat{\mathbf{z}}$ | (4$a$) | C |
| $\mathbf{B}_5$ | = | $x_2\,\mathbf{a}_1 + x_2\,\mathbf{a}_2 + x_2\,\mathbf{a}_3$ | = | $x_2a\,\hat{\mathbf{x}} + x_2a\,\hat{\mathbf{y}} + x_2a\,\hat{\mathbf{z}}$ | (8$c$) | Al |
| $\mathbf{B}_6$ | = | $\left(\frac{1}{2} - x_2\right)\,\mathbf{a}_1 - x_2\,\mathbf{a}_2 + \left(\frac{1}{2} + x_2\right)\,\mathbf{a}_3$ | = | $\left(\frac{1}{2} - x_2\right)a\,\hat{\mathbf{x}} - x_2a\,\hat{\mathbf{y}} + \left(\frac{1}{2} + x_2\right)a\,\hat{\mathbf{z}}$ | (8$c$) | Al |
| $\mathbf{B}_7$ | = | $-x_2\,\mathbf{a}_1 + \left(\frac{1}{2} + x_2\right)\,\mathbf{a}_2 + \left(\frac{1}{2} - x_2\right)\,\mathbf{a}_3$ | = | $-x_2a\,\hat{\mathbf{x}} + \left(\frac{1}{2} + x_2\right)a\,\hat{\mathbf{y}} + \left(\frac{1}{2} - x_2\right)a\,\hat{\mathbf{z}}$ | (8$c$) | Al |
| $\mathbf{B}_8$ | = | $\left(\frac{1}{2} + x_2\right)\,\mathbf{a}_1 + \left(\frac{1}{2} - x_2\right)\,\mathbf{a}_2 - x_2\,\mathbf{a}_3$ | = | $\left(\frac{1}{2} + x_2\right)a\,\hat{\mathbf{x}} + \left(\frac{1}{2} - x_2\right)a\,\hat{\mathbf{y}} - x_2a\,\hat{\mathbf{z}}$ | (8$c$) | Al |
| $\mathbf{B}_9$ | = | $\left(\frac{3}{4} + x_2\right)\,\mathbf{a}_1 + \left(\frac{1}{4} + x_2\right)\,\mathbf{a}_2 + \left(\frac{1}{4} - x_2\right)\,\mathbf{a}_3$ | = | $\left(\frac{3}{4} + x_2\right)a\,\hat{\mathbf{x}} + \left(\frac{1}{4} + x_2\right)a\,\hat{\mathbf{y}} + \left(\frac{1}{4} - x_2\right)a\,\hat{\mathbf{z}}$ | (8$c$) | Al |
| $\mathbf{B}_{10}$ | = | $\left(\frac{3}{4} - x_2\right)\,\mathbf{a}_1 + \left(\frac{3}{4} - x_2\right)\,\mathbf{a}_2 + \left(\frac{3}{4} - x_2\right)\,\mathbf{a}_3$ | = | $\left(\frac{3}{4} - x_2\right)a\,\hat{\mathbf{x}} + \left(\frac{3}{4} - x_2\right)a\,\hat{\mathbf{y}} + \left(\frac{3}{4} - x_2\right)a\,\hat{\mathbf{z}}$ | (8$c$) | Al |
| $\mathbf{B}_{11}$ | = | $\left(\frac{1}{4} + x_2\right)\,\mathbf{a}_1 + \left(\frac{1}{4} - x_2\right)\,\mathbf{a}_2 + \left(\frac{3}{4} + x_2\right)\,\mathbf{a}_3$ | = | $\left(\frac{1}{4} + x_2\right)a\,\hat{\mathbf{x}} + \left(\frac{1}{4} - x_2\right)a\,\hat{\mathbf{y}} + \left(\frac{3}{4} + x_2\right)a\,\hat{\mathbf{z}}$ | (8$c$) | Al |
| $\mathbf{B}_{12}$ | = | $\left(\frac{1}{4} - x_2\right)\,\mathbf{a}_1 + \left(\frac{3}{4} + x_2\right)\,\mathbf{a}_2 + \left(\frac{1}{4} + x_2\right)\,\mathbf{a}_3$ | = | $\left(\frac{1}{4} - x_2\right)a\,\hat{\mathbf{x}} + \left(\frac{3}{4} + x_2\right)a\,\hat{\mathbf{y}} + \left(\frac{1}{4} + x_2\right)a\,\hat{\mathbf{z}}$ | (8$c$) | Al |
| $\mathbf{B}_{13}$ | = | $\frac{1}{8}\,\mathbf{a}_1 + y_3\,\mathbf{a}_2 + \left(\frac{1}{4} + y_3\right)\,\mathbf{a}_3$ | = | $\frac{1}{8}a\,\hat{\mathbf{x}} + y_3a\,\hat{\mathbf{y}} + \left(\frac{1}{4} + y_3\right)a\,\hat{\mathbf{z}}$ | (12$d$) | Mo |
| $\mathbf{B}_{14}$ | = | $\frac{3}{8}\,\mathbf{a}_1 - y_3\,\mathbf{a}_2 + \left(\frac{3}{4} + y_3\right)\,\mathbf{a}_3$ | = | $\frac{3}{8}a\,\hat{\mathbf{x}} - y_3a\,\hat{\mathbf{y}} + \left(\frac{3}{4} + y_3\right)a\,\hat{\mathbf{z}}$ | (12$d$) | Mo |
| $\mathbf{B}_{15}$ | = | $\frac{7}{8}\,\mathbf{a}_1 + \left(\frac{1}{2} + y_3\right)\,\mathbf{a}_2 + \left(\frac{1}{4} - y_3\right)\,\mathbf{a}_3$ | = | $\frac{7}{8}a\,\hat{\mathbf{x}} + \left(\frac{1}{2} + y_3\right)a\,\hat{\mathbf{y}} + \left(\frac{1}{4} - y_3\right)a\,\hat{\mathbf{z}}$ | (12$d$) | Mo |
| $\mathbf{B}_{16}$ | = | $\frac{5}{8}\,\mathbf{a}_1 + \left(\frac{1}{2} - y_3\right)\,\mathbf{a}_2 + \left(\frac{3}{4} - y_3\right)\,\mathbf{a}_3$ | = | $\frac{5}{8}a\,\hat{\mathbf{x}} + \left(\frac{1}{2} - y_3\right)a\,\hat{\mathbf{y}} + \left(\frac{3}{4} - y_3\right)a\,\hat{\mathbf{z}}$ | (12$d$) | Mo |
| $\mathbf{B}_{17}$ | = | $\left(\frac{1}{4} + y_3\right)\,\mathbf{a}_1 + \frac{1}{8}\,\mathbf{a}_2 + y_3\,\mathbf{a}_3$ | = | $\left(\frac{1}{4} + y_3\right)a\,\hat{\mathbf{x}} + \frac{1}{8}a\,\hat{\mathbf{y}} + y_3a\,\hat{\mathbf{z}}$ | (12$d$) | Mo |
| $\mathbf{B}_{18}$ | = | $\left(\frac{3}{4} + y_3\right)\,\mathbf{a}_1 + \frac{3}{8}\,\mathbf{a}_2 - y_3\,\mathbf{a}_3$ | = | $\left(\frac{3}{4} + y_3\right)a\,\hat{\mathbf{x}} + \frac{3}{8}a\,\hat{\mathbf{y}} - y_3a\,\hat{\mathbf{z}}$ | (12$d$) | Mo |
| $\mathbf{B}_{19}$ | = | $\left(\frac{1}{4} - y_3\right)\,\mathbf{a}_1 + \frac{7}{8}\,\mathbf{a}_2 + \left(\frac{1}{2} + y_3\right)\,\mathbf{a}_3$ | = | $\left(\frac{1}{4} - y_3\right)a\,\hat{\mathbf{x}} + \frac{7}{8}a\,\hat{\mathbf{y}} + \left(\frac{1}{2} + y_3\right)a\,\hat{\mathbf{z}}$ | (12$d$) | Mo |
| $\mathbf{B}_{20}$ | = | $\left(\frac{3}{4} - y_3\right)\,\mathbf{a}_1 + \frac{5}{8}\,\mathbf{a}_2 + \left(\frac{1}{2} - y_3\right)\,\mathbf{a}_3$ | = | $\left(\frac{3}{4} - y_3\right)a\,\hat{\mathbf{x}} + \frac{5}{8}a\,\hat{\mathbf{y}} + \left(\frac{1}{2} - y_3\right)a\,\hat{\mathbf{z}}$ | (12$d$) | Mo |
| $\mathbf{B}_{21}$ | = | $y_3\,\mathbf{a}_1 + \left(\frac{1}{4} + y_3\right)\,\mathbf{a}_2 + \frac{1}{8}\,\mathbf{a}_3$ | = | $y_3a\,\hat{\mathbf{x}} + \left(\frac{1}{4} + y_3\right)a\,\hat{\mathbf{y}} + \frac{1}{8}a\,\hat{\mathbf{z}}$ | (12$d$) | Mo |
| $\mathbf{B}_{22}$ | = | $-y_3\,\mathbf{a}_1 + \left(\frac{3}{4} + y_3\right)\,\mathbf{a}_2 + \frac{3}{8}\,\mathbf{a}_3$ | = | $-y_3a\,\hat{\mathbf{x}} + \left(\frac{3}{4} + y_3\right)a\,\hat{\mathbf{y}} + \frac{3}{8}a\,\hat{\mathbf{z}}$ | (12$d$) | Mo |
| $\mathbf{B}_{23}$ | = | $\left(\frac{1}{2} + y_3\right)\,\mathbf{a}_1 + \left(\frac{1}{4} - y_3\right)\,\mathbf{a}_2 + \frac{7}{8}\,\mathbf{a}_3$ | = | $\left(\frac{1}{2} + y_3\right)a\,\hat{\mathbf{x}} + \left(\frac{1}{4} - y_3\right)a\,\hat{\mathbf{y}} + \frac{7}{8}a\,\hat{\mathbf{z}}$ | (12$d$) | Mo |
| $\mathbf{B}_{24}$ | = | $\left(\frac{1}{2} - y_3\right)\,\mathbf{a}_1 + \left(\frac{3}{4} - y_3\right)\,\mathbf{a}_2 + \frac{5}{8}\,\mathbf{a}_3$ | = | $\left(\frac{1}{2} - y_3\right)a\,\hat{\mathbf{x}} + \left(\frac{3}{4} - y_3\right)a\,\hat{\mathbf{y}} + \frac{5}{8}a\,\hat{\mathbf{z}}$ | (12$d$) | Mo |

**Geometry files:**

- CIF: pp. 1790
- POSCAR: pp. 1791

# $Mg_3Ru_2$ Structure: A3B2_cP20_213_d_c

http://aflow.org/prototype-encyclopedia/A3B2_cP20_213_d_c

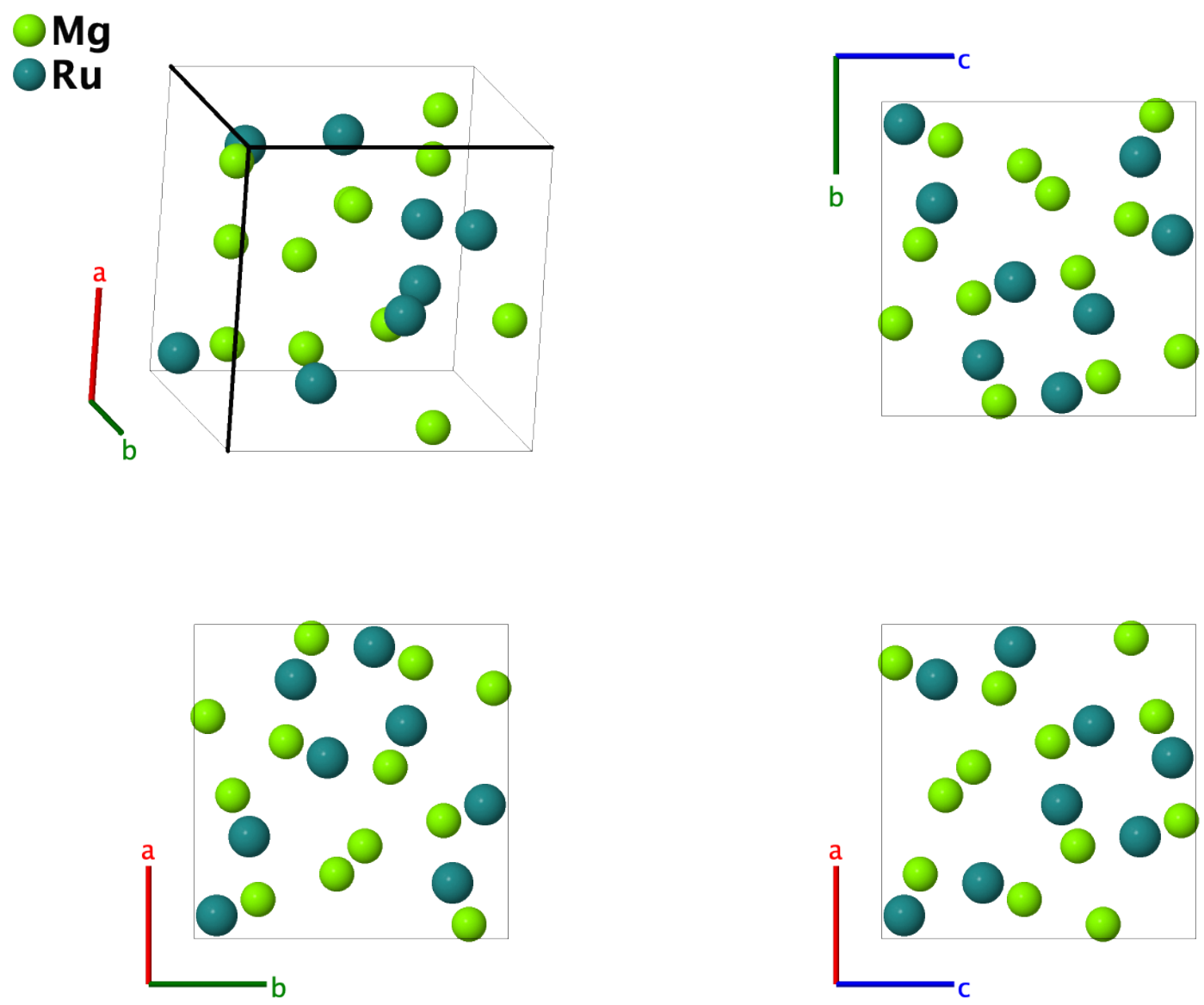

| | | |
|---|---|---|
| **Prototype** | : | $Mg_3Ru_2$ |
| **AFLOW prototype label** | : | A3B2_cP20_213_d_c |
| ***Strukturbericht* designation** | : | None |
| **Pearson symbol** | : | cP20 |
| **Space group number** | : | 213 |
| **Space group symbol** | : | $P4_132$ |
| **AFLOW prototype command** | : | `aflow --proto=A3B2_cP20_213_d_c --params=`$a, x_1, y_2$ |

- This is the binary form of the $\beta$-Mn ($A13$) structure.

**Simple Cubic primitive vectors:**

$$\mathbf{a}_1 = a\,\hat{\mathbf{x}}$$
$$\mathbf{a}_2 = a\,\hat{\mathbf{y}}$$
$$\mathbf{a}_3 = a\,\hat{\mathbf{z}}$$

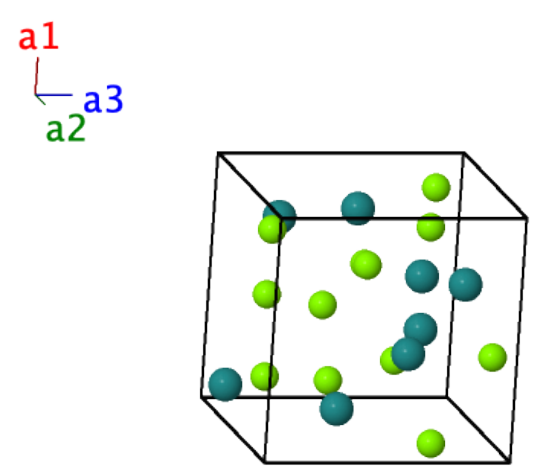

**Basis vectors:**

| | | Lattice Coordinates | | Cartesian Coordinates | Wyckoff Position | Atom Type |
|---|---|---|---|---|---|---|
| $\mathbf{B}_1$ | = | $x_1\,\mathbf{a}_1 + x_1\,\mathbf{a}_2 + x_1\,\mathbf{a}_3$ | = | $x_1 a\,\hat{\mathbf{x}} + x_1 a\,\hat{\mathbf{y}} + x_1 a\,\hat{\mathbf{z}}$ | (8$c$) | Ru |
| $\mathbf{B}_2$ | = | $\left(\frac{1}{2} - x_1\right)\mathbf{a}_1 - x_1\,\mathbf{a}_2 + \left(\frac{1}{2} + x_1\right)\mathbf{a}_3$ | = | $\left(\frac{1}{2} - x_1\right)a\,\hat{\mathbf{x}} - x_1 a\,\hat{\mathbf{y}} + \left(\frac{1}{2} + x_1\right)a\,\hat{\mathbf{z}}$ | (8$c$) | Ru |
| $\mathbf{B}_3$ | = | $-x_1\,\mathbf{a}_1 + \left(\frac{1}{2} + x_1\right)\mathbf{a}_2 + \left(\frac{1}{2} - x_1\right)\mathbf{a}_3$ | = | $-x_1 a\,\hat{\mathbf{x}} + \left(\frac{1}{2} + x_1\right)a\,\hat{\mathbf{y}} + \left(\frac{1}{2} - x_1\right)a\,\hat{\mathbf{z}}$ | (8$c$) | Ru |
| $\mathbf{B}_4$ | = | $\left(\frac{1}{2}+x_1\right)\mathbf{a}_1+\left(\frac{1}{2}-x_1\right)\mathbf{a}_2-x_1\,\mathbf{a}_3$ | = | $\left(\frac{1}{2}+x_1\right)a\,\hat{\mathbf{x}}+\left(\frac{1}{2}-x_1\right)a\,\hat{\mathbf{y}}-x_1a\,\hat{\mathbf{z}}$ | (8$c$) | Ru |
| $\mathbf{B}_5$ | = | $\left(\frac{3}{4}+x_1\right)\mathbf{a}_1+\left(\frac{1}{4}+x_1\right)\mathbf{a}_2+\left(\frac{1}{4}-x_1\right)\mathbf{a}_3$ | = | $\left(\frac{3}{4}+x_1\right)a\,\hat{\mathbf{x}}+\left(\frac{1}{4}+x_1\right)a\,\hat{\mathbf{y}}+\left(\frac{1}{4}-x_1\right)a\,\hat{\mathbf{z}}$ | (8$c$) | Ru |
| $\mathbf{B}_6$ | = | $\left(\frac{3}{4}-x_1\right)\mathbf{a}_1+\left(\frac{3}{4}-x_1\right)\mathbf{a}_2+\left(\frac{3}{4}-x_1\right)\mathbf{a}_3$ | = | $\left(\frac{3}{4}-x_1\right)a\,\hat{\mathbf{x}}+\left(\frac{3}{4}-x_1\right)a\,\hat{\mathbf{y}}+\left(\frac{3}{4}-x_1\right)a\,\hat{\mathbf{z}}$ | (8$c$) | Ru |
| $\mathbf{B}_7$ | = | $\left(\frac{1}{4}+x_1\right)\mathbf{a}_1+\left(\frac{1}{4}-x_1\right)\mathbf{a}_2+\left(\frac{3}{4}+x_1\right)\mathbf{a}_3$ | = | $\left(\frac{1}{4}+x_1\right)a\,\hat{\mathbf{x}}+\left(\frac{1}{4}-x_1\right)a\,\hat{\mathbf{y}}+\left(\frac{3}{4}+x_1\right)a\,\hat{\mathbf{z}}$ | (8$c$) | Ru |
| $\mathbf{B}_8$ | = | $\left(\frac{1}{4}-x_1\right)\mathbf{a}_1+\left(\frac{3}{4}+x_1\right)\mathbf{a}_2+\left(\frac{1}{4}+x_1\right)\mathbf{a}_3$ | = | $\left(\frac{1}{4}-x_1\right)a\,\hat{\mathbf{x}}+\left(\frac{3}{4}+x_1\right)a\,\hat{\mathbf{y}}+\left(\frac{1}{4}+x_1\right)a\,\hat{\mathbf{z}}$ | (8$c$) | Ru |
| $\mathbf{B}_9$ | = | $\frac{1}{8}\,\mathbf{a}_1+y_2\,\mathbf{a}_2+\left(\frac{1}{4}+y_2\right)\mathbf{a}_3$ | = | $\frac{1}{8}a\,\hat{\mathbf{x}}+y_2a\,\hat{\mathbf{y}}+\left(\frac{1}{4}+y_2\right)a\,\hat{\mathbf{z}}$ | (12$d$) | Mg |
| $\mathbf{B}_{10}$ | = | $\frac{3}{8}\,\mathbf{a}_1-y_2\,\mathbf{a}_2+\left(\frac{3}{4}+y_2\right)\mathbf{a}_3$ | = | $\frac{3}{8}a\,\hat{\mathbf{x}}-y_2a\,\hat{\mathbf{y}}+\left(\frac{3}{4}+y_2\right)a\,\hat{\mathbf{z}}$ | (12$d$) | Mg |
| $\mathbf{B}_{11}$ | = | $\frac{7}{8}\,\mathbf{a}_1+\left(\frac{1}{2}+y_2\right)\mathbf{a}_2+\left(\frac{1}{4}-y_2\right)\mathbf{a}_3$ | = | $\frac{7}{8}a\,\hat{\mathbf{x}}+\left(\frac{1}{2}+y_2\right)a\,\hat{\mathbf{y}}+\left(\frac{1}{4}-y_2\right)a\,\hat{\mathbf{z}}$ | (12$d$) | Mg |
| $\mathbf{B}_{12}$ | = | $\frac{5}{8}\,\mathbf{a}_1+\left(\frac{1}{2}-y_2\right)\mathbf{a}_2+\left(\frac{3}{4}-y_2\right)\mathbf{a}_3$ | = | $\frac{5}{8}a\,\hat{\mathbf{x}}+\left(\frac{1}{2}-y_2\right)a\,\hat{\mathbf{y}}+\left(\frac{3}{4}-y_2\right)a\,\hat{\mathbf{z}}$ | (12$d$) | Mg |
| $\mathbf{B}_{13}$ | = | $\left(\frac{1}{4}+y_2\right)\mathbf{a}_1+\frac{1}{8}\,\mathbf{a}_2+y_2\,\mathbf{a}_3$ | = | $\left(\frac{1}{4}+y_2\right)a\,\hat{\mathbf{x}}+\frac{1}{8}a\,\hat{\mathbf{y}}+y_2a\,\hat{\mathbf{z}}$ | (12$d$) | Mg |
| $\mathbf{B}_{14}$ | = | $\left(\frac{3}{4}+y_2\right)\mathbf{a}_1+\frac{3}{8}\,\mathbf{a}_2-y_2\,\mathbf{a}_3$ | = | $\left(\frac{3}{4}+y_2\right)a\,\hat{\mathbf{x}}+\frac{3}{8}a\,\hat{\mathbf{y}}-y_2a\,\hat{\mathbf{z}}$ | (12$d$) | Mg |
| $\mathbf{B}_{15}$ | = | $\left(\frac{1}{4}-y_2\right)\mathbf{a}_1+\frac{7}{8}\,\mathbf{a}_2+\left(\frac{1}{2}+y_2\right)\mathbf{a}_3$ | = | $\left(\frac{1}{4}-y_2\right)a\,\hat{\mathbf{x}}+\frac{7}{8}a\,\hat{\mathbf{y}}+\left(\frac{1}{2}+y_2\right)a\,\hat{\mathbf{z}}$ | (12$d$) | Mg |
| $\mathbf{B}_{16}$ | = | $\left(\frac{3}{4}-y_2\right)\mathbf{a}_1+\frac{5}{8}\,\mathbf{a}_2+\left(\frac{1}{2}-y_2\right)\mathbf{a}_3$ | = | $\left(\frac{3}{4}-y_2\right)a\,\hat{\mathbf{x}}+\frac{5}{8}a\,\hat{\mathbf{y}}+\left(\frac{1}{2}-y_2\right)a\,\hat{\mathbf{z}}$ | (12$d$) | Mg |
| $\mathbf{B}_{17}$ | = | $y_2\,\mathbf{a}_1+\left(\frac{1}{4}+y_2\right)\mathbf{a}_2+\frac{1}{8}\,\mathbf{a}_3$ | = | $y_2a\,\hat{\mathbf{x}}+\left(\frac{1}{4}+y_2\right)a\,\hat{\mathbf{y}}+\frac{1}{8}a\,\hat{\mathbf{z}}$ | (12$d$) | Mg |
| $\mathbf{B}_{18}$ | = | $-y_2\,\mathbf{a}_1+\left(\frac{3}{4}+y_2\right)\mathbf{a}_2+\frac{3}{8}\,\mathbf{a}_3$ | = | $-y_2a\,\hat{\mathbf{x}}+\left(\frac{3}{4}+y_2\right)a\,\hat{\mathbf{y}}+\frac{3}{8}a\,\hat{\mathbf{z}}$ | (12$d$) | Mg |
| $\mathbf{B}_{19}$ | = | $\left(\frac{1}{2}+y_2\right)\mathbf{a}_1+\left(\frac{1}{4}-y_2\right)\mathbf{a}_2+\frac{7}{8}\,\mathbf{a}_3$ | = | $\left(\frac{1}{2}+y_2\right)a\,\hat{\mathbf{x}}+\left(\frac{1}{4}-y_2\right)a\,\hat{\mathbf{y}}+\frac{7}{8}a\,\hat{\mathbf{z}}$ | (12$d$) | Mg |
| $\mathbf{B}_{20}$ | = | $\left(\frac{1}{2}-y_2\right)\mathbf{a}_1+\left(\frac{3}{4}-y_2\right)\mathbf{a}_2+\frac{5}{8}\,\mathbf{a}_3$ | = | $\left(\frac{1}{2}-y_2\right)a\,\hat{\mathbf{x}}+\left(\frac{3}{4}-y_2\right)a\,\hat{\mathbf{y}}+\frac{5}{8}a\,\hat{\mathbf{z}}$ | (12$d$) | Mg |

**Geometry files:**

- CIF: pp. 1791
- POSCAR: pp. 1791

# Zunyite [$Al_{13}(OH,F)_{18}Si_5O_{20}Cl$, $S0_8$] Structure: A13BC18D20E5_cF228_216_dh_b_fh_2eh_ce

http://aflow.org/prototype-encyclopedia/A13BC18D20E5_cF228_216_dh_b_fh_2eh_ce

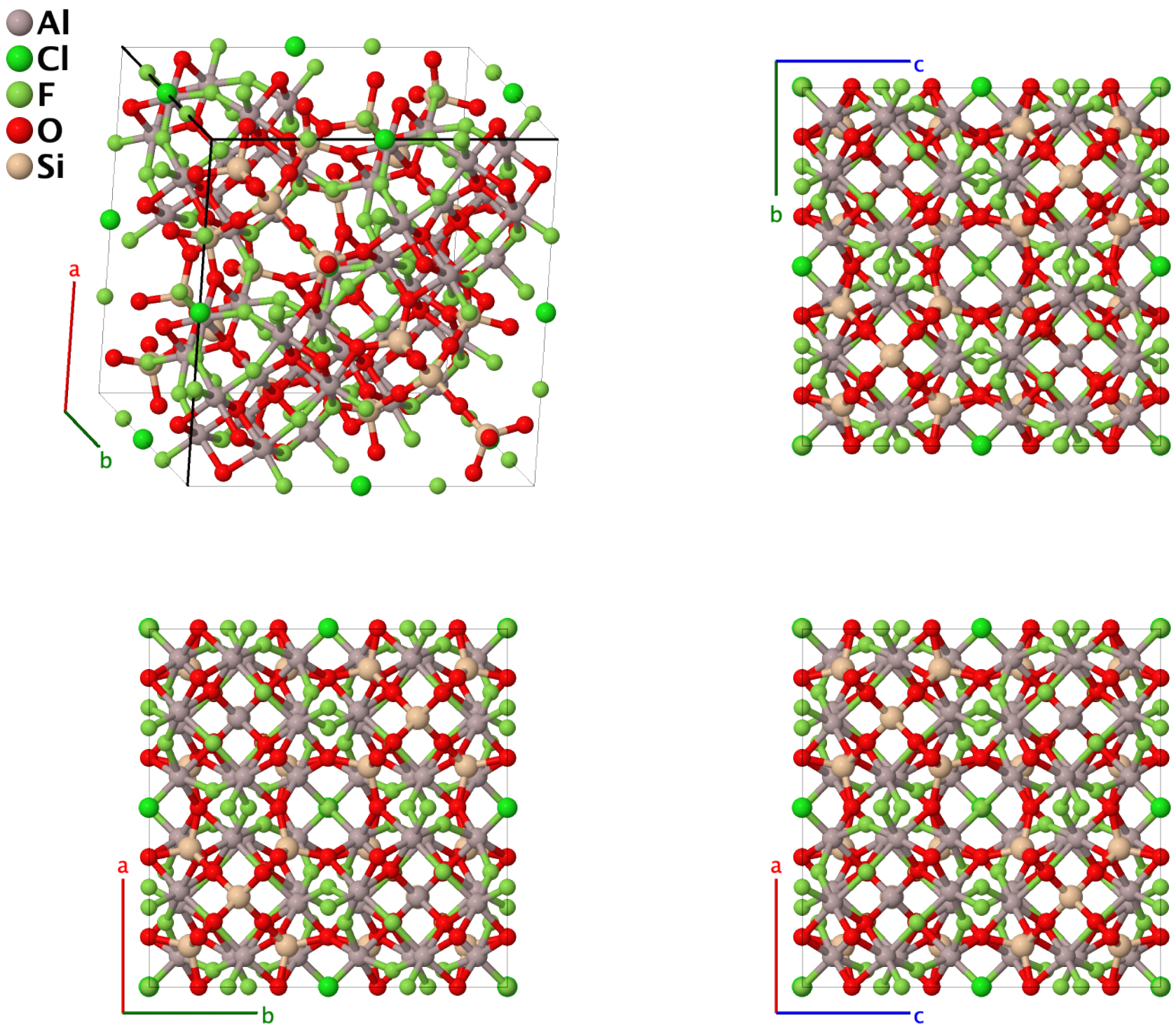

| | | |
|---|---|---|
| **Prototype** | : | $Al_{13}ClF_{18}O_{20}Si_5$ |
| **AFLOW prototype label** | : | A13BC18D20E5_cF228_216_dh_b_fh_2eh_ce |
| ***Strukturbericht* designation** | : | $S0_8$ |
| **Pearson symbol** | : | cF228 |
| **Space group number** | : | 216 |
| **Space group symbol** | : | $F\bar{4}3m$ |
| **AFLOW prototype command** | : | `aflow --proto=A13BC18D20E5_cF228_216_dh_b_fh_2eh_ce --params=`$a, x_4, x_5, x_6, x_7, x_8, z_8, x_9, z_9, x_{10}, z_{10}$ |

- We use the structure proposed by (Kamb, 1960), a refinement of the original (Pauling, 1933) $S0_8$ structure. The only major difference is the $y$ coordinate of the OH/F-I site, which is now at a more reasonable distance from the chlorine atom.
- For easy of visualization, we have used fluorine atoms to represent all of the $(OH,F)_{18}$ positions, but in reality the system is dominated by OH, not F. Kamb argues that the physics of hydrogen bonding makes it likely that the actual structure has composition $(OH)_{16}F_2$, with the fluorine atoms substituting for OH on the second (48$h$) site.

**Face-centered Cubic primitive vectors:**

$$\mathbf{a}_1 = \frac{1}{2}\,a\,\hat{\mathbf{y}} + \frac{1}{2}\,a\,\hat{\mathbf{z}}$$
$$\mathbf{a}_2 = \frac{1}{2}\,a\,\hat{\mathbf{x}} + \frac{1}{2}\,a\,\hat{\mathbf{z}}$$
$$\mathbf{a}_3 = \frac{1}{2}\,a\,\hat{\mathbf{x}} + \frac{1}{2}\,a\,\hat{\mathbf{y}}$$

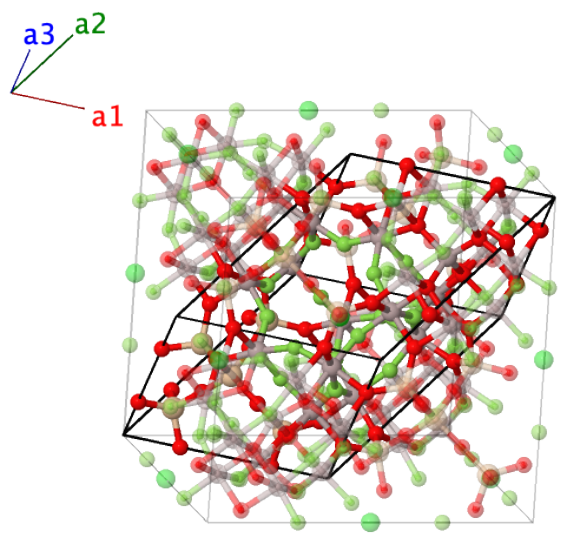

**Basis vectors:**

| | | Lattice Coordinates | | Cartesian Coordinates | Wyckoff Position | Atom Type |
|---|---|---|---|---|---|---|
| $\mathbf{B}_1$ | = | $\frac{1}{2}\,\mathbf{a}_1 + \frac{1}{2}\,\mathbf{a}_2 + \frac{1}{2}\,\mathbf{a}_3$ | = | $\frac{1}{2}a\,\hat{\mathbf{x}} + \frac{1}{2}a\,\hat{\mathbf{y}} + \frac{1}{2}a\,\hat{\mathbf{z}}$ | (4$b$) | Cl |
| $\mathbf{B}_2$ | = | $\frac{1}{4}\,\mathbf{a}_1 + \frac{1}{4}\,\mathbf{a}_2 + \frac{1}{4}\,\mathbf{a}_3$ | = | $\frac{1}{4}a\,\hat{\mathbf{x}} + \frac{1}{4}a\,\hat{\mathbf{y}} + \frac{1}{4}a\,\hat{\mathbf{z}}$ | (4$c$) | Si I |
| $\mathbf{B}_3$ | = | $\frac{3}{4}\,\mathbf{a}_1 + \frac{3}{4}\,\mathbf{a}_2 + \frac{3}{4}\,\mathbf{a}_3$ | = | $\frac{3}{4}a\,\hat{\mathbf{x}} + \frac{3}{4}a\,\hat{\mathbf{y}} + \frac{3}{4}a\,\hat{\mathbf{z}}$ | (4$d$) | Al I |
| $\mathbf{B}_4$ | = | $x_4\,\mathbf{a}_1 + x_4\,\mathbf{a}_2 + x_4\,\mathbf{a}_3$ | = | $x_4a\,\hat{\mathbf{x}} + x_4a\,\hat{\mathbf{y}} + x_4a\,\hat{\mathbf{z}}$ | (16$e$) | O I |
| $\mathbf{B}_5$ | = | $x_4\,\mathbf{a}_1 + x_4\,\mathbf{a}_2 - 3x_4\,\mathbf{a}_3$ | = | $-x_4a\,\hat{\mathbf{x}} - x_4a\,\hat{\mathbf{y}} + x_4a\,\hat{\mathbf{z}}$ | (16$e$) | O I |
| $\mathbf{B}_6$ | = | $x_4\,\mathbf{a}_1 - 3x_4\,\mathbf{a}_2 + x_4\,\mathbf{a}_3$ | = | $-x_4a\,\hat{\mathbf{x}} + x_4a\,\hat{\mathbf{y}} - x_4a\,\hat{\mathbf{z}}$ | (16$e$) | O I |
| $\mathbf{B}_7$ | = | $-3x_4\,\mathbf{a}_1 + x_4\,\mathbf{a}_2 + x_4\,\mathbf{a}_3$ | = | $x_4a\,\hat{\mathbf{x}} - x_4a\,\hat{\mathbf{y}} - x_4a\,\hat{\mathbf{z}}$ | (16$e$) | O I |
| $\mathbf{B}_8$ | = | $x_5\,\mathbf{a}_1 + x_5\,\mathbf{a}_2 + x_5\,\mathbf{a}_3$ | = | $x_5a\,\hat{\mathbf{x}} + x_5a\,\hat{\mathbf{y}} + x_5a\,\hat{\mathbf{z}}$ | (16$e$) | O II |
| $\mathbf{B}_9$ | = | $x_5\,\mathbf{a}_1 + x_5\,\mathbf{a}_2 - 3x_5\,\mathbf{a}_3$ | = | $-x_5a\,\hat{\mathbf{x}} - x_5a\,\hat{\mathbf{y}} + x_5a\,\hat{\mathbf{z}}$ | (16$e$) | O II |
| $\mathbf{B}_{10}$ | = | $x_5\,\mathbf{a}_1 - 3x_5\,\mathbf{a}_2 + x_5\,\mathbf{a}_3$ | = | $-x_5a\,\hat{\mathbf{x}} + x_5a\,\hat{\mathbf{y}} - x_5a\,\hat{\mathbf{z}}$ | (16$e$) | O II |
| $\mathbf{B}_{11}$ | = | $-3x_5\,\mathbf{a}_1 + x_5\,\mathbf{a}_2 + x_5\,\mathbf{a}_3$ | = | $x_5a\,\hat{\mathbf{x}} - x_5a\,\hat{\mathbf{y}} - x_5a\,\hat{\mathbf{z}}$ | (16$e$) | O II |
| $\mathbf{B}_{12}$ | = | $x_6\,\mathbf{a}_1 + x_6\,\mathbf{a}_2 + x_6\,\mathbf{a}_3$ | = | $x_6a\,\hat{\mathbf{x}} + x_6a\,\hat{\mathbf{y}} + x_6a\,\hat{\mathbf{z}}$ | (16$e$) | Si II |
| $\mathbf{B}_{13}$ | = | $x_6\,\mathbf{a}_1 + x_6\,\mathbf{a}_2 - 3x_6\,\mathbf{a}_3$ | = | $-x_6a\,\hat{\mathbf{x}} - x_6a\,\hat{\mathbf{y}} + x_6a\,\hat{\mathbf{z}}$ | (16$e$) | Si II |
| $\mathbf{B}_{14}$ | = | $x_6\,\mathbf{a}_1 - 3x_6\,\mathbf{a}_2 + x_6\,\mathbf{a}_3$ | = | $-x_6a\,\hat{\mathbf{x}} + x_6a\,\hat{\mathbf{y}} - x_6a\,\hat{\mathbf{z}}$ | (16$e$) | Si II |
| $\mathbf{B}_{15}$ | = | $-3x_6\,\mathbf{a}_1 + x_6\,\mathbf{a}_2 + x_6\,\mathbf{a}_3$ | = | $x_6a\,\hat{\mathbf{x}} - x_6a\,\hat{\mathbf{y}} - x_6a\,\hat{\mathbf{z}}$ | (16$e$) | Si II |
| $\mathbf{B}_{16}$ | = | $-x_7\,\mathbf{a}_1 + x_7\,\mathbf{a}_2 + x_7\,\mathbf{a}_3$ | = | $x_7a\,\hat{\mathbf{x}}$ | (24$f$) | F I |
| $\mathbf{B}_{17}$ | = | $x_7\,\mathbf{a}_1 - x_7\,\mathbf{a}_2 - x_7\,\mathbf{a}_3$ | = | $-x_7a\,\hat{\mathbf{x}}$ | (24$f$) | F I |
| $\mathbf{B}_{18}$ | = | $x_7\,\mathbf{a}_1 - x_7\,\mathbf{a}_2 + x_7\,\mathbf{a}_3$ | = | $x_7a\,\hat{\mathbf{y}}$ | (24$f$) | F I |
| $\mathbf{B}_{19}$ | = | $-x_7\,\mathbf{a}_1 + x_7\,\mathbf{a}_2 - x_7\,\mathbf{a}_3$ | = | $-x_7a\,\hat{\mathbf{y}}$ | (24$f$) | F I |
| $\mathbf{B}_{20}$ | = | $x_7\,\mathbf{a}_1 + x_7\,\mathbf{a}_2 - x_7\,\mathbf{a}_3$ | = | $x_7a\,\hat{\mathbf{z}}$ | (24$f$) | F I |
| $\mathbf{B}_{21}$ | = | $-x_7\,\mathbf{a}_1 - x_7\,\mathbf{a}_2 + x_7\,\mathbf{a}_3$ | = | $-x_7a\,\hat{\mathbf{z}}$ | (24$f$) | F I |
| $\mathbf{B}_{22}$ | = | $z_8\,\mathbf{a}_1 + z_8\,\mathbf{a}_2 + (2x_8 - z_8)\,\mathbf{a}_3$ | = | $x_8a\,\hat{\mathbf{x}} + x_8a\,\hat{\mathbf{y}} + z_8a\,\hat{\mathbf{z}}$ | (48$h$) | Al II |
| $\mathbf{B}_{23}$ | = | $z_8\,\mathbf{a}_1 + z_8\,\mathbf{a}_2 + (-2x_8 - z_8)\,\mathbf{a}_3$ | = | $-x_8a\,\hat{\mathbf{x}} - x_8a\,\hat{\mathbf{y}} + z_8a\,\hat{\mathbf{z}}$ | (48$h$) | Al II |
| $\mathbf{B}_{24}$ | = | $(2x_8 - z_8)\,\mathbf{a}_1 + (-2x_8 - z_8)\,\mathbf{a}_2 + z_8\,\mathbf{a}_3$ | = | $-x_8a\,\hat{\mathbf{x}} + x_8a\,\hat{\mathbf{y}} - z_8a\,\hat{\mathbf{z}}$ | (48$h$) | Al II |
| $\mathbf{B}_{25}$ | = | $(-2x_8 - z_8)\,\mathbf{a}_1 + (2x_8 - z_8)\,\mathbf{a}_2 + z_8\,\mathbf{a}_3$ | = | $x_8a\,\hat{\mathbf{x}} - x_8a\,\hat{\mathbf{y}} - z_8a\,\hat{\mathbf{z}}$ | (48$h$) | Al II |
| $\mathbf{B}_{26}$ | = | $(2x_8 - z_8)\,\mathbf{a}_1 + z_8\,\mathbf{a}_2 + z_8\,\mathbf{a}_3$ | = | $z_8a\,\hat{\mathbf{x}} + x_8a\,\hat{\mathbf{y}} + x_8a\,\hat{\mathbf{z}}$ | (48$h$) | Al II |
| $\mathbf{B}_{27}$ | = | $(-2x_8 - z_8)\,\mathbf{a}_1 + z_8\,\mathbf{a}_2 + z_8\,\mathbf{a}_3$ | = | $z_8a\,\hat{\mathbf{x}} - x_8a\,\hat{\mathbf{y}} - x_8a\,\hat{\mathbf{z}}$ | (48$h$) | Al II |
| $\mathbf{B}_{28}$ | = | $z_8\,\mathbf{a}_1 + (2x_8 - z_8)\,\mathbf{a}_2 + (-2x_8 - z_8)\,\mathbf{a}_3$ | = | $-z_8a\,\mathbf{\hat{x}} - x_8a\,\mathbf{\hat{y}} + x_8a\,\mathbf{\hat{z}}$ | (48$h$) | Al II |
| $\mathbf{B}_{29}$ | = | $z_8\,\mathbf{a}_1 + (-2x_8 - z_8)\,\mathbf{a}_2 + (2x_8 - z_8)\,\mathbf{a}_3$ | = | $-z_8a\,\mathbf{\hat{x}} + x_8a\,\mathbf{\hat{y}} - x_8a\,\mathbf{\hat{z}}$ | (48$h$) | Al II |
| $\mathbf{B}_{30}$ | = | $z_8\,\mathbf{a}_1 + (2x_8 - z_8)\,\mathbf{a}_2 + z_8\,\mathbf{a}_3$ | = | $x_8a\,\mathbf{\hat{x}} + z_8a\,\mathbf{\hat{y}} + x_8a\,\mathbf{\hat{z}}$ | (48$h$) | Al II |
| $\mathbf{B}_{31}$ | = | $z_8\,\mathbf{a}_1 + (-2x_8 - z_8)\,\mathbf{a}_2 + z_8\,\mathbf{a}_3$ | = | $-x_8a\,\mathbf{\hat{x}} + z_8a\,\mathbf{\hat{y}} - x_8a\,\mathbf{\hat{z}}$ | (48$h$) | Al II |
| $\mathbf{B}_{32}$ | = | $(-2x_8 - z_8)\,\mathbf{a}_1 + z_8\,\mathbf{a}_2 + (2x_8 - z_8)\,\mathbf{a}_3$ | = | $x_8a\,\mathbf{\hat{x}} - z_8a\,\mathbf{\hat{y}} - x_8a\,\mathbf{\hat{z}}$ | (48$h$) | Al II |
| $\mathbf{B}_{33}$ | = | $(2x_8 - z_8)\,\mathbf{a}_1 + z_8\,\mathbf{a}_2 + (-2x_8 - z_8)\,\mathbf{a}_3$ | = | $-x_8a\,\mathbf{\hat{x}} - z_8a\,\mathbf{\hat{y}} + x_8a\,\mathbf{\hat{z}}$ | (48$h$) | Al II |
| $\mathbf{B}_{34}$ | = | $z_9\,\mathbf{a}_1 + z_9\,\mathbf{a}_2 + (2x_9 - z_9)\,\mathbf{a}_3$ | = | $x_9a\,\mathbf{\hat{x}} + x_9a\,\mathbf{\hat{y}} + z_9a\,\mathbf{\hat{z}}$ | (48$h$) | F II |
| $\mathbf{B}_{35}$ | = | $z_9\,\mathbf{a}_1 + z_9\,\mathbf{a}_2 + (-2x_9 - z_9)\,\mathbf{a}_3$ | = | $-x_9a\,\mathbf{\hat{x}} - x_9a\,\mathbf{\hat{y}} + z_9a\,\mathbf{\hat{z}}$ | (48$h$) | F II |
| $\mathbf{B}_{36}$ | = | $(2x_9 - z_9)\,\mathbf{a}_1 + (-2x_9 - z_9)\,\mathbf{a}_2 + z_9\,\mathbf{a}_3$ | = | $-x_9a\,\mathbf{\hat{x}} + x_9a\,\mathbf{\hat{y}} - z_9a\,\mathbf{\hat{z}}$ | (48$h$) | F II |
| $\mathbf{B}_{37}$ | = | $(-2x_9 - z_9)\,\mathbf{a}_1 + (2x_9 - z_9)\,\mathbf{a}_2 + z_9\,\mathbf{a}_3$ | = | $x_9a\,\mathbf{\hat{x}} - x_9a\,\mathbf{\hat{y}} - z_9a\,\mathbf{\hat{z}}$ | (48$h$) | F II |
| $\mathbf{B}_{38}$ | = | $(2x_9 - z_9)\,\mathbf{a}_1 + z_9\,\mathbf{a}_2 + z_9\,\mathbf{a}_3$ | = | $z_9a\,\mathbf{\hat{x}} + x_9a\,\mathbf{\hat{y}} + x_9a\,\mathbf{\hat{z}}$ | (48$h$) | F II |
| $\mathbf{B}_{39}$ | = | $(-2x_9 - z_9)\,\mathbf{a}_1 + z_9\,\mathbf{a}_2 + z_9\,\mathbf{a}_3$ | = | $z_9a\,\mathbf{\hat{x}} - x_9a\,\mathbf{\hat{y}} - x_9a\,\mathbf{\hat{z}}$ | (48$h$) | F II |
| $\mathbf{B}_{40}$ | = | $z_9\,\mathbf{a}_1 + (2x_9 - z_9)\,\mathbf{a}_2 + (-2x_9 - z_9)\,\mathbf{a}_3$ | = | $-z_9a\,\mathbf{\hat{x}} - x_9a\,\mathbf{\hat{y}} + x_9a\,\mathbf{\hat{z}}$ | (48$h$) | F II |
| $\mathbf{B}_{41}$ | = | $z_9\,\mathbf{a}_1 + (-2x_9 - z_9)\,\mathbf{a}_2 + (2x_9 - z_9)\,\mathbf{a}_3$ | = | $-z_9a\,\mathbf{\hat{x}} + x_9a\,\mathbf{\hat{y}} - x_9a\,\mathbf{\hat{z}}$ | (48$h$) | F II |
| $\mathbf{B}_{42}$ | = | $z_9\,\mathbf{a}_1 + (2x_9 - z_9)\,\mathbf{a}_2 + z_9\,\mathbf{a}_3$ | = | $x_9a\,\mathbf{\hat{x}} + z_9a\,\mathbf{\hat{y}} + x_9a\,\mathbf{\hat{z}}$ | (48$h$) | F II |
| $\mathbf{B}_{43}$ | = | $z_9\,\mathbf{a}_1 + (-2x_9 - z_9)\,\mathbf{a}_2 + z_9\,\mathbf{a}_3$ | = | $-x_9a\,\mathbf{\hat{x}} + z_9a\,\mathbf{\hat{y}} - x_9a\,\mathbf{\hat{z}}$ | (48$h$) | F II |
| $\mathbf{B}_{44}$ | = | $(-2x_9 - z_9)\,\mathbf{a}_1 + z_9\,\mathbf{a}_2 + (2x_9 - z_9)\,\mathbf{a}_3$ | = | $x_9a\,\mathbf{\hat{x}} - z_9a\,\mathbf{\hat{y}} - x_9a\,\mathbf{\hat{z}}$ | (48$h$) | F II |
| $\mathbf{B}_{45}$ | = | $(2x_9 - z_9)\,\mathbf{a}_1 + z_9\,\mathbf{a}_2 + (-2x_9 - z_9)\,\mathbf{a}_3$ | = | $-x_9a\,\mathbf{\hat{x}} - z_9a\,\mathbf{\hat{y}} + x_9a\,\mathbf{\hat{z}}$ | (48$h$) | F II |
| $\mathbf{B}_{46}$ | = | $z_{10}\,\mathbf{a}_1 + z_{10}\,\mathbf{a}_2 + (2x_{10} - z_{10})\,\mathbf{a}_3$ | = | $x_{10}a\,\mathbf{\hat{x}} + x_{10}a\,\mathbf{\hat{y}} + z_{10}a\,\mathbf{\hat{z}}$ | (48$h$) | O III |
| $\mathbf{B}_{47}$ | = | $z_{10}\,\mathbf{a}_1 + z_{10}\,\mathbf{a}_2 + (-2x_{10} - z_{10})\,\mathbf{a}_3$ | = | $-x_{10}a\,\mathbf{\hat{x}} - x_{10}a\,\mathbf{\hat{y}} + z_{10}a\,\mathbf{\hat{z}}$ | (48$h$) | O III |
| $\mathbf{B}_{48}$ | = | $(2x_{10} - z_{10})\,\mathbf{a}_1 + (-2x_{10} - z_{10})\,\mathbf{a}_2 + z_{10}\,\mathbf{a}_3$ | = | $-x_{10}a\,\mathbf{\hat{x}} + x_{10}a\,\mathbf{\hat{y}} - z_{10}a\,\mathbf{\hat{z}}$ | (48$h$) | O III |
| $\mathbf{B}_{49}$ | = | $(-2x_{10} - z_{10})\,\mathbf{a}_1 + (2x_{10} - z_{10})\,\mathbf{a}_2 + z_{10}\,\mathbf{a}_3$ | = | $x_{10}a\,\mathbf{\hat{x}} - x_{10}a\,\mathbf{\hat{y}} - z_{10}a\,\mathbf{\hat{z}}$ | (48$h$) | O III |
| $\mathbf{B}_{50}$ | = | $(2x_{10} - z_{10})\,\mathbf{a}_1 + z_{10}\,\mathbf{a}_2 + z_{10}\,\mathbf{a}_3$ | = | $z_{10}a\,\mathbf{\hat{x}} + x_{10}a\,\mathbf{\hat{y}} + x_{10}a\,\mathbf{\hat{z}}$ | (48$h$) | O III |
| $\mathbf{B}_{51}$ | = | $(-2x_{10} - z_{10})\,\mathbf{a}_1 + z_{10}\,\mathbf{a}_2 + z_{10}\,\mathbf{a}_3$ | = | $z_{10}a\,\mathbf{\hat{x}} - x_{10}a\,\mathbf{\hat{y}} - x_{10}a\,\mathbf{\hat{z}}$ | (48$h$) | O III |
| $\mathbf{B}_{52}$ | = | $z_{10}\,\mathbf{a}_1 + (2x_{10} - z_{10})\,\mathbf{a}_2 + (-2x_{10} - z_{10})\,\mathbf{a}_3$ | = | $-z_{10}a\,\mathbf{\hat{x}} - x_{10}a\,\mathbf{\hat{y}} + x_{10}a\,\mathbf{\hat{z}}$ | (48$h$) | O III |
| $\mathbf{B}_{53}$ | = | $z_{10}\,\mathbf{a}_1 + (-2x_{10} - z_{10})\,\mathbf{a}_2 + (2x_{10} - z_{10})\,\mathbf{a}_3$ | = | $-z_{10}a\,\mathbf{\hat{x}} + x_{10}a\,\mathbf{\hat{y}} - x_{10}a\,\mathbf{\hat{z}}$ | (48$h$) | O III |
| $\mathbf{B}_{54}$ | = | $z_{10}\,\mathbf{a}_1 + (2x_{10} - z_{10})\,\mathbf{a}_2 + z_{10}\,\mathbf{a}_3$ | = | $x_{10}a\,\mathbf{\hat{x}} + z_{10}a\,\mathbf{\hat{y}} + x_{10}a\,\mathbf{\hat{z}}$ | (48$h$) | O III |
| $\mathbf{B}_{55}$ | = | $z_{10}\,\mathbf{a}_1 + (-2x_{10} - z_{10})\,\mathbf{a}_2 + z_{10}\,\mathbf{a}_3$ | = | $-x_{10}a\,\mathbf{\hat{x}} + z_{10}a\,\mathbf{\hat{y}} - x_{10}a\,\mathbf{\hat{z}}$ | (48$h$) | O III |
| $\mathbf{B}_{56}$ | = | $(-2x_{10} - z_{10})\,\mathbf{a}_1 + z_{10}\,\mathbf{a}_2 + (2x_{10} - z_{10})\,\mathbf{a}_3$ | = | $x_{10}a\,\mathbf{\hat{x}} - z_{10}a\,\mathbf{\hat{y}} - x_{10}a\,\mathbf{\hat{z}}$ | (48$h$) | O III |
| $\mathbf{B}_{57}$ | = | $(2x_{10} - z_{10})\,\mathbf{a}_1 + z_{10}\,\mathbf{a}_2 + (-2x_{10} - z_{10})\,\mathbf{a}_3$ | = | $-x_{10}a\,\mathbf{\hat{x}} - z_{10}a\,\mathbf{\hat{y}} + x_{10}a\,\mathbf{\hat{z}}$ | (48$h$) | O III |

**Geometry files:**

- CIF: pp. 1791

- POSCAR: pp. 1792

# Murataite $[(Y,Na)_6(Zn,Fe)_5Ti_{12}O_{29}(O,F)_{10}F_4]$ Structure: A16B40C12D6E5_cF316_216_eh_e2g2h_h_f_be

http://aflow.org/prototype-encyclopedia/A16B40C12D6E5_cF316_216_eh_e2g2h_h_f_be

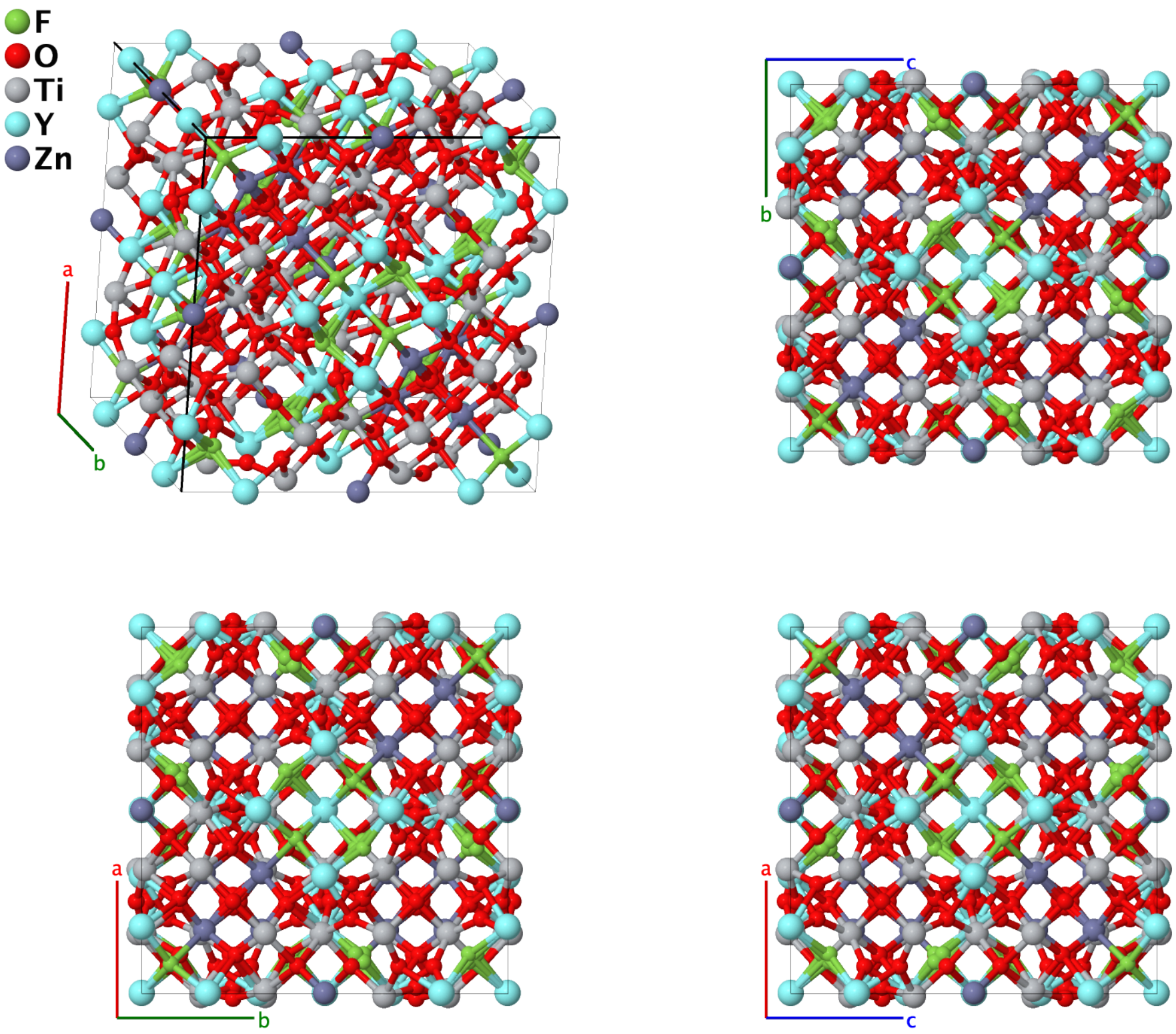

| | | |
|---|---|---|
| **Prototype** | : | $F_{16}O_{40}Ti_{12}Y_6Zn_5$ |
| **AFLOW prototype label** | : | A16B40C12D6E5_cF316_216_eh_e2g2h_h_f_be |
| ***Strukturbericht* designation** | : | None |
| **Pearson symbol** | : | cF316 |
| **Space group number** | : | 216 |
| **Space group symbol** | : | $F\bar{4}3m$ |
| **AFLOW prototype command** | : | `aflow --proto=A16B40C12D6E5_cF316_216_eh_e2g2h_h_f_be`<br>`--params=`$a, x_2, x_3, x_4, x_5, x_6, x_7, x_8, z_8, x_9, z_9, x_{10}, z_{10}, x_{11}, z_{11}$ |

- Most of the sites in this structure are somewhat disordered. The "nominal" composition is given as $F_{16}O_{40}Ti_{12}Y_6Zn_5$ by (Ercit, 1995), but as the CIF in (Downs, 2003) shows, even these labels are not quite correct. In our listing we label each Wyckoff position by the type of atom that has the largest concentration on that site. Following (Downs, 2003):
  - Site Zn-I has the composition $Zn_{0.89}Si_{0.11}$.
  - Site F-I has the composition $F_{0.55}O_{0.45}$.
  - Site O-I is pure oxygen.
  - Site Zn-II has the composition $Zn_{0.48}Fe_{0.25}Na_{0.16}Ti_{0.11}$.
  - Site Y has the composition $Y_{0.37}Na_{0.35}Mn_{0.03}HREE_{0.25}$, where "HREE" is a mixture of heavy Rare Earth elements.
  - Site O-II is pure oxygen, but only 8.3333% of the sites are occupied.
  - Site O-III has the composition $O_{0.7}F_{0.3}$.

- Site O-IV is pure oxygen.
- Site O-V is pure oxygen, but only 87% of the sites are occupied.
- Site F-II is pure fluorine, but only 33.333% of the sites are occupied.
- Site Ti has the composition $Ti_{0.76}Nb_{0.13}Na_{0.11}$.

---

**Face-centered Cubic primitive vectors:**

$$\mathbf{a}_1 = \frac{1}{2}\,a\,\hat{\mathbf{y}} + \frac{1}{2}\,a\,\hat{\mathbf{z}}$$
$$\mathbf{a}_2 = \frac{1}{2}\,a\,\hat{\mathbf{x}} + \frac{1}{2}\,a\,\hat{\mathbf{z}}$$
$$\mathbf{a}_3 = \frac{1}{2}\,a\,\hat{\mathbf{x}} + \frac{1}{2}\,a\,\hat{\mathbf{y}}$$

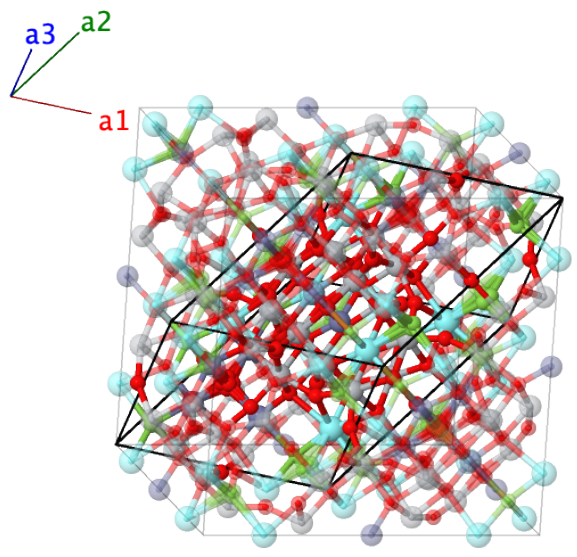

---

**Basis vectors:**

| | | Lattice Coordinates | | Cartesian Coordinates | Wyckoff Position | Atom Type |
|---|---|---|---|---|---|---|
| $\mathbf{B}_1$ | = | $\frac{1}{2}\,\mathbf{a}_1 + \frac{1}{2}\,\mathbf{a}_2 + \frac{1}{2}\,\mathbf{a}_3$ | = | $\frac{1}{2}a\,\hat{\mathbf{x}} + \frac{1}{2}a\,\hat{\mathbf{y}} + \frac{1}{2}a\,\hat{\mathbf{z}}$ | (4$b$) | Zn I |
| $\mathbf{B}_2$ | = | $x_2\,\mathbf{a}_1 + x_2\,\mathbf{a}_2 + x_2\,\mathbf{a}_3$ | = | $x_2a\,\hat{\mathbf{x}} + x_2a\,\hat{\mathbf{y}} + x_2a\,\hat{\mathbf{z}}$ | (16$e$) | F I |
| $\mathbf{B}_3$ | = | $x_2\,\mathbf{a}_1 + x_2\,\mathbf{a}_2 - 3x_2\,\mathbf{a}_3$ | = | $-x_2a\,\hat{\mathbf{x}} - x_2a\,\hat{\mathbf{y}} + x_2a\,\hat{\mathbf{z}}$ | (16$e$) | F I |
| $\mathbf{B}_4$ | = | $x_2\,\mathbf{a}_1 - 3x_2\,\mathbf{a}_2 + x_2\,\mathbf{a}_3$ | = | $-x_2a\,\hat{\mathbf{x}} + x_2a\,\hat{\mathbf{y}} - x_2a\,\hat{\mathbf{z}}$ | (16$e$) | F I |
| $\mathbf{B}_5$ | = | $-3x_2\,\mathbf{a}_1 + x_2\,\mathbf{a}_2 + x_2\,\mathbf{a}_3$ | = | $x_2a\,\hat{\mathbf{x}} - x_2a\,\hat{\mathbf{y}} - x_2a\,\hat{\mathbf{z}}$ | (16$e$) | F I |
| $\mathbf{B}_6$ | = | $x_3\,\mathbf{a}_1 + x_3\,\mathbf{a}_2 + x_3\,\mathbf{a}_3$ | = | $x_3a\,\hat{\mathbf{x}} + x_3a\,\hat{\mathbf{y}} + x_3a\,\hat{\mathbf{z}}$ | (16$e$) | O I |
| $\mathbf{B}_7$ | = | $x_3\,\mathbf{a}_1 + x_3\,\mathbf{a}_2 - 3x_3\,\mathbf{a}_3$ | = | $-x_3a\,\hat{\mathbf{x}} - x_3a\,\hat{\mathbf{y}} + x_3a\,\hat{\mathbf{z}}$ | (16$e$) | O I |
| $\mathbf{B}_8$ | = | $x_3\,\mathbf{a}_1 - 3x_3\,\mathbf{a}_2 + x_3\,\mathbf{a}_3$ | = | $-x_3a\,\hat{\mathbf{x}} + x_3a\,\hat{\mathbf{y}} - x_3a\,\hat{\mathbf{z}}$ | (16$e$) | O I |
| $\mathbf{B}_9$ | = | $-3x_3\,\mathbf{a}_1 + x_3\,\mathbf{a}_2 + x_3\,\mathbf{a}_3$ | = | $x_3a\,\hat{\mathbf{x}} - x_3a\,\hat{\mathbf{y}} - x_3a\,\hat{\mathbf{z}}$ | (16$e$) | O I |
| $\mathbf{B}_{10}$ | = | $x_4\,\mathbf{a}_1 + x_4\,\mathbf{a}_2 + x_4\,\mathbf{a}_3$ | = | $x_4a\,\hat{\mathbf{x}} + x_4a\,\hat{\mathbf{y}} + x_4a\,\hat{\mathbf{z}}$ | (16$e$) | Zn II |
| $\mathbf{B}_{11}$ | = | $x_4\,\mathbf{a}_1 + x_4\,\mathbf{a}_2 - 3x_4\,\mathbf{a}_3$ | = | $-x_4a\,\hat{\mathbf{x}} - x_4a\,\hat{\mathbf{y}} + x_4a\,\hat{\mathbf{z}}$ | (16$e$) | Zn II |
| $\mathbf{B}_{12}$ | = | $x_4\,\mathbf{a}_1 - 3x_4\,\mathbf{a}_2 + x_4\,\mathbf{a}_3$ | = | $-x_4a\,\hat{\mathbf{x}} + x_4a\,\hat{\mathbf{y}} - x_4a\,\hat{\mathbf{z}}$ | (16$e$) | Zn II |
| $\mathbf{B}_{13}$ | = | $-3x_4\,\mathbf{a}_1 + x_4\,\mathbf{a}_2 + x_4\,\mathbf{a}_3$ | = | $x_4a\,\hat{\mathbf{x}} - x_4a\,\hat{\mathbf{y}} - x_4a\,\hat{\mathbf{z}}$ | (16$e$) | Zn II |
| $\mathbf{B}_{14}$ | = | $-x_5\,\mathbf{a}_1 + x_5\,\mathbf{a}_2 + x_5\,\mathbf{a}_3$ | = | $x_5a\,\hat{\mathbf{x}}$ | (24$f$) | Y |
| $\mathbf{B}_{15}$ | = | $x_5\,\mathbf{a}_1 - x_5\,\mathbf{a}_2 - x_5\,\mathbf{a}_3$ | = | $-x_5a\,\hat{\mathbf{x}}$ | (24$f$) | Y |
| $\mathbf{B}_{16}$ | = | $x_5\,\mathbf{a}_1 - x_5\,\mathbf{a}_2 + x_5\,\mathbf{a}_3$ | = | $x_5a\,\hat{\mathbf{y}}$ | (24$f$) | Y |
| $\mathbf{B}_{17}$ | = | $-x_5\,\mathbf{a}_1 + x_5\,\mathbf{a}_2 - x_5\,\mathbf{a}_3$ | = | $-x_5a\,\hat{\mathbf{y}}$ | (24$f$) | Y |
| $\mathbf{B}_{18}$ | = | $x_5\,\mathbf{a}_1 + x_5\,\mathbf{a}_2 - x_5\,\mathbf{a}_3$ | = | $x_5a\,\hat{\mathbf{z}}$ | (24$f$) | Y |
| $\mathbf{B}_{19}$ | = | $-x_5\,\mathbf{a}_1 - x_5\,\mathbf{a}_2 + x_5\,\mathbf{a}_3$ | = | $-x_5a\,\hat{\mathbf{z}}$ | (24$f$) | Y |
| $\mathbf{B}_{20}$ | = | $\left(\frac{1}{2} - x_6\right)\,\mathbf{a}_1 + x_6\,\mathbf{a}_2 + x_6\,\mathbf{a}_3$ | = | $x_6a\,\hat{\mathbf{x}} + \frac{1}{4}a\,\hat{\mathbf{y}} + \frac{1}{4}a\,\hat{\mathbf{z}}$ | (24$g$) | O II |
| $\mathbf{B}_{21}$ | = | $x_6\,\mathbf{a}_1 + \left(\frac{1}{2} - x_6\right)\,\mathbf{a}_2 + \left(\frac{1}{2} - x_6\right)\,\mathbf{a}_3$ | = | $\left(\frac{1}{2} - x_6\right)a\,\hat{\mathbf{x}} + \frac{1}{4}a\,\hat{\mathbf{y}} + \frac{1}{4}a\,\hat{\mathbf{z}}$ | (24$g$) | O II |
| $\mathbf{B}_{22}$ | = | $x_6\,\mathbf{a}_1 + \left(\frac{1}{2} - x_6\right)\,\mathbf{a}_2 + x_6\,\mathbf{a}_3$ | = | $\frac{1}{4}a\,\hat{\mathbf{x}} + x_6a\,\hat{\mathbf{y}} + \frac{1}{4}a\,\hat{\mathbf{z}}$ | (24$g$) | O II |
| $\mathbf{B}_{23}$ | = | $\left(\frac{1}{2} - x_6\right)\,\mathbf{a}_1 + x_6\,\mathbf{a}_2 + \left(\frac{1}{2} - x_6\right)\,\mathbf{a}_3$ | = | $\frac{1}{4}a\,\hat{\mathbf{x}} + \left(\frac{1}{2} - x_6\right)a\,\hat{\mathbf{y}} + \frac{1}{4}a\,\hat{\mathbf{z}}$ | (24$g$) | O II |
| $\mathbf{B}_{24}$ | = | $x_6\,\mathbf{a}_1 + x_6\,\mathbf{a}_2 + \left(\frac{1}{2} - x_6\right)\,\mathbf{a}_3$ | = | $\frac{1}{4}a\,\mathbf{\hat{x}} + \frac{1}{4}a\,\mathbf{\hat{y}} + x_6a\,\mathbf{\hat{z}}$ | (24$g$) | O II |
| $\mathbf{B}_{25}$ | = | $\left(\frac{1}{2} - x_6\right)\,\mathbf{a}_1 + \left(\frac{1}{2} - x_6\right)\,\mathbf{a}_2 + x_6\,\mathbf{a}_3$ | = | $\frac{1}{4}a\,\mathbf{\hat{x}} + \frac{1}{4}a\,\mathbf{\hat{y}} + \left(\frac{1}{2} - x_6\right)a\,\mathbf{\hat{z}}$ | (24$g$) | O II |
| $\mathbf{B}_{26}$ | = | $\left(\frac{1}{2} - x_7\right)\,\mathbf{a}_1 + x_7\,\mathbf{a}_2 + x_7\,\mathbf{a}_3$ | = | $x_7a\,\mathbf{\hat{x}} + \frac{1}{4}a\,\mathbf{\hat{y}} + \frac{1}{4}a\,\mathbf{\hat{z}}$ | (24$g$) | O III |
| $\mathbf{B}_{27}$ | = | $x_7\,\mathbf{a}_1 + \left(\frac{1}{2} - x_7\right)\,\mathbf{a}_2 + \left(\frac{1}{2} - x_7\right)\,\mathbf{a}_3$ | = | $\left(\frac{1}{2} - x_7\right)a\,\mathbf{\hat{x}} + \frac{1}{4}a\,\mathbf{\hat{y}} + \frac{1}{4}a\,\mathbf{\hat{z}}$ | (24$g$) | O III |
| $\mathbf{B}_{28}$ | = | $x_7\,\mathbf{a}_1 + \left(\frac{1}{2} - x_7\right)\,\mathbf{a}_2 + x_7\,\mathbf{a}_3$ | = | $\frac{1}{4}a\,\mathbf{\hat{x}} + x_7a\,\mathbf{\hat{y}} + \frac{1}{4}a\,\mathbf{\hat{z}}$ | (24$g$) | O III |
| $\mathbf{B}_{29}$ | = | $\left(\frac{1}{2} - x_7\right)\,\mathbf{a}_1 + x_7\,\mathbf{a}_2 + \left(\frac{1}{2} - x_7\right)\,\mathbf{a}_3$ | = | $\frac{1}{4}a\,\mathbf{\hat{x}} + \left(\frac{1}{2} - x_7\right)a\,\mathbf{\hat{y}} + \frac{1}{4}a\,\mathbf{\hat{z}}$ | (24$g$) | O III |
| $\mathbf{B}_{30}$ | = | $x_7\,\mathbf{a}_1 + x_7\,\mathbf{a}_2 + \left(\frac{1}{2} - x_7\right)\,\mathbf{a}_3$ | = | $\frac{1}{4}a\,\mathbf{\hat{x}} + \frac{1}{4}a\,\mathbf{\hat{y}} + x_7a\,\mathbf{\hat{z}}$ | (24$g$) | O III |
| $\mathbf{B}_{31}$ | = | $\left(\frac{1}{2} - x_7\right)\,\mathbf{a}_1 + \left(\frac{1}{2} - x_7\right)\,\mathbf{a}_2 + x_7\,\mathbf{a}_3$ | = | $\frac{1}{4}a\,\mathbf{\hat{x}} + \frac{1}{4}a\,\mathbf{\hat{y}} + \left(\frac{1}{2} - x_7\right)a\,\mathbf{\hat{z}}$ | (24$g$) | O III |
| $\mathbf{B}_{32}$ | = | $z_8\,\mathbf{a}_1 + z_8\,\mathbf{a}_2 + (2x_8 - z_8)\,\mathbf{a}_3$ | = | $x_8a\,\mathbf{\hat{x}} + x_8a\,\mathbf{\hat{y}} + z_8a\,\mathbf{\hat{z}}$ | (48$h$) | F II |
| $\mathbf{B}_{33}$ | = | $z_8\,\mathbf{a}_1 + z_8\,\mathbf{a}_2 + (-2x_8 - z_8)\,\mathbf{a}_3$ | = | $-x_8a\,\mathbf{\hat{x}} - x_8a\,\mathbf{\hat{y}} + z_8a\,\mathbf{\hat{z}}$ | (48$h$) | F II |
| $\mathbf{B}_{34}$ | = | $(2x_8 - z_8)\,\mathbf{a}_1 + (-2x_8 - z_8)\,\mathbf{a}_2 + z_8\,\mathbf{a}_3$ | = | $-x_8a\,\mathbf{\hat{x}} + x_8a\,\mathbf{\hat{y}} - z_8a\,\mathbf{\hat{z}}$ | (48$h$) | F II |
| $\mathbf{B}_{35}$ | = | $(-2x_8 - z_8)\,\mathbf{a}_1 + (2x_8 - z_8)\,\mathbf{a}_2 + z_8\,\mathbf{a}_3$ | = | $x_8a\,\mathbf{\hat{x}} - x_8a\,\mathbf{\hat{y}} - z_8a\,\mathbf{\hat{z}}$ | (48$h$) | F II |
| $\mathbf{B}_{36}$ | = | $(2x_8 - z_8)\,\mathbf{a}_1 + z_8\,\mathbf{a}_2 + z_8\,\mathbf{a}_3$ | = | $z_8a\,\mathbf{\hat{x}} + x_8a\,\mathbf{\hat{y}} + x_8a\,\mathbf{\hat{z}}$ | (48$h$) | F II |
| $\mathbf{B}_{37}$ | = | $(-2x_8 - z_8)\,\mathbf{a}_1 + z_8\,\mathbf{a}_2 + z_8\,\mathbf{a}_3$ | = | $z_8a\,\mathbf{\hat{x}} - x_8a\,\mathbf{\hat{y}} - x_8a\,\mathbf{\hat{z}}$ | (48$h$) | F II |
| $\mathbf{B}_{38}$ | = | $z_8\,\mathbf{a}_1 + (2x_8 - z_8)\,\mathbf{a}_2 + (-2x_8 - z_8)\,\mathbf{a}_3$ | = | $-z_8a\,\mathbf{\hat{x}} - x_8a\,\mathbf{\hat{y}} + x_8a\,\mathbf{\hat{z}}$ | (48$h$) | F II |
| $\mathbf{B}_{39}$ | = | $z_8\,\mathbf{a}_1 + (-2x_8 - z_8)\,\mathbf{a}_2 + (2x_8 - z_8)\,\mathbf{a}_3$ | = | $-z_8a\,\mathbf{\hat{x}} + x_8a\,\mathbf{\hat{y}} - x_8a\,\mathbf{\hat{z}}$ | (48$h$) | F II |
| $\mathbf{B}_{40}$ | = | $z_8\,\mathbf{a}_1 + (2x_8 - z_8)\,\mathbf{a}_2 + z_8\,\mathbf{a}_3$ | = | $x_8a\,\mathbf{\hat{x}} + z_8a\,\mathbf{\hat{y}} + x_8a\,\mathbf{\hat{z}}$ | (48$h$) | F II |
| $\mathbf{B}_{41}$ | = | $z_8\,\mathbf{a}_1 + (-2x_8 - z_8)\,\mathbf{a}_2 + z_8\,\mathbf{a}_3$ | = | $-x_8a\,\mathbf{\hat{x}} + z_8a\,\mathbf{\hat{y}} - x_8a\,\mathbf{\hat{z}}$ | (48$h$) | F II |
| $\mathbf{B}_{42}$ | = | $(-2x_8 - z_8)\,\mathbf{a}_1 + z_8\,\mathbf{a}_2 + (2x_8 - z_8)\,\mathbf{a}_3$ | = | $x_8a\,\mathbf{\hat{x}} - z_8a\,\mathbf{\hat{y}} - x_8a\,\mathbf{\hat{z}}$ | (48$h$) | F II |
| $\mathbf{B}_{43}$ | = | $(2x_8 - z_8)\,\mathbf{a}_1 + z_8\,\mathbf{a}_2 + (-2x_8 - z_8)\,\mathbf{a}_3$ | = | $-x_8a\,\mathbf{\hat{x}} - z_8a\,\mathbf{\hat{y}} + x_8a\,\mathbf{\hat{z}}$ | (48$h$) | F II |
| $\mathbf{B}_{44}$ | = | $z_9\,\mathbf{a}_1 + z_9\,\mathbf{a}_2 + (2x_9 - z_9)\,\mathbf{a}_3$ | = | $x_9a\,\mathbf{\hat{x}} + x_9a\,\mathbf{\hat{y}} + z_9a\,\mathbf{\hat{z}}$ | (48$h$) | O IV |
| $\mathbf{B}_{45}$ | = | $z_9\,\mathbf{a}_1 + z_9\,\mathbf{a}_2 + (-2x_9 - z_9)\,\mathbf{a}_3$ | = | $-x_9a\,\mathbf{\hat{x}} - x_9a\,\mathbf{\hat{y}} + z_9a\,\mathbf{\hat{z}}$ | (48$h$) | O IV |
| $\mathbf{B}_{46}$ | = | $(2x_9 - z_9)\,\mathbf{a}_1 + (-2x_9 - z_9)\,\mathbf{a}_2 + z_9\,\mathbf{a}_3$ | = | $-x_9a\,\mathbf{\hat{x}} + x_9a\,\mathbf{\hat{y}} - z_9a\,\mathbf{\hat{z}}$ | (48$h$) | O IV |
| $\mathbf{B}_{47}$ | = | $(-2x_9 - z_9)\,\mathbf{a}_1 + (2x_9 - z_9)\,\mathbf{a}_2 + z_9\,\mathbf{a}_3$ | = | $x_9a\,\mathbf{\hat{x}} - x_9a\,\mathbf{\hat{y}} - z_9a\,\mathbf{\hat{z}}$ | (48$h$) | O IV |
| $\mathbf{B}_{48}$ | = | $(2x_9 - z_9)\,\mathbf{a}_1 + z_9\,\mathbf{a}_2 + z_9\,\mathbf{a}_3$ | = | $z_9a\,\mathbf{\hat{x}} + x_9a\,\mathbf{\hat{y}} + x_9a\,\mathbf{\hat{z}}$ | (48$h$) | O IV |
| $\mathbf{B}_{49}$ | = | $(-2x_9 - z_9)\,\mathbf{a}_1 + z_9\,\mathbf{a}_2 + z_9\,\mathbf{a}_3$ | = | $z_9a\,\mathbf{\hat{x}} - x_9a\,\mathbf{\hat{y}} - x_9a\,\mathbf{\hat{z}}$ | (48$h$) | O IV |
| $\mathbf{B}_{50}$ | = | $z_9\,\mathbf{a}_1 + (2x_9 - z_9)\,\mathbf{a}_2 + (-2x_9 - z_9)\,\mathbf{a}_3$ | = | $-z_9a\,\mathbf{\hat{x}} - x_9a\,\mathbf{\hat{y}} + x_9a\,\mathbf{\hat{z}}$ | (48$h$) | O IV |
| $\mathbf{B}_{51}$ | = | $z_9\,\mathbf{a}_1 + (-2x_9 - z_9)\,\mathbf{a}_2 + (2x_9 - z_9)\,\mathbf{a}_3$ | = | $-z_9a\,\mathbf{\hat{x}} + x_9a\,\mathbf{\hat{y}} - x_9a\,\mathbf{\hat{z}}$ | (48$h$) | O IV |
| $\mathbf{B}_{52}$ | = | $z_9\,\mathbf{a}_1 + (2x_9 - z_9)\,\mathbf{a}_2 + z_9\,\mathbf{a}_3$ | = | $x_9a\,\mathbf{\hat{x}} + z_9a\,\mathbf{\hat{y}} + x_9a\,\mathbf{\hat{z}}$ | (48$h$) | O IV |
| $\mathbf{B}_{53}$ | = | $z_9\,\mathbf{a}_1 + (-2x_9 - z_9)\,\mathbf{a}_2 + z_9\,\mathbf{a}_3$ | = | $-x_9a\,\mathbf{\hat{x}} + z_9a\,\mathbf{\hat{y}} - x_9a\,\mathbf{\hat{z}}$ | (48$h$) | O IV |
| $\mathbf{B}_{54}$ | = | $(-2x_9 - z_9)\,\mathbf{a}_1 + z_9\,\mathbf{a}_2 + (2x_9 - z_9)\,\mathbf{a}_3$ | = | $x_9a\,\mathbf{\hat{x}} - z_9a\,\mathbf{\hat{y}} - x_9a\,\mathbf{\hat{z}}$ | (48$h$) | O IV |
| $\mathbf{B}_{55}$ | = | $(2x_9 - z_9)\,\mathbf{a}_1 + z_9\,\mathbf{a}_2 + (-2x_9 - z_9)\,\mathbf{a}_3$ | = | $-x_9a\,\mathbf{\hat{x}} - z_9a\,\mathbf{\hat{y}} + x_9a\,\mathbf{\hat{z}}$ | (48$h$) | O IV |
| $\mathbf{B}_{56}$ | = | $z_{10}\,\mathbf{a}_1 + z_{10}\,\mathbf{a}_2 + (2x_{10} - z_{10})\,\mathbf{a}_3$ | = | $x_{10}a\,\mathbf{\hat{x}} + x_{10}a\,\mathbf{\hat{y}} + z_{10}a\,\mathbf{\hat{z}}$ | (48$h$) | O V |
| $\mathbf{B}_{57}$ | = | $z_{10}\,\mathbf{a}_1 + z_{10}\,\mathbf{a}_2 + (-2x_{10} - z_{10})\,\mathbf{a}_3$ | = | $-x_{10}a\,\mathbf{\hat{x}} - x_{10}a\,\mathbf{\hat{y}} + z_{10}a\,\mathbf{\hat{z}}$ | (48$h$) | O V |
| $\mathbf{B}_{58}$ | = | $(2x_{10} - z_{10})\,\mathbf{a}_1 + (-2x_{10} - z_{10})\,\mathbf{a}_2 + z_{10}\,\mathbf{a}_3$ | = | $-x_{10}a\,\mathbf{\hat{x}} + x_{10}a\,\mathbf{\hat{y}} - z_{10}a\,\mathbf{\hat{z}}$ | (48$h$) | O V |
| $\mathbf{B}_{59}$ | = | $(-2x_{10} - z_{10})\,\mathbf{a}_1 + (2x_{10} - z_{10})\,\mathbf{a}_2 + z_{10}\,\mathbf{a}_3$ | = | $x_{10}a\,\mathbf{\hat{x}} - x_{10}a\,\mathbf{\hat{y}} - z_{10}a\,\mathbf{\hat{z}}$ | (48$h$) | O V |
| $\mathbf{B}_{60}$ | = | $(2x_{10} - z_{10})\,\mathbf{a}_1 + z_{10}\,\mathbf{a}_2 + z_{10}\,\mathbf{a}_3$ | = | $z_{10}a\,\hat{\mathbf{x}} + x_{10}a\,\hat{\mathbf{y}} + x_{10}a\,\hat{\mathbf{z}}$ | (48$h$) | O V |
| $\mathbf{B}_{61}$ | = | $(-2x_{10} - z_{10})\,\mathbf{a}_1 + z_{10}\,\mathbf{a}_2 + z_{10}\,\mathbf{a}_3$ | = | $z_{10}a\,\hat{\mathbf{x}} - x_{10}a\,\hat{\mathbf{y}} - x_{10}a\,\hat{\mathbf{z}}$ | (48$h$) | O V |
| $\mathbf{B}_{62}$ | = | $z_{10}\,\mathbf{a}_1 + (2x_{10} - z_{10})\,\mathbf{a}_2 + (-2x_{10} - z_{10})\,\mathbf{a}_3$ | = | $-z_{10}a\,\hat{\mathbf{x}} - x_{10}a\,\hat{\mathbf{y}} + x_{10}a\,\hat{\mathbf{z}}$ | (48$h$) | O V |
| $\mathbf{B}_{63}$ | = | $z_{10}\,\mathbf{a}_1 + (-2x_{10} - z_{10})\,\mathbf{a}_2 + (2x_{10} - z_{10})\,\mathbf{a}_3$ | = | $-z_{10}a\,\hat{\mathbf{x}} + x_{10}a\,\hat{\mathbf{y}} - x_{10}a\,\hat{\mathbf{z}}$ | (48$h$) | O V |
| $\mathbf{B}_{64}$ | = | $z_{10}\,\mathbf{a}_1 + (2x_{10} - z_{10})\,\mathbf{a}_2 + z_{10}\,\mathbf{a}_3$ | = | $x_{10}a\,\hat{\mathbf{x}} + z_{10}a\,\hat{\mathbf{y}} + x_{10}a\,\hat{\mathbf{z}}$ | (48$h$) | O V |
| $\mathbf{B}_{65}$ | = | $z_{10}\,\mathbf{a}_1 + (-2x_{10} - z_{10})\,\mathbf{a}_2 + z_{10}\,\mathbf{a}_3$ | = | $-x_{10}a\,\hat{\mathbf{x}} + z_{10}a\,\hat{\mathbf{y}} - x_{10}a\,\hat{\mathbf{z}}$ | (48$h$) | O V |
| $\mathbf{B}_{66}$ | = | $(-2x_{10} - z_{10})\,\mathbf{a}_1 + z_{10}\,\mathbf{a}_2 + (2x_{10} - z_{10})\,\mathbf{a}_3$ | = | $x_{10}a\,\hat{\mathbf{x}} - z_{10}a\,\hat{\mathbf{y}} - x_{10}a\,\hat{\mathbf{z}}$ | (48$h$) | O V |
| $\mathbf{B}_{67}$ | = | $(2x_{10} - z_{10})\,\mathbf{a}_1 + z_{10}\,\mathbf{a}_2 + (-2x_{10} - z_{10})\,\mathbf{a}_3$ | = | $-x_{10}a\,\hat{\mathbf{x}} - z_{10}a\,\hat{\mathbf{y}} + x_{10}a\,\hat{\mathbf{z}}$ | (48$h$) | O V |
| $\mathbf{B}_{68}$ | = | $z_{11}\,\mathbf{a}_1 + z_{11}\,\mathbf{a}_2 + (2x_{11} - z_{11})\,\mathbf{a}_3$ | = | $x_{11}a\,\hat{\mathbf{x}} + x_{11}a\,\hat{\mathbf{y}} + z_{11}a\,\hat{\mathbf{z}}$ | (48$h$) | Ti |
| $\mathbf{B}_{69}$ | = | $z_{11}\,\mathbf{a}_1 + z_{11}\,\mathbf{a}_2 + (-2x_{11} - z_{11})\,\mathbf{a}_3$ | = | $-x_{11}a\,\hat{\mathbf{x}} - x_{11}a\,\hat{\mathbf{y}} + z_{11}a\,\hat{\mathbf{z}}$ | (48$h$) | Ti |
| $\mathbf{B}_{70}$ | = | $(2x_{11} - z_{11})\,\mathbf{a}_1 + (-2x_{11} - z_{11})\,\mathbf{a}_2 + z_{11}\,\mathbf{a}_3$ | = | $-x_{11}a\,\hat{\mathbf{x}} + x_{11}a\,\hat{\mathbf{y}} - z_{11}a\,\hat{\mathbf{z}}$ | (48$h$) | Ti |
| $\mathbf{B}_{71}$ | = | $(-2x_{11} - z_{11})\,\mathbf{a}_1 + (2x_{11} - z_{11})\,\mathbf{a}_2 + z_{11}\,\mathbf{a}_3$ | = | $x_{11}a\,\hat{\mathbf{x}} - x_{11}a\,\hat{\mathbf{y}} - z_{11}a\,\hat{\mathbf{z}}$ | (48$h$) | Ti |
| $\mathbf{B}_{72}$ | = | $(2x_{11} - z_{11})\,\mathbf{a}_1 + z_{11}\,\mathbf{a}_2 + z_{11}\,\mathbf{a}_3$ | = | $z_{11}a\,\hat{\mathbf{x}} + x_{11}a\,\hat{\mathbf{y}} + x_{11}a\,\hat{\mathbf{z}}$ | (48$h$) | Ti |
| $\mathbf{B}_{73}$ | = | $(-2x_{11} - z_{11})\,\mathbf{a}_1 + z_{11}\,\mathbf{a}_2 + z_{11}\,\mathbf{a}_3$ | = | $z_{11}a\,\hat{\mathbf{x}} - x_{11}a\,\hat{\mathbf{y}} - x_{11}a\,\hat{\mathbf{z}}$ | (48$h$) | Ti |
| $\mathbf{B}_{74}$ | = | $z_{11}\,\mathbf{a}_1 + (2x_{11} - z_{11})\,\mathbf{a}_2 + (-2x_{11} - z_{11})\,\mathbf{a}_3$ | = | $-z_{11}a\,\hat{\mathbf{x}} - x_{11}a\,\hat{\mathbf{y}} + x_{11}a\,\hat{\mathbf{z}}$ | (48$h$) | Ti |
| $\mathbf{B}_{75}$ | = | $z_{11}\,\mathbf{a}_1 + (-2x_{11} - z_{11})\,\mathbf{a}_2 + (2x_{11} - z_{11})\,\mathbf{a}_3$ | = | $-z_{11}a\,\hat{\mathbf{x}} + x_{11}a\,\hat{\mathbf{y}} - x_{11}a\,\hat{\mathbf{z}}$ | (48$h$) | Ti |
| $\mathbf{B}_{76}$ | = | $z_{11}\,\mathbf{a}_1 + (2x_{11} - z_{11})\,\mathbf{a}_2 + z_{11}\,\mathbf{a}_3$ | = | $x_{11}a\,\hat{\mathbf{x}} + z_{11}a\,\hat{\mathbf{y}} + x_{11}a\,\hat{\mathbf{z}}$ | (48$h$) | Ti |
| $\mathbf{B}_{77}$ | = | $z_{11}\,\mathbf{a}_1 + (-2x_{11} - z_{11})\,\mathbf{a}_2 + z_{11}\,\mathbf{a}_3$ | = | $-x_{11}a\,\hat{\mathbf{x}} + z_{11}a\,\hat{\mathbf{y}} - x_{11}a\,\hat{\mathbf{z}}$ | (48$h$) | Ti |
| $\mathbf{B}_{78}$ | = | $(-2x_{11} - z_{11})\,\mathbf{a}_1 + z_{11}\,\mathbf{a}_2 + (2x_{11} - z_{11})\,\mathbf{a}_3$ | = | $x_{11}a\,\hat{\mathbf{x}} - z_{11}a\,\hat{\mathbf{y}} - x_{11}a\,\hat{\mathbf{z}}$ | (48$h$) | Ti |
| $\mathbf{B}_{79}$ | = | $(2x_{11} - z_{11})\,\mathbf{a}_1 + z_{11}\,\mathbf{a}_2 + (-2x_{11} - z_{11})\,\mathbf{a}_3$ | = | $-x_{11}a\,\hat{\mathbf{x}} - z_{11}a\,\hat{\mathbf{y}} + x_{11}a\,\hat{\mathbf{z}}$ | (48$h$) | Ti |

**Geometry files:**

- CIF: pp. 1793
- POSCAR: pp. 1793

# $Sm_{11}Cd_{45}$ Structure: A45B11_cF448_216_bd4efg5h_ac2eh

http://aflow.org/prototype-encyclopedia/A45B11_cF448_216_bd4efg5h_ac2eh

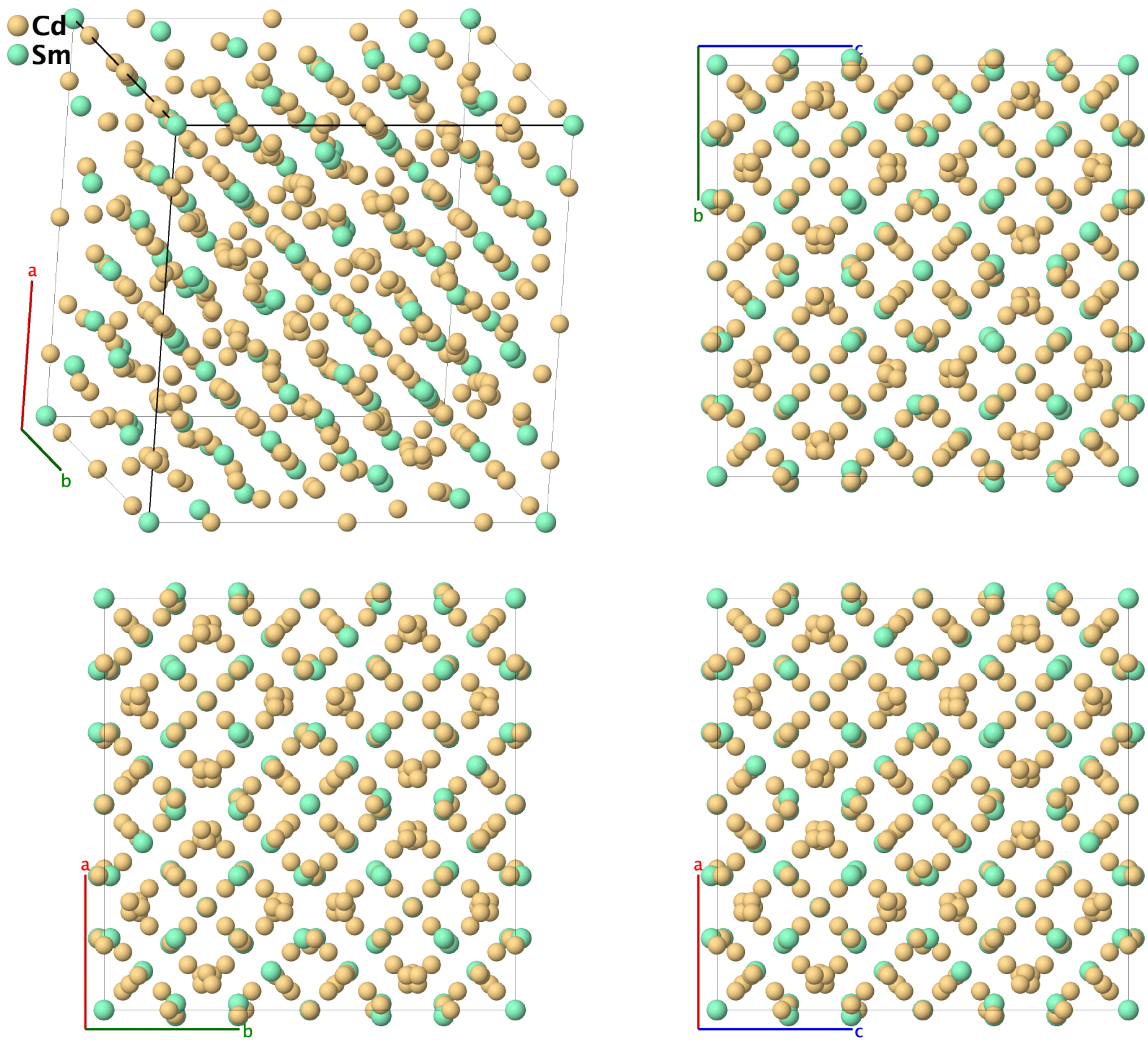

| | | |
|---|---|---|
| **Prototype** | : | $Cd_{45}Sm_{11}$ |
| **AFLOW prototype label** | : | A45B11_cF448_216_bd4efg5h_ac2eh |
| ***Strukturbericht* designation** | : | None |
| **Pearson symbol** | : | cF448 |
| **Space group number** | : | 216 |
| **Space group symbol** | : | $F\bar{4}3m$ |
| **AFLOW prototype command** | : | `aflow --proto=A45B11_cF448_216_bd4efg5h_ac2eh` `--params=`$a, x_5, x_6, x_7, x_8, x_9, x_{10}, x_{11}, x_{12}, x_{13}, z_{13}, x_{14}, z_{14}, x_{15}, z_{15}, x_{16}, z_{16}, x_{17}, z_{17}, x_{18}, z_{18}$ |

**Other compounds with this structure**

- $Dy_{11}Cd_{45}$, $Er_{11}Cd_{45}$, $Gd_{11}Cd_{45}$, $Ho_{11}Cd_{45}$, $Lu_{11}Cd_{45}$, $Nd_{11}Cd_{45}$, $Tb_{11}Cd_{45}$, $Tm_{11}Cd_{45}$, $Pr_{11}Cd_{45}$, $Y_{11}Cd_{45}$, $Ce_{11}Hg_{45}$, $Nd_{11}Hg_{45}$, $Pr_{11}Hg_{45}$, and $Sm_{11}Hg_{45}$

**Face-centered Cubic primitive vectors:**

$$\mathbf{a}_1 = \frac{1}{2}\,a\,\hat{\mathbf{y}} + \frac{1}{2}\,a\,\hat{\mathbf{z}}$$
$$\mathbf{a}_2 = \frac{1}{2}\,a\,\hat{\mathbf{x}} + \frac{1}{2}\,a\,\hat{\mathbf{z}}$$
$$\mathbf{a}_3 = \frac{1}{2}\,a\,\hat{\mathbf{x}} + \frac{1}{2}\,a\,\hat{\mathbf{y}}$$

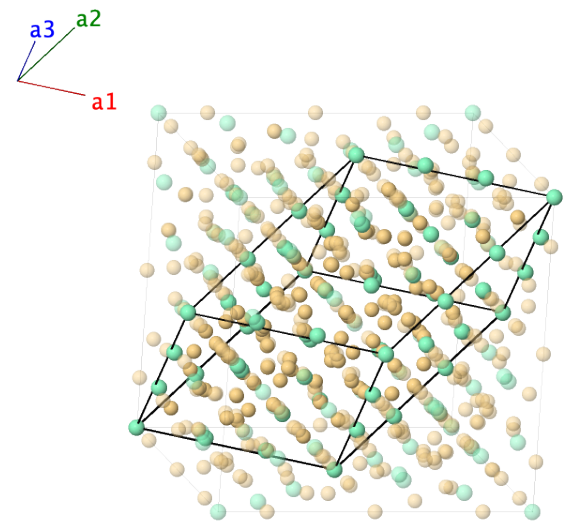

**Basis vectors:**

| | | Lattice Coordinates | | Cartesian Coordinates | Wyckoff Position | Atom Type |
|---|---|---|---|---|---|---|
| $\mathbf{B}_1$ | = | $0\,\mathbf{a}_1 + 0\,\mathbf{a}_2 + 0\,\mathbf{a}_3$ | = | $0\,\hat{\mathbf{x}} + 0\,\hat{\mathbf{y}} + 0\,\hat{\mathbf{z}}$ | (4$a$) | Sm I |
| $\mathbf{B}_2$ | = | $\frac{1}{2}\,\mathbf{a}_1 + \frac{1}{2}\,\mathbf{a}_2 + \frac{1}{2}\,\mathbf{a}_3$ | = | $\frac{1}{2}a\,\hat{\mathbf{x}} + \frac{1}{2}a\,\hat{\mathbf{y}} + \frac{1}{2}a\,\hat{\mathbf{z}}$ | (4$b$) | Cd I |
| $\mathbf{B}_3$ | = | $\frac{1}{4}\,\mathbf{a}_1 + \frac{1}{4}\,\mathbf{a}_2 + \frac{1}{4}\,\mathbf{a}_3$ | = | $\frac{1}{4}a\,\hat{\mathbf{x}} + \frac{1}{4}a\,\hat{\mathbf{y}} + \frac{1}{4}a\,\hat{\mathbf{z}}$ | (4$c$) | Sm II |
| $\mathbf{B}_4$ | = | $\frac{3}{4}\,\mathbf{a}_1 + \frac{3}{4}\,\mathbf{a}_2 + \frac{3}{4}\,\mathbf{a}_3$ | = | $\frac{3}{4}a\,\hat{\mathbf{x}} + \frac{3}{4}a\,\hat{\mathbf{y}} + \frac{3}{4}a\,\hat{\mathbf{z}}$ | (4$d$) | Cd II |
| $\mathbf{B}_5$ | = | $x_5\,\mathbf{a}_1 + x_5\,\mathbf{a}_2 + x_5\,\mathbf{a}_3$ | = | $x_5a\,\hat{\mathbf{x}} + x_5a\,\hat{\mathbf{y}} + x_5a\,\hat{\mathbf{z}}$ | (16$e$) | Cd III |
| $\mathbf{B}_6$ | = | $x_5\,\mathbf{a}_1 + x_5\,\mathbf{a}_2 - 3x_5\,\mathbf{a}_3$ | = | $-x_5a\,\hat{\mathbf{x}} - x_5a\,\hat{\mathbf{y}} + x_5a\,\hat{\mathbf{z}}$ | (16$e$) | Cd III |
| $\mathbf{B}_7$ | = | $x_5\,\mathbf{a}_1 - 3x_5\,\mathbf{a}_2 + x_5\,\mathbf{a}_3$ | = | $-x_5a\,\hat{\mathbf{x}} + x_5a\,\hat{\mathbf{y}} - x_5a\,\hat{\mathbf{z}}$ | (16$e$) | Cd III |
| $\mathbf{B}_8$ | = | $-3x_5\,\mathbf{a}_1 + x_5\,\mathbf{a}_2 + x_5\,\mathbf{a}_3$ | = | $x_5a\,\hat{\mathbf{x}} - x_5a\,\hat{\mathbf{y}} - x_5a\,\hat{\mathbf{z}}$ | (16$e$) | Cd III |
| $\mathbf{B}_9$ | = | $x_6\,\mathbf{a}_1 + x_6\,\mathbf{a}_2 + x_6\,\mathbf{a}_3$ | = | $x_6a\,\hat{\mathbf{x}} + x_6a\,\hat{\mathbf{y}} + x_6a\,\hat{\mathbf{z}}$ | (16$e$) | Cd IV |
| $\mathbf{B}_{10}$ | = | $x_6\,\mathbf{a}_1 + x_6\,\mathbf{a}_2 - 3x_6\,\mathbf{a}_3$ | = | $-x_6a\,\hat{\mathbf{x}} - x_6a\,\hat{\mathbf{y}} + x_6a\,\hat{\mathbf{z}}$ | (16$e$) | Cd IV |
| $\mathbf{B}_{11}$ | = | $x_6\,\mathbf{a}_1 - 3x_6\,\mathbf{a}_2 + x_6\,\mathbf{a}_3$ | = | $-x_6a\,\hat{\mathbf{x}} + x_6a\,\hat{\mathbf{y}} - x_6a\,\hat{\mathbf{z}}$ | (16$e$) | Cd IV |
| $\mathbf{B}_{12}$ | = | $-3x_6\,\mathbf{a}_1 + x_6\,\mathbf{a}_2 + x_6\,\mathbf{a}_3$ | = | $x_6a\,\hat{\mathbf{x}} - x_6a\,\hat{\mathbf{y}} - x_6a\,\hat{\mathbf{z}}$ | (16$e$) | Cd IV |
| $\mathbf{B}_{13}$ | = | $x_7\,\mathbf{a}_1 + x_7\,\mathbf{a}_2 + x_7\,\mathbf{a}_3$ | = | $x_7a\,\hat{\mathbf{x}} + x_7a\,\hat{\mathbf{y}} + x_7a\,\hat{\mathbf{z}}$ | (16$e$) | Cd V |
| $\mathbf{B}_{14}$ | = | $x_7\,\mathbf{a}_1 + x_7\,\mathbf{a}_2 - 3x_7\,\mathbf{a}_3$ | = | $-x_7a\,\hat{\mathbf{x}} - x_7a\,\hat{\mathbf{y}} + x_7a\,\hat{\mathbf{z}}$ | (16$e$) | Cd V |
| $\mathbf{B}_{15}$ | = | $x_7\,\mathbf{a}_1 - 3x_7\,\mathbf{a}_2 + x_7\,\mathbf{a}_3$ | = | $-x_7a\,\hat{\mathbf{x}} + x_7a\,\hat{\mathbf{y}} - x_7a\,\hat{\mathbf{z}}$ | (16$e$) | Cd V |
| $\mathbf{B}_{16}$ | = | $-3x_7\,\mathbf{a}_1 + x_7\,\mathbf{a}_2 + x_7\,\mathbf{a}_3$ | = | $x_7a\,\hat{\mathbf{x}} - x_7a\,\hat{\mathbf{y}} - x_7a\,\hat{\mathbf{z}}$ | (16$e$) | Cd V |
| $\mathbf{B}_{17}$ | = | $x_8\,\mathbf{a}_1 + x_8\,\mathbf{a}_2 + x_8\,\mathbf{a}_3$ | = | $x_8a\,\hat{\mathbf{x}} + x_8a\,\hat{\mathbf{y}} + x_8a\,\hat{\mathbf{z}}$ | (16$e$) | Cd VI |
| $\mathbf{B}_{18}$ | = | $x_8\,\mathbf{a}_1 + x_8\,\mathbf{a}_2 - 3x_8\,\mathbf{a}_3$ | = | $-x_8a\,\hat{\mathbf{x}} - x_8a\,\hat{\mathbf{y}} + x_8a\,\hat{\mathbf{z}}$ | (16$e$) | Cd VI |
| $\mathbf{B}_{19}$ | = | $x_8\,\mathbf{a}_1 - 3x_8\,\mathbf{a}_2 + x_8\,\mathbf{a}_3$ | = | $-x_8a\,\hat{\mathbf{x}} + x_8a\,\hat{\mathbf{y}} - x_8a\,\hat{\mathbf{z}}$ | (16$e$) | Cd VI |
| $\mathbf{B}_{20}$ | = | $-3x_8\,\mathbf{a}_1 + x_8\,\mathbf{a}_2 + x_8\,\mathbf{a}_3$ | = | $x_8a\,\hat{\mathbf{x}} - x_8a\,\hat{\mathbf{y}} - x_8a\,\hat{\mathbf{z}}$ | (16$e$) | Cd VI |
| $\mathbf{B}_{21}$ | = | $x_9\,\mathbf{a}_1 + x_9\,\mathbf{a}_2 + x_9\,\mathbf{a}_3$ | = | $x_9a\,\hat{\mathbf{x}} + x_9a\,\hat{\mathbf{y}} + x_9a\,\hat{\mathbf{z}}$ | (16$e$) | Sm III |
| $\mathbf{B}_{22}$ | = | $x_9\,\mathbf{a}_1 + x_9\,\mathbf{a}_2 - 3x_9\,\mathbf{a}_3$ | = | $-x_9a\,\hat{\mathbf{x}} - x_9a\,\hat{\mathbf{y}} + x_9a\,\hat{\mathbf{z}}$ | (16$e$) | Sm III |
| $\mathbf{B}_{23}$ | = | $x_9\,\mathbf{a}_1 - 3x_9\,\mathbf{a}_2 + x_9\,\mathbf{a}_3$ | = | $-x_9a\,\hat{\mathbf{x}} + x_9a\,\hat{\mathbf{y}} - x_9a\,\hat{\mathbf{z}}$ | (16$e$) | Sm III |
| $\mathbf{B}_{24}$ | = | $-3x_9\,\mathbf{a}_1 + x_9\,\mathbf{a}_2 + x_9\,\mathbf{a}_3$ | = | $x_9a\,\hat{\mathbf{x}} - x_9a\,\hat{\mathbf{y}} - x_9a\,\hat{\mathbf{z}}$ | (16$e$) | Sm III |
| $\mathbf{B}_{25}$ | = | $x_{10}\,\mathbf{a}_1 + x_{10}\,\mathbf{a}_2 + x_{10}\,\mathbf{a}_3$ | = | $x_{10}a\,\hat{\mathbf{x}} + x_{10}a\,\hat{\mathbf{y}} + x_{10}a\,\hat{\mathbf{z}}$ | (16$e$) | Sm IV |
| $\mathbf{B}_{26}$ | = | $x_{10}\,\mathbf{a}_1 + x_{10}\,\mathbf{a}_2 - 3x_{10}\,\mathbf{a}_3$ | = | $-x_{10}a\,\hat{\mathbf{x}} - x_{10}a\,\hat{\mathbf{y}} + x_{10}a\,\hat{\mathbf{z}}$ | (16$e$) | Sm IV |
| $\mathbf{B}_{27}$ | = | $x_{10}\,\mathbf{a}_1 - 3x_{10}\,\mathbf{a}_2 + x_{10}\,\mathbf{a}_3$ | = | $-x_{10}a\,\hat{\mathbf{x}} + x_{10}a\,\hat{\mathbf{y}} - x_{10}a\,\hat{\mathbf{z}}$ | $(16e)$ | Sm IV |
| $\mathbf{B}_{28}$ | = | $-3x_{10}\,\mathbf{a}_1 + x_{10}\,\mathbf{a}_2 + x_{10}\,\mathbf{a}_3$ | = | $x_{10}a\,\hat{\mathbf{x}} - x_{10}a\,\hat{\mathbf{y}} - x_{10}a\,\hat{\mathbf{z}}$ | $(16e)$ | Sm IV |
| $\mathbf{B}_{29}$ | = | $-x_{11}\,\mathbf{a}_1 + x_{11}\,\mathbf{a}_2 + x_{11}\,\mathbf{a}_3$ | = | $x_{11}a\,\hat{\mathbf{x}}$ | $(24f)$ | Cd VII |
| $\mathbf{B}_{30}$ | = | $x_{11}\,\mathbf{a}_1 - x_{11}\,\mathbf{a}_2 - x_{11}\,\mathbf{a}_3$ | = | $-x_{11}a\,\hat{\mathbf{x}}$ | $(24f)$ | Cd VII |
| $\mathbf{B}_{31}$ | = | $x_{11}\,\mathbf{a}_1 - x_{11}\,\mathbf{a}_2 + x_{11}\,\mathbf{a}_3$ | = | $x_{11}a\,\hat{\mathbf{y}}$ | $(24f)$ | Cd VII |
| $\mathbf{B}_{32}$ | = | $-x_{11}\,\mathbf{a}_1 + x_{11}\,\mathbf{a}_2 - x_{11}\,\mathbf{a}_3$ | = | $-x_{11}a\,\hat{\mathbf{y}}$ | $(24f)$ | Cd VII |
| $\mathbf{B}_{33}$ | = | $x_{11}\,\mathbf{a}_1 + x_{11}\,\mathbf{a}_2 - x_{11}\,\mathbf{a}_3$ | = | $x_{11}a\,\hat{\mathbf{z}}$ | $(24f)$ | Cd VII |
| $\mathbf{B}_{34}$ | = | $-x_{11}\,\mathbf{a}_1 - x_{11}\,\mathbf{a}_2 + x_{11}\,\mathbf{a}_3$ | = | $-x_{11}a\,\hat{\mathbf{z}}$ | $(24f)$ | Cd VII |
| $\mathbf{B}_{35}$ | = | $\left(\frac{1}{2} - x_{12}\right)\,\mathbf{a}_1 + x_{12}\,\mathbf{a}_2 + x_{12}\,\mathbf{a}_3$ | = | $x_{12}a\,\hat{\mathbf{x}} + \frac{1}{4}a\,\hat{\mathbf{y}} + \frac{1}{4}a\,\hat{\mathbf{z}}$ | $(24g)$ | Cd VIII |
| $\mathbf{B}_{36}$ | = | $x_{12}\,\mathbf{a}_1 + \left(\frac{1}{2} - x_{12}\right)\,\mathbf{a}_2 + \left(\frac{1}{2} - x_{12}\right)\,\mathbf{a}_3$ | = | $\left(\frac{1}{2} - x_{12}\right)a\,\hat{\mathbf{x}} + \frac{1}{4}a\,\hat{\mathbf{y}} + \frac{1}{4}a\,\hat{\mathbf{z}}$ | $(24g)$ | Cd VIII |
| $\mathbf{B}_{37}$ | = | $x_{12}\,\mathbf{a}_1 + \left(\frac{1}{2} - x_{12}\right)\,\mathbf{a}_2 + x_{12}\,\mathbf{a}_3$ | = | $\frac{1}{4}a\,\hat{\mathbf{x}} + x_{12}a\,\hat{\mathbf{y}} + \frac{1}{4}a\,\hat{\mathbf{z}}$ | $(24g)$ | Cd VIII |
| $\mathbf{B}_{38}$ | = | $\left(\frac{1}{2} - x_{12}\right)\,\mathbf{a}_1 + x_{12}\,\mathbf{a}_2 + \left(\frac{1}{2} - x_{12}\right)\,\mathbf{a}_3$ | = | $\frac{1}{4}a\,\hat{\mathbf{x}} + \left(\frac{1}{2} - x_{12}\right)a\,\hat{\mathbf{y}} + \frac{1}{4}a\,\hat{\mathbf{z}}$ | $(24g)$ | Cd VIII |
| $\mathbf{B}_{39}$ | = | $x_{12}\,\mathbf{a}_1 + x_{12}\,\mathbf{a}_2 + \left(\frac{1}{2} - x_{12}\right)\,\mathbf{a}_3$ | = | $\frac{1}{4}a\,\hat{\mathbf{x}} + \frac{1}{4}a\,\hat{\mathbf{y}} + x_{12}a\,\hat{\mathbf{z}}$ | $(24g)$ | Cd VIII |
| $\mathbf{B}_{40}$ | = | $\left(\frac{1}{2} - x_{12}\right)\,\mathbf{a}_1 + \left(\frac{1}{2} - x_{12}\right)\,\mathbf{a}_2 + x_{12}\,\mathbf{a}_3$ | = | $\frac{1}{4}a\,\hat{\mathbf{x}} + \frac{1}{4}a\,\hat{\mathbf{y}} + \left(\frac{1}{2} - x_{12}\right)a\,\hat{\mathbf{z}}$ | $(24g)$ | Cd VIII |
| $\mathbf{B}_{41}$ | = | $z_{13}\,\mathbf{a}_1 + z_{13}\,\mathbf{a}_2 + (2x_{13} - z_{13})\,\mathbf{a}_3$ | = | $x_{13}a\,\hat{\mathbf{x}} + x_{13}a\,\hat{\mathbf{y}} + z_{13}a\,\hat{\mathbf{z}}$ | $(48h)$ | Cd IX |
| $\mathbf{B}_{42}$ | = | $z_{13}\,\mathbf{a}_1 + z_{13}\,\mathbf{a}_2 + (-2x_{13} - z_{13})\,\mathbf{a}_3$ | = | $-x_{13}a\,\hat{\mathbf{x}} - x_{13}a\,\hat{\mathbf{y}} + z_{13}a\,\hat{\mathbf{z}}$ | $(48h)$ | Cd IX |
| $\mathbf{B}_{43}$ | = | $(2x_{13} - z_{13})\,\mathbf{a}_1 + (-2x_{13} - z_{13})\,\mathbf{a}_2 + z_{13}\,\mathbf{a}_3$ | = | $-x_{13}a\,\hat{\mathbf{x}} + x_{13}a\,\hat{\mathbf{y}} - z_{13}a\,\hat{\mathbf{z}}$ | $(48h)$ | Cd IX |
| $\mathbf{B}_{44}$ | = | $(-2x_{13} - z_{13})\,\mathbf{a}_1 + (2x_{13} - z_{13})\,\mathbf{a}_2 + z_{13}\,\mathbf{a}_3$ | = | $x_{13}a\,\hat{\mathbf{x}} - x_{13}a\,\hat{\mathbf{y}} - z_{13}a\,\hat{\mathbf{z}}$ | $(48h)$ | Cd IX |
| $\mathbf{B}_{45}$ | = | $(2x_{13} - z_{13})\,\mathbf{a}_1 + z_{13}\,\mathbf{a}_2 + z_{13}\,\mathbf{a}_3$ | = | $z_{13}a\,\hat{\mathbf{x}} + x_{13}a\,\hat{\mathbf{y}} + x_{13}a\,\hat{\mathbf{z}}$ | $(48h)$ | Cd IX |
| $\mathbf{B}_{46}$ | = | $(-2x_{13} - z_{13})\,\mathbf{a}_1 + z_{13}\,\mathbf{a}_2 + z_{13}\,\mathbf{a}_3$ | = | $z_{13}a\,\hat{\mathbf{x}} - x_{13}a\,\hat{\mathbf{y}} - x_{13}a\,\hat{\mathbf{z}}$ | $(48h)$ | Cd IX |
| $\mathbf{B}_{47}$ | = | $z_{13}\,\mathbf{a}_1 + (2x_{13} - z_{13})\,\mathbf{a}_2 + (-2x_{13} - z_{13})\,\mathbf{a}_3$ | = | $-z_{13}a\,\hat{\mathbf{x}} - x_{13}a\,\hat{\mathbf{y}} + x_{13}a\,\hat{\mathbf{z}}$ | $(48h)$ | Cd IX |
| $\mathbf{B}_{48}$ | = | $z_{13}\,\mathbf{a}_1 + (-2x_{13} - z_{13})\,\mathbf{a}_2 + (2x_{13} - z_{13})\,\mathbf{a}_3$ | = | $-z_{13}a\,\hat{\mathbf{x}} + x_{13}a\,\hat{\mathbf{y}} - x_{13}a\,\hat{\mathbf{z}}$ | $(48h)$ | Cd IX |
| $\mathbf{B}_{49}$ | = | $z_{13}\,\mathbf{a}_1 + (2x_{13} - z_{13})\,\mathbf{a}_2 + z_{13}\,\mathbf{a}_3$ | = | $x_{13}a\,\hat{\mathbf{x}} + z_{13}a\,\hat{\mathbf{y}} + x_{13}a\,\hat{\mathbf{z}}$ | $(48h)$ | Cd IX |
| $\mathbf{B}_{50}$ | = | $z_{13}\,\mathbf{a}_1 + (-2x_{13} - z_{13})\,\mathbf{a}_2 + z_{13}\,\mathbf{a}_3$ | = | $-x_{13}a\,\hat{\mathbf{x}} + z_{13}a\,\hat{\mathbf{y}} - x_{13}a\,\hat{\mathbf{z}}$ | $(48h)$ | Cd IX |
| $\mathbf{B}_{51}$ | = | $(-2x_{13} - z_{13})\,\mathbf{a}_1 + z_{13}\,\mathbf{a}_2 + (2x_{13} - z_{13})\,\mathbf{a}_3$ | = | $x_{13}a\,\hat{\mathbf{x}} - z_{13}a\,\hat{\mathbf{y}} - x_{13}a\,\hat{\mathbf{z}}$ | $(48h)$ | Cd IX |
| $\mathbf{B}_{52}$ | = | $(2x_{13} - z_{13})\,\mathbf{a}_1 + z_{13}\,\mathbf{a}_2 + (-2x_{13} - z_{13})\,\mathbf{a}_3$ | = | $-x_{13}a\,\hat{\mathbf{x}} - z_{13}a\,\hat{\mathbf{y}} + x_{13}a\,\hat{\mathbf{z}}$ | $(48h)$ | Cd IX |
| $\mathbf{B}_{53}$ | = | $z_{14}\,\mathbf{a}_1 + z_{14}\,\mathbf{a}_2 + (2x_{14} - z_{14})\,\mathbf{a}_3$ | = | $x_{14}a\,\hat{\mathbf{x}} + x_{14}a\,\hat{\mathbf{y}} + z_{14}a\,\hat{\mathbf{z}}$ | $(48h)$ | Cd X |
| $\mathbf{B}_{54}$ | = | $z_{14}\,\mathbf{a}_1 + z_{14}\,\mathbf{a}_2 + (-2x_{14} - z_{14})\,\mathbf{a}_3$ | = | $-x_{14}a\,\hat{\mathbf{x}} - x_{14}a\,\hat{\mathbf{y}} + z_{14}a\,\hat{\mathbf{z}}$ | $(48h)$ | Cd X |
| $\mathbf{B}_{55}$ | = | $(2x_{14} - z_{14})\,\mathbf{a}_1 + (-2x_{14} - z_{14})\,\mathbf{a}_2 + z_{14}\,\mathbf{a}_3$ | = | $-x_{14}a\,\hat{\mathbf{x}} + x_{14}a\,\hat{\mathbf{y}} - z_{14}a\,\hat{\mathbf{z}}$ | $(48h)$ | Cd X |
| $\mathbf{B}_{56}$ | = | $(-2x_{14} - z_{14})\,\mathbf{a}_1 + (2x_{14} - z_{14})\,\mathbf{a}_2 + z_{14}\,\mathbf{a}_3$ | = | $x_{14}a\,\hat{\mathbf{x}} - x_{14}a\,\hat{\mathbf{y}} - z_{14}a\,\hat{\mathbf{z}}$ | $(48h)$ | Cd X |
| $\mathbf{B}_{57}$ | = | $(2x_{14} - z_{14})\,\mathbf{a}_1 + z_{14}\,\mathbf{a}_2 + z_{14}\,\mathbf{a}_3$ | = | $z_{14}a\,\hat{\mathbf{x}} + x_{14}a\,\hat{\mathbf{y}} + x_{14}a\,\hat{\mathbf{z}}$ | $(48h)$ | Cd X |
| $\mathbf{B}_{58}$ | = | $(-2x_{14} - z_{14})\,\mathbf{a}_1 + z_{14}\,\mathbf{a}_2 + z_{14}\,\mathbf{a}_3$ | = | $z_{14}a\,\hat{\mathbf{x}} - x_{14}a\,\hat{\mathbf{y}} - x_{14}a\,\hat{\mathbf{z}}$ | $(48h)$ | Cd X |
| $\mathbf{B}_{59}$ | = | $z_{14}\,\mathbf{a}_1 + (2x_{14} - z_{14})\,\mathbf{a}_2 + (-2x_{14} - z_{14})\,\mathbf{a}_3$ | = | $-z_{14}a\,\hat{\mathbf{x}} - x_{14}a\,\hat{\mathbf{y}} + x_{14}a\,\hat{\mathbf{z}}$ | $(48h)$ | Cd X |
| $\mathbf{B}_{60}$ | = | $z_{14}\,\mathbf{a}_1 + (-2x_{14} - z_{14})\,\mathbf{a}_2 + (2x_{14} - z_{14})\,\mathbf{a}_3$ | = | $-z_{14}a\,\hat{\mathbf{x}} + x_{14}a\,\hat{\mathbf{y}} - x_{14}a\,\hat{\mathbf{z}}$ | $(48h)$ | Cd X |
| $\mathbf{B}_{61}$ | = | $z_{14}\,\mathbf{a}_1 + (2x_{14} - z_{14})\,\mathbf{a}_2 + z_{14}\,\mathbf{a}_3$ | = | $x_{14}a\,\hat{\mathbf{x}} + z_{14}a\,\hat{\mathbf{y}} + x_{14}a\,\hat{\mathbf{z}}$ | $(48h)$ | Cd X |
| $\mathbf{B}_{62}$ | = | $z_{14}\,\mathbf{a}_1 + (-2x_{14} - z_{14})\,\mathbf{a}_2 + z_{14}\,\mathbf{a}_3$ | = | $-x_{14}a\,\hat{\mathbf{x}} + z_{14}a\,\hat{\mathbf{y}} - x_{14}a\,\hat{\mathbf{z}}$ | $(48h)$ | Cd X |

| | | | | |
|---|---|---|---|---|
| $\mathbf{B}_{63}$ | $=\left(-2x_{14}-z_{14}\right)\mathbf{a}_1+z_{14}\mathbf{a}_2+\left(2x_{14}-z_{14}\right)\mathbf{a}_3$ | $=x_{14}a\,\hat{\mathbf{x}}-z_{14}a\,\hat{\mathbf{y}}-x_{14}a\,\hat{\mathbf{z}}$ | (48$h$) | Cd X |
| $\mathbf{B}_{64}$ | $=\left(2x_{14}-z_{14}\right)\mathbf{a}_1+z_{14}\mathbf{a}_2+\left(-2x_{14}-z_{14}\right)\mathbf{a}_3$ | $=-x_{14}a\,\hat{\mathbf{x}}-z_{14}a\,\hat{\mathbf{y}}+x_{14}a\,\hat{\mathbf{z}}$ | (48$h$) | Cd X |
| $\mathbf{B}_{65}$ | $=z_{15}\mathbf{a}_1+z_{15}\mathbf{a}_2+\left(2x_{15}-z_{15}\right)\mathbf{a}_3$ | $=x_{15}a\,\hat{\mathbf{x}}+x_{15}a\,\hat{\mathbf{y}}+z_{15}a\,\hat{\mathbf{z}}$ | (48$h$) | Cd XI |
| $\mathbf{B}_{66}$ | $=z_{15}\mathbf{a}_1+z_{15}\mathbf{a}_2+\left(-2x_{15}-z_{15}\right)\mathbf{a}_3$ | $=-x_{15}a\,\hat{\mathbf{x}}-x_{15}a\,\hat{\mathbf{y}}+z_{15}a\,\hat{\mathbf{z}}$ | (48$h$) | Cd XI |
| $\mathbf{B}_{67}$ | $=\left(2x_{15}-z_{15}\right)\mathbf{a}_1+\left(-2x_{15}-z_{15}\right)\mathbf{a}_2+z_{15}\mathbf{a}_3$ | $=-x_{15}a\,\hat{\mathbf{x}}+x_{15}a\,\hat{\mathbf{y}}-z_{15}a\,\hat{\mathbf{z}}$ | (48$h$) | Cd XI |
| $\mathbf{B}_{68}$ | $=\left(-2x_{15}-z_{15}\right)\mathbf{a}_1+\left(2x_{15}-z_{15}\right)\mathbf{a}_2+z_{15}\mathbf{a}_3$ | $=x_{15}a\,\hat{\mathbf{x}}-x_{15}a\,\hat{\mathbf{y}}-z_{15}a\,\hat{\mathbf{z}}$ | (48$h$) | Cd XI |
| $\mathbf{B}_{69}$ | $=\left(2x_{15}-z_{15}\right)\mathbf{a}_1+z_{15}\mathbf{a}_2+z_{15}\mathbf{a}_3$ | $=z_{15}a\,\hat{\mathbf{x}}+x_{15}a\,\hat{\mathbf{y}}+x_{15}a\,\hat{\mathbf{z}}$ | (48$h$) | Cd XI |
| $\mathbf{B}_{70}$ | $=\left(-2x_{15}-z_{15}\right)\mathbf{a}_1+z_{15}\mathbf{a}_2+z_{15}\mathbf{a}_3$ | $=z_{15}a\,\hat{\mathbf{x}}-x_{15}a\,\hat{\mathbf{y}}-x_{15}a\,\hat{\mathbf{z}}$ | (48$h$) | Cd XI |
| $\mathbf{B}_{71}$ | $=z_{15}\mathbf{a}_1+\left(2x_{15}-z_{15}\right)\mathbf{a}_2+\left(-2x_{15}-z_{15}\right)\mathbf{a}_3$ | $=-z_{15}a\,\hat{\mathbf{x}}-x_{15}a\,\hat{\mathbf{y}}+x_{15}a\,\hat{\mathbf{z}}$ | (48$h$) | Cd XI |
| $\mathbf{B}_{72}$ | $=z_{15}\mathbf{a}_1+\left(-2x_{15}-z_{15}\right)\mathbf{a}_2+\left(2x_{15}-z_{15}\right)\mathbf{a}_3$ | $=-z_{15}a\,\hat{\mathbf{x}}+x_{15}a\,\hat{\mathbf{y}}-x_{15}a\,\hat{\mathbf{z}}$ | (48$h$) | Cd XI |
| $\mathbf{B}_{73}$ | $=z_{15}\mathbf{a}_1+\left(2x_{15}-z_{15}\right)\mathbf{a}_2+z_{15}\mathbf{a}_3$ | $=x_{15}a\,\hat{\mathbf{x}}+z_{15}a\,\hat{\mathbf{y}}+x_{15}a\,\hat{\mathbf{z}}$ | (48$h$) | Cd XI |
| $\mathbf{B}_{74}$ | $=z_{15}\mathbf{a}_1+\left(-2x_{15}-z_{15}\right)\mathbf{a}_2+z_{15}\mathbf{a}_3$ | $=-x_{15}a\,\hat{\mathbf{x}}+z_{15}a\,\hat{\mathbf{y}}-x_{15}a\,\hat{\mathbf{z}}$ | (48$h$) | Cd XI |
| $\mathbf{B}_{75}$ | $=\left(-2x_{15}-z_{15}\right)\mathbf{a}_1+z_{15}\mathbf{a}_2+\left(2x_{15}-z_{15}\right)\mathbf{a}_3$ | $=x_{15}a\,\hat{\mathbf{x}}-z_{15}a\,\hat{\mathbf{y}}-x_{15}a\,\hat{\mathbf{z}}$ | (48$h$) | Cd XI |
| $\mathbf{B}_{76}$ | $=\left(2x_{15}-z_{15}\right)\mathbf{a}_1+z_{15}\mathbf{a}_2+\left(-2x_{15}-z_{15}\right)\mathbf{a}_3$ | $=-x_{15}a\,\hat{\mathbf{x}}-z_{15}a\,\hat{\mathbf{y}}+x_{15}a\,\hat{\mathbf{z}}$ | (48$h$) | Cd XI |
| $\mathbf{B}_{77}$ | $=z_{16}\mathbf{a}_1+z_{16}\mathbf{a}_2+\left(2x_{16}-z_{16}\right)\mathbf{a}_3$ | $=x_{16}a\,\hat{\mathbf{x}}+x_{16}a\,\hat{\mathbf{y}}+z_{16}a\,\hat{\mathbf{z}}$ | (48$h$) | Cd XII |
| $\mathbf{B}_{78}$ | $=z_{16}\mathbf{a}_1+z_{16}\mathbf{a}_2+\left(-2x_{16}-z_{16}\right)\mathbf{a}_3$ | $=-x_{16}a\,\hat{\mathbf{x}}-x_{16}a\,\hat{\mathbf{y}}+z_{16}a\,\hat{\mathbf{z}}$ | (48$h$) | Cd XII |
| $\mathbf{B}_{79}$ | $=\left(2x_{16}-z_{16}\right)\mathbf{a}_1+\left(-2x_{16}-z_{16}\right)\mathbf{a}_2+z_{16}\mathbf{a}_3$ | $=-x_{16}a\,\hat{\mathbf{x}}+x_{16}a\,\hat{\mathbf{y}}-z_{16}a\,\hat{\mathbf{z}}$ | (48$h$) | Cd XII |
| $\mathbf{B}_{80}$ | $=\left(-2x_{16}-z_{16}\right)\mathbf{a}_1+\left(2x_{16}-z_{16}\right)\mathbf{a}_2+z_{16}\mathbf{a}_3$ | $=x_{16}a\,\hat{\mathbf{x}}-x_{16}a\,\hat{\mathbf{y}}-z_{16}a\,\hat{\mathbf{z}}$ | (48$h$) | Cd XII |
| $\mathbf{B}_{81}$ | $=\left(2x_{16}-z_{16}\right)\mathbf{a}_1+z_{16}\mathbf{a}_2+z_{16}\mathbf{a}_3$ | $=z_{16}a\,\hat{\mathbf{x}}+x_{16}a\,\hat{\mathbf{y}}+x_{16}a\,\hat{\mathbf{z}}$ | (48$h$) | Cd XII |
| $\mathbf{B}_{82}$ | $=\left(-2x_{16}-z_{16}\right)\mathbf{a}_1+z_{16}\mathbf{a}_2+z_{16}\mathbf{a}_3$ | $=z_{16}a\,\hat{\mathbf{x}}-x_{16}a\,\hat{\mathbf{y}}-x_{16}a\,\hat{\mathbf{z}}$ | (48$h$) | Cd XII |
| $\mathbf{B}_{83}$ | $=z_{16}\mathbf{a}_1+\left(2x_{16}-z_{16}\right)\mathbf{a}_2+\left(-2x_{16}-z_{16}\right)\mathbf{a}_3$ | $=-z_{16}a\,\hat{\mathbf{x}}-x_{16}a\,\hat{\mathbf{y}}+x_{16}a\,\hat{\mathbf{z}}$ | (48$h$) | Cd XII |
| $\mathbf{B}_{84}$ | $=z_{16}\mathbf{a}_1+\left(-2x_{16}-z_{16}\right)\mathbf{a}_2+\left(2x_{16}-z_{16}\right)\mathbf{a}_3$ | $=-z_{16}a\,\hat{\mathbf{x}}+x_{16}a\,\hat{\mathbf{y}}-x_{16}a\,\hat{\mathbf{z}}$ | (48$h$) | Cd XII |
| $\mathbf{B}_{85}$ | $=z_{16}\mathbf{a}_1+\left(2x_{16}-z_{16}\right)\mathbf{a}_2+z_{16}\mathbf{a}_3$ | $=x_{16}a\,\hat{\mathbf{x}}+z_{16}a\,\hat{\mathbf{y}}+x_{16}a\,\hat{\mathbf{z}}$ | (48$h$) | Cd XII |
| $\mathbf{B}_{86}$ | $=z_{16}\mathbf{a}_1+\left(-2x_{16}-z_{16}\right)\mathbf{a}_2+z_{16}\mathbf{a}_3$ | $=-x_{16}a\,\hat{\mathbf{x}}+z_{16}a\,\hat{\mathbf{y}}-x_{16}a\,\hat{\mathbf{z}}$ | (48$h$) | Cd XII |
| $\mathbf{B}_{87}$ | $=\left(-2x_{16}-z_{16}\right)\mathbf{a}_1+z_{16}\mathbf{a}_2+\left(2x_{16}-z_{16}\right)\mathbf{a}_3$ | $=x_{16}a\,\hat{\mathbf{x}}-z_{16}a\,\hat{\mathbf{y}}-x_{16}a\,\hat{\mathbf{z}}$ | (48$h$) | Cd XII |
| $\mathbf{B}_{88}$ | $=\left(2x_{16}-z_{16}\right)\mathbf{a}_1+z_{16}\mathbf{a}_2+\left(-2x_{16}-z_{16}\right)\mathbf{a}_3$ | $=-x_{16}a\,\hat{\mathbf{x}}-z_{16}a\,\hat{\mathbf{y}}+x_{16}a\,\hat{\mathbf{z}}$ | (48$h$) | Cd XII |
| $\mathbf{B}_{89}$ | $=z_{17}\mathbf{a}_1+z_{17}\mathbf{a}_2+\left(2x_{17}-z_{17}\right)\mathbf{a}_3$ | $=x_{17}a\,\hat{\mathbf{x}}+x_{17}a\,\hat{\mathbf{y}}+z_{17}a\,\hat{\mathbf{z}}$ | (48$h$) | Cd XIII |
| $\mathbf{B}_{90}$ | $=z_{17}\mathbf{a}_1+z_{17}\mathbf{a}_2+\left(-2x_{17}-z_{17}\right)\mathbf{a}_3$ | $=-x_{17}a\,\hat{\mathbf{x}}-x_{17}a\,\hat{\mathbf{y}}+z_{17}a\,\hat{\mathbf{z}}$ | (48$h$) | Cd XIII |
| $\mathbf{B}_{91}$ | $=\left(2x_{17}-z_{17}\right)\mathbf{a}_1+\left(-2x_{17}-z_{17}\right)\mathbf{a}_2+z_{17}\mathbf{a}_3$ | $=-x_{17}a\,\hat{\mathbf{x}}+x_{17}a\,\hat{\mathbf{y}}-z_{17}a\,\hat{\mathbf{z}}$ | (48$h$) | Cd XIII |
| $\mathbf{B}_{92}$ | $=\left(-2x_{17}-z_{17}\right)\mathbf{a}_1+\left(2x_{17}-z_{17}\right)\mathbf{a}_2+z_{17}\mathbf{a}_3$ | $=x_{17}a\,\hat{\mathbf{x}}-x_{17}a\,\hat{\mathbf{y}}-z_{17}a\,\hat{\mathbf{z}}$ | (48$h$) | Cd XIII |
| $\mathbf{B}_{93}$ | $=\left(2x_{17}-z_{17}\right)\mathbf{a}_1+z_{17}\mathbf{a}_2+z_{17}\mathbf{a}_3$ | $=z_{17}a\,\hat{\mathbf{x}}+x_{17}a\,\hat{\mathbf{y}}+x_{17}a\,\hat{\mathbf{z}}$ | (48$h$) | Cd XIII |
| $\mathbf{B}_{94}$ | $=\left(-2x_{17}-z_{17}\right)\mathbf{a}_1+z_{17}\mathbf{a}_2+z_{17}\mathbf{a}_3$ | $=z_{17}a\,\hat{\mathbf{x}}-x_{17}a\,\hat{\mathbf{y}}-x_{17}a\,\hat{\mathbf{z}}$ | (48$h$) | Cd XIII |
| $\mathbf{B}_{95}$ | $=z_{17}\mathbf{a}_1+\left(2x_{17}-z_{17}\right)\mathbf{a}_2+\left(-2x_{17}-z_{17}\right)\mathbf{a}_3$ | $=-z_{17}a\,\hat{\mathbf{x}}-x_{17}a\,\hat{\mathbf{y}}+x_{17}a\,\hat{\mathbf{z}}$ | (48$h$) | Cd XIII |
| $\mathbf{B}_{96}$ | $=z_{17}\mathbf{a}_1+\left(-2x_{17}-z_{17}\right)\mathbf{a}_2+\left(2x_{17}-z_{17}\right)\mathbf{a}_3$ | $=-z_{17}a\,\hat{\mathbf{x}}+x_{17}a\,\hat{\mathbf{y}}-x_{17}a\,\hat{\mathbf{z}}$ | (48$h$) | Cd XIII |
| $\mathbf{B}_{97}$ | $=z_{17}\mathbf{a}_1+\left(2x_{17}-z_{17}\right)\mathbf{a}_2+z_{17}\mathbf{a}_3$ | $=x_{17}a\,\hat{\mathbf{x}}+z_{17}a\,\hat{\mathbf{y}}+x_{17}a\,\hat{\mathbf{z}}$ | (48$h$) | Cd XIII |
| $\mathbf{B}_{98}$ | $=z_{17}\mathbf{a}_1+\left(-2x_{17}-z_{17}\right)\mathbf{a}_2+z_{17}\mathbf{a}_3$ | $=-x_{17}a\,\hat{\mathbf{x}}+z_{17}a\,\hat{\mathbf{y}}-x_{17}a\,\hat{\mathbf{z}}$ | (48$h$) | Cd XIII |

| | | | | | | |
|---|---|---|---|---|---|---|
| $\mathbf{B}_{99}$ | = | $(-2x_{17} - z_{17})\,\mathbf{a}_1 + z_{17}\,\mathbf{a}_2 + (2x_{17} - z_{17})\,\mathbf{a}_3$ | = | $x_{17}a\,\hat{\mathbf{x}} - z_{17}a\,\hat{\mathbf{y}} - x_{17}a\,\hat{\mathbf{z}}$ | (48$h$) | Cd XIII |
| $\mathbf{B}_{100}$ | = | $(2x_{17} - z_{17})\,\mathbf{a}_1 + z_{17}\,\mathbf{a}_2 + (-2x_{17} - z_{17})\,\mathbf{a}_3$ | = | $-x_{17}a\,\hat{\mathbf{x}} - z_{17}a\,\hat{\mathbf{y}} + x_{17}a\,\hat{\mathbf{z}}$ | (48$h$) | Cd XIII |
| $\mathbf{B}_{101}$ | = | $z_{18}\,\mathbf{a}_1 + z_{18}\,\mathbf{a}_2 + (2x_{18} - z_{18})\,\mathbf{a}_3$ | = | $x_{18}a\,\hat{\mathbf{x}} + x_{18}a\,\hat{\mathbf{y}} + z_{18}a\,\hat{\mathbf{z}}$ | (48$h$) | Sm V |
| $\mathbf{B}_{102}$ | = | $z_{18}\,\mathbf{a}_1 + z_{18}\,\mathbf{a}_2 + (-2x_{18} - z_{18})\,\mathbf{a}_3$ | = | $-x_{18}a\,\hat{\mathbf{x}} - x_{18}a\,\hat{\mathbf{y}} + z_{18}a\,\hat{\mathbf{z}}$ | (48$h$) | Sm V |
| $\mathbf{B}_{103}$ | = | $(2x_{18} - z_{18})\,\mathbf{a}_1 + (-2x_{18} - z_{18})\,\mathbf{a}_2 + z_{18}\,\mathbf{a}_3$ | = | $-x_{18}a\,\hat{\mathbf{x}} + x_{18}a\,\hat{\mathbf{y}} - z_{18}a\,\hat{\mathbf{z}}$ | (48$h$) | Sm V |
| $\mathbf{B}_{104}$ | = | $(-2x_{18} - z_{18})\,\mathbf{a}_1 + (2x_{18} - z_{18})\,\mathbf{a}_2 + z_{18}\,\mathbf{a}_3$ | = | $x_{18}a\,\hat{\mathbf{x}} - x_{18}a\,\hat{\mathbf{y}} - z_{18}a\,\hat{\mathbf{z}}$ | (48$h$) | Sm V |
| $\mathbf{B}_{105}$ | = | $(2x_{18} - z_{18})\,\mathbf{a}_1 + z_{18}\,\mathbf{a}_2 + z_{18}\,\mathbf{a}_3$ | = | $z_{18}a\,\hat{\mathbf{x}} + x_{18}a\,\hat{\mathbf{y}} + x_{18}a\,\hat{\mathbf{z}}$ | (48$h$) | Sm V |
| $\mathbf{B}_{106}$ | = | $(-2x_{18} - z_{18})\,\mathbf{a}_1 + z_{18}\,\mathbf{a}_2 + z_{18}\,\mathbf{a}_3$ | = | $z_{18}a\,\hat{\mathbf{x}} - x_{18}a\,\hat{\mathbf{y}} - x_{18}a\,\hat{\mathbf{z}}$ | (48$h$) | Sm V |
| $\mathbf{B}_{107}$ | = | $z_{18}\,\mathbf{a}_1 + (2x_{18} - z_{18})\,\mathbf{a}_2 + (-2x_{18} - z_{18})\,\mathbf{a}_3$ | = | $-z_{18}a\,\hat{\mathbf{x}} - x_{18}a\,\hat{\mathbf{y}} + x_{18}a\,\hat{\mathbf{z}}$ | (48$h$) | Sm V |
| $\mathbf{B}_{108}$ | = | $z_{18}\,\mathbf{a}_1 + (-2x_{18} - z_{18})\,\mathbf{a}_2 + (2x_{18} - z_{18})\,\mathbf{a}_3$ | = | $-z_{18}a\,\hat{\mathbf{x}} + x_{18}a\,\hat{\mathbf{y}} - x_{18}a\,\hat{\mathbf{z}}$ | (48$h$) | Sm V |
| $\mathbf{B}_{109}$ | = | $z_{18}\,\mathbf{a}_1 + (2x_{18} - z_{18})\,\mathbf{a}_2 + z_{18}\,\mathbf{a}_3$ | = | $x_{18}a\,\hat{\mathbf{x}} + z_{18}a\,\hat{\mathbf{y}} + x_{18}a\,\hat{\mathbf{z}}$ | (48$h$) | Sm V |
| $\mathbf{B}_{110}$ | = | $z_{18}\,\mathbf{a}_1 + (-2x_{18} - z_{18})\,\mathbf{a}_2 + z_{18}\,\mathbf{a}_3$ | = | $-x_{18}a\,\hat{\mathbf{x}} + z_{18}a\,\hat{\mathbf{y}} - x_{18}a\,\hat{\mathbf{z}}$ | (48$h$) | Sm V |
| $\mathbf{B}_{111}$ | = | $(-2x_{18} - z_{18})\,\mathbf{a}_1 + z_{18}\,\mathbf{a}_2 + (2x_{18} - z_{18})\,\mathbf{a}_3$ | = | $x_{18}a\,\hat{\mathbf{x}} - z_{18}a\,\hat{\mathbf{y}} - x_{18}a\,\hat{\mathbf{z}}$ | (48$h$) | Sm V |
| $\mathbf{B}_{112}$ | = | $(2x_{18} - z_{18})\,\mathbf{a}_1 + z_{18}\,\mathbf{a}_2 + (-2x_{18} - z_{18})\,\mathbf{a}_3$ | = | $-x_{18}a\,\hat{\mathbf{x}} - z_{18}a\,\hat{\mathbf{y}} + x_{18}a\,\hat{\mathbf{z}}$ | (48$h$) | Sm V |

**Geometry files:**

- CIF: pp. 1794
- POSCAR: pp. 1795

# $Hg_2TiCu$ Inverse Heusler Structure: AB2C_cF16_216_b_ad_c

http://aflow.org/prototype-encyclopedia/AB2C_cF16_216_b_ad_c

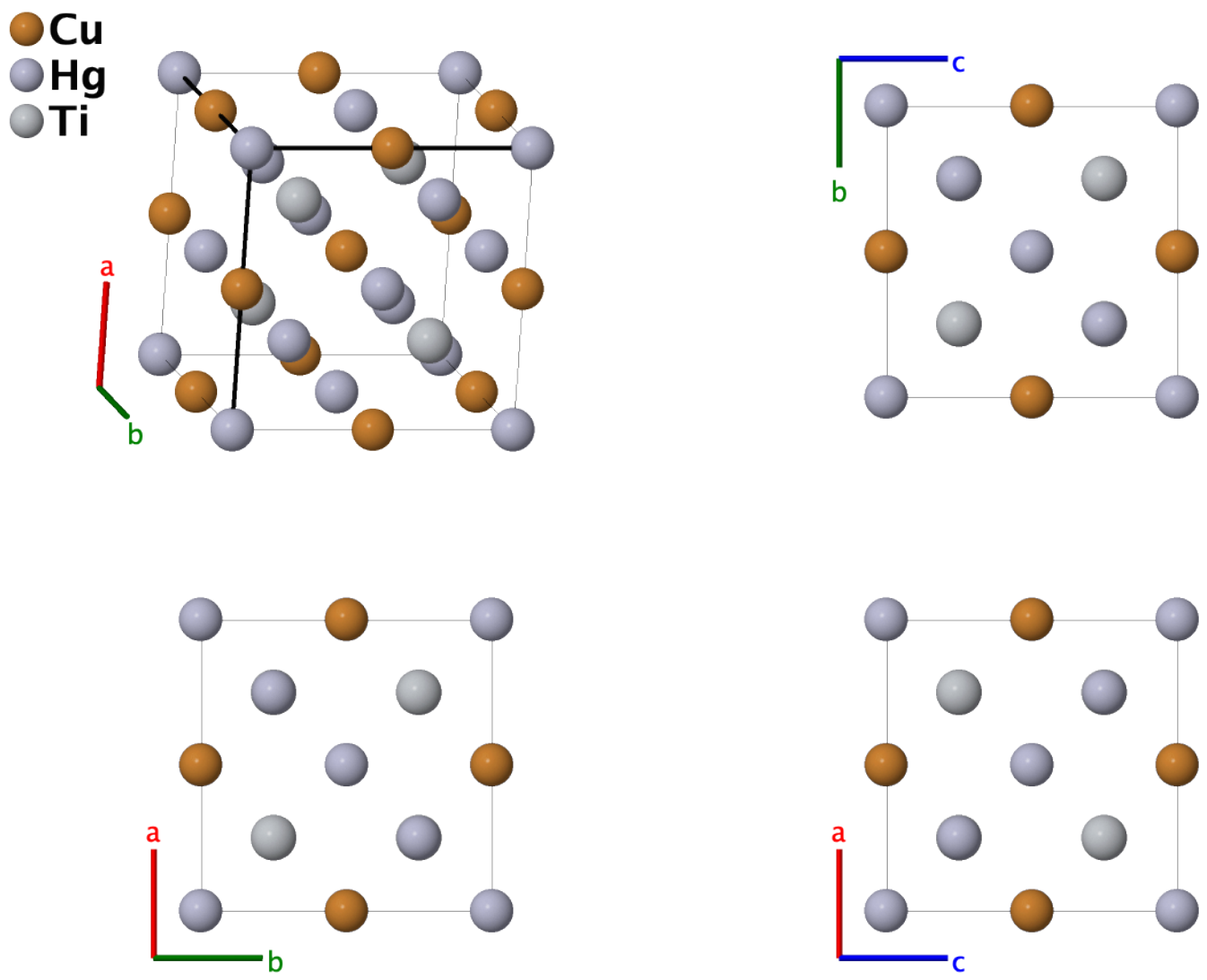

| | | |
|---|---|---|
| **Prototype** | : | $CuHg_2Ti$ |
| **AFLOW prototype label** | : | AB2C_cF16_216_b_ad_c |
| ***Strukturbericht* designation** | : | None |
| **Pearson symbol** | : | cF16 |
| **Space group number** | : | 216 |
| **Space group symbol** | : | $F\bar{4}3m$ |
| **AFLOW prototype command** | : | `aflow --proto=AB2C_cF16_216_b_ad_c --params=`$a$ |

**Other compounds with this structure**

- $Mn_2CoAl$, $Mn_2CoGa$, $Mn_2CoIn$, $Li_2AgSb$, $Li_2CuSn$, $Li_2CuSb$, $Li_2AgAl$, $Li_2AgIn$, $Li_2AgSb$, $Li_2AgSn$, $Li_2AgPb$, $Li_2AgBi$, $Li_2AuGa$, $Li_2AuIn$, $Li_2AuSb$, $Li_2AuSn$, $Li_2AuPb$, and $Li_2AuTl$

- Most of the literature on Inverse Heusler compounds identifies $Hg_2TiCu$ as the prototype structure, however (Villars, 2016) and (Villars, 2016a) use the $Li_2AgSb$ structure found in (Pauly, 1968) as the prototype. Although it is tempting to use the older paper, we follow the majority and use $Hg_2TiCu$ as the prototype.
- The inverse Heusler structure is a variation of the Quaternary Heusler structure, where the mercury atoms are on adjacent face-centered cubic sublattices, forming a diamond structure. Contrast this with the standard Heusler ($L1_2$) structure, where the like atoms are at second-neighbor positions in the fcc lattice.
- This structure is also referred to as the XA or $X_a$ structure.

### Face-centered Cubic primitive vectors:

$$\mathbf{a}_1 = \frac{1}{2}\,a\,\hat{\mathbf{y}} + \frac{1}{2}\,a\,\hat{\mathbf{z}}$$
$$\mathbf{a}_2 = \frac{1}{2}\,a\,\hat{\mathbf{x}} + \frac{1}{2}\,a\,\hat{\mathbf{z}}$$
$$\mathbf{a}_3 = \frac{1}{2}\,a\,\hat{\mathbf{x}} + \frac{1}{2}\,a\,\hat{\mathbf{y}}$$

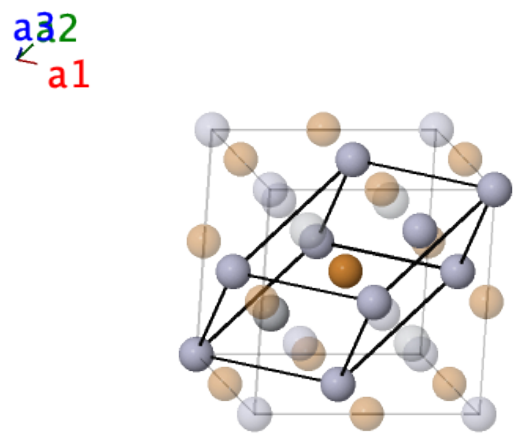

### Basis vectors:

| | | Lattice Coordinates | | Cartesian Coordinates | Wyckoff Position | Atom Type |
|---|---|---|---|---|---|---|
| $\mathbf{B}_1$ | = | $0\,\mathbf{a}_1 + 0\,\mathbf{a}_2 + 0\,\mathbf{a}_3$ | = | $0\,\hat{\mathbf{x}} + 0\,\hat{\mathbf{y}} + 0\,\hat{\mathbf{z}}$ | (4*a*) | Hg I |
| $\mathbf{B}_2$ | = | $\frac{1}{2}\,\mathbf{a}_1 + \frac{1}{2}\,\mathbf{a}_2 + \frac{1}{2}\,\mathbf{a}_3$ | = | $\frac{1}{2}a\,\hat{\mathbf{x}} + \frac{1}{2}a\,\hat{\mathbf{y}} + \frac{1}{2}a\,\hat{\mathbf{z}}$ | (4*b*) | Cu |
| $\mathbf{B}_3$ | = | $\frac{1}{4}\,\mathbf{a}_1 + \frac{1}{4}\,\mathbf{a}_2 + \frac{1}{4}\,\mathbf{a}_3$ | = | $\frac{1}{4}a\,\hat{\mathbf{x}} + \frac{1}{4}a\,\hat{\mathbf{y}} + \frac{1}{4}a\,\hat{\mathbf{z}}$ | (4*c*) | Ti |
| $\mathbf{B}_4$ | = | $\frac{3}{4}\,\mathbf{a}_1 + \frac{3}{4}\,\mathbf{a}_2 + \frac{3}{4}\,\mathbf{a}_3$ | = | $\frac{3}{4}a\,\hat{\mathbf{x}} + \frac{3}{4}a\,\hat{\mathbf{y}} + \frac{3}{4}a\,\hat{\mathbf{z}}$ | (4*d*) | Hg II |

### Found in:

### Geometry files:

- CIF: pp. 1795
- POSCAR: pp. 1796

# $GaMo_4S_8$ Structure: AB4C8_cF52_216_a_e_2e

http://aflow.org/prototype-encyclopedia/AB4C8_cF52_216_a_e_2e

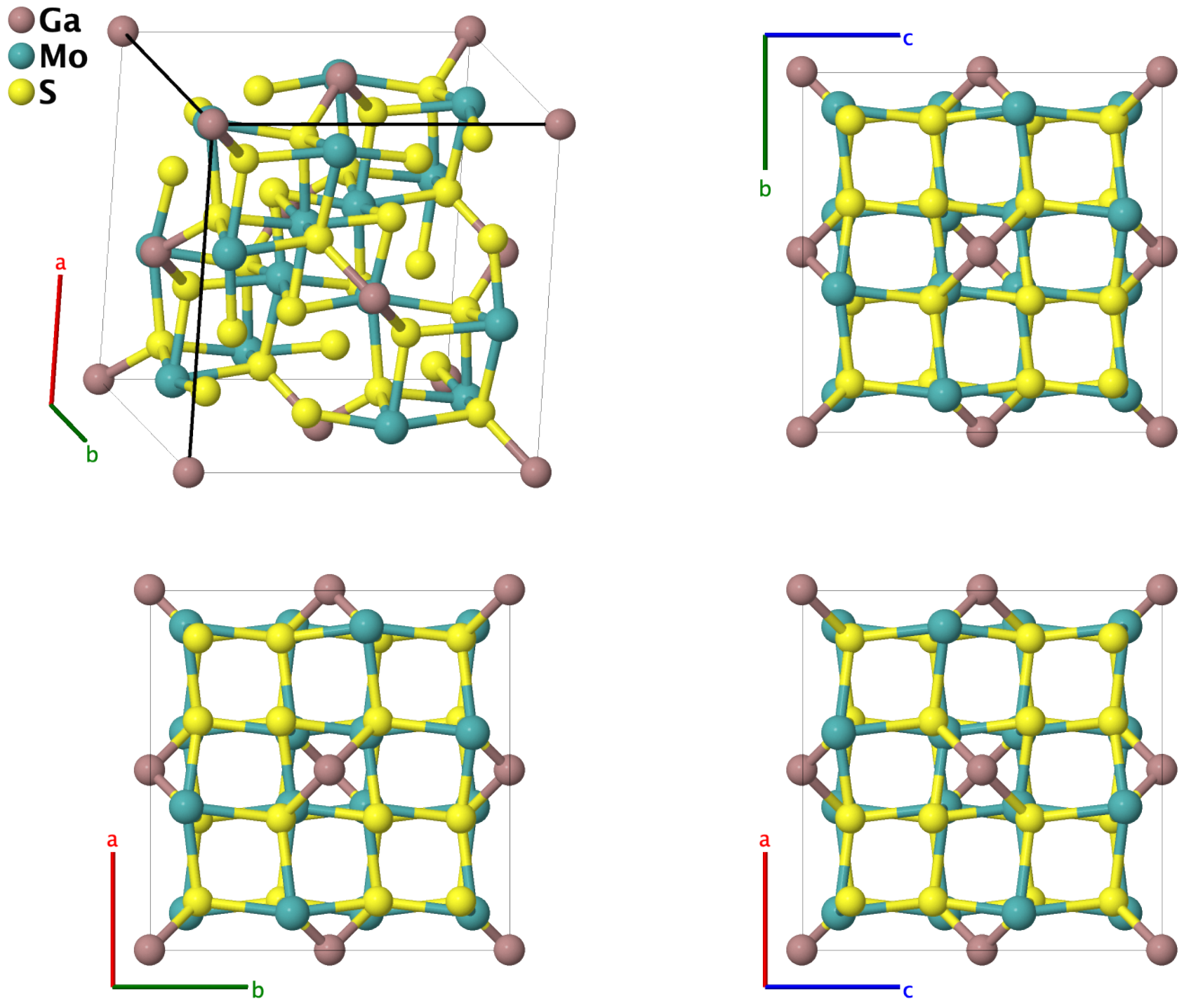

| | | |
|---|---|---|
| **Prototype** | : | $GaMo_4S_8$ |
| **AFLOW prototype label** | : | AB4C8_cF52_216_a_e_2e |
| ***Strukturbericht* designation** | : | None |
| **Pearson symbol** | : | cF52 |
| **Space group number** | : | 216 |
| **Space group symbol** | : | $F\bar{4}3m$ |
| **AFLOW prototype command** | : | `aflow --proto=AB4C8_cF52_216_a_e_2e` `--params=`$a, x_2, x_3, x_4$ |

### Other compounds with this structure

- $Co(Mo_2Re_2)S_8$, $Fe(Mo_2Re_2)S_8$, $GaMo_4S_4Te_4$, $GaMo_4S_8$, $GaMo_4Se_4Te_4$, $GaMo_4Se_8$, $GaNb_4S_8$, $GaNb_4Se_8$, $GaTa_4S_8$, $GaTa_4Se_8$, $GaV_4S_8$, $GaV_4Se_8$, $LaMo_4S_8$, $Ni(Mo_2Re_2)S_8$, and $Zn(Mo_2Re_2)S_8$

- (Ben Yaich, 1984) do not give the lattice constant for $GaMo_4S_8$. We infer it from their interatomic distances.

### Face-centered Cubic primitive vectors:

$$\mathbf{a}_1 = \frac{1}{2}\,a\,\hat{\mathbf{y}} + \frac{1}{2}\,a\,\hat{\mathbf{z}}$$
$$\mathbf{a}_2 = \frac{1}{2}\,a\,\hat{\mathbf{x}} + \frac{1}{2}\,a\,\hat{\mathbf{z}}$$
$$\mathbf{a}_3 = \frac{1}{2}\,a\,\hat{\mathbf{x}} + \frac{1}{2}\,a\,\hat{\mathbf{y}}$$

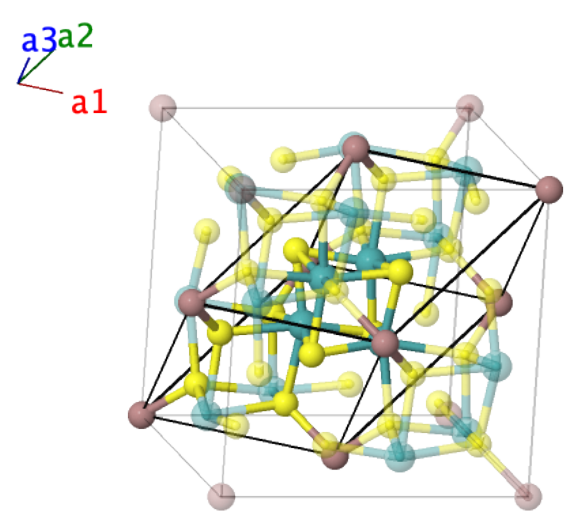

**Basis vectors:**

| | | Lattice Coordinates | | Cartesian Coordinates | Wyckoff Position | Atom Type |
|---|---|---|---|---|---|---|
| $\mathbf{B}_1$ | = | $0\,\mathbf{a}_1 + 0\,\mathbf{a}_2 + 0\,\mathbf{a}_3$ | = | $0\,\hat{\mathbf{x}} + 0\,\hat{\mathbf{y}} + 0\,\hat{\mathbf{z}}$ | (4$a$) | Ga |
| $\mathbf{B}_2$ | = | $x_2\,\mathbf{a}_1 + x_2\,\mathbf{a}_2 + x_2\,\mathbf{a}_3$ | = | $x_2a\,\hat{\mathbf{x}} + x_2a\,\hat{\mathbf{y}} + x_2a\,\hat{\mathbf{z}}$ | (16$e$) | Mo |
| $\mathbf{B}_3$ | = | $x_2\,\mathbf{a}_1 + x_2\,\mathbf{a}_2 - 3x_2\,\mathbf{a}_3$ | = | $-x_2a\,\hat{\mathbf{x}} - x_2a\,\hat{\mathbf{y}} + x_2a\,\hat{\mathbf{z}}$ | (16$e$) | Mo |
| $\mathbf{B}_4$ | = | $x_2\,\mathbf{a}_1 - 3x_2\,\mathbf{a}_2 + x_2\,\mathbf{a}_3$ | = | $-x_2a\,\hat{\mathbf{x}} + x_2a\,\hat{\mathbf{y}} - x_2a\,\hat{\mathbf{z}}$ | (16$e$) | Mo |
| $\mathbf{B}_5$ | = | $-3x_2\,\mathbf{a}_1 + x_2\,\mathbf{a}_2 + x_2\,\mathbf{a}_3$ | = | $x_2a\,\hat{\mathbf{x}} - x_2a\,\hat{\mathbf{y}} - x_2a\,\hat{\mathbf{z}}$ | (16$e$) | Mo |
| $\mathbf{B}_6$ | = | $x_3\,\mathbf{a}_1 + x_3\,\mathbf{a}_2 + x_3\,\mathbf{a}_3$ | = | $x_3a\,\hat{\mathbf{x}} + x_3a\,\hat{\mathbf{y}} + x_3a\,\hat{\mathbf{z}}$ | (16$e$) | S I |
| $\mathbf{B}_7$ | = | $x_3\,\mathbf{a}_1 + x_3\,\mathbf{a}_2 - 3x_3\,\mathbf{a}_3$ | = | $-x_3a\,\hat{\mathbf{x}} - x_3a\,\hat{\mathbf{y}} + x_3a\,\hat{\mathbf{z}}$ | (16$e$) | S I |
| $\mathbf{B}_8$ | = | $x_3\,\mathbf{a}_1 - 3x_3\,\mathbf{a}_2 + x_3\,\mathbf{a}_3$ | = | $-x_3a\,\hat{\mathbf{x}} + x_3a\,\hat{\mathbf{y}} - x_3a\,\hat{\mathbf{z}}$ | (16$e$) | S I |
| $\mathbf{B}_9$ | = | $-3x_3\,\mathbf{a}_1 + x_3\,\mathbf{a}_2 + x_3\,\mathbf{a}_3$ | = | $x_3a\,\hat{\mathbf{x}} - x_3a\,\hat{\mathbf{y}} - x_3a\,\hat{\mathbf{z}}$ | (16$e$) | S I |
| $\mathbf{B}_{10}$ | = | $x_4\,\mathbf{a}_1 + x_4\,\mathbf{a}_2 + x_4\,\mathbf{a}_3$ | = | $x_4a\,\hat{\mathbf{x}} + x_4a\,\hat{\mathbf{y}} + x_4a\,\hat{\mathbf{z}}$ | (16$e$) | S II |
| $\mathbf{B}_{11}$ | = | $x_4\,\mathbf{a}_1 + x_4\,\mathbf{a}_2 - 3x_4\,\mathbf{a}_3$ | = | $-x_4a\,\hat{\mathbf{x}} - x_4a\,\hat{\mathbf{y}} + x_4a\,\hat{\mathbf{z}}$ | (16$e$) | S II |
| $\mathbf{B}_{12}$ | = | $x_4\,\mathbf{a}_1 - 3x_4\,\mathbf{a}_2 + x_4\,\mathbf{a}_3$ | = | $-x_4a\,\hat{\mathbf{x}} + x_4a\,\hat{\mathbf{y}} - x_4a\,\hat{\mathbf{z}}$ | (16$e$) | S II |
| $\mathbf{B}_{13}$ | = | $-3x_4\,\mathbf{a}_1 + x_4\,\mathbf{a}_2 + x_4\,\mathbf{a}_3$ | = | $x_4a\,\hat{\mathbf{x}} - x_4a\,\hat{\mathbf{y}} - x_4a\,\hat{\mathbf{z}}$ | (16$e$) | S II |

**Geometry files:**

- CIF: pp. 1796
- POSCAR: pp. 1797

# High-Temperature Cubic $KClO_4$ ($H0_5$) Structure: ABC4_cF24_216_b_a_e

http://aflow.org/prototype-encyclopedia/ABC4_cF24_216_b_a_e

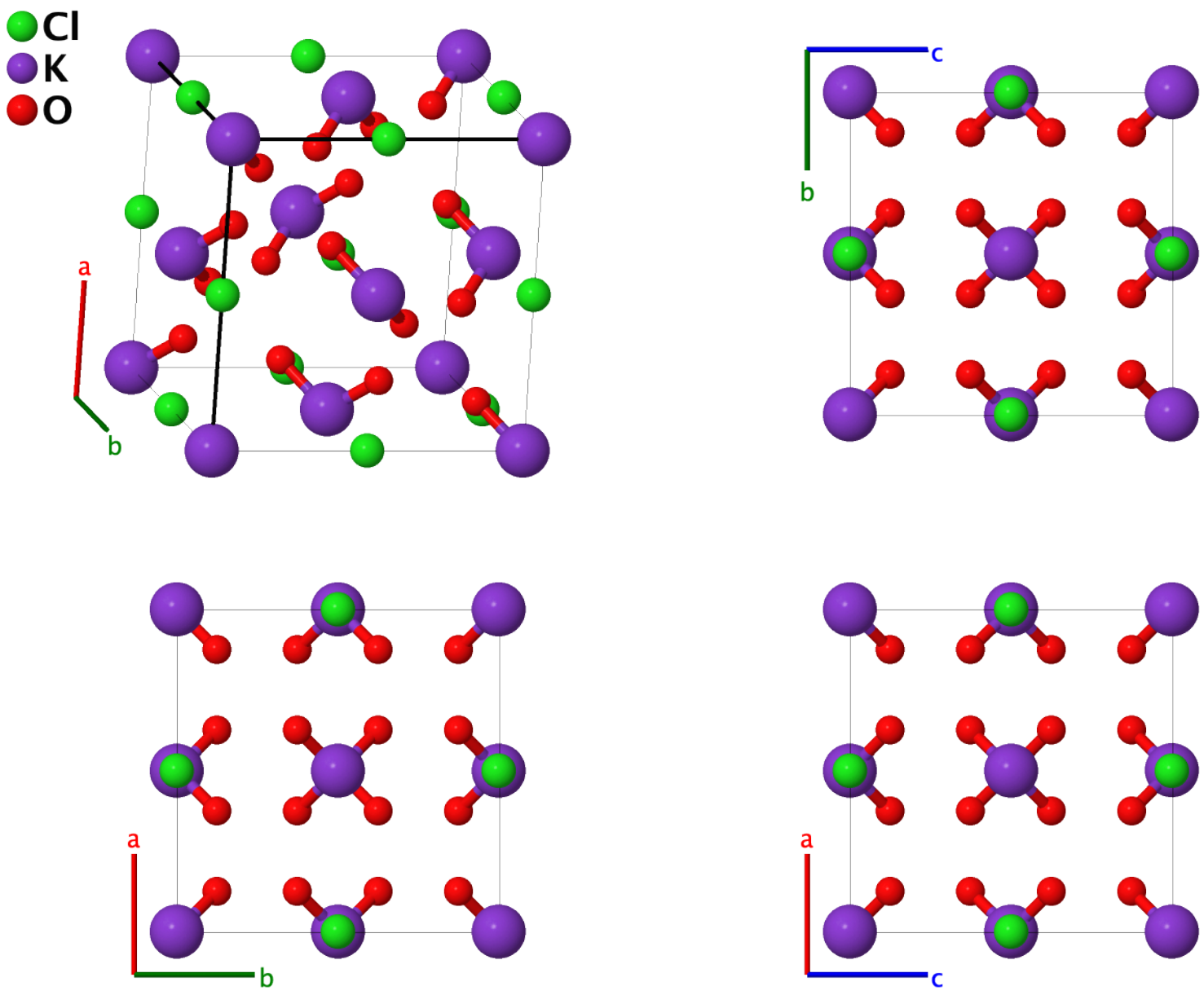

| | | |
|---|---|---|
| **Prototype** | : | $ClKO_4$ |
| **AFLOW prototype label** | : | ABC4_cF24_216_b_a_e |
| ***Strukturbericht* designation** | : | $H0_5$ |
| **Pearson symbol** | : | cF24 |
| **Space group number** | : | 216 |
| **Space group symbol** | : | $F\bar{4}3m$ |
| **AFLOW prototype command** | : | `aflow --proto=ABC4_cF24_216_b_a_e --params=`$a, x_3$ |

**Other compounds with this structure**

- $NaClO_4$, $RbClO_4$, $CsClO_4$, $NH_4ClO_4$, $AgClO_4$, and $TlClO_4$

- This is the high-temperature phase of the listed perchlorate structures. $KClO_4$ transforms from its ground-state orthorhombic structure, $H0_2$ into this structure at 299.5 °C. The transition temperature for the other compounds range from 155 °C ($AgClO_4$) to 308 °C ($NaClO_4$).
- The lattice constant for $KClO_4$ was measured at 310 °C.

**Face-centered Cubic primitive vectors:**

$$\mathbf{a}_1 = \frac{1}{2}\,a\,\hat{\mathbf{y}} + \frac{1}{2}\,a\,\hat{\mathbf{z}}$$
$$\mathbf{a}_2 = \frac{1}{2}\,a\,\hat{\mathbf{x}} + \frac{1}{2}\,a\,\hat{\mathbf{z}}$$
$$\mathbf{a}_3 = \frac{1}{2}\,a\,\hat{\mathbf{x}} + \frac{1}{2}\,a\,\hat{\mathbf{y}}$$

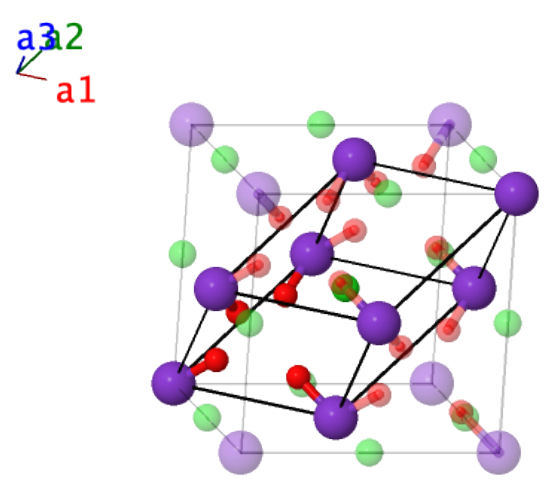

**Basis vectors:**

| | | Lattice Coordinates | | Cartesian Coordinates | Wyckoff Position | Atom Type |
|---|---|---|---|---|---|---|
| $\mathbf{B}_1$ | = | $0\,\mathbf{a}_1 + 0\,\mathbf{a}_2 + 0\,\mathbf{a}_3$ | = | $0\,\hat{\mathbf{x}} + 0\,\hat{\mathbf{y}} + 0\,\hat{\mathbf{z}}$ | $(4a)$ | K |
| $\mathbf{B}_2$ | = | $\frac{1}{2}\,\mathbf{a}_1 + \frac{1}{2}\,\mathbf{a}_2 + \frac{1}{2}\,\mathbf{a}_3$ | = | $\frac{1}{2}a\,\hat{\mathbf{x}} + \frac{1}{2}a\,\hat{\mathbf{y}} + \frac{1}{2}a\,\hat{\mathbf{z}}$ | $(4b)$ | Cl |
| $\mathbf{B}_3$ | = | $x_3\,\mathbf{a}_1 + x_3\,\mathbf{a}_2 + x_3\,\mathbf{a}_3$ | = | $x_3a\,\hat{\mathbf{x}} + x_3a\,\hat{\mathbf{y}} + x_3a\,\hat{\mathbf{z}}$ | $(16e)$ | O |
| $\mathbf{B}_4$ | = | $x_3\,\mathbf{a}_1 + x_3\,\mathbf{a}_2 - 3x_3\,\mathbf{a}_3$ | = | $-x_3a\,\hat{\mathbf{x}} - x_3a\,\hat{\mathbf{y}} + x_3a\,\hat{\mathbf{z}}$ | $(16e)$ | O |
| $\mathbf{B}_5$ | = | $x_3\,\mathbf{a}_1 - 3x_3\,\mathbf{a}_2 + x_3\,\mathbf{a}_3$ | = | $-x_3a\,\hat{\mathbf{x}} + x_3a\,\hat{\mathbf{y}} - x_3a\,\hat{\mathbf{z}}$ | $(16e)$ | O |
| $\mathbf{B}_6$ | = | $-3x_3\,\mathbf{a}_1 + x_3\,\mathbf{a}_2 + x_3\,\mathbf{a}_3$ | = | $x_3a\,\hat{\mathbf{x}} - x_3a\,\hat{\mathbf{y}} - x_3a\,\hat{\mathbf{z}}$ | $(16e)$ | O |

**Found in:**

**Geometry files:**

- CIF: pp. 1797
- POSCAR: pp. 1798

# AlN (cF40) Structure: AB_cF40_216_ce_de

http://aflow.org/prototype-encyclopedia/AB_cF40_216_ce_de

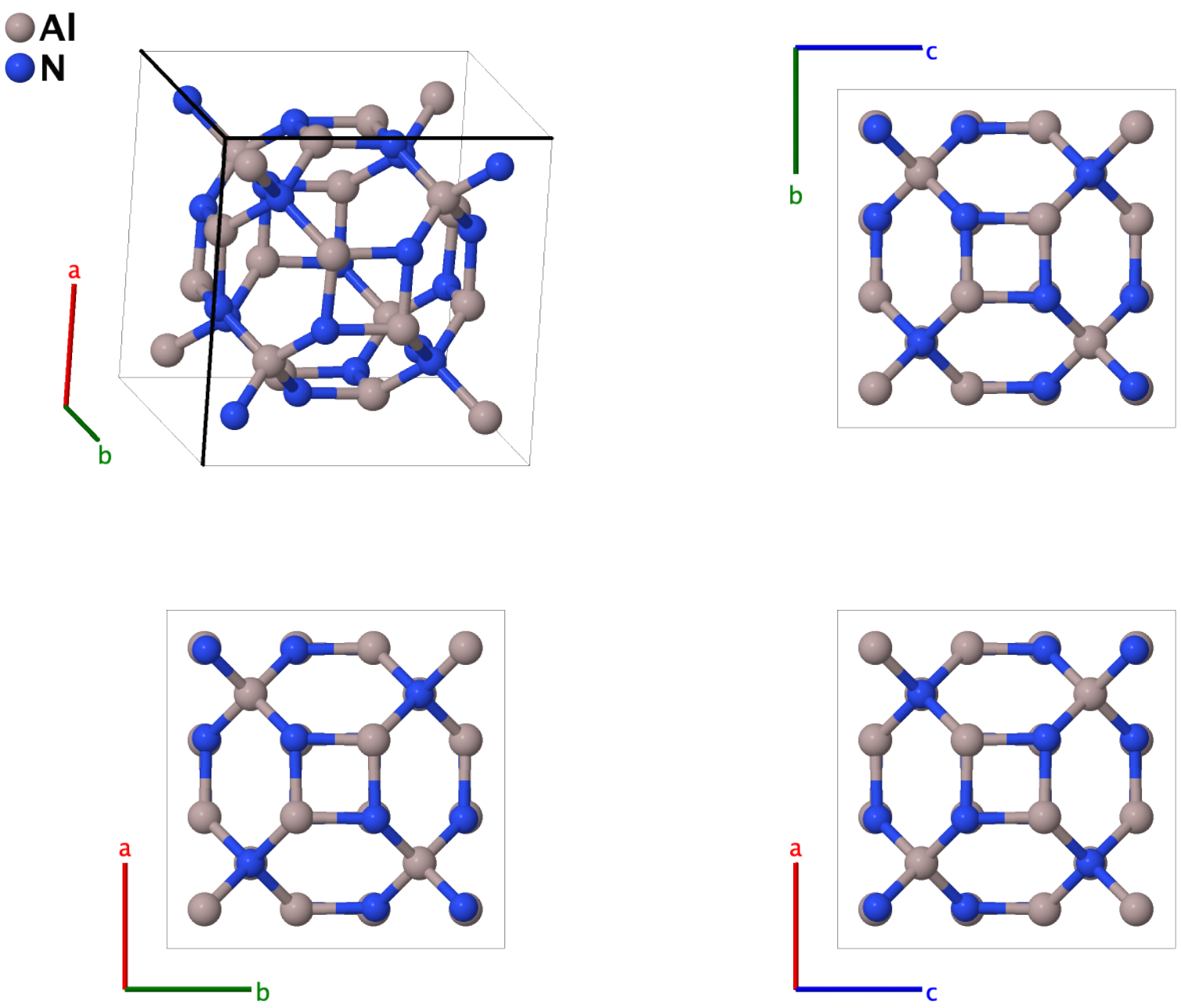

| | | |
|---|---|---|
| **Prototype** | : | AlN |
| **AFLOW prototype label** | : | AB_cF40_216_ce_de |
| ***Strukturbericht* designation** | : | None |
| **Pearson symbol** | : | cF40 |
| **Space group number** | : | 216 |
| **Space group symbol** | : | $F\bar{4}3m$ |
| **AFLOW prototype command** | : | `aflow --proto=AB_cF40_216_ce_de`<br>`--params=`$a, x_3, x_4$ |

- AlN naturally occurs in two forms (Liu, 2019): the stable wz-AlN wurtzite (*B*4) structure, and the high-pressure rs-AlN rock salt (*B*1) structure. A metastable zb-AlN zincblende (zb-AlN) structure can be synthesized via a solid-state reaction.
- (Liu, 2019) used a first-principles evolutionary technique to find four possible metastable phases: one in the SC16 structure, and three novel cubic structures, cF40, cI16, and cI24.

**Face-centered Cubic primitive vectors:**

$$
\begin{aligned}
\mathbf{a}_1 &= \tfrac{1}{2}\,a\,\mathbf{\hat{y}} + \tfrac{1}{2}\,a\,\mathbf{\hat{z}} \\
\mathbf{a}_2 &= \tfrac{1}{2}\,a\,\mathbf{\hat{x}} + \tfrac{1}{2}\,a\,\mathbf{\hat{z}} \\
\mathbf{a}_3 &= \tfrac{1}{2}\,a\,\mathbf{\hat{x}} + \tfrac{1}{2}\,a\,\mathbf{\hat{y}}
\end{aligned}
$$

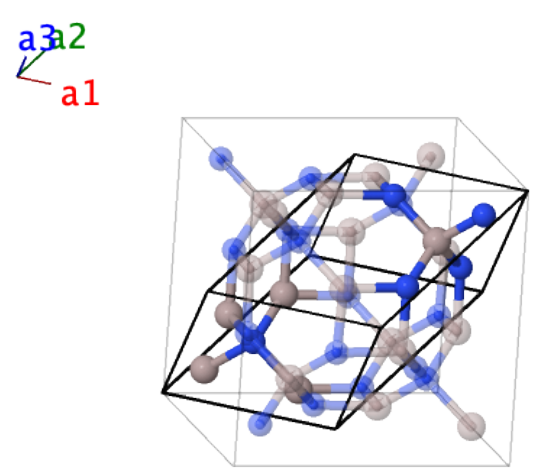

**Basis vectors:**

| | | Lattice Coordinates | | Cartesian Coordinates | Wyckoff Position | Atom Type |
|---|---|---|---|---|---|---|
| $\mathbf{B}_1$ | = | $\frac{1}{4}\,\mathbf{a}_1 + \frac{1}{4}\,\mathbf{a}_2 + \frac{1}{4}\,\mathbf{a}_3$ | = | $\frac{1}{4}a\,\hat{\mathbf{x}} + \frac{1}{4}a\,\hat{\mathbf{y}} + \frac{1}{4}a\,\hat{\mathbf{z}}$ | (4$c$) | Al I |
| $\mathbf{B}_2$ | = | $\frac{3}{4}\,\mathbf{a}_1 + \frac{3}{4}\,\mathbf{a}_2 + \frac{3}{4}\,\mathbf{a}_3$ | = | $\frac{3}{4}a\,\hat{\mathbf{x}} + \frac{3}{4}a\,\hat{\mathbf{y}} + \frac{3}{4}a\,\hat{\mathbf{z}}$ | (4$d$) | N I |
| $\mathbf{B}_3$ | = | $x_3\,\mathbf{a}_1 + x_3\,\mathbf{a}_2 + x_3\,\mathbf{a}_3$ | = | $x_3a\,\hat{\mathbf{x}} + x_3a\,\hat{\mathbf{y}} + x_3a\,\hat{\mathbf{z}}$ | (16$e$) | Al II |
| $\mathbf{B}_4$ | = | $x_3\,\mathbf{a}_1 + x_3\,\mathbf{a}_2 - 3x_3\,\mathbf{a}_3$ | = | $-x_3a\,\hat{\mathbf{x}} - x_3a\,\hat{\mathbf{y}} + x_3a\,\hat{\mathbf{z}}$ | (16$e$) | Al II |
| $\mathbf{B}_5$ | = | $x_3\,\mathbf{a}_1 - 3x_3\,\mathbf{a}_2 + x_3\,\mathbf{a}_3$ | = | $-x_3a\,\hat{\mathbf{x}} + x_3a\,\hat{\mathbf{y}} - x_3a\,\hat{\mathbf{z}}$ | (16$e$) | Al II |
| $\mathbf{B}_6$ | = | $-3x_3\,\mathbf{a}_1 + x_3\,\mathbf{a}_2 + x_3\,\mathbf{a}_3$ | = | $x_3a\,\hat{\mathbf{x}} - x_3a\,\hat{\mathbf{y}} - x_3a\,\hat{\mathbf{z}}$ | (16$e$) | Al II |
| $\mathbf{B}_7$ | = | $x_4\,\mathbf{a}_1 + x_4\,\mathbf{a}_2 + x_4\,\mathbf{a}_3$ | = | $x_4a\,\hat{\mathbf{x}} + x_4a\,\hat{\mathbf{y}} + x_4a\,\hat{\mathbf{z}}$ | (16$e$) | N II |
| $\mathbf{B}_8$ | = | $x_4\,\mathbf{a}_1 + x_4\,\mathbf{a}_2 - 3x_4\,\mathbf{a}_3$ | = | $-x_4a\,\hat{\mathbf{x}} - x_4a\,\hat{\mathbf{y}} + x_4a\,\hat{\mathbf{z}}$ | (16$e$) | N II |
| $\mathbf{B}_9$ | = | $x_4\,\mathbf{a}_1 - 3x_4\,\mathbf{a}_2 + x_4\,\mathbf{a}_3$ | = | $-x_4a\,\hat{\mathbf{x}} + x_4a\,\hat{\mathbf{y}} - x_4a\,\hat{\mathbf{z}}$ | (16$e$) | N II |
| $\mathbf{B}_{10}$ | = | $-3x_4\,\mathbf{a}_1 + x_4\,\mathbf{a}_2 + x_4\,\mathbf{a}_3$ | = | $x_4a\,\hat{\mathbf{x}} - x_4a\,\hat{\mathbf{y}} - x_4a\,\hat{\mathbf{z}}$ | (16$e$) | N II |

**Geometry files:**

- CIF: pp. 1798
- POSCAR: pp. 1799

# Tennantite ($Cu_{12}As_4S_{13}$) Structure: A4B24C13_cI82_217_c_deg_ag

http://aflow.org/prototype-encyclopedia/A4B24C13_cI82_217_c_deg_ag

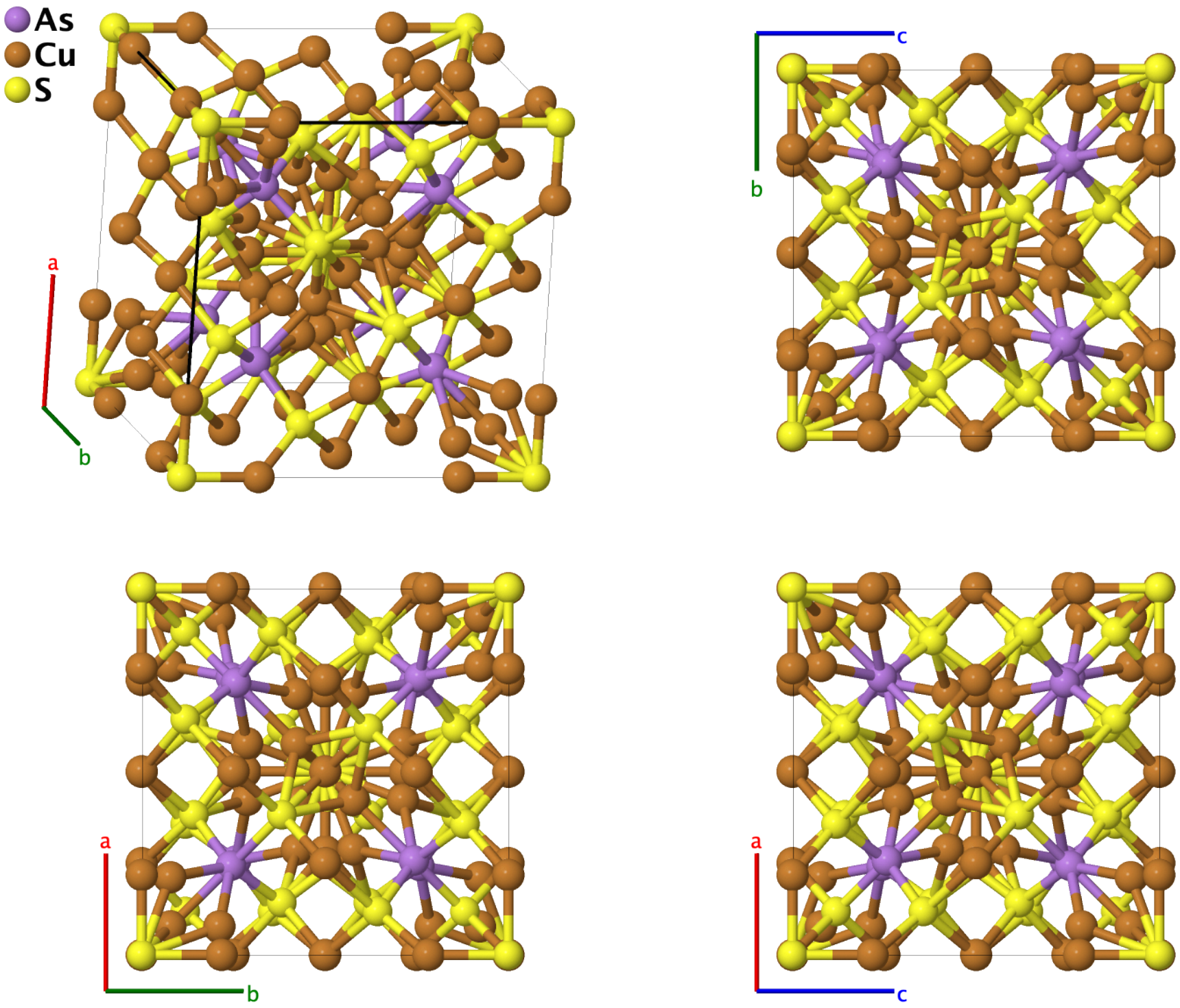

| | | |
|---|---|---|
| **Prototype** | : | $As_4Cu_{12}S_{13}$ |
| **AFLOW prototype label** | : | A4B24C13_cI82_217_c_deg_ag |
| ***Strukturbericht* designation** | : | None |
| **Pearson symbol** | : | cI82 |
| **Space group number** | : | 217 |
| **Space group symbol** | : | $I\bar{4}3m$ |
| **AFLOW prototype command** | : | `aflow --proto=A4B24C13_cI82_217_c_deg_ag`<br>`--params=`$a, x_2, x_4, x_5, z_5, x_6, z_6$ |

### Other compounds with this structure

- $Cu_{14}Sb_4S_{13}$ (tetrahedrite)

---

- The Cu-II (12$e$) site is only occupied 75.8% of the time, and the Cu-III site is occupied 12.1% of the time, so that these sites only contain twelve atoms.
- Searching (Downs, 2003) shows that natural samples often have antimony substituting for arsenic. The antimony structures (tetrahedrites) contain higher concentrations of copper.
- The AFLOW label models the structure as if the sites were fully occupied.

**Body-centered Cubic primitive vectors:**

$$\mathbf{a}_1 = -\frac{1}{2}\,a\,\hat{\mathbf{x}} + \frac{1}{2}\,a\,\hat{\mathbf{y}} + \frac{1}{2}\,a\,\hat{\mathbf{z}}$$
$$\mathbf{a}_2 = \frac{1}{2}\,a\,\hat{\mathbf{x}} - \frac{1}{2}\,a\,\hat{\mathbf{y}} + \frac{1}{2}\,a\,\hat{\mathbf{z}}$$
$$\mathbf{a}_3 = \frac{1}{2}\,a\,\hat{\mathbf{x}} + \frac{1}{2}\,a\,\hat{\mathbf{y}} - \frac{1}{2}\,a\,\hat{\mathbf{z}}$$

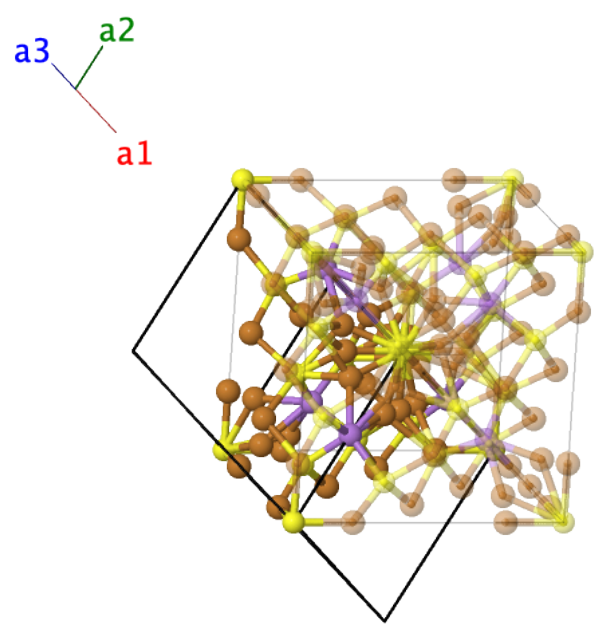

**Basis vectors:**

| | | Lattice Coordinates | | Cartesian Coordinates | Wyckoff Position | Atom Type |
|---|---|---|---|---|---|---|
| $\mathbf{B}_1$ | = | $0\,\mathbf{a}_1 + 0\,\mathbf{a}_2 + 0\,\mathbf{a}_3$ | = | $0\,\hat{\mathbf{x}} + 0\,\hat{\mathbf{y}} + 0\,\hat{\mathbf{z}}$ | (2$a$) | S I |
| $\mathbf{B}_2$ | = | $2x_2\,\mathbf{a}_1 + 2x_2\,\mathbf{a}_2 + 2x_2\,\mathbf{a}_3$ | = | $x_2a\,\hat{\mathbf{x}} + x_2a\,\hat{\mathbf{y}} + x_2a\,\hat{\mathbf{z}}$ | (8$c$) | As |
| $\mathbf{B}_3$ | = | $-2x_2\,\mathbf{a}_3$ | = | $-x_2a\,\hat{\mathbf{x}} - x_2a\,\hat{\mathbf{y}} + x_2a\,\hat{\mathbf{z}}$ | (8$c$) | As |
| $\mathbf{B}_4$ | = | $-2x_2\,\mathbf{a}_2$ | = | $-x_2a\,\hat{\mathbf{x}} + x_2a\,\hat{\mathbf{y}} - x_2a\,\hat{\mathbf{z}}$ | (8$c$) | As |
| $\mathbf{B}_5$ | = | $-2x_2\,\mathbf{a}_1$ | = | $x_2a\,\hat{\mathbf{x}} - x_2a\,\hat{\mathbf{y}} - x_2a\,\hat{\mathbf{z}}$ | (8$c$) | As |
| $\mathbf{B}_6$ | = | $\frac{1}{2}\,\mathbf{a}_1 + \frac{1}{4}\,\mathbf{a}_2 + \frac{3}{4}\,\mathbf{a}_3$ | = | $\frac{1}{4}a\,\hat{\mathbf{x}} + \frac{1}{2}a\,\hat{\mathbf{y}}$ | (12$d$) | Cu I |
| $\mathbf{B}_7$ | = | $\frac{1}{2}\,\mathbf{a}_1 + \frac{3}{4}\,\mathbf{a}_2 + \frac{1}{4}\,\mathbf{a}_3$ | = | $\frac{1}{4}a\,\hat{\mathbf{x}} + \frac{1}{2}a\,\hat{\mathbf{z}}$ | (12$d$) | Cu I |
| $\mathbf{B}_8$ | = | $\frac{3}{4}\,\mathbf{a}_1 + \frac{1}{2}\,\mathbf{a}_2 + \frac{1}{4}\,\mathbf{a}_3$ | = | $\frac{1}{4}a\,\hat{\mathbf{y}} + \frac{1}{2}a\,\hat{\mathbf{z}}$ | (12$d$) | Cu I |
| $\mathbf{B}_9$ | = | $\frac{1}{4}\,\mathbf{a}_1 + \frac{1}{2}\,\mathbf{a}_2 + \frac{3}{4}\,\mathbf{a}_3$ | = | $\frac{1}{2}a\,\hat{\mathbf{x}} + \frac{1}{4}a\,\hat{\mathbf{y}}$ | (12$d$) | Cu I |
| $\mathbf{B}_{10}$ | = | $\frac{1}{4}\,\mathbf{a}_1 + \frac{3}{4}\,\mathbf{a}_2 + \frac{1}{2}\,\mathbf{a}_3$ | = | $\frac{1}{2}a\,\hat{\mathbf{x}} + \frac{1}{4}a\,\hat{\mathbf{z}}$ | (12$d$) | Cu I |
| $\mathbf{B}_{11}$ | = | $\frac{3}{4}\,\mathbf{a}_1 + \frac{1}{4}\,\mathbf{a}_2 + \frac{1}{2}\,\mathbf{a}_3$ | = | $\frac{1}{2}a\,\hat{\mathbf{y}} + \frac{1}{4}a\,\hat{\mathbf{z}}$ | (12$d$) | Cu I |
| $\mathbf{B}_{12}$ | = | $x_4\,\mathbf{a}_2 + x_4\,\mathbf{a}_3$ | = | $x_4a\,\hat{\mathbf{x}}$ | (12$e$) | Cu II |
| $\mathbf{B}_{13}$ | = | $-x_4\,\mathbf{a}_2 - x_4\,\mathbf{a}_3$ | = | $-x_4a\,\hat{\mathbf{x}}$ | (12$e$) | Cu II |
| $\mathbf{B}_{14}$ | = | $x_4\,\mathbf{a}_1 + x_4\,\mathbf{a}_3$ | = | $x_4a\,\hat{\mathbf{y}}$ | (12$e$) | Cu II |
| $\mathbf{B}_{15}$ | = | $-x_4\,\mathbf{a}_1 - x_4\,\mathbf{a}_3$ | = | $-x_4a\,\hat{\mathbf{y}}$ | (12$e$) | Cu II |
| $\mathbf{B}_{16}$ | = | $x_4\,\mathbf{a}_1 + x_4\,\mathbf{a}_2$ | = | $x_4a\,\hat{\mathbf{z}}$ | (12$e$) | Cu II |
| $\mathbf{B}_{17}$ | = | $-x_4\,\mathbf{a}_1 - x_4\,\mathbf{a}_2$ | = | $-x_4a\,\hat{\mathbf{z}}$ | (12$e$) | Cu II |
| $\mathbf{B}_{18}$ | = | $(x_5 + z_5)\,\mathbf{a}_1 + (x_5 + z_5)\,\mathbf{a}_2 + 2x_5\,\mathbf{a}_3$ | = | $x_5a\,\hat{\mathbf{x}} + x_5a\,\hat{\mathbf{y}} + z_5a\,\hat{\mathbf{z}}$ | (24$g$) | Cu III |
| $\mathbf{B}_{19}$ | = | $(-x_5 + z_5)\,\mathbf{a}_1 + (-x_5 + z_5)\,\mathbf{a}_2 - 2x_5\,\mathbf{a}_3$ | = | $-x_5a\,\hat{\mathbf{x}} - x_5a\,\hat{\mathbf{y}} + z_5a\,\hat{\mathbf{z}}$ | (24$g$) | Cu III |
| $\mathbf{B}_{20}$ | = | $(x_5 - z_5)\,\mathbf{a}_1 + (-x_5 - z_5)\,\mathbf{a}_2$ | = | $-x_5a\,\hat{\mathbf{x}} + x_5a\,\hat{\mathbf{y}} - z_5a\,\hat{\mathbf{z}}$ | (24$g$) | Cu III |
| $\mathbf{B}_{21}$ | = | $(-x_5 - z_5)\,\mathbf{a}_1 + (x_5 - z_5)\,\mathbf{a}_2$ | = | $x_5a\,\hat{\mathbf{x}} - x_5a\,\hat{\mathbf{y}} - z_5a\,\hat{\mathbf{z}}$ | (24$g$) | Cu III |
| $\mathbf{B}_{22}$ | = | $2x_5\,\mathbf{a}_1 + (x_5 + z_5)\,\mathbf{a}_2 + (x_5 + z_5)\,\mathbf{a}_3$ | = | $z_5a\,\hat{\mathbf{x}} + x_5a\,\hat{\mathbf{y}} + x_5a\,\hat{\mathbf{z}}$ | (24$g$) | Cu III |
| $\mathbf{B}_{23}$ | = | $-2x_5\,\mathbf{a}_1 + (-x_5 + z_5)\,\mathbf{a}_2 + (-x_5 + z_5)\,\mathbf{a}_3$ | = | $z_5a\,\hat{\mathbf{x}} - x_5a\,\hat{\mathbf{y}} - x_5a\,\hat{\mathbf{z}}$ | (24$g$) | Cu III |
| $\mathbf{B}_{24}$ | = | $(x_5 - z_5)\,\mathbf{a}_2 + (-x_5 - z_5)\,\mathbf{a}_3$ | = | $-z_5a\,\hat{\mathbf{x}} - x_5a\,\hat{\mathbf{y}} + x_5a\,\hat{\mathbf{z}}$ | (24$g$) | Cu III |
| $\mathbf{B}_{25}$ | = | $(-x_5 - z_5)\,\mathbf{a}_2 + (x_5 - z_5)\,\mathbf{a}_3$ | = | $-z_5a\,\hat{\mathbf{x}} + x_5a\,\hat{\mathbf{y}} - x_5a\,\hat{\mathbf{z}}$ | (24$g$) | Cu III |
| $\mathbf{B}_{26}$ | = | $(x_5 + z_5)\,\mathbf{a}_1 + 2x_5\,\mathbf{a}_2 + (x_5 + z_5)\,\mathbf{a}_3$ | = | $x_5a\,\hat{\mathbf{x}} + z_5a\,\hat{\mathbf{y}} + x_5a\,\hat{\mathbf{z}}$ | (24$g$) | Cu III |
| $\mathbf{B}_{27}$ | = | $(-x_5 + z_5)\,\mathbf{a}_1 - 2x_5\,\mathbf{a}_2 + (-x_5 + z_5)\,\mathbf{a}_3$ | = | $-x_5a\,\hat{\mathbf{x}} + z_5a\,\hat{\mathbf{y}} - x_5a\,\hat{\mathbf{z}}$ | (24$g$) | Cu III |
| $\mathbf{B}_{28}$ | = | $(-x_5 - z_5)\,\mathbf{a}_1 + (x_5 - z_5)\,\mathbf{a}_3$ | = | $x_5 a\,\hat{\mathbf{x}} - z_5 a\,\hat{\mathbf{y}} - x_5 a\,\hat{\mathbf{z}}$ | (24$g$) | Cu III |
| $\mathbf{B}_{29}$ | = | $(x_5 - z_5)\,\mathbf{a}_1 + (-x_5 - z_5)\,\mathbf{a}_3$ | = | $-x_5 a\,\hat{\mathbf{x}} - z_5 a\,\hat{\mathbf{y}} + x_5 a\,\hat{\mathbf{z}}$ | (24$g$) | Cu III |
| $\mathbf{B}_{30}$ | = | $(x_6 + z_6)\,\mathbf{a}_1 + (x_6 + z_6)\,\mathbf{a}_2 + 2x_6\,\mathbf{a}_3$ | = | $x_6 a\,\hat{\mathbf{x}} + x_6 a\,\hat{\mathbf{y}} + z_6 a\,\hat{\mathbf{z}}$ | (24$g$) | S II |
| $\mathbf{B}_{31}$ | = | $(-x_6 + z_6)\,\mathbf{a}_1 + (-x_6 + z_6)\,\mathbf{a}_2 - 2x_6\,\mathbf{a}_3$ | = | $-x_6 a\,\hat{\mathbf{x}} - x_6 a\,\hat{\mathbf{y}} + z_6 a\,\hat{\mathbf{z}}$ | (24$g$) | S II |
| $\mathbf{B}_{32}$ | = | $(x_6 - z_6)\,\mathbf{a}_1 + (-x_6 - z_6)\,\mathbf{a}_2$ | = | $-x_6 a\,\hat{\mathbf{x}} + x_6 a\,\hat{\mathbf{y}} - z_6 a\,\hat{\mathbf{z}}$ | (24$g$) | S II |
| $\mathbf{B}_{33}$ | = | $(-x_6 - z_6)\,\mathbf{a}_1 + (x_6 - z_6)\,\mathbf{a}_2$ | = | $x_6 a\,\hat{\mathbf{x}} - x_6 a\,\hat{\mathbf{y}} - z_6 a\,\hat{\mathbf{z}}$ | (24$g$) | S II |
| $\mathbf{B}_{34}$ | = | $2x_6\,\mathbf{a}_1 + (x_6 + z_6)\,\mathbf{a}_2 + (x_6 + z_6)\,\mathbf{a}_3$ | = | $z_6 a\,\hat{\mathbf{x}} + x_6 a\,\hat{\mathbf{y}} + x_6 a\,\hat{\mathbf{z}}$ | (24$g$) | S II |
| $\mathbf{B}_{35}$ | = | $-2x_6\,\mathbf{a}_1 + (-x_6 + z_6)\,\mathbf{a}_2 + (-x_6 + z_6)\,\mathbf{a}_3$ | = | $z_6 a\,\hat{\mathbf{x}} - x_6 a\,\hat{\mathbf{y}} - x_6 a\,\hat{\mathbf{z}}$ | (24$g$) | S II |
| $\mathbf{B}_{36}$ | = | $(x_6 - z_6)\,\mathbf{a}_2 + (-x_6 - z_6)\,\mathbf{a}_3$ | = | $-z_6 a\,\hat{\mathbf{x}} - x_6 a\,\hat{\mathbf{y}} + x_6 a\,\hat{\mathbf{z}}$ | (24$g$) | S II |
| $\mathbf{B}_{37}$ | = | $(-x_6 - z_6)\,\mathbf{a}_2 + (x_6 - z_6)\,\mathbf{a}_3$ | = | $-z_6 a\,\hat{\mathbf{x}} + x_6 a\,\hat{\mathbf{y}} - x_6 a\,\hat{\mathbf{z}}$ | (24$g$) | S II |
| $\mathbf{B}_{38}$ | = | $(x_6 + z_6)\,\mathbf{a}_1 + 2x_6\,\mathbf{a}_2 + (x_6 + z_6)\,\mathbf{a}_3$ | = | $x_6 a\,\hat{\mathbf{x}} + z_6 a\,\hat{\mathbf{y}} + x_6 a\,\hat{\mathbf{z}}$ | (24$g$) | S II |
| $\mathbf{B}_{39}$ | = | $(-x_6 + z_6)\,\mathbf{a}_1 - 2x_6\,\mathbf{a}_2 + (-x_6 + z_6)\,\mathbf{a}_3$ | = | $-x_6 a\,\hat{\mathbf{x}} + z_6 a\,\hat{\mathbf{y}} - x_6 a\,\hat{\mathbf{z}}$ | (24$g$) | S II |
| $\mathbf{B}_{40}$ | = | $(-x_6 - z_6)\,\mathbf{a}_1 + (x_6 - z_6)\,\mathbf{a}_3$ | = | $x_6 a\,\hat{\mathbf{x}} - z_6 a\,\hat{\mathbf{y}} - x_6 a\,\hat{\mathbf{z}}$ | (24$g$) | S II |
| $\mathbf{B}_{41}$ | = | $(x_6 - z_6)\,\mathbf{a}_1 + (-x_6 - z_6)\,\mathbf{a}_3$ | = | $-x_6 a\,\hat{\mathbf{x}} - z_6 a\,\hat{\mathbf{y}} + x_6 a\,\hat{\mathbf{z}}$ | (24$g$) | S II |

**Geometry files:**

- CIF: pp. 1799
- POSCAR: pp. 1799

# AlN (cI16) Structure: AB_cI16_217_c_c

http://aflow.org/prototype-encyclopedia/AB_cI16_217_c_c

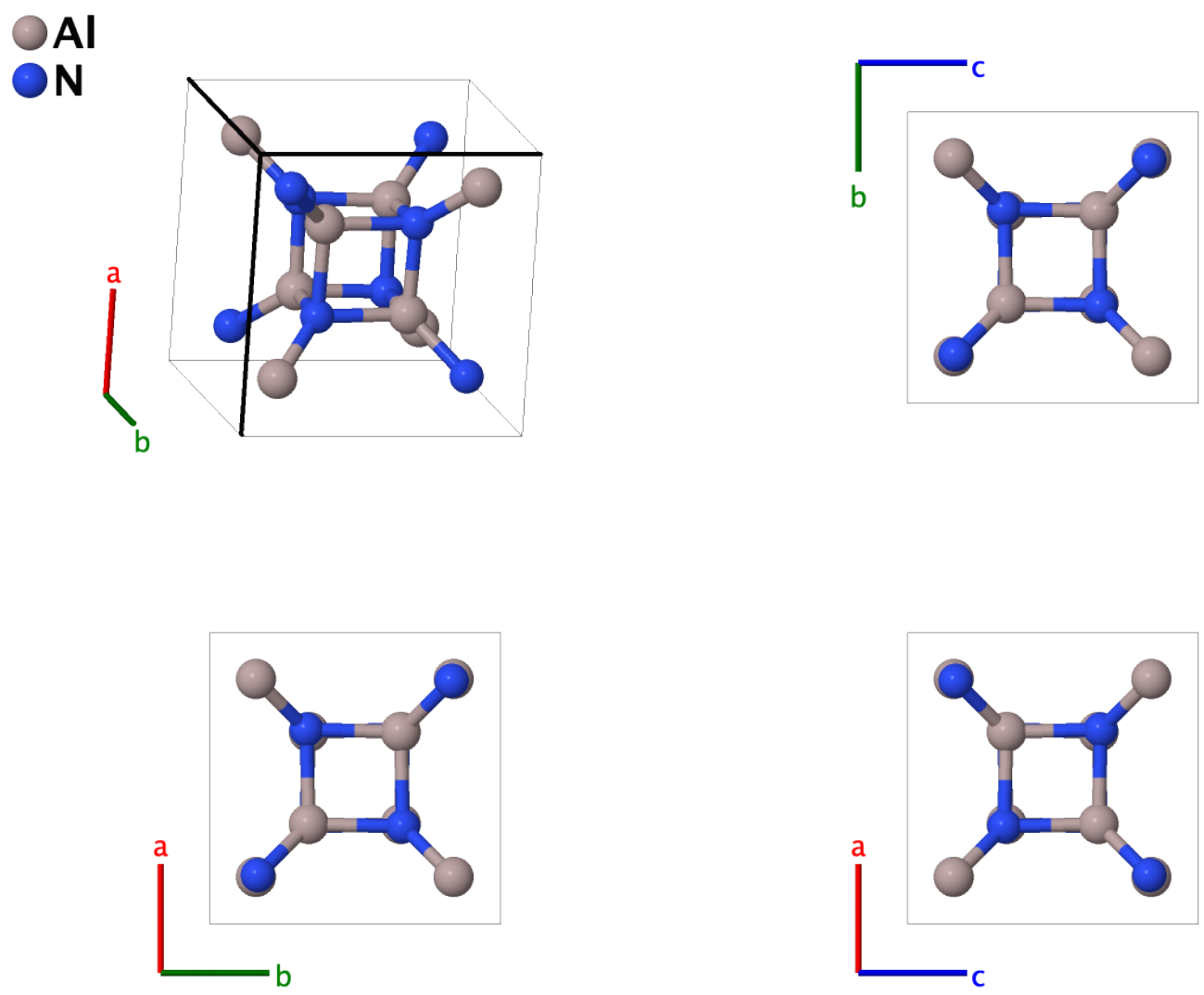

| | | |
|---|---|---|
| **Prototype** | : | AlN |
| **AFLOW prototype label** | : | AB_cI16_217_c_c |
| ***Strukturbericht* designation** | : | None |
| **Pearson symbol** | : | cI16 |
| **Space group number** | : | 217 |
| **Space group symbol** | : | $I\bar{4}3m$ |
| **AFLOW prototype command** | : | `aflow --proto=AB_cI16_217_c_c` <br> `--params=`$a, x_1, x_2$ |

- AlN naturally occurs in two forms (Liu, 2019): the stable wz-AlN wurtzite ($B4$) structure, and the high-pressure rs-AlN rock salt ($B1$) structure. A metastable zb-AlN zincblende (zb-AlN) structure can be synthesized via a solid-state reaction.
- (Liu, 2019) used a first-principles evolutionary technique to find four possible metastable phases: one in the SC16 structure, and three novel cubic structures, cF40, cI16, and cI24.

**Body-centered Cubic primitive vectors:**

$$
\begin{aligned}
\mathbf{a}_1 &= -\tfrac{1}{2}\,a\,\hat{\mathbf{x}} + \tfrac{1}{2}\,a\,\hat{\mathbf{y}} + \tfrac{1}{2}\,a\,\hat{\mathbf{z}} \\
\mathbf{a}_2 &= \tfrac{1}{2}\,a\,\hat{\mathbf{x}} - \tfrac{1}{2}\,a\,\hat{\mathbf{y}} + \tfrac{1}{2}\,a\,\hat{\mathbf{z}} \\
\mathbf{a}_3 &= \tfrac{1}{2}\,a\,\hat{\mathbf{x}} + \tfrac{1}{2}\,a\,\hat{\mathbf{y}} - \tfrac{1}{2}\,a\,\hat{\mathbf{z}}
\end{aligned}
$$

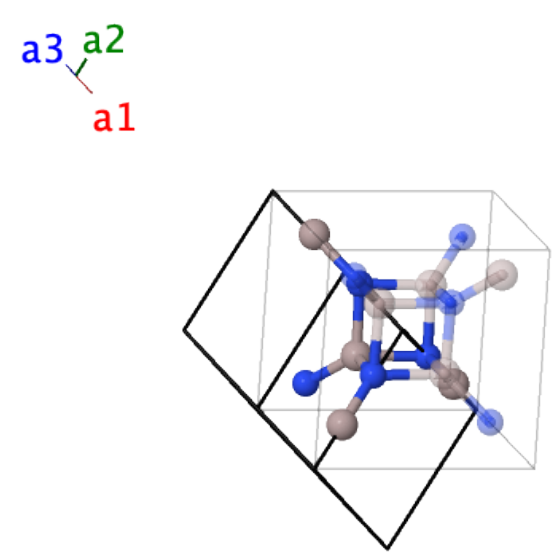

**Basis vectors:**

| | | Lattice Coordinates | | Cartesian Coordinates | Wyckoff Position | Atom Type |
|---|---|---|---|---|---|---|
| $\mathbf{B}_1$ | = | $2x_1\,\mathbf{a}_1 + 2x_1\,\mathbf{a}_2 + 2x_1\,\mathbf{a}_3$ | = | $x_1 a\,\hat{\mathbf{x}} + x_1 a\,\hat{\mathbf{y}} + x_1 a\,\hat{\mathbf{z}}$ | (8$c$) | Al |
| $\mathbf{B}_2$ | = | $-2x_1\,\mathbf{a}_3$ | = | $-x_1 a\,\hat{\mathbf{x}} - x_1 a\,\hat{\mathbf{y}} + x_1 a\,\hat{\mathbf{z}}$ | $(8c)$ | Al |
| $\mathbf{B}_3$ | = | $-2x_1\,\mathbf{a}_2$ | = | $-x_1 a\,\hat{\mathbf{x}} + x_1 a\,\hat{\mathbf{y}} - x_1 a\,\hat{\mathbf{z}}$ | $(8c)$ | Al |
| $\mathbf{B}_4$ | = | $-2x_1\,\mathbf{a}_1$ | = | $x_1 a\,\hat{\mathbf{x}} - x_1 a\,\hat{\mathbf{y}} - x_1 a\,\hat{\mathbf{z}}$ | $(8c)$ | Al |
| $\mathbf{B}_5$ | = | $2x_2\,\mathbf{a}_1 + 2x_2\,\mathbf{a}_2 + 2x_2\,\mathbf{a}_3$ | = | $x_2 a\,\hat{\mathbf{x}} + x_2 a\,\hat{\mathbf{y}} + x_2 a\,\hat{\mathbf{z}}$ | $(8c)$ | N |
| $\mathbf{B}_6$ | = | $-2x_2\,\mathbf{a}_3$ | = | $-x_2 a\,\hat{\mathbf{x}} - x_2 a\,\hat{\mathbf{y}} + x_2 a\,\hat{\mathbf{z}}$ | $(8c)$ | N |
| $\mathbf{B}_7$ | = | $-2x_2\,\mathbf{a}_2$ | = | $-x_2 a\,\hat{\mathbf{x}} + x_2 a\,\hat{\mathbf{y}} - x_2 a\,\hat{\mathbf{z}}$ | $(8c)$ | N |
| $\mathbf{B}_8$ | = | $-2x_2\,\mathbf{a}_1$ | = | $x_2 a\,\hat{\mathbf{x}} - x_2 a\,\hat{\mathbf{y}} - x_2 a\,\hat{\mathbf{z}}$ | $(8c)$ | N |

**Geometry files:**

- CIF: pp. 1800
- POSCAR: pp. 1800

# Hauyne [$(Na_{0.5}Ca_{0.3}K_{0.2})_8(Al_6Si_6O_{24})(SO_4)_{1.5}$, $S6_9$] Structure: A3B4C4D4E16F4G3_cP76_218_c_e_e_e_ei_e_d

http://aflow.org/prototype-encyclopedia/A3B4C4D4E16F4G3_cP76_218_c_e_e_e_ei_e_d

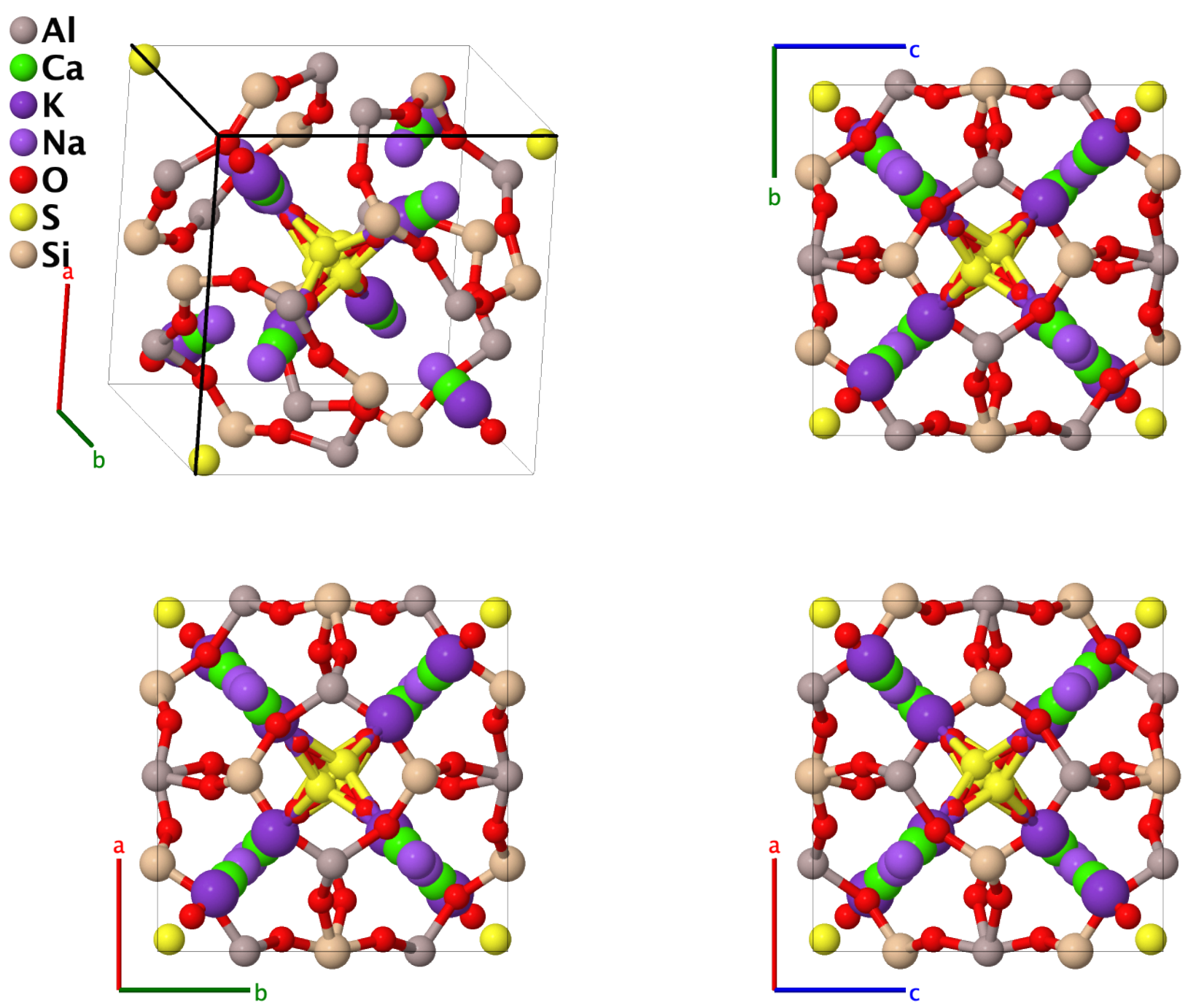

| | | |
|---|---|---|
| **Prototype** | : | $Al_6Ca_{2.4}K_{1.6}Na_4O_{30}S_{1.5}Si_6$ |
| **AFLOW prototype label** | : | A3B4C4D4E16F4G3_cP76_218_c_e_e_e_ei_e_d |
| ***Strukturbericht* designation** | : | $S6_9$ |
| **Pearson symbol** | : | cP76 |
| **Space group number** | : | 218 |
| **Space group symbol** | : | $P\bar{4}3n$ |
| **AFLOW prototype command** | : | `aflow --proto=A3B4C4D4E16F4G3_cP76_218_c_e_e_e_ei_e_d --params=`$a, x_3, x_4, x_5, x_6, x_7, x_8, y_8, z_8$ |

- (Gottfried, 1937) used the work of (Machatschki, 1934) for *Strukturbericht* label $S6_9$. We have used the updated 153 K data of (Hassan, 1991).

- (Machatschki, 1934) puts the sodium, calcium, and potassium atoms at the same (8$e$) Wyckoff position, with partial occupancies of approximately $Na_{0.75}Ca_{0.25}$ and a trace of potassium. The sample studied by (Hassan, 1991) was found to have the three atoms at slightly different (8e) positions, with 54% sodium occupation, 30% calcium occupation, and 20% potassium occupation.

- (Machatschki, 1934) put the sulfur atom at the origin, Wyckoff position (2$a$), and assumed that there were two $SO_4$ molecules per formula unit. (Hassan, 1991) found that there were statistically only 1.5 molecules per formula unit, and that the sulfur atom was slightly displaced from the origin to a (8$e$) site, where each atomic position was occupied only 19% of the time. The corresponding (8$e$) oxygen site (O-I in our notation, O2 in Hassan) is only occupied 75% of the time to maintain the $SO_4$ stoichiometry.

- We shifted the origin of (Hassan, 1991) so that the resulting atomic positions are close to those reported by (Machatschki, 1934). The later structure can be recovered from the former by moving all of the sodium, calcium, and potassium atoms to the same Wyckoff position, moving the sulfur atom to the origin, Wyckoff position (2$a$), and adjusting the occupation of each site appropriately.

- The 153K data presented in (Downs, 2003) has errors in the $z$ coordinates of the CaC2 (our Ca) and O2 (our O-I) atoms. In both cases, the $z$ coordinate should be the same as the $x$ and $y$ coordinates.
- The AFLOW label models the structure as if the sites were fully occupied.

---

**Simple Cubic primitive vectors:**

$$\mathbf{a}_1 = a\,\hat{\mathbf{x}}$$
$$\mathbf{a}_2 = a\,\hat{\mathbf{y}}$$
$$\mathbf{a}_3 = a\,\hat{\mathbf{z}}$$

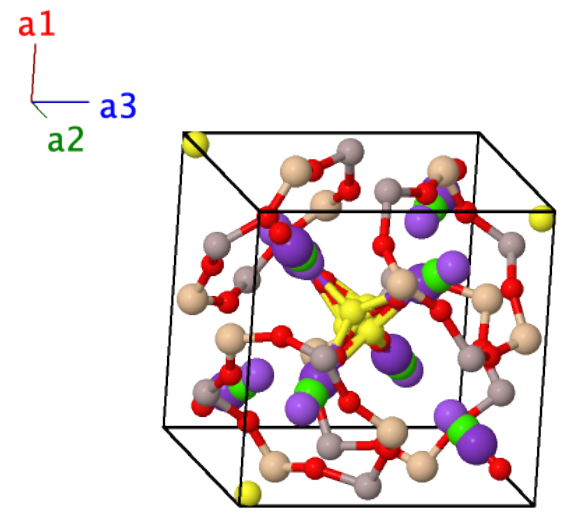

---

**Basis vectors:**

| | | Lattice Coordinates | | Cartesian Coordinates | Wyckoff Position | Atom Type |
|---|---|---|---|---|---|---|
| $\mathbf{B}_1$ | = | $\frac{1}{4}\mathbf{a}_1 + \frac{1}{2}\mathbf{a}_2$ | = | $\frac{1}{4}a\,\hat{\mathbf{x}} + \frac{1}{2}a\,\hat{\mathbf{y}}$ | (6$c$) | Al |
| $\mathbf{B}_2$ | = | $\frac{3}{4}\mathbf{a}_1 + \frac{1}{2}\mathbf{a}_2$ | = | $\frac{3}{4}a\,\hat{\mathbf{x}} + \frac{1}{2}a\,\hat{\mathbf{y}}$ | (6$c$) | Al |
| $\mathbf{B}_3$ | = | $\frac{1}{4}\mathbf{a}_2 + \frac{1}{2}\mathbf{a}_3$ | = | $\frac{1}{4}a\,\hat{\mathbf{y}} + \frac{1}{2}a\,\hat{\mathbf{z}}$ | (6$c$) | Al |
| $\mathbf{B}_4$ | = | $\frac{3}{4}\mathbf{a}_2 + \frac{1}{2}\mathbf{a}_3$ | = | $\frac{3}{4}a\,\hat{\mathbf{y}} + \frac{1}{2}a\,\hat{\mathbf{z}}$ | (6$c$) | Al |
| $\mathbf{B}_5$ | = | $\frac{1}{2}\mathbf{a}_1 + \frac{1}{4}\mathbf{a}_3$ | = | $\frac{1}{2}a\,\hat{\mathbf{x}} + \frac{1}{4}a\,\hat{\mathbf{z}}$ | (6$c$) | Al |
| $\mathbf{B}_6$ | = | $\frac{1}{2}\mathbf{a}_1 + \frac{3}{4}\mathbf{a}_3$ | = | $\frac{1}{2}a\,\hat{\mathbf{x}} + \frac{3}{4}a\,\hat{\mathbf{z}}$ | (6$c$) | Al |
| $\mathbf{B}_7$ | = | $\frac{1}{4}\mathbf{a}_1 + \frac{1}{2}\mathbf{a}_3$ | = | $\frac{1}{4}a\,\hat{\mathbf{x}} + \frac{1}{2}a\,\hat{\mathbf{z}}$ | (6$d$) | Si |
| $\mathbf{B}_8$ | = | $\frac{3}{4}\mathbf{a}_1 + \frac{1}{2}\mathbf{a}_3$ | = | $\frac{3}{4}a\,\hat{\mathbf{x}} + \frac{1}{2}a\,\hat{\mathbf{z}}$ | (6$d$) | Si |
| $\mathbf{B}_9$ | = | $\frac{1}{2}\mathbf{a}_1 + \frac{1}{4}\mathbf{a}_2$ | = | $\frac{1}{2}a\,\hat{\mathbf{x}} + \frac{1}{4}a\,\hat{\mathbf{y}}$ | (6$d$) | Si |
| $\mathbf{B}_{10}$ | = | $\frac{1}{2}\mathbf{a}_1 + \frac{3}{4}\mathbf{a}_2$ | = | $\frac{1}{2}a\,\hat{\mathbf{x}} + \frac{3}{4}a\,\hat{\mathbf{y}}$ | (6$d$) | Si |
| $\mathbf{B}_{11}$ | = | $\frac{1}{2}\mathbf{a}_2 + \frac{1}{4}\mathbf{a}_3$ | = | $\frac{1}{2}a\,\hat{\mathbf{y}} + \frac{1}{4}a\,\hat{\mathbf{z}}$ | (6$d$) | Si |
| $\mathbf{B}_{12}$ | = | $\frac{1}{2}\mathbf{a}_2 + \frac{3}{4}\mathbf{a}_3$ | = | $\frac{1}{2}a\,\hat{\mathbf{y}} + \frac{3}{4}a\,\hat{\mathbf{z}}$ | (6$d$) | Si |
| $\mathbf{B}_{13}$ | = | $x_3\,\mathbf{a}_1 + x_3\,\mathbf{a}_2 + x_3\,\mathbf{a}_3$ | = | $x_3a\,\hat{\mathbf{x}} + x_3a\,\hat{\mathbf{y}} + x_3a\,\hat{\mathbf{z}}$ | (8$e$) | Ca |
| $\mathbf{B}_{14}$ | = | $-x_3\,\mathbf{a}_1 - x_3\,\mathbf{a}_2 + x_3\,\mathbf{a}_3$ | = | $-x_3a\,\hat{\mathbf{x}} - x_3a\,\hat{\mathbf{y}} + x_3a\,\hat{\mathbf{z}}$ | (8$e$) | Ca |
| $\mathbf{B}_{15}$ | = | $-x_3\,\mathbf{a}_1 + x_3\,\mathbf{a}_2 - x_3\,\mathbf{a}_3$ | = | $-x_3a\,\hat{\mathbf{x}} + x_3a\,\hat{\mathbf{y}} - x_3a\,\hat{\mathbf{z}}$ | (8$e$) | Ca |
| $\mathbf{B}_{16}$ | = | $x_3\,\mathbf{a}_1 - x_3\,\mathbf{a}_2 - x_3\,\mathbf{a}_3$ | = | $x_3a\,\hat{\mathbf{x}} - x_3a\,\hat{\mathbf{y}} - x_3a\,\hat{\mathbf{z}}$ | (8$e$) | Ca |
| $\mathbf{B}_{17}$ | = | $\left(\frac{1}{2} + x_3\right)\mathbf{a}_1 + \left(\frac{1}{2} + x_3\right)\mathbf{a}_2 + \left(\frac{1}{2} + x_3\right)\mathbf{a}_3$ | = | $\left(\frac{1}{2} + x_3\right)a\,\hat{\mathbf{x}} + \left(\frac{1}{2} + x_3\right)a\,\hat{\mathbf{y}} + \left(\frac{1}{2} + x_3\right)a\,\hat{\mathbf{z}}$ | (8$e$) | Ca |
| $\mathbf{B}_{18}$ | = | $\left(\frac{1}{2} - x_3\right)\mathbf{a}_1 + \left(\frac{1}{2} - x_3\right)\mathbf{a}_2 + \left(\frac{1}{2} + x_3\right)\mathbf{a}_3$ | = | $\left(\frac{1}{2} - x_3\right)a\,\hat{\mathbf{x}} + \left(\frac{1}{2} - x_3\right)a\,\hat{\mathbf{y}} + \left(\frac{1}{2} + x_3\right)a\,\hat{\mathbf{z}}$ | (8$e$) | Ca |
| $\mathbf{B}_{19}$ | = | $\left(\frac{1}{2} + x_3\right)\mathbf{a}_1 + \left(\frac{1}{2} - x_3\right)\mathbf{a}_2 + \left(\frac{1}{2} - x_3\right)\mathbf{a}_3$ | = | $\left(\frac{1}{2} + x_3\right)a\,\hat{\mathbf{x}} + \left(\frac{1}{2} - x_3\right)a\,\hat{\mathbf{y}} + \left(\frac{1}{2} - x_3\right)a\,\hat{\mathbf{z}}$ | (8$e$) | Ca |
| $\mathbf{B}_{20}$ | = | $\left(\frac{1}{2} - x_3\right) \mathbf{a}_1 + \left(\frac{1}{2} + x_3\right) \mathbf{a}_2 + \left(\frac{1}{2} - x_3\right) \mathbf{a}_3$ | = | $\left(\frac{1}{2} - x_3\right) a \hat{\mathbf{x}} + \left(\frac{1}{2} + x_3\right) a \hat{\mathbf{y}} + \left(\frac{1}{2} - x_3\right) a \hat{\mathbf{z}}$ | (8$e$) | Ca |
| $\mathbf{B}_{21}$ | = | $x_4 \mathbf{a}_1 + x_4 \mathbf{a}_2 + x_4 \mathbf{a}_3$ | = | $x_4 a \hat{\mathbf{x}} + x_4 a \hat{\mathbf{y}} + x_4 a \hat{\mathbf{z}}$ | (8$e$) | K |
| $\mathbf{B}_{22}$ | = | $-x_4 \mathbf{a}_1 - x_4 \mathbf{a}_2 + x_4 \mathbf{a}_3$ | = | $-x_4 a \hat{\mathbf{x}} - x_4 a \hat{\mathbf{y}} + x_4 a \hat{\mathbf{z}}$ | (8$e$) | K |
| $\mathbf{B}_{23}$ | = | $-x_4 \mathbf{a}_1 + x_4 \mathbf{a}_2 - x_4 \mathbf{a}_3$ | = | $-x_4 a \hat{\mathbf{x}} + x_4 a \hat{\mathbf{y}} - x_4 a \hat{\mathbf{z}}$ | (8$e$) | K |
| $\mathbf{B}_{24}$ | = | $x_4 \mathbf{a}_1 - x_4 \mathbf{a}_2 - x_4 \mathbf{a}_3$ | = | $x_4 a \hat{\mathbf{x}} - x_4 a \hat{\mathbf{y}} - x_4 a \hat{\mathbf{z}}$ | (8$e$) | K |
| $\mathbf{B}_{25}$ | = | $\left(\frac{1}{2} + x_4\right) \mathbf{a}_1 + \left(\frac{1}{2} + x_4\right) \mathbf{a}_2 + \left(\frac{1}{2} + x_4\right) \mathbf{a}_3$ | = | $\left(\frac{1}{2} + x_4\right) a \hat{\mathbf{x}} + \left(\frac{1}{2} + x_4\right) a \hat{\mathbf{y}} + \left(\frac{1}{2} + x_4\right) a \hat{\mathbf{z}}$ | (8$e$) | K |
| $\mathbf{B}_{26}$ | = | $\left(\frac{1}{2} - x_4\right) \mathbf{a}_1 + \left(\frac{1}{2} - x_4\right) \mathbf{a}_2 + \left(\frac{1}{2} + x_4\right) \mathbf{a}_3$ | = | $\left(\frac{1}{2} - x_4\right) a \hat{\mathbf{x}} + \left(\frac{1}{2} - x_4\right) a \hat{\mathbf{y}} + \left(\frac{1}{2} + x_4\right) a \hat{\mathbf{z}}$ | (8$e$) | K |
| $\mathbf{B}_{27}$ | = | $\left(\frac{1}{2} + x_4\right) \mathbf{a}_1 + \left(\frac{1}{2} - x_4\right) \mathbf{a}_2 + \left(\frac{1}{2} - x_4\right) \mathbf{a}_3$ | = | $\left(\frac{1}{2} + x_4\right) a \hat{\mathbf{x}} + \left(\frac{1}{2} - x_4\right) a \hat{\mathbf{y}} + \left(\frac{1}{2} - x_4\right) a \hat{\mathbf{z}}$ | (8$e$) | K |
| $\mathbf{B}_{28}$ | = | $\left(\frac{1}{2} - x_4\right) \mathbf{a}_1 + \left(\frac{1}{2} + x_4\right) \mathbf{a}_2 + \left(\frac{1}{2} - x_4\right) \mathbf{a}_3$ | = | $\left(\frac{1}{2} - x_4\right) a \hat{\mathbf{x}} + \left(\frac{1}{2} + x_4\right) a \hat{\mathbf{y}} + \left(\frac{1}{2} - x_4\right) a \hat{\mathbf{z}}$ | (8$e$) | K |
| $\mathbf{B}_{29}$ | = | $x_5 \mathbf{a}_1 + x_5 \mathbf{a}_2 + x_5 \mathbf{a}_3$ | = | $x_5 a \hat{\mathbf{x}} + x_5 a \hat{\mathbf{y}} + x_5 a \hat{\mathbf{z}}$ | (8$e$) | Na |
| $\mathbf{B}_{30}$ | = | $-x_5 \mathbf{a}_1 - x_5 \mathbf{a}_2 + x_5 \mathbf{a}_3$ | = | $-x_5 a \hat{\mathbf{x}} - x_5 a \hat{\mathbf{y}} + x_5 a \hat{\mathbf{z}}$ | (8$e$) | Na |
| $\mathbf{B}_{31}$ | = | $-x_5 \mathbf{a}_1 + x_5 \mathbf{a}_2 - x_5 \mathbf{a}_3$ | = | $-x_5 a \hat{\mathbf{x}} + x_5 a \hat{\mathbf{y}} - x_5 a \hat{\mathbf{z}}$ | (8$e$) | Na |
| $\mathbf{B}_{32}$ | = | $x_5 \mathbf{a}_1 - x_5 \mathbf{a}_2 - x_5 \mathbf{a}_3$ | = | $x_5 a \hat{\mathbf{x}} - x_5 a \hat{\mathbf{y}} - x_5 a \hat{\mathbf{z}}$ | (8$e$) | Na |
| $\mathbf{B}_{33}$ | = | $\left(\frac{1}{2} + x_5\right) \mathbf{a}_1 + \left(\frac{1}{2} + x_5\right) \mathbf{a}_2 + \left(\frac{1}{2} + x_5\right) \mathbf{a}_3$ | = | $\left(\frac{1}{2} + x_5\right) a \hat{\mathbf{x}} + \left(\frac{1}{2} + x_5\right) a \hat{\mathbf{y}} + \left(\frac{1}{2} + x_5\right) a \hat{\mathbf{z}}$ | (8$e$) | Na |
| $\mathbf{B}_{34}$ | = | $\left(\frac{1}{2} - x_5\right) \mathbf{a}_1 + \left(\frac{1}{2} - x_5\right) \mathbf{a}_2 + \left(\frac{1}{2} + x_5\right) \mathbf{a}_3$ | = | $\left(\frac{1}{2} - x_5\right) a \hat{\mathbf{x}} + \left(\frac{1}{2} - x_5\right) a \hat{\mathbf{y}} + \left(\frac{1}{2} + x_5\right) a \hat{\mathbf{z}}$ | (8$e$) | Na |
| $\mathbf{B}_{35}$ | = | $\left(\frac{1}{2} + x_5\right) \mathbf{a}_1 + \left(\frac{1}{2} - x_5\right) \mathbf{a}_2 + \left(\frac{1}{2} - x_5\right) \mathbf{a}_3$ | = | $\left(\frac{1}{2} + x_5\right) a \hat{\mathbf{x}} + \left(\frac{1}{2} - x_5\right) a \hat{\mathbf{y}} + \left(\frac{1}{2} - x_5\right) a \hat{\mathbf{z}}$ | (8$e$) | Na |
| $\mathbf{B}_{36}$ | = | $\left(\frac{1}{2} - x_5\right) \mathbf{a}_1 + \left(\frac{1}{2} + x_5\right) \mathbf{a}_2 + \left(\frac{1}{2} - x_5\right) \mathbf{a}_3$ | = | $\left(\frac{1}{2} - x_5\right) a \hat{\mathbf{x}} + \left(\frac{1}{2} + x_5\right) a \hat{\mathbf{y}} + \left(\frac{1}{2} - x_5\right) a \hat{\mathbf{z}}$ | (8$e$) | Na |
| $\mathbf{B}_{37}$ | = | $x_6 \mathbf{a}_1 + x_6 \mathbf{a}_2 + x_6 \mathbf{a}_3$ | = | $x_6 a \hat{\mathbf{x}} + x_6 a \hat{\mathbf{y}} + x_6 a \hat{\mathbf{z}}$ | (8$e$) | O I |
| $\mathbf{B}_{38}$ | = | $-x_6 \mathbf{a}_1 - x_6 \mathbf{a}_2 + x_6 \mathbf{a}_3$ | = | $-x_6 a \hat{\mathbf{x}} - x_6 a \hat{\mathbf{y}} + x_6 a \hat{\mathbf{z}}$ | (8$e$) | O I |
| $\mathbf{B}_{39}$ | = | $-x_6 \mathbf{a}_1 + x_6 \mathbf{a}_2 - x_6 \mathbf{a}_3$ | = | $-x_6 a \hat{\mathbf{x}} + x_6 a \hat{\mathbf{y}} - x_6 a \hat{\mathbf{z}}$ | (8$e$) | O I |
| $\mathbf{B}_{40}$ | = | $x_6 \mathbf{a}_1 - x_6 \mathbf{a}_2 - x_6 \mathbf{a}_3$ | = | $x_6 a \hat{\mathbf{x}} - x_6 a \hat{\mathbf{y}} - x_6 a \hat{\mathbf{z}}$ | (8$e$) | O I |
| $\mathbf{B}_{41}$ | = | $\left(\frac{1}{2} + x_6\right) \mathbf{a}_1 + \left(\frac{1}{2} + x_6\right) \mathbf{a}_2 + \left(\frac{1}{2} + x_6\right) \mathbf{a}_3$ | = | $\left(\frac{1}{2} + x_6\right) a \hat{\mathbf{x}} + \left(\frac{1}{2} + x_6\right) a \hat{\mathbf{y}} + \left(\frac{1}{2} + x_6\right) a \hat{\mathbf{z}}$ | (8$e$) | O I |
| $\mathbf{B}_{42}$ | = | $\left(\frac{1}{2} - x_6\right) \mathbf{a}_1 + \left(\frac{1}{2} - x_6\right) \mathbf{a}_2 + \left(\frac{1}{2} + x_6\right) \mathbf{a}_3$ | = | $\left(\frac{1}{2} - x_6\right) a \hat{\mathbf{x}} + \left(\frac{1}{2} - x_6\right) a \hat{\mathbf{y}} + \left(\frac{1}{2} + x_6\right) a \hat{\mathbf{z}}$ | (8$e$) | O I |
| $\mathbf{B}_{43}$ | = | $\left(\frac{1}{2} + x_6\right) \mathbf{a}_1 + \left(\frac{1}{2} - x_6\right) \mathbf{a}_2 + \left(\frac{1}{2} - x_6\right) \mathbf{a}_3$ | = | $\left(\frac{1}{2} + x_6\right) a \hat{\mathbf{x}} + \left(\frac{1}{2} - x_6\right) a \hat{\mathbf{y}} + \left(\frac{1}{2} - x_6\right) a \hat{\mathbf{z}}$ | (8$e$) | O I |
| $\mathbf{B}_{44}$ | = | $\left(\frac{1}{2} - x_6\right) \mathbf{a}_1 + \left(\frac{1}{2} + x_6\right) \mathbf{a}_2 + \left(\frac{1}{2} - x_6\right) \mathbf{a}_3$ | = | $\left(\frac{1}{2} - x_6\right) a \hat{\mathbf{x}} + \left(\frac{1}{2} + x_6\right) a \hat{\mathbf{y}} + \left(\frac{1}{2} - x_6\right) a \hat{\mathbf{z}}$ | (8$e$) | O I |
| $\mathbf{B}_{45}$ | = | $x_7 \mathbf{a}_1 + x_7 \mathbf{a}_2 + x_7 \mathbf{a}_3$ | = | $x_7 a \hat{\mathbf{x}} + x_7 a \hat{\mathbf{y}} + x_7 a \hat{\mathbf{z}}$ | (8$e$) | S |
| $\mathbf{B}_{46}$ | = | $-x_7\,\mathbf{a}_1 - x_7\,\mathbf{a}_2 + x_7\,\mathbf{a}_3$ | = | $-x_7a\,\mathbf{\hat{x}} - x_7a\,\mathbf{\hat{y}} + x_7a\,\mathbf{\hat{z}}$ | (8$e$) | S |
| $\mathbf{B}_{47}$ | = | $-x_7\,\mathbf{a}_1 + x_7\,\mathbf{a}_2 - x_7\,\mathbf{a}_3$ | = | $-x_7a\,\mathbf{\hat{x}} + x_7a\,\mathbf{\hat{y}} - x_7a\,\mathbf{\hat{z}}$ | (8$e$) | S |
| $\mathbf{B}_{48}$ | = | $x_7\,\mathbf{a}_1 - x_7\,\mathbf{a}_2 - x_7\,\mathbf{a}_3$ | = | $x_7a\,\mathbf{\hat{x}} - x_7a\,\mathbf{\hat{y}} - x_7a\,\mathbf{\hat{z}}$ | (8$e$) | S |
| $\mathbf{B}_{49}$ | = | $\left(\frac{1}{2} + x_7\right)\mathbf{a}_1 + \left(\frac{1}{2} + x_7\right)\mathbf{a}_2 + \left(\frac{1}{2} + x_7\right)\mathbf{a}_3$ | = | $\left(\frac{1}{2} + x_7\right)a\,\mathbf{\hat{x}} + \left(\frac{1}{2} + x_7\right)a\,\mathbf{\hat{y}} + \left(\frac{1}{2} + x_7\right)a\,\mathbf{\hat{z}}$ | (8$e$) | S |
| $\mathbf{B}_{50}$ | = | $\left(\frac{1}{2} - x_7\right)\mathbf{a}_1 + \left(\frac{1}{2} - x_7\right)\mathbf{a}_2 + \left(\frac{1}{2} + x_7\right)\mathbf{a}_3$ | = | $\left(\frac{1}{2} - x_7\right)a\,\mathbf{\hat{x}} + \left(\frac{1}{2} - x_7\right)a\,\mathbf{\hat{y}} + \left(\frac{1}{2} + x_7\right)a\,\mathbf{\hat{z}}$ | (8$e$) | S |
| $\mathbf{B}_{51}$ | = | $\left(\frac{1}{2} + x_7\right)\mathbf{a}_1 + \left(\frac{1}{2} - x_7\right)\mathbf{a}_2 + \left(\frac{1}{2} - x_7\right)\mathbf{a}_3$ | = | $\left(\frac{1}{2} + x_7\right)a\,\mathbf{\hat{x}} + \left(\frac{1}{2} - x_7\right)a\,\mathbf{\hat{y}} + \left(\frac{1}{2} - x_7\right)a\,\mathbf{\hat{z}}$ | (8$e$) | S |
| $\mathbf{B}_{52}$ | = | $\left(\frac{1}{2} - x_7\right)\mathbf{a}_1 + \left(\frac{1}{2} + x_7\right)\mathbf{a}_2 + \left(\frac{1}{2} - x_7\right)\mathbf{a}_3$ | = | $\left(\frac{1}{2} - x_7\right)a\,\mathbf{\hat{x}} + \left(\frac{1}{2} + x_7\right)a\,\mathbf{\hat{y}} + \left(\frac{1}{2} - x_7\right)a\,\mathbf{\hat{z}}$ | (8$e$) | S |
| $\mathbf{B}_{53}$ | = | $x_8\,\mathbf{a}_1 + y_8\,\mathbf{a}_2 + z_8\,\mathbf{a}_3$ | = | $x_8a\,\mathbf{\hat{x}} + y_8a\,\mathbf{\hat{y}} + z_8a\,\mathbf{\hat{z}}$ | (24$i$) | O II |
| $\mathbf{B}_{54}$ | = | $-x_8\,\mathbf{a}_1 - y_8\,\mathbf{a}_2 + z_8\,\mathbf{a}_3$ | = | $-x_8a\,\mathbf{\hat{x}} - y_8a\,\mathbf{\hat{y}} + z_8a\,\mathbf{\hat{z}}$ | (24$i$) | O II |
| $\mathbf{B}_{55}$ | = | $-x_8\,\mathbf{a}_1 + y_8\,\mathbf{a}_2 - z_8\,\mathbf{a}_3$ | = | $-x_8a\,\mathbf{\hat{x}} + y_8a\,\mathbf{\hat{y}} - z_8a\,\mathbf{\hat{z}}$ | (24$i$) | O II |
| $\mathbf{B}_{56}$ | = | $x_8\,\mathbf{a}_1 - y_8\,\mathbf{a}_2 - z_8\,\mathbf{a}_3$ | = | $x_8a\,\mathbf{\hat{x}} - y_8a\,\mathbf{\hat{y}} - z_8a\,\mathbf{\hat{z}}$ | (24$i$) | O II |
| $\mathbf{B}_{57}$ | = | $z_8\,\mathbf{a}_1 + x_8\,\mathbf{a}_2 + y_8\,\mathbf{a}_3$ | = | $z_8a\,\mathbf{\hat{x}} + x_8a\,\mathbf{\hat{y}} + y_8a\,\mathbf{\hat{z}}$ | (24$i$) | O II |
| $\mathbf{B}_{58}$ | = | $z_8\,\mathbf{a}_1 - x_8\,\mathbf{a}_2 - y_8\,\mathbf{a}_3$ | = | $z_8a\,\mathbf{\hat{x}} - x_8a\,\mathbf{\hat{y}} - y_8a\,\mathbf{\hat{z}}$ | (24$i$) | O II |
| $\mathbf{B}_{59}$ | = | $-z_8\,\mathbf{a}_1 - x_8\,\mathbf{a}_2 + y_8\,\mathbf{a}_3$ | = | $-z_8a\,\mathbf{\hat{x}} - x_8a\,\mathbf{\hat{y}} + y_8a\,\mathbf{\hat{z}}$ | (24$i$) | O II |
| $\mathbf{B}_{60}$ | = | $-z_8\,\mathbf{a}_1 + x_8\,\mathbf{a}_2 - y_8\,\mathbf{a}_3$ | = | $-z_8a\,\mathbf{\hat{x}} + x_8a\,\mathbf{\hat{y}} - y_8a\,\mathbf{\hat{z}}$ | (24$i$) | O II |
| $\mathbf{B}_{61}$ | = | $y_8\,\mathbf{a}_1 + z_8\,\mathbf{a}_2 + x_8\,\mathbf{a}_3$ | = | $y_8a\,\mathbf{\hat{x}} + z_8a\,\mathbf{\hat{y}} + x_8a\,\mathbf{\hat{z}}$ | (24$i$) | O II |
| $\mathbf{B}_{62}$ | = | $-y_8\,\mathbf{a}_1 + z_8\,\mathbf{a}_2 - x_8\,\mathbf{a}_3$ | = | $-y_8a\,\mathbf{\hat{x}} + z_8a\,\mathbf{\hat{y}} - x_8a\,\mathbf{\hat{z}}$ | (24$i$) | O II |
| $\mathbf{B}_{63}$ | = | $y_8\,\mathbf{a}_1 - z_8\,\mathbf{a}_2 - x_8\,\mathbf{a}_3$ | = | $y_8a\,\mathbf{\hat{x}} - z_8a\,\mathbf{\hat{y}} - x_8a\,\mathbf{\hat{z}}$ | (24$i$) | O II |
| $\mathbf{B}_{64}$ | = | $-y_8\,\mathbf{a}_1 - z_8\,\mathbf{a}_2 + x_8\,\mathbf{a}_3$ | = | $-y_8a\,\mathbf{\hat{x}} - z_8a\,\mathbf{\hat{y}} + x_8a\,\mathbf{\hat{z}}$ | (24$i$) | O II |
| $\mathbf{B}_{65}$ | = | $\left(\frac{1}{2} + y_8\right)\mathbf{a}_1 + \left(\frac{1}{2} + x_8\right)\mathbf{a}_2 + \left(\frac{1}{2} + z_8\right)\mathbf{a}_3$ | = | $\left(\frac{1}{2} + y_8\right)a\,\mathbf{\hat{x}} + \left(\frac{1}{2} + x_8\right)a\,\mathbf{\hat{y}} + \left(\frac{1}{2} + z_8\right)a\,\mathbf{\hat{z}}$ | (24$i$) | O II |
| $\mathbf{B}_{66}$ | = | $\left(\frac{1}{2} - y_8\right)\mathbf{a}_1 + \left(\frac{1}{2} - x_8\right)\mathbf{a}_2 + \left(\frac{1}{2} + z_8\right)\mathbf{a}_3$ | = | $\left(\frac{1}{2} - y_8\right)a\,\mathbf{\hat{x}} + \left(\frac{1}{2} - x_8\right)a\,\mathbf{\hat{y}} + \left(\frac{1}{2} + z_8\right)a\,\mathbf{\hat{z}}$ | (24$i$) | O II |
| $\mathbf{B}_{67}$ | = | $\left(\frac{1}{2} + y_8\right)\mathbf{a}_1 + \left(\frac{1}{2} - x_8\right)\mathbf{a}_2 + \left(\frac{1}{2} - z_8\right)\mathbf{a}_3$ | = | $\left(\frac{1}{2} + y_8\right)a\,\mathbf{\hat{x}} + \left(\frac{1}{2} - x_8\right)a\,\mathbf{\hat{y}} + \left(\frac{1}{2} - z_8\right)a\,\mathbf{\hat{z}}$ | (24$i$) | O II |
| $\mathbf{B}_{68}$ | = | $\left(\frac{1}{2} - y_8\right)\mathbf{a}_1 + \left(\frac{1}{2} + x_8\right)\mathbf{a}_2 + \left(\frac{1}{2} - z_8\right)\mathbf{a}_3$ | = | $\left(\frac{1}{2} - y_8\right)a\,\mathbf{\hat{x}} + \left(\frac{1}{2} + x_8\right)a\,\mathbf{\hat{y}} + \left(\frac{1}{2} - z_8\right)a\,\mathbf{\hat{z}}$ | (24$i$) | O II |
| $\mathbf{B}_{69}$ | = | $\left(\frac{1}{2} + x_8\right)\mathbf{a}_1 + \left(\frac{1}{2} + z_8\right)\mathbf{a}_2 + \left(\frac{1}{2} + y_8\right)\mathbf{a}_3$ | = | $\left(\frac{1}{2} + x_8\right)a\,\mathbf{\hat{x}} + \left(\frac{1}{2} + z_8\right)a\,\mathbf{\hat{y}} + \left(\frac{1}{2} + y_8\right)a\,\mathbf{\hat{z}}$ | (24$i$) | O II |
| $\mathbf{B}_{70}$ | = | $\left(\frac{1}{2} - x_8\right)\mathbf{a}_1 + \left(\frac{1}{2} + z_8\right)\mathbf{a}_2 + \left(\frac{1}{2} - y_8\right)\mathbf{a}_3$ | = | $\left(\frac{1}{2} - x_8\right)a\,\mathbf{\hat{x}} + \left(\frac{1}{2} + z_8\right)a\,\mathbf{\hat{y}} + \left(\frac{1}{2} - y_8\right)a\,\mathbf{\hat{z}}$ | (24$i$) | O II |
| $\mathbf{B}_{71}$ | = | $\left(\frac{1}{2} - x_8\right)\mathbf{a}_1 + \left(\frac{1}{2} - z_8\right)\mathbf{a}_2 + \left(\frac{1}{2} + y_8\right)\mathbf{a}_3$ | = | $\left(\frac{1}{2} - x_8\right)a\,\mathbf{\hat{x}} + \left(\frac{1}{2} - z_8\right)a\,\mathbf{\hat{y}} + \left(\frac{1}{2} + y_8\right)a\,\mathbf{\hat{z}}$ | (24$i$) | O II |
| $\mathbf{B}_{72}$ | = | $\left(\frac{1}{2} + x_8\right)\mathbf{a}_1 + \left(\frac{1}{2} - z_8\right)\mathbf{a}_2 + \left(\frac{1}{2} - y_8\right)\mathbf{a}_3$ | = | $\left(\frac{1}{2} + x_8\right)a\,\mathbf{\hat{x}} + \left(\frac{1}{2} - z_8\right)a\,\mathbf{\hat{y}} + \left(\frac{1}{2} - y_8\right)a\,\mathbf{\hat{z}}$ | (24$i$) | O II |
| $\mathbf{B}_{73}$ | = | $\left(\frac{1}{2}+z_8\right)\mathbf{a}_1+\left(\frac{1}{2}+y_8\right)\mathbf{a}_2+\left(\frac{1}{2}+x_8\right)\mathbf{a}_3$ | = | $\left(\frac{1}{2}+z_8\right)a\,\hat{\mathbf{x}}+\left(\frac{1}{2}+y_8\right)a\,\hat{\mathbf{y}}+\left(\frac{1}{2}+x_8\right)a\,\hat{\mathbf{z}}$ | (24*i*) | O II |
| $\mathbf{B}_{74}$ | = | $\left(\frac{1}{2}+z_8\right)\mathbf{a}_1+\left(\frac{1}{2}-y_8\right)\mathbf{a}_2+\left(\frac{1}{2}-x_8\right)\mathbf{a}_3$ | = | $\left(\frac{1}{2}+z_8\right)a\,\hat{\mathbf{x}}+\left(\frac{1}{2}-y_8\right)a\,\hat{\mathbf{y}}+\left(\frac{1}{2}-x_8\right)a\,\hat{\mathbf{z}}$ | (24*i*) | O II |
| $\mathbf{B}_{75}$ | = | $\left(\frac{1}{2}-z_8\right)\mathbf{a}_1+\left(\frac{1}{2}+y_8\right)\mathbf{a}_2+\left(\frac{1}{2}-x_8\right)\mathbf{a}_3$ | = | $\left(\frac{1}{2}-z_8\right)a\,\hat{\mathbf{x}}+\left(\frac{1}{2}+y_8\right)a\,\hat{\mathbf{y}}+\left(\frac{1}{2}-x_8\right)a\,\hat{\mathbf{z}}$ | (24*i*) | O II |
| $\mathbf{B}_{76}$ | = | $\left(\frac{1}{2}-z_8\right)\mathbf{a}_1+\left(\frac{1}{2}-y_8\right)\mathbf{a}_2+\left(\frac{1}{2}+x_8\right)\mathbf{a}_3$ | = | $\left(\frac{1}{2}-z_8\right)a\,\hat{\mathbf{x}}+\left(\frac{1}{2}-y_8\right)a\,\hat{\mathbf{y}}+\left(\frac{1}{2}+x_8\right)a\,\hat{\mathbf{z}}$ | (24*i*) | O II |

**Geometry files:**

- CIF: pp. 1800
- POSCAR: pp. 1801

# Sodalite [$Na_4(AlSiO_4)_3Cl$, $S6_2$] Structure: A3BC4D12E3_cP46_218_d_a_e_i_c

http://aflow.org/prototype-encyclopedia/A3BC4D12E3_cP46_218_d_a_e_i_c

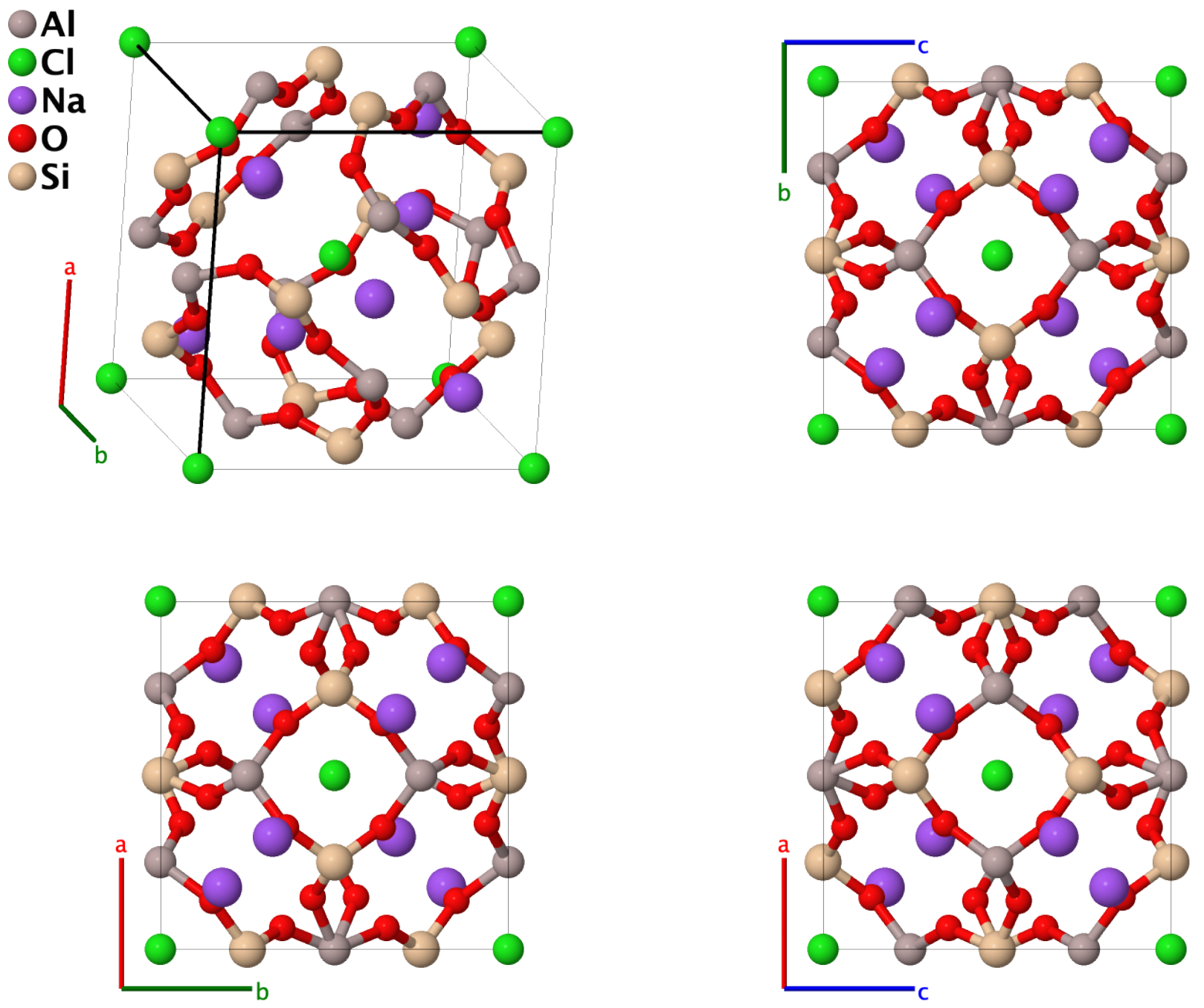

| | | |
|---|---|---|
| **Prototype** | : | $Al_3ClNa_4O_{12}Si_3$ |
| **AFLOW prototype label** | : | A3BC4D12E3_cP46_218_d_a_e_i_c |
| ***Strukturbericht* designation** | : | $S6_2$ |
| **Pearson symbol** | : | cP46 |
| **Space group number** | : | 218 |
| **Space group symbol** | : | $P\bar{4}3n$ |
| **AFLOW prototype command** | : | `aflow --proto=A3BC4D12E3_cP46_218_d_a_e_i_c --params=`$a, x_4, x_5, y_5, z_5$ |

**Other compounds with this structure**

- $(Na_xK_{1-x})_4(AlSiO_4)_3Cl$, $Li_4(AlSiO_4)_3Cl$, $(Li_xK_{1-x})_4(AlSiO_4)_3Cl$, $Na_4(AlSiO_4)_3Br$, and $Na_4(AlSiO_4)_3I$

- Sodalites of the form $A_4(AlSiO_4)_3B$ with $(A,B)$ = (Li,F), (Li,Br), (Li,I), (Na,F), (K,F), (K,Br) and (Rb,F) have been predicted to form, but not seen experimentally.

**Simple Cubic primitive vectors:**

$$\mathbf{a}_1 = a\,\hat{\mathbf{x}}$$
$$\mathbf{a}_2 = a\,\hat{\mathbf{y}}$$
$$\mathbf{a}_3 = a\,\hat{\mathbf{z}}$$

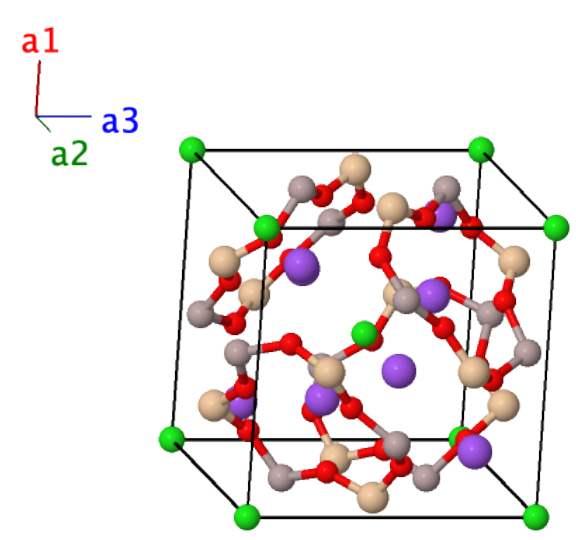

**Basis vectors:**

| | | Lattice Coordinates | | Cartesian Coordinates | Wyckoff Position | Atom Type |
|---|---|---|---|---|---|---|
| $\mathbf{B}_1$ | = | $0\,\mathbf{a}_1 + 0\,\mathbf{a}_2 + 0\,\mathbf{a}_3$ | = | $0\,\hat{\mathbf{x}} + 0\,\hat{\mathbf{y}} + 0\,\hat{\mathbf{z}}$ | (2*a*) | Cl |
| $\mathbf{B}_2$ | = | $\frac{1}{2}\,\mathbf{a}_1 + \frac{1}{2}\,\mathbf{a}_2 + \frac{1}{2}\,\mathbf{a}_3$ | = | $\frac{1}{2}a\,\hat{\mathbf{x}} + \frac{1}{2}a\,\hat{\mathbf{y}} + \frac{1}{2}a\,\hat{\mathbf{z}}$ | (2*a*) | Cl |
| $\mathbf{B}_3$ | = | $\frac{1}{4}\,\mathbf{a}_1 + \frac{1}{2}\,\mathbf{a}_2$ | = | $\frac{1}{4}a\,\hat{\mathbf{x}} + \frac{1}{2}a\,\hat{\mathbf{y}}$ | (6*c*) | Si |
| $\mathbf{B}_4$ | = | $\frac{3}{4}\,\mathbf{a}_1 + \frac{1}{2}\,\mathbf{a}_2$ | = | $\frac{3}{4}a\,\hat{\mathbf{x}} + \frac{1}{2}a\,\hat{\mathbf{y}}$ | (6*c*) | Si |
| $\mathbf{B}_5$ | = | $\frac{1}{4}\,\mathbf{a}_2 + \frac{1}{2}\,\mathbf{a}_3$ | = | $\frac{1}{4}a\,\hat{\mathbf{y}} + \frac{1}{2}a\,\hat{\mathbf{z}}$ | (6*c*) | Si |
| $\mathbf{B}_6$ | = | $\frac{3}{4}\,\mathbf{a}_2 + \frac{1}{2}\,\mathbf{a}_3$ | = | $\frac{3}{4}a\,\hat{\mathbf{y}} + \frac{1}{2}a\,\hat{\mathbf{z}}$ | (6*c*) | Si |
| $\mathbf{B}_7$ | = | $\frac{1}{2}\,\mathbf{a}_1 + \frac{1}{4}\,\mathbf{a}_3$ | = | $\frac{1}{2}a\,\hat{\mathbf{x}} + \frac{1}{4}a\,\hat{\mathbf{z}}$ | (6*c*) | Si |
| $\mathbf{B}_8$ | = | $\frac{1}{2}\,\mathbf{a}_1 + \frac{3}{4}\,\mathbf{a}_3$ | = | $\frac{1}{2}a\,\hat{\mathbf{x}} + \frac{3}{4}a\,\hat{\mathbf{z}}$ | (6*c*) | Si |
| $\mathbf{B}_9$ | = | $\frac{1}{4}\,\mathbf{a}_1 + \frac{1}{2}\,\mathbf{a}_3$ | = | $\frac{1}{4}a\,\hat{\mathbf{x}} + \frac{1}{2}a\,\hat{\mathbf{z}}$ | (6*d*) | Al |
| $\mathbf{B}_{10}$ | = | $\frac{3}{4}\,\mathbf{a}_1 + \frac{1}{2}\,\mathbf{a}_3$ | = | $\frac{3}{4}a\,\hat{\mathbf{x}} + \frac{1}{2}a\,\hat{\mathbf{z}}$ | (6*d*) | Al |
| $\mathbf{B}_{11}$ | = | $\frac{1}{2}\,\mathbf{a}_1 + \frac{1}{4}\,\mathbf{a}_2$ | = | $\frac{1}{2}a\,\hat{\mathbf{x}} + \frac{1}{4}a\,\hat{\mathbf{y}}$ | (6*d*) | Al |
| $\mathbf{B}_{12}$ | = | $\frac{1}{2}\,\mathbf{a}_1 + \frac{3}{4}\,\mathbf{a}_2$ | = | $\frac{1}{2}a\,\hat{\mathbf{x}} + \frac{3}{4}a\,\hat{\mathbf{y}}$ | (6*d*) | Al |
| $\mathbf{B}_{13}$ | = | $\frac{1}{2}\,\mathbf{a}_2 + \frac{1}{4}\,\mathbf{a}_3$ | = | $\frac{1}{2}a\,\hat{\mathbf{y}} + \frac{1}{4}a\,\hat{\mathbf{z}}$ | (6*d*) | Al |
| $\mathbf{B}_{14}$ | = | $\frac{1}{2}\,\mathbf{a}_2 + \frac{3}{4}\,\mathbf{a}_3$ | = | $\frac{1}{2}a\,\hat{\mathbf{y}} + \frac{3}{4}a\,\hat{\mathbf{z}}$ | (6*d*) | Al |
| $\mathbf{B}_{15}$ | = | $x_4\,\mathbf{a}_1 + x_4\,\mathbf{a}_2 + x_4\,\mathbf{a}_3$ | = | $x_4a\,\hat{\mathbf{x}} + x_4a\,\hat{\mathbf{y}} + x_4a\,\hat{\mathbf{z}}$ | (8*e*) | Na |
| $\mathbf{B}_{16}$ | = | $-x_4\,\mathbf{a}_1 - x_4\,\mathbf{a}_2 + x_4\,\mathbf{a}_3$ | = | $-x_4a\,\hat{\mathbf{x}} - x_4a\,\hat{\mathbf{y}} + x_4a\,\hat{\mathbf{z}}$ | (8*e*) | Na |
| $\mathbf{B}_{17}$ | = | $-x_4\,\mathbf{a}_1 + x_4\,\mathbf{a}_2 - x_4\,\mathbf{a}_3$ | = | $-x_4a\,\hat{\mathbf{x}} + x_4a\,\hat{\mathbf{y}} - x_4a\,\hat{\mathbf{z}}$ | (8*e*) | Na |
| $\mathbf{B}_{18}$ | = | $x_4\,\mathbf{a}_1 - x_4\,\mathbf{a}_2 - x_4\,\mathbf{a}_3$ | = | $x_4a\,\hat{\mathbf{x}} - x_4a\,\hat{\mathbf{y}} - x_4a\,\hat{\mathbf{z}}$ | (8*e*) | Na |
| $\mathbf{B}_{19}$ | = | $\left(\frac{1}{2} + x_4\right)\mathbf{a}_1 + \left(\frac{1}{2} + x_4\right)\mathbf{a}_2 + \left(\frac{1}{2} + x_4\right)\mathbf{a}_3$ | = | $\left(\frac{1}{2} + x_4\right)a\,\hat{\mathbf{x}} + \left(\frac{1}{2} + x_4\right)a\,\hat{\mathbf{y}} + \left(\frac{1}{2} + x_4\right)a\,\hat{\mathbf{z}}$ | (8*e*) | Na |
| $\mathbf{B}_{20}$ | = | $\left(\frac{1}{2} - x_4\right)\mathbf{a}_1 + \left(\frac{1}{2} - x_4\right)\mathbf{a}_2 + \left(\frac{1}{2} + x_4\right)\mathbf{a}_3$ | = | $\left(\frac{1}{2} - x_4\right)a\,\hat{\mathbf{x}} + \left(\frac{1}{2} - x_4\right)a\,\hat{\mathbf{y}} + \left(\frac{1}{2} + x_4\right)a\,\hat{\mathbf{z}}$ | (8*e*) | Na |
| $\mathbf{B}_{21}$ | = | $\left(\frac{1}{2} + x_4\right)\mathbf{a}_1 + \left(\frac{1}{2} - x_4\right)\mathbf{a}_2 + \left(\frac{1}{2} - x_4\right)\mathbf{a}_3$ | = | $\left(\frac{1}{2} + x_4\right)a\,\hat{\mathbf{x}} + \left(\frac{1}{2} - x_4\right)a\,\hat{\mathbf{y}} + \left(\frac{1}{2} - x_4\right)a\,\hat{\mathbf{z}}$ | (8*e*) | Na |
| $\mathbf{B}_{22}$ | = | $\left(\frac{1}{2} - x_4\right)\mathbf{a}_1 + \left(\frac{1}{2} + x_4\right)\mathbf{a}_2 + \left(\frac{1}{2} - x_4\right)\mathbf{a}_3$ | = | $\left(\frac{1}{2} - x_4\right)a\,\hat{\mathbf{x}} + \left(\frac{1}{2} + x_4\right)a\,\hat{\mathbf{y}} + \left(\frac{1}{2} - x_4\right)a\,\hat{\mathbf{z}}$ | (8*e*) | Na |
| $\mathbf{B}_{23}$ | = | $x_5\,\mathbf{a}_1 + y_5\,\mathbf{a}_2 + z_5\,\mathbf{a}_3$ | = | $x_5a\,\hat{\mathbf{x}} + y_5a\,\hat{\mathbf{y}} + z_5a\,\hat{\mathbf{z}}$ | (24*i*) | O |
| $\mathbf{B}_{24}$ | = | $-x_5\,\mathbf{a}_1 - y_5\,\mathbf{a}_2 + z_5\,\mathbf{a}_3$ | = | $-x_5a\,\hat{\mathbf{x}} - y_5a\,\hat{\mathbf{y}} + z_5a\,\hat{\mathbf{z}}$ | (24*i*) | O |
| $\mathbf{B}_{25}$ | = | $-x_5\,\mathbf{a}_1 + y_5\,\mathbf{a}_2 - z_5\,\mathbf{a}_3$ | = | $-x_5a\,\hat{\mathbf{x}} + y_5a\,\hat{\mathbf{y}} - z_5a\,\hat{\mathbf{z}}$ | (24*i*) | O |
| $\mathbf{B}_{26}$ | = | $x_5\,\mathbf{a}_1 - y_5\,\mathbf{a}_2 - z_5\,\mathbf{a}_3$ | = | $x_5a\,\hat{\mathbf{x}} - y_5a\,\hat{\mathbf{y}} - z_5a\,\hat{\mathbf{z}}$ | (24*i*) | O |
| $\mathbf{B}_{27}$ | = | $z_5\,\mathbf{a}_1 + x_5\,\mathbf{a}_2 + y_5\,\mathbf{a}_3$ | = | $z_5a\,\hat{\mathbf{x}} + x_5a\,\hat{\mathbf{y}} + y_5a\,\hat{\mathbf{z}}$ | (24*i*) | O |
| $\mathbf{B}_{28}$ | = | $z_5\,\mathbf{a}_1 - x_5\,\mathbf{a}_2 - y_5\,\mathbf{a}_3$ | = | $z_5a\,\hat{\mathbf{x}} - x_5a\,\hat{\mathbf{y}} - y_5a\,\hat{\mathbf{z}}$ | (24*i*) | O |
| $\mathbf{B}_{29}$ | = | $-z_5\,\mathbf{a}_1 - x_5\,\mathbf{a}_2 + y_5\,\mathbf{a}_3$ | = | $-z_5a\,\hat{\mathbf{x}} - x_5a\,\hat{\mathbf{y}} + y_5a\,\hat{\mathbf{z}}$ | (24*i*) | O |
| $\mathbf{B}_{30}$ | = | $-z_5\,\mathbf{a}_1 + x_5\,\mathbf{a}_2 - y_5\,\mathbf{a}_3$ | = | $-z_5a\,\hat{\mathbf{x}} + x_5a\,\hat{\mathbf{y}} - y_5a\,\hat{\mathbf{z}}$ | (24*i*) | O |
| $\mathbf{B}_{31}$ | = | $y_5\,\mathbf{a}_1 + z_5\,\mathbf{a}_2 + x_5\,\mathbf{a}_3$ | = | $y_5a\,\hat{\mathbf{x}} + z_5a\,\hat{\mathbf{y}} + x_5a\,\hat{\mathbf{z}}$ | (24*i*) | O |
| $\mathbf{B}_{32}$ | = | $-y_5\,\mathbf{a}_1 + z_5\,\mathbf{a}_2 - x_5\,\mathbf{a}_3$ | = | $-y_5 a\,\hat{\mathbf{x}} + z_5 a\,\hat{\mathbf{y}} - x_5 a\,\hat{\mathbf{z}}$ | (24$i$) | O |
| $\mathbf{B}_{33}$ | = | $y_5\,\mathbf{a}_1 - z_5\,\mathbf{a}_2 - x_5\,\mathbf{a}_3$ | = | $y_5 a\,\hat{\mathbf{x}} - z_5 a\,\hat{\mathbf{y}} - x_5 a\,\hat{\mathbf{z}}$ | (24$i$) | O |
| $\mathbf{B}_{34}$ | = | $-y_5\,\mathbf{a}_1 - z_5\,\mathbf{a}_2 + x_5\,\mathbf{a}_3$ | = | $-y_5 a\,\hat{\mathbf{x}} - z_5 a\,\hat{\mathbf{y}} + x_5 a\,\hat{\mathbf{z}}$ | (24$i$) | O |
| $\mathbf{B}_{35}$ | = | $\left(\frac{1}{2} + y_5\right)\mathbf{a}_1 + \left(\frac{1}{2} + x_5\right)\mathbf{a}_2 + \left(\frac{1}{2} + z_5\right)\mathbf{a}_3$ | = | $\left(\frac{1}{2} + y_5\right)a\,\hat{\mathbf{x}} + \left(\frac{1}{2} + x_5\right)a\,\hat{\mathbf{y}} + \left(\frac{1}{2} + z_5\right)a\,\hat{\mathbf{z}}$ | (24$i$) | O |
| $\mathbf{B}_{36}$ | = | $\left(\frac{1}{2} - y_5\right)\mathbf{a}_1 + \left(\frac{1}{2} - x_5\right)\mathbf{a}_2 + \left(\frac{1}{2} + z_5\right)\mathbf{a}_3$ | = | $\left(\frac{1}{2} - y_5\right)a\,\hat{\mathbf{x}} + \left(\frac{1}{2} - x_5\right)a\,\hat{\mathbf{y}} + \left(\frac{1}{2} + z_5\right)a\,\hat{\mathbf{z}}$ | (24$i$) | O |
| $\mathbf{B}_{37}$ | = | $\left(\frac{1}{2} + y_5\right)\mathbf{a}_1 + \left(\frac{1}{2} - x_5\right)\mathbf{a}_2 + \left(\frac{1}{2} - z_5\right)\mathbf{a}_3$ | = | $\left(\frac{1}{2} + y_5\right)a\,\hat{\mathbf{x}} + \left(\frac{1}{2} - x_5\right)a\,\hat{\mathbf{y}} + \left(\frac{1}{2} - z_5\right)a\,\hat{\mathbf{z}}$ | (24$i$) | O |
| $\mathbf{B}_{38}$ | = | $\left(\frac{1}{2} - y_5\right)\mathbf{a}_1 + \left(\frac{1}{2} + x_5\right)\mathbf{a}_2 + \left(\frac{1}{2} - z_5\right)\mathbf{a}_3$ | = | $\left(\frac{1}{2} - y_5\right)a\,\hat{\mathbf{x}} + \left(\frac{1}{2} + x_5\right)a\,\hat{\mathbf{y}} + \left(\frac{1}{2} - z_5\right)a\,\hat{\mathbf{z}}$ | (24$i$) | O |
| $\mathbf{B}_{39}$ | = | $\left(\frac{1}{2} + x_5\right)\mathbf{a}_1 + \left(\frac{1}{2} + z_5\right)\mathbf{a}_2 + \left(\frac{1}{2} + y_5\right)\mathbf{a}_3$ | = | $\left(\frac{1}{2} + x_5\right)a\,\hat{\mathbf{x}} + \left(\frac{1}{2} + z_5\right)a\,\hat{\mathbf{y}} + \left(\frac{1}{2} + y_5\right)a\,\hat{\mathbf{z}}$ | (24$i$) | O |
| $\mathbf{B}_{40}$ | = | $\left(\frac{1}{2} - x_5\right)\mathbf{a}_1 + \left(\frac{1}{2} + z_5\right)\mathbf{a}_2 + \left(\frac{1}{2} - y_5\right)\mathbf{a}_3$ | = | $\left(\frac{1}{2} - x_5\right)a\,\hat{\mathbf{x}} + \left(\frac{1}{2} + z_5\right)a\,\hat{\mathbf{y}} + \left(\frac{1}{2} - y_5\right)a\,\hat{\mathbf{z}}$ | (24$i$) | O |
| $\mathbf{B}_{41}$ | = | $\left(\frac{1}{2} - x_5\right)\mathbf{a}_1 + \left(\frac{1}{2} - z_5\right)\mathbf{a}_2 + \left(\frac{1}{2} + y_5\right)\mathbf{a}_3$ | = | $\left(\frac{1}{2} - x_5\right)a\,\hat{\mathbf{x}} + \left(\frac{1}{2} - z_5\right)a\,\hat{\mathbf{y}} + \left(\frac{1}{2} + y_5\right)a\,\hat{\mathbf{z}}$ | (24$i$) | O |
| $\mathbf{B}_{42}$ | = | $\left(\frac{1}{2} + x_5\right)\mathbf{a}_1 + \left(\frac{1}{2} - z_5\right)\mathbf{a}_2 + \left(\frac{1}{2} - y_5\right)\mathbf{a}_3$ | = | $\left(\frac{1}{2} + x_5\right)a\,\hat{\mathbf{x}} + \left(\frac{1}{2} - z_5\right)a\,\hat{\mathbf{y}} + \left(\frac{1}{2} - y_5\right)a\,\hat{\mathbf{z}}$ | (24$i$) | O |
| $\mathbf{B}_{43}$ | = | $\left(\frac{1}{2} + z_5\right)\mathbf{a}_1 + \left(\frac{1}{2} + y_5\right)\mathbf{a}_2 + \left(\frac{1}{2} + x_5\right)\mathbf{a}_3$ | = | $\left(\frac{1}{2} + z_5\right)a\,\hat{\mathbf{x}} + \left(\frac{1}{2} + y_5\right)a\,\hat{\mathbf{y}} + \left(\frac{1}{2} + x_5\right)a\,\hat{\mathbf{z}}$ | (24$i$) | O |
| $\mathbf{B}_{44}$ | = | $\left(\frac{1}{2} + z_5\right)\mathbf{a}_1 + \left(\frac{1}{2} - y_5\right)\mathbf{a}_2 + \left(\frac{1}{2} - x_5\right)\mathbf{a}_3$ | = | $\left(\frac{1}{2} + z_5\right)a\,\hat{\mathbf{x}} + \left(\frac{1}{2} - y_5\right)a\,\hat{\mathbf{y}} + \left(\frac{1}{2} - x_5\right)a\,\hat{\mathbf{z}}$ | (24$i$) | O |
| $\mathbf{B}_{45}$ | = | $\left(\frac{1}{2} - z_5\right)\mathbf{a}_1 + \left(\frac{1}{2} + y_5\right)\mathbf{a}_2 + \left(\frac{1}{2} - x_5\right)\mathbf{a}_3$ | = | $\left(\frac{1}{2} - z_5\right)a\,\hat{\mathbf{x}} + \left(\frac{1}{2} + y_5\right)a\,\hat{\mathbf{y}} + \left(\frac{1}{2} - x_5\right)a\,\hat{\mathbf{z}}$ | (24$i$) | O |
| $\mathbf{B}_{46}$ | = | $\left(\frac{1}{2} - z_5\right)\mathbf{a}_1 + \left(\frac{1}{2} - y_5\right)\mathbf{a}_2 + \left(\frac{1}{2} + x_5\right)\mathbf{a}_3$ | = | $\left(\frac{1}{2} - z_5\right)a\,\hat{\mathbf{x}} + \left(\frac{1}{2} - y_5\right)a\,\hat{\mathbf{y}} + \left(\frac{1}{2} + x_5\right)a\,\hat{\mathbf{z}}$ | (24$i$) | O |

---

---

**Geometry files:**

- CIF: pp. 1801
- POSCAR: pp. 1802

# Eulytine ($Bi_4(SiO_4)_3$, $S1_5$) Structure: A4B12C3_cI76_220_c_e_a

http://aflow.org/prototype-encyclopedia/A4B12C3_cI76_220_c_e_a

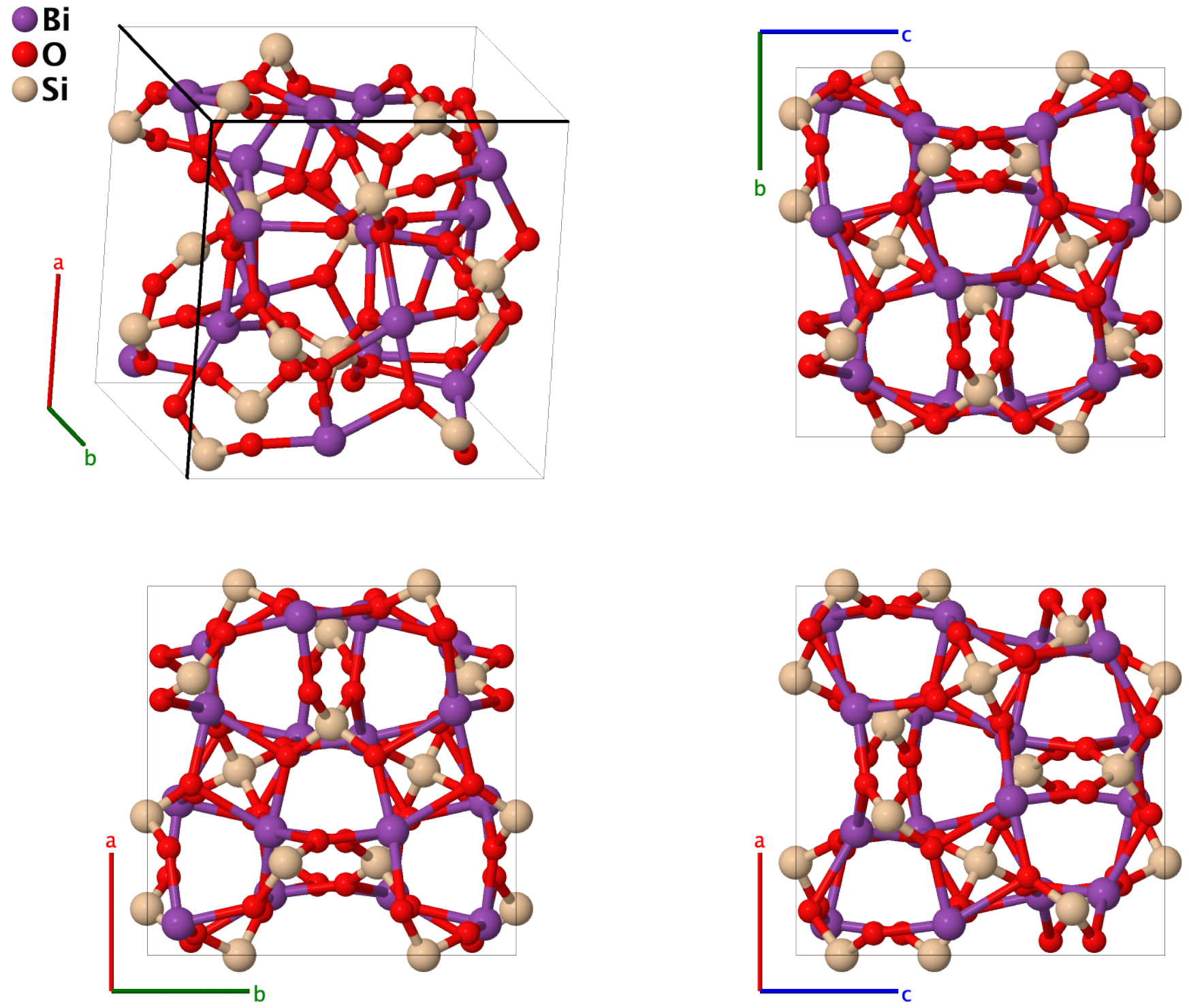

| | | |
|---|---|---|
| **Prototype** | : | $Bi_4O_{12}Si_3$ |
| **AFLOW prototype label** | : | A4B12C3_cI76_220_c_e_a |
| ***Strukturbericht* designation** | : | $S1_5$ |
| **Pearson symbol** | : | cI76 |
| **Space group number** | : | 220 |
| **Space group symbol** | : | $I\bar{4}3d$ |
| **AFLOW prototype command** | : | `aflow --proto=A4B12C3_cI76_220_c_e_a --params=`$a, x_2, x_3, y_3, z_3$ |

**Other compounds with this structure**

- $Bi_4(GeO_4)_3$, $Ca_3Bi(PO_4)_3$, and $M_3L(PO_4)_3$ ($M$ = Sr, Ba), $L$ = (La, Nd, Gd, Y, Lu, In, Bi)

**Body-centered Cubic primitive vectors:**

$$
\begin{aligned}
\mathbf{a}_1 &= -\tfrac{1}{2}\,a\,\hat{\mathbf{x}} + \tfrac{1}{2}\,a\,\hat{\mathbf{y}} + \tfrac{1}{2}\,a\,\hat{\mathbf{z}} \\
\mathbf{a}_2 &= \tfrac{1}{2}\,a\,\hat{\mathbf{x}} - \tfrac{1}{2}\,a\,\hat{\mathbf{y}} + \tfrac{1}{2}\,a\,\hat{\mathbf{z}} \\
\mathbf{a}_3 &= \tfrac{1}{2}\,a\,\hat{\mathbf{x}} + \tfrac{1}{2}\,a\,\hat{\mathbf{y}} - \tfrac{1}{2}\,a\,\hat{\mathbf{z}}
\end{aligned}
$$

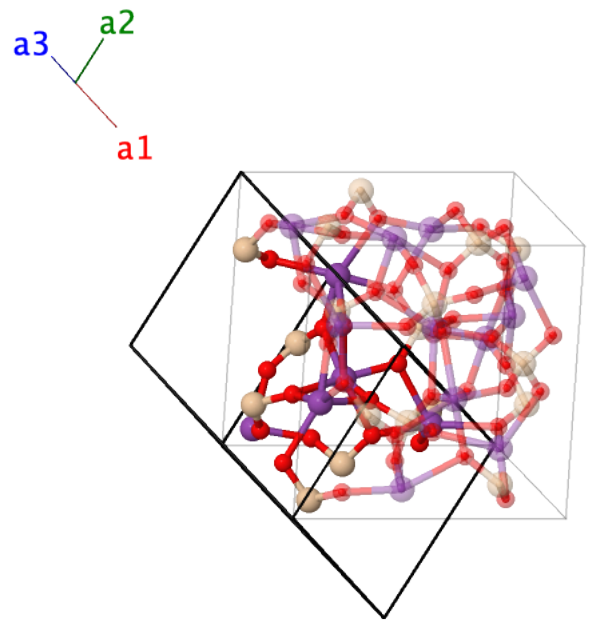

**Basis vectors:**

| | | Lattice Coordinates | | Cartesian Coordinates | Wyckoff Position | Atom Type |
|---|---|---|---|---|---|---|
| $\mathbf{B}_1$ | = | $\frac{1}{4}\,\mathbf{a}_1 + \frac{5}{8}\,\mathbf{a}_2 + \frac{3}{8}\,\mathbf{a}_3$ | = | $\frac{3}{8}a\,\hat{\mathbf{x}} + \frac{1}{4}a\,\hat{\mathbf{z}}$ | (12$a$) | Si |
| $\mathbf{B}_2$ | = | $\frac{3}{4}\,\mathbf{a}_1 + \frac{7}{8}\,\mathbf{a}_2 + \frac{1}{8}\,\mathbf{a}_3$ | = | $\frac{1}{8}a\,\hat{\mathbf{x}} + \frac{3}{4}a\,\hat{\mathbf{z}}$ | (12$a$) | Si |
| $\mathbf{B}_3$ | = | $\frac{3}{8}\,\mathbf{a}_1 + \frac{1}{4}\,\mathbf{a}_2 + \frac{5}{8}\,\mathbf{a}_3$ | = | $\frac{1}{4}a\,\hat{\mathbf{x}} + \frac{3}{8}a\,\hat{\mathbf{y}}$ | (12$a$) | Si |
| $\mathbf{B}_4$ | = | $\frac{1}{8}\,\mathbf{a}_1 + \frac{3}{4}\,\mathbf{a}_2 + \frac{7}{8}\,\mathbf{a}_3$ | = | $\frac{3}{4}a\,\hat{\mathbf{x}} + \frac{1}{8}a\,\hat{\mathbf{y}}$ | (12$a$) | Si |
| $\mathbf{B}_5$ | = | $\frac{5}{8}\,\mathbf{a}_1 + \frac{3}{8}\,\mathbf{a}_2 + \frac{1}{4}\,\mathbf{a}_3$ | = | $\frac{1}{4}a\,\hat{\mathbf{y}} + \frac{3}{8}a\,\hat{\mathbf{z}}$ | (12$a$) | Si |
| $\mathbf{B}_6$ | = | $\frac{7}{8}\,\mathbf{a}_1 + \frac{1}{8}\,\mathbf{a}_2 + \frac{3}{4}\,\mathbf{a}_3$ | = | $\frac{3}{4}a\,\hat{\mathbf{y}} + \frac{1}{8}a\,\hat{\mathbf{z}}$ | (12$a$) | Si |
| $\mathbf{B}_7$ | = | $2x_2\,\mathbf{a}_1 + 2x_2\,\mathbf{a}_2 + 2x_2\,\mathbf{a}_3$ | = | $x_2a\,\hat{\mathbf{x}} + x_2a\,\hat{\mathbf{y}} + x_2a\,\hat{\mathbf{z}}$ | (16$c$) | Bi |
| $\mathbf{B}_8$ | = | $\frac{1}{2}\,\mathbf{a}_1 + \left(\frac{1}{2} - 2x_2\right)\,\mathbf{a}_3$ | = | $-x_2a\,\hat{\mathbf{x}} + \left(\frac{1}{2} - x_2\right)a\,\hat{\mathbf{y}} + x_2a\,\hat{\mathbf{z}}$ | (16$c$) | Bi |
| $\mathbf{B}_9$ | = | $\left(\frac{1}{2} - 2x_2\right)\,\mathbf{a}_2 + \frac{1}{2}\,\mathbf{a}_3$ | = | $\left(\frac{1}{2} - x_2\right)a\,\hat{\mathbf{x}} + x_2a\,\hat{\mathbf{y}} - x_2a\,\hat{\mathbf{z}}$ | (16$c$) | Bi |
| $\mathbf{B}_{10}$ | = | $\left(\frac{1}{2} - 2x_2\right)\,\mathbf{a}_1 + \frac{1}{2}\,\mathbf{a}_2$ | = | $x_2a\,\hat{\mathbf{x}} - x_2a\,\hat{\mathbf{y}} + \left(\frac{1}{2} - x_2\right)a\,\hat{\mathbf{z}}$ | (16$c$) | Bi |
| $\mathbf{B}_{11}$ | = | $\left(\frac{1}{2} + 2x_2\right)\,\mathbf{a}_1 + \left(\frac{1}{2} + 2x_2\right)\,\mathbf{a}_2 + \left(\frac{1}{2} + 2x_2\right)\,\mathbf{a}_3$ | = | $\left(\frac{1}{4} + x_2\right)a\,\hat{\mathbf{x}} + \left(\frac{1}{4} + x_2\right)a\,\hat{\mathbf{y}} + \left(\frac{1}{4} + x_2\right)a\,\hat{\mathbf{z}}$ | (16$c$) | Bi |
| $\mathbf{B}_{12}$ | = | $\frac{1}{2}\,\mathbf{a}_1 - 2x_2\,\mathbf{a}_3$ | = | $-a\left(x_2 + \frac{1}{4}\right)\,\hat{\mathbf{x}} + \left(\frac{1}{4} - x_2\right)a\,\hat{\mathbf{y}} + \left(\frac{1}{4} + x_2\right)a\,\hat{\mathbf{z}}$ | (16$c$) | Bi |
| $\mathbf{B}_{13}$ | = | $-2x_2\,\mathbf{a}_1 + \frac{1}{2}\,\mathbf{a}_2$ | = | $\left(\frac{1}{4} + x_2\right)a\,\hat{\mathbf{x}} - a\left(x_2 + \frac{1}{4}\right)\,\hat{\mathbf{y}} + \left(\frac{1}{4} - x_2\right)a\,\hat{\mathbf{z}}$ | (16$c$) | Bi |
| $\mathbf{B}_{14}$ | = | $-2x_2\,\mathbf{a}_2 + \frac{1}{2}\,\mathbf{a}_3$ | = | $\left(\frac{1}{4} - x_2\right)a\,\hat{\mathbf{x}} + \left(\frac{1}{4} + x_2\right)a\,\hat{\mathbf{y}} - a\left(x_2 + \frac{1}{4}\right)\,\hat{\mathbf{z}}$ | (16$c$) | Bi |
| $\mathbf{B}_{15}$ | = | $(y_3 + z_3)\,\mathbf{a}_1 + (x_3 + z_3)\,\mathbf{a}_2 + (x_3 + y_3)\,\mathbf{a}_3$ | = | $x_3a\,\hat{\mathbf{x}} + y_3a\,\hat{\mathbf{y}} + z_3a\,\hat{\mathbf{z}}$ | (48$e$) | O |
| $\mathbf{B}_{16}$ | = | $\left(\frac{1}{2} - y_3 + z_3\right)\,\mathbf{a}_1 + (-x_3 + z_3)\,\mathbf{a}_2 + \left(\frac{1}{2} - x_3 - y_3\right)\,\mathbf{a}_3$ | = | $-x_3a\,\hat{\mathbf{x}} + \left(\frac{1}{2} - y_3\right)a\,\hat{\mathbf{y}} + z_3a\,\hat{\mathbf{z}}$ | (48$e$) | O |
| $\mathbf{B}_{17}$ | = | $(y_3 - z_3)\,\mathbf{a}_1 + \left(\frac{1}{2} - x_3 - z_3\right)\,\mathbf{a}_2 + \left(\frac{1}{2} - x_3 + y_3\right)\,\mathbf{a}_3$ | = | $\left(\frac{1}{2} - x_3\right)a\,\hat{\mathbf{x}} + y_3a\,\hat{\mathbf{y}} - z_3a\,\hat{\mathbf{z}}$ | (48$e$) | O |
| $\mathbf{B}_{18}$ | = | $\left(\frac{1}{2} - y_3 - z_3\right)\,\mathbf{a}_1 + \left(\frac{1}{2} + x_3 - z_3\right)\,\mathbf{a}_2 + (x_3 - y_3)\,\mathbf{a}_3$ | = | $x_3a\,\hat{\mathbf{x}} - y_3a\,\hat{\mathbf{y}} + \left(\frac{1}{2} - z_3\right)a\,\hat{\mathbf{z}}$ | (48$e$) | O |
| $\mathbf{B}_{19}$ | = | $(x_3 + y_3)\,\mathbf{a}_1 + (y_3 + z_3)\,\mathbf{a}_2 + (x_3 + z_3)\,\mathbf{a}_3$ | = | $z_3a\,\hat{\mathbf{x}} + x_3a\,\hat{\mathbf{y}} + y_3a\,\hat{\mathbf{z}}$ | (48$e$) | O |
| $\mathbf{B}_{20}$ | = | $\left(\frac{1}{2} - x_3 - y_3\right)\,\mathbf{a}_1 + \left(\frac{1}{2} - y_3 + z_3\right)\,\mathbf{a}_2 + (-x_3 + z_3)\,\mathbf{a}_3$ | = | $z_3a\,\hat{\mathbf{x}} - x_3a\,\hat{\mathbf{y}} + \left(\frac{1}{2} - y_3\right)a\,\hat{\mathbf{z}}$ | (48$e$) | O |
| $\mathbf{B}_{21}$ | = | $\left(\frac{1}{2} - x_3 + y_3\right)\,\mathbf{a}_1 + (y_3 - z_3)\,\mathbf{a}_2 + \left(\frac{1}{2} - x_3 - z_3\right)\,\mathbf{a}_3$ | = | $-z_3a\,\hat{\mathbf{x}} + \left(\frac{1}{2} - x_3\right)a\,\hat{\mathbf{y}} + y_3a\,\hat{\mathbf{z}}$ | (48$e$) | O |
| $\mathbf{B}_{22}$ | = | $(x_3 - y_3)\,\mathbf{a}_1 + \left(\frac{1}{2} - y_3 - z_3\right)\,\mathbf{a}_2 + \left(\frac{1}{2} + x_3 - z_3\right)\,\mathbf{a}_3$ | = | $\left(\frac{1}{2} - z_3\right)a\,\hat{\mathbf{x}} + x_3a\,\hat{\mathbf{y}} - y_3a\,\hat{\mathbf{z}}$ | (48$e$) | O |
| $\mathbf{B}_{23}$ | = | $(x_3 + z_3)\,\mathbf{a}_1 + (x_3 + y_3)\,\mathbf{a}_2 + (y_3 + z_3)\,\mathbf{a}_3$ | = | $y_3a\,\hat{\mathbf{x}} + z_3a\,\hat{\mathbf{y}} + x_3a\,\hat{\mathbf{z}}$ | (48$e$) | O |
| $\mathbf{B}_{24}$ | = | $(-x_3 + z_3)\,\mathbf{a}_1 + \left(\frac{1}{2} - x_3 - y_3\right)\,\mathbf{a}_2 + \left(\frac{1}{2} - y_3 + z_3\right)\,\mathbf{a}_3$ | = | $\left(\frac{1}{2} - y_3\right)a\,\hat{\mathbf{x}} + z_3a\,\hat{\mathbf{y}} - x_3a\,\hat{\mathbf{z}}$ | (48$e$) | O |
| $\mathbf{B}_{25}$ | = | $\left(\frac{1}{2} - x_3 - z_3\right)\,\mathbf{a}_1 + \left(\frac{1}{2} - x_3 + y_3\right)\,\mathbf{a}_2 + (y_3 - z_3)\,\mathbf{a}_3$ | = | $y_3a\,\hat{\mathbf{x}} - z_3a\,\hat{\mathbf{y}} + \left(\frac{1}{2} - x_3\right)a\,\hat{\mathbf{z}}$ | (48$e$) | O |
| $\mathbf{B}_{26}$ | = | $\left(\frac{1}{2}+x_3-z_3\right)\,\mathbf{a}_1+\left(x_3-y_3\right)\,\mathbf{a}_2+\left(\frac{1}{2}-y_3-z_3\right)\,\mathbf{a}_3$ | = | $-y_3a\,\hat{\mathbf{x}}+\left(\frac{1}{2}-z_3\right)a\,\hat{\mathbf{y}}+x_3a\,\hat{\mathbf{z}}$ | (48e) | O |
| $\mathbf{B}_{27}$ | = | $\left(\frac{1}{2}+x_3+z_3\right)\,\mathbf{a}_1+\left(\frac{1}{2}+y_3+z_3\right)\,\mathbf{a}_2+\left(\frac{1}{2}+x_3+y_3\right)\,\mathbf{a}_3$ | = | $\left(\frac{1}{4}+y_3\right)a\,\hat{\mathbf{x}}+\left(\frac{1}{4}+x_3\right)a\,\hat{\mathbf{y}}+\left(\frac{1}{4}+z_3\right)a\,\hat{\mathbf{z}}$ | (48e) | O |
| $\mathbf{B}_{28}$ | = | $\left(\frac{1}{2}-x_3+z_3\right)\,\mathbf{a}_1+\left(-y_3+z_3\right)\,\mathbf{a}_2+\left(-x_3-y_3\right)\,\mathbf{a}_3$ | = | $-a\left(y_3+\frac{1}{4}\right)\,\hat{\mathbf{x}}+\left(\frac{1}{4}-x_3\right)a\,\hat{\mathbf{y}}+\left(\frac{1}{4}+z_3\right)a\,\hat{\mathbf{z}}$ | (48e) | O |
| $\mathbf{B}_{29}$ | = | $\left(-x_3-z_3\right)\,\mathbf{a}_1+\left(\frac{1}{2}+y_3-z_3\right)\,\mathbf{a}_2+\left(-x_3+y_3\right)\,\mathbf{a}_3$ | = | $\left(\frac{1}{4}+y_3\right)a\,\hat{\mathbf{x}}-a\left(x_3+\frac{1}{4}\right)\,\hat{\mathbf{y}}+\left(\frac{1}{4}-z_3\right)a\,\hat{\mathbf{z}}$ | (48e) | O |
| $\mathbf{B}_{30}$ | = | $\left(x_3-z_3\right)\,\mathbf{a}_1+\left(-y_3-z_3\right)\,\mathbf{a}_2+\left(\frac{1}{2}+x_3-y_3\right)\,\mathbf{a}_3$ | = | $\left(\frac{1}{4}-y_3\right)a\,\hat{\mathbf{x}}+\left(\frac{1}{4}+x_3\right)a\,\hat{\mathbf{y}}-a\left(z_3+\frac{1}{4}\right)\,\hat{\mathbf{z}}$ | (48e) | O |
| $\mathbf{B}_{31}$ | = | $\left(\frac{1}{2}+y_3+z_3\right)\,\mathbf{a}_1+\left(\frac{1}{2}+x_3+y_3\right)\,\mathbf{a}_2+\left(\frac{1}{2}+x_3+z_3\right)\,\mathbf{a}_3$ | = | $\left(\frac{1}{4}+x_3\right)a\,\hat{\mathbf{x}}+\left(\frac{1}{4}+z_3\right)a\,\hat{\mathbf{y}}+\left(\frac{1}{4}+y_3\right)a\,\hat{\mathbf{z}}$ | (48e) | O |
| $\mathbf{B}_{32}$ | = | $\left(-y_3+z_3\right)\,\mathbf{a}_1+\left(-x_3-y_3\right)\,\mathbf{a}_2+\left(\frac{1}{2}-x_3+z_3\right)\,\mathbf{a}_3$ | = | $\left(\frac{1}{4}-x_3\right)a\,\hat{\mathbf{x}}+\left(\frac{1}{4}+z_3\right)a\,\hat{\mathbf{y}}-a\left(y_3+\frac{1}{4}\right)\,\hat{\mathbf{z}}$ | (48e) | O |
| $\mathbf{B}_{33}$ | = | $\left(\frac{1}{2}+y_3-z_3\right)\,\mathbf{a}_1+\left(-x_3+y_3\right)\,\mathbf{a}_2+\left(-x_3-z_3\right)\,\mathbf{a}_3$ | = | $-a\left(x_3+\frac{1}{4}\right)\,\hat{\mathbf{x}}+\left(\frac{1}{4}-z_3\right)a\,\hat{\mathbf{y}}+\left(\frac{1}{4}+y_3\right)a\,\hat{\mathbf{z}}$ | (48e) | O |
| $\mathbf{B}_{34}$ | = | $\left(-y_3-z_3\right)\,\mathbf{a}_1+\left(\frac{1}{2}+x_3-y_3\right)\,\mathbf{a}_2+\left(x_3-z_3\right)\,\mathbf{a}_3$ | = | $\left(\frac{1}{4}+x_3\right)a\,\hat{\mathbf{x}}-a\left(z_3+\frac{1}{4}\right)\,\hat{\mathbf{y}}+\left(\frac{1}{4}-y_3\right)a\,\hat{\mathbf{z}}$ | (48e) | O |
| $\mathbf{B}_{35}$ | = | $\left(\frac{1}{2}+x_3+y_3\right)\,\mathbf{a}_1+\left(\frac{1}{2}+x_3+z_3\right)\,\mathbf{a}_2+\left(\frac{1}{2}+y_3+z_3\right)\,\mathbf{a}_3$ | = | $\left(\frac{1}{4}+z_3\right)a\,\hat{\mathbf{x}}+\left(\frac{1}{4}+y_3\right)a\,\hat{\mathbf{y}}+\left(\frac{1}{4}+x_3\right)a\,\hat{\mathbf{z}}$ | (48e) | O |
| $\mathbf{B}_{36}$ | = | $\left(-x_3-y_3\right)\,\mathbf{a}_1+\left(\frac{1}{2}-x_3+z_3\right)\,\mathbf{a}_2+\left(-y_3+z_3\right)\,\mathbf{a}_3$ | = | $\left(\frac{1}{4}+z_3\right)a\,\hat{\mathbf{x}}-a\left(y_3+\frac{1}{4}\right)\,\hat{\mathbf{y}}+\left(\frac{1}{4}-x_3\right)a\,\hat{\mathbf{z}}$ | (48e) | O |
| $\mathbf{B}_{37}$ | = | $\left(-x_3+y_3\right)\,\mathbf{a}_1+\left(-x_3-z_3\right)\,\mathbf{a}_2+\left(\frac{1}{2}+y_3-z_3\right)\,\mathbf{a}_3$ | = | $\left(\frac{1}{4}-z_3\right)a\,\hat{\mathbf{x}}+\left(\frac{1}{4}+y_3\right)a\,\hat{\mathbf{y}}-a\left(x_3+\frac{1}{4}\right)\,\hat{\mathbf{z}}$ | (48e) | O |
| $\mathbf{B}_{38}$ | = | $\left(\frac{1}{2}+x_3-y_3\right)\,\mathbf{a}_1+\left(x_3-z_3\right)\,\mathbf{a}_2+\left(-y_3-z_3\right)\,\mathbf{a}_3$ | = | $-a\left(z_3+\frac{1}{4}\right)\,\hat{\mathbf{x}}+\left(\frac{1}{4}-y_3\right)a\,\hat{\mathbf{y}}+\left(\frac{1}{4}+x_3\right)a\,\hat{\mathbf{z}}$ | (48e) | O |

**Found in:**

**Geometry files:**

- CIF: pp. 1802
- POSCAR: pp. 1802

# Mayenite ($12CaO \cdot 7Al_2O_3$, $K7_4$, C12A7) Structure: A7B12C19_cI152_220_bc_2d_ace

http://aflow.org/prototype-encyclopedia/A7B12C19_cI152_220_bc_2d_ace

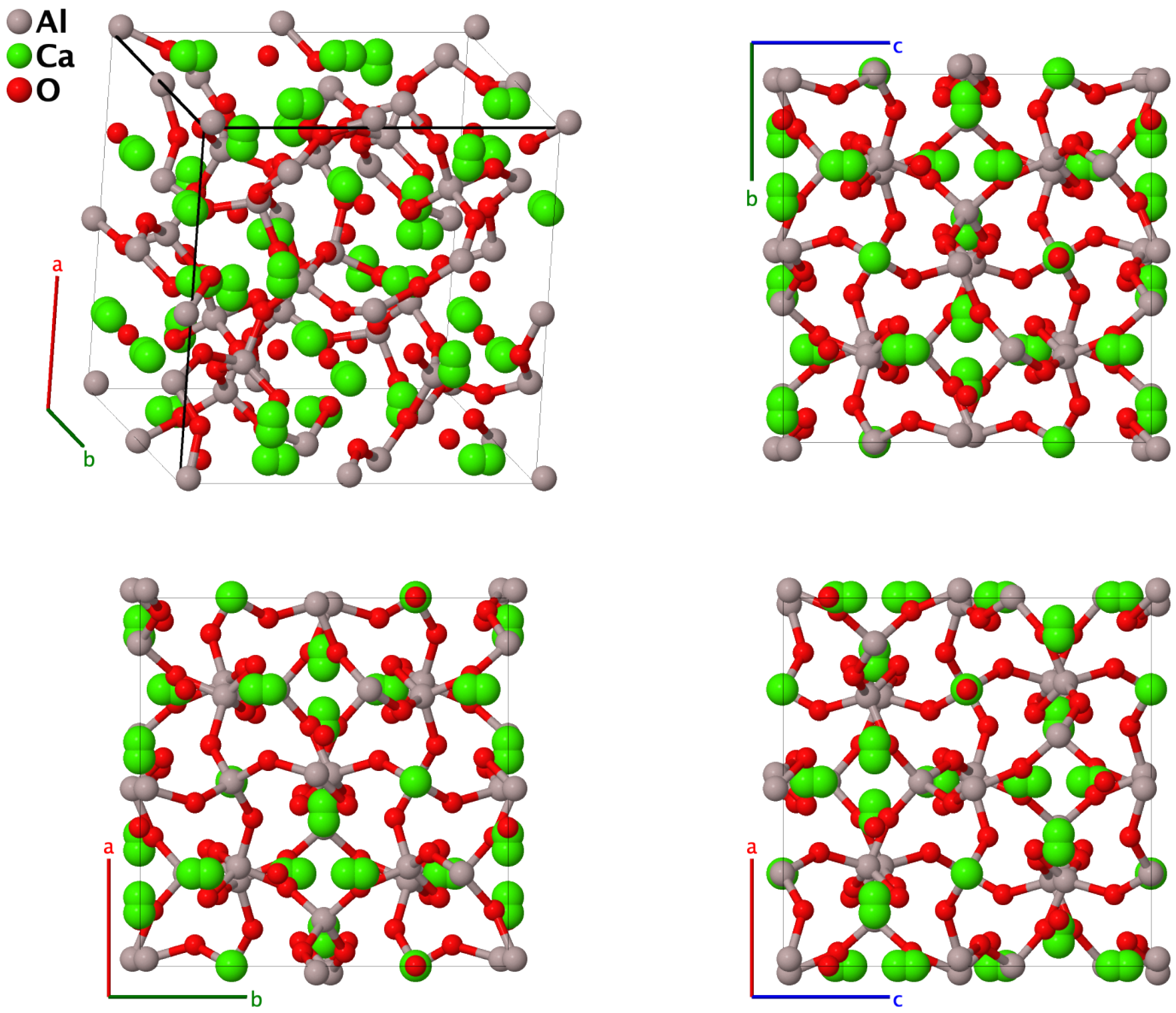

| | | |
|---|---|---|
| **Prototype** | : | $Al_{14}Ca_{12}O_{33}$ |
| **AFLOW prototype label** | : | A7B12C19_cI152_220_bc_2d_ace |
| ***Strukturbericht* designation** | : | $K7_4$ |
| **Pearson symbol** | : | cI152 |
| **Space group number** | : | 220 |
| **Space group symbol** | : | $I\bar{4}3d$ |
| **AFLOW prototype command** | : | `aflow --proto=A7B12C19_cI152_220_bc_2d_ace --params=`$a, x_3, x_4, x_5, x_6, x_7, y_7, z_7$ |

- We present the structural determined by (Boysen, 2007) with data taken at 293 K. This slightly differs from the original determination of (Büssem, 1936), which was given the $K7_4$ designation by (Gottfried, 1938). In the original work, the calcium atoms were thought to be located at a single (24*d*) site. Newer findings show that calcium is split between two (24*d*) sites, with the site we have labeled Ca-I having 87.5% of the atoms and Ca-II the remainder, although presumably only one of the two sites is occupied in any pair.

- In all works the O-I (12*a*) site is only partially occupied: if this is occupied 1/6 of the time, we get the proper stoichiometry, though (Boysen, 2007) found the occupation was 0.251 at 293 K, dropping as the temperature decreased.

- This structure is often referred to in the literature as C12A7, to distinguish it from other $CaO/Al_2O_3$ compounds.

**Body-centered Cubic primitive vectors:**

$$\mathbf{a}_1 = -\frac{1}{2}\,a\,\hat{\mathbf{x}} + \frac{1}{2}\,a\,\hat{\mathbf{y}} + \frac{1}{2}\,a\,\hat{\mathbf{z}}$$
$$\mathbf{a}_2 = \frac{1}{2}\,a\,\hat{\mathbf{x}} - \frac{1}{2}\,a\,\hat{\mathbf{y}} + \frac{1}{2}\,a\,\hat{\mathbf{z}}$$
$$\mathbf{a}_3 = \frac{1}{2}\,a\,\hat{\mathbf{x}} + \frac{1}{2}\,a\,\hat{\mathbf{y}} - \frac{1}{2}\,a\,\hat{\mathbf{z}}$$

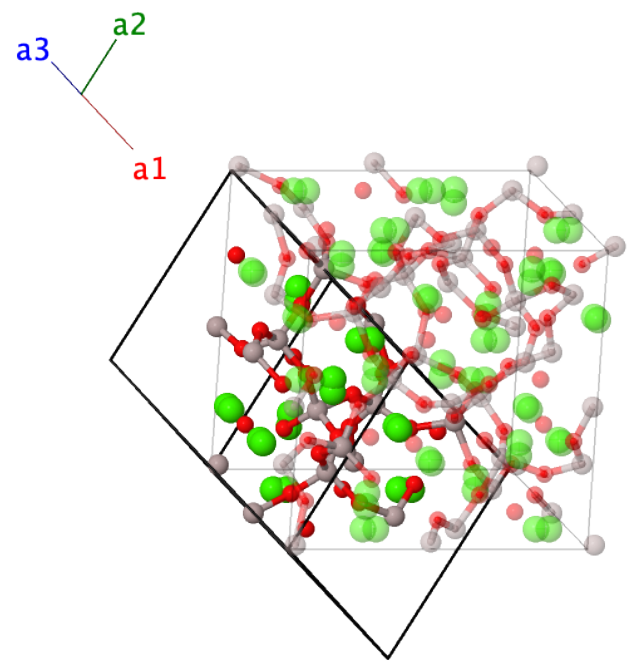

**Basis vectors:**

| | | Lattice Coordinates | | Cartesian Coordinates | Wyckoff Position | Atom Type |
|---|---|---|---|---|---|---|
| $\mathbf{B}_1$ | = | $\frac{1}{4}\,\mathbf{a}_1 + \frac{5}{8}\,\mathbf{a}_2 + \frac{3}{8}\,\mathbf{a}_3$ | = | $\frac{3}{8}a\,\hat{\mathbf{x}} + \frac{1}{4}a\,\hat{\mathbf{z}}$ | (12$a$) | O I |
| $\mathbf{B}_2$ | = | $\frac{3}{4}\,\mathbf{a}_1 + \frac{7}{8}\,\mathbf{a}_2 + \frac{1}{8}\,\mathbf{a}_3$ | = | $\frac{1}{8}a\,\hat{\mathbf{x}} + \frac{3}{4}a\,\hat{\mathbf{z}}$ | (12$a$) | O I |
| $\mathbf{B}_3$ | = | $\frac{3}{8}\,\mathbf{a}_1 + \frac{1}{4}\,\mathbf{a}_2 + \frac{5}{8}\,\mathbf{a}_3$ | = | $\frac{1}{4}a\,\hat{\mathbf{x}} + \frac{3}{8}a\,\hat{\mathbf{y}}$ | (12$a$) | O I |
| $\mathbf{B}_4$ | = | $\frac{1}{8}\,\mathbf{a}_1 + \frac{3}{4}\,\mathbf{a}_2 + \frac{7}{8}\,\mathbf{a}_3$ | = | $\frac{3}{4}a\,\hat{\mathbf{x}} + \frac{1}{8}a\,\hat{\mathbf{y}}$ | (12$a$) | O I |
| $\mathbf{B}_5$ | = | $\frac{5}{8}\,\mathbf{a}_1 + \frac{3}{8}\,\mathbf{a}_2 + \frac{1}{4}\,\mathbf{a}_3$ | = | $\frac{1}{4}a\,\hat{\mathbf{y}} + \frac{3}{8}a\,\hat{\mathbf{z}}$ | (12$a$) | O I |
| $\mathbf{B}_6$ | = | $\frac{7}{8}\,\mathbf{a}_1 + \frac{1}{8}\,\mathbf{a}_2 + \frac{3}{4}\,\mathbf{a}_3$ | = | $\frac{3}{4}a\,\hat{\mathbf{y}} + \frac{1}{8}a\,\hat{\mathbf{z}}$ | (12$a$) | O I |
| $\mathbf{B}_7$ | = | $\frac{1}{4}\,\mathbf{a}_1 + \frac{1}{8}\,\mathbf{a}_2 + \frac{7}{8}\,\mathbf{a}_3$ | = | $\frac{3}{8}a\,\hat{\mathbf{x}} + \frac{1}{2}a\,\hat{\mathbf{y}} - \frac{1}{4}a\,\hat{\mathbf{z}}$ | (12$b$) | Al I |
| $\mathbf{B}_8$ | = | $\frac{3}{4}\,\mathbf{a}_1 + \frac{3}{8}\,\mathbf{a}_2 + \frac{5}{8}\,\mathbf{a}_3$ | = | $\frac{1}{8}a\,\hat{\mathbf{x}} + \frac{1}{2}a\,\hat{\mathbf{y}} + \frac{1}{4}a\,\hat{\mathbf{z}}$ | (12$b$) | Al I |
| $\mathbf{B}_9$ | = | $\frac{7}{8}\,\mathbf{a}_1 + \frac{1}{4}\,\mathbf{a}_2 + \frac{1}{8}\,\mathbf{a}_3$ | = | $-\frac{1}{4}a\,\hat{\mathbf{x}} + \frac{3}{8}a\,\hat{\mathbf{y}} + \frac{1}{2}a\,\hat{\mathbf{z}}$ | (12$b$) | Al I |
| $\mathbf{B}_{10}$ | = | $\frac{5}{8}\,\mathbf{a}_1 + \frac{3}{4}\,\mathbf{a}_2 + \frac{3}{8}\,\mathbf{a}_3$ | = | $\frac{1}{4}a\,\hat{\mathbf{x}} + \frac{1}{8}a\,\hat{\mathbf{y}} + \frac{1}{2}a\,\hat{\mathbf{z}}$ | (12$b$) | Al I |
| $\mathbf{B}_{11}$ | = | $\frac{1}{8}\,\mathbf{a}_1 + \frac{7}{8}\,\mathbf{a}_2 + \frac{1}{4}\,\mathbf{a}_3$ | = | $\frac{1}{2}a\,\hat{\mathbf{x}} - \frac{1}{4}a\,\hat{\mathbf{y}} + \frac{3}{8}a\,\hat{\mathbf{z}}$ | (12$b$) | Al I |
| $\mathbf{B}_{12}$ | = | $\frac{3}{8}\,\mathbf{a}_1 + \frac{5}{8}\,\mathbf{a}_2 + \frac{3}{4}\,\mathbf{a}_3$ | = | $\frac{1}{2}a\,\hat{\mathbf{x}} + \frac{1}{4}a\,\hat{\mathbf{y}} + \frac{1}{8}a\,\hat{\mathbf{z}}$ | (12$b$) | Al I |
| $\mathbf{B}_{13}$ | = | $2x_3\,\mathbf{a}_1 + 2x_3\,\mathbf{a}_2 + 2x_3\,\mathbf{a}_3$ | = | $x_3a\,\hat{\mathbf{x}} + x_3a\,\hat{\mathbf{y}} + x_3a\,\hat{\mathbf{z}}$ | (16$c$) | Al II |
| $\mathbf{B}_{14}$ | = | $\frac{1}{2}\,\mathbf{a}_1 + \left(\frac{1}{2} - 2x_3\right)\,\mathbf{a}_3$ | = | $-x_3a\,\hat{\mathbf{x}} + \left(\frac{1}{2} - x_3\right)a\,\hat{\mathbf{y}} + x_3a\,\hat{\mathbf{z}}$ | (16$c$) | Al II |
| $\mathbf{B}_{15}$ | = | $\left(\frac{1}{2} - 2x_3\right)\,\mathbf{a}_2 + \frac{1}{2}\,\mathbf{a}_3$ | = | $\left(\frac{1}{2} - x_3\right)a\,\hat{\mathbf{x}} + x_3a\,\hat{\mathbf{y}} - x_3a\,\hat{\mathbf{z}}$ | (16$c$) | Al II |
| $\mathbf{B}_{16}$ | = | $\left(\frac{1}{2} - 2x_3\right)\,\mathbf{a}_1 + \frac{1}{2}\,\mathbf{a}_2$ | = | $x_3a\,\hat{\mathbf{x}} - x_3a\,\hat{\mathbf{y}} + \left(\frac{1}{2} - x_3\right)a\,\hat{\mathbf{z}}$ | (16$c$) | Al II |
| $\mathbf{B}_{17}$ | = | $\left(\frac{1}{2} + 2x_3\right)\,\mathbf{a}_1 + \left(\frac{1}{2} + 2x_3\right)\,\mathbf{a}_2 + \left(\frac{1}{2} + 2x_3\right)\,\mathbf{a}_3$ | = | $\left(\frac{1}{4} + x_3\right)a\,\hat{\mathbf{x}} + \left(\frac{1}{4} + x_3\right)a\,\hat{\mathbf{y}} + \left(\frac{1}{4} + x_3\right)a\,\hat{\mathbf{z}}$ | (16$c$) | Al II |
| $\mathbf{B}_{18}$ | = | $\frac{1}{2}\,\mathbf{a}_1 - 2x_3\,\mathbf{a}_3$ | = | $-a\left(x_3 + \frac{1}{4}\right)\,\hat{\mathbf{x}} + \left(\frac{1}{4} - x_3\right)a\,\hat{\mathbf{y}} + \left(\frac{1}{4} + x_3\right)a\,\hat{\mathbf{z}}$ | (16$c$) | Al II |
| $\mathbf{B}_{19}$ | = | $-2x_3\,\mathbf{a}_1 + \frac{1}{2}\,\mathbf{a}_2$ | = | $\left(\frac{1}{4} + x_3\right)a\,\hat{\mathbf{x}} - a\left(x_3 + \frac{1}{4}\right)\,\hat{\mathbf{y}} + \left(\frac{1}{4} - x_3\right)a\,\hat{\mathbf{z}}$ | (16$c$) | Al II |
| $\mathbf{B}_{20}$ | = | $-2x_3\,\mathbf{a}_2 + \frac{1}{2}\,\mathbf{a}_3$ | = | $\left(\frac{1}{4} - x_3\right)a\,\hat{\mathbf{x}} + \left(\frac{1}{4} + x_3\right)a\,\hat{\mathbf{y}} - a\left(x_3 + \frac{1}{4}\right)\,\hat{\mathbf{z}}$ | (16$c$) | Al II |
| $\mathbf{B}_{21}$ | = | $2x_4\,\mathbf{a}_1 + 2x_4\,\mathbf{a}_2 + 2x_4\,\mathbf{a}_3$ | = | $x_4a\,\hat{\mathbf{x}} + x_4a\,\hat{\mathbf{y}} + x_4a\,\hat{\mathbf{z}}$ | (16$c$) | O II |
| $\mathbf{B}_{22}$ | = | $\frac{1}{2}\,\mathbf{a}_1 + \left(\frac{1}{2} - 2x_4\right)\,\mathbf{a}_3$ | = | $-x_4a\,\hat{\mathbf{x}} + \left(\frac{1}{2} - x_4\right)a\,\hat{\mathbf{y}} + x_4a\,\hat{\mathbf{z}}$ | (16$c$) | O II |
| $\mathbf{B}_{23}$ | = | $\left(\frac{1}{2} - 2x_4\right)\,\mathbf{a}_2 + \frac{1}{2}\,\mathbf{a}_3$ | = | $\left(\frac{1}{2} - x_4\right)a\,\hat{\mathbf{x}} + x_4a\,\hat{\mathbf{y}} - x_4a\,\hat{\mathbf{z}}$ | (16$c$) | O II |
| $\mathbf{B}_{24}$ | = | $\left(\frac{1}{2} - 2x_4\right)\,\mathbf{a}_1 + \frac{1}{2}\,\mathbf{a}_2$ | = | $x_4a\,\hat{\mathbf{x}} - x_4a\,\hat{\mathbf{y}} + \left(\frac{1}{2} - x_4\right)a\,\hat{\mathbf{z}}$ | (16$c$) | O II |
| $\mathbf{B}_{25}$ | = | $\left(\frac{1}{2}+2x_4\right)\mathbf{a}_1+\left(\frac{1}{2}+2x_4\right)\mathbf{a}_2+\left(\frac{1}{2}+2x_4\right)\mathbf{a}_3$ | = | $\left(\frac{1}{4}+x_4\right)a\,\hat{\mathbf{x}}+\left(\frac{1}{4}+x_4\right)a\,\hat{\mathbf{y}}+\left(\frac{1}{4}+x_4\right)a\,\hat{\mathbf{z}}$ | (16*c*) | O II |
| $\mathbf{B}_{26}$ | = | $\frac{1}{2}\,\mathbf{a}_1-2x_4\,\mathbf{a}_3$ | = | $-a\left(x_4+\frac{1}{4}\right)\hat{\mathbf{x}}+\left(\frac{1}{4}-x_4\right)a\,\hat{\mathbf{y}}+\left(\frac{1}{4}+x_4\right)a\,\hat{\mathbf{z}}$ | (16*c*) | O II |
| $\mathbf{B}_{27}$ | = | $-2x_4\,\mathbf{a}_1+\frac{1}{2}\,\mathbf{a}_2$ | = | $\left(\frac{1}{4}+x_4\right)a\,\hat{\mathbf{x}}-a\left(x_4+\frac{1}{4}\right)\hat{\mathbf{y}}+\left(\frac{1}{4}-x_4\right)a\,\hat{\mathbf{z}}$ | (16*c*) | O II |
| $\mathbf{B}_{28}$ | = | $-2x_4\,\mathbf{a}_2+\frac{1}{2}\,\mathbf{a}_3$ | = | $\left(\frac{1}{4}-x_4\right)a\,\hat{\mathbf{x}}+\left(\frac{1}{4}+x_4\right)a\,\hat{\mathbf{y}}-a\left(x_4+\frac{1}{4}\right)\hat{\mathbf{z}}$ | (16*c*) | O II |
| $\mathbf{B}_{29}$ | = | $\frac{1}{4}\,\mathbf{a}_1+\left(\frac{1}{4}+x_5\right)\mathbf{a}_2+x_5\,\mathbf{a}_3$ | = | $x_5a\,\hat{\mathbf{x}}+\frac{1}{4}a\,\hat{\mathbf{z}}$ | (24*d*) | Ca I |
| $\mathbf{B}_{30}$ | = | $\frac{3}{4}\,\mathbf{a}_1+\left(\frac{1}{4}-x_5\right)\mathbf{a}_2+\left(\frac{1}{2}-x_5\right)\mathbf{a}_3$ | = | $-x_5a\,\hat{\mathbf{x}}+\frac{1}{2}a\,\hat{\mathbf{y}}+\frac{1}{4}a\,\hat{\mathbf{z}}$ | (24*d*) | Ca I |
| $\mathbf{B}_{31}$ | = | $x_5\,\mathbf{a}_1+\frac{1}{4}\,\mathbf{a}_2+\left(\frac{1}{4}+x_5\right)\mathbf{a}_3$ | = | $\frac{1}{4}a\,\hat{\mathbf{x}}+x_5a\,\hat{\mathbf{y}}$ | (24*d*) | Ca I |
| $\mathbf{B}_{32}$ | = | $\left(\frac{1}{2}-x_5\right)\mathbf{a}_1+\frac{3}{4}\,\mathbf{a}_2+\left(\frac{1}{4}-x_5\right)\mathbf{a}_3$ | = | $\frac{1}{4}a\,\hat{\mathbf{x}}-x_5a\,\hat{\mathbf{y}}+\frac{1}{2}a\,\hat{\mathbf{z}}$ | (24*d*) | Ca I |
| $\mathbf{B}_{33}$ | = | $\left(\frac{1}{4}+x_5\right)\mathbf{a}_1+x_5\,\mathbf{a}_2+\frac{1}{4}\,\mathbf{a}_3$ | = | $\frac{1}{4}a\,\hat{\mathbf{y}}+x_5a\,\hat{\mathbf{z}}$ | (24*d*) | Ca I |
| $\mathbf{B}_{34}$ | = | $\left(\frac{1}{4}-x_5\right)\mathbf{a}_1+\left(\frac{1}{2}-x_5\right)\mathbf{a}_2+\frac{3}{4}\,\mathbf{a}_3$ | = | $\frac{1}{2}a\,\hat{\mathbf{x}}+\frac{1}{4}a\,\hat{\mathbf{y}}-x_5a\,\hat{\mathbf{z}}$ | (24*d*) | Ca I |
| $\mathbf{B}_{35}$ | = | $\left(\frac{3}{4}+x_5\right)\mathbf{a}_1+\frac{3}{4}\,\mathbf{a}_2+\left(\frac{1}{2}+x_5\right)\mathbf{a}_3$ | = | $\frac{1}{4}a\,\hat{\mathbf{x}}+\left(\frac{1}{4}+x_5\right)a\,\hat{\mathbf{y}}+\frac{1}{2}a\,\hat{\mathbf{z}}$ | (24*d*) | Ca I |
| $\mathbf{B}_{36}$ | = | $\left(\frac{3}{4}-x_5\right)\mathbf{a}_1+\frac{1}{4}\,\mathbf{a}_2-x_5\,\mathbf{a}_3$ | = | $-\frac{1}{4}a\,\hat{\mathbf{x}}+\left(\frac{1}{4}-x_5\right)a\,\hat{\mathbf{y}}+\frac{1}{2}a\,\hat{\mathbf{z}}$ | (24*d*) | Ca I |
| $\mathbf{B}_{37}$ | = | $\frac{3}{4}\,\mathbf{a}_1+\left(\frac{1}{2}+x_5\right)\mathbf{a}_2+\left(\frac{3}{4}+x_5\right)\mathbf{a}_3$ | = | $\left(\frac{1}{4}+x_5\right)a\,\hat{\mathbf{x}}+\frac{1}{2}a\,\hat{\mathbf{y}}+\frac{1}{4}a\,\hat{\mathbf{z}}$ | (24*d*) | Ca I |
| $\mathbf{B}_{38}$ | = | $\frac{1}{4}\,\mathbf{a}_1-x_5\,\mathbf{a}_2+\left(\frac{3}{4}-x_5\right)\mathbf{a}_3$ | = | $\left(\frac{1}{4}-x_5\right)a\,\hat{\mathbf{x}}+\frac{1}{2}a\,\hat{\mathbf{y}}-\frac{1}{4}a\,\hat{\mathbf{z}}$ | (24*d*) | Ca I |
| $\mathbf{B}_{39}$ | = | $\left(\frac{1}{2}+x_5\right)\mathbf{a}_1+\left(\frac{3}{4}+x_5\right)\mathbf{a}_2+\frac{3}{4}\,\mathbf{a}_3$ | = | $\frac{1}{2}a\,\hat{\mathbf{x}}+\frac{1}{4}a\,\hat{\mathbf{y}}+\left(\frac{1}{4}+x_5\right)a\,\hat{\mathbf{z}}$ | (24*d*) | Ca I |
| $\mathbf{B}_{40}$ | = | $-x_5\,\mathbf{a}_1+\left(\frac{3}{4}-x_5\right)\mathbf{a}_2+\frac{1}{4}\,\mathbf{a}_3$ | = | $\frac{1}{2}a\,\hat{\mathbf{x}}-\frac{1}{4}a\,\hat{\mathbf{y}}+\left(\frac{1}{4}-x_5\right)a\,\hat{\mathbf{z}}$ | (24*d*) | Ca I |
| $\mathbf{B}_{41}$ | = | $\frac{1}{4}\,\mathbf{a}_1+\left(\frac{1}{4}+x_6\right)\mathbf{a}_2+x_6\,\mathbf{a}_3$ | = | $x_6a\,\hat{\mathbf{x}}+\frac{1}{4}a\,\hat{\mathbf{z}}$ | (24*d*) | Ca II |
| $\mathbf{B}_{42}$ | = | $\frac{3}{4}\,\mathbf{a}_1+\left(\frac{1}{4}-x_6\right)\mathbf{a}_2+\left(\frac{1}{2}-x_6\right)\mathbf{a}_3$ | = | $-x_6a\,\hat{\mathbf{x}}+\frac{1}{2}a\,\hat{\mathbf{y}}+\frac{1}{4}a\,\hat{\mathbf{z}}$ | (24*d*) | Ca II |
| $\mathbf{B}_{43}$ | = | $x_6\,\mathbf{a}_1+\frac{1}{4}\,\mathbf{a}_2+\left(\frac{1}{4}+x_6\right)\mathbf{a}_3$ | = | $\frac{1}{4}a\,\hat{\mathbf{x}}+x_6a\,\hat{\mathbf{y}}$ | (24*d*) | Ca II |
| $\mathbf{B}_{44}$ | = | $\left(\frac{1}{2}-x_6\right)\mathbf{a}_1+\frac{3}{4}\,\mathbf{a}_2+\left(\frac{1}{4}-x_6\right)\mathbf{a}_3$ | = | $\frac{1}{4}a\,\hat{\mathbf{x}}-x_6a\,\hat{\mathbf{y}}+\frac{1}{2}a\,\hat{\mathbf{z}}$ | (24*d*) | Ca II |
| $\mathbf{B}_{45}$ | = | $\left(\frac{1}{4}+x_6\right)\mathbf{a}_1+x_6\,\mathbf{a}_2+\frac{1}{4}\,\mathbf{a}_3$ | = | $\frac{1}{4}a\,\hat{\mathbf{y}}+x_6a\,\hat{\mathbf{z}}$ | (24*d*) | Ca II |
| $\mathbf{B}_{46}$ | = | $\left(\frac{1}{4}-x_6\right)\mathbf{a}_1+\left(\frac{1}{2}-x_6\right)\mathbf{a}_2+\frac{3}{4}\,\mathbf{a}_3$ | = | $\frac{1}{2}a\,\hat{\mathbf{x}}+\frac{1}{4}a\,\hat{\mathbf{y}}-x_6a\,\hat{\mathbf{z}}$ | (24*d*) | Ca II |
| $\mathbf{B}_{47}$ | = | $\left(\frac{3}{4}+x_6\right)\mathbf{a}_1+\frac{3}{4}\,\mathbf{a}_2+\left(\frac{1}{2}+x_6\right)\mathbf{a}_3$ | = | $\frac{1}{4}a\,\hat{\mathbf{x}}+\left(\frac{1}{4}+x_6\right)a\,\hat{\mathbf{y}}+\frac{1}{2}a\,\hat{\mathbf{z}}$ | (24*d*) | Ca II |
| $\mathbf{B}_{48}$ | = | $\left(\frac{3}{4}-x_6\right)\mathbf{a}_1+\frac{1}{4}\,\mathbf{a}_2-x_6\,\mathbf{a}_3$ | = | $-\frac{1}{4}a\,\hat{\mathbf{x}}+\left(\frac{1}{4}-x_6\right)a\,\hat{\mathbf{y}}+\frac{1}{2}a\,\hat{\mathbf{z}}$ | (24*d*) | Ca II |
| $\mathbf{B}_{49}$ | = | $\frac{3}{4}\,\mathbf{a}_1+\left(\frac{1}{2}+x_6\right)\mathbf{a}_2+\left(\frac{3}{4}+x_6\right)\mathbf{a}_3$ | = | $\left(\frac{1}{4}+x_6\right)a\,\hat{\mathbf{x}}+\frac{1}{2}a\,\hat{\mathbf{y}}+\frac{1}{4}a\,\hat{\mathbf{z}}$ | (24*d*) | Ca II |
| $\mathbf{B}_{50}$ | = | $\frac{1}{4}\,\mathbf{a}_1-x_6\,\mathbf{a}_2+\left(\frac{3}{4}-x_6\right)\mathbf{a}_3$ | = | $\left(\frac{1}{4}-x_6\right)a\,\hat{\mathbf{x}}+\frac{1}{2}a\,\hat{\mathbf{y}}-\frac{1}{4}a\,\hat{\mathbf{z}}$ | (24*d*) | Ca II |
| $\mathbf{B}_{51}$ | = | $\left(\frac{1}{2}+x_6\right)\mathbf{a}_1+\left(\frac{3}{4}+x_6\right)\mathbf{a}_2+\frac{3}{4}\,\mathbf{a}_3$ | = | $\frac{1}{2}a\,\hat{\mathbf{x}}+\frac{1}{4}a\,\hat{\mathbf{y}}+\left(\frac{1}{4}+x_6\right)a\,\hat{\mathbf{z}}$ | (24*d*) | Ca II |
| $\mathbf{B}_{52}$ | = | $-x_6\,\mathbf{a}_1+\left(\frac{3}{4}-x_6\right)\mathbf{a}_2+\frac{1}{4}\,\mathbf{a}_3$ | = | $\frac{1}{2}a\,\hat{\mathbf{x}}-\frac{1}{4}a\,\hat{\mathbf{y}}+\left(\frac{1}{4}-x_6\right)a\,\hat{\mathbf{z}}$ | (24*d*) | Ca II |
| $\mathbf{B}_{53}$ | = | $(y_7+z_7)\,\mathbf{a}_1+(x_7+z_7)\,\mathbf{a}_2+(x_7+y_7)\,\mathbf{a}_3$ | = | $x_7a\,\hat{\mathbf{x}}+y_7a\,\hat{\mathbf{y}}+z_7a\,\hat{\mathbf{z}}$ | (48*e*) | O III |
| $\mathbf{B}_{54}$ | = | $\left(\frac{1}{2}-y_7+z_7\right)\mathbf{a}_1+(-x_7+z_7)\,\mathbf{a}_2+\left(\frac{1}{2}-x_7-y_7\right)\mathbf{a}_3$ | = | $-x_7a\,\hat{\mathbf{x}}+\left(\frac{1}{2}-y_7\right)a\,\hat{\mathbf{y}}+z_7a\,\hat{\mathbf{z}}$ | (48*e*) | O III |
| $\mathbf{B}_{55}$ | = | $(y_7-z_7)\,\mathbf{a}_1+\left(\frac{1}{2}-x_7-z_7\right)\mathbf{a}_2+\left(\frac{1}{2}-x_7+y_7\right)\mathbf{a}_3$ | = | $\left(\frac{1}{2}-x_7\right)a\,\hat{\mathbf{x}}+y_7a\,\hat{\mathbf{y}}-z_7a\,\hat{\mathbf{z}}$ | (48*e*) | O III |
| $\mathbf{B}_{56}$ | = | $\left(\frac{1}{2} - y_7 - z_7\right)\,\mathbf{a}_1 + \left(\frac{1}{2} + x_7 - z_7\right)\,\mathbf{a}_2 + (x_7 - y_7)\,\mathbf{a}_3$ | = | $x_7 a\,\hat{\mathbf{x}} - y_7 a\,\hat{\mathbf{y}} + \left(\frac{1}{2} - z_7\right) a\,\hat{\mathbf{z}}$ | (48*e*) | O III |
| $\mathbf{B}_{57}$ | = | $(x_7 + y_7)\,\mathbf{a}_1 + (y_7 + z_7)\,\mathbf{a}_2 + (x_7 + z_7)\,\mathbf{a}_3$ | = | $z_7 a\,\hat{\mathbf{x}} + x_7 a\,\hat{\mathbf{y}} + y_7 a\,\hat{\mathbf{z}}$ | (48*e*) | O III |
| $\mathbf{B}_{58}$ | = | $\left(\frac{1}{2} - x_7 - y_7\right)\,\mathbf{a}_1 + \left(\frac{1}{2} - y_7 + z_7\right)\,\mathbf{a}_2 + (-x_7 + z_7)\,\mathbf{a}_3$ | = | $z_7 a\,\hat{\mathbf{x}} - x_7 a\,\hat{\mathbf{y}} + \left(\frac{1}{2} - y_7\right) a\,\hat{\mathbf{z}}$ | (48*e*) | O III |
| $\mathbf{B}_{59}$ | = | $\left(\frac{1}{2} - x_7 + y_7\right)\,\mathbf{a}_1 + (y_7 - z_7)\,\mathbf{a}_2 + \left(\frac{1}{2} - x_7 - z_7\right)\,\mathbf{a}_3$ | = | $-z_7 a\,\hat{\mathbf{x}} + \left(\frac{1}{2} - x_7\right) a\,\hat{\mathbf{y}} + y_7 a\,\hat{\mathbf{z}}$ | (48*e*) | O III |
| $\mathbf{B}_{60}$ | = | $(x_7 - y_7)\,\mathbf{a}_1 + \left(\frac{1}{2} - y_7 - z_7\right)\,\mathbf{a}_2 + \left(\frac{1}{2} + x_7 - z_7\right)\,\mathbf{a}_3$ | = | $\left(\frac{1}{2} - z_7\right) a\,\hat{\mathbf{x}} + x_7 a\,\hat{\mathbf{y}} - y_7 a\,\hat{\mathbf{z}}$ | (48*e*) | O III |
| $\mathbf{B}_{61}$ | = | $(x_7 + z_7)\,\mathbf{a}_1 + (x_7 + y_7)\,\mathbf{a}_2 + (y_7 + z_7)\,\mathbf{a}_3$ | = | $y_7 a\,\hat{\mathbf{x}} + z_7 a\,\hat{\mathbf{y}} + x_7 a\,\hat{\mathbf{z}}$ | (48*e*) | O III |
| $\mathbf{B}_{62}$ | = | $(-x_7 + z_7)\,\mathbf{a}_1 + \left(\frac{1}{2} - x_7 - y_7\right)\,\mathbf{a}_2 + \left(\frac{1}{2} - y_7 + z_7\right)\,\mathbf{a}_3$ | = | $\left(\frac{1}{2} - y_7\right) a\,\hat{\mathbf{x}} + z_7 a\,\hat{\mathbf{y}} - x_7 a\,\hat{\mathbf{z}}$ | (48*e*) | O III |
| $\mathbf{B}_{63}$ | = | $\left(\frac{1}{2} - x_7 - z_7\right)\,\mathbf{a}_1 + \left(\frac{1}{2} - x_7 + y_7\right)\,\mathbf{a}_2 + (y_7 - z_7)\,\mathbf{a}_3$ | = | $y_7 a\,\hat{\mathbf{x}} - z_7 a\,\hat{\mathbf{y}} + \left(\frac{1}{2} - x_7\right) a\,\hat{\mathbf{z}}$ | (48*e*) | O III |
| $\mathbf{B}_{64}$ | = | $\left(\frac{1}{2} + x_7 - z_7\right)\,\mathbf{a}_1 + (x_7 - y_7)\,\mathbf{a}_2 + \left(\frac{1}{2} - y_7 - z_7\right)\,\mathbf{a}_3$ | = | $-y_7 a\,\hat{\mathbf{x}} + \left(\frac{1}{2} - z_7\right) a\,\hat{\mathbf{y}} + x_7 a\,\hat{\mathbf{z}}$ | (48*e*) | O III |
| $\mathbf{B}_{65}$ | = | $\left(\frac{1}{2} + x_7 + z_7\right)\,\mathbf{a}_1 + \left(\frac{1}{2} + y_7 + z_7\right)\,\mathbf{a}_2 + \left(\frac{1}{2} + x_7 + y_7\right)\,\mathbf{a}_3$ | = | $\left(\frac{1}{4} + y_7\right) a\,\hat{\mathbf{x}} + \left(\frac{1}{4} + x_7\right) a\,\hat{\mathbf{y}} + \left(\frac{1}{4} + z_7\right) a\,\hat{\mathbf{z}}$ | (48*e*) | O III |
| $\mathbf{B}_{66}$ | = | $\left(\frac{1}{2} - x_7 + z_7\right)\,\mathbf{a}_1 + (-y_7 + z_7)\,\mathbf{a}_2 + (-x_7 - y_7)\,\mathbf{a}_3$ | = | $-a\left(y_7 + \frac{1}{4}\right)\,\hat{\mathbf{x}} + \left(\frac{1}{4} - x_7\right) a\,\hat{\mathbf{y}} + \left(\frac{1}{4} + z_7\right) a\,\hat{\mathbf{z}}$ | (48*e*) | O III |
| $\mathbf{B}_{67}$ | = | $(-x_7 - z_7)\,\mathbf{a}_1 + \left(\frac{1}{2} + y_7 - z_7\right)\,\mathbf{a}_2 + (-x_7 + y_7)\,\mathbf{a}_3$ | = | $\left(\frac{1}{4} + y_7\right) a\,\hat{\mathbf{x}} - a\left(x_7 + \frac{1}{4}\right)\,\hat{\mathbf{y}} + \left(\frac{1}{4} - z_7\right) a\,\hat{\mathbf{z}}$ | (48*e*) | O III |
| $\mathbf{B}_{68}$ | = | $(x_7 - z_7)\,\mathbf{a}_1 + (-y_7 - z_7)\,\mathbf{a}_2 + \left(\frac{1}{2} + x_7 - y_7\right)\,\mathbf{a}_3$ | = | $\left(\frac{1}{4} - y_7\right) a\,\hat{\mathbf{x}} + \left(\frac{1}{4} + x_7\right) a\,\hat{\mathbf{y}} - a\left(z_7 + \frac{1}{4}\right)\,\hat{\mathbf{z}}$ | (48*e*) | O III |
| $\mathbf{B}_{69}$ | = | $\left(\frac{1}{2} + y_7 + z_7\right)\,\mathbf{a}_1 + \left(\frac{1}{2} + x_7 + y_7\right)\,\mathbf{a}_2 + \left(\frac{1}{2} + x_7 + z_7\right)\,\mathbf{a}_3$ | = | $\left(\frac{1}{4} + x_7\right) a\,\hat{\mathbf{x}} + \left(\frac{1}{4} + z_7\right) a\,\hat{\mathbf{y}} + \left(\frac{1}{4} + y_7\right) a\,\hat{\mathbf{z}}$ | (48*e*) | O III |
| $\mathbf{B}_{70}$ | = | $(-y_7 + z_7)\,\mathbf{a}_1 + (-x_7 - y_7)\,\mathbf{a}_2 + \left(\frac{1}{2} - x_7 + z_7\right)\,\mathbf{a}_3$ | = | $\left(\frac{1}{4} - x_7\right) a\,\hat{\mathbf{x}} + \left(\frac{1}{4} + z_7\right) a\,\hat{\mathbf{y}} - a\left(y_7 + \frac{1}{4}\right)\,\hat{\mathbf{z}}$ | (48*e*) | O III |
| $\mathbf{B}_{71}$ | = | $\left(\frac{1}{2} + y_7 - z_7\right)\,\mathbf{a}_1 + (-x_7 + y_7)\,\mathbf{a}_2 + (-x_7 - z_7)\,\mathbf{a}_3$ | = | $-a\left(x_7 + \frac{1}{4}\right)\,\hat{\mathbf{x}} + \left(\frac{1}{4} - z_7\right) a\,\hat{\mathbf{y}} + \left(\frac{1}{4} + y_7\right) a\,\hat{\mathbf{z}}$ | (48*e*) | O III |
| $\mathbf{B}_{72}$ | = | $(-y_7 - z_7)\,\mathbf{a}_1 + \left(\frac{1}{2} + x_7 - y_7\right)\,\mathbf{a}_2 + (x_7 - z_7)\,\mathbf{a}_3$ | = | $\left(\frac{1}{4} + x_7\right) a\,\hat{\mathbf{x}} - a\left(z_7 + \frac{1}{4}\right)\,\hat{\mathbf{y}} + \left(\frac{1}{4} - y_7\right) a\,\hat{\mathbf{z}}$ | (48*e*) | O III |
| $\mathbf{B}_{73}$ | = | $\left(\frac{1}{2} + x_7 + y_7\right)\,\mathbf{a}_1 + \left(\frac{1}{2} + x_7 + z_7\right)\,\mathbf{a}_2 + \left(\frac{1}{2} + y_7 + z_7\right)\,\mathbf{a}_3$ | = | $\left(\frac{1}{4} + z_7\right) a\,\hat{\mathbf{x}} + \left(\frac{1}{4} + y_7\right) a\,\hat{\mathbf{y}} + \left(\frac{1}{4} + x_7\right) a\,\hat{\mathbf{z}}$ | (48*e*) | O III |
| $\mathbf{B}_{74}$ | = | $(-x_7 - y_7)\,\mathbf{a}_1 + \left(\frac{1}{2} - x_7 + z_7\right)\,\mathbf{a}_2 + (-y_7 + z_7)\,\mathbf{a}_3$ | = | $\left(\frac{1}{4} + z_7\right) a\,\hat{\mathbf{x}} - a\left(y_7 + \frac{1}{4}\right)\,\hat{\mathbf{y}} + \left(\frac{1}{4} - x_7\right) a\,\hat{\mathbf{z}}$ | (48*e*) | O III |
| $\mathbf{B}_{75}$ | = | $(-x_7 + y_7)\,\mathbf{a}_1 + (-x_7 - z_7)\,\mathbf{a}_2 + \left(\frac{1}{2} + y_7 - z_7\right)\,\mathbf{a}_3$ | = | $\left(\frac{1}{4} - z_7\right) a\,\hat{\mathbf{x}} + \left(\frac{1}{4} + y_7\right) a\,\hat{\mathbf{y}} - a\left(x_7 + \frac{1}{4}\right)\,\hat{\mathbf{z}}$ | (48*e*) | O III |
| $\mathbf{B}_{76}$ | = | $\left(\frac{1}{2} + x_7 - y_7\right)\,\mathbf{a}_1 + (x_7 - z_7)\,\mathbf{a}_2 + (-y_7 - z_7)\,\mathbf{a}_3$ | = | $-a\left(z_7 + \frac{1}{4}\right)\,\hat{\mathbf{x}} + \left(\frac{1}{4} - y_7\right) a\,\hat{\mathbf{y}} + \left(\frac{1}{4} + x_7\right) a\,\hat{\mathbf{z}}$ | (48*e*) | O III |

**Geometry files:**

- CIF: pp. 1803
- POSCAR: pp. 1803

# $Al(PO_3)_3$ ($G5_2$) Structure: AB9C3_cI208_220_c_3e_e

http://aflow.org/prototype-encyclopedia/AB9C3_cI208_220_c_3e_e

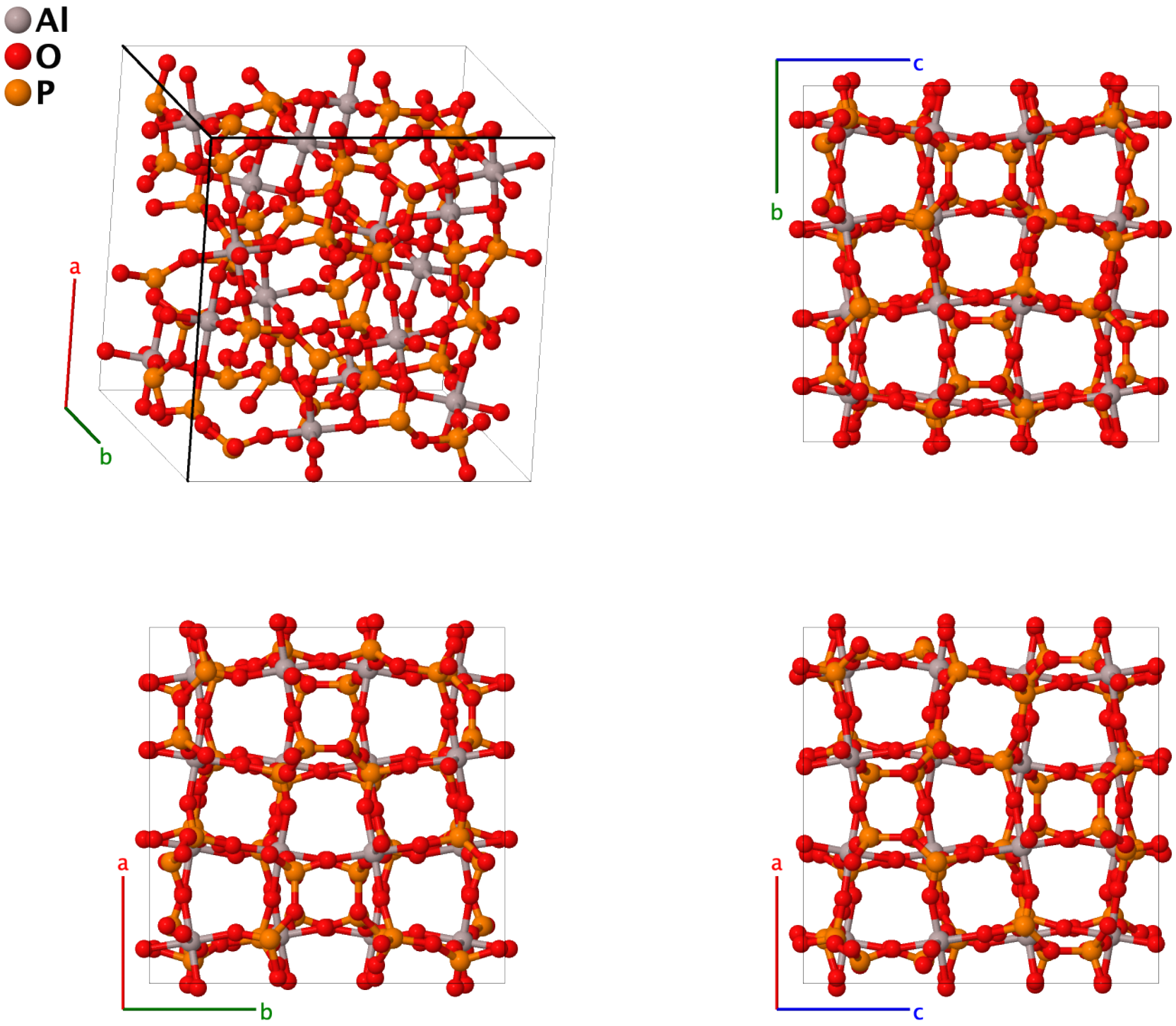

| | | |
|---|---|---|
| **Prototype** | : | $AlO_9P_3$ |
| **AFLOW prototype label** | : | AB9C3_cI208_220_c_3e_e |
| ***Strukturbericht* designation** | : | $G5_2$ |
| **Pearson symbol** | : | cI208 |
| **Space group number** | : | 220 |
| **Space group symbol** | : | $I\bar{4}3d$ |
| **AFLOW prototype command** | : | `aflow --proto=AB9C3_cI208_220_c_3e_e`<br>`--params=`$a, x_1, x_2, y_2, z_2, x_3, y_3, z_3, x_4, y_4, z_4, x_5, y_5, z_5$ |

**Body-centered Cubic primitive vectors:**

$$\mathbf{a}_1 = -\frac{1}{2}\,a\,\hat{\mathbf{x}} + \frac{1}{2}\,a\,\hat{\mathbf{y}} + \frac{1}{2}\,a\,\hat{\mathbf{z}}$$
$$\mathbf{a}_2 = \frac{1}{2}\,a\,\hat{\mathbf{x}} - \frac{1}{2}\,a\,\hat{\mathbf{y}} + \frac{1}{2}\,a\,\hat{\mathbf{z}}$$
$$\mathbf{a}_3 = \frac{1}{2}\,a\,\hat{\mathbf{x}} + \frac{1}{2}\,a\,\hat{\mathbf{y}} - \frac{1}{2}\,a\,\hat{\mathbf{z}}$$

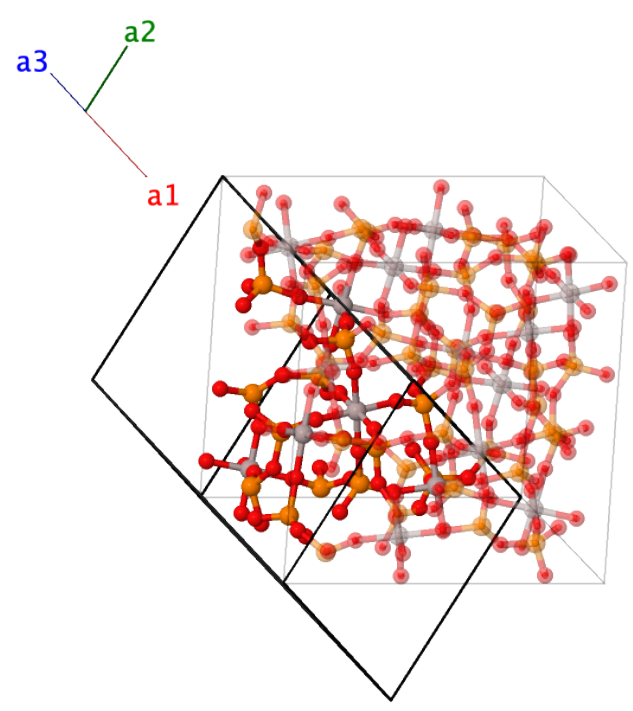

**Basis vectors:**

| | | Lattice Coordinates | | Cartesian Coordinates | Wyckoff Position | Atom Type |
|---|---|---|---|---|---|---|
| $\mathbf{B}_1$ | = | $2x_1\,\mathbf{a}_1 + 2x_1\,\mathbf{a}_2 + 2x_1\,\mathbf{a}_3$ | = | $x_1\,a\,\hat{\mathbf{x}} + x_1\,a\,\hat{\mathbf{y}} + x_1\,a\,\hat{\mathbf{z}}$ | (16c) | Al |
| $\mathbf{B}_2$ | = | $\frac{1}{2}\,\mathbf{a}_1 + \left(\frac{1}{2} - 2x_1\right)\,\mathbf{a}_3$ | = | $-x_1\,a\,\hat{\mathbf{x}} + \left(\frac{1}{2} - x_1\right)\,a\,\hat{\mathbf{y}} + x_1\,a\,\hat{\mathbf{z}}$ | (16c) | Al |
| $\mathbf{B}_{3}$ | = | $\left(\frac{1}{2} - 2x_{1}\right)\,\mathbf{a}_{2} + \frac{1}{2}\,\mathbf{a}_{3}$ | = | $\left(\frac{1}{2} - x_{1}\right)a\,\mathbf{\hat{x}} + x_{1}a\,\mathbf{\hat{y}} - x_{1}a\,\mathbf{\hat{z}}$ | (16$c$) | Al |
| $\mathbf{B}_{4}$ | = | $\left(\frac{1}{2} - 2x_{1}\right)\,\mathbf{a}_{1} + \frac{1}{2}\,\mathbf{a}_{2}$ | = | $x_{1}a\,\mathbf{\hat{x}} - x_{1}a\,\mathbf{\hat{y}} + \left(\frac{1}{2} - x_{1}\right)a\,\mathbf{\hat{z}}$ | (16$c$) | Al |
| $\mathbf{B}_{5}$ | = | $\left(\frac{1}{2} + 2x_{1}\right)\,\mathbf{a}_{1} + \left(\frac{1}{2} + 2x_{1}\right)\,\mathbf{a}_{2} + \left(\frac{1}{2} + 2x_{1}\right)\,\mathbf{a}_{3}$ | = | $\left(\frac{1}{4} + x_{1}\right)a\,\mathbf{\hat{x}} + \left(\frac{1}{4} + x_{1}\right)a\,\mathbf{\hat{y}} + \left(\frac{1}{4} + x_{1}\right)a\,\mathbf{\hat{z}}$ | (16$c$) | Al |
| $\mathbf{B}_{6}$ | = | $\frac{1}{2}\,\mathbf{a}_{1} - 2x_{1}\,\mathbf{a}_{3}$ | = | $-a\left(x_{1} + \frac{1}{4}\right)\,\mathbf{\hat{x}} + \left(\frac{1}{4} - x_{1}\right)a\,\mathbf{\hat{y}} + \left(\frac{1}{4} + x_{1}\right)a\,\mathbf{\hat{z}}$ | (16$c$) | Al |
| $\mathbf{B}_{7}$ | = | $-2x_{1}\,\mathbf{a}_{1} + \frac{1}{2}\,\mathbf{a}_{2}$ | = | $\left(\frac{1}{4} + x_{1}\right)a\,\mathbf{\hat{x}} - a\left(x_{1} + \frac{1}{4}\right)\,\mathbf{\hat{y}} + \left(\frac{1}{4} - x_{1}\right)a\,\mathbf{\hat{z}}$ | (16$c$) | Al |
| $\mathbf{B}_{8}$ | = | $-2x_{1}\,\mathbf{a}_{2} + \frac{1}{2}\,\mathbf{a}_{3}$ | = | $\left(\frac{1}{4} - x_{1}\right)a\,\mathbf{\hat{x}} + \left(\frac{1}{4} + x_{1}\right)a\,\mathbf{\hat{y}} - a\left(x_{1} + \frac{1}{4}\right)\,\mathbf{\hat{z}}$ | (16$c$) | Al |
| $\mathbf{B}_{9}$ | = | $(y_{2} + z_{2})\,\mathbf{a}_{1} + (x_{2} + z_{2})\,\mathbf{a}_{2} + (x_{2} + y_{2})\,\mathbf{a}_{3}$ | = | $x_{2}a\,\mathbf{\hat{x}} + y_{2}a\,\mathbf{\hat{y}} + z_{2}a\,\mathbf{\hat{z}}$ | (48$e$) | O I |
| $\mathbf{B}_{10}$ | = | $\left(\frac{1}{2} - y_{2} + z_{2}\right)\,\mathbf{a}_{1} + (-x_{2} + z_{2})\,\mathbf{a}_{2} + \left(\frac{1}{2} - x_{2} - y_{2}\right)\,\mathbf{a}_{3}$ | = | $-x_{2}a\,\mathbf{\hat{x}} + \left(\frac{1}{2} - y_{2}\right)a\,\mathbf{\hat{y}} + z_{2}a\,\mathbf{\hat{z}}$ | (48$e$) | O I |
| $\mathbf{B}_{11}$ | = | $(y_{2} - z_{2})\,\mathbf{a}_{1} + \left(\frac{1}{2} - x_{2} - z_{2}\right)\,\mathbf{a}_{2} + \left(\frac{1}{2} - x_{2} + y_{2}\right)\,\mathbf{a}_{3}$ | = | $\left(\frac{1}{2} - x_{2}\right)a\,\mathbf{\hat{x}} + y_{2}a\,\mathbf{\hat{y}} - z_{2}a\,\mathbf{\hat{z}}$ | (48$e$) | O I |
| $\mathbf{B}_{12}$ | = | $\left(\frac{1}{2} - y_{2} - z_{2}\right)\,\mathbf{a}_{1} + \left(\frac{1}{2} + x_{2} - z_{2}\right)\,\mathbf{a}_{2} + (x_{2} - y_{2})\,\mathbf{a}_{3}$ | = | $x_{2}a\,\mathbf{\hat{x}} - y_{2}a\,\mathbf{\hat{y}} + \left(\frac{1}{2} - z_{2}\right)a\,\mathbf{\hat{z}}$ | (48$e$) | O I |
| $\mathbf{B}_{13}$ | = | $(x_{2} + y_{2})\,\mathbf{a}_{1} + (y_{2} + z_{2})\,\mathbf{a}_{2} + (x_{2} + z_{2})\,\mathbf{a}_{3}$ | = | $z_{2}a\,\mathbf{\hat{x}} + x_{2}a\,\mathbf{\hat{y}} + y_{2}a\,\mathbf{\hat{z}}$ | (48$e$) | O I |
| $\mathbf{B}_{14}$ | = | $\left(\frac{1}{2} - x_{2} - y_{2}\right)\,\mathbf{a}_{1} + \left(\frac{1}{2} - y_{2} + z_{2}\right)\,\mathbf{a}_{2} + (-x_{2} + z_{2})\,\mathbf{a}_{3}$ | = | $z_{2}a\,\mathbf{\hat{x}} - x_{2}a\,\mathbf{\hat{y}} + \left(\frac{1}{2} - y_{2}\right)a\,\mathbf{\hat{z}}$ | (48$e$) | O I |
| $\mathbf{B}_{15}$ | = | $\left(\frac{1}{2} - x_{2} + y_{2}\right)\,\mathbf{a}_{1} + (y_{2} - z_{2})\,\mathbf{a}_{2} + \left(\frac{1}{2} - x_{2} - z_{2}\right)\,\mathbf{a}_{3}$ | = | $-z_{2}a\,\mathbf{\hat{x}} + \left(\frac{1}{2} - x_{2}\right)a\,\mathbf{\hat{y}} + y_{2}a\,\mathbf{\hat{z}}$ | (48$e$) | O I |
| $\mathbf{B}_{16}$ | = | $(x_{2} - y_{2})\,\mathbf{a}_{1} + \left(\frac{1}{2} - y_{2} - z_{2}\right)\,\mathbf{a}_{2} + \left(\frac{1}{2} + x_{2} - z_{2}\right)\,\mathbf{a}_{3}$ | = | $\left(\frac{1}{2} - z_{2}\right)a\,\mathbf{\hat{x}} + x_{2}a\,\mathbf{\hat{y}} - y_{2}a\,\mathbf{\hat{z}}$ | (48$e$) | O I |
| $\mathbf{B}_{17}$ | = | $(x_{2} + z_{2})\,\mathbf{a}_{1} + (x_{2} + y_{2})\,\mathbf{a}_{2} + (y_{2} + z_{2})\,\mathbf{a}_{3}$ | = | $y_{2}a\,\mathbf{\hat{x}} + z_{2}a\,\mathbf{\hat{y}} + x_{2}a\,\mathbf{\hat{z}}$ | (48$e$) | O I |
| $\mathbf{B}_{18}$ | = | $(-x_{2} + z_{2})\,\mathbf{a}_{1} + \left(\frac{1}{2} - x_{2} - y_{2}\right)\,\mathbf{a}_{2} + \left(\frac{1}{2} - y_{2} + z_{2}\right)\,\mathbf{a}_{3}$ | = | $\left(\frac{1}{2} - y_{2}\right)a\,\mathbf{\hat{x}} + z_{2}a\,\mathbf{\hat{y}} - x_{2}a\,\mathbf{\hat{z}}$ | (48$e$) | O I |
| $\mathbf{B}_{19}$ | = | $\left(\frac{1}{2} - x_{2} - z_{2}\right)\,\mathbf{a}_{1} + \left(\frac{1}{2} - x_{2} + y_{2}\right)\,\mathbf{a}_{2} + (y_{2} - z_{2})\,\mathbf{a}_{3}$ | = | $y_{2}a\,\mathbf{\hat{x}} - z_{2}a\,\mathbf{\hat{y}} + \left(\frac{1}{2} - x_{2}\right)a\,\mathbf{\hat{z}}$ | (48$e$) | O I |
| $\mathbf{B}_{20}$ | = | $\left(\frac{1}{2} + x_{2} - z_{2}\right)\,\mathbf{a}_{1} + (x_{2} - y_{2})\,\mathbf{a}_{2} + \left(\frac{1}{2} - y_{2} - z_{2}\right)\,\mathbf{a}_{3}$ | = | $-y_{2}a\,\mathbf{\hat{x}} + \left(\frac{1}{2} - z_{2}\right)a\,\mathbf{\hat{y}} + x_{2}a\,\mathbf{\hat{z}}$ | (48$e$) | O I |
| $\mathbf{B}_{21}$ | = | $\left(\frac{1}{2} + x_{2} + z_{2}\right)\,\mathbf{a}_{1} + \left(\frac{1}{2} + y_{2} + z_{2}\right)\,\mathbf{a}_{2} + \left(\frac{1}{2} + x_{2} + y_{2}\right)\,\mathbf{a}_{3}$ | = | $\left(\frac{1}{4} + y_{2}\right)a\,\mathbf{\hat{x}} + \left(\frac{1}{4} + x_{2}\right)a\,\mathbf{\hat{y}} + \left(\frac{1}{4} + z_{2}\right)a\,\mathbf{\hat{z}}$ | (48$e$) | O I |
| $\mathbf{B}_{22}$ | = | $\left(\frac{1}{2} - x_{2} + z_{2}\right)\,\mathbf{a}_{1} + (-y_{2} + z_{2})\,\mathbf{a}_{2} + (-x_{2} - y_{2})\,\mathbf{a}_{3}$ | = | $-a\left(y_{2} + \frac{1}{4}\right)\,\mathbf{\hat{x}} + \left(\frac{1}{4} - x_{2}\right)a\,\mathbf{\hat{y}} + \left(\frac{1}{4} + z_{2}\right)a\,\mathbf{\hat{z}}$ | (48$e$) | O I |
| $\mathbf{B}_{23}$ | = | $(-x_{2} - z_{2})\,\mathbf{a}_{1} + \left(\frac{1}{2} + y_{2} - z_{2}\right)\,\mathbf{a}_{2} + (-x_{2} + y_{2})\,\mathbf{a}_{3}$ | = | $\left(\frac{1}{4} + y_{2}\right)a\,\mathbf{\hat{x}} - a\left(x_{2} + \frac{1}{4}\right)\,\mathbf{\hat{y}} + \left(\frac{1}{4} - z_{2}\right)a\,\mathbf{\hat{z}}$ | (48$e$) | O I |
| $\mathbf{B}_{24}$ | = | $(x_{2} - z_{2})\,\mathbf{a}_{1} + (-y_{2} - z_{2})\,\mathbf{a}_{2} + \left(\frac{1}{2} + x_{2} - y_{2}\right)\,\mathbf{a}_{3}$ | = | $\left(\frac{1}{4} - y_{2}\right)a\,\mathbf{\hat{x}} + \left(\frac{1}{4} + x_{2}\right)a\,\mathbf{\hat{y}} - a\left(z_{2} + \frac{1}{4}\right)\,\mathbf{\hat{z}}$ | (48$e$) | O I |
| $\mathbf{B}_{25}$ | = | $\left(\frac{1}{2}+y_2+z_2\right)\mathbf{a}_1+\left(\frac{1}{2}+x_2+y_2\right)\mathbf{a}_2+\left(\frac{1}{2}+x_2+z_2\right)\mathbf{a}_3$ | = | $\left(\frac{1}{4}+x_2\right)a\,\hat{\mathbf{x}}+\left(\frac{1}{4}+z_2\right)a\,\hat{\mathbf{y}}+\left(\frac{1}{4}+y_2\right)a\,\hat{\mathbf{z}}$ | (48$e$) | O I |
| $\mathbf{B}_{26}$ | = | $(-y_2+z_2)\,\mathbf{a}_1+(-x_2-y_2)\,\mathbf{a}_2+\left(\frac{1}{2}-x_2+z_2\right)\mathbf{a}_3$ | = | $\left(\frac{1}{4}-x_2\right)a\,\hat{\mathbf{x}}+\left(\frac{1}{4}+z_2\right)a\,\hat{\mathbf{y}}-a\left(y_2+\frac{1}{4}\right)\hat{\mathbf{z}}$ | (48$e$) | O I |
| $\mathbf{B}_{27}$ | = | $\left(\frac{1}{2}+y_2-z_2\right)\mathbf{a}_1+(-x_2+y_2)\,\mathbf{a}_2+(-x_2-z_2)\,\mathbf{a}_3$ | = | $-a\left(x_2+\frac{1}{4}\right)\hat{\mathbf{x}}+\left(\frac{1}{4}-z_2\right)a\,\hat{\mathbf{y}}+\left(\frac{1}{4}+y_2\right)a\,\hat{\mathbf{z}}$ | (48$e$) | O I |
| $\mathbf{B}_{28}$ | = | $(-y_2-z_2)\,\mathbf{a}_1+\left(\frac{1}{2}+x_2-y_2\right)\mathbf{a}_2+(x_2-z_2)\,\mathbf{a}_3$ | = | $\left(\frac{1}{4}+x_2\right)a\,\hat{\mathbf{x}}-a\left(z_2+\frac{1}{4}\right)\hat{\mathbf{y}}+\left(\frac{1}{4}-y_2\right)a\,\hat{\mathbf{z}}$ | (48$e$) | O I |
| $\mathbf{B}_{29}$ | = | $\left(\frac{1}{2}+x_2+y_2\right)\mathbf{a}_1+\left(\frac{1}{2}+x_2+z_2\right)\mathbf{a}_2+\left(\frac{1}{2}+y_2+z_2\right)\mathbf{a}_3$ | = | $\left(\frac{1}{4}+z_2\right)a\,\hat{\mathbf{x}}+\left(\frac{1}{4}+y_2\right)a\,\hat{\mathbf{y}}+\left(\frac{1}{4}+x_2\right)a\,\hat{\mathbf{z}}$ | (48$e$) | O I |
| $\mathbf{B}_{30}$ | = | $(-x_2-y_2)\,\mathbf{a}_1+\left(\frac{1}{2}-x_2+z_2\right)\mathbf{a}_2+(-y_2+z_2)\,\mathbf{a}_3$ | = | $\left(\frac{1}{4}+z_2\right)a\,\hat{\mathbf{x}}-a\left(y_2+\frac{1}{4}\right)\hat{\mathbf{y}}+\left(\frac{1}{4}-x_2\right)a\,\hat{\mathbf{z}}$ | (48$e$) | O I |
| $\mathbf{B}_{31}$ | = | $(-x_2+y_2)\,\mathbf{a}_1+(-x_2-z_2)\,\mathbf{a}_2+\left(\frac{1}{2}+y_2-z_2\right)\mathbf{a}_3$ | = | $\left(\frac{1}{4}-z_2\right)a\,\hat{\mathbf{x}}+\left(\frac{1}{4}+y_2\right)a\,\hat{\mathbf{y}}-a\left(x_2+\frac{1}{4}\right)\hat{\mathbf{z}}$ | (48$e$) | O I |
| $\mathbf{B}_{32}$ | = | $\left(\frac{1}{2}+x_2-y_2\right)\mathbf{a}_1+(x_2-z_2)\,\mathbf{a}_2+(-y_2-z_2)\,\mathbf{a}_3$ | = | $-a\left(z_2+\frac{1}{4}\right)\hat{\mathbf{x}}+\left(\frac{1}{4}-y_2\right)a\,\hat{\mathbf{y}}+\left(\frac{1}{4}+x_2\right)a\,\hat{\mathbf{z}}$ | (48$e$) | O I |
| $\mathbf{B}_{33}$ | = | $(y_3+z_3)\,\mathbf{a}_1+(x_3+z_3)\,\mathbf{a}_2+(x_3+y_3)\,\mathbf{a}_3$ | = | $x_3a\,\hat{\mathbf{x}}+y_3a\,\hat{\mathbf{y}}+z_3a\,\hat{\mathbf{z}}$ | (48$e$) | O II |
| $\mathbf{B}_{34}$ | = | $\left(\frac{1}{2}-y_3+z_3\right)\mathbf{a}_1+(-x_3+z_3)\,\mathbf{a}_2+\left(\frac{1}{2}-x_3-y_3\right)\mathbf{a}_3$ | = | $-x_3a\,\hat{\mathbf{x}}+\left(\frac{1}{2}-y_3\right)a\,\hat{\mathbf{y}}+z_3a\,\hat{\mathbf{z}}$ | (48$e$) | O II |
| $\mathbf{B}_{35}$ | = | $(y_3-z_3)\,\mathbf{a}_1+\left(\frac{1}{2}-x_3-z_3\right)\mathbf{a}_2+\left(\frac{1}{2}-x_3+y_3\right)\mathbf{a}_3$ | = | $\left(\frac{1}{2}-x_3\right)a\,\hat{\mathbf{x}}+y_3a\,\hat{\mathbf{y}}-z_3a\,\hat{\mathbf{z}}$ | (48$e$) | O II |
| $\mathbf{B}_{36}$ | = | $\left(\frac{1}{2}-y_3-z_3\right)\mathbf{a}_1+\left(\frac{1}{2}+x_3-z_3\right)\mathbf{a}_2+(x_3-y_3)\,\mathbf{a}_3$ | = | $x_3a\,\hat{\mathbf{x}}-y_3a\,\hat{\mathbf{y}}+\left(\frac{1}{2}-z_3\right)a\,\hat{\mathbf{z}}$ | (48$e$) | O II |
| $\mathbf{B}_{37}$ | = | $(x_3+y_3)\,\mathbf{a}_1+(y_3+z_3)\,\mathbf{a}_2+(x_3+z_3)\,\mathbf{a}_3$ | = | $z_3a\,\hat{\mathbf{x}}+x_3a\,\hat{\mathbf{y}}+y_3a\,\hat{\mathbf{z}}$ | (48$e$) | O II |
| $\mathbf{B}_{38}$ | = | $\left(\frac{1}{2}-x_3-y_3\right)\mathbf{a}_1+\left(\frac{1}{2}-y_3+z_3\right)\mathbf{a}_2+(-x_3+z_3)\,\mathbf{a}_3$ | = | $z_3a\,\hat{\mathbf{x}}-x_3a\,\hat{\mathbf{y}}+\left(\frac{1}{2}-y_3\right)a\,\hat{\mathbf{z}}$ | (48$e$) | O II |
| $\mathbf{B}_{39}$ | = | $\left(\frac{1}{2}-x_3+y_3\right)\mathbf{a}_1+(y_3-z_3)\,\mathbf{a}_2+\left(\frac{1}{2}-x_3-z_3\right)\mathbf{a}_3$ | = | $-z_3a\,\hat{\mathbf{x}}+\left(\frac{1}{2}-x_3\right)a\,\hat{\mathbf{y}}+y_3a\,\hat{\mathbf{z}}$ | (48$e$) | O II |
| $\mathbf{B}_{40}$ | = | $(x_3-y_3)\,\mathbf{a}_1+\left(\frac{1}{2}-y_3-z_3\right)\mathbf{a}_2+\left(\frac{1}{2}+x_3-z_3\right)\mathbf{a}_3$ | = | $\left(\frac{1}{2}-z_3\right)a\,\hat{\mathbf{x}}+x_3a\,\hat{\mathbf{y}}-y_3a\,\hat{\mathbf{z}}$ | (48$e$) | O II |
| $\mathbf{B}_{41}$ | = | $(x_3+z_3)\,\mathbf{a}_1+(x_3+y_3)\,\mathbf{a}_2+(y_3+z_3)\,\mathbf{a}_3$ | = | $y_3a\,\hat{\mathbf{x}}+z_3a\,\hat{\mathbf{y}}+x_3a\,\hat{\mathbf{z}}$ | (48$e$) | O II |
| $\mathbf{B}_{42}$ | = | $(-x_3+z_3)\,\mathbf{a}_1+\left(\frac{1}{2}-x_3-y_3\right)\mathbf{a}_2+\left(\frac{1}{2}-y_3+z_3\right)\mathbf{a}_3$ | = | $\left(\frac{1}{2}-y_3\right)a\,\hat{\mathbf{x}}+z_3a\,\hat{\mathbf{y}}-x_3a\,\hat{\mathbf{z}}$ | (48$e$) | O II |
| $\mathbf{B}_{43}$ | = | $\left(\frac{1}{2}-x_3-z_3\right)\mathbf{a}_1+\left(\frac{1}{2}-x_3+y_3\right)\mathbf{a}_2+(y_3-z_3)\,\mathbf{a}_3$ | = | $y_3a\,\hat{\mathbf{x}}-z_3a\,\hat{\mathbf{y}}+\left(\frac{1}{2}-x_3\right)a\,\hat{\mathbf{z}}$ | (48$e$) | O II |
| $\mathbf{B}_{44}$ | = | $\left(\frac{1}{2}+x_3-z_3\right)\mathbf{a}_1+(x_3-y_3)\,\mathbf{a}_2+\left(\frac{1}{2}-y_3-z_3\right)\mathbf{a}_3$ | = | $-y_3a\,\hat{\mathbf{x}}+\left(\frac{1}{2}-z_3\right)a\,\hat{\mathbf{y}}+x_3a\,\hat{\mathbf{z}}$ | (48$e$) | O II |
| $\mathbf{B}_{45}$ | = | $\left(\frac{1}{2}+x_3+z_3\right)\mathbf{a}_1+\left(\frac{1}{2}+y_3+z_3\right)\mathbf{a}_2+\left(\frac{1}{2}+x_3+y_3\right)\mathbf{a}_3$ | = | $\left(\frac{1}{4}+y_3\right)a\,\hat{\mathbf{x}}+\left(\frac{1}{4}+x_3\right)a\,\hat{\mathbf{y}}+\left(\frac{1}{4}+z_3\right)a\,\hat{\mathbf{z}}$ | (48$e$) | O II |
| $\mathbf{B}_{46}$ | = | $\left(\frac{1}{2} - x_3 + z_3\right) \mathbf{a}_1 + (-y_3 + z_3) \mathbf{a}_2 + (-x_3 - y_3) \mathbf{a}_3$ | = | $-a\left(y_3 + \frac{1}{4}\right) \hat{\mathbf{x}} + \left(\frac{1}{4} - x_3\right) a \hat{\mathbf{y}} + \left(\frac{1}{4} + z_3\right) a \hat{\mathbf{z}}$ | (48$e$) | O II |
| $\mathbf{B}_{47}$ | = | $(-x_3 - z_3) \mathbf{a}_1 + \left(\frac{1}{2} + y_3 - z_3\right) \mathbf{a}_2 + (-x_3 + y_3) \mathbf{a}_3$ | = | $\left(\frac{1}{4} + y_3\right) a \hat{\mathbf{x}} - a\left(x_3 + \frac{1}{4}\right) \hat{\mathbf{y}} + \left(\frac{1}{4} - z_3\right) a \hat{\mathbf{z}}$ | (48$e$) | O II |
| $\mathbf{B}_{48}$ | = | $(x_3 - z_3) \mathbf{a}_1 + (-y_3 - z_3) \mathbf{a}_2 + \left(\frac{1}{2} + x_3 - y_3\right) \mathbf{a}_3$ | = | $\left(\frac{1}{4} - y_3\right) a \hat{\mathbf{x}} + \left(\frac{1}{4} + x_3\right) a \hat{\mathbf{y}} - a\left(z_3 + \frac{1}{4}\right) \hat{\mathbf{z}}$ | (48$e$) | O II |
| $\mathbf{B}_{49}$ | = | $\left(\frac{1}{2} + y_3 + z_3\right) \mathbf{a}_1 + \left(\frac{1}{2} + x_3 + y_3\right) \mathbf{a}_2 + \left(\frac{1}{2} + x_3 + z_3\right) \mathbf{a}_3$ | = | $\left(\frac{1}{4} + x_3\right) a \hat{\mathbf{x}} + \left(\frac{1}{4} + z_3\right) a \hat{\mathbf{y}} + \left(\frac{1}{4} + y_3\right) a \hat{\mathbf{z}}$ | (48$e$) | O II |
| $\mathbf{B}_{50}$ | = | $(-y_3 + z_3) \mathbf{a}_1 + (-x_3 - y_3) \mathbf{a}_2 + \left(\frac{1}{2} - x_3 + z_3\right) \mathbf{a}_3$ | = | $\left(\frac{1}{4} - x_3\right) a \hat{\mathbf{x}} + \left(\frac{1}{4} + z_3\right) a \hat{\mathbf{y}} - a\left(y_3 + \frac{1}{4}\right) \hat{\mathbf{z}}$ | (48$e$) | O II |
| $\mathbf{B}_{51}$ | = | $\left(\frac{1}{2} + y_3 - z_3\right) \mathbf{a}_1 + (-x_3 + y_3) \mathbf{a}_2 + (-x_3 - z_3) \mathbf{a}_3$ | = | $-a\left(x_3 + \frac{1}{4}\right) \hat{\mathbf{x}} + \left(\frac{1}{4} - z_3\right) a \hat{\mathbf{y}} + \left(\frac{1}{4} + y_3\right) a \hat{\mathbf{z}}$ | (48$e$) | O II |
| $\mathbf{B}_{52}$ | = | $(-y_3 - z_3) \mathbf{a}_1 + \left(\frac{1}{2} + x_3 - y_3\right) \mathbf{a}_2 + (x_3 - z_3) \mathbf{a}_3$ | = | $\left(\frac{1}{4} + x_3\right) a \hat{\mathbf{x}} - a\left(z_3 + \frac{1}{4}\right) \hat{\mathbf{y}} + \left(\frac{1}{4} - y_3\right) a \hat{\mathbf{z}}$ | (48$e$) | O II |
| $\mathbf{B}_{53}$ | = | $\left(\frac{1}{2} + x_3 + y_3\right) \mathbf{a}_1 + \left(\frac{1}{2} + x_3 + z_3\right) \mathbf{a}_2 + \left(\frac{1}{2} + y_3 + z_3\right) \mathbf{a}_3$ | = | $\left(\frac{1}{4} + z_3\right) a \hat{\mathbf{x}} + \left(\frac{1}{4} + y_3\right) a \hat{\mathbf{y}} + \left(\frac{1}{4} + x_3\right) a \hat{\mathbf{z}}$ | (48$e$) | O II |
| $\mathbf{B}_{54}$ | = | $(-x_3 - y_3) \mathbf{a}_1 + \left(\frac{1}{2} - x_3 + z_3\right) \mathbf{a}_2 + (-y_3 + z_3) \mathbf{a}_3$ | = | $\left(\frac{1}{4} + z_3\right) a \hat{\mathbf{x}} - a\left(y_3 + \frac{1}{4}\right) \hat{\mathbf{y}} + \left(\frac{1}{4} - x_3\right) a \hat{\mathbf{z}}$ | (48$e$) | O II |
| $\mathbf{B}_{55}$ | = | $(-x_3 + y_3) \mathbf{a}_1 + (-x_3 - z_3) \mathbf{a}_2 + \left(\frac{1}{2} + y_3 - z_3\right) \mathbf{a}_3$ | = | $\left(\frac{1}{4} - z_3\right) a \hat{\mathbf{x}} + \left(\frac{1}{4} + y_3\right) a \hat{\mathbf{y}} - a\left(x_3 + \frac{1}{4}\right) \hat{\mathbf{z}}$ | (48$e$) | O II |
| $\mathbf{B}_{56}$ | = | $\left(\frac{1}{2} + x_3 - y_3\right) \mathbf{a}_1 + (x_3 - z_3) \mathbf{a}_2 + (-y_3 - z_3) \mathbf{a}_3$ | = | $-a\left(z_3 + \frac{1}{4}\right) \hat{\mathbf{x}} + \left(\frac{1}{4} - y_3\right) a \hat{\mathbf{y}} + \left(\frac{1}{4} + x_3\right) a \hat{\mathbf{z}}$ | (48$e$) | O II |
| $\mathbf{B}_{57}$ | = | $(y_4 + z_4) \mathbf{a}_1 + (x_4 + z_4) \mathbf{a}_2 + (x_4 + y_4) \mathbf{a}_3$ | = | $x_4 a \hat{\mathbf{x}} + y_4 a \hat{\mathbf{y}} + z_4 a \hat{\mathbf{z}}$ | (48$e$) | O III |
| $\mathbf{B}_{58}$ | = | $\left(\frac{1}{2} - y_4 + z_4\right) \mathbf{a}_1 + (-x_4 + z_4) \mathbf{a}_2 + \left(\frac{1}{2} - x_4 - y_4\right) \mathbf{a}_3$ | = | $-x_4 a \hat{\mathbf{x}} + \left(\frac{1}{2} - y_4\right) a \hat{\mathbf{y}} + z_4 a \hat{\mathbf{z}}$ | (48$e$) | O III |
| $\mathbf{B}_{59}$ | = | $(y_4 - z_4) \mathbf{a}_1 + \left(\frac{1}{2} - x_4 - z_4\right) \mathbf{a}_2 + \left(\frac{1}{2} - x_4 + y_4\right) \mathbf{a}_3$ | = | $\left(\frac{1}{2} - x_4\right) a \hat{\mathbf{x}} + y_4 a \hat{\mathbf{y}} - z_4 a \hat{\mathbf{z}}$ | (48$e$) | O III |
| $\mathbf{B}_{60}$ | = | $\left(\frac{1}{2} - y_4 - z_4\right) \mathbf{a}_1 + \left(\frac{1}{2} + x_4 - z_4\right) \mathbf{a}_2 + (x_4 - y_4) \mathbf{a}_3$ | = | $x_4 a \hat{\mathbf{x}} - y_4 a \hat{\mathbf{y}} + \left(\frac{1}{2} - z_4\right) a \hat{\mathbf{z}}$ | (48$e$) | O III |
| $\mathbf{B}_{61}$ | = | $(x_4 + y_4) \mathbf{a}_1 + (y_4 + z_4) \mathbf{a}_2 + (x_4 + z_4) \mathbf{a}_3$ | = | $z_4 a \hat{\mathbf{x}} + x_4 a \hat{\mathbf{y}} + y_4 a \hat{\mathbf{z}}$ | (48$e$) | O III |
| $\mathbf{B}_{62}$ | = | $\left(\frac{1}{2} - x_4 - y_4\right) \mathbf{a}_1 + \left(\frac{1}{2} - y_4 + z_4\right) \mathbf{a}_2 + (-x_4 + z_4) \mathbf{a}_3$ | = | $z_4 a \hat{\mathbf{x}} - x_4 a \hat{\mathbf{y}} + \left(\frac{1}{2} - y_4\right) a \hat{\mathbf{z}}$ | (48$e$) | O III |
| $\mathbf{B}_{63}$ | = | $\left(\frac{1}{2} - x_4 + y_4\right) \mathbf{a}_1 + (y_4 - z_4) \mathbf{a}_2 + \left(\frac{1}{2} - x_4 - z_4\right) \mathbf{a}_3$ | = | $-z_4 a \hat{\mathbf{x}} + \left(\frac{1}{2} - x_4\right) a \hat{\mathbf{y}} + y_4 a \hat{\mathbf{z}}$ | (48$e$) | O III |
| $\mathbf{B}_{64}$ | = | $(x_4 - y_4) \mathbf{a}_1 + \left(\frac{1}{2} - y_4 - z_4\right) \mathbf{a}_2 + \left(\frac{1}{2} + x_4 - z_4\right) \mathbf{a}_3$ | = | $\left(\frac{1}{2} - z_4\right) a \hat{\mathbf{x}} + x_4 a \hat{\mathbf{y}} - y_4 a \hat{\mathbf{z}}$ | (48$e$) | O III |
| $\mathbf{B}_{65}$ | = | $(x_4 + z_4) \mathbf{a}_1 + (x_4 + y_4) \mathbf{a}_2 + (y_4 + z_4) \mathbf{a}_3$ | = | $y_4 a \hat{\mathbf{x}} + z_4 a \hat{\mathbf{y}} + x_4 a \hat{\mathbf{z}}$ | (48$e$) | O III |
| $\mathbf{B}_{66}$ | = | $(-x_4 + z_4) \mathbf{a}_1 + \left(\frac{1}{2} - x_4 - y_4\right) \mathbf{a}_2 + \left(\frac{1}{2} - y_4 + z_4\right) \mathbf{a}_3$ | = | $\left(\frac{1}{2} - y_4\right) a \hat{\mathbf{x}} + z_4 a \hat{\mathbf{y}} - x_4 a \hat{\mathbf{z}}$ | (48$e$) | O III |
| $\mathbf{B}_{67}$ | = | $\left(\frac{1}{2} - x_4 - z_4\right)\mathbf{a}_1 + \left(\frac{1}{2} - x_4 + y_4\right)\mathbf{a}_2 + (y_4 - z_4)\mathbf{a}_3$ | = | $y_4 a\,\hat{\mathbf{x}} - z_4 a\,\hat{\mathbf{y}} + \left(\frac{1}{2} - x_4\right) a\,\hat{\mathbf{z}}$ | (48$e$) | O III |
| $\mathbf{B}_{68}$ | = | $\left(\frac{1}{2} + x_4 - z_4\right)\mathbf{a}_1 + (x_4 - y_4)\mathbf{a}_2 + \left(\frac{1}{2} - y_4 - z_4\right)\mathbf{a}_3$ | = | $-y_4 a\,\hat{\mathbf{x}} + \left(\frac{1}{2} - z_4\right) a\,\hat{\mathbf{y}} + x_4 a\,\hat{\mathbf{z}}$ | (48$e$) | O III |
| $\mathbf{B}_{69}$ | = | $\left(\frac{1}{2} + x_4 + z_4\right)\mathbf{a}_1 + \left(\frac{1}{2} + y_4 + z_4\right)\mathbf{a}_2 + \left(\frac{1}{2} + x_4 + y_4\right)\mathbf{a}_3$ | = | $\left(\frac{1}{4} + y_4\right) a\,\hat{\mathbf{x}} + \left(\frac{1}{4} + x_4\right) a\,\hat{\mathbf{y}} + \left(\frac{1}{4} + z_4\right) a\,\hat{\mathbf{z}}$ | (48$e$) | O III |
| $\mathbf{B}_{70}$ | = | $\left(\frac{1}{2} - x_4 + z_4\right)\mathbf{a}_1 + (-y_4 + z_4)\mathbf{a}_2 + (-x_4 - y_4)\mathbf{a}_3$ | = | $-a\left(y_4 + \frac{1}{4}\right)\hat{\mathbf{x}} + \left(\frac{1}{4} - x_4\right) a\,\hat{\mathbf{y}} + \left(\frac{1}{4} + z_4\right) a\,\hat{\mathbf{z}}$ | (48$e$) | O III |
| $\mathbf{B}_{71}$ | = | $(-x_4 - z_4)\mathbf{a}_1 + \left(\frac{1}{2} + y_4 - z_4\right)\mathbf{a}_2 + (-x_4 + y_4)\mathbf{a}_3$ | = | $\left(\frac{1}{4} + y_4\right) a\,\hat{\mathbf{x}} - a\left(x_4 + \frac{1}{4}\right)\hat{\mathbf{y}} + \left(\frac{1}{4} - z_4\right) a\,\hat{\mathbf{z}}$ | (48$e$) | O III |
| $\mathbf{B}_{72}$ | = | $(x_4 - z_4)\mathbf{a}_1 + (-y_4 - z_4)\mathbf{a}_2 + \left(\frac{1}{2} + x_4 - y_4\right)\mathbf{a}_3$ | = | $\left(\frac{1}{4} - y_4\right) a\,\hat{\mathbf{x}} + \left(\frac{1}{4} + x_4\right) a\,\hat{\mathbf{y}} - a\left(z_4 + \frac{1}{4}\right)\hat{\mathbf{z}}$ | (48$e$) | O III |
| $\mathbf{B}_{73}$ | = | $\left(\frac{1}{2} + y_4 + z_4\right)\mathbf{a}_1 + \left(\frac{1}{2} + x_4 + y_4\right)\mathbf{a}_2 + \left(\frac{1}{2} + x_4 + z_4\right)\mathbf{a}_3$ | = | $\left(\frac{1}{4} + x_4\right) a\,\hat{\mathbf{x}} + \left(\frac{1}{4} + z_4\right) a\,\hat{\mathbf{y}} + \left(\frac{1}{4} + y_4\right) a\,\hat{\mathbf{z}}$ | (48$e$) | O III |
| $\mathbf{B}_{74}$ | = | $(-y_4 + z_4)\mathbf{a}_1 + (-x_4 - y_4)\mathbf{a}_2 + \left(\frac{1}{2} - x_4 + z_4\right)\mathbf{a}_3$ | = | $\left(\frac{1}{4} - x_4\right) a\,\hat{\mathbf{x}} + \left(\frac{1}{4} + z_4\right) a\,\hat{\mathbf{y}} - a\left(y_4 + \frac{1}{4}\right)\hat{\mathbf{z}}$ | (48$e$) | O III |
| $\mathbf{B}_{75}$ | = | $\left(\frac{1}{2} + y_4 - z_4\right)\mathbf{a}_1 + (-x_4 + y_4)\mathbf{a}_2 + (-x_4 - z_4)\mathbf{a}_3$ | = | $-a\left(x_4 + \frac{1}{4}\right)\hat{\mathbf{x}} + \left(\frac{1}{4} - z_4\right) a\,\hat{\mathbf{y}} + \left(\frac{1}{4} + y_4\right) a\,\hat{\mathbf{z}}$ | (48$e$) | O III |
| $\mathbf{B}_{76}$ | = | $(-y_4 - z_4)\mathbf{a}_1 + \left(\frac{1}{2} + x_4 - y_4\right)\mathbf{a}_2 + (x_4 - z_4)\mathbf{a}_3$ | = | $\left(\frac{1}{4} + x_4\right) a\,\hat{\mathbf{x}} - a\left(z_4 + \frac{1}{4}\right)\hat{\mathbf{y}} + \left(\frac{1}{4} - y_4\right) a\,\hat{\mathbf{z}}$ | (48$e$) | O III |
| $\mathbf{B}_{77}$ | = | $\left(\frac{1}{2} + x_4 + y_4\right)\mathbf{a}_1 + \left(\frac{1}{2} + x_4 + z_4\right)\mathbf{a}_2 + \left(\frac{1}{2} + y_4 + z_4\right)\mathbf{a}_3$ | = | $\left(\frac{1}{4} + z_4\right) a\,\hat{\mathbf{x}} + \left(\frac{1}{4} + y_4\right) a\,\hat{\mathbf{y}} + \left(\frac{1}{4} + x_4\right) a\,\hat{\mathbf{z}}$ | (48$e$) | O III |
| $\mathbf{B}_{78}$ | = | $(-x_4 - y_4)\mathbf{a}_1 + \left(\frac{1}{2} - x_4 + z_4\right)\mathbf{a}_2 + (-y_4 + z_4)\mathbf{a}_3$ | = | $\left(\frac{1}{4} + z_4\right) a\,\hat{\mathbf{x}} - a\left(y_4 + \frac{1}{4}\right)\hat{\mathbf{y}} + \left(\frac{1}{4} - x_4\right) a\,\hat{\mathbf{z}}$ | (48$e$) | O III |
| $\mathbf{B}_{79}$ | = | $(-x_4 + y_4)\mathbf{a}_1 + (-x_4 - z_4)\mathbf{a}_2 + \left(\frac{1}{2} + y_4 - z_4\right)\mathbf{a}_3$ | = | $\left(\frac{1}{4} - z_4\right) a\,\hat{\mathbf{x}} + \left(\frac{1}{4} + y_4\right) a\,\hat{\mathbf{y}} - a\left(x_4 + \frac{1}{4}\right)\hat{\mathbf{z}}$ | (48$e$) | O III |
| $\mathbf{B}_{80}$ | = | $\left(\frac{1}{2} + x_4 - y_4\right)\mathbf{a}_1 + (x_4 - z_4)\mathbf{a}_2 + (-y_4 - z_4)\mathbf{a}_3$ | = | $-a\left(z_4 + \frac{1}{4}\right)\hat{\mathbf{x}} + \left(\frac{1}{4} - y_4\right) a\,\hat{\mathbf{y}} + \left(\frac{1}{4} + x_4\right) a\,\hat{\mathbf{z}}$ | (48$e$) | O III |
| $\mathbf{B}_{81}$ | = | $(y_5 + z_5)\mathbf{a}_1 + (x_5 + z_5)\mathbf{a}_2 + (x_5 + y_5)\mathbf{a}_3$ | = | $x_5 a\,\hat{\mathbf{x}} + y_5 a\,\hat{\mathbf{y}} + z_5 a\,\hat{\mathbf{z}}$ | (48$e$) | P |
| $\mathbf{B}_{82}$ | = | $\left(\frac{1}{2} - y_5 + z_5\right)\mathbf{a}_1 + (-x_5 + z_5)\mathbf{a}_2 + \left(\frac{1}{2} - x_5 - y_5\right)\mathbf{a}_3$ | = | $-x_5 a\,\hat{\mathbf{x}} + \left(\frac{1}{2} - y_5\right) a\,\hat{\mathbf{y}} + z_5 a\,\hat{\mathbf{z}}$ | (48$e$) | P |
| $\mathbf{B}_{83}$ | = | $(y_5 - z_5)\mathbf{a}_1 + \left(\frac{1}{2} - x_5 - z_5\right)\mathbf{a}_2 + \left(\frac{1}{2} - x_5 + y_5\right)\mathbf{a}_3$ | = | $\left(\frac{1}{2} - x_5\right) a\,\hat{\mathbf{x}} + y_5 a\,\hat{\mathbf{y}} - z_5 a\,\hat{\mathbf{z}}$ | (48$e$) | P |
| $\mathbf{B}_{84}$ | = | $\left(\frac{1}{2} - y_5 - z_5\right)\mathbf{a}_1 + \left(\frac{1}{2} + x_5 - z_5\right)\mathbf{a}_2 + (x_5 - y_5)\mathbf{a}_3$ | = | $x_5 a\,\hat{\mathbf{x}} - y_5 a\,\hat{\mathbf{y}} + \left(\frac{1}{2} - z_5\right) a\,\hat{\mathbf{z}}$ | (48$e$) | P |
| $\mathbf{B}_{85}$ | = | $(x_5 + y_5)\mathbf{a}_1 + (y_5 + z_5)\mathbf{a}_2 + (x_5 + z_5)\mathbf{a}_3$ | = | $z_5 a\,\hat{\mathbf{x}} + x_5 a\,\hat{\mathbf{y}} + y_5 a\,\hat{\mathbf{z}}$ | (48$e$) | P |
| $\mathbf{B}_{86}$ | = | $\left(\frac{1}{2} - x_5 - y_5\right)\mathbf{a}_1 + \left(\frac{1}{2} - y_5 + z_5\right)\mathbf{a}_2 + (-x_5 + z_5)\mathbf{a}_3$ | = | $z_5 a\,\hat{\mathbf{x}} - x_5 a\,\hat{\mathbf{y}} + \left(\frac{1}{2} - y_5\right) a\,\hat{\mathbf{z}}$ | (48$e$) | P |
| $\mathbf{B}_{87}$ | = | $\left(\frac{1}{2} - x_5 + y_5\right)\mathbf{a}_1 + (y_5 - z_5)\mathbf{a}_2 + \left(\frac{1}{2} - x_5 - z_5\right)\mathbf{a}_3$ | = | $-z_5 a\,\hat{\mathbf{x}} + \left(\frac{1}{2} - x_5\right) a\,\hat{\mathbf{y}} + y_5 a\,\hat{\mathbf{z}}$ | (48$e$) | P |
| $\mathbf{B}_{88}$ | = | $\left(x_5 - y_5\right)\,\mathbf{a}_1 + \left(\frac{1}{2} - y_5 - z_5\right)\,\mathbf{a}_2 + \left(\frac{1}{2} + x_5 - z_5\right)\,\mathbf{a}_3$ | = | $\left(\frac{1}{2} - z_5\right)a\,\hat{\mathbf{x}} + x_5a\,\hat{\mathbf{y}} - y_5a\,\hat{\mathbf{z}}$ | (48$e$) | P |
| $\mathbf{B}_{89}$ | = | $\left(x_5 + z_5\right)\,\mathbf{a}_1 + \left(x_5 + y_5\right)\,\mathbf{a}_2 + \left(y_5 + z_5\right)\,\mathbf{a}_3$ | = | $y_5a\,\hat{\mathbf{x}} + z_5a\,\hat{\mathbf{y}} + x_5a\,\hat{\mathbf{z}}$ | (48$e$) | P |
| $\mathbf{B}_{90}$ | = | $\left(-x_5 + z_5\right)\,\mathbf{a}_1 + \left(\frac{1}{2} - x_5 - y_5\right)\,\mathbf{a}_2 + \left(\frac{1}{2} - y_5 + z_5\right)\,\mathbf{a}_3$ | = | $\left(\frac{1}{2} - y_5\right)a\,\hat{\mathbf{x}} + z_5a\,\hat{\mathbf{y}} - x_5a\,\hat{\mathbf{z}}$ | (48$e$) | P |
| $\mathbf{B}_{91}$ | = | $\left(\frac{1}{2} - x_5 - z_5\right)\,\mathbf{a}_1 + \left(\frac{1}{2} - x_5 + y_5\right)\,\mathbf{a}_2 + \left(y_5 - z_5\right)\,\mathbf{a}_3$ | = | $y_5a\,\hat{\mathbf{x}} - z_5a\,\hat{\mathbf{y}} + \left(\frac{1}{2} - x_5\right)a\,\hat{\mathbf{z}}$ | (48$e$) | P |
| $\mathbf{B}_{92}$ | = | $\left(\frac{1}{2} + x_5 - z_5\right)\,\mathbf{a}_1 + \left(x_5 - y_5\right)\,\mathbf{a}_2 + \left(\frac{1}{2} - y_5 - z_5\right)\,\mathbf{a}_3$ | = | $-y_5a\,\hat{\mathbf{x}} + \left(\frac{1}{2} - z_5\right)a\,\hat{\mathbf{y}} + x_5a\,\hat{\mathbf{z}}$ | (48$e$) | P |
| $\mathbf{B}_{93}$ | = | $\left(\frac{1}{2} + x_5 + z_5\right)\,\mathbf{a}_1 + \left(\frac{1}{2} + y_5 + z_5\right)\,\mathbf{a}_2 + \left(\frac{1}{2} + x_5 + y_5\right)\,\mathbf{a}_3$ | = | $\left(\frac{1}{4} + y_5\right)a\,\hat{\mathbf{x}} + \left(\frac{1}{4} + x_5\right)a\,\hat{\mathbf{y}} + \left(\frac{1}{4} + z_5\right)a\,\hat{\mathbf{z}}$ | (48$e$) | P |
| $\mathbf{B}_{94}$ | = | $\left(\frac{1}{2} - x_5 + z_5\right)\,\mathbf{a}_1 + \left(-y_5 + z_5\right)\,\mathbf{a}_2 + \left(-x_5 - y_5\right)\,\mathbf{a}_3$ | = | $-a\left(y_5 + \frac{1}{4}\right)\,\hat{\mathbf{x}} + \left(\frac{1}{4} - x_5\right)a\,\hat{\mathbf{y}} + \left(\frac{1}{4} + z_5\right)a\,\hat{\mathbf{z}}$ | (48$e$) | P |
| $\mathbf{B}_{95}$ | = | $\left(-x_5 - z_5\right)\,\mathbf{a}_1 + \left(\frac{1}{2} + y_5 - z_5\right)\,\mathbf{a}_2 + \left(-x_5 + y_5\right)\,\mathbf{a}_3$ | = | $\left(\frac{1}{4} + y_5\right)a\,\hat{\mathbf{x}} - a\left(x_5 + \frac{1}{4}\right)\,\hat{\mathbf{y}} + \left(\frac{1}{4} - z_5\right)a\,\hat{\mathbf{z}}$ | (48$e$) | P |
| $\mathbf{B}_{96}$ | = | $\left(x_5 - z_5\right)\,\mathbf{a}_1 + \left(-y_5 - z_5\right)\,\mathbf{a}_2 + \left(\frac{1}{2} + x_5 - y_5\right)\,\mathbf{a}_3$ | = | $\left(\frac{1}{4} - y_5\right)a\,\hat{\mathbf{x}} + \left(\frac{1}{4} + x_5\right)a\,\hat{\mathbf{y}} - a\left(z_5 + \frac{1}{4}\right)\,\hat{\mathbf{z}}$ | (48$e$) | P |
| $\mathbf{B}_{97}$ | = | $\left(\frac{1}{2} + y_5 + z_5\right)\,\mathbf{a}_1 + \left(\frac{1}{2} + x_5 + y_5\right)\,\mathbf{a}_2 + \left(\frac{1}{2} + x_5 + z_5\right)\,\mathbf{a}_3$ | = | $\left(\frac{1}{4} + x_5\right)a\,\hat{\mathbf{x}} + \left(\frac{1}{4} + z_5\right)a\,\hat{\mathbf{y}} + \left(\frac{1}{4} + y_5\right)a\,\hat{\mathbf{z}}$ | (48$e$) | P |
| $\mathbf{B}_{98}$ | = | $\left(-y_5 + z_5\right)\,\mathbf{a}_1 + \left(-x_5 - y_5\right)\,\mathbf{a}_2 + \left(\frac{1}{2} - x_5 + z_5\right)\,\mathbf{a}_3$ | = | $\left(\frac{1}{4} - x_5\right)a\,\hat{\mathbf{x}} + \left(\frac{1}{4} + z_5\right)a\,\hat{\mathbf{y}} - a\left(y_5 + \frac{1}{4}\right)\,\hat{\mathbf{z}}$ | (48$e$) | P |
| $\mathbf{B}_{99}$ | = | $\left(\frac{1}{2} + y_5 - z_5\right)\,\mathbf{a}_1 + \left(-x_5 + y_5\right)\,\mathbf{a}_2 + \left(-x_5 - z_5\right)\,\mathbf{a}_3$ | = | $-a\left(x_5 + \frac{1}{4}\right)\,\hat{\mathbf{x}} + \left(\frac{1}{4} - z_5\right)a\,\hat{\mathbf{y}} + \left(\frac{1}{4} + y_5\right)a\,\hat{\mathbf{z}}$ | (48$e$) | P |
| $\mathbf{B}_{100}$ | = | $\left(-y_5 - z_5\right)\,\mathbf{a}_1 + \left(\frac{1}{2} + x_5 - y_5\right)\,\mathbf{a}_2 + \left(x_5 - z_5\right)\,\mathbf{a}_3$ | = | $\left(\frac{1}{4} + x_5\right)a\,\hat{\mathbf{x}} - a\left(z_5 + \frac{1}{4}\right)\,\hat{\mathbf{y}} + \left(\frac{1}{4} - y_5\right)a\,\hat{\mathbf{z}}$ | (48$e$) | P |
| $\mathbf{B}_{101}$ | = | $\left(\frac{1}{2} + x_5 + y_5\right)\,\mathbf{a}_1 + \left(\frac{1}{2} + x_5 + z_5\right)\,\mathbf{a}_2 + \left(\frac{1}{2} + y_5 + z_5\right)\,\mathbf{a}_3$ | = | $\left(\frac{1}{4} + z_5\right)a\,\hat{\mathbf{x}} + \left(\frac{1}{4} + y_5\right)a\,\hat{\mathbf{y}} + \left(\frac{1}{4} + x_5\right)a\,\hat{\mathbf{z}}$ | (48$e$) | P |
| $\mathbf{B}_{102}$ | = | $\left(-x_5 - y_5\right)\,\mathbf{a}_1 + \left(\frac{1}{2} - x_5 + z_5\right)\,\mathbf{a}_2 + \left(-y_5 + z_5\right)\,\mathbf{a}_3$ | = | $\left(\frac{1}{4} + z_5\right)a\,\hat{\mathbf{x}} - a\left(y_5 + \frac{1}{4}\right)\,\hat{\mathbf{y}} + \left(\frac{1}{4} - x_5\right)a\,\hat{\mathbf{z}}$ | (48$e$) | P |
| $\mathbf{B}_{103}$ | = | $\left(-x_5 + y_5\right)\,\mathbf{a}_1 + \left(-x_5 - z_5\right)\,\mathbf{a}_2 + \left(\frac{1}{2} + y_5 - z_5\right)\,\mathbf{a}_3$ | = | $\left(\frac{1}{4} - z_5\right)a\,\hat{\mathbf{x}} + \left(\frac{1}{4} + y_5\right)a\,\hat{\mathbf{y}} - a\left(x_5 + \frac{1}{4}\right)\,\hat{\mathbf{z}}$ | (48$e$) | P |
| $\mathbf{B}_{104}$ | = | $\left(\frac{1}{2} + x_5 - y_5\right)\,\mathbf{a}_1 + \left(x_5 - z_5\right)\,\mathbf{a}_2 + \left(-y_5 - z_5\right)\,\mathbf{a}_3$ | = | $-a\left(z_5 + \frac{1}{4}\right)\,\hat{\mathbf{x}} + \left(\frac{1}{4} - y_5\right)a\,\hat{\mathbf{y}} + \left(\frac{1}{4} + x_5\right)a\,\hat{\mathbf{z}}$ | (48$e$) | P |

**Geometry files:**
- CIF: pp. 1804
- POSCAR: pp. 1804

# AlN (cI24) Structure: AB_cI24_220_a_b

http://aflow.org/prototype-encyclopedia/AB_cI24_220_a_b

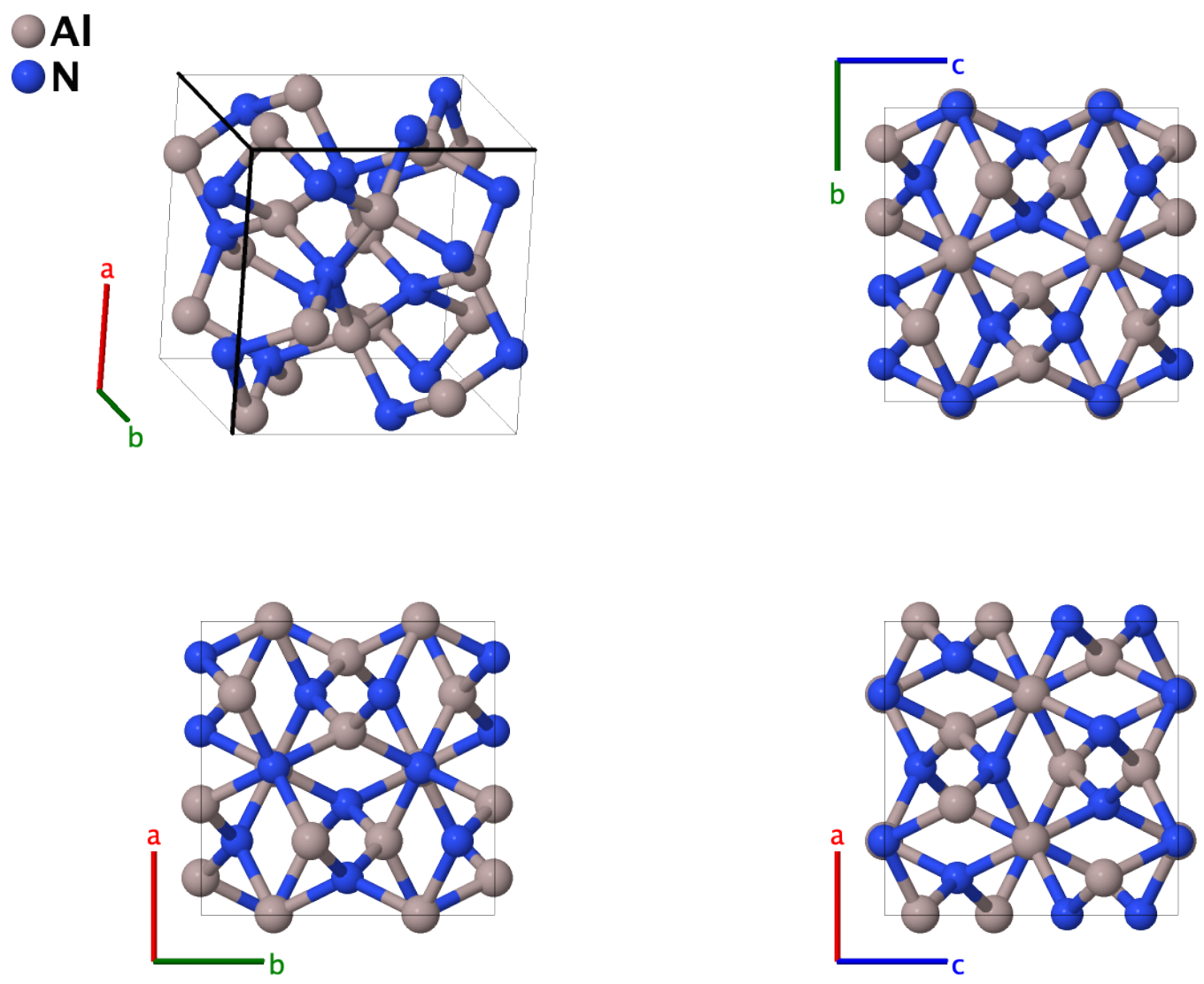

| | | |
|---|---|---|
| **Prototype** | : | AlN |
| **AFLOW prototype label** | : | AB_cI24_220_a_b |
| ***Strukturbericht* designation** | : | None |
| **Pearson symbol** | : | cI24 |
| **Space group number** | : | 220 |
| **Space group symbol** | : | $I\bar{4}3d$ |
| **AFLOW prototype command** | : | `aflow --proto=AB_cI24_220_a_b --params=`$a$ |

- AlN naturally occurs in two forms (Liu, 2019): the stable wz-AlN wurtzite ($B4$) structure, and the high-pressure rs-AlN rock salt ($B1$) structure. A metastable zb-AlN zincblende (zb-AlN) structure can be synthesized via a solid-state reaction.
- (Liu, 2019) used a first-principles evolutionary technique to find four possible metastable phases: one in the SC16 structure, and three novel cubic structures, cF40, cI16, and cI24.

**Body-centered Cubic primitive vectors:**

$$\begin{aligned}
\mathbf{a}_1 &= -\tfrac{1}{2}\,a\,\hat{\mathbf{x}} + \tfrac{1}{2}\,a\,\hat{\mathbf{y}} + \tfrac{1}{2}\,a\,\hat{\mathbf{z}} \\
\mathbf{a}_2 &= \tfrac{1}{2}\,a\,\hat{\mathbf{x}} - \tfrac{1}{2}\,a\,\hat{\mathbf{y}} + \tfrac{1}{2}\,a\,\hat{\mathbf{z}} \\
\mathbf{a}_3 &= \tfrac{1}{2}\,a\,\hat{\mathbf{x}} + \tfrac{1}{2}\,a\,\hat{\mathbf{y}} - \tfrac{1}{2}\,a\,\hat{\mathbf{z}}
\end{aligned}$$

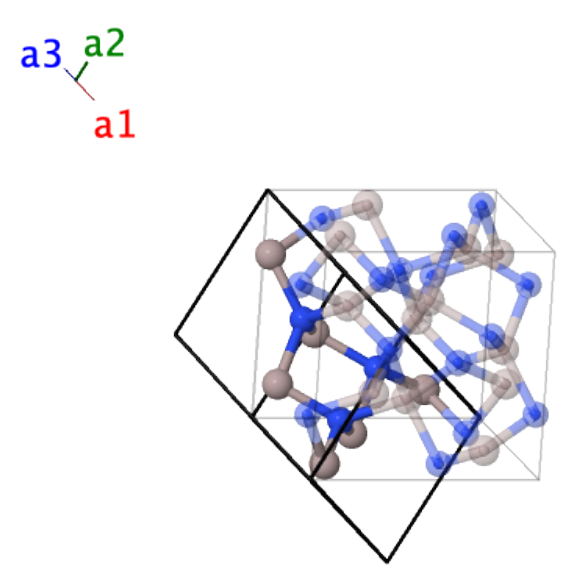

**Basis vectors:**

| | | Lattice Coordinates | | Cartesian Coordinates | Wyckoff Position | Atom Type |
|---|---|---|---|---|---|---|
| $\mathbf{B}_1$ | = | $\frac{1}{4}\,\mathbf{a}_1 + \frac{5}{8}\,\mathbf{a}_2 + \frac{3}{8}\,\mathbf{a}_3$ | = | $\frac{3}{8}a\,\hat{\mathbf{x}} + \frac{1}{4}a\,\hat{\mathbf{z}}$ | (12$a$) | Al |
| $\mathbf{B}_2$ | = | $\frac{3}{4}\,\mathbf{a}_1 + \frac{7}{8}\,\mathbf{a}_2 + \frac{1}{8}\,\mathbf{a}_3$ | = | $\frac{1}{8}a\,\mathbf{\hat{x}} + \frac{3}{4}a\,\mathbf{\hat{z}}$ | (12$a$) | Al |
| $\mathbf{B}_3$ | = | $\frac{3}{8}\,\mathbf{a}_1 + \frac{1}{4}\,\mathbf{a}_2 + \frac{5}{8}\,\mathbf{a}_3$ | = | $\frac{1}{4}a\,\mathbf{\hat{x}} + \frac{3}{8}a\,\mathbf{\hat{y}}$ | (12$a$) | Al |
| $\mathbf{B}_4$ | = | $\frac{1}{8}\,\mathbf{a}_1 + \frac{3}{4}\,\mathbf{a}_2 + \frac{7}{8}\,\mathbf{a}_3$ | = | $\frac{3}{4}a\,\mathbf{\hat{x}} + \frac{1}{8}a\,\mathbf{\hat{y}}$ | (12$a$) | Al |
| $\mathbf{B}_5$ | = | $\frac{5}{8}\,\mathbf{a}_1 + \frac{3}{8}\,\mathbf{a}_2 + \frac{1}{4}\,\mathbf{a}_3$ | = | $\frac{1}{4}a\,\mathbf{\hat{y}} + \frac{3}{8}a\,\mathbf{\hat{z}}$ | (12$a$) | Al |
| $\mathbf{B}_6$ | = | $\frac{7}{8}\,\mathbf{a}_1 + \frac{1}{8}\,\mathbf{a}_2 + \frac{3}{4}\,\mathbf{a}_3$ | = | $\frac{3}{4}a\,\mathbf{\hat{y}} + \frac{1}{8}a\,\mathbf{\hat{z}}$ | (12$a$) | Al |
| $\mathbf{B}_7$ | = | $\frac{1}{4}\,\mathbf{a}_1 + \frac{1}{8}\,\mathbf{a}_2 + \frac{7}{8}\,\mathbf{a}_3$ | = | $\frac{3}{8}a\,\mathbf{\hat{x}} + \frac{1}{2}a\,\mathbf{\hat{y}} - \frac{1}{4}a\,\mathbf{\hat{z}}$ | (12$b$) | N |
| $\mathbf{B}_8$ | = | $\frac{3}{4}\,\mathbf{a}_1 + \frac{3}{8}\,\mathbf{a}_2 + \frac{5}{8}\,\mathbf{a}_3$ | = | $\frac{1}{8}a\,\mathbf{\hat{x}} + \frac{1}{2}a\,\mathbf{\hat{y}} + \frac{1}{4}a\,\mathbf{\hat{z}}$ | (12$b$) | N |
| $\mathbf{B}_9$ | = | $\frac{7}{8}\,\mathbf{a}_1 + \frac{1}{4}\,\mathbf{a}_2 + \frac{1}{8}\,\mathbf{a}_3$ | = | $-\frac{1}{4}a\,\mathbf{\hat{x}} + \frac{3}{8}a\,\mathbf{\hat{y}} + \frac{1}{2}a\,\mathbf{\hat{z}}$ | (12$b$) | N |
| $\mathbf{B}_{10}$ | = | $\frac{5}{8}\,\mathbf{a}_1 + \frac{3}{4}\,\mathbf{a}_2 + \frac{3}{8}\,\mathbf{a}_3$ | = | $\frac{1}{4}a\,\mathbf{\hat{x}} + \frac{1}{8}a\,\mathbf{\hat{y}} + \frac{1}{2}a\,\mathbf{\hat{z}}$ | (12$b$) | N |
| $\mathbf{B}_{11}$ | = | $\frac{1}{8}\,\mathbf{a}_1 + \frac{7}{8}\,\mathbf{a}_2 + \frac{1}{4}\,\mathbf{a}_3$ | = | $\frac{1}{2}a\,\mathbf{\hat{x}} - \frac{1}{4}a\,\mathbf{\hat{y}} + \frac{3}{8}a\,\mathbf{\hat{z}}$ | (12$b$) | N |
| $\mathbf{B}_{12}$ | = | $\frac{3}{8}\,\mathbf{a}_1 + \frac{5}{8}\,\mathbf{a}_2 + \frac{3}{4}\,\mathbf{a}_3$ | = | $\frac{1}{2}a\,\mathbf{\hat{x}} + \frac{1}{4}a\,\mathbf{\hat{y}} + \frac{1}{8}a\,\mathbf{\hat{z}}$ | (12$b$) | N |

**Geometry files:**

- CIF: pp. 1805
- POSCAR: pp. 1805

# $\gamma$-$Fe_4N$ ($L'1_0$) Structure: A4B_cP5_221_bc_a

http://aflow.org/prototype-encyclopedia/A4B_cP5_221_bc_a

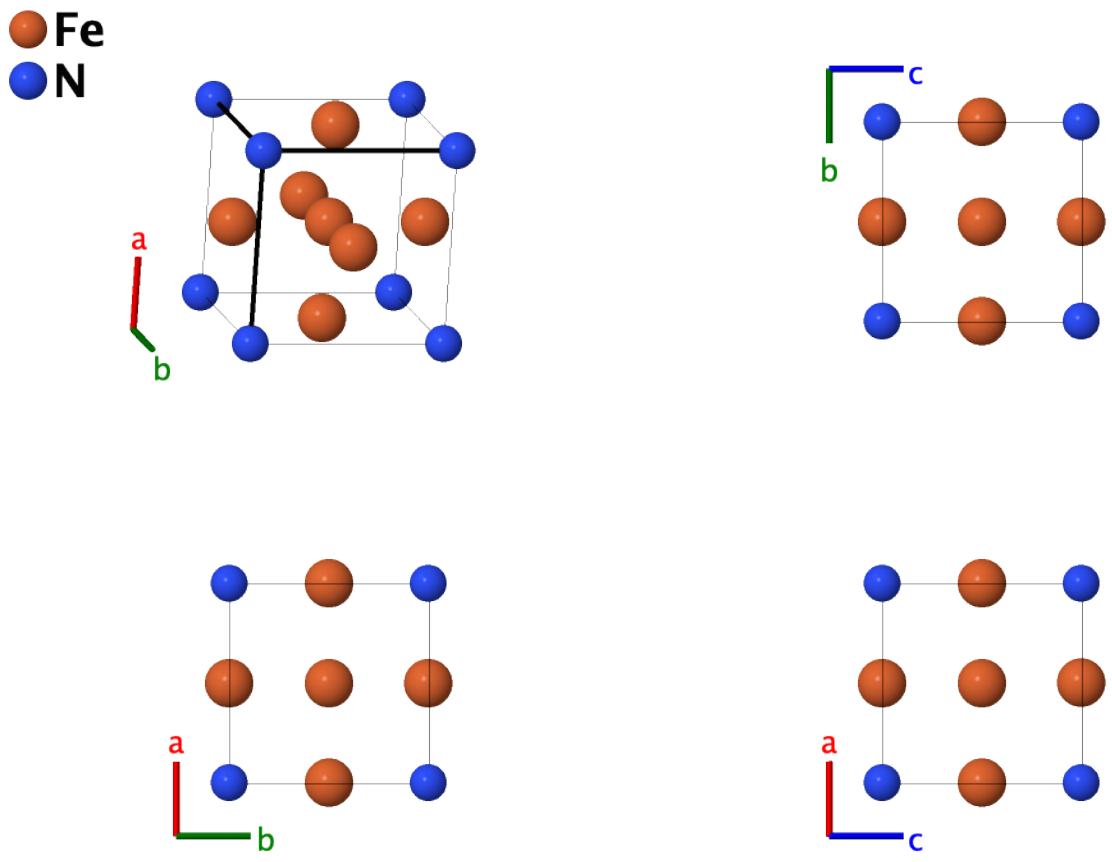

| | | |
|---|---|---|
| **Prototype** | : | $Fe_4N$ |
| **AFLOW prototype label** | : | A4B_cP5_221_bc_a |
| ***Strukturbericht* designation** | : | $L'1_0$ |
| **Pearson symbol** | : | cP5 |
| **Space group number** | : | 221 |
| **Space group symbol** | : | $Pm\bar{3}m$ |
| **AFLOW prototype command** | : | `aflow --proto=A4B_cP5_221_bc_a --params=`$a$ |

- This is a two-component form of the cubic perovskite structure (AB3C_cP5_221_a_c_b), *Strukturbericht* designation $E2_1$.

**Simple Cubic primitive vectors:**

$$\mathbf{a}_1 = a\,\hat{\mathbf{x}}$$
$$\mathbf{a}_2 = a\,\hat{\mathbf{y}}$$
$$\mathbf{a}_3 = a\,\hat{\mathbf{z}}$$

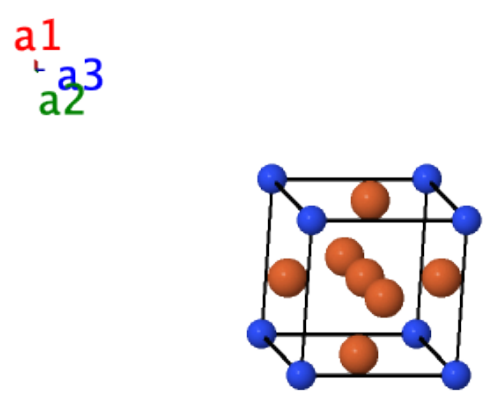

**Basis vectors:**

| | | Lattice Coordinates | | Cartesian Coordinates | Wyckoff Position | Atom Type |
|---|---|---|---|---|---|---|
| $\mathbf{B}_1$ | = | $0\,\mathbf{a}_1 + 0\,\mathbf{a}_2 + 0\,\mathbf{a}_3$ | = | $0\,\hat{\mathbf{x}} + 0\,\hat{\mathbf{y}} + 0\,\hat{\mathbf{z}}$ | (1$a$) | N |
| $\mathbf{B}_2$ | = | $\frac{1}{2}\,\mathbf{a}_1 + \frac{1}{2}\,\mathbf{a}_2 + \frac{1}{2}\,\mathbf{a}_3$ | = | $\frac{1}{2}a\,\hat{\mathbf{x}} + \frac{1}{2}a\,\hat{\mathbf{y}} + \frac{1}{2}a\,\hat{\mathbf{z}}$ | (1$b$) | Fe I |
| $\mathbf{B}_3$ | = | $\frac{1}{2}\,\mathbf{a}_2 + \frac{1}{2}\,\mathbf{a}_3$ | = | $\frac{1}{2}a\,\hat{\mathbf{y}} + \frac{1}{2}a\,\hat{\mathbf{z}}$ | (3$c$) | Fe II |
| $\mathbf{B}_4$ | = | $\frac{1}{2}\,\mathbf{a}_1 + \frac{1}{2}\,\mathbf{a}_3$ | = | $\frac{1}{2}a\,\hat{\mathbf{x}} + \frac{1}{2}a\,\hat{\mathbf{z}}$ | (3$c$) | Fe II |
| $\mathbf{B}_5$ | = | $\frac{1}{2}\,\mathbf{a}_1 + \frac{1}{2}\,\mathbf{a}_2$ | = | $\frac{1}{2}a\,\hat{\mathbf{x}} + \frac{1}{2}a\,\hat{\mathbf{y}}$ | (3$c$) | Fe II |

**Geometry files:**

- CIF: pp. 1805
- POSCAR: pp. 1806

# Predicted High-Pressure $YCaH_{12}$ Structure: AB12C_cP14_221_a_h_b

http://aflow.org/prototype-encyclopedia/AB12C_cP14_221_a_h_b

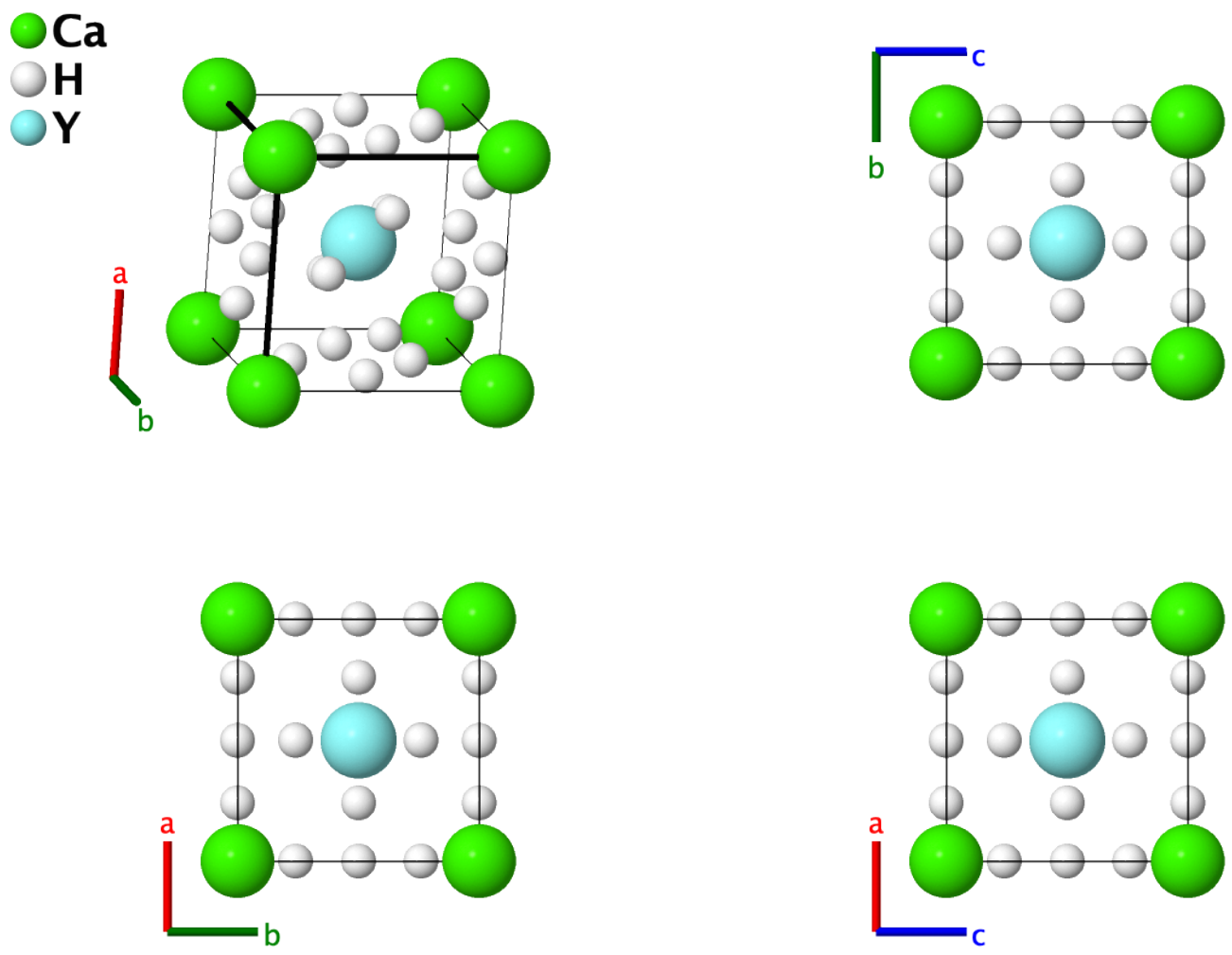

| | | |
|---|---|---|
| **Prototype** | : | $CaH_{12}Y$ |
| **AFLOW prototype label** | : | AB12C_cP14_221_a_h_b |
| ***Strukturbericht* designation** | : | None |
| **Pearson symbol** | : | cP14 |
| **Space group number** | : | 221 |
| **Space group symbol** | : | $Pm\bar{3}m$ |
| **AFLOW prototype command** | : | `aflow --proto=AB12C_cP14_221_a_h_b --params=`$a, x_3$ |

- This structure was determined by *ab initio* methods and is predicted to be stable in the pressure range 180-257 GPa, with $T_c = 230$ K at 180 GPa. We show the predicted structure at 200 GPa.

**Simple Cubic primitive vectors:**

$$\mathbf{a}_1 = a\,\hat{\mathbf{x}}$$
$$\mathbf{a}_2 = a\,\hat{\mathbf{y}}$$
$$\mathbf{a}_3 = a\,\hat{\mathbf{z}}$$

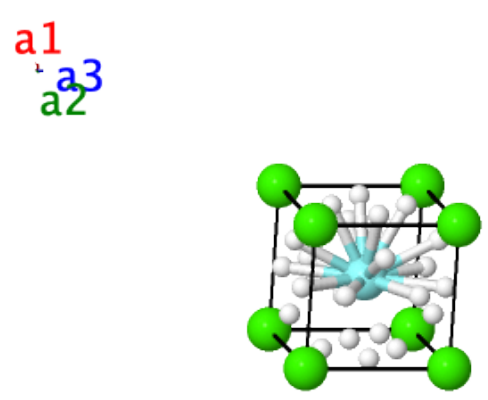

**Basis vectors:**

| | | Lattice Coordinates | | Cartesian Coordinates | Wyckoff Position | Atom Type |
|---|---|---|---|---|---|---|
| $\mathbf{B}_1$ | = | $0\,\mathbf{a}_1 + 0\,\mathbf{a}_2 + 0\,\mathbf{a}_3$ | = | $0\,\hat{\mathbf{x}} + 0\,\hat{\mathbf{y}} + 0\,\hat{\mathbf{z}}$ | (1$a$) | Ca |
| $\mathbf{B}_2$ | = | $\frac{1}{2}\,\mathbf{a}_1 + \frac{1}{2}\,\mathbf{a}_2 + \frac{1}{2}\,\mathbf{a}_3$ | = | $\frac{1}{2}a\,\hat{\mathbf{x}} + \frac{1}{2}a\,\hat{\mathbf{y}} + \frac{1}{2}a\,\hat{\mathbf{z}}$ | (1$b$) | Y |
| $\mathbf{B}_3$ | = | $x_3\,\mathbf{a}_1 + \frac{1}{2}\,\mathbf{a}_2$ | = | $x_3 a\,\hat{\mathbf{x}} + \frac{1}{2}a\,\hat{\mathbf{y}}$ | (12$h$) | H |
| $\mathbf{B}_4$ | = | $-x_3\,\mathbf{a}_1 + \frac{1}{2}\,\mathbf{a}_2$ | = | $-x_3 a\,\hat{\mathbf{x}} + \frac{1}{2} a\,\hat{\mathbf{y}}$ | $(12h)$ | H |
| $\mathbf{B}_5$ | = | $x_3\,\mathbf{a}_2 + \frac{1}{2}\,\mathbf{a}_3$ | = | $x_3 a\,\hat{\mathbf{y}} + \frac{1}{2} a\,\hat{\mathbf{z}}$ | $(12h)$ | H |
| $\mathbf{B}_6$ | = | $-x_3\,\mathbf{a}_2 + \frac{1}{2}\,\mathbf{a}_3$ | = | $-x_3 a\,\hat{\mathbf{y}} + \frac{1}{2} a\,\hat{\mathbf{z}}$ | $(12h)$ | H |
| $\mathbf{B}_7$ | = | $\frac{1}{2}\,\mathbf{a}_1 + x_3\,\mathbf{a}_3$ | = | $\frac{1}{2} a\,\hat{\mathbf{x}} + x_3 a\,\hat{\mathbf{z}}$ | $(12h)$ | H |
| $\mathbf{B}_8$ | = | $\frac{1}{2}\,\mathbf{a}_1 - x_3\,\mathbf{a}_3$ | = | $\frac{1}{2} a\,\hat{\mathbf{x}} - x_3 a\,\hat{\mathbf{z}}$ | $(12h)$ | H |
| $\mathbf{B}_9$ | = | $\frac{1}{2}\,\mathbf{a}_1 + x_3\,\mathbf{a}_2$ | = | $\frac{1}{2} a\,\hat{\mathbf{x}} + x_3 a\,\hat{\mathbf{y}}$ | $(12h)$ | H |
| $\mathbf{B}_{10}$ | = | $\frac{1}{2}\,\mathbf{a}_1 - x_3\,\mathbf{a}_2$ | = | $\frac{1}{2} a\,\hat{\mathbf{x}} - x_3 a\,\hat{\mathbf{y}}$ | $(12h)$ | H |
| $\mathbf{B}_{11}$ | = | $x_3\,\mathbf{a}_1 + \frac{1}{2}\,\mathbf{a}_3$ | = | $x_3 a\,\hat{\mathbf{x}} + \frac{1}{2} a\,\hat{\mathbf{z}}$ | $(12h)$ | H |
| $\mathbf{B}_{12}$ | = | $-x_3\,\mathbf{a}_1 + \frac{1}{2}\,\mathbf{a}_3$ | = | $-x_3 a\,\hat{\mathbf{x}} + \frac{1}{2} a\,\hat{\mathbf{z}}$ | $(12h)$ | H |
| $\mathbf{B}_{13}$ | = | $\frac{1}{2}\,\mathbf{a}_2 - x_3\,\mathbf{a}_3$ | = | $\frac{1}{2} a\,\hat{\mathbf{y}} - x_3 a\,\hat{\mathbf{z}}$ | $(12h)$ | H |
| $\mathbf{B}_{14}$ | = | $\frac{1}{2}\,\mathbf{a}_2 + x_3\,\mathbf{a}_3$ | = | $\frac{1}{2} a\,\hat{\mathbf{y}} + x_3 a\,\hat{\mathbf{z}}$ | $(12h)$ | H |

**Geometry files:**

- CIF: pp. 1806
- POSCAR: pp. 1807

# $NH_4NO_3$ I ($G0_8$) Structure: AB_cP2_221_a_b

http://aflow.org/prototype-encyclopedia/AB_cP2_221_a_b.NH4.NO3

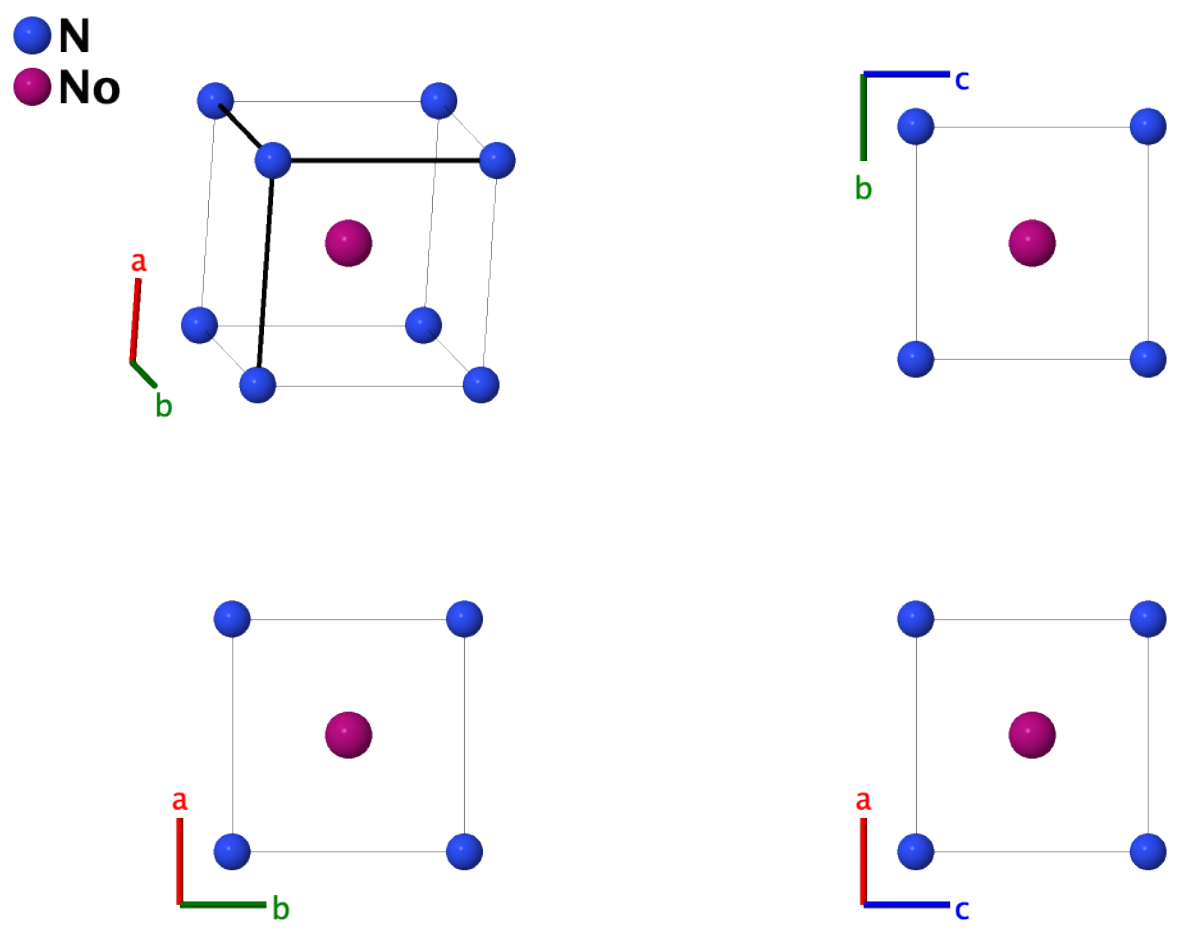

| | | |
|---|---|---|
| **Prototype** | : | $NH_4NO_3$ |
| **AFLOW prototype label** | : | AB_cP2_221_a_b |
| ***Strukturbericht* designation** | : | $G0_8$ |
| **Pearson symbol** | : | cP2 |
| **Space group number** | : | 221 |
| **Space group symbol** | : | $Pm\bar{3}m$ |
| **AFLOW prototype command** | : | `aflow --proto=AB_cP2_221_a_b --params=`$a$ |

- Ammonium Nitrate exists in a variety of forms, (Hermann, 1937) depending on the temperature:

| Phase | Temperature °C | *Strukturbericht* | Page | |
|---|---|---|---|---|
| I | 125 – 170 | $G0_8$ | AB_cP2_221_a_b.NH4.NO3 | (this structure) |
| II | 84 – 125 | $G0_9$ | ABC3_tP10_100_b_a_bc | |
| III | 32 – 84 | $G0_{10}$ | ABC3_oP20_62_c_c_cd.N.NH4.O | |
| IV | -17 – 32 | $G0_{11}$ | A4B2C3_oP18_59_ef_ab_af | |
| V | $< -17$ | Gwihabaite | A4B2C3_tP72_77_8d_ab2c2d_6d2 | |

- In the $G0_8$ structure, both the $NH_4^+$ and $NO_3^-$ ions rotate freely about their centers of mass (Kracek, 1931). The two ions sit on the same sites as atoms in the CsCl ($B2$) structure.

**Simple Cubic primitive vectors:**

$$\mathbf{a}_1 = a\,\hat{\mathbf{x}}$$
$$\mathbf{a}_2 = a\,\hat{\mathbf{y}}$$
$$\mathbf{a}_3 = a\,\hat{\mathbf{z}}$$

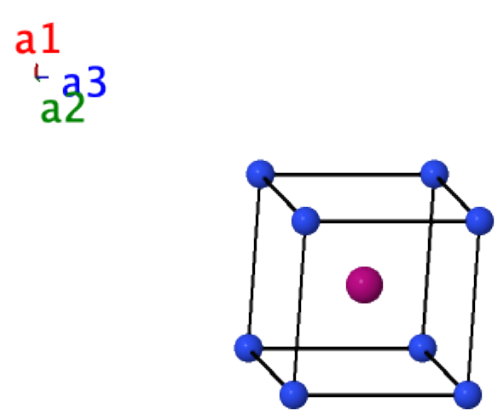

**Basis vectors:**

| | | Lattice Coordinates | | Cartesian Coordinates | Wyckoff Position | Atom Type |
|---|---|---|---|---|---|---|
| $\mathbf{B}_1$ | = | $0\,\mathbf{a}_1 + 0\,\mathbf{a}_2 + 0\,\mathbf{a}_3$ | = | $0\,\hat{\mathbf{x}} + 0\,\hat{\mathbf{y}} + 0\,\hat{\mathbf{z}}$ | $(1a)$ | $NH_4$ |
| $\mathbf{B}_2$ | = | $\frac{1}{2}\,\mathbf{a}_1 + \frac{1}{2}\,\mathbf{a}_2 + \frac{1}{2}\,\mathbf{a}_3$ | = | $\frac{1}{2}a\,\hat{\mathbf{x}} + \frac{1}{2}a\,\hat{\mathbf{y}} + \frac{1}{2}a\,\hat{\mathbf{z}}$ | $(1b)$ | $NO_3$ |

**Geometry files:**

- CIF: pp. 1807
- POSCAR: pp. 1807

# Dodecatungstophosphoric Acid Hexahydrate [$H_3PW_{12}O_{40}{\cdot}6H_2O$] Structure: A27B52CD12_cP184_224_dl_eh3k_a_k

http://aflow.org/prototype-encyclopedia/A27B52CD12_cP184_224_dl_eh3k_a_k

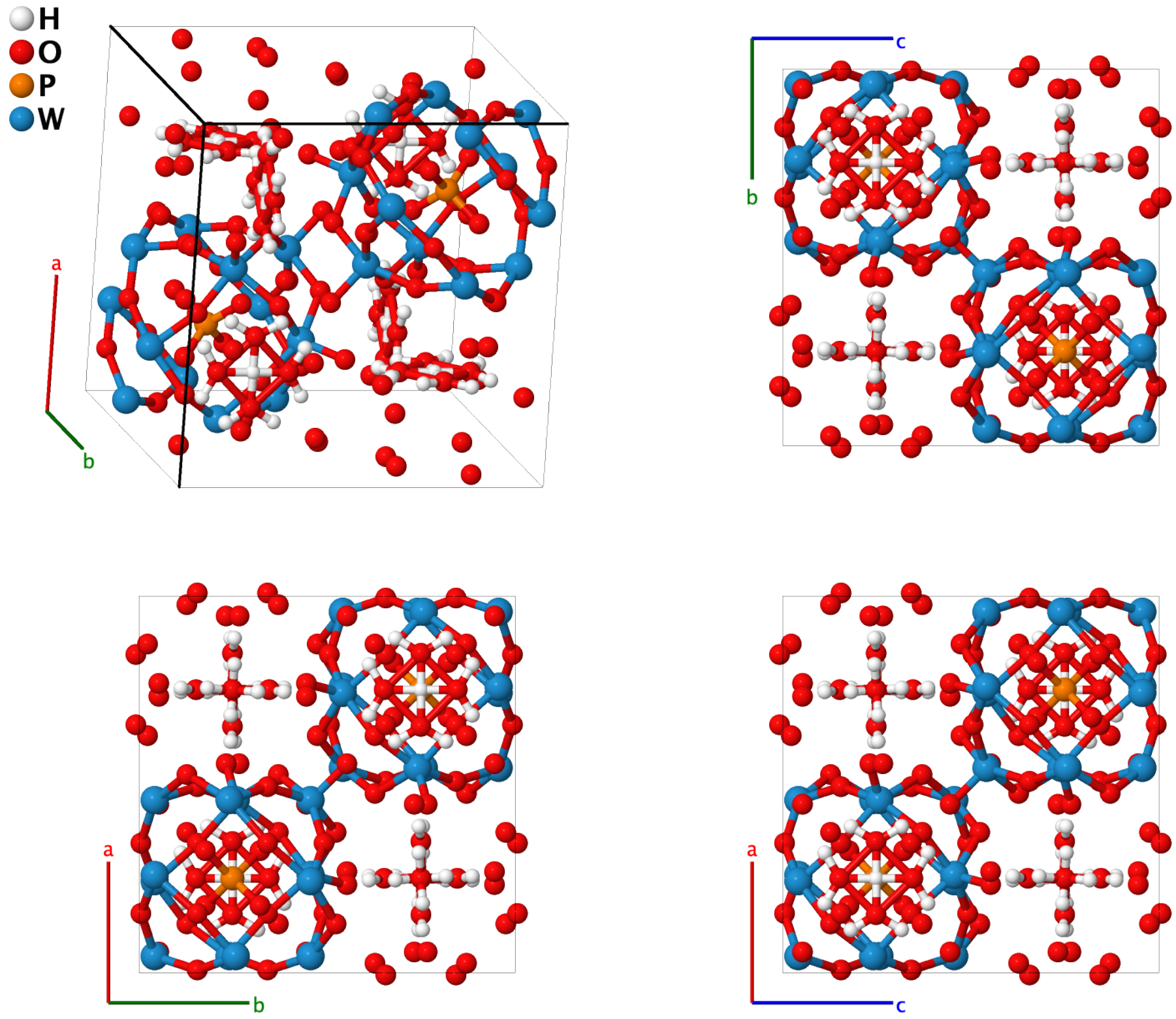

| | | |
|---|---|---|
| **Prototype** | : | $H_{15}O_{46}PW_{12}$ |
| **AFLOW prototype label** | : | A27B52CD12_cP184_224_dl_eh3k_a_k |
| ***Strukturbericht* designation** | : | None |
| **Pearson symbol** | : | cP184 |
| **Space group number** | : | 224 |
| **Space group symbol** | : | $Pn\bar{3}m$ |
| **AFLOW prototype command** | : | `aflow --proto=A27B52CD12_cP184_224_dl_eh3k_a_k --params=`$a, x_3, x_4, x_5, z_5, x_6, z_6, x_7, z_7, x_8, z_8, x_9, y_9, z_9$ |

- (Brown, 1977) presents this as an improvement on the $H4_{16}$ structure, $H_3PW_{12}O_{40}{\cdot}5H_2O$. The primary difference is the addition of a sixth water molecule and the location of the hydrogen molecules not directly attached to a water molecule.
- The water molecules are formed by the H-II and O-II atoms, and the (24$h$) (O-II) and (48$k$) (H-II) Wyckoff sites are only occupied half of the time. Presumably this means that the nearly flat H-O molecular ions in this structure actually consist of one water molecule, the central hydrogen atom (H-I), and a water molecule on the other side of the central hydrogen, with the other water positions empty. Exactly which water molecules are occupied on around each H-I atom is completely up to chance. (Brown, 1977) state that the molecule has a positive charge, and write it as $H_5O_2{}^+$.
- We use the neutron data from (Brown, 1977) to locate the non-hydrogen atoms.
- This structure is a partially dehydrated form of $H_3PW_{12}O_{40}{\cdot}29H_2O$ ($H4_{21}$). Further dehydration produces the $H_3PW_{12}O_{40}{\cdot}3H_2O$ structure.

**Simple Cubic primitive vectors:**

$$\mathbf{a}_1 = a\,\hat{\mathbf{x}}$$
$$\mathbf{a}_2 = a\,\hat{\mathbf{y}}$$
$$\mathbf{a}_3 = a\,\hat{\mathbf{z}}$$

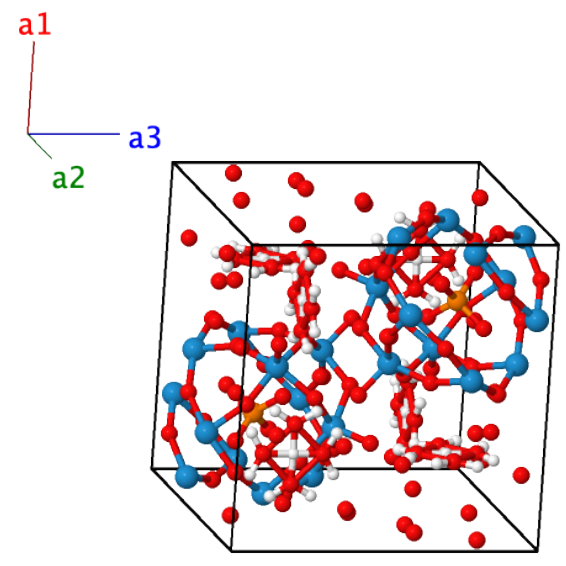

**Basis vectors:**

| | | Lattice Coordinates | | Cartesian Coordinates | Wyckoff Position | Atom Type |
|---|---|---|---|---|---|---|
| $\mathbf{B}_1$ | = | $\frac{1}{4}\,\mathbf{a}_1 + \frac{1}{4}\,\mathbf{a}_2 + \frac{1}{4}\,\mathbf{a}_3$ | = | $\frac{1}{4}a\,\hat{\mathbf{x}} + \frac{1}{4}a\,\hat{\mathbf{y}} + \frac{1}{4}a\,\hat{\mathbf{z}}$ | (2$a$) | P |
| $\mathbf{B}_2$ | = | $\frac{3}{4}\,\mathbf{a}_1 + \frac{3}{4}\,\mathbf{a}_2 + \frac{3}{4}\,\mathbf{a}_3$ | = | $\frac{3}{4}a\,\hat{\mathbf{x}} + \frac{3}{4}a\,\hat{\mathbf{y}} + \frac{3}{4}a\,\hat{\mathbf{z}}$ | (2$a$) | P |
| $\mathbf{B}_3$ | = | $\frac{1}{4}\,\mathbf{a}_1 + \frac{3}{4}\,\mathbf{a}_2 + \frac{3}{4}\,\mathbf{a}_3$ | = | $\frac{1}{4}a\,\hat{\mathbf{x}} + \frac{3}{4}a\,\hat{\mathbf{y}} + \frac{3}{4}a\,\hat{\mathbf{z}}$ | (6$d$) | H I |
| $\mathbf{B}_4$ | = | $\frac{3}{4}\,\mathbf{a}_1 + \frac{1}{4}\,\mathbf{a}_2 + \frac{3}{4}\,\mathbf{a}_3$ | = | $\frac{3}{4}a\,\hat{\mathbf{x}} + \frac{1}{4}a\,\hat{\mathbf{y}} + \frac{3}{4}a\,\hat{\mathbf{z}}$ | (6$d$) | H I |
| $\mathbf{B}_5$ | = | $\frac{3}{4}\,\mathbf{a}_1 + \frac{3}{4}\,\mathbf{a}_2 + \frac{1}{4}\,\mathbf{a}_3$ | = | $\frac{3}{4}a\,\hat{\mathbf{x}} + \frac{3}{4}a\,\hat{\mathbf{y}} + \frac{1}{4}a\,\hat{\mathbf{z}}$ | (6$d$) | H I |
| $\mathbf{B}_6$ | = | $\frac{1}{4}\,\mathbf{a}_1 + \frac{3}{4}\,\mathbf{a}_2 + \frac{1}{4}\,\mathbf{a}_3$ | = | $\frac{1}{4}a\,\hat{\mathbf{x}} + \frac{3}{4}a\,\hat{\mathbf{y}} + \frac{1}{4}a\,\hat{\mathbf{z}}$ | (6$d$) | H I |
| $\mathbf{B}_7$ | = | $\frac{3}{4}\,\mathbf{a}_1 + \frac{1}{4}\,\mathbf{a}_2 + \frac{1}{4}\,\mathbf{a}_3$ | = | $\frac{3}{4}a\,\hat{\mathbf{x}} + \frac{1}{4}a\,\hat{\mathbf{y}} + \frac{1}{4}a\,\hat{\mathbf{z}}$ | (6$d$) | H I |
| $\mathbf{B}_8$ | = | $\frac{1}{4}\,\mathbf{a}_1 + \frac{1}{4}\,\mathbf{a}_2 + \frac{3}{4}\,\mathbf{a}_3$ | = | $\frac{1}{4}a\,\hat{\mathbf{x}} + \frac{1}{4}a\,\hat{\mathbf{y}} + \frac{3}{4}a\,\hat{\mathbf{z}}$ | (6$d$) | H I |
| $\mathbf{B}_9$ | = | $x_3\,\mathbf{a}_1 + x_3\,\mathbf{a}_2 + x_3\,\mathbf{a}_3$ | = | $x_3a\,\hat{\mathbf{x}} + x_3a\,\hat{\mathbf{y}} + x_3a\,\hat{\mathbf{z}}$ | (8$e$) | O I |
| $\mathbf{B}_{10}$ | = | $\left(\frac{1}{2} - x_3\right)\mathbf{a}_1 + \left(\frac{1}{2} - x_3\right)\mathbf{a}_2 + x_3\,\mathbf{a}_3$ | = | $\left(\frac{1}{2} - x_3\right)a\,\hat{\mathbf{x}} + \left(\frac{1}{2} - x_3\right)a\,\hat{\mathbf{y}} + x_3a\,\hat{\mathbf{z}}$ | (8$e$) | O I |
| $\mathbf{B}_{11}$ | = | $\left(\frac{1}{2} - x_3\right)\mathbf{a}_1 + x_3\,\mathbf{a}_2 + \left(\frac{1}{2} - x_3\right)\mathbf{a}_3$ | = | $\left(\frac{1}{2} - x_3\right)a\,\hat{\mathbf{x}} + x_3a\,\hat{\mathbf{y}} + \left(\frac{1}{2} - x_3\right)a\,\hat{\mathbf{z}}$ | (8$e$) | O I |
| $\mathbf{B}_{12}$ | = | $x_3\,\mathbf{a}_1 + \left(\frac{1}{2} - x_3\right)\mathbf{a}_2 + \left(\frac{1}{2} - x_3\right)\mathbf{a}_3$ | = | $x_3a\,\hat{\mathbf{x}} + \left(\frac{1}{2} - x_3\right)a\,\hat{\mathbf{y}} + \left(\frac{1}{2} - x_3\right)a\,\hat{\mathbf{z}}$ | (8$e$) | O I |
| $\mathbf{B}_{13}$ | = | $\left(\frac{1}{2} + x_3\right)\mathbf{a}_1 + \left(\frac{1}{2} + x_3\right)\mathbf{a}_2 - x_3\,\mathbf{a}_3$ | = | $\left(\frac{1}{2} + x_3\right)a\,\hat{\mathbf{x}} + \left(\frac{1}{2} + x_3\right)a\,\hat{\mathbf{y}} - x_3a\,\hat{\mathbf{z}}$ | (8$e$) | O I |
| $\mathbf{B}_{14}$ | = | $-x_3\,\mathbf{a}_1 - x_3\,\mathbf{a}_2 - x_3\,\mathbf{a}_3$ | = | $-x_3a\,\hat{\mathbf{x}} - x_3a\,\hat{\mathbf{y}} - x_3a\,\hat{\mathbf{z}}$ | (8$e$) | O I |
| $\mathbf{B}_{15}$ | = | $\left(\frac{1}{2} + x_3\right)\mathbf{a}_1 - x_3\,\mathbf{a}_2 + \left(\frac{1}{2} + x_3\right)\mathbf{a}_3$ | = | $\left(\frac{1}{2} + x_3\right)a\,\hat{\mathbf{x}} - x_3a\,\hat{\mathbf{y}} + \left(\frac{1}{2} + x_3\right)a\,\hat{\mathbf{z}}$ | (8$e$) | O I |
| $\mathbf{B}_{16}$ | = | $-x_3\,\mathbf{a}_1 + \left(\frac{1}{2} + x_3\right)\mathbf{a}_2 + \left(\frac{1}{2} + x_3\right)\mathbf{a}_3$ | = | $-x_3a\,\hat{\mathbf{x}} + \left(\frac{1}{2} + x_3\right)a\,\hat{\mathbf{y}} + \left(\frac{1}{2} + x_3\right)a\,\hat{\mathbf{z}}$ | (8$e$) | O I |
| $\mathbf{B}_{17}$ | = | $x_4\,\mathbf{a}_1 + \frac{1}{4}\,\mathbf{a}_2 + \frac{3}{4}\,\mathbf{a}_3$ | = | $x_4a\,\hat{\mathbf{x}} + \frac{1}{4}a\,\hat{\mathbf{y}} + \frac{3}{4}a\,\hat{\mathbf{z}}$ | (24$h$) | O II |
| $\mathbf{B}_{18}$ | = | $\left(\frac{1}{2} - x_4\right)\mathbf{a}_1 + \frac{1}{4}\,\mathbf{a}_2 + \frac{3}{4}\,\mathbf{a}_3$ | = | $\left(\frac{1}{2} - x_4\right)a\,\hat{\mathbf{x}} + \frac{1}{4}a\,\hat{\mathbf{y}} + \frac{3}{4}a\,\hat{\mathbf{z}}$ | (24$h$) | O II |
| $\mathbf{B}_{19}$ | = | $\frac{3}{4}\,\mathbf{a}_1 + x_4\,\mathbf{a}_2 + \frac{1}{4}\,\mathbf{a}_3$ | = | $\frac{3}{4}a\,\hat{\mathbf{x}} + x_4a\,\hat{\mathbf{y}} + \frac{1}{4}a\,\hat{\mathbf{z}}$ | (24$h$) | O II |
| $\mathbf{B}_{20}$ | = | $\frac{3}{4}\,\mathbf{a}_1 + \left(\frac{1}{2} - x_4\right)\mathbf{a}_2 + \frac{1}{4}\,\mathbf{a}_3$ | = | $\frac{3}{4}a\,\hat{\mathbf{x}} + \left(\frac{1}{2} - x_4\right)a\,\hat{\mathbf{y}} + \frac{1}{4}a\,\hat{\mathbf{z}}$ | (24$h$) | O II |
| $\mathbf{B}_{21}$ | = | $\frac{1}{4}\,\mathbf{a}_1 + \frac{3}{4}\,\mathbf{a}_2 + x_4\,\mathbf{a}_3$ | = | $\frac{1}{4}a\,\hat{\mathbf{x}} + \frac{3}{4}a\,\hat{\mathbf{y}} + x_4a\,\hat{\mathbf{z}}$ | (24$h$) | O II |
| $\mathbf{B}_{22}$ | = | $\frac{1}{4}\,\mathbf{a}_1 + \frac{3}{4}\,\mathbf{a}_2 + \left(\frac{1}{2} - x_4\right)\mathbf{a}_3$ | = | $\frac{1}{4}a\,\hat{\mathbf{x}} + \frac{3}{4}a\,\hat{\mathbf{y}} + \left(\frac{1}{2} - x_4\right)a\,\hat{\mathbf{z}}$ | (24$h$) | O II |
| $\mathbf{B}_{23}$ | = | $\frac{3}{4}\,\mathbf{a}_1 + \left(\frac{1}{2} + x_4\right)\mathbf{a}_2 + \frac{1}{4}\,\mathbf{a}_3$ | = | $\frac{3}{4}a\,\hat{\mathbf{x}} + \left(\frac{1}{2} + x_4\right)a\,\hat{\mathbf{y}} + \frac{1}{4}a\,\hat{\mathbf{z}}$ | (24$h$) | O II |
| $\mathbf{B}_{24}$ | = | $\frac{3}{4}\,\mathbf{a}_1 - x_4\,\mathbf{a}_2 + \frac{1}{4}\,\mathbf{a}_3$ | = | $\frac{3}{4}a\,\hat{\mathbf{x}} - x_4a\,\hat{\mathbf{y}} + \frac{1}{4}a\,\hat{\mathbf{z}}$ | (24$h$) | O II |
| $\mathbf{B}_{25}$ | = | $\left(\frac{1}{2} + x_4\right)\mathbf{a}_1 + \frac{1}{4}\,\mathbf{a}_2 + \frac{3}{4}\,\mathbf{a}_3$ | = | $\left(\frac{1}{2} + x_4\right)a\,\hat{\mathbf{x}} + \frac{1}{4}a\,\hat{\mathbf{y}} + \frac{3}{4}a\,\hat{\mathbf{z}}$ | (24$h$) | O II |
| $\mathbf{B}_{26}$ | = | $-x_4\,\mathbf{a}_1 + \frac{1}{4}\,\mathbf{a}_2 + \frac{3}{4}\,\mathbf{a}_3$ | = | $-x_4a\,\hat{\mathbf{x}} + \frac{1}{4}a\,\hat{\mathbf{y}} + \frac{3}{4}a\,\hat{\mathbf{z}}$ | (24$h$) | O II |
| $\mathbf{B}_{27}$ | = | $\frac{1}{4}\,\mathbf{a}_1 + \frac{3}{4}\,\mathbf{a}_2 - x_4\,\mathbf{a}_3$ | = | $\frac{1}{4}a\,\hat{\mathbf{x}} + \frac{3}{4}a\,\hat{\mathbf{y}} - x_4a\,\hat{\mathbf{z}}$ | (24$h$) | O II |
| $\mathbf{B}_{28}$ | = | $\frac{1}{4}\,\mathbf{a}_1 + \frac{3}{4}\,\mathbf{a}_2 + \left(\frac{1}{2} + x_4\right)\,\mathbf{a}_3$ | = | $\frac{1}{4}a\,\hat{\mathbf{x}} + \frac{3}{4}a\,\hat{\mathbf{y}} + \left(\frac{1}{2} + x_4\right)a\,\hat{\mathbf{z}}$ | (24$h$) | O II |
| $\mathbf{B}_{29}$ | = | $-x_4\,\mathbf{a}_1 + \frac{3}{4}\,\mathbf{a}_2 + \frac{1}{4}\,\mathbf{a}_3$ | = | $-x_4a\,\hat{\mathbf{x}} + \frac{3}{4}a\,\hat{\mathbf{y}} + \frac{1}{4}a\,\hat{\mathbf{z}}$ | (24$h$) | O II |
| $\mathbf{B}_{30}$ | = | $\left(\frac{1}{2} + x_4\right)\,\mathbf{a}_1 + \frac{3}{4}\,\mathbf{a}_2 + \frac{1}{4}\,\mathbf{a}_3$ | = | $\left(\frac{1}{2} + x_4\right)a\,\hat{\mathbf{x}} + \frac{3}{4}a\,\hat{\mathbf{y}} + \frac{1}{4}a\,\hat{\mathbf{z}}$ | (24$h$) | O II |
| $\mathbf{B}_{31}$ | = | $\frac{1}{4}\,\mathbf{a}_1 - x_4\,\mathbf{a}_2 + \frac{3}{4}\,\mathbf{a}_3$ | = | $\frac{1}{4}a\,\hat{\mathbf{x}} - x_4a\,\hat{\mathbf{y}} + \frac{3}{4}a\,\hat{\mathbf{z}}$ | (24$h$) | O II |
| $\mathbf{B}_{32}$ | = | $\frac{1}{4}\,\mathbf{a}_1 + \left(\frac{1}{2} + x_4\right)\,\mathbf{a}_2 + \frac{3}{4}\,\mathbf{a}_3$ | = | $\frac{1}{4}a\,\hat{\mathbf{x}} + \left(\frac{1}{2} + x_4\right)a\,\hat{\mathbf{y}} + \frac{3}{4}a\,\hat{\mathbf{z}}$ | (24$h$) | O II |
| $\mathbf{B}_{33}$ | = | $\frac{3}{4}\,\mathbf{a}_1 + \frac{1}{4}\,\mathbf{a}_2 - x_4\,\mathbf{a}_3$ | = | $\frac{3}{4}a\,\hat{\mathbf{x}} + \frac{1}{4}a\,\hat{\mathbf{y}} - x_4a\,\hat{\mathbf{z}}$ | (24$h$) | O II |
| $\mathbf{B}_{34}$ | = | $\frac{3}{4}\,\mathbf{a}_1 + \frac{1}{4}\,\mathbf{a}_2 + \left(\frac{1}{2} + x_4\right)\,\mathbf{a}_3$ | = | $\frac{3}{4}a\,\hat{\mathbf{x}} + \frac{1}{4}a\,\hat{\mathbf{y}} + \left(\frac{1}{2} + x_4\right)a\,\hat{\mathbf{z}}$ | (24$h$) | O II |
| $\mathbf{B}_{35}$ | = | $\frac{1}{4}\,\mathbf{a}_1 + \left(\frac{1}{2} - x_4\right)\,\mathbf{a}_2 + \frac{3}{4}\,\mathbf{a}_3$ | = | $\frac{1}{4}a\,\hat{\mathbf{x}} + \left(\frac{1}{2} - x_4\right)a\,\hat{\mathbf{y}} + \frac{3}{4}a\,\hat{\mathbf{z}}$ | (24$h$) | O II |
| $\mathbf{B}_{36}$ | = | $\frac{1}{4}\,\mathbf{a}_1 + x_4\,\mathbf{a}_2 + \frac{3}{4}\,\mathbf{a}_3$ | = | $\frac{1}{4}a\,\hat{\mathbf{x}} + x_4a\,\hat{\mathbf{y}} + \frac{3}{4}a\,\hat{\mathbf{z}}$ | (24$h$) | O II |
| $\mathbf{B}_{37}$ | = | $\left(\frac{1}{2} - x_4\right)\,\mathbf{a}_1 + \frac{3}{4}\,\mathbf{a}_2 + \frac{1}{4}\,\mathbf{a}_3$ | = | $\left(\frac{1}{2} - x_4\right)a\,\hat{\mathbf{x}} + \frac{3}{4}a\,\hat{\mathbf{y}} + \frac{1}{4}a\,\hat{\mathbf{z}}$ | (24$h$) | O II |
| $\mathbf{B}_{38}$ | = | $x_4\,\mathbf{a}_1 + \frac{3}{4}\,\mathbf{a}_2 + \frac{1}{4}\,\mathbf{a}_3$ | = | $x_4a\,\hat{\mathbf{x}} + \frac{3}{4}a\,\hat{\mathbf{y}} + \frac{1}{4}a\,\hat{\mathbf{z}}$ | (24$h$) | O II |
| $\mathbf{B}_{39}$ | = | $\frac{3}{4}\,\mathbf{a}_1 + \frac{1}{4}\,\mathbf{a}_2 + x_4\,\mathbf{a}_3$ | = | $\frac{3}{4}a\,\hat{\mathbf{x}} + \frac{1}{4}a\,\hat{\mathbf{y}} + x_4a\,\hat{\mathbf{z}}$ | (24$h$) | O II |
| $\mathbf{B}_{40}$ | = | $\frac{3}{4}\,\mathbf{a}_1 + \frac{1}{4}\,\mathbf{a}_2 + \left(\frac{1}{2} - x_4\right)\,\mathbf{a}_3$ | = | $\frac{3}{4}a\,\hat{\mathbf{x}} + \frac{1}{4}a\,\hat{\mathbf{y}} + \left(\frac{1}{2} - x_4\right)a\,\hat{\mathbf{z}}$ | (24$h$) | O II |
| $\mathbf{B}_{41}$ | = | $x_5\,\mathbf{a}_1 + x_5\,\mathbf{a}_2 + z_5\,\mathbf{a}_3$ | = | $x_5a\,\hat{\mathbf{x}} + x_5a\,\hat{\mathbf{y}} + z_5a\,\hat{\mathbf{z}}$ | (24$k$) | O III |
| $\mathbf{B}_{42}$ | = | $\left(\frac{1}{2} - x_5\right)\,\mathbf{a}_1 + \left(\frac{1}{2} - x_5\right)\,\mathbf{a}_2 + z_5\,\mathbf{a}_3$ | = | $\left(\frac{1}{2} - x_5\right)a\,\hat{\mathbf{x}} + \left(\frac{1}{2} - x_5\right)a\,\hat{\mathbf{y}} + z_5a\,\hat{\mathbf{z}}$ | (24$k$) | O III |
| $\mathbf{B}_{43}$ | = | $\left(\frac{1}{2} - x_5\right)\,\mathbf{a}_1 + x_5\,\mathbf{a}_2 + \left(\frac{1}{2} - z_5\right)\,\mathbf{a}_3$ | = | $\left(\frac{1}{2} - x_5\right)a\,\hat{\mathbf{x}} + x_5a\,\hat{\mathbf{y}} + \left(\frac{1}{2} - z_5\right)a\,\hat{\mathbf{z}}$ | (24$k$) | O III |
| $\mathbf{B}_{44}$ | = | $x_5\,\mathbf{a}_1 + \left(\frac{1}{2} - x_5\right)\,\mathbf{a}_2 + \left(\frac{1}{2} - z_5\right)\,\mathbf{a}_3$ | = | $x_5a\,\hat{\mathbf{x}} + \left(\frac{1}{2} - x_5\right)a\,\hat{\mathbf{y}} + \left(\frac{1}{2} - z_5\right)a\,\hat{\mathbf{z}}$ | (24$k$) | O III |
| $\mathbf{B}_{45}$ | = | $z_5\,\mathbf{a}_1 + x_5\,\mathbf{a}_2 + x_5\,\mathbf{a}_3$ | = | $z_5a\,\hat{\mathbf{x}} + x_5a\,\hat{\mathbf{y}} + x_5a\,\hat{\mathbf{z}}$ | (24$k$) | O III |
| $\mathbf{B}_{46}$ | = | $z_5\,\mathbf{a}_1 + \left(\frac{1}{2} - x_5\right)\,\mathbf{a}_2 + \left(\frac{1}{2} - x_5\right)\,\mathbf{a}_3$ | = | $z_5a\,\hat{\mathbf{x}} + \left(\frac{1}{2} - x_5\right)a\,\hat{\mathbf{y}} + \left(\frac{1}{2} - x_5\right)a\,\hat{\mathbf{z}}$ | (24$k$) | O III |
| $\mathbf{B}_{47}$ | = | $\left(\frac{1}{2} - z_5\right)\,\mathbf{a}_1 + \left(\frac{1}{2} - x_5\right)\,\mathbf{a}_2 + x_5\,\mathbf{a}_3$ | = | $\left(\frac{1}{2} - z_5\right)a\,\hat{\mathbf{x}} + \left(\frac{1}{2} - x_5\right)a\,\hat{\mathbf{y}} + x_5a\,\hat{\mathbf{z}}$ | (24$k$) | O III |
| $\mathbf{B}_{48}$ | = | $\left(\frac{1}{2} - z_5\right)\,\mathbf{a}_1 + x_5\,\mathbf{a}_2 + \left(\frac{1}{2} - x_5\right)\,\mathbf{a}_3$ | = | $\left(\frac{1}{2} - z_5\right)a\,\hat{\mathbf{x}} + x_5a\,\hat{\mathbf{y}} + \left(\frac{1}{2} - x_5\right)a\,\hat{\mathbf{z}}$ | (24$k$) | O III |
| $\mathbf{B}_{49}$ | = | $x_5\,\mathbf{a}_1 + z_5\,\mathbf{a}_2 + x_5\,\mathbf{a}_3$ | = | $x_5a\,\hat{\mathbf{x}} + z_5a\,\hat{\mathbf{y}} + x_5a\,\hat{\mathbf{z}}$ | (24$k$) | O III |
| $\mathbf{B}_{50}$ | = | $\left(\frac{1}{2} - x_5\right)\,\mathbf{a}_1 + z_5\,\mathbf{a}_2 + \left(\frac{1}{2} - x_5\right)\,\mathbf{a}_3$ | = | $\left(\frac{1}{2} - x_5\right)a\,\hat{\mathbf{x}} + z_5a\,\hat{\mathbf{y}} + \left(\frac{1}{2} - x_5\right)a\,\hat{\mathbf{z}}$ | (24$k$) | O III |
| $\mathbf{B}_{51}$ | = | $x_5\,\mathbf{a}_1 + \left(\frac{1}{2} - z_5\right)\,\mathbf{a}_2 + \left(\frac{1}{2} - x_5\right)\,\mathbf{a}_3$ | = | $x_5a\,\hat{\mathbf{x}} + \left(\frac{1}{2} - z_5\right)a\,\hat{\mathbf{y}} + \left(\frac{1}{2} - x_5\right)a\,\hat{\mathbf{z}}$ | (24$k$) | O III |
| $\mathbf{B}_{52}$ | = | $\left(\frac{1}{2} - x_5\right)\,\mathbf{a}_1 + \left(\frac{1}{2} - z_5\right)\,\mathbf{a}_2 + x_5\,\mathbf{a}_3$ | = | $\left(\frac{1}{2} - x_5\right)a\,\hat{\mathbf{x}} + \left(\frac{1}{2} - z_5\right)a\,\hat{\mathbf{y}} + x_5a\,\hat{\mathbf{z}}$ | (24$k$) | O III |
| $\mathbf{B}_{53}$ | = | $\left(\frac{1}{2} + x_5\right)\,\mathbf{a}_1 + \left(\frac{1}{2} + x_5\right)\,\mathbf{a}_2 - z_5\,\mathbf{a}_3$ | = | $\left(\frac{1}{2} + x_5\right)a\,\hat{\mathbf{x}} + \left(\frac{1}{2} + x_5\right)a\,\hat{\mathbf{y}} - z_5a\,\hat{\mathbf{z}}$ | (24$k$) | O III |
| $\mathbf{B}_{54}$ | = | $-x_5\,\mathbf{a}_1 - x_5\,\mathbf{a}_2 - z_5\,\mathbf{a}_3$ | = | $-x_5a\,\hat{\mathbf{x}} - x_5a\,\hat{\mathbf{y}} - z_5a\,\hat{\mathbf{z}}$ | (24$k$) | O III |
| $\mathbf{B}_{55}$ | = | $\left(\frac{1}{2} + x_5\right)\,\mathbf{a}_1 - x_5\,\mathbf{a}_2 + \left(\frac{1}{2} + z_5\right)\,\mathbf{a}_3$ | = | $\left(\frac{1}{2} + x_5\right)a\,\hat{\mathbf{x}} - x_5a\,\hat{\mathbf{y}} + \left(\frac{1}{2} + z_5\right)a\,\hat{\mathbf{z}}$ | (24$k$) | O III |
| $\mathbf{B}_{56}$ | = | $-x_5\,\mathbf{a}_1 + \left(\frac{1}{2} + x_5\right)\,\mathbf{a}_2 + \left(\frac{1}{2} + z_5\right)\,\mathbf{a}_3$ | = | $-x_5a\,\hat{\mathbf{x}} + \left(\frac{1}{2} + x_5\right)a\,\hat{\mathbf{y}} + \left(\frac{1}{2} + z_5\right)a\,\hat{\mathbf{z}}$ | (24$k$) | O III |
| $\mathbf{B}_{57}$ | = | $\left(\frac{1}{2} + x_5\right)\,\mathbf{a}_1 + \left(\frac{1}{2} + z_5\right)\,\mathbf{a}_2 - x_5\,\mathbf{a}_3$ | = | $\left(\frac{1}{2} + x_5\right)a\,\hat{\mathbf{x}} + \left(\frac{1}{2} + z_5\right)a\,\hat{\mathbf{y}} - x_5a\,\hat{\mathbf{z}}$ | (24$k$) | O III |
| $\mathbf{B}_{58}$ | = | $-x_5\,\mathbf{a}_1 + \left(\frac{1}{2} + z_5\right)\,\mathbf{a}_2 + \left(\frac{1}{2} + x_5\right)\,\mathbf{a}_3$ | = | $-x_5a\,\hat{\mathbf{x}} + \left(\frac{1}{2} + z_5\right)a\,\hat{\mathbf{y}} + \left(\frac{1}{2} + x_5\right)a\,\hat{\mathbf{z}}$ | (24$k$) | O III |
| $\mathbf{B}_{59}$ | = | $-x_5\,\mathbf{a}_1 - z_5\,\mathbf{a}_2 - x_5\,\mathbf{a}_3$ | = | $-x_5a\,\hat{\mathbf{x}} - z_5a\,\hat{\mathbf{y}} - x_5a\,\hat{\mathbf{z}}$ | (24$k$) | O III |
| $\mathbf{B}_{60}$ | = | $\left(\frac{1}{2} + x_5\right)\,\mathbf{a}_1 - z_5\,\mathbf{a}_2 + \left(\frac{1}{2} + x_5\right)\,\mathbf{a}_3$ | = | $\left(\frac{1}{2} + x_5\right)a\,\hat{\mathbf{x}} - z_5a\,\hat{\mathbf{y}} + \left(\frac{1}{2} + x_5\right)a\,\hat{\mathbf{z}}$ | (24$k$) | O III |
| $\mathbf{B}_{61}$ | = | $\left(\frac{1}{2} + z_5\right)\,\mathbf{a}_1 + \left(\frac{1}{2} + x_5\right)\,\mathbf{a}_2 - x_5\,\mathbf{a}_3$ | = | $\left(\frac{1}{2} + z_5\right)a\,\hat{\mathbf{x}} + \left(\frac{1}{2} + x_5\right)a\,\hat{\mathbf{y}} - x_5a\,\hat{\mathbf{z}}$ | (24$k$) | O III |
| $\mathbf{B}_{62}$ | = | $\left(\frac{1}{2} + z_5\right)\,\mathbf{a}_1 - x_5\,\mathbf{a}_2 + \left(\frac{1}{2} + x_5\right)\,\mathbf{a}_3$ | = | $\left(\frac{1}{2} + z_5\right)a\,\hat{\mathbf{x}} - x_5a\,\hat{\mathbf{y}} + \left(\frac{1}{2} + x_5\right)a\,\hat{\mathbf{z}}$ | (24$k$) | O III |
| $\mathbf{B}_{63}$ | = | $-z_5\,\mathbf{a}_1 + \left(\frac{1}{2} + x_5\right)\,\mathbf{a}_2 + \left(\frac{1}{2} + x_5\right)\,\mathbf{a}_3$ | = | $-z_5a\,\hat{\mathbf{x}} + \left(\frac{1}{2} + x_5\right)a\,\hat{\mathbf{y}} + \left(\frac{1}{2} + x_5\right)a\,\hat{\mathbf{z}}$ | (24$k$) | O III |
| $\mathbf{B}_{64}$ | = | $-z_5\,\mathbf{a}_1 - x_5\,\mathbf{a}_2 - x_5\,\mathbf{a}_3$ | = | $-z_5 a\,\hat{\mathbf{x}} - x_5 a\,\hat{\mathbf{y}} - x_5 a\,\hat{\mathbf{z}}$ | (24$k$) | O III |
| $\mathbf{B}_{65}$ | = | $x_6\,\mathbf{a}_1 + x_6\,\mathbf{a}_2 + z_6\,\mathbf{a}_3$ | = | $x_6 a\,\hat{\mathbf{x}} + x_6 a\,\hat{\mathbf{y}} + z_6 a\,\hat{\mathbf{z}}$ | (24$k$) | O IV |
| $\mathbf{B}_{66}$ | = | $\left(\frac{1}{2} - x_6\right)\mathbf{a}_1 + \left(\frac{1}{2} - x_6\right)\mathbf{a}_2 + z_6\,\mathbf{a}_3$ | = | $\left(\frac{1}{2} - x_6\right)a\,\hat{\mathbf{x}} + \left(\frac{1}{2} - x_6\right)a\,\hat{\mathbf{y}} + z_6 a\,\hat{\mathbf{z}}$ | (24$k$) | O IV |
| $\mathbf{B}_{67}$ | = | $\left(\frac{1}{2} - x_6\right)\mathbf{a}_1 + x_6\,\mathbf{a}_2 + \left(\frac{1}{2} - z_6\right)\mathbf{a}_3$ | = | $\left(\frac{1}{2} - x_6\right)a\,\hat{\mathbf{x}} + x_6 a\,\hat{\mathbf{y}} + \left(\frac{1}{2} - z_6\right)a\,\hat{\mathbf{z}}$ | (24$k$) | O IV |
| $\mathbf{B}_{68}$ | = | $x_6\,\mathbf{a}_1 + \left(\frac{1}{2} - x_6\right)\mathbf{a}_2 + \left(\frac{1}{2} - z_6\right)\mathbf{a}_3$ | = | $x_6 a\,\hat{\mathbf{x}} + \left(\frac{1}{2} - x_6\right)a\,\hat{\mathbf{y}} + \left(\frac{1}{2} - z_6\right)a\,\hat{\mathbf{z}}$ | (24$k$) | O IV |
| $\mathbf{B}_{69}$ | = | $z_6\,\mathbf{a}_1 + x_6\,\mathbf{a}_2 + x_6\,\mathbf{a}_3$ | = | $z_6 a\,\hat{\mathbf{x}} + x_6 a\,\hat{\mathbf{y}} + x_6 a\,\hat{\mathbf{z}}$ | (24$k$) | O IV |
| $\mathbf{B}_{70}$ | = | $z_6\,\mathbf{a}_1 + \left(\frac{1}{2} - x_6\right)\mathbf{a}_2 + \left(\frac{1}{2} - x_6\right)\mathbf{a}_3$ | = | $z_6 a\,\hat{\mathbf{x}} + \left(\frac{1}{2} - x_6\right)a\,\hat{\mathbf{y}} + \left(\frac{1}{2} - x_6\right)a\,\hat{\mathbf{z}}$ | (24$k$) | O IV |
| $\mathbf{B}_{71}$ | = | $\left(\frac{1}{2} - z_6\right)\mathbf{a}_1 + \left(\frac{1}{2} - x_6\right)\mathbf{a}_2 + x_6\,\mathbf{a}_3$ | = | $\left(\frac{1}{2} - z_6\right)a\,\hat{\mathbf{x}} + \left(\frac{1}{2} - x_6\right)a\,\hat{\mathbf{y}} + x_6 a\,\hat{\mathbf{z}}$ | (24$k$) | O IV |
| $\mathbf{B}_{72}$ | = | $\left(\frac{1}{2} - z_6\right)\mathbf{a}_1 + x_6\,\mathbf{a}_2 + \left(\frac{1}{2} - x_6\right)\mathbf{a}_3$ | = | $\left(\frac{1}{2} - z_6\right)a\,\hat{\mathbf{x}} + x_6 a\,\hat{\mathbf{y}} + \left(\frac{1}{2} - x_6\right)a\,\hat{\mathbf{z}}$ | (24$k$) | O IV |
| $\mathbf{B}_{73}$ | = | $x_6\,\mathbf{a}_1 + z_6\,\mathbf{a}_2 + x_6\,\mathbf{a}_3$ | = | $x_6 a\,\hat{\mathbf{x}} + z_6 a\,\hat{\mathbf{y}} + x_6 a\,\hat{\mathbf{z}}$ | (24$k$) | O IV |
| $\mathbf{B}_{74}$ | = | $\left(\frac{1}{2} - x_6\right)\mathbf{a}_1 + z_6\,\mathbf{a}_2 + \left(\frac{1}{2} - x_6\right)\mathbf{a}_3$ | = | $\left(\frac{1}{2} - x_6\right)a\,\hat{\mathbf{x}} + z_6 a\,\hat{\mathbf{y}} + \left(\frac{1}{2} - x_6\right)a\,\hat{\mathbf{z}}$ | (24$k$) | O IV |
| $\mathbf{B}_{75}$ | = | $x_6\,\mathbf{a}_1 + \left(\frac{1}{2} - z_6\right)\mathbf{a}_2 + \left(\frac{1}{2} - x_6\right)\mathbf{a}_3$ | = | $x_6 a\,\hat{\mathbf{x}} + \left(\frac{1}{2} - z_6\right)a\,\hat{\mathbf{y}} + \left(\frac{1}{2} - x_6\right)a\,\hat{\mathbf{z}}$ | (24$k$) | O IV |
| $\mathbf{B}_{76}$ | = | $\left(\frac{1}{2} - x_6\right)\mathbf{a}_1 + \left(\frac{1}{2} - z_6\right)\mathbf{a}_2 + x_6\,\mathbf{a}_3$ | = | $\left(\frac{1}{2} - x_6\right)a\,\hat{\mathbf{x}} + \left(\frac{1}{2} - z_6\right)a\,\hat{\mathbf{y}} + x_6 a\,\hat{\mathbf{z}}$ | (24$k$) | O IV |
| $\mathbf{B}_{77}$ | = | $\left(\frac{1}{2} + x_6\right)\mathbf{a}_1 + \left(\frac{1}{2} + x_6\right)\mathbf{a}_2 - z_6\,\mathbf{a}_3$ | = | $\left(\frac{1}{2} + x_6\right)a\,\hat{\mathbf{x}} + \left(\frac{1}{2} + x_6\right)a\,\hat{\mathbf{y}} - z_6 a\,\hat{\mathbf{z}}$ | (24$k$) | O IV |
| $\mathbf{B}_{78}$ | = | $-x_6\,\mathbf{a}_1 - x_6\,\mathbf{a}_2 - z_6\,\mathbf{a}_3$ | = | $-x_6 a\,\hat{\mathbf{x}} - x_6 a\,\hat{\mathbf{y}} - z_6 a\,\hat{\mathbf{z}}$ | (24$k$) | O IV |
| $\mathbf{B}_{79}$ | = | $\left(\frac{1}{2} + x_6\right)\mathbf{a}_1 - x_6\,\mathbf{a}_2 + \left(\frac{1}{2} + z_6\right)\mathbf{a}_3$ | = | $\left(\frac{1}{2} + x_6\right)a\,\hat{\mathbf{x}} - x_6 a\,\hat{\mathbf{y}} + \left(\frac{1}{2} + z_6\right)a\,\hat{\mathbf{z}}$ | (24$k$) | O IV |
| $\mathbf{B}_{80}$ | = | $-x_6\,\mathbf{a}_1 + \left(\frac{1}{2} + x_6\right)\mathbf{a}_2 + \left(\frac{1}{2} + z_6\right)\mathbf{a}_3$ | = | $-x_6 a\,\hat{\mathbf{x}} + \left(\frac{1}{2} + x_6\right)a\,\hat{\mathbf{y}} + \left(\frac{1}{2} + z_6\right)a\,\hat{\mathbf{z}}$ | (24$k$) | O IV |
| $\mathbf{B}_{81}$ | = | $\left(\frac{1}{2} + x_6\right)\mathbf{a}_1 + \left(\frac{1}{2} + z_6\right)\mathbf{a}_2 - x_6\,\mathbf{a}_3$ | = | $\left(\frac{1}{2} + x_6\right)a\,\hat{\mathbf{x}} + \left(\frac{1}{2} + z_6\right)a\,\hat{\mathbf{y}} - x_6 a\,\hat{\mathbf{z}}$ | (24$k$) | O IV |
| $\mathbf{B}_{82}$ | = | $-x_6\,\mathbf{a}_1 + \left(\frac{1}{2} + z_6\right)\mathbf{a}_2 + \left(\frac{1}{2} + x_6\right)\mathbf{a}_3$ | = | $-x_6 a\,\hat{\mathbf{x}} + \left(\frac{1}{2} + z_6\right)a\,\hat{\mathbf{y}} + \left(\frac{1}{2} + x_6\right)a\,\hat{\mathbf{z}}$ | (24$k$) | O IV |
| $\mathbf{B}_{83}$ | = | $-x_6\,\mathbf{a}_1 - z_6\,\mathbf{a}_2 - x_6\,\mathbf{a}_3$ | = | $-x_6 a\,\hat{\mathbf{x}} - z_6 a\,\hat{\mathbf{y}} - x_6 a\,\hat{\mathbf{z}}$ | (24$k$) | O IV |
| $\mathbf{B}_{84}$ | = | $\left(\frac{1}{2} + x_6\right)\mathbf{a}_1 - z_6\,\mathbf{a}_2 + \left(\frac{1}{2} + x_6\right)\mathbf{a}_3$ | = | $\left(\frac{1}{2} + x_6\right)a\,\hat{\mathbf{x}} - z_6 a\,\hat{\mathbf{y}} + \left(\frac{1}{2} + x_6\right)a\,\hat{\mathbf{z}}$ | (24$k$) | O IV |
| $\mathbf{B}_{85}$ | = | $\left(\frac{1}{2} + z_6\right)\mathbf{a}_1 + \left(\frac{1}{2} + x_6\right)\mathbf{a}_2 - x_6\,\mathbf{a}_3$ | = | $\left(\frac{1}{2} + z_6\right)a\,\hat{\mathbf{x}} + \left(\frac{1}{2} + x_6\right)a\,\hat{\mathbf{y}} - x_6 a\,\hat{\mathbf{z}}$ | (24$k$) | O IV |
| $\mathbf{B}_{86}$ | = | $\left(\frac{1}{2} + z_6\right)\mathbf{a}_1 - x_6\,\mathbf{a}_2 + \left(\frac{1}{2} + x_6\right)\mathbf{a}_3$ | = | $\left(\frac{1}{2} + z_6\right)a\,\hat{\mathbf{x}} - x_6 a\,\hat{\mathbf{y}} + \left(\frac{1}{2} + x_6\right)a\,\hat{\mathbf{z}}$ | (24$k$) | O IV |
| $\mathbf{B}_{87}$ | = | $-z_6\,\mathbf{a}_1 + \left(\frac{1}{2} + x_6\right)\mathbf{a}_2 + \left(\frac{1}{2} + x_6\right)\mathbf{a}_3$ | = | $-z_6 a\,\hat{\mathbf{x}} + \left(\frac{1}{2} + x_6\right)a\,\hat{\mathbf{y}} + \left(\frac{1}{2} + x_6\right)a\,\hat{\mathbf{z}}$ | (24$k$) | O IV |
| $\mathbf{B}_{88}$ | = | $-z_6\,\mathbf{a}_1 - x_6\,\mathbf{a}_2 - x_6\,\mathbf{a}_3$ | = | $-z_6 a\,\hat{\mathbf{x}} - x_6 a\,\hat{\mathbf{y}} - x_6 a\,\hat{\mathbf{z}}$ | (24$k$) | O IV |
| $\mathbf{B}_{89}$ | = | $x_7\,\mathbf{a}_1 + x_7\,\mathbf{a}_2 + z_7\,\mathbf{a}_3$ | = | $x_7 a\,\hat{\mathbf{x}} + x_7 a\,\hat{\mathbf{y}} + z_7 a\,\hat{\mathbf{z}}$ | (24$k$) | O V |
| $\mathbf{B}_{90}$ | = | $\left(\frac{1}{2} - x_7\right)\mathbf{a}_1 + \left(\frac{1}{2} - x_7\right)\mathbf{a}_2 + z_7\,\mathbf{a}_3$ | = | $\left(\frac{1}{2} - x_7\right)a\,\hat{\mathbf{x}} + \left(\frac{1}{2} - x_7\right)a\,\hat{\mathbf{y}} + z_7 a\,\hat{\mathbf{z}}$ | (24$k$) | O V |
| $\mathbf{B}_{91}$ | = | $\left(\frac{1}{2} - x_7\right)\mathbf{a}_1 + x_7\,\mathbf{a}_2 + \left(\frac{1}{2} - z_7\right)\mathbf{a}_3$ | = | $\left(\frac{1}{2} - x_7\right)a\,\hat{\mathbf{x}} + x_7 a\,\hat{\mathbf{y}} + \left(\frac{1}{2} - z_7\right)a\,\hat{\mathbf{z}}$ | (24$k$) | O V |
| $\mathbf{B}_{92}$ | = | $x_7\,\mathbf{a}_1 + \left(\frac{1}{2} - x_7\right)\mathbf{a}_2 + \left(\frac{1}{2} - z_7\right)\mathbf{a}_3$ | = | $x_7 a\,\hat{\mathbf{x}} + \left(\frac{1}{2} - x_7\right)a\,\hat{\mathbf{y}} + \left(\frac{1}{2} - z_7\right)a\,\hat{\mathbf{z}}$ | (24$k$) | O V |
| $\mathbf{B}_{93}$ | = | $z_7\,\mathbf{a}_1 + x_7\,\mathbf{a}_2 + x_7\,\mathbf{a}_3$ | = | $z_7 a\,\hat{\mathbf{x}} + x_7 a\,\hat{\mathbf{y}} + x_7 a\,\hat{\mathbf{z}}$ | (24$k$) | O V |
| $\mathbf{B}_{94}$ | = | $z_7\,\mathbf{a}_1 + \left(\frac{1}{2} - x_7\right)\mathbf{a}_2 + \left(\frac{1}{2} - x_7\right)\mathbf{a}_3$ | = | $z_7 a\,\hat{\mathbf{x}} + \left(\frac{1}{2} - x_7\right)a\,\hat{\mathbf{y}} + \left(\frac{1}{2} - x_7\right)a\,\hat{\mathbf{z}}$ | (24$k$) | O V |
| $\mathbf{B}_{95}$ | = | $\left(\frac{1}{2} - z_7\right)\mathbf{a}_1 + \left(\frac{1}{2} - x_7\right)\mathbf{a}_2 + x_7\,\mathbf{a}_3$ | = | $\left(\frac{1}{2} - z_7\right)a\,\hat{\mathbf{x}} + \left(\frac{1}{2} - x_7\right)a\,\hat{\mathbf{y}} + x_7 a\,\hat{\mathbf{z}}$ | (24$k$) | O V |
| $\mathbf{B}_{96}$ | = | $\left(\frac{1}{2} - z_7\right)\mathbf{a}_1 + x_7\,\mathbf{a}_2 + \left(\frac{1}{2} - x_7\right)\mathbf{a}_3$ | = | $\left(\frac{1}{2} - z_7\right)a\,\hat{\mathbf{x}} + x_7 a\,\hat{\mathbf{y}} + \left(\frac{1}{2} - x_7\right)a\,\hat{\mathbf{z}}$ | (24$k$) | O V |
| $\mathbf{B}_{97}$ | = | $x_7\,\mathbf{a}_1 + z_7\,\mathbf{a}_2 + x_7\,\mathbf{a}_3$ | = | $x_7 a\,\hat{\mathbf{x}} + z_7 a\,\hat{\mathbf{y}} + x_7 a\,\hat{\mathbf{z}}$ | (24$k$) | O V |
| $\mathbf{B}_{98}$ | = | $\left(\frac{1}{2} - x_7\right)\mathbf{a}_1 + z_7\,\mathbf{a}_2 + \left(\frac{1}{2} - x_7\right)\mathbf{a}_3$ | = | $\left(\frac{1}{2} - x_7\right)a\,\hat{\mathbf{x}} + z_7 a\,\hat{\mathbf{y}} + \left(\frac{1}{2} - x_7\right)a\,\hat{\mathbf{z}}$ | (24$k$) | O V |
| $\mathbf{B}_{99}$ | = | $x_7\,\mathbf{a}_1 + \left(\frac{1}{2} - z_7\right)\mathbf{a}_2 + \left(\frac{1}{2} - x_7\right)\mathbf{a}_3$ | = | $x_7 a\,\hat{\mathbf{x}} + \left(\frac{1}{2} - z_7\right)a\,\hat{\mathbf{y}} + \left(\frac{1}{2} - x_7\right)a\,\hat{\mathbf{z}}$ | (24$k$) | O V |
| $\mathbf{B}_{100}$ | = | $\left(\frac{1}{2}-x_7\right)\mathbf{a}_1+\left(\frac{1}{2}-z_7\right)\mathbf{a}_2+x_7\,\mathbf{a}_3$ | = | $\left(\frac{1}{2}-x_7\right)a\,\hat{\mathbf{x}}+\left(\frac{1}{2}-z_7\right)a\,\hat{\mathbf{y}}+x_7a\,\hat{\mathbf{z}}$ | (24$k$) | O V |
| $\mathbf{B}_{101}$ | = | $\left(\frac{1}{2}+x_7\right)\mathbf{a}_1+\left(\frac{1}{2}+x_7\right)\mathbf{a}_2-z_7\,\mathbf{a}_3$ | = | $\left(\frac{1}{2}+x_7\right)a\,\hat{\mathbf{x}}+\left(\frac{1}{2}+x_7\right)a\,\hat{\mathbf{y}}-z_7a\,\hat{\mathbf{z}}$ | (24$k$) | O V |
| $\mathbf{B}_{102}$ | = | $-x_7\,\mathbf{a}_1-x_7\,\mathbf{a}_2-z_7\,\mathbf{a}_3$ | = | $-x_7a\,\hat{\mathbf{x}}-x_7a\,\hat{\mathbf{y}}-z_7a\,\hat{\mathbf{z}}$ | (24$k$) | O V |
| $\mathbf{B}_{103}$ | = | $\left(\frac{1}{2}+x_7\right)\mathbf{a}_1-x_7\,\mathbf{a}_2+\left(\frac{1}{2}+z_7\right)\mathbf{a}_3$ | = | $\left(\frac{1}{2}+x_7\right)a\,\hat{\mathbf{x}}-x_7a\,\hat{\mathbf{y}}+\left(\frac{1}{2}+z_7\right)a\,\hat{\mathbf{z}}$ | (24$k$) | O V |
| $\mathbf{B}_{104}$ | = | $-x_7\,\mathbf{a}_1+\left(\frac{1}{2}+x_7\right)\mathbf{a}_2+\left(\frac{1}{2}+z_7\right)\mathbf{a}_3$ | = | $-x_7a\,\hat{\mathbf{x}}+\left(\frac{1}{2}+x_7\right)a\,\hat{\mathbf{y}}+\left(\frac{1}{2}+z_7\right)a\,\hat{\mathbf{z}}$ | (24$k$) | O V |
| $\mathbf{B}_{105}$ | = | $\left(\frac{1}{2}+x_7\right)\mathbf{a}_1+\left(\frac{1}{2}+z_7\right)\mathbf{a}_2-x_7\,\mathbf{a}_3$ | = | $\left(\frac{1}{2}+x_7\right)a\,\hat{\mathbf{x}}+\left(\frac{1}{2}+z_7\right)a\,\hat{\mathbf{y}}-x_7a\,\hat{\mathbf{z}}$ | (24$k$) | O V |
| $\mathbf{B}_{106}$ | = | $-x_7\,\mathbf{a}_1+\left(\frac{1}{2}+z_7\right)\mathbf{a}_2+\left(\frac{1}{2}+x_7\right)\mathbf{a}_3$ | = | $-x_7a\,\hat{\mathbf{x}}+\left(\frac{1}{2}+z_7\right)a\,\hat{\mathbf{y}}+\left(\frac{1}{2}+x_7\right)a\,\hat{\mathbf{z}}$ | (24$k$) | O V |
| $\mathbf{B}_{107}$ | = | $-x_7\,\mathbf{a}_1-z_7\,\mathbf{a}_2-x_7\,\mathbf{a}_3$ | = | $-x_7a\,\hat{\mathbf{x}}-z_7a\,\hat{\mathbf{y}}-x_7a\,\hat{\mathbf{z}}$ | (24$k$) | O V |
| $\mathbf{B}_{108}$ | = | $\left(\frac{1}{2}+x_7\right)\mathbf{a}_1-z_7\,\mathbf{a}_2+\left(\frac{1}{2}+x_7\right)\mathbf{a}_3$ | = | $\left(\frac{1}{2}+x_7\right)a\,\hat{\mathbf{x}}-z_7a\,\hat{\mathbf{y}}+\left(\frac{1}{2}+x_7\right)a\,\hat{\mathbf{z}}$ | (24$k$) | O V |
| $\mathbf{B}_{109}$ | = | $\left(\frac{1}{2}+z_7\right)\mathbf{a}_1+\left(\frac{1}{2}+x_7\right)\mathbf{a}_2-x_7\,\mathbf{a}_3$ | = | $\left(\frac{1}{2}+z_7\right)a\,\hat{\mathbf{x}}+\left(\frac{1}{2}+x_7\right)a\,\hat{\mathbf{y}}-x_7a\,\hat{\mathbf{z}}$ | (24$k$) | O V |
| $\mathbf{B}_{110}$ | = | $\left(\frac{1}{2}+z_7\right)\mathbf{a}_1-x_7\,\mathbf{a}_2+\left(\frac{1}{2}+x_7\right)\mathbf{a}_3$ | = | $\left(\frac{1}{2}+z_7\right)a\,\hat{\mathbf{x}}-x_7a\,\hat{\mathbf{y}}+\left(\frac{1}{2}+x_7\right)a\,\hat{\mathbf{z}}$ | (24$k$) | O V |
| $\mathbf{B}_{111}$ | = | $-z_7\,\mathbf{a}_1+\left(\frac{1}{2}+x_7\right)\mathbf{a}_2+\left(\frac{1}{2}+x_7\right)\mathbf{a}_3$ | = | $-z_7a\,\hat{\mathbf{x}}+\left(\frac{1}{2}+x_7\right)a\,\hat{\mathbf{y}}+\left(\frac{1}{2}+x_7\right)a\,\hat{\mathbf{z}}$ | (24$k$) | O V |
| $\mathbf{B}_{112}$ | = | $-z_7\,\mathbf{a}_1-x_7\,\mathbf{a}_2-x_7\,\mathbf{a}_3$ | = | $-z_7a\,\hat{\mathbf{x}}-x_7a\,\hat{\mathbf{y}}-x_7a\,\hat{\mathbf{z}}$ | (24$k$) | O V |
| $\mathbf{B}_{113}$ | = | $x_8\,\mathbf{a}_1+x_8\,\mathbf{a}_2+z_8\,\mathbf{a}_3$ | = | $x_8a\,\hat{\mathbf{x}}+x_8a\,\hat{\mathbf{y}}+z_8a\,\hat{\mathbf{z}}$ | (24$k$) | W |
| $\mathbf{B}_{114}$ | = | $\left(\frac{1}{2}-x_8\right)\mathbf{a}_1+\left(\frac{1}{2}-x_8\right)\mathbf{a}_2+z_8\,\mathbf{a}_3$ | = | $\left(\frac{1}{2}-x_8\right)a\,\hat{\mathbf{x}}+\left(\frac{1}{2}-x_8\right)a\,\hat{\mathbf{y}}+z_8a\,\hat{\mathbf{z}}$ | (24$k$) | W |
| $\mathbf{B}_{115}$ | = | $\left(\frac{1}{2}-x_8\right)\mathbf{a}_1+x_8\,\mathbf{a}_2+\left(\frac{1}{2}-z_8\right)\mathbf{a}_3$ | = | $\left(\frac{1}{2}-x_8\right)a\,\hat{\mathbf{x}}+x_8a\,\hat{\mathbf{y}}+\left(\frac{1}{2}-z_8\right)a\,\hat{\mathbf{z}}$ | (24$k$) | W |
| $\mathbf{B}_{116}$ | = | $x_8\,\mathbf{a}_1+\left(\frac{1}{2}-x_8\right)\mathbf{a}_2+\left(\frac{1}{2}-z_8\right)\mathbf{a}_3$ | = | $x_8a\,\hat{\mathbf{x}}+\left(\frac{1}{2}-x_8\right)a\,\hat{\mathbf{y}}+\left(\frac{1}{2}-z_8\right)a\,\hat{\mathbf{z}}$ | (24$k$) | W |
| $\mathbf{B}_{117}$ | = | $z_8\,\mathbf{a}_1+x_8\,\mathbf{a}_2+x_8\,\mathbf{a}_3$ | = | $z_8a\,\hat{\mathbf{x}}+x_8a\,\hat{\mathbf{y}}+x_8a\,\hat{\mathbf{z}}$ | (24$k$) | W |
| $\mathbf{B}_{118}$ | = | $z_8\,\mathbf{a}_1+\left(\frac{1}{2}-x_8\right)\mathbf{a}_2+\left(\frac{1}{2}-x_8\right)\mathbf{a}_3$ | = | $z_8a\,\hat{\mathbf{x}}+\left(\frac{1}{2}-x_8\right)a\,\hat{\mathbf{y}}+\left(\frac{1}{2}-x_8\right)a\,\hat{\mathbf{z}}$ | (24$k$) | W |
| $\mathbf{B}_{119}$ | = | $\left(\frac{1}{2}-z_8\right)\mathbf{a}_1+\left(\frac{1}{2}-x_8\right)\mathbf{a}_2+x_8\,\mathbf{a}_3$ | = | $\left(\frac{1}{2}-z_8\right)a\,\hat{\mathbf{x}}+\left(\frac{1}{2}-x_8\right)a\,\hat{\mathbf{y}}+x_8a\,\hat{\mathbf{z}}$ | (24$k$) | W |
| $\mathbf{B}_{120}$ | = | $\left(\frac{1}{2}-z_8\right)\mathbf{a}_1+x_8\,\mathbf{a}_2+\left(\frac{1}{2}-x_8\right)\mathbf{a}_3$ | = | $\left(\frac{1}{2}-z_8\right)a\,\hat{\mathbf{x}}+x_8a\,\hat{\mathbf{y}}+\left(\frac{1}{2}-x_8\right)a\,\hat{\mathbf{z}}$ | (24$k$) | W |
| $\mathbf{B}_{121}$ | = | $x_8\,\mathbf{a}_1+z_8\,\mathbf{a}_2+x_8\,\mathbf{a}_3$ | = | $x_8a\,\hat{\mathbf{x}}+z_8a\,\hat{\mathbf{y}}+x_8a\,\hat{\mathbf{z}}$ | (24$k$) | W |
| $\mathbf{B}_{122}$ | = | $\left(\frac{1}{2}-x_8\right)\mathbf{a}_1+z_8\,\mathbf{a}_2+\left(\frac{1}{2}-x_8\right)\mathbf{a}_3$ | = | $\left(\frac{1}{2}-x_8\right)a\,\hat{\mathbf{x}}+z_8a\,\hat{\mathbf{y}}+\left(\frac{1}{2}-x_8\right)a\,\hat{\mathbf{z}}$ | (24$k$) | W |
| $\mathbf{B}_{123}$ | = | $x_8\,\mathbf{a}_1+\left(\frac{1}{2}-z_8\right)\mathbf{a}_2+\left(\frac{1}{2}-x_8\right)\mathbf{a}_3$ | = | $x_8a\,\hat{\mathbf{x}}+\left(\frac{1}{2}-z_8\right)a\,\hat{\mathbf{y}}+\left(\frac{1}{2}-x_8\right)a\,\hat{\mathbf{z}}$ | (24$k$) | W |
| $\mathbf{B}_{124}$ | = | $\left(\frac{1}{2}-x_8\right)\mathbf{a}_1+\left(\frac{1}{2}-z_8\right)\mathbf{a}_2+x_8\,\mathbf{a}_3$ | = | $\left(\frac{1}{2}-x_8\right)a\,\hat{\mathbf{x}}+\left(\frac{1}{2}-z_8\right)a\,\hat{\mathbf{y}}+x_8a\,\hat{\mathbf{z}}$ | (24$k$) | W |
| $\mathbf{B}_{125}$ | = | $\left(\frac{1}{2}+x_8\right)\mathbf{a}_1+\left(\frac{1}{2}+x_8\right)\mathbf{a}_2-z_8\,\mathbf{a}_3$ | = | $\left(\frac{1}{2}+x_8\right)a\,\hat{\mathbf{x}}+\left(\frac{1}{2}+x_8\right)a\,\hat{\mathbf{y}}-z_8a\,\hat{\mathbf{z}}$ | (24$k$) | W |
| $\mathbf{B}_{126}$ | = | $-x_8\,\mathbf{a}_1-x_8\,\mathbf{a}_2-z_8\,\mathbf{a}_3$ | = | $-x_8a\,\hat{\mathbf{x}}-x_8a\,\hat{\mathbf{y}}-z_8a\,\hat{\mathbf{z}}$ | (24$k$) | W |
| $\mathbf{B}_{127}$ | = | $\left(\frac{1}{2}+x_8\right)\mathbf{a}_1-x_8\,\mathbf{a}_2+\left(\frac{1}{2}+z_8\right)\mathbf{a}_3$ | = | $\left(\frac{1}{2}+x_8\right)a\,\hat{\mathbf{x}}-x_8a\,\hat{\mathbf{y}}+\left(\frac{1}{2}+z_8\right)a\,\hat{\mathbf{z}}$ | (24$k$) | W |
| $\mathbf{B}_{128}$ | = | $-x_8\,\mathbf{a}_1+\left(\frac{1}{2}+x_8\right)\mathbf{a}_2+\left(\frac{1}{2}+z_8\right)\mathbf{a}_3$ | = | $-x_8a\,\hat{\mathbf{x}}+\left(\frac{1}{2}+x_8\right)a\,\hat{\mathbf{y}}+\left(\frac{1}{2}+z_8\right)a\,\hat{\mathbf{z}}$ | (24$k$) | W |
| $\mathbf{B}_{129}$ | = | $\left(\frac{1}{2}+x_8\right)\mathbf{a}_1+\left(\frac{1}{2}+z_8\right)\mathbf{a}_2-x_8\,\mathbf{a}_3$ | = | $\left(\frac{1}{2}+x_8\right)a\,\hat{\mathbf{x}}+\left(\frac{1}{2}+z_8\right)a\,\hat{\mathbf{y}}-x_8a\,\hat{\mathbf{z}}$ | (24$k$) | W |
| $\mathbf{B}_{130}$ | = | $-x_8\,\mathbf{a}_1+\left(\frac{1}{2}+z_8\right)\mathbf{a}_2+\left(\frac{1}{2}+x_8\right)\mathbf{a}_3$ | = | $-x_8a\,\hat{\mathbf{x}}+\left(\frac{1}{2}+z_8\right)a\,\hat{\mathbf{y}}+\left(\frac{1}{2}+x_8\right)a\,\hat{\mathbf{z}}$ | (24$k$) | W |
| $\mathbf{B}_{131}$ | = | $-x_8\,\mathbf{a}_1-z_8\,\mathbf{a}_2-x_8\,\mathbf{a}_3$ | = | $-x_8a\,\hat{\mathbf{x}}-z_8a\,\hat{\mathbf{y}}-x_8a\,\hat{\mathbf{z}}$ | (24$k$) | W |
| $\mathbf{B}_{132}$ | = | $\left(\frac{1}{2}+x_8\right)\mathbf{a}_1-z_8\,\mathbf{a}_2+\left(\frac{1}{2}+x_8\right)\mathbf{a}_3$ | = | $\left(\frac{1}{2}+x_8\right)a\,\hat{\mathbf{x}}-z_8a\,\hat{\mathbf{y}}+\left(\frac{1}{2}+x_8\right)a\,\hat{\mathbf{z}}$ | (24$k$) | W |
| $\mathbf{B}_{133}$ | = | $\left(\frac{1}{2}+z_8\right)\mathbf{a}_1+\left(\frac{1}{2}+x_8\right)\mathbf{a}_2-x_8\,\mathbf{a}_3$ | = | $\left(\frac{1}{2}+z_8\right)a\,\hat{\mathbf{x}}+\left(\frac{1}{2}+x_8\right)a\,\hat{\mathbf{y}}-x_8a\,\hat{\mathbf{z}}$ | (24$k$) | W |
| $\mathbf{B}_{134}$ | = | $\left(\frac{1}{2}+z_8\right)\mathbf{a}_1-x_8\,\mathbf{a}_2+\left(\frac{1}{2}+x_8\right)\mathbf{a}_3$ | = | $\left(\frac{1}{2}+z_8\right)a\,\hat{\mathbf{x}}-x_8a\,\hat{\mathbf{y}}+\left(\frac{1}{2}+x_8\right)a\,\hat{\mathbf{z}}$ | (24$k$) | W |
| $\mathbf{B}_{135}$ | = | $-z_8\,\mathbf{a}_1+\left(\frac{1}{2}+x_8\right)\mathbf{a}_2+\left(\frac{1}{2}+x_8\right)\mathbf{a}_3$ | = | $-z_8a\,\hat{\mathbf{x}}+\left(\frac{1}{2}+x_8\right)a\,\hat{\mathbf{y}}+\left(\frac{1}{2}+x_8\right)a\,\hat{\mathbf{z}}$ | (24$k$) | W |
| $\mathbf{B}_{136}$ | = | $-z_8\,\mathbf{a}_1 - x_8\,\mathbf{a}_2 - x_8\,\mathbf{a}_3$ | = | $-z_8 a\,\hat{\mathbf{x}} - x_8 a\,\hat{\mathbf{y}} - x_8 a\,\hat{\mathbf{z}}$ | (24$k$) | W |
| $\mathbf{B}_{137}$ | = | $x_9\,\mathbf{a}_1 + y_9\,\mathbf{a}_2 + z_9\,\mathbf{a}_3$ | = | $x_9 a\,\hat{\mathbf{x}} + y_9 a\,\hat{\mathbf{y}} + z_9 a\,\hat{\mathbf{z}}$ | (48$l$) | H II |
| $\mathbf{B}_{138}$ | = | $\left(\frac{1}{2} - x_9\right)\mathbf{a}_1 + \left(\frac{1}{2} - y_9\right)\mathbf{a}_2 + z_9\,\mathbf{a}_3$ | = | $\left(\frac{1}{2} - x_9\right)a\,\hat{\mathbf{x}} + \left(\frac{1}{2} - y_9\right)a\,\hat{\mathbf{y}} + z_9 a\,\hat{\mathbf{z}}$ | (48$l$) | H II |
| $\mathbf{B}_{139}$ | = | $\left(\frac{1}{2} - x_9\right)\mathbf{a}_1 + y_9\,\mathbf{a}_2 + \left(\frac{1}{2} - z_9\right)\mathbf{a}_3$ | = | $\left(\frac{1}{2} - x_9\right)a\,\hat{\mathbf{x}} + y_9 a\,\hat{\mathbf{y}} + \left(\frac{1}{2} - z_9\right)a\,\hat{\mathbf{z}}$ | (48$l$) | H II |
| $\mathbf{B}_{140}$ | = | $x_9\,\mathbf{a}_1 + \left(\frac{1}{2} - y_9\right)\mathbf{a}_2 + \left(\frac{1}{2} - z_9\right)\mathbf{a}_3$ | = | $x_9 a\,\hat{\mathbf{x}} + \left(\frac{1}{2} - y_9\right)a\,\hat{\mathbf{y}} + \left(\frac{1}{2} - z_9\right)a\,\hat{\mathbf{z}}$ | (48$l$) | H II |
| $\mathbf{B}_{141}$ | = | $z_9\,\mathbf{a}_1 + x_9\,\mathbf{a}_2 + y_9\,\mathbf{a}_3$ | = | $z_9 a\,\hat{\mathbf{x}} + x_9 a\,\hat{\mathbf{y}} + y_9 a\,\hat{\mathbf{z}}$ | (48$l$) | H II |
| $\mathbf{B}_{142}$ | = | $z_9\,\mathbf{a}_1 + \left(\frac{1}{2} - x_9\right)\mathbf{a}_2 + \left(\frac{1}{2} - y_9\right)\mathbf{a}_3$ | = | $z_9 a\,\hat{\mathbf{x}} + \left(\frac{1}{2} - x_9\right)a\,\hat{\mathbf{y}} + \left(\frac{1}{2} - y_9\right)a\,\hat{\mathbf{z}}$ | (48$l$) | H II |
| $\mathbf{B}_{143}$ | = | $\left(\frac{1}{2} - z_9\right)\mathbf{a}_1 + \left(\frac{1}{2} - x_9\right)\mathbf{a}_2 + y_9\,\mathbf{a}_3$ | = | $\left(\frac{1}{2} - z_9\right)a\,\hat{\mathbf{x}} + \left(\frac{1}{2} - x_9\right)a\,\hat{\mathbf{y}} + y_9 a\,\hat{\mathbf{z}}$ | (48$l$) | H II |
| $\mathbf{B}_{144}$ | = | $\left(\frac{1}{2} - z_9\right)\mathbf{a}_1 + x_9\,\mathbf{a}_2 + \left(\frac{1}{2} - y_9\right)\mathbf{a}_3$ | = | $\left(\frac{1}{2} - z_9\right)a\,\hat{\mathbf{x}} + x_9 a\,\hat{\mathbf{y}} + \left(\frac{1}{2} - y_9\right)a\,\hat{\mathbf{z}}$ | (48$l$) | H II |
| $\mathbf{B}_{145}$ | = | $y_9\,\mathbf{a}_1 + z_9\,\mathbf{a}_2 + x_9\,\mathbf{a}_3$ | = | $y_9 a\,\hat{\mathbf{x}} + z_9 a\,\hat{\mathbf{y}} + x_9 a\,\hat{\mathbf{z}}$ | (48$l$) | H II |
| $\mathbf{B}_{146}$ | = | $\left(\frac{1}{2} - y_9\right)\mathbf{a}_1 + z_9\,\mathbf{a}_2 + \left(\frac{1}{2} - x_9\right)\mathbf{a}_3$ | = | $\left(\frac{1}{2} - y_9\right)a\,\hat{\mathbf{x}} + z_9 a\,\hat{\mathbf{y}} + \left(\frac{1}{2} - x_9\right)a\,\hat{\mathbf{z}}$ | (48$l$) | H II |
| $\mathbf{B}_{147}$ | = | $y_9\,\mathbf{a}_1 + \left(\frac{1}{2} - z_9\right)\mathbf{a}_2 + \left(\frac{1}{2} - x_9\right)\mathbf{a}_3$ | = | $y_9 a\,\hat{\mathbf{x}} + \left(\frac{1}{2} - z_9\right)a\,\hat{\mathbf{y}} + \left(\frac{1}{2} - x_9\right)a\,\hat{\mathbf{z}}$ | (48$l$) | H II |
| $\mathbf{B}_{148}$ | = | $\left(\frac{1}{2} - y_9\right)\mathbf{a}_1 + \left(\frac{1}{2} - z_9\right)\mathbf{a}_2 + x_9\,\mathbf{a}_3$ | = | $\left(\frac{1}{2} - y_9\right)a\,\hat{\mathbf{x}} + \left(\frac{1}{2} - z_9\right)a\,\hat{\mathbf{y}} + x_9 a\,\hat{\mathbf{z}}$ | (48$l$) | H II |
| $\mathbf{B}_{149}$ | = | $\left(\frac{1}{2} + y_9\right)\mathbf{a}_1 + \left(\frac{1}{2} + x_9\right)\mathbf{a}_2 - z_9\,\mathbf{a}_3$ | = | $\left(\frac{1}{2} + y_9\right)a\,\hat{\mathbf{x}} + \left(\frac{1}{2} + x_9\right)a\,\hat{\mathbf{y}} - z_9 a\,\hat{\mathbf{z}}$ | (48$l$) | H II |
| $\mathbf{B}_{150}$ | = | $-y_9\,\mathbf{a}_1 - x_9\,\mathbf{a}_2 - z_9\,\mathbf{a}_3$ | = | $-y_9 a\,\hat{\mathbf{x}} - x_9 a\,\hat{\mathbf{y}} - z_9 a\,\hat{\mathbf{z}}$ | (48$l$) | H II |
| $\mathbf{B}_{151}$ | = | $\left(\frac{1}{2} + y_9\right)\mathbf{a}_1 - x_9\,\mathbf{a}_2 + \left(\frac{1}{2} + z_9\right)\mathbf{a}_3$ | = | $\left(\frac{1}{2} + y_9\right)a\,\hat{\mathbf{x}} - x_9 a\,\hat{\mathbf{y}} + \left(\frac{1}{2} + z_9\right)a\,\hat{\mathbf{z}}$ | (48$l$) | H II |
| $\mathbf{B}_{152}$ | = | $-y_9\,\mathbf{a}_1 + \left(\frac{1}{2} + x_9\right)\mathbf{a}_2 + \left(\frac{1}{2} + z_9\right)\mathbf{a}_3$ | = | $-y_9 a\,\hat{\mathbf{x}} + \left(\frac{1}{2} + x_9\right)a\,\hat{\mathbf{y}} + \left(\frac{1}{2} + z_9\right)a\,\hat{\mathbf{z}}$ | (48$l$) | H II |
| $\mathbf{B}_{153}$ | = | $\left(\frac{1}{2} + x_9\right)\mathbf{a}_1 + \left(\frac{1}{2} + z_9\right)\mathbf{a}_2 - y_9\,\mathbf{a}_3$ | = | $\left(\frac{1}{2} + x_9\right)a\,\hat{\mathbf{x}} + \left(\frac{1}{2} + z_9\right)a\,\hat{\mathbf{y}} - y_9 a\,\hat{\mathbf{z}}$ | (48$l$) | H II |
| $\mathbf{B}_{154}$ | = | $-x_9\,\mathbf{a}_1 + \left(\frac{1}{2} + z_9\right)\mathbf{a}_2 + \left(\frac{1}{2} + y_9\right)\mathbf{a}_3$ | = | $-x_9 a\,\hat{\mathbf{x}} + \left(\frac{1}{2} + z_9\right)a\,\hat{\mathbf{y}} + \left(\frac{1}{2} + y_9\right)a\,\hat{\mathbf{z}}$ | (48$l$) | H II |
| $\mathbf{B}_{155}$ | = | $-x_9\,\mathbf{a}_1 - z_9\,\mathbf{a}_2 - y_9\,\mathbf{a}_3$ | = | $-x_9 a\,\hat{\mathbf{x}} - z_9 a\,\hat{\mathbf{y}} - y_9 a\,\hat{\mathbf{z}}$ | (48$l$) | H II |
| $\mathbf{B}_{156}$ | = | $\left(\frac{1}{2} + x_9\right)\mathbf{a}_1 - z_9\,\mathbf{a}_2 + \left(\frac{1}{2} + y_9\right)\mathbf{a}_3$ | = | $\left(\frac{1}{2} + x_9\right)a\,\hat{\mathbf{x}} - z_9 a\,\hat{\mathbf{y}} + \left(\frac{1}{2} + y_9\right)a\,\hat{\mathbf{z}}$ | (48$l$) | H II |
| $\mathbf{B}_{157}$ | = | $\left(\frac{1}{2} + z_9\right)\mathbf{a}_1 + \left(\frac{1}{2} + y_9\right)\mathbf{a}_2 - x_9\,\mathbf{a}_3$ | = | $\left(\frac{1}{2} + z_9\right)a\,\hat{\mathbf{x}} + \left(\frac{1}{2} + y_9\right)a\,\hat{\mathbf{y}} - x_9 a\,\hat{\mathbf{z}}$ | (48$l$) | H II |
| $\mathbf{B}_{158}$ | = | $\left(\frac{1}{2} + z_9\right)\mathbf{a}_1 - y_9\,\mathbf{a}_2 + \left(\frac{1}{2} + x_9\right)\mathbf{a}_3$ | = | $\left(\frac{1}{2} + z_9\right)a\,\hat{\mathbf{x}} - y_9 a\,\hat{\mathbf{y}} + \left(\frac{1}{2} + x_9\right)a\,\hat{\mathbf{z}}$ | (48$l$) | H II |
| $\mathbf{B}_{159}$ | = | $-z_9\,\mathbf{a}_1 + \left(\frac{1}{2} + y_9\right)\mathbf{a}_2 + \left(\frac{1}{2} + x_9\right)\mathbf{a}_3$ | = | $-z_9 a\,\hat{\mathbf{x}} + \left(\frac{1}{2} + y_9\right)a\,\hat{\mathbf{y}} + \left(\frac{1}{2} + x_9\right)a\,\hat{\mathbf{z}}$ | (48$l$) | H II |
| $\mathbf{B}_{160}$ | = | $-z_9\,\mathbf{a}_1 - y_9\,\mathbf{a}_2 - x_9\,\mathbf{a}_3$ | = | $-z_9 a\,\hat{\mathbf{x}} - y_9 a\,\hat{\mathbf{y}} - x_9 a\,\hat{\mathbf{z}}$ | (48$l$) | H II |
| $\mathbf{B}_{161}$ | = | $-x_9\,\mathbf{a}_1 - y_9\,\mathbf{a}_2 - z_9\,\mathbf{a}_3$ | = | $-x_9 a\,\hat{\mathbf{x}} - y_9 a\,\hat{\mathbf{y}} - z_9 a\,\hat{\mathbf{z}}$ | (48$l$) | H II |
| $\mathbf{B}_{162}$ | = | $\left(\frac{1}{2} + x_9\right)\mathbf{a}_1 + \left(\frac{1}{2} + y_9\right)\mathbf{a}_2 - z_9\,\mathbf{a}_3$ | = | $\left(\frac{1}{2} + x_9\right)a\,\hat{\mathbf{x}} + \left(\frac{1}{2} + y_9\right)a\,\hat{\mathbf{y}} - z_9 a\,\hat{\mathbf{z}}$ | (48$l$) | H II |
| $\mathbf{B}_{163}$ | = | $\left(\frac{1}{2} + x_9\right)\mathbf{a}_1 - y_9\,\mathbf{a}_2 + \left(\frac{1}{2} + z_9\right)\mathbf{a}_3$ | = | $\left(\frac{1}{2} + x_9\right)a\,\hat{\mathbf{x}} - y_9 a\,\hat{\mathbf{y}} + \left(\frac{1}{2} + z_9\right)a\,\hat{\mathbf{z}}$ | (48$l$) | H II |
| $\mathbf{B}_{164}$ | = | $-x_9\,\mathbf{a}_1 + \left(\frac{1}{2} + y_9\right)\mathbf{a}_2 + \left(\frac{1}{2} + z_9\right)\mathbf{a}_3$ | = | $-x_9 a\,\hat{\mathbf{x}} + \left(\frac{1}{2} + y_9\right)a\,\hat{\mathbf{y}} + \left(\frac{1}{2} + z_9\right)a\,\hat{\mathbf{z}}$ | (48$l$) | H II |
| $\mathbf{B}_{165}$ | = | $-z_9\,\mathbf{a}_1 - x_9\,\mathbf{a}_2 - y_9\,\mathbf{a}_3$ | = | $-z_9 a\,\hat{\mathbf{x}} - x_9 a\,\hat{\mathbf{y}} - y_9 a\,\hat{\mathbf{z}}$ | (48$l$) | H II |
| $\mathbf{B}_{166}$ | = | $-z_9\,\mathbf{a}_1 + \left(\frac{1}{2} + x_9\right)\mathbf{a}_2 + \left(\frac{1}{2} + y_9\right)\mathbf{a}_3$ | = | $-z_9 a\,\hat{\mathbf{x}} + \left(\frac{1}{2} + x_9\right)a\,\hat{\mathbf{y}} + \left(\frac{1}{2} + y_9\right)a\,\hat{\mathbf{z}}$ | (48$l$) | H II |
| $\mathbf{B}_{167}$ | = | $\left(\frac{1}{2} + z_9\right)\mathbf{a}_1 + \left(\frac{1}{2} + x_9\right)\mathbf{a}_2 - y_9\,\mathbf{a}_3$ | = | $\left(\frac{1}{2} + z_9\right)a\,\hat{\mathbf{x}} + \left(\frac{1}{2} + x_9\right)a\,\hat{\mathbf{y}} - y_9 a\,\hat{\mathbf{z}}$ | (48$l$) | H II |
| $\mathbf{B}_{168}$ | = | $\left(\frac{1}{2} + z_9\right)\mathbf{a}_1 - x_9\,\mathbf{a}_2 + \left(\frac{1}{2} + y_9\right)\mathbf{a}_3$ | = | $\left(\frac{1}{2} + z_9\right)a\,\hat{\mathbf{x}} - x_9 a\,\hat{\mathbf{y}} + \left(\frac{1}{2} + y_9\right)a\,\hat{\mathbf{z}}$ | (48$l$) | H II |
| $\mathbf{B}_{169}$ | = | $-y_9\,\mathbf{a}_1 - z_9\,\mathbf{a}_2 - x_9\,\mathbf{a}_3$ | = | $-y_9 a\,\hat{\mathbf{x}} - z_9 a\,\hat{\mathbf{y}} - x_9 a\,\hat{\mathbf{z}}$ | (48$l$) | H II |
| $\mathbf{B}_{170}$ | = | $\left(\frac{1}{2} + y_9\right)\mathbf{a}_1 - z_9\,\mathbf{a}_2 + \left(\frac{1}{2} + x_9\right)\mathbf{a}_3$ | = | $\left(\frac{1}{2} + y_9\right)a\,\hat{\mathbf{x}} - z_9 a\,\hat{\mathbf{y}} + \left(\frac{1}{2} + x_9\right)a\,\hat{\mathbf{z}}$ | (48$l$) | H II |
| $\mathbf{B}_{171}$ | = | $-y_9\,\mathbf{a}_1 + \left(\frac{1}{2} + z_9\right)\mathbf{a}_2 + \left(\frac{1}{2} + x_9\right)\mathbf{a}_3$ | = | $-y_9 a\,\hat{\mathbf{x}} + \left(\frac{1}{2} + z_9\right)a\,\hat{\mathbf{y}} + \left(\frac{1}{2} + x_9\right)a\,\hat{\mathbf{z}}$ | (48$l$) | H II |
| $\mathbf{B}_{172}$ | = | $\left(\frac{1}{2}+y_9\right)\mathbf{a}_1+\left(\frac{1}{2}+z_9\right)\mathbf{a}_2-x_9\,\mathbf{a}_3$ | = | $\left(\frac{1}{2}+y_9\right)a\,\hat{\mathbf{x}}+\left(\frac{1}{2}+z_9\right)a\,\hat{\mathbf{y}}-x_9a\,\hat{\mathbf{z}}$ | (48$l$) | H II |
| $\mathbf{B}_{173}$ | = | $\left(\frac{1}{2}-y_9\right)\mathbf{a}_1+\left(\frac{1}{2}-x_9\right)\mathbf{a}_2+z_9\,\mathbf{a}_3$ | = | $\left(\frac{1}{2}-y_9\right)a\,\hat{\mathbf{x}}+\left(\frac{1}{2}-x_9\right)a\,\hat{\mathbf{y}}+z_9a\,\hat{\mathbf{z}}$ | (48$l$) | H II |
| $\mathbf{B}_{174}$ | = | $y_9\,\mathbf{a}_1+x_9\,\mathbf{a}_2+z_9\,\mathbf{a}_3$ | = | $y_9a\,\hat{\mathbf{x}}+x_9a\,\hat{\mathbf{y}}+z_9a\,\hat{\mathbf{z}}$ | (48$l$) | H II |
| $\mathbf{B}_{175}$ | = | $\left(\frac{1}{2}-y_9\right)\mathbf{a}_1+x_9\,\mathbf{a}_2+\left(\frac{1}{2}-z_9\right)\mathbf{a}_3$ | = | $\left(\frac{1}{2}-y_9\right)a\,\hat{\mathbf{x}}+x_9a\,\hat{\mathbf{y}}+\left(\frac{1}{2}-z_9\right)a\,\hat{\mathbf{z}}$ | (48$l$) | H II |
| $\mathbf{B}_{176}$ | = | $y_9\,\mathbf{a}_1+\left(\frac{1}{2}-x_9\right)\mathbf{a}_2+\left(\frac{1}{2}-z_9\right)\mathbf{a}_3$ | = | $y_9a\,\hat{\mathbf{x}}+\left(\frac{1}{2}-x_9\right)a\,\hat{\mathbf{y}}+\left(\frac{1}{2}-z_9\right)a\,\hat{\mathbf{z}}$ | (48$l$) | H II |
| $\mathbf{B}_{177}$ | = | $\left(\frac{1}{2}-x_9\right)\mathbf{a}_1+\left(\frac{1}{2}-z_9\right)\mathbf{a}_2+y_9\,\mathbf{a}_3$ | = | $\left(\frac{1}{2}-x_9\right)a\,\hat{\mathbf{x}}+\left(\frac{1}{2}-z_9\right)a\,\hat{\mathbf{y}}+y_9a\,\hat{\mathbf{z}}$ | (48$l$) | H II |
| $\mathbf{B}_{178}$ | = | $x_9\,\mathbf{a}_1+\left(\frac{1}{2}-z_9\right)\mathbf{a}_2+\left(\frac{1}{2}-y_9\right)\mathbf{a}_3$ | = | $x_9a\,\hat{\mathbf{x}}+\left(\frac{1}{2}-z_9\right)a\,\hat{\mathbf{y}}+\left(\frac{1}{2}-y_9\right)a\,\hat{\mathbf{z}}$ | (48$l$) | H II |
| $\mathbf{B}_{179}$ | = | $x_9\,\mathbf{a}_1+z_9\,\mathbf{a}_2+y_9\,\mathbf{a}_3$ | = | $x_9a\,\hat{\mathbf{x}}+z_9a\,\hat{\mathbf{y}}+y_9a\,\hat{\mathbf{z}}$ | (48$l$) | H II |
| $\mathbf{B}_{180}$ | = | $\left(\frac{1}{2}-x_9\right)\mathbf{a}_1+z_9\,\mathbf{a}_2+\left(\frac{1}{2}-y_9\right)\mathbf{a}_3$ | = | $\left(\frac{1}{2}-x_9\right)a\,\hat{\mathbf{x}}+z_9a\,\hat{\mathbf{y}}+\left(\frac{1}{2}-y_9\right)a\,\hat{\mathbf{z}}$ | (48$l$) | H II |
| $\mathbf{B}_{181}$ | = | $\left(\frac{1}{2}-z_9\right)\mathbf{a}_1+\left(\frac{1}{2}-y_9\right)\mathbf{a}_2+x_9\,\mathbf{a}_3$ | = | $\left(\frac{1}{2}-z_9\right)a\,\hat{\mathbf{x}}+\left(\frac{1}{2}-y_9\right)a\,\hat{\mathbf{y}}+x_9a\,\hat{\mathbf{z}}$ | (48$l$) | H II |
| $\mathbf{B}_{182}$ | = | $\left(\frac{1}{2}-z_9\right)\mathbf{a}_1+y_9\,\mathbf{a}_2+\left(\frac{1}{2}-x_9\right)\mathbf{a}_3$ | = | $\left(\frac{1}{2}-z_9\right)a\,\hat{\mathbf{x}}+y_9a\,\hat{\mathbf{y}}+\left(\frac{1}{2}-x_9\right)a\,\hat{\mathbf{z}}$ | (48$l$) | H II |
| $\mathbf{B}_{183}$ | = | $z_9\,\mathbf{a}_1+\left(\frac{1}{2}-y_9\right)\mathbf{a}_2+\left(\frac{1}{2}-x_9\right)\mathbf{a}_3$ | = | $z_9a\,\hat{\mathbf{x}}+\left(\frac{1}{2}-y_9\right)a\,\hat{\mathbf{y}}+\left(\frac{1}{2}-x_9\right)a\,\hat{\mathbf{z}}$ | (48$l$) | H II |
| $\mathbf{B}_{184}$ | = | $z_9\,\mathbf{a}_1+y_9\,\mathbf{a}_2+x_9\,\mathbf{a}_3$ | = | $z_9a\,\hat{\mathbf{x}}+y_9a\,\hat{\mathbf{y}}+x_9a\,\hat{\mathbf{z}}$ | (48$l$) | H II |

**Geometry files:**

- CIF: pp. 1807
- POSCAR: pp. 1808

# $Mg_3P_2$ ($D5_5$) Structure: A3B2_cP10_224_d_b

http://aflow.org/prototype-encyclopedia/A3B2_cP10_224_d_b

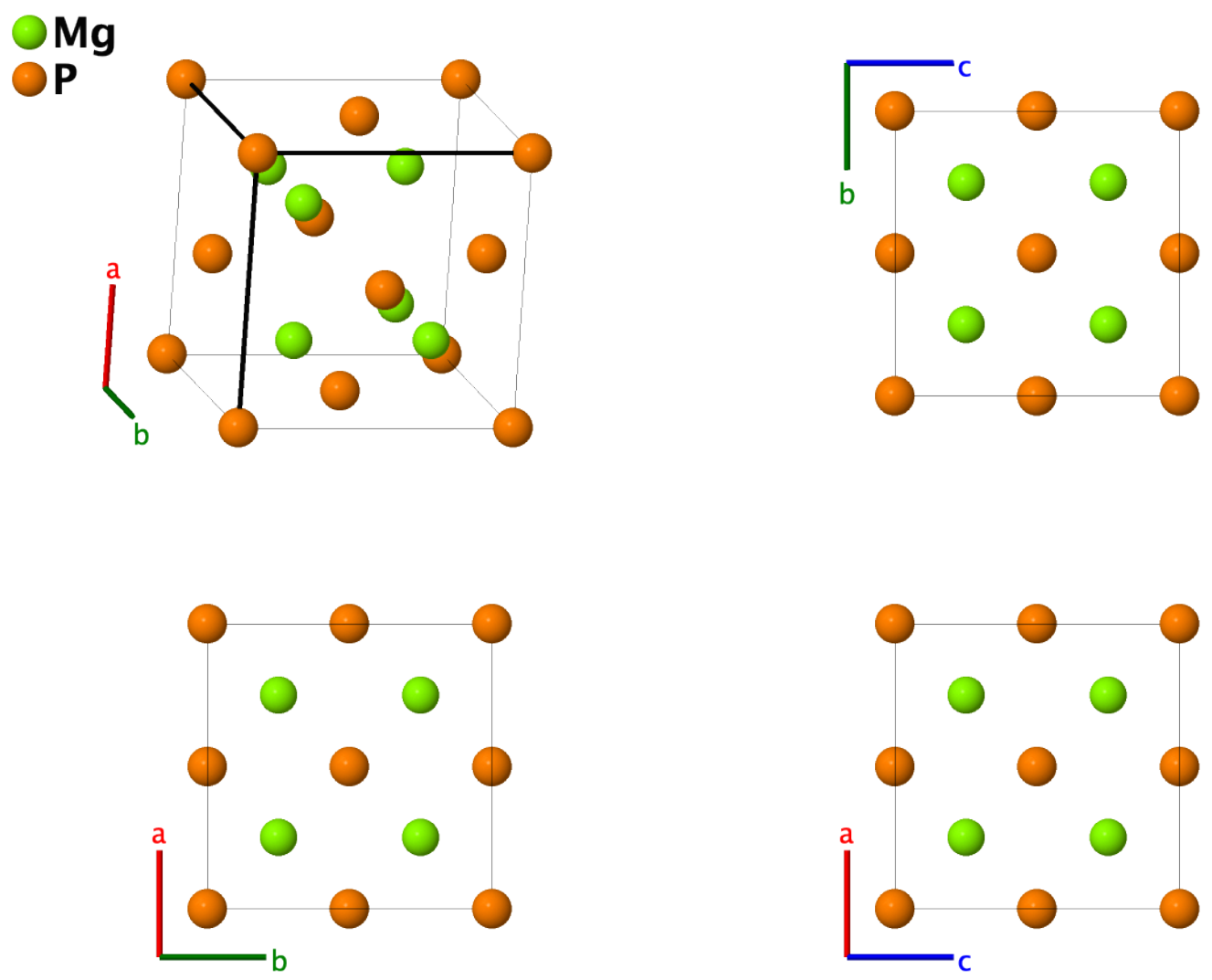

| | | |
|---|---|---|
| **Prototype** | : | $Mg_3P_2$ |
| **AFLOW prototype label** | : | A3B2_cP10_224_d_b |
| ***Strukturbericht* designation** | : | $D5_5$ |
| **Pearson symbol** | : | cP10 |
| **Space group number** | : | 224 |
| **Space group symbol** | : | $Pn\bar{3}m$ |
| **AFLOW prototype command** | : | `aflow --proto=A3B2_cP10_224_d_b --params=`$a$ |

### Other compounds with this structure

- $Al_3P_2$, $Cd_3P_2$, $Zn_3P_2$, $Mg_3As_2$, $Zn_3As_2$, and $Ag_2O_3$

---

- (Parthé, 1993) lists $As_2O_3$ as the prototype for $D5_5$.
- (Passerini, 1928) gives the atomic positions in setting 1 of space group $Pn\bar{3}m$ #224. We have shifted this to the standard, setting 2.

### Simple Cubic primitive vectors:

$$\mathbf{a}_1 = a\,\hat{\mathbf{x}}$$
$$\mathbf{a}_2 = a\,\hat{\mathbf{y}}$$
$$\mathbf{a}_3 = a\,\hat{\mathbf{z}}$$

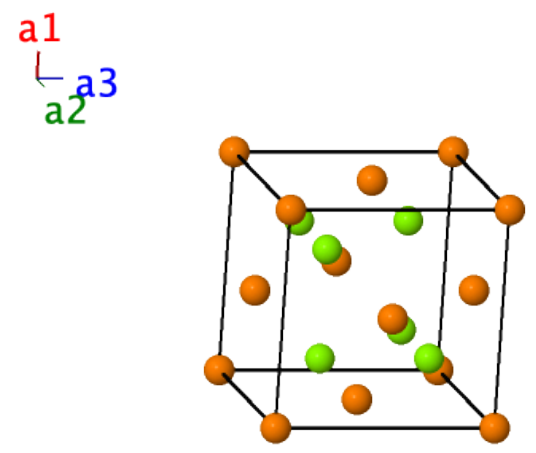

### Basis vectors:

| | | Lattice Coordinates | | Cartesian Coordinates | Wyckoff Position | Atom Type |
|---|---|---|---|---|---|---|
| $\mathbf{B}_1$ | = | $0\,\mathbf{a}_1 + 0\,\mathbf{a}_2 + 0\,\mathbf{a}_3$ | = | $0\,\hat{\mathbf{x}} + 0\,\hat{\mathbf{y}} + 0\,\hat{\mathbf{z}}$ | (4$b$) | P |
| $\mathbf{B}_2$ | = | $\frac{1}{2}\,\mathbf{a}_1 + \frac{1}{2}\,\mathbf{a}_2$ | = | $\frac{1}{2}a\,\hat{\mathbf{x}} + \frac{1}{2}a\,\hat{\mathbf{y}}$ | (4$b$) | P |
| $\mathbf{B}_3$ | = | $\frac{1}{2}\,\mathbf{a}_1 + \frac{1}{2}\,\mathbf{a}_3$ | = | $\frac{1}{2}a\,\hat{\mathbf{x}} + \frac{1}{2}a\,\hat{\mathbf{z}}$ | (4$b$) | P |
| $\mathbf{B}_4$ | = | $\frac{1}{2}\,\mathbf{a}_2 + \frac{1}{2}\,\mathbf{a}_3$ | = | $\frac{1}{2}a\,\hat{\mathbf{y}} + \frac{1}{2}a\,\hat{\mathbf{z}}$ | (4$b$) | P |
| $\mathbf{B}_5$ | = | $\frac{1}{4}\,\mathbf{a}_1 + \frac{3}{4}\,\mathbf{a}_2 + \frac{3}{4}\,\mathbf{a}_3$ | = | $\frac{1}{4}a\,\hat{\mathbf{x}} + \frac{3}{4}a\,\hat{\mathbf{y}} + \frac{3}{4}a\,\hat{\mathbf{z}}$ | (6$d$) | Mg |
| $\mathbf{B}_6$ | = | $\frac{3}{4}\,\mathbf{a}_1 + \frac{1}{4}\,\mathbf{a}_2 + \frac{3}{4}\,\mathbf{a}_3$ | = | $\frac{3}{4}a\,\hat{\mathbf{x}} + \frac{1}{4}a\,\hat{\mathbf{y}} + \frac{3}{4}a\,\hat{\mathbf{z}}$ | (6$d$) | Mg |
| $\mathbf{B}_7$ | = | $\frac{3}{4}\,\mathbf{a}_1 + \frac{3}{4}\,\mathbf{a}_2 + \frac{1}{4}\,\mathbf{a}_3$ | = | $\frac{3}{4}a\,\hat{\mathbf{x}} + \frac{3}{4}a\,\hat{\mathbf{y}} + \frac{1}{4}a\,\hat{\mathbf{z}}$ | (6$d$) | Mg |
| $\mathbf{B}_8$ | = | $\frac{1}{4}\,\mathbf{a}_1 + \frac{3}{4}\,\mathbf{a}_2 + \frac{1}{4}\,\mathbf{a}_3$ | = | $\frac{1}{4}a\,\hat{\mathbf{x}} + \frac{3}{4}a\,\hat{\mathbf{y}} + \frac{1}{4}a\,\hat{\mathbf{z}}$ | (6$d$) | Mg |
| $\mathbf{B}_9$ | = | $\frac{3}{4}\,\mathbf{a}_1 + \frac{1}{4}\,\mathbf{a}_2 + \frac{1}{4}\,\mathbf{a}_3$ | = | $\frac{3}{4}a\,\hat{\mathbf{x}} + \frac{1}{4}a\,\hat{\mathbf{y}} + \frac{1}{4}a\,\hat{\mathbf{z}}$ | (6$d$) | Mg |
| $\mathbf{B}_{10}$ | = | $\frac{1}{4}\,\mathbf{a}_1 + \frac{1}{4}\,\mathbf{a}_2 + \frac{3}{4}\,\mathbf{a}_3$ | = | $\frac{1}{4}a\,\hat{\mathbf{x}} + \frac{1}{4}a\,\hat{\mathbf{y}} + \frac{3}{4}a\,\hat{\mathbf{z}}$ | (6$d$) | Mg |

---

**Found in:**

---

**Geometry files:**

- CIF: pp. 1809
- POSCAR: pp. 1809

# $H_3PW_{12}O_{40}{\cdot}3H_2O$ Structure: A3B40CD12_cP112_224_d_e3k_a_k

http://aflow.org/prototype-encyclopedia/A3B40CD12_cP112_224_d_e3k_a_k

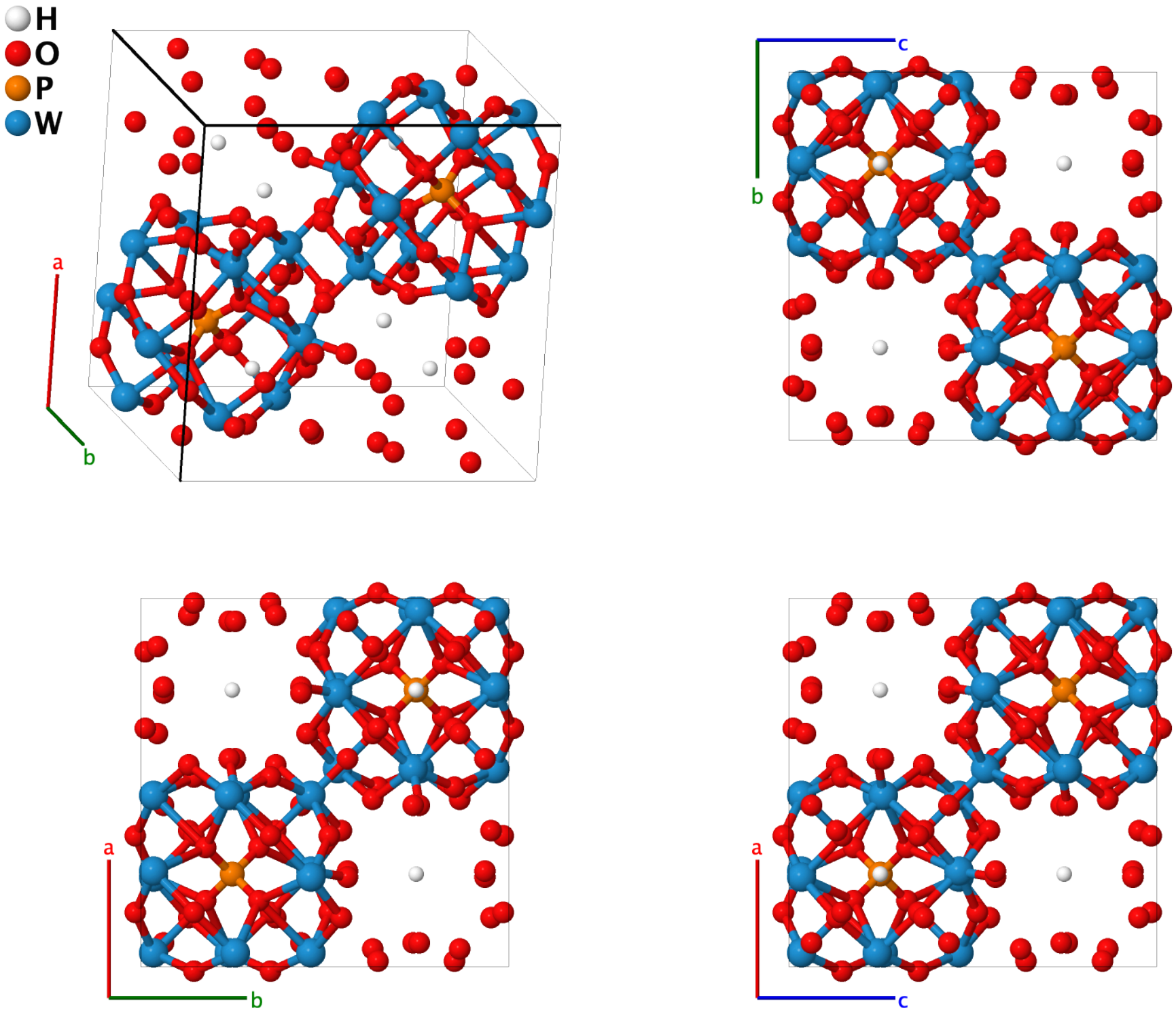

| | | |
|---|---|---|
| **Prototype** | : | $(H_3O)_3O_{40}PW_{12}$ |
| **AFLOW prototype label** | : | A3B40CD12_cP112_224_d_e3k_a_k |
| ***Strukturbericht* designation** | : | None |
| **Pearson symbol** | : | cP112 |
| **Space group number** | : | 224 |
| **Space group symbol** | : | $Pn\bar{3}m$ |
| **AFLOW prototype command** | : | `aflow --proto=A3B40CD12_cP112_224_d_e3k_a_k` `--params=`$a, x_3, x_4, z_4, x_5, z_5, x_6, z_6, x_7, z_7$ |

- This is a partially dehydrated version of $H_3PW_{12}O_{40}{\cdot}29H_2O$ ($H4_{21}$) and $H_3PW_{12}O_{40}{\cdot}5H_2O$ ($H4_{16}$) or $H_3PW_{12}O_{40}{\cdot}6H_2O$. The six $H_5O_2{}^+$ ions of the later structure have been replaced by three $H_3O^+$ ions.

**Simple Cubic primitive vectors:**

$$\mathbf{a}_1 = a\,\hat{\mathbf{x}}$$
$$\mathbf{a}_2 = a\,\hat{\mathbf{y}}$$
$$\mathbf{a}_3 = a\,\hat{\mathbf{z}}$$

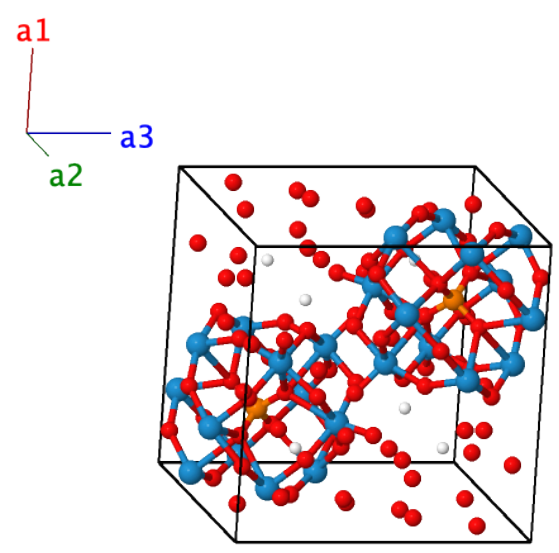

**Basis vectors:**

| | | Lattice Coordinates | | Cartesian Coordinates | Wyckoff Position | Atom Type |
|---|---|---|---|---|---|---|
| $\mathbf{B}_1$ | = | $\frac{1}{4}\,\mathbf{a}_1 + \frac{1}{4}\,\mathbf{a}_2 + \frac{1}{4}\,\mathbf{a}_3$ | = | $\frac{1}{4}a\,\hat{\mathbf{x}} + \frac{1}{4}a\,\hat{\mathbf{y}} + \frac{1}{4}a\,\hat{\mathbf{z}}$ | (2*a*) | P |
| $\mathbf{B}_2$ | = | $\frac{3}{4}\,\mathbf{a}_1 + \frac{3}{4}\,\mathbf{a}_2 + \frac{3}{4}\,\mathbf{a}_3$ | = | $\frac{3}{4}a\,\hat{\mathbf{x}} + \frac{3}{4}a\,\hat{\mathbf{y}} + \frac{3}{4}a\,\hat{\mathbf{z}}$ | (2*a*) | P |
| $\mathbf{B}_3$ | = | $\frac{1}{4}\,\mathbf{a}_1 + \frac{3}{4}\,\mathbf{a}_2 + \frac{3}{4}\,\mathbf{a}_3$ | = | $\frac{1}{4}a\,\hat{\mathbf{x}} + \frac{3}{4}a\,\hat{\mathbf{y}} + \frac{3}{4}a\,\hat{\mathbf{z}}$ | (6*d*) | $H_3O$ |
| $\mathbf{B}_4$ | = | $\frac{3}{4}\,\mathbf{a}_1 + \frac{1}{4}\,\mathbf{a}_2 + \frac{3}{4}\,\mathbf{a}_3$ | = | $\frac{3}{4}a\,\hat{\mathbf{x}} + \frac{1}{4}a\,\hat{\mathbf{y}} + \frac{3}{4}a\,\hat{\mathbf{z}}$ | (6*d*) | $H_3O$ |
| $\mathbf{B}_5$ | = | $\frac{3}{4}\,\mathbf{a}_1 + \frac{3}{4}\,\mathbf{a}_2 + \frac{1}{4}\,\mathbf{a}_3$ | = | $\frac{3}{4}a\,\hat{\mathbf{x}} + \frac{3}{4}a\,\hat{\mathbf{y}} + \frac{1}{4}a\,\hat{\mathbf{z}}$ | (6*d*) | $H_3O$ |
| $\mathbf{B}_6$ | = | $\frac{1}{4}\,\mathbf{a}_1 + \frac{3}{4}\,\mathbf{a}_2 + \frac{1}{4}\,\mathbf{a}_3$ | = | $\frac{1}{4}a\,\hat{\mathbf{x}} + \frac{3}{4}a\,\hat{\mathbf{y}} + \frac{1}{4}a\,\hat{\mathbf{z}}$ | (6*d*) | $H_3O$ |
| $\mathbf{B}_7$ | = | $\frac{3}{4}\,\mathbf{a}_1 + \frac{1}{4}\,\mathbf{a}_2 + \frac{1}{4}\,\mathbf{a}_3$ | = | $\frac{3}{4}a\,\hat{\mathbf{x}} + \frac{1}{4}a\,\hat{\mathbf{y}} + \frac{1}{4}a\,\hat{\mathbf{z}}$ | (6*d*) | $H_3O$ |
| $\mathbf{B}_8$ | = | $\frac{1}{4}\,\mathbf{a}_1 + \frac{1}{4}\,\mathbf{a}_2 + \frac{3}{4}\,\mathbf{a}_3$ | = | $\frac{1}{4}a\,\hat{\mathbf{x}} + \frac{1}{4}a\,\hat{\mathbf{y}} + \frac{3}{4}a\,\hat{\mathbf{z}}$ | (6*d*) | $H_3O$ |
| $\mathbf{B}_9$ | = | $x_3\,\mathbf{a}_1 + x_3\,\mathbf{a}_2 + x_3\,\mathbf{a}_3$ | = | $x_3a\,\hat{\mathbf{x}} + x_3a\,\hat{\mathbf{y}} + x_3a\,\hat{\mathbf{z}}$ | (8*e*) | O I |
| $\mathbf{B}_{10}$ | = | $\left(\frac{1}{2} - x_3\right)\mathbf{a}_1 + \left(\frac{1}{2} - x_3\right)\mathbf{a}_2 + x_3\,\mathbf{a}_3$ | = | $\left(\frac{1}{2} - x_3\right)a\,\hat{\mathbf{x}} + \left(\frac{1}{2} - x_3\right)a\,\hat{\mathbf{y}} + x_3a\,\hat{\mathbf{z}}$ | (8*e*) | O I |
| $\mathbf{B}_{11}$ | = | $\left(\frac{1}{2} - x_3\right)\mathbf{a}_1 + x_3\,\mathbf{a}_2 + \left(\frac{1}{2} - x_3\right)\mathbf{a}_3$ | = | $\left(\frac{1}{2} - x_3\right)a\,\hat{\mathbf{x}} + x_3a\,\hat{\mathbf{y}} + \left(\frac{1}{2} - x_3\right)a\,\hat{\mathbf{z}}$ | (8*e*) | O I |
| $\mathbf{B}_{12}$ | = | $x_3\,\mathbf{a}_1 + \left(\frac{1}{2} - x_3\right)\mathbf{a}_2 + \left(\frac{1}{2} - x_3\right)\mathbf{a}_3$ | = | $x_3a\,\hat{\mathbf{x}} + \left(\frac{1}{2} - x_3\right)a\,\hat{\mathbf{y}} + \left(\frac{1}{2} - x_3\right)a\,\hat{\mathbf{z}}$ | (8*e*) | O I |
| $\mathbf{B}_{13}$ | = | $\left(\frac{1}{2} + x_3\right)\mathbf{a}_1 + \left(\frac{1}{2} + x_3\right)\mathbf{a}_2 - x_3\,\mathbf{a}_3$ | = | $\left(\frac{1}{2} + x_3\right)a\,\hat{\mathbf{x}} + \left(\frac{1}{2} + x_3\right)a\,\hat{\mathbf{y}} - x_3a\,\hat{\mathbf{z}}$ | (8*e*) | O I |
| $\mathbf{B}_{14}$ | = | $-x_3\,\mathbf{a}_1 - x_3\,\mathbf{a}_2 - x_3\,\mathbf{a}_3$ | = | $-x_3a\,\hat{\mathbf{x}} - x_3a\,\hat{\mathbf{y}} - x_3a\,\hat{\mathbf{z}}$ | (8*e*) | O I |
| $\mathbf{B}_{15}$ | = | $\left(\frac{1}{2} + x_3\right)\mathbf{a}_1 - x_3\,\mathbf{a}_2 + \left(\frac{1}{2} + x_3\right)\mathbf{a}_3$ | = | $\left(\frac{1}{2} + x_3\right)a\,\hat{\mathbf{x}} - x_3a\,\hat{\mathbf{y}} + \left(\frac{1}{2} + x_3\right)a\,\hat{\mathbf{z}}$ | (8*e*) | O I |
| $\mathbf{B}_{16}$ | = | $-x_3\,\mathbf{a}_1 + \left(\frac{1}{2} + x_3\right)\mathbf{a}_2 + \left(\frac{1}{2} + x_3\right)\mathbf{a}_3$ | = | $-x_3a\,\hat{\mathbf{x}} + \left(\frac{1}{2} + x_3\right)a\,\hat{\mathbf{y}} + \left(\frac{1}{2} + x_3\right)a\,\hat{\mathbf{z}}$ | (8*e*) | O I |
| $\mathbf{B}_{17}$ | = | $x_4\,\mathbf{a}_1 + x_4\,\mathbf{a}_2 + z_4\,\mathbf{a}_3$ | = | $x_4a\,\hat{\mathbf{x}} + x_4a\,\hat{\mathbf{y}} + z_4a\,\hat{\mathbf{z}}$ | (24*k*) | O II |
| $\mathbf{B}_{18}$ | = | $\left(\frac{1}{2} - x_4\right)\mathbf{a}_1 + \left(\frac{1}{2} - x_4\right)\mathbf{a}_2 + z_4\,\mathbf{a}_3$ | = | $\left(\frac{1}{2} - x_4\right)a\,\hat{\mathbf{x}} + \left(\frac{1}{2} - x_4\right)a\,\hat{\mathbf{y}} + z_4a\,\hat{\mathbf{z}}$ | (24*k*) | O II |
| $\mathbf{B}_{19}$ | = | $\left(\frac{1}{2} - x_4\right)\mathbf{a}_1 + x_4\,\mathbf{a}_2 + \left(\frac{1}{2} - z_4\right)\mathbf{a}_3$ | = | $\left(\frac{1}{2} - x_4\right)a\,\hat{\mathbf{x}} + x_4a\,\hat{\mathbf{y}} + \left(\frac{1}{2} - z_4\right)a\,\hat{\mathbf{z}}$ | (24*k*) | O II |
| $\mathbf{B}_{20}$ | = | $x_4\,\mathbf{a}_1 + \left(\frac{1}{2} - x_4\right)\mathbf{a}_2 + \left(\frac{1}{2} - z_4\right)\mathbf{a}_3$ | = | $x_4a\,\hat{\mathbf{x}} + \left(\frac{1}{2} - x_4\right)a\,\hat{\mathbf{y}} + \left(\frac{1}{2} - z_4\right)a\,\hat{\mathbf{z}}$ | (24*k*) | O II |
| $\mathbf{B}_{21}$ | = | $z_4\,\mathbf{a}_1 + x_4\,\mathbf{a}_2 + x_4\,\mathbf{a}_3$ | = | $z_4a\,\hat{\mathbf{x}} + x_4a\,\hat{\mathbf{y}} + x_4a\,\hat{\mathbf{z}}$ | (24*k*) | O II |
| $\mathbf{B}_{22}$ | = | $z_4\,\mathbf{a}_1 + \left(\frac{1}{2} - x_4\right)\mathbf{a}_2 + \left(\frac{1}{2} - x_4\right)\mathbf{a}_3$ | = | $z_4a\,\hat{\mathbf{x}} + \left(\frac{1}{2} - x_4\right)a\,\hat{\mathbf{y}} + \left(\frac{1}{2} - x_4\right)a\,\hat{\mathbf{z}}$ | (24*k*) | O II |
| $\mathbf{B}_{23}$ | = | $\left(\frac{1}{2} - z_4\right)\mathbf{a}_1 + \left(\frac{1}{2} - x_4\right)\mathbf{a}_2 + x_4\,\mathbf{a}_3$ | = | $\left(\frac{1}{2} - z_4\right)a\,\hat{\mathbf{x}} + \left(\frac{1}{2} - x_4\right)a\,\hat{\mathbf{y}} + x_4a\,\hat{\mathbf{z}}$ | (24*k*) | O II |
| $\mathbf{B}_{24}$ | = | $\left(\frac{1}{2} - z_4\right)\mathbf{a}_1 + x_4\,\mathbf{a}_2 + \left(\frac{1}{2} - x_4\right)\mathbf{a}_3$ | = | $\left(\frac{1}{2} - z_4\right)a\,\hat{\mathbf{x}} + x_4a\,\hat{\mathbf{y}} + \left(\frac{1}{2} - x_4\right)a\,\hat{\mathbf{z}}$ | (24*k*) | O II |
| $\mathbf{B}_{25}$ | = | $x_4\,\mathbf{a}_1 + z_4\,\mathbf{a}_2 + x_4\,\mathbf{a}_3$ | = | $x_4a\,\hat{\mathbf{x}} + z_4a\,\hat{\mathbf{y}} + x_4a\,\hat{\mathbf{z}}$ | (24*k*) | O II |
| $\mathbf{B}_{26}$ | = | $\left(\frac{1}{2} - x_4\right)\mathbf{a}_1 + z_4\,\mathbf{a}_2 + \left(\frac{1}{2} - x_4\right)\mathbf{a}_3$ | = | $\left(\frac{1}{2} - x_4\right)a\,\hat{\mathbf{x}} + z_4a\,\hat{\mathbf{y}} + \left(\frac{1}{2} - x_4\right)a\,\hat{\mathbf{z}}$ | (24*k*) | O II |
| $\mathbf{B}_{27}$ | = | $x_4\,\mathbf{a}_1 + \left(\frac{1}{2} - z_4\right)\mathbf{a}_2 + \left(\frac{1}{2} - x_4\right)\mathbf{a}_3$ | = | $x_4a\,\hat{\mathbf{x}} + \left(\frac{1}{2} - z_4\right)a\,\hat{\mathbf{y}} + \left(\frac{1}{2} - x_4\right)a\,\hat{\mathbf{z}}$ | (24*k*) | O II |
| $\mathbf{B}_{28}$ | = | $\left(\frac{1}{2} - x_4\right)\mathbf{a}_1 + \left(\frac{1}{2} - z_4\right)\mathbf{a}_2 + x_4\,\mathbf{a}_3$ | = | $\left(\frac{1}{2} - x_4\right)a\,\hat{\mathbf{x}} + \left(\frac{1}{2} - z_4\right)a\,\hat{\mathbf{y}} + x_4a\,\hat{\mathbf{z}}$ | (24*k*) | O II |
| $\mathbf{B}_{29}$ | = | $\left(\frac{1}{2} + x_4\right)\mathbf{a}_1 + \left(\frac{1}{2} + x_4\right)\mathbf{a}_2 - z_4\,\mathbf{a}_3$ | = | $\left(\frac{1}{2} + x_4\right)a\,\hat{\mathbf{x}} + \left(\frac{1}{2} + x_4\right)a\,\hat{\mathbf{y}} - z_4a\,\hat{\mathbf{z}}$ | (24*k*) | O II |
| $\mathbf{B}_{30}$ | = | $-x_4\,\mathbf{a}_1 - x_4\,\mathbf{a}_2 - z_4\,\mathbf{a}_3$ | = | $-x_4a\,\hat{\mathbf{x}} - x_4a\,\hat{\mathbf{y}} - z_4a\,\hat{\mathbf{z}}$ | (24*k*) | O II |
| $\mathbf{B}_{31}$ | = | $\left(\frac{1}{2} + x_4\right)\mathbf{a}_1 - x_4\,\mathbf{a}_2 + \left(\frac{1}{2} + z_4\right)\mathbf{a}_3$ | = | $\left(\frac{1}{2} + x_4\right)a\,\hat{\mathbf{x}} - x_4a\,\hat{\mathbf{y}} + \left(\frac{1}{2} + z_4\right)a\,\hat{\mathbf{z}}$ | (24*k*) | O II |
| $\mathbf{B}_{32}$ | = | $-x_4\,\mathbf{a}_1 + \left(\frac{1}{2} + x_4\right)\mathbf{a}_2 + \left(\frac{1}{2} + z_4\right)\mathbf{a}_3$ | = | $-x_4a\,\hat{\mathbf{x}} + \left(\frac{1}{2} + x_4\right)a\,\hat{\mathbf{y}} + \left(\frac{1}{2} + z_4\right)a\,\hat{\mathbf{z}}$ | (24*k*) | O II |
| $\mathbf{B}_{33}$ | = | $\left(\frac{1}{2} + x_4\right)\mathbf{a}_1 + \left(\frac{1}{2} + z_4\right)\mathbf{a}_2 - x_4\,\mathbf{a}_3$ | = | $\left(\frac{1}{2} + x_4\right)a\,\hat{\mathbf{x}} + \left(\frac{1}{2} + z_4\right)a\,\hat{\mathbf{y}} - x_4a\,\hat{\mathbf{z}}$ | (24*k*) | O II |
| $\mathbf{B}_{34}$ | = | $-x_4\,\mathbf{a}_1 + \left(\frac{1}{2} + z_4\right)\mathbf{a}_2 + \left(\frac{1}{2} + x_4\right)\mathbf{a}_3$ | = | $-x_4a\,\hat{\mathbf{x}} + \left(\frac{1}{2} + z_4\right)a\,\hat{\mathbf{y}} + \left(\frac{1}{2} + x_4\right)a\,\hat{\mathbf{z}}$ | (24*k*) | O II |

$$\mathbf{B}_{35} = -x_4\,\mathbf{a}_1 - z_4\,\mathbf{a}_2 - x_4\,\mathbf{a}_3 = -x_4a\,\hat{\mathbf{x}} - z_4a\,\hat{\mathbf{y}} - x_4a\,\hat{\mathbf{z}} \qquad (24k) \qquad \text{O II}$$

$$\mathbf{B}_{36} = \left(\frac{1}{2} + x_4\right)\mathbf{a}_1 - z_4\,\mathbf{a}_2 + \left(\frac{1}{2} + x_4\right)\mathbf{a}_3 = \left(\frac{1}{2} + x_4\right)a\,\hat{\mathbf{x}} - z_4a\,\hat{\mathbf{y}} + \left(\frac{1}{2} + x_4\right)a\,\hat{\mathbf{z}} \qquad (24k) \qquad \text{O II}$$

$$\mathbf{B}_{37} = \left(\frac{1}{2} + z_4\right)\mathbf{a}_1 + \left(\frac{1}{2} + x_4\right)\mathbf{a}_2 - x_4\,\mathbf{a}_3 = \left(\frac{1}{2} + z_4\right)a\,\hat{\mathbf{x}} + \left(\frac{1}{2} + x_4\right)a\,\hat{\mathbf{y}} - x_4a\,\hat{\mathbf{z}} \qquad (24k) \qquad \text{O II}$$

$$\mathbf{B}_{38} = \left(\frac{1}{2} + z_4\right)\mathbf{a}_1 - x_4\,\mathbf{a}_2 + \left(\frac{1}{2} + x_4\right)\mathbf{a}_3 = \left(\frac{1}{2} + z_4\right)a\,\hat{\mathbf{x}} - x_4a\,\hat{\mathbf{y}} + \left(\frac{1}{2} + x_4\right)a\,\hat{\mathbf{z}} \qquad (24k) \qquad \text{O II}$$

$$\mathbf{B}_{39} = -z_4\,\mathbf{a}_1 + \left(\frac{1}{2} + x_4\right)\mathbf{a}_2 + \left(\frac{1}{2} + x_4\right)\mathbf{a}_3 = -z_4a\,\hat{\mathbf{x}} + \left(\frac{1}{2} + x_4\right)a\,\hat{\mathbf{y}} + \left(\frac{1}{2} + x_4\right)a\,\hat{\mathbf{z}} \qquad (24k) \qquad \text{O II}$$

$$\mathbf{B}_{40} = -z_4\,\mathbf{a}_1 - x_4\,\mathbf{a}_2 - x_4\,\mathbf{a}_3 = -z_4a\,\hat{\mathbf{x}} - x_4a\,\hat{\mathbf{y}} - x_4a\,\hat{\mathbf{z}} \qquad (24k) \qquad \text{O II}$$

$$\mathbf{B}_{41} = x_5\,\mathbf{a}_1 + x_5\,\mathbf{a}_2 + z_5\,\mathbf{a}_3 = x_5a\,\hat{\mathbf{x}} + x_5a\,\hat{\mathbf{y}} + z_5a\,\hat{\mathbf{z}} \qquad (24k) \qquad \text{O III}$$

$$\mathbf{B}_{42} = \left(\frac{1}{2} - x_5\right)\mathbf{a}_1 + \left(\frac{1}{2} - x_5\right)\mathbf{a}_2 + z_5\,\mathbf{a}_3 = \left(\frac{1}{2} - x_5\right)a\,\hat{\mathbf{x}} + \left(\frac{1}{2} - x_5\right)a\,\hat{\mathbf{y}} + z_5a\,\hat{\mathbf{z}} \qquad (24k) \qquad \text{O III}$$

$$\mathbf{B}_{43} = \left(\frac{1}{2} - x_5\right)\mathbf{a}_1 + x_5\,\mathbf{a}_2 + \left(\frac{1}{2} - z_5\right)\mathbf{a}_3 = \left(\frac{1}{2} - x_5\right)a\,\hat{\mathbf{x}} + x_5a\,\hat{\mathbf{y}} + \left(\frac{1}{2} - z_5\right)a\,\hat{\mathbf{z}} \qquad (24k) \qquad \text{O III}$$

$$\mathbf{B}_{44} = x_5\,\mathbf{a}_1 + \left(\frac{1}{2} - x_5\right)\mathbf{a}_2 + \left(\frac{1}{2} - z_5\right)\mathbf{a}_3 = x_5a\,\hat{\mathbf{x}} + \left(\frac{1}{2} - x_5\right)a\,\hat{\mathbf{y}} + \left(\frac{1}{2} - z_5\right)a\,\hat{\mathbf{z}} \qquad (24k) \qquad \text{O III}$$

$$\mathbf{B}_{45} = z_5\,\mathbf{a}_1 + x_5\,\mathbf{a}_2 + x_5\,\mathbf{a}_3 = z_5a\,\hat{\mathbf{x}} + x_5a\,\hat{\mathbf{y}} + x_5a\,\hat{\mathbf{z}} \qquad (24k) \qquad \text{O III}$$

$$\mathbf{B}_{46} = z_5\,\mathbf{a}_1 + \left(\frac{1}{2} - x_5\right)\mathbf{a}_2 + \left(\frac{1}{2} - x_5\right)\mathbf{a}_3 = z_5a\,\hat{\mathbf{x}} + \left(\frac{1}{2} - x_5\right)a\,\hat{\mathbf{y}} + \left(\frac{1}{2} - x_5\right)a\,\hat{\mathbf{z}} \qquad (24k) \qquad \text{O III}$$

$$\mathbf{B}_{47} = \left(\frac{1}{2} - z_5\right)\mathbf{a}_1 + \left(\frac{1}{2} - x_5\right)\mathbf{a}_2 + x_5\,\mathbf{a}_3 = \left(\frac{1}{2} - z_5\right)a\,\hat{\mathbf{x}} + \left(\frac{1}{2} - x_5\right)a\,\hat{\mathbf{y}} + x_5a\,\hat{\mathbf{z}} \qquad (24k) \qquad \text{O III}$$

$$\mathbf{B}_{48} = \left(\frac{1}{2} - z_5\right)\mathbf{a}_1 + x_5\,\mathbf{a}_2 + \left(\frac{1}{2} - x_5\right)\mathbf{a}_3 = \left(\frac{1}{2} - z_5\right)a\,\hat{\mathbf{x}} + x_5a\,\hat{\mathbf{y}} + \left(\frac{1}{2} - x_5\right)a\,\hat{\mathbf{z}} \qquad (24k) \qquad \text{O III}$$

$$\mathbf{B}_{49} = x_5\,\mathbf{a}_1 + z_5\,\mathbf{a}_2 + x_5\,\mathbf{a}_3 = x_5a\,\hat{\mathbf{x}} + z_5a\,\hat{\mathbf{y}} + x_5a\,\hat{\mathbf{z}} \qquad (24k) \qquad \text{O III}$$

$$\mathbf{B}_{50} = \left(\frac{1}{2} - x_5\right)\mathbf{a}_1 + z_5\,\mathbf{a}_2 + \left(\frac{1}{2} - x_5\right)\mathbf{a}_3 = \left(\frac{1}{2} - x_5\right)a\,\hat{\mathbf{x}} + z_5a\,\hat{\mathbf{y}} + \left(\frac{1}{2} - x_5\right)a\,\hat{\mathbf{z}} \qquad (24k) \qquad \text{O III}$$

$$\mathbf{B}_{51} = x_5\,\mathbf{a}_1 + \left(\frac{1}{2} - z_5\right)\mathbf{a}_2 + \left(\frac{1}{2} - x_5\right)\mathbf{a}_3 = x_5a\,\hat{\mathbf{x}} + \left(\frac{1}{2} - z_5\right)a\,\hat{\mathbf{y}} + \left(\frac{1}{2} - x_5\right)a\,\hat{\mathbf{z}} \qquad (24k) \qquad \text{O III}$$

$$\mathbf{B}_{52} = \left(\frac{1}{2} - x_5\right)\mathbf{a}_1 + \left(\frac{1}{2} - z_5\right)\mathbf{a}_2 + x_5\,\mathbf{a}_3 = \left(\frac{1}{2} - x_5\right)a\,\hat{\mathbf{x}} + \left(\frac{1}{2} - z_5\right)a\,\hat{\mathbf{y}} + x_5a\,\hat{\mathbf{z}} \qquad (24k) \qquad \text{O III}$$

$$\mathbf{B}_{53} = \left(\frac{1}{2} + x_5\right)\mathbf{a}_1 + \left(\frac{1}{2} + x_5\right)\mathbf{a}_2 - z_5\,\mathbf{a}_3 = \left(\frac{1}{2} + x_5\right)a\,\hat{\mathbf{x}} + \left(\frac{1}{2} + x_5\right)a\,\hat{\mathbf{y}} - z_5a\,\hat{\mathbf{z}} \qquad (24k) \qquad \text{O III}$$

$$\mathbf{B}_{54} = -x_5\,\mathbf{a}_1 - x_5\,\mathbf{a}_2 - z_5\,\mathbf{a}_3 = -x_5a\,\hat{\mathbf{x}} - x_5a\,\hat{\mathbf{y}} - z_5a\,\hat{\mathbf{z}} \qquad (24k) \qquad \text{O III}$$

$$\mathbf{B}_{55} = \left(\frac{1}{2} + x_5\right)\mathbf{a}_1 - x_5\,\mathbf{a}_2 + \left(\frac{1}{2} + z_5\right)\mathbf{a}_3 = \left(\frac{1}{2} + x_5\right)a\,\hat{\mathbf{x}} - x_5a\,\hat{\mathbf{y}} + \left(\frac{1}{2} + z_5\right)a\,\hat{\mathbf{z}} \qquad (24k) \qquad \text{O III}$$

$$\mathbf{B}_{56} = -x_5\,\mathbf{a}_1 + \left(\frac{1}{2} + x_5\right)\mathbf{a}_2 + \left(\frac{1}{2} + z_5\right)\mathbf{a}_3 = -x_5a\,\hat{\mathbf{x}} + \left(\frac{1}{2} + x_5\right)a\,\hat{\mathbf{y}} + \left(\frac{1}{2} + z_5\right)a\,\hat{\mathbf{z}} \qquad (24k) \qquad \text{O III}$$

$$\mathbf{B}_{57} = \left(\frac{1}{2} + x_5\right)\mathbf{a}_1 + \left(\frac{1}{2} + z_5\right)\mathbf{a}_2 - x_5\,\mathbf{a}_3 = \left(\frac{1}{2} + x_5\right)a\,\hat{\mathbf{x}} + \left(\frac{1}{2} + z_5\right)a\,\hat{\mathbf{y}} - x_5a\,\hat{\mathbf{z}} \qquad (24k) \qquad \text{O III}$$

$$\mathbf{B}_{58} = -x_5\,\mathbf{a}_1 + \left(\frac{1}{2} + z_5\right)\mathbf{a}_2 + \left(\frac{1}{2} + x_5\right)\mathbf{a}_3 = -x_5a\,\hat{\mathbf{x}} + \left(\frac{1}{2} + z_5\right)a\,\hat{\mathbf{y}} + \left(\frac{1}{2} + x_5\right)a\,\hat{\mathbf{z}} \qquad (24k) \qquad \text{O III}$$

$$\mathbf{B}_{59} = -x_5\,\mathbf{a}_1 - z_5\,\mathbf{a}_2 - x_5\,\mathbf{a}_3 = -x_5a\,\hat{\mathbf{x}} - z_5a\,\hat{\mathbf{y}} - x_5a\,\hat{\mathbf{z}} \qquad (24k) \qquad \text{O III}$$

$$\mathbf{B}_{60} = \left(\frac{1}{2} + x_5\right)\mathbf{a}_1 - z_5\,\mathbf{a}_2 + \left(\frac{1}{2} + x_5\right)\mathbf{a}_3 = \left(\frac{1}{2} + x_5\right)a\,\hat{\mathbf{x}} - z_5a\,\hat{\mathbf{y}} + \left(\frac{1}{2} + x_5\right)a\,\hat{\mathbf{z}} \qquad (24k) \qquad \text{O III}$$

$$\mathbf{B}_{61} = \left(\frac{1}{2} + z_5\right)\mathbf{a}_1 + \left(\frac{1}{2} + x_5\right)\mathbf{a}_2 - x_5\,\mathbf{a}_3 = \left(\frac{1}{2} + z_5\right)a\,\hat{\mathbf{x}} + \left(\frac{1}{2} + x_5\right)a\,\hat{\mathbf{y}} - x_5a\,\hat{\mathbf{z}} \qquad (24k) \qquad \text{O III}$$

$$\mathbf{B}_{62} = \left(\frac{1}{2} + z_5\right)\mathbf{a}_1 - x_5\,\mathbf{a}_2 + \left(\frac{1}{2} + x_5\right)\mathbf{a}_3 = \left(\frac{1}{2} + z_5\right)a\,\hat{\mathbf{x}} - x_5a\,\hat{\mathbf{y}} + \left(\frac{1}{2} + x_5\right)a\,\hat{\mathbf{z}} \qquad (24k) \qquad \text{O III}$$

$$\mathbf{B}_{63} = -z_5\,\mathbf{a}_1 + \left(\frac{1}{2} + x_5\right)\mathbf{a}_2 + \left(\frac{1}{2} + x_5\right)\mathbf{a}_3 = -z_5a\,\hat{\mathbf{x}} + \left(\frac{1}{2} + x_5\right)a\,\hat{\mathbf{y}} + \left(\frac{1}{2} + x_5\right)a\,\hat{\mathbf{z}} \qquad (24k) \qquad \text{O III}$$

$$\mathbf{B}_{64} = -z_5\,\mathbf{a}_1 - x_5\,\mathbf{a}_2 - x_5\,\mathbf{a}_3 = -z_5a\,\hat{\mathbf{x}} - x_5a\,\hat{\mathbf{y}} - x_5a\,\hat{\mathbf{z}} \qquad (24k) \qquad \text{O III}$$

$$\mathbf{B}_{65} = x_6\,\mathbf{a}_1 + x_6\,\mathbf{a}_2 + z_6\,\mathbf{a}_3 = x_6a\,\hat{\mathbf{x}} + x_6a\,\hat{\mathbf{y}} + z_6a\,\hat{\mathbf{z}} \qquad (24k) \qquad \text{O IV}$$

$$\mathbf{B}_{66} = \left(\frac{1}{2} - x_6\right)\mathbf{a}_1 + \left(\frac{1}{2} - x_6\right)\mathbf{a}_2 + z_6\,\mathbf{a}_3 = \left(\frac{1}{2} - x_6\right)a\,\hat{\mathbf{x}} + \left(\frac{1}{2} - x_6\right)a\,\hat{\mathbf{y}} + z_6a\,\hat{\mathbf{z}} \qquad (24k) \qquad \text{O IV}$$

$$\mathbf{B}_{67} = \left(\frac{1}{2} - x_6\right)\mathbf{a}_1 + x_6\,\mathbf{a}_2 + \left(\frac{1}{2} - z_6\right)\mathbf{a}_3 = \left(\frac{1}{2} - x_6\right)a\,\hat{\mathbf{x}} + x_6a\,\hat{\mathbf{y}} + \left(\frac{1}{2} - z_6\right)a\,\hat{\mathbf{z}} \qquad (24k) \qquad \text{O IV}$$

$$\mathbf{B}_{68} = x_6\,\mathbf{a}_1 + \left(\frac{1}{2} - x_6\right)\mathbf{a}_2 + \left(\frac{1}{2} - z_6\right)\mathbf{a}_3 = x_6a\,\hat{\mathbf{x}} + \left(\frac{1}{2} - x_6\right)a\,\hat{\mathbf{y}} + \left(\frac{1}{2} - z_6\right)a\,\hat{\mathbf{z}} \qquad (24k) \qquad \text{O IV}$$

$$\mathbf{B}_{69} = z_6\,\mathbf{a}_1 + x_6\,\mathbf{a}_2 + x_6\,\mathbf{a}_3 = z_6a\,\hat{\mathbf{x}} + x_6a\,\hat{\mathbf{y}} + x_6a\,\hat{\mathbf{z}} \qquad (24k) \qquad \text{O IV}$$

$$\mathbf{B}_{70} = z_6\,\mathbf{a}_1 + \left(\frac{1}{2} - x_6\right)\mathbf{a}_2 + \left(\frac{1}{2} - x_6\right)\mathbf{a}_3 = z_6a\,\hat{\mathbf{x}} + \left(\frac{1}{2} - x_6\right)a\,\hat{\mathbf{y}} + \left(\frac{1}{2} - x_6\right)a\,\hat{\mathbf{z}} \qquad (24k) \qquad \text{O IV}$$

| | | | | | | |
|---|---|---|---|---|---|---|
| $\mathbf{B}_{71}$ | = | $\left(\frac{1}{2} - z_6\right) \mathbf{a}_1 + \left(\frac{1}{2} - x_6\right) \mathbf{a}_2 + x_6 \mathbf{a}_3$ | = | $\left(\frac{1}{2} - z_6\right) a \hat{\mathbf{x}} + \left(\frac{1}{2} - x_6\right) a \hat{\mathbf{y}} + x_6 a \hat{\mathbf{z}}$ | (24$k$) | O IV |
| $\mathbf{B}_{72}$ | = | $\left(\frac{1}{2} - z_6\right) \mathbf{a}_1 + x_6 \mathbf{a}_2 + \left(\frac{1}{2} - x_6\right) \mathbf{a}_3$ | = | $\left(\frac{1}{2} - z_6\right) a \hat{\mathbf{x}} + x_6 a \hat{\mathbf{y}} + \left(\frac{1}{2} - x_6\right) a \hat{\mathbf{z}}$ | (24$k$) | O IV |
| $\mathbf{B}_{73}$ | = | $x_6 \mathbf{a}_1 + z_6 \mathbf{a}_2 + x_6 \mathbf{a}_3$ | = | $x_6 a \hat{\mathbf{x}} + z_6 a \hat{\mathbf{y}} + x_6 a \hat{\mathbf{z}}$ | (24$k$) | O IV |
| $\mathbf{B}_{74}$ | = | $\left(\frac{1}{2} - x_6\right) \mathbf{a}_1 + z_6 \mathbf{a}_2 + \left(\frac{1}{2} - x_6\right) \mathbf{a}_3$ | = | $\left(\frac{1}{2} - x_6\right) a \hat{\mathbf{x}} + z_6 a \hat{\mathbf{y}} + \left(\frac{1}{2} - x_6\right) a \hat{\mathbf{z}}$ | (24$k$) | O IV |
| $\mathbf{B}_{75}$ | = | $x_6 \mathbf{a}_1 + \left(\frac{1}{2} - z_6\right) \mathbf{a}_2 + \left(\frac{1}{2} - x_6\right) \mathbf{a}_3$ | = | $x_6 a \hat{\mathbf{x}} + \left(\frac{1}{2} - z_6\right) a \hat{\mathbf{y}} + \left(\frac{1}{2} - x_6\right) a \hat{\mathbf{z}}$ | (24$k$) | O IV |
| $\mathbf{B}_{76}$ | = | $\left(\frac{1}{2} - x_6\right) \mathbf{a}_1 + \left(\frac{1}{2} - z_6\right) \mathbf{a}_2 + x_6 \mathbf{a}_3$ | = | $\left(\frac{1}{2} - x_6\right) a \hat{\mathbf{x}} + \left(\frac{1}{2} - z_6\right) a \hat{\mathbf{y}} + x_6 a \hat{\mathbf{z}}$ | (24$k$) | O IV |
| $\mathbf{B}_{77}$ | = | $\left(\frac{1}{2} + x_6\right) \mathbf{a}_1 + \left(\frac{1}{2} + x_6\right) \mathbf{a}_2 - z_6 \mathbf{a}_3$ | = | $\left(\frac{1}{2} + x_6\right) a \hat{\mathbf{x}} + \left(\frac{1}{2} + x_6\right) a \hat{\mathbf{y}} - z_6 a \hat{\mathbf{z}}$ | (24$k$) | O IV |
| $\mathbf{B}_{78}$ | = | $-x_6 \mathbf{a}_1 - x_6 \mathbf{a}_2 - z_6 \mathbf{a}_3$ | = | $-x_6 a \hat{\mathbf{x}} - x_6 a \hat{\mathbf{y}} - z_6 a \hat{\mathbf{z}}$ | (24$k$) | O IV |
| $\mathbf{B}_{79}$ | = | $\left(\frac{1}{2} + x_6\right) \mathbf{a}_1 - x_6 \mathbf{a}_2 + \left(\frac{1}{2} + z_6\right) \mathbf{a}_3$ | = | $\left(\frac{1}{2} + x_6\right) a \hat{\mathbf{x}} - x_6 a \hat{\mathbf{y}} + \left(\frac{1}{2} + z_6\right) a \hat{\mathbf{z}}$ | (24$k$) | O IV |
| $\mathbf{B}_{80}$ | = | $-x_6 \mathbf{a}_1 + \left(\frac{1}{2} + x_6\right) \mathbf{a}_2 + \left(\frac{1}{2} + z_6\right) \mathbf{a}_3$ | = | $-x_6 a \hat{\mathbf{x}} + \left(\frac{1}{2} + x_6\right) a \hat{\mathbf{y}} + \left(\frac{1}{2} + z_6\right) a \hat{\mathbf{z}}$ | (24$k$) | O IV |
| $\mathbf{B}_{81}$ | = | $\left(\frac{1}{2} + x_6\right) \mathbf{a}_1 + \left(\frac{1}{2} + z_6\right) \mathbf{a}_2 - x_6 \mathbf{a}_3$ | = | $\left(\frac{1}{2} + x_6\right) a \hat{\mathbf{x}} + \left(\frac{1}{2} + z_6\right) a \hat{\mathbf{y}} - x_6 a \hat{\mathbf{z}}$ | (24$k$) | O IV |
| $\mathbf{B}_{82}$ | = | $-x_6 \mathbf{a}_1 + \left(\frac{1}{2} + z_6\right) \mathbf{a}_2 + \left(\frac{1}{2} + x_6\right) \mathbf{a}_3$ | = | $-x_6 a \hat{\mathbf{x}} + \left(\frac{1}{2} + z_6\right) a \hat{\mathbf{y}} + \left(\frac{1}{2} + x_6\right) a \hat{\mathbf{z}}$ | (24$k$) | O IV |
| $\mathbf{B}_{83}$ | = | $-x_6 \mathbf{a}_1 - z_6 \mathbf{a}_2 - x_6 \mathbf{a}_3$ | = | $-x_6 a \hat{\mathbf{x}} - z_6 a \hat{\mathbf{y}} - x_6 a \hat{\mathbf{z}}$ | (24$k$) | O IV |
| $\mathbf{B}_{84}$ | = | $\left(\frac{1}{2} + x_6\right) \mathbf{a}_1 - z_6 \mathbf{a}_2 + \left(\frac{1}{2} + x_6\right) \mathbf{a}_3$ | = | $\left(\frac{1}{2} + x_6\right) a \hat{\mathbf{x}} - z_6 a \hat{\mathbf{y}} + \left(\frac{1}{2} + x_6\right) a \hat{\mathbf{z}}$ | (24$k$) | O IV |
| $\mathbf{B}_{85}$ | = | $\left(\frac{1}{2} + z_6\right) \mathbf{a}_1 + \left(\frac{1}{2} + x_6\right) \mathbf{a}_2 - x_6 \mathbf{a}_3$ | = | $\left(\frac{1}{2} + z_6\right) a \hat{\mathbf{x}} + \left(\frac{1}{2} + x_6\right) a \hat{\mathbf{y}} - x_6 a \hat{\mathbf{z}}$ | (24$k$) | O IV |
| $\mathbf{B}_{86}$ | = | $\left(\frac{1}{2} + z_6\right) \mathbf{a}_1 - x_6 \mathbf{a}_2 + \left(\frac{1}{2} + x_6\right) \mathbf{a}_3$ | = | $\left(\frac{1}{2} + z_6\right) a \hat{\mathbf{x}} - x_6 a \hat{\mathbf{y}} + \left(\frac{1}{2} + x_6\right) a \hat{\mathbf{z}}$ | (24$k$) | O IV |
| $\mathbf{B}_{87}$ | = | $-z_6 \mathbf{a}_1 + \left(\frac{1}{2} + x_6\right) \mathbf{a}_2 + \left(\frac{1}{2} + x_6\right) \mathbf{a}_3$ | = | $-z_6 a \hat{\mathbf{x}} + \left(\frac{1}{2} + x_6\right) a \hat{\mathbf{y}} + \left(\frac{1}{2} + x_6\right) a \hat{\mathbf{z}}$ | (24$k$) | O IV |
| $\mathbf{B}_{88}$ | = | $-z_6 \mathbf{a}_1 - x_6 \mathbf{a}_2 - x_6 \mathbf{a}_3$ | = | $-z_6 a \hat{\mathbf{x}} - x_6 a \hat{\mathbf{y}} - x_6 a \hat{\mathbf{z}}$ | (24$k$) | O IV |
| $\mathbf{B}_{89}$ | = | $x_7 \mathbf{a}_1 + x_7 \mathbf{a}_2 + z_7 \mathbf{a}_3$ | = | $x_7 a \hat{\mathbf{x}} + x_7 a \hat{\mathbf{y}} + z_7 a \hat{\mathbf{z}}$ | (24$k$) | W |
| $\mathbf{B}_{90}$ | = | $\left(\frac{1}{2} - x_7\right) \mathbf{a}_1 + \left(\frac{1}{2} - x_7\right) \mathbf{a}_2 + z_7 \mathbf{a}_3$ | = | $\left(\frac{1}{2} - x_7\right) a \hat{\mathbf{x}} + \left(\frac{1}{2} - x_7\right) a \hat{\mathbf{y}} + z_7 a \hat{\mathbf{z}}$ | (24$k$) | W |
| $\mathbf{B}_{91}$ | = | $\left(\frac{1}{2} - x_7\right) \mathbf{a}_1 + x_7 \mathbf{a}_2 + \left(\frac{1}{2} - z_7\right) \mathbf{a}_3$ | = | $\left(\frac{1}{2} - x_7\right) a \hat{\mathbf{x}} + x_7 a \hat{\mathbf{y}} + \left(\frac{1}{2} - z_7\right) a \hat{\mathbf{z}}$ | (24$k$) | W |
| $\mathbf{B}_{92}$ | = | $x_7 \mathbf{a}_1 + \left(\frac{1}{2} - x_7\right) \mathbf{a}_2 + \left(\frac{1}{2} - z_7\right) \mathbf{a}_3$ | = | $x_7 a \hat{\mathbf{x}} + \left(\frac{1}{2} - x_7\right) a \hat{\mathbf{y}} + \left(\frac{1}{2} - z_7\right) a \hat{\mathbf{z}}$ | (24$k$) | W |
| $\mathbf{B}_{93}$ | = | $z_7 \mathbf{a}_1 + x_7 \mathbf{a}_2 + x_7 \mathbf{a}_3$ | = | $z_7 a \hat{\mathbf{x}} + x_7 a \hat{\mathbf{y}} + x_7 a \hat{\mathbf{z}}$ | (24$k$) | W |
| $\mathbf{B}_{94}$ | = | $z_7 \mathbf{a}_1 + \left(\frac{1}{2} - x_7\right) \mathbf{a}_2 + \left(\frac{1}{2} - x_7\right) \mathbf{a}_3$ | = | $z_7 a \hat{\mathbf{x}} + \left(\frac{1}{2} - x_7\right) a \hat{\mathbf{y}} + \left(\frac{1}{2} - x_7\right) a \hat{\mathbf{z}}$ | (24$k$) | W |
| $\mathbf{B}_{95}$ | = | $\left(\frac{1}{2} - z_7\right) \mathbf{a}_1 + \left(\frac{1}{2} - x_7\right) \mathbf{a}_2 + x_7 \mathbf{a}_3$ | = | $\left(\frac{1}{2} - z_7\right) a \hat{\mathbf{x}} + \left(\frac{1}{2} - x_7\right) a \hat{\mathbf{y}} + x_7 a \hat{\mathbf{z}}$ | (24$k$) | W |
| $\mathbf{B}_{96}$ | = | $\left(\frac{1}{2} - z_7\right) \mathbf{a}_1 + x_7 \mathbf{a}_2 + \left(\frac{1}{2} - x_7\right) \mathbf{a}_3$ | = | $\left(\frac{1}{2} - z_7\right) a \hat{\mathbf{x}} + x_7 a \hat{\mathbf{y}} + \left(\frac{1}{2} - x_7\right) a \hat{\mathbf{z}}$ | (24$k$) | W |
| $\mathbf{B}_{97}$ | = | $x_7 \mathbf{a}_1 + z_7 \mathbf{a}_2 + x_7 \mathbf{a}_3$ | = | $x_7 a \hat{\mathbf{x}} + z_7 a \hat{\mathbf{y}} + x_7 a \hat{\mathbf{z}}$ | (24$k$) | W |
| $\mathbf{B}_{98}$ | = | $\left(\frac{1}{2} - x_7\right) \mathbf{a}_1 + z_7 \mathbf{a}_2 + \left(\frac{1}{2} - x_7\right) \mathbf{a}_3$ | = | $\left(\frac{1}{2} - x_7\right) a \hat{\mathbf{x}} + z_7 a \hat{\mathbf{y}} + \left(\frac{1}{2} - x_7\right) a \hat{\mathbf{z}}$ | (24$k$) | W |
| $\mathbf{B}_{99}$ | = | $x_7 \mathbf{a}_1 + \left(\frac{1}{2} - z_7\right) \mathbf{a}_2 + \left(\frac{1}{2} - x_7\right) \mathbf{a}_3$ | = | $x_7 a \hat{\mathbf{x}} + \left(\frac{1}{2} - z_7\right) a \hat{\mathbf{y}} + \left(\frac{1}{2} - x_7\right) a \hat{\mathbf{z}}$ | (24$k$) | W |
| $\mathbf{B}_{100}$ | = | $\left(\frac{1}{2} - x_7\right) \mathbf{a}_1 + \left(\frac{1}{2} - z_7\right) \mathbf{a}_2 + x_7 \mathbf{a}_3$ | = | $\left(\frac{1}{2} - x_7\right) a \hat{\mathbf{x}} + \left(\frac{1}{2} - z_7\right) a \hat{\mathbf{y}} + x_7 a \hat{\mathbf{z}}$ | (24$k$) | W |
| $\mathbf{B}_{101}$ | = | $\left(\frac{1}{2} + x_7\right) \mathbf{a}_1 + \left(\frac{1}{2} + x_7\right) \mathbf{a}_2 - z_7 \mathbf{a}_3$ | = | $\left(\frac{1}{2} + x_7\right) a \hat{\mathbf{x}} + \left(\frac{1}{2} + x_7\right) a \hat{\mathbf{y}} - z_7 a \hat{\mathbf{z}}$ | (24$k$) | W |
| $\mathbf{B}_{102}$ | = | $-x_7 \mathbf{a}_1 - x_7 \mathbf{a}_2 - z_7 \mathbf{a}_3$ | = | $-x_7 a \hat{\mathbf{x}} - x_7 a \hat{\mathbf{y}} - z_7 a \hat{\mathbf{z}}$ | (24$k$) | W |
| $\mathbf{B}_{103}$ | = | $\left(\frac{1}{2} + x_7\right) \mathbf{a}_1 - x_7 \mathbf{a}_2 + \left(\frac{1}{2} + z_7\right) \mathbf{a}_3$ | = | $\left(\frac{1}{2} + x_7\right) a \hat{\mathbf{x}} - x_7 a \hat{\mathbf{y}} + \left(\frac{1}{2} + z_7\right) a \hat{\mathbf{z}}$ | (24$k$) | W |
| $\mathbf{B}_{104}$ | = | $-x_7 \mathbf{a}_1 + \left(\frac{1}{2} + x_7\right) \mathbf{a}_2 + \left(\frac{1}{2} + z_7\right) \mathbf{a}_3$ | = | $-x_7 a \hat{\mathbf{x}} + \left(\frac{1}{2} + x_7\right) a \hat{\mathbf{y}} + \left(\frac{1}{2} + z_7\right) a \hat{\mathbf{z}}$ | (24$k$) | W |
| $\mathbf{B}_{105}$ | = | $\left(\frac{1}{2} + x_7\right) \mathbf{a}_1 + \left(\frac{1}{2} + z_7\right) \mathbf{a}_2 - x_7 \mathbf{a}_3$ | = | $\left(\frac{1}{2} + x_7\right) a \hat{\mathbf{x}} + \left(\frac{1}{2} + z_7\right) a \hat{\mathbf{y}} - x_7 a \hat{\mathbf{z}}$ | (24$k$) | W |
| $\mathbf{B}_{106}$ | = | $-x_7 \mathbf{a}_1 + \left(\frac{1}{2} + z_7\right) \mathbf{a}_2 + \left(\frac{1}{2} + x_7\right) \mathbf{a}_3$ | = | $-x_7 a \hat{\mathbf{x}} + \left(\frac{1}{2} + z_7\right) a \hat{\mathbf{y}} + \left(\frac{1}{2} + x_7\right) a \hat{\mathbf{z}}$ | (24$k$) | W |
| $\mathbf{B}_{107}$ | = | $-x_7\,\mathbf{a}_1 - z_7\,\mathbf{a}_2 - x_7\,\mathbf{a}_3$ | = | $-x_7a\,\mathbf{\hat{x}} - z_7a\,\mathbf{\hat{y}} - x_7a\,\mathbf{\hat{z}}$ | (24$k$) | W |
| $\mathbf{B}_{108}$ | = | $\left(\frac{1}{2}+x_7\right)\mathbf{a}_1 - z_7\,\mathbf{a}_2 + \left(\frac{1}{2}+x_7\right)\mathbf{a}_3$ | = | $\left(\frac{1}{2}+x_7\right)a\,\mathbf{\hat{x}} - z_7a\,\mathbf{\hat{y}} + \left(\frac{1}{2}+x_7\right)a\,\mathbf{\hat{z}}$ | (24$k$) | W |
| $\mathbf{B}_{109}$ | = | $\left(\frac{1}{2}+z_7\right)\mathbf{a}_1 + \left(\frac{1}{2}+x_7\right)\mathbf{a}_2 - x_7\,\mathbf{a}_3$ | = | $\left(\frac{1}{2}+z_7\right)a\,\mathbf{\hat{x}} + \left(\frac{1}{2}+x_7\right)a\,\mathbf{\hat{y}} - x_7a\,\mathbf{\hat{z}}$ | (24$k$) | W |
| $\mathbf{B}_{110}$ | = | $\left(\frac{1}{2}+z_7\right)\mathbf{a}_1 - x_7\,\mathbf{a}_2 + \left(\frac{1}{2}+x_7\right)\mathbf{a}_3$ | = | $\left(\frac{1}{2}+z_7\right)a\,\mathbf{\hat{x}} - x_7a\,\mathbf{\hat{y}} + \left(\frac{1}{2}+x_7\right)a\,\mathbf{\hat{z}}$ | (24$k$) | W |
| $\mathbf{B}_{111}$ | = | $-z_7\,\mathbf{a}_1 + \left(\frac{1}{2}+x_7\right)\mathbf{a}_2 + \left(\frac{1}{2}+x_7\right)\mathbf{a}_3$ | = | $-z_7a\,\mathbf{\hat{x}} + \left(\frac{1}{2}+x_7\right)a\,\mathbf{\hat{y}} + \left(\frac{1}{2}+x_7\right)a\,\mathbf{\hat{z}}$ | (24$k$) | W |
| $\mathbf{B}_{112}$ | = | $-z_7\,\mathbf{a}_1 - x_7\,\mathbf{a}_2 - x_7\,\mathbf{a}_3$ | = | $-z_7a\,\mathbf{\hat{x}} - x_7a\,\mathbf{\hat{y}} - x_7a\,\mathbf{\hat{z}}$ | (24$k$) | W |

**Geometry files:**

- CIF: pp. 1809
- POSCAR: pp. 1810

# 12-phosphotungstic acid [$H_3PW_{12}O_{40}{\cdot}5H_2O$ ($H4_{16}$)] Structure: A5B40CD12_cP116_224_cd_e3k_a_k

http://aflow.org/prototype-encyclopedia/A5B40CD12_cP116_224_cd_e3k_a_k

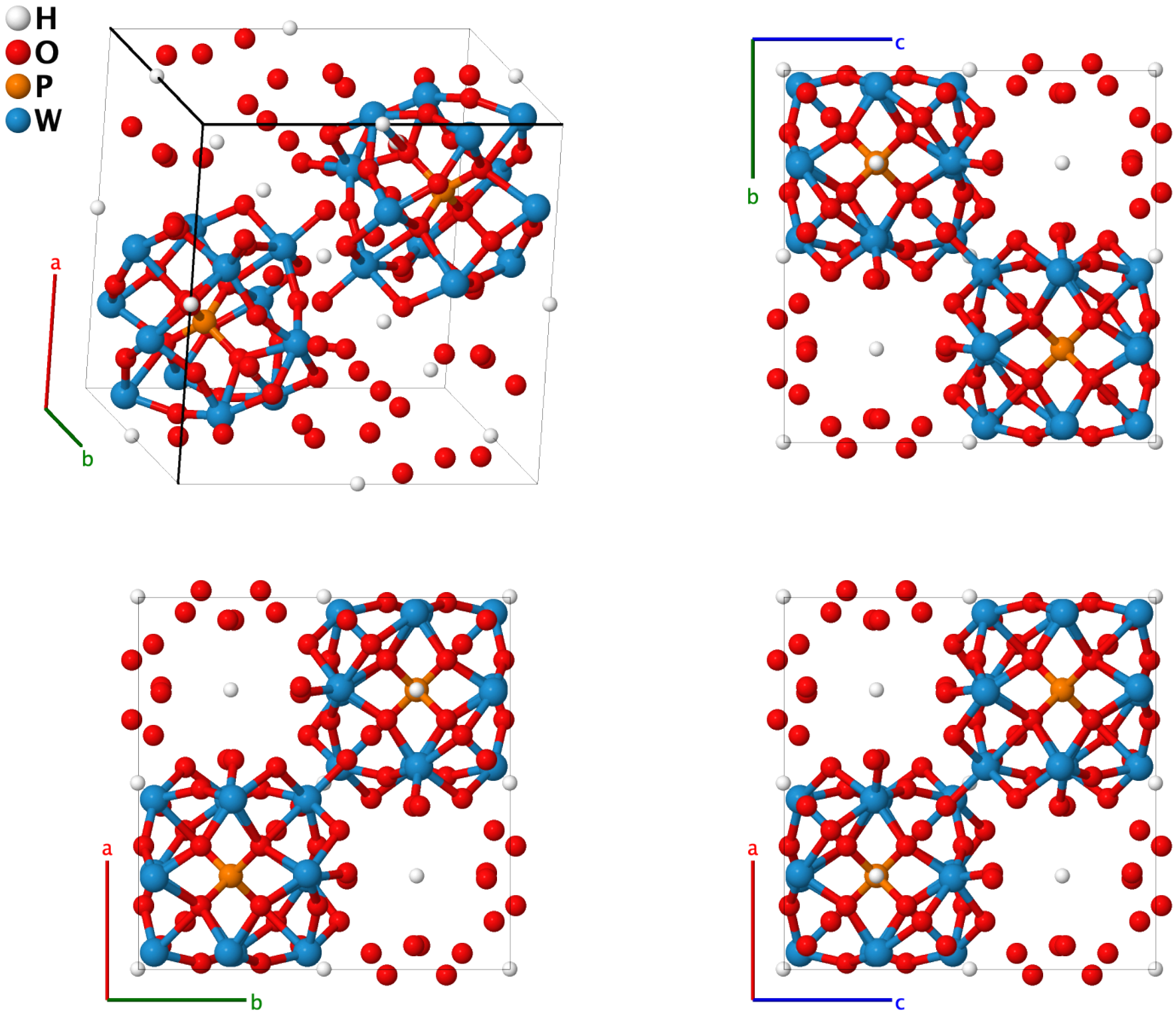

| | | |
|---|---|---|
| **Prototype** | : | $(H_{2.6}O)_5O_{40}PW_{12}$ |
| **AFLOW prototype label** | : | A5B40CD12_cP116_224_cd_e3k_a_k |
| ***Strukturbericht* designation** | : | $H4_{16}$ |
| **Pearson symbol** | : | cP116 |
| **Space group number** | : | 224 |
| **Space group symbol** | : | $Pn\bar{3}m$ |
| **AFLOW prototype command** | : | `aflow --proto=A5B40CD12_cP116_224_cd_e3k_a_k --params=`$a, x_4, x_5, z_5, x_6, z_6, x_7, z_7, x_8, z_8$ |

- This structure is a bit problematic. (Brown, 1977) offers evidence that the true structure of the hydrated phosphotungstic acid is $H_3PW_{12}O_{40}{\cdot}6H_2O$, rather than $5H_2O$ as deduced by (Keggin, 1934). The presence of the extra water molecule does change the structure somewhat, and Keggin does not give the positions of the hydrogen atoms not attached to water molecules. Ordinarily we would mark this structure obsolete, but it is possible to remove water molecules from this structure and still have a molecule that is recognizably related to the acid, as in $H_3PW_{12}O_{40}{\cdot}3H_2O$. In addition, the lattice constant reported by Keggin is 12.141 Å compared to Brown's 12.506 Å, a change consistent with the loss of a water molecule. Given this, we will not deprecate this five-water molecule structure.

- (Gottfried, 1937) does not give the positions of the water molecules, and they reverse the $x$ and $z$ coordinates for O-IV and W.

- This structure is a partially dehydrated form of $H_3PW_{12}O_{40}{\cdot}29H_2O$ ($H4_{21}$). Further dehydration produces the $H_3PW_{12}O_{40}{\cdot}3H_2O$ structure.

- The positions of the three hydrogen atoms are unknown. If we follow (Marosi, 2000), then the 'free' hydrogens are likely to be bound to some of the water molecules, giving the unusual stoichiometry in the prototype.

---

**Simple Cubic primitive vectors:**

$$\mathbf{a}_1 = a\,\hat{\mathbf{x}}$$
$$\mathbf{a}_2 = a\,\hat{\mathbf{y}}$$
$$\mathbf{a}_3 = a\,\hat{\mathbf{z}}$$

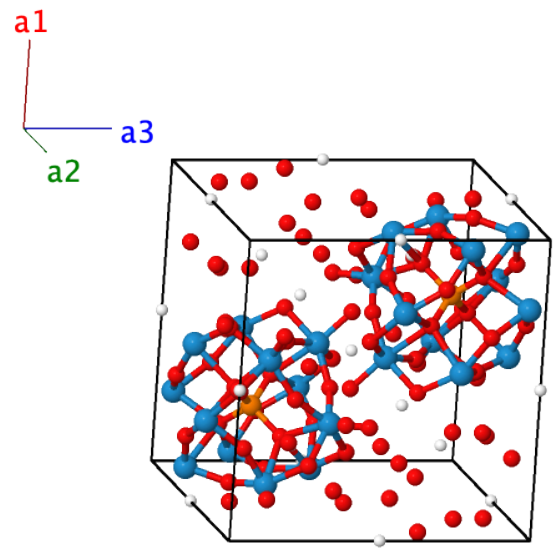

---

**Basis vectors:**

| | | Lattice Coordinates | | Cartesian Coordinates | Wyckoff Position | Atom Type |
|---|---|---|---|---|---|---|
| $\mathbf{B}_1$ | = | $\frac{1}{4}\mathbf{a}_1 + \frac{1}{4}\mathbf{a}_2 + \frac{1}{4}\mathbf{a}_3$ | = | $\frac{1}{4}a\,\hat{\mathbf{x}} + \frac{1}{4}a\,\hat{\mathbf{y}} + \frac{1}{4}a\,\hat{\mathbf{z}}$ | (2$a$) | P |
| $\mathbf{B}_2$ | = | $\frac{3}{4}\mathbf{a}_1 + \frac{3}{4}\mathbf{a}_2 + \frac{3}{4}\mathbf{a}_3$ | = | $\frac{3}{4}a\,\hat{\mathbf{x}} + \frac{3}{4}a\,\hat{\mathbf{y}} + \frac{3}{4}a\,\hat{\mathbf{z}}$ | (2$a$) | P |
| $\mathbf{B}_3$ | = | $\frac{1}{2}\mathbf{a}_1 + \frac{1}{2}\mathbf{a}_2 + \frac{1}{2}\mathbf{a}_3$ | = | $\frac{1}{2}a\,\hat{\mathbf{x}} + \frac{1}{2}a\,\hat{\mathbf{y}} + \frac{1}{2}a\,\hat{\mathbf{z}}$ | (4$c$) | $H_2O$ I |
| $\mathbf{B}_4$ | = | $\frac{1}{2}\mathbf{a}_3$ | = | $\frac{1}{2}a\,\hat{\mathbf{z}}$ | (4$c$) | $H_2O$ I |
| $\mathbf{B}_5$ | = | $\frac{1}{2}\mathbf{a}_2$ | = | $\frac{1}{2}a\,\hat{\mathbf{y}}$ | (4$c$) | $H_2O$ I |
| $\mathbf{B}_6$ | = | $\frac{1}{2}\mathbf{a}_1$ | = | $\frac{1}{2}a\,\hat{\mathbf{x}}$ | (4$c$) | $H_2O$ I |
| $\mathbf{B}_7$ | = | $\frac{1}{4}\mathbf{a}_1 + \frac{3}{4}\mathbf{a}_2 + \frac{3}{4}\mathbf{a}_3$ | = | $\frac{1}{4}a\,\hat{\mathbf{x}} + \frac{3}{4}a\,\hat{\mathbf{y}} + \frac{3}{4}a\,\hat{\mathbf{z}}$ | (6$d$) | $H_2O$ II |
| $\mathbf{B}_8$ | = | $\frac{3}{4}\mathbf{a}_1 + \frac{1}{4}\mathbf{a}_2 + \frac{3}{4}\mathbf{a}_3$ | = | $\frac{3}{4}a\,\hat{\mathbf{x}} + \frac{1}{4}a\,\hat{\mathbf{y}} + \frac{3}{4}a\,\hat{\mathbf{z}}$ | (6$d$) | $H_2O$ II |
| $\mathbf{B}_9$ | = | $\frac{3}{4}\mathbf{a}_1 + \frac{3}{4}\mathbf{a}_2 + \frac{1}{4}\mathbf{a}_3$ | = | $\frac{3}{4}a\,\hat{\mathbf{x}} + \frac{3}{4}a\,\hat{\mathbf{y}} + \frac{1}{4}a\,\hat{\mathbf{z}}$ | (6$d$) | $H_2O$ II |
| $\mathbf{B}_{10}$ | = | $\frac{1}{4}\mathbf{a}_1 + \frac{3}{4}\mathbf{a}_2 + \frac{1}{4}\mathbf{a}_3$ | = | $\frac{1}{4}a\,\hat{\mathbf{x}} + \frac{3}{4}a\,\hat{\mathbf{y}} + \frac{1}{4}a\,\hat{\mathbf{z}}$ | (6$d$) | $H_2O$ II |
| $\mathbf{B}_{11}$ | = | $\frac{3}{4}\mathbf{a}_1 + \frac{1}{4}\mathbf{a}_2 + \frac{1}{4}\mathbf{a}_3$ | = | $\frac{3}{4}a\,\hat{\mathbf{x}} + \frac{1}{4}a\,\hat{\mathbf{y}} + \frac{1}{4}a\,\hat{\mathbf{z}}$ | (6$d$) | $H_2O$ II |
| $\mathbf{B}_{12}$ | = | $\frac{1}{4}\mathbf{a}_1 + \frac{1}{4}\mathbf{a}_2 + \frac{3}{4}\mathbf{a}_3$ | = | $\frac{1}{4}a\,\hat{\mathbf{x}} + \frac{1}{4}a\,\hat{\mathbf{y}} + \frac{3}{4}a\,\hat{\mathbf{z}}$ | (6$d$) | $H_2O$ II |
| $\mathbf{B}_{13}$ | = | $x_4\,\mathbf{a}_1 + x_4\,\mathbf{a}_2 + x_4\,\mathbf{a}_3$ | = | $x_4 a\,\hat{\mathbf{x}} + x_4 a\,\hat{\mathbf{y}} + x_4 a\,\hat{\mathbf{z}}$ | (8$e$) | O I |
| $\mathbf{B}_{14}$ | = | $\left(\frac{1}{2} - x_4\right)\mathbf{a}_1 + \left(\frac{1}{2} - x_4\right)\mathbf{a}_2 + x_4\,\mathbf{a}_3$ | = | $\left(\frac{1}{2} - x_4\right)a\,\hat{\mathbf{x}} + \left(\frac{1}{2} - x_4\right)a\,\hat{\mathbf{y}} + x_4 a\,\hat{\mathbf{z}}$ | (8$e$) | O I |
| $\mathbf{B}_{15}$ | = | $\left(\frac{1}{2} - x_4\right)\mathbf{a}_1 + x_4\,\mathbf{a}_2 + \left(\frac{1}{2} - x_4\right)\mathbf{a}_3$ | = | $\left(\frac{1}{2} - x_4\right)a\,\hat{\mathbf{x}} + x_4 a\,\hat{\mathbf{y}} + \left(\frac{1}{2} - x_4\right)a\,\hat{\mathbf{z}}$ | (8$e$) | O I |
| $\mathbf{B}_{16}$ | = | $x_4\,\mathbf{a}_1 + \left(\frac{1}{2} - x_4\right)\mathbf{a}_2 + \left(\frac{1}{2} - x_4\right)\mathbf{a}_3$ | = | $x_4 a\,\hat{\mathbf{x}} + \left(\frac{1}{2} - x_4\right)a\,\hat{\mathbf{y}} + \left(\frac{1}{2} - x_4\right)a\,\hat{\mathbf{z}}$ | (8$e$) | O I |
| $\mathbf{B}_{17}$ | = | $\left(\frac{1}{2} + x_4\right)\mathbf{a}_1 + \left(\frac{1}{2} + x_4\right)\mathbf{a}_2 - x_4\,\mathbf{a}_3$ | = | $\left(\frac{1}{2} + x_4\right)a\,\hat{\mathbf{x}} + \left(\frac{1}{2} + x_4\right)a\,\hat{\mathbf{y}} - x_4 a\,\hat{\mathbf{z}}$ | (8$e$) | O I |
| $\mathbf{B}_{18}$ | = | $-x_4\,\mathbf{a}_1 - x_4\,\mathbf{a}_2 - x_4\,\mathbf{a}_3$ | = | $-x_4 a\,\hat{\mathbf{x}} - x_4 a\,\hat{\mathbf{y}} - x_4 a\,\hat{\mathbf{z}}$ | (8$e$) | O I |
| $\mathbf{B}_{19}$ | = | $\left(\frac{1}{2} + x_4\right)\mathbf{a}_1 - x_4\,\mathbf{a}_2 + \left(\frac{1}{2} + x_4\right)\mathbf{a}_3$ | = | $\left(\frac{1}{2} + x_4\right)a\,\hat{\mathbf{x}} - x_4 a\,\hat{\mathbf{y}} + \left(\frac{1}{2} + x_4\right)a\,\hat{\mathbf{z}}$ | (8$e$) | O I |
| $\mathbf{B}_{20}$ | = | $-x_4\,\mathbf{a}_1 + \left(\frac{1}{2} + x_4\right)\mathbf{a}_2 + \left(\frac{1}{2} + x_4\right)\mathbf{a}_3$ | = | $-x_4 a\,\hat{\mathbf{x}} + \left(\frac{1}{2} + x_4\right)a\,\hat{\mathbf{y}} + \left(\frac{1}{2} + x_4\right)a\,\hat{\mathbf{z}}$ | (8$e$) | O I |
| $\mathbf{B}_{21}$ | = | $x_5\,\mathbf{a}_1 + x_5\,\mathbf{a}_2 + z_5\,\mathbf{a}_3$ | = | $x_5 a\,\hat{\mathbf{x}} + x_5 a\,\hat{\mathbf{y}} + z_5 a\,\hat{\mathbf{z}}$ | (24$k$) | O II |
| $\mathbf{B}_{22}$ | = | $\left(\frac{1}{2} - x_5\right)\mathbf{a}_1 + \left(\frac{1}{2} - x_5\right)\mathbf{a}_2 + z_5\,\mathbf{a}_3$ | = | $\left(\frac{1}{2} - x_5\right)a\,\hat{\mathbf{x}} + \left(\frac{1}{2} - x_5\right)a\,\hat{\mathbf{y}} + z_5 a\,\hat{\mathbf{z}}$ | (24$k$) | O II |
| $\mathbf{B}_{23}$ | = | $\left(\frac{1}{2} - x_5\right)\mathbf{a}_1 + x_5\,\mathbf{a}_2 + \left(\frac{1}{2} - z_5\right)\mathbf{a}_3$ | = | $\left(\frac{1}{2} - x_5\right)a\,\hat{\mathbf{x}} + x_5 a\,\hat{\mathbf{y}} + \left(\frac{1}{2} - z_5\right)a\,\hat{\mathbf{z}}$ | (24$k$) | O II |
| $\mathbf{B}_{24}$ | = | $x_5\,\mathbf{a}_1 + \left(\frac{1}{2} - x_5\right)\mathbf{a}_2 + \left(\frac{1}{2} - z_5\right)\mathbf{a}_3$ | = | $x_5 a\,\hat{\mathbf{x}} + \left(\frac{1}{2} - x_5\right)a\,\hat{\mathbf{y}} + \left(\frac{1}{2} - z_5\right)a\,\hat{\mathbf{z}}$ | (24$k$) | O II |
| $\mathbf{B}_{25}$ | = | $z_5\,\mathbf{a}_1 + x_5\,\mathbf{a}_2 + x_5\,\mathbf{a}_3$ | = | $z_5 a\,\hat{\mathbf{x}} + x_5 a\,\hat{\mathbf{y}} + x_5 a\,\hat{\mathbf{z}}$ | (24$k$) | O II |
| $\mathbf{B}_{26}$ | $=$ | $z_{5}\,\mathbf{a}_{1}+\left(\frac{1}{2}-x_{5}\right)\mathbf{a}_{2}+\left(\frac{1}{2}-x_{5}\right)\mathbf{a}_{3}$ | $=$ | $z_{5}a\,\hat{\mathbf{x}}+\left(\frac{1}{2}-x_{5}\right)a\,\hat{\mathbf{y}}+\left(\frac{1}{2}-x_{5}\right)a\,\hat{\mathbf{z}}$ | (24*k*) | O II |
| $\mathbf{B}_{27}$ | $=$ | $\left(\frac{1}{2}-z_{5}\right)\mathbf{a}_{1}+\left(\frac{1}{2}-x_{5}\right)\mathbf{a}_{2}+x_{5}\,\mathbf{a}_{3}$ | $=$ | $\left(\frac{1}{2}-z_{5}\right)a\,\hat{\mathbf{x}}+\left(\frac{1}{2}-x_{5}\right)a\,\hat{\mathbf{y}}+x_{5}a\,\hat{\mathbf{z}}$ | (24*k*) | O II |
| $\mathbf{B}_{28}$ | $=$ | $\left(\frac{1}{2}-z_{5}\right)\mathbf{a}_{1}+x_{5}\,\mathbf{a}_{2}+\left(\frac{1}{2}-x_{5}\right)\mathbf{a}_{3}$ | $=$ | $\left(\frac{1}{2}-z_{5}\right)a\,\hat{\mathbf{x}}+x_{5}a\,\hat{\mathbf{y}}+\left(\frac{1}{2}-x_{5}\right)a\,\hat{\mathbf{z}}$ | (24*k*) | O II |
| $\mathbf{B}_{29}$ | $=$ | $x_{5}\,\mathbf{a}_{1}+z_{5}\,\mathbf{a}_{2}+x_{5}\,\mathbf{a}_{3}$ | $=$ | $x_{5}a\,\hat{\mathbf{x}}+z_{5}a\,\hat{\mathbf{y}}+x_{5}a\,\hat{\mathbf{z}}$ | (24*k*) | O II |
| $\mathbf{B}_{30}$ | $=$ | $\left(\frac{1}{2}-x_{5}\right)\mathbf{a}_{1}+z_{5}\,\mathbf{a}_{2}+\left(\frac{1}{2}-x_{5}\right)\mathbf{a}_{3}$ | $=$ | $\left(\frac{1}{2}-x_{5}\right)a\,\hat{\mathbf{x}}+z_{5}a\,\hat{\mathbf{y}}+\left(\frac{1}{2}-x_{5}\right)a\,\hat{\mathbf{z}}$ | (24*k*) | O II |
| $\mathbf{B}_{31}$ | $=$ | $x_{5}\,\mathbf{a}_{1}+\left(\frac{1}{2}-z_{5}\right)\mathbf{a}_{2}+\left(\frac{1}{2}-x_{5}\right)\mathbf{a}_{3}$ | $=$ | $x_{5}a\,\hat{\mathbf{x}}+\left(\frac{1}{2}-z_{5}\right)a\,\hat{\mathbf{y}}+\left(\frac{1}{2}-x_{5}\right)a\,\hat{\mathbf{z}}$ | (24*k*) | O II |
| $\mathbf{B}_{32}$ | $=$ | $\left(\frac{1}{2}-x_{5}\right)\mathbf{a}_{1}+\left(\frac{1}{2}-z_{5}\right)\mathbf{a}_{2}+x_{5}\,\mathbf{a}_{3}$ | $=$ | $\left(\frac{1}{2}-x_{5}\right)a\,\hat{\mathbf{x}}+\left(\frac{1}{2}-z_{5}\right)a\,\hat{\mathbf{y}}+x_{5}a\,\hat{\mathbf{z}}$ | (24*k*) | O II |
| $\mathbf{B}_{33}$ | $=$ | $\left(\frac{1}{2}+x_{5}\right)\mathbf{a}_{1}+\left(\frac{1}{2}+x_{5}\right)\mathbf{a}_{2}-z_{5}\,\mathbf{a}_{3}$ | $=$ | $\left(\frac{1}{2}+x_{5}\right)a\,\hat{\mathbf{x}}+\left(\frac{1}{2}+x_{5}\right)a\,\hat{\mathbf{y}}-z_{5}a\,\hat{\mathbf{z}}$ | (24*k*) | O II |
| $\mathbf{B}_{34}$ | $=$ | $-x_{5}\,\mathbf{a}_{1}-x_{5}\,\mathbf{a}_{2}-z_{5}\,\mathbf{a}_{3}$ | $=$ | $-x_{5}a\,\hat{\mathbf{x}}-x_{5}a\,\hat{\mathbf{y}}-z_{5}a\,\hat{\mathbf{z}}$ | (24*k*) | O II |
| $\mathbf{B}_{35}$ | $=$ | $\left(\frac{1}{2}+x_{5}\right)\mathbf{a}_{1}-x_{5}\,\mathbf{a}_{2}+\left(\frac{1}{2}+z_{5}\right)\mathbf{a}_{3}$ | $=$ | $\left(\frac{1}{2}+x_{5}\right)a\,\hat{\mathbf{x}}-x_{5}a\,\hat{\mathbf{y}}+\left(\frac{1}{2}+z_{5}\right)a\,\hat{\mathbf{z}}$ | (24*k*) | O II |
| $\mathbf{B}_{36}$ | $=$ | $-x_{5}\,\mathbf{a}_{1}+\left(\frac{1}{2}+x_{5}\right)\mathbf{a}_{2}+\left(\frac{1}{2}+z_{5}\right)\mathbf{a}_{3}$ | $=$ | $-x_{5}a\,\hat{\mathbf{x}}+\left(\frac{1}{2}+x_{5}\right)a\,\hat{\mathbf{y}}+\left(\frac{1}{2}+z_{5}\right)a\,\hat{\mathbf{z}}$ | (24*k*) | O II |
| $\mathbf{B}_{37}$ | $=$ | $\left(\frac{1}{2}+x_{5}\right)\mathbf{a}_{1}+\left(\frac{1}{2}+z_{5}\right)\mathbf{a}_{2}-x_{5}\,\mathbf{a}_{3}$ | $=$ | $\left(\frac{1}{2}+x_{5}\right)a\,\hat{\mathbf{x}}+\left(\frac{1}{2}+z_{5}\right)a\,\hat{\mathbf{y}}-x_{5}a\,\hat{\mathbf{z}}$ | (24*k*) | O II |
| $\mathbf{B}_{38}$ | $=$ | $-x_{5}\,\mathbf{a}_{1}+\left(\frac{1}{2}+z_{5}\right)\mathbf{a}_{2}+\left(\frac{1}{2}+x_{5}\right)\mathbf{a}_{3}$ | $=$ | $-x_{5}a\,\hat{\mathbf{x}}+\left(\frac{1}{2}+z_{5}\right)a\,\hat{\mathbf{y}}+\left(\frac{1}{2}+x_{5}\right)a\,\hat{\mathbf{z}}$ | (24*k*) | O II |
| $\mathbf{B}_{39}$ | $=$ | $-x_{5}\,\mathbf{a}_{1}-z_{5}\,\mathbf{a}_{2}-x_{5}\,\mathbf{a}_{3}$ | $=$ | $-x_{5}a\,\hat{\mathbf{x}}-z_{5}a\,\hat{\mathbf{y}}-x_{5}a\,\hat{\mathbf{z}}$ | (24*k*) | O II |
| $\mathbf{B}_{40}$ | $=$ | $\left(\frac{1}{2}+x_{5}\right)\mathbf{a}_{1}-z_{5}\,\mathbf{a}_{2}+\left(\frac{1}{2}+x_{5}\right)\mathbf{a}_{3}$ | $=$ | $\left(\frac{1}{2}+x_{5}\right)a\,\hat{\mathbf{x}}-z_{5}a\,\hat{\mathbf{y}}+\left(\frac{1}{2}+x_{5}\right)a\,\hat{\mathbf{z}}$ | (24*k*) | O II |
| $\mathbf{B}_{41}$ | $=$ | $\left(\frac{1}{2}+z_{5}\right)\mathbf{a}_{1}+\left(\frac{1}{2}+x_{5}\right)\mathbf{a}_{2}-x_{5}\,\mathbf{a}_{3}$ | $=$ | $\left(\frac{1}{2}+z_{5}\right)a\,\hat{\mathbf{x}}+\left(\frac{1}{2}+x_{5}\right)a\,\hat{\mathbf{y}}-x_{5}a\,\hat{\mathbf{z}}$ | (24*k*) | O II |
| $\mathbf{B}_{42}$ | $=$ | $\left(\frac{1}{2}+z_{5}\right)\mathbf{a}_{1}-x_{5}\,\mathbf{a}_{2}+\left(\frac{1}{2}+x_{5}\right)\mathbf{a}_{3}$ | $=$ | $\left(\frac{1}{2}+z_{5}\right)a\,\hat{\mathbf{x}}-x_{5}a\,\hat{\mathbf{y}}+\left(\frac{1}{2}+x_{5}\right)a\,\hat{\mathbf{z}}$ | (24*k*) | O II |
| $\mathbf{B}_{43}$ | $=$ | $-z_{5}\,\mathbf{a}_{1}+\left(\frac{1}{2}+x_{5}\right)\mathbf{a}_{2}+\left(\frac{1}{2}+x_{5}\right)\mathbf{a}_{3}$ | $=$ | $-z_{5}a\,\hat{\mathbf{x}}+\left(\frac{1}{2}+x_{5}\right)a\,\hat{\mathbf{y}}+\left(\frac{1}{2}+x_{5}\right)a\,\hat{\mathbf{z}}$ | (24*k*) | O II |
| $\mathbf{B}_{44}$ | $=$ | $-z_{5}\,\mathbf{a}_{1}-x_{5}\,\mathbf{a}_{2}-x_{5}\,\mathbf{a}_{3}$ | $=$ | $-z_{5}a\,\hat{\mathbf{x}}-x_{5}a\,\hat{\mathbf{y}}-x_{5}a\,\hat{\mathbf{z}}$ | (24*k*) | O II |
| $\mathbf{B}_{45}$ | $=$ | $x_{6}\,\mathbf{a}_{1}+x_{6}\,\mathbf{a}_{2}+z_{6}\,\mathbf{a}_{3}$ | $=$ | $x_{6}a\,\hat{\mathbf{x}}+x_{6}a\,\hat{\mathbf{y}}+z_{6}a\,\hat{\mathbf{z}}$ | (24*k*) | O III |
| $\mathbf{B}_{46}$ | $=$ | $\left(\frac{1}{2}-x_{6}\right)\mathbf{a}_{1}+\left(\frac{1}{2}-x_{6}\right)\mathbf{a}_{2}+z_{6}\,\mathbf{a}_{3}$ | $=$ | $\left(\frac{1}{2}-x_{6}\right)a\,\hat{\mathbf{x}}+\left(\frac{1}{2}-x_{6}\right)a\,\hat{\mathbf{y}}+z_{6}a\,\hat{\mathbf{z}}$ | (24*k*) | O III |
| $\mathbf{B}_{47}$ | $=$ | $\left(\frac{1}{2}-x_{6}\right)\mathbf{a}_{1}+x_{6}\,\mathbf{a}_{2}+\left(\frac{1}{2}-z_{6}\right)\mathbf{a}_{3}$ | $=$ | $\left(\frac{1}{2}-x_{6}\right)a\,\hat{\mathbf{x}}+x_{6}a\,\hat{\mathbf{y}}+\left(\frac{1}{2}-z_{6}\right)a\,\hat{\mathbf{z}}$ | (24*k*) | O III |
| $\mathbf{B}_{48}$ | $=$ | $x_{6}\,\mathbf{a}_{1}+\left(\frac{1}{2}-x_{6}\right)\mathbf{a}_{2}+\left(\frac{1}{2}-z_{6}\right)\mathbf{a}_{3}$ | $=$ | $x_{6}a\,\hat{\mathbf{x}}+\left(\frac{1}{2}-x_{6}\right)a\,\hat{\mathbf{y}}+\left(\frac{1}{2}-z_{6}\right)a\,\hat{\mathbf{z}}$ | (24*k*) | O III |
| $\mathbf{B}_{49}$ | $=$ | $z_{6}\,\mathbf{a}_{1}+x_{6}\,\mathbf{a}_{2}+x_{6}\,\mathbf{a}_{3}$ | $=$ | $z_{6}a\,\hat{\mathbf{x}}+x_{6}a\,\hat{\mathbf{y}}+x_{6}a\,\hat{\mathbf{z}}$ | (24*k*) | O III |
| $\mathbf{B}_{50}$ | $=$ | $z_{6}\,\mathbf{a}_{1}+\left(\frac{1}{2}-x_{6}\right)\mathbf{a}_{2}+\left(\frac{1}{2}-x_{6}\right)\mathbf{a}_{3}$ | $=$ | $z_{6}a\,\hat{\mathbf{x}}+\left(\frac{1}{2}-x_{6}\right)a\,\hat{\mathbf{y}}+\left(\frac{1}{2}-x_{6}\right)a\,\hat{\mathbf{z}}$ | (24*k*) | O III |
| $\mathbf{B}_{51}$ | $=$ | $\left(\frac{1}{2}-z_{6}\right)\mathbf{a}_{1}+\left(\frac{1}{2}-x_{6}\right)\mathbf{a}_{2}+x_{6}\,\mathbf{a}_{3}$ | $=$ | $\left(\frac{1}{2}-z_{6}\right)a\,\hat{\mathbf{x}}+\left(\frac{1}{2}-x_{6}\right)a\,\hat{\mathbf{y}}+x_{6}a\,\hat{\mathbf{z}}$ | (24*k*) | O III |
| $\mathbf{B}_{52}$ | $=$ | $\left(\frac{1}{2}-z_{6}\right)\mathbf{a}_{1}+x_{6}\,\mathbf{a}_{2}+\left(\frac{1}{2}-x_{6}\right)\mathbf{a}_{3}$ | $=$ | $\left(\frac{1}{2}-z_{6}\right)a\,\hat{\mathbf{x}}+x_{6}a\,\hat{\mathbf{y}}+\left(\frac{1}{2}-x_{6}\right)a\,\hat{\mathbf{z}}$ | (24*k*) | O III |
| $\mathbf{B}_{53}$ | $=$ | $x_{6}\,\mathbf{a}_{1}+z_{6}\,\mathbf{a}_{2}+x_{6}\,\mathbf{a}_{3}$ | $=$ | $x_{6}a\,\hat{\mathbf{x}}+z_{6}a\,\hat{\mathbf{y}}+x_{6}a\,\hat{\mathbf{z}}$ | (24*k*) | O III |
| $\mathbf{B}_{54}$ | $=$ | $\left(\frac{1}{2}-x_{6}\right)\mathbf{a}_{1}+z_{6}\,\mathbf{a}_{2}+\left(\frac{1}{2}-x_{6}\right)\mathbf{a}_{3}$ | $=$ | $\left(\frac{1}{2}-x_{6}\right)a\,\hat{\mathbf{x}}+z_{6}a\,\hat{\mathbf{y}}+\left(\frac{1}{2}-x_{6}\right)a\,\hat{\mathbf{z}}$ | (24*k*) | O III |
| $\mathbf{B}_{55}$ | $=$ | $x_{6}\,\mathbf{a}_{1}+\left(\frac{1}{2}-z_{6}\right)\mathbf{a}_{2}+\left(\frac{1}{2}-x_{6}\right)\mathbf{a}_{3}$ | $=$ | $x_{6}a\,\hat{\mathbf{x}}+\left(\frac{1}{2}-z_{6}\right)a\,\hat{\mathbf{y}}+\left(\frac{1}{2}-x_{6}\right)a\,\hat{\mathbf{z}}$ | (24*k*) | O III |
| $\mathbf{B}_{56}$ | $=$ | $\left(\frac{1}{2}-x_{6}\right)\mathbf{a}_{1}+\left(\frac{1}{2}-z_{6}\right)\mathbf{a}_{2}+x_{6}\,\mathbf{a}_{3}$ | $=$ | $\left(\frac{1}{2}-x_{6}\right)a\,\hat{\mathbf{x}}+\left(\frac{1}{2}-z_{6}\right)a\,\hat{\mathbf{y}}+x_{6}a\,\hat{\mathbf{z}}$ | (24*k*) | O III |
| $\mathbf{B}_{57}$ | $=$ | $\left(\frac{1}{2}+x_{6}\right)\mathbf{a}_{1}+\left(\frac{1}{2}+x_{6}\right)\mathbf{a}_{2}-z_{6}\,\mathbf{a}_{3}$ | $=$ | $\left(\frac{1}{2}+x_{6}\right)a\,\hat{\mathbf{x}}+\left(\frac{1}{2}+x_{6}\right)a\,\hat{\mathbf{y}}-z_{6}a\,\hat{\mathbf{z}}$ | (24*k*) | O III |
| $\mathbf{B}_{58}$ | $=$ | $-x_{6}\,\mathbf{a}_{1}-x_{6}\,\mathbf{a}_{2}-z_{6}\,\mathbf{a}_{3}$ | $=$ | $-x_{6}a\,\hat{\mathbf{x}}-x_{6}a\,\hat{\mathbf{y}}-z_{6}a\,\hat{\mathbf{z}}$ | (24*k*) | O III |
| $\mathbf{B}_{59}$ | $=$ | $\left(\frac{1}{2}+x_{6}\right)\mathbf{a}_{1}-x_{6}\,\mathbf{a}_{2}+\left(\frac{1}{2}+z_{6}\right)\mathbf{a}_{3}$ | $=$ | $\left(\frac{1}{2}+x_{6}\right)a\,\hat{\mathbf{x}}-x_{6}a\,\hat{\mathbf{y}}+\left(\frac{1}{2}+z_{6}\right)a\,\hat{\mathbf{z}}$ | (24*k*) | O III |
| $\mathbf{B}_{60}$ | $=$ | $-x_{6}\,\mathbf{a}_{1}+\left(\frac{1}{2}+x_{6}\right)\mathbf{a}_{2}+\left(\frac{1}{2}+z_{6}\right)\mathbf{a}_{3}$ | $=$ | $-x_{6}a\,\hat{\mathbf{x}}+\left(\frac{1}{2}+x_{6}\right)a\,\hat{\mathbf{y}}+\left(\frac{1}{2}+z_{6}\right)a\,\hat{\mathbf{z}}$ | (24*k*) | O III |
| $\mathbf{B}_{61}$ | $=$ | $\left(\frac{1}{2}+x_{6}\right)\mathbf{a}_{1}+\left(\frac{1}{2}+z_{6}\right)\mathbf{a}_{2}-x_{6}\,\mathbf{a}_{3}$ | $=$ | $\left(\frac{1}{2}+x_{6}\right)a\,\hat{\mathbf{x}}+\left(\frac{1}{2}+z_{6}\right)a\,\hat{\mathbf{y}}-x_{6}a\,\hat{\mathbf{z}}$ | (24*k*) | O III |
| $\mathbf{B}_{62}$ | = | $-x_6\,\mathbf{a}_1 + \left(\frac{1}{2}+z_6\right)\mathbf{a}_2 + \left(\frac{1}{2}+x_6\right)\mathbf{a}_3$ | = | $-x_6 a\,\hat{\mathbf{x}} + \left(\frac{1}{2}+z_6\right)a\,\hat{\mathbf{y}} + \left(\frac{1}{2}+x_6\right)a\,\hat{\mathbf{z}}$ | (24$k$) | O III |
| $\mathbf{B}_{63}$ | = | $-x_6\,\mathbf{a}_1 - z_6\,\mathbf{a}_2 - x_6\,\mathbf{a}_3$ | = | $-x_6 a\,\hat{\mathbf{x}} - z_6 a\,\hat{\mathbf{y}} - x_6 a\,\hat{\mathbf{z}}$ | (24$k$) | O III |
| $\mathbf{B}_{64}$ | = | $\left(\frac{1}{2}+x_6\right)\mathbf{a}_1 - z_6\,\mathbf{a}_2 + \left(\frac{1}{2}+x_6\right)\mathbf{a}_3$ | = | $\left(\frac{1}{2}+x_6\right)a\,\hat{\mathbf{x}} - z_6 a\,\hat{\mathbf{y}} + \left(\frac{1}{2}+x_6\right)a\,\hat{\mathbf{z}}$ | (24$k$) | O III |
| $\mathbf{B}_{65}$ | = | $\left(\frac{1}{2}+z_6\right)\mathbf{a}_1 + \left(\frac{1}{2}+x_6\right)\mathbf{a}_2 - x_6\,\mathbf{a}_3$ | = | $\left(\frac{1}{2}+z_6\right)a\,\hat{\mathbf{x}} + \left(\frac{1}{2}+x_6\right)a\,\hat{\mathbf{y}} - x_6 a\,\hat{\mathbf{z}}$ | (24$k$) | O III |
| $\mathbf{B}_{66}$ | = | $\left(\frac{1}{2}+z_6\right)\mathbf{a}_1 - x_6\,\mathbf{a}_2 + \left(\frac{1}{2}+x_6\right)\mathbf{a}_3$ | = | $\left(\frac{1}{2}+z_6\right)a\,\hat{\mathbf{x}} - x_6 a\,\hat{\mathbf{y}} + \left(\frac{1}{2}+x_6\right)a\,\hat{\mathbf{z}}$ | (24$k$) | O III |
| $\mathbf{B}_{67}$ | = | $-z_6\,\mathbf{a}_1 + \left(\frac{1}{2}+x_6\right)\mathbf{a}_2 + \left(\frac{1}{2}+x_6\right)\mathbf{a}_3$ | = | $-z_6 a\,\hat{\mathbf{x}} + \left(\frac{1}{2}+x_6\right)a\,\hat{\mathbf{y}} + \left(\frac{1}{2}+x_6\right)a\,\hat{\mathbf{z}}$ | (24$k$) | O III |
| $\mathbf{B}_{68}$ | = | $-z_6\,\mathbf{a}_1 - x_6\,\mathbf{a}_2 - x_6\,\mathbf{a}_3$ | = | $-z_6 a\,\hat{\mathbf{x}} - x_6 a\,\hat{\mathbf{y}} - x_6 a\,\hat{\mathbf{z}}$ | (24$k$) | O III |
| $\mathbf{B}_{69}$ | = | $x_7\,\mathbf{a}_1 + x_7\,\mathbf{a}_2 + z_7\,\mathbf{a}_3$ | = | $x_7 a\,\hat{\mathbf{x}} + x_7 a\,\hat{\mathbf{y}} + z_7 a\,\hat{\mathbf{z}}$ | (24$k$) | O IV |
| $\mathbf{B}_{70}$ | = | $\left(\frac{1}{2}-x_7\right)\mathbf{a}_1 + \left(\frac{1}{2}-x_7\right)\mathbf{a}_2 + z_7\,\mathbf{a}_3$ | = | $\left(\frac{1}{2}-x_7\right)a\,\hat{\mathbf{x}} + \left(\frac{1}{2}-x_7\right)a\,\hat{\mathbf{y}} + z_7 a\,\hat{\mathbf{z}}$ | (24$k$) | O IV |
| $\mathbf{B}_{71}$ | = | $\left(\frac{1}{2}-x_7\right)\mathbf{a}_1 + x_7\,\mathbf{a}_2 + \left(\frac{1}{2}-z_7\right)\mathbf{a}_3$ | = | $\left(\frac{1}{2}-x_7\right)a\,\hat{\mathbf{x}} + x_7 a\,\hat{\mathbf{y}} + \left(\frac{1}{2}-z_7\right)a\,\hat{\mathbf{z}}$ | (24$k$) | O IV |
| $\mathbf{B}_{72}$ | = | $x_7\,\mathbf{a}_1 + \left(\frac{1}{2}-x_7\right)\mathbf{a}_2 + \left(\frac{1}{2}-z_7\right)\mathbf{a}_3$ | = | $x_7 a\,\hat{\mathbf{x}} + \left(\frac{1}{2}-x_7\right)a\,\hat{\mathbf{y}} + \left(\frac{1}{2}-z_7\right)a\,\hat{\mathbf{z}}$ | (24$k$) | O IV |
| $\mathbf{B}_{73}$ | = | $z_7\,\mathbf{a}_1 + x_7\,\mathbf{a}_2 + x_7\,\mathbf{a}_3$ | = | $z_7 a\,\hat{\mathbf{x}} + x_7 a\,\hat{\mathbf{y}} + x_7 a\,\hat{\mathbf{z}}$ | (24$k$) | O IV |
| $\mathbf{B}_{74}$ | = | $z_7\,\mathbf{a}_1 + \left(\frac{1}{2}-x_7\right)\mathbf{a}_2 + \left(\frac{1}{2}-x_7\right)\mathbf{a}_3$ | = | $z_7 a\,\hat{\mathbf{x}} + \left(\frac{1}{2}-x_7\right)a\,\hat{\mathbf{y}} + \left(\frac{1}{2}-x_7\right)a\,\hat{\mathbf{z}}$ | (24$k$) | O IV |
| $\mathbf{B}_{75}$ | = | $\left(\frac{1}{2}-z_7\right)\mathbf{a}_1 + \left(\frac{1}{2}-x_7\right)\mathbf{a}_2 + x_7\,\mathbf{a}_3$ | = | $\left(\frac{1}{2}-z_7\right)a\,\hat{\mathbf{x}} + \left(\frac{1}{2}-x_7\right)a\,\hat{\mathbf{y}} + x_7 a\,\hat{\mathbf{z}}$ | (24$k$) | O IV |
| $\mathbf{B}_{76}$ | = | $\left(\frac{1}{2}-z_7\right)\mathbf{a}_1 + x_7\,\mathbf{a}_2 + \left(\frac{1}{2}-x_7\right)\mathbf{a}_3$ | = | $\left(\frac{1}{2}-z_7\right)a\,\hat{\mathbf{x}} + x_7 a\,\hat{\mathbf{y}} + \left(\frac{1}{2}-x_7\right)a\,\hat{\mathbf{z}}$ | (24$k$) | O IV |
| $\mathbf{B}_{77}$ | = | $x_7\,\mathbf{a}_1 + z_7\,\mathbf{a}_2 + x_7\,\mathbf{a}_3$ | = | $x_7 a\,\hat{\mathbf{x}} + z_7 a\,\hat{\mathbf{y}} + x_7 a\,\hat{\mathbf{z}}$ | (24$k$) | O IV |
| $\mathbf{B}_{78}$ | = | $\left(\frac{1}{2}-x_7\right)\mathbf{a}_1 + z_7\,\mathbf{a}_2 + \left(\frac{1}{2}-x_7\right)\mathbf{a}_3$ | = | $\left(\frac{1}{2}-x_7\right)a\,\hat{\mathbf{x}} + z_7 a\,\hat{\mathbf{y}} + \left(\frac{1}{2}-x_7\right)a\,\hat{\mathbf{z}}$ | (24$k$) | O IV |
| $\mathbf{B}_{79}$ | = | $x_7\,\mathbf{a}_1 + \left(\frac{1}{2}-z_7\right)\mathbf{a}_2 + \left(\frac{1}{2}-x_7\right)\mathbf{a}_3$ | = | $x_7 a\,\hat{\mathbf{x}} + \left(\frac{1}{2}-z_7\right)a\,\hat{\mathbf{y}} + \left(\frac{1}{2}-x_7\right)a\,\hat{\mathbf{z}}$ | (24$k$) | O IV |
| $\mathbf{B}_{80}$ | = | $\left(\frac{1}{2}-x_7\right)\mathbf{a}_1 + \left(\frac{1}{2}-z_7\right)\mathbf{a}_2 + x_7\,\mathbf{a}_3$ | = | $\left(\frac{1}{2}-x_7\right)a\,\hat{\mathbf{x}} + \left(\frac{1}{2}-z_7\right)a\,\hat{\mathbf{y}} + x_7 a\,\hat{\mathbf{z}}$ | (24$k$) | O IV |
| $\mathbf{B}_{81}$ | = | $\left(\frac{1}{2}+x_7\right)\mathbf{a}_1 + \left(\frac{1}{2}+x_7\right)\mathbf{a}_2 - z_7\,\mathbf{a}_3$ | = | $\left(\frac{1}{2}+x_7\right)a\,\hat{\mathbf{x}} + \left(\frac{1}{2}+x_7\right)a\,\hat{\mathbf{y}} - z_7 a\,\hat{\mathbf{z}}$ | (24$k$) | O IV |
| $\mathbf{B}_{82}$ | = | $-x_7\,\mathbf{a}_1 - x_7\,\mathbf{a}_2 - z_7\,\mathbf{a}_3$ | = | $-x_7 a\,\hat{\mathbf{x}} - x_7 a\,\hat{\mathbf{y}} - z_7 a\,\hat{\mathbf{z}}$ | (24$k$) | O IV |
| $\mathbf{B}_{83}$ | = | $\left(\frac{1}{2}+x_7\right)\mathbf{a}_1 - x_7\,\mathbf{a}_2 + \left(\frac{1}{2}+z_7\right)\mathbf{a}_3$ | = | $\left(\frac{1}{2}+x_7\right)a\,\hat{\mathbf{x}} - x_7 a\,\hat{\mathbf{y}} + \left(\frac{1}{2}+z_7\right)a\,\hat{\mathbf{z}}$ | (24$k$) | O IV |
| $\mathbf{B}_{84}$ | = | $-x_7\,\mathbf{a}_1 + \left(\frac{1}{2}+x_7\right)\mathbf{a}_2 + \left(\frac{1}{2}+z_7\right)\mathbf{a}_3$ | = | $-x_7 a\,\hat{\mathbf{x}} + \left(\frac{1}{2}+x_7\right)a\,\hat{\mathbf{y}} + \left(\frac{1}{2}+z_7\right)a\,\hat{\mathbf{z}}$ | (24$k$) | O IV |
| $\mathbf{B}_{85}$ | = | $\left(\frac{1}{2}+x_7\right)\mathbf{a}_1 + \left(\frac{1}{2}+z_7\right)\mathbf{a}_2 - x_7\,\mathbf{a}_3$ | = | $\left(\frac{1}{2}+x_7\right)a\,\hat{\mathbf{x}} + \left(\frac{1}{2}+z_7\right)a\,\hat{\mathbf{y}} - x_7 a\,\hat{\mathbf{z}}$ | (24$k$) | O IV |
| $\mathbf{B}_{86}$ | = | $-x_7\,\mathbf{a}_1 + \left(\frac{1}{2}+z_7\right)\mathbf{a}_2 + \left(\frac{1}{2}+x_7\right)\mathbf{a}_3$ | = | $-x_7 a\,\hat{\mathbf{x}} + \left(\frac{1}{2}+z_7\right)a\,\hat{\mathbf{y}} + \left(\frac{1}{2}+x_7\right)a\,\hat{\mathbf{z}}$ | (24$k$) | O IV |
| $\mathbf{B}_{87}$ | = | $-x_7\,\mathbf{a}_1 - z_7\,\mathbf{a}_2 - x_7\,\mathbf{a}_3$ | = | $-x_7 a\,\hat{\mathbf{x}} - z_7 a\,\hat{\mathbf{y}} - x_7 a\,\hat{\mathbf{z}}$ | (24$k$) | O IV |
| $\mathbf{B}_{88}$ | = | $\left(\frac{1}{2}+x_7\right)\mathbf{a}_1 - z_7\,\mathbf{a}_2 + \left(\frac{1}{2}+x_7\right)\mathbf{a}_3$ | = | $\left(\frac{1}{2}+x_7\right)a\,\hat{\mathbf{x}} - z_7 a\,\hat{\mathbf{y}} + \left(\frac{1}{2}+x_7\right)a\,\hat{\mathbf{z}}$ | (24$k$) | O IV |
| $\mathbf{B}_{89}$ | = | $\left(\frac{1}{2}+z_7\right)\mathbf{a}_1 + \left(\frac{1}{2}+x_7\right)\mathbf{a}_2 - x_7\,\mathbf{a}_3$ | = | $\left(\frac{1}{2}+z_7\right)a\,\hat{\mathbf{x}} + \left(\frac{1}{2}+x_7\right)a\,\hat{\mathbf{y}} - x_7 a\,\hat{\mathbf{z}}$ | (24$k$) | O IV |
| $\mathbf{B}_{90}$ | = | $\left(\frac{1}{2}+z_7\right)\mathbf{a}_1 - x_7\,\mathbf{a}_2 + \left(\frac{1}{2}+x_7\right)\mathbf{a}_3$ | = | $\left(\frac{1}{2}+z_7\right)a\,\hat{\mathbf{x}} - x_7 a\,\hat{\mathbf{y}} + \left(\frac{1}{2}+x_7\right)a\,\hat{\mathbf{z}}$ | (24$k$) | O IV |
| $\mathbf{B}_{91}$ | = | $-z_7\,\mathbf{a}_1 + \left(\frac{1}{2}+x_7\right)\mathbf{a}_2 + \left(\frac{1}{2}+x_7\right)\mathbf{a}_3$ | = | $-z_7 a\,\hat{\mathbf{x}} + \left(\frac{1}{2}+x_7\right)a\,\hat{\mathbf{y}} + \left(\frac{1}{2}+x_7\right)a\,\hat{\mathbf{z}}$ | (24$k$) | O IV |
| $\mathbf{B}_{92}$ | = | $-z_7\,\mathbf{a}_1 - x_7\,\mathbf{a}_2 - x_7\,\mathbf{a}_3$ | = | $-z_7 a\,\hat{\mathbf{x}} - x_7 a\,\hat{\mathbf{y}} - x_7 a\,\hat{\mathbf{z}}$ | (24$k$) | O IV |
| $\mathbf{B}_{93}$ | = | $x_8\,\mathbf{a}_1 + x_8\,\mathbf{a}_2 + z_8\,\mathbf{a}_3$ | = | $x_8 a\,\hat{\mathbf{x}} + x_8 a\,\hat{\mathbf{y}} + z_8 a\,\hat{\mathbf{z}}$ | (24$k$) | W |
| $\mathbf{B}_{94}$ | = | $\left(\frac{1}{2}-x_8\right)\mathbf{a}_1 + \left(\frac{1}{2}-x_8\right)\mathbf{a}_2 + z_8\,\mathbf{a}_3$ | = | $\left(\frac{1}{2}-x_8\right)a\,\hat{\mathbf{x}} + \left(\frac{1}{2}-x_8\right)a\,\hat{\mathbf{y}} + z_8 a\,\hat{\mathbf{z}}$ | (24$k$) | W |
| $\mathbf{B}_{95}$ | = | $\left(\frac{1}{2}-x_8\right)\mathbf{a}_1 + x_8\,\mathbf{a}_2 + \left(\frac{1}{2}-z_8\right)\mathbf{a}_3$ | = | $\left(\frac{1}{2}-x_8\right)a\,\hat{\mathbf{x}} + x_8 a\,\hat{\mathbf{y}} + \left(\frac{1}{2}-z_8\right)a\,\hat{\mathbf{z}}$ | (24$k$) | W |
| $\mathbf{B}_{96}$ | = | $x_8\,\mathbf{a}_1 + \left(\frac{1}{2}-x_8\right)\mathbf{a}_2 + \left(\frac{1}{2}-z_8\right)\mathbf{a}_3$ | = | $x_8 a\,\hat{\mathbf{x}} + \left(\frac{1}{2}-x_8\right)a\,\hat{\mathbf{y}} + \left(\frac{1}{2}-z_8\right)a\,\hat{\mathbf{z}}$ | (24$k$) | W |
| $\mathbf{B}_{97}$ | = | $z_8\,\mathbf{a}_1 + x_8\,\mathbf{a}_2 + x_8\,\mathbf{a}_3$ | = | $z_8 a\,\hat{\mathbf{x}} + x_8 a\,\hat{\mathbf{y}} + x_8 a\,\hat{\mathbf{z}}$ | (24$k$) | W |
| $\mathbf{B}_{98}$ | = | $z_8\,\mathbf{a}_1 + \left(\frac{1}{2} - x_8\right)\,\mathbf{a}_2 + \left(\frac{1}{2} - x_8\right)\,\mathbf{a}_3$ | = | $z_8 a\,\hat{\mathbf{x}} + \left(\frac{1}{2} - x_8\right) a\,\hat{\mathbf{y}} + \left(\frac{1}{2} - x_8\right) a\,\hat{\mathbf{z}}$ | (24k) | W |
| $\mathbf{B}_{99}$ | = | $\left(\frac{1}{2} - z_8\right)\,\mathbf{a}_1 + \left(\frac{1}{2} - x_8\right)\,\mathbf{a}_2 + x_8\,\mathbf{a}_3$ | = | $\left(\frac{1}{2} - z_8\right) a\,\hat{\mathbf{x}} + \left(\frac{1}{2} - x_8\right) a\,\hat{\mathbf{y}} + x_8 a\,\hat{\mathbf{z}}$ | (24k) | W |
| $\mathbf{B}_{100}$ | = | $\left(\frac{1}{2} - z_8\right)\,\mathbf{a}_1 + x_8\,\mathbf{a}_2 + \left(\frac{1}{2} - x_8\right)\,\mathbf{a}_3$ | = | $\left(\frac{1}{2} - z_8\right) a\,\hat{\mathbf{x}} + x_8 a\,\hat{\mathbf{y}} + \left(\frac{1}{2} - x_8\right) a\,\hat{\mathbf{z}}$ | (24k) | W |
| $\mathbf{B}_{101}$ | = | $x_8\,\mathbf{a}_1 + z_8\,\mathbf{a}_2 + x_8\,\mathbf{a}_3$ | = | $x_8 a\,\hat{\mathbf{x}} + z_8 a\,\hat{\mathbf{y}} + x_8 a\,\hat{\mathbf{z}}$ | (24k) | W |
| $\mathbf{B}_{102}$ | = | $\left(\frac{1}{2} - x_8\right)\,\mathbf{a}_1 + z_8\,\mathbf{a}_2 + \left(\frac{1}{2} - x_8\right)\,\mathbf{a}_3$ | = | $\left(\frac{1}{2} - x_8\right) a\,\hat{\mathbf{x}} + z_8 a\,\hat{\mathbf{y}} + \left(\frac{1}{2} - x_8\right) a\,\hat{\mathbf{z}}$ | (24k) | W |
| $\mathbf{B}_{103}$ | = | $x_8\,\mathbf{a}_1 + \left(\frac{1}{2} - z_8\right)\,\mathbf{a}_2 + \left(\frac{1}{2} - x_8\right)\,\mathbf{a}_3$ | = | $x_8 a\,\hat{\mathbf{x}} + \left(\frac{1}{2} - z_8\right) a\,\hat{\mathbf{y}} + \left(\frac{1}{2} - x_8\right) a\,\hat{\mathbf{z}}$ | (24k) | W |
| $\mathbf{B}_{104}$ | = | $\left(\frac{1}{2} - x_8\right)\,\mathbf{a}_1 + \left(\frac{1}{2} - z_8\right)\,\mathbf{a}_2 + x_8\,\mathbf{a}_3$ | = | $\left(\frac{1}{2} - x_8\right) a\,\hat{\mathbf{x}} + \left(\frac{1}{2} - z_8\right) a\,\hat{\mathbf{y}} + x_8 a\,\hat{\mathbf{z}}$ | (24k) | W |
| $\mathbf{B}_{105}$ | = | $\left(\frac{1}{2} + x_8\right)\,\mathbf{a}_1 + \left(\frac{1}{2} + x_8\right)\,\mathbf{a}_2 - z_8\,\mathbf{a}_3$ | = | $\left(\frac{1}{2} + x_8\right) a\,\hat{\mathbf{x}} + \left(\frac{1}{2} + x_8\right) a\,\hat{\mathbf{y}} - z_8 a\,\hat{\mathbf{z}}$ | (24k) | W |
| $\mathbf{B}_{106}$ | = | $-x_8\,\mathbf{a}_1 - x_8\,\mathbf{a}_2 - z_8\,\mathbf{a}_3$ | = | $-x_8 a\,\hat{\mathbf{x}} - x_8 a\,\hat{\mathbf{y}} - z_8 a\,\hat{\mathbf{z}}$ | (24k) | W |
| $\mathbf{B}_{107}$ | = | $\left(\frac{1}{2} + x_8\right)\,\mathbf{a}_1 - x_8\,\mathbf{a}_2 + \left(\frac{1}{2} + z_8\right)\,\mathbf{a}_3$ | = | $\left(\frac{1}{2} + x_8\right) a\,\hat{\mathbf{x}} - x_8 a\,\hat{\mathbf{y}} + \left(\frac{1}{2} + z_8\right) a\,\hat{\mathbf{z}}$ | (24k) | W |
| $\mathbf{B}_{108}$ | = | $-x_8\,\mathbf{a}_1 + \left(\frac{1}{2} + x_8\right)\,\mathbf{a}_2 + \left(\frac{1}{2} + z_8\right)\,\mathbf{a}_3$ | = | $-x_8 a\,\hat{\mathbf{x}} + \left(\frac{1}{2} + x_8\right) a\,\hat{\mathbf{y}} + \left(\frac{1}{2} + z_8\right) a\,\hat{\mathbf{z}}$ | (24k) | W |
| $\mathbf{B}_{109}$ | = | $\left(\frac{1}{2} + x_8\right)\,\mathbf{a}_1 + \left(\frac{1}{2} + z_8\right)\,\mathbf{a}_2 - x_8\,\mathbf{a}_3$ | = | $\left(\frac{1}{2} + x_8\right) a\,\hat{\mathbf{x}} + \left(\frac{1}{2} + z_8\right) a\,\hat{\mathbf{y}} - x_8 a\,\hat{\mathbf{z}}$ | (24k) | W |
| $\mathbf{B}_{110}$ | = | $-x_8\,\mathbf{a}_1 + \left(\frac{1}{2} + z_8\right)\,\mathbf{a}_2 + \left(\frac{1}{2} + x_8\right)\,\mathbf{a}_3$ | = | $-x_8 a\,\hat{\mathbf{x}} + \left(\frac{1}{2} + z_8\right) a\,\hat{\mathbf{y}} + \left(\frac{1}{2} + x_8\right) a\,\hat{\mathbf{z}}$ | (24k) | W |
| $\mathbf{B}_{111}$ | = | $-x_8\,\mathbf{a}_1 - z_8\,\mathbf{a}_2 - x_8\,\mathbf{a}_3$ | = | $-x_8 a\,\hat{\mathbf{x}} - z_8 a\,\hat{\mathbf{y}} - x_8 a\,\hat{\mathbf{z}}$ | (24k) | W |
| $\mathbf{B}_{112}$ | = | $\left(\frac{1}{2} + x_8\right)\,\mathbf{a}_1 - z_8\,\mathbf{a}_2 + \left(\frac{1}{2} + x_8\right)\,\mathbf{a}_3$ | = | $\left(\frac{1}{2} + x_8\right) a\,\hat{\mathbf{x}} - z_8 a\,\hat{\mathbf{y}} + \left(\frac{1}{2} + x_8\right) a\,\hat{\mathbf{z}}$ | (24k) | W |
| $\mathbf{B}_{113}$ | = | $\left(\frac{1}{2} + z_8\right)\,\mathbf{a}_1 + \left(\frac{1}{2} + x_8\right)\,\mathbf{a}_2 - x_8\,\mathbf{a}_3$ | = | $\left(\frac{1}{2} + z_8\right) a\,\hat{\mathbf{x}} + \left(\frac{1}{2} + x_8\right) a\,\hat{\mathbf{y}} - x_8 a\,\hat{\mathbf{z}}$ | (24k) | W |
| $\mathbf{B}_{114}$ | = | $\left(\frac{1}{2} + z_8\right)\,\mathbf{a}_1 - x_8\,\mathbf{a}_2 + \left(\frac{1}{2} + x_8\right)\,\mathbf{a}_3$ | = | $\left(\frac{1}{2} + z_8\right) a\,\hat{\mathbf{x}} - x_8 a\,\hat{\mathbf{y}} + \left(\frac{1}{2} + x_8\right) a\,\hat{\mathbf{z}}$ | (24k) | W |
| $\mathbf{B}_{115}$ | = | $-z_8\,\mathbf{a}_1 + \left(\frac{1}{2} + x_8\right)\,\mathbf{a}_2 + \left(\frac{1}{2} + x_8\right)\,\mathbf{a}_3$ | = | $-z_8 a\,\hat{\mathbf{x}} + \left(\frac{1}{2} + x_8\right) a\,\hat{\mathbf{y}} + \left(\frac{1}{2} + x_8\right) a\,\hat{\mathbf{z}}$ | (24k) | W |
| $\mathbf{B}_{116}$ | = | $-z_8\,\mathbf{a}_1 - x_8\,\mathbf{a}_2 - x_8\,\mathbf{a}_3$ | = | $-z_8 a\,\hat{\mathbf{x}} - x_8 a\,\hat{\mathbf{y}} - x_8 a\,\hat{\mathbf{z}}$ | (24k) | W |

---

---

**Geometry files:**

- CIF: pp. 1811
- POSCAR: pp. 1811

# LaH$_{10}$ High-T$_c$ Superconductor Structure: A10B_cF44_225_cf_b

http://aflow.org/prototype-encyclopedia/A10B_cF44_225_cf_b

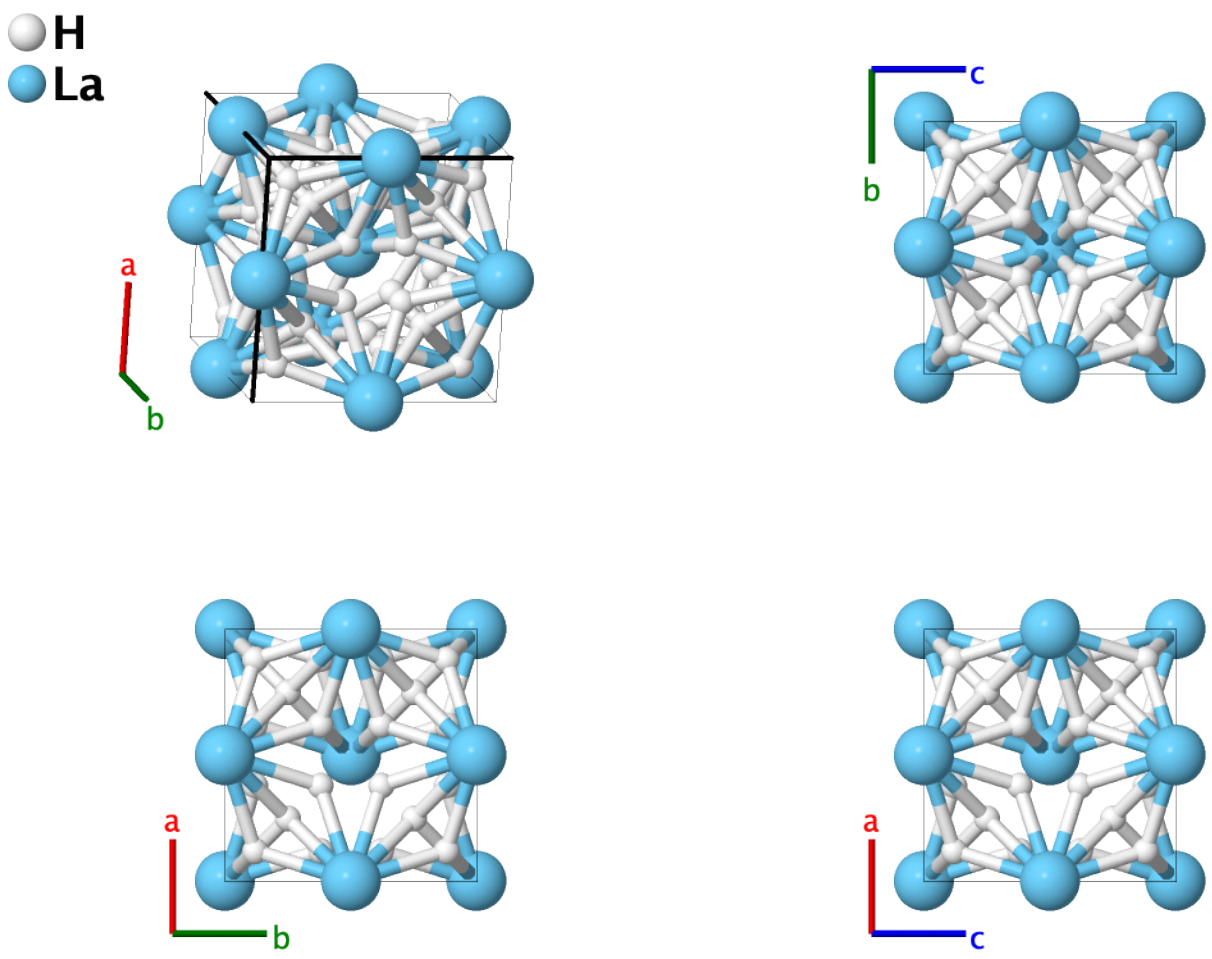

| | | |
|---|---|---|
| **Prototype** | : | $H_{10}La$ |
| **AFLOW prototype label** | : | A10B_cF44_225_cf_b |
| ***Strukturbericht* designation** | : | None |
| **Pearson symbol** | : | cF44 |
| **Space group number** | : | 225 |
| **Space group symbol** | : | $Fm\bar{3}m$ |
| **AFLOW prototype command** | : | `aflow --proto=A10B_cF44_225_cf_b --params=`$a, x_3$ |

- This structure was predicted by (Liu, 2017) as a possible near-room-temperature superconductor at pressures above 200 GPa. Its existence was confirmed by (Geballe, 2018).
- The lattice constant $a$ = 4.78 Å is taken from (Liu, 2017) prediction at 300 GPa. (Geballe, 2018) found $a$ = 5.09 Å at 172 GPa.
- (Liu, 2017) give a list of hydrogen positions which are not compatible with space group $Fm\bar{3}m$ #225. The positions we assigned are consistent with the work of (Geballe, 2018).

**Face-centered Cubic primitive vectors:**

$$\mathbf{a}_1 = \frac{1}{2}\,a\,\hat{\mathbf{y}} + \frac{1}{2}\,a\,\hat{\mathbf{z}}$$
$$\mathbf{a}_2 = \frac{1}{2}\,a\,\hat{\mathbf{x}} + \frac{1}{2}\,a\,\hat{\mathbf{z}}$$
$$\mathbf{a}_3 = \frac{1}{2}\,a\,\hat{\mathbf{x}} + \frac{1}{2}\,a\,\hat{\mathbf{y}}$$

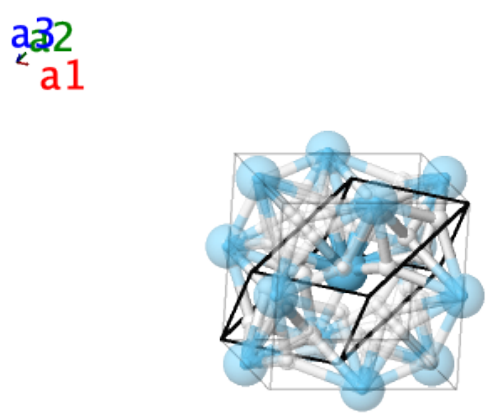

**Basis vectors:**

| | | Lattice Coordinates | | Cartesian Coordinates | Wyckoff Position | Atom Type |
|---|---|---|---|---|---|---|
| $\mathbf{B}_1$ | = | $\frac{1}{2}\,\mathbf{a}_1 + \frac{1}{2}\,\mathbf{a}_2 + \frac{1}{2}\,\mathbf{a}_3$ | = | $\frac{1}{2}a\,\hat{\mathbf{x}} + \frac{1}{2}a\,\hat{\mathbf{y}} + \frac{1}{2}a\,\hat{\mathbf{z}}$ | $(4b)$ | La |
| $\mathbf{B}_2$ | = | $\frac{1}{4}\,\mathbf{a}_1 + \frac{1}{4}\,\mathbf{a}_2 + \frac{1}{4}\,\mathbf{a}_3$ | = | $\frac{1}{4}a\,\hat{\mathbf{x}} + \frac{1}{4}a\,\hat{\mathbf{y}} + \frac{1}{4}a\,\hat{\mathbf{z}}$ | $(8c)$ | H I |
| $\mathbf{B}_3$ | = | $\frac{3}{4}\,\mathbf{a}_1 + \frac{3}{4}\,\mathbf{a}_2 + \frac{3}{4}\,\mathbf{a}_3$ | = | $\frac{3}{4}a\,\hat{\mathbf{x}} + \frac{3}{4}a\,\hat{\mathbf{y}} + \frac{3}{4}a\,\hat{\mathbf{z}}$ | $(8c)$ | H I |
| $\mathbf{B}_4$ | = | $x_3\,\mathbf{a}_1 + x_3\,\mathbf{a}_2 + x_3\,\mathbf{a}_3$ | = | $x_3a\,\hat{\mathbf{x}} + x_3a\,\hat{\mathbf{y}} + x_3a\,\hat{\mathbf{z}}$ | $(32f)$ | H II |
| $\mathbf{B}_5$ | = | $x_3\,\mathbf{a}_1 + x_3\,\mathbf{a}_2 - 3x_3\,\mathbf{a}_3$ | = | $-x_3a\,\hat{\mathbf{x}} - x_3a\,\hat{\mathbf{y}} + x_3a\,\hat{\mathbf{z}}$ | $(32f)$ | H II |
| $\mathbf{B}_6$ | = | $x_3\,\mathbf{a}_1 - 3x_3\,\mathbf{a}_2 + x_3\,\mathbf{a}_3$ | = | $-x_3a\,\hat{\mathbf{x}} + x_3a\,\hat{\mathbf{y}} - x_3a\,\hat{\mathbf{z}}$ | $(32f)$ | H II |
| $\mathbf{B}_7$ | = | $-3x_3\,\mathbf{a}_1 + x_3\,\mathbf{a}_2 + x_3\,\mathbf{a}_3$ | = | $x_3a\,\hat{\mathbf{x}} - x_3a\,\hat{\mathbf{y}} - x_3a\,\hat{\mathbf{z}}$ | $(32f)$ | H II |
| $\mathbf{B}_8$ | = | $-x_3\,\mathbf{a}_1 - x_3\,\mathbf{a}_2 + 3x_3\,\mathbf{a}_3$ | = | $x_3a\,\hat{\mathbf{x}} + x_3a\,\hat{\mathbf{y}} - x_3a\,\hat{\mathbf{z}}$ | $(32f)$ | H II |
| $\mathbf{B}_9$ | = | $-x_3\,\mathbf{a}_1 - x_3\,\mathbf{a}_2 - x_3\,\mathbf{a}_3$ | = | $-x_3a\,\hat{\mathbf{x}} - x_3a\,\hat{\mathbf{y}} - x_3a\,\hat{\mathbf{z}}$ | $(32f)$ | H II |
| $\mathbf{B}_{10}$ | = | $-x_3\,\mathbf{a}_1 + 3x_3\,\mathbf{a}_2 - x_3\,\mathbf{a}_3$ | = | $x_3a\,\hat{\mathbf{x}} - x_3a\,\hat{\mathbf{y}} + x_3a\,\hat{\mathbf{z}}$ | $(32f)$ | H II |
| $\mathbf{B}_{11}$ | = | $3x_3\,\mathbf{a}_1 - x_3\,\mathbf{a}_2 - x_3\,\mathbf{a}_3$ | = | $-x_3a\,\hat{\mathbf{x}} + x_3a\,\hat{\mathbf{y}} + x_3a\,\hat{\mathbf{z}}$ | $(32f)$ | H II |

**Geometry files:**

- CIF: pp. 1812
- POSCAR: pp. 1813

# Double Perovskite ($Ba_2MnWO_6$) Structure: A2BC6D_cF40_225_c_a_e_b

http://aflow.org/prototype-encyclopedia/A2BC6D_cF40_225_c_a_e_b

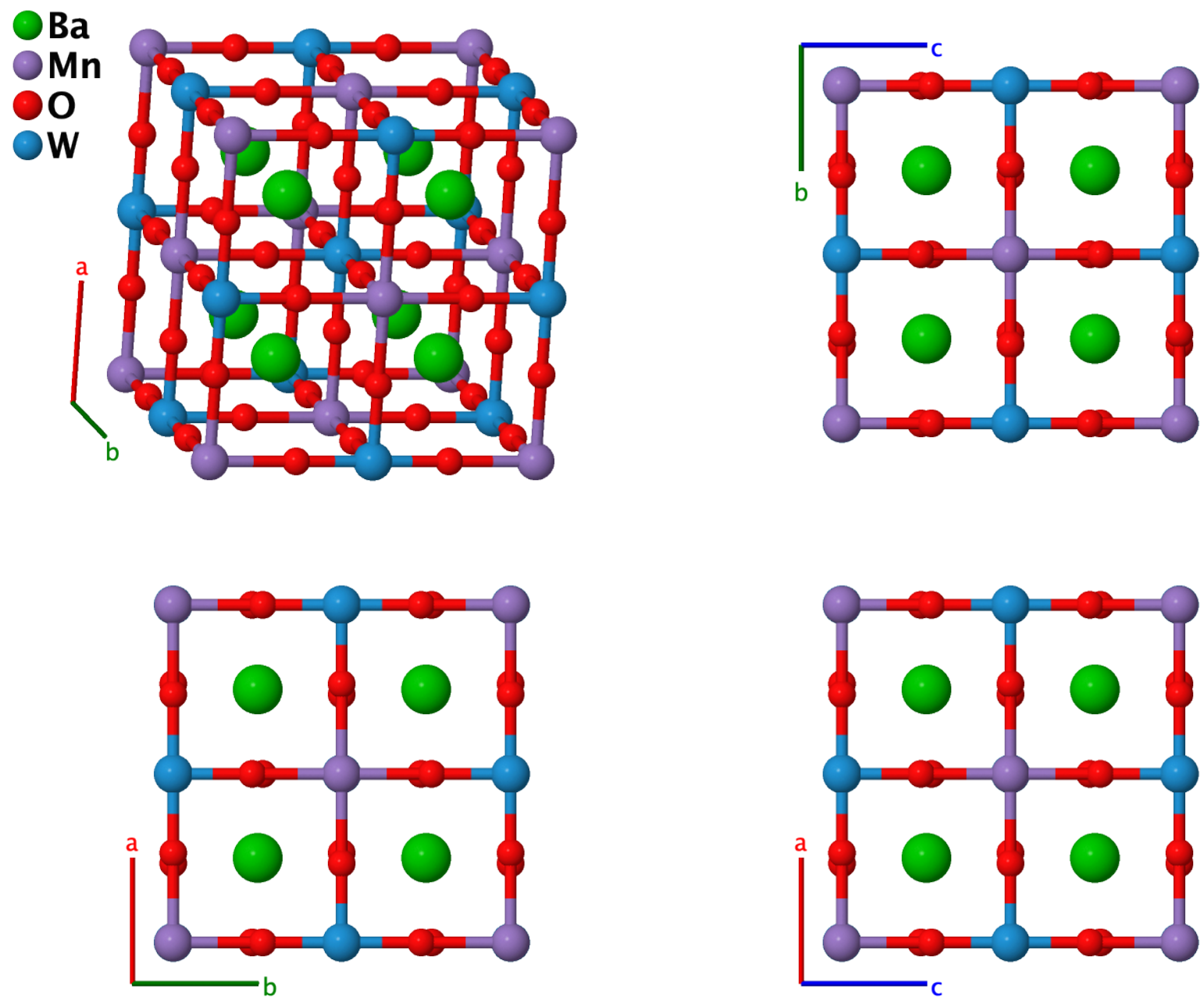

| | | |
|---|---|---|
| **Prototype** | : | $Ba_2MnO_6W$ |
| **AFLOW prototype label** | : | A2BC6D_cF40_225_c_a_e_b |
| ***Strukturbericht* designation** | : | None |
| **Pearson symbol** | : | cF40 |
| **Space group number** | : | 225 |
| **Space group symbol** | : | $Fm\bar{3}m$ |
| **AFLOW prototype command** | : | `aflow --proto=A2BC6D_cF40_225_c_a_e_b --params=`$a, x_4$ |

### Other compounds with this structure

- $Ba_2CaWO_6$, $Ba_2LiOsO_6$, $Ba_2NaOsO_6$, $Bi_2FeCrO_6$, $Ca_2MnReO_6$, $Cu_2TiSiO_6$, $Mn_2FeSbO_6$, $Sr_2FeMoO_6$, $Sr_2ReMoO_6$, $(La_{0.5}Sc_{0.5})_2MnCoO_6$, and $K_2NaAlF_6$ (elpasolite)

- We use the data taken at room temperature.

### Face-centered Cubic primitive vectors:

$$\mathbf{a}_1 = \frac{1}{2}\,a\,\hat{\mathbf{y}} + \frac{1}{2}\,a\,\hat{\mathbf{z}}$$
$$\mathbf{a}_2 = \frac{1}{2}\,a\,\hat{\mathbf{x}} + \frac{1}{2}\,a\,\hat{\mathbf{z}}$$
$$\mathbf{a}_3 = \frac{1}{2}\,a\,\hat{\mathbf{x}} + \frac{1}{2}\,a\,\hat{\mathbf{y}}$$

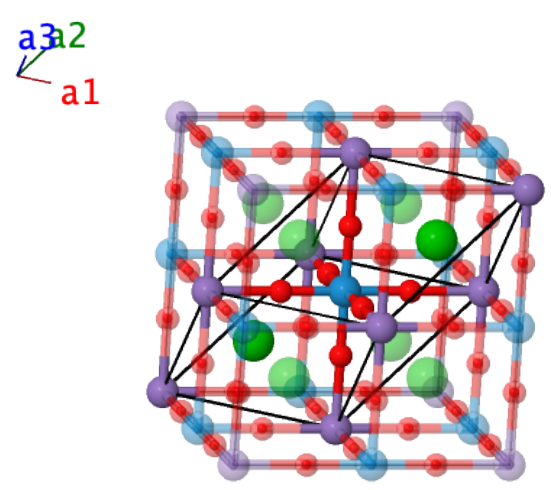

### Basis vectors:

| | | Lattice Coordinates | | Cartesian Coordinates | Wyckoff Position | Atom Type |
|---|---|---|---|---|---|---|
| $\mathbf{B}_1$ | = | $0\,\mathbf{a}_1 + 0\,\mathbf{a}_2 + 0\,\mathbf{a}_3$ | = | $0\,\hat{\mathbf{x}} + 0\,\hat{\mathbf{y}} + 0\,\hat{\mathbf{z}}$ | (4$a$) | Mn |
| $\mathbf{B}_2$ | = | $\frac{1}{2}\,\mathbf{a}_1 + \frac{1}{2}\,\mathbf{a}_2 + \frac{1}{2}\,\mathbf{a}_3$ | = | $\frac{1}{2}a\,\hat{\mathbf{x}} + \frac{1}{2}a\,\hat{\mathbf{y}} + \frac{1}{2}a\,\hat{\mathbf{z}}$ | (4$b$) | W |
| $\mathbf{B}_3$ | = | $\frac{1}{4}\,\mathbf{a}_1 + \frac{1}{4}\,\mathbf{a}_2 + \frac{1}{4}\,\mathbf{a}_3$ | = | $\frac{1}{4}a\,\hat{\mathbf{x}} + \frac{1}{4}a\,\hat{\mathbf{y}} + \frac{1}{4}a\,\hat{\mathbf{z}}$ | (8$c$) | Ba |
| $\mathbf{B}_4$ | = | $\frac{3}{4}\,\mathbf{a}_1 + \frac{3}{4}\,\mathbf{a}_2 + \frac{3}{4}\,\mathbf{a}_3$ | = | $\frac{3}{4}a\,\hat{\mathbf{x}} + \frac{3}{4}a\,\hat{\mathbf{y}} + \frac{3}{4}a\,\hat{\mathbf{z}}$ | (8$c$) | Ba |
| $\mathbf{B}_5$ | = | $-x_4\,\mathbf{a}_1 + x_4\,\mathbf{a}_2 + x_4\,\mathbf{a}_3$ | = | $x_4a\,\hat{\mathbf{x}}$ | (24$e$) | O |
| $\mathbf{B}_6$ | = | $x_4\,\mathbf{a}_1 - x_4\,\mathbf{a}_2 - x_4\,\mathbf{a}_3$ | = | $-x_4a\,\hat{\mathbf{x}}$ | (24$e$) | O |
| $\mathbf{B}_7$ | = | $x_4\,\mathbf{a}_1 - x_4\,\mathbf{a}_2 + x_4\,\mathbf{a}_3$ | = | $x_4a\,\hat{\mathbf{y}}$ | (24$e$) | O |
| $\mathbf{B}_8$ | = | $-x_4\,\mathbf{a}_1 + x_4\,\mathbf{a}_2 - x_4\,\mathbf{a}_3$ | = | $-x_4a\,\hat{\mathbf{y}}$ | (24$e$) | O |
| $\mathbf{B}_9$ | = | $x_4\,\mathbf{a}_1 + x_4\,\mathbf{a}_2 - x_4\,\mathbf{a}_3$ | = | $x_4a\,\hat{\mathbf{z}}$ | (24$e$) | O |
| $\mathbf{B}_{10}$ | = | $-x_4\,\mathbf{a}_1 - x_4\,\mathbf{a}_2 + x_4\,\mathbf{a}_3$ | = | $-x_4a\,\hat{\mathbf{z}}$ | (24$e$) | O |

**Geometry files:**

- CIF: pp. 1813
- POSCAR: pp. 1814

# $Cu_3[Fe(CN)_6]_2 \cdot xH_2O$ ($J2_5$, $x \approx 3$) Structure: A6B9CD2E6_cF96_225_e_bf_a_c_e

http://aflow.org/prototype-encyclopedia/A6B9CD2E6_cF96_225_e_bf_a_c_e

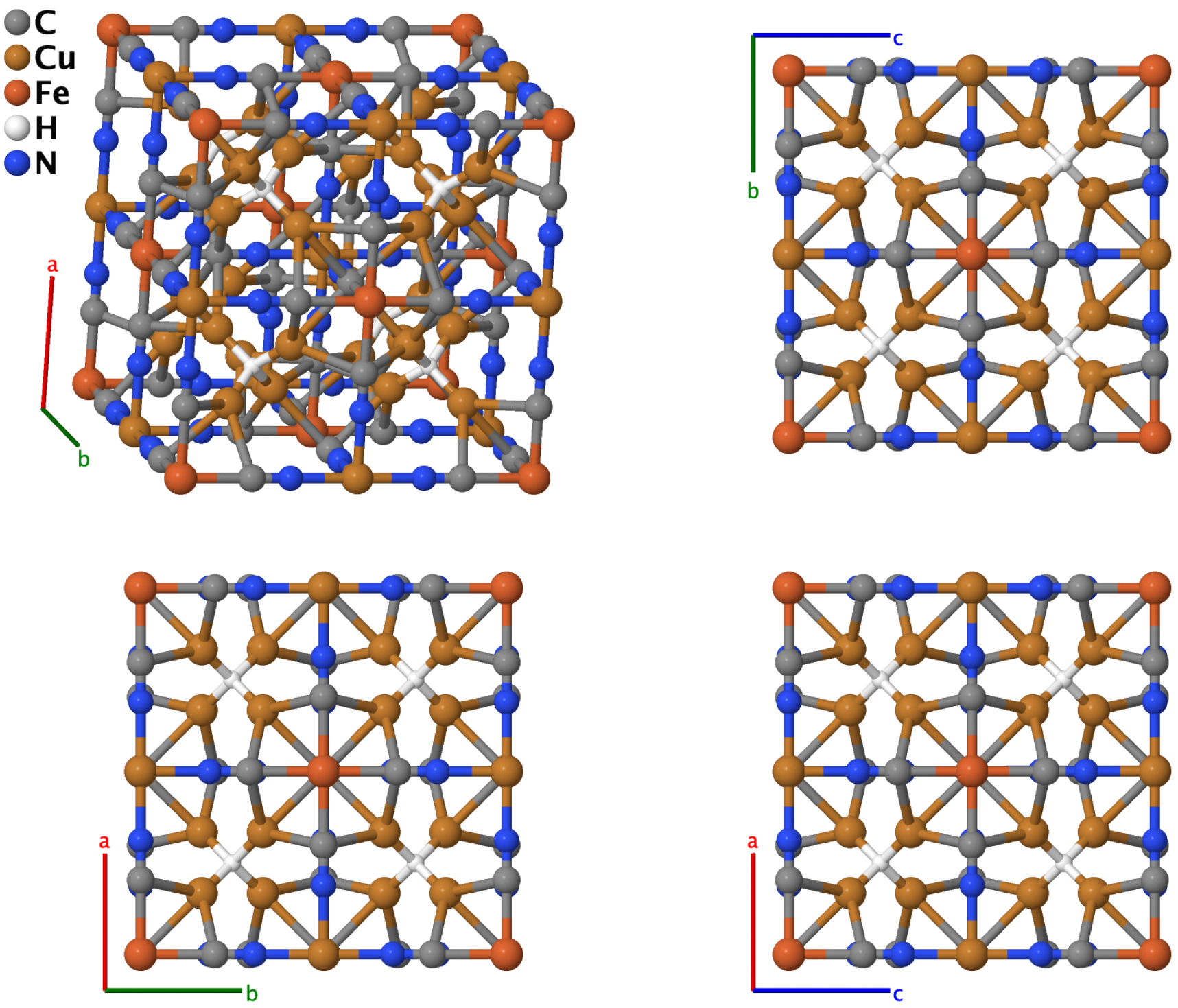

| | | |
|---|---|---|
| **Prototype** | : | $C_{12}Cu_3Fe_2(H_2O)_xN_{12}$ |
| **AFLOW prototype label** | : | A6B9CD2E6_cF96_225_e_bf_a_c_e |
| ***Strukturbericht* designation** | : | $J2_5$ |
| **Pearson symbol** | : | cF96 |
| **Space group number** | : | 225 |
| **Space group symbol** | : | $Fm\bar{3}m$ |
| **AFLOW prototype command** | : | `aflow --proto=A6B9CD2E6_cF96_225_e_bf_a_c_e` <br> `--params=`$a, x_4, x_5, x_6$ |

### Other compounds with this structure

- $Cd_3[Co(CN)_6]_2$, $Co_3[Co(CN)_6]_2$, $Cu_3[Fe(CN)_6]_2$, $Fe_3[Fe(CN)_6]_2$, $Td_3[Fe(CN)_6]_2$, and $Zn_3[Fe(CN)_6]_2$

---

- These compounds form a class called "Prussian Blue Analogs," where Prussian Blue is $Fe_3[Fe(CN)_6]_2$.
- (van Bever, 1938) studied what he believed to be the hydrated form of this structure, with $x \approx 3$. In that case, the water molecules occupy the (8*c*) sites, but each site is only occupied 75% of the time. The water sites are surrounded by a tetrahedron of copper (32*e*) sites, but only 6.25% of these sites are occupied.
- (Weiser, 1942) studied the anhydrous form. They found that in this case the copper atoms that were on the (32*e*) sites replace the water molecules on the (8*c*) site, and this site is now fully occupied with copper.
- To convert from the hydrated to anhydrous structure, remove the copper (32*e*) atoms from the (32*e*) sites in the CIF or POSCAR file, and relabel the (8*c*) site as copper.

- For a picture of the resulting structure see (Jiao, 2017).
- The AFLOW label models the structure as if the sites were fully occupied.

---

**Face-centered Cubic primitive vectors:**

$$\mathbf{a}_1 = \frac{1}{2}\,a\,\mathbf{\hat{y}} + \frac{1}{2}\,a\,\mathbf{\hat{z}}$$
$$\mathbf{a}_2 = \frac{1}{2}\,a\,\mathbf{\hat{x}} + \frac{1}{2}\,a\,\mathbf{\hat{z}}$$
$$\mathbf{a}_3 = \frac{1}{2}\,a\,\mathbf{\hat{x}} + \frac{1}{2}\,a\,\mathbf{\hat{y}}$$

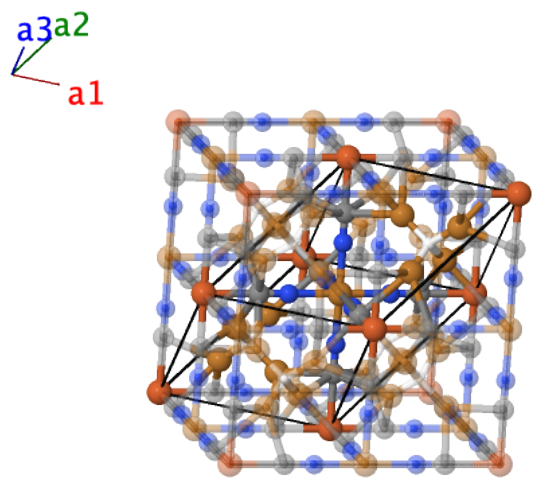

---

**Basis vectors:**

| | | Lattice Coordinates | | Cartesian Coordinates | Wyckoff Position | Atom Type |
|---|---|---|---|---|---|---|
| $\mathbf{B}_1$ | = | $0\,\mathbf{a}_1 + 0\,\mathbf{a}_2 + 0\,\mathbf{a}_3$ | = | $0\,\mathbf{\hat{x}} + 0\,\mathbf{\hat{y}} + 0\,\mathbf{\hat{z}}$ | (4$a$) | Fe |
| $\mathbf{B}_2$ | = | $\frac{1}{2}\,\mathbf{a}_1 + \frac{1}{2}\,\mathbf{a}_2 + \frac{1}{2}\,\mathbf{a}_3$ | = | $\frac{1}{2}a\,\mathbf{\hat{x}} + \frac{1}{2}a\,\mathbf{\hat{y}} + \frac{1}{2}a\,\mathbf{\hat{z}}$ | (4$b$) | Cu I |
| $\mathbf{B}_3$ | = | $\frac{1}{4}\,\mathbf{a}_1 + \frac{1}{4}\,\mathbf{a}_2 + \frac{1}{4}\,\mathbf{a}_3$ | = | $\frac{1}{4}a\,\mathbf{\hat{x}} + \frac{1}{4}a\,\mathbf{\hat{y}} + \frac{1}{4}a\,\mathbf{\hat{z}}$ | (8$c$) | $H_2O$ |
| $\mathbf{B}_4$ | = | $\frac{3}{4}\,\mathbf{a}_1 + \frac{3}{4}\,\mathbf{a}_2 + \frac{3}{4}\,\mathbf{a}_3$ | = | $\frac{3}{4}a\,\mathbf{\hat{x}} + \frac{3}{4}a\,\mathbf{\hat{y}} + \frac{3}{4}a\,\mathbf{\hat{z}}$ | (8$c$) | $H_2O$ |
| $\mathbf{B}_5$ | = | $-x_4\,\mathbf{a}_1 + x_4\,\mathbf{a}_2 + x_4\,\mathbf{a}_3$ | = | $x_4 a\,\mathbf{\hat{x}}$ | (24$e$) | C |
| $\mathbf{B}_6$ | = | $x_4\,\mathbf{a}_1 - x_4\,\mathbf{a}_2 - x_4\,\mathbf{a}_3$ | = | $-x_4 a\,\mathbf{\hat{x}}$ | (24$e$) | C |
| $\mathbf{B}_7$ | = | $x_4\,\mathbf{a}_1 - x_4\,\mathbf{a}_2 + x_4\,\mathbf{a}_3$ | = | $x_4 a\,\mathbf{\hat{y}}$ | (24$e$) | C |
| $\mathbf{B}_8$ | = | $-x_4\,\mathbf{a}_1 + x_4\,\mathbf{a}_2 - x_4\,\mathbf{a}_3$ | = | $-x_4 a\,\mathbf{\hat{y}}$ | (24$e$) | C |
| $\mathbf{B}_9$ | = | $x_4\,\mathbf{a}_1 + x_4\,\mathbf{a}_2 - x_4\,\mathbf{a}_3$ | = | $x_4 a\,\mathbf{\hat{z}}$ | (24$e$) | C |
| $\mathbf{B}_{10}$ | = | $-x_4\,\mathbf{a}_1 - x_4\,\mathbf{a}_2 + x_4\,\mathbf{a}_3$ | = | $-x_4 a\,\mathbf{\hat{z}}$ | (24$e$) | C |
| $\mathbf{B}_{11}$ | = | $-x_5\,\mathbf{a}_1 + x_5\,\mathbf{a}_2 + x_5\,\mathbf{a}_3$ | = | $x_5 a\,\mathbf{\hat{x}}$ | (24$e$) | N |
| $\mathbf{B}_{12}$ | = | $x_5\,\mathbf{a}_1 - x_5\,\mathbf{a}_2 - x_5\,\mathbf{a}_3$ | = | $-x_5 a\,\mathbf{\hat{x}}$ | (24$e$) | N |
| $\mathbf{B}_{13}$ | = | $x_5\,\mathbf{a}_1 - x_5\,\mathbf{a}_2 + x_5\,\mathbf{a}_3$ | = | $x_5 a\,\mathbf{\hat{y}}$ | (24$e$) | N |
| $\mathbf{B}_{14}$ | = | $-x_5\,\mathbf{a}_1 + x_5\,\mathbf{a}_2 - x_5\,\mathbf{a}_3$ | = | $-x_5 a\,\mathbf{\hat{y}}$ | (24$e$) | N |
| $\mathbf{B}_{15}$ | = | $x_5\,\mathbf{a}_1 + x_5\,\mathbf{a}_2 - x_5\,\mathbf{a}_3$ | = | $x_5 a\,\mathbf{\hat{z}}$ | (24$e$) | N |
| $\mathbf{B}_{16}$ | = | $-x_5\,\mathbf{a}_1 - x_5\,\mathbf{a}_2 + x_5\,\mathbf{a}_3$ | = | $-x_5 a\,\mathbf{\hat{z}}$ | (24$e$) | N |
| $\mathbf{B}_{17}$ | = | $x_6\,\mathbf{a}_1 + x_6\,\mathbf{a}_2 + x_6\,\mathbf{a}_3$ | = | $x_6 a\,\mathbf{\hat{x}} + x_6 a\,\mathbf{\hat{y}} + x_6 a\,\mathbf{\hat{z}}$ | (32$f$) | Cu II |
| $\mathbf{B}_{18}$ | = | $x_6\,\mathbf{a}_1 + x_6\,\mathbf{a}_2 - 3x_6\,\mathbf{a}_3$ | = | $-x_6 a\,\mathbf{\hat{x}} - x_6 a\,\mathbf{\hat{y}} + x_6 a\,\mathbf{\hat{z}}$ | (32$f$) | Cu II |
| $\mathbf{B}_{19}$ | = | $x_6\,\mathbf{a}_1 - 3x_6\,\mathbf{a}_2 + x_6\,\mathbf{a}_3$ | = | $-x_6 a\,\mathbf{\hat{x}} + x_6 a\,\mathbf{\hat{y}} - x_6 a\,\mathbf{\hat{z}}$ | (32$f$) | Cu II |
| $\mathbf{B}_{20}$ | = | $-3x_6\,\mathbf{a}_1 + x_6\,\mathbf{a}_2 + x_6\,\mathbf{a}_3$ | = | $x_6 a\,\mathbf{\hat{x}} - x_6 a\,\mathbf{\hat{y}} - x_6 a\,\mathbf{\hat{z}}$ | (32$f$) | Cu II |
| $\mathbf{B}_{21}$ | = | $-x_6\,\mathbf{a}_1 - x_6\,\mathbf{a}_2 + 3x_6\,\mathbf{a}_3$ | = | $x_6 a\,\mathbf{\hat{x}} + x_6 a\,\mathbf{\hat{y}} - x_6 a\,\mathbf{\hat{z}}$ | (32$f$) | Cu II |
| $\mathbf{B}_{22}$ | = | $-x_6\,\mathbf{a}_1 - x_6\,\mathbf{a}_2 - x_6\,\mathbf{a}_3$ | = | $-x_6 a\,\mathbf{\hat{x}} - x_6 a\,\mathbf{\hat{y}} - x_6 a\,\mathbf{\hat{z}}$ | (32$f$) | Cu II |
| $\mathbf{B}_{23}$ | = | $-x_6\,\mathbf{a}_1 + 3x_6\,\mathbf{a}_2 - x_6\,\mathbf{a}_3$ | = | $x_6 a\,\mathbf{\hat{x}} - x_6 a\,\mathbf{\hat{y}} + x_6 a\,\mathbf{\hat{z}}$ | (32$f$) | Cu II |
| $\mathbf{B}_{24}$ | = | $3x_6\,\mathbf{a}_1 - x_6\,\mathbf{a}_2 - x_6\,\mathbf{a}_3$ | = | $-x_6 a\,\mathbf{\hat{x}} + x_6 a\,\mathbf{\hat{y}} + x_6 a\,\mathbf{\hat{z}}$ | (32$f$) | Cu II |

**Geometry files:**

- CIF: pp. 1814
- POSCAR: pp. 1816

# $Co_9S_8$ ($D8_9$) Structure: A9B8_cF68_225_af_ce

http://aflow.org/prototype-encyclopedia/A9B8_cF68_225_af_ce

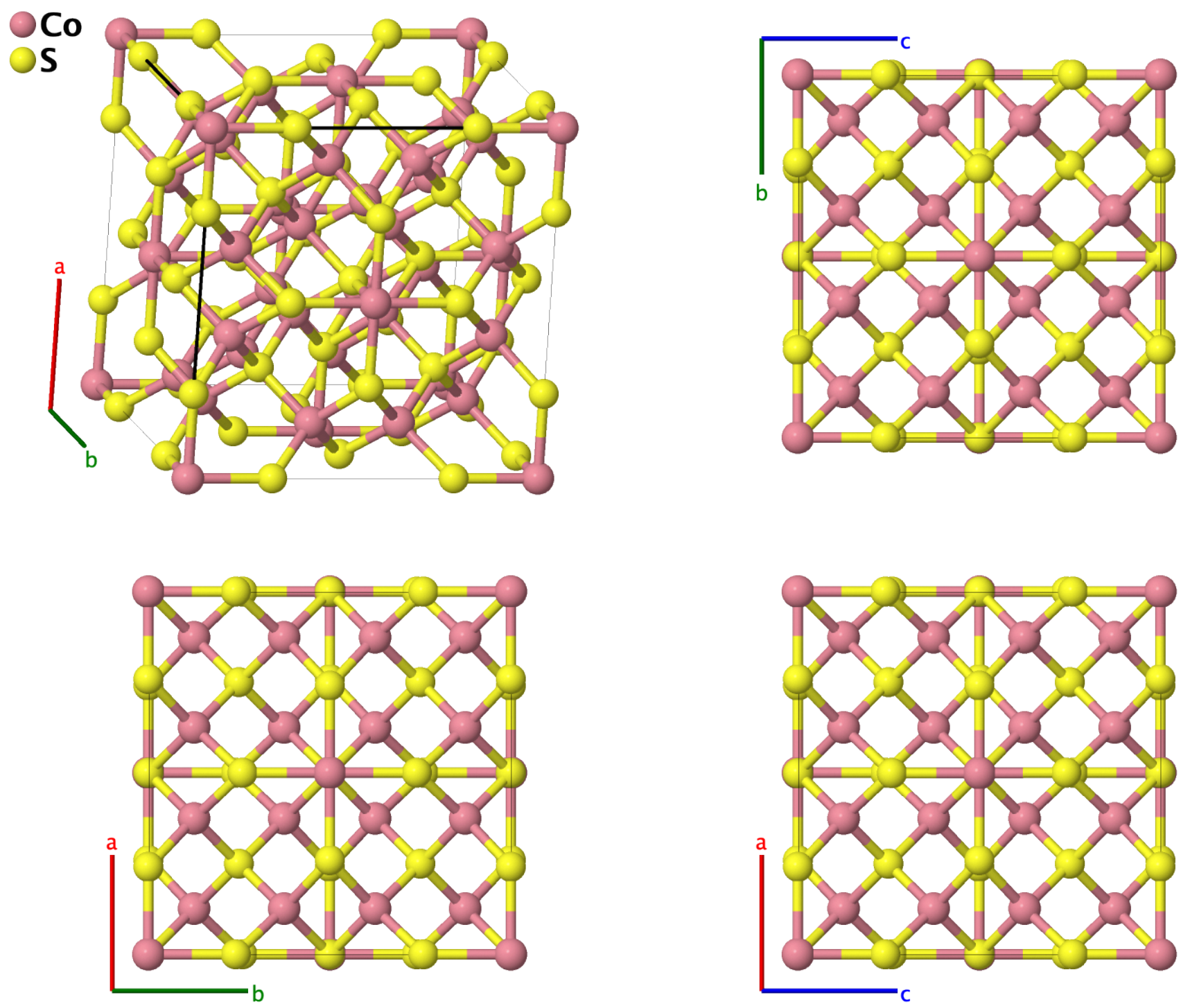

| | | |
|---|---|---|
| **Prototype** | : | $Co_9S_8$ |
| **AFLOW prototype label** | : | A9B8_cF68_225_af_ce |
| ***Strukturbericht* designation** | : | $D8_9$ |
| **Pearson symbol** | : | cF68 |
| **Space group number** | : | 225 |
| **Space group symbol** | : | $Fm\bar{3}m$ |
| **AFLOW prototype command** | : | `aflow --proto=A9B8_cF68_225_af_ce` `--params=`$a, x_3, x_4$ |

**Other compounds with this structure**

- $(Fe,Ni)_9S_8$ (pentlandite)

- (Geller,1962) placed the first Co atom at the (4*b*) Wyckoff position. We have shifted this to the origin, the (4*a*) Wyckoff position.

**Face-centered Cubic primitive vectors:**

$$\mathbf{a}_1 = \frac{1}{2}\,a\,\hat{\mathbf{y}} + \frac{1}{2}\,a\,\hat{\mathbf{z}}$$
$$\mathbf{a}_2 = \frac{1}{2}\,a\,\hat{\mathbf{x}} + \frac{1}{2}\,a\,\hat{\mathbf{z}}$$
$$\mathbf{a}_3 = \frac{1}{2}\,a\,\hat{\mathbf{x}} + \frac{1}{2}\,a\,\hat{\mathbf{y}}$$

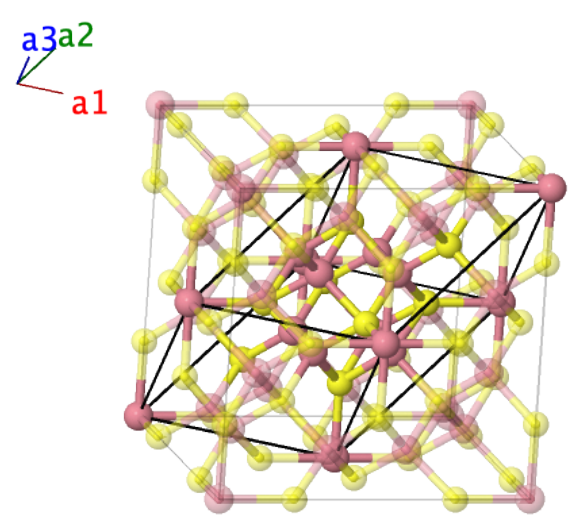

**Basis vectors:**

| | | Lattice Coordinates | | Cartesian Coordinates | Wyckoff Position | Atom Type |
|---|---|---|---|---|---|---|
| $\mathbf{B}_1$ | = | $0\,\mathbf{a}_1 + 0\,\mathbf{a}_2 + 0\,\mathbf{a}_3$ | = | $0\,\hat{\mathbf{x}} + 0\,\hat{\mathbf{y}} + 0\,\hat{\mathbf{z}}$ | $(4a)$ | Co I |
| $\mathbf{B}_2$ | = | $\frac{1}{4}\,\mathbf{a}_1 + \frac{1}{4}\,\mathbf{a}_2 + \frac{1}{4}\,\mathbf{a}_3$ | = | $\frac{1}{4}a\,\hat{\mathbf{x}} + \frac{1}{4}a\,\hat{\mathbf{y}} + \frac{1}{4}a\,\hat{\mathbf{z}}$ | $(8c)$ | S I |
| $\mathbf{B}_3$ | = | $\frac{3}{4}\,\mathbf{a}_1 + \frac{3}{4}\,\mathbf{a}_2 + \frac{3}{4}\,\mathbf{a}_3$ | = | $\frac{3}{4}a\,\hat{\mathbf{x}} + \frac{3}{4}a\,\hat{\mathbf{y}} + \frac{3}{4}a\,\hat{\mathbf{z}}$ | $(8c)$ | S I |
| $\mathbf{B}_4$ | = | $-x_3\,\mathbf{a}_1 + x_3\,\mathbf{a}_2 + x_3\,\mathbf{a}_3$ | = | $x_3a\,\hat{\mathbf{x}}$ | $(24e)$ | S II |
| $\mathbf{B}_5$ | = | $x_3\,\mathbf{a}_1 - x_3\,\mathbf{a}_2 - x_3\,\mathbf{a}_3$ | = | $-x_3a\,\hat{\mathbf{x}}$ | $(24e)$ | S II |
| $\mathbf{B}_6$ | = | $x_3\,\mathbf{a}_1 - x_3\,\mathbf{a}_2 + x_3\,\mathbf{a}_3$ | = | $x_3a\,\hat{\mathbf{y}}$ | $(24e)$ | S II |
| $\mathbf{B}_7$ | = | $-x_3\,\mathbf{a}_1 + x_3\,\mathbf{a}_2 - x_3\,\mathbf{a}_3$ | = | $-x_3a\,\hat{\mathbf{y}}$ | $(24e)$ | S II |
| $\mathbf{B}_8$ | = | $x_3\,\mathbf{a}_1 + x_3\,\mathbf{a}_2 - x_3\,\mathbf{a}_3$ | = | $x_3a\,\hat{\mathbf{z}}$ | $(24e)$ | S II |
| $\mathbf{B}_9$ | = | $-x_3\,\mathbf{a}_1 - x_3\,\mathbf{a}_2 + x_3\,\mathbf{a}_3$ | = | $-x_3a\,\hat{\mathbf{z}}$ | $(24e)$ | S II |
| $\mathbf{B}_{10}$ | = | $x_4\,\mathbf{a}_1 + x_4\,\mathbf{a}_2 + x_4\,\mathbf{a}_3$ | = | $x_4a\,\hat{\mathbf{x}} + x_4a\,\hat{\mathbf{y}} + x_4a\,\hat{\mathbf{z}}$ | $(32f)$ | Co II |
| $\mathbf{B}_{11}$ | = | $x_4\,\mathbf{a}_1 + x_4\,\mathbf{a}_2 - 3x_4\,\mathbf{a}_3$ | = | $-x_4a\,\hat{\mathbf{x}} - x_4a\,\hat{\mathbf{y}} + x_4a\,\hat{\mathbf{z}}$ | $(32f)$ | Co II |
| $\mathbf{B}_{12}$ | = | $x_4\,\mathbf{a}_1 - 3x_4\,\mathbf{a}_2 + x_4\,\mathbf{a}_3$ | = | $-x_4a\,\hat{\mathbf{x}} + x_4a\,\hat{\mathbf{y}} - x_4a\,\hat{\mathbf{z}}$ | $(32f)$ | Co II |
| $\mathbf{B}_{13}$ | = | $-3x_4\,\mathbf{a}_1 + x_4\,\mathbf{a}_2 + x_4\,\mathbf{a}_3$ | = | $x_4a\,\hat{\mathbf{x}} - x_4a\,\hat{\mathbf{y}} - x_4a\,\hat{\mathbf{z}}$ | $(32f)$ | Co II |
| $\mathbf{B}_{14}$ | = | $-x_4\,\mathbf{a}_1 - x_4\,\mathbf{a}_2 + 3x_4\,\mathbf{a}_3$ | = | $x_4a\,\hat{\mathbf{x}} + x_4a\,\hat{\mathbf{y}} - x_4a\,\hat{\mathbf{z}}$ | $(32f)$ | Co II |
| $\mathbf{B}_{15}$ | = | $-x_4\,\mathbf{a}_1 - x_4\,\mathbf{a}_2 - x_4\,\mathbf{a}_3$ | = | $-x_4a\,\hat{\mathbf{x}} - x_4a\,\hat{\mathbf{y}} - x_4a\,\hat{\mathbf{z}}$ | $(32f)$ | Co II |
| $\mathbf{B}_{16}$ | = | $-x_4\,\mathbf{a}_1 + 3x_4\,\mathbf{a}_2 - x_4\,\mathbf{a}_3$ | = | $x_4a\,\hat{\mathbf{x}} - x_4a\,\hat{\mathbf{y}} + x_4a\,\hat{\mathbf{z}}$ | $(32f)$ | Co II |
| $\mathbf{B}_{17}$ | = | $3x_4\,\mathbf{a}_1 - x_4\,\mathbf{a}_2 - x_4\,\mathbf{a}_3$ | = | $-x_4a\,\hat{\mathbf{x}} + x_4a\,\hat{\mathbf{y}} + x_4a\,\hat{\mathbf{z}}$ | $(32f)$ | Co II |

**Geometry files:**

- CIF: pp. 1816
- POSCAR: pp. 1817

# $(NH_4)_3AlF_6$ ($J2_1$) Structure: AB30C16D3_cF200_225_a_ej_2f_bc

http://aflow.org/prototype-encyclopedia/AB30C16D3_cF200_225_a_ej_2f_bc

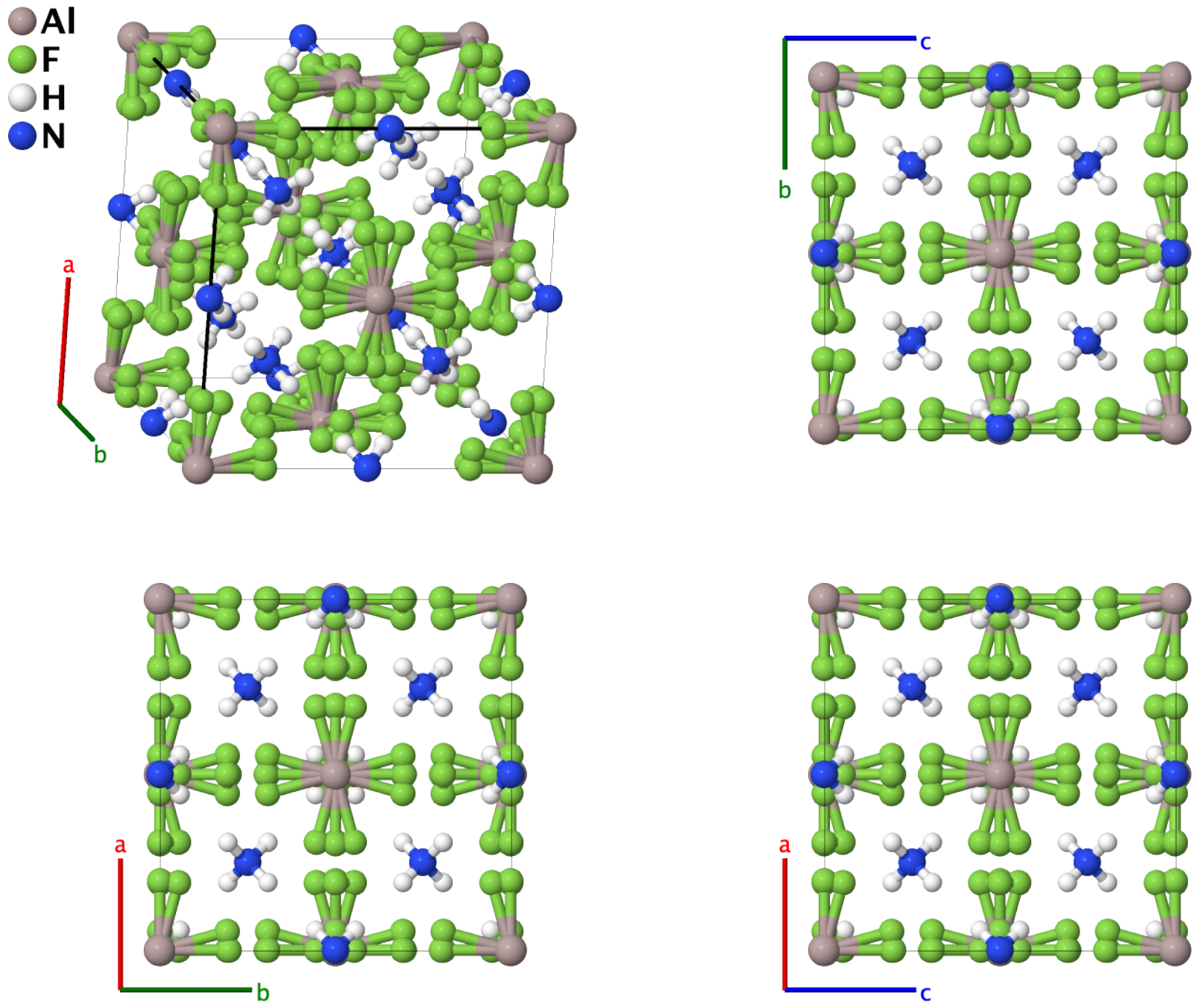

| | | |
|---|---|---|
| **Prototype** | : | $AlF_6H_{12}N_3$ |
| **AFLOW prototype label** | : | AB30C16D3_cF200_225_a_ej_2f_bc |
| ***Strukturbericht* designation** | : | $J2_1$ |
| **Pearson symbol** | : | cF200 |
| **Space group number** | : | 225 |
| **Space group symbol** | : | $Fm\bar{3}m$ |
| **AFLOW prototype command** | : | `aflow --proto=AB30C16D3_cF200_225_a_ej_2f_bc`<br>`--params=`$a, x_4, x_5, x_6, y_7, z_7$ |

### Other compounds with this structure

- $(NH_4)_3FeF_6$, $(NH_4)_3TiOF_5$, $(NH_4)_3Fe(NO_2)_6$, $K_3Ir(NO_2)_6$, $Cs_3Ir(NO_2)_6$, $Rb_3Ir(NO_2)_6$, and $Tl_3Ir(NO_2)_6$

- Early determinations of this structure placed all of the fluorine atoms on the (24$e$) site and were not able to determine the positions of the hydrogen atoms in the ammonium ion. This structure was designated $H71$ ($H7_1$) by (Ewald, 1931), renamed $I2_1$ by (Hermann, 1937) and finally given the label $J2_1$ by (Gottfried, 1937).

- The structure determined by (Udovenko, 2003) found that the fluorine atoms are split onto two sites, F-I, on Wyckoff position (24$e$) is 1/3 filled, and F-II, on (96$j$) is 1/6 filled. The positions are so close, however, that a reasonable approximation can be made by eliminating the (96$j$) site and fully occupying the (24$e$) site. The H-I (32$f$) site is fully occupied, while the H-II (32$f$) site is 50% occupied.

**Face-centered Cubic primitive vectors:**

$$\mathbf{a}_1 = \frac{1}{2}\,a\,\mathbf{\hat{y}} + \frac{1}{2}\,a\,\mathbf{\hat{z}}$$
$$\mathbf{a}_2 = \frac{1}{2}\,a\,\mathbf{\hat{x}} + \frac{1}{2}\,a\,\mathbf{\hat{z}}$$
$$\mathbf{a}_3 = \frac{1}{2}\,a\,\mathbf{\hat{x}} + \frac{1}{2}\,a\,\mathbf{\hat{y}}$$

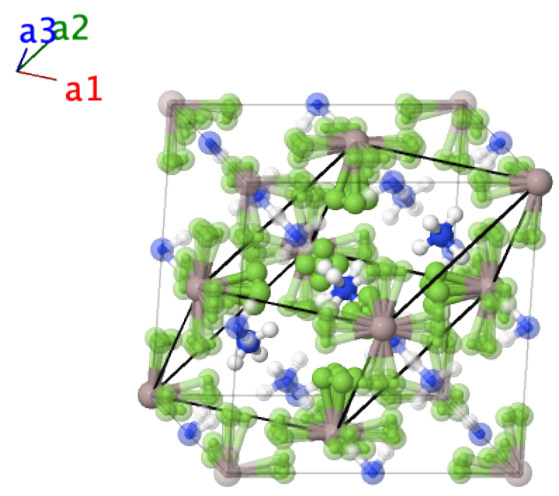

**Basis vectors:**

| | | Lattice Coordinates | | Cartesian Coordinates | Wyckoff Position | Atom Type |
|---|---|---|---|---|---|---|
| $\mathbf{B}_1$ | = | $0\,\mathbf{a}_1 + 0\,\mathbf{a}_2 + 0\,\mathbf{a}_3$ | = | $0\,\mathbf{\hat{x}} + 0\,\mathbf{\hat{y}} + 0\,\mathbf{\hat{z}}$ | $(4a)$ | Al |
| $\mathbf{B}_2$ | = | $\frac{1}{2}\,\mathbf{a}_1 + \frac{1}{2}\,\mathbf{a}_2 + \frac{1}{2}\,\mathbf{a}_3$ | = | $\frac{1}{2}a\,\mathbf{\hat{x}} + \frac{1}{2}a\,\mathbf{\hat{y}} + \frac{1}{2}a\,\mathbf{\hat{z}}$ | $(4b)$ | N I |
| $\mathbf{B}_3$ | = | $\frac{1}{4}\,\mathbf{a}_1 + \frac{1}{4}\,\mathbf{a}_2 + \frac{1}{4}\,\mathbf{a}_3$ | = | $\frac{1}{4}a\,\mathbf{\hat{x}} + \frac{1}{4}a\,\mathbf{\hat{y}} + \frac{1}{4}a\,\mathbf{\hat{z}}$ | $(8c)$ | N II |
| $\mathbf{B}_4$ | = | $\frac{3}{4}\,\mathbf{a}_1 + \frac{3}{4}\,\mathbf{a}_2 + \frac{3}{4}\,\mathbf{a}_3$ | = | $\frac{3}{4}a\,\mathbf{\hat{x}} + \frac{3}{4}a\,\mathbf{\hat{y}} + \frac{3}{4}a\,\mathbf{\hat{z}}$ | $(8c)$ | N II |
| $\mathbf{B}_5$ | = | $-x_4\,\mathbf{a}_1 + x_4\,\mathbf{a}_2 + x_4\,\mathbf{a}_3$ | = | $x_4a\,\mathbf{\hat{x}}$ | $(24e)$ | F I |
| $\mathbf{B}_6$ | = | $x_4\,\mathbf{a}_1 - x_4\,\mathbf{a}_2 - x_4\,\mathbf{a}_3$ | = | $-x_4a\,\mathbf{\hat{x}}$ | $(24e)$ | F I |
| $\mathbf{B}_7$ | = | $x_4\,\mathbf{a}_1 - x_4\,\mathbf{a}_2 + x_4\,\mathbf{a}_3$ | = | $x_4a\,\mathbf{\hat{y}}$ | $(24e)$ | F I |
| $\mathbf{B}_8$ | = | $-x_4\,\mathbf{a}_1 + x_4\,\mathbf{a}_2 - x_4\,\mathbf{a}_3$ | = | $-x_4a\,\mathbf{\hat{y}}$ | $(24e)$ | F I |
| $\mathbf{B}_9$ | = | $x_4\,\mathbf{a}_1 + x_4\,\mathbf{a}_2 - x_4\,\mathbf{a}_3$ | = | $x_4a\,\mathbf{\hat{z}}$ | $(24e)$ | F I |
| $\mathbf{B}_{10}$ | = | $-x_4\,\mathbf{a}_1 - x_4\,\mathbf{a}_2 + x_4\,\mathbf{a}_3$ | = | $-x_4a\,\mathbf{\hat{z}}$ | $(24e)$ | F I |
| $\mathbf{B}_{11}$ | = | $x_5\,\mathbf{a}_1 + x_5\,\mathbf{a}_2 + x_5\,\mathbf{a}_3$ | = | $x_5a\,\mathbf{\hat{x}} + x_5a\,\mathbf{\hat{y}} + x_5a\,\mathbf{\hat{z}}$ | $(32f)$ | H I |
| $\mathbf{B}_{12}$ | = | $x_5\,\mathbf{a}_1 + x_5\,\mathbf{a}_2 - 3x_5\,\mathbf{a}_3$ | = | $-x_5a\,\mathbf{\hat{x}} - x_5a\,\mathbf{\hat{y}} + x_5a\,\mathbf{\hat{z}}$ | $(32f)$ | H I |
| $\mathbf{B}_{13}$ | = | $x_5\,\mathbf{a}_1 - 3x_5\,\mathbf{a}_2 + x_5\,\mathbf{a}_3$ | = | $-x_5a\,\mathbf{\hat{x}} + x_5a\,\mathbf{\hat{y}} - x_5a\,\mathbf{\hat{z}}$ | $(32f)$ | H I |
| $\mathbf{B}_{14}$ | = | $-3x_5\,\mathbf{a}_1 + x_5\,\mathbf{a}_2 + x_5\,\mathbf{a}_3$ | = | $x_5a\,\mathbf{\hat{x}} - x_5a\,\mathbf{\hat{y}} - x_5a\,\mathbf{\hat{z}}$ | $(32f)$ | H I |
| $\mathbf{B}_{15}$ | = | $-x_5\,\mathbf{a}_1 - x_5\,\mathbf{a}_2 + 3x_5\,\mathbf{a}_3$ | = | $x_5a\,\mathbf{\hat{x}} + x_5a\,\mathbf{\hat{y}} - x_5a\,\mathbf{\hat{z}}$ | $(32f)$ | H I |
| $\mathbf{B}_{16}$ | = | $-x_5\,\mathbf{a}_1 - x_5\,\mathbf{a}_2 - x_5\,\mathbf{a}_3$ | = | $-x_5a\,\mathbf{\hat{x}} - x_5a\,\mathbf{\hat{y}} - x_5a\,\mathbf{\hat{z}}$ | $(32f)$ | H I |
| $\mathbf{B}_{17}$ | = | $-x_5\,\mathbf{a}_1 + 3x_5\,\mathbf{a}_2 - x_5\,\mathbf{a}_3$ | = | $x_5a\,\mathbf{\hat{x}} - x_5a\,\mathbf{\hat{y}} + x_5a\,\mathbf{\hat{z}}$ | $(32f)$ | H I |
| $\mathbf{B}_{18}$ | = | $3x_5\,\mathbf{a}_1 - x_5\,\mathbf{a}_2 - x_5\,\mathbf{a}_3$ | = | $-x_5a\,\mathbf{\hat{x}} + x_5a\,\mathbf{\hat{y}} + x_5a\,\mathbf{\hat{z}}$ | $(32f)$ | H I |
| $\mathbf{B}_{19}$ | = | $x_6\,\mathbf{a}_1 + x_6\,\mathbf{a}_2 + x_6\,\mathbf{a}_3$ | = | $x_6a\,\mathbf{\hat{x}} + x_6a\,\mathbf{\hat{y}} + x_6a\,\mathbf{\hat{z}}$ | $(32f)$ | H II |
| $\mathbf{B}_{20}$ | = | $x_6\,\mathbf{a}_1 + x_6\,\mathbf{a}_2 - 3x_6\,\mathbf{a}_3$ | = | $-x_6a\,\mathbf{\hat{x}} - x_6a\,\mathbf{\hat{y}} + x_6a\,\mathbf{\hat{z}}$ | $(32f)$ | H II |
| $\mathbf{B}_{21}$ | = | $x_6\,\mathbf{a}_1 - 3x_6\,\mathbf{a}_2 + x_6\,\mathbf{a}_3$ | = | $-x_6a\,\mathbf{\hat{x}} + x_6a\,\mathbf{\hat{y}} - x_6a\,\mathbf{\hat{z}}$ | $(32f)$ | H II |
| $\mathbf{B}_{22}$ | = | $-3x_6\,\mathbf{a}_1 + x_6\,\mathbf{a}_2 + x_6\,\mathbf{a}_3$ | = | $x_6a\,\mathbf{\hat{x}} - x_6a\,\mathbf{\hat{y}} - x_6a\,\mathbf{\hat{z}}$ | $(32f)$ | H II |
| $\mathbf{B}_{23}$ | = | $-x_6\,\mathbf{a}_1 - x_6\,\mathbf{a}_2 + 3x_6\,\mathbf{a}_3$ | = | $x_6a\,\mathbf{\hat{x}} + x_6a\,\mathbf{\hat{y}} - x_6a\,\mathbf{\hat{z}}$ | $(32f)$ | H II |
| $\mathbf{B}_{24}$ | = | $-x_6\,\mathbf{a}_1 - x_6\,\mathbf{a}_2 - x_6\,\mathbf{a}_3$ | = | $-x_6a\,\mathbf{\hat{x}} - x_6a\,\mathbf{\hat{y}} - x_6a\,\mathbf{\hat{z}}$ | $(32f)$ | H II |
| $\mathbf{B}_{25}$ | = | $-x_6\,\mathbf{a}_1 + 3x_6\,\mathbf{a}_2 - x_6\,\mathbf{a}_3$ | = | $x_6a\,\mathbf{\hat{x}} - x_6a\,\mathbf{\hat{y}} + x_6a\,\mathbf{\hat{z}}$ | $(32f)$ | H II |
| $\mathbf{B}_{26}$ | = | $3x_6\,\mathbf{a}_1 - x_6\,\mathbf{a}_2 - x_6\,\mathbf{a}_3$ | = | $-x_6a\,\mathbf{\hat{x}} + x_6a\,\mathbf{\hat{y}} + x_6a\,\mathbf{\hat{z}}$ | $(32f)$ | H II |
| $\mathbf{B}_{27}$ | = | $(y_7 + z_7)\,\mathbf{a}_1 + (-y_7 + z_7)\,\mathbf{a}_2 + (y_7 - z_7)\,\mathbf{a}_3$ | = | $y_7a\,\mathbf{\hat{y}} + z_7a\,\mathbf{\hat{z}}$ | $(96j)$ | F II |
| $\mathbf{B}_{28}$ | = | $(-y_7 + z_7)\,\mathbf{a}_1 + (y_7 + z_7)\,\mathbf{a}_2 + (-y_7 - z_7)\,\mathbf{a}_3$ | = | $-y_7 a\,\hat{\mathbf{y}} + z_7 a\,\hat{\mathbf{z}}$ | (96$j$) | F II |
| $\mathbf{B}_{29}$ | = | $(y_7 - z_7)\,\mathbf{a}_1 + (-y_7 - z_7)\,\mathbf{a}_2 + (y_7 + z_7)\,\mathbf{a}_3$ | = | $y_7 a\,\hat{\mathbf{y}} - z_7 a\,\hat{\mathbf{z}}$ | (96$j$) | F II |
| $\mathbf{B}_{30}$ | = | $(-y_7 - z_7)\,\mathbf{a}_1 + (y_7 - z_7)\,\mathbf{a}_2 + (-y_7 + z_7)\,\mathbf{a}_3$ | = | $-y_7 a\,\hat{\mathbf{y}} - z_7 a\,\hat{\mathbf{z}}$ | (96$j$) | F II |
| $\mathbf{B}_{31}$ | = | $(y_7 - z_7)\,\mathbf{a}_1 + (y_7 + z_7)\,\mathbf{a}_2 + (-y_7 + z_7)\,\mathbf{a}_3$ | = | $z_7 a\,\hat{\mathbf{x}} + y_7 a\,\hat{\mathbf{z}}$ | (96$j$) | F II |
| $\mathbf{B}_{32}$ | = | $(-y_7 - z_7)\,\mathbf{a}_1 + (-y_7 + z_7)\,\mathbf{a}_2 + (y_7 + z_7)\,\mathbf{a}_3$ | = | $z_7 a\,\hat{\mathbf{x}} - y_7 a\,\hat{\mathbf{z}}$ | (96$j$) | F II |
| $\mathbf{B}_{33}$ | = | $(y_7 + z_7)\,\mathbf{a}_1 + (y_7 - z_7)\,\mathbf{a}_2 + (-y_7 - z_7)\,\mathbf{a}_3$ | = | $-z_7 a\,\hat{\mathbf{x}} + y_7 a\,\hat{\mathbf{z}}$ | (96$j$) | F II |
| $\mathbf{B}_{34}$ | = | $(-y_7 + z_7)\,\mathbf{a}_1 + (-y_7 - z_7)\,\mathbf{a}_2 + (y_7 - z_7)\,\mathbf{a}_3$ | = | $-z_7 a\,\hat{\mathbf{x}} - y_7 a\,\hat{\mathbf{z}}$ | (96$j$) | F II |
| $\mathbf{B}_{35}$ | = | $(-y_7 + z_7)\,\mathbf{a}_1 + (y_7 - z_7)\,\mathbf{a}_2 + (y_7 + z_7)\,\mathbf{a}_3$ | = | $y_7 a\,\hat{\mathbf{x}} + z_7 a\,\hat{\mathbf{y}}$ | (96$j$) | F II |
| $\mathbf{B}_{36}$ | = | $(y_7 + z_7)\,\mathbf{a}_1 + (-y_7 - z_7)\,\mathbf{a}_2 + (-y_7 + z_7)\,\mathbf{a}_3$ | = | $-y_7 a\,\hat{\mathbf{x}} + z_7 a\,\hat{\mathbf{y}}$ | (96$j$) | F II |
| $\mathbf{B}_{37}$ | = | $(-y_7 - z_7)\,\mathbf{a}_1 + (y_7 + z_7)\,\mathbf{a}_2 + (y_7 - z_7)\,\mathbf{a}_3$ | = | $y_7 a\,\hat{\mathbf{x}} - z_7 a\,\hat{\mathbf{y}}$ | (96$j$) | F II |
| $\mathbf{B}_{38}$ | = | $(y_7 - z_7)\,\mathbf{a}_1 + (-y_7 + z_7)\,\mathbf{a}_2 + (-y_7 - z_7)\,\mathbf{a}_3$ | = | $-y_7 a\,\hat{\mathbf{x}} - z_7 a\,\hat{\mathbf{y}}$ | (96$j$) | F II |
| $\mathbf{B}_{39}$ | = | $(-y_7 - z_7)\,\mathbf{a}_1 + (y_7 - z_7)\,\mathbf{a}_2 + (y_7 + z_7)\,\mathbf{a}_3$ | = | $y_7 a\,\hat{\mathbf{x}} - z_7 a\,\hat{\mathbf{z}}$ | (96$j$) | F II |
| $\mathbf{B}_{40}$ | = | $(y_7 - z_7)\,\mathbf{a}_1 + (-y_7 - z_7)\,\mathbf{a}_2 + (-y_7 + z_7)\,\mathbf{a}_3$ | = | $-y_7 a\,\hat{\mathbf{x}} - z_7 a\,\hat{\mathbf{z}}$ | (96$j$) | F II |
| $\mathbf{B}_{41}$ | = | $(-y_7 + z_7)\,\mathbf{a}_1 + (y_7 + z_7)\,\mathbf{a}_2 + (y_7 - z_7)\,\mathbf{a}_3$ | = | $y_7 a\,\hat{\mathbf{x}} + z_7 a\,\hat{\mathbf{z}}$ | (96$j$) | F II |
| $\mathbf{B}_{42}$ | = | $(y_7 + z_7)\,\mathbf{a}_1 + (-y_7 + z_7)\,\mathbf{a}_2 + (-y_7 - z_7)\,\mathbf{a}_3$ | = | $-y_7 a\,\hat{\mathbf{x}} + z_7 a\,\hat{\mathbf{z}}$ | (96$j$) | F II |
| $\mathbf{B}_{43}$ | = | $(-y_7 + z_7)\,\mathbf{a}_1 + (-y_7 - z_7)\,\mathbf{a}_2 + (y_7 + z_7)\,\mathbf{a}_3$ | = | $z_7 a\,\hat{\mathbf{y}} - y_7 a\,\hat{\mathbf{z}}$ | (96$j$) | F II |
| $\mathbf{B}_{44}$ | = | $(y_7 + z_7)\,\mathbf{a}_1 + (y_7 - z_7)\,\mathbf{a}_2 + (-y_7 + z_7)\,\mathbf{a}_3$ | = | $z_7 a\,\hat{\mathbf{y}} + y_7 a\,\hat{\mathbf{z}}$ | (96$j$) | F II |
| $\mathbf{B}_{45}$ | = | $(-y_7 - z_7)\,\mathbf{a}_1 + (-y_7 + z_7)\,\mathbf{a}_2 + (y_7 - z_7)\,\mathbf{a}_3$ | = | $-z_7 a\,\hat{\mathbf{y}} - y_7 a\,\hat{\mathbf{z}}$ | (96$j$) | F II |
| $\mathbf{B}_{46}$ | = | $(y_7 - z_7)\,\mathbf{a}_1 + (y_7 + z_7)\,\mathbf{a}_2 + (-y_7 - z_7)\,\mathbf{a}_3$ | = | $-z_7 a\,\hat{\mathbf{y}} + y_7 a\,\hat{\mathbf{z}}$ | (96$j$) | F II |
| $\mathbf{B}_{47}$ | = | $(y_7 - z_7)\,\mathbf{a}_1 + (-y_7 + z_7)\,\mathbf{a}_2 + (y_7 + z_7)\,\mathbf{a}_3$ | = | $z_7 a\,\hat{\mathbf{x}} + y_7 a\,\hat{\mathbf{y}}$ | (96$j$) | F II |
| $\mathbf{B}_{48}$ | = | $(-y_7 - z_7)\,\mathbf{a}_1 + (y_7 + z_7)\,\mathbf{a}_2 + (-y_7 + z_7)\,\mathbf{a}_3$ | = | $z_7 a\,\hat{\mathbf{x}} - y_7 a\,\hat{\mathbf{y}}$ | (96$j$) | F II |
| $\mathbf{B}_{49}$ | = | $(y_7 + z_7)\,\mathbf{a}_1 + (-y_7 - z_7)\,\mathbf{a}_2 + (y_7 - z_7)\,\mathbf{a}_3$ | = | $-z_7 a\,\hat{\mathbf{x}} + y_7 a\,\hat{\mathbf{y}}$ | (96$j$) | F II |
| $\mathbf{B}_{50}$ | = | $(-y_7 + z_7)\,\mathbf{a}_1 + (y_7 - z_7)\,\mathbf{a}_2 + (-y_7 - z_7)\,\mathbf{a}_3$ | = | $-z_7 a\,\hat{\mathbf{x}} - y_7 a\,\hat{\mathbf{y}}$ | (96$j$) | F II |

**Geometry files:**

- CIF: pp. 1817
- POSCAR: pp. 1818

# $L1_a$ (disputed $CuPt_3$) Structure: AB7_cF32_225_b_ad

http://aflow.org/prototype-encyclopedia/AB7_cF32_225_b_ad

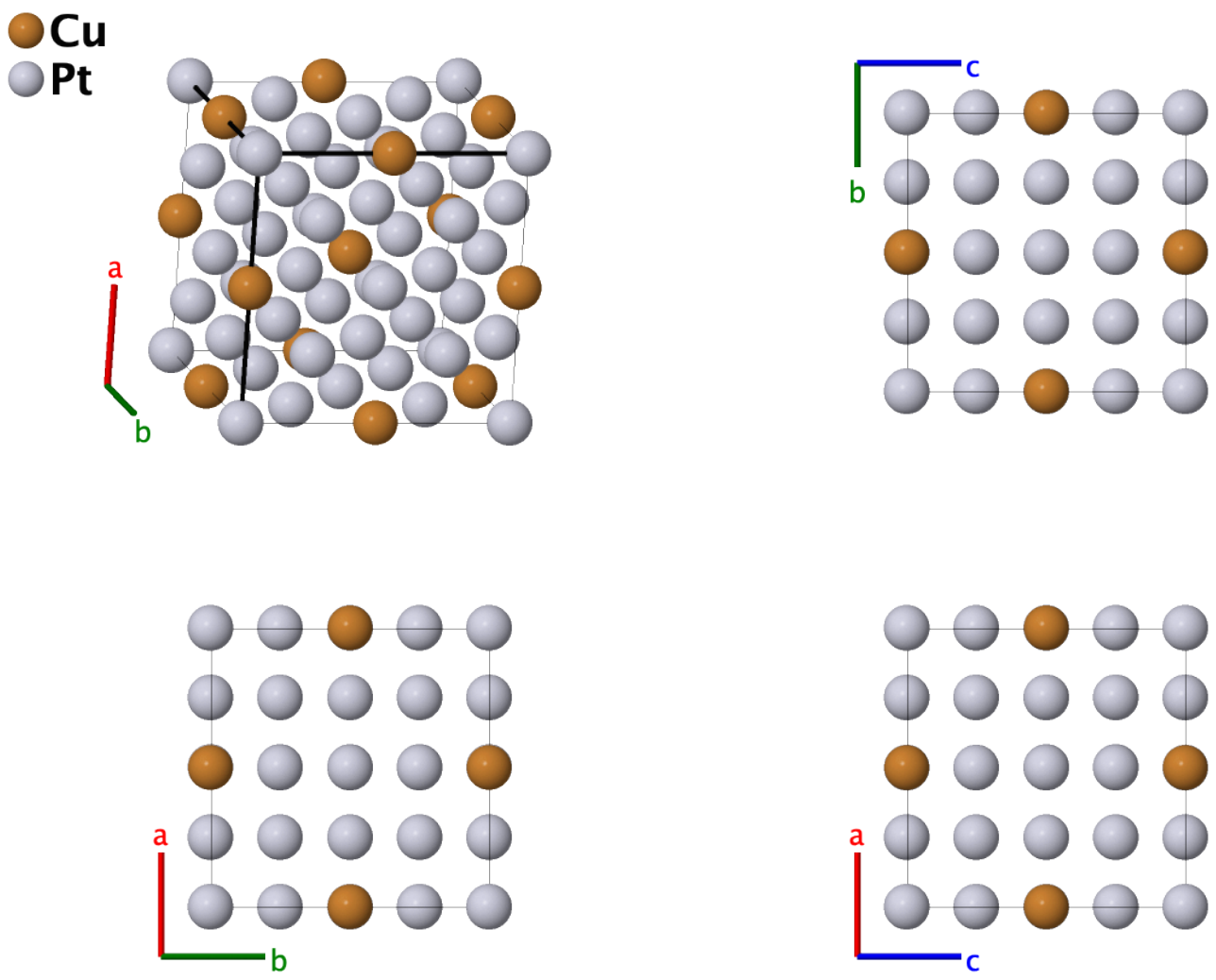

| | | |
|---|---|---|
| **Prototype** | : | $CuPt_3$ |
| **AFLOW prototype label** | : | AB7_cF32_225_b_ad |
| ***Strukturbericht* designation** | : | $L1_a$ |
| **Pearson symbol** | : | cF32 |
| **Space group number** | : | 225 |
| **Space group symbol** | : | $Fm\bar{3}m$ |
| **AFLOW prototype command** | : | `aflow --proto=AB7_cF32_225_b_ad --params=`$a$ |

- According to (Tang, 1951), the (24*d*) sites have the composition $Pt_{0.8}Cu_{0.2}$ in stoichiometric $CuPt_3$. Here we use "Pt" to specify the atoms on this site.
- (Tang, 1951) states that the crystal structure of $CuPt_3$ must be cubic, but (Mshumi, 2014) argue that it is orthorhombic, and in fact the $L1_3$ structure.
- (Smithells, 1955) gave this structure the $L1_a$ designation as part of his extension of the original *Strukturbericht* labels. He does note that an alternative orthorhombic structure had been proposed.
- (Smithells, 1955) assigns this structure to space group $F432$ #209, but the positions given by (Tang, 1951) are also consistent with $Fm\bar{3}m$ #225, so we assign this structure to the higher symmetry space group.
- (Tang, 1951) does not give the lattice constant, so we use the value estimated by (Smithells, 1955).
- The Wyckoff positions are identical to those of the $Ca_7Ge$ structure.

**Face-centered Cubic primitive vectors:**

$$
\begin{aligned}
\mathbf{a}_1 &= \tfrac{1}{2}\,a\,\hat{\mathbf{y}} + \tfrac{1}{2}\,a\,\hat{\mathbf{z}} \\
\mathbf{a}_2 &= \tfrac{1}{2}\,a\,\hat{\mathbf{x}} + \tfrac{1}{2}\,a\,\hat{\mathbf{z}} \\
\mathbf{a}_3 &= \tfrac{1}{2}\,a\,\hat{\mathbf{x}} + \tfrac{1}{2}\,a\,\hat{\mathbf{y}}
\end{aligned}
$$

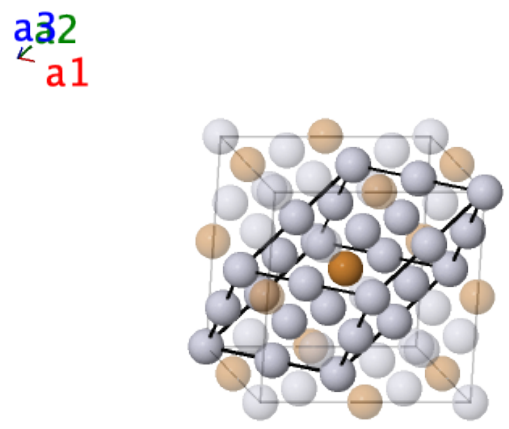

**Basis vectors:**

| | | Lattice Coordinates | | Cartesian Coordinates | Wyckoff Position | Atom Type |
|---|---|---|---|---|---|---|
| $\mathbf{B}_1$ | = | $0\,\mathbf{a}_1 + 0\,\mathbf{a}_2 + 0\,\mathbf{a}_3$ | = | $0\,\hat{\mathbf{x}} + 0\,\hat{\mathbf{y}} + 0\,\hat{\mathbf{z}}$ | (4$a$) | Pt I |
| $\mathbf{B}_2$ | = | $\frac{1}{2}\,\mathbf{a}_1 + \frac{1}{2}\,\mathbf{a}_2 + \frac{1}{2}\,\mathbf{a}_3$ | = | $\frac{1}{2}a\,\hat{\mathbf{x}} + \frac{1}{2}a\,\hat{\mathbf{y}} + \frac{1}{2}a\,\hat{\mathbf{z}}$ | (4$b$) | Cu |
| $\mathbf{B}_3$ | = | $\frac{1}{2}\,\mathbf{a}_1$ | = | $\frac{1}{4}a\,\hat{\mathbf{y}} + \frac{1}{4}a\,\hat{\mathbf{z}}$ | (24$d$) | Pt II |
| $\mathbf{B}_4$ | = | $\frac{1}{2}\,\mathbf{a}_2 + \frac{1}{2}\,\mathbf{a}_3$ | = | $\frac{1}{2}a\,\hat{\mathbf{x}} + \frac{1}{4}a\,\hat{\mathbf{y}} + \frac{1}{4}a\,\hat{\mathbf{z}}$ | (24$d$) | Pt II |
| $\mathbf{B}_5$ | = | $\frac{1}{2}\,\mathbf{a}_2$ | = | $\frac{1}{4}a\,\hat{\mathbf{x}} + \frac{1}{4}a\,\hat{\mathbf{z}}$ | (24$d$) | Pt II |
| $\mathbf{B}_6$ | = | $\frac{1}{2}\,\mathbf{a}_1 + \frac{1}{2}\,\mathbf{a}_3$ | = | $\frac{1}{4}a\,\hat{\mathbf{x}} + \frac{1}{2}a\,\hat{\mathbf{y}} + \frac{1}{4}a\,\hat{\mathbf{z}}$ | (24$d$) | Pt II |
| $\mathbf{B}_7$ | = | $\frac{1}{2}\,\mathbf{a}_3$ | = | $\frac{1}{4}a\,\hat{\mathbf{x}} + \frac{1}{4}a\,\hat{\mathbf{y}}$ | (24$d$) | Pt II |
| $\mathbf{B}_8$ | = | $\frac{1}{2}\,\mathbf{a}_1 + \frac{1}{2}\,\mathbf{a}_2$ | = | $\frac{1}{4}a\,\hat{\mathbf{x}} + \frac{1}{4}a\,\hat{\mathbf{y}} + \frac{1}{2}a\,\hat{\mathbf{z}}$ | (24$d$) | Pt II |

**Found in:**

**Geometry files:**

- CIF: pp. 1819
- POSCAR: pp. 1820

# Sulphohalite [$Na_6ClF(SO_4)_2$, $H5_8$] Structure: ABC6D8E2_cF72_225_b_a_e_f_c

http://aflow.org/prototype-encyclopedia/ABC6D8E2_cF72_225_b_a_e_f_c

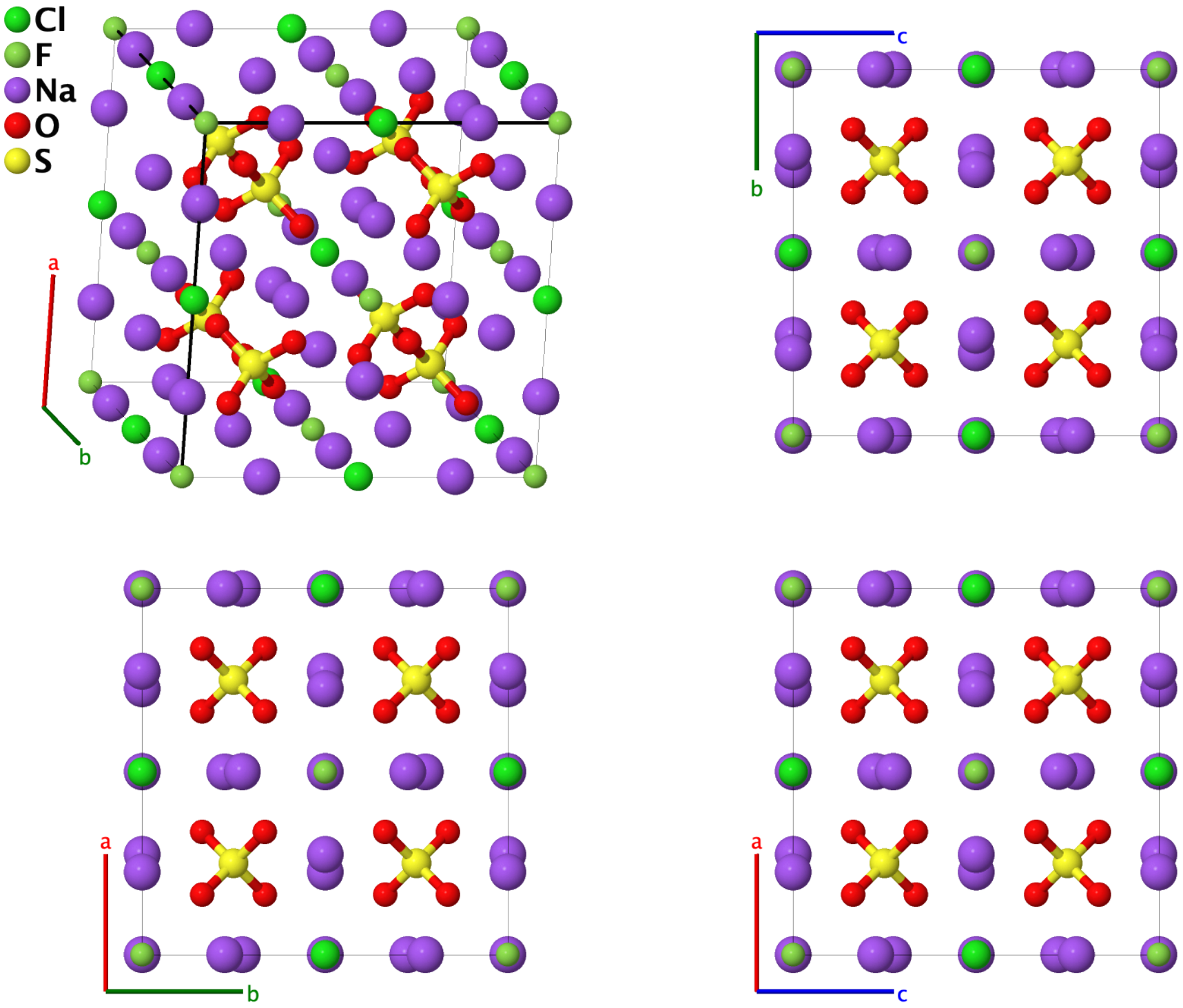

| | | |
|---|---|---|
| **Prototype** | : | $ClFNa_6O_8S_2$ |
| **AFLOW prototype label** | : | ABC6D8E2_cF72_225_b_a_e_f_c |
| ***Strukturbericht* designation** | : | $H5_8$ |
| **Pearson symbol** | : | cF72 |
| **Space group number** | : | 225 |
| **Space group symbol** | : | $Fm\bar{3}m$ |
| **AFLOW prototype command** | : | `aflow --proto=ABC6D8E2_cF72_225_b_a_e_f_c --params=`$a, x_4, x_5$ |

**Face-centered Cubic primitive vectors:**

$$\mathbf{a}_1 = \frac{1}{2}\,a\,\hat{\mathbf{y}} + \frac{1}{2}\,a\,\hat{\mathbf{z}}$$
$$\mathbf{a}_2 = \frac{1}{2}\,a\,\hat{\mathbf{x}} + \frac{1}{2}\,a\,\hat{\mathbf{z}}$$
$$\mathbf{a}_3 = \frac{1}{2}\,a\,\hat{\mathbf{x}} + \frac{1}{2}\,a\,\hat{\mathbf{y}}$$

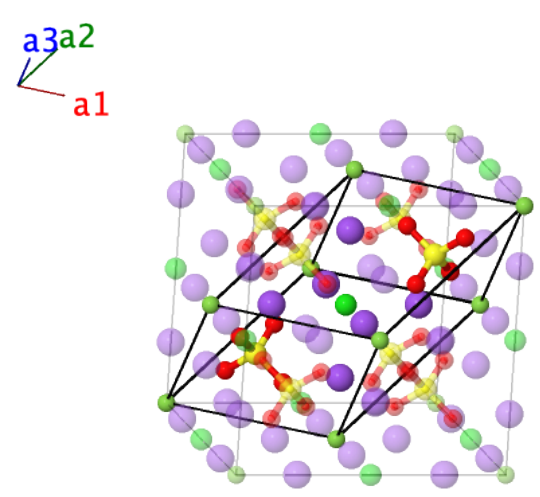

**Basis vectors:**

| | | Lattice Coordinates | | Cartesian Coordinates | Wyckoff Position | Atom Type |
|---|---|---|---|---|---|---|
| $\mathbf{B}_1$ | = | $0\,\mathbf{a}_1 + 0\,\mathbf{a}_2 + 0\,\mathbf{a}_3$ | = | $0\,\hat{\mathbf{x}} + 0\,\hat{\mathbf{y}} + 0\,\hat{\mathbf{z}}$ | (4$a$) | F |
| $\mathbf{B}_2$ | = | $\frac{1}{2}\,\mathbf{a}_1 + \frac{1}{2}\,\mathbf{a}_2 + \frac{1}{2}\,\mathbf{a}_3$ | = | $\frac{1}{2}a\,\hat{\mathbf{x}} + \frac{1}{2}a\,\hat{\mathbf{y}} + \frac{1}{2}a\,\hat{\mathbf{z}}$ | (4$b$) | Cl |
| $\mathbf{B}_3$ | = | $\frac{1}{4}\,\mathbf{a}_1 + \frac{1}{4}\,\mathbf{a}_2 + \frac{1}{4}\,\mathbf{a}_3$ | = | $\frac{1}{4}a\,\hat{\mathbf{x}} + \frac{1}{4}a\,\hat{\mathbf{y}} + \frac{1}{4}a\,\hat{\mathbf{z}}$ | (8$c$) | S |
| $\mathbf{B}_4$ | = | $\frac{3}{4}\,\mathbf{a}_1 + \frac{3}{4}\,\mathbf{a}_2 + \frac{3}{4}\,\mathbf{a}_3$ | = | $\frac{3}{4}a\,\hat{\mathbf{x}} + \frac{3}{4}a\,\hat{\mathbf{y}} + \frac{3}{4}a\,\hat{\mathbf{z}}$ | (8$c$) | S |
| $\mathbf{B}_5$ | = | $-x_4\,\mathbf{a}_1 + x_4\,\mathbf{a}_2 + x_4\,\mathbf{a}_3$ | = | $x_4a\,\hat{\mathbf{x}}$ | (24$e$) | Na |
| $\mathbf{B}_6$ | = | $x_4\,\mathbf{a}_1 - x_4\,\mathbf{a}_2 - x_4\,\mathbf{a}_3$ | = | $-x_4a\,\hat{\mathbf{x}}$ | (24$e$) | Na |
| $\mathbf{B}_7$ | = | $x_4\,\mathbf{a}_1 - x_4\,\mathbf{a}_2 + x_4\,\mathbf{a}_3$ | = | $x_4a\,\hat{\mathbf{y}}$ | (24$e$) | Na |
| $\mathbf{B}_8$ | = | $-x_4\,\mathbf{a}_1 + x_4\,\mathbf{a}_2 - x_4\,\mathbf{a}_3$ | = | $-x_4a\,\hat{\mathbf{y}}$ | (24$e$) | Na |
| $\mathbf{B}_9$ | = | $x_4\,\mathbf{a}_1 + x_4\,\mathbf{a}_2 - x_4\,\mathbf{a}_3$ | = | $x_4a\,\hat{\mathbf{z}}$ | (24$e$) | Na |
| $\mathbf{B}_{10}$ | = | $-x_4\,\mathbf{a}_1 - x_4\,\mathbf{a}_2 + x_4\,\mathbf{a}_3$ | = | $-x_4a\,\hat{\mathbf{z}}$ | (24$e$) | Na |
| $\mathbf{B}_{11}$ | = | $x_5\,\mathbf{a}_1 + x_5\,\mathbf{a}_2 + x_5\,\mathbf{a}_3$ | = | $x_5a\,\hat{\mathbf{x}} + x_5a\,\hat{\mathbf{y}} + x_5a\,\hat{\mathbf{z}}$ | (32$f$) | O |
| $\mathbf{B}_{12}$ | = | $x_5\,\mathbf{a}_1 + x_5\,\mathbf{a}_2 - 3x_5\,\mathbf{a}_3$ | = | $-x_5a\,\hat{\mathbf{x}} - x_5a\,\hat{\mathbf{y}} + x_5a\,\hat{\mathbf{z}}$ | (32$f$) | O |
| $\mathbf{B}_{13}$ | = | $x_5\,\mathbf{a}_1 - 3x_5\,\mathbf{a}_2 + x_5\,\mathbf{a}_3$ | = | $-x_5a\,\hat{\mathbf{x}} + x_5a\,\hat{\mathbf{y}} - x_5a\,\hat{\mathbf{z}}$ | (32$f$) | O |
| $\mathbf{B}_{14}$ | = | $-3x_5\,\mathbf{a}_1 + x_5\,\mathbf{a}_2 + x_5\,\mathbf{a}_3$ | = | $x_5a\,\hat{\mathbf{x}} - x_5a\,\hat{\mathbf{y}} - x_5a\,\hat{\mathbf{z}}$ | (32$f$) | O |
| $\mathbf{B}_{15}$ | = | $-x_5\,\mathbf{a}_1 - x_5\,\mathbf{a}_2 + 3x_5\,\mathbf{a}_3$ | = | $x_5a\,\hat{\mathbf{x}} + x_5a\,\hat{\mathbf{y}} - x_5a\,\hat{\mathbf{z}}$ | (32$f$) | O |
| $\mathbf{B}_{16}$ | = | $-x_5\,\mathbf{a}_1 - x_5\,\mathbf{a}_2 - x_5\,\mathbf{a}_3$ | = | $-x_5a\,\hat{\mathbf{x}} - x_5a\,\hat{\mathbf{y}} - x_5a\,\hat{\mathbf{z}}$ | (32$f$) | O |
| $\mathbf{B}_{17}$ | = | $-x_5\,\mathbf{a}_1 + 3x_5\,\mathbf{a}_2 - x_5\,\mathbf{a}_3$ | = | $x_5a\,\hat{\mathbf{x}} - x_5a\,\hat{\mathbf{y}} + x_5a\,\hat{\mathbf{z}}$ | (32$f$) | O |
| $\mathbf{B}_{18}$ | = | $3x_5\,\mathbf{a}_1 - x_5\,\mathbf{a}_2 - x_5\,\mathbf{a}_3$ | = | $-x_5a\,\hat{\mathbf{x}} + x_5a\,\hat{\mathbf{y}} + x_5a\,\hat{\mathbf{z}}$ | (32$f$) | O |

---

**Found in:**

---

**Geometry files:**

- CIF: pp. 1820
- POSCAR: pp. 1821

# $\gamma$-$Ga_2O_3$ Structure: A11B4_cF120_227_acdf_e

http://aflow.org/prototype-encyclopedia/A11B4_cF120_227_acdf_e

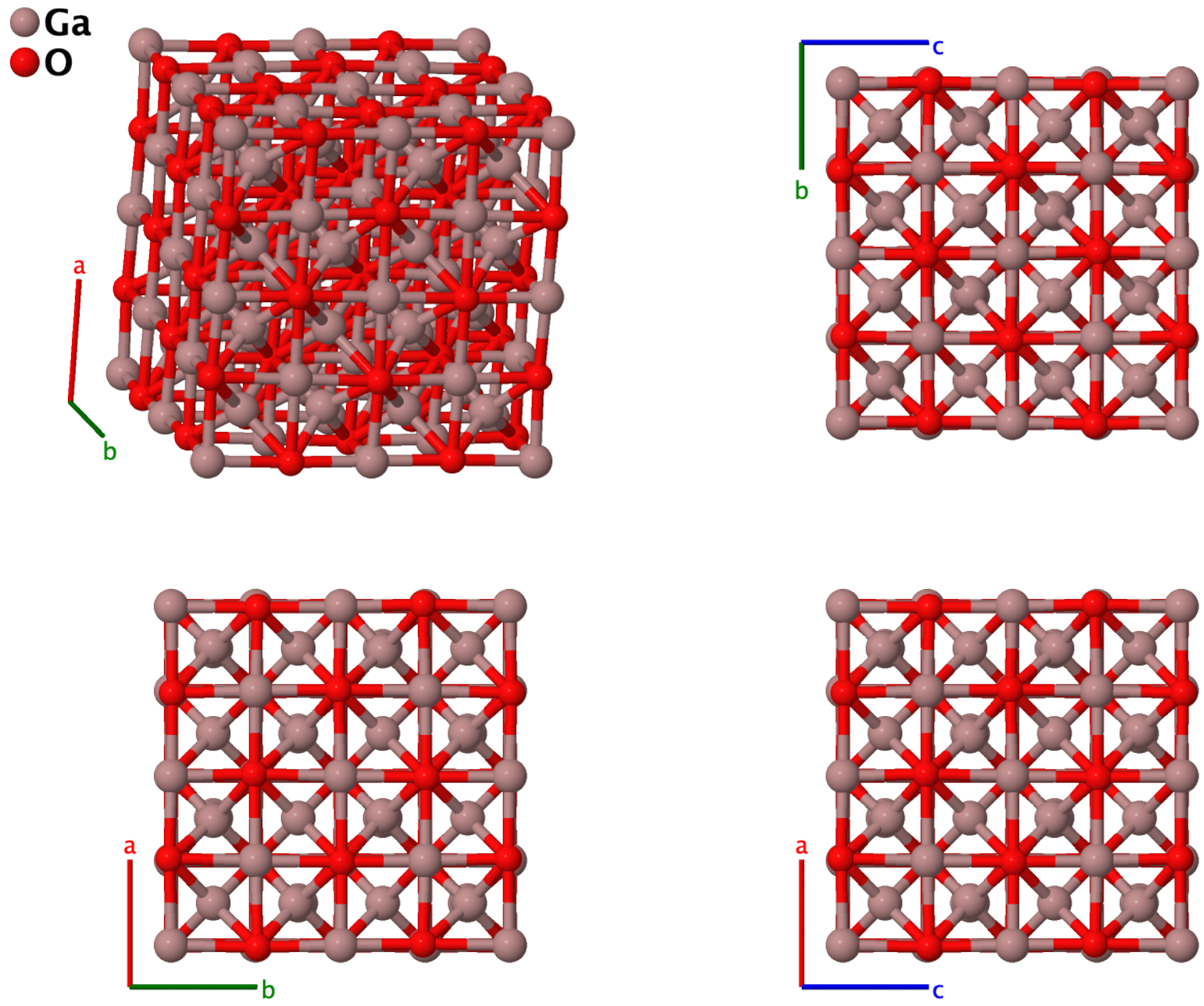

| | | |
|---|---|---|
| **Prototype** | : | $Ga_2O_3$ |
| **AFLOW prototype label** | : | A11B4_cF120_227_acdf_e |
| ***Strukturbericht* designation** | : | None |
| **Pearson symbol** | : | cF120 |
| **Space group number** | : | 227 |
| **Space group symbol** | : | $Fd\bar{3}m$ |
| **AFLOW prototype command** | : | `aflow --proto=A11B4_cF120_227_acdf_e --params=`$a, x_4, x_5$ |

- $Ga_2O_3$ takes on a variety of structures:
  - $\alpha$-$Ga_2O_3$, which has the corundum ($D5_1$) structure,
  - $\beta$-$Ga_2O_3$,
  - $\gamma$-$Ga_2O_3$, this structure, and
  - $\epsilon$-$Ga_2O_3$, a structure with many vacancies which can be approximated by the $\kappa$-alumina structure.

  Note that all of the gallium sites are only partially occupied.

**Face-centered Cubic primitive vectors:**

$$\mathbf{a}_1 = \frac{1}{2}\,a\,\hat{\mathbf{y}} + \frac{1}{2}\,a\,\hat{\mathbf{z}}$$
$$\mathbf{a}_2 = \frac{1}{2}\,a\,\hat{\mathbf{x}} + \frac{1}{2}\,a\,\hat{\mathbf{z}}$$
$$\mathbf{a}_3 = \frac{1}{2}\,a\,\hat{\mathbf{x}} + \frac{1}{2}\,a\,\hat{\mathbf{y}}$$

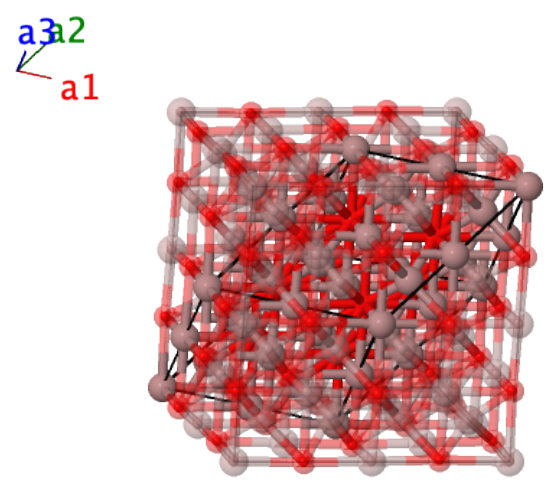

**Basis vectors:**

| | | Lattice Coordinates | | Cartesian Coordinates | Wyckoff Position | Atom Type |
|---|---|---|---|---|---|---|
| $\mathbf{B}_1$ | = | $\frac{1}{8}\,\mathbf{a}_1 + \frac{1}{8}\,\mathbf{a}_2 + \frac{1}{8}\,\mathbf{a}_3$ | = | $\frac{1}{8}a\,\hat{\mathbf{x}} + \frac{1}{8}a\,\hat{\mathbf{y}} + \frac{1}{8}a\,\hat{\mathbf{z}}$ | (8$a$) | Ga I |
| $\mathbf{B}_2$ | = | $\frac{7}{8}\,\mathbf{a}_1 + \frac{7}{8}\,\mathbf{a}_2 + \frac{7}{8}\,\mathbf{a}_3$ | = | $\frac{7}{8}a\,\hat{\mathbf{x}} + \frac{7}{8}a\,\hat{\mathbf{y}} + \frac{7}{8}a\,\hat{\mathbf{z}}$ | (8$a$) | Ga I |
| $\mathbf{B}_3$ | = | $0\,\mathbf{a}_1 + 0\,\mathbf{a}_2 + 0\,\mathbf{a}_3$ | = | $0\,\hat{\mathbf{x}} + 0\,\hat{\mathbf{y}} + 0\,\hat{\mathbf{z}}$ | (16$c$) | Ga II |
| $\mathbf{B}_4$ | = | $\frac{1}{2}\,\mathbf{a}_3$ | = | $\frac{1}{4}a\,\hat{\mathbf{x}} + \frac{1}{4}a\,\hat{\mathbf{y}}$ | (16$c$) | Ga II |
| $\mathbf{B}_5$ | = | $\frac{1}{2}\,\mathbf{a}_2$ | = | $\frac{1}{4}a\,\hat{\mathbf{x}} + \frac{1}{4}a\,\hat{\mathbf{z}}$ | (16$c$) | Ga II |
| $\mathbf{B}_6$ | = | $\frac{1}{2}\,\mathbf{a}_1$ | = | $\frac{1}{4}a\,\hat{\mathbf{y}} + \frac{1}{4}a\,\hat{\mathbf{z}}$ | (16$c$) | Ga II |
| $\mathbf{B}_7$ | = | $\frac{1}{2}\,\mathbf{a}_1 + \frac{1}{2}\,\mathbf{a}_2 + \frac{1}{2}\,\mathbf{a}_3$ | = | $\frac{1}{2}a\,\hat{\mathbf{x}} + \frac{1}{2}a\,\hat{\mathbf{y}} + \frac{1}{2}a\,\hat{\mathbf{z}}$ | (16$d$) | Ga III |
| $\mathbf{B}_8$ | = | $\frac{1}{2}\,\mathbf{a}_1 + \frac{1}{2}\,\mathbf{a}_2$ | = | $\frac{1}{4}a\,\hat{\mathbf{x}} + \frac{1}{4}a\,\hat{\mathbf{y}} + \frac{1}{2}a\,\hat{\mathbf{z}}$ | (16$d$) | Ga III |
| $\mathbf{B}_9$ | = | $\frac{1}{2}\,\mathbf{a}_1 + \frac{1}{2}\,\mathbf{a}_3$ | = | $\frac{1}{4}a\,\hat{\mathbf{x}} + \frac{1}{2}a\,\hat{\mathbf{y}} + \frac{1}{4}a\,\hat{\mathbf{z}}$ | (16$d$) | Ga III |
| $\mathbf{B}_{10}$ | = | $\frac{1}{2}\,\mathbf{a}_2 + \frac{1}{2}\,\mathbf{a}_3$ | = | $\frac{1}{2}a\,\hat{\mathbf{x}} + \frac{1}{4}a\,\hat{\mathbf{y}} + \frac{1}{4}a\,\hat{\mathbf{z}}$ | (16$d$) | Ga III |
| $\mathbf{B}_{11}$ | = | $x_4\,\mathbf{a}_1 + x_4\,\mathbf{a}_2 + x_4\,\mathbf{a}_3$ | = | $x_4a\,\hat{\mathbf{x}} + x_4a\,\hat{\mathbf{y}} + x_4a\,\hat{\mathbf{z}}$ | (32$e$) | O |
| $\mathbf{B}_{12}$ | = | $x_4\,\mathbf{a}_1 + x_4\,\mathbf{a}_2 + \left(\frac{1}{2} - 3x_4\right)\,\mathbf{a}_3$ | = | $\left(\frac{1}{4} - x_4\right)a\,\hat{\mathbf{x}} + \left(\frac{1}{4} - x_4\right)a\,\hat{\mathbf{y}} + x_4a\,\hat{\mathbf{z}}$ | (32$e$) | O |
| $\mathbf{B}_{13}$ | = | $x_4\,\mathbf{a}_1 + \left(\frac{1}{2} - 3x_4\right)\,\mathbf{a}_2 + x_4\,\mathbf{a}_3$ | = | $\left(\frac{1}{4} - x_4\right)a\,\hat{\mathbf{x}} + x_4a\,\hat{\mathbf{y}} + \left(\frac{1}{4} - x_4\right)a\,\hat{\mathbf{z}}$ | (32$e$) | O |
| $\mathbf{B}_{14}$ | = | $\left(\frac{1}{2} - 3x_4\right)\,\mathbf{a}_1 + x_4\,\mathbf{a}_2 + x_4\,\mathbf{a}_3$ | = | $x_4a\,\hat{\mathbf{x}} + \left(\frac{1}{4} - x_4\right)a\,\hat{\mathbf{y}} + \left(\frac{1}{4} - x_4\right)a\,\hat{\mathbf{z}}$ | (32$e$) | O |
| $\mathbf{B}_{15}$ | = | $-x_4\,\mathbf{a}_1 - x_4\,\mathbf{a}_2 + \left(\frac{1}{2} + 3x_4\right)\,\mathbf{a}_3$ | = | $\left(\frac{1}{4} + x_4\right)a\,\hat{\mathbf{x}} + \left(\frac{1}{4} + x_4\right)a\,\hat{\mathbf{y}} - x_4a\,\hat{\mathbf{z}}$ | (32$e$) | O |
| $\mathbf{B}_{16}$ | = | $-x_4\,\mathbf{a}_1 - x_4\,\mathbf{a}_2 - x_4\,\mathbf{a}_3$ | = | $-x_4a\,\hat{\mathbf{x}} - x_4a\,\hat{\mathbf{y}} - x_4a\,\hat{\mathbf{z}}$ | (32$e$) | O |
| $\mathbf{B}_{17}$ | = | $-x_4\,\mathbf{a}_1 + \left(\frac{1}{2} + 3x_4\right)\,\mathbf{a}_2 - x_4\,\mathbf{a}_3$ | = | $\left(\frac{1}{4} + x_4\right)a\,\hat{\mathbf{x}} - x_4a\,\hat{\mathbf{y}} + \left(\frac{1}{4} + x_4\right)a\,\hat{\mathbf{z}}$ | (32$e$) | O |
| $\mathbf{B}_{18}$ | = | $\left(\frac{1}{2} + 3x_4\right)\,\mathbf{a}_1 - x_4\,\mathbf{a}_2 - x_4\,\mathbf{a}_3$ | = | $-x_4a\,\hat{\mathbf{x}} + \left(\frac{1}{4} + x_4\right)a\,\hat{\mathbf{y}} + \left(\frac{1}{4} + x_4\right)a\,\hat{\mathbf{z}}$ | (32$e$) | O |
| $\mathbf{B}_{19}$ | = | $\left(\frac{1}{4} - x_5\right)\,\mathbf{a}_1 + x_5\,\mathbf{a}_2 + x_5\,\mathbf{a}_3$ | = | $x_5a\,\hat{\mathbf{x}} + \frac{1}{8}a\,\hat{\mathbf{y}} + \frac{1}{8}a\,\hat{\mathbf{z}}$ | (48$f$) | Ga IV |
| $\mathbf{B}_{20}$ | = | $x_5\,\mathbf{a}_1 + \left(\frac{1}{4} - x_5\right)\,\mathbf{a}_2 + \left(\frac{1}{4} - x_5\right)\,\mathbf{a}_3$ | = | $\left(\frac{1}{4} - x_5\right)a\,\hat{\mathbf{x}} + \frac{1}{8}a\,\hat{\mathbf{y}} + \frac{1}{8}a\,\hat{\mathbf{z}}$ | (48$f$) | Ga IV |
| $\mathbf{B}_{21}$ | = | $x_5\,\mathbf{a}_1 + \left(\frac{1}{4} - x_5\right)\,\mathbf{a}_2 + x_5\,\mathbf{a}_3$ | = | $\frac{1}{8}a\,\hat{\mathbf{x}} + x_5a\,\hat{\mathbf{y}} + \frac{1}{8}a\,\hat{\mathbf{z}}$ | (48$f$) | Ga IV |
| $\mathbf{B}_{22}$ | = | $\left(\frac{1}{4} - x_5\right)\,\mathbf{a}_1 + x_5\,\mathbf{a}_2 + \left(\frac{1}{4} - x_5\right)\,\mathbf{a}_3$ | = | $\frac{1}{8}a\,\hat{\mathbf{x}} + \left(\frac{1}{4} - x_5\right)a\,\hat{\mathbf{y}} + \frac{1}{8}a\,\hat{\mathbf{z}}$ | (48$f$) | Ga IV |
| $\mathbf{B}_{23}$ | = | $x_5\,\mathbf{a}_1 + x_5\,\mathbf{a}_2 + \left(\frac{1}{4} - x_5\right)\,\mathbf{a}_3$ | = | $\frac{1}{8}a\,\hat{\mathbf{x}} + \frac{1}{8}a\,\hat{\mathbf{y}} + x_5a\,\hat{\mathbf{z}}$ | (48$f$) | Ga IV |
| $\mathbf{B}_{24}$ | = | $\left(\frac{1}{4} - x_5\right)\,\mathbf{a}_1 + \left(\frac{1}{4} - x_5\right)\,\mathbf{a}_2 + x_5\,\mathbf{a}_3$ | = | $\frac{1}{8}a\,\hat{\mathbf{x}} + \frac{1}{8}a\,\hat{\mathbf{y}} + \left(\frac{1}{4} - x_5\right)a\,\hat{\mathbf{z}}$ | (48$f$) | Ga IV |
| $\mathbf{B}_{25}$ | = | $\left(\frac{3}{4} + x_5\right)\,\mathbf{a}_1 - x_5\,\mathbf{a}_2 + \left(\frac{3}{4} + x_5\right)\,\mathbf{a}_3$ | = | $\frac{3}{8}a\,\hat{\mathbf{x}} + \left(\frac{3}{4} + x_5\right)a\,\hat{\mathbf{y}} + \frac{3}{8}a\,\hat{\mathbf{z}}$ | (48$f$) | Ga IV |
| $\mathbf{B}_{26}$ | = | $-x_5\,\mathbf{a}_1 + \left(\frac{3}{4} + x_5\right)\,\mathbf{a}_2 - x_5\,\mathbf{a}_3$ | = | $\frac{3}{8}a\,\hat{\mathbf{x}} - x_5a\,\hat{\mathbf{y}} + \frac{3}{8}a\,\hat{\mathbf{z}}$ | (48$f$) | Ga IV |
| $\mathbf{B}_{27}$ | = | $-x_5\,\mathbf{a}_1 + \left(\frac{3}{4} + x_5\right)\,\mathbf{a}_2 + \left(\frac{3}{4} + x_5\right)\,\mathbf{a}_3$ | = | $\left(\frac{3}{4} + x_5\right)a\,\hat{\mathbf{x}} + \frac{3}{8}a\,\hat{\mathbf{y}} + \frac{3}{8}a\,\hat{\mathbf{z}}$ | (48$f$) | Ga IV |
| $\mathbf{B}_{28}$ | = | $\left(\frac{3}{4} + x_5\right)\,\mathbf{a}_1 - x_5\,\mathbf{a}_2 - x_5\,\mathbf{a}_3$ | = | $-x_5a\,\hat{\mathbf{x}} + \frac{3}{8}a\,\hat{\mathbf{y}} + \frac{3}{8}a\,\hat{\mathbf{z}}$ | (48$f$) | Ga IV |
| $\mathbf{B}_{29}$ | = | $-x_5\,\mathbf{a}_1 - x_5\,\mathbf{a}_2 + \left(\frac{3}{4} + x_5\right)\,\mathbf{a}_3$ | = | $\frac{3}{8}a\,\hat{\mathbf{x}} + \frac{3}{8}a\,\hat{\mathbf{y}} - x_5a\,\hat{\mathbf{z}}$ | (48$f$) | Ga IV |
| $\mathbf{B}_{30}$ | = | $\left(\frac{3}{4} + x_5\right)\,\mathbf{a}_1 + \left(\frac{3}{4} + x_5\right)\,\mathbf{a}_2 - x_5\,\mathbf{a}_3$ | = | $\frac{3}{8}a\,\hat{\mathbf{x}} + \frac{3}{8}a\,\hat{\mathbf{y}} + \left(\frac{3}{4} + x_5\right)a\,\hat{\mathbf{z}}$ | (48$f$) | Ga IV |

**Geometry files:**
- CIF: pp. 1821
- POSCAR: pp. 1823

# Predicted $Li_2MgH_{16}$ High-Temperature Superconductor (250 GPa) Structure: A16B2C_cF152_227_eg_d_a

http://aflow.org/prototype-encyclopedia/A16B2C_cF152_227_eg_d_a

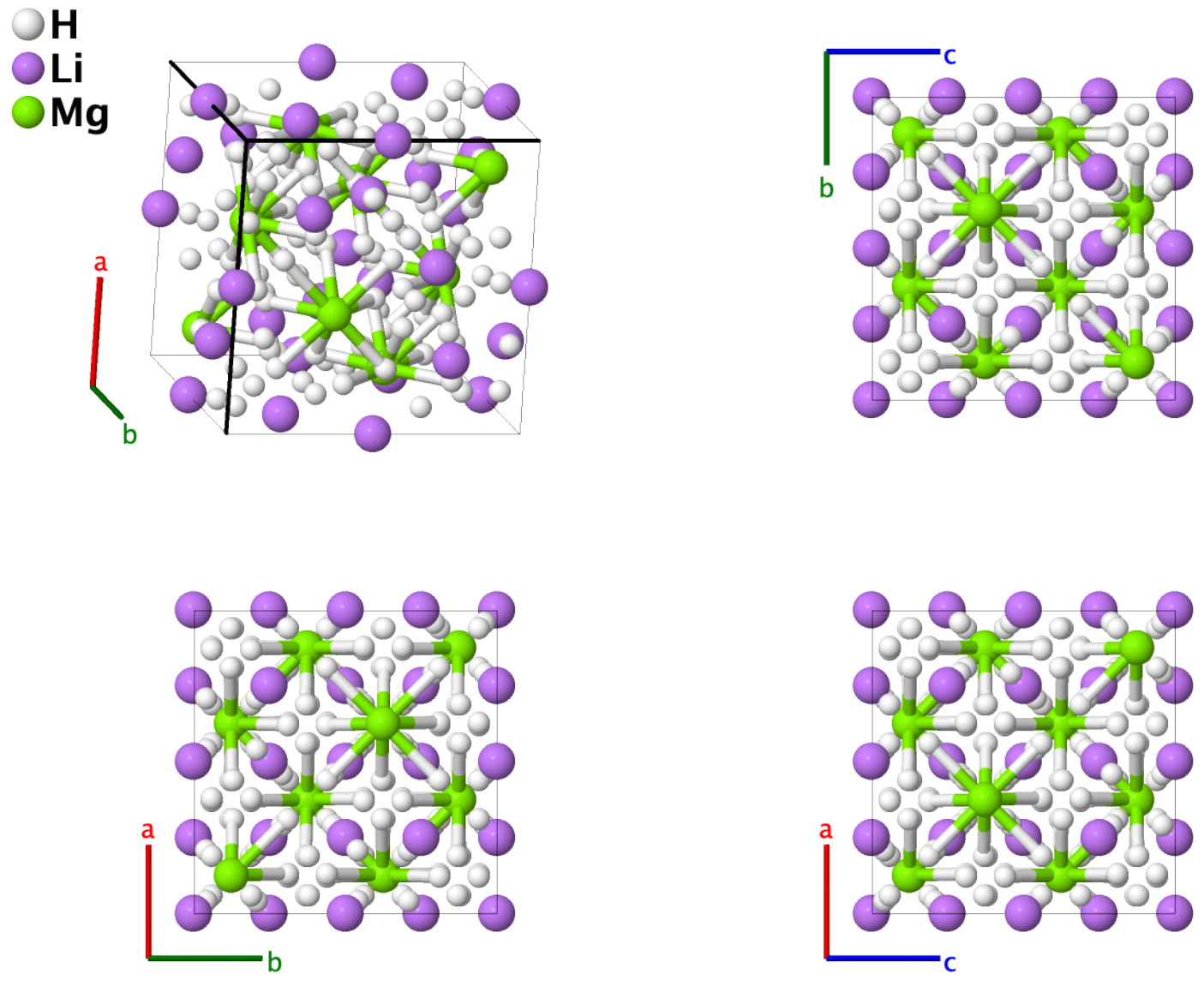

| | | |
|---|---|---|
| **Prototype** | : | $H_{16}Li_2Mg$ |
| **AFLOW prototype label** | : | A16B2C_cF152_227_eg_d_a |
| ***Strukturbericht* designation** | : | None |
| **Pearson symbol** | : | cF152 |
| **Space group number** | : | 227 |
| **Space group symbol** | : | $Fd\bar{3}m$ |
| **AFLOW prototype command** | : | `aflow --proto=A16B2C_cF152_227_eg_d_a --params=`$a, x_3, x_4, z_4$ |

- This structure was predicted by (Sun, 2019) as a metastable state of $Li_2MgH_{16}$ at 250 GPa and $T = 0$ K. If it is possible to construct this compound, or if it becomes stable due to thermodynamic considerations, it is predicted to have a superconducting transition $T_c$ between 430 and 473 K.
- The predicted $T = 0$ K ground state at 300 GPa is a $P\bar{3}m1$ #164 structure with molecular hydrogen.
- (Sun, 2019) give the Wyckoff positions in setting 1 of space group $Fd\bar{3}m$ #227. (They list the Wyckoff positions of the magnesium and lithium atoms as (8$b$) and (16$c$), respectively. They are actually at (8$a$) and (16$d$), as in their 300 GPa data.) We used FINDSYM to shift this to our standard setting 2.

**Face-centered Cubic primitive vectors:**

$$\mathbf{a}_1 = \frac{1}{2}\,a\,\hat{\mathbf{y}} + \frac{1}{2}\,a\,\hat{\mathbf{z}}$$
$$\mathbf{a}_2 = \frac{1}{2}\,a\,\hat{\mathbf{x}} + \frac{1}{2}\,a\,\hat{\mathbf{z}}$$
$$\mathbf{a}_3 = \frac{1}{2}\,a\,\hat{\mathbf{x}} + \frac{1}{2}\,a\,\hat{\mathbf{y}}$$

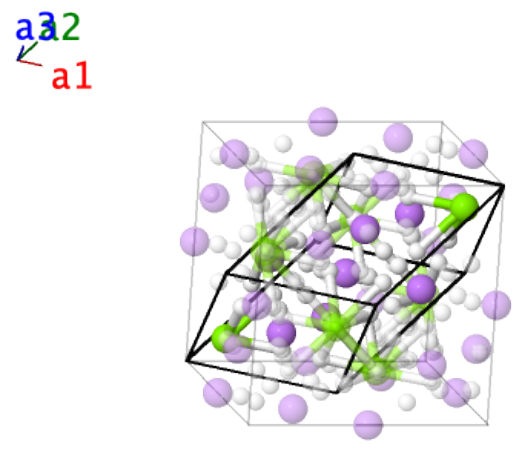

**Basis vectors:**

| | | Lattice Coordinates | | Cartesian Coordinates | Wyckoff Position | Atom Type |
|---|---|---|---|---|---|---|
| $\mathbf{B}_1$ | = | $\frac{1}{8}\,\mathbf{a}_1 + \frac{1}{8}\,\mathbf{a}_2 + \frac{1}{8}\,\mathbf{a}_3$ | = | $\frac{1}{8}a\,\hat{\mathbf{x}} + \frac{1}{8}a\,\hat{\mathbf{y}} + \frac{1}{8}a\,\hat{\mathbf{z}}$ | (8*a*) | Mg |
| $\mathbf{B}_2$ | = | $\frac{7}{8}\,\mathbf{a}_1 + \frac{7}{8}\,\mathbf{a}_2 + \frac{7}{8}\,\mathbf{a}_3$ | = | $\frac{7}{8}a\,\hat{\mathbf{x}} + \frac{7}{8}a\,\hat{\mathbf{y}} + \frac{7}{8}a\,\hat{\mathbf{z}}$ | (8*a*) | Mg |
| $\mathbf{B}_3$ | = | $\frac{1}{2}\,\mathbf{a}_1 + \frac{1}{2}\,\mathbf{a}_2 + \frac{1}{2}\,\mathbf{a}_3$ | = | $\frac{1}{2}a\,\hat{\mathbf{x}} + \frac{1}{2}a\,\hat{\mathbf{y}} + \frac{1}{2}a\,\hat{\mathbf{z}}$ | (16*d*) | Li |
| $\mathbf{B}_4$ | = | $\frac{1}{2}\,\mathbf{a}_1 + \frac{1}{2}\,\mathbf{a}_2$ | = | $\frac{1}{4}a\,\hat{\mathbf{x}} + \frac{1}{4}a\,\hat{\mathbf{y}} + \frac{1}{2}a\,\hat{\mathbf{z}}$ | (16*d*) | Li |
| $\mathbf{B}_5$ | = | $\frac{1}{2}\,\mathbf{a}_1 + \frac{1}{2}\,\mathbf{a}_3$ | = | $\frac{1}{4}a\,\hat{\mathbf{x}} + \frac{1}{2}a\,\hat{\mathbf{y}} + \frac{1}{4}a\,\hat{\mathbf{z}}$ | (16*d*) | Li |
| $\mathbf{B}_6$ | = | $\frac{1}{2}\,\mathbf{a}_2 + \frac{1}{2}\,\mathbf{a}_3$ | = | $\frac{1}{2}a\,\hat{\mathbf{x}} + \frac{1}{4}a\,\hat{\mathbf{y}} + \frac{1}{4}a\,\hat{\mathbf{z}}$ | (16*d*) | Li |
| $\mathbf{B}_7$ | = | $x_3\,\mathbf{a}_1 + x_3\,\mathbf{a}_2 + x_3\,\mathbf{a}_3$ | = | $x_3a\,\hat{\mathbf{x}} + x_3a\,\hat{\mathbf{y}} + x_3a\,\hat{\mathbf{z}}$ | (32*e*) | H I |
| $\mathbf{B}_8$ | = | $x_3\,\mathbf{a}_1 + x_3\,\mathbf{a}_2 + \left(\frac{1}{2} - 3x_3\right)\,\mathbf{a}_3$ | = | $\left(\frac{1}{4} - x_3\right)a\,\hat{\mathbf{x}} + \left(\frac{1}{4} - x_3\right)a\,\hat{\mathbf{y}} + x_3a\,\hat{\mathbf{z}}$ | (32*e*) | H I |
| $\mathbf{B}_9$ | = | $x_3\,\mathbf{a}_1 + \left(\frac{1}{2} - 3x_3\right)\,\mathbf{a}_2 + x_3\,\mathbf{a}_3$ | = | $\left(\frac{1}{4} - x_3\right)a\,\hat{\mathbf{x}} + x_3a\,\hat{\mathbf{y}} + \left(\frac{1}{4} - x_3\right)a\,\hat{\mathbf{z}}$ | (32*e*) | H I |
| $\mathbf{B}_{10}$ | = | $\left(\frac{1}{2} - 3x_3\right)\,\mathbf{a}_1 + x_3\,\mathbf{a}_2 + x_3\,\mathbf{a}_3$ | = | $x_3a\,\hat{\mathbf{x}} + \left(\frac{1}{4} - x_3\right)a\,\hat{\mathbf{y}} + \left(\frac{1}{4} - x_3\right)a\,\hat{\mathbf{z}}$ | (32*e*) | H I |
| $\mathbf{B}_{11}$ | = | $-x_3\,\mathbf{a}_1 - x_3\,\mathbf{a}_2 + \left(\frac{1}{2} + 3x_3\right)\,\mathbf{a}_3$ | = | $\left(\frac{1}{4} + x_3\right)a\,\hat{\mathbf{x}} + \left(\frac{1}{4} + x_3\right)a\,\hat{\mathbf{y}} - x_3a\,\hat{\mathbf{z}}$ | (32*e*) | H I |
| $\mathbf{B}_{12}$ | = | $-x_3\,\mathbf{a}_1 - x_3\,\mathbf{a}_2 - x_3\,\mathbf{a}_3$ | = | $-x_3a\,\hat{\mathbf{x}} - x_3a\,\hat{\mathbf{y}} - x_3a\,\hat{\mathbf{z}}$ | (32*e*) | H I |
| $\mathbf{B}_{13}$ | = | $-x_3\,\mathbf{a}_1 + \left(\frac{1}{2} + 3x_3\right)\,\mathbf{a}_2 - x_3\,\mathbf{a}_3$ | = | $\left(\frac{1}{4} + x_3\right)a\,\hat{\mathbf{x}} - x_3a\,\hat{\mathbf{y}} + \left(\frac{1}{4} + x_3\right)a\,\hat{\mathbf{z}}$ | (32*e*) | H I |
| $\mathbf{B}_{14}$ | = | $\left(\frac{1}{2} + 3x_3\right)\,\mathbf{a}_1 - x_3\,\mathbf{a}_2 - x_3\,\mathbf{a}_3$ | = | $-x_3a\,\hat{\mathbf{x}} + \left(\frac{1}{4} + x_3\right)a\,\hat{\mathbf{y}} + \left(\frac{1}{4} + x_3\right)a\,\hat{\mathbf{z}}$ | (32*e*) | H I |
| $\mathbf{B}_{15}$ | = | $z_4\,\mathbf{a}_1 + z_4\,\mathbf{a}_2 + (2x_4 - z_4)\,\mathbf{a}_3$ | = | $x_4a\,\hat{\mathbf{x}} + x_4a\,\hat{\mathbf{y}} + z_4a\,\hat{\mathbf{z}}$ | (96*g*) | H II |
| $\mathbf{B}_{16}$ | = | $z_4\,\mathbf{a}_1 + z_4\,\mathbf{a}_2 + \left(\frac{1}{2} - 2x_4 - z_4\right)\,\mathbf{a}_3$ | = | $\left(\frac{1}{4} - x_4\right)a\,\hat{\mathbf{x}} + \left(\frac{1}{4} - x_4\right)a\,\hat{\mathbf{y}} + z_4a\,\hat{\mathbf{z}}$ | (96*g*) | H II |
| $\mathbf{B}_{17}$ | = | $(2x_4 - z_4)\,\mathbf{a}_1 + \left(\frac{1}{2} - 2x_4 - z_4\right)\,\mathbf{a}_2 + z_4\,\mathbf{a}_3$ | = | $\left(\frac{1}{4} - x_4\right)a\,\hat{\mathbf{x}} + x_4a\,\hat{\mathbf{y}} + \left(\frac{1}{4} - z_4\right)a\,\hat{\mathbf{z}}$ | (96*g*) | H II |
| $\mathbf{B}_{18}$ | = | $\left(\frac{1}{2} - 2x_4 - z_4\right)\,\mathbf{a}_1 + (2x_4 - z_4)\,\mathbf{a}_2 + z_4\,\mathbf{a}_3$ | = | $x_4a\,\hat{\mathbf{x}} + \left(\frac{1}{4} - x_4\right)a\,\hat{\mathbf{y}} + \left(\frac{1}{4} - z_4\right)a\,\hat{\mathbf{z}}$ | (96*g*) | H II |
| $\mathbf{B}_{19}$ | = | $(2x_4 - z_4)\,\mathbf{a}_1 + z_4\,\mathbf{a}_2 + z_4\,\mathbf{a}_3$ | = | $z_4a\,\hat{\mathbf{x}} + x_4a\,\hat{\mathbf{y}} + x_4a\,\hat{\mathbf{z}}$ | (96*g*) | H II |
| $\mathbf{B}_{20}$ | = | $\left(\frac{1}{2} - 2x_4 - z_4\right)\,\mathbf{a}_1 + z_4\,\mathbf{a}_2 + z_4\,\mathbf{a}_3$ | = | $z_4a\,\hat{\mathbf{x}} + \left(\frac{1}{4} - x_4\right)a\,\hat{\mathbf{y}} + \left(\frac{1}{4} - x_4\right)a\,\hat{\mathbf{z}}$ | (96*g*) | H II |
| $\mathbf{B}_{21}$ | = | $z_4\,\mathbf{a}_1 + (2x_4 - z_4)\,\mathbf{a}_2 + \left(\frac{1}{2} - 2x_4 - z_4\right)\,\mathbf{a}_3$ | = | $\left(\frac{1}{4} - z_4\right)a\,\hat{\mathbf{x}} + \left(\frac{1}{4} - x_4\right)a\,\hat{\mathbf{y}} + x_4a\,\hat{\mathbf{z}}$ | (96*g*) | H II |
| $\mathbf{B}_{22}$ | = | $z_4\,\mathbf{a}_1 + \left(\frac{1}{2} - 2x_4 - z_4\right)\,\mathbf{a}_2 + (2x_4 - z_4)\,\mathbf{a}_3$ | = | $\left(\frac{1}{4} - z_4\right)a\,\hat{\mathbf{x}} + x_4a\,\hat{\mathbf{y}} + \left(\frac{1}{4} - x_4\right)a\,\hat{\mathbf{z}}$ | (96*g*) | H II |
| $\mathbf{B}_{23}$ | = | $z_4\,\mathbf{a}_1 + (2x_4 - z_4)\,\mathbf{a}_2 + z_4\,\mathbf{a}_3$ | = | $x_4a\,\hat{\mathbf{x}} + z_4a\,\hat{\mathbf{y}} + x_4a\,\hat{\mathbf{z}}$ | (96*g*) | H II |
| $\mathbf{B}_{24}$ | = | $z_4\,\mathbf{a}_1 + \left(\frac{1}{2} - 2x_4 - z_4\right)\,\mathbf{a}_2 + z_4\,\mathbf{a}_3$ | = | $\left(\frac{1}{4} - x_4\right)a\,\hat{\mathbf{x}} + z_4a\,\hat{\mathbf{y}} + \left(\frac{1}{4} - x_4\right)a\,\hat{\mathbf{z}}$ | (96*g*) | H II |
| $\mathbf{B}_{25}$ | = | $\left(\frac{1}{2} - 2x_4 - z_4\right)\,\mathbf{a}_1 + z_4\,\mathbf{a}_2 + (2x_4 - z_4)\,\mathbf{a}_3$ | = | $x_4 a\,\hat{\mathbf{x}} + \left(\frac{1}{4} - z_4\right) a\,\hat{\mathbf{y}} + \left(\frac{1}{4} - x_4\right) a\,\hat{\mathbf{z}}$ | (96$g$) | H II |
| $\mathbf{B}_{26}$ | = | $(2x_4 - z_4)\,\mathbf{a}_1 + z_4\,\mathbf{a}_2 + \left(\frac{1}{2} - 2x_4 - z_4\right)\,\mathbf{a}_3$ | = | $\left(\frac{1}{4} - x_4\right) a\,\hat{\mathbf{x}} + \left(\frac{1}{4} - z_4\right) a\,\hat{\mathbf{y}} + x_4 a\,\hat{\mathbf{z}}$ | (96$g$) | H II |
| $\mathbf{B}_{27}$ | = | $-z_4\,\mathbf{a}_1 - z_4\,\mathbf{a}_2 + \left(\frac{1}{2} + 2x_4 + z_4\right)\,\mathbf{a}_3$ | = | $\left(\frac{1}{4} + x_4\right) a\,\hat{\mathbf{x}} + \left(\frac{1}{4} + x_4\right) a\,\hat{\mathbf{y}} - z_4 a\,\hat{\mathbf{z}}$ | (96$g$) | H II |
| $\mathbf{B}_{28}$ | = | $-z_4\,\mathbf{a}_1 - z_4\,\mathbf{a}_2 + (-2x_4 + z_4)\,\mathbf{a}_3$ | = | $-x_4 a\,\hat{\mathbf{x}} - x_4 a\,\hat{\mathbf{y}} - z_4 a\,\hat{\mathbf{z}}$ | (96$g$) | H II |
| $\mathbf{B}_{29}$ | = | $(-2x_4 + z_4)\,\mathbf{a}_1 + \left(\frac{1}{2} + 2x_4 + z_4\right)\,\mathbf{a}_2 - z_4\,\mathbf{a}_3$ | = | $\left(\frac{1}{4} + x_4\right) a\,\hat{\mathbf{x}} - x_4 a\,\hat{\mathbf{y}} + \left(\frac{1}{4} + z_4\right) a\,\hat{\mathbf{z}}$ | (96$g$) | H II |
| $\mathbf{B}_{30}$ | = | $\left(\frac{1}{2} + 2x_4 + z_4\right)\,\mathbf{a}_1 + (-2x_4 + z_4)\,\mathbf{a}_2 - z_4\,\mathbf{a}_3$ | = | $-x_4 a\,\hat{\mathbf{x}} + \left(\frac{1}{4} + x_4\right) a\,\hat{\mathbf{y}} + \left(\frac{1}{4} + z_4\right) a\,\hat{\mathbf{z}}$ | (96$g$) | H II |
| $\mathbf{B}_{31}$ | = | $(-2x_4 + z_4)\,\mathbf{a}_1 - z_4\,\mathbf{a}_2 + \left(\frac{1}{2} + 2x_4 + z_4\right)\,\mathbf{a}_3$ | = | $\left(\frac{1}{4} + x_4\right) a\,\hat{\mathbf{x}} + \left(\frac{1}{4} + z_4\right) a\,\hat{\mathbf{y}} - x_4 a\,\hat{\mathbf{z}}$ | (96$g$) | H II |
| $\mathbf{B}_{32}$ | = | $\left(\frac{1}{2} + 2x_4 + z_4\right)\,\mathbf{a}_1 - z_4\,\mathbf{a}_2 + (-2x_4 + z_4)\,\mathbf{a}_3$ | = | $-x_4 a\,\hat{\mathbf{x}} + \left(\frac{1}{4} + z_4\right) a\,\hat{\mathbf{y}} + \left(\frac{1}{4} + x_4\right) a\,\hat{\mathbf{z}}$ | (96$g$) | H II |
| $\mathbf{B}_{33}$ | = | $-z_4\,\mathbf{a}_1 + (-2x_4 + z_4)\,\mathbf{a}_2 - z_4\,\mathbf{a}_3$ | = | $-x_4 a\,\hat{\mathbf{x}} - z_4 a\,\hat{\mathbf{y}} - x_4 a\,\hat{\mathbf{z}}$ | (96$g$) | H II |
| $\mathbf{B}_{34}$ | = | $-z_4\,\mathbf{a}_1 + \left(\frac{1}{2} + 2x_4 + z_4\right)\,\mathbf{a}_2 - z_4\,\mathbf{a}_3$ | = | $\left(\frac{1}{4} + x_4\right) a\,\hat{\mathbf{x}} - z_4 a\,\hat{\mathbf{y}} + \left(\frac{1}{4} + x_4\right) a\,\hat{\mathbf{z}}$ | (96$g$) | H II |
| $\mathbf{B}_{35}$ | = | $-z_4\,\mathbf{a}_1 + (-2x_4 + z_4)\,\mathbf{a}_2 + \left(\frac{1}{2} + 2x_4 + z_4\right)\,\mathbf{a}_3$ | = | $\left(\frac{1}{4} + z_4\right) a\,\hat{\mathbf{x}} + \left(\frac{1}{4} + x_4\right) a\,\hat{\mathbf{y}} - x_4 a\,\hat{\mathbf{z}}$ | (96$g$) | H II |
| $\mathbf{B}_{36}$ | = | $-z_4\,\mathbf{a}_1 + \left(\frac{1}{2} + 2x_4 + z_4\right)\,\mathbf{a}_2 + (-2x_4 + z_4)\,\mathbf{a}_3$ | = | $\left(\frac{1}{4} + z_4\right) a\,\hat{\mathbf{x}} - x_4 a\,\hat{\mathbf{y}} + \left(\frac{1}{4} + x_4\right) a\,\hat{\mathbf{z}}$ | (96$g$) | H II |
| $\mathbf{B}_{37}$ | = | $\left(\frac{1}{2} + 2x_4 + z_4\right)\,\mathbf{a}_1 - z_4\,\mathbf{a}_2 - z_4\,\mathbf{a}_3$ | = | $-z_4 a\,\hat{\mathbf{x}} + \left(\frac{1}{4} + x_4\right) a\,\hat{\mathbf{y}} + \left(\frac{1}{4} + x_4\right) a\,\hat{\mathbf{z}}$ | (96$g$) | H II |
| $\mathbf{B}_{38}$ | = | $(-2x_4 + z_4)\,\mathbf{a}_1 - z_4\,\mathbf{a}_2 - z_4\,\mathbf{a}_3$ | = | $-z_4 a\,\hat{\mathbf{x}} - x_4 a\,\hat{\mathbf{y}} - x_4 a\,\hat{\mathbf{z}}$ | (96$g$) | H II |

**Geometry files:**

- CIF: pp. 1823
- POSCAR: pp. 1824

# $Mg_3Cr_2Al_{18}$ Structure: A18B2C3_cF184_227_fg_d_ac

http://aflow.org/prototype-encyclopedia/A18B2C3_cF184_227_fg_d_ac

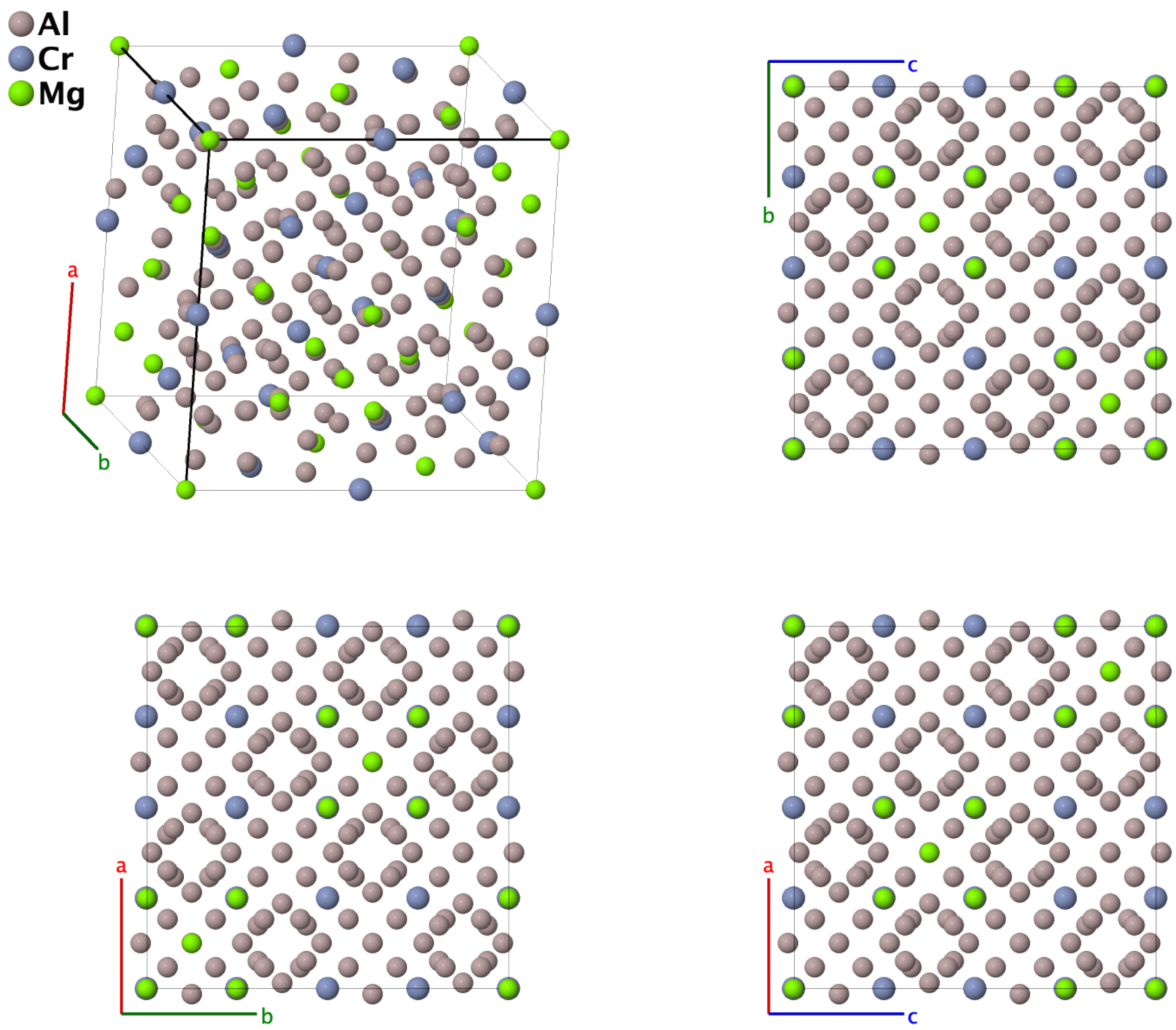

| | | |
|---|---|---|
| **Prototype** | : | $Al_{18}Cr_2Mg_3$ |
| **AFLOW prototype label** | : | A18B2C3_cF184_227_fg_d_ac |
| ***Strukturbericht* designation** | : | None |
| **Pearson symbol** | : | cF184 |
| **Space group number** | : | 227 |
| **Space group symbol** | : | $Fd\bar{3}m$ |
| **AFLOW prototype command** | : | `aflow --proto=A18B2C3_cF184_227_fg_d_ac --params=`$a, x_4, x_5, z_5$ |

### Other compounds with this structure

- $LaCr_2Al_{20}$, $CeCr_2Al_{20}$, $PrCr_2Al_{20}$, $SmCr_2Al_{20}$, $YbCr_2Al_{20}$, $CeTi_2Al_{20}$, $PrTi_2Al_{20}$, $SmTi_2Al_{20}$, $YbTi_2Al_{20}$, $CeV_2Al_{20}$, $GdV_2Al_{20}$, $LaV_2Al_{20}$, $PrV_2Al_{20}$, $SmV_2Al_{20}$, $CeNi_2Cd_{20}$, $GdNi_2Cd_{20}$, $LaNi_2Cd_{20}$, $NdNi_2Cd_{20}$, $PrNi_2Cd_{20}$, $SmNi_2Cd_{20}$, $YNi_2Cd_{20}$, $CePd_2Cd_{20}$, $PrPd_2Cd_{20}$, $SmPd_2Cd_{20}$, and $UOs_2Zn_{20}$

---

- (Samson, 1958) gives the atomic coordinates in terms of setting 1 of space group $Fd\bar{3}m$ #227. We have shifted this to the standard Setting 2, where the inversion site of the lattice is at the origin.
- If the (8*a*), (16*c*) and (16*d*) sites are occupied by the same type of atom this becomes the $Zn_{22}Zr$ structure.
- In the ternary compounds $LnMX_{20}$, the rare earth (*Ln*) metal occupies the (8*a*) site, the transition metal (*M*) the (16*d*) site, and *X*=Al,Cd,Zn occupies the (16*c*), (48*f*) and (96*g*) sites. These compounds are sometimes listed under the $CeCr_2Al_{20}$ prototype.

**Face-centered Cubic primitive vectors:**

$$
\begin{aligned}
\mathbf{a}_1 &= \frac{1}{2}\,a\,\hat{\mathbf{y}} + \frac{1}{2}\,a\,\hat{\mathbf{z}} \\
\mathbf{a}_2 &= \frac{1}{2}\,a\,\hat{\mathbf{x}} + \frac{1}{2}\,a\,\hat{\mathbf{z}} \\
\mathbf{a}_3 &= \frac{1}{2}\,a\,\hat{\mathbf{x}} + \frac{1}{2}\,a\,\hat{\mathbf{y}}
\end{aligned}
$$

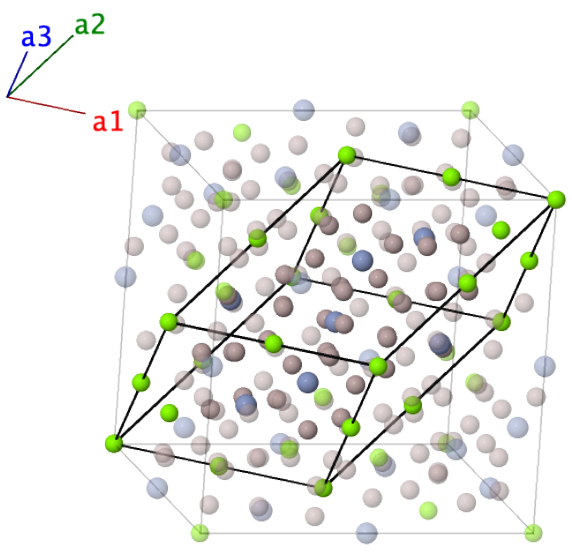

**Basis vectors:**

| | | Lattice Coordinates | | Cartesian Coordinates | Wyckoff Position | Atom Type |
|---|---|---|---|---|---|---|
| $\mathbf{B}_1$ | = | $\frac{1}{8}\,\mathbf{a}_1 + \frac{1}{8}\,\mathbf{a}_2 + \frac{1}{8}\,\mathbf{a}_3$ | = | $\frac{1}{8}a\,\hat{\mathbf{x}} + \frac{1}{8}a\,\hat{\mathbf{y}} + \frac{1}{8}a\,\hat{\mathbf{z}}$ | (8$a$) | Mg I |
| $\mathbf{B}_2$ | = | $\frac{7}{8}\,\mathbf{a}_1 + \frac{7}{8}\,\mathbf{a}_2 + \frac{7}{8}\,\mathbf{a}_3$ | = | $\frac{7}{8}a\,\hat{\mathbf{x}} + \frac{7}{8}a\,\hat{\mathbf{y}} + \frac{7}{8}a\,\hat{\mathbf{z}}$ | (8$a$) | Mg I |
| $\mathbf{B}_3$ | = | $0\,\mathbf{a}_1 + 0\,\mathbf{a}_2 + 0\,\mathbf{a}_3$ | = | $0\,\hat{\mathbf{x}} + 0\,\hat{\mathbf{y}} + 0\,\hat{\mathbf{z}}$ | (16$c$) | Mg II |
| $\mathbf{B}_4$ | = | $\frac{1}{2}\,\mathbf{a}_3$ | = | $\frac{1}{4}a\,\hat{\mathbf{x}} + \frac{1}{4}a\,\hat{\mathbf{y}}$ | (16$c$) | Mg II |
| $\mathbf{B}_5$ | = | $\frac{1}{2}\,\mathbf{a}_2$ | = | $\frac{1}{4}a\,\hat{\mathbf{x}} + \frac{1}{4}a\,\hat{\mathbf{z}}$ | (16$c$) | Mg II |
| $\mathbf{B}_6$ | = | $\frac{1}{2}\,\mathbf{a}_1$ | = | $\frac{1}{4}a\,\hat{\mathbf{y}} + \frac{1}{4}a\,\hat{\mathbf{z}}$ | (16$c$) | Mg II |
| $\mathbf{B}_7$ | = | $\frac{1}{2}\,\mathbf{a}_1 + \frac{1}{2}\,\mathbf{a}_2 + \frac{1}{2}\,\mathbf{a}_3$ | = | $\frac{1}{2}a\,\hat{\mathbf{x}} + \frac{1}{2}a\,\hat{\mathbf{y}} + \frac{1}{2}a\,\hat{\mathbf{z}}$ | (16$d$) | Cr |
| $\mathbf{B}_8$ | = | $\frac{1}{2}\,\mathbf{a}_1 + \frac{1}{2}\,\mathbf{a}_2$ | = | $\frac{1}{4}a\,\hat{\mathbf{x}} + \frac{1}{4}a\,\hat{\mathbf{y}} + \frac{1}{2}a\,\hat{\mathbf{z}}$ | (16$d$) | Cr |
| $\mathbf{B}_9$ | = | $\frac{1}{2}\,\mathbf{a}_1 + \frac{1}{2}\,\mathbf{a}_3$ | = | $\frac{1}{4}a\,\hat{\mathbf{x}} + \frac{1}{2}a\,\hat{\mathbf{y}} + \frac{1}{4}a\,\hat{\mathbf{z}}$ | (16$d$) | Cr |
| $\mathbf{B}_{10}$ | = | $\frac{1}{2}\,\mathbf{a}_2 + \frac{1}{2}\,\mathbf{a}_3$ | = | $\frac{1}{2}a\,\hat{\mathbf{x}} + \frac{1}{4}a\,\hat{\mathbf{y}} + \frac{1}{4}a\,\hat{\mathbf{z}}$ | (16$d$) | Cr |
| $\mathbf{B}_{11}$ | = | $\left(\frac{1}{4} - x_4\right)\,\mathbf{a}_1 + x_4\,\mathbf{a}_2 + x_4\,\mathbf{a}_3$ | = | $x_4a\,\hat{\mathbf{x}} + \frac{1}{8}a\,\hat{\mathbf{y}} + \frac{1}{8}a\,\hat{\mathbf{z}}$ | (48$f$) | Al I |
| $\mathbf{B}_{12}$ | = | $x_4\,\mathbf{a}_1 + \left(\frac{1}{4} - x_4\right)\,\mathbf{a}_2 + \left(\frac{1}{4} - x_4\right)\,\mathbf{a}_3$ | = | $\left(\frac{1}{4} - x_4\right)a\,\hat{\mathbf{x}} + \frac{1}{8}a\,\hat{\mathbf{y}} + \frac{1}{8}a\,\hat{\mathbf{z}}$ | (48$f$) | Al I |
| $\mathbf{B}_{13}$ | = | $x_4\,\mathbf{a}_1 + \left(\frac{1}{4} - x_4\right)\,\mathbf{a}_2 + x_4\,\mathbf{a}_3$ | = | $\frac{1}{8}a\,\hat{\mathbf{x}} + x_4a\,\hat{\mathbf{y}} + \frac{1}{8}a\,\hat{\mathbf{z}}$ | (48$f$) | Al I |
| $\mathbf{B}_{14}$ | = | $\left(\frac{1}{4} - x_4\right)\,\mathbf{a}_1 + x_4\,\mathbf{a}_2 + \left(\frac{1}{4} - x_4\right)\,\mathbf{a}_3$ | = | $\frac{1}{8}a\,\hat{\mathbf{x}} + \left(\frac{1}{4} - x_4\right)a\,\hat{\mathbf{y}} + \frac{1}{8}a\,\hat{\mathbf{z}}$ | (48$f$) | Al I |
| $\mathbf{B}_{15}$ | = | $x_4\,\mathbf{a}_1 + x_4\,\mathbf{a}_2 + \left(\frac{1}{4} - x_4\right)\,\mathbf{a}_3$ | = | $\frac{1}{8}a\,\hat{\mathbf{x}} + \frac{1}{8}a\,\hat{\mathbf{y}} + x_4a\,\hat{\mathbf{z}}$ | (48$f$) | Al I |
| $\mathbf{B}_{16}$ | = | $\left(\frac{1}{4} - x_4\right)\,\mathbf{a}_1 + \left(\frac{1}{4} - x_4\right)\,\mathbf{a}_2 + x_4\,\mathbf{a}_3$ | = | $\frac{1}{8}a\,\hat{\mathbf{x}} + \frac{1}{8}a\,\hat{\mathbf{y}} + \left(\frac{1}{4} - x_4\right)a\,\hat{\mathbf{z}}$ | (48$f$) | Al I |
| $\mathbf{B}_{17}$ | = | $\left(\frac{3}{4} + x_4\right)\,\mathbf{a}_1 - x_4\,\mathbf{a}_2 + \left(\frac{3}{4} + x_4\right)\,\mathbf{a}_3$ | = | $\frac{3}{8}a\,\hat{\mathbf{x}} + \left(\frac{3}{4} + x_4\right)a\,\hat{\mathbf{y}} + \frac{3}{8}a\,\hat{\mathbf{z}}$ | (48$f$) | Al I |
| $\mathbf{B}_{18}$ | = | $-x_4\,\mathbf{a}_1 + \left(\frac{3}{4} + x_4\right)\,\mathbf{a}_2 - x_4\,\mathbf{a}_3$ | = | $\frac{3}{8}a\,\hat{\mathbf{x}} - x_4a\,\hat{\mathbf{y}} + \frac{3}{8}a\,\hat{\mathbf{z}}$ | (48$f$) | Al I |
| $\mathbf{B}_{19}$ | = | $-x_4\,\mathbf{a}_1 + \left(\frac{3}{4} + x_4\right)\,\mathbf{a}_2 + \left(\frac{3}{4} + x_4\right)\,\mathbf{a}_3$ | = | $\left(\frac{3}{4} + x_4\right)a\,\hat{\mathbf{x}} + \frac{3}{8}a\,\hat{\mathbf{y}} + \frac{3}{8}a\,\hat{\mathbf{z}}$ | (48$f$) | Al I |
| $\mathbf{B}_{20}$ | = | $\left(\frac{3}{4} + x_4\right)\,\mathbf{a}_1 - x_4\,\mathbf{a}_2 - x_4\,\mathbf{a}_3$ | = | $-x_4a\,\hat{\mathbf{x}} + \frac{3}{8}a\,\hat{\mathbf{y}} + \frac{3}{8}a\,\hat{\mathbf{z}}$ | (48$f$) | Al I |
| $\mathbf{B}_{21}$ | = | $-x_4\,\mathbf{a}_1 - x_4\,\mathbf{a}_2 + \left(\frac{3}{4} + x_4\right)\,\mathbf{a}_3$ | = | $\frac{3}{8}a\,\hat{\mathbf{x}} + \frac{3}{8}a\,\hat{\mathbf{y}} - x_4a\,\hat{\mathbf{z}}$ | (48$f$) | Al I |
| $\mathbf{B}_{22}$ | = | $\left(\frac{3}{4} + x_4\right)\,\mathbf{a}_1 + \left(\frac{3}{4} + x_4\right)\,\mathbf{a}_2 - x_4\,\mathbf{a}_3$ | = | $\frac{3}{8}a\,\hat{\mathbf{x}} + \frac{3}{8}a\,\hat{\mathbf{y}} + \left(\frac{3}{4} + x_4\right)a\,\hat{\mathbf{z}}$ | (48$f$) | Al I |
| $\mathbf{B}_{23}$ | = | $z_5\,\mathbf{a}_1 + z_5\,\mathbf{a}_2 + (2x_5 - z_5)\,\mathbf{a}_3$ | = | $x_5a\,\hat{\mathbf{x}} + x_5a\,\hat{\mathbf{y}} + z_5a\,\hat{\mathbf{z}}$ | (96$g$) | Al II |
| $\mathbf{B}_{24}$ | = | $z_5\,\mathbf{a}_1 + z_5\,\mathbf{a}_2 + \left(\frac{1}{2} - 2x_5 - z_5\right)\,\mathbf{a}_3$ | = | $\left(\frac{1}{4} - x_5\right)a\,\hat{\mathbf{x}} + \left(\frac{1}{4} - x_5\right)a\,\hat{\mathbf{y}} + z_5a\,\hat{\mathbf{z}}$ | (96$g$) | Al II |
| $\mathbf{B}_{25}$ | = | $(2x_5 - z_5)\,\mathbf{a}_1 + \left(\frac{1}{2} - 2x_5 - z_5\right)\,\mathbf{a}_2 + z_5\,\mathbf{a}_3$ | = | $\left(\frac{1}{4} - x_5\right)a\,\hat{\mathbf{x}} + x_5a\,\hat{\mathbf{y}} + \left(\frac{1}{4} - z_5\right)a\,\hat{\mathbf{z}}$ | (96$g$) | Al II |
| $\mathbf{B}_{26}$ | = | $\left(\frac{1}{2} - 2x_5 - z_5\right)\,\mathbf{a}_1 + (2x_5 - z_5)\,\mathbf{a}_2 + z_5\,\mathbf{a}_3$ | = | $x_5a\,\hat{\mathbf{x}} + \left(\frac{1}{4} - x_5\right)a\,\hat{\mathbf{y}} + \left(\frac{1}{4} - z_5\right)a\,\hat{\mathbf{z}}$ | (96$g$) | Al II |
| $\mathbf{B}_{27}$ | = | $(2x_5 - z_5)\,\mathbf{a}_1 + z_5\,\mathbf{a}_2 + z_5\,\mathbf{a}_3$ | = | $z_5 a\,\mathbf{\hat{x}} + x_5 a\,\mathbf{\hat{y}} + x_5 a\,\mathbf{\hat{z}}$ | (96g) | Al II |
| $\mathbf{B}_{28}$ | = | $\left(\frac{1}{2} - 2x_5 - z_5\right)\mathbf{a}_1 + z_5\,\mathbf{a}_2 + z_5\,\mathbf{a}_3$ | = | $z_5 a\,\mathbf{\hat{x}} + \left(\frac{1}{4} - x_5\right) a\,\mathbf{\hat{y}} + \left(\frac{1}{4} - x_5\right) a\,\mathbf{\hat{z}}$ | (96g) | Al II |
| $\mathbf{B}_{29}$ | = | $z_5\,\mathbf{a}_1 + (2x_5 - z_5)\,\mathbf{a}_2 + \left(\frac{1}{2} - 2x_5 - z_5\right)\mathbf{a}_3$ | = | $\left(\frac{1}{4} - z_5\right) a\,\mathbf{\hat{x}} + \left(\frac{1}{4} - x_5\right) a\,\mathbf{\hat{y}} + x_5 a\,\mathbf{\hat{z}}$ | (96g) | Al II |
| $\mathbf{B}_{30}$ | = | $z_5\,\mathbf{a}_1 + \left(\frac{1}{2} - 2x_5 - z_5\right)\mathbf{a}_2 + (2x_5 - z_5)\,\mathbf{a}_3$ | = | $\left(\frac{1}{4} - z_5\right) a\,\mathbf{\hat{x}} + x_5 a\,\mathbf{\hat{y}} + \left(\frac{1}{4} - x_5\right) a\,\mathbf{\hat{z}}$ | (96g) | Al II |
| $\mathbf{B}_{31}$ | = | $z_5\,\mathbf{a}_1 + (2x_5 - z_5)\,\mathbf{a}_2 + z_5\,\mathbf{a}_3$ | = | $x_5 a\,\mathbf{\hat{x}} + z_5 a\,\mathbf{\hat{y}} + x_5 a\,\mathbf{\hat{z}}$ | (96g) | Al II |
| $\mathbf{B}_{32}$ | = | $z_5\,\mathbf{a}_1 + \left(\frac{1}{2} - 2x_5 - z_5\right)\mathbf{a}_2 + z_5\,\mathbf{a}_3$ | = | $\left(\frac{1}{4} - x_5\right) a\,\mathbf{\hat{x}} + z_5 a\,\mathbf{\hat{y}} + \left(\frac{1}{4} - x_5\right) a\,\mathbf{\hat{z}}$ | (96g) | Al II |
| $\mathbf{B}_{33}$ | = | $\left(\frac{1}{2} - 2x_5 - z_5\right)\mathbf{a}_1 + z_5\,\mathbf{a}_2 + (2x_5 - z_5)\,\mathbf{a}_3$ | = | $x_5 a\,\mathbf{\hat{x}} + \left(\frac{1}{4} - z_5\right) a\,\mathbf{\hat{y}} + \left(\frac{1}{4} - x_5\right) a\,\mathbf{\hat{z}}$ | (96g) | Al II |
| $\mathbf{B}_{34}$ | = | $(2x_5 - z_5)\,\mathbf{a}_1 + z_5\,\mathbf{a}_2 + \left(\frac{1}{2} - 2x_5 - z_5\right)\mathbf{a}_3$ | = | $\left(\frac{1}{4} - x_5\right) a\,\mathbf{\hat{x}} + \left(\frac{1}{4} - z_5\right) a\,\mathbf{\hat{y}} + x_5 a\,\mathbf{\hat{z}}$ | (96g) | Al II |
| $\mathbf{B}_{35}$ | = | $-z_5\,\mathbf{a}_1 - z_5\,\mathbf{a}_2 + \left(\frac{1}{2} + 2x_5 + z_5\right)\mathbf{a}_3$ | = | $\left(\frac{1}{4} + x_5\right) a\,\mathbf{\hat{x}} + \left(\frac{1}{4} + x_5\right) a\,\mathbf{\hat{y}} - z_5 a\,\mathbf{\hat{z}}$ | (96g) | Al II |
| $\mathbf{B}_{36}$ | = | $-z_5\,\mathbf{a}_1 - z_5\,\mathbf{a}_2 + (-2x_5 + z_5)\,\mathbf{a}_3$ | = | $-x_5 a\,\mathbf{\hat{x}} - x_5 a\,\mathbf{\hat{y}} - z_5 a\,\mathbf{\hat{z}}$ | (96g) | Al II |
| $\mathbf{B}_{37}$ | = | $(-2x_5 + z_5)\,\mathbf{a}_1 + \left(\frac{1}{2} + 2x_5 + z_5\right)\mathbf{a}_2 - z_5\,\mathbf{a}_3$ | = | $\left(\frac{1}{4} + x_5\right) a\,\mathbf{\hat{x}} - x_5 a\,\mathbf{\hat{y}} + \left(\frac{1}{4} + z_5\right) a\,\mathbf{\hat{z}}$ | (96g) | Al II |
| $\mathbf{B}_{38}$ | = | $\left(\frac{1}{2} + 2x_5 + z_5\right)\mathbf{a}_1 + (-2x_5 + z_5)\,\mathbf{a}_2 - z_5\,\mathbf{a}_3$ | = | $-x_5 a\,\mathbf{\hat{x}} + \left(\frac{1}{4} + x_5\right) a\,\mathbf{\hat{y}} + \left(\frac{1}{4} + z_5\right) a\,\mathbf{\hat{z}}$ | (96g) | Al II |
| $\mathbf{B}_{39}$ | = | $(-2x_5 + z_5)\,\mathbf{a}_1 - z_5\,\mathbf{a}_2 + \left(\frac{1}{2} + 2x_5 + z_5\right)\mathbf{a}_3$ | = | $\left(\frac{1}{4} + x_5\right) a\,\mathbf{\hat{x}} + \left(\frac{1}{4} + z_5\right) a\,\mathbf{\hat{y}} - x_5 a\,\mathbf{\hat{z}}$ | (96g) | Al II |
| $\mathbf{B}_{40}$ | = | $\left(\frac{1}{2} + 2x_5 + z_5\right)\mathbf{a}_1 - z_5\,\mathbf{a}_2 + (-2x_5 + z_5)\,\mathbf{a}_3$ | = | $-x_5 a\,\mathbf{\hat{x}} + \left(\frac{1}{4} + z_5\right) a\,\mathbf{\hat{y}} + \left(\frac{1}{4} + x_5\right) a\,\mathbf{\hat{z}}$ | (96g) | Al II |
| $\mathbf{B}_{41}$ | = | $-z_5\,\mathbf{a}_1 + (-2x_5 + z_5)\,\mathbf{a}_2 - z_5\,\mathbf{a}_3$ | = | $-x_5 a\,\mathbf{\hat{x}} - z_5 a\,\mathbf{\hat{y}} - x_5 a\,\mathbf{\hat{z}}$ | (96g) | Al II |
| $\mathbf{B}_{42}$ | = | $-z_5\,\mathbf{a}_1 + \left(\frac{1}{2} + 2x_5 + z_5\right)\mathbf{a}_2 - z_5\,\mathbf{a}_3$ | = | $\left(\frac{1}{4} + x_5\right) a\,\mathbf{\hat{x}} - z_5 a\,\mathbf{\hat{y}} + \left(\frac{1}{4} + x_5\right) a\,\mathbf{\hat{z}}$ | (96g) | Al II |
| $\mathbf{B}_{43}$ | = | $-z_5\,\mathbf{a}_1 + (-2x_5 + z_5)\,\mathbf{a}_2 + \left(\frac{1}{2} + 2x_5 + z_5\right)\mathbf{a}_3$ | = | $\left(\frac{1}{4} + z_5\right) a\,\mathbf{\hat{x}} + \left(\frac{1}{4} + x_5\right) a\,\mathbf{\hat{y}} - x_5 a\,\mathbf{\hat{z}}$ | (96g) | Al II |
| $\mathbf{B}_{44}$ | = | $-z_5\,\mathbf{a}_1 + \left(\frac{1}{2} + 2x_5 + z_5\right)\mathbf{a}_2 + (-2x_5 + z_5)\,\mathbf{a}_3$ | = | $\left(\frac{1}{4} + z_5\right) a\,\mathbf{\hat{x}} - x_5 a\,\mathbf{\hat{y}} + \left(\frac{1}{4} + x_5\right) a\,\mathbf{\hat{z}}$ | (96g) | Al II |
| $\mathbf{B}_{45}$ | = | $\left(\frac{1}{2} + 2x_5 + z_5\right)\mathbf{a}_1 - z_5\,\mathbf{a}_2 - z_5\,\mathbf{a}_3$ | = | $-z_5 a\,\mathbf{\hat{x}} + \left(\frac{1}{4} + x_5\right) a\,\mathbf{\hat{y}} + \left(\frac{1}{4} + x_5\right) a\,\mathbf{\hat{z}}$ | (96g) | Al II |
| $\mathbf{B}_{46}$ | = | $(-2x_5 + z_5)\,\mathbf{a}_1 - z_5\,\mathbf{a}_2 - z_5\,\mathbf{a}_3$ | = | $-z_5 a\,\mathbf{\hat{x}} - x_5 a\,\mathbf{\hat{y}} - x_5 a\,\mathbf{\hat{z}}$ | (96g) | Al II |

**Geometry files:**

- CIF: pp. 1824
- POSCAR: pp. 1825

# $Zn_{22}Zr$ Structure: A22B_cF184_227_cdfg_a

http://aflow.org/prototype-encyclopedia/A22B_cF184_227_cdfg_a

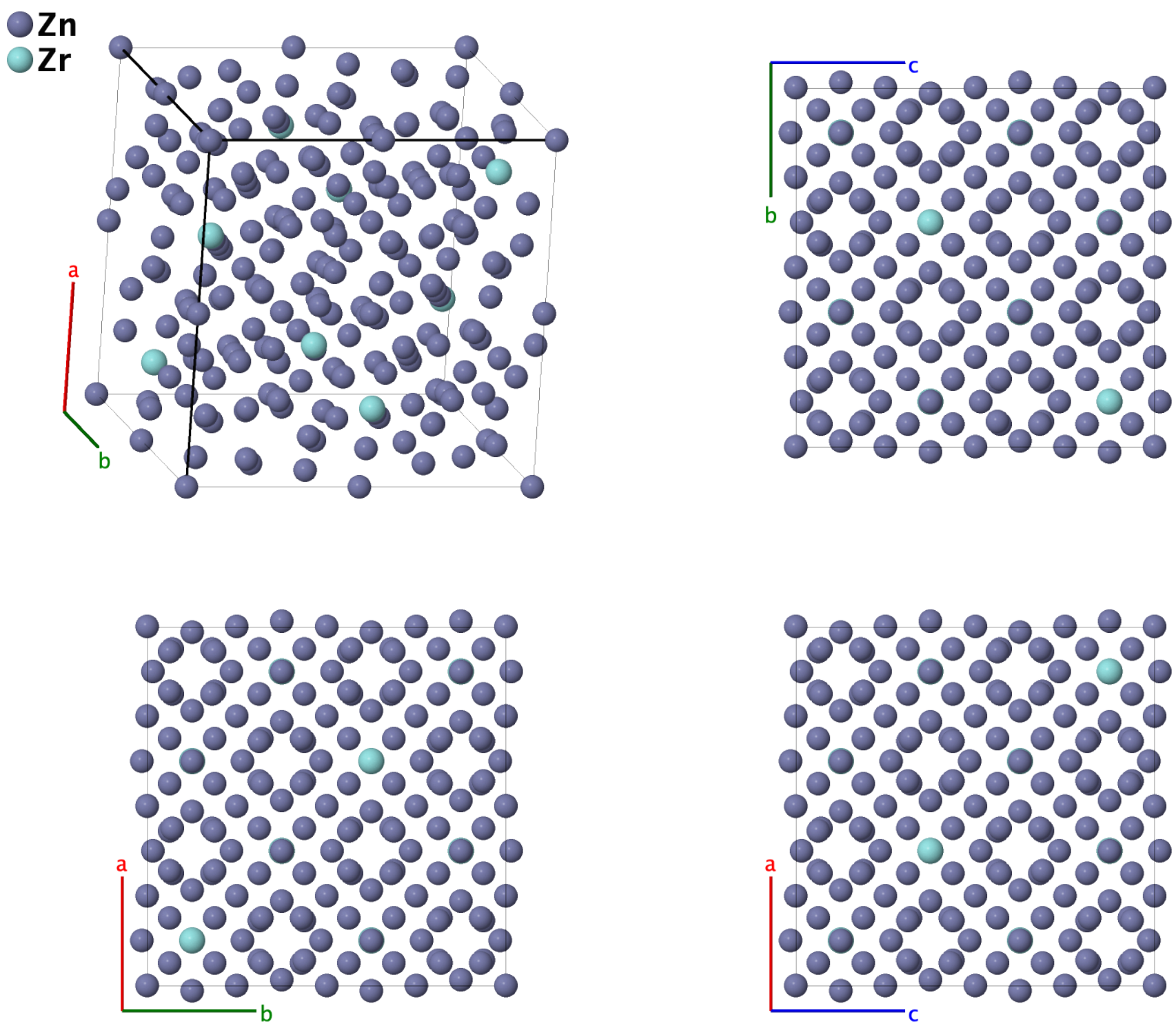

| | | |
|---|---|---|
| **Prototype** | : | $Zn_{22}Zr$ |
| **AFLOW prototype label** | : | A22B_cF184_227_cdfg_a |
| ***Strukturbericht* designation** | : | None |
| **Pearson symbol** | : | cF184 |
| **Space group number** | : | 227 |
| **Space group symbol** | : | $Fd\bar{3}m$ |
| **AFLOW prototype command** | : | `aflow --proto=A22B_cF184_227_cdfg_a`<br>`--params=`$a, x_4, x_5, z_5$ |

**Other compounds with this structure**

- $Be_{22}Mo$ and $Be_{22}W$

- (Samson, 1961) gives the atomic coordinates in terms of setting 1 of space group $F\bar{3}m$ #227. We have shifted this to the standard setting 2, where the inversion site of the lattice is at the origin.
- Samson also suggests that the "$Zn_{14}Zr$" structure is created when zirconium atoms replace some of the zinc atoms on the (16$c$) site [the (16$d$) site in Samson's orientation].
- This structure can be derived from the $Mg_3Cr_2Al_{18}$ structure.

**Face-centered Cubic primitive vectors:**

$$\mathbf{a}_1 = \frac{1}{2}\,a\,\hat{\mathbf{y}} + \frac{1}{2}\,a\,\hat{\mathbf{z}}$$
$$\mathbf{a}_2 = \frac{1}{2}\,a\,\hat{\mathbf{x}} + \frac{1}{2}\,a\,\hat{\mathbf{z}}$$
$$\mathbf{a}_3 = \frac{1}{2}\,a\,\hat{\mathbf{x}} + \frac{1}{2}\,a\,\hat{\mathbf{y}}$$

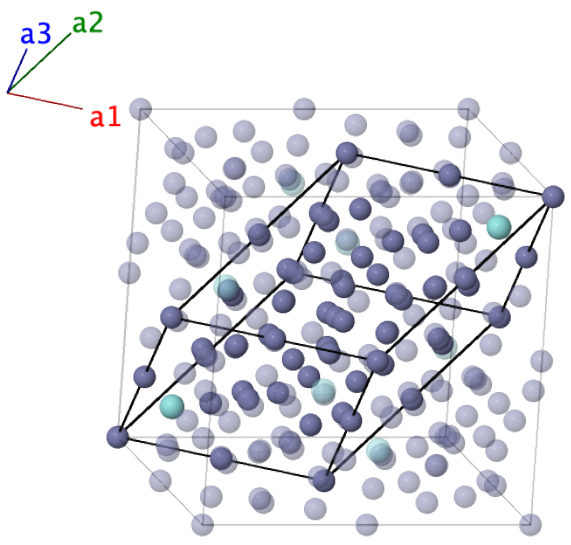

**Basis vectors:**

| | | Lattice Coordinates | | Cartesian Coordinates | Wyckoff Position | Atom Type |
|---|---|---|---|---|---|---|
| $\mathbf{B}_1$ | = | $\frac{1}{8}\,\mathbf{a}_1 + \frac{1}{8}\,\mathbf{a}_2 + \frac{1}{8}\,\mathbf{a}_3$ | = | $\frac{1}{8}a\,\hat{\mathbf{x}} + \frac{1}{8}a\,\hat{\mathbf{y}} + \frac{1}{8}a\,\hat{\mathbf{z}}$ | (8$a$) | Zr |
| $\mathbf{B}_2$ | = | $\frac{7}{8}\,\mathbf{a}_1 + \frac{7}{8}\,\mathbf{a}_2 + \frac{7}{8}\,\mathbf{a}_3$ | = | $\frac{7}{8}a\,\hat{\mathbf{x}} + \frac{7}{8}a\,\hat{\mathbf{y}} + \frac{7}{8}a\,\hat{\mathbf{z}}$ | (8$a$) | Zr |
| $\mathbf{B}_3$ | = | $0\,\mathbf{a}_1 + 0\,\mathbf{a}_2 + 0\,\mathbf{a}_3$ | = | $0\,\hat{\mathbf{x}} + 0\,\hat{\mathbf{y}} + 0\,\hat{\mathbf{z}}$ | (16$c$) | Zn I |
| $\mathbf{B}_4$ | = | $\frac{1}{2}\,\mathbf{a}_3$ | = | $\frac{1}{4}a\,\hat{\mathbf{x}} + \frac{1}{4}a\,\hat{\mathbf{y}}$ | (16$c$) | Zn I |
| $\mathbf{B}_5$ | = | $\frac{1}{2}\,\mathbf{a}_2$ | = | $\frac{1}{4}a\,\hat{\mathbf{x}} + \frac{1}{4}a\,\hat{\mathbf{z}}$ | (16$c$) | Zn I |
| $\mathbf{B}_6$ | = | $\frac{1}{2}\,\mathbf{a}_1$ | = | $\frac{1}{4}a\,\hat{\mathbf{y}} + \frac{1}{4}a\,\hat{\mathbf{z}}$ | (16$c$) | Zn I |
| $\mathbf{B}_7$ | = | $\frac{1}{2}\,\mathbf{a}_1 + \frac{1}{2}\,\mathbf{a}_2 + \frac{1}{2}\,\mathbf{a}_3$ | = | $\frac{1}{2}a\,\hat{\mathbf{x}} + \frac{1}{2}a\,\hat{\mathbf{y}} + \frac{1}{2}a\,\hat{\mathbf{z}}$ | (16$d$) | Zn II |
| $\mathbf{B}_8$ | = | $\frac{1}{2}\,\mathbf{a}_1 + \frac{1}{2}\,\mathbf{a}_2$ | = | $\frac{1}{4}a\,\hat{\mathbf{x}} + \frac{1}{4}a\,\hat{\mathbf{y}} + \frac{1}{2}a\,\hat{\mathbf{z}}$ | (16$d$) | Zn II |
| $\mathbf{B}_9$ | = | $\frac{1}{2}\,\mathbf{a}_1 + \frac{1}{2}\,\mathbf{a}_3$ | = | $\frac{1}{4}a\,\hat{\mathbf{x}} + \frac{1}{2}a\,\hat{\mathbf{y}} + \frac{1}{4}a\,\hat{\mathbf{z}}$ | (16$d$) | Zn II |
| $\mathbf{B}_{10}$ | = | $\frac{1}{2}\,\mathbf{a}_2 + \frac{1}{2}\,\mathbf{a}_3$ | = | $\frac{1}{2}a\,\hat{\mathbf{x}} + \frac{1}{4}a\,\hat{\mathbf{y}} + \frac{1}{4}a\,\hat{\mathbf{z}}$ | (16$d$) | Zn II |
| $\mathbf{B}_{11}$ | = | $\left(\frac{1}{4} - x_4\right)\,\mathbf{a}_1 + x_4\,\mathbf{a}_2 + x_4\,\mathbf{a}_3$ | = | $x_4a\,\hat{\mathbf{x}} + \frac{1}{8}a\,\hat{\mathbf{y}} + \frac{1}{8}a\,\hat{\mathbf{z}}$ | (48$f$) | Zn III |
| $\mathbf{B}_{12}$ | = | $x_4\,\mathbf{a}_1 + \left(\frac{1}{4} - x_4\right)\,\mathbf{a}_2 + \left(\frac{1}{4} - x_4\right)\,\mathbf{a}_3$ | = | $\left(\frac{1}{4} - x_4\right)a\,\hat{\mathbf{x}} + \frac{1}{8}a\,\hat{\mathbf{y}} + \frac{1}{8}a\,\hat{\mathbf{z}}$ | (48$f$) | Zn III |
| $\mathbf{B}_{13}$ | = | $x_4\,\mathbf{a}_1 + \left(\frac{1}{4} - x_4\right)\,\mathbf{a}_2 + x_4\,\mathbf{a}_3$ | = | $\frac{1}{8}a\,\hat{\mathbf{x}} + x_4a\,\hat{\mathbf{y}} + \frac{1}{8}a\,\hat{\mathbf{z}}$ | (48$f$) | Zn III |
| $\mathbf{B}_{14}$ | = | $\left(\frac{1}{4} - x_4\right)\,\mathbf{a}_1 + x_4\,\mathbf{a}_2 + \left(\frac{1}{4} - x_4\right)\,\mathbf{a}_3$ | = | $\frac{1}{8}a\,\hat{\mathbf{x}} + \left(\frac{1}{4} - x_4\right)a\,\hat{\mathbf{y}} + \frac{1}{8}a\,\hat{\mathbf{z}}$ | (48$f$) | Zn III |
| $\mathbf{B}_{15}$ | = | $x_4\,\mathbf{a}_1 + x_4\,\mathbf{a}_2 + \left(\frac{1}{4} - x_4\right)\,\mathbf{a}_3$ | = | $\frac{1}{8}a\,\hat{\mathbf{x}} + \frac{1}{8}a\,\hat{\mathbf{y}} + x_4a\,\hat{\mathbf{z}}$ | (48$f$) | Zn III |
| $\mathbf{B}_{16}$ | = | $\left(\frac{1}{4} - x_4\right)\,\mathbf{a}_1 + \left(\frac{1}{4} - x_4\right)\,\mathbf{a}_2 + x_4\,\mathbf{a}_3$ | = | $\frac{1}{8}a\,\hat{\mathbf{x}} + \frac{1}{8}a\,\hat{\mathbf{y}} + \left(\frac{1}{4} - x_4\right)a\,\hat{\mathbf{z}}$ | (48$f$) | Zn III |
| $\mathbf{B}_{17}$ | = | $\left(\frac{3}{4} + x_4\right)\,\mathbf{a}_1 - x_4\,\mathbf{a}_2 + \left(\frac{3}{4} + x_4\right)\,\mathbf{a}_3$ | = | $\frac{3}{8}a\,\hat{\mathbf{x}} + \left(\frac{3}{4} + x_4\right)a\,\hat{\mathbf{y}} + \frac{3}{8}a\,\hat{\mathbf{z}}$ | (48$f$) | Zn III |
| $\mathbf{B}_{18}$ | = | $-x_4\,\mathbf{a}_1 + \left(\frac{3}{4} + x_4\right)\,\mathbf{a}_2 - x_4\,\mathbf{a}_3$ | = | $\frac{3}{8}a\,\hat{\mathbf{x}} - x_4a\,\hat{\mathbf{y}} + \frac{3}{8}a\,\hat{\mathbf{z}}$ | (48$f$) | Zn III |
| $\mathbf{B}_{19}$ | = | $-x_4\,\mathbf{a}_1 + \left(\frac{3}{4} + x_4\right)\,\mathbf{a}_2 + \left(\frac{3}{4} + x_4\right)\,\mathbf{a}_3$ | = | $\left(\frac{3}{4} + x_4\right)a\,\hat{\mathbf{x}} + \frac{3}{8}a\,\hat{\mathbf{y}} + \frac{3}{8}a\,\hat{\mathbf{z}}$ | (48$f$) | Zn III |
| $\mathbf{B}_{20}$ | = | $\left(\frac{3}{4} + x_4\right)\,\mathbf{a}_1 - x_4\,\mathbf{a}_2 - x_4\,\mathbf{a}_3$ | = | $-x_4a\,\hat{\mathbf{x}} + \frac{3}{8}a\,\hat{\mathbf{y}} + \frac{3}{8}a\,\hat{\mathbf{z}}$ | (48$f$) | Zn III |
| $\mathbf{B}_{21}$ | = | $-x_4\,\mathbf{a}_1 - x_4\,\mathbf{a}_2 + \left(\frac{3}{4} + x_4\right)\,\mathbf{a}_3$ | = | $\frac{3}{8}a\,\hat{\mathbf{x}} + \frac{3}{8}a\,\hat{\mathbf{y}} - x_4a\,\hat{\mathbf{z}}$ | (48$f$) | Zn III |
| $\mathbf{B}_{22}$ | = | $\left(\frac{3}{4} + x_4\right)\,\mathbf{a}_1 + \left(\frac{3}{4} + x_4\right)\,\mathbf{a}_2 - x_4\,\mathbf{a}_3$ | = | $\frac{3}{8}a\,\hat{\mathbf{x}} + \frac{3}{8}a\,\hat{\mathbf{y}} + \left(\frac{3}{4} + x_4\right)a\,\hat{\mathbf{z}}$ | (48$f$) | Zn III |
| $\mathbf{B}_{23}$ | = | $z_5\,\mathbf{a}_1 + z_5\,\mathbf{a}_2 + (2x_5 - z_5)\,\mathbf{a}_3$ | = | $x_5a\,\hat{\mathbf{x}} + x_5a\,\hat{\mathbf{y}} + z_5a\,\hat{\mathbf{z}}$ | (96$g$) | Zn IV |
| $\mathbf{B}_{24}$ | = | $z_5\,\mathbf{a}_1 + z_5\,\mathbf{a}_2 + \left(\frac{1}{2} - 2x_5 - z_5\right)\,\mathbf{a}_3$ | = | $\left(\frac{1}{4} - x_5\right)a\,\hat{\mathbf{x}} + \left(\frac{1}{4} - x_5\right)a\,\hat{\mathbf{y}} + z_5a\,\hat{\mathbf{z}}$ | (96$g$) | Zn IV |
| $\mathbf{B}_{25}$ | = | $(2x_5 - z_5)\,\mathbf{a}_1 +$ $\left(\frac{1}{2} - 2x_5 - z_5\right)\,\mathbf{a}_2 + z_5\,\mathbf{a}_3$ | = | $\left(\frac{1}{4} - x_5\right)a\,\hat{\mathbf{x}} + x_5a\,\hat{\mathbf{y}} + \left(\frac{1}{4} - z_5\right)a\,\hat{\mathbf{z}}$ | (96$g$) | Zn IV |
| $\mathbf{B}_{26}$ | = | $\left(\frac{1}{2} - 2x_5 - z_5\right)\,\mathbf{a}_1 +$ $(2x_5 - z_5)\,\mathbf{a}_2 + z_5\,\mathbf{a}_3$ | = | $x_5a\,\hat{\mathbf{x}} + \left(\frac{1}{4} - x_5\right)a\,\hat{\mathbf{y}} + \left(\frac{1}{4} - z_5\right)a\,\hat{\mathbf{z}}$ | (96$g$) | Zn IV |
| $\mathbf{B}_{27}$ | = | $(2x_5 - z_5)\,\mathbf{a}_1 + z_5\,\mathbf{a}_2 + z_5\,\mathbf{a}_3$ | = | $z_5 a\,\mathbf{\hat{x}} + x_5 a\,\mathbf{\hat{y}} + x_5 a\,\mathbf{\hat{z}}$ | (96$g$) | Zn IV |
| $\mathbf{B}_{28}$ | = | $\left(\frac{1}{2} - 2x_5 - z_5\right)\mathbf{a}_1 + z_5\,\mathbf{a}_2 + z_5\,\mathbf{a}_3$ | = | $z_5 a\,\mathbf{\hat{x}} + \left(\frac{1}{4} - x_5\right) a\,\mathbf{\hat{y}} + \left(\frac{1}{4} - x_5\right) a\,\mathbf{\hat{z}}$ | (96$g$) | Zn IV |
| $\mathbf{B}_{29}$ | = | $z_5\,\mathbf{a}_1 + (2x_5 - z_5)\,\mathbf{a}_2 + \left(\frac{1}{2} - 2x_5 - z_5\right)\mathbf{a}_3$ | = | $\left(\frac{1}{4} - z_5\right) a\,\mathbf{\hat{x}} + \left(\frac{1}{4} - x_5\right) a\,\mathbf{\hat{y}} + x_5 a\,\mathbf{\hat{z}}$ | (96$g$) | Zn IV |
| $\mathbf{B}_{30}$ | = | $z_5\,\mathbf{a}_1 + \left(\frac{1}{2} - 2x_5 - z_5\right)\mathbf{a}_2 + (2x_5 - z_5)\,\mathbf{a}_3$ | = | $\left(\frac{1}{4} - z_5\right) a\,\mathbf{\hat{x}} + x_5 a\,\mathbf{\hat{y}} + \left(\frac{1}{4} - x_5\right) a\,\mathbf{\hat{z}}$ | (96$g$) | Zn IV |
| $\mathbf{B}_{31}$ | = | $z_5\,\mathbf{a}_1 + (2x_5 - z_5)\,\mathbf{a}_2 + z_5\,\mathbf{a}_3$ | = | $x_5 a\,\mathbf{\hat{x}} + z_5 a\,\mathbf{\hat{y}} + x_5 a\,\mathbf{\hat{z}}$ | (96$g$) | Zn IV |
| $\mathbf{B}_{32}$ | = | $z_5\,\mathbf{a}_1 + \left(\frac{1}{2} - 2x_5 - z_5\right)\mathbf{a}_2 + z_5\,\mathbf{a}_3$ | = | $\left(\frac{1}{4} - x_5\right) a\,\mathbf{\hat{x}} + z_5 a\,\mathbf{\hat{y}} + \left(\frac{1}{4} - x_5\right) a\,\mathbf{\hat{z}}$ | (96$g$) | Zn IV |
| $\mathbf{B}_{33}$ | = | $\left(\frac{1}{2} - 2x_5 - z_5\right)\mathbf{a}_1 + z_5\,\mathbf{a}_2 + (2x_5 - z_5)\,\mathbf{a}_3$ | = | $x_5 a\,\mathbf{\hat{x}} + \left(\frac{1}{4} - z_5\right) a\,\mathbf{\hat{y}} + \left(\frac{1}{4} - x_5\right) a\,\mathbf{\hat{z}}$ | (96$g$) | Zn IV |
| $\mathbf{B}_{34}$ | = | $(2x_5 - z_5)\,\mathbf{a}_1 + z_5\,\mathbf{a}_2 + \left(\frac{1}{2} - 2x_5 - z_5\right)\mathbf{a}_3$ | = | $\left(\frac{1}{4} - x_5\right) a\,\mathbf{\hat{x}} + \left(\frac{1}{4} - z_5\right) a\,\mathbf{\hat{y}} + x_5 a\,\mathbf{\hat{z}}$ | (96$g$) | Zn IV |
| $\mathbf{B}_{35}$ | = | $-z_5\,\mathbf{a}_1 - z_5\,\mathbf{a}_2 + \left(\frac{1}{2} + 2x_5 + z_5\right)\mathbf{a}_3$ | = | $\left(\frac{1}{4} + x_5\right) a\,\mathbf{\hat{x}} + \left(\frac{1}{4} + x_5\right) a\,\mathbf{\hat{y}} - z_5 a\,\mathbf{\hat{z}}$ | (96$g$) | Zn IV |
| $\mathbf{B}_{36}$ | = | $-z_5\,\mathbf{a}_1 - z_5\,\mathbf{a}_2 + (-2x_5 + z_5)\,\mathbf{a}_3$ | = | $-x_5 a\,\mathbf{\hat{x}} - x_5 a\,\mathbf{\hat{y}} - z_5 a\,\mathbf{\hat{z}}$ | (96$g$) | Zn IV |
| $\mathbf{B}_{37}$ | = | $(-2x_5 + z_5)\,\mathbf{a}_1 + \left(\frac{1}{2} + 2x_5 + z_5\right)\mathbf{a}_2 - z_5\,\mathbf{a}_3$ | = | $\left(\frac{1}{4} + x_5\right) a\,\mathbf{\hat{x}} - x_5 a\,\mathbf{\hat{y}} + \left(\frac{1}{4} + z_5\right) a\,\mathbf{\hat{z}}$ | (96$g$) | Zn IV |
| $\mathbf{B}_{38}$ | = | $\left(\frac{1}{2} + 2x_5 + z_5\right)\mathbf{a}_1 + (-2x_5 + z_5)\,\mathbf{a}_2 - z_5\,\mathbf{a}_3$ | = | $-x_5 a\,\mathbf{\hat{x}} + \left(\frac{1}{4} + x_5\right) a\,\mathbf{\hat{y}} + \left(\frac{1}{4} + z_5\right) a\,\mathbf{\hat{z}}$ | (96$g$) | Zn IV |
| $\mathbf{B}_{39}$ | = | $(-2x_5 + z_5)\,\mathbf{a}_1 - z_5\,\mathbf{a}_2 + \left(\frac{1}{2} + 2x_5 + z_5\right)\mathbf{a}_3$ | = | $\left(\frac{1}{4} + x_5\right) a\,\mathbf{\hat{x}} + \left(\frac{1}{4} + z_5\right) a\,\mathbf{\hat{y}} - x_5 a\,\mathbf{\hat{z}}$ | (96$g$) | Zn IV |
| $\mathbf{B}_{40}$ | = | $\left(\frac{1}{2} + 2x_5 + z_5\right)\mathbf{a}_1 - z_5\,\mathbf{a}_2 + (-2x_5 + z_5)\,\mathbf{a}_3$ | = | $-x_5 a\,\mathbf{\hat{x}} + \left(\frac{1}{4} + z_5\right) a\,\mathbf{\hat{y}} + \left(\frac{1}{4} + x_5\right) a\,\mathbf{\hat{z}}$ | (96$g$) | Zn IV |
| $\mathbf{B}_{41}$ | = | $-z_5\,\mathbf{a}_1 + (-2x_5 + z_5)\,\mathbf{a}_2 - z_5\,\mathbf{a}_3$ | = | $-x_5 a\,\mathbf{\hat{x}} - z_5 a\,\mathbf{\hat{y}} - x_5 a\,\mathbf{\hat{z}}$ | (96$g$) | Zn IV |
| $\mathbf{B}_{42}$ | = | $-z_5\,\mathbf{a}_1 + \left(\frac{1}{2} + 2x_5 + z_5\right)\mathbf{a}_2 - z_5\,\mathbf{a}_3$ | = | $\left(\frac{1}{4} + x_5\right) a\,\mathbf{\hat{x}} - z_5 a\,\mathbf{\hat{y}} + \left(\frac{1}{4} + x_5\right) a\,\mathbf{\hat{z}}$ | (96$g$) | Zn IV |
| $\mathbf{B}_{43}$ | = | $-z_5\,\mathbf{a}_1 + (-2x_5 + z_5)\,\mathbf{a}_2 + \left(\frac{1}{2} + 2x_5 + z_5\right)\mathbf{a}_3$ | = | $\left(\frac{1}{4} + z_5\right) a\,\mathbf{\hat{x}} + \left(\frac{1}{4} + x_5\right) a\,\mathbf{\hat{y}} - x_5 a\,\mathbf{\hat{z}}$ | (96$g$) | Zn IV |
| $\mathbf{B}_{44}$ | = | $-z_5\,\mathbf{a}_1 + \left(\frac{1}{2} + 2x_5 + z_5\right)\mathbf{a}_2 + (-2x_5 + z_5)\,\mathbf{a}_3$ | = | $\left(\frac{1}{4} + z_5\right) a\,\mathbf{\hat{x}} - x_5 a\,\mathbf{\hat{y}} + \left(\frac{1}{4} + x_5\right) a\,\mathbf{\hat{z}}$ | (96$g$) | Zn IV |
| $\mathbf{B}_{45}$ | = | $\left(\frac{1}{2} + 2x_5 + z_5\right)\mathbf{a}_1 - z_5\,\mathbf{a}_2 - z_5\,\mathbf{a}_3$ | = | $-z_5 a\,\mathbf{\hat{x}} + \left(\frac{1}{4} + x_5\right) a\,\mathbf{\hat{y}} + \left(\frac{1}{4} + x_5\right) a\,\mathbf{\hat{z}}$ | (96$g$) | Zn IV |
| $\mathbf{B}_{46}$ | = | $(-2x_5 + z_5)\,\mathbf{a}_1 - z_5\,\mathbf{a}_2 - z_5\,\mathbf{a}_3$ | = | $-z_5 a\,\mathbf{\hat{x}} - x_5 a\,\mathbf{\hat{y}} - x_5 a\,\mathbf{\hat{z}}$ | (96$g$) | Zn IV |

---

---

**Geometry files:**

- CIF: pp. 1826
- POSCAR: pp. 1827

# $H_3PW_{12}O_{40}\cdot 29H_2O$ ($H4_{21}$) Structure: A29B40CD12_cF656_227_ae2fg_e3g_b_g

http://aflow.org/prototype-encyclopedia/A29B40CD12_cF656_227_ae2fg_e3g_b_g

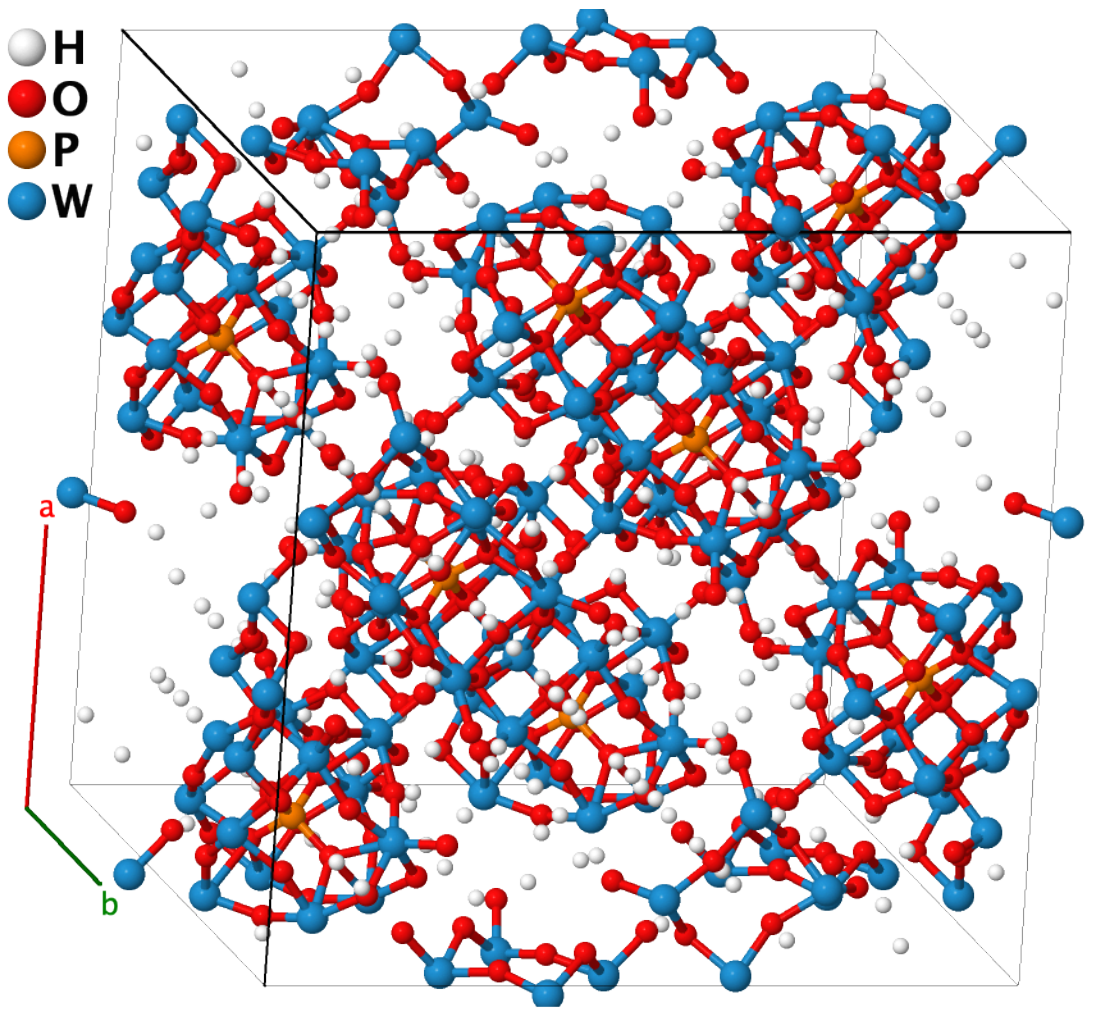

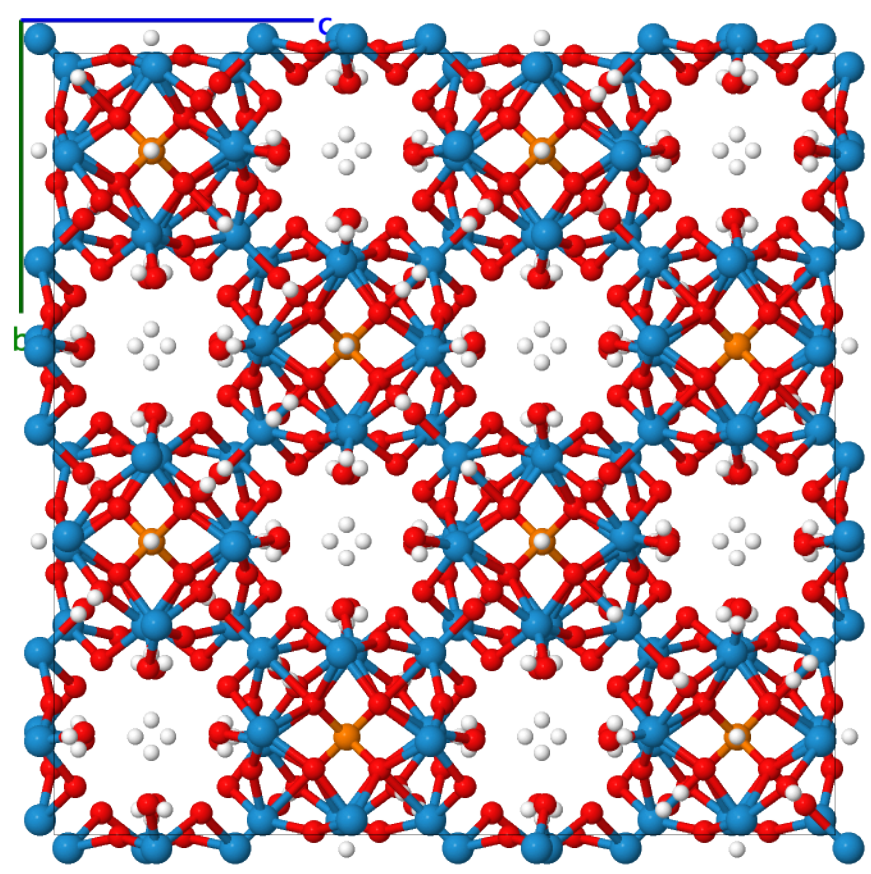

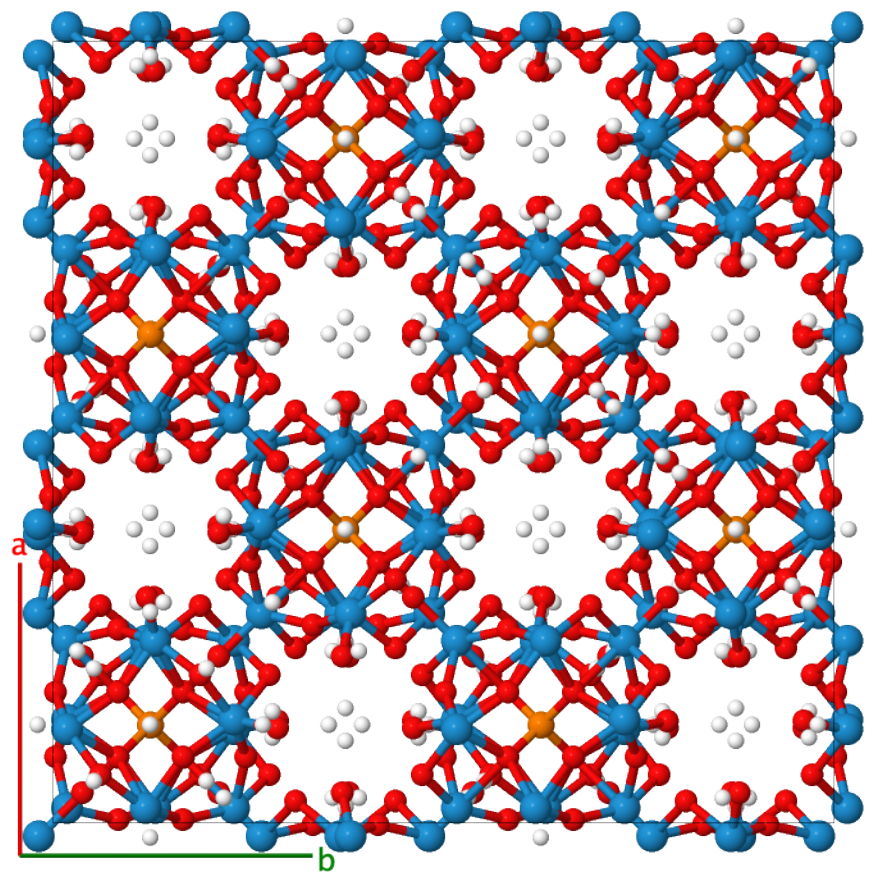

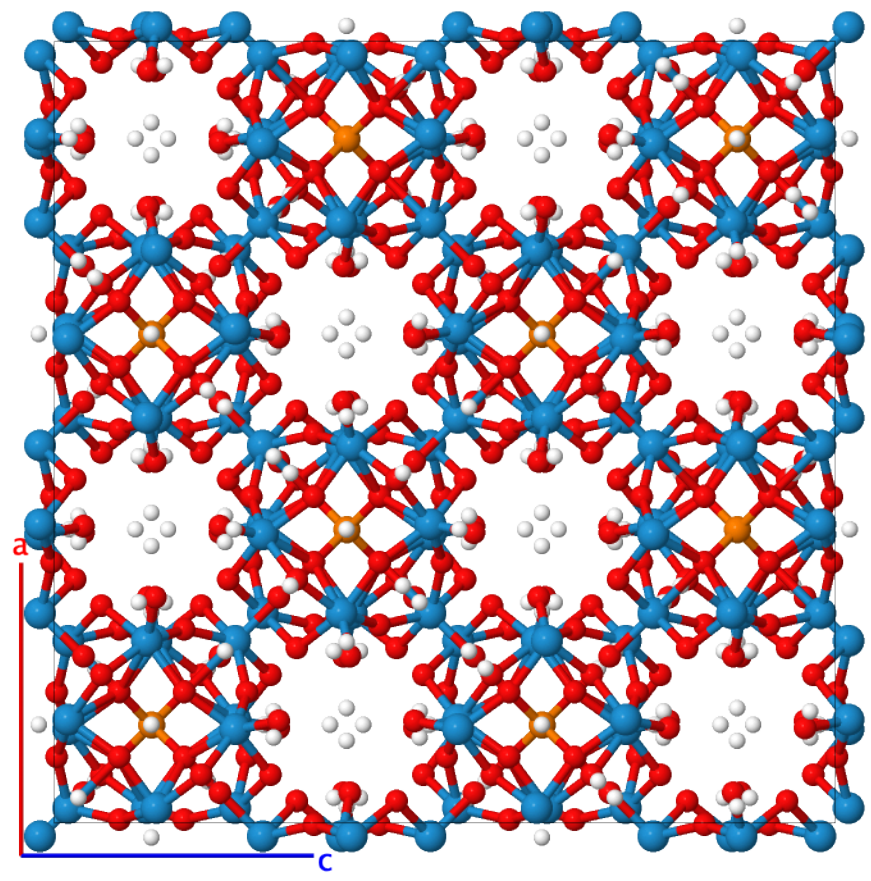

| | | |
|---|---|---|
| **Prototype** | : | $(H_2O)_{29}O_{40}PW_{12}$ |
| **AFLOW prototype label** | : | A29B40CD12_cF656_227_ae2fg_e3g_b_g |
| ***Strukturbericht* designation** | : | $H4_{21}$ |
| **Pearson symbol** | : | cF656 |
| **Space group number** | : | 227 |
| **Space group symbol** | : | $Fd\bar{3}m$ |
| **AFLOW prototype command** | : | `aflow --proto=A29B40CD12_cF656_227_ae2fg_e3g_b_g --params=`$a, x_3, x_4, x_5, x_6, x_7, z_7, x_8, z_8, x_9, z_9, x_{10}, z_{10}, x_{11}, z_{11}$ |

**Other compounds with this structure**

- $H_3PMo_{12}O_{40}\cdot 30H_2O$

- This compound is often colloquially called "PWA-29." On heating some water molecules will disassociate, leaving $H_3PW_{12}O_{40}\cdot 6H_2O$, $H_3PW_{12}O_{40}\cdot 5H_2O$ ($H4_{16}$), or $H_3PW_{12}O_{40}\cdot 3H_2O$.
- The three hydrogen atoms not formally associated with the water molecules are not located. Presumably they join with some water molecules to form $H_3O^+$ ions.

- Even the exact number and position of the water molecules is uncertain. (Clark, 1976), studying the related compound $H_3PMo_{12}O_{40}{\cdot}30H_2O$, states the the composition is approximately $30H_2O$, and that "only six of the water molecules occupy ordered sites."

---

**Face-centered Cubic primitive vectors:**

$$\mathbf{a}_1 = \frac{1}{2}\,a\,\hat{\mathbf{y}} + \frac{1}{2}\,a\,\hat{\mathbf{z}}$$
$$\mathbf{a}_2 = \frac{1}{2}\,a\,\hat{\mathbf{x}} + \frac{1}{2}\,a\,\hat{\mathbf{z}}$$
$$\mathbf{a}_3 = \frac{1}{2}\,a\,\hat{\mathbf{x}} + \frac{1}{2}\,a\,\hat{\mathbf{y}}$$

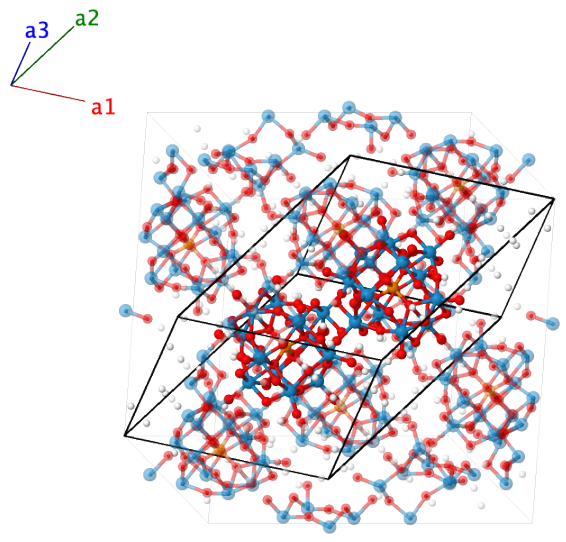

---

**Basis vectors:**

| | | Lattice Coordinates | | Cartesian Coordinates | Wyckoff Position | Atom Type |
|---|---|---|---|---|---|---|
| $\mathbf{B}_1$ | = | $\frac{1}{8}\,\mathbf{a}_1 + \frac{1}{8}\,\mathbf{a}_2 + \frac{1}{8}\,\mathbf{a}_3$ | = | $\frac{1}{8}a\,\hat{\mathbf{x}} + \frac{1}{8}a\,\hat{\mathbf{y}} + \frac{1}{8}a\,\hat{\mathbf{z}}$ | (8$a$) | $H_2O$ I |
| $\mathbf{B}_2$ | = | $\frac{7}{8}\,\mathbf{a}_1 + \frac{7}{8}\,\mathbf{a}_2 + \frac{7}{8}\,\mathbf{a}_3$ | = | $\frac{7}{8}a\,\hat{\mathbf{x}} + \frac{7}{8}a\,\hat{\mathbf{y}} + \frac{7}{8}a\,\hat{\mathbf{z}}$ | (8$a$) | $H_2O$ I |
| $\mathbf{B}_3$ | = | $\frac{3}{8}\,\mathbf{a}_1 + \frac{3}{8}\,\mathbf{a}_2 + \frac{3}{8}\,\mathbf{a}_3$ | = | $\frac{3}{8}a\,\hat{\mathbf{x}} + \frac{3}{8}a\,\hat{\mathbf{y}} + \frac{3}{8}a\,\hat{\mathbf{z}}$ | (8$b$) | P |
| $\mathbf{B}_4$ | = | $\frac{5}{8}\,\mathbf{a}_1 + \frac{5}{8}\,\mathbf{a}_2 + \frac{5}{8}\,\mathbf{a}_3$ | = | $\frac{5}{8}a\,\hat{\mathbf{x}} + \frac{5}{8}a\,\hat{\mathbf{y}} + \frac{5}{8}a\,\hat{\mathbf{z}}$ | (8$b$) | P |
| $\mathbf{B}_5$ | = | $x_3\,\mathbf{a}_1 + x_3\,\mathbf{a}_2 + x_3\,\mathbf{a}_3$ | = | $x_3a\,\hat{\mathbf{x}} + x_3a\,\hat{\mathbf{y}} + x_3a\,\hat{\mathbf{z}}$ | (32$e$) | $H_2O$ II |
| $\mathbf{B}_6$ | = | $x_3\,\mathbf{a}_1 + x_3\,\mathbf{a}_2 + \left(\frac{1}{2} - 3x_3\right)\,\mathbf{a}_3$ | = | $\left(\frac{1}{4} - x_3\right)a\,\hat{\mathbf{x}} + \left(\frac{1}{4} - x_3\right)a\,\hat{\mathbf{y}} + x_3a\,\hat{\mathbf{z}}$ | (32$e$) | $H_2O$ II |
| $\mathbf{B}_7$ | = | $x_3\,\mathbf{a}_1 + \left(\frac{1}{2} - 3x_3\right)\,\mathbf{a}_2 + x_3\,\mathbf{a}_3$ | = | $\left(\frac{1}{4} - x_3\right)a\,\hat{\mathbf{x}} + x_3a\,\hat{\mathbf{y}} + \left(\frac{1}{4} - x_3\right)a\,\hat{\mathbf{z}}$ | (32$e$) | $H_2O$ II |
| $\mathbf{B}_8$ | = | $\left(\frac{1}{2} - 3x_3\right)\,\mathbf{a}_1 + x_3\,\mathbf{a}_2 + x_3\,\mathbf{a}_3$ | = | $x_3a\,\hat{\mathbf{x}} + \left(\frac{1}{4} - x_3\right)a\,\hat{\mathbf{y}} + \left(\frac{1}{4} - x_3\right)a\,\hat{\mathbf{z}}$ | (32$e$) | $H_2O$ II |
| $\mathbf{B}_9$ | = | $-x_3\,\mathbf{a}_1 - x_3\,\mathbf{a}_2 + \left(\frac{1}{2} + 3x_3\right)\,\mathbf{a}_3$ | = | $\left(\frac{1}{4} + x_3\right)a\,\hat{\mathbf{x}} + \left(\frac{1}{4} + x_3\right)a\,\hat{\mathbf{y}} - x_3a\,\hat{\mathbf{z}}$ | (32$e$) | $H_2O$ II |
| $\mathbf{B}_{10}$ | = | $-x_3\,\mathbf{a}_1 - x_3\,\mathbf{a}_2 - x_3\,\mathbf{a}_3$ | = | $-x_3a\,\hat{\mathbf{x}} - x_3a\,\hat{\mathbf{y}} - x_3a\,\hat{\mathbf{z}}$ | (32$e$) | $H_2O$ II |
| $\mathbf{B}_{11}$ | = | $-x_3\,\mathbf{a}_1 + \left(\frac{1}{2} + 3x_3\right)\,\mathbf{a}_2 - x_3\,\mathbf{a}_3$ | = | $\left(\frac{1}{4} + x_3\right)a\,\hat{\mathbf{x}} - x_3a\,\hat{\mathbf{y}} + \left(\frac{1}{4} + x_3\right)a\,\hat{\mathbf{z}}$ | (32$e$) | $H_2O$ II |
| $\mathbf{B}_{12}$ | = | $\left(\frac{1}{2} + 3x_3\right)\,\mathbf{a}_1 - x_3\,\mathbf{a}_2 - x_3\,\mathbf{a}_3$ | = | $-x_3a\,\hat{\mathbf{x}} + \left(\frac{1}{4} + x_3\right)a\,\hat{\mathbf{y}} + \left(\frac{1}{4} + x_3\right)a\,\hat{\mathbf{z}}$ | (32$e$) | $H_2O$ II |
| $\mathbf{B}_{13}$ | = | $x_4\,\mathbf{a}_1 + x_4\,\mathbf{a}_2 + x_4\,\mathbf{a}_3$ | = | $x_4a\,\hat{\mathbf{x}} + x_4a\,\hat{\mathbf{y}} + x_4a\,\hat{\mathbf{z}}$ | (32$e$) | O I |
| $\mathbf{B}_{14}$ | = | $x_4\,\mathbf{a}_1 + x_4\,\mathbf{a}_2 + \left(\frac{1}{2} - 3x_4\right)\,\mathbf{a}_3$ | = | $\left(\frac{1}{4} - x_4\right)a\,\hat{\mathbf{x}} + \left(\frac{1}{4} - x_4\right)a\,\hat{\mathbf{y}} + x_4a\,\hat{\mathbf{z}}$ | (32$e$) | O I |
| $\mathbf{B}_{15}$ | = | $x_4\,\mathbf{a}_1 + \left(\frac{1}{2} - 3x_4\right)\,\mathbf{a}_2 + x_4\,\mathbf{a}_3$ | = | $\left(\frac{1}{4} - x_4\right)a\,\hat{\mathbf{x}} + x_4a\,\hat{\mathbf{y}} + \left(\frac{1}{4} - x_4\right)a\,\hat{\mathbf{z}}$ | (32$e$) | O I |
| $\mathbf{B}_{16}$ | = | $\left(\frac{1}{2} - 3x_4\right)\,\mathbf{a}_1 + x_4\,\mathbf{a}_2 + x_4\,\mathbf{a}_3$ | = | $x_4a\,\hat{\mathbf{x}} + \left(\frac{1}{4} - x_4\right)a\,\hat{\mathbf{y}} + \left(\frac{1}{4} - x_4\right)a\,\hat{\mathbf{z}}$ | (32$e$) | O I |
| $\mathbf{B}_{17}$ | = | $-x_4\,\mathbf{a}_1 - x_4\,\mathbf{a}_2 + \left(\frac{1}{2} + 3x_4\right)\,\mathbf{a}_3$ | = | $\left(\frac{1}{4} + x_4\right)a\,\hat{\mathbf{x}} + \left(\frac{1}{4} + x_4\right)a\,\hat{\mathbf{y}} - x_4a\,\hat{\mathbf{z}}$ | (32$e$) | O I |
| $\mathbf{B}_{18}$ | = | $-x_4\,\mathbf{a}_1 - x_4\,\mathbf{a}_2 - x_4\,\mathbf{a}_3$ | = | $-x_4a\,\hat{\mathbf{x}} - x_4a\,\hat{\mathbf{y}} - x_4a\,\hat{\mathbf{z}}$ | (32$e$) | O I |
| $\mathbf{B}_{19}$ | = | $-x_4\,\mathbf{a}_1 + \left(\frac{1}{2} + 3x_4\right)\,\mathbf{a}_2 - x_4\,\mathbf{a}_3$ | = | $\left(\frac{1}{4} + x_4\right)a\,\hat{\mathbf{x}} - x_4a\,\hat{\mathbf{y}} + \left(\frac{1}{4} + x_4\right)a\,\hat{\mathbf{z}}$ | (32$e$) | O I |
| $\mathbf{B}_{20}$ | = | $\left(\frac{1}{2} + 3x_4\right)\,\mathbf{a}_1 - x_4\,\mathbf{a}_2 - x_4\,\mathbf{a}_3$ | = | $-x_4a\,\hat{\mathbf{x}} + \left(\frac{1}{4} + x_4\right)a\,\hat{\mathbf{y}} + \left(\frac{1}{4} + x_4\right)a\,\hat{\mathbf{z}}$ | (32$e$) | O I |
| $\mathbf{B}_{21}$ | = | $\left(\frac{1}{4} - x_5\right)\,\mathbf{a}_1 + x_5\,\mathbf{a}_2 + x_5\,\mathbf{a}_3$ | = | $x_5a\,\hat{\mathbf{x}} + \frac{1}{8}a\,\hat{\mathbf{y}} + \frac{1}{8}a\,\hat{\mathbf{z}}$ | (48$f$) | $H_2O$ III |
| $\mathbf{B}_{22}$ | = | $x_5\,\mathbf{a}_1 + \left(\frac{1}{4} - x_5\right)\,\mathbf{a}_2 + \left(\frac{1}{4} - x_5\right)\,\mathbf{a}_3$ | = | $\left(\frac{1}{4} - x_5\right)a\,\hat{\mathbf{x}} + \frac{1}{8}a\,\hat{\mathbf{y}} + \frac{1}{8}a\,\hat{\mathbf{z}}$ | (48$f$) | $H_2O$ III |
| $\mathbf{B}_{23}$ | = | $x_5\,\mathbf{a}_1 + \left(\frac{1}{4} - x_5\right)\,\mathbf{a}_2 + x_5\,\mathbf{a}_3$ | = | $\frac{1}{8}a\,\hat{\mathbf{x}} + x_5a\,\hat{\mathbf{y}} + \frac{1}{8}a\,\hat{\mathbf{z}}$ | (48$f$) | $H_2O$ III |
| $\mathbf{B}_{24}$ | = | $\left(\frac{1}{4} - x_5\right)\,\mathbf{a}_1 + x_5\,\mathbf{a}_2 + \left(\frac{1}{4} - x_5\right)\,\mathbf{a}_3$ | = | $\frac{1}{8}a\,\hat{\mathbf{x}} + \left(\frac{1}{4} - x_5\right)a\,\hat{\mathbf{y}} + \frac{1}{8}a\,\hat{\mathbf{z}}$ | (48$f$) | $H_2O$ III |
| $\mathbf{B}_{25}$ | = | $x_5\,\mathbf{a}_1 + x_5\,\mathbf{a}_2 + \left(\frac{1}{4} - x_5\right)\,\mathbf{a}_3$ | = | $\frac{1}{8}a\,\hat{\mathbf{x}} + \frac{1}{8}a\,\hat{\mathbf{y}} + x_5a\,\hat{\mathbf{z}}$ | $(48f)$ | $H_2O$ III |
| $\mathbf{B}_{26}$ | = | $\left(\frac{1}{4} - x_5\right)\,\mathbf{a}_1 + \left(\frac{1}{4} - x_5\right)\,\mathbf{a}_2 + x_5\,\mathbf{a}_3$ | = | $\frac{1}{8}a\,\hat{\mathbf{x}} + \frac{1}{8}a\,\hat{\mathbf{y}} + \left(\frac{1}{4} - x_5\right)a\,\hat{\mathbf{z}}$ | $(48f)$ | $H_2O$ III |
| $\mathbf{B}_{27}$ | = | $\left(\frac{3}{4} + x_5\right)\,\mathbf{a}_1 - x_5\,\mathbf{a}_2 + \left(\frac{3}{4} + x_5\right)\,\mathbf{a}_3$ | = | $\frac{3}{8}a\,\hat{\mathbf{x}} + \left(\frac{3}{4} + x_5\right)a\,\hat{\mathbf{y}} + \frac{3}{8}a\,\hat{\mathbf{z}}$ | $(48f)$ | $H_2O$ III |
| $\mathbf{B}_{28}$ | = | $-x_5\,\mathbf{a}_1 + \left(\frac{3}{4} + x_5\right)\,\mathbf{a}_2 - x_5\,\mathbf{a}_3$ | = | $\frac{3}{8}a\,\hat{\mathbf{x}} - x_5a\,\hat{\mathbf{y}} + \frac{3}{8}a\,\hat{\mathbf{z}}$ | $(48f)$ | $H_2O$ III |
| $\mathbf{B}_{29}$ | = | $-x_5\,\mathbf{a}_1 + \left(\frac{3}{4} + x_5\right)\,\mathbf{a}_2 + \left(\frac{3}{4} + x_5\right)\,\mathbf{a}_3$ | = | $\left(\frac{3}{4} + x_5\right)a\,\hat{\mathbf{x}} + \frac{3}{8}a\,\hat{\mathbf{y}} + \frac{3}{8}a\,\hat{\mathbf{z}}$ | $(48f)$ | $H_2O$ III |
| $\mathbf{B}_{30}$ | = | $\left(\frac{3}{4} + x_5\right)\,\mathbf{a}_1 - x_5\,\mathbf{a}_2 - x_5\,\mathbf{a}_3$ | = | $-x_5a\,\hat{\mathbf{x}} + \frac{3}{8}a\,\hat{\mathbf{y}} + \frac{3}{8}a\,\hat{\mathbf{z}}$ | $(48f)$ | $H_2O$ III |
| $\mathbf{B}_{31}$ | = | $-x_5\,\mathbf{a}_1 - x_5\,\mathbf{a}_2 + \left(\frac{3}{4} + x_5\right)\,\mathbf{a}_3$ | = | $\frac{3}{8}a\,\hat{\mathbf{x}} + \frac{3}{8}a\,\hat{\mathbf{y}} - x_5a\,\hat{\mathbf{z}}$ | $(48f)$ | $H_2O$ III |
| $\mathbf{B}_{32}$ | = | $\left(\frac{3}{4} + x_5\right)\,\mathbf{a}_1 + \left(\frac{3}{4} + x_5\right)\,\mathbf{a}_2 - x_5\,\mathbf{a}_3$ | = | $\frac{3}{8}a\,\hat{\mathbf{x}} + \frac{3}{8}a\,\hat{\mathbf{y}} + \left(\frac{3}{4} + x_5\right)a\,\hat{\mathbf{z}}$ | $(48f)$ | $H_2O$ III |
| $\mathbf{B}_{33}$ | = | $\left(\frac{1}{4} - x_6\right)\,\mathbf{a}_1 + x_6\,\mathbf{a}_2 + x_6\,\mathbf{a}_3$ | = | $x_6a\,\hat{\mathbf{x}} + \frac{1}{8}a\,\hat{\mathbf{y}} + \frac{1}{8}a\,\hat{\mathbf{z}}$ | $(48f)$ | $H_2O$ IV |
| $\mathbf{B}_{34}$ | = | $x_6\,\mathbf{a}_1 + \left(\frac{1}{4} - x_6\right)\,\mathbf{a}_2 + \left(\frac{1}{4} - x_6\right)\,\mathbf{a}_3$ | = | $\left(\frac{1}{4} - x_6\right)a\,\hat{\mathbf{x}} + \frac{1}{8}a\,\hat{\mathbf{y}} + \frac{1}{8}a\,\hat{\mathbf{z}}$ | $(48f)$ | $H_2O$ IV |
| $\mathbf{B}_{35}$ | = | $x_6\,\mathbf{a}_1 + \left(\frac{1}{4} - x_6\right)\,\mathbf{a}_2 + x_6\,\mathbf{a}_3$ | = | $\frac{1}{8}a\,\hat{\mathbf{x}} + x_6a\,\hat{\mathbf{y}} + \frac{1}{8}a\,\hat{\mathbf{z}}$ | $(48f)$ | $H_2O$ IV |
| $\mathbf{B}_{36}$ | = | $\left(\frac{1}{4} - x_6\right)\,\mathbf{a}_1 + x_6\,\mathbf{a}_2 + \left(\frac{1}{4} - x_6\right)\,\mathbf{a}_3$ | = | $\frac{1}{8}a\,\hat{\mathbf{x}} + \left(\frac{1}{4} - x_6\right)a\,\hat{\mathbf{y}} + \frac{1}{8}a\,\hat{\mathbf{z}}$ | $(48f)$ | $H_2O$ IV |
| $\mathbf{B}_{37}$ | = | $x_6\,\mathbf{a}_1 + x_6\,\mathbf{a}_2 + \left(\frac{1}{4} - x_6\right)\,\mathbf{a}_3$ | = | $\frac{1}{8}a\,\hat{\mathbf{x}} + \frac{1}{8}a\,\hat{\mathbf{y}} + x_6a\,\hat{\mathbf{z}}$ | $(48f)$ | $H_2O$ IV |
| $\mathbf{B}_{38}$ | = | $\left(\frac{1}{4} - x_6\right)\,\mathbf{a}_1 + \left(\frac{1}{4} - x_6\right)\,\mathbf{a}_2 + x_6\,\mathbf{a}_3$ | = | $\frac{1}{8}a\,\hat{\mathbf{x}} + \frac{1}{8}a\,\hat{\mathbf{y}} + \left(\frac{1}{4} - x_6\right)a\,\hat{\mathbf{z}}$ | $(48f)$ | $H_2O$ IV |
| $\mathbf{B}_{39}$ | = | $\left(\frac{3}{4} + x_6\right)\,\mathbf{a}_1 - x_6\,\mathbf{a}_2 + \left(\frac{3}{4} + x_6\right)\,\mathbf{a}_3$ | = | $\frac{3}{8}a\,\hat{\mathbf{x}} + \left(\frac{3}{4} + x_6\right)a\,\hat{\mathbf{y}} + \frac{3}{8}a\,\hat{\mathbf{z}}$ | $(48f)$ | $H_2O$ IV |
| $\mathbf{B}_{40}$ | = | $-x_6\,\mathbf{a}_1 + \left(\frac{3}{4} + x_6\right)\,\mathbf{a}_2 - x_6\,\mathbf{a}_3$ | = | $\frac{3}{8}a\,\hat{\mathbf{x}} - x_6a\,\hat{\mathbf{y}} + \frac{3}{8}a\,\hat{\mathbf{z}}$ | $(48f)$ | $H_2O$ IV |
| $\mathbf{B}_{41}$ | = | $-x_6\,\mathbf{a}_1 + \left(\frac{3}{4} + x_6\right)\,\mathbf{a}_2 + \left(\frac{3}{4} + x_6\right)\,\mathbf{a}_3$ | = | $\left(\frac{3}{4} + x_6\right)a\,\hat{\mathbf{x}} + \frac{3}{8}a\,\hat{\mathbf{y}} + \frac{3}{8}a\,\hat{\mathbf{z}}$ | $(48f)$ | $H_2O$ IV |
| $\mathbf{B}_{42}$ | = | $\left(\frac{3}{4} + x_6\right)\,\mathbf{a}_1 - x_6\,\mathbf{a}_2 - x_6\,\mathbf{a}_3$ | = | $-x_6a\,\hat{\mathbf{x}} + \frac{3}{8}a\,\hat{\mathbf{y}} + \frac{3}{8}a\,\hat{\mathbf{z}}$ | $(48f)$ | $H_2O$ IV |
| $\mathbf{B}_{43}$ | = | $-x_6\,\mathbf{a}_1 - x_6\,\mathbf{a}_2 + \left(\frac{3}{4} + x_6\right)\,\mathbf{a}_3$ | = | $\frac{3}{8}a\,\hat{\mathbf{x}} + \frac{3}{8}a\,\hat{\mathbf{y}} - x_6a\,\hat{\mathbf{z}}$ | $(48f)$ | $H_2O$ IV |
| $\mathbf{B}_{44}$ | = | $\left(\frac{3}{4} + x_6\right)\,\mathbf{a}_1 + \left(\frac{3}{4} + x_6\right)\,\mathbf{a}_2 - x_6\,\mathbf{a}_3$ | = | $\frac{3}{8}a\,\hat{\mathbf{x}} + \frac{3}{8}a\,\hat{\mathbf{y}} + \left(\frac{3}{4} + x_6\right)a\,\hat{\mathbf{z}}$ | $(48f)$ | $H_2O$ IV |
| $\mathbf{B}_{45}$ | = | $z_7\,\mathbf{a}_1 + z_7\,\mathbf{a}_2 + (2x_7 - z_7)\,\mathbf{a}_3$ | = | $x_7a\,\hat{\mathbf{x}} + x_7a\,\hat{\mathbf{y}} + z_7a\,\hat{\mathbf{z}}$ | $(96g)$ | $H_2O$ V |
| $\mathbf{B}_{46}$ | = | $z_7\,\mathbf{a}_1 + z_7\,\mathbf{a}_2 + \left(\frac{1}{2} - 2x_7 - z_7\right)\,\mathbf{a}_3$ | = | $\left(\frac{1}{4} - x_7\right)a\,\hat{\mathbf{x}} + \left(\frac{1}{4} - x_7\right)a\,\hat{\mathbf{y}} + z_7a\,\hat{\mathbf{z}}$ | $(96g)$ | $H_2O$ V |
| $\mathbf{B}_{47}$ | = | $(2x_7 - z_7)\,\mathbf{a}_1 + \left(\frac{1}{2} - 2x_7 - z_7\right)\,\mathbf{a}_2 + z_7\,\mathbf{a}_3$ | = | $\left(\frac{1}{4} - x_7\right)a\,\hat{\mathbf{x}} + x_7a\,\hat{\mathbf{y}} + \left(\frac{1}{4} - z_7\right)a\,\hat{\mathbf{z}}$ | $(96g)$ | $H_2O$ V |
| $\mathbf{B}_{48}$ | = | $\left(\frac{1}{2} - 2x_7 - z_7\right)\,\mathbf{a}_1 + (2x_7 - z_7)\,\mathbf{a}_2 + z_7\,\mathbf{a}_3$ | = | $x_7a\,\hat{\mathbf{x}} + \left(\frac{1}{4} - x_7\right)a\,\hat{\mathbf{y}} + \left(\frac{1}{4} - z_7\right)a\,\hat{\mathbf{z}}$ | $(96g)$ | $H_2O$ V |
| $\mathbf{B}_{49}$ | = | $(2x_7 - z_7)\,\mathbf{a}_1 + z_7\,\mathbf{a}_2 + z_7\,\mathbf{a}_3$ | = | $z_7a\,\hat{\mathbf{x}} + x_7a\,\hat{\mathbf{y}} + x_7a\,\hat{\mathbf{z}}$ | $(96g)$ | $H_2O$ V |
| $\mathbf{B}_{50}$ | = | $\left(\frac{1}{2} - 2x_7 - z_7\right)\,\mathbf{a}_1 + z_7\,\mathbf{a}_2 + z_7\,\mathbf{a}_3$ | = | $z_7a\,\hat{\mathbf{x}} + \left(\frac{1}{4} - x_7\right)a\,\hat{\mathbf{y}} + \left(\frac{1}{4} - x_7\right)a\,\hat{\mathbf{z}}$ | $(96g)$ | $H_2O$ V |
| $\mathbf{B}_{51}$ | = | $z_7\,\mathbf{a}_1 + (2x_7 - z_7)\,\mathbf{a}_2 + \left(\frac{1}{2} - 2x_7 - z_7\right)\,\mathbf{a}_3$ | = | $\left(\frac{1}{4} - z_7\right)a\,\hat{\mathbf{x}} + \left(\frac{1}{4} - x_7\right)a\,\hat{\mathbf{y}} + x_7a\,\hat{\mathbf{z}}$ | $(96g)$ | $H_2O$ V |
| $\mathbf{B}_{52}$ | = | $z_7\,\mathbf{a}_1 + \left(\frac{1}{2} - 2x_7 - z_7\right)\,\mathbf{a}_2 + (2x_7 - z_7)\,\mathbf{a}_3$ | = | $\left(\frac{1}{4} - z_7\right)a\,\hat{\mathbf{x}} + x_7a\,\hat{\mathbf{y}} + \left(\frac{1}{4} - x_7\right)a\,\hat{\mathbf{z}}$ | $(96g)$ | $H_2O$ V |
| $\mathbf{B}_{53}$ | = | $z_7\,\mathbf{a}_1 + (2x_7 - z_7)\,\mathbf{a}_2 + z_7\,\mathbf{a}_3$ | = | $x_7a\,\hat{\mathbf{x}} + z_7a\,\hat{\mathbf{y}} + x_7a\,\hat{\mathbf{z}}$ | $(96g)$ | $H_2O$ V |
| $\mathbf{B}_{54}$ | = | $z_7\,\mathbf{a}_1 + \left(\frac{1}{2} - 2x_7 - z_7\right)\,\mathbf{a}_2 + z_7\,\mathbf{a}_3$ | = | $\left(\frac{1}{4} - x_7\right)a\,\hat{\mathbf{x}} + z_7a\,\hat{\mathbf{y}} + \left(\frac{1}{4} - x_7\right)a\,\hat{\mathbf{z}}$ | $(96g)$ | $H_2O$ V |
| $\mathbf{B}_{55}$ | = | $\left(\frac{1}{2} - 2x_7 - z_7\right)\,\mathbf{a}_1 + z_7\,\mathbf{a}_2 + (2x_7 - z_7)\,\mathbf{a}_3$ | = | $x_7a\,\hat{\mathbf{x}} + \left(\frac{1}{4} - z_7\right)a\,\hat{\mathbf{y}} + \left(\frac{1}{4} - x_7\right)a\,\hat{\mathbf{z}}$ | $(96g)$ | $H_2O$ V |
| $\mathbf{B}_{56}$ | = | $(2x_7 - z_7)\,\mathbf{a}_1 + z_7\,\mathbf{a}_2 + \left(\frac{1}{2} - 2x_7 - z_7\right)\,\mathbf{a}_3$ | = | $\left(\frac{1}{4} - x_7\right)a\,\hat{\mathbf{x}} + \left(\frac{1}{4} - z_7\right)a\,\hat{\mathbf{y}} + x_7a\,\hat{\mathbf{z}}$ | $(96g)$ | $H_2O$ V |
| $\mathbf{B}_{57}$ | = | $-z_7\,\mathbf{a}_1 - z_7\,\mathbf{a}_2 + \left(\frac{1}{2} + 2x_7 + z_7\right)\,\mathbf{a}_3$ | = | $\left(\frac{1}{4} + x_7\right)a\,\hat{\mathbf{x}} + \left(\frac{1}{4} + x_7\right)a\,\hat{\mathbf{y}} - z_7a\,\hat{\mathbf{z}}$ | $(96g)$ | $H_2O$ V |
| $\mathbf{B}_{58}$ | = | $-z_7\,\mathbf{a}_1 - z_7\,\mathbf{a}_2 + (-2x_7 + z_7)\,\mathbf{a}_3$ | = | $-x_7a\,\mathbf{\hat{x}} - x_7a\,\mathbf{\hat{y}} - z_7a\,\mathbf{\hat{z}}$ | (96g) | $H_2O$ V |
| $\mathbf{B}_{59}$ | = | $(-2x_7 + z_7)\,\mathbf{a}_1 + \left(\frac{1}{2} + 2x_7 + z_7\right)\mathbf{a}_2 - z_7\,\mathbf{a}_3$ | = | $\left(\frac{1}{4} + x_7\right)a\,\mathbf{\hat{x}} - x_7a\,\mathbf{\hat{y}} + \left(\frac{1}{4} + z_7\right)a\,\mathbf{\hat{z}}$ | (96g) | $H_2O$ V |
| $\mathbf{B}_{60}$ | = | $\left(\frac{1}{2} + 2x_7 + z_7\right)\mathbf{a}_1 + (-2x_7 + z_7)\,\mathbf{a}_2 - z_7\,\mathbf{a}_3$ | = | $-x_7a\,\mathbf{\hat{x}} + \left(\frac{1}{4} + x_7\right)a\,\mathbf{\hat{y}} + \left(\frac{1}{4} + z_7\right)a\,\mathbf{\hat{z}}$ | (96g) | $H_2O$ V |
| $\mathbf{B}_{61}$ | = | $(-2x_7 + z_7)\,\mathbf{a}_1 - z_7\,\mathbf{a}_2 + \left(\frac{1}{2} + 2x_7 + z_7\right)\mathbf{a}_3$ | = | $\left(\frac{1}{4} + x_7\right)a\,\mathbf{\hat{x}} + \left(\frac{1}{4} + z_7\right)a\,\mathbf{\hat{y}} - x_7a\,\mathbf{\hat{z}}$ | (96g) | $H_2O$ V |
| $\mathbf{B}_{62}$ | = | $\left(\frac{1}{2} + 2x_7 + z_7\right)\mathbf{a}_1 - z_7\,\mathbf{a}_2 + (-2x_7 + z_7)\,\mathbf{a}_3$ | = | $-x_7a\,\mathbf{\hat{x}} + \left(\frac{1}{4} + z_7\right)a\,\mathbf{\hat{y}} + \left(\frac{1}{4} + x_7\right)a\,\mathbf{\hat{z}}$ | (96g) | $H_2O$ V |
| $\mathbf{B}_{63}$ | = | $-z_7\,\mathbf{a}_1 + (-2x_7 + z_7)\,\mathbf{a}_2 - z_7\,\mathbf{a}_3$ | = | $-x_7a\,\mathbf{\hat{x}} - z_7a\,\mathbf{\hat{y}} - x_7a\,\mathbf{\hat{z}}$ | (96g) | $H_2O$ V |
| $\mathbf{B}_{64}$ | = | $-z_7\,\mathbf{a}_1 + \left(\frac{1}{2} + 2x_7 + z_7\right)\mathbf{a}_2 - z_7\,\mathbf{a}_3$ | = | $\left(\frac{1}{4} + x_7\right)a\,\mathbf{\hat{x}} - z_7a\,\mathbf{\hat{y}} + \left(\frac{1}{4} + x_7\right)a\,\mathbf{\hat{z}}$ | (96g) | $H_2O$ V |
| $\mathbf{B}_{65}$ | = | $-z_7\,\mathbf{a}_1 + (-2x_7 + z_7)\,\mathbf{a}_2 + \left(\frac{1}{2} + 2x_7 + z_7\right)\mathbf{a}_3$ | = | $\left(\frac{1}{4} + z_7\right)a\,\mathbf{\hat{x}} + \left(\frac{1}{4} + x_7\right)a\,\mathbf{\hat{y}} - x_7a\,\mathbf{\hat{z}}$ | (96g) | $H_2O$ V |
| $\mathbf{B}_{66}$ | = | $-z_7\,\mathbf{a}_1 + \left(\frac{1}{2} + 2x_7 + z_7\right)\mathbf{a}_2 + (-2x_7 + z_7)\,\mathbf{a}_3$ | = | $\left(\frac{1}{4} + z_7\right)a\,\mathbf{\hat{x}} - x_7a\,\mathbf{\hat{y}} + \left(\frac{1}{4} + x_7\right)a\,\mathbf{\hat{z}}$ | (96g) | $H_2O$ V |
| $\mathbf{B}_{67}$ | = | $\left(\frac{1}{2} + 2x_7 + z_7\right)\mathbf{a}_1 - z_7\,\mathbf{a}_2 - z_7\,\mathbf{a}_3$ | = | $-z_7a\,\mathbf{\hat{x}} + \left(\frac{1}{4} + x_7\right)a\,\mathbf{\hat{y}} + \left(\frac{1}{4} + x_7\right)a\,\mathbf{\hat{z}}$ | (96g) | $H_2O$ V |
| $\mathbf{B}_{68}$ | = | $(-2x_7 + z_7)\,\mathbf{a}_1 - z_7\,\mathbf{a}_2 - z_7\,\mathbf{a}_3$ | = | $-z_7a\,\mathbf{\hat{x}} - x_7a\,\mathbf{\hat{y}} - x_7a\,\mathbf{\hat{z}}$ | (96g) | $H_2O$ V |
| $\mathbf{B}_{69}$ | = | $z_8\,\mathbf{a}_1 + z_8\,\mathbf{a}_2 + (2x_8 - z_8)\,\mathbf{a}_3$ | = | $x_8a\,\mathbf{\hat{x}} + x_8a\,\mathbf{\hat{y}} + z_8a\,\mathbf{\hat{z}}$ | (96g) | O II |
| $\mathbf{B}_{70}$ | = | $z_8\,\mathbf{a}_1 + z_8\,\mathbf{a}_2 + \left(\frac{1}{2} - 2x_8 - z_8\right)\mathbf{a}_3$ | = | $\left(\frac{1}{4} - x_8\right)a\,\mathbf{\hat{x}} + \left(\frac{1}{4} - x_8\right)a\,\mathbf{\hat{y}} + z_8a\,\mathbf{\hat{z}}$ | (96g) | O II |
| $\mathbf{B}_{71}$ | = | $(2x_8 - z_8)\,\mathbf{a}_1 + \left(\frac{1}{2} - 2x_8 - z_8\right)\mathbf{a}_2 + z_8\,\mathbf{a}_3$ | = | $\left(\frac{1}{4} - x_8\right)a\,\mathbf{\hat{x}} + x_8a\,\mathbf{\hat{y}} + \left(\frac{1}{4} - z_8\right)a\,\mathbf{\hat{z}}$ | (96g) | O II |
| $\mathbf{B}_{72}$ | = | $\left(\frac{1}{2} - 2x_8 - z_8\right)\mathbf{a}_1 + (2x_8 - z_8)\,\mathbf{a}_2 + z_8\,\mathbf{a}_3$ | = | $x_8a\,\mathbf{\hat{x}} + \left(\frac{1}{4} - x_8\right)a\,\mathbf{\hat{y}} + \left(\frac{1}{4} - z_8\right)a\,\mathbf{\hat{z}}$ | (96g) | O II |
| $\mathbf{B}_{73}$ | = | $(2x_8 - z_8)\,\mathbf{a}_1 + z_8\,\mathbf{a}_2 + z_8\,\mathbf{a}_3$ | = | $z_8a\,\mathbf{\hat{x}} + x_8a\,\mathbf{\hat{y}} + x_8a\,\mathbf{\hat{z}}$ | (96g) | O II |
| $\mathbf{B}_{74}$ | = | $\left(\frac{1}{2} - 2x_8 - z_8\right)\mathbf{a}_1 + z_8\,\mathbf{a}_2 + z_8\,\mathbf{a}_3$ | = | $z_8a\,\mathbf{\hat{x}} + \left(\frac{1}{4} - x_8\right)a\,\mathbf{\hat{y}} + \left(\frac{1}{4} - x_8\right)a\,\mathbf{\hat{z}}$ | (96g) | O II |
| $\mathbf{B}_{75}$ | = | $z_8\,\mathbf{a}_1 + (2x_8 - z_8)\,\mathbf{a}_2 + \left(\frac{1}{2} - 2x_8 - z_8\right)\mathbf{a}_3$ | = | $\left(\frac{1}{4} - z_8\right)a\,\mathbf{\hat{x}} + \left(\frac{1}{4} - x_8\right)a\,\mathbf{\hat{y}} + x_8a\,\mathbf{\hat{z}}$ | (96g) | O II |
| $\mathbf{B}_{76}$ | = | $z_8\,\mathbf{a}_1 + \left(\frac{1}{2} - 2x_8 - z_8\right)\mathbf{a}_2 + (2x_8 - z_8)\,\mathbf{a}_3$ | = | $\left(\frac{1}{4} - z_8\right)a\,\mathbf{\hat{x}} + x_8a\,\mathbf{\hat{y}} + \left(\frac{1}{4} - x_8\right)a\,\mathbf{\hat{z}}$ | (96g) | O II |
| $\mathbf{B}_{77}$ | = | $z_8\,\mathbf{a}_1 + (2x_8 - z_8)\,\mathbf{a}_2 + z_8\,\mathbf{a}_3$ | = | $x_8a\,\mathbf{\hat{x}} + z_8a\,\mathbf{\hat{y}} + x_8a\,\mathbf{\hat{z}}$ | (96g) | O II |
| $\mathbf{B}_{78}$ | = | $z_8\,\mathbf{a}_1 + \left(\frac{1}{2} - 2x_8 - z_8\right)\mathbf{a}_2 + z_8\,\mathbf{a}_3$ | = | $\left(\frac{1}{4} - x_8\right)a\,\mathbf{\hat{x}} + z_8a\,\mathbf{\hat{y}} + \left(\frac{1}{4} - x_8\right)a\,\mathbf{\hat{z}}$ | (96g) | O II |
| $\mathbf{B}_{79}$ | = | $\left(\frac{1}{2} - 2x_8 - z_8\right)\mathbf{a}_1 + z_8\,\mathbf{a}_2 + (2x_8 - z_8)\,\mathbf{a}_3$ | = | $x_8a\,\mathbf{\hat{x}} + \left(\frac{1}{4} - z_8\right)a\,\mathbf{\hat{y}} + \left(\frac{1}{4} - x_8\right)a\,\mathbf{\hat{z}}$ | (96g) | O II |
| $\mathbf{B}_{80}$ | = | $(2x_8 - z_8)\,\mathbf{a}_1 + z_8\,\mathbf{a}_2 + \left(\frac{1}{2} - 2x_8 - z_8\right)\mathbf{a}_3$ | = | $\left(\frac{1}{4} - x_8\right)a\,\mathbf{\hat{x}} + \left(\frac{1}{4} - z_8\right)a\,\mathbf{\hat{y}} + x_8a\,\mathbf{\hat{z}}$ | (96g) | O II |
| $\mathbf{B}_{81}$ | = | $-z_8\,\mathbf{a}_1 - z_8\,\mathbf{a}_2 + \left(\frac{1}{2} + 2x_8 + z_8\right)\mathbf{a}_3$ | = | $\left(\frac{1}{4} + x_8\right)a\,\mathbf{\hat{x}} + \left(\frac{1}{4} + x_8\right)a\,\mathbf{\hat{y}} - z_8a\,\mathbf{\hat{z}}$ | (96g) | O II |
| $\mathbf{B}_{82}$ | = | $-z_8\,\mathbf{a}_1 - z_8\,\mathbf{a}_2 + (-2x_8 + z_8)\,\mathbf{a}_3$ | = | $-x_8a\,\mathbf{\hat{x}} - x_8a\,\mathbf{\hat{y}} - z_8a\,\mathbf{\hat{z}}$ | (96g) | O II |
| $\mathbf{B}_{83}$ | = | $(-2x_8 + z_8)\,\mathbf{a}_1 + \left(\frac{1}{2} + 2x_8 + z_8\right)\mathbf{a}_2 - z_8\,\mathbf{a}_3$ | = | $\left(\frac{1}{4} + x_8\right)a\,\mathbf{\hat{x}} - x_8a\,\mathbf{\hat{y}} + \left(\frac{1}{4} + z_8\right)a\,\mathbf{\hat{z}}$ | (96g) | O II |
| $\mathbf{B}_{84}$ | = | $\left(\frac{1}{2} + 2x_8 + z_8\right)\mathbf{a}_1 + (-2x_8 + z_8)\,\mathbf{a}_2 - z_8\,\mathbf{a}_3$ | = | $-x_8a\,\mathbf{\hat{x}} + \left(\frac{1}{4} + x_8\right)a\,\mathbf{\hat{y}} + \left(\frac{1}{4} + z_8\right)a\,\mathbf{\hat{z}}$ | (96g) | O II |
| $\mathbf{B}_{85}$ | = | $(-2x_8 + z_8)\,\mathbf{a}_1 - z_8\,\mathbf{a}_2 + \left(\frac{1}{2} + 2x_8 + z_8\right)\mathbf{a}_3$ | = | $\left(\frac{1}{4} + x_8\right)a\,\mathbf{\hat{x}} + \left(\frac{1}{4} + z_8\right)a\,\mathbf{\hat{y}} - x_8a\,\mathbf{\hat{z}}$ | (96g) | O II |
| $\mathbf{B}_{86}$ | = | $\left(\frac{1}{2} + 2x_8 + z_8\right) \mathbf{a}_1 - z_8 \mathbf{a}_2 + (-2x_8 + z_8) \mathbf{a}_3$ | = | $-x_8 a \hat{\mathbf{x}} + \left(\frac{1}{4} + z_8\right) a \hat{\mathbf{y}} + \left(\frac{1}{4} + x_8\right) a \hat{\mathbf{z}}$ | (96$g$) | O II |
| $\mathbf{B}_{87}$ | = | $-z_8 \mathbf{a}_1 + (-2x_8 + z_8) \mathbf{a}_2 - z_8 \mathbf{a}_3$ | = | $-x_8 a \hat{\mathbf{x}} - z_8 a \hat{\mathbf{y}} - x_8 a \hat{\mathbf{z}}$ | (96$g$) | O II |
| $\mathbf{B}_{88}$ | = | $-z_8 \mathbf{a}_1 + \left(\frac{1}{2} + 2x_8 + z_8\right) \mathbf{a}_2 - z_8 \mathbf{a}_3$ | = | $\left(\frac{1}{4} + x_8\right) a \hat{\mathbf{x}} - z_8 a \hat{\mathbf{y}} + \left(\frac{1}{4} + x_8\right) a \hat{\mathbf{z}}$ | (96$g$) | O II |
| $\mathbf{B}_{89}$ | = | $-z_8 \mathbf{a}_1 + (-2x_8 + z_8) \mathbf{a}_2 + \left(\frac{1}{2} + 2x_8 + z_8\right) \mathbf{a}_3$ | = | $\left(\frac{1}{4} + z_8\right) a \hat{\mathbf{x}} + \left(\frac{1}{4} + x_8\right) a \hat{\mathbf{y}} - x_8 a \hat{\mathbf{z}}$ | (96$g$) | O II |
| $\mathbf{B}_{90}$ | = | $-z_8 \mathbf{a}_1 + \left(\frac{1}{2} + 2x_8 + z_8\right) \mathbf{a}_2 + (-2x_8 + z_8) \mathbf{a}_3$ | = | $\left(\frac{1}{4} + z_8\right) a \hat{\mathbf{x}} - x_8 a \hat{\mathbf{y}} + \left(\frac{1}{4} + x_8\right) a \hat{\mathbf{z}}$ | (96$g$) | O II |
| $\mathbf{B}_{91}$ | = | $\left(\frac{1}{2} + 2x_8 + z_8\right) \mathbf{a}_1 - z_8 \mathbf{a}_2 - z_8 \mathbf{a}_3$ | = | $-z_8 a \hat{\mathbf{x}} + \left(\frac{1}{4} + x_8\right) a \hat{\mathbf{y}} + \left(\frac{1}{4} + x_8\right) a \hat{\mathbf{z}}$ | (96$g$) | O II |
| $\mathbf{B}_{92}$ | = | $(-2x_8 + z_8) \mathbf{a}_1 - z_8 \mathbf{a}_2 - z_8 \mathbf{a}_3$ | = | $-z_8 a \hat{\mathbf{x}} - x_8 a \hat{\mathbf{y}} - x_8 a \hat{\mathbf{z}}$ | (96$g$) | O II |
| $\mathbf{B}_{93}$ | = | $z_9 \mathbf{a}_1 + z_9 \mathbf{a}_2 + (2x_9 - z_9) \mathbf{a}_3$ | = | $x_9 a \hat{\mathbf{x}} + x_9 a \hat{\mathbf{y}} + z_9 a \hat{\mathbf{z}}$ | (96$g$) | O III |
| $\mathbf{B}_{94}$ | = | $z_9 \mathbf{a}_1 + z_9 \mathbf{a}_2 + \left(\frac{1}{2} - 2x_9 - z_9\right) \mathbf{a}_3$ | = | $\left(\frac{1}{4} - x_9\right) a \hat{\mathbf{x}} + \left(\frac{1}{4} - x_9\right) a \hat{\mathbf{y}} + z_9 a \hat{\mathbf{z}}$ | (96$g$) | O III |
| $\mathbf{B}_{95}$ | = | $(2x_9 - z_9) \mathbf{a}_1 + \left(\frac{1}{2} - 2x_9 - z_9\right) \mathbf{a}_2 + z_9 \mathbf{a}_3$ | = | $\left(\frac{1}{4} - x_9\right) a \hat{\mathbf{x}} + x_9 a \hat{\mathbf{y}} + \left(\frac{1}{4} - z_9\right) a \hat{\mathbf{z}}$ | (96$g$) | O III |
| $\mathbf{B}_{96}$ | = | $\left(\frac{1}{2} - 2x_9 - z_9\right) \mathbf{a}_1 + (2x_9 - z_9) \mathbf{a}_2 + z_9 \mathbf{a}_3$ | = | $x_9 a \hat{\mathbf{x}} + \left(\frac{1}{4} - x_9\right) a \hat{\mathbf{y}} + \left(\frac{1}{4} - z_9\right) a \hat{\mathbf{z}}$ | (96$g$) | O III |
| $\mathbf{B}_{97}$ | = | $(2x_9 - z_9) \mathbf{a}_1 + z_9 \mathbf{a}_2 + z_9 \mathbf{a}_3$ | = | $z_9 a \hat{\mathbf{x}} + x_9 a \hat{\mathbf{y}} + x_9 a \hat{\mathbf{z}}$ | (96$g$) | O III |
| $\mathbf{B}_{98}$ | = | $\left(\frac{1}{2} - 2x_9 - z_9\right) \mathbf{a}_1 + z_9 \mathbf{a}_2 + z_9 \mathbf{a}_3$ | = | $z_9 a \hat{\mathbf{x}} + \left(\frac{1}{4} - x_9\right) a \hat{\mathbf{y}} + \left(\frac{1}{4} - x_9\right) a \hat{\mathbf{z}}$ | (96$g$) | O III |
| $\mathbf{B}_{99}$ | = | $z_9 \mathbf{a}_1 + (2x_9 - z_9) \mathbf{a}_2 + \left(\frac{1}{2} - 2x_9 - z_9\right) \mathbf{a}_3$ | = | $\left(\frac{1}{4} - z_9\right) a \hat{\mathbf{x}} + \left(\frac{1}{4} - x_9\right) a \hat{\mathbf{y}} + x_9 a \hat{\mathbf{z}}$ | (96$g$) | O III |
| $\mathbf{B}_{100}$ | = | $z_9 \mathbf{a}_1 + \left(\frac{1}{2} - 2x_9 - z_9\right) \mathbf{a}_2 + (2x_9 - z_9) \mathbf{a}_3$ | = | $\left(\frac{1}{4} - z_9\right) a \hat{\mathbf{x}} + x_9 a \hat{\mathbf{y}} + \left(\frac{1}{4} - x_9\right) a \hat{\mathbf{z}}$ | (96$g$) | O III |
| $\mathbf{B}_{101}$ | = | $z_9 \mathbf{a}_1 + (2x_9 - z_9) \mathbf{a}_2 + z_9 \mathbf{a}_3$ | = | $x_9 a \hat{\mathbf{x}} + z_9 a \hat{\mathbf{y}} + x_9 a \hat{\mathbf{z}}$ | (96$g$) | O III |
| $\mathbf{B}_{102}$ | = | $z_9 \mathbf{a}_1 + \left(\frac{1}{2} - 2x_9 - z_9\right) \mathbf{a}_2 + z_9 \mathbf{a}_3$ | = | $\left(\frac{1}{4} - x_9\right) a \hat{\mathbf{x}} + z_9 a \hat{\mathbf{y}} + \left(\frac{1}{4} - x_9\right) a \hat{\mathbf{z}}$ | (96$g$) | O III |
| $\mathbf{B}_{103}$ | = | $\left(\frac{1}{2} - 2x_9 - z_9\right) \mathbf{a}_1 + z_9 \mathbf{a}_2 + (2x_9 - z_9) \mathbf{a}_3$ | = | $x_9 a \hat{\mathbf{x}} + \left(\frac{1}{4} - z_9\right) a \hat{\mathbf{y}} + \left(\frac{1}{4} - x_9\right) a \hat{\mathbf{z}}$ | (96$g$) | O III |
| $\mathbf{B}_{104}$ | = | $(2x_9 - z_9) \mathbf{a}_1 + z_9 \mathbf{a}_2 + \left(\frac{1}{2} - 2x_9 - z_9\right) \mathbf{a}_3$ | = | $\left(\frac{1}{4} - x_9\right) a \hat{\mathbf{x}} + \left(\frac{1}{4} - z_9\right) a \hat{\mathbf{y}} + x_9 a \hat{\mathbf{z}}$ | (96$g$) | O III |
| $\mathbf{B}_{105}$ | = | $-z_9 \mathbf{a}_1 - z_9 \mathbf{a}_2 + \left(\frac{1}{2} + 2x_9 + z_9\right) \mathbf{a}_3$ | = | $\left(\frac{1}{4} + x_9\right) a \hat{\mathbf{x}} + \left(\frac{1}{4} + x_9\right) a \hat{\mathbf{y}} - z_9 a \hat{\mathbf{z}}$ | (96$g$) | O III |
| $\mathbf{B}_{106}$ | = | $-z_9 \mathbf{a}_1 - z_9 \mathbf{a}_2 + (-2x_9 + z_9) \mathbf{a}_3$ | = | $-x_9 a \hat{\mathbf{x}} - x_9 a \hat{\mathbf{y}} - z_9 a \hat{\mathbf{z}}$ | (96$g$) | O III |
| $\mathbf{B}_{107}$ | = | $(-2x_9 + z_9) \mathbf{a}_1 + \left(\frac{1}{2} + 2x_9 + z_9\right) \mathbf{a}_2 - z_9 \mathbf{a}_3$ | = | $\left(\frac{1}{4} + x_9\right) a \hat{\mathbf{x}} - x_9 a \hat{\mathbf{y}} + \left(\frac{1}{4} + z_9\right) a \hat{\mathbf{z}}$ | (96$g$) | O III |
| $\mathbf{B}_{108}$ | = | $\left(\frac{1}{2} + 2x_9 + z_9\right) \mathbf{a}_1 + (-2x_9 + z_9) \mathbf{a}_2 - z_9 \mathbf{a}_3$ | = | $-x_9 a \hat{\mathbf{x}} + \left(\frac{1}{4} + x_9\right) a \hat{\mathbf{y}} + \left(\frac{1}{4} + z_9\right) a \hat{\mathbf{z}}$ | (96$g$) | O III |
| $\mathbf{B}_{109}$ | = | $(-2x_9 + z_9) \mathbf{a}_1 - z_9 \mathbf{a}_2 + \left(\frac{1}{2} + 2x_9 + z_9\right) \mathbf{a}_3$ | = | $\left(\frac{1}{4} + x_9\right) a \hat{\mathbf{x}} + \left(\frac{1}{4} + z_9\right) a \hat{\mathbf{y}} - x_9 a \hat{\mathbf{z}}$ | (96$g$) | O III |
| $\mathbf{B}_{110}$ | = | $\left(\frac{1}{2} + 2x_9 + z_9\right) \mathbf{a}_1 - z_9 \mathbf{a}_2 + (-2x_9 + z_9) \mathbf{a}_3$ | = | $-x_9 a \hat{\mathbf{x}} + \left(\frac{1}{4} + z_9\right) a \hat{\mathbf{y}} + \left(\frac{1}{4} + x_9\right) a \hat{\mathbf{z}}$ | (96$g$) | O III |
| $\mathbf{B}_{111}$ | = | $-z_9 \mathbf{a}_1 + (-2x_9 + z_9) \mathbf{a}_2 - z_9 \mathbf{a}_3$ | = | $-x_9 a \hat{\mathbf{x}} - z_9 a \hat{\mathbf{y}} - x_9 a \hat{\mathbf{z}}$ | (96$g$) | O III |
| $\mathbf{B}_{112}$ | = | $-z_9 \mathbf{a}_1 + \left(\frac{1}{2} + 2x_9 + z_9\right) \mathbf{a}_2 - z_9 \mathbf{a}_3$ | = | $\left(\frac{1}{4} + x_9\right) a \hat{\mathbf{x}} - z_9 a \hat{\mathbf{y}} + \left(\frac{1}{4} + x_9\right) a \hat{\mathbf{z}}$ | (96$g$) | O III |
| $\mathbf{B}_{113}$ | = | $-z_9 \mathbf{a}_1 + (-2x_9 + z_9) \mathbf{a}_2 + \left(\frac{1}{2} + 2x_9 + z_9\right) \mathbf{a}_3$ | = | $\left(\frac{1}{4} + z_9\right) a \hat{\mathbf{x}} + \left(\frac{1}{4} + x_9\right) a \hat{\mathbf{y}} - x_9 a \hat{\mathbf{z}}$ | (96$g$) | O III |
| $\mathbf{B}_{114}$ | = | $-z_9\,\mathbf{a}_1+\left(\frac{1}{2}+2x_9+z_9\right)\,\mathbf{a}_2+\left(-2x_9+z_9\right)\,\mathbf{a}_3$ | = | $\left(\frac{1}{4}+z_9\right)a\,\hat{\mathbf{x}}-x_9a\,\hat{\mathbf{y}}+\left(\frac{1}{4}+x_9\right)a\,\hat{\mathbf{z}}$ | (96$g$) | O III |
| $\mathbf{B}_{115}$ | = | $\left(\frac{1}{2}+2x_9+z_9\right)\,\mathbf{a}_1-z_9\,\mathbf{a}_2-z_9\,\mathbf{a}_3$ | = | $-z_9a\,\hat{\mathbf{x}}+\left(\frac{1}{4}+x_9\right)a\,\hat{\mathbf{y}}+\left(\frac{1}{4}+x_9\right)a\,\hat{\mathbf{z}}$ | (96$g$) | O III |
| $\mathbf{B}_{116}$ | = | $\left(-2x_9+z_9\right)\,\mathbf{a}_1-z_9\,\mathbf{a}_2-z_9\,\mathbf{a}_3$ | = | $-z_9a\,\hat{\mathbf{x}}-x_9a\,\hat{\mathbf{y}}-x_9a\,\hat{\mathbf{z}}$ | (96$g$) | O III |
| $\mathbf{B}_{117}$ | = | $z_{10}\,\mathbf{a}_1+z_{10}\,\mathbf{a}_2+\left(2x_{10}-z_{10}\right)\,\mathbf{a}_3$ | = | $x_{10}a\,\hat{\mathbf{x}}+x_{10}a\,\hat{\mathbf{y}}+z_{10}a\,\hat{\mathbf{z}}$ | (96$g$) | O IV |
| $\mathbf{B}_{118}$ | = | $z_{10}\,\mathbf{a}_1+z_{10}\,\mathbf{a}_2+\left(\frac{1}{2}-2x_{10}-z_{10}\right)\,\mathbf{a}_3$ | = | $\left(\frac{1}{4}-x_{10}\right)a\,\hat{\mathbf{x}}+\left(\frac{1}{4}-x_{10}\right)a\,\hat{\mathbf{y}}+z_{10}a\,\hat{\mathbf{z}}$ | (96$g$) | O IV |
| $\mathbf{B}_{119}$ | = | $\left(2x_{10}-z_{10}\right)\,\mathbf{a}_1+\left(\frac{1}{2}-2x_{10}-z_{10}\right)\,\mathbf{a}_2+z_{10}\,\mathbf{a}_3$ | = | $\left(\frac{1}{4}-x_{10}\right)a\,\hat{\mathbf{x}}+x_{10}a\,\hat{\mathbf{y}}+\left(\frac{1}{4}-z_{10}\right)a\,\hat{\mathbf{z}}$ | (96$g$) | O IV |
| $\mathbf{B}_{120}$ | = | $\left(\frac{1}{2}-2x_{10}-z_{10}\right)\,\mathbf{a}_1+\left(2x_{10}-z_{10}\right)\,\mathbf{a}_2+z_{10}\,\mathbf{a}_3$ | = | $x_{10}a\,\hat{\mathbf{x}}+\left(\frac{1}{4}-x_{10}\right)a\,\hat{\mathbf{y}}+\left(\frac{1}{4}-z_{10}\right)a\,\hat{\mathbf{z}}$ | (96$g$) | O IV |
| $\mathbf{B}_{121}$ | = | $\left(2x_{10}-z_{10}\right)\,\mathbf{a}_1+z_{10}\,\mathbf{a}_2+z_{10}\,\mathbf{a}_3$ | = | $z_{10}a\,\hat{\mathbf{x}}+x_{10}a\,\hat{\mathbf{y}}+x_{10}a\,\hat{\mathbf{z}}$ | (96$g$) | O IV |
| $\mathbf{B}_{122}$ | = | $\left(\frac{1}{2}-2x_{10}-z_{10}\right)\,\mathbf{a}_1+z_{10}\,\mathbf{a}_2+z_{10}\,\mathbf{a}_3$ | = | $z_{10}a\,\hat{\mathbf{x}}+\left(\frac{1}{4}-x_{10}\right)a\,\hat{\mathbf{y}}+\left(\frac{1}{4}-x_{10}\right)a\,\hat{\mathbf{z}}$ | (96$g$) | O IV |
| $\mathbf{B}_{123}$ | = | $z_{10}\,\mathbf{a}_1+\left(2x_{10}-z_{10}\right)\,\mathbf{a}_2+\left(\frac{1}{2}-2x_{10}-z_{10}\right)\,\mathbf{a}_3$ | = | $\left(\frac{1}{4}-z_{10}\right)a\,\hat{\mathbf{x}}+\left(\frac{1}{4}-x_{10}\right)a\,\hat{\mathbf{y}}+x_{10}a\,\hat{\mathbf{z}}$ | (96$g$) | O IV |
| $\mathbf{B}_{124}$ | = | $z_{10}\,\mathbf{a}_1+\left(\frac{1}{2}-2x_{10}-z_{10}\right)\,\mathbf{a}_2+\left(2x_{10}-z_{10}\right)\,\mathbf{a}_3$ | = | $\left(\frac{1}{4}-z_{10}\right)a\,\hat{\mathbf{x}}+x_{10}a\,\hat{\mathbf{y}}+\left(\frac{1}{4}-x_{10}\right)a\,\hat{\mathbf{z}}$ | (96$g$) | O IV |
| $\mathbf{B}_{125}$ | = | $z_{10}\,\mathbf{a}_1+\left(2x_{10}-z_{10}\right)\,\mathbf{a}_2+z_{10}\,\mathbf{a}_3$ | = | $x_{10}a\,\hat{\mathbf{x}}+z_{10}a\,\hat{\mathbf{y}}+x_{10}a\,\hat{\mathbf{z}}$ | (96$g$) | O IV |
| $\mathbf{B}_{126}$ | = | $z_{10}\,\mathbf{a}_1+\left(\frac{1}{2}-2x_{10}-z_{10}\right)\,\mathbf{a}_2+z_{10}\,\mathbf{a}_3$ | = | $\left(\frac{1}{4}-x_{10}\right)a\,\hat{\mathbf{x}}+z_{10}a\,\hat{\mathbf{y}}+\left(\frac{1}{4}-x_{10}\right)a\,\hat{\mathbf{z}}$ | (96$g$) | O IV |
| $\mathbf{B}_{127}$ | = | $\left(\frac{1}{2}-2x_{10}-z_{10}\right)\,\mathbf{a}_1+z_{10}\,\mathbf{a}_2+\left(2x_{10}-z_{10}\right)\,\mathbf{a}_3$ | = | $x_{10}a\,\hat{\mathbf{x}}+\left(\frac{1}{4}-z_{10}\right)a\,\hat{\mathbf{y}}+\left(\frac{1}{4}-x_{10}\right)a\,\hat{\mathbf{z}}$ | (96$g$) | O IV |
| $\mathbf{B}_{128}$ | = | $\left(2x_{10}-z_{10}\right)\,\mathbf{a}_1+z_{10}\,\mathbf{a}_2+\left(\frac{1}{2}-2x_{10}-z_{10}\right)\,\mathbf{a}_3$ | = | $\left(\frac{1}{4}-x_{10}\right)a\,\hat{\mathbf{x}}+\left(\frac{1}{4}-z_{10}\right)a\,\hat{\mathbf{y}}+x_{10}a\,\hat{\mathbf{z}}$ | (96$g$) | O IV |
| $\mathbf{B}_{129}$ | = | $-z_{10}\,\mathbf{a}_1-z_{10}\,\mathbf{a}_2+\left(\frac{1}{2}+2x_{10}+z_{10}\right)\,\mathbf{a}_3$ | = | $\left(\frac{1}{4}+x_{10}\right)a\,\hat{\mathbf{x}}+\left(\frac{1}{4}+x_{10}\right)a\,\hat{\mathbf{y}}-z_{10}a\,\hat{\mathbf{z}}$ | (96$g$) | O IV |
| $\mathbf{B}_{130}$ | = | $-z_{10}\,\mathbf{a}_1-z_{10}\,\mathbf{a}_2+\left(-2x_{10}+z_{10}\right)\,\mathbf{a}_3$ | = | $-x_{10}a\,\hat{\mathbf{x}}-x_{10}a\,\hat{\mathbf{y}}-z_{10}a\,\hat{\mathbf{z}}$ | (96$g$) | O IV |
| $\mathbf{B}_{131}$ | = | $\left(-2x_{10}+z_{10}\right)\,\mathbf{a}_1+\left(\frac{1}{2}+2x_{10}+z_{10}\right)\,\mathbf{a}_2-z_{10}\,\mathbf{a}_3$ | = | $\left(\frac{1}{4}+x_{10}\right)a\,\hat{\mathbf{x}}-x_{10}a\,\hat{\mathbf{y}}+\left(\frac{1}{4}+z_{10}\right)a\,\hat{\mathbf{z}}$ | (96$g$) | O IV |
| $\mathbf{B}_{132}$ | = | $\left(\frac{1}{2}+2x_{10}+z_{10}\right)\,\mathbf{a}_1+\left(-2x_{10}+z_{10}\right)\,\mathbf{a}_2-z_{10}\,\mathbf{a}_3$ | = | $-x_{10}a\,\hat{\mathbf{x}}+\left(\frac{1}{4}+x_{10}\right)a\,\hat{\mathbf{y}}+\left(\frac{1}{4}+z_{10}\right)a\,\hat{\mathbf{z}}$ | (96$g$) | O IV |
| $\mathbf{B}_{133}$ | = | $\left(-2x_{10}+z_{10}\right)\,\mathbf{a}_1-z_{10}\,\mathbf{a}_2+\left(\frac{1}{2}+2x_{10}+z_{10}\right)\,\mathbf{a}_3$ | = | $\left(\frac{1}{4}+x_{10}\right)a\,\hat{\mathbf{x}}+\left(\frac{1}{4}+z_{10}\right)a\,\hat{\mathbf{y}}-x_{10}a\,\hat{\mathbf{z}}$ | (96$g$) | O IV |
| $\mathbf{B}_{134}$ | = | $\left(\frac{1}{2}+2x_{10}+z_{10}\right)\,\mathbf{a}_1-z_{10}\,\mathbf{a}_2+\left(-2x_{10}+z_{10}\right)\,\mathbf{a}_3$ | = | $-x_{10}a\,\hat{\mathbf{x}}+\left(\frac{1}{4}+z_{10}\right)a\,\hat{\mathbf{y}}+\left(\frac{1}{4}+x_{10}\right)a\,\hat{\mathbf{z}}$ | (96$g$) | O IV |
| $\mathbf{B}_{135}$ | = | $-z_{10}\,\mathbf{a}_1+\left(-2x_{10}+z_{10}\right)\,\mathbf{a}_2-z_{10}\,\mathbf{a}_3$ | = | $-x_{10}a\,\hat{\mathbf{x}}-z_{10}a\,\hat{\mathbf{y}}-x_{10}a\,\hat{\mathbf{z}}$ | (96$g$) | O IV |
| $\mathbf{B}_{136}$ | = | $-z_{10}\,\mathbf{a}_1+\left(\frac{1}{2}+2x_{10}+z_{10}\right)\,\mathbf{a}_2-z_{10}\,\mathbf{a}_3$ | = | $\left(\frac{1}{4}+x_{10}\right)a\,\hat{\mathbf{x}}-z_{10}a\,\hat{\mathbf{y}}+\left(\frac{1}{4}+x_{10}\right)a\,\hat{\mathbf{z}}$ | (96$g$) | O IV |
| $\mathbf{B}_{137}$ | = | $-z_{10}\,\mathbf{a}_1+\left(-2x_{10}+z_{10}\right)\,\mathbf{a}_2+\left(\frac{1}{2}+2x_{10}+z_{10}\right)\,\mathbf{a}_3$ | = | $\left(\frac{1}{4}+z_{10}\right)a\,\hat{\mathbf{x}}+\left(\frac{1}{4}+x_{10}\right)a\,\hat{\mathbf{y}}-x_{10}a\,\hat{\mathbf{z}}$ | (96$g$) | O IV |
| $\mathbf{B}_{138}$ | = | $-z_{10}\,\mathbf{a}_1+\left(\frac{1}{2}+2x_{10}+z_{10}\right)\,\mathbf{a}_2+\left(-2x_{10}+z_{10}\right)\,\mathbf{a}_3$ | = | $\left(\frac{1}{4}+z_{10}\right)a\,\hat{\mathbf{x}}-x_{10}a\,\hat{\mathbf{y}}+\left(\frac{1}{4}+x_{10}\right)a\,\hat{\mathbf{z}}$ | (96$g$) | O IV |
| $\mathbf{B}_{139}$ | = | $\left(\frac{1}{2}+2x_{10}+z_{10}\right)\,\mathbf{a}_1-z_{10}\,\mathbf{a}_2-z_{10}\,\mathbf{a}_3$ | = | $-z_{10}a\,\hat{\mathbf{x}}+\left(\frac{1}{4}+x_{10}\right)a\,\hat{\mathbf{y}}+\left(\frac{1}{4}+x_{10}\right)a\,\hat{\mathbf{z}}$ | (96$g$) | O IV |
| $\mathbf{B}_{140}$ | = | $\left(-2x_{10}+z_{10}\right)\,\mathbf{a}_1-z_{10}\,\mathbf{a}_2-z_{10}\,\mathbf{a}_3$ | = | $-z_{10}a\,\hat{\mathbf{x}}-x_{10}a\,\hat{\mathbf{y}}-x_{10}a\,\hat{\mathbf{z}}$ | (96$g$) | O IV |
| $\mathbf{B}_{141}$ | = | $z_{11}\,\mathbf{a}_1 + z_{11}\,\mathbf{a}_2 + (2x_{11} - z_{11})\,\mathbf{a}_3$ | = | $x_{11}\,a\,\hat{\mathbf{x}} + x_{11}\,a\,\hat{\mathbf{y}} + z_{11}\,a\,\hat{\mathbf{z}}$ | (96$g$) | W |
| $\mathbf{B}_{142}$ | = | $z_{11}\,\mathbf{a}_1 + z_{11}\,\mathbf{a}_2 + \left(\frac{1}{2} - 2x_{11} - z_{11}\right)\mathbf{a}_3$ | = | $\left(\frac{1}{4} - x_{11}\right)a\,\hat{\mathbf{x}} + \left(\frac{1}{4} - x_{11}\right)a\,\hat{\mathbf{y}} + z_{11}\,a\,\hat{\mathbf{z}}$ | (96$g$) | W |
| $\mathbf{B}_{143}$ | = | $(2x_{11} - z_{11})\,\mathbf{a}_1 + \left(\frac{1}{2} - 2x_{11} - z_{11}\right)\mathbf{a}_2 + z_{11}\,\mathbf{a}_3$ | = | $\left(\frac{1}{4} - x_{11}\right)a\,\hat{\mathbf{x}} + x_{11}\,a\,\hat{\mathbf{y}} + \left(\frac{1}{4} - z_{11}\right)a\,\hat{\mathbf{z}}$ | (96$g$) | W |
| $\mathbf{B}_{144}$ | = | $\left(\frac{1}{2} - 2x_{11} - z_{11}\right)\mathbf{a}_1 + (2x_{11} - z_{11})\,\mathbf{a}_2 + z_{11}\,\mathbf{a}_3$ | = | $x_{11}\,a\,\hat{\mathbf{x}} + \left(\frac{1}{4} - x_{11}\right)a\,\hat{\mathbf{y}} + \left(\frac{1}{4} - z_{11}\right)a\,\hat{\mathbf{z}}$ | (96$g$) | W |
| $\mathbf{B}_{145}$ | = | $(2x_{11} - z_{11})\,\mathbf{a}_1 + z_{11}\,\mathbf{a}_2 + z_{11}\,\mathbf{a}_3$ | = | $z_{11}\,a\,\hat{\mathbf{x}} + x_{11}\,a\,\hat{\mathbf{y}} + x_{11}\,a\,\hat{\mathbf{z}}$ | (96$g$) | W |
| $\mathbf{B}_{146}$ | = | $\left(\frac{1}{2} - 2x_{11} - z_{11}\right)\mathbf{a}_1 + z_{11}\,\mathbf{a}_2 + z_{11}\,\mathbf{a}_3$ | = | $z_{11}\,a\,\hat{\mathbf{x}} + \left(\frac{1}{4} - x_{11}\right)a\,\hat{\mathbf{y}} + \left(\frac{1}{4} - x_{11}\right)a\,\hat{\mathbf{z}}$ | (96$g$) | W |
| $\mathbf{B}_{147}$ | = | $z_{11}\,\mathbf{a}_1 + (2x_{11} - z_{11})\,\mathbf{a}_2 + \left(\frac{1}{2} - 2x_{11} - z_{11}\right)\mathbf{a}_3$ | = | $\left(\frac{1}{4} - z_{11}\right)a\,\hat{\mathbf{x}} + \left(\frac{1}{4} - x_{11}\right)a\,\hat{\mathbf{y}} + x_{11}\,a\,\hat{\mathbf{z}}$ | (96$g$) | W |
| $\mathbf{B}_{148}$ | = | $z_{11}\,\mathbf{a}_1 + \left(\frac{1}{2} - 2x_{11} - z_{11}\right)\mathbf{a}_2 + (2x_{11} - z_{11})\,\mathbf{a}_3$ | = | $\left(\frac{1}{4} - z_{11}\right)a\,\hat{\mathbf{x}} + x_{11}\,a\,\hat{\mathbf{y}} + \left(\frac{1}{4} - x_{11}\right)a\,\hat{\mathbf{z}}$ | (96$g$) | W |
| $\mathbf{B}_{149}$ | = | $z_{11}\,\mathbf{a}_1 + (2x_{11} - z_{11})\,\mathbf{a}_2 + z_{11}\,\mathbf{a}_3$ | = | $x_{11}\,a\,\hat{\mathbf{x}} + z_{11}\,a\,\hat{\mathbf{y}} + x_{11}\,a\,\hat{\mathbf{z}}$ | (96$g$) | W |
| $\mathbf{B}_{150}$ | = | $z_{11}\,\mathbf{a}_1 + \left(\frac{1}{2} - 2x_{11} - z_{11}\right)\mathbf{a}_2 + z_{11}\,\mathbf{a}_3$ | = | $\left(\frac{1}{4} - x_{11}\right)a\,\hat{\mathbf{x}} + z_{11}\,a\,\hat{\mathbf{y}} + \left(\frac{1}{4} - x_{11}\right)a\,\hat{\mathbf{z}}$ | (96$g$) | W |
| $\mathbf{B}_{151}$ | = | $\left(\frac{1}{2} - 2x_{11} - z_{11}\right)\mathbf{a}_1 + z_{11}\,\mathbf{a}_2 + (2x_{11} - z_{11})\,\mathbf{a}_3$ | = | $x_{11}\,a\,\hat{\mathbf{x}} + \left(\frac{1}{4} - z_{11}\right)a\,\hat{\mathbf{y}} + \left(\frac{1}{4} - x_{11}\right)a\,\hat{\mathbf{z}}$ | (96$g$) | W |
| $\mathbf{B}_{152}$ | = | $(2x_{11} - z_{11})\,\mathbf{a}_1 + z_{11}\,\mathbf{a}_2 + \left(\frac{1}{2} - 2x_{11} - z_{11}\right)\mathbf{a}_3$ | = | $\left(\frac{1}{4} - x_{11}\right)a\,\hat{\mathbf{x}} + \left(\frac{1}{4} - z_{11}\right)a\,\hat{\mathbf{y}} + x_{11}\,a\,\hat{\mathbf{z}}$ | (96$g$) | W |
| $\mathbf{B}_{153}$ | = | $-z_{11}\,\mathbf{a}_1 - z_{11}\,\mathbf{a}_2 + \left(\frac{1}{2} + 2x_{11} + z_{11}\right)\mathbf{a}_3$ | = | $\left(\frac{1}{4} + x_{11}\right)a\,\hat{\mathbf{x}} + \left(\frac{1}{4} + x_{11}\right)a\,\hat{\mathbf{y}} - z_{11}\,a\,\hat{\mathbf{z}}$ | (96$g$) | W |
| $\mathbf{B}_{154}$ | = | $-z_{11}\,\mathbf{a}_1 - z_{11}\,\mathbf{a}_2 + (-2x_{11} + z_{11})\,\mathbf{a}_3$ | = | $-x_{11}\,a\,\hat{\mathbf{x}} - x_{11}\,a\,\hat{\mathbf{y}} - z_{11}\,a\,\hat{\mathbf{z}}$ | (96$g$) | W |
| $\mathbf{B}_{155}$ | = | $(-2x_{11} + z_{11})\,\mathbf{a}_1 + \left(\frac{1}{2} + 2x_{11} + z_{11}\right)\mathbf{a}_2 - z_{11}\,\mathbf{a}_3$ | = | $\left(\frac{1}{4} + x_{11}\right)a\,\hat{\mathbf{x}} - x_{11}\,a\,\hat{\mathbf{y}} + \left(\frac{1}{4} + z_{11}\right)a\,\hat{\mathbf{z}}$ | (96$g$) | W |
| $\mathbf{B}_{156}$ | = | $\left(\frac{1}{2} + 2x_{11} + z_{11}\right)\mathbf{a}_1 + (-2x_{11} + z_{11})\,\mathbf{a}_2 - z_{11}\,\mathbf{a}_3$ | = | $-x_{11}\,a\,\hat{\mathbf{x}} + \left(\frac{1}{4} + x_{11}\right)a\,\hat{\mathbf{y}} + \left(\frac{1}{4} + z_{11}\right)a\,\hat{\mathbf{z}}$ | (96$g$) | W |
| $\mathbf{B}_{157}$ | = | $(-2x_{11} + z_{11})\,\mathbf{a}_1 - z_{11}\,\mathbf{a}_2 + \left(\frac{1}{2} + 2x_{11} + z_{11}\right)\mathbf{a}_3$ | = | $\left(\frac{1}{4} + x_{11}\right)a\,\hat{\mathbf{x}} + \left(\frac{1}{4} + z_{11}\right)a\,\hat{\mathbf{y}} - x_{11}\,a\,\hat{\mathbf{z}}$ | (96$g$) | W |
| $\mathbf{B}_{158}$ | = | $\left(\frac{1}{2} + 2x_{11} + z_{11}\right)\mathbf{a}_1 - z_{11}\,\mathbf{a}_2 + (-2x_{11} + z_{11})\,\mathbf{a}_3$ | = | $-x_{11}\,a\,\hat{\mathbf{x}} + \left(\frac{1}{4} + z_{11}\right)a\,\hat{\mathbf{y}} + \left(\frac{1}{4} + x_{11}\right)a\,\hat{\mathbf{z}}$ | (96$g$) | W |
| $\mathbf{B}_{159}$ | = | $-z_{11}\,\mathbf{a}_1 + (-2x_{11} + z_{11})\,\mathbf{a}_2 - z_{11}\,\mathbf{a}_3$ | = | $-x_{11}\,a\,\hat{\mathbf{x}} - z_{11}\,a\,\hat{\mathbf{y}} - x_{11}\,a\,\hat{\mathbf{z}}$ | (96$g$) | W |
| $\mathbf{B}_{160}$ | = | $-z_{11}\,\mathbf{a}_1 + \left(\frac{1}{2} + 2x_{11} + z_{11}\right)\mathbf{a}_2 - z_{11}\,\mathbf{a}_3$ | = | $\left(\frac{1}{4} + x_{11}\right)a\,\hat{\mathbf{x}} - z_{11}\,a\,\hat{\mathbf{y}} + \left(\frac{1}{4} + x_{11}\right)a\,\hat{\mathbf{z}}$ | (96$g$) | W |
| $\mathbf{B}_{161}$ | = | $-z_{11}\,\mathbf{a}_1 + (-2x_{11} + z_{11})\,\mathbf{a}_2 + \left(\frac{1}{2} + 2x_{11} + z_{11}\right)\mathbf{a}_3$ | = | $\left(\frac{1}{4} + z_{11}\right)a\,\hat{\mathbf{x}} + \left(\frac{1}{4} + x_{11}\right)a\,\hat{\mathbf{y}} - x_{11}\,a\,\hat{\mathbf{z}}$ | (96$g$) | W |
| $\mathbf{B}_{162}$ | = | $-z_{11}\,\mathbf{a}_1 + \left(\frac{1}{2} + 2x_{11} + z_{11}\right)\mathbf{a}_2 + (-2x_{11} + z_{11})\,\mathbf{a}_3$ | = | $\left(\frac{1}{4} + z_{11}\right)a\,\hat{\mathbf{x}} - x_{11}\,a\,\hat{\mathbf{y}} + \left(\frac{1}{4} + x_{11}\right)a\,\hat{\mathbf{z}}$ | (96$g$) | W |
| $\mathbf{B}_{163}$ | = | $\left(\frac{1}{2} + 2x_{11} + z_{11}\right)\mathbf{a}_1 - z_{11}\,\mathbf{a}_2 - z_{11}\,\mathbf{a}_3$ | = | $-z_{11}\,a\,\hat{\mathbf{x}} + \left(\frac{1}{4} + x_{11}\right)a\,\hat{\mathbf{y}} + \left(\frac{1}{4} + x_{11}\right)a\,\hat{\mathbf{z}}$ | (96$g$) | W |
| $\mathbf{B}_{164}$ | = | $(-2x_{11} + z_{11})\,\mathbf{a}_1 - z_{11}\,\mathbf{a}_2 - z_{11}\,\mathbf{a}_3$ | = | $-z_{11}\,a\,\hat{\mathbf{x}} - x_{11}\,a\,\hat{\mathbf{y}} - x_{11}\,a\,\hat{\mathbf{z}}$ | (96$g$) | W |

**Found in:**

---

**Geometry files:**

- CIF: pp. 1827
- POSCAR: pp. 1828

# $G7_3$ [Northupite, $Na_3MgCl(CO_3)_2$] (*obsolete*) Structure: A2BCD3E6_cF208_227_e_c_d_f_g

http://aflow.org/prototype-encyclopedia/A2BCD3E6_cF208_227_e_c_d_f_g

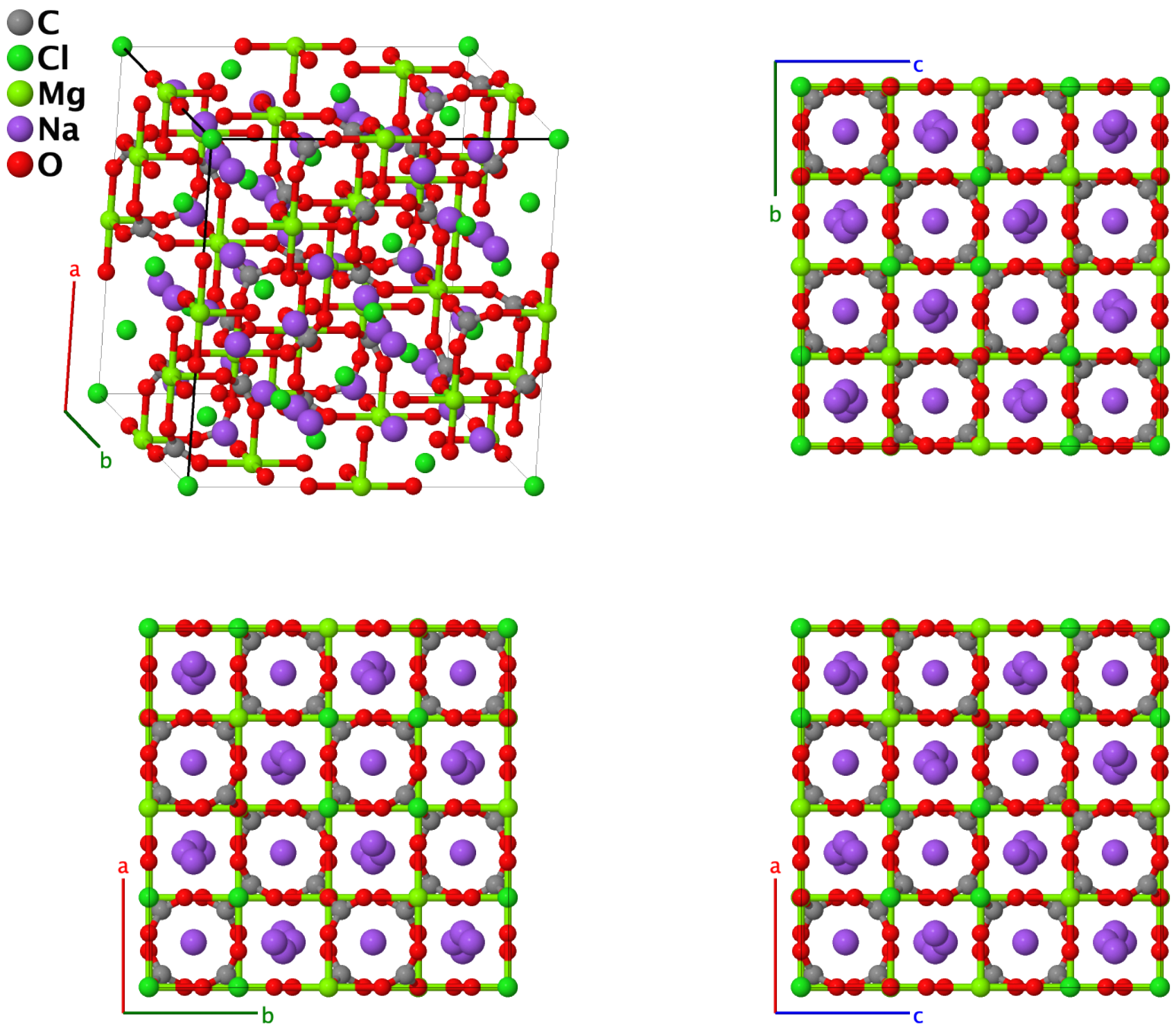

| | | |
|---|---|---|
| **Prototype** | : | $C_2ClMgNa_3O_6$ |
| **AFLOW prototype label** | : | A2BCD3E6_cF208_227_e_c_d_f_g |
| ***Strukturbericht* designation** | : | $G7_3$ |
| **Pearson symbol** | : | cF208 |
| **Space group number** | : | 227 |
| **Space group symbol** | : | $Fd\bar{3}m$ |
| **AFLOW prototype command** | : | `aflow --proto=A2BCD3E6_cF208_227_e_c_d_f_g --params=`$a, x_3, x_4, x_5, z_5$ |

- This is the original structure determined by (Shiba, 1931) and given the designation $G7_3$ in (Hermann, 1937). (Negro, 1975) showed that the correct structure was actually related to cubic pyrochlore, however the two structures are very similar, and a displacement of the oxygen atoms by less than 1 Å brings the two structures into agreement.

**Face-centered Cubic primitive vectors:**

$$\mathbf{a}_1 = \frac{1}{2}\,a\,\hat{\mathbf{y}} + \frac{1}{2}\,a\,\hat{\mathbf{z}}$$
$$\mathbf{a}_2 = \frac{1}{2}\,a\,\hat{\mathbf{x}} + \frac{1}{2}\,a\,\hat{\mathbf{z}}$$
$$\mathbf{a}_3 = \frac{1}{2}\,a\,\hat{\mathbf{x}} + \frac{1}{2}\,a\,\hat{\mathbf{y}}$$

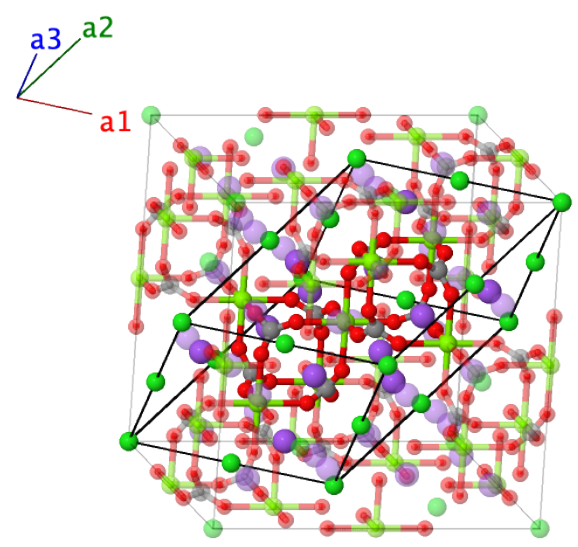

**Basis vectors:**

| | | Lattice Coordinates | | Cartesian Coordinates | Wyckoff Position | Atom Type |
|---|---|---|---|---|---|---|
| $\mathbf{B}_1$ | = | $0\,\mathbf{a}_1 + 0\,\mathbf{a}_2 + 0\,\mathbf{a}_3$ | = | $0\,\hat{\mathbf{x}} + 0\,\hat{\mathbf{y}} + 0\,\hat{\mathbf{z}}$ | $(16c)$ | Cl |
| $\mathbf{B}_2$ | = | $\frac{1}{2}\,\mathbf{a}_3$ | = | $\frac{1}{4}a\,\hat{\mathbf{x}} + \frac{1}{4}a\,\hat{\mathbf{y}}$ | $(16c)$ | Cl |
| $\mathbf{B}_3$ | = | $\frac{1}{2}\,\mathbf{a}_2$ | = | $\frac{1}{4}a\,\hat{\mathbf{x}} + \frac{1}{4}a\,\hat{\mathbf{z}}$ | $(16c)$ | Cl |
| $\mathbf{B}_4$ | = | $\frac{1}{2}\,\mathbf{a}_1$ | = | $\frac{1}{4}a\,\hat{\mathbf{y}} + \frac{1}{4}a\,\hat{\mathbf{z}}$ | $(16c)$ | Cl |
| $\mathbf{B}_5$ | = | $\frac{1}{2}\,\mathbf{a}_1 + \frac{1}{2}\,\mathbf{a}_2 + \frac{1}{2}\,\mathbf{a}_3$ | = | $\frac{1}{2}a\,\hat{\mathbf{x}} + \frac{1}{2}a\,\hat{\mathbf{y}} + \frac{1}{2}a\,\hat{\mathbf{z}}$ | $(16d)$ | Mg |
| $\mathbf{B}_6$ | = | $\frac{1}{2}\,\mathbf{a}_1 + \frac{1}{2}\,\mathbf{a}_2$ | = | $\frac{1}{4}a\,\hat{\mathbf{x}} + \frac{1}{4}a\,\hat{\mathbf{y}} + \frac{1}{2}a\,\hat{\mathbf{z}}$ | $(16d)$ | Mg |
| $\mathbf{B}_7$ | = | $\frac{1}{2}\,\mathbf{a}_1 + \frac{1}{2}\,\mathbf{a}_3$ | = | $\frac{1}{4}a\,\hat{\mathbf{x}} + \frac{1}{2}a\,\hat{\mathbf{y}} + \frac{1}{4}a\,\hat{\mathbf{z}}$ | $(16d)$ | Mg |
| $\mathbf{B}_8$ | = | $\frac{1}{2}\,\mathbf{a}_2 + \frac{1}{2}\,\mathbf{a}_3$ | = | $\frac{1}{2}a\,\hat{\mathbf{x}} + \frac{1}{4}a\,\hat{\mathbf{y}} + \frac{1}{4}a\,\hat{\mathbf{z}}$ | $(16d)$ | Mg |
| $\mathbf{B}_9$ | = | $x_3\,\mathbf{a}_1 + x_3\,\mathbf{a}_2 + x_3\,\mathbf{a}_3$ | = | $x_3a\,\hat{\mathbf{x}} + x_3a\,\hat{\mathbf{y}} + x_3a\,\hat{\mathbf{z}}$ | $(32e)$ | C |
| $\mathbf{B}_{10}$ | = | $x_3\,\mathbf{a}_1 + x_3\,\mathbf{a}_2 + \left(\frac{1}{2} - 3x_3\right)\,\mathbf{a}_3$ | = | $\left(\frac{1}{4} - x_3\right)a\,\hat{\mathbf{x}} + \left(\frac{1}{4} - x_3\right)a\,\hat{\mathbf{y}} + x_3a\,\hat{\mathbf{z}}$ | $(32e)$ | C |
| $\mathbf{B}_{11}$ | = | $x_3\,\mathbf{a}_1 + \left(\frac{1}{2} - 3x_3\right)\,\mathbf{a}_2 + x_3\,\mathbf{a}_3$ | = | $\left(\frac{1}{4} - x_3\right)a\,\hat{\mathbf{x}} + x_3a\,\hat{\mathbf{y}} + \left(\frac{1}{4} - x_3\right)a\,\hat{\mathbf{z}}$ | $(32e)$ | C |
| $\mathbf{B}_{12}$ | = | $\left(\frac{1}{2} - 3x_3\right)\,\mathbf{a}_1 + x_3\,\mathbf{a}_2 + x_3\,\mathbf{a}_3$ | = | $x_3a\,\hat{\mathbf{x}} + \left(\frac{1}{4} - x_3\right)a\,\hat{\mathbf{y}} + \left(\frac{1}{4} - x_3\right)a\,\hat{\mathbf{z}}$ | $(32e)$ | C |
| $\mathbf{B}_{13}$ | = | $-x_3\,\mathbf{a}_1 - x_3\,\mathbf{a}_2 + \left(\frac{1}{2} + 3x_3\right)\,\mathbf{a}_3$ | = | $\left(\frac{1}{4} + x_3\right)a\,\hat{\mathbf{x}} + \left(\frac{1}{4} + x_3\right)a\,\hat{\mathbf{y}} - x_3a\,\hat{\mathbf{z}}$ | $(32e)$ | C |
| $\mathbf{B}_{14}$ | = | $-x_3\,\mathbf{a}_1 - x_3\,\mathbf{a}_2 - x_3\,\mathbf{a}_3$ | = | $-x_3a\,\hat{\mathbf{x}} - x_3a\,\hat{\mathbf{y}} - x_3a\,\hat{\mathbf{z}}$ | $(32e)$ | C |
| $\mathbf{B}_{15}$ | = | $-x_3\,\mathbf{a}_1 + \left(\frac{1}{2} + 3x_3\right)\,\mathbf{a}_2 - x_3\,\mathbf{a}_3$ | = | $\left(\frac{1}{4} + x_3\right)a\,\hat{\mathbf{x}} - x_3a\,\hat{\mathbf{y}} + \left(\frac{1}{4} + x_3\right)a\,\hat{\mathbf{z}}$ | $(32e)$ | C |
| $\mathbf{B}_{16}$ | = | $\left(\frac{1}{2} + 3x_3\right)\,\mathbf{a}_1 - x_3\,\mathbf{a}_2 - x_3\,\mathbf{a}_3$ | = | $-x_3a\,\hat{\mathbf{x}} + \left(\frac{1}{4} + x_3\right)a\,\hat{\mathbf{y}} + \left(\frac{1}{4} + x_3\right)a\,\hat{\mathbf{z}}$ | $(32e)$ | C |
| $\mathbf{B}_{17}$ | = | $\left(\frac{1}{4} - x_4\right)\,\mathbf{a}_1 + x_4\,\mathbf{a}_2 + x_4\,\mathbf{a}_3$ | = | $x_4a\,\hat{\mathbf{x}} + \frac{1}{8}a\,\hat{\mathbf{y}} + \frac{1}{8}a\,\hat{\mathbf{z}}$ | $(48f)$ | Na |
| $\mathbf{B}_{18}$ | = | $x_4\,\mathbf{a}_1 + \left(\frac{1}{4} - x_4\right)\,\mathbf{a}_2 + \left(\frac{1}{4} - x_4\right)\,\mathbf{a}_3$ | = | $\left(\frac{1}{4} - x_4\right)a\,\hat{\mathbf{x}} + \frac{1}{8}a\,\hat{\mathbf{y}} + \frac{1}{8}a\,\hat{\mathbf{z}}$ | $(48f)$ | Na |
| $\mathbf{B}_{19}$ | = | $x_4\,\mathbf{a}_1 + \left(\frac{1}{4} - x_4\right)\,\mathbf{a}_2 + x_4\,\mathbf{a}_3$ | = | $\frac{1}{8}a\,\hat{\mathbf{x}} + x_4a\,\hat{\mathbf{y}} + \frac{1}{8}a\,\hat{\mathbf{z}}$ | $(48f)$ | Na |
| $\mathbf{B}_{20}$ | = | $\left(\frac{1}{4} - x_4\right)\,\mathbf{a}_1 + x_4\,\mathbf{a}_2 + \left(\frac{1}{4} - x_4\right)\,\mathbf{a}_3$ | = | $\frac{1}{8}a\,\hat{\mathbf{x}} + \left(\frac{1}{4} - x_4\right)a\,\hat{\mathbf{y}} + \frac{1}{8}a\,\hat{\mathbf{z}}$ | $(48f)$ | Na |
| $\mathbf{B}_{21}$ | = | $x_4\,\mathbf{a}_1 + x_4\,\mathbf{a}_2 + \left(\frac{1}{4} - x_4\right)\,\mathbf{a}_3$ | = | $\frac{1}{8}a\,\hat{\mathbf{x}} + \frac{1}{8}a\,\hat{\mathbf{y}} + x_4a\,\hat{\mathbf{z}}$ | $(48f)$ | Na |
| $\mathbf{B}_{22}$ | = | $\left(\frac{1}{4} - x_4\right)\,\mathbf{a}_1 + \left(\frac{1}{4} - x_4\right)\,\mathbf{a}_2 + x_4\,\mathbf{a}_3$ | = | $\frac{1}{8}a\,\hat{\mathbf{x}} + \frac{1}{8}a\,\hat{\mathbf{y}} + \left(\frac{1}{4} - x_4\right)a\,\hat{\mathbf{z}}$ | $(48f)$ | Na |
| $\mathbf{B}_{23}$ | = | $\left(\frac{3}{4} + x_4\right)\,\mathbf{a}_1 - x_4\,\mathbf{a}_2 + \left(\frac{3}{4} + x_4\right)\,\mathbf{a}_3$ | = | $\frac{3}{8}a\,\hat{\mathbf{x}} + \left(\frac{3}{4} + x_4\right)a\,\hat{\mathbf{y}} + \frac{3}{8}a\,\hat{\mathbf{z}}$ | $(48f)$ | Na |
| $\mathbf{B}_{24}$ | = | $-x_4\,\mathbf{a}_1 + \left(\frac{3}{4} + x_4\right)\,\mathbf{a}_2 - x_4\,\mathbf{a}_3$ | = | $\frac{3}{8}a\,\hat{\mathbf{x}} - x_4a\,\hat{\mathbf{y}} + \frac{3}{8}a\,\hat{\mathbf{z}}$ | $(48f)$ | Na |
| $\mathbf{B}_{25}$ | = | $-x_4\,\mathbf{a}_1 + \left(\frac{3}{4} + x_4\right)\,\mathbf{a}_2 + \left(\frac{3}{4} + x_4\right)\,\mathbf{a}_3$ | = | $\left(\frac{3}{4} + x_4\right)a\,\hat{\mathbf{x}} + \frac{3}{8}a\,\hat{\mathbf{y}} + \frac{3}{8}a\,\hat{\mathbf{z}}$ | $(48f)$ | Na |
| $\mathbf{B}_{26}$ | = | $\left(\frac{3}{4} + x_4\right)\,\mathbf{a}_1 - x_4\,\mathbf{a}_2 - x_4\,\mathbf{a}_3$ | = | $-x_4a\,\hat{\mathbf{x}} + \frac{3}{8}a\,\hat{\mathbf{y}} + \frac{3}{8}a\,\hat{\mathbf{z}}$ | $(48f)$ | Na |
| $\mathbf{B}_{27}$ | = | $-x_4\,\mathbf{a}_1 - x_4\,\mathbf{a}_2 + \left(\frac{3}{4} + x_4\right)\,\mathbf{a}_3$ | = | $\frac{3}{8}a\,\hat{\mathbf{x}} + \frac{3}{8}a\,\hat{\mathbf{y}} - x_4a\,\hat{\mathbf{z}}$ | $(48f)$ | Na |
| $\mathbf{B}_{28}$ | = | $\left(\frac{3}{4} + x_4\right)\,\mathbf{a}_1 + \left(\frac{3}{4} + x_4\right)\,\mathbf{a}_2 - x_4\,\mathbf{a}_3$ | = | $\frac{3}{8}a\,\hat{\mathbf{x}} + \frac{3}{8}a\,\hat{\mathbf{y}} + \left(\frac{3}{4} + x_4\right)a\,\hat{\mathbf{z}}$ | $(48f)$ | Na |
| $\mathbf{B}_{29}$ | = | $z_5\,\mathbf{a}_1 + z_5\,\mathbf{a}_2 + (2x_5 - z_5)\,\mathbf{a}_3$ | = | $x_5a\,\hat{\mathbf{x}} + x_5a\,\hat{\mathbf{y}} + z_5a\,\hat{\mathbf{z}}$ | $(96g)$ | O |
| $\mathbf{B}_{30}$ | = | $z_5\,\mathbf{a}_1 + z_5\,\mathbf{a}_2 + \left(\frac{1}{2} - 2x_5 - z_5\right)\,\mathbf{a}_3$ | = | $\left(\frac{1}{4} - x_5\right)a\,\hat{\mathbf{x}} + \left(\frac{1}{4} - x_5\right)a\,\hat{\mathbf{y}} + z_5a\,\hat{\mathbf{z}}$ | $(96g)$ | O |
| $\mathbf{B}_{31}$ | = | $(2x_5 - z_5)\,\mathbf{a}_1 +$ $\left(\frac{1}{2} - 2x_5 - z_5\right)\,\mathbf{a}_2 + z_5\,\mathbf{a}_3$ | = | $\left(\frac{1}{4} - x_5\right)a\,\hat{\mathbf{x}} + x_5a\,\hat{\mathbf{y}} + \left(\frac{1}{4} - z_5\right)a\,\hat{\mathbf{z}}$ | $(96g)$ | O |
| $\mathbf{B}_{32}$ | = | $\left(\frac{1}{2} - 2x_5 - z_5\right)\,\mathbf{a}_1 +$ $(2x_5 - z_5)\,\mathbf{a}_2 + z_5\,\mathbf{a}_3$ | = | $x_5a\,\hat{\mathbf{x}} + \left(\frac{1}{4} - x_5\right)a\,\hat{\mathbf{y}} + \left(\frac{1}{4} - z_5\right)a\,\hat{\mathbf{z}}$ | $(96g)$ | O |
| $\mathbf{B}_{33}$ | = | $(2x_5 - z_5)\,\mathbf{a}_1 + z_5\,\mathbf{a}_2 + z_5\,\mathbf{a}_3$ | = | $z_5a\,\hat{\mathbf{x}} + x_5a\,\hat{\mathbf{y}} + x_5a\,\hat{\mathbf{z}}$ | $(96g)$ | O |
| $\mathbf{B}_{34}$ | = | $\left(\frac{1}{2} - 2x_5 - z_5\right)\,\mathbf{a}_1 + z_5\,\mathbf{a}_2 + z_5\,\mathbf{a}_3$ | = | $z_5 a\,\mathbf{\hat{x}} + \left(\frac{1}{4} - x_5\right) a\,\mathbf{\hat{y}} + \left(\frac{1}{4} - x_5\right) a\,\mathbf{\hat{z}}$ | (96$g$) | O |
| $\mathbf{B}_{35}$ | = | $z_5\,\mathbf{a}_1 + (2x_5 - z_5)\,\mathbf{a}_2 + \left(\frac{1}{2} - 2x_5 - z_5\right)\,\mathbf{a}_3$ | = | $\left(\frac{1}{4} - z_5\right) a\,\mathbf{\hat{x}} + \left(\frac{1}{4} - x_5\right) a\,\mathbf{\hat{y}} + x_5 a\,\mathbf{\hat{z}}$ | (96$g$) | O |
| $\mathbf{B}_{36}$ | = | $z_5\,\mathbf{a}_1 + \left(\frac{1}{2} - 2x_5 - z_5\right)\,\mathbf{a}_2 + (2x_5 - z_5)\,\mathbf{a}_3$ | = | $\left(\frac{1}{4} - z_5\right) a\,\mathbf{\hat{x}} + x_5 a\,\mathbf{\hat{y}} + \left(\frac{1}{4} - x_5\right) a\,\mathbf{\hat{z}}$ | (96$g$) | O |
| $\mathbf{B}_{37}$ | = | $z_5\,\mathbf{a}_1 + (2x_5 - z_5)\,\mathbf{a}_2 + z_5\,\mathbf{a}_3$ | = | $x_5 a\,\mathbf{\hat{x}} + z_5 a\,\mathbf{\hat{y}} + x_5 a\,\mathbf{\hat{z}}$ | (96$g$) | O |
| $\mathbf{B}_{38}$ | = | $z_5\,\mathbf{a}_1 + \left(\frac{1}{2} - 2x_5 - z_5\right)\,\mathbf{a}_2 + z_5\,\mathbf{a}_3$ | = | $\left(\frac{1}{4} - x_5\right) a\,\mathbf{\hat{x}} + z_5 a\,\mathbf{\hat{y}} + \left(\frac{1}{4} - x_5\right) a\,\mathbf{\hat{z}}$ | (96$g$) | O |
| $\mathbf{B}_{39}$ | = | $\left(\frac{1}{2} - 2x_5 - z_5\right)\,\mathbf{a}_1 + z_5\,\mathbf{a}_2 + (2x_5 - z_5)\,\mathbf{a}_3$ | = | $x_5 a\,\mathbf{\hat{x}} + \left(\frac{1}{4} - z_5\right) a\,\mathbf{\hat{y}} + \left(\frac{1}{4} - x_5\right) a\,\mathbf{\hat{z}}$ | (96$g$) | O |
| $\mathbf{B}_{40}$ | = | $(2x_5 - z_5)\,\mathbf{a}_1 + z_5\,\mathbf{a}_2 + \left(\frac{1}{2} - 2x_5 - z_5\right)\,\mathbf{a}_3$ | = | $\left(\frac{1}{4} - x_5\right) a\,\mathbf{\hat{x}} + \left(\frac{1}{4} - z_5\right) a\,\mathbf{\hat{y}} + x_5 a\,\mathbf{\hat{z}}$ | (96$g$) | O |
| $\mathbf{B}_{41}$ | = | $-z_5\,\mathbf{a}_1 - z_5\,\mathbf{a}_2 + \left(\frac{1}{2} + 2x_5 + z_5\right)\,\mathbf{a}_3$ | = | $\left(\frac{1}{4} + x_5\right) a\,\mathbf{\hat{x}} + \left(\frac{1}{4} + x_5\right) a\,\mathbf{\hat{y}} - z_5 a\,\mathbf{\hat{z}}$ | (96$g$) | O |
| $\mathbf{B}_{42}$ | = | $-z_5\,\mathbf{a}_1 - z_5\,\mathbf{a}_2 + (-2x_5 + z_5)\,\mathbf{a}_3$ | = | $-x_5 a\,\mathbf{\hat{x}} - x_5 a\,\mathbf{\hat{y}} - z_5 a\,\mathbf{\hat{z}}$ | (96$g$) | O |
| $\mathbf{B}_{43}$ | = | $(-2x_5 + z_5)\,\mathbf{a}_1 + \left(\frac{1}{2} + 2x_5 + z_5\right)\,\mathbf{a}_2 - z_5\,\mathbf{a}_3$ | = | $\left(\frac{1}{4} + x_5\right) a\,\mathbf{\hat{x}} - x_5 a\,\mathbf{\hat{y}} + \left(\frac{1}{4} + z_5\right) a\,\mathbf{\hat{z}}$ | (96$g$) | O |
| $\mathbf{B}_{44}$ | = | $\left(\frac{1}{2} + 2x_5 + z_5\right)\,\mathbf{a}_1 + (-2x_5 + z_5)\,\mathbf{a}_2 - z_5\,\mathbf{a}_3$ | = | $-x_5 a\,\mathbf{\hat{x}} + \left(\frac{1}{4} + x_5\right) a\,\mathbf{\hat{y}} + \left(\frac{1}{4} + z_5\right) a\,\mathbf{\hat{z}}$ | (96$g$) | O |
| $\mathbf{B}_{45}$ | = | $(-2x_5 + z_5)\,\mathbf{a}_1 - z_5\,\mathbf{a}_2 + \left(\frac{1}{2} + 2x_5 + z_5\right)\,\mathbf{a}_3$ | = | $\left(\frac{1}{4} + x_5\right) a\,\mathbf{\hat{x}} + \left(\frac{1}{4} + z_5\right) a\,\mathbf{\hat{y}} - x_5 a\,\mathbf{\hat{z}}$ | (96$g$) | O |
| $\mathbf{B}_{46}$ | = | $\left(\frac{1}{2} + 2x_5 + z_5\right)\,\mathbf{a}_1 - z_5\,\mathbf{a}_2 + (-2x_5 + z_5)\,\mathbf{a}_3$ | = | $-x_5 a\,\mathbf{\hat{x}} + \left(\frac{1}{4} + z_5\right) a\,\mathbf{\hat{y}} + \left(\frac{1}{4} + x_5\right) a\,\mathbf{\hat{z}}$ | (96$g$) | O |
| $\mathbf{B}_{47}$ | = | $-z_5\,\mathbf{a}_1 + (-2x_5 + z_5)\,\mathbf{a}_2 - z_5\,\mathbf{a}_3$ | = | $-x_5 a\,\mathbf{\hat{x}} - z_5 a\,\mathbf{\hat{y}} - x_5 a\,\mathbf{\hat{z}}$ | (96$g$) | O |
| $\mathbf{B}_{48}$ | = | $-z_5\,\mathbf{a}_1 + \left(\frac{1}{2} + 2x_5 + z_5\right)\,\mathbf{a}_2 - z_5\,\mathbf{a}_3$ | = | $\left(\frac{1}{4} + x_5\right) a\,\mathbf{\hat{x}} - z_5 a\,\mathbf{\hat{y}} + \left(\frac{1}{4} + x_5\right) a\,\mathbf{\hat{z}}$ | (96$g$) | O |
| $\mathbf{B}_{49}$ | = | $-z_5\,\mathbf{a}_1 + (-2x_5 + z_5)\,\mathbf{a}_2 + \left(\frac{1}{2} + 2x_5 + z_5\right)\,\mathbf{a}_3$ | = | $\left(\frac{1}{4} + z_5\right) a\,\mathbf{\hat{x}} + \left(\frac{1}{4} + x_5\right) a\,\mathbf{\hat{y}} - x_5 a\,\mathbf{\hat{z}}$ | (96$g$) | O |
| $\mathbf{B}_{50}$ | = | $-z_5\,\mathbf{a}_1 + \left(\frac{1}{2} + 2x_5 + z_5\right)\,\mathbf{a}_2 + (-2x_5 + z_5)\,\mathbf{a}_3$ | = | $\left(\frac{1}{4} + z_5\right) a\,\mathbf{\hat{x}} - x_5 a\,\mathbf{\hat{y}} + \left(\frac{1}{4} + x_5\right) a\,\mathbf{\hat{z}}$ | (96$g$) | O |
| $\mathbf{B}_{51}$ | = | $\left(\frac{1}{2} + 2x_5 + z_5\right)\,\mathbf{a}_1 - z_5\,\mathbf{a}_2 - z_5\,\mathbf{a}_3$ | = | $-z_5 a\,\mathbf{\hat{x}} + \left(\frac{1}{4} + x_5\right) a\,\mathbf{\hat{y}} + \left(\frac{1}{4} + x_5\right) a\,\mathbf{\hat{z}}$ | (96$g$) | O |
| $\mathbf{B}_{52}$ | = | $(-2x_5 + z_5)\,\mathbf{a}_1 - z_5\,\mathbf{a}_2 - z_5\,\mathbf{a}_3$ | = | $-z_5 a\,\mathbf{\hat{x}} - x_5 a\,\mathbf{\hat{y}} - x_5 a\,\mathbf{\hat{z}}$ | (96$g$) | O |

**Found in:**

**Geometry files:**

- CIF: pp. 1829
- POSCAR: pp. 1831

# *D*$6_2$ ($Sb_2O_4$) (*obsolete*) Structure: A2B_cF96_227_abf_cd

http://aflow.org/prototype-encyclopedia/A2B_cF96_227_abf_cd

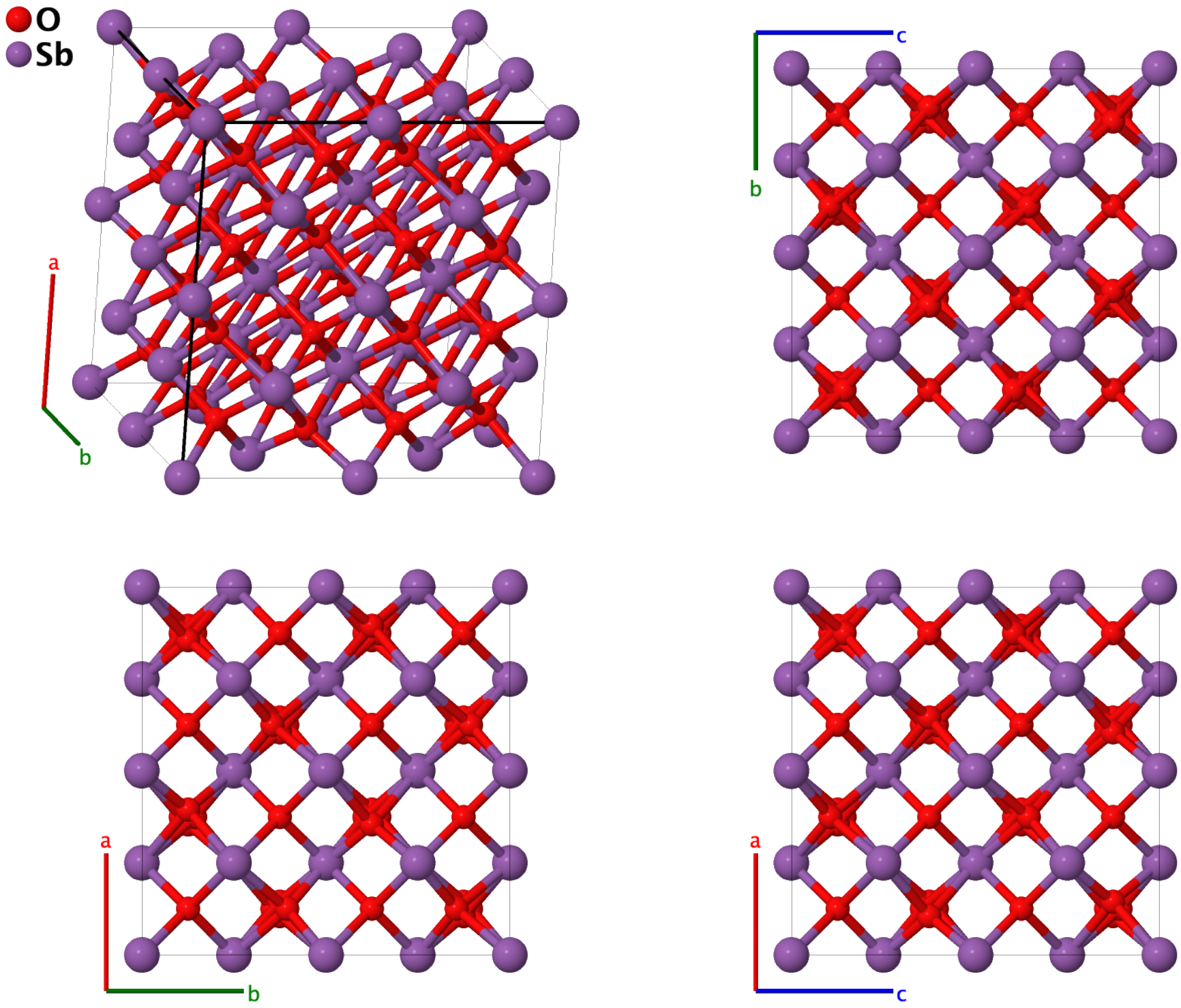

| | | |
|---|---|---|
| **Prototype** | : | $O_2Sb$ |
| **AFLOW prototype label** | : | A2B_cF96_227_abf_cd |
| ***Strukturbericht* designation** | : | *D*$6_2$ |
| **Pearson symbol** | : | cF96 |
| **Space group number** | : | 227 |
| **Space group symbol** | : | $Fd\bar{3}m$ |
| **AFLOW prototype command** | : | `aflow --proto=A2B_cF96_227_abf_cd --params=`$a, x_5$ |

- Shortly after (Gottfried, 1937) gave this compound the *Strukturbericht* designation *D*$6_2$, (Dihiström, 1937) showed that they were actually determining the structure of $Sb_3O_6OH$, making this structure obsolete. Indeed, (Herrman, 1943) formally withdraws this from the *Strukturbericht* list, saying "The type and description [in (Gottfried, 1937)] should be deleted, as the X-rays were not based on the supposed substance." We present it for its historical interest.

- Modern experiments have determined that $SbO_2$ appears as cervantite ($\alpha$-$Sb_2O_4$) or clinocervantite ($\beta$-$Sb_2O_4$).

## Face-centered Cubic primitive vectors:

$$\mathbf{a}_1 = \frac{1}{2}\,a\,\hat{\mathbf{y}} + \frac{1}{2}\,a\,\hat{\mathbf{z}}$$
$$\mathbf{a}_2 = \frac{1}{2}\,a\,\hat{\mathbf{x}} + \frac{1}{2}\,a\,\hat{\mathbf{z}}$$
$$\mathbf{a}_3 = \frac{1}{2}\,a\,\hat{\mathbf{x}} + \frac{1}{2}\,a\,\hat{\mathbf{y}}$$

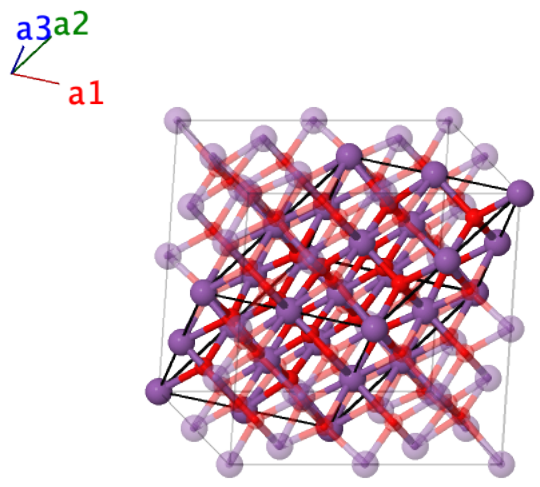

## Basis vectors:

| | | Lattice Coordinates | | Cartesian Coordinates | Wyckoff Position | Atom Type |
|---|---|---|---|---|---|---|
| $\mathbf{B}_1$ | = | $\frac{1}{8}\,\mathbf{a}_1 + \frac{1}{8}\,\mathbf{a}_2 + \frac{1}{8}\,\mathbf{a}_3$ | = | $\frac{1}{8}a\,\hat{\mathbf{x}} + \frac{1}{8}a\,\hat{\mathbf{y}} + \frac{1}{8}a\,\hat{\mathbf{z}}$ | (8$a$) | O I |
| $\mathbf{B}_2$ | = | $\frac{7}{8}\,\mathbf{a}_1 + \frac{7}{8}\,\mathbf{a}_2 + \frac{7}{8}\,\mathbf{a}_3$ | = | $\frac{7}{8}a\,\hat{\mathbf{x}} + \frac{7}{8}a\,\hat{\mathbf{y}} + \frac{7}{8}a\,\hat{\mathbf{z}}$ | (8$a$) | O I |
| $\mathbf{B}_3$ | = | $\frac{3}{8}\,\mathbf{a}_1 + \frac{3}{8}\,\mathbf{a}_2 + \frac{3}{8}\,\mathbf{a}_3$ | = | $\frac{3}{8}a\,\hat{\mathbf{x}} + \frac{3}{8}a\,\hat{\mathbf{y}} + \frac{3}{8}a\,\hat{\mathbf{z}}$ | (8$b$) | O II |
| $\mathbf{B}_4$ | = | $\frac{5}{8}\,\mathbf{a}_1 + \frac{5}{8}\,\mathbf{a}_2 + \frac{5}{8}\,\mathbf{a}_3$ | = | $\frac{5}{8}a\,\hat{\mathbf{x}} + \frac{5}{8}a\,\hat{\mathbf{y}} + \frac{5}{8}a\,\hat{\mathbf{z}}$ | (8$b$) | O II |
| $\mathbf{B}_5$ | = | $0\,\mathbf{a}_1 + 0\,\mathbf{a}_2 + 0\,\mathbf{a}_3$ | = | $0\,\hat{\mathbf{x}} + 0\,\hat{\mathbf{y}} + 0\,\hat{\mathbf{z}}$ | (16$c$) | Sb I |
| $\mathbf{B}_6$ | = | $\frac{1}{2}\,\mathbf{a}_3$ | = | $\frac{1}{4}a\,\hat{\mathbf{x}} + \frac{1}{4}a\,\hat{\mathbf{y}}$ | (16$c$) | Sb I |
| $\mathbf{B}_7$ | = | $\frac{1}{2}\,\mathbf{a}_2$ | = | $\frac{1}{4}a\,\hat{\mathbf{x}} + \frac{1}{4}a\,\hat{\mathbf{z}}$ | (16$c$) | Sb I |
| $\mathbf{B}_8$ | = | $\frac{1}{2}\,\mathbf{a}_1$ | = | $\frac{1}{4}a\,\hat{\mathbf{y}} + \frac{1}{4}a\,\hat{\mathbf{z}}$ | (16$c$) | Sb I |
| $\mathbf{B}_9$ | = | $\frac{1}{2}\,\mathbf{a}_1 + \frac{1}{2}\,\mathbf{a}_2 + \frac{1}{2}\,\mathbf{a}_3$ | = | $\frac{1}{2}a\,\hat{\mathbf{x}} + \frac{1}{2}a\,\hat{\mathbf{y}} + \frac{1}{2}a\,\hat{\mathbf{z}}$ | (16$d$) | Sb II |
| $\mathbf{B}_{10}$ | = | $\frac{1}{2}\,\mathbf{a}_1 + \frac{1}{2}\,\mathbf{a}_2$ | = | $\frac{1}{4}a\,\hat{\mathbf{x}} + \frac{1}{4}a\,\hat{\mathbf{y}} + \frac{1}{2}a\,\hat{\mathbf{z}}$ | (16$d$) | Sb II |
| $\mathbf{B}_{11}$ | = | $\frac{1}{2}\,\mathbf{a}_1 + \frac{1}{2}\,\mathbf{a}_3$ | = | $\frac{1}{4}a\,\hat{\mathbf{x}} + \frac{1}{2}a\,\hat{\mathbf{y}} + \frac{1}{4}a\,\hat{\mathbf{z}}$ | (16$d$) | Sb II |
| $\mathbf{B}_{12}$ | = | $\frac{1}{2}\,\mathbf{a}_2 + \frac{1}{2}\,\mathbf{a}_3$ | = | $\frac{1}{2}a\,\hat{\mathbf{x}} + \frac{1}{4}a\,\hat{\mathbf{y}} + \frac{1}{4}a\,\hat{\mathbf{z}}$ | (16$d$) | Sb II |
| $\mathbf{B}_{13}$ | = | $\left(\frac{1}{4} - x_5\right)\,\mathbf{a}_1 + x_5\,\mathbf{a}_2 + x_5\,\mathbf{a}_3$ | = | $x_5a\,\hat{\mathbf{x}} + \frac{1}{8}a\,\hat{\mathbf{y}} + \frac{1}{8}a\,\hat{\mathbf{z}}$ | (48$f$) | O III |
| $\mathbf{B}_{14}$ | = | $x_5\,\mathbf{a}_1 + \left(\frac{1}{4} - x_5\right)\,\mathbf{a}_2 + \left(\frac{1}{4} - x_5\right)\,\mathbf{a}_3$ | = | $\left(\frac{1}{4} - x_5\right)a\,\hat{\mathbf{x}} + \frac{1}{8}a\,\hat{\mathbf{y}} + \frac{1}{8}a\,\hat{\mathbf{z}}$ | (48$f$) | O III |
| $\mathbf{B}_{15}$ | = | $x_5\,\mathbf{a}_1 + \left(\frac{1}{4} - x_5\right)\,\mathbf{a}_2 + x_5\,\mathbf{a}_3$ | = | $\frac{1}{8}a\,\hat{\mathbf{x}} + x_5a\,\hat{\mathbf{y}} + \frac{1}{8}a\,\hat{\mathbf{z}}$ | (48$f$) | O III |
| $\mathbf{B}_{16}$ | = | $\left(\frac{1}{4} - x_5\right)\,\mathbf{a}_1 + x_5\,\mathbf{a}_2 + \left(\frac{1}{4} - x_5\right)\,\mathbf{a}_3$ | = | $\frac{1}{8}a\,\hat{\mathbf{x}} + \left(\frac{1}{4} - x_5\right)a\,\hat{\mathbf{y}} + \frac{1}{8}a\,\hat{\mathbf{z}}$ | (48$f$) | O III |
| $\mathbf{B}_{17}$ | = | $x_5\,\mathbf{a}_1 + x_5\,\mathbf{a}_2 + \left(\frac{1}{4} - x_5\right)\,\mathbf{a}_3$ | = | $\frac{1}{8}a\,\hat{\mathbf{x}} + \frac{1}{8}a\,\hat{\mathbf{y}} + x_5a\,\hat{\mathbf{z}}$ | (48$f$) | O III |
| $\mathbf{B}_{18}$ | = | $\left(\frac{1}{4} - x_5\right)\,\mathbf{a}_1 + \left(\frac{1}{4} - x_5\right)\,\mathbf{a}_2 + x_5\,\mathbf{a}_3$ | = | $\frac{1}{8}a\,\hat{\mathbf{x}} + \frac{1}{8}a\,\hat{\mathbf{y}} + \left(\frac{1}{4} - x_5\right)a\,\hat{\mathbf{z}}$ | (48$f$) | O III |
| $\mathbf{B}_{19}$ | = | $\left(\frac{3}{4} + x_5\right)\,\mathbf{a}_1 - x_5\,\mathbf{a}_2 + \left(\frac{3}{4} + x_5\right)\,\mathbf{a}_3$ | = | $\frac{3}{8}a\,\hat{\mathbf{x}} + \left(\frac{3}{4} + x_5\right)a\,\hat{\mathbf{y}} + \frac{3}{8}a\,\hat{\mathbf{z}}$ | (48$f$) | O III |
| $\mathbf{B}_{20}$ | = | $-x_5\,\mathbf{a}_1 + \left(\frac{3}{4} + x_5\right)\,\mathbf{a}_2 - x_5\,\mathbf{a}_3$ | = | $\frac{3}{8}a\,\hat{\mathbf{x}} - x_5a\,\hat{\mathbf{y}} + \frac{3}{8}a\,\hat{\mathbf{z}}$ | (48$f$) | O III |
| $\mathbf{B}_{21}$ | = | $-x_5\,\mathbf{a}_1 + \left(\frac{3}{4} + x_5\right)\,\mathbf{a}_2 + \left(\frac{3}{4} + x_5\right)\,\mathbf{a}_3$ | = | $\left(\frac{3}{4} + x_5\right)a\,\hat{\mathbf{x}} + \frac{3}{8}a\,\hat{\mathbf{y}} + \frac{3}{8}a\,\hat{\mathbf{z}}$ | (48$f$) | O III |
| $\mathbf{B}_{22}$ | = | $\left(\frac{3}{4} + x_5\right)\,\mathbf{a}_1 - x_5\,\mathbf{a}_2 - x_5\,\mathbf{a}_3$ | = | $-x_5a\,\hat{\mathbf{x}} + \frac{3}{8}a\,\hat{\mathbf{y}} + \frac{3}{8}a\,\hat{\mathbf{z}}$ | (48$f$) | O III |
| $\mathbf{B}_{23}$ | = | $-x_5\,\mathbf{a}_1 - x_5\,\mathbf{a}_2 + \left(\frac{3}{4} + x_5\right)\,\mathbf{a}_3$ | = | $\frac{3}{8}a\,\hat{\mathbf{x}} + \frac{3}{8}a\,\hat{\mathbf{y}} - x_5a\,\hat{\mathbf{z}}$ | (48$f$) | O III |
| $\mathbf{B}_{24}$ | = | $\left(\frac{3}{4} + x_5\right)\,\mathbf{a}_1 + \left(\frac{3}{4} + x_5\right)\,\mathbf{a}_2 - x_5\,\mathbf{a}_3$ | = | $\frac{3}{8}a\,\hat{\mathbf{x}} + \frac{3}{8}a\,\hat{\mathbf{y}} + \left(\frac{3}{4} + x_5\right)a\,\hat{\mathbf{z}}$ | (48$f$) | O III |

**Found in:**

---

**Geometry files:**

- CIF: pp. 1831
- POSCAR: pp. 1832

# Senarmontite ($Sb_2O_3$, $D6_1$) Structure: A3B2_cF80_227_f_e

http://aflow.org/prototype-encyclopedia/A3B2_cF80_227_f_e

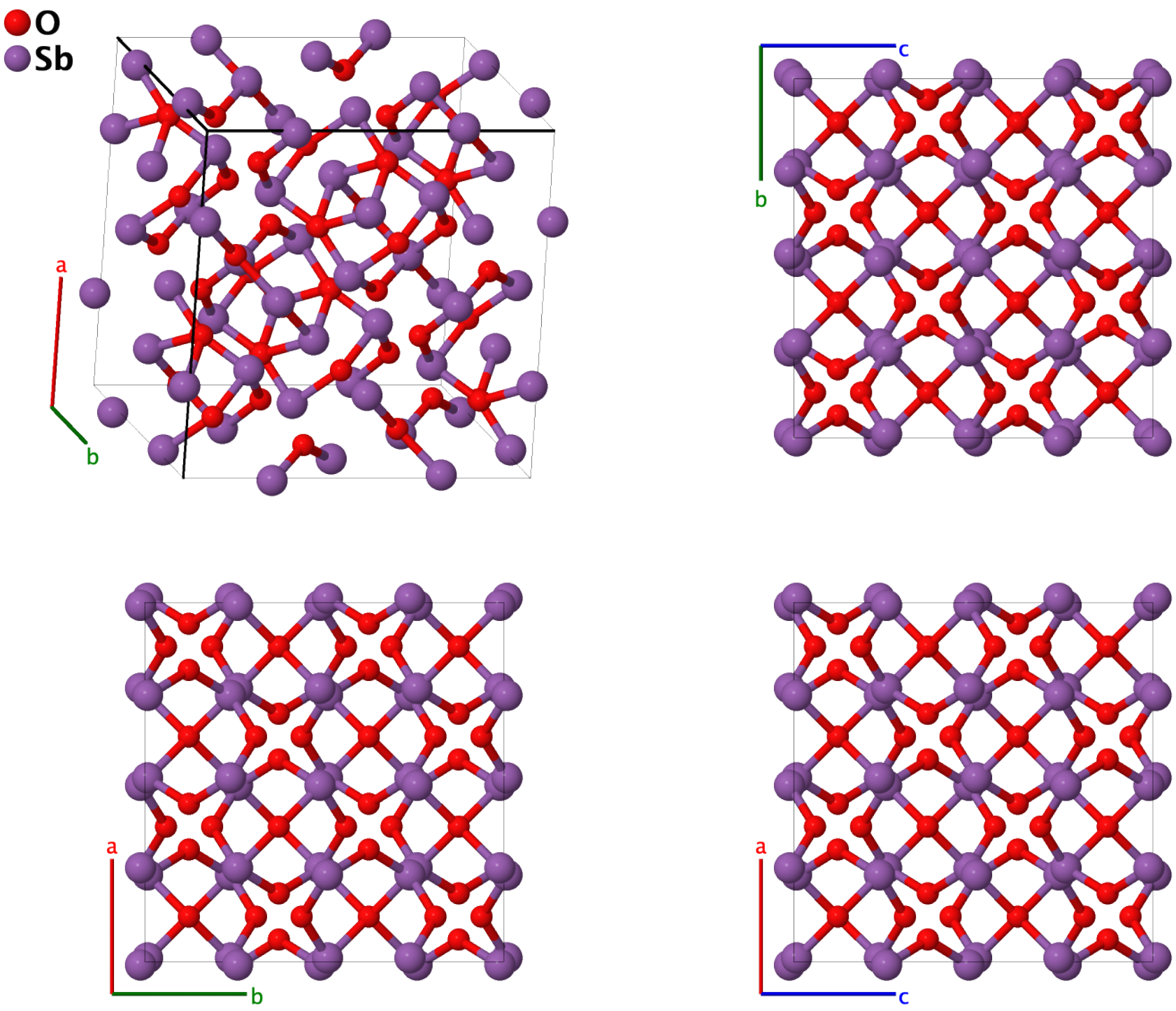

| | | |
|---|---|---|
| **Prototype** | : | $O_3Sb_2$ |
| **AFLOW prototype label** | : | A3B2_cF80_227_f_e |
| ***Strukturbericht* designation** | : | None |
| **Pearson symbol** | : | cF80 |
| **Space group number** | : | 227 |
| **Space group symbol** | : | $Fd\bar{3}m$ |
| **AFLOW prototype command** | : | `aflow --proto=A3B2_cF80_227_f_e --params=`$a, x_1, x_2$ |

**Other compounds with this structure**

- $As_2O_3$ (Arsenolite)

- (Ewald, 1931) designated this as *Strukturbericht* $D6_1$, however (Parthé, 1993) and (Villars, 1991) label this as *Strukturbericht* $D5_4$, and Parthé uses $As_2O_3$ as the prototype. While this structure obviously fits better with the $D5$ series ($A_2B_3$) than $D6$ ($A_2B_4$), the $D5_4$ structure was (inadvertently?) omitted from (Hermann, 1937), which jumps from $D5_3$ to $D5_5$. We will follow this historical record (Ewald, 1931) here.
- This is the cubic form of $Sb_2O_3$. For the orthorhombic form see the Valentinite ($D5_{11}$) structure.
- (Svensson, 1975) gave the atomic coordinates in Setting 1 of space group $Fd\bar{3}m$ #227. We used FINDSYM to shift the coordinates to the standard setting 2.

**Face-centered Cubic primitive vectors:**

$$
\begin{aligned}
\mathbf{a}_1 &= \tfrac{1}{2}\,a\,\hat{\mathbf{y}} + \tfrac{1}{2}\,a\,\hat{\mathbf{z}} \\
\mathbf{a}_2 &= \tfrac{1}{2}\,a\,\hat{\mathbf{x}} + \tfrac{1}{2}\,a\,\hat{\mathbf{z}} \\
\mathbf{a}_3 &= \tfrac{1}{2}\,a\,\hat{\mathbf{x}} + \tfrac{1}{2}\,a\,\hat{\mathbf{y}}
\end{aligned}
$$

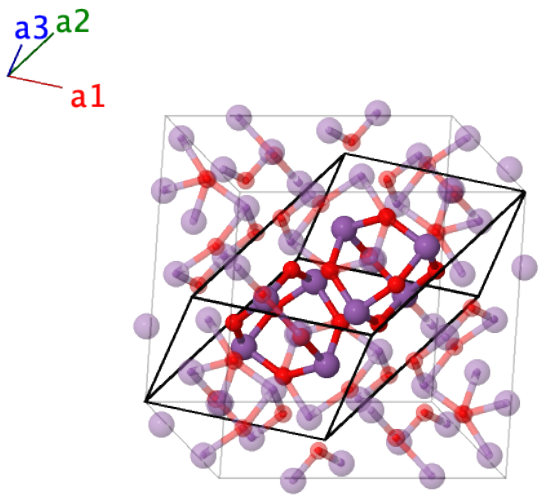

**Basis vectors:**

| | | Lattice Coordinates | | Cartesian Coordinates | Wyckoff Position | Atom Type |
|---|---|---|---|---|---|---|
| $\mathbf{B}_1$ | = | $x_1\,\mathbf{a}_1 + x_1\,\mathbf{a}_2 + x_1\,\mathbf{a}_3$ | = | $x_1 a\,\hat{\mathbf{x}} + x_1 a\,\hat{\mathbf{y}} + x_1 a\,\hat{\mathbf{z}}$ | (32$e$) | Sb |
| $\mathbf{B}_2$ | = | $x_1\,\mathbf{a}_1 + x_1\,\mathbf{a}_2 + \left(\frac{1}{2} - 3x_1\right)\mathbf{a}_3$ | = | $\left(\frac{1}{4} - x_1\right) a\,\hat{\mathbf{x}} + \left(\frac{1}{4} - x_1\right) a\,\hat{\mathbf{y}} + x_1 a\,\hat{\mathbf{z}}$ | (32$e$) | Sb |
| $\mathbf{B}_3$ | = | $x_1\,\mathbf{a}_1 + \left(\frac{1}{2} - 3x_1\right)\mathbf{a}_2 + x_1\,\mathbf{a}_3$ | = | $\left(\frac{1}{4} - x_1\right) a\,\hat{\mathbf{x}} + x_1 a\,\hat{\mathbf{y}} + \left(\frac{1}{4} - x_1\right) a\,\hat{\mathbf{z}}$ | (32$e$) | Sb |
| $\mathbf{B}_4$ | = | $\left(\frac{1}{2} - 3x_1\right)\mathbf{a}_1 + x_1\,\mathbf{a}_2 + x_1\,\mathbf{a}_3$ | = | $x_1 a\,\hat{\mathbf{x}} + \left(\frac{1}{4} - x_1\right) a\,\hat{\mathbf{y}} + \left(\frac{1}{4} - x_1\right) a\,\hat{\mathbf{z}}$ | (32$e$) | Sb |
| $\mathbf{B}_5$ | = | $-x_1\,\mathbf{a}_1 - x_1\,\mathbf{a}_2 + \left(\frac{1}{2} + 3x_1\right)\mathbf{a}_3$ | = | $\left(\frac{1}{4} + x_1\right) a\,\hat{\mathbf{x}} + \left(\frac{1}{4} + x_1\right) a\,\hat{\mathbf{y}} - x_1 a\,\hat{\mathbf{z}}$ | (32$e$) | Sb |
| $\mathbf{B}_6$ | = | $-x_1\,\mathbf{a}_1 - x_1\,\mathbf{a}_2 - x_1\,\mathbf{a}_3$ | = | $-x_1 a\,\hat{\mathbf{x}} - x_1 a\,\hat{\mathbf{y}} - x_1 a\,\hat{\mathbf{z}}$ | (32$e$) | Sb |
| $\mathbf{B}_7$ | = | $-x_1\,\mathbf{a}_1 + \left(\frac{1}{2} + 3x_1\right)\mathbf{a}_2 - x_1\,\mathbf{a}_3$ | = | $\left(\frac{1}{4} + x_1\right) a\,\hat{\mathbf{x}} - x_1 a\,\hat{\mathbf{y}} + \left(\frac{1}{4} + x_1\right) a\,\hat{\mathbf{z}}$ | (32$e$) | Sb |
| $\mathbf{B}_8$ | = | $\left(\frac{1}{2} + 3x_1\right)\mathbf{a}_1 - x_1\,\mathbf{a}_2 - x_1\,\mathbf{a}_3$ | = | $-x_1 a\,\hat{\mathbf{x}} + \left(\frac{1}{4} + x_1\right) a\,\hat{\mathbf{y}} + \left(\frac{1}{4} + x_1\right) a\,\hat{\mathbf{z}}$ | (32$e$) | Sb |
| $\mathbf{B}_9$ | = | $\left(\frac{1}{4} - x_2\right)\mathbf{a}_1 + x_2\,\mathbf{a}_2 + x_2\,\mathbf{a}_3$ | = | $x_2 a\,\hat{\mathbf{x}} + \frac{1}{8} a\,\hat{\mathbf{y}} + \frac{1}{8} a\,\hat{\mathbf{z}}$ | (48$f$) | O |
| $\mathbf{B}_{10}$ | = | $x_2\,\mathbf{a}_1 + \left(\frac{1}{4} - x_2\right)\mathbf{a}_2 + \left(\frac{1}{4} - x_2\right)\mathbf{a}_3$ | = | $\left(\frac{1}{4} - x_2\right) a\,\hat{\mathbf{x}} + \frac{1}{8} a\,\hat{\mathbf{y}} + \frac{1}{8} a\,\hat{\mathbf{z}}$ | (48$f$) | O |
| $\mathbf{B}_{11}$ | = | $x_2\,\mathbf{a}_1 + \left(\frac{1}{4} - x_2\right)\mathbf{a}_2 + x_2\,\mathbf{a}_3$ | = | $\frac{1}{8} a\,\hat{\mathbf{x}} + x_2 a\,\hat{\mathbf{y}} + \frac{1}{8} a\,\hat{\mathbf{z}}$ | (48$f$) | O |
| $\mathbf{B}_{12}$ | = | $\left(\frac{1}{4} - x_2\right)\mathbf{a}_1 + x_2\,\mathbf{a}_2 + \left(\frac{1}{4} - x_2\right)\mathbf{a}_3$ | = | $\frac{1}{8} a\,\hat{\mathbf{x}} + \left(\frac{1}{4} - x_2\right) a\,\hat{\mathbf{y}} + \frac{1}{8} a\,\hat{\mathbf{z}}$ | (48$f$) | O |
| $\mathbf{B}_{13}$ | = | $x_2\,\mathbf{a}_1 + x_2\,\mathbf{a}_2 + \left(\frac{1}{4} - x_2\right)\mathbf{a}_3$ | = | $\frac{1}{8} a\,\hat{\mathbf{x}} + \frac{1}{8} a\,\hat{\mathbf{y}} + x_2 a\,\hat{\mathbf{z}}$ | (48$f$) | O |
| $\mathbf{B}_{14}$ | = | $\left(\frac{1}{4} - x_2\right)\mathbf{a}_1 + \left(\frac{1}{4} - x_2\right)\mathbf{a}_2 + x_2\,\mathbf{a}_3$ | = | $\frac{1}{8} a\,\hat{\mathbf{x}} + \frac{1}{8} a\,\hat{\mathbf{y}} + \left(\frac{1}{4} - x_2\right) a\,\hat{\mathbf{z}}$ | (48$f$) | O |
| $\mathbf{B}_{15}$ | = | $\left(\frac{3}{4} + x_2\right)\mathbf{a}_1 - x_2\,\mathbf{a}_2 + \left(\frac{3}{4} + x_2\right)\mathbf{a}_3$ | = | $\frac{3}{8} a\,\hat{\mathbf{x}} + \left(\frac{3}{4} + x_2\right) a\,\hat{\mathbf{y}} + \frac{3}{8} a\,\hat{\mathbf{z}}$ | (48$f$) | O |
| $\mathbf{B}_{16}$ | = | $-x_2\,\mathbf{a}_1 + \left(\frac{3}{4} + x_2\right)\mathbf{a}_2 - x_2\,\mathbf{a}_3$ | = | $\frac{3}{8} a\,\hat{\mathbf{x}} - x_2 a\,\hat{\mathbf{y}} + \frac{3}{8} a\,\hat{\mathbf{z}}$ | (48$f$) | O |
| $\mathbf{B}_{17}$ | = | $-x_2\,\mathbf{a}_1 + \left(\frac{3}{4} + x_2\right)\mathbf{a}_2 + \left(\frac{3}{4} + x_2\right)\mathbf{a}_3$ | = | $\left(\frac{3}{4} + x_2\right) a\,\hat{\mathbf{x}} + \frac{3}{8} a\,\hat{\mathbf{y}} + \frac{3}{8} a\,\hat{\mathbf{z}}$ | (48$f$) | O |
| $\mathbf{B}_{18}$ | = | $\left(\frac{3}{4} + x_2\right)\mathbf{a}_1 - x_2\,\mathbf{a}_2 - x_2\,\mathbf{a}_3$ | = | $-x_2 a\,\hat{\mathbf{x}} + \frac{3}{8} a\,\hat{\mathbf{y}} + \frac{3}{8} a\,\hat{\mathbf{z}}$ | (48$f$) | O |
| $\mathbf{B}_{19}$ | = | $-x_2\,\mathbf{a}_1 - x_2\,\mathbf{a}_2 + \left(\frac{3}{4} + x_2\right)\mathbf{a}_3$ | = | $\frac{3}{8} a\,\hat{\mathbf{x}} + \frac{3}{8} a\,\hat{\mathbf{y}} - x_2 a\,\hat{\mathbf{z}}$ | (48$f$) | O |
| $\mathbf{B}_{20}$ | = | $\left(\frac{3}{4} + x_2\right)\mathbf{a}_1 + \left(\frac{3}{4} + x_2\right)\mathbf{a}_2 - x_2\,\mathbf{a}_3$ | = | $\frac{3}{8} a\,\hat{\mathbf{x}} + \frac{3}{8} a\,\hat{\mathbf{y}} + \left(\frac{3}{4} + x_2\right) a\,\hat{\mathbf{z}}$ | (48$f$) | O |

---

**Geometry files:**

- CIF: pp. 1832
- POSCAR: pp. 1833

# $H5_6$ [Tychite, $Na_6Mg_2SO_4(CO_3)_4$)] (*obsolete*) Structure: A4B2C6D16E_cF232_227_e_d_f_eg_a

http://aflow.org/prototype-encyclopedia/A4B2C6D16E_cF232_227_e_d_f_eg_a

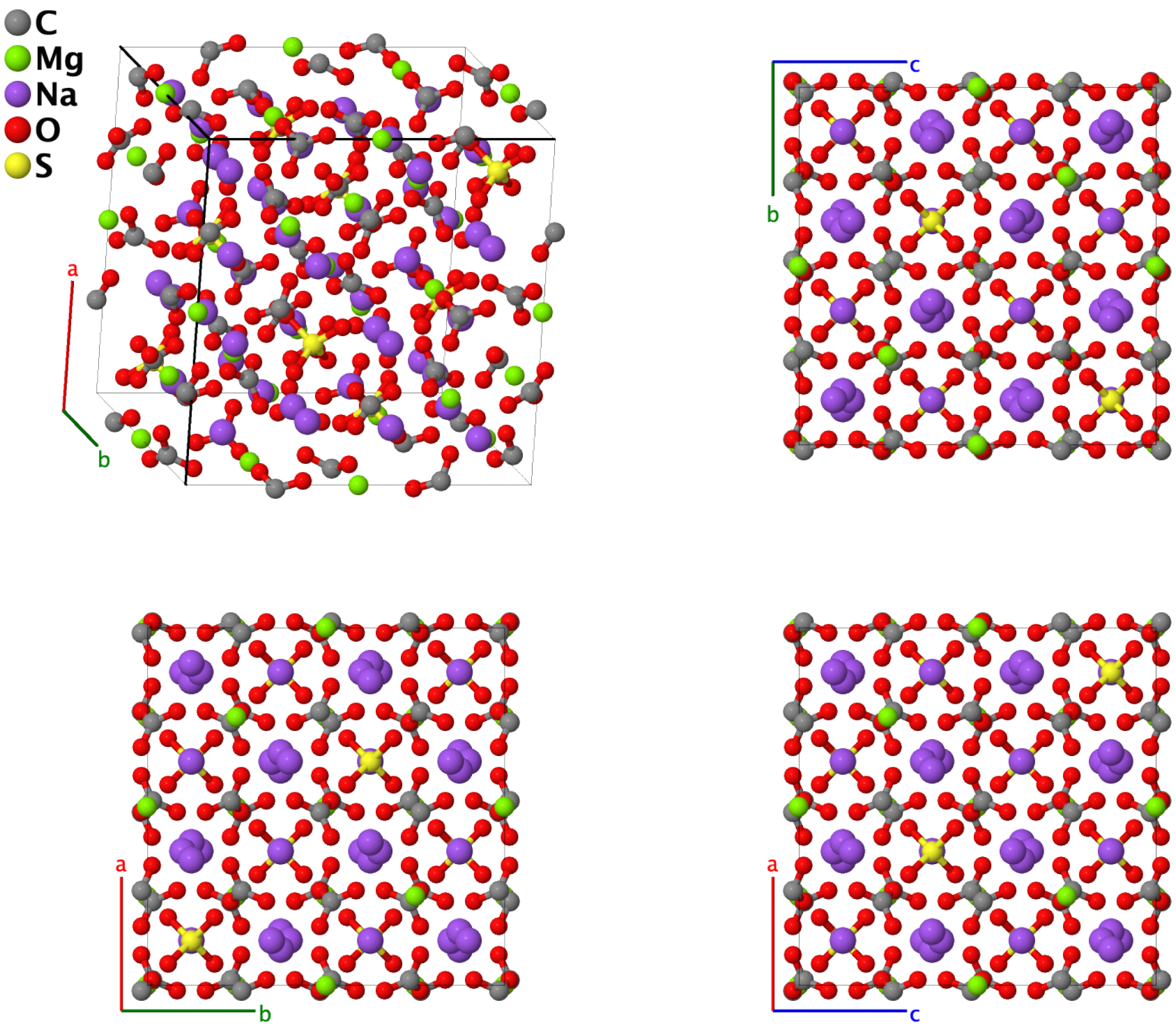

| | | |
|---|---|---|
| **Prototype** | : | $C_4Mg_2Na_6O_{16}S$ |
| **AFLOW prototype label** | : | A4B2C6D16E_cF232_227_e_d_f_eg_a |
| ***Strukturbericht* designation** | : | $H5_7$ |
| **Pearson symbol** | : | cF232 |
| **Space group number** | : | 227 |
| **Space group symbol** | : | $Fd\bar{3}m$ |
| **AFLOW prototype command** | : | `aflow --proto=A4B2C6D16E_cF232_227_e_d_f_eg_a` `--params=`$a, x_3, x_4, x_5, x_6, z_6$ |

- This is the original structure determined by (Shiba, 1931) and given the designation $H5_6$ in (Hermann, 1937). (Schmidt, 2006) showed that the true structure is in space group $Fd\bar{3}$ #203, however the two structures are very similar, and a displacement of the oxygen atoms by less than 1 Å brings the two structures into agreement.

- (Hermann, 1937) gives the chemical formula as $Na_6Mg_2SO_4(CO_3)_2$, but the given Wyckoff positions are in agreement with the correct formula, $Na_6Mg_2SO_4(CO_3)_4$.

## Face-centered Cubic primitive vectors:

$$\mathbf{a}_1 = \frac{1}{2}\,a\,\hat{\mathbf{y}} + \frac{1}{2}\,a\,\hat{\mathbf{z}}$$
$$\mathbf{a}_2 = \frac{1}{2}\,a\,\hat{\mathbf{x}} + \frac{1}{2}\,a\,\hat{\mathbf{z}}$$
$$\mathbf{a}_3 = \frac{1}{2}\,a\,\hat{\mathbf{x}} + \frac{1}{2}\,a\,\hat{\mathbf{y}}$$

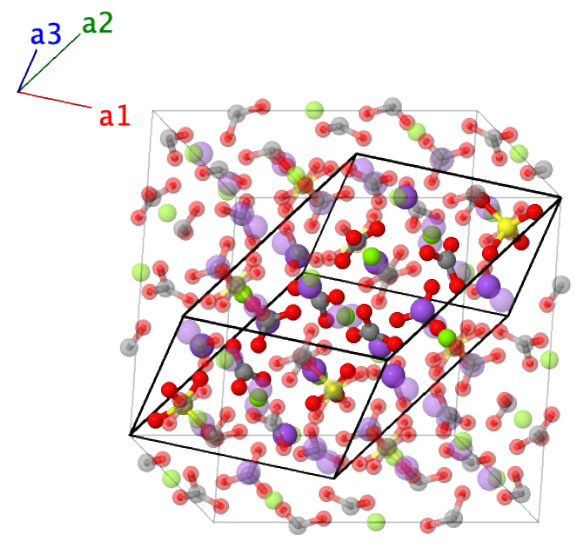

## Basis vectors:

| | | Lattice Coordinates | | Cartesian Coordinates | Wyckoff Position | Atom Type |
|---|---|---|---|---|---|---|
| $\mathbf{B}_1$ | = | $\frac{1}{8}\,\mathbf{a}_1 + \frac{1}{8}\,\mathbf{a}_2 + \frac{1}{8}\,\mathbf{a}_3$ | = | $\frac{1}{8}a\,\hat{\mathbf{x}} + \frac{1}{8}a\,\hat{\mathbf{y}} + \frac{1}{8}a\,\hat{\mathbf{z}}$ | (8$a$) | S |
| $\mathbf{B}_2$ | = | $\frac{7}{8}\,\mathbf{a}_1 + \frac{7}{8}\,\mathbf{a}_2 + \frac{7}{8}\,\mathbf{a}_3$ | = | $\frac{7}{8}a\,\hat{\mathbf{x}} + \frac{7}{8}a\,\hat{\mathbf{y}} + \frac{7}{8}a\,\hat{\mathbf{z}}$ | (8$a$) | S |
| $\mathbf{B}_3$ | = | $\frac{1}{2}\,\mathbf{a}_1 + \frac{1}{2}\,\mathbf{a}_2 + \frac{1}{2}\,\mathbf{a}_3$ | = | $\frac{1}{2}a\,\hat{\mathbf{x}} + \frac{1}{2}a\,\hat{\mathbf{y}} + \frac{1}{2}a\,\hat{\mathbf{z}}$ | (16$d$) | Mg |
| $\mathbf{B}_4$ | = | $\frac{1}{2}\,\mathbf{a}_1 + \frac{1}{2}\,\mathbf{a}_2$ | = | $\frac{1}{4}a\,\hat{\mathbf{x}} + \frac{1}{4}a\,\hat{\mathbf{y}} + \frac{1}{2}a\,\hat{\mathbf{z}}$ | (16$d$) | Mg |
| $\mathbf{B}_5$ | = | $\frac{1}{2}\,\mathbf{a}_1 + \frac{1}{2}\,\mathbf{a}_3$ | = | $\frac{1}{4}a\,\hat{\mathbf{x}} + \frac{1}{2}a\,\hat{\mathbf{y}} + \frac{1}{4}a\,\hat{\mathbf{z}}$ | (16$d$) | Mg |
| $\mathbf{B}_6$ | = | $\frac{1}{2}\,\mathbf{a}_2 + \frac{1}{2}\,\mathbf{a}_3$ | = | $\frac{1}{2}a\,\hat{\mathbf{x}} + \frac{1}{4}a\,\hat{\mathbf{y}} + \frac{1}{4}a\,\hat{\mathbf{z}}$ | (16$d$) | Mg |
| $\mathbf{B}_7$ | = | $x_3\,\mathbf{a}_1 + x_3\,\mathbf{a}_2 + x_3\,\mathbf{a}_3$ | = | $x_3a\,\hat{\mathbf{x}} + x_3a\,\hat{\mathbf{y}} + x_3a\,\hat{\mathbf{z}}$ | (32$e$) | C |
| $\mathbf{B}_8$ | = | $x_3\,\mathbf{a}_1 + x_3\,\mathbf{a}_2 + \left(\frac{1}{2} - 3x_3\right)\,\mathbf{a}_3$ | = | $\left(\frac{1}{4} - x_3\right)a\,\hat{\mathbf{x}} + \left(\frac{1}{4} - x_3\right)a\,\hat{\mathbf{y}} + x_3a\,\hat{\mathbf{z}}$ | (32$e$) | C |
| $\mathbf{B}_9$ | = | $x_3\,\mathbf{a}_1 + \left(\frac{1}{2} - 3x_3\right)\,\mathbf{a}_2 + x_3\,\mathbf{a}_3$ | = | $\left(\frac{1}{4} - x_3\right)a\,\hat{\mathbf{x}} + x_3a\,\hat{\mathbf{y}} + \left(\frac{1}{4} - x_3\right)a\,\hat{\mathbf{z}}$ | (32$e$) | C |
| $\mathbf{B}_{10}$ | = | $\left(\frac{1}{2} - 3x_3\right)\,\mathbf{a}_1 + x_3\,\mathbf{a}_2 + x_3\,\mathbf{a}_3$ | = | $x_3a\,\hat{\mathbf{x}} + \left(\frac{1}{4} - x_3\right)a\,\hat{\mathbf{y}} + \left(\frac{1}{4} - x_3\right)a\,\hat{\mathbf{z}}$ | (32$e$) | C |
| $\mathbf{B}_{11}$ | = | $-x_3\,\mathbf{a}_1 - x_3\,\mathbf{a}_2 + \left(\frac{1}{2} + 3x_3\right)\,\mathbf{a}_3$ | = | $\left(\frac{1}{4} + x_3\right)a\,\hat{\mathbf{x}} + \left(\frac{1}{4} + x_3\right)a\,\hat{\mathbf{y}} - x_3a\,\hat{\mathbf{z}}$ | (32$e$) | C |
| $\mathbf{B}_{12}$ | = | $-x_3\,\mathbf{a}_1 - x_3\,\mathbf{a}_2 - x_3\,\mathbf{a}_3$ | = | $-x_3a\,\hat{\mathbf{x}} - x_3a\,\hat{\mathbf{y}} - x_3a\,\hat{\mathbf{z}}$ | (32$e$) | C |
| $\mathbf{B}_{13}$ | = | $-x_3\,\mathbf{a}_1 + \left(\frac{1}{2} + 3x_3\right)\,\mathbf{a}_2 - x_3\,\mathbf{a}_3$ | = | $\left(\frac{1}{4} + x_3\right)a\,\hat{\mathbf{x}} - x_3a\,\hat{\mathbf{y}} + \left(\frac{1}{4} + x_3\right)a\,\hat{\mathbf{z}}$ | (32$e$) | C |
| $\mathbf{B}_{14}$ | = | $\left(\frac{1}{2} + 3x_3\right)\,\mathbf{a}_1 - x_3\,\mathbf{a}_2 - x_3\,\mathbf{a}_3$ | = | $-x_3a\,\hat{\mathbf{x}} + \left(\frac{1}{4} + x_3\right)a\,\hat{\mathbf{y}} + \left(\frac{1}{4} + x_3\right)a\,\hat{\mathbf{z}}$ | (32$e$) | C |
| $\mathbf{B}_{15}$ | = | $x_4\,\mathbf{a}_1 + x_4\,\mathbf{a}_2 + x_4\,\mathbf{a}_3$ | = | $x_4a\,\hat{\mathbf{x}} + x_4a\,\hat{\mathbf{y}} + x_4a\,\hat{\mathbf{z}}$ | (32$e$) | O I |
| $\mathbf{B}_{16}$ | = | $x_4\,\mathbf{a}_1 + x_4\,\mathbf{a}_2 + \left(\frac{1}{2} - 3x_4\right)\,\mathbf{a}_3$ | = | $\left(\frac{1}{4} - x_4\right)a\,\hat{\mathbf{x}} + \left(\frac{1}{4} - x_4\right)a\,\hat{\mathbf{y}} + x_4a\,\hat{\mathbf{z}}$ | (32$e$) | O I |
| $\mathbf{B}_{17}$ | = | $x_4\,\mathbf{a}_1 + \left(\frac{1}{2} - 3x_4\right)\,\mathbf{a}_2 + x_4\,\mathbf{a}_3$ | = | $\left(\frac{1}{4} - x_4\right)a\,\hat{\mathbf{x}} + x_4a\,\hat{\mathbf{y}} + \left(\frac{1}{4} - x_4\right)a\,\hat{\mathbf{z}}$ | (32$e$) | O I |
| $\mathbf{B}_{18}$ | = | $\left(\frac{1}{2} - 3x_4\right)\,\mathbf{a}_1 + x_4\,\mathbf{a}_2 + x_4\,\mathbf{a}_3$ | = | $x_4a\,\hat{\mathbf{x}} + \left(\frac{1}{4} - x_4\right)a\,\hat{\mathbf{y}} + \left(\frac{1}{4} - x_4\right)a\,\hat{\mathbf{z}}$ | (32$e$) | O I |
| $\mathbf{B}_{19}$ | = | $-x_4\,\mathbf{a}_1 - x_4\,\mathbf{a}_2 + \left(\frac{1}{2} + 3x_4\right)\,\mathbf{a}_3$ | = | $\left(\frac{1}{4} + x_4\right)a\,\hat{\mathbf{x}} + \left(\frac{1}{4} + x_4\right)a\,\hat{\mathbf{y}} - x_4a\,\hat{\mathbf{z}}$ | (32$e$) | O I |
| $\mathbf{B}_{20}$ | = | $-x_4\,\mathbf{a}_1 - x_4\,\mathbf{a}_2 - x_4\,\mathbf{a}_3$ | = | $-x_4a\,\hat{\mathbf{x}} - x_4a\,\hat{\mathbf{y}} - x_4a\,\hat{\mathbf{z}}$ | (32$e$) | O I |
| $\mathbf{B}_{21}$ | = | $-x_4\,\mathbf{a}_1 + \left(\frac{1}{2} + 3x_4\right)\,\mathbf{a}_2 - x_4\,\mathbf{a}_3$ | = | $\left(\frac{1}{4} + x_4\right)a\,\hat{\mathbf{x}} - x_4a\,\hat{\mathbf{y}} + \left(\frac{1}{4} + x_4\right)a\,\hat{\mathbf{z}}$ | (32$e$) | O I |
| $\mathbf{B}_{22}$ | = | $\left(\frac{1}{2} + 3x_4\right)\,\mathbf{a}_1 - x_4\,\mathbf{a}_2 - x_4\,\mathbf{a}_3$ | = | $-x_4a\,\hat{\mathbf{x}} + \left(\frac{1}{4} + x_4\right)a\,\hat{\mathbf{y}} + \left(\frac{1}{4} + x_4\right)a\,\hat{\mathbf{z}}$ | (32$e$) | O I |
| $\mathbf{B}_{23}$ | = | $\left(\frac{1}{4} - x_5\right)\,\mathbf{a}_1 + x_5\,\mathbf{a}_2 + x_5\,\mathbf{a}_3$ | = | $x_5a\,\hat{\mathbf{x}} + \frac{1}{8}a\,\hat{\mathbf{y}} + \frac{1}{8}a\,\hat{\mathbf{z}}$ | (48$f$) | Na |
| $\mathbf{B}_{24}$ | = | $x_5\,\mathbf{a}_1 + \left(\frac{1}{4} - x_5\right)\,\mathbf{a}_2 + \left(\frac{1}{4} - x_5\right)\,\mathbf{a}_3$ | = | $\left(\frac{1}{4} - x_5\right)a\,\hat{\mathbf{x}} + \frac{1}{8}a\,\hat{\mathbf{y}} + \frac{1}{8}a\,\hat{\mathbf{z}}$ | (48$f$) | Na |
| $\mathbf{B}_{25}$ | = | $x_5\,\mathbf{a}_1 + \left(\frac{1}{4} - x_5\right)\,\mathbf{a}_2 + x_5\,\mathbf{a}_3$ | = | $\frac{1}{8}a\,\hat{\mathbf{x}} + x_5a\,\hat{\mathbf{y}} + \frac{1}{8}a\,\hat{\mathbf{z}}$ | (48$f$) | Na |
| $\mathbf{B}_{26}$ | = | $\left(\frac{1}{4} - x_5\right)\,\mathbf{a}_1 + x_5\,\mathbf{a}_2 + \left(\frac{1}{4} - x_5\right)\,\mathbf{a}_3$ | = | $\frac{1}{8}a\,\hat{\mathbf{x}} + \left(\frac{1}{4} - x_5\right)a\,\hat{\mathbf{y}} + \frac{1}{8}a\,\hat{\mathbf{z}}$ | (48$f$) | Na |
| $\mathbf{B}_{27}$ | = | $x_5\,\mathbf{a}_1 + x_5\,\mathbf{a}_2 + \left(\frac{1}{4} - x_5\right)\,\mathbf{a}_3$ | = | $\frac{1}{8}a\,\hat{\mathbf{x}} + \frac{1}{8}a\,\hat{\mathbf{y}} + x_5a\,\hat{\mathbf{z}}$ | (48$f$) | Na |
| $\mathbf{B}_{28}$ | = | $\left(\frac{1}{4} - x_5\right) \mathbf{a}_1 + \left(\frac{1}{4} - x_5\right) \mathbf{a}_2 + x_5 \mathbf{a}_3$ | = | $\frac{1}{8}a\,\hat{\mathbf{x}} + \frac{1}{8}a\,\hat{\mathbf{y}} + \left(\frac{1}{4} - x_5\right) a\,\hat{\mathbf{z}}$ | $(48f)$ | Na |
| $\mathbf{B}_{29}$ | = | $\left(\frac{3}{4} + x_5\right) \mathbf{a}_1 - x_5 \mathbf{a}_2 + \left(\frac{3}{4} + x_5\right) \mathbf{a}_3$ | = | $\frac{3}{8}a\,\hat{\mathbf{x}} + \left(\frac{3}{4} + x_5\right) a\,\hat{\mathbf{y}} + \frac{3}{8}a\,\hat{\mathbf{z}}$ | $(48f)$ | Na |
| $\mathbf{B}_{30}$ | = | $-x_5 \mathbf{a}_1 + \left(\frac{3}{4} + x_5\right) \mathbf{a}_2 - x_5 \mathbf{a}_3$ | = | $\frac{3}{8}a\,\hat{\mathbf{x}} - x_5 a\,\hat{\mathbf{y}} + \frac{3}{8}a\,\hat{\mathbf{z}}$ | $(48f)$ | Na |
| $\mathbf{B}_{31}$ | = | $-x_5 \mathbf{a}_1 + \left(\frac{3}{4} + x_5\right) \mathbf{a}_2 + \left(\frac{3}{4} + x_5\right) \mathbf{a}_3$ | = | $\left(\frac{3}{4} + x_5\right) a\,\hat{\mathbf{x}} + \frac{3}{8}a\,\hat{\mathbf{y}} + \frac{3}{8}a\,\hat{\mathbf{z}}$ | $(48f)$ | Na |
| $\mathbf{B}_{32}$ | = | $\left(\frac{3}{4} + x_5\right) \mathbf{a}_1 - x_5 \mathbf{a}_2 - x_5 \mathbf{a}_3$ | = | $-x_5 a\,\hat{\mathbf{x}} + \frac{3}{8}a\,\hat{\mathbf{y}} + \frac{3}{8}a\,\hat{\mathbf{z}}$ | $(48f)$ | Na |
| $\mathbf{B}_{33}$ | = | $-x_5 \mathbf{a}_1 - x_5 \mathbf{a}_2 + \left(\frac{3}{4} + x_5\right) \mathbf{a}_3$ | = | $\frac{3}{8}a\,\hat{\mathbf{x}} + \frac{3}{8}a\,\hat{\mathbf{y}} - x_5 a\,\hat{\mathbf{z}}$ | $(48f)$ | Na |
| $\mathbf{B}_{34}$ | = | $\left(\frac{3}{4} + x_5\right) \mathbf{a}_1 + \left(\frac{3}{4} + x_5\right) \mathbf{a}_2 - x_5 \mathbf{a}_3$ | = | $\frac{3}{8}a\,\hat{\mathbf{x}} + \frac{3}{8}a\,\hat{\mathbf{y}} + \left(\frac{3}{4} + x_5\right) a\,\hat{\mathbf{z}}$ | $(48f)$ | Na |
| $\mathbf{B}_{35}$ | = | $z_6 \mathbf{a}_1 + z_6 \mathbf{a}_2 + (2x_6 - z_6) \mathbf{a}_3$ | = | $x_6 a\,\hat{\mathbf{x}} + x_6 a\,\hat{\mathbf{y}} + z_6 a\,\hat{\mathbf{z}}$ | $(96g)$ | O II |
| $\mathbf{B}_{36}$ | = | $z_6 \mathbf{a}_1 + z_6 \mathbf{a}_2 + \left(\frac{1}{2} - 2x_6 - z_6\right) \mathbf{a}_3$ | = | $\left(\frac{1}{4} - x_6\right) a\,\hat{\mathbf{x}} + \left(\frac{1}{4} - x_6\right) a\,\hat{\mathbf{y}} + z_6 a\,\hat{\mathbf{z}}$ | $(96g)$ | O II |
| $\mathbf{B}_{37}$ | = | $(2x_6 - z_6) \mathbf{a}_1 + \left(\frac{1}{2} - 2x_6 - z_6\right) \mathbf{a}_2 + z_6 \mathbf{a}_3$ | = | $\left(\frac{1}{4} - x_6\right) a\,\hat{\mathbf{x}} + x_6 a\,\hat{\mathbf{y}} + \left(\frac{1}{4} - z_6\right) a\,\hat{\mathbf{z}}$ | $(96g)$ | O II |
| $\mathbf{B}_{38}$ | = | $\left(\frac{1}{2} - 2x_6 - z_6\right) \mathbf{a}_1 + (2x_6 - z_6) \mathbf{a}_2 + z_6 \mathbf{a}_3$ | = | $x_6 a\,\hat{\mathbf{x}} + \left(\frac{1}{4} - x_6\right) a\,\hat{\mathbf{y}} + \left(\frac{1}{4} - z_6\right) a\,\hat{\mathbf{z}}$ | $(96g)$ | O II |
| $\mathbf{B}_{39}$ | = | $(2x_6 - z_6) \mathbf{a}_1 + z_6 \mathbf{a}_2 + z_6 \mathbf{a}_3$ | = | $z_6 a\,\hat{\mathbf{x}} + x_6 a\,\hat{\mathbf{y}} + x_6 a\,\hat{\mathbf{z}}$ | $(96g)$ | O II |
| $\mathbf{B}_{40}$ | = | $\left(\frac{1}{2} - 2x_6 - z_6\right) \mathbf{a}_1 + z_6 \mathbf{a}_2 + z_6 \mathbf{a}_3$ | = | $z_6 a\,\hat{\mathbf{x}} + \left(\frac{1}{4} - x_6\right) a\,\hat{\mathbf{y}} + \left(\frac{1}{4} - x_6\right) a\,\hat{\mathbf{z}}$ | $(96g)$ | O II |
| $\mathbf{B}_{41}$ | = | $z_6 \mathbf{a}_1 + (2x_6 - z_6) \mathbf{a}_2 + \left(\frac{1}{2} - 2x_6 - z_6\right) \mathbf{a}_3$ | = | $\left(\frac{1}{4} - z_6\right) a\,\hat{\mathbf{x}} + \left(\frac{1}{4} - x_6\right) a\,\hat{\mathbf{y}} + x_6 a\,\hat{\mathbf{z}}$ | $(96g)$ | O II |
| $\mathbf{B}_{42}$ | = | $z_6 \mathbf{a}_1 + \left(\frac{1}{2} - 2x_6 - z_6\right) \mathbf{a}_2 + (2x_6 - z_6) \mathbf{a}_3$ | = | $\left(\frac{1}{4} - z_6\right) a\,\hat{\mathbf{x}} + x_6 a\,\hat{\mathbf{y}} + \left(\frac{1}{4} - x_6\right) a\,\hat{\mathbf{z}}$ | $(96g)$ | O II |
| $\mathbf{B}_{43}$ | = | $z_6 \mathbf{a}_1 + (2x_6 - z_6) \mathbf{a}_2 + z_6 \mathbf{a}_3$ | = | $x_6 a\,\hat{\mathbf{x}} + z_6 a\,\hat{\mathbf{y}} + x_6 a\,\hat{\mathbf{z}}$ | $(96g)$ | O II |
| $\mathbf{B}_{44}$ | = | $z_6 \mathbf{a}_1 + \left(\frac{1}{2} - 2x_6 - z_6\right) \mathbf{a}_2 + z_6 \mathbf{a}_3$ | = | $\left(\frac{1}{4} - x_6\right) a\,\hat{\mathbf{x}} + z_6 a\,\hat{\mathbf{y}} + \left(\frac{1}{4} - x_6\right) a\,\hat{\mathbf{z}}$ | $(96g)$ | O II |
| $\mathbf{B}_{45}$ | = | $\left(\frac{1}{2} - 2x_6 - z_6\right) \mathbf{a}_1 + z_6 \mathbf{a}_2 + (2x_6 - z_6) \mathbf{a}_3$ | = | $x_6 a\,\hat{\mathbf{x}} + \left(\frac{1}{4} - z_6\right) a\,\hat{\mathbf{y}} + \left(\frac{1}{4} - x_6\right) a\,\hat{\mathbf{z}}$ | $(96g)$ | O II |
| $\mathbf{B}_{46}$ | = | $(2x_6 - z_6) \mathbf{a}_1 + z_6 \mathbf{a}_2 + \left(\frac{1}{2} - 2x_6 - z_6\right) \mathbf{a}_3$ | = | $\left(\frac{1}{4} - x_6\right) a\,\hat{\mathbf{x}} + \left(\frac{1}{4} - z_6\right) a\,\hat{\mathbf{y}} + x_6 a\,\hat{\mathbf{z}}$ | $(96g)$ | O II |
| $\mathbf{B}_{47}$ | = | $-z_6 \mathbf{a}_1 - z_6 \mathbf{a}_2 + \left(\frac{1}{2} + 2x_6 + z_6\right) \mathbf{a}_3$ | = | $\left(\frac{1}{4} + x_6\right) a\,\hat{\mathbf{x}} + \left(\frac{1}{4} + x_6\right) a\,\hat{\mathbf{y}} - z_6 a\,\hat{\mathbf{z}}$ | $(96g)$ | O II |
| $\mathbf{B}_{48}$ | = | $-z_6 \mathbf{a}_1 - z_6 \mathbf{a}_2 + (-2x_6 + z_6) \mathbf{a}_3$ | = | $-x_6 a\,\hat{\mathbf{x}} - x_6 a\,\hat{\mathbf{y}} - z_6 a\,\hat{\mathbf{z}}$ | $(96g)$ | O II |
| $\mathbf{B}_{49}$ | = | $(-2x_6 + z_6) \mathbf{a}_1 + \left(\frac{1}{2} + 2x_6 + z_6\right) \mathbf{a}_2 - z_6 \mathbf{a}_3$ | = | $\left(\frac{1}{4} + x_6\right) a\,\hat{\mathbf{x}} - x_6 a\,\hat{\mathbf{y}} + \left(\frac{1}{4} + z_6\right) a\,\hat{\mathbf{z}}$ | $(96g)$ | O II |
| $\mathbf{B}_{50}$ | = | $\left(\frac{1}{2} + 2x_6 + z_6\right) \mathbf{a}_1 + (-2x_6 + z_6) \mathbf{a}_2 - z_6 \mathbf{a}_3$ | = | $-x_6 a\,\hat{\mathbf{x}} + \left(\frac{1}{4} + x_6\right) a\,\hat{\mathbf{y}} + \left(\frac{1}{4} + z_6\right) a\,\hat{\mathbf{z}}$ | $(96g)$ | O II |
| $\mathbf{B}_{51}$ | = | $(-2x_6 + z_6) \mathbf{a}_1 - z_6 \mathbf{a}_2 + \left(\frac{1}{2} + 2x_6 + z_6\right) \mathbf{a}_3$ | = | $\left(\frac{1}{4} + x_6\right) a\,\hat{\mathbf{x}} + \left(\frac{1}{4} + z_6\right) a\,\hat{\mathbf{y}} - x_6 a\,\hat{\mathbf{z}}$ | $(96g)$ | O II |
| $\mathbf{B}_{52}$ | = | $\left(\frac{1}{2} + 2x_6 + z_6\right) \mathbf{a}_1 - z_6 \mathbf{a}_2 + (-2x_6 + z_6) \mathbf{a}_3$ | = | $-x_6 a\,\hat{\mathbf{x}} + \left(\frac{1}{4} + z_6\right) a\,\hat{\mathbf{y}} + \left(\frac{1}{4} + x_6\right) a\,\hat{\mathbf{z}}$ | $(96g)$ | O II |
| $\mathbf{B}_{53}$ | = | $-z_6 \mathbf{a}_1 + (-2x_6 + z_6) \mathbf{a}_2 - z_6 \mathbf{a}_3$ | = | $-x_6 a\,\hat{\mathbf{x}} - z_6 a\,\hat{\mathbf{y}} - x_6 a\,\hat{\mathbf{z}}$ | $(96g)$ | O II |
| $\mathbf{B}_{54}$ | = | $-z_6 \mathbf{a}_1 + \left(\frac{1}{2} + 2x_6 + z_6\right) \mathbf{a}_2 - z_6 \mathbf{a}_3$ | = | $\left(\frac{1}{4} + x_6\right) a\,\hat{\mathbf{x}} - z_6 a\,\hat{\mathbf{y}} + \left(\frac{1}{4} + x_6\right) a\,\hat{\mathbf{z}}$ | $(96g)$ | O II |
| $\mathbf{B}_{55}$ | = | $-z_6 \mathbf{a}_1 + (-2x_6 + z_6) \mathbf{a}_2 + \left(\frac{1}{2} + 2x_6 + z_6\right) \mathbf{a}_3$ | = | $\left(\frac{1}{4} + z_6\right) a\,\hat{\mathbf{x}} + \left(\frac{1}{4} + x_6\right) a\,\hat{\mathbf{y}} - x_6 a\,\hat{\mathbf{z}}$ | $(96g)$ | O II |
| $\mathbf{B}_{56}$ | = | $-z_6 \mathbf{a}_1 + \left(\frac{1}{2} + 2x_6 + z_6\right) \mathbf{a}_2 + (-2x_6 + z_6) \mathbf{a}_3$ | = | $\left(\frac{1}{4} + z_6\right) a\,\hat{\mathbf{x}} - x_6 a\,\hat{\mathbf{y}} + \left(\frac{1}{4} + x_6\right) a\,\hat{\mathbf{z}}$ | $(96g)$ | O II |
| $\mathbf{B}_{57}$ | = | $\left(\frac{1}{2} + 2x_6 + z_6\right) \mathbf{a}_1 - z_6 \mathbf{a}_2 - z_6 \mathbf{a}_3$ | = | $-z_6 a\,\hat{\mathbf{x}} + \left(\frac{1}{4} + x_6\right) a\,\hat{\mathbf{y}} + \left(\frac{1}{4} + x_6\right) a\,\hat{\mathbf{z}}$ | $(96g)$ | O II |

$$\mathbf{B}_{58} = (-2x_6 + z_6)\,\mathbf{a}_1 - z_6\,\mathbf{a}_2 - z_6\,\mathbf{a}_3 = -z_6 a\,\hat{\mathbf{x}} - x_6 a\,\hat{\mathbf{y}} - x_6 a\,\hat{\mathbf{z}} \qquad (96g) \qquad \text{O II}$$

---

**Found in:**

---

**Geometry files:**

- CIF: pp. 1834
- POSCAR: pp. 1835

# Cubic CuPt ($L1_3$ (I), $D4$) Structure: AB_cF32_227_c_d

http://aflow.org/prototype-encyclopedia/AB_cF32_227_c_d

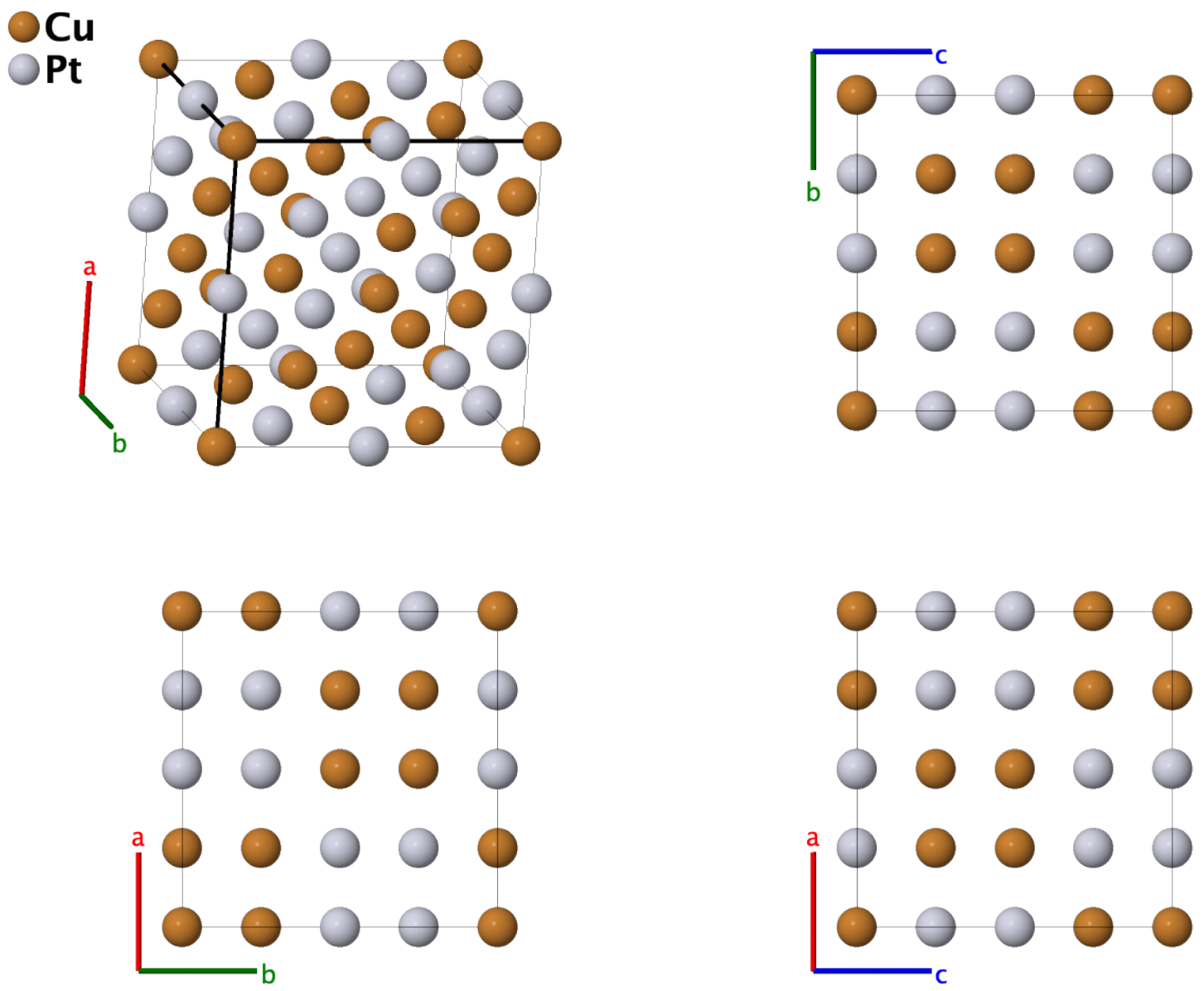

| | | |
|---|---|---|
| **Prototype** | : | CuPt |
| **AFLOW prototype label** | : | AB_cF32_227_c_d |
| ***Strukturbericht* designation** | : | $L1_3$ ($I$) |
| **Pearson symbol** | : | cF32 |
| **Space group number** | : | 227 |
| **Space group symbol** | : | $Fd\bar{3}m$ |
| **AFLOW prototype command** | : | `aflow --proto=AB_cF32_227_c_d --params=`$a$ |

- (Johansson, 1929) described two possible structures for CuPt. (Ewald, 1929) and later (Villars, 2007) used the description to determine the space group and atomic positions. This page describes the cubic structure, which (Ewald, 1929) labeled *Strukturbericht* $L1_1$. The other structure is rhombohedral, and was listed as $L1_1$. (Villars, 2007) prefers the later structure, listing the current one as "superceded."
- (Barrett, 1980) noted that even slight additions of platinum above stoichiometry will cause a change in the crystal structure.
- This structure is equivalent to the $D4$ structure of (Lu, 1991).
- This structure should not be confused with the $CuPt_3$ structure, which has also been given the $L1_3$ label, and which we will often refer to as $L1_3$ (II).

**Face-centered Cubic primitive vectors:**

$$\mathbf{a}_1 = \tfrac{1}{2}\,a\,\hat{\mathbf{y}} + \tfrac{1}{2}\,a\,\hat{\mathbf{z}}$$
$$\mathbf{a}_2 = \tfrac{1}{2}\,a\,\hat{\mathbf{x}} + \tfrac{1}{2}\,a\,\hat{\mathbf{z}}$$
$$\mathbf{a}_3 = \tfrac{1}{2}\,a\,\hat{\mathbf{x}} + \tfrac{1}{2}\,a\,\hat{\mathbf{y}}$$

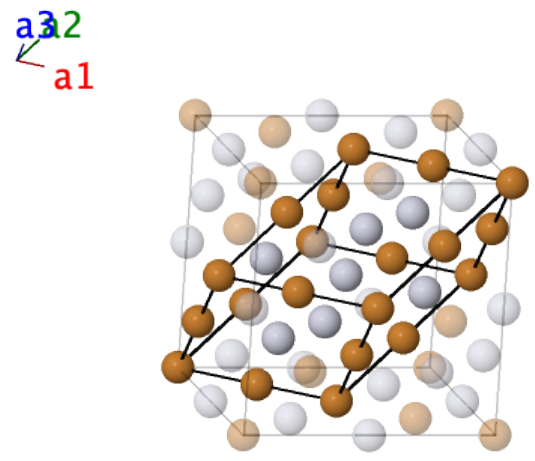

**Basis vectors:**

| | | Lattice Coordinates | | Cartesian Coordinates | Wyckoff Position | Atom Type |
|---|---|---|---|---|---|---|
| $\mathbf{B}_1$ | = | $0\,\mathbf{a}_1 + 0\,\mathbf{a}_2 + 0\,\mathbf{a}_3$ | = | $0\,\hat{\mathbf{x}} + 0\,\hat{\mathbf{y}} + 0\,\hat{\mathbf{z}}$ | (16*c*) | Cu |
| $\mathbf{B}_2$ | = | $\frac{1}{2}\,\mathbf{a}_3$ | = | $\frac{1}{4}a\,\hat{\mathbf{x}} + \frac{1}{4}a\,\hat{\mathbf{y}}$ | (16*c*) | Cu |
| $\mathbf{B}_3$ | = | $\frac{1}{2}\,\mathbf{a}_2$ | = | $\frac{1}{4}a\,\hat{\mathbf{x}} + \frac{1}{4}a\,\hat{\mathbf{z}}$ | (16*c*) | Cu |
| $\mathbf{B}_4$ | = | $\frac{1}{2}\,\mathbf{a}_1$ | = | $\frac{1}{4}a\,\hat{\mathbf{y}} + \frac{1}{4}a\,\hat{\mathbf{z}}$ | (16*c*) | Cu |
| $\mathbf{B}_5$ | = | $\frac{1}{2}\,\mathbf{a}_1 + \frac{1}{2}\,\mathbf{a}_2 + \frac{1}{2}\,\mathbf{a}_3$ | = | $\frac{1}{2}a\,\hat{\mathbf{x}} + \frac{1}{2}a\,\hat{\mathbf{y}} + \frac{1}{2}a\,\hat{\mathbf{z}}$ | (16*d*) | Pt |
| $\mathbf{B}_6$ | = | $\frac{1}{2}\,\mathbf{a}_1 + \frac{1}{2}\,\mathbf{a}_2$ | = | $\frac{1}{4}a\,\hat{\mathbf{x}} + \frac{1}{4}a\,\hat{\mathbf{y}} + \frac{1}{2}a\,\hat{\mathbf{z}}$ | (16*d*) | Pt |
| $\mathbf{B}_7$ | = | $\frac{1}{2}\,\mathbf{a}_1 + \frac{1}{2}\,\mathbf{a}_3$ | = | $\frac{1}{4}a\,\hat{\mathbf{x}} + \frac{1}{2}a\,\hat{\mathbf{y}} + \frac{1}{4}a\,\hat{\mathbf{z}}$ | (16*d*) | Pt |
| $\mathbf{B}_8$ | = | $\frac{1}{2}\,\mathbf{a}_2 + \frac{1}{2}\,\mathbf{a}_3$ | = | $\frac{1}{2}a\,\hat{\mathbf{x}} + \frac{1}{4}a\,\hat{\mathbf{y}} + \frac{1}{4}a\,\hat{\mathbf{z}}$ | (16*d*) | Pt |

**Found in:**

**Geometry files:**

- CIF: pp. 1835
- POSCAR: pp. 1836

# $\alpha$-AgI ($B23$) Structure: A21B_cI44_229_bdh_a

http://aflow.org/prototype-encyclopedia/A21B_cI44_229_bdh_a

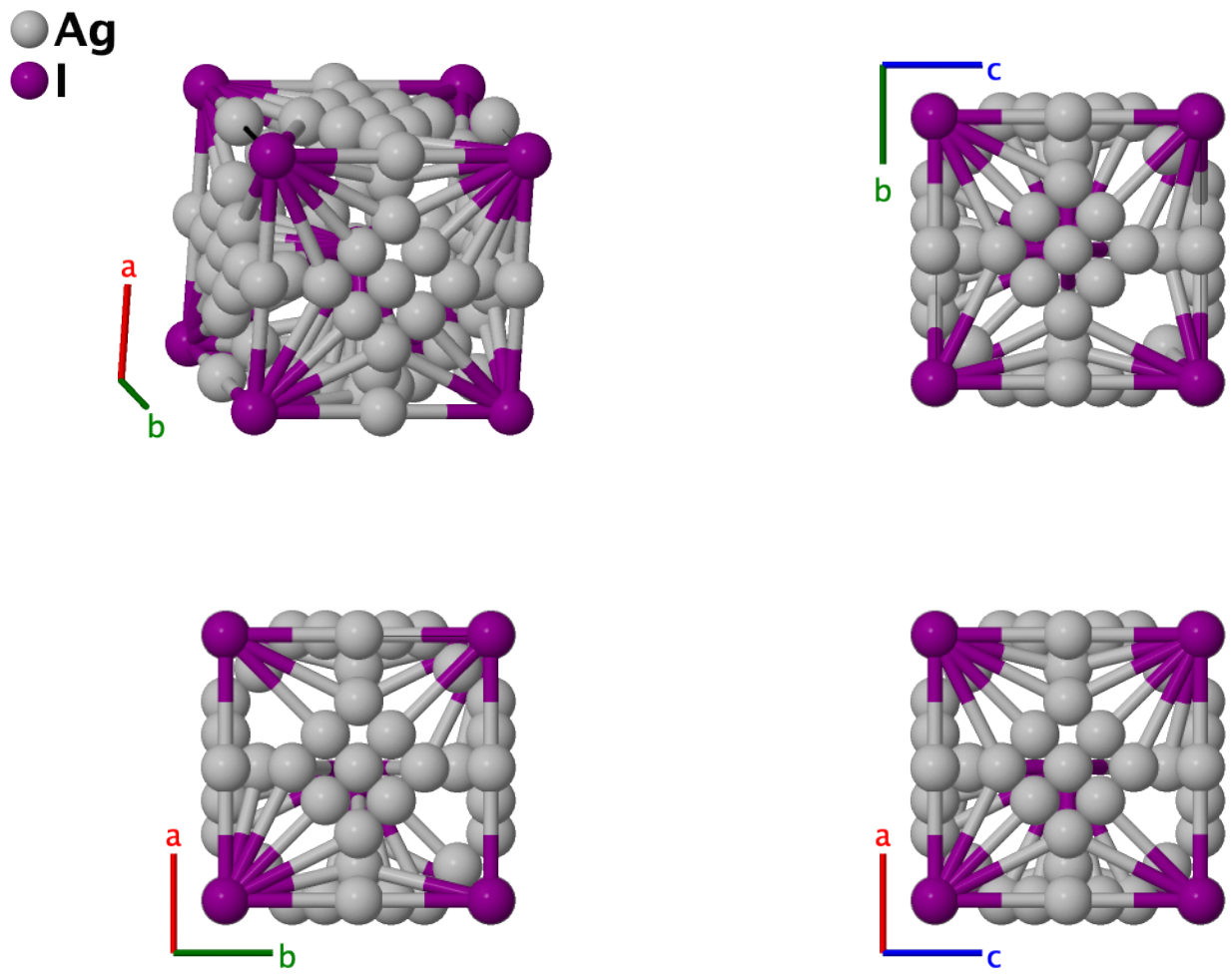

| | | |
|---|---|---|
| **Prototype** | : | AgI |
| **AFLOW prototype label** | : | A21B_cI44_229_bdh_a |
| ***Strukturbericht* designation** | : | $B23$ |
| **Pearson symbol** | : | cI44 |
| **Space group number** | : | 229 |
| **Space group symbol** | : | $Im\bar{3}m$ |
| **AFLOW prototype command** | : | `aflow --proto=A21B_cI44_229_bdh_a` `--params=`$a, y_4$ |

- Under ambient conditions, silver iodide exists as a mixture of $\beta$-AgI, which has the wurtzite ($B4$) structure, and $\gamma$-AgI, which has the zincblende ($B3$) structure (Hull, 2004). Above 420 K, AgI transforms to this superionic $\alpha$ phase. The iodine atom sits at the (2$a$) site of the bcc lattice of space group #229, while the silver atom is randomly distributed on one of the (6$b$), (12$d$), and (24$h$) Wyckoff sites in each unit cell. On average, then, each of the 21 Ag sites listed above is occupied only 4.762% of the time in any given primitive cell. This easy transport between sites drives the superionic behavior of $\alpha$-AgI.

**Body-centered Cubic primitive vectors:**

$$\mathbf{a}_1 = -\frac{1}{2}\,a\,\hat{\mathbf{x}} + \frac{1}{2}\,a\,\hat{\mathbf{y}} + \frac{1}{2}\,a\,\hat{\mathbf{z}}$$
$$\mathbf{a}_2 = \frac{1}{2}\,a\,\hat{\mathbf{x}} - \frac{1}{2}\,a\,\hat{\mathbf{y}} + \frac{1}{2}\,a\,\hat{\mathbf{z}}$$
$$\mathbf{a}_3 = \frac{1}{2}\,a\,\hat{\mathbf{x}} + \frac{1}{2}\,a\,\hat{\mathbf{y}} - \frac{1}{2}\,a\,\hat{\mathbf{z}}$$

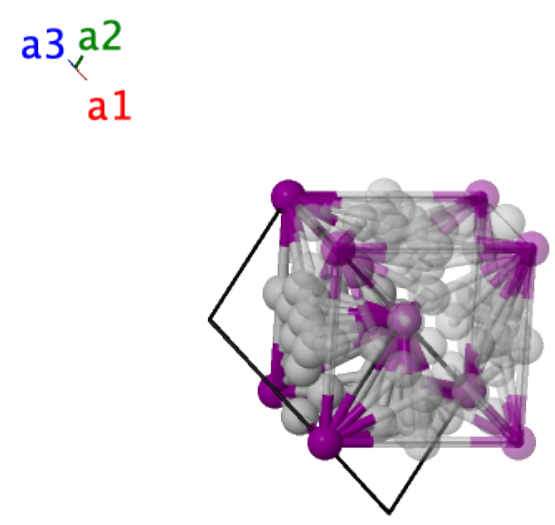

**Basis vectors:**

| | | Lattice Coordinates | | Cartesian Coordinates | Wyckoff Position | Atom Type |
|---|---|---|---|---|---|---|
| $\mathbf{B}_1$ | = | $0\,\mathbf{a}_1 + 0\,\mathbf{a}_2 + 0\,\mathbf{a}_3$ | = | $0\,\hat{\mathbf{x}} + 0\,\hat{\mathbf{y}} + 0\,\hat{\mathbf{z}}$ | (2$a$) | I |
| $\mathbf{B}_2$ | = | $\frac{1}{2}\,\mathbf{a}_2 + \frac{1}{2}\,\mathbf{a}_3$ | = | $\frac{1}{2}a\,\hat{\mathbf{x}}$ | (6$b$) | Ag I |
| $\mathbf{B}_3$ | = | $\frac{1}{2}\,\mathbf{a}_1 + \frac{1}{2}\,\mathbf{a}_3$ | = | $\frac{1}{2}a\,\mathbf{\hat{y}}$ | (6$b$) | Ag I |
| $\mathbf{B}_4$ | = | $\frac{1}{2}\,\mathbf{a}_1 + \frac{1}{2}\,\mathbf{a}_2$ | = | $\frac{1}{2}a\,\mathbf{\hat{z}}$ | (6$b$) | Ag I |
| $\mathbf{B}_5$ | = | $\frac{1}{2}\,\mathbf{a}_1 + \frac{3}{4}\,\mathbf{a}_2 + \frac{1}{4}\,\mathbf{a}_3$ | = | $\frac{1}{4}a\,\mathbf{\hat{x}} + \frac{1}{2}a\,\mathbf{\hat{z}}$ | (12$d$) | Ag II |
| $\mathbf{B}_6$ | = | $\frac{1}{2}\,\mathbf{a}_1 + \frac{1}{4}\,\mathbf{a}_2 + \frac{3}{4}\,\mathbf{a}_3$ | = | $\frac{1}{4}a\,\mathbf{\hat{x}} + \frac{1}{2}a\,\mathbf{\hat{y}}$ | (12$d$) | Ag II |
| $\mathbf{B}_7$ | = | $\frac{1}{4}\,\mathbf{a}_1 + \frac{1}{2}\,\mathbf{a}_2 + \frac{3}{4}\,\mathbf{a}_3$ | = | $\frac{1}{2}a\,\mathbf{\hat{x}} + \frac{1}{4}a\,\mathbf{\hat{y}}$ | (12$d$) | Ag II |
| $\mathbf{B}_8$ | = | $\frac{3}{4}\,\mathbf{a}_1 + \frac{1}{2}\,\mathbf{a}_2 + \frac{1}{4}\,\mathbf{a}_3$ | = | $\frac{1}{4}a\,\mathbf{\hat{y}} + \frac{1}{2}a\,\mathbf{\hat{z}}$ | (12$d$) | Ag II |
| $\mathbf{B}_9$ | = | $\frac{3}{4}\,\mathbf{a}_1 + \frac{1}{4}\,\mathbf{a}_2 + \frac{1}{2}\,\mathbf{a}_3$ | = | $\frac{1}{2}a\,\mathbf{\hat{y}} + \frac{1}{4}a\,\mathbf{\hat{z}}$ | (12$d$) | Ag II |
| $\mathbf{B}_{10}$ | = | $\frac{1}{4}\,\mathbf{a}_1 + \frac{3}{4}\,\mathbf{a}_2 + \frac{1}{2}\,\mathbf{a}_3$ | = | $\frac{1}{2}a\,\mathbf{\hat{x}} + \frac{1}{4}a\,\mathbf{\hat{z}}$ | (12$d$) | Ag II |
| $\mathbf{B}_{11}$ | = | $2y_4\,\mathbf{a}_1 + y_4\,\mathbf{a}_2 + y_4\,\mathbf{a}_3$ | = | $y_4a\,\mathbf{\hat{y}} + y_4a\,\mathbf{\hat{z}}$ | (24$h$) | Ag III |
| $\mathbf{B}_{12}$ | = | $y_4\,\mathbf{a}_2 - y_4\,\mathbf{a}_3$ | = | $-y_4a\,\mathbf{\hat{y}} + y_4a\,\mathbf{\hat{z}}$ | (24$h$) | Ag III |
| $\mathbf{B}_{13}$ | = | $-y_4\,\mathbf{a}_2 + y_4\,\mathbf{a}_3$ | = | $y_4a\,\mathbf{\hat{y}} - y_4a\,\mathbf{\hat{z}}$ | (24$h$) | Ag III |
| $\mathbf{B}_{14}$ | = | $-2y_4\,\mathbf{a}_1 - y_4\,\mathbf{a}_2 - y_4\,\mathbf{a}_3$ | = | $-y_4a\,\mathbf{\hat{y}} - y_4a\,\mathbf{\hat{z}}$ | (24$h$) | Ag III |
| $\mathbf{B}_{15}$ | = | $y_4\,\mathbf{a}_1 + 2y_4\,\mathbf{a}_2 + y_4\,\mathbf{a}_3$ | = | $y_4a\,\mathbf{\hat{x}} + y_4a\,\mathbf{\hat{z}}$ | (24$h$) | Ag III |
| $\mathbf{B}_{16}$ | = | $-y_4\,\mathbf{a}_1 + y_4\,\mathbf{a}_3$ | = | $y_4a\,\mathbf{\hat{x}} - y_4a\,\mathbf{\hat{z}}$ | (24$h$) | Ag III |
| $\mathbf{B}_{17}$ | = | $y_4\,\mathbf{a}_1 - y_4\,\mathbf{a}_3$ | = | $-y_4a\,\mathbf{\hat{x}} + y_4a\,\mathbf{\hat{z}}$ | (24$h$) | Ag III |
| $\mathbf{B}_{18}$ | = | $-y_4\,\mathbf{a}_1 - 2y_4\,\mathbf{a}_2 - y_4\,\mathbf{a}_3$ | = | $-y_4a\,\mathbf{\hat{x}} - y_4a\,\mathbf{\hat{z}}$ | (24$h$) | Ag III |
| $\mathbf{B}_{19}$ | = | $y_4\,\mathbf{a}_1 + y_4\,\mathbf{a}_2 + 2y_4\,\mathbf{a}_3$ | = | $y_4a\,\mathbf{\hat{x}} + y_4a\,\mathbf{\hat{y}}$ | (24$h$) | Ag III |
| $\mathbf{B}_{20}$ | = | $y_4\,\mathbf{a}_1 - y_4\,\mathbf{a}_2$ | = | $-y_4a\,\mathbf{\hat{x}} + y_4a\,\mathbf{\hat{y}}$ | (24$h$) | Ag III |
| $\mathbf{B}_{21}$ | = | $-y_4\,\mathbf{a}_1 + y_4\,\mathbf{a}_2$ | = | $y_4a\,\mathbf{\hat{x}} - y_4a\,\mathbf{\hat{y}}$ | (24$h$) | Ag III |
| $\mathbf{B}_{22}$ | = | $-y_4\,\mathbf{a}_1 - y_4\,\mathbf{a}_2 - 2y_4\,\mathbf{a}_3$ | = | $-y_4a\,\mathbf{\hat{x}} - y_4a\,\mathbf{\hat{y}}$ | (24$h$) | Ag III |

**Geometry files:**

- CIF: pp. 1837
- POSCAR: pp. 1837

# $Ca_3Al_2(OH)_{12}$ ($J2_3$) Structure: A2B3C12D12_cI232_230_a_c_h_h

http://aflow.org/prototype-encyclopedia/A2B3C12D12_cI232_230_a_c_h_h

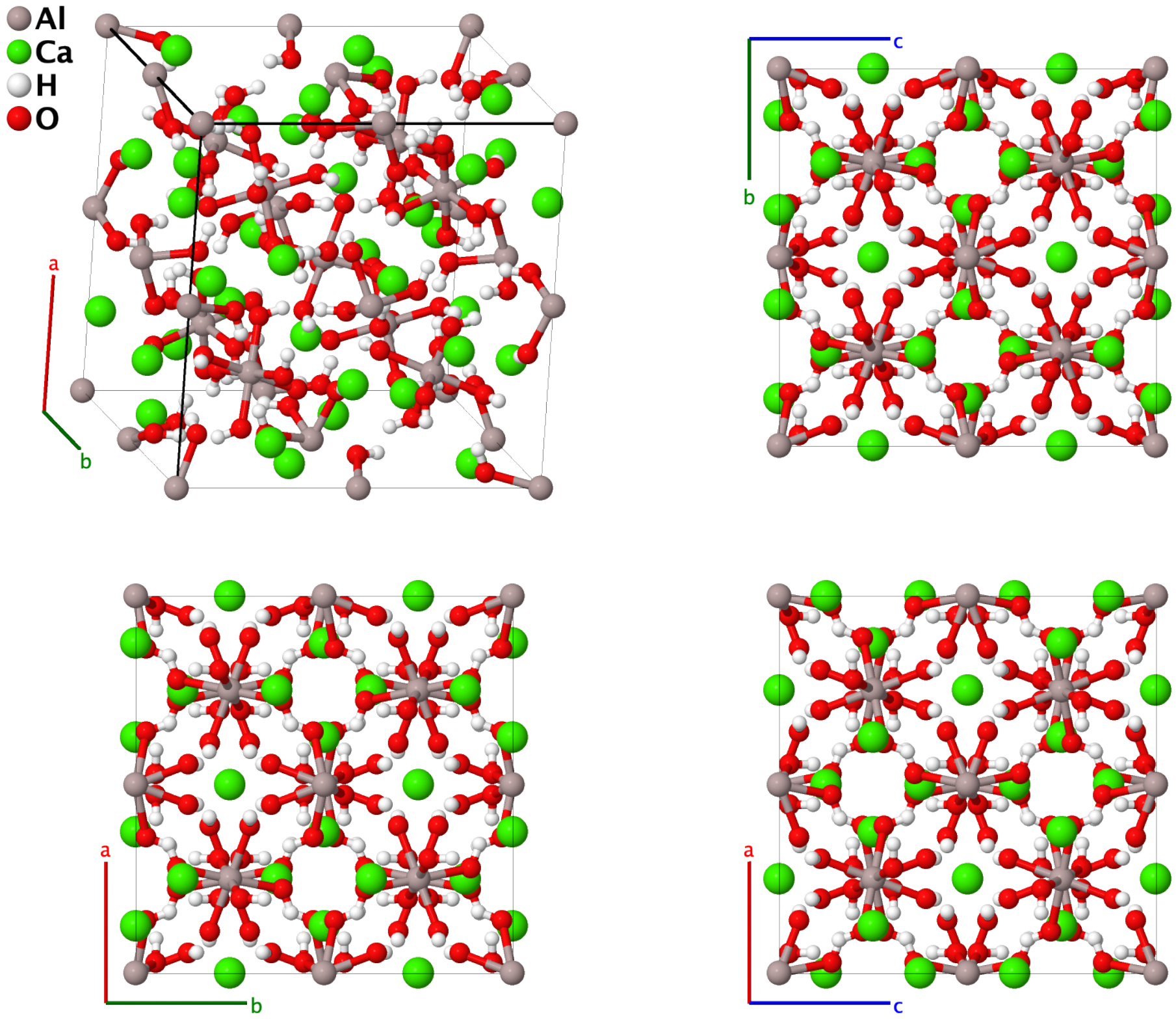

| | | |
|---|---|---|
| **Prototype** | : | $Al_2Ca_3H_{12}O_{12}$ |
| **AFLOW prototype label** | : | A2B3C12D12_cI232_230_a_c_h_h |
| ***Strukturbericht* designation** | : | $J2_3$ |
| **Pearson symbol** | : | cI232 |
| **Space group number** | : | 230 |
| **Space group symbol** | : | $Ia\bar{3}d$ |
| **AFLOW prototype command** | : | `aflow --proto=A2B3C12D12_cI232_230_a_c_h_h --params=`$a, x_3, y_3, z_3, x_4, y_4, z_4$ |

- The original determination of this structure by (Brandenberger, 1933) did not locate the hydrogen atoms, and according to (Gottfried, 1937) used the coordinates of garnet, $S1_4$. (Bartl, 1986) was able to locate the hydrogen atoms, and as they do not change the space group we include them in the $J2_3$ structure.

**Body-centered Cubic primitive vectors:**

$$\mathbf{a}_1 = -\frac{1}{2}\,a\,\hat{\mathbf{x}} + \frac{1}{2}\,a\,\hat{\mathbf{y}} + \frac{1}{2}\,a\,\hat{\mathbf{z}}$$
$$\mathbf{a}_2 = \frac{1}{2}\,a\,\hat{\mathbf{x}} - \frac{1}{2}\,a\,\hat{\mathbf{y}} + \frac{1}{2}\,a\,\hat{\mathbf{z}}$$
$$\mathbf{a}_3 = \frac{1}{2}\,a\,\hat{\mathbf{x}} + \frac{1}{2}\,a\,\hat{\mathbf{y}} - \frac{1}{2}\,a\,\hat{\mathbf{z}}$$

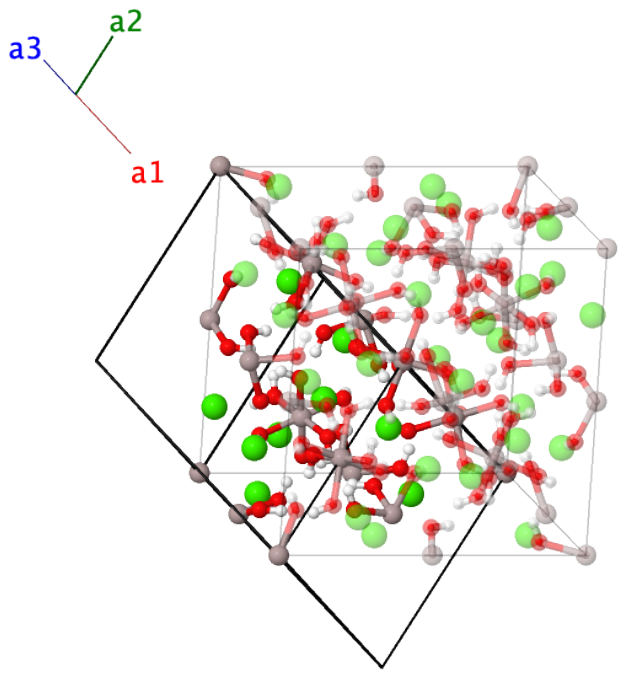

**Basis vectors:**

| | Lattice Coordinates | | Cartesian Coordinates | Wyckoff Position | Atom Type |
|---|---|---|---|---|---|
| $\mathbf{B}_1$ = | $0\,\mathbf{a}_1 + 0\,\mathbf{a}_2 + 0\,\mathbf{a}_3$ | = | $0\,\hat{\mathbf{x}} + 0\,\hat{\mathbf{y}} + 0\,\hat{\mathbf{z}}$ | (16$a$) | Al |
| $\mathbf{B}_2$ = | $\frac{1}{2}\,\mathbf{a}_1 + \frac{1}{2}\,\mathbf{a}_3$ | = | $\frac{1}{2}a\,\hat{\mathbf{y}}$ | (16$a$) | Al |
| $\mathbf{B}_3$ = | $\frac{1}{2}\,\mathbf{a}_2 + \frac{1}{2}\,\mathbf{a}_3$ | = | $\frac{1}{2}a\,\hat{\mathbf{x}}$ | (16$a$) | Al |
| $\mathbf{B}_4$ = | $\frac{1}{2}\,\mathbf{a}_1 + \frac{1}{2}\,\mathbf{a}_2$ | = | $\frac{1}{2}a\,\hat{\mathbf{z}}$ | (16$a$) | Al |
| $\mathbf{B}_5$ = | $\frac{1}{2}\,\mathbf{a}_1$ | = | $\frac{3}{4}a\,\hat{\mathbf{x}} + \frac{1}{4}a\,\hat{\mathbf{y}} + \frac{1}{4}a\,\hat{\mathbf{z}}$ | (16$a$) | Al |
| $\mathbf{B}_6$ = | $\frac{1}{2}\,\mathbf{a}_1 + \frac{1}{2}\,\mathbf{a}_2 + \frac{1}{2}\,\mathbf{a}_3$ | = | $\frac{1}{4}a\,\hat{\mathbf{x}} + \frac{1}{4}a\,\hat{\mathbf{y}} + \frac{1}{4}a\,\hat{\mathbf{z}}$ | (16$a$) | Al |
| $\mathbf{B}_7$ = | $\frac{1}{2}\,\mathbf{a}_3$ | = | $\frac{1}{4}a\,\hat{\mathbf{x}} + \frac{1}{4}a\,\hat{\mathbf{y}} + \frac{3}{4}a\,\hat{\mathbf{z}}$ | (16$a$) | Al |
| $\mathbf{B}_8$ = | $\frac{1}{2}\,\mathbf{a}_2$ | = | $\frac{1}{4}a\,\hat{\mathbf{x}} + \frac{3}{4}a\,\hat{\mathbf{y}} + \frac{1}{4}a\,\hat{\mathbf{z}}$ | (16$a$) | Al |
| $\mathbf{B}_9$ = | $\frac{1}{4}\,\mathbf{a}_1 + \frac{3}{8}\,\mathbf{a}_2 + \frac{1}{8}\,\mathbf{a}_3$ | = | $\frac{1}{8}a\,\hat{\mathbf{x}} + \frac{1}{4}a\,\hat{\mathbf{z}}$ | (24$c$) | Ca |
| $\mathbf{B}_{10}$ = | $\frac{3}{4}\,\mathbf{a}_1 + \frac{1}{8}\,\mathbf{a}_2 + \frac{3}{8}\,\mathbf{a}_3$ | = | $-\frac{1}{8}a\,\hat{\mathbf{x}} + \frac{1}{2}a\,\hat{\mathbf{y}} + \frac{1}{4}a\,\hat{\mathbf{z}}$ | (24$c$) | Ca |
| $\mathbf{B}_{11}$ = | $\frac{1}{8}\,\mathbf{a}_1 + \frac{1}{4}\,\mathbf{a}_2 + \frac{3}{8}\,\mathbf{a}_3$ | = | $\frac{1}{4}a\,\hat{\mathbf{x}} + \frac{1}{8}a\,\hat{\mathbf{y}}$ | (24$c$) | Ca |
| $\mathbf{B}_{12}$ = | $\frac{3}{8}\,\mathbf{a}_1 + \frac{3}{4}\,\mathbf{a}_2 + \frac{1}{8}\,\mathbf{a}_3$ | = | $\frac{1}{4}a\,\hat{\mathbf{x}} - \frac{1}{8}a\,\hat{\mathbf{y}} + \frac{1}{2}a\,\hat{\mathbf{z}}$ | (24$c$) | Ca |
| $\mathbf{B}_{13}$ = | $\frac{3}{8}\,\mathbf{a}_1 + \frac{1}{8}\,\mathbf{a}_2 + \frac{1}{4}\,\mathbf{a}_3$ | = | $\frac{1}{4}a\,\hat{\mathbf{y}} + \frac{1}{8}a\,\hat{\mathbf{z}}$ | (24$c$) | Ca |
| $\mathbf{B}_{14}$ = | $\frac{1}{8}\,\mathbf{a}_1 + \frac{3}{8}\,\mathbf{a}_2 + \frac{3}{4}\,\mathbf{a}_3$ | = | $\frac{1}{2}a\,\hat{\mathbf{x}} + \frac{1}{4}a\,\hat{\mathbf{y}} - \frac{1}{8}a\,\hat{\mathbf{z}}$ | (24$c$) | Ca |
| $\mathbf{B}_{15}$ = | $\frac{3}{4}\,\mathbf{a}_1 + \frac{5}{8}\,\mathbf{a}_2 + \frac{7}{8}\,\mathbf{a}_3$ | = | $\frac{3}{8}a\,\hat{\mathbf{x}} + \frac{1}{2}a\,\hat{\mathbf{y}} + \frac{1}{4}a\,\hat{\mathbf{z}}$ | (24$c$) | Ca |
| $\mathbf{B}_{16}$ = | $\frac{1}{4}\,\mathbf{a}_1 + \frac{7}{8}\,\mathbf{a}_2 + \frac{5}{8}\,\mathbf{a}_3$ | = | $\frac{5}{8}a\,\hat{\mathbf{x}} + \frac{1}{4}a\,\hat{\mathbf{z}}$ | (24$c$) | Ca |
| $\mathbf{B}_{17}$ = | $\frac{7}{8}\,\mathbf{a}_1 + \frac{3}{4}\,\mathbf{a}_2 + \frac{5}{8}\,\mathbf{a}_3$ | = | $\frac{1}{4}a\,\hat{\mathbf{x}} + \frac{3}{8}a\,\hat{\mathbf{y}} + \frac{1}{2}a\,\hat{\mathbf{z}}$ | (24$c$) | Ca |
| $\mathbf{B}_{18}$ = | $\frac{5}{8}\,\mathbf{a}_1 + \frac{1}{4}\,\mathbf{a}_2 + \frac{7}{8}\,\mathbf{a}_3$ | = | $\frac{1}{4}a\,\hat{\mathbf{x}} + \frac{5}{8}a\,\hat{\mathbf{y}}$ | (24$c$) | Ca |
| $\mathbf{B}_{19}$ = | $\frac{5}{8}\,\mathbf{a}_1 + \frac{7}{8}\,\mathbf{a}_2 + \frac{3}{4}\,\mathbf{a}_3$ | = | $\frac{1}{2}a\,\hat{\mathbf{x}} + \frac{1}{4}a\,\hat{\mathbf{y}} + \frac{3}{8}a\,\hat{\mathbf{z}}$ | (24$c$) | Ca |
| $\mathbf{B}_{20}$ = | $\frac{7}{8}\,\mathbf{a}_1 + \frac{5}{8}\,\mathbf{a}_2 + \frac{1}{4}\,\mathbf{a}_3$ | = | $\frac{1}{4}a\,\hat{\mathbf{y}} + \frac{5}{8}a\,\hat{\mathbf{z}}$ | (24$c$) | Ca |
| $\mathbf{B}_{21}$ = | $(y_3 + z_3)\,\mathbf{a}_1 + (x_3 + z_3)\,\mathbf{a}_2 + (x_3 + y_3)\,\mathbf{a}_3$ | = | $x_3a\,\hat{\mathbf{x}} + y_3a\,\hat{\mathbf{y}} + z_3a\,\hat{\mathbf{z}}$ | (96$h$) | H |
| $\mathbf{B}_{22}$ = | $\left(\frac{1}{2} - y_3 + z_3\right)\,\mathbf{a}_1 + (-x_3 + z_3)\,\mathbf{a}_2 + \left(\frac{1}{2} - x_3 - y_3\right)\,\mathbf{a}_3$ | = | $-x_3a\,\hat{\mathbf{x}} + \left(\frac{1}{2} - y_3\right)a\,\hat{\mathbf{y}} + z_3a\,\hat{\mathbf{z}}$ | (96$h$) | H |
| $\mathbf{B}_{23}$ = | $(y_3 - z_3)\,\mathbf{a}_1 + \left(\frac{1}{2} - x_3 - z_3\right)\,\mathbf{a}_2 + \left(\frac{1}{2} - x_3 + y_3\right)\,\mathbf{a}_3$ | = | $\left(\frac{1}{2} - x_3\right)a\,\hat{\mathbf{x}} + y_3a\,\hat{\mathbf{y}} - z_3a\,\hat{\mathbf{z}}$ | (96$h$) | H |
| $\mathbf{B}_{24}$ = | $\left(\frac{1}{2} - y_3 - z_3\right)\,\mathbf{a}_1 + \left(\frac{1}{2} + x_3 - z_3\right)\,\mathbf{a}_2 + (x_3 - y_3)\,\mathbf{a}_3$ | = | $x_3a\,\hat{\mathbf{x}} - y_3a\,\hat{\mathbf{y}} + \left(\frac{1}{2} - z_3\right)a\,\hat{\mathbf{z}}$ | (96$h$) | H |

| | | | | | | |
|---|---|---|---|---|---|---|
| $\mathbf{B}_{25}$ | = | $(x_3+y_3)\,\mathbf{a}_1+(y_3+z_3)\,\mathbf{a}_2+(x_3+z_3)\,\mathbf{a}_3$ | = | $z_3a\,\hat{\mathbf{x}}+x_3a\,\hat{\mathbf{y}}+y_3a\,\hat{\mathbf{z}}$ | (96$h$) | H |
| $\mathbf{B}_{26}$ | = | $\left(\frac{1}{2}-x_3-y_3\right)\mathbf{a}_1+\left(\frac{1}{2}-y_3+z_3\right)\mathbf{a}_2+(-x_3+z_3)\,\mathbf{a}_3$ | = | $z_3a\,\hat{\mathbf{x}}-x_3a\,\hat{\mathbf{y}}+\left(\frac{1}{2}-y_3\right)a\,\hat{\mathbf{z}}$ | (96$h$) | H |
| $\mathbf{B}_{27}$ | = | $\left(\frac{1}{2}-x_3+y_3\right)\mathbf{a}_1+(y_3-z_3)\,\mathbf{a}_2+\left(\frac{1}{2}-x_3-z_3\right)\mathbf{a}_3$ | = | $-z_3a\,\hat{\mathbf{x}}+\left(\frac{1}{2}-x_3\right)a\,\hat{\mathbf{y}}+y_3a\,\hat{\mathbf{z}}$ | (96$h$) | H |
| $\mathbf{B}_{28}$ | = | $(x_3-y_3)\,\mathbf{a}_1+\left(\frac{1}{2}-y_3-z_3\right)\mathbf{a}_2+\left(\frac{1}{2}+x_3-z_3\right)\mathbf{a}_3$ | = | $\left(\frac{1}{2}-z_3\right)a\,\hat{\mathbf{x}}+x_3a\,\hat{\mathbf{y}}-y_3a\,\hat{\mathbf{z}}$ | (96$h$) | H |
| $\mathbf{B}_{29}$ | = | $(x_3+z_3)\,\mathbf{a}_1+(x_3+y_3)\,\mathbf{a}_2+(y_3+z_3)\,\mathbf{a}_3$ | = | $y_3a\,\hat{\mathbf{x}}+z_3a\,\hat{\mathbf{y}}+x_3a\,\hat{\mathbf{z}}$ | (96$h$) | H |
| $\mathbf{B}_{30}$ | = | $(-x_3+z_3)\,\mathbf{a}_1+\left(\frac{1}{2}-x_3-y_3\right)\mathbf{a}_2+\left(\frac{1}{2}-y_3+z_3\right)\mathbf{a}_3$ | = | $\left(\frac{1}{2}-y_3\right)a\,\hat{\mathbf{x}}+z_3a\,\hat{\mathbf{y}}-x_3a\,\hat{\mathbf{z}}$ | (96$h$) | H |
| $\mathbf{B}_{31}$ | = | $\left(\frac{1}{2}-x_3-z_3\right)\mathbf{a}_1+\left(\frac{1}{2}-x_3+y_3\right)\mathbf{a}_2+(y_3-z_3)\,\mathbf{a}_3$ | = | $y_3a\,\hat{\mathbf{x}}-z_3a\,\hat{\mathbf{y}}+\left(\frac{1}{2}-x_3\right)a\,\hat{\mathbf{z}}$ | (96$h$) | H |
| $\mathbf{B}_{32}$ | = | $\left(\frac{1}{2}+x_3-z_3\right)\mathbf{a}_1+(x_3-y_3)\,\mathbf{a}_2+\left(\frac{1}{2}-y_3-z_3\right)\mathbf{a}_3$ | = | $-y_3a\,\hat{\mathbf{x}}+\left(\frac{1}{2}-z_3\right)a\,\hat{\mathbf{y}}+x_3a\,\hat{\mathbf{z}}$ | (96$h$) | H |
| $\mathbf{B}_{33}$ | = | $\left(\frac{1}{2}+x_3-z_3\right)\mathbf{a}_1+(y_3-z_3)\,\mathbf{a}_2+(x_3+y_3)\,\mathbf{a}_3$ | = | $\left(\frac{3}{4}+y_3\right)a\,\hat{\mathbf{x}}+\left(\frac{1}{4}+x_3\right)a\,\hat{\mathbf{y}}+\left(\frac{1}{4}-z_3\right)a\,\hat{\mathbf{z}}$ | (96$h$) | H |
| $\mathbf{B}_{34}$ | = | $\left(\frac{1}{2}-x_3-z_3\right)\mathbf{a}_1+\left(\frac{1}{2}-y_3-z_3\right)\mathbf{a}_2+\left(\frac{1}{2}-x_3-y_3\right)\mathbf{a}_3$ | = | $\left(\frac{1}{4}-y_3\right)a\,\hat{\mathbf{x}}+\left(\frac{1}{4}-x_3\right)a\,\hat{\mathbf{y}}+\left(\frac{1}{4}-z_3\right)a\,\hat{\mathbf{z}}$ | (96$h$) | H |
| $\mathbf{B}_{35}$ | = | $(-x_3+z_3)\,\mathbf{a}_1+(y_3+z_3)\,\mathbf{a}_2+\left(\frac{1}{2}-x_3+y_3\right)\mathbf{a}_3$ | = | $\left(\frac{1}{4}+y_3\right)a\,\hat{\mathbf{x}}+\left(\frac{1}{4}-x_3\right)a\,\hat{\mathbf{y}}+\left(\frac{3}{4}+z_3\right)a\,\hat{\mathbf{z}}$ | (96$h$) | H |
| $\mathbf{B}_{36}$ | = | $(x_3+z_3)\,\mathbf{a}_1+\left(\frac{1}{2}-y_3+z_3\right)\mathbf{a}_2+(x_3-y_3)\,\mathbf{a}_3$ | = | $\left(\frac{1}{4}-y_3\right)a\,\hat{\mathbf{x}}+\left(\frac{3}{4}+x_3\right)a\,\hat{\mathbf{y}}+\left(\frac{1}{4}+z_3\right)a\,\hat{\mathbf{z}}$ | (96$h$) | H |
| $\mathbf{B}_{37}$ | = | $\left(\frac{1}{2}-y_3+z_3\right)\mathbf{a}_1+(x_3-y_3)\,\mathbf{a}_2+(x_3+z_3)\,\mathbf{a}_3$ | = | $\left(\frac{3}{4}+x_3\right)a\,\hat{\mathbf{x}}+\left(\frac{1}{4}+z_3\right)a\,\hat{\mathbf{y}}+\left(\frac{1}{4}-y_3\right)a\,\hat{\mathbf{z}}$ | (96$h$) | H |
| $\mathbf{B}_{38}$ | = | $(y_3+z_3)\,\mathbf{a}_1+\left(\frac{1}{2}-x_3+y_3\right)\mathbf{a}_2+(-x_3+z_3)\,\mathbf{a}_3$ | = | $\left(\frac{1}{4}-x_3\right)a\,\hat{\mathbf{x}}+\left(\frac{3}{4}+z_3\right)a\,\hat{\mathbf{y}}+\left(\frac{1}{4}+y_3\right)a\,\hat{\mathbf{z}}$ | (96$h$) | H |
| $\mathbf{B}_{39}$ | = | $\left(\frac{1}{2}-y_3-z_3\right)\mathbf{a}_1+\left(\frac{1}{2}-x_3-y_3\right)\mathbf{a}_2+\left(\frac{1}{2}-x_3-z_3\right)\mathbf{a}_3$ | = | $\left(\frac{1}{4}-x_3\right)a\,\hat{\mathbf{x}}+\left(\frac{1}{4}-z_3\right)a\,\hat{\mathbf{y}}+\left(\frac{1}{4}-y_3\right)a\,\hat{\mathbf{z}}$ | (96$h$) | H |
| $\mathbf{B}_{40}$ | = | $(y_3-z_3)\,\mathbf{a}_1+(x_3+y_3)\,\mathbf{a}_2+\left(\frac{1}{2}+x_3-z_3\right)\mathbf{a}_3$ | = | $\left(\frac{1}{4}+x_3\right)a\,\hat{\mathbf{x}}+\left(\frac{1}{4}-z_3\right)a\,\hat{\mathbf{y}}+\left(\frac{3}{4}+y_3\right)a\,\hat{\mathbf{z}}$ | (96$h$) | H |
| $\mathbf{B}_{41}$ | = | $\left(\frac{1}{2}-x_3+y_3\right)\mathbf{a}_1+(-x_3+z_3)\,\mathbf{a}_2+(y_3+z_3)\,\mathbf{a}_3$ | = | $\left(\frac{3}{4}+z_3\right)a\,\hat{\mathbf{x}}+\left(\frac{1}{4}+y_3\right)a\,\hat{\mathbf{y}}+\left(\frac{1}{4}-x_3\right)a\,\hat{\mathbf{z}}$ | (96$h$) | H |
| $\mathbf{B}_{42}$ | = | $(x_3-y_3)\,\mathbf{a}_1+(x_3+z_3)\,\mathbf{a}_2+\left(\frac{1}{2}-y_3+z_3\right)\mathbf{a}_3$ | = | $\left(\frac{1}{4}+z_3\right)a\,\hat{\mathbf{x}}+\left(\frac{1}{4}-y_3\right)a\,\hat{\mathbf{y}}+\left(\frac{3}{4}+x_3\right)a\,\hat{\mathbf{z}}$ | (96$h$) | H |
| $\mathbf{B}_{43}$ | = | $(x_3+y_3)\,\mathbf{a}_1+\left(\frac{1}{2}+x_3-z_3\right)\mathbf{a}_2+(y_3-z_3)\,\mathbf{a}_3$ | = | $\left(\frac{1}{4}-z_3\right)a\,\hat{\mathbf{x}}+\left(\frac{3}{4}+y_3\right)a\,\hat{\mathbf{y}}+\left(\frac{1}{4}+x_3\right)a\,\hat{\mathbf{z}}$ | (96$h$) | H |
| $\mathbf{B}_{44}$ | = | $\left(\frac{1}{2}-x_3-y_3\right)\mathbf{a}_1+\left(\frac{1}{2}-x_3-z_3\right)\mathbf{a}_2+\left(\frac{1}{2}-y_3-z_3\right)\mathbf{a}_3$ | = | $\left(\frac{1}{4}-z_3\right)a\,\hat{\mathbf{x}}+\left(\frac{1}{4}-y_3\right)a\,\hat{\mathbf{y}}+\left(\frac{1}{4}-x_3\right)a\,\hat{\mathbf{z}}$ | (96$h$) | H |
| $\mathbf{B}_{45}$ | = | $(-y_3-z_3)\,\mathbf{a}_1+(-x_3-z_3)\,\mathbf{a}_2+(-x_3-y_3)\,\mathbf{a}_3$ | = | $-x_3a\,\hat{\mathbf{x}}-y_3a\,\hat{\mathbf{y}}-z_3a\,\hat{\mathbf{z}}$ | (96$h$) | H |
| $\mathbf{B}_{46}$ | = | $\left(\frac{1}{2} + y_3 - z_3\right) \mathbf{a}_1 + (x_3 - z_3) \mathbf{a}_2 + \left(\frac{1}{2} + x_3 + y_3\right) \mathbf{a}_3$ | = | $x_3 a \hat{\mathbf{x}} + \left(\frac{1}{2} + y_3\right) a \hat{\mathbf{y}} - z_3 a \hat{\mathbf{z}}$ | $(96h)$ | H |
| $\mathbf{B}_{47}$ | = | $(-y_3 + z_3) \mathbf{a}_1 + \left(\frac{1}{2} + x_3 + z_3\right) \mathbf{a}_2 + \left(\frac{1}{2} + x_3 - y_3\right) \mathbf{a}_3$ | = | $\left(\frac{1}{2} + x_3\right) a \hat{\mathbf{x}} - y_3 a \hat{\mathbf{y}} + z_3 a \hat{\mathbf{z}}$ | $(96h)$ | H |
| $\mathbf{B}_{48}$ | = | $\left(\frac{1}{2} + y_3 + z_3\right) \mathbf{a}_1 + \left(\frac{1}{2} - x_3 + z_3\right) \mathbf{a}_2 + (-x_3 + y_3) \mathbf{a}_3$ | = | $-x_3 a \hat{\mathbf{x}} + y_3 a \hat{\mathbf{y}} + \left(\frac{1}{2} + z_3\right) a \hat{\mathbf{z}}$ | $(96h)$ | H |
| $\mathbf{B}_{49}$ | = | $(-x_3 - y_3) \mathbf{a}_1 + (-y_3 - z_3) \mathbf{a}_2 + (-x_3 - z_3) \mathbf{a}_3$ | = | $-z_3 a \hat{\mathbf{x}} - x_3 a \hat{\mathbf{y}} - y_3 a \hat{\mathbf{z}}$ | $(96h)$ | H |
| $\mathbf{B}_{50}$ | = | $\left(\frac{1}{2} + x_3 + y_3\right) \mathbf{a}_1 + \left(\frac{1}{2} + y_3 - z_3\right) \mathbf{a}_2 + (x_3 - z_3) \mathbf{a}_3$ | = | $-z_3 a \hat{\mathbf{x}} + x_3 a \hat{\mathbf{y}} + \left(\frac{1}{2} + y_3\right) a \hat{\mathbf{z}}$ | $(96h)$ | H |
| $\mathbf{B}_{51}$ | = | $\left(\frac{1}{2} + x_3 - y_3\right) \mathbf{a}_1 + (-y_3 + z_3) \mathbf{a}_2 + \left(\frac{1}{2} + x_3 + z_3\right) \mathbf{a}_3$ | = | $z_3 a \hat{\mathbf{x}} + \left(\frac{1}{2} + x_3\right) a \hat{\mathbf{y}} - y_3 a \hat{\mathbf{z}}$ | $(96h)$ | H |
| $\mathbf{B}_{52}$ | = | $(-x_3 + y_3) \mathbf{a}_1 + \left(\frac{1}{2} + y_3 + z_3\right) \mathbf{a}_2 + \left(\frac{1}{2} - x_3 + z_3\right) \mathbf{a}_3$ | = | $\left(\frac{1}{2} + z_3\right) a \hat{\mathbf{x}} - x_3 a \hat{\mathbf{y}} + y_3 a \hat{\mathbf{z}}$ | $(96h)$ | H |
| $\mathbf{B}_{53}$ | = | $(-x_3 - z_3) \mathbf{a}_1 + (-x_3 - y_3) \mathbf{a}_2 + (-y_3 - z_3) \mathbf{a}_3$ | = | $-y_3 a \hat{\mathbf{x}} - z_3 a \hat{\mathbf{y}} - x_3 a \hat{\mathbf{z}}$ | $(96h)$ | H |
| $\mathbf{B}_{54}$ | = | $(x_3 - z_3) \mathbf{a}_1 + \left(\frac{1}{2} + x_3 + y_3\right) \mathbf{a}_2 + \left(\frac{1}{2} + y_3 - z_3\right) \mathbf{a}_3$ | = | $\left(\frac{1}{2} + y_3\right) a \hat{\mathbf{x}} - z_3 a \hat{\mathbf{y}} + x_3 a \hat{\mathbf{z}}$ | $(96h)$ | H |
| $\mathbf{B}_{55}$ | = | $\left(\frac{1}{2} + x_3 + z_3\right) \mathbf{a}_1 + \left(\frac{1}{2} + x_3 - y_3\right) \mathbf{a}_2 + (-y_3 + z_3) \mathbf{a}_3$ | = | $-y_3 a \hat{\mathbf{x}} + z_3 a \hat{\mathbf{y}} + \left(\frac{1}{2} + x_3\right) a \hat{\mathbf{z}}$ | $(96h)$ | H |
| $\mathbf{B}_{56}$ | = | $\left(\frac{1}{2} - x_3 + z_3\right) \mathbf{a}_1 + (-x_3 + y_3) \mathbf{a}_2 + \left(\frac{1}{2} + y_3 + z_3\right) \mathbf{a}_3$ | = | $y_3 a \hat{\mathbf{x}} + \left(\frac{1}{2} + z_3\right) a \hat{\mathbf{y}} - x_3 a \hat{\mathbf{z}}$ | $(96h)$ | H |
| $\mathbf{B}_{57}$ | = | $\left(\frac{1}{2} - x_3 + z_3\right) \mathbf{a}_1 + (-y_3 + z_3) \mathbf{a}_2 + (-x_3 - y_3) \mathbf{a}_3$ | = | $-a\left(y_3 + \frac{1}{4}\right) \hat{\mathbf{x}} + \left(\frac{1}{4} - x_3\right) a \hat{\mathbf{y}} + \left(\frac{1}{4} + z_3\right) a \hat{\mathbf{z}}$ | $(96h)$ | H |
| $\mathbf{B}_{58}$ | = | $\left(\frac{1}{2} + x_3 + z_3\right) \mathbf{a}_1 + \left(\frac{1}{2} + y_3 + z_3\right) \mathbf{a}_2 + \left(\frac{1}{2} + x_3 + y_3\right) \mathbf{a}_3$ | = | $\left(\frac{1}{4} + y_3\right) a \hat{\mathbf{x}} + \left(\frac{1}{4} + x_3\right) a \hat{\mathbf{y}} + \left(\frac{1}{4} + z_3\right) a \hat{\mathbf{z}}$ | $(96h)$ | H |
| $\mathbf{B}_{59}$ | = | $(x_3 - z_3) \mathbf{a}_1 + (-y_3 - z_3) \mathbf{a}_2 + \left(\frac{1}{2} + x_3 - y_3\right) \mathbf{a}_3$ | = | $\left(\frac{1}{4} - y_3\right) a \hat{\mathbf{x}} + \left(\frac{1}{4} + x_3\right) a \hat{\mathbf{y}} - a\left(z_3 + \frac{1}{4}\right) \hat{\mathbf{z}}$ | $(96h)$ | H |
| $\mathbf{B}_{60}$ | = | $(-x_3 - z_3) \mathbf{a}_1 + \left(\frac{1}{2} + y_3 - z_3\right) \mathbf{a}_2 + (-x_3 + y_3) \mathbf{a}_3$ | = | $\left(\frac{1}{4} + y_3\right) a \hat{\mathbf{x}} - a\left(x_3 + \frac{1}{4}\right) \hat{\mathbf{y}} + \left(\frac{1}{4} - z_3\right) a \hat{\mathbf{z}}$ | $(96h)$ | H |
| $\mathbf{B}_{61}$ | = | $\left(\frac{1}{2} + y_3 - z_3\right) \mathbf{a}_1 + (-x_3 + y_3) \mathbf{a}_2 + (-x_3 - z_3) \mathbf{a}_3$ | = | $-a\left(x_3 + \frac{1}{4}\right) \hat{\mathbf{x}} + \left(\frac{1}{4} - z_3\right) a \hat{\mathbf{y}} + \left(\frac{1}{4} + y_3\right) a \hat{\mathbf{z}}$ | $(96h)$ | H |
| $\mathbf{B}_{62}$ | = | $(-y_3 - z_3) \mathbf{a}_1 + \left(\frac{1}{2} + x_3 - y_3\right) \mathbf{a}_2 + (x_3 - z_3) \mathbf{a}_3$ | = | $\left(\frac{1}{4} + x_3\right) a \hat{\mathbf{x}} - a\left(z_3 + \frac{1}{4}\right) \hat{\mathbf{y}} + \left(\frac{1}{4} - y_3\right) a \hat{\mathbf{z}}$ | $(96h)$ | H |
| $\mathbf{B}_{63}$ | = | $\left(\frac{1}{2} + y_3 + z_3\right) \mathbf{a}_1 + \left(\frac{1}{2} + x_3 + y_3\right) \mathbf{a}_2 + \left(\frac{1}{2} + x_3 + z_3\right) \mathbf{a}_3$ | = | $\left(\frac{1}{4} + x_3\right) a \hat{\mathbf{x}} + \left(\frac{1}{4} + z_3\right) a \hat{\mathbf{y}} + \left(\frac{1}{4} + y_3\right) a \hat{\mathbf{z}}$ | $(96h)$ | H |
| $\mathbf{B}_{64}$ | = | $(-y_3 + z_3) \mathbf{a}_1 + (-x_3 - y_3) \mathbf{a}_2 + \left(\frac{1}{2} - x_3 + z_3\right) \mathbf{a}_3$ | = | $\left(\frac{1}{4} - x_3\right) a \hat{\mathbf{x}} + \left(\frac{1}{4} + z_3\right) a \hat{\mathbf{y}} - a\left(y_3 + \frac{1}{4}\right) \hat{\mathbf{z}}$ | $(96h)$ | H |
| $\mathbf{B}_{65}$ | = | $\left(\frac{1}{2} + x_3 - y_3\right) \mathbf{a}_1 + (x_3 - z_3) \mathbf{a}_2 + (-y_3 - z_3) \mathbf{a}_3$ | = | $-a\left(z_3 + \frac{1}{4}\right) \hat{\mathbf{x}} + \left(\frac{1}{4} - y_3\right) a \hat{\mathbf{y}} + \left(\frac{1}{4} + x_3\right) a \hat{\mathbf{z}}$ | $(96h)$ | H |
| $\mathbf{B}_{66}$ | = | $(-x_3 + y_3) \mathbf{a}_1 + (-x_3 - z_3) \mathbf{a}_2 + \left(\frac{1}{2} + y_3 - z_3\right) \mathbf{a}_3$ | = | $\left(\frac{1}{4} - z_3\right) a \hat{\mathbf{x}} + \left(\frac{1}{4} + y_3\right) a \hat{\mathbf{y}} - a\left(x_3 + \frac{1}{4}\right) \hat{\mathbf{z}}$ | $(96h)$ | H |
| $\mathbf{B}_{67}$ | = | $\left(-x_3 - y_3\right)\,\mathbf{a}_1 + \left(\frac{1}{2} - x_3 + z_3\right)\,\mathbf{a}_2 + \left(-y_3 + z_3\right)\,\mathbf{a}_3$ | = | $\left(\frac{1}{4} + z_3\right)a\,\mathbf{\hat{x}} - a\left(y_3 + \frac{1}{4}\right)\,\mathbf{\hat{y}} + \left(\frac{1}{4} - x_3\right)a\,\mathbf{\hat{z}}$ | (96$h$) | H |
| $\mathbf{B}_{68}$ | = | $\left(\frac{1}{2} + x_3 + y_3\right)\,\mathbf{a}_1 + \left(\frac{1}{2} + x_3 + z_3\right)\,\mathbf{a}_2 + \left(\frac{1}{2} + y_3 + z_3\right)\,\mathbf{a}_3$ | = | $\left(\frac{1}{4} + z_3\right)a\,\mathbf{\hat{x}} + \left(\frac{1}{4} + y_3\right)a\,\mathbf{\hat{y}} + \left(\frac{1}{4} + x_3\right)a\,\mathbf{\hat{z}}$ | (96$h$) | H |
| $\mathbf{B}_{69}$ | = | $\left(y_4 + z_4\right)\,\mathbf{a}_1 + \left(x_4 + z_4\right)\,\mathbf{a}_2 + \left(x_4 + y_4\right)\,\mathbf{a}_3$ | = | $x_4a\,\mathbf{\hat{x}} + y_4a\,\mathbf{\hat{y}} + z_4a\,\mathbf{\hat{z}}$ | (96$h$) | O |
| $\mathbf{B}_{70}$ | = | $\left(\frac{1}{2} - y_4 + z_4\right)\,\mathbf{a}_1 + \left(-x_4 + z_4\right)\,\mathbf{a}_2 + \left(\frac{1}{2} - x_4 - y_4\right)\,\mathbf{a}_3$ | = | $-x_4a\,\mathbf{\hat{x}} + \left(\frac{1}{2} - y_4\right)a\,\mathbf{\hat{y}} + z_4a\,\mathbf{\hat{z}}$ | (96$h$) | O |
| $\mathbf{B}_{71}$ | = | $\left(y_4 - z_4\right)\,\mathbf{a}_1 + \left(\frac{1}{2} - x_4 - z_4\right)\,\mathbf{a}_2 + \left(\frac{1}{2} - x_4 + y_4\right)\,\mathbf{a}_3$ | = | $\left(\frac{1}{2} - x_4\right)a\,\mathbf{\hat{x}} + y_4a\,\mathbf{\hat{y}} - z_4a\,\mathbf{\hat{z}}$ | (96$h$) | O |
| $\mathbf{B}_{72}$ | = | $\left(\frac{1}{2} - y_4 - z_4\right)\,\mathbf{a}_1 + \left(\frac{1}{2} + x_4 - z_4\right)\,\mathbf{a}_2 + \left(x_4 - y_4\right)\,\mathbf{a}_3$ | = | $x_4a\,\mathbf{\hat{x}} - y_4a\,\mathbf{\hat{y}} + \left(\frac{1}{2} - z_4\right)a\,\mathbf{\hat{z}}$ | (96$h$) | O |
| $\mathbf{B}_{73}$ | = | $\left(x_4 + y_4\right)\,\mathbf{a}_1 + \left(y_4 + z_4\right)\,\mathbf{a}_2 + \left(x_4 + z_4\right)\,\mathbf{a}_3$ | = | $z_4a\,\mathbf{\hat{x}} + x_4a\,\mathbf{\hat{y}} + y_4a\,\mathbf{\hat{z}}$ | (96$h$) | O |
| $\mathbf{B}_{74}$ | = | $\left(\frac{1}{2} - x_4 - y_4\right)\,\mathbf{a}_1 + \left(\frac{1}{2} - y_4 + z_4\right)\,\mathbf{a}_2 + \left(-x_4 + z_4\right)\,\mathbf{a}_3$ | = | $z_4a\,\mathbf{\hat{x}} - x_4a\,\mathbf{\hat{y}} + \left(\frac{1}{2} - y_4\right)a\,\mathbf{\hat{z}}$ | (96$h$) | O |
| $\mathbf{B}_{75}$ | = | $\left(\frac{1}{2} - x_4 + y_4\right)\,\mathbf{a}_1 + \left(y_4 - z_4\right)\,\mathbf{a}_2 + \left(\frac{1}{2} - x_4 - z_4\right)\,\mathbf{a}_3$ | = | $-z_4a\,\mathbf{\hat{x}} + \left(\frac{1}{2} - x_4\right)a\,\mathbf{\hat{y}} + y_4a\,\mathbf{\hat{z}}$ | (96$h$) | O |
| $\mathbf{B}_{76}$ | = | $\left(x_4 - y_4\right)\,\mathbf{a}_1 + \left(\frac{1}{2} - y_4 - z_4\right)\,\mathbf{a}_2 + \left(\frac{1}{2} + x_4 - z_4\right)\,\mathbf{a}_3$ | = | $\left(\frac{1}{2} - z_4\right)a\,\mathbf{\hat{x}} + x_4a\,\mathbf{\hat{y}} - y_4a\,\mathbf{\hat{z}}$ | (96$h$) | O |
| $\mathbf{B}_{77}$ | = | $\left(x_4 + z_4\right)\,\mathbf{a}_1 + \left(x_4 + y_4\right)\,\mathbf{a}_2 + \left(y_4 + z_4\right)\,\mathbf{a}_3$ | = | $y_4a\,\mathbf{\hat{x}} + z_4a\,\mathbf{\hat{y}} + x_4a\,\mathbf{\hat{z}}$ | (96$h$) | O |
| $\mathbf{B}_{78}$ | = | $\left(-x_4 + z_4\right)\,\mathbf{a}_1 + \left(\frac{1}{2} - x_4 - y_4\right)\,\mathbf{a}_2 + \left(\frac{1}{2} - y_4 + z_4\right)\,\mathbf{a}_3$ | = | $\left(\frac{1}{2} - y_4\right)a\,\mathbf{\hat{x}} + z_4a\,\mathbf{\hat{y}} - x_4a\,\mathbf{\hat{z}}$ | (96$h$) | O |
| $\mathbf{B}_{79}$ | = | $\left(\frac{1}{2} - x_4 - z_4\right)\,\mathbf{a}_1 + \left(\frac{1}{2} - x_4 + y_4\right)\,\mathbf{a}_2 + \left(y_4 - z_4\right)\,\mathbf{a}_3$ | = | $y_4a\,\mathbf{\hat{x}} - z_4a\,\mathbf{\hat{y}} + \left(\frac{1}{2} - x_4\right)a\,\mathbf{\hat{z}}$ | (96$h$) | O |
| $\mathbf{B}_{80}$ | = | $\left(\frac{1}{2} + x_4 - z_4\right)\,\mathbf{a}_1 + \left(x_4 - y_4\right)\,\mathbf{a}_2 + \left(\frac{1}{2} - y_4 - z_4\right)\,\mathbf{a}_3$ | = | $-y_4a\,\mathbf{\hat{x}} + \left(\frac{1}{2} - z_4\right)a\,\mathbf{\hat{y}} + x_4a\,\mathbf{\hat{z}}$ | (96$h$) | O |
| $\mathbf{B}_{81}$ | = | $\left(\frac{1}{2} + x_4 - z_4\right)\,\mathbf{a}_1 + \left(y_4 - z_4\right)\,\mathbf{a}_2 + \left(x_4 + y_4\right)\,\mathbf{a}_3$ | = | $\left(\frac{3}{4} + y_4\right)a\,\mathbf{\hat{x}} + \left(\frac{1}{4} + x_4\right)a\,\mathbf{\hat{y}} + \left(\frac{1}{4} - z_4\right)a\,\mathbf{\hat{z}}$ | (96$h$) | O |
| $\mathbf{B}_{82}$ | = | $\left(\frac{1}{2} - x_4 - z_4\right)\,\mathbf{a}_1 + \left(\frac{1}{2} - y_4 - z_4\right)\,\mathbf{a}_2 + \left(\frac{1}{2} - x_4 - y_4\right)\,\mathbf{a}_3$ | = | $\left(\frac{1}{4} - y_4\right)a\,\mathbf{\hat{x}} + \left(\frac{1}{4} - x_4\right)a\,\mathbf{\hat{y}} + \left(\frac{1}{4} - z_4\right)a\,\mathbf{\hat{z}}$ | (96$h$) | O |
| $\mathbf{B}_{83}$ | = | $\left(-x_4 + z_4\right)\,\mathbf{a}_1 + \left(y_4 + z_4\right)\,\mathbf{a}_2 + \left(\frac{1}{2} - x_4 + y_4\right)\,\mathbf{a}_3$ | = | $\left(\frac{1}{4} + y_4\right)a\,\mathbf{\hat{x}} + \left(\frac{1}{4} - x_4\right)a\,\mathbf{\hat{y}} + \left(\frac{3}{4} + z_4\right)a\,\mathbf{\hat{z}}$ | (96$h$) | O |
| $\mathbf{B}_{84}$ | = | $\left(x_4 + z_4\right)\,\mathbf{a}_1 + \left(\frac{1}{2} - y_4 + z_4\right)\,\mathbf{a}_2 + \left(x_4 - y_4\right)\,\mathbf{a}_3$ | = | $\left(\frac{1}{4} - y_4\right)a\,\mathbf{\hat{x}} + \left(\frac{3}{4} + x_4\right)a\,\mathbf{\hat{y}} + \left(\frac{1}{4} + z_4\right)a\,\mathbf{\hat{z}}$ | (96$h$) | O |
| $\mathbf{B}_{85}$ | = | $\left(\frac{1}{2} - y_4 + z_4\right)\,\mathbf{a}_1 + \left(x_4 - y_4\right)\,\mathbf{a}_2 + \left(x_4 + z_4\right)\,\mathbf{a}_3$ | = | $\left(\frac{3}{4} + x_4\right)a\,\mathbf{\hat{x}} + \left(\frac{1}{4} + z_4\right)a\,\mathbf{\hat{y}} + \left(\frac{1}{4} - y_4\right)a\,\mathbf{\hat{z}}$ | (96$h$) | O |
| $\mathbf{B}_{86}$ | = | $\left(y_4 + z_4\right)\,\mathbf{a}_1 + \left(\frac{1}{2} - x_4 + y_4\right)\,\mathbf{a}_2 + \left(-x_4 + z_4\right)\,\mathbf{a}_3$ | = | $\left(\frac{1}{4} - x_4\right)a\,\mathbf{\hat{x}} + \left(\frac{3}{4} + z_4\right)a\,\mathbf{\hat{y}} + \left(\frac{1}{4} + y_4\right)a\,\mathbf{\hat{z}}$ | (96$h$) | O |
| $\mathbf{B}_{87}$ | = | $\left(\frac{1}{2} - y_4 - z_4\right)\,\mathbf{a}_1 + \left(\frac{1}{2} - x_4 - y_4\right)\,\mathbf{a}_2 + \left(\frac{1}{2} - x_4 - z_4\right)\,\mathbf{a}_3$ | = | $\left(\frac{1}{4} - x_4\right)a\,\mathbf{\hat{x}} + \left(\frac{1}{4} - z_4\right)a\,\mathbf{\hat{y}} + \left(\frac{1}{4} - y_4\right)a\,\mathbf{\hat{z}}$ | (96$h$) | O |

$$\mathbf{B}_{88} = (y_4 - z_4)\,\mathbf{a}_1 + (x_4 + y_4)\,\mathbf{a}_2 + \left(\frac{1}{2} + x_4 - z_4\right)\mathbf{a}_3 = \left(\frac{1}{4} + x_4\right)a\,\hat{\mathbf{x}} + \left(\frac{1}{4} - z_4\right)a\,\hat{\mathbf{y}} + \left(\frac{3}{4} + y_4\right)a\,\hat{\mathbf{z}} \quad (96h) \quad \text{O}$$

$$\mathbf{B}_{89} = \left(\frac{1}{2} - x_4 + y_4\right)\mathbf{a}_1 + (-x_4 + z_4)\,\mathbf{a}_2 + (y_4 + z_4)\,\mathbf{a}_3 = \left(\frac{3}{4} + z_4\right)a\,\hat{\mathbf{x}} + \left(\frac{1}{4} + y_4\right)a\,\hat{\mathbf{y}} + \left(\frac{1}{4} - x_4\right)a\,\hat{\mathbf{z}} \quad (96h) \quad \text{O}$$

$$\mathbf{B}_{90} = (x_4 - y_4)\,\mathbf{a}_1 + (x_4 + z_4)\,\mathbf{a}_2 + \left(\frac{1}{2} - y_4 + z_4\right)\mathbf{a}_3 = \left(\frac{1}{4} + z_4\right)a\,\hat{\mathbf{x}} + \left(\frac{1}{4} - y_4\right)a\,\hat{\mathbf{y}} + \left(\frac{3}{4} + x_4\right)a\,\hat{\mathbf{z}} \quad (96h) \quad \text{O}$$

$$\mathbf{B}_{91} = (x_4 + y_4)\,\mathbf{a}_1 + \left(\frac{1}{2} + x_4 - z_4\right)\mathbf{a}_2 + (y_4 - z_4)\,\mathbf{a}_3 = \left(\frac{1}{4} - z_4\right)a\,\hat{\mathbf{x}} + \left(\frac{3}{4} + y_4\right)a\,\hat{\mathbf{y}} + \left(\frac{1}{4} + x_4\right)a\,\hat{\mathbf{z}} \quad (96h) \quad \text{O}$$

$$\mathbf{B}_{92} = \left(\frac{1}{2} - x_4 - y_4\right)\mathbf{a}_1 + \left(\frac{1}{2} - x_4 - z_4\right)\mathbf{a}_2 + \left(\frac{1}{2} - y_4 - z_4\right)\mathbf{a}_3 = \left(\frac{1}{4} - z_4\right)a\,\hat{\mathbf{x}} + \left(\frac{1}{4} - y_4\right)a\,\hat{\mathbf{y}} + \left(\frac{1}{4} - x_4\right)a\,\hat{\mathbf{z}} \quad (96h) \quad \text{O}$$

$$\mathbf{B}_{93} = (-y_4 - z_4)\,\mathbf{a}_1 + (-x_4 - z_4)\,\mathbf{a}_2 + (-x_4 - y_4)\,\mathbf{a}_3 = -x_4a\,\hat{\mathbf{x}} - y_4a\,\hat{\mathbf{y}} - z_4a\,\hat{\mathbf{z}} \quad (96h) \quad \text{O}$$

$$\mathbf{B}_{94} = \left(\frac{1}{2} + y_4 - z_4\right)\mathbf{a}_1 + (x_4 - z_4)\,\mathbf{a}_2 + \left(\frac{1}{2} + x_4 + y_4\right)\mathbf{a}_3 = x_4a\,\hat{\mathbf{x}} + \left(\frac{1}{2} + y_4\right)a\,\hat{\mathbf{y}} - z_4a\,\hat{\mathbf{z}} \quad (96h) \quad \text{O}$$

$$\mathbf{B}_{95} = (-y_4 + z_4)\,\mathbf{a}_1 + \left(\frac{1}{2} + x_4 + z_4\right)\mathbf{a}_2 + \left(\frac{1}{2} + x_4 - y_4\right)\mathbf{a}_3 = \left(\frac{1}{2} + x_4\right)a\,\hat{\mathbf{x}} - y_4a\,\hat{\mathbf{y}} + z_4a\,\hat{\mathbf{z}} \quad (96h) \quad \text{O}$$

$$\mathbf{B}_{96} = \left(\frac{1}{2} + y_4 + z_4\right)\mathbf{a}_1 + \left(\frac{1}{2} - x_4 + z_4\right)\mathbf{a}_2 + (-x_4 + y_4)\,\mathbf{a}_3 = -x_4a\,\hat{\mathbf{x}} + y_4a\,\hat{\mathbf{y}} + \left(\frac{1}{2} + z_4\right)a\,\hat{\mathbf{z}} \quad (96h) \quad \text{O}$$

$$\mathbf{B}_{97} = (-x_4 - y_4)\,\mathbf{a}_1 + (-y_4 - z_4)\,\mathbf{a}_2 + (-x_4 - z_4)\,\mathbf{a}_3 = -z_4a\,\hat{\mathbf{x}} - x_4a\,\hat{\mathbf{y}} - y_4a\,\hat{\mathbf{z}} \quad (96h) \quad \text{O}$$

$$\mathbf{B}_{98} = \left(\frac{1}{2} + x_4 + y_4\right)\mathbf{a}_1 + \left(\frac{1}{2} + y_4 - z_4\right)\mathbf{a}_2 + (x_4 - z_4)\,\mathbf{a}_3 = -z_4a\,\hat{\mathbf{x}} + x_4a\,\hat{\mathbf{y}} + \left(\frac{1}{2} + y_4\right)a\,\hat{\mathbf{z}} \quad (96h) \quad \text{O}$$

$$\mathbf{B}_{99} = \left(\frac{1}{2} + x_4 - y_4\right)\mathbf{a}_1 + (-y_4 + z_4)\,\mathbf{a}_2 + \left(\frac{1}{2} + x_4 + z_4\right)\mathbf{a}_3 = z_4a\,\hat{\mathbf{x}} + \left(\frac{1}{2} + x_4\right)a\,\hat{\mathbf{y}} - y_4a\,\hat{\mathbf{z}} \quad (96h) \quad \text{O}$$

$$\mathbf{B}_{100} = (-x_4 + y_4)\,\mathbf{a}_1 + \left(\frac{1}{2} + y_4 + z_4\right)\mathbf{a}_2 + \left(\frac{1}{2} - x_4 + z_4\right)\mathbf{a}_3 = \left(\frac{1}{2} + z_4\right)a\,\hat{\mathbf{x}} - x_4a\,\hat{\mathbf{y}} + y_4a\,\hat{\mathbf{z}} \quad (96h) \quad \text{O}$$

$$\mathbf{B}_{101} = (-x_4 - z_4)\,\mathbf{a}_1 + (-x_4 - y_4)\,\mathbf{a}_2 + (-y_4 - z_4)\,\mathbf{a}_3 = -y_4a\,\hat{\mathbf{x}} - z_4a\,\hat{\mathbf{y}} - x_4a\,\hat{\mathbf{z}} \quad (96h) \quad \text{O}$$

$$\mathbf{B}_{102} = (x_4 - z_4)\,\mathbf{a}_1 + \left(\frac{1}{2} + x_4 + y_4\right)\mathbf{a}_2 + \left(\frac{1}{2} + y_4 - z_4\right)\mathbf{a}_3 = \left(\frac{1}{2} + y_4\right)a\,\hat{\mathbf{x}} - z_4a\,\hat{\mathbf{y}} + x_4a\,\hat{\mathbf{z}} \quad (96h) \quad \text{O}$$

$$\mathbf{B}_{103} = \left(\frac{1}{2} + x_4 + z_4\right)\mathbf{a}_1 + \left(\frac{1}{2} + x_4 - y_4\right)\mathbf{a}_2 + (-y_4 + z_4)\,\mathbf{a}_3 = -y_4a\,\hat{\mathbf{x}} + z_4a\,\hat{\mathbf{y}} + \left(\frac{1}{2} + x_4\right)a\,\hat{\mathbf{z}} \quad (96h) \quad \text{O}$$

$$\mathbf{B}_{104} = \left(\frac{1}{2} - x_4 + z_4\right)\mathbf{a}_1 + (-x_4 + y_4)\,\mathbf{a}_2 + \left(\frac{1}{2} + y_4 + z_4\right)\mathbf{a}_3 = y_4a\,\hat{\mathbf{x}} + \left(\frac{1}{2} + z_4\right)a\,\hat{\mathbf{y}} - x_4a\,\hat{\mathbf{z}} \quad (96h) \quad \text{O}$$

$$\mathbf{B}_{105} = \left(\frac{1}{2} - x_4 + z_4\right)\mathbf{a}_1 + (-y_4 + z_4)\,\mathbf{a}_2 + (-x_4 - y_4)\,\mathbf{a}_3 = -a\left(y_4 + \frac{1}{4}\right)\hat{\mathbf{x}} + \left(\frac{1}{4} - x_4\right)a\,\hat{\mathbf{y}} + \left(\frac{1}{4} + z_4\right)a\,\hat{\mathbf{z}} \quad (96h) \quad \text{O}$$

$$\mathbf{B}_{106} = \left(\frac{1}{2} + x_4 + z_4\right)\mathbf{a}_1 + \left(\frac{1}{2} + y_4 + z_4\right)\mathbf{a}_2 + \left(\frac{1}{2} + x_4 + y_4\right)\mathbf{a}_3 = \left(\frac{1}{4} + y_4\right)a\,\hat{\mathbf{x}} + \left(\frac{1}{4} + x_4\right)a\,\hat{\mathbf{y}} + \left(\frac{1}{4} + z_4\right)a\,\hat{\mathbf{z}} \quad (96h) \quad \text{O}$$

$$\mathbf{B}_{107} = (x_4 - z_4)\,\mathbf{a}_1 + (-y_4 - z_4)\,\mathbf{a}_2 + \left(\frac{1}{2} + x_4 - y_4\right)\mathbf{a}_3 = \left(\frac{1}{4} - y_4\right)a\,\hat{\mathbf{x}} + \left(\frac{1}{4} + x_4\right)a\,\hat{\mathbf{y}} - a\left(z_4 + \frac{1}{4}\right)\hat{\mathbf{z}} \quad (96h) \quad \text{O}$$

$$\mathbf{B}_{108} = (-x_4 - z_4)\,\mathbf{a}_1 + \left(\frac{1}{2} + y_4 - z_4\right)\mathbf{a}_2 + (-x_4 + y_4)\,\mathbf{a}_3 = \left(\frac{1}{4} + y_4\right)a\,\hat{\mathbf{x}} - a\left(x_4 + \frac{1}{4}\right)\hat{\mathbf{y}} + \left(\frac{1}{4} - z_4\right)a\,\hat{\mathbf{z}} \quad (96h) \quad \text{O}$$

| | | | | | | |
|---|---|---|---|---|---|---|
| $\mathbf{B}_{109}$ | = | $\left(\frac{1}{2}+y_4-z_4\right)\mathbf{a}_1+\left(-x_4+y_4\right)\mathbf{a}_2+\left(-x_4-z_4\right)\mathbf{a}_3$ | = | $-a\left(x_4+\frac{1}{4}\right)\hat{\mathbf{x}}+\left(\frac{1}{4}-z_4\right)a\,\hat{\mathbf{y}}+\left(\frac{1}{4}+y_4\right)a\,\hat{\mathbf{z}}$ | (96$h$) | O |
| $\mathbf{B}_{110}$ | = | $\left(-y_4-z_4\right)\mathbf{a}_1+\left(\frac{1}{2}+x_4-y_4\right)\mathbf{a}_2+\left(x_4-z_4\right)\mathbf{a}_3$ | = | $\left(\frac{1}{4}+x_4\right)a\,\hat{\mathbf{x}}-a\left(z_4+\frac{1}{4}\right)\hat{\mathbf{y}}+\left(\frac{1}{4}-y_4\right)a\,\hat{\mathbf{z}}$ | (96$h$) | O |
| $\mathbf{B}_{111}$ | = | $\left(\frac{1}{2}+y_4+z_4\right)\mathbf{a}_1+\left(\frac{1}{2}+x_4+y_4\right)\mathbf{a}_2+\left(\frac{1}{2}+x_4+z_4\right)\mathbf{a}_3$ | = | $\left(\frac{1}{4}+x_4\right)a\,\hat{\mathbf{x}}+\left(\frac{1}{4}+z_4\right)a\,\hat{\mathbf{y}}+\left(\frac{1}{4}+y_4\right)a\,\hat{\mathbf{z}}$ | (96$h$) | O |
| $\mathbf{B}_{112}$ | = | $\left(-y_4+z_4\right)\mathbf{a}_1+\left(-x_4-y_4\right)\mathbf{a}_2+\left(\frac{1}{2}-x_4+z_4\right)\mathbf{a}_3$ | = | $\left(\frac{1}{4}-x_4\right)a\,\hat{\mathbf{x}}+\left(\frac{1}{4}+z_4\right)a\,\hat{\mathbf{y}}-a\left(y_4+\frac{1}{4}\right)\hat{\mathbf{z}}$ | (96$h$) | O |
| $\mathbf{B}_{113}$ | = | $\left(\frac{1}{2}+x_4-y_4\right)\mathbf{a}_1+\left(x_4-z_4\right)\mathbf{a}_2+\left(-y_4-z_4\right)\mathbf{a}_3$ | = | $-a\left(z_4+\frac{1}{4}\right)\hat{\mathbf{x}}+\left(\frac{1}{4}-y_4\right)a\,\hat{\mathbf{y}}+\left(\frac{1}{4}+x_4\right)a\,\hat{\mathbf{z}}$ | (96$h$) | O |
| $\mathbf{B}_{114}$ | = | $\left(-x_4+y_4\right)\mathbf{a}_1+\left(-x_4-z_4\right)\mathbf{a}_2+\left(\frac{1}{2}+y_4-z_4\right)\mathbf{a}_3$ | = | $\left(\frac{1}{4}-z_4\right)a\,\hat{\mathbf{x}}+\left(\frac{1}{4}+y_4\right)a\,\hat{\mathbf{y}}-a\left(x_4+\frac{1}{4}\right)\hat{\mathbf{z}}$ | (96$h$) | O |
| $\mathbf{B}_{115}$ | = | $\left(-x_4-y_4\right)\mathbf{a}_1+\left(\frac{1}{2}-x_4+z_4\right)\mathbf{a}_2+\left(-y_4+z_4\right)\mathbf{a}_3$ | = | $\left(\frac{1}{4}+z_4\right)a\,\hat{\mathbf{x}}-a\left(y_4+\frac{1}{4}\right)\hat{\mathbf{y}}+\left(\frac{1}{4}-x_4\right)a\,\hat{\mathbf{z}}$ | (96$h$) | O |
| $\mathbf{B}_{116}$ | = | $\left(\frac{1}{2}+x_4+y_4\right)\mathbf{a}_1+\left(\frac{1}{2}+x_4+z_4\right)\mathbf{a}_2+\left(\frac{1}{2}+y_4+z_4\right)\mathbf{a}_3$ | = | $\left(\frac{1}{4}+z_4\right)a\,\hat{\mathbf{x}}+\left(\frac{1}{4}+y_4\right)a\,\hat{\mathbf{y}}+\left(\frac{1}{4}+x_4\right)a\,\hat{\mathbf{z}}$ | (96$h$) | O |

---

---

**Geometry files:**

- CIF: pp. 1837
- POSCAR: pp. 1838

## CIF and POSCAR Files

$NaC_5H_{11}O_8S$: A5B11CD8E_aP26_1_5a_11a_a_8a_a - CIF

```
# CIF file
data_findsym-output
_audit_creation_method FINDSYM

_chemical_name_mineral 'C5H11NaO8S'
_chemical_formula_sum 'C5 H11 Na O8 S'

loop_
_publ_author_name
 'A. H. Haines'
 'D. L. Hughes'
_journal_name_full_name
;
 Acta Crystallographica Section E: Crystallographic Communications
;
_journal_volume 72
_journal_year 2016
_journal_page_first 628
_journal_page_last 631
_publ_Section_title
;
 Crystal structure of sodium (1S)-D-lyxit-1-yl sulfonate
;

_aflow_title 'NaC$_{5}$H$_{11}$O$_{8}$S Structure'
_aflow_proto 'A5B11CD8E_aP26_1_5a_11a_a_8a_a'
_aflow_params 'a,b/a,c/a,\alpha,\beta,\gamma,x_{1},y_{1},z_{1},x_{2},y_{
    ↪ 2},z_{2},x_{3},y_{3},z_{3},x_{4},y_{4},z_{4},x_{5},y_{5},z_{5},
    ↪ x_{6},y_{6},z_{6},x_{7},y_{7},z_{7},x_{8},y_{8},z_{8},x_{9},y_{
    ↪ 9},z_{9},x_{10},y_{10},z_{10},x_{11},y_{11},z_{11},x_{12},y_{12
    ↪ },z_{12},x_{13},y_{13},z_{13},x_{14},y_{14},z_{14},x_{15},y_{15
    ↪ },z_{15},x_{16},y_{16},z_{16},x_{17},y_{17},z_{17},x_{18},y_{18
    ↪ },z_{18},x_{19},y_{19},z_{19},x_{20},y_{20},z_{20},x_{21},y_{21
    ↪ },z_{21},x_{22},y_{22},z_{22},x_{23},y_{23},z_{23},x_{24},y_{24
    ↪ },z_{24},x_{25},y_{25},z_{25},x_{26},y_{26},z_{26}'
_aflow_params_values '4.8558,1.20466246551,1.81123604761,76.515,81.528,
    ↪ 71.392,0.1957,0.1206,0.62783,0.2966,0.2336,0.7408,0.0562,0.3094
    ↪ ,0.86872,0.1694,0.4162,-0.0197,-0.0682,0.5331,0.0947,0.158,-
    ↪ 0.022,0.679,0.45,0.117,0.794,-0.096,0.433,0.816,0.246,0.548,-
    ↪ 0.081,0.764,0.632,0.037,0.867,0.409,0.171,0.893,0.248,0.503,
    ↪ 0.552,0.42,0.642,0.781,0.158,-0.018,0.345,0.159,0.144,0.035,
    ↪ 0.799,0.113,0.14247,0.62758,0.42059,-0.0431,0.2976,0.55444,
    ↪ 0.5597,0.2447,0.39414,0.708,0.8369,0.55456,0.3305,-0.052,
    ↪ 0.37321,0.3613,0.4537,0.66271,-0.0376,0.1035,-0.05381,0.4056,
    ↪ 0.2355,0.05675,0.0303,0.6739,0.17348,0.47392,0.02789,0.47548'
_aflow_Strukturbericht 'None'
_aflow_Pearson 'aP26'

_symmetry_space_group_name_H-M "P 1"
_symmetry_Int_Tables_number 1

_cell_length_a     4.85580
_cell_length_b     5.84960
_cell_length_c     8.79500
_cell_angle_alpha 76.51500
_cell_angle_beta  81.52800
_cell_angle_gamma 71.39200

loop_
_space_group_symop_id
_space_group_symop_operation_xyz
1 x,y,z

loop_
_atom_site_label
_atom_site_type_symbol
_atom_site_symmetry_multiplicity
_atom_site_Wyckoff_label
_atom_site_fract_x
_atom_site_fract_y
_atom_site_fract_z
_atom_site_occupancy
C1 C 1 a 0.19570 0.12060 0.62783 1.00000
C2 C 1 a 0.29660 0.23360 0.74080 1.00000
C3 C 1 a 0.05620 0.30940 0.86872 1.00000
C4 C 1 a 0.16940 0.41620 -0.01970 1.00000
C5 C 1 a -0.06820 0.53310 0.09470 1.00000
H1 H 1 a 0.15800 -0.02200 0.67900 1.00000
H2 H 1 a 0.45000 0.11700 0.79400 1.00000
H3 H 1 a -0.09600 0.43300 0.81600 1.00000
H4 H 1 a 0.24600 0.54800 -0.08100 1.00000
H5 H 1 a 0.76400 0.63200 0.03700 1.00000
H6 H 1 a 0.86700 0.40900 0.17100 1.00000
H7 H 1 a 0.89300 0.24800 0.50300 1.00000
H8 H 1 a 0.55200 0.42000 0.64200 1.00000
H9 H 1 a 0.78100 0.15800 -0.01800 1.00000
H10 H 1 a 0.34500 0.15900 0.14400 1.00000
H11 H 1 a 0.03500 0.79900 0.11300 1.00000
Na1 Na 1 a 0.14247 0.62758 0.42059 1.00000
O1 O 1 a -0.04310 0.29760 0.55444 1.00000
O2 O 1 a 0.55970 0.24470 0.39414 1.00000
O3 O 1 a 0.70800 0.83690 0.55456 1.00000
O4 O 1 a 0.33050 -0.05200 0.37321 1.00000
O5 O 1 a 0.36130 0.45370 0.66271 1.00000
O6 O 1 a -0.03760 0.10350 -0.05381 1.00000
O7 O 1 a 0.40560 0.23550 0.05675 1.00000
O8 O 1 a 0.03030 0.67390 0.17348 1.00000
S1 S 1 a 0.47392 0.02789 0.47548 1.00000
```

$NaC_5H_{11}O_8S$: A5B11CD8E_aP26_1_5a_11a_a_8a_a - POSCAR

```
A5B11CD8E_aP26_1_5a_11a_a_8a_a & a,b/a,c/a,alpha,beta,gamma,x1,y1,z1,x2,
    ↪ y2,z2,x3,y3,z3,x4,y4,z4,x5,y5,z5,x6,y6,z6,x7,y7,z7,x8,y8,z8,x9,
    ↪ y9,z9,x10,y10,z10,x11,y11,z11,x12,y12,z12,x13,y13,z13,x14,y14,
    ↪ z14,x15,y15,z15,x16,y16,z16,x17,y17,z17,x18,y18,z18,x19,y19,z19
    ↪ ,x20,y20,z20,x21,y21,z21,x22,y22,z22,x23,y23,z23,x24,y24,z24,
    ↪ x25,y25,z25,x26,y26,z26 --params=4.8558,1.20466246551,
    ↪ 1.81123604761,76.515,81.528,71.392,0.1957,0.1206,0.62783,0.2966
    ↪ ,0.2336,0.7408,0.0562,0.3094,0.86872,0.1694,0.4162,-0.0197,-
    ↪ 0.0682,0.5331,0.0947,0.158,-0.022,0.679,0.45,0.117,0.794,-0.096
    ↪ ,0.433,0.816,0.246,0.548,-0.081,0.764,0.632,0.037,0.867,0.409,
    ↪ 0.171,0.893,0.248,0.503,0.552,0.42,0.642,0.781,0.158,-0.018,
    ↪ 0.345,0.159,0.144,0.035,0.799,0.113,0.14247,0.62758,0.42059,-
    ↪ 0.0431,0.2976,0.55444,0.5597,0.2447,0.39414,0.708,0.8369,
    ↪ 0.55456,0.3305,-0.052,0.37321,0.3613,0.4537,0.66271,-0.0376,
    ↪ 0.1035,-0.05381,0.4056,0.2355,0.05675,0.0303,0.6739,0.17348,
    ↪ 0.47392,0.02789,0.47548 & P1 C_{1}^{1} #1 (a^26) & aP26 & None
    ↪ & C5H11NaO8S & C5H11NaO8S & A. H. Haines and D. L. Hughes, Acta
    ↪  Crystallogr. E 72, 628-631 (2016)
   1.00000000000000
   4.85580000000000   0.00000000000000   0.00000000000000
   1.86655845513818   5.54380552450680   0.00000000000000
   1.29573277835155   1.72777700416739   8.52571921839886
     C     H    Na     O     S
     5    11     1     8     1
Direct
   0.19570000000000   0.12060000000000   0.62783000000000   C   (1a)
   0.29660000000000   0.23360000000000   0.74080000000000   C   (1a)
   0.05620000000000   0.30940000000000   0.86872000000000   C   (1a)
   0.16940000000000   0.41620000000000  -0.01970000000000   C   (1a)
  -0.06820000000000   0.53310000000000   0.09470000000000   C   (1a)
   0.15800000000000  -0.02200000000000   0.67900000000000   H   (1a)
   0.45000000000000   0.11700000000000   0.79400000000000   H   (1a)
  -0.09600000000000   0.43300000000000   0.81600000000000   H   (1a)
   0.24600000000000   0.54800000000000  -0.08100000000000   H   (1a)
   0.76400000000000   0.63200000000000   0.03700000000000   H   (1a)
   0.86700000000000   0.40900000000000   0.17100000000000   H   (1a)
   0.89300000000000   0.24800000000000   0.50300000000000   H   (1a)
   0.55200000000000   0.42000000000000   0.64200000000000   H   (1a)
   0.78100000000000   0.15800000000000  -0.01800000000000   H   (1a)
   0.34500000000000   0.15900000000000   0.14400000000000   H   (1a)
   0.03500000000000   0.79900000000000   0.11300000000000   H   (1a)
   0.14247000000000   0.62758000000000   0.42059000000000   Na  (1a)
  -0.04310000000000   0.29760000000000   0.55444000000000   O   (1a)
   0.55970000000000   0.24470000000000   0.39414000000000   O   (1a)
   0.70800000000000   0.83690000000000   0.55456000000000   O   (1a)
   0.33050000000000  -0.05200000000000   0.37321000000000   O   (1a)
   0.36130000000000   0.45370000000000   0.66271000000000   O   (1a)
  -0.03760000000000   0.10350000000000  -0.05381000000000   O   (1a)
   0.40560000000000   0.23550000000000   0.05675000000000   O   (1a)
   0.03030000000000   0.67390000000000   0.17348000000000   O   (1a)
   0.47392000000000   0.02789000000000   0.47548000000000   S   (1a)
```

$Ni(NO_3)_2(H_2O)_6$: A12B2CD12_aP54_2_12i_2i_i_12i - CIF

```
# CIF file
data_findsym-output
_audit_creation_method FINDSYM

_chemical_name_mineral 'H12N2NiO12'
_chemical_formula_sum 'H12 N2 Ni O12'

loop_
_publ_author_name
 'F. Bigoli'
 'A. Braibanti'
 'A. Tiripicchio'
 'M. T. Camellini'
_journal_name_full_name
;
 Acta Crystallographica Section B: Structural Science
;
_journal_volume 27
_journal_year 1971
_journal_page_first 1427
_journal_page_last 1434
_publ_Section_title
;
 The crystal structures of nitrates of divalent hexaaquocations. III.
    ↪ Hexaaquonickel nitrate
;

# Found in Reaction of [Ni(H$_{2}$O)$_{6}$](NO$_{3}$)$_{2}$ with gaseous
    ↪ NH$_{3}$; crystal growth {\em via} in-situ solvation, 2015

_aflow_title 'Ni(NO$_{3}$)$_{2}$(H$_{2}$O)$_{6}$ Structure'
_aflow_proto 'A12B2CD12_aP54_2_12i_2i_i_12i'
_aflow_params 'a,b/a,c/a,\alpha,\beta,\gamma,x_{1},y_{1},z_{1},x_{2},y_{
    ↪ 2},z_{2},x_{3},y_{3},z_{3},x_{4},y_{4},z_{4},x_{5},y_{5},z_{5},
    ↪ x_{6},y_{6},z_{6},x_{7},y_{7},z_{7},x_{8},y_{8},z_{8},x_{9},y_{
    ↪ 9},z_{9},x_{10},y_{10},z_{10},x_{11},y_{11},z_{11},x_{12},y_{12
    ↪ },z_{12},x_{13},y_{13},z_{13},x_{14},y_{14},z_{14},x_{15},y_{15
    ↪ },z_{15},x_{16},y_{16},z_{16},x_{17},y_{17},z_{17},x_{18},y_{18
    ↪ },z_{18},x_{19},y_{19},z_{19},x_{20},y_{20},z_{20},x_{21},y_{21
    ↪ },z_{21},x_{22},y_{22},z_{22},x_{23},y_{23},z_{23},x_{24},y_{24
    ↪ },z_{24},x_{25},y_{25},z_{25},x_{26},y_{26},z_{26},x_{27},y_{27
    ↪ },z_{27}'
_aflow_params_values '7.694,1.54873927736,0.756043670393,102.3,102.4,
    ↪ 105.9,-0.098,0.375,0.1583,0.1088,0.418,0.275,-0.0233,0.1283,
    ↪ 0.355,0.202,0.135,0.442,0.3033,0.1153,-0.095,0.07,0.04,0.7433,-
    ↪ 0.0353,0.2655,0.6483,0.1356,0.3861,0.7517,0.4,0.355,0.3917,
    ↪ 0.4704,0.3451,0.1817,0.7167,0.1417,-0.0167,0.7667,0.1167,0.7,
    ↪ 0.4207,0.873,0.6648,0.3707,0.6081,0.19,0.09,0.2325,0.0606,
    ↪ 0.4696,0.8648,0.8845,0.5323,0.8766,0.5422,0.2611,0.8757,0.5844,
    ↪ 0.5386,0.6283,0.2804,0.2587,0.6035,0.3158,0.3106,0.5904,-0.0424
```

```
    ↪ ,0.0135,0.3655,0.2694,0.1112,0.1448,0.333,0.17,0.1038,0.8533,
    ↪ 0.0773,0.3149,0.776,0.3614,0.3496,0.2179,0.8166,0.1262,0.9031'
_aflow_Strukturbericht 'None'
_aflow_Pearson 'aP54'

_symmetry_space_group_name_H-M "P -1"
_symmetry_Int_Tables_number 2

_cell_length_a     7.69400
_cell_length_b     11.91600
_cell_length_c     5.817
_cell_angle_alpha  102.30000
_cell_angle_beta   102.40000
_cell_angle_gamma  105.90000

loop_
_space_group_symop_id
_space_group_symop_operation_xyz
1 x,y,z
2 -x,-y,-z

loop_
_atom_site_label
_atom_site_type_symbol
_atom_site_symmetry_multiplicity
_atom_site_Wyckoff_label
_atom_site_fract_x
_atom_site_fract_y
_atom_site_fract_z
_atom_site_occupancy
H1 H 2 i -0.09800 0.37500 0.15830 1.00000
H2 H 2 i 0.10880 0.41800 0.27500 1.00000
H3 H 2 i -0.02330 0.12830 0.35500 1.00000
H4 H 2 i 0.20200 0.13500 0.44200 1.00000
H5 H 2 i 0.30330 0.11530 -0.09500 1.00000
H6 H 2 i 0.07000 0.04000 0.74330 1.00000
H7 H 2 i -0.03530 0.26550 0.64830 1.00000
H8 H 2 i 0.13560 0.38610 0.75170 1.00000
H9 H 2 i 0.40000 0.35500 0.39170 1.00000
H10 H 2 i 0.47040 0.34510 0.18170 1.00000
H11 H 2 i 0.71670 0.14170 -0.01670 1.00000
H12 H 2 i 0.76670 0.11670 0.70000 1.00000
N1 N 2 i 0.42070 0.87300 0.66480 1.00000
N2 N 2 i 0.37070 0.60810 0.19000 1.00000
Ni1 Ni 2 i 0.09000 0.23250 0.06060 1.00000
O1 O 2 i 0.46960 0.86480 0.88450 1.00000
O2 O 2 i 0.53230 0.87660 0.54220 1.00000
O3 O 2 i 0.26110 0.87570 0.58440 1.00000
O4 O 2 i 0.53860 0.62830 0.28040 1.00000
O5 O 2 i 0.25870 0.60350 0.31580 1.00000
O6 O 2 i 0.31060 0.59040 -0.04240 1.00000
O7 O 2 i 0.01350 0.36550 0.26940 1.00000
O8 O 2 i 0.11120 0.14480 0.33300 1.00000
O9 O 2 i 0.17000 0.10380 0.85330 1.00000
O10 O 2 i 0.07730 0.31490 0.77600 1.00000
O11 O 2 i 0.36140 0.34960 0.21790 1.00000
O12 O 2 i 0.81660 0.12620 0.90310 1.00000
```

$Ni(NO_3)_2(H_2O)_6$: A12B2CD12_aP54_2_12i_2i_i_12i - POSCAR

```
A12B2CD12_aP54_2_12i_2i_i_12i & a,b/a,c/a,alpha,beta,gamma,x1,y1,z1,x2,
    ↪ y2,z2,x3,y3,z3,x4,y4,z4,x5,y5,z5,x6,y6,z6,x7,y7,z7,x8,y8,z8,x9,
    ↪ y9,z9,x10,y10,z10,x11,y11,z11,x12,y12,z12,x13,y13,z13,x14,y14,
    ↪ z14,x15,y15,z15,x16,y16,z16,x17,y17,z17,x18,y18,z18,x19,y19,z19
    ↪ ,x20,y20,z20,x21,y21,z21,x22,y22,z22,x23,y23,z23,x24,y24,z24,
    ↪ x25,y25,z25,x26,y26,z26,x27,y27,z27 --params=7.694,
    ↪ 1.54873927736,0.756043670393,102.3,102.4,105.9,-0.098,0.375,
    ↪ 0.1583,0.1088,0.418,0.275,-0.0233,0.1283,0.355,0.202,0.135,
    ↪ 0.442,0.3033,0.1153,-0.095,0.07,0.04,0.7433,-0.0353,0.2655,
    ↪ 0.6483,0.1356,0.3861,0.7517,0.4,0.355,0.3917,0.4704,0.3451,
    ↪ 0.1817,0.7167,0.1417,-0.0167,0.7667,0.1167,0.7,0.4207,0.873,
    ↪ 0.6648,0.3707,0.6081,0.19,0.09,0.2325,0.0606,0.4696,0.8648,
    ↪ 0.8845,0.5323,0.8766,0.5422,0.2611,0.8757,0.5844,0.5386,0.6283,
    ↪ 0.2804,0.2587,0.6035,0.3158,0.3106,0.5904,-0.0424,0.0135,0.3655
    ↪ ,0.2694,0.1112,0.1448,0.333,0.17,0.1038,0.8533,0.0773,0.3149,
    ↪ 0.776,0.3614,0.3496,0.2179,0.8166,0.1262,0.9031 & P-1 C_{i}^{1}
    ↪ #2 (i^27) & aP54 & None & H12N2NiO12 & H12N2NiO12 & F. Bigoli
    ↪ et al., Acta Crystallogr. Sect. B Struct. Sci. 27, 1427-1434 (
    ↪ 1971)
   1.00000000000000
   7.69400000000000   0.00000000000000   0.00000000000000
  -3.26449804993903  11.46010944458840   0.00000000000000
  -1.24911539813081  -1.64431372531079   5.43814601632827
     H     N     Ni     O
    24     4      2    24
Direct
  -0.09800000000000   0.37500000000000   0.15830000000000   H   (2i)
   0.09800000000000  -0.37500000000000  -0.15830000000000   H   (2i)
   0.10880000000000   0.41800000000000   0.27500000000000   H   (2i)
  -0.10880000000000  -0.41800000000000  -0.27500000000000   H   (2i)
  -0.02330000000000   0.12830000000000   0.35500000000000   H   (2i)
   0.02330000000000  -0.12830000000000  -0.35500000000000   H   (2i)
   0.20200000000000   0.13500000000000   0.44200000000000   H   (2i)
  -0.20200000000000  -0.13500000000000  -0.44200000000000   H   (2i)
   0.30330000000000   0.11530000000000  -0.09500000000000   H   (2i)
  -0.30330000000000  -0.11530000000000   0.09500000000000   H   (2i)
   0.07000000000000   0.04000000000000   0.74330000000000   H   (2i)
  -0.07000000000000  -0.04000000000000  -0.74330000000000   H   (2i)
  -0.03530000000000   0.26550000000000   0.64830000000000   H   (2i)
   0.03530000000000  -0.26550000000000  -0.64830000000000   H   (2i)
   0.13560000000000   0.38610000000000   0.75170000000000   H   (2i)
  -0.13560000000000  -0.38610000000000  -0.75170000000000   H   (2i)
   0.40000000000000   0.35500000000000   0.39170000000000   H   (2i)
  -0.40000000000000  -0.35500000000000  -0.39170000000000   H   (2i)
   0.47040000000000   0.34510000000000   0.18170000000000   H   (2i)
  -0.47040000000000  -0.34510000000000  -0.18170000000000   H   (2i)
   0.71670000000000   0.14170000000000  -0.01670000000000   H   (2i)
  -0.71670000000000  -0.14170000000000   0.01670000000000   H   (2i)
   0.76670000000000   0.11670000000000   0.70000000000000   H   (2i)
  -0.76670000000000  -0.11670000000000  -0.70000000000000   H   (2i)
   0.42070000000000   0.87300000000000   0.66480000000000   N   (2i)
  -0.42070000000000  -0.87300000000000  -0.66480000000000   N   (2i)
   0.37070000000000   0.60810000000000   0.19000000000000   N   (2i)
  -0.37070000000000  -0.60810000000000  -0.19000000000000   N   (2i)
   0.09000000000000   0.23250000000000   0.06060000000000   Ni  (2i)
  -0.09000000000000  -0.23250000000000  -0.06060000000000   Ni  (2i)
   0.46960000000000   0.86480000000000   0.88450000000000   O   (2i)
  -0.46960000000000  -0.86480000000000  -0.88450000000000   O   (2i)
   0.53230000000000   0.87660000000000   0.54220000000000   O   (2i)
  -0.53230000000000  -0.87660000000000  -0.54220000000000   O   (2i)
   0.26110000000000   0.87570000000000   0.58440000000000   O   (2i)
  -0.26110000000000  -0.87570000000000  -0.58440000000000   O   (2i)
   0.53860000000000   0.62830000000000   0.28040000000000   O   (2i)
  -0.53860000000000  -0.62830000000000  -0.28040000000000   O   (2i)
   0.25870000000000   0.60350000000000   0.31580000000000   O   (2i)
  -0.25870000000000  -0.60350000000000  -0.31580000000000   O   (2i)
   0.31060000000000   0.59040000000000  -0.04240000000000   O   (2i)
  -0.31060000000000  -0.59040000000000   0.04240000000000   O   (2i)
   0.01350000000000   0.36550000000000   0.26940000000000   O   (2i)
  -0.01350000000000  -0.36550000000000  -0.26940000000000   O   (2i)
   0.11120000000000   0.14480000000000   0.33300000000000   O   (2i)
  -0.11120000000000  -0.14480000000000  -0.33300000000000   O   (2i)
   0.17000000000000   0.10380000000000   0.85330000000000   O   (2i)
  -0.17000000000000  -0.10380000000000  -0.85330000000000   O   (2i)
   0.07730000000000   0.31490000000000   0.77600000000000   O   (2i)
  -0.07730000000000  -0.31490000000000  -0.77600000000000   O   (2i)
   0.36140000000000   0.34960000000000   0.21790000000000   O   (2i)
  -0.36140000000000  -0.34960000000000  -0.21790000000000   O   (2i)
   0.81660000000000   0.12620000000000   0.90310000000000   O   (2i)
  -0.81660000000000  -0.12620000000000  -0.90310000000000   O   (2i)
```

$Co_2B_2O_5$: A2B2C5_aP18_2_2i_2i_5i - CIF

```
# CIF file
data_findsym-output
_audit_creation_method FINDSYM

_chemical_name_mineral 'B2Co2O5'
_chemical_formula_sum 'B2 Co2 O5'

loop_
_publ_author_name
 'S. V. Berger'
_journal_name_full_name
;
 Acta Chemica Scandinavica
;
_journal_volume 4
_journal_year 1950
_journal_page_first 1054
_journal_page_last 1065
_publ_Section_title
;
 The Crystal Structure of Cobaltpyroborate
;

_aflow_title 'Co$_{2}$B$_{2}$O$_{5}$ Structure'
_aflow_proto 'A2B2C5_aP18_2_2i_2i_5i'
_aflow_params 'a,b/a,c/a,\alpha,\beta,\gamma,x_{1},y_{1},z_{1},x_{2},y_{
    ↪ 2},z_{2},x_{3},y_{3},z_{3},x_{4},y_{4},z_{4},x_{5},y_{5},z_{5},
    ↪ x_{6},y_{6},z_{6},x_{7},y_{7},z_{7},x_{8},y_{8},z_{8},x_{9},y_{
    ↪ 9},z_{9}'
_aflow_params_values '3.16,1.87974683544,2.83227848101,103.9,91.0,92.0,
    ↪ 0.67,0.67,0.34,0.36,0.88,0.17,0.743,0.213,0.36,0.245,0.374,0.1,
    ↪ 0.244,0.709,0.054,0.208,0.092,0.19,0.735,0.476,0.249,0.562,
    ↪ 0.842,0.288,0.77,0.698,0.489'
_aflow_Strukturbericht 'None'
_aflow_Pearson 'aP18'

_symmetry_space_group_name_H-M "P -1"
_symmetry_Int_Tables_number 2

_cell_length_a     3.16000
_cell_length_b     5.94000
_cell_length_c     8.95000
_cell_angle_alpha  103.90000
_cell_angle_beta   91.00000
_cell_angle_gamma  92.00000

loop_
_space_group_symop_id
_space_group_symop_operation_xyz
1 x,y,z
2 -x,-y,-z

loop_
_atom_site_label
_atom_site_type_symbol
_atom_site_symmetry_multiplicity
_atom_site_Wyckoff_label
_atom_site_fract_x
_atom_site_fract_y
_atom_site_fract_z
_atom_site_occupancy
B1 B 2 i 0.67000 0.67000 0.34000 1.00000
B2 B 2 i 0.36000 0.88000 0.17000 1.00000
Co1 Co 2 i 0.74300 0.21300 0.36000 1.00000
Co2 Co 2 i 0.24500 0.37400 0.10000 1.00000
O1 O 2 i 0.24400 0.70900 0.05400 1.00000
O2 O 2 i 0.20800 0.09200 0.19000 1.00000
```

```
O3 O 2 i 0.73500 0.47600 0.24900 1.00000
O4 O 2 i 0.56200 0.84200 0.28800 1.00000
O5 O 2 i 0.77000 0.69800 0.48900 1.00000
```

$Co_2B_2O_5$: A2B2C5_aP18_2_2i_2i_5i - POSCAR

```
A2B2C5_aP18_2_2i_2i_5i & a,b/a,c/a,alpha,beta,gamma,x1,y1,z1,x2,y2,z2,x3
   ↪ ,y3,z3,x4,y4,z4,x5,y5,z5,x6,y6,z6,x7,y7,z7,x8,y8,z8,x9,y9,z9 --
   ↪ params=3.16,1.87974683544,2.83227848101,103.9,91.0,92.0,0.67,
   ↪ 0.67,0.34,0.36,0.88,0.17,0.743,0.213,0.36,0.245,0.374,0.1,0.244
   ↪ ,0.709,0.054,0.208,0.092,0.19,0.735,0.476,0.249,0.562,0.842,
   ↪ 0.288,0.77,0.698,0.489 & P-1 C_{i}^{1} #2 (i^9) & aP18 & None &
   ↪ B2Co2O5 & B2Co2O5 & S. V. Berger, Acta Chem. Scand. 4,
   ↪ 1054-1065 (1950)
   1.00000000000000
   3.16000000000000   0.00000000000000   0.00000000000000
  -0.20730301041286   5.93638151249343   0.00000000000000
  -0.15619903761369  -2.15680611598055   8.68483098504038
     B     Co     O
     4      4    10
Direct
   0.67000000000000   0.67000000000000   0.34000000000000   B   (2i)
  -0.67000000000000  -0.67000000000000  -0.34000000000000   B   (2i)
   0.36000000000000   0.88000000000000   0.17000000000000   B   (2i)
  -0.36000000000000  -0.88000000000000  -0.17000000000000   B   (2i)
   0.74300000000000   0.21300000000000   0.36000000000000   Co  (2i)
  -0.74300000000000  -0.21300000000000  -0.36000000000000   Co  (2i)
   0.24500000000000   0.37400000000000   0.10000000000000   Co  (2i)
  -0.24500000000000  -0.37400000000000  -0.10000000000000   Co  (2i)
   0.24400000000000   0.70900000000000   0.05400000000000   O   (2i)
  -0.24400000000000  -0.70900000000000  -0.05400000000000   O   (2i)
   0.20800000000000   0.09200000000000   0.19000000000000   O   (2i)
  -0.20800000000000  -0.09200000000000  -0.19000000000000   O   (2i)
   0.73500000000000   0.47600000000000   0.24900000000000   O   (2i)
  -0.73500000000000  -0.47600000000000  -0.24900000000000   O   (2i)
   0.56200000000000   0.84200000000000   0.28800000000000   O   (2i)
  -0.56200000000000  -0.84200000000000  -0.28800000000000   O   (2i)
   0.77000000000000   0.69800000000000   0.48900000000000   O   (2i)
  -0.77000000000000  -0.69800000000000  -0.48900000000000   O   (2i)
```

Kyanite ($Al_2SiO_5$, $S0_1$): A2B5C_aP32_2_4i_10i_2i - CIF

```
# CIF file
data_findsym-output
_audit_creation_method FINDSYM

_chemical_name_mineral 'Kyanite'
_chemical_formula_sum 'Al2 O5 Si'

loop_
_publ_author_name
 'H. Yang'
 'R. T. Downs'
 'L. W. Finger'
 'R. M. Hazen'
 'C. T. Prewitt'
_journal_name_full_name
;
 American Mineralogist
;
_journal_volume 82
_journal_year 1997
_journal_page_first 467
_journal_page_last 474
_publ_Section_title
;
 Compressibility and crystal structure of kyanite, Al$_{2}$SiO$_{5}$, at
   ↪ high pressure
;

_aflow_title 'Kyanite (Al$_{2}$SiO$_{5}$, $S0_{1}$) Structure'
_aflow_proto 'A2B5C_aP32_2_4i_10i_2i'
_aflow_params 'a,b/a,c/a,\alpha,\beta,\gamma,x_{1},y_{1},z_{1},x_{2},y_{
   ↪ 2},z_{2},x_{3},y_{3},z_{3},x_{4},y_{4},z_{4},x_{5},y_{5},z_{5},
   ↪ x_{6},y_{6},z_{6},x_{7},y_{7},z_{7},x_{8},y_{8},z_{8},x_{9},y_{
   ↪ 9},z_{9},x_{10},y_{10},z_{10},x_{11},y_{11},z_{11},x_{12},y_{12
   ↪ },z_{12},x_{13},y_{13},z_{13},x_{14},y_{14},z_{14},x_{15},y_{15
   ↪ },z_{15},x_{16},y_{16},z_{16}'
_aflow_params_values '7.12,1.10223314607,0.782837078652,89.974,101.177,
   ↪ 106.0,0.32533,0.70412,0.45812,0.2974,0.69882,0.9504,0.0998,
   ↪ 0.38615,0.64043,0.11205,0.9175,0.16469,0.10933,0.14685,0.12866,
   ↪ 0.12287,0.68535,0.18113,0.27507,0.45443,0.95474,0.28353,0.9357,
   ↪ 0.93567,0.10836,0.1521,0.66671,0.12192,0.63063,0.63939,0.28226,
   ↪ 0.44512,0.42868,0.29156,0.94684,0.46574,0.50074,0.27519,0.24405
   ↪ ,0.50154,0.23099,0.75595,0.29625,0.06488,0.70657,0.29102,
   ↪ 0.33168,0.18937'
_aflow_Strukturbericht '$S0_{1}$'
_aflow_Pearson 'aP32'

_symmetry_space_group_name_H-M "P -1"
_symmetry_Int_Tables_number 2

_cell_length_a     7.12000
_cell_length_b     7.84790
_cell_length_c     5.57380
_cell_angle_alpha  89.97400
_cell_angle_beta   101.17700
_cell_angle_gamma  106.00000

loop_
_space_group_symop_id
_space_group_symop_operation_xyz
1 x,y,z
2 -x,-y,-z

loop_
_atom_site_label
_atom_site_type_symbol
_atom_site_symmetry_multiplicity
_atom_site_Wyckoff_label
_atom_site_fract_x
_atom_site_fract_y
_atom_site_fract_z
_atom_site_occupancy
Al1 Al 2 i 0.32533 0.70412 0.45812 1.00000
Al2 Al 2 i 0.29740 0.69882 0.95040 1.00000
Al3 Al 2 i 0.09980 0.38615 0.64043 1.00000
Al4 Al 2 i 0.11205 0.91750 0.16469 1.00000
O1 O 2 i 0.10933 0.14685 0.12866 1.00000
O2 O 2 i 0.12287 0.68535 0.18113 1.00000
O3 O 2 i 0.27507 0.45443 0.95474 1.00000
O4 O 2 i 0.28353 0.93570 0.93567 1.00000
O5 O 2 i 0.10836 0.15210 0.66671 1.00000
O6 O 2 i 0.12192 0.63063 0.63939 1.00000
O7 O 2 i 0.28226 0.44512 0.42868 1.00000
O8 O 2 i 0.29156 0.94684 0.46574 1.00000
O9 O 2 i 0.50074 0.27519 0.24405 1.00000
O10 O 2 i 0.50154 0.23099 0.75595 1.00000
Si1 Si 2 i 0.29625 0.06488 0.70657 1.00000
Si2 Si 2 i 0.29102 0.33168 0.18937 1.00000
```

Kyanite ($Al_2SiO_5$, $S0_1$): A2B5C_aP32_2_4i_10i_2i - POSCAR

```
A2B5C_aP32_2_4i_10i_2i & a,b/a,c/a,alpha,beta,gamma,x1,y1,z1,x2,y2,z2,x3
   ↪ ,y3,z3,x4,y4,z4,x5,y5,z5,x6,y6,z6,x7,y7,z7,x8,y8,z8,x9,y9,z9,
   ↪ x10,y10,z10,x11,y11,z11,x12,y12,z12,x13,y13,z13,x14,y14,z14,x15
   ↪ ,y15,z15,x16,y16,z16 --params=7.12,1.10223314607,0.782837078652
   ↪ ,89.974,101.177,106.0,0.32533,0.70412,0.45812,0.2974,0.69882,
   ↪ 0.9504,0.0998,0.38615,0.64043,0.11205,0.9175,0.16469,0.10933,
   ↪ 0.14685,0.12866,0.12287,0.68535,0.18113,0.27507,0.45443,0.95474
   ↪ ,0.28353,0.9357,0.93567,0.10836,0.1521,0.66671,0.12192,0.63063,
   ↪ 0.63939,0.28226,0.44512,0.42868,0.29156,0.94684,0.46574,0.50074
   ↪ ,0.27519,0.24405,0.50154,0.23099,0.75595,0.29625,0.06488,
   ↪ 0.70657,0.29102,0.33168,0.18937 & P-1 C_{i}^{1} #2 (i^16) &
   ↪ aP32 & $S0_{1}$ & Al2O5Si & Kyanite & H. Yang et al., Am.
   ↪ Mineral. 82, 467-474 (1997)
   1.00000000000000
   7.12000000000000   0.00000000000000   0.00000000000000
  -2.16317440471623   7.54388566355433   0.00000000000000
  -1.08042848514409  -0.30717664286521   5.45944715503020
    Al     O    Si
     8    20     4
Direct
   0.32533000000000   0.70412000000000   0.45812000000000   Al  (2i)
  -0.32533000000000  -0.70412000000000  -0.45812000000000   Al  (2i)
   0.29740000000000   0.69882000000000   0.95040000000000   Al  (2i)
  -0.29740000000000  -0.69882000000000  -0.95040000000000   Al  (2i)
   0.09980000000000   0.38615000000000   0.64043000000000   Al  (2i)
  -0.09980000000000  -0.38615000000000  -0.64043000000000   Al  (2i)
   0.11205000000000   0.91750000000000   0.16469000000000   Al  (2i)
  -0.11205000000000  -0.91750000000000  -0.16469000000000   Al  (2i)
   0.10933000000000   0.14685000000000   0.12866000000000   O   (2i)
  -0.10933000000000  -0.14685000000000  -0.12866000000000   O   (2i)
   0.12287000000000   0.68535000000000   0.18113000000000   O   (2i)
  -0.12287000000000  -0.68535000000000  -0.18113000000000   O   (2i)
   0.27507000000000   0.45443000000000   0.95474000000000   O   (2i)
  -0.27507000000000  -0.45443000000000  -0.95474000000000   O   (2i)
   0.28353000000000   0.93570000000000   0.93567000000000   O   (2i)
  -0.28353000000000  -0.93570000000000  -0.93567000000000   O   (2i)
   0.10836000000000   0.15210000000000   0.66671000000000   O   (2i)
  -0.10836000000000  -0.15210000000000  -0.66671000000000   O   (2i)
   0.12192000000000   0.63063000000000   0.63939000000000   O   (2i)
  -0.12192000000000  -0.63063000000000  -0.63939000000000   O   (2i)
   0.28226000000000   0.44512000000000   0.42868000000000   O   (2i)
  -0.28226000000000  -0.44512000000000  -0.42868000000000   O   (2i)
   0.29156000000000   0.94684000000000   0.46574000000000   O   (2i)
  -0.29156000000000  -0.94684000000000  -0.46574000000000   O   (2i)
   0.50074000000000   0.27519000000000   0.24405000000000   O   (2i)
  -0.50074000000000  -0.27519000000000  -0.24405000000000   O   (2i)
   0.50154000000000   0.23099000000000   0.75595000000000   O   (2i)
  -0.50154000000000  -0.23099000000000  -0.75595000000000   O   (2i)
   0.29625000000000   0.06488000000000   0.70657000000000   Si  (2i)
  -0.29625000000000  -0.06488000000000  -0.70657000000000   Si  (2i)
   0.29102000000000   0.33168000000000   0.18937000000000   Si  (2i)
  -0.29102000000000  -0.33168000000000  -0.18937000000000   Si  (2i)
```

$\alpha$-$Ho_2Si_2O_7$: A2B7C2_aP44_2_4i_14i_4i - CIF

```
# CIF file
data_findsym-output
_audit_creation_method FINDSYM

_chemical_name_mineral 'Ho2O7Si2'
_chemical_formula_sum 'Ho2 O7 Si2'

loop_
_publ_author_name
 'J. Felsche'
_journal_name_full_name
;
 Naturwissenschaften
;
_journal_volume 59
_journal_year 1972
_journal_page_first 35
_journal_page_last 36
_publ_Section_title
;
 A new silicate structure containing linear [Si$_{3}$O$_{10}$] groups
;
```

```
# Found in Revision of the crystallographic data of polymorphic Y$_{2}
    ↪ $Si$_{2}$O$_{7}$ and Y$_{2}$SiO$_{5}$ compounds, 2004

_aflow_title '$\alpha$-Ho$_{2}$Si$_{2}$O$_{7}$ Structure'
_aflow_proto 'A2B7C2_aP44_2_4i_14i_4i'
_aflow_params 'a,b/a,c/a,\alpha,\beta,\gamma,x_{1},y_{1},z_{1},x_{2},y_{
    ↪ 2},z_{2},x_{3},y_{3},z_{3},x_{4},y_{4},z_{4},x_{5},y_{5},z_{5},
    ↪ x_{6},y_{6},z_{6},x_{7},y_{7},z_{7},x_{8},y_{8},z_{8},x_{9},y_{
    ↪ 9},z_{9},x_{10},y_{10},z_{10},x_{11},y_{11},z_{11},x_{12},y_{12
    ↪ },z_{12},x_{13},y_{13},z_{13},x_{14},y_{14},z_{14},x_{15},y_{15
    ↪ },z_{15},x_{16},y_{16},z_{16},x_{17},y_{17},z_{17},x_{18},y_{18
    ↪ },z_{18},x_{19},y_{19},z_{19},x_{20},y_{20},z_{20},x_{21},y_{21
    ↪ },z_{21},x_{22},y_{22},z_{22}'
_aflow_params_values '6.612,1.00862068966,1.82773744707,85.81,89.38,
    ↪ 88.57,-0.0521,0.331,0.11666,0.8845,0.0908,0.35915,0.3705,0.7756
    ↪ ,0.36947,0.6657,0.828,0.1059,0.6408,0.4893,0.1258,0.6226,0.1401
    ↪ ,0.2096,0.2968,0.2956,0.0948,0.4002,0.4255,0.304,0.5879,0.1717,
    ↪ 0.4454,0.224,0.095,0.3785,0.2937,0.422,0.5099,0.2862,0.2179,
    ↪ 0.687,-0.0374,0.2281,0.5714,0.0752,0.5789,0.6864,0.2414,-0.0859
    ↪ ,-0.001,0.3397,0.7841,0.1903,-0.0041,0.6708,0.0797,0.0023,
    ↪ 0.0137,0.1857,0.1539,0.8505,0.1168,0.4862,0.3353,0.1761,0.3781,
    ↪ 0.2726,0.4051,0.1457,0.3719,0.6179'
_aflow_Strukturbericht 'None'
_aflow_Pearson 'aP44'

_symmetry_space_group_name_H-M "P -1"
_symmetry_Int_Tables_number 2

_cell_length_a      6.61200
_cell_length_b      6.66900
_cell_length_c      12.08500
_cell_angle_alpha   85.81000
_cell_angle_beta    89.38000
_cell_angle_gamma   88.57000

loop_
_space_group_symop_id
_space_group_symop_operation_xyz
1 x,y,z
2 -x,-y,-z

loop_
_atom_site_label
_atom_site_type_symbol
_atom_site_symmetry_multiplicity
_atom_site_Wyckoff_label
_atom_site_fract_x
_atom_site_fract_y
_atom_site_fract_z
_atom_site_occupancy
Ho1 Ho 2 i -0.05210 0.33100 0.11666 1.00000
Ho2 Ho 2 i 0.88450 0.09080 0.35915 1.00000
Ho3 Ho 2 i 0.37050 0.77560 0.36947 1.00000
Ho4 Ho 2 i 0.66570 0.82800 0.10590 1.00000
O1 O 2 i 0.64080 0.48930 0.12580 1.00000
O2 O 2 i 0.62260 0.14010 0.20960 1.00000
O3 O 2 i 0.29680 0.29560 0.09480 1.00000
O4 O 2 i 0.40020 0.42550 0.30400 1.00000
O5 O 2 i 0.58790 0.17170 0.44540 1.00000
O6 O 2 i 0.22400 0.09500 0.37850 1.00000
O7 O 2 i 0.29370 0.42200 0.50990 1.00000
O8 O 2 i 0.28620 0.21790 0.68700 1.00000
O9 O 2 i -0.03740 0.22810 0.57140 1.00000
O10 O 2 i 0.07520 0.57890 0.68640 1.00000
O11 O 2 i 0.24140 -0.08590 -0.00100 1.00000
O12 O 2 i 0.33970 0.78410 0.19030 1.00000
O13 O 2 i -0.00410 0.67080 0.07970 1.00000
O14 O 2 i 0.00230 0.01370 0.18570 1.00000
Si1 Si 2 i 0.15390 0.85050 0.11680 1.00000
Si2 Si 2 i 0.48620 0.33530 0.17610 1.00000
Si3 Si 2 i 0.37810 0.27260 0.40510 1.00000
Si4 Si 2 i 0.14570 0.37190 0.61790 1.00000
```

α-$Ho_2Si_2O_7$: A2B7C2_aP44_2_4i_14i_4i - POSCAR

```
A2B7C2_aP44_2_4i_14i_4i & a,b/a,c/a,alpha,beta,gamma,x1,y1,z1,x2,y2,z2,
    ↪ x3,y3,z3,x4,y4,z4,x5,y5,z5,x6,y6,z6,x7,y7,z7,x8,y8,z8,x9,y9,z9,
    ↪ x10,y10,z10,x11,y11,z11,x12,y12,z12,x13,y13,z13,x14,y14,z14,x15
    ↪ ,y15,z15,x16,y16,z16,x17,y17,z17,x18,y18,z18,x19,y19,z19,x20,
    ↪ y20,z20,x21,y21,z21,x22,y22,z22 --params=6.612,1.00862068966,
    ↪ 1.82773744707,85.81,89.38,88.57,-0.0521,0.331,0.11666,0.8845,
    ↪ 0.0908,0.35915,0.3705,0.7756,0.36947,0.6657,0.828,0.1059,0.6408
    ↪ ,0.4893,0.1258,0.6226,0.1401,0.2096,0.2968,0.2956,0.0948,0.4002
    ↪ ,0.4255,0.304,0.5879,0.1717,0.4454,0.224,0.095,0.3785,0.2937,
    ↪ 0.422,0.5099,0.2862,0.2179,0.687,-0.0374,0.2281,0.5714,0.0752,
    ↪ 0.5789,0.6864,0.2414,-0.0859,-0.001,0.3397,0.7841,0.1903,-
    ↪ 0.0041,0.6708,0.0797,0.0023,0.0137,0.1857,0.1539,0.8505,0.1168,
    ↪ 0.4862,0.3353,0.1761,0.3781,0.2726,0.4051,0.1457,0.3719,0.6179
    ↪ & P-1 C_{i}^{1} #2 (i^22) & aP44 & None & Ho2O7Si2 & Ho2O7Si2 &
    ↪ J. Felsche, Naturwissenschaften 59, 35-36 (1972)
   1.00000000000000
   6.61200000000000   0.00000000000000   0.00000000000000
   0.16642901147786   6.66692300721544   0.00000000000000
   0.13076973275040   0.87999065761810  12.05220895601720
    Ho     O    Si
     8    28     8
Direct
  -0.05210000000000   0.33100000000000   0.11666000000000   Ho   (2i)
   0.05210000000000  -0.33100000000000  -0.11666000000000   Ho   (2i)
   0.88450000000000   0.09080000000000   0.35915000000000   Ho   (2i)
  -0.88450000000000  -0.09080000000000  -0.35915000000000   Ho   (2i)
   0.37050000000000   0.77560000000000   0.36947000000000   Ho   (2i)
  -0.37050000000000  -0.77560000000000  -0.36947000000000   Ho   (2i)
   0.66570000000000   0.82800000000000   0.10590000000000   Ho   (2i)
  -0.66570000000000  -0.82800000000000  -0.10590000000000   Ho   (2i)
   0.64080000000000   0.48930000000000   0.12580000000000   O    (2i)
  -0.64080000000000  -0.48930000000000  -0.12580000000000   O    (2i)
   0.62260000000000   0.14010000000000   0.20960000000000   O    (2i)
  -0.62260000000000  -0.14010000000000  -0.20960000000000   O    (2i)
   0.29680000000000   0.29560000000000   0.09480000000000   O    (2i)
  -0.29680000000000  -0.29560000000000  -0.09480000000000   O    (2i)
   0.40020000000000   0.42550000000000   0.30400000000000   O    (2i)
  -0.40020000000000  -0.42550000000000  -0.30400000000000   O    (2i)
   0.58790000000000   0.17170000000000   0.44540000000000   O    (2i)
  -0.58790000000000  -0.17170000000000  -0.44540000000000   O    (2i)
   0.22400000000000   0.09500000000000   0.37850000000000   O    (2i)
  -0.22400000000000  -0.09500000000000  -0.37850000000000   O    (2i)
   0.29370000000000   0.42200000000000   0.50990000000000   O    (2i)
  -0.29370000000000  -0.42200000000000  -0.50990000000000   O    (2i)
   0.28620000000000   0.21790000000000   0.68700000000000   O    (2i)
  -0.28620000000000  -0.21790000000000  -0.68700000000000   O    (2i)
  -0.03740000000000   0.22810000000000   0.57140000000000   O    (2i)
   0.03740000000000  -0.22810000000000  -0.57140000000000   O    (2i)
   0.07520000000000   0.57890000000000   0.68640000000000   O    (2i)
  -0.07520000000000  -0.57890000000000  -0.68640000000000   O    (2i)
   0.24140000000000  -0.08590000000000  -0.00100000000000   O    (2i)
  -0.24140000000000   0.08590000000000   0.00100000000000   O    (2i)
   0.33970000000000   0.78410000000000   0.19030000000000   O    (2i)
  -0.33970000000000  -0.78410000000000  -0.19030000000000   O    (2i)
  -0.00410000000000   0.67080000000000   0.07970000000000   O    (2i)
   0.00410000000000  -0.67080000000000  -0.07970000000000   O    (2i)
   0.00230000000000   0.01370000000000   0.18570000000000   O    (2i)
  -0.00230000000000  -0.01370000000000  -0.18570000000000   O    (2i)
   0.15390000000000   0.85050000000000   0.11680000000000   Si   (2i)
  -0.15390000000000  -0.85050000000000  -0.11680000000000   Si   (2i)
   0.48620000000000   0.33530000000000   0.17610000000000   Si   (2i)
  -0.48620000000000  -0.33530000000000  -0.17610000000000   Si   (2i)
   0.37810000000000   0.27260000000000   0.40510000000000   Si   (2i)
  -0.37810000000000  -0.27260000000000  -0.40510000000000   Si   (2i)
   0.14570000000000   0.37190000000000   0.61790000000000   Si   (2i)
  -0.14570000000000  -0.37190000000000  -0.61790000000000   Si   (2i)
```

$Co_3(SeO_3)_3{\cdot}H_2O$: A3B2C10D3_aP36_2_ah2i_2i_10i_3i - CIF

```
# CIF file
data_findsym-output
_audit_creation_method FINDSYM

_chemical_name_mineral 'Co3H2O10Se3'
_chemical_formula_sum 'Co3 H2 O10 Se3'

loop_
_publ_author_name
 'M. Wildner'
_journal_name_full_name
;
 Monatshefte f{\"u}r Chemie - Chemical Monthly
;
_journal_volume 122
_journal_year 1991
_journal_page_first 585
_journal_page_last 594
_publ_Section_title
;
 Crystal structures of Co$_{3}$(SeO$_{3}$)$_{3}$$\cdot$H$_{2}$O and Ni$_
    ↪ {3}$(SeO$_{3}$)$_{3}$$\cdot$H$_{2}$O, two new isotypic
    ↪ compounds
;

# Found in Tunable magnetic order in low-symmetry SeO$_{3}$ ligand
    ↪ linked $TM$$_{3}$(SeO$_{3}$)$_{3}$H$_{2}$O ($TM$ = Mn, Co and
    ↪ Ni) compounds, 2019 Found in Tunable magnetic order in
    ↪ low-symmetry SeO$_{3}$ ligand linked $TM$$_{3}$(SeO$_{3}$)$_{3}
    ↪ $H$_{2}$O ($TM$ = Mn, Co and Ni) compounds, {arXiv:1910.08175 [
    ↪ cond-mat.str-el]},

_aflow_title 'Co$_{3}$(SeO$_{3}$)$_{3}$$\cdot$H$_{2}$O Structure'
_aflow_proto 'A3B2C10D3_aP36_2_ah2i_2i_10i_3i'
_aflow_params 'a,b/a,c/a,\alpha,\beta,\gamma,x_{3},y_{3},z_{3},x_{4},y_{
    ↪ 4},z_{4},x_{5},y_{5},z_{5},x_{6},y_{6},z_{6},x_{7},y_{7},z_{7},
    ↪ x_{8},y_{8},z_{8},x_{9},y_{9},z_{9},x_{10},y_{10},z_{10},x_{11
    ↪ },y_{11},z_{11},x_{12},y_{12},z_{12},x_{13},y_{13},z_{13},x_{14
    ↪ },y_{14},z_{14},x_{15},y_{15},z_{15},x_{16},y_{16},z_{16},x_{17
    ↪ },y_{17},z_{17},x_{18},y_{18},z_{18},x_{19},y_{19},z_{19}'
_aflow_params_values '8.102,1.0144408788,1.05801036781,69.15,62.88,67.23
    ↪ ,0.65388,0.03726,0.79895,0.71664,0.85636,0.39005,0.004,0.867,
    ↪ 0.704,-0.065,0.823,0.624,0.3023,0.7124,0.3858,0.5468,0.6953,
    ↪ 0.5643,0.2128,0.8964,0.7589,0.5375,-0.0553,0.6739,-0.0504,
    ↪ 0.7241,0.4515,0.7083,0.518,0.2467,0.5988,0.8268,0.0273,0.8667,
    ↪ 0.7852,0.1424,0.1894,0.8592,0.1263,-0.0821,0.8725,0.6893,
    ↪ 0.56457,0.23365,0.23877,0.79662,0.66737,0.06492,0.17624,0.66966
    ↪ ,0.29811'
_aflow_Strukturbericht 'None'
_aflow_Pearson 'aP36'

_symmetry_space_group_name_H-M "P -1"
_symmetry_Int_Tables_number 2

_cell_length_a      8.10200
_cell_length_b      8.21900
_cell_length_c      8.57200
_cell_angle_alpha   69.15000
_cell_angle_beta    62.88000
_cell_angle_gamma   67.23000

loop_
_space_group_symop_id
_space_group_symop_operation_xyz
1 x,y,z
```

```
2 -x,-y,-z

loop_
_atom_site_label
_atom_site_type_symbol
_atom_site_symmetry_multiplicity
_atom_site_Wyckoff_label
_atom_site_fract_x
_atom_site_fract_y
_atom_site_fract_z
_atom_site_occupancy
Co1 Co 1 a 0.00000 0.00000 0.00000 1.00000
Co2 Co 1 h 0.50000 0.50000 0.50000 1.00000
Co3 Co 2 i 0.65388 0.03726 0.79895 1.00000
Co4 Co 2 i 0.71664 0.85636 0.39005 1.00000
H1 H 2 i 0.00400 0.86700 0.70400 1.00000
H2 H 2 i -0.06500 0.82300 0.62400 1.00000
O1 O 2 i 0.30230 0.71240 0.38580 1.00000
O2 O 2 i 0.54680 0.69530 0.56430 1.00000
O3 O 2 i 0.21280 0.89640 0.75890 1.00000
O4 O 2 i 0.53750 -0.05530 0.67390 1.00000
O5 O 2 i -0.05040 0.72410 0.45150 1.00000
O6 O 2 i 0.70830 0.51800 0.24670 1.00000
O7 O 2 i 0.59880 0.82680 0.02730 1.00000
O8 O 2 i 0.86670 0.78520 0.14240 1.00000
O9 O 2 i 0.18940 0.85920 0.12630 1.00000
O10 O 2 i -0.08210 0.87250 0.68930 1.00000
Se1 Se 2 i 0.56457 0.23365 0.23877 1.00000
Se2 Se 2 i 0.79662 0.66737 0.06492 1.00000
Se3 Se 2 i 0.17624 0.66966 0.29811 1.00000
```

$Co_3(SeO_3)_3 \cdot H_2O$: A3B2C10D3_aP36_2_ah2i_2i_10i_3i - POSCAR

```
A3B2C10D3_aP36_2_ah2i_2i_10i_3i & a,b/a,c/a,alpha,beta,gamma,x3,y3,z3,x4
    ↪ ,y4,z4,x5,y5,z5,x6,y6,z6,x7,y7,z7,x8,y8,z8,x9,y9,z9,x10,y10,z10
    ↪ ,x11,y11,z11,x12,y12,z12,x13,y13,z13,x14,y14,z14,x15,y15,z15,
    ↪ x16,y16,z16,x17,y17,z17,x18,y18,z18,x19,y19,z19 --params=8.102,
    ↪ 1.0144408788,1.05801036781,69.15,62.88,67.23,0.65388,0.03726,
    ↪ 0.79895,0.71664,0.85636,0.39005,0.004,0.867,0.704,-0.065,0.823,
    ↪ 0.624,0.3023,0.7124,0.3858,0.5468,0.6953,0.5643,0.2128,0.8964,
    ↪ 0.7589,0.5375,-0.0553,0.6739,-0.0504,0.7241,0.4515,0.7083,0.518
    ↪ ,0.2467,0.5988,0.8268,0.0273,0.8667,0.7852,0.1424,0.1894,0.8592
    ↪ ,0.1263,-0.0821,0.8725,0.6893,0.56457,0.23365,0.23877,0.79662,
    ↪ 0.66737,0.06492,0.17624,0.66966,0.29811 & P-1 C_{i}^{1} #2 (ahi
    ↪ ^17) & aP36 & None & Co3H2O10Se3 & Co3H2O10Se3 & M. Wildner,
    ↪ Monatshefte f{"u}r Chemie - Chemical Monthly 122, 585-594 (1991
    ↪ )
   1.00000000000000
   8.10200000000000   0.00000000000000   0.00000000000000
   3.18102296897454   7.57845986140036   0.00000000000000
   3.90759439489992   1.66864563346477   7.44483121332282
    Co      H      O     Se
     6      4     20      6
Direct
   0.00000000000000   0.00000000000000   0.00000000000000   Co   (1a)
   0.50000000000000   0.50000000000000   0.50000000000000   Co   (1h)
   0.65388000000000   0.03726000000000   0.79895000000000   Co   (2i)
  -0.65388000000000  -0.03726000000000  -0.79895000000000   Co   (2i)
   0.71664000000000   0.85636000000000   0.39005000000000   Co   (2i)
  -0.71664000000000  -0.85636000000000  -0.39005000000000   Co   (2i)
   0.00400000000000   0.86700000000000   0.70400000000000    H   (2i)
  -0.00400000000000  -0.86700000000000  -0.70400000000000    H   (2i)
  -0.06500000000000   0.82300000000000   0.62400000000000    H   (2i)
   0.06500000000000  -0.82300000000000  -0.62400000000000    H   (2i)
   0.30230000000000   0.71240000000000   0.38580000000000    O   (2i)
  -0.30230000000000  -0.71240000000000  -0.38580000000000    O   (2i)
   0.54680000000000   0.69530000000000   0.56430000000000    O   (2i)
  -0.54680000000000  -0.69530000000000  -0.56430000000000    O   (2i)
   0.21280000000000   0.89640000000000   0.75890000000000    O   (2i)
  -0.21280000000000  -0.89640000000000  -0.75890000000000    O   (2i)
   0.53750000000000  -0.05530000000000   0.67390000000000    O   (2i)
  -0.53750000000000   0.05530000000000  -0.67390000000000    O   (2i)
  -0.05040000000000   0.72410000000000   0.45150000000000    O   (2i)
   0.05040000000000  -0.72410000000000  -0.45150000000000    O   (2i)
   0.70830000000000   0.51800000000000   0.24670000000000    O   (2i)
  -0.70830000000000  -0.51800000000000  -0.24670000000000    O   (2i)
   0.59880000000000   0.82680000000000   0.02730000000000    O   (2i)
  -0.59880000000000  -0.82680000000000  -0.02730000000000    O   (2i)
   0.86670000000000   0.78520000000000   0.14240000000000    O   (2i)
  -0.86670000000000  -0.78520000000000  -0.14240000000000    O   (2i)
   0.18940000000000   0.85920000000000   0.12630000000000    O   (2i)
  -0.18940000000000  -0.85920000000000  -0.12630000000000    O   (2i)
  -0.08210000000000   0.87250000000000   0.68930000000000    O   (2i)
   0.08210000000000  -0.87250000000000  -0.68930000000000    O   (2i)
   0.56457000000000   0.23365000000000   0.23877000000000   Se   (2i)
  -0.56457000000000  -0.23365000000000  -0.23877000000000   Se   (2i)
   0.79662000000000   0.66737000000000   0.06492000000000   Se   (2i)
  -0.79662000000000  -0.66737000000000  -0.06492000000000   Se   (2i)
   0.17624000000000   0.66966000000000   0.29811000000000   Se   (2i)
  -0.17624000000000  -0.66966000000000  -0.29811000000000   Se   (2i)
```

$\delta$-$WO_3$: A3B_aP32_2_12i_4i - CIF

```
# CIF file
data_findsym-output
_audit_creation_method FINDSYM

_chemical_name_mineral 'O3W'
_chemical_formula_sum 'O3 W'

loop_
_publ_author_name
 'P. M. Woodward'
 'A. W. Sleight'
 'T. Vogt'
_journal_name_full_name
;
 Journal of Solid State Chemistry
;
_journal_volume 131
_journal_year 1997
_journal_page_first 9
_journal_page_last 17
_publ_Section_title
;
 Ferroelectric Tungsten Trioxide
;

_aflow_title '$\delta$-WO$_{3}$ Structure'
_aflow_proto 'A3B_aP32_2_12i_4i'
_aflow_params 'a,b/a,c/a,\alpha,\beta,\gamma,x_{1},y_{1},z_{1},x_{2},y_{
    ↪ 2},z_{2},x_{3},y_{3},z_{3},x_{4},y_{4},z_{4},x_{5},y_{5},z_{5},
    ↪ x_{6},y_{6},z_{6},x_{7},y_{7},z_{7},x_{8},y_{8},z_{8},x_{9},y_{
    ↪ 9},z_{9},x_{10},y_{10},z_{10},x_{11},y_{11},z_{11},x_{12},y_{12
    ↪ },z_{12},x_{13},y_{13},z_{13},x_{14},y_{14},z_{14},x_{15},y_{15
    ↪ },z_{15},x_{16},y_{16},z_{16}'
_aflow_params_values '7.309,1.02914215351,1.05048570256,88.81,89.08,
    ↪ 89.07,0.0007,0.0386,0.21,0.5038,0.5361,0.2181,0.0076,0.466,
    ↪ 0.2884,0.4972,0.9638,0.2878,0.2851,0.2574,0.287,0.2204,0.763,
    ↪ 0.2232,0.2186,0.2627,0.7258,0.284,0.7583,0.7679,0.2943,0.0422,-
    ↪ 0.0002,0.2971,0.5446,0.4982,0.2096,0.482,-0.0072,0.2088,-0.017,
    ↪ 0.5051,0.2566,0.0259,0.285,0.2502,0.528,0.2158,0.2438,0.0313,
    ↪ 0.7817,0.2499,0.5338,0.719'
_aflow_Strukturbericht 'None'
_aflow_Pearson 'aP32'

_symmetry_space_group_name_H-M "P -1"
_symmetry_Int_Tables_number 2

_cell_length_a     7.30900
_cell_length_b     7.52200
_cell_length_c     7.67800
_cell_angle_alpha  88.81000
_cell_angle_beta   89.08000
_cell_angle_gamma  89.07000

loop_
_space_group_symop_id
_space_group_symop_operation_xyz
1 x,y,z
2 -x,-y,-z

loop_
_atom_site_label
_atom_site_type_symbol
_atom_site_symmetry_multiplicity
_atom_site_Wyckoff_label
_atom_site_fract_x
_atom_site_fract_y
_atom_site_fract_z
_atom_site_occupancy
O1 O 2 i 0.00070 0.03860 0.21000 1.00000
O2 O 2 i 0.50380 0.53610 0.21810 1.00000
O3 O 2 i 0.00760 0.46600 0.28840 1.00000
O4 O 2 i 0.49720 0.96380 0.28780 1.00000
O5 O 2 i 0.28510 0.25740 0.28700 1.00000
O6 O 2 i 0.22040 0.76300 0.22320 1.00000
O7 O 2 i 0.21860 0.26270 0.72580 1.00000
O8 O 2 i 0.28400 0.75830 0.76790 1.00000
O9 O 2 i 0.29430 0.04220 -0.00020 1.00000
O10 O 2 i 0.29710 0.54460 0.49820 1.00000
O11 O 2 i 0.20960 0.48200 -0.00720 1.00000
O12 O 2 i 0.20880 -0.01700 0.50510 1.00000
W1 W 2 i 0.25660 0.02590 0.28500 1.00000
W2 W 2 i 0.25020 0.52800 0.21580 1.00000
W3 W 2 i 0.24380 0.03130 0.78170 1.00000
W4 W 2 i 0.24990 0.53380 0.71900 1.00000
```

$\delta$-$WO_3$: A3B_aP32_2_12i_4i - POSCAR

```
A3B_aP32_2_12i_4i & a,b/a,c/a,alpha,beta,gamma,x1,y1,z1,x2,y2,z2,x3,y3,
    ↪ z3,x4,y4,z4,x5,y5,z5,x6,y6,z6,x7,y7,z7,x8,y8,z8,x9,y9,z9,x10,
    ↪ y10,z10,x11,y11,z11,x12,y12,z12,x13,y13,z13,x14,y14,z14,x15,y15
    ↪ ,z15,x16,y16,z16 --params=7.309,1.02914215351,1.05048570256,
    ↪ 88.81,89.08,89.07,0.0007,0.0386,0.21,0.5038,0.5361,0.2181,
    ↪ 0.0076,0.466,0.2884,0.4972,0.9638,0.2878,0.2851,0.2574,0.287,
    ↪ 0.2204,0.763,0.2232,0.2186,0.2627,0.7258,0.284,0.7583,0.7679,
    ↪ 0.2943,0.0422,-0.0002,0.2971,0.5446,0.4982,0.2096,0.482,-0.0072
    ↪ ,0.2088,-0.017,0.5051,0.2566,0.0259,0.285,0.2502,0.528,0.2158,
    ↪ 0.2438,0.0313,0.7817,0.2499,0.5338,0.719 & P-1 C_{i}^{1} #2 (i^
    ↪ 16) & aP32 & None & O3W & O3W & P. M. Woodward and A. W.
    ↪ Sleight and T. Vogt, J. Solid State Chem. 131, 9-17 (1997)
   1.00000000000000
   7.30900000000000   0.00000000000000   0.00000000000000
   0.12208844854962   7.52100913513145   0.00000000000000
   0.12328057187699   0.15747592326685   7.67539492366282
     O      W
    24      8
Direct
   0.00070000000000   0.03860000000000   0.21000000000000    O   (2i)
  -0.00070000000000  -0.03860000000000  -0.21000000000000    O   (2i)
   0.50380000000000   0.53610000000000   0.21810000000000    O   (2i)
  -0.50380000000000  -0.53610000000000  -0.21810000000000    O   (2i)
   0.00760000000000   0.46600000000000   0.28840000000000    O   (2i)
  -0.00760000000000  -0.46600000000000  -0.28840000000000    O   (2i)
   0.49720000000000   0.96380000000000   0.28780000000000    O   (2i)
  -0.49720000000000  -0.96380000000000  -0.28780000000000    O   (2i)
   0.28510000000000   0.25740000000000   0.28700000000000    O   (2i)
  -0.28510000000000  -0.25740000000000  -0.28700000000000    O   (2i)
   0.22040000000000   0.76300000000000   0.22320000000000    O   (2i)
```

```
   -0.22040000000000   -0.76300000000000   -0.22320000000000    O    (2i)
    0.21860000000000    0.26270000000000    0.72580000000000    O    (2i)
   -0.21860000000000   -0.26270000000000   -0.72580000000000    O    (2i)
    0.28400000000000    0.75830000000000    0.76790000000000    O    (2i)
   -0.28400000000000   -0.75830000000000   -0.76790000000000    O    (2i)
    0.29430000000000    0.04220000000000   -0.00020000000000    O    (2i)
   -0.29430000000000   -0.04220000000000    0.00020000000000    O    (2i)
    0.29710000000000    0.54460000000000    0.49820000000000    O    (2i)
   -0.29710000000000   -0.54460000000000   -0.49820000000000    O    (2i)
    0.20960000000000    0.48200000000000   -0.00720000000000    O    (2i)
   -0.20960000000000   -0.48200000000000    0.00720000000000    O    (2i)
    0.20880000000000   -0.01700000000000    0.50510000000000    O    (2i)
   -0.20880000000000    0.01700000000000   -0.50510000000000    O    (2i)
    0.25660000000000    0.02590000000000    0.28500000000000    W    (2i)
   -0.25660000000000   -0.02590000000000   -0.28500000000000    W    (2i)
    0.25020000000000    0.52800000000000    0.21580000000000    W    (2i)
   -0.25020000000000   -0.52800000000000   -0.21580000000000    W    (2i)
    0.24380000000000    0.03130000000000    0.78170000000000    W    (2i)
   -0.24380000000000   -0.03130000000000   -0.78170000000000    W    (2i)
    0.24990000000000    0.53380000000000    0.71900000000000    W    (2i)
   -0.24990000000000   -0.53380000000000   -0.71900000000000    W    (2i)
```

Chalcanthite ($CuSO_4{\cdot}5H_2O$, $H4_{10}$): AB10C9D_aP42_2_ae_10i_9i_i - CIF

```
# CIF file
data_findsym-output
_audit_creation_method FINDSYM

_chemical_name_mineral 'Chalcanthite'
_chemical_formula_sum 'Cu H10 O9 S'

loop_
_publ_author_name
 'G. E. Bacon'
 'D. H. Titterton'
_journal_name_full_name
;
 Zeitschrift f{\"u}r Kristallographie - Crystalline Materials
;
_journal_volume 141
_journal_year 1975
_journal_page_first 330
_journal_page_last 341
_publ_Section_title
;
 Neutron-diffraction studies of CuSO$_{4}$$\cdot$5H$_{2}$O and CuSO$_{4}
     ↪ $$\cdot$5D$_{2}$O
;

# Found in The American Mineralogist Crystal Structure Database, 2003

_aflow_title 'Chalcanthite (CuSO$_{4}$$\cdot$5H$_{2}$O, $H4_{10}$)
     ↪ Structure'
_aflow_proto 'AB10C9D_aP42_2_ae_10i_9i_i'
_aflow_params 'a,b/a,c/a,\alpha,\beta,\gamma,x_{3},y_{3},z_{3},x_{4},y_{
     ↪ 4},z_{4},x_{5},y_{5},z_{5},x_{6},y_{6},z_{6},x_{7},y_{7},z_{7},
     ↪ x_{8},y_{8},z_{8},x_{9},y_{9},z_{9},x_{10},y_{10},z_{10},x_{11
     ↪ },y_{11},z_{11},x_{12},y_{12},z_{12},x_{13},y_{13},z_{13},x_{14
     ↪ },y_{14},z_{14},x_{15},y_{15},z_{15},x_{16},y_{16},z_{16},x_{17
     ↪ },y_{17},z_{17},x_{18},y_{18},z_{18},x_{19},y_{19},z_{19},x_{20
     ↪ },y_{20},z_{20},x_{21},y_{21},z_{21},x_{22},y_{22},z_{22}'
_aflow_params_values '6.141,1.74824947077,0.974759811106,82.26667,
     ↪ 107.43333,102.66667,0.898,0.1412,0.2547,0.7185,0.0126,0.2283,
     ↪ 0.301,0.2016,0.0667,0.3341,0.127,0.3188,0.3231,0.3785,0.3406,
     ↪ 0.6016,0.3937,0.4256,0.8012,0.4011,0.8847,0.857,0.3845,0.162,
     ↪ 0.6033,0.1321,0.6671,0.4108,0.1932,0.6922,0.9072,0.152,0.6734,
     ↪ 0.2442,0.3172,0.796,0.8061,0.3724,0.6363,0.0444,0.3022,0.3849,
     ↪ 0.8176,0.0737,0.1519,0.2887,0.1177,0.149,0.4654,0.4063,0.2975,
     ↪ 0.756,0.4161,0.0191,0.435,0.1263,0.6289,0.0133,0.2871,0.6253'
_aflow_Strukturbericht '$H4_{10}$'
_aflow_Pearson 'aP42'

_symmetry_space_group_name_H-M "P -1"
_symmetry_Int_Tables_number 2

_cell_length_a       6.14100
_cell_length_b      10.73600
_cell_length_c       5.98600
_cell_angle_alpha   82.26667
_cell_angle_beta   107.43333
_cell_angle_gamma  102.66667

loop_
_space_group_symop_id
_space_group_symop_operation_xyz
1 x,y,z
2 -x,-y,-z

loop_
_atom_site_label
_atom_site_type_symbol
_atom_site_symmetry_multiplicity
_atom_site_Wyckoff_label
_atom_site_fract_x
_atom_site_fract_y
_atom_site_fract_z
_atom_site_occupancy
Cu1 Cu 1 a 0.0000 0.0000 0.0000 1.00000
Cu2 Cu 1 e 0.5000 0.5000 0.0000 1.00000
H1 H 2 i 0.8980 0.1412 0.2547 1.00000
H2 H 2 i 0.7185 0.0126 0.2283 1.00000
H3 H 2 i 0.3010 0.2016 0.0667 1.00000
H4 H 2 i 0.3341 0.1270 0.3188 1.00000
H5 H 2 i 0.3231 0.3785 0.3406 1.00000
H6 H 2 i 0.6016 0.3937 0.4256 1.00000
H7 H 2 i 0.8012 0.4011 0.8847 1.00000
H8 H 2 i 0.8570 0.3845 0.1620 1.00000
H9 H 2 i 0.6033 0.1321 0.6671 1.00000
H10 H 2 i 0.4108 0.1932 0.6922 1.00000
O1 O 2 i 0.9072 0.1520 0.6734 1.00000
O2 O 2 i 0.2442 0.3172 0.7960 1.00000
O3 O 2 i 0.8061 0.3724 0.6363 1.00000
O4 O 2 i 0.0444 0.3022 0.3849 1.00000
O5 O 2 i 0.8176 0.0737 0.1519 1.00000
O6 O 2 i 0.2887 0.1177 0.1490 1.00000
O7 O 2 i 0.4654 0.4063 0.2975 1.00000
O8 O 2 i 0.7560 0.4161 0.0191 1.00000
O9 O 2 i 0.4350 0.1263 0.6289 1.00000
S1 S 2 i 0.0133 0.2871 0.6253 1.00000
```

Chalcanthite ($CuSO_4{\cdot}5H_2O$, $H4_{10}$): AB10C9D_aP42_2_ae_10i_9i_i - POSCAR

```
AB10C9D_aP42_2_ae_10i_9i_i & a,b/a,c/a,alpha,beta,gamma,x3,y3,z3,x4,y4,
     ↪ z4,x5,y5,z5,x6,y6,z6,x7,y7,z7,x8,y8,z8,x9,y9,z9,x10,y10,z10,x11
     ↪ ,y11,z11,x12,y12,z12,x13,y13,z13,x14,y14,z14,x15,y15,z15,x16,
     ↪ y16,z16,x17,y17,z17,x18,y18,z18,x19,y19,z19,x20,y20,z20,x21,y21
     ↪ ,z21,x22,y22,z22 --params=6.141,1.74824947077,0.974759811106,
     ↪ 82.26667,107.43333,102.66667,0.898,0.1412,0.2547,0.7185,0.0126,
     ↪ 0.2283,0.301,0.2016,0.0667,0.3341,0.127,0.3188,0.3231,0.3785,
     ↪ 0.3406,0.6016,0.3937,0.4256,0.8012,0.4011,0.8847,0.857,0.3845,
     ↪ 0.162,0.6033,0.1321,0.6671,0.4108,0.1932,0.6922,0.9072,0.152,
     ↪ 0.6734,0.2442,0.3172,0.796,0.8061,0.3724,0.6363,0.0444,0.3022,
     ↪ 0.3849,0.8176,0.0737,0.1519,0.2887,0.1177,0.149,0.4654,0.4063,
     ↪ 0.2975,0.756,0.4161,0.0191,0.435,0.1263,0.6289,0.0133,0.2871,
     ↪ 0.6253 & P-1 C_{i}^{1} #2 (aei^20) & aP42 & $H4_{10}$ &
     ↪ CuH10O9S & Chalcanthite & G. E. Bacon and D. H. Titterton,
     ↪ Zeitschrift f{"u}r Kristallographie - Crystalline Materials 141
     ↪ , 330-341 (1975)
   1.00000000000000
   6.14100000000000    0.00000000000000    0.00000000000000
  -2.35417591867527   10.47471010309730    0.00000000000000
  -1.79338070202199    0.42252508515638    5.69538885503253
    Cu       H       O       S
     2      20      18       2
Direct
   0.00000000000000    0.00000000000000    0.00000000000000    Cu    (1a)
   0.50000000000000    0.50000000000000    0.00000000000000    Cu    (1e)
   0.89800000000000    0.14120000000000    0.25470000000000    H    (2i)
  -0.89800000000000   -0.14120000000000   -0.25470000000000    H    (2i)
   0.71850000000000    0.01260000000000    0.22830000000000    H    (2i)
  -0.71850000000000   -0.01260000000000   -0.22830000000000    H    (2i)
   0.30100000000000    0.20160000000000    0.06670000000000    H    (2i)
  -0.30100000000000   -0.20160000000000   -0.06670000000000    H    (2i)
   0.33410000000000    0.12700000000000    0.31880000000000    H    (2i)
  -0.33410000000000   -0.12700000000000   -0.31880000000000    H    (2i)
   0.32310000000000    0.37850000000000    0.34060000000000    H    (2i)
  -0.32310000000000   -0.37850000000000   -0.34060000000000    H    (2i)
   0.60160000000000    0.39370000000000    0.42560000000000    H    (2i)
  -0.60160000000000   -0.39370000000000   -0.42560000000000    H    (2i)
   0.80120000000000    0.40110000000000    0.88470000000000    H    (2i)
  -0.80120000000000   -0.40110000000000   -0.88470000000000    H    (2i)
   0.85700000000000    0.38450000000000    0.16200000000000    H    (2i)
  -0.85700000000000   -0.38450000000000   -0.16200000000000    H    (2i)
   0.60330000000000    0.13210000000000    0.66710000000000    H    (2i)
  -0.60330000000000   -0.13210000000000   -0.66710000000000    H    (2i)
   0.41080000000000    0.19320000000000    0.69220000000000    H    (2i)
  -0.41080000000000   -0.19320000000000   -0.69220000000000    H    (2i)
   0.90720000000000    0.15200000000000    0.67340000000000    O    (2i)
  -0.90720000000000   -0.15200000000000   -0.67340000000000    O    (2i)
   0.24420000000000    0.31720000000000    0.79600000000000    O    (2i)
  -0.24420000000000   -0.31720000000000   -0.79600000000000    O    (2i)
   0.80610000000000    0.37240000000000    0.63630000000000    O    (2i)
  -0.80610000000000   -0.37240000000000   -0.63630000000000    O    (2i)
   0.04440000000000    0.30220000000000    0.38490000000000    O    (2i)
  -0.04440000000000   -0.30220000000000   -0.38490000000000    O    (2i)
   0.81760000000000    0.07370000000000    0.15190000000000    O    (2i)
  -0.81760000000000   -0.07370000000000   -0.15190000000000    O    (2i)
   0.28870000000000    0.11770000000000    0.14900000000000    O    (2i)
  -0.28870000000000   -0.11770000000000   -0.14900000000000    O    (2i)
   0.46540000000000    0.40630000000000    0.29750000000000    O    (2i)
  -0.46540000000000   -0.40630000000000   -0.29750000000000    O    (2i)
   0.75600000000000    0.41610000000000    0.01910000000000    O    (2i)
  -0.75600000000000   -0.41610000000000   -0.01910000000000    O    (2i)
   0.43500000000000    0.12630000000000    0.62890000000000    O    (2i)
  -0.43500000000000   -0.12630000000000   -0.62890000000000    O    (2i)
   0.01330000000000    0.28710000000000    0.62530000000000    S    (2i)
  -0.01330000000000   -0.28710000000000   -0.62530000000000    S    (2i)
```

Boric Acid ($H_3BO_3$, $G5_1$): AB3C3_aP28_2_2i_6i_6i - CIF

```
# CIF file
data_findsym-output
_audit_creation_method FINDSYM

_chemical_name_mineral 'Boric acid'
_chemical_formula_sum 'B H3 O3'

loop_
_publ_author_name
 'W. H. Zachariasen'
_journal_name_full_name
;
 Zeitschrift f{\"u}r Kristallographie - Crystalline Materials
;
_journal_volume 88
_journal_year 1934
_journal_page_first 150
_journal_page_last 161
_publ_Section_title
;
```

```
 The Crystal Lattice of Boric Acid, BO$_{3}$H$_{3}$
;

# Found in Strukturbericht Band III 1933-1935, 1937

_aflow_title 'Boric Acid (H$_{3}$BO$_{3}$, $G5_{1}$) Structure'
_aflow_proto 'AB3C3_aP28_2_2i_6i_6i'
_aflow_params 'a,b/a,c/a,\alpha,\beta,\gamma,x_{1},y_{1},z_{1},x_{2},y_{
     ↪ 2},z_{2},x_{3},y_{3},z_{3},x_{4},y_{4},z_{4},x_{5},y_{5},z_{5},
     ↪ x_{6},y_{6},z_{6},x_{7},y_{7},z_{7},x_{8},y_{8},z_{8},x_{9},y_{
     ↪ 9},z_{9},x_{10},y_{10},z_{10},x_{11},y_{11},z_{11},x_{12},y_{12
     ↪ },z_{12},x_{13},y_{13},z_{13},x_{14},y_{14},z_{14}'
_aflow_params_values '6.56,1.07317073171,1.07317073171,120.0,92.5,
     ↪ 101.16667,0.25,0.653,0.431,0.25,0.31,0.764,0.25,0.319,0.431,
     ↪ 0.25,0.319,0.097,0.25,-0.01,0.431,0.25,0.653,0.097,0.25,-0.01,
     ↪ 0.764,0.25,0.65,0.764,0.25,0.431,0.319,0.25,0.764,0.319,0.25,
     ↪ 0.764,0.653,0.25,0.21,0.542,0.25,0.21,0.875,0.25,0.542,0.875'
_aflow_Strukturbericht '$G5_{1}$'
_aflow_Pearson 'aP28'

_symmetry_space_group_name_H-M "P -1"
_symmetry_Int_Tables_number 2

_cell_length_a    6.56000
_cell_length_b    7.04000
_cell_length_c    7.04000
_cell_angle_alpha 120.00000
_cell_angle_beta  92.50000
_cell_angle_gamma 101.16667

loop_
_space_group_symop_id
_space_group_symop_operation_xyz
1 x,y,z
2 -x,-y,-z

loop_
_atom_site_label
_atom_site_type_symbol
_atom_site_symmetry_multiplicity
_atom_site_Wyckoff_label
_atom_site_fract_x
_atom_site_fract_y
_atom_site_fract_z
_atom_site_occupancy
B1 B 2 i 0.25000 0.65300 0.43100 1.00000
B2 B 2 i 0.25000 0.31000 0.76400 1.00000
H1 H 2 i 0.25000 0.31900 0.43100 1.00000
H2 H 2 i 0.25000 0.31900 0.09700 1.00000
H3 H 2 i 0.25000 -0.01000 0.43100 1.00000
H4 H 2 i 0.25000 0.65300 0.09700 1.00000
H5 H 2 i 0.25000 -0.01000 0.76400 1.00000
H6 H 2 i 0.25000 0.65000 0.76400 1.00000
O1 O 2 i 0.25000 0.43100 0.31900 1.00000
O2 O 2 i 0.25000 0.76400 0.31900 1.00000
O3 O 2 i 0.25000 0.76400 0.65300 1.00000
O4 O 2 i 0.25000 0.21000 0.54200 1.00000
O5 O 2 i 0.25000 0.21000 0.87500 1.00000
O6 O 2 i 0.25000 0.54200 0.87500 1.00000
```

Boric Acid ($H_3BO_3$, $G5_1$): AB3C3_aP28_2_2i_6i_6i - POSCAR

```
AB3C3_aP28_2_2i_6i_6i & a,b/a,c/a,alpha,beta,gamma,x1,y1,z1,x2,y2,z2,x3,
     ↪ y3,z3,x4,y4,z4,x5,y5,z5,x6,y6,z6,x7,y7,z7,x8,y8,z8,x9,y9,z9,x10
     ↪ ,y10,z10,x11,y11,z11,x12,y12,z12,x13,y13,z13,x14,y14,z14 --
     ↪ params=6.56,1.07317073171,1.07317073171,120.0,92.5,101.16667,
     ↪ 0.25,0.653,0.431,0.25,0.31,0.764,0.25,0.319,0.431,0.25,0.319,
     ↪ 0.097,0.25,-0.01,0.431,0.25,0.653,0.097,0.25,-0.01,0.764,0.25,
     ↪ 0.65,0.764,0.25,0.431,0.319,0.25,0.764,0.319,0.25,0.764,0.653,
     ↪ 0.25,0.21,0.542,0.25,0.21,0.875,0.25,0.542,0.875 & P-1 C_{i}^{1
     ↪ } #2 (i^14) & aP28 & $G5_{1}$ & BH3O3 & Boric acid & W. H.
     ↪ Zachariasen, Zeitschrift f{"u}r Kristallographie - Crystalline
     ↪ Materials 88, 150-161 (1934)
   1.00000000000000
   6.56000000000000   0.00000000000000   0.00000000000000
  -1.36339229932856   6.90671857238527   0.00000000000000
  -0.30708048705196  -3.64854466085737   6.01293800335586
     B       H       O
     4      12      12
Direct
   0.25000000000000   0.65300000000000   0.43100000000000   B   (2i)
  -0.25000000000000  -0.65300000000000  -0.43100000000000   B   (2i)
   0.25000000000000   0.31000000000000   0.76400000000000   B   (2i)
  -0.25000000000000  -0.31000000000000  -0.76400000000000   B   (2i)
   0.25000000000000   0.31900000000000   0.43100000000000   H   (2i)
  -0.25000000000000  -0.31900000000000  -0.43100000000000   H   (2i)
   0.25000000000000   0.31900000000000   0.09700000000000   H   (2i)
  -0.25000000000000  -0.31900000000000  -0.09700000000000   H   (2i)
   0.25000000000000  -0.01000000000000   0.43100000000000   H   (2i)
  -0.25000000000000   0.01000000000000  -0.43100000000000   H   (2i)
   0.25000000000000   0.65300000000000   0.09700000000000   H   (2i)
  -0.25000000000000  -0.65300000000000  -0.09700000000000   H   (2i)
   0.25000000000000  -0.01000000000000   0.76400000000000   H   (2i)
  -0.25000000000000   0.01000000000000  -0.76400000000000   H   (2i)
   0.25000000000000   0.65000000000000   0.76400000000000   H   (2i)
  -0.25000000000000  -0.65000000000000  -0.76400000000000   H   (2i)
   0.25000000000000   0.43100000000000   0.31900000000000   O   (2i)
  -0.25000000000000  -0.43100000000000  -0.31900000000000   O   (2i)
   0.25000000000000   0.76400000000000   0.31900000000000   O   (2i)
  -0.25000000000000  -0.76400000000000  -0.31900000000000   O   (2i)
   0.25000000000000   0.76400000000000   0.65300000000000   O   (2i)
  -0.25000000000000  -0.76400000000000  -0.65300000000000   O   (2i)
   0.25000000000000   0.21000000000000   0.54200000000000   O   (2i)
  -0.25000000000000  -0.21000000000000  -0.54200000000000   O   (2i)
   0.25000000000000   0.21000000000000   0.87500000000000   O   (2i)
  -0.25000000000000  -0.21000000000000  -0.87500000000000   O   (2i)
   0.25000000000000   0.54200000000000   0.87500000000000   O   (2i)
  -0.25000000000000  -0.54200000000000  -0.87500000000000   O   (2i)
```

Wollastonite ($CaSiO_3$): AB3C_aP30_2_3i_9i_3i - CIF

```
# CIF file
data_findsym-output
_audit_creation_method FINDSYM

_chemical_name_mineral 'Wollastonite'
_chemical_formula_sum 'Ca O3 Si'

loop_
_publ_author_name
 'M. J. Buerger'
 'C. T. Prewitt'
_journal_name_full_name
;
 Proceedings of the National Academy of Sciences of the United States of
     ↪ America
;
_journal_volume 47
_journal_year 1961
_journal_page_first 1884
_journal_page_last 1888
_publ_Section_title
;
 The Crystal Structures of Wollastonite and Pectolite
;

_aflow_title 'Wollastonite (CaSiO$_{3}$) Structure'
_aflow_proto 'AB3C_aP30_2_3i_9i_3i'
_aflow_params 'a,b/a,c/a,\alpha,\beta,\gamma,x_{1},y_{1},z_{1},x_{2},y_{
     ↪ 2},z_{2},x_{3},y_{3},z_{3},x_{4},y_{4},z_{4},x_{5},y_{5},z_{5},
     ↪ x_{6},y_{6},z_{6},x_{7},y_{7},z_{7},x_{8},y_{8},z_{8},x_{9},y_{
     ↪ 9},z_{9},x_{10},y_{10},z_{10},x_{11},y_{11},z_{11},x_{12},y_{12
     ↪ },z_{12},x_{13},y_{13},z_{13},x_{14},y_{14},z_{14},x_{15},y_{15
     ↪ },z_{15}'
_aflow_params_values '7.94,0.921914357683,0.890428211587,90.03333,
     ↪ 95.36667,103.43333,0.1985,0.4228,0.7608,0.2027,-0.0707,0.764,
     ↪ 0.4966,0.2495,0.472,0.4291,0.2314,0.8019,0.4008,0.7259,0.8302,
     ↪ 0.3037,0.4635,0.4641,0.3017,-0.0626,0.4655,0.0154,0.6254,0.7343
     ↪ ,0.0175,0.1319,0.7353,0.2732,0.5118,0.0919,0.2713,0.8717,0.094,
     ↪ 0.2188,0.1784,0.2228,0.1852,0.387,0.2687,0.1849,-0.0455,0.2692,
     ↪ 0.397,0.7235,0.056'
_aflow_Strukturbericht 'None'
_aflow_Pearson 'aP30'

_symmetry_space_group_name_H-M "P -1"
_symmetry_Int_Tables_number 2

_cell_length_a    7.94000
_cell_length_b    7.32000
_cell_length_c    7.07000
_cell_angle_alpha 90.03333
_cell_angle_beta  95.36667
_cell_angle_gamma 103.43333

loop_
_space_group_symop_id
_space_group_symop_operation_xyz
1 x,y,z
2 -x,-y,-z

loop_
_atom_site_label
_atom_site_type_symbol
_atom_site_symmetry_multiplicity
_atom_site_Wyckoff_label
_atom_site_fract_x
_atom_site_fract_y
_atom_site_fract_z
_atom_site_occupancy
Ca1 Ca 2 i 0.19850 0.42280 0.76080 1.00000
Ca2 Ca 2 i 0.20270 -0.07070 0.76400 1.00000
Ca3 Ca 2 i 0.49660 0.24950 0.47200 1.00000
O1 O 2 i 0.42910 0.23140 0.80190 1.00000
O2 O 2 i 0.40080 0.72590 0.83020 1.00000
O3 O 2 i 0.30370 0.46350 0.46410 1.00000
O4 O 2 i 0.30170 -0.06260 0.46550 1.00000
O5 O 2 i 0.01540 0.62540 0.73430 1.00000
O6 O 2 i 0.01750 0.13190 0.73530 1.00000
O7 O 2 i 0.27320 0.51180 0.09190 1.00000
O8 O 2 i 0.27130 0.87170 0.09400 1.00000
O9 O 2 i 0.21880 0.17840 0.22280 1.00000
Si1 Si 2 i 0.18520 0.38700 0.26870 1.00000
Si2 Si 2 i 0.18490 -0.04550 0.26920 1.00000
Si3 Si 2 i 0.39700 0.72350 0.05600 1.00000
```

Wollastonite ($CaSiO_3$): AB3C_aP30_2_3i_9i_3i - POSCAR

```
AB3C_aP30_2_3i_9i_3i & a,b/a,c/a,alpha,beta,gamma,x1,y1,z1,x2,y2,z2,x3,
     ↪ y3,z3,x4,y4,z4,x5,y5,z5,x6,y6,z6,x7,y7,z7,x8,y8,z8,x9,y9,z9,x10
     ↪ ,y10,z10,x11,y11,z11,x12,y12,z12,x13,y13,z13,x14,y14,z14,x15,
     ↪ y15,z15 --params=7.94,0.921914357683,0.890428211587,90.03333,
     ↪ 95.36667,103.43333,0.1985,0.4228,0.7608,0.2027,-0.0707,0.764,
     ↪ 0.4966,0.2495,0.472,0.4291,0.2314,0.8019,0.4008,0.7259,0.8302,
     ↪ 0.3037,0.4635,0.4641,0.3017,-0.0626,0.4655,0.0154,0.6254,0.7343
     ↪ ,0.0175,0.1319,0.7353,0.2732,0.5118,0.0919,0.2713,0.8717,0.094,
     ↪ 0.2188,0.1784,0.2228,0.1852,0.387,0.2687,0.1849,-0.0455,0.2692,
     ↪ 0.397,0.7235,0.056 & P-1 C_{i}^{1} #2 (i^15) & aP30 & None &
     ↪ CaO3Si & Wollastonite & M. J. Buerger and C. T. Prewitt, Proc.
     ↪ Natl. Acad. Sci. 47, 1884-1888 (1961)
   1.00000000000000
```

```
   7.94000000000000    0.00000000000000    0.00000000000000
  -1.70053661868821    7.11973140002490    0.00000000000000
  -0.66125116738401   -0.16216723254653    7.03714066097314
    Ca      O      Si
     6     18       6
Direct
   0.19850000000000    0.42280000000000    0.76080000000000    Ca    (2i)
  -0.19850000000000   -0.42280000000000   -0.76080000000000    Ca    (2i)
   0.20270000000000   -0.07070000000000    0.76400000000000    Ca    (2i)
  -0.20270000000000    0.07070000000000   -0.76400000000000    Ca    (2i)
   0.49660000000000    0.24950000000000    0.47200000000000    Ca    (2i)
  -0.49660000000000   -0.24950000000000   -0.47200000000000    Ca    (2i)
   0.42910000000000    0.23140000000000    0.80190000000000    O     (2i)
  -0.42910000000000   -0.23140000000000   -0.80190000000000    O     (2i)
   0.40080000000000    0.72590000000000    0.83020000000000    O     (2i)
  -0.40080000000000   -0.72590000000000   -0.83020000000000    O     (2i)
   0.30370000000000    0.46350000000000    0.46410000000000    O     (2i)
  -0.30370000000000   -0.46350000000000   -0.46410000000000    O     (2i)
   0.30170000000000   -0.06260000000000    0.46550000000000    O     (2i)
  -0.30170000000000    0.06260000000000   -0.46550000000000    O     (2i)
   0.01540000000000    0.62540000000000    0.73430000000000    O     (2i)
  -0.01540000000000   -0.62540000000000   -0.73430000000000    O     (2i)
   0.01750000000000    0.13190000000000    0.73530000000000    O     (2i)
  -0.01750000000000   -0.13190000000000   -0.73530000000000    O     (2i)
   0.27320000000000    0.51180000000000    0.09190000000000    O     (2i)
  -0.27320000000000   -0.51180000000000   -0.09190000000000    O     (2i)
   0.27130000000000    0.87170000000000    0.09400000000000    O     (2i)
  -0.27130000000000   -0.87170000000000   -0.09400000000000    O     (2i)
   0.21880000000000    0.17840000000000    0.22280000000000    O     (2i)
  -0.21880000000000   -0.17840000000000   -0.22280000000000    O     (2i)
   0.18520000000000    0.38700000000000    0.26870000000000    Si    (2i)
  -0.18520000000000   -0.38700000000000   -0.26870000000000    Si    (2i)
   0.18490000000000   -0.04550000000000    0.26920000000000    Si    (2i)
  -0.18490000000000    0.04550000000000   -0.26920000000000    Si    (2i)
   0.39700000000000    0.72350000000000    0.05600000000000    Si    (2i)
  -0.39700000000000   -0.72350000000000   -0.05600000000000    Si    (2i)
```

Albite ($NaAlSi_3O_8$, $S6_8$): ABC8D3_aP26_2_i_i_8i_3i - CIF

```
# CIF file
data_findsym-output
_audit_creation_method FINDSYM

_chemical_name_mineral 'Albite'
_chemical_formula_sum 'Al Na O8 Si3'

loop_
_publ_author_name
 'J. V. Smith'
 'G. Artioli'
 '{\AA}. Kvick'
_journal_name_full_name
;
 American Mineralogist
;
_journal_volume 71
_journal_year 1986
_journal_page_first 727
_journal_page_last 733
_publ_Section_title
;
 Low albite, NaAlSi$_{3}$O$_{8}$: Neutron diffraction study of crystal
    ↪ structure at 13~K
;

_aflow_title 'Albite (NaAlSi$_{3}$O$_{8}$, $S6_{8}$) Structure'
_aflow_proto 'ABC8D3_aP26_2_i_i_8i_3i'
_aflow_params 'a,b/a,c/a,\alpha,\beta,\gamma,x_{1},y_{1},z_{1},x_{2},y_{
    ↪ 2},z_{2},x_{3},y_{3},z_{3},x_{4},y_{4},z_{4},x_{5},y_{5},z_{5},
    ↪ x_{6},y_{6},z_{6},x_{7},y_{7},z_{7},x_{8},y_{8},z_{8},x_{9},y_{
    ↪ 9},z_{9},x_{10},y_{10},z_{10},x_{11},y_{11},z_{11},x_{12},y_{12
    ↪ },z_{12},x_{13},y_{13},z_{13}'
_aflow_params_values '7.1576,1.03716189784,1.07544707723,115.11511,
    ↪ 107.37724,100.55864,0.20773,0.84122,0.1767,0.14441,0.27518,
    ↪ 0.25422,-0.03356,0.87658,0.13472,0.72111,0.40891,0.41381,
    ↪ 0.80981,0.2953,0.07942,0.74076,0.0321,0.32928,0.26814,0.70828,
    ↪ 0.31212,0.22592,0.32857,0.71731,0.38942,0.10112,0.31914,0.43708
    ↪ ,0.31682,0.05324,0.23701,0.18228,0.8249,0.68677,0.41955,0.19963
    ↪ ,0.6404,0.203,0.4392'
_aflow_Strukturbericht '$S6_{8}$'
_aflow_Pearson 'aP26'

_symmetry_space_group_name_H-M "P -1"
_symmetry_Int_Tables_number 2

_cell_length_a    7.15760
_cell_length_b    7.42359
_cell_length_c    7.69762
_cell_angle_alpha 115.11511
_cell_angle_beta  107.37724
_cell_angle_gamma 100.55864

loop_
_space_group_symop_id
_space_group_symop_operation_xyz
1 x,y,z
2 -x,-y,-z

loop_
_atom_site_label
_atom_site_type_symbol
_atom_site_symmetry_multiplicity
_atom_site_Wyckoff_label
_atom_site_fract_x
_atom_site_fract_y
_atom_site_fract_z
_atom_site_occupancy
Al1 Al 2 i 0.20773 0.84122 0.17670 1.00000
Na1 Na 2 i 0.14441 0.27518 0.25422 1.00000
O1 O 2 i -0.03356 0.87658 0.13472 1.00000
O2 O 2 i 0.72111 0.40891 0.41381 1.00000
O3 O 2 i 0.80981 0.29530 0.07942 1.00000
O4 O 2 i 0.74076 0.03210 0.32928 1.00000
O5 O 2 i 0.26814 0.70828 0.31212 1.00000
O6 O 2 i 0.22592 0.32857 0.71731 1.00000
O7 O 2 i 0.38942 0.10112 0.31914 1.00000
O8 O 2 i 0.43708 0.31682 0.05324 1.00000
Si1 Si 2 i 0.23701 0.18228 0.82490 1.00000
Si2 Si 2 i 0.68677 0.41955 0.19963 1.00000
Si3 Si 2 i 0.64040 0.20300 0.43920 1.00000
```

Albite ($NaAlSi_3O_8$, $S6_8$): ABC8D3_aP26_2_i_i_8i_3i - POSCAR

```
ABC8D3_aP26_2_i_i_8i_3i & a,b/a,c/a,alpha,beta,gamma,x1,y1,z1,x2,y2,z2,
    ↪ x3,y3,z3,x4,y4,z4,x5,y5,z5,x6,y6,z6,x7,y7,z7,x8,y8,z8,x9,y9,z9,
    ↪ x10,y10,z10,x11,y11,z11,x12,y12,z12,x13,y13,z13 --params=7.1576
    ↪ ,1.03716189784,1.07544707723,115.11511,107.37724,100.55864,
    ↪ 0.20773,0.84122,0.1767,0.14441,0.27518,0.25422,-0.03356,0.87658
    ↪ ,0.13472,0.72111,0.40891,0.41381,0.80981,0.2953,0.07942,0.74076
    ↪ ,0.0321,0.32928,0.26814,0.70828,0.31212,0.22592,0.32857,0.71731
    ↪ ,0.38942,0.10112,0.31914,0.43708,0.31682,0.05324,0.23701,
    ↪ 0.18228,0.8249,0.68677,0.41955,0.19963,0.6404,0.203,0.4392 &
    ↪ P-1 C_{i}^{1} #2 (i^13) & aP26 & $S6_{8}$ & AlNaO8Si3 & Albite
    ↪ & J. V. Smith and G. Artioli and {\AA}. Kvick, Am. Mineral. 71,
    ↪  727-733 (1986)
   1.00000000000000
   7.15760000000000    0.00000000000000    0.00000000000000
  -1.36031164512511    7.29789289564029    0.00000000000000
  -2.29898434571175   -3.75196276024920    6.31591640922167
    Al     Na     O      Si
     2      2     16      6
Direct
   0.20773000000000    0.84122000000000    0.17670000000000    Al    (2i)
  -0.20773000000000   -0.84122000000000   -0.17670000000000    Al    (2i)
   0.14441000000000    0.27518000000000    0.25422000000000    Na    (2i)
  -0.14441000000000   -0.27518000000000   -0.25422000000000    Na    (2i)
  -0.03356000000000    0.87658000000000    0.13472000000000    O     (2i)
   0.03356000000000   -0.87658000000000   -0.13472000000000    O     (2i)
   0.72111000000000    0.40891000000000    0.41381000000000    O     (2i)
  -0.72111000000000   -0.40891000000000   -0.41381000000000    O     (2i)
   0.80981000000000    0.29530000000000    0.07942000000000    O     (2i)
  -0.80981000000000   -0.29530000000000   -0.07942000000000    O     (2i)
   0.74076000000000    0.03210000000000    0.32928000000000    O     (2i)
  -0.74076000000000   -0.03210000000000   -0.32928000000000    O     (2i)
   0.26814000000000    0.70828000000000    0.31212000000000    O     (2i)
  -0.26814000000000   -0.70828000000000   -0.31212000000000    O     (2i)
   0.22592000000000    0.32857000000000    0.71731000000000    O     (2i)
  -0.22592000000000   -0.32857000000000   -0.71731000000000    O     (2i)
   0.38942000000000    0.10112000000000    0.31914000000000    O     (2i)
  -0.38942000000000   -0.10112000000000   -0.31914000000000    O     (2i)
   0.43708000000000    0.31682000000000    0.05324000000000    O     (2i)
  -0.43708000000000   -0.31682000000000   -0.05324000000000    O     (2i)
   0.23701000000000    0.18228000000000    0.82490000000000    Si    (2i)
  -0.23701000000000   -0.18228000000000   -0.82490000000000    Si    (2i)
   0.68677000000000    0.41955000000000    0.19963000000000    Si    (2i)
  -0.68677000000000   -0.41955000000000   -0.19963000000000    Si    (2i)
   0.64040000000000    0.20300000000000    0.43920000000000    Si    (2i)
  -0.64040000000000   -0.20300000000000   -0.43920000000000    Si    (2i)
```

TaTi (BCC SQS-16): AB_aP16_2_4i_4i - CIF

```
# AFLOW.org Repositories
# TaTi/AB_aP16_2_4i_4i-001.AB params=5.4244498076,1.04446593575,
    ↪ 1.78376517004,72.976133815,87.0786711125,79.9750121379,0.375,
    ↪ 0.75,0.375,0.875,0.0,0.375,0.375,0.5,0.875,0.625,0.75,0.625,
    ↪ 0.875,0.5,0.375,0.125,0.75,0.125,0.875,0.75,0.875,0.375,-0.0,
    ↪ 0.875 SG=2 [ANRL doi: 10.1016/j.commatsci.2017.01.017 (part 1),
    ↪  doi: 10.1016/j.commatsci.2018.10.043 (part 2)]
data_TaTi
_pd_phase_name AB_aP16_2_4i_4i-001.AB

_chemical_name_mineral 'TaTi'
_chemical_formula_sum 'Ta Ti'

loop_
_publ_author_name
 'C. Jiang'
 'C. Wolverton'
 'J. Sofo'
 'L.-Q. Chen'
 'Z.-K. Liu'
_journal_name_full_name
;
 Physical Review B
;
_journal_volume 69
_journal_year 2004
_journal_page_first 214202
_journal_page_last 214202
_publ_Section_title
;
 First-principles study of binary bcc alloys using special quasirandom
    ↪ structures
;

_aflow_title 'TaTi (BCC SQS-16) Structure'
_aflow_proto 'AB_aP16_2_4i_4i'
_aflow_params 'a,b/a,c/a,\alpha,\beta,\gamma,x_{1},y_{1},z_{1},x_{2},y_{
    ↪ 2},z_{2},x_{3},y_{3},z_{3},x_{4},y_{4},z_{4},x_{5},y_{5},z_{5},
    ↪ x_{6},y_{6},z_{6},x_{7},y_{7},z_{7},x_{8},y_{8},z_{8}'
```

```
_aflow_params_values '5.4244498076,1.04446593575,1.78376517004,
    ↪ 72.976133815,87.0786711125,79.9750121379,0.375,0.75,0.375,0.875
    ↪ ,0.0,0.375,0.375,0.5,0.875,0.625,0.75,0.625,0.875,0.5,0.375,
    ↪ 0.125,0.75,0.125,0.875,0.75,0.875,0.375,0.0,0.875'
_aflow_Strukturbericht 'None'
_aflow_Pearson 'aP16'

_cell_length_a    5.4244498076
_cell_length_b    5.6656530442
_cell_length_c    9.6759446334
_cell_angle_alpha    72.9761338150
_cell_angle_beta    87.0786711125
_cell_angle_gamma    79.9750121379
_symmetry_space_group_name_H-M    'P-1'
_symmetry_Int_Tables_Number    2
loop_
_symmetry_equiv_pos_site_id
_symmetry_equiv_pos_as_xyz
  1  x,y,z
  2  -x,-y,-z
loop_
 _atom_site_label
 _atom_site_occupancy
 _atom_site_fract_x
 _atom_site_fract_y
 _atom_site_fract_z
 _atom_site_thermal_displace_type
 _atom_site_B_iso_or_equiv
 _atom_site_type_symbol
 _atom_site_symmetry_multiplicity
 _atom_site_Wyckoff_label
Ta1 1.0000000000 0.3750000000 0.7500000000 0.3750000000 Biso 1.0 Ta 2 i
Ta2 1.0000000000 0.8750000000 0.0000000000 0.3750000000 Biso 1.0 Ta 2 i
Ta3 1.0000000000 0.3750000000 0.5000000000 0.8750000000 Biso 1.0 Ta 2 i
Ta4 1.0000000000 0.6250000000 0.7500000000 0.6250000000 Biso 1.0 Ta 2 i
Ti1 1.0000000000 0.8750000000 0.5000000000 0.3750000000 Biso 1.0 Ti 2 i
Ti2 1.0000000000 0.1250000000 0.7500000000 0.1250000000 Biso 1.0 Ti 2 i
Ti3 1.0000000000 0.8750000000 0.7500000000 0.8750000000 Biso 1.0 Ti 2 i
Ti4 1.0000000000 0.3750000000 0.0000000000 0.8750000000 Biso 1.0 Ti 2 i
```

TaTi (BCC SQS-16): AB_aP16_2_4i_4i - POSCAR

```
AB_aP16_2_4i_4i & a,b/a,c/a,alpha,beta,gamma,x1,y1,z1,x2,y2,z2,x3,y3,z3,
    ↪ x4,y4,z4,x5,y5,z5,x6,y6,z6,x7,y7,z7,x8,y8,z8 --params=
    ↪ 5.4244498076,1.04446593575,1.78376517004,72.976133815,
    ↪ 87.0786711125,79.9750121379,0.375,0.75,0.375,0.875,0.0,0.375,
    ↪ 0.375,0.5,0.875,0.625,0.75,0.625,0.875,0.5,0.375,0.125,0.75,
    ↪ 0.125,0.875,0.75,0.875,0.375,0.0,0.875 & P-1 C_{i}^{1} #2 (i^8)
    ↪ & aP16 & None & TaTi & TaTi & C. Jiang et al., Phys. Rev. B 69
    ↪ , 214202(2004)
   1.00000000000000
   5.42444980760000   0.00000000000000   0.00000000000000
   0.98626360139496   5.57914944465697   0.00000000000000
   0.49313180069476   2.78957472230357   9.25197267854001
    Ta    Ti
     8     8
Direct
   0.37500000000000   0.75000000000000   0.37500000000000   Ta   (2i)
  -0.37500000000000  -0.75000000000000  -0.37500000000000   Ta   (2i)
   0.87500000000000   0.00000000000000   0.37500000000000   Ta   (2i)
  -0.87500000000000   0.00000000000000  -0.37500000000000   Ta   (2i)
   0.37500000000000   0.50000000000000   0.87500000000000   Ta   (2i)
  -0.37500000000000  -0.50000000000000  -0.87500000000000   Ta   (2i)
   0.62500000000000   0.75000000000000   0.62500000000000   Ta   (2i)
  -0.62500000000000  -0.75000000000000  -0.62500000000000   Ta   (2i)
   0.87500000000000   0.50000000000000   0.37500000000000   Ti   (2i)
  -0.87500000000000  -0.50000000000000  -0.37500000000000   Ti   (2i)
   0.12500000000000   0.75000000000000   0.12500000000000   Ti   (2i)
  -0.12500000000000  -0.75000000000000  -0.12500000000000   Ti   (2i)
   0.87500000000000   0.75000000000000   0.87500000000000   Ti   (2i)
  -0.87500000000000  -0.75000000000000  -0.87500000000000   Ti   (2i)
   0.37500000000000   0.00000000000000   0.87500000000000   Ti   (2i)
  -0.37500000000000   0.00000000000000  -0.87500000000000   Ti   (2i)
```

$W_2O_3(PO_4)_2$: A11B2C2_mP60_4_22a_4a_4a - CIF

```
# CIF file
data_findsym-output
_audit_creation_method FINDSYM

_chemical_name_mineral 'O11P2W2'
_chemical_formula_sum 'O11 P2 W2'

loop_
_publ_author_name
 'P. Kierkegaard'
 'S. {\AA}sbrink'
_journal_name_full_name
;
 Acta Chemica Scandinavica
;
_journal_volume 18
_journal_year 1964
_journal_page_first 2329
_journal_page_last 2338
_publ_Section_title
;
 The Crystal Structure of W$_{2}$O$_{3}$(PO$_{4}$)$_{2}$. Determination
    ↪ of a Superstructure by Means of Least-Squares Calculations
;

_aflow_title 'W$_{2}$O$_{3}$(PO$_{4}$)$_{2}$ Structure'
_aflow_proto 'A11B2C2_mP60_4_22a_4a_4a'
_aflow_params 'a,b/a,c/a,\beta,x_{1},y_{1},z_{1},x_{2},y_{2},z_{2},x_{3
    ↪ },y_{3},z_{3},x_{4},y_{4},z_{4},x_{5},y_{5},z_{5},x_{6},y_{6},
    ↪ z_{6},x_{7},y_{7},z_{7},x_{8},y_{8},z_{8},x_{9},y_{9},z_{9},x_{
    ↪ 10},y_{10},z_{10},x_{11},y_{11},z_{11},x_{12},y_{12},z_{12},x_{
    ↪ 13},y_{13},z_{13},x_{14},y_{14},z_{14},x_{15},y_{15},z_{15},x_{
    ↪ 16},y_{16},z_{16},x_{17},y_{17},z_{17},x_{18},y_{18},z_{18},x_{
    ↪ 19},y_{19},z_{19},x_{20},y_{20},z_{20},x_{21},y_{21},z_{21},x_{
    ↪ 22},y_{22},z_{22},x_{23},y_{23},z_{23},x_{24},y_{24},z_{24},x_{
    ↪ 25},y_{25},z_{25},x_{26},y_{26},z_{26},x_{27},y_{27},z_{27},x_{
    ↪ 28},y_{28},z_{28},x_{29},y_{29},z_{29},x_{30},y_{30},z_{30}'
_aflow_params_values '7.822,1.59805676298,0.991306571209,91.05,0.224,
    ↪ 0.099,0.084,0.796,0.889,0.811,0.258,-0.096,0.129,0.839,0.079,
    ↪ 0.842,0.193,0.009,0.442,0.852,-0.005,0.561,0.834,0.017,0.197,
    ↪ 0.086,-0.024,0.788,0.517,-0.012,0.57,0.991,0.229,0.991,0.088,
    ↪ 0.243,0.331,-0.089,0.3,0.665,0.421,0.263,0.257,0.584,0.267,
    ↪ 0.722,0.652,0.239,0.082,0.326,0.287,-0.078,0.41,0.01,0.888,
    ↪ 0.497,0.007,0.208,0.279,0.162,0.664,0.687,0.842,0.389,0.301,
    ↪ 0.841,0.669,0.64,0.155,0.343,0.3226,0.0046,0.0737,0.8943,-
    ↪ 0.0084,0.7508,0.6178,0.2582,0.2607,0.382,0.2606,0.7514,0.3082,-
    ↪ 0.0024,0.6351,0.7024,-0.0071,0.3585,0.1913,0.2501,0.1533,0.8074
    ↪ ,0.2557,0.8462'
_aflow_Strukturbericht 'None'
_aflow_Pearson 'mP60'

_symmetry_space_group_name_H-M "P 1 21 1"
_symmetry_Int_Tables_number 4

_cell_length_a     7.82200
_cell_length_b     12.50000
_cell_length_c     7.75400
_cell_angle_alpha  90.00000
_cell_angle_beta   91.05000
_cell_angle_gamma  90.00000

loop_
_space_group_symop_id
_space_group_symop_operation_xyz
1 x,y,z
2 -x,y+1/2,-z

loop_
_atom_site_label
_atom_site_type_symbol
_atom_site_symmetry_multiplicity
_atom_site_Wyckoff_label
_atom_site_fract_x
_atom_site_fract_y
_atom_site_fract_z
_atom_site_occupancy
O1 O 2 a 0.22400 0.09900 0.08400 1.00000
O2 O 2 a 0.79600 0.88900 0.81100 1.00000
O3 O 2 a 0.25800 -0.09600 0.12900 1.00000
O4 O 2 a 0.83900 0.07900 0.84200 1.00000
O5 O 2 a 0.19300 0.00900 0.44200 1.00000
O6 O 2 a 0.85200 -0.00500 0.56100 1.00000
O7 O 2 a 0.83400 0.01700 0.19700 1.00000
O8 O 2 a 0.08600 -0.02400 0.78800 1.00000
O9 O 2 a 0.51700 -0.01200 0.57000 1.00000
O10 O 2 a 0.99100 0.22900 0.99100 1.00000
O11 O 2 a 0.08800 0.24300 0.33100 1.00000
O12 O 2 a -0.08900 0.30000 0.66500 1.00000
O13 O 2 a 0.42100 0.26300 0.25700 1.00000
O14 O 2 a 0.58400 0.26700 0.72200 1.00000
O15 O 2 a 0.65200 0.23900 0.08200 1.00000
O16 O 2 a 0.32600 0.28700 -0.07800 1.00000
O17 O 2 a 0.41000 0.01000 0.88800 1.00000
O18 O 2 a 0.49700 0.00700 0.20800 1.00000
O19 O 2 a 0.27900 0.16200 0.66400 1.00000
O20 O 2 a 0.68700 0.84200 0.38900 1.00000
O21 O 2 a 0.30100 0.84100 0.66900 1.00000
O22 O 2 a 0.64000 0.15500 0.34300 1.00000
P1 P 2 a 0.32260 0.00460 0.07370 1.00000
P2 P 2 a 0.89430 -0.00840 0.75080 1.00000
P3 P 2 a 0.61780 0.25820 0.26070 1.00000
P4 P 2 a 0.38200 0.26060 0.75140 1.00000
W1 W 2 a 0.30820 -0.00240 0.63510 1.00000
W2 W 2 a 0.70240 -0.00710 0.35850 1.00000
W3 W 2 a 0.19130 0.25010 0.15330 1.00000
W4 W 2 a 0.80740 0.25570 0.84620 1.00000
```

$W_2O_3(PO_4)_2$: A11B2C2_mP60_4_22a_4a_4a - POSCAR

```
A11B2C2_mP60_4_22a_4a_4a & a,b/a,c/a,beta,x1,y1,z1,x2,y2,z2,x3,y3,z3,x4,
    ↪ y4,z4,x5,y5,z5,x6,y6,z6,x7,y7,z7,x8,y8,z8,x9,y9,z9,x10,y10,z10,
    ↪ x11,y11,z11,x12,y12,z12,x13,y13,z13,x14,y14,z14,x15,y15,z15,x16
    ↪ ,y16,z16,x17,y17,z17,x18,y18,z18,x19,y19,z19,x20,y20,z20,x21,
    ↪ y21,z21,x22,y22,z22,x23,y23,z23,x24,y24,z24,x25,y25,z25,x26,y26
    ↪ ,z26,x27,y27,z27,x28,y28,z28,x29,y29,z29,x30,y30,z30 --params=
    ↪ 7.822,1.59805676298,0.991306571209,91.05,0.224,0.099,0.084,
    ↪ 0.796,0.889,0.811,0.258,-0.096,0.129,0.839,0.079,0.842,0.193,
    ↪ 0.009,0.442,0.852,-0.005,0.561,0.834,0.017,0.197,0.086,-0.024,
    ↪ 0.788,0.517,-0.012,0.57,0.991,0.229,0.991,0.088,0.243,0.331,-
    ↪ 0.089,0.3,0.665,0.421,0.263,0.257,0.584,0.267,0.722,0.652,0.239
    ↪ ,0.082,0.326,0.287,-0.078,0.41,0.01,0.888,0.497,0.007,0.208,
    ↪ 0.279,0.162,0.664,0.687,0.842,0.389,0.301,0.841,0.669,0.64,
    ↪ 0.155,0.343,0.3226,0.0046,0.0737,0.8943,-0.0084,0.7508,0.6178,
    ↪ 0.2582,0.2607,0.382,0.2606,0.7514,0.3082,-0.0024,0.6351,0.7024
    ↪ ,-0.0071,0.3585,0.1913,0.2501,0.1533,0.8074,0.2557,0.8462 & P2_
    ↪ {1} C_{2}^{2} #4 (a^30) & mP60 & None & O11P2W2 & O11P2W2 & P.
    ↪ Kierkegaard and S. {\AA}sbrink, Acta Chem. Scand. 18, 2329-2338
    ↪ (1964)
   1.00000000000000
   7.82200000000000   0.00000000000000   0.00000000000000
   0.00000000000000  12.50000000000000   0.00000000000000
  -0.14209151804538   0.00000000000000   7.75269798202533
     O     P     W
    44     8     8
```

```
Direct
   0.22400000000000    0.09900000000000    0.08400000000000    O    (2a)
  -0.22400000000000    0.59900000000000   -0.08400000000000    O    (2a)
   0.79600000000000    0.88900000000000    0.81100000000000    O    (2a)
  -0.79600000000000    1.38900000000000   -0.81100000000000    O    (2a)
   0.25800000000000   -0.09600000000000    0.12900000000000    O    (2a)
  -0.25800000000000    0.40400000000000   -0.12900000000000    O    (2a)
   0.83900000000000    0.07900000000000    0.84200000000000    O    (2a)
  -0.83900000000000    0.57900000000000   -0.84200000000000    O    (2a)
   0.19300000000000    0.00900000000000    0.44200000000000    O    (2a)
  -0.19300000000000    0.50900000000000   -0.44200000000000    O    (2a)
   0.85200000000000   -0.00500000000000    0.56100000000000    O    (2a)
  -0.85200000000000    0.49500000000000   -0.56100000000000    O    (2a)
   0.83400000000000    0.01700000000000    0.19700000000000    O    (2a)
  -0.83400000000000    0.51700000000000   -0.19700000000000    O    (2a)
   0.08600000000000   -0.02400000000000    0.78800000000000    O    (2a)
  -0.08600000000000    0.47600000000000   -0.78800000000000    O    (2a)
   0.51700000000000   -0.01200000000000    0.57000000000000    O    (2a)
  -0.51700000000000    0.48800000000000   -0.57000000000000    O    (2a)
   0.99100000000000    0.22900000000000    0.99100000000000    O    (2a)
  -0.99100000000000    0.72900000000000   -0.99100000000000    O    (2a)
   0.08800000000000    0.24300000000000    0.33100000000000    O    (2a)
  -0.08800000000000    0.74300000000000   -0.33100000000000    O    (2a)
  -0.08900000000000    0.30000000000000    0.66500000000000    O    (2a)
   0.08900000000000    0.80000000000000   -0.66500000000000    O    (2a)
   0.42100000000000    0.26300000000000    0.25700000000000    O    (2a)
  -0.42100000000000    0.76300000000000   -0.25700000000000    O    (2a)
   0.58400000000000    0.26700000000000    0.72200000000000    O    (2a)
  -0.58400000000000    0.76700000000000   -0.72200000000000    O    (2a)
   0.65200000000000    0.23900000000000    0.08200000000000    O    (2a)
  -0.65200000000000    0.73900000000000   -0.08200000000000    O    (2a)
   0.32600000000000    0.28700000000000   -0.07800000000000    O    (2a)
  -0.32600000000000    0.78700000000000    0.07800000000000    O    (2a)
   0.41000000000000    0.01000000000000    0.88800000000000    O    (2a)
  -0.41000000000000    0.51000000000000   -0.88800000000000    O    (2a)
   0.49700000000000    0.00700000000000    0.20800000000000    O    (2a)
  -0.49700000000000    0.50700000000000   -0.20800000000000    O    (2a)
   0.27900000000000    0.16200000000000    0.66400000000000    O    (2a)
  -0.27900000000000    0.66200000000000   -0.66400000000000    O    (2a)
   0.68700000000000    0.84200000000000    0.38900000000000    O    (2a)
  -0.68700000000000    1.34200000000000   -0.38900000000000    O    (2a)
   0.30100000000000    0.84100000000000    0.66900000000000    O    (2a)
  -0.30100000000000    1.34100000000000   -0.66900000000000    O    (2a)
   0.64000000000000    0.15500000000000    0.34300000000000    O    (2a)
  -0.64000000000000    0.65500000000000   -0.34300000000000    O    (2a)
   0.32260000000000    0.00460000000000    0.07370000000000    P    (2a)
  -0.32260000000000    0.50460000000000   -0.07370000000000    P    (2a)
   0.89430000000000   -0.00840000000000    0.75080000000000    P    (2a)
  -0.89430000000000    0.49160000000000   -0.75080000000000    P    (2a)
   0.61780000000000    0.25820000000000    0.26070000000000    P    (2a)
  -0.61780000000000    0.75820000000000   -0.26070000000000    P    (2a)
   0.38200000000000    0.26060000000000    0.75140000000000    P    (2a)
  -0.38200000000000    0.76060000000000   -0.75140000000000    P    (2a)
   0.30820000000000   -0.00240000000000    0.63510000000000    W    (2a)
  -0.30820000000000    0.49760000000000   -0.63510000000000    W    (2a)
   0.70240000000000   -0.00710000000000    0.35850000000000    W    (2a)
  -0.70240000000000    0.49290000000000   -0.35850000000000    W    (2a)
   0.19130000000000    0.25010000000000    0.15330000000000    W    (2a)
  -0.19130000000000    0.75010000000000   -0.15330000000000    W    (2a)
   0.80740000000000    0.25570000000000    0.84620000000000    W    (2a)
  -0.80740000000000    0.75570000000000   -0.84620000000000    W    (2a)
```

$Li_2SO_4 \cdot H_2O$ ($H4_8$): A2B2C5D_mP20_4_2a_2a_5a_a - CIF

```
# CIF file
data_findsym-output
_audit_creation_method FINDSYM

_chemical_name_mineral 'H2Li2O5S'
_chemical_formula_sum 'H2 Li2 O5 S'

loop_
_publ_author_name
 'J.-O. Lundgren'
 '{\AA}. Kvick'
 'M. Karppinen'
 'R. Liminga'
 'S. C. Abrahams'
_journal_name_full_name
;
 Journal of Chemical Physics
;
_journal_volume 80
_journal_year 1984
_journal_page_first 423
_journal_page_last 430
_publ_Section_title
;
 Neutron diffraction structural study of pyroelectric Li$_{2}$SO$_{4}$$\
     ↪ cdot$H$_{2}$O at 293, 80, and 20 K
;

_aflow_title 'Li$_{2}$SO$_{4}$$\cdot$H$_{2}$O ($H4_{8}$) Structure'
_aflow_proto 'A2B2C5D_mP20_4_2a_2a_5a_a'
_aflow_params 'a,b/a,c/a,\beta,x_{1},y_{1},z_{1},x_{2},y_{2},z_{2},x_{3
     ↪ },y_{3},z_{3},x_{4},y_{4},z_{4},x_{5},y_{5},z_{5},x_{6},y_{6},
     ↪ z_{6},x_{7},y_{7},z_{7},x_{8},y_{8},z_{8},x_{9},y_{9},z_{9},x_{
     ↪ 10},y_{10},z_{10}'
_aflow_params_values '5.449,0.886768214351,1.49330152322,107.19,-0.02933
     ↪ ,0.38995,0.31142,0.00673,0.63149,0.44573,0.3031,0.49638,-
     ↪ 0.00741,0.56046,0.48968,0.39524,0.02087,0.07409,0.17122,0.43724
     ↪ ,0.11111,0.37857,0.39807,0.12076,0.07634,0.32462,0.69627,
     ↪ 0.20975,-0.08635,0.46436,0.40379,0.29229,0.0,0.20805'
_aflow_Strukturbericht '$H4_{8}$'
_aflow_Pearson 'mP20'

_symmetry_space_group_name_H-M "P 1 21 1"
_symmetry_Int_Tables_number 4

_cell_length_a    5.44900
_cell_length_b    4.83200
_cell_length_c    8.13700
_cell_angle_alpha 90.00000
_cell_angle_beta  107.19000
_cell_angle_gamma 90.00000

loop_
_space_group_symop_id
_space_group_symop_operation_xyz
1 x,y,z
2 -x,y+1/2,-z

loop_
_atom_site_label
_atom_site_type_symbol
_atom_site_symmetry_multiplicity
_atom_site_Wyckoff_label
_atom_site_fract_x
_atom_site_fract_y
_atom_site_fract_z
_atom_site_occupancy
H1 H 2 a -0.02933 0.38995 0.31142 1.00000
H2 H 2 a 0.00673 0.63149 0.44573 1.00000
Li1 Li 2 a 0.30310 0.49638 -0.00741 1.00000
Li2 Li 2 a 0.56046 0.48968 0.39524 1.00000
O1 O 2 a 0.02087 0.07409 0.17122 1.00000
O2 O 2 a 0.43724 0.11111 0.37857 1.00000
O3 O 2 a 0.39807 0.12076 0.07634 1.00000
O4 O 2 a 0.32462 0.69627 0.20975 1.00000
O5 O 2 a -0.08635 0.46436 0.40379 1.00000
S1 S 2 a 0.29229 0.00000 0.20805 1.00000
```

$Li_2SO_4 \cdot H_2O$ ($H4_8$): A2B2C5D_mP20_4_2a_2a_5a_a - POSCAR

```
A2B2C5D_mP20_4_2a_2a_5a_a & a,b/a,c/a,beta,x1,y1,z1,x2,y2,z2,x3,y3,z3,x4
     ↪ ,y4,z4,x5,y5,z5,x6,y6,z6,x7,y7,z7,x8,y8,z8,x9,y9,z9,x10,y10,z10
     ↪  --params=5.449,0.886768214351,1.49330152322,107.19,-0.02933,
     ↪ 0.38995,0.31142,0.00673,0.63149,0.44573,0.3031,0.49638,-0.00741
     ↪ ,0.56046,0.48968,0.39524,0.02087,0.07409,0.17122,0.43724,
     ↪ 0.11111,0.37857,0.39807,0.12076,0.07634,0.32462,0.69627,0.20975
     ↪ ,-0.08635,0.46436,0.40379,0.29229,0.0,0.20805 & P2_{1} C_{2}^{2
     ↪ } #4 (a^10) & mP20 & $H4_{8}$ & H2Li2O5S & H2Li2O5S & J.-O.
     ↪ Lundgren et al., J. Chem. Phys. 80, 423-430 (1984)
   1.00000000000000
   5.44900000000000   0.00000000000000   0.00000000000000
   0.00000000000000   4.83200000000000   0.00000000000000
  -2.40481970468058   0.00000000000000   7.77351987120249
     H     Li      O      S
     4      4     10      2
Direct
  -0.02933000000000   0.38995000000000   0.31142000000000   H    (2a)
   0.02933000000000   0.88995000000000  -0.31142000000000   H    (2a)
   0.00673000000000   0.63149000000000   0.44573000000000   H    (2a)
  -0.00673000000000   1.13149000000000  -0.44573000000000   H    (2a)
   0.30310000000000   0.49638000000000  -0.00741000000000   Li   (2a)
  -0.30310000000000   0.99638000000000   0.00741000000000   Li   (2a)
   0.56046000000000   0.48968000000000   0.39524000000000   Li   (2a)
  -0.56046000000000   0.98968000000000  -0.39524000000000   Li   (2a)
   0.02087000000000   0.07409000000000   0.17122000000000   O    (2a)
  -0.02087000000000   0.57409000000000  -0.17122000000000   O    (2a)
   0.43724000000000   0.11111000000000   0.37857000000000   O    (2a)
  -0.43724000000000   0.61111000000000  -0.37857000000000   O    (2a)
   0.39807000000000   0.12076000000000   0.07634000000000   O    (2a)
  -0.39807000000000   0.62076000000000  -0.07634000000000   O    (2a)
   0.32462000000000   0.69627000000000   0.20975000000000   O    (2a)
  -0.32462000000000   1.19627000000000  -0.20975000000000   O    (2a)
  -0.08635000000000   0.46436000000000   0.40379000000000   O    (2a)
   0.08635000000000   0.96436000000000  -0.40379000000000   O    (2a)
   0.29229000000000   0.00000000000000   0.20805000000000   S    (2a)
  -0.29229000000000   0.50000000000000  -0.20805000000000   S    (2a)
```

$Ca_3UO_6$: A3B6C_mP20_4_3a_6a_a - CIF

```
# CIF file
data_findsym-output
_audit_creation_method FINDSYM

_chemical_name_mineral 'Ca3O6U'
_chemical_formula_sum 'Ca3 O6 U'

loop_
_publ_author_name
 'B. O. Loopstra'
 'H. M. Rietveld'
_journal_name_full_name
;
 Acta Crystallographica Section B: Structural Science
;
_journal_volume 25
_journal_year 1969
_journal_page_first 787
_journal_page_last 791
_publ_Section_title
;
 The structure of some alkaline-earth metal uranates
;

_aflow_title 'Ca$_{3}$UO$_{6}$ Structure'
_aflow_proto 'A3B6C_mP20_4_3a_6a_a'
```

```
_aflow_params 'a,b/a,c/a,\beta,x_{1},y_{1},z_{1},x_{2},y_{2},z_{2},x_{3
    ↪ },y_{3},z_{3},x_{4},y_{4},z_{4},x_{5},y_{5},z_{5},x_{6},y_{6},
    ↪ z_{6},x_{7},y_{7},z_{7},x_{8},y_{8},z_{8},x_{9},y_{9},z_{9},x_{
    ↪ 10},y_{10},z_{10}'
_aflow_params_values '5.7275,1.03996508075,1.44883457006,90.568,0.2621,-
    ↪ 0.0025,0.7488,0.2723,0.4223,0.0123,0.7532,0.0288,0.4761,0.6319,
    ↪ 0.41,0.5173,0.5605,0.1618,0.1938,0.0747,0.2744,0.3323,0.1165,
    ↪ 0.0484,0.0152,0.0327,0.3053,0.6942,0.574,0.1963,0.8252,0.2459,
    ↪ 0.0,0.2519'
_aflow_Strukturbericht 'None'
_aflow_Pearson 'mP20'

_symmetry_space_group_name_H-M "P 1 21 1"
_symmetry_Int_Tables_number 4

_cell_length_a    5.72750
_cell_length_b    5.95640
_cell_length_c    8.29820
_cell_angle_alpha 90.00000
_cell_angle_beta  90.56800
_cell_angle_gamma 90.00000

loop_
_space_group_symop_id
_space_group_symop_operation_xyz
1 x,y,z
2 -x,y+1/2,-z

loop_
_atom_site_label
_atom_site_type_symbol
_atom_site_symmetry_multiplicity
_atom_site_Wyckoff_label
_atom_site_fract_x
_atom_site_fract_y
_atom_site_fract_z
_atom_site_occupancy
Ca1 Ca 2 a 0.26210 -0.00250 0.74880 1.00000
Ca2 Ca 2 a 0.27230 0.42230 0.01230 1.00000
Ca3 Ca 2 a 0.75320 0.02880 0.47610 1.00000
O1 O 2 a 0.63190 0.41000 0.51730 1.00000
O2 O 2 a 0.56050 0.16180 0.19380 1.00000
O3 O 2 a 0.07470 0.27440 0.33230 1.00000
O4 O 2 a 0.11650 0.04840 0.01520 1.00000
O5 O 2 a 0.03270 0.30530 0.69420 1.00000
O6 O 2 a 0.57400 0.19630 0.82520 1.00000
U1 U 2 a 0.24590 0.00000 0.25190 1.00000
```

$Ca_3UO_6$: A3B6C_mP20_4_3a_6a_a - POSCAR

```
A3B6C_mP20_4_3a_6a_a & a,b/a,c/a,beta,x1,y1,z1,x2,y2,z2,x3,y3,z3,x4,y4,
    ↪ z4,x5,y5,z5,x6,y6,z6,x7,y7,z7,x8,y8,z8,x9,y9,z9,x10,y10,z10 --
    ↪ params=5.7275,1.03996508075,1.44883457006,90.568,0.2621,-0.0025
    ↪ ,0.7488,0.2723,0.4223,0.0123,0.7532,0.0288,0.4761,0.6319,0.41,
    ↪ 0.5173,0.5605,0.1618,0.1938,0.0747,0.2744,0.3323,0.1165,0.0484,
    ↪ 0.0152,0.0327,0.3053,0.6942,0.574,0.1963,0.8252,0.2459,0.0,
    ↪ 0.2519 & P2_{1} C_{2}^{2} #4 (a^10) & mP20 & None & Ca3O6U &
    ↪ Ca3O6U & B. O. Loopstra and H. M. Rietveld, Acta Crystallogr.
    ↪ Sect. B Struct. Sci. 25, 787-791 (1969)
   1.0000000000000
   5.7275000000000    0.0000000000000    0.0000000000000
   0.0000000000000    5.9564000000000    0.0000000000000
  -0.0822626105753 7  0.0000000000000    8.2977922426933 1
    Ca      O      U
     6     12      2
Direct
   0.2621000000000   -0.0025000000000    0.7488000000000   Ca   (2a)
  -0.2621000000000    0.4975000000000   -0.7488000000000   Ca   (2a)
   0.2723000000000    0.4223000000000    0.0123000000000   Ca   (2a)
  -0.2723000000000    0.9223000000000   -0.0123000000000   Ca   (2a)
   0.7532000000000    0.0288000000000    0.4761000000000   Ca   (2a)
  -0.7532000000000    0.5288000000000   -0.4761000000000   Ca   (2a)
   0.6319000000000    0.4100000000000    0.5173000000000   O    (2a)
  -0.6319000000000    0.9100000000000   -0.5173000000000   O    (2a)
   0.5605000000000    0.1618000000000    0.1938000000000   O    (2a)
  -0.5605000000000    0.6618000000000   -0.1938000000000   O    (2a)
   0.0747000000000    0.2744000000000    0.3323000000000   O    (2a)
  -0.0747000000000    0.7744000000000   -0.3323000000000   O    (2a)
   0.1165000000000    0.0484000000000    0.0152000000000   O    (2a)
  -0.1165000000000    0.5484000000000   -0.0152000000000   O    (2a)
   0.0327000000000    0.3053000000000    0.6942000000000   O    (2a)
  -0.0327000000000    0.8053000000000   -0.6942000000000   O    (2a)
   0.5740000000000    0.1963000000000    0.8252000000000   O    (2a)
  -0.5740000000000    0.6963000000000   -0.8252000000000   O    (2a)
   0.2459000000000    0.0000000000000    0.2519000000000   U    (2a)
  -0.2459000000000    0.5000000000000   -0.2519000000000   U    (2a)
```

Bassanite [$CaSO_4(H_2O)_{0.5}$, $H4_7$]: A2B2C9D2_mC90_5_ab2c_3c_b13c_3c - CIF

```
# CIF file
data_findsym-output
_audit_creation_method FINDSYM

_chemical_name_mineral 'Bassanite'
_chemical_formula_sum 'Ca2 H2 O9 S2'

loop_
_publ_author_name
 'W. Abriel'
 'R. Nesper'
_journal_name_full_name
;
 Zeitschrift f{\"u}r Kristallographie - Crystalline Materials
;
_journal_volume 205
_journal_year 1993
_journal_page_first 99
_journal_page_last 113
_publ_Section_title
;
 Bestimmung der Kristallstruktur von CaSO$_{4}$(H$_{2}$O)$_{0.5}$ mit R
    ↪ \"{o}ntgenbeugungsmethoden und mit Potentialprofil-Rechnungen
;

# Found in The monoclinic $I2$ structure of bassanite, calcium sulphate
    ↪ hemihydrate (CaSO$_{4}$$\cdot$0.5H$_{2}$O), 2001

_aflow_title 'Bassanite [CaSO$_{4}$(H$_{2}$O)$_{0.5}$, $H4_{7}$]
    ↪ Structure'
_aflow_proto 'A2B2C9D2_mC90_5_ab2c_3c_b13c_3c'
_aflow_params 'a,b/a,c/a,\beta,y_{1},y_{2},y_{3},x_{4},y_{4},z_{4},x_{5
    ↪ },y_{5},z_{5},x_{6},y_{6},z_{6},x_{7},y_{7},z_{7},x_{8},y_{8},
    ↪ z_{8},x_{9},y_{9},z_{9},x_{10},y_{10},z_{10},x_{11},y_{11},z_{
    ↪ 11},x_{12},y_{12},z_{12},x_{13},y_{13},z_{13},x_{14},y_{14},z_{
    ↪ 14},x_{15},y_{15},z_{15},x_{16},y_{16},z_{16},x_{17},y_{17},z_{
    ↪ 17},x_{18},y_{18},z_{18},x_{19},y_{19},z_{19},x_{20},y_{20},z_{
    ↪ 20},x_{21},y_{21},z_{21},x_{22},y_{22},z_{22},x_{23},y_{23},z_{
    ↪ 23},x_{24},y_{24},z_{24}'
_aflow_params_values '17.45812,0.39701869388,0.688934432803,133.36555,
    ↪ 0.4553,-0.0447,0.5985,0.1667,0.2724,0.8943,0.3333,0.2724,0.6057
    ↪ ,0.5,0.171,0.572,0.333,0.307,0.783,0.833,0.023,0.712,0.5173,
    ↪ 0.226,0.8882,0.0173,0.226,0.3882,0.8507,0.4435,0.5522,0.3507,
    ↪ 0.4435,0.0522,0.184,0.3306,0.1115,0.684,0.3306,0.6115,0.6556,
    ↪ 0.1185,-0.0982,0.1556,0.1185,0.4018,-0.0111,0.3102,0.8065,
    ↪ 0.4889,0.3102,0.3065,0.3222,0.0714,0.7583,0.8222,0.0714,0.2583,
    ↪ 0.3333,0.4508,0.784,0.0833,0.2752,0.3585,0.25,0.4496,0.25,
    ↪ 0.4167,0.2752,0.1415'
_aflow_Strukturbericht '$H4_{7}$'
_aflow_Pearson 'mC90'

_symmetry_space_group_name_H-M "C 1 2 1"
_symmetry_Int_Tables_number 5

_cell_length_a    17.45812
_cell_length_b    6.93120
_cell_length_c    12.02750
_cell_angle_alpha 90.00000
_cell_angle_beta  133.36555
_cell_angle_gamma 90.00000

loop_
_space_group_symop_id
_space_group_symop_operation_xyz
1 x,y,z
2 -x,y,-z
3 x+1/2,y+1/2,z
4 -x+1/2,y+1/2,-z

loop_
_atom_site_label
_atom_site_type_symbol
_atom_site_symmetry_multiplicity
_atom_site_Wyckoff_label
_atom_site_fract_x
_atom_site_fract_y
_atom_site_fract_z
_atom_site_occupancy
Ca1 Ca 2 a 0.00000 0.45530 0.00000 1.00000
Ca2 Ca 2 b 0.00000 -0.04470 0.50000 1.00000
O1 O 2 b 0.00000 0.59850 0.50000 1.00000
Ca3 Ca 4 c 0.16670 0.27240 0.89430 1.00000
Ca4 Ca 4 c 0.33330 0.27240 0.60570 1.00000
H1 H 4 c 0.50000 0.17100 0.57200 1.00000
H2 H 4 c 0.33300 0.30700 0.78300 1.00000
H3 H 4 c 0.83300 0.02300 0.71200 1.00000
O2 O 4 c 0.51730 0.22600 0.88820 1.00000
O3 O 4 c 0.01730 0.22600 0.38820 1.00000
O4 O 4 c 0.85070 0.44350 0.55220 1.00000
O5 O 4 c 0.35070 0.44350 0.05220 1.00000
O6 O 4 c 0.18400 0.33060 0.11150 1.00000
O7 O 4 c 0.68400 0.33060 0.61150 1.00000
O8 O 4 c 0.65560 0.11850 -0.09820 1.00000
O9 O 4 c 0.15560 0.11850 0.40180 1.00000
O10 O 4 c -0.01110 0.31020 0.80650 1.00000
O11 O 4 c 0.48890 0.31020 0.30650 1.00000
O12 O 4 c 0.32220 0.07140 0.75830 1.00000
O13 O 4 c 0.82220 0.07140 0.25830 1.00000
O14 O 4 c 0.33330 0.45080 0.78400 1.00000
S1 S 4 c 0.08330 0.27520 0.35850 1.00000
S2 S 4 c 0.25000 0.44960 0.25000 1.00000
S3 S 4 c 0.41670 0.27520 0.14150 1.00000
```

Bassanite [$CaSO_4(H_2O)_{0.5}$, $H4_7$]: A2B2C9D2_mC90_5_ab2c_3c_b13c_3c - POSCAR

```
A2B2C9D2_mC90_5_ab2c_3c_b13c_3c & a,b/a,c/a,beta,y1,y2,y3,x4,y4,z4,x5,y5
    ↪ ,z5,x6,y6,z6,x7,y7,z7,x8,y8,z8,x9,y9,z9,x10,y10,z10,x11,y11,z11
    ↪ ,x12,y12,z12,x13,y13,z13,x14,y14,z14,x15,y15,z15,x16,y16,z16,
    ↪ x17,y17,z17,x18,y18,z18,x19,y19,z19,x20,y20,z20,x21,y21,z21,x22
    ↪ ,y22,z22,x23,y23,z23,x24,y24,z24 --params=17.45812,
    ↪ 0.39701869388,0.688934432803,133.36555,0.4553,-0.0447,0.5985,
    ↪ 0.1667,0.2724,0.8943,0.3333,0.2724,0.6057,0.5,0.171,0.572,0.333
    ↪ ,0.307,0.783,0.833,0.023,0.712,0.5173,0.226,0.8882,0.0173,0.226
    ↪ ,0.3882,0.8507,0.4435,0.5522,0.3507,0.4435,0.0522,0.184,0.3306,
    ↪ 0.1115,0.684,0.3306,0.6115,0.6556,0.1185,-0.0982,0.1556,0.1185,
    ↪ 0.4018,-0.0111,0.3102,0.8065,0.4889,0.3102,0.3065,0.3222,0.0714
    ↪ ,0.7583,0.8222,0.0714,0.2583,0.3333,0.4508,0.784,0.0833,0.2752,
    ↪ 0.3585,0.25,0.4496,0.25,0.4167,0.2752,0.1415 & C2 C_{2}^{3} #5
    ↪ (ab^2c^21) & mC90 & $H4_{7}$ & Ca2H2O9S2 & Bassanite & W.
    ↪ Abriel and R. Nesper, Zeitschrift f{"u}r Kristallographie -
    ↪ Crystalline Materials 205, 99-113 (1993)
```

```
   1.00000000000000
   8.72906000000000   -3.46560000000000    0.00000000000000
   8.72906000000000    3.46560000000000    0.00000000000000
  -8.25868915382723    0.00000000000000    8.74384410373701
    Ca     H     O     S
     6     6    27     6
Direct
  -0.45530000000000    0.45530000000000    0.00000000000000   Ca   (2a)
   0.04470000000000   -0.04470000000000    0.50000000000000   Ca   (2b)
  -0.10570000000000    0.43910000000000    0.89430000000000   Ca   (4c)
  -0.43910000000000    0.10570000000000   -0.89430000000000   Ca   (4c)
   0.06090000000000    0.60570000000000    0.60570000000000   Ca   (4c)
  -0.60570000000000   -0.06090000000000   -0.60570000000000   Ca   (4c)
   0.32900000000000    0.67100000000000    0.57200000000000    H   (4c)
  -0.67100000000000   -0.32900000000000   -0.57200000000000    H   (4c)
   0.02600000000000    0.64000000000000    0.78300000000000    H   (4c)
  -0.64000000000000   -0.02600000000000   -0.78300000000000    H   (4c)
   0.81000000000000    0.85600000000000    0.71200000000000    H   (4c)
  -0.85600000000000   -0.81000000000000   -0.71200000000000    H   (4c)
  -0.59850000000000    0.59850000000000    0.50000000000000    O   (2b)
   0.29130000000000    0.74330000000000    0.88820000000000    O   (4c)
  -0.74330000000000   -0.29130000000000   -0.88820000000000    O   (4c)
  -0.20870000000000    0.24330000000000    0.38820000000000    O   (4c)
  -0.24330000000000    0.20870000000000   -0.38820000000000    O   (4c)
   0.40720000000000    1.29420000000000    0.55220000000000    O   (4c)
  -1.29420000000000   -0.40720000000000   -0.55220000000000    O   (4c)
  -0.09280000000000    0.79420000000000    0.05220000000000    O   (4c)
  -0.79420000000000    0.09280000000000   -0.05220000000000    O   (4c)
  -0.14660000000000    0.51460000000000    0.11150000000000    O   (4c)
  -0.51460000000000    0.14660000000000   -0.11150000000000    O   (4c)
   0.35340000000000    1.01460000000000    0.61150000000000    O   (4c)
  -1.01460000000000   -0.35340000000000   -0.61150000000000    O   (4c)
   0.53710000000000    0.77410000000000   -0.09820000000000    O   (4c)
  -0.77410000000000   -0.53710000000000    0.09820000000000    O   (4c)
   0.03710000000000    0.27410000000000    0.40180000000000    O   (4c)
  -0.27410000000000   -0.03710000000000   -0.40180000000000    O   (4c)
  -0.32130000000000    0.29910000000000    0.80650000000000    O   (4c)
  -0.29910000000000    0.32130000000000   -0.80650000000000    O   (4c)
   0.17870000000000    0.79910000000000    0.30650000000000    O   (4c)
  -0.79910000000000   -0.17870000000000   -0.30650000000000    O   (4c)
   0.25080000000000    0.39360000000000    0.75830000000000    O   (4c)
  -0.39360000000000   -0.25080000000000   -0.75830000000000    O   (4c)
   0.75080000000000    0.89360000000000    0.25830000000000    O   (4c)
  -0.89360000000000   -0.75080000000000   -0.25830000000000    O   (4c)
  -0.11750000000000    0.78410000000000    0.78400000000000    O   (4c)
  -0.78410000000000    0.11750000000000   -0.78400000000000    O   (4c)
  -0.19190000000000    0.35850000000000    0.35850000000000    S   (4c)
  -0.35850000000000    0.19190000000000   -0.35850000000000    S   (4c)
  -0.19960000000000    0.69960000000000    0.25000000000000    S   (4c)
  -0.69960000000000    0.19960000000000   -0.25000000000000    S   (4c)
   0.14150000000000    0.69190000000000    0.14150000000000    S   (4c)
  -0.69190000000000   -0.14150000000000   -0.14150000000000    S   (4c)
```

$NbAs_2$: A2B_mC12_5_2c_c - CIF

```
# CIF file
data_findsym-output
_audit_creation_method FINDSYM

_chemical_name_mineral 'As2Nb'
_chemical_formula_sum 'As2 Nb'

loop_
_publ_author_name
 'S. Furuseth'
 'A. Kjekshus'
_journal_name_full_name
;
 Acta Cristallographica
;
_journal_volume 18
_journal_year 1965
_journal_page_first 320
_journal_page_last 324
_publ_Section_title
;
 The Crystal Structures of NbAs$_{2}$ and NbSb$_{2}$
;

_aflow_title 'NbAs$_{2}$ Structure'
_aflow_proto 'A2B_mC12_5_2c_c'
_aflow_params 'a,b/a,c/a,\beta,x_{1},y_{1},z_{1},x_{2},y_{2},z_{2},x_{3
    ↪ },y_{3},z_{3}'
_aflow_params_values '9.357,0.36147269424,0.8327455381,119.76667,0.0948,
    ↪ 0.488,0.3928,0.1399,0.067,0.0257,0.3444,0.5,0.3044'
_aflow_Strukturbericht 'None'
_aflow_Pearson 'mC12'

_symmetry_space_group_name_H-M "C 1 2 1"
_symmetry_Int_Tables_number 5

_cell_length_a     9.35700
_cell_length_b     3.38230
_cell_length_c     7.79200
_cell_angle_alpha  90.00000
_cell_angle_beta   119.76667
_cell_angle_gamma  90.00000

loop_
_space_group_symop_id
_space_group_symop_operation_xyz
1 x,y,z
2 -x,y,-z
3 x+1/2,y+1/2,z
4 -x+1/2,y+1/2,-z

loop_
_atom_site_label
_atom_site_type_symbol
_atom_site_symmetry_multiplicity
_atom_site_Wyckoff_label
_atom_site_fract_x
_atom_site_fract_y
_atom_site_fract_z
_atom_site_occupancy
As1 As 4 c 0.09480 0.48800 0.39280 1.00000
As2 As 4 c 0.13990 0.06700 0.02570 1.00000
Nb1 Nb 4 c 0.34440 0.50000 0.30440 1.00000
```

$NbAs_2$: A2B_mC12_5_2c_c - POSCAR

```
A2B_mC12_5_2c_c & a,b/a,c/a,beta,x1,y1,z1,x2,y2,z2,x3,y3,z3 --params=
    ↪ 9.357,0.36147269424,0.8327455381,119.76667,0.0948,0.488,0.3928,
    ↪ 0.1399,0.067,0.0257,0.3444,0.5,0.3044 & C2 C_{2}^{3} #5 (c^3) &
    ↪  mC12 & None & As2Nb & As2Nb & S. Furuseth and A. Kjekshus,
    ↪ Acta Cryst. 18, 320-324 (1965)
   1.00000000000000
   4.67850000000000   -1.69115000000000    0.00000000000000
   4.67850000000000    1.69115000000000    0.00000000000000
  -3.86848708676481    0.00000000000000    6.76387992645744
    As    Nb
     4     2
Direct
  -0.39320000000000    0.58280000000000    0.39280000000000   As   (4c)
  -0.58280000000000    0.39320000000000   -0.39280000000000   As   (4c)
   0.07290000000000    0.20690000000000    0.02570000000000   As   (4c)
  -0.20690000000000   -0.07290000000000   -0.02570000000000   As   (4c)
  -0.15560000000000    0.84440000000000    0.30440000000000   Nb   (4c)
  -0.84440000000000    0.15560000000000   -0.30440000000000   Nb   (4c)
```

$D0_{15}$ ($AlCl_3$) (*obsolete*): AB3_mC16_5_c_3c - CIF

```
# CIF file
data_findsym-output
_audit_creation_method FINDSYM

_chemical_name_mineral 'AlCl3'
_chemical_formula_sum 'Al Cl3'

loop_
_publ_author_name
 'J. A. A. Ketelaar'
_journal_name_full_name
;
 Zeitschrift f{\"u}r Kristallographie - Crystalline Materials
;
_journal_volume 90
_journal_year 1935
_journal_page_first 237
_journal_page_last 255
_publ_Section_title
;
 Die Kristallstruktur der Aluminiumhalogenide II
;

# Found in Strukturbericht Band III 1933-1935, 1937

_aflow_title '$D0_{15}$ (AlCl$_{3}$) ({\em{obsolete}}) Structure'
_aflow_proto 'AB3_mC16_5_c_3c'
_aflow_params 'a,b/a,c/a,\beta,x_{1},y_{1},z_{1},x_{2},y_{2},z_{2},x_{3
    ↪ },y_{3},z_{3},x_{4},y_{4},z_{4}'
_aflow_params_values '5.91,1.73205076142,1.04286294416,108.64073,0.54333
    ↪ ,0.27833,0.045,0.24,0.27778,0.22,0.74,0.44444,0.22,0.74,0.11111
    ↪ ,0.22'
_aflow_Strukturbericht '$D0_{15}$'
_aflow_Pearson 'mC16'

_symmetry_space_group_name_H-M "C 1 2 1"
_symmetry_Int_Tables_number 5

_cell_length_a     5.91000
_cell_length_b     10.23642
_cell_length_c     6.16332
_cell_angle_alpha  90.00000
_cell_angle_beta   108.64073
_cell_angle_gamma  90.00000

loop_
_space_group_symop_id
_space_group_symop_operation_xyz
1 x,y,z
2 -x,y,-z
3 x+1/2,y+1/2,z
4 -x+1/2,y+1/2,-z

loop_
_atom_site_label
_atom_site_type_symbol
_atom_site_symmetry_multiplicity
_atom_site_Wyckoff_label
_atom_site_fract_x
_atom_site_fract_y
_atom_site_fract_z
_atom_site_occupancy
Al1 Al 4 c 0.54333 0.27833 0.04500 1.00000
Cl1 Cl 4 c 0.24000 0.27778 0.22000 1.00000
Cl2 Cl 4 c 0.74000 0.44444 0.22000 1.00000
Cl3 Cl 4 c 0.74000 0.11111 0.22000 1.00000
```

$D0_{15}$ ($AlCl_3$) (*obsolete*): AB3_mC16_5_c_3c - POSCAR

```
AB3_mC16_5_c_3c & a,b/a,c/a,beta,x1,y1,z1,x2,y2,z2,x3,y3,z3,x4,y4,z4 --
    ↪ params=5.91,1.73205076142,1.04286294416,108.64073,0.54333,
    ↪ 0.27833,0.045,0.24,0.27778,0.22,0.74,0.44444,0.22,0.74,0.11111,
    ↪ 0.22 & C2 C_{2}^{3} #5 (c^4) & mC16 & $D0_{15}$ & AlCl3 & AlCl3
    ↪  & J. A. A. Ketelaar, Zeitschrift f{"u}r Kristallographie -
    ↪ Crystalline Materials 90, 237-255 (1935)
   1.00000000000000
   2.95500000000000   -5.11821000000000    0.00000000000000
   2.95500000000000    5.11821000000000    0.00000000000000
  -1.97000028435120    0.00000000000000    5.84000105325814
    Al      Cl
     2       6
Direct
   0.26500000000000    0.82166000000000    0.04500000000000   Al   (4c)
  -0.82166000000000   -0.26500000000000   -0.04500000000000   Al   (4c)
  -0.03778000000000    0.51778000000000    0.22000000000000   Cl   (4c)
  -0.51778000000000    0.03778000000000   -0.22000000000000   Cl   (4c)
   0.29556000000000    1.18444000000000    0.22000000000000   Cl   (4c)
  -1.18444000000000   -0.29556000000000   -0.22000000000000   Cl   (4c)
   0.62889000000000    0.85111000000000    0.22000000000000   Cl   (4c)
  -0.85111000000000   -0.62889000000000   -0.22000000000000   Cl   (4c)
```

$Rb_2CaCu_6(PO_4)_4O_2$: AB6C18D4E2_mC62_5_a_2b2c_9c_2c_c - CIF

```
# CIF file
data_findsym-output
_audit_creation_method FINDSYM

_chemical_name_mineral 'CaCu6O18P4Rb2'
_chemical_formula_sum 'Ca Cu6 O18 P4 Rb2'

loop_
_publ_author_name
 'S. M. Aksenov'
 'E. Y. Borovikova'
 'V. S. Mironov'
 'N. A. Yamnova'
 'A. S. Volkov'
 'D. A. Ksenofontov'
 'O. A. Gurbanova'
 'O. V. Dimitrova'
 'D. V. Deyneko'
 'E. A. Zvereva'
 'O. V. Maximova'
 'S. V. Krivovichev'
 'P. C. Burns'
 'A. N. Vasiliev'
_journal_name_full_name
;
 Acta Crystallographica Section B: Structural Science
;
_journal_volume 75
_journal_year 2019
_journal_page_first 903
_journal_page_last 913
_publ_Section_title
;
 Rb$_{2}$CaCu$_{6}$(PO$_{4}$)$_{4}$O$_{2}$, a novel oxophosphate with a
    ↪ shchurovskyite-type topology: synthesis, structure, magnetic
    ↪ properties and crystal chemistry of rubidium copper phosphates
;

_aflow_title 'Rb$_{2}$CaCu$_{6}$(PO$_{4}$)$_{4}$O$_{2}$ Structure'
_aflow_proto 'AB6C18D4E2_mC62_5_a_2b2c_9c_2c_c'
_aflow_params 'a,b/a,c/a,\beta,y_{1},y_{2},y_{3},x_{4},y_{4},z_{4},x_{5
    ↪ },y_{5},z_{5},x_{6},y_{6},z_{6},x_{7},y_{7},z_{7},x_{8},y_{8},
    ↪ z_{8},x_{9},y_{9},z_{9},x_{10},y_{10},z_{10},x_{11},y_{11},z_{
    ↪ 11},x_{12},y_{12},z_{12},x_{13},y_{13},z_{13},x_{14},y_{14},z_{
    ↪ 14},x_{15},y_{15},z_{15},x_{16},y_{16},z_{16},x_{17},y_{17},z_{
    ↪ 17}'
_aflow_params_values '16.8913,0.33393522109,0.494876060457,93.919,
    ↪ 0.67531,-0.07889,0.413,0.32143,-0.25821,0.47267,0.08396,0.18216
    ↪ ,0.22561,0.33645,-0.2696,0.2451,0.5762,0.1756,0.4677,0.07301,
    ↪ 0.1655,0.4508,0.47637,-0.1144,0.2054,0.4428,-0.5435,0.1571,
    ↪ 0.3822,-0.2162,-0.0248,0.71142,0.3504,0.4381,0.67462,-0.0515,
    ↪ 0.3216,0.61381,0.3236,0.1953,0.64484,0.2009,0.35178,0.4101,-
    ↪ 0.29307,0.14827,0.29009,0.20623,0.15752'
_aflow_Strukturbericht 'None'
_aflow_Pearson 'mC62'

_symmetry_space_group_name_H-M "C 1 2 1"
_symmetry_Int_Tables_number 5

_cell_length_a      16.89130
_cell_length_b      5.64060
_cell_length_c      8.35910
_cell_angle_alpha 90.00000
_cell_angle_beta  93.91900
_cell_angle_gamma 90.00000

loop_
_space_group_symop_id
_space_group_symop_operation_xyz
1 x,y,z
2 -x,y,-z
3 x+1/2,y+1/2,z
4 -x+1/2,y+1/2,-z

loop_
_atom_site_label
_atom_site_type_symbol
_atom_site_symmetry_multiplicity
_atom_site_Wyckoff_label
_atom_site_fract_x
_atom_site_fract_y
_atom_site_fract_z
_atom_site_occupancy
Ca1 Ca 2 a 0.00000 0.67531 0.00000 1.00000
Cu1 Cu 2 b 0.00000 -0.07889 0.50000 1.00000
Cu2 Cu 2 b 0.00000 0.41300 0.50000 1.00000
Cu3 Cu 4 c 0.32143 -0.25821 0.47267 1.00000
Cu4 Cu 4 c 0.08396 0.18216 0.22561 1.00000
O1 O 4 c 0.33645 -0.26960 0.24510 1.00000
O2 O 4 c 0.57620 0.17560 0.46770 1.00000
O3 O 4 c 0.07301 0.16550 0.45080 1.00000
O4 O 4 c 0.47637 -0.11440 0.20540 1.00000
O5 O 4 c 0.44280 -0.54350 0.15710 1.00000
O6 O 4 c 0.38220 -0.21620 -0.02480 1.00000
O7 O 4 c 0.71142 0.35040 0.43810 1.00000
O8 O 4 c 0.67462 -0.05150 0.32160 1.00000
O9 O 4 c 0.61381 0.32360 0.19530 1.00000
P1 P 4 c 0.64484 0.20090 0.35178 1.00000
P2 P 4 c 0.41010 -0.29307 0.14827 1.00000
Rb1 Rb 4 c 0.29009 0.20623 0.15752 1.00000
```

$Rb_2CaCu_6(PO_4)_4O_2$: AB6C18D4E2_mC62_5_a_2b2c_9c_2c_c - POSCAR

```
AB6C18D4E2_mC62_5_a_2b2c_9c_2c_c & a,b/a,c/a,beta,y1,y2,y3,x4,y4,z4,x5,
    ↪ y5,z5,x6,y6,z6,x7,y7,z7,x8,y8,z8,x9,y9,z9,x10,y10,z10,x11,y11,
    ↪ z11,x12,y12,z12,x13,y13,z13,x14,y14,z14,x15,y15,z15,x16,y16,z16
    ↪ ,x17,y17,z17 --params=16.8913,0.33393522109,0.494876060457,
    ↪ 93.919,0.67531,-0.07889,0.413,0.32143,-0.25821,0.47267,0.08396,
    ↪ 0.18216,0.22561,0.33645,-0.2696,0.2451,0.5762,0.1756,0.4677,
    ↪ 0.07301,0.1655,0.4508,0.47637,-0.1144,0.2054,0.4428,-0.5435,
    ↪ 0.1571,0.3822,-0.2162,-0.0248,0.71142,0.3504,0.4381,0.67462,-
    ↪ 0.0515,0.3216,0.61381,0.3236,0.1953,0.64484,0.2009,0.35178,
    ↪ 0.4101,-0.29307,0.14827,0.29009,0.20623,0.15752 & C2 C_{2}^{3}
    ↪ #5 (ab^2c^14) & mC62 & None & CaCu6O18P4Rb2 & CaCu6O18P4Rb2 &
    ↪ S. M. Aksenov et al., Acta Crystallogr. Sect. B Struct. Sci. 75
    ↪ , 903-913 (2019)
   1.00000000000000
   8.44565000000000   -2.82030000000000    0.00000000000000
   8.44565000000000    2.82030000000000    0.00000000000000
  -0.57131214828249    0.00000000000000    8.33955365947272
    Ca      Cu       O       P      Rb
     1       6      18       4       2
Direct
  -0.67531000000000    0.67531000000000    0.00000000000000   Ca   (2a)
   0.07889000000000   -0.07889000000000    0.50000000000000   Cu   (2b)
  -0.41300000000000    0.41300000000000    0.50000000000000   Cu   (2b)
   0.57964000000000    0.06322000000000    0.47267000000000   Cu   (4c)
  -0.06322000000000   -0.57964000000000   -0.47267000000000   Cu   (4c)
  -0.09820000000000    0.26612000000000    0.22561000000000   Cu   (4c)
  -0.26612000000000    0.09820000000000   -0.22561000000000   Cu   (4c)
   0.60605000000000    0.06685000000000    0.24510000000000   O    (4c)
  -0.06685000000000   -0.60605000000000   -0.24510000000000   O    (4c)
   0.40060000000000    0.75180000000000    0.46770000000000   O    (4c)
  -0.75180000000000   -0.40060000000000   -0.46770000000000   O    (4c)
  -0.09249000000000    0.23851000000000    0.45080000000000   O    (4c)
  -0.23851000000000    0.09249000000000   -0.45080000000000   O    (4c)
   0.59077000000000    0.36197000000000    0.20540000000000   O    (4c)
  -0.36197000000000   -0.59077000000000   -0.20540000000000   O    (4c)
   0.98630000000000   -0.10070000000000    0.15710000000000   O    (4c)
   0.10070000000000   -0.98630000000000   -0.15710000000000   O    (4c)
   0.59840000000000    0.16600000000000   -0.02480000000000   O    (4c)
  -0.16600000000000   -0.59840000000000    0.02480000000000   O    (4c)
   0.36102000000000    1.06182000000000    0.43810000000000   O    (4c)
  -1.06182000000000   -0.36102000000000   -0.43810000000000   O    (4c)
   0.72612000000000    0.62312000000000    0.32160000000000   O    (4c)
  -0.62312000000000   -0.72612000000000   -0.32160000000000   O    (4c)
   0.29021000000000    0.93741000000000    0.19530000000000   O    (4c)
  -0.93741000000000   -0.29021000000000   -0.19530000000000   O    (4c)
   0.44394000000000    0.84574000000000    0.35178000000000   P    (4c)
  -0.84574000000000   -0.44394000000000   -0.35178000000000   P    (4c)
   0.70317000000000    0.11703000000000    0.14827000000000   P    (4c)
  -0.11703000000000   -0.70317000000000   -0.14827000000000   P    (4c)
   0.08386000000000    0.49632000000000    0.15752000000000   Rb   (4c)
  -0.49632000000000   -0.08386000000000   -0.15752000000000   Rb   (4c)
```

*C2* $(Ba,Ca)CO_3$: ABC3_mC10_5_b_a_ac - CIF

```
# CIF file
data_findsym-output
_audit_creation_method FINDSYM

_chemical_name_mineral '(Ba,Ca)CO3'
_chemical_formula_sum 'Ba C O3'

loop_
_publ_author_name
 'D. Spahr'
 'L. Bayarjargal'
 'V. Vinograd'
 'R. Luchitskaia'
 'V. Milman'
 'B. Winkler'
_journal_name_full_name
;
 Acta Crystallographica Section B: Structural Science
;
_journal_volume 75
_journal_year 2019
_journal_page_first 291
_journal_page_last 300
_publ_Section_title
;
 A new BaCa(CO$_{3}$)$_{2}$ polymorph
;

_aflow_title '$C2$ (Ba,Ca)CO$_{3}$ Structure'
```

```
_aflow_proto 'ABC3_mC10_5_b_a_ac'
_aflow_params 'a,b/a,c/a,\beta,y_{1},y_{2},y_{3},x_{4},y_{4},z_{4}'
_aflow_params_values '6.7064,0.763241083144,0.627266491709,109.284,
    ↪ 0.6259,0.8722,0.103,0.8957,0.5037,0.7367'
_aflow_Strukturbericht 'None'
_aflow_Pearson 'mC10'

_symmetry_space_group_name_H-M "C 1 2 1"
_symmetry_Int_Tables_number 5

_cell_length_a     6.70640
_cell_length_b     5.11860
_cell_length_c     4.20670
_cell_angle_alpha  90.00000
_cell_angle_beta   109.28400
_cell_angle_gamma  90.00000

loop_
_space_group_symop_id
_space_group_symop_operation_xyz
1 x,y,z
2 -x,y,-z
3 x+1/2,y+1/2,z
4 -x+1/2,y+1/2,-z

loop_
_atom_site_label
_atom_site_type_symbol
_atom_site_symmetry_multiplicity
_atom_site_Wyckoff_label
_atom_site_fract_x
_atom_site_fract_y
_atom_site_fract_z
_atom_site_occupancy
C1 C 2 a 0.00000 0.62590 0.00000 1.00000
O1 O 2 a 0.00000 0.87220 0.00000 1.00000
Ba1 Ba 2 b 0.00000 0.10300 0.50000 1.00000
O2 O 4 c 0.89570 0.50370 0.73670 1.00000
```

*C*2 (Ba,Ca)$CO_3$: ABC3_mC10_5_b_a_ac - POSCAR

```
ABC3_mC10_5_b_a_ac & a,b/a,c/a,beta,y1,y2,y3,x4,y4,z4 --params=6.7064,
    ↪ 0.763241083144,0.627266491709,109.284,0.6259,0.8722,0.103,
    ↪ 0.8957,0.5037,0.7367 & C2 C_{2}^{3} #5 (a^2bc) & mC10 & None &
    ↪ (Ba,Ca)CO3 & (Ba,Ca)CO3 & D. Spahr et al., Acta Crystallogr.
    ↪ Sect. B Struct. Sci. 75, 291-300 (2019)
   1.00000000000000
   3.35320000000000  -2.55930000000000   0.00000000000000
   3.35320000000000   2.55930000000000   0.00000000000000
  -1.38926612821329   0.00000000000000   3.97067557413084
     Ba     C     O
      1     1     3
Direct
  -0.10300000000000   0.10300000000000   0.50000000000000   Ba   (2b)
  -0.62590000000000   0.62590000000000   0.00000000000000   C    (2a)
  -0.87220000000000   0.87220000000000   0.00000000000000   O    (2a)
   0.39200000000000   1.39940000000000   0.73670000000000   O    (4c)
  -1.39940000000000  -0.39200000000000  -0.73670000000000   O    (4c)
```

$Ta_5Ti_{11}$ (BCC SQS-16): A5B11_mP16_6_2abc_2a3b3c - CIF

```
# AFLOW.org Repositories
# TaTi/A5B11_mP16_6_2abc_2a3b3c-001.AB params=6.5421326204,1.0,1.0,90.0,
    ↪ 0.5,-0.0,0.5,0.5,0.0,-0.0,0.0,0.5,0.0,0.5,0.0,-0.0,0.5,-0.0,0.5
    ↪ ,0.5,0.25,0.25,0.25,0.75,0.25,0.25,0.75,0.25,0.75,0.25,0.25,
    ↪ 0.75 SG=6 [ANRL doi: 10.1016/j.commatsci.2017.01.017 (part 1),
    ↪ doi: 10.1016/j.commatsci.2018.10.043 (part 2)]
data_TaTi
_pd_phase_name A5B11_mP16_6_2abc_2a3b3c-001.AB

_chemical_name_mineral 'Ta5Ti11'
_chemical_formula_sum 'Ta5 Ti11'

loop_
_publ_author_name
 'T. Chakraborty'
 'J. Rogal'
 'R. Drautz'
_journal_name_full_name
;
 Physical Review B
;
_journal_volume 94
_journal_year 2016
_journal_page_first 224104
_journal_page_last 224104
_publ_Section_title
;
 Unraveling the composition dependence of the martensitic transformation
     ↪ temperature: A first-principles study of Ti-Ta alloys
;

_aflow_title 'Ta$_{5}$Ti$_{11}$ (BCC SQS-16) Structure'
_aflow_proto 'A5B11_mP16_6_2abc_2a3b3c'
_aflow_params 'a,b/a,c/a,\beta,x_{1},z_{1},x_{2},z_{2},x_{3},z_{3},x_{4
    ↪ },z_{4},x_{5},z_{5},x_{6},z_{6},x_{7},z_{7},x_{8},z_{8},x_{9},
    ↪ y_{9},z_{9},x_{10},y_{10},z_{10},x_{11},y_{11},z_{11},x_{12},y_
    ↪ {12},z_{12}'
_aflow_params_values '6.5421326204,1.0,1.0,90.0,0.0,0.5,0.5,0.5,0.0,0.0,
    ↪ 0.5,0.0,0.5,0.0,0.0,0.0,0.0,0.5,0.5,0.5,0.25,0.25,0.75,0.25,
    ↪ 0.25,0.25,0.75,0.25,0.25,0.75,0.25,0.75'
_aflow_Strukturbericht 'None'
_aflow_Pearson 'mP16'

_cell_length_a   6.5421326204
_cell_length_b   6.5421326204
_cell_length_c   6.5421326204
_cell_angle_alpha   90.0000000000
_cell_angle_beta   90.0000000000
_cell_angle_gamma   90.0000000000
_symmetry_space_group_name_H-M   'Pm'
_symmetry_Int_Tables_Number   6
loop_
_symmetry_equiv_pos_site_id
_symmetry_equiv_pos_as_xyz
  1  x,y,z
  2  x,-y,z
loop_
 _atom_site_label
 _atom_site_occupancy
 _atom_site_fract_x
 _atom_site_fract_y
 _atom_site_fract_z
 _atom_site_thermal_displace_type
 _atom_site_B_iso_or_equiv
 _atom_site_type_symbol
 _atom_site_symmetry_multiplicity
 _atom_site_Wyckoff_label
Ta1 1.0000000000 0.0000000000 0.0000000000 0.5000000000 Biso 1.0 Ta 1 a
Ta2 1.0000000000 0.5000000000 0.0000000000 0.5000000000 Biso 1.0 Ta 1 a
Ti1 1.0000000000 0.0000000000 0.0000000000 0.0000000000 Biso 1.0 Ti 1 a
Ti2 1.0000000000 0.5000000000 0.0000000000 0.0000000000 Biso 1.0 Ti 1 a
Ta3 1.0000000000 0.5000000000 0.5000000000 0.0000000000 Biso 1.0 Ta 1 b
Ti3 1.0000000000 0.0000000000 0.5000000000 0.0000000000 Biso 1.0 Ti 1 b
Ti4 1.0000000000 0.0000000000 0.5000000000 0.5000000000 Biso 1.0 Ti 1 b
Ti5 1.0000000000 0.5000000000 0.5000000000 0.5000000000 Biso 1.0 Ti 1 b
Ta4 1.0000000000 0.2500000000 0.2500000000 0.7500000000 Biso 1.0 Ta 2 c
Ti6 1.0000000000 0.2500000000 0.2500000000 0.2500000000 Biso 1.0 Ti 2 c
Ti7 1.0000000000 0.7500000000 0.2500000000 0.2500000000 Biso 1.0 Ti 2 c
Ti8 1.0000000000 0.7500000000 0.2500000000 0.7500000000 Biso 1.0 Ti 2 c
```

$Ta_5Ti_{11}$ (BCC SQS-16): A5B11_mP16_6_2abc_2a3b3c - POSCAR

```
A5B11_mP16_6_2abc_2a3b3c & a,b/a,c/a,beta,x1,z1,x2,z2,x3,z3,x4,z4,x5,z5,
    ↪ x6,z6,x7,z7,x8,z8,x9,y9,z9,x10,y10,z10,x11,y11,z11,x12,y12,z12
    ↪ --params=6.5421326204,1.0,1.0,90.0,0.0,0.5,0.5,0.5,0.0,0.0,0.5,
    ↪ 0.0,0.5,0.0,0.0,0.0,0.0,0.5,0.5,0.5,0.25,0.25,0.75,0.25,0.25,
    ↪ 0.25,0.75,0.25,0.25,0.75,0.25,0.75 & Pm C_{s}^{1} #6 (a^4b^4c^4
    ↪ ) & mP16 & None & Ta5Ti11 & Ta5Ti11 & T. Chakraborty and J.
    ↪ Rogal and R. Drautz, Phys. Rev. B 94, 224104(2016)
   1.00000000000000
   6.54213262040000   0.00000000000000   0.00000000000000
   0.00000000000000   6.54213262040000   0.00000000000000
   0.00000000000000   0.00000000000000   6.54213262040000
    Ta    Ti
     5    11
Direct
   0.00000000000000   0.00000000000000   0.50000000000000   Ta   (1a)
   0.50000000000000   0.00000000000000   0.50000000000000   Ta   (1a)
   0.50000000000000   0.50000000000000   0.00000000000000   Ta   (1b)
   0.25000000000000   0.25000000000000   0.75000000000000   Ta   (2c)
   0.25000000000000  -0.25000000000000   0.75000000000000   Ta   (2c)
   0.00000000000000   0.00000000000000   0.00000000000000   Ti   (1a)
   0.50000000000000   0.00000000000000   0.00000000000000   Ti   (1a)
   0.00000000000000   0.50000000000000   0.00000000000000   Ti   (1b)
   0.00000000000000   0.50000000000000   0.50000000000000   Ti   (1b)
   0.50000000000000   0.50000000000000   0.50000000000000   Ti   (1b)
   0.25000000000000   0.25000000000000   0.25000000000000   Ti   (2c)
   0.25000000000000  -0.25000000000000   0.25000000000000   Ti   (2c)
   0.75000000000000   0.25000000000000   0.25000000000000   Ti   (2c)
   0.75000000000000  -0.25000000000000   0.25000000000000   Ti   (2c)
   0.75000000000000   0.25000000000000   0.75000000000000   Ti   (2c)
   0.75000000000000  -0.25000000000000   0.75000000000000   Ti   (2c)
```

$Na_2Ca_6Si_4O_{15}$: A6B2C15D4_mP54_7_6a_2a_15a_4a - CIF

```
# CIF file
data_findsym-output
_audit_creation_method FINDSYM

_chemical_name_mineral 'Ca6Na2O15Si4'
_chemical_formula_sum 'Ca6 Na2 O15 Si4'

loop_
_publ_author_name
 'V. Kahlenberg'
 'M. Maier'
_journal_name_full_name
;
 Mineralogy and Petrology
;
_journal_volume 110
_journal_year 2016
_journal_page_first 905
_journal_page_last 915
_publ_Section_title
;
 On the existence of a high-temperature polymorph of Na$_{2}$Ca$_{6}
    ↪ $Si$_{4}$O$_{15}$--implications for the phase equilibria in
    ↪ the system Na$_{2}$O-CaO-SiO$_{2}$
;

_aflow_title 'Na$_{2}$Ca$_{6}$Si$_{4}$O$_{15}$ Structure'
_aflow_proto 'A6B2C15D4_mP54_7_6a_2a_15a_4a'
_aflow_params 'a,b/a,c/a,\beta,x_{1},y_{1},z_{1},x_{2},y_{2},z_{2},x_{3
    ↪ },y_{3},z_{3},x_{4},y_{4},z_{4},x_{5},y_{5},z_{5},x_{6},y_{6},
    ↪ z_{6},x_{7},y_{7},z_{7},x_{8},y_{8},z_{8},x_{9},y_{9},z_{9},x_{
    ↪ 10},y_{10},z_{10},x_{11},y_{11},z_{11},x_{12},y_{12},z_{12},x_{
    ↪ 13},y_{13},z_{13},x_{14},y_{14},z_{14},x_{15},y_{15},z_{15},x_{
    ↪ 16},y_{16},z_{16},x_{17},y_{17},z_{17},x_{18},y_{18},z_{18},x_{
```

```
    ↪ 19},y_{19},z_{19},x_{20},y_{20},z_{20},x_{21},y_{21},z_{21},x_{
    ↪ 22},y_{22},z_{22},x_{23},y_{23},z_{23},x_{24},y_{24},z_{24},x_{
    ↪ 25},y_{25},z_{25},x_{26},y_{26},z_{26},x_{27},y_{27},z_{27}'
_aflow_params_values '9.0112,0.812000621449,1.21762917259,107.72,0.3409,
    ↪ 0.2658,0.07662,0.6782,0.50471,0.1731,0.3148,0.0044,0.3329,-
    ↪ 0.0022,0.2541,0.27358,-0.0033,0.7531,0.31251,0.6554,0.2291,-
    ↪ 0.09514,0.3179,0.718,0.0741,0.6731,0.2144,0.4123,0.8207,0.4832,
    ↪ 0.8963,0.2666,0.3166,0.3293,0.2631,0.6918,0.3234,0.0643,-0.0059
    ↪ ,0.1862,0.7163,0.8148,0.1874,-0.0723,0.5119,0.1535,0.497,0.4776
    ↪ ,-0.022,0.5282,-0.0241,0.5029,0.1732,0.0159,-0.0177,0.0011,
    ↪ 0.7477,-0.0003,0.4988,0.0147,0.234,0.456,0.5016,0.2321,0.7307,
    ↪ 0.1924,0.191,0.8718,0.0697,-0.0515,0.1138,0.4252,0.0116,0.3664,
    ↪ 0.5049,0.3381,0.6271,-0.0087,0.6556,0.0263,-0.0323,0.0347,-
    ↪ 0.0333,0.5304,0.0206'
_aflow_Strukturbericht 'None'
_aflow_Pearson 'mP54'

_symmetry_space_group_name_H-M "P 1 c 1"
_symmetry_Int_Tables_number 7

_cell_length_a     9.01120
_cell_length_b     7.31710
_cell_length_c     10.97230
_cell_angle_alpha  90.00000
_cell_angle_beta   107.72000
_cell_angle_gamma  90.00000

loop_
_space_group_symop_id
_space_group_symop_operation_xyz
1 x,y,z
2 x,-y,z+1/2

loop_
_atom_site_label
_atom_site_type_symbol
_atom_site_symmetry_multiplicity
_atom_site_Wyckoff_label
_atom_site_fract_x
_atom_site_fract_y
_atom_site_fract_z
_atom_site_occupancy
Ca1 Ca 2 a 0.34090 0.26580 0.07662 1.00000
Ca2 Ca 2 a 0.67820 0.50471 0.17310 1.00000
Ca3 Ca 2 a 0.31480 0.00440 0.33290 1.00000
Ca4 Ca 2 a -0.00220 0.25410 0.27358 1.00000
Ca5 Ca 2 a -0.00330 0.75310 0.31251 1.00000
Ca6 Ca 2 a 0.65540 0.22910 -0.09514 1.00000
Na1 Na 2 a 0.31790 0.71800 0.07410 1.00000
Na2 Na 2 a 0.67310 0.21440 0.41230 1.00000
O1 O 2 a 0.82070 0.48320 0.89630 1.00000
O2 O 2 a 0.26660 0.31660 0.32930 1.00000
O3 O 2 a 0.26310 0.69180 0.32340 1.00000
O4 O 2 a 0.06430 -0.00590 0.18620 1.00000
O5 O 2 a 0.71630 0.81480 0.18740 1.00000
O6 O 2 a -0.07230 0.51190 0.15350 1.00000
O7 O 2 a 0.49700 0.47760 -0.02200 1.00000
O8 O 2 a 0.52820 -0.02410 0.50290 1.00000
O9 O 2 a 0.17320 0.01590 -0.01770 1.00000
O10 O 2 a 0.00110 0.74770 -0.00030 1.00000
O11 O 2 a 0.49880 0.01470 0.23400 1.00000
O12 O 2 a 0.45600 0.50160 0.23210 1.00000
O13 O 2 a 0.73070 0.19240 0.19100 1.00000
O14 O 2 a 0.87180 0.06970 -0.05150 1.00000
O15 O 2 a 0.11380 0.42520 0.01160 1.00000
Si1 Si 2 a 0.36640 0.50490 0.33810 1.00000
Si2 Si 2 a 0.62710 -0.00870 0.65560 1.00000
Si3 Si 2 a 0.02630 -0.03230 0.03470 1.00000
Si4 Si 2 a -0.03330 0.53040 0.02060 1.00000
```

$Na_2Ca_6Si_4O_{15}$: A6B2C15D4_mP54_7_6a_2a_15a_4a - POSCAR

```
A6B2C15D4_mP54_7_6a_2a_15a_4a & a,b/a,c/a,beta,x1,y1,z1,x2,y2,z2,x3,y3,
    ↪ z3,x4,y4,z4,x5,y5,z5,x6,y6,z6,x7,y7,z7,x8,y8,z8,x9,y9,z9,x10,
    ↪ y10,z10,x11,y11,z11,x12,y12,z12,x13,y13,z13,x14,y14,z14,x15,y15
    ↪ ,z15,x16,y16,z16,x17,y17,z17,x18,y18,z18,x19,y19,z19,x20,y20,
    ↪ z20,x21,y21,z21,x22,y22,z22,x23,y23,z23,x24,y24,z24,x25,y25,z25
    ↪ ,x26,y26,z26,x27,y27,z27 --params=9.0112,0.812000621449,
    ↪ 1.21762917259,107.72,0.3409,0.2658,0.07662,0.6782,0.50471,
    ↪ 0.1731,0.3148,0.0044,0.3329,-0.0022,0.2541,0.27358,-0.0033,
    ↪ 0.7531,0.31251,0.6554,0.2291,-0.09514,0.3179,0.718,0.0741,
    ↪ 0.6731,0.2144,0.4123,0.8207,0.4832,0.8963,0.2666,0.3166,0.3293,
    ↪ 0.2631,0.6918,0.3234,0.0643,-0.0059,0.1862,0.7163,0.8148,0.1874
    ↪ ,-0.0723,0.5119,0.1535,0.497,0.4776,-0.022,0.5282,-0.0241,
    ↪ 0.5029,0.1732,0.0159,-0.0177,0.0011,0.7477,-0.0003,0.4988,
    ↪ 0.0147,0.234,0.456,0.5016,0.2321,0.7307,0.1924,0.191,0.8718,
    ↪ 0.0697,-0.0515,0.1138,0.4252,0.0116,0.3664,0.5049,0.3381,0.6271
    ↪ ,-0.0087,0.6556,0.0263,-0.0323,0.0347,-0.0333,0.5304,0.0206 &
    ↪ Pc C_{s}^{2} #7 (a^27) & mP54 & None & Ca6Na2O15Si4 &
    ↪ Ca6Na2O15Si4 & V. Kahlenberg and M. Maier, Mineral. Petrol. 110
    ↪ , 905-915 (2016)
   1.00000000000000
   9.01120000000000   0.00000000000000   0.00000000000000
   0.00000000000000   7.31710000000000   0.00000000000000
  -3.33959049717047   0.00000000000000  10.45172247054090
    Ca    Na     O    Si
    12     4    30     8
Direct
   0.34090000000000   0.26580000000000   0.07662000000000   Ca   (2a)
   0.34090000000000  -0.26580000000000   0.57662000000000   Ca   (2a)
   0.67820000000000   0.50471000000000   0.17310000000000   Ca   (2a)
   0.67820000000000  -0.50471000000000   0.67310000000000   Ca   (2a)
   0.31480000000000   0.00440000000000   0.33290000000000   Ca   (2a)
   0.31480000000000  -0.00440000000000   0.83290000000000   Ca   (2a)
  -0.00220000000000   0.25410000000000   0.27358000000000   Ca   (2a)
  -0.00220000000000  -0.25410000000000   0.77358000000000   Ca   (2a)
  -0.00330000000000   0.75310000000000   0.31251000000000   Ca   (2a)
  -0.00330000000000  -0.75310000000000   0.81251000000000   Ca   (2a)
   0.65540000000000   0.22910000000000  -0.09514000000000   Ca   (2a)
   0.65540000000000  -0.22910000000000   0.40486000000000   Ca   (2a)
   0.31790000000000   0.71800000000000   0.07410000000000   Na   (2a)
   0.31790000000000  -0.71800000000000   0.57410000000000   Na   (2a)
   0.67310000000000   0.21440000000000   0.41230000000000   Na   (2a)
   0.67310000000000  -0.21440000000000   0.91230000000000   Na   (2a)
   0.82070000000000   0.48320000000000   0.89630000000000   O    (2a)
   0.82070000000000  -0.48320000000000   1.39630000000000   O    (2a)
   0.26660000000000   0.31660000000000   0.32930000000000   O    (2a)
   0.26660000000000  -0.31660000000000   0.82930000000000   O    (2a)
   0.26310000000000   0.69180000000000   0.32340000000000   O    (2a)
   0.26310000000000  -0.69180000000000   0.82340000000000   O    (2a)
   0.06430000000000  -0.00590000000000   0.18620000000000   O    (2a)
   0.06430000000000   0.00590000000000   0.68620000000000   O    (2a)
   0.71630000000000   0.81480000000000   0.18740000000000   O    (2a)
   0.71630000000000  -0.81480000000000   0.68740000000000   O    (2a)
  -0.07230000000000   0.51190000000000   0.15350000000000   O    (2a)
  -0.07230000000000  -0.51190000000000   0.65350000000000   O    (2a)
   0.49700000000000   0.47760000000000  -0.02200000000000   O    (2a)
   0.49700000000000  -0.47760000000000   0.47800000000000   O    (2a)
   0.52820000000000  -0.02410000000000   0.50290000000000   O    (2a)
   0.52820000000000   0.02410000000000   1.00290000000000   O    (2a)
   0.17320000000000   0.01590000000000  -0.01770000000000   O    (2a)
   0.17320000000000  -0.01590000000000   0.48230000000000   O    (2a)
   0.00110000000000   0.74770000000000  -0.00030000000000   O    (2a)
   0.00110000000000  -0.74770000000000   0.49970000000000   O    (2a)
   0.49880000000000   0.01470000000000   0.23400000000000   O    (2a)
   0.49880000000000  -0.01470000000000   0.73400000000000   O    (2a)
   0.45600000000000   0.50160000000000   0.23210000000000   O    (2a)
   0.45600000000000  -0.50160000000000   0.73210000000000   O    (2a)
   0.73070000000000   0.19240000000000   0.19100000000000   O    (2a)
   0.73070000000000  -0.19240000000000   0.69100000000000   O    (2a)
   0.87180000000000   0.06970000000000  -0.05150000000000   O    (2a)
   0.87180000000000  -0.06970000000000   0.44850000000000   O    (2a)
   0.11380000000000   0.42520000000000   0.01160000000000   O    (2a)
   0.11380000000000  -0.42520000000000   0.51160000000000   O    (2a)
   0.36640000000000   0.50490000000000   0.33810000000000   Si   (2a)
   0.36640000000000  -0.50490000000000   0.83810000000000   Si   (2a)
   0.62710000000000  -0.00870000000000   0.65560000000000   Si   (2a)
   0.62710000000000   0.00870000000000   1.15560000000000   Si   (2a)
   0.02630000000000  -0.03230000000000   0.03470000000000   Si   (2a)
   0.02630000000000   0.03230000000000   0.53470000000000   Si   (2a)
  -0.03330000000000   0.53040000000000   0.02060000000000   Si   (2a)
  -0.03330000000000  -0.53040000000000   0.52060000000000   Si   (2a)
```

Low-Temperature $Mo_8O_{23}$: A8B23_mP124_7_16a_46a - CIF

```
# CIF file
data_findsym-output
_audit_creation_method FINDSYM

_chemical_name_mineral 'Mo8O23'
_chemical_formula_sum 'Mo8 O23'

loop_
_publ_author_name
 'H. Fujishita'
 'M. Sato'
 'S. Sato'
 'S. Hoshino'
_journal_name_full_name
;
 Journal of Solid State Chemistry
;
_journal_volume 66
_journal_year 1987
_journal_page_first 40
_journal_page_last 46
_publ_Section_title
;
 Structure Determination of low-dimensional conductor Mo$_{8}$O$_{23}$
;

_aflow_title 'Low-Temperature Mo$_{8}$O$_{23}$ Structure'
_aflow_proto 'A8B23_mP124_7_16a_46a'
_aflow_params 'a,b/a,c/a,\beta,x_{1},y_{1},z_{1},x_{2},y_{2},z_{2},x_{3
    ↪ },y_{3},z_{3},x_{4},y_{4},z_{4},x_{5},y_{5},z_{5},x_{6},y_{6},
    ↪ z_{6},x_{7},y_{7},z_{7},x_{8},y_{8},z_{8},x_{9},y_{9},z_{9},x_{
    ↪ 10},y_{10},z_{10},x_{11},y_{11},z_{11},x_{12},y_{12},z_{12},x_{
    ↪ 13},y_{13},z_{13},x_{14},y_{14},z_{14},x_{15},y_{15},z_{15},x_{
    ↪ 16},y_{16},z_{16},x_{17},y_{17},z_{17},x_{18},y_{18},z_{18},x_{
    ↪ 19},y_{19},z_{19},x_{20},y_{20},z_{20},x_{21},y_{21},z_{21},x_{
    ↪ 22},y_{22},z_{22},x_{23},y_{23},z_{23},x_{24},y_{24},z_{24},x_{
    ↪ 25},y_{25},z_{25},x_{26},y_{26},z_{26},x_{27},y_{27},z_{27},x_{
    ↪ 28},y_{28},z_{28},x_{29},y_{29},z_{29},x_{30},y_{30},z_{30},x_{
    ↪ 31},y_{31},z_{31},x_{32},y_{32},z_{32},x_{33},y_{33},z_{33},x_{
    ↪ 34},y_{34},z_{34},x_{35},y_{35},z_{35},x_{36},y_{36},z_{36},x_{
    ↪ 37},y_{37},z_{37},x_{38},y_{38},z_{38},x_{39},y_{39},z_{39},x_{
    ↪ 40},y_{40},z_{40},x_{41},y_{41},z_{41},x_{42},y_{42},z_{42},x_{
    ↪ 43},y_{43},z_{43},x_{44},y_{44},z_{44},x_{45},y_{45},z_{45},x_{
    ↪ 46},y_{46},z_{46},x_{47},y_{47},z_{47},x_{48},y_{48},z_{48},x_{
    ↪ 49},y_{49},z_{49},x_{50},y_{50},z_{50},x_{51},y_{51},z_{51},x_{
    ↪ 52},y_{52},z_{52},x_{53},y_{53},z_{53},x_{54},y_{54},z_{54},x_{
    ↪ 55},y_{55},z_{55},x_{56},y_{56},z_{56},x_{57},y_{57},z_{57},x_{
    ↪ 58},y_{58},z_{58},x_{59},y_{59},z_{59},x_{60},y_{60},z_{60},x_{
    ↪ 61},y_{61},z_{61},x_{62},y_{62},z_{62}'
_aflow_params_values '13.39,0.602091112771,1.25616131441,106.02,0.0612,
    ↪ 0.2916,0.4188,-0.0652,0.2923,0.0817,0.0649,0.7928,0.417,-0.0675
    ↪ ,0.7935,0.0847,0.1842,0.2087,0.2479,0.8176,0.206,0.2543,0.1862,
    ↪ 0.707,0.2448,0.8156,0.7049,0.2563,0.3154,0.2899,0.0796,0.6851,
    ↪ 0.2954,0.4187,0.3156,0.7936,0.0774,0.6845,0.7997,0.421,0.4472,
    ↪ 0.2043,0.4015,0.5537,0.206,0.097,0.447,0.7036,0.4012,0.5546,
```

```
    ↪ 0.7049,0.0981,0.0779,-0.0011,0.405,-0.0728,0.0012,0.0891,0.0607
    ↪ ,0.499,0.4296,-0.0486,0.5048,0.0783,0.1929,0.4912,0.2339,0.7976
    ↪ ,0.504,0.2613,0.19,-0.0029,0.2574,0.8216,-0.004,0.2478,0.3214,
    ↪ 0.0082,0.067,0.6813,0.006,0.4303,0.3152,0.5,0.0882,0.6775,
    ↪ 0.5067,0.4152,0.4436,0.4978,0.4074,0.546,0.4885,0.0995,0.4521,-
    ↪ 0.0109,0.4152,0.558,-0.0026,0.0935,-0.0002,0.2505,-0.0015,
    ↪ 0.0024,0.79,-0.0015,0.0606,0.2467,0.1601,-0.0672,0.2429,0.3282,
    ↪ 0.0601,0.7542,0.1589,-0.0673,0.7428,0.3317,0.1265,0.2707,0.3274
    ↪ ,0.8668,0.2776,0.1659,0.1341,0.7186,0.3288,0.8756,0.713,0.1692,
    ↪ 0.2011,0.2464,0.4929,0.80065,0.2582,-0.001,0.1943,0.7351,0.4921
    ↪ ,0.7994,0.7476,0.0018,0.2593,0.2437,0.1578,0.7417,0.229,0.3336,
    ↪ 0.2579,0.7718,0.1605,0.7351,0.7823,0.33,0.3226,0.2545,0.3243,
    ↪ 0.6708,0.2316,0.1706,0.3225,0.7445,0.3215,0.6702,0.7512,0.1657,
    ↪ 0.4085,0.258,0.0033,0.5842,0.262,0.493,0.4024,0.7451,-0.0013,
    ↪ 0.5864,0.7398,0.4904,0.4542,0.2477,0.1492,0.5416,0.2337,0.3391,
    ↪ 0.4514,0.7444,0.1512,0.5319,0.7539,0.3435'
_aflow_Strukturbericht 'None'
_aflow_Pearson 'mP124'

_symmetry_space_group_name_H-M "P 1 c 1"
_symmetry_Int_Tables_number 7

_cell_length_a      13.39000
_cell_length_b      8.06200
_cell_length_c      16.82000
_cell_angle_alpha   90.00000
_cell_angle_beta    106.02000
_cell_angle_gamma   90.00000

loop_
_space_group_symop_id
_space_group_symop_operation_xyz
1 x,y,z
2 x,-y,z+1/2

loop_
_atom_site_label
_atom_site_type_symbol
_atom_site_symmetry_multiplicity
_atom_site_Wyckoff_label
_atom_site_fract_x
_atom_site_fract_y
_atom_site_fract_z
_atom_site_occupancy
Mo1 Mo 2 a 0.06120 0.29160 0.41880 1.00000
Mo2 Mo 2 a -0.06520 0.29230 0.08170 1.00000
Mo3 Mo 2 a 0.06490 0.79280 0.41700 1.00000
Mo4 Mo 2 a -0.06750 0.79350 0.08470 1.00000
Mo5 Mo 2 a 0.18420 0.20870 0.24790 1.00000
Mo6 Mo 2 a 0.81760 0.20600 0.25430 1.00000
Mo7 Mo 2 a 0.18620 0.70700 0.24480 1.00000
Mo8 Mo 2 a 0.81560 0.70490 0.25630 1.00000
Mo9 Mo 2 a 0.31540 0.28990 0.07960 1.00000
Mo10 Mo 2 a 0.68510 0.29540 0.41870 1.00000
Mo11 Mo 2 a 0.31560 0.79360 0.07740 1.00000
Mo12 Mo 2 a 0.68450 0.79970 0.42100 1.00000
Mo13 Mo 2 a 0.44720 0.20430 0.40150 1.00000
Mo14 Mo 2 a 0.55370 0.20600 0.09700 1.00000
Mo15 Mo 2 a 0.44700 0.70360 0.40120 1.00000
Mo16 Mo 2 a 0.55460 0.70490 0.09810 1.00000
O1 O 2 a 0.07790 -0.00110 0.40500 1.00000
O2 O 2 a -0.07280 0.00120 0.08910 1.00000
O3 O 2 a 0.06070 0.49900 0.42960 1.00000
O4 O 2 a -0.04860 0.50480 0.07830 1.00000
O5 O 2 a 0.19290 0.49120 0.23390 1.00000
O6 O 2 a 0.79760 0.50400 0.26130 1.00000
O7 O 2 a 0.19000 -0.00290 0.25740 1.00000
O8 O 2 a 0.82160 -0.00400 0.24780 1.00000
O9 O 2 a 0.32140 0.00820 0.06700 1.00000
O10 O 2 a 0.68130 0.00600 0.43030 1.00000
O11 O 2 a 0.31520 0.50000 0.08820 1.00000
O12 O 2 a 0.67750 0.50670 0.41520 1.00000
O13 O 2 a 0.44360 0.49780 0.40740 1.00000
O14 O 2 a 0.54600 0.48850 0.09950 1.00000
O15 O 2 a 0.45210 -0.01090 0.41520 1.00000
O16 O 2 a 0.55800 -0.00260 0.09350 1.00000
O17 O 2 a -0.00020 0.25050 -0.00150 1.00000
O18 O 2 a 0.00240 0.79000 -0.00150 1.00000
O19 O 2 a 0.06060 0.24670 0.16010 1.00000
O20 O 2 a -0.06720 0.24290 0.32820 1.00000
O21 O 2 a 0.06010 0.75420 0.15890 1.00000
O22 O 2 a -0.06730 0.74280 0.33170 1.00000
O23 O 2 a 0.12650 0.27070 0.32740 1.00000
O24 O 2 a 0.86680 0.27760 0.16590 1.00000
O25 O 2 a 0.13410 0.71860 0.32880 1.00000
O26 O 2 a 0.87560 0.71300 0.16920 1.00000
O27 O 2 a 0.20110 0.24640 0.49290 1.00000
O28 O 2 a 0.80065 0.25820 -0.00100 1.00000
O29 O 2 a 0.19430 0.73510 0.49210 1.00000
O30 O 2 a 0.79940 0.74760 0.00180 1.00000
O31 O 2 a 0.25930 0.24370 0.15780 1.00000
O32 O 2 a 0.74170 0.22900 0.33360 1.00000
O33 O 2 a 0.25790 0.77180 0.16050 1.00000
O34 O 2 a 0.73510 0.78230 0.33000 1.00000
O35 O 2 a 0.32260 0.25450 0.32430 1.00000
O36 O 2 a 0.67080 0.23160 0.17060 1.00000
O37 O 2 a 0.32250 0.74450 0.32150 1.00000
O38 O 2 a 0.67020 0.75120 0.16570 1.00000
O39 O 2 a 0.40850 0.25800 0.00330 1.00000
O40 O 2 a 0.58420 0.26200 0.49300 1.00000
O41 O 2 a 0.40240 0.74510 -0.00130 1.00000
O42 O 2 a 0.58640 0.73980 0.49040 1.00000
O43 O 2 a 0.45420 0.24770 0.14920 1.00000
O44 O 2 a 0.54160 0.23370 0.33910 1.00000
O45 O 2 a 0.45140 0.74440 0.15120 1.00000
O46 O 2 a 0.53190 0.75390 0.34350 1.00000
```

Low-Temperature $Mo_8O_{23}$: A8B23_mP124_7_16a_46a - POSCAR

```
A8B23_mP124_7_16a_46a & a,b/a,c/a,beta,x1,y1,z1,x2,y2,z2,x3,y3,z3,x4,y4,
    ↪ z4,x5,y5,z5,x6,y6,z6,x7,y7,z7,x8,y8,z8,x9,y9,z9,x10,y10,z10,x11
    ↪ ,y11,z11,x12,y12,z12,x13,y13,z13,x14,y14,z14,x15,y15,z15,x16,
    ↪ y16,z16,x17,y17,z17,x18,y18,z18,x19,y19,z19,x20,y20,z20,x21,y21
    ↪ ,z21,x22,y22,z22,x23,y23,z23,x24,y24,z24,x25,y25,z25,x26,y26,
    ↪ z26,x27,y27,z27,x28,y28,z28,x29,y29,z29,x30,y30,z30,x31,y31,z31
    ↪ ,x32,y32,z32,x33,y33,z33,x34,y34,z34,x35,y35,z35,x36,y36,z36,
    ↪ x37,y37,z37,x38,y38,z38,x39,y39,z39,x40,y40,z40,x41,y41,z41,x42
    ↪ ,y42,z42,x43,y43,z43,x44,y44,z44,x45,y45,z45,x46,y46,z46,x47,
    ↪ y47,z47,x48,y48,z48,x49,y49,z49,x50,y50,z50,x51,y51,z51,x52,y52
    ↪ ,z52,x53,y53,z53,x54,y54,z54,x55,y55,z55,x56,y56,z56,x57,y57,
    ↪ z57,x58,y58,z58,x59,y59,z59,x60,y60,z60,x61,y61,z61,x62,y62,z62
    ↪ --params=13.39,0.602091112771,1.25616131441,106.02,0.0612,
    ↪ 0.2916,0.4188,-0.0652,0.2923,0.0817,0.0649,0.7928,0.417,-0.0675
    ↪ ,0.7935,0.0847,0.1842,0.2087,0.2479,0.8176,0.206,0.2543,0.1862,
    ↪ 0.707,0.2448,0.8156,0.7049,0.2563,0.3154,0.2899,0.0796,0.6851,
    ↪ 0.2954,0.4187,0.3156,0.7936,0.0774,0.6845,0.7997,0.421,0.4472,
    ↪ 0.2043,0.4015,0.5537,0.206,0.097,0.447,0.7036,0.4012,0.5546,
    ↪ 0.7049,0.0981,0.0779,-0.0011,0.405,-0.0728,0.0012,0.0891,0.0607
    ↪ ,0.499,0.4296,-0.0486,0.5048,0.0783,0.1929,0.4912,0.2339,0.7976
    ↪ ,0.504,0.2613,0.19,-0.0029,0.2574,0.8216,-0.004,0.2478,0.3214,
    ↪ 0.0082,0.067,0.6813,0.006,0.4303,0.3152,0.5,0.0882,0.6775,
    ↪ 0.5067,0.4152,0.4436,0.4978,0.4074,0.546,0.4885,0.0995,0.4521,-
    ↪ 0.0109,0.4152,0.558,-0.0026,0.0935,-0.0002,0.2505,-0.0015,
    ↪ 0.0024,0.79,-0.0015,0.0606,0.2467,0.1601,-0.0672,0.2429,0.3282,
    ↪ 0.0601,0.7542,0.1589,-0.0673,0.7428,0.3317,0.1265,0.2707,0.3274
    ↪ ,0.8668,0.2776,0.1659,0.1341,0.7186,0.3288,0.8756,0.713,0.1692,
    ↪ 0.2011,0.2464,0.4929,0.80065,0.2582,-0.001,0.1943,0.7351,0.4921
    ↪ ,0.7994,0.7476,0.0018,0.2593,0.2437,0.1578,0.7417,0.229,0.3336,
    ↪ 0.2579,0.7718,0.1605,0.7351,0.7823,0.33,0.3226,0.2545,0.3243,
    ↪ 0.6708,0.2316,0.1706,0.3225,0.7445,0.3215,0.6702,0.7512,0.1657,
    ↪ 0.4085,0.258,0.0033,0.5842,0.262,0.493,0.4024,0.7451,-0.0013,
    ↪ 0.5864,0.7398,0.4904,0.4542,0.2477,0.1492,0.5416,0.2337,0.3391,
    ↪ 0.4514,0.7444,0.1512,0.5319,0.7539,0.3435 & Pc C_{s}^{2} #7 (a^
    ↪ 62) & mP124 & None & Mo8O23 & H. Fujishita et al., J.
    ↪ Solid State Chem. 66, 40-46 (1987)
   1.00000000000000
  13.39000000000000   0.00000000000000   0.00000000000000
   0.00000000000000   8.06200000000000   0.00000000000000
  -4.64186388615190   0.00000000000000  16.16680239448850
    Mo        O
    32       92
Direct
   0.06120000000000   0.29160000000000   0.41880000000000   Mo   (2a)
   0.06120000000000  -0.29160000000000   0.91880000000000   Mo   (2a)
  -0.06520000000000   0.29230000000000   0.08170000000000   Mo   (2a)
  -0.06520000000000  -0.29230000000000   0.58170000000000   Mo   (2a)
   0.06490000000000   0.79280000000000   0.41700000000000   Mo   (2a)
   0.06490000000000  -0.79280000000000   0.91700000000000   Mo   (2a)
  -0.06750000000000   0.79350000000000   0.08470000000000   Mo   (2a)
  -0.06750000000000  -0.79350000000000   0.58470000000000   Mo   (2a)
   0.18420000000000   0.20870000000000   0.24790000000000   Mo   (2a)
   0.18420000000000  -0.20870000000000   0.74790000000000   Mo   (2a)
   0.81760000000000   0.20600000000000   0.25430000000000   Mo   (2a)
   0.81760000000000  -0.20600000000000   0.75430000000000   Mo   (2a)
   0.18620000000000   0.70700000000000   0.24480000000000   Mo   (2a)
   0.18620000000000  -0.70700000000000   0.74480000000000   Mo   (2a)
   0.81560000000000   0.70490000000000   0.25630000000000   Mo   (2a)
   0.81560000000000  -0.70490000000000   0.75630000000000   Mo   (2a)
   0.31540000000000   0.28990000000000   0.07960000000000   Mo   (2a)
   0.31540000000000  -0.28990000000000   0.57960000000000   Mo   (2a)
   0.68510000000000   0.29540000000000   0.41870000000000   Mo   (2a)
   0.68510000000000  -0.29540000000000   0.91870000000000   Mo   (2a)
   0.31560000000000   0.79360000000000   0.07740000000000   Mo   (2a)
   0.31560000000000  -0.79360000000000   0.57740000000000   Mo   (2a)
   0.68450000000000   0.79970000000000   0.42100000000000   Mo   (2a)
   0.68450000000000  -0.79970000000000   0.92100000000000   Mo   (2a)
   0.44720000000000   0.20430000000000   0.40150000000000   Mo   (2a)
   0.44720000000000  -0.20430000000000   0.90150000000000   Mo   (2a)
   0.55370000000000   0.20600000000000   0.09700000000000   Mo   (2a)
   0.55370000000000  -0.20600000000000   0.59700000000000   Mo   (2a)
   0.44700000000000   0.70360000000000   0.40120000000000   Mo   (2a)
   0.44700000000000  -0.70360000000000   0.90120000000000   Mo   (2a)
   0.55460000000000   0.70490000000000   0.09810000000000   Mo   (2a)
   0.55460000000000  -0.70490000000000   0.59810000000000   Mo   (2a)
   0.07790000000000  -0.00110000000000   0.40500000000000   O    (2a)
   0.07790000000000   0.00110000000000   0.90500000000000   O    (2a)
  -0.07280000000000   0.00120000000000   0.08910000000000   O    (2a)
  -0.07280000000000  -0.00120000000000   0.58910000000000   O    (2a)
   0.06070000000000   0.49900000000000   0.42960000000000   O    (2a)
   0.06070000000000  -0.49900000000000   0.92960000000000   O    (2a)
  -0.04860000000000   0.50480000000000   0.07830000000000   O    (2a)
  -0.04860000000000  -0.50480000000000   0.57830000000000   O    (2a)
   0.19290000000000   0.49120000000000   0.23390000000000   O    (2a)
   0.19290000000000  -0.49120000000000   0.73390000000000   O    (2a)
   0.79760000000000   0.50400000000000   0.26130000000000   O    (2a)
   0.79760000000000  -0.50400000000000   0.76130000000000   O    (2a)
   0.19000000000000  -0.00290000000000   0.25740000000000   O    (2a)
   0.19000000000000   0.00290000000000   0.75740000000000   O    (2a)
   0.82160000000000  -0.00400000000000   0.24780000000000   O    (2a)
   0.82160000000000   0.00400000000000   0.74780000000000   O    (2a)
   0.32140000000000   0.00820000000000   0.06700000000000   O    (2a)
   0.32140000000000  -0.00820000000000   0.56700000000000   O    (2a)
   0.68130000000000   0.00600000000000   0.43030000000000   O    (2a)
   0.68130000000000  -0.00600000000000   0.93030000000000   O    (2a)
   0.31520000000000   0.50000000000000   0.08820000000000   O    (2a)
   0.31520000000000  -0.50000000000000   0.58820000000000   O    (2a)
   0.67750000000000   0.50670000000000   0.41520000000000   O    (2a)
   0.67750000000000  -0.50670000000000   0.91520000000000   O    (2a)
   0.44360000000000   0.49780000000000   0.40740000000000   O    (2a)
   0.44360000000000  -0.49780000000000   0.90740000000000   O    (2a)
```

```
     0.54600000000000      0.48850000000000      0.09950000000000      O      (2a)
     0.54600000000000     -0.48850000000000      0.59950000000000      O      (2a)
     0.45210000000000     -0.01090000000000      0.41520000000000      O      (2a)
     0.45210000000000      0.01090000000000      0.91520000000000      O      (2a)
     0.55800000000000     -0.00260000000000      0.09350000000000      O      (2a)
     0.55800000000000      0.00260000000000      0.59350000000000      O      (2a)
    -0.00020000000000      0.25050000000000     -0.00150000000000      O      (2a)
    -0.00020000000000     -0.25050000000000      0.49850000000000      O      (2a)
     0.00240000000000      0.79000000000000     -0.00150000000000      O      (2a)
     0.00240000000000     -0.79000000000000      0.49850000000000      O      (2a)
     0.06060000000000      0.24670000000000      0.16010000000000      O      (2a)
     0.06060000000000     -0.24670000000000      0.66010000000000      O      (2a)
    -0.06720000000000      0.24290000000000      0.32820000000000      O      (2a)
    -0.06720000000000     -0.24290000000000      0.82820000000000      O      (2a)
     0.06010000000000      0.75420000000000      0.15890000000000      O      (2a)
     0.06010000000000     -0.75420000000000      0.65890000000000      O      (2a)
    -0.06730000000000      0.74280000000000      0.33170000000000      O      (2a)
    -0.06730000000000     -0.74280000000000      0.83170000000000      O      (2a)
     0.12650000000000      0.27070000000000      0.32740000000000      O      (2a)
     0.12650000000000     -0.27070000000000      0.82740000000000      O      (2a)
     0.86680000000000      0.27760000000000      0.16590000000000      O      (2a)
     0.86680000000000     -0.27760000000000      0.66590000000000      O      (2a)
     0.13410000000000      0.71860000000000      0.32880000000000      O      (2a)
     0.13410000000000     -0.71860000000000      0.82880000000000      O      (2a)
     0.87560000000000      0.71300000000000      0.16920000000000      O      (2a)
     0.87560000000000     -0.71300000000000      0.66920000000000      O      (2a)
     0.20110000000000      0.24640000000000      0.49290000000000      O      (2a)
     0.20110000000000     -0.24640000000000      0.99290000000000      O      (2a)
     0.80065000000000      0.25820000000000     -0.00100000000000      O      (2a)
     0.80065000000000     -0.25820000000000      0.49900000000000      O      (2a)
     0.19430000000000      0.73510000000000      0.49210000000000      O      (2a)
     0.19430000000000     -0.73510000000000      0.99210000000000      O      (2a)
     0.79940000000000      0.74760000000000      0.00180000000000      O      (2a)
     0.79940000000000     -0.74760000000000      0.50180000000000      O      (2a)
     0.25930000000000      0.24370000000000      0.15780000000000      O      (2a)
     0.25930000000000     -0.24370000000000      0.65780000000000      O      (2a)
     0.74170000000000      0.22900000000000      0.33360000000000      O      (2a)
     0.74170000000000     -0.22900000000000      0.83360000000000      O      (2a)
     0.25790000000000      0.77180000000000      0.16050000000000      O      (2a)
     0.25790000000000     -0.77180000000000      0.66050000000000      O      (2a)
     0.73510000000000      0.78230000000000      0.33000000000000      O      (2a)
     0.73510000000000     -0.78230000000000      0.83000000000000      O      (2a)
     0.32260000000000      0.25450000000000      0.32430000000000      O      (2a)
     0.32260000000000     -0.25450000000000      0.82430000000000      O      (2a)
     0.67080000000000      0.23160000000000      0.17060000000000      O      (2a)
     0.67080000000000     -0.23160000000000      0.67060000000000      O      (2a)
     0.32250000000000      0.74450000000000      0.32150000000000      O      (2a)
     0.32250000000000     -0.74450000000000      0.82150000000000      O      (2a)
     0.67020000000000      0.75120000000000      0.16570000000000      O      (2a)
     0.67020000000000     -0.75120000000000      0.66570000000000      O      (2a)
     0.40850000000000      0.25800000000000      0.00330000000000      O      (2a)
     0.40850000000000     -0.25800000000000      0.50330000000000      O      (2a)
     0.58420000000000      0.26200000000000      0.49300000000000      O      (2a)
     0.58420000000000     -0.26200000000000      0.99300000000000      O      (2a)
     0.40240000000000      0.74510000000000     -0.00130000000000      O      (2a)
     0.40240000000000     -0.74510000000000      0.49870000000000      O      (2a)
     0.58640000000000      0.73980000000000      0.49040000000000      O      (2a)
     0.58640000000000     -0.73980000000000      0.99040000000000      O      (2a)
     0.45420000000000      0.24770000000000      0.14920000000000      O      (2a)
     0.45420000000000     -0.24770000000000      0.64920000000000      O      (2a)
     0.54160000000000      0.23370000000000      0.33910000000000      O      (2a)
     0.54160000000000     -0.23370000000000      0.83910000000000      O      (2a)
     0.45140000000000      0.74440000000000      0.15120000000000      O      (2a)
     0.45140000000000     -0.74440000000000      0.65120000000000      O      (2a)
     0.53190000000000      0.75390000000000      0.34350000000000      O      (2a)
     0.53190000000000     -0.75390000000000      0.84350000000000      O      (2a)
```

Calaverite ($AuTe_2$): AB2_mP12_7_2a_4a - CIF

```
# CIF file
data_findsym-output
_audit_creation_method FINDSYM

_chemical_name_mineral 'AuTe2'
_chemical_formula_sum 'Au Te2'

loop_
_publ_author_name
 'F. Pertlik'
_journal_name_full_name
;
 Zeitschrift f{\"u}r Kristallographie - Crystalline Materials
;
_journal_volume 169
_journal_year 1984
_journal_page_first 227
_journal_page_last 236
_publ_Section_title
;
 Kristallchemie nat\"{u}licher Telluride III: Die Kristallstruktur des
     ↪ Minerals Calaverit, AuTe$_2$
;

_aflow_title 'Calaverite (AuTe$_{2}$) Structure'
_aflow_proto 'AB2_mP12_7_2a_4a'
_aflow_params 'a,b/a,c/a,\beta,x_{1},y_{1},z_{1},x_{2},y_{2},z_{2},x_{3
     ↪ },y_{3},z_{3},x_{4},y_{4},z_{4},x_{5},y_{5},z_{5},x_{6},y_{6},
     ↪ z_{6}'
_aflow_params_values '8.76,0.503424657534,1.15867579909,125.2,0.7483,
     ↪ 0.4938,0.1243,0.246,-0.0059,0.8731,-0.0649,-0.0138,0.5728,
     ↪ 0.5585,0.052,0.6747,0.0575,0.4464,0.4209,0.4363,0.5585,0.3235'
_aflow_Strukturbericht 'None'
_aflow_Pearson 'mP12'

_symmetry_space_group_name_H-M "P 1 c 1"
_symmetry_Int_Tables_number 7

_cell_length_a     8.76000
_cell_length_b     4.41000
_cell_length_c     10.15000
_cell_angle_alpha  90.00000
_cell_angle_beta   125.20000
_cell_angle_gamma  90.00000

loop_
_space_group_symop_id
_space_group_symop_operation_xyz
1 x,y,z
2 x,-y,z+1/2

loop_
_atom_site_label
_atom_site_type_symbol
_atom_site_symmetry_multiplicity
_atom_site_Wyckoff_label
_atom_site_fract_x
_atom_site_fract_y
_atom_site_fract_z
_atom_site_occupancy
Au1 Au 2 a 0.74830 0.49380 0.12430 1.00000
Au2 Au 2 a 0.24600 -0.00590 0.87310 1.00000
Te1 Te 2 a -0.06490 -0.01380 0.57280 1.00000
Te2 Te 2 a 0.55850 0.05200 0.67470 1.00000
Te3 Te 2 a 0.05750 0.44640 0.42090 1.00000
Te4 Te 2 a 0.43630 0.55850 0.32350 1.00000
```

Calaverite ($AuTe_2$): AB2_mP12_7_2a_4a - POSCAR

```
AB2_mP12_7_2a_4a & a,b/a,c/a,beta,x1,y1,z1,x2,y2,z2,x3,y3,z3,x4,y4,z4,x5
     ↪ ,y5,z5,x6,y6,z6 --params=8.76,0.503424657534,1.15867579909,
     ↪ 125.2,0.7483,0.4938,0.1243,0.246,-0.0059,0.8731,-0.0649,-0.0138
     ↪ ,0.5728,0.5585,0.052,0.6747,0.0575,0.4464,0.4209,0.4363,0.5585,
     ↪ 0.3235 & Pc C_{s}^{2} #7 (a^6) & mP12 & None & AuTe2 & AuTe2 &
     ↪ F. Pertlik, Zeitschrift f{"u}r Kristallographie - Crystalline
     ↪ Materials 169, 227-236 (1984)
   1.00000000000000
   8.76000000000000    0.00000000000000    0.00000000000000
   0.00000000000000    4.41000000000000    0.00000000000000
  -5.85078800912340    0.00000000000000    8.29402071810155
    Au    Te
     4     8
Direct
   0.74830000000000    0.49380000000000    0.12430000000000    Au   (2a)
   0.74830000000000   -0.49380000000000    0.62430000000000    Au   (2a)
   0.24600000000000   -0.00590000000000    0.87310000000000    Au   (2a)
   0.24600000000000    0.00590000000000    1.37310000000000    Au   (2a)
  -0.06490000000000   -0.01380000000000    0.57280000000000    Te   (2a)
  -0.06490000000000    0.01380000000000    1.07280000000000    Te   (2a)
   0.55850000000000    0.05200000000000    0.67470000000000    Te   (2a)
   0.55850000000000   -0.05200000000000    1.17470000000000    Te   (2a)
   0.05750000000000    0.44640000000000    0.42090000000000    Te   (2a)
   0.05750000000000   -0.44640000000000    0.92090000000000    Te   (2a)
   0.43630000000000    0.55850000000000    0.32350000000000    Te   (2a)
   0.43630000000000   -0.55850000000000    0.82350000000000    Te   (2a)
```

Monoclinic $Co_4Al_{13}$: A13B4_mC102_8_17a11b_8a2b - CIF

```
# CIF file
data_findsym-output
_audit_creation_method FINDSYM

_chemical_name_mineral 'Al13Co4'
_chemical_formula_sum 'Al13 Co4'

loop_
_publ_author_name
 'R. C. Hudd'
 'W. H. Taylor'
_journal_name_full_name
;
 Acta Cristallographica
;
_journal_volume 15
_journal_year 1962
_journal_page_first 441
_journal_page_last 442
_publ_Section_title
;
 The Structure of Co$_{4}$Al$_{13}$
;

# Found in Structure investigation of the (100) surface of the
     ↪ orthorhombic Al$_{13}$Co$_{4}$ crystal, 2009

_aflow_title 'Monoclinic Co$_{4}$Al$_{13}$ Structure'
_aflow_proto 'A13B4_mC102_8_17a11b_8a2b'
_aflow_params 'a,b/a,c/a,\beta,x_{1},z_{1},x_{2},z_{2},x_{3},z_{3},x_{4
     ↪ },z_{4},x_{5},z_{5},x_{6},z_{6},x_{7},z_{7},x_{8},z_{8},x_{9},
     ↪ z_{9},x_{10},z_{10},x_{11},z_{11},x_{12},z_{12},x_{13},z_{13},
     ↪ x_{14},z_{14},x_{15},z_{15},x_{16},z_{16},x_{17},z_{17},x_{18},
     ↪ z_{18},x_{19},z_{19},x_{20},z_{20},x_{21},z_{21},x_{22},z_{22},
     ↪ x_{23},z_{23},x_{24},z_{24},x_{25},z_{25},x_{26},y_{26},z_{26},
     ↪ x_{27},y_{27},z_{27},x_{28},y_{28},z_{28},x_{29},y_{29},z_{29},
     ↪ x_{30},y_{30},z_{30},x_{31},y_{31},z_{31},x_{32},y_{32},z_{32},
     ↪ x_{33},y_{33},z_{33},x_{34},y_{34},z_{34},x_{35},y_{35},z_{35},
     ↪ x_{36},y_{36},z_{36},x_{37},y_{37},z_{37},x_{38},y_{38},z_{38}'
_aflow_params_values '15.183,0.534940393862,0.812751103208,107.9,0.055,
     ↪ 0.174,0.329,0.28,0.767,0.469,-0.077,0.424,0.754,0.031,0.522,
     ↪ 0.169,0.502,0.499,0.702,0.23,-0.095,0.212,-0.055,0.825,0.67,
     ↪ 0.72,0.232,0.532,0.077,0.576,0.245,-0.034,0.477,0.831,0.297,
```

```
    ↪ 0.769,0.095,0.788,0.086,0.382,0.604,0.373,-0.091,0.012,0.597,
    ↪ 0.018,-0.087,0.617,0.395,0.626,0.09,-0.013,0.402,-0.018,0.185,
    ↪ 0.217,0.111,0.364,0.21,0.111,0.176,0.219,0.334,0.491,0.224,
    ↪ 0.331,0.367,0.21,0.477,0.498,0.248,0.0,0.314,0.282,0.888,0.124,
    ↪ 0.271,0.888,0.322,0.28,0.666,0.008,0.279,0.669,0.131,0.289,
    ↪ 0.532,0.318,0.292,0.277,0.182,0.229,0.722'
_aflow_Strukturbericht 'None'
_aflow_Pearson 'mC102'

_symmetry_space_group_name_H-M "C 1 m 1"
_symmetry_Int_Tables_number 8

_cell_length_a    15.18300
_cell_length_b    8.12200
_cell_length_c    12.34000
_cell_angle_alpha 90.00000
_cell_angle_beta  107.90000
_cell_angle_gamma 90.00000

loop_
_space_group_symop_id
_space_group_symop_operation_xyz
1 x,y,z
2 x,-y,z
3 x+1/2,y+1/2,z
4 x+1/2,-y+1/2,z

loop_
_atom_site_label
_atom_site_type_symbol
_atom_site_symmetry_multiplicity
_atom_site_Wyckoff_label
_atom_site_fract_x
_atom_site_fract_y
_atom_site_fract_z
_atom_site_occupancy
Al1 Al 2 a 0.05500 0.00000 0.17400 1.00000
Al2 Al 2 a 0.32900 0.00000 0.28000 1.00000
Al3 Al 2 a 0.76700 0.00000 0.46900 1.00000
Al4 Al 2 a -0.07700 0.00000 0.42400 0.30000
Al5 Al 2 a 0.75400 0.00000 0.03100 0.70000
Al6 Al 2 a 0.52200 0.00000 0.16900 1.00000
Al7 Al 2 a 0.50200 0.00000 0.49900 1.00000
Al8 Al 2 a 0.70200 0.00000 0.23000 1.00000
Al9 Al 2 a -0.09500 0.00000 0.21200 0.70000
Al10 Al 2 a -0.05500 0.00000 0.82500 1.00000
Al11 Al 2 a 0.67000 0.00000 0.72000 1.00000
Al12 Al 2 a 0.23200 0.00000 0.53200 1.00000
Al13 Al 2 a 0.07700 0.00000 0.57600 0.30000
Al14 Al 2 a 0.24500 0.00000 -0.03400 0.70000
Al15 Al 2 a 0.47700 0.00000 0.83100 1.00000
Al16 Al 2 a 0.29700 0.00000 0.76900 1.00000
Al17 Al 2 a 0.09500 0.00000 0.78800 0.30000
Co1 Co 2 a 0.08600 0.00000 0.38200 1.00000
Co2 Co 2 a 0.60400 0.00000 0.37300 1.00000
Co3 Co 2 a -0.09100 0.00000 0.01200 1.00000
Co4 Co 2 a 0.59700 0.00000 0.01800 1.00000
Co5 Co 2 a -0.08700 0.00000 0.61700 1.00000
Co6 Co 2 a 0.39500 0.00000 0.62600 1.00000
Co7 Co 2 a 0.09000 0.00000 -0.01300 1.00000
Co8 Co 2 a 0.40200 0.00000 -0.01800 1.00000
Al18 Al 4 b 0.18500 0.21700 0.11100 1.00000
Al19 Al 4 b 0.36400 0.21000 0.11100 1.00000
Al20 Al 4 b 0.17600 0.21900 0.33400 1.00000
Al21 Al 4 b 0.49100 0.22400 0.33100 1.00000
Al22 Al 4 b 0.36700 0.21000 0.47700 1.00000
Al23 Al 4 b 0.49800 0.24800 0.00000 1.00000
Al24 Al 4 b 0.31400 0.28200 0.88800 1.00000
Al25 Al 4 b 0.12400 0.27100 0.88800 1.00000
Al26 Al 4 b 0.32200 0.28000 0.66600 1.00000
Al27 Al 4 b 0.00800 0.27900 0.66900 1.00000
Al28 Al 4 b 0.13100 0.28900 0.53200 1.00000
Co9 Co 4 b 0.31800 0.29200 0.27700 1.00000
Co10 Co 4 b 0.18200 0.22900 0.72200 1.00000
```

Monoclinic $Co_4Al_{13}$: A13B4_mC102_8_17a11b_8a2b - POSCAR

```
A13B4_mC102_8_17a11b_8a2b & a,b/a,c/a,beta,x1,z1,x2,z2,x3,z3,x4,z4,x5,z5
    ↪ ,x6,z6,x7,z7,x8,z8,x9,z9,x10,z10,x11,z11,x12,z12,x13,z13,x14,
    ↪ z14,x15,z15,x16,z16,x17,z17,x18,z18,x19,z19,x20,z20,x21,z21,x22
    ↪ ,z22,x23,z23,x24,z24,x25,z25,x26,y26,z26,x27,y27,z27,x28,y28,
    ↪ z28,x29,y29,z29,x30,y30,z30,x31,y31,z31,x32,y32,z32,x33,y33,z33
    ↪ ,x34,y34,z34,x35,y35,z35,x36,y36,z36,x37,y37,z37,x38,y38,z38 --
    ↪ params=15.183,0.534940393862,0.812751103208,107.9,0.055,0.174,
    ↪ 0.329,0.28,0.767,0.469,-0.077,0.424,0.754,0.031,0.522,0.169,
    ↪ 0.502,0.499,0.702,0.23,-0.095,0.212,-0.055,0.825,0.67,0.72,
    ↪ 0.232,0.532,0.077,0.576,0.245,-0.034,0.477,0.831,0.297,0.769,
    ↪ 0.095,0.788,0.086,0.382,0.604,0.373,-0.091,0.012,0.597,0.018,-
    ↪ 0.087,0.617,0.395,0.626,0.09,-0.013,0.402,-0.018,0.185,0.217,
    ↪ 0.111,0.364,0.21,0.111,0.176,0.219,0.334,0.491,0.224,0.331,
    ↪ 0.367,0.21,0.477,0.498,0.248,0.0,0.314,0.282,0.888,0.124,0.271,
    ↪ 0.888,0.322,0.28,0.666,0.008,0.279,0.669,0.131,0.289,0.532,
    ↪ 0.318,0.292,0.277,0.182,0.229,0.722 & Cm C_{s}^{3} #8 (a^25b^13
    ↪ ) & mC102 & None & Al13Co4 & Al13Co4 & R. C. Hudd and W. H.
    ↪ Taylor, Acta Cryst. 15, 441-442 (1962)
   1.00000000000000
   7.59150000000000  -4.06100000000000   0.00000000000000
   7.59150000000000   4.06100000000000   0.00000000000000
  -3.79278066364962   0.00000000000000  11.74267494387230
    Al    Co
    39    12
Direct
   0.05500000000000   0.05500000000000   0.17400000000000   Al   (2a)
   0.32900000000000   0.32900000000000   0.28000000000000   Al   (2a)
   0.76700000000000   0.76700000000000   0.46900000000000   Al   (2a)
  -0.07700000000000  -0.07700000000000   0.42400000000000   Al   (2a)
   0.75400000000000   0.75400000000000   0.03100000000000   Al   (2a)
   0.52200000000000   0.52200000000000   0.16900000000000   Al   (2a)
   0.50200000000000   0.50200000000000   0.49900000000000   Al   (2a)
   0.70200000000000   0.70200000000000   0.23000000000000   Al   (2a)
  -0.09500000000000  -0.09500000000000   0.21200000000000   Al   (2a)
  -0.05500000000000  -0.05500000000000   0.82500000000000   Al   (2a)
   0.67000000000000   0.67000000000000   0.72000000000000   Al   (2a)
   0.23200000000000   0.23200000000000   0.53200000000000   Al   (2a)
   0.07700000000000   0.07700000000000   0.57600000000000   Al   (2a)
   0.24500000000000   0.24500000000000  -0.03400000000000   Al   (2a)
   0.47700000000000   0.47700000000000   0.83100000000000   Al   (2a)
   0.29700000000000   0.29700000000000   0.76900000000000   Al   (2a)
   0.09500000000000   0.09500000000000   0.78800000000000   Al   (2a)
  -0.03200000000000   0.40200000000000   0.11100000000000   Al   (4b)
   0.40200000000000  -0.03200000000000   0.11100000000000   Al   (4b)
   0.15400000000000   0.57400000000000   0.11100000000000   Al   (4b)
   0.57400000000000   0.15400000000000   0.11100000000000   Al   (4b)
  -0.04300000000000   0.39500000000000   0.33400000000000   Al   (4b)
   0.39500000000000  -0.04300000000000   0.33400000000000   Al   (4b)
   0.26700000000000   0.71500000000000   0.33100000000000   Al   (4b)
   0.71500000000000   0.26700000000000   0.33100000000000   Al   (4b)
   0.15700000000000   0.57700000000000   0.47700000000000   Al   (4b)
   0.57700000000000   0.15700000000000   0.47700000000000   Al   (4b)
   0.25000000000000   0.74600000000000   0.00000000000000   Al   (4b)
   0.74600000000000   0.25000000000000   0.00000000000000   Al   (4b)
   0.03200000000000   0.59600000000000   0.88800000000000   Al   (4b)
   0.59600000000000   0.03200000000000   0.88800000000000   Al   (4b)
  -0.14700000000000   0.39500000000000   0.88800000000000   Al   (4b)
   0.39500000000000  -0.14700000000000   0.88800000000000   Al   (4b)
   0.04200000000000   0.60200000000000   0.66600000000000   Al   (4b)
   0.60200000000000   0.04200000000000   0.66600000000000   Al   (4b)
  -0.27100000000000   0.28700000000000   0.66900000000000   Al   (4b)
   0.28700000000000  -0.27100000000000   0.66900000000000   Al   (4b)
  -0.15800000000000   0.42000000000000   0.53200000000000   Al   (4b)
   0.42000000000000  -0.15800000000000   0.53200000000000   Al   (4b)
   0.08600000000000   0.08600000000000   0.38200000000000   Co   (2a)
   0.60400000000000   0.60400000000000   0.37300000000000   Co   (2a)
  -0.09100000000000  -0.09100000000000   0.01200000000000   Co   (2a)
   0.59700000000000   0.59700000000000   0.01800000000000   Co   (2a)
  -0.08700000000000  -0.08700000000000   0.61700000000000   Co   (2a)
   0.39500000000000   0.39500000000000   0.62600000000000   Co   (2a)
   0.09000000000000   0.09000000000000  -0.01300000000000   Co   (2a)
   0.40200000000000   0.40200000000000  -0.01800000000000   Co   (2a)
   0.02600000000000   0.61000000000000   0.27700000000000   Co   (4b)
   0.61000000000000   0.02600000000000   0.27700000000000   Co   (4b)
  -0.04700000000000   0.41100000000000   0.72200000000000   Co   (4b)
   0.41100000000000  -0.04700000000000   0.72200000000000   Co   (4b)
```

$TaTi_3$ (BCC SQS-16): AB3_mC32_8_4a_12a - CIF

```
# AFLOW.org Repositories
# TaTi/AB3_mC32_8_4a_12a-001.AB params=13.8779590179,0.333333333333,
    ↪ 0.666666666667,109.471220634,-0.0,0.5,0.5,0.75,0.375,0.9375,
    ↪ 0.125,0.8125,0.875,0.1875,0.625,0.0625,0.75,0.375,0.875,0.6875
    ↪ ,-0.0,-0.0,0.5,0.25,0.625,0.5625,0.75,0.875,0.25,0.125,0.375,
    ↪ 0.4375,0.125,0.3125,0.25,0.625 SG=8 [ANRL doi: 10.1016/
    ↪ j.commatsci.2017.01.017 (part 1), doi: 10.1016/
    ↪ j.commatsci.2018.10.043 (part 2)]
data_TaTi
_pd_phase_name AB3_mC32_8_4a_12a-001.AB

_chemical_name_mineral 'TaTi3'
_chemical_formula_sum 'Ta Ti3'

loop_
_publ_author_name
 'C. Jiang'
 'C. Wolverton'
 'J. Sofo'
 'L.-Q. Chen'
 'Z.-K. Liu'
_journal_name_full_name
;
 Physical Review B
;
_journal_volume 69
_journal_year 2004
_journal_page_first 214202
_journal_page_last 214202
_publ_Section_title
;
 First-principles study of binary bcc alloys using special quasirandom
    ↪ structures
;

_aflow_title 'TaTi$_{3}$ (BCC SQS-16) Structure'
_aflow_proto 'AB3_mC32_8_4a_12a'
_aflow_params 'a,b/a,c/a,\beta,x_{1},z_{1},x_{2},z_{2},x_{3},z_{3},x_{4
    ↪ },z_{4},x_{5},z_{5},x_{6},z_{6},x_{7},z_{7},x_{8},z_{8},x_{9},
    ↪ z_{9},x_{10},z_{10},x_{11},z_{11},x_{12},z_{12},x_{13},z_{13},
    ↪ x_{14},z_{14},x_{15},z_{15},x_{16},z_{16}'
_aflow_params_values '13.8779590179,0.333333333333,0.666666666667,
    ↪ 109.471220634,0.0,0.5,0.5,0.25,0.625,0.0625,0.875,0.1875,0.125,
    ↪ 0.8125,0.375,0.9375,0.25,0.625,0.125,0.3125,0.0,0.0,0.5,0.75,
    ↪ 0.375,0.4375,0.25,0.125,0.75,0.875,0.625,0.5625,0.875,0.6875,
    ↪ 0.75,0.375'
_aflow_Strukturbericht 'None'
_aflow_Pearson 'mC32'

_cell_length_a   13.8779590179
_cell_length_b   4.6259863393
_cell_length_c   9.2519726786
_cell_angle_alpha   90.0000000000
_cell_angle_beta   109.4712206340
```

```
_cell_angle_gamma    90.0000000000
_symmetry_space_group_name_H-M    'Cm'
_symmetry_Int_Tables_Number    8
loop_
_symmetry_equiv_pos_site_id
_symmetry_equiv_pos_as_xyz
   1  x,y,z
   2  x,-y,z
   3  x+1/2,y+1/2,z
   4  x+1/2,-y+1/2,z
loop_
 _atom_site_label
 _atom_site_occupancy
 _atom_site_fract_x
 _atom_site_fract_y
 _atom_site_fract_z
 _atom_site_thermal_displace_type
 _atom_site_B_iso_or_equiv
 _atom_site_type_symbol
 _atom_site_symmetry_multiplicity
 _atom_site_Wyckoff_label
Ta1 1.0000000000  0.0000000000  0.0000000000  0.5000000000  Biso  1.0  Ta  2  a
Ta2 1.0000000000  0.5000000000  -0.0000000000  0.2500000000  Biso  1.0  Ta  2  a
Ta3 1.0000000000  0.6250000000  0.0000000000  0.0625000000  Biso  1.0  Ta  2  a
Ta4 1.0000000000  0.8750000000  0.0000000000  0.1875000000  Biso  1.0  Ta  2  a
Ti1 1.0000000000  0.1250000000  -0.0000000000  0.8125000000  Biso  1.0  Ti  2  a
Ti2 1.0000000000  0.3750000000  -0.0000000000  0.9375000000  Biso  1.0  Ti  2  a
Ti3 1.0000000000  0.2500000000  -0.0000000000  0.6250000000  Biso  1.0  Ti  2  a
Ti4 1.0000000000  0.1250000000  -0.0000000000  0.3125000000  Biso  1.0  Ti  2  a
Ti5 1.0000000000  0.0000000000  0.0000000000  0.0000000000  Biso  1.0  Ti  2  a
Ti6 1.0000000000  0.5000000000  -0.0000000000  0.7500000000  Biso  1.0  Ti  2  a
Ti7 1.0000000000  0.3750000000  -0.0000000000  0.4375000000  Biso  1.0  Ti  2  a
Ti8 1.0000000000  0.2500000000  -0.0000000000  0.1250000000  Biso  1.0  Ti  2  a
Ti9 1.0000000000  0.7500000000  0.0000000000  0.8750000000  Biso  1.0  Ti  2  a
Ti10 1.0000000000  0.6250000000  0.0000000000  0.5625000000  Biso  1.0  Ti  2  a
Ti11 1.0000000000  0.8750000000  0.0000000000  0.6875000000  Biso  1.0  Ti  2  a
Ti12 1.0000000000  0.7500000000  0.0000000000  0.3750000000  Biso  1.0  Ti  2  a
```

$TaTi_3$ (BCC SQS-16): AB3_mC32_8_4a_12a - POSCAR

```
AB3_mC32_8_4a_12a & a,b/a,c/a,beta,x1,z1,x2,z2,x3,z3,x4,z4,x5,z5,x6,z6,
    ↪ x7,z7,x8,z8,x9,z9,x10,z10,x11,z11,x12,z12,x13,z13,x14,z14,x15,
    ↪ z15,x16,z16 --params=13.8779590179,0.333333333333,
    ↪ 0.666666666667,109.471220634,0.0,0.5,0.5,0.25,0.625,0.0625,
    ↪ 0.875,0.1875,0.125,0.8125,0.375,0.9375,0.25,0.625,0.125,0.3125,
    ↪ 0.0,0.0,0.5,0.75,0.375,0.4375,0.25,0.125,0.75,0.875,0.625,
    ↪ 0.5625,0.875,0.6875,0.75,0.375 & Cm C_{s}^{3} #8 (a^16) & mC32
    ↪ & None & TaTi3 & TaTi3 & C. Jiang et al., Phys. Rev. B 69,
    ↪ 214202(2004)
   1.00000000000000
   6.93897950895000   -2.31299316965000    0.00000000000000
   6.93897950895000    2.31299316965000    0.00000000000000
  -3.08399089279196    0.00000000000000    8.72284349388071
     Ta        Ti
      4        12
Direct
   0.00000000000000    0.00000000000000    0.50000000000000    Ta    (2a)
   0.50000000000000    0.50000000000000    0.25000000000000    Ta    (2a)
   0.62500000000000    0.62500000000000    0.06250000000000    Ta    (2a)
   0.87500000000000    0.87500000000000    0.18750000000000    Ta    (2a)
   0.12500000000000    0.12500000000000    0.81250000000000    Ti    (2a)
   0.37500000000000    0.37500000000000    0.93750000000000    Ti    (2a)
   0.25000000000000    0.25000000000000    0.62500000000000    Ti    (2a)
   0.12500000000000    0.12500000000000    0.31250000000000    Ti    (2a)
   0.00000000000000    0.00000000000000    0.00000000000000    Ti    (2a)
   0.50000000000000    0.50000000000000    0.75000000000000    Ti    (2a)
   0.37500000000000    0.37500000000000    0.43750000000000    Ti    (2a)
   0.25000000000000    0.25000000000000    0.12500000000000    Ti    (2a)
   0.75000000000000    0.75000000000000    0.87500000000000    Ti    (2a)
   0.62500000000000    0.62500000000000    0.56250000000000    Ti    (2a)
   0.87500000000000    0.87500000000000    0.68750000000000    Ti    (2a)
   0.75000000000000    0.75000000000000    0.37500000000000    Ti    (2a)
```

$TaTi_3$ (BCC SQS-16): AB3_mC32_8_4a_4a4b - CIF

```
# AFLOW.org Repositories
# TaTi/AB3_mC32_8_4a_4a4b-001.AB params=9.2519726786,1.0,0.707106781188,
    ↪ 90.0,0.0,0.0,0.0,0.5,0.25,0.25,0.5,0.5,0.25,0.75,0.5,0.0,0.75,
    ↪ 0.75,0.75,0.25,0.75,0.25,0.0,0.75,0.25,0.5,0.0,0.25,0.75,0.0,
    ↪ 0.25,0.25 SG=8 [ANRL doi: 10.1016/j.commatsci.2017.01.017 (part
    ↪ 1), doi: 10.1016/j.commatsci.2018.10.043 (part 2)]
data_TaTi
_pd_phase_name AB3_mC32_8_4a_4a4b-001.AB

_chemical_name_mineral 'TaTi3'
_chemical_formula_sum 'Ta Ti3'

loop_
_publ_author_name
 'T. Chakraborty'
 'J. Rogal'
 'R. Drautz'
_journal_name_full_name
;
 Physical Review B
;
_journal_volume 94
_journal_year 2016
_journal_page_first 224104
_journal_page_last 224104
_publ_Section_title
;
 Unraveling the composition dependence of the martensitic transformation
    ↪ temperature: A first-principles study of Ti-Ta alloys
;

_aflow_title 'TaTi$_{3}$ (BCC SQS-16) Structure'
_aflow_proto 'AB3_mC32_8_4a_4a4b'
_aflow_params 'a,b/a,c/a,\beta,x_{1},z_{1},x_{2},z_{2},x_{3},z_{3},x_{4
    ↪ },z_{4},x_{5},z_{5},x_{6},z_{6},x_{7},z_{7},x_{8},z_{8},x_{9},
    ↪ y_{9},z_{9},x_{10},y_{10},z_{10},x_{11},y_{11},z_{11},x_{12},y_
    ↪ {12},z_{12}'
_aflow_params_values '9.2519726786,1.0,0.707106781188,90.0,-0.0,0.0,-0.0
    ↪ ,0.5,0.75,0.75,0.5,0.5,0.75,0.25,0.5,0.0,0.25,0.25,0.25,0.75,
    ↪ 0.75,0.75,0.0,0.75,0.75,0.5,0.5,0.75,0.25,0.5,0.75,0.75'
_aflow_Strukturbericht 'None'
_aflow_Pearson 'mC32'

_cell_length_a    9.2519726786
_cell_length_b    9.2519726786
_cell_length_c    6.5421326204
_cell_angle_alpha    90.0000000000
_cell_angle_beta    90.0000000000
_cell_angle_gamma    90.0000000000
_symmetry_space_group_name_H-M    'Cm'
_symmetry_Int_Tables_Number    8
loop_
_symmetry_equiv_pos_site_id
_symmetry_equiv_pos_as_xyz
   1  x,y,z
   2  x,-y,z
   3  x+1/2,y+1/2,z
   4  x+1/2,-y+1/2,z
loop_
 _atom_site_label
 _atom_site_occupancy
 _atom_site_fract_x
 _atom_site_fract_y
 _atom_site_fract_z
 _atom_site_thermal_displace_type
 _atom_site_B_iso_or_equiv
 _atom_site_type_symbol
 _atom_site_symmetry_multiplicity
 _atom_site_Wyckoff_label
Ta1 1.0000000000  -0.0000000000  0.0000000000  0.0000000000  Biso  1.0  Ta  2  a
Ta2 1.0000000000  -0.0000000000  0.0000000000  0.5000000000  Biso  1.0  Ta  2  a
Ta3 1.0000000000  0.7500000000  0.0000000000  0.7500000000  Biso  1.0  Ta  2  a
Ta4 1.0000000000  0.5000000000  -0.0000000000  0.5000000000  Biso  1.0  Ta  2  a
Ti1 1.0000000000  0.7500000000  -0.0000000000  0.2500000000  Biso  1.0  Ti  2  a
Ti2 1.0000000000  0.5000000000  -0.0000000000  0.0000000000  Biso  1.0  Ti  2  a
Ti3 1.0000000000  0.2500000000  -0.0000000000  0.2500000000  Biso  1.0  Ti  2  a
Ti4 1.0000000000  0.2500000000  -0.0000000000  0.7500000000  Biso  1.0  Ti  2  a
Ti5 1.0000000000  0.7500000000  0.7500000000  0.0000000000  Biso  1.0  Ti  4  b
Ti6 1.0000000000  0.7500000000  0.7500000000  0.5000000000  Biso  1.0  Ti  4  b
Ti7 1.0000000000  0.5000000000  0.7500000000  0.2500000000  Biso  1.0  Ti  4  b
Ti8 1.0000000000  0.5000000000  0.7500000000  0.7500000000  Biso  1.0  Ti  4  b
```

$TaTi_3$ (BCC SQS-16): AB3_mC32_8_4a_4a4b - POSCAR

```
AB3_mC32_8_4a_4a4b & a,b/a,c/a,beta,x1,z1,x2,z2,x3,z3,x4,z4,x5,z5,x6,z6,
    ↪ x7,z7,x8,z8,x9,y9,z9,x10,y10,z10,x11,y11,z11,x12,y12,z12 --
    ↪ params=9.2519726786,1.0,0.707106781188,90.0,-0.0,0.0,-0.0,0.5,
    ↪ 0.75,0.75,0.5,0.5,0.75,0.25,0.5,0.0,0.25,0.25,0.25,0.75,0.75,
    ↪ 0.75,0.0,0.75,0.75,0.5,0.5,0.75,0.25,0.5,0.75,0.75 & Cm C_{s}^{
    ↪ 3} #8 (a^8b^4) & mC32 & None & TaTi3 & TaTi3 & T. Chakraborty
    ↪ and J. Rogal and R. Drautz, Phys. Rev. B 94, 224104(2016)
   1.00000000000000
   4.62598633930000   -4.62598633930000    0.00000000000000
   4.62598633930000    4.62598633930000    0.00000000000000
   0.00000000000000    0.00000000000000    6.54213262040000
     Ta        Ti
      4        12
Direct
   0.00000000000000    0.00000000000000    0.00000000000000    Ta    (2a)
   0.00000000000000    0.00000000000000    0.50000000000000    Ta    (2a)
   0.75000000000000    0.75000000000000    0.75000000000000    Ta    (2a)
   0.50000000000000    0.50000000000000    0.50000000000000    Ta    (2a)
   0.75000000000000    0.75000000000000    0.25000000000000    Ti    (2a)
   0.50000000000000    0.50000000000000    0.00000000000000    Ti    (2a)
   0.25000000000000    0.25000000000000    0.25000000000000    Ti    (2a)
   0.25000000000000    0.25000000000000    0.75000000000000    Ti    (2a)
   0.00000000000000    1.50000000000000    0.00000000000000    Ti    (4b)
   1.50000000000000    0.00000000000000    0.00000000000000    Ti    (4b)
   0.00000000000000    1.50000000000000    0.50000000000000    Ti    (4b)
   1.50000000000000    0.00000000000000    0.50000000000000    Ti    (4b)
  -0.25000000000000    1.25000000000000    0.25000000000000    Ti    (4b)
   1.25000000000000   -0.25000000000000    0.25000000000000    Ti    (4b)
  -0.25000000000000    1.25000000000000    0.75000000000000    Ti    (4b)
   1.25000000000000   -0.25000000000000    0.75000000000000    Ti    (4b)
```

$F5_{11}$ ($KNO_2$) *(obsolete)*: ABC2_mC8_8_a_a_b - CIF

```
# CIF file
data_findsym-output
_audit_creation_method FINDSYM

_chemical_name_mineral 'Potassium nitrite'
_chemical_formula_sum 'K N O2'

loop_
_publ_author_name
 'G. E. Ziegler'
_journal_name_full_name
;
 Zeitschrift f{\"u}r Kristallographie - Crystalline Materials
;
_journal_volume 94
_journal_year 1936
_journal_page_first 491
_journal_page_last 499
```

```
_publ_Section_title
;
 The Crystal Structure of Potassium Nitrite, KNO$_{2}$
;

_aflow_title '$F5_{11}$ (KNO$_{2}$) ({\em{obsolete}}) Structure'
_aflow_proto 'ABC2_mC8_8_a_a_b'
_aflow_params 'a,b/a,c/a,\beta,x_{1},z_{1},x_{2},z_{2},x_{3},y_{3},z_{3
    ↪ }'
_aflow_params_values '7.31,0.682626538988,0.608755129959,114.8333,0.0,
    ↪ 0.0,0.486,0.5,-0.083,0.306,0.444'
_aflow_Strukturbericht '$F5_{11}$'
_aflow_Pearson 'mC8'

_symmetry_space_group_name_H-M "C 1 m 1"
_symmetry_Int_Tables_number 8

_cell_length_a     7.31000
_cell_length_b     4.99000
_cell_length_c     4.45000
_cell_angle_alpha  90.00000
_cell_angle_beta   114.83330
_cell_angle_gamma  90.00000

loop_
_space_group_symop_id
_space_group_symop_operation_xyz
1 x,y,z
2 x,-y,z
3 x+1/2,y+1/2,z
4 x+1/2,-y+1/2,z

loop_
_atom_site_label
_atom_site_type_symbol
_atom_site_symmetry_multiplicity
_atom_site_Wyckoff_label
_atom_site_fract_x
_atom_site_fract_y
_atom_site_fract_z
_atom_site_occupancy
K1 K 2 a 0.00000 0.00000 0.00000 1.00000
N1 N 2 a 0.48600 0.00000 0.50000 1.00000
O1 O 4 b -0.08300 0.30600 0.44400 1.00000
```

$F5_{11}$ ($KNO_2$) (*obsolete*): ABC2_mC8_8_a_a_b - POSCAR

```
ABC2_mC8_8_a_a_b & a,b/a,c/a,beta,x1,z1,x2,z2,x3,y3,z3 --params=7.31,
    ↪ 0.682626538988,0.608755129959,114.8333,0.0,0.0,0.486,0.5,-0.083
    ↪ ,0.306,0.444 & Cm C_{s}^{3} #8 (a^2b) & mC8 & $F5_{11}$ & KNO2
    ↪ & Potassium nitrite & G. E. Ziegler, Zeitschrift f{"u}r
    ↪ Kristallographie - Crystalline Materials 94, 491-499 (1936)
   1.00000000000000
   3.65500000000000  -2.49500000000000   0.00000000000000
   3.65500000000000   2.49500000000000   0.00000000000000
  -1.86890925105819   0.00000000000000   4.03852426157243
     K      N      O
     1      1      2
Direct
   0.00000000000000   0.00000000000000   0.00000000000000   K   (2a)
   0.48600000000000   0.48600000000000   0.50000000000000   N   (2a)
  -0.38900000000000   0.22300000000000   0.44400000000000   O   (4b)
   0.22300000000000  -0.38900000000000   0.44400000000000   O   (4b)
```

Nacrite [$Al_2Si_2O_5(OH)_4$, $S5_4$]: A2B4C9D2_mC68_9_2a_4a_9a_2a - CIF

```
# CIF file
data_findsym-output
_audit_creation_method FINDSYM

_chemical_name_mineral 'Nacrite'
_chemical_formula_sum 'Al2 H4 O9 Si2'

loop_
_publ_author_name
 'A. P. Zhukhlistov'
_journal_name_full_name
;
 Crystallography Reports
;
_journal_volume 53
_journal_year 2008
_journal_page_first 76
_journal_page_last 82
_publ_Section_title
;
 Crystal structure of nacrite from the electron diffraction data
;

_aflow_title 'Nacrite [Al$_{2}$Si$_{2}$O$_{5}$(OH)$_{4}$, $S5_{4}$]
    ↪ Structure'
_aflow_proto 'A2B4C9D2_mC68_9_2a_4a_9a_2a'
_aflow_params 'a,b/a,c/a,\beta,x_{1},y_{1},z_{1},x_{2},y_{2},z_{2},x_{3
    ↪ },y_{3},z_{3},x_{4},y_{4},z_{4},x_{5},y_{5},z_{5},x_{6},y_{6},
    ↪ z_{6},x_{7},y_{7},z_{7},x_{8},y_{8},z_{8},x_{9},y_{9},z_{9},x_{
    ↪ 10},y_{10},z_{10},x_{11},y_{11},z_{11},x_{12},y_{12},z_{12},x_{
    ↪ 13},y_{13},z_{13},x_{14},y_{14},z_{14},x_{15},y_{15},z_{15},x_{
    ↪ 16},y_{16},z_{16},x_{17},y_{17},z_{17}'
_aflow_params_values '8.91,0.577328843996,1.63782267116,100.5,0.8275,
    ↪ 0.4297,-0.0002,0.1628,0.4245,0.0004,0.426,0.318,0.096,0.599,
    ↪ 0.41,0.378,0.3,0.346,0.38,-0.001,0.386,0.38,0.1349,0.0011,
    ↪ 0.2215,0.6809,0.0027,0.2362,-0.0881,0.3075,0.2371,0.0214,0.3111
    ↪ ,0.0788,0.7143,0.238,0.0794,0.3271,0.2379,0.0748,0.164,0.1153,-
    ↪ 0.066,0.7853,0.1207,-0.065,0.4772,0.0432,-0.0651,0.0632,0.2796,
    ↪ 0.1897,0.7342,0.2656,0.1907'
_aflow_Strukturbericht '$S5_{4}$'
_aflow_Pearson 'mC68'

_symmetry_space_group_name_H-M "C 1 c 1"
_symmetry_Int_Tables_number 9

_cell_length_a     8.91000
_cell_length_b     5.14400
_cell_length_c     14.59300
_cell_angle_alpha  90.00000
_cell_angle_beta   100.50000
_cell_angle_gamma  90.00000

loop_
_space_group_symop_id
_space_group_symop_operation_xyz
1 x,y,z
2 x,-y,z+1/2
3 x+1/2,y+1/2,z
4 x+1/2,-y+1/2,z+1/2

loop_
_atom_site_label
_atom_site_type_symbol
_atom_site_symmetry_multiplicity
_atom_site_Wyckoff_label
_atom_site_fract_x
_atom_site_fract_y
_atom_site_fract_z
_atom_site_occupancy
Al1 Al 4 a 0.82750 0.42970 -0.00020 1.00000
Al2 Al 4 a 0.16280 0.42450 0.00040 1.00000
H1 H 4 a 0.42600 0.31800 0.09600 1.00000
H2 H 4 a 0.59900 0.41000 0.37800 1.00000
H3 H 4 a 0.30000 0.34600 0.38000 1.00000
H4 H 4 a -0.00100 0.38600 0.38000 1.00000
O1 O 4 a 0.13490 0.00110 0.22150 1.00000
O2 O 4 a 0.68090 0.00270 0.23620 1.00000
O3 O 4 a -0.08810 0.30750 0.23710 1.00000
O4 O 4 a 0.02140 0.31110 0.07880 1.00000
O5 O 4 a 0.71430 0.23800 0.07940 1.00000
O6 O 4 a 0.32710 0.23790 0.07480 1.00000
O7 O 4 a 0.16400 0.11530 -0.06600 1.00000
O8 O 4 a 0.78530 0.12070 -0.06500 1.00000
O9 O 4 a 0.47720 0.04320 -0.06510 1.00000
Si1 Si 4 a 0.06320 0.27960 0.18970 1.00000
Si2 Si 4 a 0.73420 0.26560 0.19070 1.00000
```

Nacrite [$Al_2Si_2O_5(OH)_4$, $S5_4$]: A2B4C9D2_mC68_9_2a_4a_9a_2a - POSCAR

```
A2B4C9D2_mC68_9_2a_4a_9a_2a & a,b/a,c/a,beta,x1,y1,z1,x2,y2,z2,x3,y3,z3,
    ↪ x4,y4,z4,x5,y5,z5,x6,y6,z6,x7,y7,z7,x8,y8,z8,x9,y9,z9,x10,y10,
    ↪ z10,x11,y11,z11,x12,y12,z12,x13,y13,z13,x14,y14,z14,x15,y15,z15
    ↪ ,x16,y16,z16,x17,y17,z17 --params=8.91,0.577328843996,
    ↪ 1.63782267116,100.5,0.8275,0.4297,-0.0002,0.1628,0.4245,0.0004,
    ↪ 0.426,0.318,0.096,0.599,0.41,0.378,0.3,0.346,0.38,-0.001,0.386,
    ↪ 0.38,0.1349,0.0011,0.2215,0.6809,0.0027,0.2362,-0.0881,0.3075,
    ↪ 0.2371,0.0214,0.3111,0.0788,0.7143,0.238,0.0794,0.3271,0.2379,
    ↪ 0.0748,0.164,0.1153,-0.066,0.7853,0.1207,-0.065,0.4772,0.0432,-
    ↪ 0.0651,0.0632,0.2796,0.1897,0.7342,0.2656,0.1907 & Cc C_{s}^{4}
    ↪ #9 (a^17) & mC68 & $S5_{4}$ & Al2H4O9Si2 & Nacrite & A. P.
    ↪ Zhukhlistov, Crystal. Rep. 53, 76-82 (2008)
   1.00000000000000
   4.45500000000000  -2.57200000000000   0.00000000000000
   4.45500000000000   2.57200000000000   0.00000000000000
  -2.65936302350691   0.00000000000000  14.34863886608080
    Al     H     O    Si
     4     8    18     4
Direct
   0.39780000000000   1.25720000000000  -0.00020000000000   Al   (4a)
   1.25720000000000   0.39780000000000   0.49980000000000   Al   (4a)
  -0.26170000000000   0.58730000000000   0.00040000000000   Al   (4a)
   0.58730000000000  -0.26170000000000   0.50040000000000   Al   (4a)
   0.10800000000000   0.74400000000000   0.09600000000000   H    (4a)
   0.74400000000000   0.10800000000000   0.59600000000000   H    (4a)
   0.18900000000000   1.00900000000000   0.37800000000000   H    (4a)
   1.00900000000000   0.18900000000000   0.87800000000000   H    (4a)
  -0.04600000000000   0.64600000000000   0.38000000000000   H    (4a)
   0.64600000000000  -0.04600000000000   0.88000000000000   H    (4a)
  -0.38700000000000   0.38500000000000   0.38000000000000   H    (4a)
   0.38500000000000  -0.38700000000000   0.88000000000000   H    (4a)
   0.13380000000000   0.13600000000000   0.22150000000000   O    (4a)
   0.13600000000000   0.13380000000000   0.72150000000000   O    (4a)
   0.67820000000000   0.68360000000000   0.23620000000000   O    (4a)
   0.68360000000000   0.67820000000000   0.73620000000000   O    (4a)
  -0.39560000000000   0.21940000000000   0.23710000000000   O    (4a)
   0.21940000000000  -0.39560000000000   0.73710000000000   O    (4a)
  -0.28970000000000   0.33250000000000   0.07880000000000   O    (4a)
   0.33250000000000  -0.28970000000000   0.57880000000000   O    (4a)
   0.47630000000000   0.95230000000000   0.07940000000000   O    (4a)
   0.95230000000000   0.47630000000000   0.57940000000000   O    (4a)
   0.08920000000000   0.56500000000000   0.07480000000000   O    (4a)
   0.56500000000000   0.08920000000000   0.57480000000000   O    (4a)
   0.04870000000000   0.27930000000000  -0.06600000000000   O    (4a)
   0.27930000000000   0.04870000000000   0.43400000000000   O    (4a)
   0.66460000000000   0.90600000000000  -0.06500000000000   O    (4a)
   0.90600000000000   0.66460000000000   0.43500000000000   O    (4a)
   0.43400000000000   0.52040000000000  -0.06510000000000   O    (4a)
   0.52040000000000   0.43400000000000   0.43490000000000   O    (4a)
  -0.21640000000000   0.34280000000000   0.18970000000000   Si   (4a)
   0.34280000000000  -0.21640000000000   0.68970000000000   Si   (4a)
   0.46860000000000   0.99980000000000   0.19070000000000   Si   (4a)
   0.99980000000000   0.46860000000000   0.69070000000000   Si   (4a)
```

Chrysotile ($Mg_3Si_2O_5(OH)_4$): A3B5C4D2_mC56_9_3a_5a_4a_2a - CIF

```
# CIF file
data_findsym-output
_audit_creation_method FINDSYM

_chemical_name_mineral 'Chrysotile'
_chemical_formula_sum 'Mg3 O5 (OH)4 Si2'

loop_
_publ_author_name
 'G. Falini'
 'E. Foresti'
 'M. Gazzano'
 'A. F. Gualtieri'
 'M. Leoni'
 'I G. Lesci'
 'N. Roveri'
_journal_name_full_name
;
 Chemistry - A European Journal
;
_journal_volume 10
_journal_year 2004
_journal_page_first 3043
_journal_page_last 3049
_publ_Section_title
;
 Tubular-Shaped Stoichiometric Chrysotile Nanocrystals
;

_aflow_title 'Chrysotile (Mg$_{3}$Si$_{2}$O$_{5}$(OH)$_{4}$) Structure'
_aflow_proto 'A3B5C4D2_mC56_9_3a_5a_4a_2a'
_aflow_params 'a,b/a,c/a,\beta,x_{1},y_{1},z_{1},x_{2},y_{2},z_{2},x_{3
    ↪ },y_{3},z_{3},x_{4},y_{4},z_{4},x_{5},y_{5},z_{5},x_{6},y_{6},
    ↪ z_{6},x_{7},y_{7},z_{7},x_{8},y_{8},z_{8},x_{9},y_{9},z_{9},x_{
    ↪ 10},y_{10},z_{10},x_{11},y_{11},z_{11},x_{12},y_{12},z_{12},x_{
    ↪ 13},y_{13},z_{13},x_{14},y_{14},z_{14}'
_aflow_params_values '5.34,1.73052434457,2.75074906367,93.66,0.8852,
    ↪ 0.198,0.2303,0.3747,0.3771,0.2327,0.8818,0.5235,0.2204,0.0726,
    ↪ 0.1978,0.0,0.2387,0.475,0.0023,0.7395,0.4214,0.0063,0.0438,
    ↪ 0.3596,0.156,0.532,0.537,0.155,0.7299,0.363,0.2878,0.2395,0.197
    ↪ ,0.2791,0.5334,0.2053,0.1807,0.2025,0.5416,0.296,0.0257,0.3641,
    ↪ 0.041,0.5208,0.5335,0.04'
_aflow_Strukturbericht 'None'
_aflow_Pearson 'mC56'

_symmetry_space_group_name_H-M "C 1 c 1"
_symmetry_Int_Tables_number 9

_cell_length_a     5.34000
_cell_length_b     9.24100
_cell_length_c     14.68900
_cell_angle_alpha  90.00000
_cell_angle_beta   93.66000
_cell_angle_gamma  90.00000

loop_
_space_group_symop_id
_space_group_symop_operation_xyz
1 x,y,z
2 x,-y,z+1/2
3 x+1/2,y+1/2,z
4 x+1/2,-y+1/2,z+1/2

loop_
_atom_site_label
_atom_site_type_symbol
_atom_site_symmetry_multiplicity
_atom_site_Wyckoff_label
_atom_site_fract_x
_atom_site_fract_y
_atom_site_fract_z
_atom_site_occupancy
Mg1 Mg 4 a 0.88520 0.19800 0.23030 1.00000
Mg2 Mg 4 a 0.37470 0.37710 0.23270 1.00000
Mg3 Mg 4 a 0.88180 0.52350 0.22040 1.00000
O1 O 4 a 0.07260 0.19780 0.00000 1.00000
O2 O 4 a 0.23870 0.47500 0.00230 1.00000
O3 O 4 a 0.73950 0.42140 0.00630 1.00000
O4 O 4 a 0.04380 0.35960 0.15600 1.00000
O5 O 4 a 0.53200 0.53700 0.15500 1.00000
OH1 OH 4 a 0.72990 0.36300 0.28780 1.00000
OH2 OH 4 a 0.23950 0.19700 0.27910 1.00000
OH3 OH 4 a 0.53340 0.20530 0.18070 1.00000
OH4 OH 4 a 0.20250 0.54160 0.29600 1.00000
Si1 Si 4 a 0.02570 0.36410 0.04100 1.00000
Si2 Si 4 a 0.52080 0.53350 0.04000 1.00000
```

Chrysotile ($Mg_3Si_2O_5(OH)_4$): A3B5C4D2_mC56_9_3a_5a_4a_2a - POSCAR

```
A3B5C4D2_mC56_9_3a_5a_4a_2a & a,b/a,c/a,beta,x1,y1,z1,x2,y2,z2,x3,y3,z3,
    ↪ x4,y4,z4,x5,y5,z5,x6,y6,z6,x7,y7,z7,x8,y8,z8,x9,y9,z9,x10,y10,
    ↪ z10,x11,y11,z11,x12,y12,z12,x13,y13,z13,x14,y14,z14 --params=
    ↪ 5.34,1.73052434457,2.75074906367,93.66,0.8852,0.198,0.2303,
    ↪ 0.3747,0.3771,0.2327,0.8818,0.5235,0.2204,0.0726,0.1978,0.0,
    ↪ 0.2387,0.475,0.0023,0.7395,0.4214,0.0063,0.0438,0.3596,0.156,
    ↪ 0.532,0.537,0.155,0.7299,0.363,0.2878,0.2395,0.197,0.2791,
    ↪ 0.5334,0.2053,0.1807,0.2025,0.5416,0.296,0.0257,0.3641,0.041,
    ↪ 0.5208,0.5335,0.04 & Cc C_{s}^{4} #9 (a^14) & mC56 & None &
    ↪ Mg3O5(OH)4Si2 & Chrysotile & G. Falini et al., Chem. Eur. J. 10
    ↪ , 3043-3049 (2004)
   1.00000000000000
   2.67000000000000  -4.62050000000000   0.00000000000000
   2.67000000000000   4.62050000000000   0.00000000000000
  -0.93768136424044   0.00000000000000  14.65904071415170
    Mg     O     OH     Si
     6    10      8      4
Direct
   0.68720000000000   1.08320000000000   0.23030000000000   Mg   (4a)
   1.08320000000000   0.68720000000000   0.73030000000000   Mg   (4a)
  -0.00240000000000   0.75180000000000   0.23270000000000   Mg   (4a)
   0.75180000000000  -0.00240000000000   0.73270000000000   Mg   (4a)
   0.35830000000000   1.40530000000000   0.22040000000000   Mg   (4a)
   1.40530000000000   0.35830000000000   0.72040000000000   Mg   (4a)
  -0.12520000000000   0.27040000000000   0.00000000000000   O    (4a)
   0.27040000000000  -0.12520000000000   0.50000000000000   O    (4a)
  -0.23630000000000   0.71370000000000   0.00230000000000   O    (4a)
   0.71370000000000  -0.23630000000000   0.50230000000000   O    (4a)
   0.31810000000000   1.16090000000000   0.00630000000000   O    (4a)
   1.16090000000000   0.31810000000000   0.50630000000000   O    (4a)
  -0.31580000000000   0.40340000000000   0.15600000000000   O    (4a)
   0.40340000000000  -0.31580000000000   0.65600000000000   O    (4a)
  -0.00500000000000   1.06900000000000   0.15500000000000   O    (4a)
   1.06900000000000  -0.00500000000000   0.65500000000000   O    (4a)
   0.36690000000000   1.09290000000000   0.28780000000000   OH   (4a)
   1.09290000000000   0.36690000000000   0.78780000000000   OH   (4a)
   0.04250000000000   0.43650000000000   0.27910000000000   OH   (4a)
   0.43650000000000   0.04250000000000   0.77910000000000   OH   (4a)
   0.32810000000000   0.73870000000000   0.18070000000000   OH   (4a)
   0.73870000000000   0.32810000000000   0.68070000000000   OH   (4a)
  -0.33910000000000   0.74410000000000   0.29600000000000   OH   (4a)
   0.74410000000000  -0.33910000000000   0.79600000000000   OH   (4a)
  -0.33840000000000   0.38980000000000   0.04100000000000   Si   (4a)
   0.38980000000000  -0.33840000000000   0.54100000000000   Si   (4a)
  -0.01270000000000   1.05430000000000   0.04000000000000   Si   (4a)
   1.05430000000000  -0.01270000000000   0.54000000000000   Si   (4a)
```

$Cs_6W_{11}O_{36}$: A6B36C11_mC212_9_6a_36a_11a - CIF

```
# CIF file
data_findsym-output
_audit_creation_method FINDSYM

_chemical_name_mineral 'Cs6O36W11'
_chemical_formula_sum 'Cs6 O36 W11'

loop_
_publ_author_name
 'K. Okada'
 'F. Marumo'
 'S. Iwai'
_journal_name_full_name
;
 Acta Crystallographica Section B: Structural Science
;
_journal_volume 34
_journal_year 1978
_journal_page_first 50
_journal_page_last 54
_publ_Section_title
;
 The Crystal Structure of Cs$_{6}$W$_{11}$O$_{36}$
;

_aflow_title 'Cs$_{6}$W$_{11}$O$_{36}$ Structure'
_aflow_proto 'A6B36C11_mC212_9_6a_36a_11a'
_aflow_params 'a,b/a,c/a,\beta,x_{1},y_{1},z_{1},x_{2},y_{2},z_{2},x_{3
    ↪ },y_{3},z_{3},x_{4},y_{4},z_{4},x_{5},y_{5},z_{5},x_{6},y_{6},
    ↪ z_{6},x_{7},y_{7},z_{7},x_{8},y_{8},z_{8},x_{9},y_{9},z_{9},x_{
    ↪ 10},y_{10},z_{10},x_{11},y_{11},z_{11},x_{12},y_{12},z_{12},x_{
    ↪ 13},y_{13},z_{13},x_{14},y_{14},z_{14},x_{15},y_{15},z_{15},x_{
    ↪ 16},y_{16},z_{16},x_{17},y_{17},z_{17},x_{18},y_{18},z_{18},x_{
    ↪ 19},y_{19},z_{19},x_{20},y_{20},z_{20},x_{21},y_{21},z_{21},x_{
    ↪ 22},y_{22},z_{22},x_{23},y_{23},z_{23},x_{24},y_{24},z_{24},x_{
    ↪ 25},y_{25},z_{25},x_{26},y_{26},z_{26},x_{27},y_{27},z_{27},x_{
    ↪ 28},y_{28},z_{28},x_{29},y_{29},z_{29},x_{30},y_{30},z_{30},x_{
    ↪ 31},y_{31},z_{31},x_{32},y_{32},z_{32},x_{33},y_{33},z_{33},x_{
    ↪ 34},y_{34},z_{34},x_{35},y_{35},z_{35},x_{36},y_{36},z_{36},x_{
    ↪ 37},y_{37},z_{37},x_{38},y_{38},z_{38},x_{39},y_{39},z_{39},x_{
    ↪ 40},y_{40},z_{40},x_{41},y_{41},z_{41},x_{42},y_{42},z_{42},x_{
    ↪ 43},y_{43},z_{43},x_{44},y_{44},z_{44},x_{45},y_{45},z_{45},x_{
    ↪ 46},y_{46},z_{46},x_{47},y_{47},z_{47},x_{48},y_{48},z_{48},x_{
    ↪ 49},y_{49},z_{49},x_{50},y_{50},z_{50},x_{51},y_{51},z_{51},x_{
    ↪ 52},y_{52},z_{52},x_{53},y_{53},z_{53}'
_aflow_params_values '12.577,0.577323686094,3.00548620498,102.81,0.831,
    ↪ 0.0001,0.7206,0.2029,0.0017,0.7786,0.2482,0.4974,0.0987,0.2817,
    ↪ 0.0025,0.8992,0.4748,0.4942,-0.0638,0.058,0.0037,0.0625,0.528,
    ↪ 0.01,-0.044,0.082,0.48,0.667,0.731,0.007,0.892,0.749,0.02,0.791
    ↪ ,0.814,0.009,0.007,0.207,0.324,-0.048,0.701,0.203,-0.048,0.053,
    ↪ 0.311,0.894,0.561,0.202,0.888,0.152,0.318,0.832,0.618,0.168,
    ↪ 0.836,0.372,0.326,0.852,0.859,0.191,0.847,0.432,0.303,0.791,-
    ↪ 0.047,0.202,0.784,0.64,0.2,0.012,0.127,0.314,0.006,0.414,0.164
    ↪ ,-0.014,-0.093,0.313,-0.013,0.411,0.187,0.667,0.896,0.315,0.668
    ↪ ,0.185,0.174,0.652,0.689,0.312,0.646,0.091,0.215,0.713,0.579,
    ↪ 0.307,0.707,0.804,0.323,0.044,0.317,0.172,0.048,0.487,0.188,
    ↪ 0.611,-0.013,0.33,0.607,0.219,0.004,-0.014,-0.057,0.478,0.832,
    ↪ 0.537,0.467,0.846,0.797,0.497,0.712,0.292,0.016,0.609,0.495,
    ↪ 0.498,0.649,0.502,0.013,0.05,0.25,0.2342,0.0,0.7808,0.2354,0.0,
    ↪ 0.483,0.2476,0.8347,0.5381,0.2682,0.6654,0.0553,0.2539,0.6648,-
    ↪ 0.0022,0.2447,0.8349,0.6325,0.0051,-0.0841,-0.0894,0.495,0.0835
    ↪ ,0.7428,0.0229,0.8341,0.7913,0.4872,0.6644,0.5166,0.0298,0.0'
_aflow_Strukturbericht 'None'
_aflow_Pearson 'mC212'

_symmetry_space_group_name_H-M "C 1 c 1"
_symmetry_Int_Tables_number 9

_cell_length_a     12.57700
_cell_length_b     7.26100
_cell_length_c     37.80000
_cell_angle_alpha  90.00000
```

```
_cell_angle_beta   102.81000
_cell_angle_gamma  90.00000

loop_
_space_group_symop_id
_space_group_symop_operation_xyz
1 x,y,z
2 x,-y,z+1/2
3 x+1/2,y+1/2,z
4 x+1/2,-y+1/2,z+1/2

loop_
_atom_site_label
_atom_site_type_symbol
_atom_site_symmetry_multiplicity
_atom_site_Wyckoff_label
_atom_site_fract_x
_atom_site_fract_y
_atom_site_fract_z
_atom_site_occupancy
Cs1 Cs 4 a 0.83100 0.00010 0.72060 1.00000
Cs2 Cs 4 a 0.20290 0.00170 0.77860 1.00000
Cs3 Cs 4 a 0.24820 0.49740 0.09870 1.00000
Cs4 Cs 4 a 0.28170 0.00250 0.89920 1.00000
Cs5 Cs 4 a 0.47480 0.49420 -0.06380 1.00000
Cs6 Cs 4 a 0.05800 0.00370 0.06250 1.00000
O1 O 4 a 0.52800 0.01000 -0.04400 1.00000
O2 O 4 a 0.08200 0.48000 0.66700 1.00000
O3 O 4 a 0.73100 0.00700 0.89200 1.00000
O4 O 4 a 0.74900 0.02000 0.79100 1.00000
O5 O 4 a 0.81400 0.00900 0.00700 1.00000
O6 O 4 a 0.20700 0.32400 -0.04800 1.00000
O7 O 4 a 0.70100 0.20300 -0.04800 1.00000
O8 O 4 a 0.05300 0.31100 0.89400 1.00000
O9 O 4 a 0.56100 0.20200 0.88800 1.00000
O10 O 4 a 0.15200 0.31800 0.83200 1.00000
O11 O 4 a 0.61800 0.16800 0.83600 1.00000
O12 O 4 a 0.37200 0.32600 0.85200 1.00000
O13 O 4 a 0.85900 0.19100 0.84700 1.00000
O14 O 4 a 0.43200 0.30300 0.79100 1.00000
O15 O 4 a -0.04700 0.20200 0.78400 1.00000
O16 O 4 a 0.64000 0.20000 0.01200 1.00000
O17 O 4 a 0.12700 0.31400 0.00600 1.00000
O18 O 4 a 0.41400 0.16400 -0.01400 1.00000
O19 O 4 a -0.09300 0.31300 -0.01300 1.00000
O20 O 4 a 0.41100 0.18700 0.66700 1.00000
O21 O 4 a 0.89600 0.31500 0.66800 1.00000
O22 O 4 a 0.18500 0.17400 0.65200 1.00000
O23 O 4 a 0.68900 0.31200 0.64600 1.00000
O24 O 4 a 0.09100 0.21500 0.71300 1.00000
O25 O 4 a 0.57900 0.30700 0.70700 1.00000
O26 O 4 a 0.80400 0.32300 0.04400 1.00000
O27 O 4 a 0.31700 0.17200 0.04800 1.00000
O28 O 4 a 0.48700 0.18800 0.61100 1.00000
O29 O 4 a -0.01300 0.33000 0.60700 1.00000
O30 O 4 a 0.21900 0.00400 -0.01400 1.00000
O31 O 4 a -0.05700 0.47800 0.83200 1.00000
O32 O 4 a 0.53700 0.46700 0.84600 1.00000
O33 O 4 a 0.79700 0.49700 0.71200 1.00000
O34 O 4 a 0.29200 0.01600 0.60900 1.00000
O35 O 4 a 0.49500 0.49800 0.64900 1.00000
O36 O 4 a 0.50200 0.01300 0.05000 1.00000
W1 W 4 a 0.25000 0.23420 0.00000 1.00000
W2 W 4 a 0.78080 0.23540 0.00000 1.00000
W3 W 4 a 0.48300 0.24760 0.83470 1.00000
W4 W 4 a 0.53810 0.26820 0.66540 1.00000
W5 W 4 a 0.05530 0.25390 0.66480 1.00000
W6 W 4 a -0.00220 0.24470 0.83490 1.00000
W7 W 4 a 0.63250 0.00510 -0.08410 1.00000
W8 W 4 a -0.08940 0.49500 0.08350 1.00000
W9 W 4 a 0.74280 0.02290 0.83410 1.00000
W10 W 4 a 0.79130 0.48720 0.66440 1.00000
W11 W 4 a 0.51660 0.02980 0.00000 1.00000
```

$Cs_6W_{11}O_{36}$: A6B36C11_mC212_9_6a_36a_11a - POSCAR

```
A6B36C11_mC212_9_6a_36a_11a & a,b/a,c/a,beta,x1,y1,z1,x2,y2,z2,x3,y3,z3,
    ↪ x4,y4,z4,x5,y5,z5,x6,y6,z6,x7,y7,z7,x8,y8,z8,x9,y9,z9,x10,y10,
    ↪ z10,x11,y11,z11,x12,y12,z12,x13,y13,z13,x14,y14,z14,x15,y15,z15
    ↪ ,x16,y16,z16,x17,y17,z17,x18,y18,z18,x19,y19,z19,x20,y20,z20,
    ↪ x21,y21,z21,x22,y22,z22,x23,y23,z23,x24,y24,z24,x25,y25,z25,x26
    ↪ ,y26,z26,x27,y27,z27,x28,y28,z28,x29,y29,z29,x30,y30,z30,x31,
    ↪ y31,z31,x32,y32,z32,x33,y33,z33,x34,y34,z34,x35,y35,z35,x36,y36
    ↪ ,z36,x37,y37,z37,x38,y38,z38,x39,y39,z39,x40,y40,z40,x41,y41,
    ↪ z41,x42,y42,z42,x43,y43,z43,x44,y44,z44,x45,y45,z45,x46,y46,z46
    ↪ ,x47,y47,z47,x48,y48,z48,x49,y49,z49,x50,y50,z50,x51,y51,z51,
    ↪ x52,y52,z52,x53,y53,z53 --params=12.577,0.577323686094,
    ↪ 3.00548620498,102.81,0.831,0.0001,0.7206,0.2029,0.0017,0.7786,
    ↪ 0.2482,0.4974,0.0987,0.2817,0.0025,0.8992,0.4748,0.4942,-0.0638
    ↪ ,0.058,0.0037,0.0625,0.528,0.01,-0.044,0.082,0.48,0.667,0.731,
    ↪ 0.007,0.892,0.749,0.02,0.791,0.814,0.009,0.007,0.207,0.324,-
    ↪ 0.048,0.701,0.203,-0.048,0.053,0.311,0.894,0.561,0.202,0.888,
    ↪ 0.152,0.318,0.832,0.618,0.168,0.836,0.372,0.326,0.852,0.859,
    ↪ 0.191,0.847,0.432,0.303,0.791,-0.047,0.202,0.784,0.64,0.2,0.012
    ↪ ,0.127,0.314,0.006,0.414,0.164,-0.014,-0.093,0.313,-0.013,0.411
    ↪ ,0.187,0.667,0.896,0.315,0.668,0.185,0.174,0.652,0.689,0.312,
    ↪ 0.646,0.091,0.215,0.713,0.579,0.307,0.707,0.804,0.323,0.044,
    ↪ 0.317,0.172,0.048,0.487,0.188,0.611,-0.013,0.33,0.607,0.219,
    ↪ 0.004,-0.014,-0.057,0.478,0.832,0.537,0.467,0.846,0.797,0.497,
    ↪ 0.712,0.292,0.016,0.609,0.495,0.498,0.649,0.502,0.013,0.05,0.25
    ↪ ,0.2342,0.0,0.7808,0.2354,0.0,0.483,0.2476,0.8347,0.5381,0.2682
    ↪ ,0.6654,0.0553,0.2539,0.6648,-0.0022,0.2447,0.8349,0.6325,
    ↪ 0.0051,-0.0841,-0.0894,0.495,0.0835,0.7428,0.0229,0.8341,0.7913
    ↪ ,0.4872,0.6644,0.5166,0.0298,0.0 & Cc C_{s}^{4} #9 (a^53) &
    ↪ mC212 & None & Cs6O36W11 & Cs6O36W11 & K. Okada and F. Marumo
    ↪ and S. Iwai, Acta Crystallogr. Sect. B Struct. Sci. 34, 50-54 (
    ↪ 1978)
   1.00000000000000
   6.28850000000000  -3.63050000000000   0.00000000000000
   6.28850000000000   3.63050000000000   0.00000000000000
  -8.38096647873181   0.00000000000000  36.85918339955960
    Cs      O      W
    12     72     22
Direct
   0.83090000000000   0.83110000000000   0.72060000000000   Cs   (4a)
   0.83110000000000   0.83090000000000   1.22060000000000   Cs   (4a)
   0.20120000000000   0.20460000000000   0.77860000000000   Cs   (4a)
   0.20460000000000   0.20120000000000   1.27860000000000   Cs   (4a)
  -0.24920000000000   0.74560000000000   0.09870000000000   Cs   (4a)
   0.74560000000000  -0.24920000000000   0.59870000000000   Cs   (4a)
   0.27920000000000   0.28420000000000   0.89920000000000   Cs   (4a)
   0.28420000000000   0.27920000000000   1.39920000000000   Cs   (4a)
  -0.01940000000000   0.96900000000000  -0.06380000000000   Cs   (4a)
   0.96900000000000  -0.01940000000000   0.43620000000000   Cs   (4a)
   0.05430000000000   0.06170000000000   0.06250000000000   Cs   (4a)
   0.06170000000000   0.05430000000000   0.56250000000000   Cs   (4a)
   0.51800000000000   0.53800000000000  -0.04400000000000    O   (4a)
   0.53800000000000   0.51800000000000   0.45600000000000    O   (4a)
  -0.39800000000000   0.56200000000000   0.66700000000000    O   (4a)
   0.56200000000000  -0.39800000000000   1.16700000000000    O   (4a)
   0.72400000000000   0.73800000000000   0.89200000000000    O   (4a)
   0.73800000000000   0.72400000000000   1.39200000000000    O   (4a)
   0.72900000000000   0.76900000000000   0.79100000000000    O   (4a)
   0.76900000000000   0.72900000000000   1.29100000000000    O   (4a)
   0.80500000000000   0.82300000000000   0.00700000000000    O   (4a)
   0.82300000000000   0.80500000000000   0.50700000000000    O   (4a)
  -0.11700000000000   0.53100000000000  -0.04800000000000    O   (4a)
   0.53100000000000  -0.11700000000000   0.45200000000000    O   (4a)
   0.49800000000000   0.90400000000000  -0.04800000000000    O   (4a)
   0.90400000000000   0.49800000000000   0.45200000000000    O   (4a)
  -0.25800000000000   0.36400000000000   0.89400000000000    O   (4a)
   0.36400000000000  -0.25800000000000   1.39400000000000    O   (4a)
   0.35900000000000   0.76300000000000   0.88800000000000    O   (4a)
   0.76300000000000   0.35900000000000   1.38800000000000    O   (4a)
  -0.16600000000000   0.47000000000000   0.83200000000000    O   (4a)
   0.47000000000000  -0.16600000000000   1.33200000000000    O   (4a)
   0.45000000000000   0.78600000000000   0.83600000000000    O   (4a)
   0.78600000000000   0.45000000000000   1.33600000000000    O   (4a)
   0.04600000000000   0.69800000000000   0.85200000000000    O   (4a)
   0.69800000000000   0.04600000000000   1.35200000000000    O   (4a)
   0.66800000000000   1.05000000000000   0.84700000000000    O   (4a)
   1.05000000000000   0.66800000000000   1.34700000000000    O   (4a)
   0.12900000000000   0.73500000000000   0.79100000000000    O   (4a)
   0.73500000000000   0.12900000000000   1.29100000000000    O   (4a)
  -0.24900000000000   0.15500000000000   0.78400000000000    O   (4a)
   0.15500000000000  -0.24900000000000   1.28400000000000    O   (4a)
   0.44000000000000   0.84000000000000   0.01200000000000    O   (4a)
   0.84000000000000   0.44000000000000   0.51200000000000    O   (4a)
  -0.18700000000000   0.44100000000000   0.00600000000000    O   (4a)
   0.44100000000000  -0.18700000000000   0.50600000000000    O   (4a)
   0.25000000000000   0.57800000000000  -0.01400000000000    O   (4a)
   0.57800000000000   0.25000000000000   0.48600000000000    O   (4a)
  -0.40600000000000   0.22000000000000  -0.01300000000000    O   (4a)
   0.22000000000000  -0.40600000000000   0.48700000000000    O   (4a)
   0.22400000000000   0.59800000000000   0.66700000000000    O   (4a)
   0.59800000000000   0.22400000000000   1.16700000000000    O   (4a)
   0.58100000000000   1.21100000000000   0.66800000000000    O   (4a)
   1.21100000000000   0.58100000000000   1.16800000000000    O   (4a)
   0.01100000000000   0.35900000000000   0.65200000000000    O   (4a)
   0.35900000000000   0.01100000000000   1.15200000000000    O   (4a)
   0.37700000000000   1.00100000000000   0.64600000000000    O   (4a)
   1.00100000000000   0.37700000000000   1.14600000000000    O   (4a)
  -0.12400000000000   0.30600000000000   0.71300000000000    O   (4a)
   0.30600000000000  -0.12400000000000   1.21300000000000    O   (4a)
   0.27200000000000   0.88600000000000   0.70700000000000    O   (4a)
   0.88600000000000   0.27200000000000   1.20700000000000    O   (4a)
   0.48100000000000   1.12700000000000   0.04400000000000    O   (4a)
   1.12700000000000   0.48100000000000   0.54400000000000    O   (4a)
   0.14500000000000   0.48900000000000   0.04800000000000    O   (4a)
   0.48900000000000   0.14500000000000   0.54800000000000    O   (4a)
   0.29900000000000   0.67500000000000   0.61100000000000    O   (4a)
   0.67500000000000   0.29900000000000   1.11100000000000    O   (4a)
  -0.34300000000000   0.31700000000000   0.60700000000000    O   (4a)
   0.31700000000000  -0.34300000000000   1.10700000000000    O   (4a)
   0.21500000000000   0.22300000000000  -0.01400000000000    O   (4a)
   0.22300000000000   0.21500000000000   0.48600000000000    O   (4a)
  -0.53500000000000   0.42100000000000   0.83200000000000    O   (4a)
   0.42100000000000  -0.53500000000000   1.33200000000000    O   (4a)
   0.07000000000000   1.00400000000000   0.84600000000000    O   (4a)
   1.00400000000000   0.07000000000000   1.34600000000000    O   (4a)
   0.30000000000000   1.29400000000000   0.71200000000000    O   (4a)
   1.29400000000000   0.30000000000000   1.21200000000000    O   (4a)
   0.27600000000000   0.30800000000000   0.60900000000000    O   (4a)
   0.30800000000000   0.27600000000000   1.10900000000000    O   (4a)
  -0.00300000000000   0.99300000000000   0.64900000000000    O   (4a)
   0.99300000000000  -0.00300000000000   1.14900000000000    O   (4a)
   0.48900000000000   0.51500000000000   0.05000000000000    O   (4a)
   0.51500000000000   0.48900000000000   0.55000000000000    O   (4a)
   0.01580000000000   0.48420000000000   0.00000000000000    W   (4a)
   0.48420000000000   0.01580000000000   0.50000000000000    W   (4a)
   0.54540000000000   1.01620000000000   0.00000000000000    W   (4a)
   1.01620000000000   0.54540000000000   0.50000000000000    W   (4a)
   0.23540000000000   0.73060000000000   0.83470000000000    W   (4a)
   0.73060000000000   0.23540000000000   1.33470000000000    W   (4a)
   0.26990000000000   0.80630000000000   0.66540000000000    W   (4a)
   0.80630000000000   0.26990000000000   1.16540000000000    W   (4a)
  -0.19860000000000   0.30920000000000   0.66480000000000    W   (4a)
   0.30920000000000  -0.19860000000000   1.16480000000000    W   (4a)
  -0.24690000000000   0.24250000000000   0.83490000000000    W   (4a)
   0.24250000000000  -0.24690000000000   1.33490000000000    W   (4a)
```

```
0.62740000000000   0.63760000000000   -0.08410000000000   W   (4a)
0.63760000000000   0.62740000000000   0.41590000000000   W   (4a)
-0.58440000000000   0.40560000000000   0.08350000000000   W   (4a)
0.40560000000000   -0.58440000000000   0.58350000000000   W   (4a)
0.71990000000000   0.76570000000000   0.83410000000000   W   (4a)
0.76570000000000   0.71990000000000   1.33410000000000   W   (4a)
0.30410000000000   1.27850000000000   0.66440000000000   W   (4a)
1.27850000000000   0.30410000000000   1.16440000000000   W   (4a)
0.48680000000000   0.54640000000000   0.00000000000000   W   (4a)
0.54640000000000   0.48680000000000   0.50000000000000   W   (4a)
```

$K_2S_2O_5$ ($K0_1$): A2B5C2_mP18_11_2e_e2f_2e - CIF

```
# CIF file
data_findsym-output
_audit_creation_method FINDSYM

_chemical_name_mineral 'K2O5S2'
_chemical_formula_sum 'K2 O5 S2'

loop_
_publ_author_name
 'I-C. Chen'
 'Y. Wang'
_journal_name_full_name
;
 Acta Crystallographica Section C: Structural Chemistry
;
_journal_volume 40
_journal_year 1984
_journal_page_first 1780
_journal_page_last 1781
_publ_Section_title
;
 Reinvestigation of potassium pyrosulfite, K$_{2}$S$_{2}$O$_{5}$
;

_aflow_title 'K$_{2}$S$_{2}$O$_{5}$ ($K0_{1}$) Structure'
_aflow_proto 'A2B5C2_mP18_11_2e_e2f_2e'
_aflow_params 'a,b/a,c/a,\beta,x_{1},z_{1},x_{2},z_{2},x_{3},z_{3},x_{4
    ↪ },z_{4},x_{5},z_{5},x_{6},y_{6},z_{6},x_{7},y_{7},z_{7}'
_aflow_params_values '6.921,0.890044791215,1.08900447912,102.79,0.231,-
    ↪ 0.0647,0.6398,0.67278,0.6555,0.0407,0.0291,0.33,0.7014,0.2384,
    ↪ 0.0741,0.0515,0.2343,0.635,0.053,0.3147'
_aflow_Strukturbericht '$K0_{1}$'
_aflow_Pearson 'mP18'

_symmetry_space_group_name_H-M "P 1 21/m 1"
_symmetry_Int_Tables_number 11

_cell_length_a 6.92100
_cell_length_b 6.16000
_cell_length_c 7.53700
_cell_angle_alpha 90.00000
_cell_angle_beta 102.79000
_cell_angle_gamma 90.00000

loop_
_space_group_symop_id
_space_group_symop_operation_xyz
1 x,y,z
2 -x,y+1/2,-z
3 -x,-y,-z
4 x,-y+1/2,z

loop_
_atom_site_label
_atom_site_type_symbol
_atom_site_symmetry_multiplicity
_atom_site_Wyckoff_label
_atom_site_fract_x
_atom_site_fract_y
_atom_site_fract_z
_atom_site_occupancy
K1 K 2 e 0.23100 0.25000 -0.06470 1.00000
K2 K 2 e 0.63980 0.25000 0.67278 1.00000
O1 O 2 e 0.65550 0.25000 0.04070 1.00000
S1 S 2 e 0.02910 0.25000 0.33000 1.00000
S2 S 2 e 0.70140 0.25000 0.23840 1.00000
O2 O 4 f 0.07410 0.05150 0.23430 1.00000
O3 O 4 f 0.63500 0.05300 0.31470 1.00000
```

$K_2S_2O_5$ ($K0_1$): A2B5C2_mP18_11_2e_e2f_2e - POSCAR

```
A2B5C2_mP18_11_2e_e2f_2e & a,b/a,c/a,beta,x1,z1,x2,z2,x3,z3,x4,z4,x5,z5,
    ↪ x6,y6,z6,x7,y7,z7 --params=6.921,0.890044791215,1.08900447912,
    ↪ 102.79,0.231,-0.0647,0.6398,0.67278,0.6555,0.0407,0.0291,0.33,
    ↪ 0.7014,0.2384,0.0741,0.0515,0.2343,0.635,0.053,0.3147 & P2_{1}/
    ↪ m C_{2h}^{2} #11 (e^5f^2) & mP18 & $K0_{1}$ & K2O5S2 & K2O5S2 &
    ↪ I-C. Chen and Y. Wang, Acta Crystallogr. C 40, 1780-1781 (1984
    ↪ )
   1.00000000000000
   6.92100000000000   0.00000000000000   0.00000000000000
   0.00000000000000   6.16000000000000   0.00000000000000
  -1.66852823637215   0.00000000000000   7.34999200846020
     K     O     S
     4    10     4
Direct
   0.23100000000000   0.25000000000000   -0.06470000000000   K   (2e)
  -0.23100000000000   0.75000000000000   0.06470000000000   K   (2e)
   0.63980000000000   0.25000000000000   0.67278000000000   K   (2e)
  -0.63980000000000   0.75000000000000   -0.67278000000000   K   (2e)
   0.65550000000000   0.25000000000000   0.04070000000000   O   (2e)
  -0.65550000000000   0.75000000000000   -0.04070000000000   O   (2e)
   0.07410000000000   0.05150000000000   0.23430000000000   O   (4f)
  -0.07410000000000   0.55150000000000   -0.23430000000000   O   (4f)
  -0.07410000000000   -0.05150000000000   -0.23430000000000   O   (4f)
   0.07410000000000   0.44850000000000   0.23430000000000   O   (4f)
   0.63500000000000   0.05300000000000   0.31470000000000   O   (4f)
  -0.63500000000000   0.55300000000000   -0.31470000000000   O   (4f)
  -0.63500000000000   -0.05300000000000   -0.31470000000000   O   (4f)
   0.63500000000000   0.44700000000000   0.31470000000000   O   (4f)
   0.02910000000000   0.25000000000000   0.33000000000000   S   (2e)
  -0.02910000000000   0.75000000000000   -0.33000000000000   S   (2e)
   0.70140000000000   0.25000000000000   0.23840000000000   S   (2e)
  -0.70140000000000   0.75000000000000   -0.23840000000000   S   (2e)
```

$ZrSe_3$: A3B_mP8_11_3e_e - CIF

```
# CIF file
data_findsym-output
_audit_creation_method FINDSYM

_chemical_name_mineral 'Se3Zr'
_chemical_formula_sum 'Se3 Zr'

loop_
_publ_author_name
 'S. Furuseth'
 'L. Bratt{\aa}s'
 'A. Kjekshus'
_journal_name_full_name
;
 Acta Chemica Scandinavica
;
_journal_volume 29a
_journal_year 1975
_journal_page_first 623
_journal_page_last 631
_publ_Section_title
;
 On the Crystal Structures of TiS$_{3}$, ZrS$_{3}$, ZrSe$_{3}$, ZrTe$_{3
    ↪ }$, HfS$_{3}$, and HfSe$_{3}$
;

_aflow_title 'ZrSe$_{3}$ Structure'
_aflow_proto 'A3B_mP8_11_3e_e'
_aflow_params 'a,b/a,c/a,\beta,x_{1},z_{1},x_{2},z_{2},x_{3},z_{3},x_{4
    ↪ },z_{4}'
_aflow_params_values '5.4109,0.692823744664,1.74536583563,97.48,0.762,
    ↪ 0.554,0.456,0.174,0.888,0.169,0.285,0.656'
_aflow_Strukturbericht 'None'
_aflow_Pearson 'mP8'

_symmetry_space_group_name_H-M "P 1 21/m 1"
_symmetry_Int_Tables_number 11

_cell_length_a 5.41090
_cell_length_b 3.74880
_cell_length_c 9.44400
_cell_angle_alpha 90.00000
_cell_angle_beta 97.48000
_cell_angle_gamma 90.00000

loop_
_space_group_symop_id
_space_group_symop_operation_xyz
1 x,y,z
2 -x,y+1/2,-z
3 -x,-y,-z
4 x,-y+1/2,z

loop_
_atom_site_label
_atom_site_type_symbol
_atom_site_symmetry_multiplicity
_atom_site_Wyckoff_label
_atom_site_fract_x
_atom_site_fract_y
_atom_site_fract_z
_atom_site_occupancy
Se1 Se 2 e 0.76200 0.25000 0.55400 1.00000
Se2 Se 2 e 0.45600 0.25000 0.17400 1.00000
Se3 Se 2 e 0.88800 0.25000 0.16900 1.00000
Zr1 Zr 2 e 0.28500 0.25000 0.65600 1.00000
```

$ZrSe_3$: A3B_mP8_11_3e_e - POSCAR

```
A3B_mP8_11_3e_e & a,b/a,c/a,beta,x1,z1,x2,z2,x3,z3,x4,z4 --params=5.4109
    ↪ ,0.692823744664,1.74536583563,97.48,0.762,0.554,0.456,0.174,
    ↪ 0.888,0.169,0.285,0.656 & P2_{1}/m C_{2h}^{2} #11 (e^4) & mP8 &
    ↪ None & Se3Zr & Se3Zr & S. Furuseth and L. Bratt{\aa}s and A.
    ↪ Kjekshus, Acta Chem. Scand. 29a, 623-631 (1975)
   1.00000000000000
   5.41090000000000   0.00000000000000   0.00000000000000
   0.00000000000000   3.74880000000000   0.00000000000000
  -1.22942090908250   0.00000000000000   9.36363499012594
    Se    Zr
     6     2
Direct
   0.76200000000000   0.25000000000000   0.55400000000000   Se   (2e)
  -0.76200000000000   0.75000000000000   -0.55400000000000   Se   (2e)
   0.45600000000000   0.25000000000000   0.17400000000000   Se   (2e)
  -0.45600000000000   0.75000000000000   -0.17400000000000   Se   (2e)
   0.88800000000000   0.25000000000000   0.16900000000000   Se   (2e)
  -0.88800000000000   0.75000000000000   -0.16900000000000   Se   (2e)
   0.28500000000000   0.25000000000000   0.65600000000000   Zr   (2e)
  -0.28500000000000   0.75000000000000   -0.65600000000000   Zr   (2e)
```

$y$-$Y_2Si_2O_7$: A7B2C2_mP22_11_3e2f_2e_ab - CIF

```
# CIF file
data_findsym-output
_audit_creation_method FINDSYM

_chemical_name_mineral 'O7Si2Y2'
_chemical_formula_sum 'O7 Si2 Y2'

loop_
_publ_author_name
'N. G. Batalieva'
'Y. A. Pyatenko'
_journal_name_full_name
;
Soviet Physics Crystallography
;
_journal_volume 16
_journal_year 1972
_journal_page_first 786
_journal_page_last 789
_publ_Section_title
;
Artificial yttrialite (\'\'y-phase\'\') - a representative of a new
    ↪ structure type in the rare earth diorthosilicate series
;

# Found in Revision of the crystallographic data of polymorphic Y$_{2}
    ↪ $Si$_{2}$O$_{7}$ and Y$_{2}$SiO$_{5}$ compounds, 2004

_aflow_title '$y$-Y$_{2}$Si$_{2}$O$_{7}$ Structure'
_aflow_proto 'A7B2C2_mP22_11_3e2f_2e_ab'
_aflow_params 'a,b/a,c/a,\beta,x_{3},z_{3},x_{4},z_{4},x_{5},z_{5},x_{6
    ↪ },z_{6},x_{7},z_{7},x_{8},y_{8},z_{8},x_{9},y_{9},z_{9}'
_aflow_params_values '7.5,1.07466666667,0.669333333333,112.0,0.19,0.31,
    ↪ 0.508,0.264,0.88,0.43,0.12,0.588,0.709,0.548,0.19,0.09,0.8,
    ↪ 0.688,0.09,0.736'
_aflow_Strukturbericht 'None'
_aflow_Pearson 'mP22'

_symmetry_space_group_name_H-M "P 1 21/m 1"
_symmetry_Int_Tables_number 11

_cell_length_a      7.50
_cell_length_b      8.06000
_cell_length_c      5.02
_cell_angle_alpha   90.00000
_cell_angle_beta    112.00000
_cell_angle_gamma   90.00000

loop_
_space_group_symop_id
_space_group_symop_operation_xyz
1 x,y,z
2 -x,y+1/2,-z
3 -x,-y,-z
4 x,-y+1/2,z

loop_
_atom_site_label
_atom_site_type_symbol
_atom_site_symmetry_multiplicity
_atom_site_Wyckoff_label
_atom_site_fract_x
_atom_site_fract_y
_atom_site_fract_z
_atom_site_occupancy
Y1 Y 2 a 0.00000 0.00000 0.00000 1.00000
Y2 Y 2 b 0.50000 0.00000 0.00000 1.00000
O1 O 2 e 0.19000 0.25000 0.31000 1.00000
O2 O 2 e 0.50800 0.25000 0.26400 1.00000
O3 O 2 e 0.88000 0.25000 0.43000 1.00000
Si1 Si 2 e 0.12000 0.25000 0.58800 1.00000
Si2 Si 2 e 0.70900 0.25000 0.54800 1.00000
O4 O 4 f 0.19000 0.09000 0.80000 1.00000
O5 O 4 f 0.68800 0.09000 0.73600 1.00000
```

$y$-$Y_2Si_2O_7$: A7B2C2_mP22_11_3e2f_2e_ab - POSCAR

```
A7B2C2_mP22_11_3e2f_2e_ab & a,b/a,c/a,beta,x3,z3,x4,z4,x5,z5,x6,z6,x7,z7
    ↪ ,x8,y8,z8,x9,y9,z9 --params=7.5,1.07466666667,0.669333333333,
    ↪ 112.0,0.19,0.31,0.508,0.264,0.88,0.43,0.12,0.588,0.709,0.548,
    ↪ 0.19,0.09,0.8,0.688,0.09,0.736 & P2_{1}/m C_{2h}^{2} #11 (abe^
    ↪ 5f^2) & mP22 & None & O7Si2Y2 & O7Si2Y2 & N. G. Batalieva and
    ↪ Y. A. Pyatenko, Sov. Phys. Crystallogr. 16, 786-789 (1972)
   1.00000000000000
   7.50000000000000   0.00000000000000   0.00000000000000
   0.00000000000000   8.06000000000000   0.00000000000000
  -1.88052509894788   0.00000000000000   4.65446294992527
     O     Si      Y
    14      4      4
Direct
   0.19000000000000   0.25000000000000   0.31000000000000   O   (2e)
  -0.19000000000000   0.75000000000000  -0.31000000000000   O   (2e)
   0.50800000000000   0.25000000000000   0.26400000000000   O   (2e)
  -0.50800000000000   0.75000000000000  -0.26400000000000   O   (2e)
   0.88000000000000   0.25000000000000   0.43000000000000   O   (2e)
  -0.88000000000000   0.75000000000000  -0.43000000000000   O   (2e)
   0.19000000000000   0.09000000000000   0.80000000000000   O   (4f)
  -0.19000000000000   0.59000000000000  -0.80000000000000   O   (4f)
  -0.19000000000000  -0.09000000000000  -0.80000000000000   O   (4f)
   0.19000000000000   0.41000000000000   0.80000000000000   O   (4f)
   0.68800000000000   0.09000000000000   0.73600000000000   O   (4f)
  -0.68800000000000   0.59000000000000  -0.73600000000000   O   (4f)
  -0.68800000000000  -0.09000000000000  -0.73600000000000   O   (4f)
   0.68800000000000   0.41000000000000   0.73600000000000   O   (4f)
   0.12000000000000   0.25000000000000   0.58800000000000   Si  (2e)
  -0.12000000000000   0.75000000000000  -0.58800000000000   Si  (2e)
   0.70900000000000   0.25000000000000   0.54800000000000   Si  (2e)
  -0.70900000000000   0.75000000000000  -0.54800000000000   Si  (2e)
   0.00000000000000   0.00000000000000   0.00000000000000   Y   (2a)
   0.00000000000000   0.50000000000000   0.00000000000000   Y   (2a)
   0.50000000000000   0.00000000000000   0.00000000000000   Y   (2b)
   0.50000000000000   0.50000000000000   0.00000000000000   Y   (2b)
```

Barytocalcite ($BaCa(CO_3)_2$): AB2CD6_mP20_11_e_2e_e_2e2f - CIF

```
# CIF file
data_findsym-output
_audit_creation_method FINDSYM

_chemical_name_mineral 'Barytocalcite'
_chemical_formula_sum 'Ba C2 Ca O6'

loop_
_publ_author_name
'B. Dickens'
'J. S. Bowen'
_journal_name_full_name
;
Journal of Research of the National Bureau of Standards, Section A:
    ↪ Physics and Chemistry
;
_journal_volume 75
_journal_year 1971
_journal_page_first 197
_journal_page_last 203
_publ_Section_title
;
The Crystal Structure of BaCa(CO$_{3}$)$_{2}$ (barytocalcite)
;

# Found in A new BaCa(CO$_{3}$)$_{2}$ polymorph, 2019

_aflow_title 'Barytocalcite (BaCa(CO$_{3}$)$_{2}$) Structure'
_aflow_proto 'AB2CD6_mP20_11_e_2e_e_2e2f'
_aflow_params 'a,b/a,c/a,\beta,x_{1},z_{1},x_{2},z_{2},x_{3},z_{3},x_{4
    ↪ },z_{4},x_{5},z_{5},x_{6},z_{6},x_{7},y_{7},z_{7},x_{8},y_{8},
    ↪ z_{8}'
_aflow_params_values '5.2344,1.54592694483,1.25019104386,106.05,0.1474,
    ↪ 0.28824,0.1028,0.7517,0.6149,0.7468,0.6232,0.19855,-0.0057,
    ↪ 0.8607,0.6383,0.5644,0.8457,-0.038,0.3089,0.6066,0.4604,0.8474'
_aflow_Strukturbericht 'None'
_aflow_Pearson 'mP20'

_symmetry_space_group_name_H-M "P 1 21/m 1"
_symmetry_Int_Tables_number 11

_cell_length_a      5.23440
_cell_length_b      8.09200
_cell_length_c      6.54400
_cell_angle_alpha   90.00000
_cell_angle_beta    106.05000
_cell_angle_gamma   90.00000

loop_
_space_group_symop_id
_space_group_symop_operation_xyz
1 x,y,z
2 -x,y+1/2,-z
3 -x,-y,-z
4 x,-y+1/2,z

loop_
_atom_site_label
_atom_site_type_symbol
_atom_site_symmetry_multiplicity
_atom_site_Wyckoff_label
_atom_site_fract_x
_atom_site_fract_y
_atom_site_fract_z
_atom_site_occupancy
Ba1 Ba 2 e 0.14740 0.25000 0.28824 1.00000
C1 C 2 e 0.10280 0.25000 0.75170 1.00000
C2 C 2 e 0.61490 0.25000 0.74680 1.00000
Ca1 Ca 2 e 0.62320 0.25000 0.19855 1.00000
O1 O 2 e -0.00570 0.25000 0.86070 1.00000
O2 O 2 e 0.63830 0.25000 0.56440 1.00000
O3 O 4 f 0.84570 -0.03800 0.30890 1.00000
O4 O 4 f 0.60660 0.46040 0.84740 1.00000
```

Barytocalcite ($BaCa(CO_3)_2$): AB2CD6_mP20_11_e_2e_e_2e2f - POSCAR

```
AB2CD6_mP20_11_e_2e_e_2e2f & a,b/a,c/a,beta,x1,z1,x2,z2,x3,z3,x4,z4,x5,
    ↪ z5,x6,z6,x7,y7,z7,x8,y8,z8 --params=5.2344,1.54592694483,
    ↪ 1.25019104386,106.05,0.1474,0.28824,0.1028,0.7517,0.6149,0.7468
    ↪ ,0.6232,0.19855,-0.0057,0.8607,0.6383,0.5644,0.8457,-0.038,
    ↪ 0.3089,0.6066,0.4604,0.8474 & P2_{1}/m C_{2h}^{2} #11 (e^6f^2)
    ↪ & mP20 & None & BaC2CaO6 & Barytocalcite & B. Dickens and J. S.
    ↪ Bowen, J. Res. Nat. Stand. Sec. A 75, 197-203 (1971)
   1.00000000000000
   5.23440000000000   0.00000000000000   0.00000000000000
   0.00000000000000   8.09200000000000   0.00000000000000
  -1.80925966275357   0.00000000000000   6.28892005615693
    Ba      C     Ca      O
     2      4      2     12
Direct
   0.14740000000000   0.25000000000000   0.28824000000000   Ba  (2e)
  -0.14740000000000   0.75000000000000  -0.28824000000000   Ba  (2e)
   0.10280000000000   0.25000000000000   0.75170000000000   C   (2e)
  -0.10280000000000   0.75000000000000  -0.75170000000000   C   (2e)
   0.61490000000000   0.25000000000000   0.74680000000000   C   (2e)
```

```
−0.61490000000000   0.75000000000000  −0.74680000000000   C    (2e)
 0.62320000000000   0.25000000000000   0.19855000000000   Ca   (2e)
−0.62320000000000   0.75000000000000  −0.19855000000000   Ca   (2e)
−0.00570000000000   0.25000000000000   0.86070000000000   O    (2e)
 0.00570000000000   0.75000000000000  −0.86070000000000   O    (2e)
 0.63830000000000   0.25000000000000   0.56440000000000   O    (2e)
−0.63830000000000   0.75000000000000  −0.56440000000000   O    (2e)
 0.84570000000000  −0.03800000000000   0.30890000000000   O    (4f)
−0.84570000000000   0.46200000000000  −0.30890000000000   O    (4f)
−0.84570000000000   0.03800000000000  −0.30890000000000   O    (4f)
 0.84570000000000   0.53800000000000   0.30890000000000   O    (4f)
 0.60660000000000   0.46040000000000   0.84740000000000   O    (4f)
−0.60660000000000   0.96040000000000  −0.84740000000000   O    (4f)
−0.60660000000000  −0.46040000000000  −0.84740000000000   O    (4f)
 0.60660000000000   0.03960000000000   0.84740000000000   O    (4f)
```

O(OH)Y: ABC_mP6_11_e_e_e - CIF

```
# CIF file
data_findsym-output
_audit_creation_method FINDSYM

_chemical_name_mineral 'HO2Y'
_chemical_formula_sum 'O (OH) Y'

loop_
_publ_author_name
 'R. F. Klevtsova'
 'P. V. Klevtsov'
_journal_year 1965
_publ_Section_title
;
 The Crystal Structure of YOOH
;

_aflow_title 'O(OH)Y Structure'
_aflow_proto 'ABC_mP6_11_e_e_e'
_aflow_params 'a,b/a,c/a,\beta,x_{1},z_{1},x_{2},z_{2},x_{3},z_{3}'
_aflow_params_values '4.28,0.848130841121,1.41355140187,112.5,0.219,
    ↪ 0.448,0.69,-0.058,0.643,0.312'
_aflow_Strukturbericht 'None'
_aflow_Pearson 'mP6'

_symmetry_space_group_name_H-M "P 1 21/m 1"
_symmetry_Int_Tables_number 11

_cell_length_a    4.28000
_cell_length_b    3.63000
_cell_length_c    6.05000
_cell_angle_alpha 90.00000
_cell_angle_beta  112.50000
_cell_angle_gamma 90.00000

loop_
_space_group_symop_id
_space_group_symop_operation_xyz
1 x,y,z
2 -x,y+1/2,-z
3 -x,-y,-z
4 x,-y+1/2,z

loop_
_atom_site_label
_atom_site_type_symbol
_atom_site_symmetry_multiplicity
_atom_site_Wyckoff_label
_atom_site_fract_x
_atom_site_fract_y
_atom_site_fract_z
_atom_site_occupancy
O1 O 2 e 0.21900 0.25000 0.44800 1.00000
OH1 OH 2 e 0.69000 0.25000 -0.05800 1.00000
Y1 Y 2 e 0.64300 0.25000 0.31200 1.00000
```

O(OH)Y: ABC_mP6_11_e_e_e - POSCAR

```
ABC_mP6_11_e_e_e & a,b/a,c/a,beta,x1,z1,x2,z2,x3,z3 --params=4.28,
    ↪ 0.848130841121,1.41355140187,112.5,0.219,0.448,0.69,-0.058,
    ↪ 0.643,0.312 & P2_{1}/m C_{2h}^{2} #11 (e^3) & mP6 & None & HO2Y
    ↪ & HO2Y & R. F. Klevtsova and P. V. Klevtsov, (1965)
   1.00000000000000
   4.28000000000000   0.00000000000000   0.00000000000000
   0.00000000000000   3.63000000000000   0.00000000000000
  -2.31523476580879   0.00000000000000   5.58947117169329
     O       OH       Y
     2        2       2
Direct
   0.21900000000000   0.25000000000000   0.44800000000000   O    (2e)
  -0.21900000000000   0.75000000000000  -0.44800000000000   O    (2e)
   0.69000000000000   0.25000000000000  -0.05800000000000   OH   (2e)
  -0.69000000000000   0.75000000000000   0.05800000000000   OH   (2e)
   0.64300000000000   0.25000000000000   0.31200000000000   Y    (2e)
  -0.64300000000000   0.75000000000000  -0.31200000000000   Y    (2e)
```

$Al_{13}Fe_4$: A13B4_mC102_12_dg8i5j_4ij - CIF

```
# CIF file
data_findsym-output
_audit_creation_method FINDSYM

_chemical_name_mineral 'Al13Fe4'
_chemical_formula_sum 'Al13 Fe4'

loop_
_publ_author_name
 'P. J. Black'
_journal_name_full_name
;
 Acta Cristallographica
;
_journal_volume 8
_journal_year 1955
_journal_page_first 43
_journal_page_last 48
_publ_Section_title
;
 The Structure of FeAl$_{3}$. I
;

_aflow_title 'Al$_{13}$Fe$_{4}$ Structure'
_aflow_proto 'A13B4_mC102_12_dg8i5j_4ij'
_aflow_params 'a,b/a,c/a,\beta,y_{2},x_{3},z_{3},x_{4},z_{4},x_{5},z_{5
    ↪ },x_{6},z_{6},x_{7},z_{7},x_{8},z_{8},x_{9},z_{9},x_{10},z_{10
    ↪ },x_{11},z_{11},x_{12},z_{12},x_{13},z_{13},x_{14},z_{14},x_{15
    ↪ },y_{15},z_{15},x_{16},y_{16},z_{16},x_{17},y_{17},z_{17},x_{18
    ↪ },y_{18},z_{18},x_{19},y_{19},z_{19},x_{20},y_{20},z_{20}'
_aflow_params_values '15.489,0.521860675318,0.805474853122,107.71667,
    ↪ 0.244,0.064,0.173,0.322,0.277,0.235,0.539,0.081,0.582,0.231,-
    ↪ 0.028,0.48,0.827,0.31,0.769,0.086,0.781,0.086,0.383,0.401,0.624
    ↪ ,0.09,-0.011,0.4,-0.015,0.188,0.216,0.111,0.373,0.211,0.107,
    ↪ 0.176,0.216,0.334,0.495,0.283,0.329,0.366,0.223,0.479,0.318,
    ↪ 0.285,0.277'
_aflow_Strukturbericht 'None'
_aflow_Pearson 'mC102'

_symmetry_space_group_name_H-M "C 1 2/m 1"
_symmetry_Int_Tables_number 12

_cell_length_a    15.48900
_cell_length_b    8.08310
_cell_length_c    12.47600
_cell_angle_alpha 90.00000
_cell_angle_beta  107.71667
_cell_angle_gamma 90.00000

loop_
_space_group_symop_id
_space_group_symop_operation_xyz
1 x,y,z
2 -x,y,-z
3 -x,-y,-z
4 x,-y,z
5 x+1/2,y+1/2,z
6 -x+1/2,y+1/2,-z
7 -x+1/2,-y+1/2,-z
8 x+1/2,-y+1/2,z

loop_
_atom_site_label
_atom_site_type_symbol
_atom_site_symmetry_multiplicity
_atom_site_Wyckoff_label
_atom_site_fract_x
_atom_site_fract_y
_atom_site_fract_z
_atom_site_occupancy
Al1 Al 2 d 0.00000 0.50000 0.50000 1.00000
Al2 Al 4 g 0.00000 0.24400 0.00000 1.00000
Al3 Al 4 i 0.06400 0.00000 0.17300 1.00000
Al4 Al 4 i 0.32200 0.00000 0.27700 0.70000
Al5 Al 4 i 0.23500 0.00000 0.53900 1.00000
Al6 Al 4 i 0.08100 0.00000 0.58200 1.00000
Al7 Al 4 i 0.23100 0.00000 -0.02800 1.00000
Al8 Al 4 i 0.48000 0.00000 0.82700 1.00000
Al9 Al 4 i 0.31000 0.00000 0.76900 1.00000
Al10 Al 4 i 0.08600 0.00000 0.78100 1.00000
Fe1 Fe 4 i 0.08600 0.00000 0.38300 1.00000
Fe2 Fe 4 i 0.40100 0.00000 0.62400 1.00000
Fe3 Fe 4 i 0.09000 0.00000 -0.01100 1.00000
Fe4 Fe 4 i 0.40000 0.00000 -0.01500 1.00000
Al11 Al 8 j 0.18800 0.21600 0.11100 1.00000
Al12 Al 8 j 0.37300 0.21100 0.10700 1.00000
Al13 Al 8 j 0.17600 0.21600 0.33400 1.00000
Al14 Al 8 j 0.49500 0.28300 0.32900 1.00000
Al15 Al 8 j 0.36600 0.22300 0.47900 1.00000
Fe5 Fe 8 j 0.31800 0.28500 0.27700 1.00000
```

$Al_{13}Fe_4$: A13B4_mC102_12_dg8i5j_4ij - POSCAR

```
A13B4_mC102_12_dg8i5j_4ij & a,b/a,c/a,beta,y2,x3,z3,x4,z4,x5,z5,x6,z6,x7
    ↪ ,z7,x8,z8,x9,z9,x10,z10,x11,z11,x12,z12,x13,z13,x14,z14,x15,y15
    ↪ ,z15,x16,y16,z16,x17,y17,z17,x18,y18,z18,x19,y19,z19,x20,y20,
    ↪ z20 --params=15.489,0.521860675318,0.805474853122,107.71667,
    ↪ 0.244,0.064,0.173,0.322,0.277,0.235,0.539,0.081,0.582,0.231,-
    ↪ 0.028,0.48,0.827,0.31,0.769,0.086,0.781,0.086,0.383,0.401,0.624
    ↪ ,0.09,-0.011,0.4,-0.015,0.188,0.216,0.111,0.373,0.211,0.107,
    ↪ 0.176,0.216,0.334,0.495,0.283,0.329,0.366,0.223,0.479,0.318,
    ↪ 0.285,0.277 & C2/m C_{2h}^{3} #12 (dgi^12j^6) & mC102 & None &
    ↪ Al13Fe4 & Al13Fe4 & P. J. Black, Acta Cryst. 8, 43-48 (1955)
   1.00000000000000
   7.74450000000000  -4.04155000000000   0.00000000000000
   7.74450000000000   4.04155000000000   0.00000000000000
  -3.79657432304420   0.00000000000000  11.88430054355750
     Al      Fe
     39      12
Direct
   0.50000000000000   0.50000000000000   0.50000000000000   Al   (2d)
  -0.24400000000000   0.24400000000000   0.00000000000000   Al   (4g)
   0.24400000000000  -0.24400000000000   0.00000000000000   Al   (4g)
```

```
    0.06400000000000     0.06400000000000     0.17300000000000     Al    (4i)
   -0.06400000000000    -0.06400000000000    -0.17300000000000     Al    (4i)
    0.32200000000000     0.32200000000000     0.27700000000000     Al    (4i)
   -0.32200000000000    -0.32200000000000    -0.27700000000000     Al    (4i)
    0.23500000000000     0.23500000000000     0.53900000000000     Al    (4i)
   -0.23500000000000    -0.23500000000000    -0.53900000000000     Al    (4i)
    0.08100000000000     0.08100000000000     0.58200000000000     Al    (4i)
   -0.08100000000000    -0.08100000000000    -0.58200000000000     Al    (4i)
    0.23100000000000     0.23100000000000    -0.02800000000000     Al    (4i)
   -0.23100000000000    -0.23100000000000     0.02800000000000     Al    (4i)
    0.48000000000000     0.48000000000000     0.82700000000000     Al    (4i)
   -0.48000000000000    -0.48000000000000    -0.82700000000000     Al    (4i)
    0.31000000000000     0.31000000000000     0.76900000000000     Al    (4i)
   -0.31000000000000    -0.31000000000000    -0.76900000000000     Al    (4i)
    0.08600000000000     0.08600000000000     0.78100000000000     Al    (4i)
   -0.08600000000000    -0.08600000000000    -0.78100000000000     Al    (4i)
   -0.02800000000000     0.40400000000000     0.11100000000000     Al    (8j)
   -0.40400000000000     0.02800000000000    -0.11100000000000     Al    (8j)
    0.02800000000000    -0.40400000000000    -0.11100000000000     Al    (8j)
    0.40400000000000    -0.02800000000000     0.11100000000000     Al    (8j)
    0.16200000000000     0.58400000000000     0.10700000000000     Al    (8j)
   -0.58400000000000    -0.16200000000000    -0.10700000000000     Al    (8j)
   -0.16200000000000    -0.58400000000000    -0.10700000000000     Al    (8j)
    0.58400000000000     0.16200000000000     0.10700000000000     Al    (8j)
   -0.04000000000000     0.39200000000000     0.33400000000000     Al    (8j)
   -0.39200000000000     0.04000000000000    -0.33400000000000     Al    (8j)
    0.04000000000000    -0.39200000000000    -0.33400000000000     Al    (8j)
    0.39200000000000    -0.04000000000000     0.33400000000000     Al    (8j)
    0.21200000000000     0.77800000000000     0.32900000000000     Al    (8j)
   -0.77800000000000    -0.21200000000000    -0.32900000000000     Al    (8j)
   -0.21200000000000    -0.77800000000000    -0.32900000000000     Al    (8j)
    0.77800000000000     0.21200000000000     0.32900000000000     Al    (8j)
    0.14300000000000     0.58900000000000     0.47900000000000     Al    (8j)
   -0.58900000000000    -0.14300000000000    -0.47900000000000     Al    (8j)
   -0.14300000000000    -0.58900000000000    -0.47900000000000     Al    (8j)
    0.58900000000000     0.14300000000000     0.47900000000000     Al    (8j)
    0.08600000000000     0.08600000000000     0.38300000000000     Fe    (4i)
   -0.08600000000000    -0.08600000000000    -0.38300000000000     Fe    (4i)
    0.40100000000000     0.40100000000000     0.62400000000000     Fe    (4i)
   -0.40100000000000    -0.40100000000000    -0.62400000000000     Fe    (4i)
    0.09000000000000     0.09000000000000    -0.01100000000000     Fe    (4i)
   -0.09000000000000    -0.09000000000000     0.01100000000000     Fe    (4i)
    0.40000000000000     0.40000000000000    -0.01500000000000     Fe    (4i)
   -0.40000000000000    -0.40000000000000     0.01500000000000     Fe    (4i)
    0.03300000000000     0.60300000000000     0.27700000000000     Fe    (8j)
   -0.60300000000000    -0.03300000000000    -0.27700000000000     Fe    (8j)
   -0.03300000000000    -0.60300000000000    -0.27700000000000     Fe    (8j)
    0.60300000000000     0.03300000000000     0.27700000000000     Fe    (8j)
```

$Os_4Al_{13}$: A13B4_mC34_12_b6i_2i - CIF

```
# CIF file
data_findsym-output
_audit_creation_method FINDSYM

_chemical_name_mineral 'Al13Os4'
_chemical_formula_sum 'Al13 Os4'

loop_
_publ_author_name
 'L.-E. Edshammar'
_journal_name_full_name
;
 Acta Chemica Scandinavica
;
_journal_volume 18
_journal_year 1964
_journal_page_first 2294
_journal_page_last 2302
_publ_Section_title
;
 The Crystal Structure of Os$_{4}$Al$_{13}$
;

_aflow_title 'Os$_{4}$Al$_{13}$ Structure'
_aflow_proto 'A13B4_mC34_12_b6i_2i'
_aflow_params 'a,b/a,c/a,\beta,x_{2},z_{2},x_{3},z_{3},x_{4},z_{4},x_{5
    ↪ },z_{5},x_{6},z_{6},x_{7},z_{7},x_{8},z_{8},x_{9},z_{9}'
_aflow_params_values '17.64,0.239682539683,0.440646258503,115.15,0.587,
    ↪ 0.368,0.257,0.613,0.132,0.162,0.79,0.087,-0.086,0.432,0.409,
    ↪ 0.194,0.294,0.2915,-0.0081,0.1947'
_aflow_Strukturbericht 'None'
_aflow_Pearson 'mC34'

_symmetry_space_group_name_H-M "C 1 2/m 1"
_symmetry_Int_Tables_number 12

_cell_length_a     17.64000
_cell_length_b     4.22800
_cell_length_c     7.77300
_cell_angle_alpha  90.00000
_cell_angle_beta   115.15000
_cell_angle_gamma  90.00000

loop_
_space_group_symop_id
_space_group_symop_operation_xyz
1 x,y,z
2 -x,y,-z
3 -x,-y,-z
4 x,-y,z
5 x+1/2,y+1/2,z
6 -x+1/2,y+1/2,-z
7 -x+1/2,-y+1/2,-z
8 x+1/2,-y+1/2,z

loop_
_atom_site_label
_atom_site_type_symbol
_atom_site_symmetry_multiplicity
_atom_site_Wyckoff_label
_atom_site_fract_x
_atom_site_fract_y
_atom_site_fract_z
_atom_site_occupancy
Al1 Al 2 b 0.00000 0.50000 0.00000 1.00000
Al2 Al 4 i 0.58700 0.00000 0.36800 1.00000
Al3 Al 4 i 0.25700 0.00000 0.61300 1.00000
Al4 Al 4 i 0.13200 0.00000 0.16200 1.00000
Al5 Al 4 i 0.79000 0.00000 0.08700 1.00000
Al6 Al 4 i -0.08600 0.00000 0.43200 1.00000
Al7 Al 4 i 0.40900 0.00000 0.19400 1.00000
Os1 Os 4 i 0.29400 0.00000 0.29150 1.00000
Os2 Os 4 i -0.00810 0.00000 0.19470 1.00000
```

$Os_4Al_{13}$: A13B4_mC34_12_b6i_2i - POSCAR

```
A13B4_mC34_12_b6i_2i & a,b/a,c/a,beta,x2,z2,x3,z3,x4,z4,x5,z5,x6,z6,x7,
    ↪ z7,x8,z8,x9,z9 --params=17.64,0.239682539683,0.440646258503,
    ↪ 115.15,0.587,0.368,0.257,0.613,0.132,0.162,0.79,0.087,-0.086,
    ↪ 0.432,0.409,0.194,0.294,0.2915,-0.0081,0.1947 & C2/m C_{2h}^{3}
    ↪ #12 (bi^8) & mC34 & None & Al13Os4 & Al13Os4 & L.-E. Edshammar
    ↪ , Acta Chem. Scand. 18, 2294-2302 (1964)
   1.00000000000000
   8.82000000000000   -2.11400000000000    0.00000000000000
   8.82000000000000    2.11400000000000    0.00000000000000
  -3.30344353102559    0.00000000000000    7.03610615591644
    Al     Os
    13      4
Direct
   0.50000000000000    0.50000000000000    0.00000000000000    Al    (2b)
   0.58700000000000    0.58700000000000    0.36800000000000    Al    (4i)
  -0.58700000000000   -0.58700000000000   -0.36800000000000    Al    (4i)
   0.25700000000000    0.25700000000000    0.61300000000000    Al    (4i)
  -0.25700000000000   -0.25700000000000   -0.61300000000000    Al    (4i)
   0.13200000000000    0.13200000000000    0.16200000000000    Al    (4i)
  -0.13200000000000   -0.13200000000000   -0.16200000000000    Al    (4i)
   0.79000000000000    0.79000000000000    0.08700000000000    Al    (4i)
  -0.79000000000000   -0.79000000000000   -0.08700000000000    Al    (4i)
  -0.08600000000000   -0.08600000000000    0.43200000000000    Al    (4i)
   0.08600000000000    0.08600000000000   -0.43200000000000    Al    (4i)
   0.40900000000000    0.40900000000000    0.19400000000000    Al    (4i)
  -0.40900000000000   -0.40900000000000   -0.19400000000000    Al    (4i)
   0.29400000000000    0.29400000000000    0.29150000000000    Os    (4i)
  -0.29400000000000   -0.29400000000000   -0.29150000000000    Os    (4i)
  -0.00810000000000   -0.00810000000000    0.19470000000000    Os    (4i)
   0.00810000000000    0.00810000000000   -0.19470000000000    Os    (4i)
```

Bischofite ($MgCl_2 \cdot 6H_2O$, $J1_7$): A2B12CD6_mC42_12_i_2i2j_a_ij - CIF

```
# CIF file
data_findsym-output
_audit_creation_method FINDSYM

_chemical_name_mineral 'Bischofite'
_chemical_formula_sum 'Cl2 H12 Mg O6'

loop_
_publ_author_name
 'P. A. Agron'
 'W. R. Busing'
_journal_name_full_name
;
 Acta Crystallographica Section C: Structural Chemistry
;
_journal_volume 41
_journal_year 1985
_journal_page_first 8
_journal_page_last 10
_publ_Section_title
;
 Magnesium dichloride hexahydrate, MgCl$_{2}$$\cdot$6H$_{2}$O, by
    ↪ neutron diffraction
;

_aflow_title 'Bischofite (MgCl$_{2}$$\cdot$6H$_{2}$O, $J1_{7}$)
    ↪ Structure'
_aflow_proto 'A2B12CD6_mC42_12_i_2i2j_a_ij'
_aflow_params 'a,b/a,c/a,\beta,x_{2},z_{2},x_{3},z_{3},x_{4},z_{4},x_{5
    ↪ },z_{5},x_{6},y_{6},z_{6},x_{7},y_{7},z_{7},x_{8},y_{8},z_{8}'
_aflow_params_values '9.8607,0.720750048171,0.615950186092,93.758,0.3176
    ↪ ,0.6122,0.2372,0.2583,0.2693,0.0083,0.2019,0.1095,0.0209,0.2997
    ↪ ,0.2784,0.8839,0.1984,0.3151,-0.0429,0.2067,0.2233'
_aflow_Strukturbericht '$J1_{7}$'
_aflow_Pearson 'mC42'

_symmetry_space_group_name_H-M "C 1 2/m 1"
_symmetry_Int_Tables_number 12

_cell_length_a     9.86070
_cell_length_b     7.10710
_cell_length_c     6.07370
_cell_angle_alpha  90.00000
_cell_angle_beta   93.75800
_cell_angle_gamma  90.00000

loop_
_space_group_symop_id
_space_group_symop_operation_xyz
1 x,y,z
```

```
2 -x,y,-z
3 -x,-y,-z
4 x,-y,z
5 x+1/2,y+1/2,z
6 -x+1/2,y+1/2,-z
7 -x+1/2,-y+1/2,-z
8 x+1/2,-y+1/2,z

loop_
_atom_site_label
_atom_site_type_symbol
_atom_site_symmetry_multiplicity
_atom_site_Wyckoff_label
_atom_site_fract_x
_atom_site_fract_y
_atom_site_fract_z
_atom_site_occupancy
Mg1 Mg 2 a 0.00000 0.00000 0.00000 1.00000
Cl1 Cl 4 i 0.31760 0.00000 0.61220 1.00000
H1 H 4 i 0.23720 0.00000 0.25830 1.00000
H2 H 4 i 0.26930 0.00000 0.00830 1.00000
O1 O 4 i 0.20190 0.00000 0.10950 1.00000
H3 H 8 j 0.02090 0.29970 0.27840 1.00000
H4 H 8 j 0.88390 0.19840 0.31510 1.00000
O2 O 8 j -0.04290 0.20670 0.22330 1.00000
```

Bischofite ($MgCl_2{\cdot}6H_2O$, $J1_7$): A2B12CD6_mC42_12_i_2i2j_a_ij - POSCAR

```
A2B12CD6_mC42_12_i_2i2j_a_ij & a,b/a,c/a,beta,x2,z2,x3,z3,x4,z4,x5,z5,x6
   ↪ ,y6,z6,x7,y7,z7,x8,y8,z8 --params=9.8607,0.720750048171,
   ↪ 0.615950186092,93.758,0.3176,0.6122,0.2372,0.2583,0.2693,0.0083
   ↪ ,0.2019,0.1095,0.0209,0.2997,0.2784,0.8839,0.1984,0.3151,-
   ↪ 0.0429,0.2067,0.2233 & C2/m C_{2h}^{3} #12 (ai^4j^3) & mC42 &
   ↪ $J1_{7}$ & Cl2H12MgO6 & Bischofite & P. A. Agron and W. R.
   ↪ Busing, Acta Crystallogr. C 41, 8-10 (1985)
   1.00000000000000
   4.93035000000000  -3.55355000000000   0.00000000000000
   4.93035000000000   3.55355000000000   0.00000000000000
  -0.39808521489547   0.00000000000000   6.06064021797051
    Cl     H    Mg     O
     2    12     1     6
Direct
   0.31760000000000   0.31760000000000   0.61220000000000   Cl   (4i)
  -0.31760000000000  -0.31760000000000  -0.61220000000000   Cl   (4i)
   0.23720000000000   0.23720000000000   0.25830000000000   H    (4i)
  -0.23720000000000  -0.23720000000000  -0.25830000000000   H    (4i)
   0.26930000000000   0.26930000000000   0.00830000000000   H    (4i)
  -0.26930000000000  -0.26930000000000  -0.00830000000000   H    (4i)
  -0.27880000000000   0.32060000000000   0.27840000000000   H    (8j)
  -0.32060000000000   0.27880000000000  -0.27840000000000   H    (8j)
   0.27880000000000  -0.32060000000000  -0.27840000000000   H    (8j)
   0.32060000000000  -0.27880000000000   0.27840000000000   H    (8j)
   0.68550000000000   1.08230000000000   0.31510000000000   H    (8j)
  -1.08230000000000  -0.68550000000000  -0.31510000000000   H    (8j)
  -0.68550000000000  -1.08230000000000  -0.31510000000000   H    (8j)
   1.08230000000000   0.68550000000000   0.31510000000000   H    (8j)
   0.00000000000000   0.00000000000000   0.00000000000000   Mg   (2a)
   0.20190000000000   0.20190000000000   0.10950000000000   O    (4i)
  -0.20190000000000  -0.20190000000000  -0.10950000000000   O    (4i)
  -0.24960000000000   0.16380000000000   0.22330000000000   O    (8j)
  -0.16380000000000   0.24960000000000  -0.22330000000000   O    (8j)
   0.24960000000000  -0.16380000000000  -0.22330000000000   O    (8j)
   0.16380000000000  -0.24960000000000   0.22330000000000   O    (8j)
```

Tremolite ($Ca_2Mg_5Si_8O_{22}(OH)_2$, $S4_2$): A2B2C5D24E8_mC82_12_h_i_agh_2i5j_2j - CIF

```
# CIF file
data_findsym-output
_audit_creation_method FINDSYM

_chemical_name_mineral 'Tremolite'
_chemical_formula_sum 'Ca2 H2 Mg5 O24 Si8'

loop_
_publ_author_name
 'M. Merli'
 'L. Ungaretti'
 'R. Oberti'
_journal_name_full_name
;
 American Mineralogist
;
_journal_volume 85
_journal_year 2000
_journal_page_first 532
_journal_page_last 542
_publ_Section_title
;
 Leverage analysis and structure refinement of minerals
;

# Found in The American Mineralogist Crystal Structure Database, 2003

_aflow_title 'Tremolite (Ca$_{2}$Mg$_{5}$Si$_{8}$O$_{22}$(OH)$_{2}$,
   ↪ $S4_{2}$) Structure'
_aflow_proto 'A2B2C5D24E8_mC82_12_h_i_agh_2i5j_2j'
_aflow_params 'a,b/a,c/a,\beta,y_{2},y_{3},y_{4},x_{5},z_{5},x_{6},z_{6
   ↪ },x_{7},z_{7},x_{8},y_{8},z_{8},x_{9},y_{9},z_{9},x_{10},y_{10
   ↪ },z_{10},x_{11},y_{11},z_{11},x_{12},y_{12},z_{12},x_{13},y_{13
   ↪ },z_{13},x_{14},y_{14},z_{14}'
_aflow_params_values '9.8359,1.83460588253,0.536321028071,104.75,0.1765,
   ↪ 0.278,0.0878,0.196,0.764,0.1085,0.7155,0.3377,0.2928,0.1119,
   ↪ 0.0857,0.218,0.1187,0.1709,0.7244,0.1351,0.2519,0.2069,0.3466,
   ↪ 0.1344,0.1005,0.344,0.1188,0.5891,0.2806,0.0839,0.2972,0.2884,
   ↪ 0.1711,0.8047'
_aflow_Strukturbericht '$S4_{2}$'
_aflow_Pearson 'mC82'

_symmetry_space_group_name_H-M "C 1 2/m 1"
_symmetry_Int_Tables_number 12

_cell_length_a    9.83590
_cell_length_b    18.04500
_cell_length_c    5.27520
_cell_angle_alpha 90.00000
_cell_angle_beta  104.75000
_cell_angle_gamma 90.00000

loop_
_space_group_symop_id
_space_group_symop_operation_xyz
1 x,y,z
2 -x,y,-z
3 -x,-y,-z
4 x,-y,z
5 x+1/2,y+1/2,z
6 -x+1/2,y+1/2,-z
7 -x+1/2,-y+1/2,-z
8 x+1/2,-y+1/2,z

loop_
_atom_site_label
_atom_site_type_symbol
_atom_site_symmetry_multiplicity
_atom_site_Wyckoff_label
_atom_site_fract_x
_atom_site_fract_y
_atom_site_fract_z
_atom_site_occupancy
Mg1 Mg 2 a 0.00000 0.00000 0.00000 1.00000
Mg2 Mg 4 g 0.00000 0.17650 0.00000 1.00000
Ca1 Ca 4 h 0.00000 0.27800 0.50000 1.00000
Mg3 Mg 4 h 0.00000 0.08780 0.50000 1.00000
H1 H 4 i 0.19600 0.00000 0.76400 1.00000
O1 O 4 i 0.10850 0.00000 0.71550 1.00000
O2 O 4 i 0.33770 0.00000 0.29280 1.00000
O3 O 8 j 0.11190 0.08570 0.21800 1.00000
O4 O 8 j 0.11870 0.17090 0.72440 1.00000
O5 O 8 j 0.13510 0.25190 0.20690 1.00000
O6 O 8 j 0.34660 0.13440 0.10050 1.00000
O7 O 8 j 0.34400 0.11880 0.58910 1.00000
Si1 Si 8 j 0.28060 0.08390 0.29720 1.00000
Si2 Si 8 j 0.28840 0.17110 0.80470 1.00000
```

Tremolite ($Ca_2Mg_5Si_8O_{22}(OH)_2$, $S4_2$): A2B2C5D24E8_mC82_12_h_i_agh_2i5j_2j - POSCAR

```
A2B2C5D24E8_mC82_12_h_i_agh_2i5j_2j & a,b/a,c/a,beta,y2,y3,y4,x5,z5,x6,
   ↪ z6,x7,z7,x8,y8,z8,x9,y9,z9,x10,y10,z10,x11,y11,z11,x12,y12,z12,
   ↪ x13,y13,z13,x14,y14,z14 --params=9.8359,1.83460588253,
   ↪ 0.536321028071,104.75,0.1765,0.278,0.0878,0.196,0.764,0.1085,
   ↪ 0.7155,0.3377,0.2928,0.1119,0.0857,0.218,0.1187,0.1709,0.7244,
   ↪ 0.1351,0.2519,0.2069,0.3466,0.1344,0.1005,0.344,0.1188,0.5891,
   ↪ 0.2806,0.0839,0.2972,0.2884,0.1711,0.8047 & C2/m C_{2h}^{3} #12
   ↪ (agh^2i^3j^7) & mC82 & $S4_{2}$ & Ca2H2Mg5O24Si8 & Tremolite &
   ↪ M. Merli and L. Ungaretti and R. Oberti, Am. Mineral. 85,
   ↪ 532-542 (2000)
   1.00000000000000
   4.91795000000000  -9.02250000000000   0.00000000000000
   4.91795000000000   9.02250000000000   0.00000000000000
  -1.34307619717380   0.00000000000000   5.10136073695883
    Ca     H    Mg     O    Si
     2     2     5    24     8
Direct
  -0.27800000000000   0.27800000000000   0.50000000000000   Ca   (4h)
   0.27800000000000  -0.27800000000000   0.50000000000000   Ca   (4h)
   0.19600000000000   0.19600000000000   0.76400000000000   H    (4i)
  -0.19600000000000  -0.19600000000000  -0.76400000000000   H    (4i)
   0.00000000000000   0.00000000000000   0.00000000000000   Mg   (2a)
  -0.17650000000000   0.17650000000000   0.00000000000000   Mg   (4g)
   0.17650000000000  -0.17650000000000   0.00000000000000   Mg   (4g)
  -0.08780000000000   0.08780000000000   0.50000000000000   Mg   (4h)
   0.08780000000000  -0.08780000000000   0.50000000000000   Mg   (4h)
   0.10850000000000   0.10850000000000   0.71550000000000   O    (4i)
  -0.10850000000000  -0.10850000000000  -0.71550000000000   O    (4i)
   0.33770000000000   0.33770000000000   0.29280000000000   O    (4i)
  -0.33770000000000  -0.33770000000000  -0.29280000000000   O    (4i)
   0.02620000000000   0.19760000000000   0.21800000000000   O    (8j)
  -0.19760000000000  -0.02620000000000  -0.21800000000000   O    (8j)
  -0.02620000000000  -0.19760000000000  -0.21800000000000   O    (8j)
   0.19760000000000   0.02620000000000   0.21800000000000   O    (8j)
  -0.05220000000000   0.28960000000000   0.72440000000000   O    (8j)
  -0.28960000000000   0.05220000000000  -0.72440000000000   O    (8j)
   0.05220000000000  -0.28960000000000  -0.72440000000000   O    (8j)
   0.28960000000000  -0.05220000000000   0.72440000000000   O    (8j)
  -0.11680000000000   0.38700000000000   0.20690000000000   O    (8j)
  -0.38700000000000   0.11680000000000  -0.20690000000000   O    (8j)
   0.11680000000000  -0.38700000000000  -0.20690000000000   O    (8j)
   0.38700000000000  -0.11680000000000   0.20690000000000   O    (8j)
   0.21220000000000   0.48100000000000   0.10050000000000   O    (8j)
  -0.48100000000000  -0.21220000000000  -0.10050000000000   O    (8j)
  -0.21220000000000  -0.48100000000000  -0.10050000000000   O    (8j)
   0.48100000000000   0.21220000000000   0.10050000000000   O    (8j)
   0.22520000000000   0.46280000000000   0.58910000000000   O    (8j)
  -0.46280000000000  -0.22520000000000  -0.58910000000000   O    (8j)
  -0.22520000000000  -0.46280000000000  -0.58910000000000   O    (8j)
   0.46280000000000   0.22520000000000   0.58910000000000   O    (8j)
   0.19670000000000   0.36450000000000   0.29720000000000   Si   (8j)
  -0.36450000000000  -0.19670000000000  -0.29720000000000   Si   (8j)
  -0.19670000000000  -0.36450000000000  -0.29720000000000   Si   (8j)
   0.36450000000000   0.19670000000000   0.29720000000000   Si   (8j)
```

```
   0.11730000000000    0.45950000000000    0.80470000000000    Si    (8j)
  -0.45950000000000   -0.11730000000000   -0.80470000000000    Si    (8j)
  -0.11730000000000   -0.45950000000000   -0.80470000000000    Si    (8j)
   0.45950000000000    0.11730000000000    0.80470000000000    Si    (8j)
```

$\beta$-$Ga_2O_3$: A2B3_mC20_12_2i_3i - CIF

```
# CIF file
data_findsym-output
_audit_creation_method FINDSYM

_chemical_name_mineral 'Ga2O3'
_chemical_formula_sum 'Ga2 O3'

loop_
_publ_author_name
 'J. {\AA}hman'
 'G. Svensson'
 'J. Albertsson'
_journal_name_full_name
;
 Acta Crystallographica Section C: Structural Chemistry
;
_journal_volume 52
_journal_year 1996
_journal_page_first 1336
_journal_page_last 1338
_publ_Section_title
;
 A Reinvestigation of $\beta$-Gallium Oxide
;

_aflow_title '$\beta$-Ga$_{2}$O$_{3}$ Structure'
_aflow_proto 'A2B3_mC20_12_2i_3i'
_aflow_params 'a,b/a,c/a,\beta,x_{1},z_{1},x_{2},z_{2},x_{3},z_{3},x_{4
    ↪ },z_{4},x_{5},z_{5}'
_aflow_params_values '12.214,0.248657278533,0.474709349926,103.83,0.0905
    ↪ ,0.7946,0.65866,0.31402,0.1645,0.1098,0.1733,0.5632,0.4959,
    ↪ 0.2566'
_aflow_Strukturbericht 'None'
_aflow_Pearson 'mC20'

_symmetry_space_group_name_H-M "C 1 2/m 1"
_symmetry_Int_Tables_number 12

_cell_length_a    12.21400
_cell_length_b    3.03710
_cell_length_c    5.79810
_cell_angle_alpha 90.00000
_cell_angle_beta  103.83000
_cell_angle_gamma 90.00000

loop_
_space_group_symop_id
_space_group_symop_operation_xyz
1 x,y,z
2 -x,y,-z
3 -x,-y,-z
4 x,-y,z
5 x+1/2,y+1/2,z
6 -x+1/2,y+1/2,-z
7 -x+1/2,-y+1/2,-z
8 x+1/2,-y+1/2,z

loop_
_atom_site_label
_atom_site_type_symbol
_atom_site_symmetry_multiplicity
_atom_site_Wyckoff_label
_atom_site_fract_x
_atom_site_fract_y
_atom_site_fract_z
_atom_site_occupancy
Ga1 Ga 4 i 0.09050 0.00000 0.79460 1.00000
Ga2 Ga 4 i 0.65866 0.00000 0.31402 1.00000
O1 O 4 i 0.16450 0.00000 0.10980 1.00000
O2 O 4 i 0.17330 0.00000 0.56320 1.00000
O3 O 4 i 0.49590 0.00000 0.25660 1.00000
```

$\beta$-$Ga_2O_3$: A2B3_mC20_12_2i_3i - POSCAR

```
A2B3_mC20_12_2i_3i & a,b/a,c/a,beta,x1,z1,x2,z2,x3,z3,x4,z4,x5,z5 --
    ↪ params=12.214,0.248657278533,0.474709349926,103.83,0.0905,
    ↪ 0.7946,0.65866,0.31402,0.1645,0.1098,0.1733,0.5632,0.4959,
    ↪ 0.2566 & C2/m C_{2h}^{3} #12 (i^5) & mC20 & None & Ga2O3 &
    ↪ Ga2O3 & J. {\AA}hman and G. Svensson and J. Albertsson, Acta
    ↪ Crystallogr. C 52, 1336-1338 (1996)
   1.00000000000000
   6.10700000000000   -1.51855000000000    0.00000000000000
   6.10700000000000    1.51855000000000    0.00000000000000
  -1.38598889592188    0.00000000000000    5.63000873803774
    Ga        O
     4        6
Direct
   0.09050000000000    0.09050000000000    0.79460000000000    Ga    (4i)
  -0.09050000000000   -0.09050000000000   -0.79460000000000    Ga    (4i)
   0.65866000000000    0.65866000000000    0.31402000000000    Ga    (4i)
  -0.65866000000000   -0.65866000000000   -0.31402000000000    Ga    (4i)
   0.16450000000000    0.16450000000000    0.10980000000000    O     (4i)
  -0.16450000000000   -0.16450000000000   -0.10980000000000    O     (4i)
   0.17330000000000    0.17330000000000    0.56320000000000    O     (4i)
  -0.17330000000000   -0.17330000000000   -0.56320000000000    O     (4i)
   0.49590000000000    0.49590000000000    0.25660000000000    O     (4i)
  -0.49590000000000   -0.49590000000000   -0.25660000000000    O     (4i)
```

$K_2Ti_2O_5$: A2B5C2_mC18_12_i_a2i_i - CIF

```
# CIF file
data_findsym-output
_audit_creation_method FINDSYM

_chemical_name_mineral 'K2O5Ti2'
_chemical_formula_sum 'K2 O5 Ti2'

loop_
_publ_author_name
 'S. Andersson'
 'A. D. Wadsley'
_journal_name_full_name
;
 Acta Chemica Scandinavica
;
_journal_volume 15
_journal_year 1961
_journal_page_first 663
_journal_page_last 669
_publ_Section_title
;
 The Crystal Structure of K$_{2}$Ti$_{2}$O$_{5}$
;

_aflow_title 'K$_{2}$Ti$_{2}$O$_{5}$ Structure'
_aflow_proto 'A2B5C2_mC18_12_i_a2i_i'
_aflow_params 'a,b/a,c/a,\beta,x_{2},z_{2},x_{3},z_{3},x_{4},z_{4},x_{5
    ↪ },z_{5}'
_aflow_params_values '11.37,0.334212840809,0.582233948989,100.1,0.4022,
    ↪ 0.6439,0.128,0.664,0.325,0.008,0.1495,-0.0928'
_aflow_Strukturbericht 'None'
_aflow_Pearson 'mC18'

_symmetry_space_group_name_H-M "C 1 2/m 1"
_symmetry_Int_Tables_number 12

_cell_length_a    11.37000
_cell_length_b    3.80000
_cell_length_c    6.62000
_cell_angle_alpha 90.00000
_cell_angle_beta  100.10000
_cell_angle_gamma 90.00000

loop_
_space_group_symop_id
_space_group_symop_operation_xyz
1 x,y,z
2 -x,y,-z
3 -x,-y,-z
4 x,-y,z
5 x+1/2,y+1/2,z
6 -x+1/2,y+1/2,-z
7 -x+1/2,-y+1/2,-z
8 x+1/2,-y+1/2,z

loop_
_atom_site_label
_atom_site_type_symbol
_atom_site_symmetry_multiplicity
_atom_site_Wyckoff_label
_atom_site_fract_x
_atom_site_fract_y
_atom_site_fract_z
_atom_site_occupancy
O1 O 2 a 0.00000 0.00000 0.00000 1.00000
K1 K 4 i 0.40220 0.00000 0.64390 1.00000
O2 O 4 i 0.12800 0.00000 0.66400 1.00000
O3 O 4 i 0.32500 0.00000 0.00800 1.00000
Ti1 Ti 4 i 0.14950 0.00000 -0.09280 1.00000
```

$K_2Ti_2O_5$: A2B5C2_mC18_12_i_a2i_i - POSCAR

```
A2B5C2_mC18_12_i_a2i_i & a,b/a,c/a,beta,x2,z2,x3,z3,x4,z4,x5,z5 --params
    ↪ =11.37,0.334212840809,0.582233948989,100.1,0.4022,0.6439,0.128,
    ↪ 0.664,0.325,0.008,0.1495,-0.0928 & C2/m C_{2h}^{3} #12 (ai^4) &
    ↪  mC18 & None & K2O5Ti2 & K2O5Ti2 & S. Andersson and A. D.
    ↪ Wadsley, Acta Chem. Scand. 15, 663-669 (1961)
   1.00000000000000
   5.68500000000000   -1.90000000000000    0.00000000000000
   5.68500000000000    1.90000000000000    0.00000000000000
  -1.16092772672895    0.00000000000000    6.51741105143077
     K        O       Ti
     2        5        2
Direct
   0.40220000000000    0.40220000000000    0.64390000000000    K     (4i)
  -0.40220000000000   -0.40220000000000   -0.64390000000000    K     (4i)
   0.00000000000000    0.00000000000000    0.00000000000000    O     (2a)
   0.12800000000000    0.12800000000000    0.66400000000000    O     (4i)
  -0.12800000000000   -0.12800000000000   -0.66400000000000    O     (4i)
   0.32500000000000    0.32500000000000    0.00800000000000    O     (4i)
  -0.32500000000000   -0.32500000000000   -0.00800000000000    O     (4i)
   0.14950000000000    0.14950000000000   -0.09280000000000    Ti    (4i)
  -0.14950000000000   -0.14950000000000    0.09280000000000    Ti    (4i)
```

$CaC_2$-III: A2B_mC12_12_2i_i - CIF

```
# CIF file
data_findsym-output
_audit_creation_method FINDSYM

_chemical_name_mineral 'C2Ca'
_chemical_formula_sum 'C2 Ca'
```

```
loop_
_publ_author_name
 'M. Knapp'
 'U. Ruschewitz'
_journal_name_full_name
;
 Chemistry - A European Journal
;
_journal_volume 7
_journal_year 2001
_journal_page_first 874
_journal_page_last 880
_publ_Section_title
;
 Structural Phase Transitions in CaC$_{2}$
;

_aflow_title 'CaC$_{2}$-III Structure'
_aflow_proto 'A2B_mC12_12_2i_i'
_aflow_params 'a,b/a,c/a,\beta,x_{1},z_{1},x_{2},z_{2},x_{3},z_{3}'
_aflow_params_values '7.2286,0.532938604986,1.0204465595,107.338,0.439,
    ↪ 0.065,-0.075,0.447,0.2086,0.2486'
_aflow_Strukturbericht 'None'
_aflow_Pearson 'mC12'

_symmetry_space_group_name_H-M "C 1 2/m 1"
_symmetry_Int_Tables_number 12

_cell_length_a      7.22860
_cell_length_b      3.85240
_cell_length_c      7.37640
_cell_angle_alpha 90.00000
_cell_angle_beta  107.33800
_cell_angle_gamma 90.00000

loop_
_space_group_symop_id
_space_group_symop_operation_xyz
1 x,y,z
2 -x,y,-z
3 -x,-y,-z
4 x,-y,z
5 x+1/2,y+1/2,z
6 -x+1/2,y+1/2,-z
7 -x+1/2,-y+1/2,-z
8 x+1/2,-y+1/2,z

loop_
_atom_site_label
_atom_site_type_symbol
_atom_site_symmetry_multiplicity
_atom_site_Wyckoff_label
_atom_site_fract_x
_atom_site_fract_y
_atom_site_fract_z
_atom_site_occupancy
C1 C 4 i 0.43900 0.00000 0.06500 1.00000
C2 C 4 i -0.07500 0.00000 0.44700 1.00000
Ca1 Ca 4 i 0.20860 0.00000 0.24860 1.00000
```

$CaC_2$-III: A2B_mC12_12_2i_i - POSCAR

```
A2B_mC12_12_2i_i & a,b/a,c/a,beta,x1,z1,x2,z2,x3,z3 --params=7.2286,
    ↪ 0.532938604986,1.0204465595,107.338,0.439,0.065,-0.075,0.447,
    ↪ 0.2086,0.2486 & C2/m C_{2h}^{3} #12 (i^3) & mC12 & None & C2Ca
    ↪ & C2Ca & M. Knapp and U. Ruschewitz, Chem. Eur. J. 7, 874-880 (
    ↪ 2001)
   1.00000000000000
   3.61430000000000  -1.92620000000000   0.00000000000000
   3.61430000000000   1.92620000000000   0.00000000000000
  -2.19822643227335   0.00000000000000   7.04124119118603
     C     Ca
     4      2
Direct
   0.43900000000000   0.43900000000000   0.06500000000000   C   (4i)
  -0.43900000000000  -0.43900000000000  -0.06500000000000   C   (4i)
  -0.07500000000000  -0.07500000000000   0.44700000000000   C   (4i)
   0.07500000000000   0.07500000000000  -0.44700000000000   C   (4i)
   0.20860000000000   0.20860000000000   0.24860000000000   Ca  (4i)
  -0.20860000000000  -0.20860000000000  -0.24860000000000   Ca  (4i)
```

Tolbachite ($CuCl_2$): A2B_mC6_12_i_a - CIF

```
# CIF file
data_findsym-output
_audit_creation_method FINDSYM

_chemical_name_mineral 'Tolbachite'
_chemical_formula_sum 'Cl2 Cu'

loop_
_publ_author_name
 'P. C. Burns'
 'F. C. Hawthorne'
_journal_name_full_name
;
 American Mineralogist
;
_journal_volume 78
_journal_year 1993
_journal_page_first 187
_journal_page_last 189
_publ_Section_title
;
 Tolbachite, CuCl$_{2}$, the first example of Cu$_{2}^{+}$ octahedrally
    ↪ coordinated by Cl$^{-}$
;

# Found in The American Mineralogist Crystal Structure Database, 2003

_aflow_title 'Tolbachite (CuCl$_{2}$) Structure'
_aflow_proto 'A2B_mC6_12_i_a'
_aflow_params 'a,b/a,c/a,\beta,x_{2},z_{2}'
_aflow_params_values '6.9038,0.477925200614,0.988441148353,122.197,
    ↪ 0.5048,0.2294'
_aflow_Strukturbericht 'None'
_aflow_Pearson 'mC6'

_symmetry_space_group_name_H-M "C 1 2/m 1"
_symmetry_Int_Tables_number 12

_cell_length_a      6.90380
_cell_length_b      3.29950
_cell_length_c      6.82400
_cell_angle_alpha 90.00000
_cell_angle_beta  122.19700
_cell_angle_gamma 90.00000

loop_
_space_group_symop_id
_space_group_symop_operation_xyz
1 x,y,z
2 -x,y,-z
3 -x,-y,-z
4 x,-y,z
5 x+1/2,y+1/2,z
6 -x+1/2,y+1/2,-z
7 -x+1/2,-y+1/2,-z
8 x+1/2,-y+1/2,z

loop_
_atom_site_label
_atom_site_type_symbol
_atom_site_symmetry_multiplicity
_atom_site_Wyckoff_label
_atom_site_fract_x
_atom_site_fract_y
_atom_site_fract_z
_atom_site_occupancy
Cu1 Cu 2 a 0.00000 0.00000 0.00000 1.00000
Cl1 Cl 4 i 0.50480 0.00000 0.22940 1.00000
```

Tolbachite ($CuCl_2$): A2B_mC6_12_i_a - POSCAR

```
A2B_mC6_12_i_a & a,b/a,c/a,beta,x2,z2 --params=6.9038,0.477925200614,
    ↪ 0.988441148353,122.197,0.5048,0.2294 & C2/m C_{2h}^{3} #12 (ai)
    ↪ & mC6 & None & Cl2Cu & Tolbachite & P. C. Burns and F. C.
    ↪ Hawthorne, Am. Mineral. 78, 187-189 (1993)
   1.00000000000000
   3.45190000000000  -1.64975000000000   0.00000000000000
   3.45190000000000   1.64975000000000   0.00000000000000
  -3.63604535488460   0.00000000000000   5.77461255645971
     Cl     Cu
      2      1
Direct
   0.50480000000000   0.50480000000000   0.22940000000000   Cl  (4i)
  -0.50480000000000  -0.50480000000000  -0.22940000000000   Cl  (4i)
   0.00000000000000   0.00000000000000   0.00000000000000   Cu  (2a)
```

$\delta$-$Ni_3Sn_4$ ($D7_a$): A3B4_mC14_12_ai_2i - CIF

```
# CIF file
data_findsym-output
_audit_creation_method FINDSYM

_chemical_name_mineral 'Ni3Sn4'
_chemical_formula_sum 'Ni3 Sn4'

loop_
_publ_author_name
 'W. Jeitschko'
 'B. Jaberg'
_journal_name_full_name
;
 Acta Crystallographica Section B: Structural Science
;
_journal_volume 38
_journal_year 1982
_journal_page_first 598
_journal_page_last 600
_publ_Section_title
;
 Structure refinement of Ni$_{3}$Sn$_{4}$
;

_aflow_title '$\delta$-Ni$_{3}$Sn$_{4}$ ($D7_{a}$) Structure'
_aflow_proto 'A3B4_mC14_12_ai_2i'
_aflow_params 'a,b/a,c/a,\beta,x_{2},z_{2},x_{3},z_{3},x_{4},z_{4}'
_aflow_params_values '12.214,0.332405436384,0.427296544948,105.0,0.2147,
    ↪ 0.3369,0.4286,0.6864,0.1718,0.8123'
_aflow_Strukturbericht '$D7_{a}$'
_aflow_Pearson 'mC14'

_symmetry_space_group_name_H-M "C 1 2/m 1"
_symmetry_Int_Tables_number 12

_cell_length_a      12.21400
_cell_length_b      4.06000
_cell_length_c      5.21900
```

```
_cell_angle_alpha  90.00000
_cell_angle_beta   105.00000
_cell_angle_gamma  90.00000

loop_
_space_group_symop_id
_space_group_symop_operation_xyz
1 x,y,z
2 -x,y,-z
3 -x,-y,-z
4 x,-y,z
5 x+1/2,y+1/2,z
6 -x+1/2,y+1/2,-z
7 -x+1/2,-y+1/2,-z
8 x+1/2,-y+1/2,z

loop_
_atom_site_label
_atom_site_type_symbol
_atom_site_symmetry_multiplicity
_atom_site_Wyckoff_label
_atom_site_fract_x
_atom_site_fract_y
_atom_site_fract_z
_atom_site_occupancy
Ni1  Ni  2 a 0.00000  0.00000  0.00000  1.00000
Ni2  Ni  4 i 0.21470  0.00000  0.33690  1.00000
Sn1  Sn  4 i 0.42860  0.00000  0.68640  1.00000
Sn2  Sn  4 i 0.17180  0.00000  0.81230  1.00000
```

$\delta$-$Ni_3Sn_4$ ($D7_a$): A3B4_mC14_12_ai_2i - POSCAR

```
A3B4_mC14_12_ai_2i & a,b/a,c/a,beta,x2,z2,x3,z3,x4,z4 --params=12.214,
    ↪ 0.332405436384,0.427296544948,105.0,0.2147,0.3369,0.4286,0.6864
    ↪ ,0.1718,0.8123 & C2/m C_{2h}^{3} #12 (ai^3) & mC14 & $D7_{a}$ &
    ↪  Ni3Sn4 & Ni3Sn4 & W. Jeitschko and B. Jaberg, Acta
    ↪ Crystallogr. Sect. B Struct. Sci. 38, 598-600 (1982)
   1.00000000000000
   6.10700000000000  -2.03000000000000   0.00000000000000
   6.10700000000000   2.03000000000000   0.00000000000000
  -1.35077659639006   0.00000000000000   5.04116688740265
    Ni      Sn
     3       4
Direct
   0.00000000000000   0.00000000000000   0.00000000000000   Ni   (2a)
   0.21470000000000   0.21470000000000   0.33690000000000   Ni   (4i)
  -0.21470000000000  -0.21470000000000  -0.33690000000000   Ni   (4i)
   0.42860000000000   0.42860000000000   0.68640000000000   Sn   (4i)
  -0.42860000000000  -0.42860000000000  -0.68640000000000   Sn   (4i)
   0.17180000000000   0.17180000000000   0.81230000000000   Sn   (4i)
  -0.17180000000000  -0.17180000000000  -0.81230000000000   Sn   (4i)
```

LiOH·$H_2O$ ($B36$): A3BC2_mC24_12_ij_h_gi - CIF

```
# CIF file
data_findsym-output
_audit_creation_method FINDSYM

_chemical_name_mineral 'H3LiO2'
_chemical_formula_sum 'H3 Li O2'

loop_
_publ_author_name
 'N. W. Alcock'
_journal_name_full_name
;
 Acta Crystallographica Section B: Structural Science
;
_journal_volume 27
_journal_year 1971
_journal_page_first 1682
_journal_page_last 1683
_publ_Section_title
;
 Refinement of the crystal structure of lithium hydroxide monohydrate
;

_aflow_title 'LiOH$\cdot$H$_{2}$O ($B36$) Structure'
_aflow_proto 'A3BC2_mC24_12_ij_h_gi'
_aflow_params 'a,b/a,c/a,\beta,y_{1},y_{2},x_{3},z_{3},x_{4},z_{4},x_{5
    ↪ },y_{5},z_{5}'
_aflow_params_values '7.37,1.12075983718,0.432835820896,110.3,0.2066,
    ↪ 0.3474,0.237,0.631,0.2857,0.3952,0.107,0.118,0.004'
_aflow_Strukturbericht '$B36$'
_aflow_Pearson 'mC24'

_symmetry_space_group_name_H-M "C 1 2/m 1"
_symmetry_Int_Tables_number 12

_cell_length_a     7.37000
_cell_length_b     8.26000
_cell_length_c     3.19000
_cell_angle_alpha  90.00000
_cell_angle_beta   110.30000
_cell_angle_gamma  90.00000

loop_
_space_group_symop_id
_space_group_symop_operation_xyz
1 x,y,z
2 -x,y,-z
3 -x,-y,-z
4 x,-y,z
5 x+1/2,y+1/2,z
6 -x+1/2,y+1/2,-z
7 -x+1/2,-y+1/2,-z
8 x+1/2,-y+1/2,z

loop_
_atom_site_label
_atom_site_type_symbol
_atom_site_symmetry_multiplicity
_atom_site_Wyckoff_label
_atom_site_fract_x
_atom_site_fract_y
_atom_site_fract_z
_atom_site_occupancy
O1  O  4 g 0.00000  0.20660  0.00000  1.00000
Li1  Li  4 h 0.00000  0.34740  0.50000  1.00000
H1  H  4 i 0.23700  0.00000  0.63100  1.00000
O2  O  4 i 0.28570  0.00000  0.39520  1.00000
H2  H  8 j 0.10700  0.11800  0.00400  1.00000
```

LiOH·$H_2O$ ($B36$): A3BC2_mC24_12_ij_h_gi - POSCAR

```
A3BC2_mC24_12_ij_h_gi & a,b/a,c/a,beta,y1,y2,x3,z3,x4,z4,x5,y5,z5 --
    ↪ params=7.37,1.12075983718,0.432835820896,110.3,0.2066,0.3474,
    ↪ 0.237,0.631,0.2857,0.3952,0.107,0.118,0.004 & C2/m C_{2h}^{3} #
    ↪ 12 (ghi^2j) & mC24 & $B36$ & H3LiO2 & H3LiO2 & N. W. Alcock,
    ↪ Acta Crystallogr. Sect. B Struct. Sci. 27, 1682-1683 (1971)
   1.00000000000000
   3.68500000000000  -4.13000000000000   0.00000000000000
   3.68500000000000   4.13000000000000   0.00000000000000
  -1.10672472851869   0.00000000000000   2.99186570141195
     H       Li       O
     6        2       4
Direct
   0.23700000000000   0.23700000000000   0.63100000000000   H    (4i)
  -0.23700000000000  -0.23700000000000  -0.63100000000000   H    (4i)
  -0.01100000000000   0.22500000000000   0.00400000000000   H    (8j)
  -0.22500000000000   0.01100000000000  -0.00400000000000   H    (8j)
   0.01100000000000  -0.22500000000000  -0.00400000000000   H    (8j)
   0.22500000000000  -0.01100000000000   0.00400000000000   H    (8j)
  -0.34740000000000   0.34740000000000   0.50000000000000   Li   (4h)
   0.34740000000000  -0.34740000000000   0.50000000000000   Li   (4h)
  -0.20660000000000   0.20660000000000   0.00000000000000   O    (4g)
   0.20660000000000  -0.20660000000000   0.00000000000000   O    (4g)
   0.28570000000000   0.28570000000000   0.39520000000000   O    (4i)
  -0.28570000000000  -0.28570000000000  -0.39520000000000   O    (4i)
```

Staurolite ($Al_5Fe_2O_{10}(OH)_2Si_2$): A5B2C10D2E2_mC84_12_acghj_bdi_5j_2i_j - CIF

```
# CIF file
data_findsym-output
_audit_creation_method FINDSYM

_chemical_name_mineral 'Staurolite'
_chemical_formula_sum 'Al5 Fe2 O10 (OH)2 Si2'

loop_
_publ_author_name
 'J. V. Smith'
_journal_name_full_name
;
 American Mineralogist
;
_journal_volume 53
_journal_year 1968
_journal_page_first 1139
_journal_page_last 1155
_publ_Section_title
;
 The crystal structure of staurolite
;

_aflow_title 'Staurolite (Al$_{5}$Fe$_{2}$O$_{10}$(OH)$_{2}$Si$_{2}$)
    ↪ Structure'
_aflow_proto 'A5B2C10D2E2_mC84_12_acghj_bdi_5j_2i_j'
_aflow_params 'a,b/a,c/a,\beta,y_{5},y_{6},x_{7},z_{7},x_{8},z_{8},x_{9
    ↪ },z_{9},x_{10},y_{10},z_{10},x_{11},y_{11},z_{11},x_{12},y_{12
    ↪ },z_{12},x_{13},y_{13},z_{13},x_{14},y_{14},z_{14},x_{15},y_{15
    ↪ },z_{15},x_{16},y_{16},z_{16}'
_aflow_params_values '7.8713,2.11151906292,0.718559831286,90.0,0.67511,
    ↪ 0.67477,0.39281,0.24815,0.23274,-0.03687,0.23438,0.53428,
    ↪ 0.26356,0.41042,0.25122,0.25569,0.16153,0.01527,0.25519,0.16127
    ↪ ,0.48391,0.00143,0.08917,0.24702,0.02156,0.24925,0.25002,
    ↪ 0.52741,0.10004,0.24944,0.13414,0.16612,0.24902'
_aflow_Strukturbericht 'None'
_aflow_Pearson 'mC84'

_symmetry_space_group_name_H-M "C 1 2/m 1"
_symmetry_Int_Tables_number 12

_cell_length_a     7.87130
_cell_length_b     16.62040
_cell_length_c     5.65600
_cell_angle_alpha  90.00000
_cell_angle_beta   90.00000
_cell_angle_gamma  90.00000

loop_
_space_group_symop_id
_space_group_symop_operation_xyz
1 x,y,z
2 -x,y,-z
3 -x,-y,-z
4 x,-y,z
5 x+1/2,y+1/2,z
6 -x+1/2,y+1/2,-z
7 -x+1/2,-y+1/2,-z
```

```
8 x+1/2,-y+1/2,z

loop_
_atom_site_label
_atom_site_type_symbol
_atom_site_symmetry_multiplicity
_atom_site_Wyckoff_label
_atom_site_fract_x
_atom_site_fract_y
_atom_site_fract_z
_atom_site_occupancy
Al1  Al  2  a  0.00000  0.00000  0.00000  1.00000
Fe1  Fe  2  b  0.00000  0.50000  0.00000  1.00000
Al2  Al  2  c  0.00000  0.00000  0.50000  1.00000
Fe2  Fe  2  d  0.00000  0.50000  0.50000  1.00000
Al3  Al  4  g  0.00000  0.67511  0.00000  1.00000
Al4  Al  4  h  0.00000  0.67477  0.50000  1.00000
Fe3  Fe  4  i  0.39281  0.00000  0.24815  1.00000
OH1  OH  4  i  0.23274  0.00000  -0.03687  1.00000
OH2  OH  4  i  0.23438  0.00000  0.53428  1.00000
Al5  Al  8  j  0.26356  0.41042  0.25122  1.00000
O1  O  8  j  0.25569  0.16153  0.01527  1.00000
O2  O  8  j  0.25519  0.16127  0.48391  1.00000
O3  O  8  j  0.00143  0.08917  0.24702  1.00000
O4  O  8  j  0.02156  0.24925  0.25002  1.00000
O5  O  8  j  0.52741  0.10004  0.24944  1.00000
Si1  Si  8  j  0.13414  0.16612  0.24902  1.00000
```

Staurolite ($Al_5Fe_2O_{10}(OH)_2Si_2$): A5B2C10D2E2_mC84_12_acghj_bdi_5j_2i_j - POSCAR

```
A5B2C10D2E2_mC84_12_acghj_bdi_5j_2i_j & a,b/a,c/a,beta,y5,y6,x7,z7,x8,z8
    ↪ ,x9,z9,x10,y10,z10,x11,y11,z11,x12,y12,z12,x13,y13,z13,x14,y14,
    ↪ z14,x15,y15,z15,x16,y16,z16 --params=7.8713,2.11151906292,
    ↪ 0.718559831286,90.0,0.67511,0.67477,0.39281,0.24815,0.23274,-
    ↪ 0.03687,0.23438,0.53428,0.26356,0.41042,0.25122,0.25569,0.16153
    ↪ ,0.01527,0.25519,0.16127,0.48391,0.00143,0.08917,0.24702,
    ↪ 0.02156,0.24925,0.25002,0.52741,0.10004,0.24944,0.13414,0.16612
    ↪ ,0.24902 & C2/m C_{2h}^{3} #12 (abcdghi^3j^7) & mC84 & None &
    ↪ Al5Fe2O10(OH)2Si2 & Staurolite & J. V. Smith, Am. Mineral. 53,
    ↪ 1139-1155 (1968)
   1.00000000000000
   3.93565000000000   -8.31020000000000    0.00000000000000
   3.93565000000000    8.31020000000000    0.00000000000000
   0.00000000000000    0.00000000000000    5.65600000000000
    Al     Fe      O     OH     Si
    10      4     20      4      4
Direct
   0.00000000000000    0.00000000000000    0.00000000000000   Al   (2a)
   0.00000000000000    0.00000000000000    0.50000000000000   Al   (2c)
  -0.67511000000000    0.67511000000000    0.00000000000000   Al   (4g)
   0.67511000000000   -0.67511000000000    0.00000000000000   Al   (4g)
  -0.67477000000000    0.67477000000000    0.50000000000000   Al   (4h)
   0.67477000000000   -0.67477000000000    0.50000000000000   Al   (4h)
  -0.14686000000000    0.67398000000000    0.25122000000000   Al   (8j)
  -0.67398000000000    0.14686000000000   -0.25122000000000   Al   (8j)
   0.14686000000000   -0.67398000000000   -0.25122000000000   Al   (8j)
   0.67398000000000   -0.14686000000000    0.25122000000000   Al   (8j)
   0.50000000000000    0.50000000000000    0.00000000000000   Fe   (2b)
   0.50000000000000    0.50000000000000    0.50000000000000   Fe   (2d)
   0.39281000000000    0.39281000000000    0.24815000000000   Fe   (4i)
  -0.39281000000000   -0.39281000000000   -0.24815000000000   Fe   (4i)
   0.09416000000000    0.41722000000000    0.01527000000000   O    (8j)
  -0.41722000000000   -0.09416000000000   -0.01527000000000   O    (8j)
  -0.09416000000000   -0.41722000000000   -0.01527000000000   O    (8j)
   0.41722000000000    0.09416000000000    0.01527000000000   O    (8j)
   0.09392000000000    0.41646000000000    0.48391000000000   O    (8j)
  -0.41646000000000   -0.09392000000000   -0.48391000000000   O    (8j)
  -0.09392000000000   -0.41646000000000   -0.48391000000000   O    (8j)
   0.41646000000000    0.09392000000000    0.48391000000000   O    (8j)
  -0.08774000000000    0.09060000000000    0.24702000000000   O    (8j)
  -0.09060000000000    0.08774000000000   -0.24702000000000   O    (8j)
   0.08774000000000   -0.09060000000000   -0.24702000000000   O    (8j)
   0.09060000000000   -0.08774000000000    0.24702000000000   O    (8j)
  -0.22769000000000    0.27081000000000    0.25002000000000   O    (8j)
  -0.27081000000000    0.22769000000000   -0.25002000000000   O    (8j)
   0.22769000000000   -0.27081000000000   -0.25002000000000   O    (8j)
   0.27081000000000   -0.22769000000000    0.25002000000000   O    (8j)
   0.42737000000000    0.62745000000000    0.24944000000000   O    (8j)
  -0.62745000000000   -0.42737000000000   -0.24944000000000   O    (8j)
  -0.42737000000000   -0.62745000000000   -0.24944000000000   O    (8j)
   0.62745000000000    0.42737000000000    0.24944000000000   O    (8j)
   0.23274000000000    0.23274000000000   -0.03687000000000   OH   (4i)
  -0.23274000000000   -0.23274000000000    0.03687000000000   OH   (4i)
   0.23438000000000    0.23438000000000    0.53428000000000   OH   (4i)
  -0.23438000000000   -0.23438000000000   -0.53428000000000   OH   (4i)
  -0.03198000000000    0.30026000000000    0.24902000000000   Si   (8j)
  -0.30026000000000    0.03198000000000   -0.24902000000000   Si   (8j)
   0.03198000000000   -0.30026000000000   -0.24902000000000   Si   (8j)
   0.30026000000000   -0.03198000000000    0.24902000000000   Si   (8j)
```

Manganese-leonite [$K_2Mn(SO_4)_2 \cdot 4H_2O$, $H4_{23}$]: A8B2CD15E2_mC112_12_2i3j_j_ad_g4i5j_2i - CIF

```
# CIF file
data_findsym-output
_audit_creation_method FINDSYM

_chemical_name_mineral 'Manganese-leonite'
_chemical_formula_sum 'H8 K2 Mn O15 S2'

loop_
_publ_author_name
 'B. Hertweck'
 'G. Giester'
 'E. Libowitzky'
_journal_name_full_name
;
 American Mineralogist
;
_journal_volume 86
_journal_year 2001
_journal_page_first 1282
_journal_page_last 1292
_publ_Section_title
;
 The crystal structures of the low-temperature phases of leonite-type
    ↪ compounds, K$_{2}$$Me$(SO$_{4}$)$_{2}$$\cdot$4H$_{2}$O ($Me^{2
    ↪ +}$ = Mg, Mn, Fe)
;

_aflow_title 'Manganese-leonite [K$_{2}$Mn(SO$_{4}$)$_{2}$$\cdot$4H$_{2}
    ↪ $O, $H4_{23}$] Structure'
_aflow_proto 'A8B2CD15E2_mC112_12_2i3j_j_ad_g4i5j_2i'
_aflow_params 'a,b/a,c/a,\beta,y_{3},x_{4},z_{4},x_{5},z_{5},x_{6},z_{6
    ↪ },x_{7},z_{7},x_{8},z_{8},x_{9},z_{9},x_{10},z_{10},x_{11},z_{
    ↪ 11},x_{12},y_{12},z_{12},x_{13},y_{13},z_{13},x_{14},y_{14},z_{
    ↪ 14},x_{15},y_{15},z_{15},x_{16},y_{16},z_{16},x_{17},y_{17},z_{
    ↪ 17},x_{18},y_{18},z_{18},x_{19},y_{19},z_{19},x_{20},y_{20},z_{
    ↪ 20}'
_aflow_params_values '12.017,0.798368977282,0.827910460181,95.03,0.2267,
    ↪ 0.061,0.743,-0.041,0.728,0.1728,0.0594,0.6841,0.0965,0.6686,
    ↪ 0.4354,0.0104,0.7806,0.70695,-0.04418,0.21077,0.52021,0.072,
    ↪ 0.276,0.621,0.079,0.349,0.726,0.047,0.271,-0.027,0.17022,
    ↪ 0.26326,0.24936,0.1588,0.3743,0.8871,0.1519,0.0973,0.5987,
    ↪ 0.1723,0.1457,0.548,0.1918,0.0236,0.3754,0.0387,0.3351,0.6481'
_aflow_Strukturbericht '$H4_{23}$'
_aflow_Pearson 'mC112'

_symmetry_space_group_name_H-M "C 1 2/m 1"
_symmetry_Int_Tables_number 12

_cell_length_a     12.01700
_cell_length_b     9.59400
_cell_length_c     9.94900
_cell_angle_alpha  90.00000
_cell_angle_beta   95.03000
_cell_angle_gamma  90.00000

loop_
_space_group_symop_id
_space_group_symop_operation_xyz
1 x,y,z
2 -x,y,-z
3 -x,-y,-z
4 x,-y,z
5 x+1/2,y+1/2,z
6 -x+1/2,y+1/2,-z
7 -x+1/2,-y+1/2,-z
8 x+1/2,-y+1/2,z

loop_
_atom_site_label
_atom_site_type_symbol
_atom_site_symmetry_multiplicity
_atom_site_Wyckoff_label
_atom_site_fract_x
_atom_site_fract_y
_atom_site_fract_z
_atom_site_occupancy
Mn1  Mn  2  a  0.00000  0.00000  0.00000  1.00000
Mn2  Mn  2  d  0.00000  0.50000  0.50000  1.00000
O1  O  4  g  0.00000  0.22670  0.00000  1.00000
H1  H  4  i  0.06100  0.00000  0.74300  1.00000
H2  H  4  i  -0.04100  0.00000  0.72800  1.00000
O2  O  4  i  0.17280  0.00000  0.05940  1.00000
O3  O  4  i  0.68410  0.00000  0.09650  1.00000
O4  O  4  i  0.66860  0.00000  0.43540  1.00000
O5  O  4  i  0.01040  0.00000  0.78060  1.00000
S1  S  4  i  0.70695  0.00000  -0.04418  1.00000
S2  S  4  i  0.21077  0.00000  0.52021  1.00000
H3  H  8  j  0.07200  0.27600  0.62100  1.00000
H4  H  8  j  0.07900  0.34900  0.72600  1.00000
H5  H  8  j  0.04700  0.27100  -0.02700  1.00000
K1  K  8  j  0.17022  0.26326  0.24936  1.00000
O6  O  8  j  0.15880  0.37430  0.88710  1.00000
O7  O  8  j  0.15190  0.09730  0.59870  0.50000
O8  O  8  j  0.17230  0.14570  0.54800  0.50000
O9  O  8  j  0.19180  0.02360  0.37540  0.50000
O10  O  8  j  0.03870  0.33510  0.64810  1.00000
```

Manganese-leonite [$K_2Mn(SO_4)_2 \cdot 4H_2O$, $H4_{23}$]: A8B2CD15E2_mC112_12_2i3j_j_ad_g4i5j_2i - POSCAR

```
A8B2CD15E2_mC112_12_2i3j_j_ad_g4i5j_2i & a,b/a,c/a,beta,y3,x4,z4,x5,z5,
    ↪ x6,z6,x7,z7,x8,z8,x9,z9,x10,z10,x11,z11,x12,y12,z12,x13,y13,z13
    ↪ ,x14,y14,z14,x15,y15,z15,x16,y16,z16,x17,y17,z17,x18,y18,z18,
    ↪ x19,y19,z19,x20,y20,z20 --params=12.017,0.798368977282,
    ↪ 0.827910460181,95.03,0.2267,0.061,0.743,-0.041,0.728,0.1728,
    ↪ 0.0594,0.6841,0.0965,0.6686,0.4354,0.0104,0.7806,0.70695,-
    ↪ 0.04418,0.21077,0.52021,0.072,0.276,0.621,0.079,0.349,0.726,
    ↪ 0.047,0.271,-0.027,0.17022,0.26326,0.24936,0.1588,0.3743,0.8871
    ↪ ,0.1519,0.0973,0.5987,0.1723,0.1457,0.548,0.1918,0.0236,0.3754,
    ↪ 0.0387,0.3351,0.6481 & C2/m C_{2h}^{3} #12 (adgi^8j^9) & mC112
    ↪ & $H4_{23}$ & H8K2MnO12S2 & Manganese-leonite & B. Hertweck and
    ↪ G. Giester and E. Libowitzky, Am. Mineral. 86, 1282-1292 (2001
    ↪ )
   1.00000000000000
   6.00850000000000   -4.79700000000000    0.00000000000000
   6.00850000000000    4.79700000000000    0.00000000000000
  -0.87230182681668    0.00000000000000    9.91068567370252
     H      K     Mn      O      S
```

```
      16         4         2        30         4
Direct
   0.06100000000000    0.06100000000000    0.74300000000000    H    (4i)
  -0.06100000000000   -0.06100000000000   -0.74300000000000    H    (4i)
  -0.04100000000000   -0.04100000000000    0.72800000000000    H    (4i)
   0.04100000000000    0.04100000000000   -0.72800000000000    H    (4i)
   0.20400000000000    0.34800000000000    0.62100000000000    H    (8j)
  -0.34800000000000    0.20400000000000   -0.62100000000000    H    (8j)
   0.20400000000000   -0.34800000000000   -0.62100000000000    H    (8j)
   0.34800000000000   -0.20400000000000    0.62100000000000    H    (8j)
  -0.27000000000000    0.42800000000000    0.72600000000000    H    (8j)
  -0.42800000000000    0.27000000000000   -0.72600000000000    H    (8j)
   0.27000000000000   -0.42800000000000   -0.72600000000000    H    (8j)
   0.42800000000000   -0.27000000000000    0.72600000000000    H    (8j)
  -0.22400000000000    0.31800000000000   -0.02700000000000    H    (8j)
  -0.31800000000000    0.22400000000000    0.02700000000000    H    (8j)
   0.22400000000000   -0.31800000000000    0.02700000000000    H    (8j)
   0.31800000000000   -0.22400000000000   -0.02700000000000    H    (8j)
  -0.09304000000000    0.43348000000000    0.24936000000000    K    (8j)
  -0.43348000000000    0.09304000000000   -0.24936000000000    K    (8j)
   0.09304000000000   -0.43348000000000   -0.24936000000000    K    (8j)
   0.43348000000000   -0.09304000000000    0.24936000000000    K    (8j)
   0.00000000000000    0.00000000000000    0.00000000000000    Mn   (2a)
   0.50000000000000    0.50000000000000    0.50000000000000    Mn   (2d)
  -0.22670000000000    0.22670000000000    0.00000000000000    O    (4g)
   0.22670000000000   -0.22670000000000    0.00000000000000    O    (4g)
   0.17280000000000    0.17280000000000    0.05940000000000    O    (4i)
  -0.17280000000000   -0.17280000000000   -0.05940000000000    O    (4i)
   0.68410000000000    0.68410000000000    0.09650000000000    O    (4i)
  -0.68410000000000   -0.68410000000000   -0.09650000000000    O    (4i)
   0.66860000000000    0.66860000000000    0.43540000000000    O    (4i)
  -0.66860000000000   -0.66860000000000   -0.43540000000000    O    (4i)
   0.01040000000000    0.01040000000000    0.78060000000000    O    (4i)
  -0.01040000000000   -0.01040000000000   -0.78060000000000    O    (4i)
  -0.21550000000000    0.53310000000000    0.88710000000000    O    (8j)
  -0.53310000000000    0.21550000000000   -0.88710000000000    O    (8j)
   0.21550000000000   -0.53310000000000   -0.88710000000000    O    (8j)
   0.53310000000000   -0.21550000000000    0.88710000000000    O    (8j)
   0.05460000000000    0.24920000000000    0.59870000000000    O    (8j)
  -0.24920000000000   -0.05460000000000   -0.59870000000000    O    (8j)
  -0.05460000000000   -0.24920000000000   -0.59870000000000    O    (8j)
   0.24920000000000    0.05460000000000    0.59870000000000    O    (8j)
   0.02660000000000    0.31800000000000    0.54800000000000    O    (8j)
  -0.31800000000000   -0.02660000000000   -0.54800000000000    O    (8j)
  -0.02660000000000   -0.31800000000000   -0.54800000000000    O    (8j)
   0.31800000000000    0.02660000000000    0.54800000000000    O    (8j)
   0.16820000000000    0.21540000000000    0.37540000000000    O    (8j)
  -0.21540000000000   -0.16820000000000   -0.37540000000000    O    (8j)
  -0.16820000000000   -0.21540000000000   -0.37540000000000    O    (8j)
   0.21540000000000    0.16820000000000    0.37540000000000    O    (8j)
  -0.29640000000000    0.37380000000000    0.64810000000000    O    (8j)
  -0.37380000000000    0.29640000000000   -0.64810000000000    O    (8j)
   0.29640000000000   -0.37380000000000   -0.64810000000000    O    (8j)
   0.37380000000000   -0.29640000000000    0.64810000000000    O    (8j)
   0.70695000000000    0.70695000000000   -0.04418000000000    S    (4i)
  -0.70695000000000   -0.70695000000000    0.04418000000000    S    (4i)
   0.21077000000000    0.21077000000000    0.52021000000000    S    (4i)
  -0.21077000000000   -0.21077000000000   -0.52021000000000    S    (4i)
```

Monoclinic $FeTlSe_2$: AB2C_mC16_12_g_2i_i - CIF

```
# CIF file
data_findsym-output
_audit_creation_method FINDSYM

_chemical_name_mineral 'FeSe2Tl'
_chemical_formula_sum 'Fe Se2 Tl'

loop_
_publ_author_name
 'K. Klepp'
 'H. Boller'
_journal_name_full_name
;
 Monatshefte f{\"u}r Chemie - Chemical Monthly
;
_journal_volume 110
_journal_year 1979
_journal_page_first 1045
_journal_page_last 1055
_publ_Section_title
;
 Die Kristallstruktur von TlFeSe$_{2}$ und TlFeS$_{2}$
;

_aflow_title 'Monoclinic FeTlSe$_{2}$ Structure'
_aflow_proto 'AB2C_mC16_12_g_2i_i'
_aflow_params 'a,b/a,c/a,\beta,y_{1},x_{2},z_{2},x_{3},z_{3},x_{4},z_{4
   ↪ }'
_aflow_params_values '11.973,0.458531696317,0.593836131295,118.2,0.2493,
   ↪ 0.5357,0.2899,0.1783,0.0907,0.1752,0.6314'
_aflow_Strukturbericht 'None'
_aflow_Pearson 'mC16'

_symmetry_space_group_name_H-M "C 1 2/m 1"
_symmetry_Int_Tables_number 12

_cell_length_a     11.97300
_cell_length_b     5.49000
_cell_length_c     7.11000
_cell_angle_alpha  90.00000
_cell_angle_beta   118.20000
_cell_angle_gamma  90.00000

loop_
_space_group_symop_id
_space_group_symop_operation_xyz
1 x,y,z
2 -x,y,-z
3 -x,-y,-z
4 x,-y,z
5 x+1/2,y+1/2,z
6 -x+1/2,y+1/2,-z
7 -x+1/2,-y+1/2,-z
8 x+1/2,-y+1/2,z

loop_
_atom_site_label
_atom_site_type_symbol
_atom_site_symmetry_multiplicity
_atom_site_Wyckoff_label
_atom_site_fract_x
_atom_site_fract_y
_atom_site_fract_z
_atom_site_occupancy
Fe1 Fe 4 g 0.00000 0.24930 0.00000 1.00000
Se1 Se 4 i 0.53570 0.00000 0.28990 1.00000
Se2 Se 4 i 0.17830 0.00000 0.09070 1.00000
Tl1 Tl 4 i 0.17520 0.00000 0.63140 1.00000
```

Monoclinic $FeTlSe_2$: AB2C_mC16_12_g_2i_i - POSCAR

```
AB2C_mC16_12_g_2i_i & a,b/a,c/a,beta,y1,x2,z2,x3,z3,x4,z4 --params=
   ↪ 11.973,0.458531696317,0.593836131295,118.2,0.2493,0.5357,0.2899
   ↪ ,0.1783,0.0907,0.1752,0.6314 & C2/m C_{2h}^{3} #12 (gi^3) &
   ↪ mC16 & None & FeSe2Tl & FeSe2Tl & K. Klepp and H. Boller,
   ↪ Monatshefte f{"u}r Chemie - Chemical Monthly 110, 1045-1055 (
   ↪ 1979)
   1.00000000000000
   5.98650000000000   -2.74500000000000    0.00000000000000
   5.98650000000000    2.74500000000000    0.00000000000000
  -3.35983593821897    0.00000000000000    6.26606754418210
    Fe        Se        Tl
     2         4         2
Direct
  -0.24930000000000    0.24930000000000    0.00000000000000    Fe   (4g)
   0.24930000000000   -0.24930000000000    0.00000000000000    Fe   (4g)
   0.53570000000000    0.53570000000000    0.28990000000000    Se   (4i)
  -0.53570000000000   -0.53570000000000   -0.28990000000000    Se   (4i)
   0.17830000000000    0.17830000000000    0.09070000000000    Se   (4i)
  -0.17830000000000   -0.17830000000000   -0.09070000000000    Se   (4i)
   0.17520000000000    0.17520000000000    0.63140000000000    Tl   (4i)
  -0.17520000000000   -0.17520000000000   -0.63140000000000    Tl   (4i)
```

$NbTe_2$: AB2_mC18_12_ai_3i - CIF

```
# CIF file
data_findsym-output
_audit_creation_method FINDSYM

_chemical_name_mineral 'NbTe2'
_chemical_formula_sum 'Nb Te2'

loop_
_publ_author_name
 'B. E. Brown'
_journal_name_full_name
;
 Acta Cristallographica
;
_journal_volume 20
_journal_year 1966
_journal_page_first 264
_journal_page_last 267
_publ_Section_title
;
 The Crystal Structures of NbTe$_{2}$ and TaTe$_{2}$
;

# Found in Pearson's Handbook of Crystallographic Data for Intermetallic
   ↪ Phases, 1991

_aflow_title 'NbTe$_{2}$ Structure'
_aflow_proto 'AB2_mC18_12_ai_3i'
_aflow_params 'a,b/a,c/a,\beta,x_{2},z_{2},x_{3},z_{3},x_{4},z_{4},x_{5
   ↪ },z_{5}'
_aflow_params_values '14.73541,0.247159732915,0.636222541483,110.41211,
   ↪ 0.36025,0.7087,0.3503,-0.0096,0.703,0.6208,0.0039,0.3098'
_aflow_Strukturbericht 'None'
_aflow_Pearson 'mC18'

_symmetry_space_group_name_H-M "C 1 2/m 1"
_symmetry_Int_Tables_number 12

_cell_length_a     14.73541
_cell_length_b     3.64200
_cell_length_c     9.37500
_cell_angle_alpha  90.00000
_cell_angle_beta   110.41211
_cell_angle_gamma  90.00000

loop_
_space_group_symop_id
_space_group_symop_operation_xyz
1 x,y,z
2 -x,y,-z
3 -x,-y,-z
4 x,-y,z
5 x+1/2,y+1/2,z
6 -x+1/2,y+1/2,-z
```

```
7 -x+1/2,-y+1/2,-z
8 x+1/2,-y+1/2,z

loop_
_atom_site_label
_atom_site_type_symbol
_atom_site_symmetry_multiplicity
_atom_site_Wyckoff_label
_atom_site_fract_x
_atom_site_fract_y
_atom_site_fract_z
_atom_site_occupancy
Nb1 Nb 2 a 0.00000 0.00000 0.00000 1.00000
Nb2 Nb 4 i 0.36025 0.00000 0.70870 1.00000
Te1 Te 4 i 0.35030 0.00000 -0.00960 1.00000
Te2 Te 4 i 0.70300 0.00000 0.62080 1.00000
Te3 Te 4 i 0.00390 0.00000 0.30980 1.00000
```

$NbTe_2$: AB2_mC18_12_ai_3i - POSCAR

```
AB2_mC18_12_ai_3i & a,b/a,c/a,beta,x2,z2,x3,z3,x4,z4,x5,z5 --params=
    ↪ 14.73541,0.247159732915,0.636222541483,110.41211,0.36025,0.7087
    ↪ ,0.3503,-0.0096,0.703,0.6208,0.0039,0.3098 & C2/m C_{2h}^{3} #
    ↪ 12 (ai^4) & mC18 & None & NbTe2 & NbTe2 & B. E. Brown, Acta
    ↪ Cryst. 20, 264-267 (1966)
   1.00000000000000
   7.36770500000000  -1.82100000000000   0.00000000000000
   7.36770500000000   1.82100000000000   0.00000000000000
  -3.26972008938351   0.00000000000000   8.78632776176042
    Nb    Te
     3     6
Direct
   0.00000000000000   0.00000000000000   0.00000000000000   Nb   (2a)
   0.36025000000000   0.36025000000000   0.70870000000000   Nb   (4i)
  -0.36025000000000  -0.36025000000000  -0.70870000000000   Nb   (4i)
   0.35030000000000   0.35030000000000  -0.00960000000000   Te   (4i)
  -0.35030000000000  -0.35030000000000   0.00960000000000   Te   (4i)
   0.70300000000000   0.70300000000000   0.62080000000000   Te   (4i)
  -0.70300000000000  -0.70300000000000  -0.62080000000000   Te   (4i)
   0.00390000000000   0.00390000000000   0.30980000000000   Te   (4i)
  -0.00390000000000  -0.00390000000000  -0.30980000000000   Te   (4i)
```

*$D2_2$* ($MgZn_5$?) (*Problematic*): AB5_mC48_12_2i_ac5i2j - CIF

```
# CIF file
data_findsym-output
_audit_creation_method FINDSYM

_chemical_name_mineral 'MgZn5'
_chemical_formula_sum 'Mg Zn5'

loop_
_publ_author_name
 'C. Hermann'
 'O. Lohrmann'
 'H. Philipp'
_journal_year 1937
_publ_Section_title
;
 Strukturbericht Band II 1928-1932
;

_aflow_title '$D2_{2}$ (MgZn$_{5}$?) ({\em{Problematic}}) Structure'
_aflow_proto 'AB5_mC48_12_2i_ac5i2j'
_aflow_params 'a,b/a,c/a,\beta,x_{3},z_{3},x_{4},z_{4},x_{5},z_{5},x_{6
    ↪ },z_{6},x_{7},z_{7},x_{8},z_{8},x_{9},z_{9},x_{10},y_{10},z_{10
    ↪ },x_{11},y_{11},z_{11}'
_aflow_params_values '9.92,1.66129032258,1.0,120.0,0.66667,0.19792,
    ↪ 0.66667,0.38542,0.66667,0.66667,0.0,0.25,0.33333,0.08333,
    ↪ 0.16667,0.79167,0.16667,0.29167,0.41667,0.25,0.04167,-0.08333,
    ↪ 0.25,0.54167'
_aflow_Strukturbericht '$D2_{2}$'
_aflow_Pearson 'mC48'

_symmetry_space_group_name_H-M "C 1 2/m 1"
_symmetry_Int_Tables_number 12

_cell_length_a      9.92000
_cell_length_b      16.48000
_cell_length_c      9.92000
_cell_angle_alpha   90.00000
_cell_angle_beta    120.00000
_cell_angle_gamma   90.00000

loop_
_space_group_symop_id
_space_group_symop_operation_xyz
1 x,y,z
2 -x,y,-z
3 -x,-y,-z
4 x,-y,z
5 x+1/2,y+1/2,z
6 -x+1/2,y+1/2,-z
7 -x+1/2,-y+1/2,-z
8 x+1/2,-y+1/2,z

loop_
_atom_site_label
_atom_site_type_symbol
_atom_site_symmetry_multiplicity
_atom_site_Wyckoff_label
_atom_site_fract_x
_atom_site_fract_y
_atom_site_fract_z
_atom_site_occupancy
Zn1 Zn 2 a 0.00000 0.00000 0.00000 1.00000
Zn2 Zn 2 c 0.00000 0.00000 0.50000 1.00000
Mg1 Mg 4 i 0.66667 0.00000 0.19792 1.00000
Mg2 Mg 4 i 0.66667 0.00000 0.38542 1.00000
Zn3 Zn 4 i 0.66667 0.00000 0.66667 1.00000
Zn4 Zn 4 i 0.00000 0.00000 0.25000 1.00000
Zn5 Zn 4 i 0.33333 0.00000 0.08333 1.00000
Zn6 Zn 4 i 0.16667 0.00000 0.79167 1.00000
Zn7 Zn 4 i 0.16667 0.00000 0.29167 1.00000
Zn8 Zn 8 j 0.41667 0.25000 0.04167 1.00000
Zn9 Zn 8 j -0.08333 0.25000 0.54167 1.00000
```

*$D2_2$* ($MgZn_5$?) (*Problematic*): AB5_mC48_12_2i_ac5i2j - POSCAR

```
AB5_mC48_12_2i_ac5i2j & a,b/a,c/a,beta,x3,z3,x4,z4,x5,z5,x6,z6,x7,z7,x8,
    ↪ z8,x9,z9,x10,y10,z10,x11,y11,z11 --params=9.92,1.66129032258,
    ↪ 1.0,120.0,0.66667,0.19792,0.66667,0.38542,0.66667,0.66667,0.0,
    ↪ 0.25,0.33333,0.08333,0.16667,0.79167,0.16667,0.29167,0.41667,
    ↪ 0.25,0.04167,-0.08333,0.25,0.54167 & C2/m C_{2h}^{3} #12 (aci^
    ↪ 7j^2) & mC48 & $D2_{2}$ & MgZn5 & C. Hermann and O.
    ↪ Lohrmann and H. Philipp, (1937)
   1.00000000000000
   4.96000000000000  -8.24000000000000   0.00000000000000
   4.96000000000000   8.24000000000000   0.00000000000000
  -4.96000000000000   0.00000000000000   8.59097200554163
    Mg    Zn
     4    20
Direct
   0.66667000000000   0.66667000000000   0.19792000000000   Mg   (4i)
  -0.66667000000000  -0.66667000000000  -0.19792000000000   Mg   (4i)
   0.66667000000000   0.66667000000000   0.38542000000000   Mg   (4i)
  -0.66667000000000  -0.66667000000000  -0.38542000000000   Mg   (4i)
   0.00000000000000   0.00000000000000   0.00000000000000   Zn   (2a)
   0.00000000000000   0.00000000000000   0.50000000000000   Zn   (2c)
   0.66667000000000   0.66667000000000   0.66667000000000   Zn   (4i)
  -0.66667000000000  -0.66667000000000  -0.66667000000000   Zn   (4i)
   0.00000000000000   0.00000000000000   0.25000000000000   Zn   (4i)
   0.00000000000000   0.00000000000000  -0.25000000000000   Zn   (4i)
   0.33333000000000   0.33333000000000   0.08333000000000   Zn   (4i)
  -0.33333000000000  -0.33333000000000  -0.08333000000000   Zn   (4i)
   0.16667000000000   0.16667000000000   0.79167000000000   Zn   (4i)
  -0.16667000000000  -0.16667000000000  -0.79167000000000   Zn   (4i)
   0.16667000000000   0.16667000000000   0.29167000000000   Zn   (4i)
  -0.16667000000000  -0.16667000000000  -0.29167000000000   Zn   (4i)
   0.16667000000000   0.66667000000000   0.04167000000000   Zn   (8j)
  -0.66667000000000  -0.16667000000000  -0.04167000000000   Zn   (8j)
  -0.16667000000000  -0.66667000000000  -0.04167000000000   Zn   (8j)
   0.66667000000000   0.16667000000000   0.04167000000000   Zn   (8j)
  -0.33333000000000   0.16667000000000   0.54167000000000   Zn   (8j)
  -0.16667000000000   0.33333000000000  -0.54167000000000   Zn   (8j)
   0.33333000000000  -0.16667000000000  -0.54167000000000   Zn   (8j)
   0.16667000000000  -0.33333000000000   0.54167000000000   Zn   (8j)
```

Chrysotile ($H_4Mg_3Si_2O_9$, $S4_5$): AB6C11D6E4_mC112_12_e_gi2j_i5j_2i2j_2j - CIF

```
# CIF file
data_findsym-output
_audit_creation_method FINDSYM

_chemical_name_mineral 'Chrysotile'
_chemical_formula_sum '(H2O) Mg6 O11 (OH)6 Si4'

loop_
_publ_author_name
 'B. E. Warren'
 'W. L. Bragg'
_journal_name_full_name
;
 Zeitschrift f{\"u}r Kristallographie - Crystalline Materials
;
_journal_volume 76
_journal_year 1931
_journal_page_first 201
_journal_page_last 210
_publ_Section_title
;
 The Structure of Chrysotile H$_{4}$Mg$_{3}$Si$_{2}$O$_{9}$
;

# Found in Strukturbericht Band II 1928-1932, 1937

_aflow_title 'Chrysotile (H$_{4}$Mg$_{3}$Si$_{2}$O$_{9}$, $S4_{5}$)
    ↪ Structure'
_aflow_proto 'AB6C11D6E4_mC112_12_e_gi2j_i5j_2i2j_2j'
_aflow_params 'a,b/a,c/a,\beta,y_{2},x_{3},z_{3},x_{4},z_{4},x_{5},z_{5
    ↪ },x_{6},z_{6},x_{7},y_{7},z_{7},x_{8},y_{8},z_{8},x_{9},y_{9},
    ↪ z_{9},x_{10},y_{10},z_{10},x_{11},y_{11},z_{11},x_{12},y_{12},
    ↪ z_{12},x_{13},y_{13},z_{13},x_{14},y_{14},z_{14},x_{15},y_{15},
    ↪ z_{15},x_{16},y_{16},z_{16},x_{17},y_{17},z_{17}'
_aflow_params_values '15.31076,1.20830056771,0.348121190588,107.07121,
    ↪ 0.75,0.32,0.72,0.9,0.9,0.75,0.45,0.4,0.4,0.32,0.17,0.22,0.32,
    ↪ 0.08,0.72,0.4,0.37,0.25,0.4,0.37,0.75,0.4,0.25,0.4,0.25,0.42,-
    ↪ 0.05,0.25,0.33,0.45,0.4,0.17,0.9,0.4,0.08,0.4,0.37,0.42,-0.03,
    ↪ 0.37,0.33,0.47'
_aflow_Strukturbericht '$S4_{5}$'
_aflow_Pearson 'mC112'

_symmetry_space_group_name_H-M "C 1 2/m 1"
_symmetry_Int_Tables_number 12

_cell_length_a      15.31076
_cell_length_b      18.50000
_cell_length_c      5.33000
_cell_angle_alpha   90.00000
_cell_angle_beta    107.07121
```

```
_cell_angle_gamma 90.00000

loop_
_space_group_symop_id
_space_group_symop_operation_xyz
1 x,y,z
2 -x,y,-z
3 -x,-y,-z
4 x,-y,z
5 x+1/2,y+1/2,z
6 -x+1/2,y+1/2,-z
7 -x+1/2,-y+1/2,-z
8 x+1/2,-y+1/2,z

loop_
_atom_site_label
_atom_site_type_symbol
_atom_site_symmetry_multiplicity
_atom_site_Wyckoff_label
_atom_site_fract_x
_atom_site_fract_y
_atom_site_fract_z
_atom_site_occupancy
H2O1 H2O 4 e 0.25000 0.25000 0.00000 1.00000
Mg1 Mg 4 g 0.00000 0.75000 0.00000 1.00000
Mg2 Mg 4 i 0.32000 0.00000 0.72000 1.00000
O1 O 4 i 0.90000 0.00000 0.90000 1.00000
OH1 OH 4 i 0.75000 0.00000 0.45000 1.00000
OH2 OH 4 i 0.40000 0.00000 0.40000 1.00000
Mg3 Mg 8 j 0.32000 0.17000 0.22000 1.00000
Mg4 Mg 8 j 0.32000 0.08000 0.72000 1.00000
O2 O 8 j 0.40000 0.37000 0.25000 1.00000
O3 O 8 j 0.40000 0.37000 0.75000 1.00000
O4 O 8 j 0.40000 0.25000 0.40000 1.00000
O5 O 8 j 0.25000 0.42000 -0.05000 1.00000
O6 O 8 j 0.25000 0.33000 0.45000 1.00000
OH3 OH 8 j 0.40000 0.17000 0.90000 1.00000
OH4 OH 8 j 0.40000 0.08000 0.40000 1.00000
Si1 Si 8 j 0.37000 0.42000 -0.03000 1.00000
Si2 Si 8 j 0.37000 0.33000 0.47000 1.00000
```

Chrysotile ($H_4Mg_3Si_2O_9$, $S4_5$): AB6C11D6E4_mC112_12_e_gi2j_i5j_2i2j_2j - POSCAR

```
AB6C11D6E4_mC112_12_e_gi2j_i5j_2i2j_2j & a,b/a,c/a,beta,y2,x3,z3,x4,z4,
    ↪ x5,z5,x6,z6,x7,y7,z7,x8,y8,z8,x9,y9,z9,x10,y10,z10,x11,y11,z11,
    ↪ x12,y12,z12,x13,y13,z13,x14,y14,z14,x15,y15,z15,x16,y16,z16,x17
    ↪ ,y17,z17 --params=15.31076,1.20830056771,0.348121190588,
    ↪ 107.07121,0.75,0.32,0.72,0.9,0.9,0.75,0.45,0.4,0.4,0.32,0.17,
    ↪ 0.22,0.32,0.08,0.72,0.4,0.37,0.25,0.4,0.37,0.75,0.4,0.25,0.4,
    ↪ 0.25,0.42,-0.05,0.25,0.33,0.45,0.4,0.17,0.9,0.4,0.08,0.4,0.37,
    ↪ 0.42,-0.03,0.37,0.33,0.47 & C2/m C_{2h}^{3} #12 (egi^4j^11) &
    ↪ mC112 & $S4_{5}$ & (H2O)(OH)6O11Mg6Si4 & Chrysotile & B. E.
    ↪ Warren and W. L. Bragg, Zeitschrift f{"u}r Kristallographie -
    ↪ Crystalline Materials 76, 201-210 (1931)
   1.00000000000000
   7.65538000000000  -9.25000000000000   0.00000000000000
   7.65538000000000   9.25000000000000   0.00000000000000
  -1.56467491181852   0.00000000000000   5.09516363037790
   H2O    Mg     O    OH    Si
     2    12    22    12     8
Direct
   0.00000000000000   0.50000000000000   0.00000000000000   H2O   (4e)
   0.50000000000000   0.00000000000000   0.00000000000000   H2O   (4e)
  -0.75000000000000   0.75000000000000   0.00000000000000   Mg    (4g)
   0.75000000000000  -0.75000000000000   0.00000000000000   Mg    (4g)
   0.32000000000000   0.32000000000000   0.72000000000000   Mg    (4i)
  -0.32000000000000  -0.32000000000000  -0.72000000000000   Mg    (4i)
   0.15000000000000   0.49000000000000   0.22000000000000   Mg    (8j)
  -0.49000000000000  -0.15000000000000  -0.22000000000000   Mg    (8j)
  -0.15000000000000  -0.49000000000000  -0.22000000000000   Mg    (8j)
   0.49000000000000   0.15000000000000   0.22000000000000   Mg    (8j)
   0.24000000000000   0.40000000000000   0.72000000000000   Mg    (8j)
  -0.40000000000000  -0.24000000000000  -0.72000000000000   Mg    (8j)
  -0.24000000000000  -0.40000000000000  -0.72000000000000   Mg    (8j)
   0.40000000000000   0.24000000000000   0.72000000000000   Mg    (8j)
   0.90000000000000   0.90000000000000   0.90000000000000   O     (4i)
  -0.90000000000000  -0.90000000000000  -0.90000000000000   O     (4i)
   0.03000000000000   0.77000000000000   0.25000000000000   O     (8j)
  -0.77000000000000  -0.03000000000000  -0.25000000000000   O     (8j)
  -0.03000000000000  -0.77000000000000  -0.25000000000000   O     (8j)
   0.77000000000000   0.03000000000000   0.25000000000000   O     (8j)
   0.03000000000000   0.77000000000000   0.75000000000000   O     (8j)
  -0.77000000000000  -0.03000000000000  -0.75000000000000   O     (8j)
  -0.03000000000000  -0.77000000000000  -0.75000000000000   O     (8j)
   0.77000000000000   0.03000000000000   0.75000000000000   O     (8j)
   0.15000000000000   0.65000000000000   0.40000000000000   O     (8j)
  -0.65000000000000  -0.15000000000000  -0.40000000000000   O     (8j)
  -0.15000000000000  -0.65000000000000  -0.40000000000000   O     (8j)
   0.65000000000000   0.15000000000000   0.40000000000000   O     (8j)
  -0.17000000000000   0.67000000000000  -0.05000000000000   O     (8j)
  -0.67000000000000   0.17000000000000   0.05000000000000   O     (8j)
   0.17000000000000  -0.67000000000000   0.05000000000000   O     (8j)
   0.67000000000000  -0.17000000000000  -0.05000000000000   O     (8j)
  -0.08000000000000   0.58000000000000   0.45000000000000   O     (8j)
  -0.58000000000000   0.08000000000000  -0.45000000000000   O     (8j)
   0.08000000000000  -0.58000000000000  -0.45000000000000   O     (8j)
   0.58000000000000  -0.08000000000000   0.45000000000000   O     (8j)
   0.75000000000000   0.75000000000000   0.45000000000000   OH    (4i)
  -0.75000000000000  -0.75000000000000  -0.45000000000000   OH    (4i)
   0.40000000000000   0.40000000000000   0.40000000000000   OH    (4i)
  -0.40000000000000  -0.40000000000000  -0.40000000000000   OH    (4i)
   0.23000000000000   0.57000000000000   0.90000000000000   OH    (8j)
  -0.57000000000000  -0.23000000000000  -0.90000000000000   OH    (8j)
  -0.23000000000000  -0.57000000000000  -0.90000000000000   OH    (8j)
   0.57000000000000   0.23000000000000   0.90000000000000   OH    (8j)
   0.32000000000000   0.48000000000000   0.40000000000000   OH    (8j)
  -0.48000000000000  -0.32000000000000  -0.40000000000000   OH    (8j)
  -0.32000000000000  -0.48000000000000  -0.40000000000000   OH    (8j)
   0.48000000000000   0.32000000000000   0.40000000000000   OH    (8j)
  -0.05000000000000   0.79000000000000  -0.03000000000000   Si    (8j)
  -0.79000000000000   0.05000000000000   0.03000000000000   Si    (8j)
   0.05000000000000  -0.79000000000000   0.03000000000000   Si    (8j)
   0.79000000000000  -0.05000000000000  -0.03000000000000   Si    (8j)
   0.04000000000000   0.70000000000000   0.47000000000000   Si    (8j)
  -0.70000000000000  -0.04000000000000  -0.47000000000000   Si    (8j)
  -0.04000000000000  -0.70000000000000  -0.47000000000000   Si    (8j)
   0.70000000000000   0.04000000000000   0.47000000000000   Si    (8j)
```

$Sr_2NiTeO_6$: AB6C2D_mC40_12_ad_gh4i_j_bc - CIF

```
# CIF file
data_findsym-output
_audit_creation_method FINDSYM

_chemical_name_mineral 'NiO6Sr2Te'
_chemical_formula_sum 'Ni O6 Sr2 Te'

loop_
_publ_author_name
 'D. Iwanaga'
 'Y. Inaguma'
 'M. Itoh'
_journal_name_full_name
;
 Materials Research Bulletin
;
_journal_volume 35
_journal_year 2000
_journal_page_first 449
_journal_page_last 457
_publ_Section_title
;
 Structure and Magnetic Properties of Sr$_{2}$Ni$A$O$_{6}$ ($A$ = W, Te)
;

_aflow_title 'Sr$_{2}$NiTeO$_{6}$ Structure'
_aflow_proto 'AB6C2D_mC40_12_ad_gh4i_j_bc'
_aflow_params 'a,b/a,c/a,\beta,y_{5},y_{6},x_{7},z_{7},x_{8},z_{8},x_{9
    ↪ },z_{9},x_{10},z_{10},x_{11},y_{11},z_{11}'
_aflow_params_values '7.9174,0.99483416273,0.999823174274,90.378,0.26,
    ↪ 0.759,0.266,0.028,-0.031,0.255,0.24,0.454,0.535,0.247,0.248,
    ↪ 0.251,0.249'
_aflow_Strukturbericht 'None'
_aflow_Pearson 'mC40'

_symmetry_space_group_name_H-M "C 1 2/m 1"
_symmetry_Int_Tables_number 12

_cell_length_a    7.91740
_cell_length_b    7.87650
_cell_length_c    7.91600
_cell_angle_alpha 90.00000
_cell_angle_beta  90.37800
_cell_angle_gamma 90.00000

loop_
_space_group_symop_id
_space_group_symop_operation_xyz
1 x,y,z
2 -x,y,-z
3 -x,-y,-z
4 x,-y,z
5 x+1/2,y+1/2,z
6 -x+1/2,y+1/2,-z
7 -x+1/2,-y+1/2,-z
8 x+1/2,-y+1/2,z

loop_
_atom_site_label
_atom_site_type_symbol
_atom_site_symmetry_multiplicity
_atom_site_Wyckoff_label
_atom_site_fract_x
_atom_site_fract_y
_atom_site_fract_z
_atom_site_occupancy
Ni1 Ni 2 a 0.00000 0.00000 0.00000 1.00000
Te1 Te 2 b 0.00000 0.50000 0.00000 1.00000
Te2 Te 2 c 0.00000 0.00000 0.50000 1.00000
Ni2 Ni 2 d 0.00000 0.50000 0.50000 1.00000
O1 O 4 g 0.00000 0.26000 0.00000 1.00000
O2 O 4 h 0.00000 0.75900 0.50000 1.00000
O3 O 4 i 0.26600 0.00000 0.02800 1.00000
O4 O 4 i -0.03100 0.00000 0.25500 1.00000
O5 O 4 i 0.24000 0.00000 0.45400 1.00000
O6 O 4 i 0.53500 0.00000 0.24700 1.00000
Sr1 Sr 8 j 0.24800 0.25100 0.24900 1.00000
```

$Sr_2NiTeO_6$: AB6C2D_mC40_12_ad_gh4i_j_bc - POSCAR

```
AB6C2D_mC40_12_ad_gh4i_j_bc & a,b/a,c/a,beta,y5,y6,x7,z7,x8,z8,x9,z9,x10
    ↪ ,z10,x11,y11,z11 --params=7.9174,0.99483416273,0.999823174274,
    ↪ 90.378,0.26,0.759,0.266,0.028,-0.031,0.255,0.24,0.454,0.535,
    ↪ 0.247,0.248,0.251,0.249 & C2/m C_{2h}^{3} #12 (abcdghi^4j) &
    ↪ mC40 & None & NiO6Sr2Te & NiO6Sr2Te & D. Iwanaga and Y. Inaguma
    ↪  and M. Itoh, Mater. Res. Bull. 35, 449-457 (2000)
   1.00000000000000
   3.95870000000000  -3.93825000000000   0.00000000000000
   3.95870000000000   3.93825000000000   0.00000000000000
  -0.05222420079162   0.00000000000000   7.91582772885134
```

```
   Ni      O     Sr     Te
    2     12      4      2
Direct
   0.00000000000000    0.00000000000000    0.00000000000000    Ni    (2a)
   0.50000000000000    0.50000000000000    0.50000000000000    Ni    (2d)
  -0.26000000000000    0.26000000000000    0.00000000000000    O     (4g)
   0.26000000000000   -0.26000000000000    0.00000000000000    O     (4g)
  -0.75900000000000    0.75900000000000    0.50000000000000    O     (4h)
   0.75900000000000   -0.75900000000000    0.50000000000000    O     (4h)
   0.26600000000000    0.26600000000000    0.02800000000000    O     (4i)
  -0.26600000000000   -0.26600000000000   -0.02800000000000    O     (4i)
  -0.03100000000000   -0.03100000000000    0.25500000000000    O     (4i)
   0.03100000000000    0.03100000000000   -0.25500000000000    O     (4i)
   0.24000000000000    0.24000000000000    0.45400000000000    O     (4i)
  -0.24000000000000   -0.24000000000000   -0.45400000000000    O     (4i)
   0.53500000000000    0.53500000000000    0.24700000000000    O     (4i)
  -0.53500000000000   -0.53500000000000   -0.24700000000000    O     (4i)
  -0.00300000000000    0.49900000000000    0.24900000000000    Sr    (8j)
  -0.49900000000000    0.00300000000000   -0.24900000000000    Sr    (8j)
   0.00300000000000   -0.49900000000000   -0.24900000000000    Sr    (8j)
   0.49900000000000   -0.00300000000000    0.24900000000000    Sr    (8j)
   0.50000000000000    0.50000000000000    0.00000000000000    Te    (2b)
   0.00000000000000    0.00000000000000    0.50000000000000    Te    (2c)
```

$Ta_2PdSe_6$: AB6C2_mC18_12_a_3i_i - CIF

```
# CIF file
data_findsym-output
_audit_creation_method FINDSYM

_chemical_name_mineral 'PdSe6Ta2'
_chemical_formula_sum 'Pd Se6 Ta2'

loop_
_publ_author_name
 'D. A. Keszler'
 'P. J. Squattrito'
 'N. E. Brese'
 'J. A. Ibers'
 'M. Shang'
 'J. Lu'
_journal_year 1985
_publ_Section_title
;
 New layered ternary chalcogenides: tantalum palladium sulfide (Ta$_{2}
     ↪ $PdS$_{6}$), tantalum palladium selenide (Ta$_{2}$PdSe$_{6}$),
     ↪ niobium palladium sulfide (Nb$_{2}$PdS$_{6}$), niobium
     ↪ palladium selenide (Nb$_{2}$PdSe$_{6}$)
;

_aflow_title 'Ta$_{2}$PdSe$_{6}$ Structure'
_aflow_proto 'AB6C2_mC18_12_a_3i_i'
_aflow_params 'a,b/a,c/a,\beta,x_{2},z_{2},x_{3},z_{3},x_{4},z_{4},x_{5
     ↪ },z_{5}'
_aflow_params_values '12.46083,0.270848731585,0.836461134611,116.31902,-
     ↪ 0.00589,0.76131,0.65848,0.53954,0.21851,0.12521,0.67999,0.28976
     ↪ '
_aflow_Strukturbericht 'None'
_aflow_Pearson 'mC18'

_symmetry_space_group_name_H-M "C 1 2/m 1"
_symmetry_Int_Tables_number 12

_cell_length_a       12.46083
_cell_length_b       3.37500
_cell_length_c       10.42300
_cell_angle_alpha    90.00000
_cell_angle_beta     116.31902
_cell_angle_gamma    90.00000

loop_
_space_group_symop_id
_space_group_symop_operation_xyz
1 x,y,z
2 -x,y,-z
3 -x,-y,-z
4 x,-y,z
5 x+1/2,y+1/2,z
6 -x+1/2,y+1/2,-z
7 -x+1/2,-y+1/2,-z
8 x+1/2,-y+1/2,z

loop_
_atom_site_label
_atom_site_type_symbol
_atom_site_symmetry_multiplicity
_atom_site_Wyckoff_label
_atom_site_fract_x
_atom_site_fract_y
_atom_site_fract_z
_atom_site_occupancy
Pd1 Pd 2 a 0.00000 0.00000 0.00000 1.00000
Se1 Se 4 i -0.00589 0.00000 0.76131 1.00000
Se2 Se 4 i 0.65848 0.00000 0.53954 1.00000
Se3 Se 4 i 0.21851 0.00000 0.12521 1.00000
Ta1 Ta 4 i 0.67999 0.00000 0.28976 1.00000
```

$Ta_2PdSe_6$: AB6C2_mC18_12_a_3i_i - POSCAR

```
AB6C2_mC18_12_a_3i_i & a,b/a,c/a,beta,x2,z2,x3,z3,x4,z4,x5,z5 --params=
     ↪ 12.46083,0.270848731585,0.836461134611,116.31902,-0.00589,
     ↪ 0.76131,0.65848,0.53954,0.21851,0.12521,0.67999,0.28976 & C2/m
     ↪ C_{2h}^{3} #12 (ai^4) & mC18 & None & PdSe6Ta2 & PdSe6Ta2 & D.
     ↪ A. Keszler et al., (1985)
   1.00000000000000
   6.23041500000000   -1.68750000000000    0.00000000000000
   6.23041500000000    1.68750000000000    0.00000000000000
  -4.62123264277420    0.00000000000000    9.34254450679033
   Pd     Se     Ta
    1      6      2
Direct
   0.00000000000000    0.00000000000000    0.00000000000000    Pd    (2a)
  -0.00589000000000   -0.00589000000000    0.76131000000000    Se    (4i)
   0.00589000000000    0.00589000000000   -0.76131000000000    Se    (4i)
   0.65848000000000    0.65848000000000    0.53954000000000    Se    (4i)
  -0.65848000000000   -0.65848000000000   -0.53954000000000    Se    (4i)
   0.21851000000000    0.21851000000000    0.12521000000000    Se    (4i)
  -0.21851000000000   -0.21851000000000   -0.12521000000000    Se    (4i)
   0.67999000000000    0.67999000000000    0.28976000000000    Ta    (4i)
  -0.67999000000000   -0.67999000000000   -0.28976000000000    Ta    (4i)
```

Sanidine ($KAlSi_3O_8$, $S6_7$): AB8C4_mC52_12_i_gi3j_2j - CIF

```
# CIF file
data_findsym-output
_audit_creation_method FINDSYM

_chemical_name_mineral 'Sanidine'
_chemical_formula_sum 'K O8 Si4'

loop_
_publ_author_name
 'T. A. Scambos'
 'J. R. Smyth'
 'T. C. {McCormick}'
_journal_name_full_name
;
 American Mineralogist
;
_journal_volume 72
_journal_year 1987
_journal_page_first 973
_journal_page_last 978
_publ_Section_title
;
 Crystal-structure refinement of high sanidine from the upper mantle
;

_aflow_title 'Sanidine (KAlSi$_{3}$O$_{8}$, $S6_{7}$) Structure'
_aflow_proto 'AB8C4_mC52_12_i_gi3j_2j'
_aflow_params 'a,b/a,c/a,\beta,y_{1},x_{2},z_{2},x_{3},z_{3},x_{4},y_{4
     ↪ },z_{4},x_{5},y_{5},z_{5},x_{6},y_{6},z_{6},x_{7},y_{7},z_{7},
     ↪ x_{8},y_{8},z_{8}'
_aflow_params_values '8.595,1.51576497964,0.834787667248,115.94,0.147,
     ↪ 0.2866,0.138,0.64,0.2849,0.8302,0.1476,0.2269,0.0351,0.3105,
     ↪ 0.2569,0.1789,0.1265,0.4083,0.00991,0.1856,0.22381,0.71075,
     ↪ 0.11813,0.34438'
_aflow_Strukturbericht '$S6_{7}$'
_aflow_Pearson 'mC52'

_symmetry_space_group_name_H-M "C 1 2/m 1"
_symmetry_Int_Tables_number 12

_cell_length_a       8.59500
_cell_length_b       13.02800
_cell_length_c       7.17500
_cell_angle_alpha    90.00000
_cell_angle_beta     115.94000
_cell_angle_gamma    90.00000

loop_
_space_group_symop_id
_space_group_symop_operation_xyz
1 x,y,z
2 -x,y,-z
3 -x,-y,-z
4 x,-y,z
5 x+1/2,y+1/2,z
6 -x+1/2,y+1/2,-z
7 -x+1/2,-y+1/2,-z
8 x+1/2,-y+1/2,z

loop_
_atom_site_label
_atom_site_type_symbol
_atom_site_symmetry_multiplicity
_atom_site_Wyckoff_label
_atom_site_fract_x
_atom_site_fract_y
_atom_site_fract_z
_atom_site_occupancy
O1 O 4 g 0.00000 0.14700 0.00000 1.00000
K1 K 4 i 0.28660 0.00000 0.13800 1.00000
O2 O 4 i 0.64000 0.00000 0.28490 1.00000
O3 O 8 j 0.83020 0.14760 0.22690 1.00000
O4 O 8 j 0.03510 0.31050 0.25690 1.00000
O5 O 8 j 0.17890 0.12650 0.40830 1.00000
Si1 Si 8 j 0.00991 0.18560 0.22381 1.00000
Si2 Si 8 j 0.71075 0.11813 0.34438 1.00000
```

Sanidine ($KAlSi_3O_8$, $S6_7$): AB8C4_mC52_12_i_gi3j_2j - POSCAR

```
AB8C4_mC52_12_i_gi3j_2j & a,b/a,c/a,beta,y1,x2,z2,x3,z3,x4,y4,z4,x5,y5,
     ↪ z5,x6,y6,z6,x7,y7,z7,x8,y8,z8 --params=8.595,1.51576497964,
     ↪ 0.834787667248,115.94,0.147,0.2866,0.138,0.64,0.2849,0.8302,
     ↪ 0.1476,0.2269,0.0351,0.3105,0.2569,0.1789,0.1265,0.4083,0.00991
     ↪ ,0.1856,0.22381,0.71075,0.11813,0.34438 & C2/m C_{2h}^{3} #12 (
     ↪ gi^2j^5) & mC52 & $S6_{7}$ & AlKO8Si3 & Sanidine & T. A.
     ↪ Scambos and J. R. Smyth and T. C. {McCormick}, Am. Mineral. 72,
     ↪ 973-978 (1987)
```

```
1.00000000000000
 4.29750000000000  -6.51400000000000   0.00000000000000
 4.29750000000000   6.51400000000000   0.00000000000000
-3.13855803775329   0.00000000000000   6.45213750966718
  K    O    Si
  2   16    8
Direct
 0.28660000000000   0.28660000000000   0.13800000000000   K   (4i)
-0.28660000000000  -0.28660000000000  -0.13800000000000   K   (4i)
-0.14700000000000   0.14700000000000   0.00000000000000   O   (4g)
 0.14700000000000  -0.14700000000000   0.00000000000000   O   (4g)
 0.64000000000000   0.64000000000000   0.28490000000000   O   (4i)
-0.64000000000000  -0.64000000000000  -0.28490000000000   O   (4i)
 0.68260000000000   0.97780000000000   0.22690000000000   O   (8j)
-0.97780000000000  -0.68260000000000  -0.22690000000000   O   (8j)
-0.68260000000000  -0.97780000000000  -0.22690000000000   O   (8j)
 0.97780000000000   0.68260000000000   0.22690000000000   O   (8j)
-0.27540000000000   0.34560000000000   0.25690000000000   O   (8j)
-0.34560000000000   0.27540000000000  -0.25690000000000   O   (8j)
 0.27540000000000  -0.34560000000000  -0.25690000000000   O   (8j)
 0.34560000000000  -0.27540000000000   0.25690000000000   O   (8j)
 0.05240000000000   0.30540000000000   0.40830000000000   O   (8j)
-0.30540000000000  -0.05240000000000  -0.40830000000000   O   (8j)
-0.05240000000000  -0.30540000000000  -0.40830000000000   O   (8j)
 0.30540000000000   0.05240000000000   0.40830000000000   O   (8j)
-0.17569000000000   0.19551000000000   0.22381000000000   Si  (8j)
-0.19551000000000   0.17569000000000  -0.22381000000000   Si  (8j)
 0.17569000000000  -0.19551000000000  -0.22381000000000   Si  (8j)
 0.19551000000000  -0.17569000000000   0.22381000000000   Si  (8j)
 0.59262000000000   0.82888000000000   0.34438000000000   Si  (8j)
-0.82888000000000  -0.59262000000000  -0.34438000000000   Si  (8j)
-0.59262000000000  -0.82888000000000  -0.34438000000000   Si  (8j)
 0.82888000000000   0.59262000000000   0.34438000000000   Si  (8j)
```

$MnPS_3$: ABC3_mC20_12_g_i_ij - CIF

```
# CIF file
data_findsym-output
_audit_creation_method FINDSYM

_chemical_name_mineral 'MnPS3'
_chemical_formula_sum 'Mn P S3'

loop_
_publ_author_name
 'G. Ouvrard'
 'R. Brec'
 'J. Rouxel'
_journal_name_full_name
;
 Materials Research Bulletin
;
_journal_volume 20
_journal_year 1985
_journal_page_first 1181
_journal_page_last 1189
_publ_Section_title
;
 Structural determination of some $M$PS$_{3}$ layered phases ($M$ = Mn,
   ↪ Fe, Co, Ni and Cd)
;
# Found in Electronic band structure of the magnetic layered
   ↪ semiconductors $M$PS$_{3}$ ($M$ = Mn, Fe and Ni), 1996

_aflow_title 'MnPS$_{3}$ Structure'
_aflow_proto 'ABC3_mC20_12_g_i_ij'
_aflow_params 'a,b/a,c/a,\beta,y_{1},x_{2},z_{2},x_{3},z_{3},x_{4},y_{4
   ↪ },z_{4}'
_aflow_params_values '6.077,1.73177554714,1.11831495804,107.35,0.33258,
   ↪ 0.0556,0.1686,0.7593,0.2497,0.2438,0.1612,0.2516'
_aflow_Strukturbericht 'None'
_aflow_Pearson 'mC20'

_symmetry_space_group_name_H-M "C 1 2/m 1"
_symmetry_Int_Tables_number 12

_cell_length_a     6.07700
_cell_length_b     10.52400
_cell_length_c     6.79600
_cell_angle_alpha  90.00000
_cell_angle_beta   107.35000
_cell_angle_gamma  90.00000

loop_
_space_group_symop_id
_space_group_symop_operation_xyz
1 x,y,z
2 -x,y,-z
3 -x,-y,-z
4 x,-y,z
5 x+1/2,y+1/2,z
6 -x+1/2,y+1/2,-z
7 -x+1/2,-y+1/2,-z
8 x+1/2,-y+1/2,z

loop_
_atom_site_label
_atom_site_type_symbol
_atom_site_symmetry_multiplicity
_atom_site_Wyckoff_label
_atom_site_fract_x
_atom_site_fract_y
_atom_site_fract_z
_atom_site_occupancy
Mn1 Mn 4 g 0.00000 0.33258 0.00000 1.00000
P1 P 4 i 0.05560 0.00000 0.16860 1.00000
S1 S 4 i 0.75930 0.00000 0.24970 1.00000
S2 S 8 j 0.24380 0.16120 0.25160 1.00000
```

$MnPS_3$: ABC3_mC20_12_g_i_ij - POSCAR

```
ABC3_mC20_12_g_i_ij & a,b/a,c/a,beta,y1,x2,z2,x3,z3,x4,y4,z4 --params=
   ↪ 6.077,1.73177554714,1.11831495804,107.35,0.33258,0.0556,0.1686,
   ↪ 0.7593,0.2497,0.2438,0.1612,0.2516 & C2/m C_{2h}^{3} #12 (gi^2j
   ↪ ) & mC20 & None & MnPS3 & G. Ouvrard and R. Brec and J.
   ↪ Rouxel, Mater. Res. Bull. 20, 1181-1189 (1985)
   1.00000000000000
   3.03850000000000  -5.26200000000000   0.00000000000000
   3.03850000000000   5.26200000000000   0.00000000000000
  -2.02662120588260   0.00000000000000   6.48678830299455
    Mn    P    S
     2    2    6
Direct
  -0.33258000000000   0.33258000000000   0.00000000000000   Mn  (4g)
   0.33258000000000  -0.33258000000000   0.00000000000000   Mn  (4g)
   0.05560000000000   0.05560000000000   0.16860000000000   P   (4i)
  -0.05560000000000  -0.05560000000000  -0.16860000000000   P   (4i)
   0.75930000000000   0.75930000000000   0.24970000000000   S   (4i)
  -0.75930000000000  -0.75930000000000  -0.24970000000000   S   (4i)
   0.08260000000000   0.40500000000000   0.25160000000000   S   (8j)
  -0.40500000000000  -0.08260000000000  -0.25160000000000   S   (8j)
  -0.08260000000000  -0.40500000000000  -0.25160000000000   S   (8j)
   0.40500000000000   0.08260000000000   0.25160000000000   S   (8j)
```

$AlNbO_4$: ABC4_mC24_12_i_i_4i - CIF

```
# CIF file
data_findsym-output
_audit_creation_method FINDSYM

_chemical_name_mineral 'AlNbO4'
_chemical_formula_sum 'Al Nb O4'

loop_
_publ_author_name
 'M. Ardit'
 'M. Dondi'
 'G. Cruciani'
_journal_name_full_name
;
 American Mineralogist
;
_journal_volume 97
_journal_year 2012
_journal_page_first 910
_journal_page_last 917
_publ_Section_title
;
 Structural stability, cation ordering, and local relaxation along the
   ↪ AlNbO$_{4}$-Al$_{0.5}$Cr$_{0.5}$NbO$_{4}$ join
;

# Found in The American Mineralogist Crystal Structure Database, 2003

_aflow_title 'AlNbO$_{4}$ Structure'
_aflow_proto 'ABC4_mC24_12_i_i_4i'
_aflow_params 'a,b/a,c/a,\beta,x_{1},z_{1},x_{2},z_{2},x_{3},z_{3},x_{4
   ↪ },z_{4},x_{5},z_{5},x_{6},z_{6}'
_aflow_params_values '12.15449,0.307293847788,0.533825771382,107.6206,
   ↪ 0.19362,0.30116,0.1025,0.73176,0.1358,0.0099,0.0564,0.3636,
   ↪ 0.3615,0.2985,0.2622,0.6432'
_aflow_Strukturbericht 'None'
_aflow_Pearson 'mC24'

_symmetry_space_group_name_H-M "C 1 2/m 1"
_symmetry_Int_Tables_number 12

_cell_length_a     12.15449
_cell_length_b     3.73500
_cell_length_c     6.48838
_cell_angle_alpha  90.00000
_cell_angle_beta   107.62060
_cell_angle_gamma  90.00000

loop_
_space_group_symop_id
_space_group_symop_operation_xyz
1 x,y,z
2 -x,y,-z
3 -x,-y,-z
4 x,-y,z
5 x+1/2,y+1/2,z
6 -x+1/2,y+1/2,-z
7 -x+1/2,-y+1/2,-z
8 x+1/2,-y+1/2,z

loop_
_atom_site_label
_atom_site_type_symbol
_atom_site_symmetry_multiplicity
_atom_site_Wyckoff_label
_atom_site_fract_x
_atom_site_fract_y
_atom_site_fract_z
_atom_site_occupancy
Al1 Al 4 i 0.19362 0.00000 0.30116 1.00000
Nb1 Nb 4 i 0.10250 0.00000 0.73176 1.00000
O1 O 4 i 0.13580 0.00000 0.00990 1.00000
O2 O 4 i 0.05640 0.00000 0.36360 1.00000
```

```
O3 O 4 i 0.36150 0.00000 0.29850 1.00000
O4 O 4 i 0.26220 0.00000 0.64320 1.00000
```

$AlNbO_4$: ABC4_mC24_12_i_i_4i - POSCAR

```
ABC4_mC24_12_i_i_4i & a,b/a,c/a,beta,x1,z1,x2,z2,x3,z3,x4,z4,x5,z5,x6,z6
    ↪ --params=12.15449,0.307293847788,0.533825771382,107.6206,
    ↪ 0.19362,0.30116,0.1025,0.73176,0.1358,0.0099,0.0564,0.3636,
    ↪ 0.3615,0.2985,0.2622,0.6432 & C2/m C_{2h}^{3} #12 (i^6) & mC24
    ↪ & None & AlNbO4 & AlNbO4 & M. Ardit and M. Dondi and G.
    ↪ Cruciani, Am. Mineral. 97, 910-917 (2012)
   1.00000000000000
   6.07724500000000   -1.86750000000000    0.00000000000000
   6.07724500000000    1.86750000000000    0.00000000000000
  -1.96411424527948    0.00000000000000    6.18395749143623
     Al      Nb       O
      2       2       8
Direct
   0.19362000000000    0.19362000000000    0.30116000000000    Al    (4i)
  -0.19362000000000   -0.19362000000000   -0.30116000000000    Al    (4i)
   0.10250000000000    0.10250000000000    0.73176000000000    Nb    (4i)
  -0.10250000000000   -0.10250000000000   -0.73176000000000    Nb    (4i)
   0.13580000000000    0.13580000000000    0.00990000000000    O     (4i)
  -0.13580000000000   -0.13580000000000   -0.00990000000000    O     (4i)
   0.05640000000000    0.05640000000000    0.36360000000000    O     (4i)
  -0.05640000000000   -0.05640000000000   -0.36360000000000    O     (4i)
   0.36150000000000    0.36150000000000    0.29850000000000    O     (4i)
  -0.36150000000000   -0.36150000000000   -0.29850000000000    O     (4i)
   0.26220000000000    0.26220000000000    0.64320000000000    O     (4i)
  -0.26220000000000   -0.26220000000000   -0.64320000000000    O     (4i)
```

SiAs: AB_mC24_12_3i_3i - CIF

```
# CIF file
data_findsym-output
_audit_creation_method FINDSYM

_chemical_name_mineral 'AsSi'
_chemical_formula_sum 'As Si'

loop_
_publ_author_name
 'T. Wadsten'
_journal_name_full_name
;
 Acta Chemica Scandinavica
;
_journal_volume 19
_journal_year 1965
_journal_page_first 1232
_journal_page_last 1238
_publ_Section_title
;
 The Crystal Structure of SiAs
;

_aflow_title 'SiAs Structure'
_aflow_proto 'AB_mC24_12_3i_3i'
_aflow_params 'a,b/a,c/a,\beta,x_{1},z_{1},x_{2},z_{2},x_{3},z_{3},x_{4
    ↪ },z_{4},x_{5},z_{5},x_{6},z_{6}'
_aflow_params_values '15.98,0.229536921151,0.596307884856,106.0,0.6632,
    ↪ 0.1738,-0.0369,0.1761,0.3479,0.4543,0.2613,0.2076,0.3697,0.0838
    ↪ ,-0.0661,0.4116'
_aflow_Strukturbericht 'None'
_aflow_Pearson 'mC24'

_symmetry_space_group_name_H-M "C 1 2/m 1"
_symmetry_Int_Tables_number 12

_cell_length_a    15.98000
_cell_length_b    3.66800
_cell_length_c    9.52900
_cell_angle_alpha 90.00000
_cell_angle_beta  106.00000
_cell_angle_gamma 90.00000

loop_
_space_group_symop_id
_space_group_symop_operation_xyz
1 x,y,z
2 -x,y,-z
3 -x,-y,-z
4 x,-y,z
5 x+1/2,y+1/2,z
6 -x+1/2,y+1/2,-z
7 -x+1/2,-y+1/2,-z
8 x+1/2,-y+1/2,z

loop_
_atom_site_label
_atom_site_type_symbol
_atom_site_symmetry_multiplicity
_atom_site_Wyckoff_label
_atom_site_fract_x
_atom_site_fract_y
_atom_site_fract_z
_atom_site_occupancy
As1 As 4 i 0.66320 0.00000 0.17380 1.00000
As2 As 4 i -0.03690 0.00000 0.17610 1.00000
As3 As 4 i 0.34790 0.00000 0.45430 1.00000
Si1 Si 4 i 0.26130 0.00000 0.20760 1.00000
Si2 Si 4 i 0.36970 0.00000 0.08380 1.00000
Si3 Si 4 i -0.06610 0.00000 0.41160 1.00000
```

SiAs: AB_mC24_12_3i_3i - POSCAR

```
AB_mC24_12_3i_3i & a,b/a,c/a,beta,x1,z1,x2,z2,x3,z3,x4,z4,x5,z5,x6,z6 --
    ↪ params=15.98,0.229536921151,0.596307884856,106.0,0.6632,0.1738
    ↪ ,-0.0369,0.1761,0.3479,0.4543,0.2613,0.2076,0.3697,0.0838,-
    ↪ 0.0661,0.4116 & C2/m C_{2h}^{3} #12 (i^6) & mC24 & None & AsSi
    ↪ & AsSi & T. Wadsten, Acta Chem. Scand. 19, 1232-1238 (1965)
   1.00000000000000
   7.99000000000000   -1.83400000000000    0.00000000000000
   7.99000000000000    1.83400000000000    0.00000000000000
  -2.62654836358018    0.00000000000000    9.15986270059624
     As      Si
      6       6
Direct
   0.66320000000000    0.66320000000000    0.17380000000000    As    (4i)
  -0.66320000000000   -0.66320000000000   -0.17380000000000    As    (4i)
  -0.03690000000000   -0.03690000000000    0.17610000000000    As    (4i)
   0.03690000000000    0.03690000000000   -0.17610000000000    As    (4i)
   0.34790000000000    0.34790000000000    0.45430000000000    As    (4i)
  -0.34790000000000   -0.34790000000000   -0.45430000000000    As    (4i)
   0.26130000000000    0.26130000000000    0.20760000000000    Si    (4i)
  -0.26130000000000   -0.26130000000000   -0.20760000000000    Si    (4i)
   0.36970000000000    0.36970000000000    0.08380000000000    Si    (4i)
  -0.36970000000000   -0.36970000000000   -0.08380000000000    Si    (4i)
  -0.06610000000000   -0.06610000000000    0.41160000000000    Si    (4i)
   0.06610000000000    0.06610000000000   -0.41160000000000    Si    (4i)
```

High-Temperature $Mo_8O_{23}$: A8B23_mP62_13_4g_c11g - CIF

```
# CIF file
data_findsym-output
_audit_creation_method FINDSYM

_chemical_name_mineral 'Mo8O23'
_chemical_formula_sum 'Mo8 O23'

loop_
_publ_author_name
 'H. Fujishita'
 'M. Sato'
 'S. Sato'
 'S. Hoshino'
_journal_name_full_name
;
 Journal of Solid State Chemistry
;
_journal_volume 66
_journal_year 1987
_journal_page_first 40
_journal_page_last 46
_publ_Section_title
;
 Structure Determination of low-dimensional conductor Mo$_{8}$O$_{23}$
;

_aflow_title 'High-Temperature Mo$_{8}$O$_{23}$ Structure'
_aflow_proto 'A8B23_mP62_13_4g_c11g'
_aflow_params 'a,b/a,c/a,\beta,x_{2},y_{2},z_{2},x_{3},y_{3},z_{3},x_{4
    ↪ },y_{4},z_{4},x_{5},y_{5},z_{5},x_{6},y_{6},z_{6},x_{7},y_{7},
    ↪ z_{7},x_{8},y_{8},z_{8},x_{9},y_{9},z_{9},x_{10},y_{10},z_{10},
    ↪ x_{11},y_{11},z_{11},x_{12},y_{12},z_{12},x_{13},y_{13},z_{13},
    ↪ x_{14},y_{14},z_{14},x_{15},y_{15},z_{15},x_{16},y_{16},z_{16}'
_aflow_params_values '13.384,0.303466826061,1.26143156007,106.27,0.06397
    ↪ ,0.5858,0.41655,0.18514,0.4131,0.24568,0.31501,0.59,0.07903,
    ↪ 0.44659,0.4079,0.40204,0.0658,-0.0012,0.4164,0.1911,-0.0029,
    ↪ 0.2458,0.3191,0.0069,0.077,0.4469,-0.0082,0.4074,0.0645,0.4943,
    ↪ 0.1643,0.1293,0.493,0.3303,0.1989,0.4945,0.496,0.2612,0.5144,
    ↪ 0.1632,0.3261,0.4891,0.3282,0.4113,0.5015,0.0045,0.457,0.4908,
    ↪ 0.1532'
_aflow_Strukturbericht 'None'
_aflow_Pearson 'mP62'

_symmetry_space_group_name_H-M "P 1 2/c 1"
_symmetry_Int_Tables_number 13

_cell_length_a    13.38400
_cell_length_b    4.06160
_cell_length_c    16.88300
_cell_angle_alpha 90.00000
_cell_angle_beta  106.27000
_cell_angle_gamma 90.00000

loop_
_space_group_symop_id
_space_group_symop_operation_xyz
1 x,y,z
2 -x,y,-z+1/2
3 -x,-y,-z
4 x,-y,z+1/2

loop_
_atom_site_label
_atom_site_type_symbol
_atom_site_symmetry_multiplicity
_atom_site_Wyckoff_label
_atom_site_fract_x
_atom_site_fract_y
_atom_site_fract_z
_atom_site_occupancy
O1 O 2 c 0.00000 0.50000 0.00000 1.00000
Mo1 Mo 4 g 0.06397 0.58580 0.41655 1.00000
Mo2 Mo 4 g 0.18514 0.41310 0.24568 1.00000
Mo3 Mo 4 g 0.31501 0.59000 0.07903 1.00000
Mo4 Mo 4 g 0.44659 0.40790 0.40204 1.00000
O2 O 4 g 0.06580 -0.00120 0.41640 1.00000
O3 O 4 g 0.19110 -0.00290 0.24580 1.00000
O4 O 4 g 0.31910 0.00690 0.07700 1.00000
```

```
O5 O 4 g 0.44690 -0.00820 0.40740 1.00000
O6 O 4 g 0.06450 0.49430 0.16430 1.00000
O7 O 4 g 0.12930 0.49300 0.33030 1.00000
O8 O 4 g 0.19890 0.49450 0.49600 1.00000
O9 O 4 g 0.26120 0.51440 0.16320 1.00000
O10 O 4 g 0.32610 0.48910 0.32820 1.00000
O11 O 4 g 0.41130 0.50150 0.00450 1.00000
O12 O 4 g 0.45700 0.49080 0.15320 1.00000
```

High-Temperature $Mo_8O_{23}$: A8B23_mP62_13_4g_c11g - POSCAR

```
A8B23_mP62_13_4g_c11g & a,b/a,c/a,beta,x2,y2,z2,x3,y3,z3,x4,y4,z4,x5,y5,
    ↪ z5,x6,y6,z6,x7,y7,z7,x8,y8,z8,x9,y9,z9,x10,y10,z10,x11,y11,z11,
    ↪ x12,y12,z12,x13,y13,z13,x14,y14,z14,x15,y15,z15,x16,y16,z16 --
    ↪ params=13.384,0.303466826061,1.26143156007,106.27,0.06397,
    ↪ 0.5858,0.41655,0.18514,0.4131,0.24568,0.31501,0.59,0.07903,
    ↪ 0.44659,0.4079,0.40204,0.0658,-0.0012,0.4164,0.1911,-0.0029,
    ↪ 0.2458,0.3191,0.0069,0.077,0.4469,-0.0082,0.4074,0.0645,0.4943,
    ↪ 0.1643,0.1293,0.493,0.3303,0.1989,0.4945,0.496,0.2612,0.5144,
    ↪ 0.1632,0.3261,0.4891,0.3282,0.4113,0.5015,0.0045,0.457,0.4908,
    ↪ 0.1532 & P2/c C_{2h}^{4} #13 (cg^15) & mP62 & None & Mo8O23 &
    ↪ Mo8O23 & H. Fujishita et al., J. Solid State Chem. 66, 40-46 (
    ↪ 1987)
   1.00000000000000
  13.38400000000000    0.00000000000000    0.00000000000000
   0.00000000000000    4.06160000000000    0.00000000000000
  -4.73001079735427    0.00000000000000   16.20687159376890
    Mo        O
    16       46
Direct
   0.06397000000000    0.58580000000000    0.41655000000000   Mo   (4g)
  -0.06397000000000    0.58580000000000    0.08345000000000   Mo   (4g)
  -0.06397000000000   -0.58580000000000   -0.41655000000000   Mo   (4g)
   0.06397000000000   -0.58580000000000    0.91655000000000   Mo   (4g)
   0.18514000000000    0.41310000000000    0.24568000000000   Mo   (4g)
  -0.18514000000000    0.41310000000000    0.25432000000000   Mo   (4g)
  -0.18514000000000   -0.41310000000000   -0.24568000000000   Mo   (4g)
   0.18514000000000   -0.41310000000000    0.74568000000000   Mo   (4g)
   0.31501000000000    0.59000000000000    0.07903000000000   Mo   (4g)
  -0.31501000000000    0.59000000000000    0.42097000000000   Mo   (4g)
  -0.31501000000000   -0.59000000000000   -0.07903000000000   Mo   (4g)
   0.31501000000000   -0.59000000000000    0.57903000000000   Mo   (4g)
   0.44659000000000    0.40790000000000    0.40204000000000   Mo   (4g)
  -0.44659000000000    0.40790000000000    0.09796000000000   Mo   (4g)
  -0.44659000000000   -0.40790000000000   -0.40204000000000   Mo   (4g)
   0.44659000000000   -0.40790000000000    0.90204000000000   Mo   (4g)
   0.00000000000000    0.50000000000000    0.00000000000000   O    (2c)
   0.00000000000000    0.50000000000000    0.50000000000000   O    (2c)
   0.06580000000000   -0.00120000000000    0.41640000000000   O    (4g)
  -0.06580000000000   -0.00120000000000    0.08360000000000   O    (4g)
  -0.06580000000000    0.00120000000000   -0.41640000000000   O    (4g)
   0.06580000000000    0.00120000000000    0.91640000000000   O    (4g)
   0.19110000000000   -0.00290000000000    0.24580000000000   O    (4g)
  -0.19110000000000   -0.00290000000000    0.25420000000000   O    (4g)
  -0.19110000000000    0.00290000000000   -0.24580000000000   O    (4g)
   0.19110000000000    0.00290000000000    0.74580000000000   O    (4g)
   0.31910000000000    0.00690000000000    0.07700000000000   O    (4g)
  -0.31910000000000    0.00690000000000    0.42300000000000   O    (4g)
  -0.31910000000000   -0.00690000000000   -0.07700000000000   O    (4g)
   0.31910000000000   -0.00690000000000    0.57700000000000   O    (4g)
   0.44690000000000   -0.00820000000000    0.40740000000000   O    (4g)
  -0.44690000000000   -0.00820000000000    0.09260000000000   O    (4g)
  -0.44690000000000    0.00820000000000   -0.40740000000000   O    (4g)
   0.44690000000000    0.00820000000000    0.90740000000000   O    (4g)
   0.06450000000000    0.49430000000000    0.16430000000000   O    (4g)
  -0.06450000000000    0.49430000000000    0.33570000000000   O    (4g)
  -0.06450000000000   -0.49430000000000   -0.16430000000000   O    (4g)
   0.06450000000000   -0.49430000000000    0.66430000000000   O    (4g)
   0.12930000000000    0.49300000000000    0.33030000000000   O    (4g)
  -0.12930000000000    0.49300000000000    0.16970000000000   O    (4g)
  -0.12930000000000   -0.49300000000000   -0.33030000000000   O    (4g)
   0.12930000000000   -0.49300000000000    0.83030000000000   O    (4g)
   0.19890000000000    0.49450000000000    0.49600000000000   O    (4g)
  -0.19890000000000    0.49450000000000    0.00400000000000   O    (4g)
  -0.19890000000000   -0.49450000000000   -0.49600000000000   O    (4g)
   0.19890000000000   -0.49450000000000    0.99600000000000   O    (4g)
   0.26120000000000    0.51440000000000    0.16320000000000   O    (4g)
  -0.26120000000000    0.51440000000000    0.33680000000000   O    (4g)
  -0.26120000000000   -0.51440000000000   -0.16320000000000   O    (4g)
   0.26120000000000   -0.51440000000000    0.66320000000000   O    (4g)
   0.32610000000000    0.48910000000000    0.32820000000000   O    (4g)
  -0.32610000000000    0.48910000000000    0.17180000000000   O    (4g)
  -0.32610000000000   -0.48910000000000   -0.32820000000000   O    (4g)
   0.32610000000000   -0.48910000000000    0.82820000000000   O    (4g)
   0.41130000000000    0.50150000000000    0.00450000000000   O    (4g)
  -0.41130000000000    0.50150000000000    0.49550000000000   O    (4g)
  -0.41130000000000   -0.50150000000000   -0.00450000000000   O    (4g)
   0.41130000000000   -0.50150000000000    0.50450000000000   O    (4g)
   0.45700000000000    0.49080000000000    0.15320000000000   O    (4g)
  -0.45700000000000    0.49080000000000    0.34680000000000   O    (4g)
  -0.45700000000000   -0.49080000000000   -0.15320000000000   O    (4g)
   0.45700000000000   -0.49080000000000    0.65320000000000   O    (4g)
```

Huanzalaite ($MgWO_4$, $H0_6$): AB4C_mP12_13_f_2g_e - CIF

```
# CIF file
data_findsym-output
_audit_creation_method FINDSYM

_chemical_name_mineral 'Huanzalaite'
_chemical_formula_sum 'Mg O4 W'

loop_
_publ_author_name
 'V. B. Kravchenko'
_journal_name_full_name
;
 Journal of Structural Chemistry
;
_journal_volume 10
_journal_year 1969
_journal_page_first 139
_journal_page_last 140
_publ_Section_title
;
 Crystal structure of the monoclinic form of magnesium tungstate MgWO$_{
    ↪ 4}$
;

_aflow_title 'Huanzalaite (MgWO$_{4}$, $H0_{6}$) Structure'
_aflow_proto 'AB4C_mP12_13_f_2g_e'
_aflow_params 'a,b/a,c/a,\beta,y_{1},y_{2},x_{3},y_{3},z_{3},x_{4},y_{4
    ↪ },z_{4}'
_aflow_params_values '4.68,1.2094017094,1.05128205128,90.33333,0.1818,
    ↪ 0.6592,0.221,0.101,-0.07,0.266,0.384,0.391'
_aflow_Strukturbericht '$H0_{6}$'
_aflow_Pearson 'mP12'

_symmetry_space_group_name_H-M "P 1 2/c 1"
_symmetry_Int_Tables_number 13

_cell_length_a      4.68000
_cell_length_b      5.66000
_cell_length_c      4.92000
_cell_angle_alpha  90.00000
_cell_angle_beta   90.33333
_cell_angle_gamma  90.00000

loop_
_space_group_symop_id
_space_group_symop_operation_xyz
1 x,y,z
2 -x,y,-z+1/2
3 -x,-y,-z
4 x,-y,z+1/2

loop_
_atom_site_label
_atom_site_type_symbol
_atom_site_symmetry_multiplicity
_atom_site_Wyckoff_label
_atom_site_fract_x
_atom_site_fract_y
_atom_site_fract_z
_atom_site_occupancy
W1 W 2 e 0.00000 0.18180 0.25000 1.00000
Mg1 Mg 2 f 0.50000 0.65920 0.25000 1.00000
O1 O 4 g 0.22100 0.10100 -0.07000 1.00000
O2 O 4 g 0.26600 0.38400 0.39100 1.00000
```

Huanzalaite ($MgWO_4$, $H0_6$): AB4C_mP12_13_f_2g_e - POSCAR

```
AB4C_mP12_13_f_2g_e & a,b/a,c/a,beta,y1,y2,x3,y3,z3,x4,y4,z4 --params=
    ↪ 4.68,1.2094017094,1.05128205128,90.33333,0.1818,0.6592,0.221,
    ↪ 0.101,-0.07,0.266,0.384,0.391 & P2/c C_{2h}^{4} #13 (efg^2) &
    ↪ mP12 & $H0_{6}$ & MgO4W & Huanzalaite & V. B. Kravchenko, J.
    ↪ Struct. Chem. 10, 139-140 (1969)
   1.00000000000000
   4.68000000000000    0.00000000000000    0.00000000000000
   0.00000000000000    5.66000000000000    0.00000000000000
  -0.02862295203742    0.00000000000000    4.91991673980533
    Mg        O        W
     2        8        2
Direct
   0.50000000000000    0.65920000000000    0.25000000000000   Mg   (2f)
   0.50000000000000   -0.65920000000000    0.75000000000000   Mg   (2f)
   0.22100000000000    0.10100000000000   -0.07000000000000   O    (4g)
  -0.22100000000000    0.10100000000000    0.57000000000000   O    (4g)
  -0.22100000000000   -0.10100000000000    0.07000000000000   O    (4g)
   0.22100000000000   -0.10100000000000    0.43000000000000   O    (4g)
   0.26600000000000    0.38400000000000    0.39100000000000   O    (4g)
  -0.26600000000000    0.38400000000000    0.10900000000000   O    (4g)
  -0.26600000000000   -0.38400000000000   -0.39100000000000   O    (4g)
   0.26600000000000   -0.38400000000000    0.89100000000000   O    (4g)
   0.00000000000000    0.18180000000000    0.25000000000000   W    (2e)
   0.00000000000000   -0.18180000000000    0.75000000000000   W    (2e)
```

$Cs_{11}O_3$: A11B3_mP56_14_11e_3e - CIF

```
# CIF file
data_findsym-output
_audit_creation_method FINDSYM

_chemical_name_mineral 'Cs11O3'
_chemical_formula_sum 'Cs11 O3'

loop_
_publ_author_name
 'A. Simon'
 'E. Westerbeck'
_journal_name_full_name
;
 Zeitschrift fur Anorganische und Allgemeine Chemie
;
_journal_volume 428
_journal_year 1977
_journal_page_first 187
_journal_page_last 198
_publ_Section_title
;
```

```
 \"{U}ber Suboxide der Alkalimetalle. 10. Das ``komplexe Metall\'\' Cs$_
     ↪ {11}$O$_{3}$
;

# Found in Binary Alloy Phase Diagrams, 1990 Found in Binary Alloy Phase
     ↪ Diagrams, {Cd-Ce to Hf-Rb}}

_aflow_title 'Cs$_{11}$O$_{3}$ Structure'
_aflow_proto 'A11B3_mP56_14_11e_3e'
_aflow_params 'a,b/a,c/a,\beta,x_{1},y_{1},z_{1},x_{2},y_{2},z_{2},x_{3
     ↪ },y_{3},z_{3},x_{4},y_{4},z_{4},x_{5},y_{5},z_{5},x_{6},y_{6},
     ↪ z_{6},x_{7},y_{7},z_{7},x_{8},y_{8},z_{8},x_{9},y_{9},z_{9},x_{
     ↪ 10},y_{10},z_{10},x_{11},y_{11},z_{11},x_{12},y_{12},z_{12},x_{
     ↪ 13},y_{13},z_{13},x_{14},y_{14},z_{14}'
_aflow_params_values '17.61,0.523452583759,1.36553094832,100.14,0.3796,
     ↪ 0.7206,0.2438,0.092,-0.058,0.2834,0.2204,0.6157,0.3231,0.303,-
     ↪ 0.0205,0.3408,0.4356,0.4309,0.3821,0.5254,0.8324,0.4013,0.0663,
     ↪ 0.6974,0.4316,0.1617,0.1037,0.4538,0.3158,0.7371,0.4684,0.1517,
     ↪ 0.6716,0.1514,0.2388,0.4198,0.6692,0.2234,0.8414,0.2354,0.1795,
     ↪ 0.8554,0.3955,0.3835,0.7036,0.3654'
_aflow_Strukturbericht 'None'
_aflow_Pearson 'mP56'

_symmetry_space_group_name_H-M "P 1 21/c 1"
_symmetry_Int_Tables_number 14

_cell_length_a    17.61000
_cell_length_b    9.21800
_cell_length_c    24.04700
_cell_angle_alpha 90.00000
_cell_angle_beta  100.14000
_cell_angle_gamma 90.00000

loop_
_space_group_symop_id
_space_group_symop_operation_xyz
1 x,y,z
2 -x,y+1/2,-z+1/2
3 -x,-y,-z
4 x,-y+1/2,z+1/2

loop_
_atom_site_label
_atom_site_type_symbol
_atom_site_symmetry_multiplicity
_atom_site_Wyckoff_label
_atom_site_fract_x
_atom_site_fract_y
_atom_site_fract_z
_atom_site_occupancy
Cs1 Cs 4 e 0.37960 0.72060 0.24380 1.00000
Cs2 Cs 4 e 0.09200 -0.05800 0.28340 1.00000
Cs3 Cs 4 e 0.22040 0.61570 0.32310 1.00000
Cs4 Cs 4 e 0.30300 -0.02050 0.34080 1.00000
Cs5 Cs 4 e 0.43560 0.43090 0.38210 1.00000
Cs6 Cs 4 e 0.52540 0.83240 0.40130 1.00000
Cs7 Cs 4 e 0.06630 0.69740 0.43160 1.00000
Cs8 Cs 4 e 0.16170 0.10370 0.45380 1.00000
Cs9 Cs 4 e 0.31580 0.73710 0.46840 1.00000
Cs10 Cs 4 e 0.15170 0.67160 0.15140 1.00000
Cs11 Cs 4 e 0.23880 0.41980 0.66920 1.00000
O1 O 4 e 0.22340 0.84140 0.23540 1.00000
O2 O 4 e 0.17950 0.85540 0.39550 1.00000
O3 O 4 e 0.38350 0.70360 0.36540 1.00000
```

$Cs_{11}O_3$: A11B3_mP56_14_11e_3e - POSCAR

```
A11B3_mP56_14_11e_3e & a,b/a,c/a,beta,x1,y1,z1,x2,y2,z2,x3,y3,z3,x4,y4,
     ↪ z4,x5,y5,z5,x6,y6,z6,x7,y7,z7,x8,y8,z8,x9,y9,z9,x10,y10,z10,x11
     ↪ ,y11,z11,x12,y12,z12,x13,y13,z13,x14,y14,z14 --params=17.61,
     ↪ 0.523452583759,1.36553094832,100.14,0.3796,0.7206,0.2438,0.092
     ↪ ,-0.058,0.2834,0.2204,0.6157,0.3231,0.303,-0.0205,0.3408,0.4356
     ↪ ,0.4309,0.3821,0.5254,0.8324,0.4013,0.0663,0.6974,0.4316,0.1617
     ↪ ,0.1037,0.4538,0.3158,0.7371,0.4684,0.1517,0.6716,0.1514,0.2388
     ↪ ,0.4198,0.6692,0.2234,0.8414,0.2354,0.1795,0.8554,0.3955,0.3835
     ↪ ,0.7036,0.3654 & P2_{1}/c C_{2h}^{5} #14 (e^14) & mP56 & None &
     ↪ Cs11O3 & Cs11O3 & A. Simon and E. Westerbeck, Z. Anorg. Allg.
     ↪ Chem. 428, 187-198 (1977)
   1.00000000000000
  17.61000000000000    0.00000000000000    0.00000000000000
   0.00000000000000    9.21800000000000    0.00000000000000
  -4.23357044613628    0.00000000000000   23.67139814792530
    Cs       O
    44      12
Direct
   0.37960000000000    0.72060000000000    0.24380000000000   Cs   (4e)
  -0.37960000000000    1.22060000000000    0.25620000000000   Cs   (4e)
  -0.37960000000000   -0.72060000000000   -0.24380000000000   Cs   (4e)
   0.37960000000000   -0.22060000000000    0.74380000000000   Cs   (4e)
   0.09200000000000   -0.05800000000000    0.28340000000000   Cs   (4e)
  -0.09200000000000    0.44200000000000    0.21660000000000   Cs   (4e)
  -0.09200000000000    0.05800000000000   -0.28340000000000   Cs   (4e)
   0.09200000000000    0.55800000000000    0.78340000000000   Cs   (4e)
   0.22040000000000    0.61570000000000    0.32310000000000   Cs   (4e)
  -0.22040000000000    1.11570000000000    0.17690000000000   Cs   (4e)
  -0.22040000000000   -0.61570000000000   -0.32310000000000   Cs   (4e)
   0.22040000000000   -0.11570000000000    0.82310000000000   Cs   (4e)
   0.30300000000000   -0.02050000000000    0.34080000000000   Cs   (4e)
  -0.30300000000000    0.47950000000000    0.15920000000000   Cs   (4e)
  -0.30300000000000    0.02050000000000   -0.34080000000000   Cs   (4e)
   0.30300000000000    0.52050000000000    0.84080000000000   Cs   (4e)
   0.43560000000000    0.43090000000000    0.38210000000000   Cs   (4e)
  -0.43560000000000    0.93090000000000    0.11790000000000   Cs   (4e)
  -0.43560000000000   -0.43090000000000   -0.38210000000000   Cs   (4e)
   0.43560000000000    0.06910000000000    0.88210000000000   Cs   (4e)
   0.52540000000000    0.83240000000000    0.40130000000000   Cs   (4e)
  -0.52540000000000    1.33240000000000    0.09870000000000   Cs   (4e)
  -0.52540000000000   -0.83240000000000   -0.40130000000000   Cs   (4e)
   0.52540000000000   -0.33240000000000    0.90130000000000   Cs   (4e)
   0.06630000000000    0.69740000000000    0.43160000000000   Cs   (4e)
  -0.06630000000000    1.19740000000000    0.06840000000000   Cs   (4e)
  -0.06630000000000   -0.69740000000000   -0.43160000000000   Cs   (4e)
   0.06630000000000   -0.19740000000000    0.93160000000000   Cs   (4e)
   0.16170000000000    0.10370000000000    0.45380000000000   Cs   (4e)
  -0.16170000000000    0.60370000000000    0.04620000000000   Cs   (4e)
  -0.16170000000000   -0.10370000000000   -0.45380000000000   Cs   (4e)
   0.16170000000000    0.39630000000000    0.95380000000000   Cs   (4e)
   0.31580000000000    0.73710000000000    0.46840000000000   Cs   (4e)
  -0.31580000000000    1.23710000000000    0.03160000000000   Cs   (4e)
  -0.31580000000000   -0.73710000000000   -0.46840000000000   Cs   (4e)
   0.31580000000000   -0.23710000000000    0.96840000000000   Cs   (4e)
   0.15170000000000    0.67160000000000    0.15140000000000   Cs   (4e)
  -0.15170000000000    1.17160000000000    0.34860000000000   Cs   (4e)
  -0.15170000000000   -0.67160000000000   -0.15140000000000   Cs   (4e)
   0.15170000000000   -0.17160000000000    0.65140000000000   Cs   (4e)
   0.23880000000000    0.41980000000000    0.66920000000000   Cs   (4e)
  -0.23880000000000    0.91980000000000   -0.16920000000000   Cs   (4e)
  -0.23880000000000   -0.41980000000000   -0.66920000000000   Cs   (4e)
   0.23880000000000    0.08020000000000    1.16920000000000   Cs   (4e)
   0.22340000000000    0.84140000000000    0.23540000000000   O    (4e)
  -0.22340000000000    1.34140000000000    0.26460000000000   O    (4e)
  -0.22340000000000   -0.84140000000000   -0.23540000000000   O    (4e)
   0.22340000000000   -0.34140000000000    0.73540000000000   O    (4e)
   0.17950000000000    0.85540000000000    0.39550000000000   O    (4e)
  -0.17950000000000    1.35540000000000    0.10450000000000   O    (4e)
  -0.17950000000000   -0.85540000000000   -0.39550000000000   O    (4e)
   0.17950000000000   -0.35540000000000    0.89550000000000   O    (4e)
   0.38350000000000    0.70360000000000    0.36540000000000   O    (4e)
  -0.38350000000000    1.20360000000000    0.13460000000000   O    (4e)
  -0.38350000000000   -0.70360000000000   -0.36540000000000   O    (4e)
   0.38350000000000   -0.20360000000000    0.86540000000000   O    (4e)
```

Azurite [$Cu_3(CO_3)_2(OH)_2$, $G7_4$]: A2B3C2D8_mP30_14_e_ce_e_4e - CIF

```
# CIF file
data_findsym-output
_audit_creation_method FINDSYM

_chemical_name_mineral 'Azurite'
_chemical_formula_sum 'C2 Cu3 H2 O8'

loop_
_publ_author_name
 'K. C. Rule'
 'M. Reehuis'
 'M. C. R. Gibson'
 'B. Ouladdiaf'
 'M. J. Gutmann'
 'J.-U. Hoffmann'
 'S. Gerischer'
 'D. A. Tennant'
 'S. S\"{u}llow'
 'M. Lang'
_journal_name_full_name
;
 Physical Review B
;
_journal_volume 83
_journal_year 2011
_journal_page_first 104401
_journal_page_last 104401
_publ_Section_title
;
 Magnetic and crystal structure of azurite Cu$_{3}$(CO$_{3}$)$_{2}$(OH)
     ↪ $_{2}$ as determined by neutron diffraction
;

_aflow_title 'Azurite [Cu$_{3}$(CO$_{3}$)$_{2}$(OH)$_{2}$, $G7_{4}$]
     ↪ Structure'
_aflow_proto 'A2B3C2D8_mP30_14_e_ce_e_4e'
_aflow_params 'a,b/a,c/a,\beta,x_{2},y_{2},z_{2},x_{3},y_{3},z_{3},x_{4
     ↪ },y_{4},z_{4},x_{5},y_{5},z_{5},x_{6},y_{6},z_{6},x_{7},y_{7},
     ↪ z_{7},x_{8},y_{8},z_{8}'
_aflow_params_values '4.99995,1.16452364524,2.06746667467,92.2103,0.3308
     ↪ ,0.7994,0.3192,0.2508,-0.0033,0.0834,0.182,0.3,0.3709,0.0975,
     ↪ 0.8972,0.3318,0.0762,0.3126,0.4451,0.4518,0.7098,0.4183,0.4339,
     ↪ 0.7949,0.2065'
_aflow_Strukturbericht '$G7_{4}$'
_aflow_Pearson 'mP30'

_symmetry_space_group_name_H-M "P 1 21/c 1"
_symmetry_Int_Tables_number 14

_cell_length_a    4.99995
_cell_length_b    5.82256
_cell_length_c    10.33723
_cell_angle_alpha 90.00000
_cell_angle_beta  92.21030
_cell_angle_gamma 90.00000

loop_
_space_group_symop_id
_space_group_symop_operation_xyz
1 x,y,z
2 -x,y+1/2,-z+1/2
3 -x,-y,-z
4 x,-y+1/2,z+1/2

loop_
_atom_site_label
```

```
_atom_site_type_symbol
_atom_site_symmetry_multiplicity
_atom_site_Wyckoff_label
_atom_site_fract_x
_atom_site_fract_y
_atom_site_fract_z
_atom_site_occupancy
Cu1 Cu 2 c 0.00000 0.00000 0.50000 1.00000
C1 C 4 e 0.33080 0.79940 0.31920 1.00000
Cu2 Cu 4 e 0.25080 -0.00330 0.08340 1.00000
H1 H 4 e 0.18200 0.30000 0.37090 1.00000
O1 O 4 e 0.09750 0.89720 0.33180 1.00000
O2 O 4 e 0.07620 0.31260 0.44510 1.00000
O3 O 4 e 0.45180 0.70980 0.41830 1.00000
O4 O 4 e 0.43390 0.79490 0.20650 1.00000
```

Azurite [$Cu_3(CO_3)_2(OH)_2$, $G7_4$]: A2B3C2D8_mP30_14_e_ce_e_4e - POSCAR

```
A2B3C2D8_mP30_14_e_ce_e_4e & a,b/a,c/a,beta,x2,y2,z2,x3,y3,z3,x4,y4,z4,
    ↪ x5,y5,z5,x6,y6,z6,x7,y7,z7,x8,y8,z8 --params=4.99995,
    ↪ 1.16452364524,2.06746667467,92.2103,0.3308,0.7994,0.3192,0.2508
    ↪ ,-0.0033,0.0834,0.182,0.3,0.3709,0.0975,0.8972,0.3318,0.0762,
    ↪ 0.3126,0.4451,0.4518,0.7098,0.4183,0.4339,0.7949,0.2065 & P2_{1
    ↪ }/c C_{2h}^{5} #14 (ce^7) & mP30 & $G7_{4}$ & C2Cu3H2O8 &
    ↪ Azurite & K. C. Rule et al., Phys. Rev. B 83, 104401(2011)
   1.00000000000000
   4.99995000000000   0.00000000000000   0.00000000000000
   0.00000000000000   5.82256000000000   0.00000000000000
  -0.39868054818008   0.00000000000000  10.32953909394810
     C      Cu       H       O
     4       6       4      16
Direct
   0.33080000000000   0.79940000000000   0.31920000000000   C   (4e)
  -0.33080000000000   1.29940000000000   0.18080000000000   C   (4e)
  -0.33080000000000  -0.79940000000000  -0.31920000000000   C   (4e)
   0.33080000000000  -0.29940000000000   0.81920000000000   C   (4e)
   0.00000000000000   0.00000000000000   0.50000000000000   Cu  (2c)
   0.00000000000000   0.50000000000000   0.00000000000000   Cu  (2c)
   0.25080000000000  -0.00330000000000   0.08340000000000   Cu  (4e)
  -0.25080000000000   0.49670000000000   0.41660000000000   Cu  (4e)
  -0.25080000000000   0.00330000000000  -0.08340000000000   Cu  (4e)
   0.25080000000000   0.50330000000000   0.58340000000000   Cu  (4e)
   0.18200000000000   0.30000000000000   0.37090000000000   H   (4e)
  -0.18200000000000   0.80000000000000   0.12910000000000   H   (4e)
  -0.18200000000000  -0.30000000000000  -0.37090000000000   H   (4e)
   0.18200000000000   0.20000000000000   0.87090000000000   H   (4e)
   0.09750000000000   0.89720000000000   0.33180000000000   O   (4e)
  -0.09750000000000   1.39720000000000   0.16820000000000   O   (4e)
  -0.09750000000000  -0.89720000000000  -0.33180000000000   O   (4e)
   0.09750000000000  -0.39720000000000   0.83180000000000   O   (4e)
   0.07620000000000   0.31260000000000   0.44510000000000   O   (4e)
  -0.07620000000000   0.81260000000000   0.05490000000000   O   (4e)
  -0.07620000000000  -0.31260000000000  -0.44510000000000   O   (4e)
   0.07620000000000   0.18740000000000   0.94510000000000   O   (4e)
   0.45180000000000   0.70980000000000   0.41830000000000   O   (4e)
  -0.45180000000000   1.20980000000000   0.08170000000000   O   (4e)
  -0.45180000000000  -0.70980000000000  -0.41830000000000   O   (4e)
   0.45180000000000  -0.20980000000000   0.91830000000000   O   (4e)
   0.43390000000000   0.79490000000000   0.20650000000000   O   (4e)
  -0.43390000000000   1.29490000000000   0.29350000000000   O   (4e)
  -0.43390000000000  -0.79490000000000  -0.20650000000000   O   (4e)
   0.43390000000000  -0.29490000000000   0.70650000000000   O   (4e)
```

$HgCl_2\cdot 2HgO$: A2B3C2_mP14_14_e_ae_e - CIF

```
# CIF file
data_findsym-output
_audit_creation_method FINDSYM

_chemical_name_mineral 'Cl2Hg3O2'
_chemical_formula_sum 'Cl2 Hg3 O2'

loop_
_publ_author_name
 'S. \v{S}\\'{c}avni\\'{c}ar'
_journal_name_full_name
;
 Acta Cristallographica
;
_journal_volume 8
_journal_year 1955
_journal_page_first 379
_journal_page_last 383
_publ_Section_title
;
 The crystal structure of trimercuric oxychloride, HgCl$_{2}$$\cdot$2HgO
;

_aflow_title 'HgCl$_{2}$$\cdot$2HgO Structure'
_aflow_proto 'A2B3C2_mP14_14_e_ae_e'
_aflow_params 'a,b/a,c/a,\beta,x_{2},y_{2},z_{2},x_{3},y_{3},z_{3},x_{4
    ↪ },y_{4},z_{4}'
_aflow_params_values '7.16,0.959497206704,0.958100558659,126.16667,0.185
    ↪ ,0.62,0.233,0.412,0.365,0.151,0.397,0.115,0.331'
_aflow_Strukturbericht 'None'
_aflow_Pearson 'mP14'

_symmetry_space_group_name_H-M "P 1 21/c 1"
_symmetry_Int_Tables_number 14

_cell_length_a      7.16000
_cell_length_b      6.87000
_cell_length_c      6.86000
_cell_angle_alpha  90.00000
_cell_angle_beta  126.16667
_cell_angle_gamma  90.00000

loop_
_space_group_symop_id
_space_group_symop_operation_xyz
1 x,y,z
2 -x,y+1/2,-z+1/2
3 -x,-y,-z
4 x,-y+1/2,z+1/2

loop_
_atom_site_label
_atom_site_type_symbol
_atom_site_symmetry_multiplicity
_atom_site_Wyckoff_label
_atom_site_fract_x
_atom_site_fract_y
_atom_site_fract_z
_atom_site_occupancy
Hg1 Hg 2 a 0.00000 0.00000 0.00000 1.00000
Cl1 Cl 4 e 0.18500 0.62000 0.23300 1.00000
Hg2 Hg 4 e 0.41200 0.36500 0.15100 1.00000
O1 O 4 e 0.39700 0.11500 0.33100 1.00000
```

$HgCl_2\cdot 2HgO$: A2B3C2_mP14_14_e_ae_e - POSCAR

```
A2B3C2_mP14_14_e_ae_e & a,b/a,c/a,beta,x2,y2,z2,x3,y3,z3,x4,y4,z4 --
    ↪ params=7.16,0.959497206704,0.958100558659,126.16667,0.185,0.62,
    ↪ 0.233,0.412,0.365,0.151,0.397,0.115,0.331 & P2_{1}/c C_{2h}^{5}
    ↪ #14 (ae^3) & mP14 & None & Cl2Hg3O2 & Cl2Hg3O2 & S. \v{S}\'{c}
    ↪ avni\'{c}ar, Acta Cryst. 8, 379-383 (1955)
   1.00000000000000
   7.16000000000000   0.00000000000000   0.00000000000000
   0.00000000000000   6.87000000000000   0.00000000000000
  -4.04833394910768   0.00000000000000   5.53810366790856
    Cl      Hg       O
     4       6       4
Direct
   0.18500000000000   0.62000000000000   0.23300000000000   Cl  (4e)
  -0.18500000000000   1.12000000000000   0.26700000000000   Cl  (4e)
  -0.18500000000000  -0.62000000000000  -0.23300000000000   Cl  (4e)
   0.18500000000000  -0.12000000000000   0.73300000000000   Cl  (4e)
   0.00000000000000   0.00000000000000   0.00000000000000   Hg  (2a)
   0.00000000000000   0.50000000000000   0.50000000000000   Hg  (2a)
   0.41200000000000   0.36500000000000   0.15100000000000   Hg  (4e)
  -0.41200000000000   0.86500000000000   0.34900000000000   Hg  (4e)
  -0.41200000000000  -0.36500000000000  -0.15100000000000   Hg  (4e)
   0.41200000000000   0.13500000000000   0.65100000000000   Hg  (4e)
   0.39700000000000   0.11500000000000   0.33100000000000   O   (4e)
  -0.39700000000000   0.61500000000000   0.16900000000000   O   (4e)
  -0.39700000000000  -0.11500000000000  -0.33100000000000   O   (4e)
   0.39700000000000   0.38500000000000   0.83100000000000   O   (4e)
```

Orpiment ($As_2S_3$, $D5_f$): A2B3_mP20_14_2e_3e - CIF

```
# CIF file
data_findsym-output
_audit_creation_method FINDSYM

_chemical_name_mineral 'Orpiment'
_chemical_formula_sum 'As2 S3'

loop_
_publ_author_name
 'N. Morimoto'
_journal_name_full_name
;
 Mineralogical Journal
;
_journal_volume 1
_journal_year 1954
_journal_page_first 160
_journal_page_last 169
_publ_Section_title
;
 The Crystal Structure of Orpiment (As$_{2}$S$_{3}$) Refined
;

_aflow_title 'Orpiment (As$_{2}$S$_{3}$, $D5_{f}$) Structure'
_aflow_proto 'A2B3_mP20_14_2e_3e'
_aflow_params 'a,b/a,c/a,\beta,x_{1},y_{1},z_{1},x_{2},y_{2},z_{2},x_{3
    ↪ },y_{3},z_{3},x_{4},y_{4},z_{4},x_{5},y_{5},z_{5}'
_aflow_params_values '4.22,2.26777251185,2.88570616114,109.77475,0.124,
    ↪ 0.19,0.267,0.841,0.323,0.484,0.895,0.12,0.395,0.342,0.397,0.355
    ↪ ,0.715,0.293,0.125'
_aflow_Strukturbericht '$D5_{f}$'
_aflow_Pearson 'mP20'

_symmetry_space_group_name_H-M "P 1 21/c 1"
_symmetry_Int_Tables_number 14

_cell_length_a      4.22000
_cell_length_b      9.57000
_cell_length_c     12.17768
_cell_angle_alpha  90.00000
_cell_angle_beta  109.77475
_cell_angle_gamma  90.00000

loop_
_space_group_symop_id
_space_group_symop_operation_xyz
1 x,y,z
2 -x,y+1/2,-z+1/2
3 -x,-y,-z
4 x,-y+1/2,z+1/2
```

```
loop_
_atom_site_label
_atom_site_type_symbol
_atom_site_symmetry_multiplicity
_atom_site_Wyckoff_label
_atom_site_fract_x
_atom_site_fract_y
_atom_site_fract_z
_atom_site_occupancy
As1 As 4 e 0.12400 0.19000 0.26700 1.00000
As2 As 4 e 0.84100 0.32300 0.48400 1.00000
S1 S 4 e 0.89500 0.12000 0.39500 1.00000
S2 S 4 e 0.34200 0.39700 0.35500 1.00000
S3 S 4 e 0.71500 0.29300 0.12500 1.00000
```

Orpiment ($As_2S_3$, $D5_f$): A2B3_mP20_14_2e_3e - POSCAR

```
A2B3_mP20_14_2e_3e & a,b/a,c/a,beta,x1,y1,z1,x2,y2,z2,x3,y3,z3,x4,y4,z4,
    ↪ x5,y5,z5 --params=4.22,2.26777251185,2.88570616114,109.77475,
    ↪ 0.124,0.19,0.267,0.841,0.323,0.484,0.895,0.12,0.395,0.342,0.397
    ↪ ,0.355,0.715,0.293,0.125 & P2_{1}/c C_{2h}^{5} #14 (e^5) & mP20
    ↪ & $D5_{f}$ & As2S3 & Orpiment & N. Morimoto, Mineral. J. 1,
    ↪ 160-169 (1954)
   1.00000000000000
   4.22000000000000    0.00000000000000    0.00000000000000
   0.00000000000000    9.57000000000000    0.00000000000000
  -4.11999221802126    0.00000000000000   11.45956169780700
    As        S
     8       12
Direct
   0.12400000000000    0.19000000000000    0.26700000000000    As    (4e)
  -0.12400000000000    0.69000000000000    0.23300000000000    As    (4e)
  -0.12400000000000   -0.19000000000000   -0.26700000000000    As    (4e)
   0.12400000000000    0.31000000000000    0.76700000000000    As    (4e)
   0.84100000000000    0.32300000000000    0.48400000000000    As    (4e)
  -0.84100000000000    0.82300000000000    0.01600000000000    As    (4e)
  -0.84100000000000   -0.32300000000000   -0.48400000000000    As    (4e)
   0.84100000000000    0.17700000000000    0.98400000000000    As    (4e)
   0.89500000000000    0.12000000000000    0.39500000000000    S     (4e)
  -0.89500000000000    0.62000000000000    0.10500000000000    S     (4e)
  -0.89500000000000   -0.12000000000000   -0.39500000000000    S     (4e)
   0.89500000000000    0.38000000000000    0.89500000000000    S     (4e)
   0.34200000000000    0.39700000000000    0.35500000000000    S     (4e)
  -0.34200000000000    0.89700000000000    0.14500000000000    S     (4e)
  -0.34200000000000   -0.39700000000000   -0.35500000000000    S     (4e)
   0.34200000000000    0.10300000000000    0.85500000000000    S     (4e)
   0.71500000000000    0.29300000000000    0.12500000000000    S     (4e)
  -0.71500000000000    0.79300000000000    0.37500000000000    S     (4e)
  -0.71500000000000   -0.29300000000000   -0.12500000000000    S     (4e)
   0.71500000000000    0.20700000000000    0.62500000000000    S     (4e)
```

Monoclinic $Cu_2OSeO_3$: A2B4C_mP28_14_abe_4e_e - CIF

```
# CIF file
data_findsym-output
_audit_creation_method FINDSYM

_chemical_name_mineral 'Cu2O4Se'
_chemical_formula_sum 'Cu2 O4 Se'

loop_
_publ_author_name
 'H. Effenberger'
 'F. Pertlik'
_journal_name_full_name
;
 Monatshefte f{\"u}r Chemie - Chemical Monthly
;
_journal_volume 117
_journal_year 1986
_journal_page_first 887
_journal_page_last 896
_publ_Section_title
;
 Die Kristallstrukturen der Kupfer(II)-oxo-selenite Cu$_{2}$O(SeO$_{3}$)
    ↪ (kubisch und monoklin) und Cu$_{4}$O(SeO$_{3}$)$_{3}$ (
    ↪ monoklin und triklin)
;

# Found in Magnon spectrum of the helimagnetic insulator Cu$_{2}$OSeO$_{
    ↪ 3}$, 2016

_aflow_title 'Monoclinic Cu$_{2}$OSeO$_{3}$ Structure'
_aflow_proto 'A2B4C_mP28_14_abe_4e_e'
_aflow_params 'a,b/a,c/a,\beta,x_{3},y_{3},z_{3},x_{4},y_{4},z_{4},x_{5
    ↪ },y_{5},z_{5},x_{6},y_{6},z_{6},x_{7},y_{7},z_{7},x_{8},y_{8},
    ↪ z_{8}'
_aflow_params_values '6.987,0.852010877344,1.53753112924,128.36609,
    ↪ 0.5082,0.705,0.7605,0.6549,-0.031,-0.0981,0.408,0.5641,0.0995,-
    ↪ 0.051,0.717,0.896,0.261,0.7826,0.8405,0.1649,0.5773,-0.0999'
_aflow_Strukturbericht 'None'
_aflow_Pearson 'mP28'

_symmetry_space_group_name_H-M "P 1 21/c 1"
_symmetry_Int_Tables_number 14

_cell_length_a      6.98700
_cell_length_b      5.95300
_cell_length_c      10.74273
_cell_angle_alpha   90.00000
_cell_angle_beta    128.36609
_cell_angle_gamma   90.00000

loop_
_space_group_symop_id
_space_group_symop_operation_xyz
1 x,y,z
2 -x,y+1/2,-z+1/2
3 -x,-y,-z
4 x,-y+1/2,z+1/2

loop_
_atom_site_label
_atom_site_type_symbol
_atom_site_symmetry_multiplicity
_atom_site_Wyckoff_label
_atom_site_fract_x
_atom_site_fract_y
_atom_site_fract_z
_atom_site_occupancy
Cu1 Cu 2 a 0.00000 0.00000 0.00000 1.00000
Cu2 Cu 2 b 0.50000 0.00000 0.00000 1.00000
Cu3 Cu 4 e 0.50820 0.70500 0.76050 1.00000
O1 O 4 e 0.65490 -0.03100 -0.09810 1.00000
O2 O 4 e 0.40800 0.56410 0.09950 1.00000
O3 O 4 e -0.05100 0.71700 0.89600 1.00000
O4 O 4 e 0.26100 0.78260 0.84050 1.00000
Se1 Se 4 e 0.16490 0.57730 -0.09990 1.00000
```

Monoclinic $Cu_2OSeO_3$: A2B4C_mP28_14_abe_4e_e - POSCAR

```
A2B4C_mP28_14_abe_4e_e & a,b/a,c/a,beta,x3,y3,z3,x4,y4,z4,x5,y5,z5,x6,y6
    ↪ ,z6,x7,y7,z7,x8,y8,z8 --params=6.987,0.852010877344,
    ↪ 1.53753112924,128.36609,0.5082,0.705,0.7605,0.6549,-0.031,-
    ↪ 0.0981,0.408,0.5641,0.0995,-0.051,0.717,0.896,0.261,0.7826,
    ↪ 0.8405,0.1649,0.5773,-0.0999 & P2_{1}/c C_{2h}^{5} #14 (abe^6)
    ↪ & mP28 & None & Cu2O4Se & Cu2O4Se & H. Effenberger and F.
    ↪ Pertlik, Monatshefte f{"u}r Chemie - Chemical Monthly 117,
    ↪ 887-896 (1986)
   1.00000000000000
   6.98700000000000    0.00000000000000    0.00000000000000
   0.00000000000000    5.95300000000000    0.00000000000000
  -6.66783901034905    0.00000000000000    8.42295499127043
    Cu       O       Se
     8      16        4
Direct
   0.00000000000000    0.00000000000000    0.00000000000000    Cu    (2a)
   0.00000000000000    0.50000000000000    0.50000000000000    Cu    (2a)
   0.50000000000000    0.00000000000000    0.00000000000000    Cu    (2b)
   0.50000000000000    0.50000000000000    0.50000000000000    Cu    (2b)
   0.50820000000000    0.70500000000000    0.76050000000000    Cu    (4e)
  -0.50820000000000    1.20500000000000   -0.26050000000000    Cu    (4e)
  -0.50820000000000   -0.70500000000000   -0.76050000000000    Cu    (4e)
   0.50820000000000   -0.20500000000000    1.26050000000000    Cu    (4e)
   0.65490000000000   -0.03100000000000   -0.09810000000000    O     (4e)
  -0.65490000000000    0.46900000000000    0.59810000000000    O     (4e)
  -0.65490000000000    0.03100000000000    0.09810000000000    O     (4e)
   0.65490000000000    0.53100000000000    0.40190000000000    O     (4e)
   0.40800000000000    0.56410000000000    0.09950000000000    O     (4e)
  -0.40800000000000    1.06410000000000    0.40050000000000    O     (4e)
  -0.40800000000000   -0.56410000000000   -0.09950000000000    O     (4e)
   0.40800000000000   -0.06410000000000    0.59950000000000    O     (4e)
  -0.05100000000000    0.71700000000000    0.89600000000000    O     (4e)
   0.05100000000000    1.21700000000000   -0.39600000000000    O     (4e)
   0.05100000000000   -0.71700000000000   -0.89600000000000    O     (4e)
  -0.05100000000000   -0.21700000000000    1.39600000000000    O     (4e)
   0.26100000000000    0.78260000000000    0.84050000000000    O     (4e)
  -0.26100000000000    1.28260000000000   -0.34050000000000    O     (4e)
  -0.26100000000000   -0.78260000000000   -0.84050000000000    O     (4e)
   0.26100000000000   -0.28260000000000    1.34050000000000    O     (4e)
   0.16490000000000    0.57730000000000   -0.09990000000000    Se    (4e)
  -0.16490000000000    1.07730000000000    0.59990000000000    Se    (4e)
  -0.16490000000000   -0.57730000000000    0.09990000000000    Se    (4e)
   0.16490000000000   -0.07730000000000    0.40010000000000    Se    (4e)
```

$Sb_4O_5Cl_2$: A2B5C4_mP22_14_e_c2e_2e - CIF

```
# CIF file
data_findsym-output
_audit_creation_method FINDSYM

_chemical_name_mineral 'Cl2O5Sb4'
_chemical_formula_sum 'Cl2 O5 Sb4'

loop_
_publ_author_name
 'C. S\"{a}rnstrand'
_journal_name_full_name
;
 Acta Crystallographica Section B: Structural Science
;
_journal_volume 34
_journal_year 1978
_journal_page_first 2402
_journal_page_last 2407
_publ_Section_title
;
 The crystal structure of antimony(III) chloride oxide Sb$_{4}$O$_{5}
    ↪ $Cl$_{2}$
;

_aflow_title 'Sb$_{4}$O$_{5}$Cl$_{2}$ Structure'
_aflow_proto 'A2B5C4_mP22_14_e_c2e_2e'
_aflow_params 'a,b/a,c/a,\beta,x_{2},y_{2},z_{2},x_{3},y_{3},z_{3},x_{4
    ↪ },y_{4},z_{4},x_{5},y_{5},z_{5},x_{6},y_{6},z_{6}'
_aflow_params_values '6.238,0.819365181148,2.170246874,97.217,0.51,0.703
    ↪ ,0.115,0.13,0.35,0.185,0.07,0.05,-0.085,0.186,0.225,0.049,0.796
    ↪ ,0.113,0.203'
_aflow_Strukturbericht 'None'
```

```
_aflow_Pearson 'mP22'

_symmetry_space_group_name_H-M "P 1 21/c 1"
_symmetry_Int_Tables_number 14

_cell_length_a      6.23800
_cell_length_b      5.11120
_cell_length_c      13.53800
_cell_angle_alpha 90.00000
_cell_angle_beta   97.21700
_cell_angle_gamma 90.00000

loop_
_space_group_symop_id
_space_group_symop_operation_xyz
1 x,y,z
2 -x,y+1/2,-z+1/2
3 -x,-y,-z
4 x,-y+1/2,z+1/2

loop_
_atom_site_label
_atom_site_type_symbol
_atom_site_symmetry_multiplicity
_atom_site_Wyckoff_label
_atom_site_fract_x
_atom_site_fract_y
_atom_site_fract_z
_atom_site_occupancy
O1 O 2 c 0.00000 0.00000 0.50000 1.00000
Cl1 Cl 4 e 0.51000 0.70300 0.11500 1.00000
O2 O 4 e 0.13000 0.35000 0.18500 1.00000
O3 O 4 e 0.07000 0.05000 -0.08500 1.00000
Sb1 Sb 4 e 0.18600 0.22500 0.04900 1.00000
Sb2 Sb 4 e 0.79600 0.11300 0.20300 1.00000
```

$Sb_4O_5Cl_2$: A2B5C4_mP22_14_e_c2e_2e - POSCAR

```
A2B5C4_mP22_14_e_c2e_2e & a,b/a,c/a,beta,x2,y2,z2,x3,y3,z3,x4,y4,z4,x5,
    ↪ y5,z5,x6,y6,z6 --params=6.238,0.819365181148,2.170246874,97.217
    ↪ ,0.51,0.703,0.115,0.13,0.35,0.185,0.07,0.05,-0.085,0.186,0.225,
    ↪ 0.049,0.796,0.113,0.203 & P2_{1}/c C_{2h}^{5} #14 (ce^5) & mP22
    ↪ & None & Cl2O5Sb4 & Cl2O5Sb4 & C. S\"{a}rnstrand, Acta
    ↪ Crystallogr. Sect. B Struct. Sci. 34, 2402-2407 (1978)
   1.00000000000000
   6.23800000000000    0.00000000000000    0.00000000000000
   0.00000000000000    5.11120000000000    0.00000000000000
  -1.70074637299785    0.00000000000000   13.43074479597970
    Cl       O      Sb
     4      10       8
Direct
   0.51000000000000    0.70300000000000    0.11500000000000   Cl   (4e)
  -0.51000000000000    1.20300000000000    0.38500000000000   Cl   (4e)
  -0.51000000000000   -0.70300000000000   -0.11500000000000   Cl   (4e)
   0.51000000000000   -0.20300000000000    0.61500000000000   Cl   (4e)
   0.00000000000000    0.00000000000000    0.50000000000000   O    (2c)
   0.00000000000000    0.50000000000000    0.00000000000000   O    (2c)
   0.13000000000000    0.35000000000000    0.18500000000000   O    (4e)
  -0.13000000000000    0.85000000000000    0.31500000000000   O    (4e)
  -0.13000000000000   -0.35000000000000   -0.18500000000000   O    (4e)
   0.13000000000000    0.15000000000000    0.68500000000000   O    (4e)
   0.07000000000000    0.05000000000000   -0.08500000000000   O    (4e)
  -0.07000000000000    0.55000000000000    0.58500000000000   O    (4e)
  -0.07000000000000   -0.05000000000000    0.08500000000000   O    (4e)
   0.07000000000000    0.45000000000000    0.41500000000000   O    (4e)
   0.18600000000000    0.22500000000000    0.04900000000000   Sb   (4e)
  -0.18600000000000    0.72500000000000    0.45100000000000   Sb   (4e)
  -0.18600000000000   -0.22500000000000   -0.04900000000000   Sb   (4e)
   0.18600000000000    0.27500000000000    0.54900000000000   Sb   (4e)
   0.79600000000000    0.11300000000000    0.20300000000000   Sb   (4e)
  -0.79600000000000    0.61300000000000    0.29700000000000   Sb   (4e)
  -0.79600000000000   -0.11300000000000   -0.20300000000000   Sb   (4e)
   0.79600000000000    0.38700000000000    0.70300000000000   Sb   (4e)
```

$Ca_2UO_5$: A2B5C_mP32_14_2e_5e_ab - CIF

```
# CIF file
data_findsym-output
_audit_creation_method FINDSYM

_chemical_name_mineral 'Ca2O5U'
_chemical_formula_sum 'Ca2 O5 U'

loop_
_publ_author_name
 'B. O. Loopstra'
 'H. M. Rietveld'
_journal_name_full_name
;
 Acta Crystallographica Section B: Structural Science
;
_journal_volume 25
_journal_year 1969
_journal_page_first 787
_journal_page_last 791
_publ_Section_title
;
 The structure of some alkaline-earth metal uranates
;

_aflow_title 'Ca$_{2}$UO$_{5}$ Structure'
_aflow_proto 'A2B5C_mP32_14_2e_5e_ab'
_aflow_params 'a,b/a,c/a,\beta,x_{3},y_{3},z_{3},x_{4},y_{4},z_{4},x_{5
    ↪ },y_{5},z_{5},x_{6},y_{6},z_{6},x_{7},y_{7},z_{7},x_{8},y_{8},
    ↪ z_{8},x_{9},y_{9},z_{9}'
_aflow_params_values '7.9137,0.687529221477,1.44663052681,108.803,0.1592
    ↪ ,0.0621,0.32,0.3462,-0.0015,0.6414,0.4447,0.3202,0.3267,0.8802,
    ↪ 0.2831,0.3532,0.6197,0.3043,0.072,0.0163,0.261,0.1295,0.2507,
    ↪ 0.1873,-0.0047'
_aflow_Strukturbericht 'None'
_aflow_Pearson 'mP32'

_symmetry_space_group_name_H-M "P 1 21/c 1"
_symmetry_Int_Tables_number 14

_cell_length_a      7.91370
_cell_length_b      5.44090
_cell_length_c      11.44820
_cell_angle_alpha 90.00000
_cell_angle_beta   108.80300
_cell_angle_gamma 90.00000

loop_
_space_group_symop_id
_space_group_symop_operation_xyz
1 x,y,z
2 -x,y+1/2,-z+1/2
3 -x,-y,-z
4 x,-y+1/2,z+1/2

loop_
_atom_site_label
_atom_site_type_symbol
_atom_site_symmetry_multiplicity
_atom_site_Wyckoff_label
_atom_site_fract_x
_atom_site_fract_y
_atom_site_fract_z
_atom_site_occupancy
U1 U 2 a 0.00000 0.00000 0.00000 1.00000
U2 U 2 b 0.50000 0.00000 0.00000 1.00000
Ca1 Ca 4 e 0.15920 0.06210 0.32000 1.00000
Ca2 Ca 4 e 0.34620 -0.00150 0.64140 1.00000
O1 O 4 e 0.44470 0.32020 0.32670 1.00000
O2 O 4 e 0.88020 0.28310 0.35320 1.00000
O3 O 4 e 0.61970 0.30430 0.07200 1.00000
O4 O 4 e 0.01630 0.26100 0.12950 1.00000
O5 O 4 e 0.25070 0.18730 -0.00470 1.00000
```

$Ca_2UO_5$: A2B5C_mP32_14_2e_5e_ab - POSCAR

```
A2B5C_mP32_14_2e_5e_ab & a,b/a,c/a,beta,x3,y3,z3,x4,y4,z4,x5,y5,z5,x6,y6
    ↪ ,z6,x7,y7,z7,x8,y8,z8,x9,y9,z9 --params=7.9137,0.687529221477,
    ↪ 1.44663052681,108.803,0.1592,0.0621,0.32,0.3462,-0.0015,0.6414,
    ↪ 0.4447,0.3202,0.3267,0.8802,0.2831,0.3532,0.6197,0.3043,0.072,
    ↪ 0.0163,0.261,0.1295,0.2507,0.1873,-0.0047 & P2_{1}/c C_{2h}^{5}
    ↪ #14 (abe^7) & mP32 & None & Ca2O5U & Ca2O5U & B. O. Loopstra
    ↪ and H. M. Rietveld, Acta Crystallogr. Sect. B Struct. Sci. 25,
    ↪ 787-791 (1969)
   1.00000000000000
   7.91370000000000    0.00000000000000    0.00000000000000
   0.00000000000000    5.44090000000000    0.00000000000000
  -3.68992957359137    0.00000000000000   10.83723687025140
    Ca       O       U
     8      20       4
Direct
   0.15920000000000    0.06210000000000    0.32000000000000   Ca   (4e)
  -0.15920000000000    0.56210000000000    0.18000000000000   Ca   (4e)
  -0.15920000000000   -0.06210000000000   -0.32000000000000   Ca   (4e)
   0.15920000000000    0.43790000000000    0.82000000000000   Ca   (4e)
   0.34620000000000   -0.00150000000000    0.64140000000000   Ca   (4e)
  -0.34620000000000    0.49850000000000   -0.14140000000000   Ca   (4e)
  -0.34620000000000    0.00150000000000   -0.64140000000000   Ca   (4e)
   0.34620000000000    0.50150000000000    1.14140000000000   Ca   (4e)
   0.44470000000000    0.32020000000000    0.32670000000000   O    (4e)
  -0.44470000000000    0.82020000000000    0.17330000000000   O    (4e)
  -0.44470000000000   -0.32020000000000   -0.32670000000000   O    (4e)
   0.44470000000000    0.17980000000000    0.82670000000000   O    (4e)
   0.88020000000000    0.28310000000000    0.35320000000000   O    (4e)
  -0.88020000000000    0.78310000000000    0.14680000000000   O    (4e)
  -0.88020000000000   -0.28310000000000   -0.35320000000000   O    (4e)
   0.88020000000000    0.21690000000000    0.85320000000000   O    (4e)
   0.61970000000000    0.30430000000000    0.07200000000000   O    (4e)
  -0.61970000000000    0.80430000000000    0.42800000000000   O    (4e)
  -0.61970000000000   -0.30430000000000   -0.07200000000000   O    (4e)
   0.61970000000000    0.19570000000000    0.57200000000000   O    (4e)
   0.01630000000000    0.26100000000000    0.12950000000000   O    (4e)
  -0.01630000000000    0.76100000000000    0.37050000000000   O    (4e)
  -0.01630000000000   -0.26100000000000   -0.12950000000000   O    (4e)
   0.01630000000000    0.23900000000000    0.62950000000000   O    (4e)
   0.25070000000000    0.18730000000000   -0.00470000000000   O    (4e)
  -0.25070000000000    0.68730000000000    0.50470000000000   O    (4e)
  -0.25070000000000   -0.18730000000000    0.00470000000000   O    (4e)
   0.25070000000000    0.31270000000000    0.49530000000000   O    (4e)
   0.00000000000000    0.00000000000000    0.00000000000000   U    (2a)
   0.00000000000000    0.50000000000000    0.50000000000000   U    (2a)
   0.50000000000000    0.00000000000000    0.00000000000000   U    (2b)
   0.50000000000000    0.50000000000000    0.50000000000000   U    (2b)
```

$Gd_2SiO_5$ ($RE_2SiO_5$ X1): A2B5C_mP32_14_2e_5e_e - CIF

```
# CIF file
data_findsym-output
_audit_creation_method FINDSYM

_chemical_name_mineral 'Gd2O5Si'
_chemical_formula_sum 'Gd2 O5 Si'

loop_
_publ_author_name
```

```
'G. V. Anan\'eva'
'A. M. Korovkin'
'T. I. Merkulyaeva'
'A. M. Morozova'
'M. V. Petrov'
'I. R. Savinova'
'V. R. Startsev'
'P. P. Feofilov'
_journal_name_full_name
;
 Inorganic Materials
;
_journal_volume 17
_journal_year 1981
_journal_page_first 754
_journal_page_last 758
_publ_Section_title
;
 Growth of lanthanide oxyorthosilicate single crystals, and their
     ↪ structural and optical characteristics
;

# Found in Gd$_{2}$SiO$_{5}$ (Gd$_{2}$[SiO$_{4}$]O) Crystal Structure,
     ↪ 2016 Found in Gd$_{2}$SiO$_{5}$ (Gd$_{2}$[SiO$_{4}$]O) Crystal
     ↪ Structure, {PAULING FILE in: Inorganic Solid Phases,
     ↪ SpringerMaterials (online database), Springer, Heidelberg (ed.)
     ↪ SpringerMaterials },

_aflow_title 'Gd$_{2}$SiO$_{5}$ ($RE$$_{2}$SiO$_{5}$ X1) Structure'
_aflow_proto 'A2B5C_mP32_14_2e_5e_e'
_aflow_params 'a,b/a,c/a,\beta,x_{1},y_{1},z_{1},x_{2},y_{2},z_{2},x_{3
     ↪ },y_{3},z_{3},x_{4},y_{4},z_{4},x_{5},y_{5},z_{5},x_{6},y_{6},
     ↪ z_{6},x_{7},y_{7},z_{7},x_{8},y_{8},z_{8}'
_aflow_params_values '9.16,0.770742358079,0.745633187773,107.58,0.02458,
     ↪ 0.12451,0.23428,0.38547,0.146,0.08372,0.1163,0.3782,0.4513,
     ↪ 0.2968,0.0698,0.3547,0.3683,0.4587,0.248,0.5941,0.2681,0.4507,
     ↪ 0.8839,0.3639,0.0059,0.298,0.5876,0.0402'
_aflow_Strukturbericht 'None'
_aflow_Pearson 'mP32'

_symmetry_space_group_name_H-M "P 1 21/c 1"
_symmetry_Int_Tables_number 14

_cell_length_a     9.16000
_cell_length_b     7.06000
_cell_length_c     6.83000
_cell_angle_alpha  90.00000
_cell_angle_beta   107.58000
_cell_angle_gamma  90.00000

loop_
_space_group_symop_id
_space_group_symop_operation_xyz
1 x,y,z
2 -x,y+1/2,-z+1/2
3 -x,-y,-z
4 x,-y+1/2,z+1/2

loop_
_atom_site_label
_atom_site_type_symbol
_atom_site_symmetry_multiplicity
_atom_site_Wyckoff_label
_atom_site_fract_x
_atom_site_fract_y
_atom_site_fract_z
_atom_site_occupancy
Gd1 Gd 4 e 0.02458 0.12451 0.23428 1.00000
Gd2 Gd 4 e 0.38547 0.14600 0.08372 1.00000
O1 O 4 e 0.11630 0.37820 0.45130 1.00000
O2 O 4 e 0.29680 0.06980 0.35470 1.00000
O3 O 4 e 0.36830 0.45870 0.24800 1.00000
O4 O 4 e 0.59410 0.26810 0.45070 1.00000
O5 O 4 e 0.88390 0.36390 0.00590 1.00000
Si1 Si 4 e 0.29800 0.58760 0.04020 1.00000
```

$Gd_2SiO_5$ ($RE_2SiO_5$ X1): A2B5C_mP32_14_2e_5e_e - POSCAR

```
A2B5C_mP32_14_2e_5e_e & a,b/a,c/a,beta,x1,y1,z1,x2,y2,z2,x3,y3,z3,x4,y4,
     ↪ z4,x5,y5,z5,x6,y6,z6,x7,y7,z7,x8,y8,z8 --params=9.16,
     ↪ 0.770742358079,0.745633187773,107.58,0.02458,0.12451,0.23428,
     ↪ 0.38547,0.146,0.08372,0.1163,0.3782,0.4513,0.2968,0.0698,0.3547
     ↪ ,0.3683,0.4587,0.248,0.5941,0.2681,0.4507,0.8839,0.3639,0.0059,
     ↪ 0.298,0.5876,0.0402 & P2_{1}/c C_{2h}^{5} #14 (e^8) & mP32 &
     ↪ None & Gd2O5Si & Gd2O5Si & G. V. Anan'eva et al., Inorg. Mat.
     ↪ 17, 754-758 (1981)
   1.00000000000000
   9.16000000000000   0.00000000000000   0.00000000000000
   0.00000000000000   7.06000000000000   0.00000000000000
  -2.06291370734890   0.00000000000000   6.51101275041234
    Gd      O     Si
     8     20      4
Direct
   0.02458000000000   0.12451000000000   0.23428000000000   Gd   (4e)
  -0.02458000000000   0.62451000000000   0.26572000000000   Gd   (4e)
  -0.02458000000000  -0.12451000000000  -0.23428000000000   Gd   (4e)
   0.02458000000000   0.37549000000000   0.73428000000000   Gd   (4e)
   0.38547000000000   0.14600000000000   0.08372000000000   Gd   (4e)
  -0.38547000000000   0.64600000000000   0.41628000000000   Gd   (4e)
  -0.38547000000000  -0.14600000000000  -0.08372000000000   Gd   (4e)
   0.38547000000000   0.35400000000000   0.58372000000000   Gd   (4e)
   0.11630000000000   0.37820000000000   0.45130000000000   O    (4e)
  -0.11630000000000   0.87820000000000   0.04870000000000   O    (4e)
  -0.11630000000000  -0.37820000000000  -0.45130000000000   O    (4e)
   0.11630000000000   0.12180000000000   0.95130000000000   O    (4e)
   0.29680000000000   0.06980000000000   0.35470000000000   O    (4e)
  -0.29680000000000   0.56980000000000   0.14530000000000   O    (4e)
  -0.29680000000000  -0.06980000000000  -0.35470000000000   O    (4e)
   0.29680000000000   0.43020000000000   0.85470000000000   O    (4e)
   0.36830000000000   0.45870000000000   0.24800000000000   O    (4e)
  -0.36830000000000   0.95870000000000   0.25200000000000   O    (4e)
  -0.36830000000000  -0.45870000000000  -0.24800000000000   O    (4e)
   0.36830000000000   0.04130000000000   0.74800000000000   O    (4e)
   0.59410000000000   0.26810000000000   0.45070000000000   O    (4e)
  -0.59410000000000   0.76810000000000   0.04930000000000   O    (4e)
  -0.59410000000000  -0.26810000000000  -0.45070000000000   O    (4e)
   0.59410000000000   0.23190000000000   0.95070000000000   O    (4e)
   0.88390000000000   0.36390000000000   0.00590000000000   O    (4e)
  -0.88390000000000   0.86390000000000   0.49410000000000   O    (4e)
  -0.88390000000000  -0.36390000000000  -0.00590000000000   O    (4e)
   0.88390000000000   0.13610000000000   0.50590000000000   O    (4e)
   0.29800000000000   0.58760000000000   0.04020000000000   Si   (4e)
  -0.29800000000000   1.08760000000000   0.45980000000000   Si   (4e)
  -0.29800000000000  -0.58760000000000  -0.04020000000000   Si   (4e)
   0.29800000000000  -0.08760000000000   0.54020000000000   Si   (4e)
```

Sanguite ($KCuCl_3$): A3BC_mP20_14_3e_e_e - CIF

```
# CIF file
data_findsym-output
_audit_creation_method FINDSYM

_chemical_name_mineral 'Sanguite'
_chemical_formula_sum 'Cl3 Cu K'

loop_
_publ_author_name
 'R. D. Willett'
 'C. {Dwiggins, Jr.}'
 'R. F. Kruh'
 'R. E. Rundle'
_journal_name_full_name
;
 Journal of Chemical Physics
;
_journal_volume 38
_journal_year 1963
_journal_page_first 2429
_journal_page_last 2436
_publ_Section_title
;
 Crystal Structures of KCuCl$_{3}$ and NH$_{4}$CuCl$_{3}$
;

_aflow_title 'Sanguite (KCuCl$_{3}$) Structure'
_aflow_proto 'A3BC_mP20_14_3e_e_e'
_aflow_params 'a,b/a,c/a,\beta,x_{1},y_{1},z_{1},x_{2},y_{2},z_{2},x_{3
     ↪ },y_{3},z_{3},x_{4},y_{4},z_{4},x_{5},y_{5},z_{5}'
_aflow_params_values '4.029,3.42144452718,2.16827997022,97.33333,0.2754,
     ↪ 0.19875,0.263,0.6782,-0.00745,0.32171,0.8203,0.09875,-0.03369,
     ↪ 0.2408,0.04976,0.1575,0.7825,0.17081,0.55692'
_aflow_Strukturbericht 'None'
_aflow_Pearson 'mP20'

_symmetry_space_group_name_H-M "P 1 21/c 1"
_symmetry_Int_Tables_number 14

_cell_length_a     4.02900
_cell_length_b     13.78500
_cell_length_c     8.73600
_cell_angle_alpha  90.00000
_cell_angle_beta   97.33333
_cell_angle_gamma  90.00000

loop_
_space_group_symop_id
_space_group_symop_operation_xyz
1 x,y,z
2 -x,y+1/2,-z+1/2
3 -x,-y,-z
4 x,-y+1/2,z+1/2

loop_
_atom_site_label
_atom_site_type_symbol
_atom_site_symmetry_multiplicity
_atom_site_Wyckoff_label
_atom_site_fract_x
_atom_site_fract_y
_atom_site_fract_z
_atom_site_occupancy
Cl1 Cl 4 e 0.27540 0.19875 0.26300 1.00000
Cl2 Cl 4 e 0.67820 -0.00745 0.32171 1.00000
Cl3 Cl 4 e 0.82030 0.09875 -0.03369 1.00000
Cu1 Cu 4 e 0.24080 0.04976 0.15750 1.00000
K1 K 4 e 0.78250 0.17081 0.55692 1.00000
```

Sanguite ($KCuCl_3$): A3BC_mP20_14_3e_e_e - POSCAR

```
A3BC_mP20_14_3e_e_e & a,b/a,c/a,beta,x1,y1,z1,x2,y2,z2,x3,y3,z3,x4,y4,z4
     ↪ ,x5,y5,z5 --params=4.029,3.42144452718,2.16827997022,97.33333,
     ↪ 0.2754,0.19875,0.263,0.6782,-0.00745,0.32171,0.8203,0.09875,-
     ↪ 0.03369,0.2408,0.04976,0.1575,0.7825,0.17081,0.55692 & P2_{1}/c
     ↪ C_{2h}^{5} #14 (e^5) & mP20 & None & Cl3CuK & Sanguite & R. D.
     ↪ Willett et al., J. Chem. Phys. 38, 2429-2436 (1963)
   1.00000000000000
   4.02900000000000   0.00000000000000   0.00000000000000
   0.00000000000000  13.78500000000000   0.00000000000000
  -1.11507693162750   0.00000000000000   8.66454265593702
    Cl     Cu      K
```

```
      12        4        4
Direct
   0.27540000000000    0.19875000000000    0.26300000000000    Cl    (4e)
  -0.27540000000000    0.69875000000000    0.23700000000000    Cl    (4e)
  -0.27540000000000   -0.19875000000000   -0.26300000000000    Cl    (4e)
   0.27540000000000    0.30125000000000    0.76300000000000    Cl    (4e)
   0.67820000000000   -0.00745000000000    0.32171000000000    Cl    (4e)
  -0.67820000000000    0.49255000000000    0.17829000000000    Cl    (4e)
  -0.67820000000000    0.00745000000000   -0.32171000000000    Cl    (4e)
   0.67820000000000    0.50745000000000    0.82171000000000    Cl    (4e)
   0.82030000000000    0.09875000000000   -0.03369000000000    Cl    (4e)
  -0.82030000000000    0.59875000000000    0.53369000000000    Cl    (4e)
  -0.82030000000000   -0.09875000000000    0.03369000000000    Cl    (4e)
   0.82030000000000    0.40125000000000    0.46631000000000    Cl    (4e)
   0.24080000000000    0.04976000000000    0.15750000000000    Cu    (4e)
  -0.24080000000000    0.54976000000000    0.34250000000000    Cu    (4e)
  -0.24080000000000   -0.04976000000000   -0.15750000000000    Cu    (4e)
   0.24080000000000    0.45024000000000    0.65750000000000    Cu    (4e)
   0.78250000000000    0.17081000000000    0.55692000000000    K     (4e)
  -0.78250000000000    0.67081000000000   -0.05692000000000    K     (4e)
  -0.78250000000000   -0.17081000000000   -0.55692000000000    K     (4e)
   0.78250000000000    0.32919000000000    1.05692000000000    K     (4e)
```

$\gamma$-$WO_3$: A3B_mP32_14_6e_2e - CIF

```
# CIF file
data_findsym-output
_audit_creation_method FINDSYM

_chemical_name_mineral 'O3W'
_chemical_formula_sum 'O3 W'

loop_
_publ_author_name
 'P. M. Woodward'
 'A. W. Sleight'
 'T. Vogt'
_journal_name_full_name
;
 Journal of Solid State Chemistry
;
_journal_volume 131
_journal_year 1997
_journal_page_first 9
_journal_page_last 17
_publ_Section_title
;
 Ferroelectric Tungsten Trioxide
;

_aflow_title '$\gamma$-WO$_{3}$ Structure'
_aflow_proto 'A3B_mP32_14_6e_2e'
_aflow_params 'a,b/a,c/a,\beta,x_{1},y_{1},z_{1},x_{2},y_{2},z_{2},x_{3
    ↪ },y_{3},z_{3},x_{4},y_{4},z_{4},x_{5},y_{5},z_{5},x_{6},y_{6},
    ↪ z_{6},x_{7},y_{7},z_{7},x_{8},y_{8},z_{8}'
_aflow_params_values '7.3271,1.03238661954,1.44716463539,133.21984,0.782
    ↪ ,-0.03,0.782,0.781,0.536,0.779,0.005,0.736,0.723,0.472,0.743,
    ↪ 0.258,0.2768,-0.028,-0.0002,0.288,0.502,0.0,-0.03,-0.026,0.717,
    ↪ 0.4649,-0.033,0.2189'
_aflow_Strukturbericht 'None'
_aflow_Pearson 'mP32'

_symmetry_space_group_name_H-M "P 1 21/c 1"
_symmetry_Int_Tables_number 14

_cell_length_a      7.32710
_cell_length_b      7.56440
_cell_length_c      10.60352
_cell_angle_alpha 90.00000
_cell_angle_beta  133.21984
_cell_angle_gamma 90.00000

loop_
_space_group_symop_id
_space_group_symop_operation_xyz
1 x,y,z
2 -x,y+1/2,-z+1/2
3 -x,-y,-z
4 x,-y+1/2,z+1/2

loop_
_atom_site_label
_atom_site_type_symbol
_atom_site_symmetry_multiplicity
_atom_site_Wyckoff_label
_atom_site_fract_x
_atom_site_fract_y
_atom_site_fract_z
_atom_site_occupancy
O1 O 4 e 0.78200 -0.03000 0.78200 1.00000
O2 O 4 e 0.78100 0.53600 0.77900 1.00000
O3 O 4 e 0.00500 0.73600 0.72300 1.00000
O4 O 4 e 0.47200 0.74300 0.25800 1.00000
O5 O 4 e 0.27680 -0.02800 -0.00020 1.00000
O6 O 4 e 0.28800 0.50200 0.00000 1.00000
W1 W 4 e -0.03000 -0.02600 0.71700 1.00000
W2 W 4 e 0.46490 -0.03300 0.21890 1.00000
```

$\gamma$-$WO_3$: A3B_mP32_14_6e_2e - POSCAR

```
A3B_mP32_14_6e_2e & a,b/a,c/a,beta,x1,y1,z1,x2,y2,z2,x3,y3,z3,x4,y4,z4,
    ↪ x5,y5,z5,x6,y6,z6,x7,y7,z7,x8,y8,z8 --params=7.3271,
    ↪ 1.03238661954,1.44716463539,133.21984,0.782,-0.03,0.782,0.781,
    ↪ 0.536,0.779,0.005,0.736,0.723,0.472,0.743,0.258,0.2768,-0.028,-
    ↪ 0.0002,0.288,0.502,0.0,-0.03,-0.026,0.717,0.4649,-0.033,0.2189
    ↪ & P2_{1}/c C_{2h}^{5} #14 (e^8) & mP32 & None & O3W & O3W & P.
    ↪ M. Woodward and A. W. Sleight and T. Vogt, J. Solid State Chem.
    ↪ 131, 9-17 (1997)
   1.00000000000000
   7.32710000000000    0.00000000000000    0.00000000000000
   0.00000000000000    7.56440000000000    0.00000000000000
  -7.26128505928556    0.00000000000000    7.72711949423563
     O        W
    24        8
Direct
   0.78200000000000   -0.03000000000000    0.78200000000000    O    (4e)
  -0.78200000000000    0.47000000000000   -0.28200000000000    O    (4e)
  -0.78200000000000    0.03000000000000   -0.78200000000000    O    (4e)
   0.78200000000000    0.53000000000000    1.28200000000000    O    (4e)
   0.78100000000000    0.53600000000000    0.77900000000000    O    (4e)
  -0.78100000000000    1.03600000000000   -0.27900000000000    O    (4e)
  -0.78100000000000   -0.53600000000000   -0.77900000000000    O    (4e)
   0.78100000000000   -0.03600000000000    1.27900000000000    O    (4e)
   0.00500000000000    0.73600000000000    0.72300000000000    O    (4e)
  -0.00500000000000    1.23600000000000   -0.22300000000000    O    (4e)
  -0.00500000000000   -0.73600000000000   -0.72300000000000    O    (4e)
   0.00500000000000   -0.23600000000000    1.22300000000000    O    (4e)
   0.47200000000000    0.74300000000000    0.25800000000000    O    (4e)
  -0.47200000000000    1.24300000000000    0.24200000000000    O    (4e)
  -0.47200000000000   -0.74300000000000   -0.25800000000000    O    (4e)
   0.47200000000000   -0.24300000000000    0.75800000000000    O    (4e)
   0.27680000000000   -0.02800000000000   -0.00020000000000    O    (4e)
  -0.27680000000000    0.47200000000000    0.50020000000000    O    (4e)
  -0.27680000000000    0.02800000000000    0.00020000000000    O    (4e)
   0.27680000000000    0.52800000000000    0.49980000000000    O    (4e)
   0.28800000000000    0.50200000000000    0.00000000000000    O    (4e)
  -0.28800000000000    1.00200000000000    0.50000000000000    O    (4e)
  -0.28800000000000   -0.50200000000000    0.00000000000000    O    (4e)
   0.28800000000000   -0.00200000000000    0.50000000000000    O    (4e)
  -0.03000000000000   -0.02600000000000    0.71700000000000    W    (4e)
   0.03000000000000    0.47400000000000   -0.21700000000000    W    (4e)
   0.03000000000000    0.02600000000000   -0.71700000000000    W    (4e)
  -0.03000000000000    0.52600000000000    1.21700000000000    W    (4e)
   0.46490000000000   -0.03300000000000    0.21890000000000    W    (4e)
  -0.46490000000000    0.46700000000000    0.28110000000000    W    (4e)
  -0.46490000000000    0.03300000000000   -0.21890000000000    W    (4e)
   0.46490000000000    0.53300000000000    0.71890000000000    W    (4e)
```

$K_2Ni(CN)_4$: A4B2C4D_mP22_14_2e_e_2e_a - CIF

```
# CIF file
data_findsym-output
_audit_creation_method FINDSYM

_chemical_name_mineral 'C4K2N4Ni'
_chemical_formula_sum 'C4 K2 N4 Ni'

loop_
_publ_author_name
 'N.-G. Vannerberg'
_journal_name_full_name
;
 Acta Chemica Scandinavica
;
_journal_volume 18
_journal_year 1964
_journal_page_first 2385
_journal_page_last 2391
_publ_Section_title
;
 The Crystal Structure of K$_{2}$Ni(CN)$_{4}$
;

_aflow_title 'K$_{2}$Ni(CN)$_{4}$ Structure'
_aflow_proto 'A4B2C4D_mP22_14_2e_e_2e_a'
_aflow_params 'a,b/a,c/a,\beta,x_{2},y_{2},z_{2},x_{3},y_{3},z_{3},x_{4
    ↪ },y_{4},z_{4},x_{5},y_{5},z_{5},x_{6},y_{6},z_{6}'
_aflow_params_values '4.294,1.78854215184,3.03446669772,87.26667,0.167,
    ↪ 0.207,0.055,0.849,0.382,0.39,0.281,0.046,0.341,0.261,0.338,
    ↪ 0.082,0.764,0.3,0.327'
_aflow_Strukturbericht 'None'
_aflow_Pearson 'mP22'

_symmetry_space_group_name_H-M "P 1 21/c 1"
_symmetry_Int_Tables_number 14

_cell_length_a      4.29400
_cell_length_b      7.68000
_cell_length_c      13.03000
_cell_angle_alpha 90.00000
_cell_angle_beta  87.26667
_cell_angle_gamma 90.00000

loop_
_space_group_symop_id
_space_group_symop_operation_xyz
1 x,y,z
2 -x,y+1/2,-z+1/2
3 -x,-y,-z
4 x,-y+1/2,z+1/2

loop_
_atom_site_label
_atom_site_type_symbol
_atom_site_symmetry_multiplicity
_atom_site_Wyckoff_label
_atom_site_fract_x
_atom_site_fract_y
_atom_site_fract_z
_atom_site_occupancy
```

```
Ni1 Ni 2 a 0.00000 0.00000 0.00000 1.00000
C1 C 4 e 0.16700 0.20700 0.05500 1.00000
C2 C 4 e 0.84900 0.38200 0.39000 1.00000
K1 K 4 e 0.28100 0.04600 0.34100 1.00000
N1 N 4 e 0.26100 0.33800 0.08200 1.00000
N2 N 4 e 0.76400 0.30000 0.32700 1.00000
```

$K_2Ni(CN)_4$: A4B2C4D_mP22_14_2e_e_2e_a - POSCAR

```
A4B2C4D_mP22_14_2e_e_2e_a & a,b/a,c/a,beta,x2,y2,z2,x3,y3,z3,x4,y4,z4,x5
    ↪ ,y5,z5,x6,y6,z6 --params=4.294,1.78854215184,3.03446669772,
    ↪ 87.26667,0.167,0.207,0.055,0.849,0.382,0.39,0.281,0.046,0.341,
    ↪ 0.261,0.338,0.082,0.764,0.3,0.327 & P2_{1}/c C_{2h}^{5} #14 (ae
    ↪ ^5) & mP22 & None & C4K2N4Ni & C4K2N4Ni & N.-G. Vannerberg,
    ↪ Acta Chem. Scand. 18, 2385-2391 (1964)
   1.00000000000000
   4.29400000000000   0.00000000000000   0.00000000000000
   0.00000000000000   7.68000000000000   0.00000000000000
   0.62136832261634   0.00000000000000  13.01517581163040
     C      K      N     Ni
     8      4      8      2
Direct
   0.16700000000000   0.20700000000000   0.05500000000000   C   (4e)
  -0.16700000000000   0.70700000000000   0.44500000000000   C   (4e)
  -0.16700000000000  -0.20700000000000  -0.05500000000000   C   (4e)
   0.16700000000000   0.29300000000000   0.55500000000000   C   (4e)
   0.84900000000000   0.38200000000000   0.39000000000000   C   (4e)
  -0.84900000000000   0.88200000000000   0.11000000000000   C   (4e)
  -0.84900000000000  -0.38200000000000  -0.39000000000000   C   (4e)
   0.84900000000000   0.11800000000000   0.89000000000000   C   (4e)
   0.28100000000000   0.04600000000000   0.34100000000000   K   (4e)
  -0.28100000000000   0.54600000000000   0.15900000000000   K   (4e)
  -0.28100000000000  -0.04600000000000  -0.34100000000000   K   (4e)
   0.28100000000000   0.45400000000000   0.84100000000000   K   (4e)
   0.26100000000000   0.33800000000000   0.08200000000000   N   (4e)
  -0.26100000000000   0.83800000000000   0.41800000000000   N   (4e)
  -0.26100000000000  -0.33800000000000  -0.08200000000000   N   (4e)
   0.26100000000000   0.16200000000000   0.58200000000000   N   (4e)
   0.76400000000000   0.30000000000000   0.32700000000000   N   (4e)
  -0.76400000000000   0.80000000000000   0.17300000000000   N   (4e)
  -0.76400000000000  -0.30000000000000  -0.32700000000000   N   (4e)
   0.76400000000000   0.20000000000000   0.82700000000000   N   (4e)
   0.00000000000000   0.00000000000000   0.00000000000000   Ni  (2a)
   0.00000000000000   0.50000000000000   0.50000000000000   Ni  (2a)
```

$KICl_4 \cdot H_2O$ ($H0_{10}$): A4BCD_mP28_14_4e_e_e_e - CIF

```
# CIF file
data_findsym-output
_audit_creation_method FINDSYM

_chemical_name_mineral 'Cl4IK \cdot H2O'
_chemical_formula_sum 'Cl4 (H2O) I K'

loop_
_publ_author_name
 'R. J. Elema'
 'J. L. {de Boer}'
 'A. Vos'
_journal_name_full_name
;
 Acta Cristallographica
;
_journal_volume 16
_journal_year 1963
_journal_page_first 243
_journal_page_last 247
_publ_Section_title
;
 The refinement of the crystal structure of KICl$_{4}$ $\cdot$ H$_{2}$O
;

_aflow_title 'KICl$_{4}$$\cdot$H$_{2}$O ($H0_{10}$) Structure'
_aflow_proto 'A4BCD_mP28_14_4e_e_e_e'
_aflow_params 'a,b/a,c/a,\beta,x_{1},y_{1},z_{1},x_{2},y_{2},z_{2},x_{3
    ↪ },y_{3},z_{3},x_{4},y_{4},z_{4},x_{5},y_{5},z_{5},x_{6},y_{6},
    ↪ z_{6},x_{7},y_{7},z_{7}'
_aflow_params_values '4.284,3.34990662932,3.16137721755,102.62529,0.4857
    ↪ ,0.7807,0.2292,-0.06,0.0955,0.1866,0.625,-0.0116,0.3824,0.7875,
    ↪ 0.8953,0.0398,0.264,0.708,-0.043,0.7052,-0.0592,0.2041,0.1442,
    ↪ 0.3409,-0.0791'
_aflow_Strukturbericht '$H0_{10}$'
_aflow_Pearson 'mP28'

_symmetry_space_group_name_H-M "P 1 21/c 1"
_symmetry_Int_Tables_number 14

_cell_length_a    4.28400
_cell_length_b    14.35100
_cell_length_c    13.54334
_cell_angle_alpha 90.00000
_cell_angle_beta  102.62529
_cell_angle_gamma 90.00000

loop_
_space_group_symop_id
_space_group_symop_operation_xyz
1 x,y,z
2 -x,y+1/2,-z+1/2
3 -x,-y,-z
4 x,-y+1/2,z+1/2

loop_
_atom_site_label
_atom_site_type_symbol
_atom_site_symmetry_multiplicity
_atom_site_Wyckoff_label
_atom_site_fract_x
_atom_site_fract_y
_atom_site_fract_z
_atom_site_occupancy
Cl1 Cl 4 e 0.48570 0.78070 0.22920 1.00000
Cl2 Cl 4 e -0.06000 0.09550 0.18660 1.00000
Cl3 Cl 4 e 0.62500 -0.01160 0.38240 1.00000
Cl4 Cl 4 e 0.78750 0.89530 0.03980 1.00000
H2O1 H2O 4 e 0.26400 0.70800 -0.04300 1.00000
I1 I 4 e 0.70520 -0.05920 0.20410 1.00000
K1 K 4 e 0.14420 0.34090 -0.07910 1.00000
```

$KICl_4 \cdot H_2O$ ($H0_{10}$): A4BCD_mP28_14_4e_e_e_e - POSCAR

```
A4BCD_mP28_14_4e_e_e_e & a,b/a,c/a,beta,x1,y1,z1,x2,y2,z2,x3,y3,z3,x4,y4
    ↪ ,z4,x5,y5,z5,x6,y6,z6,x7,y7,z7 --params=4.284,3.34990662932,
    ↪ 3.16137721755,102.62529,0.4857,0.7807,0.2292,-0.06,0.0955,
    ↪ 0.1866,0.625,-0.0116,0.3824,0.7875,0.8953,0.0398,0.264,0.708,-
    ↪ 0.043,0.7052,-0.0592,0.2041,0.1442,0.3409,-0.0791 & P2_{1}/c C_
    ↪ {2h}^{5} #14 (e^7) & mP28 & $H0_{10}$ & Cl4IK \cdot H2O & Cl4IK
    ↪ \cdot H2O & R. J. Elema and J. L. {de Boer} and A. Vos, Acta
    ↪ Cryst. 16, 243-247 (1963)
   1.00000000000000
   4.28400000000000   0.00000000000000   0.00000000000000
   0.00000000000000  14.35100000000000   0.00000000000000
  -2.96022177652463   0.00000000000000  13.21586718264790
    Cl    H2O     I      K
    16     4      4      4
Direct
   0.48570000000000   0.78070000000000   0.22920000000000   Cl   (4e)
  -0.48570000000000   1.28070000000000   0.27080000000000   Cl   (4e)
  -0.48570000000000  -0.78070000000000  -0.22920000000000   Cl   (4e)
   0.48570000000000  -0.28070000000000   0.72920000000000   Cl   (4e)
  -0.06000000000000   0.09550000000000   0.18660000000000   Cl   (4e)
   0.06000000000000   0.59550000000000   0.31340000000000   Cl   (4e)
   0.06000000000000  -0.09550000000000  -0.18660000000000   Cl   (4e)
  -0.06000000000000   0.40450000000000   0.68660000000000   Cl   (4e)
   0.62500000000000  -0.01160000000000   0.38240000000000   Cl   (4e)
  -0.62500000000000   0.48840000000000   0.11760000000000   Cl   (4e)
  -0.62500000000000   0.01160000000000  -0.38240000000000   Cl   (4e)
   0.62500000000000   0.51160000000000   0.88240000000000   Cl   (4e)
   0.78750000000000   0.89530000000000   0.03980000000000   Cl   (4e)
  -0.78750000000000   1.39530000000000   0.46020000000000   Cl   (4e)
  -0.78750000000000  -0.89530000000000  -0.03980000000000   Cl   (4e)
   0.78750000000000  -0.39530000000000   0.53980000000000   Cl   (4e)
   0.26400000000000   0.70800000000000  -0.04300000000000   H2O  (4e)
  -0.26400000000000   1.20800000000000   0.54300000000000   H2O  (4e)
  -0.26400000000000  -0.70800000000000   0.04300000000000   H2O  (4e)
   0.26400000000000  -0.20800000000000   0.45700000000000   H2O  (4e)
   0.70520000000000  -0.05920000000000   0.20410000000000   I    (4e)
  -0.70520000000000   0.44080000000000   0.29590000000000   I    (4e)
  -0.70520000000000   0.05920000000000  -0.20410000000000   I    (4e)
   0.70520000000000   0.55920000000000   0.70410000000000   I    (4e)
   0.14420000000000   0.34090000000000  -0.07910000000000   K    (4e)
  -0.14420000000000   0.84090000000000   0.57910000000000   K    (4e)
  -0.14420000000000  -0.34090000000000   0.07910000000000   K    (4e)
   0.14420000000000   0.15910000000000   0.42090000000000   K    (4e)
```

$\gamma$-$Y_2Si_2O_7$: A4BC_mP24_14_4e_e_e - CIF

```
# CIF file
data_findsym-output
_audit_creation_method FINDSYM

_chemical_name_mineral 'O7Si2Y2'
_chemical_formula_sum 'O4 Si Y'

loop_
_publ_author_name
 'A. N. Christensen'
 'R. G. Hazell'
 'A. W. Hewat'
_journal_name_full_name
;
 Acta Chemica Scandinavica
;
_journal_volume 51
_journal_year 1997
_journal_page_first 37
_journal_page_last 43
_publ_Section_title
;
 Synthesis, Crystal Growth and Structure Investigations of Rare-Earth
    ↪ Disilicates and Rare-Earth Oxyapatites
;

# Found in Revision of the crystallographic data of polymorphic Y$_{2}
    ↪ $Si$_{2}$O$_{7}$ and Y$_{2}$SiO$_{5}$ compounds, 2004

_aflow_title '$\gamma$-Y$_{2}$Si$_{2}$O$_{7}$ Structure'
_aflow_proto 'A4BC_mP24_14_4e_e_e'
_aflow_params 'a,b/a,c/a,\beta,x_{1},y_{1},z_{1},x_{2},y_{2},z_{2},x_{3
    ↪ },y_{3},z_{3},x_{4},y_{4},z_{4},x_{5},y_{5},z_{5},x_{6},y_{6},
    ↪ z_{6}'
_aflow_params_values '4.6916,2.31309148265,1.1908943644,96.04,0.7933,
    ↪ 0.0502,0.1358,0.867,0.1822,0.5371,0.534,-0.0106,0.4976,0.3792,
    ↪ 0.2017,0.2513,0.6445,0.1136,0.3696,0.8904,0.8495,0.0941'
_aflow_Strukturbericht 'None'
_aflow_Pearson 'mP24'

_symmetry_space_group_name_H-M "P 1 21/c 1"
_symmetry_Int_Tables_number 14
```

```
_cell_length_a      4.69160
_cell_length_b      10.85210
_cell_length_c      5.58720
_cell_angle_alpha  90.00000
_cell_angle_beta   96.04000
_cell_angle_gamma  90.00000

loop_
_space_group_symop_id
_space_group_symop_operation_xyz
1 x,y,z
2 -x,y+1/2,-z+1/2
3 -x,-y,-z
4 x,-y+1/2,z+1/2

loop_
_atom_site_label
_atom_site_type_symbol
_atom_site_symmetry_multiplicity
_atom_site_Wyckoff_label
_atom_site_fract_x
_atom_site_fract_y
_atom_site_fract_z
_atom_site_occupancy
O1 O 4 e 0.79330 0.05020 0.13580 1.00000
O2 O 4 e 0.86700 0.18220 0.53710 1.00000
O3 O 4 e 0.53400 -0.01060 0.49760 0.50000
O4 O 4 e 0.37920 0.20170 0.25130 1.00000
Si1 Si 4 e 0.64450 0.11360 0.36960 1.00000
Y1 Y 4 e 0.89040 0.84950 0.09410 1.00000
```

$\gamma$-$Y_2Si_2O_7$: A4BC_mP24_14_4e_e_e - POSCAR

```
A4BC_mP24_14_4e_e_e & a,b/a,c/a,beta,x1,y1,z1,x2,y2,z2,x3,y3,z3,x4,y4,z4
    ↪ ,x5,y5,z5,x6,y6,z6 --params=4.6916,2.31309148265,1.1908943644,
    ↪ 96.04,0.7933,0.0502,0.1358,0.867,0.1822,0.5371,0.534,-0.0106,
    ↪ 0.4976,0.3792,0.2017,0.2513,0.6445,0.1136,0.3696,0.8904,0.8495,
    ↪ 0.0941 & P2_{1}/c C_{2h}^{5} #14 (e^6) & mP24 & None & O7Si2Y2
    ↪ & O7Si2Y2 & A. N. Christensen and R. G. Hazell and A. W. Hewat,
    ↪  Acta Chem. Scand. 51, 37-43 (1997)
   1.00000000000000
   4.69160000000000    0.00000000000000    0.00000000000000
   0.00000000000000   10.85210000000000    0.00000000000000
  -0.58790052086733    0.00000000000000    5.55618365585263
    O      Si      Y
   16       4      4
Direct
   0.79330000000000    0.05020000000000    0.13580000000000    O   (4e)
  -0.79330000000000    0.55020000000000    0.36420000000000    O   (4e)
  -0.79330000000000   -0.05020000000000   -0.13580000000000    O   (4e)
   0.79330000000000    0.44980000000000    0.63580000000000    O   (4e)
   0.86700000000000    0.18220000000000    0.53710000000000    O   (4e)
  -0.86700000000000    0.68220000000000   -0.03710000000000    O   (4e)
  -0.86700000000000   -0.18220000000000   -0.53710000000000    O   (4e)
   0.86700000000000    0.31780000000000    1.03710000000000    O   (4e)
   0.53400000000000   -0.01060000000000    0.49760000000000    O   (4e)
  -0.53400000000000    0.48940000000000    0.00240000000000    O   (4e)
  -0.53400000000000    0.01060000000000   -0.49760000000000    O   (4e)
   0.53400000000000    0.51060000000000    0.99760000000000    O   (4e)
   0.37920000000000    0.20170000000000    0.25130000000000    O   (4e)
  -0.37920000000000    0.70170000000000    0.24870000000000    O   (4e)
  -0.37920000000000   -0.20170000000000   -0.25130000000000    O   (4e)
   0.37920000000000    0.29830000000000    0.75130000000000    O   (4e)
   0.64450000000000    0.11360000000000    0.36960000000000    Si  (4e)
  -0.64450000000000    0.61360000000000    0.13040000000000    Si  (4e)
  -0.64450000000000   -0.11360000000000   -0.36960000000000    Si  (4e)
   0.64450000000000    0.38640000000000    0.86960000000000    Si  (4e)
   0.89040000000000    0.84950000000000    0.09410000000000    Y   (4e)
  -0.89040000000000    1.34950000000000    0.40590000000000    Y   (4e)
  -0.89040000000000   -0.84950000000000   -0.09410000000000    Y   (4e)
   0.89040000000000   -0.34950000000000    0.59410000000000    Y   (4e)
```

$K_2Pt(SCN)_6 \cdot 2H_2O$: A6B4C2D6E2FG6_mP54_14_3e_2e_e_3e_e_a_3e - CIF

```
# CIF file
data_findsym-output
_audit_creation_method FINDSYM

_chemical_name_mineral 'C6H4K2N6O2PtS6'
_chemical_formula_sum 'C6 H4 K2 N6 O2 Pt S6'

loop_
_publ_author_name
 'J. Arpalahti'
 'J. H\"{o}ls\"{a}'
 'R. Sillanp\"{a}\"{a}'
_journal_name_full_name
;
 Acta Chemica Scandinavica
;
_journal_volume 47
_journal_year 1993
_journal_page_first 1078
_journal_page_last 1082
_publ_Section_title
;
 Studies on Potassium Thiocyanatoplatinates. II. Crystal Structure of
    ↪ Potassium Hexathiocyanatoplatinate(IV) Dihydrate, K$_{2}$Pt(
    ↪ SCN)$_{6}$ $\cdot$ 2H$_{2}$O
;

_aflow_title 'K$_{2}$Pt(SCN)$_{6}$$\cdot$2H$_{2}$O Structure'
_aflow_proto 'A6B4C2D6E2FG6_mP54_14_3e_2e_e_3e_e_a_3e'
_aflow_params 'a,b/a,c/a,\beta,x_{2},y_{2},z_{2},x_{3},y_{3},z_{3},x_{4
    ↪ },y_{4},z_{4},x_{5},y_{5},z_{5},x_{6},y_{6},z_{6},x_{7},y_{7},
    ↪ z_{7},x_{8},y_{8},z_{8},x_{9},y_{9},z_{9},x_{10},y_{10},z_{10},
    ↪ x_{11},y_{11},z_{11},x_{12},y_{12},z_{12},x_{13},y_{13},z_{13},
    ↪ x_{14},y_{14},z_{14}'
_aflow_params_values '11.33,0.983495145631,0.644571932921,94.76,0.7648,
    ↪ 0.8251,-0.0619,0.2402,-0.0953,0.8473,0.2213,-0.0589,0.3316,
    ↪ 0.4514,-0.0915,0.1233,0.5826,-0.0741,0.2021,0.4986,0.8473,
    ↪ 0.6064,0.6695,0.8204,-0.0303,0.327,0.8561,0.8963,0.2991,-0.0101
    ↪ ,0.4229,0.5087,-0.0876,0.2358,-0.0988,0.8216,0.8743,0.1151,-
    ↪ 0.0193,0.7552,0.1129,0.8611,0.2058'
_aflow_Strukturbericht 'None'
_aflow_Pearson 'mP54'

_symmetry_space_group_name_H-M "P 1 21/c 1"
_symmetry_Int_Tables_number 14

_cell_length_a      11.33000
_cell_length_b      11.14300
_cell_length_c      7.30300
_cell_angle_alpha  90.00000
_cell_angle_beta   94.76000
_cell_angle_gamma  90.00000

loop_
_space_group_symop_id
_space_group_symop_operation_xyz
1 x,y,z
2 -x,y+1/2,-z+1/2
3 -x,-y,-z
4 x,-y+1/2,z+1/2

loop_
_atom_site_label
_atom_site_type_symbol
_atom_site_symmetry_multiplicity
_atom_site_Wyckoff_label
_atom_site_fract_x
_atom_site_fract_y
_atom_site_fract_z
_atom_site_occupancy
Pt1 Pt 2 a 0.00000 0.00000 0.00000 1.00000
C1 C 4 e 0.76480 0.82510 -0.06190 1.00000
C2 C 4 e 0.24020 -0.09530 0.84730 1.00000
C3 C 4 e 0.22130 -0.05890 0.33160 1.00000
H1 H 4 e 0.45140 -0.09150 0.12330 1.00000
H2 H 4 e 0.58260 -0.07410 0.20210 1.00000
K1 K 4 e 0.49860 0.84730 0.60640 1.00000
N1 N 4 e 0.66950 0.82040 -0.03030 1.00000
N2 N 4 e 0.32700 0.85610 0.89630 1.00000
N3 N 4 e 0.29910 -0.01010 0.42290 1.00000
O1 O 4 e 0.50870 -0.08760 0.23580 1.00000
S1 S 4 e -0.09880 0.82160 0.87430 1.00000
S2 S 4 e 0.11510 -0.01930 0.75520 1.00000
S3 S 4 e 0.11290 0.86110 0.20580 1.00000
```

$K_2Pt(SCN)_6 \cdot 2H_2O$: A6B4C2D6E2FG6_mP54_14_3e_2e_e_3e_e_a_3e - POSCAR

```
A6B4C2D6E2FG6_mP54_14_3e_2e_e_3e_e_a_3e & a,b/a,c/a,beta,x2,y2,z2,x3,y3,
    ↪ z3,x4,y4,z4,x5,y5,z5,x6,y6,z6,x7,y7,z7,x8,y8,z8,x9,y9,z9,x10,
    ↪ y10,z10,x11,y11,z11,x12,y12,z12,x13,y13,z13,x14,y14,z14 --
    ↪ params=11.33,0.983495145631,0.644571932921,94.76,0.7648,0.8251
    ↪ ,-0.0619,0.2402,-0.0953,0.8473,0.2213,-0.0589,0.3316,0.4514,-
    ↪ 0.0915,0.1233,0.5826,-0.0741,0.2021,0.4986,0.8473,0.6064,0.6695
    ↪ ,0.8204,-0.0303,0.327,0.8561,0.8963,0.2991,-0.0101,0.4229,
    ↪ 0.5087,-0.0876,0.2358,-0.0988,0.8216,0.8743,0.1151,-0.0193,
    ↪ 0.7552,0.1129,0.8611,0.2058 & P2_{1}/c C_{2h}^{5} #14 (ae^13) &
    ↪  mP54 & None & C6H4K2N6O2PtS6 & C6H4K2N6O2PtS6 & J. Arpalahti
    ↪ and J. H\"{o}ls\"{a} and R. Sillanp\"{a}\"{a}, Acta Chem.
    ↪ Scand. 47, 1078-1082 (1993)
   1.00000000000000
  11.33000000000000    0.00000000000000    0.00000000000000
   0.00000000000000   11.14300000000000    0.00000000000000
  -0.60601856653863    0.00000000000000    7.27781220539596
    C       H       K       N       O      Pt       S
   12       8       4      12       4       2      12
Direct
   0.76480000000000    0.82510000000000   -0.06190000000000    C   (4e)
  -0.76480000000000    1.32510000000000    0.56190000000000    C   (4e)
  -0.76480000000000   -0.82510000000000    0.06190000000000    C   (4e)
   0.76480000000000   -0.32510000000000    0.43810000000000    C   (4e)
   0.24020000000000   -0.09530000000000    0.84730000000000    C   (4e)
  -0.24020000000000    0.40470000000000   -0.34730000000000    C   (4e)
  -0.24020000000000    0.09530000000000   -0.84730000000000    C   (4e)
   0.24020000000000    0.59530000000000    1.34730000000000    C   (4e)
   0.22130000000000   -0.05890000000000    0.33160000000000    C   (4e)
  -0.22130000000000    0.44110000000000    0.16840000000000    C   (4e)
  -0.22130000000000    0.05890000000000   -0.33160000000000    C   (4e)
   0.22130000000000    0.55890000000000    0.83160000000000    C   (4e)
   0.45140000000000   -0.09150000000000    0.12330000000000    H   (4e)
  -0.45140000000000    0.40850000000000    0.37670000000000    H   (4e)
  -0.45140000000000    0.09150000000000   -0.12330000000000    H   (4e)
   0.45140000000000    0.59150000000000    0.62330000000000    H   (4e)
   0.58260000000000   -0.07410000000000    0.20210000000000    H   (4e)
  -0.58260000000000    0.42590000000000    0.29790000000000    H   (4e)
  -0.58260000000000    0.07410000000000   -0.20210000000000    H   (4e)
   0.58260000000000    0.57410000000000    0.70210000000000    H   (4e)
   0.49860000000000    0.84730000000000    0.60640000000000    K   (4e)
  -0.49860000000000    1.34730000000000   -0.10640000000000    K   (4e)
  -0.49860000000000   -0.84730000000000   -0.60640000000000    K   (4e)
   0.49860000000000   -0.34730000000000    1.10640000000000    K   (4e)
   0.66950000000000    0.82040000000000   -0.03030000000000    N   (4e)
  -0.66950000000000    1.32040000000000    0.53030000000000    N   (4e)
  -0.66950000000000   -0.82040000000000    0.03030000000000    N   (4e)
   0.66950000000000   -0.32040000000000    0.46970000000000    N   (4e)
   0.32700000000000    0.85610000000000    0.89630000000000    N   (4e)
  -0.32700000000000    1.35610000000000   -0.39630000000000    N   (4e)
```

```
  -0.32700000000000   -0.85610000000000   -0.89630000000000    N    (4e)
   0.32700000000000   -0.35610000000000    1.39630000000000    N    (4e)
   0.29910000000000   -0.01010000000000    0.42290000000000    N    (4e)
  -0.29910000000000    0.48990000000000    0.07710000000000    N    (4e)
  -0.29910000000000    0.01010000000000   -0.42290000000000    N    (4e)
   0.29910000000000    0.51010000000000    0.92290000000000    N    (4e)
   0.50870000000000   -0.08760000000000    0.23580000000000    O    (4e)
  -0.50870000000000    0.41240000000000    0.26420000000000    O    (4e)
  -0.50870000000000    0.08760000000000   -0.23580000000000    O    (4e)
   0.50870000000000    0.58760000000000    0.73580000000000    O    (4e)
   0.00000000000000    0.00000000000000    0.00000000000000    Pt   (2a)
   0.00000000000000    0.50000000000000    0.50000000000000    Pt   (2a)
  -0.09880000000000    0.82160000000000    0.87430000000000    S    (4e)
   0.09880000000000    1.32160000000000   -0.37430000000000    S    (4e)
   0.09880000000000   -0.82160000000000   -0.87430000000000    S    (4e)
  -0.09880000000000   -0.32160000000000    1.37430000000000    S    (4e)
   0.11510000000000   -0.01930000000000    0.75520000000000    S    (4e)
  -0.11510000000000    0.48070000000000   -0.25520000000000    S    (4e)
  -0.11510000000000    0.01930000000000   -0.75520000000000    S    (4e)
   0.11510000000000    0.51930000000000    1.25520000000000    S    (4e)
   0.11290000000000    0.86110000000000    0.20580000000000    S    (4e)
  -0.11290000000000    1.36110000000000    0.29420000000000    S    (4e)
  -0.11290000000000   -0.86110000000000   -0.20580000000000    S    (4e)
   0.11290000000000   -0.36110000000000    0.70580000000000    S    (4e)
```

$K_2NbF_7$ ($K6_2$): A7B2C_mP40_14_7e_2e_e - CIF

```
# CIF file
data_findsym-output
_audit_creation_method FINDSYM

_chemical_name_mineral 'F7K2Nb'
_chemical_formula_sum 'F7 K2 Nb'

loop_
_publ_author_name
 'G. M. Brown'
 'L. A. Walker'
_journal_name_full_name
;
 Acta Cristallographica
;
_journal_volume 20
_journal_year 1966
_journal_page_first 220
_journal_page_last 229
_publ_Section_title
;
 Refinement of the structure of potassium heptafluoroniobate, K$_{2}
    ↪ $NbF$_{7}$, from neutron-diffraction data
;

# Found in Structure and luminescence of K$_{2}$TaF$_{7}$ and K$_{2}
    ↪ $NbF$_{7}$, 1987

_aflow_title 'K$_{2}$NbF$_{7}$ ($K6_{2}$) Structure'
_aflow_proto 'A7B2C_mP40_14_7e_2e_e'
_aflow_params 'a,b/a,c/a,\beta,x_{1},y_{1},z_{1},x_{2},y_{2},z_{2},x_{3
    ↪ },y_{3},z_{3},x_{4},y_{4},z_{4},x_{5},y_{5},z_{5},x_{6},y_{6},
    ↪ z_{6},x_{7},y_{7},z_{7},x_{8},y_{8},z_{8},x_{9},y_{9},z_{9},x_{
    ↪ 10},y_{10},z_{10}'
_aflow_params_values '5.846,2.17122819022,1.4565514882,90.0,0.0463,
    ↪ 0.2431,0.1994,0.4573,0.2567,0.2114,0.0448,0.1072,0.388,0.4599,
    ↪ 0.1107,0.4101,0.2166,-0.0207,0.2159,0.1875,0.1216,0.0063,0.5572
    ↪ ,0.0792,0.1317,0.2397,0.4404,0.1833,0.7626,0.2845,0.4446,0.2718
    ↪ ,0.1288,0.2229'
_aflow_Strukturbericht '$K6_{2}$'
_aflow_Pearson 'mP40'

_symmetry_space_group_name_H-M "P 1 21/c 1"
_symmetry_Int_Tables_number 14

_cell_length_a      5.84600
_cell_length_b      12.69300
_cell_length_c      8.51500
_cell_angle_alpha   90.00000
_cell_angle_beta    90.00000
_cell_angle_gamma   90.00000

loop_
_space_group_symop_id
_space_group_symop_operation_xyz
1 x,y,z
2 -x,y+1/2,-z+1/2
3 -x,-y,-z
4 x,-y+1/2,z+1/2

loop_
_atom_site_label
_atom_site_type_symbol
_atom_site_symmetry_multiplicity
_atom_site_Wyckoff_label
_atom_site_fract_x
_atom_site_fract_y
_atom_site_fract_z
_atom_site_occupancy
F1 F 4 e 0.04630 0.24310 0.19940 1.00000
F2 F 4 e 0.45730 0.25670 0.21140 1.00000
F3 F 4 e 0.04480 0.10720 0.38800 1.00000
F4 F 4 e 0.45990 0.11070 0.41010 1.00000
F5 F 4 e 0.21660 -0.02070 0.21590 1.00000
F6 F 4 e 0.18750 0.12160 0.00630 1.00000
F7 F 4 e 0.55720 0.07920 0.13170 1.00000
K1 K 4 e 0.23970 0.44040 0.18330 1.00000
K2 K 4 e 0.76260 0.28450 0.44460 1.00000
Nb1 Nb 4 e 0.27180 0.12880 0.22290 1.00000
```

$K_2NbF_7$ ($K6_2$): A7B2C_mP40_14_7e_2e_e - POSCAR

```
A7B2C_mP40_14_7e_2e_e & a,b/a,c/a,beta,x1,y1,z1,x2,y2,z2,x3,y3,z3,x4,y4,
    ↪ z4,x5,y5,z5,x6,y6,z6,x7,y7,z7,x8,y8,z8,x9,y9,z9,x10,y10,z10 --
    ↪ params=5.846,2.17122819022,1.4565514882,90.0,0.0463,0.2431,
    ↪ 0.1994,0.4573,0.2567,0.2114,0.0448,0.1072,0.388,0.4599,0.1107,
    ↪ 0.4101,0.2166,-0.0207,0.2159,0.1875,0.1216,0.0063,0.5572,0.0792
    ↪ ,0.1317,0.2397,0.4404,0.1833,0.7626,0.2845,0.4446,0.2718,0.1288
    ↪ ,0.2229 & P2_{1}/c C_{2h}^{5} #14 (e^10) & mP40 & $K6_{2}$ &
    ↪ F7K2Nb & F7K2Nb & G. M. Brown and L. A. Walker, Acta Cryst. 20,
    ↪ 220-229 (1966)
   1.00000000000000
   5.84600000000000    0.00000000000000    0.00000000000000
   0.00000000000000   12.69300000000000    0.00000000000000
   0.00000000000000    0.00000000000000    8.51500000000000
     F       K      Nb
    28       8       4
Direct
   0.04630000000000    0.24310000000000    0.19940000000000    F    (4e)
  -0.04630000000000    0.74310000000000    0.30060000000000    F    (4e)
  -0.04630000000000   -0.24310000000000   -0.19940000000000    F    (4e)
   0.04630000000000    0.25690000000000    0.69940000000000    F    (4e)
   0.45730000000000    0.25670000000000    0.21140000000000    F    (4e)
  -0.45730000000000    0.75670000000000    0.28860000000000    F    (4e)
  -0.45730000000000   -0.25670000000000   -0.21140000000000    F    (4e)
   0.45730000000000    0.24330000000000    0.71140000000000    F    (4e)
   0.04480000000000    0.10720000000000    0.38800000000000    F    (4e)
  -0.04480000000000    0.60720000000000    0.11200000000000    F    (4e)
  -0.04480000000000   -0.10720000000000   -0.38800000000000    F    (4e)
   0.04480000000000    0.39280000000000    0.88800000000000    F    (4e)
   0.45990000000000    0.11070000000000    0.41010000000000    F    (4e)
  -0.45990000000000    0.61070000000000    0.08990000000000    F    (4e)
  -0.45990000000000   -0.11070000000000   -0.41010000000000    F    (4e)
   0.45990000000000    0.38930000000000    0.91010000000000    F    (4e)
   0.21660000000000   -0.02070000000000    0.21590000000000    F    (4e)
  -0.21660000000000    0.47930000000000    0.28410000000000    F    (4e)
  -0.21660000000000    0.02070000000000   -0.21590000000000    F    (4e)
   0.21660000000000    0.52070000000000    0.71590000000000    F    (4e)
   0.18750000000000    0.12160000000000    0.00630000000000    F    (4e)
  -0.18750000000000    0.62160000000000    0.49370000000000    F    (4e)
  -0.18750000000000   -0.12160000000000   -0.00630000000000    F    (4e)
   0.18750000000000    0.37840000000000    0.50630000000000    F    (4e)
   0.55720000000000    0.07920000000000    0.13170000000000    F    (4e)
  -0.55720000000000    0.57920000000000    0.36830000000000    F    (4e)
  -0.55720000000000   -0.07920000000000   -0.13170000000000    F    (4e)
   0.55720000000000    0.42080000000000    0.63170000000000    F    (4e)
   0.23970000000000    0.44040000000000    0.18330000000000    K    (4e)
  -0.23970000000000    0.94040000000000    0.31670000000000    K    (4e)
  -0.23970000000000   -0.44040000000000   -0.18330000000000    K    (4e)
   0.23970000000000    0.05960000000000    0.68330000000000    K    (4e)
   0.76260000000000    0.28450000000000    0.44460000000000    K    (4e)
  -0.76260000000000    0.78450000000000    0.05540000000000    K    (4e)
  -0.76260000000000   -0.28450000000000   -0.44460000000000    K    (4e)
   0.76260000000000    0.21550000000000    0.94460000000000    K    (4e)
   0.27180000000000    0.12880000000000    0.22290000000000    Nb   (4e)
  -0.27180000000000    0.62880000000000    0.27710000000000    Nb   (4e)
  -0.27180000000000   -0.12880000000000   -0.22290000000000    Nb   (4e)
   0.27180000000000    0.37120000000000    0.72290000000000    Nb   (4e)
```

Manganese-leonite 110 K [$K_2Mn(SO_4)_2{\cdot}4H_2O$]: A8B2CD12E2_mP100_14_8e_2e_ad_12e_2e - CIF

```
# CIF file
data_findsym-output
_audit_creation_method FINDSYM

_chemical_name_mineral 'Manganese-leonite'
_chemical_formula_sum 'H8 K2 Mn O12 S2'

loop_
_publ_author_name
 'B. Hertweck'
 'G. Giester'
 'E. Libowitzky'
_journal_name_full_name
;
 American Mineralogist
;
_journal_volume 86
_journal_year 2001
_journal_page_first 1282
_journal_page_last 1292
_publ_Section_title
;
 The crystal structures of the low-temperature phases of leonite-type
    ↪ compounds, K$_{2}$$Me$(SO$_{4}$)$_{2}$$\cdot$4H$_{2}$O ($Me^{2
    ↪ +}$ = Mg, Mn, Fe)
;

_aflow_title 'Manganese-leonite 110~K [K$_{2}$Mn(SO$_{4}$)$_{2}$$\
    ↪ cdot$4H$_{2}$O] Structure'
_aflow_proto 'A8B2CD12E2_mP100_14_8e_2e_ad_12e_2e'
_aflow_params 'a,b/a,c/a,\beta,x_{3},y_{3},z_{3},x_{4},y_{4},z_{4},x_{5
    ↪ },y_{5},z_{5},x_{6},y_{6},z_{6},x_{7},y_{7},z_{7},x_{8},y_{8},
    ↪ z_{8},x_{9},y_{9},z_{9},x_{10},y_{10},z_{10},x_{11},y_{11},z_{
    ↪ 11},x_{12},y_{12},z_{12},x_{13},y_{13},z_{13},x_{14},y_{14},z_{
    ↪ 14},x_{15},y_{15},z_{15},x_{16},y_{16},z_{16},x_{17},y_{17},z_{
    ↪ 17},x_{18},y_{18},z_{18},x_{19},y_{19},z_{19},x_{20},y_{20},z_{
    ↪ 20},x_{21},y_{21},z_{21},x_{22},y_{22},z_{22},x_{23},y_{23},z_{
    ↪ 23},x_{24},y_{24},z_{24},x_{25},y_{25},z_{25},x_{26},y_{26},z_{
    ↪ 26}'
_aflow_params_values '9.902,0.962532821652,1.21500706928,95.02,0.387,
    ↪ 0.729,-0.07,0.275,0.649,-0.074,0.62,0.735,0.074,0.721,0.648,
    ↪ 0.081,0.025,0.72,-0.049,-0.037,0.726,0.043,0.733,-0.008,0.06,
```

```
  ↪ 0.729,0.014,−0.044,0.29373,0.73232,0.16965,0.25852,0.25782,
  ↪ 0.16992,0.1163,0.6276,0.8466,0.8888,0.6261,0.1633,0.5998,−
  ↪ 0.0908,0.1493,0.5455,0.1528,0.1791,0.0611,−0.0127,0.1725,0.0973
  ↪ ,0.4938,0.1853,0.4342,0.4911,0.1674,0.3741,−0.0227,0.1931,
  ↪ 0.3537,0.6684,−0.0342,0.6484,0.6653,0.0431,−0.0131,0.7721,−
  ↪ 0.0051,0.7802,0.0103,0.0111,−0.04519,0.49523,0.20732,0.51973,
  ↪ 0.00747,0.21252'
_aflow_Strukturbericht 'None'
_aflow_Pearson 'mP100'

_symmetry_space_group_name_H−M "P 1 21/c 1"
_symmetry_Int_Tables_number 14

_cell_length_a     9.90200
_cell_length_b     9.53100
_cell_length_c     12.03100
_cell_angle_alpha  90.00000
_cell_angle_beta   95.02000
_cell_angle_gamma  90.00000

loop_
_space_group_symop_id
_space_group_symop_operation_xyz
1 x,y,z
2 −x,y+1/2,−z+1/2
3 −x,−y,−z
4 x,−y+1/2,z+1/2

loop_
_atom_site_label
_atom_site_type_symbol
_atom_site_symmetry_multiplicity
_atom_site_Wyckoff_label
_atom_site_fract_x
_atom_site_fract_y
_atom_site_fract_z
_atom_site_occupancy
Mn1 Mn 2 a 0.00000 0.00000 0.00000 1.00000
Mn2 Mn 2 d 0.50000 0.00000 0.50000 1.00000
H1 H 4 e 0.38700 0.72900 −0.07000 1.00000
H2 H 4 e 0.27500 0.64900 −0.07400 1.00000
H3 H 4 e 0.62000 0.73500 0.07400 1.00000
H4 H 4 e 0.72100 0.64800 0.08100 1.00000
H5 H 4 e 0.02500 0.72000 −0.04900 1.00000
H6 H 4 e −0.03700 0.72600 0.04300 1.00000
H7 H 4 e 0.73300 −0.00800 0.06000 1.00000
H8 H 4 e 0.72900 0.01400 −0.04400 1.00000
K1 K 4 e 0.29373 0.73232 0.16965 1.00000
K2 K 4 e 0.25852 0.25782 0.16992 1.00000
O1 O 4 e 0.11630 0.62760 0.84660 1.00000
O2 O 4 e 0.88880 0.62610 0.16330 1.00000
O3 O 4 e 0.59980 −0.09080 0.14930 1.00000
O4 O 4 e 0.54550 0.15280 0.17910 1.00000
O5 O 4 e 0.06110 −0.01270 0.17250 1.00000
O6 O 4 e 0.09730 0.49380 0.18530 1.00000
O7 O 4 e 0.43420 0.49110 0.16740 1.00000
O8 O 4 e 0.37410 −0.02270 0.19310 1.00000
O9 O 4 e 0.35370 0.66840 −0.03420 1.00000
O10 O 4 e 0.64840 0.66530 0.04310 1.00000
O11 O 4 e −0.01310 0.77210 −0.00510 1.00000
O12 O 4 e 0.78020 0.01030 0.01110 1.00000
S1 S 4 e −0.04519 0.49523 0.20732 1.00000
S2 S 4 e 0.51973 0.00747 0.21252 1.00000
```

Manganese-leonite 110 K [$K_2Mn(SO_4)_2{\cdot}4H_2O$]: A8B2CD12E2_mP100_14_8e_2e_ad_12e_2e - POSCAR

```
A8B2CD12E2_mP100_14_8e_2e_ad_12e_2e & a,b/a,c/a,beta,x3,y3,z3,x4,y4,z4,
  ↪ x5,y5,z5,x6,y6,z6,x7,y7,z7,x8,y8,z8,x9,y9,z9,x10,y10,z10,x11,
  ↪ y11,z11,x12,y12,z12,x13,y13,z13,x14,y14,z14,x15,y15,z15,x16,y16
  ↪ ,z16,x17,y17,z17,x18,y18,z18,x19,y19,z19,x20,y20,z20,x21,y21,
  ↪ z21,x22,y22,z22,x23,y23,z23,x24,y24,z24,x25,y25,z25,x26,y26,z26
  ↪  −−params=9.902,0.962532821652,1.21500706928,95.02,0.387,0.729
  ↪ ,−0.07,0.275,0.649,−0.074,0.62,0.735,0.074,0.721,0.648,0.081,
  ↪ 0.025,0.72,−0.049,−0.037,0.726,0.043,0.733,−0.008,0.06,0.729,
  ↪ 0.014,−0.044,0.29373,0.73232,0.16965,0.25852,0.25782,0.16992,
  ↪ 0.1163,0.6276,0.8466,0.8888,0.6261,0.1633,0.5998,−0.0908,0.1493
  ↪ ,0.5455,0.1528,0.1791,0.0611,−0.0127,0.1725,0.0973,0.4938,
  ↪ 0.1853,0.4342,0.4911,0.1674,0.3741,−0.0227,0.1931,0.3537,0.6684
  ↪ ,−0.0342,0.6484,0.6653,0.0431,−0.0131,0.7721,−0.0051,0.7802,
  ↪ 0.0103,0.0111,−0.04519,0.49523,0.20732,0.51973,0.00747,0.21252
  ↪ & P2_{1}/c C_{2h}^{5} #14 (ade^24) & mP100 & None & H8K2MnO12S2
  ↪  & Manganese−leonite & B. Hertweck and G. Giester and E.
  ↪ Libowitzky, Am. Mineral. 86, 1282−1292 (2001)
   1.00000000000000
   9.90200000000000   0.00000000000000   0.00000000000000
   0.00000000000000   9.53100000000000   0.00000000000000
  −1.05275430748699   0.00000000000000  11.98485166233060
      H     K    Mn     O     S
     32     8     4    48     8
Direct
   0.38700000000000   0.72900000000000  −0.07000000000000   H   (4e)
  −0.38700000000000   1.22900000000000   0.57000000000000   H   (4e)
  −0.38700000000000  −0.72900000000000   0.07000000000000   H   (4e)
   0.38700000000000  −0.22900000000000   0.43000000000000   H   (4e)
   0.27500000000000   0.64900000000000  −0.07400000000000   H   (4e)
  −0.27500000000000   1.14900000000000   0.57400000000000   H   (4e)
  −0.27500000000000  −0.64900000000000   0.07400000000000   H   (4e)
   0.27500000000000  −0.14900000000000   0.42600000000000   H   (4e)
   0.62000000000000   0.73500000000000   0.07400000000000   H   (4e)
  −0.62000000000000   1.23500000000000   0.42600000000000   H   (4e)
  −0.62000000000000  −0.73500000000000  −0.07400000000000   H   (4e)
   0.62000000000000  −0.23500000000000   0.57400000000000   H   (4e)
   0.72100000000000   0.64800000000000   0.08100000000000   H   (4e)
  −0.72100000000000   1.14800000000000   0.41900000000000   H   (4e)
  −0.72100000000000  −0.64800000000000  −0.08100000000000   H   (4e)
   0.72100000000000  −0.14800000000000   0.58100000000000   H   (4e)
   0.02500000000000   0.72000000000000  −0.04900000000000   H   (4e)
  −0.02500000000000   1.22000000000000   0.54900000000000   H   (4e)
  −0.02500000000000  −0.72000000000000   0.04900000000000   H   (4e)
   0.02500000000000  −0.22000000000000   0.45100000000000   H   (4e)
  −0.03700000000000   0.72600000000000   0.04300000000000   H   (4e)
   0.03700000000000   1.22600000000000   0.45700000000000   H   (4e)
   0.03700000000000  −0.72600000000000  −0.04300000000000   H   (4e)
  −0.03700000000000  −0.22600000000000   0.54300000000000   H   (4e)
   0.73300000000000  −0.00800000000000   0.06000000000000   H   (4e)
  −0.73300000000000   0.49200000000000   0.44000000000000   H   (4e)
  −0.73300000000000   0.00800000000000  −0.06000000000000   H   (4e)
   0.73300000000000   0.50800000000000   0.56000000000000   H   (4e)
   0.72900000000000   0.01400000000000  −0.04400000000000   H   (4e)
  −0.72900000000000   0.51400000000000   0.54400000000000   H   (4e)
  −0.72900000000000  −0.01400000000000   0.04400000000000   H   (4e)
   0.72900000000000   0.48600000000000   0.45600000000000   H   (4e)
   0.29373000000000   0.73232000000000   0.16965000000000   K   (4e)
  −0.29373000000000   1.23232000000000   0.33035000000000   K   (4e)
  −0.29373000000000  −0.73232000000000  −0.16965000000000   K   (4e)
   0.29373000000000  −0.23232000000000   0.66965000000000   K   (4e)
   0.25852000000000   0.25782000000000   0.16992000000000   K   (4e)
  −0.25852000000000   0.75782000000000   0.33008000000000   K   (4e)
  −0.25852000000000  −0.25782000000000  −0.16992000000000   K   (4e)
   0.25852000000000   0.24218000000000   0.66992000000000   K   (4e)
   0.00000000000000   0.00000000000000   0.00000000000000   Mn  (2a)
   0.00000000000000   0.50000000000000   0.50000000000000   Mn  (2a)
   0.50000000000000   0.00000000000000   0.50000000000000   Mn  (2d)
   0.50000000000000   0.50000000000000   0.00000000000000   Mn  (2d)
   0.11630000000000   0.62760000000000   0.84660000000000   O   (4e)
  −0.11630000000000   1.12760000000000  −0.34660000000000   O   (4e)
  −0.11630000000000  −0.62760000000000  −0.84660000000000   O   (4e)
   0.11630000000000  −0.12760000000000   1.34660000000000   O   (4e)
   0.88880000000000   0.62610000000000   0.16330000000000   O   (4e)
  −0.88880000000000   1.12610000000000   0.33670000000000   O   (4e)
  −0.88880000000000  −0.62610000000000  −0.16330000000000   O   (4e)
   0.88880000000000  −0.12610000000000   0.66330000000000   O   (4e)
   0.59980000000000  −0.09080000000000   0.14930000000000   O   (4e)
  −0.59980000000000   0.40920000000000   0.35070000000000   O   (4e)
  −0.59980000000000   0.09080000000000  −0.14930000000000   O   (4e)
   0.59980000000000   0.59080000000000   0.64930000000000   O   (4e)
   0.54550000000000   0.15280000000000   0.17910000000000   O   (4e)
  −0.54550000000000   0.65280000000000   0.32090000000000   O   (4e)
  −0.54550000000000  −0.15280000000000  −0.17910000000000   O   (4e)
   0.54550000000000   0.34720000000000   0.67910000000000   O   (4e)
   0.06110000000000  −0.01270000000000   0.17250000000000   O   (4e)
  −0.06110000000000   0.48730000000000   0.32750000000000   O   (4e)
  −0.06110000000000   0.01270000000000  −0.17250000000000   O   (4e)
   0.06110000000000   0.51270000000000   0.67250000000000   O   (4e)
   0.09730000000000   0.49380000000000   0.18530000000000   O   (4e)
  −0.09730000000000   0.99380000000000   0.31470000000000   O   (4e)
  −0.09730000000000  −0.49380000000000  −0.18530000000000   O   (4e)
   0.09730000000000   0.00620000000000   0.68530000000000   O   (4e)
   0.43420000000000   0.49110000000000   0.16740000000000   O   (4e)
  −0.43420000000000   0.99110000000000   0.33260000000000   O   (4e)
  −0.43420000000000  −0.49110000000000  −0.16740000000000   O   (4e)
   0.43420000000000   0.00890000000000   0.66740000000000   O   (4e)
   0.37410000000000  −0.02270000000000   0.19310000000000   O   (4e)
  −0.37410000000000   0.47730000000000   0.30690000000000   O   (4e)
  −0.37410000000000   0.02270000000000  −0.19310000000000   O   (4e)
   0.37410000000000   0.52270000000000   0.69310000000000   O   (4e)
   0.35370000000000   0.66840000000000  −0.03420000000000   O   (4e)
  −0.35370000000000   1.16840000000000   0.53420000000000   O   (4e)
  −0.35370000000000  −0.66840000000000   0.03420000000000   O   (4e)
   0.35370000000000  −0.16840000000000   0.46580000000000   O   (4e)
   0.64840000000000   0.66530000000000   0.04310000000000   O   (4e)
  −0.64840000000000   1.16530000000000   0.45690000000000   O   (4e)
  −0.64840000000000  −0.66530000000000  −0.04310000000000   O   (4e)
   0.64840000000000  −0.16530000000000   0.54310000000000   O   (4e)
  −0.01310000000000   0.77210000000000  −0.00510000000000   O   (4e)
   0.01310000000000   1.27210000000000   0.50510000000000   O   (4e)
   0.01310000000000  −0.77210000000000   0.00510000000000   O   (4e)
  −0.01310000000000  −0.27210000000000   0.49490000000000   O   (4e)
   0.78020000000000   0.01030000000000   0.01110000000000   O   (4e)
  −0.78020000000000   0.51030000000000   0.48890000000000   O   (4e)
  −0.78020000000000  −0.01030000000000  −0.01110000000000   O   (4e)
   0.78020000000000   0.48970000000000   0.51110000000000   O   (4e)
  −0.04519000000000   0.49523000000000   0.20732000000000   S   (4e)
   0.04519000000000   0.99523000000000   0.29268000000000   S   (4e)
   0.04519000000000  −0.49523000000000  −0.20732000000000   S   (4e)
  −0.04519000000000   0.00477000000000   0.70732000000000   S   (4e)
   0.51973000000000   0.00747000000000   0.21252000000000   S   (4e)
  −0.51973000000000   0.50747000000000   0.28748000000000   S   (4e)
  −0.51973000000000  −0.00747000000000  −0.21252000000000   S   (4e)
   0.51973000000000   0.49253000000000   0.71252000000000   S   (4e)
```

$Co_2Al_9$ ($D8_d$): A9B2_mP22_14_a4e_e - CIF

```
# CIF file
data_findsym−output
_audit_creation_method FINDSYM

_chemical_name_mineral 'Co2Al9'
_chemical_formula_sum 'Al9 Co2'

loop_
_publ_author_name
 'A. M. B. Douglas'
_journal_name_full_name
;
 Acta Crystallographica Section B: Structural Science
;
_journal_volume 3
_journal_year 1950
_journal_page_first 19
```

```
_journal_page_last 24
_publ_Section_title
;
 The Structure of Co$_{2}$Al$_{9}$
;

_aflow_title 'Co$_{2}$Al$_{9}$ ($D8_{d}$) Structure'
_aflow_proto 'A9B2_mP22_14_a4e_e'
_aflow_params 'a,b/a,c/a,\beta,x_{2},y_{2},z_{2},x_{3},y_{3},z_{3},x_{4}
    ↪ },y_{4},z_{4},x_{5},y_{5},z_{5},x_{6},y_{6},z_{6}'
_aflow_params_values '6.213,1.01239336874,1.37719298246,94.76,0.4044,
    ↪ 0.5381,0.2682,0.0889,0.2101,0.293,0.3891,0.3069,-0.0014,0.2159,
    ↪ 0.8852,0.0417,0.2646,0.8851,0.3335'
_aflow_Strukturbericht '$D8_{d}$'
_aflow_Pearson 'mP22'

_symmetry_space_group_name_H-M "P 1 21/c 1"
_symmetry_Int_Tables_number 14

_cell_length_a     6.21300
_cell_length_b     6.29000
_cell_length_c     8.55650
_cell_angle_alpha  90.00000
_cell_angle_beta   94.76000
_cell_angle_gamma  90.00000

loop_
_space_group_symop_id
_space_group_symop_operation_xyz
1 x,y,z
2 -x,y+1/2,-z+1/2
3 -x,-y,-z
4 x,-y+1/2,z+1/2

loop_
_atom_site_label
_atom_site_type_symbol
_atom_site_symmetry_multiplicity
_atom_site_Wyckoff_label
_atom_site_fract_x
_atom_site_fract_y
_atom_site_fract_z
_atom_site_occupancy
Al1 Al 2 a 0.00000 0.00000 0.00000 1.00000
Al2 Al 4 e 0.40440 0.53810 0.26820 1.00000
Al3 Al 4 e 0.08890 0.21010 0.29300 1.00000
Al4 Al 4 e 0.38910 0.30690 -0.00140 1.00000
Al5 Al 4 e 0.21590 0.88520 0.04170 1.00000
Co1 Co 4 e 0.26460 0.88510 0.33350 1.00000
```

$Co_2Al_9$ ($D8_d$): A9B2_mP22_14_a4e_e - POSCAR

```
A9B2_mP22_14_a4e_e & a,b/a,c/a,beta,x2,y2,z2,x3,y3,z3,x4,y4,z4,x5,y5,z5,
    ↪ x6,y6,z6 --params=6.213,1.01239336874,1.37719298246,94.76,
    ↪ 0.4044,0.5381,0.2682,0.0889,0.2101,0.293,0.3891,0.3069,-0.0014,
    ↪ 0.2159,0.8852,0.0417,0.2646,0.8851,0.3335 & P2_{1}/c C_{2h}^{5}
    ↪ #14 (ae^5) & mP22 & $D8_{d}$ & Co2Al9 & Co2Al9 & A. M. B.
    ↪ Douglas, Acta Crystallogr. Sect. B Struct. Sci. 3, 19-24 (1950)
   1.00000000000000
   6.21300000000000   0.00000000000000   0.00000000000000
   0.00000000000000   6.29000000000000   0.00000000000000
  -0.71003667870571   0.00000000000000   8.52698892721766
    Al       Co
    18        4
Direct
   0.00000000000000   0.00000000000000   0.00000000000000   Al   (2a)
   0.00000000000000   0.50000000000000   0.50000000000000   Al   (2a)
   0.40440000000000   0.53810000000000   0.26820000000000   Al   (4e)
  -0.40440000000000   1.03810000000000   0.23180000000000   Al   (4e)
  -0.40440000000000  -0.53810000000000  -0.26820000000000   Al   (4e)
   0.40440000000000  -0.03810000000000   0.76820000000000   Al   (4e)
   0.08890000000000   0.21010000000000   0.29300000000000   Al   (4e)
  -0.08890000000000   0.71010000000000   0.20700000000000   Al   (4e)
  -0.08890000000000  -0.21010000000000  -0.29300000000000   Al   (4e)
   0.08890000000000   0.28990000000000   0.79300000000000   Al   (4e)
   0.38910000000000   0.30690000000000  -0.00140000000000   Al   (4e)
  -0.38910000000000   0.80690000000000   0.50140000000000   Al   (4e)
  -0.38910000000000  -0.30690000000000   0.00140000000000   Al   (4e)
   0.38910000000000   0.19310000000000   0.49860000000000   Al   (4e)
   0.21590000000000   0.88520000000000   0.04170000000000   Al   (4e)
  -0.21590000000000   1.38520000000000   0.45830000000000   Al   (4e)
  -0.21590000000000  -0.88520000000000  -0.04170000000000   Al   (4e)
   0.21590000000000  -0.38520000000000   0.54170000000000   Al   (4e)
   0.26460000000000   0.88510000000000   0.33350000000000   Co   (4e)
  -0.26460000000000   1.38510000000000   0.16650000000000   Co   (4e)
  -0.26460000000000  -0.88510000000000  -0.33350000000000   Co   (4e)
   0.26460000000000  -0.38510000000000   0.83350000000000   Co   (4e)
```

Tutton salt [Cu(NH$_4$)$_2$(SO$_4$)$_2$·6H$_2$O, $H4_4$]: AB20C2D14E2_mP78_14_a_10e_e_7e_e - CIF

```
# CIF file
data_findsym-output
_audit_creation_method FINDSYM

_chemical_name_mineral 'Tutton salt'
_chemical_formula_sum 'Cu H20 N2 O14 S2'

loop_
_publ_author_name
 'G. M. Brown'
 'R. Chidambaram'
_journal_name_full_name
;
 Acta Crystallographica Section B: Structural Science
;
_journal_volume 25
_journal_year 1969
_journal_page_first 676
_journal_page_last 687
_publ_Section_title
;
 The structure of copper ammonium sulfate hexahydrate from neutron
    ↪ diffraction data
;

_aflow_title 'Tutton salt [Cu(NH$_{4}$)$_{2}$(SO$_{4}$)$_{2}$$\cdot$6H$_
    ↪ {2}$O, $H4_{4}$] Structure'
_aflow_proto 'AB20C2D14E2_mP78_14_a_10e_e_7e_e'
_aflow_params 'a,b/a,c/a,\beta,x_{2},y_{2},z_{2},x_{3},y_{3},z_{3},x_{4
    ↪ },y_{4},z_{4},x_{5},y_{5},z_{5},x_{6},y_{6},z_{6},x_{7},y_{7},
    ↪ z_{7},x_{8},y_{8},z_{8},x_{9},y_{9},z_{9},x_{10},y_{10},z_{10},
    ↪ x_{11},y_{11},z_{11},x_{12},y_{12},z_{12},x_{13},y_{13},z_{13},
    ↪ x_{14},y_{14},z_{14},x_{15},y_{15},z_{15},x_{16},y_{16},z_{16},
    ↪ x_{17},y_{17},z_{17},x_{18},y_{18},z_{18},x_{19},y_{19},z_{19},
    ↪ x_{20},y_{20},z_{20}'
_aflow_params_values '6.3016,1.96450107909,1.4616129237,106.112,0.20966,
    ↪ 0.66679,0.06554,0.40162,0.70084,0.22515,0.46879,0.66573,0.07617
    ↪ ,0.36184,0.57533,0.16861,0.32899,-0.0962,0.22103,0.1068,0.87132
    ↪ ,0.25555,-0.0597,-0.09535,0.73282,0.00217,0.81595,0.85798,
    ↪ 0.31563,0.05877,0.89729,0.3095,0.14082,0.02457,0.35953,0.65214,
    ↪ 0.13458,0.60057,0.76827,0.41576,0.78118,-0.0758,0.54888,0.63349
    ↪ ,-0.07032,0.28057,-0.04314,0.82138,0.39147,0.17726,0.88327,
    ↪ 0.17658,0.03043,0.89108,0.83608,0.2821,0.06534,-0.00531,0.74517
    ↪ ,0.86102,0.41069'
_aflow_Strukturbericht '$H4_{4}$'
_aflow_Pearson 'mP78'

_symmetry_space_group_name_H-M "P 1 21/c 1"
_symmetry_Int_Tables_number 14

_cell_length_a     6.30160
_cell_length_b     12.37950
_cell_length_c     9.21050
_cell_angle_alpha  90.00000
_cell_angle_beta   106.11200
_cell_angle_gamma  90.00000

loop_
_space_group_symop_id
_space_group_symop_operation_xyz
1 x,y,z
2 -x,y+1/2,-z+1/2
3 -x,-y,-z
4 x,-y+1/2,z+1/2

loop_
_atom_site_label
_atom_site_type_symbol
_atom_site_symmetry_multiplicity
_atom_site_Wyckoff_label
_atom_site_fract_x
_atom_site_fract_y
_atom_site_fract_z
_atom_site_occupancy
Cu1 Cu 2 a 0.00000 0.00000 0.00000 1.00000
H1 H 4 e 0.20966 0.66679 0.06554 1.00000
H2 H 4 e 0.40162 0.70084 0.22515 1.00000
H3 H 4 e 0.46879 0.66573 0.07617 1.00000
H4 H 4 e 0.36184 0.57533 0.16861 1.00000
H5 H 4 e 0.32899 -0.09620 0.22103 1.00000
H6 H 4 e 0.10680 0.87132 0.25555 1.00000
H7 H 4 e -0.05970 -0.09535 0.73282 1.00000
H8 H 4 e 0.00217 0.81595 0.85798 1.00000
H9 H 4 e 0.31563 0.05877 0.89729 1.00000
H10 H 4 e 0.30950 0.14082 0.02457 1.00000
N1 N 4 e 0.35953 0.65214 0.13458 1.00000
O1 O 4 e 0.60057 0.76827 0.41576 1.00000
O2 O 4 e 0.78118 -0.07580 0.54888 1.00000
O3 O 4 e 0.63349 -0.07032 0.28057 1.00000
O4 O 4 e -0.04314 0.82138 0.39147 1.00000
O5 O 4 e 0.17726 0.88327 0.17658 1.00000
O6 O 4 e 0.03043 0.89108 0.83608 1.00000
O7 O 4 e 0.28210 0.06534 -0.00531 1.00000
S1 S 4 e 0.74517 0.86102 0.41069 1.00000
```

Tutton salt [Cu(NH$_4$)$_2$(SO$_4$)$_2$·6H$_2$O, $H4_4$]: AB20C2D14E2_mP78_14_a_10e_e_7e_e - POSCAR

```
AB20C2D14E2_mP78_14_a_10e_e_7e_e & a,b/a,c/a,beta,x2,y2,z2,x3,y3,z3,x4,
    ↪ y4,z4,x5,y5,z5,x6,y6,z6,x7,y7,z7,x8,y8,z8,x9,y9,z9,x10,y10,z10,
    ↪ x11,y11,z11,x12,y12,z12,x13,y13,z13,x14,y14,z14,x15,y15,z15,x16
    ↪ ,y16,z16,x17,y17,z17,x18,y18,z18,x19,y19,z19,x20,y20,z20 --
    ↪ params=6.3016,1.96450107909,1.4616129237,106.112,0.20966,
    ↪ 0.66679,0.06554,0.40162,0.70084,0.22515,0.46879,0.66573,0.07617
    ↪ ,0.36184,0.57533,0.16861,0.32899,-0.0962,0.22103,0.1068,0.87132
    ↪ ,0.25555,-0.0597,-0.09535,0.73282,0.00217,0.81595,0.85798,
    ↪ 0.31563,0.05877,0.89729,0.3095,0.14082,0.02457,0.35953,0.65214,
    ↪ 0.13458,0.60057,0.76827,0.41576,0.78118,-0.0758,0.54888,0.63349
    ↪ ,-0.07032,0.28057,-0.04314,0.82138,0.39147,0.17726,0.88327,
    ↪ 0.17658,0.03043,0.89108,0.83608,0.2821,0.06534,-0.00531,0.74517
    ↪ ,0.86102,0.41069 & P2_{1}/c C_{2h}^{5} #14 (ae^19) & mP78 &
    ↪ $H4_{4}$ & CuH20N2O14S2 & Tutton salt & G. M. Brown and R.
    ↪ Chidambaram, Acta Crystallogr. Sect. B Struct. Sci. 25, 676-687
    ↪ (1969)
   1.00000000000000
   6.30160000000000   0.00000000000000   0.00000000000000
   0.00000000000000  12.37950000000000   0.00000000000000
  -2.55605994213425   0.00000000000000   8.84872125350419
    Cu     H      N      O      S
     2    40      4     28      4
Direct
```

```
  0.00000000000000   0.00000000000000   0.00000000000000   Cu   (2a)
  0.00000000000000   0.50000000000000   0.50000000000000   Cu   (2a)
  0.20966000000000   0.66679000000000   0.06554000000000   H   (4e)
 -0.20966000000000   1.16679000000000   0.43446000000000   H   (4e)
 -0.20966000000000  -0.66679000000000  -0.06554000000000   H   (4e)
  0.20966000000000  -0.16679000000000   0.56554000000000   H   (4e)
  0.40162000000000   0.70084000000000   0.22515000000000   H   (4e)
 -0.40162000000000   1.20084000000000   0.27485000000000   H   (4e)
 -0.40162000000000  -0.70084000000000  -0.22515000000000   H   (4e)
  0.40162000000000  -0.20084000000000   0.72515000000000   H   (4e)
  0.46879000000000   0.66573000000000   0.07617000000000   H   (4e)
 -0.46879000000000   1.16573000000000   0.42383000000000   H   (4e)
 -0.46879000000000  -0.66573000000000  -0.07617000000000   H   (4e)
  0.46879000000000  -0.16573000000000   0.57617000000000   H   (4e)
  0.36184000000000   0.57533000000000   0.16861000000000   H   (4e)
 -0.36184000000000   1.07533000000000   0.33139000000000   H   (4e)
 -0.36184000000000  -0.57533000000000  -0.16861000000000   H   (4e)
  0.36184000000000  -0.07533000000000   0.66861000000000   H   (4e)
  0.32899000000000  -0.09620000000000   0.22103000000000   H   (4e)
 -0.32899000000000   0.40380000000000   0.27897000000000   H   (4e)
 -0.32899000000000   0.09620000000000  -0.22103000000000   H   (4e)
  0.32899000000000   0.59620000000000   0.72103000000000   H   (4e)
  0.10680000000000   0.87132000000000   0.25555000000000   H   (4e)
 -0.10680000000000   1.37132000000000   0.24445000000000   H   (4e)
 -0.10680000000000  -0.87132000000000  -0.25555000000000   H   (4e)
  0.10680000000000  -0.37132000000000   0.75555000000000   H   (4e)
 -0.05970000000000  -0.09535000000000   0.73282000000000   H   (4e)
  0.05970000000000   0.40465000000000  -0.23282000000000   H   (4e)
  0.05970000000000   0.09535000000000  -0.73282000000000   H   (4e)
 -0.05970000000000   0.59535000000000   1.23282000000000   H   (4e)
  0.00217000000000   0.81595000000000   0.85798000000000   H   (4e)
 -0.00217000000000   1.31595000000000  -0.35798000000000   H   (4e)
 -0.00217000000000  -0.81595000000000  -0.85798000000000   H   (4e)
  0.00217000000000  -0.31595000000000   1.35798000000000   H   (4e)
  0.31563000000000   0.05877000000000   0.89729000000000   H   (4e)
 -0.31563000000000   0.55877000000000  -0.39729000000000   H   (4e)
 -0.31563000000000  -0.05877000000000  -0.89729000000000   H   (4e)
  0.31563000000000   0.44123000000000   1.39729000000000   H   (4e)
  0.30950000000000   0.14082000000000   0.02457000000000   H   (4e)
 -0.30950000000000   0.64082000000000   0.47543000000000   H   (4e)
 -0.30950000000000  -0.14082000000000  -0.02457000000000   H   (4e)
  0.30950000000000   0.35918000000000   0.52457000000000   H   (4e)
  0.35953000000000   0.65214000000000   0.13458000000000   N   (4e)
 -0.35953000000000   1.15214000000000   0.36542000000000   N   (4e)
 -0.35953000000000  -0.65214000000000  -0.13458000000000   N   (4e)
  0.35953000000000  -0.15214000000000   0.63458000000000   N   (4e)
  0.60057000000000   0.76827000000000   0.41576000000000   O   (4e)
 -0.60057000000000   1.26827000000000   0.08424000000000   O   (4e)
 -0.60057000000000  -0.76827000000000  -0.41576000000000   O   (4e)
  0.60057000000000  -0.26827000000000   0.91576000000000   O   (4e)
  0.78118000000000  -0.07580000000000   0.54888000000000   O   (4e)
 -0.78118000000000   0.42420000000000  -0.04888000000000   O   (4e)
 -0.78118000000000   0.07580000000000  -0.54888000000000   O   (4e)
  0.78118000000000   0.57580000000000   1.04888000000000   O   (4e)
  0.63349000000000  -0.07032000000000   0.28057000000000   O   (4e)
 -0.63349000000000   0.42968000000000   0.21943000000000   O   (4e)
 -0.63349000000000   0.07032000000000  -0.28057000000000   O   (4e)
  0.63349000000000   0.57032000000000   0.78057000000000   O   (4e)
 -0.04314000000000   0.82138000000000   0.39147000000000   O   (4e)
  0.04314000000000   1.32138000000000   0.10853000000000   O   (4e)
  0.04314000000000  -0.82138000000000  -0.39147000000000   O   (4e)
 -0.04314000000000  -0.32138000000000   0.89147000000000   O   (4e)
  0.17726000000000   0.88327000000000   0.17658000000000   O   (4e)
 -0.17726000000000   1.38327000000000   0.32342000000000   O   (4e)
 -0.17726000000000  -0.88327000000000  -0.17658000000000   O   (4e)
  0.17726000000000  -0.38327000000000   0.67658000000000   O   (4e)
  0.03043000000000   0.89108000000000   0.83608000000000   O   (4e)
 -0.03043000000000   1.39108000000000  -0.33608000000000   O   (4e)
 -0.03043000000000  -0.89108000000000  -0.83608000000000   O   (4e)
  0.03043000000000  -0.39108000000000   1.33608000000000   O   (4e)
  0.28210000000000   0.06534000000000  -0.00531000000000   O   (4e)
 -0.28210000000000   0.56534000000000   0.50531000000000   O   (4e)
 -0.28210000000000  -0.06534000000000   0.00531000000000   O   (4e)
  0.28210000000000   0.43466000000000   0.49469000000000   O   (4e)
  0.74517000000000   0.86102000000000   0.41069000000000   S   (4e)
 -0.74517000000000   1.36102000000000   0.08931000000000   S   (4e)
 -0.74517000000000  -0.86102000000000  -0.41069000000000   S   (4e)
  0.74517000000000  -0.36102000000000   0.91069000000000   S   (4e)
```

Parawollastonite ($CaSiO_3$, $S3_3$ (II)): AB3C_mP60_14_3e_9e_3e - CIF

```
# CIF file
data_findsym-output
_audit_creation_method FINDSYM

_chemical_name_mineral 'Parawollastonite'
_chemical_formula_sum 'Ca O3 Si'

loop_
_publ_author_name
 'F. J. Trojer'
_journal_name_full_name
;
 Zeitschrift f{\"u}r Kristallographie - Crystalline Materials
;
_journal_volume 127
_journal_year 1968
_journal_page_first 291
_journal_page_last 308
_publ_Section_title
;
 The crystal structure of parawollastonite
;

_aflow_title 'Parawollastonite (CaSiO$_{3}$, $S3_{3}$ (II)) Structure'
_aflow_proto 'AB3C_mP60_14_3e_9e_3e'
_aflow_params 'a,b/a,c/a,\beta,x_{1},y_{1},z_{1},x_{2},y_{2},z_{2},x_{3
    ↪ },y_{3},z_{3},x_{4},y_{4},z_{4},x_{5},y_{5},z_{5},x_{6},y_{6},
    ↪ z_{6},x_{7},y_{7},z_{7},x_{8},y_{8},z_{8},x_{9},y_{9},z_{9},x_{
    ↪ 10},y_{10},z_{10},x_{11},y_{11},z_{11},x_{12},y_{12},z_{12},x_{
    ↪ 13},y_{13},z_{13},x_{14},y_{14},z_{14},x_{15},y_{15},z_{15}'
_aflow_params_values '7.066,1.03594678743,2.18313048401,95.40417,-0.0288
    ↪ ,0.6242,0.2482,0.7397,0.3735,0.4011,0.7364,0.8791,0.3987,0.6685
    ↪ ,0.6253,0.3,0.3031,0.6241,0.2156,0.0328,0.8603,0.349,0.0348,
    ↪ 0.3843,0.3473,0.2388,0.8774,0.5086,0.2347,0.3824,0.5078,0.406,
    ↪ 0.8038,0.3642,0.4067,0.4467,0.3633,0.2767,0.1245,0.3906,0.2313
    ↪ ,-0.0907,0.4076,0.2313,0.3402,0.4075,0.4432,0.6239,0.3016'
_aflow_Strukturbericht '$S3_{3}$ (II)'
_aflow_Pearson 'mP60'

_symmetry_space_group_name_H-M "P 1 21/c 1"
_symmetry_Int_Tables_number 14

_cell_length_a    7.06600
_cell_length_b    7.32000
_cell_length_c    15.42600
_cell_angle_alpha 90.00000
_cell_angle_beta  95.40417
_cell_angle_gamma 90.00000

loop_
_space_group_symop_id
_space_group_symop_operation_xyz
1 x,y,z
2 -x,y+1/2,-z+1/2
3 -x,-y,-z
4 x,-y+1/2,z+1/2

loop_
_atom_site_label
_atom_site_type_symbol
_atom_site_symmetry_multiplicity
_atom_site_Wyckoff_label
_atom_site_fract_x
_atom_site_fract_y
_atom_site_fract_z
_atom_site_occupancy
Ca1 Ca 4 e -0.02880 0.62420 0.24820 1.00000
Ca2 Ca 4 e 0.73970 0.37350 0.40110 1.00000
Ca3 Ca 4 e 0.73640 0.87910 0.39870 1.00000
O1 O 4 e 0.66850 0.62530 0.30000 1.00000
O2 O 4 e 0.30310 0.62410 0.21560 1.00000
O3 O 4 e 0.03280 0.86030 0.34900 1.00000
O4 O 4 e 0.03480 0.38430 0.34730 1.00000
O5 O 4 e 0.23880 0.87740 0.50860 1.00000
O6 O 4 e 0.23470 0.38240 0.50780 1.00000
O7 O 4 e 0.40600 0.80380 0.36420 1.00000
O8 O 4 e 0.40670 0.44670 0.36330 1.00000
O9 O 4 e 0.27670 0.12450 0.39060 1.00000
Si1 Si 4 e 0.23130 -0.09070 0.40760 1.00000
Si2 Si 4 e 0.23130 0.34020 0.40750 1.00000
Si3 Si 4 e 0.44320 0.62390 0.30160 1.00000
```

Parawollastonite ($CaSiO_3$, $S3_3$ (II)): AB3C_mP60_14_3e_9e_3e - POSCAR

```
AB3C_mP60_14_3e_9e_3e & a,b/a,c/a,beta,x1,y1,z1,x2,y2,z2,x3,y3,z3,x4,y4,
    ↪ z4,x5,y5,z5,x6,y6,z6,x7,y7,z7,x8,y8,z8,x9,y9,z9,x10,y10,z10,x11
    ↪ ,y11,z11,x12,y12,z12,x13,y13,z13,x14,y14,z14,x15,y15,z15 --
    ↪ params=7.066,1.03594678743,2.18313048401,95.40417,-0.0288,
    ↪ 0.6242,0.2482,0.7397,0.3735,0.4011,0.7364,0.8791,0.3987,0.6685,
    ↪ 0.6253,0.3,0.3031,0.6241,0.2156,0.0328,0.8603,0.349,0.0348,
    ↪ 0.3843,0.3473,0.2388,0.8774,0.5086,0.2347,0.3824,0.5078,0.406,
    ↪ 0.8038,0.3642,0.4067,0.4467,0.3633,0.2767,0.1245,0.3906,0.2313
    ↪ ,-0.0907,0.4076,0.2313,0.3402,0.4075,0.4432,0.6239,0.3016 & P2_
    ↪ {1}/c C_{2h}^{5} #14 (e^15) & mP60 & $S3_{3}$ (II) & CaO3Si &
    ↪ Parawollastonite & F. J. Trojer, Zeitschrift f{"u}r
    ↪ Kristallographie - Crystalline Materials 127, 291-308 (1968)
   1.00000000000000
   7.06600000000000   0.00000000000000   0.00000000000000
   0.00000000000000   7.32000000000000   0.00000000000000
  -1.45283256261353   0.00000000000000  15.35743316915330
    Ca     O    Si
    12    36    12
Direct
  -0.02880000000000   0.62420000000000   0.24820000000000   Ca   (4e)
   0.02880000000000   1.12420000000000   0.25180000000000   Ca   (4e)
   0.02880000000000  -0.62420000000000  -0.24820000000000   Ca   (4e)
  -0.02880000000000  -0.12420000000000   0.74820000000000   Ca   (4e)
   0.73970000000000   0.37350000000000   0.40110000000000   Ca   (4e)
  -0.73970000000000   0.87350000000000   0.09890000000000   Ca   (4e)
  -0.73970000000000  -0.37350000000000  -0.40110000000000   Ca   (4e)
   0.73970000000000   0.12650000000000   0.90110000000000   Ca   (4e)
   0.73640000000000   0.87910000000000   0.39870000000000   Ca   (4e)
  -0.73640000000000   1.37910000000000   0.10130000000000   Ca   (4e)
  -0.73640000000000  -0.87910000000000  -0.39870000000000   Ca   (4e)
   0.73640000000000  -0.37910000000000   0.89870000000000   Ca   (4e)
   0.66850000000000   0.62530000000000   0.30000000000000   O   (4e)
  -0.66850000000000   1.12530000000000   0.20000000000000   O   (4e)
  -0.66850000000000  -0.62530000000000  -0.30000000000000   O   (4e)
   0.66850000000000  -0.12530000000000   0.80000000000000   O   (4e)
   0.30310000000000   0.62410000000000   0.21560000000000   O   (4e)
  -0.30310000000000   1.12410000000000   0.28440000000000   O   (4e)
  -0.30310000000000  -0.62410000000000  -0.21560000000000   O   (4e)
   0.30310000000000  -0.12410000000000   0.71560000000000   O   (4e)
   0.03280000000000   0.86030000000000   0.34900000000000   O   (4e)
  -0.03280000000000   1.36030000000000   0.15100000000000   O   (4e)
  -0.03280000000000  -0.86030000000000  -0.34900000000000   O   (4e)
   0.03280000000000  -0.36030000000000   0.84900000000000   O   (4e)
   0.03480000000000   0.38430000000000   0.34730000000000   O   (4e)
  -0.03480000000000   0.88430000000000   0.15270000000000   O   (4e)
```

```
   -0.03480000000000   -0.38430000000000   -0.34730000000000     O    (4e)
    0.03480000000000    0.11570000000000    0.84730000000000     O    (4e)
    0.23880000000000    0.87740000000000    0.50860000000000     O    (4e)
   -0.23880000000000    1.37740000000000   -0.00860000000000     O    (4e)
   -0.23880000000000   -0.87740000000000   -0.50860000000000     O    (4e)
    0.23880000000000   -0.37740000000000    1.00860000000000     O    (4e)
    0.23470000000000    0.38240000000000    0.50780000000000     O    (4e)
   -0.23470000000000    0.88240000000000   -0.00780000000000     O    (4e)
   -0.23470000000000   -0.38240000000000   -0.50780000000000     O    (4e)
    0.23470000000000    0.11760000000000    1.00780000000000     O    (4e)
    0.40600000000000    0.80380000000000    0.36420000000000     O    (4e)
   -0.40600000000000    1.30380000000000    0.13580000000000     O    (4e)
   -0.40600000000000   -0.80380000000000   -0.36420000000000     O    (4e)
    0.40600000000000   -0.30380000000000    0.86420000000000     O    (4e)
    0.40670000000000    0.44670000000000    0.36330000000000     O    (4e)
   -0.40670000000000    0.94670000000000    0.13670000000000     O    (4e)
   -0.40670000000000   -0.44670000000000   -0.36330000000000     O    (4e)
    0.40670000000000    0.05330000000000    0.86330000000000     O    (4e)
    0.27670000000000    0.12450000000000    0.39060000000000     O    (4e)
   -0.27670000000000    0.62450000000000    0.10940000000000     O    (4e)
   -0.27670000000000   -0.12450000000000   -0.39060000000000     O    (4e)
    0.27670000000000    0.37550000000000    0.89060000000000     O    (4e)
    0.23130000000000   -0.09070000000000    0.40760000000000    Si    (4e)
   -0.23130000000000    0.40930000000000    0.09240000000000    Si    (4e)
   -0.23130000000000    0.09070000000000   -0.40760000000000    Si    (4e)
    0.23130000000000    0.59070000000000    0.90760000000000    Si    (4e)
    0.23130000000000    0.34020000000000    0.40750000000000    Si    (4e)
   -0.23130000000000    0.84020000000000    0.09250000000000    Si    (4e)
   -0.23130000000000   -0.34020000000000   -0.40750000000000    Si    (4e)
    0.23130000000000    0.15980000000000    0.90750000000000    Si    (4e)
    0.44320000000000    0.62390000000000    0.30160000000000    Si    (4e)
   -0.44320000000000    1.12390000000000    0.19840000000000    Si    (4e)
   -0.44320000000000   -0.62390000000000   -0.30160000000000    Si    (4e)
    0.44320000000000   -0.12390000000000    0.80160000000000    Si    (4e)
```

$\beta$-$B_2H_6$: AB3_mP16_14_e_3e - CIF

```
# CIF file
data_findsym-output
_audit_creation_method FINDSYM

_chemical_name_mineral 'beta-B2H6'
_chemical_formula_sum 'B H3'

loop_
_publ_author_name
 'H. W. Smith'
 'W. N. Lipscomb'
_journal_name_full_name
;
 Journal of Chemical Physics
;
_journal_volume 43
_journal_year 1965
_journal_page_first 1060
_journal_page_last 1064
_publ_Section_title
;
 Single-Crystal X-Ray Diffraction Study of $\beta$-Diborane
;

_aflow_title '$\beta$-B$_{2}$H$_{6}$ Structure'
_aflow_proto 'AB3_mP16_14_e_3e'
_aflow_params 'a,b/a,c/a,\beta,x_{1},y_{1},z_{1},x_{2},y_{2},z_{2},x_{3
    ↪ },y_{3},z_{3},x_{4},y_{4},z_{4}'
_aflow_params_values '4.4,1.47727272727,1.41869318182,117.78562,0.144,
    ↪ 0.042,0.146,0.36,0.14,0.166,0.098,-0.005,0.294,0.877,0.112,-
    ↪ 0.019'
_aflow_Strukturbericht 'None'
_aflow_Pearson 'mP16'

_symmetry_space_group_name_H-M "P 1 21/c 1"
_symmetry_Int_Tables_number 14

_cell_length_a    4.40000
_cell_length_b    6.50000
_cell_length_c    6.24225
_cell_angle_alpha 90.00000
_cell_angle_beta  117.78562
_cell_angle_gamma 90.00000

loop_
_space_group_symop_id
_space_group_symop_operation_xyz
1 x,y,z
2 -x,y+1/2,-z+1/2
3 -x,-y,-z
4 x,-y+1/2,z+1/2

loop_
_atom_site_label
_atom_site_type_symbol
_atom_site_symmetry_multiplicity
_atom_site_Wyckoff_label
_atom_site_fract_x
_atom_site_fract_y
_atom_site_fract_z
_atom_site_occupancy
B1 B 4 e 0.14400 0.04200 0.14600 1.00000
H1 H 4 e 0.36000 0.14000 0.16600 1.00000
H2 H 4 e 0.09800 -0.00500 0.29400 1.00000
H3 H 4 e 0.87700 0.11200 -0.01900 1.00000
```

$\beta$-$B_2H_6$: AB3_mP16_14_e_3e - POSCAR

```
AB3_mP16_14_e_3e & a,b/a,c/a,beta,x1,y1,z1,x2,y2,z2,x3,y3,z3,x4,y4,z4 --
    ↪ params=4.4,1.47727272727,1.41869318182,117.78562,0.144,0.042,
    ↪ 0.146,0.36,0.14,0.166,0.098,-0.005,0.294,0.877,0.112,-0.019 &
    ↪ P2_{1}/c C_{2h}^{5} #14 (e^4) & mP16 & None & B2H6 & beta & H.
    ↪ W. Smith and W. N. Lipscomb, J. Chem. Phys. 43, 1060-1064 (1965
    ↪ )
   1.00000000000000
   4.40000000000000    0.00000000000000    0.00000000000000
   0.00000000000000    6.50000000000000    0.00000000000000
  -2.90991606787698    0.00000000000000    5.52250609238337
     B      H
     4     12
Direct
   0.14400000000000    0.04200000000000    0.14600000000000     B    (4e)
  -0.14400000000000    0.54200000000000    0.35400000000000     B    (4e)
  -0.14400000000000   -0.04200000000000   -0.14600000000000     B    (4e)
   0.14400000000000    0.45800000000000    0.64600000000000     B    (4e)
   0.36000000000000    0.14000000000000    0.16600000000000     H    (4e)
  -0.36000000000000    0.64000000000000    0.33400000000000     H    (4e)
  -0.36000000000000   -0.14000000000000   -0.16600000000000     H    (4e)
   0.36000000000000    0.36000000000000    0.66600000000000     H    (4e)
   0.09800000000000   -0.00500000000000    0.29400000000000     H    (4e)
  -0.09800000000000    0.49500000000000    0.20600000000000     H    (4e)
  -0.09800000000000    0.00500000000000   -0.29400000000000     H    (4e)
   0.09800000000000    0.50500000000000    0.79400000000000     H    (4e)
   0.87700000000000    0.11200000000000   -0.01900000000000     H    (4e)
  -0.87700000000000    0.61200000000000    0.51900000000000     H    (4e)
  -0.87700000000000   -0.11200000000000    0.01900000000000     H    (4e)
   0.87700000000000    0.38800000000000    0.48100000000000     H    (4e)
```

$B_2H_6$ ($P2_1/c$): AB3_mP16_14_e_3e - CIF

```
# CIF file
data_findsym-output
_audit_creation_method FINDSYM

_chemical_name_mineral 'B2H6'
_chemical_formula_sum 'B H3'

loop_
_publ_author_name
 'Y. Yao'
 'R. Hoffmann'
_journal_name_full_name
;
 Journal of the American Chemical Society
;
_journal_volume 133
_journal_year 2011
_journal_page_first 21002
_journal_page_last 21009
_publ_Section_title
;
 BH$_{3}$ under Pressure: Leaving the Molecular Diborane Motif
;

_aflow_title 'B$_{2}$H$_{6}$ ($P2_{1}/c$) Structure'
_aflow_proto 'AB3_mP16_14_e_3e'
_aflow_params 'a,b/a,c/a,\beta,x_{1},y_{1},z_{1},x_{2},y_{2},z_{2},x_{3
    ↪ },y_{3},z_{3},x_{4},y_{4},z_{4}'
_aflow_params_values '4.46,1.94618834081,1.0201793722,120.5,0.3698,
    ↪ 0.4285,0.0119,0.0865,0.4057,0.7725,0.4789,0.1446,0.7686,0.5981,
    ↪ 0.4237,-0.0903'
_aflow_Strukturbericht 'None'
_aflow_Pearson 'mP16'

_symmetry_space_group_name_H-M "P 1 21/c 1"
_symmetry_Int_Tables_number 14

_cell_length_a    4.46000
_cell_length_b    8.68000
_cell_length_c    4.55000
_cell_angle_alpha 90.00000
_cell_angle_beta  120.50000
_cell_angle_gamma 90.00000

loop_
_space_group_symop_id
_space_group_symop_operation_xyz
1 x,y,z
2 -x,y+1/2,-z+1/2
3 -x,-y,-z
4 x,-y+1/2,z+1/2

loop_
_atom_site_label
_atom_site_type_symbol
_atom_site_symmetry_multiplicity
_atom_site_Wyckoff_label
_atom_site_fract_x
_atom_site_fract_y
_atom_site_fract_z
_atom_site_occupancy
B1 B 4 e 0.36980 0.42850 0.01190 1.00000
H1 H 4 e 0.08650 0.40570 0.77250 1.00000
H2 H 4 e 0.47890 0.14460 0.76860 1.00000
H3 H 4 e 0.59810 0.42370 -0.09030 1.00000
```

$B_2H_6$ ($P2_1/c$): AB3_mP16_14_e_3e - POSCAR

```
AB3_mP16_14_e_3e & a,b/a,c/a,beta,x1,y1,z1,x2,y2,z2,x3,y3,z3,x4,y4,z4 --
    ↪ params=4.46,1.94618834081,1.0201793722,120.5,0.3698,0.4285,
    ↪ 0.0119,0.0865,0.4057,0.7725,0.4789,0.1446,0.7686,0.5981,0.4237
    ↪ ,-0.0903 & P2_{1}/c C_{2h}^{5} #14 (e^4) & mP16 & None & B2H6 &
```

```
     ↪ B2H6 & Y. Yao and R. Hoffmann, J. Am. Chem. Soc. 133,
     ↪ 21002–21009 (2011)
   1.00000000000000
   4.46000000000000    0.00000000000000    0.00000000000000
   0.00000000000000    8.68000000000000    0.00000000000000
  -2.30929955147120    0.00000000000000    3.92041268000894
     B       H
     4      12
Direct
   0.36980000000000    0.42850000000000    0.01190000000000    B    (4e)
  -0.36980000000000    0.92850000000000    0.48810000000000    B    (4e)
  -0.36980000000000   -0.42850000000000   -0.01190000000000    B    (4e)
   0.36980000000000    0.07150000000000    0.51190000000000    B    (4e)
   0.08650000000000    0.40570000000000    0.77250000000000    H    (4e)
  -0.08650000000000    0.90570000000000   -0.27250000000000    H    (4e)
  -0.08650000000000   -0.40570000000000   -0.77250000000000    H    (4e)
   0.08650000000000    0.09430000000000    1.27250000000000    H    (4e)
   0.47890000000000    0.14460000000000    0.76860000000000    H    (4e)
  -0.47890000000000    0.64460000000000   -0.26860000000000    H    (4e)
  -0.47890000000000   -0.14460000000000   -0.76860000000000    H    (4e)
   0.47890000000000    0.35540000000000    1.26860000000000    H    (4e)
   0.59810000000000    0.42370000000000   -0.09030000000000    H    (4e)
  -0.59810000000000    0.92370000000000    0.59030000000000    H    (4e)
  -0.59810000000000   -0.42370000000000    0.09030000000000    H    (4e)
   0.59810000000000    0.07630000000000    0.40970000000000    H    (4e)
```

KAuBr$_4$·2H$_2$O ($H4_{19}$): AB4C2D_mP32_14_e_4e_2e_e - CIF

```
# CIF file
data_findsym-output
_audit_creation_method FINDSYM

_chemical_name_mineral 'AuBr4(H2O)2K'
_chemical_formula_sum 'Au Br4 (H2O)2 K'

loop_
_publ_author_name
 'H. Omrani'
 'F. Th\\'{e}obald'
 'H. Vivier'
_journal_name_full_name
;
 Acta Crystallographica Section C: Structural Chemistry
;
_journal_volume 42
_journal_year 1986
_journal_page_first 1091
_journal_page_last 1092
_publ_Section_title
;
 Structure of potassium tetrabromoaurate(III) dihydrate
;

# Found in Sodium tetrabromoaurate(III) dihydrate, 2001

_aflow_title 'KAuBr$_{4}$$\cdot$2H$_{2}$O ($H4_{19}$) Structure'
_aflow_proto 'AB4C2D_mP32_14_e_4e_2e_e'
_aflow_params 'a,b/a,c/a,\beta,x_{1},y_{1},z_{1},x_{2},y_{2},z_{2},x_{3
    ↪ },y_{3},z_{3},x_{4},y_{4},z_{4},x_{5},y_{5},z_{5},x_{6},y_{6},
    ↪ z_{6},x_{7},y_{7},z_{7},x_{8},y_{8},z_{8}'
_aflow_params_values '8.48,1.41580188679,1.45127830189,129.09091,0.2507,
    ↪ 0.0067,0.5026,0.3935,0.8371,0.6395,0.114,0.1778,0.3679,0.2208,-
    ↪ 0.0834,0.313,0.2767,0.0954,0.6901,0.5843,0.1353,0.0613,0.0171,
    ↪ 0.364,0.5512,0.2671,0.4723,0.5181'
_aflow_Strukturbericht '$H4_{19}$'
_aflow_Pearson 'mP32'

_symmetry_space_group_name_H-M "P 1 21/c 1"
_symmetry_Int_Tables_number 14

_cell_length_a     8.48000
_cell_length_b     12.00600
_cell_length_c     12.30684
_cell_angle_alpha  90.00000
_cell_angle_beta   129.09091
_cell_angle_gamma  90.00000

loop_
_space_group_symop_id
_space_group_symop_operation_xyz
1 x,y,z
2 -x,y+1/2,-z+1/2
3 -x,-y,-z
4 x,-y+1/2,z+1/2

loop_
_atom_site_label
_atom_site_type_symbol
_atom_site_symmetry_multiplicity
_atom_site_Wyckoff_label
_atom_site_fract_x
_atom_site_fract_y
_atom_site_fract_z
_atom_site_occupancy
Au1 Au 4 e 0.25070 0.00670 0.50260 1.00000
Br1 Br 4 e 0.39350 0.83710 0.63950 1.00000
Br2 Br 4 e 0.11400 0.17780 0.36790 1.00000
Br3 Br 4 e 0.22080 -0.08340 0.31300 1.00000
Br4 Br 4 e 0.27670 0.09540 0.69010 1.00000
H2O1 H2O 4 e 0.58430 0.13530 0.06130 1.00000
H2O2 H2O 4 e 0.01710 0.36400 0.55120 1.00000
K1 K 4 e 0.26710 0.47230 0.51810 1.00000
```

KAuBr$_4$·2H$_2$O ($H4_{19}$): AB4C2D_mP32_14_e_4e_2e_e - POSCAR

```
AB4C2D_mP32_14_e_4e_2e_e & a,b/a,c/a,beta,x1,y1,z1,x2,y2,z2,x3,y3,z3,x4,
     ↪ y4,z4,x5,y5,z5,x6,y6,z6,x7,y7,z7,x8,y8,z8 --params=8.48,
     ↪ 1.41580188679,1.45127830189,129.09091,0.2507,0.0067,0.5026,
     ↪ 0.3935,0.8371,0.6395,0.114,0.1778,0.3679,0.2208,-0.0834,0.313,
     ↪ 0.2767,0.0954,0.6901,0.5843,0.1353,0.0613,0.0171,0.364,0.5512,
     ↪ 0.2671,0.4723,0.5181 & P2_{1}/c C_{2h}^{5} #14 (e^8) & mP32 &
     ↪ $H4_{19}$ & AuBr4(H2O)2K & AuBr4(H2O)2K & H. Omrani and F. Th
     ↪ \'{e}obald and H. Vivier, Acta Crystallogr. C 42, 1091–1092 (
     ↪ 1986)
   1.00000000000000
   8.48000000000000    0.00000000000000    0.00000000000000
   0.00000000000000   12.00600000000000    0.00000000000000
  -7.76011093693911    0.00000000000000    9.55191022947756
    Au     Br    H2O      K
     4     16      8      4
Direct
   0.25070000000000    0.00670000000000    0.50260000000000    Au    (4e)
  -0.25070000000000    0.50670000000000   -0.00260000000000    Au    (4e)
  -0.25070000000000   -0.00670000000000   -0.50260000000000    Au    (4e)
   0.25070000000000    0.49330000000000    1.00260000000000    Au    (4e)
   0.39350000000000    0.83710000000000    0.63950000000000    Br    (4e)
  -0.39350000000000    1.33710000000000   -0.13950000000000    Br    (4e)
  -0.39350000000000   -0.83710000000000   -0.63950000000000    Br    (4e)
   0.39350000000000   -0.33710000000000    1.13950000000000    Br    (4e)
   0.11400000000000    0.17780000000000    0.36790000000000    Br    (4e)
  -0.11400000000000    0.67780000000000    0.13210000000000    Br    (4e)
  -0.11400000000000   -0.17780000000000   -0.36790000000000    Br    (4e)
   0.11400000000000    0.32220000000000    0.86790000000000    Br    (4e)
   0.22080000000000   -0.08340000000000    0.31300000000000    Br    (4e)
  -0.22080000000000    0.41660000000000    0.18700000000000    Br    (4e)
  -0.22080000000000    0.08340000000000   -0.31300000000000    Br    (4e)
   0.22080000000000    0.58340000000000    0.81300000000000    Br    (4e)
   0.27670000000000    0.09540000000000    0.69010000000000    Br    (4e)
  -0.27670000000000    0.59540000000000   -0.19010000000000    Br    (4e)
  -0.27670000000000   -0.09540000000000   -0.69010000000000    Br    (4e)
   0.27670000000000    0.40460000000000    1.19010000000000    Br    (4e)
   0.58430000000000    0.13530000000000    0.06130000000000   H2O    (4e)
  -0.58430000000000    0.63530000000000    0.43870000000000   H2O    (4e)
  -0.58430000000000   -0.13530000000000   -0.06130000000000   H2O    (4e)
   0.58430000000000    0.36470000000000    0.56130000000000   H2O    (4e)
   0.01710000000000    0.36400000000000    0.55120000000000   H2O    (4e)
  -0.01710000000000    0.86400000000000   -0.05120000000000   H2O    (4e)
  -0.01710000000000   -0.36400000000000   -0.55120000000000   H2O    (4e)
   0.01710000000000    0.13600000000000    1.05120000000000   H2O    (4e)
   0.26710000000000    0.47230000000000    0.51810000000000     K    (4e)
  -0.26710000000000    0.97230000000000   -0.01810000000000     K    (4e)
  -0.26710000000000   -0.47230000000000   -0.51810000000000     K    (4e)
   0.26710000000000    0.02770000000000    1.01810000000000     K    (4e)
```

Anhydrous KAuBr$_4$: AB4C_mP24_14_ab_4e_e - CIF

```
# CIF file
data_findsym-output
_audit_creation_method FINDSYM

_chemical_name_mineral 'AuBr4K'
_chemical_formula_sum 'Au Br4 K'

loop_
_publ_author_name
 'H. Omrani'
 'R. Welter'
 'R. Vangelisti'
_journal_name_full_name
;
 Acta Crystallographica Section C: Structural Chemistry
;
_journal_volume 55
_journal_year 1999
_journal_page_first 13
_journal_page_last 14
_publ_Section_title
;
 Potassium tetrabromoaurate(III)
;

# Found in Sodium tetrabromoaurate(III) dihydrate, 2001

_aflow_title 'Anhydrous KAuBr$_{4}$ Structure'
_aflow_proto 'AB4C_mP24_14_ab_4e_e'
_aflow_params 'a,b/a,c/a,\beta,x_{3},y_{3},z_{3},x_{4},y_{4},z_{4},x_{5
    ↪ },y_{5},z_{5},x_{6},y_{6},z_{6},x_{7},y_{7},z_{7}'
_aflow_params_values '9.0306,0.736208003898,1.41740305185,96.88,0.86659,
    ↪ 0.2934,0.04766,0.09492,-0.0538,0.18234,0.5056,0.1517,0.1707,
    ↪ 0.6635,0.7336,0.07207,0.2128,0.4492,0.1813'
_aflow_Strukturbericht 'None'
_aflow_Pearson 'mP24'

_symmetry_space_group_name_H-M "P 1 21/c 1"
_symmetry_Int_Tables_number 14

_cell_length_a     9.03060
_cell_length_b     6.64840
_cell_length_c     12.80000
_cell_angle_alpha  90.00000
_cell_angle_beta   96.88000
_cell_angle_gamma  90.00000

loop_
_space_group_symop_id
_space_group_symop_operation_xyz
1 x,y,z
2 -x,y+1/2,-z+1/2
3 -x,-y,-z
4 x,-y+1/2,z+1/2
```

```
loop_
_atom_site_label
_atom_site_type_symbol
_atom_site_symmetry_multiplicity
_atom_site_Wyckoff_label
_atom_site_fract_x
_atom_site_fract_y
_atom_site_fract_z
_atom_site_occupancy
Au1 Au 2 a 0.00000 0.00000 0.00000 1.00000
Au2 Au 2 b 0.50000 0.00000 0.00000 1.00000
Br1 Br 4 e 0.86659 0.29340 0.04766 1.00000
Br2 Br 4 e 0.09492 -0.05380 0.18234 1.00000
Br3 Br 4 e 0.50560 0.15170 0.17070 1.00000
Br4 Br 4 e 0.66350 0.73360 0.07207 1.00000
K1 K 4 e 0.21280 0.44920 0.18130 1.00000
```

Anhydrous $KAuBr_4$: AB4C_mP24_14_ab_4e_e - POSCAR

```
AB4C_mP24_14_ab_4e_e & a,b/a,c/a,beta,x3,y3,z3,x4,y4,z4,x5,y5,z5,x6,y6,
    ↪ z6,x7,y7,z7 --params=9.0306,0.736208003898,1.41740305185,96.88,
    ↪ 0.86659,0.2934,0.04766,0.09492,-0.0538,0.18234,0.5056,0.1517,
    ↪ 0.1707,0.6635,0.7336,0.07207,0.2128,0.4492,0.1813 & P2_{1}/c C_
    ↪ {2h}^{5} #14 (abe^5) & mP24 & None & AuBr4K & AuBr4K & H.
    ↪ Omrani and R. Welter and R. Vangelisti, Acta Crystallogr. C 55,
    ↪ 13-14 (1999)
   1.00000000000000
   9.03060000000000   0.00000000000000   0.00000000000000
   0.00000000000000   6.64840000000000   0.00000000000000
  -1.53331576111014   0.00000000000000  12.70782997906140
    Au      Br       K
     4      16       4
Direct
   0.00000000000000   0.00000000000000   0.00000000000000   Au   (2a)
   0.00000000000000   0.50000000000000   0.50000000000000   Au   (2a)
   0.50000000000000   0.00000000000000   0.00000000000000   Au   (2b)
   0.50000000000000   0.50000000000000   0.50000000000000   Au   (2b)
   0.86659000000000   0.29340000000000   0.04766000000000   Br   (4e)
  -0.86659000000000   0.79340000000000   0.45234000000000   Br   (4e)
  -0.86659000000000  -0.29340000000000  -0.04766000000000   Br   (4e)
   0.86659000000000   0.20660000000000   0.54766000000000   Br   (4e)
   0.09492000000000  -0.05380000000000   0.18234000000000   Br   (4e)
  -0.09492000000000   0.44620000000000   0.31766000000000   Br   (4e)
  -0.09492000000000   0.05380000000000  -0.18234000000000   Br   (4e)
   0.09492000000000   0.55380000000000   0.68234000000000   Br   (4e)
   0.50560000000000   0.15170000000000   0.17070000000000   Br   (4e)
  -0.50560000000000   0.65170000000000   0.32930000000000   Br   (4e)
  -0.50560000000000  -0.15170000000000  -0.17070000000000   Br   (4e)
   0.50560000000000   0.34830000000000   0.67070000000000   Br   (4e)
   0.66350000000000   0.73360000000000   0.07207000000000   Br   (4e)
  -0.66350000000000   1.23360000000000   0.42793000000000   Br   (4e)
  -0.66350000000000  -0.73360000000000  -0.07207000000000   Br   (4e)
   0.66350000000000  -0.23360000000000   0.57207000000000   Br   (4e)
   0.21280000000000   0.44920000000000   0.18130000000000    K   (4e)
  -0.21280000000000   0.94920000000000   0.31870000000000    K   (4e)
  -0.21280000000000  -0.44920000000000  -0.18130000000000    K   (4e)
   0.21280000000000   0.05080000000000   0.68130000000000    K   (4e)
```

Ammonium Persulfate $[(NH_4)_2S_2O_8$, $K4_1]$: AB4C_mP24_14_e_4e_e - CIF

```
# CIF file
data_findsym-output
_audit_creation_method FINDSYM

_chemical_name_mineral 'Ammonium persulfate'
_chemical_formula_sum '(NH4) O4 S'

loop_
_publ_author_name
 'B. K. Sivertsen'
 'H. Sorum'
_journal_name_full_name
;
 Zeitschrift f{\"u}r Kristallographie - Crystalline Materials
;
_journal_volume 130
_journal_year 1969
_journal_page_first 449
_journal_page_last 460
_publ_Section_title
;
 A reinvestigation of the crystal structure of ammonium persulfate, (
    ↪ NH$_{4}$)$_{2}$S$_{2}$O$_{8}$
;

_aflow_title 'Ammonium Persulfate [(NH$_{4}$)$_{2}$S$_{2}$O$_{8}$, $K4_{
    ↪ 1}$] Structure'
_aflow_proto 'AB4C_mP24_14_e_4e_e'
_aflow_params 'a,b/a,c/a,\beta,x_{1},y_{1},z_{1},x_{2},y_{2},z_{2},x_{3
    ↪ },y_{3},z_{3},x_{4},y_{4},z_{4},x_{5},y_{5},z_{5},x_{6},y_{6},
    ↪ z_{6}'
_aflow_params_values '7.83,1.02273307791,1.21392081737,139.39,0.0899,
    ↪ 0.8839,0.2331,0.4059,0.5186,0.3818,0.343,0.217,0.3158,0.2414,
    ↪ 0.4109,0.0667,0.6985,0.351,0.4181,0.4242,0.3572,0.2869'
_aflow_Strukturbericht '$K4_{1}$'
_aflow_Pearson 'mP24'

_symmetry_space_group_name_H-M "P 1 21/c 1"
_symmetry_Int_Tables_number 14

_cell_length_a      7.83000
_cell_length_b      8.00800
_cell_length_c      9.50500
_cell_angle_alpha  90.00000
_cell_angle_beta  139.39000
_cell_angle_gamma  90.00000

loop_
_space_group_symop_id
_space_group_symop_operation_xyz
1 x,y,z
2 -x,y+1/2,-z+1/2
3 -x,-y,-z
4 x,-y+1/2,z+1/2

loop_
_atom_site_label
_atom_site_type_symbol
_atom_site_symmetry_multiplicity
_atom_site_Wyckoff_label
_atom_site_fract_x
_atom_site_fract_y
_atom_site_fract_z
_atom_site_occupancy
NH41 NH4 4 e 0.08990 0.88390 0.23310 1.00000
O1 O 4 e 0.40590 0.51860 0.38180 1.00000
O2 O 4 e 0.34300 0.21700 0.31580 1.00000
O3 O 4 e 0.24140 0.41090 0.06670 1.00000
O4 O 4 e 0.69850 0.35100 0.41810 1.00000
S1 S 4 e 0.42420 0.35720 0.28690 1.00000
```

Ammonium Persulfate $[(NH_4)_2S_2O_8$, $K4_1]$: AB4C_mP24_14_e_4e_e - POSCAR

```
AB4C_mP24_14_e_4e_e & a,b/a,c/a,beta,x1,y1,z1,x2,y2,z2,x3,y3,z3,x4,y4,z4
    ↪ ,x5,y5,z5,x6,y6,z6 --params=7.83,1.02273307791,1.21392081737,
    ↪ 139.39,0.0899,0.8839,0.2331,0.4059,0.5186,0.3818,0.343,0.217,
    ↪ 0.3158,0.2414,0.4109,0.0667,0.6985,0.351,0.4181,0.4242,0.3572,
    ↪ 0.2869 & P2_{1}/c C_{2h}^{5} #14 (e^6) & mP24 & $K4_{1}$ & (NH4
    ↪ )O4S & Ammonium persulfate & B. K. Sivertsen and H. Sorum,
    ↪ Zeitschrift f{"u}r Kristallographie - Crystalline Materials 130
    ↪ , 449-460 (1969)
   1.00000000000000
   7.83000000000000   0.00000000000000   0.00000000000000
   0.00000000000000   8.00800000000000   0.00000000000000
  -7.21579407388266   0.00000000000000   6.18686842298427
   NH4       O       S
     4      16       4
Direct
   0.08990000000000   0.88390000000000   0.23310000000000  NH4   (4e)
  -0.08990000000000   1.38390000000000   0.26690000000000  NH4   (4e)
  -0.08990000000000  -0.88390000000000  -0.23310000000000  NH4   (4e)
   0.08990000000000  -0.38390000000000   0.73310000000000  NH4   (4e)
   0.40590000000000   0.51860000000000   0.38180000000000    O   (4e)
  -0.40590000000000   1.01860000000000   0.11820000000000    O   (4e)
  -0.40590000000000  -0.51860000000000  -0.38180000000000    O   (4e)
   0.40590000000000  -0.01860000000000   0.88180000000000    O   (4e)
   0.34300000000000   0.21700000000000   0.31580000000000    O   (4e)
  -0.34300000000000   0.71700000000000   0.18420000000000    O   (4e)
  -0.34300000000000  -0.21700000000000  -0.31580000000000    O   (4e)
   0.34300000000000   0.28300000000000   0.81580000000000    O   (4e)
   0.24140000000000   0.41090000000000   0.06670000000000    O   (4e)
  -0.24140000000000   0.91090000000000   0.43330000000000    O   (4e)
  -0.24140000000000  -0.41090000000000  -0.06670000000000    O   (4e)
   0.24140000000000   0.08910000000000   0.56670000000000    O   (4e)
   0.69850000000000   0.35100000000000   0.41810000000000    O   (4e)
  -0.69850000000000   0.85100000000000   0.08190000000000    O   (4e)
  -0.69850000000000  -0.35100000000000  -0.41810000000000    O   (4e)
   0.69850000000000   0.14900000000000   0.91810000000000    O   (4e)
   0.42420000000000   0.35720000000000   0.28690000000000    S   (4e)
  -0.42420000000000   0.85720000000000   0.21310000000000    S   (4e)
  -0.42420000000000  -0.35720000000000  -0.28690000000000    S   (4e)
   0.42420000000000   0.14280000000000   0.78690000000000    S   (4e)
```

Monasite ($LaPO_4$): AB4C_mP24_14_e_4e_e - CIF

```
# CIF file
data_findsym-output
_audit_creation_method FINDSYM

_chemical_name_mineral 'Monasite'
_chemical_formula_sum 'La O4 P'

loop_
_publ_author_name
 'Y. Ni'
 'J. M. Hughes'
 'A. N. Mariano'
_journal_name_full_name
;
 American Mineralogist
;
_journal_volume 80
_journal_year 1995
_journal_page_first 21
_journal_page_last 26
_publ_Section_title
;
 Crystal chemistry of the monazite and xenotime structures
;

# Found in Gasparite-(La), La(AsO$_{4}$), a new mineral from Mn ores of
    ↪ the Ushkatyn-III deposit, Central Kazakhstan, and metamorphic
    ↪ rocks of the Wanni glacier, Switzerland, 2019

_aflow_title 'Monasite (LaPO$_{4}$) Structure'
_aflow_proto 'AB4C_mP24_14_e_4e_e'
_aflow_params 'a,b/a,c/a,\beta,x_{1},y_{1},z_{1},x_{2},y_{2},z_{2},x_{3
    ↪ },y_{3},z_{3},x_{4},y_{4},z_{4},x_{5},y_{5},z_{5},x_{6},y_{6},
    ↪ z_{6}'
```

```
_aflow_params_values '6.5034,1.08720054126,1.27324630193,126.58575,
    ↪ 0.18086,0.16033,0.28154,0.8026,0.0077,0.2503,0.8835,0.3315,
    ↪ 0.3799,0.673,0.1071,0.4748,0.4176,0.2168,0.1277,0.6926,0.1639,
    ↪ 0.3047'
_aflow_Strukturbericht 'None'
_aflow_Pearson 'mP24'

_symmetry_space_group_name_H-M "P 1 21/c 1"
_symmetry_Int_Tables_number 14

_cell_length_a    6.50340
_cell_length_b    7.07050
_cell_length_c    8.28043
_cell_angle_alpha 90.00000
_cell_angle_beta  126.58575
_cell_angle_gamma 90.00000

loop_
_space_group_symop_id
_space_group_symop_operation_xyz
1 x,y,z
2 -x,y+1/2,-z+1/2
3 -x,-y,-z
4 x,-y+1/2,z+1/2

loop_
_atom_site_label
_atom_site_type_symbol
_atom_site_symmetry_multiplicity
_atom_site_Wyckoff_label
_atom_site_fract_x
_atom_site_fract_y
_atom_site_fract_z
_atom_site_occupancy
La1 La 4 e 0.18086 0.16033 0.28154 1.00000
O1 O 4 e 0.80260 0.00770 0.25030 1.00000
O2 O 4 e 0.88350 0.33150 0.37990 1.00000
O3 O 4 e 0.67300 0.10710 0.47480 1.00000
O4 O 4 e 0.41760 0.21680 0.12770 1.00000
P1 P 4 e 0.69260 0.16390 0.30470 1.00000
```

Monasite ($LaPO_4$): AB4C_mP24_14_e_4e_e - POSCAR

```
AB4C_mP24_14_e_4e_e & a,b/a,c/a,beta,x1,y1,z1,x2,y2,z2,x3,y3,z3,x4,y4,z4
    ↪ ,x5,y5,z5,x6,y6,z6 --params=6.5034,1.08720054126,1.27324630193,
    ↪ 126.58575,0.18086,0.16033,0.28154,0.8026,0.0077,0.2503,0.8835,
    ↪ 0.3315,0.3799,0.673,0.1071,0.4748,0.4176,0.2168,0.1277,0.6926,
    ↪ 0.1639,0.3047 & P2_{1}/c C_{2h}^{5} #14 (e^6) & mP24 & None &
    ↪ LaO4P & Monasite & Y. Ni and J. M. Hughes and A. N. Mariano,
    ↪ Am. Mineral. 80, 21-26 (1995)
   1.00000000000000
   6.50340000000000   0.00000000000000   0.00000000000000
   0.00000000000000   7.07050000000000   0.00000000000000
  -4.93534484962521   0.00000000000000   6.64890157846978
    La     O     P
     4    16     4
Direct
   0.18086000000000   0.16033000000000   0.28154000000000   La   (4e)
  -0.18086000000000   0.66033000000000   0.21846000000000   La   (4e)
  -0.18086000000000  -0.16033000000000  -0.28154000000000   La   (4e)
   0.18086000000000   0.33967000000000   0.78154000000000   La   (4e)
   0.80260000000000   0.00770000000000   0.25030000000000   O    (4e)
  -0.80260000000000   0.50770000000000   0.24970000000000   O    (4e)
  -0.80260000000000  -0.00770000000000  -0.25030000000000   O    (4e)
   0.80260000000000   0.49230000000000   0.75030000000000   O    (4e)
   0.88350000000000   0.33150000000000   0.37990000000000   O    (4e)
  -0.88350000000000   0.83150000000000   0.12010000000000   O    (4e)
  -0.88350000000000  -0.33150000000000  -0.37990000000000   O    (4e)
   0.88350000000000   0.16850000000000   0.87990000000000   O    (4e)
   0.67300000000000   0.10710000000000   0.47480000000000   O    (4e)
  -0.67300000000000   0.60710000000000   0.02520000000000   O    (4e)
  -0.67300000000000  -0.10710000000000  -0.47480000000000   O    (4e)
   0.67300000000000   0.39290000000000   0.97480000000000   O    (4e)
   0.41760000000000   0.21680000000000   0.12770000000000   O    (4e)
  -0.41760000000000   0.71680000000000   0.37230000000000   O    (4e)
  -0.41760000000000  -0.21680000000000  -0.12770000000000   O    (4e)
   0.41760000000000   0.28320000000000   0.62770000000000   O    (4e)
   0.69260000000000   0.16390000000000   0.30470000000000   P    (4e)
  -0.69260000000000   0.66390000000000   0.19530000000000   P    (4e)
  -0.69260000000000  -0.16390000000000  -0.30470000000000   P    (4e)
   0.69260000000000   0.33610000000000   0.80470000000000   P    (4e)
```

$Sr_2MnTeO_6$: AB6C2D_mP20_14_a_3e_e_d - CIF

```
# CIF file
data_findsym-output
_audit_creation_method FINDSYM

_chemical_name_mineral 'MnO6Sr2Te'
_chemical_formula_sum 'Mn O6 Sr2 Te'

loop_
_publ_author_name
 'L. {Ortega-San Martin}'
 'J. P. Chapman'
 'E. {Hern\\'{a}ndez-Bocanegra}'
 'M. Insausti'
 'M. I. Arriortua'
 'T. Rojo'
_journal_name_full_name
;
 Journal of Physics: Condensed Matter
;
_journal_volume 16
_journal_year 2004
_journal_page_first 3879
_journal_page_last 3888
_publ_Section_title
;
 Structural phase transitions in the ordered double perovskite Sr${_2}
    ↪ $MnTeO$_{6}$
;

_aflow_title 'Sr$_{2}$MnTeO$_{6}$ Structure'
_aflow_proto 'AB6C2D_mP20_14_a_3e_e_d'
_aflow_params 'a,b/a,c/a,\beta,x_{3},y_{3},z_{3},x_{4},y_{4},z_{4},x_{5
    ↪ },y_{5},z_{5},x_{6},y_{6},z_{6}'
_aflow_params_values '5.7009,0.995807679489,1.72670455542,125.30487,
    ↪ 0.2103,-0.0065,0.265,0.3193,0.7621,0.0287,0.7308,0.7126,-0.0283
    ↪ ,0.2523,-0.01532,0.7508'
_aflow_Strukturbericht 'None'
_aflow_Pearson 'mP20'

_symmetry_space_group_name_H-M "P 1 21/c 1"
_symmetry_Int_Tables_number 14

_cell_length_a    5.70090
_cell_length_b    5.67700
_cell_length_c    9.84377
_cell_angle_alpha 90.00000
_cell_angle_beta  125.30487
_cell_angle_gamma 90.00000

loop_
_space_group_symop_id
_space_group_symop_operation_xyz
1 x,y,z
2 -x,y+1/2,-z+1/2
3 -x,-y,-z
4 x,-y+1/2,z+1/2

loop_
_atom_site_label
_atom_site_type_symbol
_atom_site_symmetry_multiplicity
_atom_site_Wyckoff_label
_atom_site_fract_x
_atom_site_fract_y
_atom_site_fract_z
_atom_site_occupancy
Mn1 Mn 2 a 0.00000 0.00000 0.00000 1.00000
Te1 Te 2 d 0.50000 0.00000 0.50000 1.00000
O1 O 4 e 0.21030 -0.00650 0.26500 1.00000
O2 O 4 e 0.31930 0.76210 0.02870 1.00000
O3 O 4 e 0.73080 0.71260 -0.02830 1.00000
Sr1 Sr 4 e 0.25230 -0.01532 0.75080 1.00000
```

$Sr_2MnTeO_6$: AB6C2D_mP20_14_a_3e_e_d - POSCAR

```
AB6C2D_mP20_14_a_3e_e_d & a,b/a,c/a,beta,x3,y3,z3,x4,y4,z4,x5,y5,z5,x6,
    ↪ y6,z6 --params=5.7009,0.995807679489,1.72670455542,125.30487,
    ↪ 0.2103,-0.0065,0.265,0.3193,0.7621,0.0287,0.7308,0.7126,-0.0283
    ↪ ,0.2523,-0.01532,0.7508 & P2_{1}/c C_{2h}^{5} #14 (ade^4) &
    ↪ mP20 & None & MnO6Sr2Te & MnO6Sr2Te & L. {Ortega-San Martin} et
    ↪ al., J. Phys.: Condens. Matter 16, 3879-3888 (2004)
   1.00000000000000
   5.70090000000000   0.00000000000000   0.00000000000000
   0.00000000000000   5.67700000000000   0.00000000000000
  -5.68898038583372   0.00000000000000   8.03338720481586
    Mn     O    Sr    Te
     2    12     4     2
Direct
   0.00000000000000   0.00000000000000   0.00000000000000   Mn   (2a)
   0.00000000000000   0.50000000000000   0.50000000000000   Mn   (2a)
   0.21030000000000  -0.00650000000000   0.26500000000000   O    (4e)
  -0.21030000000000   0.49350000000000   0.23500000000000   O    (4e)
  -0.21030000000000   0.00650000000000  -0.26500000000000   O    (4e)
   0.21030000000000   0.50650000000000   0.76500000000000   O    (4e)
   0.31930000000000   0.76210000000000   0.02870000000000   O    (4e)
  -0.31930000000000   1.26210000000000   0.47130000000000   O    (4e)
  -0.31930000000000  -0.76210000000000  -0.02870000000000   O    (4e)
   0.31930000000000  -0.26210000000000   0.52870000000000   O    (4e)
   0.73080000000000   0.71260000000000  -0.02830000000000   O    (4e)
  -0.73080000000000   1.21260000000000   0.52830000000000   O    (4e)
  -0.73080000000000  -0.71260000000000   0.02830000000000   O    (4e)
   0.73080000000000  -0.21260000000000   0.47170000000000   O    (4e)
   0.25230000000000  -0.01532000000000   0.75080000000000   Sr   (4e)
  -0.25230000000000   0.48468000000000  -0.25080000000000   Sr   (4e)
  -0.25230000000000   0.01532000000000  -0.75080000000000   Sr   (4e)
   0.25230000000000   0.51532000000000   1.25080000000000   Sr   (4e)
   0.50000000000000   0.00000000000000   0.50000000000000   Te   (2d)
   0.50000000000000   0.50000000000000   0.00000000000000   Te   (2d)
```

Cryolite ($Na_3AlF_6$, $J2_6$): AB6C3_mP20_14_a_3e_de - CIF

```
# CIF file
data_findsym-output
_audit_creation_method FINDSYM

_chemical_name_mineral 'Cryolite'
_chemical_formula_sum 'Al F6 Na3'

loop_
_publ_author_name
 'H. Yang'
 'S. Ghose'
 'D. M. Hatch'
_journal_name_full_name
;
 Physics and Chemistry of Minerals
```

```
;
_journal_volume 19
_journal_year 1993
_journal_page_first 528
_journal_page_last 544
_publ_Section_title
;
 Ferroelastic phase transition in cryolite, Na$_{3}$AlF$_{6}$, a mixed
     ↪ fluoride perovskite: High temperature single crystal X-ray
     ↪ diffraction study and symmetry analysis of the transition
     ↪ mechanism
;

# Found in The American Mineralogist Crystal Structure Database, 2003

_aflow_title 'Cryolite (Na$_{3}$AlF$_{6}$, $J2_{6}$) Structure'
_aflow_proto 'AB6C3_mP20_14_a_3e_de'
_aflow_params 'a,b/a,c/a,\beta,x_{3},y_{3},z_{3},x_{4},y_{4},z_{4},x_{5
     ↪ },y_{5},z_{5},x_{6},y_{6},z_{6}'
_aflow_params_values '5.4139,1.03459613218,1.74764587451,124.72045,
     ↪ 0.8828,-0.0442,0.7812,0.6828,0.6733,0.455,0.2239,0.7313,0.0617,
     ↪ 0.7353,0.5508,0.7474'
_aflow_Strukturbericht '$J2_{6}$'
_aflow_Pearson 'mP20'

_symmetry_space_group_name_H-M "P 1 21/c 1"
_symmetry_Int_Tables_number 14

_cell_length_a    5.41390
_cell_length_b    5.60120
_cell_length_c    9.46158
_cell_angle_alpha 90.00000
_cell_angle_beta  124.72045
_cell_angle_gamma 90.00000

loop_
_space_group_symop_id
_space_group_symop_operation_xyz
1 x,y,z
2 -x,y+1/2,-z+1/2
3 -x,-y,-z
4 x,-y+1/2,z+1/2

loop_
_atom_site_label
_atom_site_type_symbol
_atom_site_symmetry_multiplicity
_atom_site_Wyckoff_label
_atom_site_fract_x
_atom_site_fract_y
_atom_site_fract_z
_atom_site_occupancy
Al1 Al 2 a 0.00000 0.00000 0.00000 1.00000
Na1 Na 2 d 0.50000 0.00000 0.50000 1.00000
F1 F 4 e 0.88280 -0.04420 0.78120 1.00000
F2 F 4 e 0.68280 0.67330 0.45500 1.00000
F3 F 4 e 0.22390 0.73130 0.06170 1.00000
Na2 Na 4 e 0.73530 0.55080 0.74740 1.00000
```

Cryolite ($Na_3AlF_6$, $J2_6$): AB6C3_mP20_14_a_3e_de - POSCAR

```
AB6C3_mP20_14_a_3e_de & a,b/a,c/a,beta,x3,y3,z3,x4,y4,z4,x5,y5,z5,x6,y6,
     ↪ z6 --params=5.4139,1.03459613218,1.74764587451,124.72045,0.8828
     ↪ ,-0.0442,0.7812,0.6828,0.6733,0.455,0.2239,0.7313,0.0617,0.7353
     ↪ ,0.5508,0.7474 & P2_{1}/c C_{2h}^{5} #14 (ade^4) & mP20 & $J2_{
     ↪ 6}$ & AlF6Na3 & Cryolite & H. Yang and S. Ghose and D. M. Hatch
     ↪ , Phys. Chem. Miner. 19, 528-544 (1993)
   1.00000000000000
   5.41390000000000   0.00000000000000   0.00000000000000
   0.00000000000000   5.60120000000000   0.00000000000000
  -5.38905981158747   0.00000000000000   7.77685864880755
     Al      F      Na
      2     12       6
Direct
   0.00000000000000   0.00000000000000   0.00000000000000   Al   (2a)
   0.00000000000000   0.50000000000000   0.50000000000000   Al   (2a)
   0.88280000000000  -0.04420000000000   0.78120000000000   F    (4e)
  -0.88280000000000   0.45580000000000  -0.28120000000000   F    (4e)
  -0.88280000000000   0.04420000000000  -0.78120000000000   F    (4e)
   0.88280000000000   0.54420000000000   1.28120000000000   F    (4e)
   0.68280000000000   0.67330000000000   0.45500000000000   F    (4e)
  -0.68280000000000   1.17330000000000   0.04500000000000   F    (4e)
  -0.68280000000000  -0.67330000000000  -0.45500000000000   F    (4e)
   0.68280000000000  -0.17330000000000   0.95500000000000   F    (4e)
   0.22390000000000   0.73130000000000   0.06170000000000   F    (4e)
  -0.22390000000000   1.23130000000000   0.43830000000000   F    (4e)
  -0.22390000000000  -0.73130000000000  -0.06170000000000   F    (4e)
   0.22390000000000  -0.23130000000000   0.56170000000000   F    (4e)
   0.50000000000000   0.00000000000000   0.50000000000000   Na   (2d)
   0.50000000000000   0.50000000000000   0.00000000000000   Na   (2d)
   0.73530000000000   0.55080000000000   0.74740000000000   Na   (4e)
  -0.73530000000000   1.05080000000000  -0.24740000000000   Na   (4e)
  -0.73530000000000  -0.55080000000000  -0.74740000000000   Na   (4e)
   0.73530000000000  -0.05080000000000   1.24740000000000   Na   (4e)
```

$KNO_2$ III: ABC2_mP16_14_e_e_2e - CIF

```
# CIF file
data_findsym-output
_audit_creation_method FINDSYM

_chemical_name_mineral 'Potassium nitrite'
_chemical_formula_sum 'K N O2'

loop_
_publ_author_name
 'N. {Onoda-Yamamuro}'
 'H. Honda'
 'R. Ikeda'
 'O. Yamamuro'
 'T. Matsuo'
 'K. Oikawa'
 'T. Kamiyama'
 'F. Izumi'
_journal_name_full_name
;
 Journal of Physics: Condensed Matter
;
_journal_volume 10
_journal_year 1998
_journal_page_first 3341
_journal_page_last 3351
_publ_Section_title
;
 Neutron powder diffraction study of the low-temperature phases of KNO$_
     ↪ {2}$
;

# Found in Order-disorder phase transitions in KNO$_{2}$, CsNO$_{2}$,
     ↪ and TlNO$_{2}$ crystals: A molecular dynamics study, 2001

_aflow_title 'KNO$_{2}$ III Structure'
_aflow_proto 'ABC2_mP16_14_e_e_2e'
_aflow_params 'a,b/a,c/a,\beta,x_{1},y_{1},z_{1},x_{2},y_{2},z_{2},x_{3
     ↪ },y_{3},z_{3},x_{4},y_{4},z_{4}'
_aflow_params_values '4.401,2.18057259714,1.58638945694,108.274,0.4105,
     ↪ 0.1407,0.2101,-0.0732,0.4107,0.1839,0.2034,0.4127,0.3067,0.8385
     ↪ ,0.2999,0.0865'
_aflow_Strukturbericht 'None'
_aflow_Pearson 'mP16'

_symmetry_space_group_name_H-M "P 1 21/c 1"
_symmetry_Int_Tables_number 14

_cell_length_a    4.40100
_cell_length_b    9.59670
_cell_length_c    6.98170
_cell_angle_alpha 90.00000
_cell_angle_beta  108.27400
_cell_angle_gamma 90.00000

loop_
_space_group_symop_id
_space_group_symop_operation_xyz
1 x,y,z
2 -x,y+1/2,-z+1/2
3 -x,-y,-z
4 x,-y+1/2,z+1/2

loop_
_atom_site_label
_atom_site_type_symbol
_atom_site_symmetry_multiplicity
_atom_site_Wyckoff_label
_atom_site_fract_x
_atom_site_fract_y
_atom_site_fract_z
_atom_site_occupancy
K1 K 4 e 0.41050 0.14070 0.21010 1.00000
N1 N 4 e -0.07320 0.41070 0.18390 1.00000
O1 O 4 e 0.20340 0.41270 0.30670 1.00000
O2 O 4 e 0.83850 0.29990 0.08650 1.00000
```

$KNO_2$ III: ABC2_mP16_14_e_e_2e - POSCAR

```
ABC2_mP16_14_e_e_2e & a,b/a,c/a,beta,x1,y1,z1,x2,y2,z2,x3,y3,z3,x4,y4,z4
     ↪ --params=4.401,2.18057259714,1.58638945694,108.274,0.4105,
     ↪ 0.1407,0.2101,-0.0732,0.4107,0.1839,0.2034,0.4127,0.3067,0.8385
     ↪ ,0.2999,0.0865 & P2_{1}/c C_{2h}^{5} #14 (e^4) & mP16 & None &
     ↪ KNO2 & Potassium nitrite & N. {Onoda-Yamamuro} et al., J. Phys.
     ↪ : Condens. Matter 10, 3341-3351 (1998)
   1.00000000000000
   4.40100000000000   0.00000000000000   0.00000000000000
   0.00000000000000   9.59670000000000   0.00000000000000
  -2.18919293920523   0.00000000000000   6.62959796404986
     K       N       O
     4       4       8
Direct
   0.41050000000000   0.14070000000000   0.21010000000000   K   (4e)
  -0.41050000000000   0.64070000000000   0.28990000000000   K   (4e)
  -0.41050000000000  -0.14070000000000  -0.21010000000000   K   (4e)
   0.41050000000000   0.35930000000000   0.71010000000000   K   (4e)
  -0.07320000000000   0.41070000000000   0.18390000000000   N   (4e)
   0.07320000000000   0.91070000000000   0.31610000000000   N   (4e)
   0.07320000000000  -0.41070000000000  -0.18390000000000   N   (4e)
  -0.07320000000000   0.08930000000000   0.68390000000000   N   (4e)
   0.20340000000000   0.41270000000000   0.30670000000000   O   (4e)
  -0.20340000000000   0.91270000000000   0.19330000000000   O   (4e)
  -0.20340000000000  -0.41270000000000  -0.30670000000000   O   (4e)
   0.20340000000000   0.08730000000000   0.80670000000000   O   (4e)
   0.83850000000000   0.29990000000000   0.08650000000000   O   (4e)
  -0.83850000000000   0.79990000000000   0.41350000000000   O   (4e)
  -0.83850000000000  -0.29990000000000  -0.08650000000000   O   (4e)
   0.83850000000000   0.20010000000000   0.58650000000000   O   (4e)
```

Manganite ($\gamma$-MnO(OH), $E0_6$): ABC2_mP16_14_e_e_2e - CIF

```
# CIF file
data_findsym-output
_audit_creation_method FINDSYM
```

```
_chemical_name_mineral 'Manganite'
_chemical_formula_sum 'H Mn O2'

loop_
_publ_author_name
 'T. Kohler'
 'T. Armbruster'
 'E. Libowitzky'
_journal_name_full_name
;
 Journal of Solid State Chemistry
;
_journal_volume 133
_journal_year 1997
_journal_page_first 486
_journal_page_last 500
_publ_Section_title
;
 Hydrogen Bonding and Jahn-Teller Distortion in Groutite, $\alpha$-MnOOH
    ↪ , and Manganite, $\gamma$-MnOOH, and Their Relations to the
    ↪ Manganese Dioxides Ramsdellite and Pyrolusite
;

_aflow_title 'Manganite ($\gamma$-MnO(OH), $E0_{6}$) Structure'
_aflow_proto 'ABC2_mP16_14_e_e_2e'
_aflow_params 'a,b/a,c/a,\beta,x_{1},y_{1},z_{1},x_{2},y_{2},z_{2},x_{3
    ↪ },y_{3},z_{3},x_{4},y_{4},z_{4}'
_aflow_params_values '5.304,0.994909502262,1.0,114.38,0.284,0.027,0.725,
    ↪ 0.76316,0.01033,0.75464,0.3749,0.1238,0.6279,0.8752,0.1256,
    ↪ 0.1206'
_aflow_Strukturbericht '$E0_{6}$'
_aflow_Pearson 'mP16'

_symmetry_space_group_name_H-M "P 1 21/c 1"
_symmetry_Int_Tables_number 14

_cell_length_a    5.30400
_cell_length_b    5.27700
_cell_length_c    5.30400
_cell_angle_alpha 90.00000
_cell_angle_beta  114.38000
_cell_angle_gamma 90.00000

loop_
_space_group_symop_id
_space_group_symop_operation_xyz
1 x,y,z
2 -x,y+1/2,-z+1/2
3 -x,-y,-z
4 x,-y+1/2,z+1/2

loop_
_atom_site_label
_atom_site_type_symbol
_atom_site_symmetry_multiplicity
_atom_site_Wyckoff_label
_atom_site_fract_x
_atom_site_fract_y
_atom_site_fract_z
_atom_site_occupancy
H1 H 4 e 0.28400 0.02700 0.72500 1.00000
Mn1 Mn 4 e 0.76316 0.01033 0.75464 1.00000
O1 O 4 e 0.37490 0.12380 0.62790 1.00000
O2 O 4 e 0.87520 0.12560 0.12060 1.00000
```

Manganite ($\gamma$-MnO(OH), $E0_6$): ABC2_mP16_14_e_e_2e - POSCAR

```
ABC2_mP16_14_e_e_2e & a,b/a,c/a,beta,x1,y1,z1,x2,y2,z2,x3,y3,z3,x4,y4,z4
    ↪ --params=5.304,0.994909502262,1.0,114.38,0.284,0.027,0.725,
    ↪ 0.76316,0.01033,0.75464,0.3749,0.1238,0.6279,0.8752,0.1256,
    ↪ 0.1206 & P2_{1}/c C_{2h}^{5} #14 (e^4) & mP16 & $E0_{6}$ &
    ↪ MnHO2 & Manganite & T. Kohler and T. Armbruster and E.
    ↪ Libowitzky, J. Solid State Chem. 133, 486-500 (1997)
   1.00000000000000
   5.30400000000000   0.00000000000000   0.00000000000000
   0.00000000000000   5.27700000000000   0.00000000000000
  -2.18941968137707   0.00000000000000   4.83103068286662
     H     Mn     O
     4      4     8
Direct
   0.28400000000000   0.02700000000000   0.72500000000000   H   (4e)
  -0.28400000000000   0.52700000000000  -0.22500000000000   H   (4e)
  -0.28400000000000  -0.02700000000000  -0.72500000000000   H   (4e)
   0.28400000000000   0.47300000000000   1.22500000000000   H   (4e)
   0.76316000000000   0.01033000000000   0.75464000000000   Mn  (4e)
  -0.76316000000000   0.51033000000000  -0.25464000000000   Mn  (4e)
  -0.76316000000000  -0.01033000000000  -0.75464000000000   Mn  (4e)
   0.76316000000000   0.48967000000000   1.25464000000000   Mn  (4e)
   0.37490000000000   0.12380000000000   0.62790000000000   O   (4e)
  -0.37490000000000   0.62380000000000  -0.12790000000000   O   (4e)
  -0.37490000000000  -0.12380000000000  -0.62790000000000   O   (4e)
   0.37490000000000   0.37620000000000   1.12790000000000   O   (4e)
   0.87520000000000   0.12560000000000   0.12060000000000   O   (4e)
  -0.87520000000000   0.62560000000000   0.37940000000000   O   (4e)
  -0.87520000000000  -0.12560000000000  -0.12060000000000   O   (4e)
   0.87520000000000   0.37440000000000   0.62060000000000   O   (4e)
```

$AgMnO_4$ ($H0_9$): ABC4_mP24_14_e_e_4e - CIF

```
# CIF file
data_findsym-output
_audit_creation_method FINDSYM

_chemical_name_mineral 'AgMnO4'
_chemical_formula_sum 'Ag Mn O4'

loop_
_publ_author_name
 'E. G. Boonstra'
_journal_name_full_name
;
 Acta Crystallographica Section B: Structural Science
;
_journal_volume 24
_journal_year 1968
_journal_page_first 1053
_journal_page_last 1062
_publ_Section_title
;
 The Crystal Structure of Silver Permanganate
;

_aflow_title 'AgMnO$_{4}$ ($H0_{9}$) Structure'
_aflow_proto 'ABC4_mP24_14_e_e_4e'
_aflow_params 'a,b/a,c/a,\beta,x_{1},y_{1},z_{1},x_{2},y_{2},z_{2},x_{3
    ↪ },y_{3},z_{3},x_{4},y_{4},z_{4},x_{5},y_{5},z_{5},x_{6},y_{6},
    ↪ z_{6}'
_aflow_params_values '5.64,1.47695035461,1.57941666667,126.99641,0.0792,
    ↪ 0.4672,0.8353,0.5965,0.3132,0.3392,0.7908,0.1508,0.4079,0.2708,
    ↪ 0.2767,0.2847,0.5599,0.381,0.1555,0.7557,0.441,0.5037'
_aflow_Strukturbericht '$H0_{9}$'
_aflow_Pearson 'mP24'

_symmetry_space_group_name_H-M "P 1 21/c 1"
_symmetry_Int_Tables_number 14

_cell_length_a    5.64000
_cell_length_b    8.33000
_cell_length_c    8.90791
_cell_angle_alpha 90.00000
_cell_angle_beta  126.99641
_cell_angle_gamma 90.00000

loop_
_space_group_symop_id
_space_group_symop_operation_xyz
1 x,y,z
2 -x,y+1/2,-z+1/2
3 -x,-y,-z
4 x,-y+1/2,z+1/2

loop_
_atom_site_label
_atom_site_type_symbol
_atom_site_symmetry_multiplicity
_atom_site_Wyckoff_label
_atom_site_fract_x
_atom_site_fract_y
_atom_site_fract_z
_atom_site_occupancy
Ag1 Ag 4 e 0.07920 0.46720 0.83530 1.00000
Mn1 Mn 4 e 0.59650 0.31320 0.33920 1.00000
O1 O 4 e 0.79080 0.15080 0.40790 1.00000
O2 O 4 e 0.27080 0.27670 0.28470 1.00000
O3 O 4 e 0.55990 0.38100 0.15550 1.00000
O4 O 4 e 0.75570 0.44100 0.50370 1.00000
```

$AgMnO_4$ ($H0_9$): ABC4_mP24_14_e_e_4e - POSCAR

```
ABC4_mP24_14_e_e_4e & a,b/a,c/a,beta,x1,y1,z1,x2,y2,z2,x3,y3,z3,x4,y4,z4
    ↪ ,x5,y5,z5,x6,y6,z6 --params=5.64,1.47695035461,1.57941666667,
    ↪ 126.99641,0.0792,0.4672,0.8353,0.5965,0.3132,0.3392,0.7908,
    ↪ 0.1508,0.4079,0.2708,0.2767,0.2847,0.5599,0.381,0.1555,0.7557,
    ↪ 0.441,0.5037 & P2_{1}/c C_{2h}^{5} #14 (e^6) & mP24 & $H0_{9}$
    ↪ & AgMnO4 & AgMnO4 & E. G. Boonstra, Acta Crystallogr. Sect. B
    ↪ Struct. Sci. 24, 1053-1062 (1968)
   1.00000000000000
   5.64000000000000   0.00000000000000   0.00000000000000
   0.00000000000000   8.33000000000000   0.00000000000000
  -5.36046829733267   0.00000000000000   7.11450913284897
    Ag     Mn     O
     4      4    16
Direct
   0.07920000000000   0.46720000000000   0.83530000000000   Ag  (4e)
  -0.07920000000000   0.96720000000000  -0.33530000000000   Ag  (4e)
  -0.07920000000000  -0.46720000000000  -0.83530000000000   Ag  (4e)
   0.07920000000000   0.03280000000000   1.33530000000000   Ag  (4e)
   0.59650000000000   0.31320000000000   0.33920000000000   Mn  (4e)
  -0.59650000000000   0.81320000000000   0.16080000000000   Mn  (4e)
  -0.59650000000000  -0.31320000000000  -0.33920000000000   Mn  (4e)
   0.59650000000000   0.18680000000000   0.83920000000000   Mn  (4e)
   0.79080000000000   0.15080000000000   0.40790000000000   O   (4e)
  -0.79080000000000   0.65080000000000   0.09210000000000   O   (4e)
  -0.79080000000000  -0.15080000000000  -0.40790000000000   O   (4e)
   0.79080000000000   0.34920000000000   0.90790000000000   O   (4e)
   0.27080000000000   0.27670000000000   0.28470000000000   O   (4e)
  -0.27080000000000   0.77670000000000   0.21530000000000   O   (4e)
  -0.27080000000000  -0.27670000000000  -0.28470000000000   O   (4e)
   0.27080000000000   0.22330000000000   0.78470000000000   O   (4e)
   0.55990000000000   0.38100000000000   0.15550000000000   O   (4e)
  -0.55990000000000   0.88100000000000   0.34450000000000   O   (4e)
  -0.55990000000000  -0.38100000000000  -0.15550000000000   O   (4e)
   0.55990000000000   0.11900000000000   0.65550000000000   O   (4e)
   0.75570000000000   0.44100000000000   0.50370000000000   O   (4e)
  -0.75570000000000   0.94100000000000  -0.00370000000000   O   (4e)
  -0.75570000000000  -0.44100000000000  -0.50370000000000   O   (4e)
   0.75570000000000   0.05900000000000   1.00370000000000   O   (4e)
```

Nahcolite ($NaHCO_3$, $G0_{12}$): ABCD3_mP24_14_e_e_e_3e - CIF

```
# CIF file
data_findsym-output
_audit_creation_method FINDSYM

_chemical_name_mineral 'Nacolite'
_chemical_formula_sum 'C H Na O3'

loop_
_publ_author_name
 'R. L. Sass'
 'R. F. Scheuerman'
_journal_name_full_name
;
 Acta Cristallographica
;
_journal_volume 15
_journal_year 1962
_journal_page_first 77
_journal_page_last 81
_publ_Section_title
;
 The Crystal Structure of Sodium Bicarbonate
;

# Found in The American Mineralogist Crystal Structure Database, 2003

_aflow_title 'Nahcolite (NaHCO$_{3}$, $G0_{12}$) Structure'
_aflow_proto 'ABCD3_mP24_14_e_e_e_3e'
_aflow_params 'a,b/a,c/a,\beta,x_{1},y_{1},z_{1},x_{2},y_{2},z_{2},x_{3
    ↪ },y_{3},z_{3},x_{4},y_{4},z_{4},x_{5},y_{5},z_{5},x_{6},y_{6},
    ↪ z_{6}'
_aflow_params_values '3.51,2.76638176638,2.29344729345,111.85,0.2098,
    ↪ 0.237,-0.0768,0.74205,0.2539,0.1773,0.4274,0.0047,0.7145,0.1896
    ↪ ,0.3668,-0.0709,-0.0117,0.1629,0.7946,0.4958,0.1707,0.06'
_aflow_Strukturbericht '$G0_{12}$'
_aflow_Pearson 'mP24'

_symmetry_space_group_name_H-M "P 1 21/c 1"
_symmetry_Int_Tables_number 14

_cell_length_a      3.51000
_cell_length_b      9.71000
_cell_length_c      8.05000
_cell_angle_alpha 90.00000
_cell_angle_beta  111.85000
_cell_angle_gamma 90.00000

loop_
_space_group_symop_id
_space_group_symop_operation_xyz
1 x,y,z
2 -x,y+1/2,-z+1/2
3 -x,-y,-z
4 x,-y+1/2,z+1/2

loop_
_atom_site_label
_atom_site_type_symbol
_atom_site_symmetry_multiplicity
_atom_site_Wyckoff_label
_atom_site_fract_x
_atom_site_fract_y
_atom_site_fract_z
_atom_site_occupancy
C1 C 4 e 0.20980 0.23700 -0.07680 1.00000
H1 H 4 e 0.74205 0.25390 0.17730 1.00000
Na1 Na 4 e 0.42740 0.00470 0.71450 1.00000
O1 O 4 e 0.18960 0.36680 -0.07090 1.00000
O2 O 4 e -0.01170 0.16290 0.79460 1.00000
O3 O 4 e 0.49580 0.17070 0.06000 1.00000
```

Nahcolite (NaHCO$_3$, $G0_{12}$): ABCD3_mP24_14_e_e_e_3e - POSCAR

```
ABCD3_mP24_14_e_e_e_3e & a,b/a,c/a,beta,x1,y1,z1,x2,y2,z2,x3,y3,z3,x4,y4
    ↪ ,z4,x5,y5,z5,x6,y6,z6 --params=3.51,2.76638176638,2.29344729345
    ↪ ,111.85,0.2098,0.237,-0.0768,0.74205,0.2539,0.1773,0.4274,
    ↪ 0.0047,0.7145,0.1896,0.3668,-0.0709,-0.0117,0.1629,0.7946,
    ↪ 0.4958,0.1707,0.06 & P2_{1}/c C_{2h}^{5} #14 (e^6) & mP24 &
    ↪ $G0_{12}$ & CHNaO3 & Nacolite & R. L. Sass and R. F. Scheuerman
    ↪ , Acta Cryst. 15, 77-81 (1962)
   1.00000000000000
   3.51000000000000   0.00000000000000   0.00000000000000
   0.00000000000000   9.71000000000000   0.00000000000000
  -2.99603250376187   0.00000000000000   7.47169922015082
     C      H     Na      O
     4      4      4     12
Direct
   0.20980000000000   0.23700000000000  -0.07680000000000   C   (4e)
  -0.20980000000000   0.73700000000000   0.57680000000000   C   (4e)
  -0.20980000000000  -0.23700000000000   0.07680000000000   C   (4e)
   0.20980000000000   0.26300000000000   0.42320000000000   C   (4e)
   0.74205000000000   0.25390000000000   0.17730000000000   H   (4e)
  -0.74205000000000   0.75390000000000   0.32270000000000   H   (4e)
  -0.74205000000000  -0.25390000000000  -0.17730000000000   H   (4e)
   0.74205000000000   0.24610000000000   0.67730000000000   H   (4e)
   0.42740000000000   0.00470000000000   0.71450000000000   Na  (4e)
  -0.42740000000000   0.50470000000000  -0.21450000000000   Na  (4e)
  -0.42740000000000  -0.00470000000000  -0.71450000000000   Na  (4e)
   0.42740000000000   0.49530000000000   1.21450000000000   Na  (4e)
   0.18960000000000   0.36680000000000  -0.07090000000000   O   (4e)
  -0.18960000000000   0.86680000000000   0.57090000000000   O   (4e)
  -0.18960000000000  -0.36680000000000   0.07090000000000   O   (4e)
   0.18960000000000   0.13320000000000   0.42910000000000   O   (4e)
  -0.01170000000000   0.16290000000000   0.79460000000000   O   (4e)
   0.01170000000000   0.66290000000000  -0.29460000000000   O   (4e)
   0.01170000000000  -0.16290000000000  -0.79460000000000   O   (4e)
  -0.01170000000000   0.33710000000000   1.29460000000000   O   (4e)
   0.49580000000000   0.17070000000000   0.06000000000000   O   (4e)
  -0.49580000000000   0.67070000000000   0.44000000000000   O   (4e)
  -0.49580000000000  -0.17070000000000  -0.06000000000000   O   (4e)
   0.49580000000000   0.32930000000000   0.56000000000000   O   (4e)
```

Cu(OH)Cl: ABCD_mP16_14_e_e_e_e - CIF

```
# CIF file
data_findsym-output
_audit_creation_method FINDSYM

_chemical_name_mineral 'ClCuHO'
_chemical_formula_sum 'Cl Cu H O'

loop_
_publ_author_name
 'Y. Cudennec'
 'A. Riou'
 'Y. G\\'{e}rault'
 'A. Lecerf'
_journal_name_full_name
;
 Journal of Solid State Chemistry
;
_journal_volume 151
_journal_year 2000
_journal_page_first 308
_journal_page_last 312
_publ_Section_title
;
 Synthesis and Crystal Structures of Cd(OH)Cl and Cu(OH)Cl and
    ↪ Relationship to Brucite Type
;

_aflow_title 'Cu(OH)Cl Structure'
_aflow_proto 'ABCD_mP16_14_e_e_e_e'
_aflow_params 'a,b/a,c/a,\beta,x_{1},y_{1},z_{1},x_{2},y_{2},z_{2},x_{3
    ↪ },y_{3},z_{3},x_{4},y_{4},z_{4}'
_aflow_params_values '6.2953,1.05871046654,0.882880879386,118.138,0.3115
    ↪ ,0.0907,0.1334,0.03201,0.11772,0.28577,0.674,0.128,0.528,0.8807
    ↪ ,0.1478,0.5318'
_aflow_Strukturbericht 'None'
_aflow_Pearson 'mP16'

_symmetry_space_group_name_H-M "P 1 21/c 1"
_symmetry_Int_Tables_number 14

_cell_length_a      6.29530
_cell_length_b      6.66490
_cell_length_c      5.55800
_cell_angle_alpha 90.00000
_cell_angle_beta  118.13800
_cell_angle_gamma 90.00000

loop_
_space_group_symop_id
_space_group_symop_operation_xyz
1 x,y,z
2 -x,y+1/2,-z+1/2
3 -x,-y,-z
4 x,-y+1/2,z+1/2

loop_
_atom_site_label
_atom_site_type_symbol
_atom_site_symmetry_multiplicity
_atom_site_Wyckoff_label
_atom_site_fract_x
_atom_site_fract_y
_atom_site_fract_z
_atom_site_occupancy
Cl1 Cl 4 e 0.31150 0.09070 0.13340 1.00000
Cu1 Cu 4 e 0.03201 0.11772 0.28577 1.00000
H1 H 4 e 0.67400 0.12800 0.52800 1.00000
O1 O 4 e 0.88070 0.14780 0.53180 1.00000
```

Cu(OH)Cl: ABCD_mP16_14_e_e_e_e - POSCAR

```
ABCD_mP16_14_e_e_e_e & a,b/a,c/a,beta,x1,y1,z1,x2,y2,z2,x3,y3,z3,x4,y4,
    ↪ z4 --params=6.2953,1.05871046654,0.882880879386,118.138,0.3115,
    ↪ 0.0907,0.1334,0.03201,0.11772,0.28577,0.674,0.128,0.528,0.8807,
    ↪ 0.1478,0.5318 & P2_{1}/c C_{2h}^{5} #14 (e^4) & mP16 & None &
    ↪ ClCuHO & ClCuHO & Y. Cudennec et al., J. Solid State Chem. 151,
    ↪ 308-312 (2000)
   1.00000000000000
   6.29530000000000   0.00000000000000   0.00000000000000
   0.00000000000000   6.66490000000000   0.00000000000000
  -2.62113516045273   0.00000000000000   4.90112379670606
    Cl     Cu      H      O
     4      4      4      4
Direct
   0.31150000000000   0.09070000000000   0.13340000000000   Cl  (4e)
  -0.31150000000000   0.59070000000000   0.36660000000000   Cl  (4e)
  -0.31150000000000  -0.09070000000000  -0.13340000000000   Cl  (4e)
   0.31150000000000   0.40930000000000   0.63340000000000   Cl  (4e)
   0.03201000000000   0.11772000000000   0.28577000000000   Cu  (4e)
  -0.03201000000000   0.61772000000000   0.21423000000000   Cu  (4e)
  -0.03201000000000  -0.11772000000000  -0.28577000000000   Cu  (4e)
   0.03201000000000   0.38228000000000   0.78577000000000   Cu  (4e)
   0.67400000000000   0.12800000000000   0.52800000000000   H   (4e)
  -0.67400000000000   0.62800000000000  -0.02800000000000   H   (4e)
  -0.67400000000000  -0.12800000000000  -0.52800000000000   H   (4e)
   0.67400000000000   0.37200000000000   1.02800000000000   H   (4e)
```

```
   0.88070000000000    0.14780000000000    0.53180000000000     O    (4e)
  -0.88070000000000    0.64780000000000   -0.03180000000000     O    (4e)
  -0.88070000000000   -0.14780000000000   -0.53180000000000     O    (4e)
   0.88070000000000    0.35220000000000    1.03180000000000     O    (4e)
```

Arsenopyrite (FeAsS, $E0_7$): ABC_mP12_14_e_e_e - CIF

```
# CIF file
data_findsym-output
_audit_creation_method FINDSYM

_chemical_name_mineral 'Arsenopyrite'
_chemical_formula_sum 'As Fe S'

loop_
_publ_author_name
 'L. Bindi'
 'Y. Mo\"{e}lo'
 'P. L\\'eone'
 'M. Suchaud'
_journal_name_full_name
;
 Canadian Mineralogist
;
_journal_volume 50
_journal_year 2012
_journal_page_first 471
_journal_page_last 479
_publ_Section_title
;
 Stoichiometric Arsenopyrite, FeAsS, from La Roche-Balue Quarry,
       ↪ Loire-Atlantique, France: Crystal Structure And M\"{o}ssbauer
       ↪ Study
;

_aflow_title 'Arsenopyrite (FeAsS, $E0_{7}$) Structure'
_aflow_proto 'ABC_mP12_14_e_e_e'
_aflow_params 'a,b/a,c/a,\beta,x_{1},y_{1},z_{1},x_{2},y_{2},z_{2},x_{3
       ↪ },y_{3},z_{3}'
_aflow_params_values '5.7612,0.986617371381,1.00107616469,111.721,
       ↪ 0.14746,0.13055,0.86937,0.28353,-0.00643,0.29429,0.6551,0.1311,
       ↪ 0.3211'
_aflow_Strukturbericht '$E0_{7}$'
_aflow_Pearson 'mP12'

_symmetry_space_group_name_H-M "P 1 21/c 1"
_symmetry_Int_Tables_number 14

_cell_length_a       5.76120
_cell_length_b       5.68410
_cell_length_c       5.76740
_cell_angle_alpha  90.00000
_cell_angle_beta   111.72100
_cell_angle_gamma  90.00000

loop_
_space_group_symop_id
_space_group_symop_operation_xyz
1 x,y,z
2 -x,y+1/2,-z+1/2
3 -x,-y,-z
4 x,-y+1/2,z+1/2

loop_
_atom_site_label
_atom_site_type_symbol
_atom_site_symmetry_multiplicity
_atom_site_Wyckoff_label
_atom_site_fract_x
_atom_site_fract_y
_atom_site_fract_z
_atom_site_occupancy
As1 As 4 e 0.14746 0.13055 0.86937 1.00000
Fe1 Fe 4 e 0.28353 -0.00643 0.29429 1.00000
S1 S 4 e 0.65510 0.13110 0.32110 1.00000
```

Arsenopyrite (FeAsS, $E0_7$): ABC_mP12_14_e_e_e - POSCAR

```
ABC_mP12_14_e_e_e & a,b/a,c/a,beta,x1,y1,z1,x2,y2,z2,x3,y3,z3 --params=
       ↪ 5.7612,0.986617371381,1.00107616469,111.721,0.14746,0.13055,
       ↪ 0.86937,0.28353,-0.00643,0.29429,0.6551,0.1311,0.3211 & P2_{1}/
       ↪ c C_{2h}^{5} #14 (e^3) & mP12 & $E0_{7}$ & AsFeS & Arsenopyrite
       ↪ & L. Bindi et al., Can. Mineral. 50, 471-479 (2012)
   1.00000000000000
   5.76120000000000    0.00000000000000    0.00000000000000
   0.00000000000000    5.68410000000000    0.00000000000000
  -2.13444136312758    0.00000000000000    5.35789723934400
     As       Fe        S
      4        4        4
Direct
   0.14746000000000    0.13055000000000    0.86937000000000    As    (4e)
  -0.14746000000000    0.63055000000000   -0.36937000000000    As    (4e)
  -0.14746000000000   -0.13055000000000   -0.86937000000000    As    (4e)
   0.14746000000000    0.36945000000000    1.36937000000000    As    (4e)
   0.28353000000000   -0.00643000000000    0.29429000000000    Fe    (4e)
  -0.28353000000000    0.49357000000000    0.20571000000000    Fe    (4e)
  -0.28353000000000    0.00643000000000   -0.29429000000000    Fe    (4e)
   0.28353000000000    0.50643000000000    0.79429000000000    Fe    (4e)
   0.65510000000000    0.13110000000000    0.32110000000000     S    (4e)
  -0.65510000000000    0.63110000000000    0.17890000000000     S    (4e)
  -0.65510000000000   -0.13110000000000   -0.32110000000000     S    (4e)
   0.65510000000000    0.36890000000000    0.82110000000000     S    (4e)
```

$\alpha$-ICl: AB_mP16_14_2e_2e - CIF

```
# CIF file
data_findsym-output
_audit_creation_method FINDSYM

_chemical_name_mineral 'ClI'
_chemical_formula_sum 'Cl I'

loop_
_publ_author_name
 'K. H. Boswijk'
 'J. {van der Heide}'
 'A. Vos'
 'E. H. Wiebenga'
_journal_name_full_name
;
 Acta Cristallographica
;
_journal_volume 9
_journal_year 1956
_journal_page_first 274
_journal_page_last 277
_publ_Section_title
;
 The crystal structure of $\alpha$-ICl
;

_aflow_title '$\alpha$-ICl Structure'
_aflow_proto 'AB_mP16_14_2e_2e'
_aflow_params 'a,b/a,c/a,\beta,x_{1},y_{1},z_{1},x_{2},y_{2},z_{2},x_{3
       ↪ },y_{3},z_{3},x_{4},y_{4},z_{4}'
_aflow_params_values '12.6,0.347619047619,0.944444444444,119.5,0.084,
       ↪ 0.152,0.706,0.462,0.858,0.62,0.179,0.366,0.588,0.297,0.632,
       ↪ 0.436'
_aflow_Strukturbericht 'None'
_aflow_Pearson 'mP16'

_symmetry_space_group_name_H-M "P 1 21/c 1"
_symmetry_Int_Tables_number 14

_cell_length_a       12.60000
_cell_length_b       4.38000
_cell_length_c       11.90000
_cell_angle_alpha  90.00000
_cell_angle_beta   119.50000
_cell_angle_gamma  90.00000

loop_
_space_group_symop_id
_space_group_symop_operation_xyz
1 x,y,z
2 -x,y+1/2,-z+1/2
3 -x,-y,-z
4 x,-y+1/2,z+1/2

loop_
_atom_site_label
_atom_site_type_symbol
_atom_site_symmetry_multiplicity
_atom_site_Wyckoff_label
_atom_site_fract_x
_atom_site_fract_y
_atom_site_fract_z
_atom_site_occupancy
Cl1 Cl 4 e 0.08400 0.15200 0.70600 1.00000
Cl2 Cl 4 e 0.46200 0.85800 0.62000 1.00000
I1 I 4 e 0.17900 0.36600 0.58800 1.00000
I2 I 4 e 0.29700 0.63200 0.43600 1.00000
```

$\alpha$-ICl: AB_mP16_14_2e_2e - POSCAR

```
AB_mP16_14_2e_2e & a,b/a,c/a,beta,x1,y1,z1,x2,y2,z2,x3,y3,z3,x4,y4,z4 --
       ↪ params=12.6,0.347619047619,0.944444444444,119.5,0.084,0.152,
       ↪ 0.706,0.462,0.858,0.62,0.179,0.366,0.588,0.297,0.632,0.436 &
       ↪ P2_{1}/c C_{2h}^{5} #14 (e^4) & mP16 & None & ClI & ClI & K. H.
       ↪ Boswijk et al., Acta Cryst. 9, 274-277 (1956)
   1.00000000000000
  12.60000000000000    0.00000000000000    0.00000000000000
   0.00000000000000    4.38000000000000    0.00000000000000
  -5.85984036523126    0.00000000000000   10.35723278168480
     Cl        I
      8        8
Direct
   0.08400000000000    0.15200000000000    0.70600000000000    Cl    (4e)
  -0.08400000000000    0.65200000000000   -0.20600000000000    Cl    (4e)
  -0.08400000000000   -0.15200000000000   -0.70600000000000    Cl    (4e)
   0.08400000000000    0.34800000000000    1.20600000000000    Cl    (4e)
   0.46200000000000    0.85800000000000    0.62000000000000    Cl    (4e)
  -0.46200000000000    1.35800000000000   -0.12000000000000    Cl    (4e)
  -0.46200000000000   -0.85800000000000   -0.62000000000000    Cl    (4e)
   0.46200000000000   -0.35800000000000    1.12000000000000    Cl    (4e)
   0.17900000000000    0.36600000000000    0.58800000000000     I    (4e)
  -0.17900000000000    0.86600000000000   -0.08800000000000     I    (4e)
  -0.17900000000000   -0.36600000000000   -0.58800000000000     I    (4e)
   0.17900000000000    0.13400000000000    1.08800000000000     I    (4e)
   0.29700000000000    0.63200000000000    0.43600000000000     I    (4e)
  -0.29700000000000    1.13200000000000    0.06400000000000     I    (4e)
  -0.29700000000000   -0.63200000000000   -0.43600000000000     I    (4e)
   0.29700000000000   -0.13200000000000    0.93600000000000     I    (4e)
```

LiAs: AB_mP16_14_2e_2e - CIF

```
# CIF file
data_findsym-output
_audit_creation_method FINDSYM
```

```
_chemical_name_mineral 'AsLi'
_chemical_formula_sum 'As Li'

loop_
_publ_author_name
 'D. T. Cromer'
_journal_name_full_name
;
 Acta Cristallographica
;
_journal_volume 12
_journal_year 1959
_journal_page_first 36
_journal_page_last 41
_publ_Section_title
;
 The Crystal Structure of LiAs
;

# Found in Pearson's Handbook of Crystallographic Data for Intermetallic
    ↪ Phases, 1991

_aflow_title 'LiAs Structure'
_aflow_proto 'AB_mP16_14_2e_2e'
_aflow_params 'a,b/a,c/a,\beta,x_{1},y_{1},z_{1},x_{2},y_{2},z_{2},x_{3
    ↪ },y_{3},z_{3},x_{4},y_{4},z_{4}'
_aflow_params_values '5.79,0.905008635579,1.84801381693,117.4,0.3042,
    ↪ 0.9143,0.2992,0.2891,0.1626,0.1011,0.235,0.402,0.329,0.232,
    ↪ 0.669,0.045'
_aflow_Strukturbericht 'None'
_aflow_Pearson 'mP16'

_symmetry_space_group_name_H-M "P 1 21/c 1"
_symmetry_Int_Tables_number 14

_cell_length_a    5.79000
_cell_length_b    5.24000
_cell_length_c    10.70000
_cell_angle_alpha 90.00000
_cell_angle_beta  117.40000
_cell_angle_gamma 90.00000

loop_
_space_group_symop_id
_space_group_symop_operation_xyz
1 x,y,z
2 -x,y+1/2,-z+1/2
3 -x,-y,-z
4 x,-y+1/2,z+1/2

loop_
_atom_site_label
_atom_site_type_symbol
_atom_site_symmetry_multiplicity
_atom_site_Wyckoff_label
_atom_site_fract_x
_atom_site_fract_y
_atom_site_fract_z
_atom_site_occupancy
As1 As 4 e 0.30420 0.91430 0.29920 1.00000
As2 As 4 e 0.28910 0.16260 0.10110 1.00000
Li1 Li 4 e 0.23500 0.40200 0.32900 1.00000
Li2 Li 4 e 0.23200 0.66900 0.04500 1.00000
```

LiAs: AB_mP16_14_2e_2e - POSCAR

```
AB_mP16_14_2e_2e & a,b/a,c/a,beta,x1,y1,z1,x2,y2,z2,x3,y3,z3,x4,y4,z4 --
    ↪ params=5.79,0.905008635579,1.84801381693,117.4,0.3042,0.9143,
    ↪ 0.2992,0.2891,0.1626,0.1011,0.235,0.402,0.329,0.232,0.669,0.045
    ↪ & P2_{1}/c C_{2h}^{5} #14 (e^4) & mP16 & None & AsLi & AsLi &
    ↪ D. T. Cromer, Acta Cryst. 12, 36-41 (1959)
   1.00000000000000
   5.79000000000000   0.00000000000000   0.00000000000000
   0.00000000000000   5.24000000000000   0.00000000000000
  -4.92413769718721   0.00000000000000   9.49962462095949
    As    Li
     8     8
Direct
   0.30420000000000   0.91430000000000   0.29920000000000   As   (4e)
  -0.30420000000000   1.41430000000000   0.20080000000000   As   (4e)
  -0.30420000000000  -0.91430000000000  -0.29920000000000   As   (4e)
   0.30420000000000  -0.41430000000000   0.79920000000000   As   (4e)
   0.28910000000000   0.16260000000000   0.10110000000000   As   (4e)
  -0.28910000000000   0.66260000000000   0.39890000000000   As   (4e)
  -0.28910000000000  -0.16260000000000  -0.10110000000000   As   (4e)
   0.28910000000000   0.33740000000000   0.60110000000000   As   (4e)
   0.23500000000000   0.40200000000000   0.32900000000000   Li   (4e)
  -0.23500000000000   0.90200000000000   0.17100000000000   Li   (4e)
  -0.23500000000000  -0.40200000000000  -0.32900000000000   Li   (4e)
   0.23500000000000   0.09800000000000   0.82900000000000   Li   (4e)
   0.23200000000000   0.66900000000000   0.04500000000000   Li   (4e)
  -0.23200000000000   1.16900000000000   0.45500000000000   Li   (4e)
  -0.23200000000000  -0.66900000000000  -0.04500000000000   Li   (4e)
   0.23200000000000  -0.16900000000000   0.54500000000000   Li   (4e)
```

$\epsilon$-1,2,3,4,5,6-Hexachlorocyclohexane ($C_6Cl_6$): AB_mP24_14_3e_3e - CIF

```
# CIF file
data_findsym-output
_audit_creation_method FINDSYM

_chemical_name_mineral '$\epsilon$-1,2,3,4,5,6-hexachlorocyclohexane'
_chemical_formula_sum 'C Cl'

loop_
_publ_author_name
 'N. Norman'
_journal_name_full_name
;
 Acta Chemica Scandinavica
;
_journal_volume 4
_journal_year 1950
_journal_page_first 251
_journal_page_last 259
_publ_Section_title
;
 The Crystal Structure of the Epsilon Isomer of 1,2,3,4,5,6-Hexachloro\
    ↪ em{cyclo}hexane
;

_aflow_title '$\epsilon$-1,2,3,4,5,6-Hexachlorocyclohexane (C$_{6}$Cl$_{
    ↪ 6}$) Structure'
_aflow_proto 'AB_mP24_14_3e_3e'
_aflow_params 'a,b/a,c/a,\beta,x_{1},y_{1},z_{1},x_{2},y_{2},z_{2},x_{3
    ↪ },y_{3},z_{3},x_{4},y_{4},z_{4},x_{5},y_{5},z_{5},x_{6},y_{6},
    ↪ z_{6}'
_aflow_params_values '7.02,1.67948717949,0.968660968661,112.0,0.206,
    ↪ 0.013,0.005,0.089,0.115,0.095,-0.035,0.041,0.177,0.369,-0.071,
    ↪ 0.212,0.27,0.188,0.312,0.81,0.13,0.298'
_aflow_Strukturbericht 'None'
_aflow_Pearson 'mP24'

_symmetry_space_group_name_H-M "P 1 21/c 1"
_symmetry_Int_Tables_number 14

_cell_length_a    7.02000
_cell_length_b    11.79000
_cell_length_c    6.80000
_cell_angle_alpha 90.00000
_cell_angle_beta  112.00000
_cell_angle_gamma 90.00000

loop_
_space_group_symop_id
_space_group_symop_operation_xyz
1 x,y,z
2 -x,y+1/2,-z+1/2
3 -x,-y,-z
4 x,-y+1/2,z+1/2

loop_
_atom_site_label
_atom_site_type_symbol
_atom_site_symmetry_multiplicity
_atom_site_Wyckoff_label
_atom_site_fract_x
_atom_site_fract_y
_atom_site_fract_z
_atom_site_occupancy
C1 C 4 e 0.20600 0.01300 0.00500 1.00000
C2 C 4 e 0.08900 0.11500 0.09500 1.00000
C3 C 4 e -0.03500 0.04100 0.17700 1.00000
Cl1 Cl 4 e 0.36900 -0.07100 0.21200 1.00000
Cl2 Cl 4 e 0.27000 0.18800 0.31200 1.00000
Cl3 Cl 4 e 0.81000 0.13000 0.29800 1.00000
```

$\epsilon$-1,2,3,4,5,6-Hexachlorocyclohexane ($C_6Cl_6$): AB_mP24_14_3e_3e - POSCAR

```
AB_mP24_14_3e_3e & a,b/a,c/a,beta,x1,y1,z1,x2,y2,z2,x3,y3,z3,x4,y4,z4,x5
    ↪ ,y5,z5,x6,y6,z6 --params=7.02,1.67948717949,0.968660968661,
    ↪ 112.0,0.206,0.013,0.005,0.089,0.115,0.095,-0.035,0.041,0.177,
    ↪ 0.369,-0.071,0.212,0.27,0.188,0.312,0.81,0.13,0.298 & P2_{1}/c
    ↪ C_{2h}^{5} #14 (e^6) & mP24 & None & C6Cl6 & $\epsilon$-1,2,3,4
    ↪ ,5,6-hexachlorocyclohexane & N. Norman, Acta Chem. Scand. 4,
    ↪ 251-259 (1950)
   1.00000000000000
   7.02000000000000   0.00000000000000   0.00000000000000
   0.00000000000000  11.79000000000000   0.00000000000000
  -2.54732483522820   0.00000000000000   6.30485021105415
     C    Cl
    12    12
Direct
   0.20600000000000   0.01300000000000   0.00500000000000   C   (4e)
  -0.20600000000000   0.51300000000000   0.49500000000000   C   (4e)
  -0.20600000000000  -0.01300000000000  -0.00500000000000   C   (4e)
   0.20600000000000   0.48700000000000   0.50500000000000   C   (4e)
   0.08900000000000   0.11500000000000   0.09500000000000   C   (4e)
  -0.08900000000000   0.61500000000000   0.40500000000000   C   (4e)
  -0.08900000000000  -0.11500000000000  -0.09500000000000   C   (4e)
   0.08900000000000   0.38500000000000   0.59500000000000   C   (4e)
  -0.03500000000000   0.04100000000000   0.17700000000000   C   (4e)
   0.03500000000000   0.54100000000000   0.32300000000000   C   (4e)
   0.03500000000000  -0.04100000000000  -0.17700000000000   C   (4e)
  -0.03500000000000   0.45900000000000   0.67700000000000   C   (4e)
   0.36900000000000  -0.07100000000000   0.21200000000000   Cl  (4e)
  -0.36900000000000   0.42900000000000   0.28800000000000   Cl  (4e)
  -0.36900000000000   0.07100000000000  -0.21200000000000   Cl  (4e)
   0.36900000000000   0.57100000000000   0.71200000000000   Cl  (4e)
   0.27000000000000   0.18800000000000   0.31200000000000   Cl  (4e)
  -0.27000000000000   0.68800000000000   0.18800000000000   Cl  (4e)
  -0.27000000000000  -0.18800000000000  -0.31200000000000   Cl  (4e)
   0.27000000000000   0.31200000000000   0.81200000000000   Cl  (4e)
   0.81000000000000   0.13000000000000   0.29800000000000   Cl  (4e)
  -0.81000000000000   0.63000000000000   0.20200000000000   Cl  (4e)
  -0.81000000000000  -0.13000000000000  -0.29800000000000   Cl  (4e)
   0.81000000000000   0.37000000000000   0.79800000000000   Cl  (4e)
```

Pararealgar (AsS): AB_mP32_14_4e_4e - CIF

```
# CIF file
data_findsym-output
_audit_creation_method FINDSYM

_chemical_name_mineral 'Pararealgar'
_chemical_formula_sum 'As S'

loop_
_publ_author_name
 'P. Bonazzi'
 'S. Menchetti'
 'G. Pratesi'
_journal_name_full_name
;
 American Mineralogist
;
_journal_volume 80
_journal_year 1995
_journal_page_first 400
_journal_page_last 403
_publ_Section_title
;
 The crystal structure of pararealgar, As$_{4}$S$_{4}$
;

_aflow_title 'Pararealgar (AsS) Structure'
_aflow_proto 'AB_mP32_14_4e_4e'
_aflow_params 'a,b/a,c/a,\beta,x_{1},y_{1},z_{1},x_{2},y_{2},z_{2},x_{3
    ↪ },y_{3},z_{3},x_{4},y_{4},z_{4},x_{5},y_{5},z_{5},x_{6},y_{6},
    ↪ z_{6},x_{7},y_{7},z_{7},x_{8},y_{8},z_{8}'
_aflow_params_values '9.909,0.97436673731,0.858007871632,97.29,0.3187,
    ↪ 0.6355,0.0432,0.0819,0.5427,0.3252,0.3698,0.3607,0.3431,0.1455,
    ↪ 0.3439,0.1643,0.1645,0.7187,0.1923,0.2537,0.4782,0.5099,0.4703,
    ↪ 0.5276,0.2192,0.1964,0.4483,-0.0492'
_aflow_Strukturbericht 'None'
_aflow_Pearson 'mP32'

_symmetry_space_group_name_H-M "P 1 21/c 1"
_symmetry_Int_Tables_number 14

_cell_length_a    9.90900
_cell_length_b    9.65500
_cell_length_c    8.50200
_cell_angle_alpha 90.00000
_cell_angle_beta  97.29000
_cell_angle_gamma 90.00000

loop_
_space_group_symop_id
_space_group_symop_operation_xyz
1 x,y,z
2 -x,y+1/2,-z+1/2
3 -x,-y,-z
4 x,-y+1/2,z+1/2

loop_
_atom_site_label
_atom_site_type_symbol
_atom_site_symmetry_multiplicity
_atom_site_Wyckoff_label
_atom_site_fract_x
_atom_site_fract_y
_atom_site_fract_z
_atom_site_occupancy
As1 As 4 e 0.31870 0.63550 0.04320 1.00000
As2 As 4 e 0.08190 0.54270 0.32520 1.00000
As3 As 4 e 0.36980 0.36070 0.34310 1.00000
As4 As 4 e 0.14550 0.34390 0.16430 1.00000
S1 S 4 e 0.16450 0.71870 0.19230 1.00000
S2 S 4 e 0.25370 0.47820 0.50990 1.00000
S3 S 4 e 0.47030 0.52760 0.21920 1.00000
S4 S 4 e 0.19640 0.44830 -0.04920 1.00000
```

Pararealgar (AsS): AB_mP32_14_4e_4e - POSCAR

```
AB_mP32_14_4e_4e & a,b/a,c/a,beta,x1,y1,z1,x2,y2,z2,x3,y3,z3,x4,y4,z4,x5
    ↪ ,y5,z5,x6,y6,z6,x7,y7,z7,x8,y8,z8 --params=9.909,0.97436673731,
    ↪ 0.858007871632,97.29,0.3187,0.6355,0.0432,0.0819,0.5427,0.3252,
    ↪ 0.3698,0.3607,0.3431,0.1455,0.3439,0.1643,0.1645,0.7187,0.1923,
    ↪ 0.2537,0.4782,0.5099,0.4703,0.5276,0.2192,0.1964,0.4483,-0.0492
    ↪ & P2_{1}/c C_{2h}^{5} #14 (e^8) & mP32 & None & AsS &
    ↪ Pararealgar & P. Bonazzi and S. Menchetti and G. Pratesi, Am.
    ↪ Mineral. 80, 400-403 (1995)
   1.00000000000000
   9.90900000000000   0.00000000000000   0.00000000000000
   0.00000000000000   9.65500000000000   0.00000000000000
  -1.07883143461801   0.00000000000000   8.43327497095168
     As        S
     16       16
Direct
   0.31870000000000   0.63550000000000   0.04320000000000   As   (4e)
  -0.31870000000000   1.13550000000000   0.45680000000000   As   (4e)
  -0.31870000000000  -0.63550000000000  -0.04320000000000   As   (4e)
   0.31870000000000  -0.13550000000000   0.54320000000000   As   (4e)
   0.08190000000000   0.54270000000000   0.32520000000000   As   (4e)
  -0.08190000000000   1.04270000000000   0.17480000000000   As   (4e)
  -0.08190000000000  -0.54270000000000  -0.32520000000000   As   (4e)
   0.08190000000000  -0.04270000000000   0.82520000000000   As   (4e)
   0.36980000000000   0.36070000000000   0.34310000000000   As   (4e)
  -0.36980000000000   0.86070000000000   0.15690000000000   As   (4e)
  -0.36980000000000  -0.36070000000000  -0.34310000000000   As   (4e)
   0.36980000000000   0.13930000000000   0.84310000000000   As   (4e)
   0.14550000000000   0.34390000000000   0.16430000000000   As   (4e)
  -0.14550000000000   0.84390000000000   0.33570000000000   As   (4e)
  -0.14550000000000  -0.34390000000000  -0.16430000000000   As   (4e)
   0.14550000000000   0.15610000000000   0.66430000000000   As   (4e)
   0.16450000000000   0.71870000000000   0.19230000000000   S    (4e)
  -0.16450000000000   1.21870000000000   0.30770000000000   S    (4e)
  -0.16450000000000  -0.71870000000000  -0.19230000000000   S    (4e)
   0.16450000000000  -0.21870000000000   0.69230000000000   S    (4e)
   0.25370000000000   0.47820000000000   0.50990000000000   S    (4e)
  -0.25370000000000   0.97820000000000  -0.00990000000000   S    (4e)
  -0.25370000000000  -0.47820000000000  -0.50990000000000   S    (4e)
   0.25370000000000   0.02180000000000   1.00990000000000   S    (4e)
   0.47030000000000   0.52760000000000   0.21920000000000   S    (4e)
  -0.47030000000000   1.02760000000000   0.28080000000000   S    (4e)
  -0.47030000000000  -0.52760000000000  -0.21920000000000   S    (4e)
   0.47030000000000  -0.02760000000000   0.71920000000000   S    (4e)
   0.19640000000000   0.44830000000000  -0.04920000000000   S    (4e)
  -0.19640000000000   0.94830000000000   0.54920000000000   S    (4e)
  -0.19640000000000  -0.44830000000000   0.04920000000000   S    (4e)
   0.19640000000000   0.05170000000000   0.45080000000000   S    (4e)
```

Realgar (AsS, $B_l$): AB_mP32_14_4e_4e - CIF

```
# CIF file
data_findsym-output
_audit_creation_method FINDSYM

_chemical_name_mineral 'Realgar'
_chemical_formula_sum 'As S'

loop_
_publ_author_name
 'T. Ito'
 'N. Morimoto'
 'R. Sadanaga'
_journal_name_full_name
;
 Acta Cristallographica
;
_journal_volume 5
_journal_year 1952
_journal_page_first 755
_journal_page_last 782
_publ_Section_title
;
 The Crystal Structure of Realgar
;

_aflow_title 'Realgar (AsS, $B_{l}$) Structure'
_aflow_proto 'AB_mP32_14_4e_4e'
_aflow_params 'a,b/a,c/a,\beta,x_{1},y_{1},z_{1},x_{2},y_{2},z_{2},x_{3
    ↪ },y_{3},z_{3},x_{4},y_{4},z_{4},x_{5},y_{5},z_{5},x_{6},y_{6},
    ↪ z_{6},x_{7},y_{7},z_{7},x_{8},y_{8},z_{8}'
_aflow_params_values '6.56,2.05792682927,1.47941615854,113.75283,0.359,
    ↪ 0.524,0.118,0.567,0.36,0.425,0.137,0.373,0.318,0.328,0.339,
    ↪ 0.038,0.641,0.508,0.346,0.093,0.524,0.213,0.608,0.275,0.245,
    ↪ 0.067,0.285,0.115'
_aflow_Strukturbericht '$B_{l}$'
_aflow_Pearson 'mP32'

_symmetry_space_group_name_H-M "P 1 21/c 1"
_symmetry_Int_Tables_number 14

_cell_length_a    6.56000
_cell_length_b    13.50000
_cell_length_c    9.70497
_cell_angle_alpha 90.00000
_cell_angle_beta  113.75283
_cell_angle_gamma 90.00000

loop_
_space_group_symop_id
_space_group_symop_operation_xyz
1 x,y,z
2 -x,y+1/2,-z+1/2
3 -x,-y,-z
4 x,-y+1/2,z+1/2

loop_
_atom_site_label
_atom_site_type_symbol
_atom_site_symmetry_multiplicity
_atom_site_Wyckoff_label
_atom_site_fract_x
_atom_site_fract_y
_atom_site_fract_z
_atom_site_occupancy
As1 As 4 e 0.35900 0.52400 0.11800 1.00000
As2 As 4 e 0.56700 0.36000 0.42500 1.00000
As3 As 4 e 0.13700 0.37300 0.31800 1.00000
As4 As 4 e 0.32800 0.33900 0.03800 1.00000
S1 S 4 e 0.64100 0.50800 0.34600 1.00000
S2 S 4 e 0.09300 0.52400 0.21300 1.00000
S3 S 4 e 0.60800 0.27500 0.24500 1.00000
S4 S 4 e 0.06700 0.28500 0.11500 1.00000
```

Realgar (AsS, $B_l$): AB_mP32_14_4e_4e - POSCAR

```
AB_mP32_14_4e_4e & a,b/a,c/a,beta,x1,y1,z1,x2,y2,z2,x3,y3,z3,x4,y4,z4,x5
    ↪ ,y5,z5,x6,y6,z6,x7,y7,z7,x8,y8,z8 --params=6.56,2.05792682927,
    ↪ 1.47941615854,113.75283,0.359,0.524,0.118,0.567,0.36,0.425,
    ↪ 0.137,0.373,0.318,0.328,0.339,0.038,0.641,0.508,0.346,0.093,
    ↪ 0.524,0.213,0.608,0.275,0.245,0.067,0.285,0.115 & P2_{1}/c C_{
    ↪ 2h}^{5} #14 (e^8) & mP32 & $B_{1}$ & AsS & Realgar & T. Ito and
    ↪  N. Morimoto and R. Sadanaga, Acta Cryst. 5, 755-782 (1952)
   1.00000000000000
   6.56000000000000   0.00000000000000   0.00000000000000
   0.00000000000000  13.50000000000000   0.00000000000000
```

```
   -3.90908329779061     0.00000000000000     8.88287737570628
      As       S
      16      16
Direct
    0.35900000000000     0.52400000000000     0.11800000000000    As    (4e)
   -0.35900000000000     1.02400000000000     0.38200000000000    As    (4e)
   -0.35900000000000    -0.52400000000000    -0.11800000000000    As    (4e)
    0.35900000000000    -0.02400000000000     0.61800000000000    As    (4e)
    0.56700000000000     0.36000000000000     0.42500000000000    As    (4e)
   -0.56700000000000     0.86000000000000     0.07500000000000    As    (4e)
   -0.56700000000000    -0.36000000000000    -0.42500000000000    As    (4e)
    0.56700000000000     0.14000000000000     0.92500000000000    As    (4e)
    0.13700000000000     0.37300000000000     0.31800000000000    As    (4e)
   -0.13700000000000     0.87300000000000     0.18200000000000    As    (4e)
   -0.13700000000000    -0.37300000000000    -0.31800000000000    As    (4e)
    0.13700000000000     0.12700000000000     0.81800000000000    As    (4e)
    0.32800000000000     0.33900000000000     0.03800000000000    As    (4e)
   -0.32800000000000     0.83900000000000     0.46200000000000    As    (4e)
   -0.32800000000000    -0.33900000000000    -0.03800000000000    As    (4e)
    0.32800000000000     0.16100000000000     0.53800000000000    As    (4e)
    0.64100000000000     0.50800000000000     0.34600000000000     S    (4e)
   -0.64100000000000     1.00800000000000     0.15400000000000     S    (4e)
   -0.64100000000000    -0.50800000000000    -0.34600000000000     S    (4e)
    0.64100000000000    -0.00800000000000     0.84600000000000     S    (4e)
    0.09300000000000     0.52400000000000     0.21300000000000     S    (4e)
   -0.09300000000000     1.02400000000000     0.28700000000000     S    (4e)
   -0.09300000000000    -0.52400000000000    -0.21300000000000     S    (4e)
    0.09300000000000    -0.02400000000000     0.71300000000000     S    (4e)
    0.60800000000000     0.27500000000000     0.24500000000000     S    (4e)
   -0.60800000000000     0.77500000000000     0.25500000000000     S    (4e)
   -0.60800000000000    -0.27500000000000    -0.24500000000000     S    (4e)
    0.60800000000000     0.22500000000000     0.74500000000000     S    (4e)
    0.06700000000000     0.28500000000000     0.11500000000000     S    (4e)
   -0.06700000000000     0.78500000000000     0.38500000000000     S    (4e)
   -0.06700000000000    -0.28500000000000    -0.11500000000000     S    (4e)
    0.06700000000000     0.21500000000000     0.61500000000000     S    (4e)
```

$Ag_2PbO_2$: A2B2C_mC20_15_ad_f_e - CIF

```
# CIF file
data_findsym-output
_audit_creation_method FINDSYM

_chemical_name_mineral 'Ag2O2Pb'
_chemical_formula_sum 'Ag2 O2 Pb'

loop_
_publ_author_name
 'A. Bystr\"{o}m'
 'L. Evers'
_journal_name_full_name
;
 Acta Chemica Scandinavica
;
_journal_volume 4
_journal_year 1950
_journal_page_first 613
_journal_page_last 627
_publ_Section_title
;
 The Crystal Structures of Ag$_{2}$PbO$_{2}$ and Ag5Pb$_{2}$O$_{6}$
;

_aflow_title 'Ag$_{2}$PbO$_{2}$ Structure'
_aflow_proto 'A2B2C_mC20_15_ad_f_e'
_aflow_params 'a,b/a,c/a,\beta,y_{3},x_{4},y_{4},z_{4}'
_aflow_params_values '8.65,1.00751445087,0.703121387283,130.86424,0.625,
    ↪ 0.804,0.445,0.865'
_aflow_Strukturbericht 'None'
_aflow_Pearson 'mC20'

_symmetry_space_group_name_H-M "C 1 2/c 1"
_symmetry_Int_Tables_number 15

_cell_length_a      8.65000
_cell_length_b      8.71500
_cell_length_c      6.08200
_cell_angle_alpha  90.00000
_cell_angle_beta  130.86424
_cell_angle_gamma  90.00000

loop_
_space_group_symop_id
_space_group_symop_operation_xyz
1 x,y,z
2 -x,y,-z+1/2
3 -x,-y,-z
4 x,-y,z+1/2
5 x+1/2,y+1/2,z
6 -x+1/2,y+1/2,-z+1/2
7 -x+1/2,-y+1/2,-z
8 x+1/2,-y+1/2,z+1/2

loop_
_atom_site_label
_atom_site_type_symbol
_atom_site_symmetry_multiplicity
_atom_site_Wyckoff_label
_atom_site_fract_x
_atom_site_fract_y
_atom_site_fract_z
_atom_site_occupancy
Ag1 Ag 4 a 0.00000 0.00000 0.00000 1.00000
Ag2 Ag 4 d 0.25000 0.25000 0.50000 1.00000
Pb1 Pb 4 e 0.00000 0.62500 0.25000 1.00000
O1 O 8 f 0.80400 0.44500 0.86500 1.00000
```

$Ag_2PbO_2$: A2B2C_mC20_15_ad_f_e - POSCAR

```
A2B2C_mC20_15_ad_f_e & a,b/a,c/a,beta,y3,x4,y4,z4 --params=8.65,
    ↪ 1.00751445087,0.703121387283,130.86424,0.625,0.804,0.445,0.865
    ↪ & C2/c C_{2h}^{6} #15 (adef) & mC20 & None & Ag2O2Pb & Ag2O2Pb
    ↪ & A. Bystr\"{o}m and L. Evers, Acta Chem. Scand. 4, 613-627 (
    ↪ 1950)
   1.00000000000000
   4.32500000000000    -4.35750000000000     0.00000000000000
   4.32500000000000     4.35750000000000     0.00000000000000
  -3.97926366629002     0.00000000000000     4.59958527197226
     Ag       O      Pb
      4       4       2
Direct
   0.00000000000000     0.00000000000000     0.00000000000000    Ag    (4a)
   0.00000000000000     0.00000000000000     0.50000000000000    Ag    (4a)
   0.00000000000000     0.50000000000000     0.50000000000000    Ag    (4d)
   0.50000000000000     0.00000000000000     0.00000000000000    Ag    (4d)
   0.35900000000000     1.24900000000000     0.86500000000000     O    (8f)
  -1.24900000000000    -0.35900000000000    -0.36500000000000     O    (8f)
  -0.35900000000000    -1.24900000000000    -0.86500000000000     O    (8f)
   1.24900000000000     0.35900000000000     1.36500000000000     O    (8f)
  -0.62500000000000     0.62500000000000     0.25000000000000    Pb    (4e)
   0.62500000000000    -0.62500000000000     0.75000000000000    Pb    (4e)
```

Catapleiite ($Na_2ZrSi_3O_9{\cdot}2H_2O$): A2B3C9D3E_mC144_15_2f_bcdef_9f_3f_ae - CIF

```
# CIF file
data_findsym-output
_audit_creation_method FINDSYM

_chemical_name_mineral 'Catapleiite'
_chemical_formula_sum '(H2O)2 Na3 O9 Si3 Zr'

loop_
_publ_author_name
 'G. D. Ilyushin'
 'A. A. Voronkov'
 'V. V. Ilyukhin'
 'N. N. Nevskii'
 'N. V. Belov'
_journal_name_full_name
;
 Doklady Akademii Nauk SSSR
;
_journal_volume 260
_journal_year 1981
_journal_page_first 623
_journal_page_last 627
_publ_Section_title
;
 Crystal structure of natural monoclinic catapleiite, Na$_{2}$ZrSi$_{3}
    ↪ $O$_{9} \cdot$ 2H$_{2}$O
;

# Found in The American Mineralogist Crystal Structure Database, 2003

_aflow_title 'Catapleiite (Na$_{2}$ZrSi$_{3}$O$_{9}\cdot$2H$_{2}$O)
    ↪ Structure'
_aflow_proto 'A2B3C9D3E_mC144_15_2f_bcdef_9f_3f_ae'
_aflow_params 'a,b/a,c/a,\beta,y_{5},y_{6},x_{7},y_{7},z_{7},x_{8},y_{8
    ↪ },z_{8},x_{9},y_{9},z_{9},x_{10},y_{10},z_{10},x_{11},y_{11},z_
    ↪ {11},x_{12},y_{12},z_{12},x_{13},y_{13},z_{13},x_{14},y_{14},z_
    ↪ {14},x_{15},y_{15},z_{15},x_{16},y_{16},z_{16},x_{17},y_{17},z_
    ↪ {17},x_{18},y_{18},z_{18},x_{19},y_{19},z_{19},x_{20},y_{20},z_
    ↪ {20},x_{21},y_{21},z_{21}'
_aflow_params_values '23.8927,0.311057352246,0.843270120162,147.42281,
    ↪ 0.4941,-0.0074,0.8315,0.4922,0.0362,0.3371,0.0092,0.8816,0.2531
    ↪ ,0.2432,0.255,0.8661,0.0088,-0.0767,0.3666,0.49,0.0586,0.0493,
    ↪ 0.4955,0.1747,0.0711,0.1973,0.1285,0.0641,0.1942,0.2558,0.2265,
    ↪ 0.3221,0.3544,0.562,0.298,0.1203,0.569,0.2938,0.2625,0.7239,
    ↪ 0.1738,0.3455,0.5976,0.2004,0.2221,0.0995,0.2923,0.2252,0.1972,
    ↪ 0.0015,0.5723'
_aflow_Strukturbericht 'None'
_aflow_Pearson 'mC144'

_symmetry_space_group_name_H-M "C 1 2/c 1"
_symmetry_Int_Tables_number 15

_cell_length_a     23.89270
_cell_length_b      7.43200
_cell_length_c     20.14800
_cell_angle_alpha  90.00000
_cell_angle_beta  147.42281
_cell_angle_gamma  90.00000

loop_
_space_group_symop_id
_space_group_symop_operation_xyz
1 x,y,z
2 -x,y,-z+1/2
3 -x,-y,-z
4 x,-y,z+1/2
5 x+1/2,y+1/2,z
6 -x+1/2,y+1/2,-z+1/2
7 -x+1/2,-y+1/2,-z
8 x+1/2,-y+1/2,z+1/2

loop_
_atom_site_label
_atom_site_type_symbol
_atom_site_symmetry_multiplicity
_atom_site_Wyckoff_label
```

```
_atom_site_fract_x
_atom_site_fract_y
_atom_site_fract_z
_atom_site_occupancy
Zr1  Zr  4  a  0.00000  0.00000  0.00000  1.00000
Na1  Na  4  b  0.00000  0.50000  0.00000  1.00000
Na2  Na  4  c  0.25000  0.25000  0.00000  1.00000
Na3  Na  4  d  0.25000  0.25000  0.50000  1.00000
Na4  Na  4  e  0.00000  0.49410  0.25000  1.00000
Zr2  Zr  4  e  0.00000  -0.00740  0.25000  1.00000
H2O1  H2O  8  f  0.83150  0.49220  0.03620  1.00000
H2O2  H2O  8  f  0.33710  0.00920  0.88160  1.00000
Na5  Na  8  f  0.25310  0.24320  0.25500  1.00000
O1  O  8  f  0.86610  0.00880  -0.07670  1.00000
O2  O  8  f  0.36660  0.49000  0.05860  1.00000
O3  O  8  f  0.04930  0.49550  0.17470  1.00000
O4  O  8  f  0.07110  0.19730  0.12850  1.00000
O5  O  8  f  0.06410  0.19420  0.25580  1.00000
O6  O  8  f  0.22650  0.32210  0.35440  1.00000
O7  O  8  f  0.56200  0.29800  0.12030  1.00000
O8  O  8  f  0.56900  0.29380  0.26250  1.00000
O9  O  8  f  0.72390  0.17380  0.34550  1.00000
Si1  Si  8  f  0.59760  0.20040  0.22210  1.00000
Si2  Si  8  f  0.09950  0.29230  0.22520  1.00000
Si3  Si  8  f  0.19720  0.00150  0.57230  1.00000
```

Catapleiite ($Na_2ZrSi_3O_9 \cdot 2H_2O$): A2B3C9D3E_mC144_15_2f_bcdef_9f_3f_ae - POSCAR

```
A2B3C9D3E_mC144_15_2f_bcdef_9f_3f_ae & a,b/a,c/a,beta,y5,y6,x7,y7,z7,x8,
    ↪ y8,z8,x9,y9,z9,x10,y10,z10,x11,y11,z11,x12,y12,z12,x13,y13,z13,
    ↪ x14,y14,z14,x15,y15,z15,x16,y16,z16,x17,y17,z17,x18,y18,z18,x19
    ↪ ,y19,z19,x20,y20,z20,x21,y21,z21 --params=23.8927,
    ↪ 0.311057352246,0.843270120162,147.42281,0.4941,-0.0074,0.8315,
    ↪ 0.4922,0.0362,0.3371,0.0092,0.8816,0.2531,0.2432,0.255,0.8661,
    ↪ 0.0088,-0.0767,0.3666,0.49,0.0586,0.0493,0.4955,0.1747,0.0711,
    ↪ 0.1973,0.1285,0.0641,0.1942,0.2558,0.2265,0.3221,0.3544,0.562,
    ↪ 0.298,0.1203,0.569,0.2938,0.2625,0.7239,0.1738,0.3455,0.5976,
    ↪ 0.2004,0.2221,0.0995,0.2923,0.2252,0.1972,0.0015,0.5723 & C2/c
    ↪ C_{2h}^{6} #15 (abcde^2f^15) & mC144 & None & (H2O)2Na2O9Si3Zr
    ↪ & Catapleiite & G. D. Ilyushin et al., Doklady Akademii Nauk
    ↪ SSSR 260, 623-627 (1981)
   1.00000000000000
  11.94635000000000   -3.71600000000000    0.00000000000000
  11.94635000000000    3.71600000000000    0.00000000000000
 -16.97805109064560    0.00000000000000   10.84839551101580
    H2O      Na       O       Si      Zr
     8       12      36       12       4
Direct
   0.33930000000000    1.32370000000000    0.03620000000000   H2O   (8f)
  -1.32370000000000   -0.33930000000000    0.46380000000000   H2O   (8f)
  -0.33930000000000   -1.32370000000000   -0.03620000000000   H2O   (8f)
   1.32370000000000    0.33930000000000    0.53620000000000   H2O   (8f)
   0.32790000000000    0.34630000000000    0.88160000000000   H2O   (8f)
  -0.34630000000000   -0.32790000000000   -0.38160000000000   H2O   (8f)
  -0.32790000000000   -0.34630000000000   -0.88160000000000   H2O   (8f)
   0.34630000000000    0.32790000000000    1.38160000000000   H2O   (8f)
   0.50000000000000    0.50000000000000    0.00000000000000    Na   (4b)
   0.50000000000000    0.50000000000000    0.50000000000000    Na   (4b)
   0.00000000000000    0.50000000000000    0.00000000000000    Na   (4c)
   0.50000000000000    0.00000000000000    0.50000000000000    Na   (4c)
   0.00000000000000    0.50000000000000    0.50000000000000    Na   (4d)
   0.50000000000000    0.00000000000000    0.00000000000000    Na   (4d)
  -0.49410000000000    0.49410000000000    0.25000000000000    Na   (4e)
   0.49410000000000   -0.49410000000000    0.75000000000000    Na   (4e)
   0.00990000000000    0.49630000000000    0.25500000000000    Na   (8f)
  -0.49630000000000   -0.00990000000000    0.24500000000000    Na   (8f)
  -0.00990000000000   -0.49630000000000   -0.25500000000000    Na   (8f)
   0.49630000000000    0.00990000000000    0.75500000000000    Na   (8f)
   0.85730000000000    0.87490000000000   -0.07670000000000     O   (8f)
  -0.87490000000000   -0.85730000000000    0.57670000000000     O   (8f)
  -0.85730000000000   -0.87490000000000    0.07670000000000     O   (8f)
   0.87490000000000    0.85730000000000    0.42330000000000     O   (8f)
  -0.12340000000000    0.85660000000000    0.05860000000000     O   (8f)
  -0.85660000000000    0.12340000000000    0.44140000000000     O   (8f)
   0.12340000000000   -0.85660000000000   -0.05860000000000     O   (8f)
   0.85660000000000   -0.12340000000000    0.55860000000000     O   (8f)
  -0.44620000000000    0.54480000000000    0.17470000000000     O   (8f)
  -0.54480000000000    0.44620000000000    0.32530000000000     O   (8f)
   0.44620000000000   -0.54480000000000   -0.17470000000000     O   (8f)
   0.54480000000000   -0.44620000000000    0.67470000000000     O   (8f)
  -0.12620000000000    0.26840000000000    0.12850000000000     O   (8f)
  -0.26840000000000    0.12620000000000    0.37150000000000     O   (8f)
   0.12620000000000   -0.26840000000000   -0.12850000000000     O   (8f)
   0.26840000000000   -0.12620000000000    0.62850000000000     O   (8f)
  -0.13010000000000    0.25830000000000    0.25580000000000     O   (8f)
  -0.25830000000000    0.13010000000000    0.24420000000000     O   (8f)
   0.13010000000000   -0.25830000000000   -0.25580000000000     O   (8f)
   0.25830000000000   -0.13010000000000    0.75580000000000     O   (8f)
  -0.09560000000000    0.54860000000000    0.35440000000000     O   (8f)
  -0.54860000000000    0.09560000000000    0.14560000000000     O   (8f)
   0.09560000000000   -0.54860000000000   -0.35440000000000     O   (8f)
   0.54860000000000   -0.09560000000000    0.85440000000000     O   (8f)
   0.26400000000000    0.86000000000000    0.12030000000000     O   (8f)
  -0.86000000000000   -0.26400000000000    0.37970000000000     O   (8f)
  -0.26400000000000   -0.86000000000000   -0.12030000000000     O   (8f)
   0.86000000000000    0.26400000000000    0.62030000000000     O   (8f)
   0.27520000000000    0.86280000000000    0.26250000000000     O   (8f)
  -0.86280000000000   -0.27520000000000    0.23750000000000     O   (8f)
  -0.27520000000000   -0.86280000000000   -0.26250000000000     O   (8f)
   0.86280000000000    0.27520000000000    0.76250000000000     O   (8f)
   0.55010000000000    0.89770000000000    0.34550000000000     O   (8f)
  -0.89770000000000   -0.55010000000000    0.15450000000000     O   (8f)
  -0.55010000000000   -0.89770000000000   -0.34550000000000     O   (8f)
   0.89770000000000    0.55010000000000    0.84550000000000     O   (8f)
   0.39720000000000    0.79800000000000    0.22210000000000    Si   (8f)
  -0.79800000000000   -0.39720000000000    0.27790000000000    Si   (8f)
  -0.39720000000000   -0.79800000000000   -0.22210000000000    Si   (8f)
   0.79800000000000    0.39720000000000    0.72210000000000    Si   (8f)
  -0.19280000000000    0.39180000000000    0.22520000000000    Si   (8f)
  -0.39180000000000    0.19280000000000    0.27480000000000    Si   (8f)
   0.19280000000000   -0.39180000000000   -0.22520000000000    Si   (8f)
   0.39180000000000   -0.19280000000000    0.72520000000000    Si   (8f)
   0.19570000000000    0.19870000000000    0.57230000000000    Si   (8f)
  -0.19870000000000   -0.19570000000000   -0.07230000000000    Si   (8f)
  -0.19570000000000   -0.19870000000000   -0.57230000000000    Si   (8f)
   0.19870000000000    0.19570000000000    1.07230000000000    Si   (8f)
   0.00000000000000    0.00000000000000    0.00000000000000    Zr   (4a)
   0.00000000000000    0.00000000000000    0.50000000000000    Zr   (4a)
   0.00740000000000   -0.00740000000000    0.25000000000000    Zr   (4e)
  -0.00740000000000    0.00740000000000    0.75000000000000    Zr   (4e)
```

$Na_2PrO_3$: A2B3C_mC48_15_aef_3f_2e - CIF

```
# CIF file
data_findsym-output
_audit_creation_method FINDSYM

_chemical_name_mineral 'Na2O3Pr'
_chemical_formula_sum 'Na2 O3 Pr'

loop_
_publ_author_name
 'Y. Hinatsu'
 'Y. Doi'
_journal_name_full_name
;
 Journal of Alloys and Compounds
;
_journal_volume 418
_journal_year 2006
_journal_page_first 155
_journal_page_last 160
_publ_Section_title
;
 Crystal structures and magnetic properties of alkali-metal lanthanide
    ↪ oxides $A_{2}$$Ln$O$_{3}$ ($A$ = Li, Na; $Ln$ = Ce, Pr, Tb)
;

_aflow_title 'Na$_{2}$PrO$_{3}$ Structure'
_aflow_proto 'A2B3C_mC48_15_aef_3f_2e'
_aflow_params 'a,b/a,c/a,\beta,y_{2},y_{3},y_{4},x_{5},y_{5},z_{5},x_{6
    ↪ },y_{6},z_{6},x_{7},y_{7},z_{7},x_{8},y_{8},z_{8}'
_aflow_params_values '5.9649,1.72911532465,1.96968934936,110.09,0.8146,
    ↪ 0.1681,0.4994,-0.04,0.325,-0.004,0.241,0.501,0.141,0.251,0.135,
    ↪ 0.144,0.301,0.861,0.143'
_aflow_Strukturbericht 'None'
_aflow_Pearson 'mC48'

_symmetry_space_group_name_H-M "C 1 2/c 1"
_symmetry_Int_Tables_number 15

_cell_length_a     5.96490
_cell_length_b     10.31400
_cell_length_c     11.74900
_cell_angle_alpha  90.00000
_cell_angle_beta   110.09000
_cell_angle_gamma  90.00000

loop_
_space_group_symop_id
_space_group_symop_operation_xyz
1 x,y,z
2 -x,y,-z+1/2
3 -x,-y,-z
4 x,-y,z+1/2
5 x+1/2,y+1/2,z
6 -x+1/2,y+1/2,-z+1/2
7 -x+1/2,-y+1/2,-z
8 x+1/2,-y+1/2,z+1/2

loop_
_atom_site_label
_atom_site_type_symbol
_atom_site_symmetry_multiplicity
_atom_site_Wyckoff_label
_atom_site_fract_x
_atom_site_fract_y
_atom_site_fract_z
_atom_site_occupancy
Na1  Na  4  a  0.00000  0.00000  0.00000  1.00000
Na2  Na  4  e  0.00000  0.81460  0.25000  1.00000
Pr1  Pr  4  e  0.00000  0.16810  0.25000  1.00000
Pr2  Pr  4  e  0.00000  0.49940  0.25000  1.00000
Na3  Na  8  f  -0.04000  0.32500  -0.00400  1.00000
O1  O  8  f  0.24100  0.50100  0.14100  1.00000
O2  O  8  f  0.25100  0.13500  0.14400  1.00000
O3  O  8  f  0.30100  0.86100  0.14300  1.00000
```

$Na_2PrO_3$: A2B3C_mC48_15_aef_3f_2e - POSCAR

```
A2B3C_mC48_15_aef_3f_2e & a,b/a,c/a,beta,y2,y3,y4,x5,y5,z5,x6,y6,z6,x7,
    ↪ y7,z7,x8,y8,z8 --params=5.9649,1.72911532465,1.96968934936,
    ↪ 110.09,0.8146,0.1681,0.4994,-0.04,0.325,-0.004,0.241,0.501,
    ↪ 0.141,0.251,0.135,0.144,0.301,0.861,0.143 & C2/c C_{2h}^{6} #15
    ↪ (ae^3f^4) & mC48 & None & Na2O3Pr & Na2O3Pr & Y. Hinatsu and
    ↪ Y. Doi, J. Alloys Compd. 418, 155-160 (2006)
   1.00000000000000
   2.98245000000000   -5.15700000000000    0.00000000000000
   2.98245000000000    5.15700000000000    0.00000000000000
  -4.03573199541645    0.00000000000000   11.03412290402690
```

```
   Na     O    Pr
    8    12     4
Direct
   0.00000000000000    0.00000000000000    0.00000000000000   Na   (4a)
   0.00000000000000    0.00000000000000    0.50000000000000   Na   (4a)
  -0.81460000000000    0.81460000000000    0.25000000000000   Na   (4e)
   0.81460000000000   -0.81460000000000    0.75000000000000   Na   (4e)
  -0.36500000000000    0.28500000000000   -0.00400000000000   Na   (8f)
  -0.28500000000000    0.36500000000000    0.50400000000000   Na   (8f)
   0.36500000000000   -0.28500000000000    0.00400000000000   Na   (8f)
   0.28500000000000   -0.36500000000000    0.49600000000000   Na   (8f)
  -0.26000000000000    0.74200000000000    0.14100000000000    O   (8f)
  -0.74200000000000    0.26000000000000    0.35900000000000    O   (8f)
   0.26000000000000   -0.74200000000000   -0.14100000000000    O   (8f)
   0.74200000000000   -0.26000000000000    0.64100000000000    O   (8f)
   0.11600000000000    0.38600000000000    0.14400000000000    O   (8f)
  -0.38600000000000   -0.11600000000000    0.35600000000000    O   (8f)
  -0.11600000000000   -0.38600000000000   -0.14400000000000    O   (8f)
   0.38600000000000    0.11600000000000    0.64400000000000    O   (8f)
  -0.56000000000000    1.16200000000000    0.14300000000000    O   (8f)
  -1.16200000000000    0.56000000000000    0.35700000000000    O   (8f)
   0.56000000000000   -1.16200000000000   -0.14300000000000    O   (8f)
   1.16200000000000   -0.56000000000000    0.64300000000000    O   (8f)
  -0.16810000000000    0.16810000000000    0.25000000000000   Pr   (4e)
   0.16810000000000   -0.16810000000000    0.75000000000000   Pr   (4e)
  -0.49940000000000    0.49940000000000    0.25000000000000   Pr   (4e)
   0.49940000000000   -0.49940000000000    0.75000000000000   Pr   (4e)
```

Eudidymite (BeHNa$O_8$$Si_3$: A2B4C2D17E6_mC124_15_f_2f_f_e8f_3f - CIF

```
# CIF file
data_findsym-output
_audit_creation_method FINDSYM

_chemical_name_mineral 'Eudidymite'
_chemical_formula_sum 'Be2 H4 Na2 O17 Si6'

loop_
_publ_author_name
 'G. {Diego Gatta}'
 'N. Rotiroti'
 'G. J. {McIntyre}'
 'A. Guastoni'
 'F. Nestola'
_journal_name_full_name
;
 American Mineralogist
;
_journal_volume 93
_journal_year 2008
_journal_page_first 1158
_journal_page_last 1165
_publ_Section_title
;
 New insights into the crystal chemistry of epididymite and eudidymite
     ↪ from Malosa, Malawi: A single-crystal neutron diffraction
     ↪ study
;

_aflow_title 'Eudidymite (BeHNaO$_{8}$Si$_{3}$ Structure'
_aflow_proto 'A2B4C2D17E6_mC124_15_f_2f_f_e8f_3f'
_aflow_params 'a,b/a,c/a,\beta,y_{1},x_{2},y_{2},z_{2},x_{3},y_{3},z_{3
     ↪ },x_{4},y_{4},z_{4},x_{5},y_{5},z_{5},x_{6},y_{6},z_{6},x_{7},
     ↪ y_{7},z_{7},x_{8},y_{8},z_{8},x_{9},y_{9},z_{9},x_{10},y_{10},
     ↪ z_{10},x_{11},y_{11},z_{11},x_{12},y_{12},z_{12},x_{13},y_{13},
     ↪ z_{13},x_{14},y_{14},z_{14},x_{15},y_{15},z_{15},x_{16},y_{16},
     ↪ z_{16}'
_aflow_params_values '12.6188,0.584691095825,1.10898025169,103.762,
     ↪ 0.17321,0.16817,0.32317,0.4971,0.084,0.4645,0.7449,0.4773,
     ↪ 0.0419,0.7486,0.02899,0.34078,-0.07147,0.0727,0.03247,0.62614,
     ↪ 0.24036,0.1803,0.1154,0.24045,0.16471,0.57054,0.23677,0.43859,
     ↪ 0.24484,0.1548,0.49864,0.05623,0.12854,0.05189,0.87541,0.05237,
     ↪ 0.25694,0.43729,0.01209,0.43098,0.74466,0.20415,0.01965,0.64127
     ↪ ,0.25166,0.1025,0.86648,0.02964,0.09397,0.36195'
_aflow_Strukturbericht 'None'
_aflow_Pearson 'mC124'

_symmetry_space_group_name_H-M "C 1 2/c 1"
_symmetry_Int_Tables_number 15

_cell_length_a     12.61880
_cell_length_b     7.37810
_cell_length_c     13.99400
_cell_angle_alpha  90.00000
_cell_angle_beta   103.76200
_cell_angle_gamma  90.00000

loop_
_space_group_symop_id
_space_group_symop_operation_xyz
1 x,y,z
2 -x,y,-z+1/2
3 -x,-y,-z
4 x,-y,z+1/2
5 x+1/2,y+1/2,z
6 -x+1/2,y+1/2,-z+1/2
7 -x+1/2,-y+1/2,-z
8 x+1/2,-y+1/2,z+1/2

loop_
_atom_site_label
_atom_site_type_symbol
_atom_site_symmetry_multiplicity
_atom_site_Wyckoff_label
_atom_site_fract_x
_atom_site_fract_y
_atom_site_fract_z
_atom_site_occupancy
O1 O 4 e 0.00000 0.17321 0.25000 1.00000
Be1 Be 8 f 0.16817 0.32317 0.49710 1.00000
H1 H 8 f 0.08400 0.46450 0.74490 1.00000
H2 H 8 f 0.47730 0.04190 0.74860 1.00000
Na1 Na 8 f 0.02899 0.34078 -0.07147 1.00000
O2 O 8 f 0.07270 0.03247 0.62614 1.00000
O3 O 8 f 0.24036 0.18030 0.11540 1.00000
O4 O 8 f 0.24045 0.16471 0.57054 1.00000
O5 O 8 f 0.23677 0.43859 0.24484 1.00000
O6 O 8 f 0.15480 0.49864 0.05623 1.00000
O7 O 8 f 0.12854 0.05189 0.87541 1.00000
O8 O 8 f 0.05237 0.25694 0.43729 1.00000
O9 O 8 f 0.01209 0.43098 0.74466 1.00000
Si1 Si 8 f 0.20415 0.01965 0.64127 1.00000
Si2 Si 8 f 0.25166 0.10250 0.86648 1.00000
Si3 Si 8 f 0.02964 0.09397 0.36195 1.00000
```

Eudidymite (BeHNa$O_8$$Si_3$: A2B4C2D17E6_mC124_15_f_2f_f_e8f_3f - POSCAR

```
A2B4C2D17E6_mC124_15_f_2f_f_e8f_3f & a,b/a,c/a,beta,y1,x2,y2,z2,x3,y3,z3
     ↪ ,x4,y4,z4,x5,y5,z5,x6,y6,z6,x7,y7,z7,x8,y8,z8,x9,y9,z9,x10,y10,
     ↪ z10,x11,y11,z11,x12,y12,z12,x13,y13,z13,x14,y14,z14,x15,y15,z15
     ↪ ,x16,y16,z16 --params=12.6188,0.584691095825,1.10898025169,
     ↪ 103.762,0.17321,0.16817,0.32317,0.4971,0.084,0.4645,0.7449,
     ↪ 0.4773,0.0419,0.7486,0.02899,0.34078,-0.07147,0.0727,0.03247,
     ↪ 0.62614,0.24036,0.1803,0.1154,0.24045,0.16471,0.57054,0.23677,
     ↪ 0.43859,0.24484,0.1548,0.49864,0.05623,0.12854,0.05189,0.87541,
     ↪ 0.05237,0.25694,0.43729,0.01209,0.43098,0.74466,0.20415,0.01965
     ↪ ,0.64127,0.25166,0.1025,0.86648,0.02964,0.09397,0.36195 & C2/c
     ↪ C_{2h}^{6} #15 (ef^15) & mC124 & None & BeHNaO8Si3 & Eudidymite
     ↪  & G. {Diego Gatta} et al., Am. Mineral. 93, 1158-1165 (2008)
   1.00000000000000
   6.30940000000000   -3.68905000000000    0.00000000000000
   6.30940000000000    3.68905000000000    0.00000000000000
  -3.32902320731927    0.00000000000000   13.59226399409350
    Be     H    Na     O    Si
     4     8     4    34    12
Direct
  -0.15500000000000    0.49134000000000    0.49710000000000   Be   (8f)
  -0.49134000000000    0.15500000000000    0.00290000000000   Be   (8f)
   0.15500000000000   -0.49134000000000   -0.49710000000000   Be   (8f)
   0.49134000000000   -0.15500000000000    0.99710000000000   Be   (8f)
  -0.38050000000000    0.54850000000000    0.74490000000000    H   (8f)
  -0.54850000000000    0.38050000000000   -0.24490000000000    H   (8f)
   0.38050000000000   -0.54850000000000   -0.74490000000000    H   (8f)
   0.54850000000000   -0.38050000000000    1.24490000000000    H   (8f)
   0.43540000000000    0.51920000000000    0.74860000000000    H   (8f)
  -0.51920000000000   -0.43540000000000   -0.24860000000000    H   (8f)
  -0.43540000000000   -0.51920000000000   -0.74860000000000    H   (8f)
   0.51920000000000    0.43540000000000    1.24860000000000    H   (8f)
  -0.31179000000000    0.36977000000000   -0.07147000000000   Na   (8f)
  -0.36977000000000    0.31179000000000    0.57147000000000   Na   (8f)
   0.31179000000000   -0.36977000000000    0.07147000000000   Na   (8f)
   0.36977000000000   -0.31179000000000    0.42853000000000   Na   (8f)
  -0.17321000000000    0.17321000000000    0.25000000000000    O   (4e)
   0.17321000000000   -0.17321000000000    0.75000000000000    O   (4e)
   0.04023000000000    0.10517000000000    0.62614000000000    O   (8f)
  -0.10517000000000   -0.04023000000000   -0.12614000000000    O   (8f)
  -0.04023000000000   -0.10517000000000   -0.62614000000000    O   (8f)
   0.10517000000000    0.04023000000000    1.12614000000000    O   (8f)
   0.06006000000000    0.42066000000000    0.11540000000000    O   (8f)
  -0.42066000000000   -0.06006000000000    0.38460000000000    O   (8f)
  -0.06006000000000   -0.42066000000000   -0.11540000000000    O   (8f)
   0.42066000000000    0.06006000000000    0.61540000000000    O   (8f)
   0.07574000000000    0.40516000000000    0.57054000000000    O   (8f)
  -0.40516000000000   -0.07574000000000   -0.07054000000000    O   (8f)
  -0.07574000000000   -0.40516000000000   -0.57054000000000    O   (8f)
   0.40516000000000    0.07574000000000    1.07054000000000    O   (8f)
  -0.20182000000000    0.67536000000000    0.24484000000000    O   (8f)
  -0.67536000000000    0.20182000000000    0.25516000000000    O   (8f)
   0.20182000000000   -0.67536000000000   -0.24484000000000    O   (8f)
   0.67536000000000   -0.20182000000000    0.74484000000000    O   (8f)
  -0.34384000000000    0.65344000000000    0.05623000000000    O   (8f)
  -0.65344000000000    0.34384000000000    0.44377000000000    O   (8f)
   0.34384000000000   -0.65344000000000   -0.05623000000000    O   (8f)
   0.65344000000000   -0.34384000000000    0.55623000000000    O   (8f)
   0.07665000000000    0.18043000000000    0.87541000000000    O   (8f)
  -0.18043000000000   -0.07665000000000   -0.37541000000000    O   (8f)
  -0.07665000000000   -0.18043000000000   -0.87541000000000    O   (8f)
   0.18043000000000    0.07665000000000    1.37541000000000    O   (8f)
  -0.20457000000000    0.30931000000000    0.43729000000000    O   (8f)
  -0.30931000000000    0.20457000000000    0.06271000000000    O   (8f)
   0.20457000000000   -0.30931000000000   -0.43729000000000    O   (8f)
   0.30931000000000   -0.20457000000000    0.93729000000000    O   (8f)
  -0.41889000000000    0.44307000000000    0.74466000000000    O   (8f)
  -0.44307000000000    0.41889000000000   -0.24466000000000    O   (8f)
   0.41889000000000   -0.44307000000000   -0.74466000000000    O   (8f)
   0.44307000000000   -0.41889000000000    1.24466000000000    O   (8f)
   0.18450000000000    0.22380000000000    0.64127000000000   Si   (8f)
  -0.22380000000000   -0.18450000000000   -0.14127000000000   Si   (8f)
  -0.18450000000000   -0.22380000000000   -0.64127000000000   Si   (8f)
   0.22380000000000    0.18450000000000    1.14127000000000   Si   (8f)
   0.14916000000000    0.35416000000000    0.86648000000000   Si   (8f)
  -0.35416000000000   -0.14916000000000   -0.36648000000000   Si   (8f)
  -0.14916000000000   -0.35416000000000   -0.86648000000000   Si   (8f)
   0.35416000000000    0.14916000000000    1.36648000000000   Si   (8f)
  -0.06433000000000    0.12361000000000    0.36195000000000   Si   (8f)
  -0.12361000000000    0.06433000000000    0.13805000000000   Si   (8f)
   0.06433000000000   -0.12361000000000   -0.36195000000000   Si   (8f)
   0.12361000000000   -0.06433000000000    0.86195000000000   Si   (8f)
```

$\zeta$-$Nb_2O_5$ (B-$Nb_2O_5$): A2B5_mC28_15_f_e2f - CIF

```
# CIF file
data_findsym-output
_audit_creation_method FINDSYM

_chemical_name_mineral 'Nb2O5'
_chemical_formula_sum 'Nb2 O5'

loop_
_publ_author_name
 'T. S. Ercit'
_journal_name_full_name
;
 Mineralogy and Petrology
;
_journal_volume 43
_journal_year 1991
_journal_page_first 217
_journal_page_last 223
_publ_Section_title
;
 Refinement of the structure of $\zeta$-Nb$_{2}$O$_{5}$ and its
      ↪ relationship to the rutile and thoreaulite structures
;

# Found in ResearchGate, {~},

_aflow_title '$\zeta$-Nb$_{2}$O$_{5}$ (B-Nb$_{2}$O$_{5}$) Structure'
_aflow_proto 'A2B5_mC28_15_f_e2f'
_aflow_params 'a,b/a,c/a,\beta,y_{1},x_{2},y_{2},z_{2},x_{3},y_{3},z_{3
      ↪ },x_{4},y_{4},z_{4}'
_aflow_params_values '12.74,0.38328100471,0.436491365777,105.02,0.6465,
      ↪ 0.3598,0.25874,0.2512,0.1088,0.4421,0.4709,0.2055,0.0723,0.1256
      ↪ '
_aflow_Strukturbericht 'None'
_aflow_Pearson 'mC28'

_symmetry_space_group_name_H-M "C 1 2/c 1"
_symmetry_Int_Tables_number 15

_cell_length_a    12.74000
_cell_length_b    4.88300
_cell_length_c    5.56090
_cell_angle_alpha 90.00000
_cell_angle_beta  105.02000
_cell_angle_gamma 90.00000

loop_
_space_group_symop_id
_space_group_symop_operation_xyz
1 x,y,z
2 -x,y,-z+1/2
3 -x,-y,-z
4 x,-y,z+1/2
5 x+1/2,y+1/2,z
6 -x+1/2,y+1/2,-z+1/2
7 -x+1/2,-y+1/2,-z
8 x+1/2,-y+1/2,z+1/2

loop_
_atom_site_label
_atom_site_type_symbol
_atom_site_symmetry_multiplicity
_atom_site_Wyckoff_label
_atom_site_fract_x
_atom_site_fract_y
_atom_site_fract_z
_atom_site_occupancy
O1 O 4 e 0.00000 0.64650 0.25000 1.00000
Nb1 Nb 8 f 0.35980 0.25874 0.25120 1.00000
O2 O 8 f 0.10880 0.44210 0.47090 1.00000
O3 O 8 f 0.20550 0.07230 0.12560 1.00000
```

$\zeta$-$Nb_2O_5$ (B-$Nb_2O_5$): A2B5_mC28_15_f_e2f - POSCAR

```
A2B5_mC28_15_f_e2f & a,b/a,c/a,beta,y1,x2,y2,z2,x3,y3,z3,x4,y4,z4 --
      ↪ params=12.74,0.38328100471,0.436491365777,105.02,0.6465,0.3598,
      ↪ 0.25874,0.2512,0.1088,0.4421,0.4709,0.2055,0.0723,0.1256 & C2/c
      ↪ C_{2h}^{6} #15 (ef^3) & mC28 & None & Nb2O5 & Nb2O5 & T. S.
      ↪ Ercit, Mineral. Petrol. 43, 217-223 (1991)
   1.00000000000000
   6.37000000000000  -2.44150000000000   0.00000000000000
   6.37000000000000   2.44150000000000   0.00000000000000
  -1.44114171840500   0.00000000000000   5.37091420127642
    Nb     O
     4    10
Direct
   0.10106000000000   0.61854000000000   0.25120000000000   Nb   (8f)
  -0.61854000000000  -0.10106000000000   0.24880000000000   Nb   (8f)
  -0.10106000000000  -0.61854000000000  -0.25120000000000   Nb   (8f)
   0.61854000000000   0.10106000000000   0.75120000000000   Nb   (8f)
  -0.64650000000000   0.64650000000000   0.25000000000000    O   (4e)
   0.64650000000000  -0.64650000000000   0.75000000000000    O   (4e)
  -0.33330000000000   0.55090000000000   0.47090000000000    O   (8f)
  -0.55090000000000   0.33330000000000   0.02910000000000    O   (8f)
   0.33330000000000  -0.55090000000000  -0.47090000000000    O   (8f)
   0.55090000000000  -0.33330000000000   0.97090000000000    O   (8f)
   0.13320000000000   0.27780000000000   0.12560000000000    O   (8f)
  -0.27780000000000  -0.13320000000000   0.37440000000000    O   (8f)
  -0.13320000000000  -0.27780000000000  -0.12560000000000    O   (8f)
   0.27780000000000   0.13320000000000   0.62560000000000    O   (8f)
```

Muscovite ($KH_2Al_3Si_3O_{12}$, $S5_1$): A2BC10D2E4_mC76_15_f_e_5f_f_2f - CIF

```
# CIF file
data_findsym-output
_audit_creation_method FINDSYM

_chemical_name_mineral 'Muscovite'
_chemical_formula_sum 'Al2 K O10 (OH)2 Si4'

loop_
_publ_author_name
 'S. M. Richardson'
 'J. W. {Richardson, Jr.}'
_journal_name_full_name
;
 American Mineralogist
;
_journal_volume 67
_journal_year 1982
_journal_page_first 69
_journal_page_last 75
_publ_Section_title
;
 Crystal structure of a pink muscovite from Archer\'s Post, Kenya:
      ↪ Implications for reverse pleochroism in dioctahedral micas
;

# Found in The American Mineralogist Crystal Structure Database, 2003

_aflow_title 'Muscovite (KH$_{2}$Al$_{3}$Si$_{3}$O$_{12}$, $S5_{1}$)
      ↪ Structure'
_aflow_proto 'A2BC10D2E4_mC76_15_f_e_5f_f_2f'
_aflow_params 'a,b/a,c/a,\beta,y_{1},x_{2},y_{2},z_{2},x_{3},y_{3},z_{3
      ↪ },x_{4},y_{4},z_{4},x_{5},y_{5},z_{5},x_{6},y_{6},z_{6},x_{7},
      ↪ y_{7},z_{7},x_{8},y_{8},z_{8},x_{9},y_{9},z_{9},x_{10},y_{10},
      ↪ z_{10}'
_aflow_params_values '5.1988,1.73628529661,3.86739247519,95.782,0.0992,
      ↪ 0.2506,0.0838,0.0002,0.3872,0.2525,0.0543,0.0366,0.4431,0.4459,
      ↪ 0.4178,0.0931,0.1685,0.2475,0.3712,0.1685,0.2509,0.3132,0.3424,
      ↪ 0.0422,0.0622,0.4492,0.451,0.2587,0.1355,0.0354,0.4298,0.3646'
_aflow_Strukturbericht '$S5_{1}$'
_aflow_Pearson 'mC76'

_symmetry_space_group_name_H-M "C 1 2/c 1"
_symmetry_Int_Tables_number 15

_cell_length_a    5.19880
_cell_length_b    9.02660
_cell_length_c    20.10580
_cell_angle_alpha 90.00000
_cell_angle_beta  95.78200
_cell_angle_gamma 90.00000

loop_
_space_group_symop_id
_space_group_symop_operation_xyz
1 x,y,z
2 -x,y,-z+1/2
3 -x,-y,-z
4 x,-y,z+1/2
5 x+1/2,y+1/2,z
6 -x+1/2,y+1/2,-z+1/2
7 -x+1/2,-y+1/2,-z
8 x+1/2,-y+1/2,z+1/2

loop_
_atom_site_label
_atom_site_type_symbol
_atom_site_symmetry_multiplicity
_atom_site_Wyckoff_label
_atom_site_fract_x
_atom_site_fract_y
_atom_site_fract_z
_atom_site_occupancy
K1 K 4 e 0.00000 0.09920 0.25000 1.00000
Al1 Al 8 f 0.25060 0.08380 0.00020 1.00000
O1 O 8 f 0.38720 0.25250 0.05430 1.00000
O2 O 8 f 0.03660 0.44310 0.44590 1.00000
O3 O 8 f 0.41780 0.09310 0.16850 1.00000
O4 O 8 f 0.24750 0.37120 0.16850 1.00000
O5 O 8 f 0.25090 0.31320 0.34240 1.00000
OH1 OH 8 f 0.04220 0.06220 0.44920 1.00000
Si1 Si 8 f 0.45100 0.25870 0.13550 1.00000
Si2 Si 8 f 0.03540 0.42980 0.36460 1.00000
```

Muscovite ($KH_2Al_3Si_3O_{12}$, $S5_1$): A2BC10D2E4_mC76_15_f_e_5f_f_2f - POSCAR

```
A2BC10D2E4_mC76_15_f_e_5f_f_2f & a,b/a,c/a,beta,y1,x2,y2,z2,x3,y3,z3,x4,
      ↪ y4,z4,x5,y5,z5,x6,y6,z6,x7,y7,z7,x8,y8,z8,x9,y9,z9,x10,y10,z10
      ↪ --params=5.1988,1.73628529661,3.86739247519,95.782,0.0992,
      ↪ 0.2506,0.0838,0.0002,0.3872,0.2525,0.0543,0.0366,0.4431,0.4459,
      ↪ 0.4178,0.0931,0.1685,0.2475,0.3712,0.1685,0.2509,0.3132,0.3424,
      ↪ 0.0422,0.0622,0.4492,0.451,0.2587,0.1355,0.0354,0.4298,0.3646 &
      ↪ C2/c C_{2h}^{6} #15 (ef^9) & mC76 & $S5_{1}$ & Al3KO10(OH)2Si3
      ↪ & Muscovite & S. M. Richardson and J. W. {Richardson, Jr.},
      ↪ Am. Mineral. 67, 69-75 (1982)
   1.00000000000000
   2.59940000000000  -4.51330000000000   0.00000000000000
   2.59940000000000   4.51330000000000   0.00000000000000
  -2.02553351199094   0.00000000000000  20.00350988281310
    Al     K     O    OH    Si
     4     2    20     4     8
Direct
   0.16680000000000   0.33440000000000   0.00020000000000   Al   (8f)
  -0.33440000000000  -0.16680000000000   0.49980000000000   Al   (8f)
  -0.16680000000000  -0.33440000000000  -0.00020000000000   Al   (8f)
   0.33440000000000   0.16680000000000   0.50020000000000   Al   (8f)
  -0.09920000000000   0.09920000000000   0.25000000000000    K   (4e)
```

```
    0.09920000000000  -0.09920000000000   0.75000000000000   K   (4e)
    0.13470000000000   0.63970000000000   0.05430000000000   O   (8f)
   -0.63970000000000  -0.13470000000000   0.44570000000000   O   (8f)
   -0.13470000000000  -0.63970000000000  -0.05430000000000   O   (8f)
    0.63970000000000   0.13470000000000   0.55430000000000   O   (8f)
   -0.40650000000000   0.47970000000000   0.44590000000000   O   (8f)
   -0.47970000000000   0.40650000000000   0.05410000000000   O   (8f)
    0.40650000000000  -0.47970000000000  -0.44590000000000   O   (8f)
    0.47970000000000  -0.40650000000000   0.94590000000000   O   (8f)
    0.32470000000000   0.51090000000000   0.16850000000000   O   (8f)
   -0.51090000000000  -0.32470000000000   0.33150000000000   O   (8f)
   -0.32470000000000  -0.51090000000000  -0.16850000000000   O   (8f)
    0.51090000000000   0.32470000000000   0.66850000000000   O   (8f)
   -0.12370000000000   0.61870000000000   0.16850000000000   O   (8f)
   -0.61870000000000   0.12370000000000   0.33150000000000   O   (8f)
    0.12370000000000  -0.61870000000000  -0.16850000000000   O   (8f)
    0.61870000000000  -0.12370000000000   0.66850000000000   O   (8f)
   -0.06230000000000   0.56410000000000   0.34240000000000   O   (8f)
   -0.56410000000000   0.06230000000000   0.15760000000000   O   (8f)
    0.06230000000000  -0.56410000000000  -0.34240000000000   O   (8f)
    0.56410000000000  -0.06230000000000   0.84240000000000   O   (8f)
   -0.02000000000000   0.10440000000000   0.44920000000000   OH  (8f)
   -0.10440000000000   0.02000000000000   0.05080000000000   OH  (8f)
    0.02000000000000  -0.10440000000000  -0.44920000000000   OH  (8f)
    0.10440000000000  -0.02000000000000   0.94920000000000   OH  (8f)
    0.19230000000000   0.70970000000000   0.13550000000000   Si  (8f)
   -0.70970000000000  -0.19230000000000   0.36450000000000   Si  (8f)
   -0.19230000000000  -0.70970000000000  -0.13550000000000   Si  (8f)
    0.70970000000000   0.19230000000000   0.63550000000000   Si  (8f)
   -0.39440000000000   0.46520000000000   0.36460000000000   Si  (8f)
   -0.46520000000000   0.39440000000000   0.13540000000000   Si  (8f)
    0.39440000000000  -0.46520000000000  -0.36460000000000   Si  (8f)
    0.46520000000000  -0.39440000000000   0.86460000000000   Si  (8f)
```

$Rb_2C_2O_4 \cdot H_2O$: A2BC4D2_mC36_15_f_e_2f_f - CIF

```
# CIF file
data_findsym-output
_audit_creation_method FINDSYM

_chemical_name_mineral 'C2(H2O)O4Rb2'
_chemical_formula_sum 'C2 (H2O) O4 Rb2'

loop_
_publ_author_name
 'B. F. Pedersen'
_journal_name_full_name
;
 Acta Chemica Scandinavica
;
_journal_volume 19
_journal_year 1965
_journal_page_first 1815
_journal_page_last 1818
_publ_Section_title
;
 The Crystal Structure of Rubidium. Oxalate Monohydrate, Rb$_{2}$C$_{2}
    ↪ $O$_{4}\cdot$H$_{2}$O
;

_aflow_title 'Rb$_{2}$C$_{2}$O$_{4}\cdot$H$_{2}$O Structure'
_aflow_proto 'A2BC4D2_mC36_15_f_e_2f_f'
_aflow_params 'a,b/a,c/a,\beta,y_{1},x_{2},y_{2},z_{2},x_{3},y_{3},z_{3
    ↪ },x_{4},y_{4},z_{4},x_{5},y_{5},z_{5}'
_aflow_params_values '9.662,0.657213827365,1.147588491,109.4,0.4733,
    ↪ 0.2396,0.3216,0.0546,0.1355,0.2748,0.0933,0.3262,0.4768,0.094,
    ↪ 0.129,0.8156,0.1297'
_aflow_Strukturbericht 'None'
_aflow_Pearson 'mC36'

_symmetry_space_group_name_H-M "C 1 2/c 1"
_symmetry_Int_Tables_number 15

_cell_length_a    9.66200
_cell_length_b    6.35000
_cell_length_c    11.08800
_cell_angle_alpha 90.00000
_cell_angle_beta  109.40000
_cell_angle_gamma 90.00000

loop_
_space_group_symop_id
_space_group_symop_operation_xyz
1 x,y,z
2 -x,y,-z+1/2
3 -x,-y,-z
4 x,-y,z+1/2
5 x+1/2,y+1/2,z
6 -x+1/2,y+1/2,-z+1/2
7 -x+1/2,-y+1/2,-z
8 x+1/2,-y+1/2,z+1/2

loop_
_atom_site_label
_atom_site_type_symbol
_atom_site_symmetry_multiplicity
_atom_site_Wyckoff_label
_atom_site_fract_x
_atom_site_fract_y
_atom_site_fract_z
_atom_site_occupancy
H2O1 H2O 4 e 0.00000 0.47330 0.25000 1.00000
C1 C 8 f 0.23960 0.32160 0.05460 1.00000
O1 O 8 f 0.13550 0.27480 0.09330 1.00000
O2 O 8 f 0.32620 0.47680 0.09400 1.00000
Rb1 Rb 8 f 0.12900 0.81560 0.12970 1.00000
```

$Rb_2C_2O_4 \cdot H_2O$: A2BC4D2_mC36_15_f_e_2f_f - POSCAR

```
A2BC4D2_mC36_15_f_e_2f_f & a,b/a,c/a,beta,y1,x2,y2,z2,x3,y3,z3,x4,y4,z4,
    ↪ x5,y5,z5 --params=9.662,0.657213827365,1.147588491,109.4,0.4733
    ↪ ,0.2396,0.3216,0.0546,0.1355,0.2748,0.0933,0.3262,0.4768,0.094,
    ↪ 0.129,0.8156,0.1297 & C2/c C_{2h}^{6} #15 (ef^4) & mC36 & None
    ↪ & C2(H2O)O4Rb2 & C2(H2O)O4Rb2 & B. F. Pedersen, Acta Chem.
    ↪ Scand. 19, 1815-1818 (1965)
   1.00000000000000
   4.83100000000000  -3.17500000000000   0.00000000000000
   4.83100000000000   3.17500000000000   0.00000000000000
  -3.68300263032650   0.00000000000000  10.45845283132300
     C    H2O    O    Rb
     4     2     8     4
Direct
  -0.08200000000000   0.56120000000000   0.05460000000000   C    (8f)
  -0.56120000000000   0.08200000000000   0.44540000000000   C    (8f)
   0.08200000000000  -0.56120000000000  -0.05460000000000   C    (8f)
   0.56120000000000  -0.08200000000000   0.55460000000000   C    (8f)
  -0.47330000000000   0.47330000000000   0.25000000000000   H2O  (4e)
   0.47330000000000  -0.47330000000000   0.75000000000000   H2O  (4e)
  -0.13930000000000   0.41030000000000   0.09330000000000   O    (8f)
  -0.41030000000000   0.13930000000000   0.40670000000000   O    (8f)
   0.13930000000000  -0.41030000000000  -0.09330000000000   O    (8f)
   0.41030000000000  -0.13930000000000   0.59330000000000   O    (8f)
  -0.15060000000000   0.80300000000000   0.09400000000000   O    (8f)
  -0.80300000000000   0.15060000000000   0.40600000000000   O    (8f)
   0.15060000000000  -0.80300000000000  -0.09400000000000   O    (8f)
   0.80300000000000  -0.15060000000000   0.59400000000000   O    (8f)
  -0.68660000000000   0.94460000000000   0.12970000000000   Rb   (8f)
  -0.94460000000000   0.68660000000000   0.37030000000000   Rb   (8f)
   0.68660000000000  -0.94460000000000  -0.12970000000000   Rb   (8f)
   0.94460000000000  -0.68660000000000   0.62970000000000   Rb   (8f)
```

Alluaudite [$NaMnFe_2(PO_4)_3$]: A2BCD12E3_mC76_15_f_e_b_6f_ef - CIF

```
# CIF file
data_findsym-output
_audit_creation_method FINDSYM

_chemical_name_mineral 'Alluaudite'
_chemical_formula_sum 'Fe2 Mn Na O12 P3'

loop_
_publ_author_name
 'P. B. Moore'
_journal_name_full_name
;
 American Mineralogist
;
_journal_volume 56
_journal_year 1971
_journal_page_first 1955
_journal_page_last 1975
_publ_Section_title
;
 Crystal Chemistry of the Alluaudite Structure Type: Contribution to the
    ↪ Paragenesis of Pegmatite Phosphate Giant Crystals
;

# Found in The American Mineralogist Crystal Structure Database, 2003

_aflow_title 'Alluaudite [NaMnFe$_{2}$(PO$_{4}$)$_{3}$] Structure'
_aflow_proto 'A2BCD12E3_mC76_15_f_e_b_6f_ef'
_aflow_params 'a,b/a,c/a,\beta,y_{2},y_{3},x_{4},y_{4},z_{4},x_{5},y_{5
    ↪ },z_{5},x_{6},y_{6},z_{6},x_{7},y_{7},z_{7},x_{8},y_{8},z_{8},
    ↪ x_{9},y_{9},z_{9},x_{10},y_{10},z_{10},x_{11},y_{11},z_{11}'
_aflow_params_values '12.004,1.04406864379,0.533488837054,114.4,0.2599,
    ↪ 0.7145,0.2812,0.6525,0.3713,0.4533,0.7152,0.5342,0.0988,0.6375,
    ↪ 0.2401,0.3272,0.6633,0.102,0.1213,0.3974,0.3119,0.2251,0.822,
    ↪ 0.3172,0.3102,0.5021,0.3735,0.2424,0.8911,0.1325'
_aflow_Strukturbericht 'None'
_aflow_Pearson 'mC76'

_symmetry_space_group_name_H-M "C 1 2/c 1"
_symmetry_Int_Tables_number 15

_cell_length_a    12.00400
_cell_length_b    12.53300
_cell_length_c    6.40400
_cell_angle_alpha 90.00000
_cell_angle_beta  114.40000
_cell_angle_gamma 90.00000

loop_
_space_group_symop_id
_space_group_symop_operation_xyz
1 x,y,z
2 -x,y,-z+1/2
3 -x,-y,-z
4 x,-y,z+1/2
5 x+1/2,y+1/2,z
6 -x+1/2,y+1/2,-z+1/2
7 -x+1/2,-y+1/2,-z
8 x+1/2,-y+1/2,z+1/2

loop_
_atom_site_label
_atom_site_type_symbol
_atom_site_symmetry_multiplicity
_atom_site_Wyckoff_label
_atom_site_fract_x
_atom_site_fract_y
```

```
_atom_site_fract_z
_atom_site_occupancy
Na1 Na 4 b 0.00000 0.50000 0.00000 1.00000
Mn1 Mn 4 e 0.00000 0.25990 0.25000 1.00000
P1 P 4 e 0.00000 0.71450 0.25000 1.00000
Fe1 Fe 8 f 0.28120 0.65250 0.37130 1.00000
O1 O 8 f 0.45330 0.71520 0.53420 1.00000
O2 O 8 f 0.09880 0.63750 0.24010 1.00000
O3 O 8 f 0.32720 0.66330 0.10200 1.00000
O4 O 8 f 0.12130 0.39740 0.31190 1.00000
O5 O 8 f 0.22510 0.82200 0.31720 1.00000
O6 O 8 f 0.31020 0.50210 0.37350 1.00000
P2 P 8 f 0.24240 0.89110 0.13250 1.00000
```

Alluaudite [$NaMnFe_2(PO_4)_3$]: A2BCD12E3_mC76_15_f_e_b_6f_ef - POSCAR

```
A2BCD12E3_mC76_15_f_e_b_6f_ef & a,b/a,c/a,beta,y2,y3,x4,y4,z4,x5,y5,z5,
    ↪ x6,y6,z6,x7,y7,z7,x8,y8,z8,x9,y9,z9,x10,y10,z10,x11,y11,z11 --
    ↪ params=12.004,1.04406864379,0.533488837054,114.4,0.2599,0.7145,
    ↪ 0.2812,0.6525,0.3713,0.4533,0.7152,0.5342,0.0988,0.6375,0.2401,
    ↪ 0.3272,0.6633,0.102,0.1213,0.3974,0.3119,0.2251,0.822,0.3172,
    ↪ 0.3102,0.5021,0.3735,0.2424,0.8911,0.1325 & C2/c C_{2h}^{6} #15
    ↪ (be^2f^8) & mC76 & None & Fe2MnNaO12P3 & Alluaudite & P. B.
    ↪ Moore, Am. Mineral. 56, 1955-1975 (1971)
   1.00000000000000
   6.00200000000000  -6.26650000000000   0.00000000000000
   6.00200000000000   6.26650000000000   0.00000000000000
  -2.64552076859637   0.00000000000000   5.83201816380276
    Fe     Mn     Na     O      P
     4      2      2     24     6
Direct
  -0.37130000000000   0.93370000000000   0.37130000000000   Fe   (8f)
  -0.93370000000000   0.37130000000000   0.12870000000000   Fe   (8f)
   0.37130000000000  -0.93370000000000  -0.37130000000000   Fe   (8f)
   0.93370000000000  -0.37130000000000   0.87130000000000   Fe   (8f)
  -0.25990000000000   0.25990000000000   0.25000000000000   Mn   (4e)
   0.25990000000000  -0.25990000000000   0.75000000000000   Mn   (4e)
   0.50000000000000   0.50000000000000   0.00000000000000   Na   (4b)
   0.50000000000000   0.50000000000000   0.50000000000000   Na   (4b)
  -0.26190000000000   1.16850000000000   0.53420000000000   O    (8f)
  -1.16850000000000   0.26190000000000  -0.03420000000000   O    (8f)
   0.26190000000000  -1.16850000000000  -0.53420000000000   O    (8f)
   1.16850000000000  -0.26190000000000   1.03420000000000   O    (8f)
  -0.53870000000000   0.73630000000000   0.24010000000000   O    (8f)
  -0.73630000000000   0.53870000000000   0.25990000000000   O    (8f)
   0.53870000000000  -0.73630000000000  -0.24010000000000   O    (8f)
   0.73630000000000  -0.53870000000000   0.74010000000000   O    (8f)
  -0.33610000000000   0.99050000000000   0.10200000000000   O    (8f)
  -0.99050000000000   0.33610000000000   0.39800000000000   O    (8f)
   0.33610000000000  -0.99050000000000  -0.10200000000000   O    (8f)
   0.99050000000000  -0.33610000000000   0.60200000000000   O    (8f)
  -0.27610000000000   0.51870000000000   0.31190000000000   O    (8f)
  -0.51870000000000   0.27610000000000   0.18810000000000   O    (8f)
   0.27610000000000  -0.51870000000000  -0.31190000000000   O    (8f)
   0.51870000000000  -0.27610000000000   0.81190000000000   O    (8f)
  -0.59690000000000   1.04710000000000   0.31720000000000   O    (8f)
  -1.04710000000000   0.59690000000000   0.18280000000000   O    (8f)
   0.59690000000000  -1.04710000000000  -0.31720000000000   O    (8f)
   1.04710000000000  -0.59690000000000   0.81720000000000   O    (8f)
  -0.19190000000000   0.81230000000000   0.37350000000000   O    (8f)
  -0.81230000000000   0.19190000000000   0.12650000000000   O    (8f)
   0.19190000000000  -0.81230000000000  -0.37350000000000   O    (8f)
   0.81230000000000  -0.19190000000000   0.87350000000000   O    (8f)
  -0.71450000000000   0.71450000000000   0.25000000000000   P    (4e)
   0.71450000000000  -0.71450000000000   0.75000000000000   P    (4e)
  -0.64870000000000   1.13350000000000   0.13250000000000   P    (8f)
  -1.13350000000000   0.64870000000000   0.36750000000000   P    (8f)
   0.64870000000000  -1.13350000000000  -0.13250000000000   P    (8f)
   1.13350000000000  -0.64870000000000   0.63250000000000   P    (8f)
```

$ThC_2$ ($C_g$): A2B_mC12_15_f_e - CIF

```
# CIF file
data_findsym-output
_audit_creation_method FINDSYM

_chemical_name_mineral 'C2Th'
_chemical_formula_sum 'C2 Th'

loop_
_publ_author_name
 'A. L. Bowman'
 'N. H. Krikorian'
 'G. P. Arnold'
 'T. C. Wallace'
 'N. G. Nereson'
_journal_name_full_name
;
 Acta Crystallographica Section B: Structural Science
;
_journal_volume 24
_journal_year 1968
_journal_page_first 1121
_journal_page_last 1123
_publ_Section_title
;
 The Crystal Structure of ThC$_{2}$
;

_aflow_title 'ThC$_{2}$ (C$_{g}$) Structure'
_aflow_proto 'A2B_mC12_15_f_e'
_aflow_params 'a,b/a,c/a,\beta,y_{1},x_{2},y_{2},z_{2}'
_aflow_params_values '6.692,0.631052002391,1.00777047221,103.12,0.2074,
    ↪ 0.2992,0.1326,0.054'
_aflow_Strukturbericht 'None'
_aflow_Pearson 'mC12'

_symmetry_space_group_name_H-M "C 1 2/c 1"
_symmetry_Int_Tables_number 15

_cell_length_a     6.69200
_cell_length_b     4.22300
_cell_length_c     6.74400
_cell_angle_alpha 90.00000
_cell_angle_beta  103.12000
_cell_angle_gamma 90.00000

loop_
_space_group_symop_id
_space_group_symop_operation_xyz
1 x,y,z
2 -x,y,-z+1/2
3 -x,-y,-z
4 x,-y,z+1/2
5 x+1/2,y+1/2,z
6 -x+1/2,y+1/2,-z+1/2
7 -x+1/2,-y+1/2,-z
8 x+1/2,-y+1/2,z+1/2

loop_
_atom_site_label
_atom_site_type_symbol
_atom_site_symmetry_multiplicity
_atom_site_Wyckoff_label
_atom_site_fract_x
_atom_site_fract_y
_atom_site_fract_z
_atom_site_occupancy
Th1 Th 4 e 0.00000 0.20740 0.25000 1.00000
C1 C 8 f 0.29920 0.13260 0.05400 1.00000
```

$ThC_2$ ($C_g$): A2B_mC12_15_f_e - POSCAR

```
A2B_mC12_15_f_e & a,b/a,c/a,beta,y1,x2,y2,z2 --params=6.692,
    ↪ 0.631052002391,1.00777047221,103.12,0.2074,0.2992,0.1326,0.054
    ↪ & C2/c C_{2h}^{6} #15 (ef) & mC12 & None & C2Th & C2Th & A. L.
    ↪ Bowman et al., Acta Crystallogr. Sect. B Struct. Sci. 24,
    ↪ 1121-1123 (1968)
   1.00000000000000
   3.34600000000000  -2.11150000000000   0.00000000000000
   3.34600000000000   2.11150000000000   0.00000000000000
  -1.53082916110101   0.00000000000000   6.56795996330084
     C     Th
     4      2
Direct
   0.16660000000000   0.43180000000000   0.05400000000000   C    (8f)
  -0.43180000000000  -0.16660000000000   0.44600000000000   C    (8f)
  -0.16660000000000  -0.43180000000000  -0.05400000000000   C    (8f)
   0.43180000000000   0.16660000000000   0.55400000000000   C    (8f)
  -0.20740000000000   0.20740000000000   0.25000000000000   Th   (4e)
   0.20740000000000  -0.20740000000000   0.75000000000000   Th   (4e)
```

Clinocervantite ($\beta$-$Sb_2O_4$): A2B_mC24_15_2f_ce - CIF

```
# CIF file
data_findsym-output
_audit_creation_method FINDSYM

_chemical_name_mineral 'Clinocervantite'
_chemical_formula_sum 'O2 Sb'

loop_
_publ_author_name
 'R. Basso'
 'G. Lucchetti'
 'L. Zefiro'
 'A. Palenzona'
_journal_name_full_name
;
 European Journal of Mineralogy
;
_journal_volume 11
_journal_year 1999
_journal_page_first 95
_journal_page_last 100
_publ_Section_title
;
 Clinocervantite, $\beta$-Sb$_{2}$O$_{4}$, the natural monoclinic
    ↪ polymorph of cervantite from the Cetine mine, Siena, Italy
;

# Found in The American Mineralogist Crystal Structure Database, 2003

_aflow_title 'Clinocervantite ($\beta$-Sb$_{2}$O$_{4}$) Structure'
_aflow_proto 'A2B_mC24_15_2f_ce'
_aflow_params 'a,b/a,c/a,\beta,y_{2},x_{3},y_{3},z_{3},x_{4},y_{4},z_{4
    ↪ }'
_aflow_params_values '12.061,0.40096177763,0.446314567615,103.12,0.2851,
    ↪ 0.1918,0.0517,0.6746,0.0939,0.4122,-0.0351'
_aflow_Strukturbericht 'None'
_aflow_Pearson 'mC24'

_symmetry_space_group_name_H-M "C 1 2/c 1"
_symmetry_Int_Tables_number 15

_cell_length_a     12.06100
_cell_length_b     4.83600
_cell_length_c     5.38300
_cell_angle_alpha 90.00000
_cell_angle_beta  103.12000
```

```
_cell_angle_gamma 90.00000

loop_
_space_group_symop_id
_space_group_symop_operation_xyz
1 x,y,z
2 -x,y,-z+1/2
3 -x,-y,-z
4 x,-y,z+1/2
5 x+1/2,y+1/2,z
6 -x+1/2,y+1/2,-z+1/2
7 -x+1/2,-y+1/2,-z
8 x+1/2,-y+1/2,z+1/2

loop_
_atom_site_label
_atom_site_type_symbol
_atom_site_symmetry_multiplicity
_atom_site_Wyckoff_label
_atom_site_fract_x
_atom_site_fract_y
_atom_site_fract_z
_atom_site_occupancy
Sb1 Sb 4 c 0.25000 0.25000 0.00000 1.00000
Sb2 Sb 4 e 0.00000 0.28510 0.25000 1.00000
O1 O 8 f 0.19180 0.05170 0.67460 1.00000
O2 O 8 f 0.09390 0.41220 -0.03510 1.00000
```

Clinocervantite ($\beta$-$Sb_2O_4$): A2B_mC24_15_2f_ce - POSCAR

```
A2B_mC24_15_2f_ce & a,b/a,c/a,beta,y2,x3,y3,z3,x4,y4,z4 --params=12.061,
    ↪ 0.40096177763,0.446314567615,103.12,0.2851,0.1918,0.0517,0.6746
    ↪ ,0.0939,0.4122,-0.0351 & C2/c C_{2h}^{6} #15 (cef^2) & mC24 &
    ↪ None & O2Sb & Clinocervantite & R. Basso et al., Eur. J.
    ↪ Mineral. 11, 95-100 (1999)
   1.00000000000000
   6.03050000000000  -2.41800000000000   0.00000000000000
   6.03050000000000   2.41800000000000   0.00000000000000
  -1.22189403532129   0.00000000000000   5.24248642978179
     O    Sb
     8     4
Direct
   0.14010000000000   0.24350000000000   0.67460000000000   O   (8f)
  -0.24350000000000  -0.14010000000000  -0.17460000000000   O   (8f)
  -0.14010000000000  -0.24350000000000  -0.67460000000000   O   (8f)
   0.24350000000000   0.14010000000000   1.17460000000000   O   (8f)
  -0.31830000000000   0.50610000000000  -0.03510000000000   O   (8f)
  -0.50610000000000   0.31830000000000   0.53510000000000   O   (8f)
   0.31830000000000  -0.50610000000000   0.03510000000000   O   (8f)
   0.50610000000000  -0.31830000000000   0.46490000000000   O   (8f)
   0.00000000000000   0.50000000000000   0.00000000000000   Sb  (4c)
   0.50000000000000   0.00000000000000   0.50000000000000   Sb  (4c)
  -0.28510000000000   0.28510000000000   0.25000000000000   Sb  (4e)
   0.28510000000000  -0.28510000000000   0.75000000000000   Sb  (4e)
```

$(CdSO_4)_3{\cdot}8H_2O$ ($H4_{20}$): A3B16C20D3_mC168_15_ef_8f_10f_ef - CIF

```
# CIF file
data_findsym-output
_audit_creation_method FINDSYM

_chemical_name_mineral 'Cd3H16O20S3'
_chemical_formula_sum 'Cd3 H16 O20 S3'

loop_
_publ_author_name
 'R. Caminiti'
 'G. Johansson'
_journal_name_full_name
;
 Acta Chemica Scandinavica
;
_journal_volume 35a
_journal_year 1981
_journal_page_first 451
_journal_page_last 455
_publ_Section_title
;
 A refinement of the Crystal Structure of the Cadmium Sulfate 3CdSO$_{4}
    ↪ $$\cdot$8H$_{2}$O
;

_aflow_title '(CdSO$_{4}$)$_{3}$$\cdot$8H$_{2}$O ($H4_{20}$) Structure'
_aflow_proto 'A3B16C20D3_mC168_15_ef_8f_10f_ef'
_aflow_params 'a,b/a,c/a,\beta,y_{1},y_{2},x_{3},y_{3},z_{3},x_{4},y_{4
    ↪ },z_{4},x_{5},y_{5},z_{5},x_{6},y_{6},z_{6},x_{7},y_{7},z_{7},
    ↪ x_{8},y_{8},z_{8},x_{9},y_{9},z_{9},x_{10},y_{10},z_{10},x_{11
    ↪ },y_{11},z_{11},x_{12},y_{12},z_{12},x_{13},y_{13},z_{13},x_{14
    ↪ },y_{14},z_{14},x_{15},y_{15},z_{15},x_{16},y_{16},z_{16},x_{17
    ↪ },y_{17},z_{17},x_{18},y_{18},z_{18},x_{19},y_{19},z_{19},x_{20
    ↪ },y_{20},z_{20},x_{21},y_{21},z_{21},x_{22},y_{22},z_{22}'
_aflow_params_values '14.818,0.803279794844,0.638952625186,97.39,-
    ↪ 0.05363,0.50828,0.15443,0.40511,0.035,0.083,0.252,0.716,0.095,
    ↪ 0.209,0.601,0.251,0.079,0.406,0.271,0.075,0.309,0.195,0.326,
    ↪ 0.288,0.273,0.347,0.273,0.094,0.203,0.346,0.06,0.3,0.402,0.0258
    ↪ ,0.4358,0.1357,0.4224,0.0799,0.1936,0.3962,0.4709,0.0521,0.1008
    ↪ ,0.2269,0.0033,0.0848,0.0936,0.1832,0.2296,0.1152,0.0982,0.0936
    ↪ ,0.1928,0.6827,0.2796,0.0746,0.4019,0.2214,0.3584,0.2573,0.0955
    ↪ ,0.2506,0.4077,0.12978,0.11705,0.05632'
_aflow_Strukturbericht '$H4_{20}$'
_aflow_Pearson 'mC168'

_symmetry_space_group_name_H-M "C 1 2/c 1"
_symmetry_Int_Tables_number 15
_cell_length_a    14.81800
_cell_length_b    11.90300
_cell_length_c    9.46800
_cell_angle_alpha 90.00000
_cell_angle_beta  97.39000
_cell_angle_gamma 90.00000

loop_
_space_group_symop_id
_space_group_symop_operation_xyz
1 x,y,z
2 -x,y,-z+1/2
3 -x,-y,-z
4 x,-y,z+1/2
5 x+1/2,y+1/2,z
6 -x+1/2,y+1/2,-z+1/2
7 -x+1/2,-y+1/2,-z
8 x+1/2,-y+1/2,z+1/2

loop_
_atom_site_label
_atom_site_type_symbol
_atom_site_symmetry_multiplicity
_atom_site_Wyckoff_label
_atom_site_fract_x
_atom_site_fract_y
_atom_site_fract_z
_atom_site_occupancy
Cd1 Cd 4 e 0.00000 -0.05363 0.25000 1.00000
S1 S 4 e 0.00000 0.50828 0.25000 1.00000
Cd2 Cd 8 f 0.15443 0.40511 0.03500 1.00000
H1 H 8 f 0.08300 0.25200 0.71600 1.00000
H2 H 8 f 0.09500 0.20900 0.60100 1.00000
H3 H 8 f 0.25100 0.07900 0.40600 1.00000
H4 H 8 f 0.27100 0.07500 0.30900 1.00000
H5 H 8 f 0.19500 0.32600 0.28800 1.00000
H6 H 8 f 0.27300 0.34700 0.27300 1.00000
H7 H 8 f 0.09400 0.20300 0.34600 1.00000
H8 H 8 f 0.06000 0.30000 0.40200 1.00000
O1 O 8 f 0.02580 0.43580 0.13570 1.00000
O2 O 8 f 0.42240 0.07990 0.19360 1.00000
O3 O 8 f 0.39620 0.47090 0.05210 1.00000
O4 O 8 f 0.10080 0.22690 0.00330 1.00000
O5 O 8 f 0.08480 0.09360 0.18320 1.00000
O6 O 8 f 0.22960 0.11520 0.09820 1.00000
O7 O 8 f 0.09360 0.19280 0.68270 1.00000
O8 O 8 f 0.27960 0.07460 0.40190 1.00000
O9 O 8 f 0.22140 0.35840 0.25730 1.00000
O10 O 8 f 0.09550 0.25060 0.40770 1.00000
S2 S 8 f 0.12978 0.11705 0.05632 1.00000
```

$(CdSO_4)_3{\cdot}8H_2O$ ($H4_{20}$): A3B16C20D3_mC168_15_ef_8f_10f_ef - POSCAR

```
A3B16C20D3_mC168_15_ef_8f_10f_ef & a,b/a,c/a,beta,y1,y2,x3,y3,z3,x4,y4,
    ↪ z4,x5,y5,z5,x6,y6,z6,x7,y7,z7,x8,y8,z8,x9,y9,z9,x10,y10,z10,x11
    ↪ ,y11,z11,x12,y12,z12,x13,y13,z13,x14,y14,z14,x15,y15,z15,x16,
    ↪ y16,z16,x17,y17,z17,x18,y18,z18,x19,y19,z19,x20,y20,z20,x21,y21
    ↪ ,z21,x22,y22,z22 --params=14.818,0.803279794844,0.638952625186,
    ↪ 97.39,-0.05363,0.50828,0.15443,0.40511,0.035,0.083,0.252,0.716,
    ↪ 0.095,0.209,0.601,0.251,0.079,0.406,0.271,0.075,0.309,0.195,
    ↪ 0.326,0.288,0.273,0.347,0.273,0.094,0.203,0.346,0.06,0.3,0.402,
    ↪ 0.0258,0.4358,0.1357,0.4224,0.0799,0.1936,0.3962,0.4709,0.0521,
    ↪ 0.1008,0.2269,0.0033,0.0848,0.0936,0.1832,0.2296,0.1152,0.0982,
    ↪ 0.0936,0.1928,0.6827,0.2796,0.0746,0.4019,0.2214,0.3584,0.2573,
    ↪ 0.0955,0.2506,0.4077,0.12978,0.11705,0.05632 & C2/c C_{2h}^{6}
    ↪ #15 (e^2f^20) & mC168 & $H4_{20}$ & Cd3H16O20S3 & Cd3H16O20S3 &
    ↪  R. Caminiti and G. Johansson, Acta Chem. Scand. 35a, 451-455 (
    ↪ 1981)
   1.00000000000000
   7.40900000000000  -5.95150000000000   0.00000000000000
   7.40900000000000   5.95150000000000   0.00000000000000
  -1.21779797531454   0.00000000000000   9.38935525429301
    Cd     H     O     S
     6    32    40     6
Direct
   0.05363000000000  -0.05363000000000   0.25000000000000   Cd  (4e)
  -0.05363000000000   0.05363000000000   0.75000000000000   Cd  (4e)
  -0.25068000000000   0.55954000000000   0.03500000000000   Cd  (8f)
  -0.55954000000000   0.25068000000000   0.46500000000000   Cd  (8f)
   0.25068000000000  -0.55954000000000  -0.03500000000000   Cd  (8f)
   0.55954000000000  -0.25068000000000   0.53500000000000   Cd  (8f)
  -0.16900000000000   0.33500000000000   0.71600000000000   H   (8f)
  -0.33500000000000   0.16900000000000  -0.21600000000000   H   (8f)
   0.16900000000000  -0.33500000000000  -0.71600000000000   H   (8f)
   0.33500000000000  -0.16900000000000   1.21600000000000   H   (8f)
  -0.11400000000000   0.30400000000000   0.60100000000000   H   (8f)
  -0.30400000000000   0.11400000000000  -0.10100000000000   H   (8f)
   0.11400000000000  -0.30400000000000  -0.60100000000000   H   (8f)
   0.30400000000000  -0.11400000000000   1.10100000000000   H   (8f)
   0.17200000000000   0.33000000000000   0.40600000000000   H   (8f)
  -0.33000000000000  -0.17200000000000   0.09400000000000   H   (8f)
  -0.17200000000000  -0.33000000000000  -0.40600000000000   H   (8f)
   0.33000000000000   0.17200000000000   0.90600000000000   H   (8f)
   0.19600000000000   0.34600000000000   0.30900000000000   H   (8f)
  -0.34600000000000  -0.19600000000000   0.19100000000000   H   (8f)
  -0.19600000000000  -0.34600000000000  -0.30900000000000   H   (8f)
   0.34600000000000   0.19600000000000   0.80900000000000   H   (8f)
  -0.13100000000000   0.52100000000000   0.28800000000000   H   (8f)
  -0.52100000000000   0.13100000000000   0.21200000000000   H   (8f)
   0.13100000000000  -0.52100000000000  -0.28800000000000   H   (8f)
   0.52100000000000  -0.13100000000000   0.78800000000000   H   (8f)
  -0.07400000000000   0.62000000000000   0.27300000000000   H   (8f)
  -0.62000000000000   0.07400000000000   0.22700000000000   H   (8f)
   0.07400000000000  -0.62000000000000  -0.27300000000000   H   (8f)
   0.62000000000000  -0.07400000000000   0.77300000000000   H   (8f)
```

```
 -0.10900000000000   0.29700000000000   0.34600000000000   H   (8f)
 -0.29700000000000   0.10900000000000   0.15400000000000   H   (8f)
  0.10900000000000  -0.29700000000000  -0.34600000000000   H   (8f)
  0.29700000000000  -0.10900000000000   0.84600000000000   H   (8f)
 -0.24000000000000   0.36000000000000   0.40200000000000   H   (8f)
 -0.36000000000000   0.24000000000000   0.09800000000000   H   (8f)
  0.24000000000000  -0.36000000000000  -0.40200000000000   H   (8f)
  0.36000000000000  -0.24000000000000   0.90200000000000   H   (8f)
 -0.41000000000000   0.46160000000000   0.13570000000000   O   (8f)
 -0.46160000000000   0.41000000000000   0.36430000000000   O   (8f)
  0.41000000000000  -0.46160000000000  -0.13570000000000   O   (8f)
  0.46160000000000  -0.41000000000000   0.63570000000000   O   (8f)
  0.34250000000000   0.50230000000000   0.19360000000000   O   (8f)
 -0.50230000000000  -0.34250000000000   0.30640000000000   O   (8f)
 -0.34250000000000  -0.50230000000000  -0.19360000000000   O   (8f)
  0.50230000000000   0.34250000000000   0.69360000000000   O   (8f)
 -0.07470000000000   0.86710000000000   0.05210000000000   O   (8f)
 -0.86710000000000   0.07470000000000   0.44790000000000   O   (8f)
  0.07470000000000  -0.86710000000000  -0.05210000000000   O   (8f)
  0.86710000000000  -0.07470000000000   0.55210000000000   O   (8f)
 -0.12610000000000   0.32770000000000   0.00330000000000   O   (8f)
 -0.32770000000000   0.12610000000000   0.49670000000000   O   (8f)
  0.12610000000000  -0.32770000000000  -0.00330000000000   O   (8f)
  0.32770000000000  -0.12610000000000   0.50330000000000   O   (8f)
 -0.00880000000000   0.17840000000000   0.18320000000000   O   (8f)
 -0.17840000000000   0.00880000000000   0.31680000000000   O   (8f)
  0.00880000000000  -0.17840000000000  -0.18320000000000   O   (8f)
  0.17840000000000  -0.00880000000000   0.68320000000000   O   (8f)
  0.11440000000000   0.34480000000000   0.09820000000000   O   (8f)
 -0.34480000000000  -0.11440000000000   0.40180000000000   O   (8f)
 -0.11440000000000  -0.34480000000000  -0.09820000000000   O   (8f)
  0.34480000000000   0.11440000000000   0.59820000000000   O   (8f)
 -0.09920000000000   0.28640000000000   0.68270000000000   O   (8f)
 -0.28640000000000   0.09920000000000  -0.18270000000000   O   (8f)
  0.09920000000000  -0.28640000000000  -0.68270000000000   O   (8f)
  0.28640000000000  -0.09920000000000   1.18270000000000   O   (8f)
  0.20500000000000   0.35420000000000   0.40190000000000   O   (8f)
 -0.35420000000000  -0.20500000000000   0.09810000000000   O   (8f)
 -0.20500000000000  -0.35420000000000  -0.40190000000000   O   (8f)
  0.35420000000000   0.20500000000000   0.90190000000000   O   (8f)
 -0.13700000000000   0.57980000000000   0.25730000000000   O   (8f)
 -0.57980000000000   0.13700000000000   0.24270000000000   O   (8f)
  0.13700000000000  -0.57980000000000  -0.25730000000000   O   (8f)
  0.57980000000000  -0.13700000000000   0.75730000000000   O   (8f)
 -0.15510000000000   0.34610000000000   0.40770000000000   O   (8f)
 -0.34610000000000   0.15510000000000   0.09230000000000   O   (8f)
  0.15510000000000  -0.34610000000000  -0.40770000000000   O   (8f)
  0.34610000000000  -0.15510000000000   0.90770000000000   O   (8f)
 -0.50828000000000   0.50828000000000   0.25000000000000   S   (4e)
  0.50828000000000  -0.50828000000000   0.75000000000000   S   (4e)
  0.01273000000000   0.24683000000000   0.05632000000000   S   (8f)
 -0.24683000000000  -0.01273000000000   0.44368000000000   S   (8f)
 -0.01273000000000  -0.24683000000000  -0.05632000000000   S   (8f)
  0.24683000000000   0.01273000000000   0.55632000000000   S   (8f)
```

$Al_2Mg_5Si_3O_{10}(OH)_8$ ($S5_5$): A5B10C8D4_mC108_15_a2ef_5f_4f_2f - CIF

```
# CIF file
data_findsym-output
_audit_creation_method FINDSYM

_chemical_name_mineral 'Al2H8Mg5O18Si3'
_chemical_formula_sum 'Mg5 O10 (OH)8 Si4'

loop_
_publ_author_name
 'R. C. {McMurchy}'
_journal_name_full_name
;
 Zeitschrift f{\"u}r Kristallographie - Crystalline Materials
;
_journal_volume 88
_journal_year 1934
_journal_page_first 420
_journal_page_last 432
_publ_Section_title
;
 The Crystal Structure of the Chlorite Minerals
;

# Found in Strukturbericht Band III 1933-1935, 1937

_aflow_title 'Al$_{2}$Mg$_{5}$Si$_{3}$O$_{10}$(OH)$_{8}$ ($S5_{5}$)
    ↪ Structure'
_aflow_proto 'A5B10C8D4_mC108_15_a2ef_5f_4f_2f'
_aflow_params 'a,b/a,c/a,\beta,y_{2},y_{3},x_{4},y_{4},z_{4},x_{5},y_{5
    ↪ },z_{5},x_{6},y_{6},z_{6},x_{7},y_{7},z_{7},x_{8},y_{8},z_{8},
    ↪ x_{9},y_{9},z_{9},x_{10},y_{10},z_{10},x_{11},y_{11},z_{11},x_{
    ↪ 12},y_{12},z_{12},x_{13},y_{13},z_{13},x_{14},y_{14},z_{14},x_{
    ↪ 15},y_{15},z_{15}'
_aflow_params_values '5.31,1.73822975518,5.3615819209,97.14444,0.5,0.167
    ↪ ,0.75,0.333,0.0,0.692,0.667,0.039,0.692,0.0,0.039,-0.006,0.0835
    ↪ ,0.886,-0.006,0.583,0.117,0.744,0.167,0.114,0.692,0.667,0.039,
    ↪ 0.142,0.0,0.211,0.142,0.333,0.211,0.142,0.667,0.211,0.731,0.0,
    ↪ 0.094,0.731,0.667,0.094'
_aflow_Strukturbericht '$S5_{5}$'
_aflow_Pearson 'mC108'

_symmetry_space_group_name_H-M "C 1 2/c 1"
_symmetry_Int_Tables_number 15

_cell_length_a      5.31000
_cell_length_b      9.23000
_cell_length_c      28.47000
_cell_angle_alpha   90.00000
_cell_angle_beta    97.14444
_cell_angle_gamma   90.00000

loop_
_space_group_symop_id
_space_group_symop_operation_xyz
1 x,y,z
2 -x,y,-z+1/2
3 -x,-y,-z
4 x,-y,z+1/2
5 x+1/2,y+1/2,z
6 -x+1/2,y+1/2,-z+1/2
7 -x+1/2,-y+1/2,-z
8 x+1/2,-y+1/2,z+1/2

loop_
_atom_site_label
_atom_site_type_symbol
_atom_site_symmetry_multiplicity
_atom_site_Wyckoff_label
_atom_site_fract_x
_atom_site_fract_y
_atom_site_fract_z
_atom_site_occupancy
Mg1 Mg 4 a 0.00000 0.00000 0.00000 1.00000
Mg2 Mg 4 e 0.00000 0.50000 0.25000 1.00000
Mg3 Mg 4 e 0.00000 0.16700 0.25000 1.00000
Mg4 Mg 8 f 0.75000 0.33300 0.00000 1.00000
O1 O 8 f 0.69200 0.66700 0.03900 1.00000
O2 O 8 f 0.69200 0.00000 0.03900 1.00000
O3 O 8 f -0.00600 0.08350 0.88600 1.00000
O4 O 8 f -0.00600 0.58300 0.11700 1.00000
O5 O 8 f 0.74400 0.16700 0.11400 1.00000
OH1 OH 8 f 0.69200 0.66700 0.03900 1.00000
OH2 OH 8 f 0.14200 0.00000 0.21100 1.00000
OH3 OH 8 f 0.14200 0.33300 0.21100 1.00000
OH4 OH 8 f 0.14200 0.66700 0.21100 1.00000
Si1 Si 8 f 0.73100 0.00000 0.09400 1.00000
Si2 Si 8 f 0.73100 0.66700 0.09400 1.00000
```

$Al_2Mg_5Si_3O_{10}(OH)_8$ ($S5_5$): A5B10C8D4_mC108_15_a2ef_5f_4f_2f - POSCAR

```
A5B10C8D4_mC108_15_a2ef_5f_4f_2f & a,b/a,c/a,beta,y2,y3,x4,y4,z4,x5,y5,
    ↪ z5,x6,y6,z6,x7,y7,z7,x8,y8,z8,x9,y9,z9,x10,y10,z10,x11,y11,z11,
    ↪ x12,y12,z12,x13,y13,z13,x14,y14,z14,x15,y15,z15 --params=5.31,
    ↪ 1.73822975518,5.3615819209,97.14444,0.5,0.167,0.75,0.333,0.0,
    ↪ 0.692,0.667,0.039,0.692,0.0,0.039,-0.006,0.0835,0.886,-0.006,
    ↪ 0.583,0.117,0.744,0.167,0.114,0.692,0.667,0.039,0.142,0.0,0.211
    ↪ ,0.142,0.333,0.211,0.142,0.667,0.211,0.731,0.0,0.094,0.731,
    ↪ 0.667,0.094 & C2/c C_{2h}^{6} #15 (ae^2f^12) & mC108 & $S5_{5}$
    ↪ & Al2H8Mg5O18Si3 & Al2H8Mg5O18Si3 & R. C. {McMurchy},
    ↪ Zeitschrift f{"u}r Kristallographie - Crystalline Materials 88,
    ↪ 420-432 (1934)
   1.00000000000000
   2.65500000000000  -4.61500000000000   0.00000000000000
   2.65500000000000   4.61500000000000   0.00000000000000
  -3.54084568016567   0.00000000000000  28.24895240304060
    Mg     O    OH    Si
    10    20    16     8
Direct
   0.00000000000000   0.00000000000000   0.00000000000000   Mg   (4a)
   0.00000000000000   0.00000000000000   0.50000000000000   Mg   (4a)
  -0.50000000000000   0.50000000000000   0.25000000000000   Mg   (4e)
   0.50000000000000  -0.50000000000000   0.75000000000000   Mg   (4e)
  -0.16700000000000   0.16700000000000   0.25000000000000   Mg   (4e)
   0.16700000000000  -0.16700000000000   0.75000000000000   Mg   (4e)
   0.41700000000000   1.08300000000000   0.00000000000000   Mg   (8f)
  -1.08300000000000  -0.41700000000000   0.50000000000000   Mg   (8f)
  -0.41700000000000  -1.08300000000000   0.00000000000000   Mg   (8f)
   1.08300000000000   0.41700000000000   0.50000000000000   Mg   (8f)
   0.02500000000000   1.35900000000000   0.03900000000000   O    (8f)
  -1.35900000000000  -0.02500000000000   0.46100000000000   O    (8f)
  -0.02500000000000  -1.35900000000000  -0.03900000000000   O    (8f)
   1.35900000000000   0.02500000000000   0.53900000000000   O    (8f)
   0.69200000000000   0.69200000000000   0.03900000000000   O    (8f)
  -0.69200000000000  -0.69200000000000   0.46100000000000   O    (8f)
  -0.69200000000000  -0.69200000000000  -0.03900000000000   O    (8f)
   0.69200000000000   0.69200000000000   0.53900000000000   O    (8f)
  -0.08950000000000   0.07750000000000   0.88600000000000   O    (8f)
  -0.07750000000000   0.08950000000000  -0.38600000000000   O    (8f)
   0.08950000000000  -0.07750000000000  -0.88600000000000   O    (8f)
   0.07750000000000  -0.08950000000000   1.38600000000000   O    (8f)
  -0.58900000000000   0.57700000000000   0.11700000000000   O    (8f)
  -0.57700000000000   0.58900000000000   0.38300000000000   O    (8f)
   0.58900000000000  -0.57700000000000  -0.11700000000000   O    (8f)
   0.57700000000000  -0.58900000000000   0.61700000000000   O    (8f)
   0.57700000000000   0.91100000000000   0.11400000000000   O    (8f)
  -0.91100000000000  -0.57700000000000   0.38600000000000   O    (8f)
  -0.57700000000000  -0.91100000000000  -0.11400000000000   O    (8f)
   0.91100000000000   0.57700000000000   0.61400000000000   O    (8f)
   0.02500000000000   1.35900000000000   0.03900000000000   OH   (8f)
  -1.35900000000000  -0.02500000000000   0.46100000000000   OH   (8f)
  -0.02500000000000  -1.35900000000000  -0.03900000000000   OH   (8f)
   1.35900000000000   0.02500000000000   0.53900000000000   OH   (8f)
   0.14200000000000   0.14200000000000   0.21100000000000   OH   (8f)
  -0.14200000000000  -0.14200000000000   0.28900000000000   OH   (8f)
  -0.14200000000000  -0.14200000000000  -0.21100000000000   OH   (8f)
   0.14200000000000   0.14200000000000   0.71100000000000   OH   (8f)
  -0.19100000000000   0.47500000000000   0.21100000000000   OH   (8f)
  -0.47500000000000   0.19100000000000   0.28900000000000   OH   (8f)
   0.19100000000000  -0.47500000000000  -0.21100000000000   OH   (8f)
   0.47500000000000  -0.19100000000000   0.71100000000000   OH   (8f)
  -0.52500000000000   0.80900000000000   0.21100000000000   OH   (8f)
  -0.80900000000000   0.52500000000000   0.28900000000000   OH   (8f)
   0.52500000000000  -0.80900000000000  -0.21100000000000   OH   (8f)
```

```
   0.80900000000000  -0.52500000000000   0.71100000000000   OH   (8f)
   0.73100000000000   0.73100000000000   0.09400000000000   Si   (8f)
  -0.73100000000000  -0.73100000000000   0.40600000000000   Si   (8f)
  -0.73100000000000  -0.73100000000000  -0.09400000000000   Si   (8f)
   0.73100000000000   0.73100000000000   0.59400000000000   Si   (8f)
   0.06400000000000   1.39800000000000   0.09400000000000   Si   (8f)
  -1.39800000000000  -0.06400000000000   0.40600000000000   Si   (8f)
  -0.06400000000000  -1.39800000000000  -0.09400000000000   Si   (8f)
   1.39800000000000   0.06400000000000   0.59400000000000   Si   (8f)
```

$Y_2SiO_5$ ($RE_2SiO_5$ X2): A5BC2_mC64_15_5f_f_2f - CIF

```
# CIF file
data_findsym-output
_audit_creation_method FINDSYM

_chemical_name_mineral 'O5SiY2'
_chemical_formula_sum 'O5 Si Y2'

loop_
_publ_author_name
 'G. V. Anan\'eva'
 'A. M. Korovkin'
 'T. I. Merkulyaeva'
 'A. M. Morozova'
 'M. V. Petrov'
 'I. R. Savinova'
 'V. R. Startsev'
 'P. P. Feofilov'
_journal_name_full_name
;
 Inorganic Materials
;
_journal_volume 17
_journal_year 1981
_journal_page_first 754
_journal_page_last 758
_publ_Section_title
;
 Growth of lanthanide oxyorthosilicate single crystals, and their
    ↪ structural and optical characteristics
;

# Found in Y$_{2}$SiO$_{5}$ (Y$_{2}$[SiO$_{4}$]O ht) Crystal Structure,
    ↪ 2016 Found in Y$_{2}$SiO$_{5}$ (Y$_{2}$[SiO$_{4}$]O ht) Crystal
    ↪ Structure, {PAULING FILE in: Inorganic Solid Phases,
    ↪ SpringerMaterials (online database), Springer, Heidelberg (ed.)
    ↪ SpringerMaterials },

_aflow_title 'Y$_{2}$SiO$_{5}$ ($RE$$_{2}$SiO$_{5}$ X2) Structure'
_aflow_proto 'A5BC2_mC64_15_5f_f_2f'
_aflow_params 'a,b/a,c/a,\beta,x_{1},y_{1},z_{1},x_{2},y_{2},z_{2},x_{3
    ↪ },y_{3},z_{3},x_{4},y_{4},z_{4},x_{5},y_{5},z_{5},x_{6},y_{6},
    ↪ z_{6},x_{7},y_{7},z_{7},x_{8},y_{8},z_{8}'
_aflow_params_values '14.43,0.466597366597,0.721413721414,122.13,0.089,
    ↪ 0.002,0.143,0.118,0.287,0.318,0.297,0.429,0.06,0.298,0.157,0.33
    ↪ ,0.485,0.102,0.103,0.181,0.093,0.308,0.037,0.257,0.466,0.359,
    ↪ 0.122,0.165'
_aflow_Strukturbericht 'None'
_aflow_Pearson 'mC64'

_symmetry_space_group_name_H-M "C 1 2/c 1"
_symmetry_Int_Tables_number 15

_cell_length_a    14.43000
_cell_length_b    6.73300
_cell_length_c    10.41000
_cell_angle_alpha 90.00000
_cell_angle_beta  122.13000
_cell_angle_gamma 90.00000

loop_
_space_group_symop_id
_space_group_symop_operation_xyz
1 x,y,z
2 -x,y,-z+1/2
3 -x,-y,-z
4 x,-y,z+1/2
5 x+1/2,y+1/2,z
6 -x+1/2,y+1/2,-z+1/2
7 -x+1/2,-y+1/2,-z
8 x+1/2,-y+1/2,z+1/2

loop_
_atom_site_label
_atom_site_type_symbol
_atom_site_symmetry_multiplicity
_atom_site_Wyckoff_label
_atom_site_fract_x
_atom_site_fract_y
_atom_site_fract_z
_atom_site_occupancy
O1 O 8 f 0.08900 0.00200 0.14300 1.00000
O2 O 8 f 0.11800 0.28700 0.31800 1.00000
O3 O 8 f 0.29700 0.42900 0.06000 1.00000
O4 O 8 f 0.29800 0.15700 0.33000 1.00000
O5 O 8 f 0.48500 0.10200 0.10300 1.00000
Si1 Si 8 f 0.18100 0.09300 0.30800 1.00000
Y1 Y 8 f 0.03700 0.25700 0.46600 1.00000
Y2 Y 8 f 0.35900 0.12200 0.16500 1.00000
```

$Y_2SiO_5$ ($RE_2SiO_5$ X2): A5BC2_mC64_15_5f_f_2f - POSCAR

```
A5BC2_mC64_15_5f_f_2f & a,b/a,c/a,beta,x1,y1,z1,x2,y2,z2,x3,y3,z3,x4,y4,
    ↪ z4,x5,y5,z5,x6,y6,z6,x7,y7,z7,x8,y8,z8 --params=14.43,
    ↪ 0.466597366597,0.721413721414,122.13,0.089,0.002,0.143,0.118,
    ↪ 0.287,0.318,0.297,0.429,0.06,0.298,0.157,0.33,0.485,0.102,0.103
    ↪ ,0.181,0.093,0.308,0.037,0.257,0.466,0.359,0.122,0.165 & C2/c
    ↪ C_{2h}^{6} #15 (f^8) & mC64 & None & O5SiY2 & O5SiY2 & G. V.
    ↪ Anan'eva et al., Inorg. Mat. 17, 754-758 (1981)
   1.00000000000000
   7.21500000000000  -3.36650000000000   0.00000000000000
   7.21500000000000   3.36650000000000   0.00000000000000
  -5.53647583060549   0.00000000000000   8.81564151818353
     O    Si     Y
    20     4     8
Direct
   0.08700000000000   0.09100000000000   0.14300000000000   O   (8f)
  -0.09100000000000  -0.08700000000000   0.35700000000000   O   (8f)
  -0.08700000000000  -0.09100000000000  -0.14300000000000   O   (8f)
   0.09100000000000   0.08700000000000   0.64300000000000   O   (8f)
  -0.16900000000000   0.40500000000000   0.31800000000000   O   (8f)
  -0.40500000000000   0.16900000000000   0.18200000000000   O   (8f)
   0.16900000000000  -0.40500000000000  -0.31800000000000   O   (8f)
   0.40500000000000  -0.16900000000000   0.81800000000000   O   (8f)
  -0.13200000000000   0.72600000000000   0.06000000000000   O   (8f)
  -0.72600000000000   0.13200000000000   0.44000000000000   O   (8f)
   0.13200000000000  -0.72600000000000  -0.06000000000000   O   (8f)
   0.72600000000000  -0.13200000000000   0.56000000000000   O   (8f)
   0.14100000000000   0.45500000000000   0.33000000000000   O   (8f)
  -0.45500000000000  -0.14100000000000   0.17000000000000   O   (8f)
  -0.14100000000000  -0.45500000000000  -0.33000000000000   O   (8f)
   0.45500000000000   0.14100000000000   0.83000000000000   O   (8f)
   0.38300000000000   0.58700000000000   0.10300000000000   O   (8f)
  -0.58700000000000  -0.38300000000000   0.39700000000000   O   (8f)
  -0.38300000000000  -0.58700000000000  -0.10300000000000   O   (8f)
   0.58700000000000   0.38300000000000   0.60300000000000   O   (8f)
   0.08800000000000   0.27400000000000   0.30800000000000   Si  (8f)
  -0.27400000000000  -0.08800000000000   0.19200000000000   Si  (8f)
  -0.08800000000000  -0.27400000000000  -0.30800000000000   Si  (8f)
   0.27400000000000   0.08800000000000   0.80800000000000   Si  (8f)
  -0.22000000000000   0.29400000000000   0.46600000000000   Y   (8f)
  -0.29400000000000   0.22000000000000   0.03400000000000   Y   (8f)
   0.22000000000000  -0.29400000000000  -0.46600000000000   Y   (8f)
   0.29400000000000  -0.22000000000000   0.96600000000000   Y   (8f)
   0.23700000000000   0.48100000000000   0.16500000000000   Y   (8f)
  -0.48100000000000  -0.23700000000000   0.33500000000000   Y   (8f)
  -0.23700000000000  -0.48100000000000  -0.16500000000000   Y   (8f)
   0.48100000000000   0.23700000000000   0.66500000000000   Y   (8f)
```

$\alpha$-$Zn_2V_2O_7$: A7B2C2_mC44_15_e3f_f_f - CIF

```
# CIF file
data_findsym-output
_audit_creation_method FINDSYM

_chemical_name_mineral 'O7V2Zn2'
_chemical_formula_sum 'O7 V2 Zn2'

loop_
_publ_author_name
 'R. Gopal'
 'C. Calvo'
_journal_name_full_name
;
 Canadian Journal of Chemistry
;
_journal_volume 51
_journal_year 1973
_journal_page_first 1004
_journal_page_last 1009
_publ_Section_title
;
 Crystal Structure of $\alpha$-Zn$_{2}$V$_{2}$O$_{7}$
;

# Found in $\alpha$ Cupric Divanadate, 1975

_aflow_title '$\alpha$-Zn$_{2}$V$_{2}$O$_{7}$ Structure'
_aflow_proto 'A7B2C2_mC44_15_e3f_f_f'
_aflow_params 'a,b/a,c/a,\beta,y_{1},x_{2},y_{2},z_{2},x_{3},y_{3},z_{3
    ↪ },x_{4},y_{4},z_{4},x_{5},y_{5},z_{5},x_{6},y_{6},z_{6}'
_aflow_params_values '7.429,1.12262754072,1.35926773455,111.37,0.0612,
    ↪ 0.3984,-0.0189,0.362,0.244,0.1541,0.1056,0.1531,0.8353,0.1138,
    ↪ 0.2016,0.0049,0.2058,-0.04958,0.3239,0.51955'
_aflow_Strukturbericht 'None'
_aflow_Pearson 'mC44'

_symmetry_space_group_name_H-M "C 1 2/c 1"
_symmetry_Int_Tables_number 15

_cell_length_a    7.42900
_cell_length_b    8.34000
_cell_length_c    10.09800
_cell_angle_alpha 90.00000
_cell_angle_beta  111.37000
_cell_angle_gamma 90.00000

loop_
_space_group_symop_id
_space_group_symop_operation_xyz
1 x,y,z
2 -x,y,-z+1/2
3 -x,-y,-z
4 x,-y,z+1/2
5 x+1/2,y+1/2,z
6 -x+1/2,y+1/2,-z+1/2
7 -x+1/2,-y+1/2,-z
8 x+1/2,-y+1/2,z+1/2
```

```
loop_
_atom_site_label
_atom_site_type_symbol
_atom_site_symmetry_multiplicity
_atom_site_Wyckoff_label
_atom_site_fract_x
_atom_site_fract_y
_atom_site_fract_z
_atom_site_occupancy
O1 O 4 e 0.00000 0.06120 0.25000 1.00000
O2 O 8 f 0.39840 -0.01890 0.36200 1.00000
O3 O 8 f 0.24400 0.15410 0.10560 1.00000
O4 O 8 f 0.15310 0.83530 0.11380 1.00000
V1 V 8 f 0.20160 0.00490 0.20580 1.00000
Zn1 Zn 8 f -0.04958 0.32390 0.51955 1.00000
```

$\alpha$-$Zn_2V_2O_7$: A7B2C2_mC44_15_e3f_f_f - POSCAR

```
A7B2C2_mC44_15_e3f_f_f & a,b/a,c/a,beta,y1,x2,y2,z2,x3,y3,z3,x4,y4,z4,x5
   ↪ ,y5,z5,x6,y6,z6 --params=7.429,1.12262754072,1.35926773455,
   ↪ 111.37,0.0612,0.3984,-0.0189,0.362,0.244,0.1541,0.1056,0.1531,
   ↪ 0.8353,0.1138,0.2016,0.0049,0.2058,-0.04958,0.3239,0.51955 & C2
   ↪ /c C_{2h}^{6} #15 (ef^5) & mC44 & None & O7V2Zn2 & O7V2Zn2 & R.
   ↪ Gopal and C. Calvo, Can. J. Chem. 51, 1004-1009 (1973)
   1.00000000000000
   3.71450000000000  -4.17000000000000   0.00000000000000
   3.71450000000000   4.17000000000000   0.00000000000000
  -3.67960249157826   0.00000000000000   9.40372955289395
     O       V      Zn
    14       4       4
Direct
  -0.06120000000000   0.06120000000000   0.25000000000000   O   (4e)
   0.06120000000000  -0.06120000000000   0.75000000000000   O   (4e)
   0.41730000000000   0.37950000000000   0.36200000000000   O   (8f)
  -0.37950000000000  -0.41730000000000   0.13800000000000   O   (8f)
  -0.41730000000000  -0.37950000000000  -0.36200000000000   O   (8f)
   0.37950000000000   0.41730000000000   0.86200000000000   O   (8f)
   0.08990000000000   0.39810000000000   0.10560000000000   O   (8f)
  -0.39810000000000  -0.08990000000000   0.39440000000000   O   (8f)
  -0.08990000000000  -0.39810000000000  -0.10560000000000   O   (8f)
   0.39810000000000   0.08990000000000   0.60560000000000   O   (8f)
  -0.68220000000000   0.98840000000000   0.11380000000000   O   (8f)
  -0.98840000000000   0.68220000000000   0.38620000000000   O   (8f)
   0.68220000000000  -0.98840000000000  -0.11380000000000   O   (8f)
   0.98840000000000  -0.68220000000000   0.61380000000000   O   (8f)
   0.19670000000000   0.20650000000000   0.20580000000000   V   (8f)
  -0.20650000000000  -0.19670000000000   0.29420000000000   V   (8f)
  -0.19670000000000  -0.20650000000000  -0.20580000000000   V   (8f)
   0.20650000000000   0.19670000000000   0.70580000000000   V   (8f)
  -0.37348000000000   0.27432000000000   0.51955000000000   Zn  (8f)
  -0.27432000000000   0.37348000000000  -0.01955000000000   Zn  (8f)
   0.37348000000000  -0.27432000000000  -0.51955000000000   Zn  (8f)
   0.27432000000000  -0.37348000000000   1.01955000000000   Zn  (8f)
```

Manganese-leonite 185 K [$K_2Mn(SO_4)_2{\cdot}4H_2O$]: A8B2CD12E2_mC200_15_8f_2f_ce_2e11f_2f - CIF

```
# CIF file
data_findsym-output
_audit_creation_method FINDSYM

_chemical_name_mineral 'Manganese-leonite'
_chemical_formula_sum 'H8 K2 Mn O12 S2'

loop_
_publ_author_name
 'B. Hertweck'
 'G. Giester'
 'E. Libowitzky'
_journal_name_full_name
;
 American Mineralogist
;
_journal_volume 86
_journal_year 2001
_journal_page_first 1282
_journal_page_last 1292
_publ_Section_title
;
 The crystal structures of the low-temperature phases of leonite-type
   ↪ compounds, K$_{2}$$Me$(SO$_{4}$)$_{2}$$\cdot$4H$_{2}$O ($Me^{2
   ↪ +}$ = Mg, Mn, Fe)
;

_aflow_title 'Manganese-leonite 185~K [K$_{2}$Mn(SO$_{4}$)$_{2}$$\
   ↪ cdot$4H$_{2}$O] Structure'
_aflow_proto 'A8B2CD12E2_mC200_15_8f_2f_ce_2e11f_2f'
_aflow_params 'a,b/a,c/a,\beta,y_{2},y_{3},y_{4},x_{5},y_{5},z_{5},x_{6
   ↪ },y_{6},z_{6},x_{7},y_{7},z_{7},x_{8},y_{8},z_{8},x_{9},y_{9},
   ↪ z_{9},x_{10},y_{10},z_{10},x_{11},y_{11},z_{11},x_{12},y_{12},
   ↪ z_{12},x_{13},y_{13},z_{13},x_{14},y_{14},z_{14},x_{15},y_{15},
   ↪ z_{15},x_{16},y_{16},z_{16},x_{17},y_{17},z_{17},x_{18},y_{18},
   ↪ z_{18},x_{19},y_{19},z_{19},x_{20},y_{20},z_{20},x_{21},y_{21},
   ↪ z_{21},x_{22},y_{22},z_{22},x_{23},y_{23},z_{23},x_{24},y_{24},
   ↪ z_{24},x_{25},y_{25},z_{25},x_{26},y_{26},z_{26},x_{27},y_{27},
   ↪ z_{27}'
_aflow_params_values '22.291,0.42837916648,0.539903997129,117.548,0.2496
   ↪ ,0.478,0.0215,0.1942,0.4916,0.023,0.1939,0.0004,0.5205,0.3664,
   ↪ 0.1105,0.0332,0.3651,0.3925,0.0298,0.4808,0.036,0.1753,0.481,
   ↪ 0.4628,0.1753,0.1379,0.2306,0.4589,0.1408,0.258,0.334,0.3794,
   ↪ 0.017,0.2991,0.372,0.4906,0.292,0.4444,0.1248,0.036,0.4419,
   ↪ 0.3781,0.0326,0.2991,0.3418,0.3983,0.2731,0.0985,0.3462,0.0301,
   ↪ 0.2498,0.1074,0.5483,0.257,0.1144,0.2827,0.2554,0.2006,0.1872,
   ↪ 0.2697,0.2437,0.3248,0.0844,0.0375,0.3231,0.417,0.0324,0.1102,
   ↪ 0.2518,0.3703,0.4779,0.2528,0.0207,0.2598,0.2429,0.2977'
_aflow_Strukturbericht 'None'
_aflow_Pearson 'mC200'

_symmetry_space_group_name_H-M "C 1 2/c 1"
_symmetry_Int_Tables_number 15

_cell_length_a    22.29100
_cell_length_b    9.54900
_cell_length_c    12.03500
_cell_angle_alpha 90.00000
_cell_angle_beta  117.54800
_cell_angle_gamma 90.00000

loop_
_space_group_symop_id
_space_group_symop_operation_xyz
1 x,y,z
2 -x,y,-z+1/2
3 -x,-y,-z
4 x,-y,z+1/2
5 x+1/2,y+1/2,z
6 -x+1/2,y+1/2,-z+1/2
7 -x+1/2,-y+1/2,-z
8 x+1/2,-y+1/2,z+1/2

loop_
_atom_site_label
_atom_site_type_symbol
_atom_site_symmetry_multiplicity
_atom_site_Wyckoff_label
_atom_site_fract_x
_atom_site_fract_y
_atom_site_fract_z
_atom_site_occupancy
Mn1 Mn 4 c 0.25000 0.25000 0.00000 1.00000
Mn2 Mn 4 e 0.00000 0.24960 0.25000 1.00000
O1 O 4 e 0.00000 0.47800 0.25000 1.00000
O2 O 4 e 0.00000 0.02150 0.25000 1.00000
H1 H 8 f 0.19420 0.49160 0.02300 1.00000
H2 H 8 f 0.19390 0.00040 0.52050 1.00000
H3 H 8 f 0.36640 0.11050 0.03320 1.00000
H4 H 8 f 0.36510 0.39250 0.02980 1.00000
H5 H 8 f 0.48080 0.03600 0.17530 1.00000
H6 H 8 f 0.48100 0.46280 0.17530 1.00000
H7 H 8 f 0.13790 0.23060 0.45890 1.00000
H8 H 8 f 0.14080 0.25800 0.33400 1.00000
K1 K 8 f 0.37940 0.01700 0.29910 1.00000
K2 K 8 f 0.37200 0.49060 0.29200 1.00000
O3 O 8 f 0.44440 0.12480 0.03600 1.00000
O4 O 8 f 0.44190 0.37810 0.03260 1.00000
O5 O 8 f 0.29910 0.34180 0.39830 1.00000
O6 O 8 f 0.27310 0.09850 0.34620 1.00000
O7 O 8 f 0.03010 0.24980 0.10740 1.00000
O8 O 8 f 0.54830 0.25700 0.11440 1.00000
O9 O 8 f 0.28270 0.25540 0.20060 1.00000
O10 O 8 f 0.18720 0.26970 0.24370 1.00000
O11 O 8 f 0.32480 0.08440 0.03750 1.00000
O12 O 8 f 0.32310 0.41700 0.03240 1.00000
O13 O 8 f 0.11020 0.25180 0.37030 1.00000
S1 S 8 f 0.47790 0.25280 0.02070 1.00000
S2 S 8 f 0.25980 0.24290 0.29770 1.00000
```

Manganese-leonite 185 K [$K_2Mn(SO_4)_2{\cdot}4H_2O$]: A8B2CD12E2_mC200_15_8f_2f_ce_2e11f_2f - POSCAR

```
A8B2CD12E2_mC200_15_8f_2f_ce_2e11f_2f & a,b/a,c/a,beta,y2,y3,y4,x5,y5,z5
   ↪ ,x6,y6,z6,x7,y7,z7,x8,y8,z8,x9,y9,z9,x10,y10,z10,x11,y11,z11,
   ↪ x12,y12,z12,x13,y13,z13,x14,y14,z14,x15,y15,z15,x16,y16,z16,x17
   ↪ ,y17,z17,x18,y18,z18,x19,y19,z19,x20,y20,z20,x21,y21,z21,x22,
   ↪ y22,z22,x23,y23,z23,x24,y24,z24,x25,y25,z25,x26,y26,z26,x27,y27
   ↪ ,z27 --params=22.291,0.42837916648,0.539903997129,117.548,
   ↪ 0.2496,0.478,0.0215,0.1942,0.4916,0.023,0.1939,0.0004,0.5205,
   ↪ 0.3664,0.1105,0.0332,0.3651,0.3925,0.0298,0.4808,0.036,0.1753,
   ↪ 0.481,0.4628,0.1753,0.1379,0.2306,0.4589,0.1408,0.258,0.334,
   ↪ 0.3794,0.017,0.2991,0.372,0.4906,0.292,0.4444,0.1248,0.036,
   ↪ 0.4419,0.3781,0.0326,0.2991,0.3418,0.3983,0.2731,0.0985,0.3462,
   ↪ 0.0301,0.2498,0.1074,0.5483,0.257,0.1144,0.2827,0.2554,0.2006,
   ↪ 0.1872,0.2697,0.2437,0.3248,0.0844,0.0375,0.3231,0.417,0.0324,
   ↪ 0.1102,0.2518,0.3703,0.4779,0.2528,0.0207,0.2598,0.2429,0.2977
   ↪ & C2/c C_{2h}^{6} #15 (ce^3f^23) & mC200 & None & H8K2MnO12S2 &
   ↪ Manganese-leonite & B. Hertweck and G. Giester and E.
   ↪ Libowitzky, Am. Mineral. 86, 1282-1292 (2001)
   1.00000000000000
  11.14550000000000  -4.77450000000000   0.00000000000000
  11.14550000000000   4.77450000000000   0.00000000000000
  -5.56608582313949   0.00000000000000  10.67051608917980
     H       K      Mn       O       S
    32       8       4      48       8
Direct
  -0.29740000000000   0.68580000000000   0.02300000000000   H   (8f)
  -0.68580000000000   0.29740000000000   0.47700000000000   H   (8f)
   0.29740000000000  -0.68580000000000  -0.02300000000000   H   (8f)
   0.68580000000000  -0.29740000000000   0.52300000000000   H   (8f)
   0.19350000000000   0.19430000000000   0.52050000000000   H   (8f)
  -0.19430000000000  -0.19350000000000  -0.02050000000000   H   (8f)
  -0.19350000000000  -0.19430000000000  -0.52050000000000   H   (8f)
   0.19430000000000   0.19350000000000   1.02050000000000   H   (8f)
   0.25590000000000   0.47690000000000   0.03320000000000   H   (8f)
  -0.47690000000000  -0.25590000000000   0.46680000000000   H   (8f)
  -0.25590000000000  -0.47690000000000  -0.03320000000000   H   (8f)
   0.47690000000000   0.25590000000000   0.53320000000000   H   (8f)
  -0.02740000000000   0.75760000000000   0.02980000000000   H   (8f)
  -0.75760000000000   0.02740000000000   0.47020000000000   H   (8f)
   0.02740000000000  -0.75760000000000  -0.02980000000000   H   (8f)
   0.75760000000000  -0.02740000000000   0.52980000000000   H   (8f)
   0.44480000000000   0.51680000000000   0.17530000000000   H   (8f)
```

```
   -0.51680000000000   -0.44480000000000    0.32470000000000    H    (8f)
   -0.44480000000000   -0.51680000000000   -0.17530000000000    H    (8f)
    0.51680000000000    0.44480000000000    0.67530000000000    H    (8f)
    0.01820000000000    0.94380000000000    0.17530000000000    H    (8f)
   -0.94380000000000   -0.01820000000000    0.32470000000000    H    (8f)
   -0.01820000000000   -0.94380000000000   -0.17530000000000    H    (8f)
    0.94380000000000    0.01820000000000    0.67530000000000    H    (8f)
   -0.09270000000000    0.36850000000000    0.45890000000000    H    (8f)
   -0.36850000000000    0.09270000000000    0.04110000000000    H    (8f)
    0.09270000000000   -0.36850000000000   -0.45890000000000    H    (8f)
    0.36850000000000   -0.09270000000000    0.95890000000000    H    (8f)
   -0.11720000000000    0.39880000000000    0.33400000000000    H    (8f)
   -0.39880000000000    0.11720000000000    0.16600000000000    H    (8f)
    0.11720000000000   -0.39880000000000   -0.33400000000000    H    (8f)
    0.39880000000000   -0.11720000000000    0.83400000000000    H    (8f)
    0.36240000000000    0.39640000000000    0.29910000000000    K    (8f)
   -0.39640000000000   -0.36240000000000    0.20090000000000    K    (8f)
   -0.36240000000000   -0.39640000000000   -0.29910000000000    K    (8f)
    0.39640000000000    0.36240000000000    0.79910000000000    K    (8f)
   -0.11860000000000    0.86260000000000    0.29200000000000    K    (8f)
   -0.86260000000000    0.11860000000000    0.20800000000000    K    (8f)
    0.11860000000000   -0.86260000000000   -0.29200000000000    K    (8f)
    0.86260000000000   -0.11860000000000    0.79200000000000    K    (8f)
    0.00000000000000    0.50000000000000    0.00000000000000    Mn   (4c)
    0.50000000000000    0.00000000000000    0.50000000000000    Mn   (4c)
   -0.24960000000000    0.24960000000000    0.25000000000000    Mn   (4e)
    0.24960000000000   -0.24960000000000    0.75000000000000    Mn   (4e)
   -0.47800000000000    0.47800000000000    0.25000000000000    O    (4e)
    0.47800000000000   -0.47800000000000    0.75000000000000    O    (4e)
   -0.02150000000000    0.02150000000000    0.25000000000000    O    (4e)
    0.02150000000000   -0.02150000000000    0.75000000000000    O    (4e)
    0.31960000000000    0.56920000000000    0.03600000000000    O    (8f)
   -0.56920000000000   -0.31960000000000    0.46400000000000    O    (8f)
   -0.31960000000000   -0.56920000000000   -0.03600000000000    O    (8f)
    0.56920000000000    0.31960000000000    0.53600000000000    O    (8f)
    0.06380000000000    0.82000000000000    0.03260000000000    O    (8f)
   -0.82000000000000   -0.06380000000000    0.46740000000000    O    (8f)
   -0.06380000000000   -0.82000000000000   -0.03260000000000    O    (8f)
    0.82000000000000    0.06380000000000    0.53260000000000    O    (8f)
   -0.04270000000000    0.64090000000000    0.39830000000000    O    (8f)
   -0.64090000000000    0.04270000000000    0.10170000000000    O    (8f)
    0.04270000000000   -0.64090000000000   -0.39830000000000    O    (8f)
    0.64090000000000   -0.04270000000000    0.89830000000000    O    (8f)
    0.17460000000000    0.37160000000000    0.34620000000000    O    (8f)
   -0.37160000000000   -0.17460000000000    0.15380000000000    O    (8f)
   -0.17460000000000   -0.37160000000000   -0.34620000000000    O    (8f)
    0.37160000000000    0.17460000000000    0.84620000000000    O    (8f)
   -0.21970000000000    0.27990000000000    0.10740000000000    O    (8f)
   -0.27990000000000    0.21970000000000    0.39260000000000    O    (8f)
    0.21970000000000   -0.27990000000000   -0.10740000000000    O    (8f)
    0.27990000000000   -0.21970000000000    0.60740000000000    O    (8f)
    0.29130000000000    0.80530000000000    0.11440000000000    O    (8f)
   -0.80530000000000   -0.29130000000000    0.38560000000000    O    (8f)
   -0.29130000000000   -0.80530000000000   -0.11440000000000    O    (8f)
    0.80530000000000    0.29130000000000    0.61440000000000    O    (8f)
    0.02730000000000    0.53810000000000    0.20060000000000    O    (8f)
   -0.53810000000000   -0.02730000000000    0.29940000000000    O    (8f)
   -0.02730000000000   -0.53810000000000   -0.20060000000000    O    (8f)
    0.53810000000000    0.02730000000000    0.70060000000000    O    (8f)
   -0.08250000000000    0.45690000000000    0.24370000000000    O    (8f)
   -0.45690000000000    0.08250000000000    0.25630000000000    O    (8f)
    0.08250000000000   -0.45690000000000   -0.24370000000000    O    (8f)
    0.45690000000000   -0.08250000000000    0.74370000000000    O    (8f)
    0.24040000000000    0.40920000000000    0.03750000000000    O    (8f)
   -0.40920000000000   -0.24040000000000    0.46250000000000    O    (8f)
   -0.24040000000000   -0.40920000000000   -0.03750000000000    O    (8f)
    0.40920000000000    0.24040000000000    0.53750000000000    O    (8f)
   -0.09390000000000    0.74010000000000    0.03240000000000    O    (8f)
   -0.74010000000000    0.09390000000000    0.46760000000000    O    (8f)
    0.09390000000000   -0.74010000000000   -0.03240000000000    O    (8f)
    0.74010000000000   -0.09390000000000    0.53240000000000    O    (8f)
   -0.14160000000000    0.36200000000000    0.37030000000000    O    (8f)
   -0.36200000000000    0.14160000000000    0.12970000000000    O    (8f)
    0.14160000000000   -0.36200000000000   -0.37030000000000    O    (8f)
    0.36200000000000   -0.14160000000000    0.87030000000000    O    (8f)
    0.22510000000000    0.73070000000000    0.02070000000000    S    (8f)
   -0.73070000000000   -0.22510000000000    0.47930000000000    S    (8f)
   -0.22510000000000   -0.73070000000000   -0.02070000000000    S    (8f)
    0.73070000000000    0.22510000000000    0.52070000000000    S    (8f)
    0.01690000000000    0.50270000000000    0.29770000000000    S    (8f)
   -0.50270000000000   -0.01690000000000    0.20230000000000    S    (8f)
   -0.01690000000000   -0.50270000000000   -0.29770000000000    S    (8f)
    0.50270000000000    0.01690000000000    0.79770000000000    S    (8f)
```

$BaNi(CN)_4 \cdot 4H_2O$ ($H4_{22}$): AB4C4D4E_mC56_15_e_2f_2f_2f_a - CIF

```
# CIF file
data_findsym-output
_audit_creation_method FINDSYM

_chemical_name_mineral 'BaC4(H2O)4N4Ni'
_chemical_formula_sum 'Ba C4 (H2O)4 N4 Ni'

loop_
_publ_author_name
 'F. K. Larsen'
 'R. G. Hazell'
 'S. E. Rasmussen'
_journal_name_full_name
;
 Acta Chemica Scandinavica
;
_journal_volume 23
_journal_year 1969
_journal_page_first 61
_journal_page_last 69
_publ_Section_title
;
 The Crystal Structure of Barium Tetracyanonickelate(II) Tetrahydrate
;

_aflow_title 'BaNi(CN)$_{4}$$\cdot$4H$_{2}$O ($H4_{22}$) Structure'
_aflow_proto 'AB4C4D4E_mC56_15_e_2f_2f_2f_a'
_aflow_params 'a,b/a,c/a,\beta,y_{2},x_{3},y_{3},z_{3},x_{4},y_{4},z_{4
    ↪ },x_{5},y_{5},z_{5},x_{6},y_{6},z_{6},x_{7},y_{7},z_{7},x_{8},
    ↪ y_{8},z_{8}'
_aflow_params_values '12.07,1.12758906379,0.556752278376,107.54,0.37049,
    ↪ 0.0641,0.125,0.0557,0.6475,0.4437,0.0905,0.3588,0.1896,0.3431,
    ↪ 0.4018,0.0551,0.0562,0.0997,0.2032,0.0914,0.7393,0.4076,0.1439'
_aflow_Strukturbericht '$H4_{22}$'
_aflow_Pearson 'mC56'

_symmetry_space_group_name_H-M "C 1 2/c 1"
_symmetry_Int_Tables_number 15

_cell_length_a    12.07000
_cell_length_b    13.61000
_cell_length_c    6.72000
_cell_angle_alpha 90.00000
_cell_angle_beta  107.54000
_cell_angle_gamma 90.00000

loop_
_space_group_symop_id
_space_group_symop_operation_xyz
1 x,y,z
2 -x,y,-z+1/2
3 -x,-y,-z
4 x,-y,z+1/2
5 x+1/2,y+1/2,z
6 -x+1/2,y+1/2,-z+1/2
7 -x+1/2,-y+1/2,-z
8 x+1/2,-y+1/2,z+1/2

loop_
_atom_site_label
_atom_site_type_symbol
_atom_site_symmetry_multiplicity
_atom_site_Wyckoff_label
_atom_site_fract_x
_atom_site_fract_y
_atom_site_fract_z
_atom_site_occupancy
Ni1 Ni 4 a 0.00000 0.00000 0.00000 1.00000
Ba1 Ba 4 e 0.00000 0.37049 0.25000 1.00000
C1 C 8 f 0.06410 0.12500 0.05570 1.00000
C2 C 8 f 0.64750 0.44370 0.09050 1.00000
H2O1 H2O 8 f 0.35880 0.18960 0.34310 1.00000
H2O2 H2O 8 f 0.40180 0.05510 0.05620 1.00000
N1 N 8 f 0.09970 0.20320 0.09140 1.00000
N2 N 8 f 0.73930 0.40760 0.14390 1.00000
```

$BaNi(CN)_4 \cdot 4H_2O$ ($H4_{22}$): AB4C4D4E_mC56_15_e_2f_2f_2f_a - POSCAR

```
AB4C4D4E_mC56_15_e_2f_2f_2f_a & a,b/a,c/a,beta,y2,x3,y3,z3,x4,y4,z4,x5,
    ↪ y5,z5,x6,y6,z6,x7,y7,z7,x8,y8,z8 --params=12.07,1.12758906379,
    ↪ 0.556752278376,107.54,0.37049,0.0641,0.125,0.0557,0.6475,0.4437
    ↪ ,0.0905,0.3588,0.1896,0.3431,0.4018,0.0551,0.0562,0.0997,0.2032
    ↪ ,0.0914,0.7393,0.4076,0.1439 & C2/c C_{2h}^{6} #15 (aef^6) &
    ↪ mC56 & $H4_{22}$ & BaC4(H2O)4N4Ni & BaC4(H2O)4N4Ni & F. K.
    ↪ Larsen and R. G. Hazell and S. E. Rasmussen, Acta Chem. Scand.
    ↪ 23, 61-69 (1969)
   1.00000000000000
   6.03500000000000   -6.80500000000000    0.00000000000000
   6.03500000000000    6.80500000000000    0.00000000000000
  -2.02521679051048    0.00000000000000    6.40756560258531
     Ba        C      H2O        N       Ni
      2        8        8        8        2
Direct
  -0.37049000000000    0.37049000000000    0.25000000000000    Ba    (4e)
   0.37049000000000   -0.37049000000000    0.75000000000000    Ba    (4e)
  -0.06090000000000    0.18910000000000    0.05570000000000    C     (8f)
  -0.18910000000000    0.06090000000000    0.44430000000000    C     (8f)
   0.06090000000000   -0.18910000000000   -0.05570000000000    C     (8f)
   0.18910000000000   -0.06090000000000    0.55570000000000    C     (8f)
   0.20380000000000    1.09120000000000    0.09050000000000    C     (8f)
  -1.09120000000000   -0.20380000000000    0.40950000000000    C     (8f)
  -0.20380000000000   -1.09120000000000   -0.09050000000000    C     (8f)
   1.09120000000000    0.20380000000000    0.59050000000000    C     (8f)
   0.16920000000000    0.54840000000000    0.34310000000000    H2O   (8f)
  -0.54840000000000   -0.16920000000000    0.15690000000000    H2O   (8f)
  -0.16920000000000   -0.54840000000000   -0.34310000000000    H2O   (8f)
   0.54840000000000    0.16920000000000    0.84310000000000    H2O   (8f)
   0.34670000000000    0.45690000000000    0.05620000000000    H2O   (8f)
  -0.45690000000000   -0.34670000000000    0.44380000000000    H2O   (8f)
  -0.34670000000000   -0.45690000000000   -0.05620000000000    H2O   (8f)
   0.45690000000000    0.34670000000000    0.55620000000000    H2O   (8f)
  -0.10350000000000    0.30290000000000    0.09140000000000    N     (8f)
  -0.30290000000000    0.10350000000000    0.40860000000000    N     (8f)
   0.10350000000000   -0.30290000000000   -0.09140000000000    N     (8f)
   0.30290000000000   -0.10350000000000    0.59140000000000    N     (8f)
   0.33170000000000    1.14690000000000    0.14390000000000    N     (8f)
  -1.14690000000000   -0.33170000000000    0.35610000000000    N     (8f)
  -0.33170000000000   -1.14690000000000   -0.14390000000000    N     (8f)
   1.14690000000000    0.33170000000000    0.64390000000000    N     (8f)
   0.00000000000000    0.00000000000000    0.00000000000000    Ni    (4a)
   0.00000000000000    0.00000000000000    0.50000000000000    Ni    (4a)
```

Gypsum ($CaSO_4 \cdot 2H_2O$, $H4_6$): AB4C6D_mC48_15_e_2f_3f_e - CIF

```
# CIF file
data_findsym-output
_audit_creation_method FINDSYM

_chemical_name_mineral 'Gypsum'
_chemical_formula_sum 'Ca H4 O6 S'

loop_
_publ_author_name
 'P. Comodi'
 'S. Nazzareni'
 'P. F. Zanazzi'
 'S. Speziale'
_journal_name_full_name
;
 American Mineralogist
;
_journal_volume 93
_journal_year 2008
_journal_page_first 1530
_journal_page_last 1537
_publ_Section_title
;
 High-pressure behavior of gypsum: A single-crystal X-ray study
;

# Found in The American Mineralogist Crystal Structure Database, 2003

_aflow_title 'Gypsum (CaSO$_{4}$$\cdot$2H$_{2}$O, $H4_{6}$) Structure'
_aflow_proto 'AB4C6D_mC48_15_e_2f_3f_e'
_aflow_params 'a,b/a,c/a,\beta,y_{1},y_{2},x_{3},y_{3},z_{3},x_{4},y_{4
    ↪ },z_{4},x_{5},y_{5},z_{5},x_{6},y_{6},z_{6},x_{7},y_{7},z_{7}'
_aflow_params_values '6.277,2.41851202804,0.90361637725,114.11,0.1705,
    ↪ 0.67273,0.742,0.087,0.766,0.756,0.02,-0.077,0.08319,0.27218,
    ↪ 0.59103,0.19997,0.38195,-0.08702,0.79177,0.06826,-0.07831'
_aflow_Strukturbericht '$H4_{6}$'
_aflow_Pearson 'mC48'

_symmetry_space_group_name_H-M "C 1 2/c 1"
_symmetry_Int_Tables_number 15

_cell_length_a      6.27700
_cell_length_b      15.18100
_cell_length_c      5.67200
_cell_angle_alpha   90.00000
_cell_angle_beta    114.11000
_cell_angle_gamma   90.00000

loop_
_space_group_symop_id
_space_group_symop_operation_xyz
1 x,y,z
2 -x,y,-z+1/2
3 -x,-y,-z
4 x,-y,z+1/2
5 x+1/2,y+1/2,z
6 -x+1/2,y+1/2,-z+1/2
7 -x+1/2,-y+1/2,-z
8 x+1/2,-y+1/2,z+1/2

loop_
_atom_site_label
_atom_site_type_symbol
_atom_site_symmetry_multiplicity
_atom_site_Wyckoff_label
_atom_site_fract_x
_atom_site_fract_y
_atom_site_fract_z
_atom_site_occupancy
Ca1 Ca 4 e 0.00000 0.17050 0.25000 1.00000
S1 S 4 e 0.00000 0.67273 0.25000 1.00000
H1 H 8 f 0.74200 0.08700 0.76600 1.00000
H2 H 8 f 0.75600 0.02000 -0.07700 1.00000
O1 O 8 f 0.08319 0.27218 0.59103 1.00000
O2 O 8 f 0.19997 0.38195 -0.08702 1.00000
O3 O 8 f 0.79177 0.06826 -0.07831 1.00000
```

Gypsum ($CaSO_4 \cdot 2H_2O$, $H4_6$): AB4C6D_mC48_15_e_2f_3f_e - POSCAR

```
AB4C6D_mC48_15_e_2f_3f_e & a,b/a,c/a,beta,y1,y2,x3,y3,z3,x4,y4,z4,x5,y5,
    ↪ z5,x6,y6,z6,x7,y7,z7 --params=6.277,2.41851202804,0.90361637725
    ↪ ,114.11,0.1705,0.67273,0.742,0.087,0.766,0.756,0.02,-0.077,
    ↪ 0.08319,0.27218,0.59103,0.19997,0.38195,-0.08702,0.79177,
    ↪ 0.06826,-0.07831 & C2/c C_{2h}^{6} #15 (e^2f^5) & mC48 & $H4_{6
    ↪ }$ & CaH4O6S & Gypsum & P. Comodi et al., Am. Mineral. 93,
    ↪ 1530-1537 (2008)
   1.00000000000000
   3.13850000000000  -7.59050000000000   0.00000000000000
   3.13850000000000   7.59050000000000   0.00000000000000
  -2.31695399688312   0.00000000000000   5.17719114736237
    Ca     H     O     S
     2     8    12     2
Direct
  -0.17050000000000   0.17050000000000   0.25000000000000   Ca   (4e)
   0.17050000000000  -0.17050000000000   0.75000000000000   Ca   (4e)
   0.65500000000000   0.82900000000000   0.76600000000000   H    (8f)
  -0.82900000000000  -0.65500000000000  -0.26600000000000   H    (8f)
  -0.65500000000000  -0.82900000000000  -0.76600000000000   H    (8f)
   0.82900000000000   0.65500000000000   1.26600000000000   H    (8f)
   0.73600000000000   0.77600000000000  -0.07700000000000   H    (8f)
  -0.77600000000000  -0.73600000000000   0.57700000000000   H    (8f)
  -0.73600000000000  -0.77600000000000   0.07700000000000   H    (8f)
   0.77600000000000   0.73600000000000   0.42300000000000   H    (8f)
  -0.18899000000000   0.35537000000000   0.59103000000000   O    (8f)
  -0.35537000000000   0.18899000000000  -0.09103000000000   O    (8f)
   0.18899000000000  -0.35537000000000  -0.59103000000000   O    (8f)
   0.35537000000000  -0.18899000000000   1.09103000000000   O    (8f)
  -0.18198000000000   0.58192000000000  -0.08702000000000   O    (8f)
  -0.58192000000000   0.18198000000000   0.58702000000000   O    (8f)
   0.18198000000000  -0.58192000000000   0.08702000000000   O    (8f)
   0.58192000000000  -0.18198000000000   0.41298000000000   O    (8f)
   0.72351000000000   0.86003000000000  -0.07831000000000   O    (8f)
  -0.86003000000000  -0.72351000000000   0.57831000000000   O    (8f)
  -0.72351000000000  -0.86003000000000   0.07831000000000   O    (8f)
   0.86003000000000   0.72351000000000   0.42169000000000   O    (8f)
  -0.67273000000000   0.67273000000000   0.25000000000000   S    (4e)
   0.67273000000000  -0.67273000000000   0.75000000000000   S    (4e)
```

$Ta_2NiSe_5$: AB5C2_mC32_15_e_e2f_f - CIF

```
# CIF file
data_findsym-output
_audit_creation_method FINDSYM

_chemical_name_mineral 'NiSe5Ta2'
_chemical_formula_sum 'Ni Se5 Ta2'

loop_
_publ_author_name
 'S. A. Sunshine'
 'J. A. Ibers'
_journal_name_full_name
;
 Inorganic Chemistry
;
_journal_volume 24
_journal_year 1985
_journal_page_first 3611
_journal_page_last 3614
_publ_Section_title
;
 Structure and physical properties of the new layered ternary
    ↪ chalcogenides tantalum nickel sulfide (Ta$_{2}$NiS$_{5}$) and
    ↪ tantalum nickel selenide (Ta$_{2}$NiSe$_{5}$)
;

# Found in Physical and structural properties of the new layered
    ↪ compounds Ta$_{2}$NiS$_{5}$ and Ta$_{2}$NiSe$_{5}$, 1986

_aflow_title 'Ta$_{2}$NiSe$_{5}$ Structure'
_aflow_proto 'AB5C2_mC32_15_e_e2f_f'
_aflow_params 'a,b/a,c/a,\beta,y_{1},y_{2},x_{3},y_{3},z_{3},x_{4},y_{4
    ↪ },z_{4},x_{5},y_{5},z_{5}'
_aflow_params_values '3.496,3.66962242563,4.47397025172,90.53,0.70113,
    ↪ 0.32714,0.5053,0.08039,0.13798,-0.00513,0.14565,-0.04913,-
    ↪ 0.00793,0.22135,0.11044'
_aflow_Strukturbericht 'None'
_aflow_Pearson 'mC32'

_symmetry_space_group_name_H-M "C 1 2/c 1"
_symmetry_Int_Tables_number 15

_cell_length_a      3.49600
_cell_length_b      12.82900
_cell_length_c      15.64100
_cell_angle_alpha   90.00000
_cell_angle_beta    90.53000
_cell_angle_gamma   90.00000

loop_
_space_group_symop_id
_space_group_symop_operation_xyz
1 x,y,z
2 -x,y,-z+1/2
3 -x,-y,-z
4 x,-y,z+1/2
5 x+1/2,y+1/2,z
6 -x+1/2,y+1/2,-z+1/2
7 -x+1/2,-y+1/2,-z
8 x+1/2,-y+1/2,z+1/2

loop_
_atom_site_label
_atom_site_type_symbol
_atom_site_symmetry_multiplicity
_atom_site_Wyckoff_label
_atom_site_fract_x
_atom_site_fract_y
_atom_site_fract_z
_atom_site_occupancy
Ni1 Ni 4 e 0.00000 0.70113 0.25000 1.00000
Se1 Se 4 e 0.00000 0.32714 0.25000 1.00000
Se2 Se 8 f 0.50530 0.08039 0.13798 1.00000
Se3 Se 8 f -0.00513 0.14565 -0.04913 1.00000
Ta1 Ta 8 f -0.00793 0.22135 0.11044 1.00000
```

$Ta_2NiSe_5$: AB5C2_mC32_15_e_e2f_f - POSCAR

```
AB5C2_mC32_15_e_e2f_f & a,b/a,c/a,beta,y1,y2,x3,y3,z3,x4,y4,z4,x5,y5,z5
    ↪ --params=3.496,3.66962242563,4.47397025172,90.53,0.70113,
    ↪ 0.32714,0.5053,0.08039,0.13798,-0.00513,0.14565,-0.04913,-
    ↪ 0.00793,0.22135,0.11044 & C2/c C_{2h}^{6} #15 (e^2f^3) & mC32 &
    ↪ None & NiSe5Ta2 & NiSe5Ta2 & S. A. Sunshine and J. A. Ibers,
    ↪ Inorg. Chem. 24, 3611-3614 (1985)
   1.00000000000000
   1.74800000000000  -6.41450000000000   0.00000000000000
   1.74800000000000   6.41450000000000   0.00000000000000
  -0.14468101925982   0.00000000000000  15.64033082778830
    Ni    Se    Ta
     2    10     4
```

```
Direct
  -0.70113000000000    0.70113000000000    0.25000000000000    Ni    (4e)
   0.70113000000000   -0.70113000000000    0.75000000000000    Ni    (4e)
  -0.32714000000000    0.32714000000000    0.25000000000000    Se    (4e)
   0.32714000000000   -0.32714000000000    0.75000000000000    Se    (4e)
   0.42491000000000    0.58569000000000    0.13798000000000    Se    (8f)
  -0.58569000000000   -0.42491000000000    0.36202000000000    Se    (8f)
  -0.42491000000000   -0.58569000000000   -0.13798000000000    Se    (8f)
   0.58569000000000    0.42491000000000    0.63798000000000    Se    (8f)
  -0.15078000000000    0.14052000000000   -0.04913000000000    Se    (8f)
  -0.14052000000000    0.15078000000000    0.54913000000000    Se    (8f)
   0.15078000000000   -0.14052000000000    0.04913000000000    Se    (8f)
   0.14052000000000   -0.15078000000000    0.45087000000000    Se    (8f)
  -0.22928000000000    0.21342000000000    0.11044000000000    Ta    (8f)
  -0.21342000000000    0.22928000000000    0.38956000000000    Ta    (8f)
   0.22928000000000   -0.21342000000000   -0.11044000000000    Ta    (8f)
   0.21342000000000   -0.22928000000000    0.61044000000000    Ta    (8f)
```

Pyrophyllite [$AlSi_2O_5(OH)$, $S5_6$]: AB5CD2_mC72_15_f_5f_f_2f - CIF

```
# CIF file
data_findsym-output
_audit_creation_method FINDSYM

_chemical_name_mineral 'Pyrophyllite'
_chemical_formula_sum 'Al O5 (OH) Si2'

loop_
_publ_author_name
 'J. W. Gruner'
_journal_name_full_name
;
 Zeitschrift f{\"u}r Kristallographie - Crystalline Materials
;
_journal_volume 88
_journal_year 1934
_journal_page_first 412
_journal_page_last 419
_publ_Section_title
;
 The Crystal Structures of Talc and Pyrophyllite
;

# Found in Strukturbericht Band III 1933-1935, 1937

_aflow_title 'Pyrophyllite [AlSi$_{2}$O$_{5}$(OH), $S5_{6}$] Structure'
_aflow_proto 'AB5CD2_mC72_15_f_5f_f_2f'
_aflow_params 'a,b/a,c/a,\beta,x_{1},y_{1},z_{1},x_{2},y_{2},z_{2},x_{3
    ↪ },y_{3},z_{3},x_{4},y_{4},z_{4},x_{5},y_{5},z_{5},x_{6},y_{6},
    ↪ z_{6},x_{7},y_{7},z_{7},x_{8},y_{8},z_{8},x_{9},y_{9},z_{9}'
_aflow_params_values '5.14,1.73151750973,3.60894941634,99.91667,0.0,
    ↪ 0.33333,0.0,0.20278,0.5,0.05833,0.20278,0.16667,0.05833,0.025,
    ↪ 0.08333,0.17639,0.525,0.08333,0.17639,0.275,0.33333,0.17639,
    ↪ 0.20278,0.83333,0.05833,0.76111,0.0,0.14306,0.26111,0.16667,
    ↪ 0.14306'
_aflow_Strukturbericht '$S5_{6}$'
_aflow_Pearson 'mC72'

_symmetry_space_group_name_H-M "C 1 2/c 1"
_symmetry_Int_Tables_number 15

_cell_length_a     5.14000
_cell_length_b     8.90000
_cell_length_c     18.55000
_cell_angle_alpha  90.00000
_cell_angle_beta   99.91667
_cell_angle_gamma  90.00000

loop_
_space_group_symop_id
_space_group_symop_operation_xyz
1 x,y,z
2 -x,y,-z+1/2
3 -x,-y,-z
4 x,-y,z+1/2
5 x+1/2,y+1/2,z
6 -x+1/2,y+1/2,-z+1/2
7 -x+1/2,-y+1/2,-z
8 x+1/2,-y+1/2,z+1/2

loop_
_atom_site_label
_atom_site_type_symbol
_atom_site_symmetry_multiplicity
_atom_site_Wyckoff_label
_atom_site_fract_x
_atom_site_fract_y
_atom_site_fract_z
_atom_site_occupancy
Al1 Al 8 f 0.00000 0.33333 0.00000 1.00000
O1 O 8 f 0.20278 0.50000 0.05833 1.00000
O2 O 8 f 0.20278 0.16667 0.05833 1.00000
O3 O 8 f 0.02500 0.08333 0.17639 1.00000
O4 O 8 f 0.52500 0.08333 0.17639 1.00000
O5 O 8 f 0.27500 0.33333 0.17639 1.00000
OH1 OH 8 f 0.20278 0.83333 0.05833 1.00000
Si1 Si 8 f 0.76111 0.00000 0.14306 1.00000
Si2 Si 8 f 0.26111 0.16667 0.14306 1.00000
```

Pyrophyllite [$AlSi_2O_5(OH)$, $S5_6$]: AB5CD2_mC72_15_f_5f_f_2f - POSCAR

```
AB5CD2_mC72_15_f_5f_f_2f & a,b/a,c/a,beta,x1,y1,z1,x2,y2,z2,x3,y3,z3,x4,
    ↪ y4,z4,x5,y5,z5,x6,y6,z6,x7,y7,z7,x8,y8,z8,x9,y9,z9 --params=
    ↪ 5.14,1.73151750973,3.60894941634,99.91667,0.0,0.33333,0.0,
    ↪ 0.20278,0.5,0.05833,0.20278,0.16667,0.05833,0.025,0.08333,
    ↪ 0.17639,0.525,0.08333,0.17639,0.275,0.33333,0.17639,0.20278,
    ↪ 0.83333,0.05833,0.76111,0.0,0.14306,0.26111,0.16667,0.14306 &
    ↪ C2/c C_{2h}^{6} #15 (f^9) & mC72 & $S5_{6}$ & AlO5(OH)Si2 &
    ↪ Pyrophyllite & J. W. Gruner, Zeitschrift f{"u}r
    ↪ Kristallographie - Crystalline Materials 88, 412-419 (1934)
   1.00000000000000
   2.57000000000000   -4.45000000000000    0.00000000000000
   2.57000000000000    4.45000000000000    0.00000000000000
  -3.19460136479521    0.00000000000000   18.27284931585790
    Al     O    OH    Si
     4    20     4     8
Direct
  -0.33333000000000    0.33333000000000    0.00000000000000    Al    (8f)
  -0.33333000000000    0.33333000000000    0.50000000000000    Al    (8f)
   0.33333000000000   -0.33333000000000    0.00000000000000    Al    (8f)
   0.33333000000000   -0.33333000000000    0.50000000000000    Al    (8f)
  -0.29722000000000    0.70278000000000    0.05833000000000    O     (8f)
  -0.70278000000000    0.29722000000000    0.44167000000000    O     (8f)
   0.29722000000000   -0.70278000000000   -0.05833000000000    O     (8f)
   0.70278000000000   -0.29722000000000    0.55833000000000    O     (8f)
   0.03611000000000    0.36945000000000    0.05833000000000    O     (8f)
  -0.36945000000000   -0.03611000000000    0.44167000000000    O     (8f)
  -0.03611000000000   -0.36945000000000   -0.05833000000000    O     (8f)
   0.36945000000000    0.03611000000000    0.55833000000000    O     (8f)
  -0.05833000000000    0.10833000000000    0.17639000000000    O     (8f)
  -0.10833000000000    0.05833000000000    0.32361000000000    O     (8f)
   0.05833000000000   -0.10833000000000   -0.17639000000000    O     (8f)
   0.10833000000000   -0.05833000000000    0.67639000000000    O     (8f)
   0.44167000000000    0.60833000000000    0.17639000000000    O     (8f)
  -0.60833000000000   -0.44167000000000    0.32361000000000    O     (8f)
  -0.44167000000000   -0.60833000000000   -0.17639000000000    O     (8f)
   0.60833000000000    0.44167000000000    0.67639000000000    O     (8f)
  -0.05833000000000    0.60833000000000    0.17639000000000    O     (8f)
  -0.60833000000000    0.05833000000000    0.32361000000000    O     (8f)
   0.05833000000000   -0.60833000000000   -0.17639000000000    O     (8f)
   0.60833000000000   -0.05833000000000    0.67639000000000    O     (8f)
  -0.63055000000000    1.03611000000000    0.05833000000000    OH    (8f)
  -1.03611000000000    0.63055000000000    0.44167000000000    OH    (8f)
   0.63055000000000   -1.03611000000000   -0.05833000000000    OH    (8f)
   1.03611000000000   -0.63055000000000    0.55833000000000    OH    (8f)
   0.76111000000000    0.76111000000000    0.14306000000000    Si    (8f)
  -0.76111000000000   -0.76111000000000    0.35694000000000    Si    (8f)
  -0.76111000000000   -0.76111000000000   -0.14306000000000    Si    (8f)
   0.76111000000000    0.76111000000000    0.64306000000000    Si    (8f)
   0.09444000000000    0.42778000000000    0.14306000000000    Si    (8f)
  -0.42778000000000   -0.09444000000000    0.35694000000000    Si    (8f)
  -0.09444000000000   -0.42778000000000   -0.14306000000000    Si    (8f)
   0.42778000000000    0.09444000000000    0.64306000000000    Si    (8f)
```

Titanite ($CaTiSiO_5$, $S0_6$): AB5CD_mC32_15_e_e2f_e_b - CIF

```
# CIF file
data_findsym-output
_audit_creation_method FINDSYM

_chemical_name_mineral 'Titanite'
_chemical_formula_sum 'Ca O5 Si Ti'

loop_
_publ_author_name
 'F. C. Hawthorne'
 'L. A. Groat'
 'M. Raudsepp'
 'N. A. Ball'
 'M. Kimata'
 'F. D. Spike'
 'R. Gaba'
 'N. M. Halden'
 'G. R. Lumpkin'
 'R. C. Ewing'
 'R. B. Greegor'
 'F. W. Lytle'
 'T. {Scott Ercit}'
 'G. R. Rossman'
 'F. J. Wicks'
 'R. A. Ramik'
 'B. L. Sherriff'
 'M. E. Fleet'
 'C. {McCammon}'
_journal_name_full_name
;
 American Mineralogist
;
_journal_volume 76
_journal_year 1991
_journal_page_first 370
_journal_page_last 396
_publ_Section_title
;
 Alpha-decay damage in titanite
;

_aflow_title 'Titanite (CaTiSiO$_{5}$, $S0_{6}$) Structure'
_aflow_proto 'AB5CD_mC32_15_e_e2f_e_b'
_aflow_params 'a,b/a,c/a,\beta,y_{2},y_{3},y_{4},x_{5},y_{5},z_{5},x_{6
    ↪ },y_{6},z_{6}'
_aflow_params_values '6.549,1.32768361582,1.07802717972,113.87,0.8323,
    ↪ 0.5714,0.1828,0.1855,0.0663,0.4102,0.1025,0.2893,0.1185'
_aflow_Strukturbericht '$S0_{6}$'
_aflow_Pearson 'mC32'

_symmetry_space_group_name_H-M "C 1 2/c 1"
_symmetry_Int_Tables_number 15

_cell_length_a     6.54900
```

```
_cell_length_b       8.69500
_cell_length_c       7.06000
_cell_angle_alpha  90.00000
_cell_angle_beta   113.87000
_cell_angle_gamma  90.00000

loop_
_space_group_symop_id
_space_group_symop_operation_xyz
1 x,y,z
2 -x,y,-z+1/2
3 -x,-y,-z
4 x,-y,z+1/2
5 x+1/2,y+1/2,z
6 -x+1/2,y+1/2,-z+1/2
7 -x+1/2,-y+1/2,-z
8 x+1/2,-y+1/2,z+1/2

loop_
_atom_site_label
_atom_site_type_symbol
_atom_site_symmetry_multiplicity
_atom_site_Wyckoff_label
_atom_site_fract_x
_atom_site_fract_y
_atom_site_fract_z
_atom_site_occupancy
Ti1  Ti  4 b  0.00000  0.50000  0.00000  1.00000
Ca1  Ca  4 e  0.00000  0.83230  0.25000  1.00000
O1  O  4 e  0.00000  0.57140  0.25000  1.00000
Si1  Si  4 e  0.00000  0.18280  0.25000  1.00000
O2  O  8 f  0.18550  0.06630  0.41020  1.00000
O3  O  8 f  0.10250  0.28930  0.11850  1.00000
```

Titanite ($CaTiSiO_5$, $S0_6$): AB5CD_mC32_15_e_e2f_e_b - POSCAR

```
AB5CD_mC32_15_e_e2f_e_b & a,b/a,c/a,beta,y2,y3,y4,x5,y5,z5,x6,y6,z6 --
    ↪ params=6.549,1.32768361582,1.07802717972,113.87,0.8323,0.5714,
    ↪ 0.1828,0.1855,0.0663,0.4102,0.1025,0.2893,0.1185 & C2/c C_{2h
    ↪ }^{6} #15 (be^3f^2) & mC32 & $S0_{6}$ & CaO5SiTi & Titanite &
    ↪ F. C. Hawthorne et al., Am. Mineral. 76, 370-396 (1991)
   1.0000000000000000
   3.27450000000000  -4.34750000000000   0.00000000000000
   3.27450000000000   4.34750000000000   0.00000000000000
  -2.85691957284066   0.00000000000000   6.45612968846816
     Ca     O    Si    Ti
      2    10     2     2
Direct
  -0.83230000000000   0.83230000000000   0.25000000000000   Ca   (4e)
   0.83230000000000  -0.83230000000000   0.75000000000000   Ca   (4e)
  -0.57140000000000   0.57140000000000   0.25000000000000   O    (4e)
   0.57140000000000  -0.57140000000000   0.75000000000000   O    (4e)
   0.11920000000000   0.25180000000000   0.41020000000000   O    (8f)
  -0.25180000000000  -0.11920000000000   0.08980000000000   O    (8f)
  -0.11920000000000  -0.25180000000000  -0.41020000000000   O    (8f)
   0.25180000000000   0.11920000000000   0.91020000000000   O    (8f)
  -0.18680000000000   0.39180000000000   0.11850000000000   O    (8f)
  -0.39180000000000   0.18680000000000   0.38150000000000   O    (8f)
   0.18680000000000  -0.39180000000000  -0.11850000000000   O    (8f)
   0.39180000000000  -0.18680000000000   0.61850000000000   O    (8f)
  -0.18280000000000   0.18280000000000   0.25000000000000   Si   (4e)
   0.18280000000000  -0.18280000000000   0.75000000000000   Si   (4e)
   0.50000000000000   0.50000000000000   0.00000000000000   Ti   (4b)
   0.50000000000000   0.50000000000000   0.50000000000000   Ti   (4b)
```

$KFeS_2$ ($F5_a$): ABC2_mC16_15_e_e_f - CIF

```
# CIF file
data_findsym-output
_audit_creation_method FINDSYM

_chemical_name_mineral 'FeKS2'
_chemical_formula_sum 'Fe K S2'

loop_
_publ_author_name
 'W. Bronger'
 'A. Kyas'
 'P. M\"{u}ller'
_journal_name_full_name
;
 Journal of Solid State Chemistry
;
_journal_volume 70
_journal_year 1987
_journal_page_first 262
_journal_page_last 270
_publ_Section_title
;
 The antiferromagnetic structures of KFeS$_{2}$, RbFeS$_{2}$, KFeSe$_{2}
    ↪ $, and RbFeSe$_{2}$ and the correlation between magnetic
    ↪ moments and crystal field calculations
;

_aflow_title 'KFeS$_{2}$ ($F5_{a}$) Structure'
_aflow_proto 'ABC2_mC16_15_e_e_f'
_aflow_params 'a,b/a,c/a,\beta,y_{1},y_{2},x_{3},y_{3},z_{3}'
_aflow_params_values '7.084,1.595567476,0.761434217956,113.2,-0.00332,
    ↪ 0.3572,0.196,0.1098,0.1068'
_aflow_Strukturbericht '$F5_{a}$'
_aflow_Pearson 'mC16'

_symmetry_space_group_name_H-M "C 1 2/c 1"
_symmetry_Int_Tables_number 15
```

```
_cell_length_a       7.08400
_cell_length_b       11.30300
_cell_length_c       5.39400
_cell_angle_alpha  90.00000
_cell_angle_beta   113.20000
_cell_angle_gamma  90.00000

loop_
_space_group_symop_id
_space_group_symop_operation_xyz
1 x,y,z
2 -x,y,-z+1/2
3 -x,-y,-z
4 x,-y,z+1/2
5 x+1/2,y+1/2,z
6 -x+1/2,y+1/2,-z+1/2
7 -x+1/2,-y+1/2,-z
8 x+1/2,-y+1/2,z+1/2

loop_
_atom_site_label
_atom_site_type_symbol
_atom_site_symmetry_multiplicity
_atom_site_Wyckoff_label
_atom_site_fract_x
_atom_site_fract_y
_atom_site_fract_z
_atom_site_occupancy
Fe1  Fe  4 e  0.00000  -0.00332  0.25000  1.00000
K1  K  4 e  0.00000  0.35720  0.25000  1.00000
S1  S  8 f  0.19600  0.10980  0.10680  1.00000
```

$KFeS_2$ ($F5_a$): ABC2_mC16_15_e_e_f - POSCAR

```
ABC2_mC16_15_e_e_f & a,b/a,c/a,beta,y1,y2,x3,y3,z3 --params=7.084,
    ↪ 1.595567476,0.761434217956,113.2,-0.00332,0.3572,0.196,0.1098,
    ↪ 0.1068 & C2/c C_{2h}^{6} #15 (e^2f) & mC16 & $F5_{a}$ & FeKS2 &
    ↪ FeKS2 & W. Bronger and A. Kyas and P. M\"{u}ller, J. Solid
    ↪ State Chem. 70, 262-270 (1987)
   1.0000000000000000
   3.54200000000000  -5.65150000000000   0.00000000000000
   3.54200000000000   5.65150000000000   0.00000000000000
  -2.12492266033359   0.00000000000000   4.95781601994273
     Fe     K     S
      2     2     4
Direct
   0.00332000000000  -0.00332000000000   0.25000000000000   Fe   (4e)
  -0.00332000000000   0.00332000000000   0.75000000000000   Fe   (4e)
  -0.35720000000000   0.35720000000000   0.25000000000000   K    (4e)
   0.35720000000000  -0.35720000000000   0.75000000000000   K    (4e)
   0.08620000000000   0.30580000000000   0.10680000000000   S    (8f)
  -0.30580000000000  -0.08620000000000   0.39320000000000   S    (8f)
  -0.08620000000000  -0.30580000000000  -0.10680000000000   S    (8f)
   0.30580000000000   0.08620000000000   0.60680000000000   S    (8f)
```

Diopside [$CaMg(SiO_3)_2$, $S4_1$]: ABC6D2_mC40_15_e_e_3f_f - CIF

```
# CIF file
data_findsym-output
_audit_creation_method FINDSYM

_chemical_name_mineral 'Diopside'
_chemical_formula_sum 'Ca Mg O6 Si2'

loop_
_publ_author_name
 'L. W. Finger'
 'Y. Ohashi'
_journal_name_full_name
;
 American Mineralogist
;
_journal_volume 61
_journal_year 1976
_journal_page_first 303
_journal_page_last 310
_publ_Section_title
;
 The thermal expansion of diopside to 800$^{\circ}$C and a refinement of
    ↪ the crystal structure at 700$^{\circ}$C
;

_aflow_title 'Diopside [CaMg(SiO$_{3}$)$_{2}$, $S4_{1}$] Structure'
_aflow_proto 'ABC6D2_mC40_15_e_e_3f_f'
_aflow_params 'a,b/a,c/a,\beta,y_{1},y_{2},x_{3},y_{3},z_{3},x_{4},y_{4
    ↪ },z_{4},x_{5},y_{5},z_{5},x_{6},y_{6},z_{6}'
_aflow_params_values '9.804,0.921052631579,0.538249694002,106.02,-0.0933
    ↪ ,0.3003,0.1167,0.0872,0.1417,0.3617,0.246,0.3166,0.3493,0.0155
    ↪ ,-0.0022,0.2864,0.0923,0.2299'
_aflow_Strukturbericht '$S4_{1}$'
_aflow_Pearson 'mC40'

_symmetry_space_group_name_H-M "C 1 2/c 1"
_symmetry_Int_Tables_number 15

_cell_length_a       9.80400
_cell_length_b       9.03000
_cell_length_c       5.27700
_cell_angle_alpha  90.00000
_cell_angle_beta   106.02000
_cell_angle_gamma  90.00000

loop_
_space_group_symop_id
_space_group_symop_operation_xyz
```

```
1 x,y,z
2 -x,y,-z+1/2
3 -x,-y,-z
4 x,-y,z+1/2
5 x+1/2,y+1/2,z
6 -x+1/2,y+1/2,-z+1/2
7 -x+1/2,-y+1/2,-z
8 x+1/2,-y+1/2,z+1/2

loop_
_atom_site_label
_atom_site_type_symbol
_atom_site_symmetry_multiplicity
_atom_site_Wyckoff_label
_atom_site_fract_x
_atom_site_fract_y
_atom_site_fract_z
_atom_site_occupancy
Ca1 Ca 4 e 0.00000 -0.09330 0.25000 1.00000
Mg1 Mg 4 e 0.00000 0.30030 0.25000 1.00000
O1 O 8 f 0.11670 0.08720 0.14170 1.00000
O2 O 8 f 0.36170 0.24600 0.31660 1.00000
O3 O 8 f 0.34930 0.01550 -0.00220 1.00000
Si1 Si 8 f 0.28640 0.09230 0.22990 1.00000
```

Diopside [$CaMg(SiO_3)_2$, $S4_1$]: ABC6D2_mC40_15_e_e_3f_f - POSCAR

```
ABC6D2_mC40_15_e_e_3f_f & a,b/a,c/a,beta,y1,y2,x3,y3,z3,x4,y4,z4,x5,y5,
    ↪ z5,x6,y6,z6 --params=9.804,0.921052631579,0.538249694002,106.02
    ↪ ,-0.0933,0.3003,0.1167,0.0872,0.1417,0.3617,0.246,0.3166,0.3493
    ↪ ,0.0155,-0.0022,0.2864,0.0923,0.2299 & C2/c C_{2h}^{6} #15 (e^
    ↪ 2f^4) & mC40 & $S4_{1}$ & CaMgO6Si2 & Diopside & L. W. Finger
    ↪ and Y. Ohashi, Am. Mineral. 61, 303-310 (1976)
   1.00000000000000
   4.90200000000000   -4.51500000000000    0.00000000000000
   4.90200000000000    4.51500000000000    0.00000000000000
  -1.45630890173743    0.00000000000000    5.07206993077977
    Ca     Mg     O     Si
     2      2    12      4
Direct
   0.09330000000000   -0.09330000000000    0.25000000000000    Ca    (4e)
  -0.09330000000000    0.09330000000000    0.75000000000000    Ca    (4e)
  -0.30030000000000    0.30030000000000    0.25000000000000    Mg    (4e)
   0.30030000000000   -0.30030000000000    0.75000000000000    Mg    (4e)
   0.02950000000000    0.20390000000000    0.14170000000000    O     (8f)
  -0.20390000000000   -0.02950000000000    0.35830000000000    O     (8f)
  -0.02950000000000   -0.20390000000000   -0.14170000000000    O     (8f)
   0.20390000000000    0.02950000000000    0.64170000000000    O     (8f)
   0.11570000000000    0.60770000000000    0.31660000000000    O     (8f)
  -0.60770000000000   -0.11570000000000    0.18340000000000    O     (8f)
  -0.11570000000000   -0.60770000000000   -0.31660000000000    O     (8f)
   0.60770000000000    0.11570000000000    0.81660000000000    O     (8f)
   0.33380000000000    0.36480000000000   -0.00220000000000    O     (8f)
  -0.36480000000000   -0.33380000000000    0.50220000000000    O     (8f)
  -0.33380000000000   -0.36480000000000    0.00220000000000    O     (8f)
   0.36480000000000    0.33380000000000    0.49780000000000    O     (8f)
   0.19410000000000    0.37870000000000    0.22990000000000    Si    (8f)
  -0.37870000000000   -0.19410000000000    0.27010000000000    Si    (8f)
  -0.19410000000000   -0.37870000000000   -0.22990000000000    Si    (8f)
   0.37870000000000    0.19410000000000    0.72990000000000    Si    (8f)
```

$\beta$-Ga (*obsolete*): A_mC4_15_e - CIF

```
# CIF file
data_findsym-output
_audit_creation_method FINDSYM

_chemical_name_mineral 'Ga'
_chemical_formula_sum 'Ga'

loop_
_publ_author_name
 'L. Bosio'
 'A. Defrain'
 'H. Curien'
 'A. Rimsky'
_journal_name_full_name
;
 Acta Crystallographica Section B: Structural Science
;
_journal_volume 25
_journal_year 1969
_journal_page_first 995
_journal_page_last 995
_publ_Section_title
;
 Structure cristalline du gallium $\beta$
;

# Found in The Structures of the Elements, 1982 Found in The Structures
    ↪ of the Elements, {Reprint of the 1974 John Wiley \& Sons
    ↪ edition},

_aflow_title '$\beta$-Ga ({\em{obsolete}}) Structure'
_aflow_proto 'A_mC4_15_e'
_aflow_params 'a,b/a,c/a,\beta,y_{1}'
_aflow_params_values '2.766,2.91142443962,1.20462762111,92.03333,0.131'
_aflow_Strukturbericht 'None'
_aflow_Pearson 'mC4'

_symmetry_space_group_name_H-M "C 1 2/c 1"
_symmetry_Int_Tables_number 15

_cell_length_a     2.76600
_cell_length_b     8.05300
_cell_length_c     3.33200
_cell_angle_alpha  90.00000
_cell_angle_beta   92.03333
_cell_angle_gamma  90.00000

loop_
_space_group_symop_id
_space_group_symop_operation_xyz
1 x,y,z
2 -x,y,-z+1/2
3 -x,-y,-z
4 x,-y,z+1/2
5 x+1/2,y+1/2,z
6 -x+1/2,y+1/2,-z+1/2
7 -x+1/2,-y+1/2,-z
8 x+1/2,-y+1/2,z+1/2

loop_
_atom_site_label
_atom_site_type_symbol
_atom_site_symmetry_multiplicity
_atom_site_Wyckoff_label
_atom_site_fract_x
_atom_site_fract_y
_atom_site_fract_z
_atom_site_occupancy
Ga1 Ga 4 e 0.00000 0.13100 0.25000 1.00000
```

$\beta$-Ga (*obsolete*): A_mC4_15_e - POSCAR

```
A_mC4_15_e & a,b/a,c/a,beta,y1 --params=2.766,2.91142443962,
    ↪ 1.20462762111,92.03333,0.131 & C2/c C_{2h}^{6} #15 (e) & mC4 &
    ↪ None & Ga & Ga & L. Bosio et al., Acta Crystallogr. Sect. B
    ↪ Struct. Sci. 25, 995(1969)
   1.00000000000000
   1.38300000000000   -4.02650000000000    0.00000000000000
   1.38300000000000    4.02650000000000    0.00000000000000
  -0.11822220765196    0.00000000000000    3.32990202702991
    Ga
     2
Direct
  -0.13100000000000    0.13100000000000    0.25000000000000    Ga    (4e)
   0.13100000000000   -0.13100000000000    0.75000000000000    Ga    (4e)
```

$NaNbO_3$: ABC3_oP40_17_abcd_2e_abcd4e - CIF

```
# CIF file
data_findsym-output
_audit_creation_method FINDSYM

_chemical_name_mineral 'NaNbO3'
_chemical_formula_sum 'Na Nb O3'

loop_
_publ_author_name
 'P. Vousden'
_journal_name_full_name
;
 Acta Cristallographica
;
_journal_volume 4
_journal_year 1951
_journal_page_first 545
_journal_page_last 551
_publ_Section_title
;
 The Structure of Ferroelectric Sodium Niobate at Room Temperature
;

# Found in The American Mineralogist Crystal Structure Database, 2003

_aflow_title 'NaNbO$_{3}$ Structure'
_aflow_proto 'ABC3_oP40_17_abcd_2e_abcd4e'
_aflow_params 'a,b/a,c/a,x_{1},x_{2},x_{3},x_{4},y_{5},y_{6},y_{7},y_{8
    ↪ },x_{9},y_{9},z_{9},x_{10},y_{10},z_{10},x_{11},y_{11},z_{11},
    ↪ x_{12},y_{12},z_{12},x_{13},y_{13},z_{13},x_{14},y_{14},z_{14}'
_aflow_params_values '5.504,1.01162790698,2.81976744186,0.481,-0.025,
    ↪ 0.019,0.525,0.519,0.025,-0.019,0.475,0.014,0.005,0.375,0.514,
    ↪ 0.505,0.375,0.25,0.25,0.375,0.75,0.25,0.375,0.25,0.75,0.375,
    ↪ 0.75,0.75,0.375'
_aflow_Strukturbericht 'None'
_aflow_Pearson 'oP40'

_symmetry_space_group_name_H-M "P 2 2 21"
_symmetry_Int_Tables_number 17

_cell_length_a     5.50400
_cell_length_b     5.56800
_cell_length_c     15.52000
_cell_angle_alpha  90.00000
_cell_angle_beta   90.00000
_cell_angle_gamma  90.00000

loop_
_space_group_symop_id
_space_group_symop_operation_xyz
1 x,y,z
2 x,-y,-z
3 -x,y,-z+1/2
4 -x,-y,z+1/2

loop_
_atom_site_label
_atom_site_type_symbol
_atom_site_symmetry_multiplicity
```

```
_atom_site_Wyckoff_label
_atom_site_fract_x
_atom_site_fract_y
_atom_site_fract_z
_atom_site_occupancy
Na1 Na 2 a 0.48100 0.00000 0.00000 1.00000
O1 O 2 a -0.02500 0.00000 0.00000 1.00000
Na2 Na 2 b 0.01900 0.50000 0.00000 1.00000
O2 O 2 b 0.52500 0.50000 0.00000 1.00000
Na3 Na 2 c 0.00000 0.51900 0.25000 1.00000
O3 O 2 c 0.00000 0.02500 0.25000 1.00000
Na4 Na 2 d 0.50000 -0.01900 0.25000 1.00000
O4 O 2 d 0.50000 0.47500 0.25000 1.00000
Nb1 Nb 4 e 0.01400 0.00500 0.37500 1.00000
Nb2 Nb 4 e 0.51400 0.50500 0.37500 1.00000
O5 O 4 e 0.25000 0.25000 0.37500 1.00000
O6 O 4 e 0.75000 0.25000 0.37500 1.00000
O7 O 4 e 0.25000 0.75000 0.37500 1.00000
O8 O 4 e 0.75000 0.75000 0.37500 1.00000
```

$NaNbO_3$: ABC3_oP40_17_abcd_2e_abcd4e - POSCAR

```
ABC3_oP40_17_abcd_2e_abcd4e & a,b/a,c/a,x1,x2,x3,x4,y5,y6,y7,y8,x9,y9,z9
    ↪ ,x10,y10,z10,x11,y11,z11,x12,y12,z12,x13,y13,z13,x14,y14,z14 --
    ↪ params=5.504,1.01162790698,2.81976744186,0.481,-0.025,0.019,
    ↪ 0.525,0.519,0.025,-0.019,0.475,0.014,0.005,0.375,0.514,0.505,
    ↪ 0.375,0.25,0.25,0.375,0.75,0.25,0.375,0.25,0.75,0.375,0.75,0.75
    ↪ ,0.375 & P222_{1} D_{2}^{2} #17 (a^2b^2c^2d^2e^6) & oP40 & None
    ↪ & NaNbO3 & NaNbO3 & P. Vousden, Acta Cryst. 4, 545-551 (1951)
   1.00000000000000
   5.50400000000000   0.00000000000000   0.00000000000000
   0.00000000000000   5.56800000000000   0.00000000000000
   0.00000000000000   0.00000000000000  15.52000000000000
    Na    Nb    O
     8     8   24
Direct
   0.48100000000000   0.00000000000000   0.00000000000000   Na   (2a)
  -0.48100000000000   0.00000000000000   0.50000000000000   Na   (2a)
   0.01900000000000   0.50000000000000   0.00000000000000   Na   (2b)
  -0.01900000000000   0.50000000000000   0.50000000000000   Na   (2b)
   0.00000000000000   0.51900000000000   0.25000000000000   Na   (2c)
   0.00000000000000  -0.51900000000000   0.75000000000000   Na   (2c)
   0.50000000000000  -0.01900000000000   0.25000000000000   Na   (2d)
   0.50000000000000   0.01900000000000   0.75000000000000   Na   (2d)
   0.01400000000000   0.00500000000000   0.37500000000000   Nb   (4e)
  -0.01400000000000  -0.00500000000000   0.87500000000000   Nb   (4e)
  -0.01400000000000   0.00500000000000   0.12500000000000   Nb   (4e)
   0.01400000000000  -0.00500000000000  -0.37500000000000   Nb   (4e)
   0.51400000000000   0.50500000000000   0.37500000000000   Nb   (4e)
  -0.51400000000000  -0.50500000000000   0.87500000000000   Nb   (4e)
  -0.51400000000000   0.50500000000000   0.12500000000000   Nb   (4e)
   0.51400000000000  -0.50500000000000  -0.37500000000000   Nb   (4e)
  -0.02500000000000   0.00000000000000   0.00000000000000   O    (2a)
   0.02500000000000   0.00000000000000   0.50000000000000   O    (2a)
   0.52500000000000   0.50000000000000   0.00000000000000   O    (2b)
  -0.52500000000000   0.50000000000000   0.50000000000000   O    (2b)
   0.00000000000000   0.02500000000000   0.25000000000000   O    (2c)
   0.00000000000000  -0.02500000000000   0.75000000000000   O    (2c)
   0.50000000000000   0.47500000000000   0.25000000000000   O    (2d)
   0.50000000000000  -0.47500000000000   0.75000000000000   O    (2d)
   0.25000000000000   0.25000000000000   0.37500000000000   O    (4e)
  -0.25000000000000  -0.25000000000000   0.87500000000000   O    (4e)
  -0.25000000000000   0.25000000000000   0.12500000000000   O    (4e)
   0.25000000000000  -0.25000000000000  -0.37500000000000   O    (4e)
   0.75000000000000   0.25000000000000   0.37500000000000   O    (4e)
  -0.75000000000000  -0.25000000000000   0.87500000000000   O    (4e)
  -0.75000000000000   0.25000000000000   0.12500000000000   O    (4e)
   0.75000000000000  -0.25000000000000  -0.37500000000000   O    (4e)
   0.25000000000000   0.75000000000000   0.37500000000000   O    (4e)
  -0.25000000000000  -0.75000000000000   0.87500000000000   O    (4e)
  -0.25000000000000   0.75000000000000   0.12500000000000   O    (4e)
   0.25000000000000  -0.75000000000000  -0.37500000000000   O    (4e)
   0.75000000000000   0.75000000000000   0.37500000000000   O    (4e)
  -0.75000000000000  -0.75000000000000   0.87500000000000   O    (4e)
  -0.75000000000000   0.75000000000000   0.12500000000000   O    (4e)
   0.75000000000000  -0.75000000000000  -0.37500000000000   O    (4e)
```

$\gamma$-$TeO_2$: A2B_oP12_18_2c_c - CIF

```
# CIF file
data_findsym-output
_audit_creation_method FINDSYM

_chemical_name_mineral 'O2Te'
_chemical_formula_sum 'O2 Te'

loop_
_publ_author_name
 'J. C. {Champarnaud-Mesjard}'
 'S. Blanchandin'
 'P. Thomas'
 'A. Mirgorodsky'
 'T. {Merle-M\\'{e}jean}'
 'B. Frit'
_journal_name_full_name
;
 Journal of Physics and Chemistry of Solids
;
_journal_volume 61
_journal_year 2000
_journal_page_first 1499
_journal_page_last 1507
_publ_Section_title
;
 Crystal structure, Raman spectrum and lattice dynamics of a new
    ↪ metastable form of tellurium dioxide: $\gamma$-TeO$_{2}$
;

# Found in {\em Ab initio} study of the vibrational properties of
    ↪ crystalline TeO$_{2}$: The $\alpha$, $\beta$, and $\gamma$
    ↪ phases, 2006

_aflow_title '$\gamma$-TeO$_{2}$ Structure'
_aflow_proto 'A2B_oP12_18_2c_c'
_aflow_params 'a,b/a,c/a,x_{1},y_{1},z_{1},x_{2},y_{2},z_{2},x_{3},y_{3
    ↪ },z_{3}'
_aflow_params_values '4.898,1.75091874234,0.888321763985,0.759,0.281,
    ↪ 0.173,0.855,0.036,0.727,-0.0304,0.1016,0.1358'
_aflow_Strukturbericht 'None'
_aflow_Pearson 'oP12'

_symmetry_space_group_name_H-M "P 21 21 2"
_symmetry_Int_Tables_number 18

_cell_length_a     4.89800
_cell_length_b     8.57600
_cell_length_c     4.35100
_cell_angle_alpha 90.00000
_cell_angle_beta  90.00000
_cell_angle_gamma 90.00000

loop_
_space_group_symop_id
_space_group_symop_operation_xyz
1 x,y,z
2 x+1/2,-y+1/2,-z
3 -x+1/2,y+1/2,-z
4 -x,-y,z

loop_
_atom_site_label
_atom_site_type_symbol
_atom_site_symmetry_multiplicity
_atom_site_Wyckoff_label
_atom_site_fract_x
_atom_site_fract_y
_atom_site_fract_z
_atom_site_occupancy
O1 O 4 c 0.75900 0.28100 0.17300 1.00000
O2 O 4 c 0.85500 0.03600 0.72700 1.00000
Te1 Te 4 c -0.03040 0.10160 0.13580 1.00000
```

$\gamma$-$TeO_2$: A2B_oP12_18_2c_c - POSCAR

```
A2B_oP12_18_2c_c & a,b/a,c/a,x1,y1,z1,x2,y2,z2,x3,y3,z3 --params=4.898,
    ↪ 1.75091874234,0.888321763985,0.759,0.281,0.173,0.855,0.036,
    ↪ 0.727,-0.0304,0.1016,0.1358 & P2_{1}2_{1}2 D_{2}^{3} #18 (c^3)
    ↪ & oP12 & None & O2Te & O2Te & J. C. {Champarnaud-Mesjard} et
    ↪ al., J. Phys. Chem. Solids 61, 1499-1507 (2000)
   1.00000000000000
   4.89800000000000   0.00000000000000   0.00000000000000
   0.00000000000000   8.57600000000000   0.00000000000000
   0.00000000000000   0.00000000000000   4.35100000000000
     O    Te
     8     4
Direct
   0.75900000000000   0.28100000000000   0.17300000000000   O    (4c)
  -0.75900000000000  -0.28100000000000   0.17300000000000   O    (4c)
  -0.25900000000000   0.78100000000000  -0.17300000000000   O    (4c)
   1.25900000000000   0.21900000000000  -0.17300000000000   O    (4c)
   0.85500000000000   0.03600000000000   0.72700000000000   O    (4c)
  -0.85500000000000  -0.03600000000000   0.72700000000000   O    (4c)
  -0.35500000000000   0.53600000000000  -0.72700000000000   O    (4c)
   1.35500000000000   0.46400000000000  -0.72700000000000   O    (4c)
  -0.03040000000000   0.10160000000000   0.13580000000000   Te   (4c)
   0.03040000000000  -0.10160000000000   0.13580000000000   Te   (4c)
   0.53040000000000   0.60160000000000  -0.13580000000000   Te   (4c)
   0.46960000000000   0.39840000000000  -0.13580000000000   Te   (4c)
```

Diamminetriamidodizinc Chloride ($[Zn_2(NH_3)_2(NH_2)_3]Cl$): AB12C5D2_oP40_18_a_6c_b2c_c - CIF

```
# CIF file
data_findsym-output
_audit_creation_method FINDSYM

_chemical_name_mineral 'Diamminetriamidodizinc chloride'
_chemical_formula_sum 'Cl H12 N5 Zn2'

loop_
_publ_author_name
 'T. M. M. Richter'
 'S. Strobel'
 'N. S. A. Alt'
 'E. Schl\"{u}cker'
 'R. Niewa'
_journal_name_full_name
;
 Inorganics
;
_journal_volume 4
_journal_year 2016
_journal_page_first 41
_journal_page_last 41
_publ_Section_title
;
 Ammonothermal Synthesis and Crystal Structures of
    ↪ Diamminetriamidodizinc Chloride [Zn$_{2}$(NH$_{3}$)$_{2}$(NH$_
    ↪ {2}$)$_{3}$]Cl and Diamminemonoamidozinc Bromide [Zn(NH$_{3}$)
    ↪ $_{2}$(NH$_{2}$)]Br
```

```
;

_aflow_title 'Diamminetriamidodizinc Chloride ([Zn$_{2}$(NH$_{3}$)$_{2}$
      ↪ (NH$_{2}$)$_{3}$]Cl) Structure'
_aflow_proto 'AB12C5D2_oP40_18_a_6c_b2c_c'
_aflow_params 'a,b/a,c/a,z_{1},z_{2},x_{3},y_{3},z_{3},x_{4},y_{4},z_{4
      ↪ },x_{5},y_{5},z_{5},x_{6},y_{6},z_{6},x_{7},y_{7},z_{7},x_{8},
      ↪ y_{8},z_{8},x_{9},y_{9},z_{9},x_{10},y_{10},z_{10},x_{11},y_{11
      ↪ },z_{11}'
_aflow_params_values '5.7715,1.77352508014,1.13412457767,-0.0434,0.395,
      ↪ 0.35,0.248,0.67,0.5,0.354,0.67,0.0,0.259,-0.09,0.09,0.36,-0.04,
      ↪ 0.41,0.118,0.08,0.354,0.01,0.65,0.449,0.3291,0.553,0.007,0.3377
      ↪ ,0.853,0.1009,0.3376,0.5374'
_aflow_Strukturbericht 'None'
_aflow_Pearson 'oP40'

_symmetry_space_group_name_H-M "P 21 21 2"
_symmetry_Int_Tables_number 18

_cell_length_a     5.77150
_cell_length_b     10.23590
_cell_length_c     6.54560
_cell_angle_alpha  90.00000
_cell_angle_beta   90.00000
_cell_angle_gamma  90.00000

loop_
_space_group_symop_id
_space_group_symop_operation_xyz
1 x,y,z
2 x+1/2,-y+1/2,-z
3 -x+1/2,y+1/2,-z
4 -x,-y,z

loop_
_atom_site_label
_atom_site_type_symbol
_atom_site_symmetry_multiplicity
_atom_site_Wyckoff_label
_atom_site_fract_x
_atom_site_fract_y
_atom_site_fract_z
_atom_site_occupancy
Cl1 Cl 2 a 0.00000 0.00000 -0.04340 1.00000
N1 N 2 b 0.00000 0.50000 0.39500 1.00000
H1 H 4 c 0.35000 0.24800 0.67000 1.00000
H2 H 4 c 0.50000 0.35400 0.67000 1.00000
H3 H 4 c 0.00000 0.25900 -0.09000 1.00000
H4 H 4 c 0.09000 0.36000 -0.04000 1.00000
H5 H 4 c 0.41000 0.11800 0.08000 1.00000
H6 H 4 c 0.35400 0.01000 0.65000 1.00000
N2 N 4 c 0.44900 0.32910 0.55300 1.00000
N3 N 4 c 0.00700 0.33770 0.85300 1.00000
Zn1 Zn 4 c 0.10090 0.33760 0.53740 1.00000
```

Diamminetriamidodizinc Chloride ($[Zn_2(NH_3)_2(NH_2)_3]Cl$): AB12C5D2_oP40_18_a_6c_b2c_c - POSCAR

```
AB12C5D2_oP40_18_a_6c_b2c_c & a,b/a,c/a,z1,z2,x3,y3,z3,x4,y4,z4,x5,y5,z5
      ↪ ,x6,y6,z6,x7,y7,z7,x8,y8,z8,x9,y9,z9,x10,y10,z10,x11,y11,z11 --
      ↪ params=5.7715,1.77352508014,1.13412457767,-0.0434,0.395,0.35,
      ↪ 0.248,0.67,0.5,0.354,0.67,0.0,0.259,-0.09,0.09,0.36,-0.04,0.41,
      ↪ 0.118,0.08,0.354,0.01,0.65,0.449,0.3291,0.553,0.007,0.3377,
      ↪ 0.853,0.1009,0.3376,0.5374 & P2_{1}2_{1}2 D_{2}^{3} #18 (abc^9)
      ↪  & oP40 & None & ClH12N5Zn2 & Diamminetriamidodizinc chloride &
      ↪  T. M. M. Richter et al., Inorganics 4, 41(2016)
   1.00000000000000
   5.77150000000000   0.00000000000000   0.00000000000000
   0.00000000000000  10.23590000000000   0.00000000000000
   0.00000000000000   0.00000000000000   6.54560000000000
    Cl       H       N      Zn
     2      24      10       4
Direct
   0.00000000000000   0.00000000000000  -0.04340000000000   Cl   (2a)
   0.50000000000000   0.50000000000000   0.04340000000000   Cl   (2a)
   0.35000000000000   0.24800000000000   0.67000000000000    H   (4c)
  -0.35000000000000  -0.24800000000000   0.67000000000000    H   (4c)
   0.15000000000000   0.74800000000000  -0.67000000000000    H   (4c)
   0.85000000000000   0.25200000000000  -0.67000000000000    H   (4c)
   0.50000000000000   0.35400000000000   0.67000000000000    H   (4c)
  -0.50000000000000  -0.35400000000000   0.67000000000000    H   (4c)
   0.00000000000000   0.85400000000000  -0.67000000000000    H   (4c)
   1.00000000000000   0.14600000000000  -0.67000000000000    H   (4c)
   0.00000000000000   0.25900000000000  -0.09000000000000    H   (4c)
   0.00000000000000  -0.25900000000000  -0.09000000000000    H   (4c)
   0.50000000000000   0.75900000000000   0.09000000000000    H   (4c)
   0.50000000000000   0.24100000000000   0.09000000000000    H   (4c)
   0.09000000000000   0.36000000000000  -0.04000000000000    H   (4c)
  -0.09000000000000  -0.36000000000000  -0.04000000000000    H   (4c)
   0.41000000000000   0.86000000000000   0.04000000000000    H   (4c)
   0.59000000000000   0.14000000000000   0.04000000000000    H   (4c)
   0.41000000000000   0.11800000000000   0.08000000000000    H   (4c)
  -0.41000000000000  -0.11800000000000   0.08000000000000    H   (4c)
   0.09000000000000   0.61800000000000  -0.08000000000000    H   (4c)
   0.91000000000000   0.38200000000000  -0.08000000000000    H   (4c)
   0.35400000000000   0.01000000000000   0.65000000000000    H   (4c)
  -0.35400000000000  -0.01000000000000   0.65000000000000    H   (4c)
   0.14600000000000   0.51000000000000  -0.65000000000000    H   (4c)
   0.85400000000000   0.49000000000000  -0.65000000000000    H   (4c)
   0.00000000000000   0.50000000000000   0.39500000000000    N   (2b)
   0.50000000000000   0.00000000000000  -0.39500000000000    N   (2b)
   0.44900000000000   0.32910000000000   0.55300000000000    N   (4c)
  -0.44900000000000  -0.32910000000000   0.55300000000000    N   (4c)
   0.05100000000000   0.82910000000000  -0.55300000000000    N   (4c)
   0.94900000000000   0.17090000000000  -0.55300000000000    N   (4c)
   0.00700000000000   0.33770000000000   0.85300000000000    N   (4c)
  -0.00700000000000  -0.33770000000000   0.85300000000000    N   (4c)
   0.49300000000000   0.83770000000000  -0.85300000000000    N   (4c)
   0.50700000000000   0.16230000000000  -0.85300000000000    N   (4c)
   0.10090000000000   0.33760000000000   0.53740000000000   Zn   (4c)
  -0.10090000000000  -0.33760000000000   0.53740000000000   Zn   (4c)
   0.39910000000000   0.83760000000000  -0.53740000000000   Zn   (4c)
   0.60090000000000   0.16240000000000  -0.53740000000000   Zn   (4c)
```

Morenosite ($NiSO_4{\cdot}7H_2O$, $H4_{12}$): A14BC11D_oP108_19_14a_a_11a_a - CIF

```
# CIF file
data_findsym-output
_audit_creation_method FINDSYM

_chemical_name_mineral 'Morenosite'
_chemical_formula_sum 'H14 Ni O11 S'

loop_
_publ_author_name
 'H. Ptasiewicz-Bak'
 'I. Olovsson'
 'G. J. McIntyre'
_journal_name_full_name
;
 Acta Crystallographica Section B: Structural Science
;
_journal_volume 53
_journal_year 1997
_journal_page_first 325
_journal_page_last 336
_publ_Section_title
;
 Charge Density in Orthorhombic NiSO$_{4}$$\cdot$7H$_{2}$O at Room
      ↪ Temperature and 25 K
;

_aflow_title 'Morenosite (NiSO$_{4}$$\cdot$7H$_{2}$O, $H4_{12}$)
      ↪ Structure'
_aflow_proto 'A14BC11D_oP108_19_14a_a_11a_a'
_aflow_params 'a,b/a,c/a,x_{1},y_{1},z_{1},x_{2},y_{2},z_{2},x_{3},y_{3
      ↪ },z_{3},x_{4},y_{4},z_{4},x_{5},y_{5},z_{5},x_{6},y_{6},z_{6},
      ↪ x_{7},y_{7},z_{7},x_{8},y_{8},z_{8},x_{9},y_{9},z_{9},x_{10},y_
      ↪ {10},z_{10},x_{11},y_{11},z_{11},x_{12},y_{12},z_{12},x_{13},y_
      ↪ {13},z_{13},x_{14},y_{14},z_{14},x_{15},y_{15},z_{15},x_{16},y_
      ↪ {16},z_{16},x_{17},y_{17},z_{17},x_{18},y_{18},z_{18},x_{19},y_
      ↪ {19},z_{19},x_{20},y_{20},z_{20},x_{21},y_{21},z_{21},x_{22},y_
      ↪ {22},z_{22},x_{23},y_{23},z_{23},x_{24},y_{24},z_{24},x_{25},y_
      ↪ {25},z_{25},x_{26},y_{26},z_{26},x_{27},y_{27},z_{27}'
_aflow_params_values '6.706,1.75902177155,1.78183716075,0.109,0.7344,
      ↪ 0.7161,0.875,0.74,0.6932,0.298,-0.0819,0.7675,0.251,0.0439,
      ↪ 0.7498,0.689,-0.0769,0.7175,0.74,0.0473,0.6778,0.197,0.1238,
      ↪ 0.5537,0.007,0.1119,0.48,-0.023,0.8655,0.3955,0.779,-0.0807,
      ↪ 0.4437,0.319,0.7843,0.5251,0.396,-0.0866,0.4995,-0.05,-0.0782,-
      ↪ 0.0187,-0.002,-0.0288,0.8664,0.04106,-0.07911,0.60442,0.42704,
      ↪ 0.18381,0.57465,0.48169,0.35142,0.68668,0.69542,0.18889,0.70501
      ↪ ,0.36068,0.17853,0.77191,0.00529,0.7607,0.66624,0.18986,-
      ↪ 0.03077,0.74842,0.7752,-0.03247,0.66931,0.07373,0.08217,0.54556
      ↪ ,0.8931,0.87229,0.46088,0.29961,0.8655,0.537,-0.0694,-0.01048,-
      ↪ 0.0641,0.49094,0.2263,0.68354'
_aflow_Strukturbericht '$H4_{12}$'
_aflow_Pearson 'oP108'

_symmetry_space_group_name_H-M "P 21 21 21"
_symmetry_Int_Tables_number 19

_cell_length_a     6.70600
_cell_length_b     11.79600
_cell_length_c     11.94900
_cell_angle_alpha  90.00000
_cell_angle_beta   90.00000
_cell_angle_gamma  90.00000

loop_
_space_group_symop_id
_space_group_symop_operation_xyz
1 x,y,z
2 x+1/2,-y+1/2,-z
3 -x,y+1/2,-z+1/2
4 -x+1/2,-y,z+1/2

loop_
_atom_site_label
_atom_site_type_symbol
_atom_site_symmetry_multiplicity
_atom_site_Wyckoff_label
_atom_site_fract_x
_atom_site_fract_y
_atom_site_fract_z
_atom_site_occupancy
H1 H 4 a 0.10900 0.73440 0.71610 1.00000
H2 H 4 a 0.87500 0.74000 0.69320 1.00000
H3 H 4 a 0.29800 -0.08190 0.76750 1.00000
H4 H 4 a 0.25100 0.04390 0.74980 1.00000
H5 H 4 a 0.68900 -0.07690 0.71750 1.00000
H6 H 4 a 0.74000 0.04730 0.67780 1.00000
H7 H 4 a 0.19700 0.12380 0.55370 1.00000
H8 H 4 a 0.00700 0.11190 0.48000 1.00000
H9 H 4 a -0.02300 0.86550 0.39550 1.00000
H10 H 4 a 0.77900 -0.08070 0.44370 1.00000
H11 H 4 a 0.31900 0.78430 0.52510 1.00000
H12 H 4 a 0.39600 -0.08660 0.49950 1.00000
H13 H 4 a -0.05000 -0.07820 -0.01870 1.00000
H14 H 4 a -0.00200 -0.02880 0.86640 1.00000
Ni1 Ni 4 a 0.04106 -0.07911 0.60442 1.00000
```

```
O1 O 4 a 0.42704 0.18381 0.57465 1.00000
O2 O 4 a 0.48169 0.35142 0.68668 1.00000
O3 O 4 a 0.69542 0.18889 0.70501 1.00000
O4 O 4 a 0.36068 0.17853 0.77191 1.00000
O5 O 4 a 0.00529 0.76070 0.66624 1.00000
O6 O 4 a 0.18986 -0.03077 0.74842 1.00000
O7 O 4 a 0.77520 -0.03247 0.66931 1.00000
O8 O 4 a 0.07373 0.08217 0.54556 1.00000
O9 O 4 a 0.89310 0.87229 0.46088 1.00000
O10 O 4 a 0.29961 0.86550 0.53700 1.00000
O11 O 4 a -0.06940 -0.01048 -0.06410 1.00000
S1 S 4 a 0.49094 0.22630 0.68354 1.00000
```

Morenosite (NiSO$_4$·7H$_2$O, $H4_{12}$): A14BC11D_oP108_19_14a_a_11a_a - POSCAR

```
A14BC11D_oP108_19_14a_a_11a_a & a,b/a,c/a,x1,y1,z1,x2,y2,z2,x3,y3,z3,x4,
    ↪ y4,z4,x5,y5,z5,x6,y6,z6,x7,y7,z7,x8,y8,z8,x9,y9,z9,x10,y10,z10,
    ↪ x11,y11,z11,x12,y12,z12,x13,y13,z13,x14,y14,z14,x15,y15,z15,x16
    ↪ ,y16,z16,x17,y17,z17,x18,y18,z18,x19,y19,z19,x20,y20,z20,x21,
    ↪ y21,z21,x22,y22,z22,x23,y23,z23,x24,y24,z24,x25,y25,z25,x26,y26
    ↪ ,z26,x27,y27,z27 --params=6.706,1.75902177155,1.78183716075,
    ↪ 0.109,0.7344,0.7161,0.875,0.74,0.6932,0.298,-0.0819,0.7675,
    ↪ 0.251,0.0439,0.7498,0.689,-0.0769,0.7175,0.74,0.0473,0.6778,
    ↪ 0.197,0.1238,0.5537,0.007,0.1119,0.48,-0.023,0.8655,0.3955,
    ↪ 0.779,-0.0807,0.4437,0.319,0.7843,0.5251,0.396,-0.0866,0.4995,-
    ↪ 0.05,-0.0782,-0.0187,-0.002,-0.0288,0.8664,0.04106,-0.07911,
    ↪ 0.60442,0.42704,0.18381,0.57465,0.48169,0.35142,0.68668,0.69542
    ↪ ,0.18889,0.70501,0.36068,0.17853,0.77191,0.00529,0.7607,0.66624
    ↪ ,0.18986,-0.03077,0.74842,0.7752,-0.03247,0.66931,0.07373,
    ↪ 0.08217,0.54556,0.8931,0.87229,0.46088,0.29961,0.8655,0.537,-
    ↪ 0.0694,-0.01048,-0.0641,0.49094,0.2263,0.68354 & P2_{1}2_{1}2_{
    ↪ 1} D_{2}^{4} #19 (a^27) & oP108 & $H4_{12}$ & H14NiO11S &
    ↪ Morenosite & H. Ptasiewicz-Bak and I. Olovsson and G. J.
    ↪ McIntyre, Acta Crystallogr. Sect. B Struct. Sci. 53, 325-336 (
    ↪ 1997)
   1.00000000000000
   6.70600000000000   0.00000000000000   0.00000000000000
   0.00000000000000  11.79600000000000   0.00000000000000
   0.00000000000000   0.00000000000000  11.94900000000000
     H    Ni     O     S
    56     4    44     4
Direct
   0.10900000000000   0.73440000000000   0.71610000000000   H   (4a)
   0.39100000000000  -0.73440000000000   1.21610000000000   H   (4a)
  -0.10900000000000   1.23440000000000  -0.21610000000000   H   (4a)
   0.60900000000000  -0.23440000000000  -0.71610000000000   H   (4a)
   0.87500000000000   0.74000000000000   0.69320000000000   H   (4a)
  -0.37500000000000  -0.74000000000000   1.19320000000000   H   (4a)
  -0.87500000000000   1.24000000000000  -0.19320000000000   H   (4a)
   1.37500000000000  -0.24000000000000  -0.69320000000000   H   (4a)
   0.29800000000000  -0.08190000000000   0.76750000000000   H   (4a)
   0.20200000000000   0.08190000000000   1.26750000000000   H   (4a)
  -0.29800000000000   0.41810000000000  -0.26750000000000   H   (4a)
   0.79800000000000   0.58190000000000  -0.76750000000000   H   (4a)
   0.25100000000000   0.04390000000000   0.74980000000000   H   (4a)
   0.24900000000000  -0.04390000000000   1.24980000000000   H   (4a)
  -0.25100000000000   0.54390000000000  -0.24980000000000   H   (4a)
   0.75100000000000   0.45610000000000  -0.74980000000000   H   (4a)
   0.68900000000000  -0.07690000000000   0.71750000000000   H   (4a)
  -0.18900000000000   0.07690000000000   1.21750000000000   H   (4a)
  -0.68900000000000   0.42310000000000  -0.21750000000000   H   (4a)
   1.18900000000000   0.57690000000000  -0.71750000000000   H   (4a)
   0.74000000000000   0.04730000000000   0.67780000000000   H   (4a)
  -0.24000000000000  -0.04730000000000   1.17780000000000   H   (4a)
  -0.74000000000000   0.54730000000000  -0.17780000000000   H   (4a)
   1.24000000000000   0.45270000000000  -0.67780000000000   H   (4a)
   0.19700000000000   0.12380000000000   0.55370000000000   H   (4a)
   0.30300000000000  -0.12380000000000   1.05370000000000   H   (4a)
  -0.19700000000000   0.62380000000000  -0.05370000000000   H   (4a)
   0.69700000000000   0.37620000000000  -0.55370000000000   H   (4a)
   0.00700000000000   0.11190000000000   0.48000000000000   H   (4a)
   0.49300000000000  -0.11190000000000   0.98000000000000   H   (4a)
  -0.00700000000000   0.61190000000000   0.02000000000000   H   (4a)
   0.50700000000000   0.38810000000000  -0.48000000000000   H   (4a)
  -0.02300000000000   0.86550000000000   0.39550000000000   H   (4a)
   0.52300000000000  -0.86550000000000   0.89550000000000   H   (4a)
   0.02300000000000   1.36550000000000   0.10450000000000   H   (4a)
   0.47700000000000  -0.36550000000000  -0.39550000000000   H   (4a)
   0.77900000000000  -0.08070000000000   0.44370000000000   H   (4a)
  -0.27900000000000   0.08070000000000   0.94370000000000   H   (4a)
  -0.77900000000000   0.41930000000000   0.05630000000000   H   (4a)
   1.27900000000000   0.58070000000000  -0.44370000000000   H   (4a)
   0.31900000000000   0.78430000000000   0.52510000000000   H   (4a)
   0.18100000000000  -0.78430000000000   1.02510000000000   H   (4a)
  -0.31900000000000   1.28430000000000  -0.02510000000000   H   (4a)
   0.81900000000000  -0.28430000000000  -0.52510000000000   H   (4a)
   0.39600000000000  -0.08660000000000   0.49950000000000   H   (4a)
   0.10400000000000   0.08660000000000   0.99950000000000   H   (4a)
  -0.39600000000000   0.41340000000000   0.00050000000000   H   (4a)
   0.89600000000000   0.58660000000000  -0.49950000000000   H   (4a)
  -0.05000000000000  -0.07820000000000  -0.01870000000000   H   (4a)
   0.55000000000000   0.07820000000000   0.48130000000000   H   (4a)
   0.05000000000000   0.42180000000000   0.51870000000000   H   (4a)
   0.45000000000000   0.57820000000000   0.01870000000000   H   (4a)
  -0.00200000000000  -0.02880000000000   0.86640000000000   H   (4a)
   0.50200000000000   0.02880000000000   1.36640000000000   H   (4a)
   0.00200000000000   0.47120000000000  -0.36640000000000   H   (4a)
   0.49800000000000   0.52880000000000  -0.86640000000000   H   (4a)
   0.04106000000000  -0.07911000000000   0.60442000000000   Ni  (4a)
   0.45894000000000   0.07911000000000   1.10442000000000   Ni  (4a)
  -0.04106000000000   0.42089000000000  -0.10442000000000   Ni  (4a)
   0.54106000000000   0.57911000000000  -0.60442000000000   Ni  (4a)
   0.42704000000000   0.18381000000000   0.57465000000000   O   (4a)
   0.07296000000000  -0.18381000000000   1.07465000000000   O   (4a)
  -0.42704000000000   0.68381000000000  -0.07465000000000   O   (4a)
   0.92704000000000   0.31619000000000  -0.57465000000000   O   (4a)
   0.48169000000000   0.35142000000000   0.68668000000000   O   (4a)
   0.01831000000000  -0.35142000000000   1.18668000000000   O   (4a)
  -0.48169000000000   0.85142000000000  -0.18668000000000   O   (4a)
   0.98169000000000   0.14858000000000  -0.68668000000000   O   (4a)
   0.69542000000000   0.18889000000000   0.70501000000000   O   (4a)
  -0.19542000000000  -0.18889000000000   1.20501000000000   O   (4a)
  -0.69542000000000   0.68889000000000  -0.20501000000000   O   (4a)
   1.19542000000000   0.31111000000000  -0.70501000000000   O   (4a)
   0.36068000000000   0.17853000000000   0.77191000000000   O   (4a)
   0.13932000000000  -0.17853000000000   1.27191000000000   O   (4a)
  -0.36068000000000   0.67853000000000  -0.27191000000000   O   (4a)
   0.86068000000000   0.32147000000000  -0.77191000000000   O   (4a)
   0.00529000000000   0.76070000000000   0.66624000000000   O   (4a)
   0.49471000000000  -0.76070000000000   1.16624000000000   O   (4a)
  -0.00529000000000   1.26070000000000  -0.16624000000000   O   (4a)
   0.50529000000000  -0.26070000000000  -0.66624000000000   O   (4a)
   0.18986000000000  -0.03077000000000   0.74842000000000   O   (4a)
   0.31014000000000   0.03077000000000   1.24842000000000   O   (4a)
  -0.18986000000000   0.46923000000000  -0.24842000000000   O   (4a)
   0.68986000000000   0.53077000000000  -0.74842000000000   O   (4a)
   0.77520000000000  -0.03247000000000   0.66931000000000   O   (4a)
  -0.27520000000000   0.03247000000000   1.16931000000000   O   (4a)
  -0.77520000000000   0.46753000000000  -0.16931000000000   O   (4a)
   1.27520000000000   0.53247000000000  -0.66931000000000   O   (4a)
   0.07373000000000   0.08217000000000   0.54556000000000   O   (4a)
   0.42627000000000  -0.08217000000000   1.04556000000000   O   (4a)
  -0.07373000000000   0.58217000000000  -0.04556000000000   O   (4a)
   0.57373000000000   0.41783000000000  -0.54556000000000   O   (4a)
   0.89310000000000   0.87229000000000   0.46088000000000   O   (4a)
  -0.39310000000000  -0.87229000000000   0.96088000000000   O   (4a)
  -0.89310000000000   1.37229000000000   0.03912000000000   O   (4a)
   1.39310000000000  -0.37229000000000  -0.46088000000000   O   (4a)
   0.29961000000000   0.86550000000000   0.53700000000000   O   (4a)
   0.20039000000000  -0.86550000000000   1.03700000000000   O   (4a)
  -0.29961000000000   1.36550000000000  -0.03700000000000   O   (4a)
   0.79961000000000  -0.36550000000000  -0.53700000000000   O   (4a)
  -0.06940000000000  -0.01048000000000  -0.06410000000000   O   (4a)
   0.56940000000000   0.01048000000000   0.43590000000000   O   (4a)
   0.06940000000000   0.48952000000000   0.56410000000000   O   (4a)
   0.43060000000000   0.51048000000000   0.06410000000000   O   (4a)
   0.49094000000000   0.22630000000000   0.68354000000000   S   (4a)
   0.00906000000000  -0.22630000000000   1.18354000000000   S   (4a)
  -0.49094000000000   0.72630000000000  -0.18354000000000   S   (4a)
   0.99094000000000   0.27370000000000  -0.68354000000000   S   (4a)
```

Wülfingite ($\epsilon$-Zn(OH)$_2$, $C31$): A2B2C_oP20_19_2a_2a_a - CIF

```
# CIF file
data_findsym-output
_audit_creation_method FINDSYM

_chemical_name_mineral 'W\"{u}lfingite'
_chemical_formula_sum 'H2 O2 Zn'

loop_
_publ_author_name
 'R. Stahl'
 'C. Jung'
 'H. D. Lutz'
 'W. Kockelmann'
 'H. Jacobs'
_journal_name_full_name
;
 Zeitschrift fur Anorganische und Allgemeine Chemie
;
_journal_volume 624
_journal_year 1998
_journal_page_first 1130
_journal_page_last 1136
_publ_Section_title
;
 Kristallstrukturen und Wasserstoffbr\"{u}ckenbindungen bei $\beta$-Be(
     ↪ OH)$_{2}$ und $\epsilon$-Zn(OH)$_{2}$
;

# Found in The American Mineralogist Crystal Structure Database, 2003

_aflow_title 'W\"{u}lfingite ($\epsilon$-Zn(OH)$_{2}$, $C31$) Structure'
_aflow_proto 'A2B2C_oP20_19_2a_2a_a'
_aflow_params 'a,b/a,c/a,x_{1},y_{1},z_{1},x_{2},y_{2},z_{2},x_{3},y_{3
    ↪ },z_{3},x_{4},y_{4},z_{4},x_{5},y_{5},z_{5}'
_aflow_params_values '4.905,1.04852191641,1.72742099898,0.501,0.36,0.857
    ↪ ,0.28,0.79,0.17,0.1155,0.1256,0.0795,0.167,0.3172,0.7198,
    ↪ 0.07096,0.64832,0.6242'
_aflow_Strukturbericht '$C31$'
_aflow_Pearson 'oP20'

_symmetry_space_group_name_H-M "P 21 21 21"
_symmetry_Int_Tables_number 19

_cell_length_a    4.90500
_cell_length_b    5.14300
_cell_length_c    8.47300
_cell_angle_alpha 90.00000
_cell_angle_beta  90.00000
_cell_angle_gamma 90.00000

loop_
_space_group_symop_id
_space_group_symop_operation_xyz
1 x,y,z
2 x+1/2,-y+1/2,-z
3 -x,y+1/2,-z+1/2
4 -x+1/2,-y,z+1/2
```

```
loop_
_atom_site_label
_atom_site_type_symbol
_atom_site_symmetry_multiplicity
_atom_site_Wyckoff_label
_atom_site_fract_x
_atom_site_fract_y
_atom_site_fract_z
_atom_site_occupancy
H1 H 4 a 0.50100 0.36000 0.85700 1.00000
H2 H 4 a 0.28000 0.79000 0.17000 1.00000
O1 O 4 a 0.11550 0.12560 0.07950 1.00000
O2 O 4 a 0.16700 0.31720 0.71980 1.00000
Zn1 Zn 4 a 0.07096 0.64832 0.62420 1.00000
```

Wülfingite ($\epsilon$-$Zn(OH)_2$, $C31$): A2B2C_oP20_19_2a_2a_a - POSCAR

```
A2B2C_oP20_19_2a_2a_a & a,b/a,c/a,x1,y1,z1,x2,y2,z2,x3,y3,z3,x4,y4,z4,x5
    ↪ ,y5,z5 --params=4.905,1.04852191641,1.72742099898,0.501,0.36,
    ↪ 0.857,0.28,0.79,0.17,0.1155,0.1256,0.0795,0.167,0.3172,0.7198,
    ↪ 0.07096,0.64832,0.6242 & P2_{1}2_{1}2_{1} D_{2}^{4} #19 (a^5) &
    ↪ oP20 & $C31$ & H2O2Zn & W\"{u}lfingite & R. Stahl et al., Z.
    ↪ Anorg. Allg. Chem. 624, 1130-1136 (1998)
   1.00000000000000
   4.90500000000000   0.00000000000000   0.00000000000000
   0.00000000000000   5.14300000000000   0.00000000000000
   0.00000000000000   0.00000000000000   8.47300000000000
     H     O     Zn
     8     8     4
Direct
   0.50100000000000   0.36000000000000   0.85700000000000   H   (4a)
  -0.00100000000000  -0.36000000000000   1.35700000000000   H   (4a)
  -0.50100000000000   0.86000000000000  -0.35700000000000   H   (4a)
   1.00100000000000   0.14000000000000  -0.85700000000000   H   (4a)
   0.28000000000000   0.79000000000000   0.17000000000000   H   (4a)
   0.22000000000000  -0.79000000000000   0.67000000000000   H   (4a)
  -0.28000000000000   1.29000000000000   0.33000000000000   H   (4a)
   0.78000000000000  -0.29000000000000  -0.17000000000000   H   (4a)
   0.11550000000000   0.12560000000000   0.07950000000000   O   (4a)
   0.38450000000000  -0.12560000000000   0.57950000000000   O   (4a)
  -0.11550000000000   0.62560000000000   0.42050000000000   O   (4a)
   0.61550000000000   0.37440000000000  -0.07950000000000   O   (4a)
   0.16700000000000   0.31720000000000   0.71980000000000   O   (4a)
   0.33300000000000  -0.31720000000000   1.21980000000000   O   (4a)
  -0.16700000000000   0.81720000000000  -0.21980000000000   O   (4a)
   0.66700000000000   0.18280000000000  -0.71980000000000   O   (4a)
   0.07096000000000   0.64832000000000   0.62420000000000   Zn  (4a)
   0.42904000000000  -0.64832000000000   1.12420000000000   Zn  (4a)
  -0.07096000000000   1.14832000000000  -0.12420000000000   Zn  (4a)
   0.57096000000000  -0.14832000000000  -0.62420000000000   Zn  (4a)
```

Ferroelectric $NH_4H_2PO_4$: A6BC4D_oP48_19_6a_a_4a_a - CIF

```
# CIF file
data_findsym-output
_audit_creation_method FINDSYM

_chemical_name_mineral 'H6NO4P'
_chemical_formula_sum 'H6 N O4 P'

loop_
_publ_author_name
 'T. Fukami'
 'S. Akahoshi'
 'K. Hukuda'
 'T. Yagi'
_journal_name_full_name
;
 Journal of the Physical Society of Japan
;
_journal_volume 56
_journal_year 1987
_journal_page_first 2223
_journal_page_last 2224
_publ_Section_title
;
 Refinement of the Crystal Structure of NH$_{4}$H$_{2}$PO$_{4}$ above
    ↪ and below Antiferroelectric Phase Transition Temperature
;

_aflow_title 'Ferroelectric NH$_{4}$H$_{2}$PO$_{4}$ Structure'
_aflow_proto 'A6BC4D_oP48_19_6a_a_4a_a'
_aflow_params 'a,b/a,c/a,x_{1},y_{1},z_{1},x_{2},y_{2},z_{2},x_{3},y_{3
    ↪ },z_{3},x_{4},y_{4},z_{4},x_{5},y_{5},z_{5},x_{6},y_{6},z_{6},
    ↪ x_{7},y_{7},z_{7},x_{8},y_{8},z_{8},x_{9},y_{9},z_{9},x_{10},y_
    ↪ {10},z_{10},x_{11},y_{11},z_{11},x_{12},y_{12},z_{12}'
_aflow_params_values '7.503,1.00119952019,0.99800079968,0.012,0.574,
    ↪ 0.587,0.0,0.388,0.547,-0.068,0.504,0.421,0.113,0.487,0.453,
    ↪ 0.789,0.348,0.122,0.154,0.298,0.892,0.011,0.4837,0.5016,0.082,
    ↪ 0.6443,0.1192,-0.0853,0.3498,0.1145,0.8628,0.5773,0.872,0.1571,
    ↪ 0.41,0.8974,0.0,0.5019,0.0004'
_aflow_Strukturbericht 'None'
_aflow_Pearson 'oP48'

_symmetry_space_group_name_H-M "P 21 21 21"
_symmetry_Int_Tables_number 19

_cell_length_a    7.50300
_cell_length_b    7.51200
_cell_length_c    7.48800
_cell_angle_alpha 90.00000
_cell_angle_beta  90.00000
_cell_angle_gamma 90.00000

loop_
_space_group_symop_id
_space_group_symop_operation_xyz
1 x,y,z
2 x+1/2,-y+1/2,-z
3 -x,y+1/2,-z+1/2
4 -x+1/2,-y,z+1/2

loop_
_atom_site_label
_atom_site_type_symbol
_atom_site_symmetry_multiplicity
_atom_site_Wyckoff_label
_atom_site_fract_x
_atom_site_fract_y
_atom_site_fract_z
_atom_site_occupancy
H1 H 4 a 0.01200 0.57400 0.58700 1.00000
H2 H 4 a 0.00000 0.38800 0.54700 1.00000
H3 H 4 a -0.06800 0.50400 0.42100 1.00000
H4 H 4 a 0.11300 0.48700 0.45300 1.00000
H5 H 4 a 0.78900 0.34800 0.12200 1.00000
H6 H 4 a 0.15400 0.29800 0.89200 1.00000
N1 N 4 a 0.01100 0.48370 0.50160 1.00000
O1 O 4 a 0.08200 0.64430 0.11920 1.00000
O2 O 4 a -0.08530 0.34980 0.11450 1.00000
O3 O 4 a 0.86280 0.57730 0.87200 1.00000
O4 O 4 a 0.15710 0.41000 0.89740 1.00000
P1 P 4 a 0.00000 0.50190 0.00040 1.00000
```

Ferroelectric $NH_4H_2PO_4$: A6BC4D_oP48_19_6a_a_4a_a - POSCAR

```
A6BC4D_oP48_19_6a_a_4a_a & a,b/a,c/a,x1,y1,z1,x2,y2,z2,x3,y3,z3,x4,y4,z4
    ↪ ,x5,y5,z5,x6,y6,z6,x7,y7,z7,x8,y8,z8,x9,y9,z9,x10,y10,z10,x11,
    ↪ y11,z11,x12,y12,z12 --params=7.503,1.00119952019,0.99800079968,
    ↪ 0.012,0.574,0.587,0.0,0.388,0.547,-0.068,0.504,0.421,0.113,
    ↪ 0.487,0.453,0.789,0.348,0.122,0.154,0.298,0.892,0.011,0.4837,
    ↪ 0.5016,0.082,0.6443,0.1192,-0.0853,0.3498,0.1145,0.8628,0.5773,
    ↪ 0.872,0.1571,0.41,0.8974,0.0,0.5019,0.0004 & P2_{1}2_{1}2_{1}
    ↪ D_{2}^{4} #19 (a^12) & oP48 & None & H6NO4P & H6NO4P & T.
    ↪ Fukami et al., J. Phys. Soc. Jpn. 56, 2223-2224 (1987)
   1.00000000000000
   7.50300000000000   0.00000000000000   0.00000000000000
   0.00000000000000   7.51200000000000   0.00000000000000
   0.00000000000000   0.00000000000000   7.48800000000000
     H     N     O     P
    24     4    16     4
Direct
   0.01200000000000   0.57400000000000   0.58700000000000   H   (4a)
   0.48800000000000  -0.57400000000000   1.08700000000000   H   (4a)
  -0.01200000000000   1.07400000000000  -0.08700000000000   H   (4a)
   0.51200000000000  -0.07400000000000  -0.58700000000000   H   (4a)
   0.00000000000000   0.38800000000000   0.54700000000000   H   (4a)
   0.50000000000000  -0.38800000000000   1.04700000000000   H   (4a)
   0.00000000000000   0.88800000000000  -0.04700000000000   H   (4a)
   0.50000000000000   0.11200000000000  -0.54700000000000   H   (4a)
  -0.06800000000000   0.50400000000000   0.42100000000000   H   (4a)
   0.56800000000000  -0.50400000000000   0.92100000000000   H   (4a)
   0.06800000000000   1.00400000000000   0.07900000000000   H   (4a)
   0.43200000000000  -0.00400000000000  -0.42100000000000   H   (4a)
   0.11300000000000   0.48700000000000   0.45300000000000   H   (4a)
   0.38700000000000  -0.48700000000000   0.95300000000000   H   (4a)
  -0.11300000000000   0.98700000000000   0.04700000000000   H   (4a)
   0.61300000000000   0.01300000000000  -0.45300000000000   H   (4a)
   0.78900000000000   0.34800000000000   0.12200000000000   H   (4a)
  -0.28900000000000  -0.34800000000000   0.62200000000000   H   (4a)
  -0.78900000000000   0.84800000000000   0.37800000000000   H   (4a)
   1.28900000000000   0.15200000000000  -0.12200000000000   H   (4a)
   0.15400000000000   0.29800000000000   0.89200000000000   H   (4a)
   0.34600000000000  -0.29800000000000   1.39200000000000   H   (4a)
  -0.15400000000000   0.79800000000000  -0.39200000000000   H   (4a)
   0.65400000000000   0.20200000000000  -0.89200000000000   H   (4a)
   0.01100000000000   0.48370000000000   0.50160000000000   N   (4a)
   0.48900000000000  -0.48370000000000   1.00160000000000   N   (4a)
  -0.01100000000000   0.98370000000000  -0.00160000000000   N   (4a)
   0.51100000000000   0.01630000000000  -0.50160000000000   N   (4a)
   0.08200000000000   0.64430000000000   0.11920000000000   O   (4a)
   0.41800000000000  -0.64430000000000   0.61920000000000   O   (4a)
  -0.08200000000000   1.14430000000000   0.38080000000000   O   (4a)
   0.58200000000000  -0.14430000000000  -0.11920000000000   O   (4a)
  -0.08530000000000   0.34980000000000   0.11450000000000   O   (4a)
   0.58530000000000  -0.34980000000000   0.61450000000000   O   (4a)
   0.08530000000000   0.84980000000000   0.38550000000000   O   (4a)
   0.41470000000000   0.15020000000000  -0.11450000000000   O   (4a)
   0.86280000000000   0.57730000000000   0.87200000000000   O   (4a)
  -0.36280000000000  -0.57730000000000   1.37200000000000   O   (4a)
  -0.86280000000000   1.07730000000000  -0.37200000000000   O   (4a)
   1.36280000000000  -0.07730000000000  -0.87200000000000   O   (4a)
   0.15710000000000   0.41000000000000   0.89740000000000   O   (4a)
   0.34290000000000  -0.41000000000000   1.39740000000000   O   (4a)
  -0.15710000000000   0.91000000000000  -0.39740000000000   O   (4a)
   0.65710000000000   0.09000000000000  -0.89740000000000   O   (4a)
   0.00000000000000   0.50190000000000   0.00040000000000   P   (4a)
   0.50000000000000  -0.50190000000000   0.50040000000000   P   (4a)
   0.00000000000000   1.00190000000000   0.49960000000000   P   (4a)
   0.50000000000000  -0.00190000000000  -0.00040000000000   P   (4a)
```

$\beta$-Arabinose [$(CH_2O)_{20}$]: AB2C_oP80_19_5a_10a_5a - CIF

```
# CIF file
data_findsym-output
_audit_creation_method FINDSYM

_chemical_name_mineral 'CH2O'
_chemical_formula_sum 'C H2 O'
```

```
loop_
_publ_author_name
 'A. Hordvik'
_journal_name_full_name
;
 Acta Chemica Scandinavica
;
_journal_volume 15
_journal_year 1961
_journal_page_first 16
_journal_page_last 30
_publ_Section_title
;
 Refinement of the Crystal Structure of $\beta$-Arabinose
;

_aflow_title '$\beta$-Arabinose [(CH$_{2}$O)$_{20}$] Structure'
_aflow_proto 'AB2C_oP80_19_5a_10a_5a'
_aflow_params 'a,b/a,c/a,x_{1},y_{1},z_{1},x_{2},y_{2},z_{2},x_{3},y_{3
    ↪ },z_{3},x_{4},y_{4},z_{4},x_{5},y_{5},z_{5},x_{6},y_{6},z_{6},
    ↪ x_{7},y_{7},z_{7},x_{8},y_{8},z_{8},x_{9},y_{9},z_{9},x_{10},y_
    ↪ {10},z_{10},x_{11},y_{11},z_{11},x_{12},y_{12},z_{12},x_{13},y_
    ↪ {13},z_{13},x_{14},y_{14},z_{14},x_{15},y_{15},z_{15},x_{16},y_
    ↪ {16},z_{16},x_{17},y_{17},z_{17},x_{18},y_{18},z_{18},x_{19},y_
    ↪ {19},z_{19},x_{20},y_{20},z_{20}'
_aflow_params_values '6.535,2.97888293803,0.74078041316,0.493,-0.0828,
    ↪ 0.688,0.431,0.8469,0.7985,0.251,0.8198,0.631,0.0695,0.8712,
    ↪ 0.649,0.15,-0.0586,0.553,0.628,-0.058,0.79,0.376,0.847,0.0,
    ↪ 0.303,0.811,0.422,-0.043,0.843,0.507,0.024,-0.028,0.564,0.22,-
    ↪ 0.061,0.351,0.6,-0.046,0.38,0.539,0.767,0.892,0.173,0.737,0.567
    ↪ ,0.875,0.852,-0.056,0.5625,-0.0901,0.42,0.6075,0.8019,0.771,
    ↪ 0.191,0.7545,0.737,-0.004,0.8781,-0.074,0.325,-0.0368,0.711'
_aflow_Strukturbericht 'None'
_aflow_Pearson 'oP80'

_symmetry_space_group_name_H-M "P 21 21 21"
_symmetry_Int_Tables_number 19

_cell_length_a     6.53500
_cell_length_b     19.46700
_cell_length_c     4.84100
_cell_angle_alpha  90.00000
_cell_angle_beta   90.00000
_cell_angle_gamma  90.00000

loop_
_space_group_symop_id
_space_group_symop_operation_xyz
1 x,y,z
2 x+1/2,-y+1/2,-z
3 -x,y+1/2,-z+1/2
4 -x+1/2,-y,z+1/2

loop_
_atom_site_label
_atom_site_type_symbol
_atom_site_symmetry_multiplicity
_atom_site_Wyckoff_label
_atom_site_fract_x
_atom_site_fract_y
_atom_site_fract_z
_atom_site_occupancy
C1 C 4 a 0.49300 -0.08280 0.68800 1.00000
C2 C 4 a 0.43100 0.84690 0.79850 1.00000
C3 C 4 a 0.25100 0.81980 0.63100 1.00000
C4 C 4 a 0.06950 0.87120 0.64900 1.00000
C5 C 4 a 0.15000 -0.05860 0.55300 1.00000
H1 H 4 a 0.62800 -0.05800 0.79000 1.00000
H2 H 4 a 0.37600 0.84700 0.00000 1.00000
H3 H 4 a 0.30300 0.81100 0.42200 1.00000
H4 H 4 a -0.04300 0.84300 0.50700 1.00000
H5 H 4 a 0.02400 -0.02800 0.56400 1.00000
H6 H 4 a 0.22000 -0.06100 0.35100 1.00000
H7 H 4 a 0.60000 -0.04600 0.38000 1.00000
H8 H 4 a 0.53900 0.76700 0.89200 1.00000
H9 H 4 a 0.17300 0.73700 0.56700 1.00000
H10 H 4 a 0.87500 0.85200 -0.05600 1.00000
O1 O 4 a 0.56250 -0.09010 0.42000 1.00000
O2 O 4 a 0.60750 0.80190 0.77100 1.00000
O3 O 4 a 0.19100 0.75450 0.73700 1.00000
O4 O 4 a -0.00400 0.87810 -0.07400 1.00000
O5 O 4 a 0.32500 -0.03680 0.71100 1.00000
```

$\beta$-Arabinose [$(CH_2O)_{20}$]: AB2C_oP80_19_5a_10a_5a - POSCAR

```
AB2C_oP80_19_5a_10a_5a & a,b/a,c/a,x1,y1,z1,x2,y2,z2,x3,y3,z3,x4,y4,z4,
    ↪ x5,y5,z5,x6,y6,z6,x7,y7,z7,x8,y8,z8,x9,y9,z9,x10,y10,z10,x11,
    ↪ y11,z11,x12,y12,z12,x13,y13,z13,x14,y14,z14,x15,y15,z15,x16,y16
    ↪ ,z16,x17,y17,z17,x18,y18,z18,x19,y19,z19,x20,y20,z20 --params=
    ↪ 6.535,2.97888293803,0.74078041316,0.493,-0.0828,0.688,0.431,
    ↪ 0.8469,0.7985,0.251,0.8198,0.631,0.0695,0.8712,0.649,0.15,-
    ↪ 0.0586,0.553,0.628,-0.058,0.79,0.376,0.847,0.0,0.303,0.811,
    ↪ 0.422,-0.043,0.843,0.507,0.024,-0.028,0.564,0.22,-0.061,0.351,
    ↪ 0.6,-0.046,0.38,0.539,0.767,0.892,0.173,0.737,0.567,0.875,0.852
    ↪ ,-0.056,0.5625,-0.0901,0.42,0.6075,0.8019,0.771,0.191,0.7545,
    ↪ 0.737,-0.004,0.8781,-0.074,0.325,-0.0368,0.711 & P2_{1}2_{1}2_{
    ↪ 1} D_{2}^{4} #19 (a^20) & oP80 & CH2O & CH2O & A.
    ↪ Hordvik, Acta Chem. Scand. 15, 16-30 (1961)
   1.00000000000000
   6.53500000000000   0.00000000000000   0.00000000000000
   0.00000000000000  19.46700000000000   0.00000000000000
   0.00000000000000   0.00000000000000   4.84100000000000
     C      H      O
    20     40     20
Direct
   0.49300000000000  -0.08280000000000   0.68800000000000   C   (4a)
   0.00700000000000   0.08280000000000   1.18800000000000   C   (4a)
  -0.49300000000000   0.41720000000000  -0.18800000000000   C   (4a)
   0.99300000000000   0.58280000000000  -0.68800000000000   C   (4a)
   0.43100000000000   0.84690000000000   0.79850000000000   C   (4a)
   0.06900000000000  -0.84690000000000   1.29850000000000   C   (4a)
  -0.43100000000000   1.34690000000000  -0.29850000000000   C   (4a)
   0.93100000000000  -0.34690000000000  -0.79850000000000   C   (4a)
   0.25100000000000   0.81980000000000   0.63100000000000   C   (4a)
   0.24900000000000  -0.81980000000000   1.13100000000000   C   (4a)
  -0.25100000000000   1.31980000000000  -0.13100000000000   C   (4a)
   0.75100000000000  -0.31980000000000  -0.63100000000000   C   (4a)
   0.06950000000000   0.87120000000000   0.64900000000000   C   (4a)
   0.43050000000000  -0.87120000000000   1.14900000000000   C   (4a)
  -0.06950000000000   1.37120000000000  -0.14900000000000   C   (4a)
   0.56950000000000  -0.37120000000000  -0.64900000000000   C   (4a)
   0.15000000000000  -0.05860000000000   0.55300000000000   C   (4a)
   0.35000000000000   0.05860000000000   1.05300000000000   C   (4a)
  -0.15000000000000   0.44140000000000  -0.05300000000000   C   (4a)
   0.65000000000000   0.55860000000000  -0.55300000000000   C   (4a)
   0.62800000000000  -0.05800000000000   0.79000000000000   H   (4a)
  -0.12800000000000   0.05800000000000   1.29000000000000   H   (4a)
  -0.62800000000000   0.44200000000000  -0.29000000000000   H   (4a)
   1.12800000000000   0.55800000000000  -0.79000000000000   H   (4a)
   0.37600000000000   0.84700000000000   0.00000000000000   H   (4a)
   0.12400000000000  -0.84700000000000   0.50000000000000   H   (4a)
  -0.37600000000000   1.34700000000000   0.50000000000000   H   (4a)
   0.87600000000000  -0.34700000000000   0.00000000000000   H   (4a)
   0.30300000000000   0.81100000000000   0.42200000000000   H   (4a)
   0.19700000000000  -0.81100000000000   0.92200000000000   H   (4a)
  -0.30300000000000   1.31100000000000   0.07800000000000   H   (4a)
   0.80300000000000  -0.31100000000000  -0.42200000000000   H   (4a)
  -0.04300000000000   0.84300000000000   0.50700000000000   H   (4a)
   0.54300000000000  -0.84300000000000   1.00700000000000   H   (4a)
   0.04300000000000   1.34300000000000  -0.00700000000000   H   (4a)
   0.45700000000000  -0.34300000000000  -0.50700000000000   H   (4a)
   0.02400000000000  -0.02800000000000   0.56400000000000   H   (4a)
   0.47600000000000   0.02800000000000   1.06400000000000   H   (4a)
  -0.02400000000000   0.47200000000000  -0.06400000000000   H   (4a)
   0.52400000000000   0.52800000000000  -0.56400000000000   H   (4a)
   0.22000000000000  -0.06100000000000   0.35100000000000   H   (4a)
   0.28000000000000   0.06100000000000   0.85100000000000   H   (4a)
  -0.22000000000000   0.43900000000000   0.14900000000000   H   (4a)
   0.72000000000000   0.56100000000000  -0.35100000000000   H   (4a)
   0.60000000000000  -0.04600000000000   0.38000000000000   H   (4a)
  -0.10000000000000   0.04600000000000   0.88000000000000   H   (4a)
  -0.60000000000000   0.45400000000000   0.12000000000000   H   (4a)
   1.10000000000000   0.54600000000000  -0.38000000000000   H   (4a)
   0.53900000000000   0.76700000000000   0.89200000000000   H   (4a)
  -0.03900000000000  -0.76700000000000   1.39200000000000   H   (4a)
  -0.53900000000000   1.26700000000000  -0.39200000000000   H   (4a)
   1.03900000000000  -0.26700000000000  -0.89200000000000   H   (4a)
   0.17300000000000   0.73700000000000   0.56700000000000   H   (4a)
   0.32700000000000  -0.73700000000000   1.06700000000000   H   (4a)
  -0.17300000000000   1.23700000000000  -0.06700000000000   H   (4a)
   0.67300000000000  -0.23700000000000  -0.56700000000000   H   (4a)
   0.87500000000000   0.85200000000000  -0.05600000000000   H   (4a)
  -0.37500000000000  -0.85200000000000   0.44400000000000   H   (4a)
  -0.87500000000000   1.35200000000000   0.55600000000000   H   (4a)
   1.37500000000000  -0.35200000000000   0.05600000000000   H   (4a)
   0.56250000000000  -0.09010000000000   0.42000000000000   O   (4a)
  -0.06250000000000   0.09010000000000   0.92000000000000   O   (4a)
  -0.56250000000000   0.40990000000000   0.08000000000000   O   (4a)
   1.06250000000000   0.59010000000000  -0.42000000000000   O   (4a)
   0.60750000000000   0.80190000000000   0.77100000000000   O   (4a)
  -0.10750000000000  -0.80190000000000   1.27100000000000   O   (4a)
  -0.60750000000000   1.30190000000000  -0.27100000000000   O   (4a)
   1.10750000000000  -0.30190000000000  -0.77100000000000   O   (4a)
   0.19100000000000   0.75450000000000   0.73700000000000   O   (4a)
   0.30900000000000  -0.75450000000000   1.23700000000000   O   (4a)
  -0.19100000000000   1.25450000000000  -0.23700000000000   O   (4a)
   0.69100000000000  -0.25450000000000  -0.73700000000000   O   (4a)
  -0.00400000000000   0.87810000000000  -0.07400000000000   O   (4a)
   0.50400000000000  -0.87810000000000   0.42600000000000   O   (4a)
   0.00400000000000   1.37810000000000   0.57400000000000   O   (4a)
   0.49600000000000  -0.37810000000000   0.07400000000000   O   (4a)
   0.32500000000000  -0.03680000000000   0.71100000000000   O   (4a)
   0.17500000000000   0.03680000000000   1.21100000000000   O   (4a)
  -0.32500000000000   0.46320000000000  -0.21100000000000   O   (4a)
   0.82500000000000   0.53680000000000  -0.71100000000000   O   (4a)
```

$NaAlCl_4$: AB4C_oP24_19_a_4a_a - CIF

```
# CIF file
data_findsym-output
_audit_creation_method FINDSYM

_chemical_name_mineral 'AlCl4Na'
_chemical_formula_sum 'Al Cl4 Na'

loop_
_publ_author_name
 'G. Mairesse'
 'P. Barbier'
 'J.-P. Wignacourt'
_journal_name_full_name
;
 Acta Crystallographica Section B: Structural Science
;
_journal_volume 35
_journal_year 1979
_journal_page_first 1573
_journal_page_last 1580
_publ_Section_title
```

```
;
Comparison of the crystal structures of alkaline ($M$ = Li, Na, K, Rb,
    ↪ Cs) and pseudo-alkaline ($M$ = NO,NH$_{4}$)
    ↪ tetrachloroaluminates, $M$AlCl$_{4}$
;

# Found in The American Mineralogist Crystal Structure Database, 2003

_aflow_title 'NaAlCl$_{4}$ Structure'
_aflow_proto 'AB4C_oP24_19_a_4a_a'
_aflow_params 'a,b/a,c/a,x_{1},y_{1},z_{1},x_{2},y_{2},z_{2},x_{3},y_{3
    ↪ },z_{3},x_{4},y_{4},z_{4},x_{5},y_{5},z_{5},x_{6},y_{6},z_{6}'
_aflow_params_values '9.886,0.623811450536,1.0441027716,0.51429,0.20707
    ↪ ,-0.03774,0.6855,0.10957,0.85152,0.33503,0.07337,0.87726,
    ↪ 0.52257,0.07458,0.15367,0.50874,0.55281,-0.03214,0.71343,
    ↪ 0.31122,0.37466'
_aflow_Strukturbericht 'None'
_aflow_Pearson 'oP24'

_symmetry_space_group_name_H-M "P 21 21 21"
_symmetry_Int_Tables_number 19

_cell_length_a      9.88600
_cell_length_b      6.16700
_cell_length_c      10.32200
_cell_angle_alpha   90.00000
_cell_angle_beta    90.00000
_cell_angle_gamma   90.00000

loop_
_space_group_symop_id
_space_group_symop_operation_xyz
1 x,y,z
2 x+1/2,-y+1/2,-z
3 -x,y+1/2,-z+1/2
4 -x+1/2,-y,z+1/2

loop_
_atom_site_label
_atom_site_type_symbol
_atom_site_symmetry_multiplicity
_atom_site_Wyckoff_label
_atom_site_fract_x
_atom_site_fract_y
_atom_site_fract_z
_atom_site_occupancy
Al1 Al 4 a 0.51429 0.20707 -0.03774 1.00000
Cl1 Cl 4 a 0.68550 0.10957 0.85152 1.00000
Cl2 Cl 4 a 0.33503 0.07337 0.87726 1.00000
Cl3 Cl 4 a 0.52257 0.07458 0.15367 1.00000
Cl4 Cl 4 a 0.50874 0.55281 -0.03214 1.00000
Na1 Na 4 a 0.71343 0.31122 0.37466 1.00000
```

NaAlCl$_4$: AB4C_oP24_19_a_4a_a - POSCAR

```
AB4C_oP24_19_a_4a_a & a,b/a,c/a,x1,y1,z1,x2,y2,z2,x3,y3,z3,x4,y4,z4,x5,
    ↪ y5,z5,x6,y6,z6 --params=9.886,0.623811450536,1.0441027716,
    ↪ 0.51429,0.20707,-0.03774,0.6855,0.10957,0.85152,0.33503,0.07337
    ↪ ,0.87726,0.52257,0.07458,0.15367,0.50874,0.55281,-0.03214,
    ↪ 0.71343,0.31122,0.37466 & P2_{1}2_{1}2_{1} D_{2}^{4} #19 (a^6)
    ↪ & oP24 & None & AlCl4Na & AlCl4Na & G. Mairesse and P. Barbier
    ↪ and J.-P. Wignacourt, Acta Crystallogr. Sect. B Struct. Sci. 35
    ↪ , 1573-1580 (1979)
   1.00000000000000
   9.88600000000000   0.00000000000000   0.00000000000000
   0.00000000000000   6.16700000000000   0.00000000000000
   0.00000000000000   0.00000000000000  10.32200000000000
    Al     Cl     Na
     4     16      4
Direct
   0.51429000000000   0.20707000000000  -0.03774000000000   Al   (4a)
  -0.01429000000000  -0.20707000000000   0.46226000000000   Al   (4a)
  -0.51429000000000   0.70707000000000   0.53774000000000   Al   (4a)
   1.01429000000000   0.29293000000000   0.03774000000000   Al   (4a)
   0.68550000000000   0.10957000000000   0.85152000000000   Cl   (4a)
  -0.18550000000000  -0.10957000000000   1.35152000000000   Cl   (4a)
  -0.68550000000000   0.60957000000000  -0.35152000000000   Cl   (4a)
   1.18550000000000   0.39043000000000  -0.85152000000000   Cl   (4a)
   0.33503000000000   0.07337000000000   0.87726000000000   Cl   (4a)
   0.16497000000000  -0.07337000000000   1.37726000000000   Cl   (4a)
  -0.33503000000000   0.57337000000000  -0.37726000000000   Cl   (4a)
   0.83503000000000   0.42663000000000  -0.87726000000000   Cl   (4a)
   0.52257000000000   0.07458000000000   0.15367000000000   Cl   (4a)
  -0.02257000000000  -0.07458000000000   0.65367000000000   Cl   (4a)
  -0.52257000000000   0.57458000000000   0.34633000000000   Cl   (4a)
   1.02257000000000   0.42542000000000  -0.15367000000000   Cl   (4a)
   0.50874000000000   0.55281000000000  -0.03214000000000   Cl   (4a)
  -0.00874000000000  -0.55281000000000   0.46786000000000   Cl   (4a)
  -0.50874000000000   1.05281000000000   0.53214000000000   Cl   (4a)
   1.00874000000000  -0.05281000000000   0.03214000000000   Cl   (4a)
   0.71343000000000   0.31122000000000   0.37466000000000   Na   (4a)
  -0.21343000000000  -0.31122000000000   0.87466000000000   Na   (4a)
  -0.71343000000000   0.81122000000000   0.12534000000000   Na   (4a)
   1.21343000000000   0.18878000000000  -0.37466000000000   Na   (4a)
```

NaP: AB_oP16_19_2a_2a - CIF

```
# CIF file
data_findsym-output
_audit_creation_method FINDSYM

_chemical_name_mineral 'NaP'
_chemical_formula_sum 'Na P'

loop_
_publ_author_name
'H. G. {von Schnering}'
'W. H\"{o}nle'
_journal_name_full_name
;
Zeitschrift fur Anorganische und Allgemeine Chemie
;
_journal_volume 456
_journal_year 1979
_journal_page_first 194
_journal_page_last 206
_publ_Section_title
;
Zur Chemie und Strukturchemie der Phosphide und Polyphosphide. 20.
    ↪ Darstellung, Struktur und Eigenschaften der
    ↪ Alkalimetallmonophosphide NaP und KP
;

# Found in NaP Crystal Structure, 2016 Found in NaP Crystal Structure, {
    ↪ PAULING FILE in: Inorganic Solid Phases, SpringerMaterials (
    ↪ online database)},

_aflow_title 'NaP Structure'
_aflow_proto 'AB_oP16_19_2a_2a'
_aflow_params 'a,b/a,c/a,x_{1},y_{1},z_{1},x_{2},y_{2},z_{2},x_{3},y_{3
    ↪ },z_{3},x_{4},y_{4},z_{4}'
_aflow_params_values '6.038,0.934580987082,1.67969526333,0.4174,-0.0911,
    ↪ 0.0318,0.1338,0.6367,0.3313,0.3086,0.1404,0.2838,0.4287,0.402,
    ↪ 0.1341'
_aflow_Strukturbericht 'None'
_aflow_Pearson 'oP16'

_symmetry_space_group_name_H-M "P 21 21 21"
_symmetry_Int_Tables_number 19

_cell_length_a      6.03800
_cell_length_b      5.64300
_cell_length_c      10.14200
_cell_angle_alpha   90.00000
_cell_angle_beta    90.00000
_cell_angle_gamma   90.00000

loop_
_space_group_symop_id
_space_group_symop_operation_xyz
1 x,y,z
2 x+1/2,-y+1/2,-z
3 -x,y+1/2,-z+1/2
4 -x+1/2,-y,z+1/2

loop_
_atom_site_label
_atom_site_type_symbol
_atom_site_symmetry_multiplicity
_atom_site_Wyckoff_label
_atom_site_fract_x
_atom_site_fract_y
_atom_site_fract_z
_atom_site_occupancy
Na1 Na 4 a 0.41740 -0.09110 0.03180 1.00000
Na2 Na 4 a 0.13380 0.63670 0.33130 1.00000
P1 P 4 a 0.30860 0.14040 0.28380 1.00000
P2 P 4 a 0.42870 0.40200 0.13410 1.00000
```

NaP: AB_oP16_19_2a_2a - POSCAR

```
AB_oP16_19_2a_2a & a,b/a,c/a,x1,y1,z1,x2,y2,z2,x3,y3,z3,x4,y4,z4 --
    ↪ params=6.038,0.934580987082,1.67969526333,0.4174,-0.0911,0.0318
    ↪ ,0.1338,0.6367,0.3313,0.3086,0.1404,0.2838,0.4287,0.402,0.1341
    ↪ & P2_{1}2_{1}2_{1} D_{2}^{4} #19 (a^4) & oP16 & None & NaP &
    ↪ NaP & H. G. {von Schnering} and W. H\"{o}nle, Z. Anorg. Allg.
    ↪ Chem. 456, 194-206 (1979)
   1.00000000000000
   6.03800000000000   0.00000000000000   0.00000000000000
   0.00000000000000   5.64300000000000   0.00000000000000
   0.00000000000000   0.00000000000000  10.14200000000000
    Na     P
     8     8
Direct
   0.41740000000000  -0.09110000000000   0.03180000000000   Na   (4a)
   0.08260000000000   0.09110000000000   0.53180000000000   Na   (4a)
  -0.41740000000000   0.40890000000000   0.46820000000000   Na   (4a)
   0.91740000000000   0.59110000000000  -0.03180000000000   Na   (4a)
   0.13380000000000   0.63670000000000   0.33130000000000   Na   (4a)
   0.36620000000000  -0.63670000000000   0.83130000000000   Na   (4a)
  -0.13380000000000   1.13670000000000   0.16870000000000   Na   (4a)
   0.63380000000000  -0.13670000000000  -0.33130000000000   Na   (4a)
   0.30860000000000   0.14040000000000   0.28380000000000    P   (4a)
   0.19140000000000  -0.14040000000000   0.78380000000000    P   (4a)
  -0.30860000000000   0.64040000000000   0.21620000000000    P   (4a)
   0.80860000000000   0.35960000000000  -0.28380000000000    P   (4a)
   0.42870000000000   0.40200000000000   0.13410000000000    P   (4a)
   0.07130000000000  -0.40200000000000   0.63410000000000    P   (4a)
  -0.42870000000000   0.90200000000000   0.36590000000000    P   (4a)
   0.92870000000000   0.09800000000000  -0.13410000000000    P   (4a)
```

*D*$0_7$ (CrO$_3$) (*obsolete*): AB3_oC16_20_a_bc - CIF

```
# CIF file
data_findsym-output
_audit_creation_method FINDSYM

_chemical_name_mineral 'CrO3'
_chemical_formula_sum 'Cr O3'
```

```
loop_
_publ_author_name
 'H. Br\"{a}kken'
_journal_name_full_name
;
 Zeitschrift f{\"u}r Kristallographie - Crystalline Materials
;
_journal_volume 78
_journal_year 1931
_journal_page_first 484
_journal_page_last 488
_publ_Section_title
;
 Die Kristallstrukturen der Trioxyde von Chrom, Molybd\"{a}n und Wolfram
;

# Found in The Crystal Structure of Chromium Trioxide, 1950

_aflow_title '$D0_{7}$ (CrO$_{3}$) ({\em{obsolete}}) Structure'
_aflow_proto 'AB3_oC16_20_a_bc'
_aflow_params 'a,b/a,c/a,x_{1},y_{2},x_{3},y_{3},z_{3}'
_aflow_params_values '8.46,0.563829787234,0.673758865248,0.33333,0.33333
    ↪ ,0.16667,-0.16667,0.25'
_aflow_Strukturbericht '$D0_{7}$'
_aflow_Pearson 'oC16'

_symmetry_space_group_name_H-M "C 2 2 21"
_symmetry_Int_Tables_number 20

_cell_length_a    8.46000
_cell_length_b    4.77000
_cell_length_c    5.70000
_cell_angle_alpha 90.00000
_cell_angle_beta  90.00000
_cell_angle_gamma 90.00000

loop_
_space_group_symop_id
_space_group_symop_operation_xyz
1 x,y,z
2 x,-y,-z
3 -x,y,-z+1/2
4 -x,-y,z+1/2
5 x+1/2,y+1/2,z
6 x+1/2,-y+1/2,-z
7 -x+1/2,y+1/2,-z+1/2
8 -x+1/2,-y+1/2,z+1/2

loop_
_atom_site_label
_atom_site_type_symbol
_atom_site_symmetry_multiplicity
_atom_site_Wyckoff_label
_atom_site_fract_x
_atom_site_fract_y
_atom_site_fract_z
_atom_site_occupancy
Cr1 Cr 4 a 0.33333 0.00000 0.00000 1.00000
O1 O 4 b 0.00000 0.33333 0.25000 1.00000
O2 O 8 c 0.16667 -0.16667 0.25000 1.00000
```

$D0_7$ ($CrO_3$) (*obsolete*): AB3_oC16_20_a_bc - POSCAR

```
AB3_oC16_20_a_bc & a,b/a,c/a,x1,y2,x3,y3,z3 --params=8.46,0.563829787234
    ↪ ,0.673758865248,0.33333,0.33333,0.16667,-0.16667,0.25 & C222_{1
    ↪ } D_{2}^{5} #20 (abc) & oC16 & $D0_{7}$ & CrO3 & CrO3 & H. Br
    ↪ \"{a}kken, Zeitschrift f{"u}r Kristallographie - Crystalline
    ↪ Materials 78, 484-488 (1931)
   1.00000000000000
   4.23000000000000  -2.38500000000000   0.00000000000000
   4.23000000000000   2.38500000000000   0.00000000000000
   0.00000000000000   0.00000000000000   5.70000000000000
    Cr     O
     2     6
Direct
   0.33333000000000   0.33333000000000   0.00000000000000   Cr   (4a)
  -0.33333000000000  -0.33333000000000   0.50000000000000   Cr   (4a)
  -0.33333000000000   0.33333000000000   0.25000000000000    O   (4b)
   0.33333000000000  -0.33333000000000   0.75000000000000    O   (4b)
   0.33334000000000   0.00000000000000   0.25000000000000    O   (8c)
  -0.33334000000000   0.00000000000000   0.75000000000000    O   (8c)
   0.00000000000000  -0.33334000000000   0.25000000000000    O   (8c)
   0.00000000000000   0.33334000000000  -0.25000000000000    O   (8c)
```

$AlPO_4$ “low cristobalite type”: AB4C_oC24_20_b_2c_a - CIF

```
# CIF file
data_findsym-output
_audit_creation_method FINDSYM

_chemical_name_mineral 'Low cristobalite type'
_chemical_formula_sum 'Al O4 P'

loop_
_publ_author_name
 'R. C. L. Mooney'
_journal_name_full_name
;
 Acta Cristallographica
;
_journal_volume 9
_journal_year 1956
_journal_page_first 728
_journal_page_last 734
_publ_Section_title
;
 The crystal structure of aluminium phosphate and gallium phosphate,
    ↪ low-cristobalite type
;

# Found in The $\alpha-\beta$ phase transition in AlPO$_{4}$
    ↪ cristobalite: Symmetry analysis, domain structure and
    ↪ transition dynamics, 1994

_aflow_title 'AlPO$_{4}$ ``low cristobalite type\'\' Structure'
_aflow_proto 'AB4C_oC24_20_b_2c_a'
_aflow_params 'a,b/a,c/a,x_{1},y_{2},x_{3},y_{3},z_{3},x_{4},y_{4},z_{4
    ↪ }'
_aflow_params_values '7.009,1.0,0.999571978884,0.306,0.198,0.179,0.058,
    ↪ 0.172,0.433,0.17,0.941'
_aflow_Strukturbericht 'None'
_aflow_Pearson 'oC24'

_symmetry_space_group_name_H-M "C 2 2 21"
_symmetry_Int_Tables_number 20

_cell_length_a    7.00900
_cell_length_b    7.00900
_cell_length_c    7.00600
_cell_angle_alpha 90.00000
_cell_angle_beta  90.00000
_cell_angle_gamma 90.00000

loop_
_space_group_symop_id
_space_group_symop_operation_xyz
1 x,y,z
2 x,-y,-z
3 -x,y,-z+1/2
4 -x,-y,z+1/2
5 x+1/2,y+1/2,z
6 x+1/2,-y+1/2,-z
7 -x+1/2,y+1/2,-z+1/2
8 -x+1/2,-y+1/2,z+1/2

loop_
_atom_site_label
_atom_site_type_symbol
_atom_site_symmetry_multiplicity
_atom_site_Wyckoff_label
_atom_site_fract_x
_atom_site_fract_y
_atom_site_fract_z
_atom_site_occupancy
P1 P 4 a 0.30600 0.00000 0.00000 1.00000
Al1 Al 4 b 0.00000 0.19800 0.25000 1.00000
O1 O 8 c 0.17900 0.05800 0.17200 1.00000
O2 O 8 c 0.43300 0.17000 0.94100 1.00000
```

$AlPO_4$ “low cristobalite type”: AB4C_oC24_20_b_2c_a - POSCAR

```
AB4C_oC24_20_b_2c_a & a,b/a,c/a,x1,y2,x3,y3,z3,x4,y4,z4 --params=7.009,
    ↪ 1.0,0.999571978884,0.306,0.198,0.179,0.058,0.172,0.433,0.17,
    ↪ 0.941 & C222_{1} D_{2}^{5} #20 (abc^2) & oC24 & None & AlO4P &
    ↪ Low cristobalite type & R. C. L. Mooney, Acta Cryst. 9, 728-734
    ↪ (1956)
   1.00000000000000
   3.50450000000000  -3.50450000000000   0.00000000000000
   3.50450000000000   3.50450000000000   0.00000000000000
   0.00000000000000   0.00000000000000   7.00600000000000
    Al     O     P
     2     8     2
Direct
  -0.19800000000000   0.19800000000000   0.25000000000000   Al   (4b)
   0.19800000000000  -0.19800000000000   0.75000000000000   Al   (4b)
   0.12100000000000   0.23700000000000   0.17200000000000    O   (8c)
  -0.12100000000000  -0.23700000000000   0.67200000000000    O   (8c)
  -0.23700000000000  -0.12100000000000   0.32800000000000    O   (8c)
   0.23700000000000   0.12100000000000  -0.17200000000000    O   (8c)
   0.26300000000000   0.60300000000000   0.94100000000000    O   (8c)
  -0.26300000000000  -0.60300000000000   1.44100000000000    O   (8c)
  -0.60300000000000  -0.26300000000000  -0.44100000000000    O   (8c)
   0.60300000000000   0.26300000000000  -0.94100000000000    O   (8c)
   0.30600000000000   0.30600000000000   0.00000000000000    P   (4a)
  -0.30600000000000  -0.30600000000000   0.50000000000000    P   (4a)
```

$Tl_2AlF_5$ ($K3_3$): AB5C2_oC32_20_b_a2bc_c - CIF

```
# CIF file
data_findsym-output
_audit_creation_method FINDSYM

_chemical_name_mineral 'AlF5Tl2'
_chemical_formula_sum 'Al F5 Tl2'

loop_
_publ_author_name
 'C. Brosset'
_journal_name_full_name
;
 Zeitschrift fur Anorganische und Allgemeine Chemie
;
_journal_volume 235
_journal_year 1937
_journal_page_first 139
_journal_page_last 147
_publ_Section_title
;
 Herstellung und Kristallbau der Verbindungen TlAlF$_{4}$ und Tl$_{2}
    ↪ $AlF$_{5}$
```

```
;

# Found in A Structural Classification of Fluoaluminates, 1950

_aflow_title 'Tl$_{2}$AlF$_{5}$ ($K3_{3}$) Structure'
_aflow_proto 'AB5C2_oC32_20_b_a2bc_c'
_aflow_params 'a,b/a,c/a,x_{1},y_{2},y_{3},y_{4},x_{5},y_{5},z_{5},x_{6}
    ↪ },y_{6},z_{6}'
_aflow_params_values '10.06,0.819085487078,0.741550695825,0.033,0.0,0.23
    ↪ ,0.78,0.19,0.0,0.29,0.29,0.2,0.0'
_aflow_Strukturbericht 'None'
_aflow_Pearson 'oC32'

_symmetry_space_group_name_H-M "C 2 2 21"
_symmetry_Int_Tables_number 20

_cell_length_a    10.06000
_cell_length_b    8.24000
_cell_length_c    7.46000
_cell_angle_alpha 90.00000
_cell_angle_beta  90.00000
_cell_angle_gamma 90.00000

loop_
_space_group_symop_id
_space_group_symop_operation_xyz
1 x,y,z
2 x,-y,-z
3 -x,y,-z+1/2
4 -x,-y,z+1/2
5 x+1/2,y+1/2,z
6 x+1/2,-y+1/2,-z
7 -x+1/2,y+1/2,-z+1/2
8 -x+1/2,-y+1/2,z+1/2

loop_
_atom_site_label
_atom_site_type_symbol
_atom_site_symmetry_multiplicity
_atom_site_Wyckoff_label
_atom_site_fract_x
_atom_site_fract_y
_atom_site_fract_z
_atom_site_occupancy
F1 F 4 a 0.03300 0.00000 0.00000 1.00000
Al1 Al 4 b 0.00000 0.00000 0.25000 1.00000
F2 F 4 b 0.00000 0.23000 0.25000 1.00000
F3 F 4 b 0.00000 0.78000 0.25000 1.00000
F4 F 8 c 0.19000 0.00000 0.29000 1.00000
Tl1 Tl 8 c 0.29000 0.20000 0.00000 1.00000
```

$Tl_2AlF_5$ ($K3_3$): AB5C2_oC32_20_b_a2bc_c - POSCAR

```
AB5C2_oC32_20_b_a2bc_c & a,b/a,c/a,x1,y2,y3,y4,x5,y5,z5,x6,y6,z6 --
    ↪ params=10.06,0.819085487078,0.741550695825,0.033,0.0,0.23,0.78,
    ↪ 0.19,0.0,0.29,0.29,0.2,0.0 & C222_{1} D_{2}^{5} #20 (ab^3c^2) &
    ↪ oC32 & None & AlF5Tl2 & AlF5Tl2 & C. Brosset, Z. Anorg. Allg.
    ↪ Chem. 235, 139-147 (1937)
   1.00000000000000
   5.03000000000000  -4.12000000000000   0.00000000000000
   5.03000000000000   4.12000000000000   0.00000000000000
   0.00000000000000   0.00000000000000   7.46000000000000
    Al      F      Tl
     2     10       4
Direct
   0.00000000000000   0.00000000000000   0.25000000000000   Al   (4b)
   0.00000000000000   0.00000000000000   0.75000000000000   Al   (4b)
   0.03300000000000   0.03300000000000   0.00000000000000   F    (4a)
  -0.03300000000000  -0.03300000000000   0.50000000000000   F    (4a)
  -0.23000000000000   0.23000000000000   0.25000000000000   F    (4b)
   0.23000000000000  -0.23000000000000   0.75000000000000   F    (4b)
  -0.78000000000000   0.78000000000000   0.25000000000000   F    (4b)
   0.78000000000000  -0.78000000000000   0.75000000000000   F    (4b)
   0.19000000000000   0.19000000000000   0.29000000000000   F    (8c)
  -0.19000000000000  -0.19000000000000   0.79000000000000   F    (8c)
  -0.19000000000000  -0.19000000000000   0.21000000000000   F    (8c)
   0.19000000000000   0.19000000000000  -0.29000000000000   F    (8c)
   0.09000000000000   0.49000000000000   0.00000000000000   Tl   (8c)
  -0.09000000000000  -0.49000000000000   0.50000000000000   Tl   (8c)
  -0.49000000000000  -0.09000000000000   0.50000000000000   Tl   (8c)
   0.49000000000000   0.09000000000000   0.00000000000000   Tl   (8c)
```

$HoSb_2$: AB2_oC6_21_a_k - CIF

```
# CIF file
data_findsym-output
_audit_creation_method FINDSYM

_chemical_name_mineral 'HoSb2'
_chemical_formula_sum 'Ho Sb2'

loop_
_publ_author_name
 'Q. Johnson'
_journal_name_full_name
;
 Inorganic Chemistry
;
_journal_volume 10
_journal_year 1971
_journal_page_first 2089
_journal_page_last 2090
_publ_Section_title
;
 The Crystal Structure of High-Pressure Synthesized Holmium Diantimonde
;

# Found in The Ho-Sb Alloy System, 1984

_aflow_title 'HoSb$_{2}$ Structure'
_aflow_proto 'AB2_oC6_21_a_k'
_aflow_params 'a,b/a,c/a,z_{2}'
_aflow_params_values '3.343,1.73197726593,2.34519892312,0.34'
_aflow_Strukturbericht 'None'
_aflow_Pearson 'oC6'

_symmetry_space_group_name_H-M "C 2 2 2"
_symmetry_Int_Tables_number 21

_cell_length_a    3.34300
_cell_length_b    5.79000
_cell_length_c    7.84000
_cell_angle_alpha 90.00000
_cell_angle_beta  90.00000
_cell_angle_gamma 90.00000

loop_
_space_group_symop_id
_space_group_symop_operation_xyz
1 x,y,z
2 x,-y,-z
3 -x,y,-z
4 -x,-y,z
5 x+1/2,y+1/2,z
6 x+1/2,-y+1/2,-z
7 -x+1/2,y+1/2,-z
8 -x+1/2,-y+1/2,z

loop_
_atom_site_label
_atom_site_type_symbol
_atom_site_symmetry_multiplicity
_atom_site_Wyckoff_label
_atom_site_fract_x
_atom_site_fract_y
_atom_site_fract_z
_atom_site_occupancy
Ho1 Ho 2 a 0.00000 0.00000 0.00000 1.00000
Sb1 Sb 4 k 0.25000 0.25000 0.34000 1.00000
```

$HoSb_2$: AB2_oC6_21_a_k - POSCAR

```
AB2_oC6_21_a_k & a,b/a,c/a,z2 --params=3.343,1.73197726593,2.34519892312
    ↪ ,0.34 & C222 D_{2}^{6} #21 (ak) & oC6 & None & HoSb2 & HoSb2 &
    ↪ Q. Johnson, Inorg. Chem. 10, 2089-2090 (1971)
   1.00000000000000
   1.67150000000000  -2.89500000000000   0.00000000000000
   1.67150000000000   2.89500000000000   0.00000000000000
   0.00000000000000   0.00000000000000   7.84000000000000
    Ho      Sb
     1       2
Direct
   0.00000000000000   0.00000000000000   0.00000000000000   Ho   (2a)
   0.00000000000000   0.50000000000000   0.34000000000000   Sb   (4k)
   0.50000000000000   0.00000000000000  -0.34000000000000   Sb   (4k)
```

Predicted Phase IV $Cd_2Re_2O_7$: A2B7C2_oF88_22_k_bdefghij_k - CIF

```
# CIF file
data_findsym-output
_audit_creation_method FINDSYM

_chemical_name_mineral 'Cd2O7Re2'
_chemical_formula_sum 'Cd2 O7 Re2'

loop_
_publ_author_name
 'K. J. Kapcia'
 'M. Reedyk'
 'M. Hajialamdari'
 'A. Ptok'
 'P. Piekarz'
 'F. S. Razavi'
 'A. M. Ole\\'{s}'
 'R. K. Kremer'
_journal_year 2019
_publ_Section_title
;
 Low-Temperature Phase of the Cd$_{2}$Re$_{2}$O$_{7}$ Superconductor: {\
    ↪ em Ab initio} Phonon Calculations and Raman Scattering
;

_aflow_title 'Predicted Phase IV Cd$_{2}$Re$_{2}$O$_{7}$ Structure'
_aflow_proto 'A2B7C2_oF88_22_k_bdefghij_k'
_aflow_params 'a,b/a,c/a,x_{3},y_{4},z_{5},z_{6},y_{7},x_{8},x_{9},y_{9
    ↪ },z_{9},x_{10},y_{10},z_{10}'
_aflow_params_values '10.3832,0.999961476231,1.00259072348,0.2,0.2,
    ↪ 0.1689,0.0447,0.4295,0.0706,0.63141,0.63137,0.6114,0.1245,
    ↪ 0.124502,0.126'
_aflow_Strukturbericht 'None'
_aflow_Pearson 'oF88'

_symmetry_space_group_name_H-M "F 2 2 2"
_symmetry_Int_Tables_number 22

_cell_length_a    10.38320
_cell_length_b    10.38280
_cell_length_c    10.41010
_cell_angle_alpha 90.00000
_cell_angle_beta  90.00000
```

```
_cell_angle_gamma 90.00000

loop_
_space_group_symop_id
_space_group_symop_operation_xyz
1 x,y,z
2 x,-y,-z
3 -x,y,-z
4 -x,-y,z
5 x,y+1/2,z+1/2
6 x,-y+1/2,-z+1/2
7 -x,y+1/2,-z+1/2
8 -x,-y+1/2,z+1/2
9 x+1/2,y,z+1/2
10 x+1/2,-y,-z+1/2
11 -x+1/2,y,-z+1/2
12 -x+1/2,-y,z+1/2
13 x+1/2,y+1/2,z
14 x+1/2,-y+1/2,-z
15 -x+1/2,y+1/2,-z
16 -x+1/2,-y+1/2,z

loop_
_atom_site_label
_atom_site_type_symbol
_atom_site_symmetry_multiplicity
_atom_site_Wyckoff_label
_atom_site_fract_x
_atom_site_fract_y
_atom_site_fract_z
_atom_site_occupancy
O1 O 4 b 0.000000 0.000000 0.500000 1.00000
O2 O 4 d 0.250000 0.250000 0.750000 1.00000
O3 O 8 e 0.200000 0.000000 0.000000 1.00000
O4 O 8 f 0.000000 0.200000 0.000000 1.00000
O5 O 8 g 0.000000 0.000000 0.168900 1.00000
O6 O 8 h 0.250000 0.250000 0.044700 1.00000
O7 O 8 i 0.250000 0.429500 0.250000 1.00000
O8 O 8 j 0.070600 0.250000 0.250000 1.00000
Cd1 Cd 16 k 0.631410 0.631370 0.611400 1.00000
Re1 Re 16 k 0.124500 0.124502 0.126000 1.00000
```

Predicted Phase IV $Cd_2Re_2O_7$: A2B7C2_oF88_22_k_bdefghij_k - POSCAR

```
A2B7C2_oF88_22_k_bdefghij_k & a,b/a,c/a,x3,y4,z5,z6,y7,x8,x9,y9,z9,x10,
    ↪ y10,z10 --params=10.3832,0.999961476231,1.00259072348,0.2,0.2,
    ↪ 0.1689,0.0447,0.4295,0.0706,0.63141,0.63137,0.6114,0.1245,
    ↪ 0.124502,0.126 & F222 D_{2}^{7} #22 (bdefghijk^2) & oF88 & None
    ↪ & Cd2O7Re2 & Cd2O7Re2 & K. J. Kapcia et al., (2019)
   1.00000000000000
   0.00000000000000   5.19140000000000   5.20505000000000
   5.19160000000000   0.00000000000000   5.20505000000000
   5.19160000000000   5.19140000000000   0.00000000000000
   Cd      O      Re
    4     14       4
Direct
   0.61136000000000   0.61144000000000   0.65138000000000   Cd   (16k)
   0.61144000000000   0.61136000000000  -1.87418000000000   Cd   (16k)
   0.65138000000000  -1.87418000000000   0.61136000000000   Cd   (16k)
  -1.87418000000000   0.65138000000000   0.61144000000000   Cd   (16k)
   0.50000000000000   0.50000000000000   0.50000000000000   O    (4b)
   0.75000000000000   0.75000000000000   0.75000000000000   O    (4d)
  -0.20000000000000   0.20000000000000   0.20000000000000   O    (8e)
   0.20000000000000  -0.20000000000000  -0.20000000000000   O    (8e)
   0.20000000000000  -0.20000000000000   0.20000000000000   O    (8f)
  -0.20000000000000   0.20000000000000  -0.20000000000000   O    (8f)
   0.16890000000000   0.16890000000000  -0.16890000000000   O    (8g)
  -0.16890000000000  -0.16890000000000   0.16890000000000   O    (8g)
   0.04470000000000   0.04470000000000   0.45530000000000   O    (8h)
   0.45530000000000   0.45530000000000   0.04470000000000   O    (8h)
   0.42950000000000   0.07050000000000   0.42950000000000   O    (8i)
   0.07050000000000   0.42950000000000   0.07050000000000   O    (8i)
   0.42940000000000   0.07060000000000   0.07060000000000   O    (8j)
   0.07060000000000   0.42940000000000   0.42940000000000   O    (8j)
   0.12600200000000   0.12599800000000   0.12300200000000   Re   (16k)
   0.12599800000000   0.12600200000000  -0.37500200000000   Re   (16k)
   0.12300200000000  -0.37500200000000   0.12600200000000   Re   (16k)
  -0.37500200000000   0.12300200000000   0.12599800000000   Re   (16k)
```

Mercury (II) Azide $[Hg(N_3)_2]$: AB6_oP28_29_a_6a - CIF

```
# CIF file
data_findsym-output
_audit_creation_method FINDSYM

_chemical_name_mineral 'Mercury (ii) azide'
_chemical_formula_sum 'Hg N6'

loop_
_publ_author_name
 'U. M\"{u}ller'
_journal_name_full_name
;
 Zeitschrift fur Anorganische und Allgemeine Chemie
;
_journal_volume 399
_journal_year 1973
_journal_page_first 183
_journal_page_last 192
_publ_Section_title
;
 Die Kristallstruktur von $\alpha$-Quecksilber(II)-Azid
;

# Found in Binary Alloy Phase Diagrams, 1990 Found in Binary Alloy Phase
    ↪ Diagrams, {Hf-Re to Zn-Zr}

_aflow_title 'Mercury (II) Azide [Hg(N$_{3}$)$_{2}$] Structure'
_aflow_proto 'AB6_oP28_29_a_6a'
_aflow_params 'a,b/a,c/a,x_{1},y_{1},z_{1},x_{2},y_{2},z_{2},x_{3},y_{3
    ↪ },z_{3},x_{4},y_{4},z_{4},x_{5},y_{5},z_{5},x_{6},y_{6},z_{6},
    ↪ x_{7},y_{7},z_{7}'
_aflow_params_values '10.632,0.589164785553,0.59471407073,0.03183,
    ↪ 0.23877,0.25,0.082,-0.043,0.106,0.188,0.893,0.121,0.284,0.82,
    ↪ 0.112,0.962,0.529,0.39,0.883,0.609,0.28,0.807,0.7,0.172'
_aflow_Strukturbericht 'None'
_aflow_Pearson 'oP28'

_symmetry_space_group_name_H-M "P c a 21"
_symmetry_Int_Tables_number 29

_cell_length_a     10.63200
_cell_length_b     6.26400
_cell_length_c     6.32300
_cell_angle_alpha  90.00000
_cell_angle_beta   90.00000
_cell_angle_gamma  90.00000

loop_
_space_group_symop_id
_space_group_symop_operation_xyz
1 x,y,z
2 -x,-y,z+1/2
3 -x+1/2,y,z+1/2
4 x+1/2,-y,z

loop_
_atom_site_label
_atom_site_type_symbol
_atom_site_symmetry_multiplicity
_atom_site_Wyckoff_label
_atom_site_fract_x
_atom_site_fract_y
_atom_site_fract_z
_atom_site_occupancy
Hg1 Hg 4 a 0.03183 0.23877 0.25000 1.00000
N1 N 4 a 0.08200 -0.04300 0.10600 1.00000
N2 N 4 a 0.18800 0.89300 0.12100 1.00000
N3 N 4 a 0.28400 0.82000 0.11200 1.00000
N4 N 4 a 0.96200 0.52900 0.39000 1.00000
N5 N 4 a 0.88300 0.60900 0.28000 1.00000
N6 N 4 a 0.80700 0.70000 0.17200 1.00000
```

Mercury (II) Azide $[Hg(N_3)_2]$: AB6_oP28_29_a_6a - POSCAR

```
AB6_oP28_29_a_6a & a,b/a,c/a,x1,y1,z1,x2,y2,z2,x3,y3,z3,x4,y4,z4,x5,y5,
    ↪ z5,x6,y6,z6,x7,y7,z7 --params=10.632,0.589164785553,
    ↪ 0.59471407073,0.03183,0.23877,0.25,0.082,-0.043,0.106,0.188,
    ↪ 0.893,0.121,0.284,0.82,0.112,0.962,0.529,0.39,0.883,0.609,0.28,
    ↪ 0.807,0.7,0.172 & Pca2_{1} C_{2v}^{5} #29 (a^7) & oP28 & None &
    ↪ HgN6 & Mercury (ii) azide & U. M\"{u}ller, Z. Anorg. Allg.
    ↪ Chem. 399, 183-192 (1973)
   1.00000000000000
  10.63200000000000   0.00000000000000   0.00000000000000
   0.00000000000000   6.26400000000000   0.00000000000000
   0.00000000000000   0.00000000000000   6.32300000000000
   Hg      N
    4     24
Direct
   0.03183000000000   0.23877000000000   0.25000000000000   Hg   (4a)
  -0.03183000000000  -0.23877000000000   0.75000000000000   Hg   (4a)
   0.53183000000000  -0.23877000000000   0.25000000000000   Hg   (4a)
   0.46817000000000   0.23877000000000   0.75000000000000   Hg   (4a)
   0.08200000000000  -0.04300000000000   0.10600000000000   N    (4a)
  -0.08200000000000   0.04300000000000   0.60600000000000   N    (4a)
   0.58200000000000   0.04300000000000   0.10600000000000   N    (4a)
   0.41800000000000  -0.04300000000000   0.60600000000000   N    (4a)
   0.18800000000000   0.89300000000000   0.12100000000000   N    (4a)
  -0.18800000000000  -0.89300000000000   0.62100000000000   N    (4a)
   0.68800000000000  -0.89300000000000   0.12100000000000   N    (4a)
   0.31200000000000   0.89300000000000   0.62100000000000   N    (4a)
   0.28400000000000   0.82000000000000   0.11200000000000   N    (4a)
  -0.28400000000000  -0.82000000000000   0.61200000000000   N    (4a)
   0.78400000000000  -0.82000000000000   0.11200000000000   N    (4a)
   0.21600000000000   0.82000000000000   0.61200000000000   N    (4a)
   0.96200000000000   0.52900000000000   0.39000000000000   N    (4a)
  -0.96200000000000  -0.52900000000000   0.89000000000000   N    (4a)
   1.46200000000000  -0.52900000000000   0.39000000000000   N    (4a)
  -0.46200000000000   0.52900000000000   0.89000000000000   N    (4a)
   0.88300000000000   0.60900000000000   0.28000000000000   N    (4a)
  -0.88300000000000  -0.60900000000000   0.78000000000000   N    (4a)
   1.38300000000000  -0.60900000000000   0.28000000000000   N    (4a)
  -0.38300000000000   0.60900000000000   0.78000000000000   N    (4a)
   0.80700000000000   0.70000000000000   0.17200000000000   N    (4a)
  -0.80700000000000  -0.70000000000000   0.67200000000000   N    (4a)
   1.30700000000000  -0.70000000000000   0.17200000000000   N    (4a)
  -0.30700000000000   0.70000000000000   0.67200000000000   N    (4a)
```

Low-Temperature $(NH_3CH_3)Al(SO_4)_2{\cdot}12H_2O$: ABC30DE20F2_oP220_29_a_a_30a_a_20a_2a - CIF

```
# CIF file
data_findsym-output
_audit_creation_method FINDSYM

_chemical_name_mineral 'AlCH30NO20S2'
_chemical_formula_sum 'Al C H30 N O20 S2'

loop_
_publ_author_name
```

```
 'R. O. W. Fletcher'
 'H. Steeple'
_journal_name_full_name
;
 Acta Cristallographica
;
_journal_volume 17
_journal_year 1964
_journal_page_first 290
_journal_page_last 294
_publ_Section_title
;
 The crystal structure of the low-temperature phase of methylammonium
    ↪ alum
;

_aflow_title 'Low-Temperature (NH$_{3}$CH$_{3}$)Al(SO$_{4}$)$_{2}$$\
    ↪ cdot$12H$_{2}$O Structure'
_aflow_proto 'ABC30DE20F2_oP220_29_a_a_30a_a_20a_2a'
_aflow_params 'a,b/a,c/a,x_{1},y_{1},z_{1},x_{2},y_{2},z_{2},x_{3},y_{3
    ↪ },z_{3},x_{4},y_{4},z_{4},x_{5},y_{5},z_{5},x_{6},y_{6},z_{6},
    ↪ x_{7},y_{7},z_{7},x_{8},y_{8},z_{8},x_{9},y_{9},z_{9},x_{10},y_
    ↪ {10},z_{10},x_{11},y_{11},z_{11},x_{12},y_{12},z_{12},x_{13},y_
    ↪ {13},z_{13},x_{14},y_{14},z_{14},x_{15},y_{15},z_{15},x_{16},y_
    ↪ {16},z_{16},x_{17},y_{17},z_{17},x_{18},y_{18},z_{18},x_{19},y_
    ↪ {19},z_{19},x_{20},y_{20},z_{20},x_{21},y_{21},z_{21},x_{22},y_
    ↪ {22},z_{22},x_{23},y_{23},z_{23},x_{24},y_{24},z_{24},x_{25},y_
    ↪ {25},z_{25},x_{26},y_{26},z_{26},x_{27},y_{27},z_{27},x_{28},y_
    ↪ {28},z_{28},x_{29},y_{29},z_{29},x_{30},y_{30},z_{30},x_{31},y_
    ↪ {31},z_{31},x_{32},y_{32},z_{32},x_{33},y_{33},z_{33},x_{34},y_
    ↪ {34},z_{34},x_{35},y_{35},z_{35},x_{36},y_{36},z_{36},x_{37},y_
    ↪ {37},z_{37},x_{38},y_{38},z_{38},x_{39},y_{39},z_{39},x_{40},y_
    ↪ {40},z_{40},x_{41},y_{41},z_{41},x_{42},y_{42},z_{42},x_{43},y_
    ↪ {43},z_{43},x_{44},y_{44},z_{44},x_{45},y_{45},z_{45},x_{46},y_
    ↪ {46},z_{46},x_{47},y_{47},z_{47},x_{48},y_{48},z_{48},x_{49},y_
    ↪ {49},z_{49},x_{50},y_{50},z_{50},x_{51},y_{51},z_{51},x_{52},y_
    ↪ {52},z_{52},x_{53},y_{53},z_{53},x_{54},y_{54},z_{54},x_{55},y_
    ↪ {55},z_{55}'
_aflow_params_values '12.57,0.980906921241,0.984884645982,0.006,0.259,
    ↪ 0.256,0.047,0.77,0.28,0.21,0.237,0.305,0.199,0.277,0.175,0.196,
    ↪ 0.722,0.693,0.193,0.774,0.825,0.076,0.057,0.235,0.048,0.953,
    ↪ 0.781,0.07,0.455,0.22,0.059,0.542,0.768,0.026,0.18,0.45,0.024,
    ↪ 0.698,0.96,0.025,0.192,0.048,0.021,0.68,0.547,0.006,0.012,0.502
    ↪ ,0.136,0.06,0.507,0.247,0.385,0.557,0.243,0.264,0.012,0.225,
    ↪ 0.488,0.752,0.183,0.48,0.885,0.004,0.503,0.502,0.139,0.552,0.49
    ↪ ,0.249,0.757,0.49,0.244,0.893,0.445,0.249,0.01,0.743,0.202,
    ↪ 0.002,0.108,0.12,0.784,0.34,0.033,0.84,0.245,0.083,0.71,0.225,
    ↪ 0.119,0.275,0.665,0.017,0.33,0.755,0.08,0.205,0.775,0.045,0.268
    ↪ ,0.711,0.136,0.04,0.811,0.24,0.022,0.978,0.064,0.096,0.976,
    ↪ 0.199,0.202,0.904,0.169,0.528,0.19,0.225,0.51,0.011,0.061,0.61,
    ↪ 0.046,0.231,0.684,0.097,0.153,0.257,0.236,0.145,0.748,0.759,
    ↪ 0.012,0.105,0.258,0.003,0.405,0.247,0.007,0.242,0.397,0.0,0.256
    ↪ ,0.098,0.228,0.302,0.591,0.154,0.456,0.801,0.235,0.808,0.425,
    ↪ 0.174,0.988,0.182,0.057,0.086,0.521,0.053,0.577,0.472,0.163,
    ↪ 0.088,0.915,0.171,0.581,0.085'
_aflow_Strukturbericht 'None'
_aflow_Pearson 'oP220'

_symmetry_space_group_name_H-M "P c a 21"
_symmetry_Int_Tables_number 29

_cell_length_a      12.57000
_cell_length_b      12.33000
_cell_length_c      12.38000
_cell_angle_alpha   90.00000
_cell_angle_beta    90.00000
_cell_angle_gamma   90.00000

loop_
_space_group_symop_id
_space_group_symop_operation_xyz
1 x,y,z
2 -x,-y,z+1/2
3 -x+1/2,y,z+1/2
4 x+1/2,-y,z

loop_
_atom_site_label
_atom_site_type_symbol
_atom_site_symmetry_multiplicity
_atom_site_Wyckoff_label
_atom_site_fract_x
_atom_site_fract_y
_atom_site_fract_z
_atom_site_occupancy
Al1 Al 4 a 0.00600 0.25900 0.25600 1.00000
C1 C 4 a 0.04700 0.77000 0.28000 1.00000
H1 H 4 a 0.21000 0.23700 0.30500 1.00000
H2 H 4 a 0.19900 0.27700 0.17500 1.00000
H3 H 4 a 0.19600 0.72200 0.69300 1.00000
H4 H 4 a 0.19300 0.77400 0.82500 1.00000
H5 H 4 a 0.07600 0.05700 0.23500 1.00000
H6 H 4 a 0.04800 0.95300 0.78100 1.00000
H7 H 4 a 0.07000 0.45500 0.22000 1.00000
H8 H 4 a 0.05900 0.54200 0.76800 1.00000
H9 H 4 a 0.02600 0.18000 0.45000 1.00000
H10 H 4 a 0.02400 0.69800 0.96000 1.00000
H11 H 4 a 0.02500 0.19200 0.04800 1.00000
H12 H 4 a 0.02100 0.68000 0.54700 1.00000
H13 H 4 a 0.00600 0.01200 0.50200 1.00000
H14 H 4 a 0.13600 0.06000 0.50700 1.00000
H15 H 4 a 0.24700 0.38500 0.55700 1.00000
H16 H 4 a 0.24300 0.26400 0.01200 1.00000
H17 H 4 a 0.22500 0.48800 0.75200 1.00000
H18 H 4 a 0.18300 0.48000 0.88500 1.00000
H19 H 4 a 0.00400 0.50300 0.50200 1.00000
H20 H 4 a 0.13900 0.55200 0.49000 1.00000
H21 H 4 a 0.24900 0.75700 0.49000 1.00000
H22 H 4 a 0.24400 0.89300 0.44500 1.00000
H23 H 4 a 0.24900 0.01000 0.74300 1.00000
H24 H 4 a 0.20200 0.00200 0.10800 1.00000
H25 H 4 a 0.12000 0.78400 0.34000 1.00000
H26 H 4 a 0.03300 0.84000 0.24500 1.00000
H27 H 4 a 0.08300 0.71000 0.22500 1.00000
H28 H 4 a 0.11900 0.27500 0.66500 1.00000
H29 H 4 a 0.01700 0.33000 0.75500 1.00000
H30 H 4 a 0.08000 0.20500 0.77500 1.00000
N1 N 4 a 0.04500 0.26800 0.71100 1.00000
O1 O 4 a 0.13600 0.04000 0.81100 1.00000
O2 O 4 a 0.24000 0.02200 0.97800 1.00000
O3 O 4 a 0.06400 0.09600 0.97600 1.00000
O4 O 4 a 0.19900 0.20200 0.90400 1.00000
O5 O 4 a 0.16900 0.52800 0.19000 1.00000
O6 O 4 a 0.22500 0.51000 0.01100 1.00000
O7 O 4 a 0.06100 0.61000 0.04600 1.00000
O8 O 4 a 0.23100 0.68400 0.09700 1.00000
O9 O 4 a 0.15300 0.25700 0.23600 1.00000
O10 O 4 a 0.14500 0.74800 0.75900 1.00000
O11 O 4 a 0.01200 0.10500 0.25800 1.00000
O12 O 4 a 0.00300 0.40500 0.24700 1.00000
O13 O 4 a 0.00700 0.24200 0.39700 1.00000
O14 O 4 a 0.00000 0.25600 0.09800 1.00000
O15 O 4 a 0.22800 0.30200 0.59100 1.00000
O16 O 4 a 0.15400 0.45600 0.80100 1.00000
O17 O 4 a 0.23500 0.80800 0.42500 1.00000
O18 O 4 a 0.17400 0.98800 0.18200 1.00000
O19 O 4 a 0.05700 0.08600 0.52100 1.00000
O20 O 4 a 0.05300 0.57700 0.47200 1.00000
S1 S 4 a 0.16300 0.08800 0.91500 1.00000
S2 S 4 a 0.17100 0.58100 0.08500 1.00000
```

Low-Temperature $(NH_3CH_3)Al(SO_4)_2 \cdot 12H_2O$: ABC30DE20F2_oP220_29_a_a_30a_a_20a_2a - POSCAR

```
ABC30DE20F2_oP220_29_a_a_30a_a_20a_2a & a,b/a,c/a,x1,y1,z1,x2,y2,z2,x3,
    ↪ y3,z3,x4,y4,z4,x5,y5,z5,x6,y6,z6,x7,y7,z7,x8,y8,z8,x9,y9,z9,x10
    ↪ ,y10,z10,x11,y11,z11,x12,y12,z12,x13,y13,z13,x14,y14,z14,x15,
    ↪ y15,z15,x16,y16,z16,x17,y17,z17,x18,y18,z18,x19,y19,z19,x20,y20
    ↪ ,z20,x21,y21,z21,x22,y22,z22,x23,y23,z23,x24,y24,z24,x25,y25,
    ↪ z25,x26,y26,z26,x27,y27,z27,x28,y28,z28,x29,y29,z29,x30,y30,z30
    ↪ ,x31,y31,z31,x32,y32,z32,x33,y33,z33,x34,y34,z34,x35,y35,z35,
    ↪ x36,y36,z36,x37,y37,z37,x38,y38,z38,x39,y39,z39,x40,y40,z40,x41
    ↪ ,y41,z41,x42,y42,z42,x43,y43,z43,x44,y44,z44,x45,y45,z45,x46,
    ↪ y46,z46,x47,y47,z47,x48,y48,z48,x49,y49,z49,x50,y50,z50,x51,y51
    ↪ ,z51,x52,y52,z52,x53,y53,z53,x54,y54,z54,x55,y55,z55 --params=
    ↪ 12.57,0.980906921241,0.984884645982,0.006,0.259,0.256,0.047,
    ↪ 0.77,0.28,0.21,0.237,0.305,0.199,0.277,0.175,0.196,0.722,0.693,
    ↪ 0.193,0.774,0.825,0.076,0.057,0.235,0.048,0.953,0.781,0.07,
    ↪ 0.455,0.22,0.059,0.542,0.768,0.026,0.18,0.45,0.024,0.698,0.96,
    ↪ 0.025,0.192,0.048,0.021,0.68,0.547,0.006,0.012,0.502,0.136,0.06
    ↪ ,0.507,0.247,0.385,0.557,0.243,0.264,0.012,0.225,0.488,0.752,
    ↪ 0.183,0.48,0.885,0.004,0.503,0.502,0.139,0.552,0.49,0.249,0.757
    ↪ ,0.49,0.244,0.893,0.445,0.249,0.01,0.743,0.202,0.002,0.108,0.12
    ↪ ,0.784,0.34,0.033,0.84,0.245,0.083,0.71,0.225,0.119,0.275,0.665
    ↪ ,0.017,0.33,0.755,0.08,0.205,0.775,0.045,0.268,0.711,0.136,0.04
    ↪ ,0.811,0.24,0.022,0.978,0.064,0.096,0.976,0.199,0.202,0.904,
    ↪ 0.169,0.528,0.19,0.225,0.51,0.011,0.061,0.61,0.046,0.231,0.684,
    ↪ 0.097,0.153,0.257,0.236,0.145,0.748,0.759,0.012,0.105,0.258,
    ↪ 0.003,0.405,0.247,0.007,0.242,0.397,0.0,0.256,0.098,0.228,0.302
    ↪ ,0.591,0.154,0.456,0.801,0.235,0.808,0.425,0.174,0.988,0.182,
    ↪ 0.057,0.086,0.521,0.053,0.577,0.472,0.163,0.088,0.915,0.171,
    ↪ 0.581,0.085 & Pca2_{1} C_{2v}^{5} #29 (a^55) & oP220 & None &
    ↪ AlCH30NO20S2 & AlCH30NO20S2 & R. O. W. Fletcher and H. Steeple,
    ↪ Acta Cryst. 17, 290-294 (1964)
   1.00000000000000
  12.57000000000000   0.00000000000000   0.00000000000000
   0.00000000000000  12.33000000000000   0.00000000000000
   0.00000000000000   0.00000000000000  12.38000000000000
    Al     C     H     N     O     S
     4     4   120     4    80     8
Direct
   0.00600000000000   0.25900000000000   0.25600000000000   Al   (4a)
  -0.00600000000000  -0.25900000000000   0.75600000000000   Al   (4a)
   0.50600000000000  -0.25900000000000   0.25600000000000   Al   (4a)
   0.49400000000000   0.25900000000000   0.75600000000000   Al   (4a)
   0.04700000000000   0.77000000000000   0.28000000000000   C    (4a)
  -0.04700000000000  -0.77000000000000   0.78000000000000   C    (4a)
   0.54700000000000  -0.77000000000000   0.28000000000000   C    (4a)
   0.45300000000000   0.77000000000000   0.78000000000000   C    (4a)
   0.21000000000000   0.23700000000000   0.30500000000000   H    (4a)
  -0.21000000000000  -0.23700000000000   0.80500000000000   H    (4a)
   0.71000000000000  -0.23700000000000   0.30500000000000   H    (4a)
   0.29000000000000   0.23700000000000   0.80500000000000   H    (4a)
   0.19900000000000   0.27700000000000   0.17500000000000   H    (4a)
  -0.19900000000000  -0.27700000000000   0.67500000000000   H    (4a)
   0.69900000000000  -0.27700000000000   0.17500000000000   H    (4a)
   0.30100000000000   0.27700000000000   0.67500000000000   H    (4a)
   0.19600000000000   0.72200000000000   0.69300000000000   H    (4a)
  -0.19600000000000  -0.72200000000000   1.19300000000000   H    (4a)
   0.69600000000000  -0.72200000000000   0.69300000000000   H    (4a)
   0.30400000000000   0.72200000000000   1.19300000000000   H    (4a)
   0.19300000000000   0.77400000000000   0.82500000000000   H    (4a)
  -0.19300000000000  -0.77400000000000   1.32500000000000   H    (4a)
   0.69300000000000  -0.77400000000000   0.82500000000000   H    (4a)
   0.30700000000000   0.77400000000000   1.32500000000000   H    (4a)
   0.07600000000000   0.05700000000000   0.23500000000000   H    (4a)
  -0.07600000000000  -0.05700000000000   0.73500000000000   H    (4a)
   0.57600000000000  -0.05700000000000   0.23500000000000   H    (4a)
   0.42400000000000   0.05700000000000   0.73500000000000   H    (4a)
   0.04800000000000   0.95300000000000   0.78100000000000   H    (4a)
  -0.04800000000000  -0.95300000000000   1.28100000000000   H    (4a)
```

| | | | | |
|---|---|---|---|---|
| 0.54800000000000 | −0.95300000000000 | 0.78100000000000 | H | (4a) |
| 0.45200000000000 | 0.95300000000000 | 1.28100000000000 | H | (4a) |
| 0.07000000000000 | 0.45500000000000 | 0.22000000000000 | H | (4a) |
| −0.07000000000000 | −0.45500000000000 | 0.72000000000000 | H | (4a) |
| 0.57000000000000 | −0.45500000000000 | 0.22000000000000 | H | (4a) |
| 0.43000000000000 | 0.45500000000000 | 0.72000000000000 | H | (4a) |
| 0.05900000000000 | 0.54200000000000 | 0.76800000000000 | H | (4a) |
| −0.05900000000000 | −0.54200000000000 | 1.26800000000000 | H | (4a) |
| 0.55900000000000 | −0.54200000000000 | 0.76800000000000 | H | (4a) |
| 0.44100000000000 | 0.54200000000000 | 1.26800000000000 | H | (4a) |
| 0.02600000000000 | 0.18000000000000 | 0.45000000000000 | H | (4a) |
| −0.02600000000000 | −0.18000000000000 | 0.95000000000000 | H | (4a) |
| 0.52600000000000 | −0.18000000000000 | 0.45000000000000 | H | (4a) |
| 0.47400000000000 | 0.18000000000000 | 0.95000000000000 | H | (4a) |
| 0.02400000000000 | 0.69800000000000 | 0.96000000000000 | H | (4a) |
| −0.02400000000000 | −0.69800000000000 | 1.46000000000000 | H | (4a) |
| 0.52400000000000 | −0.69800000000000 | 0.96000000000000 | H | (4a) |
| 0.47600000000000 | 0.69800000000000 | 1.46000000000000 | H | (4a) |
| 0.02500000000000 | 0.19200000000000 | 0.04800000000000 | H | (4a) |
| −0.02500000000000 | −0.19200000000000 | 0.54800000000000 | H | (4a) |
| 0.52500000000000 | −0.19200000000000 | 0.04800000000000 | H | (4a) |
| 0.47500000000000 | 0.19200000000000 | 0.54800000000000 | H | (4a) |
| 0.02100000000000 | 0.68000000000000 | 0.54700000000000 | H | (4a) |
| −0.02100000000000 | −0.68000000000000 | 1.04700000000000 | H | (4a) |
| 0.52100000000000 | −0.68000000000000 | 0.54700000000000 | H | (4a) |
| 0.47900000000000 | 0.68000000000000 | 1.04700000000000 | H | (4a) |
| 0.00600000000000 | 0.01200000000000 | 0.50200000000000 | H | (4a) |
| −0.00600000000000 | −0.01200000000000 | 1.00200000000000 | H | (4a) |
| 0.50600000000000 | −0.01200000000000 | 0.50200000000000 | H | (4a) |
| 0.49400000000000 | 0.01200000000000 | 1.00200000000000 | H | (4a) |
| 0.13600000000000 | 0.06000000000000 | 0.50700000000000 | H | (4a) |
| −0.13600000000000 | −0.06000000000000 | 1.00700000000000 | H | (4a) |
| 0.63600000000000 | −0.06000000000000 | 0.50700000000000 | H | (4a) |
| 0.36400000000000 | 0.06000000000000 | 1.00700000000000 | H | (4a) |
| 0.24700000000000 | 0.38500000000000 | 0.55700000000000 | H | (4a) |
| −0.24700000000000 | −0.38500000000000 | 1.05700000000000 | H | (4a) |
| 0.74700000000000 | −0.38500000000000 | 0.55700000000000 | H | (4a) |
| 0.25300000000000 | 0.38500000000000 | 1.05700000000000 | H | (4a) |
| 0.24300000000000 | 0.26400000000000 | 0.01200000000000 | H | (4a) |
| −0.24300000000000 | −0.26400000000000 | 0.51200000000000 | H | (4a) |
| 0.74300000000000 | −0.26400000000000 | 0.01200000000000 | H | (4a) |
| 0.25700000000000 | 0.26400000000000 | 0.51200000000000 | H | (4a) |
| 0.22500000000000 | 0.48800000000000 | 0.75200000000000 | H | (4a) |
| −0.22500000000000 | −0.48800000000000 | 1.25200000000000 | H | (4a) |
| 0.72500000000000 | −0.48800000000000 | 0.75200000000000 | H | (4a) |
| 0.27500000000000 | 0.48800000000000 | 1.25200000000000 | H | (4a) |
| 0.18300000000000 | 0.48000000000000 | 0.88500000000000 | H | (4a) |
| −0.18300000000000 | −0.48000000000000 | 1.38500000000000 | H | (4a) |
| 0.68300000000000 | −0.48000000000000 | 0.88500000000000 | H | (4a) |
| 0.31700000000000 | 0.48000000000000 | 1.38500000000000 | H | (4a) |
| 0.00400000000000 | 0.50300000000000 | 0.50200000000000 | H | (4a) |
| −0.00400000000000 | −0.50300000000000 | 1.00200000000000 | H | (4a) |
| 0.50400000000000 | −0.50300000000000 | 0.50200000000000 | H | (4a) |
| 0.49600000000000 | 0.50300000000000 | 1.00200000000000 | H | (4a) |
| 0.13900000000000 | 0.55200000000000 | 0.49000000000000 | H | (4a) |
| −0.13900000000000 | −0.55200000000000 | 0.99000000000000 | H | (4a) |
| 0.63900000000000 | −0.55200000000000 | 0.49000000000000 | H | (4a) |
| 0.36100000000000 | 0.55200000000000 | 0.99000000000000 | H | (4a) |
| 0.24900000000000 | 0.75700000000000 | 0.49000000000000 | H | (4a) |
| −0.24900000000000 | −0.75700000000000 | 0.99000000000000 | H | (4a) |
| 0.74900000000000 | −0.75700000000000 | 0.49000000000000 | H | (4a) |
| 0.25100000000000 | 0.75700000000000 | 0.99000000000000 | H | (4a) |
| 0.24400000000000 | 0.89300000000000 | 0.44500000000000 | H | (4a) |
| −0.24400000000000 | −0.89300000000000 | 0.94500000000000 | H | (4a) |
| 0.74400000000000 | −0.89300000000000 | 0.44500000000000 | H | (4a) |
| 0.25600000000000 | 0.89300000000000 | 0.94500000000000 | H | (4a) |
| 0.24900000000000 | 0.01000000000000 | 0.74300000000000 | H | (4a) |
| −0.24900000000000 | −0.01000000000000 | 1.24300000000000 | H | (4a) |
| 0.74900000000000 | −0.01000000000000 | 0.74300000000000 | H | (4a) |
| 0.25100000000000 | 0.01000000000000 | 1.24300000000000 | H | (4a) |
| 0.20200000000000 | 0.00200000000000 | 0.10800000000000 | H | (4a) |
| −0.20200000000000 | −0.00200000000000 | 0.60800000000000 | H | (4a) |
| 0.70200000000000 | −0.00200000000000 | 0.10800000000000 | H | (4a) |
| 0.29800000000000 | 0.00200000000000 | 0.60800000000000 | H | (4a) |
| 0.12000000000000 | 0.78400000000000 | 0.34000000000000 | H | (4a) |
| −0.12000000000000 | −0.78400000000000 | 0.84000000000000 | H | (4a) |
| 0.62000000000000 | −0.78400000000000 | 0.34000000000000 | H | (4a) |
| 0.38000000000000 | 0.78400000000000 | 0.84000000000000 | H | (4a) |
| 0.03300000000000 | 0.84000000000000 | 0.24500000000000 | H | (4a) |
| −0.03300000000000 | −0.84000000000000 | 0.74500000000000 | H | (4a) |
| 0.53300000000000 | −0.84000000000000 | 0.24500000000000 | H | (4a) |
| 0.46700000000000 | 0.84000000000000 | 0.74500000000000 | H | (4a) |
| 0.08300000000000 | 0.71000000000000 | 0.22500000000000 | H | (4a) |
| −0.08300000000000 | −0.71000000000000 | 0.72500000000000 | H | (4a) |
| 0.58300000000000 | −0.71000000000000 | 0.22500000000000 | H | (4a) |
| 0.41700000000000 | 0.71000000000000 | 0.72500000000000 | H | (4a) |
| 0.11900000000000 | 0.27500000000000 | 0.66500000000000 | H | (4a) |
| −0.11900000000000 | −0.27500000000000 | 1.16500000000000 | H | (4a) |
| 0.61900000000000 | −0.27500000000000 | 0.66500000000000 | H | (4a) |
| 0.38100000000000 | 0.27500000000000 | 1.16500000000000 | H | (4a) |
| 0.01700000000000 | 0.33000000000000 | 0.75500000000000 | H | (4a) |
| −0.01700000000000 | −0.33000000000000 | 1.25500000000000 | H | (4a) |
| 0.51700000000000 | −0.33000000000000 | 0.75500000000000 | H | (4a) |
| 0.48300000000000 | 0.33000000000000 | 1.25500000000000 | H | (4a) |
| 0.08000000000000 | 0.20500000000000 | 0.77500000000000 | H | (4a) |
| −0.08000000000000 | −0.20500000000000 | 1.27500000000000 | H | (4a) |
| 0.58000000000000 | −0.20500000000000 | 0.77500000000000 | H | (4a) |
| 0.42000000000000 | 0.20500000000000 | 1.27500000000000 | H | (4a) |
| 0.04500000000000 | 0.26800000000000 | 0.71100000000000 | N | (4a) |
| −0.04500000000000 | −0.26800000000000 | 1.21100000000000 | N | (4a) |
| 0.54500000000000 | −0.26800000000000 | 0.71100000000000 | N | (4a) |
| 0.45500000000000 | 0.26800000000000 | 1.21100000000000 | N | (4a) |
| 0.13600000000000 | 0.04000000000000 | 0.81100000000000 | O | (4a) |
| −0.13600000000000 | −0.04000000000000 | 1.31100000000000 | O | (4a) |
| 0.63600000000000 | −0.04000000000000 | 0.81100000000000 | O | (4a) |
| 0.36400000000000 | 0.04000000000000 | 1.31100000000000 | O | (4a) |
| 0.24000000000000 | 0.02200000000000 | 0.97800000000000 | O | (4a) |
| −0.24000000000000 | −0.02200000000000 | 1.47800000000000 | O | (4a) |
| 0.74000000000000 | −0.02200000000000 | 0.97800000000000 | O | (4a) |
| 0.26000000000000 | 0.02200000000000 | 1.47800000000000 | O | (4a) |
| 0.06400000000000 | 0.09600000000000 | 0.97600000000000 | O | (4a) |
| −0.06400000000000 | −0.09600000000000 | 1.47600000000000 | O | (4a) |
| 0.56400000000000 | −0.09600000000000 | 0.97600000000000 | O | (4a) |
| 0.43600000000000 | 0.09600000000000 | 1.47600000000000 | O | (4a) |
| 0.19900000000000 | 0.20200000000000 | 0.90400000000000 | O | (4a) |
| −0.19900000000000 | −0.20200000000000 | 1.40400000000000 | O | (4a) |
| 0.69900000000000 | −0.20200000000000 | 0.90400000000000 | O | (4a) |
| 0.30100000000000 | 0.20200000000000 | 1.40400000000000 | O | (4a) |
| 0.16900000000000 | 0.52800000000000 | 0.19000000000000 | O | (4a) |
| −0.16900000000000 | −0.52800000000000 | 0.69000000000000 | O | (4a) |
| 0.66900000000000 | −0.52800000000000 | 0.19000000000000 | O | (4a) |
| 0.33100000000000 | 0.52800000000000 | 0.69000000000000 | O | (4a) |
| 0.22500000000000 | 0.51000000000000 | 0.01100000000000 | O | (4a) |
| −0.22500000000000 | −0.51000000000000 | 0.51100000000000 | O | (4a) |
| 0.72500000000000 | −0.51000000000000 | 0.01100000000000 | O | (4a) |
| 0.27500000000000 | 0.51000000000000 | 0.51100000000000 | O | (4a) |
| 0.06100000000000 | 0.61000000000000 | 0.04600000000000 | O | (4a) |
| −0.06100000000000 | −0.61000000000000 | 0.54600000000000 | O | (4a) |
| 0.56100000000000 | −0.61000000000000 | 0.04600000000000 | O | (4a) |
| 0.43900000000000 | 0.61000000000000 | 0.54600000000000 | O | (4a) |
| 0.23100000000000 | 0.68400000000000 | 0.09700000000000 | O | (4a) |
| −0.23100000000000 | −0.68400000000000 | 0.59700000000000 | O | (4a) |
| 0.73100000000000 | −0.68400000000000 | 0.09700000000000 | O | (4a) |
| 0.26900000000000 | 0.68400000000000 | 0.59700000000000 | O | (4a) |
| 0.15300000000000 | 0.25700000000000 | 0.23600000000000 | O | (4a) |
| −0.15300000000000 | −0.25700000000000 | 0.73600000000000 | O | (4a) |
| 0.65300000000000 | −0.25700000000000 | 0.23600000000000 | O | (4a) |
| 0.34700000000000 | 0.25700000000000 | 0.73600000000000 | O | (4a) |
| 0.14500000000000 | 0.74800000000000 | 0.75900000000000 | O | (4a) |
| −0.14500000000000 | −0.74800000000000 | 1.25900000000000 | O | (4a) |
| 0.64500000000000 | −0.74800000000000 | 0.75900000000000 | O | (4a) |
| 0.35500000000000 | 0.74800000000000 | 1.25900000000000 | O | (4a) |
| 0.01200000000000 | 0.10500000000000 | 0.25800000000000 | O | (4a) |
| −0.01200000000000 | −0.10500000000000 | 0.75800000000000 | O | (4a) |
| 0.51200000000000 | −0.10500000000000 | 0.25800000000000 | O | (4a) |
| 0.48800000000000 | 0.10500000000000 | 0.75800000000000 | O | (4a) |
| 0.00300000000000 | 0.40500000000000 | 0.24700000000000 | O | (4a) |
| −0.00300000000000 | −0.40500000000000 | 0.74700000000000 | O | (4a) |
| 0.50300000000000 | −0.40500000000000 | 0.24700000000000 | O | (4a) |
| 0.49700000000000 | 0.40500000000000 | 0.74700000000000 | O | (4a) |
| 0.00700000000000 | 0.24200000000000 | 0.39700000000000 | O | (4a) |
| −0.00700000000000 | −0.24200000000000 | 0.89700000000000 | O | (4a) |
| 0.50700000000000 | −0.24200000000000 | 0.39700000000000 | O | (4a) |
| 0.49300000000000 | 0.24200000000000 | 0.89700000000000 | O | (4a) |
| 0.00000000000000 | 0.25600000000000 | 0.09800000000000 | O | (4a) |
| 0.00000000000000 | −0.25600000000000 | 0.59800000000000 | O | (4a) |
| 0.50000000000000 | −0.25600000000000 | 0.09800000000000 | O | (4a) |
| 0.50000000000000 | 0.25600000000000 | 0.59800000000000 | O | (4a) |
| 0.22800000000000 | 0.30200000000000 | 0.59100000000000 | O | (4a) |
| −0.22800000000000 | −0.30200000000000 | 1.09100000000000 | O | (4a) |
| 0.72800000000000 | −0.30200000000000 | 0.59100000000000 | O | (4a) |
| 0.27200000000000 | 0.30200000000000 | 1.09100000000000 | O | (4a) |
| 0.15400000000000 | 0.45600000000000 | 0.80100000000000 | O | (4a) |
| −0.15400000000000 | −0.45600000000000 | 1.30100000000000 | O | (4a) |
| 0.65400000000000 | −0.45600000000000 | 0.80100000000000 | O | (4a) |
| 0.34600000000000 | 0.45600000000000 | 1.30100000000000 | O | (4a) |
| 0.23500000000000 | 0.80800000000000 | 0.42500000000000 | O | (4a) |
| −0.23500000000000 | −0.80800000000000 | 0.92500000000000 | O | (4a) |
| 0.73500000000000 | −0.80800000000000 | 0.42500000000000 | O | (4a) |
| 0.26500000000000 | 0.80800000000000 | 0.92500000000000 | O | (4a) |
| 0.17400000000000 | 0.98800000000000 | 0.18200000000000 | O | (4a) |
| −0.17400000000000 | −0.98800000000000 | 0.68200000000000 | O | (4a) |
| 0.67400000000000 | −0.98800000000000 | 0.18200000000000 | O | (4a) |
| 0.32600000000000 | 0.98800000000000 | 0.68200000000000 | O | (4a) |
| 0.05700000000000 | 0.08600000000000 | 0.52100000000000 | O | (4a) |
| −0.05700000000000 | −0.08600000000000 | 1.02100000000000 | O | (4a) |
| 0.55700000000000 | −0.08600000000000 | 0.52100000000000 | O | (4a) |
| 0.44300000000000 | 0.08600000000000 | 1.02100000000000 | O | (4a) |
| 0.05300000000000 | 0.57700000000000 | 0.47200000000000 | O | (4a) |
| −0.05300000000000 | −0.57700000000000 | 0.97200000000000 | O | (4a) |
| 0.55300000000000 | −0.57700000000000 | 0.47200000000000 | O | (4a) |
| 0.44700000000000 | 0.57700000000000 | 0.97200000000000 | O | (4a) |
| 0.16300000000000 | 0.08800000000000 | 0.91500000000000 | S | (4a) |
| −0.16300000000000 | −0.08800000000000 | 1.41500000000000 | S | (4a) |
| 0.66300000000000 | −0.08800000000000 | 0.91500000000000 | S | (4a) |
| 0.33700000000000 | 0.08800000000000 | 1.41500000000000 | S | (4a) |
| 0.17100000000000 | 0.58100000000000 | 0.08500000000000 | S | (4a) |
| −0.17100000000000 | −0.58100000000000 | 0.58500000000000 | S | (4a) |
| 0.67100000000000 | −0.58100000000000 | 0.08500000000000 | S | (4a) |
| 0.32900000000000 | 0.58100000000000 | 0.58500000000000 | S | (4a) |

Orthorhombic $Co_4Al_{13}$: A13B4_oP102_31_17a11b_8a2b - CIF

```
# CIF file
data_findsym-output
_audit_creation_method FINDSYM

_chemical_name_mineral 'Al13Co4'
_chemical_formula_sum 'Al13 Co4'

loop_
_publ_author_name
 'J. Grin'
 'U. Burkhard'
 'M. Ellner'
 'K. Peters'
_journal_name_full_name
;
 Journal of Alloys and Compounds
;
```

```
_journal_volume 206
_journal_year 1994
_journal_page_first 243
_journal_page_last 247
_publ_Section_title
;
 Crystal structure of orthorhombic Co$_{4}$Al$_{13}$
;

# Found in Structure investigation of the (100) surface of the
    ↪ orthorhombic Al$_{13}$Co$_{4}$ crystal, 2009

_aflow_title 'Orthorhombic Co$_{4}$Al$_{13}$ Structure'
_aflow_proto 'A13B4_oP102_31_17a11b_8a2b'
_aflow_params 'a,b/a,c/a,y_{1},z_{1},y_{2},z_{2},y_{3},z_{3},y_{4},z_{4
    ↪ },y_{5},z_{5},y_{6},z_{6},y_{7},z_{7},y_{8},z_{8},y_{9},z_{9},
    ↪ y_{10},z_{10},y_{11},z_{11},y_{12},z_{12},y_{13},z_{13},y_{14},
    ↪ z_{14},y_{15},z_{15},y_{16},z_{16},y_{17},z_{17},y_{18},z_{18},
    ↪ y_{19},z_{19},y_{20},z_{20},y_{21},z_{21},y_{22},z_{22},y_{23},
    ↪ z_{23},y_{24},z_{24},y_{25},z_{25},x_{26},y_{26},z_{26},x_{27},
    ↪ y_{27},z_{27},x_{28},y_{28},z_{28},x_{29},y_{29},z_{29},x_{30},
    ↪ y_{30},z_{30},x_{31},y_{31},z_{31},x_{32},y_{32},z_{32},x_{33},
    ↪ y_{33},z_{33},x_{34},y_{34},z_{34},x_{35},y_{35},z_{35},x_{36},
    ↪ y_{36},z_{36},x_{37},y_{37},z_{37},x_{38},y_{38},z_{38}'
_aflow_params_values '8.158,1.51287080167,1.77151262564,-0.0055,0.8151,
    ↪ 0.0914,-0.0431,-0.0947,0.6226,-0.0915,0.4431,0.1394,0.6728,
    ↪ 0.8113,0.1564,0.8092,0.836,0.252,0.414,0.407,0.852,0.401,0.1491
    ↪ ,0.6837,0.6749,0.59,0.396,0.5271,0.0054,0.5986,0.2149,0.7111,-
    ↪ 0.0332,0.1047,0.238,0.4179,0.5881,0.8978,0.0,0.0901,0.5114,
    ↪ 0.1977,0.825,0.772,0.3167,0.5986,0.8248,0.4119,0.3137,0.7309,
    ↪ 0.518,0.2877,0.0076,0.2135,0.2138,0.0996,0.2138,-0.098,0.2831,
    ↪ 0.2401,0.0829,0.4088,0.2534,0.255,-0.0828,0.2251,-0.019,0.0935,
    ↪ 0.2214,0.2229,0.55,0.2259,0.5885,0.5422,0.2244,0.2946,0.7352,
    ↪ 0.2291,0.422,0.4252,0.2133,0.5174,0.7338,0.2319,0.2763,0.2849,
    ↪ 0.2257,-0.0901,0.7346,0.2181,0.5969,0.0984'
_aflow_Strukturbericht 'None'
_aflow_Pearson 'oP102'

_symmetry_space_group_name_H-M "P m n 21"
_symmetry_Int_Tables_number 31

_cell_length_a    8.15800
_cell_length_b    12.34200
_cell_length_c    14.45200
_cell_angle_alpha 90.00000
_cell_angle_beta  90.00000
_cell_angle_gamma 90.00000

loop_
_space_group_symop_id
_space_group_symop_operation_xyz
1 x,y,z
2 -x+1/2,-y,z+1/2
3 -x,y,z
4 x+1/2,-y,z+1/2

loop_
_atom_site_label
_atom_site_type_symbol
_atom_site_symmetry_multiplicity
_atom_site_Wyckoff_label
_atom_site_fract_x
_atom_site_fract_y
_atom_site_fract_z
_atom_site_occupancy
Al1 Al 2 a 0.00000 -0.00550 0.81510 1.00000
Al2 Al 2 a 0.00000 0.09140 -0.04310 1.00000
Al3 Al 2 a 0.00000 -0.09470 0.62260 1.00000
Al4 Al 2 a 0.00000 -0.09150 0.44310 1.00000
Al5 Al 2 a 0.00000 0.13940 0.67280 1.00000
Al6 Al 2 a 0.00000 0.81130 0.15640 1.00000
Al7 Al 2 a 0.00000 0.80920 0.83600 1.00000
Al8 Al 2 a 0.00000 0.25200 0.41400 1.00000
Al9 Al 2 a 0.00000 0.40700 0.85200 1.00000
Al10 Al 2 a 0.00000 0.40100 0.14910 1.00000
Al11 Al 2 a 0.00000 0.68370 0.67490 1.00000
Al12 Al 2 a 0.00000 0.59000 0.39600 1.00000
Al13 Al 2 a 0.00000 0.52710 0.00540 1.00000
Al14 Al 2 a 0.00000 0.59860 0.21490 1.00000
Al15 Al 2 a 0.00000 0.71110 -0.03320 1.00000
Al16 Al 2 a 0.00000 0.10470 0.23800 1.00000
Al17 Al 2 a 0.00000 0.41790 0.58810 1.00000
Co1 Co 2 a 0.00000 0.89780 0.00000 1.00000
Co2 Co 2 a 0.00000 0.09010 0.51140 1.00000
Co3 Co 2 a 0.00000 0.19770 0.82500 1.00000
Co4 Co 2 a 0.00000 0.77200 0.31670 1.00000
Co5 Co 2 a 0.00000 0.59860 0.82480 1.00000
Co6 Co 2 a 0.00000 0.41190 0.31370 1.00000
Co7 Co 2 a 0.00000 0.73090 0.51800 1.00000
Co8 Co 2 a 0.00000 0.28770 0.00760 1.00000
Al18 Al 4 b 0.21350 0.21380 0.09960 1.00000
Al19 Al 4 b 0.21380 -0.09800 0.28310 1.00000
Al20 Al 4 b 0.24010 0.08290 0.40880 1.00000
Al21 Al 4 b 0.25340 0.25500 -0.08280 1.00000
Al22 Al 4 b 0.22510 -0.01900 0.09350 1.00000
Al23 Al 4 b 0.22140 0.22290 0.55000 1.00000
Al24 Al 4 b 0.22590 0.58850 0.54220 1.00000
Al25 Al 4 b 0.22440 0.29460 0.73520 1.00000
Al26 Al 4 b 0.22910 0.42200 0.42520 1.00000
Al27 Al 4 b 0.21330 0.51740 0.73380 1.00000
Al28 Al 4 b 0.23190 0.27630 0.28490 1.00000
Co9 Co 4 b 0.22570 -0.09010 0.73460 1.00000
Co10 Co 4 b 0.21810 0.59690 0.09840 1.00000
```

Orthorhombic $Co_4Al_{13}$: A13B4_oP102_31_17a11b_8a2b - POSCAR

```
A13B4_oP102_31_17a11b_8a2b & a,b/a,c/a,y1,z1,y2,z2,y3,z3,y4,z4,y5,z5,y6,
    ↪ z6,y7,z7,y8,z8,y9,z9,y10,z10,y11,z11,y12,z12,y13,z13,y14,z14,
    ↪ y15,z15,y16,z16,y17,z17,y18,z18,y19,z19,y20,z20,y21,z21,y22,z22
    ↪ ,y23,z23,y24,z24,y25,z25,x26,y26,z26,x27,y27,z27,x28,y28,z28,
    ↪ x29,y29,z29,x30,y30,z30,x31,y31,z31,x32,y32,z32,x33,y33,z33,x34
    ↪ ,y34,z34,x35,y35,z35,x36,y36,z36,x37,y37,z37,x38,y38,z38 --
    ↪ params=8.158,1.51287080167,1.77151262564,-0.0055,0.8151,0.0914
    ↪ ,-0.0431,-0.0947,0.6226,-0.0915,0.4431,0.1394,0.6728,0.8113,
    ↪ 0.1564,0.8092,0.836,0.252,0.414,0.407,0.852,0.401,0.1491,0.6837
    ↪ ,0.6749,0.59,0.396,0.5271,0.0054,0.5986,0.2149,0.7111,-0.0332,
    ↪ 0.1047,0.238,0.4179,0.5881,0.8978,0.0,0.0901,0.5114,0.1977,
    ↪ 0.825,0.772,0.3167,0.5986,0.8248,0.4119,0.3137,0.7309,0.518,
    ↪ 0.2877,0.0076,0.2135,0.2138,0.0996,0.2138,-0.098,0.2831,0.2401,
    ↪ 0.0829,0.4088,0.2534,0.255,-0.0828,0.2251,-0.019,0.0935,0.2214,
    ↪ 0.2229,0.55,0.2259,0.5885,0.5422,0.2244,0.2946,0.7352,0.2291,
    ↪ 0.422,0.4252,0.2133,0.5174,0.7338,0.2319,0.2763,0.2849,0.2257,-
    ↪ 0.0901,0.7346,0.2181,0.5969,0.0984 & Pmn2_{1} C_{2v}^{7} #31 (a
    ↪ ^25b^13) & oP102 & None & Al13Co4 & Al13Co4 & J. Grin et al.,
    ↪ J. Alloys Compd. 206, 243-247 (1994)
   1.00000000000000
   8.15800000000000    0.00000000000000    0.00000000000000
   0.00000000000000   12.34200000000000    0.00000000000000
   0.00000000000000    0.00000000000000   14.45200000000000
    Al    Co
    78    24
Direct
   0.00000000000000   -0.00550000000000    0.81510000000000   Al   (2a)
   0.50000000000000    0.00550000000000    1.31510000000000   Al   (2a)
   0.00000000000000    0.09140000000000   -0.04310000000000   Al   (2a)
   0.50000000000000   -0.09140000000000    0.45690000000000   Al   (2a)
   0.00000000000000   -0.09470000000000    0.62260000000000   Al   (2a)
   0.50000000000000    0.09470000000000    1.12260000000000   Al   (2a)
   0.00000000000000   -0.09150000000000    0.44310000000000   Al   (2a)
   0.50000000000000    0.09150000000000    0.94310000000000   Al   (2a)
   0.00000000000000    0.13940000000000    0.67280000000000   Al   (2a)
   0.50000000000000   -0.13940000000000    1.17280000000000   Al   (2a)
   0.00000000000000    0.81130000000000    0.15640000000000   Al   (2a)
   0.50000000000000   -0.81130000000000    0.65640000000000   Al   (2a)
   0.00000000000000    0.80920000000000    0.83600000000000   Al   (2a)
   0.50000000000000   -0.80920000000000    1.33600000000000   Al   (2a)
   0.00000000000000    0.25200000000000    0.41400000000000   Al   (2a)
   0.50000000000000   -0.25200000000000    0.91400000000000   Al   (2a)
   0.00000000000000    0.40700000000000    0.85200000000000   Al   (2a)
   0.50000000000000   -0.40700000000000    1.35200000000000   Al   (2a)
   0.00000000000000    0.40100000000000    0.14910000000000   Al   (2a)
   0.50000000000000   -0.40100000000000    0.64910000000000   Al   (2a)
   0.00000000000000    0.68370000000000    0.67490000000000   Al   (2a)
   0.50000000000000   -0.68370000000000    1.17490000000000   Al   (2a)
   0.00000000000000    0.59000000000000    0.39600000000000   Al   (2a)
   0.50000000000000   -0.59000000000000    0.89600000000000   Al   (2a)
   0.00000000000000    0.52710000000000    0.00540000000000   Al   (2a)
   0.50000000000000   -0.52710000000000    0.50540000000000   Al   (2a)
   0.00000000000000    0.59860000000000    0.21490000000000   Al   (2a)
   0.50000000000000   -0.59860000000000    0.71490000000000   Al   (2a)
   0.00000000000000    0.71110000000000   -0.03320000000000   Al   (2a)
   0.50000000000000   -0.71110000000000    0.46680000000000   Al   (2a)
   0.00000000000000    0.10470000000000    0.23800000000000   Al   (2a)
   0.50000000000000   -0.10470000000000    0.73800000000000   Al   (2a)
   0.00000000000000    0.41790000000000    0.58810000000000   Al   (2a)
   0.50000000000000   -0.41790000000000    1.08810000000000   Al   (2a)
   0.21350000000000    0.21380000000000    0.09960000000000   Al   (4b)
   0.28650000000000   -0.21380000000000    0.59960000000000   Al   (4b)
   0.71350000000000   -0.21380000000000    0.59960000000000   Al   (4b)
  -0.21350000000000    0.21380000000000    0.09960000000000   Al   (4b)
   0.21380000000000   -0.09800000000000    0.28310000000000   Al   (4b)
   0.28620000000000    0.09800000000000    0.78310000000000   Al   (4b)
   0.71380000000000    0.09800000000000    0.78310000000000   Al   (4b)
  -0.21380000000000   -0.09800000000000    0.28310000000000   Al   (4b)
   0.24010000000000    0.08290000000000    0.40880000000000   Al   (4b)
   0.25990000000000   -0.08290000000000    0.90880000000000   Al   (4b)
   0.74010000000000   -0.08290000000000    0.90880000000000   Al   (4b)
  -0.24010000000000    0.08290000000000    0.40880000000000   Al   (4b)
   0.25340000000000    0.25500000000000   -0.08280000000000   Al   (4b)
   0.24660000000000   -0.25500000000000    0.41720000000000   Al   (4b)
   0.75340000000000   -0.25500000000000    0.41720000000000   Al   (4b)
  -0.25340000000000    0.25500000000000   -0.08280000000000   Al   (4b)
   0.22510000000000   -0.01900000000000    0.09350000000000   Al   (4b)
   0.27490000000000    0.01900000000000    0.59350000000000   Al   (4b)
   0.72510000000000    0.01900000000000    0.59350000000000   Al   (4b)
  -0.22510000000000   -0.01900000000000    0.09350000000000   Al   (4b)
   0.22140000000000    0.22290000000000    0.55000000000000   Al   (4b)
   0.27860000000000   -0.22290000000000    1.05000000000000   Al   (4b)
   0.72140000000000   -0.22290000000000    1.05000000000000   Al   (4b)
  -0.22140000000000    0.22290000000000    0.55000000000000   Al   (4b)
   0.22590000000000    0.58850000000000    0.54220000000000   Al   (4b)
   0.27410000000000   -0.58850000000000    1.04220000000000   Al   (4b)
   0.72590000000000   -0.58850000000000    1.04220000000000   Al   (4b)
  -0.22590000000000    0.58850000000000    0.54220000000000   Al   (4b)
   0.22440000000000    0.29460000000000    0.73520000000000   Al   (4b)
   0.27560000000000   -0.29460000000000    1.23520000000000   Al   (4b)
   0.72440000000000   -0.29460000000000    1.23520000000000   Al   (4b)
  -0.22440000000000    0.29460000000000    0.73520000000000   Al   (4b)
   0.22910000000000    0.42200000000000    0.42520000000000   Al   (4b)
   0.27090000000000   -0.42200000000000    0.92520000000000   Al   (4b)
   0.72910000000000   -0.42200000000000    0.92520000000000   Al   (4b)
  -0.22910000000000    0.42200000000000    0.42520000000000   Al   (4b)
   0.21330000000000    0.51740000000000    0.73380000000000   Al   (4b)
   0.28670000000000   -0.51740000000000    1.23380000000000   Al   (4b)
   0.71330000000000   -0.51740000000000    1.23380000000000   Al   (4b)
  -0.21330000000000    0.51740000000000    0.73380000000000   Al   (4b)
   0.23190000000000    0.27630000000000    0.28490000000000   Al   (4b)
   0.26810000000000   -0.27630000000000    0.78490000000000   Al   (4b)
   0.73190000000000   -0.27630000000000    0.78490000000000   Al   (4b)
  -0.23190000000000    0.27630000000000    0.28490000000000   Al   (4b)
   0.00000000000000    0.89780000000000    0.00000000000000   Co   (2a)
```

```
0.50000000000000   -0.89780000000000   0.50000000000000   Co   (2a)
0.00000000000000    0.09010000000000   0.51140000000000   Co   (2a)
0.50000000000000   -0.09010000000000   1.01140000000000   Co   (2a)
0.00000000000000    0.19770000000000   0.82500000000000   Co   (2a)
0.50000000000000   -0.19770000000000   1.32500000000000   Co   (2a)
0.00000000000000    0.77200000000000   0.31670000000000   Co   (2a)
0.50000000000000   -0.77200000000000   0.81670000000000   Co   (2a)
0.00000000000000    0.59860000000000   0.82480000000000   Co   (2a)
0.50000000000000   -0.59860000000000   1.32480000000000   Co   (2a)
0.00000000000000    0.41190000000000   0.31370000000000   Co   (2a)
0.50000000000000   -0.41190000000000   0.81370000000000   Co   (2a)
0.00000000000000    0.73090000000000   0.51800000000000   Co   (2a)
0.50000000000000   -0.73090000000000   1.01800000000000   Co   (2a)
0.00000000000000    0.28770000000000   0.00760000000000   Co   (2a)
0.50000000000000   -0.28770000000000   0.50760000000000   Co   (2a)
0.22570000000000   -0.09010000000000   0.73460000000000   Co   (4b)
0.27430000000000    0.09010000000000   1.23460000000000   Co   (4b)
0.72570000000000    0.09010000000000   1.23460000000000   Co   (4b)
-0.22570000000000  -0.09010000000000   0.73460000000000   Co   (4b)
0.21810000000000    0.59690000000000   0.09840000000000   Co   (4b)
0.28190000000000   -0.59690000000000   0.59840000000000   Co   (4b)
0.71810000000000   -0.59690000000000   0.59840000000000   Co   (4b)
-0.21810000000000   0.59690000000000   0.09840000000000   Co   (4b)
```

Mg(ClO$_4$)$_2$·6H$_2$O ($H4_{11}$): A2B6CD8_oP34_31_2a_2a2b_a_4a2b - CIF

```
# CIF file
data_findsym-output
_audit_creation_method FINDSYM

_chemical_name_mineral 'Cl2(H2O)6MgO8'
_chemical_formula_sum 'Cl2 (H2O)6 Mg O8'

loop_
_publ_author_name
 'C. D. West'
_journal_name_full_name
;
 Zeitschrift f{\"u}r Kristallographie - Crystalline Materials
;
_journal_volume 91
_journal_year 1935
_journal_page_first 480
_journal_page_last 493
_publ_Section_title
;
 Crystal Structures of Hydrated Compounds II. Structure Type Mg(ClO$_{4}
    ↪ $)$_{2}$$\cdot$6H$_{2}$O
;

# Found in Stability of phases in the Mg(ClO$_{4}$)$_{2}$$\cdot$$n$H$_{2
    ↪ }$O system and implications for perchlorate occurrences on Mars
    ↪ , 2011

_aflow_title 'Mg(ClO$_{4}$)$_{2}$$\cdot$6H$_{2}$O ($H4_{11}$) Structure'
_aflow_proto 'A2B6CD8_oP34_31_2a_2a2b_a_4a2b'
_aflow_params 'a,b/a,c/a,y_{1},z_{1},y_{2},z_{2},y_{3},z_{3},y_{4},z_{4
    ↪ },y_{5},z_{5},y_{6},z_{6},y_{7},z_{7},y_{8},z_{8},y_{9},z_{9},
    ↪ x_{10},y_{10},z_{10},x_{11},y_{11},z_{11},x_{12},y_{12},z_{12},
    ↪ x_{13},y_{13},z_{13}'
_aflow_params_values '7.76,1.73453608247,0.677835051546,-0.08333,0.5,
    ↪ 0.58333,0.0,0.125,0.0,0.375,0.5,0.25,0.75,-0.08333,0.77778,
    ↪ 0.58333,0.27778,0.81111,0.40833,0.68889,-0.09167,0.8125,0.1875,
    ↪ 0.5,0.8125,0.3125,0.0,0.84167,-0.03056,0.40833,0.84167,0.53056
    ↪ ,-0.09167'
_aflow_Strukturbericht '$H4_{11}$'
_aflow_Pearson 'oP34'

_symmetry_space_group_name_H-M "P m n 21"
_symmetry_Int_Tables_number 31

_cell_length_a    7.76000
_cell_length_b    13.46000
_cell_length_c    5.26000
_cell_angle_alpha 90.00000
_cell_angle_beta  90.00000
_cell_angle_gamma 90.00000

loop_
_space_group_symop_id
_space_group_symop_operation_xyz
1 x,y,z
2 -x+1/2,-y,z+1/2
3 -x,y,z
4 x+1/2,-y,z+1/2

loop_
_atom_site_label
_atom_site_type_symbol
_atom_site_symmetry_multiplicity
_atom_site_Wyckoff_label
_atom_site_fract_x
_atom_site_fract_y
_atom_site_fract_z
_atom_site_occupancy
Cl1 Cl 2 a 0.00000 -0.08333 0.50000 1.00000
Cl2 Cl 2 a 0.00000 0.58333 0.00000 1.00000
H2O1 H2O 2 a 0.00000 0.12500 0.00000 1.00000
H2O2 H2O 2 a 0.00000 0.37500 0.50000 1.00000
Mg1 Mg 2 a 0.00000 0.25000 0.75000 1.00000
O1 O 2 a 0.00000 -0.08333 0.77778 1.00000
O2 O 2 a 0.00000 0.58333 0.27778 1.00000
O3 O 2 a 0.00000 0.81111 0.40833 1.00000
O4 O 2 a 0.00000 0.68889 -0.09167 1.00000
H2O3 H2O 4 b 0.81250 0.18750 0.50000 1.00000
H2O4 H2O 4 b 0.81250 0.31250 0.00000 1.00000
O5 O 4 b 0.84167 -0.03056 0.40833 1.00000
O6 O 4 b 0.84167 0.53056 -0.09167 1.00000
```

Mg(ClO$_4$)$_2$·6H$_2$O ($H4_{11}$): A2B6CD8_oP34_31_2a_2a2b_a_4a2b - POSCAR

```
A2B6CD8_oP34_31_2a_2a2b_a_4a2b & a,b/a,c/a,y1,z1,y2,z2,y3,z3,y4,z4,y5,z5
    ↪ ,y6,z6,y7,z7,y8,z8,y9,z9,x10,y10,z10,x11,y11,z11,x12,y12,z12,
    ↪ x13,y13,z13 --params=7.76,1.73453608247,0.677835051546,-0.08333
    ↪ ,0.5,0.58333,0.0,0.125,0.0,0.375,0.5,0.25,0.75,-0.08333,0.77778
    ↪ ,0.58333,0.27778,0.81111,0.40833,0.68889,-0.09167,0.8125,0.1875
    ↪ ,0.5,0.8125,0.3125,0.0,0.84167,-0.03056,0.40833,0.84167,0.53056
    ↪ ,-0.09167 & Pmn2_{1} C_{2v}^{7} #31 (a^9b^4) & oP34 & $H4_{11}$
    ↪ & Cl2(H2O)6MgO8 & Cl2(H2O)6MgO8 & C. D. West, Zeitschrift f{"u
    ↪ }r Kristallographie - Crystalline Materials 91, 480-493 (1935)
   1.00000000000000
   7.76000000000000    0.00000000000000    0.00000000000000
   0.00000000000000   13.46000000000000    0.00000000000000
   0.00000000000000    0.00000000000000    5.26000000000000
    Cl    H2O    Mg     O
     4     12     2    16
Direct
   0.00000000000000   -0.08333000000000   0.50000000000000   Cl    (2a)
   0.50000000000000    0.08333000000000   1.00000000000000   Cl    (2a)
   0.00000000000000    0.58333000000000   0.00000000000000   Cl    (2a)
   0.50000000000000   -0.58333000000000   0.50000000000000   Cl    (2a)
   0.00000000000000    0.12500000000000   0.00000000000000   H2O   (2a)
   0.50000000000000   -0.12500000000000   0.50000000000000   H2O   (2a)
   0.00000000000000    0.37500000000000   0.50000000000000   H2O   (2a)
   0.50000000000000   -0.37500000000000   1.00000000000000   H2O   (2a)
   0.81250000000000    0.18750000000000   0.50000000000000   H2O   (4b)
  -0.31250000000000   -0.18750000000000   1.00000000000000   H2O   (4b)
   1.31250000000000   -0.18750000000000   1.00000000000000   H2O   (4b)
  -0.81250000000000    0.18750000000000   0.50000000000000   H2O   (4b)
   0.81250000000000    0.31250000000000   0.00000000000000   H2O   (4b)
  -0.31250000000000   -0.31250000000000   0.50000000000000   H2O   (4b)
   1.31250000000000   -0.31250000000000   0.50000000000000   H2O   (4b)
  -0.81250000000000    0.31250000000000   0.00000000000000   H2O   (4b)
   0.00000000000000    0.25000000000000   0.75000000000000   Mg    (2a)
   0.50000000000000   -0.25000000000000   1.25000000000000   Mg    (2a)
   0.00000000000000   -0.08333000000000   0.77778000000000   O     (2a)
   0.50000000000000    0.08333000000000   1.27778000000000   O     (2a)
   0.00000000000000    0.58333000000000   0.27778000000000   O     (2a)
   0.50000000000000   -0.58333000000000   0.77778000000000   O     (2a)
   0.00000000000000    0.81111000000000   0.40833000000000   O     (2a)
   0.50000000000000   -0.81111000000000   0.90833000000000   O     (2a)
   0.00000000000000    0.68889000000000  -0.09167000000000   O     (2a)
   0.50000000000000   -0.68889000000000   0.40833000000000   O     (2a)
   0.84167000000000   -0.03056000000000   0.40833000000000   O     (4b)
  -0.34167000000000    0.03056000000000   0.90833000000000   O     (4b)
   1.34167000000000    0.03056000000000   0.90833000000000   O     (4b)
  -0.84167000000000   -0.03056000000000   0.40833000000000   O     (4b)
   0.84167000000000    0.53056000000000  -0.09167000000000   O     (4b)
  -0.34167000000000   -0.53056000000000   0.40833000000000   O     (4b)
   1.34167000000000   -0.53056000000000   0.40833000000000   O     (4b)
  -0.84167000000000    0.53056000000000  -0.09167000000000   O     (4b)
```

B$_4$SrO$_7$: A4B7C_oP24_31_2b_a3b_a - CIF

```
# CIF file
data_findsym-output
_audit_creation_method FINDSYM

_chemical_name_mineral 'B4O7Sr'
_chemical_formula_sum 'B4 O7 Sr'

loop_
_publ_author_name
 'J. {Krogh-Moe}'
_journal_name_full_name
;
 Acta Chemica Scandinavica
;
_journal_volume 18
_journal_year 1964
_journal_page_first 2055
_journal_page_last 2060
_publ_Section_title
;
 The Crystal Structure of Strontium Diborate, SrO$\cdot$2B$_{2}$O$_{3}$
;

_aflow_title 'B$_{4}$SrO$_{7}$ Structure'
_aflow_proto 'A4B7C_oP24_31_2b_a3b_a'
_aflow_params 'a,b/a,c/a,y_{1},z_{1},y_{2},z_{2},x_{3},y_{3},z_{3},x_{4
    ↪ },y_{4},z_{4},x_{5},y_{5},z_{5},x_{6},y_{6},z_{6},x_{7},y_{7},
    ↪ z_{7}'
_aflow_params_values '10.711,0.413313416114,0.395387918962,0.728,0.454,
    ↪ 0.289,0.0,0.379,0.174,-0.024,0.246,0.671,-0.037,0.359,0.857,
    ↪ 0.064,0.221,0.631,0.335,0.365,0.226,0.335'
_aflow_Strukturbericht 'None'
_aflow_Pearson 'oP24'

_symmetry_space_group_name_H-M "P m n 21"
_symmetry_Int_Tables_number 31

_cell_length_a    10.71100
_cell_length_b    4.42700
_cell_length_c    4.23500
_cell_angle_alpha 90.00000
_cell_angle_beta  90.00000
_cell_angle_gamma 90.00000

loop_
_space_group_symop_id
```

```
_space_group_symop_operation_xyz
1 x,y,z
2 -x+1/2,-y,z+1/2
3 -x,y,z
4 x+1/2,-y,z+1/2

loop_
_atom_site_label
_atom_site_type_symbol
_atom_site_symmetry_multiplicity
_atom_site_Wyckoff_label
_atom_site_fract_x
_atom_site_fract_y
_atom_site_fract_z
_atom_site_occupancy
O1 O 2 a 0.00000 0.72800 0.45400 1.00000
Sr1 Sr 2 a 0.00000 0.28900 0.00000 1.00000
B1 B 4 b 0.37900 0.17400 -0.02400 1.00000
B2 B 4 b 0.24600 0.67100 -0.03700 1.00000
O2 O 4 b 0.35900 0.85700 0.06400 1.00000
O3 O 4 b 0.22100 0.63100 0.33500 1.00000
O4 O 4 b 0.36500 0.22600 0.33500 1.00000
```

$B_4SrO_7$: A4B7C_oP24_31_2b_a3b_a - POSCAR

```
A4B7C_oP24_31_2b_a3b_a & a,b/a,c/a,y1,z1,y2,z2,x3,y3,z3,x4,y4,z4,x5,y5,
    ↪ z5,x6,y6,z6,x7,y7,z7 --params=10.711,0.413313416114,
    ↪ 0.395387918962,0.728,0.454,0.289,0.0,0.379,0.174,-0.024,0.246,
    ↪ 0.671,-0.037,0.359,0.857,0.064,0.221,0.631,0.335,0.365,0.226,
    ↪ 0.335 & Pmn2_{1} C_{2v}^{7} #31 (a^2b^5) & oP24 & None & B4O7Sr
    ↪ & B4O7Sr & J. {Krogh-Moe}, Acta Chem. Scand. 18, 2055-2060 (
    ↪ 1964)
   1.00000000000000
  10.71100000000000   0.00000000000000   0.00000000000000
   0.00000000000000   4.42700000000000   0.00000000000000
   0.00000000000000   0.00000000000000   4.23500000000000
     B     O    Sr
     8    14     2
Direct
   0.37900000000000   0.17400000000000  -0.02400000000000   B   (4b)
   0.12100000000000  -0.17400000000000   0.47600000000000   B   (4b)
   0.87900000000000  -0.17400000000000   0.47600000000000   B   (4b)
  -0.37900000000000   0.17400000000000  -0.02400000000000   B   (4b)
   0.24600000000000   0.67100000000000  -0.03700000000000   B   (4b)
   0.25400000000000  -0.67100000000000   0.46300000000000   B   (4b)
   0.74600000000000  -0.67100000000000   0.46300000000000   B   (4b)
  -0.24600000000000   0.67100000000000  -0.03700000000000   B   (4b)
   0.00000000000000   0.72800000000000   0.45400000000000   O   (2a)
   0.50000000000000  -0.72800000000000   0.95400000000000   O   (2a)
   0.35900000000000   0.85700000000000   0.06400000000000   O   (4b)
   0.14100000000000  -0.85700000000000   0.56400000000000   O   (4b)
   0.85900000000000  -0.85700000000000   0.56400000000000   O   (4b)
  -0.35900000000000   0.85700000000000   0.06400000000000   O   (4b)
   0.22100000000000   0.63100000000000   0.33500000000000   O   (4b)
   0.27900000000000  -0.63100000000000   0.83500000000000   O   (4b)
   0.72100000000000  -0.63100000000000   0.83500000000000   O   (4b)
  -0.22100000000000   0.63100000000000   0.33500000000000   O   (4b)
   0.36500000000000   0.22600000000000   0.33500000000000   O   (4b)
   0.13500000000000  -0.22600000000000   0.83500000000000   O   (4b)
   0.86500000000000  -0.22600000000000   0.83500000000000   O   (4b)
  -0.36500000000000   0.22600000000000   0.33500000000000   O   (4b)
   0.00000000000000   0.28900000000000   0.00000000000000   Sr  (2a)
   0.50000000000000  -0.28900000000000   0.50000000000000   Sr  (2a)
```

*D*8$_7$ (Shcherbinaite, $V_2O_5$) (*obsolete*): A5B2_oP14_31_a2b_b - CIF

```
# CIF file
data_findsym-output
_audit_creation_method FINDSYM

_chemical_name_mineral 'Shcherbinaite'
_chemical_formula_sum 'O5 V2'

loop_
_publ_author_name
 'J. A. A. Ketelaar'
_journal_name_full_name
;
 Zeitschrift f{\"u}r Kristallographie - Crystalline Materials
;
_journal_volume 95
_journal_year 1936
_journal_page_first 9
_journal_page_last 27
_publ_Section_title
;
 Die Kristallstruktur des Vanadinpentoxyds
;

# Found in A Refinement of the Structure of V$_{2}$O$_{5}$, 1986

_aflow_title '$D8_{7}$ (Shcherbinaite, V$_{2}$O$_{5}$) ({\em{obsolete
    ↪ }}) Structure'
_aflow_proto 'A5B2_oP14_31_a2b_b'
_aflow_params 'a,b/a,c/a,y_{1},z_{1},x_{2},y_{2},z_{2},x_{3},y_{3},z_{3
    ↪ },x_{4},y_{4},z_{4}'
_aflow_params_values '11.48,0.379790940767,0.309233449477,0.08,0.89,
    ↪ 0.148,0.45,-0.08,0.2,0.03,0.46,0.148,0.097,0.0'
_aflow_Strukturbericht '$D8_{7}$'
_aflow_Pearson 'oP14'

_symmetry_space_group_name_H-M "P m n 21"
_symmetry_Int_Tables_number 31

_cell_length_a    11.48000
_cell_length_b    4.36000
_cell_length_c    3.55000
_cell_angle_alpha 90.00000
_cell_angle_beta  90.00000
_cell_angle_gamma 90.00000

loop_
_space_group_symop_id
_space_group_symop_operation_xyz
1 x,y,z
2 -x+1/2,-y,z+1/2
3 -x,y,z
4 x+1/2,-y,z+1/2

loop_
_atom_site_label
_atom_site_type_symbol
_atom_site_symmetry_multiplicity
_atom_site_Wyckoff_label
_atom_site_fract_x
_atom_site_fract_y
_atom_site_fract_z
_atom_site_occupancy
O1 O 2 a 0.00000 0.08000 0.89000 1.00000
O2 O 4 b 0.14800 0.45000 -0.08000 1.00000
O3 O 4 b 0.20000 0.03000 0.46000 1.00000
V1 V 4 b 0.14800 0.09700 0.00000 1.00000
```

*D*8$_7$ (Shcherbinaite, $V_2O_5$) (*obsolete*): A5B2_oP14_31_a2b_b - POSCAR

```
A5B2_oP14_31_a2b_b & a,b/a,c/a,y1,z1,x2,y2,z2,x3,y3,z3,x4,y4,z4 --params
    ↪ =11.48,0.379790940767,0.309233449477,0.08,0.89,0.148,0.45,-0.08
    ↪ ,0.2,0.03,0.46,0.148,0.097,0.0 & Pmn2_{1} C_{2v}^{7} #31 (ab^3)
    ↪ & oP14 & $D8_{7}$ & O5V2 & Shcherbinaite & J. A. A. Ketelaar,
    ↪ Zeitschrift f{"u}r Kristallographie - Crystalline Materials 95,
    ↪ 9-27 (1936)
   1.00000000000000
  11.48000000000000   0.00000000000000   0.00000000000000
   0.00000000000000   4.36000000000000   0.00000000000000
   0.00000000000000   0.00000000000000   3.55000000000000
     O     V
    10     4
Direct
   0.00000000000000   0.08000000000000   0.89000000000000   O   (2a)
   0.50000000000000  -0.08000000000000   1.39000000000000   O   (2a)
   0.14800000000000   0.45000000000000  -0.08000000000000   O   (4b)
   0.35200000000000  -0.45000000000000   0.42000000000000   O   (4b)
   0.64800000000000  -0.45000000000000   0.42000000000000   O   (4b)
  -0.14800000000000   0.45000000000000  -0.08000000000000   O   (4b)
   0.20000000000000   0.03000000000000   0.46000000000000   O   (4b)
   0.30000000000000  -0.03000000000000   0.96000000000000   O   (4b)
   0.70000000000000  -0.03000000000000   0.96000000000000   O   (4b)
  -0.20000000000000   0.03000000000000   0.46000000000000   O   (4b)
   0.14800000000000   0.09700000000000   0.00000000000000   V   (4b)
   0.35200000000000  -0.09700000000000   0.50000000000000   V   (4b)
   0.64800000000000  -0.09700000000000   0.50000000000000   V   (4b)
  -0.14800000000000   0.09700000000000   0.00000000000000   V   (4b)
```

$Mo_{17}O_{47}$: A17B47_oP128_32_a8c_a23c - CIF

```
# CIF file
data_findsym-output
_audit_creation_method FINDSYM

_chemical_name_mineral 'Mo17O47'
_chemical_formula_sum 'Mo17 O47'

loop_
_publ_author_name
 'L. Kihlborg'
_journal_name_full_name
;
 Acta Chemica Scandinavica
;
_journal_volume 17
_journal_year 1963
_journal_page_first 1485
_journal_page_last 1487
_publ_Section_title
;
 Least Squares Refinement of the Structure of Mo$_{17}$O$_{47}$
;

_aflow_title 'Mo$_{17}$O$_{47}$ Structure'
_aflow_proto 'A17B47_oP128_32_a8c_a23c'
_aflow_params 'a,b/a,c/a,z_{1},z_{2},x_{3},y_{3},z_{3},x_{4},y_{4},z_{4
    ↪ },x_{5},y_{5},z_{5},x_{6},y_{6},z_{6},x_{7},y_{7},z_{7},x_{8},
    ↪ y_{8},z_{8},x_{9},y_{9},z_{9},x_{10},y_{10},z_{10},x_{11},y_{11
    ↪ },z_{11},x_{12},y_{12},z_{12},x_{13},y_{13},z_{13},x_{14},y_{14
    ↪ },z_{14},x_{15},y_{15},z_{15},x_{16},y_{16},z_{16},x_{17},y_{17
    ↪ },z_{17},x_{18},y_{18},z_{18},x_{19},y_{19},z_{19},x_{20},y_{20
    ↪ },z_{20},x_{21},y_{21},z_{21},x_{22},y_{22},z_{22},x_{23},y_{23
    ↪ },z_{23},x_{24},y_{24},z_{24},x_{25},y_{25},z_{25},x_{26},y_{26
    ↪ },z_{26},x_{27},y_{27},z_{27},x_{28},y_{28},z_{28},x_{29},y_{29
    ↪ },z_{29},x_{30},y_{30},z_{30},x_{31},y_{31},z_{31},x_{32},y_{32
    ↪ },z_{32},x_{33},y_{33},z_{33}'
_aflow_params_values '21.61,0.908375751967,0.182832022212,0.579,0.033,
    ↪ 0.02398,0.26126,0.586,0.12899,0.11845,0.4177,0.13648,0.39897,
    ↪ 0.4329,0.2426,0.25684,0.579,0.28825,0.06556,0.5848,0.38285,
    ↪ 0.19333,0.4334,0.38519,0.36637,0.4205,0.46561,0.05514,0.5551,
    ↪ 0.0225,0.263,0.027,0.131,0.1206,-0.052,0.1349,0.4038,-0.011,
    ↪ 0.2465,0.2532,0.007,0.2882,0.0716,0.003,0.381,0.1938,-0.008,
    ↪ 0.3919,0.3568,-0.047,0.4649,0.0579,-0.001,0.0767,0.0429,0.505,
    ↪ 0.0612,0.179,0.514,0.093,0.3224,0.521,0.0585,0.4581,0.483,
    ↪ 0.2048,0.0683,0.486,0.1788,0.1999,0.503,0.2064,0.3388,0.503,
```

```
    ↪ 0.1892,0.4792,0.538,0.2969,0.1672,0.49,0.3297,0.2778,0.481,
    ↪ 0.3266,0.4217,0.465,0.3764,0.0925,0.494,0.4697,0.1567,0.477,
    ↪ 0.4396,0.2746,0.483,0.4595,0.4132,0.481'
_aflow_Strukturbericht 'None'
_aflow_Pearson 'oP128'

_symmetry_space_group_name_H-M "P b a 2"
_symmetry_Int_Tables_number 32

_cell_length_a     21.61000
_cell_length_b     19.63000
_cell_length_c     3.95100
_cell_angle_alpha  90.00000
_cell_angle_beta   90.00000
_cell_angle_gamma  90.00000

loop_
_space_group_symop_id
_space_group_symop_operation_xyz
1 x,y,z
2 -x,-y,z
3 -x+1/2,y+1/2,z
4 x+1/2,-y+1/2,z

loop_
_atom_site_label
_atom_site_type_symbol
_atom_site_symmetry_multiplicity
_atom_site_Wyckoff_label
_atom_site_fract_x
_atom_site_fract_y
_atom_site_fract_z
_atom_site_occupancy
Mo1 Mo 2 a 0.00000 0.00000 0.57900 1.00000
O1 O 2 a 0.00000 0.00000 0.03300 1.00000
Mo2 Mo 4 c 0.02398 0.26126 0.58600 1.00000
Mo3 Mo 4 c 0.12899 0.11845 0.41770 1.00000
Mo4 Mo 4 c 0.13648 0.39897 0.43290 1.00000
Mo5 Mo 4 c 0.24260 0.25684 0.57900 1.00000
Mo6 Mo 4 c 0.28825 0.06556 0.58480 1.00000
Mo7 Mo 4 c 0.38285 0.19333 0.43340 1.00000
Mo8 Mo 4 c 0.38519 0.36637 0.42050 1.00000
Mo9 Mo 4 c 0.46561 0.05514 0.55510 1.00000
O2 O 4 c 0.02250 0.26300 0.02700 1.00000
O3 O 4 c 0.13100 0.12060 -0.05200 1.00000
O4 O 4 c 0.13490 0.40380 -0.01100 1.00000
O5 O 4 c 0.24650 0.25320 0.00700 1.00000
O6 O 4 c 0.28820 0.07160 0.00300 1.00000
O7 O 4 c 0.38100 0.19380 -0.00800 1.00000
O8 O 4 c 0.39190 0.35680 -0.04700 1.00000
O9 O 4 c 0.46490 0.05790 -0.00100 1.00000
O10 O 4 c 0.07670 0.04290 0.50500 1.00000
O11 O 4 c 0.06120 0.17900 0.51400 1.00000
O12 O 4 c 0.09300 0.32240 0.52100 1.00000
O13 O 4 c 0.05850 0.45810 0.48300 1.00000
O14 O 4 c 0.20480 0.06830 0.48600 1.00000
O15 O 4 c 0.17880 0.19990 0.50300 1.00000
O16 O 4 c 0.20640 0.33880 0.50300 1.00000
O17 O 4 c 0.18920 0.47920 0.53800 1.00000
O18 O 4 c 0.29690 0.16720 0.49000 1.00000
O19 O 4 c 0.32970 0.27780 0.48100 1.00000
O20 O 4 c 0.32660 0.42170 0.46500 1.00000
O21 O 4 c 0.37640 0.09250 0.49400 1.00000
O22 O 4 c 0.46970 0.15670 0.47700 1.00000
O23 O 4 c 0.43960 0.27460 0.48300 1.00000
O24 O 4 c 0.45950 0.41320 0.48100 1.00000
```

$Mo_{17}O_{47}$: A17B47_oP128_32_a8c_a23c - POSCAR

```
A17B47_oP128_32_a8c_a23c & a,b/a,c/a,z1,z2,x3,y3,z3,x4,y4,z4,x5,y5,z5,x6
    ↪ ,y6,z6,x7,y7,z7,x8,y8,z8,x9,y9,z9,x10,y10,z10,x11,y11,z11,x12,
    ↪ y12,z12,x13,y13,z13,x14,y14,z14,x15,y15,z15,x16,y16,z16,x17,y17
    ↪ ,z17,x18,y18,z18,x19,y19,z19,x20,y20,z20,x21,y21,z21,x22,y22,
    ↪ z22,x23,y23,z23,x24,y24,z24,x25,y25,z25,x26,y26,z26,x27,y27,z27
    ↪ ,x28,y28,z28,x29,y29,z29,x30,y30,z30,x31,y31,z31,x32,y32,z32,
    ↪ x33,y33,z33 --params=21.61,0.908375751967,0.182832022212,0.579,
    ↪ 0.033,0.02398,0.26126,0.586,0.12899,0.11845,0.4177,0.13648,
    ↪ 0.39897,0.4329,0.2426,0.25684,0.579,0.28825,0.06556,0.5848,
    ↪ 0.38285,0.19333,0.4334,0.38519,0.36637,0.4205,0.46561,0.05514,
    ↪ 0.5551,0.0225,0.263,0.027,0.131,0.1206,-0.052,0.1349,0.4038,-
    ↪ 0.011,0.2465,0.2532,0.007,0.2882,0.0716,0.003,0.381,0.1938,-
    ↪ 0.008,0.3919,0.3568,-0.047,0.4649,0.0579,-0.001,0.0767,0.0429,
    ↪ 0.505,0.0612,0.179,0.514,0.093,0.3224,0.521,0.0585,0.4581,0.483
    ↪ ,0.2048,0.0683,0.486,0.1788,0.1999,0.503,0.2064,0.3388,0.503,
    ↪ 0.1892,0.4792,0.538,0.2969,0.1672,0.49,0.3297,0.2778,0.481,
    ↪ 0.3266,0.4217,0.465,0.3764,0.0925,0.494,0.4697,0.1567,0.477,
    ↪ 0.4396,0.2746,0.483,0.4595,0.4132,0.481 & Pba2 C_{2v}^{8} #32 (
    ↪ a^2c^31) & oP128 & None & Mo17O47 & Mo17O47 & L. Kihlborg, Acta
    ↪ Chem. Scand. 17, 1485-1487 (1963)
   1.00000000000000
  21.61000000000000    0.00000000000000    0.00000000000000
   0.00000000000000   19.63000000000000    0.00000000000000
   0.00000000000000    0.00000000000000    3.95100000000000
    Mo        O
    34       94
Direct
   0.00000000000000    0.00000000000000    0.57900000000000   Mo   (2a)
   0.50000000000000    0.50000000000000    0.57900000000000   Mo   (2a)
   0.02398000000000    0.26126000000000    0.58600000000000   Mo   (4c)
  -0.02398000000000   -0.26126000000000    0.58600000000000   Mo   (4c)
   0.52398000000000    0.23874000000000    0.58600000000000   Mo   (4c)
   0.47602000000000    0.76126000000000    0.58600000000000   Mo   (4c)
   0.12899000000000    0.11845000000000    0.41770000000000   Mo   (4c)
  -0.12899000000000   -0.11845000000000    0.41770000000000   Mo   (4c)
   0.62899000000000    0.38155000000000    0.41770000000000   Mo   (4c)
   0.37101000000000    0.61845000000000    0.41770000000000   Mo   (4c)
   0.13648000000000    0.39897000000000    0.43290000000000   Mo   (4c)
  -0.13648000000000   -0.39897000000000    0.43290000000000   Mo   (4c)
   0.63648000000000    0.10103000000000    0.43290000000000   Mo   (4c)
   0.36352000000000    0.89897000000000    0.43290000000000   Mo   (4c)
   0.24260000000000    0.25684000000000    0.57900000000000   Mo   (4c)
  -0.24260000000000   -0.25684000000000    0.57900000000000   Mo   (4c)
   0.74260000000000    0.24316000000000    0.57900000000000   Mo   (4c)
   0.25740000000000    0.75684000000000    0.57900000000000   Mo   (4c)
   0.28825000000000    0.06556000000000    0.58480000000000   Mo   (4c)
  -0.28825000000000   -0.06556000000000    0.58480000000000   Mo   (4c)
   0.78825000000000    0.43444000000000    0.58480000000000   Mo   (4c)
   0.21175000000000    0.56556000000000    0.58480000000000   Mo   (4c)
   0.38285000000000    0.19333000000000    0.43340000000000   Mo   (4c)
  -0.38285000000000   -0.19333000000000    0.43340000000000   Mo   (4c)
   0.88285000000000    0.30667000000000    0.43340000000000   Mo   (4c)
   0.11715000000000    0.69333000000000    0.43340000000000   Mo   (4c)
   0.38519000000000    0.36637000000000    0.42050000000000   Mo   (4c)
  -0.38519000000000   -0.36637000000000    0.42050000000000   Mo   (4c)
   0.88519000000000    0.13363000000000    0.42050000000000   Mo   (4c)
   0.11481000000000    0.86637000000000    0.42050000000000   Mo   (4c)
   0.46561000000000    0.05514000000000    0.55510000000000   Mo   (4c)
  -0.46561000000000   -0.05514000000000    0.55510000000000   Mo   (4c)
   0.96561000000000    0.44486000000000    0.55510000000000   Mo   (4c)
   0.03439000000000    0.55514000000000    0.55510000000000   Mo   (4c)
   0.00000000000000    0.00000000000000    0.03300000000000   O    (2a)
   0.50000000000000    0.50000000000000    0.03300000000000   O    (2a)
   0.02250000000000    0.26300000000000    0.02700000000000   O    (4c)
  -0.02250000000000   -0.26300000000000    0.02700000000000   O    (4c)
   0.52250000000000    0.23700000000000    0.02700000000000   O    (4c)
   0.47750000000000    0.76300000000000    0.02700000000000   O    (4c)
   0.13100000000000    0.12060000000000   -0.05200000000000   O    (4c)
  -0.13100000000000   -0.12060000000000   -0.05200000000000   O    (4c)
   0.63100000000000    0.37940000000000   -0.05200000000000   O    (4c)
   0.36900000000000    0.62060000000000   -0.05200000000000   O    (4c)
   0.13490000000000    0.40380000000000   -0.01100000000000   O    (4c)
  -0.13490000000000   -0.40380000000000   -0.01100000000000   O    (4c)
   0.63490000000000    0.09620000000000   -0.01100000000000   O    (4c)
   0.36510000000000    0.90380000000000   -0.01100000000000   O    (4c)
   0.24650000000000    0.25320000000000    0.00700000000000   O    (4c)
  -0.24650000000000   -0.25320000000000    0.00700000000000   O    (4c)
   0.74650000000000    0.24680000000000    0.00700000000000   O    (4c)
   0.25350000000000    0.75320000000000    0.00700000000000   O    (4c)
   0.28820000000000    0.07160000000000    0.00300000000000   O    (4c)
  -0.28820000000000   -0.07160000000000    0.00300000000000   O    (4c)
   0.78820000000000    0.42840000000000    0.00300000000000   O    (4c)
   0.21180000000000    0.57160000000000    0.00300000000000   O    (4c)
   0.38100000000000    0.19380000000000   -0.00800000000000   O    (4c)
  -0.38100000000000   -0.19380000000000   -0.00800000000000   O    (4c)
   0.88100000000000    0.30620000000000   -0.00800000000000   O    (4c)
   0.11900000000000    0.69380000000000   -0.00800000000000   O    (4c)
   0.39190000000000    0.35680000000000   -0.04700000000000   O    (4c)
  -0.39190000000000   -0.35680000000000   -0.04700000000000   O    (4c)
   0.89190000000000    0.14320000000000   -0.04700000000000   O    (4c)
   0.10810000000000    0.85680000000000   -0.04700000000000   O    (4c)
   0.46490000000000    0.05790000000000   -0.00100000000000   O    (4c)
  -0.46490000000000   -0.05790000000000   -0.00100000000000   O    (4c)
   0.96490000000000    0.44210000000000   -0.00100000000000   O    (4c)
   0.03510000000000    0.55790000000000   -0.00100000000000   O    (4c)
   0.07670000000000    0.04290000000000    0.50500000000000   O    (4c)
  -0.07670000000000   -0.04290000000000    0.50500000000000   O    (4c)
   0.57670000000000    0.45710000000000    0.50500000000000   O    (4c)
   0.42330000000000    0.54290000000000    0.50500000000000   O    (4c)
   0.06120000000000    0.17900000000000    0.51400000000000   O    (4c)
  -0.06120000000000   -0.17900000000000    0.51400000000000   O    (4c)
   0.56120000000000    0.32100000000000    0.51400000000000   O    (4c)
   0.43880000000000    0.67900000000000    0.51400000000000   O    (4c)
   0.09300000000000    0.32240000000000    0.52100000000000   O    (4c)
  -0.09300000000000   -0.32240000000000    0.52100000000000   O    (4c)
   0.59300000000000    0.17760000000000    0.52100000000000   O    (4c)
   0.40700000000000    0.82240000000000    0.52100000000000   O    (4c)
   0.05850000000000    0.45810000000000    0.48300000000000   O    (4c)
  -0.05850000000000   -0.45810000000000    0.48300000000000   O    (4c)
   0.55850000000000    0.04190000000000    0.48300000000000   O    (4c)
   0.44150000000000    0.95810000000000    0.48300000000000   O    (4c)
   0.20480000000000    0.06830000000000    0.48600000000000   O    (4c)
  -0.20480000000000   -0.06830000000000    0.48600000000000   O    (4c)
   0.70480000000000    0.43170000000000    0.48600000000000   O    (4c)
   0.29520000000000    0.56830000000000    0.48600000000000   O    (4c)
   0.17880000000000    0.19990000000000    0.50300000000000   O    (4c)
  -0.17880000000000   -0.19990000000000    0.50300000000000   O    (4c)
   0.67880000000000    0.30010000000000    0.50300000000000   O    (4c)
   0.32120000000000    0.69990000000000    0.50300000000000   O    (4c)
   0.20640000000000    0.33880000000000    0.50300000000000   O    (4c)
  -0.20640000000000   -0.33880000000000    0.50300000000000   O    (4c)
   0.70640000000000    0.16120000000000    0.50300000000000   O    (4c)
   0.29360000000000    0.83880000000000    0.50300000000000   O    (4c)
   0.18920000000000    0.47920000000000    0.53800000000000   O    (4c)
  -0.18920000000000   -0.47920000000000    0.53800000000000   O    (4c)
   0.68920000000000    0.02080000000000    0.53800000000000   O    (4c)
   0.31080000000000    0.97920000000000    0.53800000000000   O    (4c)
   0.29690000000000    0.16720000000000    0.49000000000000   O    (4c)
  -0.29690000000000   -0.16720000000000    0.49000000000000   O    (4c)
   0.79690000000000    0.33280000000000    0.49000000000000   O    (4c)
   0.20310000000000    0.66720000000000    0.49000000000000   O    (4c)
   0.32970000000000    0.27780000000000    0.48100000000000   O    (4c)
  -0.32970000000000   -0.27780000000000    0.48100000000000   O    (4c)
   0.82970000000000    0.22220000000000    0.48100000000000   O    (4c)
   0.17030000000000    0.77780000000000    0.48100000000000   O    (4c)
   0.32660000000000    0.42170000000000    0.46500000000000   O    (4c)
  -0.32660000000000   -0.42170000000000    0.46500000000000   O    (4c)
   0.82660000000000    0.07830000000000    0.46500000000000   O    (4c)
   0.17340000000000    0.92170000000000    0.46500000000000   O    (4c)
   0.37640000000000    0.09250000000000    0.49400000000000   O    (4c)
  -0.37640000000000   -0.09250000000000    0.49400000000000   O    (4c)
```

```
   0.87640000000000    0.40750000000000    0.49400000000000    O    (4c)
   0.12360000000000    0.59250000000000    0.49400000000000    O    (4c)
   0.46970000000000    0.15670000000000    0.47700000000000    O    (4c)
  -0.46970000000000   -0.15670000000000    0.47700000000000    O    (4c)
   0.96970000000000    0.34330000000000    0.47700000000000    O    (4c)
   0.03030000000000    0.65670000000000    0.47700000000000    O    (4c)
   0.43960000000000    0.27460000000000    0.48300000000000    O    (4c)
  -0.43960000000000   -0.27460000000000    0.48300000000000    O    (4c)
   0.93960000000000    0.22540000000000    0.48300000000000    O    (4c)
   0.06040000000000    0.77460000000000    0.48300000000000    O    (4c)
   0.45950000000000    0.41320000000000    0.48100000000000    O    (4c)
  -0.45950000000000   -0.41320000000000    0.48100000000000    O    (4c)
   0.95950000000000    0.08680000000000    0.48100000000000    O    (4c)
   0.04050000000000    0.91320000000000    0.48100000000000    O    (4c)
```

Possible $\delta$-$Gd_2Si_2O_7$: A2B7C2_oP44_33_2a_7a_2a - CIF

```
# CIF file
data_findsym-output
_audit_creation_method FINDSYM

_chemical_name_mineral 'Gd2O7Si2'
_chemical_formula_sum 'Gd2 O7 Si2'

loop_
_publ_author_name
 'Y. I. Smolin'
 'Y. F. Shepelev'
_journal_name_full_name
;
 Acta Crystallographica Section B: Structural Science
;
_journal_volume 26
_journal_year 1970
_journal_page_first 484
_journal_page_last 492
_publ_Section_title
;
 The Crystal Structures of the Rare Earth Pyrosilicates
;

# Found in The crystal structure of $\delta$-yttrium pyrosilicate, $\
    ↪ delta$-Y$_{2}$Si$_{2}$O$_{7}$, 1990

_aflow_title 'Possible $\delta$-Gd$_{2}$Si$_{2}$O$_{7}$ Structure'
_aflow_proto 'A2B7C2_oP44_33_2a_7a_2a'
_aflow_params 'a,b/a,c/a,x_{1},y_{1},z_{1},x_{2},y_{2},z_{2},x_{3},y_{3
    ↪ },z_{3},x_{4},y_{4},z_{4},x_{5},y_{5},z_{5},x_{6},y_{6},z_{6},
    ↪ x_{7},y_{7},z_{7},x_{8},y_{8},z_{8},x_{9},y_{9},z_{9},x_{10},y_
    ↪ {10},z_{10},x_{11},y_{11},z_{11}'
_aflow_params_values '13.87,0.365753424658,0.600576784427,0.12551,0.3373
    ↪ ,-0.00169,0.12564,0.33739,0.51409,0.2715,0.4769,0.0876,0.2658,
    ↪ 0.4857,0.413,0.3457,0.0706,0.2465,0.4211,0.5557,0.2448,0.5472,
    ↪ 0.7858,0.0866,0.5456,0.7882,0.4206,0.5988,0.3526,0.256,0.3205,
    ↪ 0.3744,0.2505,0.539,0.6253,0.2498'
_aflow_Strukturbericht 'None'
_aflow_Pearson 'oP44'

_symmetry_space_group_name_H-M "P n a 21"
_symmetry_Int_Tables_number 33

_cell_length_a      13.87000
_cell_length_b      5.07300
_cell_length_c      8.33000
_cell_angle_alpha   90.00000
_cell_angle_beta    90.00000
_cell_angle_gamma   90.00000

loop_
_space_group_symop_id
_space_group_symop_operation_xyz
1 x,y,z
2 -x,-y,z+1/2
3 -x+1/2,y+1/2,z+1/2
4 x+1/2,-y+1/2,z

loop_
_atom_site_label
_atom_site_type_symbol
_atom_site_symmetry_multiplicity
_atom_site_Wyckoff_label
_atom_site_fract_x
_atom_site_fract_y
_atom_site_fract_z
_atom_site_occupancy
Gd1 Gd 4 a 0.12551 0.33730 -0.00169 1.00000
Gd2 Gd 4 a 0.12564 0.33739 0.51409 1.00000
O1 O 4 a 0.27150 0.47690 0.08760 1.00000
O2 O 4 a 0.26580 0.48570 0.41300 1.00000
O3 O 4 a 0.34570 0.07060 0.24650 1.00000
O4 O 4 a 0.42110 0.55570 0.24480 1.00000
O5 O 4 a 0.54720 0.78580 0.08660 1.00000
O6 O 4 a 0.54560 0.78820 0.42060 1.00000
O7 O 4 a 0.59880 0.35260 0.25600 1.00000
Si1 Si 4 a 0.32050 0.37440 0.25050 1.00000
Si2 Si 4 a 0.53900 0.62530 0.24980 1.00000
```

Possible $\delta$-$Gd_2Si_2O_7$: A2B7C2_oP44_33_2a_7a_2a - POSCAR

```
A2B7C2_oP44_33_2a_7a_2a & a,b/a,c/a,x1,y1,z1,x2,y2,z2,x3,y3,z3,x4,y4,z4,
    ↪ x5,y5,z5,x6,y6,z6,x7,y7,z7,x8,y8,z8,x9,y9,z9,x10,y10,z10,x11,
    ↪ y11,z11 --params=13.87,0.365753424658,0.600576784427,0.12551,
    ↪ 0.3373,-0.00169,0.12564,0.33739,0.51409,0.2715,0.4769,0.0876,
    ↪ 0.2658,0.4857,0.413,0.3457,0.0706,0.2465,0.4211,0.5557,0.2448,
    ↪ 0.5472,0.7858,0.0866,0.5456,0.7882,0.4206,0.5988,0.3526,0.256,
    ↪ 0.3205,0.3744,0.2505,0.539,0.6253,0.2498 & Pna2_{1} C_{2v}^{9}
    ↪ #33 (a^11) & oP44 & None & Gd2O7Si2 & Gd2O7Si2 & Y. I. Smolin
    ↪ and Y. F. Shepelev, Acta Crystallogr. Sect. B Struct. Sci. 26,
    ↪ 484-492 (1970)
   1.00000000000000
  13.87000000000000    0.00000000000000    0.00000000000000
   0.00000000000000    5.07300000000000    0.00000000000000
   0.00000000000000    0.00000000000000    8.33000000000000
    Gd     O    Si
     8    28     8
Direct
   0.12551000000000    0.33730000000000   -0.00169000000000   Gd   (4a)
  -0.12551000000000   -0.33730000000000    0.49831000000000   Gd   (4a)
   0.62551000000000    0.16270000000000   -0.00169000000000   Gd   (4a)
   0.37449000000000    0.83730000000000    0.49831000000000   Gd   (4a)
   0.12564000000000    0.33739000000000    0.51409000000000   Gd   (4a)
  -0.12564000000000   -0.33739000000000    1.01409000000000   Gd   (4a)
   0.62564000000000    0.16261000000000    0.51409000000000   Gd   (4a)
   0.37436000000000    0.83739000000000    1.01409000000000   Gd   (4a)
   0.27150000000000    0.47690000000000    0.08760000000000   O    (4a)
  -0.27150000000000   -0.47690000000000    0.58760000000000   O    (4a)
   0.77150000000000    0.02310000000000    0.08760000000000   O    (4a)
   0.22850000000000    0.97690000000000    0.58760000000000   O    (4a)
   0.26580000000000    0.48570000000000    0.41300000000000   O    (4a)
  -0.26580000000000   -0.48570000000000    0.91300000000000   O    (4a)
   0.76580000000000    0.01430000000000    0.41300000000000   O    (4a)
   0.23420000000000    0.98570000000000    0.91300000000000   O    (4a)
   0.34570000000000    0.07060000000000    0.24650000000000   O    (4a)
  -0.34570000000000   -0.07060000000000    0.74650000000000   O    (4a)
   0.84570000000000    0.42940000000000    0.24650000000000   O    (4a)
   0.15430000000000    0.57060000000000    0.74650000000000   O    (4a)
   0.42110000000000    0.55570000000000    0.24480000000000   O    (4a)
  -0.42110000000000   -0.55570000000000    0.74480000000000   O    (4a)
   0.92110000000000   -0.05570000000000    0.24480000000000   O    (4a)
   0.07890000000000    1.05570000000000    0.74480000000000   O    (4a)
   0.54720000000000    0.78580000000000    0.08660000000000   O    (4a)
  -0.54720000000000   -0.78580000000000    0.58660000000000   O    (4a)
   1.04720000000000   -0.28580000000000    0.08660000000000   O    (4a)
  -0.04720000000000    1.28580000000000    0.58660000000000   O    (4a)
   0.54560000000000    0.78820000000000    0.42060000000000   O    (4a)
  -0.54560000000000   -0.78820000000000    0.92060000000000   O    (4a)
   1.04560000000000   -0.28820000000000    0.42060000000000   O    (4a)
  -0.04560000000000    1.28820000000000    0.92060000000000   O    (4a)
   0.59880000000000    0.35260000000000    0.25600000000000   O    (4a)
  -0.59880000000000   -0.35260000000000    0.75600000000000   O    (4a)
   1.09880000000000    0.14740000000000    0.25600000000000   O    (4a)
  -0.09880000000000    0.85260000000000    0.75600000000000   O    (4a)
   0.32050000000000    0.37440000000000    0.25050000000000   Si   (4a)
  -0.32050000000000   -0.37440000000000    0.75050000000000   Si   (4a)
   0.82050000000000    0.12560000000000    0.25050000000000   Si   (4a)
   0.17950000000000    0.87440000000000    0.75050000000000   Si   (4a)
   0.53900000000000    0.62530000000000    0.24980000000000   Si   (4a)
  -0.53900000000000   -0.62530000000000    0.74980000000000   Si   (4a)
   1.03900000000000   -0.12530000000000    0.24980000000000   Si   (4a)
  -0.03900000000000    1.12530000000000    0.74980000000000   Si   (4a)
```

$CaB_2O_4$ (III): A2BC4_oP84_33_6a_3a_12a - CIF

```
# CIF file
data_findsym-output
_audit_creation_method FINDSYM

_chemical_name_mineral 'B2CaO4'
_chemical_formula_sum 'B2 Ca O4'

loop_
_publ_author_name
 'M. Marezio'
 'J. P. Remeika'
 'P. D. Dernier'
_journal_name_full_name
;
 Acta Crystallographica Section B: Structural Science
;
_journal_volume 25
_journal_year 1969
_journal_page_first 955
_journal_page_last 964
_publ_Section_title
;
 The crystal structure of the high-pressure phase CaB$_{2}$O$_{4}$(III)
;

_aflow_title 'CaB$_{2}$O$_{4}$ (III) Structure'
_aflow_proto 'A2BC4_oP84_33_6a_3a_12a'
_aflow_params 'a,b/a,c/a,x_{1},y_{1},z_{1},x_{2},y_{2},z_{2},x_{3},y_{3
    ↪ },z_{3},x_{4},y_{4},z_{4},x_{5},y_{5},z_{5},x_{6},y_{6},z_{6},
    ↪ x_{7},y_{7},z_{7},x_{8},y_{8},z_{8},x_{9},y_{9},z_{9},x_{10},y_
    ↪ {10},z_{10},x_{11},y_{11},z_{11},x_{12},y_{12},z_{12},x_{13},y_
    ↪ {13},z_{13},x_{14},y_{14},z_{14},x_{15},y_{15},z_{15},x_{16},y_
    ↪ {16},z_{16},x_{17},y_{17},z_{17},x_{18},y_{18},z_{18},x_{19},y_
    ↪ {19},z_{19},x_{20},y_{20},z_{20},x_{21},y_{21},z_{21}'
_aflow_params_values '11.38,0.554217926186,0.993321616872,0.1401,0.2529,
    ↪ 0.539,0.251,0.4193,0.0515,0.2131,0.1191,0.7403,0.3222,0.1006,
    ↪ 0.9442,0.4658,0.0183,0.48,0.4748,0.3758,0.2477,0.07351,0.05043,
    ↪ 0.0,0.03257,0.65995,0.23386,0.25309,0.11288,0.27955,0.2657,
    ↪ 0.1944,0.0459,0.2171,0.4704,0.1734,0.0717,0.0297,0.2009,0.4547,
    ↪ 0.1715,0.2567,0.2009,0.375,0.4486,0.1456,0.0322,0.5116,0.0061,
    ↪ 0.3009,0.5305,0.1734,0.2946,0.6635,0.1069,0.022,0.7981,0.287,
    ↪ 0.2046,0.8336,0.4563,0.137,0.9458,0.1504,0.4768,0.9668'
_aflow_Strukturbericht 'None'
_aflow_Pearson 'oP84'

_symmetry_space_group_name_H-M "P n a 21"
_symmetry_Int_Tables_number 33
```

```
_cell_length_a      11.38000
_cell_length_b      6.30700
_cell_length_c      11.30400
_cell_angle_alpha  90.00000
_cell_angle_beta   90.00000
_cell_angle_gamma  90.00000

loop_
_space_group_symop_id
_space_group_symop_operation_xyz
1 x,y,z
2 -x,-y,z+1/2
3 -x+1/2,y+1/2,z+1/2
4 x+1/2,-y+1/2,z

loop_
_atom_site_label
_atom_site_type_symbol
_atom_site_symmetry_multiplicity
_atom_site_Wyckoff_label
_atom_site_fract_x
_atom_site_fract_y
_atom_site_fract_z
_atom_site_occupancy
B1 B 4 a 0.14010 0.25290 0.53900 1.00000
B2 B 4 a 0.25100 0.41930 0.05150 1.00000
B3 B 4 a 0.21310 0.11910 0.74030 1.00000
B4 B 4 a 0.32220 0.10060 0.94420 1.00000
B5 B 4 a 0.46580 0.01830 0.48000 1.00000
B6 B 4 a 0.47480 0.37580 0.24770 1.00000
Ca1 Ca 4 a 0.07351 0.05043 0.00000 1.00000
Ca2 Ca 4 a 0.03257 0.65995 0.23386 1.00000
Ca3 Ca 4 a 0.25309 0.11288 0.27955 1.00000
O1 O 4 a 0.26570 0.19440 0.04590 1.00000
O2 O 4 a 0.21710 0.47040 0.17340 1.00000
O3 O 4 a 0.07170 0.02970 0.20090 1.00000
O4 O 4 a 0.45470 0.17150 0.25670 1.00000
O5 O 4 a 0.20090 0.37500 0.44860 1.00000
O6 O 4 a 0.14560 0.03220 0.51160 1.00000
O7 O 4 a 0.00610 0.30090 0.53050 1.00000
O8 O 4 a 0.17340 0.29460 0.66350 1.00000
O9 O 4 a 0.10690 0.02200 0.79810 1.00000
O10 O 4 a 0.28700 0.20460 0.83360 1.00000
O11 O 4 a 0.45630 0.13700 0.94580 1.00000
O12 O 4 a 0.15040 0.47680 0.96680 1.00000
```

$CaB_2O_4$ (III): A2BC4_oP84_33_6a_3a_12a - POSCAR

```
A2BC4_oP84_33_6a_3a_12a & a,b/a,c/a,x1,y1,z1,x2,y2,z2,x3,y3,z3,x4,y4,z4,
    ↪ x5,y5,z5,x6,y6,z6,x7,y7,z7,x8,y8,z8,x9,y9,z9,x10,y10,z10,x11,
    ↪ y11,z11,x12,y12,z12,x13,y13,z13,x14,y14,z14,x15,y15,z15,x16,y16
    ↪ ,z16,x17,y17,z17,x18,y18,z18,x19,y19,z19,x20,y20,z20,x21,y21,
    ↪ z21 --params=11.38,0.554217926186,0.993321616872,0.1401,0.2529,
    ↪ 0.539,0.251,0.4193,0.0515,0.2131,0.1191,0.7403,0.3222,0.1006,
    ↪ 0.9442,0.4658,0.0183,0.48,0.4748,0.3758,0.2477,0.07351,0.05043,
    ↪ 0.0,0.03257,0.65995,0.23386,0.25309,0.11288,0.27955,0.2657,
    ↪ 0.1944,0.0459,0.2171,0.4704,0.1734,0.0717,0.0297,0.2009,0.4547,
    ↪ 0.1715,0.2567,0.2009,0.375,0.4486,0.1456,0.0322,0.5116,0.0061,
    ↪ 0.3009,0.5305,0.1734,0.2946,0.6635,0.1069,0.022,0.7981,0.287,
    ↪ 0.2046,0.8336,0.4563,0.137,0.9458,0.1504,0.4768,0.9668 & Pna2_{
    ↪ 1} C_{2v}^{9} #33 (a^21) & oP84 & None & B2CaO4 & B2CaO4 & M.
    ↪ Marezio and J. P. Remeika and P. D. Dernier, Acta Crystallogr.
    ↪ Sect. B Struct. Sci. 25, 955-964 (1969)
   1.00000000000000
  11.38000000000000   0.00000000000000   0.00000000000000
   0.00000000000000   6.30700000000000   0.00000000000000
   0.00000000000000   0.00000000000000  11.30400000000000
     B     Ca     O
    24     12    48
Direct
   0.14010000000000   0.25290000000000   0.53900000000000   B   (4a)
  -0.14010000000000  -0.25290000000000   1.03900000000000   B   (4a)
   0.64010000000000   0.24710000000000   0.53900000000000   B   (4a)
   0.35990000000000   0.75290000000000   1.03900000000000   B   (4a)
   0.25100000000000   0.41930000000000   0.05150000000000   B   (4a)
  -0.25100000000000  -0.41930000000000   0.55150000000000   B   (4a)
   0.75100000000000   0.08070000000000   0.05150000000000   B   (4a)
   0.24900000000000   0.91930000000000   0.55150000000000   B   (4a)
   0.21310000000000   0.11910000000000   0.74030000000000   B   (4a)
  -0.21310000000000  -0.11910000000000   1.24030000000000   B   (4a)
   0.71310000000000   0.38090000000000   0.74030000000000   B   (4a)
   0.28690000000000   0.61910000000000   1.24030000000000   B   (4a)
   0.32220000000000   0.10060000000000   0.94420000000000   B   (4a)
  -0.32220000000000  -0.10060000000000   1.44420000000000   B   (4a)
   0.82220000000000   0.39940000000000   0.94420000000000   B   (4a)
   0.17780000000000   0.60060000000000   1.44420000000000   B   (4a)
   0.46580000000000   0.01830000000000   0.48000000000000   B   (4a)
  -0.46580000000000  -0.01830000000000   0.98000000000000   B   (4a)
   0.96580000000000   0.48170000000000   0.48000000000000   B   (4a)
   0.03420000000000   0.51830000000000   0.98000000000000   B   (4a)
   0.47480000000000   0.37580000000000   0.24770000000000   B   (4a)
  -0.47480000000000  -0.37580000000000   0.74770000000000   B   (4a)
   0.97480000000000   0.12420000000000   0.24770000000000   B   (4a)
   0.02520000000000   0.87580000000000   0.74770000000000   B   (4a)
   0.07351000000000   0.05043000000000   0.00000000000000   Ca  (4a)
  -0.07351000000000  -0.05043000000000   0.50000000000000   Ca  (4a)
   0.57351000000000   0.44957000000000   0.00000000000000   Ca  (4a)
   0.42649000000000   0.55043000000000   0.50000000000000   Ca  (4a)
   0.03257000000000   0.65995000000000   0.23386000000000   Ca  (4a)
  -0.03257000000000  -0.65995000000000   0.73386000000000   Ca  (4a)
   0.53257000000000  -0.15995000000000   0.23386000000000   Ca  (4a)
   0.46743000000000   1.15995000000000   0.73386000000000   Ca  (4a)
   0.25309000000000   0.11288000000000   0.27955000000000   Ca  (4a)
  -0.25309000000000  -0.11288000000000   0.77955000000000   Ca  (4a)
   0.75309000000000   0.38712000000000   0.27955000000000   Ca  (4a)
   0.24691000000000   0.61288000000000   0.77955000000000   Ca  (4a)
   0.26570000000000   0.19440000000000   0.04590000000000   O   (4a)
  -0.26570000000000  -0.19440000000000   0.54590000000000   O   (4a)
   0.76570000000000   0.30560000000000   0.04590000000000   O   (4a)
   0.23430000000000   0.69440000000000   0.54590000000000   O   (4a)
   0.21710000000000   0.47040000000000   0.17340000000000   O   (4a)
  -0.21710000000000  -0.47040000000000   0.67340000000000   O   (4a)
   0.71710000000000   0.02960000000000   0.17340000000000   O   (4a)
   0.28290000000000   0.97040000000000   0.67340000000000   O   (4a)
   0.07170000000000   0.02970000000000   0.20090000000000   O   (4a)
  -0.07170000000000  -0.02970000000000   0.70090000000000   O   (4a)
   0.57170000000000   0.47030000000000   0.20090000000000   O   (4a)
   0.42830000000000   0.52970000000000   0.70090000000000   O   (4a)
   0.45470000000000   0.17150000000000   0.25670000000000   O   (4a)
  -0.45470000000000  -0.17150000000000   0.75670000000000   O   (4a)
   0.95470000000000   0.32850000000000   0.25670000000000   O   (4a)
   0.04530000000000   0.67150000000000   0.75670000000000   O   (4a)
   0.20090000000000   0.37500000000000   0.44860000000000   O   (4a)
  -0.20090000000000  -0.37500000000000   0.94860000000000   O   (4a)
   0.70090000000000   0.12500000000000   0.44860000000000   O   (4a)
   0.29910000000000   0.87500000000000   0.94860000000000   O   (4a)
   0.14560000000000   0.03220000000000   0.51160000000000   O   (4a)
  -0.14560000000000  -0.03220000000000   1.01160000000000   O   (4a)
   0.64560000000000   0.46780000000000   0.51160000000000   O   (4a)
   0.35440000000000   0.53220000000000   1.01160000000000   O   (4a)
   0.00610000000000   0.30090000000000   0.53050000000000   O   (4a)
  -0.00610000000000  -0.30090000000000   1.03050000000000   O   (4a)
   0.50610000000000   0.19910000000000   0.53050000000000   O   (4a)
   0.49390000000000   0.80090000000000   1.03050000000000   O   (4a)
   0.17340000000000   0.29460000000000   0.66350000000000   O   (4a)
  -0.17340000000000  -0.29460000000000   1.16350000000000   O   (4a)
   0.67340000000000   0.20540000000000   0.66350000000000   O   (4a)
   0.32660000000000   0.79460000000000   1.16350000000000   O   (4a)
   0.10690000000000   0.02200000000000   0.79810000000000   O   (4a)
  -0.10690000000000  -0.02200000000000   1.29810000000000   O   (4a)
   0.60690000000000   0.47800000000000   0.79810000000000   O   (4a)
   0.39310000000000   0.52200000000000   1.29810000000000   O   (4a)
   0.28700000000000   0.20460000000000   0.83360000000000   O   (4a)
  -0.28700000000000  -0.20460000000000   1.33360000000000   O   (4a)
   0.78700000000000   0.29540000000000   0.83360000000000   O   (4a)
   0.21300000000000   0.70460000000000   1.33360000000000   O   (4a)
   0.45630000000000   0.13700000000000   0.94580000000000   O   (4a)
  -0.45630000000000  -0.13700000000000   1.44580000000000   O   (4a)
   0.95630000000000   0.36300000000000   0.94580000000000   O   (4a)
   0.04370000000000   0.63700000000000   1.44580000000000   O   (4a)
   0.15040000000000   0.47680000000000   0.96680000000000   O   (4a)
  -0.15040000000000  -0.47680000000000   1.46680000000000   O   (4a)
   0.65040000000000   0.02320000000000   0.96680000000000   O   (4a)
   0.34960000000000   0.97680000000000   1.46680000000000   O   (4a)
```

Cervantite ($\alpha$-$Sb_2O_4$): A2B_oP24_33_4a_2a - CIF

```
# CIF file
data_findsym-output
_audit_creation_method FINDSYM

_chemical_name_mineral 'Cervantite'
_chemical_formula_sum 'O2 Sb'

loop_
_publ_author_name
 'G. Thornton'
_journal_name_full_name
;
 Acta Crystallographica Section B: Structural Science
;
_journal_volume 33
_journal_year 1977
_journal_page_first 1271
_journal_page_last 1273
_publ_Section_title
;
 A Neutron Diffraction Study of $\alpha$-Sb$_{2}$O$_{3}$
;

# Found in The American Mineralogist Crystal Structure Database, 2003

_aflow_title 'Cervantite ($\alpha$-Sb$_{2}$O$_{4}$) Structure'
_aflow_proto 'A2B_oP24_33_4a_2a'
_aflow_params 'a,b/a,c/a,x_{1},y_{1},z_{1},x_{2},y_{2},z_{2},x_{3},y_{3
    ↪ },z_{3},x_{4},y_{4},z_{4},x_{5},y_{5},z_{5},x_{6},y_{6},z_{6}'
_aflow_params_values '5.456,0.882331378299,2.16037390029,0.34,0.177,
    ↪ 0.0962,0.159,0.71,0.195,0.086,0.208,0.312,0.33,0.84,0.41,-0.032
    ↪ ,0.036,0.009,0.366,0.016,0.253'
_aflow_Strukturbericht 'None'
_aflow_Pearson 'oP24'

_symmetry_space_group_name_H-M "P n a 21"
_symmetry_Int_Tables_number 33

_cell_length_a      5.45600
_cell_length_b      4.81400
_cell_length_c      11.78700
_cell_angle_alpha  90.00000
_cell_angle_beta   90.00000
_cell_angle_gamma  90.00000

loop_
_space_group_symop_id
_space_group_symop_operation_xyz
1 x,y,z
2 -x,-y,z+1/2
3 -x+1/2,y+1/2,z+1/2
4 x+1/2,-y+1/2,z
```

```
loop_
_atom_site_label
_atom_site_type_symbol
_atom_site_symmetry_multiplicity
_atom_site_Wyckoff_label
_atom_site_fract_x
_atom_site_fract_y
_atom_site_fract_z
_atom_site_occupancy
O1 O 4 a 0.34000 0.17700 0.09620 1.00000
O2 O 4 a 0.15900 0.71000 0.19500 1.00000
O3 O 4 a 0.08600 0.20800 0.31200 1.00000
O4 O 4 a 0.33000 0.84000 0.41000 1.00000
Sb1 Sb 4 a -0.03200 0.03600 0.00900 1.00000
Sb2 Sb 4 a 0.36600 0.01600 0.25300 1.00000
```

Cervantite ($\alpha$-$Sb_2O_4$): A2B_oP24_33_4a_2a - POSCAR

```
A2B_oP24_33_4a_2a & a,b/a,c/a,x1,y1,z1,x2,y2,z2,x3,y3,z3,x4,y4,z4,x5,y5,
    ↪ z5,x6,y6,z6 --params=5.456,0.882331378299,2.16037390029,0.34,
    ↪ 0.177,0.0962,0.159,0.71,0.195,0.086,0.208,0.312,0.33,0.84,0.41
    ↪ ,-0.032,0.036,0.009,0.366,0.016,0.253 & Pna2_{1} C_{2v}^{9} #33
    ↪ (a^6) & oP24 & None & O2Sb & Cervantite & G. Thornton, Acta
    ↪ Crystallogr. Sect. B Struct. Sci. 33, 1271-1273 (1977)
   1.00000000000000
   5.45600000000000   0.00000000000000   0.00000000000000
   0.00000000000000   4.81400000000000   0.00000000000000
   0.00000000000000   0.00000000000000  11.78700000000000
     O     Sb
    16      8
Direct
   0.34000000000000   0.17700000000000   0.09620000000000   O   (4a)
  -0.34000000000000  -0.17700000000000   0.59620000000000   O   (4a)
   0.84000000000000   0.32300000000000   0.09620000000000   O   (4a)
   0.16000000000000   0.67700000000000   0.59620000000000   O   (4a)
   0.15900000000000   0.71000000000000   0.19500000000000   O   (4a)
  -0.15900000000000  -0.71000000000000   0.69500000000000   O   (4a)
   0.65900000000000  -0.21000000000000   0.19500000000000   O   (4a)
   0.34100000000000   1.21000000000000   0.69500000000000   O   (4a)
   0.08600000000000   0.20800000000000   0.31200000000000   O   (4a)
  -0.08600000000000  -0.20800000000000   0.81200000000000   O   (4a)
   0.58600000000000   0.29200000000000   0.31200000000000   O   (4a)
   0.41400000000000   0.70800000000000   0.81200000000000   O   (4a)
   0.33000000000000   0.84000000000000   0.41000000000000   O   (4a)
  -0.33000000000000  -0.84000000000000   0.91000000000000   O   (4a)
   0.83000000000000  -0.34000000000000   0.41000000000000   O   (4a)
   0.17000000000000   1.34000000000000   0.91000000000000   O   (4a)
  -0.03200000000000   0.03600000000000   0.00900000000000   Sb  (4a)
   0.03200000000000  -0.03600000000000   0.50900000000000   Sb  (4a)
   0.46800000000000   0.46400000000000   0.00900000000000   Sb  (4a)
   0.53200000000000   0.53600000000000   0.50900000000000   Sb  (4a)
   0.36600000000000   0.01600000000000   0.25300000000000   Sb  (4a)
  -0.36600000000000  -0.01600000000000   0.75300000000000   Sb  (4a)
   0.86600000000000   0.48400000000000   0.25300000000000   Sb  (4a)
   0.13400000000000   0.51600000000000   0.75300000000000   Sb  (4a)
```

$CsB_4O_6F$: A4BCD6_oP48_33_4a_a_a_6a - CIF

```
# CIF file
data_findsym-output
_audit_creation_method FINDSYM

_chemical_name_mineral 'B4CsFO6'
_chemical_formula_sum 'B4 Cs F O6'

loop_
_publ_author_name
 'X. Wang'
 'Y. Wang'
 'B. Zhang'
 'F. Zhang'
 'Z. Yang'
 'S. Pan'
_journal_name_full_name
;
 Angewandte Chemie
;
_journal_volume 129
_journal_year 2017
_journal_page_first 14307
_journal_page_last 14311
_publ_Section_title
;
 CsB$_{4}$O$_{6}$F: A Congruent-Melting Deep-Ultraviolet Nonlinear
    ↪ Optical Material by Combining Superior Functional Units
;

_aflow_title 'CsB$_{4}$O$_{6}$F Structure'
_aflow_proto 'A4BCD6_oP48_33_4a_a_a_6a'
_aflow_params 'a,b/a,c/a,x_{1},y_{1},z_{1},x_{2},y_{2},z_{2},x_{3},y_{3
    ↪ },z_{3},x_{4},y_{4},z_{4},x_{5},y_{5},z_{5},x_{6},y_{6},z_{6},
    ↪ x_{7},y_{7},z_{7},x_{8},y_{8},z_{8},x_{9},y_{9},z_{9},x_{10},y_
    ↪ {10},z_{10},x_{11},y_{11},z_{11},x_{12},y_{12},z_{12}'
_aflow_params_values '7.9241,1.43859870522,0.840953546775,0.2971,0.5446,
    ↪ 0.6555,0.2977,0.5006,0.0044,0.2304,0.2003,0.3942,0.2327,0.4189,
    ↪ 0.348,0.49584,0.22117,0.852,0.0566,0.4371,0.3269,0.306,0.4658,
    ↪ 0.8049,0.2888,0.6655,0.701,0.3,0.5194,0.4588,0.2626,0.3109,
    ↪ 0.4526,0.3038,0.4165,0.1445,0.2784,0.6167,0.0497'
_aflow_Strukturbericht 'None'
_aflow_Pearson 'oP48'

_symmetry_space_group_name_H-M "P n a 21"
_symmetry_Int_Tables_number 33

_cell_length_a    7.92410
_cell_length_b    11.39960
_cell_length_c    6.66380
_cell_angle_alpha 90.00000
_cell_angle_beta  90.00000
_cell_angle_gamma 90.00000

loop_
_space_group_symop_id
_space_group_symop_operation_xyz
1 x,y,z
2 -x,-y,z+1/2
3 -x+1/2,y+1/2,z+1/2
4 x+1/2,-y+1/2,z

loop_
_atom_site_label
_atom_site_type_symbol
_atom_site_symmetry_multiplicity
_atom_site_Wyckoff_label
_atom_site_fract_x
_atom_site_fract_y
_atom_site_fract_z
_atom_site_occupancy
B1 B 4 a 0.29710 0.54460 0.65550 1.00000
B2 B 4 a 0.29770 0.50060 0.00440 1.00000
B3 B 4 a 0.23040 0.20030 0.39420 1.00000
B4 B 4 a 0.23270 0.41890 0.34800 1.00000
Cs1 Cs 4 a 0.49584 0.22117 0.85200 1.00000
F1 F 4 a 0.05660 0.43710 0.32690 1.00000
O1 O 4 a 0.30600 0.46580 0.80490 1.00000
O2 O 4 a 0.28880 0.66550 0.70100 1.00000
O3 O 4 a 0.30000 0.51940 0.45880 1.00000
O4 O 4 a 0.26260 0.31090 0.45260 1.00000
O5 O 4 a 0.30380 0.41650 0.14450 1.00000
O6 O 4 a 0.27840 0.61670 0.04970 1.00000
```

$CsB_4O_6F$: A4BCD6_oP48_33_4a_a_a_6a - POSCAR

```
A4BCD6_oP48_33_4a_a_a_6a & a,b/a,c/a,x1,y1,z1,x2,y2,z2,x3,y3,z3,x4,y4,z4
    ↪ ,x5,y5,z5,x6,y6,z6,x7,y7,z7,x8,y8,z8,x9,y9,z9,x10,y10,z10,x11,
    ↪ y11,z11,x12,y12,z12 --params=7.9241,1.43859870522,
    ↪ 0.840953546775,0.2971,0.5446,0.6555,0.2977,0.5006,0.0044,0.2304
    ↪ ,0.2003,0.3942,0.2327,0.4189,0.348,0.49584,0.22117,0.852,0.0566
    ↪ ,0.4371,0.3269,0.306,0.4658,0.8049,0.2888,0.6655,0.701,0.3,
    ↪ 0.5194,0.4588,0.2626,0.3109,0.4526,0.3038,0.4165,0.1445,0.2784,
    ↪ 0.6167,0.0497 & Pna2_{1} C_{2v}^{9} #33 (a^12) & oP48 & None &
    ↪ B4CsFO6 & B4CsFO6 & X. Wang et al., Angew. Chem. 129,
    ↪ 14307-14311 (2017)
   1.00000000000000
   7.92410000000000   0.00000000000000   0.00000000000000
   0.00000000000000  11.39960000000000   0.00000000000000
   0.00000000000000   0.00000000000000   6.66380000000000
     B     Cs     F     O
    16      4     4    24
Direct
   0.29710000000000   0.54460000000000   0.65550000000000   B    (4a)
  -0.29710000000000  -0.54460000000000   1.15550000000000   B    (4a)
   0.79710000000000  -0.04460000000000   0.65550000000000   B    (4a)
   0.20290000000000   1.04460000000000   1.15550000000000   B    (4a)
   0.29770000000000   0.50060000000000   0.00440000000000   B    (4a)
  -0.29770000000000  -0.50060000000000   0.50440000000000   B    (4a)
   0.79770000000000  -0.00060000000000   0.00440000000000   B    (4a)
   0.20230000000000   1.00060000000000   0.50440000000000   B    (4a)
   0.23040000000000   0.20030000000000   0.39420000000000   B    (4a)
  -0.23040000000000  -0.20030000000000   0.89420000000000   B    (4a)
   0.73040000000000   0.29970000000000   0.39420000000000   B    (4a)
   0.26960000000000   0.70030000000000   0.89420000000000   B    (4a)
   0.23270000000000   0.41890000000000   0.34800000000000   B    (4a)
  -0.23270000000000  -0.41890000000000   0.84800000000000   B    (4a)
   0.73270000000000   0.08110000000000   0.34800000000000   B    (4a)
   0.26730000000000   0.91890000000000   0.84800000000000   B    (4a)
   0.49584000000000   0.22117000000000   0.85200000000000   Cs   (4a)
  -0.49584000000000  -0.22117000000000   1.35200000000000   Cs   (4a)
   0.99584000000000   0.27883000000000   0.85200000000000   Cs   (4a)
   0.00416000000000   0.72117000000000   1.35200000000000   Cs   (4a)
   0.05660000000000   0.43710000000000   0.32690000000000   F    (4a)
  -0.05660000000000  -0.43710000000000   0.82690000000000   F    (4a)
   0.55660000000000   0.06290000000000   0.32690000000000   F    (4a)
   0.44340000000000   0.93710000000000   0.82690000000000   F    (4a)
   0.30600000000000   0.46580000000000   0.80490000000000   O    (4a)
  -0.30600000000000  -0.46580000000000   1.30490000000000   O    (4a)
   0.80600000000000   0.03420000000000   0.80490000000000   O    (4a)
   0.19400000000000   0.96580000000000   1.30490000000000   O    (4a)
   0.28880000000000   0.66550000000000   0.70100000000000   O    (4a)
  -0.28880000000000  -0.66550000000000   1.20100000000000   O    (4a)
   0.78880000000000  -0.16550000000000   0.70100000000000   O    (4a)
   0.21120000000000   1.16550000000000   1.20100000000000   O    (4a)
   0.30000000000000   0.51940000000000   0.45880000000000   O    (4a)
  -0.30000000000000  -0.51940000000000   0.95880000000000   O    (4a)
   0.80000000000000  -0.01940000000000   0.45880000000000   O    (4a)
   0.20000000000000   1.01940000000000   0.95880000000000   O    (4a)
   0.26260000000000   0.31090000000000   0.45260000000000   O    (4a)
  -0.26260000000000  -0.31090000000000   0.95260000000000   O    (4a)
   0.76260000000000   0.18910000000000   0.45260000000000   O    (4a)
   0.23740000000000   0.81090000000000   0.95260000000000   O    (4a)
   0.30380000000000   0.41650000000000   0.14450000000000   O    (4a)
  -0.30380000000000  -0.41650000000000   0.64450000000000   O    (4a)
   0.80380000000000   0.08350000000000   0.14450000000000   O    (4a)
   0.19620000000000   0.91650000000000   0.64450000000000   O    (4a)
   0.27840000000000   0.61670000000000   0.04970000000000   O    (4a)
  -0.27840000000000  -0.61670000000000   0.54970000000000   O    (4a)
   0.77840000000000  -0.11670000000000   0.04970000000000   O    (4a)
   0.22160000000000   1.11670000000000   0.54970000000000   O    (4a)
```

LiGaO$_2$: ABC2_oP16_33_a_a_2a - CIF

```
# CIF file
data_findsym-output
_audit_creation_method FINDSYM

_chemical_name_mineral 'GaLiO2'
_chemical_formula_sum 'Ga Li O2'

loop_
_publ_author_name
 'M. Marezio'
_journal_name_full_name
;
 Acta Cristallographica
;
_journal_volume 18
_journal_year 1965
_journal_page_first 481
_journal_page_last 484
_publ_Section_title
;
 The Crystal Structure of LiGaO$_{2}$
;

_aflow_title 'LiGaO$_{2}$ Structure'
_aflow_proto 'ABC2_oP16_33_a_a_2a'
_aflow_params 'a,b/a,c/a,x_{1},y_{1},z_{1},x_{2},y_{2},z_{2},x_{3},y_{3
    ↪ },z_{3},x_{4},y_{4},z_{4}'
_aflow_params_values '5.402,1.17956312477,0.926878933728,0.0821,0.1263,
    ↪ 0.0,0.4207,0.1267,0.4936,0.4066,0.1388,0.8972,0.0697,0.1121,
    ↪ 0.3708'
_aflow_Strukturbericht 'None'
_aflow_Pearson 'oP16'

_symmetry_space_group_name_H-M "P n a 21"
_symmetry_Int_Tables_number 33

_cell_length_a    5.40200
_cell_length_b    6.37200
_cell_length_c    5.00700
_cell_angle_alpha 90.00000
_cell_angle_beta  90.00000
_cell_angle_gamma 90.00000

loop_
_space_group_symop_id
_space_group_symop_operation_xyz
1 x,y,z
2 -x,-y,z+1/2
3 -x+1/2,y+1/2,z+1/2
4 x+1/2,-y+1/2,z

loop_
_atom_site_label
_atom_site_type_symbol
_atom_site_symmetry_multiplicity
_atom_site_Wyckoff_label
_atom_site_fract_x
_atom_site_fract_y
_atom_site_fract_z
_atom_site_occupancy
Ga1 Ga 4 a 0.08210 0.12630 0.00000 1.00000
Li1 Li 4 a 0.42070 0.12670 0.49360 1.00000
O1 O 4 a 0.40660 0.13880 0.89720 1.00000
O2 O 4 a 0.06970 0.11210 0.37080 1.00000
```

LiGaO$_2$: ABC2_oP16_33_a_a_2a - POSCAR

```
ABC2_oP16_33_a_a_2a & a,b/a,c/a,x1,y1,z1,x2,y2,z2,x3,y3,z3,x4,y4,z4 --
    ↪ params=5.402,1.17956312477,0.926878933728,0.0821,0.1263,0.0,
    ↪ 0.4207,0.1267,0.4936,0.4066,0.1388,0.8972,0.0697,0.1121,0.3708
    ↪ & Pna2_{1} C_{2v}^{9} #33 (a^4) & oP16 & None & GaLiO2 & GaLiO2
    ↪ & M. Marezio, Acta Cryst. 18, 481-484 (1965)
   1.00000000000000
   5.40200000000000   0.00000000000000   0.00000000000000
   0.00000000000000   6.37200000000000   0.00000000000000
   0.00000000000000   0.00000000000000   5.00700000000000
    Ga    Li    O
     4     4    8
Direct
   0.08210000000000   0.12630000000000   0.00000000000000   Ga   (4a)
  -0.08210000000000  -0.12630000000000   0.50000000000000   Ga   (4a)
   0.58210000000000   0.37370000000000   0.00000000000000   Ga   (4a)
   0.41790000000000   0.62630000000000   0.50000000000000   Ga   (4a)
   0.42070000000000   0.12670000000000   0.49360000000000   Li   (4a)
  -0.42070000000000  -0.12670000000000   0.99360000000000   Li   (4a)
   0.92070000000000   0.37330000000000   0.49360000000000   Li   (4a)
   0.07930000000000   0.62670000000000   0.99360000000000   Li   (4a)
   0.40660000000000   0.13880000000000   0.89720000000000   O    (4a)
  -0.40660000000000  -0.13880000000000   1.39720000000000   O    (4a)
   0.90660000000000   0.36120000000000   0.89720000000000   O    (4a)
   0.09340000000000   0.63880000000000   1.39720000000000   O    (4a)
   0.06970000000000   0.11210000000000   0.37080000000000   O    (4a)
  -0.06970000000000  -0.11210000000000   0.87080000000000   O    (4a)
   0.56970000000000   0.38790000000000   0.37080000000000   O    (4a)
   0.43030000000000   0.61210000000000   0.87080000000000   O    (4a)
```

$\gamma$-LiIO$_3$: ABC3_oP20_33_a_a_3a - CIF

```
# CIF file
data_findsym-output
_audit_creation_method FINDSYM

_chemical_name_mineral 'ILiO3'
_chemical_formula_sum 'I Li O3'

loop_
_publ_author_name
 'R. Liminga'
 'C. Svensson'
 'J. Albertsson'
 'S. C. Abrahams'
_journal_name_full_name
;
 Journal of Chemical Physics
;
_journal_volume 77
_journal_year 1982
_journal_page_first 4222
_journal_page_last 4226
_publ_Section_title
;
 Gamma-lithium iodate structure at 515~K and the $\alpha$-LiIO$_{3}$ to
    ↪ $\gamma$-LiIO$_{3}$, $\gamma$-LiIO$_{3}$ to $\beta$-LiIO$_{3}$
    ↪ phase transitions
;

_aflow_title '$\gamma$-LiIO$_{3}$ Structure'
_aflow_proto 'ABC3_oP20_33_a_a_3a'
_aflow_params 'a,b/a,c/a,x_{1},y_{1},z_{1},x_{2},y_{2},z_{2},x_{3},y_{3
    ↪ },z_{3},x_{4},y_{4},z_{4},x_{5},y_{5},z_{5}'
_aflow_params_values '9.422,0.622054765443,0.562619401401,0.3179,0.0751,
    ↪ 0.0,0.0,0.0,0.0706,0.1298,0.0556,0.8087,0.4117,0.8806,0.8382,
    ↪ 0.3663,0.3439,0.8147'
_aflow_Strukturbericht 'None'
_aflow_Pearson 'oP20'

_symmetry_space_group_name_H-M "P n a 21"
_symmetry_Int_Tables_number 33

_cell_length_a    9.42200
_cell_length_b    5.86100
_cell_length_c    5.30100
_cell_angle_alpha 90.00000
_cell_angle_beta  90.00000
_cell_angle_gamma 90.00000

loop_
_space_group_symop_id
_space_group_symop_operation_xyz
1 x,y,z
2 -x,-y,z+1/2
3 -x+1/2,y+1/2,z+1/2
4 x+1/2,-y+1/2,z

loop_
_atom_site_label
_atom_site_type_symbol
_atom_site_symmetry_multiplicity
_atom_site_Wyckoff_label
_atom_site_fract_x
_atom_site_fract_y
_atom_site_fract_z
_atom_site_occupancy
I1 I 4 a 0.31790 0.07510 0.00000 1.00000
Li1 Li 4 a 0.00000 0.00000 0.07060 1.00000
O1 O 4 a 0.12980 0.05560 0.80870 1.00000
O2 O 4 a 0.41170 0.88060 0.83820 1.00000
O3 O 4 a 0.36630 0.34390 0.81470 1.00000
```

$\gamma$-LiIO$_3$: ABC3_oP20_33_a_a_3a - POSCAR

```
ABC3_oP20_33_a_a_3a & a,b/a,c/a,x1,y1,z1,x2,y2,z2,x3,y3,z3,x4,y4,z4,x5,
    ↪ y5,z5 --params=9.422,0.622054765443,0.562619401401,0.3179,
    ↪ 0.0751,0.0,0.0,0.0,0.0706,0.1298,0.0556,0.8087,0.4117,0.8806,
    ↪ 0.8382,0.3663,0.3439,0.8147 & Pna2_{1} C_{2v}^{9} #33 (a^5) &
    ↪ oP20 & None & ILiO3 & ILiO3 & R. Liminga et al., J. Chem. Phys.
    ↪ 77, 4222-4226 (1982)
   1.00000000000000
   9.42200000000000   0.00000000000000   0.00000000000000
   0.00000000000000   5.86100000000000   0.00000000000000
   0.00000000000000   0.00000000000000   5.30100000000000
     I    Li    O
     4     4   12
Direct
   0.31790000000000   0.07510000000000   0.00000000000000   I    (4a)
  -0.31790000000000  -0.07510000000000   0.50000000000000   I    (4a)
   0.81790000000000   0.42490000000000   0.00000000000000   I    (4a)
   0.18210000000000   0.57510000000000   0.50000000000000   I    (4a)
   0.00000000000000   0.00000000000000   0.07060000000000   Li   (4a)
   0.00000000000000   0.00000000000000   0.57060000000000   Li   (4a)
   0.50000000000000   0.50000000000000   0.07060000000000   Li   (4a)
   0.50000000000000   0.50000000000000   0.57060000000000   Li   (4a)
   0.12980000000000   0.05560000000000   0.80870000000000   O    (4a)
  -0.12980000000000  -0.05560000000000   1.30870000000000   O    (4a)
   0.62980000000000   0.44440000000000   0.80870000000000   O    (4a)
   0.37020000000000   0.55560000000000   1.30870000000000   O    (4a)
   0.41170000000000   0.88060000000000   0.83820000000000   O    (4a)
  -0.41170000000000  -0.88060000000000   1.33820000000000   O    (4a)
   0.91170000000000  -0.38060000000000   0.83820000000000   O    (4a)
   0.08830000000000   1.38060000000000   1.33820000000000   O    (4a)
   0.36630000000000   0.34390000000000   0.81470000000000   O    (4a)
  -0.36630000000000  -0.34390000000000   1.31470000000000   O    (4a)
   0.86630000000000   0.15610000000000   0.81470000000000   O    (4a)
   0.13370000000000   0.84390000000000   1.31470000000000   O    (4a)
```

MnF$_{2-x}$(OH)$_x$: A2B2CD2_oP14_34_c_c_a_c - CIF

```
# CIF file
```

```
data_findsym-output
_audit_creation_method FINDSYM

_chemical_name_mineral 'F_{2-xHxMnOx'
_chemical_formula_sum 'F2 H2 Mn O2'

loop_
_publ_author_name
 'H. B. Yahia'
 'M. Shikano'
 'H. Kobayashi'
 'M. Avdeev'
 'S. Liu'
 'C. D. Ling'
_journal_name_full_name
;
 Physical Chemistry Chemical Physics
;
_journal_volume 15
_journal_year 2013
_journal_page_first 13061
_journal_page_last 13069
_publ_Section_title
;
 Synthesis and characterization of the crystal structure and magnetic
    ↪ properties of the hydroxyfluoride MnF$_{2-x}$(OH)$_{x}$ ($x \
    ↪ approx 0.8$)
;

_aflow_title 'MnF$_{2-x}$(OH)$_{x}$ Structure'
_aflow_proto 'A2B2CD2_oP14_34_c_c_a_c'
_aflow_params 'a,b/a,c/a,z_{1},x_{2},y_{2},z_{2},x_{3},y_{3},z_{3},x_{4
    ↪ },y_{4},z_{4}'
_aflow_params_values '4.71143,1.11290415012,0.689459463475,0.5,0.2511,
    ↪ 0.1533,0.504,0.425,0.0287,0.543,0.2511,0.1533,0.504'
_aflow_Strukturbericht 'None'
_aflow_Pearson 'oP14'

_symmetry_space_group_name_H-M "P n n 2"
_symmetry_Int_Tables_number 34

_cell_length_a      4.71143
_cell_length_b      5.24337
_cell_length_c      3.24834
_cell_angle_alpha 90.00000
_cell_angle_beta  90.00000
_cell_angle_gamma 90.00000

loop_
_space_group_symop_id
_space_group_symop_operation_xyz
1 x,y,z
2 -x,-y,z
3 -x+1/2,y+1/2,z+1/2
4 x+1/2,-y+1/2,z+1/2

loop_
_atom_site_label
_atom_site_type_symbol
_atom_site_symmetry_multiplicity
_atom_site_Wyckoff_label
_atom_site_fract_x
_atom_site_fract_y
_atom_site_fract_z
_atom_site_occupancy
Mn1 Mn 2 a 0.00000 0.00000 0.50000 1.00000
F1 F 4 c 0.25110 0.15330 0.50400 0.60100
H1 H 4 c 0.42500 0.02870 0.54300 0.39900
O1 O 4 c 0.25110 0.15330 0.50400 0.39900
```

$MnF_{2-x}(OH)_x$: A2B2CD2_oP14_34_c_c_a_c - POSCAR

```
A2B2CD2_oP14_34_c_c_a_c & a,b/a,c/a,z1,x2,y2,z2,x3,y3,z3,x4,y4,z4 --
    ↪ params=4.71143,1.11290415012,0.689459463475,0.5,0.2511,0.1533,
    ↪ 0.504,0.425,0.0287,0.543,0.2511,0.1533,0.504 & Pnn2 C_{2v}^{10}
    ↪ #34 (ac^3) & oP14 & None & xHxMnOx & F_{2 & H. B. Yahia et al.
    ↪ , Phys. Chem. Chem. Phys. 15, 13061-13069 (2013)
   1.00000000000000
   4.71143000000000   0.00000000000000   0.00000000000000
   0.00000000000000   5.24337000000000   0.00000000000000
   0.00000000000000   0.00000000000000   3.24834000000000
     F     H    Mn     O
     4     4     2     4
Direct
   0.25110000000000   0.15330000000000   0.50400000000000   F   (4c)
  -0.25110000000000  -0.15330000000000   0.50400000000000   F   (4c)
   0.75110000000000   0.34670000000000   1.00400000000000   F   (4c)
   0.24890000000000   0.65330000000000   1.00400000000000   F   (4c)
   0.42500000000000   0.02870000000000   0.54300000000000   H   (4c)
  -0.42500000000000  -0.02870000000000   0.54300000000000   H   (4c)
   0.92500000000000   0.47130000000000   1.04300000000000   H   (4c)
   0.07500000000000   0.52870000000000   1.04300000000000   H   (4c)
   0.00000000000000   0.00000000000000   0.50000000000000   Mn  (2a)
   0.50000000000000   0.50000000000000   1.00000000000000   Mn  (2a)
   0.25110000000000   0.15330000000000   0.50400000000000   O   (4c)
  -0.25110000000000  -0.15330000000000   0.50400000000000   O   (4c)
   0.75110000000000   0.34670000000000   1.00400000000000   O   (4c)
   0.24890000000000   0.65330000000000   1.00400000000000   O   (4c)
```

$Si_2N_2O$: A2BC2_oC20_36_b_a_b - CIF

```
# CIF file
data_findsym-output
_audit_creation_method FINDSYM

_chemical_name_mineral 'N2OSi2'
_chemical_formula_sum 'N2 O Si2'

loop_
_publ_author_name
 'I. Idrestedt'
 'C. Brosset'
_journal_name_full_name
;
 Acta Chemica Scandinavica
;
_journal_volume 18
_journal_year 1964
_journal_page_first 1879
_journal_page_last 1886
_publ_Section_title
;
 Structure of Si$_{2}$N$_{2}$O
;

_aflow_title 'Si$_{2}$N$_{2}$O Structure'
_aflow_proto 'A2BC2_oC20_36_b_a_b'
_aflow_params 'a,b/a,c/a,y_{1},z_{1},x_{2},y_{2},z_{2},x_{3},y_{3},z_{3
    ↪ }'
_aflow_params_values '8.843,0.618907610539,0.546760149271,0.214,0.23,
    ↪ 0.218,0.121,0.642,0.1763,0.1509,0.2898'
_aflow_Strukturbericht 'None'
_aflow_Pearson 'oC20'

_symmetry_space_group_name_H-M "C m c 21"
_symmetry_Int_Tables_number 36

_cell_length_a      8.84300
_cell_length_b      5.47300
_cell_length_c      4.83500
_cell_angle_alpha 90.00000
_cell_angle_beta  90.00000
_cell_angle_gamma 90.00000

loop_
_space_group_symop_id
_space_group_symop_operation_xyz
1 x,y,z
2 -x,-y,z+1/2
3 -x,y,z
4 x,-y,z+1/2
5 x+1/2,y+1/2,z
6 -x+1/2,-y+1/2,z+1/2
7 -x+1/2,y+1/2,z
8 x+1/2,-y+1/2,z+1/2

loop_
_atom_site_label
_atom_site_type_symbol
_atom_site_symmetry_multiplicity
_atom_site_Wyckoff_label
_atom_site_fract_x
_atom_site_fract_y
_atom_site_fract_z
_atom_site_occupancy
O1 O 4 a 0.00000 0.21400 0.23000 1.00000
N1 N 8 b 0.21800 0.12100 0.64200 1.00000
Si1 Si 8 b 0.17630 0.15090 0.28980 1.00000
```

$Si_2N_2O$: A2BC2_oC20_36_b_a_b - POSCAR

```
A2BC2_oC20_36_b_a_b & a,b/a,c/a,y1,z1,x2,y2,z2,x3,y3,z3 --params=8.843,
    ↪ 0.618907610539,0.546760149271,0.214,0.23,0.218,0.121,0.642,
    ↪ 0.1763,0.1509,0.2898 & Cmc2_{1} C_{2v}^{12} #36 (ab^2) & oC20 &
    ↪ None & N2OSi2 & N2OSi2 & I. Idrestedt and C. Brosset, Acta
    ↪ Chem. Scand. 18, 1879-1886 (1964)
   1.00000000000000
   4.42150000000000  -2.73650000000000   0.00000000000000
   4.42150000000000   2.73650000000000   0.00000000000000
   0.00000000000000   0.00000000000000   4.83500000000000
     N     O    Si
     4     2     4
Direct
   0.09700000000000   0.33900000000000   0.64200000000000   N   (8b)
  -0.09700000000000  -0.33900000000000   1.14200000000000   N   (8b)
   0.33900000000000   0.09700000000000   1.14200000000000   N   (8b)
  -0.33900000000000  -0.09700000000000   0.64200000000000   N   (8b)
  -0.21400000000000   0.21400000000000   0.23000000000000   O   (4a)
   0.21400000000000  -0.21400000000000   0.73000000000000   O   (4a)
   0.02540000000000   0.32720000000000   0.28980000000000   Si  (8b)
  -0.02540000000000  -0.32720000000000   0.78980000000000   Si  (8b)
   0.32720000000000   0.02540000000000   0.78980000000000   Si  (8b)
  -0.32720000000000  -0.02540000000000   0.28980000000000   Si  (8b)
```

$Bi_2GeO_5$: A2BC5_oC32_36_b_a_a2b - CIF

```
# CIF file
data_findsym-output
_audit_creation_method FINDSYM

_chemical_name_mineral 'Bi2GeO5'
_chemical_formula_sum 'Bi2 Ge O5'

loop_
_publ_author_name
 'B. Aurivillius'
 'C.-I. Lindblom'
 'P. St\\'{e}nson'
_journal_name_full_name
;
```

```
 Acta Chemica Scandinavica
;
_journal_volume 18
_journal_year 1964
_journal_page_first 1555
_journal_page_last 1557
_publ_Section_title
;
 The Crystal Structure of Bi$_{2}$GeO$_{5}$
;

_aflow_title 'Bi$_{2}$GeO$_{5}$ Structure'
_aflow_proto 'A2BC5_oC32_36_b_a_a2b'
_aflow_params 'a,b/a,c/a,y_{1},z_{1},y_{2},z_{2},x_{3},y_{3},z_{3},x_{4
    ↪ },y_{4},z_{4},x_{5},y_{5},z_{5}'
_aflow_params_values '15.69,0.350031867431,0.343084767368,0.68572,
    ↪ 0.20003,0.5803,0.4908,0.16758,0.21738,0.25,0.0949,0.1368,0.6482
    ↪ ,0.2541,0.4777,0.4621'
_aflow_Strukturbericht 'None'
_aflow_Pearson 'oC32'

_symmetry_space_group_name_H-M "C m c 21"
_symmetry_Int_Tables_number 36

_cell_length_a      15.69000
_cell_length_b      5.49200
_cell_length_c      5.38300
_cell_angle_alpha 90.00000
_cell_angle_beta  90.00000
_cell_angle_gamma 90.00000

loop_
_space_group_symop_id
_space_group_symop_operation_xyz
1 x,y,z
2 -x,-y,z+1/2
3 -x,y,z
4 x,-y,z+1/2
5 x+1/2,y+1/2,z
6 -x+1/2,-y+1/2,z+1/2
7 -x+1/2,y+1/2,z
8 x+1/2,-y+1/2,z+1/2

loop_
_atom_site_label
_atom_site_type_symbol
_atom_site_symmetry_multiplicity
_atom_site_Wyckoff_label
_atom_site_fract_x
_atom_site_fract_y
_atom_site_fract_z
_atom_site_occupancy
Ge1 Ge 4 a 0.00000 0.68572 0.20003 1.00000
O1 O 4 a 0.00000 0.58030 0.49080 1.00000
Bi1 Bi 8 b 0.16758 0.21738 0.25000 1.00000
O2 O 8 b 0.09490 0.13680 0.64820 1.00000
O3 O 8 b 0.25410 0.47770 0.46210 1.00000
```

$Bi_2GeO_5$: A2BC5_oC32_36_b_a_a2b - POSCAR

```
A2BC5_oC32_36_b_a_a2b & a,b/a,c/a,y1,z1,y2,z2,x3,y3,z3,x4,y4,z4,x5,y5,z5
    ↪ --params=15.69,0.350031867431,0.343084767368,0.68572,0.20003,
    ↪ 0.5803,0.4908,0.16758,0.21738,0.25,0.0949,0.1368,0.6482,0.2541,
    ↪ 0.4777,0.4621 & Cmc2_{1} C_{2v}^{12} #36 (a^2b^3) & oC32 & None
    ↪ & Bi2GeO5 & Bi2GeO5 & B. Aurivillius and C.-I. Lindblom and P.
    ↪ St\'{e}nson, Acta Chem. Scand. 18, 1555-1557 (1964)
   1.00000000000000
   7.84500000000000  -2.74600000000000   0.00000000000000
   7.84500000000000   2.74600000000000   0.00000000000000
   0.00000000000000   0.00000000000000   5.38300000000000
    Bi     Ge      O
     4      2     10
Direct
  -0.04980000000000   0.38496000000000   0.25000000000000   Bi   (8b)
   0.04980000000000  -0.38496000000000   0.75000000000000   Bi   (8b)
   0.38496000000000  -0.04980000000000   0.75000000000000   Bi   (8b)
  -0.38496000000000   0.04980000000000   0.25000000000000   Bi   (8b)
  -0.68572000000000   0.68572000000000   0.20003000000000   Ge   (4a)
   0.68572000000000  -0.68572000000000   0.70003000000000   Ge   (4a)
  -0.58030000000000   0.58030000000000   0.49080000000000    O   (4a)
   0.58030000000000  -0.58030000000000   0.99080000000000    O   (4a)
  -0.04190000000000   0.23170000000000   0.64820000000000    O   (8b)
   0.04190000000000  -0.23170000000000   1.14820000000000    O   (8b)
   0.23170000000000  -0.04190000000000   1.14820000000000    O   (8b)
  -0.23170000000000   0.04190000000000   0.64820000000000    O   (8b)
  -0.22360000000000   0.73180000000000   0.46210000000000    O   (8b)
   0.22360000000000  -0.73180000000000   0.96210000000000    O   (8b)
   0.73180000000000  -0.22360000000000   0.96210000000000    O   (8b)
  -0.73180000000000   0.22360000000000   0.46210000000000    O   (8b)
```

$Ni_3Si_2$: A3B2_oC80_36_4a4b_2a3b - CIF

```
# CIF file
data_findsym-output
_audit_creation_method FINDSYM

_chemical_name_mineral 'Ni3Si2'
_chemical_formula_sum 'Ni3 Si2'

loop_
_publ_author_name
 'G. Pilstr\"{o}m'
_journal_name_full_name
;
 Acta Chemica Scandinavica
;
_journal_volume 15
_journal_year 1961
_journal_page_first 893
_journal_page_last 902
_publ_Section_title
;
 The Crystal Structure of Ni$_{3}$Si$_{2}$ with some Notes on Ni$_{5}
    ↪ $Si$_{2}$
;

_aflow_title 'Ni$_{3}$Si$_{2}$ Structure'
_aflow_proto 'A3B2_oC80_36_4a4b_2a3b'
_aflow_params 'a,b/a,c/a,y_{1},z_{1},y_{2},z_{2},y_{3},z_{3},y_{4},z_{4
    ↪ },y_{5},z_{5},y_{6},z_{6},x_{7},y_{7},z_{7},x_{8},y_{8},z_{8},
    ↪ x_{9},y_{9},z_{9},x_{10},y_{10},z_{10},x_{11},y_{11},z_{11},x_{
    ↪ 12},y_{12},z_{12},x_{13},y_{13},z_{13}'
_aflow_params_values '12.22901,0.883554760361,0.566194646991,0.0,0.0,
    ↪ 0.2345,0.024,0.233,0.4,0.3814,0.714,0.157,0.712,0.409,0.218,
    ↪ 0.1732,0.1177,0.518,0.1723,0.1189,-0.1,0.1972,0.2467,0.217,
    ↪ 0.1824,0.4975,0.225,0.12,0.059,0.214,0.152,0.344,0.506,0.151,
    ↪ 0.343,-0.08'
_aflow_Strukturbericht 'None'
_aflow_Pearson 'oC80'

_symmetry_space_group_name_H-M "C m c 21"
_symmetry_Int_Tables_number 36

_cell_length_a      12.22901
_cell_length_b      10.80500
_cell_length_c      6.92400
_cell_angle_alpha 90.00000
_cell_angle_beta  90.00000
_cell_angle_gamma 90.00000

loop_
_space_group_symop_id
_space_group_symop_operation_xyz
1 x,y,z
2 -x,-y,z+1/2
3 -x,y,z
4 x,-y,z+1/2
5 x+1/2,y+1/2,z
6 -x+1/2,-y+1/2,z+1/2
7 -x+1/2,y+1/2,z
8 x+1/2,-y+1/2,z+1/2

loop_
_atom_site_label
_atom_site_type_symbol
_atom_site_symmetry_multiplicity
_atom_site_Wyckoff_label
_atom_site_fract_x
_atom_site_fract_y
_atom_site_fract_z
_atom_site_occupancy
Ni1  Ni  4  a  0.00000  0.00000  0.00000  1.00000
Ni2  Ni  4  a  0.00000  0.23450  0.02400  1.00000
Ni3  Ni  4  a  0.00000  0.23300  0.40000  1.00000
Ni4  Ni  4  a  0.00000  0.38140  0.71400  1.00000
Si1  Si  4  a  0.00000  0.15700  0.71200  1.00000
Si2  Si  4  a  0.00000  0.40900  0.21800  1.00000
Ni5  Ni  8  b  0.17320  0.11770  0.51800  1.00000
Ni6  Ni  8  b  0.17230  0.11890  -0.10000  1.00000
Ni7  Ni  8  b  0.19720  0.24670  0.21700  1.00000
Ni8  Ni  8  b  0.18240  0.49750  0.22500  1.00000
Si3  Si  8  b  0.12000  0.05900  0.21400  1.00000
Si4  Si  8  b  0.15200  0.34400  0.50600  1.00000
Si5  Si  8  b  0.15100  0.34300  -0.08000  1.00000
```

$Ni_3Si_2$: A3B2_oC80_36_4a4b_2a3b - POSCAR

```
A3B2_oC80_36_4a4b_2a3b & a,b/a,c/a,y1,z1,y2,z2,y3,z3,y4,z4,y5,z5,y6,z6,
    ↪ x7,y7,z7,x8,y8,z8,x9,y9,z9,x10,y10,z10,x11,y11,z11,x12,y12,z12,
    ↪ x13,y13,z13 --params=12.22901,0.883554760361,0.566194646991,0.0
    ↪ ,0.0,0.2345,0.024,0.233,0.4,0.3814,0.714,0.157,0.712,0.409,
    ↪ 0.218,0.1732,0.1177,0.518,0.1723,0.1189,-0.1,0.1972,0.2467,
    ↪ 0.217,0.1824,0.4975,0.225,0.12,0.059,0.214,0.152,0.344,0.506,
    ↪ 0.151,0.343,-0.08 & Cmc2_{1} C_{2v}^{12} #36 (a^6b^7) & oC80 &
    ↪ None & Ni3Si2 & Ni3Si2 & G. Pilstr\"{o}m, Acta Chem. Scand. 15,
    ↪ 893-902 (1961)
   1.00000000000000
   6.11450500000000  -5.40250000000000   0.00000000000000
   6.11450500000000   5.40250000000000   0.00000000000000
   0.00000000000000   0.00000000000000   6.92400000000000
    Ni     Si
    24     16
Direct
   0.00000000000000   0.00000000000000   0.00000000000000   Ni   (4a)
   0.00000000000000   0.00000000000000   0.50000000000000   Ni   (4a)
  -0.23450000000000   0.23450000000000   0.02400000000000   Ni   (4a)
   0.23450000000000  -0.23450000000000   0.52400000000000   Ni   (4a)
  -0.23300000000000   0.23300000000000   0.40000000000000   Ni   (4a)
   0.23300000000000  -0.23300000000000   0.90000000000000   Ni   (4a)
  -0.38140000000000   0.38140000000000   0.71400000000000   Ni   (4a)
   0.38140000000000  -0.38140000000000   1.21400000000000   Ni   (4a)
   0.05550000000000   0.29090000000000   0.51800000000000   Ni   (8b)
  -0.05550000000000  -0.29090000000000   1.01800000000000   Ni   (8b)
   0.29090000000000   0.05550000000000   1.01800000000000   Ni   (8b)
  -0.29090000000000  -0.05550000000000   0.51800000000000   Ni   (8b)
   0.05340000000000   0.29120000000000  -0.10000000000000   Ni   (8b)
  -0.05340000000000  -0.29120000000000   0.40000000000000   Ni   (8b)
   0.29120000000000   0.05340000000000   0.40000000000000   Ni   (8b)
  -0.29120000000000  -0.05340000000000  -0.10000000000000   Ni   (8b)
  -0.04950000000000   0.44390000000000   0.21700000000000   Ni   (8b)
```

```
   0.04950000000000  -0.44390000000000   0.71700000000000   Ni   (8b)
   0.44390000000000  -0.04950000000000   0.71700000000000   Ni   (8b)
  -0.44390000000000   0.04950000000000   0.21700000000000   Ni   (8b)
  -0.31510000000000   0.67990000000000   0.22500000000000   Ni   (8b)
   0.31510000000000  -0.67990000000000   0.72500000000000   Ni   (8b)
   0.67990000000000  -0.31510000000000   0.72500000000000   Ni   (8b)
  -0.67990000000000   0.31510000000000   0.22500000000000   Ni   (8b)
  -0.15700000000000   0.15700000000000   0.71200000000000   Si   (4a)
   0.15700000000000  -0.15700000000000   1.21200000000000   Si   (4a)
  -0.40900000000000   0.40900000000000   0.21800000000000   Si   (4a)
   0.40900000000000  -0.40900000000000   0.71800000000000   Si   (4a)
   0.06100000000000   0.17900000000000   0.21400000000000   Si   (8b)
  -0.06100000000000  -0.17900000000000   0.71400000000000   Si   (8b)
   0.17900000000000   0.06100000000000   0.71400000000000   Si   (8b)
  -0.17900000000000  -0.06100000000000   0.21400000000000   Si   (8b)
  -0.19200000000000   0.49600000000000   0.50600000000000   Si   (8b)
   0.19200000000000  -0.49600000000000   1.00600000000000   Si   (8b)
   0.49600000000000  -0.19200000000000   1.00600000000000   Si   (8b)
  -0.49600000000000   0.19200000000000   0.50600000000000   Si   (8b)
  -0.19200000000000   0.49400000000000  -0.08000000000000   Si   (8b)
   0.19200000000000  -0.49400000000000   0.42000000000000   Si   (8b)
   0.49400000000000  -0.19200000000000   0.42000000000000   Si   (8b)
  -0.49400000000000   0.19200000000000  -0.08000000000000   Si   (8b)
```

Bertrandite ($Be_4Si_2O_7(OH)_2$, $S4_6$): A4B7C2D2_oC60_36_2b_a3b_2a_b - CIF

```
# CIF file
data_findsym-output
_audit_creation_method FINDSYM

_chemical_name_mineral 'Bertrandite'
_chemical_formula_sum 'Be4 O7 (OH)2 Si2'

loop_
_publ_author_name
 'R. M. Hazen'
 'A. Y. Au'
_journal_name_full_name
;
 Physics and Chemistry of Minerals
;
_journal_volume 13
_journal_year 1986
_journal_page_first 69
_journal_page_last 78
_publ_Section_title
;
 High-pressure crystal chemistry of phenakite (Be$_{2}$SiO$_{4}$) and
     ↪ bertrandite (Be$_{4}$Si$_{2}$O$_{7}$(OH)$_{2}$)
;

# Found in The power of databases: the RRUFF project, 2015

_aflow_title 'Bertrandite (Be$_{4}$Si$_{2}$O$_{7}$(OH)$_{2}$, $S4_{6}$)
     ↪ Structure'
_aflow_proto 'A4B7C2D2_oC60_36_2b_a3b_2a_b'
_aflow_params 'a,b/a,c/a,y_{1},z_{1},y_{2},z_{2},y_{3},z_{3},x_{4},y_{4
     ↪ },z_{4},x_{5},y_{5},z_{5},x_{6},y_{6},z_{6},x_{7},y_{7},z_{7},
     ↪ x_{8},y_{8},z_{8},x_{9},y_{9},z_{9}'
_aflow_params_values '8.7135,1.75222356114,0.524278418546,0.5852,0.5951,
     ↪ 0.7553,0.0886,0.0872,0.0993,0.1735,0.0527,0.1628,0.3264,0.2203,
     ↪ 0.1517,0.2897,0.1244,0.0,0.2101,0.0431,0.5074,0.2934,0.2093,
     ↪ 0.5024,0.3254,0.1144,0.6523'
_aflow_Strukturbericht '$S4_{6}$'
_aflow_Pearson 'oC60'

_symmetry_space_group_name_H-M "C m c 21"
_symmetry_Int_Tables_number 36

_cell_length_a       8.71350
_cell_length_b       15.26800
_cell_length_c       4.56830
_cell_angle_alpha    90.00000
_cell_angle_beta     90.00000
_cell_angle_gamma    90.00000

loop_
_space_group_symop_id
_space_group_symop_operation_xyz
1 x,y,z
2 -x,-y,z+1/2
3 -x,y,z
4 x,-y,z+1/2
5 x+1/2,y+1/2,z
6 -x+1/2,-y+1/2,z+1/2
7 -x+1/2,y+1/2,z
8 x+1/2,-y+1/2,z+1/2

loop_
_atom_site_label
_atom_site_type_symbol
_atom_site_symmetry_multiplicity
_atom_site_Wyckoff_label
_atom_site_fract_x
_atom_site_fract_y
_atom_site_fract_z
_atom_site_occupancy
O1 O 4 a 0.00000 0.58520 0.59510 1.00000
OH1 OH 4 a 0.00000 0.75530 0.08860 1.00000
OH2 OH 4 a 0.00000 0.08720 0.09930 1.00000
Be1 Be 8 b 0.17350 0.05270 0.16280 1.00000
Be2 Be 8 b 0.32640 0.22030 0.15170 1.00000
O2 O 8 b 0.28970 0.12440 0.00000 1.00000
O3 O 8 b 0.21010 0.04310 0.50740 1.00000
O4 O 8 b 0.29340 0.20930 0.50240 1.00000
Si1 Si 8 b 0.32540 0.11440 0.65230 1.00000
```

Bertrandite ($Be_4Si_2O_7(OH)_2$, $S4_6$): A4B7C2D2_oC60_36_2b_a3b_2a_b - POSCAR

```
A4B7C2D2_oC60_36_2b_a3b_2a_b & a,b/a,c/a,y1,z1,y2,z2,y3,z3,x4,y4,z4,x5,
     ↪ y5,z5,x6,y6,z6,x7,y7,z7,x8,y8,z8,x9,y9,z9 --params=8.7135,
     ↪ 1.75222356114,0.524278418546,0.5852,0.5951,0.7553,0.0886,0.0872
     ↪ ,0.0993,0.1735,0.0527,0.1628,0.3264,0.2203,0.1517,0.2897,0.1244
     ↪ ,0.0,0.2101,0.0431,0.5074,0.2934,0.2093,0.5024,0.3254,0.1144,
     ↪ 0.6523 & Cmc2_{1} C_{2v}^{12} #36 (a^3b^6) & oC60 & $S4_{6}$ &
     ↪ Be4O7(OH)2Si2 & Bertrandite & R. M. Hazen and A. Y. Au, Phys.
     ↪ Chem. Miner. 13, 69-78 (1986)
   1.00000000000000
   4.35675000000000  -7.63400000000000   0.00000000000000
   4.35675000000000   7.63400000000000   0.00000000000000
   0.00000000000000   0.00000000000000   4.56830000000000
    Be     O    OH    Si
     8    14     4     4
Direct
   0.12080000000000   0.22620000000000   0.16280000000000   Be   (8b)
  -0.12080000000000  -0.22620000000000   0.66280000000000   Be   (8b)
   0.22620000000000   0.12080000000000   0.66280000000000   Be   (8b)
  -0.22620000000000  -0.12080000000000   0.16280000000000   Be   (8b)
   0.10610000000000   0.54670000000000   0.15170000000000   Be   (8b)
  -0.10610000000000  -0.54670000000000   0.65170000000000   Be   (8b)
   0.54670000000000   0.10610000000000   0.65170000000000   Be   (8b)
  -0.54670000000000  -0.10610000000000   0.15170000000000   Be   (8b)
  -0.58520000000000   0.58520000000000   0.59510000000000   O    (4a)
   0.58520000000000  -0.58520000000000   1.09510000000000   O    (4a)
   0.16530000000000   0.41410000000000   0.00000000000000   O    (8b)
  -0.16530000000000  -0.41410000000000   0.50000000000000   O    (8b)
   0.41410000000000   0.16530000000000   0.50000000000000   O    (8b)
  -0.41410000000000  -0.16530000000000   0.00000000000000   O    (8b)
   0.16700000000000   0.25320000000000   0.50740000000000   O    (8b)
  -0.16700000000000  -0.25320000000000   1.00740000000000   O    (8b)
   0.25320000000000   0.16700000000000   1.00740000000000   O    (8b)
  -0.25320000000000  -0.16700000000000   0.50740000000000   O    (8b)
   0.08410000000000   0.50270000000000   0.50240000000000   O    (8b)
  -0.08410000000000  -0.50270000000000   1.00240000000000   O    (8b)
   0.50270000000000   0.08410000000000   1.00240000000000   O    (8b)
  -0.50270000000000  -0.08410000000000   0.50240000000000   O    (8b)
  -0.75530000000000   0.75530000000000   0.08860000000000   OH   (4a)
   0.75530000000000  -0.75530000000000   0.58860000000000   OH   (4a)
  -0.08720000000000   0.08720000000000   0.09930000000000   OH   (4a)
   0.08720000000000  -0.08720000000000   0.59930000000000   OH   (4a)
   0.21100000000000   0.43980000000000   0.65230000000000   Si   (8b)
  -0.21100000000000  -0.43980000000000   1.15230000000000   Si   (8b)
   0.43980000000000   0.21100000000000   1.15230000000000   Si   (8b)
  -0.43980000000000  -0.21100000000000   0.65230000000000   Si   (8b)
```

$MoP_2$: AB2_oC12_36_a_2a - CIF

```
# CIF file
data_findsym-output
_audit_creation_method FINDSYM

_chemical_name_mineral 'MoP2'
_chemical_formula_sum 'Mo P2'

loop_
_publ_author_name
 'S. Rundqvist'
 'T. Lundstr\"{o}m'
_journal_name_full_name
;
 Acta Chemica Scandinavica
;
_journal_volume 17
_journal_year 1963
_journal_page_first 37
_journal_page_last 46
_publ_Section_title
;
 X-Ray Studies of Molybdenum and Tungsten Phosphides
;

_aflow_title 'MoP$_{2}$ Structure'
_aflow_proto 'AB2_oC12_36_a_2a'
_aflow_params 'a,b/a,c/a,y_{1},z_{1},y_{2},z_{2},y_{3},z_{3}'
_aflow_params_values '3.145,3.55612082671,1.84702384738,0.0934,0.0,0.294
     ↪ ,0.803,0.426,0.121'
_aflow_Strukturbericht 'None'
_aflow_Pearson 'oC12'

_symmetry_space_group_name_H-M "C m c 21"
_symmetry_Int_Tables_number 36

_cell_length_a       3.14500
_cell_length_b       11.18400
_cell_length_c       5.80889
_cell_angle_alpha    90.00000
_cell_angle_beta     90.00000
_cell_angle_gamma    90.00000

loop_
_space_group_symop_id
_space_group_symop_operation_xyz
1 x,y,z
2 -x,-y,z+1/2
3 -x,y,z
4 x,-y,z+1/2
5 x+1/2,y+1/2,z
6 -x+1/2,-y+1/2,z+1/2
7 -x+1/2,y+1/2,z
8 x+1/2,-y+1/2,z+1/2
```

```
loop_
_atom_site_label
_atom_site_type_symbol
_atom_site_symmetry_multiplicity
_atom_site_Wyckoff_label
_atom_site_fract_x
_atom_site_fract_y
_atom_site_fract_z
_atom_site_occupancy
Mo1 Mo 4 a 0.00000 0.09340 0.00000 1.00000
P1 P 4 a 0.00000 0.29400 0.80300 1.00000
P2 P 4 a 0.00000 0.42600 0.12100 1.00000
```

$MoP_2$: AB2_oC12_36_a_2a - POSCAR

```
AB2_oC12_36_a_2a & a,b/a,c/a,y1,z1,y2,z2,y3,z3 --params=3.145,
    ↪ 3.55612082671,1.84702384738,0.0934,0.0,0.294,0.803,0.426,0.121
    ↪ & Cmc2_{1} C_{2v}^{12} #36 (a^3) & oC12 & None & MoP2 & MoP2 &
    ↪ S. Rundqvist and T. Lundstr\"{o}m, Acta Chem. Scand. 17, 37-46
    ↪ (1963)
   1.00000000000000
   1.57250000000000  -5.59200000000000   0.00000000000000
   1.57250000000000   5.59200000000000   0.00000000000000
   0.00000000000000   0.00000000000000   5.80889000000000
    Mo     P
     2     4
Direct
  -0.09340000000000   0.09340000000000   0.00000000000000   Mo   (4a)
   0.09340000000000  -0.09340000000000   0.50000000000000   Mo   (4a)
  -0.29400000000000   0.29400000000000   0.80300000000000   P    (4a)
   0.29400000000000  -0.29400000000000   1.30300000000000   P    (4a)
  -0.42600000000000   0.42600000000000   0.12100000000000   P    (4a)
   0.42600000000000  -0.42600000000000   0.62100000000000   P    (4a)
```

α-Potassium Nitrate ($KNO_3$) II: ABC3_oC80_36_2ab_2ab_2a5b - CIF

```
# CIF file
data_findsym-output
_audit_creation_method FINDSYM

_chemical_name_mineral 'KNO3'
_chemical_formula_sum 'K N O3'

loop_
_publ_author_name
 'G. Adiwidjaja'
 'D. Pohl'
_journal_name_full_name
;
 Acta Crystallographica Section C: Structural Chemistry
;
_journal_volume 59
_journal_year 2003
_journal_page_first i139
_journal_page_last i140
_publ_Section_title
;
 Superstructure of $\alpha$-phase potassium nitrate
;

_aflow_title '$\alpha$-Potassium Nitrate (KNO$_{3}$) II Structure'
_aflow_proto 'ABC3_oC80_36_2ab_2ab_2a5b'
_aflow_params 'a,b/a,c/a,y_{1},z_{1},y_{2},z_{2},y_{3},z_{3},y_{4},z_{4
    ↪ },y_{5},z_{5},y_{6},z_{6},x_{7},y_{7},z_{7},x_{8},y_{8},z_{8},
    ↪ x_{9},y_{9},z_{9},x_{10},y_{10},z_{10},x_{11},y_{11},z_{11},x_{
    ↪ 12},y_{12},z_{12},x_{13},y_{13},z_{13}'
_aflow_params_values '10.825,1.69524249423,0.594457274827,0.41703,0.2604
    ↪ ,-0.08321,0.2614,0.2483,0.4215,0.7469,0.4208,0.1807,0.414,
    ↪ 0.6788,0.4138,0.2498,0.1667,0.25104,0.2497,0.4979,0.0911,0.0999
    ↪ ,0.283,0.4206,0.3987,0.2809,0.4248,0.6506,0.0313,0.0984,0.8479,
    ↪ 0.0321,0.0896,0.2478,0.4296,0.0937'
_aflow_Strukturbericht 'None'
_aflow_Pearson 'oC80'

_symmetry_space_group_name_H-M "C m c 21"
_symmetry_Int_Tables_number 36

_cell_length_a     10.82500
_cell_length_b     18.35100
_cell_length_c     6.43500
_cell_angle_alpha  90.00000
_cell_angle_beta   90.00000
_cell_angle_gamma  90.00000

loop_
_space_group_symop_id
_space_group_symop_operation_xyz
1 x,y,z
2 -x,-y,z+1/2
3 -x,y,z
4 x,-y,z+1/2
5 x+1/2,y+1/2,z
6 -x+1/2,-y+1/2,z+1/2
7 -x+1/2,y+1/2,z
8 x+1/2,-y+1/2,z+1/2

loop_
_atom_site_label
_atom_site_type_symbol
_atom_site_symmetry_multiplicity
_atom_site_Wyckoff_label
_atom_site_fract_x
_atom_site_fract_y
_atom_site_fract_z
_atom_site_occupancy
K1 K 4 a 0.00000 0.41703 0.26040 1.00000
K2 K 4 a 0.00000 -0.08321 0.26140 1.00000
N1 N 4 a 0.00000 0.24830 0.42150 1.00000
N2 N 4 a 0.00000 0.74690 0.42080 1.00000
O1 O 4 a 0.00000 0.18070 0.41400 1.00000
O2 O 4 a 0.00000 0.67880 0.41380 1.00000
K3 K 8 b 0.24980 0.16670 0.25104 1.00000
N3 N 8 b 0.24970 0.49790 0.09110 1.00000
O3 O 8 b 0.09990 0.28300 0.42060 1.00000
O4 O 8 b 0.39870 0.28090 0.42480 1.00000
O5 O 8 b 0.65060 0.03130 0.09840 1.00000
O6 O 8 b 0.84790 0.03210 0.08960 1.00000
O7 O 8 b 0.24780 0.42960 0.09370 1.00000
```

α-Potassium Nitrate ($KNO_3$) II: ABC3_oC80_36_2ab_2ab_2a5b - POSCAR

```
ABC3_oC80_36_2ab_2ab_2a5b & a,b/a,c/a,y1,z1,y2,z2,y3,z3,y4,z4,y5,z5,y6,
    ↪ z6,x7,y7,z7,x8,y8,z8,x9,y9,z9,x10,y10,z10,x11,y11,z11,x12,y12,
    ↪ z12,x13,y13,z13 --params=10.825,1.69524249423,0.594457274827,
    ↪ 0.41703,0.2604,-0.08321,0.2614,0.2483,0.4215,0.7469,0.4208,
    ↪ 0.1807,0.414,0.6788,0.4138,0.2498,0.1667,0.25104,0.2497,0.4979,
    ↪ 0.0911,0.0999,0.283,0.4206,0.3987,0.2809,0.4248,0.6506,0.0313,
    ↪ 0.0984,0.8479,0.0321,0.0896,0.2478,0.4296,0.0937 & Cmc2_{1} C_{
    ↪ 2v}^{12} #36 (a^6b^7) & oC80 & None & KNO3 & KNO3 & G.
    ↪ Adiwidjaja and D. Pohl, Acta Crystallogr. C 59, i139-i140 (2003
    ↪ )
   1.00000000000000
   5.41250000000000  -9.17550000000000   0.00000000000000
   5.41250000000000   9.17550000000000   0.00000000000000
   0.00000000000000   0.00000000000000   6.43500000000000
     K      N      O
     8      8     24
Direct
  -0.41703000000000   0.41703000000000   0.26040000000000   K   (4a)
   0.41703000000000  -0.41703000000000   0.76040000000000   K   (4a)
   0.08321000000000  -0.08321000000000   0.26140000000000   K   (4a)
  -0.08321000000000   0.08321000000000   0.76140000000000   K   (4a)
   0.08310000000000   0.41650000000000   0.25104000000000   K   (8b)
  -0.08310000000000  -0.41650000000000   0.75104000000000   K   (8b)
   0.41650000000000   0.08310000000000   0.75104000000000   K   (8b)
  -0.41650000000000  -0.08310000000000   0.25104000000000   K   (8b)
  -0.24830000000000   0.24830000000000   0.42150000000000   N   (4a)
   0.24830000000000  -0.24830000000000   0.92150000000000   N   (4a)
  -0.74690000000000   0.74690000000000   0.42080000000000   N   (4a)
   0.74690000000000  -0.74690000000000   0.92080000000000   N   (4a)
  -0.24820000000000   0.74760000000000   0.09110000000000   N   (8b)
   0.24820000000000  -0.74760000000000   0.59110000000000   N   (8b)
   0.74760000000000  -0.24820000000000   0.59110000000000   N   (8b)
  -0.74760000000000   0.24820000000000   0.09110000000000   N   (8b)
  -0.18070000000000   0.18070000000000   0.41400000000000   O   (4a)
   0.18070000000000  -0.18070000000000   0.91400000000000   O   (4a)
  -0.67880000000000   0.67880000000000   0.41380000000000   O   (4a)
   0.67880000000000  -0.67880000000000   0.91380000000000   O   (4a)
  -0.18310000000000   0.38290000000000   0.42060000000000   O   (8b)
   0.18310000000000  -0.38290000000000   0.92060000000000   O   (8b)
   0.38290000000000  -0.18310000000000   0.92060000000000   O   (8b)
  -0.38290000000000   0.18310000000000   0.42060000000000   O   (8b)
   0.11780000000000   0.67960000000000   0.42480000000000   O   (8b)
  -0.11780000000000  -0.67960000000000   0.92480000000000   O   (8b)
   0.67960000000000   0.11780000000000   0.92480000000000   O   (8b)
  -0.67960000000000  -0.11780000000000   0.42480000000000   O   (8b)
   0.61930000000000   0.68190000000000   0.09840000000000   O   (8b)
  -0.61930000000000  -0.68190000000000   0.59840000000000   O   (8b)
   0.68190000000000   0.61930000000000   0.59840000000000   O   (8b)
  -0.68190000000000  -0.61930000000000   0.09840000000000   O   (8b)
   0.81580000000000   0.88000000000000   0.08960000000000   O   (8b)
  -0.81580000000000  -0.88000000000000   0.58960000000000   O   (8b)
   0.88000000000000   0.81580000000000   0.58960000000000   O   (8b)
  -0.88000000000000  -0.81580000000000   0.08960000000000   O   (8b)
  -0.18180000000000   0.67740000000000   0.09370000000000   O   (8b)
   0.18180000000000  -0.67740000000000   0.59370000000000   O   (8b)
   0.67740000000000  -0.18180000000000   0.59370000000000   O   (8b)
  -0.67740000000000   0.18180000000000   0.09370000000000   O   (8b)
```

$Ta_3Ti_{13}$ (BCC SQS-16): A3B13_oC32_38_ac_a2bcdef - CIF

```
# AFLOW.org Repositories
# TaTi/A3B13_oC32_38_ac_a2bcdef-001.AB params=6.5421326204,1.41421356237
    ↪ ,1.41421356237,0.5,0.0,0.5,0.0,0.75,0.25,0.75,0.75,0.25,0.25,
    ↪ 0.25,0.25,0.75,0.75,0.0 SG=38 [ANRL doi: 10.1016/
    ↪ j.commatsci.2017.01.017 (part 1), doi: 10.1016/
    ↪ j.commatsci.2018.10.043 (part 2)]
data_TaTi
_pd_phase_name A3B13_oC32_38_ac_a2bcdef-001.AB

_chemical_name_mineral 'Ta3Ti13'
_chemical_formula_sum 'Ta3 Ti13'

loop_
_publ_author_name
 'T. Chakraborty'
 'J. Rogal'
 'R. Drautz'
_journal_name_full_name
;
 Physical Review B
;
_journal_volume 94
_journal_year 2016
_journal_page_first 224104
_journal_page_last 224104
_publ_Section_title
;
```

```
 Unraveling the composition dependence of the martensitic transformation
      ↪ temperature: A first-principles study of Ti-Ta alloys
;

_aflow_title 'Ta$_{3}$Ti$_{13}$ (BCC SQS-16) Structure'
_aflow_proto 'A3B13_oC32_38_ac_a2bcdef'
_aflow_params 'a,b/a,c/a,z_{1},z_{2},z_{3},z_{4},x_{5},z_{5},x_{6},z_{6
      ↪ },y_{7},z_{7},y_{8},z_{8},x_{9},y_{9},z_{9}'
_aflow_params_values '6.5421326204,1.41421356237,1.41421356237,0.5,0.0,
      ↪ 0.5,0.0,0.75,0.75,0.25,0.75,0.75,0.75,0.75,0.75,0.25,0.0'
_aflow_Strukturbericht 'None'
_aflow_Pearson 'oC32'

_cell_length_a    6.5421326204
_cell_length_b    9.2519726786
_cell_length_c    9.2519726786
_cell_angle_alpha 90.0000000000
_cell_angle_beta  90.0000000000
_cell_angle_gamma 90.0000000000
_symmetry_space_group_name_H-M 'Amm2'
_symmetry_Int_Tables_Number 38
loop_
_symmetry_equiv_pos_site_id
_symmetry_equiv_pos_as_xyz
  1  x,y,z
  2  -x,-y,z
  3  x,-y,z
  4  -x,y,z
  5  x,y+1/2,z+1/2
  6  -x,-y+1/2,z+1/2
  7  x,-y+1/2,z+1/2
  8  -x,y+1/2,z+1/2
loop_
 _atom_site_label
 _atom_site_occupancy
 _atom_site_fract_x
 _atom_site_fract_y
 _atom_site_fract_z
 _atom_site_thermal_displace_type
 _atom_site_B_iso_or_equiv
 _atom_site_type_symbol
 _atom_site_symmetry_multiplicity
 _atom_site_Wyckoff_label
Ta1 1.0000000000 0.0000000000 -0.0000000000 0.5000000000 Biso 1.0 Ta 2 a
Ti1 1.0000000000 0.0000000000 0.0000000000 0.0000000000 Biso 1.0 Ti 2 a
Ti2 1.0000000000 0.5000000000 -0.0000000000 0.5000000000 Biso 1.0 Ti 2 b
Ti3 1.0000000000 0.5000000000 0.0000000000 0.0000000000 Biso 1.0 Ti 2 b
Ta2 1.0000000000 0.7500000000 -0.0000000000 0.7500000000 Biso 1.0 Ta 4 c
Ti4 1.0000000000 0.7500000000 -0.0000000000 0.2500000000 Biso 1.0 Ti 4 c
Ti5 1.0000000000 0.0000000000 0.7500000000 0.2500000000 Biso 1.0 Ti 4 d
Ti6 1.0000000000 0.5000000000 0.7500000000 0.7500000000 Biso 1.0 Ti 4 e
Ti7 1.0000000000 0.7500000000 0.2500000000 0.0000000000 Biso 1.0 Ti 8 f
```

$Ta_3Ti_{13}$ (BCC SQS-16): A3B13_oC32_38_ac_a2bcdef - POSCAR

```
A3B13_oC32_38_ac_a2bcdef & a,b/a,c/a,z1,z2,z3,z4,x5,z5,x6,z6,y7,z7,y8,z8
      ↪ ,x9,y9,z9 --params=6.5421326204,1.41421356237,1.41421356237,0.5
      ↪ ,0.0,0.5,0.0,0.75,0.75,0.75,0.25,0.75,0.75,0.75,0.75,0.75,0.25,
      ↪ 0.0 & Amm2 C_{2v}^{14} #38 (a^2b^2c^2def) & oC32 & None &
      ↪ Ta3Ti13 & Ta3Ti13 & T. Chakraborty and J. Rogal and R. Drautz,
      ↪ Phys. Rev. B 94, 224104(2016)
   1.00000000000000
   6.54213262040000   0.00000000000000   0.00000000000000
   0.00000000000000   4.62598633930000  -4.62598633930000
   0.00000000000000   4.62598633930000   4.62598633930000
    Ta     Ti
     3     13
Direct
   0.00000000000000  -0.50000000000000   0.50000000000000   Ta   (2a)
   0.75000000000000  -0.75000000000000   0.75000000000000   Ta   (4c)
  -0.75000000000000  -0.75000000000000   0.75000000000000   Ta   (4c)
   0.00000000000000   0.00000000000000   0.00000000000000   Ti   (2a)
   0.50000000000000  -0.50000000000000   0.50000000000000   Ti   (2b)
   0.50000000000000   0.00000000000000   0.00000000000000   Ti   (2b)
   0.75000000000000  -0.25000000000000   0.25000000000000   Ti   (4c)
  -0.75000000000000  -0.25000000000000   0.25000000000000   Ti   (4c)
   0.00000000000000   0.00000000000000   1.50000000000000   Ti   (4d)
   0.00000000000000  -1.50000000000000   0.00000000000000   Ti   (4d)
   0.50000000000000   0.00000000000000   1.50000000000000   Ti   (4e)
   0.50000000000000  -1.50000000000000   0.00000000000000   Ti   (4e)
   0.75000000000000   0.25000000000000   0.25000000000000   Ti   (8f)
  -0.75000000000000  -0.25000000000000  -0.25000000000000   Ti   (8f)
   0.75000000000000  -0.25000000000000  -0.25000000000000   Ti   (8f)
  -0.75000000000000   0.25000000000000   0.25000000000000   Ti   (8f)
```

$Ta_3Ti_5$ (BCC SQS-16): A3B5_oC32_38_abce_abcdf - CIF

```
# AFLOW.org Repositories
# TaTi/A3B5_oC32_38_abce_abcdf-001.AB params=6.5421326204,1.41421356237,
      ↪ 1.41421356237,0.5,-0.0,-0.0,0.5,0.25,0.75,0.25,0.25,0.75,0.25,
      ↪ 0.75,0.25,0.25,0.75,-0.0 SG=38 [ANRL doi: 10.1016/
      ↪ j.commatsci.2017.01.017 (part 1), doi: 10.1016/
      ↪ j.commatsci.2018.10.043 (part 2)]
data_TaTi
_pd_phase_name A3B5_oC32_38_abce_abcdf-001.AB

_chemical_name_mineral 'Ta3Ti5'
_chemical_formula_sum 'Ta3 Ti5'

loop_
_publ_author_name
 'T. Chakraborty'
 'J. Rogal'
 'R. Drautz'
_journal_name_full_name
;
 Physical Review B
;
_journal_volume 94
_journal_year 2016
_journal_page_first 224104
_journal_page_last 224104
_publ_Section_title
;
 Unraveling the composition dependence of the martensitic transformation
      ↪ temperature: A first-principles study of Ti-Ta alloys
;

_aflow_title 'Ta$_{3}$Ti$_{5}$ (BCC SQS-16) Structure'
_aflow_proto 'A3B5_oC32_38_abce_abcdf'
_aflow_params 'a,b/a,c/a,z_{1},z_{2},z_{3},z_{4},x_{5},z_{5},x_{6},z_{6
      ↪ },y_{7},z_{7},y_{8},z_{8},x_{9},y_{9},z_{9}'
_aflow_params_values '6.5421326204,1.41421356237,1.41421356237,0.5,0.0,
      ↪ 0.0,0.5,0.25,0.25,0.25,0.75,0.75,0.25,0.75,0.25,0.25,0.25,0.0'
_aflow_Strukturbericht 'None'
_aflow_Pearson 'oC32'

_cell_length_a    6.5421326204
_cell_length_b    9.2519726786
_cell_length_c    9.2519726786
_cell_angle_alpha 90.0000000000
_cell_angle_beta  90.0000000000
_cell_angle_gamma 90.0000000000
_symmetry_space_group_name_H-M 'Amm2'
_symmetry_Int_Tables_Number 38
loop_
_symmetry_equiv_pos_site_id
_symmetry_equiv_pos_as_xyz
  1  x,y,z
  2  -x,-y,z
  3  x,-y,z
  4  -x,y,z
  5  x,y+1/2,z+1/2
  6  -x,-y+1/2,z+1/2
  7  x,-y+1/2,z+1/2
  8  -x,y+1/2,z+1/2
loop_
 _atom_site_label
 _atom_site_occupancy
 _atom_site_fract_x
 _atom_site_fract_y
 _atom_site_fract_z
 _atom_site_thermal_displace_type
 _atom_site_B_iso_or_equiv
 _atom_site_type_symbol
 _atom_site_symmetry_multiplicity
 _atom_site_Wyckoff_label
Ta1 1.0000000000 0.0000000000 -0.0000000000 0.5000000000 Biso 1.0 Ta 2 a
Ti1 1.0000000000 0.0000000000 0.0000000000 0.0000000000 Biso 1.0 Ti 2 a
Ta2 1.0000000000 0.5000000000 0.0000000000 0.0000000000 Biso 1.0 Ta 2 b
Ti2 1.0000000000 0.5000000000 -0.0000000000 0.5000000000 Biso 1.0 Ti 2 b
Ta3 1.0000000000 0.2500000000 -0.0000000000 0.2500000000 Biso 1.0 Ta 4 c
Ti3 1.0000000000 0.2500000000 -0.0000000000 0.7500000000 Biso 1.0 Ti 4 c
Ti4 1.0000000000 0.0000000000 0.7500000000 0.2500000000 Biso 1.0 Ti 4 d
Ta4 1.0000000000 0.5000000000 0.7500000000 0.2500000000 Biso 1.0 Ta 4 e
Ti5 1.0000000000 0.2500000000 0.2500000000 0.0000000000 Biso 1.0 Ti 8 f
```

$Ta_3Ti_5$ (BCC SQS-16): A3B5_oC32_38_abce_abcdf - POSCAR

```
A3B5_oC32_38_abce_abcdf & a,b/a,c/a,z1,z2,z3,z4,x5,z5,x6,z6,y7,z7,y8,z8,
      ↪ x9,y9,z9 --params=6.5421326204,1.41421356237,1.41421356237,0.5,
      ↪ 0.0,0.0,0.5,0.25,0.25,0.25,0.75,0.75,0.25,0.75,0.25,0.25,0.25,
      ↪ 0.0 & Amm2 C_{2v}^{14} #38 (a^2b^2c^2def) & oC32 & None &
      ↪ Ta3Ti5 & Ta3Ti5 & T. Chakraborty and J. Rogal and R. Drautz,
      ↪ Phys. Rev. B 94, 224104(2016)
   1.00000000000000
   6.54213262040000   0.00000000000000   0.00000000000000
   0.00000000000000   4.62598633930000  -4.62598633930000
   0.00000000000000   4.62598633930000   4.62598633930000
    Ta     Ti
     6     10
Direct
   0.00000000000000  -0.50000000000000   0.50000000000000   Ta   (2a)
   0.50000000000000   0.00000000000000   0.00000000000000   Ta   (2b)
   0.25000000000000  -0.25000000000000   0.25000000000000   Ta   (4c)
  -0.25000000000000  -0.25000000000000   0.25000000000000   Ta   (4c)
   0.50000000000000   0.50000000000000   1.00000000000000   Ta   (4e)
   0.50000000000000  -1.00000000000000  -0.50000000000000   Ta   (4e)
   0.00000000000000   0.00000000000000   0.00000000000000   Ti   (2a)
   0.50000000000000  -0.50000000000000   0.50000000000000   Ti   (2b)
   0.25000000000000  -0.75000000000000   0.75000000000000   Ti   (4c)
  -0.25000000000000  -0.75000000000000   0.75000000000000   Ti   (4c)
   0.00000000000000   0.50000000000000   1.00000000000000   Ti   (4d)
   0.00000000000000  -1.00000000000000  -0.50000000000000   Ti   (4d)
   0.25000000000000   0.25000000000000   0.25000000000000   Ti   (8f)
  -0.25000000000000  -0.25000000000000  -0.25000000000000   Ti   (8f)
   0.25000000000000  -0.25000000000000  -0.25000000000000   Ti   (8f)
  -0.25000000000000   0.25000000000000   0.25000000000000   Ti   (8f)
```

$NaNb_6O_{15}F$: ABC6D15_oC46_38_b_b_2a2d_2ab4d2e - CIF

```
# CIF file
data_findsym-output
_audit_creation_method FINDSYM

_chemical_name_mineral 'FNaNb6O15'
_chemical_formula_sum 'F Na Nb6 O15'

loop_
_publ_author_name
```

```
'S. Andersson'
_journal_name_full_name
;
 Acta Chemica Scandinavica
;
_journal_volume 19
_journal_year 1965
_journal_page_first 2285
_journal_page_last 2290
_publ_Section_title
;
 The Crystal Structure of NaNb$_{6}$O$_{15}$F and NaNb$_{6}$O$_{15}$OH
;

_aflow_title 'NaNb$_{6}$O$_{15}$F Structure'
_aflow_proto 'ABC6D15_oC46_38_b_b_2a2d_2ab4d2e'
_aflow_params 'a,b/a,c/a,z_{1},z_{2},z_{3},z_{4},z_{5},z_{6},z_{7},y_{8
    ↪ },z_{8},y_{9},z_{9},y_{10},z_{10},y_{11},z_{11},y_{12},z_{12},
    ↪ y_{13},z_{13},y_{14},z_{14},y_{15},z_{15}'
_aflow_params_values '3.949,2.58090655862,3.7277791846,0.0,0.2142,0.856,
    ↪ 0.519,0.0,0.404,0.226,0.3186,0.062,0.3153,0.3154,0.119,0.105,
    ↪ 0.133,0.311,0.305,0.447,0.351,0.178,0.324,0.05,0.323,0.307'
_aflow_Strukturbericht 'None'
_aflow_Pearson 'oC46'

_symmetry_space_group_name_H-M "A m m 2"
_symmetry_Int_Tables_number 38

_cell_length_a      3.94900
_cell_length_b      10.19200
_cell_length_c      14.72100
_cell_angle_alpha   90.00000
_cell_angle_beta    90.00000
_cell_angle_gamma   90.00000

loop_
_space_group_symop_id
_space_group_symop_operation_xyz
1 x,y,z
2 -x,-y,z
3 x,-y,z
4 -x,y,z
5 x,y+1/2,z+1/2
6 -x,-y+1/2,z+1/2
7 x,-y+1/2,z+1/2
8 -x,y+1/2,z+1/2

loop_
_atom_site_label
_atom_site_type_symbol
_atom_site_symmetry_multiplicity
_atom_site_Wyckoff_label
_atom_site_fract_x
_atom_site_fract_y
_atom_site_fract_z
_atom_site_occupancy
Nb1 Nb 2 a 0.00000 0.00000 0.00000 1.00000
Nb2 Nb 2 a 0.00000 0.00000 0.21420 1.00000
O1 O 2 a 0.00000 0.00000 0.85600 1.00000
O2 O 2 a 0.00000 0.00000 0.51900 1.00000
F1 F 2 b 0.50000 0.00000 0.00000 1.00000
Na1 Na 2 b 0.50000 0.00000 0.40400 1.00000
O3 O 2 b 0.50000 0.00000 0.22600 1.00000
Nb3 Nb 4 d 0.00000 0.31860 0.06200 1.00000
Nb4 Nb 4 d 0.00000 0.31530 0.31540 1.00000
O4 O 4 d 0.00000 0.11900 0.10500 1.00000
O5 O 4 d 0.00000 0.13300 0.31100 1.00000
O6 O 4 d 0.00000 0.30500 0.44700 1.00000
O7 O 4 d 0.00000 0.35100 0.17800 1.00000
O8 O 4 e 0.50000 0.32400 0.05000 1.00000
O9 O 4 e 0.50000 0.32300 0.30700 1.00000
```

$NaNb_6O_{15}F$: ABC6D15_oC46_38_b_b_2a2d_2ab4d2e - POSCAR

```
ABC6D15_oC46_38_b_b_2a2d_2ab4d2e & a,b/a,c/a,z1,z2,z3,z4,z5,z6,z7,y8,z8,
    ↪ y9,z9,y10,z10,y11,z11,y12,z12,y13,z13,y14,z14,y15,z15 --params=
    ↪ 3.949,2.58090655862,3.7277791846,0.0,0.2142,0.856,0.519,0.0,
    ↪ 0.404,0.226,0.3186,0.062,0.3153,0.3154,0.119,0.105,0.133,0.311,
    ↪ 0.305,0.447,0.351,0.178,0.324,0.05,0.323,0.307 & Amm2 C_{2v}^{
    ↪ 14} #38 (a^4b^3d^6e^2) & oC46 & None & FNaNb6O15 & FNaNb6O15 &
    ↪ S. Andersson, Acta Chem. Scand. 19, 2285-2290 (1965)
   1.00000000000000
   3.94900000000000   0.00000000000000   0.00000000000000
   0.00000000000000   5.09600000000000  -7.36050000000000
   0.00000000000000   5.09600000000000   7.36050000000000
     F     Na     Nb     O
     1      1      6    15
Direct
   0.50000000000000   0.00000000000000   0.00000000000000   F   (2b)
   0.50000000000000  -0.40400000000000   0.40400000000000   Na  (2b)
   0.00000000000000   0.00000000000000   0.00000000000000   Nb  (2a)
   0.00000000000000  -0.21420000000000   0.21420000000000   Nb  (2a)
   0.00000000000000   0.25660000000000   0.38060000000000   Nb  (4d)
   0.00000000000000  -0.38060000000000  -0.25660000000000   Nb  (4d)
   0.00000000000000  -0.00010000000000   0.63070000000000   Nb  (4d)
   0.00000000000000  -0.63070000000000   0.00010000000000   Nb  (4d)
   0.00000000000000  -0.85600000000000   0.85600000000000   O   (2a)
   0.00000000000000  -0.51900000000000   0.51900000000000   O   (2a)
   0.50000000000000  -0.22600000000000   0.22600000000000   O   (2b)
   0.00000000000000   0.01400000000000   0.22400000000000   O   (4d)
   0.00000000000000  -0.22400000000000  -0.01400000000000   O   (4d)
   0.00000000000000  -0.17800000000000   0.44400000000000   O   (4d)
   0.00000000000000  -0.44400000000000   0.17800000000000   O   (4d)
   0.00000000000000  -0.14200000000000   0.75200000000000   O   (4d)
   0.00000000000000  -0.75200000000000   0.14200000000000   O   (4d)
   0.00000000000000   0.17300000000000   0.52900000000000   O   (4d)
   0.00000000000000  -0.52900000000000  -0.17300000000000   O   (4d)
   0.50000000000000   0.27400000000000   0.37400000000000   O   (4e)
   0.50000000000000  -0.37400000000000  -0.27400000000000   O   (4e)
   0.50000000000000   0.01600000000000   0.63000000000000   O   (4e)
   0.50000000000000  -0.63000000000000  -0.01600000000000   O   (4e)
```

$Rb_2Mo_2O_7$: A2B7C2_oC88_40_abc_2b6c_a3b - CIF

```
# CIF file
data_findsym-output
_audit_creation_method FINDSYM

_chemical_name_mineral 'Mo2O7Rb2'
_chemical_formula_sum 'Mo2 O7 Rb2'

loop_
_publ_author_name
 'Z. A. Solodovnikova'
 'S. F. Solodovnikov'
_journal_name_full_name
;
 Acta Crystallographica Section C: Structural Chemistry
;
_journal_volume 62
_journal_year 2006
_journal_page_first i53
_journal_page_last i56
_publ_Section_title
;
 Rubidium dimolybdate, Rb$_{2}$Mo$_{2}$O$_{7}$, and caesium dimolybdate,
    ↪ Cs$_{2}$Mo$_{2}$O$_{7}$
;

_aflow_title 'Rb$_{2}$Mo$_{2}$O$_{7}$ Structure'
_aflow_proto 'A2B7C2_oC88_40_abc_2b6c_a3b'
_aflow_params 'a,b/a,c/a,z_{1},z_{2},y_{3},z_{3},y_{4},z_{4},y_{5},z_{5
    ↪ },y_{6},z_{6},y_{7},z_{7},y_{8},z_{8},x_{9},y_{9},z_{9},x_{10},
    ↪ y_{10},z_{10},x_{11},y_{11},z_{11},x_{12},y_{12},z_{12},x_{13},
    ↪ y_{13},z_{13},x_{14},y_{14},z_{14},x_{15},y_{15},z_{15}'
_aflow_params_values '11.8887,1.07920125834,0.861860422081,0.56117,0.0,
    ↪ 0.47105,0.28712,0.3413,0.244,0.4846,0.4559,0.48386,0.7361,
    ↪ 0.21144,0.4933,0.2554,-0.0226,0.01169,0.25276,0.22891,0.1103,
    ↪ 0.5171,-0.0426,0.0319,0.3571,0.1036,0.1236,0.5306,0.228,0.8759,
    ↪ 0.205,0.2201,0.1013,0.1507,0.1898,0.0325,0.2957,0.3879'
_aflow_Strukturbericht 'None'
_aflow_Pearson 'oC88'

_symmetry_space_group_name_H-M "A m a 2"
_symmetry_Int_Tables_number 40

_cell_length_a      11.88870
_cell_length_b      12.83030
_cell_length_c      10.24640
_cell_angle_alpha   90.00000
_cell_angle_beta    90.00000
_cell_angle_gamma   90.00000

loop_
_space_group_symop_id
_space_group_symop_operation_xyz
1 x,y,z
2 -x,-y,z
3 x+1/2,-y,z
4 -x+1/2,y,z
5 x,y+1/2,z+1/2
6 -x,-y+1/2,z+1/2
7 x+1/2,-y+1/2,z+1/2
8 -x+1/2,y+1/2,z+1/2

loop_
_atom_site_label
_atom_site_type_symbol
_atom_site_symmetry_multiplicity
_atom_site_Wyckoff_label
_atom_site_fract_x
_atom_site_fract_y
_atom_site_fract_z
_atom_site_occupancy
Mo1 Mo 4 a 0.00000 0.00000 0.56117 1.00000
Rb1 Rb 4 a 0.00000 0.00000 0.00000 1.00000
Mo2 Mo 4 b 0.25000 0.47105 0.28712 1.00000
O1 O 4 b 0.25000 0.34130 0.24400 1.00000
O2 O 4 b 0.25000 0.48460 0.45590 1.00000
Rb2 Rb 4 b 0.25000 0.48386 0.73610 1.00000
Rb3 Rb 4 b 0.25000 0.21144 0.49330 1.00000
Rb4 Rb 4 b 0.25000 0.25540 -0.02260 1.00000
Mo3 Mo 8 c 0.01169 0.25276 0.22891 1.00000
O3 O 8 c 0.11030 0.51710 -0.04260 1.00000
O4 O 8 c 0.03190 0.35710 0.10360 1.00000
O5 O 8 c 0.12360 0.53060 0.22800 1.00000
O6 O 8 c 0.87590 0.20500 0.22010 1.00000
O7 O 8 c 0.10130 0.15070 0.18980 1.00000
O8 O 8 c 0.03250 0.29570 0.38790 1.00000
```

$Rb_2Mo_2O_7$: A2B7C2_oC88_40_abc_2b6c_a3b - POSCAR

```
A2B7C2_oC88_40_abc_2b6c_a3b & a,b/a,c/a,z1,z2,y3,z3,y4,z4,y5,z5,y6,z6,y7
    ↪ ,z7,y8,z8,x9,y9,z9,x10,y10,z10,x11,y11,z11,x12,y12,z12,x13,y13,
    ↪ z13,x14,y14,z14,x15,y15,z15 --params=11.8887,1.07920125834,
    ↪ 0.861860422081,0.56117,0.0,0.47105,0.28712,0.3413,0.244,0.4846,
    ↪ 0.4559,0.48386,0.7361,0.21144,0.4933,0.2554,-0.0226,0.01169,
    ↪ 0.25276,0.22891,0.1103,0.5171,-0.0426,0.0319,0.3571,0.1036,
    ↪ 0.1236,0.5306,0.228,0.8759,0.205,0.2201,0.1013,0.1507,0.1898,
    ↪ 0.0325,0.2957,0.3879 & Ama2 C_{2v}^{16} #40 (a^2b^6c^7) & oC88
```

```
    ↪ & None & Mo2O7Rb2 & Mo2O7Rb2 & Z. A. Solodovnikova and S. F.
    ↪ Solodovnikov, Acta Crystallogr. C 62, i53-i56 (2006)
  1.00000000000000
 11.88870000000000   0.00000000000000   0.00000000000000
  0.00000000000000   6.41515000000000  -5.12320000000000
  0.00000000000000   6.41515000000000   5.12320000000000
   Mo        O        Rb
    8       28         8
Direct
  0.00000000000000  -0.56117000000000   0.56117000000000   Mo   (4a)
  0.50000000000000  -0.56117000000000   0.56117000000000   Mo   (4a)
  0.25000000000000   0.18393000000000   0.75817000000000   Mo   (4b)
  0.75000000000000  -0.75817000000000  -0.18393000000000   Mo   (4b)
  0.01169000000000   0.02385000000000   0.48167000000000   Mo   (8c)
 -0.01169000000000  -0.48167000000000  -0.02385000000000   Mo   (8c)
  0.51169000000000  -0.48167000000000  -0.02385000000000   Mo   (8c)
  0.48831000000000   0.02385000000000   0.48167000000000   Mo   (8c)
  0.25000000000000   0.09730000000000   0.58530000000000    O   (4b)
  0.75000000000000  -0.58530000000000  -0.09730000000000    O   (4b)
  0.25000000000000   0.02870000000000   0.94050000000000    O   (4b)
  0.75000000000000  -0.94050000000000  -0.02870000000000    O   (4b)
  0.11030000000000   0.55970000000000   0.47450000000000    O   (8c)
 -0.11030000000000  -0.47450000000000  -0.55970000000000    O   (8c)
  0.61030000000000  -0.47450000000000  -0.55970000000000    O   (8c)
  0.38970000000000   0.55970000000000   0.47450000000000    O   (8c)
  0.03190000000000   0.25350000000000   0.46070000000000    O   (8c)
 -0.03190000000000  -0.46070000000000  -0.25350000000000    O   (8c)
  0.53190000000000  -0.46070000000000  -0.25350000000000    O   (8c)
  0.46810000000000   0.25350000000000   0.46070000000000    O   (8c)
  0.12360000000000   0.30260000000000   0.75860000000000    O   (8c)
 -0.12360000000000  -0.75860000000000  -0.30260000000000    O   (8c)
  0.62360000000000  -0.75860000000000  -0.30260000000000    O   (8c)
  0.37640000000000   0.30260000000000   0.75860000000000    O   (8c)
  0.87590000000000  -0.01510000000000   0.42510000000000    O   (8c)
 -0.87590000000000  -0.42510000000000   0.01510000000000    O   (8c)
  1.37590000000000  -0.42510000000000   0.01510000000000    O   (8c)
 -0.37590000000000  -0.01510000000000   0.42510000000000    O   (8c)
  0.10130000000000  -0.03910000000000   0.34050000000000    O   (8c)
 -0.10130000000000  -0.34050000000000   0.03910000000000    O   (8c)
  0.60130000000000  -0.34050000000000   0.03910000000000    O   (8c)
  0.39870000000000  -0.03910000000000   0.34050000000000    O   (8c)
  0.03250000000000  -0.09220000000000   0.68360000000000    O   (8c)
 -0.03250000000000  -0.68360000000000   0.09220000000000    O   (8c)
  0.53250000000000  -0.68360000000000   0.09220000000000    O   (8c)
  0.46750000000000  -0.09220000000000   0.68360000000000    O   (8c)
  0.00000000000000   0.00000000000000   0.00000000000000   Rb   (4a)
  0.50000000000000   0.00000000000000   0.00000000000000   Rb   (4a)
  0.25000000000000  -0.25224000000000   1.21996000000000   Rb   (4b)
  0.75000000000000  -1.21996000000000   0.25224000000000   Rb   (4b)
  0.25000000000000  -0.28186000000000   0.70474000000000   Rb   (4b)
  0.75000000000000  -0.70474000000000   0.28186000000000   Rb   (4b)
  0.25000000000000   0.27800000000000   0.23280000000000   Rb   (4b)
  0.75000000000000  -0.23280000000000  -0.27800000000000   Rb   (4b)
```

Orthorhombic $CrO_3$: AB3_oC16_40_b_a2b - CIF

```
# CIF file
data_findsym-output
_audit_creation_method FINDSYM

_chemical_name_mineral 'CrO3'
_chemical_formula_sum 'Cr O3'

loop_
_publ_author_name
 'A. Bystr\"{o}m'
 'K.-A. Wilhelmi'
_journal_name_full_name
;
 Acta Chemica Scandinavica
;
_journal_volume 4
_journal_year 1950
_journal_page_first 1131
_journal_page_last 1141
_publ_Section_title
;
 The Crystal Structure of Chromium Trioxide
;

_aflow_title 'Orthorhombic CrO$_{3}$ Structure'
_aflow_proto 'AB3_oC16_40_b_a2b'
_aflow_params 'a,b/a,c/a,z_{1},y_{2},z_{2},y_{3},z_{3},y_{4},z_{4}'
_aflow_params_values '5.743,1.54199895525,0.833884729236,0.444,0.097,
    ↪ 0.306,0.778,0.0,0.222,0.0'
_aflow_Strukturbericht 'None'
_aflow_Pearson 'oC16'

_symmetry_space_group_name_H-M "A m a 2"
_symmetry_Int_Tables_number 40

_cell_length_a      5.74300
_cell_length_b      8.85570
_cell_length_c      4.78900
_cell_angle_alpha  90.00000
_cell_angle_beta   90.00000
_cell_angle_gamma  90.00000

loop_
_space_group_symop_id
_space_group_symop_operation_xyz
1 x,y,z
2 -x,-y,z
3 x+1/2,-y,z
4 -x+1/2,y,z
5 x,y+1/2,z+1/2
6 -x,-y+1/2,z+1/2
7 x+1/2,-y+1/2,z+1/2
8 -x+1/2,y+1/2,z+1/2

loop_
_atom_site_label
_atom_site_type_symbol
_atom_site_symmetry_multiplicity
_atom_site_Wyckoff_label
_atom_site_fract_x
_atom_site_fract_y
_atom_site_fract_z
_atom_site_occupancy
O1 O 4 a 0.00000 0.00000 0.44400 1.00000
Cr1 Cr 4 b 0.25000 0.09700 0.30600 1.00000
O2 O 4 b 0.25000 0.77800 0.00000 1.00000
O3 O 4 b 0.25000 0.22200 0.00000 1.00000
```

Orthorhombic $CrO_3$: AB3_oC16_40_b_a2b - POSCAR

```
AB3_oC16_40_b_a2b & a,b/a,c/a,z1,y2,z2,y3,z3,y4,z4 --params=5.743,
    ↪ 1.54199895525,0.833884729236,0.444,0.097,0.306,0.778,0.0,0.222,
    ↪ 0.0 & Ama2 C_{2v}^{16} #40 (ab^3) & oC16 & None & CrO3 & CrO3 &
    ↪  A. Bystr\"{o}m and K.-A. Wilhelmi, Acta Chem. Scand. 4,
    ↪ 1131-1141 (1950)
  1.00000000000000
  5.74300000000000   0.00000000000000   0.00000000000000
  0.00000000000000   4.42785000000000  -2.39450000000000
  0.00000000000000   4.42785000000000   2.39450000000000
   Cr        O
    2        6
Direct
  0.25000000000000  -0.20900000000000   0.40300000000000   Cr   (4b)
  0.75000000000000  -0.40300000000000   0.20900000000000   Cr   (4b)
  0.00000000000000  -0.44400000000000   0.44400000000000    O   (4a)
  0.50000000000000  -0.44400000000000   0.44400000000000    O   (4a)
  0.25000000000000   0.77800000000000   0.77800000000000    O   (4b)
  0.75000000000000  -0.77800000000000  -0.77800000000000    O   (4b)
  0.25000000000000   0.22200000000000   0.22200000000000    O   (4b)
  0.75000000000000  -0.22200000000000  -0.22200000000000    O   (4b)
```

Santite ($KB_5O_8 \cdot 4H_2O$, $K3_5$): A5B8CD12_oC104_41_a2b_4b_a_6b - CIF

```
# CIF file
data_findsym-output
_audit_creation_method FINDSYM

_chemical_name_mineral 'Santite'
_chemical_formula_sum 'B5 H8 K O12'

loop_
_publ_author_name
 'W. H. Zachariasen'
 'H. A. Plettinger'
_journal_name_full_name
;
 Acta Cristallographica
;
_journal_volume 16
_journal_year 1963
_journal_page_first 376
_journal_page_last 379
_publ_Section_title
;
 Refinement of the structure of potassium pentaborate tetrahydrate
;

_aflow_title 'Santite (KB$_{5}$O$_{8}$$\cdot$4H$_{2}$O, $K3_{5}$)
    ↪ Structure'
_aflow_proto 'A5B8CD12_oC104_41_a2b_4b_a_6b'
_aflow_params 'a,b/a,c/a,z_{1},z_{2},x_{3},y_{3},z_{3},x_{4},y_{4},z_{4
    ↪ },x_{5},y_{5},z_{5},x_{6},y_{6},z_{6},x_{7},y_{7},z_{7},x_{8},
    ↪ y_{8},z_{8},x_{9},y_{9},z_{9},x_{10},y_{10},z_{10},x_{11},y_{11
    ↪ },z_{11},x_{12},y_{12},z_{12},x_{13},y_{13},z_{13},x_{14},y_{14
    ↪ },z_{14}'
_aflow_params_values '11.062,1.01021605496,0.817302476948,0.4041,0.0,
    ↪ 0.189,0.0943,0.3126,0.2042,0.9432,0.4952,0.978,0.247,0.804,
    ↪ 0.967,0.111,0.649,0.161,0.159,0.741,0.251,0.316,0.193,0.0672,
    ↪ 0.0843,0.3082,0.0832,0.9313,0.4991,0.2591,0.0254,0.4033,0.2448,
    ↪ 0.1753,0.2235,0.2816,0.8801,0.5802,0.0151,0.1707,0.7588'
_aflow_Strukturbericht '$K3_{5}$'
_aflow_Pearson 'oC104'

_symmetry_space_group_name_H-M "A b a 2"
_symmetry_Int_Tables_number 41

_cell_length_a     11.06200
_cell_length_b     11.17501
_cell_length_c      9.04100
_cell_angle_alpha  90.00000
_cell_angle_beta   90.00000
_cell_angle_gamma  90.00000

loop_
_space_group_symop_id
_space_group_symop_operation_xyz
1 x,y,z
2 -x,-y,z
3 x+1/2,-y,z+1/2
4 -x+1/2,y,z+1/2
5 x,y+1/2,z+1/2
6 -x,-y+1/2,z+1/2
7 x+1/2,-y+1/2,z
8 -x+1/2,y+1/2,z
```

```
loop_
_atom_site_label
_atom_site_type_symbol
_atom_site_symmetry_multiplicity
_atom_site_Wyckoff_label
_atom_site_fract_x
_atom_site_fract_y
_atom_site_fract_z
_atom_site_occupancy
B1 B 4 a 0.00000 0.00000 0.40410 1.00000
K1 K 4 a 0.00000 0.00000 0.00000 1.00000
B2 B 8 b 0.18900 0.09430 0.31260 1.00000
B3 B 8 b 0.20420 0.94320 0.49520 1.00000
H1 H 8 b 0.97800 0.24700 0.80400 1.00000
H2 H 8 b 0.96700 0.11100 0.64900 1.00000
H3 H 8 b 0.16100 0.15900 0.74100 1.00000
H4 H 8 b 0.25100 0.31600 0.19300 1.00000
O1 O 8 b 0.06720 0.08430 0.30820 1.00000
O2 O 8 b 0.08320 0.93130 0.49910 1.00000
O3 O 8 b 0.25910 0.02540 0.40330 1.00000
O4 O 8 b 0.24480 0.17530 0.22350 1.00000
O5 O 8 b 0.28160 0.88010 0.58020 1.00000
O6 O 8 b 0.01510 0.17070 0.75880 1.00000
```

Santite ($KB_5O_8 \cdot 4H_2O$, $K3_5$): A5B8CD12_oC104_41_a2b_4b_a_6b - POSCAR

```
A5B8CD12_oC104_41_a2b_4b_a_6b & a,b/a,c/a,z1,z2,x3,y3,z3,x4,y4,z4,x5,y5,
    ↪ z5,x6,y6,z6,x7,y7,z7,x8,y8,z8,x9,y9,z9,x10,y10,z10,x11,y11,z11,
    ↪ x12,y12,z12,x13,y13,z13,x14,y14,z14 --params=11.062,
    ↪ 1.01021605496,0.817302476948,0.4041,0.0,0.189,0.0943,0.3126,
    ↪ 0.2042,0.9432,0.4952,0.978,0.247,0.804,0.967,0.111,0.649,0.161,
    ↪ 0.159,0.741,0.251,0.316,0.193,0.0672,0.0843,0.3082,0.0832,
    ↪ 0.9313,0.4991,0.2591,0.0254,0.4033,0.2448,0.1753,0.2235,0.2816,
    ↪ 0.8801,0.5802,0.0151,0.1707,0.7588 & Aba2 C_{2v}^{17} #41 (a^2b
    ↪ ^12) & oC104 & $K3_{5}$ & B5H8KO12 & Santite & W. H.
    ↪ Zachariasen and H. A. Plettinger, Acta Cryst. 16, 376-379 (1963
    ↪ )
   1.00000000000000
  11.06200000000000   0.00000000000000   0.00000000000000
   0.00000000000000   5.58750500000000  -4.52050000000000
   0.00000000000000   5.58750500000000   4.52050000000000
     B      H      K      O
    10     16      2     24
Direct
   0.00000000000000  -0.40410000000000   0.40410000000000   B   (4a)
   0.50000000000000   0.09590000000000   0.90410000000000   B   (4a)
   0.18900000000000  -0.21830000000000   0.40690000000000   B   (8b)
  -0.18900000000000  -0.40690000000000   0.21830000000000   B   (8b)
   0.68900000000000   0.09310000000000   0.71830000000000   B   (8b)
   0.31100000000000   0.28170000000000   0.90690000000000   B   (8b)
   0.20420000000000   0.44800000000000   1.43840000000000   B   (8b)
  -0.20420000000000  -1.43840000000000  -0.44800000000000   B   (8b)
   0.70420000000000  -0.93840000000000   0.05200000000000   B   (8b)
   0.29580000000000   0.94800000000000   1.93840000000000   B   (8b)
   0.97800000000000  -0.55700000000000   1.05100000000000   H   (8b)
  -0.97800000000000  -1.05100000000000   0.55700000000000   H   (8b)
   1.47800000000000  -0.55100000000000   1.05700000000000   H   (8b)
  -0.47800000000000  -0.05700000000000   1.55100000000000   H   (8b)
   0.96700000000000  -0.53800000000000   0.76000000000000   H   (8b)
  -0.96700000000000  -0.76000000000000   0.53800000000000   H   (8b)
   1.46700000000000  -0.26000000000000   1.03800000000000   H   (8b)
  -0.46700000000000  -0.03800000000000   1.26000000000000   H   (8b)
   0.16100000000000  -0.58200000000000   0.90000000000000   H   (8b)
  -0.16100000000000  -0.90000000000000   0.58200000000000   H   (8b)
   0.66100000000000  -0.40000000000000   1.08200000000000   H   (8b)
   0.33900000000000  -0.08200000000000   1.40000000000000   H   (8b)
   0.25100000000000   0.12300000000000   0.50900000000000   H   (8b)
  -0.25100000000000  -0.50900000000000  -0.12300000000000   H   (8b)
   0.75100000000000  -0.00900000000000   0.37700000000000   H   (8b)
   0.24900000000000   0.62300000000000   1.00900000000000   H   (8b)
   0.00000000000000   0.00000000000000   0.00000000000000   K   (4a)
   0.50000000000000   0.50000000000000   0.50000000000000   K   (4a)
   0.06720000000000  -0.22390000000000   0.39250000000000   O   (8b)
  -0.06720000000000  -0.39250000000000   0.22390000000000   O   (8b)
   0.56720000000000   0.10750000000000   0.72390000000000   O   (8b)
   0.43280000000000   0.27610000000000   0.89250000000000   O   (8b)
   0.08320000000000   0.43220000000000   1.43040000000000   O   (8b)
  -0.08320000000000  -1.43040000000000  -0.43220000000000   O   (8b)
   0.58320000000000  -0.93040000000000   0.06780000000000   O   (8b)
   0.41680000000000   0.93220000000000   1.93040000000000   O   (8b)
   0.25910000000000  -0.37790000000000   0.42870000000000   O   (8b)
  -0.25910000000000  -0.42870000000000   0.37790000000000   O   (8b)
   0.75910000000000   0.07130000000000   0.87790000000000   O   (8b)
   0.24090000000000   0.12210000000000   0.92870000000000   O   (8b)
   0.24480000000000  -0.04820000000000   0.39880000000000   O   (8b)
  -0.24480000000000  -0.39880000000000   0.04820000000000   O   (8b)
   0.74480000000000   0.10120000000000   0.54820000000000   O   (8b)
   0.25520000000000   0.45180000000000   0.89880000000000   O   (8b)
   0.28160000000000   0.29990000000000   1.46030000000000   O   (8b)
  -0.28160000000000  -1.46030000000000  -0.29990000000000   O   (8b)
   0.78160000000000  -0.96030000000000   0.20010000000000   O   (8b)
   0.21840000000000   0.79990000000000   1.96030000000000   O   (8b)
   0.01510000000000  -0.58810000000000   0.92950000000000   O   (8b)
  -0.01510000000000  -0.92950000000000   0.58810000000000   O   (8b)
   0.51510000000000  -0.42950000000000   1.08810000000000   O   (8b)
   0.48490000000000  -0.08810000000000   1.42950000000000   O   (8b)
```

$Ag_2O_3$: A2B3_oF40_43_b_ab - CIF

```
# CIF file
data_findsym-output
_audit_creation_method FINDSYM

_chemical_name_mineral 'Ag2O3'
_chemical_formula_sum 'Ag2 O3'

loop_
_publ_author_name
 'B. Standke'
 'M. Jansen'
_journal_name_full_name
;
 Zeitschrift fur Anorganische und Allgemeine Chemie
;
_journal_volume 535
_journal_year 1986
_journal_page_first 39
_journal_page_last 46
_publ_Section_title
;
 Darstellung und Kristallstruktur von Ag$_{2}$O$_{3}$
;

_aflow_title 'Ag$_{2}$O$_{3}$ Structure'
_aflow_proto 'A2B3_oF40_43_b_ab'
_aflow_params 'a,b/a,c/a,z_{1},x_{2},y_{2},z_{2},x_{3},y_{3},z_{3}'
_aflow_params_values '12.869,0.815137151294,0.284699665864,0.8402,-
    ↪ 0.0467,0.1325,0.5,0.083,0.2333,0.6153'
_aflow_Strukturbericht 'None'
_aflow_Pearson 'oF40'

_symmetry_space_group_name_H-M "F d d 2"
_symmetry_Int_Tables_number 43

_cell_length_a      12.86900
_cell_length_b      10.49000
_cell_length_c       3.66380
_cell_angle_alpha   90.00000
_cell_angle_beta    90.00000
_cell_angle_gamma   90.00000

loop_
_space_group_symop_id
_space_group_symop_operation_xyz
1 x,y,z
2 -x,-y,z
3 -x+1/4,y+1/4,z+1/4
4 x+1/4,-y+1/4,z+1/4
5 x,y+1/2,z+1/2
6 -x,-y+1/2,z+1/2
7 -x+1/4,y+3/4,z+3/4
8 x+1/4,-y+3/4,z+3/4
9 x+1/2,y,z+1/2
10 -x+1/2,-y,z+1/2
11 -x+3/4,y+1/4,z+3/4
12 x+3/4,-y+1/4,z+3/4
13 x+1/2,y+1/2,z
14 -x+1/2,-y+1/2,z
15 -x+3/4,y+3/4,z+1/4
16 x+3/4,-y+3/4,z+1/4

loop_
_atom_site_label
_atom_site_type_symbol
_atom_site_symmetry_multiplicity
_atom_site_Wyckoff_label
_atom_site_fract_x
_atom_site_fract_y
_atom_site_fract_z
_atom_site_occupancy
O1 O 8 a 0.00000 0.00000 0.84020 1.00000
Ag1 Ag 16 b -0.04670 0.13250 0.50000 1.00000
O2 O 16 b 0.08300 0.23330 0.61530 1.00000
```

$Ag_2O_3$: A2B3_oF40_43_b_ab - POSCAR

```
A2B3_oF40_43_b_ab & a,b/a,c/a,z1,x2,y2,z2,x3,y3,z3 --params=12.869,
    ↪ 0.815137151294,0.284699665864,0.8402,-0.0467,0.1325,0.5,0.083,
    ↪ 0.2333,0.6153 & Fdd2 C_{2v}^{19} #43 (ab^2) & oF40 & None &
    ↪ Ag2O3 & Ag2O3 & B. Standke and M. Jansen, Z. Anorg. Allg. Chem.
    ↪ 535, 39-46 (1986)
   1.00000000000000
   0.00000000000000   5.24500000000000   1.83190000000000
   6.43450000000000   0.00000000000000   1.83190000000000
   6.43450000000000   5.24500000000000   0.00000000000000
    Ag      O
     4      6
Direct
   0.67920000000000   0.32080000000000  -0.41420000000000   Ag   (16b)
   0.32080000000000   0.67920000000000  -0.58580000000000   Ag   (16b)
   0.66420000000000   0.83580000000000  -0.42920000000000   Ag   (16b)
   0.83580000000000   0.66420000000000  -0.07080000000000   Ag   (16b)
   0.84020000000000   0.84020000000000  -0.84020000000000   O    (8a)
   1.09020000000000   1.09020000000000  -0.59020000000000   O    (8a)
   0.76560000000000   0.46500000000000  -0.29900000000000   O   (16b)
   0.46500000000000   0.76560000000000  -0.93160000000000   O   (16b)
   0.54900000000000   1.18160000000000  -0.51560000000000   O   (16b)
   1.18160000000000   0.54900000000000  -0.21500000000000   O   (16b)
```

Natrolite ($Na_2Al_2Si_3O_{10} \cdot 2H_2O$, $S6_{10}$): A2B4C2D12E3_oF184_43_b_2b_b_6b_ab - CIF

```
# CIF file
data_findsym-output
_audit_creation_method FINDSYM

_chemical_name_mineral 'Natrolite'
_chemical_formula_sum 'Al2 H4 Na2 O12 Si3'

loop_
```

```
_publ_author_name
 'A. Kirfel'
 'M. Orthen'
 'G. Will'
_journal_name_full_name
;
 Zeolites
;
_journal_volume 4
_journal_year 1984
_journal_page_first 140
_journal_page_last 146
_publ_Section_title
;
 Natrolite: refinement of the crystal structure of two samples from
     ↪ Marienberg (Usti nad Labem, CSSR)
;

_aflow_title 'Natrolite (Na$_{2}$Al$_{2}$Si$_{3}$O$_{10}$$\cdot$2H$_{2}
     ↪ $O, $S6_{10}$) Structure'
_aflow_proto 'A2B4C2D12E3_oF184_43_b_2b_b_6b_ab'
_aflow_params 'a,b/a,c/a,z_{1},x_{2},y_{2},z_{2},x_{3},y_{3},z_{3},x_{4
     ↪ },y_{4},z_{4},x_{5},y_{5},z_{5},x_{6},y_{6},z_{6},x_{7},y_{7},
     ↪ z_{7},x_{8},y_{8},z_{8},x_{9},y_{9},z_{9},x_{10},y_{10},z_{10},
     ↪ x_{11},y_{11},z_{11},x_{12},y_{12},z_{12}'
_aflow_params_values '18.296,1.01918452121,0.359914735461,0.0,0.0374,
     ↪ 0.0937,0.6152,0.0533,0.1604,0.0616,0.0848,0.1932,0.1731,0.2208,
     ↪ 0.0307,0.6179,0.0226,0.0686,0.866,0.0701,0.1818,0.6099,0.0983,
     ↪ 0.0351,0.5003,0.2065,0.1529,0.7261,0.1804,0.2273,0.3903,0.0561,
     ↪ 0.1896,0.1112,0.1533,0.2113,0.6231'
_aflow_Strukturbericht '$S6_{10}$'
_aflow_Pearson 'oF184'

_symmetry_space_group_name_H-M "F d d 2"
_symmetry_Int_Tables_number 43

_cell_length_a      18.29600
_cell_length_b      18.64700
_cell_length_c      6.58500
_cell_angle_alpha   90.00000
_cell_angle_beta    90.00000
_cell_angle_gamma   90.00000

loop_
_space_group_symop_id
_space_group_symop_operation_xyz
1 x,y,z
2 -x,-y,z
3 -x+1/4,y+1/4,z+1/4
4 x+1/4,-y+1/4,z+1/4
5 x,y+1/2,z+1/2
6 -x,-y+1/2,z+1/2
7 -x+1/4,y+3/4,z+3/4
8 x+1/4,-y+3/4,z+3/4
9 x+1/2,y,z+1/2
10 -x+1/2,-y,z+1/2
11 -x+3/4,y+1/4,z+3/4
12 x+3/4,-y+1/4,z+3/4
13 x+1/2,y+1/2,z
14 -x+1/2,-y+1/2,z
15 -x+3/4,y+3/4,z+1/4
16 x+3/4,-y+3/4,z+1/4

loop_
_atom_site_label
_atom_site_type_symbol
_atom_site_symmetry_multiplicity
_atom_site_Wyckoff_label
_atom_site_fract_x
_atom_site_fract_y
_atom_site_fract_z
_atom_site_occupancy
Si1 Si 8 a 0.00000 0.00000 0.00000 1.00000
Al1 Al 16 b 0.03740 0.09370 0.61520 1.00000
H1 H 16 b 0.05330 0.16040 0.06160 1.00000
H2 H 16 b 0.08480 0.19320 0.17310 1.00000
Na1 Na 16 b 0.22080 0.03070 0.61790 1.00000
O1 O 16 b 0.02260 0.06860 0.86600 1.00000
O2 O 16 b 0.07010 0.18180 0.60990 1.00000
O3 O 16 b 0.09830 0.03510 0.50030 1.00000
O4 O 16 b 0.20650 0.15290 0.72610 1.00000
O5 O 16 b 0.18040 0.22730 0.39030 1.00000
O6 O 16 b 0.05610 0.18960 0.11120 1.00000
Si2 Si 16 b 0.15330 0.21130 0.62310 1.00000
```

Natrolite ($Na_2Al_2Si_3O_{10}{\cdot}2H_2O$, $S6_{10}$): A2B4C2D12E3_oF184_43_b_2b_b_6b_ab - POSCAR

```
A2B4C2D12E3_oF184_43_b_2b_b_6b_ab & a,b/a,c/a,z1,x2,y2,z2,x3,y3,z3,x4,y4
     ↪ ,z4,x5,y5,z5,x6,y6,z6,x7,y7,z7,x8,y8,z8,x9,y9,z9,x10,y10,z10,
     ↪ x11,y11,z11,x12,y12,z12 --params=18.296,1.01918452121,
     ↪ 0.359914735461,0.0,0.0374,0.0937,0.6152,0.0533,0.1604,0.0616,
     ↪ 0.0848,0.1932,0.1731,0.2208,0.0307,0.6179,0.0226,0.0686,0.866,
     ↪ 0.0701,0.1818,0.6099,0.0983,0.0351,0.5003,0.2065,0.1529,0.7261,
     ↪ 0.1804,0.2273,0.3903,0.0561,0.1896,0.1112,0.1533,0.2113,0.6231
     ↪ & Fdd2 C_{2v}^{19} #43 (ab^11) & oF184 & $S6_{10}$ &
     ↪ Al2H4Na2O12Si3 & Natrolite & A. Kirfel and M. Orthen and G.
     ↪ Will, Zeolites 4, 140-146 (1984)
   1.00000000000000
   0.00000000000000   9.32350000000000   3.29250000000000
   9.14800000000000   0.00000000000000   3.29250000000000
   9.14800000000000   9.32350000000000   0.00000000000000
     Al       H      Na       O      Si
      4       8       4      24       6
Direct
   0.67150000000000   0.55890000000000  -0.48410000000000   Al   (16b)
   0.55890000000000   0.67150000000000  -0.74630000000000   Al   (16b)
   0.73410000000000   0.99630000000000  -0.42150000000000   Al   (16b)
   0.99630000000000   0.73410000000000  -0.30890000000000   Al   (16b)
   0.16870000000000  -0.04550000000000   0.15210000000000   H    (16b)
  -0.04550000000000   0.16870000000000  -0.27530000000000   H    (16b)
   0.09790000000000   0.52530000000000   0.08130000000000   H    (16b)
   0.52530000000000   0.09790000000000   0.29550000000000   H    (16b)
   0.28150000000000   0.06470000000000   0.10490000000000   H    (16b)
   0.06470000000000   0.28150000000000  -0.45110000000000   H    (16b)
   0.14510000000000   0.70110000000000  -0.03150000000000   H    (16b)
   0.70110000000000   0.14510000000000   0.18530000000000   H    (16b)
   0.42780000000000   0.80800000000000  -0.36640000000000   Na   (16b)
   0.80800000000000   0.42780000000000  -0.86940000000000   Na   (16b)
   0.61640000000000   1.11940000000000  -0.17780000000000   Na   (16b)
   1.11940000000000   0.61640000000000  -0.55800000000000   Na   (16b)
   0.91200000000000   0.82000000000000  -0.77480000000000   O    (16b)
   0.82000000000000   0.91200000000000  -0.95720000000000   O    (16b)
   1.02480000000000   1.20720000000000  -0.66200000000000   O    (16b)
   1.20720000000000   1.02480000000000  -0.57000000000000   O    (16b)
   0.72160000000000   0.49820000000000  -0.35800000000000   O    (16b)
   0.49820000000000   0.72160000000000  -0.86180000000000   O    (16b)
   0.60800000000000   1.11180000000000  -0.47160000000000   O    (16b)
   1.11180000000000   0.60800000000000  -0.24820000000000   O    (16b)
   0.43710000000000   0.56350000000000  -0.36690000000000   O    (16b)
   0.56350000000000   0.43710000000000  -0.63370000000000   O    (16b)
   0.61690000000000   0.88370000000000  -0.18710000000000   O    (16b)
   0.88370000000000   0.61690000000000  -0.31350000000000   O    (16b)
   0.67250000000000   0.77970000000000  -0.36670000000000   O    (16b)
   0.77970000000000   0.67250000000000  -1.08550000000000   O    (16b)
   0.61670000000000   1.33550000000000  -0.42250000000000   O    (16b)
   1.33550000000000   0.61670000000000  -0.52970000000000   O    (16b)
   0.43720000000000   0.34340000000000   0.01740000000000   O    (16b)
   0.34340000000000   0.43720000000000  -0.79800000000000   O    (16b)
   0.23260000000000   1.04800000000000  -0.18720000000000   O    (16b)
   1.04800000000000   0.23260000000000  -0.09340000000000   O    (16b)
   0.24470000000000  -0.02230000000000   0.13450000000000   O    (16b)
  -0.02230000000000   0.24470000000000  -0.35690000000000   O    (16b)
   0.11550000000000   0.60690000000000   0.00530000000000   O    (16b)
   0.60690000000000   0.11550000000000   0.27230000000000   O    (16b)
   0.00000000000000   0.00000000000000   0.00000000000000   Si    (8a)
   0.25000000000000   0.25000000000000   0.25000000000000   Si    (8a)
   0.68110000000000   0.56510000000000  -0.25850000000000   Si   (16b)
   0.56510000000000   0.68110000000000  -0.98770000000000   Si   (16b)
   0.50850000000000   1.23770000000000  -0.43110000000000   Si   (16b)
   1.23770000000000   0.50850000000000  -0.31510000000000   Si   (16b)
```

Blossite ($\alpha$-$Cu_2V_2O_7$): A2B7C2_oF88_43_b_a3b_b - CIF

```
# CIF file
data_findsym-output
_audit_creation_method FINDSYM

_chemical_name_mineral 'Blossite'
_chemical_formula_sum 'Cu2 O7 V2'

loop_
_publ_author_name
 'C. Calvo'
 'R. Faggiani'
_journal_name_full_name
;
 Acta Crystallographica Section B: Structural Science
;
_journal_volume 31
_journal_year 1975
_journal_page_first 603
_journal_page_last 605
_publ_Section_title
;
 $\alpha$ Cupric Divanadate
;

# Found in The American Mineralogist Crystal Structure Database, 2003

_aflow_title 'Blossite ($\alpha$-Cu$_{2}$V$_{2}$O$_{7}$) Structure'
_aflow_proto 'A2B7C2_oF88_43_b_a3b_b'
_aflow_params 'a,b/a,c/a,z_{1},x_{2},y_{2},z_{2},x_{3},y_{3},z_{3},x_{4
     ↪ },y_{4},z_{4},x_{5},y_{5},z_{5},x_{6},y_{6},z_{6}'
_aflow_params_values '20.645,0.406054734803,0.312036812788,-0.0908,
     ↪ 0.1658,0.3646,0.75,0.2453,0.5622,0.2774,0.1446,0.4368,0.0332,
     ↪ 0.1619,0.3477,0.4592,0.19906,0.4054,0.2343'
_aflow_Strukturbericht 'None'
_aflow_Pearson 'oF88'

_symmetry_space_group_name_H-M "F d d 2"
_symmetry_Int_Tables_number 43

_cell_length_a      20.64500
_cell_length_b      8.38300
_cell_length_c      6.44200
_cell_angle_alpha   90.00000
_cell_angle_beta    90.00000
_cell_angle_gamma   90.00000

loop_
_space_group_symop_id
_space_group_symop_operation_xyz
1 x,y,z
2 -x,-y,z
3 -x+1/4,y+1/4,z+1/4
4 x+1/4,-y+1/4,z+1/4
5 x,y+1/2,z+1/2
6 -x,-y+1/2,z+1/2
7 -x+1/4,y+3/4,z+3/4
8 x+1/4,-y+3/4,z+3/4
```

```
9 x+1/2,y,z+1/2
10 -x+1/2,-y,z+1/2
11 -x+3/4,y+1/4,z+3/4
12 x+3/4,-y+1/4,z+3/4
13 x+1/2,y+1/2,z
14 -x+1/2,-y+1/2,z
15 -x+3/4,y+3/4,z+1/4
16 x+3/4,-y+3/4,z+1/4

loop_
_atom_site_label
_atom_site_type_symbol
_atom_site_symmetry_multiplicity
_atom_site_Wyckoff_label
_atom_site_fract_x
_atom_site_fract_y
_atom_site_fract_z
_atom_site_occupancy
O1 O 8 a 0.00000 0.00000 -0.09080 1.00000
Cu1 Cu 16 b 0.16580 0.36460 0.75000 1.00000
O2 O 16 b 0.24530 0.56220 0.27740 1.00000
O3 O 16 b 0.14460 0.43680 0.03320 1.00000
O4 O 16 b 0.16190 0.34770 0.45920 1.00000
V1 V 16 b 0.19906 0.40540 0.23430 1.00000
```

Blossite ($\alpha$-$Cu_2V_2O_7$): A2B7C2_oF88_43_b_a3b_b - POSCAR

```
A2B7C2_oF88_43_b_a3b_b & a,b/a,c/a,z1,x2,y2,z2,x3,y3,z3,x4,y4,z4,x5,y5,
    ↪ z5,x6,y6,z6 --params=20.645,0.406054734803,0.312036812788,-
    ↪ 0.0908,0.1658,0.3646,0.75,0.2453,0.5622,0.2774,0.1446,0.4368,
    ↪ 0.0332,0.1619,0.3477,0.4592,0.19906,0.4054,0.2343 & Fdd2 C_{2v
    ↪ }^{19} #43 (ab^5) & oF88 & None & Cu2O7V2 & Blossite & C. Calvo
    ↪  and R. Faggiani, Acta Crystallogr. Sect. B Struct. Sci. 31,
    ↪ 603-605 (1975)
   1.00000000000000
   0.00000000000000   4.19150000000000   3.22100000000000
  10.32250000000000   0.00000000000000   3.22100000000000
  10.32250000000000   4.19150000000000   0.00000000000000
    Cu        O        V
     4       14        4
Direct
   0.94880000000000   0.55120000000000  -0.21960000000000   Cu   (16b)
   0.55120000000000   0.94880000000000  -1.28040000000000   Cu   (16b)
   0.46960000000000   1.53040000000000  -0.69880000000000   Cu   (16b)
   1.53040000000000   0.46960000000000  -0.30120000000000   Cu   (16b)
  -0.09080000000000  -0.09080000000000   0.09080000000000   O    (8a)
   0.15920000000000   0.15920000000000   0.34080000000000   O    (8a)
   0.59430000000000  -0.03950000000000   0.53010000000000   O   (16b)
  -0.03950000000000   0.59430000000000  -1.08490000000000   O   (16b)
  -0.28010000000000   1.33490000000000  -0.34430000000000   O   (16b)
   1.33490000000000  -0.28010000000000   0.28950000000000   O   (16b)
   0.32540000000000  -0.25900000000000   0.54820000000000   O   (16b)
  -0.25900000000000   0.32540000000000  -0.61460000000000   O   (16b)
  -0.29820000000000   0.86460000000000  -0.07540000000000   O   (16b)
   0.86460000000000  -0.29820000000000   0.50900000000000   O   (16b)
   0.64500000000000   0.27340000000000   0.05040000000000   O   (16b)
   0.27340000000000   0.64500000000000  -0.96880000000000   O   (16b)
   0.19960000000000   1.21880000000000  -0.39500000000000   O   (16b)
   1.21880000000000   0.19960000000000  -0.02340000000000   O   (16b)
   0.44064000000000   0.02796000000000   0.37016000000000   V   (16b)
   0.02796000000000   0.44064000000000  -0.83876000000000   V   (16b)
  -0.12016000000000   1.08876000000000  -0.19064000000000   V   (16b)
   1.08876000000000  -0.12016000000000   0.22204000000000   V   (16b)
```

Archerite ($KH_2PO_4$): A2BC4D_oF64_43_b_a_2b_a - CIF

```
# CIF file
data_findsym-output
_audit_creation_method FINDSYM

_chemical_name_mineral 'Archerite'
_chemical_formula_sum 'H2 K O4 P'

loop_
_publ_author_name
 'H. A. Levy'
 'S. W. Peterson'
 'S. H. Simonsen'
_journal_name_full_name
;
 Physical Review
;
_journal_volume 93
_journal_year 1954
_journal_page_first 1120
_journal_page_last 1121
_publ_Section_title
;
 Neutron Diffraction Study of the Ferroelectric Modification of
    ↪ Potassium Dihydrogen Phosphate
;

# Found in The American Mineralogist Crystal Structure Database, 2003

_aflow_title 'Archerite (KH$_{2}$PO$_{4}$) Structure'
_aflow_proto 'A2BC4D_oF64_43_b_a_2b_a'
_aflow_params 'a,b/a,c/a,z_{1},z_{2},x_{3},y_{3},z_{3},x_{4},y_{4},z_{4
    ↪ },x_{5},y_{5},z_{5}'
_aflow_params_values '10.53,0.991452991453,0.655270655271,0.512,0.0,
    ↪ 0.188,-0.0375,0.1355,0.116,0.0345,0.131,-0.0345,0.116,0.8765'
_aflow_Strukturbericht 'None'
_aflow_Pearson 'oF64'

_symmetry_space_group_name_H-M "F d d 2"
_symmetry_Int_Tables_number 43
```

```
_cell_length_a     10.53000
_cell_length_b     10.44000
_cell_length_c     6.90000
_cell_angle_alpha  90.00000
_cell_angle_beta   90.00000
_cell_angle_gamma  90.00000

loop_
_space_group_symop_id
_space_group_symop_operation_xyz
1 x,y,z
2 -x,-y,z
3 -x+1/4,y+1/4,z+1/4
4 x+1/4,-y+1/4,z+1/4
5 x,y+1/2,z+1/2
6 -x,-y+1/2,z+1/2
7 -x+1/4,y+3/4,z+3/4
8 x+1/4,-y+3/4,z+3/4
9 x+1/2,y,z+1/2
10 -x+1/2,-y,z+1/2
11 -x+3/4,y+1/4,z+3/4
12 x+3/4,-y+1/4,z+3/4
13 x+1/2,y+1/2,z
14 -x+1/2,-y+1/2,z
15 -x+3/4,y+3/4,z+1/4
16 x+3/4,-y+3/4,z+1/4

loop_
_atom_site_label
_atom_site_type_symbol
_atom_site_symmetry_multiplicity
_atom_site_Wyckoff_label
_atom_site_fract_x
_atom_site_fract_y
_atom_site_fract_z
_atom_site_occupancy
K1 K 8 a 0.00000 0.00000 0.51200 1.00000
P1 P 8 a 0.00000 0.00000 0.00000 1.00000
H1 H 16 b 0.18800 -0.03750 0.13550 1.00000
O1 O 16 b 0.11600 0.03450 0.13100 1.00000
O2 O 16 b -0.03450 0.11600 0.87650 1.00000
```

Archerite ($KH_2PO_4$): A2BC4D_oF64_43_b_a_2b_a - POSCAR

```
A2BC4D_oF64_43_b_a_2b_a & a,b/a,c/a,z1,z2,x3,y3,z3,x4,y4,z4,x5,y5,z5 --
    ↪ params=10.53,0.991452991453,0.655270655271,0.512,0.0,0.188,-
    ↪ 0.0375,0.1355,0.116,0.0345,0.131,-0.0345,0.116,0.8765 & Fdd2 C_
    ↪ {2v}^{19} #43 (a^2b^3) & oF64 & None & H2KO4P & Archerite & H.
    ↪ A. Levy and S. W. Peterson and S. H. Simonsen, Phys. Rev. 93,
    ↪ 1120-1121 (1954)
   1.00000000000000
   0.00000000000000   5.22000000000000   3.45000000000000
   5.26500000000000   0.00000000000000   3.45000000000000
   5.26500000000000   5.22000000000000   0.00000000000000
     H        K        O        P
     4        2        8        2
Direct
  -0.09000000000000   0.36100000000000   0.01500000000000   H   (16b)
   0.36100000000000  -0.09000000000000  -0.28600000000000   H   (16b)
   0.23500000000000   0.53600000000000   0.34000000000000   H   (16b)
   0.53600000000000   0.23500000000000  -0.11100000000000   H   (16b)
   0.51200000000000   0.51200000000000  -0.51200000000000   K    (8a)
   0.76200000000000   0.76200000000000  -0.26200000000000   K    (8a)
   0.04950000000000   0.21250000000000   0.01950000000000   O   (16b)
   0.21250000000000   0.04950000000000  -0.28150000000000   O   (16b)
   0.23050000000000   0.53150000000000   0.20050000000000   O   (16b)
   0.53150000000000   0.23050000000000   0.03750000000000   O   (16b)
   1.02700000000000   0.72600000000000  -0.79500000000000   O   (16b)
   0.72600000000000   1.02700000000000  -0.95800000000000   O   (16b)
   1.04500000000000   1.20800000000000  -0.77700000000000   O   (16b)
   1.20800000000000   1.04500000000000  -0.47600000000000   O   (16b)
   0.00000000000000   0.00000000000000   0.00000000000000   P    (8a)
   0.25000000000000   0.25000000000000   0.25000000000000   P    (8a)
```

$Cs_2Se$: A2B_oF24_43_b_a - CIF

```
# CIF file
data_findsym-output
_audit_creation_method FINDSYM

_chemical_name_mineral 'Cs2Se'
_chemical_formula_sum 'Cs2 Se'

loop_
_publ_author_name
 'P. B\"{o}ttcher'
_journal_name_full_name
;
 Journal of the Less-Common Metals
;
_journal_volume 76
_journal_year 1980
_journal_page_first 271
_journal_page_last 277
_publ_Section_title
;
 Zur Kenntnis von Cs$_{2}$Se
;

# Found in Pearson's Handbook of Crystallographic Data, 1991

_aflow_title 'Cs$_{2}$Se Structure'
_aflow_proto 'A2B_oF24_43_b_a'
_aflow_params 'a,b/a,c/a,z_{1},x_{2},y_{2},z_{2}'
```

```
_aflow_params_values '16.49,0.712553062462,0.410855063675,0.0,0.0749,
    ↪ 0.202,0.8118'
_aflow_Strukturbericht 'None'
_aflow_Pearson 'oF24'

_symmetry_space_group_name_H-M "F d d 2"
_symmetry_Int_Tables_number 43

_cell_length_a     16.49000
_cell_length_b     11.75000
_cell_length_c     6.77500
_cell_angle_alpha  90.00000
_cell_angle_beta   90.00000
_cell_angle_gamma  90.00000

loop_
_space_group_symop_id
_space_group_symop_operation_xyz
1 x,y,z
2 -x,-y,z
3 -x+1/4,y+1/4,z+1/4
4 x+1/4,-y+1/4,z+1/4
5 x,y+1/2,z+1/2
6 -x,-y+1/2,z+1/2
7 -x+1/4,y+3/4,z+3/4
8 x+1/4,-y+3/4,z+3/4
9 x+1/2,y,z+1/2
10 -x+1/2,-y,z+1/2
11 -x+3/4,y+1/4,z+3/4
12 x+3/4,-y+1/4,z+3/4
13 x+1/2,y+1/2,z
14 -x+1/2,-y+1/2,z
15 -x+3/4,y+3/4,z+1/4
16 x+3/4,-y+3/4,z+1/4

loop_
_atom_site_label
_atom_site_type_symbol
_atom_site_symmetry_multiplicity
_atom_site_Wyckoff_label
_atom_site_fract_x
_atom_site_fract_y
_atom_site_fract_z
_atom_site_occupancy
Se1 Se 8 a 0.00000 0.00000 0.00000 1.00000
Cs1 Cs 16 b 0.07490 0.20200 0.81180 1.00000
```

$Cs_2Se$: A2B_oF24_43_b_a - POSCAR

```
A2B_oF24_43_b_a & a,b/a,c/a,z1,x2,y2,z2 --params=16.49,0.712553062462,
    ↪ 0.410855063675,0.0,0.0749,0.202,0.8118 & Fdd2 C_{2v}^{19} #43 (
    ↪ ab) & oF24 & None & Cs2Se & Cs2Se & P. B\"{o}ttcher, J.
    ↪ Less-Common Met. 76, 271-277 (1980)
   1.00000000000000
   0.00000000000000   5.87500000000000   3.38750000000000
   8.24500000000000   0.00000000000000   3.38750000000000
   8.24500000000000   5.87500000000000   0.00000000000000
    Cs      Se
     4       2
Direct
   0.93890000000000   0.68470000000000  -0.53490000000000   Cs   (16b)
   0.68470000000000   0.93890000000000  -1.08870000000000   Cs   (16b)
   0.78490000000000   1.33870000000000  -0.68890000000000   Cs   (16b)
   1.33870000000000   0.78490000000000  -0.43470000000000   Cs   (16b)
   0.00000000000000   0.00000000000000   0.00000000000000   Se    (8a)
   0.25000000000000   0.25000000000000   0.25000000000000   Se    (8a)
```

$Zr_2Al_3$: A3B2_oF40_43_ab_b - CIF

```
# CIF file
data_findsym-output
_audit_creation_method FINDSYM

_chemical_name_mineral 'Al3Zr2'
_chemical_formula_sum 'Al3 Zr2'

loop_
_publ_author_name
 'T. J. Renouf'
 'C. A. Beevers'
_journal_name_full_name
;
 Acta Cristallographica
;
_journal_volume 14
_journal_year 1961
_journal_page_first 469
_journal_page_last 472
_publ_Section_title
;
 The Crystal Structure of Zr$_{2}$Al$_{3}$
;

# Found in Crystal Structure Investigations on the Zr-Al and Hf-Al
    ↪ Systems, 1962

_aflow_title 'Zr$_{2}$Al$_{3}$ Structure'
_aflow_proto 'A3B2_oF40_43_ab_b'
_aflow_params 'a,b/a,c/a,z_{1},x_{2},y_{2},z_{2},x_{3},y_{3},z_{3}'
_aflow_params_values '9.601,1.44839079263,0.58014790126,0.125,0.065,
    ↪ 0.116,0.5,0.068,0.196,0.0'
_aflow_Strukturbericht 'None'
_aflow_Pearson 'oF40'

_symmetry_space_group_name_H-M "F d d 2"
_symmetry_Int_Tables_number 43

_cell_length_a     9.60100
_cell_length_b     13.90600
_cell_length_c     5.57000
_cell_angle_alpha  90.00000
_cell_angle_beta   90.00000
_cell_angle_gamma  90.00000

loop_
_space_group_symop_id
_space_group_symop_operation_xyz
1 x,y,z
2 -x,-y,z
3 -x+1/4,y+1/4,z+1/4
4 x+1/4,-y+1/4,z+1/4
5 x,y+1/2,z+1/2
6 -x,-y+1/2,z+1/2
7 -x+1/4,y+3/4,z+3/4
8 x+1/4,-y+3/4,z+3/4
9 x+1/2,y,z+1/2
10 -x+1/2,-y,z+1/2
11 -x+3/4,y+1/4,z+3/4
12 x+3/4,-y+1/4,z+3/4
13 x+1/2,y+1/2,z
14 -x+1/2,-y+1/2,z
15 -x+3/4,y+3/4,z+1/4
16 x+3/4,-y+3/4,z+1/4

loop_
_atom_site_label
_atom_site_type_symbol
_atom_site_symmetry_multiplicity
_atom_site_Wyckoff_label
_atom_site_fract_x
_atom_site_fract_y
_atom_site_fract_z
_atom_site_occupancy
Al1 Al 8 a 0.00000 0.00000 0.12500 1.00000
Al2 Al 16 b 0.06500 0.11600 0.50000 1.00000
Zr1 Zr 16 b 0.06800 0.19600 0.00000 1.00000
```

$Zr_2Al_3$: A3B2_oF40_43_ab_b - POSCAR

```
A3B2_oF40_43_ab_b & a,b/a,c/a,z1,x2,y2,z2,x3,y3,z3 --params=9.601,
    ↪ 1.44839079263,0.58014790126,0.125,0.065,0.116,0.5,0.068,0.196,
    ↪ 0.0 & Fdd2 C_{2v}^{19} #43 (ab^2) & oF40 & None & Al3Zr2 &
    ↪ Al3Zr2 & T. J. Renouf and C. A. Beevers, Acta Cryst. 14,
    ↪ 469-472 (1961)
   1.00000000000000
   0.00000000000000   6.95300000000000   2.78500000000000
   4.80050000000000   0.00000000000000   2.78500000000000
   4.80050000000000   6.95300000000000   0.00000000000000
    Al      Zr
     6       4
Direct
   0.12500000000000   0.12500000000000  -0.12500000000000   Al    (8a)
   0.37500000000000   0.37500000000000  -0.37500000000000   Al    (8a)
   0.55100000000000   0.44900000000000  -0.31900000000000   Al   (16b)
   0.44900000000000   0.55100000000000  -0.68100000000000   Al   (16b)
   0.56900000000000   0.93100000000000  -0.30100000000000   Al   (16b)
   0.93100000000000   0.56900000000000  -0.19900000000000   Al   (16b)
   0.12800000000000  -0.12800000000000   0.26400000000000   Zr   (16b)
  -0.12800000000000   0.12800000000000  -0.26400000000000   Zr   (16b)
  -0.01400000000000   0.51400000000000   0.12200000000000   Zr   (16b)
   0.51400000000000  -0.01400000000000   0.37800000000000   Zr   (16b)
```

Hemimorphite ($Zn_4Si_2O_7(OH)_2{\cdot}H_2O$, $S2_2$): A2B5CD2_oI40_44_2c_abcde_d_e - CIF

```
# CIF file
data_findsym-output
_audit_creation_method FINDSYM

_chemical_name_mineral 'Hemimorphite'
_chemical_formula_sum 'H2 O5 Si Zn2'

loop_
_publ_author_name
 'R. J. Hill'
 'G. V. Gibbs'
 'J. R. Craig'
 'F. K. Ross'
 'J. M. Williams'
_journal_name_full_name
;
 Zeitschrift f{\"u}r Kristallographie - Crystalline Materials
;
_journal_volume 146
_journal_year 1977
_journal_page_first 241
_journal_page_last 259
_publ_Section_title
;
 A neutron-diffraction study of hemimorphite
;

# Found in The American Mineralogist Crystal Structure Database, 2003

_aflow_title 'Hemimorphite (Zn$_{4}$Si$_{2}$O$_{7}$(OH)$_{2}\cdot$H$_{2}
    ↪ $O, $S2_{2}$) Structure'
_aflow_proto 'A2B5CD2_oI40_44_2c_abcde_d_e'
_aflow_params 'a,b/a,c/a,z_{1},z_{2},x_{3},z_{3},x_{4},z_{4},x_{5},z_{5
    ↪ },y_{6},z_{6},y_{7},z_{7},x_{8},y_{8},z_{8},x_{9},y_{9},z_{9}'
```

```
_aflow_params_values '8.367,1.28241902713,0.611330225887,0.5912,0.0195,
    ↪ 0.374,0.19,0.4256,0.643,0.305,0.041,0.1669,0.1938,0.1465,0.5076
    ↪ ,0.1602,0.2055,0.6362,0.2047,0.1613,0.0'
_aflow_Strukturbericht '$S2_{2}$'
_aflow_Pearson 'oI40'

_symmetry_space_group_name_H-M "I m m 2"
_symmetry_Int_Tables_number 44

_cell_length_a    8.36700
_cell_length_b    10.73000
_cell_length_c    5.11500
_cell_angle_alpha 90.00000
_cell_angle_beta  90.00000
_cell_angle_gamma 90.00000

loop_
_space_group_symop_id
_space_group_symop_operation_xyz
1 x,y,z
2 -x,-y,z
3 -x,y,z
4 x,-y,z
5 x+1/2,y+1/2,z+1/2
6 -x+1/2,-y+1/2,z+1/2
7 -x+1/2,y+1/2,z+1/2
8 x+1/2,-y+1/2,z+1/2

loop_
_atom_site_label
_atom_site_type_symbol
_atom_site_symmetry_multiplicity
_atom_site_Wyckoff_label
_atom_site_fract_x
_atom_site_fract_y
_atom_site_fract_z
_atom_site_occupancy
O1 O 2 a 0.00000 0.00000 0.59120 1.00000
O2 O 2 b 0.00000 0.50000 0.01950 1.00000
H1 H 4 c 0.37400 0.00000 0.19000 1.00000
H2 H 4 c 0.42560 0.00000 0.64300 1.00000
O3 O 4 c 0.30500 0.00000 0.04100 1.00000
O4 O 4 d 0.00000 0.16690 0.19380 1.00000
Si1 Si 4 d 0.00000 0.14650 0.50760 1.00000
O5 O 8 e 0.16020 0.20550 0.63620 1.00000
Zn1 Zn 8 e 0.20470 0.16130 0.00000 1.00000
```

Hemimorphite ($Zn_4Si_2O_7(OH)_2 \cdot H_2O$, $S2_2$): A2B5CD2_oI40_44_2c_abcde_d_e - POSCAR

```
A2B5CD2_oI40_44_2c_abcde_d_e & a,b/a,c/a,z1,z2,x3,z3,x4,z4,x5,z5,y6,z6,
    ↪ y7,z7,x8,y8,z8,x9,y9,z9 --params=8.367,1.28241902713,
    ↪ 0.611330225887,0.5912,0.0195,0.374,0.19,0.4256,0.643,0.305,
    ↪ 0.041,0.1669,0.1938,0.1465,0.5076,0.1602,0.2055,0.6362,0.2047,
    ↪ 0.1613,0.0 & Imm2 C_{2v}^{20} #44 (abc^3d^2e^2) & oI40 & $S2_{2
    ↪ }$ & H2O5SiZn2 & Hemimorphite & R. J. Hill et al., Zeitschrift
    ↪ f{"u}r Kristallographie - Crystalline Materials 146, 241-259 (
    ↪ 1977)
   1.00000000000000
  -4.18350000000000   5.36500000000000   2.55750000000000
   4.18350000000000  -5.36500000000000   2.55750000000000
   4.18350000000000   5.36500000000000  -2.55750000000000
     H      O     Si     Zn
     4     10      2      4
Direct
   0.19000000000000   0.56400000000000   0.37400000000000   H   (4c)
   0.19000000000000  -0.18400000000000  -0.37400000000000   H   (4c)
   0.64300000000000   1.06860000000000   0.42560000000000   H   (4c)
   0.64300000000000   0.21740000000000  -0.42560000000000   H   (4c)
   0.59120000000000   0.59120000000000   0.00000000000000   O   (2a)
   0.51950000000000   0.01950000000000   0.50000000000000   O   (2b)
   0.04100000000000   0.34600000000000   0.30500000000000   O   (4c)
   0.04100000000000  -0.26400000000000  -0.30500000000000   O   (4c)
   0.36070000000000   0.19380000000000   0.16690000000000   O   (4d)
   0.02690000000000   0.19380000000000  -0.16690000000000   O   (4d)
   0.84170000000000   0.79640000000000   0.36570000000000   O   (8e)
   0.43070000000000   0.47600000000000  -0.36570000000000   O   (8e)
   0.43070000000000   0.79640000000000  -0.04530000000000   O   (8e)
   0.84170000000000   0.47600000000000   0.04530000000000   O   (8e)
   0.65410000000000   0.50760000000000   0.14650000000000   Si  (4d)
   0.36110000000000   0.50760000000000  -0.14650000000000   Si  (4d)
   0.16130000000000   0.20470000000000   0.36600000000000   Zn  (8e)
  -0.16130000000000  -0.20470000000000  -0.36600000000000   Zn  (8e)
  -0.16130000000000   0.20470000000000   0.04340000000000   Zn  (8e)
   0.16130000000000  -0.20470000000000  -0.04340000000000   Zn  (8e)
```

Ferroelectric $NaNO_2$ ($F5_5$): ABC2_oI8_44_a_a_c - CIF

```
# CIF file
data_findsym-output
_audit_creation_method FINDSYM

_chemical_name_mineral 'Sodium nitrite'
_chemical_formula_sum 'N Na O2'

loop_
_publ_author_name
 'M. I. Kay'
 'B. C. Frazier'
_journal_name_full_name
;
 Acta Cristallographica
;
_journal_volume 14
_journal_year 1961
_journal_page_first 56
_journal_page_last 57
_publ_Section_title
;
 A neutron diffraction refinement of the low temperature phase of NaNO$_
    ↪ {2}$
;

_aflow_title 'Ferroelectric NaNO$_{2}$ ($F5_{5}$) Structure'
_aflow_proto 'ABC2_oI8_44_a_a_c'
_aflow_params 'a,b/a,c/a,z_{1},z_{2},x_{3},z_{3}'
_aflow_params_values '5.384,0.661218424963,1.03324665676,0.88,0.4147,
    ↪ 0.8059,0.0'
_aflow_Strukturbericht '$F5_{5}$'
_aflow_Pearson 'oI8'

_symmetry_space_group_name_H-M "I m m 2"
_symmetry_Int_Tables_number 44

_cell_length_a    5.38400
_cell_length_b    3.56000
_cell_length_c    5.56300
_cell_angle_alpha 90.00000
_cell_angle_beta  90.00000
_cell_angle_gamma 90.00000

loop_
_space_group_symop_id
_space_group_symop_operation_xyz
1 x,y,z
2 -x,-y,z
3 -x,y,z
4 x,-y,z
5 x+1/2,y+1/2,z+1/2
6 -x+1/2,-y+1/2,z+1/2
7 -x+1/2,y+1/2,z+1/2
8 x+1/2,-y+1/2,z+1/2

loop_
_atom_site_label
_atom_site_type_symbol
_atom_site_symmetry_multiplicity
_atom_site_Wyckoff_label
_atom_site_fract_x
_atom_site_fract_y
_atom_site_fract_z
_atom_site_occupancy
N1 N 2 a 0.00000 0.00000 0.88000 1.00000
Na1 Na 2 a 0.00000 0.00000 0.41470 1.00000
O1 O 4 c 0.80590 0.00000 0.00000 1.00000
```

Ferroelectric $NaNO_2$ ($F5_5$): ABC2_oI8_44_a_a_c - POSCAR

```
ABC2_oI8_44_a_a_c & a,b/a,c/a,z1,z2,x3,z3 --params=5.384,0.661218424963,
    ↪ 1.03324665676,0.88,0.4147,0.8059,0.0 & Imm2 C_{2v}^{20} #44 (a^
    ↪ 2c) & oI8 & $F5_{5}$ & NaNO2 & Sodium nitrite & M. I. Kay and
    ↪ B. C. Frazier, Acta Cryst. 14, 56-57 (1961)
   1.00000000000000
  -2.69200000000000   1.78000000000000   2.78150000000000
   2.69200000000000  -1.78000000000000   2.78150000000000
   2.69200000000000   1.78000000000000  -2.78150000000000
     N     Na      O
     1      1      2
Direct
   0.88000000000000   0.88000000000000   0.00000000000000   N    (2a)
   0.41470000000000   0.41470000000000   0.00000000000000   Na   (2a)
   0.00000000000000   0.80590000000000   0.80590000000000   O    (4c)
   0.00000000000000  -0.80590000000000  -0.80590000000000   O    (4c)
```

$AgNO_2$ ($F5_{12}$): ABC2_oI8_44_a_a_d - CIF

```
# CIF file
data_findsym-output
_audit_creation_method FINDSYM

_chemical_name_mineral 'Silver nitrite'
_chemical_formula_sum 'Ag N O2'

loop_
_publ_author_name
 'S. Ohba'
 'Y. Saito'
_journal_name_full_name
;
 Acta Crystallographica Section B: Structural Science
;
_journal_volume 37
_journal_year 1981
_journal_page_first 1911
_journal_page_last 1913
_publ_Section_title
;
 Structure of silver(I) nitrite, a redetermination
;

_aflow_title 'AgNO$_{2}$ ($F5_{12}$) Structure'
_aflow_proto 'ABC2_oI8_44_a_a_d'
_aflow_params 'a,b/a,c/a,z_{1},z_{2},y_{3},z_{3}'
_aflow_params_values '3.528,1.74943310658,1.46853741497,0.0,0.4446,
    ↪ 0.1701,0.5747'
_aflow_Strukturbericht '$F5_{12}$'
_aflow_Pearson 'oI8'

_symmetry_space_group_name_H-M "I m m 2"
_symmetry_Int_Tables_number 44
```

```
_cell_length_a      3.52800
_cell_length_b      6.17200
_cell_length_c      5.18100
_cell_angle_alpha   90.00000
_cell_angle_beta    90.00000
_cell_angle_gamma   90.00000

loop_
_space_group_symop_id
_space_group_symop_operation_xyz
1 x,y,z
2 -x,-y,z
3 -x,y,z
4 x,-y,z
5 x+1/2,y+1/2,z+1/2
6 -x+1/2,-y+1/2,z+1/2
7 -x+1/2,y+1/2,z+1/2
8 x+1/2,-y+1/2,z+1/2

loop_
_atom_site_label
_atom_site_type_symbol
_atom_site_symmetry_multiplicity
_atom_site_Wyckoff_label
_atom_site_fract_x
_atom_site_fract_y
_atom_site_fract_z
_atom_site_occupancy
Ag1 Ag 2 a 0.00000 0.00000 0.00000 1.00000
N1 N 2 a 0.00000 0.00000 0.44460 1.00000
O1 O 4 d 0.00000 0.17010 0.57470 1.00000
```

$AgNO_2$ ($F5_{12}$): ABC2_oI8_44_a_a_d - POSCAR

```
ABC2_oI8_44_a_a_d & a,b/a,c/a,z1,z2,y3,z3 --params=3.528,1.74943310658,
    ↪ 1.46853741497,0.0,0.4446,0.1701,0.5747 & Imm2 C_{2v}^{20} #44 (
    ↪ a^2d) & oI8 & $F5_{12}$ & AgNO2 & Silver nitrite & S. Ohba and
    ↪ Y. Saito, Acta Crystallogr. Sect. B Struct. Sci. 37, 1911-1913
    ↪ (1981)
   1.00000000000000
  -1.76400000000000    3.08600000000000    2.59050000000000
   1.76400000000000   -3.08600000000000    2.59050000000000
   1.76400000000000    3.08600000000000   -2.59050000000000
     Ag        N        O
      1        1        2
Direct
   0.00000000000000    0.00000000000000    0.00000000000000   Ag   (2a)
   0.44460000000000    0.44460000000000    0.00000000000000   N    (2a)
   0.74480000000000    0.57470000000000    0.17010000000000   O    (4d)
   0.40460000000000    0.57470000000000   -0.17010000000000   O    (4d)
```

$B30$ (MgZn?): AB_oI48_44_6d_ab2cde - CIF

```
# CIF file
data_findsym-output
_audit_creation_method FINDSYM

_chemical_name_mineral 'MgZn'
_chemical_formula_sum 'Mg Zn'

loop_
_publ_author_name
 'C. Hermann'
 'O. Lohrmann'
 'H. Philipp'
_journal_year 1937
_publ_Section_title
;
 Strukturbericht Band II 1928-1932
;

_aflow_title '$B30$ (MgZn?) Structure'
_aflow_proto 'AB_oI48_44_6d_ab2cde'
_aflow_params 'a,b/a,c/a,z_{1},z_{2},x_{3},z_{3},x_{4},z_{4},y_{5},z_{5
    ↪ },y_{6},z_{6},y_{7},z_{7},y_{8},z_{8},y_{9},z_{9},y_{10},z_{10
    ↪ },y_{11},z_{11},x_{12},y_{12},z_{12}'
_aflow_params_values '7.53776,2.27653838806,1.0,-0.04167,-0.04167,0.25,
    ↪ 0.70833,0.75,0.20833,0.09375,0.45833,0.59375,0.45833,0.84375,
    ↪ 0.79167,0.34375,0.79167,0.59375,0.125,0.15625,0.125,0.75,
    ↪ 0.29167,0.75,0.25,0.04167'
_aflow_Strukturbericht '$B30$'
_aflow_Pearson 'oI48'

_symmetry_space_group_name_H-M "I m m 2"
_symmetry_Int_Tables_number 44

_cell_length_a      7.53776
_cell_length_b      17.16000
_cell_length_c      7.53776
_cell_angle_alpha   90.00000
_cell_angle_beta    90.00000
_cell_angle_gamma   90.00000

loop_
_space_group_symop_id
_space_group_symop_operation_xyz
1 x,y,z
2 -x,-y,z
3 -x,y,z
4 x,-y,z
5 x+1/2,y+1/2,z+1/2
6 -x+1/2,-y+1/2,z+1/2
7 -x+1/2,y+1/2,z+1/2
8 x+1/2,-y+1/2,z+1/2

loop_
_atom_site_label
_atom_site_type_symbol
_atom_site_symmetry_multiplicity
_atom_site_Wyckoff_label
_atom_site_fract_x
_atom_site_fract_y
_atom_site_fract_z
_atom_site_occupancy
Zn1 Zn 2 a 0.00000 0.00000 -0.04167 1.00000
Zn2 Zn 2 b 0.00000 0.50000 -0.04167 1.00000
Zn3 Zn 4 c 0.25000 0.00000 0.70833 1.00000
Zn4 Zn 4 c 0.75000 0.00000 0.20833 1.00000
Mg1 Mg 4 d 0.00000 0.09375 0.45833 1.00000
Mg2 Mg 4 d 0.00000 0.59375 0.45833 1.00000
Mg3 Mg 4 d 0.00000 0.84375 0.79167 1.00000
Mg4 Mg 4 d 0.00000 0.34375 0.79167 1.00000
Mg5 Mg 4 d 0.00000 0.59375 0.12500 1.00000
Mg6 Mg 4 d 0.00000 0.15625 0.12500 1.00000
Zn5 Zn 4 d 0.00000 0.75000 0.29167 1.00000
Zn6 Zn 8 e 0.75000 0.25000 0.04167 1.00000
```

$B30$ (MgZn?): AB_oI48_44_6d_ab2cde - POSCAR

```
AB_oI48_44_6d_ab2cde & a,b/a,c/a,z1,z2,x3,z3,x4,z4,y5,z5,y6,z6,y7,z7,y8,
    ↪ z8,y9,z9,y10,z10,y11,z11,x12,y12,z12 --params=7.53776,
    ↪ 2.27653838806,1.0,-0.04167,-0.04167,0.25,0.70833,0.75,0.20833,
    ↪ 0.09375,0.45833,0.59375,0.45833,0.84375,0.79167,0.34375,0.79167
    ↪ ,0.59375,0.125,0.15625,0.125,0.75,0.29167,0.75,0.25,0.04167 &
    ↪ Imm2 C_{2v}^{20} #44 (abc^2d^7e) & oI48 & $B30$ & MgZn & MgZn &
    ↪  C. Hermann and O. Lohrmann and H. Philipp, (1937)
   1.00000000000000
  -3.76888000000000    8.58000000000000    3.76888000000000
   3.76888000000000   -8.58000000000000    3.76888000000000
   3.76888000000000    8.58000000000000   -3.76888000000000
     Mg       Zn
     12       12
Direct
   0.55208000000000    0.45833000000000    0.09375000000000   Mg   (4d)
   0.36458000000000    0.45833000000000   -0.09375000000000   Mg   (4d)
   1.05208000000000    0.45833000000000    0.59375000000000   Mg   (4d)
  -0.13542000000000    0.45833000000000   -0.59375000000000   Mg   (4d)
   1.63542000000000    0.79167000000000    0.84375000000000   Mg   (4d)
  -0.05208000000000    0.79167000000000   -0.84375000000000   Mg   (4d)
   1.13542000000000    0.79167000000000    0.34375000000000   Mg   (4d)
   0.44792000000000    0.79167000000000   -0.34375000000000   Mg   (4d)
   0.71875000000000    0.12500000000000    0.59375000000000   Mg   (4d)
  -0.46875000000000    0.12500000000000   -0.59375000000000   Mg   (4d)
   0.28125000000000    0.12500000000000    0.15625000000000   Mg   (4d)
  -0.03125000000000    0.12500000000000   -0.15625000000000   Mg   (4d)
  -0.04167000000000   -0.04167000000000    0.00000000000000   Zn   (2a)
   0.45833000000000   -0.04167000000000    0.50000000000000   Zn   (2b)
   0.70833000000000    0.95833000000000    0.25000000000000   Zn   (4c)
   0.70833000000000    0.45833000000000   -0.25000000000000   Zn   (4c)
   0.20833000000000    0.95833000000000    0.75000000000000   Zn   (4c)
   0.20833000000000   -0.54167000000000   -0.75000000000000   Zn   (4c)
   1.04167000000000    0.29167000000000    0.75000000000000   Zn   (4d)
  -0.45833000000000    0.29167000000000   -0.75000000000000   Zn   (4d)
   0.29167000000000    0.79167000000000    1.00000000000000   Zn   (8e)
  -0.20833000000000   -0.70833000000000   -1.00000000000000   Zn   (8e)
  -0.20833000000000    0.79167000000000    0.50000000000000   Zn   (8e)
   0.29167000000000   -0.70833000000000   -0.50000000000000   Zn   (8e)
```

$Nb_2Zr_6O_{17}$: A2B17C6_oI100_46_ab_b8c_3c - CIF

```
# CIF file
data_findsym-output
_audit_creation_method FINDSYM

_chemical_name_mineral 'Nb2O17Zr6'
_chemical_formula_sum 'Nb2 O17 Zr6'

loop_
_publ_author_name
 'J. Galy'
 'R. S. Roth'
_journal_name_full_name
;
 Journal of Solid State Chemistry
;
_journal_volume 7
_journal_year 1973
_journal_page_first 277
_journal_page_last 285
_publ_Section_title
;
 The Crystal Structure of Nb$_{2}$Zr$_{6}$O$_{17}$
;

# Found in Crystal structure solution for the $A_{6}$$B_{2}$O$_{17}$ (
    ↪ $A$ = Zr, Hf; $B$ = Nb, Ta) superstructure, 2019

_aflow_title 'Nb$_{2}$Zr$_{6}$O$_{17}$ Structure'
_aflow_proto 'A2B17C6_oI100_46_ab_b8c_3c'
_aflow_params 'a,b/a,c/a,z_{1},y_{2},z_{2},y_{3},z_{3},x_{4},y_{4},z_{4
    ↪ },x_{5},y_{5},z_{5},x_{6},y_{6},z_{6},x_{7},y_{7},z_{7},x_{8},
    ↪ y_{8},z_{8},x_{9},y_{9},z_{9},x_{10},y_{10},z_{10},x_{11},y_{11
    ↪ },z_{11},x_{12},y_{12},z_{12},x_{13},y_{13},z_{13},x_{14},y_{14
    ↪ },z_{14}'
_aflow_params_values '40.92,0.120478983382,0.128787878788,0.0,0.5151,
    ↪ 0.5087,0.3564,0.1537,0.1976,0.3485,0.4426,0.2195,0.8082,0.311,
    ↪ 0.1589,0.7925,0.3251,0.0956,0.7915,0.2785,0.0362,0.7505,0.3797,
    ↪ 0.0827,0.2994,0.4251,0.0249,0.2973,0.1839,0.1419,0.3421,0.1826,
    ↪ 0.065,-0.0152,0.5894,0.1278,0.5138,0.5136,0.18965,-0.0151,
    ↪ 0.5861'
```

```
_aflow_Strukturbericht 'None'
_aflow_Pearson 'oI100'

_symmetry_space_group_name_H-M "I m a 2"
_symmetry_Int_Tables_number 46

_cell_length_a     40.92000
_cell_length_b     4.93000
_cell_length_c     5.27000
_cell_angle_alpha  90.00000
_cell_angle_beta   90.00000
_cell_angle_gamma  90.00000

loop_
_space_group_symop_id
_space_group_symop_operation_xyz
1 x,y,z
2 -x,-y,z
3 -x+1/2,y,z
4 x+1/2,-y,z
5 x+1/2,y+1/2,z+1/2
6 -x+1/2,-y+1/2,z+1/2
7 -x,y+1/2,z+1/2
8 x,-y+1/2,z+1/2

loop_
_atom_site_label
_atom_site_type_symbol
_atom_site_symmetry_multiplicity
_atom_site_Wyckoff_label
_atom_site_fract_x
_atom_site_fract_y
_atom_site_fract_z
_atom_site_occupancy
Nb1 Nb 4 a 0.00000 0.00000 0.00000 1.00000
Nb2 Nb 4 b 0.25000 0.51510 0.50870 1.00000
O1 O 4 b 0.25000 0.35640 0.15370 1.00000
O2 O 8 c 0.19760 0.34850 0.44260 1.00000
O3 O 8 c 0.21950 0.80820 0.31100 1.00000
O4 O 8 c 0.15890 0.79250 0.32510 1.00000
O5 O 8 c 0.09560 0.79150 0.27850 1.00000
O6 O 8 c 0.03620 0.75050 0.37970 1.00000
O7 O 8 c 0.08270 0.29940 0.42510 1.00000
O8 O 8 c 0.02490 0.29730 0.18390 1.00000
O9 O 8 c 0.14190 0.34210 0.18260 1.00000
Zr1 Zr 8 c 0.06500 -0.01520 0.58940 1.00000
Zr2 Zr 8 c 0.12780 0.51380 0.51360 1.00000
Zr3 Zr 8 c 0.18965 -0.01510 0.58610 1.00000
```

$Nb_2Zr_6O_{17}$: A2B17C6_oI100_46_ab_b8c_3c - POSCAR

```
A2B17C6_oI100_46_ab_b8c_3c & a,b/a,c/a,z1,y2,z2,y3,z3,x4,y4,z4,x5,y5,z5,
    ↪ x6,y6,z6,x7,y7,z7,x8,y8,z8,x9,y9,z9,x10,y10,z10,x11,y11,z11,x12
    ↪ ,y12,z12,x13,y13,z13,x14,y14,z14 --params=40.92,0.120478983382,
    ↪ 0.128787878788,0.0,0.5151,0.5087,0.3564,0.1537,0.1976,0.3485,
    ↪ 0.4426,0.2195,0.8082,0.311,0.1589,0.7925,0.3251,0.0956,0.7915,
    ↪ 0.2785,0.0362,0.7505,0.3797,0.0827,0.2994,0.4251,0.0249,0.2973,
    ↪ 0.1839,0.1419,0.3421,0.1826,0.065,-0.0152,0.5894,0.1278,0.5138,
    ↪ 0.5136,0.18965,-0.0151,0.5861 & Ima2 C_{2v}^{22} #46 (ab^2c^11)
    ↪ & oI100 & None & Nb2O17Zr6 & Nb2O17Zr6 & J. Galy and R. S.
    ↪ Roth, J. Solid State Chem. 7, 277-285 (1973)
   1.00000000000000
 -20.46000000000000   2.46500000000000   2.63500000000000
  20.46000000000000  -2.46500000000000   2.63500000000000
  20.46000000000000   2.46500000000000  -2.63500000000000
    Nb     O    Zr
     4    34    12
Direct
   0.00000000000000   0.00000000000000   0.00000000000000   Nb   (4a)
   0.00000000000000   0.50000000000000   0.50000000000000   Nb   (4a)
   1.02380000000000   0.75870000000000   0.76510000000000   Nb   (4b)
  -0.00640000000000   1.25870000000000   0.23490000000000   Nb   (4b)
   0.51010000000000   0.40370000000000   0.60640000000000   O    (4b)
  -0.20270000000000   0.90370000000000   0.39360000000000   O    (4b)
   0.79110000000000   0.64020000000000   0.54610000000000   O    (8c)
   0.09410000000000   0.24500000000000  -0.54610000000000   O    (8c)
   0.09410000000000   1.14020000000000   0.34910000000000   O    (8c)
   0.79110000000000   0.74500000000000   0.65090000000000   O    (8c)
   1.11920000000000   0.53050000000000   1.02770000000000   O    (8c)
  -0.49720000000000   0.09150000000000  -1.02770000000000   O    (8c)
  -0.49720000000000   1.03050000000000  -0.08870000000000   O    (8c)
   1.11920000000000   0.59150000000000   1.08870000000000   O    (8c)
   1.11760000000000   0.48400000000000   0.95140000000000   O    (8c)
  -0.46740000000000   0.16620000000000  -0.95140000000000   O    (8c)
  -0.46740000000000   0.98400000000000  -0.13360000000000   O    (8c)
   1.11760000000000   0.66620000000000   1.13360000000000   O    (8c)
   1.07000000000000   0.37410000000000   0.88710000000000   O    (8c)
  -0.51300000000000   0.18290000000000  -0.88710000000000   O    (8c)
  -0.51300000000000   0.87410000000000  -0.19590000000000   O    (8c)
   1.07000000000000   0.68290000000000   1.19590000000000   O    (8c)
   1.13020000000000   0.41590000000000   0.78670000000000   O    (8c)
  -0.37080000000000   0.34350000000000  -0.78670000000000   O    (8c)
  -0.37080000000000   0.91590000000000  -0.21430000000000   O    (8c)
   1.13020000000000   0.84350000000000   1.21430000000000   O    (8c)
   0.72450000000000   0.50780000000000   0.38210000000000   O    (8c)
   0.12570000000000   0.34240000000000  -0.38210000000000   O    (8c)
   0.12570000000000   1.00780000000000   0.28330000000000   O    (8c)
   0.72450000000000   0.84240000000000   0.71670000000000   O    (8c)
   0.48120000000000   0.20880000000000   0.32220000000000   O    (8c)
  -0.11340000000000   0.15900000000000  -0.32220000000000   O    (8c)
  -0.11340000000000   0.70880000000000   0.22760000000000   O    (8c)
   0.48120000000000   0.65900000000000   0.77240000000000   O    (8c)
   0.52470000000000   0.32450000000000   0.48400000000000   O    (8c)
  -0.15950000000000   0.04070000000000  -0.48400000000000   O    (8c)
  -0.15950000000000   0.82450000000000   0.29980000000000   O    (8c)
   0.52470000000000   0.54070000000000   0.70020000000000   O    (8c)
   0.57420000000000   0.65440000000000   0.04980000000000   Zr   (8c)
   0.60460000000000   0.52440000000000  -0.04980000000000   Zr   (8c)
   0.60460000000000   1.15440000000000   0.58020000000000   Zr   (8c)
   0.57420000000000   1.02440000000000   0.41980000000000   Zr   (8c)
   1.02740000000000   0.64140000000000   0.64160000000000   Zr   (8c)
  -0.00020000000000   0.38580000000000  -0.64160000000000   Zr   (8c)
  -0.00020000000000   1.14140000000000   0.11400000000000   Zr   (8c)
   1.02740000000000   0.88580000000000   0.88600000000000   Zr   (8c)
   0.57100000000000   0.77575000000000   0.17455000000000   Zr   (8c)
   0.60120000000000   0.39645000000000  -0.17455000000000   Zr   (8c)
   0.60120000000000   1.27575000000000   0.70475000000000   Zr   (8c)
   0.57100000000000   0.89645000000000   0.29525000000000   Zr   (8c)
```

Parkerite ($Ni_3Bi_2S_2$): AB2C_oP8_51_e_be_f - CIF

```
# CIF file
data_findsym-output
_audit_creation_method FINDSYM

_chemical_name_mineral 'Parkerite'
_chemical_formula_sum 'Bi Ni2 S'

loop_
_publ_author_name
 'M. E. Fleet'
_journal_name_full_name
;
 American Mineralogist
;
_journal_volume 58
_journal_year 1973
_journal_page_first 435
_journal_page_last 439
_publ_Section_title
;
 The Crystal Structure of Parkerite (Ni$_{3}$Bi$_{2}$S$_{2}$)
;

_aflow_title 'Parkerite (Ni$_{3}$Bi$_{2}$S$_{2}$) Structure'
_aflow_proto 'AB2C_oP8_51_e_be_f'
_aflow_params 'a,b/a,c/a,z_{2},z_{3},z_{4}'
_aflow_params_values '5.545,0.730748422002,1.03354373309,0.7423,0.204,
    ↪ 0.256'
_aflow_Strukturbericht 'None'
_aflow_Pearson 'oP8'

_symmetry_space_group_name_H-M "P 21/m 2/m 2/a"
_symmetry_Int_Tables_number 51

_cell_length_a     5.54500
_cell_length_b     4.05200
_cell_length_c     5.73100
_cell_angle_alpha  90.00000
_cell_angle_beta   90.00000
_cell_angle_gamma  90.00000

loop_
_space_group_symop_id
_space_group_symop_operation_xyz
1 x,y,z
2 x+1/2,-y,-z
3 -x,y,-z
4 -x+1/2,-y,z
5 -x,-y,-z
6 -x+1/2,y,z
7 x,-y,z
8 x+1/2,y,-z

loop_
_atom_site_label
_atom_site_type_symbol
_atom_site_symmetry_multiplicity
_atom_site_Wyckoff_label
_atom_site_fract_x
_atom_site_fract_y
_atom_site_fract_z
_atom_site_occupancy
Ni1 Ni 2 b 0.00000 0.50000 0.00000 1.00000
Bi1 Bi 2 e 0.25000 0.00000 0.74230 1.00000
Ni2 Ni 2 e 0.25000 0.00000 0.20400 0.50000
S1 S 2 f 0.25000 0.50000 0.25600 1.00000
```

Parkerite ($Ni_3Bi_2S_2$): AB2C_oP8_51_e_be_f - POSCAR

```
AB2C_oP8_51_e_be_f & a,b/a,c/a,z2,z3,z4 --params=5.545,0.730748422002,
    ↪ 1.03354373309,0.7423,0.204,0.256 & Pmma D_{2h}^{5} #51 (be^2f)
    ↪ & oP8 & None & Bi2Ni3S2 & Parkerite & M. E. Fleet, Am. Mineral.
    ↪ 58, 435-439 (1973)
   1.00000000000000
   5.54500000000000   0.00000000000000   0.00000000000000
   0.00000000000000   4.05200000000000   0.00000000000000
   0.00000000000000   0.00000000000000   5.73100000000000
    Bi    Ni     S
     2     4     2
Direct
   0.25000000000000   0.00000000000000   0.74230000000000   Bi   (2e)
   0.75000000000000   0.00000000000000  -0.74230000000000   Bi   (2e)
   0.00000000000000   0.50000000000000   0.00000000000000   Ni   (2b)
   0.50000000000000   0.50000000000000   0.00000000000000   Ni   (2b)
   0.25000000000000   0.00000000000000   0.20400000000000   Ni   (2e)
   0.75000000000000   0.00000000000000  -0.20400000000000   Ni   (2e)
   0.25000000000000   0.50000000000000   0.25600000000000   S    (2f)
   0.75000000000000   0.50000000000000  -0.25600000000000   S    (2f)
```

$LiNb_6O_{15}F$: ABC6D15_oP46_51_f_d_2e2i_aef4i2j - CIF

```
# CIF file
data_findsym-output
_audit_creation_method FINDSYM

_chemical_name_mineral 'FLiNb6O15'
_chemical_formula_sum 'F Li Nb6 O15'

loop_
_publ_author_name
 'M. Lundberg'
_journal_name_full_name
;
 Acta Chemica Scandinavica
;
_journal_volume 19
_journal_year 1965
_journal_page_first 2274
_journal_page_last 2284
_publ_Section_title
;
 The Crystal Structure of LiNb$_{6}$O$_{15}$F
;

_aflow_title 'LiNb$_{6}$O$_{15}$F Structure'
_aflow_proto 'ABC6D15_oP46_51_f_d_2e2i_aef4i2j'
_aflow_params 'a,b/a,c/a,z_{3},z_{4},z_{5},z_{6},z_{7},x_{8},z_{8},x_{9
    ↪ },z_{9},x_{10},z_{10},x_{11},z_{11},x_{12},z_{12},x_{13},z_{13
    ↪ },x_{14},z_{14},x_{15},z_{15}'
_aflow_params_values '16.635,0.238292756237,0.534295160806,0.0578,0.6874
    ↪ ,0.4582,0.6922,0.0524,0.1305,0.3973,0.0593,0.8146,0.1309,0.6149
    ↪ ,0.1744,0.8823,0.1636,0.1963,0.0206,0.3386,0.1286,0.3998,0.0583
    ↪ ,0.817'
_aflow_Strukturbericht 'None'
_aflow_Pearson 'oP46'

_symmetry_space_group_name_H-M "P 21/m 2/m 2/a"
_symmetry_Int_Tables_number 51

_cell_length_a 16.63500
_cell_length_b 3.96400
_cell_length_c 8.88800
_cell_angle_alpha 90.00000
_cell_angle_beta 90.00000
_cell_angle_gamma 90.00000

loop_
_space_group_symop_id
_space_group_symop_operation_xyz
1 x,y,z
2 x+1/2,-y,-z
3 -x,y,-z
4 -x+1/2,-y,z
5 -x,-y,-z
6 -x+1/2,y,z
7 x,-y,z
8 x+1/2,y,-z

loop_
_atom_site_label
_atom_site_type_symbol
_atom_site_symmetry_multiplicity
_atom_site_Wyckoff_label
_atom_site_fract_x
_atom_site_fract_y
_atom_site_fract_z
_atom_site_occupancy
O1 O 2 a 0.00000 0.00000 0.00000 1.00000
Li1 Li 2 d 0.00000 0.50000 0.50000 1.00000
Nb1 Nb 2 e 0.25000 0.00000 0.05780 1.00000
Nb2 Nb 2 e 0.25000 0.00000 0.68740 1.00000
O2 O 2 e 0.25000 0.00000 0.45820 1.00000
F1 F 2 f 0.25000 0.50000 0.69220 1.00000
O3 O 2 f 0.25000 0.50000 0.05240 1.00000
Nb3 Nb 4 i 0.13050 0.00000 0.39730 1.00000
Nb4 Nb 4 i 0.05930 0.00000 0.81460 1.00000
O4 O 4 i 0.13090 0.00000 0.61490 1.00000
O5 O 4 i 0.17440 0.00000 0.88230 1.00000
O6 O 4 i 0.16360 0.00000 0.19630 1.00000
O7 O 4 i 0.02060 0.00000 0.33860 1.00000
O8 O 4 j 0.12860 0.50000 0.39980 1.00000
O9 O 4 j 0.05830 0.50000 0.81700 1.00000
```

$LiNb_6O_{15}F$: ABC6D15_oP46_51_f_d_2e2i_aef4i2j - POSCAR

```
ABC6D15_oP46_51_f_d_2e2i_aef4i2j & a,b/a,c/a,z3,z4,z5,z6,z7,x8,z8,x9,z9,
    ↪ x10,z10,x11,z11,x12,z12,x13,z13,x14,z14,x15,z15 --params=16.635
    ↪ ,0.238292756237,0.534295160806,0.0578,0.6874,0.4582,0.6922,
    ↪ 0.0524,0.1305,0.3973,0.0593,0.8146,0.1309,0.6149,0.1744,0.8823,
    ↪ 0.1636,0.1963,0.0206,0.3386,0.1286,0.3998,0.0583,0.817 & Pmma
    ↪ D_{2h}^{5} #51 (ade^3f^2i^6j^2) & oP46 & None & FLiNb6O15 &
    ↪ FLiNb6O15 & M. Lundberg, Acta Chem. Scand. 19, 2274-2284 (1965)
   1.00000000000000
  16.63500000000000   0.00000000000000   0.00000000000000
   0.00000000000000   3.96400000000000   0.00000000000000
   0.00000000000000   0.00000000000000   8.88800000000000
    F    Li    Nb    O
    2     2    12   30
Direct
   0.25000000000000   0.50000000000000   0.69220000000000   F   (2f)
   0.75000000000000   0.50000000000000  -0.69220000000000   F   (2f)
   0.00000000000000   0.50000000000000   0.50000000000000   Li  (2d)
   0.50000000000000   0.50000000000000   0.50000000000000   Li  (2d)
   0.25000000000000   0.00000000000000   0.05780000000000   Nb  (2e)
   0.75000000000000   0.00000000000000  -0.05780000000000   Nb  (2e)
   0.25000000000000   0.00000000000000   0.68740000000000   Nb  (2e)
   0.75000000000000   0.00000000000000  -0.68740000000000   Nb  (2e)
   0.13050000000000   0.00000000000000   0.39730000000000   Nb  (4i)
   0.36950000000000   0.00000000000000   0.39730000000000   Nb  (4i)
  -0.13050000000000   0.00000000000000  -0.39730000000000   Nb  (4i)
   0.63050000000000   0.00000000000000  -0.39730000000000   Nb  (4i)
   0.05930000000000   0.00000000000000   0.81460000000000   Nb  (4i)
   0.44070000000000   0.00000000000000   0.81460000000000   Nb  (4i)
  -0.05930000000000   0.00000000000000  -0.81460000000000   Nb  (4i)
   0.55930000000000   0.00000000000000  -0.81460000000000   Nb  (4i)
   0.00000000000000   0.00000000000000   0.00000000000000   O   (2a)
   0.50000000000000   0.00000000000000   0.00000000000000   O   (2a)
   0.25000000000000   0.00000000000000   0.45820000000000   O   (2e)
   0.75000000000000   0.00000000000000  -0.45820000000000   O   (2e)
   0.25000000000000   0.50000000000000   0.05240000000000   O   (2f)
   0.75000000000000   0.50000000000000  -0.05240000000000   O   (2f)
   0.13090000000000   0.00000000000000   0.61490000000000   O   (4i)
   0.36910000000000   0.00000000000000   0.61490000000000   O   (4i)
  -0.13090000000000   0.00000000000000  -0.61490000000000   O   (4i)
   0.63090000000000   0.00000000000000  -0.61490000000000   O   (4i)
   0.17440000000000   0.00000000000000   0.88230000000000   O   (4i)
   0.32560000000000   0.00000000000000   0.88230000000000   O   (4i)
  -0.17440000000000   0.00000000000000  -0.88230000000000   O   (4i)
   0.67440000000000   0.00000000000000  -0.88230000000000   O   (4i)
   0.16360000000000   0.00000000000000   0.19630000000000   O   (4i)
   0.33640000000000   0.00000000000000   0.19630000000000   O   (4i)
  -0.16360000000000   0.00000000000000  -0.19630000000000   O   (4i)
   0.66360000000000   0.00000000000000  -0.19630000000000   O   (4i)
   0.02060000000000   0.00000000000000   0.33860000000000   O   (4i)
   0.47940000000000   0.00000000000000   0.33860000000000   O   (4i)
  -0.02060000000000   0.00000000000000  -0.33860000000000   O   (4i)
   0.52060000000000   0.00000000000000  -0.33860000000000   O   (4i)
   0.12860000000000   0.50000000000000   0.39980000000000   O   (4j)
   0.37140000000000   0.50000000000000   0.39980000000000   O   (4j)
  -0.12860000000000   0.50000000000000  -0.39980000000000   O   (4j)
   0.62860000000000   0.50000000000000  -0.39980000000000   O   (4j)
   0.05830000000000   0.50000000000000   0.81700000000000   O   (4j)
   0.44170000000000   0.50000000000000   0.81700000000000   O   (4j)
  -0.05830000000000   0.50000000000000  -0.81700000000000   O   (4j)
   0.55830000000000   0.50000000000000  -0.81700000000000   O   (4j)
```

Carnallite $[Mg(H_2O)_6KCl_3]$: A3B12CDE6_oP276_52_d4e_18e_ce_de_2d8e - CIF

```
# CIF file
data_findsym-output
_audit_creation_method FINDSYM

_chemical_name_mineral 'Carnallite'
_chemical_formula_sum 'Cl3 H12 K Mg O6'

loop_
_publ_author_name
 'E. O. Schlemper'
 'P. K. {Sen Gupta}'
 'T. Zoltai'
_journal_name_full_name
;
 American Mineralogist
;
_journal_volume 70
_journal_year 1985
_journal_page_first 1309
_journal_page_last 1313
_publ_Section_title
;
 Refinement of the structure of carnallite, Mg(H$_{2}$O)$_{6}$KCl$_{3}$
;

_aflow_title 'Carnallite [Mg(H$_{2}$O)$_{6}$KCl$_{3}$] Structure'
_aflow_proto 'A3B12CDE6_oP276_52_d4e_18e_ce_de_2d8e'
_aflow_params 'a,b/a,c/a,z_{1},x_{2},x_{3},x_{4},x_{5},x_{6},y_{6},z_{6
    ↪ },x_{7},y_{7},z_{7},x_{8},y_{8},z_{8},x_{9},y_{9},z_{9},x_{10},
    ↪ y_{10},z_{10},x_{11},y_{11},z_{11},x_{12},y_{12},z_{12},x_{13},
    ↪ y_{13},z_{13},x_{14},y_{14},z_{14},x_{15},y_{15},z_{15},x_{16},
    ↪ y_{16},z_{16},x_{17},y_{17},z_{17},x_{18},y_{18},z_{18},x_{19},
    ↪ y_{19},z_{19},x_{20},y_{20},z_{20},x_{21},y_{21},z_{21},x_{22},
    ↪ y_{22},z_{22},x_{23},y_{23},z_{23},x_{24},y_{24},z_{24},x_{25},
    ↪ y_{25},z_{25},x_{26},y_{26},z_{26},x_{27},y_{27},z_{27},x_{28},
    ↪ y_{28},z_{28},x_{29},y_{29},z_{29},x_{30},y_{30},z_{30},x_{31},
    ↪ y_{31},z_{31},x_{32},y_{32},z_{32},x_{33},y_{33},z_{33},x_{34},
    ↪ y_{34},z_{34},x_{35},y_{35},z_{35},x_{36},y_{36},z_{36},x_{37},
    ↪ y_{37},z_{37}'
_aflow_params_values '16.119,1.39413114958,0.592530554005,0.2514,0.76061
    ↪ ,0.24343,0.3692,0.1162,0.16583,0.57525,-0.01474,0.16928,0.58113
    ↪ ,0.48834,0.41462,0.58176,0.25235,0.01987,0.74957,-0.02314,0.604
    ↪ ,0.75,0.579,0.412,0.751,0.313,0.737,0.673,0.543,0.25,0.722,
    ↪ 0.505,0.734,0.848,0.587,0.303,0.845,0.335,0.514,0.509,0.326,
    ↪ 0.417,0.494,0.606,0.768,0.572,0.345,0.264,-0.066,0.682,0.399,-
    ↪ 0.01,0.568,0.422,-0.058,0.5,0.359,0.817,0.673,0.429,0.811,0.623
    ↪ ,0.565,-0.099,0.679,0.573,0.604,0.695,0.438,0.66,0.561,0.411,
    ↪ 0.615,0.475,0.0886,0.65668,0.74995,0.42094,0.58981,0.7477,
    ↪ 0.2554,0.70704,0.4384,0.2673,0.66936,0.1506,0.4466,0.51041,0.65
    ↪ ,0.2967,0.57117,0.7349,0.4259,0.54662,-0.0646,0.3956,0.66912,
    ↪ 0.8444,0.5448,0.60769,0.7631,0.4179,0.63155,0.558'
_aflow_Strukturbericht 'None'
_aflow_Pearson 'oP276'

_symmetry_space_group_name_H-M "P 2/n 21/n 2/a"
_symmetry_Int_Tables_number 52

_cell_length_a 16.11900
_cell_length_b 22.47200
_cell_length_c 9.55100
_cell_angle_alpha 90.00000
```

```
_cell_angle_beta   90.00000
_cell_angle_gamma  90.00000

loop_
_space_group_symop_id
_space_group_symop_operation_xyz
1 x,y,z
2 x,−y+1/2,−z+1/2
3 −x+1/2,y+1/2,−z+1/2
4 −x+1/2,−y,z
5 −x,−y,−z
6 −x,y+1/2,z+1/2
7 x+1/2,−y+1/2,z+1/2
8 x+1/2,y,−z

loop_
_atom_site_label
_atom_site_type_symbol
_atom_site_symmetry_multiplicity
_atom_site_Wyckoff_label
_atom_site_fract_x
_atom_site_fract_y
_atom_site_fract_z
_atom_site_occupancy
K1 K 4 c 0.25000 0.00000 0.25140 1.00000
Cl1 Cl 4 d 0.76061 0.25000 0.25000 1.00000
Mg1 Mg 4 d 0.24343 0.25000 0.25000 1.00000
O1 O 4 d 0.36920 0.25000 0.25000 1.00000
O2 O 4 d 0.11620 0.25000 0.25000 1.00000
Cl2 Cl 8 e 0.16583 0.57525 −0.01474 1.00000
Cl3 Cl 8 e 0.16928 0.58113 0.48834 1.00000
Cl4 Cl 8 e 0.41462 0.58176 0.25235 1.00000
Cl5 Cl 8 e 0.01987 0.74957 −0.02314 1.00000
H1 H 8 e 0.60400 0.75000 0.57900 1.00000
H2 H 8 e 0.41200 0.75100 0.31300 1.00000
H3 H 8 e 0.73700 0.67300 0.54300 1.00000
H4 H 8 e 0.25000 0.72200 0.50500 1.00000
H5 H 8 e 0.73400 0.84800 0.58700 1.00000
H6 H 8 e 0.30300 0.84500 0.33500 1.00000
H7 H 8 e 0.51400 0.50900 0.32600 1.00000
H8 H 8 e 0.41700 0.49400 0.60600 1.00000
H9 H 8 e 0.76800 0.57200 0.34500 1.00000
H10 H 8 e 0.26400 −0.06600 0.68200 1.00000
H11 H 8 e 0.39900 −0.01000 0.56800 1.00000
H12 H 8 e 0.42200 −0.05800 0.50000 1.00000
H13 H 8 e 0.35900 0.81700 0.67300 1.00000
H14 H 8 e 0.42900 0.81100 0.62300 1.00000
H15 H 8 e 0.56500 −0.09900 0.67900 1.00000
H16 H 8 e 0.57300 0.60400 0.69500 1.00000
H17 H 8 e 0.43800 0.66000 0.56100 1.00000
H18 H 8 e 0.41100 0.61500 0.47500 1.00000
K2 K 8 e 0.08860 0.65668 0.74995 1.00000
Mg2 Mg 8 e 0.42094 0.58981 0.74770 1.00000
O3 O 8 e 0.25540 0.70704 0.43840 1.00000
O4 O 8 e 0.26730 0.66936 0.15060 1.00000
O5 O 8 e 0.44660 0.51041 0.65000 1.00000
O6 O 8 e 0.29670 0.57117 0.73490 1.00000
O7 O 8 e 0.42590 0.54662 −0.06460 1.00000
O8 O 8 e 0.39560 0.66912 0.84440 1.00000
O9 O 8 e 0.54480 0.60769 0.76310 1.00000
O10 O 8 e 0.41790 0.63155 0.55800 1.00000
```

Carnallite [$Mg(H_2O)_6KCl_3$]: A3B12CDE6_oP276_52_d4e_18e_ce_de_2d8e - POSCAR

```
A3B12CDE6_oP276_52_d4e_18e_ce_de_2d8e & a,b/a,c/a,z1,x2,x3,x4,x5,x6,y6,
    ↪ z6,x7,y7,z7,x8,y8,z8,x9,y9,z9,x10,y10,z10,x11,y11,z11,x12,y12,
    ↪ z12,x13,y13,z13,x14,y14,z14,x15,y15,z15,x16,y16,z16,x17,y17,z17
    ↪ ,x18,y18,z18,x19,y19,z19,x20,y20,z20,x21,y21,z21,x22,y22,z22,
    ↪ x23,y23,z23,x24,y24,z24,x25,y25,z25,x26,y26,z26,x27,y27,z27,x28
    ↪ ,y28,z28,x29,y29,z29,x30,y30,z30,x31,y31,z31,x32,y32,z32,x33,
    ↪ y33,z33,x34,y34,z34,x35,y35,z35,x36,y36,z36,x37,y37,z37 --
    ↪ params=16.119,1.39413114958,0.592530554005,0.2514,0.76061,
    ↪ 0.24343,0.3692,0.1162,0.16583,0.57525,−0.01474,0.16928,0.58113,
    ↪ 0.48834,0.41462,0.58176,0.25235,0.01987,0.74957,−0.02314,0.604,
    ↪ 0.75,0.579,0.412,0.751,0.313,0.737,0.673,0.543,0.25,0.722,0.505
    ↪ ,0.734,0.848,0.587,0.303,0.845,0.335,0.514,0.509,0.326,0.417,
    ↪ 0.494,0.606,0.768,0.572,0.345,0.264,−0.066,0.682,0.399,−0.01,
    ↪ 0.568,0.422,−0.058,0.5,0.359,0.817,0.673,0.429,0.811,0.623,
    ↪ 0.565,−0.099,0.679,0.573,0.604,0.695,0.438,0.66,0.561,0.411,
    ↪ 0.615,0.475,0.0886,0.65668,0.74995,0.42094,0.58981,0.7477,
    ↪ 0.2554,0.70704,0.4384,0.2673,0.66936,0.1506,0.4466,0.51041,0.65
    ↪ ,0.2967,0.57117,0.7349,0.4259,0.54662,−0.0646,0.3956,0.66912,
    ↪ 0.8444,0.5448,0.60769,0.7631,0.4179,0.63155,0.558 & Pnna D_{2h
    ↪ }^{6} #52 (cd^4e^32) & oP276 & None & Cl3H12KMgO6 & Carnallite
    ↪ & E. O. Schlemper and P. K. {Sen Gupta} and T. Zoltai, Am.
    ↪ Mineral. 70, 1309−1313 (1985)
   1.00000000000000
  16.11900000000000   0.00000000000000   0.00000000000000
   0.00000000000000  22.47200000000000   0.00000000000000
   0.00000000000000   0.00000000000000   9.55100000000000
     Cl      H      K     Mg      O
     36    144     12     12     72
Direct
   0.76061000000000   0.25000000000000   0.25000000000000  Cl  (4d)
  −0.26061000000000   0.75000000000000   0.25000000000000  Cl  (4d)
  −0.76061000000000   0.75000000000000   0.75000000000000  Cl  (4d)
   1.26061000000000   0.25000000000000   0.75000000000000  Cl  (4d)
   0.16583000000000   0.57525000000000  −0.01474000000000  Cl  (8e)
   0.33417000000000  −0.57525000000000  −0.01474000000000  Cl  (8e)
   0.33417000000000   1.07525000000000   0.51474000000000  Cl  (8e)
   0.16583000000000  −0.07525000000000   0.51474000000000  Cl  (8e)
  −0.16583000000000  −0.57525000000000   0.01474000000000  Cl  (8e)
   0.66583000000000   0.57525000000000   0.01474000000000  Cl  (8e)
   0.66583000000000  −0.07525000000000   0.48526000000000  Cl  (8e)
  −0.16583000000000   1.07525000000000   0.48526000000000  Cl  (8e)
   0.16928000000000   0.58113000000000   0.48834000000000  Cl  (8e)
   0.33072000000000  −0.58113000000000   0.48834000000000  Cl  (8e)
   0.33072000000000   1.08113000000000   0.01166000000000  Cl  (8e)
   0.16928000000000  −0.08113000000000   0.01166000000000  Cl  (8e)
  −0.16928000000000  −0.58113000000000  −0.48834000000000  Cl  (8e)
   0.66928000000000   0.58113000000000  −0.48834000000000  Cl  (8e)
   0.66928000000000  −0.08113000000000   0.98834000000000  Cl  (8e)
  −0.16928000000000   1.08113000000000   0.98834000000000  Cl  (8e)
   0.41462000000000   0.58176000000000   0.25235000000000  Cl  (8e)
   0.08538000000000  −0.58176000000000   0.25235000000000  Cl  (8e)
   0.08538000000000   1.08176000000000   0.24765000000000  Cl  (8e)
   0.41462000000000  −0.08176000000000   0.24765000000000  Cl  (8e)
  −0.41462000000000  −0.58176000000000  −0.25235000000000  Cl  (8e)
   0.91462000000000   0.58176000000000  −0.25235000000000  Cl  (8e)
   0.91462000000000  −0.08176000000000   0.75235000000000  Cl  (8e)
  −0.41462000000000   1.08176000000000   0.75235000000000  Cl  (8e)
   0.01987000000000   0.74957000000000  −0.02314000000000  Cl  (8e)
   0.48013000000000  −0.74957000000000  −0.02314000000000  Cl  (8e)
   0.48013000000000   1.24957000000000   0.52314000000000  Cl  (8e)
   0.01987000000000  −0.24957000000000   0.52314000000000  Cl  (8e)
  −0.01987000000000  −0.74957000000000   0.02314000000000  Cl  (8e)
   0.51987000000000   0.74957000000000   0.02314000000000  Cl  (8e)
   0.51987000000000  −0.24957000000000   0.47686000000000  Cl  (8e)
  −0.01987000000000   1.24957000000000   0.47686000000000  Cl  (8e)
   0.60400000000000   0.75000000000000   0.57900000000000   H  (8e)
  −0.10400000000000  −0.75000000000000   0.57900000000000   H  (8e)
  −0.10400000000000   1.25000000000000  −0.07900000000000   H  (8e)
   0.60400000000000  −0.25000000000000  −0.07900000000000   H  (8e)
  −0.60400000000000  −0.75000000000000  −0.57900000000000   H  (8e)
   1.10400000000000   0.75000000000000  −0.57900000000000   H  (8e)
   1.10400000000000  −0.25000000000000   1.07900000000000   H  (8e)
  −0.60400000000000   1.25000000000000   1.07900000000000   H  (8e)
   0.41200000000000   0.75100000000000   0.31300000000000   H  (8e)
   0.08800000000000  −0.75100000000000   0.31300000000000   H  (8e)
   0.08800000000000   1.25100000000000   0.18700000000000   H  (8e)
   0.41200000000000  −0.25100000000000   0.18700000000000   H  (8e)
  −0.41200000000000  −0.75100000000000  −0.31300000000000   H  (8e)
   0.91200000000000   0.75100000000000  −0.31300000000000   H  (8e)
   0.91200000000000  −0.25100000000000   0.81300000000000   H  (8e)
  −0.41200000000000   1.25100000000000   0.81300000000000   H  (8e)
   0.73700000000000   0.67300000000000   0.54300000000000   H  (8e)
  −0.23700000000000  −0.67300000000000   0.54300000000000   H  (8e)
  −0.23700000000000   1.17300000000000  −0.04300000000000   H  (8e)
   0.73700000000000  −0.17300000000000  −0.04300000000000   H  (8e)
  −0.73700000000000  −0.67300000000000  −0.54300000000000   H  (8e)
   1.23700000000000   0.67300000000000  −0.54300000000000   H  (8e)
   1.23700000000000  −0.17300000000000   1.04300000000000   H  (8e)
  −0.73700000000000   1.17300000000000   1.04300000000000   H  (8e)
   0.25000000000000   0.72200000000000   0.50500000000000   H  (8e)
   0.25000000000000  −0.72200000000000   0.50500000000000   H  (8e)
   0.25000000000000   1.22200000000000  −0.00500000000000   H  (8e)
   0.25000000000000  −0.22200000000000  −0.00500000000000   H  (8e)
  −0.25000000000000  −0.72200000000000  −0.50500000000000   H  (8e)
   0.75000000000000   0.72200000000000  −0.50500000000000   H  (8e)
   0.75000000000000  −0.22200000000000   1.00500000000000   H  (8e)
  −0.25000000000000   1.22200000000000   1.00500000000000   H  (8e)
   0.73400000000000   0.84800000000000   0.58700000000000   H  (8e)
  −0.23400000000000  −0.84800000000000   0.58700000000000   H  (8e)
  −0.23400000000000   1.34800000000000  −0.08700000000000   H  (8e)
   0.73400000000000  −0.34800000000000  −0.08700000000000   H  (8e)
  −0.73400000000000  −0.84800000000000  −0.58700000000000   H  (8e)
   1.23400000000000   0.84800000000000  −0.58700000000000   H  (8e)
   1.23400000000000  −0.34800000000000   1.08700000000000   H  (8e)
  −0.73400000000000   1.34800000000000   1.08700000000000   H  (8e)
   0.30300000000000   0.84500000000000   0.33500000000000   H  (8e)
   0.19700000000000  −0.84500000000000   0.33500000000000   H  (8e)
   0.19700000000000   1.34500000000000   0.16500000000000   H  (8e)
   0.30300000000000  −0.34500000000000   0.16500000000000   H  (8e)
  −0.30300000000000  −0.84500000000000  −0.33500000000000   H  (8e)
   0.80300000000000   0.84500000000000  −0.33500000000000   H  (8e)
   0.80300000000000  −0.34500000000000   0.83500000000000   H  (8e)
  −0.30300000000000   1.34500000000000   0.83500000000000   H  (8e)
   0.51400000000000   0.50900000000000   0.32600000000000   H  (8e)
  −0.01400000000000  −0.50900000000000   0.32600000000000   H  (8e)
  −0.01400000000000   1.00900000000000   0.17400000000000   H  (8e)
   0.51400000000000  −0.00900000000000   0.17400000000000   H  (8e)
  −0.51400000000000  −0.50900000000000  −0.32600000000000   H  (8e)
   1.01400000000000   0.50900000000000  −0.32600000000000   H  (8e)
   1.01400000000000  −0.00900000000000   0.82600000000000   H  (8e)
  −0.51400000000000   1.00900000000000   0.82600000000000   H  (8e)
   0.41700000000000   0.49400000000000   0.60600000000000   H  (8e)
   0.08300000000000  −0.49400000000000   0.60600000000000   H  (8e)
   0.08300000000000   0.99400000000000  −0.10600000000000   H  (8e)
   0.41700000000000   0.00600000000000  −0.10600000000000   H  (8e)
  −0.41700000000000  −0.49400000000000  −0.60600000000000   H  (8e)
   0.91700000000000   0.49400000000000  −0.60600000000000   H  (8e)
   0.91700000000000   0.00600000000000   1.10600000000000   H  (8e)
  −0.41700000000000   0.99400000000000   1.10600000000000   H  (8e)
   0.76800000000000   0.57200000000000   0.34500000000000   H  (8e)
  −0.26800000000000  −0.57200000000000   0.34500000000000   H  (8e)
  −0.26800000000000   1.07200000000000   0.15500000000000   H  (8e)
   0.76800000000000  −0.07200000000000   0.15500000000000   H  (8e)
  −0.76800000000000  −0.57200000000000  −0.34500000000000   H  (8e)
   1.26800000000000   0.57200000000000  −0.34500000000000   H  (8e)
   1.26800000000000  −0.07200000000000   0.84500000000000   H  (8e)
  −0.76800000000000   1.07200000000000   0.84500000000000   H  (8e)
   0.26400000000000  −0.06600000000000   0.68200000000000   H  (8e)
   0.23600000000000   0.06600000000000   0.68200000000000   H  (8e)
   0.23600000000000   0.43400000000000  −0.18200000000000   H  (8e)
   0.26400000000000   0.56600000000000  −0.18200000000000   H  (8e)
  −0.26400000000000   0.06600000000000  −0.68200000000000   H  (8e)
   0.76400000000000  −0.06600000000000  −0.68200000000000   H  (8e)
   0.76400000000000   0.56600000000000   1.18200000000000   H  (8e)
  −0.26400000000000   0.43400000000000   1.18200000000000   H  (8e)
   0.39900000000000  −0.01000000000000   0.56800000000000   H  (8e)
```

```
0.10100000000000   0.01000000000000   0.56800000000000   H   (8e)
0.10100000000000   0.49000000000000  -0.06800000000000   H   (8e)
0.39900000000000   0.51000000000000  -0.06800000000000   H   (8e)
-0.39900000000000   0.01000000000000  -0.56800000000000   H   (8e)
0.89900000000000  -0.01000000000000  -0.56800000000000   H   (8e)
0.89900000000000   0.51000000000000   1.06800000000000   H   (8e)
-0.39900000000000   0.49000000000000   1.06800000000000   H   (8e)
0.42200000000000  -0.05800000000000   0.50000000000000   H   (8e)
0.07800000000000   0.05800000000000   0.50000000000000   H   (8e)
0.07800000000000   0.44200000000000   0.00000000000000   H   (8e)
0.42200000000000   0.55800000000000   0.00000000000000   H   (8e)
-0.42200000000000   0.05800000000000  -0.50000000000000   H   (8e)
0.92200000000000  -0.05800000000000  -0.50000000000000   H   (8e)
0.92200000000000   0.55800000000000   1.00000000000000   H   (8e)
-0.42200000000000   0.44200000000000   1.00000000000000   H   (8e)
0.35900000000000   0.81700000000000   0.67300000000000   H   (8e)
0.14100000000000  -0.81700000000000   0.67300000000000   H   (8e)
0.14100000000000   1.31700000000000  -0.17300000000000   H   (8e)
0.35900000000000  -0.31700000000000  -0.17300000000000   H   (8e)
-0.35900000000000  -0.81700000000000  -0.67300000000000   H   (8e)
0.85900000000000   0.81700000000000  -0.67300000000000   H   (8e)
0.85900000000000  -0.31700000000000   1.17300000000000   H   (8e)
-0.35900000000000   1.31700000000000   1.17300000000000   H   (8e)
0.42900000000000   0.81100000000000   0.62300000000000   H   (8e)
0.07100000000000  -0.81100000000000   0.62300000000000   H   (8e)
0.07100000000000   1.31100000000000  -0.12300000000000   H   (8e)
0.42900000000000  -0.31100000000000  -0.12300000000000   H   (8e)
-0.42900000000000  -0.81100000000000  -0.62300000000000   H   (8e)
0.92900000000000   0.81100000000000  -0.62300000000000   H   (8e)
0.92900000000000  -0.31100000000000   1.12300000000000   H   (8e)
-0.42900000000000   1.31100000000000   1.12300000000000   H   (8e)
0.56500000000000  -0.09900000000000   0.67900000000000   H   (8e)
-0.06500000000000   0.09900000000000   0.67900000000000   H   (8e)
-0.06500000000000   0.40100000000000  -0.17900000000000   H   (8e)
0.56500000000000   0.59900000000000  -0.17900000000000   H   (8e)
-0.56500000000000   0.09900000000000  -0.67900000000000   H   (8e)
1.06500000000000  -0.09900000000000  -0.67900000000000   H   (8e)
1.06500000000000   0.59900000000000   1.17900000000000   H   (8e)
-0.56500000000000   0.40100000000000   1.17900000000000   H   (8e)
0.57300000000000   0.60400000000000   0.69500000000000   H   (8e)
-0.07300000000000  -0.60400000000000   0.69500000000000   H   (8e)
-0.07300000000000   1.10400000000000  -0.19500000000000   H   (8e)
0.57300000000000  -0.10400000000000  -0.19500000000000   H   (8e)
-0.57300000000000  -0.60400000000000  -0.69500000000000   H   (8e)
1.07300000000000   0.60400000000000  -0.69500000000000   H   (8e)
1.07300000000000  -0.10400000000000   1.19500000000000   H   (8e)
-0.57300000000000   1.10400000000000   1.19500000000000   H   (8e)
0.43800000000000   0.66000000000000   0.56100000000000   H   (8e)
0.06200000000000  -0.66000000000000   0.56100000000000   H   (8e)
0.06200000000000   1.16000000000000  -0.06100000000000   H   (8e)
0.43800000000000  -0.16000000000000  -0.06100000000000   H   (8e)
-0.43800000000000  -0.66000000000000  -0.56100000000000   H   (8e)
0.93800000000000   0.66000000000000  -0.56100000000000   H   (8e)
0.93800000000000  -0.16000000000000   1.06100000000000   H   (8e)
-0.43800000000000   1.16000000000000   1.06100000000000   H   (8e)
0.41100000000000   0.61500000000000   0.47500000000000   H   (8e)
0.08900000000000  -0.61500000000000   0.47500000000000   H   (8e)
0.08900000000000   1.11500000000000   0.02500000000000   H   (8e)
0.41100000000000  -0.11500000000000   0.02500000000000   H   (8e)
-0.41100000000000  -0.61500000000000  -0.47500000000000   H   (8e)
0.91100000000000   0.61500000000000  -0.47500000000000   H   (8e)
0.91100000000000  -0.11500000000000   0.97500000000000   H   (8e)
-0.41100000000000   1.11500000000000   0.97500000000000   H   (8e)
0.25000000000000   0.00000000000000   0.25140000000000   K   (4c)
0.25000000000000   0.50000000000000   0.24860000000000   K   (4c)
0.75000000000000   0.00000000000000  -0.25140000000000   K   (4c)
0.75000000000000   0.50000000000000   0.75140000000000   K   (4c)
0.08860000000000   0.65668000000000   0.74995000000000   K   (8e)
0.41140000000000  -0.65668000000000   0.74995000000000   K   (8e)
0.41140000000000   1.15668000000000  -0.24995000000000   K   (8e)
0.08860000000000  -0.15668000000000  -0.24995000000000   K   (8e)
-0.08860000000000  -0.65668000000000  -0.74995000000000   K   (8e)
0.58860000000000   0.65668000000000  -0.74995000000000   K   (8e)
0.58860000000000  -0.15668000000000   1.24995000000000   K   (8e)
-0.08860000000000   1.15668000000000   1.24995000000000   K   (8e)
0.24343000000000   0.25000000000000   0.25000000000000   Mg  (4d)
0.25657000000000   0.75000000000000   0.25000000000000   Mg  (4d)
-0.24343000000000   0.75000000000000   0.75000000000000   Mg  (4d)
0.74343000000000   0.25000000000000   0.75000000000000   Mg  (4d)
0.42094000000000   0.58981000000000   0.74770000000000   Mg  (8e)
0.07906000000000  -0.58981000000000   0.74770000000000   Mg  (8e)
0.07906000000000   1.08981000000000  -0.24770000000000   Mg  (8e)
0.42094000000000  -0.08981000000000  -0.24770000000000   Mg  (8e)
-0.42094000000000  -0.58981000000000  -0.74770000000000   Mg  (8e)
0.92094000000000   0.58981000000000  -0.74770000000000   Mg  (8e)
0.92094000000000  -0.08981000000000   1.24770000000000   Mg  (8e)
-0.42094000000000   1.08981000000000   1.24770000000000   Mg  (8e)
0.36920000000000   0.25000000000000   0.25000000000000   O   (4d)
0.13080000000000   0.75000000000000   0.25000000000000   O   (4d)
-0.36920000000000   0.75000000000000   0.75000000000000   O   (4d)
0.86920000000000   0.25000000000000   0.75000000000000   O   (4d)
0.11620000000000   0.25000000000000   0.25000000000000   O   (4d)
0.38380000000000   0.75000000000000   0.25000000000000   O   (4d)
-0.11620000000000   0.75000000000000   0.75000000000000   O   (4d)
0.61620000000000   0.25000000000000   0.75000000000000   O   (4d)
0.25540000000000   0.70704000000000   0.43840000000000   O   (8e)
0.24460000000000  -0.70704000000000   0.43840000000000   O   (8e)
0.24460000000000   1.20704000000000   0.06160000000000   O   (8e)
0.25540000000000  -0.20704000000000   0.06160000000000   O   (8e)
-0.25540000000000  -0.70704000000000  -0.43840000000000   O   (8e)
0.75540000000000   0.70704000000000  -0.43840000000000   O   (8e)
0.75540000000000  -0.20704000000000   0.93840000000000   O   (8e)
-0.25540000000000   1.20704000000000   0.93840000000000   O   (8e)
0.26730000000000   0.66936000000000   0.15060000000000   O   (8e)
0.23270000000000  -0.66936000000000   0.15060000000000   O   (8e)
```

```
0.23270000000000   1.16936000000000   0.34940000000000   O   (8e)
0.26730000000000  -0.16936000000000   0.34940000000000   O   (8e)
-0.26730000000000  -0.66936000000000  -0.15060000000000   O   (8e)
0.76730000000000   0.66936000000000  -0.15060000000000   O   (8e)
0.76730000000000  -0.16936000000000   0.65060000000000   O   (8e)
-0.26730000000000   1.16936000000000   0.65060000000000   O   (8e)
0.44660000000000   0.51041000000000   0.65000000000000   O   (8e)
0.05340000000000  -0.51041000000000   0.65000000000000   O   (8e)
0.05340000000000   1.01041000000000  -0.15000000000000   O   (8e)
0.44660000000000  -0.01041000000000  -0.15000000000000   O   (8e)
-0.44660000000000  -0.51041000000000  -0.65000000000000   O   (8e)
0.94660000000000   0.51041000000000  -0.65000000000000   O   (8e)
0.94660000000000  -0.01041000000000   1.15000000000000   O   (8e)
-0.44660000000000   1.01041000000000   1.15000000000000   O   (8e)
0.29670000000000   0.57117000000000   0.73490000000000   O   (8e)
0.20330000000000  -0.57117000000000   0.73490000000000   O   (8e)
0.20330000000000   1.07117000000000  -0.23490000000000   O   (8e)
0.29670000000000  -0.07117000000000  -0.23490000000000   O   (8e)
-0.29670000000000  -0.57117000000000  -0.73490000000000   O   (8e)
0.79670000000000   0.57117000000000  -0.73490000000000   O   (8e)
0.79670000000000  -0.07117000000000   1.23490000000000   O   (8e)
-0.29670000000000   1.07117000000000   1.23490000000000   O   (8e)
0.42590000000000   0.54662000000000  -0.06460000000000   O   (8e)
0.07410000000000  -0.54662000000000  -0.06460000000000   O   (8e)
0.07410000000000   1.04662000000000   0.56460000000000   O   (8e)
0.42590000000000  -0.04662000000000   0.56460000000000   O   (8e)
-0.42590000000000  -0.54662000000000   0.06460000000000   O   (8e)
0.92590000000000   0.54662000000000   0.06460000000000   O   (8e)
0.92590000000000  -0.04662000000000   0.43540000000000   O   (8e)
-0.42590000000000   1.04662000000000   0.43540000000000   O   (8e)
0.39560000000000   0.66912000000000   0.84440000000000   O   (8e)
0.10440000000000  -0.66912000000000   0.84440000000000   O   (8e)
0.10440000000000   1.16912000000000  -0.34440000000000   O   (8e)
0.39560000000000  -0.16912000000000  -0.34440000000000   O   (8e)
-0.39560000000000  -0.66912000000000  -0.84440000000000   O   (8e)
0.89560000000000   0.66912000000000  -0.84440000000000   O   (8e)
0.89560000000000  -0.16912000000000   1.34440000000000   O   (8e)
-0.39560000000000   1.16912000000000   1.34440000000000   O   (8e)
0.54480000000000   0.60769000000000   0.76310000000000   O   (8e)
-0.04480000000000  -0.60769000000000   0.76310000000000   O   (8e)
-0.04480000000000   1.10769000000000  -0.26310000000000   O   (8e)
0.54480000000000  -0.10769000000000  -0.26310000000000   O   (8e)
-0.54480000000000  -0.60769000000000  -0.76310000000000   O   (8e)
1.04480000000000   0.60769000000000  -0.76310000000000   O   (8e)
1.04480000000000  -0.10769000000000   1.26310000000000   O   (8e)
-0.54480000000000   1.10769000000000   1.26310000000000   O   (8e)
0.41790000000000   0.63155000000000   0.55800000000000   O   (8e)
0.08210000000000  -0.63155000000000   0.55800000000000   O   (8e)
0.08210000000000   1.13155000000000  -0.05800000000000   O   (8e)
0.41790000000000  -0.13155000000000  -0.05800000000000   O   (8e)
-0.41790000000000  -0.63155000000000  -0.55800000000000   O   (8e)
0.91790000000000   0.63155000000000  -0.55800000000000   O   (8e)
0.91790000000000  -0.13155000000000   1.05800000000000   O   (8e)
-0.41790000000000   1.13155000000000   1.05800000000000   O   (8e)
```

Eriochalcite ($CuCl_2 \cdot 2H_2O$, $C45$): A2BC4D2_oP18_53_h_a_i_e - CIF

```
# CIF file
data_findsym-output
_audit_creation_method FINDSYM

_chemical_name_mineral 'Eriochalcite'
_chemical_formula_sum 'Cl2 Cu H4 O2'

loop_
_publ_author_name
 'S. Brownstein'
 'N. F. Han'
 'E. Gabe'
 'Y. {LePage}'
_journal_name_full_name
;
 Zeitschrift f{\"u}r Kristallographie - Crystalline Materials
;
_journal_volume 189
_journal_year 1989
_journal_page_first 13
_journal_page_last 15
_publ_Section_title
;
 A redetermination of the crystal structure of cupric chloride dihydrate
;

_aflow_title 'Eriochalcite (CuCl$_{2} \cdot 2$H$_{2}$O, $C45$) Structure
    ↪ '
_aflow_proto 'A2BC4D2_oP18_53_h_a_i_e'
_aflow_params 'a,b/a,c/a,x_{2},y_{3},z_{3},x_{4},y_{4},z_{4}'
_aflow_params_values '8.0886,0.46309620948,0.916611032812,0.2402,0.3798,
    ↪ 0.23998,0.275,0.064,0.099'
_aflow_Strukturbericht '$C45$'
_aflow_Pearson 'oP18'

_symmetry_space_group_name_H-M "P 2/m 2/n 21/a"
_symmetry_Int_Tables_number 53

_cell_length_a    8.08860
_cell_length_b    3.74580
_cell_length_c    7.41410
_cell_angle_alpha 90.00000
_cell_angle_beta  90.00000
_cell_angle_gamma 90.00000

loop_
_space_group_symop_id
_space_group_symop_operation_xyz
```

```
1 x,y,z
2 x,-y,-z
3 -x+1/2,y,-z+1/2
4 -x+1/2,-y,z+1/2
5 -x,-y,-z
6 -x,y,z
7 x+1/2,-y,z+1/2
8 x+1/2,y,-z+1/2

loop_
_atom_site_label
_atom_site_type_symbol
_atom_site_symmetry_multiplicity
_atom_site_Wyckoff_label
_atom_site_fract_x
_atom_site_fract_y
_atom_site_fract_z
_atom_site_occupancy
Cu1 Cu 2 a 0.00000 0.00000 0.00000 1.00000
O1 O 4 e 0.24020 0.00000 0.00000 1.00000
Cl1 Cl 4 h 0.00000 0.37980 0.23998 1.00000
H1 H 8 i 0.27500 0.06400 0.09900 1.00000
```

Eriochalcite ($CuCl_2 \cdot 2H_2O$, *C*45): A2BC4D2_oP18_53_h_a_i_e - POSCAR

```
A2BC4D2_oP18_53_h_a_i_e & a,b/a,c/a,x2,y3,z3,x4,y4,z4 --params=8.0886,
    ↪ 0.46309620948,0.916611032812,0.2402,0.3798,0.23998,0.275,0.064,
    ↪ 0.099 & Pmna D_{2h}^{7} #53 (aehi) & oP18 & $C45$ & Cl2CuH4O2 &
    ↪ Eriochalcite & S. Brownstein et al., Zeitschrift f{"u}r
    ↪ Kristallographie - Crystalline Materials 189, 13-15 (1989)
   1.00000000000000
   8.08860000000000   0.00000000000000   0.00000000000000
   0.00000000000000   3.74580000000000   0.00000000000000
   0.00000000000000   0.00000000000000   7.41410000000000
     Cl      Cu       H       O
      4       2       8       4
Direct
   0.00000000000000   0.37980000000000   0.23998000000000   Cl   (4h)
   0.50000000000000  -0.73998000000000   0.73998000000000   Cl   (4h)
   0.50000000000000   0.37980000000000   0.26002000000000   Cl   (4h)
   0.00000000000000  -0.37980000000000  -0.23998000000000   Cl   (4h)
   0.00000000000000   0.00000000000000   0.00000000000000   Cu   (2a)
   0.50000000000000   0.00000000000000   0.50000000000000   Cu   (2a)
   0.27500000000000   0.06400000000000   0.09900000000000    H   (8i)
   0.22500000000000  -0.06400000000000   0.59900000000000    H   (8i)
   0.22500000000000   0.06400000000000   0.40100000000000    H   (8i)
   0.27500000000000  -0.06400000000000  -0.09900000000000    H   (8i)
  -0.27500000000000  -0.06400000000000  -0.09900000000000    H   (8i)
   0.77500000000000   0.06400000000000   0.40100000000000    H   (8i)
   0.77500000000000  -0.06400000000000   0.59900000000000    H   (8i)
  -0.27500000000000   0.06400000000000   0.09900000000000    H   (8i)
   0.24020000000000   0.00000000000000   0.00000000000000    O   (4e)
   0.25980000000000   0.00000000000000   0.50000000000000    O   (4e)
  -0.24020000000000   0.00000000000000   0.00000000000000    O   (4e)
   0.74020000000000   0.00000000000000   0.50000000000000    O   (4e)
```

$NH_4HF_2$ ($F5_8$): A2BC_oP16_53_eh_ab_g - CIF

```
# CIF file
data_findsym-output
_audit_creation_method FINDSYM

_chemical_name_mineral 'F2H5N'
_chemical_formula_sum 'F2 H (NH4)'

loop_
_publ_author_name
 'M. T. Rogers'
 'L. Helmholz'
_journal_name_full_name
;
 Journal of the American Chemical Society
;
_journal_volume 62
_journal_year 1940
_journal_page_first 1533
_journal_page_last 1536
_publ_Section_title
;
 A Redetermination of the Parameters in Ammonium Bifluoride
;

# Found in The American Mineralogist Crystal Structure Database, 2003

_aflow_title 'NH$_{4}$HF$_{2}$ ($F5_{8}$) Structure'
_aflow_proto 'A2BC_oP16_53_eh_ab_g'
_aflow_params 'a,b/a,c/a,x_{3},y_{4},y_{5},z_{5}'
_aflow_params_values '8.426,0.437930215998,0.970804652267,0.863,0.46,
    ↪ 0.869,0.371'
_aflow_Strukturbericht '$F5_{8}$'
_aflow_Pearson 'oP16'

_symmetry_space_group_name_H-M "P 2/m 2/n 21/a"
_symmetry_Int_Tables_number 53

_cell_length_a    8.42600
_cell_length_b    3.69000
_cell_length_c    8.18000
_cell_angle_alpha 90.00000
_cell_angle_beta  90.00000
_cell_angle_gamma 90.00000

loop_
_space_group_symop_id
_space_group_symop_operation_xyz
1 x,y,z
2 x,-y,-z
3 -x+1/2,y,-z+1/2
4 -x+1/2,-y,z+1/2
5 -x,-y,-z
6 -x,y,z
7 x+1/2,-y,z+1/2
8 x+1/2,y,-z+1/2

loop_
_atom_site_label
_atom_site_type_symbol
_atom_site_symmetry_multiplicity
_atom_site_Wyckoff_label
_atom_site_fract_x
_atom_site_fract_y
_atom_site_fract_z
_atom_site_occupancy
H1 H 2 a 0.00000 0.00000 0.00000 1.00000
H2 H 2 b 0.50000 0.00000 0.00000 1.00000
F1 F 4 e 0.86300 0.00000 0.00000 1.00000
NH41 NH4 4 g 0.25000 0.46000 0.25000 1.00000
F2 F 4 h 0.00000 0.86900 0.37100 1.00000
```

$NH_4HF_2$ ($F5_8$): A2BC_oP16_53_eh_ab_g - POSCAR

```
A2BC_oP16_53_eh_ab_g & a,b/a,c/a,x3,y4,y5,z5 --params=8.426,
    ↪ 0.437930215998,0.970804652267,0.863,0.46,0.869,0.371 & Pmna D_{
    ↪ 2h}^{7} #53 (abegh) & oP16 & $F5_{8}$ & F2H5N & F2H5N & M. T.
    ↪ Rogers and L. Helmholz, J. Am. Chem. Soc. 62, 1533-1536 (1940)
   1.00000000000000
   8.42600000000000   0.00000000000000   0.00000000000000
   0.00000000000000   3.69000000000000   0.00000000000000
   0.00000000000000   0.00000000000000   8.18000000000000
      F       H     NH4
      8       4       4
Direct
   0.86300000000000   0.00000000000000   0.00000000000000    F   (4e)
  -0.36300000000000   0.00000000000000   0.50000000000000    F   (4e)
  -0.86300000000000   0.00000000000000   0.00000000000000    F   (4e)
   1.36300000000000   0.00000000000000   0.50000000000000    F   (4e)
   0.00000000000000   0.86900000000000   0.37100000000000    F   (4h)
   0.50000000000000  -0.86900000000000   0.87100000000000    F   (4h)
   0.50000000000000   0.86900000000000   0.12900000000000    F   (4h)
   0.00000000000000  -0.86900000000000  -0.37100000000000    F   (4h)
   0.00000000000000   0.00000000000000   0.00000000000000    H   (2a)
   0.50000000000000   0.00000000000000   0.50000000000000    H   (2a)
   0.50000000000000   0.00000000000000   0.00000000000000    H   (2b)
   0.00000000000000   0.00000000000000   0.50000000000000    H   (2b)
   0.25000000000000   0.46000000000000   0.25000000000000  NH4   (4g)
   0.25000000000000  -0.46000000000000   0.75000000000000  NH4   (4g)
   0.75000000000000  -0.46000000000000   0.75000000000000  NH4   (4g)
   0.75000000000000   0.46000000000000   0.25000000000000  NH4   (4g)
```

Orthorhombic $Sr_4Ru_3O_{10}$: A10B3C4_oP68_55_2e2fgh2i_adef_2e2f - CIF

```
# CIF file
data_findsym-output
_audit_creation_method FINDSYM

_chemical_name_mineral 'O10Ru3Sr4'
_chemical_formula_sum 'O10 Ru3 Sr4'

loop_
_publ_author_name
 'M. K. Crawford'
 'R. L. Harlow'
 'W. Marshall'
 'Z. Li'
 'G. Cao'
 'R. L. Lindstrom'
 'Q. Huang'
 'J. W. Lynn'
_journal_name_full_name
;
 Physical Review B
;
_journal_volume 65
_journal_year 2002
_journal_page_first 214412
_journal_page_last 214412
_publ_Section_title
;
 Structure and magnetism of single crystal Sr$_{4}$Ru$_{3}$O$_{10}$: A
    ↪ ferromagnetic triple-layer ruthenate
;

_aflow_title 'Orthorhombic Sr$_{4}$Ru$_{3}$O$_{10}$ Structure'
_aflow_proto 'A10B3C4_oP68_55_2e2fgh2i_adef_2e2f'
_aflow_params 'a,b/a,c/a,z_{3},z_{4},z_{5},z_{6},z_{7},z_{8},z_{9},z_{10
    ↪ },z_{11},z_{12},x_{13},y_{13},x_{14},y_{14},x_{15},y_{15},z_{15
    ↪ },x_{16},y_{16},z_{16}'
_aflow_params_values '3.9001,1.0,7.32622240455,0.0695,0.213,0.1402,
    ↪ 0.2961,0.4301,0.2871,0.4303,0.3598,0.0699,0.2038,0.2028,0.2971,
    ↪ 0.2966,0.2964,0.2721,0.2271,0.1392,0.2266,0.2275,0.3608'
_aflow_Strukturbericht 'None'
_aflow_Pearson 'oP68'

_symmetry_space_group_name_H-M "P 21/b 21/a 2/m"
_symmetry_Int_Tables_number 55

_cell_length_a    3.90010
_cell_length_b    3.90010
_cell_length_c    28.57300
_cell_angle_alpha 90.00000
```

```
_cell_angle_beta   90.00000
_cell_angle_gamma  90.00000

loop_
_space_group_symop_id
_space_group_symop_operation_xyz
1 x,y,z
2 x+1/2,-y+1/2,-z
3 -x+1/2,y+1/2,-z
4 -x,-y,z
5 -x,-y,-z
6 -x+1/2,y+1/2,z
7 x+1/2,-y+1/2,z
8 x,y,-z

loop_
_atom_site_label
_atom_site_type_symbol
_atom_site_symmetry_multiplicity
_atom_site_Wyckoff_label
_atom_site_fract_x
_atom_site_fract_y
_atom_site_fract_z
_atom_site_occupancy
Ru1 Ru 2 a 0.00000 0.00000 0.00000 1.00000
Ru2 Ru 2 d 0.00000 0.50000 0.50000 1.00000
O1 O 4 e 0.00000 0.00000 0.06950 1.00000
O2 O 4 e 0.00000 0.00000 0.21300 1.00000
Ru3 Ru 4 e 0.00000 0.00000 0.14020 1.00000
Sr1 Sr 4 e 0.00000 0.00000 0.29610 1.00000
Sr2 Sr 4 e 0.00000 0.00000 0.43010 1.00000
O3 O 4 f 0.00000 0.50000 0.28710 1.00000
O4 O 4 f 0.00000 0.50000 0.43030 1.00000
Ru4 Ru 4 f 0.00000 0.50000 0.35980 1.00000
Sr3 Sr 4 f 0.00000 0.50000 0.06990 1.00000
Sr4 Sr 4 f 0.00000 0.50000 0.20380 1.00000
O5 O 4 g 0.20280 0.29710 0.00000 1.00000
O6 O 4 h 0.29660 0.29640 0.50000 1.00000
O7 O 8 i 0.27210 0.22710 0.13920 1.00000
O8 O 8 i 0.22660 0.22750 0.36080 1.00000
```

Orthorhombic $Sr_4Ru_3O_{10}$: A10B3C4_oP68_55_2e2fgh2i_adef_2e2f - POSCAR

```
A10B3C4_oP68_55_2e2fgh2i_adef_2e2f & a,b/a,c/a,z3,z4,z5,z6,z7,z8,z9,z10,
    ↪ z11,z12,x13,y13,x14,y14,x15,y15,z15,x16,y16,z16 --params=3.9001
    ↪ ,1.0,7.32622240455,0.0695,0.213,0.1402,0.2961,0.4301,0.2871,
    ↪ 0.4303,0.3598,0.0699,0.2038,0.2028,0.2971,0.2966,0.2964,0.2721,
    ↪ 0.2271,0.1392,0.2266,0.2275,0.3608 & Pbam D_{2h}^{9} #55 (ade^
    ↪ 5f^5ghi^2) & oP68 & None & O10Ru3Sr4 & O10Ru3Sr4 & M. K.
    ↪ Crawford et al., Phys. Rev. B 65, 214412(2002)
   1.00000000000000
   3.90010000000000   0.00000000000000   0.00000000000000
   0.00000000000000   3.90010000000000   0.00000000000000
   0.00000000000000   0.00000000000000  28.57300000000000
     O     Ru     Sr
    40     12     16
Direct
   0.00000000000000   0.00000000000000   0.06950000000000   O   (4e)
   0.50000000000000   0.50000000000000  -0.06950000000000   O   (4e)
   0.00000000000000   0.00000000000000  -0.06950000000000   O   (4e)
   0.50000000000000   0.50000000000000   0.06950000000000   O   (4e)
   0.00000000000000   0.00000000000000   0.21300000000000   O   (4e)
   0.50000000000000   0.50000000000000  -0.21300000000000   O   (4e)
   0.00000000000000   0.00000000000000  -0.21300000000000   O   (4e)
   0.50000000000000   0.50000000000000   0.21300000000000   O   (4e)
   0.00000000000000   0.50000000000000   0.28710000000000   O   (4f)
   0.50000000000000   0.00000000000000  -0.28710000000000   O   (4f)
   0.00000000000000   0.50000000000000  -0.28710000000000   O   (4f)
   0.50000000000000   0.00000000000000   0.28710000000000   O   (4f)
   0.00000000000000   0.50000000000000   0.43030000000000   O   (4f)
   0.50000000000000   0.00000000000000  -0.43030000000000   O   (4f)
   0.00000000000000   0.50000000000000  -0.43030000000000   O   (4f)
   0.50000000000000   0.00000000000000   0.43030000000000   O   (4f)
   0.20280000000000   0.29710000000000   0.00000000000000   O   (4g)
  -0.20280000000000  -0.29710000000000   0.00000000000000   O   (4g)
   0.29720000000000   0.79710000000000   0.00000000000000   O   (4g)
   0.70280000000000   0.20290000000000   0.00000000000000   O   (4g)
   0.29660000000000   0.29640000000000   0.50000000000000   O   (4h)
  -0.29660000000000  -0.29640000000000   0.50000000000000   O   (4h)
   0.20340000000000   0.79640000000000   0.50000000000000   O   (4h)
   0.79660000000000   0.20360000000000   0.50000000000000   O   (4h)
   0.27210000000000   0.22710000000000   0.13920000000000   O   (8i)
  -0.27210000000000  -0.22710000000000   0.13920000000000   O   (8i)
   0.22790000000000   0.72710000000000  -0.13920000000000   O   (8i)
   0.77210000000000   0.27290000000000  -0.13920000000000   O   (8i)
  -0.27210000000000  -0.22710000000000  -0.13920000000000   O   (8i)
   0.27210000000000   0.22710000000000  -0.13920000000000   O   (8i)
   0.77210000000000   0.27290000000000   0.13920000000000   O   (8i)
   0.22790000000000   0.72710000000000   0.13920000000000   O   (8i)
   0.22660000000000   0.22750000000000   0.36080000000000   O   (8i)
  -0.22660000000000  -0.22750000000000   0.36080000000000   O   (8i)
   0.27340000000000   0.72750000000000  -0.36080000000000   O   (8i)
   0.72660000000000   0.27250000000000  -0.36080000000000   O   (8i)
  -0.22660000000000  -0.22750000000000  -0.36080000000000   O   (8i)
   0.22660000000000   0.22750000000000  -0.36080000000000   O   (8i)
   0.72660000000000   0.27250000000000   0.36080000000000   O   (8i)
   0.27340000000000   0.72750000000000   0.36080000000000   O   (8i)
   0.00000000000000   0.00000000000000   0.00000000000000   Ru  (2a)
   0.50000000000000   0.50000000000000   0.00000000000000   Ru  (2a)
   0.00000000000000   0.50000000000000   0.50000000000000   Ru  (2d)
   0.50000000000000   0.00000000000000   0.50000000000000   Ru  (2d)
   0.00000000000000   0.00000000000000   0.14020000000000   Ru  (4e)
   0.50000000000000   0.50000000000000  -0.14020000000000   Ru  (4e)
   0.00000000000000   0.00000000000000  -0.14020000000000   Ru  (4e)
   0.50000000000000   0.50000000000000   0.14020000000000   Ru  (4e)
   0.00000000000000   0.50000000000000   0.35980000000000   Ru  (4f)
   0.50000000000000   0.00000000000000  -0.35980000000000   Ru  (4f)
   0.00000000000000   0.50000000000000  -0.35980000000000   Ru  (4f)
   0.50000000000000   0.00000000000000   0.35980000000000   Ru  (4f)
   0.00000000000000   0.00000000000000   0.29610000000000   Sr  (4e)
   0.50000000000000   0.50000000000000  -0.29610000000000   Sr  (4e)
   0.00000000000000   0.00000000000000  -0.29610000000000   Sr  (4e)
   0.50000000000000   0.50000000000000   0.29610000000000   Sr  (4e)
   0.00000000000000   0.00000000000000   0.43010000000000   Sr  (4e)
   0.50000000000000   0.50000000000000  -0.43010000000000   Sr  (4e)
   0.00000000000000   0.00000000000000  -0.43010000000000   Sr  (4e)
   0.50000000000000   0.50000000000000   0.43010000000000   Sr  (4e)
   0.00000000000000   0.50000000000000   0.06990000000000   Sr  (4f)
   0.50000000000000   0.00000000000000  -0.06990000000000   Sr  (4f)
   0.00000000000000   0.50000000000000  -0.06990000000000   Sr  (4f)
   0.50000000000000   0.00000000000000   0.06990000000000   Sr  (4f)
   0.00000000000000   0.50000000000000   0.20380000000000   Sr  (4f)
   0.50000000000000   0.00000000000000  -0.20380000000000   Sr  (4f)
   0.00000000000000   0.50000000000000  -0.20380000000000   Sr  (4f)
   0.50000000000000   0.00000000000000   0.20380000000000   Sr  (4f)
```

$Nb_2Pd_3Se_8$: A2B3C8_oP26_55_h_ag_2g2h - CIF

```
# CIF file
data_findsym-output
_audit_creation_method FINDSYM

_chemical_name_mineral 'Nb2Pd3Se8'
_chemical_formula_sum 'Nb2 Pd3 Se8'

loop_
_publ_author_name
 'D. A. Keszler'
 'J. A. Ibers'
_journal_name_full_name
;
 Journal of Solid State Chemistry
;
_journal_volume 52
_journal_year 1984
_journal_page_first 73
_journal_page_last 79
_publ_Section_title
;
 A new structural type in ternary chalcogenide chemistry: Structure and
    ↪ properties of Nb$_{2}$Pd$_{3}$Se$_{8}$
;

_aflow_title 'Nb$_{2}$Pd$_{3}$Se$_{8}$ Structure'
_aflow_proto 'A2B3C8_oP26_55_h_ag_2g2h'
_aflow_params 'a,b/a,c/a,x_{2},y_{2},x_{3},y_{3},x_{4},y_{4},x_{5},y_{5
    ↪ },x_{6},y_{6},x_{7},y_{7}'
_aflow_params_values '15.074,0.701406395117,0.235305824599,0.21616,
    ↪ 0.3812,-0.01109,0.23185,0.15734,0.04391,0.11591,0.21532,0.28335
    ↪ ,0.25032,0.11562,0.45742'
_aflow_Strukturbericht 'None'
_aflow_Pearson 'oP26'

_symmetry_space_group_name_H-M "P 21/b 21/a 2/m"
_symmetry_Int_Tables_number 55

_cell_length_a     15.07400
_cell_length_b     10.57300
_cell_length_c     3.54700
_cell_angle_alpha  90.00000
_cell_angle_beta   90.00000
_cell_angle_gamma  90.00000

loop_
_space_group_symop_id
_space_group_symop_operation_xyz
1 x,y,z
2 x+1/2,-y+1/2,-z
3 -x+1/2,y+1/2,-z
4 -x,-y,z
5 -x,-y,-z
6 -x+1/2,y+1/2,z
7 x+1/2,-y+1/2,z
8 x,y,-z

loop_
_atom_site_label
_atom_site_type_symbol
_atom_site_symmetry_multiplicity
_atom_site_Wyckoff_label
_atom_site_fract_x
_atom_site_fract_y
_atom_site_fract_z
_atom_site_occupancy
Pd1 Pd 2 a 0.00000 0.00000 0.00000 1.00000
Pd2 Pd 4 g 0.21616 0.38120 0.00000 1.00000
Se1 Se 4 g -0.01109 0.23185 0.00000 1.00000
Se2 Se 4 g 0.15734 0.04391 0.00000 1.00000
Nb1 Nb 4 h 0.11591 0.21532 0.50000 1.00000
Se3 Se 4 h 0.28335 0.25032 0.50000 1.00000
Se4 Se 4 h 0.11562 0.45742 0.50000 1.00000
```

$Nb_2Pd_3Se_8$: A2B3C8_oP26_55_h_ag_2g2h - POSCAR

```
A2B3C8_oP26_55_h_ag_2g2h & a,b/a,c/a,x2,y2,x3,y3,x4,y4,x5,y5,x6,y6,x7,y7
    ↪ --params=15.074,0.701406395117,0.235305824599,0.21616,0.3812,-
    ↪ 0.01109,0.23185,0.15734,0.04391,0.11591,0.21532,0.28335,0.25032
    ↪ ,0.11562,0.45742 & Pbam D_{2h}^{9} #55 (ag^3h^3) & oP26 & None
    ↪ & Nb2Pd3Se8 & Nb2Pd3Se8 & D. A. Keszler and J. A. Ibers, J.
    ↪ Solid State Chem. 52, 73-79 (1984)
```

```
   1.00000000000000
  15.07400000000000    0.00000000000000    0.00000000000000
   0.00000000000000   10.57300000000000    0.00000000000000
   0.00000000000000    0.00000000000000    3.54700000000000
    Nb      Pd      Se
     4       6      16
Direct
   0.11591000000000    0.21532000000000    0.50000000000000   Nb   (4h)
  -0.11591000000000   -0.21532000000000    0.50000000000000   Nb   (4h)
   0.38409000000000    0.71532000000000    0.50000000000000   Nb   (4h)
   0.61591000000000    0.28468000000000    0.50000000000000   Nb   (4h)
   0.00000000000000    0.00000000000000    0.00000000000000   Pd   (2a)
   0.50000000000000    0.50000000000000    0.00000000000000   Pd   (2a)
   0.21616000000000    0.38120000000000    0.00000000000000   Pd   (4g)
  -0.21616000000000   -0.38120000000000    0.00000000000000   Pd   (4g)
   0.28384000000000    0.88120000000000    0.00000000000000   Pd   (4g)
   0.71616000000000    0.11880000000000    0.00000000000000   Pd   (4g)
  -0.01109000000000    0.23185000000000    0.00000000000000   Se   (4g)
   0.01109000000000   -0.23185000000000    0.00000000000000   Se   (4g)
   0.51109000000000    0.73185000000000    0.00000000000000   Se   (4g)
   0.48891000000000    0.26815000000000    0.00000000000000   Se   (4g)
   0.15734000000000    0.04391000000000    0.00000000000000   Se   (4g)
  -0.15734000000000   -0.04391000000000    0.00000000000000   Se   (4g)
   0.34266000000000    0.54391000000000    0.00000000000000   Se   (4g)
   0.65734000000000    0.45609000000000    0.00000000000000   Se   (4g)
   0.28335000000000    0.25032000000000    0.50000000000000   Se   (4h)
  -0.28335000000000   -0.25032000000000    0.50000000000000   Se   (4h)
   0.21665000000000    0.75032000000000    0.50000000000000   Se   (4h)
   0.78335000000000    0.24968000000000    0.50000000000000   Se   (4h)
   0.11562000000000    0.45742000000000    0.50000000000000   Se   (4h)
  -0.11562000000000   -0.45742000000000    0.50000000000000   Se   (4h)
   0.38438000000000    0.95742000000000    0.50000000000000   Se   (4h)
   0.61562000000000    0.04258000000000    0.50000000000000   Se   (4h)
```

$K_2HgCl_4 \cdot H_2O$ ($E3_4$): A4BCD2_oP32_55_ghi_f_e_gh - CIF

```
# CIF file
data_findsym-output
_audit_creation_method FINDSYM

_chemical_name_mineral 'Cl4(H2O)HgK2'
_chemical_formula_sum 'Cl4 (H2O) Hg K2'

loop_
_publ_author_name
 'K. Aurivillius'
 'C. St{\aa}lhandske'
_journal_name_full_name
;
 Acta Chemica Scandinavica
;
_journal_volume 27
_journal_year 1973
_journal_page_first 1086
_journal_page_last 1088
_publ_Section_title
;
 An X-Ray Single Crystal Study of K$_{2}$HgCl$_{4}$$\cdot$H$_{2}$O
;

_aflow_title 'K$_{2}$HgCl$_{4}$$\cdot$H$_{2}$O ($E3_{4}$) Structure'
_aflow_proto 'A4BCD2_oP32_55_ghi_f_e_gh'
_aflow_params 'a,b/a,c/a,z_{1},z_{2},x_{3},y_{3},x_{4},y_{4},x_{5},y_{5
    ↪ },x_{6},y_{6},x_{7},y_{7},z_{7}'
_aflow_params_values '8.258,1.41220634536,1.08077016227,0.22925,0.2312,
    ↪ 0.2042,0.0763,0.0819,0.341,0.2489,0.0601,0.1047,0.3047,0.8831,
    ↪ 0.186,0.252'
_aflow_Strukturbericht '$E3_{4}$'
_aflow_Pearson 'oP32'

_symmetry_space_group_name_H-M "P 21/b 21/a 2/m"
_symmetry_Int_Tables_number 55

_cell_length_a      8.25800
_cell_length_b      11.66200
_cell_length_c      8.92500
_cell_angle_alpha   90.00000
_cell_angle_beta    90.00000
_cell_angle_gamma   90.00000

loop_
_space_group_symop_id
_space_group_symop_operation_xyz
1 x,y,z
2 x+1/2,-y+1/2,-z
3 -x+1/2,y+1/2,-z
4 -x,-y,z
5 -x,-y,-z
6 -x+1/2,y+1/2,z
7 x+1/2,-y+1/2,z
8 x,y,-z

loop_
_atom_site_label
_atom_site_type_symbol
_atom_site_symmetry_multiplicity
_atom_site_Wyckoff_label
_atom_site_fract_x
_atom_site_fract_y
_atom_site_fract_z
_atom_site_occupancy
Hg1 Hg 4 e 0.00000 0.00000 0.22925 1.00000
H2O1 H2O 4 f 0.00000 0.50000 0.23120 1.00000
Cl1 Cl 4 g 0.20420 0.07630 0.00000 1.00000
K1 K 4 g 0.08190 0.34100 0.00000 1.00000
Cl2 Cl 4 h 0.24890 0.06010 0.50000 1.00000
K2 K 4 h 0.10470 0.30470 0.50000 1.00000
Cl3 Cl 8 i 0.88310 0.18600 0.25200 1.00000
```

$K_2HgCl_4 \cdot H_2O$ ($E3_4$): A4BCD2_oP32_55_ghi_f_e_gh - POSCAR

```
A4BCD2_oP32_55_ghi_f_e_gh & a,b/a,c/a,z1,z2,x3,y3,x4,y4,x5,y5,x6,y6,x7,
    ↪ y7,z7 --params=8.258,1.41220634536,1.08077016227,0.22925,0.2312
    ↪ ,0.2042,0.0763,0.0819,0.341,0.2489,0.0601,0.1047,0.3047,0.8831,
    ↪ 0.186,0.252 & Pbam D_{2h}^{9} #55 (efg^2h^2i) & oP32 & $E3_{4}$
    ↪ & Cl4(H2O)HgK2 & Cl4(H2O)HgK2 & K. Aurivillius and C. St{\aa}
    ↪ lhandske, Acta Chem. Scand. 27, 1086-1088 (1973)
   1.00000000000000
   8.25800000000000    0.00000000000000    0.00000000000000
   0.00000000000000   11.66200000000000    0.00000000000000
   0.00000000000000    0.00000000000000    8.92500000000000
    Cl     H2O      Hg       K
    16       4       4       8
Direct
   0.20420000000000    0.07630000000000    0.00000000000000   Cl    (4g)
  -0.20420000000000   -0.07630000000000    0.00000000000000   Cl    (4g)
   0.29580000000000    0.57630000000000    0.00000000000000   Cl    (4g)
   0.70420000000000    0.42370000000000    0.00000000000000   Cl    (4g)
   0.24890000000000    0.06010000000000    0.50000000000000   Cl    (4h)
  -0.24890000000000   -0.06010000000000    0.50000000000000   Cl    (4h)
   0.25110000000000    0.56010000000000    0.50000000000000   Cl    (4h)
   0.74890000000000    0.43990000000000    0.50000000000000   Cl    (4h)
   0.88310000000000    0.18600000000000    0.25200000000000   Cl    (8i)
  -0.88310000000000   -0.18600000000000    0.25200000000000   Cl    (8i)
  -0.38310000000000    0.68600000000000   -0.25200000000000   Cl    (8i)
   1.38310000000000    0.31400000000000   -0.25200000000000   Cl    (8i)
  -0.88310000000000   -0.18600000000000   -0.25200000000000   Cl    (8i)
   0.88310000000000    0.18600000000000   -0.25200000000000   Cl    (8i)
   1.38310000000000    0.31400000000000    0.25200000000000   Cl    (8i)
  -0.38310000000000    0.68600000000000    0.25200000000000   Cl    (8i)
   0.00000000000000    0.50000000000000    0.23120000000000   H2O   (4f)
   0.50000000000000    0.00000000000000   -0.23120000000000   H2O   (4f)
   0.00000000000000    0.50000000000000   -0.23120000000000   H2O   (4f)
   0.50000000000000    0.00000000000000    0.23120000000000   H2O   (4f)
   0.00000000000000    0.00000000000000    0.22925000000000   Hg    (4e)
   0.50000000000000    0.50000000000000   -0.22925000000000   Hg    (4e)
   0.00000000000000    0.00000000000000   -0.22925000000000   Hg    (4e)
   0.50000000000000    0.50000000000000    0.22925000000000   Hg    (4e)
   0.08190000000000    0.34100000000000    0.00000000000000   K     (4g)
  -0.08190000000000   -0.34100000000000    0.00000000000000   K     (4g)
   0.41810000000000    0.84100000000000    0.00000000000000   K     (4g)
   0.58190000000000    0.15900000000000    0.00000000000000   K     (4g)
   0.10470000000000    0.30470000000000    0.50000000000000   K     (4h)
  -0.10470000000000   -0.30470000000000    0.50000000000000   K     (4h)
   0.39530000000000    0.80470000000000    0.50000000000000   K     (4h)
   0.60470000000000    0.19530000000000    0.50000000000000   K     (4h)
```

$Ru_{11}B_8$: A8B11_oP38_55_g3h_a3g2h - CIF

```
# CIF file
data_findsym-output
_audit_creation_method FINDSYM

_chemical_name_mineral 'B8Ru11'
_chemical_formula_sum 'B8 Ru11'

loop_
_publ_author_name
 'J. {\AA}selius'
_journal_name_full_name
;
 Acta Chemica Scandinavica
;
_journal_volume 14
_journal_year 1960
_journal_page_first 2169
_journal_page_last 2176
_publ_Section_title
;
 The Crystal Structure of Ru$_{11}$B$_{8}$
;

_aflow_title 'Ru$_{11}$B$_{8}$ Structure'
_aflow_proto 'A8B11_oP38_55_g3h_a3g2h'
_aflow_params 'a,b/a,c/a,x_{2},y_{2},x_{3},y_{3},x_{4},y_{4},x_{5},y_{5
    ↪ },x_{6},y_{6},x_{7},y_{7},x_{8},y_{8},x_{9},y_{9},x_{10},y_{10
    ↪ }'
_aflow_params_values '11.609,0.97700060298,0.244293220777,0.34,0.21,
    ↪ 0.2844,0.3913,0.0429,0.3952,0.1686,0.174,0.13,0.01,0.15,0.32,
    ↪ 0.27,0.25,0.4636,0.2962,0.3404,0.0616'
_aflow_Strukturbericht 'None'
_aflow_Pearson 'oP38'

_symmetry_space_group_name_H-M "P 21/b 21/a 2/m"
_symmetry_Int_Tables_number 55

_cell_length_a      11.60900
_cell_length_b      11.34200
_cell_length_c      2.83600
_cell_angle_alpha   90.00000
_cell_angle_beta    90.00000
_cell_angle_gamma   90.00000

loop_
_space_group_symop_id
_space_group_symop_operation_xyz
1 x,y,z
2 x+1/2,-y+1/2,-z
3 -x+1/2,y+1/2,-z
4 -x,-y,z
```

```
5 -x,-y,-z
6 -x+1/2,y+1/2,z
7 x+1/2,-y+1/2,z
8 x,y,-z

loop_
_atom_site_label
_atom_site_type_symbol
_atom_site_symmetry_multiplicity
_atom_site_Wyckoff_label
_atom_site_fract_x
_atom_site_fract_y
_atom_site_fract_z
_atom_site_occupancy
Ru1 Ru 2 a 0.00000 0.00000 0.00000 1.00000
B1 B 4 g 0.34000 0.21000 0.00000 1.00000
Ru2 Ru 4 g 0.28440 0.39130 0.00000 1.00000
Ru3 Ru 4 g 0.04290 0.39520 0.00000 1.00000
Ru4 Ru 4 g 0.16860 0.17400 0.00000 1.00000
B2 B 4 h 0.13000 0.01000 0.50000 1.00000
B3 B 4 h 0.15000 0.32000 0.50000 1.00000
B4 B 4 h 0.27000 0.25000 0.50000 1.00000
Ru5 Ru 4 h 0.46360 0.29620 0.50000 1.00000
Ru6 Ru 4 h 0.34040 0.06160 0.50000 1.00000
```

$Ru_{11}B_8$: A8B11_oP38_55_g3h_a3g2h - POSCAR

```
A8B11_oP38_55_g3h_a3g2h & a,b/a,c/a,x2,y2,x3,y3,x4,y4,x5,y5,x6,y6,x7,y7,
    ↪ x8,y8,x9,y9,x10,y10 --params=11.609,0.97700060298,
    ↪ 0.244293220777,0.34,0.21,0.2844,0.3913,0.0429,0.3952,0.1686,
    ↪ 0.174,0.13,0.01,0.15,0.32,0.27,0.25,0.4636,0.2962,0.3404,0.0616
    ↪ & Pbam D_{2h}^{9} #55 (ag^4h^5) & oP38 & None & B8Ru11 &
    ↪ B8Ru11 & J. {\AA}selius, Acta Chem. Scand. 14, 2169-2176 (1960)
   1.00000000000000
  11.60900000000000   0.00000000000000   0.00000000000000
   0.00000000000000  11.34200000000000   0.00000000000000
   0.00000000000000   0.00000000000000   2.83600000000000
     B      Ru
    16      22
Direct
   0.34000000000000   0.21000000000000   0.00000000000000   B   (4g)
  -0.34000000000000  -0.21000000000000   0.00000000000000   B   (4g)
   0.16000000000000   0.71000000000000   0.00000000000000   B   (4g)
   0.84000000000000   0.29000000000000   0.00000000000000   B   (4g)
   0.13000000000000   0.01000000000000   0.50000000000000   B   (4h)
  -0.13000000000000  -0.01000000000000   0.50000000000000   B   (4h)
   0.37000000000000   0.51000000000000   0.50000000000000   B   (4h)
   0.63000000000000   0.49000000000000   0.50000000000000   B   (4h)
   0.15000000000000   0.32000000000000   0.50000000000000   B   (4h)
  -0.15000000000000  -0.32000000000000   0.50000000000000   B   (4h)
   0.35000000000000   0.82000000000000   0.50000000000000   B   (4h)
   0.65000000000000   0.18000000000000   0.50000000000000   B   (4h)
   0.27000000000000   0.25000000000000   0.50000000000000   B   (4h)
  -0.27000000000000  -0.25000000000000   0.50000000000000   B   (4h)
   0.23000000000000   0.75000000000000   0.50000000000000   B   (4h)
   0.77000000000000   0.25000000000000   0.50000000000000   B   (4h)
   0.00000000000000   0.00000000000000   0.00000000000000   Ru  (2a)
   0.50000000000000   0.50000000000000   0.00000000000000   Ru  (2a)
   0.28440000000000   0.39130000000000   0.00000000000000   Ru  (4g)
  -0.28440000000000  -0.39130000000000   0.00000000000000   Ru  (4g)
   0.21560000000000   0.89130000000000   0.00000000000000   Ru  (4g)
   0.78440000000000   0.10870000000000   0.00000000000000   Ru  (4g)
   0.04290000000000   0.39520000000000   0.00000000000000   Ru  (4g)
  -0.04290000000000  -0.39520000000000   0.00000000000000   Ru  (4g)
   0.45710000000000   0.89520000000000   0.00000000000000   Ru  (4g)
   0.54290000000000   0.10480000000000   0.00000000000000   Ru  (4g)
   0.16860000000000   0.17400000000000   0.00000000000000   Ru  (4g)
  -0.16860000000000  -0.17400000000000   0.00000000000000   Ru  (4g)
   0.33140000000000   0.67400000000000   0.00000000000000   Ru  (4g)
   0.66860000000000   0.32600000000000   0.00000000000000   Ru  (4g)
   0.46360000000000   0.29620000000000   0.50000000000000   Ru  (4h)
  -0.46360000000000  -0.29620000000000   0.50000000000000   Ru  (4h)
   0.03640000000000   0.79620000000000   0.50000000000000   Ru  (4h)
   0.96360000000000   0.20380000000000   0.50000000000000   Ru  (4h)
   0.34040000000000   0.06160000000000   0.50000000000000   Ru  (4h)
  -0.34040000000000  -0.06160000000000   0.50000000000000   Ru  (4h)
   0.15960000000000   0.56160000000000   0.50000000000000   Ru  (4h)
   0.84040000000000   0.43840000000000   0.50000000000000   Ru  (4h)
```

$HoMn_2O_5$: AB2C5_oP32_55_g_fh_eghi - CIF

```
# CIF file
data_findsym-output
_audit_creation_method FINDSYM

_chemical_name_mineral 'HoMn2O5'
_chemical_formula_sum 'Ho Mn2 O5'

loop_
_publ_author_name
 'S. {Quezel-Ambrunaz}'
 'F. Bertaut'
 'G. Buisson'
_journal_name_full_name
;
 Comptes rendus de l'{A}cad\'{e}mie des Sciences
;
_journal_volume 258
_journal_year 1964
_journal_page_first 3025
_journal_page_last 3027
_publ_Section_title
;
 Structure des compos\\'{e}s d\'oxydes de terres rares et de mangan\`{e}
    ↪ se de formule TMn$_{2}$O$_{5}$
;

# Found in Structure of NdMn$_{2}$O$_{5}$, 1993

_aflow_title 'HoMn$_{2}$O$_{5}$ Structure'
_aflow_proto 'AB2C5_oP32_55_g_fh_eghi'
_aflow_params 'a,b/a,c/a,z_{1},z_{2},x_{3},y_{3},x_{4},y_{4},x_{5},y_{5
    ↪ },x_{6},y_{6},x_{7},y_{7},z_{7}'
_aflow_params_values '7.36,1.1535326087,0.773097826087,0.25,0.25,0.143,
    ↪ 0.172,0.14,0.44,0.09,0.848,0.14,0.44,0.1,0.72,0.25'
_aflow_Strukturbericht 'None'
_aflow_Pearson 'oP32'

_symmetry_space_group_name_H-M "P 21/b 21/a 2/m"
_symmetry_Int_Tables_number 55

_cell_length_a     7.36000
_cell_length_b     8.49000
_cell_length_c     5.69000
_cell_angle_alpha  90.00000
_cell_angle_beta   90.00000
_cell_angle_gamma  90.00000

loop_
_space_group_symop_id
_space_group_symop_operation_xyz
1 x,y,z
2 x+1/2,-y+1/2,-z
3 -x+1/2,y+1/2,-z
4 -x,-y,z
5 -x,-y,-z
6 -x+1/2,y+1/2,z
7 x+1/2,-y+1/2,z
8 x,y,-z

loop_
_atom_site_label
_atom_site_type_symbol
_atom_site_symmetry_multiplicity
_atom_site_Wyckoff_label
_atom_site_fract_x
_atom_site_fract_y
_atom_site_fract_z
_atom_site_occupancy
O1 O 4 e 0.00000 0.00000 0.25000 1.00000
Mn1 Mn 4 f 0.00000 0.50000 0.25000 1.00000
Ho1 Ho 4 g 0.14300 0.17200 0.00000 1.00000
O2 O 4 g 0.14000 0.44000 0.00000 1.00000
Mn2 Mn 4 h 0.09000 0.84800 0.50000 1.00000
O3 O 4 h 0.14000 0.44000 0.50000 1.00000
O4 O 8 i 0.10000 0.72000 0.25000 1.00000
```

$HoMn_2O_5$: AB2C5_oP32_55_g_fh_eghi - POSCAR

```
AB2C5_oP32_55_g_fh_eghi & a,b/a,c/a,z1,z2,x3,y3,x4,y4,x5,y5,x6,y6,x7,y7,
    ↪ z7 --params=7.36,1.1535326087,0.773097826087,0.25,0.25,0.143,
    ↪ 0.172,0.14,0.44,0.09,0.848,0.14,0.44,0.1,0.72,0.25 & Pbam D_{2h
    ↪ }^{9} #55 (efg^2h^2i) & oP32 & None & HoMn2O5 & HoMn2O5 & S. {
    ↪ Quezel-Ambrunaz} and F. Bertaut and G. Buisson, {C. R. Acad.
    ↪ Sci. 258, 3025-3027 (1964)
   1.00000000000000
   7.36000000000000   0.00000000000000   0.00000000000000
   0.00000000000000   8.49000000000000   0.00000000000000
   0.00000000000000   0.00000000000000   5.69000000000000
    Ho     Mn      O
     4      8     20
Direct
   0.14300000000000   0.17200000000000   0.00000000000000   Ho  (4g)
  -0.14300000000000  -0.17200000000000   0.00000000000000   Ho  (4g)
   0.35700000000000   0.67200000000000   0.00000000000000   Ho  (4g)
   0.64300000000000   0.32800000000000   0.00000000000000   Ho  (4g)
   0.00000000000000   0.50000000000000   0.25000000000000   Mn  (4f)
   0.50000000000000   0.00000000000000  -0.25000000000000   Mn  (4f)
   0.00000000000000   0.50000000000000  -0.25000000000000   Mn  (4f)
   0.50000000000000   0.00000000000000   0.25000000000000   Mn  (4f)
   0.09000000000000   0.84800000000000   0.50000000000000   Mn  (4h)
  -0.09000000000000  -0.84800000000000   0.50000000000000   Mn  (4h)
   0.41000000000000   1.34800000000000   0.50000000000000   Mn  (4h)
   0.59000000000000  -0.34800000000000   0.50000000000000   Mn  (4h)
   0.00000000000000   0.00000000000000   0.25000000000000   O   (4e)
   0.50000000000000   0.50000000000000  -0.25000000000000   O   (4e)
   0.00000000000000   0.00000000000000  -0.25000000000000   O   (4e)
   0.50000000000000   0.50000000000000   0.25000000000000   O   (4e)
   0.14000000000000   0.44000000000000   0.00000000000000   O   (4g)
  -0.14000000000000  -0.44000000000000   0.00000000000000   O   (4g)
   0.36000000000000   0.94000000000000   0.00000000000000   O   (4g)
   0.64000000000000   0.06000000000000   0.00000000000000   O   (4g)
   0.14000000000000   0.44000000000000   0.50000000000000   O   (4h)
  -0.14000000000000  -0.44000000000000   0.50000000000000   O   (4h)
   0.36000000000000   0.94000000000000   0.50000000000000   O   (4h)
   0.64000000000000   0.06000000000000   0.50000000000000   O   (4h)
   0.10000000000000   0.72000000000000   0.25000000000000   O   (8i)
  -0.10000000000000  -0.72000000000000   0.25000000000000   O   (8i)
   0.40000000000000   1.22000000000000  -0.25000000000000   O   (8i)
   0.60000000000000  -0.22000000000000  -0.25000000000000   O   (8i)
  -0.10000000000000  -0.72000000000000  -0.25000000000000   O   (8i)
   0.10000000000000   0.72000000000000  -0.25000000000000   O   (8i)
   0.60000000000000  -0.22000000000000   0.25000000000000   O   (8i)
   0.40000000000000   1.22000000000000   0.25000000000000   O   (8i)
```

Calciborite ($CaB_2O_4$ II): A2BC4_oP56_56_2e_e_4e - CIF

```
# CIF file
data_findsym-output
_audit_creation_method FINDSYM
```

```
_chemical_name_mineral 'Calciborite'
_chemical_formula_sum 'B2 Ca O4'

loop_
_publ_author_name
 'D. N. Shashkin'
 'M. A. Simonov'
 'N. V. Belov'
_journal_name_full_name
;
 Doklady Akademii Nauk SSSR
;
_journal_volume 195
_journal_year 1970
_journal_page_first 345
_journal_page_last 348
_publ_Section_title
;
 Crystal structure of calciborite CaB$_{2}$O$_{4}$=Ca$_{2}$[BO$_{3}$BO]
    ↪ $_{2}$
;

# Found in The American Mineralogist Crystal Structure Database, 2003

_aflow_title 'Calciborite (CaB$_{2}$O$_{4}$ II) Structure'
_aflow_proto 'A2BC4_oP56_56_2e_e_4e'
_aflow_params 'a,b/a,c/a,x_{1},y_{1},z_{1},x_{2},y_{2},z_{2},x_{3},y_{3
    ↪ },z_{3},x_{4},y_{4},z_{4},x_{5},y_{5},z_{5},x_{6},y_{6},z_{6},
    ↪ x_{7},y_{7},z_{7}'
_aflow_params_values '8.38,1.6491646778,0.597374701671,0.537,0.139,0.624
    ↪ ,0.742,0.052,0.365,0.386,0.143,0.123,0.391,0.185,0.633,0.742,-
    ↪ 0.009,0.114,0.596,0.112,0.365,0.885,0.112,0.378'
_aflow_Strukturbericht 'None'
_aflow_Pearson 'oP56'

_symmetry_space_group_name_H-M "P 21/c 21/c 2/n"
_symmetry_Int_Tables_number 56

_cell_length_a     8.38000
_cell_length_b     13.82000
_cell_length_c     5.00600
_cell_angle_alpha  90.00000
_cell_angle_beta   90.00000
_cell_angle_gamma  90.00000

loop_
_space_group_symop_id
_space_group_symop_operation_xyz
1 x,y,z
2 x+1/2,-y,-z+1/2
3 -x,y+1/2,-z+1/2
4 -x+1/2,-y+1/2,z
5 -x,-y,-z
6 -x+1/2,y,z+1/2
7 x,-y+1/2,z+1/2
8 x+1/2,y+1/2,-z

loop_
_atom_site_label
_atom_site_type_symbol
_atom_site_symmetry_multiplicity
_atom_site_Wyckoff_label
_atom_site_fract_x
_atom_site_fract_y
_atom_site_fract_z
_atom_site_occupancy
B1 B 8 e 0.53700 0.13900 0.62400 1.00000
B2 B 8 e 0.74200 0.05200 0.36500 1.00000
Ca1 Ca 8 e 0.38600 0.14300 0.12300 1.00000
O1 O 8 e 0.39100 0.18500 0.63300 1.00000
O2 O 8 e 0.74200 -0.00900 0.11400 1.00000
O3 O 8 e 0.59600 0.11200 0.36500 1.00000
O4 O 8 e 0.88500 0.11200 0.37800 1.00000
```

Calciborite ($CaB_2O_4$ II): A2BC4_oP56_56_2e_e_4e - POSCAR

```
A2BC4_oP56_56_2e_e_4e & a,b/a,c/a,x1,y1,z1,x2,y2,z2,x3,y3,z3,x4,y4,z4,x5
    ↪ ,y5,z5,x6,y6,z6,x7,y7,z7 --params=8.38,1.6491646778,
    ↪ 0.597374701671,0.537,0.139,0.624,0.742,0.052,0.365,0.386,0.143,
    ↪ 0.123,0.391,0.185,0.633,0.742,-0.009,0.114,0.596,0.112,0.365,
    ↪ 0.885,0.112,0.378 & Pccn D_{2h}^{10} #56 (e^7) & oP56 & None &
    ↪ B2CaO4 & Calciborite & D. N. Shashkin and M. A. Simonov and N.
    ↪ V. Belov, Doklady Akademii Nauk SSSR 195, 345-348 (1970)
   1.00000000000000
   8.38000000000000   0.00000000000000   0.00000000000000
   0.00000000000000  13.82000000000000   0.00000000000000
   0.00000000000000   0.00000000000000   5.00600000000000
     B      Ca      O
    16       8     32
Direct
   0.53700000000000   0.13900000000000   0.62400000000000   B   (8e)
  -0.03700000000000   0.36100000000000   0.62400000000000   B   (8e)
  -0.53700000000000   0.63900000000000  -0.12400000000000   B   (8e)
   1.03700000000000  -0.13900000000000  -0.12400000000000   B   (8e)
  -0.53700000000000  -0.13900000000000  -0.62400000000000   B   (8e)
   1.03700000000000   0.63900000000000  -0.62400000000000   B   (8e)
   0.53700000000000   0.36100000000000   1.12400000000000   B   (8e)
  -0.03700000000000   0.13900000000000   1.12400000000000   B   (8e)
   0.74200000000000   0.05200000000000   0.36500000000000   B   (8e)
  -0.24200000000000   0.44800000000000   0.36500000000000   B   (8e)
  -0.74200000000000   0.55200000000000   0.13500000000000   B   (8e)
   1.24200000000000  -0.05200000000000   0.13500000000000   B   (8e)
  -0.74200000000000  -0.05200000000000  -0.36500000000000   B   (8e)
   1.24200000000000   0.55200000000000  -0.36500000000000   B   (8e)
   0.74200000000000   0.44800000000000   0.86500000000000   B   (8e)
  -0.24200000000000   0.05200000000000   0.86500000000000   B   (8e)
   0.38600000000000   0.14300000000000   0.12300000000000   Ca  (8e)
   0.11400000000000   0.35700000000000   0.12300000000000   Ca  (8e)
  -0.38600000000000   0.64300000000000   0.37700000000000   Ca  (8e)
   0.88600000000000  -0.14300000000000   0.37700000000000   Ca  (8e)
  -0.38600000000000  -0.14300000000000  -0.12300000000000   Ca  (8e)
   0.88600000000000   0.64300000000000  -0.12300000000000   Ca  (8e)
   0.38600000000000   0.35700000000000   0.62300000000000   Ca  (8e)
   0.11400000000000   0.14300000000000   0.62300000000000   Ca  (8e)
   0.39100000000000   0.18500000000000   0.63300000000000   O   (8e)
   0.10900000000000   0.31500000000000   0.63300000000000   O   (8e)
  -0.39100000000000   0.68500000000000  -0.13300000000000   O   (8e)
   0.89100000000000  -0.18500000000000  -0.13300000000000   O   (8e)
  -0.39100000000000  -0.18500000000000  -0.63300000000000   O   (8e)
   0.89100000000000   0.68500000000000  -0.63300000000000   O   (8e)
   0.39100000000000   0.31500000000000   1.13300000000000   O   (8e)
   0.10900000000000   0.18500000000000   1.13300000000000   O   (8e)
   0.74200000000000  -0.00900000000000   0.11400000000000   O   (8e)
  -0.24200000000000   0.50900000000000   0.11400000000000   O   (8e)
  -0.74200000000000   0.49100000000000   0.38600000000000   O   (8e)
   1.24200000000000   0.00900000000000   0.38600000000000   O   (8e)
  -0.74200000000000   0.00900000000000  -0.11400000000000   O   (8e)
   1.24200000000000   0.49100000000000  -0.11400000000000   O   (8e)
   0.74200000000000   0.50900000000000   0.61400000000000   O   (8e)
  -0.24200000000000  -0.00900000000000   0.61400000000000   O   (8e)
   0.59600000000000   0.11200000000000   0.36500000000000   O   (8e)
  -0.09600000000000   0.38800000000000   0.36500000000000   O   (8e)
  -0.59600000000000   0.61200000000000   0.13500000000000   O   (8e)
   1.09600000000000  -0.11200000000000   0.13500000000000   O   (8e)
  -0.59600000000000  -0.11200000000000  -0.36500000000000   O   (8e)
   1.09600000000000   0.61200000000000  -0.36500000000000   O   (8e)
   0.59600000000000   0.38800000000000   0.86500000000000   O   (8e)
  -0.09600000000000   0.11200000000000   0.86500000000000   O   (8e)
   0.88500000000000   0.11200000000000   0.37800000000000   O   (8e)
  -0.38500000000000   0.38800000000000   0.37800000000000   O   (8e)
  -0.88500000000000   0.61200000000000   0.12200000000000   O   (8e)
   1.38500000000000  -0.11200000000000   0.12200000000000   O   (8e)
  -0.88500000000000  -0.11200000000000  -0.37800000000000   O   (8e)
   1.38500000000000   0.61200000000000  -0.37800000000000   O   (8e)
   0.88500000000000   0.38800000000000   0.87800000000000   O   (8e)
  -0.38500000000000   0.11200000000000   0.87800000000000   O   (8e)
```

$D0_{10}$ ($WO_3$) (*obsolete*): A3B_oP16_57_a2d_d - CIF

```
# CIF file
data_findsym-output
_audit_creation_method FINDSYM

_chemical_name_mineral 'O3W'
_chemical_formula_sum 'O3 W'

loop_
_publ_author_name
 'P. M. Woodward'
 'A. W. Sleight'
 'T. Vogt'
_journal_name_full_name
;
 Journal of Solid State Chemistry
;
_journal_volume 131
_journal_year 1997
_journal_page_first 9
_journal_page_last 17
_publ_Section_title
;
 Ferroelectric Tungsten Trioxide
;

_aflow_title '$D0_{10}$ (WO$_{3}$) ({\em obsolete}) Structure'
_aflow_proto 'A3B_oP16_57_a2d_d'
_aflow_params 'a,b/a,c/a,x_{2},y_{2},x_{3},y_{3},x_{4},y_{4}'
_aflow_params_values '3.8,1.94736842105,1.89473684211,0.0,0.78125,0.5625
    ↪ ,-0.03125,0.0625,-0.03125'
_aflow_Strukturbericht '$D0_{10}$'
_aflow_Pearson 'oP16'

_symmetry_space_group_name_H-M "P 2/b 21/c 21/m"
_symmetry_Int_Tables_number 57

_cell_length_a     3.80000
_cell_length_b     7.40000
_cell_length_c     7.20000
_cell_angle_alpha  90.00000
_cell_angle_beta   90.00000
_cell_angle_gamma  90.00000

loop_
_space_group_symop_id
_space_group_symop_operation_xyz
1 x,y,z
2 x,-y+1/2,-z
3 -x,y+1/2,-z+1/2
4 -x,-y,z+1/2
5 -x,-y,-z
6 -x,y+1/2,z
7 x,-y+1/2,z+1/2
8 x,y,-z+1/2

loop_
_atom_site_label
_atom_site_type_symbol
_atom_site_symmetry_multiplicity
_atom_site_Wyckoff_label
```

```
_atom_site_fract_x
_atom_site_fract_y
_atom_site_fract_z
_atom_site_occupancy
O1 O 4 a 0.00000 0.00000 0.00000 1.00000
O2 O 4 d 0.00000 0.78125 0.25000 1.00000
O3 O 4 d 0.56250 -0.03125 0.25000 1.00000
W1 W 4 d 0.06250 -0.03125 0.25000 1.00000
```

$D0_{10}$ ($WO_3$) *(obsolete)*: A3B_oP16_57_a2d_d - POSCAR

```
A3B_oP16_57_a2d_d & a,b/a,c/a,x2,y2,x3,y3,x4,y4 --params=3.8 ,
    ↪ 1.94736842105,1.89473684211,0.0,0.78125,0.5625,-0.03125,0.0625
    ↪ ,-0.03125 & Pbcm D_{2h}^{11} #57 (ad^3) & oP16 & $D0_{10}$ &
    ↪ O3W & O3W & P. M. Woodward and A. W. Sleight and T. Vogt, J.
    ↪ Solid State Chem. 131, 9-17 (1997)
   1.00000000000000
   3.80000000000000   0.00000000000000   0.00000000000000
   0.00000000000000   7.40000000000000   0.00000000000000
   0.00000000000000   0.00000000000000   7.20000000000000
     O     W
    12     4
Direct
   0.00000000000000   0.00000000000000   0.00000000000000   O   (4a)
   0.00000000000000   0.00000000000000   0.50000000000000   O   (4a)
   0.00000000000000   0.50000000000000   0.50000000000000   O   (4a)
   0.00000000000000   0.50000000000000   0.00000000000000   O   (4a)
   0.00000000000000   0.78125000000000   0.25000000000000   O   (4d)
   0.00000000000000  -0.78125000000000   0.75000000000000   O   (4d)
   0.00000000000000   1.28125000000000   0.25000000000000   O   (4d)
   0.00000000000000  -0.28125000000000   0.75000000000000   O   (4d)
   0.56250000000000  -0.03125000000000   0.25000000000000   O   (4d)
  -0.56250000000000   0.03125000000000   0.75000000000000   O   (4d)
  -0.56250000000000   0.46875000000000   0.25000000000000   O   (4d)
   0.56250000000000   0.53125000000000   0.75000000000000   O   (4d)
   0.06250000000000  -0.03125000000000   0.25000000000000   W   (4d)
  -0.06250000000000   0.03125000000000   0.75000000000000   W   (4d)
  -0.06250000000000   0.46875000000000   0.25000000000000   W   (4d)
   0.06250000000000   0.53125000000000   0.75000000000000   W   (4d)
```

$SrUO_4$: A4BC_oP24_57_cde_d_a - CIF

```
# CIF file
data_findsym-output
_audit_creation_method FINDSYM

_chemical_name_mineral 'O4SrU'
_chemical_formula_sum 'O4 Sr U'

loop_
_publ_author_name
 'B. O. Loopstra'
 'H. M. Rietveld'
_journal_name_full_name
;
 Acta Crystallographica Section B: Structural Science
;
_journal_volume 25
_journal_year 1969
_journal_page_first 787
_journal_page_last 791
_publ_Section_title
;
 The structure of some alkaline-earth metal uranates
;

_aflow_title 'SrUO$_{4}$ Structure'
_aflow_proto 'A4BC_oP24_57_cde_d_a'
_aflow_params 'a,b/a,c/a,x_{2},x_{3},y_{3},x_{4},y_{4},x_{5},y_{5},z_{5
    ↪ }'
_aflow_params_values '5.4896,1.4531113378,1.48092757214,0.1726,0.8602,
    ↪ 0.03,0.471,0.2013,0.2998,-0.0807,0.081'
_aflow_Strukturbericht 'None'
_aflow_Pearson 'oP24'

_symmetry_space_group_name_H-M "P 2/b 21/c 21/m"
_symmetry_Int_Tables_number 57

_cell_length_a      5.48960
_cell_length_b      7.97700
_cell_length_c      8.12970
_cell_angle_alpha 90.00000
_cell_angle_beta   90.00000
_cell_angle_gamma 90.00000

loop_
_space_group_symop_id
_space_group_symop_operation_xyz
1 x,y,z
2 x,-y+1/2,-z
3 -x,y+1/2,-z+1/2
4 -x,-y,z+1/2
5 -x,-y,-z
6 -x,y+1/2,z
7 x,-y+1/2,z+1/2
8 x,y,-z+1/2

loop_
_atom_site_label
_atom_site_type_symbol
_atom_site_symmetry_multiplicity
_atom_site_Wyckoff_label
_atom_site_fract_x
_atom_site_fract_y
_atom_site_fract_z
_atom_site_occupancy
U1 U 4 a 0.00000 0.00000 0.00000 1.00000
O1 O 4 c 0.17260 0.25000 0.00000 1.00000
O2 O 4 d 0.86020 0.03000 0.25000 1.00000
Sr1 Sr 4 d 0.47100 0.20130 0.25000 1.00000
O3 O 8 e 0.29980 -0.08070 0.08100 1.00000
```

$SrUO_4$: A4BC_oP24_57_cde_d_a - POSCAR

```
A4BC_oP24_57_cde_d_a & a,b/a,c/a,x2,x3,y3,x4,y4,x5,y5,z5 --params=5.4896
    ↪ ,1.4531113378,1.48092757214,0.1726,0.8602,0.03,0.471,0.2013,
    ↪ 0.2998,-0.0807,0.081 & Pbcm D_{2h}^{11} #57 (acd^2e) & oP24 &
    ↪ None & O4SrU & O4SrU & B. O. Loopstra and H. M. Rietveld, Acta
    ↪ Crystallogr. Sect. B Struct. Sci. 25, 787-791 (1969)
   1.00000000000000
   5.48960000000000   0.00000000000000   0.00000000000000
   0.00000000000000   7.97700000000000   0.00000000000000
   0.00000000000000   0.00000000000000   8.12970000000000
     O    Sr     U
    16     4     4
Direct
   0.17260000000000   0.25000000000000   0.00000000000000   O   (4c)
  -0.17260000000000   0.75000000000000   0.50000000000000   O   (4c)
  -0.17260000000000   0.75000000000000   0.00000000000000   O   (4c)
   0.17260000000000   0.25000000000000   0.50000000000000   O   (4c)
   0.86020000000000   0.03000000000000   0.25000000000000   O   (4d)
  -0.86020000000000  -0.03000000000000   0.75000000000000   O   (4d)
  -0.86020000000000   0.53000000000000   0.25000000000000   O   (4d)
   0.86020000000000   0.47000000000000   0.75000000000000   O   (4d)
   0.29980000000000  -0.08070000000000   0.08100000000000   O   (8e)
  -0.29980000000000   0.08070000000000   0.58100000000000   O   (8e)
  -0.29980000000000   0.41930000000000   0.41900000000000   O   (8e)
   0.29980000000000   0.58070000000000  -0.08100000000000   O   (8e)
  -0.29980000000000   0.08070000000000  -0.08100000000000   O   (8e)
   0.29980000000000  -0.08070000000000   0.41900000000000   O   (8e)
   0.29980000000000   0.58070000000000   0.58100000000000   O   (8e)
  -0.29980000000000   0.41930000000000   0.08100000000000   O   (8e)
   0.47100000000000   0.20130000000000   0.25000000000000  Sr   (4d)
  -0.47100000000000  -0.20130000000000   0.75000000000000  Sr   (4d)
  -0.47100000000000   0.70130000000000   0.25000000000000  Sr   (4d)
   0.47100000000000   0.29870000000000   0.75000000000000  Sr   (4d)
   0.00000000000000   0.00000000000000   0.00000000000000   U   (4a)
   0.00000000000000   0.00000000000000   0.50000000000000   U   (4a)
   0.00000000000000   0.50000000000000   0.50000000000000   U   (4a)
   0.00000000000000   0.50000000000000   0.00000000000000   U   (4a)
```

Lueshite ($NaNbO_3$): ABC3_oP40_57_cd_e_cd2e - CIF

```
# CIF file
data_findsym-output
_audit_creation_method FINDSYM

_chemical_name_mineral 'Lueshite'
_chemical_formula_sum 'Na Nb O3'

loop_
_publ_author_name
 'A. C. {Sakowski-Cowley}'
 'K. Lukaszewicz'
 'H. D. Megaw'
_journal_name_full_name
;
 Acta Crystallographica Section B: Structural Science
;
_journal_volume 25
_journal_year 1969
_journal_page_first 851
_journal_page_last 865
_publ_Section_title
;
 The structure of sodium niobate at room temperature, and the problem of
    ↪ reliability in pseudosymmetric structures
;

# Found in The American Mineralogist Crystal Structure Database, 2003

_aflow_title 'Lueshite (NaNbO$_{3}$) Structure'
_aflow_proto 'ABC3_oP40_57_cd_e_cd2e'
_aflow_params 'a,b/a,c/a,x_{1},x_{2},x_{3},y_{3},x_{4},y_{4},x_{5},y_{5
    ↪ },z_{5},x_{6},y_{6},z_{6},x_{7},y_{7},z_{7}'
_aflow_params_values '5.506,1.01089720305,2.81874318925,0.243,0.696,
    ↪ 0.239,0.218,0.191,0.767,0.2566,0.7722,0.6262,0.536,0.532,0.64,-
    ↪ 0.034,-0.033,0.61'
_aflow_Strukturbericht 'None'
_aflow_Pearson 'oP40'

_symmetry_space_group_name_H-M "P 2/b 21/c 21/m"
_symmetry_Int_Tables_number 57

_cell_length_a      5.50600
_cell_length_b      5.56600
_cell_length_c      15.52000
_cell_angle_alpha 90.00000
_cell_angle_beta   90.00000
_cell_angle_gamma 90.00000

loop_
_space_group_symop_id
_space_group_symop_operation_xyz
1 x,y,z
2 x,-y+1/2,-z
3 -x,y+1/2,-z+1/2
4 -x,-y,z+1/2
5 -x,-y,-z
6 -x,y+1/2,z
```

```
7 x,-y+1/2,z+1/2
8 x,y,-z+1/2

loop_
_atom_site_label
_atom_site_type_symbol
_atom_site_symmetry_multiplicity
_atom_site_Wyckoff_label
_atom_site_fract_x
_atom_site_fract_y
_atom_site_fract_z
_atom_site_occupancy
Na1 Na 4 c 0.24300 0.25000 0.00000 1.00000
O1 O 4 c 0.69600 0.25000 0.00000 1.00000
Na2 Na 4 d 0.23900 0.21800 0.25000 1.00000
O2 O 4 d 0.19100 0.76700 0.25000 1.00000
Nb1 Nb 8 e 0.25660 0.77220 0.62620 1.00000
O3 O 8 e 0.53600 0.53200 0.64000 1.00000
O4 O 8 e -0.03400 -0.03300 0.61000 1.00000
```

Lueshite ($NaNbO_3$): ABC3_oP40_57_cd_e_cd2e - POSCAR

```
ABC3_oP40_57_cd_e_cd2e & a,b/a,c/a,x1,x2,x3,y3,x4,y4,x5,y5,z5,x6,y6,z6,
    ↪ x7,y7,z7 --params=5.506,1.01089720305,2.81874318925,0.243,0.696
    ↪ ,0.239,0.218,0.191,0.767,0.2566,0.7722,0.6262,0.536,0.532,0.64
    ↪ ,-0.034,-0.033,0.61 & Pbcm D_{2h}^{11} #57 (c^2d^2e^3) & oP40 &
    ↪ None & NaNbO3 & Lueshite & A. C. {Sakowski-Cowley} and K.
    ↪ Lukaszewicz and H. D. Megaw, Acta Crystallogr. Sect. B Struct.
    ↪ Sci. 25, 851-865 (1969)
   1.00000000000000
   5.50600000000000   0.00000000000000   0.00000000000000
   0.00000000000000   5.56600000000000   0.00000000000000
   0.00000000000000   0.00000000000000  15.52000000000000
    Na     Nb     O
     8      8    24
Direct
   0.24300000000000   0.25000000000000   0.00000000000000   Na   (4c)
  -0.24300000000000   0.75000000000000   0.50000000000000   Na   (4c)
  -0.24300000000000   0.75000000000000   0.00000000000000   Na   (4c)
   0.24300000000000   0.25000000000000   0.50000000000000   Na   (4c)
   0.23900000000000   0.21800000000000   0.25000000000000   Na   (4d)
  -0.23900000000000  -0.21800000000000   0.75000000000000   Na   (4d)
  -0.23900000000000   0.71800000000000   0.25000000000000   Na   (4d)
   0.23900000000000   0.28200000000000   0.75000000000000   Na   (4d)
   0.25660000000000   0.77220000000000   0.62620000000000   Nb   (8e)
  -0.25660000000000  -0.77220000000000   1.12620000000000   Nb   (8e)
  -0.25660000000000   1.27220000000000  -0.12620000000000   Nb   (8e)
   0.25660000000000  -0.27220000000000  -0.62620000000000   Nb   (8e)
  -0.25660000000000  -0.77220000000000  -0.62620000000000   Nb   (8e)
   0.25660000000000   0.77220000000000  -0.12620000000000   Nb   (8e)
   0.25660000000000  -0.27220000000000   1.12620000000000   Nb   (8e)
  -0.25660000000000   1.27220000000000   0.62620000000000   Nb   (8e)
   0.69600000000000   0.25000000000000   0.00000000000000   O    (4c)
  -0.69600000000000   0.75000000000000   0.50000000000000   O    (4c)
  -0.69600000000000   0.75000000000000   0.00000000000000   O    (4c)
   0.69600000000000   0.25000000000000   0.50000000000000   O    (4c)
   0.19100000000000   0.76700000000000   0.25000000000000   O    (4d)
  -0.19100000000000  -0.76700000000000   0.75000000000000   O    (4d)
  -0.19100000000000   1.26700000000000   0.25000000000000   O    (4d)
   0.19100000000000  -0.26700000000000   0.75000000000000   O    (4d)
   0.53600000000000   0.53200000000000   0.64000000000000   O    (8e)
  -0.53600000000000  -0.53200000000000   1.14000000000000   O    (8e)
  -0.53600000000000   1.03200000000000  -0.14000000000000   O    (8e)
   0.53600000000000  -0.03200000000000  -0.64000000000000   O    (8e)
  -0.53600000000000  -0.53200000000000  -0.64000000000000   O    (8e)
   0.53600000000000   0.53200000000000  -0.14000000000000   O    (8e)
   0.53600000000000  -0.03200000000000   1.14000000000000   O    (8e)
  -0.53600000000000   1.03200000000000   0.64000000000000   O    (8e)
  -0.03400000000000  -0.03300000000000   0.61000000000000   O    (8e)
   0.03400000000000   0.03300000000000   1.11000000000000   O    (8e)
   0.03400000000000   0.46700000000000  -0.11000000000000   O    (8e)
  -0.03400000000000   0.53300000000000  -0.61000000000000   O    (8e)
   0.03400000000000   0.03300000000000  -0.61000000000000   O    (8e)
  -0.03400000000000  -0.03300000000000  -0.11000000000000   O    (8e)
  -0.03400000000000   0.53300000000000   1.11000000000000   O    (8e)
   0.03400000000000   0.46700000000000   0.61000000000000   O    (8e)
```

Kotoite ($Mg_3(BO_3)_2$): A2B3C6_oP22_58_g_af_gh - CIF

```
# CIF file
data_findsym-output
_audit_creation_method FINDSYM

_chemical_name_mineral 'Kotoite'
_chemical_formula_sum 'B2 Mg3 O6'

loop_
_publ_author_name
 'S. V. Berger'
_journal_name_full_name
;
 Acta Chemica Scandinavica
;
_journal_volume 3
_journal_year 1949
_journal_page_first 660
_journal_page_last 675
_publ_Section_title
;
 The Crystal Structure of the Isomorphous Orthoborates of Cobalt and
    ↪ Magnesium
;

_aflow_title 'Kotoite (Mg$_{3}$(BO$_{3}$)$_{2}$) Structure'
_aflow_proto 'A2B3C6_oP22_58_g_af_gh'
_aflow_params 'a,b/a,c/a,z_{2},x_{3},y_{3},x_{4},y_{4},x_{5},y_{5},z_{5
    ↪ }'
_aflow_params_values '4.497,1.20035579275,1.8714698688,0.179,0.56,0.25,
    ↪ 0.258,0.316,0.705,0.218,0.139'
_aflow_Strukturbericht 'None'
_aflow_Pearson 'oP22'

_symmetry_space_group_name_H-M "P 21/n 21/n 2/m"
_symmetry_Int_Tables_number 58

_cell_length_a     4.49700
_cell_length_b     5.39800
_cell_length_c     8.41600
_cell_angle_alpha  90.00000
_cell_angle_beta   90.00000
_cell_angle_gamma  90.00000

loop_
_space_group_symop_id
_space_group_symop_operation_xyz
1 x,y,z
2 x+1/2,-y+1/2,-z+1/2
3 -x+1/2,y+1/2,-z+1/2
4 -x,-y,z
5 -x,-y,-z
6 -x+1/2,y+1/2,z+1/2
7 x+1/2,-y+1/2,z+1/2
8 x,y,-z

loop_
_atom_site_label
_atom_site_type_symbol
_atom_site_symmetry_multiplicity
_atom_site_Wyckoff_label
_atom_site_fract_x
_atom_site_fract_y
_atom_site_fract_z
_atom_site_occupancy
Mg1 Mg 2 a 0.00000 0.00000 0.00000 1.00000
Mg2 Mg 4 f 0.00000 0.50000 0.17900 1.00000
B1 B 4 g 0.56000 0.25000 0.00000 1.00000
O1 O 4 g 0.25800 0.31600 0.00000 1.00000
O2 O 8 h 0.70500 0.21800 0.13900 1.00000
```

Kotoite ($Mg_3(BO_3)_2$): A2B3C6_oP22_58_g_af_gh - POSCAR

```
A2B3C6_oP22_58_g_af_gh & a,b/a,c/a,z2,x3,y3,x4,y4,x5,y5,z5 --params=
    ↪ 4.497,1.20035579275,1.8714698688,0.179,0.56,0.25,0.258,0.316,
    ↪ 0.705,0.218,0.139 & Pnnm D_{2h}^{12} #58 (afg^2h) & oP22 & None
    ↪ & B2Mg3O6 & Kotoite & S. V. Berger, Acta Chem. Scand. 3,
    ↪ 660-675 (1949)
   1.00000000000000
   4.49700000000000   0.00000000000000   0.00000000000000
   0.00000000000000   5.39800000000000   0.00000000000000
   0.00000000000000   0.00000000000000   8.41600000000000
    B     Mg     O
    4      6    12
Direct
   0.56000000000000   0.25000000000000   0.00000000000000   B    (4g)
  -0.56000000000000  -0.25000000000000   0.00000000000000   B    (4g)
  -0.06000000000000   0.75000000000000   0.50000000000000   B    (4g)
   1.06000000000000   0.25000000000000   0.50000000000000   B    (4g)
   0.00000000000000   0.00000000000000   0.00000000000000   Mg   (2a)
   0.50000000000000   0.50000000000000   0.50000000000000   Mg   (2a)
   0.00000000000000   0.50000000000000   0.17900000000000   Mg   (4f)
   0.50000000000000   0.00000000000000   0.32100000000000   Mg   (4f)
   0.00000000000000   0.50000000000000  -0.17900000000000   Mg   (4f)
   0.50000000000000   0.00000000000000   0.67900000000000   Mg   (4f)
   0.25800000000000   0.31600000000000   0.00000000000000   O    (4g)
  -0.25800000000000  -0.31600000000000   0.00000000000000   O    (4g)
   0.24200000000000   0.81600000000000   0.50000000000000   O    (4g)
   0.75800000000000   0.18400000000000   0.50000000000000   O    (4g)
   0.70500000000000   0.21800000000000   0.13900000000000   O    (8h)
  -0.70500000000000  -0.21800000000000   0.13900000000000   O    (8h)
  -0.20500000000000   0.71800000000000   0.36100000000000   O    (8h)
   1.20500000000000   0.28200000000000   0.36100000000000   O    (8h)
  -0.70500000000000  -0.21800000000000  -0.13900000000000   O    (8h)
   0.70500000000000   0.21800000000000  -0.13900000000000   O    (8h)
   1.20500000000000   0.28200000000000   0.63900000000000   O    (8h)
  -0.20500000000000   0.71800000000000   0.63900000000000   O    (8h)
```

Andalusite ($Al_2SiO_5$, $S0_2$): A2B5C_oP32_58_eg_3gh_g - CIF

```
# CIF file
data_findsym-output
_audit_creation_method FINDSYM

_chemical_name_mineral 'Andalusite'
_chemical_formula_sum 'Al2 O5 S'

loop_
_publ_author_name
 'J. K. Winter'
 'S. Ghose'
_journal_name_full_name
;
 American Mineralogist
;
_journal_volume 64
_journal_year 1979
_journal_page_first 573
_journal_page_last 586
_publ_Section_title
;
```

```
 Thermal expansion and high-temperature crystal chemistry of the Al$_{2
       ↪ }$SiO$_{4}$ polymorphs
;

# Found in The American Mineralogist Crystal Structure Database, 2003

_aflow_title 'Andalusite (Al$_{2}$SiO$_{5}$, $S0_{2}$) Structure'
_aflow_proto 'A2B5C_oP32_58_eg_3gh_g'
_aflow_params 'a,b/a,c/a,z_{1},x_{2},y_{2},x_{3},y_{3},x_{4},y_{4},x_{5
       ↪ },y_{5},x_{6},y_{6},x_{7},y_{7},z_{7}'
_aflow_params_values '7.798,1.01347781482,0.712567324955,0.2419,0.1295,
       ↪ 0.6391,0.0767,0.8629,0.4246,0.3629,0.103,0.4003,0.246,0.252,
       ↪ 0.2305,0.1339,0.2394'
_aflow_Strukturbericht '$S0_{2}$'
_aflow_Pearson 'oP32'

_symmetry_space_group_name_H-M "P 21/n 21/n 2/m"
_symmetry_Int_Tables_number 58

_cell_length_a     7.79800
_cell_length_b     7.90310
_cell_length_c     5.55660
_cell_angle_alpha  90.00000
_cell_angle_beta   90.00000
_cell_angle_gamma  90.00000

loop_
_space_group_symop_id
_space_group_symop_operation_xyz
1 x,y,z
2 x+1/2,-y+1/2,-z+1/2
3 -x+1/2,y+1/2,-z+1/2
4 -x,-y,z
5 -x,-y,-z
6 -x+1/2,y+1/2,z+1/2
7 x+1/2,-y+1/2,z+1/2
8 x,y,-z

loop_
_atom_site_label
_atom_site_type_symbol
_atom_site_symmetry_multiplicity
_atom_site_Wyckoff_label
_atom_site_fract_x
_atom_site_fract_y
_atom_site_fract_z
_atom_site_occupancy
Al1 Al 4 e 0.00000 0.00000 0.24190 1.00000
Al2 Al 4 g 0.12950 0.63910 0.00000 1.00000
O1 O 4 g 0.07670 0.86290 0.00000 1.00000
O2 O 4 g 0.42460 0.36290 0.00000 1.00000
O3 O 4 g 0.10300 0.40030 0.00000 1.00000
S1 S 4 g 0.24600 0.25200 0.00000 1.00000
O4 O 8 h 0.23050 0.13390 0.23940 1.00000
```

Andalusite ($Al_2SiO_5$, $S0_2$): A2B5C_oP32_58_eg_3gh_g - POSCAR

```
A2B5C_oP32_58_eg_3gh_g & a,b/a,c/a,z1,x2,y2,x3,y3,x4,y4,x5,y5,x6,y6,x7,
       ↪ y7,z7 --params=7.798,1.01347781482,0.712567324955,0.2419,0.1295
       ↪ ,0.6391,0.0767,0.8629,0.4246,0.3629,0.103,0.4003,0.246,0.252,
       ↪ 0.2305,0.1339,0.2394 & Pnnm D_{2h}^{12} #58 (eg^5h) & oP32 &
       ↪ $S0_{2}$ & Al2O5Si & Andalusite & J. K. Winter and S. Ghose,
       ↪ Am. Mineral. 64, 573-586 (1979)
   1.00000000000000
   7.79800000000000   0.00000000000000   0.00000000000000
   0.00000000000000   7.90310000000000   0.00000000000000
   0.00000000000000   0.00000000000000   5.55660000000000
    Al      O      S
     8     20      4
Direct
   0.00000000000000   0.00000000000000   0.24190000000000   Al   (4e)
   0.50000000000000   0.50000000000000   0.25810000000000   Al   (4e)
   0.00000000000000   0.00000000000000  -0.24190000000000   Al   (4e)
   0.50000000000000   0.50000000000000   0.74190000000000   Al   (4e)
   0.12950000000000   0.63910000000000   0.00000000000000   Al   (4g)
  -0.12950000000000  -0.63910000000000   0.00000000000000   Al   (4g)
   0.37050000000000   1.13910000000000   0.50000000000000   Al   (4g)
   0.62950000000000  -0.13910000000000   0.50000000000000   Al   (4g)
   0.07670000000000   0.86290000000000   0.00000000000000    O   (4g)
  -0.07670000000000  -0.86290000000000   0.00000000000000    O   (4g)
   0.42330000000000   1.36290000000000   0.50000000000000    O   (4g)
   0.57670000000000  -0.36290000000000   0.50000000000000    O   (4g)
   0.42460000000000   0.36290000000000   0.00000000000000    O   (4g)
  -0.42460000000000  -0.36290000000000   0.00000000000000    O   (4g)
   0.07540000000000   0.86290000000000   0.50000000000000    O   (4g)
   0.92460000000000   0.13710000000000   0.50000000000000    O   (4g)
   0.10300000000000   0.40030000000000   0.00000000000000    O   (4g)
  -0.10300000000000  -0.40030000000000   0.00000000000000    O   (4g)
   0.39700000000000   0.90030000000000   0.50000000000000    O   (4g)
   0.60300000000000   0.09970000000000   0.50000000000000    O   (4g)
   0.23050000000000   0.13390000000000   0.23940000000000    O   (8h)
  -0.23050000000000  -0.13390000000000   0.23940000000000    O   (8h)
   0.26950000000000   0.63390000000000   0.26060000000000    O   (8h)
   0.73050000000000   0.36610000000000   0.26060000000000    O   (8h)
  -0.23050000000000  -0.13390000000000  -0.23940000000000    O   (8h)
   0.23050000000000   0.13390000000000  -0.23940000000000    O   (8h)
   0.73050000000000   0.36610000000000   0.73940000000000    O   (8h)
   0.26950000000000   0.63390000000000   0.73940000000000    O   (8h)
   0.24600000000000   0.25200000000000   0.00000000000000    S   (4g)
  -0.24600000000000  -0.25200000000000   0.00000000000000    S   (4g)
   0.25400000000000   0.75200000000000   0.50000000000000    S   (4g)
   0.74600000000000   0.24800000000000   0.50000000000000    S   (4g)
```

Protoanthophyllite ($H_2Mg_7Si_8O_{24}$): A2B7C24D8_oP82_58_g_ae2f_2g5h_2h - CIF

```
# CIF file
data_findsym-output
_audit_creation_method FINDSYM

_chemical_name_mineral 'Protoanthophyllite'
_chemical_formula_sum 'H2 Mg7 O24 Si8'

loop_
_publ_author_name
 'H. Konishi'
 'T. L. Groy'
 'I. D\\'{o}dony'
 'R. Miyawaki'
 'S. Matsubara'
 'P. R. Buseck'
_journal_name_full_name
;
 American Mineralogist
;
_journal_volume 88
_journal_year 2003
_journal_page_first 1718
_journal_page_last 1723
_publ_Section_title
;
 Crystal structure of protoanthophyllite: A new mineral from the Takase
       ↪ ultramafic complex, Japan
;

_aflow_title 'Protoanthophyllite (H$_{2}$Mg$_{7}$Si$_{8}$O$_{24}$)
       ↪ Structure'
_aflow_proto 'A2B7C24D8_oP82_58_g_ae2f_2g5h_2h'
_aflow_params 'a,b/a,c/a,z_{2},z_{3},z_{4},x_{5},y_{5},x_{6},y_{6},x_{7
       ↪ },y_{7},x_{8},y_{8},z_{8},x_{9},y_{9},z_{9},x_{10},y_{10},z_{10
       ↪ },x_{11},y_{11},z_{11},x_{12},y_{12},z_{12},x_{13},y_{13},z_{13
       ↪ },x_{14},y_{14},z_{14}'
_aflow_params_values '5.3117,1.76126287253,3.37571775514,0.17748,0.41276
       ↪ ,0.24009,0.682,0.225,0.6656,0.1128,0.1502,0.3414,0.1666,0.11279
       ↪ ,0.08868,0.6699,0.12137,0.17464,0.1834,0.25265,0.433,
       ↪ 0.3436,0.11987,-0.0626,0.34888,0.13244,0.17327,0.28464,0.08483,
       ↪ 0.67034,0.29399,0.17095'
_aflow_Strukturbericht 'None'
_aflow_Pearson 'oP82'

_symmetry_space_group_name_H-M "P 21/n 21/n 2/m"
_symmetry_Int_Tables_number 58

_cell_length_a     5.31170
_cell_length_b     9.35530
_cell_length_c     17.93080
_cell_angle_alpha  90.00000
_cell_angle_beta   90.00000
_cell_angle_gamma  90.00000

loop_
_space_group_symop_id
_space_group_symop_operation_xyz
1 x,y,z
2 x+1/2,-y+1/2,-z+1/2
3 -x+1/2,y+1/2,-z+1/2
4 -x,-y,z
5 -x,-y,-z
6 -x+1/2,y+1/2,z+1/2
7 x+1/2,-y+1/2,z+1/2
8 x,y,-z

loop_
_atom_site_label
_atom_site_type_symbol
_atom_site_symmetry_multiplicity
_atom_site_Wyckoff_label
_atom_site_fract_x
_atom_site_fract_y
_atom_site_fract_z
_atom_site_occupancy
Mg1 Mg 2 a 0.00000 0.00000 0.00000 1.00000
Mg2 Mg 4 e 0.00000 0.00000 0.17748 1.00000
Mg3 Mg 4 f 0.00000 0.50000 0.41276 1.00000
Mg4 Mg 4 f 0.00000 0.50000 0.24009 1.00000
H1 H 4 g 0.68200 0.22500 0.00000 1.00000
O1 O 4 g 0.66560 0.11280 0.00000 1.00000
O2 O 4 g 0.15020 0.34140 0.00000 1.00000
O3 O 8 h 0.16660 0.11279 0.08868 1.00000
O4 O 8 h 0.66990 0.12137 0.17464 1.00000
O5 O 8 h 0.18340 0.12110 0.25265 1.00000
O6 O 8 h 0.43300 0.34360 0.11987 1.00000
O7 O 8 h -0.06260 0.34888 0.13244 1.00000
Si1 Si 8 h 0.17327 0.28464 0.08483 1.00000
Si2 Si 8 h 0.67034 0.29399 0.17095 1.00000
```

Protoanthophyllite ($H_2Mg_7Si_8O_{24}$): A2B7C24D8_oP82_58_g_ae2f_2g5h_2h - POSCAR

```
A2B7C24D8_oP82_58_g_ae2f_2g5h_2h & a,b/a,c/a,z2,z3,z4,x5,y5,x6,y6,x7,y7,
       ↪ x8,y8,z8,x9,y9,z9,x10,y10,z10,x11,y11,z11,x12,y12,z12,x13,y13,
       ↪ z13,x14,y14,z14 --params=5.3117,1.76126287253,3.37571775514,
       ↪ 0.17748,0.41276,0.24009,0.682,0.225,0.6656,0.1128,0.1502,0.3414
       ↪ ,0.1666,0.11279,0.08868,0.6699,0.12137,0.17464,0.1834,0.1211,
       ↪ 0.25265,0.433,0.3436,0.11987,-0.0626,0.34888,0.13244,0.17327,
       ↪ 0.28464,0.08483,0.67034,0.29399,0.17095 & Pnnm D_{2h}^{12} #58
       ↪ (aef^2g^3h^7) & oP82 & None & H2Mg7O24Si8 & Protoanthophyllite
       ↪ & H. Konishi et al., Am. Mineral. 88, 1718-1723 (2003)
   1.00000000000000
   5.31170000000000   0.00000000000000   0.00000000000000
   0.00000000000000   9.35530000000000   0.00000000000000
   0.00000000000000   0.00000000000000  17.93080000000000
```

```
     H    Mg     O    Si
     4    14    48    16
Direct
   0.68200000000000   0.22500000000000   0.00000000000000   H   (4g)
  -0.68200000000000  -0.22500000000000   0.00000000000000   H   (4g)
  -0.18200000000000   0.72500000000000   0.50000000000000   H   (4g)
   1.18200000000000   0.27500000000000   0.50000000000000   H   (4g)
   0.00000000000000   0.00000000000000   0.00000000000000   Mg  (2a)
   0.50000000000000   0.50000000000000   0.50000000000000   Mg  (2a)
   0.00000000000000   0.00000000000000   0.17748000000000   Mg  (4e)
   0.50000000000000   0.50000000000000   0.32252000000000   Mg  (4e)
   0.00000000000000   0.00000000000000  -0.17748000000000   Mg  (4e)
   0.50000000000000   0.50000000000000   0.67748000000000   Mg  (4e)
   0.00000000000000   0.50000000000000   0.41276000000000   Mg  (4f)
   0.50000000000000   0.00000000000000   0.08724000000000   Mg  (4f)
   0.00000000000000   0.50000000000000  -0.41276000000000   Mg  (4f)
   0.50000000000000   0.00000000000000   0.91276000000000   Mg  (4f)
   0.00000000000000   0.50000000000000   0.24009000000000   Mg  (4f)
   0.50000000000000   0.00000000000000   0.25991000000000   Mg  (4f)
   0.00000000000000   0.50000000000000  -0.24009000000000   Mg  (4f)
   0.50000000000000   0.00000000000000   0.74009000000000   Mg  (4f)
   0.66560000000000   0.11280000000000   0.00000000000000   O   (4g)
  -0.66560000000000  -0.11280000000000   0.00000000000000   O   (4g)
  -0.16560000000000   0.61280000000000   0.50000000000000   O   (4g)
   1.16560000000000   0.38720000000000   0.50000000000000   O   (4g)
   0.15020000000000   0.34140000000000   0.00000000000000   O   (4g)
  -0.15020000000000  -0.34140000000000   0.00000000000000   O   (4g)
   0.34980000000000   0.84140000000000   0.50000000000000   O   (4g)
   0.65020000000000   0.15860000000000   0.50000000000000   O   (4g)
   0.16660000000000   0.11279000000000   0.08868000000000   O   (8h)
  -0.16660000000000  -0.11279000000000   0.08868000000000   O   (8h)
   0.33340000000000   0.61279000000000   0.41132000000000   O   (8h)
   0.66660000000000   0.38721000000000   0.41132000000000   O   (8h)
  -0.16660000000000  -0.11279000000000  -0.08868000000000   O   (8h)
   0.16660000000000   0.11279000000000  -0.08868000000000   O   (8h)
   0.66660000000000   0.38721000000000   0.58868000000000   O   (8h)
   0.33340000000000   0.61279000000000   0.58868000000000   O   (8h)
   0.66990000000000   0.12137000000000   0.17464000000000   O   (8h)
  -0.66990000000000  -0.12137000000000   0.17464000000000   O   (8h)
  -0.16990000000000   0.62137000000000   0.32536000000000   O   (8h)
   1.16990000000000   0.37863000000000   0.32536000000000   O   (8h)
  -0.66990000000000  -0.12137000000000  -0.17464000000000   O   (8h)
   0.66990000000000   0.12137000000000  -0.17464000000000   O   (8h)
   1.16990000000000   0.37863000000000   0.67464000000000   O   (8h)
  -0.16990000000000   0.62137000000000   0.67464000000000   O   (8h)
   0.18340000000000   0.12110000000000   0.25265000000000   O   (8h)
  -0.18340000000000  -0.12110000000000   0.25265000000000   O   (8h)
   0.31660000000000   0.62110000000000   0.24735000000000   O   (8h)
   0.68340000000000   0.37890000000000   0.24735000000000   O   (8h)
  -0.18340000000000  -0.12110000000000  -0.25265000000000   O   (8h)
   0.18340000000000   0.12110000000000  -0.25265000000000   O   (8h)
   0.68340000000000   0.37890000000000   0.75265000000000   O   (8h)
   0.31660000000000   0.62110000000000   0.75265000000000   O   (8h)
   0.43300000000000   0.34360000000000   0.11987000000000   O   (8h)
  -0.43300000000000  -0.34360000000000   0.11987000000000   O   (8h)
   0.06700000000000   0.84360000000000   0.38013000000000   O   (8h)
   0.93300000000000   0.15640000000000   0.38013000000000   O   (8h)
  -0.43300000000000  -0.34360000000000  -0.11987000000000   O   (8h)
   0.43300000000000   0.34360000000000  -0.11987000000000   O   (8h)
   0.93300000000000   0.15640000000000   0.61987000000000   O   (8h)
   0.06700000000000   0.84360000000000   0.61987000000000   O   (8h)
  -0.06260000000000   0.34888000000000   0.13244000000000   O   (8h)
   0.06260000000000  -0.34888000000000   0.13244000000000   O   (8h)
   0.56260000000000   0.84888000000000   0.36756000000000   O   (8h)
   0.43740000000000   0.15112000000000   0.36756000000000   O   (8h)
   0.06260000000000  -0.34888000000000  -0.13244000000000   O   (8h)
  -0.06260000000000   0.34888000000000  -0.13244000000000   O   (8h)
   0.43740000000000   0.15112000000000   0.63244000000000   O   (8h)
   0.56260000000000   0.84888000000000   0.63244000000000   O   (8h)
   0.17327000000000   0.28464000000000   0.08483000000000   Si  (8h)
  -0.17327000000000  -0.28464000000000   0.08483000000000   Si  (8h)
   0.32673000000000   0.78464000000000   0.41517000000000   Si  (8h)
   0.67327000000000   0.21536000000000   0.41517000000000   Si  (8h)
  -0.17327000000000  -0.28464000000000  -0.08483000000000   Si  (8h)
   0.17327000000000   0.28464000000000  -0.08483000000000   Si  (8h)
   0.67327000000000   0.21536000000000   0.58483000000000   Si  (8h)
   0.32673000000000   0.78464000000000   0.58483000000000   Si  (8h)
   0.67034000000000   0.29399000000000   0.17095000000000   Si  (8h)
  -0.67034000000000  -0.29399000000000   0.17095000000000   Si  (8h)
  -0.17034000000000   0.79399000000000   0.32905000000000   Si  (8h)
   1.17034000000000   0.20601000000000   0.32905000000000   Si  (8h)
  -0.67034000000000  -0.29399000000000  -0.17095000000000   Si  (8h)
   0.67034000000000   0.29399000000000  -0.17095000000000   Si  (8h)
   1.17034000000000   0.20601000000000   0.67095000000000   Si  (8h)
  -0.17034000000000   0.79399000000000   0.67095000000000   Si  (8h)
```

$In_4Se_3$: A4B3_oP28_58_4g_3g - CIF

```
# CIF file
data_findsym-output
_audit_creation_method FINDSYM

_chemical_name_mineral 'In4Se3'
_chemical_formula_sum 'In4 Se3'

loop_
_publ_author_name
 'J. H. C. Hogg'
 'H. H. Sutherland'
 'D. J. Williams'
_journal_name_full_name
;
 Acta Crystallographica Section B: Structural Science
;
_journal_volume 29
_journal_year 1973
_journal_page_first 1590
_journal_page_last 1593
_publ_Section_title
;
 The crystal structure of tetraindium triselenide
;

# Found in Structural and optical properties of In$_{4}$Se$_{3}$ thin
     ↪ films obtained by flash evaporation, 1995

_aflow_title 'In$_{4}$Se$_{3}$ Structure'
_aflow_proto 'A4B3_oP28_58_4g_3g'
_aflow_params 'a,b/a,c/a,x_{1},y_{1},x_{2},y_{2},x_{3},y_{3},x_{4},y_{4
     ↪ },x_{5},y_{5},x_{6},y_{6},x_{7},y_{7}'
_aflow_params_values '15.297,0.804602209584,0.266784336798,0.7111,0.3393
     ↪ ,0.8157,0.5236,-0.0325,0.6442,0.4238,0.3974,-0.0967,0.8493,
     ↪ 0.7688,0.1386,0.4239,0.156'
_aflow_Strukturbericht 'None'
_aflow_Pearson 'oP28'

_symmetry_space_group_name_H-M "P 21/n 21/n 2/m"
_symmetry_Int_Tables_number 58

_cell_length_a      15.29700
_cell_length_b      12.30800
_cell_length_c      4.08100
_cell_angle_alpha   90.00000
_cell_angle_beta    90.00000
_cell_angle_gamma   90.00000

loop_
_space_group_symop_id
_space_group_symop_operation_xyz
1 x,y,z
2 x+1/2,-y+1/2,-z+1/2
3 -x+1/2,y+1/2,-z+1/2
4 -x,-y,z
5 -x,-y,-z
6 -x+1/2,y+1/2,z+1/2
7 x+1/2,-y+1/2,z+1/2
8 x,y,-z

loop_
_atom_site_label
_atom_site_type_symbol
_atom_site_symmetry_multiplicity
_atom_site_Wyckoff_label
_atom_site_fract_x
_atom_site_fract_y
_atom_site_fract_z
_atom_site_occupancy
In1 In 4 g 0.71110 0.33930 0.00000 1.00000
In2 In 4 g 0.81570 0.52360 0.00000 1.00000
In3 In 4 g -0.03250 0.64420 0.00000 1.00000
In4 In 4 g 0.42380 0.39740 0.00000 1.00000
Se1 Se 4 g -0.09670 0.84930 0.00000 1.00000
Se2 Se 4 g 0.76880 0.13860 0.00000 1.00000
Se3 Se 4 g 0.42390 0.15600 0.00000 1.00000
```

$In_4Se_3$: A4B3_oP28_58_4g_3g - POSCAR

```
A4B3_oP28_58_4g_3g & a,b/a,c/a,x1,y1,x2,y2,x3,y3,x4,y4,x5,y5,x6,y6,x7,y7
     ↪ --params=15.297,0.804602209584,0.266784336798,0.7111,0.3393,
     ↪ 0.8157,0.5236,-0.0325,0.6442,0.4238,0.3974,-0.0967,0.8493,
     ↪ 0.7688,0.1386,0.4239,0.156 & Pnnm D_{2h}^{12} #58 (g^7) & oP28
     ↪ & None & In4Se3 & J. H. C. Hogg and H. H. Sutherland
     ↪ and D. J. Williams, Acta Crystallogr. Sect. B Struct. Sci. 29,
     ↪ 1590-1593 (1973)
   1.00000000000000
  15.29700000000000   0.00000000000000   0.00000000000000
   0.00000000000000  12.30800000000000   0.00000000000000
   0.00000000000000   0.00000000000000   4.08100000000000
    In    Se
    16    12
Direct
   0.71110000000000   0.33930000000000   0.00000000000000   In   (4g)
  -0.71110000000000  -0.33930000000000   0.00000000000000   In   (4g)
  -0.21110000000000   0.83930000000000   0.50000000000000   In   (4g)
   1.21110000000000   0.16070000000000   0.50000000000000   In   (4g)
   0.81570000000000   0.52360000000000   0.00000000000000   In   (4g)
  -0.81570000000000  -0.52360000000000   0.00000000000000   In   (4g)
  -0.31570000000000   1.02360000000000   0.50000000000000   In   (4g)
   1.31570000000000  -0.02360000000000   0.50000000000000   In   (4g)
  -0.03250000000000   0.64420000000000   0.00000000000000   In   (4g)
   0.03250000000000  -0.64420000000000   0.00000000000000   In   (4g)
   0.53250000000000   1.14420000000000   0.50000000000000   In   (4g)
   0.46750000000000  -0.14420000000000   0.50000000000000   In   (4g)
   0.42380000000000   0.39740000000000   0.00000000000000   In   (4g)
  -0.42380000000000  -0.39740000000000   0.00000000000000   In   (4g)
   0.07620000000000   0.89740000000000   0.50000000000000   In   (4g)
   0.92380000000000   0.10260000000000   0.50000000000000   In   (4g)
  -0.09670000000000   0.84930000000000   0.00000000000000   Se   (4g)
   0.09670000000000  -0.84930000000000   0.00000000000000   Se   (4g)
   0.59670000000000   1.34930000000000   0.50000000000000   Se   (4g)
   0.40330000000000  -0.34930000000000   0.50000000000000   Se   (4g)
   0.76880000000000   0.13860000000000   0.00000000000000   Se   (4g)
  -0.76880000000000  -0.13860000000000   0.00000000000000   Se   (4g)
  -0.26880000000000   0.63860000000000   0.50000000000000   Se   (4g)
   1.26880000000000   0.36140000000000   0.50000000000000   Se   (4g)
   0.42390000000000   0.15600000000000   0.00000000000000   Se   (4g)
  -0.42390000000000  -0.15600000000000   0.00000000000000   Se   (4g)
   0.07610000000000   0.65600000000000   0.50000000000000   Se   (4g)
   0.92390000000000   0.34400000000000   0.50000000000000   Se   (4g)
```

Adamite [$Zn_2(AsO_4)(OH)$, $H2_7$]: ABC5D2_oP36_58_g_g_3gh_eg - CIF

```
# CIF file
data_findsym-output
_audit_creation_method FINDSYM

_chemical_name_mineral 'Adamite'
_chemical_formula_sum 'As H O5 Zn2'

loop_
_publ_author_name
 'R. J. Hill'
_journal_name_full_name
;
 American Mineralogist
;
_journal_volume 61
_journal_year 1976
_journal_page_first 979
_journal_page_last 986
_publ_Section_title
;
 The crystal structure and infrared properties of adamite
;

_aflow_title 'Adamite [Zn$_{2}$(AsO$_{4}$)(OH), $H2_{7}$] Structure'
_aflow_proto 'ABC5D2_oP36_58_g_g_3gh_eg'
_aflow_params 'a,b/a,c/a,z_{1},x_{2},y_{2},x_{3},y_{3},x_{4},y_{4},x_{5
    ↪ },y_{5},x_{6},y_{6},x_{7},y_{7},x_{8},y_{8},z_{8}'
_aflow_params_values '8.306,1.02624608717,0.727546352035,0.24737,0.24952
    ↪ ,0.74394,0.2,0.13,0.424,0.6447,0.1079,0.1268,0.104,0.6063,
    ↪ 0.13482,0.36423,0.2685,0.3615,0.2778'
_aflow_Strukturbericht '$H2_{7}$'
_aflow_Pearson 'oP36'

_symmetry_space_group_name_H-M "P 21/n 21/n 2/m"
_symmetry_Int_Tables_number 58

_cell_length_a     8.30600
_cell_length_b     8.52400
_cell_length_c     6.04300
_cell_angle_alpha 90.00000
_cell_angle_beta  90.00000
_cell_angle_gamma 90.00000

loop_
_space_group_symop_id
_space_group_symop_operation_xyz
1 x,y,z
2 x+1/2,-y+1/2,-z+1/2
3 -x+1/2,y+1/2,-z+1/2
4 -x,-y,z
5 -x,-y,-z
6 -x+1/2,y+1/2,z+1/2
7 x+1/2,-y+1/2,z+1/2
8 x,y,-z

loop_
_atom_site_label
_atom_site_type_symbol
_atom_site_symmetry_multiplicity
_atom_site_Wyckoff_label
_atom_site_fract_x
_atom_site_fract_y
_atom_site_fract_z
_atom_site_occupancy
Zn1 Zn 4 e 0.00000 0.00000 0.24737 1.00000
As1 As 4 g 0.24952 0.74394 0.00000 1.00000
H1 H 4 g 0.20000 0.13000 0.00000 1.00000
O1 O 4 g 0.42400 0.64470 0.00000 1.00000
O2 O 4 g 0.10790 0.12680 0.00000 1.00000
O3 O 4 g 0.10400 0.60630 0.00000 1.00000
Zn2 Zn 4 g 0.13482 0.36423 0.00000 1.00000
O4 O 8 h 0.26850 0.36150 0.27780 1.00000
```

Adamite [$Zn_2(AsO_4)(OH)$, $H2_7$]: ABC5D2_oP36_58_g_g_3gh_eg - POSCAR

```
ABC5D2_oP36_58_g_g_3gh_eg & a,b/a,c/a,z1,x2,y2,x3,y3,x4,y4,x5,y5,x6,y6,
    ↪ x7,y7,x8,y8,z8 --params=8.306,1.02624608717,0.727546352035,
    ↪ 0.24737,0.24952,0.74394,0.2,0.13,0.424,0.6447,0.1079,0.1268,
    ↪ 0.104,0.6063,0.13482,0.36423,0.2685,0.3615,0.2778 & Pnnm D_{2h
    ↪ }^{12} #58 (eg^6h) & oP36 & $H2_{7}$ & AsHO5Zn2 & Adamite & R.
    ↪ J. Hill, Am. Mineral. 61, 979-986 (1976)
   1.00000000000000
   8.30600000000000   0.00000000000000   0.00000000000000
   0.00000000000000   8.52400000000000   0.00000000000000
   0.00000000000000   0.00000000000000   6.04300000000000
     As       H       O      Zn
      4       4      20       8
Direct
   0.24952000000000   0.74394000000000   0.00000000000000   As   (4g)
  -0.24952000000000  -0.74394000000000   0.00000000000000   As   (4g)
   0.25048000000000   1.24394000000000   0.50000000000000   As   (4g)
   0.74952000000000  -0.24394000000000   0.50000000000000   As   (4g)
   0.20000000000000   0.13000000000000   0.00000000000000    H   (4g)
  -0.20000000000000  -0.13000000000000   0.00000000000000    H   (4g)
   0.30000000000000   0.63000000000000   0.50000000000000    H   (4g)
   0.70000000000000   0.37000000000000   0.50000000000000    H   (4g)
   0.42400000000000   0.64470000000000   0.00000000000000    O   (4g)
  -0.42400000000000  -0.64470000000000   0.00000000000000    O   (4g)
   0.07600000000000   1.14470000000000   0.50000000000000    O   (4g)
   0.92400000000000  -0.14470000000000   0.50000000000000    O   (4g)
   0.10790000000000   0.12680000000000   0.00000000000000    O   (4g)
  -0.10790000000000  -0.12680000000000   0.00000000000000    O   (4g)
   0.39210000000000   0.62680000000000   0.50000000000000    O   (4g)
   0.60790000000000   0.37320000000000   0.50000000000000    O   (4g)
   0.10400000000000   0.60630000000000   0.00000000000000    O   (4g)
  -0.10400000000000  -0.60630000000000   0.00000000000000    O   (4g)
   0.39600000000000   1.10630000000000   0.50000000000000    O   (4g)
   0.60400000000000  -0.10630000000000   0.50000000000000    O   (4g)
   0.26850000000000   0.36150000000000   0.27780000000000    O   (8h)
  -0.26850000000000  -0.36150000000000   0.27780000000000    O   (8h)
   0.23150000000000   0.86150000000000   0.22220000000000    O   (8h)
   0.76850000000000   0.13850000000000   0.22220000000000    O   (8h)
  -0.26850000000000  -0.36150000000000  -0.27780000000000    O   (8h)
   0.26850000000000   0.36150000000000  -0.27780000000000    O   (8h)
   0.76850000000000   0.13850000000000   0.77780000000000    O   (8h)
   0.23150000000000   0.86150000000000   0.77780000000000    O   (8h)
   0.00000000000000   0.00000000000000   0.24737000000000   Zn   (4e)
   0.50000000000000   0.50000000000000   0.25263000000000   Zn   (4e)
   0.00000000000000   0.00000000000000  -0.24737000000000   Zn   (4e)
   0.50000000000000   0.50000000000000   0.74737000000000   Zn   (4e)
   0.13482000000000   0.36423000000000   0.00000000000000   Zn   (4g)
  -0.13482000000000  -0.36423000000000   0.00000000000000   Zn   (4g)
   0.36518000000000   0.86423000000000   0.50000000000000   Zn   (4g)
   0.63482000000000   0.13577000000000   0.50000000000000   Zn   (4g)
```

InS: AB_oP8_58_g_g - CIF

```
# CIF file
data_findsym-output
_audit_creation_method FINDSYM

_chemical_name_mineral 'InS'
_chemical_formula_sum 'In S'

loop_
_publ_author_name
 'K. Schubert'
 'E. D\"{o}rre'
 'E. G\"{u}nzel'
_journal_name_full_name
;
 Naturwissenschaften
;
_journal_volume 41
_journal_year 1954
_journal_page_first 448
_journal_page_last 448
_publ_Section_title
;
 Kristallchemische Ergebnisse an Phasen aus B-Elementen
;

# Found in Pearson's Handbook of Crystallographic Data for Intermetallic
    ↪ Phases, 1991

_aflow_title 'InS Structure'
_aflow_proto 'AB_oP8_58_g_g'
_aflow_params 'a,b/a,c/a,x_{1},y_{1},x_{2},y_{2}'
_aflow_params_values '4.443,2.39522844925,0.886788206167,0.125,0.121,-
    ↪ 0.005,0.355'
_aflow_Strukturbericht 'None'
_aflow_Pearson 'oP8'

_symmetry_space_group_name_H-M "P 21/n 21/n 2/m"
_symmetry_Int_Tables_number 58

_cell_length_a     4.44300
_cell_length_b     10.64200
_cell_length_c     3.94000
_cell_angle_alpha 90.00000
_cell_angle_beta  90.00000
_cell_angle_gamma 90.00000

loop_
_space_group_symop_id
_space_group_symop_operation_xyz
1 x,y,z
2 x+1/2,-y+1/2,-z+1/2
3 -x+1/2,y+1/2,-z+1/2
4 -x,-y,z
5 -x,-y,-z
6 -x+1/2,y+1/2,z+1/2
7 x+1/2,-y+1/2,z+1/2
8 x,y,-z

loop_
_atom_site_label
_atom_site_type_symbol
_atom_site_symmetry_multiplicity
_atom_site_Wyckoff_label
_atom_site_fract_x
_atom_site_fract_y
_atom_site_fract_z
_atom_site_occupancy
In1 In 4 g 0.12500 0.12100 0.00000 1.00000
S1 S 4 g -0.00500 0.35500 0.00000 1.00000
```

InS: AB_oP8_58_g_g - POSCAR

```
AB_oP8_58_g_g & a,b/a,c/a,x1,y1,x2,y2 --params=4.443,2.39522844925,
    ↪ 0.886788206167,0.125,0.121,-0.005,0.355 & Pnnm D_{2h}^{12} #58
    ↪ (g^2) & oP8 & None & InS & InS & K. Schubert and E. D\"{o}rre
    ↪ and E. G\"{u}nzel, Naturwissenschaften 41, 448(1954)
   1.00000000000000
   4.44300000000000   0.00000000000000   0.00000000000000
   0.00000000000000  10.64200000000000   0.00000000000000
   0.00000000000000   0.00000000000000   3.94000000000000
    In       S
```

```
        4        4
Direct
   0.12500000000000    0.12100000000000    0.00000000000000    In    (4g)
  -0.12500000000000   -0.12100000000000    0.00000000000000    In    (4g)
   0.37500000000000    0.62100000000000    0.50000000000000    In    (4g)
   0.62500000000000    0.37900000000000    0.50000000000000    In    (4g)
  -0.00500000000000    0.35500000000000    0.00000000000000     S    (4g)
   0.00500000000000   -0.35500000000000    0.00000000000000     S    (4g)
   0.50500000000000    0.85500000000000    0.50000000000000     S    (4g)
   0.49500000000000    0.14500000000000    0.50000000000000     S    (4g)
```

$RuB_2$: A2B_oP6_59_f_a - CIF

```
# CIF file
data_findsym-output
_audit_creation_method FINDSYM

_chemical_name_mineral 'B2Ru'
_chemical_formula_sum 'B2 Ru'

loop_
_publ_author_name
 'B. Aronsson'
_journal_name_full_name
;
 Acta Chemica Scandinavica
;
_journal_volume 17
_journal_year 1963
_journal_page_first 2036
_journal_page_last 2050
_publ_Section_title
;
 The Crystal Structure of RuB$_{2}$, OsB$_{2}$, and IrB$_{1.35}$ and
     ↪ Some General Comments on the Crystal Chemistry of Borides in
     ↪ the Composition Range MeB - MeB$_{3}$
;

_aflow_title 'RuB$_{2}$ Structure'
_aflow_proto 'A2B_oP6_59_f_a'
_aflow_params 'a,b/a,c/a,z_{1},x_{2},z_{2}'
_aflow_params_values '4.645,0.616792249731,0.870828848224,0.1508,0.059,
     ↪ 0.639'
_aflow_Strukturbericht 'None'
_aflow_Pearson 'oP6'

_symmetry_space_group_name_H-M "P 21/m 21/m 2/n (origin choice 2)"
_symmetry_Int_Tables_number 59

_cell_length_a      4.64500
_cell_length_b      2.86500
_cell_length_c      4.04500
_cell_angle_alpha 90.00000
_cell_angle_beta  90.00000
_cell_angle_gamma 90.00000

loop_
_space_group_symop_id
_space_group_symop_operation_xyz
1 x,y,z
2 x+1/2,-y,-z
3 -x,y+1/2,-z
4 -x+1/2,-y+1/2,z
5 -x,-y,-z
6 -x+1/2,y,z
7 x,-y+1/2,z
8 x+1/2,y+1/2,-z

loop_
_atom_site_label
_atom_site_type_symbol
_atom_site_symmetry_multiplicity
_atom_site_Wyckoff_label
_atom_site_fract_x
_atom_site_fract_y
_atom_site_fract_z
_atom_site_occupancy
Ru1 Ru 2 a 0.25000 0.25000 0.15080 1.00000
B1 B 4 f 0.05900 0.25000 0.63900 1.00000
```

$RuB_2$: A2B_oP6_59_f_a - POSCAR

```
A2B_oP6_59_f_a & a,b/a,c/a,z1,x2,z2 --params=4.645,0.616792249731,
     ↪ 0.870828848224,0.1508,0.059,0.639 & Pmmn D_{2h}^{13} #59 (af) &
     ↪ oP6 & None & B2Ru & B2Ru & B. Aronsson, Acta Chem. Scand. 17,
     ↪ 2036-2050 (1963)
   1.00000000000000
   4.64500000000000    0.00000000000000    0.00000000000000
   0.00000000000000    2.86500000000000    0.00000000000000
   0.00000000000000    0.00000000000000    4.04500000000000
     B      Ru
     4       2
Direct
   0.05900000000000    0.25000000000000    0.63900000000000     B    (4f)
   0.44100000000000    0.25000000000000    0.63900000000000     B    (4f)
  -0.05900000000000    0.75000000000000   -0.63900000000000     B    (4f)
   0.55900000000000    0.75000000000000   -0.63900000000000     B    (4f)
   0.25000000000000    0.25000000000000    0.15080000000000    Ru    (2a)
   0.75000000000000    0.75000000000000   -0.15080000000000    Ru    (2a)
```

$NH_4NO_3$ IV ($G0_{11}$): A4B2C3_oP18_59_ef_ab_af - CIF

```
# CIF file
data_findsym-output
_audit_creation_method FINDSYM

_chemical_name_mineral 'H4N2O3'
_chemical_formula_sum 'H4 N2 O3'

loop_
_publ_author_name
 'C. S. Choi'
 'J. E. Mapes'
 'E. Prince'
_journal_name_full_name
;
 Acta Crystallographica Section B: Structural Science
;
_journal_volume 28
_journal_year 1972
_journal_page_first 1357
_journal_page_last 1361
_publ_Section_title
;
 The structure of ammonium nitrate (IV)
;

_aflow_title 'NH$_{4}$NO$_{3}$ IV ($G0_{11}$) Structure'
_aflow_proto 'A4B2C3_oP18_59_ef_ab_af'
_aflow_params 'a,b/a,c/a,z_{1},z_{2},z_{3},y_{4},z_{4},x_{5},z_{5},x_{6
     ↪ },z_{6}'
_aflow_params_values '5.745,0.946562228024,0.860226283725,0.5067,0.7629,
     ↪ 0.0836,0.8989,-0.0324,0.6045,0.8102,0.43442,0.3832'
_aflow_Strukturbericht '$G0_{11}$'
_aflow_Pearson 'oP18'

_symmetry_space_group_name_H-M "P 21/m 21/m 2/n (origin choice 2)"
_symmetry_Int_Tables_number 59

_cell_length_a      5.74500
_cell_length_b      5.43800
_cell_length_c      4.94200
_cell_angle_alpha 90.00000
_cell_angle_beta  90.00000
_cell_angle_gamma 90.00000

loop_
_space_group_symop_id
_space_group_symop_operation_xyz
1 x,y,z
2 x+1/2,-y,-z
3 -x,y+1/2,-z
4 -x+1/2,-y+1/2,z
5 -x,-y,-z
6 -x+1/2,y,z
7 x,-y+1/2,z
8 x+1/2,y+1/2,-z

loop_
_atom_site_label
_atom_site_type_symbol
_atom_site_symmetry_multiplicity
_atom_site_Wyckoff_label
_atom_site_fract_x
_atom_site_fract_y
_atom_site_fract_z
_atom_site_occupancy
N1 N 2 a 0.25000 0.25000 0.50670 1.00000
O1 O 2 a 0.25000 0.25000 0.76290 1.00000
N2 N 2 b 0.25000 0.75000 0.08360 1.00000
H1 H 4 e 0.25000 0.89890 -0.03240 1.00000
H2 H 4 f 0.60450 0.25000 0.81020 1.00000
O2 O 4 f 0.43442 0.25000 0.38320 1.00000
```

$NH_4NO_3$ IV ($G0_{11}$): A4B2C3_oP18_59_ef_ab_af - POSCAR

```
A4B2C3_oP18_59_ef_ab_af & a,b/a,c/a,z1,z2,z3,y4,z4,x5,z5,x6,z6 --params=
     ↪ 5.745,0.946562228024,0.860226283725,0.5067,0.7629,0.0836,0.8989
     ↪ ,-0.0324,0.6045,0.8102,0.43442,0.3832 & Pmmn D_{2h}^{13} #59 (a
     ↪ ^2bef^2) & oP18 & $G0_{11}$ & H4N2O3 & H4N2O3 & C. S. Choi and
     ↪ J. E. Mapes and E. Prince, Acta Crystallogr. Sect. B Struct.
     ↪ Sci. 28, 1357-1361 (1972)
   1.00000000000000
   5.74500000000000    0.00000000000000    0.00000000000000
   0.00000000000000    5.43800000000000    0.00000000000000
   0.00000000000000    0.00000000000000    4.94200000000000
     H       N       O
     8       4       6
Direct
   0.25000000000000    0.89890000000000   -0.03240000000000     H    (4e)
   0.25000000000000   -0.39890000000000   -0.03240000000000     H    (4e)
   0.75000000000000    1.39890000000000    0.03240000000000     H    (4e)
   0.75000000000000   -0.89890000000000    0.03240000000000     H    (4e)
   0.60450000000000    0.25000000000000    0.81020000000000     H    (4f)
  -0.10450000000000    0.25000000000000    0.81020000000000     H    (4f)
  -0.60450000000000    0.75000000000000   -0.81020000000000     H    (4f)
   1.10450000000000    0.75000000000000   -0.81020000000000     H    (4f)
   0.25000000000000    0.25000000000000    0.50670000000000     N    (2a)
   0.75000000000000    0.75000000000000   -0.50670000000000     N    (2a)
   0.25000000000000    0.75000000000000    0.08360000000000     N    (2b)
   0.75000000000000    0.25000000000000   -0.08360000000000     N    (2b)
   0.25000000000000    0.25000000000000    0.76290000000000     O    (2a)
   0.75000000000000    0.75000000000000   -0.76290000000000     O    (2a)
   0.43442000000000    0.25000000000000    0.38320000000000     O    (4f)
   0.06558000000000    0.25000000000000    0.38320000000000     O    (4f)
  -0.43442000000000    0.75000000000000   -0.38320000000000     O    (4f)
   0.93442000000000    0.75000000000000   -0.38320000000000     O    (4f)
```

Shcherbinaite ($V_2O_5$) (*Revised*): A5B2_oP14_59_a2f_f - CIF

```
# CIF file
data_findsym-output
_audit_creation_method FINDSYM

_chemical_name_mineral 'Shcherbinaite'
_chemical_formula_sum 'O5 V2'

loop_
_publ_author_name
 'R. Enjalbert'
 'J. Galy'
_journal_name_full_name
;
 Acta Crystallographica Section C: Structural Chemistry
;
_journal_volume 42
_journal_year 1986
_journal_page_first 1467
_journal_page_last 1469
_publ_Section_title
;
 A Refinement of the Structure of V$_{2}$O$_{5}$
;

_aflow_title 'Shcherbinaite (V$_{2}$O$_{5}$) ({\em{Revised}}) Structure
    ↪ '
_aflow_proto 'A5B2_oP14_59_a2f_f'
_aflow_params 'a,b/a,c/a,z_{1},x_{2},z_{2},x_{3},z_{3},x_{4},z_{4}'
_aflow_params_values '11.512,0.309589993051,0.379430159833,0.001,0.1043,
    ↪ 0.531,-0.0689,0.003,0.10118,0.891'
_aflow_Strukturbericht 'None'
_aflow_Pearson 'oP14'

_symmetry_space_group_name_H-M "P 21/m 21/m 2/n (origin choice 2)"
_symmetry_Int_Tables_number 59

_cell_length_a     11.51200
_cell_length_b     3.56400
_cell_length_c     4.36800
_cell_angle_alpha  90.00000
_cell_angle_beta   90.00000
_cell_angle_gamma  90.00000

loop_
_space_group_symop_id
_space_group_symop_operation_xyz
1 x,y,z
2 x+1/2,-y,-z
3 -x,y+1/2,-z
4 -x+1/2,-y+1/2,z
5 -x,-y,-z
6 -x+1/2,y,z
7 x,-y+1/2,z
8 x+1/2,y+1/2,-z

loop_
_atom_site_label
_atom_site_type_symbol
_atom_site_symmetry_multiplicity
_atom_site_Wyckoff_label
_atom_site_fract_x
_atom_site_fract_y
_atom_site_fract_z
_atom_site_occupancy
O1 O 2 a 0.25000 0.25000 0.00100 1.00000
O2 O 4 f 0.10430 0.25000 0.53100 1.00000
O3 O 4 f -0.06890 0.25000 0.00300 1.00000
V1 V 4 f 0.10118 0.25000 0.89100 1.00000
```

Shcherbinaite ($V_2O_5$) (*Revised*): A5B2_oP14_59_a2f_f - POSCAR

```
A5B2_oP14_59_a2f_f & a,b/a,c/a,z1,x2,z2,x3,z3,x4,z4 --params=11.512,
    ↪ 0.309589993051,0.379430159833,0.001,0.1043,0.531,-0.0689,0.003,
    ↪ 0.10118,0.891 & Pmmn D_{2h}^{13} #59 (af^3) & oP14 & None &
    ↪ O5V2 & Shcherbinaite & R. Enjalbert and J. Galy, Acta
    ↪ Crystallogr. C 42, 1467-1469 (1986)
   1.00000000000000
  11.51200000000000   0.00000000000000   0.00000000000000
   0.00000000000000   3.56400000000000   0.00000000000000
   0.00000000000000   0.00000000000000   4.36800000000000
     O        V
    10        4
Direct
   0.25000000000000   0.25000000000000   0.00100000000000   O   (2a)
   0.75000000000000   0.75000000000000  -0.00100000000000   O   (2a)
   0.10430000000000   0.25000000000000   0.53100000000000   O   (4f)
   0.39570000000000   0.25000000000000   0.53100000000000   O   (4f)
  -0.10430000000000   0.75000000000000  -0.53100000000000   O   (4f)
   0.60430000000000   0.75000000000000  -0.53100000000000   O   (4f)
  -0.06890000000000   0.25000000000000   0.00300000000000   O   (4f)
   0.56890000000000   0.25000000000000   0.00300000000000   O   (4f)
   0.06890000000000   0.75000000000000  -0.00300000000000   O   (4f)
   0.43110000000000   0.75000000000000  -0.00300000000000   O   (4f)
   0.10118000000000   0.25000000000000   0.89100000000000   V   (4f)
   0.39882000000000   0.25000000000000   0.89100000000000   V   (4f)
  -0.10118000000000   0.75000000000000  -0.89100000000000   V   (4f)
   0.60118000000000   0.75000000000000  -0.89100000000000   V   (4f)
```

$CaB_2O_4$ I ($E3_2$): A2BC4_oP28_60_d_c_2d - CIF

```
# CIF file
data_findsym-output
_audit_creation_method FINDSYM

_chemical_name_mineral 'B2CaO4'
_chemical_formula_sum 'B2 Ca O4'

loop_
_publ_author_name
 'M. Marezio'
 'H. A. Plettinger'
 'W. H. Zachariasen'
_journal_name_full_name
;
 Acta Cristallographica
;
_journal_volume 16
_journal_year 1963
_journal_page_first 390
_journal_page_last 392
_publ_Section_title
;
 Refinement of the calcium metaborate structure
;

# Found in The crystal structure of the high-pressure phase CaB$_{2}$O$_
    ↪ {4}$(IV), and polymorphism in CaB$_{2}$O$_{4}$, 1969

_aflow_title 'CaB$_{2}$O$_{4}$ I ($E3_{2}$) Structure'
_aflow_proto 'A2BC4_oP28_60_d_c_2d'
_aflow_params 'a,b/a,c/a,y_{1},x_{2},y_{2},z_{2},x_{3},y_{3},z_{3},x_{4
    ↪ },y_{4},z_{4}'
_aflow_params_values '11.604,0.369269217511,0.535504998276,0.2726,0.1924
    ↪ ,0.8296,0.1258,0.0862,0.7268,0.0917,0.2078,0.1485,0.1478'
_aflow_Strukturbericht '$E3_{2}$'
_aflow_Pearson 'oP28'

_symmetry_space_group_name_H-M "P 21/b 2/c 21/n"
_symmetry_Int_Tables_number 60

_cell_length_a     11.60400
_cell_length_b     4.28500
_cell_length_c     6.21400
_cell_angle_alpha  90.00000
_cell_angle_beta   90.00000
_cell_angle_gamma  90.00000

loop_
_space_group_symop_id
_space_group_symop_operation_xyz
1 x,y,z
2 x+1/2,-y+1/2,-z
3 -x,y,-z+1/2
4 -x+1/2,-y+1/2,z+1/2
5 -x,-y,-z
6 -x+1/2,y+1/2,z
7 x,-y,z+1/2
8 x+1/2,y+1/2,-z+1/2

loop_
_atom_site_label
_atom_site_type_symbol
_atom_site_symmetry_multiplicity
_atom_site_Wyckoff_label
_atom_site_fract_x
_atom_site_fract_y
_atom_site_fract_z
_atom_site_occupancy
Ca1 Ca 4 c 0.00000 0.27260 0.25000 1.00000
B1 B 8 d 0.19240 0.82960 0.12580 1.00000
O1 O 8 d 0.08620 0.72680 0.09170 1.00000
O2 O 8 d 0.20780 0.14850 0.14780 1.00000
```

$CaB_2O_4$ I ($E3_2$): A2BC4_oP28_60_d_c_2d - POSCAR

```
A2BC4_oP28_60_d_c_2d & a,b/a,c/a,y1,x2,y2,z2,x3,y3,z3,x4,y4,z4 --params=
    ↪ 11.604,0.369269217511,0.535504998276,0.2726,0.1924,0.8296,
    ↪ 0.1258,0.0862,0.7268,0.0917,0.2078,0.1485,0.1478 & Pbcn D_{2h
    ↪ }^{14} #60 (cd^3) & oP28 & $E3_{2}$ & B2CaO4 & B2CaO4 & M.
    ↪ Marezio and H. A. Plettinger and W. H. Zachariasen, Acta Cryst.
    ↪ 16, 390-392 (1963)
   1.00000000000000
  11.60400000000000   0.00000000000000   0.00000000000000
   0.00000000000000   4.28500000000000   0.00000000000000
   0.00000000000000   0.00000000000000   6.21400000000000
     B       Ca        O
     8        4       16
Direct
   0.19240000000000   0.82960000000000   0.12580000000000   B   (8d)
   0.30760000000000  -0.32960000000000   0.62580000000000   B   (8d)
  -0.19240000000000   0.82960000000000   0.37420000000000   B   (8d)
   0.69240000000000  -0.32960000000000  -0.12580000000000   B   (8d)
  -0.19240000000000  -0.82960000000000  -0.12580000000000   B   (8d)
   0.69240000000000   1.32960000000000   0.37420000000000   B   (8d)
   0.19240000000000  -0.82960000000000   0.62580000000000   B   (8d)
   0.30760000000000   1.32960000000000   0.12580000000000   B   (8d)
   0.00000000000000   0.27260000000000   0.25000000000000   Ca  (4c)
   0.50000000000000   0.22740000000000   0.75000000000000   Ca  (4c)
   0.00000000000000  -0.27260000000000   0.75000000000000   Ca  (4c)
   0.50000000000000   0.77260000000000   0.25000000000000   Ca  (4c)
   0.08620000000000   0.72680000000000   0.09170000000000   O   (8d)
   0.41380000000000  -0.22680000000000   0.59170000000000   O   (8d)
  -0.08620000000000   0.72680000000000   0.40830000000000   O   (8d)
   0.58620000000000  -0.22680000000000  -0.09170000000000   O   (8d)
  -0.08620000000000  -0.72680000000000  -0.09170000000000   O   (8d)
   0.58620000000000   1.22680000000000   0.40830000000000   O   (8d)
   0.08620000000000  -0.72680000000000   0.59170000000000   O   (8d)
   0.41380000000000   1.22680000000000   0.09170000000000   O   (8d)
   0.20780000000000   0.14850000000000   0.14780000000000   O   (8d)
```

```
    0.29220000000000    0.35150000000000    0.64780000000000    O    (8d)
   -0.20780000000000    0.14850000000000    0.35220000000000    O    (8d)
    0.70780000000000    0.35150000000000   -0.14780000000000    O    (8d)
   -0.20780000000000   -0.14850000000000   -0.14780000000000    O    (8d)
    0.70780000000000    0.64850000000000    0.35220000000000    O    (8d)
    0.20780000000000   -0.14850000000000    0.64780000000000    O    (8d)
    0.29220000000000    0.64850000000000    0.14780000000000    O    (8d)
```

$\zeta$-$Fe_2N$: A2B_oP12_60_d_c - CIF

```
# CIF file
data_findsym-output
_audit_creation_method FINDSYM

_chemical_name_mineral 'Fe2N'
_chemical_formula_sum 'Fe2 N'

loop_
_publ_author_name
 'D. Rechenbach'
 'H. Jacobs'
_journal_name_full_name
;
 Journal of Alloys and Compounds
;
_journal_volume 235
_journal_year 1996
_journal_page_first 15
_journal_page_last 22
_publ_Section_title
;
 Structure determination of $\zeta$-Fe$_{2}$N by neutron and synchrotron
     ↪ powder diffraction
;

_aflow_title '$\zeta$-Fe$_{2}$N Structure'
_aflow_proto 'A2B_oP12_60_d_c'
_aflow_params 'a,b/a,c/a,y_{1},x_{2},y_{2},z_{2}'
_aflow_params_values '4.4373,1.24879994591,1.09140693665,0.364,0.249,
     ↪ 0.128,0.0827'
_aflow_Strukturbericht 'None'
_aflow_Pearson 'oP12'

_symmetry_space_group_name_H-M "P 21/b 2/c 21/n"
_symmetry_Int_Tables_number 60

_cell_length_a      4.43730
_cell_length_b      5.54130
_cell_length_c      4.84290
_cell_angle_alpha  90.00000
_cell_angle_beta   90.00000
_cell_angle_gamma  90.00000

loop_
_space_group_symop_id
_space_group_symop_operation_xyz
1 x,y,z
2 x+1/2,-y+1/2,-z
3 -x,y,-z+1/2
4 -x+1/2,-y+1/2,z+1/2
5 -x,-y,-z
6 -x+1/2,y+1/2,z
7 x,-y,z+1/2
8 x+1/2,y+1/2,-z+1/2

loop_
_atom_site_label
_atom_site_type_symbol
_atom_site_symmetry_multiplicity
_atom_site_Wyckoff_label
_atom_site_fract_x
_atom_site_fract_y
_atom_site_fract_z
_atom_site_occupancy
N1 N 4 c 0.00000 0.36400 0.25000 1.00000
Fe1 Fe 8 d 0.24900 0.12800 0.08270 1.00000
```

$\zeta$-$Fe_2N$: A2B_oP12_60_d_c - POSCAR

```
A2B_oP12_60_d_c & a,b/a,c/a,y1,x2,y2,z2 --params=4.4373,1.24879994591,
     ↪ 1.09140693665,0.364,0.249,0.128,0.0827 & Pbcn D_{2h}^{14} #60 (
     ↪ cd) & oP12 & None & Fe2N & Fe2N & D. Rechenbach and H. Jacobs,
     ↪ J. Alloys Compd. 235, 15-22 (1996)
    1.00000000000000
    4.43730000000000    0.00000000000000    0.00000000000000
    0.00000000000000    5.54130000000000    0.00000000000000
    0.00000000000000    0.00000000000000    4.84290000000000
     Fe         N
      8         4
Direct
    0.24900000000000    0.12800000000000    0.08270000000000    Fe    (8d)
    0.25100000000000    0.37200000000000    0.58270000000000    Fe    (8d)
   -0.24900000000000    0.12800000000000    0.41730000000000    Fe    (8d)
    0.74900000000000    0.37200000000000   -0.08270000000000    Fe    (8d)
   -0.24900000000000   -0.12800000000000   -0.08270000000000    Fe    (8d)
    0.74900000000000    0.62800000000000    0.41730000000000    Fe    (8d)
    0.24900000000000   -0.12800000000000    0.58270000000000    Fe    (8d)
    0.25100000000000    0.62800000000000    0.08270000000000    Fe    (8d)
    0.00000000000000    0.36400000000000    0.25000000000000    N     (4c)
    0.50000000000000    0.13600000000000    0.75000000000000    N     (4c)
    0.00000000000000   -0.36400000000000    0.75000000000000    N     (4c)
    0.50000000000000    0.86400000000000    0.25000000000000    N     (4c)
```

$\alpha$-$PbO_2$: A2B_oP12_60_d_c - CIF

```
# CIF file
data_findsym-output
_audit_creation_method FINDSYM

_chemical_name_mineral 'O2Pb'
_chemical_formula_sum 'O2 Pb'

loop_
_publ_author_name
 'R. J. Hill'
_journal_name_full_name
;
 Materials Research Bulletin
;
_journal_volume 17
_journal_year 1982
_journal_page_first 769
_journal_page_last 784
_publ_Section_title
;
 The Crystal Structures of Lead Dioxides from the Positive Plate of the
     ↪ Lead/Acid Battery
;

# Found in Pearson's Handbook of Crystallographic Data for Intermetallic
     ↪ Phases, 1991

_aflow_title '$\alpha$-PbO$_{2}$ Structure'
_aflow_proto 'A2B_oP12_60_d_c'
_aflow_params 'a,b/a,c/a,y_{1},x_{2},y_{2},z_{2}'
_aflow_params_values '4.9898,1.19191149946,1.09535452323,0.1779,0.2685,
     ↪ 0.401,0.0248'
_aflow_Strukturbericht 'None'
_aflow_Pearson 'oP12'

_symmetry_space_group_name_H-M "P 21/b 2/c 21/n"
_symmetry_Int_Tables_number 60

_cell_length_a      4.98980
_cell_length_b      5.94740
_cell_length_c      5.46560
_cell_angle_alpha  90.00000
_cell_angle_beta   90.00000
_cell_angle_gamma  90.00000

loop_
_space_group_symop_id
_space_group_symop_operation_xyz
1 x,y,z
2 x+1/2,-y+1/2,-z
3 -x,y,-z+1/2
4 -x+1/2,-y+1/2,z+1/2
5 -x,-y,-z
6 -x+1/2,y+1/2,z
7 x,-y,z+1/2
8 x+1/2,y+1/2,-z+1/2

loop_
_atom_site_label
_atom_site_type_symbol
_atom_site_symmetry_multiplicity
_atom_site_Wyckoff_label
_atom_site_fract_x
_atom_site_fract_y
_atom_site_fract_z
_atom_site_occupancy
Pb1 Pb 4 c 0.00000 0.17790 0.25000 0.49000
O1 O 8 d 0.26850 0.40100 0.02480 1.00000
```

$\alpha$-$PbO_2$: A2B_oP12_60_d_c - POSCAR

```
A2B_oP12_60_d_c & a,b/a,c/a,y1,x2,y2,z2 --params=4.9898,1.19191149946,
     ↪ 1.09535452323,0.1779,0.2685,0.401,0.0248 & Pbcn D_{2h}^{14} #60
     ↪ (cd) & oP12 & None & O2Pb & O2Pb & R. J. Hill, Mater. Res.
     ↪ Bull. 17, 769-784 (1982)
    1.00000000000000
    4.98980000000000    0.00000000000000    0.00000000000000
    0.00000000000000    5.94740000000000    0.00000000000000
    0.00000000000000    0.00000000000000    5.46560000000000
     O         Pb
     8          4
Direct
    0.26850000000000    0.40100000000000    0.02480000000000    O     (8d)
    0.23150000000000    0.09900000000000    0.52480000000000    O     (8d)
   -0.26850000000000    0.40100000000000    0.47520000000000    O     (8d)
    0.76850000000000    0.09900000000000   -0.02480000000000    O     (8d)
   -0.26850000000000   -0.40100000000000   -0.02480000000000    O     (8d)
    0.76850000000000    0.90100000000000    0.47520000000000    O     (8d)
    0.26850000000000   -0.40100000000000    0.52480000000000    O     (8d)
    0.23150000000000    0.90100000000000    0.02480000000000    O     (8d)
    0.00000000000000    0.17790000000000    0.25000000000000    Pb    (4c)
    0.50000000000000    0.32210000000000    0.75000000000000    Pb    (4c)
    0.00000000000000   -0.17790000000000    0.75000000000000    Pb    (4c)
    0.50000000000000    0.67790000000000    0.25000000000000    Pb    (4c)
```

$Cr_5O_{12}$: A5B12_oP68_60_c2d_6d - CIF

```
# CIF file
data_findsym-output
_audit_creation_method FINDSYM

_chemical_name_mineral 'Cr5O12'
_chemical_formula_sum 'Cr5 O12'
```

```
loop_
_publ_author_name
 'K.-A. Wilhelmi'
_journal_name_full_name
;
 Acta Chemica Scandinavica
;
_journal_volume 19
_journal_year 1965
_journal_page_first 165
_journal_page_last 176
_publ_Section_title
;
 The Crystal Structure of Cr$_{5}$O$_{12}$
;

_aflow_title 'Cr$_{5}$O$_{12}$ Structure'
_aflow_proto 'A5B12_oP68_60_c2d_6d'
_aflow_params 'a,b/a,c/a,y_{1},x_{2},y_{2},z_{2},x_{3},y_{3},z_{3},x_{4
    ↪ },y_{4},z_{4},x_{5},y_{5},z_{5},x_{6},y_{6},z_{6},x_{7},y_{7},
    ↪ z_{7},x_{8},y_{8},z_{8},x_{9},y_{9},z_{9}'
_aflow_params_values '12.044,0.681833277981,0.678927266689,0.3939,0.1719
    ↪ ,0.8878,0.4768,0.0857,0.7517,0.1119,0.2454,0.7314,0.0906,0.075,
    ↪ 0.7736,0.3535,0.1003,-0.0101,0.1032,0.0806,0.508,0.1356,0.0707,
    ↪ 0.27,0.3741,0.2538,0.4894,0.3625'
_aflow_Strukturbericht 'None'
_aflow_Pearson 'oP68'

_symmetry_space_group_name_H-M "P 21/b 2/c 21/n"
_symmetry_Int_Tables_number 60

_cell_length_a    12.04400
_cell_length_b    8.21200
_cell_length_c    8.17700
_cell_angle_alpha 90.00000
_cell_angle_beta  90.00000
_cell_angle_gamma 90.00000

loop_
_space_group_symop_id
_space_group_symop_operation_xyz
1 x,y,z
2 x+1/2,-y+1/2,-z
3 -x,y,-z+1/2
4 -x+1/2,-y+1/2,z+1/2
5 -x,-y,-z
6 -x+1/2,y+1/2,z
7 x,-y,z+1/2
8 x+1/2,y+1/2,-z+1/2

loop_
_atom_site_label
_atom_site_type_symbol
_atom_site_symmetry_multiplicity
_atom_site_Wyckoff_label
_atom_site_fract_x
_atom_site_fract_y
_atom_site_fract_z
_atom_site_occupancy
Cr1 Cr 4 c 0.00000 0.39390 0.25000 1.00000
Cr2 Cr 8 d 0.17190 0.88780 0.47680 1.00000
Cr3 Cr 8 d 0.08570 0.75170 0.11190 1.00000
O1 O 8 d 0.24540 0.73140 0.09060 1.00000
O2 O 8 d 0.07500 0.77360 0.35350 1.00000
O3 O 8 d 0.10030 -0.01010 0.10320 1.00000
O4 O 8 d 0.08060 0.50800 0.13560 1.00000
O5 O 8 d 0.07070 0.27000 0.37410 1.00000
O6 O 8 d 0.25380 0.48940 0.36250 1.00000
```

$Cr_5O_{12}$: A5B12_oP68_60_c2d_6d - POSCAR

```
A5B12_oP68_60_c2d_6d & a,b/a,c/a,y1,x2,y2,z2,x3,y3,z3,x4,y4,z4,x5,y5,z5,
    ↪ x6,y6,z6,x7,y7,z7,x8,y8,z8,x9,y9,z9 --params=12.044,
    ↪ 0.681833277981,0.678927266689,0.3939,0.1719,0.8878,0.4768,
    ↪ 0.0857,0.7517,0.1119,0.2454,0.7314,0.0906,0.075,0.7736,0.3535,
    ↪ 0.1003,-0.0101,0.1032,0.0806,0.508,0.1356,0.0707,0.27,0.3741,
    ↪ 0.2538,0.4894,0.3625 & Pbcn D_{2h}^{14} #60 (cd^8) & oP68 &
    ↪ None & Cr5O12 & Cr5O12 & K.-A. Wilhelmi, Acta Chem. Scand. 19,
    ↪ 165-176 (1965)
   1.00000000000000
  12.04400000000000   0.00000000000000   0.00000000000000
   0.00000000000000   8.21200000000000   0.00000000000000
   0.00000000000000   0.00000000000000   8.17700000000000
    Cr     O
    20    48
Direct
   0.00000000000000   0.39390000000000   0.25000000000000   Cr   (4c)
   0.50000000000000   0.10610000000000   0.75000000000000   Cr   (4c)
   0.00000000000000  -0.39390000000000   0.75000000000000   Cr   (4c)
   0.50000000000000   0.89390000000000   0.25000000000000   Cr   (4c)
   0.17190000000000   0.88780000000000   0.47680000000000   Cr   (8d)
   0.32810000000000  -0.38780000000000   0.97680000000000   Cr   (8d)
  -0.17190000000000   0.88780000000000   0.02320000000000   Cr   (8d)
   0.67190000000000  -0.38780000000000  -0.47680000000000   Cr   (8d)
  -0.17190000000000  -0.88780000000000  -0.47680000000000   Cr   (8d)
   0.67190000000000   1.38780000000000   0.02320000000000   Cr   (8d)
   0.17190000000000  -0.88780000000000   0.97680000000000   Cr   (8d)
   0.32810000000000   1.38780000000000   0.47680000000000   Cr   (8d)
   0.08570000000000   0.75170000000000   0.11190000000000   Cr   (8d)
   0.41430000000000  -0.25170000000000   0.61190000000000   Cr   (8d)
  -0.08570000000000   0.75170000000000   0.38810000000000   Cr   (8d)
   0.58570000000000  -0.25170000000000  -0.11190000000000   Cr   (8d)
  -0.08570000000000  -0.75170000000000  -0.11190000000000   Cr   (8d)
   0.58570000000000   1.25170000000000   0.38810000000000   Cr   (8d)
   0.08570000000000  -0.75170000000000   0.61190000000000   Cr   (8d)
   0.41430000000000   1.25170000000000   0.11190000000000   Cr   (8d)
   0.24540000000000   0.73140000000000   0.09060000000000   O    (8d)
   0.25460000000000  -0.23140000000000   0.59060000000000   O    (8d)
  -0.24540000000000   0.73140000000000   0.40940000000000   O    (8d)
   0.74540000000000  -0.23140000000000  -0.09060000000000   O    (8d)
  -0.24540000000000  -0.73140000000000  -0.09060000000000   O    (8d)
   0.74540000000000   1.23140000000000   0.40940000000000   O    (8d)
   0.24540000000000  -0.73140000000000   0.59060000000000   O    (8d)
   0.25460000000000   1.23140000000000   0.09060000000000   O    (8d)
   0.07500000000000   0.77360000000000   0.35350000000000   O    (8d)
   0.42500000000000  -0.27360000000000   0.85350000000000   O    (8d)
  -0.07500000000000   0.77360000000000   0.14650000000000   O    (8d)
   0.57500000000000  -0.27360000000000  -0.35350000000000   O    (8d)
  -0.07500000000000  -0.77360000000000  -0.35350000000000   O    (8d)
   0.57500000000000   1.27360000000000   0.14650000000000   O    (8d)
   0.07500000000000  -0.77360000000000   0.85350000000000   O    (8d)
   0.42500000000000   1.27360000000000   0.35350000000000   O    (8d)
   0.10030000000000  -0.01010000000000   0.10320000000000   O    (8d)
   0.39970000000000   0.51010000000000   0.60320000000000   O    (8d)
  -0.10030000000000  -0.01010000000000   0.39680000000000   O    (8d)
   0.60030000000000   0.51010000000000  -0.10320000000000   O    (8d)
  -0.10030000000000   0.01010000000000  -0.10320000000000   O    (8d)
   0.60030000000000   0.48990000000000   0.39680000000000   O    (8d)
   0.10030000000000   0.01010000000000   0.60320000000000   O    (8d)
   0.39970000000000   0.48990000000000   0.10320000000000   O    (8d)
   0.08060000000000   0.50800000000000   0.13560000000000   O    (8d)
   0.41940000000000  -0.00800000000000   0.63560000000000   O    (8d)
  -0.08060000000000   0.50800000000000   0.36440000000000   O    (8d)
   0.58060000000000  -0.00800000000000  -0.13560000000000   O    (8d)
  -0.08060000000000  -0.50800000000000  -0.13560000000000   O    (8d)
   0.58060000000000   1.00800000000000   0.36440000000000   O    (8d)
   0.08060000000000  -0.50800000000000   0.63560000000000   O    (8d)
   0.41940000000000   1.00800000000000   0.13560000000000   O    (8d)
   0.07070000000000   0.27000000000000   0.37410000000000   O    (8d)
   0.42930000000000   0.23000000000000   0.87410000000000   O    (8d)
  -0.07070000000000   0.27000000000000   0.12590000000000   O    (8d)
   0.57070000000000   0.23000000000000  -0.37410000000000   O    (8d)
  -0.07070000000000  -0.27000000000000  -0.37410000000000   O    (8d)
   0.57070000000000   0.77000000000000   0.12590000000000   O    (8d)
   0.07070000000000  -0.27000000000000   0.87410000000000   O    (8d)
   0.42930000000000   0.77000000000000   0.37410000000000   O    (8d)
   0.25380000000000   0.48940000000000   0.36250000000000   O    (8d)
   0.24620000000000   0.01060000000000   0.86250000000000   O    (8d)
  -0.25380000000000   0.48940000000000   0.13750000000000   O    (8d)
   0.75380000000000   0.01060000000000  -0.36250000000000   O    (8d)
  -0.25380000000000  -0.48940000000000  -0.36250000000000   O    (8d)
   0.75380000000000   0.98940000000000   0.13750000000000   O    (8d)
   0.25380000000000  -0.48940000000000   0.86250000000000   O    (8d)
   0.24620000000000   0.98940000000000   0.36250000000000   O    (8d)
```

Columbite ($FeNb_2O_4$, $E5_1$): AB2C6_oP36_60_c_d_3d - CIF

```
# CIF file
data_findsym-output
_audit_creation_method FINDSYM

_chemical_name_mineral 'Columbite'
_chemical_formula_sum 'Fe Nb2 O6'

loop_
_publ_author_name
 'P. Bordet'
 'A. {McHale}'
 'A. Santoro'
 'R. S. Roth'
_journal_name_full_name
;
 Journal of Solid State Chemistry
;
_journal_volume 64
_journal_year 1986
_journal_page_first 30
_journal_page_last 46
_publ_Section_title
;
 Powder neutron diffraction study of ZrTiO$_{4}$, Zr$_{5}$Ti$_{7}$O$_{24
    ↪ }$, and FeNb$_{2}$O$_{6}$
;

# Found in The American Mineralogist Crystal Structure Database, 2003

_aflow_title 'Columbite (FeNb$_{2}$O$_{4}$, $E5_{1}$) Structure'
_aflow_proto 'AB2C6_oP36_60_c_d_3d'
_aflow_params 'a,b/a,c/a,y_{1},x_{2},y_{2},z_{2},x_{3},y_{3},z_{3},x_{4
    ↪ },y_{4},z_{4},x_{5},y_{5},z_{5}'
_aflow_params_values '14.2661,0.401889794688,0.353950974688,0.3311,
    ↪ 0.3389,0.3191,0.2506,0.0963,0.1041,0.0727,0.4189,0.1163,0.099,
    ↪ 0.756,0.1236,0.0793'
_aflow_Strukturbericht '$E5_{1}$'
_aflow_Pearson 'oP36'

_symmetry_space_group_name_H-M "P 21/b 2/c 21/n"
_symmetry_Int_Tables_number 60

_cell_length_a    14.26610
_cell_length_b    5.73340
_cell_length_c    5.04950
_cell_angle_alpha 90.00000
_cell_angle_beta  90.00000
_cell_angle_gamma 90.00000

loop_
_space_group_symop_id
_space_group_symop_operation_xyz
1 x,y,z
```

```
2 x+1/2,-y+1/2,-z
3 -x,y,-z+1/2
4 -x+1/2,-y+1/2,z+1/2
5 -x,-y,-z
6 -x+1/2,y+1/2,z
7 x,-y,z+1/2
8 x+1/2,y+1/2,-z+1/2

loop_
_atom_site_label
_atom_site_type_symbol
_atom_site_symmetry_multiplicity
_atom_site_Wyckoff_label
_atom_site_fract_x
_atom_site_fract_y
_atom_site_fract_z
_atom_site_occupancy
Fe1 Fe 4 c 0.00000 0.33110 0.25000 1.00000
Nb1 Nb 8 d 0.33890 0.31910 0.25060 1.00000
O1 O 8 d 0.09630 0.10410 0.07270 1.00000
O2 O 8 d 0.41890 0.11630 0.09900 1.00000
O3 O 8 d 0.75600 0.12360 0.07930 1.00000
```

Columbite ($FeNb_2O_4$, $E5_1$): AB2C6_oP36_60_c_d_3d - POSCAR

```
AB2C6_oP36_60_c_d_3d & a,b/a,c/a,y1,x2,y2,z2,x3,y3,z3,x4,y4,z4,x5,y5,z5
    ↪ --params=14.2661,0.401889794688,0.353950974688,0.3311,0.3389,
    ↪ 0.3191,0.2506,0.0963,0.1041,0.0727,0.4189,0.1163,0.099,0.756,
    ↪ 0.1236,0.0793 & Pbcn D_{2h}^{14} #60 (cd^4) & oP36 & $E5_{1}$ &
    ↪ FeNb2O6 & Columbite & P. Bordet et al., J. Solid State Chem.
    ↪ 64, 30-46 (1986)
   1.00000000000000
  14.26610000000000   0.00000000000000   0.00000000000000
   0.00000000000000   5.73340000000000   0.00000000000000
   0.00000000000000   0.00000000000000   5.04950000000000
    Fe     Nb      O
     4      8     24
Direct
   0.00000000000000   0.33110000000000   0.25000000000000   Fe   (4c)
   0.50000000000000   0.16890000000000   0.75000000000000   Fe   (4c)
   0.00000000000000  -0.33110000000000   0.75000000000000   Fe   (4c)
   0.50000000000000   0.83110000000000   0.25000000000000   Fe   (4c)
   0.33890000000000   0.31910000000000   0.25060000000000   Nb   (8d)
   0.16110000000000   0.18090000000000   0.75060000000000   Nb   (8d)
  -0.33890000000000   0.31910000000000   0.24940000000000   Nb   (8d)
   0.83890000000000   0.18090000000000  -0.25060000000000   Nb   (8d)
  -0.33890000000000  -0.31910000000000  -0.25060000000000   Nb   (8d)
   0.83890000000000   0.81910000000000   0.24940000000000   Nb   (8d)
   0.33890000000000  -0.31910000000000   0.75060000000000   Nb   (8d)
   0.16110000000000   0.81910000000000   0.25060000000000   Nb   (8d)
   0.09630000000000   0.10410000000000   0.07270000000000    O   (8d)
   0.40370000000000   0.39590000000000   0.57270000000000    O   (8d)
  -0.09630000000000   0.10410000000000   0.42730000000000    O   (8d)
   0.59630000000000   0.39590000000000  -0.07270000000000    O   (8d)
  -0.09630000000000  -0.10410000000000  -0.07270000000000    O   (8d)
   0.59630000000000   0.60410000000000   0.42730000000000    O   (8d)
   0.09630000000000  -0.10410000000000   0.57270000000000    O   (8d)
   0.40370000000000   0.60410000000000   0.07270000000000    O   (8d)
   0.41890000000000   0.11630000000000   0.09900000000000    O   (8d)
   0.08110000000000   0.38370000000000   0.59900000000000    O   (8d)
  -0.41890000000000   0.11630000000000   0.40100000000000    O   (8d)
   0.91890000000000   0.38370000000000  -0.09900000000000    O   (8d)
  -0.41890000000000  -0.11630000000000  -0.09900000000000    O   (8d)
   0.91890000000000   0.61630000000000   0.40100000000000    O   (8d)
   0.41890000000000  -0.11630000000000   0.59900000000000    O   (8d)
   0.08110000000000   0.61630000000000   0.09900000000000    O   (8d)
   0.75600000000000   0.12360000000000   0.07930000000000    O   (8d)
  -0.25600000000000   0.37640000000000   0.57930000000000    O   (8d)
  -0.75600000000000   0.12360000000000   0.42070000000000    O   (8d)
   1.25600000000000   0.37640000000000  -0.07930000000000    O   (8d)
  -0.75600000000000  -0.12360000000000  -0.07930000000000    O   (8d)
   1.25600000000000   0.62360000000000   0.42070000000000    O   (8d)
   0.75600000000000  -0.12360000000000   0.57930000000000    O   (8d)
  -0.25600000000000   0.62360000000000   0.07930000000000    O   (8d)
```

$Ca_2RuO_4$: A2B4C_oP28_61_c_2c_a - CIF

```
# CIF file
data_findsym-output
_audit_creation_method FINDSYM

_chemical_name_mineral 'Ca2O4Ru'
_chemical_formula_sum 'Ca2 O4 Ru'

loop_
_publ_author_name
 'O. Friedt'
 'M. Braden'
 'G. Andr\\'{e}'
 'P. Adelmann'
 'S. Nakatsuji'
 'Y. Maeno'
_journal_name_full_name
;
 Physical Review B
;
_journal_volume 63
_journal_year 2001
_journal_page_first 174432
_journal_page_last 174432
_publ_Section_title
;
 Structural and magnetic aspects of the metal-insulator transition in
    ↪ Ca$_{2-x}$Sr$_{x}$RuO$_{4}$
;

# Found in Unique Crystal Structure of Ca$_{2}$RuO$_{4}$ in the Current
    ↪ Stabilized Semimetallic State, 2019

_aflow_title 'Ca$_{2}$RuO$_{4}$ Structure'
_aflow_proto 'A2B4C_oP28_61_c_2c_a'
_aflow_params 'a,b/a,c/a,x_{2},y_{2},z_{2},x_{3},y_{3},z_{3},x_{4},y_{4
    ↪ },z_{4}'
_aflow_params_values '5.3945,1.03807581796,2.18098062842,0.0042,0.0559,
    ↪ 0.3524,0.1961,0.3018,0.0264,0.0673,0.4782,0.3355'
_aflow_Strukturbericht 'None'
_aflow_Pearson 'oP28'

_symmetry_space_group_name_H-M "P 21/b 21/c 21/a"
_symmetry_Int_Tables_number 61

_cell_length_a      5.39450
_cell_length_b      5.59990
_cell_length_c      11.76530
_cell_angle_alpha  90.00000
_cell_angle_beta   90.00000
_cell_angle_gamma  90.00000

loop_
_space_group_symop_id
_space_group_symop_operation_xyz
1 x,y,z
2 x+1/2,-y+1/2,-z
3 -x,y+1/2,-z+1/2
4 -x+1/2,-y,z+1/2
5 -x,-y,-z
6 -x+1/2,y+1/2,z
7 x,-y+1/2,z+1/2
8 x+1/2,y,-z+1/2

loop_
_atom_site_label
_atom_site_type_symbol
_atom_site_symmetry_multiplicity
_atom_site_Wyckoff_label
_atom_site_fract_x
_atom_site_fract_y
_atom_site_fract_z
_atom_site_occupancy
Ru1 Ru 4 a 0.00000 0.00000 0.00000 1.00000
Ca1 Ca 8 c 0.00420 0.05590 0.35240 1.00000
O1 O 8 c 0.19610 0.30180 0.02640 1.00000
O2 O 8 c 0.06730 0.47820 0.33550 1.00000
```

$Ca_2RuO_4$: A2B4C_oP28_61_c_2c_a - POSCAR

```
A2B4C_oP28_61_c_2c_a & a,b/a,c/a,x2,y2,z2,x3,y3,z3,x4,y4,z4 --params=
    ↪ 5.3945,1.03807581796,2.18098062842,0.0042,0.0559,0.3524,0.1961,
    ↪ 0.3018,0.0264,0.0673,0.4782,0.3355 & Pbca D_{2h}^{15} #61 (ac^3
    ↪ ) & oP28 & None & Ca2O4Ru & Ca2O4Ru & O. Friedt et al., Phys.
    ↪ Rev. B 63, 174432(2001)
   1.00000000000000
   5.39450000000000   0.00000000000000   0.00000000000000
   0.00000000000000   5.59990000000000   0.00000000000000
   0.00000000000000   0.00000000000000  11.76530000000000
    Ca      O     Ru
     8     16      4
Direct
   0.00420000000000   0.05590000000000   0.35240000000000   Ca   (8c)
   0.49580000000000  -0.05590000000000   0.85240000000000   Ca   (8c)
  -0.00420000000000   0.55590000000000   0.14760000000000   Ca   (8c)
   0.50420000000000   0.44410000000000  -0.35240000000000   Ca   (8c)
  -0.00420000000000  -0.05590000000000  -0.35240000000000   Ca   (8c)
   0.50420000000000   0.05590000000000   0.14760000000000   Ca   (8c)
   0.00420000000000   0.44410000000000   0.85240000000000   Ca   (8c)
   0.49580000000000   0.55590000000000   0.35240000000000   Ca   (8c)
   0.19610000000000   0.30180000000000   0.02640000000000    O   (8c)
   0.30390000000000  -0.30180000000000   0.52640000000000    O   (8c)
  -0.19610000000000   0.80180000000000   0.47360000000000    O   (8c)
   0.69610000000000   0.19820000000000  -0.02640000000000    O   (8c)
  -0.19610000000000  -0.30180000000000  -0.02640000000000    O   (8c)
   0.69610000000000   0.30180000000000   0.47360000000000    O   (8c)
   0.19610000000000   0.19820000000000   0.52640000000000    O   (8c)
   0.30390000000000   0.80180000000000   0.02640000000000    O   (8c)
   0.06730000000000   0.47820000000000   0.33550000000000    O   (8c)
   0.43270000000000  -0.47820000000000   0.83550000000000    O   (8c)
  -0.06730000000000   0.97820000000000   0.16450000000000    O   (8c)
   0.56730000000000   0.02180000000000  -0.33550000000000    O   (8c)
  -0.06730000000000  -0.47820000000000  -0.33550000000000    O   (8c)
   0.56730000000000   0.47820000000000   0.16450000000000    O   (8c)
   0.06730000000000   0.02180000000000   0.83550000000000    O   (8c)
   0.43270000000000   0.97820000000000   0.33550000000000    O   (8c)
   0.00000000000000   0.00000000000000   0.00000000000000   Ru   (4a)
   0.50000000000000   0.00000000000000   0.50000000000000   Ru   (4a)
   0.00000000000000   0.50000000000000   0.50000000000000   Ru   (4a)
   0.50000000000000   0.50000000000000   0.00000000000000   Ru   (4a)
```

Tellurite ($\beta$-$TeO_2$, $C52$): A2B_oP24_61_2c_c - CIF

```
# CIF file
data_findsym-output
_audit_creation_method FINDSYM

_chemical_name_mineral 'Tellurite'
_chemical_formula_sum 'O2 Te'

loop_
_publ_author_name
 'H. Beyer'
_journal_name_full_name
```

```
;
 Zeitschrift f{\"u}r Kristallographie - Crystalline Materials
;
_journal_volume 124
_journal_year 1967
_journal_page_first 228
_journal_page_last 237
_publ_Section_title
;
 Verfeinerung der Kristallstruktur von Tellurit, dem rhombischen TeO$_{2
    ↪ }$
;

# Found in {\em Ab initio} study of the vibrational properties of
    ↪ crystalline TeO$_{2}$: The $\alpha$, $\beta$, and $\gamma$
    ↪ phases, 2006

_aflow_title 'Tellurite ($\beta$-TeO$_{2}$, $C52$) Structure'
_aflow_proto 'A2B_oP24_61_2c_c'
_aflow_params 'a,b/a,c/a,x_{1},y_{1},z_{1},x_{2},y_{2},z_{2},x_{3},y_{3
    ↪ },z_{3}'
_aflow_params_values '15.035,0.363418689724,0.372929830396,0.028,0.634,
    ↪ 0.171,0.168,0.221,0.081,0.1181,0.0252,0.3378'
_aflow_Strukturbericht '$C52$'
_aflow_Pearson 'oP24'

_symmetry_space_group_name_H-M "P 21/b 21/c 21/a"
_symmetry_Int_Tables_number 61

_cell_length_a     15.03500
_cell_length_b     5.46400
_cell_length_c     5.60700
_cell_angle_alpha  90.00000
_cell_angle_beta   90.00000
_cell_angle_gamma  90.00000

loop_
_space_group_symop_id
_space_group_symop_operation_xyz
1 x,y,z
2 x+1/2,-y+1/2,-z
3 -x,y+1/2,-z+1/2
4 -x+1/2,-y,z+1/2
5 -x,-y,-z
6 -x+1/2,y+1/2,z
7 x,-y+1/2,z+1/2
8 x+1/2,y,-z+1/2

loop_
_atom_site_label
_atom_site_type_symbol
_atom_site_symmetry_multiplicity
_atom_site_Wyckoff_label
_atom_site_fract_x
_atom_site_fract_y
_atom_site_fract_z
_atom_site_occupancy
O1 O 8 c 0.02800 0.63400 0.17100 1.00000
O2 O 8 c 0.16800 0.22100 0.08100 1.00000
Te1 Te 8 c 0.11810 0.02520 0.33780 1.00000
```

Tellurite ($\beta$-$TeO_2$, $C52$): A2B_oP24_61_2c_c - POSCAR

```
A2B_oP24_61_2c_c & a,b/a,c/a,x1,y1,z1,x2,y2,z2,x3,y3,z3 --params=15.035,
    ↪ 0.363418689724,0.372929830396,0.028,0.634,0.171,0.168,0.221,
    ↪ 0.081,0.1181,0.0252,0.3378 & Pbca D_{2h}^{15} #61 (c^3) & oP24
    ↪ & $C52$ & O2Te & Tellurite & H. Beyer, Zeitschrift f{"u}r
    ↪ Kristallographie - Crystalline Materials 124, 228-237 (1967)
   1.00000000000000
  15.03500000000000   0.00000000000000   0.00000000000000
   0.00000000000000   5.46400000000000   0.00000000000000
   0.00000000000000   0.00000000000000   5.60700000000000
     O      Te
    16       8
Direct
   0.02800000000000   0.63400000000000   0.17100000000000   O    (8c)
   0.47200000000000  -0.63400000000000   0.67100000000000   O    (8c)
  -0.02800000000000   1.13400000000000   0.32900000000000   O    (8c)
   0.52800000000000  -0.13400000000000  -0.17100000000000   O    (8c)
  -0.02800000000000  -0.63400000000000  -0.17100000000000   O    (8c)
   0.52800000000000   0.63400000000000   0.32900000000000   O    (8c)
   0.02800000000000  -0.13400000000000   0.67100000000000   O    (8c)
   0.47200000000000   1.13400000000000   0.17100000000000   O    (8c)
   0.16800000000000   0.22100000000000   0.08100000000000   O    (8c)
   0.33200000000000  -0.22100000000000   0.58100000000000   O    (8c)
  -0.16800000000000   0.72100000000000   0.41900000000000   O    (8c)
   0.66800000000000   0.27900000000000  -0.08100000000000   O    (8c)
  -0.16800000000000  -0.22100000000000  -0.08100000000000   O    (8c)
   0.66800000000000   0.22100000000000   0.41900000000000   O    (8c)
   0.16800000000000   0.27900000000000   0.58100000000000   O    (8c)
   0.33200000000000   0.72100000000000   0.08100000000000   O    (8c)
   0.11810000000000   0.02520000000000   0.33780000000000   Te   (8c)
   0.38190000000000  -0.02520000000000   0.83780000000000   Te   (8c)
  -0.11810000000000   0.52520000000000   0.16220000000000   Te   (8c)
   0.61810000000000   0.47480000000000  -0.33780000000000   Te   (8c)
  -0.11810000000000  -0.02520000000000  -0.33780000000000   Te   (8c)
   0.61810000000000   0.02520000000000   0.16220000000000   Te   (8c)
   0.11810000000000   0.47480000000000   0.83780000000000   Te   (8c)
   0.38190000000000   0.52520000000000   0.33780000000000   Te   (8c)
```

($TiCl_4$·$POCl_3$)$_2$: A7BCD_oP80_61_7c_c_c_c - CIF

```
# CIF file
data_findsym-output
_audit_creation_method FINDSYM

_chemical_name_mineral 'Cl7OPTi'
_chemical_formula_sum 'Cl7 O P Ti'

loop_
_publ_author_name
 'C.-I. Br\"{a}nd\\'{e}n'
 'I. Lindqvist'
_journal_name_full_name
;
 Acta Chemica Scandinavica
;
_journal_volume 14
_journal_year 1960
_journal_page_first 726
_journal_page_last 732
_publ_Section_title
;
 The Crystal Structure of (TiCl$_{4}$$\cdot$POCl$_{3}$)$_{2}$
;

_aflow_title '(TiCl$_{4}$$\cdot$POCl$_{3}$)$_{2}$ Structure'
_aflow_proto 'A7BCD_oP80_61_7c_c_c_c'
_aflow_params 'a,b/a,c/a,x_{1},y_{1},z_{1},x_{2},y_{2},z_{2},x_{3},y_{3
    ↪ },z_{3},x_{4},y_{4},z_{4},x_{5},y_{5},z_{5},x_{6},y_{6},z_{6},
    ↪ x_{7},y_{7},z_{7},x_{8},y_{8},z_{8},x_{9},y_{9},z_{9},x_{10},y_
    ↪ {10},z_{10}'
_aflow_params_values '12.42,1.02737520129,1.08776167472,0.1062,0.5507,
    ↪ 0.5424,0.2364,0.3205,0.5231,0.0098,0.2107,0.4306,0.0092,0.3285,
    ↪ 0.6506,0.2852,0.5483,0.3103,0.2769,0.3103,0.2522,0.1207,0.4706,
    ↪ 0.1548,0.1075,0.4124,0.3564,0.1875,0.4353,0.2819,0.0629,0.3616,
    ↪ 0.4986'
_aflow_Strukturbericht 'None'
_aflow_Pearson 'oP80'

_symmetry_space_group_name_H-M "P 21/b 21/c 21/a"
_symmetry_Int_Tables_number 61

_cell_length_a     12.42000
_cell_length_b     12.76000
_cell_length_c     13.51000
_cell_angle_alpha  90.00000
_cell_angle_beta   90.00000
_cell_angle_gamma  90.00000

loop_
_space_group_symop_id
_space_group_symop_operation_xyz
1 x,y,z
2 x+1/2,-y+1/2,-z
3 -x,y+1/2,-z+1/2
4 -x+1/2,-y,z+1/2
5 -x,-y,-z
6 -x+1/2,y+1/2,z
7 x,-y+1/2,z+1/2
8 x+1/2,y,-z+1/2

loop_
_atom_site_label
_atom_site_type_symbol
_atom_site_symmetry_multiplicity
_atom_site_Wyckoff_label
_atom_site_fract_x
_atom_site_fract_y
_atom_site_fract_z
_atom_site_occupancy
Cl1 Cl 8 c 0.10620 0.55070 0.54240 1.00000
Cl2 Cl 8 c 0.23640 0.32050 0.52310 1.00000
Cl3 Cl 8 c 0.00980 0.21070 0.43060 1.00000
Cl4 Cl 8 c 0.00920 0.32850 0.65060 1.00000
Cl5 Cl 8 c 0.28520 0.54830 0.31030 1.00000
Cl6 Cl 8 c 0.27690 0.31030 0.25220 1.00000
Cl7 Cl 8 c 0.12070 0.47060 0.15480 1.00000
O1 O 8 c 0.10750 0.41240 0.35640 1.00000
P1 P 8 c 0.18750 0.43530 0.28190 1.00000
Ti1 Ti 8 c 0.06290 0.36160 0.49860 1.00000
```

($TiCl_4$·$POCl_3$)$_2$: A7BCD_oP80_61_7c_c_c_c - POSCAR

```
A7BCD_oP80_61_7c_c_c_c & a,b/a,c/a,x1,y1,z1,x2,y2,z2,x3,y3,z3,x4,y4,z4,
    ↪ x5,y5,z5,x6,y6,z6,x7,y7,z7,x8,y8,z8,x9,y9,z9,x10,y10,z10 --
    ↪ params=12.42,1.02737520129,1.08776167472,0.1062,0.5507,0.5424,
    ↪ 0.2364,0.3205,0.5231,0.0098,0.2107,0.4306,0.0092,0.3285,0.6506,
    ↪ 0.2852,0.5483,0.3103,0.2769,0.3103,0.2522,0.1207,0.4706,0.1548,
    ↪ 0.1075,0.4124,0.3564,0.1875,0.4353,0.2819,0.0629,0.3616,0.4986
    ↪ & Pbca D_{2h}^{15} #61 (c^10) & oP80 & None & Cl7OPTi & Cl7OPTi
    ↪ & C.-I. Br\"{a}nd\'{e}n and I. Lindqvist, Acta Chem. Scand. 14
    ↪ , 726-732 (1960)
   1.00000000000000
  12.42000000000000   0.00000000000000   0.00000000000000
   0.00000000000000  12.76000000000000   0.00000000000000
   0.00000000000000   0.00000000000000  13.51000000000000
    Cl      O       P       Ti
    56      8       8       8
Direct
   0.10620000000000   0.55070000000000   0.54240000000000   Cl   (8c)
   0.39380000000000  -0.55070000000000   1.04240000000000   Cl   (8c)
  -0.10620000000000   1.05070000000000  -0.04240000000000   Cl   (8c)
   0.60620000000000  -0.05070000000000  -0.54240000000000   Cl   (8c)
  -0.10620000000000  -0.55070000000000  -0.54240000000000   Cl   (8c)
   0.60620000000000   0.55070000000000  -0.04240000000000   Cl   (8c)
   0.10620000000000  -0.05070000000000   1.04240000000000   Cl   (8c)
   0.39380000000000   1.05070000000000   0.54240000000000   Cl   (8c)
   0.23640000000000   0.32050000000000   0.52310000000000   Cl   (8c)
   0.26360000000000  -0.32050000000000   1.02310000000000   Cl   (8c)
```

```
   -0.23640000000000    0.82050000000000   -0.02310000000000    Cl    (8c)
    0.73640000000000    0.17950000000000   -0.52310000000000    Cl    (8c)
   -0.23640000000000   -0.32050000000000   -0.52310000000000    Cl    (8c)
    0.73640000000000    0.32050000000000   -0.02310000000000    Cl    (8c)
    0.23640000000000    0.17950000000000    1.02310000000000    Cl    (8c)
    0.26360000000000    0.82050000000000    0.52310000000000    Cl    (8c)
    0.00980000000000    0.21070000000000    0.43060000000000    Cl    (8c)
    0.49020000000000   -0.21070000000000    0.93060000000000    Cl    (8c)
   -0.00980000000000    0.71070000000000    0.06940000000000    Cl    (8c)
    0.50980000000000    0.28930000000000   -0.43060000000000    Cl    (8c)
   -0.00980000000000   -0.21070000000000   -0.43060000000000    Cl    (8c)
    0.50980000000000    0.21070000000000    0.06940000000000    Cl    (8c)
    0.00980000000000    0.28930000000000    0.93060000000000    Cl    (8c)
    0.49020000000000    0.71070000000000    0.43060000000000    Cl    (8c)
    0.00920000000000    0.32850000000000    0.65060000000000    Cl    (8c)
    0.49080000000000   -0.32850000000000    1.15060000000000    Cl    (8c)
   -0.00920000000000    0.82850000000000   -0.15060000000000    Cl    (8c)
    0.50920000000000    0.17150000000000   -0.65060000000000    Cl    (8c)
   -0.00920000000000   -0.32850000000000   -0.65060000000000    Cl    (8c)
    0.50920000000000    0.32850000000000   -0.15060000000000    Cl    (8c)
    0.00920000000000    0.17150000000000    1.15060000000000    Cl    (8c)
    0.49080000000000    0.82850000000000    0.65060000000000    Cl    (8c)
    0.28520000000000    0.54830000000000    0.31030000000000    Cl    (8c)
    0.21480000000000   -0.54830000000000    0.81030000000000    Cl    (8c)
   -0.28520000000000    1.04830000000000    0.18970000000000    Cl    (8c)
    0.78520000000000   -0.04830000000000   -0.31030000000000    Cl    (8c)
   -0.28520000000000   -0.54830000000000   -0.31030000000000    Cl    (8c)
    0.78520000000000    0.54830000000000    0.18970000000000    Cl    (8c)
    0.28520000000000   -0.04830000000000    0.81030000000000    Cl    (8c)
    0.21480000000000    1.04830000000000    0.31030000000000    Cl    (8c)
    0.27690000000000    0.31030000000000    0.25220000000000    Cl    (8c)
    0.22310000000000   -0.31030000000000    0.75220000000000    Cl    (8c)
   -0.27690000000000    0.81030000000000    0.24780000000000    Cl    (8c)
    0.77690000000000    0.18970000000000   -0.25220000000000    Cl    (8c)
   -0.27690000000000   -0.31030000000000   -0.25220000000000    Cl    (8c)
    0.77690000000000    0.31030000000000    0.24780000000000    Cl    (8c)
    0.27690000000000    0.18970000000000    0.75220000000000    Cl    (8c)
    0.22310000000000    0.81030000000000    0.25220000000000    Cl    (8c)
    0.12070000000000    0.47060000000000    0.15480000000000    Cl    (8c)
    0.37930000000000   -0.47060000000000    0.65480000000000    Cl    (8c)
   -0.12070000000000    0.97060000000000    0.34520000000000    Cl    (8c)
    0.62070000000000    0.02940000000000   -0.15480000000000    Cl    (8c)
   -0.12070000000000   -0.47060000000000   -0.15480000000000    Cl    (8c)
    0.62070000000000    0.47060000000000    0.34520000000000    Cl    (8c)
    0.12070000000000    0.02940000000000    0.65480000000000    Cl    (8c)
    0.37930000000000    0.97060000000000    0.15480000000000    Cl    (8c)
    0.10750000000000    0.41240000000000    0.35640000000000    O     (8c)
    0.39250000000000   -0.41240000000000    0.85640000000000    O     (8c)
   -0.10750000000000    0.91240000000000    0.14360000000000    O     (8c)
    0.60750000000000    0.08760000000000   -0.35640000000000    O     (8c)
   -0.10750000000000   -0.41240000000000   -0.35640000000000    O     (8c)
    0.60750000000000    0.41240000000000    0.14360000000000    O     (8c)
    0.10750000000000    0.08760000000000    0.85640000000000    O     (8c)
    0.39250000000000    0.91240000000000    0.35640000000000    O     (8c)
    0.18750000000000    0.43530000000000    0.28190000000000    P     (8c)
    0.31250000000000   -0.43530000000000    0.78190000000000    P     (8c)
   -0.18750000000000    0.93530000000000    0.21810000000000    P     (8c)
    0.68750000000000    0.06470000000000   -0.28190000000000    P     (8c)
   -0.18750000000000   -0.43530000000000   -0.28190000000000    P     (8c)
    0.68750000000000    0.43530000000000    0.21810000000000    P     (8c)
    0.18750000000000    0.06470000000000    0.78190000000000    P     (8c)
    0.31250000000000    0.93530000000000    0.28190000000000    P     (8c)
    0.06290000000000    0.36160000000000    0.49860000000000    Ti    (8c)
    0.43710000000000   -0.36160000000000    0.99860000000000    Ti    (8c)
   -0.06290000000000    0.86160000000000    0.00140000000000    Ti    (8c)
    0.56290000000000    0.13840000000000   -0.49860000000000    Ti    (8c)
   -0.06290000000000   -0.36160000000000   -0.49860000000000    Ti    (8c)
    0.56290000000000    0.36160000000000    0.00140000000000    Ti    (8c)
    0.06290000000000    0.13840000000000    0.99860000000000    Ti    (8c)
    0.43710000000000    0.86160000000000    0.49860000000000    Ti    (8c)
```

Hambergite [$Be_2BO_3(OH)$ ($G7_2$)]: AB2CD4_oP64_61_c_2c_c_4c - CIF

```
# CIF file
data_findsym-output
_audit_creation_method FINDSYM

_chemical_name_mineral 'Hambergite'
_chemical_formula_sum 'B Be2 H O4'

loop_
_publ_author_name
 'G. D. Gatta'
 'G. J. {McIntyre}'
 'G. Bromiley'
 'A. Guastoni'
 'F. Nestola'
_journal_name_full_name
;
 American Mineralogist
;
_journal_volume 97
_journal_year 2012
_journal_page_first 1891
_journal_page_last 1897
_publ_Section_title
;
 A single-crystal neutron diffraction study of hambergite, Be$_{2}$BO$_{
    ↪ 3}$(OH,F)
;

_aflow_title 'Hambergite [Be$_{2}$BO$_{3}$(OH) ($G7_{2}$)] Structure'
_aflow_proto 'AB2CD4_oP64_61_c_2c_c_4c'
_aflow_params 'a,b/a,c/a,x_{1},y_{1},z_{1},x_{2},y_{2},z_{2},x_{3},y_{3
    ↪ },z_{3},x_{4},y_{4},z_{4},x_{5},y_{5},z_{5},x_{6},y_{6},z_{6},
    ↪ x_{7},y_{7},z_{7},x_{8},y_{8},z_{8}'
_aflow_params_values '9.762,1.24984634296,0.453800450727,0.10617,0.60704
    ↪ ,0.77298,0.00261,0.68871,0.26018,0.23724,0.56757,0.27717,0.3138
    ↪ ,0.7228,0.4574,0.0376,0.68766,0.61914,0.1012,0.60302,0.08204,
    ↪ 0.18691,0.5345,0.61701,0.33976,0.67302,0.296'
_aflow_Strukturbericht '$G7_{2}$'
_aflow_Pearson 'oP64'

_symmetry_space_group_name_H-M "P 21/b 21/c 21/a"
_symmetry_Int_Tables_number 61

_cell_length_a     9.76200
_cell_length_b     12.20100
_cell_length_c     4.43000
_cell_angle_alpha 90.00000
_cell_angle_beta  90.00000
_cell_angle_gamma 90.00000

loop_
_space_group_symop_id
_space_group_symop_operation_xyz
1 x,y,z
2 x+1/2,-y+1/2,-z
3 -x,y+1/2,-z+1/2
4 -x+1/2,-y,z+1/2
5 -x,-y,-z
6 -x+1/2,y+1/2,z
7 x,-y+1/2,z+1/2
8 x+1/2,y,-z+1/2

loop_
_atom_site_label
_atom_site_type_symbol
_atom_site_symmetry_multiplicity
_atom_site_Wyckoff_label
_atom_site_fract_x
_atom_site_fract_y
_atom_site_fract_z
_atom_site_occupancy
B1  B  8 c 0.10617 0.60704 0.77298 1.00000
Be1 Be 8 c 0.00261 0.68871 0.26018 1.00000
Be2 Be 8 c 0.23724 0.56757 0.27717 1.00000
H1  H  8 c 0.31380 0.72280 0.45740 1.00000
O1  O  8 c 0.03760 0.68766 0.61914 1.00000
O2  O  8 c 0.10120 0.60302 0.08204 1.00000
O3  O  8 c 0.18691 0.53450 0.61701 1.00000
O4  O  8 c 0.33976 0.67302 0.29600 1.00000
```

Hambergite [$Be_2BO_3(OH)$ ($G7_2$)]: AB2CD4_oP64_61_c_2c_c_4c - POSCAR

```
AB2CD4_oP64_61_c_2c_c_4c & a,b/a,c/a,x1,y1,z1,x2,y2,z2,x3,y3,z3,x4,y4,z4
    ↪ ,x5,y5,z5,x6,y6,z6,x7,y7,z7,x8,y8,z8 --params=9.762,
    ↪ 1.24984634296,0.453800450727,0.10617,0.60704,0.77298,0.00261,
    ↪ 0.68871,0.26018,0.23724,0.56757,0.27717,0.3138,0.7228,0.4574,
    ↪ 0.0376,0.68766,0.61914,0.1012,0.60302,0.08204,0.18691,0.5345,
    ↪ 0.61701,0.33976,0.67302,0.296 & Pbca D_{2h}^{15} #61 (c^8) &
    ↪ oP64 & $G7_{2}$ & BBe2HO4 & Hambergite & G. D. Gatta et al.,
    ↪ Am. Mineral. 97, 1891-1897 (2012)
   1.00000000000000
   9.76200000000000    0.00000000000000    0.00000000000000
   0.00000000000000   12.20100000000000    0.00000000000000
   0.00000000000000    0.00000000000000    4.43000000000000
     B     Be     H     O
     8     16     8     32
Direct
    0.10617000000000    0.60704000000000    0.77298000000000    B     (8c)
    0.39383000000000   -0.60704000000000    1.27298000000000    B     (8c)
   -0.10617000000000    1.10704000000000   -0.27298000000000    B     (8c)
    0.60617000000000   -0.10704000000000   -0.77298000000000    B     (8c)
   -0.10617000000000   -0.60704000000000   -0.77298000000000    B     (8c)
    0.60617000000000    0.60704000000000   -0.27298000000000    B     (8c)
    0.10617000000000   -0.10704000000000    1.27298000000000    B     (8c)
    0.39383000000000    1.10704000000000    0.77298000000000    B     (8c)
    0.00261000000000    0.68871000000000    0.26018000000000    Be    (8c)
    0.49739000000000   -0.68871000000000    0.76018000000000    Be    (8c)
   -0.00261000000000    1.18871000000000    0.23982000000000    Be    (8c)
    0.50261000000000   -0.18871000000000   -0.26018000000000    Be    (8c)
   -0.00261000000000   -0.68871000000000   -0.26018000000000    Be    (8c)
    0.50261000000000    0.68871000000000    0.23982000000000    Be    (8c)
    0.00261000000000   -0.18871000000000    0.76018000000000    Be    (8c)
    0.49739000000000    1.18871000000000    0.26018000000000    Be    (8c)
    0.23724000000000    0.56757000000000    0.27717000000000    Be    (8c)
    0.26276000000000   -0.56757000000000    0.77717000000000    Be    (8c)
   -0.23724000000000    1.06757000000000    0.22283000000000    Be    (8c)
    0.73724000000000   -0.06757000000000   -0.27717000000000    Be    (8c)
   -0.23724000000000   -0.56757000000000   -0.27717000000000    Be    (8c)
    0.73724000000000    0.56757000000000    0.22283000000000    Be    (8c)
    0.23724000000000   -0.06757000000000    0.77717000000000    Be    (8c)
    0.26276000000000    1.06757000000000    0.27717000000000    Be    (8c)
    0.31380000000000    0.72280000000000    0.45740000000000    H     (8c)
    0.18620000000000   -0.72280000000000    0.95740000000000    H     (8c)
   -0.31380000000000    1.22280000000000    0.04260000000000    H     (8c)
    0.81380000000000   -0.22280000000000   -0.45740000000000    H     (8c)
   -0.31380000000000   -0.72280000000000   -0.45740000000000    H     (8c)
    0.81380000000000    0.72280000000000    0.04260000000000    H     (8c)
    0.31380000000000   -0.22280000000000    0.95740000000000    H     (8c)
    0.18620000000000    1.22280000000000    0.45740000000000    H     (8c)
    0.03760000000000    0.68766000000000    0.61914000000000    O     (8c)
    0.46240000000000   -0.68766000000000    1.11914000000000    O     (8c)
   -0.03760000000000    1.18766000000000   -0.11914000000000    O     (8c)
    0.53760000000000   -0.18766000000000   -0.61914000000000    O     (8c)
   -0.03760000000000   -0.68766000000000   -0.61914000000000    O     (8c)
    0.53760000000000    0.68766000000000   -0.11914000000000    O     (8c)
    0.03760000000000   -0.18766000000000    1.11914000000000    O     (8c)
    0.46240000000000    1.18766000000000    0.61914000000000    O     (8c)
```

```
  0.10120000000000   0.60302000000000   0.08204000000000   O   (8c)
  0.39880000000000  -0.60302000000000   0.58204000000000   O   (8c)
 -0.10120000000000   1.10302000000000   0.41796000000000   O   (8c)
  0.60120000000000  -0.10302000000000  -0.08204000000000   O   (8c)
 -0.10120000000000  -0.60302000000000  -0.08204000000000   O   (8c)
  0.60120000000000   0.60302000000000   0.41796000000000   O   (8c)
  0.10120000000000  -0.10302000000000   0.58204000000000   O   (8c)
  0.39880000000000   1.10302000000000   0.08204000000000   O   (8c)
  0.18691000000000   0.53450000000000   0.61701000000000   O   (8c)
  0.31309000000000  -0.53450000000000   1.11701000000000   O   (8c)
 -0.18691000000000   1.03450000000000  -0.11701000000000   O   (8c)
  0.68691000000000  -0.03450000000000  -0.61701000000000   O   (8c)
 -0.18691000000000  -0.53450000000000  -0.61701000000000   O   (8c)
  0.68691000000000   0.53450000000000  -0.11701000000000   O   (8c)
  0.18691000000000  -0.03450000000000   1.11701000000000   O   (8c)
  0.31309000000000   1.03450000000000   0.61701000000000   O   (8c)
  0.33976000000000   0.67302000000000   0.29600000000000   O   (8c)
  0.16024000000000  -0.67302000000000   0.79600000000000   O   (8c)
 -0.33976000000000   1.17302000000000   0.20400000000000   O   (8c)
  0.83976000000000  -0.17302000000000  -0.29600000000000   O   (8c)
 -0.33976000000000  -0.67302000000000  -0.29600000000000   O   (8c)
  0.83976000000000   0.67302000000000   0.20400000000000   O   (8c)
  0.33976000000000  -0.17302000000000   0.79600000000000   O   (8c)
  0.16024000000000   1.17302000000000   0.29600000000000   O   (8c)
```

Enstatite ($MgSiO_3$, $S4_3$): AB3C_oP80_61_2c_6c_2c - CIF

```
# CIF file
data_findsym-output
_audit_creation_method FINDSYM

_chemical_name_mineral 'Enstatite'
_chemical_formula_sum 'Mg O3 Si'

loop_
_publ_author_name
 'B. E. Warren'
 'D. I. Modell'
_journal_name_full_name
;
 Zeitschrift f{\"u}r Kristallographie - Crystalline Materials
;
_journal_volume 75
_journal_year 1930
_journal_page_first 1
_journal_page_last 14
_publ_Section_title
;
 The Structure of Enstatite MgSiO$_{3}$
;

_aflow_title 'Enstatite (MgSiO$_{3}$, $S4_{3}$) Structure'
_aflow_proto 'AB3C_oP80_61_2c_6c_2c'
_aflow_params 'a,b/a,c/a,x_{1},y_{1},z_{1},x_{2},y_{2},z_{2},x_{3},y_{3
   ↪ },z_{3},x_{4},y_{4},z_{4},x_{5},y_{5},z_{5},x_{6},y_{6},z_{6},
   ↪ x_{7},y_{7},z_{7},x_{8},y_{8},z_{8},x_{9},y_{9},z_{9},x_{10},y_
   ↪ {10},z_{10}'
_aflow_params_values '18.2,0.486813186813,0.285714285714,0.13,0.33,0.37,
   ↪ 0.13,0.96,0.37,0.06,0.14,0.2,0.06,0.5,0.2,0.05,0.75,0.05,0.19,
   ↪ 0.35,0.06,0.19,0.01,0.05,0.2,0.75,0.3,0.03,0.65,0.29,0.22,0.85,
   ↪ 0.04'
_aflow_Strukturbericht '$S4_{3}$'
_aflow_Pearson 'oP80'

_symmetry_space_group_name_H-M "P 21/b 21/c 21/a"
_symmetry_Int_Tables_number 61

_cell_length_a    18.20000
_cell_length_b    8.86000
_cell_length_c    5.20000
_cell_angle_alpha 90.00000
_cell_angle_beta  90.00000
_cell_angle_gamma 90.00000

loop_
_space_group_symop_id
_space_group_symop_operation_xyz
1 x,y,z
2 x+1/2,-y+1/2,-z
3 -x,y+1/2,-z+1/2
4 -x+1/2,-y,z+1/2
5 -x,-y,-z
6 -x+1/2,y+1/2,z
7 x,-y+1/2,z+1/2
8 x+1/2,y,-z+1/2

loop_
_atom_site_label
_atom_site_type_symbol
_atom_site_symmetry_multiplicity
_atom_site_Wyckoff_label
_atom_site_fract_x
_atom_site_fract_y
_atom_site_fract_z
_atom_site_occupancy
Mg1 Mg 8 c 0.13000 0.33000 0.37000 1.00000
Mg2 Mg 8 c 0.13000 0.96000 0.37000 1.00000
O1 O 8 c 0.06000 0.14000 0.20000 1.00000
O2 O 8 c 0.06000 0.50000 0.20000 1.00000
O3 O 8 c 0.05000 0.75000 0.05000 1.00000
O4 O 8 c 0.19000 0.35000 0.06000 1.00000
O5 O 8 c 0.19000 0.01000 0.05000 1.00000
O6 O 8 c 0.20000 0.75000 0.30000 1.00000
Si1 Si 8 c 0.03000 0.65000 0.29000 1.00000
Si2 Si 8 c 0.22000 0.85000 0.04000 1.00000
```

Enstatite ($MgSiO_3$, $S4_3$): AB3C_oP80_61_2c_6c_2c - POSCAR

```
AB3C_oP80_61_2c_6c_2c & a,b/a,c/a,x1,y1,z1,x2,y2,z2,x3,y3,z3,x4,y4,z4,x5
   ↪ ,y5,z5,x6,y6,z6,x7,y7,z7,x8,y8,z8,x9,y9,z9,x10,y10,z10 --params
   ↪ =18.2,0.486813186813,0.285714285714,0.13,0.33,0.37,0.13,0.96,
   ↪ 0.37,0.06,0.14,0.2,0.06,0.5,0.2,0.05,0.75,0.05,0.19,0.35,0.06,
   ↪ 0.19,0.01,0.05,0.2,0.75,0.3,0.03,0.65,0.29,0.22,0.85,0.04 &
   ↪ Pbca D_{2h}^{15} #61 (c^10) & oP80 & $S4_{3}$ & MgO3Si &
   ↪ Enstatite & B. E. Warren and D. I. Modell, Zeitschrift f{"u}r
   ↪ Kristallographie - Crystalline Materials 75, 1-14 (1930)
   1.00000000000000
  18.20000000000000   0.00000000000000   0.00000000000000
   0.00000000000000   8.86000000000000   0.00000000000000
   0.00000000000000   0.00000000000000   5.20000000000000
    Mg     O    Si
    16    48    16
Direct
   0.13000000000000   0.33000000000000   0.37000000000000   Mg   (8c)
   0.37000000000000  -0.33000000000000   0.87000000000000   Mg   (8c)
  -0.13000000000000   0.83000000000000   0.13000000000000   Mg   (8c)
   0.63000000000000   0.17000000000000  -0.37000000000000   Mg   (8c)
  -0.13000000000000  -0.33000000000000  -0.37000000000000   Mg   (8c)
   0.63000000000000   0.33000000000000   0.13000000000000   Mg   (8c)
   0.13000000000000   0.17000000000000   0.87000000000000   Mg   (8c)
   0.37000000000000   0.83000000000000   0.37000000000000   Mg   (8c)
   0.13000000000000   0.96000000000000   0.37000000000000   Mg   (8c)
   0.37000000000000  -0.96000000000000   0.87000000000000   Mg   (8c)
  -0.13000000000000   1.46000000000000   0.13000000000000   Mg   (8c)
   0.63000000000000  -0.46000000000000  -0.37000000000000   Mg   (8c)
  -0.13000000000000  -0.96000000000000  -0.37000000000000   Mg   (8c)
   0.63000000000000   0.96000000000000   0.13000000000000   Mg   (8c)
   0.13000000000000  -0.46000000000000   0.87000000000000   Mg   (8c)
   0.37000000000000   1.46000000000000   0.37000000000000   Mg   (8c)
   0.06000000000000   0.14000000000000   0.20000000000000    O   (8c)
   0.44000000000000  -0.14000000000000   0.70000000000000    O   (8c)
  -0.06000000000000   0.64000000000000   0.30000000000000    O   (8c)
   0.56000000000000   0.36000000000000  -0.20000000000000    O   (8c)
  -0.06000000000000  -0.14000000000000  -0.20000000000000    O   (8c)
   0.56000000000000   0.14000000000000   0.30000000000000    O   (8c)
   0.06000000000000   0.36000000000000   0.70000000000000    O   (8c)
   0.44000000000000   0.64000000000000   0.20000000000000    O   (8c)
   0.06000000000000   0.50000000000000   0.20000000000000    O   (8c)
   0.44000000000000  -0.50000000000000   0.70000000000000    O   (8c)
  -0.06000000000000   1.00000000000000   0.30000000000000    O   (8c)
   0.56000000000000   0.00000000000000  -0.20000000000000    O   (8c)
  -0.06000000000000  -0.50000000000000  -0.20000000000000    O   (8c)
   0.56000000000000   0.50000000000000   0.30000000000000    O   (8c)
   0.06000000000000   0.00000000000000   0.70000000000000    O   (8c)
   0.44000000000000   1.00000000000000   0.20000000000000    O   (8c)
   0.05000000000000   0.75000000000000   0.05000000000000    O   (8c)
   0.45000000000000  -0.75000000000000   0.55000000000000    O   (8c)
  -0.05000000000000   1.25000000000000   0.45000000000000    O   (8c)
   0.55000000000000  -0.25000000000000  -0.05000000000000    O   (8c)
  -0.05000000000000  -0.75000000000000  -0.05000000000000    O   (8c)
   0.55000000000000   0.75000000000000   0.45000000000000    O   (8c)
   0.05000000000000  -0.25000000000000   0.55000000000000    O   (8c)
   0.45000000000000   1.25000000000000   0.05000000000000    O   (8c)
   0.19000000000000   0.35000000000000   0.06000000000000    O   (8c)
   0.31000000000000  -0.35000000000000   0.56000000000000    O   (8c)
  -0.19000000000000   0.85000000000000   0.44000000000000    O   (8c)
   0.69000000000000   0.15000000000000  -0.06000000000000    O   (8c)
  -0.19000000000000  -0.35000000000000  -0.06000000000000    O   (8c)
   0.69000000000000   0.35000000000000   0.44000000000000    O   (8c)
   0.19000000000000   0.15000000000000   0.56000000000000    O   (8c)
   0.31000000000000   0.85000000000000   0.06000000000000    O   (8c)
   0.19000000000000   0.01000000000000   0.05000000000000    O   (8c)
   0.31000000000000  -0.01000000000000   0.55000000000000    O   (8c)
  -0.19000000000000   0.51000000000000   0.45000000000000    O   (8c)
   0.69000000000000   0.49000000000000  -0.05000000000000    O   (8c)
  -0.19000000000000  -0.01000000000000  -0.05000000000000    O   (8c)
   0.69000000000000   0.01000000000000   0.45000000000000    O   (8c)
   0.19000000000000   0.49000000000000   0.55000000000000    O   (8c)
   0.31000000000000   0.51000000000000   0.05000000000000    O   (8c)
   0.20000000000000   0.75000000000000   0.30000000000000    O   (8c)
   0.30000000000000  -0.75000000000000   0.80000000000000    O   (8c)
  -0.20000000000000   1.25000000000000   0.20000000000000    O   (8c)
   0.70000000000000  -0.25000000000000  -0.30000000000000    O   (8c)
  -0.20000000000000  -0.75000000000000  -0.30000000000000    O   (8c)
   0.70000000000000   0.75000000000000   0.20000000000000    O   (8c)
   0.20000000000000  -0.25000000000000   0.80000000000000    O   (8c)
   0.30000000000000   1.25000000000000   0.30000000000000    O   (8c)
   0.03000000000000   0.65000000000000   0.29000000000000   Si   (8c)
   0.47000000000000  -0.65000000000000   0.79000000000000   Si   (8c)
  -0.03000000000000   1.15000000000000   0.21000000000000   Si   (8c)
   0.53000000000000  -0.15000000000000  -0.29000000000000   Si   (8c)
  -0.03000000000000  -0.65000000000000  -0.29000000000000   Si   (8c)
   0.53000000000000   0.65000000000000   0.21000000000000   Si   (8c)
   0.03000000000000  -0.15000000000000   0.79000000000000   Si   (8c)
   0.47000000000000   1.15000000000000   0.29000000000000   Si   (8c)
   0.22000000000000   0.85000000000000   0.04000000000000   Si   (8c)
   0.28000000000000  -0.85000000000000   0.54000000000000   Si   (8c)
  -0.22000000000000   1.35000000000000   0.46000000000000   Si   (8c)
   0.72000000000000  -0.35000000000000  -0.04000000000000   Si   (8c)
  -0.22000000000000  -0.85000000000000  -0.04000000000000   Si   (8c)
   0.72000000000000   0.85000000000000   0.46000000000000   Si   (8c)
   0.22000000000000  -0.35000000000000   0.54000000000000   Si   (8c)
   0.28000000000000   1.35000000000000   0.04000000000000   Si   (8c)
```

COCl: ABC_oP24_61_c_c_c - CIF

```
# CIF file
data_findsym-output
_audit_creation_method FINDSYM

_chemical_name_mineral 'CClO'
```

```
_chemical_formula_sum 'C Cl O'

loop_
_publ_author_name
 'P. Groth'
 'O. Hassel'
_journal_name_full_name
;
 Acta Chemica Scandinavica
;
_journal_volume 16
_journal_year 1962
_journal_page_first 2311
_journal_page_last 2317
_publ_Section_title
;
 Crystal Structures of Oxalyl Bromide and Oxalyl Chloride
;

_aflow_title 'COCl Structure'
_aflow_proto 'ABC_oP24_61_c_c_c'
_aflow_params 'a,b/a,c/a,x_{1},y_{1},z_{1},x_{2},y_{2},z_{2},x_{3},y_{3
    ↪ },z_{3}'
_aflow_params_values '6.44,0.944099378882,1.85248447205,-0.006,0.062,
    ↪ 0.053,-0.034,-0.083,0.167,0.118,0.23,0.05'
_aflow_Strukturbericht 'None'
_aflow_Pearson 'oP24'

_symmetry_space_group_name_H-M "P 21/b 21/c 21/a"
_symmetry_Int_Tables_number 61

_cell_length_a    6.44000
_cell_length_b    6.08000
_cell_length_c    11.93000
_cell_angle_alpha 90.00000
_cell_angle_beta  90.00000
_cell_angle_gamma 90.00000

loop_
_space_group_symop_id
_space_group_symop_operation_xyz
1 x,y,z
2 x+1/2,-y+1/2,-z
3 -x,y+1/2,-z+1/2
4 -x+1/2,-y,z+1/2
5 -x,-y,-z
6 -x+1/2,y+1/2,z
7 x,-y+1/2,z+1/2
8 x+1/2,y,-z+1/2

loop_
_atom_site_label
_atom_site_type_symbol
_atom_site_symmetry_multiplicity
_atom_site_Wyckoff_label
_atom_site_fract_x
_atom_site_fract_y
_atom_site_fract_z
_atom_site_occupancy
C1 C 8 c -0.00600 0.06200 0.05300 1.00000
Cl1 Cl 8 c -0.03400 -0.08300 0.16700 1.00000
O1 O 8 c 0.11800 0.23000 0.05000 1.00000
```

COCl: ABC_oP24_61_c_c_c - POSCAR

```
ABC_oP24_61_c_c_c & a,b/a,c/a,x1,y1,z1,x2,y2,z2,x3,y3,z3 --params=6.44,
    ↪ 0.944099378882,1.85248447205,-0.006,0.062,0.053,-0.034,-0.083,
    ↪ 0.167,0.118,0.23,0.05 & Pbca D_{2h}^{15} #61 (c^3) & oP24 &
    ↪ None & CClO & CClO & P. Groth and O. Hassel, Acta Chem. Scand.
    ↪ 16, 2311-2317 (1962)
   1.00000000000000
   6.44000000000000   0.00000000000000   0.00000000000000
   0.00000000000000   6.08000000000000   0.00000000000000
   0.00000000000000   0.00000000000000  11.93000000000000
     C     Cl      O
     8      8      8
Direct
  -0.00600000000000   0.06200000000000   0.05300000000000   C   (8c)
   0.50600000000000  -0.06200000000000   0.55300000000000   C   (8c)
   0.00600000000000   0.56200000000000   0.44700000000000   C   (8c)
   0.49400000000000   0.43800000000000  -0.05300000000000   C   (8c)
   0.00600000000000  -0.06200000000000  -0.05300000000000   C   (8c)
   0.49400000000000   0.06200000000000   0.44700000000000   C   (8c)
  -0.00600000000000   0.43800000000000   0.55300000000000   C   (8c)
   0.50600000000000   0.56200000000000   0.05300000000000   C   (8c)
  -0.03400000000000  -0.08300000000000   0.16700000000000   Cl  (8c)
   0.53400000000000   0.08300000000000   0.66700000000000   Cl  (8c)
   0.03400000000000   0.41700000000000   0.33300000000000   Cl  (8c)
   0.46600000000000   0.58300000000000  -0.16700000000000   Cl  (8c)
   0.03400000000000   0.08300000000000  -0.16700000000000   Cl  (8c)
   0.46600000000000  -0.08300000000000   0.33300000000000   Cl  (8c)
  -0.03400000000000   0.58300000000000   0.66700000000000   Cl  (8c)
   0.53400000000000   0.41700000000000   0.16700000000000   Cl  (8c)
   0.11800000000000   0.23000000000000   0.05000000000000   O   (8c)
   0.38200000000000  -0.23000000000000   0.55000000000000   O   (8c)
  -0.11800000000000   0.73000000000000   0.45000000000000   O   (8c)
   0.61800000000000   0.27000000000000  -0.05000000000000   O   (8c)
  -0.11800000000000  -0.23000000000000  -0.05000000000000   O   (8c)
   0.61800000000000   0.23000000000000   0.45000000000000   O   (8c)
   0.11800000000000   0.27000000000000   0.55000000000000   O   (8c)
   0.38200000000000   0.73000000000000   0.05000000000000   O   (8c)
```

Topaz ($Al_2SiO_4F_2$, $S0_5$): A2B2C4D_oP36_62_d_d_2cd_c - CIF

```
# CIF file
data_findsym-output
_audit_creation_method FINDSYM

_chemical_name_mineral 'Topaz'
_chemical_formula_sum 'Al2 F2 O4 Si'

loop_
_publ_author_name
 'K. Komatsu'
 'T. Kuribayashi'
 'Y. Kudoh'
_journal_name_full_name
;
 Journal of Mineralogical and Petrological Sciences
;
_journal_volume 98
_journal_year 2003
_journal_page_first 167
_journal_page_last 180
_publ_Section_title
;
 Effect of temperature and pressure on the crystal structure of topaz,
     ↪ Al$_{2}$SiO$_{4}$(OH,F)$_{2}$
;

_aflow_title 'Topaz (Al$_{2}$SiO$_{4}$F$_{2}$, $S0_{5}$) Structure'
_aflow_proto 'A2B2C4D_oP36_62_d_d_2cd_c'
_aflow_params 'a,b/a,c/a,x_{1},z_{1},x_{2},z_{2},x_{3},z_{3},x_{4},y_{4
     ↪ },z_{4},x_{5},y_{5},z_{5},x_{6},y_{6},z_{6}'
_aflow_params_values '8.7935,0.954762040143,0.528674589185,0.468,0.7957,
     ↪ 0.2436,0.4566,0.55945,0.1022,0.63102,0.5824,0.40358,0.75265,
     ↪ 0.55741,0.0981,0.4893,0.5923,0.711'
_aflow_Strukturbericht '$S0_{5}$'
_aflow_Pearson 'oP36'

_symmetry_space_group_name_H-M "P 21/n 21/m 21/a"
_symmetry_Int_Tables_number 62

_cell_length_a    8.79350
_cell_length_b    8.39570
_cell_length_c    4.64890
_cell_angle_alpha 90.00000
_cell_angle_beta  90.00000
_cell_angle_gamma 90.00000

loop_
_space_group_symop_id
_space_group_symop_operation_xyz
1 x,y,z
2 x+1/2,-y+1/2,-z+1/2
3 -x,y+1/2,-z
4 -x+1/2,-y,z+1/2
5 -x,-y,-z
6 -x+1/2,y+1/2,z+1/2
7 x,-y+1/2,z
8 x+1/2,y,-z+1/2

loop_
_atom_site_label
_atom_site_type_symbol
_atom_site_symmetry_multiplicity
_atom_site_Wyckoff_label
_atom_site_fract_x
_atom_site_fract_y
_atom_site_fract_z
_atom_site_occupancy
O1 O 4 c 0.46800 0.25000 0.79570 1.00000
O2 O 4 c 0.24360 0.25000 0.45660 1.00000
Si1 Si 4 c 0.55945 0.25000 0.10220 1.00000
Al1 Al 8 d 0.63102 0.58240 0.40358 1.00000
F1 F 8 d 0.75265 0.55741 0.09810 1.00000
O3 O 8 d 0.48930 0.59230 0.71100 1.00000
```

Topaz ($Al_2SiO_4F_2$, $S0_5$): A2B2C4D_oP36_62_d_d_2cd_c - POSCAR

```
A2B2C4D_oP36_62_d_d_2cd_c & a,b/a,c/a,x1,z1,x2,z2,x3,z3,x4,y4,z4,x5,y5,
     ↪ z5,x6,y6,z6 --params=8.7935,0.954762040143,0.528674589185,0.468
     ↪ ,0.7957,0.2436,0.4566,0.55945,0.1022,0.63102,0.5824,0.40358,
     ↪ 0.75265,0.55741,0.0981,0.4893,0.5923,0.711 & Pnma D_{2h}^{16} #
     ↪ 62 (c^3d^3) & oP36 & $S0_{5}$ & Al2F2O4Si & Topaz & K. Komatsu
     ↪ and T. Kuribayashi and Y. Kudoh, J. Miner. Petrol. Sci. 98,
     ↪ 167-180 (2003)
   1.00000000000000
   8.79350000000000   0.00000000000000   0.00000000000000
   0.00000000000000   8.39570000000000   0.00000000000000
   0.00000000000000   0.00000000000000   4.64890000000000
    Al      F      O     Si
     8      8     16      4
Direct
   0.63102000000000   0.58240000000000   0.40358000000000   Al  (8d)
  -0.13102000000000  -0.58240000000000   0.90358000000000   Al  (8d)
  -0.63102000000000   1.08240000000000  -0.40358000000000   Al  (8d)
   1.13102000000000  -0.08240000000000   0.09642000000000   Al  (8d)
  -0.63102000000000  -0.58240000000000  -0.40358000000000   Al  (8d)
   1.13102000000000   0.58240000000000   0.09642000000000   Al  (8d)
   0.63102000000000  -0.08240000000000   0.40358000000000   Al  (8d)
  -0.13102000000000   1.08240000000000   0.90358000000000   Al  (8d)
   0.75265000000000   0.55741000000000   0.09810000000000   F   (8d)
  -0.25265000000000  -0.55741000000000   0.59810000000000   F   (8d)
  -0.75265000000000   1.05741000000000  -0.09810000000000   F   (8d)
   1.25265000000000  -0.05741000000000   0.40190000000000   F   (8d)
  -0.75265000000000  -0.55741000000000  -0.09810000000000   F   (8d)
   1.25265000000000   0.55741000000000   0.40190000000000   F   (8d)
   0.75265000000000  -0.05741000000000   0.09810000000000   F   (8d)
  -0.25265000000000   1.05741000000000   0.59810000000000   F   (8d)
```

```
    0.46800000000000     0.25000000000000     0.79570000000000     O    (4c)
    0.03200000000000     0.75000000000000     1.29570000000000     O    (4c)
   -0.46800000000000     0.75000000000000    -0.79570000000000     O    (4c)
    0.96800000000000     0.25000000000000    -0.29570000000000     O    (4c)
    0.24360000000000     0.25000000000000     0.45660000000000     O    (4c)
    0.25640000000000     0.75000000000000     0.95660000000000     O    (4c)
   -0.24360000000000     0.75000000000000    -0.45660000000000     O    (4c)
    0.74360000000000     0.25000000000000     0.04340000000000     O    (4c)
    0.48930000000000     0.59230000000000     0.71100000000000     O    (8d)
    0.01070000000000    -0.59230000000000     1.21100000000000     O    (8d)
   -0.48930000000000     1.09230000000000    -0.71100000000000     O    (8d)
    0.98930000000000    -0.09230000000000    -0.21100000000000     O    (8d)
   -0.48930000000000    -0.59230000000000    -0.71100000000000     O    (8d)
    0.98930000000000     0.59230000000000    -0.21100000000000     O    (8d)
    0.48930000000000    -0.09230000000000     0.71100000000000     O    (8d)
    0.01070000000000     1.09230000000000     1.21100000000000     O    (8d)
    0.55945000000000     0.25000000000000     0.10220000000000     Si   (4c)
   -0.05945000000000     0.75000000000000     0.60220000000000     Si   (4c)
   -0.55945000000000     0.75000000000000    -0.10220000000000     Si   (4c)
    1.05945000000000     0.25000000000000     0.39780000000000     Si   (4c)
```

Norbergite [Mg(F,OH)$_2$ · Mg$_2$SiO$_4$, $S0_7$]: A2B3C4D_oP40_62_d_cd_2cd_c - CIF

```
# CIF file
data_findsym-output
_audit_creation_method FINDSYM

_chemical_name_mineral 'Norbergite'
_chemical_formula_sum 'F2 Mg3 O4 Si'

loop_
_publ_author_name
 'G. V. Gibbs'
 'P. H. Ribbe'
_journal_name_full_name
;
 American Mineralogist
;
_journal_volume 54
_journal_year 1969
_journal_page_first 376
_journal_page_last 390
_publ_Section_title
;
 The crystal structures of the humite minerals: I. Norbergite
;

# Found in The RRUFF$^\mathroman{TM}$ Project, {Norbergite, RRUFF ID:
    ↪ R050280},

_aflow_title 'Norbergite [Mg(F,OH)$_{2}$ $\cdot$ Mg$_{2}$SiO$_{4}$, $S0_
    ↪ {7}$] Structure'
_aflow_proto 'A2B3C4D_oP40_62_d_cd_2cd_c'
_aflow_params 'a,b/a,c/a,x_{1},z_{1},x_{2},z_{2},x_{3},z_{3},x_{4},z_{4
    ↪ },x_{5},y_{5},z_{5},x_{6},y_{6},z_{6},x_{7},y_{7},z_{7}'
_aflow_params_values '10.2718,0.851613154462,0.458575906852,-0.0923,-
    ↪ 0.0076,0.7204,0.7617,0.574,0.2793,0.7196,0.4195,-0.0318,0.0834,
    ↪ 0.7295,0.633,0.4305,-0.011,0.7907,0.1034,0.269'
_aflow_Strukturbericht '$S0_{7}$'
_aflow_Pearson 'oP40'

_symmetry_space_group_name_H-M "P 21/n 21/m 21/a"
_symmetry_Int_Tables_number 62

_cell_length_a      10.27180
_cell_length_b      8.74760
_cell_length_c      4.71040
_cell_angle_alpha 90.00000
_cell_angle_beta  90.00000
_cell_angle_gamma 90.00000

loop_
_space_group_symop_id
_space_group_symop_operation_xyz
1 x,y,z
2 x+1/2,-y+1/2,-z+1/2
3 -x,y+1/2,-z
4 -x+1/2,-y,z+1/2
5 -x,-y,-z
6 -x+1/2,y+1/2,z+1/2
7 x,-y+1/2,z
8 x+1/2,y,-z+1/2

loop_
_atom_site_label
_atom_site_type_symbol
_atom_site_symmetry_multiplicity
_atom_site_Wyckoff_label
_atom_site_fract_x
_atom_site_fract_y
_atom_site_fract_z
_atom_site_occupancy
Mg1 Mg 4 c -0.09230 0.25000 -0.00760 1.00000
O1 O 4 c 0.72040 0.25000 0.76170 1.00000
O2 O 4 c 0.57400 0.25000 0.27930 1.00000
Si1 Si 4 c 0.71960 0.25000 0.41950 1.00000
F1 F 8 d -0.03180 0.08340 0.72950 1.00000
Mg2 Mg 8 d 0.63300 0.43050 -0.01100 1.00000
O3 O 8 d 0.79070 0.10340 0.26900 1.00000
```

Norbergite [Mg(F,OH)$_2$ · Mg$_2$SiO$_4$, $S0_7$]: A2B3C4D_oP40_62_d_cd_2cd_c - POSCAR

```
A2B3C4D_oP40_62_d_cd_2cd_c & a,b/a,c/a,x1,z1,x2,z2,x3,z3,x4,z4,x5,y5,z5,
    ↪ x6,y6,z6,x7,y7,z7 --params=10.2718,0.851613154462,
    ↪ 0.458575906852,-0.0923,-0.0076,0.7204,0.7617,0.574,0.2793,
    ↪ 0.7196,0.4195,-0.0318,0.0834,0.7295,0.633,0.4305,-0.011,0.7907,
    ↪ 0.1034,0.269 & Pnma D_{2h}^{16} #62 (c^4d^3) & oP40 & $S0_{7}$
    ↪ & F2Mg3O4Si & Norbergite & G. V. Gibbs and P. H. Ribbe, Am.
    ↪ Mineral. 54, 376-390 (1969)
   1.00000000000000
  10.27180000000000     0.00000000000000     0.00000000000000
   0.00000000000000     8.74760000000000     0.00000000000000
   0.00000000000000     0.00000000000000     4.71040000000000
     F     Mg      O     Si
     8     12     16      4
Direct
  -0.03180000000000     0.08340000000000     0.72950000000000     F    (8d)
   0.53180000000000    -0.08340000000000     1.22950000000000     F    (8d)
   0.03180000000000     0.58340000000000    -0.72950000000000     F    (8d)
   0.46820000000000     0.41660000000000    -0.22950000000000     F    (8d)
   0.03180000000000    -0.08340000000000    -0.72950000000000     F    (8d)
   0.46820000000000     0.08340000000000    -0.22950000000000     F    (8d)
  -0.03180000000000     0.41660000000000     0.72950000000000     F    (8d)
   0.53180000000000     0.58340000000000     1.22950000000000     F    (8d)
  -0.09230000000000     0.25000000000000    -0.00760000000000     Mg   (4c)
   0.59230000000000     0.75000000000000     0.49240000000000     Mg   (4c)
   0.09230000000000     0.75000000000000     0.00760000000000     Mg   (4c)
   0.40770000000000     0.25000000000000     0.50760000000000     Mg   (4c)
   0.63300000000000     0.43050000000000    -0.01100000000000     Mg   (8d)
  -0.13300000000000    -0.43050000000000     0.48900000000000     Mg   (8d)
  -0.63300000000000     0.93050000000000     0.01100000000000     Mg   (8d)
   1.13300000000000     0.06950000000000     0.51100000000000     Mg   (8d)
  -0.63300000000000    -0.43050000000000     0.01100000000000     Mg   (8d)
   1.13300000000000     0.43050000000000     0.51100000000000     Mg   (8d)
   0.63300000000000     0.06950000000000    -0.01100000000000     Mg   (8d)
  -0.13300000000000     0.93050000000000     0.48900000000000     Mg   (8d)
   0.72040000000000     0.25000000000000     0.76170000000000     O    (4c)
  -0.22040000000000     0.75000000000000     1.26170000000000     O    (4c)
  -0.72040000000000     0.75000000000000    -0.76170000000000     O    (4c)
   1.22040000000000     0.25000000000000    -0.26170000000000     O    (4c)
   0.57400000000000     0.25000000000000     0.27930000000000     O    (4c)
  -0.07400000000000     0.75000000000000     0.77930000000000     O    (4c)
  -0.57400000000000     0.75000000000000    -0.27930000000000     O    (4c)
   1.07400000000000     0.25000000000000     0.22070000000000     O    (4c)
   0.79070000000000     0.10340000000000     0.26900000000000     O    (8d)
  -0.29070000000000    -0.10340000000000     0.76900000000000     O    (8d)
  -0.79070000000000     0.60340000000000    -0.26900000000000     O    (8d)
   1.29070000000000     0.39660000000000     0.23100000000000     O    (8d)
  -0.79070000000000    -0.10340000000000    -0.26900000000000     O    (8d)
   1.29070000000000     0.10340000000000     0.23100000000000     O    (8d)
   0.79070000000000     0.39660000000000     0.26900000000000     O    (8d)
  -0.29070000000000     0.60340000000000     0.76900000000000     O    (8d)
   0.71960000000000     0.25000000000000     0.41950000000000     Si   (4c)
  -0.21960000000000     0.75000000000000     0.91950000000000     Si   (4c)
  -0.71960000000000     0.75000000000000    -0.41950000000000     Si   (4c)
   1.21960000000000     0.25000000000000     0.08050000000000     Si   (4c)
```

Arcanite (K$_2$SO$_4$, $H1_6$): A2B4C_oP28_62_2c_2cd_c - CIF

```
# CIF file
data_findsym-output
_audit_creation_method FINDSYM

_chemical_name_mineral 'Arcanite'
_chemical_formula_sum 'K2 O4 S'

loop_
_publ_author_name
 'J. A. {McGinnety}'
_journal_name_full_name
;
 Acta Crystallographica Section B: Structural Science
;
_journal_volume 28
_journal_year 1972
_journal_page_first 2845
_journal_page_last 2852
_publ_Section_title
;
 Redetermination of the structures of potassium sulphate and potassium
      ↪ chromate: the effect of electrostatic crystal forces upon
      ↪ observed bond lengths
;

# Found in The American Mineralogist Crystal Structure Database, 2003

_aflow_title 'Arcanite (K$_{2}$SO$_{4}$, $H1_{6}$) Structure'
_aflow_proto 'A2B4C_oP28_62_2c_2cd_c'
_aflow_params 'a,b/a,c/a,x_{1},z_{1},x_{2},z_{2},x_{3},z_{3},x_{4},z_{4
    ↪ },x_{5},z_{5},x_{6},y_{6},z_{6}'
_aflow_params_values '7.476,0.770866773676,1.34711075441,0.82623,0.21062
    ↪ ,0.51104,0.20406,0.4621,-0.0834,0.2037,0.0582,0.26702,-0.08029,
    ↪ 0.1991,0.0412,0.8522'
_aflow_Strukturbericht '$H1_{6}$'
_aflow_Pearson 'oP28'

_symmetry_space_group_name_H-M "P 21/n 21/m 21/a"
_symmetry_Int_Tables_number 62

_cell_length_a      7.47600
_cell_length_b      5.76300
_cell_length_c      10.07100
_cell_angle_alpha 90.00000
_cell_angle_beta  90.00000
_cell_angle_gamma 90.00000

loop_
_space_group_symop_id
_space_group_symop_operation_xyz
1 x,y,z
```

```
2 x+1/2,-y+1/2,-z+1/2
3 -x,y+1/2,-z
4 -x+1/2,-y,z+1/2
5 -x,-y,-z
6 -x+1/2,y+1/2,z+1/2
7 x,-y+1/2,z
8 x+1/2,y,-z+1/2

loop_
_atom_site_label
_atom_site_type_symbol
_atom_site_symmetry_multiplicity
_atom_site_Wyckoff_label
_atom_site_fract_x
_atom_site_fract_y
_atom_site_fract_z
_atom_site_occupancy
K1 K 4 c 0.82623 0.25000 0.21062 1.00000
K2 K 4 c 0.51104 0.25000 0.20406 1.00000
O1 O 4 c 0.46210 0.25000 -0.08340 1.00000
O2 O 4 c 0.20370 0.25000 0.05820 1.00000
S1 S 4 c 0.26702 0.25000 -0.08029 1.00000
O3 O 8 d 0.19910 0.04120 0.85220 1.00000
```

Arcanite ($K_2SO_4$, $H1_6$): A2B4C_oP28_62_2c_2cd_c - POSCAR

```
A2B4C_oP28_62_2c_2cd_c & a,b/a,c/a,x1,z1,x2,z2,x3,z3,x4,z4,x5,z5,x6,y6,
  ↪ z6 --params=7.476,0.770866773676,1.34711075441,0.82623,0.21062,
  ↪ 0.51104,0.20406,0.4621,-0.0834,0.2037,0.0582,0.26702,-0.08029,
  ↪ 0.1991,0.0412,0.8522 & Pnma D_{2h}^{16} #62 (c^5d) & oP28 &
  ↪ $H1_{6}$ & K2O4S & Arcanite & J. A. {McGinnety}, Acta
  ↪ Crystallogr. Sect. B Struct. Sci. 28, 2845-2852 (1972)
   1.0000000000000
   7.4760000000000    0.0000000000000    0.0000000000000
   0.0000000000000    5.7630000000000    0.0000000000000
   0.0000000000000    0.0000000000000   10.0710000000000
     K     O     S
     8    16     4
Direct
   0.8262300000000    0.2500000000000    0.2106200000000   K   (4c)
  -0.3262300000000    0.7500000000000    0.7106200000000   K   (4c)
  -0.8262300000000    0.7500000000000   -0.2106200000000   K   (4c)
   1.3262300000000    0.2500000000000    0.2893800000000   K   (4c)
   0.5110400000000    0.2500000000000    0.2040600000000   K   (4c)
  -0.0110400000000    0.7500000000000    0.7040600000000   K   (4c)
  -0.5110400000000    0.7500000000000   -0.2040600000000   K   (4c)
   1.0110400000000    0.2500000000000    0.2959400000000   K   (4c)
   0.4621000000000    0.2500000000000   -0.0834000000000   O   (4c)
   0.0379000000000    0.7500000000000    0.4166000000000   O   (4c)
  -0.4621000000000    0.7500000000000    0.0834000000000   O   (4c)
   0.9621000000000    0.2500000000000    0.5834000000000   O   (4c)
   0.2037000000000    0.2500000000000    0.0582000000000   O   (4c)
   0.2963000000000    0.7500000000000    0.5582000000000   O   (4c)
  -0.2037000000000    0.7500000000000   -0.0582000000000   O   (4c)
   0.7037000000000    0.2500000000000    0.4418000000000   O   (4c)
   0.1991000000000    0.0412000000000    0.8522000000000   O   (8d)
   0.3009000000000   -0.0412000000000    1.3522000000000   O   (8d)
  -0.1991000000000    0.5412000000000   -0.8522000000000   O   (8d)
   0.6991000000000    0.4588000000000   -0.3522000000000   O   (8d)
  -0.1991000000000   -0.0412000000000   -0.8522000000000   O   (8d)
   0.6991000000000    0.0412000000000   -0.3522000000000   O   (8d)
   0.1991000000000    0.4588000000000    0.8522000000000   O   (8d)
   0.3009000000000    0.5412000000000    1.3522000000000   O   (8d)
   0.2670200000000    0.2500000000000   -0.0802900000000   S   (4c)
   0.2329800000000    0.7500000000000    0.4197100000000   S   (4c)
  -0.2670200000000    0.7500000000000    0.0802900000000   S   (4c)
   0.7670200000000    0.2500000000000    0.5802900000000   S   (4c)
```

Anthophyllite ($Mg_5Fe_2Si_8O_{22}(OH)_2$, $S4_4$): A2B5C22D2E8_oP156_62_d_c2d_2c10d_2c_4d - CIF

```
# CIF file
data_findsym-output
_audit_creation_method FINDSYM

_chemical_name_mineral 'Anthophyllite'
_chemical_formula_sum 'Fe2 Mg5 O22 (OH)2 Si8'

loop_
_publ_author_name
 'E. M. Walitzi'
 'F. Walter'
 'K. Ettinger'
_journal_name_full_name
;
 Zeitschrift f{\"u}r Kristallographie - Crystalline Materials
;
_journal_volume 188
_journal_year 1989
_journal_page_first 237
_journal_page_last 244
_publ_Section_title
;
 Verfeinerung der Kristallstruktur von Anthophyllit vom Ochsenkogel/
    ↪ Gleinalpe, \"{O}sterreich
;

# Found in The American Mineralogist Crystal Structure Database, 2003

_aflow_title 'Anthophyllite (Mg$_{5}$Fe$_{2}$Si$_{8}$O$_{22}$(OH)$_{2}$,
  ↪ $S4_{4}$) Structure'
_aflow_proto 'A2B5C22D2E8_oP156_62_d_c2d_2c10d_2c_4d'
_aflow_params 'a,b/a,c/a,x_{1},z_{1},x_{2},z_{2},x_{3},z_{3},x_{4},z_{4
  ↪ },x_{5},z_{5},x_{6},y_{6},z_{6},x_{7},y_{7},z_{7},x_{8},y_{8},
  ↪ z_{8},x_{9},y_{9},z_{9},x_{10},y_{10},z_{10},x_{11},y_{11},z_{
  ↪ 11},x_{12},y_{12},z_{12},x_{13},y_{13},z_{13},x_{14},y_{14},z_{
  ↪ 14},x_{15},y_{15},z_{15},x_{16},y_{16},z_{16},x_{17},y_{17},z_{
  ↪ 17},x_{18},y_{18},z_{18},x_{19},y_{19},z_{19},x_{20},y_{20},z_{
  ↪ 20},x_{21},y_{21},z_{21},x_{22},y_{22},z_{22}'
_aflow_params_values '18.544,0.972066436583,0.284836065574,0.1255,0.893,
  ↪ 0.7975,0.4595,0.9536,0.778,0.1831,0.5573,0.0697,0.227,0.1239,-
  ↪ 0.0094,0.3897,0.1251,0.1634,0.393,0.125,0.0731,0.8196,0.1825,
  ↪ 0.1635,0.0599,0.0682,0.1636,0.7275,0.1857,0.0773,0.563,0.0631,
  ↪ 0.0772,0.2209,0.1868,-0.0012,0.0743,0.0668,-0.0068,0.7092,
  ↪ 0.1973,0.8822,0.3321,0.0504,0.8882,0.0565,0.2003,0.869,0.8288,
  ↪ 0.0486,0.8596,0.5504,0.2305,0.8345,0.567,0.0186,0.8336,0.2743,
  ↪ 0.2271,-0.0798,0.0637,0.0245,-0.0817,0.7754'
_aflow_Strukturbericht '$S4_{4}$'
_aflow_Pearson 'oP156'

_symmetry_space_group_name_H-M "P 21/n 21/m 21/a"
_symmetry_Int_Tables_number 62

_cell_length_a     18.54400
_cell_length_b     18.02600
_cell_length_c     5.28200
_cell_angle_alpha  90.00000
_cell_angle_beta   90.00000
_cell_angle_gamma  90.00000

loop_
_space_group_symop_id
_space_group_symop_operation_xyz
1 x,y,z
2 x+1/2,-y+1/2,-z+1/2
3 -x,y+1/2,-z
4 -x+1/2,-y,z+1/2
5 -x,-y,-z
6 -x+1/2,y+1/2,z+1/2
7 x,-y+1/2,z
8 x+1/2,y,-z+1/2

loop_
_atom_site_label
_atom_site_type_symbol
_atom_site_symmetry_multiplicity
_atom_site_Wyckoff_label
_atom_site_fract_x
_atom_site_fract_y
_atom_site_fract_z
_atom_site_occupancy
Mg1 Mg 4 c 0.12550 0.25000 0.89300 1.00000
O1 O 4 c 0.79750 0.25000 0.45950 1.00000
O2 O 4 c 0.95360 0.25000 0.77800 1.00000
OH1 OH 4 c 0.18310 0.25000 0.55730 1.00000
OH2 OH 4 c 0.06970 0.25000 0.22700 1.00000
Fe1 Fe 8 d 0.12390 -0.00940 0.38970 1.00000
Mg2 Mg 8 d 0.12510 0.16340 0.39300 1.00000
Mg3 Mg 8 d 0.12500 0.07310 0.81960 1.00000
O3 O 8 d 0.18250 0.16350 0.05990 1.00000
O4 O 8 d 0.06820 0.16360 0.72750 1.00000
O5 O 8 d 0.18570 0.07730 0.56300 1.00000
O6 O 8 d 0.06310 0.07720 0.22090 1.00000
O7 O 8 d 0.18680 -0.00120 0.07430 1.00000
O8 O 8 d 0.06680 -0.00680 0.70920 1.00000
O9 O 8 d 0.19730 0.88220 0.33210 1.00000
O10 O 8 d 0.05040 0.88820 0.05650 1.00000
O11 O 8 d 0.20030 0.86900 0.82880 1.00000
O12 O 8 d 0.04860 0.85960 0.55040 1.00000
Si1 Si 8 d 0.23050 0.83450 0.56700 1.00000
Si2 Si 8 d 0.01860 0.83360 0.27430 1.00000
Si3 Si 8 d 0.22710 -0.07980 0.06370 1.00000
Si4 Si 8 d 0.02450 -0.08170 0.77540 1.00000
```

Anthophyllite ($Mg_5Fe_2Si_8O_{22}(OH)_2$, $S4_4$): A2B5C22D2E8_oP156_62_d_c2d_2c10d_2c_4d - POSCAR

```
A2B5C22D2E8_oP156_62_d_c2d_2c10d_2c_4d & a,b/a,c/a,x1,z1,x2,z2,x3,z3,x4,
  ↪ z4,x5,z5,x6,y6,z6,x7,y7,z7,x8,y8,z8,x9,y9,z9,x10,y10,z10,x11,
  ↪ y11,z11,x12,y12,z12,x13,y13,z13,x14,y14,z14,x15,y15,z15,x16,y16
  ↪ ,z16,x17,y17,z17,x18,y18,z18,x19,y19,z19,x20,y20,z20,x21,y21,
  ↪ z21,x22,y22,z22 --params=18.544,0.972066436583,0.284836065574,
  ↪ 0.1255,0.893,0.7975,0.4595,0.9536,0.778,0.1831,0.5573,0.0697,
  ↪ 0.227,0.1239,-0.0094,0.3897,0.1251,0.1634,0.393,0.125,0.0731,
  ↪ 0.8196,0.1825,0.1635,0.0599,0.0682,0.1636,0.7275,0.1857,0.0773,
  ↪ 0.563,0.0631,0.0772,0.2209,0.1868,-0.0012,0.0743,0.0668,-0.0068
  ↪ ,0.7092,0.1973,0.8822,0.3321,0.0504,0.8882,0.0565,0.2003,0.869,
  ↪ 0.8288,0.0486,0.8596,0.5504,0.2305,0.8345,0.567,0.0186,0.8336,
  ↪ 0.2743,0.2271,-0.0798,0.0637,0.0245,-0.0817,0.7754 & Pnma D_{2h
  ↪ }^{16} #62 (c^5d^17) & oP156 & $S4_{4}$ & Fe2Mg5O22(OH)2Si8 &
  ↪ Anthophyllite & E. M. Walitzi and F. Walter and K. Ettinger,
  ↪ Zeitschrift f{"u}r Kristallographie - Crystalline Materials 188
  ↪ , 237-244 (1989)
   1.0000000000000
  18.5440000000000    0.0000000000000    0.0000000000000
   0.0000000000000   18.0260000000000    0.0000000000000
   0.0000000000000    0.0000000000000    5.2820000000000
    Fe    Mg     O    OH    Si
     8    20    88     8    32
Direct
   0.1239000000000   -0.0094000000000    0.3897000000000   Fe   (8d)
   0.3761000000000    0.0094000000000    0.8897000000000   Fe   (8d)
  -0.1239000000000    0.4906000000000   -0.3897000000000   Fe   (8d)
   0.6239000000000    0.5094000000000    0.1103000000000   Fe   (8d)
  -0.1239000000000    0.0094000000000   -0.3897000000000   Fe   (8d)
   0.6239000000000   -0.0094000000000    0.1103000000000   Fe   (8d)
   0.1239000000000    0.5094000000000    0.3897000000000   Fe   (8d)
   0.3761000000000    0.4906000000000    0.8897000000000   Fe   (8d)
   0.1255000000000    0.2500000000000    0.8930000000000   Mg   (4c)
   0.3745000000000    0.7500000000000    1.3930000000000   Mg   (4c)
  -0.1255000000000    0.7500000000000   -0.8930000000000   Mg   (4c)
   0.6255000000000    0.2500000000000   -0.3930000000000   Mg   (4c)
```

```
 0.12510000000000   0.16340000000000   0.39300000000000  Mg   (8d)
 0.37490000000000  -0.16340000000000   0.89300000000000  Mg   (8d)
-0.12510000000000   0.66340000000000  -0.39300000000000  Mg   (8d)
 0.62510000000000   0.33660000000000   0.10700000000000  Mg   (8d)
-0.12510000000000  -0.16340000000000  -0.39300000000000  Mg   (8d)
 0.62510000000000   0.16340000000000   0.10700000000000  Mg   (8d)
 0.12510000000000   0.33660000000000   0.39300000000000  Mg   (8d)
 0.37490000000000   0.66340000000000   0.89300000000000  Mg   (8d)
 0.12500000000000   0.07310000000000   0.81960000000000  Mg   (8d)
 0.37500000000000  -0.07310000000000   1.31960000000000  Mg   (8d)
-0.12500000000000   0.57310000000000  -0.81960000000000  Mg   (8d)
 0.62500000000000   0.42690000000000  -0.31960000000000  Mg   (8d)
-0.12500000000000  -0.07310000000000  -0.81960000000000  Mg   (8d)
 0.62500000000000   0.07310000000000  -0.31960000000000  Mg   (8d)
 0.12500000000000   0.42690000000000   0.81960000000000  Mg   (8d)
 0.37500000000000   0.57310000000000   1.31960000000000  Mg   (8d)
 0.79750000000000   0.25000000000000   0.45950000000000  O    (4c)
-0.29750000000000   0.75000000000000   0.95950000000000  O    (4c)
-0.79750000000000   0.75000000000000  -0.45950000000000  O    (4c)
 1.29750000000000   0.25000000000000   0.04050000000000  O    (4c)
 0.95360000000000   0.25000000000000   0.77800000000000  O    (4c)
-0.45360000000000   0.75000000000000   1.27800000000000  O    (4c)
-0.95360000000000   0.75000000000000  -0.77800000000000  O    (4c)
 1.45360000000000   0.25000000000000  -0.27800000000000  O    (4c)
 0.18250000000000   0.16350000000000   0.05990000000000  O    (8d)
 0.31750000000000  -0.16350000000000   0.55990000000000  O    (8d)
-0.18250000000000   0.66350000000000  -0.05990000000000  O    (8d)
 0.68250000000000   0.33650000000000   0.44010000000000  O    (8d)
-0.18250000000000  -0.16350000000000  -0.05990000000000  O    (8d)
 0.68250000000000   0.16350000000000   0.44010000000000  O    (8d)
 0.18250000000000   0.33650000000000   0.05990000000000  O    (8d)
 0.31750000000000   0.66350000000000   0.55990000000000  O    (8d)
 0.06820000000000   0.16360000000000   0.72750000000000  O    (8d)
 0.43180000000000  -0.16360000000000   1.22750000000000  O    (8d)
-0.06820000000000   0.66360000000000  -0.72750000000000  O    (8d)
 0.56820000000000   0.33640000000000  -0.22750000000000  O    (8d)
-0.06820000000000  -0.16360000000000  -0.72750000000000  O    (8d)
 0.56820000000000   0.16360000000000  -0.22750000000000  O    (8d)
 0.06820000000000   0.33640000000000   0.72750000000000  O    (8d)
 0.43180000000000   0.66360000000000   1.22750000000000  O    (8d)
 0.18570000000000   0.07730000000000   0.56300000000000  O    (8d)
 0.31430000000000  -0.07730000000000   1.06300000000000  O    (8d)
-0.18570000000000   0.57730000000000  -0.56300000000000  O    (8d)
 0.68570000000000   0.42270000000000  -0.06300000000000  O    (8d)
-0.18570000000000  -0.07730000000000  -0.56300000000000  O    (8d)
 0.68570000000000   0.07730000000000  -0.06300000000000  O    (8d)
 0.18570000000000   0.42270000000000   0.56300000000000  O    (8d)
 0.31430000000000   0.57730000000000   1.06300000000000  O    (8d)
 0.06310000000000   0.07720000000000   0.22090000000000  O    (8d)
 0.43690000000000  -0.07720000000000   0.72090000000000  O    (8d)
-0.06310000000000   0.57720000000000  -0.22090000000000  O    (8d)
 0.56310000000000   0.42280000000000   0.27910000000000  O    (8d)
-0.06310000000000  -0.07720000000000  -0.22090000000000  O    (8d)
 0.56310000000000   0.07720000000000   0.27910000000000  O    (8d)
 0.06310000000000   0.42280000000000   0.22090000000000  O    (8d)
 0.43690000000000   0.57720000000000   0.72090000000000  O    (8d)
 0.18680000000000  -0.00120000000000   0.07430000000000  O    (8d)
 0.31320000000000   0.00120000000000   0.57430000000000  O    (8d)
-0.18680000000000   0.49880000000000  -0.07430000000000  O    (8d)
 0.68680000000000   0.50120000000000   0.42570000000000  O    (8d)
-0.18680000000000   0.00120000000000  -0.07430000000000  O    (8d)
 0.68680000000000  -0.00120000000000   0.42570000000000  O    (8d)
 0.18680000000000   0.50120000000000   0.07430000000000  O    (8d)
 0.31320000000000   0.49880000000000   0.57430000000000  O    (8d)
 0.06680000000000  -0.00680000000000   0.70920000000000  O    (8d)
 0.43320000000000   0.00680000000000   1.20920000000000  O    (8d)
-0.06680000000000   0.49320000000000  -0.70920000000000  O    (8d)
 0.56680000000000   0.50680000000000  -0.20920000000000  O    (8d)
-0.06680000000000   0.00680000000000  -0.70920000000000  O    (8d)
 0.56680000000000  -0.00680000000000  -0.20920000000000  O    (8d)
 0.06680000000000   0.50680000000000   0.70920000000000  O    (8d)
 0.43320000000000   0.49320000000000   1.20920000000000  O    (8d)
 0.19730000000000   0.88220000000000   0.33210000000000  O    (8d)
 0.30270000000000  -0.88220000000000   0.83210000000000  O    (8d)
-0.19730000000000   1.38220000000000  -0.33210000000000  O    (8d)
 0.69730000000000  -0.38220000000000   0.16790000000000  O    (8d)
-0.19730000000000  -0.88220000000000  -0.33210000000000  O    (8d)
 0.69730000000000   0.88220000000000   0.16790000000000  O    (8d)
 0.19730000000000  -0.38220000000000   0.33210000000000  O    (8d)
 0.30270000000000   1.38220000000000   0.83210000000000  O    (8d)
 0.05040000000000   0.88820000000000   0.05650000000000  O    (8d)
 0.44960000000000  -0.88820000000000   0.55650000000000  O    (8d)
-0.05040000000000   1.38820000000000  -0.05650000000000  O    (8d)
 0.55040000000000  -0.38820000000000   0.44350000000000  O    (8d)
-0.05040000000000  -0.88820000000000  -0.05650000000000  O    (8d)
 0.55040000000000   0.88820000000000   0.44350000000000  O    (8d)
 0.05040000000000  -0.38820000000000   0.05650000000000  O    (8d)
 0.44960000000000   1.38820000000000   0.55650000000000  O    (8d)
 0.20030000000000   0.86900000000000   0.82880000000000  O    (8d)
 0.29970000000000  -0.86900000000000   1.32880000000000  O    (8d)
-0.20030000000000   1.36900000000000  -0.82880000000000  O    (8d)
 0.70030000000000  -0.36900000000000  -0.32880000000000  O    (8d)
-0.20030000000000  -0.86900000000000  -0.82880000000000  O    (8d)
 0.70030000000000   0.86900000000000  -0.32880000000000  O    (8d)
 0.20030000000000  -0.36900000000000   0.82880000000000  O    (8d)
 0.29970000000000   1.36900000000000   1.32880000000000  O    (8d)
 0.04860000000000   0.85960000000000   0.55040000000000  O    (8d)
 0.45140000000000  -0.85960000000000   1.05040000000000  O    (8d)
-0.04860000000000   1.35960000000000  -0.55040000000000  O    (8d)
 0.54860000000000  -0.35960000000000  -0.05040000000000  O    (8d)
-0.04860000000000  -0.85960000000000  -0.55040000000000  O    (8d)
 0.54860000000000   0.85960000000000  -0.05040000000000  O    (8d)
 0.04860000000000  -0.35960000000000   0.55040000000000  O    (8d)
 0.45140000000000   1.35960000000000   1.05040000000000  O    (8d)
 0.18310000000000   0.25000000000000   0.55730000000000  OH   (4c)
 0.31690000000000   0.75000000000000   1.05730000000000  OH   (4c)
-0.18310000000000   0.75000000000000  -0.55730000000000  OH   (4c)
 0.68310000000000   0.25000000000000  -0.05730000000000  OH   (4c)
 0.06970000000000   0.25000000000000   0.22700000000000  OH   (4c)
 0.43030000000000   0.75000000000000   0.72700000000000  OH   (4c)
-0.06970000000000   0.75000000000000  -0.22700000000000  OH   (4c)
 0.56970000000000   0.25000000000000   0.27300000000000  OH   (4c)
 0.23050000000000   0.83450000000000   0.56700000000000  Si   (8d)
 0.26950000000000  -0.83450000000000   1.06700000000000  Si   (8d)
-0.23050000000000   1.33450000000000  -0.56700000000000  Si   (8d)
 0.73050000000000  -0.33450000000000  -0.06700000000000  Si   (8d)
-0.23050000000000  -0.83450000000000  -0.56700000000000  Si   (8d)
 0.73050000000000   0.83450000000000  -0.06700000000000  Si   (8d)
 0.23050000000000  -0.33450000000000   0.56700000000000  Si   (8d)
 0.26950000000000   1.33450000000000   1.06700000000000  Si   (8d)
 0.01860000000000   0.83360000000000   0.27430000000000  Si   (8d)
 0.48140000000000  -0.83360000000000   0.77430000000000  Si   (8d)
-0.01860000000000   1.33360000000000  -0.27430000000000  Si   (8d)
 0.51860000000000  -0.33360000000000   0.22570000000000  Si   (8d)
-0.01860000000000  -0.83360000000000  -0.27430000000000  Si   (8d)
 0.51860000000000   0.83360000000000   0.22570000000000  Si   (8d)
 0.01860000000000  -0.33360000000000   0.27430000000000  Si   (8d)
 0.48140000000000   1.33360000000000   0.77430000000000  Si   (8d)
 0.22710000000000  -0.07980000000000   0.06370000000000  Si   (8d)
 0.27290000000000   0.07980000000000   0.56370000000000  Si   (8d)
-0.22710000000000   0.42020000000000  -0.06370000000000  Si   (8d)
 0.72710000000000   0.57980000000000   0.43630000000000  Si   (8d)
-0.22710000000000   0.07980000000000  -0.06370000000000  Si   (8d)
 0.72710000000000  -0.07980000000000   0.43630000000000  Si   (8d)
 0.22710000000000   0.57980000000000   0.06370000000000  Si   (8d)
 0.27290000000000   0.42020000000000   0.56370000000000  Si   (8d)
 0.02450000000000  -0.08170000000000   0.77540000000000  Si   (8d)
 0.47550000000000   0.08170000000000   1.27540000000000  Si   (8d)
-0.02450000000000   0.41830000000000  -0.77540000000000  Si   (8d)
 0.52450000000000   0.58170000000000  -0.27540000000000  Si   (8d)
-0.02450000000000   0.08170000000000  -0.77540000000000  Si   (8d)
 0.52450000000000  -0.08170000000000  -0.27540000000000  Si   (8d)
 0.02450000000000   0.58170000000000   0.77540000000000  Si   (8d)
 0.47550000000000   0.41830000000000   1.27540000000000  Si   (8d)
```

Sillimanite ($Al_2SiO_5$, $S0_3$): A2B5C_oP32_62_bc_3cd_c - CIF

```
# CIF file
data_findsym-output
_audit_creation_method FINDSYM

_chemical_name_mineral 'Sillimanite'
_chemical_formula_sum 'Al2 O5 Si'

loop_
_publ_author_name
 'H. Yang'
 'R. M. Hazen'
 'L. W. Finger'
 'C. T. Prewitt'
 'R. T. Downs'
_journal_name_full_name
;
 Physics and Chemistry of Minerals
;
_journal_volume 25
_journal_year 1997
_journal_page_first 39
_journal_page_last 47
_publ_Section_title
;
 Compressibility and crystal structure of sillimanite, Al$_{2}$SiO$_{5}$
   ↪ , at high pressure
;

_aflow_title 'Sillimanite (Al$_{2}$SiO$_{5}$, $S0_{3}$) Structure'
_aflow_proto 'A2B5C_oP32_62_bc_3cd_c'
_aflow_params 'a,b/a,c/a,x_{2},z_{2},x_{3},z_{3},x_{4},z_{4},x_{5},z_{5
   ↪ },x_{6},z_{6},x_{7},y_{7},z_{7}'
_aflow_params_values '7.675,0.752456026059,0.975335504886,0.1548,0.8583
   ↪ ,-0.0911,0.3602,0.0661,0.6434,0.5009,0.4767,0.8404,0.1532,
   ↪ 0.7232,0.0145,0.1255'
_aflow_Strukturbericht '$S0_{3}$'
_aflow_Pearson 'oP32'

_symmetry_space_group_name_H-M "P 21/n 21/m 21/a"
_symmetry_Int_Tables_number 62

_cell_length_a     7.67500
_cell_length_b     5.77510
_cell_length_c     7.48570
_cell_angle_alpha  90.00000
_cell_angle_beta   90.00000
_cell_angle_gamma  90.00000

loop_
_space_group_symop_id
_space_group_symop_operation_xyz
1 x,y,z
2 x+1/2,-y+1/2,-z+1/2
3 -x,y+1/2,-z
4 -x+1/2,-y,z+1/2
5 -x,-y,-z
6 -x+1/2,y+1/2,z+1/2
7 x,-y+1/2,z
8 x+1/2,y,-z+1/2

loop_
_atom_site_label
_atom_site_type_symbol
```

```
_atom_site_symmetry_multiplicity
_atom_site_Wyckoff_label
_atom_site_fract_x
_atom_site_fract_y
_atom_site_fract_z
_atom_site_occupancy
Al1 Al 4 b 0.00000 0.00000 0.50000 1.00000
Al2 Al 4 c 0.15480 0.25000 0.85830 1.00000
O1 O 4 c -0.09110 0.25000 0.36020 1.00000
O2 O 4 c 0.06610 0.25000 0.64340 1.00000
O3 O 4 c 0.50090 0.25000 0.47670 1.00000
Si1 Si 4 c 0.84040 0.25000 0.15320 1.00000
O4 O 8 d 0.72320 0.01450 0.12550 1.00000
```

Sillimanite ($Al_2SiO_5$, $S0_3$): A2B5C_oP32_62_bc_3cd_c - POSCAR

```
A2B5C_oP32_62_bc_3cd_c & a,b/a,c/a,x2,z2,x3,z3,x4,z4,x5,z5,x6,z6,x7,y7,
    ↪ z7 --params=7.675,0.752456026059,0.975335504886,0.1548,0.8583,-
    ↪ 0.0911,0.3602,0.0661,0.6434,0.5009,0.4767,0.8404,0.1532,0.7232,
    ↪ 0.0145,0.1255 & Pnma D_{2h}^{16} #62 (bc^5d) & oP32 & $S0_{3}$
    ↪ & Al2O5Si & Sillimanite & H. Yang et al., Phys. Chem. Miner. 25
    ↪ , 39-47 (1997)
   1.00000000000000
   7.67500000000000   0.00000000000000   0.00000000000000
   0.00000000000000   5.77510000000000   0.00000000000000
   0.00000000000000   0.00000000000000   7.48570000000000
    Al     O    Si
     8    20     4
Direct
   0.00000000000000   0.00000000000000   0.50000000000000   Al   (4b)
   0.50000000000000   0.00000000000000   0.00000000000000   Al   (4b)
   0.00000000000000   0.50000000000000   0.50000000000000   Al   (4b)
   0.50000000000000   0.50000000000000   0.00000000000000   Al   (4b)
   0.15480000000000   0.25000000000000   0.85830000000000   Al   (4c)
   0.34520000000000   0.75000000000000   1.35830000000000   Al   (4c)
  -0.15480000000000   0.75000000000000  -0.85830000000000   Al   (4c)
   0.65480000000000   0.25000000000000  -0.35830000000000   Al   (4c)
  -0.09110000000000   0.25000000000000   0.36020000000000    O   (4c)
   0.59110000000000   0.75000000000000   0.86020000000000    O   (4c)
   0.09110000000000   0.75000000000000  -0.36020000000000    O   (4c)
   0.40890000000000   0.25000000000000   0.13980000000000    O   (4c)
   0.06610000000000   0.25000000000000   0.64340000000000    O   (4c)
   0.43390000000000   0.75000000000000   1.14340000000000    O   (4c)
  -0.06610000000000   0.75000000000000  -0.64340000000000    O   (4c)
   0.56610000000000   0.25000000000000  -0.14340000000000    O   (4c)
   0.50090000000000   0.25000000000000   0.47670000000000    O   (4c)
  -0.00090000000000   0.75000000000000   0.97670000000000    O   (4c)
  -0.50090000000000   0.75000000000000  -0.47670000000000    O   (4c)
   1.00090000000000   0.25000000000000   0.02330000000000    O   (4c)
   0.72320000000000   0.01450000000000   0.12550000000000    O   (8d)
  -0.22320000000000  -0.01450000000000   0.62550000000000    O   (8d)
  -0.72320000000000   0.51450000000000  -0.12550000000000    O   (8d)
   1.22320000000000   0.48550000000000   0.37450000000000    O   (8d)
  -0.72320000000000  -0.01450000000000  -0.12550000000000    O   (8d)
   1.22320000000000   0.01450000000000   0.37450000000000    O   (8d)
   0.72320000000000   0.48550000000000   0.12550000000000    O   (8d)
  -0.22320000000000   0.51450000000000   0.62550000000000    O   (8d)
   0.84040000000000   0.25000000000000   0.15320000000000   Si   (4c)
  -0.34040000000000   0.75000000000000   0.65320000000000   Si   (4c)
  -0.84040000000000   0.75000000000000  -0.15320000000000   Si   (4c)
   1.34040000000000   0.25000000000000   0.34680000000000   Si   (4c)
```

$K_2S_3O_6$ ($K5_1$): A2B6C3_oP44_62_2c_2c2d_3c - CIF

```
# CIF file
data_findsym-output
_audit_creation_method FINDSYM

_chemical_name_mineral 'K2O6S3'
_chemical_formula_sum 'K2 O6 S3'

loop_
_publ_author_name
 'J. M. Stewart'
 'J. T. Szyma\\'{n}ski'
_journal_name_full_name
;
 Acta Crystallographica Section B: Structural Science
;
_journal_volume 35
_journal_year 1979
_journal_page_first 1967
_journal_page_last 1970
_publ_Section_title
;
 A redetermination of the crystal structure of potassium trithionate,
    ↪ K$_{2}$S$_{3}$O$_{6}$
;

_aflow_title 'K$_{2}$S$_{3}$O$_{6}$ ($K5_{1}$) Structure'
_aflow_proto 'A2B6C3_oP44_62_2c_2c2d_3c'
_aflow_params 'a,b/a,c/a,x_{1},z_{1},x_{2},z_{2},x_{3},z_{3},x_{4},z_{4
    ↪ },x_{5},z_{5},x_{6},z_{6},x_{7},z_{7},x_{8},y_{8},z_{8},x_{9},
    ↪ y_{9},z_{9}'
_aflow_params_values '9.794,0.58627731264,1.3977945681,0.1324,0.4084,
    ↪ 0.1792,0.736,0.8837,0.1626,0.3229,-0.0759,0.0298,0.1757,0.0907,
    ↪ 0.0289,0.3015,0.0293,-0.0807,0.5411,0.779,0.6492,0.541,-0.076'
_aflow_Strukturbericht '$K5_{1}$'
_aflow_Pearson 'oP44'

_symmetry_space_group_name_H-M "P 21/n 21/m 21/a"
_symmetry_Int_Tables_number 62

_cell_length_a     9.79400
_cell_length_b     5.74200
_cell_length_c     13.69000
_cell_angle_alpha  90.00000
_cell_angle_beta   90.00000
_cell_angle_gamma  90.00000

loop_
_space_group_symop_id
_space_group_symop_operation_xyz
1 x,y,z
2 x+1/2,-y+1/2,-z+1/2
3 -x,y+1/2,-z
4 -x+1/2,-y,z+1/2
5 -x,-y,-z
6 -x+1/2,y+1/2,z+1/2
7 x,-y+1/2,z
8 x+1/2,y,-z+1/2

loop_
_atom_site_label
_atom_site_type_symbol
_atom_site_symmetry_multiplicity
_atom_site_Wyckoff_label
_atom_site_fract_x
_atom_site_fract_y
_atom_site_fract_z
_atom_site_occupancy
K1 K 4 c 0.13240 0.25000 0.40840 1.00000
K2 K 4 c 0.17920 0.25000 0.73600 1.00000
O1 O 4 c 0.88370 0.25000 0.16260 1.00000
O2 O 4 c 0.32290 0.25000 -0.07590 1.00000
S1 S 4 c 0.02980 0.25000 0.17570 1.00000
S2 S 4 c 0.09070 0.25000 0.02890 1.00000
S3 S 4 c 0.30150 0.25000 0.02930 1.00000
O3 O 8 d -0.08070 0.54110 0.77900 1.00000
O4 O 8 d 0.64920 0.54100 -0.07600 1.00000
```

$K_2S_3O_6$ ($K5_1$): A2B6C3_oP44_62_2c_2c2d_3c - POSCAR

```
A2B6C3_oP44_62_2c_2c2d_3c & a,b/a,c/a,x1,z1,x2,z2,x3,z3,x4,z4,x5,z5,x6,
    ↪ z6,x7,z7,x8,y8,z8,x9,y9,z9 --params=9.794,0.58627731264,
    ↪ 1.3977945681,0.1324,0.4084,0.1792,0.736,0.8837,0.1626,0.3229,-
    ↪ 0.0759,0.0298,0.1757,0.0907,0.0289,0.3015,0.0293,-0.0807,0.5411
    ↪ ,0.779,0.6492,0.541,-0.076 & Pnma D_{2h}^{16} #62 (c^7d^2) &
    ↪ oP44 & $K5_{1}$ & K2O6S3 & K2O6S3 & J. M. Stewart and J. T.
    ↪ Szyma\'{n}ski, Acta Crystallogr. Sect. B Struct. Sci. 35,
    ↪ 1967-1970 (1979)
   1.00000000000000
   9.79400000000000   0.00000000000000   0.00000000000000
   0.00000000000000   5.74200000000000   0.00000000000000
   0.00000000000000   0.00000000000000  13.69000000000000
     K     O     S
     8    24    12
Direct
   0.13240000000000   0.25000000000000   0.40840000000000    K   (4c)
   0.36760000000000   0.75000000000000   0.90840000000000    K   (4c)
  -0.13240000000000   0.75000000000000  -0.40840000000000    K   (4c)
   0.63240000000000   0.25000000000000   0.09160000000000    K   (4c)
   0.17920000000000   0.25000000000000   0.73600000000000    K   (4c)
   0.32080000000000   0.75000000000000   1.23600000000000    K   (4c)
  -0.17920000000000   0.75000000000000  -0.73600000000000    K   (4c)
   0.67920000000000   0.25000000000000  -0.23600000000000    K   (4c)
   0.88370000000000   0.25000000000000   0.16260000000000    O   (4c)
  -0.38370000000000   0.75000000000000   0.66260000000000    O   (4c)
  -0.88370000000000   0.75000000000000  -0.16260000000000    O   (4c)
   1.38370000000000   0.25000000000000   0.33740000000000    O   (4c)
   0.32290000000000   0.25000000000000  -0.07590000000000    O   (4c)
   0.17710000000000   0.75000000000000   0.42410000000000    O   (4c)
  -0.32290000000000   0.75000000000000   0.07590000000000    O   (4c)
   0.82290000000000   0.25000000000000   0.57590000000000    O   (4c)
  -0.08070000000000   0.54110000000000   0.77900000000000    O   (8d)
   0.58070000000000  -0.54110000000000   1.27900000000000    O   (8d)
   0.08070000000000   1.04110000000000  -0.77900000000000    O   (8d)
   0.41930000000000  -0.04110000000000  -0.27900000000000    O   (8d)
   0.08070000000000  -0.54110000000000  -0.77900000000000    O   (8d)
   0.41930000000000   0.54110000000000  -0.27900000000000    O   (8d)
  -0.08070000000000  -0.04110000000000   0.77900000000000    O   (8d)
   0.58070000000000   1.04110000000000   1.27900000000000    O   (8d)
   0.64920000000000   0.54100000000000  -0.07600000000000    O   (8d)
  -0.14920000000000  -0.54100000000000   0.42400000000000    O   (8d)
  -0.64920000000000   1.04100000000000   0.07600000000000    O   (8d)
   1.14920000000000  -0.04100000000000   0.57600000000000    O   (8d)
  -0.64920000000000  -0.54100000000000   0.07600000000000    O   (8d)
   1.14920000000000   0.54100000000000   0.57600000000000    O   (8d)
   0.64920000000000  -0.04100000000000  -0.07600000000000    O   (8d)
  -0.14920000000000   1.04100000000000   0.42400000000000    O   (8d)
   0.02980000000000   0.25000000000000   0.17570000000000    S   (4c)
   0.47020000000000   0.75000000000000   0.67570000000000    S   (4c)
  -0.02980000000000   0.75000000000000  -0.17570000000000    S   (4c)
   0.52980000000000   0.25000000000000   0.32430000000000    S   (4c)
   0.09070000000000   0.25000000000000   0.02890000000000    S   (4c)
   0.40930000000000   0.75000000000000   0.52890000000000    S   (4c)
  -0.09070000000000   0.75000000000000  -0.02890000000000    S   (4c)
   0.59070000000000   0.25000000000000   0.47110000000000    S   (4c)
   0.30150000000000   0.25000000000000   0.02930000000000    S   (4c)
   0.19850000000000   0.75000000000000   0.52930000000000    S   (4c)
  -0.30150000000000   0.75000000000000  -0.02930000000000    S   (4c)
   0.80150000000000   0.25000000000000   0.47070000000000    S   (4c)
```

Danburite ($CaB_2Si_2O_8$, $S6_3$): A2BC8D2_oP52_62_d_c_2c3d_d - CIF

```
# CIF file
data_findsym-output
_audit_creation_method FINDSYM

_chemical_name_mineral 'Danburite'
```

```
_chemical_formula_sum 'B2 Ca O8 Si2'

loop_
_publ_author_name
 'K. Sugiyama'
 'Y. Tak\\'{e}uchi'
_journal_name_full_name
;
 Zeitschrift f{\"u}r Kristallographie - Crystalline Materials
;
_journal_volume 173
_journal_year 1985
_journal_page_first 293
_journal_page_last 304
_publ_Section_title
;
 Unusual thermal expansion of a B-O bond in the structure of danburite
    ↪ CaB$_{2}$Si$_{2}$O$_{8}$
;

# Found in The American Mineralogist Crystal Structure Database, 2003

_aflow_title 'Danburite (CaB$_{2}$Si$_{2}$O$_{8}$, $S6_{3}$) Structure'
_aflow_proto 'A2BC8D2_oP52_62_d_c_2c3d_d'
_aflow_params 'a,b/a,c/a,x_{1},z_{1},x_{2},z_{2},x_{3},z_{3},x_{4},y_{4
    ↪ },z_{4},x_{5},y_{5},z_{5},x_{6},y_{6},z_{6},x_{7},y_{7},z_{7},
    ↪ x_{8},y_{8},z_{8}'
_aflow_params_values '8.037,0.960781386089,1.09120318527,0.61445,0.38657
    ↪ ,0.4863,0.664,0.8162,0.4282,0.7411,0.4206,0.4192,0.80709,-
    ↪ 0.00324,0.06797,0.87368,-0.04233,0.36496,0.60035,0.0782,0.31351
    ↪ ,-0.05333,-0.05574,0.1925'
_aflow_Strukturbericht '$S6_{3}$'
_aflow_Pearson 'oP52'

_symmetry_space_group_name_H-M "P 21/n 21/m 21/a"
_symmetry_Int_Tables_number 62

_cell_length_a     8.03700
_cell_length_b     7.72180
_cell_length_c     8.77000
_cell_angle_alpha 90.00000
_cell_angle_beta  90.00000
_cell_angle_gamma 90.00000

loop_
_space_group_symop_id
_space_group_symop_operation_xyz
1 x,y,z
2 x+1/2,-y+1/2,-z+1/2
3 -x,y+1/2,-z
4 -x+1/2,-y,z+1/2
5 -x,-y,-z
6 -x+1/2,y+1/2,z+1/2
7 x,-y+1/2,z
8 x+1/2,y,-z+1/2

loop_
_atom_site_label
_atom_site_type_symbol
_atom_site_symmetry_multiplicity
_atom_site_Wyckoff_label
_atom_site_fract_x
_atom_site_fract_y
_atom_site_fract_z
_atom_site_occupancy
Ca1 Ca 4 c 0.61445 0.25000 0.38657 1.00000
O1 O 4 c 0.48630 0.25000 0.66400 1.00000
O2 O 4 c 0.81620 0.25000 0.42820 1.00000
B1 B 8 d 0.74110 0.42060 0.41920 1.00000
O3 O 8 d 0.80709 -0.00324 0.06797 1.00000
O4 O 8 d 0.87368 -0.04233 0.36496 1.00000
O5 O 8 d 0.60035 0.07820 0.31351 1.00000
Si1 Si 8 d -0.05333 -0.05574 0.19250 1.00000
```

Danburite ($CaB_2Si_2O_8$, $S6_3$): A2BC8D2_oP52_62_d_c_2c3d_d - POSCAR

```
A2BC8D2_oP52_62_d_c_2c3d_d & a,b/a,c/a,x1,z1,x2,z2,x3,z3,x4,y4,z4,x5,y5,
    ↪ z5,x6,y6,z6,x7,y7,z7,x8,y8,z8 --params=8.037,0.960781386089,
    ↪ 1.09120318527,0.61445,0.38657,0.4863,0.664,0.8162,0.4282,0.7411
    ↪ ,0.4206,0.4192,0.80709,-0.00324,0.06797,0.87368,-0.04233,
    ↪ 0.36496,0.60035,0.0782,0.31351,-0.05333,-0.05574,0.1925 & Pnma
    ↪ D_{2h}^{16} #62 (c^3d^5) & oP52 & $S6_{3}$ & B2CaO8Si2 &
    ↪ Danburite & K. Sugiyama and Y. Tak\'{e}uchi, Zeitschrift f{"u}r
    ↪  Kristallographie - Crystalline Materials 173, 293-304 (1985)
   1.00000000000000
   8.03700000000000   0.00000000000000   0.00000000000000
   0.00000000000000   7.72180000000000   0.00000000000000
   0.00000000000000   0.00000000000000   8.77000000000000
     B     Ca      O     Si
     8      4     32      8
Direct
   0.74110000000000   0.42060000000000   0.41920000000000   B   (8d)
  -0.24110000000000  -0.42060000000000   0.91920000000000   B   (8d)
  -0.74110000000000   0.92060000000000  -0.41920000000000   B   (8d)
   1.24110000000000   0.07940000000000   0.08080000000000   B   (8d)
  -0.74110000000000  -0.42060000000000  -0.41920000000000   B   (8d)
   1.24110000000000   0.42060000000000   0.08080000000000   B   (8d)
   0.74110000000000   0.07940000000000   0.41920000000000   B   (8d)
  -0.24110000000000   0.92060000000000   0.91920000000000   B   (8d)
   0.61445000000000   0.25000000000000   0.38657000000000   Ca  (4c)
  -0.11445000000000   0.75000000000000   0.88657000000000   Ca  (4c)
  -0.61445000000000   0.75000000000000  -0.38657000000000   Ca  (4c)
   1.11445000000000   0.25000000000000   0.11343000000000   Ca  (4c)
   0.48630000000000   0.25000000000000   0.66400000000000   O   (4c)
   0.01370000000000   0.75000000000000   1.16400000000000   O   (4c)
  -0.48630000000000   0.75000000000000  -0.66400000000000   O   (4c)
   0.98630000000000   0.25000000000000  -0.16400000000000   O   (4c)
   0.81620000000000   0.25000000000000   0.42820000000000   O   (4c)
  -0.31620000000000   0.75000000000000   0.92820000000000   O   (4c)
  -0.81620000000000   0.75000000000000  -0.42820000000000   O   (4c)
   1.31620000000000   0.25000000000000   0.07180000000000   O   (4c)
   0.80709000000000  -0.00324000000000   0.06797000000000   O   (8d)
  -0.30709000000000   0.00324000000000   0.56797000000000   O   (8d)
  -0.80709000000000   0.49676000000000  -0.06797000000000   O   (8d)
   1.30709000000000   0.50324000000000   0.43203000000000   O   (8d)
  -0.80709000000000   0.00324000000000  -0.06797000000000   O   (8d)
   1.30709000000000  -0.00324000000000   0.43203000000000   O   (8d)
   0.80709000000000   0.50324000000000   0.06797000000000   O   (8d)
  -0.30709000000000   0.49676000000000   0.56797000000000   O   (8d)
   0.87368000000000  -0.04233000000000   0.36496000000000   O   (8d)
  -0.37368000000000   0.04233000000000   0.86496000000000   O   (8d)
  -0.87368000000000   0.45767000000000  -0.36496000000000   O   (8d)
   1.37368000000000   0.54233000000000   0.13504000000000   O   (8d)
  -0.87368000000000   0.04233000000000  -0.36496000000000   O   (8d)
   1.37368000000000  -0.04233000000000   0.13504000000000   O   (8d)
   0.87368000000000   0.54233000000000   0.36496000000000   O   (8d)
  -0.37368000000000   0.45767000000000   0.86496000000000   O   (8d)
   0.60035000000000   0.07820000000000   0.31351000000000   O   (8d)
  -0.10035000000000  -0.07820000000000   0.81351000000000   O   (8d)
  -0.60035000000000   0.57820000000000  -0.31351000000000   O   (8d)
   1.10035000000000   0.42180000000000   0.18649000000000   O   (8d)
  -0.60035000000000  -0.07820000000000  -0.31351000000000   O   (8d)
   1.10035000000000   0.07820000000000   0.18649000000000   O   (8d)
   0.60035000000000   0.42180000000000   0.31351000000000   O   (8d)
  -0.10035000000000   0.57820000000000   0.81351000000000   O   (8d)
  -0.05333000000000  -0.05574000000000   0.19250000000000   Si  (8d)
   0.55333000000000   0.05574000000000   0.69250000000000   Si  (8d)
   0.05333000000000   0.44426000000000  -0.19250000000000   Si  (8d)
   0.44667000000000   0.55574000000000   0.30750000000000   Si  (8d)
   0.05333000000000   0.05574000000000  -0.19250000000000   Si  (8d)
   0.44667000000000  -0.05574000000000   0.30750000000000   Si  (8d)
  -0.05333000000000   0.55574000000000   0.19250000000000   Si  (8d)
   0.55333000000000   0.44426000000000   0.69250000000000   Si  (8d)
```

C53 ($SrBr_2$) (*obsolete*): A2B_oP12_62_2c_c - CIF

```
# CIF file
data_findsym-output
_audit_creation_method FINDSYM

_chemical_name_mineral 'Br2Sr'
_chemical_formula_sum 'Br2 Sr'

loop_
_publ_author_name
 'M. A. Kamermans'
_journal_name_full_name
;
 Zeitschrift f{\"u}r Kristallographie - Crystalline Materials
;
_journal_volume 101
_journal_year 1939
_journal_page_first 406
_journal_page_last 411
_publ_Section_title
;
 The Crystal Structure of SrBr$_{2}$
;

# Found in Strukturbericht Band VII 1939, 1943

_aflow_title '$C53$ (SrBr$_{2}$) ({\em{obsolete}}) Structure'
_aflow_proto 'A2B_oP12_62_2c_c'
_aflow_params 'a,b/a,c/a,x_{1},z_{1},x_{2},z_{2},x_{3},z_{3}'
_aflow_params_values '11.42,0.376532399299,0.805604203152,0.103,0.119,
    ↪ 0.614,0.842,0.811,0.108'
_aflow_Strukturbericht '$C53$'
_aflow_Pearson 'oP12'

_symmetry_space_group_name_H-M "P 21/n 21/m 21/a"
_symmetry_Int_Tables_number 62

_cell_length_a    11.42000
_cell_length_b     4.30000
_cell_length_c     9.20000
_cell_angle_alpha 90.00000
_cell_angle_beta  90.00000
_cell_angle_gamma 90.00000

loop_
_space_group_symop_id
_space_group_symop_operation_xyz
1 x,y,z
2 x+1/2,-y+1/2,-z+1/2
3 -x,y+1/2,-z
4 -x+1/2,-y,z+1/2
5 -x,-y,-z
6 -x+1/2,y+1/2,z+1/2
7 x,-y+1/2,z
8 x+1/2,y,-z+1/2

loop_
_atom_site_label
_atom_site_type_symbol
_atom_site_symmetry_multiplicity
_atom_site_Wyckoff_label
_atom_site_fract_x
_atom_site_fract_y
_atom_site_fract_z
_atom_site_occupancy
```

```
Br1  Br  4  c  0.10300  0.25000  0.11900  1.00000
Br2  Br  4  c  0.61400  0.25000  0.84200  1.00000
Sr1  Sr  4  c  0.81100  0.25000  0.10800  1.00000
```

C53 ($SrBr_2$) (*obsolete*): A2B_oP12_62_2c_c - POSCAR

```
A2B_oP12_62_2c_c & a,b/a,c/a,x1,z1,x2,z2,x3,z3 --params=11.42,
    ↪ 0.376532399299,0.805604203152,0.103,0.119,0.614,0.842,0.811,
    ↪ 0.108 & Pnma D_{2h}^{16} #62 (c^3) & oP12 & $C53$ & Br2Sr &
    ↪ Br2Sr & M. A. Kamermans, Zeitschrift f{"u}r Kristallographie -
    ↪ Crystalline Materials 101, 406-411 (1939)
   1.00000000000000
  11.42000000000000   0.00000000000000   0.00000000000000
   0.00000000000000   4.30000000000000   0.00000000000000
   0.00000000000000   0.00000000000000   9.20000000000000
    Br     Sr
     8      4
Direct
   0.10300000000000   0.25000000000000   0.11900000000000   Br   (4c)
   0.39700000000000   0.75000000000000   0.61900000000000   Br   (4c)
  -0.10300000000000   0.75000000000000  -0.11900000000000   Br   (4c)
   0.60300000000000   0.25000000000000   0.38100000000000   Br   (4c)
   0.61400000000000   0.25000000000000   0.84200000000000   Br   (4c)
  -0.11400000000000   0.75000000000000   1.34200000000000   Br   (4c)
  -0.61400000000000   0.75000000000000  -0.84200000000000   Br   (4c)
   1.11400000000000   0.25000000000000  -0.34200000000000   Br   (4c)
   0.81100000000000   0.25000000000000   0.10800000000000   Sr   (4c)
  -0.31100000000000   0.75000000000000   0.60800000000000   Sr   (4c)
  -0.81100000000000   0.75000000000000  -0.10800000000000   Sr   (4c)
   1.31100000000000   0.25000000000000   0.39200000000000   Sr   (4c)
```

$Cs_2Sb$: A2B_oP24_62_4c_2c - CIF

```
# CIF file
data_findsym-output
_audit_creation_method FINDSYM

_chemical_name_mineral 'Cs2Sb'
_chemical_formula_sum 'Cs2 Sb'

loop_
_publ_author_name
 'C. Hirschle'
 'C. R\"{o}hr'
_journal_name_full_name
;
 Zeitschrift fur Anorganische und Allgemeine Chemie
;
_journal_volume 626
_journal_year 2000
_journal_page_first 1992
_journal_page_last 1998
_publ_Section_title
;
 Darstellung und Kristallstruktur der bekannten Zintl-Phasen Cs$_{3}
    ↪ $Sb$_{7}$ und Cs$_{4}$Sb$_{2}$
;

# Found in PAULING FILE in: Inorganic Solid Phases (online database),
    ↪ 2016 Found in PAULING FILE in: Inorganic Solid Phases (online
    ↪ database), {Cs$_{4}$Sb$_{2}$ (Cs$_{2}$Sb) Crystal Structure},

_aflow_title 'Cs$_{2}$Sb Structure'
_aflow_proto 'A2B_oP24_62_4c_2c'
_aflow_params 'a,b/a,c/a,x_{1},z_{1},x_{2},z_{2},x_{3},z_{3},x_{4},z_{4
    ↪ },x_{5},z_{5},x_{6},z_{6}'
_aflow_params_values '15.985,0.395308101345,0.687832342821,0.00065,
    ↪ 0.6743,0.20442,0.40591,0.25609,0.73979,0.43929,0.41364,0.16182,
    ↪ 0.06288,0.34361,0.09158'
_aflow_Strukturbericht 'None'
_aflow_Pearson 'oP24'

_symmetry_space_group_name_H-M "P 21/n 21/m 21/a"
_symmetry_Int_Tables_number 62

_cell_length_a      15.98500
_cell_length_b      6.31900
_cell_length_c      10.99500
_cell_angle_alpha  90.00000
_cell_angle_beta   90.00000
_cell_angle_gamma  90.00000

loop_
_space_group_symop_id
_space_group_symop_operation_xyz
1 x,y,z
2 x+1/2,-y+1/2,-z+1/2
3 -x,y+1/2,-z
4 -x+1/2,-y,z+1/2
5 -x,-y,-z
6 -x+1/2,y+1/2,z+1/2
7 x,-y+1/2,z
8 x+1/2,y,-z+1/2

loop_
_atom_site_label
_atom_site_type_symbol
_atom_site_symmetry_multiplicity
_atom_site_Wyckoff_label
_atom_site_fract_x
_atom_site_fract_y
_atom_site_fract_z
_atom_site_occupancy
Cs1  Cs  4  c  0.00065  0.25000  0.67430  1.00000
Cs2  Cs  4  c  0.20442  0.25000  0.40591  1.00000
Cs3  Cs  4  c  0.25609  0.25000  0.73979  1.00000
Cs4  Cs  4  c  0.43929  0.25000  0.41364  1.00000
Sb1  Sb  4  c  0.16182  0.25000  0.06288  1.00000
Sb2  Sb  4  c  0.34361  0.25000  0.09158  1.00000
```

$Cs_2Sb$: A2B_oP24_62_4c_2c - POSCAR

```
A2B_oP24_62_4c_2c & a,b/a,c/a,x1,z1,x2,z2,x3,z3,x4,z4,x5,z5,x6,z6 --
    ↪ params=15.985,0.395308101345,0.687832342821,0.00065,0.6743,
    ↪ 0.20442,0.40591,0.25609,0.73979,0.43929,0.41364,0.16182,0.06288
    ↪ ,0.34361,0.09158 & Pnma D_{2h}^{16} #62 (c^6) & oP24 & None &
    ↪ Cs2Sb & Cs2Sb & C. Hirschle and C. R\"{o}hr, Z. Anorg. Allg.
    ↪ Chem. 626, 1992-1998 (2000)
   1.00000000000000
  15.98500000000000   0.00000000000000   0.00000000000000
   0.00000000000000   6.31900000000000   0.00000000000000
   0.00000000000000   0.00000000000000  10.99500000000000
    Cs     Sb
    16      8
Direct
   0.00065000000000   0.25000000000000   0.67430000000000   Cs   (4c)
   0.49935000000000   0.75000000000000   1.17430000000000   Cs   (4c)
  -0.00065000000000   0.75000000000000  -0.67430000000000   Cs   (4c)
   0.50065000000000   0.25000000000000  -0.17430000000000   Cs   (4c)
   0.20442000000000   0.25000000000000   0.40591000000000   Cs   (4c)
   0.29558000000000   0.75000000000000   0.90591000000000   Cs   (4c)
  -0.20442000000000   0.75000000000000  -0.40591000000000   Cs   (4c)
   0.70442000000000   0.25000000000000   0.09409000000000   Cs   (4c)
   0.25609000000000   0.25000000000000   0.73979000000000   Cs   (4c)
   0.24391000000000   0.75000000000000   1.23979000000000   Cs   (4c)
  -0.25609000000000   0.75000000000000  -0.73979000000000   Cs   (4c)
   0.75609000000000   0.25000000000000  -0.23979000000000   Cs   (4c)
   0.43929000000000   0.25000000000000   0.41364000000000   Cs   (4c)
   0.06071000000000   0.75000000000000   0.91364000000000   Cs   (4c)
  -0.43929000000000   0.75000000000000  -0.41364000000000   Cs   (4c)
   0.93929000000000   0.25000000000000   0.08636000000000   Cs   (4c)
   0.16182000000000   0.25000000000000   0.06288000000000   Sb   (4c)
   0.33818000000000   0.75000000000000   0.56288000000000   Sb   (4c)
  -0.16182000000000   0.75000000000000  -0.06288000000000   Sb   (4c)
   0.66182000000000   0.25000000000000   0.43712000000000   Sb   (4c)
   0.34361000000000   0.25000000000000   0.09158000000000   Sb   (4c)
   0.15639000000000   0.75000000000000   0.59158000000000   Sb   (4c)
  -0.34361000000000   0.75000000000000  -0.09158000000000   Sb   (4c)
   0.84361000000000   0.25000000000000   0.40842000000000   Sb   (4c)
```

$RhCl_2(NH_3)_5Cl$ ($J1_8$): A3B15C5D_oP96_62_cd_3c6d_3cd_c - CIF

```
# CIF file
data_findsym-output
_audit_creation_method FINDSYM

_chemical_name_mineral 'Cl3H15N5Rh'
_chemical_formula_sum 'Cl3 H15 N5 Rh'

loop_
_publ_author_name
 'R. S. Evans'
 'E. A. Hopcus'
 'J. Bordner'
 'A. F. Schreiner'
_journal_name_full_name
;
 Journal of Crystal and Molecular Structure
;
_journal_volume 3
_journal_year 1973
_journal_page_first 235
_journal_page_last 245
_publ_Section_title
;
 Molecular and crystal structures of halopentaamminerhodium-(III)
    ↪ complexes, [Rh(NH$_{3}$)$_{5}$Cl]Cl$_{2}$ and [Rh(NH$_{3}$)$_{
    ↪ 5}$Br]Br$_{2}$
;

# Found in Comparisons of $\pi$-bonding and hydrogen bonding in
    ↪ isomorphous compounds: [$M$(NH$_{3}$)$_{5}$Cl]Cl$_{2}$ ($M$=Cr,
    ↪ Co, Rh, Ir, Ru, Os), 1986

_aflow_title 'RhCl$_{2}$(NH$_{3}$)$_{5}$Cl ($J1_{8}$) Structure'
_aflow_proto 'A3B15C5D_oP96_62_cd_3c6d_3cd_c'
_aflow_params 'a,b/a,c/a,x_{1},z_{1},x_{2},z_{2},x_{3},z_{3},x_{4},z_{4
    ↪ },x_{5},z_{5},x_{6},z_{6},x_{7},z_{7},x_{8},z_{8},x_{9},y_{9},
    ↪ z_{9},x_{10},y_{10},z_{10},x_{11},y_{11},z_{11},x_{12},y_{12},
    ↪ z_{12},x_{13},y_{13},z_{13},x_{14},y_{14},z_{14},x_{15},y_{15},
    ↪ z_{15},x_{16},y_{16},z_{16}'
_aflow_params_values '13.36,0.782934131737,0.504565868263,0.4705,-0.0482
    ↪ ,0.771,-0.003,0.527,0.527,0.778,0.337,0.7207,0.3777,0.7041,-
    ↪ 0.0519,0.5046,0.4129,0.6038,0.1805,0.1484,-0.0018,0.8394,0.396,
    ↪ 0.509,-0.06,0.357,0.52,0.739,0.452,0.528,0.798,0.297,0.696,
    ↪ 0.125,0.522,0.807,0.567,0.274,0.809,0.541,0.3975,0.5534,0.8224'
_aflow_Strukturbericht '$J1_{8}$'
_aflow_Pearson 'oP96'

_symmetry_space_group_name_H-M "P 21/n 21/m 21/a"
_symmetry_Int_Tables_number 62

_cell_length_a      13.36000
_cell_length_b      10.46000
_cell_length_c      6.74100
_cell_angle_alpha  90.00000
_cell_angle_beta   90.00000
_cell_angle_gamma  90.00000

loop_
```

```
_space_group_symop_id
_space_group_symop_operation_xyz
1 x,y,z
2 x+1/2,-y+1/2,-z+1/2
3 -x,y+1/2,-z
4 -x+1/2,-y,z+1/2
5 -x,-y,-z
6 -x+1/2,y+1/2,z+1/2
7 x,-y+1/2,z
8 x+1/2,y,-z+1/2

loop_
_atom_site_label
_atom_site_type_symbol
_atom_site_symmetry_multiplicity
_atom_site_Wyckoff_label
_atom_site_fract_x
_atom_site_fract_y
_atom_site_fract_z
_atom_site_occupancy
Cl1 Cl 4 c 0.47050 0.25000 -0.04820 1.00000
H1 H 4 c 0.77100 0.25000 -0.00300 1.00000
H2 H 4 c 0.52700 0.25000 0.52700 1.00000
H3 H 4 c 0.77800 0.25000 0.33700 1.00000
N1 N 4 c 0.72070 0.25000 0.37770 1.00000
N2 N 4 c 0.70410 0.25000 -0.05190 1.00000
N3 N 4 c 0.50460 0.25000 0.41290 1.00000
Rh1 Rh 4 c 0.60380 0.25000 0.18050 1.00000
Cl2 Cl 8 d 0.14840 -0.00180 0.83940 1.00000
H4 H 8 d 0.39600 0.50900 -0.06000 1.00000
H5 H 8 d 0.35700 0.52000 0.73900 1.00000
H6 H 8 d 0.45200 0.52800 0.79800 1.00000
H7 H 8 d 0.29700 0.69600 0.12500 1.00000
H8 H 8 d 0.52200 0.80700 0.56700 1.00000
H9 H 8 d 0.27400 0.80900 0.54100 1.00000
N4 N 8 d 0.39750 0.55340 0.82240 1.00000
```

$RhCl_2(NH_3)_5Cl$ ($J1_8$): A3B15C5D_oP96_62_cd_3c6d_3cd_c - POSCAR

```
A3B15C5D_oP96_62_cd_3c6d_3cd_c & a,b/a,c/a,x1,z1,x2,z2,x3,z3,x4,z4,x5,z5
    ↪ ,x6,z6,x7,z7,x8,z8,x9,y9,z9,x10,y10,z10,x11,y11,z11,x12,y12,z12
    ↪ ,x13,y13,z13,x14,y14,z14,x15,y15,z15,x16,y16,z16 --params=13.36
    ↪ ,0.782934131737,0.504565868263,0.4705,-0.0482,0.771,-0.003,
    ↪ 0.527,0.527,0.778,0.337,0.7207,0.3777,0.7041,-0.0519,0.5046,
    ↪ 0.4129,0.6038,0.1805,0.1484,-0.0018,0.8394,0.396,0.509,-0.06,
    ↪ 0.357,0.52,0.739,0.452,0.528,0.798,0.297,0.696,0.125,0.522,
    ↪ 0.807,0.567,0.274,0.809,0.541,0.3975,0.5534,0.8224 & Pnma D_{2h
    ↪ }^{16} #62 (c^8d^8) & oP96 & $J1_{8}$ & Cl3H15N5Rh & Cl3H15N5Rh
    ↪ & R. S. Evans et al., J. Cryst. Mol. Struct. 3, 235-245 (1973)
   1.00000000000000
  13.36000000000000   0.00000000000000   0.00000000000000
   0.00000000000000  10.46000000000000   0.00000000000000
   0.00000000000000   0.00000000000000   6.74100000000000
    Cl     H     N    Rh
    12    60    20     4
Direct
   0.47050000000000   0.25000000000000  -0.04820000000000   Cl   (4c)
   0.02950000000000   0.75000000000000   0.45180000000000   Cl   (4c)
  -0.47050000000000   0.75000000000000   0.04820000000000   Cl   (4c)
   0.97050000000000   0.25000000000000   0.54820000000000   Cl   (4c)
   0.14840000000000  -0.00180000000000   0.83940000000000   Cl   (8d)
   0.35160000000000   0.00180000000000   1.33940000000000   Cl   (8d)
  -0.14840000000000   0.49820000000000  -0.83940000000000   Cl   (8d)
   0.64840000000000   0.50180000000000  -0.33940000000000   Cl   (8d)
  -0.14840000000000   0.00180000000000  -0.83940000000000   Cl   (8d)
   0.64840000000000  -0.00180000000000  -0.33940000000000   Cl   (8d)
   0.14840000000000   0.50180000000000   0.83940000000000   Cl   (8d)
   0.35160000000000   0.49820000000000   1.33940000000000   Cl   (8d)
   0.77100000000000   0.25000000000000  -0.00300000000000   H    (4c)
  -0.27100000000000   0.75000000000000   0.49700000000000   H    (4c)
  -0.77100000000000   0.75000000000000   0.00300000000000   H    (4c)
   1.27100000000000   0.25000000000000   0.50300000000000   H    (4c)
   0.52700000000000   0.25000000000000   0.52700000000000   H    (4c)
  -0.02700000000000   0.75000000000000   1.02700000000000   H    (4c)
  -0.52700000000000   0.75000000000000  -0.52700000000000   H    (4c)
   1.02700000000000   0.25000000000000  -0.02700000000000   H    (4c)
   0.77800000000000   0.25000000000000   0.33700000000000   H    (4c)
  -0.27800000000000   0.75000000000000   0.83700000000000   H    (4c)
  -0.77800000000000   0.75000000000000  -0.33700000000000   H    (4c)
   1.27800000000000   0.25000000000000   0.16300000000000   H    (4c)
   0.39600000000000   0.50900000000000  -0.06000000000000   H    (8d)
   0.10400000000000  -0.50900000000000   0.44000000000000   H    (8d)
  -0.39600000000000   1.00900000000000   0.06000000000000   H    (8d)
   0.89600000000000  -0.00900000000000   0.56000000000000   H    (8d)
  -0.39600000000000  -0.50900000000000   0.06000000000000   H    (8d)
   0.89600000000000   0.50900000000000   0.56000000000000   H    (8d)
   0.39600000000000  -0.00900000000000  -0.06000000000000   H    (8d)
   0.10400000000000   1.00900000000000   0.44000000000000   H    (8d)
   0.35700000000000   0.52000000000000   0.73900000000000   H    (8d)
   0.14300000000000  -0.52000000000000   1.23900000000000   H    (8d)
  -0.35700000000000   1.02000000000000  -0.73900000000000   H    (8d)
   0.85700000000000  -0.02000000000000  -0.23900000000000   H    (8d)
  -0.35700000000000  -0.52000000000000  -0.73900000000000   H    (8d)
   0.85700000000000   0.52000000000000  -0.23900000000000   H    (8d)
   0.35700000000000  -0.02000000000000   0.73900000000000   H    (8d)
   0.14300000000000   1.02000000000000   1.23900000000000   H    (8d)
   0.45200000000000   0.52800000000000   0.79800000000000   H    (8d)
   0.04800000000000  -0.52800000000000   1.29800000000000   H    (8d)
  -0.45200000000000   1.02800000000000  -0.79800000000000   H    (8d)
   0.95200000000000  -0.02800000000000  -0.29800000000000   H    (8d)
  -0.45200000000000  -0.52800000000000  -0.79800000000000   H    (8d)
   0.95200000000000   0.52800000000000  -0.29800000000000   H    (8d)
   0.45200000000000  -0.02800000000000   0.79800000000000   H    (8d)
   0.04800000000000   1.02800000000000   1.29800000000000   H    (8d)
   0.29700000000000   0.69600000000000   0.12500000000000   H    (8d)
   0.20300000000000  -0.69600000000000   0.62500000000000   H    (8d)
  -0.29700000000000   1.19600000000000  -0.12500000000000   H    (8d)
   0.79700000000000  -0.19600000000000   0.37500000000000   H    (8d)
  -0.29700000000000  -0.69600000000000  -0.12500000000000   H    (8d)
   0.79700000000000   0.69600000000000   0.37500000000000   H    (8d)
   0.29700000000000  -0.19600000000000   0.12500000000000   H    (8d)
   0.20300000000000   1.19600000000000   0.62500000000000   H    (8d)
   0.52200000000000   0.80700000000000   0.56700000000000   H    (8d)
  -0.02200000000000  -0.80700000000000   1.06700000000000   H    (8d)
  -0.52200000000000   1.30700000000000  -0.56700000000000   H    (8d)
   1.02200000000000  -0.30700000000000  -0.06700000000000   H    (8d)
  -0.52200000000000  -0.80700000000000  -0.56700000000000   H    (8d)
   1.02200000000000   0.80700000000000  -0.06700000000000   H    (8d)
   0.52200000000000  -0.30700000000000   0.56700000000000   H    (8d)
  -0.02200000000000   1.30700000000000   1.06700000000000   H    (8d)
   0.27400000000000   0.80900000000000   0.54100000000000   H    (8d)
   0.22600000000000  -0.80900000000000   1.04100000000000   H    (8d)
  -0.27400000000000   1.30900000000000  -0.54100000000000   H    (8d)
   0.77400000000000  -0.30900000000000  -0.04100000000000   H    (8d)
  -0.27400000000000  -0.80900000000000  -0.54100000000000   H    (8d)
   0.77400000000000   0.80900000000000  -0.04100000000000   H    (8d)
   0.27400000000000  -0.30900000000000   0.54100000000000   H    (8d)
   0.22600000000000   1.30900000000000   1.04100000000000   H    (8d)
   0.72070000000000   0.25000000000000   0.37770000000000   N    (4c)
  -0.22070000000000   0.75000000000000   0.87770000000000   N    (4c)
  -0.72070000000000   0.75000000000000  -0.37770000000000   N    (4c)
   1.22070000000000   0.25000000000000   0.12230000000000   N    (4c)
   0.70410000000000   0.25000000000000  -0.05190000000000   N    (4c)
  -0.20410000000000   0.75000000000000   0.44810000000000   N    (4c)
  -0.70410000000000   0.75000000000000   0.05190000000000   N    (4c)
   1.20410000000000   0.25000000000000   0.55190000000000   N    (4c)
   0.50460000000000   0.25000000000000   0.41290000000000   N    (4c)
  -0.00460000000000   0.75000000000000   0.91290000000000   N    (4c)
  -0.50460000000000   0.75000000000000  -0.41290000000000   N    (4c)
   1.00460000000000   0.25000000000000   0.08710000000000   N    (4c)
   0.39750000000000   0.55340000000000   0.82240000000000   N    (8d)
   0.10250000000000  -0.55340000000000   1.32240000000000   N    (8d)
  -0.39750000000000   1.05340000000000  -0.82240000000000   N    (8d)
   0.89750000000000  -0.05340000000000  -0.32240000000000   N    (8d)
  -0.39750000000000  -0.55340000000000  -0.82240000000000   N    (8d)
   0.89750000000000   0.55340000000000  -0.32240000000000   N    (8d)
   0.39750000000000  -0.05340000000000   0.82240000000000   N    (8d)
   0.10250000000000   1.05340000000000   1.32240000000000   N    (8d)
   0.60380000000000   0.25000000000000   0.18050000000000   Rh   (4c)
  -0.10380000000000   0.75000000000000   0.68050000000000   Rh   (4c)
  -0.60380000000000   0.75000000000000  -0.18050000000000   Rh   (4c)
   1.10380000000000   0.25000000000000   0.31950000000000   Rh   (4c)
```

$NH_4I_3$ ($D0_{16}$): A3B_oP16_62_3c_c - CIF

```
# CIF file
data_findsym-output
_audit_creation_method FINDSYM

_chemical_name_mineral 'I3(NH4)'
_chemical_formula_sum 'I3 (NH4)'

loop_
_publ_author_name
 'G. H. Cheesman'
 'A. J. T. Finney'
_journal_name_full_name
;
 Acta Crystallographica Section B: Structural Science
;
_journal_volume 26
_journal_year 1970
_journal_page_first 904
_journal_page_last 906
_publ_Section_title
;
 Refinement of the structure of ammonium triiodide, NH$_{4}$I$_{3}$
;

# Found in The Structure and Stability of Simple Tri-Iodides, 1973 Found
    ↪ in The Structure and Stability of Simple Tri-Iodides, {Ph.D.
    ↪ Thesis, University of Tasmania},

_aflow_title 'NH$_{4}$I$_{3}$ ($D0_{16}$) Structure'
_aflow_proto 'A3B_oP16_62_3c_c'
_aflow_params 'a,b/a,c/a,x_{1},z_{1},x_{2},z_{2},x_{3},z_{3},x_{4},z_{4
    ↪ }'
_aflow_params_values '10.819,0.613735095665,0.89305850818,0.84315,
    ↪ 0.15287,0.6188,-0.04894,0.42157,0.76482,0.16484,0.02957'
_aflow_Strukturbericht '$D0_{16}$'
_aflow_Pearson 'oP16'

_symmetry_space_group_name_H-M "P 21/n 21/m 21/a"
_symmetry_Int_Tables_number 62

_cell_length_a    10.81900
_cell_length_b    6.64000
_cell_length_c    9.66200
_cell_angle_alpha 90.00000
_cell_angle_beta  90.00000
_cell_angle_gamma 90.00000

loop_
_space_group_symop_id
_space_group_symop_operation_xyz
1 x,y,z
2 x+1/2,-y+1/2,-z+1/2
3 -x,y+1/2,-z
4 -x+1/2,-y,z+1/2
5 -x,-y,-z
```

```
6 -x+1/2,y+1/2,z+1/2
7 x,-y+1/2,z
8 x+1/2,y,-z+1/2

loop_
_atom_site_label
_atom_site_type_symbol
_atom_site_symmetry_multiplicity
_atom_site_Wyckoff_label
_atom_site_fract_x
_atom_site_fract_y
_atom_site_fract_z
_atom_site_occupancy
I1 I 4 c 0.84315 0.25000 0.15287 1.00000
I2 I 4 c 0.61880 0.25000 -0.04894 1.00000
I3 I 4 c 0.42157 0.25000 0.76482 1.00000
NH41 NH4 4 c 0.16484 0.25000 0.02957 1.00000
```

$NH_4I_3$ ($D0_{16}$): A3B_oP16_62_3c_c - POSCAR

```
A3B_oP16_62_3c_c & a,b/a,c/a,x1,z1,x2,z2,x3,z3,x4,z4 --params=10.819,
    ↪ 0.613735095665,0.89305850818,0.84315,0.15287,0.6188,-0.04894,
    ↪ 0.42157,0.76482,0.16484,0.02957 & Pnma D_{2h}^{16} #62 (c^4) &
    ↪ oP16 & $D0_{16}$ & I3(NH4) & I3(NH4) & G. H. Cheesman and A. J.
    ↪ T. Finney, Acta Crystallogr. Sect. B Struct. Sci. 26, 904-906
    ↪ (1970)
   1.00000000000000
  10.81900000000000   0.00000000000000   0.00000000000000
   0.00000000000000   6.64000000000000   0.00000000000000
   0.00000000000000   0.00000000000000   9.66200000000000
     I     NH4
    12       4
Direct
   0.84315000000000   0.25000000000000   0.15287000000000   I    (4c)
  -0.34315000000000   0.75000000000000   0.65287000000000   I    (4c)
  -0.84315000000000   0.75000000000000  -0.15287000000000   I    (4c)
   1.34315000000000   0.25000000000000   0.34713000000000   I    (4c)
   0.61880000000000   0.25000000000000  -0.04894000000000   I    (4c)
  -0.11880000000000   0.75000000000000   0.45106000000000   I    (4c)
  -0.61880000000000   0.75000000000000   0.04894000000000   I    (4c)
   1.11880000000000   0.25000000000000   0.54894000000000   I    (4c)
   0.42157000000000   0.25000000000000   0.76482000000000   I    (4c)
   0.07843000000000   0.75000000000000   1.26482000000000   I    (4c)
  -0.42157000000000   0.75000000000000  -0.76482000000000   I    (4c)
   0.92157000000000   0.25000000000000  -0.26482000000000   I    (4c)
   0.16484000000000   0.25000000000000   0.02957000000000   NH4  (4c)
   0.33516000000000   0.75000000000000   0.52957000000000   NH4  (4c)
  -0.16484000000000   0.75000000000000  -0.02957000000000   NH4  (4c)
   0.66484000000000   0.25000000000000   0.47043000000000   NH4  (4c)
```

Original $\beta$-$WO_3$ (*obsolete*): A3B_oP32_62_ab4c_2c - CIF

```
# CIF file
data_findsym-output
_audit_creation_method FINDSYM

_chemical_name_mineral 'O3W'
_chemical_formula_sum 'O3 W'

loop_
_publ_author_name
 'P. M. Woodward'
 'A. W. Sleight'
 'T. Vogt'
_journal_name_full_name
;
 Journal of Solid State Chemistry
;
_journal_volume 131
_journal_year 1997
_journal_page_first 9
_journal_page_last 17
_publ_Section_title
;
 Ferroelectric Tungsten Trioxide
;

_aflow_title 'Original $\beta$-WO$_{3}$ ({\em obsolete}) Structure'
_aflow_proto 'A3B_oP32_62_ab4c_2c'
_aflow_params 'a,b/a,c/a,x_{3},z_{3},x_{4},z_{4},x_{5},z_{5},x_{6},z_{6
    ↪ },x_{7},z_{7},x_{8},z_{8}'
_aflow_params_values '7.57,0.969749009247,1.02430647292,0.231,0.473,
    ↪ 0.222,0.029,0.496,0.238,0.485,0.724,0.471,0.469,0.47,-0.032'
_aflow_Strukturbericht 'None'
_aflow_Pearson 'oP32'

_symmetry_space_group_name_H-M "P 21/n 21/m 21/a"
_symmetry_Int_Tables_number 62

_cell_length_a     7.57000
_cell_length_b     7.34100
_cell_length_c     7.75400
_cell_angle_alpha 90.00000
_cell_angle_beta  90.00000
_cell_angle_gamma 90.00000

loop_
_space_group_symop_id
_space_group_symop_operation_xyz
1 x,y,z
2 x+1/2,-y+1/2,-z+1/2
3 -x,y+1/2,-z
4 -x+1/2,-y,z+1/2
5 -x,-y,-z
6 -x+1/2,y+1/2,z+1/2
7 x,-y+1/2,z
8 x+1/2,y,-z+1/2

loop_
_atom_site_label
_atom_site_type_symbol
_atom_site_symmetry_multiplicity
_atom_site_Wyckoff_label
_atom_site_fract_x
_atom_site_fract_y
_atom_site_fract_z
_atom_site_occupancy
O1 O 4 a 0.00000 0.00000 0.00000 1.00000
O2 O 4 b 0.00000 0.00000 0.50000 1.00000
O3 O 4 c 0.23100 0.25000 0.47300 1.00000
O4 O 4 c 0.22200 0.25000 0.02900 1.00000
O5 O 4 c 0.49600 0.25000 0.23800 1.00000
O6 O 4 c 0.48500 0.25000 0.72400 1.00000
W1 W 4 c 0.47100 0.25000 0.46900 1.00000
W2 W 4 c 0.47000 0.25000 -0.03200 1.00000
```

Original $\beta$-$WO_3$ (*obsolete*): A3B_oP32_62_ab4c_2c - POSCAR

```
A3B_oP32_62_ab4c_2c & a,b/a,c/a,x3,z3,x4,z4,x5,z5,x6,z6,x7,z7,x8,z8 --
    ↪ params=7.57,0.969749009247,1.02430647292,0.231,0.473,0.222,
    ↪ 0.029,0.496,0.238,0.485,0.724,0.471,0.469,0.47,-0.032 & Pnma D_
    ↪ {2h}^{16} #62 (abc^6) & oP32 & None & O3W & O3W & P. M.
    ↪ Woodward and A. W. Sleight and T. Vogt, J. Solid State Chem.
    ↪ 131, 9-17 (1997)
   1.00000000000000
   7.57000000000000   0.00000000000000   0.00000000000000
   0.00000000000000   7.34100000000000   0.00000000000000
   0.00000000000000   0.00000000000000   7.75400000000000
     O      W
    24      8
Direct
   0.00000000000000   0.00000000000000   0.00000000000000   O   (4a)
   0.50000000000000   0.00000000000000   0.50000000000000   O   (4a)
   0.00000000000000   0.50000000000000   0.00000000000000   O   (4a)
   0.50000000000000   0.50000000000000   0.50000000000000   O   (4a)
   0.00000000000000   0.00000000000000   0.50000000000000   O   (4b)
   0.50000000000000   0.00000000000000   0.00000000000000   O   (4b)
   0.00000000000000   0.50000000000000   0.50000000000000   O   (4b)
   0.50000000000000   0.50000000000000   0.00000000000000   O   (4b)
   0.23100000000000   0.25000000000000   0.47300000000000   O   (4c)
   0.26900000000000   0.75000000000000   0.97300000000000   O   (4c)
  -0.23100000000000   0.75000000000000  -0.47300000000000   O   (4c)
   0.73100000000000   0.25000000000000   0.02700000000000   O   (4c)
   0.22200000000000   0.25000000000000   0.02900000000000   O   (4c)
   0.27800000000000   0.75000000000000   0.52900000000000   O   (4c)
  -0.22200000000000   0.75000000000000  -0.02900000000000   O   (4c)
   0.72200000000000   0.25000000000000   0.47100000000000   O   (4c)
   0.49600000000000   0.25000000000000   0.23800000000000   O   (4c)
   0.00400000000000   0.75000000000000   0.73800000000000   O   (4c)
  -0.49600000000000   0.75000000000000  -0.23800000000000   O   (4c)
   0.99600000000000   0.25000000000000   0.26200000000000   O   (4c)
   0.48500000000000   0.25000000000000   0.72400000000000   O   (4c)
   0.01500000000000   0.75000000000000   1.22400000000000   O   (4c)
  -0.48500000000000   0.75000000000000  -0.72400000000000   O   (4c)
   0.98500000000000   0.25000000000000  -0.22400000000000   O   (4c)
   0.47100000000000   0.25000000000000   0.46900000000000   W   (4c)
   0.02900000000000   0.75000000000000   0.96900000000000   W   (4c)
  -0.47100000000000   0.75000000000000  -0.46900000000000   W   (4c)
   0.97100000000000   0.25000000000000   0.03100000000000   W   (4c)
   0.47000000000000   0.25000000000000  -0.03200000000000   W   (4c)
   0.03000000000000   0.75000000000000   0.46800000000000   W   (4c)
  -0.47000000000000   0.75000000000000   0.03200000000000   W   (4c)
   0.97000000000000   0.25000000000000   0.53200000000000   W   (4c)
```

$P_4Se_3$: A4B3_oP112_62_8c4d_4c4d - CIF

```
# CIF file
data_findsym-output
_audit_creation_method FINDSYM

_chemical_name_mineral 'P4Se3'
_chemical_formula_sum 'P4 Se3'

loop_
_publ_author_name
 'E. Keulen'
 'A. Vos'
_journal_name_full_name
;
 Acta Cristallographica
;
_journal_volume 12
_journal_year 1959
_journal_page_first 323
_journal_page_last 329
_publ_Section_title
;
 The Crystal Structure of P$_{4}$Se$_{3}$
;

_aflow_title 'P$_{4}$Se$_{3}$ Structure'
_aflow_proto 'A4B3_oP112_62_8c4d_4c4d'
_aflow_params 'a,b/a,c/a,x_{1},z_{1},x_{2},z_{2},x_{3},z_{3},x_{4},z_{4
    ↪ },x_{5},z_{5},x_{6},z_{6},x_{7},z_{7},x_{8},z_{8},x_{9},z_{9},
    ↪ x_{10},z_{10},x_{11},z_{11},x_{12},z_{12},x_{13},y_{13},z_{13},
    ↪ x_{14},y_{14},z_{14},x_{15},y_{15},z_{15},x_{16},y_{16},z_{16},
    ↪ x_{17},y_{17},z_{17},x_{18},y_{18},z_{18},x_{19},y_{19},z_{19},
    ↪ x_{20},y_{20},z_{20}'
_aflow_params_values '11.797,0.825548868356,2.22683733152,0.086,0.627,
    ↪ 0.041,0.5,-0.034,0.282,0.238,0.331,-0.048,0.825,0.239,0.807,
```

```
    ↪ 0.192,0.104,−0.032,0.015,−0.057,0.571,0.0148,0.2567,0.107,0.87,
    ↪ 0.163,0.018,0.227,0.134,0.5875,0.304,0.634,0.6443,−0.01,0.134,
    ↪ 0.7527,−0.039,0.634,0.8663,0.1592,0.072,0.5139,0.8527,0.572,
    ↪ 0.6298,0.1723,0.072,0.7612,0.0602,0.572,−0.0645'
_aflow_Strukturbericht 'None'
_aflow_Pearson 'oP112'

_symmetry_space_group_name_H−M "P 21/n 21/m 21/a"
_symmetry_Int_Tables_number 62

_cell_length_a      11.79700
_cell_length_b      9.73900
_cell_length_c      26.27000
_cell_angle_alpha 90.00000
_cell_angle_beta   90.00000
_cell_angle_gamma 90.00000

loop_
_space_group_symop_id
_space_group_symop_operation_xyz
1 x,y,z
2 x+1/2,−y+1/2,−z+1/2
3 −x,y+1/2,−z
4 −x+1/2,−y,z+1/2
5 −x,−y,−z
6 −x+1/2,y+1/2,z+1/2
7 x,−y+1/2,z
8 x+1/2,y,−z+1/2

loop_
_atom_site_label
_atom_site_type_symbol
_atom_site_symmetry_multiplicity
_atom_site_Wyckoff_label
_atom_site_fract_x
_atom_site_fract_y
_atom_site_fract_z
_atom_site_occupancy
P1 P 4 c 0.08600 0.25000 0.62700 1.00000
P2 P 4 c 0.04100 0.25000 0.50000 1.00000
P3 P 4 c −0.03400 0.25000 0.28200 1.00000
P4 P 4 c 0.23800 0.25000 0.33100 1.00000
P5 P 4 c −0.04800 0.25000 0.82500 1.00000
P6 P 4 c 0.23900 0.25000 0.80700 1.00000
P7 P 4 c 0.19200 0.25000 0.10400 1.00000
P8 P 4 c −0.03200 0.25000 0.01500 1.00000
Se1 Se 4 c −0.05700 0.25000 0.57100 1.00000
Se2 Se 4 c 0.01480 0.25000 0.25670 1.00000
Se3 Se 4 c 0.10700 0.25000 0.87000 1.00000
Se4 Se 4 c 0.16300 0.25000 0.01800 1.00000
P9 P 8 d 0.22700 0.13400 0.58750 1.00000
P10 P 8 d 0.30400 0.63400 0.64430 1.00000
P11 P 8 d −0.01000 0.13400 0.75270 1.00000
P12 P 8 d −0.03900 0.63400 0.86630 1.00000
Se5 Se 8 d 0.15920 0.07200 0.51390 1.00000
Se6 Se 8 d 0.85270 0.57200 0.62980 1.00000
Se7 Se 8 d 0.17230 0.07200 0.76120 1.00000
Se8 Se 8 d 0.06020 0.57200 −0.06450 1.00000
```

$P_4Se_3$: A4B3_oP112_62_8c4d_4c4d - POSCAR

```
A4B3_oP112_62_8c4d_4c4d & a,b/a,c/a,x1,z1,x2,z2,x3,z3,x4,z4,x5,z5,x6,z6,
    ↪ x7,z7,x8,z8,x9,z9,x10,z10,x11,z11,x12,z12,x13,y13,z13,x14,y14,
    ↪ z14,x15,y15,z15,x16,y16,z16,x17,y17,z17,x18,y18,z18,x19,y19,z19
    ↪ ,x20,y20,z20 −−params=11.797,0.825548868356,2.22683733152,0.086
    ↪ ,0.627,0.041,0.5,−0.034,0.282,0.238,0.331,−0.048,0.825,0.239,
    ↪ 0.807,0.192,0.104,−0.032,0.015,−0.057,0.571,0.0148,0.2567,0.107
    ↪ ,0.87,0.163,0.018,0.227,0.134,0.5875,0.304,0.634,0.6443,−0.01,
    ↪ 0.134,0.7527,−0.039,0.634,0.8663,0.1592,0.072,0.5139,0.8527,
    ↪ 0.572,0.6298,0.1723,0.072,0.7612,0.0602,0.572,−0.0645 & Pnma D_
    ↪ {2h}^{16} #62 (c^12d^8) & oP112 & None & P4Se3 & P4Se3 & E.
    ↪ Keulen and A. Vos, Acta Cryst. 12, 323−329 (1959)
   1.00000000000000
  11.79700000000000    0.00000000000000    0.00000000000000
   0.00000000000000    9.73900000000000    0.00000000000000
   0.00000000000000    0.00000000000000   26.27000000000000
     P      Se
    64      48
Direct
   0.08600000000000    0.25000000000000    0.62700000000000    P   (4c)
   0.41400000000000    0.75000000000000    1.12700000000000    P   (4c)
  −0.08600000000000    0.75000000000000   −0.62700000000000    P   (4c)
   0.58600000000000    0.25000000000000   −0.12700000000000    P   (4c)
   0.04100000000000    0.25000000000000    0.50000000000000    P   (4c)
   0.45900000000000    0.75000000000000    1.00000000000000    P   (4c)
  −0.04100000000000    0.75000000000000   −0.50000000000000    P   (4c)
   0.54100000000000    0.25000000000000    0.00000000000000    P   (4c)
  −0.03400000000000    0.25000000000000    0.28200000000000    P   (4c)
   0.53400000000000    0.75000000000000    0.78200000000000    P   (4c)
   0.03400000000000    0.75000000000000   −0.28200000000000    P   (4c)
   0.46600000000000    0.25000000000000    0.21800000000000    P   (4c)
   0.23800000000000    0.25000000000000    0.33100000000000    P   (4c)
   0.26200000000000    0.75000000000000    0.83100000000000    P   (4c)
  −0.23800000000000    0.75000000000000   −0.33100000000000    P   (4c)
   0.73800000000000    0.25000000000000    0.16900000000000    P   (4c)
  −0.04800000000000    0.25000000000000    0.82500000000000    P   (4c)
   0.54800000000000    0.75000000000000    1.32500000000000    P   (4c)
   0.04800000000000    0.75000000000000   −0.82500000000000    P   (4c)
   0.45200000000000    0.25000000000000   −0.32500000000000    P   (4c)
   0.23900000000000    0.25000000000000    0.80700000000000    P   (4c)
   0.26100000000000    0.75000000000000    1.30700000000000    P   (4c)
  −0.23900000000000    0.75000000000000   −0.80700000000000    P   (4c)
   0.73900000000000    0.25000000000000   −0.30700000000000    P   (4c)
   0.19200000000000    0.25000000000000    0.10400000000000    P   (4c)
   0.30800000000000    0.75000000000000    0.60400000000000    P   (4c)
  −0.19200000000000    0.75000000000000   −0.10400000000000    P   (4c)
   0.69200000000000    0.25000000000000    0.39600000000000    P   (4c)
  −0.03200000000000    0.25000000000000    0.01500000000000    P   (4c)
   0.53200000000000    0.75000000000000    0.51500000000000    P   (4c)
   0.03200000000000    0.75000000000000   −0.01500000000000    P   (4c)
   0.46800000000000    0.25000000000000    0.48500000000000    P   (4c)
   0.22700000000000    0.13400000000000    0.58750000000000    P   (8d)
   0.27300000000000   −0.13400000000000    1.08750000000000    P   (8d)
  −0.22700000000000    0.63400000000000   −0.58750000000000    P   (8d)
   0.72700000000000    0.36600000000000   −0.08750000000000    P   (8d)
  −0.22700000000000   −0.13400000000000   −0.58750000000000    P   (8d)
   0.72700000000000    0.13400000000000   −0.08750000000000    P   (8d)
   0.22700000000000    0.36600000000000    0.58750000000000    P   (8d)
   0.27300000000000    0.63400000000000    1.08750000000000    P   (8d)
   0.30400000000000    0.63400000000000    0.64430000000000    P   (8d)
   0.19600000000000   −0.63400000000000    1.14430000000000    P   (8d)
  −0.30400000000000    1.13400000000000   −0.64430000000000    P   (8d)
   0.80400000000000   −0.13400000000000   −0.14430000000000    P   (8d)
  −0.30400000000000   −0.63400000000000   −0.64430000000000    P   (8d)
   0.80400000000000    0.63400000000000   −0.14430000000000    P   (8d)
   0.30400000000000   −0.13400000000000    0.64430000000000    P   (8d)
   0.19600000000000    1.13400000000000    1.14430000000000    P   (8d)
  −0.01000000000000    0.13400000000000    0.75270000000000    P   (8d)
   0.51000000000000   −0.13400000000000    1.25270000000000    P   (8d)
   0.01000000000000    0.63400000000000   −0.75270000000000    P   (8d)
   0.49000000000000    0.36600000000000   −0.25270000000000    P   (8d)
   0.01000000000000   −0.13400000000000   −0.75270000000000    P   (8d)
   0.49000000000000    0.13400000000000   −0.25270000000000    P   (8d)
  −0.01000000000000    0.36600000000000    0.75270000000000    P   (8d)
   0.51000000000000    0.63400000000000    1.25270000000000    P   (8d)
  −0.03900000000000    0.63400000000000    0.86630000000000    P   (8d)
   0.53900000000000   −0.63400000000000    1.36630000000000    P   (8d)
   0.03900000000000    1.13400000000000   −0.86630000000000    P   (8d)
   0.46100000000000   −0.13400000000000   −0.36630000000000    P   (8d)
   0.03900000000000   −0.63400000000000   −0.86630000000000    P   (8d)
   0.46100000000000    0.63400000000000   −0.36630000000000    P   (8d)
  −0.03900000000000   −0.13400000000000    0.86630000000000    P   (8d)
   0.53900000000000    1.13400000000000    1.36630000000000    P   (8d)
  −0.05700000000000    0.25000000000000    0.57100000000000    Se  (4c)
   0.55700000000000    0.75000000000000    1.07100000000000    Se  (4c)
   0.05700000000000    0.75000000000000   −0.57100000000000    Se  (4c)
   0.44300000000000    0.25000000000000   −0.07100000000000    Se  (4c)
   0.01480000000000    0.25000000000000    0.25670000000000    Se  (4c)
   0.48520000000000    0.75000000000000    0.75670000000000    Se  (4c)
  −0.01480000000000    0.75000000000000   −0.25670000000000    Se  (4c)
   0.51480000000000    0.25000000000000    0.24330000000000    Se  (4c)
   0.10700000000000    0.25000000000000    0.87000000000000    Se  (4c)
   0.39300000000000    0.75000000000000    1.37000000000000    Se  (4c)
  −0.10700000000000    0.75000000000000   −0.87000000000000    Se  (4c)
   0.60700000000000    0.25000000000000   −0.37000000000000    Se  (4c)
   0.16300000000000    0.25000000000000    0.01800000000000    Se  (4c)
   0.33700000000000    0.75000000000000    0.51800000000000    Se  (4c)
  −0.16300000000000    0.75000000000000   −0.01800000000000    Se  (4c)
   0.66300000000000    0.25000000000000    0.48200000000000    Se  (4c)
   0.15920000000000    0.07200000000000    0.51390000000000    Se  (8d)
   0.34080000000000   −0.07200000000000    1.01390000000000    Se  (8d)
  −0.15920000000000    0.57200000000000   −0.51390000000000    Se  (8d)
   0.65920000000000    0.42800000000000   −0.01390000000000    Se  (8d)
  −0.15920000000000   −0.07200000000000   −0.51390000000000    Se  (8d)
   0.65920000000000    0.07200000000000   −0.01390000000000    Se  (8d)
   0.15920000000000    0.42800000000000    0.51390000000000    Se  (8d)
   0.34080000000000    0.57200000000000    1.01390000000000    Se  (8d)
   0.85270000000000    0.57200000000000    0.62980000000000    Se  (8d)
  −0.35270000000000   −0.57200000000000    1.12980000000000    Se  (8d)
  −0.85270000000000    1.07200000000000   −0.62980000000000    Se  (8d)
   1.35270000000000   −0.07200000000000   −0.12980000000000    Se  (8d)
  −0.85270000000000   −0.57200000000000   −0.62980000000000    Se  (8d)
   1.35270000000000    0.57200000000000   −0.12980000000000    Se  (8d)
   0.85270000000000   −0.07200000000000    0.62980000000000    Se  (8d)
  −0.35270000000000    1.07200000000000    1.12980000000000    Se  (8d)
   0.17230000000000    0.07200000000000    0.76120000000000    Se  (8d)
   0.32770000000000   −0.07200000000000    1.26120000000000    Se  (8d)
  −0.17230000000000    0.57200000000000   −0.76120000000000    Se  (8d)
   0.67230000000000    0.42800000000000   −0.26120000000000    Se  (8d)
  −0.17230000000000   −0.07200000000000   −0.76120000000000    Se  (8d)
   0.67230000000000    0.07200000000000   −0.26120000000000    Se  (8d)
   0.17230000000000    0.42800000000000    0.76120000000000    Se  (8d)
   0.32770000000000    0.57200000000000    1.26120000000000    Se  (8d)
   0.06020000000000    0.57200000000000   −0.06450000000000    Se  (8d)
   0.43980000000000   −0.57200000000000    0.43550000000000    Se  (8d)
  −0.06020000000000    1.07200000000000    0.06450000000000    Se  (8d)
   0.56020000000000   −0.07200000000000    0.56450000000000    Se  (8d)
  −0.06020000000000   −0.57200000000000    0.06450000000000    Se  (8d)
   0.56020000000000    0.57200000000000    0.56450000000000    Se  (8d)
   0.06020000000000   −0.07200000000000   −0.06450000000000    Se  (8d)
   0.43980000000000    1.07200000000000    0.43550000000000    Se  (8d)
```

$Mo_4P_3$: A4B3_oP56_62_8c_6c - CIF

```
# CIF file
data_findsym−output
_audit_creation_method FINDSYM

_chemical_name_mineral 'Mo4P3'
_chemical_formula_sum 'Mo4 P3'

loop_
_publ_author_name
 'S. Rundqvist'
_journal_name_full_name
;
 Acta Chemica Scandinavica
;
_journal_volume 19
_journal_year 1965
```

```
_journal_page_first 393
_journal_page_last 400
_publ_Section_title
;
 The Crystal Structure of Mo$_{4}$P$_{3}$
;

_aflow_title 'Mo$_{4}$P$_{3}$ Structure'
_aflow_proto 'A4B3_oP56_62_8c_6c'
_aflow_params 'a,b/a,c/a,x_{1},z_{1},x_{2},z_{2},x_{3},z_{3},x_{4},z_{4}
    ↪ },x_{5},z_{5},x_{6},z_{6},x_{7},z_{7},x_{8},z_{8},x_{9},z_{9},
    ↪ x_{10},z_{10},x_{11},z_{11},x_{12},z_{12},x_{13},z_{13},x_{14},
    ↪ z_{14}'
_aflow_params_values '12.428,0.254103636949,1.64467331831,0.1939,0.2262,
    ↪ 0.3843,0.8518,0.1218,0.6891,0.298,0.0835,0.4007,0.5269,0.0952,
    ↪ 0.4665,0.2977,0.3863,0.0095,0.8272,0.2464,-0.027,0.4824,0.4186,
    ↪ 0.3206,0.6371,0.0382,0.5805,0.0155,0.2745,0.1994,0.7941'
_aflow_Strukturbericht 'None'
_aflow_Pearson 'oP56'

_symmetry_space_group_name_H-M "P 21/n 21/m 21/a"
_symmetry_Int_Tables_number 62

_cell_length_a    12.42800
_cell_length_b    3.15800
_cell_length_c    20.44000
_cell_angle_alpha 90.00000
_cell_angle_beta  90.00000
_cell_angle_gamma 90.00000

loop_
_space_group_symop_id
_space_group_symop_operation_xyz
1 x,y,z
2 x+1/2,-y+1/2,-z+1/2
3 -x,y+1/2,-z
4 -x+1/2,-y,z+1/2
5 -x,-y,-z
6 -x+1/2,y+1/2,z+1/2
7 x,-y+1/2,z
8 x+1/2,y,-z+1/2

loop_
_atom_site_label
_atom_site_type_symbol
_atom_site_symmetry_multiplicity
_atom_site_Wyckoff_label
_atom_site_fract_x
_atom_site_fract_y
_atom_site_fract_z
_atom_site_occupancy
Mo1 Mo 4 c 0.19390 0.25000 0.22620 1.00000
Mo2 Mo 4 c 0.38430 0.25000 0.85180 1.00000
Mo3 Mo 4 c 0.12180 0.25000 0.68910 1.00000
Mo4 Mo 4 c 0.29800 0.25000 0.08350 1.00000
Mo5 Mo 4 c 0.40070 0.25000 0.52690 1.00000
Mo6 Mo 4 c 0.09520 0.25000 0.46650 1.00000
Mo7 Mo 4 c 0.29770 0.25000 0.38630 1.00000
Mo8 Mo 4 c 0.00950 0.25000 0.82720 1.00000
P1 P 4 c 0.24640 0.25000 -0.02700 1.00000
P2 P 4 c 0.48240 0.25000 0.41860 1.00000
P3 P 4 c 0.32060 0.25000 0.63710 1.00000
P4 P 4 c 0.03820 0.25000 0.58050 1.00000
P5 P 4 c 0.01550 0.25000 0.27450 1.00000
P6 P 4 c 0.19940 0.25000 0.79410 1.00000
```

$Mo_4P_3$: A4B3_oP56_62_8c_6c - POSCAR

```
A4B3_oP56_62_8c_6c & a,b/a,c/a,x1,z1,x2,z2,x3,z3,x4,z4,x5,z5,x6,z6,x7,z7
    ↪ ,x8,z8,x9,z9,x10,z10,x11,z11,x12,z12,x13,z13,x14,z14 --params=
    ↪ 12.428,0.254103636949,1.64467331831,0.1939,0.2262,0.3843,0.8518
    ↪ ,0.1218,0.6891,0.298,0.0835,0.4007,0.5269,0.0952,0.4665,0.2977,
    ↪ 0.3863,0.0095,0.8272,0.2464,-0.027,0.4824,0.4186,0.3206,0.6371,
    ↪ 0.0382,0.5805,0.0155,0.2745,0.1994,0.7941 & Pnma D_{2h}^{16} #
    ↪ 62 (c^14) & oP56 & None & Mo4P3 & Mo4P3 & S. Rundqvist, Acta
    ↪ Chem. Scand. 19, 393-400 (1965)
   1.00000000000000
  12.42800000000000   0.00000000000000   0.00000000000000
   0.00000000000000   3.15800000000000   0.00000000000000
   0.00000000000000   0.00000000000000  20.44000000000000
    Mo     P
    32    24
Direct
   0.19390000000000   0.25000000000000   0.22620000000000   Mo   (4c)
   0.30610000000000   0.75000000000000   0.72620000000000   Mo   (4c)
  -0.19390000000000   0.75000000000000  -0.22620000000000   Mo   (4c)
   0.69390000000000   0.25000000000000   0.27380000000000   Mo   (4c)
   0.38430000000000   0.25000000000000   0.85180000000000   Mo   (4c)
   0.11570000000000   0.75000000000000   1.35180000000000   Mo   (4c)
  -0.38430000000000   0.75000000000000  -0.85180000000000   Mo   (4c)
   0.88430000000000   0.25000000000000  -0.35180000000000   Mo   (4c)
   0.12180000000000   0.25000000000000   0.68910000000000   Mo   (4c)
   0.37820000000000   0.75000000000000   1.18910000000000   Mo   (4c)
  -0.12180000000000   0.75000000000000  -0.68910000000000   Mo   (4c)
   0.62180000000000   0.25000000000000  -0.18910000000000   Mo   (4c)
   0.29800000000000   0.25000000000000   0.08350000000000   Mo   (4c)
   0.20200000000000   0.75000000000000   0.58350000000000   Mo   (4c)
  -0.29800000000000   0.75000000000000  -0.08350000000000   Mo   (4c)
   0.79800000000000   0.25000000000000   0.41650000000000   Mo   (4c)
   0.40070000000000   0.25000000000000   0.52690000000000   Mo   (4c)
   0.09930000000000   0.75000000000000   1.02690000000000   Mo   (4c)
  -0.40070000000000   0.75000000000000  -0.52690000000000   Mo   (4c)
   0.90070000000000   0.25000000000000  -0.02690000000000   Mo   (4c)
   0.09520000000000   0.25000000000000   0.46650000000000   Mo   (4c)
   0.40480000000000   0.75000000000000   0.96650000000000   Mo   (4c)
  -0.09520000000000   0.75000000000000  -0.46650000000000   Mo   (4c)
   0.59520000000000   0.25000000000000   0.03350000000000   Mo   (4c)
   0.29770000000000   0.25000000000000   0.38630000000000   Mo   (4c)
   0.20230000000000   0.75000000000000   0.88630000000000   Mo   (4c)
  -0.29770000000000   0.75000000000000  -0.38630000000000   Mo   (4c)
   0.79770000000000   0.25000000000000   0.11370000000000   Mo   (4c)
   0.00950000000000   0.25000000000000   0.82720000000000   Mo   (4c)
   0.49050000000000   0.75000000000000   1.32720000000000   Mo   (4c)
  -0.00950000000000   0.75000000000000  -0.82720000000000   Mo   (4c)
   0.50950000000000   0.25000000000000  -0.32720000000000   Mo   (4c)
   0.24640000000000   0.25000000000000  -0.02700000000000   P    (4c)
   0.25360000000000   0.75000000000000   0.47300000000000   P    (4c)
  -0.24640000000000   0.75000000000000   0.02700000000000   P    (4c)
   0.74640000000000   0.25000000000000   0.52700000000000   P    (4c)
   0.48240000000000   0.25000000000000   0.41860000000000   P    (4c)
   0.01760000000000   0.75000000000000   0.91860000000000   P    (4c)
  -0.48240000000000   0.75000000000000  -0.41860000000000   P    (4c)
   0.98240000000000   0.25000000000000   0.08140000000000   P    (4c)
   0.32060000000000   0.25000000000000   0.63710000000000   P    (4c)
   0.17940000000000   0.75000000000000   1.13710000000000   P    (4c)
  -0.32060000000000   0.75000000000000  -0.63710000000000   P    (4c)
   0.82060000000000   0.25000000000000  -0.13710000000000   P    (4c)
   0.03820000000000   0.25000000000000   0.58050000000000   P    (4c)
   0.46180000000000   0.75000000000000   1.08050000000000   P    (4c)
  -0.03820000000000   0.75000000000000  -0.58050000000000   P    (4c)
   0.53820000000000   0.25000000000000  -0.08050000000000   P    (4c)
   0.01550000000000   0.25000000000000   0.27450000000000   P    (4c)
   0.48450000000000   0.75000000000000   0.77450000000000   P    (4c)
  -0.01550000000000   0.75000000000000  -0.27450000000000   P    (4c)
   0.51550000000000   0.25000000000000   0.22550000000000   P    (4c)
   0.19940000000000   0.25000000000000   0.79410000000000   P    (4c)
   0.30060000000000   0.75000000000000   1.29410000000000   P    (4c)
  -0.19940000000000   0.75000000000000  -0.79410000000000   P    (4c)
   0.69940000000000   0.25000000000000  -0.29410000000000   P    (4c)
```

$K_2SnCl_4{\cdot}H_2O$ ($E3_5$): A4BC2D_oP32_62_2cd_b_2c_a - CIF

```
# CIF file
data_findsym-output
_audit_creation_method FINDSYM

_chemical_name_mineral 'Cl4(H2O)K2Sn'
_chemical_formula_sum 'Cl4 (H2O) K2 Sn'

loop_
_publ_author_name
 'H. Brasseur'
 'A. {de Rassenfosse}'
_journal_name_full_name
;
 Zeitschrift f{\"u}r Kristallographie - Crystalline Materials
;
_journal_volume 101
_journal_year 1939
_journal_page_first 389
_journal_page_last 395
_publ_Section_title
;
 The Crystal Structure of Hydrated Potassium Chlorostannite
;

# Found in Strukturbericht Band VII 1939, 1943

_aflow_title 'K$_{2}$SnCl$_{4}$\cdot$H$_{2}$O ($E3_{5}$) Structure'
_aflow_proto 'A4BC2D_oP32_62_2cd_b_2c_a'
_aflow_params 'a,b/a,c/a,x_{3},z_{3},x_{4},z_{4},x_{5},z_{5},x_{6},z_{6
    ↪ },x_{7},y_{7},z_{7}'
_aflow_params_values '12.05,0.755186721992,0.684647302905,-0.05,0.21,
    ↪ 0.05,0.79,0.2,0.34,0.825,0.6,0.25,-0.04,0.12'
_aflow_Strukturbericht '$E3_{5}$'
_aflow_Pearson 'oP32'

_symmetry_space_group_name_H-M "P 21/n 21/m 21/a"
_symmetry_Int_Tables_number 62

_cell_length_a    12.05000
_cell_length_b    9.10000
_cell_length_c    8.25000
_cell_angle_alpha 90.00000
_cell_angle_beta  90.00000
_cell_angle_gamma 90.00000

loop_
_space_group_symop_id
_space_group_symop_operation_xyz
1 x,y,z
2 x+1/2,-y+1/2,-z+1/2
3 -x,y+1/2,-z
4 -x+1/2,-y,z+1/2
5 -x,-y,-z
6 -x+1/2,y+1/2,z+1/2
7 x,-y+1/2,z
8 x+1/2,y,-z+1/2

loop_
_atom_site_label
_atom_site_type_symbol
_atom_site_symmetry_multiplicity
_atom_site_Wyckoff_label
_atom_site_fract_x
_atom_site_fract_y
_atom_site_fract_z
_atom_site_occupancy
Sn1 Sn 4 a 0.00000 0.00000 0.00000 1.00000
H2O1 H2O 4 b 0.00000 0.00000 0.50000 1.00000
```

```
Cl1  Cl  4  c  -0.05000  0.25000  0.21000  1.00000
Cl2  Cl  4  c  0.05000  0.25000  0.79000  1.00000
K1  K  4  c  0.20000  0.25000  0.34000  1.00000
K2  K  4  c  0.82500  0.25000  0.60000  1.00000
Cl3  Cl  8  d  0.25000  -0.04000  0.12000  1.00000
```

$K_2SnCl_4{\cdot}H_2O$ ($E3_5$): A4BC2D_oP32_62_2cd_b_2c_a - POSCAR

```
A4BC2D_oP32_62_2cd_b_2c_a & a,b/a,c/a,x3,z3,x4,z4,x5,z5,x6,z6,x7,y7,z7
    ↪ --params=12.05,0.755186721992,0.684647302905,-0.05,0.21,0.05,
    ↪ 0.79,0.2,0.34,0.825,0.6,0.25,-0.04,0.12 & Pnma D_{2h}^{16} #62
    ↪ (abc^4d) & oP32 & $E3_{5}$ & Cl4(H2O)K2Sn & Cl4(H2O)K2Sn & H.
    ↪ Brasseur and A. {de Rassenfosse}, Zeitschrift f{"u}r
    ↪ Kristallographie - Crystalline Materials 101, 389-395 (1939)
   1.00000000000000
  12.05000000000000    0.00000000000000    0.00000000000000
   0.00000000000000    9.10000000000000    0.00000000000000
   0.00000000000000    0.00000000000000    8.25000000000000
    Cl      H2O       K       Sn
    16       4        8        4
Direct
  -0.05000000000000    0.25000000000000    0.21000000000000    Cl    (4c)
   0.55000000000000    0.75000000000000    0.71000000000000    Cl    (4c)
   0.05000000000000    0.75000000000000   -0.21000000000000    Cl    (4c)
   0.45000000000000    0.25000000000000    0.29000000000000    Cl    (4c)
   0.05000000000000    0.25000000000000    0.79000000000000    Cl    (4c)
   0.45000000000000    0.75000000000000    1.29000000000000    Cl    (4c)
  -0.05000000000000    0.75000000000000   -0.79000000000000    Cl    (4c)
   0.55000000000000    0.25000000000000   -0.29000000000000    Cl    (4c)
   0.25000000000000   -0.04000000000000    0.12000000000000    Cl    (8d)
   0.25000000000000    0.04000000000000    0.62000000000000    Cl    (8d)
  -0.25000000000000    0.46000000000000   -0.12000000000000    Cl    (8d)
   0.75000000000000    0.54000000000000    0.38000000000000    Cl    (8d)
  -0.25000000000000    0.04000000000000   -0.12000000000000    Cl    (8d)
   0.75000000000000   -0.04000000000000    0.38000000000000    Cl    (8d)
   0.25000000000000    0.54000000000000    0.12000000000000    Cl    (8d)
   0.25000000000000    0.46000000000000    0.62000000000000    Cl    (8d)
   0.00000000000000    0.00000000000000    0.50000000000000   H2O    (4b)
   0.50000000000000    0.00000000000000    0.00000000000000   H2O    (4b)
   0.00000000000000    0.50000000000000    0.50000000000000   H2O    (4b)
   0.50000000000000    0.50000000000000    0.00000000000000   H2O    (4b)
   0.20000000000000    0.25000000000000    0.34000000000000    K     (4c)
   0.30000000000000    0.75000000000000    0.84000000000000    K     (4c)
  -0.20000000000000    0.75000000000000   -0.34000000000000    K     (4c)
   0.70000000000000    0.25000000000000    0.16000000000000    K     (4c)
   0.82500000000000    0.25000000000000    0.60000000000000    K     (4c)
  -0.32500000000000    0.75000000000000    1.10000000000000    K     (4c)
  -0.82500000000000    0.75000000000000   -0.60000000000000    K     (4c)
   1.32500000000000    0.25000000000000   -0.10000000000000    K     (4c)
   0.00000000000000    0.00000000000000    0.00000000000000    Sn    (4a)
   0.50000000000000    0.00000000000000    0.50000000000000    Sn    (4a)
   0.00000000000000    0.50000000000000    0.00000000000000    Sn    (4a)
   0.50000000000000    0.50000000000000    0.50000000000000    Sn    (4a)
```

$K_2SnCl_4{\cdot}H_2O$: A4BC2D_oP32_62_2cd_c_d_c - CIF

```
# CIF file
data_findsym-output
_audit_creation_method FINDSYM

_chemical_name_mineral 'Cl4(H2O)K2Sn'
_chemical_formula_sum 'Cl4 (H2O) K2 Sn'

loop_
_publ_author_name
 'B. Kamenar'
 'D. Grdeni\\'{c}'
_journal_name_full_name
;
 Journal of Inorganic and Nuclear Chemistry
;
_journal_volume 24
_journal_year 1962
_journal_page_first 1039
_journal_page_last 1045
_publ_Section_title
;
 The crystal structure of potassium chloride trichlorostannite hydrate,
    ↪ KCl, KSnCl$_{3}$, H$_{2}$O
;

_aflow_title 'K$_{2}$SnCl$_{4}$$\cdot$H$_{2}$O Structure'
_aflow_proto 'A4BC2D_oP32_62_2cd_c_d_c'
_aflow_params 'a,b/a,c/a,x_{1},z_{1},x_{2},z_{2},x_{3},z_{3},x_{4},z_{4
    ↪ },x_{5},y_{5},z_{5},x_{6},y_{6},z_{6}'
_aflow_params_values '12.05,0.758506224066,0.683817427386,0.308,0.888,
    ↪ 0.766,0.099,0.54,0.441,0.512,0.0,0.444,-0.052,0.207,0.685,0.482
    ↪ ,0.37'
_aflow_Strukturbericht 'None'
_aflow_Pearson 'oP32'

_symmetry_space_group_name_H-M "P 21/n 21/m 21/a"
_symmetry_Int_Tables_number 62

_cell_length_a      12.05000
_cell_length_b      9.14000
_cell_length_c      8.24000
_cell_angle_alpha 90.00000
_cell_angle_beta  90.00000
_cell_angle_gamma 90.00000

loop_
_space_group_symop_id
_space_group_symop_operation_xyz
1 x,y,z
2 x+1/2,-y+1/2,-z+1/2
3 -x,y+1/2,-z
4 -x+1/2,-y,z+1/2
5 -x,-y,-z
6 -x+1/2,y+1/2,z+1/2
7 x,-y+1/2,z
8 x+1/2,y,-z+1/2

loop_
_atom_site_label
_atom_site_type_symbol
_atom_site_symmetry_multiplicity
_atom_site_Wyckoff_label
_atom_site_fract_x
_atom_site_fract_y
_atom_site_fract_z
_atom_site_occupancy
Cl1  Cl  4  c  0.30800  0.25000  0.88800  1.00000
Cl2  Cl  4  c  0.76600  0.25000  0.09900  1.00000
H2O1  H2O  4  c  0.54000  0.25000  0.44100  1.00000
Sn1  Sn  4  c  0.51200  0.25000  0.00000  1.00000
Cl3  Cl  8  d  0.44400  -0.05200  0.20700  1.00000
K1  K  8  d  0.68500  0.48200  0.37000  1.00000
```

$K_2SnCl_4{\cdot}H_2O$: A4BC2D_oP32_62_2cd_c_d_c - POSCAR

```
A4BC2D_oP32_62_2cd_c_d_c & a,b/a,c/a,x1,z1,x2,z2,x3,z3,x4,z4,x5,y5,z5,x6
    ↪ ,y6,z6 --params=12.05,0.758506224066,0.683817427386,0.308,0.888
    ↪ ,0.766,0.099,0.54,0.441,0.512,0.0,0.444,-0.052,0.207,0.685,
    ↪ 0.482,0.37 & Pnma D_{2h}^{16} #62 (c^4d^2) & oP32 & None & Cl4(
    ↪ H2O)K2Sn & Cl4(H2O)K2Sn & B. Kamenar and D. Grdeni\'{c}, J.
    ↪ Inorg. Nucl. Chem. 24, 1039-1045 (1962)
   1.00000000000000
  12.05000000000000    0.00000000000000    0.00000000000000
   0.00000000000000    9.14000000000000    0.00000000000000
   0.00000000000000    0.00000000000000    8.24000000000000
    Cl      H2O       K       Sn
    16       4        8        4
Direct
   0.30800000000000    0.25000000000000    0.88800000000000    Cl    (4c)
   0.19200000000000    0.75000000000000    1.38800000000000    Cl    (4c)
  -0.30800000000000    0.75000000000000   -0.88800000000000    Cl    (4c)
   0.80800000000000    0.25000000000000   -0.38800000000000    Cl    (4c)
   0.76600000000000    0.25000000000000    0.09900000000000    Cl    (4c)
  -0.26600000000000    0.75000000000000    0.59900000000000    Cl    (4c)
  -0.76600000000000    0.75000000000000   -0.09900000000000    Cl    (4c)
   1.26600000000000    0.25000000000000    0.40100000000000    Cl    (4c)
   0.44400000000000   -0.05200000000000    0.20700000000000    Cl    (8d)
   0.05600000000000    0.05200000000000    0.70700000000000    Cl    (8d)
  -0.44400000000000    0.44800000000000   -0.20700000000000    Cl    (8d)
   0.94400000000000    0.55200000000000    0.29300000000000    Cl    (8d)
  -0.44400000000000    0.05200000000000   -0.20700000000000    Cl    (8d)
   0.94400000000000   -0.05200000000000    0.29300000000000    Cl    (8d)
   0.44400000000000    0.55200000000000    0.20700000000000    Cl    (8d)
   0.05600000000000    0.44800000000000    0.70700000000000    Cl    (8d)
   0.54000000000000    0.25000000000000    0.44100000000000   H2O    (4c)
  -0.04000000000000    0.75000000000000    0.94100000000000   H2O    (4c)
  -0.54000000000000    0.75000000000000   -0.44100000000000   H2O    (4c)
   1.04000000000000    0.25000000000000    0.05900000000000   H2O    (4c)
   0.68500000000000    0.48200000000000    0.37000000000000    K     (8d)
  -0.18500000000000   -0.48200000000000    0.87000000000000    K     (8d)
  -0.68500000000000    0.98200000000000   -0.37000000000000    K     (8d)
   1.18500000000000    0.01800000000000    0.13000000000000    K     (8d)
  -0.68500000000000   -0.48200000000000   -0.37000000000000    K     (8d)
   1.18500000000000    0.48200000000000    0.13000000000000    K     (8d)
   0.68500000000000    0.01800000000000    0.37000000000000    K     (8d)
  -0.18500000000000    0.98200000000000    0.87000000000000    K     (8d)
   0.51200000000000    0.25000000000000    0.00000000000000    Sn    (4c)
  -0.01200000000000    0.75000000000000    0.50000000000000    Sn    (4c)
  -0.51200000000000    0.75000000000000    0.00000000000000    Sn    (4c)
   1.01200000000000    0.25000000000000    0.50000000000000    Sn    (4c)
```

$VOSO_4$: A5BC_oP28_62_3cd_c_c - CIF

```
# CIF file
data_findsym-output
_audit_creation_method FINDSYM

_chemical_name_mineral 'O5SV'
_chemical_formula_sum 'O5 S V'

loop_
_publ_author_name
 'P. Kierkegaard'
 'J. M. Longo'
 'B.-O. Marinder'
_journal_name_full_name
;
 Acta Chemica Scandinavica
;
_journal_volume 19
_journal_year 1965
_journal_page_first 763
_journal_page_last 764
_publ_Section_title
;
 Note on the Crystal Structure of VOSO$_{4}$
;

_aflow_title 'VOSO$_{4}$ Structure'
_aflow_proto 'A5BC_oP28_62_3cd_c_c'
_aflow_params 'a,b/a,c/a,x_{1},z_{1},x_{2},z_{2},x_{3},z_{3},x_{4},z_{4
    ↪ },x_{5},z_{5},x_{6},y_{6},z_{6}'
```

```
_aflow_params_values '7.371,0.850495183829,0.960792294126,0.7187,-0.0104
    ↪ ,0.0399,-0.0201,0.3719,0.1647,0.8761,0.8669,0.1658,0.2327,
    ↪ 0.1251,0.5733,0.2569'
_aflow_Strukturbericht 'None'
_aflow_Pearson 'oP28'

_symmetry_space_group_name_H-M "P 21/n 21/m 21/a"
_symmetry_Int_Tables_number 62

_cell_length_a    7.37100
_cell_length_b    6.26900
_cell_length_c    7.08200
_cell_angle_alpha 90.00000
_cell_angle_beta  90.00000
_cell_angle_gamma 90.00000

loop_
_space_group_symop_id
_space_group_symop_operation_xyz
1 x,y,z
2 x+1/2,-y+1/2,-z+1/2
3 -x,y+1/2,-z
4 -x+1/2,-y,z+1/2
5 -x,-y,-z
6 -x+1/2,y+1/2,z+1/2
7 x,-y+1/2,z
8 x+1/2,y,-z+1/2

loop_
_atom_site_label
_atom_site_type_symbol
_atom_site_symmetry_multiplicity
_atom_site_Wyckoff_label
_atom_site_fract_x
_atom_site_fract_y
_atom_site_fract_z
_atom_site_occupancy
O1 O 4 c 0.71870 0.25000 -0.01040 1.00000
O2 O 4 c 0.03990 0.25000 -0.02010 1.00000
O3 O 4 c 0.37190 0.25000 0.16470 1.00000
S1 S 4 c 0.87610 0.25000 0.86690 1.00000
V1 V 4 c 0.16580 0.25000 0.23270 1.00000
O4 O 8 d 0.12510 0.57330 0.25690 1.00000
```

$VOSO_4$: A5BC_oP28_62_3cd_c_c - POSCAR

```
A5BC_oP28_62_3cd_c_c & a,b/a,c/a,x1,z1,x2,z2,x3,z3,x4,z4,x5,z5,x6,y6,z6
    ↪ --params=7.371,0.850495183829,0.960792294126,0.7187,-0.0104,
    ↪ 0.0399,-0.0201,0.3719,0.1647,0.8761,0.8669,0.1658,0.2327,0.1251
    ↪ ,0.5733,0.2569 & Pnma D_{2h}^{16} #62 (c^5d) & oP28 & None &
    ↪ O5SV & O5SV & P. Kierkegaard and J. M. Longo and B.-O. Marinder
    ↪ , Acta Chem. Scand. 19, 763-764 (1965)
   1.00000000000000
   7.37100000000000   0.00000000000000   0.00000000000000
   0.00000000000000   6.26900000000000   0.00000000000000
   0.00000000000000   0.00000000000000   7.08200000000000
     O     S     V
    20     4     4
Direct
   0.71870000000000   0.25000000000000  -0.01040000000000   O   (4c)
  -0.21870000000000   0.75000000000000   0.48960000000000   O   (4c)
  -0.71870000000000   0.75000000000000   0.01040000000000   O   (4c)
   1.21870000000000   0.25000000000000   0.51040000000000   O   (4c)
   0.03990000000000   0.25000000000000  -0.02010000000000   O   (4c)
   0.46010000000000   0.75000000000000   0.47990000000000   O   (4c)
  -0.03990000000000   0.75000000000000   0.02010000000000   O   (4c)
   0.53990000000000   0.25000000000000   0.52010000000000   O   (4c)
   0.37190000000000   0.25000000000000   0.16470000000000   O   (4c)
   0.12810000000000   0.75000000000000   0.66470000000000   O   (4c)
  -0.37190000000000   0.75000000000000  -0.16470000000000   O   (4c)
   0.87190000000000   0.25000000000000   0.33530000000000   O   (4c)
   0.12510000000000   0.57330000000000   0.25690000000000   O   (8d)
   0.37490000000000  -0.57330000000000   0.75690000000000   O   (8d)
  -0.12510000000000   1.07330000000000  -0.25690000000000   O   (8d)
   0.62510000000000  -0.07330000000000   0.24310000000000   O   (8d)
  -0.12510000000000  -0.57330000000000  -0.25690000000000   O   (8d)
   0.62510000000000   0.57330000000000   0.24310000000000   O   (8d)
   0.12510000000000  -0.07330000000000   0.25690000000000   O   (8d)
   0.37490000000000   1.07330000000000   0.75690000000000   O   (8d)
   0.87610000000000   0.25000000000000   0.86690000000000   S   (4c)
  -0.37610000000000   0.75000000000000   1.36690000000000   S   (4c)
  -0.87610000000000   0.75000000000000  -0.86690000000000   S   (4c)
   1.37610000000000   0.25000000000000  -0.36690000000000   S   (4c)
   0.16580000000000   0.25000000000000   0.23270000000000   V   (4c)
   0.33420000000000   0.75000000000000   0.73270000000000   V   (4c)
  -0.16580000000000   0.75000000000000  -0.23270000000000   V   (4c)
   0.66580000000000   0.25000000000000   0.26730000000000   V   (4c)
```

Possible $\delta$-$Y_2Si_2O_7$: A7B2C2_oP44_62_3c2d_2c_d - CIF

```
# CIF file
data_findsym-output
_audit_creation_method FINDSYM

_chemical_name_mineral 'O7Si2Y2'
_chemical_formula_sum 'O7 Si2 Y2'

loop_
_publ_author_name
 'H. W. Dias'
 'F. P. Glasser'
 'R. P. Gunwardane'
 'R. A. Howie'
_journal_name_full_name
;
 Zeitschrift f{\"u}r Kristallographie - Crystalline Materials
;
_journal_volume 191
_journal_year 1990
_journal_page_first 117
_journal_page_last 124
_publ_Section_title
;
 The crystal structure of $\delta$-yttrium pyrosilicate, $\delta$-Y$_{2}
    ↪ $Si$_{2}$O$_{7}$
;

# Found in Revision of the crystallographic data of polymorphic Y$_{2}
    ↪ $Si$_{2}$O$_{7}$ and Y$_{2}$SiO$_{5}$ compounds, 2004

_aflow_title 'Possible $\delta$-Y$_{2}$Si$_{2}$O$_{7}$ Structure'
_aflow_proto 'A7B2C2_oP44_62_3c2d_2c_d'
_aflow_params 'a,b/a,c/a,x_{1},z_{1},x_{2},z_{2},x_{3},z_{3},x_{4},z_{4
    ↪ },x_{5},z_{5},x_{6},y_{6},z_{6},x_{7},y_{7},z_{7},x_{8},y_{8},
    ↪ z_{8}'
_aflow_params_values '13.655,0.596045404614,0.367337971439,0.1524,0.0613
    ↪ ,0.0782,0.5588,0.8969,0.3386,0.1821,0.3711,-0.041,0.6214,0.2329
    ↪ ,0.0861,0.4875,-0.0477,0.0817,0.7994,0.3745,-0.0096,0.3395'
_aflow_Strukturbericht 'None'
_aflow_Pearson 'oP44'

_symmetry_space_group_name_H-M "P 21/n 21/m 21/a"
_symmetry_Int_Tables_number 62

_cell_length_a    13.65500
_cell_length_b    8.13900
_cell_length_c    5.01600
_cell_angle_alpha 90.00000
_cell_angle_beta  90.00000
_cell_angle_gamma 90.00000

loop_
_space_group_symop_id
_space_group_symop_operation_xyz
1 x,y,z
2 x+1/2,-y+1/2,-z+1/2
3 -x,y+1/2,-z
4 -x+1/2,-y,z+1/2
5 -x,-y,-z
6 -x+1/2,y+1/2,z+1/2
7 x,-y+1/2,z
8 x+1/2,y,-z+1/2

loop_
_atom_site_label
_atom_site_type_symbol
_atom_site_symmetry_multiplicity
_atom_site_Wyckoff_label
_atom_site_fract_x
_atom_site_fract_y
_atom_site_fract_z
_atom_site_occupancy
O1 O 4 c 0.15240 0.25000 0.06130 1.00000
O2 O 4 c 0.07820 0.25000 0.55880 1.00000
O3 O 4 c 0.89690 0.25000 0.33860 1.00000
Si1 Si 4 c 0.18210 0.25000 0.37110 1.00000
Si2 Si 4 c -0.04100 0.25000 0.62140 1.00000
O4 O 8 d 0.23290 0.08610 0.48750 1.00000
O5 O 8 d -0.04770 0.08170 0.79940 1.00000
Y1 Y 8 d 0.37450 -0.00960 0.33950 1.00000
```

Possible $\delta$-$Y_2Si_2O_7$: A7B2C2_oP44_62_3c2d_2c_d - POSCAR

```
A7B2C2_oP44_62_3c2d_2c_d & a,b/a,c/a,x1,z1,x2,z2,x3,z3,x4,z4,x5,z5,x6,y6
    ↪ ,z6,x7,y7,z7,x8,y8,z8 --params=13.655,0.596045404614,
    ↪ 0.367337971439,0.1524,0.0613,0.0782,0.5588,0.8969,0.3386,0.1821
    ↪ ,0.3711,-0.041,0.6214,0.2329,0.0861,0.4875,-0.0477,0.0817,
    ↪ 0.7994,0.3745,-0.0096,0.3395 & Pnma D_{2h}^{16} #62 (c^5d^3) &
    ↪ oP44 & None & O7Si2Y2 & O7Si2Y2 & H. W. Dias et al.,
    ↪ Zeitschrift f{"u}r Kristallographie - Crystalline Materials 191
    ↪ , 117-124 (1990)
   1.00000000000000
  13.65500000000000   0.00000000000000   0.00000000000000
   0.00000000000000   8.13900000000000   0.00000000000000
   0.00000000000000   0.00000000000000   5.01600000000000
     O    Si     Y
    28     8     8
Direct
   0.15240000000000   0.25000000000000   0.06130000000000   O   (4c)
   0.34760000000000   0.75000000000000   0.56130000000000   O   (4c)
  -0.15240000000000   0.75000000000000  -0.06130000000000   O   (4c)
   0.65240000000000   0.25000000000000   0.43870000000000   O   (4c)
   0.07820000000000   0.25000000000000   0.55880000000000   O   (4c)
   0.42180000000000   0.75000000000000   1.05880000000000   O   (4c)
  -0.07820000000000   0.75000000000000  -0.55880000000000   O   (4c)
   0.57820000000000   0.25000000000000  -0.05880000000000   O   (4c)
   0.89690000000000   0.25000000000000   0.33860000000000   O   (4c)
  -0.39690000000000   0.75000000000000   0.83860000000000   O   (4c)
  -0.89690000000000   0.75000000000000  -0.33860000000000   O   (4c)
   1.39690000000000   0.25000000000000   0.16140000000000   O   (4c)
   0.23290000000000   0.08610000000000   0.48750000000000   O   (8d)
   0.26710000000000  -0.08610000000000   0.98750000000000   O   (8d)
  -0.23290000000000   0.58610000000000  -0.48750000000000   O   (8d)
   0.73290000000000   0.41390000000000   0.01250000000000   O   (8d)
  -0.23290000000000  -0.08610000000000  -0.48750000000000   O   (8d)
   0.73290000000000   0.08610000000000   0.01250000000000   O   (8d)
   0.23290000000000   0.41390000000000   0.48750000000000   O   (8d)
   0.26710000000000   0.58610000000000   0.98750000000000   O   (8d)
  -0.04770000000000   0.08170000000000   0.79940000000000   O   (8d)
   0.54770000000000  -0.08170000000000   1.29940000000000   O   (8d)
```

```
   0.04770000000000    0.58170000000000   -0.79940000000000     O    (8d)
   0.45230000000000    0.41830000000000   -0.29940000000000     O    (8d)
   0.04770000000000   -0.08170000000000   -0.79940000000000     O    (8d)
   0.45230000000000    0.08170000000000   -0.29940000000000     O    (8d)
  -0.04770000000000    0.41830000000000    0.79940000000000     O    (8d)
   0.54770000000000    0.58170000000000    1.29940000000000     O    (8d)
   0.18210000000000    0.25000000000000    0.37110000000000     Si   (4c)
   0.31790000000000    0.75000000000000    0.87110000000000     Si   (4c)
  -0.18210000000000    0.75000000000000   -0.37110000000000     Si   (4c)
   0.68210000000000    0.25000000000000    0.12890000000000     Si   (4c)
  -0.04100000000000    0.25000000000000    0.62140000000000     Si   (4c)
   0.54100000000000    0.75000000000000    1.12140000000000     Si   (4c)
   0.04100000000000    0.75000000000000   -0.62140000000000     Si   (4c)
   0.45900000000000    0.25000000000000   -0.12140000000000     Si   (4c)
   0.37450000000000   -0.00960000000000    0.33950000000000     Y    (8d)
   0.12550000000000    0.00960000000000    0.83950000000000     Y    (8d)
  -0.37450000000000    0.49040000000000   -0.33950000000000     Y    (8d)
   0.87450000000000    0.50960000000000    0.16050000000000     Y    (8d)
  -0.37450000000000    0.00960000000000   -0.33950000000000     Y    (8d)
   0.87450000000000   -0.00960000000000    0.16050000000000     Y    (8d)
   0.37450000000000    0.50960000000000    0.33950000000000     Y    (8d)
   0.12550000000000    0.49040000000000    0.83950000000000     Y    (8d)
```

$K_4[Mo(CN)_8]\cdot 2H_2O$ ($F2_1$): A8B4C4DE8F2_oP108_62_4c2d_2d_2cd_c_4c2d_d - CIF

```
# CIF file
data_findsym-output
_audit_creation_method FINDSYM

_chemical_name_mineral 'C8H4K4MoN8O2'
_chemical_formula_sum 'C8 H4 K4 Mo N8 O2'

loop_
_publ_author_name
 'I. Typilo'
 'O. Sereda'
 'H. {Stoeckli-Evans}'
 'R. Gladyshevskii'
 'D. Semenyshyn'
_journal_name_full_name
;
 Chemistry of Metals and Alloys
;
_journal_volume 3
_journal_year 2010
_journal_page_first 49
_journal_page_last 52
_publ_Section_title
;
 Refinement of the crystal structure of potassium octacyanomolybdate(IV)
     ↪ dihydrate
;

_aflow_title 'K$_{4}$[Mo(CN)$_{8}$]$\cdot$2H$_{2}$O ($F2_{1}$) Structure
     ↪ '
_aflow_proto 'A8B4C4DE8F2_oP108_62_4c2d_2d_2cd_c_4c2d_d'
_aflow_params 'a,b/a,c/a,x_{1},z_{1},x_{2},z_{2},x_{3},z_{3},x_{4},z_{4
     ↪ },x_{5},z_{5},x_{6},z_{6},x_{7},z_{7},x_{8},z_{8},x_{9},z_{9},
     ↪ x_{10},z_{10},x_{11},z_{11},x_{12},y_{12},z_{12},x_{13},y_{13},
     ↪ z_{13},x_{14},y_{14},z_{14},x_{15},y_{15},z_{15},x_{16},y_{16},
     ↪ z_{16},x_{17},y_{17},z_{17},x_{18},y_{18},z_{18},x_{19},y_{19},
     ↪ z_{19}'
_aflow_params_values '16.6959,0.695320408004,0.519864158266,0.0619,
     ↪ 0.3937,0.1418,0.8446,0.2168,0.3978,0.2602,0.6753,0.36535,0.025,
     ↪ 0.46252,0.52249,0.1375,0.59643,0.021,0.2866,0.2614,0.2957,0.326
     ↪ ,0.7133,0.6401,0.5221,0.0391,0.1388,0.6648,0.1731,0.0725,0.5689
     ↪ ,0.349,0.08,0.367,0.412,0.017,0.347,0.14835,0.04756,0.18788,
     ↪ 0.0123,0.5809,0.2893,0.3068,0.0216,0.05,0.3888,0.0593,0.409'
_aflow_Strukturbericht '$F2_{1}$'
_aflow_Pearson 'oP108'

_symmetry_space_group_name_H-M "P 21/n 21/m 21/a"
_symmetry_Int_Tables_number 62

_cell_length_a    16.69590
_cell_length_b    11.60900
_cell_length_c    8.67960
_cell_angle_alpha 90.00000
_cell_angle_beta  90.00000
_cell_angle_gamma 90.00000

loop_
_space_group_symop_id
_space_group_symop_operation_xyz
1 x,y,z
2 x+1/2,-y+1/2,-z+1/2
3 -x,y+1/2,-z
4 -x+1/2,-y,z+1/2
5 -x,-y,-z
6 -x+1/2,y+1/2,z+1/2
7 x,-y+1/2,z
8 x+1/2,y,-z+1/2

loop_
_atom_site_label
_atom_site_type_symbol
_atom_site_symmetry_multiplicity
_atom_site_Wyckoff_label
_atom_site_fract_x
_atom_site_fract_y
_atom_site_fract_z
_atom_site_occupancy
C1 C 4 c 0.06190 0.25000 0.39370 1.00000
C2 C 4 c 0.14180 0.25000 0.84460 1.00000
C3 C 4 c 0.21680 0.25000 0.39780 1.00000
C4 C 4 c 0.26020 0.25000 0.67530 1.00000
K1 K 4 c 0.36535 0.25000 0.02500 1.00000
K2 K 4 c 0.46252 0.25000 0.52249 1.00000
Mo1 Mo 4 c 0.13750 0.25000 0.59643 1.00000
N1 N 4 c 0.02100 0.25000 0.28660 1.00000
N2 N 4 c 0.26140 0.25000 0.29570 1.00000
N3 N 4 c 0.32600 0.25000 0.71330 1.00000
N4 N 4 c 0.64010 0.25000 0.52210 1.00000
C5 C 8 d 0.03910 0.13880 0.66480 1.00000
C6 C 8 d 0.17310 0.07250 0.56890 1.00000
H1 H 8 d 0.34900 0.08000 0.36700 1.00000
H2 H 8 d 0.41200 0.01700 0.34700 1.00000
K3 K 8 d 0.14835 0.04756 0.18788 1.00000
N5 N 8 d 0.01230 0.58090 0.28930 1.00000
N6 N 8 d 0.30680 0.02160 0.05000 1.00000
O1 O 8 d 0.38880 0.05930 0.40900 1.00000
```

$K_4[Mo(CN)_8]\cdot 2H_2O$ ($F2_1$): A8B4C4DE8F2_oP108_62_4c2d_2d_2cd_c_4c2d_d - POSCAR

```
A8B4C4DE8F2_oP108_62_4c2d_2d_2cd_c_4c2d_d & a,b/a,c/a,x1,z1,x2,z2,x3,z3,
     ↪ x4,z4,x5,z5,x6,z6,x7,z7,x8,z8,x9,z9,x10,z10,x11,z11,x12,y12,z12
     ↪ ,x13,y13,z13,x14,y14,z14,x15,y15,z15,x16,y16,z16,x17,y17,z17,
     ↪ x18,y18,z18,x19,y19,z19 --params=16.6959,0.695320408004,
     ↪ 0.519864158266,0.0619,0.3937,0.1418,0.8446,0.2168,0.3978,0.2602
     ↪ ,0.6753,0.36535,0.025,0.46252,0.52249,0.1375,0.59643,0.021,
     ↪ 0.2866,0.2614,0.2957,0.326,0.7133,0.6401,0.5221,0.0391,0.1388,
     ↪ 0.6648,0.1731,0.0725,0.5689,0.349,0.08,0.367,0.412,0.017,0.347,
     ↪ 0.14835,0.04756,0.18788,0.0123,0.5809,0.2893,0.3068,0.0216,0.05
     ↪ ,0.3888,0.0593,0.409 & Pnma D_{2h}^{16} #62 (c^11d^8) & oP108 &
     ↪ $F2_{1}$ & C8H4K4MoN8O2 & C8H4K4MoN8O2 & I. Typilo et al.,
     ↪ Chem. Met. Alloys 3, 49-52 (2010)
   1.00000000000000
  16.69590000000000    0.00000000000000    0.00000000000000
   0.00000000000000   11.60900000000000    0.00000000000000
   0.00000000000000    0.00000000000000    8.67960000000000
     C      H      K     Mo      N      O
    32     16     16      4     32      8
Direct
   0.06190000000000    0.25000000000000    0.39370000000000     C    (4c)
   0.43810000000000    0.75000000000000    0.89370000000000     C    (4c)
  -0.06190000000000    0.75000000000000   -0.39370000000000     C    (4c)
   0.56190000000000    0.25000000000000    0.10630000000000     C    (4c)
   0.14180000000000    0.25000000000000    0.84460000000000     C    (4c)
   0.35820000000000    0.75000000000000    1.34460000000000     C    (4c)
  -0.14180000000000    0.75000000000000   -0.84460000000000     C    (4c)
   0.64180000000000    0.25000000000000   -0.34460000000000     C    (4c)
   0.21680000000000    0.25000000000000    0.39780000000000     C    (4c)
   0.28320000000000    0.75000000000000    0.89780000000000     C    (4c)
  -0.21680000000000    0.75000000000000   -0.39780000000000     C    (4c)
   0.71680000000000    0.25000000000000    0.10220000000000     C    (4c)
   0.26020000000000    0.25000000000000    0.67530000000000     C    (4c)
   0.23980000000000    0.75000000000000    1.17530000000000     C    (4c)
  -0.26020000000000    0.75000000000000   -0.67530000000000     C    (4c)
   0.76020000000000    0.25000000000000   -0.17530000000000     C    (4c)
   0.03910000000000    0.13880000000000    0.66480000000000     C    (8d)
   0.46090000000000   -0.13880000000000    1.16480000000000     C    (8d)
  -0.03910000000000    0.63880000000000   -0.66480000000000     C    (8d)
   0.53910000000000    0.36120000000000   -0.16480000000000     C    (8d)
  -0.03910000000000   -0.13880000000000   -0.66480000000000     C    (8d)
   0.53910000000000    0.13880000000000   -0.16480000000000     C    (8d)
   0.03910000000000    0.36120000000000    0.66480000000000     C    (8d)
   0.46090000000000    0.63880000000000    1.16480000000000     C    (8d)
   0.17310000000000    0.07250000000000    0.56890000000000     C    (8d)
   0.32690000000000   -0.07250000000000    1.06890000000000     C    (8d)
  -0.17310000000000    0.57250000000000   -0.56890000000000     C    (8d)
   0.67310000000000    0.42750000000000   -0.06890000000000     C    (8d)
  -0.17310000000000   -0.07250000000000   -0.56890000000000     C    (8d)
   0.67310000000000    0.07250000000000   -0.06890000000000     C    (8d)
   0.17310000000000    0.42750000000000    0.56890000000000     C    (8d)
   0.32690000000000    0.57250000000000    1.06890000000000     C    (8d)
   0.34900000000000    0.08000000000000    0.36700000000000     H    (8d)
   0.15100000000000   -0.08000000000000    0.86700000000000     H    (8d)
  -0.34900000000000    0.58000000000000   -0.36700000000000     H    (8d)
   0.84900000000000    0.42000000000000    0.13300000000000     H    (8d)
  -0.34900000000000   -0.08000000000000   -0.36700000000000     H    (8d)
   0.84900000000000    0.08000000000000    0.13300000000000     H    (8d)
   0.34900000000000    0.42000000000000    0.36700000000000     H    (8d)
   0.15100000000000    0.58000000000000    0.86700000000000     H    (8d)
   0.41200000000000    0.01700000000000    0.34700000000000     H    (8d)
   0.08800000000000   -0.01700000000000    0.84700000000000     H    (8d)
  -0.41200000000000    0.51700000000000   -0.34700000000000     H    (8d)
   0.91200000000000    0.48300000000000    0.15300000000000     H    (8d)
  -0.41200000000000   -0.01700000000000   -0.34700000000000     H    (8d)
   0.91200000000000    0.01700000000000    0.15300000000000     H    (8d)
   0.41200000000000    0.48300000000000    0.34700000000000     H    (8d)
   0.08800000000000    0.51700000000000    0.84700000000000     H    (8d)
   0.36535000000000    0.25000000000000    0.02500000000000     K    (4c)
   0.13465000000000    0.75000000000000    0.52500000000000     K    (4c)
  -0.36535000000000    0.75000000000000   -0.02500000000000     K    (4c)
   0.86535000000000    0.25000000000000    0.47500000000000     K    (4c)
   0.46252000000000    0.25000000000000    0.52249000000000     K    (4c)
   0.03748000000000    0.75000000000000    1.02249000000000     K    (4c)
  -0.46252000000000    0.75000000000000   -0.52249000000000     K    (4c)
   0.96252000000000    0.25000000000000   -0.02249000000000     K    (4c)
   0.14835000000000    0.04756000000000    0.18788000000000     K    (8d)
   0.35165000000000   -0.04756000000000    0.68788000000000     K    (8d)
  -0.14835000000000    0.54756000000000   -0.18788000000000     K    (8d)
   0.64835000000000    0.45244000000000    0.31212000000000     K    (8d)
  -0.14835000000000   -0.04756000000000   -0.18788000000000     K    (8d)
   0.64835000000000    0.04756000000000    0.31212000000000     K    (8d)
   0.14835000000000    0.45244000000000    0.18788000000000     K    (8d)
   0.35165000000000    0.54756000000000    0.68788000000000     K    (8d)
   0.13750000000000    0.25000000000000    0.59643000000000     Mo   (4c)
   0.36250000000000    0.75000000000000    1.09643000000000     Mo   (4c)
  -0.13750000000000    0.75000000000000   -0.59643000000000     Mo   (4c)
```

```
   0.63750000000000    0.25000000000000   -0.09643000000000    Mo    (4c)
   0.02100000000000    0.25000000000000    0.28660000000000    N     (4c)
   0.47900000000000    0.75000000000000    0.78660000000000    N     (4c)
  -0.02100000000000    0.75000000000000   -0.28660000000000    N     (4c)
   0.52100000000000    0.25000000000000    0.21340000000000    N     (4c)
   0.26140000000000    0.25000000000000    0.29570000000000    N     (4c)
   0.23860000000000    0.75000000000000    0.79570000000000    N     (4c)
  -0.26140000000000    0.75000000000000   -0.29570000000000    N     (4c)
   0.76140000000000    0.25000000000000    0.20430000000000    N     (4c)
   0.32600000000000    0.25000000000000    0.71330000000000    N     (4c)
   0.17400000000000    0.75000000000000    1.21330000000000    N     (4c)
  -0.32600000000000    0.75000000000000   -0.71330000000000    N     (4c)
   0.82600000000000    0.25000000000000   -0.21330000000000    N     (4c)
   0.64010000000000    0.25000000000000    0.52210000000000    N     (4c)
  -0.14010000000000    0.75000000000000    1.02210000000000    N     (4c)
  -0.64010000000000    0.75000000000000   -0.52210000000000    N     (4c)
   1.14010000000000    0.25000000000000   -0.02210000000000    N     (4c)
   0.01230000000000    0.58090000000000    0.28930000000000    N     (8d)
   0.48770000000000   -0.58090000000000    0.78930000000000    N     (8d)
  -0.01230000000000    1.08090000000000   -0.28930000000000    N     (8d)
   0.51230000000000   -0.08090000000000    0.21070000000000    N     (8d)
  -0.01230000000000   -0.58090000000000   -0.28930000000000    N     (8d)
   0.51230000000000    0.58090000000000    0.21070000000000    N     (8d)
   0.01230000000000   -0.08090000000000    0.28930000000000    N     (8d)
   0.48770000000000    1.08090000000000    0.78930000000000    N     (8d)
   0.30680000000000    0.02160000000000    0.05000000000000    N     (8d)
   0.19320000000000   -0.02160000000000    0.55000000000000    N     (8d)
  -0.30680000000000    0.52160000000000   -0.05000000000000    N     (8d)
   0.80680000000000    0.47840000000000    0.45000000000000    N     (8d)
  -0.30680000000000   -0.02160000000000   -0.05000000000000    N     (8d)
   0.80680000000000    0.02160000000000    0.45000000000000    N     (8d)
   0.30680000000000    0.47840000000000    0.05000000000000    N     (8d)
   0.19320000000000    0.52160000000000    0.55000000000000    N     (8d)
   0.38880000000000    0.05930000000000    0.40900000000000    O     (8d)
   0.11120000000000   -0.05930000000000    0.90900000000000    O     (8d)
  -0.38880000000000    0.55930000000000   -0.40900000000000    O     (8d)
   0.88880000000000    0.44070000000000    0.09100000000000    O     (8d)
  -0.38880000000000   -0.05930000000000   -0.40900000000000    O     (8d)
   0.88880000000000    0.05930000000000    0.09100000000000    O     (8d)
   0.38880000000000    0.44070000000000    0.40900000000000    O     (8d)
   0.11120000000000    0.55930000000000    0.90900000000000    O     (8d)
```

$SbCl_5 \cdot POCl_3$: A8BCD_oP44_62_4c2d_c_c_c - CIF

```
# CIF file
data_findsym-output
_audit_creation_method FINDSYM

_chemical_name_mineral 'Cl8OPSb'
_chemical_formula_sum 'Cl8 O P Sb'

loop_
_publ_author_name
 'I. Lindqvist'
 'C.-I. Br\"{a}nd\\'{e}n'
_journal_name_full_name
;
 Acta Cristallographica
;
_journal_volume 12
_journal_year 1959
_journal_page_first 642
_journal_page_last 645
_publ_Section_title
;
 The Crystal Structure of SbCl$_{5}$$\cdot$POCl$_{3}$
;

# Found in The Crystal Structure of SbCl$_{5}$.PO(CH$_{3}$)$_{3}$, 1961

_aflow_title 'SbCl$_{5}$$\cdot$POCl$_{3}$ Structure'
_aflow_proto 'A8BCD_oP44_62_4c2d_c_c_c'
_aflow_params 'a,b/a,c/a,x_{1},z_{1},x_{2},z_{2},x_{3},z_{3},x_{4},z_{4
    ↪ },x_{5},z_{5},x_{6},z_{6},x_{7},z_{7},x_{8},y_{8},z_{8},x_{9},
    ↪ y_{9},z_{9}'
_aflow_params_values '16.42,0.490864799026,0.543848964677,0.2585,-0.0805
    ↪ ,0.0202,0.2048,0.2249,0.2938,0.4644,0.8699,0.0706,0.8774,0.0742
    ↪ ,0.7142,0.145,0.0797,0.1314,0.4442,0.6327,0.1405,0.5383,0.0597'
_aflow_Strukturbericht 'None'
_aflow_Pearson 'oP44'

_symmetry_space_group_name_H-M "P 21/n 21/m 21/a"
_symmetry_Int_Tables_number 62

_cell_length_a    16.42000
_cell_length_b    8.06000
_cell_length_c    8.93000
_cell_angle_alpha 90.00000
_cell_angle_beta  90.00000
_cell_angle_gamma 90.00000

loop_
_space_group_symop_id
_space_group_symop_operation_xyz
1 x,y,z
2 x+1/2,-y+1/2,-z+1/2
3 -x,y+1/2,-z
4 -x+1/2,-y,z+1/2
5 -x,-y,-z
6 -x+1/2,y+1/2,z+1/2
7 x,-y+1/2,z
8 x+1/2,y,-z+1/2

loop_
_atom_site_label
_atom_site_type_symbol
_atom_site_symmetry_multiplicity
_atom_site_Wyckoff_label
_atom_site_fract_x
_atom_site_fract_y
_atom_site_fract_z
_atom_site_occupancy
Cl1 Cl 4 c 0.25850 0.25000 -0.08050 1.00000
Cl2 Cl 4 c 0.02020 0.25000 0.20480 1.00000
Cl3 Cl 4 c 0.22490 0.25000 0.29380 1.00000
Cl4 Cl 4 c 0.46440 0.25000 0.86990 1.00000
O1 O 4 c 0.07060 0.25000 0.87740 1.00000
P1 P 4 c 0.07420 0.25000 0.71420 1.00000
Sb1 Sb 4 c 0.14500 0.25000 0.07970 1.00000
Cl5 Cl 8 d 0.13140 0.44420 0.63270 1.00000
Cl6 Cl 8 d 0.14050 0.53830 0.05970 1.00000
```

$SbCl_5 \cdot POCl_3$: A8BCD_oP44_62_4c2d_c_c_c - POSCAR

```
A8BCD_oP44_62_4c2d_c_c_c & a,b/a,c/a,x1,z1,x2,z2,x3,z3,x4,z4,x5,z5,x6,z6
    ↪ ,x7,z7,x8,y8,z8,x9,y9,z9 --params=16.42,0.490864799026,
    ↪ 0.543848964677,0.2585,-0.0805,0.0202,0.2048,0.2249,0.2938,
    ↪ 0.4644,0.8699,0.0706,0.8774,0.0742,0.7142,0.145,0.0797,0.1314,
    ↪ 0.4442,0.6327,0.1405,0.5383,0.0597 & Pnma D_{2h}^{16} #62 (c^7d
    ↪ ^2) & oP44 & None & Cl8OPSb & Cl8OPSb & I. Lindqvist and C.-I.
    ↪ Br\"{a}nd\'{e}n, Acta Cryst. 12, 642-645 (1959)
   1.00000000000000
  16.42000000000000    0.00000000000000    0.00000000000000
   0.00000000000000    8.06000000000000    0.00000000000000
   0.00000000000000    0.00000000000000    8.93000000000000
    Cl     O     P     Sb
    32     4     4      4
Direct
   0.25850000000000    0.25000000000000   -0.08050000000000    Cl    (4c)
   0.24150000000000    0.75000000000000    0.41950000000000    Cl    (4c)
  -0.25850000000000    0.75000000000000    0.08050000000000    Cl    (4c)
   0.75850000000000    0.25000000000000    0.58050000000000    Cl    (4c)
   0.02020000000000    0.25000000000000    0.20480000000000    Cl    (4c)
   0.47980000000000    0.75000000000000    0.70480000000000    Cl    (4c)
  -0.02020000000000    0.75000000000000   -0.20480000000000    Cl    (4c)
   0.52020000000000    0.25000000000000    0.29520000000000    Cl    (4c)
   0.22490000000000    0.25000000000000    0.29380000000000    Cl    (4c)
   0.27510000000000    0.75000000000000    0.79380000000000    Cl    (4c)
  -0.22490000000000    0.75000000000000   -0.29380000000000    Cl    (4c)
   0.72490000000000    0.25000000000000    0.20620000000000    Cl    (4c)
   0.46440000000000    0.25000000000000    0.86990000000000    Cl    (4c)
   0.03560000000000    0.75000000000000    1.36990000000000    Cl    (4c)
  -0.46440000000000    0.75000000000000   -0.86990000000000    Cl    (4c)
   0.96440000000000    0.25000000000000   -0.36990000000000    Cl    (4c)
   0.13140000000000    0.44420000000000    0.63270000000000    Cl    (8d)
   0.36860000000000   -0.44420000000000    1.13270000000000    Cl    (8d)
  -0.13140000000000    0.94420000000000   -0.63270000000000    Cl    (8d)
   0.63140000000000    0.05580000000000   -0.13270000000000    Cl    (8d)
  -0.13140000000000   -0.44420000000000   -0.63270000000000    Cl    (8d)
   0.63140000000000    0.44420000000000   -0.13270000000000    Cl    (8d)
   0.13140000000000    0.05580000000000    0.63270000000000    Cl    (8d)
   0.36860000000000    0.94420000000000    1.13270000000000    Cl    (8d)
   0.14050000000000    0.53830000000000    0.05970000000000    Cl    (8d)
   0.35950000000000   -0.53830000000000    0.55970000000000    Cl    (8d)
  -0.14050000000000    1.03830000000000   -0.05970000000000    Cl    (8d)
   0.64050000000000   -0.03830000000000    0.44030000000000    Cl    (8d)
  -0.14050000000000   -0.53830000000000   -0.05970000000000    Cl    (8d)
   0.64050000000000    0.53830000000000    0.44030000000000    Cl    (8d)
   0.14050000000000   -0.03830000000000    0.05970000000000    Cl    (8d)
   0.35950000000000    1.03830000000000    0.55970000000000    Cl    (8d)
   0.07060000000000    0.25000000000000    0.87740000000000    O     (4c)
   0.42940000000000    0.75000000000000    1.37740000000000    O     (4c)
  -0.07060000000000    0.75000000000000   -0.87740000000000    O     (4c)
   0.57060000000000    0.25000000000000   -0.37740000000000    O     (4c)
   0.07420000000000    0.25000000000000    0.71420000000000    P     (4c)
   0.42580000000000    0.75000000000000    1.21420000000000    P     (4c)
  -0.07420000000000    0.75000000000000   -0.71420000000000    P     (4c)
   0.57420000000000    0.25000000000000   -0.21420000000000    P     (4c)
   0.14500000000000    0.25000000000000    0.07970000000000    Sb    (4c)
   0.35500000000000    0.75000000000000    0.57970000000000    Sb    (4c)
  -0.14500000000000    0.75000000000000   -0.07970000000000    Sb    (4c)
   0.64500000000000    0.25000000000000    0.42030000000000    Sb    (4c)
```

Autunite $\{Ca[(UO_2)(PO_4)]_2(H_2O)_{11}\}$: AB22C23D2E2_oP200_62_c_11d_3c10d_d_d - CIF

```
# CIF file
data_findsym-output
_audit_creation_method FINDSYM

_chemical_name_mineral 'Autunite'
_chemical_formula_sum 'Ca H22 O23 P2 U2'

loop_
_publ_author_name
 'A. J. Locock'
 'P. C. Burns'
_journal_name_full_name
;
 American Mineralogist
;
_journal_volume 88
_journal_year 2003
_journal_page_first 240
_journal_page_last 244
_publ_Section_title
;
 The crystal structure of synthetic autunite, Ca[(UO$_{2}$)(PO$_{4}$)]$_
    ↪ {2}$(H$_{2}$O)$_{11}$
;
```

```
# Found in The American Mineralogist Crystal Structure Database, 2003

_aflow_title 'Autunite \{Ca[(UO$_{2}$)(PO$_{4}$)]$_{2}$(H$_{2}$O)$_{11}$
    ↪ \} Structure'
_aflow_proto 'AB22C23D2E2_oP200_62_c_11d_3c10d_d_d'
_aflow_params 'a,b/a,c/a,x_{1},z_{1},x_{2},z_{2},x_{3},z_{3},x_{4},z_{4
    ↪ },x_{5},y_{5},z_{5},x_{6},y_{6},z_{6},x_{7},y_{7},z_{7},x_{8},
    ↪ y_{8},z_{8},x_{9},y_{9},z_{9},x_{10},y_{10},z_{10},x_{11},y_{11
    ↪ },z_{11},x_{12},y_{12},z_{12},x_{13},y_{13},z_{13},x_{14},y_{14
    ↪ },z_{14},x_{15},y_{15},z_{15},x_{16},y_{16},z_{16},x_{17},y_{17
    ↪ },z_{17},x_{18},y_{18},z_{18},x_{19},y_{19},z_{19},x_{20},y_{20
    ↪ },z_{20},x_{21},y_{21},z_{21},x_{22},y_{22},z_{22},x_{23},y_{23
    ↪ },z_{23},x_{24},y_{24},z_{24},x_{25},y_{25},z_{25},x_{26},y_{26
    ↪ },z_{26},x_{27},y_{27},z_{27}'
_aflow_params_values '14.0135,1.47800335391,0.499225746601,0.6251,-
    ↪ 0.0499,0.7409,0.2149,0.4998,0.2167,0.6263,0.5757,0.226,0.783,
    ↪ 0.722,0.697,0.133,-0.088,0.721,0.154,0.715,0.533,0.1191,0.835,
    ↪ 0.508,0.165,0.701,0.537,0.7851,0.733,0.63,0.2056,0.544,0.772,
    ↪ 0.174,0.558,0.8972,0.155,0.473,0.568,0.8819,0.575,0.6483,0.836,
    ↪ 0.522,0.6235,-0.0449,0.2482,0.625,0.1278,0.2472,0.6252,0.044,
    ↪ 0.5733,0.5373,-0.044,0.7451,0.7118,-0.0443,0.7553,0.6206,0.0452
    ↪ ,-0.0774,0.7368,0.1647,0.851,0.5126,0.1645,0.8438,0.8257,0.1614
    ↪ ,0.4745,0.5759,0.8379,0.5243,0.6245,0.0003,0.7478,0.625,0.0412,
    ↪ 0.2498'
_aflow_Strukturbericht 'None'
_aflow_Pearson 'oP200'

_symmetry_space_group_name_H-M "P 21/n 21/m 21/a"
_symmetry_Int_Tables_number 62

_cell_length_a 14.01350
_cell_length_b 20.71200
_cell_length_c 6.99590
_cell_angle_alpha 90.00000
_cell_angle_beta 90.00000
_cell_angle_gamma 90.00000

loop_
_space_group_symop_id
_space_group_symop_operation_xyz
1 x,y,z
2 x+1/2,-y+1/2,-z+1/2
3 -x,y+1/2,-z
4 -x+1/2,-y,z+1/2
5 -x,-y,-z
6 -x+1/2,y+1/2,z+1/2
7 x,-y+1/2,z
8 x+1/2,y,-z+1/2

loop_
_atom_site_label
_atom_site_type_symbol
_atom_site_symmetry_multiplicity
_atom_site_Wyckoff_label
_atom_site_fract_x
_atom_site_fract_y
_atom_site_fract_z
_atom_site_occupancy
Ca1 Ca 4 c 0.62510 0.25000 -0.04990 1.00000
O1 O 4 c 0.74090 0.25000 0.21490 1.00000
O2 O 4 c 0.49980 0.25000 0.21670 1.00000
O3 O 4 c 0.62630 0.25000 0.57570 1.00000
H1 H 8 d 0.22600 0.78300 0.72200 1.00000
H2 H 8 d 0.69700 0.13300 -0.08800 1.00000
H3 H 8 d 0.72100 0.15400 0.71500 1.00000
H4 H 8 d 0.53300 0.11910 0.83500 1.00000
H5 H 8 d 0.50800 0.16500 0.70100 1.00000
H6 H 8 d 0.53700 0.78510 0.73300 1.00000
H7 H 8 d 0.63000 0.20560 0.54400 1.00000
H8 H 8 d 0.77200 0.17400 0.55800 1.00000
H9 H 8 d 0.89720 0.15500 0.47300 1.00000
H10 H 8 d 0.56800 0.88190 0.57500 1.00000
H11 H 8 d 0.64830 0.83600 0.52200 1.00000
O4 O 8 d 0.62350 -0.04490 0.24820 1.00000
O5 O 8 d 0.62500 0.12780 0.24720 1.00000
O6 O 8 d 0.62520 0.04400 0.57330 1.00000
O7 O 8 d 0.53730 -0.04400 0.74510 1.00000
O8 O 8 d 0.71180 -0.04430 0.75530 1.00000
O9 O 8 d 0.62060 0.04520 -0.07740 1.00000
O10 O 8 d 0.73680 0.16470 0.85100 1.00000
O11 O 8 d 0.51260 0.16450 0.84380 1.00000
O12 O 8 d 0.82570 0.16140 0.47450 1.00000
O13 O 8 d 0.57590 0.83790 0.52430 1.00000
P1 P 8 d 0.62450 0.00030 0.74780 1.00000
U1 U 8 d 0.62500 0.04120 0.24980 1.00000
```

Autunite {$Ca[(UO_2)(PO_4)]_2(H_2O)_{11}$}: AB22C23D2E2_oP200_62_c_11d_3c10d_d_d - POSCAR

```
AB22C23D2E2_oP200_62_c_11d_3c10d_d_d & a,b/a,c/a,x1,z1,x2,z2,x3,z3,x4,z4
    ↪ ,x5,y5,z5,x6,y6,z6,x7,y7,z7,x8,y8,z8,x9,y9,z9,x10,y10,z10,x11,
    ↪ y11,z11,x12,y12,z12,x13,y13,z13,x14,y14,z14,x15,y15,z15,x16,y16
    ↪ ,z16,x17,y17,z17,x18,y18,z18,x19,y19,z19,x20,y20,z20,x21,y21,
    ↪ z21,x22,y22,z22,x23,y23,z23,x24,y24,z24,x25,y25,z25,x26,y26,z26
    ↪ ,x27,y27,z27 --params=14.0135,1.47800335391,0.499225746601,
    ↪ 0.6251,-0.0499,0.7409,0.2149,0.4998,0.2167,0.6263,0.5757,0.226,
    ↪ 0.783,0.722,0.697,0.133,-0.088,0.721,0.154,0.715,0.533,0.1191,
    ↪ 0.835,0.508,0.165,0.701,0.537,0.7851,0.733,0.63,0.2056,0.544,
    ↪ 0.772,0.174,0.558,0.8972,0.155,0.473,0.568,0.8819,0.575,0.6483,
    ↪ 0.836,0.522,0.6235,-0.0449,0.2482,0.625,0.1278,0.2472,0.6252,
    ↪ 0.044,0.5733,0.5373,-0.044,0.7451,0.7118,-0.0443,0.7553,0.6206,
    ↪ 0.0452,-0.0774,0.7368,0.1647,0.851,0.5126,0.1645,0.8438,0.8257,
    ↪ 0.1614,0.4745,0.5759,0.8379,0.5243,0.6245,0.0003,0.7478,0.625,
    ↪ 0.0412,0.2498 & Pnma D_{2h}^{16} #62 (c^4d^23) & oP200 & None &
    ↪  CaH22O23P2U2 & Autunite & A. J. Locock and P. C. Burns, Am.
    ↪ Mineral. 88, 240-244 (2003)
```

```
1.00000000000000
14.01350000000000 0.00000000000000 0.00000000000000
0.00000000000000 20.71200000000000 0.00000000000000
0.00000000000000 0.00000000000000 6.99590000000000
Ca H O P U
4 88 92 8 8
Direct
0.62510000000000 0.25000000000000 -0.04990000000000 Ca (4c)
-0.12510000000000 0.75000000000000 0.45010000000000 Ca (4c)
-0.62510000000000 0.75000000000000 0.04990000000000 Ca (4c)
1.12510000000000 0.25000000000000 0.54990000000000 Ca (4c)
0.22600000000000 0.78300000000000 0.72200000000000 H (8d)
0.27400000000000 -0.78300000000000 1.22200000000000 H (8d)
-0.22600000000000 1.28300000000000 -0.72200000000000 H (8d)
0.72600000000000 -0.28300000000000 -0.22200000000000 H (8d)
-0.22600000000000 -0.78300000000000 -0.72200000000000 H (8d)
0.72600000000000 0.78300000000000 -0.22200000000000 H (8d)
0.22600000000000 -0.28300000000000 0.72200000000000 H (8d)
0.27400000000000 1.28300000000000 1.22200000000000 H (8d)
0.69700000000000 0.13300000000000 -0.08800000000000 H (8d)
-0.19700000000000 -0.13300000000000 0.41200000000000 H (8d)
-0.69700000000000 0.63300000000000 0.08800000000000 H (8d)
1.19700000000000 0.36700000000000 0.58800000000000 H (8d)
-0.69700000000000 -0.13300000000000 0.08800000000000 H (8d)
1.19700000000000 0.13300000000000 0.58800000000000 H (8d)
0.69700000000000 0.36700000000000 -0.08800000000000 H (8d)
-0.19700000000000 0.63300000000000 0.41200000000000 H (8d)
0.72100000000000 0.15400000000000 0.71500000000000 H (8d)
-0.22100000000000 -0.15400000000000 1.21500000000000 H (8d)
-0.72100000000000 0.65400000000000 -0.71500000000000 H (8d)
1.22100000000000 0.34600000000000 -0.21500000000000 H (8d)
-0.72100000000000 -0.15400000000000 -0.71500000000000 H (8d)
1.22100000000000 0.15400000000000 -0.21500000000000 H (8d)
0.72100000000000 0.34600000000000 0.71500000000000 H (8d)
-0.22100000000000 0.65400000000000 1.21500000000000 H (8d)
0.53300000000000 0.11910000000000 0.83500000000000 H (8d)
-0.03300000000000 -0.11910000000000 1.33500000000000 H (8d)
-0.53300000000000 0.61910000000000 -0.83500000000000 H (8d)
1.03300000000000 0.38090000000000 -0.33500000000000 H (8d)
-0.53300000000000 -0.11910000000000 -0.83500000000000 H (8d)
1.03300000000000 0.11910000000000 -0.33500000000000 H (8d)
0.53300000000000 0.38090000000000 0.83500000000000 H (8d)
-0.03300000000000 0.61910000000000 1.33500000000000 H (8d)
0.50800000000000 0.16500000000000 0.70100000000000 H (8d)
-0.00800000000000 -0.16500000000000 1.20100000000000 H (8d)
-0.50800000000000 0.66500000000000 -0.70100000000000 H (8d)
1.00800000000000 0.33500000000000 -0.20100000000000 H (8d)
-0.50800000000000 -0.16500000000000 -0.70100000000000 H (8d)
1.00800000000000 0.16500000000000 -0.20100000000000 H (8d)
0.50800000000000 0.33500000000000 0.70100000000000 H (8d)
-0.00800000000000 0.66500000000000 1.20100000000000 H (8d)
0.53700000000000 0.78510000000000 0.73300000000000 H (8d)
-0.03700000000000 -0.78510000000000 1.23300000000000 H (8d)
-0.53700000000000 1.28510000000000 -0.73300000000000 H (8d)
1.03700000000000 -0.28510000000000 -0.23300000000000 H (8d)
-0.53700000000000 -0.78510000000000 -0.73300000000000 H (8d)
1.03700000000000 0.78510000000000 -0.23300000000000 H (8d)
0.53700000000000 -0.28510000000000 0.73300000000000 H (8d)
-0.03700000000000 1.28510000000000 1.23300000000000 H (8d)
0.63000000000000 0.20560000000000 0.54400000000000 H (8d)
-0.13000000000000 -0.20560000000000 1.04400000000000 H (8d)
-0.63000000000000 0.70560000000000 -0.54400000000000 H (8d)
1.13000000000000 0.29440000000000 -0.04400000000000 H (8d)
-0.63000000000000 -0.20560000000000 -0.54400000000000 H (8d)
1.13000000000000 0.20560000000000 -0.04400000000000 H (8d)
0.63000000000000 0.29440000000000 0.54400000000000 H (8d)
-0.13000000000000 0.70560000000000 1.04400000000000 H (8d)
0.77200000000000 0.17400000000000 0.55800000000000 H (8d)
-0.27200000000000 -0.17400000000000 1.05800000000000 H (8d)
-0.77200000000000 0.67400000000000 -0.55800000000000 H (8d)
1.27200000000000 0.32600000000000 -0.05800000000000 H (8d)
-0.77200000000000 -0.17400000000000 -0.55800000000000 H (8d)
1.27200000000000 0.17400000000000 -0.05800000000000 H (8d)
0.77200000000000 0.32600000000000 0.55800000000000 H (8d)
-0.27200000000000 0.67400000000000 1.05800000000000 H (8d)
0.89720000000000 0.15500000000000 0.47300000000000 H (8d)
-0.39720000000000 -0.15500000000000 0.97300000000000 H (8d)
-0.89720000000000 0.65500000000000 -0.47300000000000 H (8d)
1.39720000000000 0.34500000000000 0.02700000000000 H (8d)
-0.89720000000000 -0.15500000000000 -0.47300000000000 H (8d)
1.39720000000000 0.15500000000000 0.02700000000000 H (8d)
0.89720000000000 0.34500000000000 0.47300000000000 H (8d)
-0.39720000000000 0.65500000000000 0.97300000000000 H (8d)
0.56800000000000 0.88190000000000 0.57500000000000 H (8d)
-0.06800000000000 -0.88190000000000 1.07500000000000 H (8d)
-0.56800000000000 1.38190000000000 -0.57500000000000 H (8d)
1.06800000000000 -0.38190000000000 -0.07500000000000 H (8d)
-0.56800000000000 -0.88190000000000 -0.57500000000000 H (8d)
1.06800000000000 0.88190000000000 -0.07500000000000 H (8d)
0.56800000000000 -0.38190000000000 0.57500000000000 H (8d)
-0.06800000000000 1.38190000000000 1.07500000000000 H (8d)
0.64830000000000 0.83600000000000 0.52200000000000 H (8d)
-0.14830000000000 -0.83600000000000 1.02200000000000 H (8d)
-0.64830000000000 1.33600000000000 -0.52200000000000 H (8d)
1.14830000000000 -0.33600000000000 -0.02200000000000 H (8d)
-0.64830000000000 -0.83600000000000 -0.52200000000000 H (8d)
1.14830000000000 0.83600000000000 -0.02200000000000 H (8d)
0.64830000000000 -0.33600000000000 0.52200000000000 H (8d)
-0.14830000000000 1.33600000000000 1.02200000000000 H (8d)
0.74090000000000 0.25000000000000 0.21490000000000 O (4c)
-0.24090000000000 0.75000000000000 0.71490000000000 O (4c)
-0.74090000000000 0.75000000000000 -0.21490000000000 O (4c)
1.24090000000000 0.25000000000000 0.28510000000000 O (4c)
0.49980000000000 0.25000000000000 0.21670000000000 O (4c)
0.00020000000000 0.75000000000000 0.71670000000000 O (4c)
```

```
-0.49980000000000   0.75000000000000  -0.21670000000000   O   (4c)
 0.99980000000000   0.25000000000000   0.28330000000000   O   (4c)
 0.62630000000000   0.25000000000000   0.57570000000000   O   (4c)
-0.12630000000000   0.75000000000000   1.07570000000000   O   (4c)
-0.62630000000000   0.75000000000000  -0.57570000000000   O   (4c)
 1.12630000000000   0.25000000000000  -0.07570000000000   O   (4c)
 0.62350000000000  -0.04490000000000   0.24820000000000   O   (8d)
-0.12350000000000   0.04490000000000   0.74820000000000   O   (8d)
-0.62350000000000   0.45510000000000  -0.24820000000000   O   (8d)
 1.12350000000000   0.54490000000000   0.25180000000000   O   (8d)
-0.62350000000000   0.04490000000000  -0.24820000000000   O   (8d)
 1.12350000000000  -0.04490000000000   0.25180000000000   O   (8d)
 0.62350000000000   0.54490000000000   0.24820000000000   O   (8d)
-0.12350000000000   0.45510000000000   0.74820000000000   O   (8d)
 0.62500000000000   0.12780000000000   0.24720000000000   O   (8d)
-0.12500000000000  -0.12780000000000   0.74720000000000   O   (8d)
-0.62500000000000   0.62780000000000  -0.24720000000000   O   (8d)
 1.12500000000000   0.37220000000000   0.25280000000000   O   (8d)
-0.62500000000000  -0.12780000000000  -0.24720000000000   O   (8d)
 1.12500000000000   0.12780000000000   0.25280000000000   O   (8d)
 0.62500000000000   0.37220000000000   0.24720000000000   O   (8d)
-0.12500000000000   0.62780000000000   0.74720000000000   O   (8d)
 0.62520000000000   0.04400000000000   0.57330000000000   O   (8d)
-0.12520000000000  -0.04400000000000   1.07330000000000   O   (8d)
-0.62520000000000   0.54400000000000  -0.57330000000000   O   (8d)
 1.12520000000000   0.45600000000000  -0.07330000000000   O   (8d)
-0.62520000000000  -0.04400000000000  -0.57330000000000   O   (8d)
 1.12520000000000   0.04400000000000  -0.07330000000000   O   (8d)
 0.62520000000000   0.45600000000000   0.57330000000000   O   (8d)
-0.12520000000000   0.54400000000000   1.07330000000000   O   (8d)
 0.53730000000000  -0.04400000000000   0.74510000000000   O   (8d)
-0.03730000000000   0.04400000000000   1.24510000000000   O   (8d)
-0.53730000000000   0.45600000000000  -0.74510000000000   O   (8d)
 1.03730000000000   0.54400000000000  -0.24510000000000   O   (8d)
-0.53730000000000   0.04400000000000  -0.74510000000000   O   (8d)
 1.03730000000000  -0.04400000000000  -0.24510000000000   O   (8d)
 0.53730000000000   0.54400000000000   0.74510000000000   O   (8d)
-0.03730000000000   0.45600000000000   1.24510000000000   O   (8d)
 0.71180000000000  -0.04430000000000   0.75530000000000   O   (8d)
-0.21180000000000   0.04430000000000   1.25530000000000   O   (8d)
-0.71180000000000   0.45570000000000  -0.75530000000000   O   (8d)
 1.21180000000000   0.54430000000000  -0.25530000000000   O   (8d)
-0.71180000000000   0.04430000000000  -0.75530000000000   O   (8d)
 1.21180000000000  -0.04430000000000  -0.25530000000000   O   (8d)
 0.71180000000000   0.54430000000000   0.75530000000000   O   (8d)
-0.21180000000000   0.45570000000000   1.25530000000000   O   (8d)
 0.62060000000000   0.04520000000000  -0.07740000000000   O   (8d)
-0.12060000000000  -0.04520000000000   0.42260000000000   O   (8d)
-0.62060000000000   0.54520000000000   0.07740000000000   O   (8d)
 1.12060000000000   0.45480000000000   0.57740000000000   O   (8d)
-0.62060000000000  -0.04520000000000   0.07740000000000   O   (8d)
 1.12060000000000   0.04520000000000   0.57740000000000   O   (8d)
 0.62060000000000   0.45480000000000  -0.07740000000000   O   (8d)
-0.12060000000000   0.54520000000000   0.42260000000000   O   (8d)
 0.73680000000000   0.16470000000000   0.85100000000000   O   (8d)
-0.23680000000000  -0.16470000000000   1.35100000000000   O   (8d)
-0.73680000000000   0.66470000000000  -0.85100000000000   O   (8d)
 1.23680000000000   0.33530000000000  -0.35100000000000   O   (8d)
-0.73680000000000  -0.16470000000000  -0.85100000000000   O   (8d)
 1.23680000000000   0.16470000000000  -0.35100000000000   O   (8d)
 0.73680000000000   0.33530000000000   0.85100000000000   O   (8d)
-0.23680000000000   0.66470000000000   1.35100000000000   O   (8d)
 0.51260000000000   0.16450000000000   0.84380000000000   O   (8d)
-0.01260000000000  -0.16450000000000   1.34380000000000   O   (8d)
-0.51260000000000   0.66450000000000  -0.84380000000000   O   (8d)
 1.01260000000000   0.33550000000000  -0.34380000000000   O   (8d)
-0.51260000000000  -0.16450000000000  -0.84380000000000   O   (8d)
 1.01260000000000   0.16450000000000  -0.34380000000000   O   (8d)
 0.51260000000000   0.33550000000000   0.84380000000000   O   (8d)
-0.01260000000000   0.66450000000000   1.34380000000000   O   (8d)
 0.82570000000000   0.16140000000000   0.47450000000000   O   (8d)
-0.32570000000000  -0.16140000000000   0.97450000000000   O   (8d)
-0.82570000000000   0.66140000000000  -0.47450000000000   O   (8d)
 1.32570000000000   0.33860000000000   0.02550000000000   O   (8d)
-0.82570000000000  -0.16140000000000  -0.47450000000000   O   (8d)
 1.32570000000000   0.16140000000000   0.02550000000000   O   (8d)
 0.82570000000000   0.33860000000000   0.47450000000000   O   (8d)
-0.32570000000000   0.66140000000000   0.97450000000000   O   (8d)
 0.57590000000000   0.83790000000000   0.52430000000000   O   (8d)
-0.07590000000000  -0.83790000000000   1.02430000000000   O   (8d)
-0.57590000000000   1.33790000000000  -0.52430000000000   O   (8d)
 1.07590000000000  -0.33790000000000  -0.02430000000000   O   (8d)
-0.57590000000000  -0.83790000000000  -0.52430000000000   O   (8d)
 1.07590000000000   0.83790000000000  -0.02430000000000   O   (8d)
 0.57590000000000  -0.33790000000000   0.52430000000000   O   (8d)
-0.07590000000000   1.33790000000000   1.02430000000000   O   (8d)
 0.62450000000000   0.00030000000000   0.74780000000000   P   (8d)
-0.12450000000000  -0.00030000000000   1.24780000000000   P   (8d)
-0.62450000000000   0.50030000000000  -0.74780000000000   P   (8d)
 1.12450000000000   0.49970000000000  -0.24780000000000   P   (8d)
-0.62450000000000  -0.00030000000000  -0.74780000000000   P   (8d)
 1.12450000000000   0.00030000000000  -0.24780000000000   P   (8d)
 0.62450000000000   0.49970000000000   0.74780000000000   P   (8d)
-0.12450000000000   0.50030000000000   1.24780000000000   P   (8d)
 0.62500000000000   0.04120000000000   0.24980000000000   U   (8d)
-0.12500000000000  -0.04120000000000   0.74980000000000   U   (8d)
-0.62500000000000   0.54120000000000  -0.24980000000000   U   (8d)
 1.12500000000000   0.45880000000000   0.25020000000000   U   (8d)
-0.62500000000000  -0.04120000000000  -0.24980000000000   U   (8d)
 1.12500000000000   0.04120000000000   0.25020000000000   U   (8d)
 0.62500000000000   0.45880000000000   0.24980000000000   U   (8d)
-0.12500000000000   0.54120000000000   0.74980000000000   U   (8d)
```

Atacamite ($Cu_2(OH)_3Cl$): AB2C3D3_oP36_62_c_ac_cd_cd - CIF

```
# CIF file
data_findsym-output
_audit_creation_method FINDSYM

_chemical_name_mineral 'Atacamite'
_chemical_formula_sum 'Cl Cu2 H3 O3'

loop_
_publ_author_name
 'J. B. Parise'
 'B. G. Hyde'
_journal_name_full_name
;
 Acta Crystallographica Section C: Structural Chemistry
;
_journal_volume 42
_journal_year 1986
_journal_page_first 1277
_journal_page_last 1280
_publ_Section_title
;
 The structure of atacamite and its relationship to spinel
;

# Found in The American Mineralogist Crystal Structure Database, 2003

_aflow_title 'Atacamite (Cu$_{2}$(OH)$_{3}$Cl) Structure'
_aflow_proto 'AB2C3D3_oP36_62_c_ac_cd_cd'
_aflow_params 'a,b/a,c/a,x_{2},z_{2},x_{3},z_{3},x_{4},z_{4},x_{5},z_{5
    ↪ },x_{6},y_{6},z_{6},x_{7},y_{7},z_{7}'
_aflow_params_values '6.03,1.13847429519,1.51243781095,0.8518,0.5556,
    ↪ 0.3094,0.2447,0.1951,0.5148,0.3502,0.5018,-0.0669,0.4666,0.7279
    ↪ ,-0.0594,0.5651,0.7879'
_aflow_Strukturbericht 'None'
_aflow_Pearson 'oP36'

_symmetry_space_group_name_H-M "P 21/n 21/m 21/a"
_symmetry_Int_Tables_number 62

_cell_length_a    6.03000
_cell_length_b    6.86500
_cell_length_c    9.12000
_cell_angle_alpha 90.00000
_cell_angle_beta  90.00000
_cell_angle_gamma 90.00000

loop_
_space_group_symop_id
_space_group_symop_operation_xyz
1 x,y,z
2 x+1/2,-y+1/2,-z+1/2
3 -x,y+1/2,-z
4 -x+1/2,-y,z+1/2
5 -x,-y,-z
6 -x+1/2,y+1/2,z+1/2
7 x,-y+1/2,z
8 x+1/2,y,-z+1/2

loop_
_atom_site_label
_atom_site_type_symbol
_atom_site_symmetry_multiplicity
_atom_site_Wyckoff_label
_atom_site_fract_x
_atom_site_fract_y
_atom_site_fract_z
_atom_site_occupancy
Cu1 Cu 4 a 0.00000 0.00000 0.00000 1.00000
Cl1 Cl 4 c 0.85180 0.25000 0.55560 1.00000
Cu2 Cu 4 c 0.30940 0.25000 0.24470 1.00000
H1 H 4 c 0.19510 0.25000 0.51480 1.00000
O1 O 4 c 0.35020 0.25000 0.50180 1.00000
H2 H 8 d -0.06690 0.46660 0.72790 1.00000
O2 O 8 d -0.05940 0.56510 0.78790 1.00000
```

Atacamite ($Cu_2(OH)_3Cl$): AB2C3D3_oP36_62_c_ac_cd_cd - POSCAR

```
AB2C3D3_oP36_62_c_ac_cd_cd & a,b/a,c/a,x2,z2,x3,z3,x4,z4,x5,z5,x6,y6,z6,
    ↪ x7,y7,z7 --params=6.03,1.13847429519,1.51243781095,0.8518,
    ↪ 0.5556,0.3094,0.2447,0.1951,0.5148,0.3502,0.5018,-0.0669,0.4666
    ↪ ,0.7279,-0.0594,0.5651,0.7879 & Pnma D_{2h}^{16} #62 (ac^4d^2)
    ↪ & oP36 & None & ClCu2H3O3 & Atacamite & J. B. Parise and B. G.
    ↪ Hyde, Acta Crystallogr. C 42, 1277-1280 (1986)
   1.00000000000000
   6.03000000000000   0.00000000000000   0.00000000000000
   0.00000000000000   6.86500000000000   0.00000000000000
   0.00000000000000   0.00000000000000   9.12000000000000
    Cl      Cu      H      O
     4       8     12     12
Direct
   0.85180000000000   0.25000000000000   0.55560000000000   Cl   (4c)
  -0.35180000000000   0.75000000000000   1.05560000000000   Cl   (4c)
  -0.85180000000000   0.75000000000000  -0.55560000000000   Cl   (4c)
   1.35180000000000   0.25000000000000  -0.05560000000000   Cl   (4c)
   0.00000000000000   0.00000000000000   0.00000000000000   Cu   (4a)
   0.50000000000000   0.00000000000000   0.50000000000000   Cu   (4a)
   0.00000000000000   0.50000000000000   0.00000000000000   Cu   (4a)
   0.50000000000000   0.50000000000000   0.50000000000000   Cu   (4a)
   0.30940000000000   0.25000000000000   0.24470000000000   Cu   (4c)
   0.19060000000000   0.75000000000000   0.74470000000000   Cu   (4c)
  -0.30940000000000   0.75000000000000  -0.24470000000000   Cu   (4c)
   0.80940000000000   0.25000000000000   0.25530000000000   Cu   (4c)
   0.19510000000000   0.25000000000000   0.51480000000000    H   (4c)
   0.30490000000000   0.75000000000000   1.01480000000000    H   (4c)
  -0.19510000000000   0.75000000000000  -0.51480000000000    H   (4c)
```

```
   0.69510000000000    0.25000000000000   -0.01480000000000    H    (4c)
  -0.06690000000000    0.46660000000000    0.72790000000000    H    (8d)
   0.56690000000000   -0.46660000000000    1.22790000000000    H    (8d)
   0.06690000000000    0.96660000000000   -0.72790000000000    H    (8d)
   0.43310000000000    0.03340000000000   -0.22790000000000    H    (8d)
   0.06690000000000   -0.46660000000000   -0.72790000000000    H    (8d)
   0.43310000000000    0.46660000000000   -0.22790000000000    H    (8d)
  -0.06690000000000    0.03340000000000    0.72790000000000    H    (8d)
   0.56690000000000    0.96660000000000    1.22790000000000    H    (8d)
   0.35020000000000    0.25000000000000    0.50180000000000    O    (4c)
   0.14980000000000    0.75000000000000    1.00180000000000    O    (4c)
  -0.35020000000000    0.75000000000000   -0.50180000000000    O    (4c)
   0.85020000000000    0.25000000000000   -0.00180000000000    O    (4c)
  -0.05940000000000    0.56510000000000    0.78790000000000    O    (8d)
   0.55940000000000   -0.56510000000000    1.28790000000000    O    (8d)
   0.05940000000000    1.06510000000000   -0.78790000000000    O    (8d)
   0.44060000000000   -0.06510000000000   -0.28790000000000    O    (8d)
   0.05940000000000   -0.56510000000000   -0.78790000000000    O    (8d)
   0.44060000000000    0.56510000000000   -0.28790000000000    O    (8d)
  -0.05940000000000   -0.06510000000000    0.78790000000000    O    (8d)
   0.55940000000000    1.06510000000000    1.28790000000000    O    (8d)
```

NH$_4$CdCl$_3$ ($E2_4$): AB3C_oP20_62_c_3c_c - CIF

```
# CIF file
data_findsym-output
_audit_creation_method FINDSYM

_chemical_name_mineral 'CdCl3(NH4)'
_chemical_formula_sum 'Cd Cl3 (NH4)'

loop_
_publ_author_name
 'H. Brasseur'
 'L. Pauling'
_journal_name_full_name
;
 Journal of the American Chemical Society
;
_journal_volume 60
_journal_year 1938
_journal_page_first 2886
_journal_page_last 2890
_publ_Section_title
;
 The Crystal Structure of Ammonium Cadmium Chloride, NH$_{4}$CdCl$_{3}$
;

_aflow_title 'NH$_{4}$CdCl$_{3}$ ($E2_{4}$) Structure'
_aflow_proto 'AB3C_oP20_62_c_3c_c'
_aflow_params 'a,b/a,c/a,x_{1},z_{1},x_{2},z_{2},x_{3},z_{3},x_{4},z_{4
    ↪ },x_{5},z_{5}'
_aflow_params_values '8.96,0.443080357143,1.65959821429,0.665,-0.054,
    ↪ 0.784,0.785,0.667,0.504,0.526,0.102,-0.07,0.18'
_aflow_Strukturbericht '$E2_{4}$'
_aflow_Pearson 'oP20'

_symmetry_space_group_name_H-M "P 21/n 21/m 21/a"
_symmetry_Int_Tables_number 62

_cell_length_a      8.96000
_cell_length_b      3.97000
_cell_length_c      14.87000
_cell_angle_alpha  90.00000
_cell_angle_beta   90.00000
_cell_angle_gamma  90.00000

loop_
_space_group_symop_id
_space_group_symop_operation_xyz
1 x,y,z
2 x+1/2,-y+1/2,-z+1/2
3 -x,y+1/2,-z
4 -x+1/2,-y,z+1/2
5 -x,-y,-z
6 -x+1/2,y+1/2,z+1/2
7 x,-y+1/2,z
8 x+1/2,y,-z+1/2

loop_
_atom_site_label
_atom_site_type_symbol
_atom_site_symmetry_multiplicity
_atom_site_Wyckoff_label
_atom_site_fract_x
_atom_site_fract_y
_atom_site_fract_z
_atom_site_occupancy
Cd1 Cd 4 c 0.66500 0.25000 -0.05400 1.00000
Cl1 Cl 4 c 0.78400 0.25000 0.78500 1.00000
Cl2 Cl 4 c 0.66700 0.25000 0.50400 1.00000
Cl3 Cl 4 c 0.52600 0.25000 0.10200 1.00000
NH41 NH4 4 c -0.07000 0.25000 0.18000 1.00000
```

NH$_4$CdCl$_3$ ($E2_4$): AB3C_oP20_62_c_3c_c - POSCAR

```
AB3C_oP20_62_c_3c_c & a,b/a,c/a,x1,z1,x2,z2,x3,z3,x4,z4,x5,z5 --params=
    ↪ 8.96,0.443080357143,1.65959821429,0.665,-0.054,0.784,0.785,
    ↪ 0.667,0.504,0.526,0.102,-0.07,0.18 & Pnma D_{2h}^{16} #62 (c^5)
    ↪  & oP20 & $E2_{4}$ & CdCl3(NH4) & CdCl3(NH4) & H. Brasseur and
    ↪ L. Pauling, J. Am. Chem. Soc. 60, 2886-2890 (1938)
   1.00000000000000
   8.96000000000000    0.00000000000000    0.00000000000000
   0.00000000000000    3.97000000000000    0.00000000000000
   0.00000000000000    0.00000000000000   14.87000000000000
    Cd     Cl     NH4
     4     12      4
Direct
   0.66500000000000    0.25000000000000   -0.05400000000000    Cd    (4c)
  -0.16500000000000    0.75000000000000    0.44600000000000    Cd    (4c)
  -0.66500000000000    0.75000000000000    0.05400000000000    Cd    (4c)
   1.16500000000000    0.25000000000000    0.55400000000000    Cd    (4c)
   0.78400000000000    0.25000000000000    0.78500000000000    Cl    (4c)
  -0.28400000000000    0.75000000000000    1.28500000000000    Cl    (4c)
  -0.78400000000000    0.75000000000000   -0.78500000000000    Cl    (4c)
   1.28400000000000    0.25000000000000   -0.28500000000000    Cl    (4c)
   0.66700000000000    0.25000000000000    0.50400000000000    Cl    (4c)
  -0.16700000000000    0.75000000000000    1.00400000000000    Cl    (4c)
  -0.66700000000000    0.75000000000000   -0.50400000000000    Cl    (4c)
   1.16700000000000    0.25000000000000   -0.00400000000000    Cl    (4c)
   0.52600000000000    0.25000000000000    0.10200000000000    Cl    (4c)
  -0.02600000000000    0.75000000000000    0.60200000000000    Cl    (4c)
  -0.52600000000000    0.75000000000000   -0.10200000000000    Cl    (4c)
   1.02600000000000    0.25000000000000    0.39800000000000    Cl    (4c)
  -0.07000000000000    0.25000000000000    0.18000000000000    NH4   (4c)
   0.57000000000000    0.75000000000000    0.68000000000000    NH4   (4c)
   0.07000000000000    0.75000000000000   -0.18000000000000    NH4   (4c)
   0.43000000000000    0.25000000000000    0.32000000000000    NH4   (4c)
```

Berthierite (FeSb$_2$S$_4$, $E3_3$): AB4C2_oP28_62_c_4c_2c - CIF

```
# CIF file
data_findsym-output
_audit_creation_method FINDSYM

_chemical_name_mineral 'Berthierite'
_chemical_formula_sum 'Fe S4 Sb2'

loop_
_publ_author_name
 'M. J. Buerger'
 'T. Hahn'
_journal_year 1953
_publ_Section_title
;
 The Crystal Structure of Berthierite, FeSb$_{2}$S$_{4}$
;

_aflow_title 'Berthierite (FeSb$_{2}$S$_{4}$, $E3_{3}$) Structure'
_aflow_proto 'AB4C2_oP28_62_c_4c_2c'
_aflow_params 'a,b/a,c/a,x_{1},z_{1},x_{2},z_{2},x_{3},z_{3},x_{4},z_{4
    ↪ },x_{5},z_{5},x_{6},z_{6},x_{7},z_{7}'
_aflow_params_values '11.44,0.328671328671,1.23426573427,0.184,0.334,
    ↪ 0.695,0.728,0.076,0.184,0.274,0.492,-0.049,0.595,0.355,0.062,
    ↪ 0.537,0.614'
_aflow_Strukturbericht '$E3_{3}$'
_aflow_Pearson 'oP28'

_symmetry_space_group_name_H-M "P 21/n 21/m 21/a"
_symmetry_Int_Tables_number 62

_cell_length_a      11.44000
_cell_length_b      3.76000
_cell_length_c      14.12000
_cell_angle_alpha  90.00000
_cell_angle_beta   90.00000
_cell_angle_gamma  90.00000

loop_
_space_group_symop_id
_space_group_symop_operation_xyz
1 x,y,z
2 x+1/2,-y+1/2,-z+1/2
3 -x,y+1/2,-z
4 -x+1/2,-y,z+1/2
5 -x,-y,-z
6 -x+1/2,y+1/2,z+1/2
7 x,-y+1/2,z
8 x+1/2,y,-z+1/2

loop_
_atom_site_label
_atom_site_type_symbol
_atom_site_symmetry_multiplicity
_atom_site_Wyckoff_label
_atom_site_fract_x
_atom_site_fract_y
_atom_site_fract_z
_atom_site_occupancy
Fe1 Fe 4 c 0.18400 0.25000 0.33400 1.00000
S1 S 4 c 0.69500 0.25000 0.72800 1.00000
S2 S 4 c 0.07600 0.25000 0.18400 1.00000
S3 S 4 c 0.27400 0.25000 0.49200 1.00000
S4 S 4 c -0.04900 0.25000 0.59500 1.00000
Sb1 Sb 4 c 0.35500 0.25000 0.06200 1.00000
Sb2 Sb 4 c 0.53700 0.25000 0.61400 1.00000
```

Berthierite (FeSb$_2$S$_4$, $E3_3$): AB4C2_oP28_62_c_4c_2c - POSCAR

```
AB4C2_oP28_62_c_4c_2c & a,b/a,c/a,x1,z1,x2,z2,x3,z3,x4,z4,x5,z5,x6,z6,x7
    ↪ ,z7 --params=11.44,0.328671328671,1.23426573427,0.184,0.334,
    ↪ 0.695,0.728,0.076,0.184,0.274,0.492,-0.049,0.595,0.355,0.062,
    ↪ 0.537,0.614 & Pnma D_{2h}^{16} #62 (c^7) & oP28 & $E3_{3}$ &
    ↪ FeS4Sb2 & Berthierite & M. J. Buerger and T. Hahn, (1953)
   1.00000000000000
  11.44000000000000    0.00000000000000    0.00000000000000
   0.00000000000000    3.76000000000000    0.00000000000000
   0.00000000000000    0.00000000000000   14.12000000000000
    Fe      S     Sb
     4     16      8
```

```
Direct
    0.18400000000000    0.25000000000000    0.33400000000000    Fe    (4c)
    0.31600000000000    0.75000000000000    0.83400000000000    Fe    (4c)
   -0.18400000000000    0.75000000000000   -0.33400000000000    Fe    (4c)
    0.68400000000000    0.25000000000000    0.16600000000000    Fe    (4c)
    0.69500000000000    0.25000000000000    0.72800000000000    S     (4c)
   -0.19500000000000    0.75000000000000    1.22800000000000    S     (4c)
   -0.69500000000000    0.75000000000000   -0.72800000000000    S     (4c)
    1.19500000000000    0.25000000000000   -0.22800000000000    S     (4c)
    0.07600000000000    0.25000000000000    0.18400000000000    S     (4c)
    0.42400000000000    0.75000000000000    0.68400000000000    S     (4c)
   -0.07600000000000    0.75000000000000   -0.18400000000000    S     (4c)
    0.57600000000000    0.25000000000000    0.31600000000000    S     (4c)
    0.27400000000000    0.25000000000000    0.49200000000000    S     (4c)
    0.22600000000000    0.75000000000000    0.99200000000000    S     (4c)
   -0.27400000000000    0.75000000000000   -0.49200000000000    S     (4c)
    0.77400000000000    0.25000000000000    0.00800000000000    S     (4c)
   -0.04900000000000    0.25000000000000    0.59500000000000    S     (4c)
    0.54900000000000    0.75000000000000    1.09500000000000    S     (4c)
    0.04900000000000    0.75000000000000   -0.59500000000000    S     (4c)
    0.45100000000000    0.25000000000000   -0.09500000000000    S     (4c)
    0.35500000000000    0.25000000000000    0.06200000000000    Sb    (4c)
    0.14500000000000    0.75000000000000    0.56200000000000    Sb    (4c)
   -0.35500000000000    0.75000000000000   -0.06200000000000    Sb    (4c)
    0.85500000000000    0.25000000000000    0.43800000000000    Sb    (4c)
    0.53700000000000    0.25000000000000    0.61400000000000    Sb    (4c)
   -0.03700000000000    0.75000000000000    1.11400000000000    Sb    (4c)
   -0.53700000000000    0.75000000000000   -0.61400000000000    Sb    (4c)
    1.03700000000000    0.25000000000000   -0.11400000000000    Sb    (4c)
```

Chalcocyanite ($CuSO_4$): AB4C_oP24_62_a_2cd_c - CIF

```
# CIF file
data_findsym-output
_audit_creation_method FINDSYM

_chemical_name_mineral 'Chalcocyanite'
_chemical_formula_sum 'Cu O4 S'

loop_
_publ_author_name
 'M. Wildner'
 'G. Giester'
_journal_name_full_name
;
 Mineralogy and Petrology
;
_journal_volume 39
_journal_year 1988
_journal_page_first 201
_journal_page_last 209
_publ_Section_title
;
 Crystal structure refinements of synthetic chalcocyanite (CuSO$_{4}$)
    ↪ and zincosite (ZnSO$_{4}$)
;

_aflow_title 'Chalcocyanite (CuSO$_{4}$) Structure'
_aflow_proto 'AB4C_oP24_62_a_2cd_c'
_aflow_params 'a,b/a,c/a,x_{2},z_{2},x_{3},z_{3},x_{4},z_{4},x_{5},y_{5
    ↪ },z_{5}'
_aflow_params_values '8.409,0.797835652277,0.574741348555,0.1293,0.7353,
    ↪ 0.3646,0.4385,0.18363,0.44979,0.1328,0.0674,0.3083'
_aflow_Strukturbericht 'None'
_aflow_Pearson 'oP24'

_symmetry_space_group_name_H-M "P 21/n 21/m 21/a"
_symmetry_Int_Tables_number 62

_cell_length_a     8.40900
_cell_length_b     6.70900
_cell_length_c     4.83300
_cell_angle_alpha  90.00000
_cell_angle_beta   90.00000
_cell_angle_gamma  90.00000

loop_
_space_group_symop_id
_space_group_symop_operation_xyz
1 x,y,z
2 x+1/2,-y+1/2,-z+1/2
3 -x,y+1/2,-z
4 -x+1/2,-y,z+1/2
5 -x,-y,-z
6 -x+1/2,y+1/2,z+1/2
7 x,-y+1/2,z
8 x+1/2,y,-z+1/2

loop_
_atom_site_label
_atom_site_type_symbol
_atom_site_symmetry_multiplicity
_atom_site_Wyckoff_label
_atom_site_fract_x
_atom_site_fract_y
_atom_site_fract_z
_atom_site_occupancy
Cu1 Cu 4 a 0.00000 0.00000 0.00000 1.00000
O1 O 4 c 0.12930 0.25000 0.73530 1.00000
O2 O 4 c 0.36460 0.25000 0.43850 1.00000
S1 S 4 c 0.18363 0.25000 0.44979 1.00000
O3 O 8 d 0.13280 0.06740 0.30830 1.00000
```

Chalcocyanite ($CuSO_4$): AB4C_oP24_62_a_2cd_c - POSCAR

```
AB4C_oP24_62_a_2cd_c & a,b/a,c/a,x2,z2,x3,z3,x4,z4,x5,y5,z5 --params=
    ↪ 8.409,0.797835652277,0.574741348555,0.1293,0.7353,0.3646,0.4385
    ↪ ,0.18363,0.44979,0.1328,0.0674,0.3083 & Pnma D_{2h}^{16} #62 (
    ↪ ac^3d) & oP24 & None & CuO4S & Chalcocyanite & M. Wildner and
    ↪ G. Giester, Mineral. Petrol. 39, 201-209 (1988)
   1.00000000000000
   8.40900000000000    0.00000000000000    0.00000000000000
   0.00000000000000    6.70900000000000    0.00000000000000
   0.00000000000000    0.00000000000000    4.83300000000000
    Cu     O     S
     4    16     4
Direct
   0.00000000000000    0.00000000000000    0.00000000000000    Cu    (4a)
   0.50000000000000    0.00000000000000    0.50000000000000    Cu    (4a)
   0.00000000000000    0.50000000000000    0.00000000000000    Cu    (4a)
   0.50000000000000    0.50000000000000    0.50000000000000    Cu    (4a)
   0.12930000000000    0.25000000000000    0.73530000000000    O     (4c)
   0.37070000000000    0.75000000000000    1.23530000000000    O     (4c)
  -0.12930000000000    0.75000000000000   -0.73530000000000    O     (4c)
   0.62930000000000    0.25000000000000   -0.23530000000000    O     (4c)
   0.36460000000000    0.25000000000000    0.43850000000000    O     (4c)
   0.13540000000000    0.75000000000000    0.93850000000000    O     (4c)
  -0.36460000000000    0.75000000000000   -0.43850000000000    O     (4c)
   0.86460000000000    0.25000000000000    0.06150000000000    O     (4c)
   0.13280000000000    0.06740000000000    0.30830000000000    O     (8d)
   0.36720000000000   -0.06740000000000    0.80830000000000    O     (8d)
  -0.13280000000000    0.56740000000000   -0.30830000000000    O     (8d)
   0.63280000000000    0.43260000000000    0.19170000000000    O     (8d)
  -0.13280000000000   -0.06740000000000   -0.30830000000000    O     (8d)
   0.63280000000000    0.06740000000000    0.19170000000000    O     (8d)
   0.13280000000000    0.43260000000000    0.30830000000000    O     (8d)
   0.36720000000000    0.56740000000000    0.80830000000000    O     (8d)
   0.18363000000000    0.25000000000000    0.44979000000000    S     (4c)
   0.31637000000000    0.75000000000000    0.94979000000000    S     (4c)
  -0.18363000000000    0.75000000000000   -0.44979000000000    S     (4c)
   0.68363000000000    0.25000000000000    0.05021000000000    S     (4c)
```

Rynersonite (Orthorhombic $CaTa_2O_6$): AB6C2_oP36_62_c_2c2d_d - CIF

```
# CIF file
data_findsym-output
_audit_creation_method FINDSYM

_chemical_name_mineral 'Rynersonite'
_chemical_formula_sum 'Ca O6 Ta2'

loop_
_publ_author_name
 'L. Jahnberg'
_journal_name_full_name
;
 Acta Chemica Scandinavica
;
_journal_volume 71
_journal_year 1963
_journal_page_first 2548
_journal_page_last 2559
_publ_Section_title
;
 Crystal Structure of Orthorhombic CaTa$_{2}$O$_{6}$
;

_aflow_title 'Rynersonite (Orthorhombic CaTa$_{2}$O$_{6}$) Structure'
_aflow_proto 'AB6C2_oP36_62_c_2c2d_d'
_aflow_params 'a,b/a,c/a,x_{1},z_{1},x_{2},z_{2},x_{3},z_{3},x_{4},y_{4
    ↪ },z_{4},x_{5},y_{5},z_{5},x_{6},y_{6},z_{6}'
_aflow_params_values '11.068,0.678080954102,0.485905312613,0.042,0.54,
    ↪ 0.146,-0.033,0.878,0.838,-0.024,0.035,0.225,0.213,0.049,0.383,
    ↪ 0.1412,-0.0056,0.0376'
_aflow_Strukturbericht 'None'
_aflow_Pearson 'oP36'

_symmetry_space_group_name_H-M "P 21/n 21/m 21/a"
_symmetry_Int_Tables_number 62

_cell_length_a     11.06800
_cell_length_b     7.50500
_cell_length_c     5.37800
_cell_angle_alpha  90.00000
_cell_angle_beta   90.00000
_cell_angle_gamma  90.00000

loop_
_space_group_symop_id
_space_group_symop_operation_xyz
1 x,y,z
2 x+1/2,-y+1/2,-z+1/2
3 -x,y+1/2,-z
4 -x+1/2,-y,z+1/2
5 -x,-y,-z
6 -x+1/2,y+1/2,z+1/2
7 x,-y+1/2,z
8 x+1/2,y,-z+1/2

loop_
_atom_site_label
_atom_site_type_symbol
_atom_site_symmetry_multiplicity
_atom_site_Wyckoff_label
_atom_site_fract_x
_atom_site_fract_y
_atom_site_fract_z
_atom_site_occupancy
Ca1 Ca 4 c 0.04200 0.25000 0.54000 1.00000
O1 O 4 c 0.14600 0.25000 -0.03300 1.00000
```

```
O2 O 4 c 0.87800 0.25000 0.83800 1.00000
O3 O 8 d -0.02400 0.03500 0.22500 1.00000
O4 O 8 d 0.21300 0.04900 0.38300 1.00000
Ta1 Ta 8 d 0.14120 -0.00560 0.03760 1.00000
```

Rynersonite (Orthorhombic $CaTa_2O_6$): AB6C2_oP36_62_c_2c2d_d - POSCAR

```
AB6C2_oP36_62_c_2c2d_d & a,b/a,c/a,x1,z1,x2,z2,x3,z3,x4,y4,z4,x5,y5,z5,
   ↪ x6,y6,z6 --params=11.068,0.678080954102,0.485905312613,0.042,
   ↪ 0.54,0.146,-0.033,0.878,0.838,-0.024,0.035,0.225,0.213,0.049,
   ↪ 0.383,0.1412,-0.0056,0.0376 & Pnma D_{2h}^{16} #62 (c^3d^3) &
   ↪ oP36 & None & CaO6Ta2 & Rynersonite & L. Jahnberg, Acta Chem.
   ↪ Scand. 71, 2548-2559 (1963)
   1.00000000000000
  11.06800000000000   0.00000000000000   0.00000000000000
   0.00000000000000   7.50500000000000   0.00000000000000
   0.00000000000000   0.00000000000000   5.37800000000000
    Ca     O     Ta
     4    24      8
Direct
   0.04200000000000   0.25000000000000   0.54000000000000   Ca   (4c)
   0.45800000000000   0.75000000000000   1.04000000000000   Ca   (4c)
  -0.04200000000000   0.75000000000000  -0.54000000000000   Ca   (4c)
   0.54200000000000   0.25000000000000  -0.04000000000000   Ca   (4c)
   0.14600000000000   0.25000000000000  -0.03300000000000   O    (4c)
   0.35400000000000   0.75000000000000   0.46700000000000   O    (4c)
  -0.14600000000000   0.75000000000000   0.03300000000000   O    (4c)
   0.64600000000000   0.25000000000000   0.53300000000000   O    (4c)
   0.87800000000000   0.25000000000000   0.83800000000000   O    (4c)
  -0.37800000000000   0.75000000000000   1.33800000000000   O    (4c)
  -0.87800000000000   0.75000000000000  -0.83800000000000   O    (4c)
   1.37800000000000   0.25000000000000  -0.33800000000000   O    (4c)
  -0.02400000000000   0.03500000000000   0.22500000000000   O    (8d)
   0.52400000000000  -0.03500000000000   0.72500000000000   O    (8d)
   0.02400000000000   0.53500000000000  -0.22500000000000   O    (8d)
   0.47600000000000   0.46500000000000   0.27500000000000   O    (8d)
   0.02400000000000  -0.03500000000000  -0.22500000000000   O    (8d)
   0.47600000000000   0.03500000000000   0.27500000000000   O    (8d)
  -0.02400000000000   0.46500000000000   0.22500000000000   O    (8d)
   0.52400000000000   0.53500000000000   0.72500000000000   O    (8d)
   0.21300000000000   0.04900000000000   0.38300000000000   O    (8d)
   0.28700000000000  -0.04900000000000   0.88300000000000   O    (8d)
  -0.21300000000000   0.54900000000000  -0.38300000000000   O    (8d)
   0.71300000000000   0.45100000000000   0.11700000000000   O    (8d)
  -0.21300000000000  -0.04900000000000  -0.38300000000000   O    (8d)
   0.71300000000000   0.04900000000000   0.11700000000000   O    (8d)
   0.21300000000000   0.45100000000000   0.38300000000000   O    (8d)
   0.28700000000000   0.54900000000000   0.88300000000000   O    (8d)
   0.14120000000000  -0.00560000000000   0.03760000000000   Ta   (8d)
   0.35880000000000   0.00560000000000   0.53760000000000   Ta   (8d)
  -0.14120000000000   0.49440000000000  -0.03760000000000   Ta   (8d)
   0.64120000000000   0.50560000000000   0.46240000000000   Ta   (8d)
  -0.14120000000000   0.00560000000000  -0.03760000000000   Ta   (8d)
   0.64120000000000  -0.00560000000000   0.46240000000000   Ta   (8d)
   0.14120000000000   0.50560000000000   0.03760000000000   Ta   (8d)
   0.35880000000000   0.49440000000000   0.53760000000000   Ta   (8d)
```

Copper (II) Azide $[Cu(N_3)_2]$: AB6_oP28_62_c_6c - CIF

```
# CIF file
data_findsym-output
_audit_creation_method FINDSYM

_chemical_name_mineral 'Copper (ii) azide'
_chemical_formula_sum 'Cu N6'

loop_
_publ_author_name
 'I. Agrell'
_journal_name_full_name
;
 Acta Chemica Scandinavica
;
_journal_volume 21
_journal_year 1967
_journal_page_first 2647
_journal_page_last 2658
_publ_Section_title
;
 The Crystal Structure of Cu(N$_{3}$)$_{2}$
;

_aflow_title 'Copper (II) Azide [Cu(N$_{3}$)$_{2}$] Structure'
_aflow_proto 'AB6_oP28_62_c_6c'
_aflow_params 'a,b/a,c/a,x_{1},z_{1},x_{2},z_{2},x_{3},z_{3},x_{4},z_{4
   ↪ },x_{5},z_{5},x_{6},z_{6},x_{7},z_{7}'
_aflow_params_values '13.481,0.228766411987,0.673243824642,0.39692,
   ↪ 0.58022,0.81158,0.66343,0.7465,0.59084,0.6767,0.50317,0.52408,
   ↪ 0.35102,0.47071,0.24672,0.42026,0.15152'
_aflow_Strukturbericht 'None'
_aflow_Pearson 'oP28'

_symmetry_space_group_name_H-M "P 21/n 21/m 21/a"
_symmetry_Int_Tables_number 62

_cell_length_a    13.48100
_cell_length_b    3.08400
_cell_length_c    9.07600
_cell_angle_alpha 90.00000
_cell_angle_beta  90.00000
_cell_angle_gamma 90.00000

loop_
_space_group_symop_id
_space_group_symop_operation_xyz
1 x,y,z
2 x+1/2,-y+1/2,-z+1/2
3 -x,y+1/2,-z
4 -x+1/2,-y,z+1/2
5 -x,-y,-z
6 -x+1/2,y+1/2,z+1/2
7 x,-y+1/2,z
8 x+1/2,y,-z+1/2

loop_
_atom_site_label
_atom_site_type_symbol
_atom_site_symmetry_multiplicity
_atom_site_Wyckoff_label
_atom_site_fract_x
_atom_site_fract_y
_atom_site_fract_z
_atom_site_occupancy
Cu1 Cu 4 c 0.39692 0.25000 0.58022 1.00000
N1 N 4 c 0.81158 0.25000 0.66343 1.00000
N2 N 4 c 0.74650 0.25000 0.59084 1.00000
N3 N 4 c 0.67670 0.25000 0.50317 1.00000
N4 N 4 c 0.52408 0.25000 0.35102 1.00000
N5 N 4 c 0.47071 0.25000 0.24672 1.00000
N6 N 4 c 0.42026 0.25000 0.15152 1.00000
```

Copper (II) Azide $[Cu(N_3)_2]$: AB6_oP28_62_c_6c - POSCAR

```
AB6_oP28_62_c_6c & a,b/a,c/a,x1,z1,x2,z2,x3,z3,x4,z4,x5,z5,x6,z6,x7,z7
   ↪ --params=13.481,0.228766411987,0.673243824642,0.39692,0.58022,
   ↪ 0.81158,0.66343,0.7465,0.59084,0.6767,0.50317,0.52408,0.35102,
   ↪ 0.47071,0.24672,0.42026,0.15152 & Pnma D_{2h}^{16} #62 (c^7) &
   ↪ oP28 & None & CuN6 & Copper (ii) azide & I. Agrell, Acta Chem.
   ↪ Scand. 21, 2647-2658 (1967)
   1.00000000000000
  13.48100000000000   0.00000000000000   0.00000000000000
   0.00000000000000   3.08400000000000   0.00000000000000
   0.00000000000000   0.00000000000000   9.07600000000000
    Cu     N
     4    24
Direct
   0.39692000000000   0.25000000000000   0.58022000000000   Cu   (4c)
   0.10308000000000   0.75000000000000   1.08022000000000   Cu   (4c)
  -0.39692000000000   0.75000000000000  -0.58022000000000   Cu   (4c)
   0.89692000000000   0.25000000000000  -0.08022000000000   Cu   (4c)
   0.81158000000000   0.25000000000000   0.66343000000000   N    (4c)
  -0.31158000000000   0.75000000000000   1.16343000000000   N    (4c)
  -0.81158000000000   0.75000000000000  -0.66343000000000   N    (4c)
   1.31158000000000   0.25000000000000  -0.16343000000000   N    (4c)
   0.74650000000000   0.25000000000000   0.59084000000000   N    (4c)
  -0.24650000000000   0.75000000000000   1.09084000000000   N    (4c)
  -0.74650000000000   0.75000000000000  -0.59084000000000   N    (4c)
   1.24650000000000   0.25000000000000  -0.09084000000000   N    (4c)
   0.67670000000000   0.25000000000000   0.50317000000000   N    (4c)
  -0.17670000000000   0.75000000000000   1.00317000000000   N    (4c)
  -0.67670000000000   0.75000000000000  -0.50317000000000   N    (4c)
   1.17670000000000   0.25000000000000  -0.00317000000000   N    (4c)
   0.52408000000000   0.25000000000000   0.35102000000000   N    (4c)
  -0.02408000000000   0.75000000000000   0.85102000000000   N    (4c)
  -0.52408000000000   0.75000000000000  -0.35102000000000   N    (4c)
   1.02408000000000   0.25000000000000   0.14898000000000   N    (4c)
   0.47071000000000   0.25000000000000   0.24672000000000   N    (4c)
   0.02929000000000   0.75000000000000   0.74672000000000   N    (4c)
  -0.47071000000000   0.75000000000000  -0.24672000000000   N    (4c)
   0.97071000000000   0.25000000000000   0.25328000000000   N    (4c)
   0.42026000000000   0.25000000000000   0.15152000000000   N    (4c)
   0.07974000000000   0.75000000000000   0.65152000000000   N    (4c)
  -0.42026000000000   0.75000000000000  -0.15152000000000   N    (4c)
   0.92026000000000   0.25000000000000   0.34848000000000   N    (4c)
```

Diaspore (AlOOH, $E0_2$): ABC2_oP16_62_c_c_2c - CIF

```
# CIF file
data_findsym-output
_audit_creation_method FINDSYM

_chemical_name_mineral 'Diaspore'
_chemical_formula_sum 'Al H O2'

loop_
_publ_author_name
 'R. J. Hill'
_journal_name_full_name
;
 Physics and Chemistry of Minerals
;
_journal_volume 5
_journal_year 1979
_journal_page_first 179
_journal_page_last 200
_publ_Section_title
;
 Crystal Structure Refinement and Electron Density Distribution in
   ↪ Diaspore
;

# Found in The American Mineralogist Crystal Structure Database, 2003

_aflow_title 'Diaspore (AlOOH, $E0_{2}$) Structure'
_aflow_proto 'ABC2_oP16_62_c_c_2c'
_aflow_params 'a,b/a,c/a,x_{1},z_{1},x_{2},z_{2},x_{3},z_{3},x_{4},z_{4
   ↪ }'
_aflow_params_values '9.4253,0.301868375542,0.466902910252,0.1445,-
   ↪ 0.04472,0.80108,0.28766,-0.05338,0.80286,-0.0876,0.5905'
_aflow_Strukturbericht '$E0_{2}$'
```

```
_aflow_Pearson 'oP16'

_symmetry_space_group_name_H-M "P 21/n 21/m 21/a"
_symmetry_Int_Tables_number 62

_cell_length_a    9.42530
_cell_length_b    2.84520
_cell_length_c    4.40070
_cell_angle_alpha 90.00000
_cell_angle_beta  90.00000
_cell_angle_gamma 90.00000

loop_
_space_group_symop_id
_space_group_symop_operation_xyz
1 x,y,z
2 x+1/2,-y+1/2,-z+1/2
3 -x,y+1/2,-z
4 -x+1/2,-y,z+1/2
5 -x,-y,-z
6 -x+1/2,y+1/2,z+1/2
7 x,-y+1/2,z
8 x+1/2,y,-z+1/2

loop_
_atom_site_label
_atom_site_type_symbol
_atom_site_symmetry_multiplicity
_atom_site_Wyckoff_label
_atom_site_fract_x
_atom_site_fract_y
_atom_site_fract_z
_atom_site_occupancy
Al1 Al 4 c 0.14450 0.25000 -0.04472 1.00000
H1 H 4 c 0.80108 0.25000 0.28766 1.00000
O1 O 4 c -0.05338 0.25000 0.80286 1.00000
O2 O 4 c -0.08760 0.25000 0.59050 1.00000
```

Diaspore (AlOOH, $E0_2$): ABC2_oP16_62_c_c_2c - POSCAR

```
ABC2_oP16_62_c_c_2c & a,b/a,c/a,x1,z1,x2,z2,x3,z3,x4,z4 --params=9.4253,
    ↪ 0.301868375542,0.466902910252,0.1445,-0.04472,0.80108,0.28766,-
    ↪ 0.05338,0.80286,-0.0876,0.5905 & Pnma D_{2h}^{16} #62 (c^4) &
    ↪ oP16 & $E0_{2}$ & AlHO2 & Diaspore & R. J. Hill, Phys. Chem.
    ↪ Miner. 5, 179-200 (1979)
   1.00000000000000
   9.42530000000000   0.00000000000000   0.00000000000000
   0.00000000000000   2.84520000000000   0.00000000000000
   0.00000000000000   0.00000000000000   4.40070000000000
    Al    H    O
     4    4    8
Direct
   0.14450000000000   0.25000000000000  -0.04472000000000   Al   (4c)
   0.35550000000000   0.75000000000000   0.45528000000000   Al   (4c)
  -0.14450000000000   0.75000000000000   0.04472000000000   Al   (4c)
   0.64450000000000   0.25000000000000   0.54472000000000   Al   (4c)
   0.80108000000000   0.25000000000000   0.28766000000000   H    (4c)
  -0.30108000000000   0.75000000000000   0.78766000000000   H    (4c)
  -0.80108000000000   0.75000000000000  -0.28766000000000   H    (4c)
   1.30108000000000   0.25000000000000   0.21234000000000   H    (4c)
  -0.05338000000000   0.25000000000000   0.80286000000000   O    (4c)
   0.55338000000000   0.75000000000000   1.30286000000000   O    (4c)
   0.05338000000000   0.75000000000000  -0.80286000000000   O    (4c)
   0.44662000000000   0.25000000000000  -0.30286000000000   O    (4c)
  -0.08760000000000   0.25000000000000   0.59050000000000   O    (4c)
   0.58760000000000   0.75000000000000   1.09050000000000   O    (4c)
   0.08760000000000   0.75000000000000  -0.59050000000000   O    (4c)
   0.41240000000000   0.25000000000000  -0.09050000000000   O    (4c)
```

$\alpha$-Potassium Nitrate ($KNO_3$) I: ABC3_oP20_62_c_c_cd - CIF

```
# CIF file
data_findsym-output
_audit_creation_method FINDSYM

_chemical_name_mineral 'KNO3'
_chemical_formula_sum 'K N O3'

loop_
_publ_author_name
 'J. K. Nimmo'
 'B. W. Lucas'
_journal_name_full_name
;
 Journal of Physics C: Solid State Physics
;
_journal_volume 6
_journal_year 1973
_journal_page_first 201
_journal_page_last 211
_publ_Section_title
;
 A neutron diffraction determination of the crystal structure of $\
    ↪ alpha$-phase potassium nitrate at 25$^{\circ}$C and 100$^{\
    ↪ circ}$C
;

# Found in The American Mineralogist Crystal Structure Database, 2003

_aflow_title '$\alpha$-Potassium Nitrate (KNO$_{3}$) I Structure'
_aflow_proto 'ABC3_oP20_62_c_c_cd'
_aflow_params 'a,b/a,c/a,x_{1},z_{1},x_{2},z_{2},x_{3},z_{3},x_{4},y_{4
    ↪ },z_{4}'
_aflow_params_values '6.4213,0.842804416551,1.425988507,0.7568,0.4166,-
    ↪ 0.0848,0.7548,-0.0893,0.8902,-0.0849,0.4492,0.6866'
```

```
_aflow_Strukturbericht 'None'
_aflow_Pearson 'oP20'

_symmetry_space_group_name_H-M "P 21/n 21/m 21/a"
_symmetry_Int_Tables_number 62

_cell_length_a    6.42130
_cell_length_b    5.41190
_cell_length_c    9.15670
_cell_angle_alpha 90.00000
_cell_angle_beta  90.00000
_cell_angle_gamma 90.00000

loop_
_space_group_symop_id
_space_group_symop_operation_xyz
1 x,y,z
2 x+1/2,-y+1/2,-z+1/2
3 -x,y+1/2,-z
4 -x+1/2,-y,z+1/2
5 -x,-y,-z
6 -x+1/2,y+1/2,z+1/2
7 x,-y+1/2,z
8 x+1/2,y,-z+1/2

loop_
_atom_site_label
_atom_site_type_symbol
_atom_site_symmetry_multiplicity
_atom_site_Wyckoff_label
_atom_site_fract_x
_atom_site_fract_y
_atom_site_fract_z
_atom_site_occupancy
K1 K 4 c 0.75680 0.25000 0.41660 1.00000
N1 N 4 c -0.08480 0.25000 0.75480 1.00000
O1 O 4 c -0.08930 0.25000 0.89020 1.00000
O2 O 8 d -0.08490 0.44920 0.68660 1.00000
```

$\alpha$-Potassium Nitrate ($KNO_3$) I: ABC3_oP20_62_c_c_cd - POSCAR

```
ABC3_oP20_62_c_c_cd & a,b/a,c/a,x1,z1,x2,z2,x3,z3,x4,y4,z4 --params=
    ↪ 6.4213,0.842804416551,1.425988507,0.7568,0.4166,-0.0848,0.7548
    ↪ ,-0.0893,0.8902,-0.0849,0.4492,0.6866 & Pnma D_{2h}^{16} #62 (c
    ↪ ^3d) & oP20 & None & KNO3 & KNO3 & J. K. Nimmo and B. W. Lucas,
    ↪ J. Phys. C: Solid State Phys. 6, 201-211 (1973)
   1.00000000000000
   6.42130000000000   0.00000000000000   0.00000000000000
   0.00000000000000   5.41190000000000   0.00000000000000
   0.00000000000000   0.00000000000000   9.15670000000000
     K    N    O
     4    4   12
Direct
   0.75680000000000   0.25000000000000   0.41660000000000   K    (4c)
  -0.25680000000000   0.75000000000000   0.91660000000000   K    (4c)
  -0.75680000000000   0.75000000000000  -0.41660000000000   K    (4c)
   1.25680000000000   0.25000000000000   0.08340000000000   K    (4c)
  -0.08480000000000   0.25000000000000   0.75480000000000   N    (4c)
   0.58480000000000   0.75000000000000   1.25480000000000   N    (4c)
   0.08480000000000   0.75000000000000  -0.75480000000000   N    (4c)
   0.41520000000000   0.25000000000000  -0.25480000000000   N    (4c)
  -0.08930000000000   0.25000000000000   0.89020000000000   O    (4c)
   0.58930000000000   0.75000000000000   1.39020000000000   O    (4c)
   0.08930000000000   0.75000000000000  -0.89020000000000   O    (4c)
   0.41070000000000   0.25000000000000  -0.39020000000000   O    (4c)
  -0.08490000000000   0.44920000000000   0.68660000000000   O    (8d)
   0.58490000000000  -0.44920000000000   1.18660000000000   O    (8d)
   0.08490000000000   0.94920000000000  -0.68660000000000   O    (8d)
   0.41510000000000   0.05080000000000  -0.18660000000000   O    (8d)
   0.08490000000000  -0.44920000000000  -0.68660000000000   O    (8d)
   0.41510000000000   0.44920000000000  -0.18660000000000   O    (8d)
  -0.08490000000000   0.05080000000000   0.68660000000000   O    (8d)
   0.58490000000000   0.94920000000000   1.18660000000000   O    (8d)
```

$NH_4NO_3$ III ($G0_{10}$): ABC3_oP20_62_c_c_cd - CIF

```
# CIF file
data_findsym-output
_audit_creation_method FINDSYM

_chemical_name_mineral 'N(NH4)O3'
_chemical_formula_sum 'N (NH4) O3'

loop_
_publ_author_name
 'T. H. Goodwin'
 'J. Whetstone'
_journal_name_full_name
;
 Journal of the Chemical Society
;
_journal_volume
_journal_year 1947
_journal_page_first 1455
_journal_page_last 1461
_publ_Section_title
;
 The crystal structure of ammonium nitrate III, and atomic scattering
    ↪ factors in ionic crystals
;

_aflow_title 'NH$_{4}$NO$_{3}$ III ($G0_{10}$) Structure'
_aflow_proto 'ABC3_oP20_62_c_c_cd'
_aflow_params 'a,b/a,c/a,x_{1},z_{1},x_{2},z_{2},x_{3},z_{3},x_{4},y_{4
    ↪ },z_{4}'
```

```
_aflow_params_values '7.65,0.762091503268,0.933333333333,0.145,0.14,0.02
    ↪ ,0.68,0.085,0.3,0.175,0.06,0.06'
_aflow_Strukturbericht '$G0_{10}$'
_aflow_Pearson 'oP20'

_symmetry_space_group_name_H-M "P 21/n 21/m 21/a"
_symmetry_Int_Tables_number 62

_cell_length_a    7.65000
_cell_length_b    5.83000
_cell_length_c    7.14000
_cell_angle_alpha 90.00000
_cell_angle_beta  90.00000
_cell_angle_gamma 90.00000

loop_
_space_group_symop_id
_space_group_symop_operation_xyz
1 x,y,z
2 x+1/2,-y+1/2,-z+1/2
3 -x,y+1/2,-z
4 -x+1/2,-y,z+1/2
5 -x,-y,-z
6 -x+1/2,y+1/2,z+1/2
7 x,-y+1/2,z
8 x+1/2,y,-z+1/2

loop_
_atom_site_label
_atom_site_type_symbol
_atom_site_symmetry_multiplicity
_atom_site_Wyckoff_label
_atom_site_fract_x
_atom_site_fract_y
_atom_site_fract_z
_atom_site_occupancy
N1 N 4 c 0.14500 0.25000 0.14000 1.00000
NH41 NH4 4 c 0.02000 0.25000 0.68000 1.00000
O1 O 4 c 0.08500 0.25000 0.30000 1.00000
O2 O 8 d 0.17500 0.06000 0.06000 1.00000
```

$NH_4NO_3$ III ($G0_{10}$): ABC3_oP20_62_c_c_cd - POSCAR

```
ABC3_oP20_62_c_c_cd & a,b/a,c/a,x1,z1,x2,z2,x3,z3,x4,y4,z4 --params=7.65
    ↪ ,0.762091503268,0.933333333333,0.145,0.14,0.02,0.68,0.085,0.3,
    ↪ 0.175,0.06,0.06 & Pnma D_{2h}^{16} #62 (c^3d) & oP20 & $G0_{10}
    ↪ $ & N(NH4)O3 & N(NH4)O3 & T. H. Goodwin and J. Whetstone, J.
    ↪ Chem. Soc. , 1455-1461 (1947)
   1.00000000000000
   7.65000000000000   0.00000000000000   0.00000000000000
   0.00000000000000   5.83000000000000   0.00000000000000
   0.00000000000000   0.00000000000000   7.14000000000000
     N    NH4     O
     4     4     12
Direct
   0.14500000000000   0.25000000000000   0.14000000000000   N    (4c)
   0.35500000000000   0.75000000000000   0.64000000000000   N    (4c)
  -0.14500000000000   0.75000000000000  -0.14000000000000   N    (4c)
   0.64500000000000   0.25000000000000   0.36000000000000   N    (4c)
   0.02000000000000   0.25000000000000   0.68000000000000   NH4  (4c)
   0.48000000000000   0.75000000000000   1.18000000000000   NH4  (4c)
  -0.02000000000000   0.75000000000000  -0.68000000000000   NH4  (4c)
   0.52000000000000   0.25000000000000  -0.18000000000000   NH4  (4c)
   0.08500000000000   0.25000000000000   0.30000000000000   O    (4c)
   0.41500000000000   0.75000000000000   0.80000000000000   O    (4c)
  -0.08500000000000   0.75000000000000  -0.30000000000000   O    (4c)
   0.58500000000000   0.25000000000000   0.20000000000000   O    (4c)
   0.17500000000000   0.06000000000000   0.06000000000000   O    (8d)
   0.32500000000000  -0.06000000000000   0.56000000000000   O    (8d)
  -0.17500000000000   0.56000000000000  -0.06000000000000   O    (8d)
   0.67500000000000   0.44000000000000   0.44000000000000   O    (8d)
  -0.17500000000000  -0.06000000000000  -0.06000000000000   O    (8d)
   0.67500000000000   0.06000000000000   0.44000000000000   O    (8d)
   0.17500000000000   0.44000000000000   0.06000000000000   O    (8d)
   0.32500000000000   0.56000000000000   0.56000000000000   O    (8d)
```

Aragonite ($CaCO_3$, $G0_2$): ABC3_oP20_62_c_c_cd - CIF

```
# CIF file
data_findsym-output
_audit_creation_method FINDSYM

_chemical_name_mineral 'Aragonite'
_chemical_formula_sum 'C Ca O3'

loop_
_publ_author_name
 'J. P. R. {de Villiers}'
_journal_name_full_name
;
 American Mineralogist
;
_journal_volume 56
_journal_year 1971
_journal_page_first 758
_journal_page_last 767
_publ_Section_title
;
 Crystal Structures of Aragonite, Strontianite, and Witherite
;

# Found in Aragonite, {Mineral Database},

_aflow_title 'Aragonite (CaCO$_{3}$, $G0_{2}$) Structure'
_aflow_proto 'ABC3_oP20_62_c_c_cd'
_aflow_params 'a,b/a,c/a,x_{1},z_{1},x_{2},z_{2},x_{3},z_{3},x_{4},y_{4
    ↪ },z_{4}'
_aflow_params_values '5.7404,0.864295171068,1.38789979792,0.4138,0.2622,
    ↪ 0.2597,-0.085,0.4038,0.4225,0.414,0.4736,0.181'
_aflow_Strukturbericht '$G0_{2}$'
_aflow_Pearson 'oP20'

_symmetry_space_group_name_H-M "P 21/n 21/m 21/a"
_symmetry_Int_Tables_number 62

_cell_length_a    5.74040
_cell_length_b    4.96140
_cell_length_c    7.96710
_cell_angle_alpha 90.00000
_cell_angle_beta  90.00000
_cell_angle_gamma 90.00000

loop_
_space_group_symop_id
_space_group_symop_operation_xyz
1 x,y,z
2 x+1/2,-y+1/2,-z+1/2
3 -x,y+1/2,-z
4 -x+1/2,-y,z+1/2
5 -x,-y,-z
6 -x+1/2,y+1/2,z+1/2
7 x,-y+1/2,z
8 x+1/2,y,-z+1/2

loop_
_atom_site_label
_atom_site_type_symbol
_atom_site_symmetry_multiplicity
_atom_site_Wyckoff_label
_atom_site_fract_x
_atom_site_fract_y
_atom_site_fract_z
_atom_site_occupancy
C1 C 4 c 0.41380 0.25000 0.26220 1.00000
Ca1 Ca 4 c 0.25970 0.25000 -0.08500 1.00000
O1 O 4 c 0.40380 0.25000 0.42250 1.00000
O2 O 8 d 0.41400 0.47360 0.18100 1.00000
```

Aragonite ($CaCO_3$, $G0_2$): ABC3_oP20_62_c_c_cd - POSCAR

```
ABC3_oP20_62_c_c_cd & a,b/a,c/a,x1,z1,x2,z2,x3,z3,x4,y4,z4 --params=
    ↪ 5.7404,0.864295171068,1.38789979792,0.4138,0.2622,0.2597,-0.085
    ↪ ,0.4038,0.4225,0.414,0.4736,0.181 & Pnma D_{2h}^{16} #62 (c^3d)
    ↪ & oP20 & $G0_{2}$ & CaCO3 & Aragonite & J. P. R. {de Villiers
    ↪ }, Am. Mineral. 56, 758-767 (1971)
   1.00000000000000
   5.74040000000000   0.00000000000000   0.00000000000000
   0.00000000000000   4.96140000000000   0.00000000000000
   0.00000000000000   0.00000000000000   7.96710000000000
     C    Ca     O
     4     4    12
Direct
   0.41380000000000   0.25000000000000   0.26220000000000   C    (4c)
   0.08620000000000   0.75000000000000   0.76220000000000   C    (4c)
  -0.41380000000000   0.75000000000000  -0.26220000000000   C    (4c)
   0.91380000000000   0.25000000000000   0.23780000000000   C    (4c)
   0.25970000000000   0.25000000000000  -0.08500000000000   Ca   (4c)
   0.24030000000000   0.75000000000000   0.41500000000000   Ca   (4c)
  -0.25970000000000   0.75000000000000   0.08500000000000   Ca   (4c)
   0.75970000000000   0.25000000000000   0.58500000000000   Ca   (4c)
   0.40380000000000   0.25000000000000   0.42250000000000   O    (4c)
   0.09620000000000   0.75000000000000   0.92250000000000   O    (4c)
  -0.40380000000000   0.75000000000000  -0.42250000000000   O    (4c)
   0.90380000000000   0.25000000000000   0.07750000000000   O    (4c)
   0.41400000000000   0.47360000000000   0.18100000000000   O    (8d)
   0.08600000000000  -0.47360000000000   0.68100000000000   O    (8d)
  -0.41400000000000   0.97360000000000  -0.18100000000000   O    (8d)
   0.91400000000000   0.02640000000000   0.31900000000000   O    (8d)
  -0.41400000000000  -0.47360000000000  -0.18100000000000   O    (8d)
   0.91400000000000   0.47360000000000   0.31900000000000   O    (8d)
   0.41400000000000   0.02640000000000   0.18100000000000   O    (8d)
   0.08600000000000   0.97360000000000   0.68100000000000   O    (8d)
```

Epididymite ($BeHNaO_8Si_3$, $S4_7$): ABCD8E3_oP112_62_d_2c_d_4c6d_3d - CIF

```
# CIF file
data_findsym-output
_audit_creation_method FINDSYM

_chemical_name_mineral 'Epididymite'
_chemical_formula_sum 'Be H Na O8 Si3'

loop_
_publ_author_name
 'G. {Diego Gatta}'
 'N. Rotiroti'
 'G. J. {McIntyre}'
 'A. Guastoni'
 'F. Nestola'
_journal_name_full_name
;
 American Mineralogist
;
_journal_volume 93
_journal_year 2008
_journal_page_first 1158
_journal_page_last 1165
_publ_Section_title
;
```

```
New insights into the crystal chemistry of epididymite and eudidymite
    ↪ from Malosa, Malawi: A single-crystal neutron diffraction
    ↪ study
;

# Found in The American Mineralogist Crystal Structure Database, 2003

_aflow_title 'Epididymite (BeHNaO$_{8}$Si$_{3}$, $S4_{7}$) Structure'
_aflow_proto 'ABCD8E3_oP112_62_d_2c_d_4c6d_3d'
_aflow_params 'a,b/a,c/a,x_{1},z_{1},x_{2},z_{2},x_{3},z_{3},x_{4},z_{4
    ↪ },x_{5},z_{5},x_{6},z_{6},x_{7},y_{7},z_{7},x_{8},y_{8},z_{8},
    ↪ x_{9},y_{9},z_{9},x_{10},y_{10},z_{10},x_{11},y_{11},z_{11},x_{
    ↪ 12},y_{12},z_{12},x_{13},y_{13},z_{13},x_{14},y_{14},z_{14},x_{
    ↪ 15},y_{15},z_{15},x_{16},y_{16},z_{16},x_{17},y_{17},z_{17}'
_aflow_params_values '12.7334,1.07039753719,0.576962947838,0.00647,
    ↪ 0.58561,0.0134,0.38486,0.12005,1e-05,0.37761,0.76529,0.37,
    ↪ 0.26883,0.05713,0.48974,0.49472,-0.00042,0.34273,0.10285,
    ↪ 0.06934,0.50192,0.21704,0.1224,0.21004,0.06068,0.06612,0.00658,
    ↪ 0.23777,0.11405,0.84938,0.43489,0.06457,0.78946,0.30497,0.1234,
    ↪ 0.52074,0.41552,0.063,0.22624,0.15908,0.1378,0.01543,0.3426,
    ↪ 0.13635,0.73115,0.33173,0.13741,0.30525'
_aflow_Strukturbericht '$S4_{7}$'
_aflow_Pearson 'oP112'

_symmetry_space_group_name_H-M "P 21/n 21/m 21/a"
_symmetry_Int_Tables_number 62

_cell_length_a    12.73340
_cell_length_b    13.62980
_cell_length_c    7.34670
_cell_angle_alpha 90.00000
_cell_angle_beta  90.00000
_cell_angle_gamma 90.00000

loop_
_space_group_symop_id
_space_group_symop_operation_xyz
1 x,y,z
2 x+1/2,-y+1/2,-z+1/2
3 -x,y+1/2,-z
4 -x+1/2,-y,z+1/2
5 -x,-y,-z
6 -x+1/2,y+1/2,z+1/2
7 x,-y+1/2,z
8 x+1/2,y,-z+1/2

loop_
_atom_site_label
_atom_site_type_symbol
_atom_site_symmetry_multiplicity
_atom_site_Wyckoff_label
_atom_site_fract_x
_atom_site_fract_y
_atom_site_fract_z
_atom_site_occupancy
H1 H 4 c 0.00647 0.25000 0.58561 1.00000
H2 H 4 c 0.01340 0.25000 0.38486 1.00000
O1 O 4 c 0.12005 0.25000 0.00001 1.00000
O2 O 4 c 0.37761 0.25000 0.76529 1.00000
O3 O 4 c 0.37000 0.25000 0.26883 1.00000
O4 O 4 c 0.05713 0.25000 0.48974 1.00000
Be1 Be 8 d 0.49472 -0.00042 0.34273 1.00000
Na1 Na 8 d 0.10285 0.06934 0.50192 1.00000
O5 O 8 d 0.21704 0.12240 0.21004 1.00000
O6 O 8 d 0.06068 0.06612 0.00658 1.00000
O7 O 8 d 0.23777 0.11405 0.84938 1.00000
O8 O 8 d 0.43489 0.06457 0.78946 1.00000
O9 O 8 d 0.30497 0.12340 0.52074 1.00000
O10 O 8 d 0.41552 0.06300 0.22624 1.00000
Si1 Si 8 d 0.15908 0.13780 0.01543 1.00000
Si2 Si 8 d 0.34260 0.13635 0.73115 1.00000
Si3 Si 8 d 0.33173 0.13741 0.30525 1.00000
```

Epididymite (BeHNaO$_8$Si$_3$, $S4_7$): ABCD8E3_oP112_62_d_2c_d_4c6d_3d - POSCAR

```
ABCD8E3_oP112_62_d_2c_d_4c6d_3d & a,b/a,c/a,x1,z1,x2,z2,x3,z3,x4,z4,x5,
    ↪ z5,x6,z6,x7,y7,z7,x8,y8,z8,x9,y9,z9,x10,y10,z10,x11,y11,z11,x12
    ↪ ,y12,z12,x13,y13,z13,x14,y14,z14,x15,y15,z15,x16,y16,z16,x17,
    ↪ y17,z17 --params=12.7334,1.07039753719,0.576962947838,0.00647,
    ↪ 0.58561,0.0134,0.38486,0.12005,1e-05,0.37761,0.76529,0.37,
    ↪ 0.26883,0.05713,0.48974,0.49472,-0.00042,0.34273,0.10285,
    ↪ 0.06934,0.50192,0.21704,0.1224,0.21004,0.06068,0.06612,0.00658,
    ↪ 0.23777,0.11405,0.84938,0.43489,0.06457,0.78946,0.30497,0.1234,
    ↪ 0.52074,0.41552,0.063,0.22624,0.15908,0.1378,0.01543,0.3426,
    ↪ 0.13635,0.73115,0.33173,0.13741,0.30525 & Pnma D_{2h}^{16} #62
    ↪ (c^6d^11) & oP112 & $S4_{7}$ & BeHNaO8Si3 & Epididymite & G. {
    ↪ Diego Gatta} et al., Am. Mineral. 93, 1158-1165 (2008)
   1.00000000000000
  12.73340000000000   0.00000000000000   0.00000000000000
   0.00000000000000  13.62980000000000   0.00000000000000
   0.00000000000000   0.00000000000000   7.34670000000000
    Be     H    Na     O    Si
     8     8     8    64    24
Direct
   0.49472000000000  -0.00042000000000   0.34273000000000   Be   (8d)
   0.00528000000000   0.00042000000000   0.84273000000000   Be   (8d)
  -0.49472000000000   0.49958000000000  -0.34273000000000   Be   (8d)
   0.99472000000000   0.50042000000000   0.15727000000000   Be   (8d)
  -0.49472000000000   0.00042000000000  -0.34273000000000   Be   (8d)
   0.99472000000000  -0.00042000000000   0.15727000000000   Be   (8d)
   0.49472000000000   0.50042000000000   0.34273000000000   Be   (8d)
   0.00528000000000   0.49958000000000   0.84273000000000   Be   (8d)
   0.00647000000000   0.25000000000000   0.58561000000000    H   (4c)
   0.49353000000000   0.75000000000000   1.08561000000000    H   (4c)
  -0.00647000000000   0.75000000000000  -0.58561000000000    H   (4c)
   0.50647000000000   0.25000000000000  -0.08561000000000    H   (4c)
   0.01340000000000   0.25000000000000   0.38486000000000    H   (4c)
   0.48660000000000   0.75000000000000   0.88486000000000    H   (4c)
  -0.01340000000000   0.75000000000000  -0.38486000000000    H   (4c)
   0.51340000000000   0.25000000000000   0.11514000000000    H   (4c)
   0.10285000000000   0.06934000000000   0.50192000000000   Na   (8d)
   0.39715000000000  -0.06934000000000   1.00192000000000   Na   (8d)
  -0.10285000000000   0.56934000000000  -0.50192000000000   Na   (8d)
   0.60285000000000   0.43066000000000  -0.00192000000000   Na   (8d)
  -0.10285000000000  -0.06934000000000  -0.50192000000000   Na   (8d)
   0.60285000000000   0.06934000000000  -0.00192000000000   Na   (8d)
   0.10285000000000   0.43066000000000   0.50192000000000   Na   (8d)
   0.39715000000000   0.56934000000000   1.00192000000000   Na   (8d)
   0.12005000000000   0.25000000000000   0.00001000000000    O   (4c)
   0.37995000000000   0.75000000000000   0.50001000000000    O   (4c)
  -0.12005000000000   0.75000000000000  -0.00001000000000    O   (4c)
   0.62005000000000   0.25000000000000   0.49999000000000    O   (4c)
   0.37761000000000   0.25000000000000   0.76529000000000    O   (4c)
   0.12239000000000   0.75000000000000   1.26529000000000    O   (4c)
  -0.37761000000000   0.75000000000000  -0.76529000000000    O   (4c)
   0.87761000000000   0.25000000000000  -0.26529000000000    O   (4c)
   0.37000000000000   0.25000000000000   0.26883000000000    O   (4c)
   0.13000000000000   0.75000000000000   0.76883000000000    O   (4c)
  -0.37000000000000   0.75000000000000  -0.26883000000000    O   (4c)
   0.87000000000000   0.25000000000000   0.23117000000000    O   (4c)
   0.05713000000000   0.25000000000000   0.48974000000000    O   (4c)
   0.44287000000000   0.75000000000000   0.98974000000000    O   (4c)
  -0.05713000000000   0.75000000000000  -0.48974000000000    O   (4c)
   0.55713000000000   0.25000000000000   0.01026000000000    O   (4c)
   0.21704000000000   0.12240000000000   0.21004000000000    O   (8d)
   0.28296000000000  -0.12240000000000   0.71004000000000    O   (8d)
  -0.21704000000000   0.62240000000000  -0.21004000000000    O   (8d)
   0.71704000000000   0.37760000000000   0.28996000000000    O   (8d)
  -0.21704000000000  -0.12240000000000  -0.21004000000000    O   (8d)
   0.71704000000000   0.12240000000000   0.28996000000000    O   (8d)
   0.21704000000000   0.37760000000000   0.21004000000000    O   (8d)
   0.28296000000000   0.62240000000000   0.71004000000000    O   (8d)
   0.06068000000000   0.06612000000000   0.00658000000000    O   (8d)
   0.43932000000000  -0.06612000000000   0.50658000000000    O   (8d)
  -0.06068000000000   0.56612000000000  -0.00658000000000    O   (8d)
   0.56068000000000   0.43388000000000   0.49342000000000    O   (8d)
  -0.06068000000000  -0.06612000000000  -0.00658000000000    O   (8d)
   0.56068000000000   0.06612000000000   0.49342000000000    O   (8d)
   0.06068000000000   0.43388000000000   0.00658000000000    O   (8d)
   0.43932000000000   0.56612000000000   0.50658000000000    O   (8d)
   0.23777000000000   0.11405000000000   0.84938000000000    O   (8d)
   0.26223000000000  -0.11405000000000   1.34938000000000    O   (8d)
  -0.23777000000000   0.61405000000000  -0.84938000000000    O   (8d)
   0.73777000000000   0.38595000000000  -0.34938000000000    O   (8d)
  -0.23777000000000  -0.11405000000000  -0.84938000000000    O   (8d)
   0.73777000000000   0.11405000000000  -0.34938000000000    O   (8d)
   0.23777000000000   0.38595000000000   0.84938000000000    O   (8d)
   0.26223000000000   0.61405000000000   1.34938000000000    O   (8d)
   0.43489000000000   0.06457000000000   0.78946000000000    O   (8d)
   0.06511000000000  -0.06457000000000   1.28946000000000    O   (8d)
  -0.43489000000000   0.56457000000000  -0.78946000000000    O   (8d)
   0.93489000000000   0.43543000000000  -0.28946000000000    O   (8d)
  -0.43489000000000  -0.06457000000000  -0.78946000000000    O   (8d)
   0.93489000000000   0.06457000000000  -0.28946000000000    O   (8d)
   0.43489000000000   0.43543000000000   0.78946000000000    O   (8d)
   0.06511000000000   0.56457000000000   1.28946000000000    O   (8d)
   0.30497000000000   0.12340000000000   0.52074000000000    O   (8d)
   0.19503000000000  -0.12340000000000   1.02074000000000    O   (8d)
  -0.30497000000000   0.62340000000000  -0.52074000000000    O   (8d)
   0.80497000000000   0.37660000000000  -0.02074000000000    O   (8d)
  -0.30497000000000  -0.12340000000000  -0.52074000000000    O   (8d)
   0.80497000000000   0.12340000000000  -0.02074000000000    O   (8d)
   0.30497000000000   0.37660000000000   0.52074000000000    O   (8d)
   0.19503000000000   0.62340000000000   1.02074000000000    O   (8d)
   0.41552000000000   0.06300000000000   0.22624000000000    O   (8d)
   0.08448000000000  -0.06300000000000   0.72624000000000    O   (8d)
  -0.41552000000000   0.56300000000000  -0.22624000000000    O   (8d)
   0.91552000000000   0.43700000000000   0.27376000000000    O   (8d)
  -0.41552000000000  -0.06300000000000  -0.22624000000000    O   (8d)
   0.91552000000000   0.06300000000000   0.27376000000000    O   (8d)
   0.41552000000000   0.43700000000000   0.22624000000000    O   (8d)
   0.08448000000000   0.56300000000000   0.72624000000000    O   (8d)
   0.15908000000000   0.13780000000000   0.01543000000000   Si   (8d)
   0.34092000000000  -0.13780000000000   0.51543000000000   Si   (8d)
  -0.15908000000000   0.63780000000000  -0.01543000000000   Si   (8d)
   0.65908000000000   0.36220000000000   0.48457000000000   Si   (8d)
  -0.15908000000000  -0.13780000000000  -0.01543000000000   Si   (8d)
   0.65908000000000   0.13780000000000   0.48457000000000   Si   (8d)
   0.15908000000000   0.36220000000000   0.01543000000000   Si   (8d)
   0.34092000000000   0.63780000000000   0.51543000000000   Si   (8d)
   0.34260000000000   0.13635000000000   0.73115000000000   Si   (8d)
   0.15740000000000  -0.13635000000000   1.23115000000000   Si   (8d)
  -0.34260000000000   0.63635000000000  -0.73115000000000   Si   (8d)
   0.84260000000000   0.36365000000000  -0.23115000000000   Si   (8d)
  -0.34260000000000  -0.13635000000000  -0.73115000000000   Si   (8d)
   0.84260000000000   0.13635000000000  -0.23115000000000   Si   (8d)
   0.34260000000000   0.36365000000000   0.73115000000000   Si   (8d)
   0.15740000000000   0.63635000000000   1.23115000000000   Si   (8d)
   0.33173000000000   0.13741000000000   0.30525000000000   Si   (8d)
   0.16827000000000  -0.13741000000000   0.80525000000000   Si   (8d)
  -0.33173000000000   0.63741000000000  -0.30525000000000   Si   (8d)
   0.83173000000000   0.36259000000000   0.19475000000000   Si   (8d)
  -0.33173000000000  -0.13741000000000  -0.30525000000000   Si   (8d)
   0.83173000000000   0.13741000000000   0.19475000000000   Si   (8d)
   0.33173000000000   0.36259000000000   0.30525000000000   Si   (8d)
   0.16827000000000   0.63741000000000   0.80525000000000   Si   (8d)
```

$NH_4ClBrI$ ($F5_{14}$): ABCD_oP16_62_c_c_c_c - CIF

```
# CIF file
```

```
data_findsym-output
_audit_creation_method FINDSYM

_chemical_name_mineral 'ClBrI(NH4)'
_chemical_formula_sum 'Br Cl I (NH4)'

loop_
_publ_author_name
 'R. C. L. Mooney'
_journal_name_full_name
;
 Zeitschrift f{\"u}r Kristallographie - Crystalline Materials
;
_journal_volume 98
_journal_year 1938
_journal_page_first 324
_journal_page_last 333
_publ_Section_title
;
 The Crystal Structure of Ammonium Chlorobromoiodide and the
     ↪ Configuration of the Chlorobromoiodide Group
;

# Found in Strukturbericht Band V 1937, 1940

_aflow_title 'NH$_{4}$ClBrI ($F5_{14}$) Structure'
_aflow_proto 'ABCD_oP16_62_c_c_c_c'
_aflow_params 'a,b/a,c/a,x_{1},z_{1},x_{2},z_{2},x_{3},z_{3},x_{4},z_{4
     ↪ }'
_aflow_params_values '9.94,0.616700201207,0.855130784708,0.576,0.748,
     ↪ 0.192,0.381,0.376,0.561,0.372,0.074'
_aflow_Strukturbericht '$F5_{14}$'
_aflow_Pearson 'oP16'

_symmetry_space_group_name_H-M "P 21/n 21/m 21/a"
_symmetry_Int_Tables_number 62

_cell_length_a     9.94000
_cell_length_b     6.13000
_cell_length_c     8.50000
_cell_angle_alpha 90.00000
_cell_angle_beta  90.00000
_cell_angle_gamma 90.00000

loop_
_space_group_symop_id
_space_group_symop_operation_xyz
1 x,y,z
2 x+1/2,-y+1/2,-z+1/2
3 -x,y+1/2,-z
4 -x+1/2,-y,z+1/2
5 -x,-y,-z
6 -x+1/2,y+1/2,z+1/2
7 x,-y+1/2,z
8 x+1/2,y,-z+1/2

loop_
_atom_site_label
_atom_site_type_symbol
_atom_site_symmetry_multiplicity
_atom_site_Wyckoff_label
_atom_site_fract_x
_atom_site_fract_y
_atom_site_fract_z
_atom_site_occupancy
Br1 Br 4 c 0.57600 0.25000 0.74800 1.00000
Cl1 Cl 4 c 0.19200 0.25000 0.38100 1.00000
I1 I 4 c 0.37600 0.25000 0.56100 1.00000
NH41 NH4 4 c 0.37200 0.25000 0.07400 1.00000
```

NH$_4$ClBrI ($F5_{14}$): ABCD_oP16_62_c_c_c_c - POSCAR

```
ABCD_oP16_62_c_c_c_c & a,b/a,c/a,x1,z1,x2,z2,x3,z3,x4,z4 --params=9.94,
     ↪ 0.616700201207,0.855130784708,0.576,0.748,0.192,0.381,0.376,
     ↪ 0.561,0.372,0.074 & Pnma D_{2h}^{16} #62 (c^4) & oP16 & $F5_{14
     ↪ }$ & ClBrI(NH4) & ClBrI(NH4) & R. C. L. Mooney, Zeitschrift f{"
     ↪ u}r Kristallographie - Crystalline Materials 98, 324-333 (1938)
   1.00000000000000
   9.94000000000000   0.00000000000000   0.00000000000000
   0.00000000000000   6.13000000000000   0.00000000000000
   0.00000000000000   0.00000000000000   8.50000000000000
    Br    Cl     I   NH4
     4     4     4     4
Direct
   0.57600000000000   0.25000000000000   0.74800000000000   Br   (4c)
  -0.07600000000000   0.75000000000000   1.24800000000000   Br   (4c)
  -0.57600000000000   0.75000000000000  -0.74800000000000   Br   (4c)
   1.07600000000000   0.25000000000000  -0.24800000000000   Br   (4c)
   0.19200000000000   0.25000000000000   0.38100000000000   Cl   (4c)
   0.30800000000000   0.75000000000000   0.88100000000000   Cl   (4c)
  -0.19200000000000   0.75000000000000  -0.38100000000000   Cl   (4c)
   0.69200000000000   0.25000000000000   0.11900000000000   Cl   (4c)
   0.37600000000000   0.25000000000000   0.56100000000000    I   (4c)
   0.12400000000000   0.75000000000000   1.06100000000000    I   (4c)
  -0.37600000000000   0.75000000000000  -0.56100000000000    I   (4c)
   0.87600000000000   0.25000000000000  -0.06100000000000    I   (4c)
   0.37200000000000   0.25000000000000   0.07400000000000  NH4   (4c)
   0.12800000000000   0.75000000000000   0.57400000000000  NH4   (4c)
  -0.37200000000000   0.75000000000000  -0.07400000000000  NH4   (4c)
   0.87200000000000   0.25000000000000   0.42600000000000  NH4   (4c)
```

MnCuP: ABC_oP12_62_c_c_c - CIF

```
# CIF file
data_findsym-output
_audit_creation_method FINDSYM

_chemical_name_mineral 'CuMnP'
_chemical_formula_sum 'Cu Mn P'

loop_
_publ_author_name
 'J. M\"{u}ndelein'
 'H.-U. Schuster'
_journal_name_full_name
;
 Zeitschrift f{\"u}r Naturforschung B
;
_journal_volume 47
_journal_year 1992
_journal_page_first 925
_journal_page_last 928
_publ_Section_title
;
 Darstellung und Kristallstruktur der Verbindungen MnCu$X$ ($X$ = P, As,
     ↪ P$_{x}$As$_{1-x}$)
;

# Found in Room-temperature antiferromagnetism in CuMnAs, 2012

_aflow_title 'MnCuP Structure'
_aflow_proto 'ABC_oP12_62_c_c_c'
_aflow_params 'a,b/a,c/a,x_{1},z_{1},x_{2},z_{2},x_{3},z_{3}'
_aflow_params_values '6.3187,0.58934590976,1.12178137908,0.6282,0.061,
     ↪ 0.5376,0.6709,0.2492,0.1245'
_aflow_Strukturbericht 'None'
_aflow_Pearson 'oP12'

_symmetry_space_group_name_H-M "P 21/n 21/m 21/a"
_symmetry_Int_Tables_number 62

_cell_length_a     6.31870
_cell_length_b     3.72390
_cell_length_c     7.08820
_cell_angle_alpha 90.00000
_cell_angle_beta  90.00000
_cell_angle_gamma 90.00000

loop_
_space_group_symop_id
_space_group_symop_operation_xyz
1 x,y,z
2 x+1/2,-y+1/2,-z+1/2
3 -x,y+1/2,-z
4 -x+1/2,-y,z+1/2
5 -x,-y,-z
6 -x+1/2,y+1/2,z+1/2
7 x,-y+1/2,z
8 x+1/2,y,-z+1/2

loop_
_atom_site_label
_atom_site_type_symbol
_atom_site_symmetry_multiplicity
_atom_site_Wyckoff_label
_atom_site_fract_x
_atom_site_fract_y
_atom_site_fract_z
_atom_site_occupancy
Cu1 Cu 4 c 0.62820 0.25000 0.06100 1.00000
Mn1 Mn 4 c 0.53760 0.25000 0.67090 1.00000
P1 P 4 c 0.24920 0.25000 0.12450 1.00000
```

MnCuP: ABC_oP12_62_c_c_c - POSCAR

```
ABC_oP12_62_c_c_c & a,b/a,c/a,x1,z1,x2,z2,x3,z3 --params=6.3187,
     ↪ 0.58934590976,1.12178137908,0.6282,0.061,0.5376,0.6709,0.2492,
     ↪ 0.1245 & Pnma D_{2h}^{16} #62 (c^3) & oP12 & None & CuMnP &
     ↪ CuMnP & J. M\"{u}ndelein and H.-U. Schuster, Z. Naturforsch. B
     ↪ 47, 925-928 (1992)
   1.00000000000000
   6.31870000000000   0.00000000000000   0.00000000000000
   0.00000000000000   3.72390000000000   0.00000000000000
   0.00000000000000   0.00000000000000   7.08820000000000
    Cu    Mn     P
     4     4     4
Direct
   0.62820000000000   0.25000000000000   0.06100000000000   Cu   (4c)
  -0.12820000000000   0.75000000000000   0.56100000000000   Cu   (4c)
  -0.62820000000000   0.75000000000000  -0.06100000000000   Cu   (4c)
   1.12820000000000   0.25000000000000   0.43900000000000   Cu   (4c)
   0.53760000000000   0.25000000000000   0.67090000000000   Mn   (4c)
  -0.03760000000000   0.75000000000000   1.17090000000000   Mn   (4c)
  -0.53760000000000   0.75000000000000  -0.67090000000000   Mn   (4c)
   1.03760000000000   0.25000000000000  -0.17090000000000   Mn   (4c)
   0.24920000000000   0.25000000000000   0.12450000000000    P   (4c)
   0.25080000000000   0.75000000000000   0.62450000000000    P   (4c)
  -0.24920000000000   0.75000000000000  -0.12450000000000    P   (4c)
   0.74920000000000   0.25000000000000   0.37550000000000    P   (4c)
```

$\eta$-NiSi ($B_d$): AB_oP8_62_c_c - CIF

```
# CIF file
data_findsym-output
_audit_creation_method FINDSYM

_chemical_name_mineral '$\eta$-nisi'
_chemical_formula_sum 'Ni Si'

loop_
```

```
_publ_author_name
 'K. Toman'
_journal_name_full_name
;
 Acta Cristallographica
;
_journal_volume 4
_journal_year 1951
_journal_page_first 462
_journal_page_last 464
_publ_Section_title
;
 The structure of NiSi
;

_aflow_title '$\eta$-NiSi ($B_{d}$) Structure'
_aflow_proto 'AB_oP8_62_c_c'
_aflow_params 'a,b/a,c/a,x_{1},z_{1},x_{2},z_{2}'
_aflow_params_values '5.18,0.644787644788,1.08494208494,0.006,0.184,0.67
    ↪ ,-0.08'
_aflow_Strukturbericht '$B_{d}$'
_aflow_Pearson 'oP8'

_symmetry_space_group_name_H-M "P 21/n 21/m 21/a"
_symmetry_Int_Tables_number 62

_cell_length_a      5.18000
_cell_length_b      3.34000
_cell_length_c      5.62000
_cell_angle_alpha 90.00000
_cell_angle_beta  90.00000
_cell_angle_gamma 90.00000

loop_
_space_group_symop_id
_space_group_symop_operation_xyz
1 x,y,z
2 x+1/2,-y+1/2,-z+1/2
3 -x,y+1/2,-z
4 -x+1/2,-y,z+1/2
5 -x,-y,-z
6 -x+1/2,y+1/2,z+1/2
7 x,-y+1/2,z
8 x+1/2,y,-z+1/2

loop_
_atom_site_label
_atom_site_type_symbol
_atom_site_symmetry_multiplicity
_atom_site_Wyckoff_label
_atom_site_fract_x
_atom_site_fract_y
_atom_site_fract_z
_atom_site_occupancy
Ni1 Ni 4 c 0.00600 0.25000 0.18400 1.00000
Si1 Si 4 c 0.67000 0.25000 -0.08000 1.00000
```

$\eta$-NiSi ($B_d$): AB_oP8_62_c_c - POSCAR

```
AB_oP8_62_c_c & a,b/a,c/a,x1,z1,x2,z2 --params=5.18,0.644787644788,
    ↪ 1.08494208494,0.006,0.184,0.67,-0.08 & Pnma D_{2h}^{16} #62 (c^
    ↪ 2) & oP8 & $B_{d}$ & NiSi & $\eta$-nisi & K. Toman, Acta Cryst.
    ↪ 4, 462-464 (1951)
   1.00000000000000
   5.18000000000000   0.00000000000000   0.00000000000000
   0.00000000000000   3.34000000000000   0.00000000000000
   0.00000000000000   0.00000000000000   5.62000000000000
    Ni      Si
     4       4
Direct
   0.00600000000000   0.25000000000000   0.18400000000000   Ni   (4c)
   0.49400000000000   0.75000000000000   0.68400000000000   Ni   (4c)
  -0.00600000000000   0.75000000000000  -0.18400000000000   Ni   (4c)
   0.50600000000000   0.25000000000000   0.31600000000000   Ni   (4c)
   0.67000000000000   0.25000000000000  -0.08000000000000   Si   (4c)
  -0.17000000000000   0.75000000000000   0.42000000000000   Si   (4c)
  -0.67000000000000   0.75000000000000   0.08000000000000   Si   (4c)
   1.17000000000000   0.25000000000000   0.58000000000000   Si   (4c)
```

$Cu_2Pb(SeO_3)_2Br_2$: A2B2C6DE2_oC52_63_g_e_fh_c_f - CIF

```
# CIF file
data_findsym-output
_audit_creation_method FINDSYM

_chemical_name_mineral 'Br2Cu2O6PbSe2'
_chemical_formula_sum 'Br2 Cu2 O6 Pb Se2'

loop_
_publ_author_name
 'O. I. Siidra'
 'M. S. Kozin'
 'W. Depmeier'
 'R. A. Kayukov'
 'V. M. Kovrugin'
_journal_name_full_name
;
 Acta Crystallographica Section B: Structural Science
;
_journal_volume 74
_journal_year 2018
_journal_page_first 712
_journal_page_last 724
_publ_Section_title
;
 Copper-lead selenite bromides: a new large family of compounds partly
    ↪ having Cu$^{2+}$ substructures derivable from kagome nets
;

_aflow_title 'Cu$_{2}$Pb(SeO$_{3}$)$_{2}$Br$_{2}$ Structure'
_aflow_proto 'A2B2C6DE2_oC52_63_g_e_fh_c_f'
_aflow_params 'a,b/a,c/a,y_{1},x_{2},y_{3},z_{3},y_{4},z_{4},x_{5},y_{5
    ↪ },x_{6},y_{6},z_{6}'
_aflow_params_values '8.275,1.12422960725,1.43697885196,0.33496,0.1862,
    ↪ 0.1287,0.5177,0.31221,0.54036,0.22773,0.07717,0.1531,0.3602,
    ↪ 0.4531'
_aflow_Strukturbericht 'None'
_aflow_Pearson 'oC52'

_symmetry_space_group_name_H-M "C 2/m 2/c 21/m"
_symmetry_Int_Tables_number 63

_cell_length_a      8.27500
_cell_length_b      9.30300
_cell_length_c     11.89100
_cell_angle_alpha 90.00000
_cell_angle_beta  90.00000
_cell_angle_gamma 90.00000

loop_
_space_group_symop_id
_space_group_symop_operation_xyz
1 x,y,z
2 x,-y,-z
3 -x,y,-z+1/2
4 -x,-y,z+1/2
5 -x,-y,-z
6 -x,y,z
7 x,-y,z+1/2
8 x,y,-z+1/2
9 x+1/2,y+1/2,z
10 x+1/2,-y+1/2,-z
11 -x+1/2,y+1/2,-z+1/2
12 -x+1/2,-y+1/2,z+1/2
13 -x+1/2,-y+1/2,-z
14 -x+1/2,y+1/2,z
15 x+1/2,-y+1/2,z+1/2
16 x+1/2,y+1/2,-z+1/2

loop_
_atom_site_label
_atom_site_type_symbol
_atom_site_symmetry_multiplicity
_atom_site_Wyckoff_label
_atom_site_fract_x
_atom_site_fract_y
_atom_site_fract_z
_atom_site_occupancy
Pb1 Pb 4 c 0.00000 0.33496 0.25000 1.00000
Cu1 Cu 8 e 0.18620 0.00000 0.00000 1.00000
O1 O 8 f 0.00000 0.12870 0.51770 1.00000
Se1 Se 8 f 0.00000 0.31221 0.54036 1.00000
Br1 Br 8 g 0.22773 0.07717 0.25000 1.00000
O2 O 16 h 0.15310 0.36020 0.45310 1.00000
```

$Cu_2Pb(SeO_3)_2Br_2$: A2B2C6DE2_oC52_63_g_e_fh_c_f - POSCAR

```
A2B2C6DE2_oC52_63_g_e_fh_c_f & a,b/a,c/a,y1,x2,y3,z3,y4,z4,x5,y5,x6,y6,
    ↪ z6 --params=8.275,1.12422960725,1.43697885196,0.33496,0.1862,
    ↪ 0.1287,0.5177,0.31221,0.54036,0.22773,0.07717,0.1531,0.3602,
    ↪ 0.4531 & Cmcm D_{2h}^{17} #63 (cef^2gh) & oC52 & None &
    ↪ Br2Cu2O6PbSe2 & Br2Cu2O6PbSe2 & O. I. Siidra et al., Acta
    ↪ Crystallogr. Sect. B Struct. Sci. 74, 712-724 (2018)
   1.00000000000000
   4.13750000000000  -4.65150000000000   0.00000000000000
   4.13750000000000   4.65150000000000   0.00000000000000
   0.00000000000000   0.00000000000000  11.89100000000000
    Br      Cu      O      Pb      Se
     4       4     12       2       4
Direct
   0.15056000000000   0.30490000000000   0.25000000000000   Br   (8g)
  -0.15056000000000  -0.30490000000000   0.75000000000000   Br   (8g)
  -0.30490000000000  -0.15056000000000   0.25000000000000   Br   (8g)
   0.30490000000000   0.15056000000000   0.75000000000000   Br   (8g)
   0.18620000000000   0.18620000000000   0.00000000000000   Cu   (8e)
  -0.18620000000000  -0.18620000000000   0.50000000000000   Cu   (8e)
  -0.18620000000000  -0.18620000000000   0.00000000000000   Cu   (8e)
   0.18620000000000   0.18620000000000   0.50000000000000   Cu   (8e)
  -0.12870000000000   0.12870000000000   0.51770000000000   O    (8f)
   0.12870000000000  -0.12870000000000   1.01770000000000   O    (8f)
  -0.12870000000000   0.12870000000000  -0.01770000000000   O    (8f)
   0.12870000000000  -0.12870000000000  -0.51770000000000   O    (8f)
  -0.20710000000000   0.51330000000000   0.45310000000000   O   (16h)
   0.20710000000000  -0.51330000000000   0.95310000000000   O   (16h)
  -0.51330000000000   0.20710000000000   0.04690000000000   O   (16h)
   0.51330000000000  -0.20710000000000  -0.45310000000000   O   (16h)
   0.20710000000000  -0.51330000000000  -0.45310000000000   O   (16h)
  -0.20710000000000   0.51330000000000   0.04690000000000   O   (16h)
   0.51330000000000  -0.20710000000000   0.95310000000000   O   (16h)
  -0.51330000000000   0.20710000000000   0.45310000000000   O   (16h)
  -0.33496000000000   0.33496000000000   0.25000000000000   Pb   (4c)
   0.33496000000000  -0.33496000000000   0.75000000000000   Pb   (4c)
  -0.31221000000000   0.31221000000000   0.54036000000000   Se   (8f)
   0.31221000000000  -0.31221000000000   1.04036000000000   Se   (8f)
  -0.31221000000000   0.31221000000000  -0.04036000000000   Se   (8f)
   0.31221000000000  -0.31221000000000  -0.54036000000000   Se   (8f)
```

Pseudobrookite ($Fe_2TiO_5$, $E4_1$): A2B5C_oC32_63_f_c2f_c - CIF

```
# CIF file
```

```
data_findsym-output
_audit_creation_method FINDSYM

_chemical_name_mineral 'Pseudobrookite'
_chemical_formula_sum 'Fe2 O5 Ti'

loop_
_publ_author_name
 'W. Q. Guo'
 'S. Malus'
 'D. H. Ryan'
 'Z. Altounian'
_journal_name_full_name
;
 Journal of Physics: Condensed Matter
;
_journal_volume 11
_journal_year 1999
_journal_page_first 6337
_journal_page_last 6346
_publ_Section_title
;
 Crystal structure and cation distributions in the FeTi$_{2}$O$_{5}
    ↪ $-Fe$_{2}$TiO$_{5}$ solid solution series
;

_aflow_title 'Pseudobrookite (Fe$_{2}$TiO$_{5}$, $E4_{1}$) Structure'
_aflow_proto 'A2B5C_oC32_63_f_c2f_c'
_aflow_params 'a,b/a,c/a,y_{1},y_{2},y_{3},z_{3},y_{4},z_{4},y_{5},z_{5
    ↪ }'
_aflow_params_values '3.7318,2.62428318774,2.67393751005,0.7617,0.18848,
    ↪ 0.1367,0.5649,0.048,0.1167,0.3108,0.0709'
_aflow_Strukturbericht '$E4_{1}$'
_aflow_Pearson 'oC32'

_symmetry_space_group_name_H-M "C 2/m 2/c 21/m"
_symmetry_Int_Tables_number 63

_cell_length_a    3.73180
_cell_length_b    9.79330
_cell_length_c    9.97860
_cell_angle_alpha 90.00000
_cell_angle_beta  90.00000
_cell_angle_gamma 90.00000

loop_
_space_group_symop_id
_space_group_symop_operation_xyz
1 x,y,z
2 x,-y,-z
3 -x,y,-z+1/2
4 -x,-y,z+1/2
5 -x,-y,-z
6 -x,y,z
7 x,-y,z+1/2
8 x,y,-z+1/2
9 x+1/2,y+1/2,z
10 x+1/2,-y+1/2,-z
11 -x+1/2,y+1/2,-z+1/2
12 -x+1/2,-y+1/2,z+1/2
13 -x+1/2,-y+1/2,-z
14 -x+1/2,y+1/2,z
15 x+1/2,-y+1/2,z+1/2
16 x+1/2,y+1/2,-z+1/2

loop_
_atom_site_label
_atom_site_type_symbol
_atom_site_symmetry_multiplicity
_atom_site_Wyckoff_label
_atom_site_fract_x
_atom_site_fract_y
_atom_site_fract_z
_atom_site_occupancy
O1 O 4 c 0.00000 0.76170 0.25000 1.00000
Ti1 Ti 4 c 0.00000 0.18848 0.25000 1.00000
Fe1 Fe 8 f 0.00000 0.13670 0.56490 1.00000
O2 O 8 f 0.00000 0.04800 0.11670 1.00000
O3 O 8 f 0.00000 0.31080 0.07090 1.00000
```

Pseudobrookite ($Fe_2TiO_5$, $E4_1$): A2B5C_oC32_63_f_c2f_c - POSCAR

```
A2B5C_oC32_63_f_c2f_c & a,b/a,c/a,y1,y2,y3,z3,y4,z4,y5,z5 --params=
    ↪ 3.7318,2.62428318774,2.67393751005,0.7617,0.18848,0.1367,0.5649
    ↪ ,0.048,0.1167,0.3108,0.0709 & Cmcm D_{2h}^{17} #63 (c^2f^3) &
    ↪ oC32 & $E4_{1}$ & Fe2O5Ti & Pseudobrookite & W. Q. Guo et al.,
    ↪ J. Phys.: Condens. Matter 11, 6337-6346 (1999)
   1.00000000000000
   1.86590000000000  -4.89665000000000   0.00000000000000
   1.86590000000000   4.89665000000000   0.00000000000000
   0.00000000000000   0.00000000000000   9.97860000000000
    Fe      O     Ti
     4     10      2
Direct
  -0.13670000000000   0.13670000000000   0.56490000000000   Fe   (8f)
   0.13670000000000  -0.13670000000000   1.06490000000000   Fe   (8f)
  -0.13670000000000   0.13670000000000  -0.06490000000000   Fe   (8f)
   0.13670000000000  -0.13670000000000  -0.56490000000000   Fe   (8f)
  -0.76170000000000   0.76170000000000   0.25000000000000   O    (4c)
   0.76170000000000  -0.76170000000000   0.75000000000000   O    (4c)
  -0.04800000000000   0.04800000000000   0.11670000000000   O    (8f)
   0.04800000000000  -0.04800000000000   0.61670000000000   O    (8f)
  -0.04800000000000   0.04800000000000   0.38330000000000   O    (8f)
   0.04800000000000  -0.04800000000000  -0.11670000000000   O    (8f)
  -0.31080000000000   0.31080000000000   0.07090000000000   O    (8f)
   0.31080000000000  -0.31080000000000   0.57090000000000   O    (8f)
  -0.31080000000000   0.31080000000000   0.42910000000000   O    (8f)
   0.31080000000000  -0.31080000000000  -0.07090000000000   O    (8f)
  -0.18848000000000   0.18848000000000   0.25000000000000   Ti   (4c)
   0.18848000000000  -0.18848000000000   0.75000000000000   Ti   (4c)
```

$MgCuAl_2$ ($E1_a$): A2BC_oC16_63_f_c_c - CIF

```
# CIF file
data_findsym-output
_audit_creation_method FINDSYM

_chemical_name_mineral 'Al2CuMg'
_chemical_formula_sum 'Al2 Cu Mg'

loop_
_publ_author_name
 'B. Heying'
 'R.-D. Hoffmann'
 'R. P\"{o}ttgen'
_journal_name_full_name
;
 Zeitschrift f{\"u}r Naturforschung B
;
_journal_volume 60
_journal_year 2005
_journal_page_first 491
_journal_page_last 494
_publ_Section_title
;
 Structure Refinement of the S-Phase Precipitate MgCuAl$_{2}$
;

_aflow_title 'MgCuAl$_{2}$ ($E1_{a}$) Structure'
_aflow_proto 'A2BC_oC16_63_f_c_c'
_aflow_params 'a,b/a,c/a,y_{1},y_{2},y_{3},z_{3}'
_aflow_params_values '4.0119,2.3093795957,1.77571724121,0.7801,0.0651,
    ↪ 0.3558,0.0556'
_aflow_Strukturbericht 'None'
_aflow_Pearson 'oC16'

_symmetry_space_group_name_H-M "C 2/m 2/c 21/m"
_symmetry_Int_Tables_number 63

_cell_length_a    4.01190
_cell_length_b    9.26500
_cell_length_c    7.12400
_cell_angle_alpha 90.00000
_cell_angle_beta  90.00000
_cell_angle_gamma 90.00000

loop_
_space_group_symop_id
_space_group_symop_operation_xyz
1 x,y,z
2 x,-y,-z
3 -x,y,-z+1/2
4 -x,-y,z+1/2
5 -x,-y,-z
6 -x,y,z
7 x,-y,z+1/2
8 x,y,-z+1/2
9 x+1/2,y+1/2,z
10 x+1/2,-y+1/2,-z
11 -x+1/2,y+1/2,-z+1/2
12 -x+1/2,-y+1/2,z+1/2
13 -x+1/2,-y+1/2,-z
14 -x+1/2,y+1/2,z
15 x+1/2,-y+1/2,z+1/2
16 x+1/2,y+1/2,-z+1/2

loop_
_atom_site_label
_atom_site_type_symbol
_atom_site_symmetry_multiplicity
_atom_site_Wyckoff_label
_atom_site_fract_x
_atom_site_fract_y
_atom_site_fract_z
_atom_site_occupancy
Cu1 Cu 4 c 0.00000 0.78010 0.25000 1.00000
Mg1 Mg 4 c 0.00000 0.06510 0.25000 1.00000
Al1 Al 8 f 0.00000 0.35580 0.05560 1.00000
```

$MgCuAl_2$ ($E1_a$): A2BC_oC16_63_f_c_c - POSCAR

```
A2BC_oC16_63_f_c_c & a,b/a,c/a,y1,y2,y3,z3 --params=4.0119,2.3093795957,
    ↪ 1.77571724121,0.7801,0.0651,0.3558,0.0556 & Cmcm D_{2h}^{17} #
    ↪ 63 (c^2f) & oC16 & None & Al2CuMg & Al2CuMg & B. Heying and
    ↪ R.-D. Hoffmann and R. P\"{o}ttgen, Z. Naturforsch. B 60,
    ↪ 491-494 (2005)
   1.00000000000000
   2.00595000000000  -4.63250000000000   0.00000000000000
   2.00595000000000   4.63250000000000   0.00000000000000
   0.00000000000000   0.00000000000000   7.12400000000000
    Al     Cu     Mg
     4      2      2
Direct
  -0.35580000000000   0.35580000000000   0.05560000000000   Al   (8f)
   0.35580000000000  -0.35580000000000   0.55560000000000   Al   (8f)
  -0.35580000000000   0.35580000000000   0.44440000000000   Al   (8f)
   0.35580000000000  -0.35580000000000  -0.05560000000000   Al   (8f)
  -0.78010000000000   0.78010000000000   0.25000000000000   Cu   (4c)
   0.78010000000000  -0.78010000000000   0.75000000000000   Cu   (4c)
  -0.06510000000000   0.06510000000000   0.25000000000000   Mg   (4c)
```

```
    0.06510000000000   -0.06510000000000    0.75000000000000    Mg    (4c)
```

$S0_4$ (Staurolite, $Fe(OH)_2Al_4Si_2O_{10}$) (*obsolete*): A4BC12D2_oC76_63_eg_c_f3gh_g - CIF

```
# CIF file
data_findsym-output
_audit_creation_method FINDSYM

_chemical_name_mineral 'Staurolite'
_chemical_formula_sum 'Al4 Fe O12 Si2'

loop_
_publ_author_name
 'C. Hermann'
 'O. Lohrmann'
 'H. Philipp'
_journal_year 1937
_publ_Section_title
;
 Strukturbericht Band II 1928-1932
;

_aflow_title '$S0_{4}$ (Staurolite, Fe(OH)$_{2}$Al$_{4}$Si$_{2}$O$_{10}$
    ↪ ) ({\em{obsolete}}) Structure'
_aflow_proto 'A4BC12D2_oC76_63_eg_c_f3gh_g'
_aflow_params 'a,b/a,c/a,y_{1},x_{2},y_{3},z_{3},x_{4},y_{4},x_{5},y_{5
    ↪ },x_{6},y_{6},x_{7},y_{7},x_{8},y_{8},x_{9},y_{9},z_{9}'
_aflow_params_values '16.52,0.473365617433,0.340799031477,0.611,0.328,
    ↪ 0.778,0.0,-0.092,0.25,0.094,0.0,0.25,0.0,0.406,0.0,0.667,0.375,
    ↪ 0.658,0.173,0.0'
_aflow_Strukturbericht '$S0_{4}$'
_aflow_Pearson 'oC76'

_symmetry_space_group_name_H-M "C 2/m 2/c 21/m"
_symmetry_Int_Tables_number 63

_cell_length_a    16.52000
_cell_length_b    7.82000
_cell_length_c    5.63000
_cell_angle_alpha 90.00000
_cell_angle_beta  90.00000
_cell_angle_gamma 90.00000

loop_
_space_group_symop_id
_space_group_symop_operation_xyz
1 x,y,z
2 x,-y,-z
3 -x,y,-z+1/2
4 -x,-y,z+1/2
5 -x,-y,-z
6 -x,y,z
7 x,-y,z+1/2
8 x,y,-z+1/2
9 x+1/2,y+1/2,z
10 x+1/2,-y+1/2,-z
11 -x+1/2,y+1/2,-z+1/2
12 -x+1/2,-y+1/2,z+1/2
13 -x+1/2,-y+1/2,-z
14 -x+1/2,y+1/2,z
15 x+1/2,-y+1/2,z+1/2
16 x+1/2,y+1/2,-z+1/2

loop_
_atom_site_label
_atom_site_type_symbol
_atom_site_symmetry_multiplicity
_atom_site_Wyckoff_label
_atom_site_fract_x
_atom_site_fract_y
_atom_site_fract_z
_atom_site_occupancy
Fe1 Fe 4 c 0.00000 0.61100 0.25000 1.00000
Al1 Al 8 e 0.32800 0.00000 0.00000 1.00000
O1 O 8 f 0.00000 0.77800 0.00000 1.00000
Al2 Al 8 g -0.09200 0.25000 0.25000 1.00000
O2 O 8 g 0.09400 0.00000 0.25000 1.00000
O3 O 8 g 0.25000 0.00000 0.25000 1.00000
O4 O 8 g 0.40600 0.00000 0.25000 1.00000
Si1 Si 8 g 0.66700 0.37500 0.25000 1.00000
O5 O 16 h 0.65800 0.17300 0.00000 1.00000
```

$S0_4$ (Staurolite, $Fe(OH)_2Al_4Si_2O_{10}$) (*obsolete*): A4BC12D2_oC76_63_eg_c_f3gh_g - POSCAR

```
A4BC12D2_oC76_63_eg_c_f3gh_g & a,b/a,c/a,y1,x2,y3,z3,x4,y4,x5,y5,x6,y6,
    ↪ x7,y7,x8,y8,x9,y9,z9 --params=16.52,0.473365617433,
    ↪ 0.340799031477,0.611,0.328,0.778,0.0,-0.092,0.25,0.094,0.0,0.25
    ↪ ,0.0,0.406,0.0,0.667,0.375,0.658,0.173,0.0 & Cmcm D_{2h}^{17} #
    ↪ 63 (cefg^5h) & oC76 & $S0_{4}$ & Al4FeO12Si2 & Staurolite & C.
    ↪ Hermann and O. Lohrmann and H. Philipp, (1937)
   1.00000000000000
   8.26000000000000   -3.91000000000000    0.00000000000000
   8.26000000000000    3.91000000000000    0.00000000000000
   0.00000000000000    0.00000000000000    5.63000000000000
    Al     Fe      O     Si
     8      2     24      4
Direct
   0.32800000000000    0.32800000000000    0.00000000000000    Al    (8e)
  -0.32800000000000   -0.32800000000000    0.50000000000000    Al    (8e)
  -0.32800000000000   -0.32800000000000    0.00000000000000    Al    (8e)
   0.32800000000000    0.32800000000000    0.50000000000000    Al    (8e)
  -0.34200000000000    0.15800000000000    0.25000000000000    Al    (8g)
   0.34200000000000   -0.15800000000000    0.75000000000000    Al    (8g)
  -0.15800000000000    0.34200000000000    0.25000000000000    Al    (8g)
   0.15800000000000   -0.34200000000000    0.75000000000000    Al    (8g)
```

```
  -0.61100000000000    0.61100000000000    0.25000000000000    Fe    (4c)
   0.61100000000000   -0.61100000000000    0.75000000000000    Fe    (4c)
  -0.77800000000000    0.77800000000000    0.00000000000000    O     (8f)
   0.77800000000000   -0.77800000000000    0.50000000000000    O     (8f)
  -0.77800000000000    0.77800000000000    0.50000000000000    O     (8f)
   0.77800000000000   -0.77800000000000    0.00000000000000    O     (8f)
   0.09400000000000    0.09400000000000    0.25000000000000    O     (8g)
  -0.09400000000000   -0.09400000000000    0.75000000000000    O     (8g)
  -0.09400000000000   -0.09400000000000    0.25000000000000    O     (8g)
   0.09400000000000    0.09400000000000    0.75000000000000    O     (8g)
   0.25000000000000    0.25000000000000    0.25000000000000    O     (8g)
  -0.25000000000000   -0.25000000000000    0.75000000000000    O     (8g)
  -0.25000000000000   -0.25000000000000    0.25000000000000    O     (8g)
   0.25000000000000    0.25000000000000    0.75000000000000    O     (8g)
   0.40600000000000    0.40600000000000    0.25000000000000    O     (8g)
  -0.40600000000000   -0.40600000000000    0.75000000000000    O     (8g)
  -0.40600000000000   -0.40600000000000    0.25000000000000    O     (8g)
   0.40600000000000    0.40600000000000    0.75000000000000    O     (8g)
   0.48500000000000    0.83100000000000    0.00000000000000    O    (16h)
  -0.48500000000000   -0.83100000000000    0.50000000000000    O    (16h)
  -0.83100000000000   -0.48500000000000    0.50000000000000    O    (16h)
   0.83100000000000    0.48500000000000    0.00000000000000    O    (16h)
  -0.48500000000000   -0.83100000000000    0.00000000000000    O    (16h)
   0.48500000000000    0.83100000000000    0.50000000000000    O    (16h)
   0.83100000000000    0.48500000000000    0.50000000000000    O    (16h)
  -0.83100000000000   -0.48500000000000    0.00000000000000    O    (16h)
   0.29200000000000    1.04200000000000    0.25000000000000    Si    (8g)
  -0.29200000000000   -1.04200000000000    0.75000000000000    Si    (8g)
  -1.04200000000000   -0.29200000000000    0.25000000000000    Si    (8g)
   1.04200000000000    0.29200000000000    0.75000000000000    Si    (8g)
```

$Pd_5Pu_3$: A5B3_oC32_63_cfg_ce - CIF

```
# CIF file
data_findsym-output
_audit_creation_method FINDSYM

_chemical_name_mineral 'Pd5Pu3'
_chemical_formula_sum 'Pd5 Pu3'

loop_
_publ_author_name
 'D. T. Cromer'
_journal_name_full_name
;
 Acta Crystallographica Section B: Structural Science
;
_journal_volume 32
_journal_year 1976
_journal_page_first 1930
_journal_page_last 1932
_publ_Section_title
;
 Plutonium-palladium Pu$_{3}$Pd$_{5}$
;

# Found in New $R$$_{3}$Pd$_{5}$ Compounds ($R$ = Sc, Y, Gd-Lu):
    ↪ Formation and Stability, Crystal Structure, and
    ↪ Antiferromagnetism, 2016

_aflow_title 'Pd$_{5}$Pu$_{3}$ Structure'
_aflow_proto 'A5B3_oC32_63_cfg_ce'
_aflow_params 'a,b/a,c/a,y_{1},y_{2},x_{3},y_{4},z_{4},x_{5},y_{5}'
_aflow_params_values '9.20101,0.781979369656,1.0619486339,0.0254,0.6251,
    ↪ 0.2018,0.3147,0.451,0.2219,0.2863'
_aflow_Strukturbericht 'None'
_aflow_Pearson 'oC32'

_symmetry_space_group_name_H-M "C 2/m 2/c 21/m"
_symmetry_Int_Tables_number 63

_cell_length_a    9.20101
_cell_length_b    7.19500
_cell_length_c    9.77100
_cell_angle_alpha 90.00000
_cell_angle_beta  90.00000
_cell_angle_gamma 90.00000

loop_
_space_group_symop_id
_space_group_symop_operation_xyz
1 x,y,z
2 x,-y,-z
3 -x,y,-z+1/2
4 -x,-y,z+1/2
5 -x,-y,-z
6 -x,y,z
7 x,-y,z+1/2
8 x,y,-z+1/2
9 x+1/2,y+1/2,z
10 x+1/2,-y+1/2,-z
11 -x+1/2,y+1/2,-z+1/2
12 -x+1/2,-y+1/2,z+1/2
13 -x+1/2,-y+1/2,-z
14 -x+1/2,y+1/2,z
15 x+1/2,-y+1/2,z+1/2
16 x+1/2,y+1/2,-z+1/2

loop_
_atom_site_label
_atom_site_type_symbol
_atom_site_symmetry_multiplicity
_atom_site_Wyckoff_label
_atom_site_fract_x
_atom_site_fract_y
```

```
_atom_site_fract_z
_atom_site_occupancy
Pd1 Pd 4 c 0.00000 0.02540 0.25000 1.00000
Pu1 Pu 4 c 0.00000 0.62510 0.25000 1.00000
Pu2 Pu 8 e 0.20180 0.00000 0.00000 1.00000
Pd2 Pd 8 f 0.00000 0.31470 0.45100 1.00000
Pd3 Pd 8 g 0.22190 0.28630 0.25000 1.00000
```

$Pd_5Pu_3$: A5B3_oC32_63_cfg_ce - POSCAR

```
A5B3_oC32_63_cfg_ce & a,b/a,c/a,y1,y2,x3,y4,z4,x5,y5 --params=9.20101,
    ↪ 0.781979369656,1.0619486339,0.0254,0.6251,0.2018,0.3147,0.451,
    ↪ 0.2219,0.2863 & Cmcm D_{2h}^{17} #63 (c^2efg) & oC32 & None &
    ↪ Pd5Pu3 & Pd5Pu3 & D. T. Cromer, Acta Crystallogr. Sect. B
    ↪ Struct. Sci. 32, 1930-1932 (1976)
   1.00000000000000
   4.60050500000000  -3.59750000000000   0.00000000000000
   4.60050500000000   3.59750000000000   0.00000000000000
   0.00000000000000   0.00000000000000   9.77100000000000
    Pd       Pu
    10        6
Direct
  -0.02540000000000   0.02540000000000   0.25000000000000   Pd   (4c)
   0.02540000000000  -0.02540000000000   0.75000000000000   Pd   (4c)
  -0.31470000000000   0.31470000000000   0.45100000000000   Pd   (8f)
   0.31470000000000  -0.31470000000000   0.95100000000000   Pd   (8f)
  -0.31470000000000   0.31470000000000   0.04900000000000   Pd   (8f)
   0.31470000000000  -0.31470000000000  -0.45100000000000   Pd   (8f)
  -0.06440000000000   0.50820000000000   0.25000000000000   Pd   (8g)
   0.06440000000000  -0.50820000000000   0.75000000000000   Pd   (8g)
  -0.50820000000000   0.06440000000000   0.25000000000000   Pd   (8g)
   0.50820000000000  -0.06440000000000   0.75000000000000   Pd   (8g)
  -0.62510000000000   0.62510000000000   0.25000000000000   Pu   (4c)
   0.62510000000000  -0.62510000000000   0.75000000000000   Pu   (4c)
   0.20180000000000   0.20180000000000   0.00000000000000   Pu   (8e)
  -0.20180000000000  -0.20180000000000   0.50000000000000   Pu   (8e)
  -0.20180000000000  -0.20180000000000   0.00000000000000   Pu   (8e)
   0.20180000000000   0.20180000000000   0.50000000000000   Pu   (8e)
```

$ZrTe_5$: A5B_oC24_63_c2f_c - CIF

```
# CIF file
data_findsym-output
_audit_creation_method FINDSYM

_chemical_name_mineral 'Te5Zr'
_chemical_formula_sum 'Te5 Zr'

loop_
_publ_author_name
 'H. Fjellv{\aa}g'
 'A. Kjekshus'
_journal_name_full_name
;
 Solid State Communications
;
_journal_volume 60
_journal_year 1986
_journal_page_first 91
_journal_page_last 93
_publ_Section_title
;
 Structural Properties of ZrTe$_{5}$ and HfTe$_{5}$ as Seen by Power
    ↪ Diffraction
;

_aflow_title 'ZrTe$_{5}$ Structure'
_aflow_proto 'A5B_oC24_63_c2f_c'
_aflow_params 'a,b/a,c/a,y_{1},y_{2},y_{3},z_{3},y_{4},z_{4}'
_aflow_params_values '3.9875,3.64388714734,3.4407523511,0.316,0.633,-
    ↪ 0.067,0.151,0.209,0.434'
_aflow_Strukturbericht 'None'
_aflow_Pearson 'oC24'

_symmetry_space_group_name_H-M "C 2/m 2/c 21/m"
_symmetry_Int_Tables_number 63

_cell_length_a     3.98750
_cell_length_b     14.53000
_cell_length_c     13.72000
_cell_angle_alpha 90.00000
_cell_angle_beta  90.00000
_cell_angle_gamma 90.00000

loop_
_space_group_symop_id
_space_group_symop_operation_xyz
1 x,y,z
2 x,-y,-z
3 -x,y,-z+1/2
4 -x,-y,z+1/2
5 -x,-y,-z
6 -x,y,z
7 x,-y,z+1/2
8 x,y,-z+1/2
9 x+1/2,y+1/2,z
10 x+1/2,-y+1/2,-z
11 -x+1/2,y+1/2,-z+1/2
12 -x+1/2,-y+1/2,z+1/2
13 -x+1/2,-y+1/2,-z
14 -x+1/2,y+1/2,z
15 x+1/2,-y+1/2,z+1/2
16 x+1/2,y+1/2,-z+1/2

loop_
```

```
_atom_site_label
_atom_site_type_symbol
_atom_site_symmetry_multiplicity
_atom_site_Wyckoff_label
_atom_site_fract_x
_atom_site_fract_y
_atom_site_fract_z
_atom_site_occupancy
Te1 Te 4 c 0.00000 0.31600 0.25000 1.00000
Zr1 Zr 4 c 0.00000 0.63300 0.25000 1.00000
Te2 Te 8 f 0.00000 -0.06700 0.15100 1.00000
Te3 Te 8 f 0.00000 0.20900 0.43400 1.00000
```

$ZrTe_5$: A5B_oC24_63_c2f_c - POSCAR

```
A5B_oC24_63_c2f_c & a,b/a,c/a,y1,y2,y3,z3,y4,z4 --params=3.9875,
    ↪ 3.64388714734,3.4407523511,0.316,0.633,-0.067,0.151,0.209,0.434
    ↪ & Cmcm D_{2h}^{17} #63 (c^2f^2) & oC24 & None & Te5Zr & Te5Zr
    ↪ & H. Fjellv{\aa}g and A. Kjekshus, Solid State Commun. 60,
    ↪ 91-93 (1986)
   1.00000000000000
   1.99375000000000  -7.26500000000000   0.00000000000000
   1.99375000000000   7.26500000000000   0.00000000000000
   0.00000000000000   0.00000000000000  13.72000000000000
    Te       Zr
    10        2
Direct
  -0.31600000000000   0.31600000000000   0.25000000000000   Te   (4c)
   0.31600000000000  -0.31600000000000   0.75000000000000   Te   (4c)
   0.06700000000000  -0.06700000000000   0.15100000000000   Te   (8f)
  -0.06700000000000   0.06700000000000   0.65100000000000   Te   (8f)
   0.06700000000000  -0.06700000000000   0.34900000000000   Te   (8f)
  -0.06700000000000   0.06700000000000  -0.15100000000000   Te   (8f)
  -0.20900000000000   0.20900000000000   0.43400000000000   Te   (8f)
   0.20900000000000  -0.20900000000000   0.93400000000000   Te   (8f)
  -0.20900000000000   0.20900000000000   0.06600000000000   Te   (8f)
   0.20900000000000  -0.20900000000000  -0.43400000000000   Te   (8f)
  -0.63300000000000   0.63300000000000   0.25000000000000   Zr   (4c)
   0.63300000000000  -0.63300000000000   0.75000000000000   Zr   (4c)
```

Lepidocrocite ($\gamma$-FeO(OH), $E0_4$): AB2C2_oC20_63_c_f_2c - CIF

```
# CIF file
data_findsym-output
_audit_creation_method FINDSYM

_chemical_name_mineral 'Lepidocrocite'
_chemical_formula_sum 'Fe H2 O2'

loop_
_publ_author_name
 'A. {N{\o}rlund Christensen}'
 'M. S. Lehmann'
 'P. Convert'
_journal_name_full_name
;
 Acta Chemica Scandinavica
;
_journal_volume 36a
_journal_year 1982
_journal_page_first 303
_journal_page_last 308
_publ_Section_title
;
 Deuteration of Crystalline Hydroxides, Hydrogen Bonds of $\gamma$-AlOO(
    ↪ H,D) and $\gamma$-FeOO(H,D)
;

# Found in The American Mineralogist Crystal Structure Database, 2003

_aflow_title 'Lepidocrocite ($\gamma$-FeO(OH), $E0_{4}$) Structure'
_aflow_proto 'AB2C2_oC20_63_c_f_2c'
_aflow_params 'a,b/a,c/a,y_{1},y_{2},y_{3},y_{4},z_{4}'
_aflow_params_values '3.07,4.08143322476,1.22475570033,-0.3137,0.2842,
    ↪ 0.0724,0.0143,0.3663'
_aflow_Strukturbericht '$E0_{4}$'
_aflow_Pearson 'oC20'

_symmetry_space_group_name_H-M "C 2/m 2/c 21/m"
_symmetry_Int_Tables_number 63

_cell_length_a     3.07000
_cell_length_b     12.53000
_cell_length_c     3.76000
_cell_angle_alpha 90.00000
_cell_angle_beta  90.00000
_cell_angle_gamma 90.00000

loop_
_space_group_symop_id
_space_group_symop_operation_xyz
1 x,y,z
2 x,-y,-z
3 -x,y,-z+1/2
4 -x,-y,z+1/2
5 -x,-y,-z
6 -x,y,z
7 x,-y,z+1/2
8 x,y,-z+1/2
9 x+1/2,y+1/2,z
10 x+1/2,-y+1/2,-z
11 -x+1/2,y+1/2,-z+1/2
12 -x+1/2,-y+1/2,z+1/2
13 -x+1/2,-y+1/2,-z
14 -x+1/2,y+1/2,z
```

```
15 x+1/2,-y+1/2,z+1/2
16 x+1/2,y+1/2,-z+1/2

loop_
_atom_site_label
_atom_site_type_symbol
_atom_site_symmetry_multiplicity
_atom_site_Wyckoff_label
_atom_site_fract_x
_atom_site_fract_y
_atom_site_fract_z
_atom_site_occupancy
Fe1 Fe 4 c 0.00000 -0.31370 0.25000 1.00000
O1 O 4 c 0.00000 0.28420 0.25000 1.00000
O2 O 4 c 0.00000 0.07240 0.25000 1.00000
H1 H 8 f 0.00000 0.01430 0.36630 0.47100
```

Lepidocrocite (γ-FeO(OH), $E0_4$): AB2C2_oC20_63_c_f_2c - POSCAR

```
AB2C2_oC20_63_c_f_2c & a,b/a,c/a,y1,y2,y3,y4,z4 --params=3.07,
    ↪ 4.08143322476,1.22475570033,-0.3137,0.2842,0.0724,0.0143,0.3663
    ↪ & Cmcm D_{2h}^{17} #63 (c^3f) & oC20 & $E0_{4}$ & FeHO2 &
    ↪ Lepidocrocite & A. {N{\o}rlund Christensen} and M. S. Lehmann
    ↪ and P. Convert, Acta Chem. Scand. 36a, 303-308 (1982)
   1.00000000000000
   1.53500000000000  -6.26500000000000   0.00000000000000
   1.53500000000000   6.26500000000000   0.00000000000000
   0.00000000000000   0.00000000000000   3.76000000000000
    Fe      H      O
     2      4      4
Direct
   0.31370000000000  -0.31370000000000   0.25000000000000   Fe   (4c)
  -0.31370000000000   0.31370000000000   0.75000000000000   Fe   (4c)
  -0.01430000000000   0.01430000000000   0.36630000000000   H    (8f)
   0.01430000000000  -0.01430000000000   0.86630000000000   H    (8f)
  -0.01430000000000   0.01430000000000   0.13370000000000   H    (8f)
   0.01430000000000  -0.01430000000000  -0.36630000000000   H    (8f)
  -0.28420000000000   0.28420000000000   0.25000000000000   O    (4c)
   0.28420000000000  -0.28420000000000   0.75000000000000   O    (4c)
  -0.07240000000000   0.07240000000000   0.25000000000000   O    (4c)
   0.07240000000000  -0.07240000000000   0.75000000000000   O    (4c)
```

$Na_2CrO_4$ ($H1_8$): AB2C4_oC28_63_c_bc_fg - CIF

```
# CIF file
data_findsym-output
_audit_creation_method FINDSYM

_chemical_name_mineral 'CrNa2O4'
_chemical_formula_sum 'Cr Na2 O4'

loop_
_publ_author_name
 'A. Niggli'
_journal_name_full_name
;
 Acta Cristallographica
;
_journal_volume 7
_journal_year 1954
_journal_page_first 776
_journal_page_last 776
_publ_Section_title
;
 Die Raumgruppe von Na$_{2}$CrO$_{4}$
;

_aflow_title 'Na$_{2}$CrO$_{4}$ ($H1_{8}$) Structure'
_aflow_proto 'AB2C4_oC28_63_c_bc_fg'
_aflow_params 'a,b/a,c/a,y_{2},y_{3},y_{4},z_{4},x_{5},y_{5}'
_aflow_params_values '5.91,1.56175972927,1.21827411168,0.8472,0.2,0.75,
    ↪ 0.0694,0.2778,0.45'
_aflow_Strukturbericht '$H1_{8}$'
_aflow_Pearson 'oC28'

_symmetry_space_group_name_H-M "C 2/m 2/c 21/m"
_symmetry_Int_Tables_number 63

_cell_length_a      5.91000
_cell_length_b      9.23000
_cell_length_c      7.20000
_cell_angle_alpha  90.00000
_cell_angle_beta   90.00000
_cell_angle_gamma  90.00000

loop_
_space_group_symop_id
_space_group_symop_operation_xyz
1 x,y,z
2 x,-y,-z
3 -x,y,-z+1/2
4 -x,-y,z+1/2
5 -x,-y,-z
6 -x,y,z
7 x,-y,z+1/2
8 x,y,-z+1/2
9 x+1/2,y+1/2,z
10 x+1/2,-y+1/2,-z
11 -x+1/2,y+1/2,-z+1/2
12 -x+1/2,-y+1/2,z+1/2
13 -x+1/2,-y+1/2,-z
14 -x+1/2,y+1/2,z
15 x+1/2,-y+1/2,z+1/2
16 x+1/2,y+1/2,-z+1/2
```

```
loop_
_atom_site_label
_atom_site_type_symbol
_atom_site_symmetry_multiplicity
_atom_site_Wyckoff_label
_atom_site_fract_x
_atom_site_fract_y
_atom_site_fract_z
_atom_site_occupancy
Na1 Na 4 b 0.00000 0.50000 0.00000 1.00000
Cr1 Cr 4 c 0.00000 0.84720 0.25000 1.00000
Na2 Na 4 c 0.00000 0.20000 0.25000 1.00000
O1 O 8 f 0.00000 0.75000 0.06940 1.00000
O2 O 8 g 0.27780 0.45000 0.25000 1.00000
```

$Na_2CrO_4$ ($H1_8$): AB2C4_oC28_63_c_bc_fg - POSCAR

```
AB2C4_oC28_63_c_bc_fg & a,b/a,c/a,y2,y3,y4,z4,x5,y5 --params=5.91,
    ↪ 1.56175972927,1.21827411168,0.8472,0.2,0.75,0.0694,0.2778,0.45
    ↪ & Cmcm D_{2h}^{17} #63 (bc^2fg) & oC28 & $H1_{8}$ & CrNa2O4 &
    ↪ CrNa2O4 & A. Niggli, Acta Cryst. 7, 776(1954)
   1.00000000000000
   2.95500000000000  -4.61500000000000   0.00000000000000
   2.95500000000000   4.61500000000000   0.00000000000000
   0.00000000000000   0.00000000000000   7.20000000000000
    Cr     Na      O
     2      4      8
Direct
  -0.84720000000000   0.84720000000000   0.25000000000000   Cr   (4c)
   0.84720000000000  -0.84720000000000   0.75000000000000   Cr   (4c)
   0.50000000000000   0.50000000000000   0.00000000000000   Na   (4b)
   0.50000000000000   0.50000000000000   0.50000000000000   Na   (4b)
  -0.20000000000000   0.20000000000000   0.25000000000000   Na   (4c)
   0.20000000000000  -0.20000000000000   0.75000000000000   Na   (4c)
  -0.75000000000000   0.75000000000000   0.06940000000000   O    (8f)
   0.75000000000000  -0.75000000000000   0.56940000000000   O    (8f)
  -0.75000000000000   0.75000000000000   0.43060000000000   O    (8f)
   0.75000000000000  -0.75000000000000  -0.06940000000000   O    (8f)
  -0.17220000000000   0.72780000000000   0.25000000000000   O    (8g)
   0.17220000000000  -0.72780000000000   0.75000000000000   O    (8g)
  -0.72780000000000   0.17220000000000   0.25000000000000   O    (8g)
   0.72780000000000  -0.17220000000000   0.75000000000000   O    (8g)
```

$ThFe_2SiC$: AB2CD_oC20_63_b_f_c_c - CIF

```
# CIF file
data_findsym-output
_audit_creation_method FINDSYM

_chemical_name_mineral 'CFe2SiTh'
_chemical_formula_sum 'C Fe2 Si Th'

loop_
_publ_author_name
 'A. M. Witte'
 'W. Jeitschko'
_journal_name_full_name
;
 Journal of Solid State Chemistry
;
_journal_volume 112
_journal_year 1994
_journal_page_first 232
_journal_page_last 236
_publ_Section_title
;
 Carbides with Filled Re$_{3}$B-Type Structure
;

_aflow_title 'ThFe$_{2}$SiC Structure'
_aflow_proto 'AB2CD_oC20_63_b_f_c_c'
_aflow_params 'a,b/a,c/a,y_{2},y_{3},y_{4},z_{4}'
_aflow_params_values '3.863,2.79730779187,1.79911985503,0.7719,0.05275,
    ↪ 0.66585,0.5611'
_aflow_Strukturbericht 'None'
_aflow_Pearson 'oC20'

_symmetry_space_group_name_H-M "C 2/m 2/c 21/m"
_symmetry_Int_Tables_number 63

_cell_length_a      3.86300
_cell_length_b      10.80600
_cell_length_c      6.95000
_cell_angle_alpha  90.00000
_cell_angle_beta   90.00000
_cell_angle_gamma  90.00000

loop_
_space_group_symop_id
_space_group_symop_operation_xyz
1 x,y,z
2 x,-y,-z
3 -x,y,-z+1/2
4 -x,-y,z+1/2
5 -x,-y,-z
6 -x,y,z
7 x,-y,z+1/2
8 x,y,-z+1/2
9 x+1/2,y+1/2,z
10 x+1/2,-y+1/2,-z
11 -x+1/2,y+1/2,-z+1/2
12 -x+1/2,-y+1/2,z+1/2
13 -x+1/2,-y+1/2,-z
14 -x+1/2,y+1/2,z
15 x+1/2,-y+1/2,z+1/2
```

```
16 x+1/2,y+1/2,-z+1/2

loop_
_atom_site_label
_atom_site_type_symbol
_atom_site_symmetry_multiplicity
_atom_site_Wyckoff_label
_atom_site_fract_x
_atom_site_fract_y
_atom_site_fract_z
_atom_site_occupancy
C1 C 4 b 0.00000 0.50000 0.00000 1.00000
Si1 Si 4 c 0.00000 0.77190 0.25000 1.00000
Th1 Th 4 c 0.00000 0.05275 0.25000 1.00000
Fe1 Fe 8 f 0.00000 0.66585 0.56110 1.00000
```

$ThFe_2SiC$: AB2CD_oC20_63_b_f_c_c - POSCAR

```
AB2CD_oC20_63_b_f_c_c & a,b/a,c/a,y2,y3,y4,z4 --params=3.863,
    ↪ 2.79730779187,1.79911985503,0.7719,0.05275,0.66585,0.5611 &
    ↪ Cmcm D_{2h}^{17} #63 (bc^2f) & oC20 & None & CFe2SiTh &
    ↪ CFe2SiTh & A. M. Witte and W. Jeitschko, J. Solid State Chem.
    ↪ 112, 232-236 (1994)
   1.00000000000000
   1.93150000000000  -5.40300000000000   0.00000000000000
   1.93150000000000   5.40300000000000   0.00000000000000
   0.00000000000000   0.00000000000000   6.95000000000000
     C     Fe     Si     Th
     2      4      2      2
Direct
   0.50000000000000   0.50000000000000   0.00000000000000    C   (4b)
   0.50000000000000   0.50000000000000   0.50000000000000    C   (4b)
  -0.66585000000000   0.66585000000000   0.56110000000000    Fe  (8f)
   0.66585000000000  -0.66585000000000   1.06110000000000    Fe  (8f)
  -0.66585000000000   0.66585000000000  -0.06110000000000    Fe  (8f)
   0.66585000000000  -0.66585000000000  -0.56110000000000    Fe  (8f)
  -0.77190000000000   0.77190000000000   0.25000000000000    Si  (4c)
   0.77190000000000  -0.77190000000000   0.75000000000000    Si  (4c)
  -0.05275000000000   0.05275000000000   0.25000000000000    Th  (4c)
   0.05275000000000  -0.05275000000000   0.75000000000000    Th  (4c)
```

$Mn_3As$ ($D0_d$): AB3_oC16_63_c_3c - CIF

```
# CIF file
data_findsym-output
_audit_creation_method FINDSYM

_chemical_name_mineral 'AsMn3'
_chemical_formula_sum 'As Mn3'

loop_
_publ_author_name
 'W. {Carrillo-Cabrera}'
_journal_name_full_name
;
 Acta Chemica Scandinavica
;
_journal_volume 37a
_journal_year 1983
_journal_page_first 93
_journal_page_last 98
_publ_Section_title
;
 The Crystal Structure of TiCu$_{2}$P and Its Relationship to the
    ↪ Structure of Mn$_{3}$As
;

_aflow_title 'Mn$_{3}$As ($D0_{d}$) Structure'
_aflow_proto 'AB3_oC16_63_c_3c'
_aflow_params 'a,b/a,c/a,y_{1},y_{2},y_{3},y_{4}'
_aflow_params_values '3.78779,4.28999759754,1.0,0.159,-0.0565,0.5565,
    ↪ 0.3155'
_aflow_Strukturbericht '$D0_{d}$'
_aflow_Pearson 'oC16'

_symmetry_space_group_name_H-M "C 2/m 2/c 21/m"
_symmetry_Int_Tables_number 63

_cell_length_a    3.78779
_cell_length_b    16.24961
_cell_length_c    3.78779
_cell_angle_alpha 90.00000
_cell_angle_beta  90.00000
_cell_angle_gamma 90.00000

loop_
_space_group_symop_id
_space_group_symop_operation_xyz
1 x,y,z
2 x,-y,-z
3 -x,y,-z+1/2
4 -x,-y,z+1/2
5 -x,-y,-z
6 -x,y,z
7 x,-y,z+1/2
8 x,y,-z+1/2
9 x+1/2,y+1/2,z
10 x+1/2,-y+1/2,-z
11 -x+1/2,y+1/2,-z+1/2
12 -x+1/2,-y+1/2,z+1/2
13 -x+1/2,-y+1/2,-z
14 -x+1/2,y+1/2,z
15 x+1/2,-y+1/2,z+1/2
16 x+1/2,y+1/2,-z+1/2

loop_
_atom_site_label
_atom_site_type_symbol
_atom_site_symmetry_multiplicity
_atom_site_Wyckoff_label
_atom_site_fract_x
_atom_site_fract_y
_atom_site_fract_z
_atom_site_occupancy
As1 As 4 c 0.00000 0.15900 0.25000 1.00000
Mn1 Mn 4 c 0.00000 -0.05650 0.25000 1.00000
Mn2 Mn 4 c 0.00000 0.55650 0.25000 1.00000
Mn3 Mn 4 c 0.00000 0.31550 0.25000 1.00000
```

$Mn_3As$ ($D0_d$): AB3_oC16_63_c_3c - POSCAR

```
AB3_oC16_63_c_3c & a,b/a,c/a,y1,y2,y3,y4 --params=3.78779,4.28999759754,
    ↪ 1.0,0.159,-0.0565,0.5565,0.3155 & Cmcm D_{2h}^{17} #63 (c^4) &
    ↪ oC16 & $D0_{d}$ & AsMn3 & AsMn3 & W. {Carrillo-Cabrera}, Acta
    ↪ Chem. Scand. 37a, 93-98 (1983)
   1.00000000000000
   1.89389500000000  -8.12480500000000   0.00000000000000
   1.89389500000000   8.12480500000000   0.00000000000000
   0.00000000000000   0.00000000000000   3.78779000000000
    As    Mn
     2     6
Direct
  -0.15900000000000   0.15900000000000   0.25000000000000    As  (4c)
   0.15900000000000  -0.15900000000000   0.75000000000000    As  (4c)
   0.05650000000000  -0.05650000000000   0.25000000000000    Mn  (4c)
  -0.05650000000000   0.05650000000000   0.75000000000000    Mn  (4c)
  -0.55650000000000   0.55650000000000   0.25000000000000    Mn  (4c)
   0.55650000000000  -0.55650000000000   0.75000000000000    Mn  (4c)
  -0.31550000000000   0.31550000000000   0.25000000000000    Mn  (4c)
   0.31550000000000  -0.31550000000000   0.75000000000000    Mn  (4c)
```

$Re_3B$: AB3_oC16_63_c_cf - CIF

```
# CIF file
data_findsym-output
_audit_creation_method FINDSYM

_chemical_name_mineral 'BRe3'
_chemical_formula_sum 'B Re3'

loop_
_publ_author_name
 'B. Aronsson'
 'M. B\"{a}ckman'
 'S. Rundqvist'
_journal_name_full_name
;
 Acta Chemica Scandinavica
;
_journal_volume 14
_journal_year 1960
_journal_page_first 1001
_journal_page_last 1005
_publ_Section_title
;
 The Crystal Structure of Re$_{3}$B
;

_aflow_title 'Re$_{3}$B Structure'
_aflow_proto 'AB3_oC16_63_c_cf'
_aflow_params 'a,b/a,c/a,y_{1},y_{2},y_{3},z_{3}'
_aflow_params_values '2.89,3.22249134948,2.51141868512,0.744,0.4262,
    ↪ 0.1345,0.062'
_aflow_Strukturbericht 'None'
_aflow_Pearson 'oC16'

_symmetry_space_group_name_H-M "C 2/m 2/c 21/m"
_symmetry_Int_Tables_number 63

_cell_length_a    2.89000
_cell_length_b    9.31300
_cell_length_c    7.25800
_cell_angle_alpha 90.00000
_cell_angle_beta  90.00000
_cell_angle_gamma 90.00000

loop_
_space_group_symop_id
_space_group_symop_operation_xyz
1 x,y,z
2 x,-y,-z
3 -x,y,-z+1/2
4 -x,-y,z+1/2
5 -x,-y,-z
6 -x,y,z
7 x,-y,z+1/2
8 x,y,-z+1/2
9 x+1/2,y+1/2,z
10 x+1/2,-y+1/2,-z
11 -x+1/2,y+1/2,-z+1/2
12 -x+1/2,-y+1/2,z+1/2
13 -x+1/2,-y+1/2,-z
14 -x+1/2,y+1/2,z
15 x+1/2,-y+1/2,z+1/2
16 x+1/2,y+1/2,-z+1/2

loop_
_atom_site_label
_atom_site_type_symbol
_atom_site_symmetry_multiplicity
```

```
_atom_site_Wyckoff_label
_atom_site_fract_x
_atom_site_fract_y
_atom_site_fract_z
_atom_site_occupancy
B1 B 4 c 0.00000 0.74400 0.25000 1.00000
Re1 Re 4 c 0.00000 0.42620 0.25000 1.00000
Re2 Re 8 f 0.00000 0.13450 0.06200 1.00000
```

$Re_3B$: AB3_oC16_63_c_cf - POSCAR

```
AB3_oC16_63_c_cf & a,b/a,c/a,y1,y2,y3,z3 --params=2.89,3.22249134948,
    ↪ 2.51141868512,0.744,0.4262,0.1345,0.062 & Cmcm D_{2h}^{17} #63
    ↪ (c^2f) & oC16 & None & BRe3 & BRe3 & B. Aronsson and M. B\"{a}
    ↪ ckman and S. Rundqvist, Acta Chem. Scand. 14, 1001-1005 (1960)
   1.00000000000000
   1.44500000000000  -4.65650000000000   0.00000000000000
   1.44500000000000   4.65650000000000   0.00000000000000
   0.00000000000000   0.00000000000000   7.25800000000000
     B      Re
     2       6
Direct
  -0.74400000000000   0.74400000000000   0.25000000000000   B    (4c)
   0.74400000000000  -0.74400000000000   0.75000000000000   B    (4c)
  -0.42620000000000   0.42620000000000   0.25000000000000   Re   (4c)
   0.42620000000000  -0.42620000000000   0.75000000000000   Re   (4c)
  -0.13450000000000   0.13450000000000   0.06200000000000   Re   (8f)
   0.13450000000000  -0.13450000000000   0.56200000000000   Re   (8f)
  -0.13450000000000   0.13450000000000   0.43800000000000   Re   (8f)
   0.13450000000000  -0.13450000000000  -0.06200000000000   Re   (8f)
```

$Ta_2NiS_5$: AB5C2_oC32_63_c_c2f_f - CIF

```
# CIF file
data_findsym-output
_audit_creation_method FINDSYM

_chemical_name_mineral 'NiS5Ta2'
_chemical_formula_sum 'Ni S5 Ta2'

loop_
_publ_author_name
 'S. A. Sunshine'
 'J. A. Ibers'
_journal_name_full_name
;
 Inorganic Chemistry
;
_journal_volume 24
_journal_year 1985
_journal_page_first 3611
_journal_page_last 3614
_publ_Section_title
;
 Structure and physical properties of the new layered ternary
    ↪ chalcogenides tantalum nickel sulfide (Ta$_{2}$NiS$_{5}$) and
    ↪ tantalum nickel selenide (Ta$_{2}$NiSe$_{5}$)
;

# Found in Physical and structural properties of the new layered
    ↪ compounds Ta$_{2}$NiS$_{5}$ and Ta$_{2}$NiSe$_{5}$, 1986

_aflow_title 'Ta$_{2}$NiS$_{5}$ Structure'
_aflow_proto 'AB5C2_oC32_63_c_c2f_f'
_aflow_params 'a,b/a,c/a,y_{1},y_{2},y_{3},z_{3},y_{4},z_{4},y_{5},z_{5
    ↪ }'
_aflow_params_values '3.415,3.55666471449,4.42079062958,0.69692,0.31968,
    ↪ 0.58282,0.13527,0.1485,-0.05026,0.22082,0.10879'
_aflow_Strukturbericht 'None'
_aflow_Pearson 'oC32'

_symmetry_space_group_name_H-M "C 2/m 2/c 21/m"
_symmetry_Int_Tables_number 63

_cell_length_a    3.41500
_cell_length_b    12.14601
_cell_length_c    15.09700
_cell_angle_alpha 90.00000
_cell_angle_beta  90.00000
_cell_angle_gamma 90.00000

loop_
_space_group_symop_id
_space_group_symop_operation_xyz
1 x,y,z
2 x,-y,-z
3 -x,y,-z+1/2
4 -x,-y,z+1/2
5 -x,-y,-z
6 -x,y,z
7 x,-y,z+1/2
8 x,y,-z+1/2
9 x+1/2,y+1/2,z
10 x+1/2,-y+1/2,-z
11 -x+1/2,y+1/2,-z+1/2
12 -x+1/2,-y+1/2,z+1/2
13 -x+1/2,-y+1/2,-z
14 -x+1/2,y+1/2,z
15 x+1/2,-y+1/2,z+1/2
16 x+1/2,y+1/2,-z+1/2

loop_
_atom_site_label
_atom_site_type_symbol
_atom_site_symmetry_multiplicity
_atom_site_Wyckoff_label
_atom_site_fract_x
_atom_site_fract_y
_atom_site_fract_z
_atom_site_occupancy
Ni1 Ni 4 c 0.00000 0.69692 0.25000 1.00000
S1 S 4 c 0.00000 0.31968 0.25000 1.00000
S2 S 8 f 0.00000 0.58282 0.13527 1.00000
S3 S 8 f 0.00000 0.14850 -0.05026 1.00000
Ta1 Ta 8 f 0.00000 0.22082 0.10879 1.00000
```

$Ta_2NiS_5$: AB5C2_oC32_63_c_c2f_f - POSCAR

```
AB5C2_oC32_63_c_c2f_f & a,b/a,c/a,y1,y2,y3,z3,y4,z4,y5,z5 --params=3.415
    ↪ ,3.55666471449,4.42079062958,0.69692,0.31968,0.58282,0.13527,
    ↪ 0.1485,-0.05026,0.22082,0.10879 & Cmcm D_{2h}^{17} #63 (c^2f^3)
    ↪ & oC32 & None & NiS5Ta2 & NiS5Ta2 & S. A. Sunshine and J. A.
    ↪ Ibers, Inorg. Chem. 24, 3611-3614 (1985)
   1.00000000000000
   1.70750000000000  -6.07300500000000   0.00000000000000
   1.70750000000000   6.07300500000000   0.00000000000000
   0.00000000000000   0.00000000000000  15.09700000000000
    Ni      S      Ta
     2     10       4
Direct
  -0.69692000000000   0.69692000000000   0.25000000000000   Ni   (4c)
   0.69692000000000  -0.69692000000000   0.75000000000000   Ni   (4c)
  -0.31968000000000   0.31968000000000   0.25000000000000   S    (4c)
   0.31968000000000  -0.31968000000000   0.75000000000000   S    (4c)
  -0.58282000000000   0.58282000000000   0.13527000000000   S    (8f)
   0.58282000000000  -0.58282000000000   0.63527000000000   S    (8f)
  -0.58282000000000   0.58282000000000   0.36473000000000   S    (8f)
   0.58282000000000  -0.58282000000000  -0.13527000000000   S    (8f)
  -0.14850000000000   0.14850000000000  -0.05026000000000   S    (8f)
   0.14850000000000  -0.14850000000000   0.44974000000000   S    (8f)
  -0.14850000000000   0.14850000000000   0.55026000000000   S    (8f)
   0.14850000000000  -0.14850000000000   0.05026000000000   S    (8f)
  -0.22082000000000   0.22082000000000   0.10879000000000   Ta   (8f)
   0.22082000000000  -0.22082000000000   0.60879000000000   Ta   (8f)
  -0.22082000000000   0.22082000000000   0.39121000000000   Ta   (8f)
   0.22082000000000  -0.22082000000000  -0.10879000000000   Ta   (8f)
```

$V_3AsC$: ABC3_oC20_63_c_b_cf - CIF

```
# CIF file
data_findsym-output
_audit_creation_method FINDSYM

_chemical_name_mineral 'AsCV3'
_chemical_formula_sum 'As C V3'

loop_
_publ_author_name
 'H. Boller'
 'H. Nowotny'
_journal_name_full_name
;
 Monatshefte f{\"u}r Chemie und verwandte Teile anderer Wissenschaften
;
_journal_volume 98
_journal_year 1967
_journal_page_first 2127
_journal_page_last 2132
_publ_Section_title
;
 Zum Dreistoff: Vanadin-Arsen-Kohlenstoff
;

# Found in Carbides with Filled Re$_{3}$B-Type Structure, 1994

_aflow_title 'V$_{3}$AsC Structure'
_aflow_proto 'ABC3_oC20_63_c_b_cf'
_aflow_params 'a,b/a,c/a,y_{2},y_{3},y_{4},z_{4}'
_aflow_params_values '3.128,3.24168797954,2.46131713555,0.756,0.452,
    ↪ 0.1283,0.052'
_aflow_Strukturbericht 'None'
_aflow_Pearson 'oC20'

_symmetry_space_group_name_H-M "C 2/m 2/c 21/m"
_symmetry_Int_Tables_number 63

_cell_length_a    3.12800
_cell_length_b    10.14000
_cell_length_c    7.69900
_cell_angle_alpha 90.00000
_cell_angle_beta  90.00000
_cell_angle_gamma 90.00000

loop_
_space_group_symop_id
_space_group_symop_operation_xyz
1 x,y,z
2 x,-y,-z
3 -x,y,-z+1/2
4 -x,-y,z+1/2
5 -x,-y,-z
6 -x,y,z
7 x,-y,z+1/2
8 x,y,-z+1/2
9 x+1/2,y+1/2,z
10 x+1/2,-y+1/2,-z
11 -x+1/2,y+1/2,-z+1/2
12 -x+1/2,-y+1/2,z+1/2
13 -x+1/2,-y+1/2,-z
14 -x+1/2,y+1/2,z
```

```
15 x+1/2,-y+1/2,z+1/2
16 x+1/2,y+1/2,-z+1/2

loop_
_atom_site_label
_atom_site_type_symbol
_atom_site_symmetry_multiplicity
_atom_site_Wyckoff_label
_atom_site_fract_x
_atom_site_fract_y
_atom_site_fract_z
_atom_site_occupancy
C1 C 4 b 0.00000 0.50000 0.00000 1.00000
As1 As 4 c 0.00000 0.75600 0.25000 1.00000
V1 V 4 c 0.00000 0.45200 0.25000 1.00000
V2 V 8 f 0.00000 0.12830 0.05200 1.00000
```

$V_3AsC$: ABC3_oC20_63_c_b_cf - POSCAR

```
ABC3_oC20_63_c_b_cf & a,b/a,c/a,y2,y3,y4,z4 --params=3.128,3.24168797954
    ↪ ,2.46131713555,0.756,0.452,0.1283,0.052 & Cmcm D_{2h}^{17} #63
    ↪ (bc^2f) & oC20 & None & AsCV3 & AsCV3 & H. Boller and H.
    ↪ Nowotny, Monatsh. Chem. Verw. Teile Anderer Wiss. 98, 2127-2132
    ↪ (1967)
   1.00000000000000
   1.56400000000000  -5.07000000000000   0.00000000000000
   1.56400000000000   5.07000000000000   0.00000000000000
   0.00000000000000   0.00000000000000   7.69900000000000
    As    C    V
     2    2    6
Direct
  -0.75600000000000   0.75600000000000   0.25000000000000   As   (4c)
   0.75600000000000  -0.75600000000000   0.75000000000000   As   (4c)
   0.50000000000000   0.50000000000000   0.00000000000000   C    (4b)
   0.50000000000000   0.50000000000000   0.50000000000000   C    (4b)
  -0.45200000000000   0.45200000000000   0.25000000000000   V    (4c)
   0.45200000000000  -0.45200000000000   0.75000000000000   V    (4c)
  -0.12830000000000   0.12830000000000   0.05200000000000   V    (8f)
   0.12830000000000  -0.12830000000000   0.55200000000000   V    (8f)
  -0.12830000000000   0.12830000000000   0.44800000000000   V    (8f)
   0.12830000000000  -0.12830000000000  -0.05200000000000   V    (8f)
```

$Si_{24}$ Clathrate: A_oC24_63_3f - CIF

```
# CIF file
data_findsym-output
_audit_creation_method FINDSYM

_chemical_name_mineral 'Clathrate'
_chemical_formula_sum 'Si'

loop_
_publ_author_name
 'D. Y. Kim'
 'S. Stefanoski'
 'O. O. Kurakevych'
 'T. A. Strobel'
_journal_name_full_name
;
 Nature Materials
;
_journal_volume 14
_journal_year 2015
_journal_page_first 169
_journal_page_last 173
_publ_Section_title
;
 Synthesis of an open-framework allotrope of silicon
;

_aflow_title 'Si$_{24}$ Clathrate Structure'
_aflow_proto 'A_oC24_63_3f'
_aflow_params 'a,b/a,c/a,y_{1},z_{1},y_{2},z_{2},y_{3},z_{3}'
_aflow_params_values '3.82236,2.79950083195,3.30314256114,0.0284,0.5903,
    ↪ 0.2435,0.5551,0.5705,0.3412'
_aflow_Strukturbericht 'None'
_aflow_Pearson 'oC24'

_symmetry_space_group_name_H-M "C 2/m 2/c 21/m"
_symmetry_Int_Tables_number 63

_cell_length_a    3.82236
_cell_length_b    10.70070
_cell_length_c    12.62580
_cell_angle_alpha 90.00000
_cell_angle_beta  90.00000
_cell_angle_gamma 90.00000

loop_
_space_group_symop_id
_space_group_symop_operation_xyz
1 x,y,z
2 x,-y,-z
3 -x,y,-z+1/2
4 -x,-y,z+1/2
5 -x,-y,-z
6 -x,y,z
7 x,-y,z+1/2
8 x,y,-z+1/2
9 x+1/2,y+1/2,z
10 x+1/2,-y+1/2,-z
11 -x+1/2,y+1/2,-z+1/2
12 -x+1/2,-y+1/2,z+1/2
13 -x+1/2,-y+1/2,-z
14 -x+1/2,y+1/2,z
```

```
15 x+1/2,-y+1/2,z+1/2
16 x+1/2,y+1/2,-z+1/2

loop_
_atom_site_label
_atom_site_type_symbol
_atom_site_symmetry_multiplicity
_atom_site_Wyckoff_label
_atom_site_fract_x
_atom_site_fract_y
_atom_site_fract_z
_atom_site_occupancy
Si1 Si 8 f 0.00000 0.02840 0.59030 1.00000
Si2 Si 8 f 0.00000 0.24350 0.55510 1.00000
Si3 Si 8 f 0.00000 0.57050 0.34120 1.00000
```

$Si_{24}$ Clathrate: A_oC24_63_3f - POSCAR

```
A_oC24_63_3f & a,b/a,c/a,y1,z1,y2,z2,y3,z3 --params=3.82236,
    ↪ 2.79950083195,3.30314256114,0.0284,0.5903,0.2435,0.5551,0.5705,
    ↪ 0.3412 & Cmcm D_{2h}^{17} #63 (f^3) & oC24 & None & Si24 &
    ↪ Clathrate & D. Y. Kim et al., Nat. Mater. 14, 169-173 (2015)
   1.00000000000000
   1.91118000000000  -5.35035000000000   0.00000000000000
   1.91118000000000   5.35035000000000   0.00000000000000
   0.00000000000000   0.00000000000000  12.62580000000000
    Si
    12
Direct
  -0.02840000000000   0.02840000000000   0.59030000000000   Si   (8f)
   0.02840000000000  -0.02840000000000   1.09030000000000   Si   (8f)
  -0.02840000000000   0.02840000000000  -0.09030000000000   Si   (8f)
   0.02840000000000  -0.02840000000000  -0.59030000000000   Si   (8f)
  -0.24350000000000   0.24350000000000   0.55510000000000   Si   (8f)
   0.24350000000000  -0.24350000000000   1.05510000000000   Si   (8f)
  -0.24350000000000   0.24350000000000  -0.05510000000000   Si   (8f)
   0.24350000000000  -0.24350000000000  -0.55510000000000   Si   (8f)
  -0.57050000000000   0.57050000000000   0.34120000000000   Si   (8f)
   0.57050000000000  -0.57050000000000   0.84120000000000   Si   (8f)
  -0.57050000000000   0.57050000000000   0.15880000000000   Si   (8f)
   0.57050000000000  -0.57050000000000  -0.34120000000000   Si   (8f)
```

Base-centered orthorhombic $Sr_4Ru_3O_{10}$: A10B3C4_oC68_64_2dfg_ad_2d - CIF

```
# CIF file
data_findsym-output
_audit_creation_method FINDSYM

_chemical_name_mineral 'O10Ru3Sr4'
_chemical_formula_sum 'O10 Ru3 Sr4'

loop_
_publ_author_name
 'M. K. Crawford'
 'R. L. Harlow'
 'W. Marshall'
 'Z. Li'
 'G. Cao'
 'R. L. Lindstrom'
 'Q. Huang'
 'J. W. Lynn'
_journal_name_full_name
;
 Physical Review B
;
_journal_volume 65
_journal_year 2002
_journal_page_first 214412
_journal_page_last 214412
_publ_Section_title
;
 Structure and magnetism of single crystal Sr$_{4}$Ru$_{3}$O$_{10}$: A
    ↪ ferromagnetic triple-layer ruthenate
;

_aflow_title 'Base-centered orthorhombic Sr$_{4}$Ru$_{3}$O$_{10}$
    ↪ Structure'
_aflow_proto 'A10B3C4_oC68_64_2dfg_ad_2d'
_aflow_params 'a,b/a,c/a,x_{2},x_{3},x_{4},x_{5},x_{6},y_{7},z_{7},x_{8
    ↪ },y_{8},z_{8}'
_aflow_params_values '28.573,0.13649599272,0.13649599272,-0.0695,0.787,
    ↪ 0.8598,0.7039,0.5699,0.7971,0.2972,0.6392,0.2271,0.2279'
_aflow_Strukturbericht 'None'
_aflow_Pearson 'oC68'

_symmetry_space_group_name_H-M "C 2/m 2/c 21/a"
_symmetry_Int_Tables_number 64

_cell_length_a    28.57300
_cell_length_b    3.90010
_cell_length_c    3.90010
_cell_angle_alpha 90.00000
_cell_angle_beta  90.00000
_cell_angle_gamma 90.00000

loop_
_space_group_symop_id
_space_group_symop_operation_xyz
1 x,y,z
2 x,-y,-z
3 -x+1/2,y,-z+1/2
4 -x+1/2,-y,z+1/2
5 -x,-y,-z
6 -x,y,z
7 x+1/2,-y,z+1/2
```

```
8 x+1/2,y,-z+1/2
9 x+1/2,y+1/2,z
10 x+1/2,-y+1/2,-z
11 -x,y+1/2,-z+1/2
12 -x,-y+1/2,z+1/2
13 -x+1/2,-y+1/2,-z
14 -x+1/2,y+1/2,z
15 x,-y+1/2,z+1/2
16 x,y+1/2,-z+1/2

loop_
_atom_site_label
_atom_site_type_symbol
_atom_site_symmetry_multiplicity
_atom_site_Wyckoff_label
_atom_site_fract_x
_atom_site_fract_y
_atom_site_fract_z
_atom_site_occupancy
Ru1 Ru 4 a 0.00000 0.00000 0.00000 1.00000
O1 O 8 d -0.06950 0.00000 0.00000 1.00000
O2 O 8 d 0.78700 0.00000 0.00000 1.00000
Ru2 Ru 8 d 0.85980 0.00000 0.00000 1.00000
Sr1 Sr 8 d 0.70390 0.00000 0.00000 1.00000
Sr2 Sr 8 d 0.56990 0.00000 0.00000 1.00000
O3 O 8 f 0.00000 0.79710 0.29720 1.00000
O4 O 16 g 0.63920 0.22710 0.22790 1.00000
```

Base-centered orthorhombic $Sr_4Ru_3O_{10}$: A10B3C4_oC68_64_2dfg_ad_2d - POSCAR

```
A10B3C4_oC68_64_2dfg_ad_2d & a,b/a,c/a,x2,x3,x4,x5,x6,y7,z7,x8,y8,z8 --
    ↪ params=28.573,0.13649599272,0.13649599272,-0.0695,0.787,0.8598,
    ↪ 0.7039,0.5699,0.7971,0.2972,0.6392,0.2271,0.2279 & Cmca D_{2h
    ↪ }^{18} #64 (ad^5fg) & oC68 & None & O10Ru3Sr4 & O10Ru3Sr4 & M.
    ↪ K. Crawford et al., Phys. Rev. B 65, 214412(2002)
   1.00000000000000
  14.28650000000000  -1.95005000000000   0.00000000000000
  14.28650000000000   1.95005000000000   0.00000000000000
   0.00000000000000   0.00000000000000   3.90010000000000
     O      Ru     Sr
    20       6      8
Direct
  -0.06950000000000  -0.06950000000000   0.00000000000000   O   (8d)
   0.56950000000000   0.56950000000000   0.50000000000000   O   (8d)
   0.06950000000000   0.06950000000000   0.00000000000000   O   (8d)
   0.43050000000000   0.43050000000000   0.50000000000000   O   (8d)
   0.78700000000000   0.78700000000000   0.00000000000000   O   (8d)
  -0.28700000000000  -0.28700000000000   0.50000000000000   O   (8d)
  -0.78700000000000  -0.78700000000000   0.00000000000000   O   (8d)
   1.28700000000000   1.28700000000000   0.50000000000000   O   (8d)
  -0.79710000000000   0.79710000000000   0.29720000000000   O   (8f)
   1.29710000000000  -0.29710000000000   0.79720000000000   O   (8f)
  -0.29710000000000   1.29710000000000   0.20280000000000   O   (8f)
   0.79710000000000  -0.79710000000000  -0.29720000000000   O   (8f)
   0.41210000000000   0.86630000000000   0.22790000000000   O  (16g)
   0.08790000000000  -0.36630000000000   0.72790000000000   O  (16g)
  -0.36630000000000   0.08790000000000   0.27210000000000   O  (16g)
   0.86630000000000   0.41210000000000  -0.22790000000000   O  (16g)
  -0.41210000000000  -0.86630000000000  -0.22790000000000   O  (16g)
   0.91210000000000   1.36630000000000   0.27210000000000   O  (16g)
   1.36630000000000   0.91210000000000   0.72790000000000   O  (16g)
  -0.86630000000000  -0.41210000000000   0.22790000000000   O  (16g)
   0.00000000000000   0.00000000000000   0.00000000000000   Ru  (4a)
   0.50000000000000   0.50000000000000   0.50000000000000   Ru  (4a)
   0.85980000000000   0.85980000000000   0.00000000000000   Ru  (8d)
  -0.35980000000000  -0.35980000000000   0.50000000000000   Ru  (8d)
  -0.85980000000000  -0.85980000000000   0.00000000000000   Ru  (8d)
   1.35980000000000   1.35980000000000   0.50000000000000   Ru  (8d)
   0.70390000000000   0.70390000000000   0.00000000000000   Sr  (8d)
  -0.20390000000000  -0.20390000000000   0.50000000000000   Sr  (8d)
  -0.70390000000000  -0.70390000000000   0.00000000000000   Sr  (8d)
   1.20390000000000   1.20390000000000   0.50000000000000   Sr  (8d)
   0.56990000000000   0.56990000000000   0.00000000000000   Sr  (8d)
  -0.06990000000000  -0.06990000000000   0.50000000000000   Sr  (8d)
  -0.56990000000000  -0.56990000000000   0.00000000000000   Sr  (8d)
   1.06990000000000   1.06990000000000   0.50000000000000   Sr  (8d)
```

$Na_2Mo_2O_7$: A2B2C7_oC88_64_ef_df_3f2g - CIF

```
# CIF file
data_findsym-output
_audit_creation_method FINDSYM

_chemical_name_mineral 'Mo2Na2O7'
_chemical_formula_sum 'Mo2 Na2 O7'

loop_
_publ_author_name
 'I. Lindqvist'
_journal_name_full_name
;
 Acta Chemica Scandinavica
;
_journal_volume 4
_journal_year 1950
_journal_page_first 1066
_journal_page_last 1074
_publ_Section_title
;
 Crystal Structure Studies on Anhydrous Sodium Molybdates and Tungstates
;

_aflow_title 'Na$_{2}$Mo$_{2}$O$_{7}$ Structure'
_aflow_proto 'A2B2C7_oC88_64_ef_df_3f2g'
_aflow_params 'a,b/a,c/a,x_{1},y_{2},y_{3},z_{3},y_{4},z_{4},y_{5},z_{5
    ↪ },y_{6},z_{6},y_{7},z_{7},x_{8},y_{8},z_{8},x_{9},y_{9},z_{9}'
_aflow_params_values '7.17,1.64993026499,2.05020920502,0.25,0.08,0.25,
    ↪ 0.08,0.36,0.3,0.08,0.27,0.41,0.13,0.35,0.49,0.21,0.21,0.16,0.21
    ↪ ,0.49,0.34'
_aflow_Strukturbericht 'None'
_aflow_Pearson 'oC88'

_symmetry_space_group_name_H-M "C 2/m 2/c 21/a"
_symmetry_Int_Tables_number 64

_cell_length_a     7.17000
_cell_length_b     11.83000
_cell_length_c     14.70000
_cell_angle_alpha  90.00000
_cell_angle_beta   90.00000
_cell_angle_gamma  90.00000

loop_
_space_group_symop_id
_space_group_symop_operation_xyz
1 x,y,z
2 x,-y,-z
3 -x+1/2,y,-z+1/2
4 -x+1/2,-y,z+1/2
5 -x,-y,-z
6 -x,y,z
7 x+1/2,-y,z+1/2
8 x+1/2,y,-z+1/2
9 x+1/2,y+1/2,z
10 x+1/2,-y+1/2,-z
11 -x,y+1/2,-z+1/2
12 -x,-y+1/2,z+1/2
13 -x+1/2,-y+1/2,-z
14 -x+1/2,y+1/2,z
15 x,-y+1/2,z+1/2
16 x,y+1/2,-z+1/2

loop_
_atom_site_label
_atom_site_type_symbol
_atom_site_symmetry_multiplicity
_atom_site_Wyckoff_label
_atom_site_fract_x
_atom_site_fract_y
_atom_site_fract_z
_atom_site_occupancy
Na1 Na 8 d 0.25000 0.00000 0.00000 1.00000
Mo1 Mo 8 e 0.25000 0.08000 0.25000 1.00000
Mo2 Mo 8 f 0.00000 0.25000 0.08000 1.00000
Na2 Na 8 f 0.00000 0.36000 0.30000 1.00000
O1 O 8 f 0.00000 0.08000 0.27000 1.00000
O2 O 8 f 0.00000 0.41000 0.13000 1.00000
O3 O 8 f 0.00000 0.35000 0.49000 1.00000
O4 O 16 g 0.21000 0.21000 0.16000 1.00000
O5 O 16 g 0.21000 0.49000 0.34000 1.00000
```

$Na_2Mo_2O_7$: A2B2C7_oC88_64_ef_df_3f2g - POSCAR

```
A2B2C7_oC88_64_ef_df_3f2g & a,b/a,c/a,x1,y2,y3,z3,y4,z4,y5,z5,y6,z6,y7,
    ↪ z7,x8,y8,z8,x9,y9,z9 --params=7.17,1.64993026499,2.05020920502,
    ↪ 0.25,0.08,0.25,0.08,0.36,0.3,0.08,0.27,0.41,0.13,0.35,0.49,0.21
    ↪ ,0.21,0.16,0.21,0.49,0.34 & Cmca D_{2h}^{18} #64 (def^5g^2) &
    ↪ oC88 & None & Mo2Na2O7 & Mo2Na2O7 & I. Lindqvist, Acta Chem.
    ↪ Scand. 4, 1066-1074 (1950)
   1.00000000000000
   3.58500000000000  -5.91500000000000   0.00000000000000
   3.58500000000000   5.91500000000000   0.00000000000000
   0.00000000000000   0.00000000000000  14.70000000000000
     Mo      Na      O
      8       8     28
Direct
   0.17000000000000   0.33000000000000   0.25000000000000   Mo  (8e)
   0.33000000000000   0.17000000000000   0.75000000000000   Mo  (8e)
   0.83000000000000   0.67000000000000   0.75000000000000   Mo  (8e)
   0.67000000000000   0.83000000000000   0.25000000000000   Mo  (8e)
  -0.25000000000000   0.25000000000000   0.08000000000000   Mo  (8f)
   0.75000000000000   0.25000000000000   0.58000000000000   Mo  (8f)
   0.25000000000000   0.75000000000000   0.42000000000000   Mo  (8f)
   0.25000000000000  -0.25000000000000  -0.08000000000000   Mo  (8f)
   0.25000000000000   0.25000000000000   0.00000000000000   Na  (8d)
   0.25000000000000   0.25000000000000   0.50000000000000   Na  (8d)
  -0.25000000000000  -0.25000000000000   0.00000000000000   Na  (8d)
   0.75000000000000   0.75000000000000   0.50000000000000   Na  (8d)
  -0.36000000000000   0.36000000000000   0.30000000000000   Na  (8f)
   0.86000000000000   0.14000000000000   0.80000000000000   Na  (8f)
   0.14000000000000   0.86000000000000   0.20000000000000   Na  (8f)
   0.36000000000000  -0.36000000000000  -0.30000000000000   Na  (8f)
  -0.08000000000000   0.08000000000000   0.27000000000000   O   (8f)
   0.58000000000000   0.42000000000000   0.77000000000000   O   (8f)
   0.42000000000000   0.58000000000000   0.23000000000000   O   (8f)
   0.08000000000000  -0.08000000000000  -0.27000000000000   O   (8f)
  -0.41000000000000   0.41000000000000   0.13000000000000   O   (8f)
   0.91000000000000   0.09000000000000   0.63000000000000   O   (8f)
   0.09000000000000   0.91000000000000   0.37000000000000   O   (8f)
   0.41000000000000  -0.41000000000000  -0.13000000000000   O   (8f)
  -0.35000000000000   0.35000000000000   0.49000000000000   O   (8f)
   0.85000000000000   0.15000000000000   0.99000000000000   O   (8f)
   0.15000000000000   0.85000000000000   0.01000000000000   O   (8f)
   0.35000000000000  -0.35000000000000  -0.49000000000000   O   (8f)
   0.00000000000000   0.42000000000000   0.16000000000000   O  (16g)
   0.50000000000000   0.08000000000000   0.66000000000000   O  (16g)
   0.08000000000000   0.50000000000000   0.34000000000000   O  (16g)
   0.42000000000000   0.00000000000000  -0.16000000000000   O  (16g)
   0.00000000000000  -0.42000000000000  -0.16000000000000   O  (16g)
```

```
   0.50000000000000    0.92000000000000    0.34000000000000    O   (16g)
   0.92000000000000    0.50000000000000    0.66000000000000    O   (16g)
  -0.42000000000000    0.00000000000000    0.16000000000000    O   (16g)
  -0.28000000000000    0.70000000000000    0.34000000000000    O   (16g)
   0.78000000000000   -0.20000000000000    0.84000000000000    O   (16g)
  -0.20000000000000    0.78000000000000    0.16000000000000    O   (16g)
   0.70000000000000   -0.28000000000000   -0.34000000000000    O   (16g)
   0.28000000000000   -0.70000000000000   -0.34000000000000    O   (16g)
   0.22000000000000    1.20000000000000    0.16000000000000    O   (16g)
   1.20000000000000    0.22000000000000    0.84000000000000    O   (16g)
  -0.70000000000000    0.28000000000000    0.34000000000000    O   (16g)
```

$Li_2PrO_3$: A2B3C_oC12_65_h_bh_a - CIF

```
# CIF file
data_findsym-output
_audit_creation_method FINDSYM

_chemical_name_mineral 'Li2O3Pr'
_chemical_formula_sum 'Li2 O3 Pr'

loop_
_publ_author_name
 'Y. Hinatsu'
 'Y. Doi'
_journal_name_full_name
;
 Journal of Alloys and Compounds
;
_journal_volume 418
_journal_year 2006
_journal_page_first 155
_journal_page_last 160
_publ_Section_title
;
 Crystal structures and magnetic properties of alkali-metal lanthanide
      ↪ oxides $A_{2}$$Ln$O$_{3}$ ($A$ = Li, Na; $Ln$ = Ce, Pr, Tb)
;

_aflow_title 'Li$_{2}$PrO$_{3}$ Structure'
_aflow_proto 'A2B3C_oC12_65_h_bh_a'
_aflow_params 'a,b/a,c/a,x_{3},x_{4}'
_aflow_params_values '9.5198,0.467551839324,0.36581650875,0.6556,0.8557'
_aflow_Strukturbericht 'None'
_aflow_Pearson 'oC12'

_symmetry_space_group_name_H-M "C 2/m 2/m 2/m"
_symmetry_Int_Tables_number 65

_cell_length_a      9.51980
_cell_length_b      4.45100
_cell_length_c      3.48250
_cell_angle_alpha  90.00000
_cell_angle_beta   90.00000
_cell_angle_gamma  90.00000

loop_
_space_group_symop_id
_space_group_symop_operation_xyz
1 x,y,z
2 x,-y,-z
3 -x,y,-z
4 -x,-y,z
5 -x,-y,-z
6 -x,y,z
7 x,-y,z
8 x,y,-z
9 x+1/2,y+1/2,z
10 x+1/2,-y+1/2,-z
11 -x+1/2,y+1/2,-z
12 -x+1/2,-y+1/2,z
13 -x+1/2,-y+1/2,-z
14 -x+1/2,y+1/2,z
15 x+1/2,-y+1/2,z
16 x+1/2,y+1/2,-z

loop_
_atom_site_label
_atom_site_type_symbol
_atom_site_symmetry_multiplicity
_atom_site_Wyckoff_label
_atom_site_fract_x
_atom_site_fract_y
_atom_site_fract_z
_atom_site_occupancy
Pr1 Pr 2 a 0.00000 0.00000 0.00000 1.00000
O1 O 2 b 0.50000 0.00000 0.00000 1.00000
Li1 Li 4 h 0.65560 0.00000 0.50000 1.00000
O2 O 4 h 0.85570 0.00000 0.50000 1.00000
```

$Li_2PrO_3$: A2B3C_oC12_65_h_bh_a - POSCAR

```
A2B3C_oC12_65_h_bh_a & a,b/a,c/a,x3,x4 --params=9.5198,0.467551839324,
     ↪ 0.36581650875,0.6556,0.8557 & Cmmm D_{2h}^{19} #65 (abh^2) &
     ↪ oC12 & None & Li2O3Pr & Li2O3Pr & Y. Hinatsu and Y. Doi, J.
     ↪ Alloys Compd. 418, 155-160 (2006)
   1.00000000000000
   4.75990000000000   -2.22550000000000    0.00000000000000
   4.75990000000000    2.22550000000000    0.00000000000000
   0.00000000000000    0.00000000000000    3.48250000000000
    Li     O    Pr
     2     3     1
Direct
   0.65560000000000    0.65560000000000    0.50000000000000    Li   (4h)
  -0.65560000000000   -0.65560000000000    0.50000000000000    Li   (4h)
   0.50000000000000    0.50000000000000    0.00000000000000    O    (2b)
   0.85570000000000    0.85570000000000    0.50000000000000    O    (4h)
  -0.85570000000000   -0.85570000000000    0.50000000000000    O    (4h)
   0.00000000000000    0.00000000000000    0.00000000000000    Pr   (2a)
```

$Mg(NH_3)_2Cl_2$ ($E1_3$): A2B8CD2_oC26_65_h_r_a_i - CIF

```
# CIF file
data_findsym-output
_audit_creation_method FINDSYM

_chemical_name_mineral 'Cl2H6MgN2'
_chemical_formula_sum 'Cl2 H8 Mg N2'

loop_
_publ_author_name
 'A. Leineweber'
 'M. W. Friedriszik'
 'H. Jacobs'
_journal_name_full_name
;
 Journal of Solid State Chemistry
;
_journal_volume 147
_journal_year 1999
_journal_page_first 229
_journal_page_last 234
_publ_Section_title
;
 Preparation and Crystal Structures of Mg(NH$_{3}$)$_{2}$Cl$_{2}$, Mg(
      ↪ NH$_{3}$)$_{2}$Br$_{2}$, and Mg(NH$_{3}$)$_{2}$I$_{2}$
;

_aflow_title 'Mg(NH$_{3}$)$_{2}$Cl$_{2}$ ($E1_{3}$) Structure'
_aflow_proto 'A2B8CD2_oC26_65_h_r_a_i'
_aflow_params 'a,b/a,c/a,x_{2},y_{3},x_{4},y_{4},z_{4}'
_aflow_params_values '8.18099,1.00314142909,0.458990904524,0.2133,0.2595
      ↪ ,0.045,0.312,0.158'
_aflow_Strukturbericht '$E1_{3}$'
_aflow_Pearson 'oC26'

_symmetry_space_group_name_H-M "C 2/m 2/m 2/m"
_symmetry_Int_Tables_number 65

_cell_length_a      8.18099
_cell_length_b      8.20669
_cell_length_c      3.75500
_cell_angle_alpha  90.00000
_cell_angle_beta   90.00000
_cell_angle_gamma  90.00000

loop_
_space_group_symop_id
_space_group_symop_operation_xyz
1 x,y,z
2 x,-y,-z
3 -x,y,-z
4 -x,-y,z
5 -x,-y,-z
6 -x,y,z
7 x,-y,z
8 x,y,-z
9 x+1/2,y+1/2,z
10 x+1/2,-y+1/2,-z
11 -x+1/2,y+1/2,-z
12 -x+1/2,-y+1/2,z
13 -x+1/2,-y+1/2,-z
14 -x+1/2,y+1/2,z
15 x+1/2,-y+1/2,z
16 x+1/2,y+1/2,-z

loop_
_atom_site_label
_atom_site_type_symbol
_atom_site_symmetry_multiplicity
_atom_site_Wyckoff_label
_atom_site_fract_x
_atom_site_fract_y
_atom_site_fract_z
_atom_site_occupancy
Mg1 Mg 2 a 0.00000 0.00000 0.00000 1.00000
Cl1 Cl 4 h 0.21330 0.00000 0.50000 1.00000
N1 N 4 i 0.00000 0.25950 0.00000 1.00000
H1 H 16 r 0.04500 0.31200 0.15800 0.75000
```

$Mg(NH_3)_2Cl_2$ ($E1_3$): A2B8CD2_oC26_65_h_r_a_i - POSCAR

```
A2B8CD2_oC26_65_h_r_a_i & a,b/a,c/a,x2,y3,x4,y4,z4 --params=8.18099,
     ↪ 1.00314142909,0.458990904524,0.2133,0.2595,0.045,0.312,0.158 &
     ↪ Cmmm D_{2h}^{19} #65 (ahir) & oC26 & $E1_{3}$ & Cl2H6MgN2 &
     ↪ Cl2H6MgN2 & A. Leineweber and M. W. Friedriszik and H. Jacobs,
     ↪ J. Solid State Chem. 147, 229-234 (1999)
   1.00000000000000
   4.09049500000000   -4.10334500000000    0.00000000000000
   4.09049500000000    4.10334500000000    0.00000000000000
   0.00000000000000    0.00000000000000    3.75500000000000
    Cl     H    Mg     N
     2     8     1     2
Direct
   0.21330000000000    0.21330000000000    0.50000000000000    Cl   (4h)
  -0.21330000000000   -0.21330000000000    0.50000000000000    Cl   (4h)
  -0.26700000000000    0.35700000000000    0.15800000000000    H   (16r)
   0.26700000000000   -0.35700000000000    0.15800000000000    H   (16r)
  -0.35700000000000    0.26700000000000   -0.15800000000000    H   (16r)
   0.35700000000000   -0.26700000000000   -0.15800000000000    H   (16r)
```

```
   0.26700000000000   -0.35700000000000   -0.15800000000000    H   (16r)
  -0.26700000000000    0.35700000000000   -0.15800000000000    H   (16r)
   0.35700000000000   -0.26700000000000    0.15800000000000    H   (16r)
  -0.35700000000000    0.26700000000000    0.15800000000000    H   (16r)
   0.00000000000000    0.00000000000000    0.00000000000000   Mg   (2a)
  -0.25950000000000    0.25950000000000    0.00000000000000    N   (4i)
   0.25950000000000   -0.25950000000000    0.00000000000000    N   (4i)
```

$Nb_3O_7F$: A3B8_oC22_65_ag_bd2gh - CIF

```
# CIF file
data_findsym-output
_audit_creation_method FINDSYM

_chemical_name_mineral 'FNb3O7'
_chemical_formula_sum 'Nb3 O8'

loop_
_publ_author_name
 'S. Andersson'
_journal_name_full_name
;
 Acta Chemica Scandinavica
;
_journal_volume 18
_journal_year 1964
_journal_page_first 2339
_journal_page_last 2344
_publ_Section_title
;
 The Crystal Structure of Nb$_{3}$O$_{7}$F
;

_aflow_title 'Nb$_{3}$O$_{7}$F Structure'
_aflow_proto 'A3B8_oC22_65_ag_bd2gh'
_aflow_params 'a,b/a,c/a,x_{4},x_{5},x_{6},x_{7}'
_aflow_params_values '20.66999,0.185437922321,0.189985578126,0.1836,
    ↪ 0.094,0.71,0.189'
_aflow_Strukturbericht 'None'
_aflow_Pearson 'oC22'

_symmetry_space_group_name_H-M "C 2/m 2/m 2/m"
_symmetry_Int_Tables_number 65

_cell_length_a    20.66999
_cell_length_b    3.83300
_cell_length_c    3.92700
_cell_angle_alpha 90.00000
_cell_angle_beta  90.00000
_cell_angle_gamma 90.00000

loop_
_space_group_symop_id
_space_group_symop_operation_xyz
1 x,y,z
2 x,-y,-z
3 -x,y,-z
4 -x,-y,z
5 -x,-y,-z
6 -x,y,z
7 x,-y,z
8 x,y,-z
9 x+1/2,y+1/2,z
10 x+1/2,-y+1/2,-z
11 -x+1/2,y+1/2,-z
12 -x+1/2,-y+1/2,z
13 -x+1/2,-y+1/2,-z
14 -x+1/2,y+1/2,z
15 x+1/2,-y+1/2,z
16 x+1/2,y+1/2,-z

loop_
_atom_site_label
_atom_site_type_symbol
_atom_site_symmetry_multiplicity
_atom_site_Wyckoff_label
_atom_site_fract_x
_atom_site_fract_y
_atom_site_fract_z
_atom_site_occupancy
Nb1 Nb 2 a 0.00000 0.00000 0.00000 1.00000
O1 O 2 b 0.50000 0.00000 0.00000 1.00000
O2 O 2 d 0.00000 0.00000 0.50000 1.00000
Nb2 Nb 4 g 0.18360 0.00000 0.00000 1.00000
O3 O 4 g 0.09400 0.00000 0.00000 1.00000
O4 O 4 g 0.71000 0.00000 0.00000 1.00000
O5 O 4 h 0.18900 0.00000 0.50000 1.00000
```

$Nb_3O_7F$: A3B8_oC22_65_ag_bd2gh - POSCAR

```
A3B8_oC22_65_ag_bd2gh & a,b/a,c/a,x4,x5,x6,x7 --params=20.66999,
    ↪ 0.185437922321,0.189985578126,0.1836,0.094,0.71,0.189 & Cmmm D_
    ↪ {2h}^{19} #65 (abdg^3h) & oC22 & None & FNb3O7 & FNb3O7 & S.
    ↪ Andersson, Acta Chem. Scand. 18, 2339-2344 (1964)
   1.00000000000000
  10.33499500000000   -1.91650000000000    0.00000000000000
  10.33499500000000    1.91650000000000    0.00000000000000
   0.00000000000000    0.00000000000000    3.92700000000000
    Nb      O
     3      8
Direct
   0.00000000000000    0.00000000000000    0.00000000000000   Nb   (2a)
   0.18360000000000    0.18360000000000    0.00000000000000   Nb   (4g)
  -0.18360000000000   -0.18360000000000    0.00000000000000   Nb   (4g)
   0.50000000000000    0.50000000000000    0.00000000000000    O   (2b)
```

```
   0.00000000000000    0.00000000000000    0.50000000000000    O   (2d)
   0.09400000000000    0.09400000000000    0.00000000000000    O   (4g)
  -0.09400000000000   -0.09400000000000    0.00000000000000    O   (4g)
   0.71000000000000    0.71000000000000    0.00000000000000    O   (4g)
  -0.71000000000000   -0.71000000000000    0.00000000000000    O   (4g)
   0.18900000000000    0.18900000000000    0.50000000000000    O   (4h)
  -0.18900000000000   -0.18900000000000    0.50000000000000    O   (4h)
```

$NH_4H_2PO_2$ ($F5_7$): A2BC2D_oC24_67_m_a_n_g - CIF

```
# CIF file
data_findsym-output
_audit_creation_method FINDSYM

_chemical_name_mineral 'H2(NH4)O2P'
_chemical_formula_sum 'H2 (NH4) O2 P'

loop_
_publ_author_name
 'W. H. Zachariasen'
 'R. C. L. Mooney'
_journal_name_full_name
;
 Journal of Chemical Physics
;
_journal_volume 2
_journal_year 1934
_journal_page_first 34
_journal_page_last 37
_publ_Section_title
;
 The Structure of the Hypophosphite Group as Determined from the Crystal
    ↪ Lattice of Ammonium Hypophosphite
;

# Found in Strukturbericht Band III 1933-1935, 1937

_aflow_title 'NH$_{4}$H$_{2}$PO$_{2}$ ($F5_{7}$) Structure'
_aflow_proto 'A2BC2D_oC24_67_m_a_n_g'
_aflow_params 'a,b/a,c/a,z_{2},y_{3},z_{3},x_{4},z_{4}'
_aflow_params_values '11.47,0.659982563208,0.346992153444,0.458,0.608,
    ↪ 0.806,0.386,0.348'
_aflow_Strukturbericht '$F5_{7}$'
_aflow_Pearson 'oC24'

_symmetry_space_group_name_H-M "C 2/m 2/m 2/a"
_symmetry_Int_Tables_number 67

_cell_length_a    11.47000
_cell_length_b    7.57000
_cell_length_c    3.98000
_cell_angle_alpha 90.00000
_cell_angle_beta  90.00000
_cell_angle_gamma 90.00000

loop_
_space_group_symop_id
_space_group_symop_operation_xyz
1 x,y,z
2 x,-y,-z
3 -x+1/2,y,-z
4 -x+1/2,-y,z
5 -x,-y,-z
6 -x,y,z
7 x+1/2,-y,z
8 x+1/2,y,-z
9 x+1/2,y+1/2,z
10 x+1/2,-y+1/2,-z
11 -x,y+1/2,-z
12 -x,-y+1/2,z
13 -x+1/2,-y+1/2,-z
14 -x+1/2,y+1/2,z
15 x,-y+1/2,z
16 x,y+1/2,-z

loop_
_atom_site_label
_atom_site_type_symbol
_atom_site_symmetry_multiplicity
_atom_site_Wyckoff_label
_atom_site_fract_x
_atom_site_fract_y
_atom_site_fract_z
_atom_site_occupancy
NH41 NH4 4 a 0.25000 0.00000 0.00000 1.00000
P1 P 4 g 0.00000 0.25000 0.45800 1.00000
H1 H 8 m 0.00000 0.60800 0.80600 1.00000
O1 O 8 n 0.38600 0.25000 0.34800 1.00000
```

$NH_4H_2PO_2$ ($F5_7$): A2BC2D_oC24_67_m_a_n_g - POSCAR

```
A2BC2D_oC24_67_m_a_n_g & a,b/a,c/a,z2,y3,z3,x4,z4 --params=11.47,
    ↪ 0.659982563208,0.346992153444,0.458,0.608,0.806,0.386,0.348 &
    ↪ Cmma D_{2h}^{21} #67 (agmn) & oC24 & $F5_{7}$ & H2(NH4)O2P & H2
    ↪ (NH4)O2P & W. H. Zachariasen and R. C. L. Mooney, J. Chem.
    ↪ Phys. 2, 34-37 (1934)
   1.00000000000000
   5.73500000000000   -3.78500000000000    0.00000000000000
   5.73500000000000    3.78500000000000    0.00000000000000
   0.00000000000000    0.00000000000000    3.98000000000000
     H    NH4      O      P
     4      2      4      2
Direct
  -0.60800000000000    0.60800000000000    0.80600000000000    H   (8m)
   1.10800000000000   -0.10800000000000    0.80600000000000    H   (8m)
```

```
 -0.10800000000000   1.10800000000000  -0.80600000000000    H    (8m)
  0.60800000000000  -0.60800000000000  -0.80600000000000    H    (8m)
  0.25000000000000   0.25000000000000   0.00000000000000  NH4    (4a)
  0.75000000000000   0.75000000000000   0.00000000000000  NH4    (4a)
  1.13600000000000   0.63600000000000   0.34800000000000    O    (8n)
  0.36400000000000  -0.13600000000000   0.34800000000000    O    (8n)
 -0.13600000000000   0.36400000000000  -0.34800000000000    O    (8n)
  0.63600000000000   1.13600000000000  -0.34800000000000    O    (8n)
  0.75000000000000   0.25000000000000   0.45800000000000    P    (4g)
  0.25000000000000   0.75000000000000  -0.45800000000000    P    (4g)
```

Thenardite [$Na_2SO_4$ (V), $H1_7$]: A2B4C_oF56_70_g_h_a - CIF

```
# CIF file
data_findsym-output
_audit_creation_method FINDSYM

_chemical_name_mineral 'Thenardite'
_chemical_formula_sum 'Na2 O4 S'

loop_
_publ_author_name
 'A. G. Nord'
_journal_name_full_name
;
 Acta Chemica Scandinavica
;
_journal_volume 27
_journal_year 1973
_journal_page_first 814
_journal_page_last 822
_publ_Section_title
;
 Refinement of the Crystal Structure of Thenardite, Na$_{2}$SO$_{4}$(V)
;

_aflow_title 'Thenardite [Na$_{2}$SO$_{4}$ (V), $H1_{7}$] Structure'
_aflow_proto 'A2B4C_oF56_70_g_h_a'
_aflow_params 'a,b/a,c/a,z_{2},x_{3},y_{3},z_{3}'
_aflow_params_values '5.8596,2.09987029831,1.67537033245,0.4414,-0.0203,
    ↪ 0.0572,0.2137'
_aflow_Strukturbericht '$H1_{7}$'
_aflow_Pearson 'oF56'

_symmetry_space_group_name_H-M "F 2/d 2/d 2/d (origin choice 2)"
_symmetry_Int_Tables_number 70

_cell_length_a      5.85960
_cell_length_b      12.30440
_cell_length_c      9.81700
_cell_angle_alpha 90.00000
_cell_angle_beta   90.00000
_cell_angle_gamma 90.00000

loop_
_space_group_symop_id
_space_group_symop_operation_xyz
1 x,y,z
2 x,-y+3/4,-z+3/4
3 -x+3/4,y,-z+3/4
4 -x+3/4,-y+3/4,z
5 -x,-y,-z
6 -x,y+1/4,z+1/4
7 x+1/4,-y,z+1/4
8 x+1/4,y+1/4,-z
9 x,y+1/2,z+1/2
10 x,-y+1/4,-z+1/4
11 -x+3/4,y+1/2,-z+1/4
12 -x+3/4,-y+1/4,z+1/2
13 -x,-y+1/2,-z+1/2
14 -x,y+3/4,z+3/4
15 x+1/4,-y+1/2,z+3/4
16 x+1/4,y+3/4,-z+1/2
17 x+1/2,y,z+1/2
18 x+1/2,-y+3/4,-z+1/4
19 -x+1/4,y,-z+1/4
20 -x+1/4,-y+3/4,z+1/2
21 -x+1/2,-y,-z+1/2
22 -x+1/2,y+1/4,z+3/4
23 x+3/4,-y,z+3/4
24 x+3/4,y+1/4,-z+1/2
25 x+1/2,y+1/2,z
26 x+1/2,-y+1/4,-z+3/4
27 -x+1/4,y+1/2,-z+3/4
28 -x+1/4,-y+1/4,z
29 -x+1/2,-y+1/2,-z
30 -x+1/2,y+3/4,z+1/4
31 x+3/4,-y+1/2,z+1/4
32 x+3/4,y+3/4,-z

loop_
_atom_site_label
_atom_site_type_symbol
_atom_site_symmetry_multiplicity
_atom_site_Wyckoff_label
_atom_site_fract_x
_atom_site_fract_y
_atom_site_fract_z
_atom_site_occupancy
S1 S 8 a 0.12500 0.12500 0.12500 1.00000
Na1 Na 16 g 0.12500 0.12500 0.44140 1.00000
O1 O 32 h -0.02030 0.05720 0.21370 1.00000
```

Thenardite [$Na_2SO_4$ (V), $H1_7$]: A2B4C_oF56_70_g_h_a - POSCAR

```
A2B4C_oF56_70_g_h_a & a,b/a,c/a,z2,x3,y3,z3 --params=5.8596,
    ↪ 2.09987029831,1.67537033245,0.4414,-0.0203,0.0572,0.2137 & Fddd
    ↪ D_{2h}^{24} #70 (agh) & oF56 & $H1_{7}$ & Na2O4S & Thenardite
    ↪ & A. G. Nord, Acta Chem. Scand. 27, 814-822 (1973)
   1.00000000000000
   0.00000000000000   6.15220000000000   4.90850000000000
   2.92980000000000   0.00000000000000   4.90850000000000
   2.92980000000000   6.15220000000000   0.00000000000000
    Na        O        S
     4        8        2
Direct
   0.44140000000000   0.44140000000000  -0.19140000000000   Na   (16g)
  -0.19140000000000  -0.19140000000000   0.44140000000000   Na   (16g)
  -0.44140000000000  -0.44140000000000   1.19140000000000   Na   (16g)
   1.19140000000000   1.19140000000000  -0.44140000000000   Na   (16g)
   0.29120000000000   0.13620000000000  -0.17680000000000    O   (32h)
   0.13620000000000   0.29120000000000   0.24940000000000    O   (32h)
  -0.17680000000000   0.24940000000000   0.29120000000000    O   (32h)
   0.24940000000000  -0.17680000000000   0.13620000000000    O   (32h)
  -0.29120000000000  -0.13620000000000   0.17680000000000    O   (32h)
  -0.13620000000000  -0.29120000000000   0.75060000000000    O   (32h)
   0.17680000000000   0.75060000000000  -0.29120000000000    O   (32h)
   0.75060000000000   0.17680000000000  -0.13620000000000    O   (32h)
   0.12500000000000   0.12500000000000   0.12500000000000    S    (8a)
   0.87500000000000   0.87500000000000   0.87500000000000    S    (8a)
```

$Mg_2Cu$ ($C_b$): AB2_oF48_70_g_fg - CIF

```
# CIF file
data_findsym-output
_audit_creation_method FINDSYM

_chemical_name_mineral 'CuMg2'
_chemical_formula_sum 'Cu Mg2'

loop_
_publ_author_name
 'F. Gingl'
 'P. Selvam'
 'K. Yvon'
_journal_name_full_name
;
 Acta Crystallographica Section B: Structural Science
;
_journal_volume 49
_journal_year 1993
_journal_page_first 201
_journal_page_last 203
_publ_Section_title
;
 Structure refinement of Mg$_{2}$Cu and a comparison of the Mg$_{2}$Cu,
    ↪ Mg$_{2}$Ni and Al$_{2}$Cu structure types
;

# Found in Crystal Structure of CuMg$_{2}$, 2018 Found in Crystal
    ↪ Structure of CuMg$_{2}$, {Crystallography online.com},

_aflow_title 'Mg$_{2}$Cu ($C_{b}$) Structure'
_aflow_proto 'AB2_oF48_70_g_fg'
_aflow_params 'a,b/a,c/a,y_{1},z_{2},z_{3}'
_aflow_params_values '5.275,1.71450236967,3.47450236967,0.4586,0.49819,
    ↪ 0.0415'
_aflow_Strukturbericht '$C_{b}$'
_aflow_Pearson 'oF48'

_symmetry_space_group_name_H-M "F 2/d 2/d 2/d (origin choice 2)"
_symmetry_Int_Tables_number 70

_cell_length_a      5.27500
_cell_length_b      9.04400
_cell_length_c      18.32800
_cell_angle_alpha 90.00000
_cell_angle_beta   90.00000
_cell_angle_gamma 90.00000

loop_
_space_group_symop_id
_space_group_symop_operation_xyz
1 x,y,z
2 x,-y+3/4,-z+3/4
3 -x+3/4,y,-z+3/4
4 -x+3/4,-y+3/4,z
5 -x,-y,-z
6 -x,y+1/4,z+1/4
7 x+1/4,-y,z+1/4
8 x+1/4,y+1/4,-z
9 x,y+1/2,z+1/2
10 x,-y+1/4,-z+1/4
11 -x+3/4,y+1/2,-z+1/4
12 -x+3/4,-y+1/4,z+1/2
13 -x,-y+1/2,-z+1/2
14 -x,y+3/4,z+3/4
15 x+1/4,-y+1/2,z+3/4
16 x+1/4,y+3/4,-z+1/2
17 x+1/2,y,z+1/2
18 x+1/2,-y+3/4,-z+1/4
19 -x+1/4,y,-z+1/4
20 -x+1/4,-y+3/4,z+1/2
21 -x+1/2,-y,-z+1/2
22 -x+1/2,y+1/4,z+3/4
23 x+3/4,-y,z+3/4
24 x+3/4,y+1/4,-z+1/2
25 x+1/2,y+1/2,z
26 x+1/2,-y+1/4,-z+3/4
27 -x+1/4,y+1/2,-z+3/4
```

```
28 -x+1/4,-y+1/4,z
29 -x+1/2,-y+1/2,-z
30 -x+1/2,y+3/4,z+1/4
31 x+3/4,-y+1/2,z+1/4
32 x+3/4,y+3/4,-z

loop_
_atom_site_label
_atom_site_type_symbol
_atom_site_symmetry_multiplicity
_atom_site_Wyckoff_label
_atom_site_fract_x
_atom_site_fract_y
_atom_site_fract_z
_atom_site_occupancy
Mg1 Mg 16 f 0.12500 0.45860 0.12500 1.00000
Cu1 Cu 16 g 0.12500 0.12500 0.49819 1.00000
Mg2 Mg 16 g 0.12500 0.12500 0.04150 1.00000
```

$Mg_2Cu$ ($C_b$): AB2_oF48_70_g_fg - POSCAR

```
AB2_oF48_70_g_fg & a,b/a,c/a,y1,z2,z3 --params=5.275,1.71450236967,
    ↪ 3.47450236967,0.4586,0.49819,0.0415 & Fddd D_{2h}^{24} #70 (fg^
    ↪ 2) & oF48 & $C_{b}$ & CuMg2 & CuMg2 & F. Gingl and P. Selvam
    ↪ and K. Yvon, Acta Crystallogr. Sect. B Struct. Sci. 49, 201-203
    ↪ (1993)
   1.00000000000000
   0.00000000000000   4.52200000000000   9.16400000000000
   2.63750000000000   0.00000000000000   9.16400000000000
   2.63750000000000   4.52200000000000   0.00000000000000
    Cu      Mg
     4       8
Direct
   0.49819000000000   0.49819000000000  -0.24819000000000   Cu   (16g)
  -0.24819000000000  -0.24819000000000   0.49819000000000   Cu   (16g)
  -0.49819000000000  -0.49819000000000   0.24819000000000   Cu   (16g)
   1.24819000000000   1.24819000000000  -0.49819000000000   Cu   (16g)
   0.45860000000000  -0.20860000000000   0.45860000000000   Mg   (16f)
  -0.20860000000000   0.45860000000000  -0.20860000000000   Mg   (16f)
  -0.45860000000000   1.20860000000000  -0.45860000000000   Mg   (16f)
   1.20860000000000  -0.45860000000000   1.20860000000000   Mg   (16f)
   0.04150000000000   0.04150000000000   0.20850000000000   Mg   (16g)
   0.20850000000000   0.20850000000000   0.04150000000000   Mg   (16g)
  -0.04150000000000  -0.04150000000000   0.79150000000000   Mg   (16g)
   0.79150000000000   0.79150000000000  -0.04150000000000   Mg   (16g)
```

High-Temperature Cryolite ($Na_3AlF_6$): AB6C3_oI20_71_a_in_cj - CIF

```
# CIF file
data_findsym-output
_audit_creation_method FINDSYM

_chemical_name_mineral 'Cryolite'
_chemical_formula_sum 'Al F6 Na3'

loop_
_publ_author_name
 'H. Yang'
 'S. Ghose'
 'D. M. Hatch'
_journal_name_full_name
;
 Physics and Chemistry of Minerals
;
_journal_volume 19
_journal_year 1993
_journal_page_first 528
_journal_page_last 544
_publ_Section_title
;
 Ferroelastic phase transition in cryolite, Na$_{3}$AlF$_{6}$, a mixed
    ↪ fluoride perovskite: High temperature single crystal X-ray
    ↪ diffraction study and symmetry analysis of the transition
    ↪ mechanism
;

# Found in The American Mineralogist Crystal Structure Database, 2003

_aflow_title 'High-Temperature Cryolite (Na$_{3}$AlF$_{6}$) Structure'
_aflow_proto 'AB6C3_oI20_71_a_in_cj'
_aflow_params 'a,b/a,c/a,z_{3},z_{4},x_{5},y_{5}'
_aflow_params_values '5.6333,0.99893490494,1.41336339268,0.2192,0.2485,
    ↪ 0.2335,0.2166'
_aflow_Strukturbericht 'None'
_aflow_Pearson 'oI20'

_symmetry_space_group_name_H-M "I 2/m 2/m 2/m"
_symmetry_Int_Tables_number 71

_cell_length_a    5.63330
_cell_length_b    5.62730
_cell_length_c    7.96190
_cell_angle_alpha 90.00000
_cell_angle_beta  90.00000
_cell_angle_gamma 90.00000

loop_
_space_group_symop_id
_space_group_symop_operation_xyz
1 x,y,z
2 x,-y,-z
3 -x,y,-z
4 -x,-y,z
5 -x,-y,-z
6 -x,y,z
7 x,-y,z
8 x,y,-z
9 x+1/2,y+1/2,z+1/2
10 x+1/2,-y+1/2,-z+1/2
11 -x+1/2,y+1/2,-z+1/2
12 -x+1/2,-y+1/2,z+1/2
13 -x+1/2,-y+1/2,-z+1/2
14 -x+1/2,y+1/2,z+1/2
15 x+1/2,-y+1/2,z+1/2
16 x+1/2,y+1/2,-z+1/2

loop_
_atom_site_label
_atom_site_type_symbol
_atom_site_symmetry_multiplicity
_atom_site_Wyckoff_label
_atom_site_fract_x
_atom_site_fract_y
_atom_site_fract_z
_atom_site_occupancy
Al1 Al 2 a 0.00000 0.00000 0.00000 1.00000
Na1 Na 2 c 0.50000 0.50000 0.00000 1.00000
F1  F  4 i 0.00000 0.00000 0.21920 1.00000
Na2 Na 4 j 0.50000 0.00000 0.24850 1.00000
F2  F  8 n 0.23350 0.21660 0.00000 1.00000
```

High-Temperature Cryolite ($Na_3AlF_6$): AB6C3_oI20_71_a_in_cj - POSCAR

```
AB6C3_oI20_71_a_in_cj & a,b/a,c/a,z3,z4,x5,y5 --params=5.6333,
    ↪ 0.99893490494,1.41336339268,0.2192,0.2485,0.2335,0.2166 & Immm
    ↪ D_{2h}^{25} #71 (acijn) & oI20 & None & AlF6Na3 & Cryolite & H.
    ↪ Yang and S. Ghose and D. M. Hatch, Phys. Chem. Miner. 19,
    ↪ 528-544 (1993)
   1.00000000000000
  -2.81665000000000   2.81365000000000   3.98095000000000
   2.81665000000000  -2.81365000000000   3.98095000000000
   2.81665000000000   2.81365000000000  -3.98095000000000
    Al       F       Na
     1       6        3
Direct
   0.00000000000000   0.00000000000000   0.00000000000000   Al   (2a)
   0.21920000000000   0.21920000000000   0.00000000000000   F    (4i)
  -0.21920000000000  -0.21920000000000   0.00000000000000   F    (4i)
   0.21660000000000   0.23350000000000   0.45010000000000   F    (8n)
  -0.21660000000000  -0.23350000000000  -0.45010000000000   F    (8n)
   0.21660000000000  -0.23350000000000  -0.01690000000000   F    (8n)
  -0.21660000000000   0.23350000000000   0.01690000000000   F    (8n)
   0.50000000000000   0.50000000000000   0.00000000000000   Na   (2c)
   0.24850000000000   0.74850000000000   0.50000000000000   Na   (4j)
  -0.24850000000000   0.25150000000000   0.50000000000000   Na   (4j)
```

$CsFeS_2$ (100 K): ABC2_oI16_71_g_i_eh - CIF

```
# CIF file
data_findsym-output
_audit_creation_method FINDSYM

_chemical_name_mineral 'CsFeS2'
_chemical_formula_sum 'Cs Fe S2'

loop_
_publ_author_name
 'Y. Ito'
 'M. Nishi'
 'C. F. Majkrzak'
 'L. Passell'
_journal_name_full_name
;
 Journal of the Physical Society of Japan
;
_journal_volume 54
_journal_year 1985
_journal_page_first 348
_journal_page_last 357
_publ_Section_title
;
 Low Temperature Powder Neutron Diffraction Studies of CsFeS$_{2}$
;

# Found in CsFeS2 (100K) Crystal Structure, 2016 Found in CsFeS2 (100K)
    ↪ Crystal Structure, {PAULING FILE in: Inorganic Solid Phases,
    ↪ SpringerMaterials (online database), Springer, Heidelberg (ed.)
    ↪ SpringerMaterials },

_aflow_title 'CsFeS$_{2}$ (100~K) Structure'
_aflow_proto 'ABC2_oI16_71_g_i_eh'
_aflow_params 'a,b/a,c/a,x_{1},y_{2},y_{3},z_{4}'
_aflow_params_values '7.09,1.67136812412,0.764456981664,0.256,0.33,0.148
    ↪ ,0.249'
_aflow_Strukturbericht 'None'
_aflow_Pearson 'oI16'

_symmetry_space_group_name_H-M "I 2/m 2/m 2/m"
_symmetry_Int_Tables_number 71

_cell_length_a    7.09000
_cell_length_b    11.85000
_cell_length_c    5.42000
_cell_angle_alpha 90.00000
_cell_angle_beta  90.00000
_cell_angle_gamma 90.00000

loop_
_space_group_symop_id
_space_group_symop_operation_xyz
```

```
1 x,y,z
2 x,-y,-z
3 -x,y,-z
4 -x,-y,z
5 -x,-y,-z
6 -x,y,z
7 x,-y,z
8 x,y,-z
9 x+1/2,y+1/2,z+1/2
10 x+1/2,-y+1/2,-z+1/2
11 -x+1/2,y+1/2,-z+1/2
12 -x+1/2,-y+1/2,z+1/2
13 -x+1/2,-y+1/2,-z+1/2
14 -x+1/2,y+1/2,z+1/2
15 x+1/2,-y+1/2,z+1/2
16 x+1/2,y+1/2,-z+1/2

loop_
_atom_site_label
_atom_site_type_symbol
_atom_site_symmetry_multiplicity
_atom_site_Wyckoff_label
_atom_site_fract_x
_atom_site_fract_y
_atom_site_fract_z
_atom_site_occupancy
S1 S 4 e 0.25600 0.00000 0.00000 1.00000
Cs1 Cs 4 g 0.00000 0.33000 0.00000 1.00000
S2 S 4 h 0.00000 0.14800 0.50000 1.00000
Fe1 Fe 4 i 0.00000 0.00000 0.24900 1.00000
```

$CsFeS_2$ (100 K): ABC2_oI16_71_g_i_eh - POSCAR

```
ABC2_oI16_71_g_i_eh & a,b/a,c/a,x1,y2,y3,z4 --params=7.09,1.67136812412,
    ↪ 0.764456981664,0.256,0.33,0.148,0.249 & Immm D_{2h}^{25} #71 (
    ↪ eghi) & oI16 & None & CsFeS2 & CsFeS2 & Y. Ito et al., J. Phys.
    ↪ Soc. Jpn. 54, 348-357 (1985)
   1.00000000000000
  -3.54500000000000   5.92500000000000   2.71000000000000
   3.54500000000000  -5.92500000000000   2.71000000000000
   3.54500000000000   5.92500000000000  -2.71000000000000
    Cs      Fe      S
     2       2      4
Direct
   0.33000000000000   0.00000000000000   0.33000000000000   Cs   (4g)
  -0.33000000000000   0.00000000000000  -0.33000000000000   Cs   (4g)
   0.24900000000000   0.24900000000000   0.00000000000000   Fe   (4i)
  -0.24900000000000  -0.24900000000000   0.00000000000000   Fe   (4i)
   0.00000000000000   0.25600000000000   0.25600000000000   S    (4e)
   0.00000000000000  -0.25600000000000  -0.25600000000000   S    (4e)
   0.64800000000000   0.50000000000000   0.14800000000000   S    (4h)
   0.35200000000000   0.50000000000000  -0.14800000000000   S    (4h)
```

CsO: AB_oI8_71_g_i - CIF

```
# CIF file
data_findsym-output
_audit_creation_method FINDSYM

_chemical_name_mineral 'CsO'
_chemical_formula_sum 'Cs O'

loop_
_publ_author_name
 'M. E. Rengade'
_journal_name_full_name
;
 Comptes Rendus de l'Acad{\'e}mie des Sciences
;
_journal_volume 148
_journal_year 1909
_journal_page_first 1199
_journal_page_last 1202
_publ_Section_title
;
 Sur les Sous-Oxydes de Caesium
;

# Found in Binary Alloy Phase Diagrams, 1990 Found in Binary Alloy Phase
    ↪ Diagrams, {Cd-Ce to Hf-Rb}}

_aflow_title 'CsO Structure'
_aflow_proto 'AB_oI8_71_g_i'
_aflow_params 'a,b/a,c/a,y_{1},z_{2}'
_aflow_params_values '4.322,1.73924109209,1.48773715872,0.25,0.38'
_aflow_Strukturbericht 'None'
_aflow_Pearson 'oI8'

_symmetry_space_group_name_H-M "I 2/m 2/m 2/m"
_symmetry_Int_Tables_number 71

_cell_length_a      4.32200
_cell_length_b      7.51700
_cell_length_c      6.43000
_cell_angle_alpha  90.00000
_cell_angle_beta   90.00000
_cell_angle_gamma  90.00000

loop_
_space_group_symop_id
_space_group_symop_operation_xyz
1 x,y,z
2 x,-y,-z
3 -x,y,-z
4 -x,-y,z
5 -x,-y,-z
6 -x,y,z
7 x,-y,z
8 x,y,-z
9 x+1/2,y+1/2,z+1/2
10 x+1/2,-y+1/2,-z+1/2
11 -x+1/2,y+1/2,-z+1/2
12 -x+1/2,-y+1/2,z+1/2
13 -x+1/2,-y+1/2,-z+1/2
14 -x+1/2,y+1/2,z+1/2
15 x+1/2,-y+1/2,z+1/2
16 x+1/2,y+1/2,-z+1/2

loop_
_atom_site_label
_atom_site_type_symbol
_atom_site_symmetry_multiplicity
_atom_site_Wyckoff_label
_atom_site_fract_x
_atom_site_fract_y
_atom_site_fract_z
_atom_site_occupancy
Cs1 Cs 4 g 0.00000 0.25000 0.00000 1.00000
O1 O 4 i 0.00000 0.00000 0.38000 1.00000
```

CsO: AB_oI8_71_g_i - POSCAR

```
AB_oI8_71_g_i & a,b/a,c/a,y1,z2 --params=4.322,1.73924109209,
    ↪ 1.48773715872,0.25,0.38 & Immm D_{2h}^{25} #71 (gi) & oI8 &
    ↪ None & CsO & CsO & M. E. Rengade, {C. R. Acad. Sci. C 148,
    ↪ 1199-1202 (1909)
   1.00000000000000
  -2.16100000000000   3.75850000000000   3.21500000000000
   2.16100000000000  -3.75850000000000   3.21500000000000
   2.16100000000000   3.75850000000000  -3.21500000000000
    Cs      O
     2      2
Direct
   0.25000000000000   0.00000000000000   0.25000000000000   Cs   (4g)
  -0.25000000000000   0.00000000000000  -0.25000000000000   Cs   (4g)
   0.38000000000000   0.38000000000000   0.00000000000000   O    (4i)
  -0.38000000000000  -0.38000000000000   0.00000000000000   O    (4i)
```

$Ga_2Mg_5$ ($D8_g$): A2B5_oI28_72_j_bfj - CIF

```
# CIF file
data_findsym-output
_audit_creation_method FINDSYM

_chemical_name_mineral 'Ga2Mg5'
_chemical_formula_sum 'Ga2 Mg5'

loop_
_publ_author_name
 'K. Schubert'
 'K. Frank'
 'R. Gohle'
 'A. Maldonado'
 'H. G. Meissner'
 'A. Raman'
 'W. Rossteutscher'
_journal_name_full_name
;
 Naturwissenschaften
;
_journal_volume 50
_journal_year 1963
_journal_page_first 41
_journal_page_last 41
_publ_Section_title
;
 Einige Strukturdaten metallischer Phasen (8)
;

# Found in Crystal Structure of Ga$_{2}$Mg$_{5}$, 2018 Found in Crystal
    ↪ Structure of Ga$_{2}$Mg$_{5}$, {Crystallography online.com},

_aflow_title 'Ga$_{2}$Mg$_{5}$ ($D8_{g}$) Structure'
_aflow_proto 'A2B5_oI28_72_j_bfj'
_aflow_params 'a,b/a,c/a,x_{2},x_{3},y_{3},x_{4},y_{4}'
_aflow_params_values '13.7,0.511678832117,0.439416058394,0.26,0.378,
    ↪ 0.255,0.27,0.345'
_aflow_Strukturbericht '$D8_{g}$'
_aflow_Pearson 'oI28'

_symmetry_space_group_name_H-M "I 2/b 2/a 2/m"
_symmetry_Int_Tables_number 72

_cell_length_a     13.70000
_cell_length_b      7.01000
_cell_length_c      6.02000
_cell_angle_alpha  90.00000
_cell_angle_beta   90.00000
_cell_angle_gamma  90.00000

loop_
_space_group_symop_id
_space_group_symop_operation_xyz
1 x,y,z
2 x,-y,-z+1/2
3 -x,y,-z+1/2
4 -x,-y,z
5 -x,-y,-z
6 -x,y,z+1/2
7 x,-y,z+1/2
```

```
8 x,y,-z
9 x+1/2,y+1/2,z+1/2
10 x+1/2,-y+1/2,-z
11 -x+1/2,y+1/2,-z
12 -x+1/2,-y+1/2,z+1/2
13 -x+1/2,-y+1/2,-z+1/2
14 -x+1/2,y+1/2,z
15 x+1/2,-y+1/2,z
16 x+1/2,y+1/2,-z+1/2

loop_
_atom_site_label
_atom_site_type_symbol
_atom_site_symmetry_multiplicity
_atom_site_Wyckoff_label
_atom_site_fract_x
_atom_site_fract_y
_atom_site_fract_z
_atom_site_occupancy
Mg1 Mg 4 b 0.50000 0.00000 0.25000 1.00000
Mg2 Mg 8 f 0.26000 0.00000 0.25000 1.00000
Ga1 Ga 8 j 0.37800 0.25500 0.00000 1.00000
Mg3 Mg 8 j 0.27000 0.34500 0.00000 1.00000
```

$Ga_2Mg_5$ ($D8_g$): A2B5_oI28_72_j_bfj - POSCAR

```
A2B5_oI28_72_j_bfj & a,b/a,c/a,x2,x3,y3,x4,y4 --params=13.7,
    ↪ 0.511678832117,0.439416058394,0.26,0.378,0.255,0.27,0.345 &
    ↪ Ibam D_{2h}^{26} #72 (bfj^2) & oI28 & $D8_{g}$ & Ga2Mg5 &
    ↪ Ga2Mg5 & K. Schubert et al., Naturwissenschaften 50, 41(1963)
   1.00000000000000
  -6.85000000000000    3.50500000000000    3.01000000000000
   6.85000000000000   -3.50500000000000    3.01000000000000
   6.85000000000000    3.50500000000000   -3.01000000000000
    Ga       Mg
     4       10
Direct
   0.25500000000000    0.37800000000000    0.63300000000000    Ga    (8j)
  -0.25500000000000   -0.37800000000000   -0.63300000000000    Ga    (8j)
   0.75500000000000    0.12200000000000   -0.12300000000000    Ga    (8j)
   0.24500000000000    0.87800000000000    0.12300000000000    Ga    (8j)
   0.25000000000000    0.75000000000000    0.50000000000000    Mg    (4b)
   0.75000000000000    0.25000000000000    0.50000000000000    Mg    (4b)
   0.25000000000000    0.51000000000000    0.26000000000000    Mg    (8f)
   0.25000000000000   -0.01000000000000   -0.26000000000000    Mg    (8f)
   0.75000000000000    0.49000000000000   -0.26000000000000    Mg    (8f)
   0.75000000000000    1.01000000000000    0.26000000000000    Mg    (8f)
   0.34500000000000    0.27000000000000    0.61500000000000    Mg    (8j)
  -0.34500000000000   -0.27000000000000   -0.61500000000000    Mg    (8j)
   0.84500000000000    0.23000000000000    0.07500000000000    Mg    (8j)
   0.15500000000000    0.77000000000000   -0.07500000000000    Mg    (8j)
```

$Zn(NH_3)_2Cl_2$ ($E1_2$): A2B6C2D_oI44_74_h_ij_i_e - CIF

```
# CIF file
data_findsym-output
_audit_creation_method FINDSYM

_chemical_name_mineral 'Cl2H6N2Zn'
_chemical_formula_sum 'Cl2 H6 N2 Zn'

loop_
_publ_author_name
 'T. Iv\v{s}i\'{c}'
 'D. W. Bi'
 'A. Magrez'
_journal_name_full_name
;
 Acta Crystallographica Section E: Crystallographic Communications
;
_journal_volume 75
_journal_year 2019
_journal_page_first 1386
_journal_page_last 1388
_publ_Section_title
;
 New refinement of the crystal structure of Zn(NH$_{3}$)$_{2}$Cl$_{2}$
    ↪ at 100K
;

_aflow_title 'Zn(NH$_{3}$)$_{2}$Cl$_{2}$ ($E1_{2}$) Structure'
_aflow_proto 'A2B6C2D_oI44_74_h_ij_i_e'
_aflow_params 'a,b/a,c/a,z_{1},y_{2},z_{2},x_{3},z_{3},x_{4},z_{4},x_{5
    ↪ },y_{5},z_{5}'
_aflow_params_values '7.7077,1.04085524865,1.09664361612,0.88832,0.47954
    ↪ ,0.73085,0.206,0.533,0.28237,0.47949,0.274,0.3281,0.416'
_aflow_Strukturbericht '$E1_{2}$'
_aflow_Pearson 'oI44'

_symmetry_space_group_name_H-M "I 21/m 21/m 21/a"
_symmetry_Int_Tables_number 74

_cell_length_a     7.70770
_cell_length_b     8.02260
_cell_length_c     8.45260
_cell_angle_alpha 90.00000
_cell_angle_beta  90.00000
_cell_angle_gamma 90.00000

loop_
_space_group_symop_id
_space_group_symop_operation_xyz
1 x,y,z
2 x,-y,-z
3 -x,y+1/2,-z
4 -x,-y+1/2,z
5 -x,-y,-z
6 -x,y,z
7 x,-y+1/2,z
8 x,y+1/2,-z
9 x+1/2,y+1/2,z+1/2
10 x+1/2,-y+1/2,-z+1/2
11 -x+1/2,y,-z+1/2
12 -x+1/2,-y,z+1/2
13 -x+1/2,-y+1/2,-z+1/2
14 -x+1/2,y+1/2,z+1/2
15 x+1/2,-y,z+1/2
16 x+1/2,y,-z+1/2

loop_
_atom_site_label
_atom_site_type_symbol
_atom_site_symmetry_multiplicity
_atom_site_Wyckoff_label
_atom_site_fract_x
_atom_site_fract_y
_atom_site_fract_z
_atom_site_occupancy
Zn1 Zn 4 e 0.00000 0.25000 0.88832 1.00000
Cl1 Cl 8 h 0.00000 0.47954 0.73085 1.00000
H1 H 8 i 0.20600 0.25000 0.53300 1.00000
N1 N 8 i 0.28237 0.25000 0.47949 1.00000
H2 H 16 j 0.27400 0.32810 0.41600 1.00000
```

$Zn(NH_3)_2Cl_2$ ($E1_2$): A2B6C2D_oI44_74_h_ij_i_e - POSCAR

```
A2B6C2D_oI44_74_h_ij_i_e & a,b/a,c/a,z1,y2,z2,x3,z3,x4,z4,x5,y5,z5 --
    ↪ params=7.7077,1.04085524865,1.09664361612,0.88832,0.47954,
    ↪ 0.73085,0.206,0.533,0.28237,0.47949,0.274,0.3281,0.416 & Imma
    ↪ D_{2h}^{28} #74 (ehi^2j) & oI44 & $E1_{2}$ & Cl2H6N2Zn &
    ↪ Cl2H6N2Zn & T. Iv\v{s}i\'{c} and D. W. Bi and A. Magrez, Acta
    ↪ Crystallogr. E 75, 1386-1388 (2019)
   1.00000000000000
  -3.85385000000000    4.01130000000000    4.22630000000000
   3.85385000000000   -4.01130000000000    4.22630000000000
   3.85385000000000    4.01130000000000   -4.22630000000000
    Cl       H        N       Zn
     4      12        4        2
Direct
   1.21039000000000    0.73085000000000    0.47954000000000    Cl    (8h)
   0.75131000000000    0.73085000000000    0.02046000000000    Cl    (8h)
   0.24869000000000   -0.73085000000000    0.97954000000000    Cl    (8h)
  -1.21039000000000   -0.73085000000000   -0.47954000000000    Cl    (8h)
   0.78300000000000    0.73900000000000    0.45600000000000     H    (8i)
   0.78300000000000    0.32700000000000    0.04400000000000     H    (8i)
   0.21700000000000   -0.73900000000000    0.54400000000000     H    (8i)
   0.21700000000000   -0.32700000000000    0.95600000000000     H    (8i)
   0.74410000000000    0.69000000000000    0.60210000000000     H   (16j)
   0.58790000000000    0.14200000000000   -0.10210000000000     H   (16j)
   0.41210000000000   -0.69000000000000    0.55410000000000     H   (16j)
  -0.74410000000000   -0.14200000000000   -0.05410000000000     H   (16j)
  -0.74410000000000   -0.69000000000000   -0.60210000000000     H   (16j)
   0.41210000000000   -0.14200000000000    1.10210000000000     H   (16j)
   0.58790000000000    0.69000000000000    0.44590000000000     H   (16j)
   0.74410000000000    0.14200000000000    0.05410000000000     H   (16j)
   0.72949000000000    0.76186000000000    0.53237000000000     N    (8i)
   0.72949000000000    0.19712000000000   -0.03237000000000     N    (8i)
   0.27051000000000   -0.76186000000000    0.46763000000000     N    (8i)
   0.27051000000000   -0.19712000000000    1.03237000000000     N    (8i)
   1.13832000000000    0.88832000000000    0.25000000000000    Zn    (4e)
  -0.13832000000000   -0.88832000000000    0.75000000000000    Zn    (4e)
```

$CeCu_2$: AB2_oI12_74_e_h - CIF

```
# CIF file
data_findsym-output
_audit_creation_method FINDSYM

_chemical_name_mineral 'CeCu2'
_chemical_formula_sum 'Ce Cu2'

loop_
_publ_author_name
 'A. C. Larson'
 'D. T. Cromer'
_journal_name_full_name
;
 Acta Cristallographica
;
_journal_volume 14
_journal_year 1961
_journal_page_first 73
_journal_page_last 74
_publ_Section_title
;
 The crystal structure of CeCu$_{2}$
;

_aflow_title 'CeCu$_{2}$ Structure'
_aflow_proto 'AB2_oI12_74_e_h'
_aflow_params 'a,b/a,c/a,z_{1},y_{2},z_{2}'
_aflow_params_values '4.43,1.5914221219,1.68171557562,0.5377,0.051,
    ↪ 0.1648'
_aflow_Strukturbericht 'None'
_aflow_Pearson 'oI12'

_symmetry_space_group_name_H-M "I 21/m 21/m 21/a"
_symmetry_Int_Tables_number 74

_cell_length_a     4.43000
```

```
_cell_length_b      7.05000
_cell_length_c      7.45000
_cell_angle_alpha  90.00000
_cell_angle_beta   90.00000
_cell_angle_gamma  90.00000

loop_
_space_group_symop_id
_space_group_symop_operation_xyz
1 x,y,z
2 x,-y,-z
3 -x,y+1/2,-z
4 -x,-y+1/2,z
5 -x,-y,-z
6 -x,y,z
7 x,-y+1/2,z
8 x,y+1/2,-z
9 x+1/2,y+1/2,z+1/2
10 x+1/2,-y+1/2,-z+1/2
11 -x+1/2,y,-z+1/2
12 -x+1/2,-y,z+1/2
13 -x+1/2,-y+1/2,-z+1/2
14 -x+1/2,y+1/2,z+1/2
15 x+1/2,-y,z+1/2
16 x+1/2,y,-z+1/2

loop_
_atom_site_label
_atom_site_type_symbol
_atom_site_symmetry_multiplicity
_atom_site_Wyckoff_label
_atom_site_fract_x
_atom_site_fract_y
_atom_site_fract_z
_atom_site_occupancy
Ce1 Ce 4 e 0.00000 0.25000 0.53770 1.00000
Cu1 Cu 8 h 0.00000 0.05100 0.16480 1.00000
```

CeCu$_2$: AB2_oI12_74_e_h - POSCAR

```
AB2_oI12_74_e_h & a,b/a,c/a,z1,y2,z2 --params=4.43,1.5914221219,
    ↪ 1.68171557562,0.5377,0.051,0.1648 & Imma D_{2h}^{28} #74 (eh) &
    ↪ oI12 & None & CeCu2 & CeCu2 & A. C. Larson and D. T. Cromer,
    ↪ Acta Cryst. 14, 73-74 (1961)
   1.00000000000000
  -2.21500000000000   3.52500000000000   3.72500000000000
   2.21500000000000  -3.52500000000000   3.72500000000000
   2.21500000000000   3.52500000000000  -3.72500000000000
    Ce    Cu
     2     4
Direct
   0.78770000000000   0.53770000000000   0.25000000000000   Ce   (4e)
   0.21230000000000  -0.53770000000000   0.75000000000000   Ce   (4e)
   0.21580000000000   0.16480000000000   0.05100000000000   Cu   (8h)
   0.61380000000000   0.16480000000000   0.44900000000000   Cu   (8h)
   0.38620000000000  -0.16480000000000   0.55100000000000   Cu   (8h)
  -0.21580000000000  -0.16480000000000  -0.05100000000000   Cu   (8h)
```

LiCuVO$_4$: ABC4D_oI28_74_a_d_hi_e - CIF

```
# CIF file
data_findsym-output
_audit_creation_method FINDSYM

_chemical_name_mineral 'CuLiO4V'
_chemical_formula_sum 'Cu Li O4 V'

loop_
_publ_author_name
 'M. A. Lafontaine'
 'M. Leblanc'
 'G. Ferey'
_journal_name_full_name
;
 Acta Crystallographica Section C: Structural Chemistry
;
_journal_volume 45
_journal_year 1989
_journal_page_first 1205
_journal_page_last 1206
_publ_Section_title
;
 New refinement of the room-temperature structure of LiCuVO$_{4}$
;

# Found in Structure, stoichiometry and magnetic properties of the
    ↪ low-dimensional structure phase LiCuVO$_{4}$, 2004

_aflow_title 'LiCuVO$_{4}$ Structure'
_aflow_proto 'ABC4D_oI28_74_a_d_hi_e'
_aflow_params 'a,b/a,c/a,z_{3},y_{4},z_{4},x_{5},z_{5}'
_aflow_params_values '5.662,1.0259625574,1.54680324974,0.386,0.0164,
    ↪ 0.2748,0.2352,-0.0007'
_aflow_Strukturbericht 'None'
_aflow_Pearson 'oI28'

_symmetry_space_group_name_H-M "I 21/m 21/m 21/a"
_symmetry_Int_Tables_number 74

_cell_length_a      5.66200
_cell_length_b      5.80900
_cell_length_c      8.75800
_cell_angle_alpha  90.00000
_cell_angle_beta   90.00000
_cell_angle_gamma  90.00000

loop_
_space_group_symop_id
_space_group_symop_operation_xyz
1 x,y,z
2 x,-y,-z
3 -x,y+1/2,-z
4 -x,-y+1/2,z
5 -x,-y,-z
6 -x,y,z
7 x,-y+1/2,z
8 x,y+1/2,-z
9 x+1/2,y+1/2,z+1/2
10 x+1/2,-y+1/2,-z+1/2
11 -x+1/2,y,-z+1/2
12 -x+1/2,-y,z+1/2
13 -x+1/2,-y+1/2,-z+1/2
14 -x+1/2,y+1/2,z+1/2
15 x+1/2,-y,z+1/2
16 x+1/2,y,-z+1/2

loop_
_atom_site_label
_atom_site_type_symbol
_atom_site_symmetry_multiplicity
_atom_site_Wyckoff_label
_atom_site_fract_x
_atom_site_fract_y
_atom_site_fract_z
_atom_site_occupancy
Cu1 Cu 4 a 0.00000 0.00000 0.00000 1.00000
Li1 Li 4 d 0.25000 0.25000 0.75000 1.00000
V1 V 4 e 0.00000 0.25000 0.38600 1.00000
O1 O 8 h 0.00000 0.01640 0.27480 1.00000
O2 O 8 i 0.23520 0.25000 -0.00070 1.00000
```

LiCuVO$_4$: ABC4D_oI28_74_a_d_hi_e - POSCAR

```
ABC4D_oI28_74_a_d_hi_e & a,b/a,c/a,z3,y4,z4,x5,z5 --params=5.662,
    ↪ 1.0259625574,1.54680324974,0.386,0.0164,0.2748,0.2352,-0.0007 &
    ↪ Imma D_{2h}^{28} #74 (adehi) & oI28 & None & CuLiO4V & CuLiO4V
    ↪ & M. A. Lafontaine and M. Leblanc and G. Ferey, Acta
    ↪ Crystallogr. C 45, 1205-1206 (1989)
   1.00000000000000
  -2.83100000000000   2.90450000000000   4.37900000000000
   2.83100000000000  -2.90450000000000   4.37900000000000
   2.83100000000000   2.90450000000000  -4.37900000000000
    Cu    Li    O    V
     2     2    8    2
Direct
   0.00000000000000   0.00000000000000   0.00000000000000   Cu   (4a)
   0.50000000000000   0.00000000000000   0.50000000000000   Cu   (4a)
   0.00000000000000   0.00000000000000   0.50000000000000   Li   (4d)
   0.00000000000000   0.50000000000000   0.00000000000000   Li   (4d)
   0.29120000000000   0.27480000000000   0.01640000000000   O    (8h)
   0.75840000000000   0.27480000000000   0.48360000000000   O    (8h)
   0.24160000000000  -0.27480000000000   0.51640000000000   O    (8h)
  -0.29120000000000  -0.27480000000000  -0.01640000000000   O    (8h)
   0.24930000000000   0.23450000000000   0.48520000000000   O    (8i)
   0.24930000000000  -0.23590000000000   0.01480000000000   O    (8i)
   0.75070000000000  -0.23450000000000   0.51480000000000   O    (8i)
   0.75070000000000   0.23590000000000   0.98520000000000   O    (8i)
   0.63600000000000   0.38600000000000   0.25000000000000   V    (4e)
   0.36400000000000  -0.38600000000000   0.75000000000000   V    (4e)
```

Gwihabaite [NH$_4$NO$_3$ (V)]: A4B2C3_tP72_77_8d_ab2c2d_6d - CIF

```
# CIF file
data_findsym-output
_audit_creation_method FINDSYM

_chemical_name_mineral 'Gwihabaite'
_chemical_formula_sum 'H4 N2 O3'

loop_
_publ_author_name
 'J. L. Amor\\'{o}s'
 'F. Arrese'
 'M. Canut'
_journal_name_full_name
;
 Zeitschrift f{\"u}r Kristallographie - Crystalline Materials
;
_journal_volume 117
_journal_year 1962
_journal_page_first 92
_journal_page_last 107
_publ_Section_title
;
 The crystal structure of the low-temperature phase of NH$_{4}$NO$_{3}$
    ↪ (V) at -- 150$^{\circ}$C
;

# Found in The American Mineralogist Crystal Structure Database, 2003

_aflow_title 'Gwihabaite [NH$_{4}$NO$_{3}$ (V)] Structure'
_aflow_proto 'A4B2C3_tP72_77_8d_ab2c2d_6d'
_aflow_params 'a,c/a,z_{1},z_{2},z_{3},z_{4},x_{5},y_{5},z_{5},x_{6},y_{
    ↪ 6},z_{6},x_{7},y_{7},z_{7},x_{8},y_{8},z_{8},x_{9},y_{9},z_{9},
    ↪ x_{10},y_{10},z_{10},x_{11},y_{11},z_{11},x_{12},y_{12},z_{12},
    ↪ x_{13},y_{13},z_{13},x_{14},y_{14},z_{14},x_{15},y_{15},z_{15},
    ↪ x_{16},y_{16},z_{16},x_{17},y_{17},z_{17},x_{18},y_{18},z_{18},
    ↪ x_{19},y_{19},z_{19},x_{20},y_{20},z_{20}'
_aflow_params_values '7.98,1.22556390977,0.25,0.25,0.75,0.25,0.076,0.076
    ↪ ,0.19,0.076,-0.076,0.31,0.576,0.076,0.19,0.424,0.076,0.31,0.424
```

```
      ↪ ,0.424,0.31,0.424,0.576,0.19,-0.076,0.424,0.31,-0.076,0.576,
      ↪ 0.19,0.25,0.25,0.518,0.25,0.25,0.018,0.12,0.27,0.456,0.33,0.23,
      ↪ 0.456,0.25,0.25,0.642,0.12,0.23,-0.044,0.38,0.27,-0.044,0.25,
      ↪ 0.25,0.142'
_aflow_Strukturbericht 'None'
_aflow_Pearson 'tP72'

_symmetry_space_group_name_H-M "P 42"
_symmetry_Int_Tables_number 77

_cell_length_a      7.98000
_cell_length_b      7.98000
_cell_length_c      9.78000
_cell_angle_alpha  90.00000
_cell_angle_beta   90.00000
_cell_angle_gamma  90.00000

loop_
_space_group_symop_id
_space_group_symop_operation_xyz
1 x,y,z
2 -x,-y,z
3 -y,x,z+1/2
4 y,-x,z+1/2

loop_
_atom_site_label
_atom_site_type_symbol
_atom_site_symmetry_multiplicity
_atom_site_Wyckoff_label
_atom_site_fract_x
_atom_site_fract_y
_atom_site_fract_z
_atom_site_occupancy
N1 N 2 a 0.00000 0.00000 0.25000 1.00000
N2 N 2 b 0.50000 0.50000 0.25000 1.00000
N3 N 2 c 0.00000 0.50000 0.75000 1.00000
N4 N 2 c 0.00000 0.50000 0.25000 1.00000
H1 H 4 d 0.07600 0.07600 0.19000 1.00000
H2 H 4 d 0.07600 -0.07600 0.31000 1.00000
H3 H 4 d 0.57600 0.07600 0.19000 1.00000
H4 H 4 d 0.42400 0.07600 0.31000 1.00000
H5 H 4 d 0.42400 0.42400 0.31000 1.00000
H6 H 4 d 0.42400 0.57600 0.19000 1.00000
H7 H 4 d -0.07600 0.42400 0.31000 1.00000
H8 H 4 d -0.07600 0.57600 0.19000 1.00000
N5 N 4 d 0.25000 0.25000 0.51800 1.00000
N6 N 4 d 0.25000 0.25000 0.01800 1.00000
O1 O 4 d 0.12000 0.27000 0.45600 1.00000
O2 O 4 d 0.33000 0.23000 0.45600 1.00000
O3 O 4 d 0.25000 0.25000 0.64200 1.00000
O4 O 4 d 0.12000 0.23000 -0.04400 1.00000
O5 O 4 d 0.38000 0.27000 -0.04400 1.00000
O6 O 4 d 0.25000 0.25000 0.14200 1.00000
```

Gwihabaite [$NH_4NO_3$ (V)]: A4B2C3_tP72_77_8d_ab2c2d_6d - POSCAR

```
A4B2C3_tP72_77_8d_ab2c2d_6d & a,c/a,z1,z2,z3,z4,x5,y5,z5,x6,y6,z6,x7,y7,
      ↪ z7,x8,y8,z8,x9,y9,z9,x10,y10,z10,x11,y11,z11,x12,y12,z12,x13,
      ↪ y13,z13,x14,y14,z14,x15,y15,z15,x16,y16,z16,x17,y17,z17,x18,y18
      ↪ ,z18,x19,y19,z19,x20,y20,z20 --params=7.98,1.22556390977,0.25,
      ↪ 0.25,0.75,0.25,0.076,0.076,0.19,0.076,-0.076,0.31,0.576,0.076,
      ↪ 0.19,0.424,0.076,0.31,0.424,0.424,0.31,0.424,0.576,0.19,-0.076,
      ↪ 0.424,0.31,-0.076,0.576,0.19,0.25,0.25,0.518,0.25,0.25,0.018,
      ↪ 0.12,0.27,0.456,0.33,0.23,0.456,0.25,0.25,0.642,0.12,0.23,-
      ↪ 0.044,0.38,0.27,-0.044,0.25,0.25,0.142 & P4_{2} C_{4}^{3} #77 (
      ↪ abc^2d^16) & tP72 & None & H4N2O3 & Gwihabaite & J. L. Amor\'{o
      ↪ }s and F. Arrese and M. Canut, Zeitschrift f{"u}r
      ↪ Kristallographie - Crystalline Materials 117, 92-107 (1962)
   1.00000000000000
   7.98000000000000   0.00000000000000   0.00000000000000
   0.00000000000000   7.98000000000000   0.00000000000000
   0.00000000000000   0.00000000000000   9.78000000000000
    H     N     O
   32    16    24
Direct
   0.07600000000000   0.07600000000000   0.19000000000000   H   (4d)
  -0.07600000000000  -0.07600000000000   0.19000000000000   H   (4d)
  -0.07600000000000   0.07600000000000   0.69000000000000   H   (4d)
   0.07600000000000  -0.07600000000000   0.69000000000000   H   (4d)
   0.07600000000000  -0.07600000000000   0.31000000000000   H   (4d)
  -0.07600000000000   0.07600000000000   0.31000000000000   H   (4d)
   0.07600000000000   0.07600000000000   0.81000000000000   H   (4d)
  -0.07600000000000  -0.07600000000000   0.81000000000000   H   (4d)
   0.57600000000000   0.07600000000000   0.19000000000000   H   (4d)
  -0.57600000000000  -0.07600000000000   0.19000000000000   H   (4d)
  -0.07600000000000   0.57600000000000   0.69000000000000   H   (4d)
   0.07600000000000  -0.57600000000000   0.69000000000000   H   (4d)
   0.42400000000000   0.07600000000000   0.31000000000000   H   (4d)
  -0.42400000000000  -0.07600000000000   0.31000000000000   H   (4d)
  -0.07600000000000   0.42400000000000   0.81000000000000   H   (4d)
   0.07600000000000  -0.42400000000000   0.81000000000000   H   (4d)
   0.42400000000000   0.42400000000000   0.31000000000000   H   (4d)
  -0.42400000000000  -0.42400000000000   0.31000000000000   H   (4d)
  -0.42400000000000   0.42400000000000   0.81000000000000   H   (4d)
   0.42400000000000  -0.42400000000000   0.81000000000000   H   (4d)
   0.42400000000000   0.57600000000000   0.19000000000000   H   (4d)
  -0.42400000000000  -0.57600000000000   0.19000000000000   H   (4d)
  -0.57600000000000   0.42400000000000   0.69000000000000   H   (4d)
   0.57600000000000  -0.42400000000000   0.69000000000000   H   (4d)
  -0.07600000000000   0.42400000000000   0.31000000000000   H   (4d)
   0.07600000000000  -0.42400000000000   0.31000000000000   H   (4d)
  -0.42400000000000  -0.07600000000000   0.81000000000000   H   (4d)
   0.42400000000000   0.07600000000000   0.81000000000000   H   (4d)
  -0.07600000000000   0.57600000000000   0.19000000000000   H   (4d)
   0.07600000000000  -0.57600000000000   0.19000000000000   H   (4d)
  -0.57600000000000  -0.07600000000000   0.69000000000000   H   (4d)
   0.57600000000000   0.07600000000000   0.69000000000000   H   (4d)
   0.00000000000000   0.00000000000000   0.25000000000000   N   (2a)
   0.00000000000000   0.00000000000000   0.75000000000000   N   (2a)
   0.50000000000000   0.50000000000000   0.25000000000000   N   (2b)
   0.50000000000000   0.50000000000000   0.75000000000000   N   (2b)
   0.00000000000000   0.50000000000000   0.75000000000000   N   (2c)
   0.50000000000000   0.00000000000000   1.25000000000000   N   (2c)
   0.00000000000000   0.50000000000000   0.25000000000000   N   (2c)
   0.50000000000000   0.00000000000000   0.75000000000000   N   (2c)
   0.25000000000000   0.25000000000000   0.51800000000000   N   (4d)
  -0.25000000000000  -0.25000000000000   0.51800000000000   N   (4d)
  -0.25000000000000   0.25000000000000   1.01800000000000   N   (4d)
   0.25000000000000  -0.25000000000000   1.01800000000000   N   (4d)
   0.25000000000000   0.25000000000000   0.01800000000000   N   (4d)
  -0.25000000000000  -0.25000000000000   0.01800000000000   N   (4d)
  -0.25000000000000   0.25000000000000   0.51800000000000   N   (4d)
   0.25000000000000  -0.25000000000000   0.51800000000000   N   (4d)
   0.12000000000000   0.27000000000000   0.45600000000000   O   (4d)
  -0.12000000000000  -0.27000000000000   0.45600000000000   O   (4d)
  -0.27000000000000   0.12000000000000   0.95600000000000   O   (4d)
   0.27000000000000  -0.12000000000000   0.95600000000000   O   (4d)
   0.33000000000000   0.23000000000000   0.45600000000000   O   (4d)
  -0.33000000000000  -0.23000000000000   0.45600000000000   O   (4d)
  -0.23000000000000   0.33000000000000   0.95600000000000   O   (4d)
   0.23000000000000  -0.33000000000000   0.95600000000000   O   (4d)
   0.25000000000000   0.25000000000000   0.64200000000000   O   (4d)
  -0.25000000000000  -0.25000000000000   0.64200000000000   O   (4d)
  -0.25000000000000   0.25000000000000   1.14200000000000   O   (4d)
   0.25000000000000  -0.25000000000000   1.14200000000000   O   (4d)
   0.12000000000000   0.23000000000000  -0.04400000000000   O   (4d)
  -0.12000000000000  -0.23000000000000  -0.04400000000000   O   (4d)
  -0.23000000000000   0.12000000000000   0.45600000000000   O   (4d)
   0.23000000000000  -0.12000000000000   0.45600000000000   O   (4d)
   0.38000000000000   0.27000000000000  -0.04400000000000   O   (4d)
  -0.38000000000000  -0.27000000000000  -0.04400000000000   O   (4d)
  -0.27000000000000   0.38000000000000   0.45600000000000   O   (4d)
   0.27000000000000  -0.38000000000000   0.45600000000000   O   (4d)
   0.25000000000000   0.25000000000000   0.14200000000000   O   (4d)
  -0.25000000000000  -0.25000000000000   0.14200000000000   O   (4d)
  -0.25000000000000   0.25000000000000   0.64200000000000   O   (4d)
   0.25000000000000  -0.25000000000000   0.64200000000000   O   (4d)
```

Kesterite [$Cu_2(Zn,Fe)SnS_4$]: A2BCD4_tI16_82_ac_b_d_g - CIF

```
# AFLOW.org Repositories
# CuFeSSn/A2BCD4_tI16_82_ac_b_d_g-001.ABCD params=-1,2.00313248572,
      ↪ 0.7434,0.256,0.6278 SG=82 [ANRL doi: 10.1016/
      ↪ j.commatsci.2017.01.017 (part 1), doi: 10.1016/
      ↪ j.commatsci.2018.10.043 (part 2)]
data_CuFeSSn
_pd_phase_name A2BCD4_tI16_82_ac_b_d_g-001.ABCD

_chemical_name_mineral 'Kesterite'
_chemical_formula_sum 'Cu2 Fe S Sn4'

loop_
_publ_author_name
 'S. R. Hall'
 'J. T. Szyma{\\'n}ski'
 'J. M. Stewart'
_journal_name_full_name
;
 Canadian Mineralogist
;
_journal_volume 16
_journal_year 1978
_journal_page_first 131
_journal_page_last 137
_publ_Section_title
;
 Kesterite, Cu$_{2}$(Zn,Fe)SnS$_{4}$, and stannite, Cu$_{2}$(Fe,Zn)SnS$_
      ↪ {4}$, structurally similar but distinct minerals
;

_aflow_title 'Kesterite [Cu$_{2}$(Zn,Fe)SnS$_{4}$] Structure'
_aflow_proto 'A2BCD4_tI16_82_ac_b_d_g'
_aflow_params 'a,c/a,x_{5},y_{5},z_{5}'
_aflow_params_values '5.428437725,2.00313248571,0.7434,0.256,0.6278'
_aflow_Strukturbericht 'None'
_aflow_Pearson 'tI16'

_cell_length_a    5.4284377250
_cell_length_b    5.4284377250
_cell_length_c    10.8738799536
_cell_angle_alpha   90.0000000000
_cell_angle_beta    90.0000000000
_cell_angle_gamma   90.0000000000
_symmetry_space_group_name_H-M   'I-4'
_symmetry_Int_Tables_Number   82
loop_
_symmetry_equiv_pos_site_id
_symmetry_equiv_pos_as_xyz
  1  x,y,z
  2  -x,-y,z
  3  y,-x,-z
  4  -y,x,-z
  5  x+1/2,y+1/2,z+1/2
  6  -x+1/2,-y+1/2,z+1/2
  7  y+1/2,-x+1/2,-z+1/2
  8  -y+1/2,x+1/2,-z+1/2
loop_
 _atom_site_label
 _atom_site_occupancy
```

```
_atom_site_fract_x
_atom_site_fract_y
_atom_site_fract_z
_atom_site_thermal_displace_type
_atom_site_B_iso_or_equiv
_atom_site_type_symbol
_atom_site_symmetry_multiplicity
_atom_site_Wyckoff_label
Cu1 1.0000000000 -0.0000000000 0.0000000000 0.0000000000 Biso 1.0 Cu 2 a
Fe1 1.0000000000 0.0000000000 0.0000000000 0.5000000000 Biso 1.0 Fe 2 b
Cu2 1.0000000000 -0.0000000000 0.5000000000 0.2500000000 Biso 1.0 Cu 2 c
S1 1.0000000000 0.0000000000 0.5000000000 0.7500000000 Biso 1.0 S 2 d
Sn1 1.0000000000 0.7434000000 0.2560000000 0.6278000000 Biso 1.0 Sn 8 g
```

Kesterite [$Cu_2$(Zn,Fe)$SnS_4$]: A2BCD4_tI16_82_ac_b_d_g - POSCAR

```
A2BCD4_tI16_82_ac_b_d_g & a,c/a,x5,y5,z5 --params=5.428437725,
   ↪ 2.00313248571,0.7434,0.256,0.6278 & I-4 S_{4}^{2} #82 (abcdg) &
   ↪ tI16 & None & Cu2(Fe,Zn)SnS4 & Kesterite & S. R. Hall and J.
   ↪ T. Szyma{\'n}ski and J. M. Stewart, Can. Mineral. 16, 131-137 (
   ↪ 1978)
   1.00000000000000
  -2.71421886250000   2.71421886250000   5.43693997680000
   2.71421886250000  -2.71421886250000   5.43693997680000
   2.71421886250000   2.71421886250000  -5.43693997680000
    Cu    Fe     S    Sn
     2     1     1     4
Direct
   0.00000000000000   0.00000000000000   0.00000000000000   Cu   (2a)
   0.75000000000000   0.25000000000000   0.50000000000000   Cu   (2c)
   0.50000000000000   0.50000000000000   0.00000000000000   Fe   (2b)
   0.25000000000000   0.75000000000000   0.50000000000000    S   (2d)
   0.88380000000000   1.37120000000000   0.99940000000000   Sn   (8g)
   0.37180000000000  -0.11560000000000  -0.99940000000000   Sn   (8g)
  -1.37120000000000  -0.37180000000000  -0.48740000000000   Sn   (8g)
   0.11560000000000  -0.88380000000000   0.48740000000000   Sn   (8g)
```

Bromocarnallite (KMg($H_2O$)$_6$(Cl,Br)$_3$, $E2_6$): A3B6CD_tP44_85_bcg_3g_ac_e - CIF

```
# CIF file
data_findsym-output
_audit_creation_method FINDSYM

_chemical_name_mineral 'Bromocarnallite'
_chemical_formula_sum 'Br3 (H2O)6 K Mg'

loop_
_publ_author_name
 'K. R. Andre{\ss}'
 'O. Saffe'
_journal_name_full_name
;
 Zeitschrift f{\"u}r Kristallographie - Crystalline Materials
;
_journal_volume 101
_journal_year 1939
_journal_page_first 451
_journal_page_last 469
_publ_Section_title
;
 R\"{o}ntgenographische Untersuchung der Mischkristallreihe
     ↪ Karnallit-Bromkarnallit
;

# Found in Strukturbericht Band VII 1939, 1943

_aflow_title 'Bromocarnallite (KMg(H$_{2}$O)$_{6}$(Cl,Br)$_{3}$, $E2_{6}
   ↪ $) Structure'
_aflow_proto 'A3B6CD_tP44_85_bcg_3g_ac_e'
_aflow_params 'a,c/a,z_{3},z_{4},x_{6},y_{6},z_{6},x_{7},y_{7},z_{7},x_{
   ↪ 8},y_{8},z_{8},x_{9},y_{9},z_{9}'
_aflow_params_values '13.51,0.501110288675,0.58,0.08,0.25,0.0,0.008,0.55
   ↪ ,0.05,0.78,0.54,0.85,0.58,0.34,-0.03,0.58'
_aflow_Strukturbericht '$E2_{6}$'
_aflow_Pearson 'tP44'

_symmetry_space_group_name_H-M "P 4/n (origin choice 2)"
_symmetry_Int_Tables_number 85

_cell_length_a     13.51000
_cell_length_b     13.51000
_cell_length_c     6.77000
_cell_angle_alpha  90.00000
_cell_angle_beta   90.00000
_cell_angle_gamma  90.00000

loop_
_space_group_symop_id
_space_group_symop_operation_xyz
1 x,y,z
2 -x+1/2,-y+1/2,z
3 -y+1/2,x,z
4 y,-x+1/2,z
5 -x,-y,-z
6 x+1/2,y+1/2,-z
7 y+1/2,-x,-z
8 -y,x+1/2,-z

loop_
_atom_site_label
_atom_site_type_symbol
_atom_site_symmetry_multiplicity
_atom_site_Wyckoff_label
_atom_site_fract_x
_atom_site_fract_y
_atom_site_fract_z
_atom_site_occupancy
K1 K 2 a 0.25000 0.75000 0.00000 1.00000
Br1 Br 2 b 0.25000 0.75000 0.50000 1.00000
Br2 Br 2 c 0.25000 0.25000 0.58000 1.00000
K2 K 2 c 0.25000 0.25000 0.08000 1.00000
Mg1 Mg 4 e 0.00000 0.00000 0.50000 1.00000
Br3 Br 8 g 0.25000 0.00000 0.00800 1.00000
H2O1 H2O 8 g 0.55000 0.05000 0.78000 1.00000
H2O2 H2O 8 g 0.54000 0.85000 0.58000 1.00000
H2O3 H2O 8 g 0.34000 -0.03000 0.58000 1.00000
```

Bromocarnallite (KMg($H_2O$)$_6$(Cl,Br)$_3$, $E2_6$): A3B6CD_tP44_85_bcg_3g_ac_e - POSCAR

```
A3B6CD_tP44_85_bcg_3g_ac_e & a,c/a,z3,z4,x6,y6,z6,x7,y7,z7,x8,y8,z8,x9,
   ↪ y9,z9 --params=13.51,0.501110288675,0.58,0.08,0.25,0.0,0.008,
   ↪ 0.55,0.05,0.78,0.54,0.85,0.58,0.34,-0.03,0.58 & P4/n C_{4h}^{3}
   ↪ #85 (abc^2eg^4) & tP44 & $E2_{6}$ & (Br,Cl)3(H2O)6KMg &
   ↪ Bromocarnallite & K. R. Andre{\ss} and O. Saffe, Zeitschrift f
   ↪ {"u}r Kristallographie - Crystalline Materials 101, 451-469 (
   ↪ 1939)
   1.00000000000000
  13.51000000000000   0.00000000000000   0.00000000000000
   0.00000000000000  13.51000000000000   0.00000000000000
   0.00000000000000   0.00000000000000   6.77000000000000
    Br   H2O     K    Mg
    12    24     4     4
Direct
   0.25000000000000   0.75000000000000   0.50000000000000   Br   (2b)
   0.75000000000000   0.25000000000000   0.50000000000000   Br   (2b)
   0.25000000000000   0.25000000000000   0.58000000000000   Br   (2c)
   0.75000000000000   0.75000000000000  -0.58000000000000   Br   (2c)
   0.25000000000000   0.00000000000000   0.00800000000000   Br   (8g)
   0.25000000000000   0.50000000000000   0.00800000000000   Br   (8g)
   0.50000000000000   0.25000000000000   0.00800000000000   Br   (8g)
   0.00000000000000   0.25000000000000   0.00800000000000   Br   (8g)
  -0.25000000000000   0.00000000000000  -0.00800000000000   Br   (8g)
   0.75000000000000   0.50000000000000  -0.00800000000000   Br   (8g)
   0.50000000000000  -0.25000000000000  -0.00800000000000   Br   (8g)
   0.00000000000000   0.75000000000000  -0.00800000000000   Br   (8g)
   0.55000000000000   0.05000000000000   0.78000000000000  H2O   (8g)
  -0.05000000000000   0.45000000000000   0.78000000000000  H2O   (8g)
   0.45000000000000   0.55000000000000   0.78000000000000  H2O   (8g)
   0.05000000000000  -0.05000000000000   0.78000000000000  H2O   (8g)
  -0.55000000000000  -0.05000000000000  -0.78000000000000  H2O   (8g)
   1.05000000000000   0.55000000000000  -0.78000000000000  H2O   (8g)
   0.55000000000000  -0.55000000000000  -0.78000000000000  H2O   (8g)
  -0.05000000000000   1.05000000000000  -0.78000000000000  H2O   (8g)
   0.54000000000000   0.85000000000000   0.58000000000000  H2O   (8g)
  -0.04000000000000  -0.35000000000000   0.58000000000000  H2O   (8g)
  -0.35000000000000   0.54000000000000   0.58000000000000  H2O   (8g)
   0.85000000000000  -0.04000000000000   0.58000000000000  H2O   (8g)
  -0.54000000000000  -0.85000000000000  -0.58000000000000  H2O   (8g)
   1.04000000000000   1.35000000000000  -0.58000000000000  H2O   (8g)
   1.35000000000000  -0.54000000000000  -0.58000000000000  H2O   (8g)
  -0.85000000000000   1.04000000000000  -0.58000000000000  H2O   (8g)
   0.34000000000000  -0.03000000000000   0.58000000000000  H2O   (8g)
   0.16000000000000   0.53000000000000   0.58000000000000  H2O   (8g)
   0.53000000000000   0.34000000000000   0.58000000000000  H2O   (8g)
  -0.03000000000000   0.16000000000000   0.58000000000000  H2O   (8g)
  -0.34000000000000   0.03000000000000  -0.58000000000000  H2O   (8g)
   0.84000000000000   0.47000000000000  -0.58000000000000  H2O   (8g)
   0.47000000000000  -0.34000000000000  -0.58000000000000  H2O   (8g)
   0.03000000000000   0.84000000000000  -0.58000000000000  H2O   (8g)
   0.25000000000000   0.75000000000000   0.00000000000000    K   (2a)
   0.75000000000000   0.25000000000000   0.00000000000000    K   (2a)
   0.25000000000000   0.25000000000000   0.08000000000000    K   (2c)
   0.75000000000000   0.75000000000000  -0.08000000000000    K   (2c)
   0.00000000000000   0.00000000000000   0.50000000000000   Mg   (4e)
   0.50000000000000   0.50000000000000   0.50000000000000   Mg   (4e)
   0.50000000000000   0.00000000000000   0.50000000000000   Mg   (4e)
   0.00000000000000   0.50000000000000   0.50000000000000   Mg   (4e)
```

$MoPO_5$: AB5C_tP14_85_c_cg_b - CIF

```
# CIF file
data_findsym-output
_audit_creation_method FINDSYM

_chemical_name_mineral 'MoO5P'
_chemical_formula_sum 'Mo O5 P'

loop_
_publ_author_name
 'P. Kierkegaard'
 'M. Westerlund'
_journal_name_full_name
;
 Acta Chemica Scandinavica
;
_journal_volume 18
_journal_year 1964
_journal_page_first 2217
_journal_page_last 2225
_publ_Section_title
;
 The Crystal Structure of MoOPO$_{4}$
;

_aflow_title 'MoPO$_{5}$ Structure'
_aflow_proto 'AB5C_tP14_85_c_cg_b'
_aflow_params 'a,c/a,z_{2},z_{3},x_{4},y_{4},z_{4}'
_aflow_params_values '6.1768,0.695052454345,0.1975,0.8102,0.3125,-0.0554
   ↪ ,0.2994'
_aflow_Strukturbericht 'None'
```

```
_aflow_Pearson 'tP14'

_symmetry_space_group_name_H-M "P 4/n (origin choice 2)"
_symmetry_Int_Tables_number 85

_cell_length_a     6.17680
_cell_length_b     6.17680
_cell_length_c     4.29320
_cell_angle_alpha 90.00000
_cell_angle_beta  90.00000
_cell_angle_gamma 90.00000

loop_
_space_group_symop_id
_space_group_symop_operation_xyz
1 x,y,z
2 -x+1/2,-y+1/2,z
3 -y+1/2,x,z
4 y,-x+1/2,z
5 -x,-y,-z
6 x+1/2,y+1/2,-z
7 y+1/2,-x,-z
8 -y,x+1/2,-z

loop_
_atom_site_label
_atom_site_type_symbol
_atom_site_symmetry_multiplicity
_atom_site_Wyckoff_label
_atom_site_fract_x
_atom_site_fract_y
_atom_site_fract_z
_atom_site_occupancy
P1 P 2 b 0.25000 0.75000 0.50000 1.00000
Mo1 Mo 2 c 0.25000 0.25000 0.19750 1.00000
O1 O 2 c 0.25000 0.25000 0.81020 1.00000
O2 O 8 g 0.31250 -0.05540 0.29940 1.00000
```

$MoPO_5$: AB5C_tP14_85_c_cg_b - POSCAR

```
AB5C_tP14_85_c_cg_b & a,c/a,z2,z3,x4,y4,z4 --params=6.1768,
    ↪ 0.695052454345,0.1975,0.8102,0.3125,-0.0554,0.2994 & P4/n C_{4h
    ↪ }^{3} #85 (bc^2g) & tP14 & None & MoO5P & MoO5P & P.
    ↪ Kierkegaard and M. Westerlund, Acta Chem. Scand. 18, 2217-2225
    ↪ (1964)
   1.00000000000000
   6.17680000000000   0.00000000000000   0.00000000000000
   0.00000000000000   6.17680000000000   0.00000000000000
   0.00000000000000   0.00000000000000   4.29320000000000
    Mo      O      P
     2     10      2
Direct
   0.25000000000000   0.25000000000000   0.19750000000000   Mo   (2c)
   0.75000000000000   0.75000000000000  -0.19750000000000   Mo   (2c)
   0.25000000000000   0.25000000000000   0.81020000000000   O    (2c)
   0.75000000000000   0.75000000000000  -0.81020000000000   O    (2c)
   0.31250000000000  -0.05540000000000   0.29940000000000   O    (8g)
   0.18750000000000   0.55540000000000   0.29940000000000   O    (8g)
   0.55540000000000   0.31250000000000   0.29940000000000   O    (8g)
  -0.05540000000000   0.18750000000000   0.29940000000000   O    (8g)
  -0.31250000000000   0.05540000000000  -0.29940000000000   O    (8g)
   0.81250000000000   0.44460000000000  -0.29940000000000   O    (8g)
   0.44460000000000  -0.31250000000000  -0.29940000000000   O    (8g)
   0.05540000000000   0.81250000000000  -0.29940000000000   O    (8g)
   0.25000000000000   0.75000000000000   0.50000000000000   P    (2b)
   0.75000000000000   0.25000000000000   0.50000000000000   P    (2b)
```

$PNCl_2$ ($E1_4$): A2BC_tP32_86_2g_g_g - CIF

```
# CIF file
data_findsym-output
_audit_creation_method FINDSYM

_chemical_name_mineral 'Cl2NP'
_chemical_formula_sum 'Cl2 N P'

loop_
_publ_author_name
 'J. A. A. Ketelaar'
 'T. A. {de Vries}'
_journal_name_full_name
;
 Recueil des Travaux Chimiques des Pays-Bas
;
_journal_volume 58
_journal_year 1939
_journal_page_first 1081
_journal_page_last 1099
_publ_Section_title
;
 The crystal structure of tetra phosphonitrile chloride, P$_{4}$N$_{4}
    ↪ $Cl$_{8}$
;

_aflow_title 'PNCl$_{2}$ ($E1_{4}$) Structure'
_aflow_proto 'A2BC_tP32_86_2g_g_g'
_aflow_params 'a,c/a,x_{1},y_{1},z_{1},x_{2},y_{2},z_{2},x_{3},y_{3},z_{
    ↪ 3},x_{4},y_{4},z_{4}'
_aflow_params_values '10.82,0.549907578558,0.551,0.852,0.86,0.594,0.198,
    ↪ 0.69,0.625,0.645,0.14,0.574,0.791,0.17'
_aflow_Strukturbericht '$E1_{4}$'
_aflow_Pearson 'tP32'

_symmetry_space_group_name_H-M "P 42/n (origin choice 2)"
_symmetry_Int_Tables_number 86

_cell_length_a     10.82000
_cell_length_b     10.82000
_cell_length_c     5.95000
_cell_angle_alpha 90.00000
_cell_angle_beta  90.00000
_cell_angle_gamma 90.00000

loop_
_space_group_symop_id
_space_group_symop_operation_xyz
1 x,y,z
2 -x+1/2,-y+1/2,z
3 -y,x+1/2,z+1/2
4 y+1/2,-x,z+1/2
5 -x,-y,-z
6 x+1/2,y+1/2,-z
7 y,-x+1/2,-z+1/2
8 -y+1/2,x,-z+1/2

loop_
_atom_site_label
_atom_site_type_symbol
_atom_site_symmetry_multiplicity
_atom_site_Wyckoff_label
_atom_site_fract_x
_atom_site_fract_y
_atom_site_fract_z
_atom_site_occupancy
Cl1 Cl 8 g 0.55100 0.85200 0.86000 1.00000
Cl2 Cl 8 g 0.59400 0.19800 0.69000 1.00000
N1 N 8 g 0.62500 0.64500 0.14000 1.00000
P1 P 8 g 0.57400 0.79100 0.17000 1.00000
```

$PNCl_2$ ($E1_4$): A2BC_tP32_86_2g_g_g - POSCAR

```
A2BC_tP32_86_2g_g_g & a,c/a,x1,y1,z1,x2,y2,z2,x3,y3,z3,x4,y4,z4 --params
    ↪ =10.82,0.549907578558,0.551,0.852,0.86,0.594,0.198,0.69,0.625,
    ↪ 0.645,0.14,0.574,0.791,0.17 & P4_{2}/n C_{4h}^{4} #86 (g^4) &
    ↪ tP32 & $E1_{4}$ & Cl2NP & Cl2NP & J. A. A. Ketelaar and T. A. {
    ↪ de Vries}, Rec. Trav. Chim. Pays-Bas 58, 1081-1099 (1939)
   1.00000000000000
  10.82000000000000   0.00000000000000   0.00000000000000
   0.00000000000000  10.82000000000000   0.00000000000000
   0.00000000000000   0.00000000000000   5.95000000000000
    Cl       N       P
    16       8       8
Direct
   0.55100000000000   0.85200000000000   0.86000000000000   Cl   (8g)
  -0.05100000000000  -0.35200000000000   0.86000000000000   Cl   (8g)
  -0.85200000000000   1.05100000000000   1.36000000000000   Cl   (8g)
   1.35200000000000  -0.55100000000000   1.36000000000000   Cl   (8g)
  -0.55100000000000  -0.85200000000000  -0.86000000000000   Cl   (8g)
   1.05100000000000   1.35200000000000  -0.86000000000000   Cl   (8g)
   0.85200000000000  -0.05100000000000  -0.36000000000000   Cl   (8g)
  -0.35200000000000   0.55100000000000  -0.36000000000000   Cl   (8g)
   0.59400000000000   0.19800000000000   0.69000000000000   Cl   (8g)
  -0.09400000000000   0.30200000000000   0.69000000000000   Cl   (8g)
  -0.19800000000000   1.09400000000000   1.19000000000000   Cl   (8g)
   0.69800000000000  -0.59400000000000   1.19000000000000   Cl   (8g)
  -0.59400000000000  -0.19800000000000  -0.69000000000000   Cl   (8g)
   1.09400000000000   0.69800000000000  -0.69000000000000   Cl   (8g)
   0.19800000000000  -0.09400000000000  -0.19000000000000   Cl   (8g)
   0.30200000000000   0.59400000000000  -0.19000000000000   Cl   (8g)
   0.62500000000000   0.64500000000000   0.14000000000000   N    (8g)
  -0.12500000000000  -0.14500000000000   0.14000000000000   N    (8g)
  -0.64500000000000   1.12500000000000   0.64000000000000   N    (8g)
   1.14500000000000  -0.62500000000000   0.64000000000000   N    (8g)
  -0.62500000000000  -0.64500000000000  -0.14000000000000   N    (8g)
   1.12500000000000   1.14500000000000  -0.14000000000000   N    (8g)
   0.64500000000000  -0.12500000000000   0.36000000000000   N    (8g)
  -0.14500000000000   0.62500000000000   0.36000000000000   N    (8g)
   0.57400000000000   0.79100000000000   0.17000000000000   P    (8g)
  -0.07400000000000  -0.29100000000000   0.17000000000000   P    (8g)
  -0.79100000000000   1.07400000000000   0.67000000000000   P    (8g)
   1.29100000000000  -0.57400000000000   0.67000000000000   P    (8g)
  -0.57400000000000  -0.79100000000000  -0.17000000000000   P    (8g)
   1.07400000000000   1.29100000000000  -0.17000000000000   P    (8g)
   0.79100000000000  -0.07400000000000   0.33000000000000   P    (8g)
  -0.29100000000000   0.57400000000000   0.33000000000000   P    (8g)
```

$Nd_4Re_2O_{11}$: A4B11C2_tP68_86_2g_ab5g_g - CIF

```
# CIF file
data_findsym-output
_audit_creation_method FINDSYM

_chemical_name_mineral 'Nd4O11Re2'
_chemical_formula_sum 'Nd4 O11 Re2'

loop_
_publ_author_name
 'K.-A. Wilhelmi'
 'E. Lagervall'
 'O. Muller'
_journal_name_full_name
;
 Acta Chemica Scandinavica
;
_journal_volume 24
_journal_year 1970
_journal_page_first 3406
_journal_page_last 3408
_publ_Section_title
;
```

```
On the Crystal Structure of Nd$_{4}$Re$_{2}$O$_{11}$
;

# Found in The American Mineralogist Crystal Structure Database, 2003

_aflow_title 'Nd$_{4}$Re$_{2}$O$_{11}$ Structure'
_aflow_proto 'A4B11C2_tP68_86_2g_ab5g_g'
_aflow_params 'a,c/a,x_{3},y_{3},z_{3},x_{4},y_{4},z_{4},x_{5},y_{5},z_{
    ↪ 5},x_{6},y_{6},z_{6},x_{7},y_{7},z_{7},x_{8},y_{8},z_{8},x_{9},
    ↪ y_{9},z_{9},x_{10},y_{10},z_{10}'
_aflow_params_values '12.676,0.441858630483,0.184,0.1182,0.4979,0.1107,
    ↪ 0.8064,0.603,0.0301,0.1938,0.7009,0.0013,0.4086,0.6937,0.0415,
    ↪ 0.6291,0.7905,0.1484,0.8032,-0.0057,0.4559,0.34,0.1868,0.0247,-
    ↪ 0.087,0.0707'
_aflow_Strukturbericht 'None'
_aflow_Pearson 'tP68'

_symmetry_space_group_name_H-M "P 42/n (origin choice 2)"
_symmetry_Int_Tables_number 86

_cell_length_a     12.67600
_cell_length_b     12.67600
_cell_length_c     5.60100
_cell_angle_alpha  90.00000
_cell_angle_beta   90.00000
_cell_angle_gamma  90.00000

loop_
_space_group_symop_id
_space_group_symop_operation_xyz
1 x,y,z
2 -x+1/2,-y+1/2,z
3 -y,x+1/2,z+1/2
4 y+1/2,-x,z+1/2
5 -x,-y,-z
6 x+1/2,y+1/2,-z
7 y,-x+1/2,-z+1/2
8 -y+1/2,x,-z+1/2

loop_
_atom_site_label
_atom_site_type_symbol
_atom_site_symmetry_multiplicity
_atom_site_Wyckoff_label
_atom_site_fract_x
_atom_site_fract_y
_atom_site_fract_z
_atom_site_occupancy
O1 O 2 a 0.25000 0.25000 0.25000 1.00000
O2 O 2 b 0.25000 0.25000 0.75000 1.00000
Nd1 Nd 8 g 0.18400 0.11820 0.49790 1.00000
Nd2 Nd 8 g 0.11070 0.80640 0.60300 1.00000
O3 O 8 g 0.03010 0.19380 0.70090 1.00000
O4 O 8 g 0.00130 0.40860 0.69370 1.00000
O5 O 8 g 0.04150 0.62910 0.79050 1.00000
O6 O 8 g 0.14840 0.80320 -0.00570 1.00000
O7 O 8 g 0.45590 0.34000 0.18680 1.00000
Re1 Re 8 g 0.02470 -0.08700 0.07070 1.00000
```

$Nd_4Re_2O_{11}$: A4B11C2_tP68_86_2g_ab5g_g - POSCAR

```
A4B11C2_tP68_86_2g_ab5g_g & a,c/a,x3,y3,z3,x4,y4,z4,x5,y5,z5,x6,y6,z6,x7
    ↪ ,y7,z7,x8,y8,z8,x9,y9,z9,x10,y10,z10 --params=12.676,
    ↪ 0.441858630483,0.184,0.1182,0.4979,0.1107,0.8064,0.603,0.0301,
    ↪ 0.1938,0.7009,0.0013,0.4086,0.6937,0.0415,0.6291,0.7905,0.1484,
    ↪ 0.8032,-0.0057,0.4559,0.34,0.1868,0.0247,-0.087,0.0707 & P4_{2
    ↪ }/n C_{4h}^{4} #86 (abg^8) & tP68 & None & Nd4O11Re2 &
    ↪ Nd4O11Re2 & K.-A. Wilhelmi and E. Lagervall and O. Muller, Acta
    ↪ Chem. Scand. 24, 3406-3408 (1970)
   1.00000000000000
  12.67600000000000   0.00000000000000   0.00000000000000
   0.00000000000000  12.67600000000000   0.00000000000000
   0.00000000000000   0.00000000000000   5.60100000000000
    Nd     O    Re
    16    44     8
Direct
   0.18400000000000   0.11820000000000   0.49790000000000   Nd   (8g)
   0.31600000000000   0.38180000000000   0.49790000000000   Nd   (8g)
  -0.11820000000000   0.68400000000000   0.99790000000000   Nd   (8g)
   0.61820000000000  -0.18400000000000   0.99790000000000   Nd   (8g)
  -0.18400000000000  -0.11820000000000  -0.49790000000000   Nd   (8g)
   0.68400000000000   0.61820000000000  -0.49790000000000   Nd   (8g)
   0.11820000000000   0.31600000000000   0.00210000000000   Nd   (8g)
   0.38180000000000   0.18400000000000   0.00210000000000   Nd   (8g)
   0.11070000000000   0.80640000000000   0.60300000000000   Nd   (8g)
   0.38930000000000  -0.30640000000000   0.60300000000000   Nd   (8g)
  -0.80640000000000   0.61070000000000   1.10300000000000   Nd   (8g)
   1.30640000000000  -0.11070000000000   1.10300000000000   Nd   (8g)
  -0.11070000000000  -0.80640000000000  -0.60300000000000   Nd   (8g)
   0.61070000000000   1.30640000000000  -0.60300000000000   Nd   (8g)
   0.80640000000000   0.38930000000000  -0.10300000000000   Nd   (8g)
  -0.30640000000000   0.11070000000000  -0.10300000000000   Nd   (8g)
   0.25000000000000   0.25000000000000   0.25000000000000    O   (2a)
   0.75000000000000   0.75000000000000   0.75000000000000    O   (2a)
   0.25000000000000   0.25000000000000   0.75000000000000    O   (2b)
   0.75000000000000   0.75000000000000   0.25000000000000    O   (2b)
   0.03010000000000   0.19380000000000   0.70090000000000    O   (8g)
   0.46990000000000   0.30620000000000   0.70090000000000    O   (8g)
  -0.19380000000000   0.53010000000000   1.20090000000000    O   (8g)
   0.69380000000000  -0.03010000000000   1.20090000000000    O   (8g)
  -0.03010000000000  -0.19380000000000  -0.70090000000000    O   (8g)
   0.53010000000000   0.69380000000000  -0.70090000000000    O   (8g)
   0.19380000000000   0.46990000000000  -0.20090000000000    O   (8g)
   0.30620000000000   0.03010000000000  -0.20090000000000    O   (8g)
   0.00130000000000   0.40860000000000   0.69370000000000    O   (8g)
   0.49870000000000   0.09140000000000   0.69370000000000    O   (8g)
  -0.40860000000000   0.50130000000000   1.19370000000000    O   (8g)
   0.90860000000000  -0.00130000000000   1.19370000000000    O   (8g)
  -0.00130000000000  -0.40860000000000  -0.69370000000000    O   (8g)
   0.50130000000000   0.90860000000000  -0.69370000000000    O   (8g)
   0.40860000000000   0.49870000000000  -0.19370000000000    O   (8g)
   0.09140000000000   0.00130000000000  -0.19370000000000    O   (8g)
   0.04150000000000   0.62910000000000   0.79050000000000    O   (8g)
   0.45850000000000  -0.12910000000000   0.79050000000000    O   (8g)
  -0.62910000000000   0.54150000000000   1.29050000000000    O   (8g)
   1.12910000000000  -0.04150000000000   1.29050000000000    O   (8g)
  -0.04150000000000  -0.62910000000000  -0.79050000000000    O   (8g)
   0.54150000000000   1.12910000000000  -0.79050000000000    O   (8g)
   0.62910000000000   0.45850000000000  -0.29050000000000    O   (8g)
  -0.12910000000000   0.04150000000000  -0.29050000000000    O   (8g)
   0.14840000000000   0.80320000000000  -0.00570000000000    O   (8g)
   0.35160000000000  -0.30320000000000  -0.00570000000000    O   (8g)
  -0.80320000000000   0.64840000000000   0.49430000000000    O   (8g)
   1.30320000000000  -0.14840000000000   0.49430000000000    O   (8g)
  -0.14840000000000  -0.80320000000000   0.00570000000000    O   (8g)
   0.64840000000000   1.30320000000000   0.00570000000000    O   (8g)
   0.80320000000000   0.35160000000000   0.50570000000000    O   (8g)
  -0.30320000000000   0.14840000000000   0.50570000000000    O   (8g)
   0.45590000000000   0.34000000000000   0.18680000000000    O   (8g)
   0.04410000000000   0.16000000000000   0.18680000000000    O   (8g)
  -0.34000000000000   0.95590000000000   0.68680000000000    O   (8g)
   0.84000000000000  -0.45590000000000   0.68680000000000    O   (8g)
  -0.45590000000000  -0.34000000000000  -0.18680000000000    O   (8g)
   0.95590000000000   0.84000000000000  -0.18680000000000    O   (8g)
   0.34000000000000   0.04410000000000   0.31320000000000    O   (8g)
   0.16000000000000   0.45590000000000   0.31320000000000    O   (8g)
   0.02470000000000  -0.08700000000000   0.07070000000000   Re   (8g)
   0.47530000000000   0.58700000000000   0.07070000000000   Re   (8g)
   0.08700000000000   0.52470000000000   0.57070000000000   Re   (8g)
   0.41300000000000  -0.02470000000000   0.57070000000000   Re   (8g)
  -0.02470000000000   0.08700000000000  -0.07070000000000   Re   (8g)
   0.52470000000000   0.41300000000000  -0.07070000000000   Re   (8g)
  -0.08700000000000   0.47530000000000   0.42930000000000   Re   (8g)
   0.58700000000000   0.02470000000000   0.42930000000000   Re   (8g)
```

$NaSb(OH)_6$ ($J1_{11}$): AB6C_tP32_86_d_3g_c - CIF

```
# CIF file
data_findsym-output
_audit_creation_method FINDSYM

_chemical_name_mineral 'Na(OH)6Sb'
_chemical_formula_sum 'Na O6 Sb'

loop_
_publ_author_name
 'T. Asai'
_journal_name_full_name
;
 Bulletin of the Chemical Society of Japan
;
_journal_volume 48
_journal_year 1975
_journal_page_first 2677
_journal_page_last 2679
_publ_Section_title
;
 Refinement of the Crystal Structure of Sodium Hexahydroxoantimonate(V),
    ↪ NaSb(OH)$_{6}$
;

# Found in The Fascination of Crystals and Symmetry, 2014 Found in The
    ↪ Fascination of Crystals and Symmetry, {NaSb(OH)$_{6}$},

_aflow_title 'NaSb(OH)$_{6}$ ($J1_{11}$) Structure'
_aflow_proto 'AB6C_tP32_86_d_3g_c'
_aflow_params 'a,c/a,x_{3},y_{3},z_{3},x_{4},y_{4},z_{4},x_{5},y_{5},z_{
    ↪ 5}'
_aflow_params_values '8.029,0.983185950928,0.0558,0.2226,-0.0982,0.2792,
    ↪ 0.582,-0.0828,0.0895,0.0685,0.2219'
_aflow_Strukturbericht '$J1_{11}$'
_aflow_Pearson 'tP32'

_symmetry_space_group_name_H-M "P 42/n (origin choice 2)"
_symmetry_Int_Tables_number 86

_cell_length_a     8.02900
_cell_length_b     8.02900
_cell_length_c     7.89400
_cell_angle_alpha  90.00000
_cell_angle_beta   90.00000
_cell_angle_gamma  90.00000

loop_
_space_group_symop_id
_space_group_symop_operation_xyz
1 x,y,z
2 -x+1/2,-y+1/2,z
3 -y,x+1/2,z+1/2
4 y+1/2,-x,z+1/2
5 -x,-y,-z
6 x+1/2,y+1/2,-z
7 y,-x+1/2,-z+1/2
8 -y+1/2,x,-z+1/2

loop_
_atom_site_label
_atom_site_type_symbol
_atom_site_symmetry_multiplicity
_atom_site_Wyckoff_label
```

```
_atom_site_fract_x
_atom_site_fract_y
_atom_site_fract_z
_atom_site_occupancy
Sb1 Sb 4 c 0.00000 0.00000 0.00000 1.00000
Na1 Na 4 d 0.00000 0.00000 0.50000 1.00000
O1 O 8 g 0.05580 0.22260 -0.09820 1.00000
O2 O 8 g 0.27920 0.58200 -0.08280 1.00000
O3 O 8 g 0.08950 0.06850 0.22190 1.00000
```

NaSb(OH)$_6$ ($J1_{11}$): AB6C_tP32_86_d_3g_c - POSCAR

```
AB6C_tP32_86_d_3g_c & a,c/a,x3,y3,z3,x4,y4,z4,x5,y5,z5 --params=8.029,
    ↪ 0.983185950928,0.0558,0.2226,-0.0982,0.2792,0.582,-0.0828,
    ↪ 0.0895,0.0685,0.2219 & P4_{2}/n C_{4h}^{4} #86 (cdg^3) & tP32 &
    ↪ $J1_{11}$ & Na(OH)6Sb & Na(OH)6Sb & T. Asai, Bull. Chem. Soc.
    ↪ Jpn. 48, 2677-2679 (1975)
   1.00000000000000
   8.02900000000000   0.00000000000000   0.00000000000000
   0.00000000000000   8.02900000000000   0.00000000000000
   0.00000000000000   0.00000000000000   7.89400000000000
    Na         O        Sb
     4        24         4
Direct
   0.00000000000000   0.00000000000000   0.50000000000000   Na   (4d)
   0.50000000000000   0.50000000000000   0.50000000000000   Na   (4d)
   0.00000000000000   0.50000000000000   0.00000000000000   Na   (4d)
   0.50000000000000   0.00000000000000   0.00000000000000   Na   (4d)
   0.05580000000000   0.22260000000000  -0.09820000000000   O    (8g)
   0.44420000000000   0.27740000000000  -0.09820000000000   O    (8g)
  -0.22260000000000   0.55580000000000   0.40180000000000   O    (8g)
   0.72260000000000  -0.05580000000000   0.40180000000000   O    (8g)
  -0.05580000000000  -0.22260000000000   0.09820000000000   O    (8g)
   0.55580000000000   0.72260000000000   0.09820000000000   O    (8g)
   0.22260000000000   0.44420000000000   0.59820000000000   O    (8g)
   0.27740000000000   0.05580000000000   0.59820000000000   O    (8g)
   0.27920000000000   0.58200000000000  -0.08280000000000   O    (8g)
   0.22080000000000  -0.08200000000000  -0.08280000000000   O    (8g)
  -0.58200000000000   0.77920000000000   0.41720000000000   O    (8g)
   1.08200000000000  -0.27920000000000   0.41720000000000   O    (8g)
  -0.27920000000000  -0.58200000000000   0.08280000000000   O    (8g)
   0.77920000000000   1.08200000000000   0.08280000000000   O    (8g)
   0.58200000000000   0.22080000000000   0.58280000000000   O    (8g)
  -0.08200000000000   0.27920000000000   0.58280000000000   O    (8g)
   0.08950000000000   0.06850000000000   0.22190000000000   O    (8g)
   0.41050000000000   0.43150000000000   0.22190000000000   O    (8g)
  -0.06850000000000   0.58950000000000   0.72190000000000   O    (8g)
   0.56850000000000  -0.08950000000000   0.72190000000000   O    (8g)
  -0.08950000000000  -0.06850000000000  -0.22190000000000   O    (8g)
   0.58950000000000   0.56850000000000  -0.22190000000000   O    (8g)
   0.06850000000000   0.41050000000000   0.27810000000000   O    (8g)
   0.43150000000000   0.08950000000000   0.27810000000000   O    (8g)
   0.00000000000000   0.00000000000000   0.00000000000000   Sb   (4c)
   0.50000000000000   0.50000000000000   0.00000000000000   Sb   (4c)
   0.00000000000000   0.50000000000000   0.50000000000000   Sb   (4c)
   0.50000000000000   0.00000000000000   0.50000000000000   Sb   (4c)
```

$\beta$-LiIO$_3$: ABC3_tP40_86_g_g_3g - CIF

```
# CIF file
data_findsym-output
_audit_creation_method FINDSYM

_chemical_name_mineral 'ILiO3'
_chemical_formula_sum 'I Li O3'

loop_
_publ_author_name
 'H. Schulz'
_journal_name_full_name
;
 Acta Crystallographica Section B: Structural Science
;
_journal_volume 29
_journal_year 1973
_journal_page_first 2285
_journal_page_last 2289
_publ_Section_title
;
 The structure of $\beta$-LiIO$_{3}$
;

_aflow_title '$\beta$-LiIO$_{3}$ Structure'
_aflow_proto 'ABC3_tP40_86_g_g_3g'
_aflow_params 'a,c/a,x_{1},y_{1},z_{1},x_{2},y_{2},z_{2},x_{3},y_{3},z_{
    ↪ 3},x_{4},y_{4},z_{4},x_{5},y_{5},z_{5}'
_aflow_params_values '9.7329,0.63255555898,0.0343,0.7576,0.115,0.426,
    ↪ 0.236,0.426,0.836,0.121,-0.05,0.094,0.209,0.094,0.847,0.552,
    ↪ 0.172'
_aflow_Strukturbericht 'None'
_aflow_Pearson 'tP40'

_symmetry_space_group_name_H-M "P 42/n (origin choice 2)"
_symmetry_Int_Tables_number 86

_cell_length_a     9.73290
_cell_length_b     9.73290
_cell_length_c     6.15660
_cell_angle_alpha 90.00000
_cell_angle_beta  90.00000
_cell_angle_gamma 90.00000

loop_
_space_group_symop_id
_space_group_symop_operation_xyz
1 x,y,z
2 -x+1/2,-y+1/2,z
3 -y,x+1/2,z+1/2
4 y+1/2,-x,z+1/2
5 -x,-y,-z
6 x+1/2,y+1/2,-z
7 y,-x+1/2,-z+1/2
8 -y+1/2,x,-z+1/2

loop_
_atom_site_label
_atom_site_type_symbol
_atom_site_symmetry_multiplicity
_atom_site_Wyckoff_label
_atom_site_fract_x
_atom_site_fract_y
_atom_site_fract_z
_atom_site_occupancy
I1 I 8 g 0.03430 0.75760 0.11500 1.00000
Li1 Li 8 g 0.42600 0.23600 0.42600 1.00000
O1 O 8 g 0.83600 0.12100 -0.05000 1.00000
O2 O 8 g 0.09400 0.20900 0.09400 1.00000
O3 O 8 g 0.84700 0.55200 0.17200 1.00000
```

$\beta$-LiIO$_3$: ABC3_tP40_86_g_g_3g - POSCAR

```
ABC3_tP40_86_g_g_3g & a,c/a,x1,y1,z1,x2,y2,z2,x3,y3,z3,x4,y4,z4,x5,y5,z5
    ↪ --params=9.7329,0.63255555898,0.0343,0.7576,0.115,0.426,0.236,
    ↪ 0.426,0.836,0.121,-0.05,0.094,0.209,0.094,0.847,0.552,0.172 &
    ↪ P4_{2}/n C_{4h}^{4} #86 (g^5) & tP40 & None & ILiO3 & ILiO3 &
    ↪ H. Schulz, Acta Crystallogr. Sect. B Struct. Sci. 29, 2285-2289
    ↪ (1973)
   1.00000000000000
   9.73290000000000   0.00000000000000   0.00000000000000
   0.00000000000000   9.73290000000000   0.00000000000000
   0.00000000000000   0.00000000000000   6.15660000000000
     I        Li         O
     8         8        24
Direct
   0.03430000000000   0.75760000000000   0.11500000000000   I    (8g)
   0.46570000000000  -0.25760000000000   0.11500000000000   I    (8g)
  -0.75760000000000   0.53430000000000   0.61500000000000   I    (8g)
   1.25760000000000  -0.03430000000000   0.61500000000000   I    (8g)
  -0.03430000000000  -0.75760000000000  -0.11500000000000   I    (8g)
   0.53430000000000   1.25760000000000  -0.11500000000000   I    (8g)
   0.75760000000000   0.46570000000000   0.38500000000000   I    (8g)
  -0.25760000000000   0.03430000000000   0.38500000000000   I    (8g)
   0.42600000000000   0.23600000000000   0.42600000000000   Li   (8g)
   0.07400000000000   0.26400000000000   0.42600000000000   Li   (8g)
  -0.23600000000000   0.92600000000000   0.92600000000000   Li   (8g)
   0.73600000000000  -0.42600000000000   0.92600000000000   Li   (8g)
  -0.42600000000000  -0.23600000000000  -0.42600000000000   Li   (8g)
   0.92600000000000   0.73600000000000  -0.42600000000000   Li   (8g)
   0.23600000000000   0.07400000000000   0.07400000000000   Li   (8g)
   0.26400000000000   0.42600000000000   0.07400000000000   Li   (8g)
   0.83600000000000   0.12100000000000  -0.05000000000000   O    (8g)
  -0.33600000000000   0.37900000000000  -0.05000000000000   O    (8g)
  -0.12100000000000   1.33600000000000   0.45000000000000   O    (8g)
   0.62100000000000  -0.83600000000000   0.45000000000000   O    (8g)
  -0.83600000000000  -0.12100000000000   0.05000000000000   O    (8g)
   1.33600000000000   0.62100000000000   0.05000000000000   O    (8g)
   0.12100000000000  -0.33600000000000   0.55000000000000   O    (8g)
   0.37900000000000   0.83600000000000   0.55000000000000   O    (8g)
   0.09400000000000   0.20900000000000   0.09400000000000   O    (8g)
   0.40600000000000   0.29100000000000   0.09400000000000   O    (8g)
  -0.20900000000000   0.59400000000000   0.59400000000000   O    (8g)
   0.70900000000000  -0.09400000000000   0.59400000000000   O    (8g)
  -0.09400000000000  -0.20900000000000  -0.09400000000000   O    (8g)
   0.59400000000000   0.70900000000000  -0.09400000000000   O    (8g)
   0.20900000000000   0.40600000000000   0.40600000000000   O    (8g)
   0.29100000000000   0.09400000000000   0.40600000000000   O    (8g)
   0.84700000000000   0.55200000000000   0.17200000000000   O    (8g)
  -0.34700000000000  -0.05200000000000   0.17200000000000   O    (8g)
  -0.55200000000000   1.34700000000000   0.67200000000000   O    (8g)
   1.05200000000000  -0.84700000000000   0.67200000000000   O    (8g)
  -0.84700000000000  -0.55200000000000  -0.17200000000000   O    (8g)
   1.34700000000000   1.05200000000000  -0.17200000000000   O    (8g)
   0.55200000000000  -0.34700000000000   0.32800000000000   O    (8g)
  -0.05200000000000   0.84700000000000   0.32800000000000   O    (8g)
```

Marialite Scapolite [$Na_4Cl(AlSi_3)_3O_{24}$, $S6_4$]: AB4C24D12_tI82_87_a_h_2h2i_hi - CIF

```
# CIF file
data_findsym-output
_audit_creation_method FINDSYM

_chemical_name_mineral 'Marialite scapolite'
_chemical_formula_sum 'Cl Na4 O24 Si12'

loop_
_publ_author_name
 'J. J. Papike'
 'T. Zoltai'
_journal_name_full_name
;
 American Mineralogist
;
_journal_volume 50
_journal_year 1965
_journal_page_first 641
_journal_page_last 655
_publ_Section_title
;
 The crystal structure of a marialite scapolite
;
```

```
_aflow_title 'Marialite Scapolite [Na$_{4}$Cl(AlSi$_{3}$)$_{3}$O$_{24}$,
      ↪  $S6_{4}$] Structure'
_aflow_proto 'AB4C24D12_tI82_87_a_h_2h2i_hi'
_aflow_params 'a,c/a,x_{2},y_{2},x_{3},y_{3},x_{4},y_{4},x_{5},y_{5},x_{
      ↪ 6},y_{6},z_{6},x_{7},y_{7},z_{7},x_{8},y_{8},z_{8}'
_aflow_params_values '12.06,0.627860696517,0.134,0.2113,0.4587,0.3483,
      ↪ 0.3066,0.1206,0.3388,0.4104,0.0517,0.35,0.2148,0.2293,0.1289,
      ↪ 0.3281,0.3374,0.0851,0.206'
_aflow_Strukturbericht '$S6_{4}$'
_aflow_Pearson 'tI82'

_symmetry_space_group_name_H-M "I 4/m"
_symmetry_Int_Tables_number 87

_cell_length_a    12.06000
_cell_length_b    12.06000
_cell_length_c    7.57200
_cell_angle_alpha 90.00000
_cell_angle_beta  90.00000
_cell_angle_gamma 90.00000

loop_
_space_group_symop_id
_space_group_symop_operation_xyz
1 x,y,z
2 -x,-y,z
3 -y,x,z
4 y,-x,z
5 -x,-y,-z
6 x,y,-z
7 y,-x,-z
8 -y,x,-z
9 x+1/2,y+1/2,z+1/2
10 -x+1/2,-y+1/2,z+1/2
11 -y+1/2,x+1/2,z+1/2
12 y+1/2,-x+1/2,z+1/2
13 -x+1/2,-y+1/2,-z+1/2
14 x+1/2,y+1/2,-z+1/2
15 y+1/2,-x+1/2,-z+1/2
16 -y+1/2,x+1/2,-z+1/2

loop_
_atom_site_label
_atom_site_type_symbol
_atom_site_symmetry_multiplicity
_atom_site_Wyckoff_label
_atom_site_fract_x
_atom_site_fract_y
_atom_site_fract_z
_atom_site_occupancy
Cl1 Cl 2 a 0.00000 0.00000 0.00000 1.00000
Na1 Na 8 h 0.13400 0.21130 0.00000 1.00000
O1 O 8 h 0.45870 0.34830 0.00000 1.00000
O2 O 8 h 0.30660 0.12060 0.00000 1.00000
Si1 Si 8 h 0.33880 0.41040 0.00000 1.00000
O3 O 16 i 0.05170 0.35000 0.21480 1.00000
O4 O 16 i 0.22930 0.12890 0.32810 1.00000
Si2 Si 16 i 0.33740 0.08510 0.20600 1.00000
```

Marialite Scapolite [$Na_4Cl(AlSi_3)_3O_{24}$, $S6_4$]: AB4C24D12_tI82_87_a_h_2h2i_hi - POSCAR

```
AB4C24D12_tI82_87_a_h_2h2i_hi & a,c/a,x2,y2,x3,y3,x4,y4,x5,y5,x6,y6,z6,
      ↪ x7,y7,z7,x8,y8,z8 --params=12.06,0.627860696517,0.134,0.2113,
      ↪ 0.4587,0.3483,0.3066,0.1206,0.3388,0.4104,0.0517,0.35,0.2148,
      ↪ 0.2293,0.1289,0.3281,0.3374,0.0851,0.206 & I4/m C_{4h}^{5} #87
      ↪ (ah^4i^3) & tI82 & $S6_{4}$ & ClNa4O24(Al3Si9) & Marialite
      ↪ scapolite & J. J. Papike and T. Zoltai, Am. Mineral. 50,
      ↪ 641-655 (1965)
   1.00000000000000
  -6.03000000000000     6.03000000000000     3.78600000000000
   6.03000000000000    -6.03000000000000     3.78600000000000
   6.03000000000000     6.03000000000000    -3.78600000000000
    Cl      Na       O      Si
     1       4      24      12
Direct
   0.00000000000000     0.00000000000000     0.00000000000000    Cl    (2a)
   0.21130000000000     0.13400000000000     0.34530000000000    Na    (8h)
  -0.21130000000000    -0.13400000000000    -0.34530000000000    Na    (8h)
   0.13400000000000    -0.21130000000000    -0.07730000000000    Na    (8h)
  -0.13400000000000     0.21130000000000     0.07730000000000    Na    (8h)
   0.34830000000000     0.45870000000000     0.80700000000000     O    (8h)
  -0.34830000000000    -0.45870000000000    -0.80700000000000     O    (8h)
   0.45870000000000    -0.34830000000000     0.11040000000000     O    (8h)
  -0.45870000000000     0.34830000000000    -0.11040000000000     O    (8h)
   0.12060000000000     0.30660000000000     0.42720000000000     O    (8h)
  -0.12060000000000    -0.30660000000000    -0.42720000000000     O    (8h)
   0.30660000000000    -0.12060000000000     0.18600000000000     O    (8h)
  -0.30660000000000     0.12060000000000    -0.18600000000000     O    (8h)
   0.56480000000000     0.26650000000000     0.40170000000000     O   (16i)
  -0.13520000000000     0.16310000000000    -0.40170000000000     O   (16i)
   0.26650000000000    -0.13520000000000    -0.29830000000000     O   (16i)
   0.16310000000000     0.56480000000000     0.29830000000000     O   (16i)
  -0.56480000000000    -0.26650000000000    -0.40170000000000     O   (16i)
   0.13520000000000    -0.16310000000000     0.40170000000000     O   (16i)
  -0.26650000000000     0.13520000000000     0.29830000000000     O   (16i)
  -0.16310000000000    -0.56480000000000    -0.29830000000000     O   (16i)
   0.45700000000000     0.55740000000000     0.35820000000000     O   (16i)
   0.19920000000000     0.09880000000000    -0.35820000000000     O   (16i)
   0.55740000000000     0.19920000000000     0.10040000000000     O   (16i)
   0.09880000000000     0.45700000000000    -0.10040000000000     O   (16i)
  -0.45700000000000    -0.55740000000000    -0.35820000000000     O   (16i)
  -0.19920000000000    -0.09880000000000     0.35820000000000     O   (16i)
  -0.55740000000000    -0.19920000000000    -0.10040000000000     O   (16i)
  -0.09880000000000    -0.45700000000000     0.10040000000000     O   (16i)
   0.41040000000000     0.33880000000000     0.74920000000000    Si    (8h)
  -0.41040000000000    -0.33880000000000    -0.74920000000000    Si    (8h)
   0.33880000000000    -0.41040000000000    -0.07160000000000    Si    (8h)
  -0.33880000000000     0.41040000000000     0.07160000000000    Si    (8h)
   0.29110000000000     0.54340000000000     0.42250000000000    Si   (16i)
   0.12090000000000    -0.13140000000000    -0.42250000000000    Si   (16i)
   0.54340000000000     0.12090000000000     0.25230000000000    Si   (16i)
  -0.13140000000000     0.29110000000000    -0.25230000000000    Si   (16i)
  -0.29110000000000    -0.54340000000000    -0.42250000000000    Si   (16i)
  -0.12090000000000     0.13140000000000     0.42250000000000    Si   (16i)
  -0.54340000000000    -0.12090000000000    -0.25230000000000    Si   (16i)
   0.13140000000000    -0.29110000000000     0.25230000000000    Si   (16i)
```

$Sr_2NiWO_6$: AB6C2D_tI20_87_a_eh_d_b - CIF

```
# CIF file
data_findsym-output
_audit_creation_method FINDSYM

_chemical_name_mineral 'NiO6Sr2W'
_chemical_formula_sum 'Ni O6 Sr2 W'

loop_
_publ_author_name
 'D. Iwanaga'
 'Y. Inaguma'
 'M. Itoh'
_journal_name_full_name
;
 Materials Research Bulletin
;
_journal_volume 35
_journal_year 2000
_journal_page_first 449
_journal_page_last 457
_publ_Section_title
;
 Structure and Magnetic Properties of Sr$_{2}$Ni$A$O$_{6}$ ($A$ = W, Te)
;

_aflow_title 'Sr$_{2}$NiWO$_{6}$ Structure'
_aflow_proto 'AB6C2D_tI20_87_a_eh_d_b'
_aflow_params 'a,c/a,z_{4},x_{5},y_{5}'
_aflow_params_values '5.5571,1.42396213853,0.255,0.289,0.227'
_aflow_Strukturbericht 'None'
_aflow_Pearson 'tI20'

_symmetry_space_group_name_H-M "I 4/m"
_symmetry_Int_Tables_number 87

_cell_length_a    5.55710
_cell_length_b    5.55710
_cell_length_c    7.91310
_cell_angle_alpha 90.00000
_cell_angle_beta  90.00000
_cell_angle_gamma 90.00000

loop_
_space_group_symop_id
_space_group_symop_operation_xyz
1 x,y,z
2 -x,-y,z
3 -y,x,z
4 y,-x,z
5 -x,-y,-z
6 x,y,-z
7 y,-x,-z
8 -y,x,-z
9 x+1/2,y+1/2,z+1/2
10 -x+1/2,-y+1/2,z+1/2
11 -y+1/2,x+1/2,z+1/2
12 y+1/2,-x+1/2,z+1/2
13 -x+1/2,-y+1/2,-z+1/2
14 x+1/2,y+1/2,-z+1/2
15 y+1/2,-x+1/2,-z+1/2
16 -y+1/2,x+1/2,-z+1/2

loop_
_atom_site_label
_atom_site_type_symbol
_atom_site_symmetry_multiplicity
_atom_site_Wyckoff_label
_atom_site_fract_x
_atom_site_fract_y
_atom_site_fract_z
_atom_site_occupancy
Ni1 Ni 2 a 0.00000 0.00000 0.00000 1.00000
W1 W 2 b 0.00000 0.00000 0.50000 1.00000
Sr1 Sr 4 d 0.00000 0.50000 0.25000 1.00000
O1 O 4 e 0.00000 0.00000 0.25500 1.00000
O2 O 8 h 0.28900 0.22700 0.00000 1.00000
```

$Sr_2NiWO_6$: AB6C2D_tI20_87_a_eh_d_b - POSCAR

```
AB6C2D_tI20_87_a_eh_d_b & a,c/a,z4,x5,y5 --params=5.5571,1.42396213853,
      ↪ 0.255,0.289,0.227 & I4/m C_{4h}^{5} #87 (abdeh) & tI20 & None &
      ↪ NiO6Sr2W & NiO6Sr2W & D. Iwanaga and Y. Inaguma and M. Itoh,
      ↪ Mater. Res. Bull. 35, 449-457 (2000)
   1.00000000000000
  -2.77855000000000     2.77855000000000     3.95655000000000
   2.77855000000000    -2.77855000000000     3.95655000000000
   2.77855000000000     2.77855000000000    -3.95655000000000
    Ni       O      Sr       W
     1       6       2       1
Direct
```

```
   0.00000000000000    0.00000000000000    0.00000000000000    Ni    (2a)
   0.25500000000000    0.25500000000000    0.00000000000000     O    (4e)
  -0.25500000000000   -0.25500000000000    0.00000000000000     O    (4e)
   0.22700000000000    0.28900000000000    0.51600000000000     O    (8h)
  -0.22700000000000   -0.28900000000000   -0.51600000000000     O    (8h)
   0.28900000000000   -0.22700000000000    0.06200000000000     O    (8h)
  -0.28900000000000    0.22700000000000   -0.06200000000000     O    (8h)
   0.75000000000000    0.25000000000000    0.50000000000000    Sr    (4d)
   0.25000000000000    0.75000000000000    0.50000000000000    Sr    (4d)
   0.50000000000000    0.50000000000000    0.00000000000000     W    (2b)
```

$Na_4Ge_9O_{20}$: A9B4C20_tI132_88_a2f_f_5f - CIF

```
# CIF file
data_findsym-output
_audit_creation_method FINDSYM

_chemical_name_mineral 'Ge9Na4O20'
_chemical_formula_sum 'Ge9 Na4 O20'

loop_
_publ_author_name
 'N. Ingri'
 'G. Lundgren'
_journal_name_full_name
;
 Acta Chemica Scandinavica
;
_journal_volume 17
_journal_year 1963
_journal_page_first 617
_journal_page_last 633
_publ_Section_title
;
 The Crystal Structure of Na$_{4}$Ge$_{9}$O$_{20}$
;

_aflow_title 'Na$_{4}$Ge$_{9}$O$_{20}$ Structure'
_aflow_proto 'A9B4C20_tI132_88_a2f_f_5f'
_aflow_params 'a,c/a,x_{2},y_{2},z_{2},x_{3},y_{3},z_{3},x_{4},y_{4},z_{
    ↪ 4},x_{5},y_{5},z_{5},x_{6},y_{6},z_{6},x_{7},y_{7},z_{7},x_{8},
    ↪ y_{8},z_{8},x_{9},y_{9},z_{9}'
_aflow_params_values '14.97999,0.492924227586,0.1366,0.046,0.7009,0.0956
    ↪ ,0.218,0.4912,0.0878,0.0544,0.1702,0.0928,0.2127,0.2448,0.0824,
    ↪ 0.2227,0.7476,0.1889,0.0739,-0.0926,0.0253,0.0446,0.7555,0.1554
    ↪ ,0.1088,0.507'
_aflow_Strukturbericht 'None'
_aflow_Pearson 'tI132'

_symmetry_space_group_name_H-M "I 41/a (origin choice 2)"
_symmetry_Int_Tables_number 88

_cell_length_a    14.97999
_cell_length_b    14.97999
_cell_length_c    7.38400
_cell_angle_alpha 90.00000
_cell_angle_beta  90.00000
_cell_angle_gamma 90.00000

loop_
_space_group_symop_id
_space_group_symop_operation_xyz
1 x,y,z
2 -x,-y+1/2,z
3 -y+3/4,x+1/4,z+1/4
4 y+1/4,-x+1/4,z+1/4
5 -x,-y,-z
6 x,y+1/2,-z
7 y+1/4,-x+3/4,-z+3/4
8 -y+3/4,x+3/4,-z+3/4
9 x+1/2,y+1/2,z+1/2
10 -x+1/2,-y,z+1/2
11 -y+1/4,x+3/4,z+3/4
12 y+3/4,-x+3/4,z+3/4
13 -x+1/2,-y+1/2,-z+1/2
14 x+1/2,y,-z+1/2
15 y+3/4,-x+1/4,-z+1/4
16 -y+1/4,x+1/4,-z+1/4

loop_
_atom_site_label
_atom_site_type_symbol
_atom_site_symmetry_multiplicity
_atom_site_Wyckoff_label
_atom_site_fract_x
_atom_site_fract_y
_atom_site_fract_z
_atom_site_occupancy
Ge1 Ge 4 a 0.00000 0.25000 0.12500 1.00000
Ge2 Ge 16 f 0.13660 0.04600 0.70090 1.00000
Ge3 Ge 16 f 0.09560 0.21800 0.49120 1.00000
Na1 Na 16 f 0.08780 0.05440 0.17020 1.00000
O1 O 16 f 0.09280 0.21270 0.24480 1.00000
O2 O 16 f 0.08240 0.22270 0.74760 1.00000
O3 O 16 f 0.18890 0.07390 -0.09260 1.00000
O4 O 16 f 0.02530 0.04460 0.75550 1.00000
O5 O 16 f 0.15540 0.10880 0.50700 1.00000
```

$Na_4Ge_9O_{20}$: A9B4C20_tI132_88_a2f_f_5f - POSCAR

```
A9B4C20_tI132_88_a2f_f_5f & a,c/a,x2,y2,z2,x3,y3,z3,x4,y4,z4,x5,y5,z5,x6
    ↪ ,y6,z6,x7,y7,z7,x8,y8,z8,x9,y9,z9 --params=14.97999,
    ↪ 0.492924227586,0.1366,0.046,0.7009,0.0956,0.218,0.4912,0.0878,
    ↪ 0.0544,0.1702,0.0928,0.2127,0.2448,0.0824,0.2227,0.7476,0.1889,
    ↪ 0.0739,-0.0926,0.0253,0.0446,0.7555,0.1554,0.1088,0.507 & I4_{1
    ↪ }/a C_{4h}^{6} #88 (af^8) & tI132 & None & Ge9Na4O20 &
    ↪ Ge9Na4O20 & N. Ingri and G. Lundgren, Acta Chem. Scand. 17,
    ↪ 617-633 (1963)
   1.00000000000000
  -7.48999500000000    7.48999500000000    3.69200000000000
   7.48999500000000   -7.48999500000000    3.69200000000000
   7.48999500000000    7.48999500000000   -3.69200000000000
    Ge    Na     O
    18     8    40
Direct
   0.37500000000000    0.12500000000000    0.25000000000000    Ge    (4a)
   0.62500000000000    0.87500000000000    0.75000000000000    Ge    (4a)
   0.74690000000000    0.83750000000000    0.18260000000000    Ge   (16f)
   1.15490000000000    0.56430000000000    0.31740000000000    Ge   (16f)
   1.33750000000000    0.65490000000000    0.09060000000000    Ge   (16f)
   1.06430000000000    1.24690000000000    0.40940000000000    Ge   (16f)
  -0.74690000000000   -0.83750000000000   -0.18260000000000    Ge   (16f)
  -0.15490000000000   -0.56430000000000    0.68260000000000    Ge   (16f)
  -0.33750000000000   -0.65490000000000   -0.09060000000000    Ge   (16f)
  -0.06430000000000   -0.24690000000000    0.59060000000000    Ge   (16f)
   0.70920000000000    0.58680000000000    0.31360000000000    Ge   (16f)
   0.77320000000000    0.39560000000000    0.18640000000000    Ge   (16f)
   1.08680000000000    0.27320000000000   -0.12240000000000    Ge   (16f)
   0.89560000000000    1.20920000000000    0.62240000000000    Ge   (16f)
  -0.70920000000000   -0.58680000000000   -0.31360000000000    Ge   (16f)
   0.22680000000000   -0.39560000000000    0.81360000000000    Ge   (16f)
  -0.08680000000000   -0.27320000000000    0.12240000000000    Ge   (16f)
   0.10440000000000   -0.20920000000000    0.37760000000000    Ge   (16f)
   0.22460000000000    0.25800000000000    0.14220000000000    Na   (16f)
   0.61580000000000    0.08240000000000    0.35780000000000    Na   (16f)
   0.75800000000000    0.11580000000000    0.03340000000000    Na   (16f)
   0.58240000000000    0.72460000000000    0.46660000000000    Na   (16f)
  -0.22460000000000   -0.25800000000000   -0.14220000000000    Na   (16f)
   0.38420000000000   -0.08240000000000    0.64220000000000    Na   (16f)
   0.24200000000000   -0.11580000000000   -0.03340000000000    Na   (16f)
   0.41760000000000    0.27540000000000    0.53340000000000    Na   (16f)
   0.45750000000000    0.33760000000000    0.30550000000000     O   (16f)
   0.53210000000000    0.15200000000000    0.19450000000000     O   (16f)
   0.83760000000000    0.03210000000000   -0.11990000000000     O   (16f)
   0.65200000000000    0.95750000000000    0.61990000000000     O   (16f)
  -0.45750000000000   -0.33760000000000   -0.30550000000000     O   (16f)
   0.46790000000000   -0.15200000000000    0.80550000000000     O   (16f)
   0.16240000000000   -0.03210000000000    0.11990000000000     O   (16f)
   0.34800000000000    0.04250000000000    0.38010000000000     O   (16f)
   0.97030000000000    0.83000000000000    0.30510000000000     O   (16f)
   1.02490000000000    0.66520000000000    0.19490000000000     O   (16f)
   1.33000000000000    0.52490000000000   -0.14030000000000     O   (16f)
   1.16520000000000    1.47030000000000    0.64030000000000     O   (16f)
  -0.97030000000000   -0.83000000000000   -0.30510000000000     O   (16f)
  -0.02490000000000   -0.66520000000000    0.80510000000000     O   (16f)
  -0.33000000000000   -0.52490000000000    0.14030000000000     O   (16f)
  -0.16520000000000   -0.47030000000000    0.35970000000000     O   (16f)
  -0.01870000000000    0.09630000000000    0.26280000000000     O   (16f)
   0.33350000000000   -0.28150000000000    0.23720000000000     O   (16f)
   0.59630000000000   -0.16650000000000    0.11500000000000     O   (16f)
   0.21850000000000    0.48130000000000    0.38500000000000     O   (16f)
   0.01870000000000   -0.09630000000000   -0.26280000000000     O   (16f)
   0.66650000000000    0.28150000000000    0.76280000000000     O   (16f)
   0.40370000000000    0.16650000000000   -0.11500000000000     O   (16f)
   0.78150000000000    0.51870000000000    0.61500000000000     O   (16f)
   0.80010000000000    0.78080000000000    0.06990000000000     O   (16f)
   1.21090000000000    0.73020000000000    0.43010000000000     O   (16f)
   1.28080000000000    0.71090000000000   -0.01930000000000     O   (16f)
   1.23020000000000    1.30010000000000    0.51930000000000     O   (16f)
  -0.80010000000000   -0.78080000000000   -0.06990000000000     O   (16f)
  -0.21090000000000   -0.73020000000000    0.56990000000000     O   (16f)
  -0.28080000000000   -0.71090000000000    0.01930000000000     O   (16f)
  -0.23020000000000   -0.30010000000000    0.48070000000000     O   (16f)
   0.61580000000000    0.66240000000000    0.26420000000000     O   (16f)
   0.89820000000000    0.35160000000000    0.23580000000000     O   (16f)
   1.16240000000000    0.39820000000000    0.04660000000000     O   (16f)
   0.85160000000000    1.11580000000000    0.45340000000000     O   (16f)
  -0.61580000000000   -0.66240000000000   -0.26420000000000     O   (16f)
   0.10180000000000   -0.35160000000000    0.76420000000000     O   (16f)
  -0.16240000000000   -0.39820000000000   -0.04660000000000     O   (16f)
   0.14840000000000   -0.11580000000000    0.54660000000000     O   (16f)
```

Copper (I) Azide ($CuN_3$): AB3_tI32_88_d_cf - CIF

```
# CIF file
data_findsym-output
_audit_creation_method FINDSYM

_chemical_name_mineral 'Copper (i) azide'
_chemical_formula_sum 'Cu N3'

loop_
_publ_author_name
 'H. Wilsdorf'
_journal_name_full_name
;
 Acta Crystallographica
;
_journal_volume 1
_journal_year 1948
_journal_page_first 115
_journal_page_last 118
_publ_Section_title
;
 Die Kristallstruktur des einwertigen Kupferazids, CuN$_{3}$
;

# Found in {\em Ab initio} study of electronic structure and optical
    ↪ properties of heavy-metal azides: TlN$_{3}$, AgN$_{3}$, and
    ↪ CuN$_{3}$, 2008
```

```
_aflow_title 'Copper (I) Azide (CuN$_{3}$) Structure'
_aflow_proto 'AB3_tI32_88_d_cf'
_aflow_params 'a,c/a,x_{3},y_{3},z_{3}'
_aflow_params_values '8.65001,0.646242027466,0.173,0.173,0.375'
_aflow_Strukturbericht 'None'
_aflow_Pearson 'tI32'

_symmetry_space_group_name_H-M "I 41/a (origin choice 2)"
_symmetry_Int_Tables_number 88

_cell_length_a    8.65001
_cell_length_b    8.65001
_cell_length_c    5.59000
_cell_angle_alpha 90.00000
_cell_angle_beta  90.00000
_cell_angle_gamma 90.00000

loop_
_space_group_symop_id
_space_group_symop_operation_xyz
1 x,y,z
2 -x,-y+1/2,z
3 -y+3/4,x+1/4,z+1/4
4 y+1/4,-x+1/4,z+1/4
5 -x,-y,-z
6 x,y+1/2,-z
7 y+1/4,-x+3/4,-z+3/4
8 -y+3/4,x+3/4,-z+3/4
9 x+1/2,y+1/2,z+1/2
10 -x+1/2,-y,z+1/2
11 -y+1/4,x+3/4,z+3/4
12 y+3/4,-x+3/4,z+3/4
13 -x+1/2,-y+1/2,-z+1/2
14 x+1/2,y,-z+1/2
15 y+3/4,-x+1/4,-z+1/4
16 -y+1/4,x+1/4,-z+1/4

loop_
_atom_site_label
_atom_site_type_symbol
_atom_site_symmetry_multiplicity
_atom_site_Wyckoff_label
_atom_site_fract_x
_atom_site_fract_y
_atom_site_fract_z
_atom_site_occupancy
N1 N 8 c 0.00000 0.00000 0.00000 1.00000
Cu1 Cu 8 d 0.00000 0.00000 0.50000 1.00000
N2 N 16 f 0.17300 0.17300 0.37500 1.00000
```

Copper (I) Azide (CuN$_3$): AB3_tI32_88_d_cf - POSCAR

```
AB3_tI32_88_d_cf & a,c/a,x3,y3,z3 --params=8.65001,0.646242027466,0.173,
    ↪ 0.173,0.375 & I4_{1}/a C_{4h}^{6} #88 (cdf) & tI32 & None &
    ↪ CuN3 & Copper (i) azide & H. Wilsdorf, Acta Cryst. 1, 115-118 (
    ↪ 1948)
   1.00000000000000
  -4.32500500000000   4.32500500000000   2.79500000000000
   4.32500500000000  -4.32500500000000   2.79500000000000
   4.32500500000000   4.32500500000000  -2.79500000000000
    Cu        N
     4       12
Direct
   0.50000000000000   0.50000000000000   0.00000000000000   Cu   (8d)
   0.00000000000000   0.50000000000000   0.50000000000000   Cu   (8d)
   0.00000000000000   0.50000000000000   0.00000000000000   Cu   (8d)
   0.00000000000000   0.00000000000000   0.50000000000000   Cu   (8d)
   0.00000000000000   0.00000000000000   0.00000000000000    N   (8c)
   0.50000000000000   0.00000000000000   0.50000000000000    N   (8c)
   0.50000000000000   0.00000000000000   0.00000000000000    N   (8c)
   0.50000000000000   0.50000000000000   0.50000000000000    N   (8c)
   0.54800000000000   0.54800000000000   0.34600000000000    N  (16f)
   0.70200000000000   0.20200000000000   0.15400000000000    N  (16f)
   1.04800000000000   0.20200000000000   0.00000000000000    N  (16f)
   0.70200000000000   1.04800000000000   0.50000000000000    N  (16f)
  -0.54800000000000  -0.54800000000000  -0.34600000000000    N  (16f)
   0.29800000000000  -0.20200000000000   0.84600000000000    N  (16f)
  -0.04800000000000  -0.20200000000000   0.00000000000000    N  (16f)
   0.29800000000000  -0.04800000000000   0.50000000000000    N  (16f)
```

Scheelite (CaWO$_4$, $H0_4$): AB4C_tI24_88_b_f_a - CIF

```
# CIF file
data_findsym-output
_audit_creation_method FINDSYM

_chemical_name_mineral 'Scheelite'
_chemical_formula_sum 'Ca O4 W'

loop_
_publ_author_name
 'R. M. Hazen'
 'L. W. Finger'
 'J. W. E. Mariathasan'
_journal_name_full_name
;
 Journal of Physics and Chemistry of Solids
;
_journal_volume 46
_journal_year 1985
_journal_page_first 253
_journal_page_last 263
_publ_Section_title
;
 High-pressure crystal chemistry of scheelite-type tungstates and
    ↪ molybdates
;

# Found in Electronic band structures of the scheelite materials CaMoO$_
    ↪ {4}$, CaWO$_{4}$, PbMoO$_{4}$, and PbWO$_{4}$, 1988

_aflow_title 'Scheelite (CaWO$_{4}$, $H0_{4}$) Structure'
_aflow_proto 'AB4C_tI24_88_b_f_a'
_aflow_params 'a,c/a,x_{3},y_{3},z_{3}'
_aflow_params_values '5.2429,2.16935283908,0.1488,0.0038,0.2159'
_aflow_Strukturbericht '$H0_{4}$'
_aflow_Pearson 'tI24'

_symmetry_space_group_name_H-M "I 41/a (origin choice 2)"
_symmetry_Int_Tables_number 88

_cell_length_a    5.24290
_cell_length_b    5.24290
_cell_length_c    11.37370
_cell_angle_alpha 90.00000
_cell_angle_beta  90.00000
_cell_angle_gamma 90.00000

loop_
_space_group_symop_id
_space_group_symop_operation_xyz
1 x,y,z
2 -x,-y+1/2,z
3 -y+3/4,x+1/4,z+1/4
4 y+1/4,-x+1/4,z+1/4
5 -x,-y,-z
6 x,y+1/2,-z
7 y+1/4,-x+3/4,-z+3/4
8 -y+3/4,x+3/4,-z+3/4
9 x+1/2,y+1/2,z+1/2
10 -x+1/2,-y,z+1/2
11 -y+1/4,x+3/4,z+3/4
12 y+3/4,-x+3/4,z+3/4
13 -x+1/2,-y+1/2,-z+1/2
14 x+1/2,y,-z+1/2
15 y+3/4,-x+1/4,-z+1/4
16 -y+1/4,x+1/4,-z+1/4

loop_
_atom_site_label
_atom_site_type_symbol
_atom_site_symmetry_multiplicity
_atom_site_Wyckoff_label
_atom_site_fract_x
_atom_site_fract_y
_atom_site_fract_z
_atom_site_occupancy
W1 W 4 a 0.00000 0.25000 0.12500 1.00000
Ca1 Ca 4 b 0.00000 0.25000 0.62500 1.00000
O1 O 16 f 0.14880 0.00380 0.21590 1.00000
```

Scheelite (CaWO$_4$, $H0_4$): AB4C_tI24_88_b_f_a - POSCAR

```
AB4C_tI24_88_b_f_a & a,c/a,x3,y3,z3 --params=5.2429,2.16935283908,0.1488
    ↪ ,0.0038,0.2159 & I4_{1}/a C_{4h}^{6} #88 (abf) & tI24 & $H0_{4}
    ↪ $ & CaO4W & Scheelite & R. M. Hazen and L. W. Finger and J. W.
    ↪ E. Mariathasan, J. Phys. Chem. Solids 46, 253-263 (1985)
   1.00000000000000
  -2.62145000000000   2.62145000000000   5.68685000000000
   2.62145000000000  -2.62145000000000   5.68685000000000
   2.62145000000000   2.62145000000000  -5.68685000000000
    Ca       O       W
     2       8       2
Direct
   0.87500000000000   0.62500000000000   0.25000000000000   Ca   (4b)
   0.12500000000000   0.37500000000000   0.75000000000000   Ca   (4b)
   0.21970000000000   0.36470000000000   0.15260000000000    O  (16f)
   0.71210000000000   0.06710000000000   0.34740000000000    O  (16f)
   0.86470000000000   0.21210000000000   0.14500000000000    O  (16f)
   0.56710000000000   0.71970000000000   0.35500000000000    O  (16f)
  -0.21970000000000  -0.36470000000000  -0.15260000000000    O  (16f)
   0.28790000000000  -0.06710000000000   0.65260000000000    O  (16f)
   0.13530000000000  -0.21210000000000  -0.14500000000000    O  (16f)
   0.43290000000000   0.28030000000000   0.64500000000000    O  (16f)
   0.37500000000000   0.12500000000000   0.25000000000000    W   (4a)
   0.62500000000000   0.87500000000000   0.75000000000000    W   (4a)
```

$G7_5$ (PbCO$_3$ · PbCl$_2$, Phosgenite) (*obsolete*): AB2C3D2_tP16_90_c_f_ce_e - CIF

```
# CIF file
data_findsym-output
_audit_creation_method FINDSYM

_chemical_name_mineral 'Phosgenite'
_chemical_formula_sum 'C Cl2 O3 Pb2'

loop_
_publ_author_name
 'E. Onorato'
_journal_name_full_name
;
 Periodico di Mineralogia
;
_journal_volume 5
_journal_year 1934
_journal_page_first 1
_journal_page_last 27
_publ_Section_title
;
```

```
 La struttura della Fosgenite
;

# Found in Reexamination of the crystal structure of phosgenite, Pb$_{2}
     ↪ $Cl$_{2}$(CO$_{3}$), 1974

_aflow_title '$G7_{5}$ (PbCO$_{3}$ $\cdot$ PbCl$_{2}$, Phosgenite) ({\em
     ↪ {obsolete}}) Structure'
_aflow_proto 'AB2C3D2_tP16_90_c_f_ce_e'
_aflow_params 'a,c/a,z_{1},z_{2},x_{3},x_{4},x_{5}'
_aflow_params_values '8.12,0.541871921182,0.55,0.84,0.11,0.34,0.24'
_aflow_Strukturbericht '$G7_{5}$'
_aflow_Pearson 'tP16'

_symmetry_space_group_name_H-M "P 4 21 2"
_symmetry_Int_Tables_number 90

_cell_length_a    8.12000
_cell_length_b    8.12000
_cell_length_c    4.40000
_cell_angle_alpha 90.00000
_cell_angle_beta  90.00000
_cell_angle_gamma 90.00000

loop_
_space_group_symop_id
_space_group_symop_operation_xyz
1 x,y,z
2 x+1/2,-y+1/2,-z
3 -x+1/2,y+1/2,-z
4 -x,-y,z
5 -y,-x,-z
6 -y+1/2,x+1/2,z
7 y+1/2,-x+1/2,z
8 y,x,-z

loop_
_atom_site_label
_atom_site_type_symbol
_atom_site_symmetry_multiplicity
_atom_site_Wyckoff_label
_atom_site_fract_x
_atom_site_fract_y
_atom_site_fract_z
_atom_site_occupancy
C1 C 2 c 0.00000 0.50000 0.55000 1.00000
O1 O 2 c 0.00000 0.50000 0.84000 1.00000
O2 O 4 e 0.11000 0.11000 0.00000 1.00000
Pb1 Pb 4 e 0.34000 0.34000 0.00000 1.00000
Cl1 Cl 4 f 0.24000 0.24000 0.50000 1.00000
```

$G7_5$ (PbCO$_3$ · PbCl$_2$, Phosgenite) (*obsolete*): AB2C3D2_tP16_90_c_f_ce_e - POSCAR

```
AB2C3D2_tP16_90_c_f_ce_e & a,c/a,z1,z2,x3,x4,x5 --params=8.12,
     ↪ 0.541871921182,0.55,0.84,0.11,0.34,0.24 & P42_{1}2 D_{4}^{2} #
     ↪ 90 (c^2e^2f) & tP16 & $G7_{5}$ & CCl2O3Pb2 & Phosgenite & E.
     ↪ Onorato, Period. Mineral. 5, 1-27 (1934)
   1.00000000000000
   8.12000000000000   0.00000000000000   0.00000000000000
   0.00000000000000   8.12000000000000   0.00000000000000
   0.00000000000000   0.00000000000000   4.40000000000000
     C    Cl     O    Pb
     2     4     6     4
Direct
   0.00000000000000   0.50000000000000   0.55000000000000   C   (2c)
   0.50000000000000   0.00000000000000  -0.55000000000000   C   (2c)
   0.24000000000000   0.24000000000000   0.50000000000000   Cl  (4f)
  -0.24000000000000  -0.24000000000000   0.50000000000000   Cl  (4f)
   0.26000000000000   0.74000000000000   0.50000000000000   Cl  (4f)
   0.74000000000000   0.26000000000000   0.50000000000000   Cl  (4f)
   0.00000000000000   0.50000000000000   0.84000000000000   O   (2c)
   0.50000000000000   0.00000000000000  -0.84000000000000   O   (2c)
   0.11000000000000   0.11000000000000   0.00000000000000   O   (4e)
  -0.11000000000000  -0.11000000000000   0.00000000000000   O   (4e)
   0.39000000000000   0.61000000000000   0.00000000000000   O   (4e)
   0.61000000000000   0.39000000000000   0.00000000000000   O   (4e)
   0.34000000000000   0.34000000000000   0.00000000000000   Pb  (4e)
  -0.34000000000000  -0.34000000000000   0.00000000000000   Pb  (4e)
   0.16000000000000   0.84000000000000   0.00000000000000   Pb  (4e)
   0.84000000000000   0.16000000000000   0.00000000000000   Pb  (4e)
```

Retgersite ($\alpha$-NiSO$_4$·6H$_2$O, $H4_5$): A12BC10D_tP96_92_6b_a_5b_a - CIF

```
# CIF file
data_findsym-output
_audit_creation_method FINDSYM

_chemical_name_mineral 'Retgersite'
_chemical_formula_sum 'H12 Ni O10 S'

loop_
_publ_author_name
 'K. Stadnicka'
 'A. M. Glazer'
 'M. Koralewski'
_journal_name_full_name
;
 Acta Crystallographica Section B: Structural Science
;
_journal_volume 43
_journal_year 1987
_journal_page_first 319
_journal_page_last 325
_publ_Section_title
;
 Structure, absolute configuration and optical activity of $\
     ↪ alpha$-nickel sulfate hexahydrate
;

# Found in Handbook of Mineralogy, 2004

_aflow_title 'Retgersite ($\alpha$-NiSO$_{4}$$\cdot$6H$_{2}$O, $H4_{5}$)
     ↪ Structure'
_aflow_proto 'A12BC10D_tP96_92_6b_a_5b_a'
_aflow_params 'a,c/a,x_{1},x_{2},x_{3},y_{3},z_{3},x_{4},y_{4},z_{4},x_{
     ↪ 5},y_{5},z_{5},x_{6},y_{6},z_{6},x_{7},y_{7},z_{7},x_{8},y_{8},
     ↪ z_{8},x_{9},y_{9},z_{9},x_{10},y_{10},z_{10},x_{11},y_{11},z_{
     ↪ 11},x_{12},y_{12},z_{12},x_{13},y_{13},z_{13}'
_aflow_params_values '6.783,2.69615214507,0.2106,0.70943,0.1097,0.8667,
     ↪ 0.0394,0.2215,-0.0746,0.0867,0.5702,0.1559,0.0507,0.5371,0.3532
     ↪ ,0.0601,-0.0051,0.4453,0.0744,-0.0101,0.2941,0.1149,0.1727,-
     ↪ 0.047,0.0528,0.4705,0.2449,0.0561,0.0658,0.3599,0.085,0.6209,
     ↪ 0.6203,0.0658,0.9237,0.6731,0.0003'
_aflow_Strukturbericht '$H4_{5}$'
_aflow_Pearson 'tP96'

_symmetry_space_group_name_H-M "P 41 21 2"
_symmetry_Int_Tables_number 92

_cell_length_a    6.78300
_cell_length_b    6.78300
_cell_length_c    18.28800
_cell_angle_alpha 90.00000
_cell_angle_beta  90.00000
_cell_angle_gamma 90.00000

loop_
_space_group_symop_id
_space_group_symop_operation_xyz
1 x,y,z
2 x+1/2,-y+1/2,-z+3/4
3 -x+1/2,y+1/2,-z+1/4
4 -x,-y,z+1/2
5 -y,-x,-z+1/2
6 -y+1/2,x+1/2,z+1/4
7 y+1/2,-x+1/2,z+3/4
8 y,x,-z

loop_
_atom_site_label
_atom_site_type_symbol
_atom_site_symmetry_multiplicity
_atom_site_Wyckoff_label
_atom_site_fract_x
_atom_site_fract_y
_atom_site_fract_z
_atom_site_occupancy
Ni1 Ni 4 a 0.21060 0.21060 0.00000 1.00000
S1 S 4 a 0.70943 0.70943 0.00000 1.00000
H1 H 8 b 0.10970 0.86670 0.03940 1.00000
H2 H 8 b 0.22150 -0.07460 0.08670 1.00000
H3 H 8 b 0.57020 0.15590 0.05070 1.00000
H4 H 8 b 0.53710 0.35320 0.06010 1.00000
H5 H 8 b -0.00510 0.44530 0.07440 1.00000
H6 H 8 b -0.01010 0.29410 0.11490 1.00000
O1 O 8 b 0.17270 -0.04700 0.05280 1.00000
O2 O 8 b 0.47050 0.24490 0.05610 1.00000
O3 O 8 b 0.06580 0.35990 0.08500 1.00000
O4 O 8 b 0.62090 0.62030 0.06580 1.00000
O5 O 8 b 0.92370 0.67310 0.00030 1.00000
```

Retgersite ($\alpha$-NiSO$_4$·6H$_2$O, $H4_5$): A12BC10D_tP96_92_6b_a_5b_a - POSCAR

```
A12BC10D_tP96_92_6b_a_5b_a & a,c/a,x1,x2,x3,y3,z3,x4,y4,z4,x5,y5,z5,x6,
     ↪ y6,z6,x7,y7,z7,x8,y8,z8,x9,y9,z9,x10,y10,z10,x11,y11,z11,x12,
     ↪ y12,z12,x13,y13,z13 --params=6.783,2.69615214507,0.2106,0.70943
     ↪ ,0.1097,0.8667,0.0394,0.2215,-0.0746,0.0867,0.5702,0.1559,
     ↪ 0.0507,0.5371,0.3532,0.0601,-0.0051,0.4453,0.0744,-0.0101,
     ↪ 0.2941,0.1149,0.1727,-0.047,0.0528,0.4705,0.2449,0.0561,0.0658,
     ↪ 0.3599,0.085,0.6209,0.6203,0.0658,0.9237,0.6731,0.0003 & P4_{1}
     ↪ 2_{1}2 D_{4}^{4} #92 (a^2b^11) & tP96 & $H4_{5}$ & H12NiO10S &
     ↪ Retgersite & K. Stadnicka and A. M. Glazer and M. Koralewski,
     ↪ Acta Crystallogr. Sect. B Struct. Sci. 43, 319-325 (1987)
   1.00000000000000
   6.78300000000000   0.00000000000000   0.00000000000000
   0.00000000000000   6.78300000000000   0.00000000000000
   0.00000000000000   0.00000000000000  18.28800000000000
     H    Ni     O     S
    48     4    40     4
Direct
   0.10970000000000   0.86670000000000   0.03940000000000   H   (8b)
  -0.10970000000000  -0.86670000000000   0.53940000000000   H   (8b)
  -0.36670000000000   0.60970000000000   0.28940000000000   H   (8b)
   1.36670000000000   0.39030000000000   0.78940000000000   H   (8b)
   0.39030000000000   1.36670000000000   0.21060000000000   H   (8b)
   0.60970000000000  -0.36670000000000   0.71060000000000   H   (8b)
   0.86670000000000   0.10970000000000  -0.03940000000000   H   (8b)
  -0.86670000000000  -0.10970000000000   0.46060000000000   H   (8b)
   0.22150000000000  -0.07460000000000   0.08670000000000   H   (8b)
  -0.22150000000000   0.07460000000000   0.58670000000000   H   (8b)
   0.57460000000000   0.72150000000000   0.33670000000000   H   (8b)
   0.42540000000000   0.27850000000000   0.83670000000000   H   (8b)
   0.27850000000000   0.42540000000000   0.16330000000000   H   (8b)
   0.72150000000000   0.57460000000000   0.66330000000000   H   (8b)
  -0.07460000000000   0.22150000000000  -0.08670000000000   H   (8b)
   0.07460000000000  -0.22150000000000   0.41330000000000   H   (8b)
   0.57020000000000   0.15590000000000   0.05070000000000   H   (8b)
  -0.57020000000000  -0.15590000000000   0.55070000000000   H   (8b)
   0.34410000000000   1.07020000000000   0.30070000000000   H   (8b)
   0.65590000000000  -0.07020000000000   0.80070000000000   H   (8b)
```

```
  -0.07020000000000   0.65590000000000   0.19930000000000   H   (8b)
   1.07020000000000   0.34410000000000   0.69930000000000   H   (8b)
   0.15590000000000   0.57020000000000  -0.05070000000000   H   (8b)
  -0.15590000000000  -0.57020000000000   0.44930000000000   H   (8b)
   0.53710000000000   0.35320000000000   0.06010000000000   H   (8b)
  -0.53710000000000  -0.35320000000000   0.56010000000000   H   (8b)
   0.14680000000000   1.03710000000000   0.31010000000000   H   (8b)
   0.85320000000000  -0.03710000000000   0.81010000000000   H   (8b)
  -0.03710000000000   0.85320000000000   0.18990000000000   H   (8b)
   1.03710000000000   0.14680000000000   0.68990000000000   H   (8b)
   0.35320000000000   0.53710000000000  -0.06010000000000   H   (8b)
  -0.35320000000000  -0.53710000000000   0.43990000000000   H   (8b)
  -0.00510000000000   0.44530000000000   0.07440000000000   H   (8b)
   0.00510000000000  -0.44530000000000   0.57440000000000   H   (8b)
   0.05470000000000   0.49490000000000   0.32440000000000   H   (8b)
   0.94530000000000   0.50510000000000   0.82440000000000   H   (8b)
   0.50510000000000   0.94530000000000   0.17560000000000   H   (8b)
   0.49490000000000   0.05470000000000   0.67560000000000   H   (8b)
   0.44530000000000  -0.00510000000000  -0.07440000000000   H   (8b)
  -0.44530000000000   0.00510000000000   0.42560000000000   H   (8b)
  -0.01010000000000   0.29410000000000   0.11490000000000   H   (8b)
   0.01010000000000  -0.29410000000000   0.61490000000000   H   (8b)
   0.20590000000000   0.48990000000000   0.36490000000000   H   (8b)
   0.79410000000000   0.51010000000000   0.86490000000000   H   (8b)
   0.51010000000000   0.79410000000000   0.13510000000000   H   (8b)
   0.48990000000000   0.20590000000000   0.63510000000000   H   (8b)
   0.29410000000000  -0.01010000000000  -0.11490000000000   H   (8b)
  -0.29410000000000   0.01010000000000   0.38510000000000   H   (8b)
   0.21060000000000   0.21060000000000   0.00000000000000   Ni  (4a)
  -0.21060000000000  -0.21060000000000   0.50000000000000   Ni  (4a)
   0.28940000000000   0.71060000000000   0.25000000000000   Ni  (4a)
   0.71060000000000   0.28940000000000   0.75000000000000   Ni  (4a)
   0.17270000000000  -0.04700000000000   0.05280000000000   O   (8b)
  -0.17270000000000   0.04700000000000   0.55280000000000   O   (8b)
   0.54700000000000   0.67270000000000   0.30280000000000   O   (8b)
   0.45300000000000   0.32730000000000   0.80280000000000   O   (8b)
   0.32730000000000   0.45300000000000   0.19720000000000   O   (8b)
   0.67270000000000   0.54700000000000   0.69720000000000   O   (8b)
  -0.04700000000000   0.17270000000000  -0.05280000000000   O   (8b)
   0.04700000000000  -0.17270000000000   0.44720000000000   O   (8b)
   0.47050000000000   0.24490000000000   0.05610000000000   O   (8b)
  -0.47050000000000  -0.24490000000000   0.55610000000000   O   (8b)
   0.25510000000000   0.97050000000000   0.30610000000000   O   (8b)
   0.74490000000000   0.02950000000000   0.80610000000000   O   (8b)
   0.02950000000000   0.74490000000000   0.19390000000000   O   (8b)
   0.97050000000000   0.25510000000000   0.69390000000000   O   (8b)
   0.24490000000000   0.47050000000000  -0.05610000000000   O   (8b)
  -0.24490000000000  -0.47050000000000   0.44390000000000   O   (8b)
   0.06580000000000   0.35990000000000   0.08500000000000   O   (8b)
  -0.06580000000000  -0.35990000000000   0.58500000000000   O   (8b)
   0.14010000000000   0.56580000000000   0.33500000000000   O   (8b)
   0.85990000000000   0.43420000000000   0.83500000000000   O   (8b)
   0.43420000000000   0.85990000000000   0.16500000000000   O   (8b)
   0.56580000000000   0.14010000000000   0.66500000000000   O   (8b)
   0.35990000000000   0.06580000000000  -0.08500000000000   O   (8b)
  -0.35990000000000  -0.06580000000000   0.41500000000000   O   (8b)
   0.62090000000000   0.62030000000000   0.06580000000000   O   (8b)
  -0.62090000000000  -0.62030000000000   0.56580000000000   O   (8b)
  -0.12030000000000   1.12090000000000   0.31580000000000   O   (8b)
   1.12030000000000  -0.12090000000000   0.81580000000000   O   (8b)
  -0.12090000000000   1.12030000000000   0.18420000000000   O   (8b)
   1.12090000000000  -0.12030000000000   0.68420000000000   O   (8b)
   0.62030000000000   0.62090000000000  -0.06580000000000   O   (8b)
  -0.62030000000000  -0.62090000000000   0.43420000000000   O   (8b)
   0.92370000000000   0.67310000000000   0.00030000000000   O   (8b)
  -0.92370000000000  -0.67310000000000   0.50030000000000   O   (8b)
  -0.17310000000000   1.42370000000000   0.25030000000000   O   (8b)
   1.17310000000000  -0.42370000000000   0.75030000000000   O   (8b)
  -0.42370000000000   1.17310000000000   0.24970000000000   O   (8b)
   1.42370000000000  -0.17310000000000   0.74970000000000   O   (8b)
   0.67310000000000   0.92370000000000  -0.00030000000000   O   (8b)
  -0.67310000000000  -0.92370000000000   0.49970000000000   O   (8b)
   0.70943000000000   0.70943000000000   0.00000000000000   S   (4a)
  -0.70943000000000  -0.70943000000000   0.50000000000000   S   (4a)
  -0.20943000000000   1.20943000000000   0.25000000000000   S   (4a)
   1.20943000000000  -0.20943000000000   0.75000000000000   S   (4a)
```

Paratellurite ($\alpha$-$TeO_2$): A2B_tP12_92_b_a - CIF

```
# CIF file
data_findsym-output
_audit_creation_method FINDSYM

_chemical_name_mineral 'Paratellurite'
_chemical_formula_sum 'O2 Te'

loop_
_publ_author_name
 'P. A. Thomas'
_journal_name_full_name
;
 Journal of Physics C: Solid State Physics
;
_journal_volume 21
_journal_year 1988
_journal_page_first 4611
_journal_page_last 4627
_publ_Section_title
;
 The crystal structure and absolute optical chirality of paratellurite,
    ↪ $\alpha$-TeO$_{2}$
;

# Found in {\em Ab initio} study of the vibrational properties of
    ↪ crystalline TeO$_{2}$: The $\alpha$, $\beta$, and $\gamma$
    ↪ phases, 2006

_aflow_title 'Paratellurite ($\alpha$-TeO$_{2}$) Structure'
_aflow_proto 'A2B_tP12_92_b_a'
_aflow_params 'a,c/a,x_{1},x_{2},y_{2},z_{2}'
_aflow_params_values '4.8082,1.60392662535,0.0268,0.1386,0.2576,0.1862'
_aflow_Strukturbericht 'None'
_aflow_Pearson 'tP12'

_symmetry_space_group_name_H-M "P 41 21 2"
_symmetry_Int_Tables_number 92

_cell_length_a    4.80820
_cell_length_b    4.80820
_cell_length_c    7.71200
_cell_angle_alpha 90.00000
_cell_angle_beta  90.00000
_cell_angle_gamma 90.00000

loop_
_space_group_symop_id
_space_group_symop_operation_xyz
1 x,y,z
2 x+1/2,-y+1/2,-z+3/4
3 -x+1/2,y+1/2,-z+1/4
4 -x,-y,z+1/2
5 -y,-x,-z+1/2
6 -y+1/2,x+1/2,z+1/4
7 y+1/2,-x+1/2,z+3/4
8 y,x,-z

loop_
_atom_site_label
_atom_site_type_symbol
_atom_site_symmetry_multiplicity
_atom_site_Wyckoff_label
_atom_site_fract_x
_atom_site_fract_y
_atom_site_fract_z
_atom_site_occupancy
Te1 Te 4 a 0.02680 0.02680 0.00000 1.00000
O1 O 8 b 0.13860 0.25760 0.18620 1.00000
```

Paratellurite ($\alpha$-$TeO_2$): A2B_tP12_92_b_a - POSCAR

```
A2B_tP12_92_b_a & a,c/a,x1,x2,y2,z2 --params=4.8082,1.60392662535,0.0268
    ↪ ,0.1386,0.2576,0.1862 & P4_{1}2_{1}2 D_{4}^{4} #92 (ab) & tP12
    ↪ & None & O2Te & Paratellurite & P. A. Thomas, J. Phys. C: Solid
    ↪ State Phys. 21, 4611-4627 (1988)
   1.00000000000000
   4.80820000000000   0.00000000000000   0.00000000000000
   0.00000000000000   4.80820000000000   0.00000000000000
   0.00000000000000   0.00000000000000   7.71200000000000
     O     Te
     8      4
Direct
   0.13860000000000   0.25760000000000   0.18620000000000   O    (8b)
  -0.13860000000000  -0.25760000000000   0.68620000000000   O    (8b)
   0.24240000000000   0.63860000000000   0.43620000000000   O    (8b)
   0.75760000000000   0.36140000000000   0.93620000000000   O    (8b)
   0.36140000000000   0.75760000000000   0.06380000000000   O    (8b)
   0.63860000000000   0.24240000000000   0.56380000000000   O    (8b)
   0.25760000000000   0.13860000000000  -0.18620000000000   O    (8b)
  -0.25760000000000  -0.13860000000000   0.31380000000000   O    (8b)
   0.02680000000000   0.02680000000000   0.00000000000000   Te   (4a)
  -0.02680000000000  -0.02680000000000   0.50000000000000   Te   (4a)
   0.47320000000000   0.52680000000000   0.25000000000000   Te   (4a)
   0.52680000000000   0.47320000000000   0.75000000000000   Te   (4a)
```

Phase III $Cd_2Re_2O_7$: A2B7C2_tI44_98_f_bcde_f - CIF

```
# CIF file
data_findsym-output
_audit_creation_method FINDSYM

_chemical_name_mineral 'Cd2O7Re2'
_chemical_formula_sum 'Cd2 O7 Re2'

loop_
_publ_author_name
 'S.-W. Huang'
 'H.-T. Jeng'
 'J.-Y. Lin'
 'W. J. Chang'
 'J. M. Chen'
 'G. H. Lee'
 'H. Berger'
 'H. D. Yang'
 'K. S. Liang'
_journal_name_full_name
;
 Journal of Physics: Condensed Matter
;
_journal_volume 21
_journal_year 2009
_journal_page_first 195602
_journal_page_last 195602
_publ_Section_title
;
 Electronic structure of pyrochlore Cd$_{2}$Re$_{2}$O$_{7}$
;

# Found in The crystal structure of the inversion breaking metal Cd$_{2}
    ↪ $Re$_{2}$O$_{7}$, 2019 Found in The crystal structure of the
```

```
↪ inversion breaking metal Cd$_{2}$Re$_{2}$O$_{7}$, {arXiv:
↪ 1911.10141 [cond-mat.str-el]},

_aflow_title 'Phase III Cd$_{2}$Re$_{2}$O$_{7}$ Structure'
_aflow_proto 'A2B7C2_tI44_98_f_bcde_f'
_aflow_params 'a,c/a,z_{2},x_{3},x_{4},x_{5},x_{6}'
_aflow_params_values '7.2313,1.41443447236,0.81048,0.188,0.803,0.5041,-
↪ 0.0033'
_aflow_Strukturbericht 'None'
_aflow_Pearson 'tI44'

_symmetry_space_group_name_H-M "I 41 2 2"
_symmetry_Int_Tables_number 98

_cell_length_a      7.23130
_cell_length_b      7.23130
_cell_length_c      10.22820
_cell_angle_alpha   90.00000
_cell_angle_beta    90.00000
_cell_angle_gamma   90.00000

loop_
_space_group_symop_id
_space_group_symop_operation_xyz
1 x,y,z
2 x,-y+1/2,-z+1/4
3 -x,y+1/2,-z+1/4
4 -x,-y,z
5 -y,-x,-z
6 -y,x+1/2,z+1/4
7 y,-x+1/2,z+1/4
8 y,x,-z
9 x+1/2,y+1/2,z+1/2
10 x+1/2,-y,-z+3/4
11 -x+1/2,y,-z+3/4
12 -x+1/2,-y+1/2,z+1/2
13 -y+1/2,-x+1/2,-z+1/2
14 -y+1/2,x,z+3/4
15 y+1/2,-x,z+3/4
16 y+1/2,x+1/2,-z+1/2

loop_
_atom_site_label
_atom_site_type_symbol
_atom_site_symmetry_multiplicity
_atom_site_Wyckoff_label
_atom_site_fract_x
_atom_site_fract_y
_atom_site_fract_z
_atom_site_occupancy
O1 O 4 b 0.00000 0.00000 0.50000 1.00000
O2 O 8 c 0.00000 0.00000 0.81048 1.00000
O3 O 8 d 0.18800 0.18800 0.00000 1.00000
O4 O 8 e 0.80300 0.19700 0.00000 1.00000
Cd1 Cd 8 f 0.50410 0.25000 0.12500 1.00000
Re1 Re 8 f -0.00330 0.25000 0.12500 1.00000
```

Phase III $Cd_2Re_2O_7$: A2B7C2_tI44_98_f_bcde_f - POSCAR

```
A2B7C2_tI44_98_f_bcde_f & a,c/a,z2,x3,x4,x5,x6 --params=7.2313,
↪ 1.41443447236,0.81048,0.188,0.803,0.5041,-0.0033 & I4_{1}22 D_{
↪ 4}^{10} #98 (bcdef^2) & tI44 & None & Cd2O7Re2 & Cd2O7Re2 &
↪ S.-W. Huang et al., J. Phys.: Condens. Matter 21, 195602(2009)
   1.00000000000000
  -3.61565000000000   3.61565000000000   5.11410000000000
   3.61565000000000  -3.61565000000000   5.11410000000000
   3.61565000000000   3.61565000000000  -5.11410000000000
    Cd     O    Re
     4    14     4
Direct
   0.37500000000000   0.62910000000000   0.75410000000000   Cd   (8f)
   0.87500000000000  -0.37910000000000   0.24590000000000   Cd   (8f)
   1.37910000000000   0.12500000000000   0.75410000000000   Cd   (8f)
   0.37090000000000   0.62500000000000   0.24590000000000   Cd   (8f)
   0.50000000000000   0.50000000000000   0.00000000000000    O   (4b)
   0.25000000000000   0.75000000000000   0.50000000000000    O   (4b)
   0.81048000000000   0.81048000000000   0.00000000000000    O   (8c)
   1.56048000000000   1.06048000000000   0.50000000000000    O   (8c)
  -0.06048000000000  -0.56048000000000   0.50000000000000    O   (8c)
  -0.81048000000000  -0.81048000000000   0.00000000000000    O   (8c)
   0.18800000000000   0.18800000000000   0.37600000000000    O   (8d)
  -0.18800000000000  -0.18800000000000  -0.37600000000000    O   (8d)
   0.93800000000000   0.06200000000000   0.50000000000000    O   (8d)
   0.56200000000000   0.43800000000000   0.50000000000000    O   (8d)
   0.80300000000000  -0.80300000000000   0.00000000000000    O   (8e)
  -0.80300000000000   0.80300000000000   0.00000000000000    O   (8e)
  -0.05300000000000  -0.55300000000000  -1.10600000000000    O   (8e)
   1.55300000000000   1.05300000000000   2.10600000000000    O   (8e)
   0.37500000000000   0.12170000000000   0.24670000000000   Re   (8f)
   0.87500000000000   0.12830000000000   0.75330000000000   Re   (8f)
   0.87170000000000   0.12500000000000   0.24670000000000   Re   (8f)
   0.87830000000000   0.62500000000000   0.75330000000000   Re   (8f)
```

*F*$5_4$ ($NH_4ClO_2$) (*obsolete*): ABC2_tP8_100_b_a_c - CIF

```
# CIF file
data_findsym-output
_audit_creation_method FINDSYM

_chemical_name_mineral 'Cl(NH4)O2'
_chemical_formula_sum 'Cl (NH4) O2'

loop_
_publ_author_name
 'G. R. Levi'
 'A. Scherillo'
_journal_name_full_name
;
 Zeitschrift f{\"u}r Kristallographie - Crystalline Materials
;
_journal_volume 76
_journal_year 1931
_journal_page_first 431
_journal_page_last 452
_publ_Section_title
;
 Ricerche cristallografiche sui sali dell\'acido cloroso
;

# Found in Strukturbericht Band II 1928-1932, 1937

_aflow_title '$F5_{4}$ (NH$_{4}$ClO$_{2}$) ({\em{obsolete}}) Structure'
_aflow_proto 'ABC2_tP8_100_b_a_c'
_aflow_params 'a,c/a,z_{1},z_{2},x_{3},z_{3}'
_aflow_params_values '6.3,0.587301587302,0.0,0.25,0.65,0.5'
_aflow_Strukturbericht '$F5_{4}$'
_aflow_Pearson 'tP8'

_symmetry_space_group_name_H-M "P 4 b m"
_symmetry_Int_Tables_number 100

_cell_length_a      6.30000
_cell_length_b      6.30000
_cell_length_c      3.70000
_cell_angle_alpha   90.00000
_cell_angle_beta    90.00000
_cell_angle_gamma   90.00000

loop_
_space_group_symop_id
_space_group_symop_operation_xyz
1 x,y,z
2 -x,-y,z
3 -y,x,z
4 y,-x,z
5 -x+1/2,y+1/2,z
6 x+1/2,-y+1/2,z
7 y+1/2,x+1/2,z
8 -y+1/2,-x+1/2,z

loop_
_atom_site_label
_atom_site_type_symbol
_atom_site_symmetry_multiplicity
_atom_site_Wyckoff_label
_atom_site_fract_x
_atom_site_fract_y
_atom_site_fract_z
_atom_site_occupancy
NH41 NH4 2 a 0.00000 0.00000 0.00000 1.00000
Cl1 Cl 2 b 0.50000 0.00000 0.25000 1.00000
O1 O 4 c 0.65000 0.15000 0.50000 1.00000
```

*F*$5_4$ ($NH_4ClO_2$) (*obsolete*): ABC2_tP8_100_b_a_c - POSCAR

```
ABC2_tP8_100_b_a_c & a,c/a,z1,z2,x3,z3 --params=6.3,0.587301587302,0.0,
↪ 0.25,0.65,0.5 & P4bm C_{4v}^{2} #100 (abc) & tP8 & $F5_{4}$ &
↪ Cl(NH4)O2 & Cl(NH4)O2 & G. R. Levi and A. Scherillo,
↪ Zeitschrift f{"u}r Kristallographie - Crystalline Materials 76,
↪ 431-452 (1931)
   1.00000000000000
   6.30000000000000   0.00000000000000   0.00000000000000
   0.00000000000000   6.30000000000000   0.00000000000000
   0.00000000000000   0.00000000000000   3.70000000000000
    Cl   NH4     O
     2     2     4
Direct
   0.50000000000000   0.00000000000000   0.25000000000000   Cl    (2b)
   0.00000000000000   0.50000000000000   0.25000000000000   Cl    (2b)
   0.00000000000000   0.00000000000000   0.00000000000000   NH4   (2a)
   0.50000000000000   0.50000000000000   0.00000000000000   NH4   (2a)
   0.65000000000000   1.15000000000000   0.50000000000000    O    (4c)
  -0.65000000000000  -0.15000000000000   0.50000000000000    O    (4c)
  -0.15000000000000   0.65000000000000   0.50000000000000    O    (4c)
   1.15000000000000  -0.65000000000000   0.50000000000000    O    (4c)
```

$NH_4NO_3$ II (*G*$0_9$): ABC3_tP10_100_b_a_bc - CIF

```
# CIF file
data_findsym-output
_audit_creation_method FINDSYM

_chemical_name_mineral 'N(NH4)O3'
_chemical_formula_sum 'N (NH4) O3'

loop_
_publ_author_name
 'Y. Shinnaka'
_journal_name_full_name
;
 Journal of the Physical Society of Japan
;
_journal_volume 11
_journal_year 1956
_journal_page_first 393
_journal_page_last 396
_publ_Section_title
;
 On the Metastable Transition and the Crystal Structure of Ammonium
↪ Nitrate (Tetragonal Modification)
```

```
;

# Found in The structure of ammonium nitrate (IV), 1972

_aflow_title 'NH$_{4}$NO$_{3}$ II ($G0_{9}$) Structure'
_aflow_proto 'ABC3_tP10_100_b_a_bc'
_aflow_params 'a,c/a,z_{1},z_{2},z_{3},x_{4},z_{4}'
_aflow_params_values '5.74,0.862369337979,0.0,0.53,0.75,0.62,0.42'
_aflow_Strukturbericht '$G0_{9}$'
_aflow_Pearson 'tP10'

_symmetry_space_group_name_H-M "P 4 b m"
_symmetry_Int_Tables_number 100

_cell_length_a    5.74000
_cell_length_b    5.74000
_cell_length_c    4.95000
_cell_angle_alpha 90.00000
_cell_angle_beta  90.00000
_cell_angle_gamma 90.00000

loop_
_space_group_symop_id
_space_group_symop_operation_xyz
1 x,y,z
2 -x,-y,z
3 -y,x,z
4 y,-x,z
5 -x+1/2,y+1/2,z
6 x+1/2,-y+1/2,z
7 y+1/2,x+1/2,z
8 -y+1/2,-x+1/2,z

loop_
_atom_site_label
_atom_site_type_symbol
_atom_site_symmetry_multiplicity
_atom_site_Wyckoff_label
_atom_site_fract_x
_atom_site_fract_y
_atom_site_fract_z
_atom_site_occupancy
NH41 NH4 2 a 0.00000 0.00000 0.00000 1.00000
N1 N 2 b 0.50000 0.00000 0.53000 1.00000
O1 O 2 b 0.50000 0.00000 0.75000 1.00000
O2 O 4 c 0.62000 0.12000 0.42000 1.00000
```

$NH_4NO_3$ II ($G0_9$): ABC3_tP10_100_b_a_bc - POSCAR

```
ABC3_tP10_100_b_a_bc & a,c/a,z1,z2,z3,x4,z4 --params=5.74,0.862369337979
    ↪ ,0.0,0.53,0.75,0.62,0.42 & P4bm C_{4v}^{2} #100 (ab^2c) & tP10
    ↪ & $G0_{9}$ & N(NH4)O3 & N(NH4)O3 & Y. Shinnaka, J. Phys. Soc.
    ↪ Jpn. 11, 393-396 (1956)
   1.00000000000000
   5.74000000000000   0.00000000000000   0.00000000000000
   0.00000000000000   5.74000000000000   0.00000000000000
   0.00000000000000   0.00000000000000   4.95000000000000
     N     NH4     O
     2      2      6
Direct
   0.50000000000000   0.00000000000000   0.53000000000000   N    (2b)
   0.00000000000000   0.50000000000000   0.53000000000000   N    (2b)
   0.00000000000000   0.00000000000000   0.00000000000000   NH4  (2a)
   0.50000000000000   0.50000000000000   0.00000000000000   NH4  (2a)
   0.50000000000000   0.00000000000000   0.75000000000000   O    (2b)
   0.00000000000000   0.50000000000000   0.75000000000000   O    (2b)
   0.62000000000000   1.12000000000000   0.42000000000000   O    (4c)
  -0.62000000000000  -0.12000000000000   0.42000000000000   O    (4c)
  -0.12000000000000   0.62000000000000   0.42000000000000   O    (4c)
   1.12000000000000  -0.62000000000000   0.42000000000000   O    (4c)
```

$VSe_2O_6$: A6B2C_tP72_103_abc5d_2d_abc - CIF

```
# CIF file
data_findsym-output
_audit_creation_method FINDSYM

_chemical_name_mineral 'O6Se2V'
_chemical_formula_sum 'O6 Se2 V'

loop_
_publ_author_name
 'G. Meunier'
 'M. Bertaud'
 'J. Galy'
_journal_name_full_name
;
 Acta Crystallographica Section B: Structural Science
;
_journal_volume 30
_journal_year 1974
_journal_page_first 2834
_journal_page_last 2839
_publ_Section_title
;
 Cristallochimie du s\\'{e}l\\'{e}nium(+IV). I. VSe$_{2}$O$_{6}$, une
    ↪ structure \`{a} trois cha\^{i}nes parall\`{e}les (VO$_{5}$)$^{
    ↪ 6n-}_{n}$ ind\\'{e}pendantes pont\\'{e}es par des groupements
    ↪ (Se$_{2}$O)$^{6+}$
;

_aflow_title 'VSe$_{2}$O$_{6}$ Structure'
_aflow_proto 'A6B2C_tP72_103_abc5d_2d_abc'
_aflow_params 'a,c/a,z_{1},z_{2},z_{3},z_{4},z_{5},z_{6},x_{7},y_{7},z_{
    ↪ 7},x_{8},y_{8},z_{8},x_{9},y_{9},z_{9},x_{10},y_{10},z_{10},x_{
    ↪ 11},y_{11},z_{11},x_{12},y_{12},z_{12},x_{13},y_{13},z_{13}'
_aflow_params_values '11.22,0.700534759358,0.377,0.1715,0.223,0.4274,
    ↪ 0.391,0.0972,0.1233,0.12,0.13,0.2546,0.255,0.44,0.0919,0.3511,
    ↪ 0.146,0.3271,0.4759,0.468,0.4059,0.1432,0.15,0.1033,0.2405,0.0,
    ↪ 0.3671,0.2818,0.1031'
_aflow_Strukturbericht 'None'
_aflow_Pearson 'tP72'

_symmetry_space_group_name_H-M "P 4 c c"
_symmetry_Int_Tables_number 103

_cell_length_a    11.22000
_cell_length_b    11.22000
_cell_length_c    7.86000
_cell_angle_alpha 90.00000
_cell_angle_beta  90.00000
_cell_angle_gamma 90.00000

loop_
_space_group_symop_id
_space_group_symop_operation_xyz
1 x,y,z
2 -x,-y,z
3 -y,x,z
4 y,-x,z
5 -x,y,z+1/2
6 x,-y,z+1/2
7 y,x,z+1/2
8 -y,-x,z+1/2

loop_
_atom_site_label
_atom_site_type_symbol
_atom_site_symmetry_multiplicity
_atom_site_Wyckoff_label
_atom_site_fract_x
_atom_site_fract_y
_atom_site_fract_z
_atom_site_occupancy
O1 O 2 a 0.00000 0.00000 0.37700 1.00000
V1 V 2 a 0.00000 0.00000 0.17150 1.00000
O2 O 2 b 0.50000 0.50000 0.22300 1.00000
V2 V 2 b 0.50000 0.50000 0.42740 1.00000
O3 O 4 c 0.00000 0.50000 0.39100 1.00000
V3 V 4 c 0.00000 0.50000 0.09720 1.00000
O4 O 8 d 0.12330 0.12000 0.13000 1.00000
O5 O 8 d 0.25460 0.25500 0.44000 1.00000
O6 O 8 d 0.09190 0.35110 0.14600 1.00000
O7 O 8 d 0.32710 0.47590 0.46800 1.00000
O8 O 8 d 0.40590 0.14320 0.15000 1.00000
Se1 Se 8 d 0.10330 0.24050 0.00000 1.00000
Se2 Se 8 d 0.36710 0.28180 0.10310 1.00000
```

$VSe_2O_6$: A6B2C_tP72_103_abc5d_2d_abc - POSCAR

```
A6B2C_tP72_103_abc5d_2d_abc & a,c/a,z1,z2,z3,z4,z5,z6,x7,y7,z7,x8,y8,z8,
    ↪ x9,y9,z9,x10,y10,z10,x11,y11,z11,x12,y12,z12,x13,y13,z13 --
    ↪ params=11.22,0.700534759358,0.377,0.1715,0.223,0.4274,0.391,
    ↪ 0.0972,0.1233,0.12,0.13,0.2546,0.255,0.44,0.0919,0.3511,0.146,
    ↪ 0.3271,0.4759,0.468,0.4059,0.1432,0.15,0.1033,0.2405,0.0,0.3671
    ↪ ,0.2818,0.1031 & P4cc C_{4v}^{5} #103 (a^2b^2c^2d^7) & tP72 &
    ↪ None & O6Se2V & O6Se2V & G. Meunier and M. Bertaud and J. Galy,
    ↪ Acta Crystallogr. Sect. B Struct. Sci. 30, 2834-2839 (1974)
   1.00000000000000
  11.22000000000000   0.00000000000000   0.00000000000000
   0.00000000000000  11.22000000000000   0.00000000000000
   0.00000000000000   0.00000000000000   7.86000000000000
     O     Se     V
    48     16     8
Direct
   0.00000000000000   0.00000000000000   0.37700000000000   O   (2a)
   0.00000000000000   0.00000000000000   0.87700000000000   O   (2a)
   0.50000000000000   0.50000000000000   0.22300000000000   O   (2b)
   0.50000000000000   0.50000000000000   0.72300000000000   O   (2b)
   0.00000000000000   0.50000000000000   0.39100000000000   O   (4c)
   0.50000000000000   0.00000000000000   0.39100000000000   O   (4c)
   0.00000000000000   0.50000000000000   0.89100000000000   O   (4c)
   0.50000000000000   0.00000000000000   0.89100000000000   O   (4c)
   0.12330000000000   0.12000000000000   0.13000000000000   O   (8d)
  -0.12330000000000  -0.12000000000000   0.13000000000000   O   (8d)
  -0.12000000000000   0.12330000000000   0.13000000000000   O   (8d)
   0.12000000000000  -0.12330000000000   0.13000000000000   O   (8d)
   0.12330000000000  -0.12000000000000   0.63000000000000   O   (8d)
  -0.12330000000000   0.12000000000000   0.63000000000000   O   (8d)
  -0.12000000000000  -0.12330000000000   0.63000000000000   O   (8d)
   0.12000000000000   0.12330000000000   0.63000000000000   O   (8d)
   0.25460000000000   0.25500000000000   0.44000000000000   O   (8d)
  -0.25460000000000  -0.25500000000000   0.44000000000000   O   (8d)
  -0.25500000000000   0.25460000000000   0.44000000000000   O   (8d)
   0.25500000000000  -0.25460000000000   0.44000000000000   O   (8d)
   0.25460000000000  -0.25500000000000   0.94000000000000   O   (8d)
  -0.25460000000000   0.25500000000000   0.94000000000000   O   (8d)
  -0.25500000000000  -0.25460000000000   0.94000000000000   O   (8d)
   0.25500000000000   0.25460000000000   0.94000000000000   O   (8d)
   0.09190000000000   0.35110000000000   0.14600000000000   O   (8d)
  -0.09190000000000  -0.35110000000000   0.14600000000000   O   (8d)
  -0.35110000000000   0.09190000000000   0.14600000000000   O   (8d)
   0.35110000000000  -0.09190000000000   0.14600000000000   O   (8d)
   0.09190000000000  -0.35110000000000   0.64600000000000   O   (8d)
  -0.09190000000000   0.35110000000000   0.64600000000000   O   (8d)
  -0.35110000000000  -0.09190000000000   0.64600000000000   O   (8d)
   0.35110000000000   0.09190000000000   0.64600000000000   O   (8d)
   0.32710000000000   0.47590000000000   0.46800000000000   O   (8d)
```

```
 -0.32710000000000  -0.47590000000000   0.46800000000000   O   (8d)
 -0.47590000000000   0.32710000000000   0.46800000000000   O   (8d)
  0.47590000000000  -0.32710000000000   0.46800000000000   O   (8d)
  0.32710000000000  -0.47590000000000   0.96800000000000   O   (8d)
 -0.32710000000000   0.47590000000000   0.96800000000000   O   (8d)
 -0.47590000000000  -0.32710000000000   0.96800000000000   O   (8d)
  0.47590000000000   0.32710000000000   0.96800000000000   O   (8d)
  0.40590000000000   0.14320000000000   0.15000000000000   O   (8d)
 -0.40590000000000  -0.14320000000000   0.15000000000000   O   (8d)
 -0.14320000000000   0.40590000000000   0.15000000000000   O   (8d)
  0.14320000000000  -0.40590000000000   0.15000000000000   O   (8d)
  0.40590000000000  -0.14320000000000   0.65000000000000   O   (8d)
 -0.40590000000000   0.14320000000000   0.65000000000000   O   (8d)
 -0.14320000000000  -0.40590000000000   0.65000000000000   O   (8d)
  0.14320000000000   0.40590000000000   0.65000000000000   O   (8d)
  0.10330000000000   0.24050000000000   0.00000000000000   Se  (8d)
 -0.10330000000000  -0.24050000000000   0.00000000000000   Se  (8d)
 -0.24050000000000   0.10330000000000   0.00000000000000   Se  (8d)
  0.24050000000000  -0.10330000000000   0.00000000000000   Se  (8d)
  0.10330000000000  -0.24050000000000   0.50000000000000   Se  (8d)
 -0.10330000000000   0.24050000000000   0.50000000000000   Se  (8d)
 -0.24050000000000  -0.10330000000000   0.50000000000000   Se  (8d)
  0.24050000000000   0.10330000000000   0.50000000000000   Se  (8d)
  0.36710000000000   0.28180000000000   0.10310000000000   Se  (8d)
 -0.36710000000000  -0.28180000000000   0.10310000000000   Se  (8d)
 -0.28180000000000   0.36710000000000   0.10310000000000   Se  (8d)
  0.28180000000000  -0.36710000000000   0.10310000000000   Se  (8d)
  0.36710000000000  -0.28180000000000   0.60310000000000   Se  (8d)
 -0.36710000000000   0.28180000000000   0.60310000000000   Se  (8d)
 -0.28180000000000  -0.36710000000000   0.60310000000000   Se  (8d)
  0.28180000000000   0.36710000000000   0.60310000000000   Se  (8d)
  0.00000000000000   0.00000000000000   0.17150000000000   V   (2a)
  0.00000000000000   0.00000000000000   0.67150000000000   V   (2a)
  0.50000000000000   0.50000000000000   0.42740000000000   V   (2b)
  0.50000000000000   0.50000000000000   0.92740000000000   V   (2b)
  0.00000000000000   0.50000000000000   0.09720000000000   V   (4c)
  0.50000000000000   0.00000000000000   0.09720000000000   V   (4c)
  0.00000000000000   0.50000000000000   0.59720000000000   V   (4c)
  0.50000000000000   0.00000000000000   0.59720000000000   V   (4c)
```

$BaNiSn_3$: ABC3_tI10_107_a_a_ab - CIF

```
# CIF file
data_findsym-output
_audit_creation_method FINDSYM

_chemical_name_mineral 'BaNiSn3'
_chemical_formula_sum 'Ba Ni Sn3'

loop_
_publ_author_name
 'W. D\"{o}rrscheidt'
 'H. Sch\"{a}fer'
_journal_name_full_name
;
 Journal of the Less-Common Metals
;
_journal_volume 58
_journal_year 1978
_journal_page_first 209
_journal_page_last 216
_publ_Section_title
;
 Die Struktur des BaPtSn$_{3}$, BaNiSn$_{3}$ und SrNiSn$_{3}$ und ihre
   ↪ Verwandtschaft zum ThCr$_{2}$Si$_{2}$-Strukturtyp
;

_aflow_title 'BaNiSn$_{3}$ Structure'
_aflow_proto 'ABC3_tI10_107_a_a_ab'
_aflow_params 'a,c/a,z_{1},z_{2},z_{3},z_{4}'
_aflow_params_values '4.82,2.26763692946,0.0,0.6554,0.4241,0.2516'
_aflow_Strukturbericht 'None'
_aflow_Pearson 'tI10'

_symmetry_space_group_name_H-M "I 4 m m"
_symmetry_Int_Tables_number 107

_cell_length_a      4.82000
_cell_length_b      4.82000
_cell_length_c      10.93001
_cell_angle_alpha   90.00000
_cell_angle_beta    90.00000
_cell_angle_gamma   90.00000

loop_
_space_group_symop_id
_space_group_symop_operation_xyz
1 x,y,z
2 -x,-y,z
3 -y,x,z
4 y,-x,z
5 -x,y,z
6 x,-y,z
7 y,x,z
8 -y,-x,z
9 x+1/2,y+1/2,z+1/2
10 -x+1/2,-y+1/2,z+1/2
11 -y+1/2,x+1/2,z+1/2
12 y+1/2,-x+1/2,z+1/2
13 -x+1/2,y+1/2,z+1/2
14 x+1/2,-y+1/2,z+1/2
15 y+1/2,x+1/2,z+1/2
16 -y+1/2,-x+1/2,z+1/2

loop_
_atom_site_label
_atom_site_type_symbol
_atom_site_symmetry_multiplicity
_atom_site_Wyckoff_label
_atom_site_fract_x
_atom_site_fract_y
_atom_site_fract_z
_atom_site_occupancy
Ba1 Ba 2 a 0.00000 0.00000 0.00000 1.00000
Ni1 Ni 2 a 0.00000 0.00000 0.65540 1.00000
Sn1 Sn 2 a 0.00000 0.00000 0.42410 1.00000
Sn2 Sn 4 b 0.00000 0.50000 0.25160 1.00000
```

$BaNiSn_3$: ABC3_tI10_107_a_a_ab - POSCAR

```
ABC3_tI10_107_a_a_ab & a,c/a,z1,z2,z3,z4 --params=4.82,2.26763692946,0.0
   ↪ ,0.6554,0.4241,0.2516 & I4mm C_{4v}^{9} #107 (a^3b) & tI10 &
   ↪ None & BaNiSn3 & BaNiSn3 & W. D\"{o}rrscheidt and H. Sch\"{a}
   ↪ fer, J. Less-Common Met. 58, 209-216 (1978)
   1.00000000000000
  -2.41000000000000   2.41000000000000   5.46500500000000
   2.41000000000000  -2.41000000000000   5.46500500000000
   2.41000000000000   2.41000000000000  -5.46500500000000
    Ba     Ni     Sn
     1      1      3
Direct
   0.00000000000000   0.00000000000000   0.00000000000000   Ba   (2a)
   0.65540000000000   0.65540000000000   0.00000000000000   Ni   (2a)
   0.42410000000000   0.42410000000000   0.00000000000000   Sn   (2a)
   0.75160000000000   0.25160000000000   0.50000000000000   Sn   (4b)
   0.25160000000000   0.75160000000000   0.50000000000000   Sn   (4b)
```

$E3_1$ ($\beta$-$Ag_2HgI_4$) (*obsolete*): A2BC4_tP7_111_f_a_n - CIF

```
# CIF file
data_findsym-output
_audit_creation_method FINDSYM

_chemical_name_mineral 'Ag2HgI4'
_chemical_formula_sum 'Ag2 Hg I4'

loop_
_publ_author_name
 'J. A. A. Ketelaar'
_journal_name_full_name
;
 Zeitschrift f{\"u}r Kristallographie - Crystalline Materials
;
_journal_volume 80
_journal_year 1931
_journal_page_first 190
_journal_page_last 203
_publ_Section_title
;
 Strukturbestimmung der komplexen Quecksilberverbindungen Ag$_{2}$HgJ$_{
   ↪ 4}$ und Cu$_{2}$HgJ$_{4}$
;

# Found in Single-crystal studies of $\beta$-Ag$_{2}$HgI$_{4}$, 1974

_aflow_title '$E3_{1}$ ($\beta$-Ag$_{2}$HgI$_{4}$) ({\em{obsolete}})
   ↪ Structure'
_aflow_proto 'A2BC4_tP7_111_f_a_n'
_aflow_params 'a,c/a,x_{3},z_{3}'
_aflow_params_values '6.34,1.0,0.27,0.225'
_aflow_Strukturbericht '$E3_{1}$'
_aflow_Pearson 'tP7'

_symmetry_space_group_name_H-M "P -4 2 m"
_symmetry_Int_Tables_number 111

_cell_length_a      6.34000
_cell_length_b      6.34000
_cell_length_c      6.34000
_cell_angle_alpha   90.00000
_cell_angle_beta    90.00000
_cell_angle_gamma   90.00000

loop_
_space_group_symop_id
_space_group_symop_operation_xyz
1 x,y,z
2 x,-y,-z
3 -x,y,-z
4 -x,-y,z
5 y,x,z
6 y,-x,-z
7 -y,x,-z
8 -y,-x,z

loop_
_atom_site_label
_atom_site_type_symbol
_atom_site_symmetry_multiplicity
_atom_site_Wyckoff_label
_atom_site_fract_x
_atom_site_fract_y
_atom_site_fract_z
_atom_site_occupancy
Hg1 Hg 1 a 0.00000 0.00000 0.00000 1.00000
Ag1 Ag 2 f 0.50000 0.00000 0.50000 1.00000
I1 I 4 n 0.27000 0.27000 0.22500 1.00000
```

$E3_1$ ($\beta$-$Ag_2HgI_4$) (*obsolete*): A2BC4_tP7_111_f_a_n - POSCAR

```
A2BC4_tP7_111_f_a_n & a,c/a,x3,z3 --params=6.34,1.0,0.27,0.225 & P-42m
    ↪ D_{2d}^{1} #111 (afn) & tP7 & $E3_{1}$ & Ag2HgI4 & Ag2HgI4 & J.
    ↪ A. A. Ketelaar, Zeitschrift f{"u}r Kristallographie -
    ↪ Crystalline Materials 80, 190-203 (1931)
   1.00000000000000
   6.34000000000000   0.00000000000000   0.00000000000000
   0.00000000000000   6.34000000000000   0.00000000000000
   0.00000000000000   0.00000000000000   6.34000000000000
    Ag    Hg    I
     2     1    4
Direct
   0.50000000000000   0.00000000000000   0.50000000000000   Ag   (2f)
   0.00000000000000   0.50000000000000   0.50000000000000   Ag   (2f)
   0.00000000000000   0.00000000000000   0.00000000000000   Hg   (1a)
   0.27000000000000   0.27000000000000   0.22500000000000   I    (4n)
  -0.27000000000000  -0.27000000000000   0.22500000000000   I    (4n)
   0.27000000000000  -0.27000000000000  -0.22500000000000   I    (4n)
  -0.27000000000000   0.27000000000000  -0.22500000000000   I    (4n)
```

Ammonium Chlorite ($NH_4ClO_2$): AB4CD2_tP16_113_c_f_a_e - CIF

```
# CIF file
data_findsym-output
_audit_creation_method FINDSYM

_chemical_name_mineral 'ClH4NO2'
_chemical_formula_sum 'Cl H4 N O2'

loop_
_publ_author_name
 'A. I. Smolentsev'
 'D. Y. Naumov'
_journal_name_full_name
;
 Acta Crystallographica Section E: Crystallographic Communications
;
_journal_volume 61
_journal_year 2005
_journal_page_first i38
_journal_page_last i40
_publ_Section_title
;
 Ammonium chlorite, NH$_{4}$ClO$_{2}$, at 150~K
;

_aflow_title 'Ammonium Chlorite (NH$_{4}$ClO$_{2}$) Structure'
_aflow_proto 'AB4CD2_tP16_113_c_f_a_e'
_aflow_params 'a,c/a,z_{2},x_{3},z_{3},x_{4},y_{4},z_{4}'
_aflow_params_values '6.3397,0.592457056328,0.37123,0.14562,0.6062,0.111
    ↪ ,0.03,0.125'
_aflow_Strukturbericht 'None'
_aflow_Pearson 'tP16'

_symmetry_space_group_name_H-M "P -4 21 m"
_symmetry_Int_Tables_number 113

_cell_length_a      6.33970
_cell_length_b      6.33970
_cell_length_c      3.75600
_cell_angle_alpha  90.00000
_cell_angle_beta   90.00000
_cell_angle_gamma  90.00000

loop_
_space_group_symop_id
_space_group_symop_operation_xyz
1 x,y,z
2 x+1/2,-y+1/2,-z
3 -x+1/2,y+1/2,-z
4 -x,-y,z
5 y+1/2,x+1/2,z
6 y,-x,-z
7 -y,x,-z
8 -y+1/2,-x+1/2,z

loop_
_atom_site_label
_atom_site_type_symbol
_atom_site_symmetry_multiplicity
_atom_site_Wyckoff_label
_atom_site_fract_x
_atom_site_fract_y
_atom_site_fract_z
_atom_site_occupancy
N1 N 2 a 0.00000 0.00000 0.00000 1.00000
Cl1 Cl 2 c 0.00000 0.50000 0.37123 1.00000
O1 O 4 e 0.14562 0.64562 0.60620 1.00000
H1 H 8 f 0.11100 0.03000 0.12500 1.00000
```

Ammonium Chlorite ($NH_4ClO_2$): AB4CD2_tP16_113_c_f_a_e - POSCAR

```
AB4CD2_tP16_113_c_f_a_e & a,c/a,z2,x3,z3,x4,y4,z4 --params=6.3397,
    ↪ 0.592457056328,0.37123,0.14562,0.6062,0.111,0.03,0.125 & P-42_{
    ↪ 1}m D_{2d}^{3} #113 (acef) & tP16 & None & ClH4NO2 & ClH4NO2 &
    ↪ A. I. Smolentsev and D. Y. Naumov, Acta Crystallogr. E 61,
    ↪ i38-i40 (2005)
   1.00000000000000
   6.33970000000000   0.00000000000000   0.00000000000000
   0.00000000000000   6.33970000000000   0.00000000000000
   0.00000000000000   0.00000000000000   3.75600000000000
    Cl     H     N     O
     2     8     2     4
Direct
   0.00000000000000   0.50000000000000   0.37123000000000   Cl   (2c)
   0.50000000000000   0.00000000000000  -0.37123000000000   Cl   (2c)
   0.11100000000000   0.03000000000000   0.12500000000000   H    (8f)
  -0.11100000000000  -0.03000000000000   0.12500000000000   H    (8f)
   0.03000000000000  -0.11100000000000  -0.12500000000000   H    (8f)
  -0.03000000000000   0.11100000000000  -0.12500000000000   H    (8f)
   0.38900000000000   0.53000000000000  -0.12500000000000   H    (8f)
   0.61100000000000   0.47000000000000  -0.12500000000000   H    (8f)
   0.47000000000000   0.38900000000000   0.12500000000000   H    (8f)
   0.53000000000000   0.61100000000000   0.12500000000000   H    (8f)
   0.00000000000000   0.00000000000000   0.00000000000000   N    (2a)
   0.50000000000000   0.50000000000000   0.00000000000000   N    (2a)
   0.14562000000000   0.64562000000000   0.60620000000000   O    (4e)
  -0.14562000000000   0.35438000000000   0.60620000000000   O    (4e)
   0.64562000000000  -0.14562000000000  -0.60620000000000   O    (4e)
   0.35438000000000   0.14562000000000  -0.60620000000000   O    (4e)
```

$C_{19}Sc_{15}$: A19B15_tP68_114_bc4e_ac3e - CIF

```
# CIF file
data_findsym-output
_audit_creation_method FINDSYM

_chemical_name_mineral 'C19Sc15'
_chemical_formula_sum 'C19 Sc15'

loop_
_publ_author_name
 'H. Jedlicka'
 'H. Nowotny'
 'F. Benesovsky'
_journal_name_full_name
;
 Monatshefte f{\"u}r Chemie - Chemical Monthly
;
_journal_volume 102
_journal_year 1971
_journal_page_first 389
_journal_page_last 403
_publ_Section_title
;
 Zum System Scandium-Kohlenstoff, 2. Mitt.: Kristallstruktur des
    ↪ C-reichen Carbids
;

_aflow_title 'C$_{19}$Sc$_{15}$ Structure'
_aflow_proto 'A19B15_tP68_114_bc4e_ac3e'
_aflow_params 'a,c/a,z_{3},z_{4},x_{5},y_{5},z_{5},x_{6},y_{6},z_{6},x_{
    ↪ 7},y_{7},z_{7},x_{8},y_{8},z_{8},x_{9},y_{9},z_{9},x_{10},y_{10
    ↪ },z_{10},x_{11},y_{11},z_{11}'
_aflow_params_values '7.5,2.0,0.165,0.3198,0.712,0.118,-0.008,0.384,0.2,
    ↪ 0.1333,0.436,0.208,0.212,-0.194,0.375,0.179,0.4021,0.1993,-0.02
    ↪ ,0.0982,0.2823,0.1593,0.7157,0.0982,0.1423'
_aflow_Strukturbericht 'None'
_aflow_Pearson 'tP68'

_symmetry_space_group_name_H-M "P -4 21 c"
_symmetry_Int_Tables_number 114

_cell_length_a      7.50000
_cell_length_b      7.50000
_cell_length_c     15.00000
_cell_angle_alpha  90.00000
_cell_angle_beta   90.00000
_cell_angle_gamma  90.00000

loop_
_space_group_symop_id
_space_group_symop_operation_xyz
1 x,y,z
2 x+1/2,-y+1/2,-z+1/2
3 -x+1/2,y+1/2,-z+1/2
4 -x,-y,z
5 y+1/2,x+1/2,z+1/2
6 y,-x,-z
7 -y,x,-z
8 -y+1/2,-x+1/2,z+1/2

loop_
_atom_site_label
_atom_site_type_symbol
_atom_site_symmetry_multiplicity
_atom_site_Wyckoff_label
_atom_site_fract_x
_atom_site_fract_y
_atom_site_fract_z
_atom_site_occupancy
Sc1 Sc 2 a 0.00000 0.00000 0.00000 1.00000
C1 C 2 b 0.00000 0.00000 0.50000 1.00000
C2 C 4 c 0.00000 0.00000 0.16500 1.00000
Sc2 Sc 4 c 0.00000 0.00000 0.31980 1.00000
C3 C 8 e 0.71200 0.11800 -0.00800 1.00000
C4 C 8 e 0.38400 0.20000 0.13330 1.00000
C5 C 8 e 0.43600 0.20800 0.21200 1.00000
C6 C 8 e -0.19400 0.37500 0.17900 1.00000
Sc3 Sc 8 e 0.40210 0.19930 -0.02000 1.00000
Sc4 Sc 8 e 0.09820 0.28230 0.15930 1.00000
Sc5 Sc 8 e 0.71570 0.09820 0.14230 1.00000
```

$C_{19}Sc_{15}$: A19B15_tP68_114_bc4e_ac3e - POSCAR

```
A19B15_tP68_114_bc4e_ac3e & a,c/a,z3,z4,x5,y5,z5,x6,y6,z6,x7,y7,z7,x8,y8
    ↪ ,z8,x9,y9,z9,x10,y10,z10,x11,y11,z11 --params=7.5,2.0,0.165,
    ↪ 0.3198,0.712,0.118,-0.008,0.384,0.2,0.1333,0.436,0.208,0.212,-
    ↪ 0.194,0.375,0.179,0.4021,0.1993,-0.02,0.0982,0.2823,0.1593,
    ↪ 0.7157,0.0982,0.1423 & P-42_{1}c D_{2d}^{4} #114 (abc^2e^7) &
```

```
      ↪ tP68 & None & C19Sc15 & C19Sc15 & H. Jedlicka and H. Nowotny
      ↪ and F. Benesovsky, Monatshefte f{"u}r Chemie - Chemical Monthly
      ↪ 102, 389-403 (1971)
   1.00000000000000
   7.50000000000000    0.00000000000000    0.00000000000000
   0.00000000000000    7.50000000000000    0.00000000000000
   0.00000000000000    0.00000000000000   15.00000000000000
     C      Sc
    38      30
Direct
   0.00000000000000    0.00000000000000    0.50000000000000   C    (2b)
   0.50000000000000    0.50000000000000    0.00000000000000   C    (2b)
   0.00000000000000    0.00000000000000    0.16500000000000   C    (4c)
   0.00000000000000    0.00000000000000   -0.16500000000000   C    (4c)
   0.50000000000000    0.50000000000000    0.33500000000000   C    (4c)
   0.50000000000000    0.50000000000000    0.66500000000000   C    (4c)
   0.71200000000000    0.11800000000000   -0.00800000000000   C    (8e)
  -0.71200000000000   -0.11800000000000   -0.00800000000000   C    (8e)
   0.11800000000000   -0.71200000000000    0.00800000000000   C    (8e)
  -0.11800000000000    0.71200000000000    0.00800000000000   C    (8e)
  -0.21200000000000    0.61800000000000    0.50800000000000   C    (8e)
   1.21200000000000    0.38200000000000    0.50800000000000   C    (8e)
   0.38200000000000   -0.21200000000000    0.49200000000000   C    (8e)
   0.61800000000000    1.21200000000000    0.49200000000000   C    (8e)
   0.38400000000000    0.20000000000000    0.13330000000000   C    (8e)
  -0.38400000000000   -0.20000000000000    0.13330000000000   C    (8e)
   0.20000000000000   -0.38400000000000   -0.13330000000000   C    (8e)
  -0.20000000000000    0.38400000000000   -0.13330000000000   C    (8e)
   0.11600000000000    0.70000000000000    0.36670000000000   C    (8e)
   0.88400000000000    0.30000000000000    0.36670000000000   C    (8e)
   0.30000000000000    0.11600000000000    0.63330000000000   C    (8e)
   0.70000000000000    0.88400000000000    0.63330000000000   C    (8e)
   0.43600000000000    0.20800000000000    0.21200000000000   C    (8e)
  -0.43600000000000   -0.20800000000000    0.21200000000000   C    (8e)
   0.20800000000000   -0.43600000000000   -0.21200000000000   C    (8e)
  -0.20800000000000    0.43600000000000   -0.21200000000000   C    (8e)
   0.06400000000000    0.70800000000000    0.28800000000000   C    (8e)
   0.93600000000000    0.29200000000000    0.28800000000000   C    (8e)
   0.29200000000000    0.06400000000000    0.71200000000000   C    (8e)
   0.70800000000000    0.93600000000000    0.71200000000000   C    (8e)
  -0.19400000000000    0.37500000000000    0.17900000000000   C    (8e)
   0.19400000000000   -0.37500000000000    0.17900000000000   C    (8e)
   0.37500000000000    0.19400000000000   -0.17900000000000   C    (8e)
  -0.37500000000000   -0.19400000000000   -0.17900000000000   C    (8e)
   0.69400000000000    0.87500000000000    0.32100000000000   C    (8e)
   0.30600000000000    0.12500000000000    0.32100000000000   C    (8e)
   0.12500000000000    0.69400000000000    0.67900000000000   C    (8e)
   0.87500000000000    0.30600000000000    0.67900000000000   C    (8e)
   0.00000000000000    0.00000000000000    0.00000000000000   Sc   (2a)
   0.50000000000000    0.50000000000000    0.50000000000000   Sc   (2a)
   0.00000000000000    0.00000000000000    0.31980000000000   Sc   (4c)
   0.00000000000000    0.00000000000000   -0.31980000000000   Sc   (4c)
   0.50000000000000    0.50000000000000    0.18020000000000   Sc   (4c)
   0.50000000000000    0.50000000000000    0.81980000000000   Sc   (4c)
   0.40210000000000    0.19930000000000   -0.02000000000000   Sc   (8e)
  -0.40210000000000   -0.19930000000000   -0.02000000000000   Sc   (8e)
   0.19930000000000   -0.40210000000000    0.02000000000000   Sc   (8e)
  -0.19930000000000    0.40210000000000    0.02000000000000   Sc   (8e)
   0.09790000000000    0.69930000000000    0.52000000000000   Sc   (8e)
   0.90210000000000    0.30070000000000    0.52000000000000   Sc   (8e)
   0.30070000000000    0.09790000000000    0.48000000000000   Sc   (8e)
   0.69930000000000    0.90210000000000    0.48000000000000   Sc   (8e)
   0.09820000000000    0.28230000000000    0.15930000000000   Sc   (8e)
  -0.09820000000000   -0.28230000000000    0.15930000000000   Sc   (8e)
   0.28230000000000   -0.09820000000000   -0.15930000000000   Sc   (8e)
  -0.28230000000000    0.09820000000000   -0.15930000000000   Sc   (8e)
   0.40180000000000    0.78230000000000    0.34070000000000   Sc   (8e)
   0.59820000000000    0.21770000000000    0.34070000000000   Sc   (8e)
   0.21770000000000    0.40180000000000    0.65930000000000   Sc   (8e)
   0.78230000000000    0.59820000000000    0.65930000000000   Sc   (8e)
   0.71570000000000    0.09820000000000    0.14230000000000   Sc   (8e)
  -0.71570000000000   -0.09820000000000    0.14230000000000   Sc   (8e)
   0.09820000000000   -0.71570000000000   -0.14230000000000   Sc   (8e)
  -0.09820000000000    0.71570000000000   -0.14230000000000   Sc   (8e)
  -0.21570000000000    0.59820000000000    0.35770000000000   Sc   (8e)
   1.21570000000000    0.40180000000000    0.35770000000000   Sc   (8e)
   0.40180000000000   -0.21570000000000    0.64230000000000   Sc   (8e)
   0.59820000000000    1.21570000000000    0.64230000000000   Sc   (8e)
```

$Ag_2SO_4 \cdot 4NH_3$ ($H4_{17}$): A2B12C4D4E_tP46_114_d_3e_e_e_a - CIF

```
# CIF file
data_findsym-output
_audit_creation_method FINDSYM

_chemical_name_mineral 'Ag2H12N4O4S'
_chemical_formula_sum 'Ag2 H12 N4 O4 S'

loop_
_publ_author_name
 'U. Zachwieja'
 'H. Jacobs'
_journal_name_full_name
;
 Zeitschrift f{\"u}r Kristallographie - Crystalline Materials
;
_journal_volume 201
_journal_year 1992
_journal_page_first 207
_journal_page_last 212
_publ_Section_title
;
 Redetermination of the crystal structure of diammine silver(I)-sulfate,
      ↪ [Ag(NH$_{3}$)$_{2}$]$_{2}$SO$_{4}$
;

_aflow_title 'Ag$_{2}$SO$_{4}$$\cdot$4NH$_{3}$ ($H4_{17}$) Structure'
_aflow_proto 'A2B12C4D4E_tP46_114_d_3e_e_e_a'
_aflow_params 'a,c/a,z_{2},x_{3},y_{3},z_{3},x_{4},y_{4},z_{4},x_{5},y_{
      ↪ 5},z_{5},x_{6},y_{6},z_{6},x_{7},y_{7},z_{7}'
_aflow_params_values '8.442,0.757995735608,0.49194,0.082,0.21,0.416,
      ↪ 0.183,0.287,0.52,0.063,0.227,0.603,0.1081,0.275,0.5083,0.1301,
      ↪ 0.0595,0.1309'
_aflow_Strukturbericht '$H4_{17}$'
_aflow_Pearson 'tP46'

_symmetry_space_group_name_H-M "P -4 21 c"
_symmetry_Int_Tables_number 114

_cell_length_a     8.44200
_cell_length_b     8.44200
_cell_length_c     6.39900
_cell_angle_alpha 90.00000
_cell_angle_beta  90.00000
_cell_angle_gamma 90.00000

loop_
_space_group_symop_id
_space_group_symop_operation_xyz
1 x,y,z
2 x+1/2,-y+1/2,-z+1/2
3 -x+1/2,y+1/2,-z+1/2
4 -x,-y,z
5 y+1/2,x+1/2,z+1/2
6 y,-x,-z
7 -y,x,-z
8 -y+1/2,-x+1/2,z+1/2

loop_
_atom_site_label
_atom_site_type_symbol
_atom_site_symmetry_multiplicity
_atom_site_Wyckoff_label
_atom_site_fract_x
_atom_site_fract_y
_atom_site_fract_z
_atom_site_occupancy
S1 S 2 a 0.00000 0.00000 0.00000 1.00000
Ag1 Ag 4 d 0.00000 0.50000 0.49194 1.00000
H1 H 8 e 0.08200 0.21000 0.41600 1.00000
H2 H 8 e 0.18300 0.28700 0.52000 1.00000
H3 H 8 e 0.06300 0.22700 0.60300 1.00000
N1 N 8 e 0.10810 0.27500 0.50830 1.00000
O1 O 8 e 0.13010 0.05950 0.13090 1.00000
```

$Ag_2SO_4 \cdot 4NH_3$ ($H4_{17}$): A2B12C4D4E_tP46_114_d_3e_e_e_a - POSCAR

```
A2B12C4D4E_tP46_114_d_3e_e_e_a & a,c/a,z2,x3,y3,z3,x4,y4,z4,x5,y5,z5,x6,
      ↪ y6,z6,x7,y7,z7 --params=8.442,0.757995735608,0.49194,0.082,0.21
      ↪ ,0.416,0.183,0.287,0.52,0.063,0.227,0.603,0.1081,0.275,0.5083,
      ↪ 0.1301,0.0595,0.1309 & P-42_{1}c D_{2d}^{4} #114 (ade^5) & tP46
      ↪ & $H4_{17}$ & Ag2H12N4O4S & Ag2H12N4O4S & U. Zachwieja and H.
      ↪ Jacobs, Zeitschrift f{"u}r Kristallographie - Crystalline
      ↪ Materials 201, 207-212 (1992)
   1.00000000000000
   8.44200000000000    0.00000000000000    0.00000000000000
   0.00000000000000    8.44200000000000    0.00000000000000
   0.00000000000000    0.00000000000000    6.39900000000000
    Ag       H       N       O       S
     4      24       8       8       2
Direct
   0.00000000000000    0.50000000000000    0.49194000000000   Ag   (4d)
   0.50000000000000    0.00000000000000   -0.49194000000000   Ag   (4d)
   0.50000000000000    0.00000000000000    0.00806000000000   Ag   (4d)
   0.00000000000000    0.50000000000000    0.99194000000000   Ag   (4d)
   0.08200000000000    0.21000000000000    0.41600000000000   H    (8e)
  -0.08200000000000   -0.21000000000000    0.41600000000000   H    (8e)
   0.21000000000000   -0.08200000000000   -0.41600000000000   H    (8e)
  -0.21000000000000    0.08200000000000   -0.41600000000000   H    (8e)
   0.41800000000000    0.71000000000000    0.08400000000000   H    (8e)
   0.58200000000000    0.29000000000000    0.08400000000000   H    (8e)
   0.29000000000000    0.41800000000000    0.91600000000000   H    (8e)
   0.71000000000000    0.58200000000000    0.91600000000000   H    (8e)
   0.18300000000000    0.28700000000000    0.52000000000000   H    (8e)
  -0.18300000000000   -0.28700000000000    0.52000000000000   H    (8e)
   0.28700000000000   -0.18300000000000   -0.52000000000000   H    (8e)
  -0.28700000000000    0.18300000000000   -0.52000000000000   H    (8e)
   0.31700000000000    0.78700000000000   -0.02000000000000   H    (8e)
   0.68300000000000    0.21300000000000   -0.02000000000000   H    (8e)
   0.21300000000000    0.31700000000000    1.02000000000000   H    (8e)
   0.78700000000000    0.68300000000000    1.02000000000000   H    (8e)
   0.06300000000000    0.22700000000000    0.60300000000000   H    (8e)
  -0.06300000000000   -0.22700000000000    0.60300000000000   H    (8e)
   0.22700000000000   -0.06300000000000   -0.60300000000000   H    (8e)
  -0.22700000000000    0.06300000000000   -0.60300000000000   H    (8e)
   0.43700000000000    0.72700000000000   -0.10300000000000   H    (8e)
   0.56300000000000    0.27300000000000   -0.10300000000000   H    (8e)
   0.27300000000000    0.43700000000000    1.10300000000000   H    (8e)
   0.72700000000000    0.56300000000000    1.10300000000000   H    (8e)
   0.10810000000000    0.27500000000000    0.50830000000000   N    (8e)
  -0.10810000000000   -0.27500000000000    0.50830000000000   N    (8e)
   0.27500000000000   -0.10810000000000   -0.50830000000000   N    (8e)
  -0.27500000000000    0.10810000000000   -0.50830000000000   N    (8e)
   0.39190000000000    0.77500000000000   -0.00830000000000   N    (8e)
   0.60810000000000    0.22500000000000   -0.00830000000000   N    (8e)
   0.22500000000000    0.39190000000000    1.00830000000000   N    (8e)
   0.77500000000000    0.60810000000000    1.00830000000000   N    (8e)
   0.13010000000000    0.05950000000000    0.13090000000000   O    (8e)
  -0.13010000000000   -0.05950000000000    0.13090000000000   O    (8e)
   0.05950000000000   -0.13010000000000   -0.13090000000000   O    (8e)
```

```
   -0.05950000000000    0.13010000000000   -0.13090000000000     O    (8e)
    0.36990000000000    0.55950000000000    0.36910000000000     O    (8e)
    0.63010000000000    0.44050000000000    0.36910000000000     O    (8e)
    0.44050000000000    0.36990000000000    0.63090000000000     O    (8e)
    0.55950000000000    0.63010000000000    0.63090000000000     O    (8e)
    0.00000000000000    0.00000000000000    0.00000000000000     S    (2a)
    0.50000000000000    0.50000000000000    0.50000000000000     S    (2a)
```

$F6_1$ (Chalcopyrite, $CuFeS_2$) (*obsolete*): ABC2_tP4_115_a_c_g - CIF

```
# CIF file
data_findsym-output
_audit_creation_method FINDSYM

_chemical_name_mineral 'CuFeS2'
_chemical_formula_sum 'Cu Fe S2'

loop_
_publ_author_name
 'R. Gro{\ss}'
 'N. Gro{\ss}'
_journal_name_full_name
;
 Neues Jahrbuch fur Mineralogie, Monatshefte
;
_journal_volume 48
_journal_year 1923
_journal_page_first 113
_journal_page_last 135
_publ_Section_title
;
 ~
;

# Found in Strukturbericht 1913-1928, 1931

_aflow_title '$F6_{1}$ (Chalcopyrite, CuFeS$_{2}$) ({\em{obsolete}})
     ↪ Structure'
_aflow_proto 'ABC2_tP4_115_a_c_g'
_aflow_params 'a,c/a,z_{3}'
_aflow_params_values '3.72645,1.39381985536,0.19'
_aflow_Strukturbericht '$F6_{1}$'
_aflow_Pearson 'tP4'

_symmetry_space_group_name_H-M "P -4 m 2"
_symmetry_Int_Tables_number 115

_cell_length_a      3.72645
_cell_length_b      3.72645
_cell_length_c      5.19400
_cell_angle_alpha  90.00000
_cell_angle_beta   90.00000
_cell_angle_gamma  90.00000

loop_
_space_group_symop_id
_space_group_symop_operation_xyz
1 x,y,z
2 -x,-y,z
3 -y,-x,-z
4 y,x,-z
5 -x,y,z
6 x,-y,z
7 y,-x,-z
8 -y,x,-z

loop_
_atom_site_label
_atom_site_type_symbol
_atom_site_symmetry_multiplicity
_atom_site_Wyckoff_label
_atom_site_fract_x
_atom_site_fract_y
_atom_site_fract_z
_atom_site_occupancy
Cu1 Cu 1 a 0.00000 0.00000 0.00000 1.00000
Fe1 Fe 1 c 0.50000 0.50000 0.50000 1.00000
S1 S 2 g 0.00000 0.50000 0.19000 1.00000
```

$F6_1$ (Chalcopyrite, $CuFeS_2$) (*obsolete*): ABC2_tP4_115_a_c_g - POSCAR

```
ABC2_tP4_115_a_c_g & a,c/a,z3 --params=3.72645,1.39381985536,0.19 &
     ↪ P-4m2 D_{2d}^{5} #115 (acg) & tP4 & $F6_{1}$ & CuFeS2 & CuFeS2
     ↪ & R. Gro{\ss} and N. Gro{\ss}, Neues Jahrb. Mineral. Monatsh.
     ↪ 48, 113-135 (1923)
   1.00000000000000
   3.72645000000000    0.00000000000000    0.00000000000000
   0.00000000000000    3.72645000000000    0.00000000000000
   0.00000000000000    0.00000000000000    5.19400000000000
     Cu        Fe        S
      1         1        2
Direct
   0.00000000000000    0.00000000000000    0.00000000000000    Cu    (1a)
   0.50000000000000    0.50000000000000    0.50000000000000    Fe    (1c)
   0.00000000000000    0.50000000000000    0.19000000000000    S     (2g)
   0.50000000000000    0.00000000000000   -0.19000000000000    S     (2g)
```

Phase II $Cd_2Re_2O_7$: A2B7C2_tI44_119_i_bdefgh_i - CIF

```
# CIF file
data_findsym-output
_audit_creation_method FINDSYM

_chemical_name_mineral 'Cd2O7Re2'
_chemical_formula_sum 'Cd2 O7 Re2'

loop_
_publ_author_name
 'S.-W. Huang'
 'H.-T. Jeng'
 'J.-Y. Lin'
 'W. J. Chang'
 'J. M. Chen'
 'G. H. Lee'
 'H. Berger'
 'H. D. Yang'
 'K. S. Liang'
_journal_name_full_name
;
 Journal of Physics: Condensed Matter
;
_journal_volume 21
_journal_year 2009
_journal_page_first 195602
_journal_page_last 195602
_publ_Section_title
;
 Electronic structure of pyrochlore Cd$_{2}$Re$_{2}$O$_{7}$
;

# Found in The crystal structure of the inversion breaking metal Cd$_{2}
     ↪ $Re$_{2}$O$_{7}$, 2019 Found in The crystal structure of the
     ↪ inversion breaking metal Cd$_{2}$Re$_{2}$O$_{7}$, {arXiv:
     ↪ 1911.10141 [cond-mat.str-el]},

_aflow_title 'Phase II Cd$_{2}$Re$_{2}$O$_{7}$ Structure'
_aflow_proto 'A2B7C2_tI44_119_i_bdefgh_i'
_aflow_params 'a,c/a,z_{3},z_{4},x_{5},x_{6},x_{7},z_{7},x_{8},z_{8}'
_aflow_params_values '7.2312,1.41410830844,0.1974,0.0683,0.1889,0.1941,
     ↪ 0.2471,0.3741,0.2471,0.87294'
_aflow_Strukturbericht 'None'
_aflow_Pearson 'tI44'

_symmetry_space_group_name_H-M "I -4 m 2"
_symmetry_Int_Tables_number 119

_cell_length_a      7.23120
_cell_length_b      7.23120
_cell_length_c     10.22570
_cell_angle_alpha  90.00000
_cell_angle_beta   90.00000
_cell_angle_gamma  90.00000

loop_
_space_group_symop_id
_space_group_symop_operation_xyz
1 x,y,z
2 -x,-y,z
3 -y,-x,-z
4 y,x,-z
5 -x,y,z
6 x,-y,z
7 y,-x,-z
8 -y,x,-z
9 x+1/2,y+1/2,z+1/2
10 -x+1/2,-y+1/2,z+1/2
11 -y+1/2,-x+1/2,-z+1/2
12 y+1/2,x+1/2,-z+1/2
13 -x+1/2,y+1/2,z+1/2
14 x+1/2,-y+1/2,z+1/2
15 y+1/2,-x+1/2,-z+1/2
16 -y+1/2,x+1/2,-z+1/2

loop_
_atom_site_label
_atom_site_type_symbol
_atom_site_symmetry_multiplicity
_atom_site_Wyckoff_label
_atom_site_fract_x
_atom_site_fract_y
_atom_site_fract_z
_atom_site_occupancy
O1 O 2 b 0.00000 0.00000 0.50000 1.00000
O2 O 2 d 0.00000 0.50000 0.75000 1.00000
O3 O 4 e 0.00000 0.00000 0.19740 1.00000
O4 O 4 f 0.00000 0.50000 0.06830 1.00000
O5 O 8 g 0.18890 0.18890 0.00000 1.00000
O6 O 8 h 0.19410 0.69410 0.25000 1.00000
Cd1 Cd 8 i 0.24710 0.00000 0.37410 1.00000
Re1 Re 8 i 0.24710 0.00000 0.87294 1.00000
```

Phase II $Cd_2Re_2O_7$: A2B7C2_tI44_119_i_bdefgh_i - POSCAR

```
A2B7C2_tI44_119_i_bdefgh_i & a,c/a,z3,z4,x5,x6,x7,z7,x8,z8 --params=
     ↪ 7.2312,1.41410830844,0.1974,0.0683,0.1889,0.1941,0.2471,0.3741,
     ↪ 0.2471,0.87294 & I-4m2 D_{2d}^{9} #119 (bdefghi^2) & tI44 &
     ↪ None & Cd2O7Re2 & Cd2O7Re2 & S.-W. Huang et al., J. Phys.:
     ↪ Condens. Matter 21, 195602(2009)
   1.00000000000000
  -3.61560000000000    3.61560000000000    5.11285000000000
   3.61560000000000   -3.61560000000000    5.11285000000000
   3.61560000000000    3.61560000000000   -5.11285000000000
     Cd        O        Re
      4       14         4
Direct
   0.37410000000000    0.62120000000000    0.24710000000000    Cd    (8i)
   0.37410000000000    0.12700000000000   -0.24710000000000    Cd    (8i)
  -0.62120000000000   -0.37410000000000   -0.24710000000000    Cd    (8i)
  -0.12700000000000   -0.37410000000000    0.24710000000000    Cd    (8i)
```

```
    0.50000000000000    0.50000000000000    0.00000000000000     O    (2b)
    0.25000000000000    0.75000000000000    0.50000000000000     O    (2d)
    0.19740000000000    0.19740000000000    0.00000000000000     O    (4e)
   -0.19740000000000   -0.19740000000000    0.00000000000000     O    (4e)
    0.56830000000000    0.06830000000000    0.50000000000000     O    (4f)
   -0.06830000000000    0.43170000000000    0.50000000000000     O    (4f)
    0.18890000000000    0.18890000000000    0.37780000000000     O    (8g)
   -0.18890000000000   -0.18890000000000   -0.37780000000000     O    (8g)
   -0.18890000000000    0.18890000000000    0.00000000000000     O    (8g)
    0.18890000000000   -0.18890000000000    0.00000000000000     O    (8g)
    0.94410000000000    0.44410000000000    0.88820000000000     O    (8h)
    0.55590000000000    0.05590000000000    0.11180000000000     O    (8h)
    0.55590000000000    0.44410000000000    0.50000000000000     O    (8h)
    0.94410000000000    0.05590000000000    0.50000000000000     O    (8h)
    0.87294000000000    1.12004000000000    0.24710000000000    Re    (8i)
    0.87294000000000    0.62584000000000   -0.24710000000000    Re    (8i)
   -1.12004000000000   -0.87294000000000   -0.24710000000000    Re    (8i)
   -0.62584000000000   -0.87294000000000    0.24710000000000    Re    (8i)
```

Tetragonal $TlFeS_2$: AB2C_tI8_119_c_e_a - CIF

```
# CIF file
data_findsym-output
_audit_creation_method FINDSYM

_chemical_name_mineral 'FeS2Tl'
_chemical_formula_sum 'Fe S2 Tl'

loop_
_publ_author_name
 'A. Kutoglu'
_journal_name_full_name
;
 Naturwissenschaften
;
_journal_volume 61
_journal_year 1974
_journal_page_first 125
_journal_page_last 126
_publ_Section_title
;
 Synthese und Kristallstrukturen von TlFeS$_{2}$ und TlFeSe$_{2}$
;

# Found in Neutron-Diffraction Study in TlFeS$_{2}$ and TlFeSe$_{2}$ at
    ↪ Low Temperatures, 2014

_aflow_title 'Tetragonal TlFeS$_{2}$ Structure'
_aflow_proto 'AB2C_tI8_119_c_e_a'
_aflow_params 'a,c/a,z_{3}'
_aflow_params_values '3.753,3.55502264855,0.35126'
_aflow_Strukturbericht 'None'
_aflow_Pearson 'tI8'

_symmetry_space_group_name_H-M "I -4 m 2"
_symmetry_Int_Tables_number 119

_cell_length_a      3.75300
_cell_length_b      3.75300
_cell_length_c      13.34200
_cell_angle_alpha   90.00000
_cell_angle_beta    90.00000
_cell_angle_gamma   90.00000

loop_
_space_group_symop_id
_space_group_symop_operation_xyz
1 x,y,z
2 -x,-y,z
3 -y,-x,-z
4 y,x,-z
5 -x,y,z
6 x,-y,z
7 y,-x,-z
8 -y,x,-z
9 x+1/2,y+1/2,z+1/2
10 -x+1/2,-y+1/2,z+1/2
11 -y+1/2,-x+1/2,-z+1/2
12 y+1/2,x+1/2,-z+1/2
13 -x+1/2,y+1/2,z+1/2
14 x+1/2,-y+1/2,z+1/2
15 y+1/2,-x+1/2,-z+1/2
16 -y+1/2,x+1/2,-z+1/2

loop_
_atom_site_label
_atom_site_type_symbol
_atom_site_symmetry_multiplicity
_atom_site_Wyckoff_label
_atom_site_fract_x
_atom_site_fract_y
_atom_site_fract_z
_atom_site_occupancy
Tl1 Tl 2 a 0.00000 0.00000 0.00000 1.00000
Fe1 Fe 2 c 0.00000 0.50000 0.25000 1.00000
S1 S 4 e 0.00000 0.00000 0.35126 1.00000
```

Tetragonal $TlFeS_2$: AB2C_tI8_119_c_e_a - POSCAR

```
AB2C_tI8_119_c_e_a & a,c/a,z3 --params=3.753,3.55502264855,0.35126 &
    ↪ I-4m2 D_{2d}^{9} #119 (ace) & tI8 & None & FeS2Tl & FeS2Tl & A.
    ↪ Kutoglu, Naturwissenschaften 61, 125-126 (1974)
   1.00000000000000
  -1.87650000000000    1.87650000000000    6.67100000000000
   1.87650000000000   -1.87650000000000    6.67100000000000
   1.87650000000000    1.87650000000000   -6.67100000000000
    Fe     S    Tl
     1     2     1
Direct
   0.75000000000000    0.25000000000000    0.50000000000000    Fe    (2c)
   0.35126000000000    0.35126000000000    0.00000000000000     S    (4e)
  -0.35126000000000   -0.35126000000000    0.00000000000000     S    (4e)
   0.00000000000000    0.00000000000000    0.00000000000000    Tl    (2a)
```

$BeSO_4{\cdot}4H_2O$ ($H4_3$): AB8C8D_tI72_120_c_2i_2i_b - CIF

```
# CIF file
data_findsym-output
_audit_creation_method FINDSYM

_chemical_name_mineral 'BeH8O8S'
_chemical_formula_sum 'Be H8 O8 S'

loop_
_publ_author_name
 'S. K. Sikka'
 'R. Chidambaram'
_journal_name_full_name
;
 Acta Crystallographica Section B: Structural Science
;
_journal_volume 25
_journal_year 1969
_journal_page_first 310
_journal_page_last 315
_publ_Section_title
;
 A neutron diffraction determination of the structure of beryllium
    ↪ sulphate tetrahydrate, BeSO$_{4}$$\cdot$4H$_{2}$O
;

_aflow_title 'BeSO$_{4}$$\cdot$4H$_{2}$O ($H4_{3}$) Structure'
_aflow_proto 'AB8C8D_tI72_120_c_2i_2i_b'
_aflow_params 'a,c/a,x_{3},y_{3},z_{3},x_{4},y_{4},z_{4},x_{5},y_{5},z_{
    ↪ 5},x_{6},y_{6},z_{6}'
_aflow_params_values '7.99,1.33767209011,0.22729,0.45696,0.13106,0.1229,
    ↪ 0.28418,0.13628,0.12563,0.08408,0.07733,0.13758,0.3956,0.17093'
_aflow_Strukturbericht '$H4_{3}$'
_aflow_Pearson 'tI72'

_symmetry_space_group_name_H-M "I -4 c 2"
_symmetry_Int_Tables_number 120

_cell_length_a      7.99000
_cell_length_b      7.99000
_cell_length_c      10.68800
_cell_angle_alpha   90.00000
_cell_angle_beta    90.00000
_cell_angle_gamma   90.00000

loop_
_space_group_symop_id
_space_group_symop_operation_xyz
1 x,y,z
2 -x,-y,z
3 -y,-x,-z+1/2
4 y,x,-z+1/2
5 -x,y,z+1/2
6 x,-y,z+1/2
7 y,-x,-z
8 -y,x,-z
9 x+1/2,y+1/2,z+1/2
10 -x+1/2,-y+1/2,z+1/2
11 -y+1/2,-x+1/2,-z
12 y+1/2,x+1/2,-z
13 -x+1/2,y+1/2,z
14 x+1/2,-y+1/2,z
15 y+1/2,-x+1/2,-z+1/2
16 -y+1/2,x+1/2,-z+1/2

loop_
_atom_site_label
_atom_site_type_symbol
_atom_site_symmetry_multiplicity
_atom_site_Wyckoff_label
_atom_site_fract_x
_atom_site_fract_y
_atom_site_fract_z
_atom_site_occupancy
S1 S 4 b 0.00000 0.00000 0.00000 1.00000
Be1 Be 4 c 0.00000 0.50000 0.25000 1.00000
H1 H 16 i 0.22729 0.45696 0.13106 1.00000
H2 H 16 i 0.12290 0.28418 0.13628 1.00000
O1 O 16 i 0.12563 0.08408 0.07733 1.00000
O2 O 16 i 0.13758 0.39560 0.17093 1.00000
```

$BeSO_4{\cdot}4H_2O$ ($H4_3$): AB8C8D_tI72_120_c_2i_2i_b - POSCAR

```
AB8C8D_tI72_120_c_2i_2i_b & a,c/a,x3,y3,z3,x4,y4,z4,x5,y5,z5,x6,y6,z6 --
    ↪ params=7.99,1.33767209011,0.22729,0.45696,0.13106,0.1229,
    ↪ 0.28418,0.13628,0.12563,0.08408,0.07733,0.13758,0.3956,0.17093
    ↪ & I-4c2 D_{2d}^{10} #120 (bci^4) & tI72 & $H4_{3}$ & BeH8O8S &
    ↪ BeH8O8S & S. K. Sikka and R. Chidambaram, Acta Crystallogr.
    ↪ Sect. B Struct. Sci. 25, 310-315 (1969)
   1.00000000000000
  -3.99500000000000    3.99500000000000    5.34400000000000
   3.99500000000000   -3.99500000000000    5.34400000000000
   3.99500000000000    3.99500000000000   -5.34400000000000
    Be     H     O     S
     2    16    16     2
```

```
Direct
   0.75000000000000     0.25000000000000     0.50000000000000    Be     (4c)
   0.25000000000000     0.75000000000000     0.50000000000000    Be     (4c)
   0.58802000000000     0.35835000000000     0.68425000000000     H   (16i)
  -0.32590000000000    -0.09623000000000    -0.68425000000000     H   (16i)
  -0.35835000000000     0.32590000000000     0.22967000000000     H   (16i)
   0.09623000000000    -0.58802000000000    -0.22967000000000     H   (16i)
   0.17410000000000     0.85835000000000    -0.22967000000000     H   (16i)
   1.08802000000000     0.40377000000000     0.22967000000000     H   (16i)
   0.59623000000000     0.82590000000000     0.68425000000000     H   (16i)
   0.14165000000000    -0.08802000000000    -0.68425000000000     H   (16i)
   0.42046000000000     0.25918000000000     0.40708000000000     H   (16i)
  -0.14790000000000     0.01338000000000    -0.40708000000000     H   (16i)
  -0.25918000000000     0.14790000000000     0.16128000000000     H   (16i)
  -0.01338000000000    -0.42046000000000    -0.16128000000000     H   (16i)
   0.35210000000000     0.75918000000000    -0.16128000000000     H   (16i)
   0.92046000000000     0.51338000000000     0.16128000000000     H   (16i)
   0.48662000000000     0.64790000000000     0.40708000000000     H   (16i)
   0.24082000000000     0.07954000000000    -0.40708000000000     H   (16i)
   0.16141000000000     0.20296000000000     0.20971000000000     O   (16i)
  -0.00675000000000    -0.04830000000000    -0.20971000000000     O   (16i)
  -0.20296000000000     0.00675000000000    -0.04155000000000     O   (16i)
   0.04830000000000    -0.16141000000000     0.04155000000000     O   (16i)
   0.49325000000000     0.70296000000000     0.04155000000000     O   (16i)
   0.66141000000000     0.45170000000000    -0.04155000000000     O   (16i)
   0.54830000000000     0.50675000000000     0.20971000000000     O   (16i)
   0.29704000000000     0.33859000000000    -0.20971000000000     O   (16i)
   0.56653000000000     0.30851000000000     0.53318000000000     O   (16i)
  -0.22467000000000     0.03335000000000    -0.53318000000000     O   (16i)
  -0.30851000000000     0.22467000000000     0.25802000000000     O   (16i)
  -0.03335000000000    -0.56653000000000    -0.25802000000000     O   (16i)
   0.27533000000000     0.80851000000000    -0.25802000000000     O   (16i)
   1.06653000000000     0.53335000000000     0.25802000000000     O   (16i)
   0.46665000000000     0.72467000000000     0.53318000000000     O   (16i)
   0.19149000000000    -0.06653000000000    -0.53318000000000     O   (16i)
   0.00000000000000     0.00000000000000     0.00000000000000     S    (4b)
   0.50000000000000     0.50000000000000     0.00000000000000     S    (4b)
```

$SrCu_2(BO_3)_2$: A2B2C6D_tI44_121_i_i_ij_c - CIF

```
# CIF file
data_findsym-output
_audit_creation_method FINDSYM

_chemical_name_mineral 'B2Cu2O6Sr'
_chemical_formula_sum 'B2 Cu2 O6 Sr'

loop_
_publ_author_name
 'R. W. Smith'
 'D. A. Keszler'
_journal_name_full_name
;
 Journal of Solid State Chemistry
;
_journal_volume 93
_journal_year 1991
_journal_page_first 430
_journal_page_last 435
_publ_Section_title
;
 Synthesis, structure, and properties of the orthoborate SrCu$_{2}$(BO$_
      ↪ {3}$)$_{2}$
;

# Found in Exact Dimer Ground State and Quantized Magnetization Plateaus
      ↪ in the Two-Dimensional Spin System SrCu$_{2}$(BO$_{3}$)$_{2}$,
      ↪ 1999

_aflow_title 'SrCu$_{2}$(BO$_{3}$)$_{2}$ Structure'
_aflow_proto 'A2B2C6D_tI44_121_i_i_ij_c'
_aflow_params 'a,c/a,x_{2},z_{2},x_{3},z_{3},x_{4},z_{4},x_{5},y_{5},z_{
      ↪ 5}'
_aflow_params_values '8.995,0.739188438021,0.2953,0.243,0.11412,0.2783,
      ↪ 0.4004,0.212,0.3276,0.1456,0.254'
_aflow_Strukturbericht 'None'
_aflow_Pearson 'tI44'

_symmetry_space_group_name_H-M "I -4 2 m"
_symmetry_Int_Tables_number 121

_cell_length_a      8.99500
_cell_length_b      8.99500
_cell_length_c      6.64900
_cell_angle_alpha 90.00000
_cell_angle_beta  90.00000
_cell_angle_gamma 90.00000

loop_
_space_group_symop_id
_space_group_symop_operation_xyz
1 x,y,z
2 x,-y,-z
3 -x,y,-z
4 -x,-y,z
5 y,x,z
6 y,-x,-z
7 -y,x,-z
8 -y,-x,z
9 x+1/2,y+1/2,z+1/2
10 x+1/2,-y+1/2,-z+1/2
11 -x+1/2,y+1/2,-z+1/2
12 -x+1/2,-y+1/2,z+1/2
13 y+1/2,x+1/2,z+1/2
14 y+1/2,-x+1/2,-z+1/2
15 -y+1/2,x+1/2,-z+1/2
16 -y+1/2,-x+1/2,z+1/2

loop_
_atom_site_label
_atom_site_type_symbol
_atom_site_symmetry_multiplicity
_atom_site_Wyckoff_label
_atom_site_fract_x
_atom_site_fract_y
_atom_site_fract_z
_atom_site_occupancy
Sr1 Sr 4 c 0.00000 0.50000 0.00000 1.00000
B1 B 8 i 0.29530 0.29530 0.24300 1.00000
Cu1 Cu 8 i 0.11412 0.11412 0.27830 1.00000
O1 O 8 i 0.40040 0.40040 0.21200 1.00000
O2 O 16 j 0.32760 0.14560 0.25400 1.00000
```

$SrCu_2(BO_3)_2$: A2B2C6D_tI44_121_i_i_ij_c - POSCAR

```
A2B2C6D_tI44_121_i_i_ij_c & a,c/a,x2,z2,x3,z3,x4,z4,x5,y5,z5 --params=
      ↪ 8.995,0.739188438021,0.2953,0.243,0.11412,0.2783,0.4004,0.212,
      ↪ 0.3276,0.1456,0.254 & I-42m D_{2d}^{11} #121 (ci^3j) & tI44 &
      ↪ None & B2Cu2O6Sr & B2Cu2O6Sr & R. W. Smith and D. A. Keszler,
      ↪ J. Solid State Chem. 93, 430-435 (1991)
   1.00000000000000
  -4.49750000000000     4.49750000000000     3.32450000000000
   4.49750000000000    -4.49750000000000     3.32450000000000
   4.49750000000000     4.49750000000000    -3.32450000000000
     B      Cu       O      Sr
     4       4      12       2
Direct
   0.53830000000000     0.53830000000000     0.59060000000000     B    (8i)
  -0.05230000000000    -0.05230000000000    -0.59060000000000     B    (8i)
  -0.53830000000000     0.05230000000000     0.00000000000000     B    (8i)
   0.05230000000000    -0.53830000000000     0.00000000000000     B    (8i)
   0.39242000000000     0.39242000000000     0.22824000000000    Cu    (8i)
   0.16418000000000     0.16418000000000    -0.22824000000000    Cu    (8i)
  -0.39242000000000    -0.16418000000000     0.00000000000000    Cu    (8i)
  -0.16418000000000    -0.39242000000000     0.00000000000000    Cu    (8i)
   0.61240000000000     0.61240000000000     0.80080000000000     O    (8i)
  -0.18840000000000    -0.18840000000000    -0.80080000000000     O    (8i)
  -0.61240000000000     0.18840000000000     0.00000000000000     O    (8i)
   0.18840000000000    -0.61240000000000     0.00000000000000     O    (8i)
   0.39960000000000     0.58160000000000     0.47320000000000     O   (16j)
   0.10840000000000    -0.07360000000000    -0.47320000000000     O   (16j)
  -0.58160000000000    -0.10840000000000    -0.18200000000000     O   (16j)
   0.07360000000000    -0.39960000000000     0.18200000000000     O   (16j)
  -0.10840000000000    -0.58160000000000    -0.18200000000000     O   (16j)
  -0.39960000000000     0.07360000000000     0.18200000000000     O   (16j)
  -0.07360000000000     0.10840000000000    -0.47320000000000     O   (16j)
   0.58160000000000     0.39960000000000     0.47320000000000     O   (16j)
   0.50000000000000     0.00000000000000     0.50000000000000    Sr    (4c)
   0.00000000000000     0.50000000000000     0.50000000000000    Sr    (4c)
```

*C*17 ($Fe_2B$) (*obsolete*): AB2_tI12_121_ab_i - CIF

```
# CIF file
data_findsym-output
_audit_creation_method FINDSYM

_chemical_name_mineral 'BFe2'
_chemical_formula_sum 'B Fe2'

loop_
_publ_author_name
 'F. Wever'
 'A. M\"{u}ller'
_journal_name_full_name
;
 Zeitschrift fur Anorganische und Allgemeine Chemie
;
_journal_volume 192
_journal_year 1930
_journal_page_first 317
_journal_page_last 336
_publ_Section_title
;
 \"{U}ber das Zweistoffsystem Eisen-Bor und \"{u}ber die Struktur des
      ↪ Eisenborides Fe$_{4}$B$_{2}$
;

_aflow_title '$C17$ (Fe$_{2}$B) ({\em{obsolete}}) Structure'
_aflow_proto 'AB2_tI12_121_ab_i'
_aflow_params 'a,c/a,x_{3},z_{3}'
_aflow_params_values '5.078,0.831626624655,0.16667,0.2'
_aflow_Strukturbericht '$C17$'
_aflow_Pearson 'tI12'

_symmetry_space_group_name_H-M "I -4 2 m"
_symmetry_Int_Tables_number 121

_cell_length_a      5.07800
_cell_length_b      5.07800
_cell_length_c      4.22300
_cell_angle_alpha 90.00000
_cell_angle_beta  90.00000
_cell_angle_gamma 90.00000

loop_
_space_group_symop_id
_space_group_symop_operation_xyz
1 x,y,z
2 x,-y,-z
3 -x,y,-z
```

```
4 -x,-y,z
5 y,x,z
6 y,-x,-z
7 -y,x,-z
8 -y,-x,z
9 x+1/2,y+1/2,z+1/2
10 x+1/2,-y+1/2,-z+1/2
11 -x+1/2,y+1/2,-z+1/2
12 -x+1/2,-y+1/2,z+1/2
13 y+1/2,x+1/2,z+1/2
14 y+1/2,-x+1/2,-z+1/2
15 -y+1/2,x+1/2,-z+1/2
16 -y+1/2,-x+1/2,z+1/2

loop_
_atom_site_label
_atom_site_type_symbol
_atom_site_symmetry_multiplicity
_atom_site_Wyckoff_label
_atom_site_fract_x
_atom_site_fract_y
_atom_site_fract_z
_atom_site_occupancy
B1 B 2 a 0.00000 0.00000 0.00000 1.00000
B2 B 2 b 0.00000 0.00000 0.50000 1.00000
Fe1 Fe 8 i 0.16667 0.16667 0.20000 1.00000
```

C17 ($Fe_2B$) (*obsolete*): AB2_tI12_121_ab_i - POSCAR

```
AB2_tI12_121_ab_i & a,c/a,x3,z3 --params=5.078,0.831626624655,0.16667,
    ↪ 0.2 & I-42m D_{2d}^{11} #121 (abi) & tI12 & $C17$ & BFe2 & BFe2
    ↪ & F. Wever and A. M\"{u}ller, Z. Anorg. Allg. Chem. 192,
    ↪ 317-336 (1930)
  1.00000000000000
 -2.53900000000000   2.53900000000000   2.11150000000000
  2.53900000000000  -2.53900000000000   2.11150000000000
  2.53900000000000   2.53900000000000  -2.11150000000000
    B    Fe
    2     4
Direct
  0.00000000000000   0.00000000000000   0.00000000000000   B   (2a)
  0.50000000000000   0.50000000000000   0.00000000000000   B   (2b)
  0.36667000000000   0.36667000000000   0.33334000000000   Fe  (8i)
  0.03333000000000   0.03333000000000  -0.33334000000000   Fe  (8i)
 -0.36667000000000  -0.03333000000000   0.00000000000000   Fe  (8i)
 -0.03333000000000  -0.36667000000000   0.00000000000000   Fe  (8i)
```

$K_3CrO_8$: AB3C8_tI24_121_a_bd_2i - CIF

```
# CIF file
data_findsym-output
_audit_creation_method FINDSYM

_chemical_name_mineral 'CrK3O8'
_chemical_formula_sum 'Cr K3 O8'

loop_
_publ_author_name
 'R. Stomberg'
_journal_name_full_name
;
 Acta Chemica Scandinavica
;
_journal_volume 17
_journal_year 1963
_journal_page_first 1563
_journal_page_last 1566
_publ_Section_title
;
 Least-Squares Refinement of the Crystal Structure of Potassium
    ↪ Peroxochromate
;

_aflow_title 'K$_{3}$CrO$_{8}$ Structure'
_aflow_proto 'AB3C8_tI24_121_a_bd_2i'
_aflow_params 'a,c/a,x_{4},z_{4},x_{5},z_{5}'
_aflow_params_values '6.703,1.13859465911,0.1355,0.1788,0.2079,0.0082'
_aflow_Strukturbericht 'None'
_aflow_Pearson 'tI24'

_symmetry_space_group_name_H-M "I -4 2 m"
_symmetry_Int_Tables_number 121

_cell_length_a      6.70300
_cell_length_b      6.70300
_cell_length_c      7.63200
_cell_angle_alpha  90.00000
_cell_angle_beta   90.00000
_cell_angle_gamma  90.00000

loop_
_space_group_symop_id
_space_group_symop_operation_xyz
1 x,y,z
2 x,-y,-z
3 -x,y,-z
4 -x,-y,z
5 y,x,z
6 y,-x,-z
7 -y,x,-z
8 -y,-x,z
9 x+1/2,y+1/2,z+1/2
10 x+1/2,-y+1/2,-z+1/2
11 -x+1/2,y+1/2,-z+1/2
12 -x+1/2,-y+1/2,z+1/2
13 y+1/2,x+1/2,z+1/2
14 y+1/2,-x+1/2,-z+1/2
15 -y+1/2,x+1/2,-z+1/2
16 -y+1/2,-x+1/2,z+1/2

loop_
_atom_site_label
_atom_site_type_symbol
_atom_site_symmetry_multiplicity
_atom_site_Wyckoff_label
_atom_site_fract_x
_atom_site_fract_y
_atom_site_fract_z
_atom_site_occupancy
Cr1 Cr 2 a 0.00000 0.00000 0.00000 1.00000
K1 K 2 b 0.00000 0.00000 0.50000 1.00000
K2 K 4 d 0.00000 0.50000 0.25000 1.00000
O1 O 8 i 0.13550 0.13550 0.17880 1.00000
O2 O 8 i 0.20790 0.20790 0.00820 1.00000
```

$K_3CrO_8$: AB3C8_tI24_121_a_bd_2i - POSCAR

```
AB3C8_tI24_121_a_bd_2i & a,c/a,x4,z4,x5,z5 --params=6.703,1.13859465911,
    ↪ 0.1355,0.1788,0.2079,0.0082 & I-42m D_{2d}^{11} #121 (abdi^2) &
    ↪ tI24 & None & CrK3O8 & CrK3O8 & R. Stomberg, Acta Chem. Scand.
    ↪ 17, 1563-1566 (1963)
  1.00000000000000
 -3.35150000000000   3.35150000000000   3.81600000000000
  3.35150000000000  -3.35150000000000   3.81600000000000
  3.35150000000000   3.35150000000000  -3.81600000000000
    Cr     K     O
     1     3     8
Direct
  0.00000000000000   0.00000000000000   0.00000000000000   Cr  (2a)
  0.50000000000000   0.50000000000000   0.00000000000000   K   (2b)
  0.75000000000000   0.25000000000000   0.50000000000000   K   (4d)
  0.25000000000000   0.75000000000000   0.50000000000000   K   (4d)
  0.31430000000000   0.31430000000000   0.27100000000000   O   (8i)
  0.04330000000000   0.04330000000000  -0.27100000000000   O   (8i)
 -0.31430000000000  -0.04330000000000   0.00000000000000   O   (8i)
 -0.04330000000000  -0.31430000000000   0.00000000000000   O   (8i)
  0.21610000000000   0.21610000000000   0.41580000000000   O   (8i)
 -0.19970000000000  -0.19970000000000  -0.41580000000000   O   (8i)
 -0.21610000000000   0.19970000000000   0.00000000000000   O   (8i)
  0.19970000000000  -0.21610000000000   0.00000000000000   O   (8i)
```

$\alpha$-$V_3S$: AB3_tI32_121_g_f2i - CIF

```
# CIF file
data_findsym-output
_audit_creation_method FINDSYM

_chemical_name_mineral 'SV3'
_chemical_formula_sum 'S V3'

loop_
_publ_author_name
 'B. Pedersen'
 'F. Gr{\o}nvold'
_journal_name_full_name
;
 Acta Cristallographica
;
_journal_volume 12
_journal_year 1959
_journal_page_first 1022
_journal_page_last 1027
_publ_Section_title
;
 The Crystal Structures of $\alpha$-V$_{3}$S and $\beta$-V$_{3}$S
;

_aflow_title '$\alpha$-V$_{3}$S Structure'
_aflow_proto 'AB3_tI32_121_g_f2i'
_aflow_params 'a,c/a,x_{1},x_{2},x_{3},z_{3},x_{4},z_{4}'
_aflow_params_values '9.47,0.484582893347,0.355,0.2851,0.5932,0.25,0.2,
    ↪ 0.25'
_aflow_Strukturbericht 'None'
_aflow_Pearson 'tI32'

_symmetry_space_group_name_H-M "I -4 2 m"
_symmetry_Int_Tables_number 121

_cell_length_a      9.47000
_cell_length_b      9.47000
_cell_length_c      4.58900
_cell_angle_alpha  90.00000
_cell_angle_beta   90.00000
_cell_angle_gamma  90.00000

loop_
_space_group_symop_id
_space_group_symop_operation_xyz
1 x,y,z
2 x,-y,-z
3 -x,y,-z
4 -x,-y,z
5 y,x,z
6 y,-x,-z
7 -y,x,-z
8 -y,-x,z
9 x+1/2,y+1/2,z+1/2
10 x+1/2,-y+1/2,-z+1/2
11 -x+1/2,y+1/2,-z+1/2
12 -x+1/2,-y+1/2,z+1/2
```

```
13 y+1/2,x+1/2,z+1/2
14 y+1/2,-x+1/2,-z+1/2
15 -y+1/2,x+1/2,-z+1/2
16 -y+1/2,-x+1/2,z+1/2

loop_
_atom_site_label
_atom_site_type_symbol
_atom_site_symmetry_multiplicity
_atom_site_Wyckoff_label
_atom_site_fract_x
_atom_site_fract_y
_atom_site_fract_z
_atom_site_occupancy
V1 V 8 f 0.35500 0.00000 0.00000 1.00000
S1 S 8 g 0.28510 0.00000 0.50000 1.00000
V2 V 8 i 0.59320 0.59320 0.25000 1.00000
V3 V 8 i 0.20000 0.20000 0.25000 1.00000
```

$\alpha$-$V_3S$: AB3_tI32_121_g_f2i - POSCAR

```
AB3_tI32_121_g_f2i & a,c/a,x1,x2,x3,z3,x4,z4 --params=9.47,
    ↪ 0.484582893347,0.355,0.2851,0.5932,0.25,0.2,0.25 & I-42m D_{2d
    ↪ }^{11} #121 (fgi^2) & tI32 & None & SV3 & SV3 & B. Pedersen and
    ↪ F. Gr{\o}nvold, Acta Cryst. 12, 1022-1027 (1959)
   1.00000000000000
  -4.73500000000000   4.73500000000000   2.29450000000000
   4.73500000000000  -4.73500000000000   2.29450000000000
   4.73500000000000   4.73500000000000  -2.29450000000000
     S     V
     4    12
Direct
   0.50000000000000   0.78510000000000   0.28510000000000   S   (8g)
   0.50000000000000   0.21490000000000  -0.28510000000000   S   (8g)
   0.21490000000000   0.50000000000000  -0.28510000000000   S   (8g)
   0.78510000000000   0.50000000000000   0.28510000000000   S   (8g)
   0.00000000000000   0.35500000000000   0.35500000000000   V   (8f)
   0.00000000000000  -0.35500000000000  -0.35500000000000   V   (8f)
  -0.35500000000000   0.00000000000000  -0.35500000000000   V   (8f)
   0.35500000000000   0.00000000000000   0.35500000000000   V   (8f)
   0.84320000000000   0.84320000000000   1.18640000000000   V   (8i)
  -0.34320000000000  -0.34320000000000  -1.18640000000000   V   (8i)
  -0.84320000000000   0.34320000000000   0.00000000000000   V   (8i)
   0.34320000000000  -0.84320000000000   0.00000000000000   V   (8i)
   0.45000000000000   0.45000000000000   0.40000000000000   V   (8i)
   0.05000000000000   0.05000000000000  -0.40000000000000   V   (8i)
  -0.45000000000000  -0.05000000000000   0.00000000000000   V   (8i)
  -0.05000000000000  -0.45000000000000   0.00000000000000   V   (8i)
```

Mercury Cyanide [$Hg(CN)_2$, $F1_1$]: A2BC2_tI40_122_e_d_e - CIF

```
# CIF file
data_findsym-output
_audit_creation_method FINDSYM

_chemical_name_mineral 'Mercury cyanide'
_chemical_formula_sum 'C2 Hg N2'

loop_
_publ_author_name
 'O. Reckeweg'
 'A. Simon'
_journal_name_full_name
;
 Zeitschrift f{\"u}r Naturforschung B
;
_journal_volume 57
_journal_year 2002
_journal_page_first 895
_journal_page_last 900
_publ_Section_title
;
 X-Ray and Raman Investigations on Cyanides of Mono- and Divalent Metals
    ↪ and Synthesis, Crystal Structure and Raman Spectrum of Tl$_{5
    ↪ }$(CO$_{3}$)$_{2}$(CN)
;

# Found in Hg(CN)$_{2}$ (Hg[CN]$_{2}$) Crystal Structure, 2016 Found in
    ↪ Hg(CN)$_{2}$ (Hg[CN]$_{2}$) Crystal Structure, {PAULING FILE in
    ↪ : Inorganic Solid Phases, SpringerMaterials (online database)},

_aflow_title 'Mercury Cyanide [Hg(CN)$_{2}$, $F1_{1}$] Structure'
_aflow_proto 'A2BC2_tI40_122_e_d_e'
_aflow_params 'a,c/a,x_{1},x_{2},y_{2},z_{2},x_{3},y_{3},z_{3}'
_aflow_params_values '9.6922,0.918418934814,0.21203,0.2035,-0.0729,
    ↪ 0.1785,0.205,0.0464,0.1639'
_aflow_Strukturbericht '$F1_{1}$'
_aflow_Pearson 'tI40'

_symmetry_space_group_name_H-M "I -4 2 d"
_symmetry_Int_Tables_number 122

_cell_length_a      9.69220
_cell_length_b      9.69220
_cell_length_c      8.90150
_cell_angle_alpha  90.00000
_cell_angle_beta   90.00000
_cell_angle_gamma  90.00000

loop_
_space_group_symop_id
_space_group_symop_operation_xyz
1 x,y,z
2 x,-y+1/2,-z+1/4
3 -x,y+1/2,-z+1/4
4 -x,-y,z
5 y,x+1/2,z+1/4
6 y,-x,-z
7 -y,x,-z
8 -y,-x+1/2,z+1/4
9 x+1/2,y+1/2,z+1/2
10 x+1/2,-y,-z+3/4
11 -x+1/2,y,-z+3/4
12 -x+1/2,-y+1/2,z+1/2
13 y+1/2,x,z+3/4
14 y+1/2,-x+1/2,-z+1/2
15 -y+1/2,x+1/2,-z+1/2
16 -y+1/2,-x,z+3/4

loop_
_atom_site_label
_atom_site_type_symbol
_atom_site_symmetry_multiplicity
_atom_site_Wyckoff_label
_atom_site_fract_x
_atom_site_fract_y
_atom_site_fract_z
_atom_site_occupancy
Hg1 Hg 8 d 0.21203 0.25000 0.12500 1.00000
C1 C 16 e 0.20350 -0.07290 0.17850 1.00000
N1 N 16 e 0.20500 0.04640 0.16390 1.00000
```

Mercury Cyanide [$Hg(CN)_2$, $F1_1$]: A2BC2_tI40_122_e_d_e - POSCAR

```
A2BC2_tI40_122_e_d_e & a,c/a,x1,x2,y2,z2,x3,y3,z3 --params=9.6922,
    ↪ 0.918418934814,0.21203,0.2035,-0.0729,0.1785,0.205,0.0464,
    ↪ 0.1639 & I-42d D_{2d}^{12} #122 (de^2) & tI40 & $F1_{1}$ &
    ↪ C2HgN2 & Mercury cyanide & O. Reckeweg and A. Simon, Z.
    ↪ Naturforsch. B 57, 895-900 (2002)
   1.00000000000000
  -4.84610000000000   4.84610000000000   4.45075000000000
   4.84610000000000  -4.84610000000000   4.45075000000000
   4.84610000000000   4.84610000000000  -4.45075000000000
     C    Hg     N
     8     4     8
Direct
   0.10560000000000   0.38200000000000   0.13060000000000   C   (16e)
   0.25140000000000  -0.02500000000000  -0.13060000000000   C   (16e)
  -0.38200000000000  -0.25140000000000  -0.27640000000000   C   (16e)
   0.02500000000000  -0.10560000000000   0.27640000000000   C   (16e)
   0.49860000000000  -0.13200000000000   0.22360000000000   C   (16e)
   0.64440000000000   0.27500000000000   0.77640000000000   C   (16e)
   0.72500000000000   0.50140000000000   0.36940000000000   C   (16e)
   1.13200000000000   0.35560000000000   0.63060000000000   C   (16e)
   0.37500000000000   0.33703000000000   0.46203000000000   Hg   (8d)
   0.87500000000000  -0.08703000000000   0.53797000000000   Hg   (8d)
   0.66297000000000   0.12500000000000   0.03797000000000   Hg   (8d)
   1.08703000000000   0.62500000000000   0.96203000000000   Hg   (8d)
   0.21030000000000   0.36890000000000   0.25140000000000   N   (16e)
   0.11750000000000  -0.04110000000000  -0.25140000000000   N   (16e)
  -0.36890000000000  -0.11750000000000  -0.15860000000000   N   (16e)
   0.04110000000000  -0.21030000000000   0.15860000000000   N   (16e)
   0.63250000000000  -0.11890000000000   0.34140000000000   N   (16e)
   0.53970000000000   0.29110000000000   0.65860000000000   N   (16e)
   0.70890000000000   0.36750000000000   0.24860000000000   N   (16e)
   1.11890000000000   0.46030000000000   0.75140000000000   N   (16e)
```

$KH_2PO_4$ ($H2_2$): A4BC4D_tI40_122_e_b_e_a - CIF

```
# CIF file
data_findsym-output
_audit_creation_method FINDSYM

_chemical_name_mineral 'KH2PO4'
_chemical_formula_sum 'H4 K O4 P'

loop_
_publ_author_name
 'R. J. Nelmes'
 'G. M. Meyer'
 'J. E. Tibballs'
_journal_name_full_name
;
 Journal of Physics C: Solid State Physics
;
_journal_volume 15
_journal_year 1982
_journal_page_first 59
_journal_page_last 75
_publ_Section_title
;
 The crystal structure of tetragonal KH$_{2}$PO$_{4}$ and KD$_{2}$PO$_{4
    ↪ }$ as a function of temperature
;

_aflow_title 'KH$_{2}$PO$_{4}$ ($H2_{2}$) Structure'
_aflow_proto 'A4BC4D_tI40_122_e_b_e_a'
_aflow_params 'a,c/a,x_{3},y_{3},z_{3},x_{4},y_{4},z_{4}'
_aflow_params_values '7.4264,0.933292039211,0.14867,0.22713,0.12266,
    ↪ 0.1933,0.08283,0.12675'
_aflow_Strukturbericht '$H2_{2}$'
_aflow_Pearson 'tI40'

_symmetry_space_group_name_H-M "I -4 2 d"
_symmetry_Int_Tables_number 122

_cell_length_a      7.42640
_cell_length_b      7.42640
_cell_length_c      6.93100
_cell_angle_alpha  90.00000
```

```
_cell_angle_beta   90.00000
_cell_angle_gamma  90.00000

loop_
_space_group_symop_id
_space_group_symop_operation_xyz
1 x,y,z
2 x,-y+1/2,-z+1/4
3 -x,y+1/2,-z+1/4
4 -x,-y,z
5 y,x+1/2,z+1/4
6 y,-x,-z
7 -y,x,-z
8 -y,-x+1/2,z+1/4
9 x+1/2,y+1/2,z+1/2
10 x+1/2,-y,-z+3/4
11 -x+1/2,y,-z+3/4
12 -x+1/2,-y+1/2,z+1/2
13 y+1/2,x,z+3/4
14 y+1/2,-x+1/2,-z+1/2
15 -y+1/2,x+1/2,-z+1/2
16 -y+1/2,-x,z+3/4

loop_
_atom_site_label
_atom_site_type_symbol
_atom_site_symmetry_multiplicity
_atom_site_Wyckoff_label
_atom_site_fract_x
_atom_site_fract_y
_atom_site_fract_z
_atom_site_occupancy
P1 P 4 a 0.00000 0.00000 0.00000 1.00000
K1 K 4 b 0.00000 0.00000 0.50000 1.00000
H1 H 16 e 0.14867 0.22713 0.12266 0.50000
O1 O 16 e 0.19330 0.08283 0.12675 1.00000
```

$KH_2PO_4$ ($H2_2$): A4BC4D_tI40_122_e_b_e_a - POSCAR

```
A4BC4D_tI40_122_e_b_e_a & a,c/a,x3,y3,z3,x4,y4,z4 --params=7.4264,
    ↪ 0.933292039211,0.14867,0.22713,0.12266,0.1933,0.08283,0.12675 &
    ↪ I-42d D_{2d}^{12} #122 (abe^2) & tI40 & $H2_{2}$ & KH2PO4 &
    ↪ KH2PO4 & R. J. Nelmes and G. M. Meyer and J. E. Tibballs, J.
    ↪ Phys. C: Solid State Phys. 15, 59-75 (1982)
   1.00000000000000
  -3.71320000000000   3.71320000000000   3.46550000000000
   3.71320000000000  -3.71320000000000   3.46550000000000
   3.71320000000000   3.71320000000000  -3.46550000000000
     H     K     O     P
     8     2     8     2
Direct
   0.34979000000000   0.27133000000000   0.37580000000000   H   (16e)
  -0.10447000000000  -0.02601000000000  -0.37580000000000   H   (16e)
  -0.27133000000000   0.10447000000000   0.07846000000000   H   (16e)
   0.02601000000000  -0.34979000000000  -0.07846000000000   H   (16e)
   0.85447000000000  -0.02133000000000   0.57846000000000   H   (16e)
   0.40021000000000   0.27601000000000   0.42154000000000   H   (16e)
   0.72399000000000   0.14553000000000   0.12420000000000   H   (16e)
   1.02133000000000   0.59979000000000   0.87580000000000   H   (16e)
   0.50000000000000   0.50000000000000   0.00000000000000   K    (4b)
   0.25000000000000   0.75000000000000   0.50000000000000   K    (4b)
   0.20958000000000   0.32005000000000   0.27613000000000   O   (16e)
   0.04392000000000  -0.06655000000000  -0.27613000000000   O   (16e)
  -0.32005000000000  -0.04392000000000  -0.11047000000000   O   (16e)
   0.06655000000000  -0.20958000000000   0.11047000000000   O   (16e)
   0.70608000000000  -0.07005000000000   0.38953000000000   O   (16e)
   0.54042000000000   0.31655000000000   0.61047000000000   O   (16e)
   0.68345000000000   0.29392000000000   0.22387000000000   O   (16e)
   1.07005000000000   0.45958000000000   0.77613000000000   O   (16e)
   0.00000000000000   0.00000000000000   0.00000000000000   P    (4a)
   0.75000000000000   0.25000000000000   0.50000000000000   P    (4a)
```

$NH_4H_2PO_4$: A8BC4D_tI56_122_2e_b_e_a - CIF

```
# CIF file
data_findsym-output
_audit_creation_method FINDSYM

_chemical_name_mineral 'H6NO4P'
_chemical_formula_sum 'H8 N O4 P'

loop_
_publ_author_name
 'A. A. Khan'
 'W. H. Baur'
_journal_name_full_name
;
 Acta Crystallographica Section B: Structural Science
;
_journal_volume 29
_journal_year 1973
_journal_page_first 2721
_journal_page_last 2726
_publ_Section_title
;
 Refinement of the crystal structures of ammonium dihydrogen phosphate
     ↪ and ammonium dihydrogen arsenate
;

# Found in Refinement of the Crystal Structure of NH$_{4}$H$_{2}$PO$_{4}
    ↪ $ above and below Antiferroelectric Phase Transition
    ↪ Temperature, 1987

_aflow_title 'NH$_{4}$H$_{2}$PO$_{4}$ Structure'
_aflow_proto 'A8BC4D_tI56_122_2e_b_e_a'
_aflow_params 'a,c/a,x_{3},y_{3},z_{3},x_{4},y_{4},z_{4},x_{5},y_{5},z_{
    ↪ 5}'
_aflow_params_values '7.4997,1.00662693174,0.25,0.15,0.125,0.498,0.589,
    ↪ 0.063,0.0843,0.1466,0.1151'
_aflow_Strukturbericht 'None'
_aflow_Pearson 'tI56'

_symmetry_space_group_name_H-M "I -4 2 d"
_symmetry_Int_Tables_number 122

_cell_length_a      7.49970
_cell_length_b      7.49970
_cell_length_c      7.54940
_cell_angle_alpha  90.00000
_cell_angle_beta   90.00000
_cell_angle_gamma  90.00000

loop_
_space_group_symop_id
_space_group_symop_operation_xyz
1 x,y,z
2 x,-y+1/2,-z+1/4
3 -x,y+1/2,-z+1/4
4 -x,-y,z
5 y,x+1/2,z+1/4
6 y,-x,-z
7 -y,x,-z
8 -y,-x+1/2,z+1/4
9 x+1/2,y+1/2,z+1/2
10 x+1/2,-y,-z+3/4
11 -x+1/2,y,-z+3/4
12 -x+1/2,-y+1/2,z+1/2
13 y+1/2,x,z+3/4
14 y+1/2,-x+1/2,-z+1/2
15 -y+1/2,x+1/2,-z+1/2
16 -y+1/2,-x,z+3/4

loop_
_atom_site_label
_atom_site_type_symbol
_atom_site_symmetry_multiplicity
_atom_site_Wyckoff_label
_atom_site_fract_x
_atom_site_fract_y
_atom_site_fract_z
_atom_site_occupancy
P1 P 4 a 0.00000 0.00000 0.00000 1.00000
N1 N 4 b 0.00000 0.00000 0.50000 1.00000
H1 H 16 e 0.25000 0.15000 0.12500 0.50000
H2 H 16 e 0.49800 0.58900 0.06300 1.00000
O1 O 16 e 0.08430 0.14660 0.11510 1.00000
```

$NH_4H_2PO_4$: A8BC4D_tI56_122_2e_b_e_a - POSCAR

```
A8BC4D_tI56_122_2e_b_e_a & a,c/a,x3,y3,z3,x4,y4,z4,x5,y5,z5 --params=
    ↪ 7.4997,1.00662693174,0.25,0.15,0.125,0.498,0.589,0.063,0.0843,
    ↪ 0.1466,0.1151 & I-42d D_{2d}^{12} #122 (abe^3) & tI56 & None &
    ↪ H6NO4P & H6NO4P & A. A. Khan and W. H. Baur, Acta Crystallogr.
    ↪ Sect. B Struct. Sci. 29, 2721-2726 (1973)
   1.00000000000000
  -3.74985000000000   3.74985000000000   3.77470000000000
   3.74985000000000  -3.74985000000000   3.77470000000000
   3.74985000000000   3.74985000000000  -3.77470000000000
     H     N     O     P
    16     2     8     2
Direct
   0.27500000000000   0.37500000000000   0.40000000000000   H   (16e)
  -0.02500000000000  -0.12500000000000  -0.40000000000000   H   (16e)
  -0.37500000000000   0.02500000000000  -0.10000000000000   H   (16e)
   0.12500000000000  -0.27500000000000   0.10000000000000   H   (16e)
   0.77500000000000  -0.12500000000000   0.40000000000000   H   (16e)
   0.47500000000000   0.37500000000000   0.60000000000000   H   (16e)
   0.62500000000000   0.22500000000000   0.10000000000000   H   (16e)
   1.12500000000000   0.52500000000000   0.90000000000000   H   (16e)
   0.65200000000000   0.56100000000000   1.08700000000000   H   (16e)
  -0.52600000000000  -0.43500000000000  -1.08700000000000   H   (16e)
  -0.56100000000000   0.52600000000000   0.09100000000000   H   (16e)
   0.43500000000000  -0.65200000000000  -0.09100000000000   H   (16e)
   1.27600000000000  -0.31100000000000   0.59100000000000   H   (16e)
   0.09800000000000   0.68500000000000   0.40900000000000   H   (16e)
   0.31500000000000  -0.27600000000000  -0.58700000000000   H   (16e)
   1.31100000000000   0.90200000000000   1.58700000000000   H   (16e)
   0.50000000000000   0.50000000000000   0.00000000000000   N    (4b)
   0.25000000000000   0.75000000000000   0.50000000000000   N    (4b)
   0.26170000000000   0.19940000000000   0.23090000000000   O   (16e)
  -0.03150000000000   0.03080000000000  -0.23090000000000   O   (16e)
  -0.19940000000000   0.03150000000000   0.06230000000000   O   (16e)
  -0.03080000000000  -0.26170000000000  -0.06230000000000   O   (16e)
   0.78150000000000   0.05060000000000   0.56230000000000   O   (16e)
   0.48830000000000   0.21920000000000   0.43770000000000   O   (16e)
   0.78080000000000   0.21850000000000   0.26910000000000   O   (16e)
   0.94940000000000   0.51170000000000   0.73090000000000   O   (16e)
   0.00000000000000   0.00000000000000   0.00000000000000   P    (4a)
   0.75000000000000   0.25000000000000   0.50000000000000   P    (4a)
```

$NaS_2$: AB2_tI48_122_cd_2e - CIF

```
# CIF file
data_findsym-output
_audit_creation_method FINDSYM

_chemical_name_mineral 'NaS2'
_chemical_formula_sum 'Na S2'

loop_
```

```
_publ_author_name
 'R. Tegman'
_journal_name_full_name
;
 Acta Crystallographica Section B: Structural Science
;
_journal_volume 29
_journal_year 1973
_journal_page_first 1463
_journal_page_last 1469
_publ_Section_title
;
 The Crystal Structure of Sodium Tetrasulphide, Na$_{2}$S$_{4}$
;

# Found in Pearson's Handbook of Crystallographic Data for Intermetallic
    ↪ Phases, 1991

_aflow_title 'NaS$_{2}$ Structure'
_aflow_proto 'AB2_tI48_122_cd_2e'
_aflow_params 'a,c/a,z_{1},x_{2},x_{3},y_{3},z_{3},x_{4},y_{4},z_{4}'
_aflow_params_values '9.5965,1.22841661022,0.16953,0.27647,0.03373,
    ↪ 0.24952,0.29644,0.14693,0.116,0.39956'
_aflow_Strukturbericht 'None'
_aflow_Pearson 'tI48'

_symmetry_space_group_name_H-M "I -4 2 d"
_symmetry_Int_Tables_number 122

_cell_length_a     9.59650
_cell_length_b     9.59650
_cell_length_c     11.78850
_cell_angle_alpha  90.00000
_cell_angle_beta   90.00000
_cell_angle_gamma  90.00000

loop_
_space_group_symop_id
_space_group_symop_operation_xyz
1 x,y,z
2 x,-y+1/2,-z+1/4
3 -x,y+1/2,-z+1/4
4 -x,-y,z
5 y,x+1/2,z+1/4
6 y,-x,-z
7 -y,x,-z
8 -y,-x+1/2,z+1/4
9 x+1/2,y+1/2,z+1/2
10 x+1/2,-y,-z+3/4
11 -x+1/2,y,-z+3/4
12 -x+1/2,-y+1/2,z+1/2
13 y+1/2,x,z+3/4
14 y+1/2,-x+1/2,-z+1/2
15 -y+1/2,x+1/2,-z+1/2
16 -y+1/2,-x,z+3/4

loop_
_atom_site_label
_atom_site_type_symbol
_atom_site_symmetry_multiplicity
_atom_site_Wyckoff_label
_atom_site_fract_x
_atom_site_fract_y
_atom_site_fract_z
_atom_site_occupancy
Na1 Na 8 c 0.00000 0.00000 0.16953 1.00000
Na2 Na 8 d 0.27647 0.25000 0.12500 1.00000
S1 S 16 e 0.03373 0.24952 0.29644 1.00000
S2 S 16 e 0.14693 0.11600 0.39956 1.00000
```

$NaS_2$: AB2_tI48_122_cd_2e - POSCAR

```
AB2_tI48_122_cd_2e & a,c/a,z1,x2,x3,y3,z3,x4,y4,z4 --params=9.5965,
    ↪ 1.22841661022,0.16953,0.27647,0.03373,0.24952,0.29644,0.14693,
    ↪ 0.116,0.39956 & I-42d D_{2d}^{12} #122 (cde^2) & tI48 & None &
    ↪ NaS2 & NaS2 & R. Tegman, Acta Crystallogr. Sect. B Struct. Sci.
    ↪ 29, 1463-1469 (1973)
   1.00000000000000
  -4.79825000000000   4.79825000000000   5.89425000000000
   4.79825000000000  -4.79825000000000   5.89425000000000
   4.79825000000000   4.79825000000000  -5.89425000000000
    Na     S
     8    16
Direct
   0.16953000000000   0.16953000000000   0.00000000000000   Na   (8c)
  -0.16953000000000  -0.16953000000000   0.00000000000000   Na   (8c)
   0.58047000000000   0.08047000000000   0.50000000000000   Na   (8c)
   0.91953000000000   0.41953000000000   0.50000000000000   Na   (8c)
   0.37500000000000   0.40147000000000   0.52647000000000   Na   (8d)
   0.87500000000000  -0.15147000000000   0.47353000000000   Na   (8d)
   0.59853000000000   0.12500000000000  -0.02647000000000   Na   (8d)
   1.15147000000000   0.62500000000000   1.02647000000000   Na   (8d)
   0.54596000000000   0.33017000000000   0.28325000000000   S   (16e)
   0.04692000000000   0.26271000000000  -0.28325000000000   S   (16e)
  -0.33017000000000  -0.04692000000000   0.21579000000000   S   (16e)
  -0.26271000000000  -0.54596000000000  -0.21579000000000   S   (16e)
   0.70308000000000  -0.08017000000000   0.71579000000000   S   (16e)
   0.20404000000000  -0.01271000000000   0.28421000000000   S   (16e)
   1.01271000000000   0.29692000000000   0.21675000000000   S   (16e)
   1.08017000000000   0.79596000000000   0.78325000000000   S   (16e)
   0.51556000000000   0.54649000000000   0.26293000000000   S   (16e)
   0.28356000000000   0.25263000000000  -0.26293000000000   S   (16e)
  -0.54649000000000  -0.28356000000000  -0.03093000000000   S   (16e)
  -0.25263000000000  -0.51556000000000   0.03093000000000   S   (16e)
   0.46644000000000  -0.29649000000000   0.46907000000000   S   (16e)
   0.23444000000000  -0.00263000000000   0.53093000000000   S   (16e)
   1.00263000000000   0.53356000000000   0.23707000000000   S   (16e)
   1.29649000000000   0.76556000000000   0.76293000000000   S   (16e)
```

$NH_4HgCl_3$ ($E2_5$): A3BC_tP5_123_cg_a_d - CIF

```
# CIF file
data_findsym-output
_audit_creation_method FINDSYM

_chemical_name_mineral 'Cl3Hg(NH4)'
_chemical_formula_sum 'Cl3 Hg (NH4)'

loop_
_publ_author_name
 'E. J. Harmsen'
_journal_name_full_name
;
 Zeitschrift f{\"u}r Kristallographie - Crystalline Materials
;
_journal_volume 100
_journal_year 1939
_journal_page_first 208
_journal_page_last 211
_publ_Section_title
;
 The Crystal Structure of NH$_{4}$HgCl$_{3}$
;

# Found in Strukturbericht Band VII 1939, 1943

_aflow_title 'NH$_{4}$HgCl$_{3}$ ($E2_{5}$) Structure'
_aflow_proto 'A3BC_tP5_123_cg_a_d'
_aflow_params 'a,c/a,z_{4}'
_aflow_params_values '4.19,1.89498806683,0.294'
_aflow_Strukturbericht '$E2_{5}$'
_aflow_Pearson 'tP5'

_symmetry_space_group_name_H-M "P 4/m 2/m 2/m"
_symmetry_Int_Tables_number 123

_cell_length_a     4.19000
_cell_length_b     4.19000
_cell_length_c     7.94000
_cell_angle_alpha  90.00000
_cell_angle_beta   90.00000
_cell_angle_gamma  90.00000

loop_
_space_group_symop_id
_space_group_symop_operation_xyz
1 x,y,z
2 x,-y,-z
3 -x,y,-z
4 -x,-y,z
5 -y,-x,-z
6 -y,x,z
7 y,-x,z
8 y,x,-z
9 -x,-y,-z
10 -x,y,z
11 x,-y,z
12 x,y,-z
13 y,x,z
14 y,-x,-z
15 -y,x,-z
16 -y,-x,z

loop_
_atom_site_label
_atom_site_type_symbol
_atom_site_symmetry_multiplicity
_atom_site_Wyckoff_label
_atom_site_fract_x
_atom_site_fract_y
_atom_site_fract_z
_atom_site_occupancy
Hg1 Hg 1 a 0.00000 0.00000 0.00000 1.00000
Cl1 Cl 1 c 0.50000 0.50000 0.00000 1.00000
NH41 NH4 1 d 0.50000 0.50000 0.50000 1.00000
Cl2 Cl 2 g 0.00000 0.00000 0.29400 1.00000
```

$NH_4HgCl_3$ ($E2_5$): A3BC_tP5_123_cg_a_d - POSCAR

```
A3BC_tP5_123_cg_a_d & a,c/a,z4 --params=4.19,1.89498806683,0.294 & P4/
    ↪ mmm D_{4h}^{1} #123 (acdg) & tP5 & $E2_{5}$ & Cl3Hg(NH4) &
    ↪ Cl3Hg(NH4) & E. J. Harmsen, Zeitschrift f{"u}r Kristallographie
    ↪ - Crystalline Materials 100, 208-211 (1939)
   1.00000000000000
   4.19000000000000   0.00000000000000   0.00000000000000
   0.00000000000000   4.19000000000000   0.00000000000000
   0.00000000000000   0.00000000000000   7.94000000000000
    Cl    Hg   NH4
     3     1     1
Direct
   0.50000000000000   0.50000000000000   0.00000000000000   Cl   (1c)
   0.00000000000000   0.00000000000000   0.29400000000000   Cl   (2g)
   0.00000000000000   0.00000000000000  -0.29400000000000   Cl   (2g)
   0.00000000000000   0.00000000000000   0.00000000000000   Hg   (1a)
   0.50000000000000   0.50000000000000   0.50000000000000   NH4  (1d)
```

$K_2PtCl_4$ ($H1_5$): A4B2C_tP7_123_j_e_a - CIF

```
# CIF file
```

```
data_findsym-output
_audit_creation_method FINDSYM

_chemical_name_mineral 'Cl4K2Pt'
_chemical_formula_sum 'Cl4 K2 Pt'

loop_
_publ_author_name
 'R. H. B. Mais'
 'P. G. Owston'
 'A. Wood'
_journal_name_full_name
;
 Acta Crystallographica Section B: Structural Science
;
_journal_volume 28
_journal_year 1972
_journal_page_first 393
_journal_page_last 399
_publ_Section_title
;
 The crystal structure of K$_{2}$PtCl$_{4}$ and K$_{2}$PdCl$_{4}$ with
   ↪ estimates of the factors affecting accuracy
;

_aflow_title 'K$_{2}$PtCl$_{4}$ ($H1_{5}$) Structure'
_aflow_proto 'A4B2C_tP7_123_j_e_a'
_aflow_params 'a,c/a,x_{3}'
_aflow_params_values '7.025,0.589893238434,0.2324'
_aflow_Strukturbericht '$H1_{5}$'
_aflow_Pearson 'tP7'

_symmetry_space_group_name_H-M "P 4/m 2/m 2/m"
_symmetry_Int_Tables_number 123

_cell_length_a      7.02500
_cell_length_b      7.02500
_cell_length_c      4.14400
_cell_angle_alpha  90.00000
_cell_angle_beta   90.00000
_cell_angle_gamma  90.00000

loop_
_space_group_symop_id
_space_group_symop_operation_xyz
1 x,y,z
2 x,-y,-z
3 -x,y,-z
4 -x,-y,z
5 -y,-x,-z
6 -y,x,z
7 y,-x,z
8 y,x,-z
9 -x,-y,-z
10 -x,y,z
11 x,-y,z
12 x,y,-z
13 y,x,z
14 y,-x,-z
15 -y,x,-z
16 -y,-x,z

loop_
_atom_site_label
_atom_site_type_symbol
_atom_site_symmetry_multiplicity
_atom_site_Wyckoff_label
_atom_site_fract_x
_atom_site_fract_y
_atom_site_fract_z
_atom_site_occupancy
Pt1 Pt 1 a 0.00000 0.00000 0.00000 1.00000
K1 K 2 e 0.00000 0.50000 0.50000 1.00000
Cl1 Cl 4 j 0.23240 0.23240 0.00000 1.00000
```

$K_2PtCl_4$ (*$H1_5$*): A4B2C_tP7_123_j_e_a - POSCAR

```
A4B2C_tP7_123_j_e_a & a,c/a,x3 --params=7.025,0.589893238434,0.2324 & P4
   ↪ /mmm D_{4h}^{1} #123 (aej) & tP7 & $H1_{5}$ & Cl4K2Pt & Cl4K2Pt
   ↪  & R. H. B. Mais and P. G. Owston and A. Wood, Acta
   ↪ Crystallogr. Sect. B Struct. Sci. 28, 393-399 (1972)
   1.00000000000000
   7.02500000000000   0.00000000000000   0.00000000000000
   0.00000000000000   7.02500000000000   0.00000000000000
   0.00000000000000   0.00000000000000   4.14400000000000
    Cl      K      Pt
     4      2      1
Direct
   0.23240000000000   0.23240000000000   0.00000000000000   Cl   (4j)
  -0.23240000000000  -0.23240000000000   0.00000000000000   Cl   (4j)
  -0.23240000000000   0.23240000000000   0.00000000000000   Cl   (4j)
   0.23240000000000  -0.23240000000000   0.00000000000000   Cl   (4j)
   0.00000000000000   0.50000000000000   0.50000000000000    K   (2e)
   0.50000000000000   0.00000000000000   0.50000000000000    K   (2e)
   0.00000000000000   0.00000000000000   0.00000000000000   Pt   (1a)
```

*$E6_1$* ($Sr(OH)_2(H_2O)_8$) (*Obsolete*): A8B2C_tP11_123_r_f_a - CIF

```
# CIF file
data_findsym-output
_audit_creation_method FINDSYM

_chemical_name_mineral 'Sr(OH)2(H2O)8'
_chemical_formula_sum '(H2O)8 (OH)2 Sr'

loop_
_publ_author_name
 'G. Natta'
_journal_name_full_name
;
 Gazzetta Chimica Italiana
;
_journal_volume 58
_journal_year 1928
_journal_page_first 870
_journal_page_last 872
_publ_Section_title
;
 Constitution of hydroxides and of hydrates. III. Octahydrated strontium
   ↪ hydroxide
;

# Found in Strukturbericht Band II 1928-1932, 1937

_aflow_title '$E6_{1}$ (Sr(OH)$_{2}$(H$_{2}$O)$_{8}$) ({\em{Obsolete}})
   ↪ Structure'
_aflow_proto 'A8B2C_tP11_123_r_f_a'
_aflow_params 'a,c/a,x_{3},z_{3}'
_aflow_params_values '6.41,0.905928237129,0.29,0.25'
_aflow_Strukturbericht '$E6_{1}$'
_aflow_Pearson 'tP11'

_symmetry_space_group_name_H-M "P 4/m 2/m 2/m"
_symmetry_Int_Tables_number 123

_cell_length_a      6.41000
_cell_length_b      6.41000
_cell_length_c      5.80700
_cell_angle_alpha  90.00000
_cell_angle_beta   90.00000
_cell_angle_gamma  90.00000

loop_
_space_group_symop_id
_space_group_symop_operation_xyz
1 x,y,z
2 x,-y,-z
3 -x,y,-z
4 -x,-y,z
5 -y,-x,-z
6 -y,x,z
7 y,-x,z
8 y,x,-z
9 -x,-y,-z
10 -x,y,z
11 x,-y,z
12 x,y,-z
13 y,x,z
14 y,-x,-z
15 -y,x,-z
16 -y,-x,z

loop_
_atom_site_label
_atom_site_type_symbol
_atom_site_symmetry_multiplicity
_atom_site_Wyckoff_label
_atom_site_fract_x
_atom_site_fract_y
_atom_site_fract_z
_atom_site_occupancy
Sr1 Sr 1 a 0.00000 0.00000 0.00000 1.00000
OH1 OH 2 f 0.00000 0.50000 0.00000 1.00000
H2O1 H2O 8 r 0.29000 0.29000 0.25000 1.00000
```

*$E6_1$* ($Sr(OH)_2(H_2O)_8$) (*Obsolete*): A8B2C_tP11_123_r_f_a - POSCAR

```
A8B2C_tP11_123_r_f_a & a,c/a,x3,z3 --params=6.41,0.905928237129,0.29,
   ↪ 0.25 & P4/mmm D_{4h}^{1} #123 (afr) & tP11 & $E6_{1}$ & Sr(OH)2
   ↪ (H2O)8 & Sr(OH)2(H2O)8 & G. Natta, Gazz. Chim. Ital. 58,
   ↪ 870-872 (1928)
   1.00000000000000
   6.41000000000000   0.00000000000000   0.00000000000000
   0.00000000000000   6.41000000000000   0.00000000000000
   0.00000000000000   0.00000000000000   5.80700000000000
   H2O     OH     Sr
     8      2      1
Direct
   0.29000000000000   0.29000000000000   0.25000000000000  H2O   (8r)
  -0.29000000000000  -0.29000000000000   0.25000000000000  H2O   (8r)
  -0.29000000000000   0.29000000000000   0.25000000000000  H2O   (8r)
   0.29000000000000  -0.29000000000000   0.25000000000000  H2O   (8r)
  -0.29000000000000   0.29000000000000  -0.25000000000000  H2O   (8r)
   0.29000000000000  -0.29000000000000  -0.25000000000000  H2O   (8r)
   0.29000000000000   0.29000000000000  -0.25000000000000  H2O   (8r)
  -0.29000000000000  -0.29000000000000  -0.25000000000000  H2O   (8r)
   0.00000000000000   0.50000000000000   0.00000000000000   OH   (2f)
   0.50000000000000   0.00000000000000   0.00000000000000   OH   (2f)
   0.00000000000000   0.00000000000000   0.00000000000000   Sr   (1a)
```

*$E6_2$* [$SrO_2(H_2O)_8$] (*possibly obsolete*): A8B2C_tP11_123_r_h_a - CIF

```
# CIF file
data_findsym-output
_audit_creation_method FINDSYM

_chemical_name_mineral '(H2O)8SrO2'
_chemical_formula_sum '(H2O)8 O2 Sr'

loop_
```

```
_publ_author_name
 'G. Natta'
_journal_name_full_name
;
 Gazzetta Chimica Italiana
;
_journal_volume 62
_journal_year 1932
_journal_page_first 444
_journal_page_last 444
_publ_Section_title
;
 ~
;

# Found in The space group of calcium peroxide octahydrate, 1951

_aflow_title '$E6_{2}$ [SrO$_{2}$(H$_{2}$O)$_{8}$] ({\em{possibly
    ↪ obsolete}}) Structure'
_aflow_proto 'A8B2C_tP11_123_r_h_a'
_aflow_params 'a,c/a,z_{2},x_{3},z_{3}'
_aflow_params_values '6.32,0.879746835443,0.1,0.2,0.25'
_aflow_Strukturbericht '$E6_{2}$'
_aflow_Pearson 'tP11'

_symmetry_space_group_name_H-M "P 4/m 2/m 2/m"
_symmetry_Int_Tables_number 123

_cell_length_a     6.32000
_cell_length_b     6.32000
_cell_length_c     5.56000
_cell_angle_alpha 90.00000
_cell_angle_beta  90.00000
_cell_angle_gamma 90.00000

loop_
_space_group_symop_id
_space_group_symop_operation_xyz
1 x,y,z
2 x,-y,-z
3 -x,y,-z
4 -x,-y,z
5 -y,-x,-z
6 -y,x,z
7 y,-x,z
8 y,x,-z
9 -x,-y,-z
10 -x,y,z
11 x,-y,z
12 x,y,-z
13 y,x,z
14 y,-x,-z
15 -y,x,-z
16 -y,-x,z

loop_
_atom_site_label
_atom_site_type_symbol
_atom_site_symmetry_multiplicity
_atom_site_Wyckoff_label
_atom_site_fract_x
_atom_site_fract_y
_atom_site_fract_z
_atom_site_occupancy
Sr1 Sr 1 a 0.00000 0.00000 0.00000 1.00000
O1 O 2 h 0.50000 0.50000 0.10000 1.00000
H2O1 H2O 8 r 0.20000 0.20000 0.25000 1.00000
```

$E6_2$ [$SrO_2(H_2O)_8$] (*possibly obsolete*): A8B2C_tP11_123_r_h_a - POSCAR

```
A8B2C_tP11_123_r_h_a & a,c/a,z2,x3,z3 --params=6.32,0.879746835443,0.1,
    ↪ 0.2,0.25 & P4/mmm D_{4h}^{1} #123 (ahr) & tP11 & $E6_{2}$ & (
    ↪ H2O)8SrO2 & (H2O)8SrO2 & G. Natta, Gazz. Chim. Ital. 62, 444(
    ↪ 1932)
   1.00000000000000
   6.32000000000000   0.00000000000000   0.00000000000000
   0.00000000000000   6.32000000000000   0.00000000000000
   0.00000000000000   0.00000000000000   5.56000000000000
   H2O      O     Sr
     8      2      1
Direct
   0.20000000000000   0.20000000000000   0.25000000000000   H2O   (8r)
  -0.20000000000000  -0.20000000000000   0.25000000000000   H2O   (8r)
  -0.20000000000000   0.20000000000000   0.25000000000000   H2O   (8r)
   0.20000000000000  -0.20000000000000   0.25000000000000   H2O   (8r)
  -0.20000000000000   0.20000000000000  -0.25000000000000   H2O   (8r)
   0.20000000000000  -0.20000000000000  -0.25000000000000   H2O   (8r)
   0.20000000000000   0.20000000000000  -0.25000000000000   H2O   (8r)
  -0.20000000000000  -0.20000000000000  -0.25000000000000   H2O   (8r)
   0.50000000000000   0.50000000000000   0.10000000000000     O   (2h)
   0.50000000000000   0.50000000000000  -0.10000000000000     O   (2h)
   0.00000000000000   0.00000000000000   0.00000000000000    Sr   (1a)
```

$TlAlF_4$ ($H0_8$): AB4C_tP6_123_d_eh_a - CIF

```
# CIF file
data_findsym-output
_audit_creation_method FINDSYM

_chemical_name_mineral 'AlF4Tl'
_chemical_formula_sum 'Al F4 Tl'

loop_
_publ_author_name
 'C. Brosset'
_journal_name_full_name
;
 Zeitschrift fur Anorganische und Allgemeine Chemie
;
_journal_volume 235
_journal_year 1937
_journal_page_first 139
_journal_page_last 147
_publ_Section_title
;
 Herstellung und Kristallbau der Verbindungen TlAlF$_{4}$ und Tl$_{2}
    ↪ $AlF$_{5}$
;

# Found in A Structural Classification of Fluoaluminates, 1950

_aflow_title 'TlAlF$_{4}$ ($H0_8$) Structure'
_aflow_proto 'AB4C_tP6_123_d_eh_a'
_aflow_params 'a,c/a,z_{4}'
_aflow_params_values '3.61,1.76454293629,0.215'
_aflow_Strukturbericht '$H0_{8}$'
_aflow_Pearson 'tP6'

_symmetry_space_group_name_H-M "P 4/m 2/m 2/m"
_symmetry_Int_Tables_number 123

_cell_length_a     3.61000
_cell_length_b     3.61000
_cell_length_c     6.37000
_cell_angle_alpha 90.00000
_cell_angle_beta  90.00000
_cell_angle_gamma 90.00000

loop_
_space_group_symop_id
_space_group_symop_operation_xyz
1 x,y,z
2 x,-y,-z
3 -x,y,-z
4 -x,-y,z
5 -y,-x,-z
6 -y,x,z
7 y,-x,z
8 y,x,-z
9 -x,-y,-z
10 -x,y,z
11 x,-y,z
12 x,y,-z
13 y,x,z
14 y,-x,-z
15 -y,x,-z
16 -y,-x,z

loop_
_atom_site_label
_atom_site_type_symbol
_atom_site_symmetry_multiplicity
_atom_site_Wyckoff_label
_atom_site_fract_x
_atom_site_fract_y
_atom_site_fract_z
_atom_site_occupancy
Tl1 Tl 1 a 0.00000 0.00000 0.00000 1.00000
Al1 Al 1 d 0.50000 0.50000 0.50000 1.00000
F1 F 2 e 0.00000 0.50000 0.50000 1.00000
F2 F 2 h 0.50000 0.50000 0.21500 1.00000
```

$TlAlF_4$ ($H0_8$): AB4C_tP6_123_d_eh_a - POSCAR

```
AB4C_tP6_123_d_eh_a & a,c/a,z4 --params=3.61,1.76454293629,0.215 & P4/
    ↪ mmm D_{4h}^{1} #123 (adeh) & tP6 & $H0_{8}$ & AlF4Tl & AlF4Tl &
    ↪ C. Brosset, Z. Anorg. Allg. Chem. 235, 139-147 (1937)
   1.00000000000000
   3.61000000000000   0.00000000000000   0.00000000000000
   0.00000000000000   3.61000000000000   0.00000000000000
   0.00000000000000   0.00000000000000   6.37000000000000
    Al      F     Tl
     1      4      1
Direct
   0.50000000000000   0.50000000000000   0.50000000000000    Al   (1d)
   0.00000000000000   0.50000000000000   0.50000000000000     F   (2e)
   0.50000000000000   0.00000000000000   0.50000000000000     F   (2e)
   0.50000000000000   0.50000000000000   0.21500000000000     F   (2h)
   0.50000000000000   0.50000000000000  -0.21500000000000     F   (2h)
   0.00000000000000   0.00000000000000   0.00000000000000    Tl   (1a)
```

$\delta$-CuTi ($L2_a$): AB_tP2_123_a_d - CIF

```
# CIF file
data_findsym-output
_audit_creation_method FINDSYM

_chemical_name_mineral 'CuTi'
_chemical_formula_sum 'Cu Ti'

loop_
_publ_author_name
 'N. Karlsson'
_journal_name_full_name
;
 Journal of the Institute of Metals
;
_journal_volume 79
_journal_year 1951
_journal_page_first 391
```

```
_journal_page_last 405
_publ_Section_title
;
 An X-Ray Study of the Phases in the Copper-Titanium System
;

# Found in A Handbook of Lattice Spacings and Structures of Metals and
      ↪ Alloys, 1958 Found in A Handbook of Lattice Spacings and
      ↪ Structures of Metals and Alloys, {N.~R.~C. No. 4303},

_aflow_title '$\delta$-CuTi ($L2_{a}$) Structure'
_aflow_proto 'AB_tP2_123_a_d'
_aflow_params 'a,c/a'
_aflow_params_values '4.44,0.643243243243'
_aflow_Strukturbericht '$L2_{a}$'
_aflow_Pearson 'tP2'

_symmetry_space_group_name_H-M "P 4/m 2/m 2/m"
_symmetry_Int_Tables_number 123

_cell_length_a      4.44000
_cell_length_b      4.44000
_cell_length_c      2.85600
_cell_angle_alpha  90.00000
_cell_angle_beta   90.00000
_cell_angle_gamma  90.00000

loop_
_space_group_symop_id
_space_group_symop_operation_xyz
1 x,y,z
2 x,-y,-z
3 -x,y,-z
4 -x,-y,z
5 -y,-x,-z
6 -y,x,z
7 y,-x,z
8 y,x,-z
9 -x,-y,-z
10 -x,y,z
11 x,-y,z
12 x,y,-z
13 y,x,z
14 y,-x,-z
15 -y,x,-z
16 -y,-x,z

loop_
_atom_site_label
_atom_site_type_symbol
_atom_site_symmetry_multiplicity
_atom_site_Wyckoff_label
_atom_site_fract_x
_atom_site_fract_y
_atom_site_fract_z
_atom_site_occupancy
Cu1 Cu 1 a 0.00000 0.00000 0.00000 1.00000
Ti1 Ti 1 d 0.50000 0.50000 0.50000 1.00000
```

$\delta$-CuTi ($L2_a$): AB_tP2_123_a_d - POSCAR

```
AB_tP2_123_a_d & a,c/a --params=4.44,0.643243243243 & P4/mmm D_{4h}^{1}
      ↪ #123 (ad) & tP2 & $L2_{a}$ & CuTi & CuTi & N. Karlsson, J.
      ↪ Inst. Met. 79, 391-405 (1951)
   1.00000000000000
   4.44000000000000   0.00000000000000   0.00000000000000
   0.00000000000000   4.44000000000000   0.00000000000000
   0.00000000000000   0.00000000000000   2.85600000000000
    Cu    Ti
     1     1
Direct
   0.00000000000000   0.00000000000000   0.00000000000000   Cu   (1a)
   0.50000000000000   0.50000000000000   0.50000000000000   Ti   (1d)
```

$CaO_2(H_2O)_8$: AB8C2_tP22_124_a_n_h - CIF

```
# CIF file
data_findsym-output
_audit_creation_method FINDSYM

_chemical_name_mineral 'Ca(H2O)8O2'
_chemical_formula_sum 'Ca (H2O)8 O2'

loop_
_publ_author_name
 'R. S. Shineman'
 'A. J. King'
_journal_name_full_name
;
 Acta Cristallographica
;
_journal_volume 4
_journal_year 1951
_journal_page_first 67
_journal_page_last 68
_publ_Section_title
;
 The space group of calcium peroxide octahydrate
;

_aflow_title 'CaO$_{2}$(H$_{2}$O)$_{8}$ Structure'
_aflow_proto 'AB8C2_tP22_124_a_n_h'
_aflow_params 'a,c/a,z_{2},x_{3},y_{3},z_{3}'
_aflow_params_values '6.21,1.77133655395,0.19,0.3,0.11,0.13'
_aflow_Strukturbericht 'None'
_aflow_Pearson 'tP22'

_symmetry_space_group_name_H-M "P 4/m 2/c 2/c"
_symmetry_Int_Tables_number 124

_cell_length_a      6.21000
_cell_length_b      6.21000
_cell_length_c     11.00000
_cell_angle_alpha  90.00000
_cell_angle_beta   90.00000
_cell_angle_gamma  90.00000

loop_
_space_group_symop_id
_space_group_symop_operation_xyz
1 x,y,z
2 x,-y,-z+1/2
3 -x,y,-z+1/2
4 -x,-y,z
5 -y,-x,-z+1/2
6 -y,x,z
7 y,-x,z
8 y,x,-z+1/2
9 -x,-y,-z
10 -x,y,z+1/2
11 x,-y,z+1/2
12 x,y,-z
13 y,x,z+1/2
14 y,-x,-z
15 -y,x,-z
16 -y,-x,z+1/2

loop_
_atom_site_label
_atom_site_type_symbol
_atom_site_symmetry_multiplicity
_atom_site_Wyckoff_label
_atom_site_fract_x
_atom_site_fract_y
_atom_site_fract_z
_atom_site_occupancy
Ca1 Ca 2 a 0.00000 0.00000 0.25000 1.00000
O1 O 4 h 0.50000 0.50000 0.19000 1.00000
H2O1 H2O 16 n 0.30000 0.11000 0.13000 1.00000
```

$CaO_2(H_2O)_8$: AB8C2_tP22_124_a_n_h - POSCAR

```
AB8C2_tP22_124_a_n_h & a,c/a,z2,x3,y3,z3 --params=6.21,1.77133655395,
      ↪ 0.19,0.3,0.11,0.13 & P4/mcc D_{4h}^{2} #124 (ahn) & tP22 & None
      ↪ & Ca(H2O)8O2 & Ca(H2O)8O2 & R. S. Shineman and A. J. King,
      ↪ Acta Cryst. 4, 67-68 (1951)
   1.00000000000000
   6.21000000000000   0.00000000000000   0.00000000000000
   0.00000000000000   6.21000000000000   0.00000000000000
   0.00000000000000   0.00000000000000  11.00000000000000
    Ca    H2O    O
     2     16    4
Direct
   0.00000000000000   0.00000000000000   0.25000000000000   Ca   (2a)
   0.00000000000000   0.00000000000000   0.75000000000000   Ca   (2a)
   0.30000000000000   0.11000000000000   0.13000000000000   H2O  (16n)
  -0.30000000000000  -0.11000000000000   0.13000000000000   H2O  (16n)
  -0.11000000000000   0.30000000000000   0.13000000000000   H2O  (16n)
   0.11000000000000  -0.30000000000000   0.13000000000000   H2O  (16n)
  -0.30000000000000   0.11000000000000   0.37000000000000   H2O  (16n)
   0.30000000000000  -0.11000000000000   0.37000000000000   H2O  (16n)
   0.11000000000000   0.30000000000000   0.37000000000000   H2O  (16n)
  -0.11000000000000  -0.30000000000000   0.37000000000000   H2O  (16n)
  -0.30000000000000  -0.11000000000000  -0.13000000000000   H2O  (16n)
   0.30000000000000   0.11000000000000  -0.13000000000000   H2O  (16n)
   0.11000000000000  -0.30000000000000  -0.13000000000000   H2O  (16n)
  -0.11000000000000   0.30000000000000  -0.13000000000000   H2O  (16n)
   0.30000000000000  -0.11000000000000   0.63000000000000   H2O  (16n)
  -0.30000000000000   0.11000000000000   0.63000000000000   H2O  (16n)
  -0.11000000000000  -0.30000000000000   0.63000000000000   H2O  (16n)
   0.11000000000000   0.30000000000000   0.63000000000000   H2O  (16n)
   0.50000000000000   0.50000000000000   0.19000000000000    O   (4h)
   0.50000000000000   0.50000000000000   0.31000000000000    O   (4h)
   0.50000000000000   0.50000000000000  -0.19000000000000    O   (4h)
   0.50000000000000   0.50000000000000   0.69000000000000    O   (4h)
```

Vesuvianite ($Ca_{10}Al_4(Mg,Fe)_2Si_9O_{34}(OH)_4$, $S2_3$): A4B10C2D34E4F9_tP252_126_k_ce2k_f_h8k_k_d2k - CIF

```
# CIF file
data_findsym-output
_audit_creation_method FINDSYM

_chemical_name_mineral 'Vesuvianite'
_chemical_formula_sum 'Ag4 Ca10 Mg2 O34 (OH)4 Si9'

loop_
_publ_author_name
 'B. E. Warren'
 'D. I. Modell'
_journal_name_full_name
;
 Zeitschrift f{\"u}r Kristallographie - Crystalline Materials
;
_journal_volume 78
_journal_year 1931
_journal_page_first 422
_journal_page_last 432
_publ_Section_title
;
```

```
 The Structure of Vesuvianite Ca$_{10}$Al$_{4}$(Mg,Fe)$_{2}$Si$_{9}$O$_{
      ↪ 34}$(OH)$_{4}$
;

# Found in Strukturbericht Band II 1928-1932, 1937

_aflow_title 'Vesuvianite (Ca$_{10}$Al$_{4}$(Mg,Fe)$_{2}$Si$_{9}$O$_{34}
      ↪ $(OH)$_{4}$, $S2_{3}$) Structure'
_aflow_proto 'A4B10C2D34E4F9_tP252_126_k_ce2k_f_h8k_k_d2k'
_aflow_params 'a,c/a,z_{3},x_{5},x_{6},y_{6},z_{6},x_{7},y_{7},z_{7},x_{
      ↪ 8},y_{8},z_{8},x_{9},y_{9},z_{9},x_{10},y_{10},z_{10},x_{11},y_
      ↪ {11},z_{11},x_{12},y_{12},z_{12},x_{13},y_{13},z_{13},x_{14},y_
      ↪ {14},z_{14},x_{15},y_{15},z_{15},x_{16},y_{16},z_{16},x_{17},y_
      ↪ {17},z_{17},x_{18},y_{18},z_{18},x_{19},y_{19},z_{19}'
_aflow_params_values '15.63,0.756877799104,0.13,0.055,0.11,0.89,0.87,
      ↪ 0.81,0.05,0.86,0.17,0.09,0.88,0.83,0.22,0.08,0.87,0.16,0.78,
      ↪ 0.22,-0.06,-0.08,-0.07,0.13,-0.02,0.01,0.83,0.82,0.78,0.87,-
      ↪ 0.07,-0.05,0.82,0.18,-0.08,-0.1,-0.07,0.06,-0.01,0.83,-0.05,
      ↪ 0.19,0.87,-0.09,0.83,0.87'
_aflow_Strukturbericht '$S2_{3}$'
_aflow_Pearson 'tP252'

_symmetry_space_group_name_H-M "P 4/n 2/n 2/c (origin choice 2)"
_symmetry_Int_Tables_number 126

_cell_length_a      15.63000
_cell_length_b      15.63000
_cell_length_c      11.83000
_cell_angle_alpha 90.00000
_cell_angle_beta   90.00000
_cell_angle_gamma 90.00000

loop_
_space_group_symop_id
_space_group_symop_operation_xyz
1 x,y,z
2 x,-y+1/2,-z+1/2
3 -x+1/2,y,-z+1/2
4 -x+1/2,-y+1/2,z
5 -y+1/2,-x+1/2,-z+1/2
6 -y+1/2,x,z
7 y,-x+1/2,z
8 y,x,-z+1/2
9 -x,-y,-z
10 -x,y+1/2,z+1/2
11 x+1/2,-y,z+1/2
12 x+1/2,y+1/2,-z
13 y+1/2,x+1/2,z+1/2
14 y+1/2,-x,-z
15 -y,x+1/2,-z
16 -y,-x,z+1/2

loop_
_atom_site_label
_atom_site_type_symbol
_atom_site_symmetry_multiplicity
_atom_site_Wyckoff_label
_atom_site_fract_x
_atom_site_fract_y
_atom_site_fract_z
_atom_site_occupancy
Ca1 Ca 4 c 0.25000 0.75000 0.75000 1.00000
Si1 Si 4 d 0.25000 0.75000 0.00000 1.00000
Ca2 Ca 4 e 0.25000 0.25000 0.13000 1.00000
Mg1 Mg 8 f 0.00000 0.00000 0.00000 1.00000
O1 O 8 h 0.05500 0.05500 0.25000 1.00000
Ag1 Ag 16 k 0.11000 0.89000 0.87000 1.00000
Ca3 Ca 16 k 0.81000 0.05000 0.86000 1.00000
Ca4 Ca 16 k 0.17000 0.09000 0.88000 1.00000
O2 O 16 k 0.83000 0.22000 0.08000 1.00000
O3 O 16 k 0.87000 0.16000 0.78000 1.00000
O4 O 16 k 0.22000 -0.06000 -0.08000 1.00000
O5 O 16 k -0.07000 0.13000 -0.02000 1.00000
O6 O 16 k 0.01000 0.83000 0.82000 1.00000
O7 O 16 k 0.78000 0.87000 -0.07000 1.00000
O8 O 16 k -0.05000 0.82000 0.18000 1.00000
O9 O 16 k -0.08000 -0.10000 -0.07000 1.00000
OH1 OH 16 k 0.06000 -0.01000 0.83000 1.00000
Si2 Si 16 k -0.05000 0.19000 0.87000 1.00000
Si3 Si 16 k -0.09000 0.83000 0.87000 1.00000
```

Vesuvianite ($Ca_{10}Al_4(Mg,Fe)_2Si_9O_{34}(OH)_4$, $S2_3$): A4B10C2D34E4F9_tP252_126_k_ce2k_f_h8k_k_d2k - POSCAR

```
A4B10C2D34E4F9_tP252_126_k_ce2k_f_h8k_k_d2k & a,c/a,z3,x5,x6,y6,z6,x7,y7
      ↪ ,z7,x8,y8,z8,x9,y9,z9,x10,y10,z10,x11,y11,z11,x12,y12,z12,x13,
      ↪ y13,z13,x14,y14,z14,x15,y15,z15,x16,y16,z16,x17,y17,z17,x18,y18
      ↪ ,z18,x19,y19,z19 --params=15.63,0.756877799104,0.13,0.055,0.11,
      ↪ 0.89,0.87,0.81,0.05,0.86,0.17,0.09,0.88,0.83,0.22,0.08,0.87,
      ↪ 0.16,0.78,0.22,-0.06,-0.08,-0.07,0.13,-0.02,0.01,0.83,0.82,0.78
      ↪ ,0.87,-0.07,-0.05,0.82,0.18,-0.08,-0.1,-0.07,0.06,-0.01,0.83,-
      ↪ 0.05,0.19,0.87,-0.09,0.83,0.87 & P4/nnc D_{4h}^{4} #126 (cdefhk
      ↪ ^14) & tP252 & $S2_{3}$ & Al4Ca10(Mg,Fe)2O34(OH)4Si9 &
      ↪ Vesuvianite & B. E. Warren and D. I. Modell, Zeitschrift f"{u}r
      ↪ Kristallographie - Crystalline Materials 78, 422-432 (1931)
   1.00000000000000
  15.63000000000000   0.00000000000000   0.00000000000000
   0.00000000000000  15.63000000000000   0.00000000000000
   0.00000000000000   0.00000000000000  11.83000000000000
    Ag     Ca     Mg      O     OH     Si
    16     40      8    136     16     36
Direct
   0.11000000000000   0.89000000000000   0.87000000000000   Ag   (16k)
   0.39000000000000  -0.39000000000000   0.87000000000000   Ag   (16k)
  -0.39000000000000   0.11000000000000   0.87000000000000   Ag   (16k)
   0.89000000000000   0.39000000000000   0.87000000000000   Ag   (16k)
   0.39000000000000   0.89000000000000  -0.37000000000000   Ag   (16k)
   0.11000000000000  -0.39000000000000  -0.37000000000000   Ag   (16k)
   0.89000000000000   0.11000000000000  -0.37000000000000   Ag   (16k)
  -0.39000000000000   0.39000000000000  -0.37000000000000   Ag   (16k)
  -0.11000000000000  -0.89000000000000  -0.87000000000000   Ag   (16k)
   0.61000000000000   1.39000000000000  -0.87000000000000   Ag   (16k)
   1.39000000000000  -0.11000000000000  -0.87000000000000   Ag   (16k)
  -0.89000000000000   0.61000000000000  -0.87000000000000   Ag   (16k)
   0.61000000000000  -0.89000000000000   1.37000000000000   Ag   (16k)
  -0.11000000000000   1.39000000000000   1.37000000000000   Ag   (16k)
  -0.89000000000000  -0.11000000000000   1.37000000000000   Ag   (16k)
   1.39000000000000   0.61000000000000   1.37000000000000   Ag   (16k)
   0.25000000000000   0.75000000000000   0.75000000000000   Ca    (4c)
   0.75000000000000   0.25000000000000   0.75000000000000   Ca    (4c)
   0.75000000000000   0.25000000000000   0.25000000000000   Ca    (4c)
   0.25000000000000   0.75000000000000   0.25000000000000   Ca    (4c)
   0.25000000000000   0.25000000000000   0.13000000000000   Ca    (4e)
   0.25000000000000   0.25000000000000   0.37000000000000   Ca    (4e)
   0.75000000000000   0.75000000000000  -0.13000000000000   Ca    (4e)
   0.75000000000000   0.75000000000000   0.63000000000000   Ca    (4e)
   0.81000000000000   0.05000000000000   0.86000000000000   Ca   (16k)
  -0.31000000000000   0.45000000000000   0.86000000000000   Ca   (16k)
   0.45000000000000   0.81000000000000   0.86000000000000   Ca   (16k)
   0.05000000000000  -0.31000000000000   0.86000000000000   Ca   (16k)
  -0.31000000000000   0.05000000000000  -0.36000000000000   Ca   (16k)
   0.81000000000000   0.45000000000000  -0.36000000000000   Ca   (16k)
   0.05000000000000   0.81000000000000  -0.36000000000000   Ca   (16k)
   0.45000000000000  -0.31000000000000  -0.36000000000000   Ca   (16k)
  -0.81000000000000  -0.05000000000000  -0.86000000000000   Ca   (16k)
   1.31000000000000   0.55000000000000  -0.86000000000000   Ca   (16k)
   0.55000000000000  -0.81000000000000  -0.86000000000000   Ca   (16k)
  -0.05000000000000   1.31000000000000  -0.86000000000000   Ca   (16k)
   1.31000000000000  -0.05000000000000   1.36000000000000   Ca   (16k)
  -0.81000000000000   0.55000000000000   1.36000000000000   Ca   (16k)
  -0.05000000000000  -0.81000000000000   1.36000000000000   Ca   (16k)
   0.55000000000000   1.31000000000000   1.36000000000000   Ca   (16k)
   0.17000000000000   0.09000000000000   0.88000000000000   Ca   (16k)
   0.33000000000000   0.41000000000000   0.88000000000000   Ca   (16k)
   0.41000000000000   0.17000000000000   0.88000000000000   Ca   (16k)
   0.09000000000000   0.33000000000000   0.88000000000000   Ca   (16k)
   0.33000000000000   0.09000000000000  -0.38000000000000   Ca   (16k)
   0.17000000000000   0.41000000000000  -0.38000000000000   Ca   (16k)
   0.09000000000000   0.17000000000000  -0.38000000000000   Ca   (16k)
   0.41000000000000   0.33000000000000  -0.38000000000000   Ca   (16k)
  -0.17000000000000  -0.09000000000000  -0.88000000000000   Ca   (16k)
   0.67000000000000   0.59000000000000  -0.88000000000000   Ca   (16k)
   0.59000000000000  -0.17000000000000  -0.88000000000000   Ca   (16k)
  -0.09000000000000   0.67000000000000  -0.88000000000000   Ca   (16k)
   0.67000000000000  -0.09000000000000   1.38000000000000   Ca   (16k)
  -0.17000000000000   0.59000000000000   1.38000000000000   Ca   (16k)
  -0.09000000000000  -0.17000000000000   1.38000000000000   Ca   (16k)
   0.59000000000000   0.67000000000000   1.38000000000000   Ca   (16k)
   0.00000000000000   0.00000000000000   0.00000000000000   Mg    (8f)
   0.50000000000000   0.50000000000000   0.00000000000000   Mg    (8f)
   0.50000000000000   0.00000000000000   0.00000000000000   Mg    (8f)
   0.00000000000000   0.50000000000000   0.00000000000000   Mg    (8f)
   0.50000000000000   0.00000000000000   0.50000000000000   Mg    (8f)
   0.00000000000000   0.50000000000000   0.50000000000000   Mg    (8f)
   0.00000000000000   0.00000000000000   0.50000000000000   Mg    (8f)
   0.50000000000000   0.50000000000000   0.50000000000000   Mg    (8f)
   0.05500000000000   0.05500000000000   0.25000000000000   O     (8h)
   0.44500000000000   0.44500000000000   0.25000000000000   O     (8h)
   0.44500000000000   0.05500000000000   0.25000000000000   O     (8h)
   0.05500000000000   0.44500000000000   0.25000000000000   O     (8h)
  -0.05500000000000  -0.05500000000000   0.75000000000000   O     (8h)
   0.55500000000000   0.55500000000000   0.75000000000000   O     (8h)
   0.55500000000000  -0.05500000000000   0.75000000000000   O     (8h)
  -0.05500000000000   0.55500000000000   0.75000000000000   O     (8h)
   0.83000000000000   0.22000000000000   0.08000000000000   O    (16k)
  -0.33000000000000   0.28000000000000   0.08000000000000   O    (16k)
   0.28000000000000   0.83000000000000   0.08000000000000   O    (16k)
   0.22000000000000  -0.33000000000000   0.08000000000000   O    (16k)
  -0.33000000000000   0.22000000000000   0.42000000000000   O    (16k)
   0.83000000000000   0.28000000000000   0.42000000000000   O    (16k)
   0.22000000000000   0.83000000000000   0.42000000000000   O    (16k)
   0.28000000000000  -0.33000000000000   0.42000000000000   O    (16k)
  -0.83000000000000  -0.22000000000000  -0.08000000000000   O    (16k)
   1.33000000000000   0.72000000000000  -0.08000000000000   O    (16k)
   0.72000000000000  -0.83000000000000  -0.08000000000000   O    (16k)
  -0.22000000000000   1.33000000000000  -0.08000000000000   O    (16k)
   1.33000000000000  -0.22000000000000   0.58000000000000   O    (16k)
  -0.83000000000000   0.72000000000000   0.58000000000000   O    (16k)
  -0.22000000000000  -0.83000000000000   0.58000000000000   O    (16k)
   0.72000000000000   1.33000000000000   0.58000000000000   O    (16k)
   0.87000000000000   0.16000000000000   0.78000000000000   O    (16k)
  -0.37000000000000   0.34000000000000   0.78000000000000   O    (16k)
   0.34000000000000   0.87000000000000   0.78000000000000   O    (16k)
   0.16000000000000  -0.37000000000000   0.78000000000000   O    (16k)
  -0.37000000000000   0.16000000000000  -0.28000000000000   O    (16k)
   0.87000000000000   0.34000000000000  -0.28000000000000   O    (16k)
   0.16000000000000   0.87000000000000  -0.28000000000000   O    (16k)
   0.34000000000000  -0.37000000000000  -0.28000000000000   O    (16k)
  -0.87000000000000  -0.16000000000000  -0.78000000000000   O    (16k)
   1.37000000000000   0.66000000000000  -0.78000000000000   O    (16k)
   0.66000000000000  -0.87000000000000  -0.78000000000000   O    (16k)
  -0.16000000000000   1.37000000000000  -0.78000000000000   O    (16k)
   1.37000000000000  -0.16000000000000   1.28000000000000   O    (16k)
  -0.87000000000000   0.66000000000000   1.28000000000000   O    (16k)
  -0.16000000000000  -0.87000000000000   1.28000000000000   O    (16k)
   0.66000000000000   1.37000000000000   1.28000000000000   O    (16k)
   0.22000000000000  -0.06000000000000  -0.08000000000000   O    (16k)
   0.28000000000000   0.56000000000000  -0.08000000000000   O    (16k)
   0.56000000000000   0.22000000000000  -0.08000000000000   O    (16k)
  -0.06000000000000   0.28000000000000  -0.08000000000000   O    (16k)
```

| | | | | |
|---|---|---|---|---|
| 0.28000000000000 | −0.06000000000000 | 0.58000000000000 | O | (16k) |
| 0.22000000000000 | 0.56000000000000 | 0.58000000000000 | O | (16k) |
| −0.06000000000000 | 0.22000000000000 | 0.58000000000000 | O | (16k) |
| 0.56000000000000 | 0.28000000000000 | 0.58000000000000 | O | (16k) |
| −0.22000000000000 | 0.06000000000000 | 0.08000000000000 | O | (16k) |
| 0.72000000000000 | 0.44000000000000 | 0.08000000000000 | O | (16k) |
| 0.44000000000000 | −0.22000000000000 | 0.08000000000000 | O | (16k) |
| 0.06000000000000 | 0.72000000000000 | 0.08000000000000 | O | (16k) |
| 0.72000000000000 | 0.06000000000000 | 0.42000000000000 | O | (16k) |
| −0.22000000000000 | 0.44000000000000 | 0.42000000000000 | O | (16k) |
| 0.06000000000000 | −0.22000000000000 | 0.42000000000000 | O | (16k) |
| 0.44000000000000 | 0.72000000000000 | 0.42000000000000 | O | (16k) |
| −0.07000000000000 | 0.13000000000000 | −0.02000000000000 | O | (16k) |
| 0.57000000000000 | 0.37000000000000 | −0.02000000000000 | O | (16k) |
| 0.37000000000000 | −0.07000000000000 | −0.02000000000000 | O | (16k) |
| 0.13000000000000 | 0.57000000000000 | −0.02000000000000 | O | (16k) |
| 0.57000000000000 | 0.13000000000000 | 0.52000000000000 | O | (16k) |
| −0.07000000000000 | 0.37000000000000 | 0.52000000000000 | O | (16k) |
| 0.13000000000000 | −0.07000000000000 | 0.52000000000000 | O | (16k) |
| 0.37000000000000 | 0.57000000000000 | 0.52000000000000 | O | (16k) |
| 0.07000000000000 | −0.13000000000000 | 0.02000000000000 | O | (16k) |
| 0.43000000000000 | 0.63000000000000 | 0.02000000000000 | O | (16k) |
| 0.63000000000000 | 0.07000000000000 | 0.02000000000000 | O | (16k) |
| −0.13000000000000 | 0.43000000000000 | 0.02000000000000 | O | (16k) |
| 0.43000000000000 | −0.13000000000000 | 0.48000000000000 | O | (16k) |
| 0.07000000000000 | 0.63000000000000 | 0.48000000000000 | O | (16k) |
| −0.13000000000000 | 0.07000000000000 | 0.48000000000000 | O | (16k) |
| 0.63000000000000 | 0.43000000000000 | 0.48000000000000 | O | (16k) |
| 0.01000000000000 | 0.83000000000000 | 0.82000000000000 | O | (16k) |
| 0.49000000000000 | −0.33000000000000 | 0.82000000000000 | O | (16k) |
| −0.33000000000000 | 0.01000000000000 | 0.82000000000000 | O | (16k) |
| 0.83000000000000 | 0.49000000000000 | 0.82000000000000 | O | (16k) |
| 0.49000000000000 | 0.83000000000000 | −0.32000000000000 | O | (16k) |
| 0.01000000000000 | −0.33000000000000 | −0.32000000000000 | O | (16k) |
| 0.83000000000000 | 0.01000000000000 | −0.32000000000000 | O | (16k) |
| −0.33000000000000 | 0.49000000000000 | −0.32000000000000 | O | (16k) |
| −0.01000000000000 | −0.83000000000000 | −0.82000000000000 | O | (16k) |
| 0.51000000000000 | 1.33000000000000 | −0.82000000000000 | O | (16k) |
| 1.33000000000000 | −0.01000000000000 | −0.82000000000000 | O | (16k) |
| −0.83000000000000 | 0.51000000000000 | −0.82000000000000 | O | (16k) |
| 0.51000000000000 | −0.83000000000000 | 1.32000000000000 | O | (16k) |
| −0.01000000000000 | 1.33000000000000 | 1.32000000000000 | O | (16k) |
| −0.83000000000000 | −0.01000000000000 | 1.32000000000000 | O | (16k) |
| 1.33000000000000 | 0.51000000000000 | 1.32000000000000 | O | (16k) |
| 0.78000000000000 | 0.87000000000000 | −0.07000000000000 | O | (16k) |
| −0.28000000000000 | −0.37000000000000 | −0.07000000000000 | O | (16k) |
| −0.37000000000000 | 0.78000000000000 | −0.07000000000000 | O | (16k) |
| 0.87000000000000 | −0.28000000000000 | −0.07000000000000 | O | (16k) |
| −0.28000000000000 | 0.87000000000000 | 0.57000000000000 | O | (16k) |
| 0.78000000000000 | −0.37000000000000 | 0.57000000000000 | O | (16k) |
| 0.87000000000000 | 0.78000000000000 | 0.57000000000000 | O | (16k) |
| −0.37000000000000 | −0.28000000000000 | 0.57000000000000 | O | (16k) |
| −0.78000000000000 | −0.87000000000000 | 0.07000000000000 | O | (16k) |
| 1.28000000000000 | 1.37000000000000 | 0.07000000000000 | O | (16k) |
| 1.37000000000000 | −0.78000000000000 | 0.07000000000000 | O | (16k) |
| −0.87000000000000 | 1.28000000000000 | 0.07000000000000 | O | (16k) |
| 1.28000000000000 | −0.87000000000000 | 0.43000000000000 | O | (16k) |
| −0.78000000000000 | 1.37000000000000 | 0.43000000000000 | O | (16k) |
| −0.87000000000000 | −0.78000000000000 | 0.43000000000000 | O | (16k) |
| 1.37000000000000 | 1.28000000000000 | 0.43000000000000 | O | (16k) |
| −0.05000000000000 | 0.82000000000000 | 0.18000000000000 | O | (16k) |
| 0.55000000000000 | −0.32000000000000 | 0.18000000000000 | O | (16k) |
| −0.32000000000000 | −0.05000000000000 | 0.18000000000000 | O | (16k) |
| 0.82000000000000 | 0.55000000000000 | 0.18000000000000 | O | (16k) |
| 0.55000000000000 | 0.82000000000000 | 0.32000000000000 | O | (16k) |
| −0.05000000000000 | −0.32000000000000 | 0.32000000000000 | O | (16k) |
| 0.82000000000000 | −0.05000000000000 | 0.32000000000000 | O | (16k) |
| −0.32000000000000 | 0.55000000000000 | 0.32000000000000 | O | (16k) |
| 0.05000000000000 | −0.82000000000000 | −0.18000000000000 | O | (16k) |
| 0.45000000000000 | 1.32000000000000 | −0.18000000000000 | O | (16k) |
| 1.32000000000000 | 0.05000000000000 | −0.18000000000000 | O | (16k) |
| −0.82000000000000 | 0.45000000000000 | −0.18000000000000 | O | (16k) |
| 0.45000000000000 | −0.82000000000000 | 0.68000000000000 | O | (16k) |
| 0.05000000000000 | 1.32000000000000 | 0.68000000000000 | O | (16k) |
| −0.82000000000000 | 0.05000000000000 | 0.68000000000000 | O | (16k) |
| 1.32000000000000 | 0.45000000000000 | 0.68000000000000 | O | (16k) |
| −0.08000000000000 | −0.10000000000000 | −0.07000000000000 | O | (16k) |
| 0.58000000000000 | 0.60000000000000 | −0.07000000000000 | O | (16k) |
| 0.60000000000000 | −0.08000000000000 | −0.07000000000000 | O | (16k) |
| −0.10000000000000 | 0.58000000000000 | −0.07000000000000 | O | (16k) |
| 0.58000000000000 | −0.10000000000000 | 0.57000000000000 | O | (16k) |
| −0.08000000000000 | 0.60000000000000 | 0.57000000000000 | O | (16k) |
| −0.10000000000000 | −0.08000000000000 | 0.57000000000000 | O | (16k) |
| 0.60000000000000 | 0.58000000000000 | 0.57000000000000 | O | (16k) |
| 0.08000000000000 | 0.10000000000000 | 0.07000000000000 | O | (16k) |
| 0.42000000000000 | 0.40000000000000 | 0.07000000000000 | O | (16k) |
| 0.40000000000000 | 0.08000000000000 | 0.07000000000000 | O | (16k) |
| 0.10000000000000 | 0.42000000000000 | 0.07000000000000 | O | (16k) |
| 0.42000000000000 | 0.10000000000000 | 0.43000000000000 | O | (16k) |
| 0.08000000000000 | 0.40000000000000 | 0.43000000000000 | O | (16k) |
| 0.10000000000000 | 0.08000000000000 | 0.43000000000000 | O | (16k) |
| 0.40000000000000 | 0.42000000000000 | 0.43000000000000 | O | (16k) |
| 0.06000000000000 | −0.01000000000000 | 0.83000000000000 | OH | (16k) |
| 0.44000000000000 | 0.51000000000000 | 0.83000000000000 | OH | (16k) |
| 0.51000000000000 | 0.06000000000000 | 0.83000000000000 | OH | (16k) |
| −0.01000000000000 | 0.44000000000000 | 0.83000000000000 | OH | (16k) |
| 0.44000000000000 | −0.01000000000000 | −0.33000000000000 | OH | (16k) |
| 0.06000000000000 | 0.51000000000000 | −0.33000000000000 | OH | (16k) |
| −0.01000000000000 | 0.06000000000000 | −0.33000000000000 | OH | (16k) |
| 0.51000000000000 | 0.44000000000000 | −0.33000000000000 | OH | (16k) |
| −0.06000000000000 | 0.01000000000000 | −0.83000000000000 | OH | (16k) |
| 0.56000000000000 | 0.49000000000000 | −0.83000000000000 | OH | (16k) |
| 0.49000000000000 | −0.06000000000000 | −0.83000000000000 | OH | (16k) |
| 0.01000000000000 | 0.56000000000000 | −0.83000000000000 | OH | (16k) |
| 0.56000000000000 | 0.01000000000000 | 1.33000000000000 | OH | (16k) |
| −0.06000000000000 | 0.49000000000000 | 1.33000000000000 | OH | (16k) |
| 0.01000000000000 | −0.06000000000000 | 1.33000000000000 | OH | (16k) |
| 0.49000000000000 | 0.56000000000000 | 1.33000000000000 | OH | (16k) |
| 0.25000000000000 | 0.75000000000000 | 0.00000000000000 | Si | (4d) |
| 0.75000000000000 | 0.25000000000000 | 0.00000000000000 | Si | (4d) |
| 0.25000000000000 | 0.75000000000000 | 0.50000000000000 | Si | (4d) |
| 0.75000000000000 | 0.25000000000000 | 0.50000000000000 | Si | (4d) |
| −0.05000000000000 | 0.19000000000000 | 0.87000000000000 | Si | (16k) |
| 0.55000000000000 | 0.31000000000000 | 0.87000000000000 | Si | (16k) |
| 0.31000000000000 | −0.05000000000000 | 0.87000000000000 | Si | (16k) |
| 0.19000000000000 | 0.55000000000000 | 0.87000000000000 | Si | (16k) |
| 0.55000000000000 | 0.19000000000000 | −0.37000000000000 | Si | (16k) |
| −0.05000000000000 | 0.31000000000000 | −0.37000000000000 | Si | (16k) |
| 0.19000000000000 | −0.05000000000000 | −0.37000000000000 | Si | (16k) |
| 0.31000000000000 | 0.55000000000000 | −0.37000000000000 | Si | (16k) |
| 0.05000000000000 | −0.19000000000000 | −0.87000000000000 | Si | (16k) |
| 0.45000000000000 | 0.69000000000000 | −0.87000000000000 | Si | (16k) |
| 0.69000000000000 | 0.05000000000000 | −0.87000000000000 | Si | (16k) |
| −0.19000000000000 | 0.45000000000000 | −0.87000000000000 | Si | (16k) |
| 0.45000000000000 | −0.19000000000000 | 1.37000000000000 | Si | (16k) |
| 0.05000000000000 | 0.69000000000000 | 1.37000000000000 | Si | (16k) |
| −0.19000000000000 | 0.05000000000000 | 1.37000000000000 | Si | (16k) |
| 0.69000000000000 | 0.45000000000000 | 1.37000000000000 | Si | (16k) |
| −0.09000000000000 | 0.83000000000000 | 0.87000000000000 | Si | (16k) |
| 0.59000000000000 | −0.33000000000000 | 0.87000000000000 | Si | (16k) |
| −0.33000000000000 | −0.09000000000000 | 0.87000000000000 | Si | (16k) |
| 0.83000000000000 | 0.59000000000000 | 0.87000000000000 | Si | (16k) |
| 0.59000000000000 | 0.83000000000000 | −0.37000000000000 | Si | (16k) |
| −0.09000000000000 | −0.33000000000000 | −0.37000000000000 | Si | (16k) |
| 0.83000000000000 | −0.09000000000000 | −0.37000000000000 | Si | (16k) |
| −0.33000000000000 | 0.59000000000000 | −0.37000000000000 | Si | (16k) |
| 0.09000000000000 | −0.83000000000000 | −0.87000000000000 | Si | (16k) |
| 0.41000000000000 | 1.33000000000000 | −0.87000000000000 | Si | (16k) |
| 1.33000000000000 | 0.09000000000000 | −0.87000000000000 | Si | (16k) |
| −0.83000000000000 | 0.41000000000000 | −0.87000000000000 | Si | (16k) |
| 0.41000000000000 | −0.83000000000000 | 1.37000000000000 | Si | (16k) |
| 0.09000000000000 | 1.33000000000000 | 1.37000000000000 | Si | (16k) |
| −0.83000000000000 | 0.09000000000000 | 1.37000000000000 | Si | (16k) |
| 1.33000000000000 | 0.41000000000000 | 1.37000000000000 | Si | (16k) |

$Ag[Co(NH_3)_2(NO_2)_4]$ ($J1_9$): ABC4D2E8_tP32_126_a_b_h_e_k - CIF

```
# CIF file
data_findsym-output
_audit_creation_method FINDSYM

_chemical_name_mineral 'AgCoN4(NH3)2O8'
_chemical_formula_sum 'Ag Co N4 (NH3)2 O8'

loop_
_publ_author_name
 'A. F. Wells'
_journal_name_full_name
;
 Zeitschrift f{\"u}r Kristallographie - Crystalline Materials
;
_journal_volume 95
_journal_year 1936
_journal_page_first 74
_journal_page_last 82
_publ_Section_title
;
 The Crystal Structure of Silver Diamminotetranitro-cobaltiate Ag[Co(
    ↪ NH$_{3}$)$_{2}$(NO$_{2}$)$_{4}$]
;

# Found in Strukturbericht Band IV 1936, 1938

_aflow_title 'Ag[Co(NH$_{3}$)$_{2}$(NO$_{2}$)$_{4}$] ($J1_{9}$)
    ↪ Structure'
_aflow_proto 'ABC4D2E8_tP32_126_a_b_h_e_k'
_aflow_params 'a,c/a,z_{3},x_{4},x_{5},y_{5},z_{5}'
_aflow_params_values '6.67,1.56371814093,0.53,-0.055,0.06,-0.03,0.34'
_aflow_Strukturbericht '$J1_{9}$'
_aflow_Pearson 'tP32'

_symmetry_space_group_name_H-M "P 4/n 2/n 2/c (origin choice 2)"
_symmetry_Int_Tables_number 126

_cell_length_a 6.67000
_cell_length_b 6.67000
_cell_length_c 10.43000
_cell_angle_alpha 90.00000
_cell_angle_beta 90.00000
_cell_angle_gamma 90.00000

loop_
_space_group_symop_id
_space_group_symop_operation_xyz
1 x,y,z
2 x,-y+1/2,-z+1/2
3 -x+1/2,y,-z+1/2
4 -x+1/2,-y+1/2,z
5 -y+1/2,-x+1/2,-z+1/2
6 -y+1/2,x,z
7 y,-x+1/2,z
8 y,x,-z+1/2
9 -x,-y,-z
10 -x,y+1/2,z+1/2
11 x+1/2,-y,z+1/2
12 x+1/2,y+1/2,-z
13 y+1/2,x+1/2,z+1/2
14 y+1/2,-x,-z
15 -y,x+1/2,-z
16 -y,-x,z+1/2
```

```
loop_
_atom_site_label
_atom_site_type_symbol
_atom_site_symmetry_multiplicity
_atom_site_Wyckoff_label
_atom_site_fract_x
_atom_site_fract_y
_atom_site_fract_z
_atom_site_occupancy
Ag1 Ag 2 a 0.25000 0.25000 0.25000 1.00000
Co1 Co 2 b 0.25000 0.25000 0.75000 1.00000
NH31 NH3 4 e 0.25000 0.25000 0.53000 1.00000
N1 N 8 h -0.05500 -0.05500 0.25000 1.00000
O1 O 16 k 0.06000 -0.03000 0.34000 1.00000
```

Ag[Co($NH_3$)$_2$($NO_2$)$_4$] ($J1_9$): ABC4D2E8_tP32_126_a_b_h_e_k - POSCAR

```
ABC4D2E8_tP32_126_a_b_h_e_k & a,c/a,z3,x4,x5,y5,z5 --params=6.67,
    ↪ 1.56371814093,0.53,-0.055,0.06,-0.03,0.34 & P4/nnc D_{4h}^{4} #
    ↪ 126 (abehk) & tP32 & $J1_{9}$ & AgCoN4(NH3)2O8 & AgCoN4(NH3)2O8
    ↪ & A. F. Wells, Zeitschrift f{"u}r Kristallographie -
    ↪ Crystalline Materials 95, 74-82 (1936)
   1.00000000000000
   6.67000000000000   0.00000000000000   0.00000000000000
   0.00000000000000   6.67000000000000   0.00000000000000
   0.00000000000000   0.00000000000000  10.43000000000000
    Ag    Co     N   NH3     O
     2     2     8     4    16
Direct
   0.25000000000000   0.25000000000000   0.25000000000000   Ag   (2a)
   0.75000000000000   0.75000000000000   0.75000000000000   Ag   (2a)
   0.25000000000000   0.25000000000000   0.75000000000000   Co   (2b)
   0.75000000000000   0.75000000000000   0.25000000000000   Co   (2b)
  -0.05500000000000  -0.05500000000000   0.25000000000000   N    (8h)
   0.55500000000000   0.55500000000000   0.25000000000000   N    (8h)
   0.55500000000000  -0.05500000000000   0.25000000000000   N    (8h)
  -0.05500000000000   0.55500000000000   0.25000000000000   N    (8h)
   0.05500000000000   0.05500000000000   0.75000000000000   N    (8h)
   0.44500000000000   0.44500000000000   0.75000000000000   N    (8h)
   0.44500000000000   0.05500000000000   0.75000000000000   N    (8h)
   0.05500000000000   0.44500000000000   0.75000000000000   N    (8h)
   0.25000000000000   0.25000000000000   0.53000000000000   NH3  (4e)
   0.25000000000000   0.25000000000000  -0.03000000000000   NH3  (4e)
   0.75000000000000   0.75000000000000  -0.53000000000000   NH3  (4e)
   0.75000000000000   0.75000000000000   1.03000000000000   NH3  (4e)
   0.06000000000000  -0.03000000000000   0.34000000000000   O    (16k)
   0.44000000000000   0.53000000000000   0.34000000000000   O    (16k)
   0.53000000000000   0.06000000000000   0.34000000000000   O    (16k)
  -0.03000000000000   0.44000000000000   0.34000000000000   O    (16k)
   0.44000000000000  -0.03000000000000   0.16000000000000   O    (16k)
   0.06000000000000   0.53000000000000   0.16000000000000   O    (16k)
  -0.03000000000000   0.06000000000000   0.16000000000000   O    (16k)
   0.53000000000000   0.44000000000000   0.16000000000000   O    (16k)
  -0.06000000000000   0.03000000000000  -0.34000000000000   O    (16k)
   0.56000000000000   0.47000000000000  -0.34000000000000   O    (16k)
   0.47000000000000  -0.06000000000000  -0.34000000000000   O    (16k)
   0.03000000000000   0.56000000000000  -0.34000000000000   O    (16k)
   0.56000000000000   0.03000000000000   0.84000000000000   O    (16k)
  -0.06000000000000   0.47000000000000   0.84000000000000   O    (16k)
   0.03000000000000  -0.06000000000000   0.84000000000000   O    (16k)
   0.47000000000000   0.56000000000000   0.84000000000000   O    (16k)
```

Pd($NH_3$)$_4$$Cl_2$·$H_2O$ ($H4_9$): A2BC4D_tP16_127_h_d_i_a - CIF

```
# CIF file
data_findsym-output
_audit_creation_method FINDSYM

_chemical_name_mineral 'Cl2(H2O)(NH3)4Pd'
_chemical_formula_sum 'Cl2 (H2O) (NH3)4 Pd'

loop_
_publ_author_name
 'B. N. Dickinson'
_journal_name_full_name
;
 Zeitschrift f{\"u}r Kristallographie - Crystalline Materials
;
_journal_volume 88
_journal_year 1934
_journal_page_first 281
_journal_page_last 297
_publ_Section_title
;
 The Crystal Structure of Tetramminopalladous Chloride Pd(NH$_{3}$)$_{4}
    ↪ $Cl$_{2}$$\cdot$H$_{2}$O
;

_aflow_title 'Pd(NH$_{3}$)$_{4}$Cl$_{2}$$\cdot$H$_{2}$O ($H4_{9}$)
    ↪ Structure'
_aflow_proto 'A2BC4D_tP16_127_h_d_i_a'
_aflow_params 'a,c/a,x_{3},x_{4},y_{4}'
_aflow_params_values '10.302,0.42127742186,0.285,0.194,0.027'
_aflow_Strukturbericht '$H4_{9}$'
_aflow_Pearson 'tP16'

_symmetry_space_group_name_H-M "P 4/m 21/b 2/m"
_symmetry_Int_Tables_number 127

_cell_length_a     10.30200
_cell_length_b     10.30200
_cell_length_c     4.34000
_cell_angle_alpha  90.00000
_cell_angle_beta   90.00000
_cell_angle_gamma  90.00000

loop_
_space_group_symop_id
_space_group_symop_operation_xyz
1 x,y,z
2 x+1/2,-y+1/2,-z
3 -x+1/2,y+1/2,-z
4 -x,-y,z
5 -y+1/2,-x+1/2,-z
6 -y,x,z
7 y,-x,z
8 y+1/2,x+1/2,-z
9 -x,-y,-z
10 -x+1/2,y+1/2,z
11 x+1/2,-y+1/2,z
12 x,y,-z
13 y+1/2,x+1/2,z
14 y,-x,-z
15 -y,x,-z
16 -y+1/2,-x+1/2,z

loop_
_atom_site_label
_atom_site_type_symbol
_atom_site_symmetry_multiplicity
_atom_site_Wyckoff_label
_atom_site_fract_x
_atom_site_fract_y
_atom_site_fract_z
_atom_site_occupancy
Pd1 Pd 2 a 0.00000 0.00000 0.00000 1.00000
H2O1 H2O 2 d 0.00000 0.50000 0.00000 1.00000
Cl1 Cl 4 h 0.28500 0.78500 0.50000 1.00000
NH31 NH3 8 i 0.19400 0.02700 0.00000 1.00000
```

Pd($NH_3$)$_4$$Cl_2$·$H_2O$ ($H4_9$): A2BC4D_tP16_127_h_d_i_a - POSCAR

```
A2BC4D_tP16_127_h_d_i_a & a,c/a,x3,x4,y4 --params=10.302,0.42127742186,
    ↪ 0.285,0.194,0.027 & P4/mbm D_{4h}^{5} #127 (adhi) & tP16 & $H4_
    ↪ {9}$ & Cl2(H2O)(NH3)4Pd & Cl2(H2O)(NH3)4Pd & B. N. Dickinson,
    ↪ Zeitschrift f{"u}r Kristallographie - Crystalline Materials 88,
    ↪ 281-297 (1934)
   1.00000000000000
  10.30200000000000   0.00000000000000   0.00000000000000
   0.00000000000000  10.30200000000000   0.00000000000000
   0.00000000000000   0.00000000000000   4.34000000000000
    Cl   H2O   NH3    Pd
     4     2     8     2
Direct
   0.28500000000000   0.78500000000000   0.50000000000000   Cl   (4h)
  -0.28500000000000   0.21500000000000   0.50000000000000   Cl   (4h)
   0.21500000000000   0.28500000000000   0.50000000000000   Cl   (4h)
   0.78500000000000  -0.28500000000000   0.50000000000000   Cl   (4h)
   0.00000000000000   0.50000000000000   0.00000000000000   H2O  (2d)
   0.50000000000000   0.00000000000000   0.00000000000000   H2O  (2d)
   0.19400000000000   0.02700000000000   0.00000000000000   NH3  (8i)
  -0.19400000000000  -0.02700000000000   0.00000000000000   NH3  (8i)
  -0.02700000000000   0.19400000000000   0.00000000000000   NH3  (8i)
   0.02700000000000  -0.19400000000000   0.00000000000000   NH3  (8i)
   0.30600000000000   0.52700000000000   0.00000000000000   NH3  (8i)
   0.69400000000000   0.47300000000000   0.00000000000000   NH3  (8i)
   0.52700000000000   0.69400000000000   0.00000000000000   NH3  (8i)
   0.47300000000000   0.30600000000000   0.00000000000000   NH3  (8i)
   0.00000000000000   0.00000000000000   0.00000000000000   Pd   (2a)
   0.50000000000000   0.50000000000000   0.00000000000000   Pd   (2a)
```

Phosgenite [$Pb_2Cl_2(CO_3)$]: AB2C3D2_tP32_127_g_eh_gk_k - CIF

```
# CIF file
data_findsym-output
_audit_creation_method FINDSYM

_chemical_name_mineral 'Phosgenite'
_chemical_formula_sum 'C Cl2 O3 Pb2'

loop_
_publ_author_name
 'G. Giuseppetti'
 'C. Tadini'
_journal_name_full_name
;
 Tschermaks mineralogische und petrographische Mitteilungen
;
_journal_volume 21
_journal_year 1974
_journal_page_first 101
_journal_page_last 109
_publ_Section_title
;
 Reexamination of the crystal structure of phosgenite, Pb$_{2}$Cl$_{2}$(
    ↪ CO$_{3}$)
;

_aflow_title 'Phosgenite [Pb$_{2}$Cl$_{2}$(CO$_{3}$)] Structure'
_aflow_proto 'AB2C3D2_tP32_127_g_eh_gk_k'
_aflow_params 'a,c/a,z_{1},x_{2},x_{3},x_{4},x_{5},z_{5},x_{6},z_{6}'
_aflow_params_values '8.16,1.08860294118,0.7572,0.3257,0.211,0.3521,
    ↪ 0.3726,0.1269,0.1659,0.2594'
_aflow_Strukturbericht 'None'
_aflow_Pearson 'tP32'

_symmetry_space_group_name_H-M "P 4/m 21/b 2/m"
_symmetry_Int_Tables_number 127
```

```
_cell_length_a       8.16000
_cell_length_b       8.16000
_cell_length_c       8.88300
_cell_angle_alpha  90.00000
_cell_angle_beta   90.00000
_cell_angle_gamma  90.00000

loop_
_space_group_symop_id
_space_group_symop_operation_xyz
1 x,y,z
2 x+1/2,-y+1/2,-z
3 -x+1/2,y+1/2,-z
4 -x,-y,z
5 -y+1/2,-x+1/2,-z
6 -y,x,z
7 y,-x,z
8 y+1/2,x+1/2,-z
9 -x,-y,-z
10 -x+1/2,y+1/2,z
11 x+1/2,-y+1/2,z
12 x,y,-z
13 y+1/2,x+1/2,z
14 y,-x,-z
15 -y,x,-z
16 -y+1/2,-x+1/2,z

loop_
_atom_site_label
_atom_site_type_symbol
_atom_site_symmetry_multiplicity
_atom_site_Wyckoff_label
_atom_site_fract_x
_atom_site_fract_y
_atom_site_fract_z
_atom_site_occupancy
Cl1  Cl  4  e  0.00000  0.00000  0.75720  1.00000
C1  C  4  g  0.32570  0.82570  0.00000  1.00000
O1  O  4  g  0.21100  0.71100  0.00000  1.00000
Cl2  Cl  4  h  0.35210  0.85210  0.50000  1.00000
O2  O  8  k  0.37260  0.87260  0.12690  1.00000
Pb1  Pb  8  k  0.16590  0.66590  0.25940  1.00000
```

Phosgenite [$Pb_2Cl_2(CO_3)$]: AB2C3D2_tP32_127_g_eh_gk_k - POSCAR

```
AB2C3D2_tP32_127_g_eh_gk_k & a,c/a,z1,x2,x3,x4,x5,z5,x6,z6 --params=8.16
   ↪ ,1.08860294118,0.7572,0.3257,0.211,0.3521,0.3726,0.1269,0.1659,
   ↪ 0.2594 & P4/mbm D_{4h}^{5} #127 (eg^2hk^2) & tP32 & None &
   ↪ CCl2O3Pb2 & Phosgenite & G. Giuseppetti and C. Tadini,
   ↪ Tschermaks Min. Petr. Mitt. 21, 101-109 (1974)
   1.00000000000000
   8.16000000000000    0.00000000000000    0.00000000000000
   0.00000000000000    8.16000000000000    0.00000000000000
   0.00000000000000    0.00000000000000    8.88300000000000
      C      Cl      O      Pb
      4       8     12       8
Direct
   0.32570000000000    0.82570000000000    0.00000000000000    C    (4g)
  -0.32570000000000    0.17430000000000    0.00000000000000    C    (4g)
   0.17430000000000    0.32570000000000    0.00000000000000    C    (4g)
   0.82570000000000   -0.32570000000000    0.00000000000000    C    (4g)
   0.00000000000000    0.00000000000000    0.75720000000000    Cl   (4e)
   0.50000000000000    0.50000000000000   -0.75720000000000    Cl   (4e)
   0.00000000000000    0.00000000000000   -0.75720000000000    Cl   (4e)
   0.50000000000000    0.50000000000000    0.75720000000000    Cl   (4e)
   0.35210000000000    0.85210000000000    0.50000000000000    Cl   (4h)
  -0.35210000000000    0.14790000000000    0.50000000000000    Cl   (4h)
   0.14790000000000    0.35210000000000    0.50000000000000    Cl   (4h)
   0.85210000000000   -0.35210000000000    0.50000000000000    Cl   (4h)
   0.21100000000000    0.71100000000000    0.00000000000000    O    (4g)
  -0.21100000000000    0.28900000000000    0.00000000000000    O    (4g)
   0.28900000000000    0.21100000000000    0.00000000000000    O    (4g)
   0.71100000000000   -0.21100000000000    0.00000000000000    O    (4g)
   0.37260000000000    0.87260000000000    0.12690000000000    O    (8k)
  -0.37260000000000    0.12740000000000    0.12690000000000    O    (8k)
   0.12740000000000    0.37260000000000    0.12690000000000    O    (8k)
   0.87260000000000   -0.37260000000000    0.12690000000000    O    (8k)
   0.12740000000000    0.37260000000000   -0.12690000000000    O    (8k)
   0.87260000000000   -0.37260000000000   -0.12690000000000    O    (8k)
   0.37260000000000    0.87260000000000   -0.12690000000000    O    (8k)
  -0.37260000000000    0.12740000000000   -0.12690000000000    O    (8k)
   0.16590000000000    0.66590000000000    0.25940000000000    Pb   (8k)
  -0.16590000000000    0.33410000000000    0.25940000000000    Pb   (8k)
   0.33410000000000    0.16590000000000    0.25940000000000    Pb   (8k)
   0.66590000000000   -0.16590000000000    0.25940000000000    Pb   (8k)
   0.33410000000000    0.16590000000000   -0.25940000000000    Pb   (8k)
   0.66590000000000   -0.16590000000000   -0.25940000000000    Pb   (8k)
   0.16590000000000    0.66590000000000   -0.25940000000000    Pb   (8k)
  -0.16590000000000    0.33410000000000   -0.25940000000000    Pb   (8k)
```

Chiolite ($Na_5Al_3F_{14}$, $K7_5$): A3B14C5_tP44_128_ac_ehi_bg - CIF

```
# CIF file
data_findsym-output
_audit_creation_method FINDSYM

_chemical_name_mineral 'Chiolite'
_chemical_formula_sum 'Al3 F14 Na5'

loop_
_publ_author_name
 'C. Jacoboni'
 'A. Leble'
 'J. J. Rousseau'
_journal_name_full_name
;
 Journal of Solid State Chemistry
;
_journal_volume 36
_journal_year 1981
_journal_page_first 297
_journal_page_last 304
_publ_Section_title
;
 D\\'{e}termination pr\\'{e}cise de la structure de la chiolite Na$_{5}
   ↪ $Al$_{3}$F$_{14}$ et \\'{e}tude par R.P.E. de Na$_{5}$Al$_{3}
   ↪ $F$_{14}$:Cr$^{3+}$
;

# Found in The American Mineralogist Crystal Structure Database, 2003

_aflow_title 'Chiolite (Na$_{5}$Al$_{3}$F$_{14}$, $K7_{5}$) Structure'
_aflow_proto 'A3B14C5_tP44_128_ac_ehi_bg'
_aflow_params 'a,c/a,z_{4},x_{5},x_{6},y_{6},x_{7},y_{7},z_{7}'
_aflow_params_values '7.0138,1.48307622116,0.1711,0.2768,0.0642,0.2477,
   ↪ 0.1794,0.5364,0.1198'
_aflow_Strukturbericht '$K7_{5}$'
_aflow_Pearson 'tP44'

_symmetry_space_group_name_H-M "P 4/m 21/n 2/c"
_symmetry_Int_Tables_number 128

_cell_length_a       7.01380
_cell_length_b       7.01380
_cell_length_c      10.40200
_cell_angle_alpha  90.00000
_cell_angle_beta   90.00000
_cell_angle_gamma  90.00000

loop_
_space_group_symop_id
_space_group_symop_operation_xyz
1 x,y,z
2 x+1/2,-y+1/2,-z+1/2
3 -x+1/2,y+1/2,-z+1/2
4 -x,-y,z
5 -y+1/2,-x+1/2,-z+1/2
6 -y,x,z
7 y,-x,z
8 y+1/2,x+1/2,-z+1/2
9 -x,-y,-z
10 -x+1/2,y+1/2,z+1/2
11 x+1/2,-y+1/2,z+1/2
12 x,y,-z
13 y+1/2,x+1/2,z+1/2
14 y,-x,-z
15 -y,x,-z
16 -y+1/2,-x+1/2,z+1/2

loop_
_atom_site_label
_atom_site_type_symbol
_atom_site_symmetry_multiplicity
_atom_site_Wyckoff_label
_atom_site_fract_x
_atom_site_fract_y
_atom_site_fract_z
_atom_site_occupancy
Al1  Al  2  a  0.00000  0.00000  0.00000  1.00000
Na1  Na  2  b  0.00000  0.00000  0.50000  1.00000
Al2  Al  4  c  0.00000  0.50000  0.00000  1.00000
F1  F  4  e  0.00000  0.00000  0.17110  1.00000
Na2  Na  8  g  0.27680  0.77680  0.25000  1.00000
F2  F  8  h  0.06420  0.24770  0.00000  1.00000
F3  F  16  i  0.17940  0.53640  0.11980  1.00000
```

Chiolite ($Na_5Al_3F_{14}$, $K7_5$): A3B14C5_tP44_128_ac_ehi_bg - POSCAR

```
A3B14C5_tP44_128_ac_ehi_bg & a,c/a,z4,x5,x6,y6,x7,y7,z7 --params=7.0138,
   ↪ 1.48307622116,0.1711,0.2768,0.0642,0.2477,0.1794,0.5364,0.1198
   ↪ & P4/mnc D_{4h}^{6} #128 (abceghi) & tP44 & $K7_{5}$ &
   ↪ Al3F14Na5 & Chiolite & C. Jacoboni and A. Leble and J. J.
   ↪ Rousseau, J. Solid State Chem. 36, 297-304 (1981)
   1.00000000000000
   7.01380000000000    0.00000000000000    0.00000000000000
   0.00000000000000    7.01380000000000    0.00000000000000
   0.00000000000000    0.00000000000000   10.40200000000000
      Al      F      Na
       6     28      10
Direct
   0.00000000000000    0.00000000000000    0.00000000000000    Al   (2a)
   0.50000000000000    0.50000000000000    0.50000000000000    Al   (2a)
   0.00000000000000    0.50000000000000    0.00000000000000    Al   (4c)
   0.50000000000000    0.00000000000000    0.00000000000000    Al   (4c)
   0.50000000000000    0.00000000000000    0.50000000000000    Al   (4c)
   0.00000000000000    0.50000000000000    0.50000000000000    Al   (4c)
   0.00000000000000    0.00000000000000    0.17110000000000    F    (4e)
   0.50000000000000    0.50000000000000    0.32890000000000    F    (4e)
   0.00000000000000    0.00000000000000   -0.17110000000000    F    (4e)
   0.50000000000000    0.50000000000000    0.67110000000000    F    (4e)
   0.06420000000000    0.24770000000000    0.00000000000000    F    (8h)
  -0.06420000000000   -0.24770000000000    0.00000000000000    F    (8h)
  -0.24770000000000    0.06420000000000    0.00000000000000    F    (8h)
   0.24770000000000   -0.06420000000000    0.00000000000000    F    (8h)
   0.43580000000000    0.74770000000000    0.50000000000000    F    (8h)
   0.56420000000000    0.25230000000000    0.50000000000000    F    (8h)
   0.74770000000000    0.56420000000000    0.50000000000000    F    (8h)
   0.25230000000000    0.43580000000000    0.50000000000000    F    (8h)
   0.17940000000000    0.53640000000000    0.11980000000000    F   (16i)
  -0.17940000000000   -0.53640000000000    0.11980000000000    F   (16i)
```

```
  -0.53640000000000    0.17940000000000    0.11980000000000    F    (16i)
   0.53640000000000   -0.17940000000000    0.11980000000000    F    (16i)
   0.32060000000000    1.03640000000000    0.38020000000000    F    (16i)
   0.67940000000000   -0.03640000000000    0.38020000000000    F    (16i)
   1.03640000000000    0.67940000000000    0.38020000000000    F    (16i)
  -0.03640000000000    0.32060000000000    0.38020000000000    F    (16i)
  -0.17940000000000   -0.53640000000000   -0.11980000000000    F    (16i)
   0.17940000000000    0.53640000000000   -0.11980000000000    F    (16i)
   0.53640000000000   -0.17940000000000   -0.11980000000000    F    (16i)
  -0.53640000000000    0.17940000000000   -0.11980000000000    F    (16i)
   0.67940000000000   -0.03640000000000    0.61980000000000    F    (16i)
   0.32060000000000    1.03640000000000    0.61980000000000    F    (16i)
  -0.03640000000000    0.32060000000000    0.61980000000000    F    (16i)
   1.03640000000000    0.67940000000000    0.61980000000000    F    (16i)
   0.00000000000000    0.00000000000000    0.50000000000000    Na    (2b)
   0.50000000000000    0.50000000000000    0.00000000000000    Na    (2b)
   0.27680000000000    0.77680000000000    0.25000000000000    Na    (8g)
  -0.27680000000000    0.22320000000000    0.25000000000000    Na    (8g)
   0.22320000000000    0.27680000000000    0.25000000000000    Na    (8g)
   0.77680000000000   -0.27680000000000    0.25000000000000    Na    (8g)
  -0.27680000000000    0.22320000000000    0.75000000000000    Na    (8g)
   0.27680000000000    0.77680000000000    0.75000000000000    Na    (8g)
   0.77680000000000   -0.27680000000000    0.75000000000000    Na    (8g)
   0.22320000000000    0.27680000000000    0.75000000000000    Na    (8g)
```

Apophyllite ($KCa_4Si_8O_{20}F{\cdot}8H_2O$, $S5_2$): A4BC16DE28F8_tP116_128_h_a_2i_b_g3i_i - CIF

```
# CIF file
data_findsym-output
_audit_creation_method FINDSYM

_chemical_name_mineral 'Apophyllite'
_chemical_formula_sum 'Ca4 F H16 K O28 Si8'

loop_
_publ_author_name
 'G. Y. Chao'
_journal_name_full_name
;
 American Mineralogist
;
_journal_volume 56
_journal_year 1971
_journal_page_first 1234
_journal_page_last 1242
_publ_Section_title
;
 The refinement of the crystal structure of apophyllite: II.
    ↪ Determination of the hydrogen positions by X-ray diffraction
;

_aflow_title 'Apophyllite (KCa$_{4}$Si$_{8}$O$_{20}$F$\cdot$8H$_{2}$O,
    ↪ $S5_{2}$) Structure'
_aflow_proto 'A4BC16DE28F8_tP116_128_h_a_2i_b_g3i_i'
_aflow_params 'a,c/a,x_{3},x_{4},y_{4},x_{5},y_{5},z_{5},x_{6},y_{6},z_{
    ↪ 6},x_{7},y_{7},z_{7},x_{8},y_{8},z_{8},x_{9},y_{9},z_{9},x_{10
    ↪ },y_{10},z_{10}'
_aflow_params_values '8.965,1.75872838818,0.1369,0.1094,0.2466,0.4515,
    ↪ 0.177,0.0775,0.2362,0.4706,0.1198,0.0846,0.1891,0.2178,0.2636,
    ↪ 0.1026,0.0923,0.2131,0.4491,0.0898,0.2256,0.0865,0.19'
_aflow_Strukturbericht '$S5_{2}$'
_aflow_Pearson 'tP116'

_symmetry_space_group_name_H-M "P 4/m 21/n 2/c"
_symmetry_Int_Tables_number 128

_cell_length_a      8.96500
_cell_length_b      8.96500
_cell_length_c      15.76700
_cell_angle_alpha 90.00000
_cell_angle_beta  90.00000
_cell_angle_gamma 90.00000

loop_
_space_group_symop_id
_space_group_symop_operation_xyz
1 x,y,z
2 x+1/2,-y+1/2,-z+1/2
3 -x+1/2,y+1/2,-z+1/2
4 -x,-y,z
5 -y+1/2,-x+1/2,-z+1/2
6 -y,x,z
7 y,-x,z
8 y+1/2,x+1/2,-z+1/2
9 -x,-y,-z
10 -x+1/2,y+1/2,z+1/2
11 x+1/2,-y+1/2,z+1/2
12 x,y,-z
13 y+1/2,x+1/2,z+1/2
14 y,-x,-z
15 -y,x,-z
16 -y+1/2,-x+1/2,z+1/2

loop_
_atom_site_label
_atom_site_type_symbol
_atom_site_symmetry_multiplicity
_atom_site_Wyckoff_label
_atom_site_fract_x
_atom_site_fract_y
_atom_site_fract_z
_atom_site_occupancy
F1 F 2 a 0.00000 0.00000 0.00000 1.00000
K1 K 2 b 0.00000 0.00000 0.50000 1.00000
O1 O 8 g 0.13690 0.63690 0.25000 1.00000
Ca1 Ca 8 h 0.10940 0.24660 0.00000 1.00000
H1 H 16 i 0.45150 0.17700 0.07750 1.00000
H2 H 16 i 0.23620 0.47060 0.11980 1.00000
O2 O 16 i 0.08460 0.18910 0.21780 1.00000
O3 O 16 i 0.26360 0.10260 0.09230 1.00000
O4 O 16 i 0.21310 0.44910 0.08980 1.00000
Si1 Si 16 i 0.22560 0.08650 0.19000 1.00000
```

Apophyllite ($KCa_4Si_8O_{20}F{\cdot}8H_2O$, $S5_2$): A4BC16DE28F8_tP116_128_h_a_2i_b_g3i_i - POSCAR

```
A4BC16DE28F8_tP116_128_h_a_2i_b_g3i_i & a,c/a,x3,x4,y4,x5,y5,z5,x6,y6,z6
    ↪ ,x7,y7,z7,x8,y8,z8,x9,y9,z9,x10,y10,z10 --params=8.965,
    ↪ 1.75872838818,0.1369,0.1094,0.2466,0.4515,0.177,0.0775,0.2362,
    ↪ 0.4706,0.1198,0.0846,0.1891,0.2178,0.2636,0.1026,0.0923,0.2131,
    ↪ 0.4491,0.0898,0.2256,0.0865,0.19 & P4/mnc D_{4h}^{6} #128 (
    ↪ abghi^6) & tP116 & $S5_{2}$ & Ca4FH16KO28Si8 & Apophyllite & G.
    ↪ Y. Chao, Am. Mineral. 56, 1234-1242 (1971)
   1.00000000000000
   8.96500000000000    0.00000000000000    0.00000000000000
   0.00000000000000    8.96500000000000    0.00000000000000
   0.00000000000000    0.00000000000000   15.76700000000000
    Ca     F     H     K     O    Si
     8     2    32     2    56    16
Direct
   0.10940000000000    0.24660000000000    0.00000000000000    Ca    (8h)
  -0.10940000000000   -0.24660000000000    0.00000000000000    Ca    (8h)
  -0.24660000000000    0.10940000000000    0.00000000000000    Ca    (8h)
   0.24660000000000   -0.10940000000000    0.00000000000000    Ca    (8h)
   0.39060000000000    0.74660000000000    0.50000000000000    Ca    (8h)
   0.60940000000000    0.25340000000000    0.50000000000000    Ca    (8h)
   0.74660000000000    0.60940000000000    0.50000000000000    Ca    (8h)
   0.25340000000000    0.39060000000000    0.50000000000000    Ca    (8h)
   0.00000000000000    0.00000000000000    0.00000000000000    F    (2a)
   0.50000000000000    0.50000000000000    0.50000000000000    F    (2a)
   0.45150000000000    0.17700000000000    0.07750000000000    H   (16i)
  -0.45150000000000   -0.17700000000000    0.07750000000000    H   (16i)
  -0.17700000000000    0.45150000000000    0.07750000000000    H   (16i)
   0.17700000000000   -0.45150000000000    0.07750000000000    H   (16i)
   0.04850000000000    0.67700000000000    0.42250000000000    H   (16i)
   0.95150000000000    0.32300000000000    0.42250000000000    H   (16i)
   0.67700000000000    0.95150000000000    0.42250000000000    H   (16i)
   0.32300000000000    0.04850000000000    0.42250000000000    H   (16i)
  -0.45150000000000   -0.17700000000000   -0.07750000000000    H   (16i)
   0.45150000000000    0.17700000000000   -0.07750000000000    H   (16i)
   0.17700000000000   -0.45150000000000   -0.07750000000000    H   (16i)
  -0.17700000000000    0.45150000000000   -0.07750000000000    H   (16i)
   0.95150000000000    0.32300000000000    0.57750000000000    H   (16i)
   0.04850000000000    0.67700000000000    0.57750000000000    H   (16i)
   0.32300000000000    0.04850000000000    0.57750000000000    H   (16i)
   0.67700000000000    0.95150000000000    0.57750000000000    H   (16i)
   0.23620000000000    0.47060000000000    0.11980000000000    H   (16i)
  -0.23620000000000   -0.47060000000000    0.11980000000000    H   (16i)
  -0.47060000000000    0.23620000000000    0.11980000000000    H   (16i)
   0.47060000000000   -0.23620000000000    0.11980000000000    H   (16i)
   0.26380000000000    0.97060000000000    0.38020000000000    H   (16i)
   0.73620000000000    0.02940000000000    0.38020000000000    H   (16i)
   0.97060000000000    0.73620000000000    0.38020000000000    H   (16i)
   0.02940000000000    0.26380000000000    0.38020000000000    H   (16i)
  -0.23620000000000   -0.47060000000000   -0.11980000000000    H   (16i)
   0.23620000000000    0.47060000000000   -0.11980000000000    H   (16i)
   0.47060000000000   -0.23620000000000   -0.11980000000000    H   (16i)
  -0.47060000000000    0.23620000000000   -0.11980000000000    H   (16i)
   0.73620000000000    0.02940000000000    0.61980000000000    H   (16i)
   0.26380000000000    0.97060000000000    0.61980000000000    H   (16i)
   0.02940000000000    0.26380000000000    0.61980000000000    H   (16i)
   0.97060000000000    0.73620000000000    0.61980000000000    H   (16i)
   0.00000000000000    0.00000000000000    0.50000000000000    K    (2b)
   0.50000000000000    0.50000000000000    0.00000000000000    K    (2b)
   0.13690000000000    0.63690000000000    0.25000000000000    O    (8g)
  -0.13690000000000    0.36310000000000    0.25000000000000    O    (8g)
   0.36310000000000    0.13690000000000    0.25000000000000    O    (8g)
   0.63690000000000   -0.13690000000000    0.25000000000000    O    (8g)
  -0.13690000000000    0.36310000000000    0.75000000000000    O    (8g)
   0.13690000000000    0.63690000000000    0.75000000000000    O    (8g)
   0.63690000000000   -0.13690000000000    0.75000000000000    O    (8g)
   0.36310000000000    0.13690000000000    0.75000000000000    O    (8g)
   0.08460000000000    0.18910000000000    0.21780000000000    O   (16i)
  -0.08460000000000   -0.18910000000000    0.21780000000000    O   (16i)
  -0.18910000000000    0.08460000000000    0.21780000000000    O   (16i)
   0.18910000000000   -0.08460000000000    0.21780000000000    O   (16i)
   0.41540000000000    0.68910000000000    0.28220000000000    O   (16i)
   0.58460000000000    0.31090000000000    0.28220000000000    O   (16i)
   0.68910000000000    0.58460000000000    0.28220000000000    O   (16i)
   0.31090000000000    0.41540000000000    0.28220000000000    O   (16i)
  -0.08460000000000   -0.18910000000000   -0.21780000000000    O   (16i)
   0.08460000000000    0.18910000000000   -0.21780000000000    O   (16i)
   0.18910000000000   -0.08460000000000   -0.21780000000000    O   (16i)
  -0.18910000000000    0.08460000000000   -0.21780000000000    O   (16i)
   0.58460000000000    0.31090000000000    0.71780000000000    O   (16i)
   0.41540000000000    0.68910000000000    0.71780000000000    O   (16i)
   0.31090000000000    0.41540000000000    0.71780000000000    O   (16i)
   0.68910000000000    0.58460000000000    0.71780000000000    O   (16i)
   0.26360000000000    0.10260000000000    0.09230000000000    O   (16i)
  -0.26360000000000   -0.10260000000000    0.09230000000000    O   (16i)
  -0.10260000000000    0.26360000000000    0.09230000000000    O   (16i)
   0.10260000000000   -0.26360000000000    0.09230000000000    O   (16i)
   0.23640000000000    0.60260000000000    0.40770000000000    O   (16i)
   0.76360000000000    0.39740000000000    0.40770000000000    O   (16i)
   0.60260000000000    0.76360000000000    0.40770000000000    O   (16i)
   0.39740000000000    0.23640000000000    0.40770000000000    O   (16i)
  -0.26360000000000   -0.10260000000000   -0.09230000000000    O   (16i)
   0.26360000000000    0.10260000000000   -0.09230000000000    O   (16i)
   0.10260000000000   -0.26360000000000   -0.09230000000000    O   (16i)
  -0.10260000000000    0.26360000000000   -0.09230000000000    O   (16i)
   0.76360000000000    0.39740000000000    0.59230000000000    O   (16i)
```

```
0.23640000000000   0.60260000000000   0.59230000000000   O   (16i)
0.39740000000000   0.23640000000000   0.59230000000000   O   (16i)
0.60260000000000   0.76360000000000   0.59230000000000   O   (16i)
0.21310000000000   0.44910000000000   0.08980000000000   O   (16i)
-0.21310000000000  -0.44910000000000   0.08980000000000   O   (16i)
-0.44910000000000   0.21310000000000   0.08980000000000   O   (16i)
0.44910000000000  -0.21310000000000   0.08980000000000   O   (16i)
0.28690000000000   0.94910000000000   0.41020000000000   O   (16i)
0.71310000000000   0.05090000000000   0.41020000000000   O   (16i)
0.94910000000000   0.71310000000000   0.41020000000000   O   (16i)
0.05090000000000   0.28690000000000   0.41020000000000   O   (16i)
-0.21310000000000  -0.44910000000000  -0.08980000000000   O   (16i)
0.21310000000000   0.44910000000000  -0.08980000000000   O   (16i)
0.44910000000000  -0.21310000000000  -0.08980000000000   O   (16i)
-0.44910000000000   0.21310000000000  -0.08980000000000   O   (16i)
0.71310000000000   0.05090000000000   0.58980000000000   O   (16i)
0.28690000000000   0.94910000000000   0.58980000000000   O   (16i)
0.05090000000000   0.28690000000000   0.58980000000000   O   (16i)
0.94910000000000   0.71310000000000   0.58980000000000   O   (16i)
0.22560000000000   0.08650000000000   0.19000000000000   Si  (16i)
-0.22560000000000  -0.08650000000000   0.19000000000000   Si  (16i)
-0.08650000000000   0.22560000000000   0.19000000000000   Si  (16i)
0.08650000000000  -0.22560000000000   0.19000000000000   Si  (16i)
0.27440000000000   0.58650000000000   0.31000000000000   Si  (16i)
0.72560000000000   0.41350000000000   0.31000000000000   Si  (16i)
0.58650000000000   0.72560000000000   0.31000000000000   Si  (16i)
0.41350000000000   0.27440000000000   0.31000000000000   Si  (16i)
-0.22560000000000  -0.08650000000000  -0.19000000000000   Si  (16i)
0.22560000000000   0.08650000000000  -0.19000000000000   Si  (16i)
0.08650000000000  -0.22560000000000  -0.19000000000000   Si  (16i)
-0.08650000000000   0.22560000000000  -0.19000000000000   Si  (16i)
0.72560000000000   0.41350000000000   0.69000000000000   Si  (16i)
0.27440000000000   0.58650000000000   0.69000000000000   Si  (16i)
0.41350000000000   0.27440000000000   0.69000000000000   Si  (16i)
0.58650000000000   0.72560000000000   0.69000000000000   Si  (16i)
```

$CaBe_2Ge_2$: A2BC2_tP10_129_ac_c_bc - CIF

```
# CIF file
data_findsym-output
_audit_creation_method FINDSYM

_chemical_name_mineral 'Be2CaGe2'
_chemical_formula_sum 'Be2 Ca Ge2'

loop_
_publ_author_name
 'B. Eisenmann'
 'N. May'
 'W. M\"{u}ller'
 'H. Sch\"{a}fer'
_journal_year 1972
_publ_Section_title
;
 Eine neue strukturelle Variante des BaAl$_{4}$-Typs: Der CaBe$_{2}$Ge$_
    ↪ {2}$-Typ
;

_aflow_title 'CaBe$_{2}$Ge$_{2}$ Structure'
_aflow_proto 'A2BC2_tP10_129_ac_c_bc'
_aflow_params 'a,c/a,z_{3},z_{4},z_{5}'
_aflow_params_values '4.02,2.46766169154,0.608,0.249,0.868'
_aflow_Strukturbericht 'None'
_aflow_Pearson 'tP10'

_symmetry_space_group_name_H-M "P 4/n 21/m 2/m (origin choice 2)"
_symmetry_Int_Tables_number 129

_cell_length_a      4.02000
_cell_length_b      4.02000
_cell_length_c      9.92000
_cell_angle_alpha 90.00000
_cell_angle_beta  90.00000
_cell_angle_gamma 90.00000

loop_
_space_group_symop_id
_space_group_symop_operation_xyz
1 x,y,z
2 x+1/2,-y,-z
3 -x,y+1/2,-z
4 -x+1/2,-y+1/2,z
5 -y,-x,-z
6 -y+1/2,x,z
7 y,-x+1/2,z
8 y+1/2,x+1/2,-z
9 -x,-y,-z
10 -x+1/2,y,z
11 x,-y+1/2,z
12 x+1/2,y+1/2,-z
13 y,x,z
14 y+1/2,-x,-z
15 -y,x+1/2,-z
16 -y+1/2,-x+1/2,z

loop_
_atom_site_label
_atom_site_type_symbol
_atom_site_symmetry_multiplicity
_atom_site_Wyckoff_label
_atom_site_fract_x
_atom_site_fract_y
_atom_site_fract_z
_atom_site_occupancy
Be1 Be 2 a 0.75000 0.25000 0.00000 1.00000
Ge1 Ge 2 b 0.75000 0.25000 0.50000 1.00000
Be2 Be 2 c 0.25000 0.25000 0.60800 1.00000
Ca1 Ca 2 c 0.25000 0.25000 0.24900 1.00000
Ge2 Ge 2 c 0.25000 0.25000 0.86800 1.00000
```

$CaBe_2Ge_2$: A2BC2_tP10_129_ac_c_bc - POSCAR

```
A2BC2_tP10_129_ac_c_bc & a,c/a,z3,z4,z5 --params=4.02,2.46766169154,
    ↪ 0.608,0.249,0.868 & P4/nmm D_{4h}^{7} #129 (abc^3) & tP10 &
    ↪ None & Be2CaGe2 & Be2CaGe2 & B. Eisenmann et al., (1972)
   1.00000000000000
   4.02000000000000   0.00000000000000   0.00000000000000
   0.00000000000000   4.02000000000000   0.00000000000000
   0.00000000000000   0.00000000000000   9.92000000000000
    Be     Ca     Ge
     4      2      4
Direct
   0.75000000000000   0.25000000000000   0.00000000000000   Be   (2a)
   0.25000000000000   0.75000000000000   0.00000000000000   Be   (2a)
   0.25000000000000   0.25000000000000   0.60800000000000   Be   (2c)
   0.75000000000000   0.75000000000000  -0.60800000000000   Be   (2c)
   0.25000000000000   0.25000000000000   0.24900000000000   Ca   (2c)
   0.75000000000000   0.75000000000000  -0.24900000000000   Ca   (2c)
   0.75000000000000   0.25000000000000   0.50000000000000   Ge   (2b)
   0.25000000000000   0.75000000000000   0.50000000000000   Ge   (2b)
   0.25000000000000   0.25000000000000   0.86800000000000   Ge   (2c)
   0.75000000000000   0.75000000000000  -0.86800000000000   Ge   (2c)
```

Meta-autunite (I) $[Ca(UO_2)_2(PO_4)_2 \cdot 6H_2O$, $H5_{10}]$: AB4C6DE_tP26_129_c_j_2ci_a_c - CIF

```
# CIF file
data_findsym-output
_audit_creation_method FINDSYM

_chemical_name_mineral 'Meta-autunite (i)'
_chemical_formula_sum 'Ca (H2O)4 O6 P U'

loop_
_publ_author_name
 'Y. S. Makarov'
 'V. I. Ivanov'
_journal_name_full_name
;
 Doklady Akademii Nauk SSSR
;
_journal_volume 132
_journal_year 1960
_journal_page_first 601
_journal_page_last 603
_publ_Section_title
;
 The crystal structure of meta-autunite, Ca(UO$_{2}$)$_{2}$(PO$_{4}$)$_{
    ↪ 2}$*6H$_{2}$O
;

# Found in The American Mineralogist Crystal Structure Database, 2003

_aflow_title 'Meta-autunite (I) [Ca(UO$_{2}$)$_{2}$(PO$_{4}$)$_{2}$$\
    ↪ cdot$6H$_{2}$O, $H5_{10}$] Structure'
_aflow_proto 'AB4C6DE_tP26_129_c_j_2ci_a_c'
_aflow_params 'a,c/a,z_{2},z_{3},z_{4},z_{5},y_{6},z_{6},x_{7},z_{7}'
_aflow_params_values '6.98,1.20630372493,0.612,0.343,0.893,0.106,0.584,
    ↪ 0.106,0.486,0.392'
_aflow_Strukturbericht '$H5_{10}$'
_aflow_Pearson 'tP26'

_symmetry_space_group_name_H-M "P 4/n 21/m 2/m (origin choice 2)"
_symmetry_Int_Tables_number 129

_cell_length_a      6.98000
_cell_length_b      6.98000
_cell_length_c      8.42000
_cell_angle_alpha 90.00000
_cell_angle_beta  90.00000
_cell_angle_gamma 90.00000

loop_
_space_group_symop_id
_space_group_symop_operation_xyz
1 x,y,z
2 x+1/2,-y,-z
3 -x,y+1/2,-z
4 -x+1/2,-y+1/2,z
5 -y,-x,-z
6 -y+1/2,x,z
7 y,-x+1/2,z
8 y+1/2,x+1/2,-z
9 -x,-y,-z
10 -x+1/2,y,z
11 x,-y+1/2,z
12 x+1/2,y+1/2,-z
13 y,x,z
14 y+1/2,-x,-z
15 -y,x+1/2,-z
16 -y+1/2,-x+1/2,z

loop_
_atom_site_label
_atom_site_type_symbol
_atom_site_symmetry_multiplicity
_atom_site_Wyckoff_label
_atom_site_fract_x
_atom_site_fract_y
_atom_site_fract_z
_atom_site_occupancy
```

```
P1 P 2 a 0.75000 0.25000 0.00000 1.00000
Ca1 Ca 2 c 0.25000 0.25000 0.61200 0.50000
O1 O 2 c 0.25000 0.25000 0.34300 1.00000
O2 O 2 c 0.25000 0.25000 0.89300 1.00000
U1 U 2 c 0.25000 0.25000 0.10600 1.00000
O3 O 8 i 0.25000 0.58400 0.10600 1.00000
H2O1 H2O 8 j 0.48600 0.48600 0.39200 0.75000
```

Meta-autunite (I) [$Ca(UO_2)_2(PO_4)_2{\cdot}6H_2O$, $H5_{10}$]: AB4C6DE_tP26_129_c_j_2ci_a_c - POSCAR

```
AB4C6DE_tP26_129_c_j_2ci_a_c & a,c/a,z2,z3,z4,z5,y6,z6,x7,z7 --params=
   ↪ 6.98,1.20630372493,0.612,0.343,0.893,0.106,0.584,0.106,0.486,
   ↪ 0.392 & P4/nmm D_{4h}^{7} #129 (ac^4ij) & tP26 & $H5_{10}$ &
   ↪ CaH12O18P2U2 & Meta-autunite (i) & Y. S. Makarov and V. I.
   ↪ Ivanov, Doklady Akademii Nauk SSSR 132, 601-603 (1960)
   1.00000000000000
   6.98000000000000   0.00000000000000   0.00000000000000
   0.00000000000000   6.98000000000000   0.00000000000000
   0.00000000000000   0.00000000000000   8.42000000000000
    Ca    H2O    O     P     U
     2     8    12     2     2
Direct
   0.25000000000000   0.25000000000000   0.61200000000000   Ca   (2c)
   0.75000000000000   0.75000000000000  -0.61200000000000   Ca   (2c)
   0.48600000000000   0.48600000000000   0.39200000000000   H2O  (8j)
   0.01400000000000   0.01400000000000   0.39200000000000   H2O  (8j)
   0.01400000000000   0.48600000000000   0.39200000000000   H2O  (8j)
   0.48600000000000   0.01400000000000   0.39200000000000   H2O  (8j)
  -0.48600000000000   0.98600000000000  -0.39200000000000   H2O  (8j)
   0.98600000000000  -0.48600000000000  -0.39200000000000   H2O  (8j)
   0.98600000000000   0.98600000000000  -0.39200000000000   H2O  (8j)
  -0.48600000000000  -0.48600000000000  -0.39200000000000   H2O  (8j)
   0.25000000000000   0.25000000000000   0.34300000000000   O    (2c)
   0.75000000000000   0.75000000000000  -0.34300000000000   O    (2c)
   0.25000000000000   0.25000000000000   0.89300000000000   O    (2c)
   0.75000000000000   0.75000000000000  -0.89300000000000   O    (2c)
   0.25000000000000   0.58400000000000   0.10600000000000   O    (8i)
   0.25000000000000  -0.08400000000000   0.10600000000000   O    (8i)
  -0.08400000000000   0.25000000000000   0.10600000000000   O    (8i)
   0.58400000000000   0.25000000000000   0.10600000000000   O    (8i)
   0.75000000000000   1.08400000000000  -0.10600000000000   O    (8i)
   0.75000000000000  -0.58400000000000  -0.10600000000000   O    (8i)
   1.08400000000000   0.75000000000000  -0.10600000000000   O    (8i)
  -0.58400000000000   0.75000000000000  -0.10600000000000   O    (8i)
   0.75000000000000   0.25000000000000   0.00000000000000   P    (2a)
   0.25000000000000   0.75000000000000   0.00000000000000   P    (2a)
   0.25000000000000   0.25000000000000   0.10600000000000   U    (2c)
   0.75000000000000   0.75000000000000  -0.10600000000000   U    (2c)
```

$NH_4Br$ ($B25$): AB4C_tP12_129_c_i_a - CIF

```
# CIF file
data_findsym-output
_audit_creation_method FINDSYM

_chemical_name_mineral 'BrH4N'
_chemical_formula_sum 'Br H4 N'

loop_
_publ_author_name
 'H. A. Levy'
 'S. W. Peterson'
_journal_name_full_name
;
 Journal of the American Chemical Society
;
_journal_volume 75
_journal_year 1953
_journal_page_first 1536
_journal_page_last 1542
_publ_Section_title
;
 Neutron Diffraction Determination of the Crystal Structure of Ammonium
    ↪ Bromide in Four Phases
;

_aflow_title 'NH$_{4}$Br ($B25$) Structure'
_aflow_proto 'AB4C_tP12_129_c_i_a'
_aflow_params 'a,c/a,z_{2},y_{3},z_{3}'
_aflow_params_values '5.82,0.710652920962,0.48,0.897,0.147'
_aflow_Strukturbericht '$B25$'
_aflow_Pearson 'tP12'

_symmetry_space_group_name_H-M "P 4/n 21/m 2/m (origin choice 2)"
_symmetry_Int_Tables_number 129

_cell_length_a     5.82000
_cell_length_b     5.82000
_cell_length_c     4.13600
_cell_angle_alpha 90.00000
_cell_angle_beta  90.00000
_cell_angle_gamma 90.00000

loop_
_space_group_symop_id
_space_group_symop_operation_xyz
1 x,y,z
2 x+1/2,-y,-z
3 -x,y+1/2,-z
4 -x+1/2,-y+1/2,z
5 -y,-x,-z
6 -y+1/2,x,z
7 y,-x+1/2,z
8 y+1/2,x+1/2,-z
9 -x,-y,-z
10 -x+1/2,y,z
11 x,-y+1/2,z
12 x+1/2,y+1/2,-z
13 y,x,z
14 y+1/2,-x,-z
15 -y,x+1/2,-z
16 -y+1/2,-x+1/2,z

loop_
_atom_site_label
_atom_site_type_symbol
_atom_site_symmetry_multiplicity
_atom_site_Wyckoff_label
_atom_site_fract_x
_atom_site_fract_y
_atom_site_fract_z
_atom_site_occupancy
N1 N 2 a 0.25000 0.75000 0.00000 1.00000
Br1 Br 2 c 0.25000 0.25000 0.48000 1.00000
H1 H 8 i 0.25000 0.89700 0.14700 1.00000
```

$NH_4Br$ ($B25$): AB4C_tP12_129_c_i_a - POSCAR

```
AB4C_tP12_129_c_i_a & a,c/a,z2,y3,z3 --params=5.82,0.710652920962,0.48,
   ↪ 0.897,0.147 & P4/nmm D_{4h}^{7} #129 (aci) & tP12 & $B25$ &
   ↪ BrH4N & BrH4N & H. A. Levy and S. W. Peterson, J. Am. Chem.
   ↪ Soc. 75, 1536-1542 (1953)
   1.00000000000000
   5.82000000000000   0.00000000000000   0.00000000000000
   0.00000000000000   5.82000000000000   0.00000000000000
   0.00000000000000   0.00000000000000   4.13600000000000
    Br     H     N
     2     8     2
Direct
   0.25000000000000   0.25000000000000   0.48000000000000   Br   (2c)
   0.75000000000000   0.75000000000000  -0.48000000000000   Br   (2c)
   0.25000000000000   0.89700000000000   0.14700000000000   H    (8i)
   0.25000000000000  -0.39700000000000   0.14700000000000   H    (8i)
  -0.39700000000000   0.25000000000000   0.14700000000000   H    (8i)
   0.89700000000000   0.25000000000000   0.14700000000000   H    (8i)
   0.75000000000000   1.39700000000000  -0.14700000000000   H    (8i)
   0.75000000000000  -0.89700000000000  -0.14700000000000   H    (8i)
   1.39700000000000   0.75000000000000  -0.14700000000000   H    (8i)
  -0.89700000000000   0.75000000000000  -0.14700000000000   H    (8i)
   0.75000000000000   0.25000000000000   0.00000000000000   N    (2a)
   0.25000000000000   0.75000000000000   0.00000000000000   N    (2a)
```

LaOAgS: ABCD_tP8_129_b_c_a_c - CIF

```
# CIF file
data_findsym-output
_audit_creation_method FINDSYM

_chemical_name_mineral 'AgLaOS'
_chemical_formula_sum 'Ag La O S'

loop_
_publ_author_name
 'M. Palazzi'
 'S. Jaulmes'
_journal_name_full_name
;
 Acta Crystallographica Section B: Structural Science
;
_journal_volume 37
_journal_year 1981
_journal_page_first 1337
_journal_page_last 1339
_publ_Section_title
;
 Structure du Conducteur Ionique (LaO)AgS
;

# Found in Pnictides and Chalcogenides II, 2003

_aflow_title 'LaOAgS Structure'
_aflow_proto 'ABCD_tP8_129_b_c_a_c'
_aflow_params 'a,c/a,z_{3},z_{4}'
_aflow_params_values '4.05,2.23185185185,0.1356,0.6929'
_aflow_Strukturbericht 'None'
_aflow_Pearson 'tP8'

_symmetry_space_group_name_H-M "P 4/n 21/m 2/m (origin choice 2)"
_symmetry_Int_Tables_number 129

_cell_length_a     4.05000
_cell_length_b     4.05000
_cell_length_c     9.03900
_cell_angle_alpha 90.00000
_cell_angle_beta  90.00000
_cell_angle_gamma 90.00000

loop_
_space_group_symop_id
_space_group_symop_operation_xyz
1 x,y,z
2 x+1/2,-y,-z
3 -x,y+1/2,-z
4 -x+1/2,-y+1/2,z
5 -y,-x,-z
6 -y+1/2,x,z
7 y,-x+1/2,z
8 y+1/2,x+1/2,-z
9 -x,-y,-z
10 -x+1/2,y,z
```

```
11 x,-y+1/2,z
12 x+1/2,y+1/2,-z
13 y,x,z
14 y+1/2,-x,-z
15 -y,x+1/2,-z
16 -y+1/2,-x+1/2,z

loop_
_atom_site_label
_atom_site_type_symbol
_atom_site_symmetry_multiplicity
_atom_site_Wyckoff_label
_atom_site_fract_x
_atom_site_fract_y
_atom_site_fract_z
_atom_site_occupancy
O1 O 2 a 0.75000 0.25000 0.00000 1.00000
Ag1 Ag 2 b 0.75000 0.25000 0.50000 1.00000
La1 La 2 c 0.25000 0.25000 0.13560 1.00000
S1 S 2 c 0.25000 0.25000 0.69290 1.00000
```

LaOAgS: ABCD_tP8_129_b_c_a_c - POSCAR

```
ABCD_tP8_129_b_c_a_c & a,c/a,z3,z4 --params=4.05,2.23185185185,0.1356,
    ↪ 0.6929 & P4/nmm D_{4h}^{7} #129 (abc^2) & tP8 & None & AgLaOS &
    ↪ AgLaOS & M. Palazzi and S. Jaulmes, Acta Crystallogr. Sect. B
    ↪ Struct. Sci. 37, 1337-1339 (1981)
   1.00000000000000
   4.05000000000000   0.00000000000000   0.00000000000000
   0.00000000000000   4.05000000000000   0.00000000000000
   0.00000000000000   0.00000000000000   9.03900000000000
    Ag    La     O     S
     2     2     2     2
Direct
   0.75000000000000   0.25000000000000   0.50000000000000   Ag   (2b)
   0.25000000000000   0.75000000000000   0.50000000000000   Ag   (2b)
   0.25000000000000   0.25000000000000   0.13560000000000   La   (2c)
   0.75000000000000   0.75000000000000  -0.13560000000000   La   (2c)
   0.75000000000000   0.25000000000000   0.00000000000000    O   (2a)
   0.25000000000000   0.75000000000000   0.00000000000000    O   (2a)
   0.25000000000000   0.25000000000000   0.69290000000000    S   (2c)
   0.75000000000000   0.75000000000000  -0.69290000000000    S   (2c)
```

$Sr(OH)_2(H_2O)_8$: A18B10C_tP116_130_2c4g_2c2g_a - CIF

```
# CIF file
data_findsym-output
_audit_creation_method FINDSYM

_chemical_name_mineral 'H18O10Sr'
_chemical_formula_sum 'H18 O10 Sr'

loop_
_publ_author_name
 'J. S. Ricci'
 'R. C. Stevens'
 'R. K. {McMullan}'
 'W. T. Klooster'
_journal_name_full_name
;
 Acta Crystallographica Section B: Structural Science
;
_journal_volume 61
_journal_year 2005
_journal_page_first 381
_journal_page_last 386
_publ_Section_title
;
 Structure of strontium hydroxide octahydrate, Sr(OH)$_{2} \cdot $8H$_{2
    ↪ }$O, at 20, 100 and 200 K from neutron diffraction
;

_aflow_title 'Sr(OH)$_{2}$(H$_{2}$O)$_{8}$ Structure'
_aflow_proto 'A18B10C_tP116_130_2c4g_2c2g_a'
_aflow_params 'a,c/a,z_{2},z_{3},z_{4},z_{5},x_{6},y_{6},z_{6},x_{7},y_{
    ↪ 7},z_{7},x_{8},y_{8},z_{8},x_{9},y_{9},z_{9},x_{10},y_{10},z_{
    ↪ 10},x_{11},y_{11},z_{11}'
_aflow_params_values '8.984,1.28194568121,0.4888,0.2426,0.40482,0.15904,
    ↪ 0.8044,0.5759,0.1201,-0.047,0.4874,0.1327,0.8054,0.5766,0.8623,
    ↪ 0.8392,0.4596,-0.0398,0.84543,0.47219,0.12549,0.84587,0.47366,
    ↪ 0.87655'
_aflow_Strukturbericht 'None'
_aflow_Pearson 'tP116'

_symmetry_space_group_name_H-M "P 4/n 21/c 2/c (origin choice 2)"
_symmetry_Int_Tables_number 130

_cell_length_a     8.98400
_cell_length_b     8.98400
_cell_length_c     11.51700
_cell_angle_alpha  90.00000
_cell_angle_beta   90.00000
_cell_angle_gamma  90.00000

loop_
_space_group_symop_id
_space_group_symop_operation_xyz
1 x,y,z
2 x+1/2,-y,-z+1/2
3 -x,y+1/2,-z+1/2
4 -x+1/2,-y+1/2,z
5 -y,-x,-z+1/2
6 -y+1/2,x,z
7 y,-x+1/2,z
8 y+1/2,x+1/2,-z+1/2
9 -x,-y,-z
10 -x+1/2,y,z+1/2
11 x,-y+1/2,z+1/2
12 x+1/2,y+1/2,-z
13 y,x,z+1/2
14 y+1/2,-x,-z
15 -y,x+1/2,-z
16 -y+1/2,-x+1/2,z+1/2

loop_
_atom_site_label
_atom_site_type_symbol
_atom_site_symmetry_multiplicity
_atom_site_Wyckoff_label
_atom_site_fract_x
_atom_site_fract_y
_atom_site_fract_z
_atom_site_occupancy
Sr1 Sr 4 a 0.75000 0.25000 0.25000 1.00000
H1 H 4 c 0.25000 0.25000 0.48880 1.00000
H2 H 4 c 0.25000 0.25000 0.24260 1.00000
O1 O 4 c 0.25000 0.25000 0.40482 1.00000
O2 O 4 c 0.25000 0.25000 0.15904 1.00000
H3 H 16 g 0.80440 0.57590 0.12010 1.00000
H4 H 16 g -0.04700 0.48740 0.13270 1.00000
H5 H 16 g 0.80540 0.57660 0.86230 1.00000
H6 H 16 g 0.83920 0.45960 -0.03980 1.00000
O3 O 16 g 0.84543 0.47219 0.12549 1.00000
O4 O 16 g 0.84587 0.47366 0.87655 1.00000
```

$Sr(OH)_2(H_2O)_8$: A18B10C_tP116_130_2c4g_2c2g_a - POSCAR

```
A18B10C_tP116_130_2c4g_2c2g_a & a,c/a,z2,z3,z4,z5,x6,y6,z6,x7,y7,z7,x8,
    ↪ y8,z8,x9,y9,z9,x10,y10,z10,x11,y11,z11 --params=8.984,
    ↪ 1.28194568121,0.4888,0.2426,0.40482,0.15904,0.8044,0.5759,
    ↪ 0.1201,-0.047,0.4874,0.1327,0.8054,0.5766,0.8623,0.8392,0.4596
    ↪ ,-0.0398,0.84543,0.47219,0.12549,0.84587,0.47366,0.87655 & P4/
    ↪ ncc D_{4h}^{8} #130 (ac^4g^6) & tP116 & None & H18O10Sr &
    ↪ H18O10Sr & J. S. Ricci et al., Acta Crystallogr. Sect. B
    ↪ Struct. Sci. 61, 381-386 (2005)
   1.00000000000000
   8.98400000000000   0.00000000000000   0.00000000000000
   0.00000000000000   8.98400000000000   0.00000000000000
   0.00000000000000   0.00000000000000  11.51700000000000
     H     O    Sr
    72    40     4
Direct
   0.25000000000000   0.25000000000000   0.48880000000000    H   (4c)
   0.75000000000000   0.75000000000000   0.01120000000000    H   (4c)
   0.75000000000000   0.75000000000000  -0.48880000000000    H   (4c)
   0.25000000000000   0.25000000000000   0.98880000000000    H   (4c)
   0.25000000000000   0.25000000000000   0.24260000000000    H   (4c)
   0.75000000000000   0.75000000000000   0.25740000000000    H   (4c)
   0.75000000000000   0.75000000000000  -0.24260000000000    H   (4c)
   0.25000000000000   0.25000000000000   0.74260000000000    H   (4c)
   0.80440000000000   0.57590000000000   0.12010000000000    H  (16g)
  -0.30440000000000  -0.07590000000000   0.12010000000000    H  (16g)
  -0.07590000000000   0.80440000000000   0.12010000000000    H  (16g)
   0.57590000000000  -0.30440000000000   0.12010000000000    H  (16g)
  -0.80440000000000   1.07590000000000   0.37990000000000    H  (16g)
   1.30440000000000  -0.57590000000000   0.37990000000000    H  (16g)
   1.07590000000000   1.30440000000000   0.37990000000000    H  (16g)
  -0.57590000000000  -0.80440000000000   0.37990000000000    H  (16g)
  -0.80440000000000  -0.57590000000000  -0.12010000000000    H  (16g)
   1.30440000000000   1.07590000000000  -0.12010000000000    H  (16g)
   1.07590000000000  -0.80440000000000  -0.12010000000000    H  (16g)
  -0.57590000000000   1.30440000000000  -0.12010000000000    H  (16g)
   0.80440000000000  -0.07590000000000   0.62010000000000    H  (16g)
  -0.30440000000000   0.57590000000000   0.62010000000000    H  (16g)
  -0.07590000000000  -0.30440000000000   0.62010000000000    H  (16g)
   0.57590000000000   0.80440000000000   0.62010000000000    H  (16g)
  -0.04700000000000   0.48740000000000   0.13270000000000    H  (16g)
   0.54700000000000   0.01260000000000   0.13270000000000    H  (16g)
   0.01260000000000  -0.04700000000000   0.13270000000000    H  (16g)
   0.48740000000000   0.54700000000000   0.13270000000000    H  (16g)
   0.04700000000000   0.98740000000000   0.36730000000000    H  (16g)
   0.45300000000000  -0.48740000000000   0.36730000000000    H  (16g)
   0.98740000000000   0.45300000000000   0.36730000000000    H  (16g)
  -0.48740000000000   0.04700000000000   0.36730000000000    H  (16g)
   0.04700000000000  -0.48740000000000  -0.13270000000000    H  (16g)
   0.45300000000000   0.98740000000000  -0.13270000000000    H  (16g)
   0.98740000000000   0.04700000000000  -0.13270000000000    H  (16g)
  -0.48740000000000   0.45300000000000  -0.13270000000000    H  (16g)
  -0.04700000000000   0.01260000000000   0.63270000000000    H  (16g)
   0.54700000000000   0.48740000000000   0.63270000000000    H  (16g)
   0.01260000000000   0.54700000000000   0.63270000000000    H  (16g)
   0.48740000000000  -0.04700000000000   0.63270000000000    H  (16g)
   0.80540000000000   0.57660000000000   0.86230000000000    H  (16g)
  -0.30540000000000  -0.07660000000000   0.86230000000000    H  (16g)
  -0.07660000000000   0.80540000000000   0.86230000000000    H  (16g)
   0.57660000000000  -0.30540000000000   0.86230000000000    H  (16g)
  -0.80540000000000   1.07660000000000  -0.36230000000000    H  (16g)
   1.30540000000000  -0.57660000000000  -0.36230000000000    H  (16g)
   1.07660000000000   1.30540000000000  -0.36230000000000    H  (16g)
  -0.57660000000000  -0.80540000000000  -0.36230000000000    H  (16g)
  -0.80540000000000  -0.57660000000000  -0.86230000000000    H  (16g)
   1.30540000000000   1.07660000000000  -0.86230000000000    H  (16g)
   1.07660000000000  -0.80540000000000  -0.86230000000000    H  (16g)
  -0.57660000000000   1.30540000000000  -0.86230000000000    H  (16g)
   0.80540000000000  -0.07660000000000   1.36230000000000    H  (16g)
  -0.30540000000000   0.57660000000000   1.36230000000000    H  (16g)
  -0.07660000000000  -0.30540000000000   1.36230000000000    H  (16g)
   0.57660000000000   0.80540000000000   1.36230000000000    H  (16g)
   0.83920000000000   0.45960000000000  -0.03980000000000    H  (16g)
  -0.33920000000000   0.04040000000000  -0.03980000000000    H  (16g)
```

```
0.04040000000000    0.83920000000000   -0.03980000000000    H   (16g)
0.45960000000000   -0.33920000000000   -0.03980000000000    H   (16g)
-0.83920000000000    0.95960000000000    0.53980000000000    H   (16g)
1.33920000000000   -0.45960000000000    0.53980000000000    H   (16g)
0.95960000000000    1.33920000000000    0.53980000000000    H   (16g)
-0.45960000000000   -0.83920000000000    0.53980000000000    H   (16g)
-0.83920000000000   -0.45960000000000    0.03980000000000    H   (16g)
1.33920000000000    0.95960000000000    0.03980000000000    H   (16g)
0.95960000000000   -0.83920000000000    0.03980000000000    H   (16g)
-0.45960000000000    1.33920000000000    0.03980000000000    H   (16g)
0.83920000000000    0.04040000000000    0.46020000000000    H   (16g)
-0.33920000000000    0.45960000000000    0.46020000000000    H   (16g)
0.04040000000000   -0.33920000000000    0.46020000000000    H   (16g)
0.45960000000000    0.83920000000000    0.46020000000000    H   (16g)
0.25000000000000    0.25000000000000    0.40482000000000    O    (4c)
0.75000000000000    0.75000000000000    0.09518000000000    O    (4c)
0.75000000000000    0.75000000000000   -0.40482000000000    O    (4c)
0.25000000000000    0.25000000000000    0.90482000000000    O    (4c)
0.25000000000000    0.25000000000000    0.15904000000000    O    (4c)
0.75000000000000    0.75000000000000    0.34096000000000    O    (4c)
0.75000000000000    0.75000000000000   -0.15904000000000    O    (4c)
0.25000000000000    0.25000000000000    0.65904000000000    O    (4c)
0.84543000000000    0.47219000000000    0.12549000000000    O   (16g)
-0.34543000000000    0.02781000000000    0.12549000000000    O   (16g)
0.02781000000000    0.84543000000000    0.12549000000000    O   (16g)
0.47219000000000   -0.34543000000000    0.12549000000000    O   (16g)
-0.84543000000000    0.97219000000000    0.37451000000000    O   (16g)
1.34543000000000   -0.47219000000000    0.37451000000000    O   (16g)
0.97219000000000    1.34543000000000    0.37451000000000    O   (16g)
-0.47219000000000   -0.84543000000000    0.37451000000000    O   (16g)
-0.84543000000000   -0.47219000000000   -0.12549000000000    O   (16g)
1.34543000000000    0.97219000000000   -0.12549000000000    O   (16g)
0.97219000000000   -0.84543000000000   -0.12549000000000    O   (16g)
-0.47219000000000    1.34543000000000   -0.12549000000000    O   (16g)
0.84543000000000    0.02781000000000    0.62549000000000    O   (16g)
-0.34543000000000    0.47219000000000    0.62549000000000    O   (16g)
0.02781000000000   -0.34543000000000    0.62549000000000    O   (16g)
0.47219000000000    0.84543000000000    0.62549000000000    O   (16g)
0.84587000000000    0.47366000000000    0.87655000000000    O   (16g)
-0.34587000000000    0.02634000000000    0.87655000000000    O   (16g)
0.02634000000000    0.84587000000000    0.87655000000000    O   (16g)
0.47366000000000   -0.34587000000000    0.87655000000000    O   (16g)
-0.84587000000000    0.97366000000000   -0.37655000000000    O   (16g)
1.34587000000000   -0.47366000000000   -0.37655000000000    O   (16g)
0.97366000000000    1.34587000000000   -0.37655000000000    O   (16g)
-0.47366000000000   -0.84587000000000   -0.37655000000000    O   (16g)
-0.84587000000000   -0.47366000000000   -0.87655000000000    O   (16g)
1.34587000000000    0.97366000000000   -0.87655000000000    O   (16g)
0.97366000000000   -0.84587000000000   -0.87655000000000    O   (16g)
-0.47366000000000    1.34587000000000   -0.87655000000000    O   (16g)
0.84587000000000    0.02634000000000    1.37655000000000    O   (16g)
-0.34587000000000    0.47366000000000    1.37655000000000    O   (16g)
0.02634000000000   -0.34587000000000    1.37655000000000    O   (16g)
0.47366000000000    0.84587000000000    1.37655000000000    O   (16g)
0.75000000000000    0.25000000000000    0.25000000000000   Sr    (4a)
0.25000000000000    0.75000000000000    0.25000000000000   Sr    (4a)
0.25000000000000    0.75000000000000    0.75000000000000   Sr    (4a)
0.75000000000000    0.25000000000000    0.75000000000000   Sr    (4a)
```

$\alpha$-$WO_3$: A3B_tP16_130_cf_c - CIF

```
# CIF file
data_findsym-output
_audit_creation_method FINDSYM

_chemical_name_mineral 'O3W'
_chemical_formula_sum 'O3 W'

loop_
_publ_author_name
 'P. M. Woodward'
 'A. W. Sleight'
 'T. Vogt'
_journal_name_full_name
;
 Journal of Solid State Chemistry
;
_journal_volume 131
_journal_year 1997
_journal_page_first 9
_journal_page_last 17
_publ_Section_title
;
 Ferroelectric Tungsten Trioxide
;

_aflow_title '$\alpha$-WO$_{3}$ Structure'
_aflow_proto 'A3B_tP16_130_cf_c'
_aflow_params 'a,c/a,z_{1},z_{2},x_{3}'
_aflow_params_values '5.2759,1.48717754317,0.0028,0.2847,0.0287'
_aflow_Strukturbericht 'None'
_aflow_Pearson 'tP16'

_symmetry_space_group_name_H-M "P 4/n 21/c 2/c (origin choice 2)"
_symmetry_Int_Tables_number 130

_cell_length_a    5.27590
_cell_length_b    5.27590
_cell_length_c    7.84620
_cell_angle_alpha 90.00000
_cell_angle_beta  90.00000
_cell_angle_gamma 90.00000

loop_
_space_group_symop_id
_space_group_symop_operation_xyz
1 x,y,z
2 x+1/2,-y,-z+1/2
3 -x,y+1/2,-z+1/2
4 -x+1/2,-y+1/2,z
5 -y,-x,-z+1/2
6 -y+1/2,x,z
7 y,-x+1/2,z
8 y+1/2,x+1/2,-z+1/2
9 -x,-y,-z
10 -x+1/2,y,z+1/2
11 x,-y+1/2,z+1/2
12 x+1/2,y+1/2,-z
13 y,x,z+1/2
14 y+1/2,-x,-z
15 -y,x+1/2,-z
16 -y+1/2,-x+1/2,z+1/2

loop_
_atom_site_label
_atom_site_type_symbol
_atom_site_symmetry_multiplicity
_atom_site_Wyckoff_label
_atom_site_fract_x
_atom_site_fract_y
_atom_site_fract_z
_atom_site_occupancy
O1 O 4 c 0.25000 0.25000 0.00280 1.00000
W1 W 4 c 0.25000 0.25000 0.28470 1.00000
O2 O 8 f 0.02870 -0.02870 0.25000 1.00000
```

$\alpha$-$WO_3$: A3B_tP16_130_cf_c - POSCAR

```
A3B_tP16_130_cf_c & a,c/a,z1,z2,x3 --params=5.2759,1.48717754317,0.0028,
    ↪ 0.2847,0.0287 & P4/ncc D_{4h}^{8} #130 (c^2f) & tP16 & None &
    ↪ O3W & O3W & P. M. Woodward and A. W. Sleight and T. Vogt, J.
    ↪ Solid State Chem. 131, 9-17 (1997)
   1.00000000000000
   5.27590000000000    0.00000000000000    0.00000000000000
   0.00000000000000    5.27590000000000    0.00000000000000
   0.00000000000000    0.00000000000000    7.84620000000000
     O       W
    12       4
Direct
   0.25000000000000    0.25000000000000    0.00280000000000    O    (4c)
   0.75000000000000    0.75000000000000    0.49720000000000    O    (4c)
   0.75000000000000    0.75000000000000   -0.00280000000000    O    (4c)
   0.25000000000000    0.25000000000000    0.50280000000000    O    (4c)
   0.02870000000000   -0.02870000000000    0.25000000000000    O    (8f)
   0.47130000000000    0.52870000000000    0.25000000000000    O    (8f)
   0.52870000000000    0.02870000000000    0.25000000000000    O    (8f)
  -0.02870000000000    0.47130000000000    0.25000000000000    O    (8f)
  -0.02870000000000    0.02870000000000    0.75000000000000    O    (8f)
   0.52870000000000    0.47130000000000    0.75000000000000    O    (8f)
   0.47130000000000   -0.02870000000000    0.75000000000000    O    (8f)
   0.02870000000000    0.52870000000000    0.75000000000000    O    (8f)
   0.25000000000000    0.25000000000000    0.28470000000000    W    (4c)
   0.75000000000000    0.75000000000000    0.21530000000000    W    (4c)
   0.75000000000000    0.75000000000000   -0.28470000000000    W    (4c)
   0.25000000000000    0.25000000000000    0.78470000000000    W    (4c)
```

$Zr_3Al_2$: A2B3_tP20_136_j_dfg - CIF

```
# CIF file
data_findsym-output
_audit_creation_method FINDSYM

_chemical_name_mineral 'Al2Zr3'
_chemical_formula_sum 'Al2 Zr3'

loop_
_publ_author_name
 'C. G. Wilson'
 'F. J. Spooner'
_journal_name_full_name
;
 Acta Cristallographica
;
_journal_volume 13
_journal_year 1960
_journal_page_first 358
_journal_page_last 359
_publ_Section_title
;
 The Crystal Structure of Zr$_{3}$Al$_{2}$
;

# Found in Crystal Structure Investigations on the Zr-Al and Hf-Al
    ↪ Systems, 1962

_aflow_title 'Zr$_{3}$Al$_{2}$ Structure'
_aflow_proto 'A2B3_tP20_136_j_dfg'
_aflow_params 'a,c/a,x_{2},x_{3},x_{4},z_{4}'
_aflow_params_values '7.63,0.917169069463,0.34,0.2,0.125,0.21'
_aflow_Strukturbericht 'None'
_aflow_Pearson 'tP20'

_symmetry_space_group_name_H-M "P 42/m 21/n 2/m"
_symmetry_Int_Tables_number 136

_cell_length_a    7.63000
_cell_length_b    7.63000
_cell_length_c    6.99800
_cell_angle_alpha 90.00000
_cell_angle_beta  90.00000
```

```
_cell_angle_gamma  90.00000

loop_
_space_group_symop_id
_space_group_symop_operation_xyz
1 x,y,z
2 x+1/2,-y+1/2,-z+1/2
3 -x+1/2,y+1/2,-z+1/2
4 -x,-y,z
5 -y,-x,-z
6 -y+1/2,x+1/2,z+1/2
7 y+1/2,-x+1/2,z+1/2
8 y,x,-z
9 -x,-y,-z
10 -x+1/2,y+1/2,z+1/2
11 x+1/2,-y+1/2,z+1/2
12 x,y,-z
13 y,x,z
14 y+1/2,-x+1/2,-z+1/2
15 -y+1/2,x+1/2,-z+1/2
16 -y,-x,z

loop_
_atom_site_label
_atom_site_type_symbol
_atom_site_symmetry_multiplicity
_atom_site_Wyckoff_label
_atom_site_fract_x
_atom_site_fract_y
_atom_site_fract_z
_atom_site_occupancy
Zr1 Zr 4 d 0.00000 0.50000 0.25000 1.00000
Zr2 Zr 4 f 0.34000 0.34000 0.00000 1.00000
Zr3 Zr 4 g 0.20000 0.80000 0.00000 1.00000
Al1 Al 8 j 0.12500 0.12500 0.21000 1.00000
```

$Zr_3Al_2$: A2B3_tP20_136_j_dfg - POSCAR

```
A2B3_tP20_136_j_dfg & a,c/a,x2,x3,x4,z4 --params=7.63,0.917169069463,
    ↪ 0.34,0.2,0.125,0.21 & P4_{2}/mnm D_{4h}^{14} #136 (dfgj) & tP20
    ↪ & None & Al2Zr3 & Al2Zr3 & C. G. Wilson and F. J. Spooner,
    ↪ Acta Cryst. 13, 358-359 (1960)
   1.00000000000000
   7.63000000000000    0.00000000000000    0.00000000000000
   0.00000000000000    7.63000000000000    0.00000000000000
   0.00000000000000    0.00000000000000    6.99800000000000
    Al     Zr
     8     12
Direct
   0.12500000000000    0.12500000000000    0.21000000000000   Al   (8j)
  -0.12500000000000   -0.12500000000000    0.21000000000000   Al   (8j)
   0.37500000000000    0.62500000000000    0.71000000000000   Al   (8j)
   0.62500000000000    0.37500000000000    0.71000000000000   Al   (8j)
   0.37500000000000    0.62500000000000    0.29000000000000   Al   (8j)
   0.62500000000000    0.37500000000000    0.29000000000000   Al   (8j)
   0.12500000000000    0.12500000000000   -0.21000000000000   Al   (8j)
  -0.12500000000000   -0.12500000000000   -0.21000000000000   Al   (8j)
   0.00000000000000    0.50000000000000    0.25000000000000   Zr   (4d)
   0.00000000000000    0.50000000000000    0.75000000000000   Zr   (4d)
   0.50000000000000    0.00000000000000    0.25000000000000   Zr   (4d)
   0.50000000000000    0.00000000000000    0.75000000000000   Zr   (4d)
   0.34000000000000    0.34000000000000    0.00000000000000   Zr   (4f)
  -0.34000000000000   -0.34000000000000    0.00000000000000   Zr   (4f)
   0.16000000000000    0.84000000000000    0.50000000000000   Zr   (4f)
   0.84000000000000    0.16000000000000    0.50000000000000   Zr   (4f)
   0.20000000000000   -0.20000000000000    0.00000000000000   Zr   (4g)
  -0.20000000000000    0.20000000000000    0.00000000000000   Zr   (4g)
   0.70000000000000    0.70000000000000    0.50000000000000   Zr   (4g)
   0.30000000000000    0.30000000000000    0.50000000000000   Zr   (4g)
```

$ZrFe_4Si_2$: A4B2C_tP14_136_i_g_b - CIF

```
# CIF file
data_findsym-output
_audit_creation_method FINDSYM

_chemical_name_mineral 'Fe4Si2Zr'
_chemical_formula_sum 'Fe4 Si2 Zr'

loop_
_publ_author_name
 'Y. P. Yarmolyuk'
 'L. A. Lysenko'
 'E. I. Gladyshevsky'
_journal_name_full_name
;
 Dopov. Akad. Nauk Ukr. RSR, Ser. A
;
_journal_volume 37
_journal_year 1975
_journal_page_first 281
_journal_page_last 284
_publ_Section_title
;
 Crystal Structure of ZrFe$_{4}$Si$_{2}$ -- A New Structure Type of
    ↪ Ternary Silicides
;

# Found in Simultaneous structural and magnetic transitions in YFe$_{4}
    ↪ $Ge$_{2}$ studied by neutron diffraction and magnetic
    ↪ measurements, 2001

_aflow_title 'ZrFe$_{4}$Si$_{2}$ Structure'
_aflow_proto 'A4B2C_tP14_136_i_g_b'
_aflow_params 'a,c/a,x_{2},x_{3},y_{3}'
_aflow_params_values '7.004,0.536122215877,0.2201,0.092,0.3468'
_aflow_Strukturbericht 'None'
_aflow_Pearson 'tP14'

_symmetry_space_group_name_H-M "P 42/m 21/n 2/m"
_symmetry_Int_Tables_number 136

_cell_length_a     7.00400
_cell_length_b     7.00400
_cell_length_c     3.75500
_cell_angle_alpha  90.00000
_cell_angle_beta   90.00000
_cell_angle_gamma  90.00000

loop_
_space_group_symop_id
_space_group_symop_operation_xyz
1 x,y,z
2 x+1/2,-y+1/2,-z+1/2
3 -x+1/2,y+1/2,-z+1/2
4 -x,-y,z
5 -y,-x,-z
6 -y+1/2,x+1/2,z+1/2
7 y+1/2,-x+1/2,z+1/2
8 y,x,-z
9 -x,-y,-z
10 -x+1/2,y+1/2,z+1/2
11 x+1/2,-y+1/2,z+1/2
12 x,y,-z
13 y,x,z
14 y+1/2,-x+1/2,-z+1/2
15 -y+1/2,x+1/2,-z+1/2
16 -y,-x,z

loop_
_atom_site_label
_atom_site_type_symbol
_atom_site_symmetry_multiplicity
_atom_site_Wyckoff_label
_atom_site_fract_x
_atom_site_fract_y
_atom_site_fract_z
_atom_site_occupancy
Zr1 Zr 2 b 0.00000 0.00000 0.50000 1.00000
Si1 Si 4 g 0.22010 0.77990 0.00000 1.00000
Fe1 Fe 8 i 0.09200 0.34680 0.00000 1.00000
```

$ZrFe_4Si_2$: A4B2C_tP14_136_i_g_b - POSCAR

```
A4B2C_tP14_136_i_g_b & a,c/a,x2,x3,y3 --params=7.004,0.536122215877,
    ↪ 0.2201,0.092,0.3468 & P4_{2}/mnm D_{4h}^{14} #136 (bgi) & tP14
    ↪ & None & Fe4Si2Zr & Fe4Si2Zr & Y. P. Yarmolyuk and L. A.
    ↪ Lysenko and E. I. Gladyshevsky, Dopov. Akad. Nauk Ukr. RSR,
    ↪ Ser. A 37, 281-284 (1975)
   1.00000000000000
   7.00400000000000    0.00000000000000    0.00000000000000
   0.00000000000000    7.00400000000000    0.00000000000000
   0.00000000000000    0.00000000000000    3.75500000000000
    Fe     Si     Zr
     8      4      2
Direct
   0.09200000000000    0.34680000000000    0.00000000000000   Fe   (8i)
  -0.09200000000000   -0.34680000000000    0.00000000000000   Fe   (8i)
   0.15320000000000    0.59200000000000    0.50000000000000   Fe   (8i)
   0.84680000000000    0.40800000000000    0.50000000000000   Fe   (8i)
   0.40800000000000    0.84680000000000    0.50000000000000   Fe   (8i)
   0.59200000000000    0.15320000000000    0.50000000000000   Fe   (8i)
   0.34680000000000    0.09200000000000    0.00000000000000   Fe   (8i)
  -0.34680000000000   -0.09200000000000    0.00000000000000   Fe   (8i)
   0.22010000000000   -0.22010000000000    0.00000000000000   Si   (4g)
  -0.22010000000000    0.22010000000000    0.00000000000000   Si   (4g)
   0.72010000000000    0.72010000000000    0.50000000000000   Si   (4g)
   0.27990000000000    0.27990000000000    0.50000000000000   Si   (4g)
   0.00000000000000    0.00000000000000    0.50000000000000   Zr   (2b)
   0.50000000000000    0.50000000000000    0.00000000000000   Zr   (2b)
```

$K_2CuCl_4 \cdot 2H_2O$ ($H4_1$): A4BC4D2E2_tP26_136_fg_a_j_d_e - CIF

```
# CIF file
data_findsym-output
_audit_creation_method FINDSYM

_chemical_name_mineral 'Cl4CuH4K2O2'
_chemical_formula_sum 'Cl4 Cu H4 K2 O2'

loop_
_publ_author_name
 'R. Chidambaram'
 'Q. O. Navarro'
 'A. Garcia'
 'K. Linggoatmodjo'
 'L. {Shi-Chien}'
 'I.-H. Suh'
_journal_name_full_name
;
 Acta Crystallographica Section B: Structural Science
;
_journal_volume 26
_journal_year 1970
_journal_page_first 827
_journal_page_last 830
_publ_Section_title
;
 Neutron diffraction refinement of the crystal structure of potassium
    ↪ copper chloride dihydrate, K$_{2}$CuCl$_{4}$ $\cdot$ 2H$_{2}$O
```

```
;
_aflow_title 'K$_{2}$CuCl$_{4}$$\cdot$2H$_{2}$O ($H4_{1}$) Structure'
_aflow_proto 'A4BC4D2E2_tP26_136_fg_a_j_d_e'
_aflow_params 'a,c/a,z_{3},x_{4},x_{5},x_{6},z_{6}'
_aflow_params_values '7.477,1.06125451384,0.2484,0.2161,0.7262,0.0739,
    ↪ 0.3178'
_aflow_Strukturbericht '$H4_{1}$'
_aflow_Pearson 'tP26'

_symmetry_space_group_name_H-M "P 42/m 21/n 2/m"
_symmetry_Int_Tables_number 136

_cell_length_a      7.47700
_cell_length_b      7.47700
_cell_length_c      7.93500
_cell_angle_alpha 90.00000
_cell_angle_beta  90.00000
_cell_angle_gamma 90.00000

loop_
_space_group_symop_id
_space_group_symop_operation_xyz
1 x,y,z
2 x+1/2,-y+1/2,-z+1/2
3 -x+1/2,y+1/2,-z+1/2
4 -x,-y,z
5 -y,-x,-z
6 -y+1/2,x+1/2,z+1/2
7 y+1/2,-x+1/2,z+1/2
8 y,x,-z
9 -x,-y,-z
10 -x+1/2,y+1/2,z+1/2
11 x+1/2,-y+1/2,z+1/2
12 x,y,-z
13 y,x,z
14 y+1/2,-x+1/2,-z+1/2
15 -y+1/2,x+1/2,-z+1/2
16 -y,-x,z

loop_
_atom_site_label
_atom_site_type_symbol
_atom_site_symmetry_multiplicity
_atom_site_Wyckoff_label
_atom_site_fract_x
_atom_site_fract_y
_atom_site_fract_z
_atom_site_occupancy
Cu1 Cu 2 a 0.00000 0.00000 0.00000 1.00000
K1 K 4 d 0.00000 0.50000 0.25000 1.00000
O1 O 4 e 0.00000 0.00000 0.24840 1.00000
Cl1 Cl 4 f 0.21610 0.21610 0.00000 1.00000
Cl2 Cl 4 g 0.72620 0.27380 0.00000 1.00000
H1 H 8 j 0.07390 0.07390 0.31780 1.00000
```

$K_2CuCl_4 \cdot 2H_2O$ ($H4_1$): A4BC4D2E2_tP26_136_fg_a_j_d_e - POSCAR

```
A4BC4D2E2_tP26_136_fg_a_j_d_e & a,c/a,z3,x4,x5,x6,z6 --params=7.477,
    ↪ 1.06125451384,0.2484,0.2161,0.7262,0.0739,0.3178 & P4_{2}/mnm
    ↪ D_{4h}^{14} #136 (adefgj) & tP26 & $H4_{1}$ & Cl4CuH4K2O2 &
    ↪ Cl4CuH4K2O2 & R. Chidambaram et al., Acta Crystallogr. Sect. B
    ↪ Struct. Sci. 26, 827-830 (1970)
   1.00000000000000
   7.47700000000000   0.00000000000000   0.00000000000000
   0.00000000000000   7.47700000000000   0.00000000000000
   0.00000000000000   0.00000000000000   7.93500000000000
     Cl    Cu     H     K     O
      8     2     8     4     4
Direct
   0.21610000000000   0.21610000000000   0.00000000000000   Cl   (4f)
  -0.21610000000000  -0.21610000000000   0.00000000000000   Cl   (4f)
   0.28390000000000   0.71610000000000   0.50000000000000   Cl   (4f)
   0.71610000000000   0.28390000000000   0.50000000000000   Cl   (4f)
   0.72620000000000  -0.72620000000000   0.00000000000000   Cl   (4g)
  -0.72620000000000   0.72620000000000   0.00000000000000   Cl   (4g)
   1.22620000000000   1.22620000000000   0.50000000000000   Cl   (4g)
  -0.22620000000000  -0.22620000000000   0.50000000000000   Cl   (4g)
   0.00000000000000   0.00000000000000   0.00000000000000   Cu   (2a)
   0.50000000000000   0.50000000000000   0.50000000000000   Cu   (2a)
   0.07390000000000   0.07390000000000   0.31780000000000    H   (8j)
  -0.07390000000000  -0.07390000000000   0.31780000000000    H   (8j)
   0.42610000000000   0.57390000000000   0.81780000000000    H   (8j)
   0.57390000000000   0.42610000000000   0.81780000000000    H   (8j)
   0.42610000000000   0.57390000000000   0.18220000000000    H   (8j)
   0.57390000000000   0.42610000000000   0.18220000000000    H   (8j)
   0.07390000000000   0.07390000000000  -0.31780000000000    H   (8j)
  -0.07390000000000  -0.07390000000000  -0.31780000000000    H   (8j)
   0.00000000000000   0.50000000000000   0.25000000000000    K   (4d)
   0.00000000000000   0.50000000000000   0.75000000000000    K   (4d)
   0.50000000000000   0.00000000000000   0.25000000000000    K   (4d)
   0.50000000000000   0.00000000000000   0.75000000000000    K   (4d)
   0.00000000000000   0.00000000000000   0.24840000000000    O   (4e)
   0.50000000000000   0.50000000000000   0.74840000000000    O   (4e)
   0.50000000000000   0.50000000000000   0.25160000000000    O   (4e)
   0.00000000000000   0.00000000000000  -0.24840000000000    O   (4e)
```

$Nd_2Fe_{14}B$: AB14C2_tP68_136_f_ce2j2k_fg - CIF

```
# CIF file
data_findsym-output
_audit_creation_method FINDSYM

_chemical_name_mineral 'BFe14Nd2'
_chemical_formula_sum 'B Fe14 Nd2'

loop_
_publ_author_name
 'D. Givord'
 'H. S. Li'
 'J. M. Moreau'
_journal_name_full_name
;
 Solid State Communications
;
_journal_volume 50
_journal_year 1984
_journal_page_first 497
_journal_page_last 499
_publ_Section_title
;
 Magnetic properties and crystal structure of Nd$_{2}$Fe$_{14}$B
;

_aflow_title 'Nd$_{2}$Fe$_{14}$B Structure'
_aflow_proto 'AB14C2_tP68_136_f_ce2j2k_fg'
_aflow_params 'a,c/a,z_{2},x_{3},x_{4},x_{5},x_{6},z_{6},x_{7},z_{7},x_{
    ↪ 8},y_{8},z_{8},x_{9},y_{9},z_{9}'
_aflow_params_values '8.792,1.38648771611,0.116,0.124,0.3572,0.7698,
    ↪ 0.0978,0.2942,0.3184,0.255,0.567,0.2245,0.3735,0.1397,0.537,
    ↪ 0.1759'
_aflow_Strukturbericht 'None'
_aflow_Pearson 'tP68'

_symmetry_space_group_name_H-M "P 42/m 21/n 2/m"
_symmetry_Int_Tables_number 136

_cell_length_a      8.79200
_cell_length_b      8.79200
_cell_length_c      12.19000
_cell_angle_alpha 90.00000
_cell_angle_beta  90.00000
_cell_angle_gamma 90.00000

loop_
_space_group_symop_id
_space_group_symop_operation_xyz
1 x,y,z
2 x+1/2,-y+1/2,-z+1/2
3 -x+1/2,y+1/2,-z+1/2
4 -x,-y,z
5 -y,-x,-z
6 -y+1/2,x+1/2,z+1/2
7 y+1/2,-x+1/2,z+1/2
8 y,x,-z
9 -x,-y,-z
10 -x+1/2,y+1/2,z+1/2
11 x+1/2,-y+1/2,z+1/2
12 x,y,-z
13 y,x,z
14 y+1/2,-x+1/2,-z+1/2
15 -y+1/2,x+1/2,-z+1/2
16 -y,-x,z

loop_
_atom_site_label
_atom_site_type_symbol
_atom_site_symmetry_multiplicity
_atom_site_Wyckoff_label
_atom_site_fract_x
_atom_site_fract_y
_atom_site_fract_z
_atom_site_occupancy
Fe1 Fe 4 c 0.00000 0.50000 0.00000 1.00000
Fe2 Fe 4 e 0.00000 0.00000 0.11600 1.00000
B1 B 4 f 0.12400 0.12400 0.00000 1.00000
Nd1 Nd 4 f 0.35720 0.35720 0.00000 1.00000
Nd2 Nd 4 g 0.76980 0.23020 0.00000 1.00000
Fe3 Fe 8 j 0.09780 0.09780 0.29420 1.00000
Fe4 Fe 8 j 0.31840 0.31840 0.25500 1.00000
Fe5 Fe 16 k 0.56700 0.22450 0.37350 1.00000
Fe6 Fe 16 k 0.13970 0.53700 0.17590 1.00000
```

$Nd_2Fe_{14}B$: AB14C2_tP68_136_f_ce2j2k_fg - POSCAR

```
AB14C2_tP68_136_f_ce2j2k_fg & a,c/a,z2,x3,x4,x5,x6,z6,x7,z7,x8,y8,z8,x9,
    ↪ y9,z9 --params=8.792,1.38648771611,0.116,0.124,0.3572,0.7698,
    ↪ 0.0978,0.2942,0.3184,0.255,0.567,0.2245,0.3735,0.1397,0.537,
    ↪ 0.1759 & P4_{2}/mnm D_{4h}^{14} #136 (cef^2gj^2k^2) & tP68 &
    ↪ None & BFe14Nd2 & BFe14Nd2 & D. Givord and H. S. Li and J. M.
    ↪ Moreau, Solid State Commun. 50, 497-499 (1984)
   1.00000000000000
   8.79200000000000   0.00000000000000   0.00000000000000
   0.00000000000000   8.79200000000000   0.00000000000000
   0.00000000000000   0.00000000000000  12.19000000000000
      B    Fe    Nd
      4    56     8
Direct
   0.12400000000000   0.12400000000000   0.00000000000000    B   (4f)
  -0.12400000000000  -0.12400000000000   0.00000000000000    B   (4f)
   0.37600000000000   0.62400000000000   0.50000000000000    B   (4f)
   0.62400000000000   0.37600000000000   0.50000000000000    B   (4f)
   0.00000000000000   0.50000000000000   0.00000000000000   Fe   (4c)
   0.00000000000000   0.50000000000000   0.50000000000000   Fe   (4c)
   0.50000000000000   0.00000000000000   0.50000000000000   Fe   (4c)
   0.50000000000000   0.00000000000000   0.00000000000000   Fe   (4c)
   0.00000000000000   0.00000000000000   0.11600000000000   Fe   (4e)
   0.50000000000000   0.50000000000000   0.61600000000000   Fe   (4e)
   0.50000000000000   0.50000000000000   0.38400000000000   Fe   (4e)
   0.00000000000000   0.00000000000000  -0.11600000000000   Fe   (4e)
```

```
   0.09780000000000    0.09780000000000    0.29420000000000    Fe    (8j)
  -0.09780000000000   -0.09780000000000    0.29420000000000    Fe    (8j)
   0.40220000000000    0.59780000000000    0.79420000000000    Fe    (8j)
   0.59780000000000    0.40220000000000    0.79420000000000    Fe    (8j)
   0.40220000000000    0.59780000000000    0.20580000000000    Fe    (8j)
   0.59780000000000    0.40220000000000    0.20580000000000    Fe    (8j)
   0.09780000000000    0.09780000000000   -0.29420000000000    Fe    (8j)
  -0.09780000000000   -0.09780000000000   -0.29420000000000    Fe    (8j)
   0.31840000000000    0.31840000000000    0.25500000000000    Fe    (8j)
  -0.31840000000000   -0.31840000000000    0.25500000000000    Fe    (8j)
   0.18160000000000    0.81840000000000    0.75500000000000    Fe    (8j)
   0.81840000000000    0.18160000000000    0.75500000000000    Fe    (8j)
   0.18160000000000    0.81840000000000    0.24500000000000    Fe    (8j)
   0.81840000000000    0.18160000000000    0.24500000000000    Fe    (8j)
   0.31840000000000    0.31840000000000   -0.25500000000000    Fe    (8j)
  -0.31840000000000   -0.31840000000000   -0.25500000000000    Fe    (8j)
   0.56700000000000    0.22450000000000    0.37350000000000    Fe   (16k)
  -0.56700000000000   -0.22450000000000    0.37350000000000    Fe   (16k)
   0.27550000000000    1.06700000000000    0.87350000000000    Fe   (16k)
   0.72450000000000   -0.06700000000000    0.87350000000000    Fe   (16k)
  -0.06700000000000    0.72450000000000    0.12650000000000    Fe   (16k)
   1.06700000000000    0.27550000000000    0.12650000000000    Fe   (16k)
   0.22450000000000    0.56700000000000   -0.37350000000000    Fe   (16k)
  -0.22450000000000   -0.56700000000000   -0.37350000000000    Fe   (16k)
  -0.56700000000000   -0.22450000000000   -0.37350000000000    Fe   (16k)
   0.56700000000000    0.22450000000000   -0.37350000000000    Fe   (16k)
   0.72450000000000   -0.06700000000000    0.12650000000000    Fe   (16k)
   0.27550000000000    1.06700000000000    0.12650000000000    Fe   (16k)
   1.06700000000000    0.27550000000000    0.87350000000000    Fe   (16k)
  -0.06700000000000    0.72450000000000    0.87350000000000    Fe   (16k)
  -0.22450000000000   -0.56700000000000    0.37350000000000    Fe   (16k)
   0.22450000000000    0.56700000000000    0.37350000000000    Fe   (16k)
   0.13970000000000    0.53700000000000    0.17590000000000    Fe   (16k)
  -0.13970000000000   -0.53700000000000    0.17590000000000    Fe   (16k)
  -0.03700000000000    0.63970000000000    0.67590000000000    Fe   (16k)
   1.03700000000000    0.36030000000000    0.67590000000000    Fe   (16k)
   0.36030000000000    1.03700000000000    0.32410000000000    Fe   (16k)
   0.63970000000000   -0.03700000000000    0.32410000000000    Fe   (16k)
   0.53700000000000    0.13970000000000   -0.17590000000000    Fe   (16k)
  -0.53700000000000   -0.13970000000000   -0.17590000000000    Fe   (16k)
  -0.13970000000000   -0.53700000000000   -0.17590000000000    Fe   (16k)
   0.13970000000000    0.53700000000000   -0.17590000000000    Fe   (16k)
   1.03700000000000    0.36030000000000    0.32410000000000    Fe   (16k)
  -0.03700000000000    0.63970000000000    0.32410000000000    Fe   (16k)
   0.63970000000000   -0.03700000000000    0.67590000000000    Fe   (16k)
   0.36030000000000    1.03700000000000    0.67590000000000    Fe   (16k)
  -0.53700000000000   -0.13970000000000    0.17590000000000    Fe   (16k)
   0.53700000000000    0.13970000000000    0.17590000000000    Fe   (16k)
   0.35720000000000    0.35720000000000    0.00000000000000    Nd    (4f)
  -0.35720000000000   -0.35720000000000    0.00000000000000    Nd    (4f)
   0.14280000000000    0.85720000000000    0.50000000000000    Nd    (4f)
   0.85720000000000    0.14280000000000    0.50000000000000    Nd    (4f)
   0.76980000000000   -0.76980000000000    0.00000000000000    Nd    (4g)
  -0.76980000000000    0.76980000000000    0.00000000000000    Nd    (4g)
   1.26980000000000    1.26980000000000    0.50000000000000    Nd    (4g)
  -0.26980000000000   -0.26980000000000    0.50000000000000    Nd    (4g)
```

$Sr_4Ti_3O_{10}$: A10B4C3_tI34_139_c2eg_2e_ae - CIF

```
# CIF file
data_findsym-output
_audit_creation_method FINDSYM

_chemical_name_mineral 'O10Sr4Ti3'
_chemical_formula_sum 'O10 Sr4 Ti3'

loop_
_publ_author_name
 'S. N. Ruddlesden'
 'P. Popper'
_journal_name_full_name
;
 Acta Cristallographica
;
_journal_volume 11
_journal_year 1958
_journal_page_first 54
_journal_page_last 55
_publ_Section_title
;
 The compound Sr$_{3}$Ti$_{2}$O$_{7}$ and its structure
;

# Found in Ruddlesden-Popper phase, {A$_{3}$B$_{2}$X$_{7}$ series},

_aflow_title 'Sr$_{4}$Ti$_{3}$O$_{10}$ Structure'
_aflow_proto 'A10B4C3_tI34_139_c2eg_2e_ae'
_aflow_params 'a,c/a,z_{3},z_{4},z_{5},z_{6},z_{7},z_{8}'
_aflow_params_values '3.9,7.20512820513,0.068,0.204,0.432,0.296,0.136,
    ↪ 0.136'
_aflow_Strukturbericht 'None'
_aflow_Pearson 'tI34'

_symmetry_space_group_name_H-M "I 4/m 2/m 2/m"
_symmetry_Int_Tables_number 139

_cell_length_a      3.90000
_cell_length_b      3.90000
_cell_length_c      28.10000
_cell_angle_alpha   90.00000
_cell_angle_beta    90.00000
_cell_angle_gamma   90.00000

loop_
_space_group_symop_id
_space_group_symop_operation_xyz
1 x,y,z
2 x,-y,-z
3 -x,y,-z
4 -x,-y,z
5 -y,-x,-z
6 -y,x,z
7 y,-x,z
8 y,x,-z
9 -x,-y,-z
10 -x,y,z
11 x,-y,z
12 x,y,-z
13 y,x,z
14 y,-x,-z
15 -y,x,-z
16 -y,-x,z
17 x+1/2,y+1/2,z+1/2
18 x+1/2,-y+1/2,-z+1/2
19 -x+1/2,y+1/2,-z+1/2
20 -x+1/2,-y+1/2,z+1/2
21 -y+1/2,-x+1/2,-z+1/2
22 -y+1/2,x+1/2,z+1/2
23 y+1/2,-x+1/2,z+1/2
24 y+1/2,x+1/2,-z+1/2
25 -x+1/2,-y+1/2,-z+1/2
26 -x+1/2,y+1/2,z+1/2
27 x+1/2,-y+1/2,z+1/2
28 x+1/2,y+1/2,-z+1/2
29 y+1/2,x+1/2,z+1/2
30 y+1/2,-x+1/2,-z+1/2
31 -y+1/2,x+1/2,-z+1/2
32 -y+1/2,-x+1/2,z+1/2

loop_
_atom_site_label
_atom_site_type_symbol
_atom_site_symmetry_multiplicity
_atom_site_Wyckoff_label
_atom_site_fract_x
_atom_site_fract_y
_atom_site_fract_z
_atom_site_occupancy
Ti1 Ti 2 a 0.00000 0.00000 0.00000 1.00000
O1 O 4 c 0.00000 0.50000 0.00000 1.00000
O2 O 4 e 0.00000 0.00000 0.06800 1.00000
O3 O 4 e 0.00000 0.00000 0.20400 1.00000
Sr1 Sr 4 e 0.00000 0.00000 0.43200 1.00000
Sr2 Sr 4 e 0.00000 0.00000 0.29600 1.00000
Ti2 Ti 4 e 0.00000 0.00000 0.13600 1.00000
O4 O 8 g 0.00000 0.50000 0.13600 1.00000
```

$Sr_4Ti_3O_{10}$: A10B4C3_tI34_139_c2eg_2e_ae - POSCAR

```
A10B4C3_tI34_139_c2eg_2e_ae & a,c/a,z3,z4,z5,z6,z7,z8 --params=3.9,
    ↪ 7.20512820513,0.068,0.204,0.432,0.296,0.136,0.136 & I4/mmm D_{
    ↪ 4h}^{17} #139 (ace^5g) & tI34 & None & O10Sr4Ti3 & O10Sr4Ti3 &
    ↪ S. N. Ruddlesden and P. Popper, Acta Cryst. 11, 54-55 (1958)
   1.00000000000000
  -1.95000000000000    1.95000000000000   14.05000000000000
   1.95000000000000   -1.95000000000000   14.05000000000000
   1.95000000000000    1.95000000000000  -14.05000000000000
     O     Sr     Ti
    10      4      3
Direct
   0.50000000000000    0.00000000000000    0.50000000000000    O    (4c)
   0.00000000000000    0.50000000000000    0.50000000000000    O    (4c)
   0.06800000000000    0.06800000000000    0.00000000000000    O    (4e)
  -0.06800000000000   -0.06800000000000    0.00000000000000    O    (4e)
   0.20400000000000    0.20400000000000    0.00000000000000    O    (4e)
  -0.20400000000000   -0.20400000000000    0.00000000000000    O    (4e)
   0.63600000000000    0.13600000000000    0.50000000000000    O    (8g)
   0.13600000000000    0.63600000000000    0.50000000000000    O    (8g)
   0.36400000000000   -0.13600000000000    0.50000000000000    O    (8g)
  -0.13600000000000    0.36400000000000    0.50000000000000    O    (8g)
   0.43200000000000    0.43200000000000    0.00000000000000    Sr   (4e)
  -0.43200000000000   -0.43200000000000    0.00000000000000    Sr   (4e)
   0.29600000000000    0.29600000000000    0.00000000000000    Sr   (4e)
  -0.29600000000000   -0.29600000000000    0.00000000000000    Sr   (4e)
   0.00000000000000    0.00000000000000    0.00000000000000    Ti   (2a)
   0.13600000000000    0.13600000000000    0.00000000000000    Ti   (4e)
  -0.13600000000000   -0.13600000000000    0.00000000000000    Ti   (4e)
```

$TlCo_2S_2$: A2B2C_tI10_139_d_e_a - CIF

```
# CIF file
data_findsym-output
_audit_creation_method FINDSYM

_chemical_name_mineral 'Co2S2Tl'
_chemical_formula_sum 'Co2 S2 Tl'

loop_
_publ_author_name
 'K. Klepp'
 'H. Boller'
_journal_name_full_name
;
 Monatshefte f{\"u}r Chemie - Chemical Monthly
;
_journal_volume 109
_journal_year 1978
_journal_page_first 1049
_journal_page_last 1057
_publ_Section_title
```

```
;
 Tern\"{a}re Thallium-\"{U}bergangsmetall-Chalkogenide mit ThCr$_{2}
      ↪ $Si$_{2}$-Struktur
;

# Found in TlCu$_{2}$Se$_{2}$: A p-type metal with a layer structure,
      ↪ 1984

_aflow_title 'TlCo$_{2}$S$_{2}$ Structure'
_aflow_proto 'A2B2C_tI10_139_d_e_a'
_aflow_params 'a,c/a,z_{3}'
_aflow_params_values '3.741,3.46324779471,0.3474'
_aflow_Strukturbericht 'None'
_aflow_Pearson 'tI10'

_symmetry_space_group_name_H-M "I 4/m 2/m 2/m"
_symmetry_Int_Tables_number 139

_cell_length_a      3.74100
_cell_length_b      3.74100
_cell_length_c      12.95601
_cell_angle_alpha   90.00000
_cell_angle_beta    90.00000
_cell_angle_gamma   90.00000

loop_
_space_group_symop_id
_space_group_symop_operation_xyz
1 x,y,z
2 x,-y,-z
3 -x,y,-z
4 -x,-y,z
5 -y,-x,-z
6 -y,x,z
7 y,-x,z
8 y,x,-z
9 -x,-y,-z
10 -x,y,z
11 x,-y,z
12 x,y,-z
13 y,x,z
14 y,-x,-z
15 -y,x,-z
16 -y,-x,z
17 x+1/2,y+1/2,z+1/2
18 x+1/2,-y+1/2,-z+1/2
19 -x+1/2,y+1/2,-z+1/2
20 -x+1/2,-y+1/2,z+1/2
21 -y+1/2,-x+1/2,-z+1/2
22 -y+1/2,x+1/2,z+1/2
23 y+1/2,-x+1/2,z+1/2
24 y+1/2,x+1/2,-z+1/2
25 -x+1/2,-y+1/2,-z+1/2
26 -x+1/2,y+1/2,z+1/2
27 x+1/2,-y+1/2,z+1/2
28 x+1/2,y+1/2,-z+1/2
29 y+1/2,x+1/2,z+1/2
30 y+1/2,-x+1/2,-z+1/2
31 -y+1/2,x+1/2,-z+1/2
32 -y+1/2,-x+1/2,z+1/2

loop_
_atom_site_label
_atom_site_type_symbol
_atom_site_symmetry_multiplicity
_atom_site_Wyckoff_label
_atom_site_fract_x
_atom_site_fract_y
_atom_site_fract_z
_atom_site_occupancy
Tl1 Tl 2 a 0.00000 0.00000 0.00000 1.00000
Co1 Co 4 d 0.00000 0.50000 0.25000 1.00000
S1 S 4 e 0.00000 0.00000 0.34740 1.00000
```

$TlCo_2S_2$: A2B2C_tI10_139_d_e_a - POSCAR

```
A2B2C_tI10_139_d_e_a & a,c/a,z3 --params=3.741,3.46324779471,0.3474 & I4
      ↪ /mmm D_{4h}^{17} #139 (ade) & tI10 & None & Co2S2Tl & Co2S2Tl &
      ↪ K. Klepp and H. Boller, Monatshefte f{"u}r Chemie - Chemical
      ↪ Monthly 109, 1049-1057 (1978)
   1.00000000000000
  -1.87050000000000   1.87050000000000   6.47800500000000
   1.87050000000000  -1.87050000000000   6.47800500000000
   1.87050000000000   1.87050000000000  -6.47800500000000
    Co     S     Tl
     2     2      1
Direct
   0.75000000000000   0.25000000000000   0.50000000000000   Co   (4d)
   0.25000000000000   0.75000000000000   0.50000000000000   Co   (4d)
   0.34740000000000   0.34740000000000   0.00000000000000    S   (4e)
  -0.34740000000000  -0.34740000000000   0.00000000000000    S   (4e)
   0.00000000000000   0.00000000000000   0.00000000000000   Tl   (2a)
```

$ThCr_2Si_2$: A2B2C_tI10_139_d_e_a - CIF

```
# CIF file
data_findsym-output
_audit_creation_method FINDSYM

_chemical_name_mineral 'Cr2Si2Th'
_chemical_formula_sum 'Cr2 Si2 Th'

loop_
_publ_author_name
 'Z. Ban'
 'M. Sikirica'
_journal_name_full_name
;
 Acta Cristallographica
;
_journal_volume 18
_journal_year 1965
_journal_page_first 594
_journal_page_last 599
_publ_Section_title
;
 The crystal structure of ternary silicides Th$M$$_{2}$Si$_{2}$ ($M$ =
      ↪ Cr, Mn, Fe, Co, Ni and Cu)
;

_aflow_title 'ThCr$_{2}$Si$_{2}$ Structure'
_aflow_proto 'A2B2C_tI10_139_d_e_a'
_aflow_params 'a,c/a,z_{3}'
_aflow_params_values '4.043,2.61612663863,0.374'
_aflow_Strukturbericht 'None'
_aflow_Pearson 'tI10'

_symmetry_space_group_name_H-M "I 4/m 2/m 2/m"
_symmetry_Int_Tables_number 139

_cell_length_a      4.04300
_cell_length_b      4.04300
_cell_length_c      10.57700
_cell_angle_alpha   90.00000
_cell_angle_beta    90.00000
_cell_angle_gamma   90.00000

loop_
_space_group_symop_id
_space_group_symop_operation_xyz
1 x,y,z
2 x,-y,-z
3 -x,y,-z
4 -x,-y,z
5 -y,-x,-z
6 -y,x,z
7 y,-x,z
8 y,x,-z
9 -x,-y,-z
10 -x,y,z
11 x,-y,z
12 x,y,-z
13 y,x,z
14 y,-x,-z
15 -y,x,-z
16 -y,-x,z
17 x+1/2,y+1/2,z+1/2
18 x+1/2,-y+1/2,-z+1/2
19 -x+1/2,y+1/2,-z+1/2
20 -x+1/2,-y+1/2,z+1/2
21 -y+1/2,-x+1/2,-z+1/2
22 -y+1/2,x+1/2,z+1/2
23 y+1/2,-x+1/2,z+1/2
24 y+1/2,x+1/2,-z+1/2
25 -x+1/2,-y+1/2,-z+1/2
26 -x+1/2,y+1/2,z+1/2
27 x+1/2,-y+1/2,z+1/2
28 x+1/2,y+1/2,-z+1/2
29 y+1/2,x+1/2,z+1/2
30 y+1/2,-x+1/2,-z+1/2
31 -y+1/2,x+1/2,-z+1/2
32 -y+1/2,-x+1/2,z+1/2

loop_
_atom_site_label
_atom_site_type_symbol
_atom_site_symmetry_multiplicity
_atom_site_Wyckoff_label
_atom_site_fract_x
_atom_site_fract_y
_atom_site_fract_z
_atom_site_occupancy
Th1 Th 2 a 0.00000 0.00000 0.00000 1.00000
Cr1 Cr 4 d 0.00000 0.50000 0.25000 1.00000
Si1 Si 4 e 0.00000 0.00000 0.37400 1.00000
```

$ThCr_2Si_2$: A2B2C_tI10_139_d_e_a - POSCAR

```
A2B2C_tI10_139_d_e_a & a,c/a,z3 --params=4.043,2.61612663863,0.374 & I4/
      ↪ mmm D_{4h}^{17} #139 (ade) & tI10 & None & Cr2Si2Th & Cr2Si2Th
      ↪ & Z. Ban and M. Sikirica, Acta Cryst. 18, 594-599 (1965)
   1.00000000000000
  -2.02150000000000   2.02150000000000   5.28850000000000
   2.02150000000000  -2.02150000000000   5.28850000000000
   2.02150000000000   2.02150000000000  -5.28850000000000
    Cr    Si    Th
     2     2     1
Direct
   0.75000000000000   0.25000000000000   0.50000000000000   Cr   (4d)
   0.25000000000000   0.75000000000000   0.50000000000000   Cr   (4d)
   0.37400000000000   0.37400000000000   0.00000000000000   Si   (4e)
  -0.37400000000000  -0.37400000000000   0.00000000000000   Si   (4e)
   0.00000000000000   0.00000000000000   0.00000000000000   Th   (2a)
```

$Au_2Nb_3$: A2B3_tI10_139_e_ae - CIF

```
# CIF file
data_findsym-output
_audit_creation_method FINDSYM
```

```
_chemical_name_mineral 'Au2Nb3'
_chemical_formula_sum 'Au2 Nb3'

loop_
_publ_author_name
 'K. Schubert'
 'T. R. Anantharaman'
 'H. O. K. Ata'
 'H. G. Meissner'
 'M. P\"{o}tzschke'
 'W. Rossteutscher'
 'E. Stolz'
_journal_name_full_name
;
 Naturwissenschaften
;
_journal_volume 47
_journal_year 1960
_journal_page_first 512
_journal_page_last 512
_publ_Section_title
;
 Einige strukturelle Ergebnisse an metallischen Phasen (6)
;

# Found in The Crystal Structures of Os$_{2}$Al$_{3}$ and OsAl$_{2}$,
    ↪ 1965

_aflow_title 'Au$_{2}$Nb$_{3}$ Structure'
_aflow_proto 'A2B3_tI10_139_e_ae'
_aflow_params 'a,c/a,z_{2},z_{3}'
_aflow_params_values '3.36301,4.49555309083,0.2,0.4'
_aflow_Strukturbericht 'None'
_aflow_Pearson 'tI10'

_symmetry_space_group_name_H-M "I 4/m 2/m 2/m"
_symmetry_Int_Tables_number 139

_cell_length_a     3.36301
_cell_length_b     3.36301
_cell_length_c     15.11859
_cell_angle_alpha  90.00000
_cell_angle_beta   90.00000
_cell_angle_gamma  90.00000

loop_
_space_group_symop_id
_space_group_symop_operation_xyz
1 x,y,z
2 x,-y,-z
3 -x,y,-z
4 -x,-y,z
5 -y,-x,-z
6 -y,x,z
7 y,-x,z
8 y,x,-z
9 -x,-y,-z
10 -x,y,z
11 x,-y,z
12 x,y,-z
13 y,x,z
14 y,-x,-z
15 -y,x,-z
16 -y,-x,z
17 x+1/2,y+1/2,z+1/2
18 x+1/2,-y+1/2,-z+1/2
19 -x+1/2,y+1/2,-z+1/2
20 -x+1/2,-y+1/2,z+1/2
21 -y+1/2,-x+1/2,-z+1/2
22 -y+1/2,x+1/2,z+1/2
23 y+1/2,-x+1/2,z+1/2
24 y+1/2,x+1/2,-z+1/2
25 -x+1/2,-y+1/2,-z+1/2
26 -x+1/2,y+1/2,z+1/2
27 x+1/2,-y+1/2,z+1/2
28 x+1/2,y+1/2,-z+1/2
29 y+1/2,x+1/2,z+1/2
30 y+1/2,-x+1/2,-z+1/2
31 -y+1/2,x+1/2,-z+1/2
32 -y+1/2,-x+1/2,z+1/2

loop_
_atom_site_label
_atom_site_type_symbol
_atom_site_symmetry_multiplicity
_atom_site_Wyckoff_label
_atom_site_fract_x
_atom_site_fract_y
_atom_site_fract_z
_atom_site_occupancy
Nb1 Nb 2 a 0.00000 0.00000 0.00000 1.00000
Au1 Au 4 e 0.00000 0.00000 0.20000 1.00000
Nb2 Nb 4 e 0.00000 0.00000 0.40000 1.00000
```

$Au_2Nb_3$: A2B3_tI10_139_e_ae - POSCAR

```
A2B3_tI10_139_e_ae & a,c/a,z2,z3 --params=3.36301,4.49555309083,0.2,0.4
    ↪ & I4/mmm D_{4h}^{17} #139 (ae^2) & tI10 & None & Au2Nb3 &
    ↪ Au2Nb3 & K. Schubert et al., Naturwissenschaften 47, 512(1960)
   1.00000000000000
  -1.68150500000000   1.68150500000000   7.55929500000000
   1.68150500000000  -1.68150500000000   7.55929500000000
   1.68150500000000   1.68150500000000  -7.55929500000000
    Au     Nb
     2      3
Direct
   0.20000000000000   0.20000000000000   0.00000000000000   Au   (4e)
  -0.20000000000000  -0.20000000000000   0.00000000000000   Au   (4e)
   0.00000000000000   0.00000000000000   0.00000000000000   Nb   (2a)
   0.40000000000000   0.40000000000000   0.00000000000000   Nb   (4e)
  -0.40000000000000  -0.40000000000000   0.00000000000000   Nb   (4e)
```

$CaC_2$-I ($C11_a$): A2B_tI6_139_e_a - CIF

```
# CIF file
data_findsym-output
_audit_creation_method FINDSYM

_chemical_name_mineral 'C2Ca'
_chemical_formula_sum 'C2 Ca'

loop_
_publ_author_name
 'M. {von Stackelberg}'
_journal_name_full_name
;
 Naturwissenschaften
;
_journal_volume 18
_journal_year 1930
_journal_page_first 305
_journal_page_last 306
_publ_Section_title
;
 Die Krystallstruktur des CaC$_{2}$
;

# Found in The Crystal Structure of Calcium Carbide III, 1961

_aflow_title 'CaC$_{2}$-I ($C11_{a}$) Structure'
_aflow_proto 'A2B_tI6_139_e_a'
_aflow_params 'a,c/a,z_{2}'
_aflow_params_values '3.87,1.64599483204,0.38'
_aflow_Strukturbericht '$C11_{a}$'
_aflow_Pearson 'tI6'

_symmetry_space_group_name_H-M "I 4/m 2/m 2/m"
_symmetry_Int_Tables_number 139

_cell_length_a     3.87000
_cell_length_b     3.87000
_cell_length_c     6.37000
_cell_angle_alpha  90.00000
_cell_angle_beta   90.00000
_cell_angle_gamma  90.00000

loop_
_space_group_symop_id
_space_group_symop_operation_xyz
1 x,y,z
2 x,-y,-z
3 -x,y,-z
4 -x,-y,z
5 -y,-x,-z
6 -y,x,z
7 y,-x,z
8 y,x,-z
9 -x,-y,-z
10 -x,y,z
11 x,-y,z
12 x,y,-z
13 y,x,z
14 y,-x,-z
15 -y,x,-z
16 -y,-x,z
17 x+1/2,y+1/2,z+1/2
18 x+1/2,-y+1/2,-z+1/2
19 -x+1/2,y+1/2,-z+1/2
20 -x+1/2,-y+1/2,z+1/2
21 -y+1/2,-x+1/2,-z+1/2
22 -y+1/2,x+1/2,z+1/2
23 y+1/2,-x+1/2,z+1/2
24 y+1/2,x+1/2,-z+1/2
25 -x+1/2,-y+1/2,-z+1/2
26 -x+1/2,y+1/2,z+1/2
27 x+1/2,-y+1/2,z+1/2
28 x+1/2,y+1/2,-z+1/2
29 y+1/2,x+1/2,z+1/2
30 y+1/2,-x+1/2,-z+1/2
31 -y+1/2,x+1/2,-z+1/2
32 -y+1/2,-x+1/2,z+1/2

loop_
_atom_site_label
_atom_site_type_symbol
_atom_site_symmetry_multiplicity
_atom_site_Wyckoff_label
_atom_site_fract_x
_atom_site_fract_y
_atom_site_fract_z
_atom_site_occupancy
Ca1 Ca 2 a 0.00000 0.00000 0.00000 1.00000
C1 C 4 e 0.00000 0.00000 0.38000 1.00000
```

$CaC_2$-I ($C11_a$): A2B_tI6_139_e_a - POSCAR

```
A2B_tI6_139_e_a & a,c/a,z2 --params=3.87,1.64599483204,0.38 & I4/mmm D_{
    ↪ 4h}^{17} #139 (ae) & tI6 & $C11_{a}$ & C2Ca & C2Ca & M. {von
    ↪ Stackelberg}, Naturwissenschaften 18, 305-306 (1930)
   1.00000000000000
```

```
  -1.93500000000000    1.93500000000000    3.18500000000000
   1.93500000000000   -1.93500000000000    3.18500000000000
   1.93500000000000    1.93500000000000   -3.18500000000000
     C      Ca
     2       1
Direct
   0.38000000000000    0.38000000000000    0.00000000000000    C    (4e)
  -0.38000000000000   -0.38000000000000    0.00000000000000    C    (4e)
   0.00000000000000    0.00000000000000    0.00000000000000    Ca   (2a)
```

$K_2OsO_2Cl_4$ ($J1_5$): A4B2C2D_tI18_139_h_d_e_a - CIF

```
# CIF file
data_findsym-output
_audit_creation_method FINDSYM

_chemical_name_mineral 'Cl4K2O2Os'
_chemical_formula_sum 'Cl4 K2 O2 Os'

loop_
_publ_author_name
 'J. L. Hoard'
 'J. D. Grenko'
_journal_name_full_name
;
 Zeitschrift f{\"u}r Kristallographie - Crystalline Materials
;
_journal_volume 87
_journal_year 1934
_journal_page_first 100
_journal_page_last 109
_publ_Section_title
;
 The Crystal Structure of Potassium Osmyl Chloride, K$_{2}$OsO$_{2}$Cl$_
     ↪ {4}$
;

# Found in The American Mineralogist Crystal Structure Database, 2003

_aflow_title 'K$_{2}$OsO$_{2}$Cl$_{4}$ ($J1_{5}$) Structure'
_aflow_proto 'A4B2C2D_tI18_139_h_d_e_a'
_aflow_params 'a,c/a,z_{3},x_{4}'
_aflow_params_values '6.99,1.25178826896,0.212,0.23'
_aflow_Strukturbericht '$J1_{5}$'
_aflow_Pearson 'tI18'

_symmetry_space_group_name_H-M "I 4/m 2/m 2/m"
_symmetry_Int_Tables_number 139

_cell_length_a      6.99000
_cell_length_b      6.99000
_cell_length_c      8.75000
_cell_angle_alpha  90.00000
_cell_angle_beta   90.00000
_cell_angle_gamma  90.00000

loop_
_space_group_symop_id
_space_group_symop_operation_xyz
1 x,y,z
2 x,-y,-z
3 -x,y,-z
4 -x,-y,z
5 -y,-x,-z
6 -y,x,z
7 y,-x,z
8 y,x,-z
9 -x,-y,-z
10 -x,y,z
11 x,-y,z
12 x,y,-z
13 y,x,z
14 y,-x,-z
15 -y,x,-z
16 -y,-x,z
17 x+1/2,y+1/2,z+1/2
18 x+1/2,-y+1/2,-z+1/2
19 -x+1/2,y+1/2,-z+1/2
20 -x+1/2,-y+1/2,z+1/2
21 -y+1/2,-x+1/2,-z+1/2
22 -y+1/2,x+1/2,z+1/2
23 y+1/2,-x+1/2,z+1/2
24 y+1/2,x+1/2,-z+1/2
25 -x+1/2,-y+1/2,-z+1/2
26 -x+1/2,y+1/2,z+1/2
27 x+1/2,-y+1/2,z+1/2
28 x+1/2,y+1/2,-z+1/2
29 y+1/2,x+1/2,z+1/2
30 y+1/2,-x+1/2,-z+1/2
31 -y+1/2,x+1/2,-z+1/2
32 -y+1/2,-x+1/2,z+1/2

loop_
_atom_site_label
_atom_site_type_symbol
_atom_site_symmetry_multiplicity
_atom_site_Wyckoff_label
_atom_site_fract_x
_atom_site_fract_y
_atom_site_fract_z
_atom_site_occupancy
Os1 Os 2 a 0.00000 0.00000 0.00000 1.00000
K1 K 4 d 0.00000 0.50000 0.25000 1.00000
O1 O 4 e 0.00000 0.00000 0.21200 1.00000
Cl1 Cl 8 h 0.23000 0.23000 0.00000 1.00000
```

$K_2OsO_2Cl_4$ ($J1_5$): A4B2C2D_tI18_139_h_d_e_a - POSCAR

```
A4B2C2D_tI18_139_h_d_e_a & a,c/a,z3,x4 --params=6.99,1.25178826896,0.212
     ↪ ,0.23 & I4/mmm D_{4h}^{17} #139 (adeh) & tI18 & $J1_{5}$ &
     ↪ Cl4K2O2Os & Cl4K2O2Os & J. L. Hoard and J. D. Grenko,
     ↪ Zeitschrift f{"u}r Kristallographie - Crystalline Materials 87,
     ↪ 100-109 (1934)
   1.00000000000000
  -3.49500000000000    3.49500000000000    4.37500000000000
   3.49500000000000   -3.49500000000000    4.37500000000000
   3.49500000000000    3.49500000000000   -4.37500000000000
     Cl      K       O      Os
      4      2       2       1
Direct
   0.23000000000000    0.23000000000000    0.46000000000000    Cl   (8h)
  -0.23000000000000   -0.23000000000000   -0.46000000000000    Cl   (8h)
   0.23000000000000   -0.23000000000000    0.00000000000000    Cl   (8h)
  -0.23000000000000    0.23000000000000    0.00000000000000    Cl   (8h)
   0.75000000000000    0.25000000000000    0.50000000000000    K    (4d)
   0.25000000000000    0.75000000000000    0.50000000000000    K    (4d)
   0.21200000000000    0.21200000000000    0.00000000000000    O    (4e)
  -0.21200000000000   -0.21200000000000    0.00000000000000    O    (4e)
   0.00000000000000    0.00000000000000    0.00000000000000    Os   (2a)
```

$K_2NiF_4$: A4B2C_tI14_139_ce_e_a - CIF

```
# CIF file
data_findsym-output
_audit_creation_method FINDSYM

_chemical_name_mineral 'F4K2Ni'
_chemical_formula_sum 'F4 K2 Ni'

loop_
_publ_author_name
 'S. N. Ruddlesden'
 'P. Popper'
_journal_name_full_name
;
 Acta Cristallographica
;
_journal_volume 10
_journal_year 1957
_journal_page_first 538
_journal_page_last 539
_publ_Section_title
;
 New compounds of the K$_{2}$NiF$_{4}$ type
;

# Found in Ruddlesden-Popper phase, {A$_{2}$BX$_{4}$ series},

_aflow_title 'K$_{2}$NiF$_{4}$ Structure'
_aflow_proto 'A4B2C_tI14_139_ce_e_a'
_aflow_params 'a,c/a,z_{3},z_{4}'
_aflow_params_values '4.0,3.2675,0.151,0.352'
_aflow_Strukturbericht 'None'
_aflow_Pearson 'tI14'

_symmetry_space_group_name_H-M "I 4/m 2/m 2/m"
_symmetry_Int_Tables_number 139

_cell_length_a      4.00000
_cell_length_b      4.00000
_cell_length_c     13.07000
_cell_angle_alpha  90.00000
_cell_angle_beta   90.00000
_cell_angle_gamma  90.00000

loop_
_space_group_symop_id
_space_group_symop_operation_xyz
1 x,y,z
2 x,-y,-z
3 -x,y,-z
4 -x,-y,z
5 -y,-x,-z
6 -y,x,z
7 y,-x,z
8 y,x,-z
9 -x,-y,-z
10 -x,y,z
11 x,-y,z
12 x,y,-z
13 y,x,z
14 y,-x,-z
15 -y,x,-z
16 -y,-x,z
17 x+1/2,y+1/2,z+1/2
18 x+1/2,-y+1/2,-z+1/2
19 -x+1/2,y+1/2,-z+1/2
20 -x+1/2,-y+1/2,z+1/2
21 -y+1/2,-x+1/2,-z+1/2
22 -y+1/2,x+1/2,z+1/2
23 y+1/2,-x+1/2,z+1/2
24 y+1/2,x+1/2,-z+1/2
25 -x+1/2,-y+1/2,-z+1/2
26 -x+1/2,y+1/2,z+1/2
27 x+1/2,-y+1/2,z+1/2
28 x+1/2,y+1/2,-z+1/2
29 y+1/2,x+1/2,z+1/2
30 y+1/2,-x+1/2,-z+1/2
31 -y+1/2,x+1/2,-z+1/2
32 -y+1/2,-x+1/2,z+1/2
```

```
loop_
_atom_site_label
_atom_site_type_symbol
_atom_site_symmetry_multiplicity
_atom_site_Wyckoff_label
_atom_site_fract_x
_atom_site_fract_y
_atom_site_fract_z
_atom_site_occupancy
Ni1 Ni 2 a 0.00000 0.00000 0.00000 1.00000
F1 F 4 c 0.00000 0.50000 0.00000 1.00000
F2 F 4 e 0.00000 0.00000 0.15100 1.00000
K1 K 4 e 0.00000 0.00000 0.35200 1.00000
```

$K_2NiF_4$: A4B2C_tI14_139_ce_e_a - POSCAR

```
A4B2C_tI14_139_ce_e_a & a,c/a,z3,z4 --params=4.0,3.2675,0.151,0.352 & I4
    ↪ /mmm D_{4h}^{17} #139 (ace^2) & tI14 & None & F4K2Ni & F4K2Ni &
    ↪ S. N. Ruddlesden and P. Popper, Acta Cryst. 10, 538-539 (1957)
  1.00000000000000
 -2.00000000000000   2.00000000000000   6.53500000000000
  2.00000000000000  -2.00000000000000   6.53500000000000
  2.00000000000000   2.00000000000000  -6.53500000000000
     F      K     Ni
     4      2      1
Direct
  0.50000000000000   0.00000000000000   0.50000000000000   F   (4c)
  0.00000000000000   0.50000000000000   0.50000000000000   F   (4c)
  0.15100000000000   0.15100000000000   0.00000000000000   F   (4e)
 -0.15100000000000  -0.15100000000000   0.00000000000000   F   (4e)
  0.35200000000000   0.35200000000000   0.00000000000000   K   (4e)
 -0.35200000000000  -0.35200000000000   0.00000000000000   K   (4e)
  0.00000000000000   0.00000000000000   0.00000000000000   Ni  (2a)
```

$K_3TlCl_6{\cdot}2H_2O$ ($J3_1$): A6B2C3D_tI168_139_egikl2m_ejn_bh2n_acf - CIF

```
# CIF file
data_findsym-output
_audit_creation_method FINDSYM

_chemical_name_mineral 'Cl6(H2O)2K3Tl'
_chemical_formula_sum 'Cl6 (H2O)2 K3 Tl'

loop_
_publ_author_name
 'J. L. Hoard'
 'L. Goldstein'
_journal_name_full_name
;
 Journal of Chemical Physics
;
_journal_volume 3
_journal_year 1935
_journal_page_first 645
_journal_page_last 649
_publ_Section_title
;
 The Structure of Potassium Hexachlorothalliate Dihydrate
;

_aflow_title 'K$_{3}$TlCl$_{6}$$\cdot$2H$_{2}$O ($J3_{1}$) Structure'
_aflow_proto 'A6B2C3D_tI168_139_egikl2m_ejn_bh2n_acf'
_aflow_params 'a,c/a,z_{4},z_{5},z_{7},x_{8},x_{9},x_{10},x_{11},x_{12},
    ↪ y_{12},x_{13},z_{13},x_{14},z_{14},y_{15},z_{15},y_{16},z_{16},
    ↪ y_{17},z_{17}'
_aflow_params_values '15.841,1.1366075374,0.142,0.347,0.142,0.214,0.161,
    ↪ 0.673,0.364,0.386,0.114,0.181,0.362,0.16,0.163,0.157,0.276,
    ↪ 0.295,0.132,0.293,0.376'
_aflow_Strukturbericht '$J3_{1}$'
_aflow_Pearson 'tI168'

_symmetry_space_group_name_H-M "I 4/m 2/m 2/m"
_symmetry_Int_Tables_number 139

_cell_length_a     15.84100
_cell_length_b     15.84100
_cell_length_c     18.00500
_cell_angle_alpha  90.00000
_cell_angle_beta   90.00000
_cell_angle_gamma  90.00000

loop_
_space_group_symop_id
_space_group_symop_operation_xyz
1 x,y,z
2 x,-y,-z
3 -x,y,-z
4 -x,-y,z
5 -y,-x,-z
6 -y,x,z
7 y,-x,z
8 y,x,-z
9 -x,-y,-z
10 -x,y,z
11 x,-y,z
12 x,y,-z
13 y,x,z
14 y,-x,-z
15 -y,x,-z
16 -y,-x,z
17 x+1/2,y+1/2,z+1/2
18 x+1/2,-y+1/2,-z+1/2
19 -x+1/2,y+1/2,-z+1/2
20 -x+1/2,-y+1/2,z+1/2
21 -y+1/2,-x+1/2,-z+1/2
22 -y+1/2,x+1/2,z+1/2
23 y+1/2,-x+1/2,z+1/2
24 y+1/2,x+1/2,-z+1/2
25 -x+1/2,-y+1/2,-z+1/2
26 -x+1/2,y+1/2,z+1/2
27 x+1/2,-y+1/2,z+1/2
28 x+1/2,y+1/2,-z+1/2
29 y+1/2,x+1/2,z+1/2
30 y+1/2,-x+1/2,-z+1/2
31 -y+1/2,x+1/2,-z+1/2
32 -y+1/2,-x+1/2,z+1/2

loop_
_atom_site_label
_atom_site_type_symbol
_atom_site_symmetry_multiplicity
_atom_site_Wyckoff_label
_atom_site_fract_x
_atom_site_fract_y
_atom_site_fract_z
_atom_site_occupancy
Tl1 Tl 2 a 0.00000 0.00000 0.00000 1.00000
K1 K 2 b 0.00000 0.00000 0.50000 1.00000
Tl2 Tl 4 c 0.00000 0.50000 0.00000 1.00000
Cl1 Cl 4 e 0.00000 0.00000 0.14200 1.00000
H2O1 H2O 4 e 0.00000 0.00000 0.34700 1.00000
Tl3 Tl 8 f 0.25000 0.25000 0.25000 1.00000
Cl2 Cl 8 g 0.00000 0.50000 0.14200 1.00000
K2 K 8 h 0.21400 0.21400 0.00000 1.00000
Cl3 Cl 8 i 0.16100 0.00000 0.00000 1.00000
H2O2 H2O 8 j 0.67300 0.50000 0.00000 1.00000
Cl4 Cl 16 k 0.36400 0.86400 0.25000 1.00000
Cl5 Cl 16 l 0.38600 0.11400 0.00000 1.00000
Cl6 Cl 16 m 0.18100 0.18100 0.36200 1.00000
Cl7 Cl 16 m 0.16000 0.16000 0.16300 1.00000
H2O3 H2O 16 n 0.00000 0.15700 0.27600 1.00000
K3 K 16 n 0.00000 0.29500 0.13200 1.00000
K4 K 16 n 0.00000 0.29300 0.37600 1.00000
```

$K_3TlCl_6{\cdot}2H_2O$ ($J3_1$): A6B2C3D_tI168_139_egikl2m_ejn_bh2n_acf - POSCAR

```
A6B2C3D_tI168_139_egikl2m_ejn_bh2n_acf & a,c/a,z4,z5,z7,x8,x9,x10,x11,
    ↪ x12,y12,x13,z13,x14,z14,y15,z15,y16,z16,y17,z17 --params=15.841
    ↪ ,1.1366075374,0.142,0.347,0.142,0.214,0.161,0.673,0.364,0.386,
    ↪ 0.114,0.181,0.362,0.16,0.163,0.157,0.276,0.295,0.132,0.293,
    ↪ 0.376 & I4/mmm D_{4h}^{17} #139 (abce^2fghijklm^2n^3) & tI168 &
    ↪ $J3_{1}$ & Cl6(H2O)2K3Tl & Cl6(H2O)2K3Tl & J. L. Hoard and L.
    ↪ Goldstein, J. Chem. Phys. 3, 645-649 (1935)
  1.00000000000000
 -7.92050000000000   7.92050000000000   9.00250000000000
  7.92050000000000  -7.92050000000000   9.00250000000000
  7.92050000000000   7.92050000000000  -9.00250000000000
    Cl    H2O      K     Tl
    42     14     21      7
Direct
  0.14200000000000   0.14200000000000   0.00000000000000   Cl   (4e)
 -0.14200000000000  -0.14200000000000   0.00000000000000   Cl   (4e)
  0.64200000000000   0.14200000000000   0.50000000000000   Cl   (8g)
  0.14200000000000   0.64200000000000   0.50000000000000   Cl   (8g)
  0.35800000000000  -0.14200000000000   0.50000000000000   Cl   (8g)
 -0.14200000000000   0.35800000000000   0.50000000000000   Cl   (8g)
  0.00000000000000   0.16100000000000   0.16100000000000   Cl   (8i)
  0.00000000000000  -0.16100000000000  -0.16100000000000   Cl   (8i)
  0.16100000000000   0.00000000000000   0.16100000000000   Cl   (8i)
 -0.16100000000000   0.00000000000000  -0.16100000000000   Cl   (8i)
  1.11400000000000   0.61400000000000   1.22800000000000   Cl  (16k)
  0.38600000000000  -0.11400000000000  -0.22800000000000   Cl  (16k)
  0.61400000000000   0.38600000000000   0.50000000000000   Cl  (16k)
 -0.11400000000000   1.11400000000000   0.50000000000000   Cl  (16k)
 -0.11400000000000   0.38600000000000  -0.22800000000000   Cl  (16k)
  0.61400000000000   1.11400000000000   1.22800000000000   Cl  (16k)
  0.38600000000000   0.61400000000000   0.50000000000000   Cl  (16k)
  1.11400000000000  -0.11400000000000   0.50000000000000   Cl  (16k)
  0.11400000000000   0.38600000000000   0.50000000000000   Cl  (16l)
 -0.11400000000000  -0.38600000000000  -0.50000000000000   Cl  (16l)
  0.38600000000000  -0.11400000000000   0.27200000000000   Cl  (16l)
 -0.38600000000000   0.11400000000000  -0.27200000000000   Cl  (16l)
  0.11400000000000  -0.38600000000000  -0.27200000000000   Cl  (16l)
 -0.11400000000000   0.38600000000000   0.27200000000000   Cl  (16l)
  0.38600000000000   0.11400000000000   0.50000000000000   Cl  (16l)
 -0.38600000000000  -0.11400000000000  -0.50000000000000   Cl  (16l)
  0.54300000000000   0.54300000000000   0.36200000000000   Cl  (16m)
  0.18100000000000   0.18100000000000  -0.36200000000000   Cl  (16m)
  0.54300000000000   0.18100000000000   0.00000000000000   Cl  (16m)
  0.18100000000000   0.54300000000000   0.00000000000000   Cl  (16m)
 -0.18100000000000  -0.54300000000000   0.00000000000000   Cl  (16m)
 -0.54300000000000  -0.18100000000000   0.00000000000000   Cl  (16m)
 -0.18100000000000  -0.18100000000000   0.36200000000000   Cl  (16m)
 -0.54300000000000  -0.54300000000000  -0.36200000000000   Cl  (16m)
  0.32300000000000   0.32300000000000   0.32000000000000   Cl  (16m)
  0.00300000000000   0.00300000000000  -0.32000000000000   Cl  (16m)
  0.32300000000000   0.00300000000000   0.00000000000000   Cl  (16m)
  0.00300000000000   0.32300000000000   0.00000000000000   Cl  (16m)
 -0.00300000000000  -0.32300000000000   0.00000000000000   Cl  (16m)
 -0.32300000000000  -0.00300000000000   0.00000000000000   Cl  (16m)
 -0.00300000000000  -0.00300000000000   0.32000000000000   Cl  (16m)
 -0.32300000000000  -0.32300000000000  -0.32000000000000   Cl  (16m)
  0.34700000000000   0.34700000000000   0.00000000000000   H2O   (4e)
 -0.34700000000000  -0.34700000000000   0.00000000000000   H2O   (4e)
  0.50000000000000   0.67300000000000   1.17300000000000   H2O   (8j)
  0.50000000000000  -0.67300000000000  -0.17300000000000   H2O   (8j)
  0.67300000000000   0.50000000000000   1.17300000000000   H2O   (8j)
 -0.67300000000000   0.50000000000000  -0.17300000000000   H2O   (8j)
  0.43300000000000   0.27600000000000   0.15700000000000   H2O  (16n)
  0.11900000000000   0.27600000000000  -0.15700000000000   H2O  (16n)
```

```
   0.27600000000000    0.11900000000000   -0.15700000000000   H2O   (16n)
   0.27600000000000    0.43300000000000    0.15700000000000   H2O   (16n)
  -0.11900000000000   -0.27600000000000    0.15700000000000   H2O   (16n)
  -0.43300000000000   -0.27600000000000   -0.15700000000000   H2O   (16n)
  -0.27600000000000   -0.11900000000000    0.15700000000000   H2O   (16n)
  -0.27600000000000   -0.43300000000000   -0.15700000000000   H2O   (16n)
   0.50000000000000    0.50000000000000    0.00000000000000    K    (2b)
   0.21400000000000    0.21400000000000    0.42800000000000    K    (8h)
  -0.21400000000000   -0.21400000000000   -0.42800000000000    K    (8h)
   0.21400000000000   -0.21400000000000    0.00000000000000    K    (8h)
  -0.21400000000000    0.21400000000000    0.00000000000000    K    (8h)
   0.42700000000000    0.13200000000000    0.29500000000000    K   (16n)
  -0.16300000000000    0.13200000000000   -0.29500000000000    K   (16n)
   0.13200000000000   -0.16300000000000   -0.29500000000000    K   (16n)
   0.13200000000000    0.42700000000000    0.29500000000000    K   (16n)
   0.16300000000000   -0.13200000000000    0.29500000000000    K   (16n)
  -0.42700000000000   -0.13200000000000   -0.29500000000000    K   (16n)
  -0.13200000000000    0.16300000000000    0.29500000000000    K   (16n)
  -0.13200000000000   -0.42700000000000   -0.29500000000000    K   (16n)
   0.66900000000000    0.37600000000000    0.29300000000000    K   (16n)
   0.08300000000000    0.37600000000000   -0.29300000000000    K   (16n)
   0.37600000000000    0.08300000000000   -0.29300000000000    K   (16n)
   0.37600000000000    0.66900000000000    0.29300000000000    K   (16n)
  -0.08300000000000   -0.37600000000000    0.29300000000000    K   (16n)
  -0.66900000000000   -0.37600000000000   -0.29300000000000    K   (16n)
  -0.37600000000000   -0.08300000000000    0.29300000000000    K   (16n)
  -0.37600000000000   -0.66900000000000   -0.29300000000000    K   (16n)
   0.00000000000000    0.00000000000000    0.00000000000000    Tl   (2a)
   0.50000000000000    0.00000000000000    0.50000000000000    Tl   (4c)
   0.00000000000000    0.50000000000000    0.50000000000000    Tl   (4c)
   0.50000000000000    0.50000000000000    0.50000000000000    Tl   (8f)
   0.00000000000000    0.00000000000000    0.50000000000000    Tl   (8f)
   0.50000000000000    0.00000000000000    0.00000000000000    Tl   (8f)
   0.00000000000000    0.50000000000000    0.00000000000000    Tl   (8f)
```

$Sr_3Ti_2O_7$: A7B3C2_tI24_139_aeg_be_e - CIF

```
# CIF file
data_findsym-output
_audit_creation_method FINDSYM

_chemical_name_mineral 'O7Sr3Ti2'
_chemical_formula_sum 'O7 Sr3 Ti2'

loop_
_publ_author_name
 'S. N. Ruddlesden'
 'P. Popper'
_journal_name_full_name
;
 Acta Cristallographica
;
_journal_volume 11
_journal_year 1958
_journal_page_first 54
_journal_page_last 55
_publ_Section_title
;
 The compound Sr$_{3}$Ti$_{2}$O$_{7}$ and its structure
;

# Found in Ruddlesden-Popper phase, {A$_{3}$B$_{2}$X$_{7}$ series},

_aflow_title 'Sr$_{3}$Ti$_{2}$O$_{7}$ Structure'
_aflow_proto 'A7B3C2_tI24_139_aeg_be_e'
_aflow_params 'a,c/a,z_{3},z_{4},z_{5},z_{6}'
_aflow_params_values '3.9,5.22564102564,0.188,0.312,0.094,0.094'
_aflow_Strukturbericht 'None'
_aflow_Pearson 'tI24'

_symmetry_space_group_name_H-M "I 4/m 2/m 2/m"
_symmetry_Int_Tables_number 139

_cell_length_a    3.90000
_cell_length_b    3.90000
_cell_length_c    20.38000
_cell_angle_alpha 90.00000
_cell_angle_beta  90.00000
_cell_angle_gamma 90.00000

loop_
_space_group_symop_id
_space_group_symop_operation_xyz
1 x,y,z
2 x,-y,-z
3 -x,y,-z
4 -x,-y,z
5 -y,-x,-z
6 -y,x,z
7 y,-x,z
8 y,x,-z
9 -x,-y,-z
10 -x,y,z
11 x,-y,z
12 x,y,-z
13 y,x,z
14 y,-x,-z
15 -y,x,-z
16 -y,-x,z
17 x+1/2,y+1/2,z+1/2
18 x+1/2,-y+1/2,-z+1/2
19 -x+1/2,y+1/2,-z+1/2
20 -x+1/2,-y+1/2,z+1/2
21 -y+1/2,-x+1/2,-z+1/2
22 -y+1/2,x+1/2,z+1/2
23 y+1/2,-x+1/2,z+1/2
24 y+1/2,x+1/2,-z+1/2
25 -x+1/2,-y+1/2,-z+1/2
26 -x+1/2,y+1/2,z+1/2
27 x+1/2,-y+1/2,z+1/2
28 x+1/2,y+1/2,-z+1/2
29 y+1/2,x+1/2,z+1/2
30 y+1/2,-x+1/2,-z+1/2
31 -y+1/2,x+1/2,-z+1/2
32 -y+1/2,-x+1/2,z+1/2

loop_
_atom_site_label
_atom_site_type_symbol
_atom_site_symmetry_multiplicity
_atom_site_Wyckoff_label
_atom_site_fract_x
_atom_site_fract_y
_atom_site_fract_z
_atom_site_occupancy
O1 O 2 a 0.00000 0.00000 0.00000 1.00000
Sr1 Sr 2 b 0.00000 0.00000 0.50000 1.00000
O2 O 4 e 0.00000 0.00000 0.18800 1.00000
Sr2 Sr 4 e 0.00000 0.00000 0.31200 1.00000
Ti1 Ti 4 e 0.00000 0.00000 0.09400 1.00000
O3 O 8 g 0.00000 0.50000 0.09400 1.00000
```

$Sr_3Ti_2O_7$: A7B3C2_tI24_139_aeg_be_e - POSCAR

```
A7B3C2_tI24_139_aeg_be_e & a,c/a,z3,z4,z5,z6 --params=3.9,5.22564102564,
    ↪ 0.188,0.312,0.094,0.094 & I4/mmm D_{4h}^{17} #139 (abe^3g) &
    ↪ tI24 & None & O7Sr3Ti2 & O7Sr3Ti2 & S. N. Ruddlesden and P.
    ↪ Popper, Acta Cryst. 11, 54-55 (1958)
   1.00000000000000
  -1.95000000000000    1.95000000000000   10.19000000000000
   1.95000000000000   -1.95000000000000   10.19000000000000
   1.95000000000000    1.95000000000000  -10.19000000000000
     O     Sr     Ti
     7      3      2
Direct
   0.00000000000000    0.00000000000000    0.00000000000000    O    (2a)
   0.18800000000000    0.18800000000000    0.00000000000000    O    (4e)
  -0.18800000000000   -0.18800000000000    0.00000000000000    O    (4e)
   0.59400000000000    0.09400000000000    0.50000000000000    O    (8g)
   0.09400000000000    0.59400000000000    0.50000000000000    O    (8g)
   0.40600000000000   -0.09400000000000    0.50000000000000    O    (8g)
  -0.09400000000000    0.40600000000000    0.50000000000000    O    (8g)
   0.50000000000000    0.50000000000000    0.00000000000000    Sr   (2b)
   0.31200000000000    0.31200000000000    0.00000000000000    Sr   (4e)
  -0.31200000000000   -0.31200000000000    0.00000000000000    Sr   (4e)
   0.09400000000000    0.09400000000000    0.00000000000000    Ti   (4e)
  -0.09400000000000   -0.09400000000000    0.00000000000000    Ti   (4e)
```

$Fe_8N$ ($D2_g$): A8B_tI18_139_deh_a - CIF

```
# CIF file
data_findsym-output
_audit_creation_method FINDSYM

_chemical_name_mineral 'Fe8N'
_chemical_formula_sum 'Fe8 N'

loop_
_publ_author_name
 'S. Yamashita'
 'Y. Masubuchi'
 'Y. Nakazawa'
 'T. Okayama'
 'M. Tsuchiya'
 'S. Kikkawa'
_journal_name_full_name
;
 Journal of Solid State Chemistry
;
_journal_volume 194
_journal_year 2012
_journal_page_first 76
_journal_page_last 79
_publ_Section_title
;
 Crystal structure and magnetic properties of ``$\alpha$$\'\'$-Fe$_{16}
    ↪ $N$_{2}$\'\' containing reisdual $\alpha$-Fe prepared by
    ↪ low-temperature ammonia nitridation
;

_aflow_title 'Fe$_{8}$N ($D2_{g}$) Structure'
_aflow_proto 'A8B_tI18_139_deh_a'
_aflow_params 'a,c/a,z_{3},x_{4}'
_aflow_params_values '5.71361,1.10055114017,0.2922,0.24433'
_aflow_Strukturbericht '$D2_{g}$'
_aflow_Pearson 'tI18'

_symmetry_space_group_name_H-M "I 4/m 2/m 2/m"
_symmetry_Int_Tables_number 139

_cell_length_a    5.71361
_cell_length_b    5.71361
_cell_length_c    6.28812
_cell_angle_alpha 90.00000
_cell_angle_beta  90.00000
_cell_angle_gamma 90.00000

loop_
_space_group_symop_id
_space_group_symop_operation_xyz
```

```
1 x,y,z
2 x,-y,-z
3 -x,y,-z
4 -x,-y,z
5 -y,-x,-z
6 -y,x,z
7 y,-x,z
8 y,x,-z
9 -x,-y,-z
10 -x,y,z
11 x,-y,z
12 x,y,-z
13 y,x,z
14 y,-x,-z
15 -y,x,-z
16 -y,-x,z
17 x+1/2,y+1/2,z+1/2
18 x+1/2,-y+1/2,-z+1/2
19 -x+1/2,y+1/2,-z+1/2
20 -x+1/2,-y+1/2,z+1/2
21 -y+1/2,-x+1/2,-z+1/2
22 -y+1/2,x+1/2,z+1/2
23 y+1/2,-x+1/2,z+1/2
24 y+1/2,x+1/2,-z+1/2
25 -x+1/2,-y+1/2,-z+1/2
26 -x+1/2,y+1/2,z+1/2
27 x+1/2,-y+1/2,z+1/2
28 x+1/2,y+1/2,-z+1/2
29 y+1/2,x+1/2,z+1/2
30 y+1/2,-x+1/2,-z+1/2
31 -y+1/2,x+1/2,-z+1/2
32 -y+1/2,-x+1/2,z+1/2

loop_
_atom_site_label
_atom_site_type_symbol
_atom_site_symmetry_multiplicity
_atom_site_Wyckoff_label
_atom_site_fract_x
_atom_site_fract_y
_atom_site_fract_z
_atom_site_occupancy
N1 N 2 a 0.00000 0.00000 0.00000 1.00000
Fe1 Fe 4 d 0.00000 0.50000 0.25000 1.00000
Fe2 Fe 4 e 0.00000 0.00000 0.29220 1.00000
Fe3 Fe 8 h 0.24433 0.24433 0.00000 1.00000
```

$Fe_8N$ ($D2_g$): A8B_tI18_139_deh_a - POSCAR

```
A8B_tI18_139_deh_a & a,c/a,z3,x4 --params=5.71361,1.10055114017,0.2922,
    ↪ 0.24433 & I4/mmm D_{4h}^{17} #139 (adeh) & tI18 & $D2_{g}$ &
    ↪ Fe8N & Fe8N & S. Yamashita et al., J. Solid State Chem. 194,
    ↪ 76-79 (2012)
   1.00000000000000
  -2.85680500000000   2.85680500000000   3.14406000000000
   2.85680500000000  -2.85680500000000   3.14406000000000
   2.85680500000000   2.85680500000000  -3.14406000000000
    Fe     N
     8     1
Direct
   0.75000000000000   0.25000000000000   0.50000000000000   Fe   (4d)
   0.25000000000000   0.75000000000000   0.50000000000000   Fe   (4d)
   0.29220000000000   0.29220000000000   0.00000000000000   Fe   (4e)
  -0.29220000000000  -0.29220000000000   0.00000000000000   Fe   (4e)
   0.24433000000000   0.24433000000000   0.48866000000000   Fe   (8h)
  -0.24433000000000  -0.24433000000000  -0.48866000000000   Fe   (8h)
   0.24433000000000  -0.24433000000000   0.00000000000000   Fe   (8h)
  -0.24433000000000   0.24433000000000   0.00000000000000   Fe   (8h)
   0.00000000000000   0.00000000000000   0.00000000000000    N   (2a)
```

$Li_2CN_2$: AB2C2_tI10_139_a_d_e - CIF

```
# CIF file
data_findsym-output
_audit_creation_method FINDSYM

_chemical_name_mineral 'CLi2N2'
_chemical_formula_sum 'C Li2 N2'

loop_
_publ_author_name
 'M. G. Down'
 'M. J. Haley'
 'P. Hubberstey'
 'R. J. Pulham'
 'A. E. Thunder'
_journal_name_full_name
;
 Journal of the Chemical Society, Dalton Transactions
;
_journal_volume
_journal_year 1978
_journal_page_first 1407
_journal_page_last 1411
_publ_Section_title
;
 Solutions of lithium salts in liquid lithium: preparation and X-ray
    ↪ crystal structure of the dilithium salt of carbodi-imide (
    ↪ cyanamide)
;

# Found in {Commentary: The Materials Project: A materials genome
    ↪ approach to accelerating materials innovation}, 2013

_aflow_title 'Li$_{2}$CN$_{2}$ Structure'
_aflow_proto 'AB2C2_tI10_139_a_d_e'
_aflow_params 'a,c/a,z_{3}'
_aflow_params_values '3.687,2.35096284242,0.1419'
_aflow_Strukturbericht 'None'
_aflow_Pearson 'tI10'

_symmetry_space_group_name_H-M "I 4/m 2/m 2/m"
_symmetry_Int_Tables_number 139

_cell_length_a     3.68700
_cell_length_b     3.68700
_cell_length_c     8.66800
_cell_angle_alpha 90.00000
_cell_angle_beta  90.00000
_cell_angle_gamma 90.00000

loop_
_space_group_symop_id
_space_group_symop_operation_xyz
1 x,y,z
2 x,-y,-z
3 -x,y,-z
4 -x,-y,z
5 -y,-x,-z
6 -y,x,z
7 y,-x,z
8 y,x,-z
9 -x,-y,-z
10 -x,y,z
11 x,-y,z
12 x,y,-z
13 y,x,z
14 y,-x,-z
15 -y,x,-z
16 -y,-x,z
17 x+1/2,y+1/2,z+1/2
18 x+1/2,-y+1/2,-z+1/2
19 -x+1/2,y+1/2,-z+1/2
20 -x+1/2,-y+1/2,z+1/2
21 -y+1/2,-x+1/2,-z+1/2
22 -y+1/2,x+1/2,z+1/2
23 y+1/2,-x+1/2,z+1/2
24 y+1/2,x+1/2,-z+1/2
25 -x+1/2,-y+1/2,-z+1/2
26 -x+1/2,y+1/2,z+1/2
27 x+1/2,-y+1/2,z+1/2
28 x+1/2,y+1/2,-z+1/2
29 y+1/2,x+1/2,z+1/2
30 y+1/2,-x+1/2,-z+1/2
31 -y+1/2,x+1/2,-z+1/2
32 -y+1/2,-x+1/2,z+1/2

loop_
_atom_site_label
_atom_site_type_symbol
_atom_site_symmetry_multiplicity
_atom_site_Wyckoff_label
_atom_site_fract_x
_atom_site_fract_y
_atom_site_fract_z
_atom_site_occupancy
C1 C 2 a 0.00000 0.00000 0.00000 1.00000
Li1 Li 4 d 0.00000 0.50000 0.25000 1.00000
N1 N 4 e 0.00000 0.00000 0.14190 1.00000
```

$Li_2CN_2$: AB2C2_tI10_139_a_d_e - POSCAR

```
AB2C2_tI10_139_a_d_e & a,c/a,z3 --params=3.687,2.35096284242,0.1419 & I4
    ↪ /mmm D_{4h}^{17} #139 (ade) & tI10 & None & CLi2N2 & CLi2N2 &
    ↪ M. G. Down et al., {J. Chem. Soc. Dalton Trans., 1407-1411 (
    ↪ 1978)
   1.00000000000000
  -1.84350000000000   1.84350000000000   4.33400000000000
   1.84350000000000  -1.84350000000000   4.33400000000000
   1.84350000000000   1.84350000000000  -4.33400000000000
     C    Li     N
     1     2     2
Direct
   0.00000000000000   0.00000000000000   0.00000000000000    C   (2a)
   0.75000000000000   0.25000000000000   0.50000000000000   Li   (4d)
   0.25000000000000   0.75000000000000   0.50000000000000   Li   (4d)
   0.14190000000000   0.14190000000000   0.00000000000000    N   (4e)
  -0.14190000000000  -0.14190000000000   0.00000000000000    N   (4e)
```

$H5_9$ [Autunite, $Ca(UO_2)_2(PO_4)_2 \cdot 10\frac{1}{2}H_2O$] (*obsolete*): AB2C2_tI10_139_a_d_e - CIF

```
# CIF file
data_findsym-output
_audit_creation_method FINDSYM

_chemical_name_mineral 'Autunite'
_chemical_formula_sum 'Ca P2 U2'

loop_
_publ_author_name
 'J. Beintema'
_journal_name_full_name
;
 Recueil des Travaux Chimiques des Pays-Bas
;
_journal_volume 57
_journal_year 1938
_journal_page_first 155
_journal_page_last 175
_publ_Section_title
```

```
;
 On the composition and the crystallography of autunite and the
       ↪ meta−autunites
;

_aflow_title '$H5_{9}$ [Autunite, Ca(UO$_{2}$)$_{2}$(PO$_{4}$)$_{2}$$\
       ↪ cdot$10$\frac{1}{2}$H$_{2}$O] ({\em{obsolete}}) Structure'
_aflow_proto 'AB2C2_tI10_139_a_d_e'
_aflow_params 'a,c/a,z_{3}'
_aflow_params_values '6.98,2.95558739255,0.208'
_aflow_Strukturbericht '$H5_{9}$'
_aflow_Pearson 'tI10'

_symmetry_space_group_name_H−M "I 4/m 2/m 2/m"
_symmetry_Int_Tables_number 139

_cell_length_a      6.98000
_cell_length_b      6.98000
_cell_length_c      20.63000
_cell_angle_alpha  90.00000
_cell_angle_beta   90.00000
_cell_angle_gamma  90.00000

loop_
_space_group_symop_id
_space_group_symop_operation_xyz
1 x,y,z
2 x,−y,−z
3 −x,y,−z
4 −x,−y,z
5 −y,−x,−z
6 −y,x,z
7 y,−x,z
8 y,x,−z
9 −x,−y,−z
10 −x,y,z
11 x,−y,z
12 x,y,−z
13 y,x,z
14 y,−x,−z
15 −y,x,−z
16 −y,−x,z
17 x+1/2,y+1/2,z+1/2
18 x+1/2,−y+1/2,−z+1/2
19 −x+1/2,y+1/2,−z+1/2
20 −x+1/2,−y+1/2,z+1/2
21 −y+1/2,−x+1/2,−z+1/2
22 −y+1/2,x+1/2,z+1/2
23 y+1/2,−x+1/2,z+1/2
24 y+1/2,x+1/2,−z+1/2
25 −x+1/2,−y+1/2,−z+1/2
26 −x+1/2,y+1/2,z+1/2
27 x+1/2,−y+1/2,z+1/2
28 x+1/2,y+1/2,−z+1/2
29 y+1/2,x+1/2,z+1/2
30 y+1/2,−x+1/2,−z+1/2
31 −y+1/2,x+1/2,−z+1/2
32 −y+1/2,−x+1/2,z+1/2

loop_
_atom_site_label
_atom_site_type_symbol
_atom_site_symmetry_multiplicity
_atom_site_Wyckoff_label
_atom_site_fract_x
_atom_site_fract_y
_atom_site_fract_z
_atom_site_occupancy
Ca1 Ca 2 a 0.00000 0.00000 0.00000 1.00000
P1 P 4 d 0.00000 0.50000 0.25000 1.00000
U1 U 4 e 0.00000 0.00000 0.20800 1.00000
```

$H5_9$ [Autunite, $Ca(UO_2)_2(PO_4)_2{\cdot}10\frac{1}{2}H_2O$] (*obsolete*): AB2C2_tI10_139_a_d_e - POSCAR

```
AB2C2_tI10_139_a_d_e & a,c/a,z3 −−params=6.98,2.95558739255,0.208 & I4/
       ↪ mmm D_{4h}^{17} #139 (ade) & tI10 & $H5_{9}$ & Ca(UO2)2(PO4)2\
       ↪ cdot10\frac{1{2H2O & Autunite & J. Beintema, Rec. Trav. Chim.
       ↪ Pays−Bas 57, 155−175 (1938)
   1.00000000000000
  −3.49000000000000    3.49000000000000   10.31500000000000
   3.49000000000000   −3.49000000000000   10.31500000000000
   3.49000000000000    3.49000000000000  −10.31500000000000
     Ca         P         U
      1         2         2
Direct
   0.00000000000000    0.00000000000000    0.00000000000000    Ca    (2a)
   0.75000000000000    0.25000000000000    0.50000000000000    P     (4d)
   0.25000000000000    0.75000000000000    0.50000000000000    P     (4d)
   0.20800000000000    0.20800000000000    0.00000000000000    U     (4e)
  −0.20800000000000   −0.20800000000000    0.00000000000000    U     (4e)
```

$AuCsCl_3$ ($K7_6$): AB3C_tI20_139_ab_eh_d - CIF

```
# CIF file
data_findsym−output
_audit_creation_method FINDSYM

_chemical_name_mineral 'AuCl3Cs'
_chemical_formula_sum 'Au Cl3 Cs'

loop_
_publ_author_name
 'N. Elliott'
 'L. Pauling'
_journal_name_full_name
;
 Journal of the American Chemical Society
;
_journal_volume 60
_journal_year 1938
_journal_page_first 1846
_journal_page_last 1851
_publ_Section_title
;
 The Crystal Structure of Cesium Aurous Auric Chloride, Cs$_{2}$AuAuCl$_
       ↪ {6}$, and Cesium Argentous Auric Chloride, Cs$_{2}$AgAuCl$_{6}
       ↪ $
;

# Found in The American Mineralogist Crystal Structure Database, 2003

_aflow_title 'AuCsCl$_{3}$ ($K7_{6}$) Structure'
_aflow_proto 'AB3C_tI20_139_ab_eh_d'
_aflow_params 'a,c/a,z_{4},x_{5}'
_aflow_params_values '7.49,1.45126835781,0.288,0.228'
_aflow_Strukturbericht '$K7_{6}$'
_aflow_Pearson 'tI20'

_symmetry_space_group_name_H−M "I 4/m 2/m 2/m"
_symmetry_Int_Tables_number 139

_cell_length_a      7.49000
_cell_length_b      7.49000
_cell_length_c      10.87000
_cell_angle_alpha  90.00000
_cell_angle_beta   90.00000
_cell_angle_gamma  90.00000

loop_
_space_group_symop_id
_space_group_symop_operation_xyz
1 x,y,z
2 x,−y,−z
3 −x,y,−z
4 −x,−y,z
5 −y,−x,−z
6 −y,x,z
7 y,−x,z
8 y,x,−z
9 −x,−y,−z
10 −x,y,z
11 x,−y,z
12 x,y,−z
13 y,x,z
14 y,−x,−z
15 −y,x,−z
16 −y,−x,z
17 x+1/2,y+1/2,z+1/2
18 x+1/2,−y+1/2,−z+1/2
19 −x+1/2,y+1/2,−z+1/2
20 −x+1/2,−y+1/2,z+1/2
21 −y+1/2,−x+1/2,−z+1/2
22 −y+1/2,x+1/2,z+1/2
23 y+1/2,−x+1/2,z+1/2
24 y+1/2,x+1/2,−z+1/2
25 −x+1/2,−y+1/2,−z+1/2
26 −x+1/2,y+1/2,z+1/2
27 x+1/2,−y+1/2,z+1/2
28 x+1/2,y+1/2,−z+1/2
29 y+1/2,x+1/2,z+1/2
30 y+1/2,−x+1/2,−z+1/2
31 −y+1/2,x+1/2,−z+1/2
32 −y+1/2,−x+1/2,z+1/2

loop_
_atom_site_label
_atom_site_type_symbol
_atom_site_symmetry_multiplicity
_atom_site_Wyckoff_label
_atom_site_fract_x
_atom_site_fract_y
_atom_site_fract_z
_atom_site_occupancy
Au1 Au 2 a 0.00000 0.00000 0.00000 1.00000
Au2 Au 2 b 0.00000 0.00000 0.50000 1.00000
Cs1 Cs 4 d 0.00000 0.50000 0.25000 1.00000
Cl1 Cl 4 e 0.00000 0.00000 0.28800 1.00000
Cl2 Cl 8 h 0.22800 0.22800 0.00000 1.00000
```

$AuCsCl_3$ ($K7_6$): AB3C_tI20_139_ab_eh_d - POSCAR

```
AB3C_tI20_139_ab_eh_d & a,c/a,z4,x5 −−params=7.49,1.45126835781,0.288,
       ↪ 0.228 & I4/mmm D_{4h}^{17} #139 (abdeh) & tI20 & $K7_{6}$ &
       ↪ AuCl3Cs & AuCl3Cs & N. Elliott and L. Pauling, J. Am. Chem.
       ↪ Soc. 60, 1846−1851 (1938)
   1.00000000000000
  −3.74500000000000    3.74500000000000    5.43500000000000
   3.74500000000000   −3.74500000000000    5.43500000000000
   3.74500000000000    3.74500000000000   −5.43500000000000
     Au        Cl        Cs
      2         6         2
Direct
   0.00000000000000    0.00000000000000    0.00000000000000    Au    (2a)
   0.50000000000000    0.50000000000000    0.00000000000000    Au    (2b)
   0.28800000000000    0.28800000000000    0.00000000000000    Cl    (4e)
  −0.28800000000000   −0.28800000000000    0.00000000000000    Cl    (4e)
   0.22800000000000    0.22800000000000    0.45600000000000    Cl    (8h)
  −0.22800000000000   −0.22800000000000   −0.45600000000000    Cl    (8h)
   0.22800000000000   −0.22800000000000    0.00000000000000    Cl    (8h)
  −0.22800000000000    0.22800000000000    0.00000000000000    Cl    (8h)
```

```
    0.75000000000000     0.25000000000000     0.50000000000000     Cs     (4d)
    0.25000000000000     0.75000000000000     0.50000000000000     Cs     (4d)
```

"Martensite Type" $FeC_x$ ($x \le 0.06$) ($L2_0$): AB_tI4_139_b_a - CIF

```
# CIF file
data_findsym-output
_audit_creation_method FINDSYM

_chemical_name_mineral '``martensite type'''
_chemical_formula_sum 'C Fe'

loop_
_publ_author_name
 'E. A. Brandes'
 'G. B. Brook'
_journal_year 1992
_publ_Section_title
;
 Smithells Metals Reference Book
;

# Found in Strukturbericht 1913-1928, 1931

_aflow_title '``Martensite Type'' FeC$_{x}$ ($x \le 0.06$) ($L'2_{0}$)
    ↪ Structure'
_aflow_proto 'AB_tI4_139_b_a'
_aflow_params 'a,c/a'
_aflow_params_values '2.84,1.04577464789'
_aflow_Strukturbericht '$L'2_{0}$'
_aflow_Pearson 'tI4'

_symmetry_space_group_name_H-M "I 4/m 2/m 2/m"
_symmetry_Int_Tables_number 139

_cell_length_a      2.84000
_cell_length_b      2.84000
_cell_length_c      2.97000
_cell_angle_alpha  90.00000
_cell_angle_beta   90.00000
_cell_angle_gamma  90.00000

loop_
_space_group_symop_id
_space_group_symop_operation_xyz
1 x,y,z
2 x,-y,-z
3 -x,y,-z
4 -x,-y,z
5 -y,-x,-z
6 -y,x,z
7 y,-x,z
8 y,x,-z
9 -x,-y,-z
10 -x,y,z
11 x,-y,z
12 x,y,-z
13 y,x,z
14 y,-x,-z
15 -y,x,-z
16 -y,-x,z
17 x+1/2,y+1/2,z+1/2
18 x+1/2,-y+1/2,-z+1/2
19 -x+1/2,y+1/2,-z+1/2
20 -x+1/2,-y+1/2,z+1/2
21 -y+1/2,-x+1/2,-z+1/2
22 -y+1/2,x+1/2,z+1/2
23 y+1/2,-x+1/2,z+1/2
24 y+1/2,x+1/2,-z+1/2
25 -x+1/2,-y+1/2,-z+1/2
26 -x+1/2,y+1/2,z+1/2
27 x+1/2,-y+1/2,z+1/2
28 x+1/2,y+1/2,-z+1/2
29 y+1/2,x+1/2,z+1/2
30 y+1/2,-x+1/2,-z+1/2
31 -y+1/2,x+1/2,-z+1/2
32 -y+1/2,-x+1/2,z+1/2

loop_
_atom_site_label
_atom_site_type_symbol
_atom_site_symmetry_multiplicity
_atom_site_Wyckoff_label
_atom_site_fract_x
_atom_site_fract_y
_atom_site_fract_z
_atom_site_occupancy
Fe1 Fe 2 a 0.00000 0.00000 0.00000 1.00000
C1 C 2 b 0.00000 0.00000 0.50000 0.06000
```

"Martensite Type" $FeC_x$ ($x \le 0.06$) ($L2_0$): AB_tI4_139_b_a - POSCAR

```
AB_tI4_139_b_a & a,c/a --params=2.84,1.04577464789 & I4/mmm D_{4h}^{17}
    ↪ #139 (ab) & tI4 & $L'2_{0}$ & CxFe & ``martensite type'' & E.
    ↪ A. Brandes and G. B. Brook, (1992)
   1.00000000000000
  -1.42000000000000     1.42000000000000     1.48500000000000
   1.42000000000000    -1.42000000000000     1.48500000000000
   1.42000000000000     1.42000000000000    -1.48500000000000
     C       Fe
     1        1
Direct
    0.50000000000000     0.50000000000000     0.00000000000000     C     (2b)
    0.00000000000000     0.00000000000000     0.00000000000000     Fe    (2a)
```

$V_4SiSb_2$: A2BC4_tI28_140_h_a_k - CIF

```
# CIF file
data_findsym-output
_audit_creation_method FINDSYM

_chemical_name_mineral 'Sb2SiV4'
_chemical_formula_sum 'Sb2 Si V4'

loop_
_publ_author_name
 'P. Wollesen'
 'W. Jeitschko'
_journal_name_full_name
;
 Journal of Alloys and Compounds
;
_journal_volume 243
_journal_year 1996
_journal_page_first 67
_journal_page_last 69
_publ_Section_title
;
 V$_{4}$SiSb$_{2}$, a vanadium silicide antimonide crystallizing with a
    ↪ defect variant of the W$_{5}$Si$_{3}$-type structure
;

_aflow_title 'V$_{4}$SiSb$_{2}$ Structure'
_aflow_proto 'A2BC4_tI28_140_h_a_k'
_aflow_params 'a,c/a,x_{2},x_{3},y_{3}'
_aflow_params_values '9.87199,0.476803562402,0.14094,0.0852,0.20674'
_aflow_Strukturbericht 'None'
_aflow_Pearson 'tI28'

_symmetry_space_group_name_H-M "I 4/m 2/c 2/m"
_symmetry_Int_Tables_number 140

_cell_length_a      9.87199
_cell_length_b      9.87199
_cell_length_c      4.70700
_cell_angle_alpha  90.00000
_cell_angle_beta   90.00000
_cell_angle_gamma  90.00000

loop_
_space_group_symop_id
_space_group_symop_operation_xyz
1 x,y,z
2 x,-y,-z+1/2
3 -x,y,-z+1/2
4 -x,-y,z
5 -y,-x,-z+1/2
6 -y,x,z
7 y,-x,z
8 y,x,-z+1/2
9 -x,-y,-z
10 -x,y,z+1/2
11 x,-y,z+1/2
12 x,y,-z
13 y,x,z+1/2
14 y,-x,-z
15 -y,x,-z
16 -y,-x,z+1/2
17 x+1/2,y+1/2,z+1/2
18 x+1/2,-y+1/2,-z
19 -x+1/2,y+1/2,-z
20 -x+1/2,-y+1/2,z+1/2
21 -y+1/2,-x+1/2,-z
22 -y+1/2,x+1/2,z+1/2
23 y+1/2,-x+1/2,z+1/2
24 y+1/2,x+1/2,-z
25 -x+1/2,-y+1/2,-z+1/2
26 -x+1/2,y+1/2,z
27 x+1/2,-y+1/2,z
28 x+1/2,y+1/2,-z+1/2
29 y+1/2,x+1/2,z
30 y+1/2,-x+1/2,-z+1/2
31 -y+1/2,x+1/2,-z+1/2
32 -y+1/2,-x+1/2,z

loop_
_atom_site_label
_atom_site_type_symbol
_atom_site_symmetry_multiplicity
_atom_site_Wyckoff_label
_atom_site_fract_x
_atom_site_fract_y
_atom_site_fract_z
_atom_site_occupancy
Si1 Si 4 a 0.00000 0.00000 0.25000 1.00000
Sb1 Sb 8 h 0.14094 0.64094 0.00000 1.00000
V1 V 16 k 0.08520 0.20674 0.00000 1.00000
```

$V_4SiSb_2$: A2BC4_tI28_140_h_a_k - POSCAR

```
A2BC4_tI28_140_h_a_k & a,c/a,x2,x3,y3 --params=9.87199,0.476803562402,
    ↪ 0.14094,0.0852,0.20674 & I4/mcm D_{4h}^{18} #140 (ahk) & tI28 &
    ↪ None & Sb2SiV4 & Sb2SiV4 & P. Wollesen and W. Jeitschko, J.
    ↪ Alloys Compd. 243, 67-69 (1996)
   1.00000000000000
  -4.93599500000000     4.93599500000000     2.35350000000000
   4.93599500000000    -4.93599500000000     2.35350000000000
   4.93599500000000     4.93599500000000    -2.35350000000000
    Sb       Si       V
     4        2       8
```

```
Direct
    0.64094000000000    0.14094000000000    0.78188000000000    Sb    (8h)
    0.35906000000000   -0.14094000000000    0.21812000000000    Sb    (8h)
    0.14094000000000    0.35906000000000    0.50000000000000    Sb    (8h)
   -0.14094000000000    0.64094000000000    0.50000000000000    Sb    (8h)
    0.25000000000000    0.25000000000000    0.00000000000000    Si    (4a)
    0.75000000000000    0.75000000000000    0.00000000000000    Si    (4a)
    0.20674000000000    0.08520000000000    0.29194000000000    V    (16k)
   -0.20674000000000   -0.08520000000000   -0.29194000000000    V    (16k)
    0.08520000000000   -0.20674000000000   -0.12154000000000    V    (16k)
   -0.08520000000000    0.20674000000000    0.12154000000000    V    (16k)
    0.70674000000000    0.41480000000000    0.12154000000000    V    (16k)
    0.29326000000000    0.58520000000000   -0.12154000000000    V    (16k)
    0.58520000000000    0.70674000000000    0.29194000000000    V    (16k)
    0.41480000000000    0.29326000000000   -0.29194000000000    V    (16k)
```

$KHF_2$ ($F5_2$): A2BC_tI16_140_h_d_a - CIF

```
# CIF file
data_findsym-output
_audit_creation_method FINDSYM

_chemical_name_mineral 'F2HK'
_chemical_formula_sum 'F2 H K'

loop_
_publ_author_name
 'J. A. Ibers'
_journal_name_full_name
;
 Journal of Chemical Physics
;
_journal_volume 40
_journal_year 1964
_journal_page_first 402
_journal_page_last 404
_publ_Section_title
;
 Refinement of Peterson and Levy\'s Neutron Diffraction Data on KHF$_{2}
   ↪ $
;

_aflow_title 'KHF$_{2}$ ($F5_{2}$) Structure'
_aflow_proto 'A2BC_tI16_140_h_d_a'
_aflow_params 'a,c/a,x_{3}'
_aflow_params_values '5.67,1.20105820106,0.142'
_aflow_Strukturbericht '$F5_{2}$'
_aflow_Pearson 'tI16'

_symmetry_space_group_name_H-M "I 4/m 2/c 2/m"
_symmetry_Int_Tables_number 140

_cell_length_a      5.67000
_cell_length_b      5.67000
_cell_length_c      6.81000
_cell_angle_alpha  90.00000
_cell_angle_beta   90.00000
_cell_angle_gamma  90.00000

loop_
_space_group_symop_id
_space_group_symop_operation_xyz
1 x,y,z
2 x,-y,-z+1/2
3 -x,y,-z+1/2
4 -x,-y,z
5 -y,-x,-z+1/2
6 -y,x,z
7 y,-x,z
8 y,x,-z+1/2
9 -x,-y,-z
10 -x,y,z+1/2
11 x,-y,z+1/2
12 x,y,-z
13 y,x,z+1/2
14 y,-x,-z
15 -y,x,-z
16 -y,-x,z+1/2
17 x+1/2,y+1/2,z+1/2
18 x+1/2,-y+1/2,-z
19 -x+1/2,y+1/2,-z
20 -x+1/2,-y+1/2,z+1/2
21 -y+1/2,-x+1/2,-z
22 -y+1/2,x+1/2,z+1/2
23 y+1/2,-x+1/2,z+1/2
24 y+1/2,x+1/2,-z
25 -x+1/2,-y+1/2,-z+1/2
26 -x+1/2,y+1/2,z
27 x+1/2,-y+1/2,z
28 x+1/2,y+1/2,-z+1/2
29 y+1/2,x+1/2,z
30 y+1/2,-x+1/2,-z+1/2
31 -y+1/2,x+1/2,-z+1/2
32 -y+1/2,-x+1/2,z

loop_
_atom_site_label
_atom_site_type_symbol
_atom_site_symmetry_multiplicity
_atom_site_Wyckoff_label
_atom_site_fract_x
_atom_site_fract_y
_atom_site_fract_z
_atom_site_occupancy
K1 K 4 a 0.00000 0.00000 0.25000 1.00000
H1 H 4 d 0.00000 0.50000 0.00000 1.00000
F1 F 8 h 0.14200 0.64200 0.00000 1.00000
```

$KHF_2$ ($F5_2$): A2BC_tI16_140_h_d_a - POSCAR

```
A2BC_tI16_140_h_d_a & a,c/a,x3 --params=5.67,1.20105820106,0.142 & I4/
   ↪ mcm D_{4h}^{18} #140 (adh) & tI16 & $F5_{2}$ & F2HK & F2HK & J.
   ↪ A. Ibers, J. Chem. Phys. 40, 402-404 (1964)
   1.00000000000000
  -2.83500000000000    2.83500000000000    3.40500000000000
   2.83500000000000   -2.83500000000000    3.40500000000000
   2.83500000000000    2.83500000000000   -3.40500000000000
     F      H      K
     4      2      2
Direct
   0.64200000000000    0.14200000000000    0.78400000000000    F    (8h)
   0.35800000000000   -0.14200000000000    0.21600000000000    F    (8h)
   0.14200000000000    0.35800000000000    0.50000000000000    F    (8h)
  -0.14200000000000    0.64200000000000    0.50000000000000    F    (8h)
   0.50000000000000    0.00000000000000    0.50000000000000    H    (4d)
   0.00000000000000    0.50000000000000    0.50000000000000    H    (4d)
   0.25000000000000    0.25000000000000    0.00000000000000    K    (4a)
   0.75000000000000    0.75000000000000    0.00000000000000    K    (4a)
```

$Pu_{31}Rh_{20}$: A31B20_tI204_140_b2gh3m_ac2fh3l - CIF

```
# CIF file
data_findsym-output
_audit_creation_method FINDSYM

_chemical_name_mineral 'Pu31Rh20'
_chemical_formula_sum 'Pu31 Rh20'

loop_
_publ_author_name
 'D. T. Cromer'
 'A. C. Larson'
_journal_name_full_name
;
 Acta Crystallographica Section B: Structural Science
;
_journal_volume 33
_journal_year 1977
_journal_page_first 2620
_journal_page_last 2627
_publ_Section_title
;
 The Crystal Structure of Pu$_{31}$Pt$_{20}$ and Pu$_{31}$Rh$_{20}$
;

_aflow_title 'Pu$_{31}$Rh$_{20}$ Structure'
_aflow_proto 'A31B20_tI204_140_b2gh3m_ac2fh3l'
_aflow_params 'a,c/a,z_{4},z_{5},z_{6},z_{7},x_{8},x_{9},x_{10},z_{10},
   ↪ x_{11},z_{11},x_{12},z_{12},x_{13},y_{13},z_{13},x_{14},y_{14},
   ↪ z_{14},x_{15},y_{15},z_{15}'
_aflow_params_values '11.076,3.33450794511,0.094,0.1734,0.0756,0.1656,
   ↪ 0.1586,0.4035,0.1812,0.0726,0.3417,0.1269,0.1536,0.2109,0.2947,
   ↪ 0.4299,0.05,0.2855,0.5774,0.1346,0.2819,0.4125,0.2114'
_aflow_Strukturbericht 'None'
_aflow_Pearson 'tI204'

_symmetry_space_group_name_H-M "I 4/m 2/c 2/m"
_symmetry_Int_Tables_number 140

_cell_length_a     11.07600
_cell_length_b     11.07600
_cell_length_c     36.93301
_cell_angle_alpha  90.00000
_cell_angle_beta   90.00000
_cell_angle_gamma  90.00000

loop_
_space_group_symop_id
_space_group_symop_operation_xyz
1 x,y,z
2 x,-y,-z+1/2
3 -x,y,-z+1/2
4 -x,-y,z
5 -y,-x,-z+1/2
6 -y,x,z
7 y,-x,z
8 y,x,-z+1/2
9 -x,-y,-z
10 -x,y,z+1/2
11 x,-y,z+1/2
12 x,y,-z
13 y,x,z+1/2
14 y,-x,-z
15 -y,x,-z
16 -y,-x,z+1/2
17 x+1/2,y+1/2,z+1/2
18 x+1/2,-y+1/2,-z
19 -x+1/2,y+1/2,-z
20 -x+1/2,-y+1/2,z+1/2
21 -y+1/2,-x+1/2,-z
22 -y+1/2,x+1/2,z+1/2
23 y+1/2,-x+1/2,z+1/2
24 y+1/2,x+1/2,-z
25 -x+1/2,-y+1/2,-z+1/2
26 -x+1/2,y+1/2,z
27 x+1/2,-y+1/2,z
28 x+1/2,y+1/2,-z+1/2
29 y+1/2,x+1/2,z
30 y+1/2,-x+1/2,-z+1/2
31 -y+1/2,x+1/2,-z+1/2
```

```
32 -y+1/2,-x+1/2,z

loop_
_atom_site_label
_atom_site_type_symbol
_atom_site_symmetry_multiplicity
_atom_site_Wyckoff_label
_atom_site_fract_x
_atom_site_fract_y
_atom_site_fract_z
_atom_site_occupancy
Rh1 Rh 4 a 0.00000 0.00000 0.25000 1.00000
Pu1 Pu 4 b 0.00000 0.50000 0.25000 1.00000
Rh2 Rh 4 c 0.00000 0.00000 0.00000 1.00000
Rh3 Rh 8 f 0.00000 0.00000 0.09400 1.00000
Rh4 Rh 8 f 0.00000 0.00000 0.17340 1.00000
Pu2 Pu 8 g 0.00000 0.50000 0.07560 1.00000
Pu3 Pu 8 g 0.00000 0.50000 0.16560 1.00000
Pu4 Pu 8 h 0.15860 0.65860 0.00000 1.00000
Rh5 Rh 8 h 0.40350 0.90350 0.00000 1.00000
Rh6 Rh 16 l 0.18120 0.68120 0.07260 1.00000
Rh7 Rh 16 l 0.34170 0.84170 0.12690 1.00000
Rh8 Rh 16 l 0.15360 0.65360 0.21090 1.00000
Pu5 Pu 32 m 0.29470 0.42990 0.05000 1.00000
Pu6 Pu 32 m 0.28550 0.57740 0.13460 1.00000
Pu7 Pu 32 m 0.28190 0.41250 0.21140 1.00000
```

$Pu_{31}Rh_{20}$: A31B20_tI204_140_b2gh3m_ac2fh3l - POSCAR

```
A31B20_tI204_140_b2gh3m_ac2fh3l & a,c/a,z4,z5,z6,z7,x8,x9,x10,z10,x11,
    ↪ z11,x12,z12,x13,y13,z13,x14,y14,z14,x15,y15,z15 --params=11.076
    ↪ ,3.33450794511,0.094,0.1734,0.0756,0.1656,0.1586,0.4035,0.1812,
    ↪ 0.0726,0.3417,0.1269,0.1536,0.2109,0.2947,0.4299,0.05,0.2855,
    ↪ 0.5774,0.1346,0.2819,0.4125,0.2114 & I4/mcm D_{4h}^{18} #140 (
    ↪ abcf^2g^2h^21^3m^3) & tI204 & None & Pu31Rh20 & Pu31Rh20 & D.
    ↪ T. Cromer and A. C. Larson, Acta Crystallogr. Sect. B Struct.
    ↪ Sci. 33, 2620-2627 (1977)
   1.00000000000000
  -5.53800000000000   5.53800000000000  18.46650500000000
   5.53800000000000  -5.53800000000000  18.46650500000000
   5.53800000000000   5.53800000000000 -18.46650500000000
    Pu      Rh
    62      40
Direct
   0.75000000000000   0.25000000000000   0.50000000000000   Pu   (4b)
   0.25000000000000   0.75000000000000   0.50000000000000   Pu   (4b)
   0.57560000000000   0.07560000000000   0.50000000000000   Pu   (8g)
   0.07560000000000   0.57560000000000   0.50000000000000   Pu   (8g)
  -0.07560000000000   0.42440000000000   0.50000000000000   Pu   (8g)
   0.42440000000000  -0.07560000000000   0.50000000000000   Pu   (8g)
   0.66560000000000   0.16560000000000   0.50000000000000   Pu   (8g)
   0.16560000000000   0.66560000000000   0.50000000000000   Pu   (8g)
  -0.16560000000000   0.33440000000000   0.50000000000000   Pu   (8g)
   0.33440000000000  -0.16560000000000   0.50000000000000   Pu   (8g)
   0.65860000000000   0.15860000000000   0.81720000000000   Pu   (8h)
   0.34140000000000  -0.15860000000000   0.18280000000000   Pu   (8h)
   0.15860000000000   0.34140000000000   0.50000000000000   Pu   (8h)
  -0.15860000000000   0.65860000000000   0.50000000000000   Pu   (8h)
   0.47990000000000   0.34470000000000   0.72460000000000   Pu  (32m)
  -0.37990000000000  -0.24470000000000  -0.72460000000000   Pu  (32m)
   0.34470000000000  -0.37990000000000  -0.13520000000000   Pu  (32m)
  -0.24470000000000   0.47990000000000   0.13520000000000   Pu  (32m)
   0.87990000000000   0.15530000000000   0.13520000000000   Pu  (32m)
   0.02010000000000   0.74470000000000  -0.13520000000000   Pu  (32m)
   0.74470000000000   0.87990000000000   0.72460000000000   Pu  (32m)
   0.15530000000000   0.02010000000000  -0.72460000000000   Pu  (32m)
  -0.47990000000000  -0.34470000000000  -0.72460000000000   Pu  (32m)
   0.37990000000000   0.24470000000000   0.72460000000000   Pu  (32m)
  -0.34470000000000   0.37990000000000   0.13520000000000   Pu  (32m)
   0.24470000000000  -0.47990000000000  -0.13520000000000   Pu  (32m)
   0.12010000000000   0.84470000000000  -0.13520000000000   Pu  (32m)
   0.97990000000000   0.25530000000000   0.13520000000000   Pu  (32m)
   0.25530000000000   0.12010000000000  -0.72460000000000   Pu  (32m)
   0.84470000000000   0.97990000000000   0.72460000000000   Pu  (32m)
   0.71200000000000   0.42010000000000   0.86290000000000   Pu  (32m)
  -0.44280000000000  -0.15090000000000  -0.86290000000000   Pu  (32m)
   0.42010000000000  -0.44280000000000  -0.29190000000000   Pu  (32m)
  -0.15090000000000   0.71200000000000   0.29190000000000   Pu  (32m)
   0.94280000000000   0.07990000000000   0.29190000000000   Pu  (32m)
  -0.21200000000000   0.65090000000000  -0.29190000000000   Pu  (32m)
   0.65090000000000   0.94280000000000   0.86290000000000   Pu  (32m)
   0.07990000000000  -0.21200000000000  -0.86290000000000   Pu  (32m)
  -0.71200000000000  -0.42010000000000  -0.86290000000000   Pu  (32m)
   0.44280000000000   0.15090000000000   0.86290000000000   Pu  (32m)
  -0.42010000000000   0.44280000000000   0.29190000000000   Pu  (32m)
   0.15090000000000  -0.71200000000000  -0.29190000000000   Pu  (32m)
   0.05720000000000   0.92010000000000  -0.29190000000000   Pu  (32m)
   1.21200000000000   0.34910000000000   0.29190000000000   Pu  (32m)
   0.34910000000000   0.05720000000000  -0.86290000000000   Pu  (32m)
   0.92010000000000   1.21200000000000   0.86290000000000   Pu  (32m)
   0.62390000000000   0.49330000000000   0.69440000000000   Pu  (32m)
  -0.20110000000000  -0.07050000000000  -0.69440000000000   Pu  (32m)
   0.49330000000000  -0.20110000000000  -0.13060000000000   Pu  (32m)
  -0.07050000000000   0.62390000000000   0.13060000000000   Pu  (32m)
   0.70110000000000   0.00670000000000   0.13060000000000   Pu  (32m)
  -0.12390000000000   0.57050000000000  -0.13060000000000   Pu  (32m)
   0.57050000000000   0.70110000000000   0.69440000000000   Pu  (32m)
   0.00670000000000  -0.12390000000000  -0.69440000000000   Pu  (32m)
  -0.62390000000000  -0.49330000000000  -0.69440000000000   Pu  (32m)
   0.20110000000000   0.07050000000000   0.69440000000000   Pu  (32m)
  -0.49330000000000   0.20110000000000   0.13060000000000   Pu  (32m)
   0.07050000000000  -0.62390000000000  -0.13060000000000   Pu  (32m)
   0.29890000000000   0.99330000000000  -0.13060000000000   Pu  (32m)
   1.12390000000000   0.42950000000000   0.13060000000000   Pu  (32m)
   0.42950000000000   0.29890000000000  -0.69440000000000   Pu  (32m)
   0.99330000000000   1.12390000000000   0.69440000000000   Pu  (32m)
   0.25000000000000   0.25000000000000   0.00000000000000   Rh   (4a)
   0.75000000000000   0.75000000000000   0.00000000000000   Rh   (4a)
   0.00000000000000   0.00000000000000   0.00000000000000   Rh   (4c)
   0.50000000000000   0.50000000000000   0.00000000000000   Rh   (4c)
   0.09400000000000   0.09400000000000   0.00000000000000   Rh   (8f)
   0.40600000000000   0.40600000000000   0.00000000000000   Rh   (8f)
  -0.09400000000000  -0.09400000000000   0.00000000000000   Rh   (8f)
   0.59400000000000   0.59400000000000   0.00000000000000   Rh   (8f)
   0.17340000000000   0.17340000000000   0.00000000000000   Rh   (8f)
   0.32660000000000   0.32660000000000   0.00000000000000   Rh   (8f)
  -0.17340000000000  -0.17340000000000   0.00000000000000   Rh   (8f)
   0.67340000000000   0.67340000000000   0.00000000000000   Rh   (8f)
   0.90350000000000   0.40350000000000   1.30700000000000   Rh   (8h)
   0.09650000000000  -0.40350000000000  -0.30700000000000   Rh   (8h)
   0.40350000000000   0.09650000000000   0.50000000000000   Rh   (8h)
  -0.40350000000000   0.90350000000000   0.50000000000000   Rh   (8h)
   0.75380000000000   0.25380000000000   0.86240000000000   Rh  (16l)
   0.39140000000000  -0.10860000000000   0.13760000000000   Rh  (16l)
   0.25380000000000   0.39140000000000   0.50000000000000   Rh  (16l)
  -0.10860000000000   0.75380000000000   0.50000000000000   Rh  (16l)
   0.10860000000000   0.24620000000000   0.50000000000000   Rh  (16l)
  -0.25380000000000   0.60860000000000   0.50000000000000   Rh  (16l)
   0.60860000000000   0.10860000000000   0.86240000000000   Rh  (16l)
   0.24620000000000  -0.25380000000000   0.13760000000000   Rh  (16l)
   0.96860000000000   0.46860000000000   1.18340000000000   Rh  (16l)
   0.28520000000000  -0.21480000000000  -0.18340000000000   Rh  (16l)
   0.46860000000000   0.28520000000000   0.50000000000000   Rh  (16l)
  -0.21480000000000   0.96860000000000   0.50000000000000   Rh  (16l)
   0.21480000000000   0.03140000000000   0.50000000000000   Rh  (16l)
  -0.46860000000000   0.71480000000000   0.50000000000000   Rh  (16l)
   0.71480000000000   0.21480000000000   1.18340000000000   Rh  (16l)
   0.03140000000000  -0.46860000000000  -0.18340000000000   Rh  (16l)
   0.86450000000000   0.36450000000000   0.80720000000000   Rh  (16l)
   0.55730000000000   0.05730000000000   0.19280000000000   Rh  (16l)
   0.36450000000000   0.55730000000000   0.50000000000000   Rh  (16l)
   0.05730000000000   0.86450000000000   0.50000000000000   Rh  (16l)
  -0.05730000000000   0.13550000000000   0.50000000000000   Rh  (16l)
  -0.36450000000000   0.44270000000000   0.50000000000000   Rh  (16l)
   0.44270000000000  -0.05730000000000   0.80720000000000   Rh  (16l)
   0.13550000000000  -0.36450000000000   0.19280000000000   Rh  (16l)
```

$NH_4Pb_2Br_5$ ($K3_4$): A5BC2_tI32_140_bl_a_h - CIF

```
# CIF file
data_findsym-output
_audit_creation_method FINDSYM

_chemical_name_mineral 'Br5(NH4)Pb2'
_chemical_formula_sum 'Br5 (NH4) Pb2'

loop_
_publ_author_name
 'H. M. Powell'
 'H. S. Tasker'
_journal_year 1937
_publ_Section_title
;
 The valency angle of bivalent lead: the crystal structure of ammonium,
     ↪ rubidium, and potassium pentabromodiplumbites
;

# Found in Strukturbericht Band V 1937, 1940

_aflow_title 'NH$_{4}$Pb$_{2}$Br$_{5}$ ($K3_{4}$) Structure'
_aflow_proto 'A5BC2_tI32_140_bl_a_h'
_aflow_params 'a,c/a,x_{3},x_{4},z_{4}'
_aflow_params_values '8.39,1.70917759237,0.158,0.163,0.363'
_aflow_Strukturbericht '$K3_{4}$'
_aflow_Pearson 'tI32'

_symmetry_space_group_name_H-M "I 4/m 2/c 2/m"
_symmetry_Int_Tables_number 140

_cell_length_a      8.39000
_cell_length_b      8.39000
_cell_length_c      14.34000
_cell_angle_alpha  90.00000
_cell_angle_beta   90.00000
_cell_angle_gamma  90.00000

loop_
_space_group_symop_id
_space_group_symop_operation_xyz
1 x,y,z
2 x,-y,-z+1/2
3 -x,y,-z+1/2
4 -x,-y,z
5 -y,-x,-z+1/2
6 -y,x,z
7 y,-x,z
8 y,x,-z+1/2
9 -x,-y,-z
10 -x,y,z+1/2
11 x,-y,z+1/2
12 x,y,-z
13 y,x,z+1/2
14 y,-x,-z
15 -y,x,-z
16 -y,-x,z+1/2
17 x+1/2,y+1/2,z+1/2
18 x+1/2,-y+1/2,-z
19 -x+1/2,y+1/2,-z
20 -x+1/2,-y+1/2,z+1/2
21 -y+1/2,-x+1/2,-z
```

```
22 -y+1/2,x+1/2,z+1/2
23 y+1/2,-x+1/2,z+1/2
24 y+1/2,x+1/2,-z
25 -x+1/2,-y+1/2,-z+1/2
26 -x+1/2,y+1/2,z
27 x+1/2,-y+1/2,z
28 x+1/2,y+1/2,-z+1/2
29 y+1/2,x+1/2,z
30 y+1/2,-x+1/2,-z+1/2
31 -y+1/2,x+1/2,-z+1/2
32 -y+1/2,-x+1/2,z

loop_
_atom_site_label
_atom_site_type_symbol
_atom_site_symmetry_multiplicity
_atom_site_Wyckoff_label
_atom_site_fract_x
_atom_site_fract_y
_atom_site_fract_z
_atom_site_occupancy
NH41 NH4 4 a 0.00000 0.00000 0.25000 1.00000
Br1 Br 4 b 0.00000 0.50000 0.25000 1.00000
Pb1 Pb 8 h 0.15800 0.65800 0.00000 1.00000
Br2 Br 16 l 0.16300 0.66300 0.36300 1.00000
```

$NH_4Pb_2Br_5$ ($K3_4$): A5BC2_tI32_140_bl_a_h - POSCAR

```
A5BC2_tI32_140_bl_a_h & a,c/a,x3,x4,z4 --params=8.39,1.70917759237,0.158
   ↪ ,0.163,0.363 & I4/mcm D_{4h}^{18} #140 (abhl) & tI32 & $K3_{4}$
   ↪ & Br5(NH4)Pb2 & Br5(NH4)Pb2 & H. M. Powell and H. S. Tasker, (
   ↪ 1937)
   1.00000000000000
  -4.19500000000000   4.19500000000000   7.17000000000000
   4.19500000000000  -4.19500000000000   7.17000000000000
   4.19500000000000   4.19500000000000  -7.17000000000000
    Br    NH4    Pb
    10     2      4
Direct
   0.75000000000000   0.25000000000000   0.50000000000000   Br    (4b)
   0.25000000000000   0.75000000000000   0.50000000000000   Br    (4b)
   1.02600000000000   0.52600000000000   0.82600000000000   Br   (16l)
   0.70000000000000   0.20000000000000   0.17400000000000   Br   (16l)
   0.52600000000000   0.70000000000000   0.50000000000000   Br   (16l)
   0.20000000000000   1.02600000000000   0.50000000000000   Br   (16l)
  -0.20000000000000  -0.02600000000000   0.50000000000000   Br   (16l)
  -0.52600000000000   0.30000000000000   0.50000000000000   Br   (16l)
   0.30000000000000  -0.20000000000000   0.82600000000000   Br   (16l)
  -0.02600000000000  -0.52600000000000   0.17400000000000   Br   (16l)
   0.25000000000000   0.25000000000000   0.00000000000000   NH4   (4a)
   0.75000000000000   0.75000000000000   0.00000000000000   NH4   (4a)
   0.65800000000000   0.15800000000000   0.81600000000000   Pb    (8h)
   0.34200000000000  -0.15800000000000   0.18400000000000   Pb    (8h)
   0.15800000000000   0.34200000000000   0.50000000000000   Pb    (8h)
  -0.15800000000000   0.65800000000000   0.50000000000000   Pb    (8h)
```

$Cs_3CoCl_5$ ($K3_1$): A5BC3_tI36_140_cl_b_ah - CIF

```
# CIF file
data_findsym-output
_audit_creation_method FINDSYM

_chemical_name_mineral 'Cl5CoCs3'
_chemical_formula_sum 'Cl5 Co Cs3'

loop_
_publ_author_name
 'B. N. Figgis'
 'R. Mason'
 'A. R. P. Smith'
 'G. A. Williams'
_journal_name_full_name
;
 Acta Crystallographica Section B: Structural Science
;
_journal_volume 36
_journal_year 1980
_journal_page_first 509
_journal_page_last 512
_publ_Section_title
;
 Neutron Diffraction Structure of Cs$_{3}$CoCl$_{5}$ at 4.2~K
;

_aflow_title 'Cs$_{3}$CoCl$_{5}$ ($K3_{1}$) Structure'
_aflow_proto 'A5BC3_tI36_140_cl_b_ah'
_aflow_params 'a,c/a,x_{4},x_{5},z_{5}'
_aflow_params_values '9.063,1.59439479201,0.66225,0.1421,0.15711'
_aflow_Strukturbericht '$K3_{1}$'
_aflow_Pearson 'tI36'

_symmetry_space_group_name_H-M "I 4/m 2/c 2/m"
_symmetry_Int_Tables_number 140

_cell_length_a     9.06300
_cell_length_b     9.06300
_cell_length_c     14.45000
_cell_angle_alpha 90.00000
_cell_angle_beta  90.00000
_cell_angle_gamma 90.00000

loop_
_space_group_symop_id
_space_group_symop_operation_xyz
1 x,y,z
2 x,-y,-z+1/2
3 -x,y,-z+1/2
4 -x,-y,z
5 -y,-x,-z+1/2
6 -y,x,z
7 y,-x,z
8 y,x,-z+1/2
9 -x,-y,-z
10 -x,y,z+1/2
11 x,-y,z+1/2
12 x,y,-z
13 y,x,z+1/2
14 y,-x,-z
15 -y,x,-z
16 -y,-x,z+1/2
17 x+1/2,y+1/2,z+1/2
18 x+1/2,-y+1/2,-z
19 -x+1/2,y+1/2,-z
20 -x+1/2,-y+1/2,z+1/2
21 -y+1/2,-x+1/2,-z
22 -y+1/2,x+1/2,z+1/2
23 y+1/2,-x+1/2,z+1/2
24 y+1/2,x+1/2,-z
25 -x+1/2,-y+1/2,-z+1/2
26 -x+1/2,y+1/2,z
27 x+1/2,-y+1/2,z
28 x+1/2,y+1/2,-z+1/2
29 y+1/2,x+1/2,z
30 y+1/2,-x+1/2,-z+1/2
31 -y+1/2,x+1/2,-z+1/2
32 -y+1/2,-x+1/2,z

loop_
_atom_site_label
_atom_site_type_symbol
_atom_site_symmetry_multiplicity
_atom_site_Wyckoff_label
_atom_site_fract_x
_atom_site_fract_y
_atom_site_fract_z
_atom_site_occupancy
Cs1 Cs 4 a 0.00000 0.00000 0.25000 1.00000
Co1 Co 4 b 0.00000 0.50000 0.25000 1.00000
Cl1 Cl 4 c 0.00000 0.00000 0.00000 1.00000
Cs2 Cs 8 h 0.66225 0.16225 0.00000 1.00000
Cl2 Cl 16 l 0.14210 0.64210 0.15711 1.00000
```

$Cs_3CoCl_5$ ($K3_1$): A5BC3_tI36_140_cl_b_ah - POSCAR

```
A5BC3_tI36_140_cl_b_ah & a,c/a,x4,x5,z5 --params=9.063,1.59439479201,
   ↪ 0.66225,0.1421,0.15711 & I4/mcm D_{4h}^{18} #140 (abchl) & tI36
   ↪ & $K3_{1}$ & Cl5CoCs3 & Cl5CoCs3 & B. N. Figgis et al., Acta
   ↪ Crystallogr. Sect. B Struct. Sci. 36, 509-512 (1980)
   1.00000000000000
  -4.53150000000000   4.53150000000000   7.22500000000000
   4.53150000000000  -4.53150000000000   7.22500000000000
   4.53150000000000   4.53150000000000  -7.22500000000000
    Cl    Co    Cs
    10     2     6
Direct
   0.00000000000000   0.00000000000000   0.00000000000000   Cl    (4c)
   0.50000000000000   0.50000000000000   0.00000000000000   Cl    (4c)
   0.79921000000000   0.29921000000000   0.78420000000000   Cl   (16l)
   0.51501000000000   0.01501000000000   0.21580000000000   Cl   (16l)
   0.29921000000000   0.51501000000000   0.50000000000000   Cl   (16l)
   0.01501000000000   0.79921000000000   0.50000000000000   Cl   (16l)
  -0.01501000000000   0.20079000000000   0.50000000000000   Cl   (16l)
  -0.29921000000000   0.48499000000000   0.50000000000000   Cl   (16l)
   0.48499000000000  -0.01501000000000   0.78420000000000   Cl   (16l)
   0.20079000000000  -0.29921000000000   0.21580000000000   Cl   (16l)
   0.75000000000000   0.25000000000000   0.50000000000000   Co    (4b)
   0.25000000000000   0.75000000000000   0.50000000000000   Co    (4b)
   0.25000000000000   0.25000000000000   0.00000000000000   Cs    (4a)
   0.75000000000000   0.75000000000000   0.00000000000000   Cs    (4a)
   1.16225000000000   0.66225000000000   1.82450000000000   Cs    (8h)
  -0.16225000000000  -0.66225000000000  -0.82450000000000   Cs    (8h)
   0.66225000000000  -0.16225000000000   0.50000000000000   Cs    (8h)
  -0.66225000000000   1.16225000000000   0.50000000000000   Cs    (8h)
```

$U_6Mn$ ($D2_c$): AB6_tI28_140_a_hk - CIF

```
# CIF file
data_findsym-output
_audit_creation_method FINDSYM

_chemical_name_mineral 'MnU6'
_chemical_formula_sum 'Mn U6'

loop_
_publ_author_name
 'N. C. Baenziger'
 'R. E. Rundle'
 'A. I. Snow'
 'A. S. Wilson'
_journal_name_full_name
;
 Acta Cristallographica
;
_journal_volume 3
_journal_year 1950
_journal_page_first 34
_journal_page_last 40
_publ_Section_title
;
```

```
 Compounds of uranium with the transition metals of the first long
      ↪ period
;

# Found in PAULING FILE, 2016 Found in PAULING FILE, {in: Inorganic
      ↪ Solid Phases, SpringerMaterials (online database), Springer,
      ↪ Heidelberg},

_aflow_title 'U$_{6}$Mn ($D2_{c}$) Structure'
_aflow_proto 'AB6_tI28_140_a_hk'
_aflow_params 'a,c/a,x_{2},x_{3},y_{3}'
_aflow_params_values '10.29,0.509232264334,0.4068,0.2141,0.1021'
_aflow_Strukturbericht '$D2_{c}$'
_aflow_Pearson 'tI28'

_symmetry_space_group_name_H-M "I 4/m 2/c 2/m"
_symmetry_Int_Tables_number 140

_cell_length_a     10.29000
_cell_length_b     10.29000
_cell_length_c     5.24000
_cell_angle_alpha 90.00000
_cell_angle_beta  90.00000
_cell_angle_gamma 90.00000

loop_
_space_group_symop_id
_space_group_symop_operation_xyz
1 x,y,z
2 x,-y,-z+1/2
3 -x,y,-z+1/2
4 -x,-y,z
5 -y,-x,-z+1/2
6 -y,x,z
7 y,-x,z
8 y,x,-z+1/2
9 -x,-y,-z
10 -x,y,z+1/2
11 x,-y,z+1/2
12 x,y,-z
13 y,x,z+1/2
14 y,-x,-z
15 -y,x,-z
16 -y,-x,z+1/2
17 x+1/2,y+1/2,z+1/2
18 x+1/2,-y+1/2,-z
19 -x+1/2,y+1/2,-z
20 -x+1/2,-y+1/2,z+1/2
21 -y+1/2,-x+1/2,-z
22 -y+1/2,x+1/2,z+1/2
23 y+1/2,-x+1/2,z+1/2
24 y+1/2,x+1/2,-z
25 -x+1/2,-y+1/2,-z+1/2
26 -x+1/2,y+1/2,z
27 x+1/2,-y+1/2,z
28 x+1/2,y+1/2,-z+1/2
29 y+1/2,x+1/2,z
30 y+1/2,-x+1/2,-z+1/2
31 -y+1/2,x+1/2,-z+1/2
32 -y+1/2,-x+1/2,z

loop_
_atom_site_label
_atom_site_type_symbol
_atom_site_symmetry_multiplicity
_atom_site_Wyckoff_label
_atom_site_fract_x
_atom_site_fract_y
_atom_site_fract_z
_atom_site_occupancy
Mn1 Mn 4 a 0.00000 0.00000 0.25000 1.00000
U1 U 8 h 0.40680 -0.09320 0.00000 1.00000
U2 U 16 k 0.21410 0.10210 0.00000 1.00000
```

$U_6$Mn ($D2_c$): AB6_tI28_140_a_hk - POSCAR

```
AB6_tI28_140_a_hk & a,c/a,x2,x3,y3 --params=10.29,0.509232264334,0.4068,
      ↪ 0.2141,0.1021 & I4/mcm D_{4h}^{18} #140 (ahk) & tI28 & $D2_{c}$
      ↪ & MnU6 & MnU6 & N. C. Baenziger et al., Acta Cryst. 3, 34-40 (
      ↪ 1950)
   1.00000000000000
  -5.14500000000000   5.14500000000000   2.62000000000000
   5.14500000000000  -5.14500000000000   2.62000000000000
   5.14500000000000   5.14500000000000  -2.62000000000000
    Mn       U
     2      12
Direct
   0.25000000000000   0.25000000000000   0.00000000000000   Mn   (4a)
   0.75000000000000   0.75000000000000   0.00000000000000   Mn   (4a)
   0.90680000000000   0.40680000000000   1.31360000000000   U    (8h)
   0.09320000000000  -0.40680000000000  -0.31360000000000   U    (8h)
   0.40680000000000   0.09320000000000   0.50000000000000   U    (8h)
  -0.40680000000000   0.90680000000000   0.50000000000000   U    (8h)
   0.10210000000000   0.21410000000000   0.31620000000000   U   (16k)
  -0.10210000000000  -0.21410000000000  -0.31620000000000   U   (16k)
   0.21410000000000  -0.10210000000000   0.11200000000000   U   (16k)
  -0.21410000000000   0.10210000000000  -0.11200000000000   U   (16k)
   0.60210000000000   0.28590000000000  -0.11200000000000   U   (16k)
   0.39790000000000   0.71410000000000   0.11200000000000   U   (16k)
   0.71410000000000   0.60210000000000   0.31620000000000   U   (16k)
   0.28590000000000   0.39790000000000  -0.31620000000000   U   (16k)
```

$BaCd_{11}$: AB11_tI48_141_a_bdi - CIF

```
# CIF file
data_findsym-output
_audit_creation_method FINDSYM

_chemical_name_mineral 'BaCd11'
_chemical_formula_sum 'Ba Cd11'

loop_
_publ_author_name
 'M. J. Sanderson'
 'N. C. Baenziger'
_journal_name_full_name
;
 Acta Cristallographica
;
_journal_volume 6
_journal_year 1953
_journal_page_first 627
_journal_page_last 631
_publ_Section_title
;
 The Crystal Structure of BaCd$_{11}$
;

_aflow_title 'BaCd$_{11}$ Structure'
_aflow_proto 'AB11_tI48_141_a_bdi'
_aflow_params 'a,c/a,x_{4},y_{4},z_{4}'
_aflow_params_values '12.02,0.643926788686,0.123,0.455,0.183'
_aflow_Strukturbericht 'None'
_aflow_Pearson 'tI48'

_symmetry_space_group_name_H-M "I 41/a 2/m 2/d (origin choice 2)"
_symmetry_Int_Tables_number 141

_cell_length_a     12.02000
_cell_length_b     12.02000
_cell_length_c     7.74000
_cell_angle_alpha 90.00000
_cell_angle_beta  90.00000
_cell_angle_gamma 90.00000

loop_
_space_group_symop_id
_space_group_symop_operation_xyz
1 x,y,z
2 x,-y,-z
3 -x,y+1/2,-z
4 -x,-y+1/2,z
5 -y+1/4,-x+1/4,-z+3/4
6 -y+1/4,x+3/4,z+1/4
7 y+3/4,-x+3/4,z+1/4
8 y+3/4,x+1/4,-z+3/4
9 -x,-y,-z
10 -x,y,z
11 x,-y+1/2,z
12 x,y+1/2,-z
13 y+3/4,x+3/4,z+1/4
14 y+3/4,-x+1/4,-z+3/4
15 -y+1/4,x+1/4,-z+3/4
16 -y+1/4,-x+3/4,z+1/4
17 x+1/2,y+1/2,z+1/2
18 x+1/2,-y+1/2,-z+1/2
19 -x+1/2,y,-z+1/2
20 -x+1/2,-y,z+1/2
21 -y+3/4,-x+3/4,-z+1/4
22 -y+3/4,x+1/4,z+3/4
23 y+1/4,-x+1/4,z+3/4
24 y+1/4,x+3/4,-z+1/4
25 -x+1/2,-y+1/2,-z+1/2
26 -x+1/2,y+1/2,z+1/2
27 x+1/2,-y,z+1/2
28 x+1/2,y,-z+1/2
29 y+1/4,x+1/4,z+3/4
30 y+1/4,-x+3/4,-z+1/4
31 -y+3/4,x+3/4,-z+1/4
32 -y+3/4,-x+1/4,z+3/4

loop_
_atom_site_label
_atom_site_type_symbol
_atom_site_symmetry_multiplicity
_atom_site_Wyckoff_label
_atom_site_fract_x
_atom_site_fract_y
_atom_site_fract_z
_atom_site_occupancy
Ba1 Ba 4 a 0.00000 0.75000 0.12500 1.00000
Cd1 Cd 4 b 0.00000 0.25000 0.37500 1.00000
Cd2 Cd 8 d 0.00000 0.00000 0.50000 1.00000
Cd3 Cd 32 i 0.12300 0.45500 0.18300 1.00000
```

$BaCd_{11}$: AB11_tI48_141_a_bdi - POSCAR

```
AB11_tI48_141_a_bdi & a,c/a,x4,y4,z4 --params=12.02,0.643926788686,0.123
      ↪ ,0.455,0.183 & I4_{1}/amd D_{4h}^{19} #141 (abdi) & tI48 & None
      ↪ & BaCd11 & BaCd11 & M. J. Sanderson and N. C. Baenziger, Acta
      ↪ Cryst. 6, 627-631 (1953)
   1.00000000000000
  -6.01000000000000   6.01000000000000   3.87000000000000
   6.01000000000000  -6.01000000000000   3.87000000000000
   6.01000000000000   6.01000000000000  -3.87000000000000
    Ba      Cd
     2      22
Direct
   0.87500000000000   0.12500000000000   0.75000000000000   Ba   (4a)
   0.12500000000000   0.87500000000000   0.25000000000000   Ba   (4a)
```

```
  0.62500000000000     0.37500000000000     0.25000000000000    Cd     (4b)
  0.37500000000000     0.62500000000000     0.75000000000000    Cd     (4b)
  0.50000000000000     0.50000000000000     0.00000000000000    Cd     (8d)
  0.00000000000000     0.50000000000000     0.50000000000000    Cd     (8d)
  0.50000000000000     0.00000000000000     0.00000000000000    Cd     (8d)
  0.50000000000000     0.50000000000000     0.50000000000000    Cd     (8d)
  0.63800000000000     0.30600000000000     0.57800000000000    Cd    (32i)
  0.22800000000000     0.06000000000000    -0.07800000000000    Cd    (32i)
  0.30600000000000     0.22800000000000    -0.33200000000000    Cd    (32i)
  0.06000000000000     0.63800000000000     0.83200000000000    Cd    (32i)
  0.77200000000000    -0.30600000000000     0.83200000000000    Cd    (32i)
 -0.63800000000000    -0.06000000000000    -0.33200000000000    Cd    (32i)
 -0.06000000000000     0.77200000000000     0.57800000000000    Cd    (32i)
 -0.30600000000000    -0.63800000000000    -0.07800000000000    Cd    (32i)
 -0.63800000000000    -0.30600000000000    -0.57800000000000    Cd    (32i)
  0.77200000000000    -0.06000000000000     1.07800000000000    Cd    (32i)
 -0.30600000000000     0.77200000000000     0.33200000000000    Cd    (32i)
 -0.06000000000000    -0.63800000000000     0.16800000000000    Cd    (32i)
  0.22800000000000     0.30600000000000     0.16800000000000    Cd    (32i)
  0.63800000000000     0.06000000000000     0.33200000000000    Cd    (32i)
  0.06000000000000     0.22800000000000    -0.57800000000000    Cd    (32i)
  0.30600000000000     0.63800000000000     1.07800000000000    Cd    (32i)
```

Analcime ($NaAlSi_2O_6 \cdot H_2O$, $S6_1$): A2B2C3D12E4_tI184_142_f_f_be_3g_g - CIF

```
# CIF file
data_findsym-output
_audit_creation_method FINDSYM

_chemical_name_mineral 'Analcime'
_chemical_formula_sum 'Al2 (H2O)2 Na3 O12 Si4'

loop_
_publ_author_name
 'F. Mazzi'
 'E. Galli'
_journal_name_full_name
;
 American Mineralogist
;
_journal_volume 63
_journal_year 1978
_journal_page_first 448
_journal_page_last 460
_publ_Section_title
;
 Is each analcime different?
;

# Found in The crystal structure of natural monoclinic analcime (
    ↪ NaAlSi$_{2}$O$_{6}$$\cdot$H$_{2}$O), 1988

_aflow_title 'Analcime (NaAlSi$_{2}$O$_{6}$$\cdot$H$_{2}$O, $S6_{1}$)
    ↪ Structure'
_aflow_proto 'A2B2C3D12E4_tI184_142_f_f_be_3g_g'
_aflow_params 'a,c/a,x_{2},y_{2},x_{3},x_{4},x_{5},y_{5},z_{5},x_{6},y_{
    ↪ 6},z_{6},x_{7},y_{7},z_{7},x_{8},y_{8},z_{8}'
_aflow_params_values '13.723,0.997303796546,0.1242,0.0,0.1631,0.3805,
    ↪ 0.105,0.3705,0.2186,0.2218,0.1028,0.3633,0.3631,0.2182,0.1051,
    ↪ 0.1264,0.1617,0.4118'
_aflow_Strukturbericht '$S6_{1}$'
_aflow_Pearson 'tI184'

_symmetry_space_group_name_H-M "I 41/a 2/c 2/d (origin choice 2)"
_symmetry_Int_Tables_number 142

_cell_length_a      13.72300
_cell_length_b      13.72300
_cell_length_c      13.68600
_cell_angle_alpha   90.00000
_cell_angle_beta    90.00000
_cell_angle_gamma   90.00000

loop_
_space_group_symop_id
_space_group_symop_operation_xyz
1 x,y,z
2 x+1/2,-y+1/2,-z
3 -x+1/2,y,-z
4 -x,-y+1/2,z
5 -y+1/4,-x+1/4,-z+1/4
6 -y+1/4,x+3/4,z+1/4
7 y+3/4,-x+3/4,z+1/4
8 y+3/4,x+1/4,-z+1/4
9 -x,-y,-z
10 -x,y,z+1/2
11 x,-y+1/2,z+1/2
12 x,y+1/2,-z
13 y+1/4,x+1/4,z+1/4
14 y+3/4,-x+1/4,-z+3/4
15 -y+1/4,x+1/4,-z+3/4
16 -y+3/4,-x+1/4,z+1/4
17 x+1/2,y+1/2,z+1/2
18 x,-y,-z+1/2
19 -x,y+1/2,-z+1/2
20 -x+1/2,-y,z+1/2
21 -y+3/4,-x+3/4,-z+3/4
22 -y+3/4,x+1/4,z+3/4
23 y+1/4,-x+1/4,z+3/4
24 y+1/4,x+3/4,-z+3/4
25 -x+1/2,-y+1/2,-z+1/2
26 -x+1/2,y+1/2,z
27 x+1/2,-y,z
28 x+1/2,y,-z+1/2
29 y+3/4,x+3/4,z+3/4
30 y+1/4,-x+3/4,-z+1/4
31 -y+3/4,x+3/4,-z+1/4
32 -y+1/4,-x+3/4,z+3/4

loop_
_atom_site_label
_atom_site_type_symbol
_atom_site_symmetry_multiplicity
_atom_site_Wyckoff_label
_atom_site_fract_x
_atom_site_fract_y
_atom_site_fract_z
_atom_site_occupancy
Na1 Na 8 b 0.00000 0.25000 0.12500 0.23000
Na2 Na 16 e 0.12420 0.00000 0.25000 0.82000
Al1 Al 16 f 0.16310 0.41310 0.12500 1.00000
H2O1 H2O 16 f 0.38050 0.63050 0.12500 1.00000
O1 O 32 g 0.10500 0.37050 0.21860 1.00000
O2 O 32 g 0.22180 0.10280 0.36330 1.00000
O3 O 32 g 0.36310 0.21820 0.10510 1.00000
Si1 Si 32 g 0.12640 0.16170 0.41180 1.00000
```

Analcime ($NaAlSi_2O_6 \cdot H_2O$, $S6_1$): A2B2C3D12E4_tI184_142_f_f_be_3g_g - POSCAR

```
A2B2C3D12E4_tI184_142_f_f_be_3g_g & a,c/a,x2,y2,x3,x4,x5,y5,z5,x6,y6,z6,
    ↪ x7,y7,z7,x8,y8,z8 --params=13.723,0.997303796546,0.1242,0.0,
    ↪ 0.1631,0.3805,0.105,0.3705,0.2186,0.2218,0.1028,0.3633,0.3631,
    ↪ 0.2182,0.1051,0.1264,0.1617,0.4118 & I4_{1}/acd D_{4h}^{20} #
    ↪ 142 (bef^2g^4) & tI184 & $S6_{1}$ & AlH2NaO7Si2 & Analcime & F.
    ↪ Mazzi and E. Galli, Am. Mineral. 63, 448-460 (1978)
   1.00000000000000
  -6.86150000000000     6.86150000000000     6.84300000000000
   6.86150000000000    -6.86150000000000     6.84300000000000
   6.86150000000000     6.86150000000000    -6.84300000000000
     Al     H2O      Na       O      Si
      8       8      12      48      16
Direct
   0.53810000000000     0.28810000000000     0.57620000000000    Al    (16f)
   0.21190000000000    -0.03810000000000    -0.07620000000000    Al    (16f)
   0.28810000000000     0.21190000000000     0.75000000000000    Al    (16f)
  -0.03810000000000     0.53810000000000     0.75000000000000    Al    (16f)
   0.46190000000000     0.71190000000000     0.42380000000000    Al    (16f)
   0.78810000000000     1.03810000000000     1.07620000000000    Al    (16f)
   0.71190000000000     0.78810000000000     0.25000000000000    Al    (16f)
   1.03810000000000     0.46190000000000     0.25000000000000    Al    (16f)
   0.75550000000000     0.50550000000000     1.01100000000000   H2O    (16f)
  -0.00550000000000    -0.25550000000000    -0.51100000000000   H2O    (16f)
   0.50550000000000    -0.00550000000000     0.75000000000000   H2O    (16f)
  -0.25550000000000     0.75550000000000     0.75000000000000   H2O    (16f)
   0.24450000000000     0.49450000000000    -0.01100000000000   H2O    (16f)
   1.00550000000000     1.25550000000000     1.51100000000000   H2O    (16f)
   0.49450000000000     1.00550000000000     0.25000000000000   H2O    (16f)
   1.25550000000000     0.24450000000000     0.25000000000000   H2O    (16f)
   0.37500000000000     0.12500000000000     0.25000000000000    Na     (8b)
   0.12500000000000     0.37500000000000     0.75000000000000    Na     (8b)
   0.62500000000000     0.87500000000000     0.75000000000000    Na     (8b)
   0.87500000000000     0.62500000000000     0.25000000000000    Na     (8b)
   0.25000000000000     0.37420000000000     0.12420000000000    Na    (16e)
   0.75000000000000     0.12580000000000     0.37580000000000    Na    (16e)
   0.37420000000000     0.75000000000000     0.12420000000000    Na    (16e)
   0.12580000000000     0.25000000000000     0.37580000000000    Na    (16e)
   0.75000000000000     0.62580000000000    -0.12420000000000    Na    (16e)
   0.25000000000000     0.87420000000000     0.62420000000000    Na    (16e)
   0.12580000000000     0.75000000000000    -0.12420000000000    Na    (16e)
   0.87420000000000     0.75000000000000     0.62420000000000    Na    (16e)
   0.58910000000000     0.32360000000000     0.47550000000000     O    (32g)
   0.34810000000000     0.11360000000000     0.02450000000000     O    (32g)
   0.32360000000000     0.34810000000000    -0.26550000000000     O    (32g)
   0.11360000000000     0.58910000000000     0.76550000000000     O    (32g)
   0.15190000000000     0.17640000000000     0.76550000000000     O    (32g)
  -0.08910000000000     0.38640000000000    -0.26550000000000     O    (32g)
   0.38640000000000     0.15190000000000     0.47550000000000     O    (32g)
   0.17640000000000    -0.08910000000000     0.02450000000000     O    (32g)
  -0.58910000000000    -0.32360000000000    -0.47550000000000     O    (32g)
   0.65190000000000    -0.11360000000000     0.97550000000000     O    (32g)
  -0.32360000000000     0.65190000000000     0.26550000000000     O    (32g)
  -0.11360000000000    -0.58910000000000     0.23450000000000     O    (32g)
  -0.15190000000000     0.82360000000000     0.23450000000000     O    (32g)
   1.08910000000000     0.61360000000000     0.26550000000000     O    (32g)
   0.61360000000000    -0.15190000000000    -0.47550000000000     O    (32g)
   0.82360000000000     1.08910000000000     0.97550000000000     O    (32g)
   0.46610000000000     0.58510000000000     0.32460000000000     O    (32g)
   0.76050000000000     0.14150000000000     0.17540000000000     O    (32g)
   0.58510000000000     0.76050000000000     0.11900000000000     O    (32g)
   0.14150000000000     0.46610000000000     0.38100000000000     O    (32g)
  -0.26050000000000    -0.08510000000000     0.38100000000000     O    (32g)
   0.03390000000000     0.35850000000000     0.11900000000000     O    (32g)
   0.35850000000000    -0.26050000000000     0.32460000000000     O    (32g)
  -0.08510000000000     0.03390000000000     0.17540000000000     O    (32g)
  -0.46610000000000    -0.58510000000000    -0.32460000000000     O    (32g)
   0.23950000000000    -0.14150000000000     0.82460000000000     O    (32g)
  -0.58510000000000     0.23950000000000    -0.11900000000000     O    (32g)
  -0.14150000000000    -0.46610000000000     0.61900000000000     O    (32g)
   0.26050000000000     1.08510000000000     0.61900000000000     O    (32g)
   0.96610000000000     0.64150000000000    -0.11900000000000     O    (32g)
   0.64150000000000     0.26050000000000    -0.32460000000000     O    (32g)
   1.08510000000000     0.96610000000000     0.82460000000000     O    (32g)
   0.32330000000000     0.46820000000000     0.58130000000000     O    (32g)
   0.38690000000000    -0.25800000000000    -0.08130000000000     O    (32g)
   0.46820000000000     0.38690000000000     0.14490000000000     O    (32g)
  -0.25800000000000     0.32330000000000     0.35510000000000     O    (32g)
   0.11310000000000     0.03180000000000     0.35510000000000     O    (32g)
   0.17670000000000     0.75800000000000     0.14490000000000     O    (32g)
   0.75800000000000     0.11310000000000     0.58130000000000     O    (32g)
   0.03180000000000     0.17670000000000    -0.08130000000000     O    (32g)
```

```
 -0.32330000000000  -0.46820000000000  -0.58130000000000   O  (32g)
  0.61310000000000   0.25800000000000   1.08130000000000   O  (32g)
 -0.46820000000000   0.61310000000000  -0.14490000000000   O  (32g)
  0.25800000000000  -0.32330000000000   0.64490000000000   O  (32g)
 -0.11310000000000   0.96820000000000   0.64490000000000   O  (32g)
  0.82330000000000   0.24200000000000  -0.14490000000000   O  (32g)
  0.24200000000000  -0.11310000000000  -0.58130000000000   O  (32g)
  0.96820000000000   0.82330000000000   1.08130000000000   O  (32g)
  0.57350000000000   0.53820000000000   0.28810000000000   Si (32g)
  0.75010000000000   0.28540000000000   0.21190000000000   Si (32g)
  0.53820000000000   0.75010000000000  -0.03530000000000   Si (32g)
  0.28540000000000   0.57350000000000   0.53530000000000   Si (32g)
 -0.25010000000000  -0.03820000000000   0.53530000000000   Si (32g)
 -0.07350000000000   0.21460000000000  -0.03530000000000   Si (32g)
  0.21460000000000  -0.25010000000000   0.28810000000000   Si (32g)
 -0.03820000000000  -0.07350000000000   0.21190000000000   Si (32g)
 -0.57350000000000  -0.53820000000000  -0.28810000000000   Si (32g)
  0.24990000000000  -0.28540000000000   0.78810000000000   Si (32g)
 -0.53820000000000   0.24990000000000   0.03530000000000   Si (32g)
 -0.28540000000000  -0.57350000000000   0.46470000000000   Si (32g)
  0.25010000000000   1.03820000000000   0.46470000000000   Si (32g)
  1.07350000000000   0.78540000000000   0.03530000000000   Si (32g)
  0.78540000000000   0.25010000000000  -0.28810000000000   Si (32g)
  1.03820000000000   1.07350000000000   0.78810000000000   Si (32g)
```

$Cd_3As_2$: A2B3_tI160_142_deg_3g - CIF

```
# CIF file
data_findsym-output
_audit_creation_method FINDSYM

_chemical_name_mineral 'As2Cd3'
_chemical_formula_sum 'As2 Cd3'

loop_
_publ_author_name
 'M. N. Ali'
 'Q. Gibson'
 'S. Jeon'
 'B. B. Zhou'
 'A. Yazdani'
 'R. J. Cava'
_journal_name_full_name
;
 Inorganic Chemistry
;
_journal_volume 53
_journal_year 2014
_journal_page_first 4062
_journal_page_last 4067
_publ_Section_title
;
 The Crystal and Electronic Structures of Cd$_{3}$As$_{2}$, the
    ↪ Three-Dimensional Electronic Analogue of Graphene
;

_aflow_title 'Cd$_{3}$As$_{2}$ Structure'
_aflow_proto 'A2B3_tI160_142_deg_3g'
_aflow_params 'a,c/a,z_{1},x_{2},y_{2},x_{3},y_{3},z_{3},x_{4},y_{4},z_{
    ↪ 4},x_{5},y_{5},z_{5},x_{6},y_{6},z_{6}'
_aflow_params_values '12.633,2.01274439959,-0.00073,0.73924,0.0,0.24597,
    ↪ 0.25789,0.12315,0.13951,0.36959,0.05249,0.11169,0.64224,0.0725,
    ↪ 0.11879,0.1061,0.06251'
_aflow_Strukturbericht 'None'
_aflow_Pearson 'tI160'

_symmetry_space_group_name_H-M "I 41/a 2/c 2/d (origin choice 2)"
_symmetry_Int_Tables_number 142

_cell_length_a    12.63300
_cell_length_b    12.63300
_cell_length_c    25.42700
_cell_angle_alpha 90.00000
_cell_angle_beta  90.00000
_cell_angle_gamma 90.00000

loop_
_space_group_symop_id
_space_group_symop_operation_xyz
1 x,y,z
2 x+1/2,-y+1/2,-z
3 -x+1/2,y,-z
4 -x,-y+1/2,z
5 -y+1/4,-x+1/4,-z+1/4
6 -y+1/4,x+3/4,z+1/4
7 y+3/4,-x+3/4,z+1/4
8 y+3/4,x+1/4,-z+1/4
9 -x,-y,-z
10 -x,y,z+1/2
11 x,-y+1/2,z+1/2
12 x,y+1/2,-z
13 y+1/4,x+1/4,z+1/4
14 y+3/4,-x+1/4,-z+3/4
15 -y+1/4,x+1/4,-z+3/4
16 -y+3/4,-x+1/4,z+1/4
17 x+1/2,y+1/2,z+1/2
18 x,-y,-z+1/2
19 -x,y+1/2,-z+1/2
20 -x+1/2,-y,z+1/2
21 -y+3/4,-x+3/4,-z+3/4
22 -y+3/4,x+1/4,z+3/4
23 y+1/4,-x+1/4,z+3/4
24 y+1/4,x+3/4,-z+3/4
25 -x+1/2,-y+1/2,-z+1/2
26 -x+1/2,y+1/2,z
27 x+1/2,-y,z
28 x+1/2,y,-z+1/2
29 y+3/4,x+3/4,z+3/4
30 y+1/4,-x+3/4,-z+1/4
31 -y+3/4,x+3/4,-z+1/4
32 -y+1/4,-x+3/4,z+3/4

loop_
_atom_site_label
_atom_site_type_symbol
_atom_site_symmetry_multiplicity
_atom_site_Wyckoff_label
_atom_site_fract_x
_atom_site_fract_y
_atom_site_fract_z
_atom_site_occupancy
As1 As 16 d 0.00000 0.25000 -0.00073 1.00000
As2 As 16 e 0.73924 0.00000 0.25000 1.00000
As3 As 32 g 0.24597 0.25789 0.12315 1.00000
Cd1 Cd 32 g 0.13951 0.36959 0.05249 1.00000
Cd2 Cd 32 g 0.11169 0.64224 0.07250 1.00000
Cd3 Cd 32 g 0.11879 0.10610 0.06251 1.00000
```

$Cd_3As_2$: A2B3_tI160_142_deg_3g - POSCAR

```
A2B3_tI160_142_deg_3g & a,c/a,z1,x2,y2,x3,y3,z3,x4,y4,z4,x5,y5,z5,x6,y6,
    ↪ z6 --params=12.633,2.01274439959,-0.00073,0.73924,0.0,0.24597,
    ↪ 0.25789,0.12315,0.13951,0.36959,0.05249,0.11169,0.64224,0.0725,
    ↪ 0.11879,0.1061,0.06251 & I4_{1}/acd D_{4h}^{20} #142 (deg^4) &
    ↪ tI160 & None & As2Cd3 & M. N. Ali et al., Inorg. Chem.
    ↪ 53, 4062-4067 (2014)
   1.00000000000000
  -6.31650000000000   6.31650000000000  12.71350000000000
   6.31650000000000  -6.31650000000000  12.71350000000000
   6.31650000000000   6.31650000000000 -12.71350000000000
    As    Cd
    32    48
Direct
   0.24927000000000  -0.00073000000000   0.25000000000000   As  (16d)
  -0.00073000000000   0.24927000000000   0.75000000000000   As  (16d)
   0.25073000000000   0.50073000000000   0.75000000000000   As  (16d)
   0.50073000000000   0.25073000000000   0.25000000000000   As  (16d)
   0.75073000000000   0.00073000000000   0.75000000000000   As  (16d)
   0.00073000000000   0.75073000000000   0.25000000000000   As  (16d)
   0.74927000000000   0.49927000000000   0.25000000000000   As  (16d)
   0.49927000000000   0.74927000000000   0.75000000000000   As  (16d)
   0.25000000000000   0.98924000000000   0.73924000000000   As  (16e)
   0.75000000000000  -0.48924000000000  -0.23924000000000   As  (16e)
   0.98924000000000   0.75000000000000   0.73924000000000   As  (16e)
  -0.48924000000000   0.25000000000000  -0.23924000000000   As  (16e)
   0.75000000000000   0.01076000000000  -0.73924000000000   As  (16e)
   0.25000000000000   1.48924000000000   1.23924000000000   As  (16e)
  -0.48924000000000   0.75000000000000  -0.73924000000000   As  (16e)
   1.48924000000000   0.75000000000000   1.23924000000000   As  (16e)
   0.38104000000000   0.36912000000000   0.50386000000000   As  (32g)
   0.36526000000000  -0.12282000000000  -0.00386000000000   As  (32g)
   0.36912000000000   0.36526000000000  -0.01192000000000   As  (32g)
  -0.12282000000000   0.38104000000000   0.51192000000000   As  (32g)
   0.13474000000000   0.13088000000000   0.51192000000000   As  (32g)
   0.11896000000000   0.62282000000000  -0.01192000000000   As  (32g)
   0.62282000000000   0.13474000000000   0.50386000000000   As  (32g)
   0.13088000000000   0.11896000000000  -0.00386000000000   As  (32g)
  -0.38104000000000  -0.36912000000000  -0.50386000000000   As  (32g)
   0.63474000000000   0.12282000000000   1.00386000000000   As  (32g)
  -0.36912000000000   0.63474000000000   0.01192000000000   As  (32g)
   0.12282000000000  -0.38104000000000   0.48808000000000   As  (32g)
  -0.13474000000000   0.86912000000000   0.48808000000000   As  (32g)
   0.88104000000000   0.37718000000000   0.01192000000000   As  (32g)
   0.37718000000000  -0.13474000000000  -0.50386000000000   As  (32g)
   0.86912000000000   0.88104000000000   1.00386000000000   As  (32g)
   0.42208000000000   0.19200000000000   0.50910000000000   Cd  (32g)
   0.18290000000000  -0.08702000000000  -0.00910000000000   Cd  (32g)
   0.19200000000000   0.18290000000000  -0.23008000000000   Cd  (32g)
  -0.08702000000000   0.42208000000000   0.73008000000000   Cd  (32g)
   0.31710000000000   0.30800000000000   0.73008000000000   Cd  (32g)
   0.07792000000000   0.58702000000000  -0.23008000000000   Cd  (32g)
   0.58702000000000   0.31710000000000   0.50910000000000   Cd  (32g)
   0.30800000000000   0.07792000000000  -0.00910000000000   Cd  (32g)
  -0.42208000000000  -0.19200000000000  -0.50910000000000   Cd  (32g)
   0.81710000000000   0.08702000000000   1.00910000000000   Cd  (32g)
  -0.19200000000000   0.81710000000000   0.23008000000000   Cd  (32g)
   0.08702000000000  -0.42208000000000   0.26992000000000   Cd  (32g)
  -0.31710000000000   0.69200000000000   0.26992000000000   Cd  (32g)
   0.92208000000000   0.41298000000000   0.23008000000000   Cd  (32g)
   0.41298000000000  -0.31710000000000  -0.50910000000000   Cd  (32g)
   0.69200000000000   0.92208000000000   1.00910000000000   Cd  (32g)
   0.71474000000000   0.18419000000000   0.75393000000000   Cd  (32g)
  -0.06974000000000  -0.03919000000000  -0.25393000000000   Cd  (32g)
   0.18419000000000  -0.06974000000000  -0.53055000000000   Cd  (32g)
  -0.03919000000000   0.71474000000000   1.03055000000000   Cd  (32g)
   0.56974000000000   0.31581000000000   1.03055000000000   Cd  (32g)
  -0.21474000000000   0.53919000000000  -0.53055000000000   Cd  (32g)
   0.53919000000000   0.56974000000000   0.75393000000000   Cd  (32g)
   0.31581000000000  -0.21474000000000  -0.25393000000000   Cd  (32g)
  -0.71474000000000  -0.18419000000000  -0.75393000000000   Cd  (32g)
   1.06974000000000   0.03919000000000   1.25393000000000   Cd  (32g)
  -0.18419000000000   1.06974000000000   0.53055000000000   Cd  (32g)
   0.03919000000000  -0.71474000000000  -0.03055000000000   Cd  (32g)
  -0.56974000000000   0.68419000000000  -0.03055000000000   Cd  (32g)
   1.21474000000000   0.46081000000000   0.53055000000000   Cd  (32g)
   0.46081000000000  -0.56974000000000  -0.75393000000000   Cd  (32g)
   0.68419000000000   1.21474000000000   1.25393000000000   Cd  (32g)
   0.16861000000000   0.18130000000000   0.22489000000000   Cd  (32g)
   0.45641000000000  -0.05628000000000   0.27511000000000   Cd  (32g)
   0.18130000000000   0.45641000000000   0.01269000000000   Cd  (32g)
```

```
   -0.05628000000000    0.16861000000000    0.48731000000000    Cd   (32g)
    0.04359000000000    0.31870000000000    0.48731000000000    Cd   (32g)
    0.33139000000000    0.55628000000000    0.01269000000000    Cd   (32g)
    0.55628000000000    0.04359000000000    0.22489000000000    Cd   (32g)
    0.31870000000000    0.33139000000000    0.27511000000000    Cd   (32g)
   -0.16861000000000   -0.18130000000000   -0.22489000000000    Cd   (32g)
    0.54359000000000    0.05628000000000    0.72489000000000    Cd   (32g)
   -0.18130000000000    0.54359000000000   -0.01269000000000    Cd   (32g)
    0.05628000000000   -0.16861000000000    0.51269000000000    Cd   (32g)
   -0.04359000000000    0.68130000000000    0.51269000000000    Cd   (32g)
    0.66861000000000    0.44372000000000   -0.01269000000000    Cd   (32g)
    0.44372000000000   -0.04359000000000   -0.22489000000000    Cd   (32g)
    0.68130000000000    0.66861000000000    0.72489000000000    Cd   (32g)
```

$La_3BWO_9$ ($P3$): AB3C9D_hP28_143_2a_2d_6d_bc - CIF

```
# CIF file
data_findsym-output
_audit_creation_method FINDSYM

_chemical_name_mineral 'BaLi3O9W'
_chemical_formula_sum 'B La3 O9 W'

loop_
_publ_author_name
 'J. Han'
 'F. Pan'
 'M. S. Molokeev'
 'J. Dai'
 'M. Peng'
 'W. Zhou'
 'J. Wang'
_journal_name_full_name
;
 ACS Applied Materials and Interfaces
;
_journal_volume 10
_journal_year 2018
_journal_page_first 13660
_journal_page_last 13668
_publ_Section_title
;
 Redefinition of Crystal Structure and Bi$^{3+}$ Yellow Luminescence
   ↪ with Strong Near-Ultraviolet Excitation in La$_{3}$BWO$_{9}$:
   ↪ Bi$^{3+}$ Phosphor for White Light-Emitting Diodes
;

_aflow_title 'La$_{3}$BWO$_{9}$ ($P3$) Structure'
_aflow_proto 'AB3C9D_hP28_143_2a_2d_6d_bc'
_aflow_params 'a,c/a,z_{1},z_{2},z_{3},z_{4},x_{5},y_{5},z_{5},x_{6},y_{
   ↪ 6},z_{6},x_{7},y_{7},z_{7},x_{8},y_{8},z_{8},x_{9},y_{9},z_{9},
   ↪ x_{10},y_{10},z_{10},x_{11},y_{11},z_{11},x_{12},y_{12},z_{12}'
_aflow_params_values '8.84326,0.630483554707,0.36,0.84,0.25,0.77,0.363,
   ↪ 0.084,0.229,0.278,0.365,0.737,0.172,0.056,0.839,0.138,0.169,
   ↪ 0.418,0.181,0.498,-0.017,0.728,0.198,0.594,0.134,0.502,0.427,
   ↪ 0.617,0.143,-0.004'
_aflow_Strukturbericht 'None'
_aflow_Pearson 'hP28'

_symmetry_space_group_name_H-M "P 3"
_symmetry_Int_Tables_number 143

_cell_length_a     8.84326
_cell_length_b     8.84326
_cell_length_c     5.57553
_cell_angle_alpha  90.00000
_cell_angle_beta   90.00000
_cell_angle_gamma  120.00000

loop_
_space_group_symop_id
_space_group_symop_operation_xyz
1 x,y,z
2 -y,x-y,z
3 -x+y,-x,z

loop_
_atom_site_label
_atom_site_type_symbol
_atom_site_symmetry_multiplicity
_atom_site_Wyckoff_label
_atom_site_fract_x
_atom_site_fract_y
_atom_site_fract_z
_atom_site_occupancy
B1 B 1 a 0.00000 0.00000 0.36000 1.00000
B2 B 1 a 0.00000 0.00000 0.84000 1.00000
W1 W 1 b 0.33333 0.66667 0.25000 1.00000
W2 W 1 c 0.66667 0.33333 0.77000 1.00000
La1 La 3 d 0.36300 0.08400 0.22900 1.00000
La2 La 3 d 0.27800 0.36500 0.73700 1.00000
O1 O 3 d 0.17200 0.05600 0.83900 1.00000
O2 O 3 d 0.13800 0.16900 0.41800 1.00000
O3 O 3 d 0.18100 0.49800 -0.01700 1.00000
O4 O 3 d 0.72800 0.19800 0.59400 1.00000
O5 O 3 d 0.13400 0.50200 0.42700 1.00000
O6 O 3 d 0.61700 0.14300 -0.00400 1.00000
```

$La_3BWO_9$ ($P3$): AB3C9D_hP28_143_2a_2d_6d_bc - POSCAR

```
AB3C9D_hP28_143_2a_2d_6d_bc & a,c/a,z1,z2,z3,z4,x5,y5,z5,x6,y6,z6,x7,y7,
   ↪ z7,x8,y8,z8,x9,y9,z9,x10,y10,z10,x11,y11,z11,x12,y12,z12 --
   ↪ params=8.84326,0.630483554707,0.36,0.84,0.25,0.77,0.363,0.084,
   ↪ 0.229,0.278,0.365,0.737,0.172,0.056,0.839,0.138,0.169,0.418,
   ↪ 0.181,0.498,-0.017,0.728,0.198,0.594,0.134,0.502,0.427,0.617,
   ↪ 0.143,-0.004 & P3 C_{3}^{1} #143 (a^2bcd^8) & hP28 & None &
   ↪ BaLi3O9W & BaLi3O9W & J. Han et al., ACS Appl. Mater.
   ↪ Interfaces 10, 13660-13668 (2018)
   1.00000000000000
   4.42163000000000   -7.65848781227078    0.00000000000000
   4.42163000000000    7.65848781227078    0.00000000000000
   0.00000000000000    0.00000000000000    5.57553000000000
     B     La      O      W
     2      6     18      2
Direct
   0.00000000000000    0.00000000000000    0.36000000000000    B   (1a)
   0.00000000000000    0.00000000000000    0.84000000000000    B   (1a)
   0.36300000000000    0.08400000000000    0.22900000000000    La  (3d)
  -0.08400000000000    0.27900000000000    0.22900000000000    La  (3d)
  -0.27900000000000   -0.36300000000000    0.22900000000000    La  (3d)
   0.27800000000000    0.36500000000000    0.73700000000000    La  (3d)
  -0.36500000000000   -0.08700000000000    0.73700000000000    La  (3d)
   0.08700000000000   -0.27800000000000    0.73700000000000    La  (3d)
   0.17200000000000    0.05600000000000    0.83900000000000    O   (3d)
  -0.05600000000000    0.11600000000000    0.83900000000000    O   (3d)
  -0.11600000000000   -0.17200000000000    0.83900000000000    O   (3d)
   0.13800000000000    0.16900000000000    0.41800000000000    O   (3d)
  -0.16900000000000   -0.03100000000000    0.41800000000000    O   (3d)
   0.03100000000000   -0.13800000000000    0.41800000000000    O   (3d)
   0.18100000000000    0.49800000000000   -0.01700000000000    O   (3d)
  -0.49800000000000   -0.31700000000000   -0.01700000000000    O   (3d)
   0.31700000000000   -0.18100000000000   -0.01700000000000    O   (3d)
   0.72800000000000    0.19800000000000    0.59400000000000    O   (3d)
  -0.19800000000000    0.53000000000000    0.59400000000000    O   (3d)
  -0.53000000000000   -0.72800000000000    0.59400000000000    O   (3d)
   0.13400000000000    0.50200000000000    0.42700000000000    O   (3d)
  -0.50200000000000   -0.36800000000000    0.42700000000000    O   (3d)
   0.36800000000000   -0.13400000000000    0.42700000000000    O   (3d)
   0.61700000000000    0.14300000000000   -0.00400000000000    O   (3d)
  -0.14300000000000    0.47400000000000   -0.00400000000000    O   (3d)
  -0.47400000000000   -0.61700000000000   -0.00400000000000    O   (3d)
   0.33333333333333    0.66666666666667    0.25000000000000    W   (1b)
   0.66666666666667    0.33333333333333    0.77000000000000    W   (1c)
```

$RbNO_3$ (IV): AB3C_hP45_144_3a_9a_3a - CIF

```
# CIF file
data_findsym-output
_audit_creation_method FINDSYM

_chemical_name_mineral 'NO3Rb'
_chemical_formula_sum 'N O3 Rb'

loop_
_publ_author_name
 'M. Shamsuzzoha'
 'B. W. Lucas'
_journal_name_full_name
;
 Acta Crystallographica Section B: Structural Science
;
_journal_volume 38
_journal_year 1982
_journal_page_first 2353
_journal_page_last 2357
_publ_Section_title
;
 Structure (neutron) of phase IV rubidium nitrate at 298 and 403~K
;

_aflow_title 'RbNO$_{3}$ (IV) Structure'
_aflow_proto 'AB3C_hP45_144_3a_9a_3a'
_aflow_params 'a,c/a,x_{1},y_{1},z_{1},x_{2},y_{2},z_{2},x_{3},y_{3},z_{
   ↪ 3},x_{4},y_{4},z_{4},x_{5},y_{5},z_{5},x_{6},y_{6},z_{6},x_{7},
   ↪ y_{7},z_{7},x_{8},y_{8},z_{8},x_{9},y_{9},z_{9},x_{10},y_{10},
   ↪ z_{10},x_{11},y_{11},z_{11},x_{12},y_{12},z_{12},x_{13},y_{13},
   ↪ z_{13},x_{14},y_{14},z_{14},x_{15},y_{15},z_{15}'
_aflow_params_values '10.55,0.708056872038,0.4388,0.5668,0.1053,0.0962,
   ↪ 0.2042,0.5318,0.742,0.2057,0.1168,0.3371,0.5582,0.0094,0.3964,
   ↪ 0.4704,0.228,0.5646,0.6512,0.0664,-0.0011,0.1089,0.6245,0.2309,
   ↪ 0.2495,0.5637,0.0581,0.2465,0.3949,0.6922,0.1208,0.2503,0.8721,
   ↪ 0.2774,0.0897,0.6541,0.2174,0.014,0.4566,0.5691,0.6236,0.1184,
   ↪ 0.2192,0.0,0.7772,0.2214,0.6381'
_aflow_Strukturbericht 'None'
_aflow_Pearson 'hP45'

_symmetry_space_group_name_H-M "P 31"
_symmetry_Int_Tables_number 144

_cell_length_a     10.55000
_cell_length_b     10.55000
_cell_length_c     7.47000
_cell_angle_alpha  90.00000
_cell_angle_beta   90.00000
_cell_angle_gamma  120.00000

loop_
_space_group_symop_id
_space_group_symop_operation_xyz
1 x,y,z
2 -y,x-y,z+1/3
3 -x+y,-x,z+2/3

loop_
_atom_site_label
_atom_site_type_symbol
_atom_site_symmetry_multiplicity
_atom_site_Wyckoff_label
_atom_site_fract_x
_atom_site_fract_y
```

```
_atom_site_fract_z
_atom_site_occupancy
N1 N 3 a 0.43880 0.56680 0.10530 1.00000
N2 N 3 a 0.09620 0.20420 0.53180 1.00000
N3 N 3 a 0.74200 0.20570 0.11680 1.00000
O1 O 3 a 0.33710 0.55820 0.00940 1.00000
O2 O 3 a 0.39640 0.47040 0.22800 1.00000
O3 O 3 a 0.56460 0.65120 0.06640 1.00000
O4 O 3 a -0.00110 0.10890 0.62450 1.00000
O5 O 3 a 0.23090 0.24950 0.56370 1.00000
O6 O 3 a 0.05810 0.24650 0.39490 1.00000
O7 O 3 a 0.69220 0.12080 0.25030 1.00000
O8 O 3 a 0.87210 0.27740 0.08970 1.00000
O9 O 3 a 0.65410 0.21740 0.01400 1.00000
Rb1 Rb 3 a 0.45660 0.56910 0.62360 1.00000
Rb2 Rb 3 a 0.11840 0.21920 0.00000 1.00000
Rb3 Rb 3 a 0.77720 0.22140 0.63810 1.00000
```

$RbNO_3$ (IV): AB3C_hP45_144_3a_9a_3a - POSCAR

```
AB3C_hP45_144_3a_9a_3a & a,c/a,x1,y1,z1,x2,y2,z2,x3,y3,z3,x4,y4,z4,x5,y5
    ↪ ,z5,x6,y6,z6,x7,y7,z7,x8,y8,z8,x9,y9,z9,x10,y10,z10,x11,y11,z11
    ↪ ,x12,y12,z12,x13,y13,z13,x14,y14,z14,x15,y15,z15 --params=10.55
    ↪ ,0.708056872038,0.4388,0.5668,0.1053,0.0962,0.2042,0.5318,0.742
    ↪ ,0.2057,0.1168,0.3371,0.5582,0.0094,0.3964,0.4704,0.228,0.5646,
    ↪ 0.6512,0.0664,-0.0011,0.1089,0.6245,0.2309,0.2495,0.5637,0.0581
    ↪ ,0.2465,0.3949,0.6922,0.1208,0.2503,0.8721,0.2774,0.0897,0.6541
    ↪ ,0.2174,0.014,0.4566,0.5691,0.6236,0.1184,0.2192,0.0,0.7772,
    ↪ 0.2214,0.6381 & P3_{1} C_{3}^{2} #144 (a^15) & hP45 & None &
    ↪ NO3Rb & NO3Rb & M. Shamsuzzoha and B. W. Lucas, Acta
    ↪ Crystallogr. Sect. B Struct. Sci. 38, 2353-2357 (1982)
   1.00000000000000
   5.27500000000000  -9.13656800992583   0.00000000000000
   5.27500000000000   9.13656800992583   0.00000000000000
   0.00000000000000   0.00000000000000   7.47000000000000
    N     O    Rb
    9    27     9
Direct
   0.43880000000000   0.56680000000000   0.10530000000000   N   (3a)
  -0.56680000000000  -0.12800000000000   0.43863333333333   N   (3a)
   0.12800000000000  -0.43880000000000   0.77196666666667   N   (3a)
   0.09620000000000   0.20420000000000   0.53180000000000   N   (3a)
  -0.20420000000000  -0.10800000000000   0.86513333333333   N   (3a)
   0.10800000000000  -0.09620000000000   1.19846666666667   N   (3a)
   0.74200000000000   0.20570000000000   0.11680000000000   N   (3a)
  -0.20570000000000   0.53630000000000   0.45013333333333   N   (3a)
  -0.53630000000000  -0.74200000000000   0.78346666666667   N   (3a)
   0.33710000000000   0.55820000000000   0.00940000000000   O   (3a)
  -0.55820000000000  -0.22110000000000   0.34273333333333   O   (3a)
   0.22110000000000  -0.33710000000000   0.67606666666667   O   (3a)
   0.39640000000000   0.47040000000000   0.22800000000000   O   (3a)
  -0.47040000000000  -0.07400000000000   0.56133333333333   O   (3a)
   0.07400000000000  -0.39640000000000   0.89466666666667   O   (3a)
   0.56460000000000   0.65120000000000   0.06640000000000   O   (3a)
  -0.65120000000000  -0.08660000000000   0.39973333333333   O   (3a)
   0.08660000000000  -0.56460000000000   0.73306666666667   O   (3a)
  -0.00110000000000   0.10890000000000   0.62450000000000   O   (3a)
  -0.10890000000000  -0.11000000000000   0.95783333333333   O   (3a)
   0.11000000000000   0.00110000000000   1.29116666666667   O   (3a)
   0.23090000000000   0.24950000000000   0.56370000000000   O   (3a)
  -0.24950000000000  -0.01860000000000   0.89703333333333   O   (3a)
   0.01860000000000  -0.23090000000000   1.23036666666667   O   (3a)
   0.05810000000000   0.24650000000000   0.39490000000000   O   (3a)
  -0.24650000000000  -0.18840000000000   0.72823333333333   O   (3a)
   0.18840000000000  -0.05810000000000   1.06156666666667   O   (3a)
   0.69220000000000   0.12080000000000   0.25030000000000   O   (3a)
  -0.12080000000000   0.57140000000000   0.58363333333333   O   (3a)
  -0.57140000000000  -0.69220000000000   0.91696666666667   O   (3a)
   0.87210000000000   0.27740000000000   0.08970000000000   O   (3a)
  -0.27740000000000   0.59470000000000   0.42303333333333   O   (3a)
  -0.59470000000000  -0.87210000000000   0.75636666666667   O   (3a)
   0.65410000000000   0.21740000000000   0.01400000000000   O   (3a)
  -0.21740000000000   0.43670000000000   0.34733333333333   O   (3a)
  -0.43670000000000  -0.65410000000000   0.68066666666667   O   (3a)
   0.45660000000000   0.56910000000000   0.62360000000000   Rb  (3a)
  -0.56910000000000  -0.11250000000000   0.95693333333333   Rb  (3a)
   0.11250000000000  -0.45660000000000   1.29026666666667   Rb  (3a)
   0.11840000000000   0.21920000000000   0.00000000000000   Rb  (3a)
  -0.21920000000000  -0.10080000000000   0.33333333333333   Rb  (3a)
   0.10080000000000  -0.11840000000000   0.66666666666667   Rb  (3a)
   0.77720000000000   0.22140000000000   0.63810000000000   Rb  (3a)
  -0.22140000000000   0.55580000000000   0.97143333333333   Rb  (3a)
  -0.55580000000000  -0.77720000000000   1.30476666666667   Rb  (3a)
```

$Na_2SO_3$ ($G3_2$): A2B3C_hP12_147_abd_g_d - CIF

```
# CIF file
data_findsym-output
_audit_creation_method FINDSYM

_chemical_name_mineral 'Na2O3S'
_chemical_formula_sum 'Na2 O3 S'

loop_
_publ_author_name
 'W. H. Zachariasen'
 'H. E. Buckley'
_journal_name_full_name
;
 Physical Review
;
_journal_volume 37
_journal_year 1931
_journal_page_first 1295
_journal_page_last 1305
_publ_Section_title
;
 The Crystal Lattice of Anhydrous Sodium Sulphite, Na$_{2}$SO$_{3}$
;

_aflow_title 'Na$_{2}$SO$_{3}$ ($G3_{2}$) Structure'
_aflow_proto 'A2B3C_hP12_147_abd_g_d'
_aflow_params 'a,c/a,z_{3},z_{4},x_{5},y_{5},z_{5}'
_aflow_params_values '5.441,1.12718250322,0.67,0.17,0.14,0.4,0.25'
_aflow_Strukturbericht '$G3_{2}$'
_aflow_Pearson 'hP12'

_symmetry_space_group_name_H-M "P -3"
_symmetry_Int_Tables_number 147

_cell_length_a     5.44100
_cell_length_b     5.44100
_cell_length_c     6.13300
_cell_angle_alpha  90.00000
_cell_angle_beta   90.00000
_cell_angle_gamma  120.00000

loop_
_space_group_symop_id
_space_group_symop_operation_xyz
1 x,y,z
2 -y,x-y,z
3 -x+y,-x,z
4 -x,-y,-z
5 y,-x+y,-z
6 x-y,x,-z

loop_
_atom_site_label
_atom_site_type_symbol
_atom_site_symmetry_multiplicity
_atom_site_Wyckoff_label
_atom_site_fract_x
_atom_site_fract_y
_atom_site_fract_z
_atom_site_occupancy
Na1 Na 1 a 0.00000 0.00000 0.00000 1.00000
Na2 Na 1 b 0.00000 0.00000 0.50000 1.00000
Na3 Na 2 d 0.33333 0.66667 0.67000 1.00000
S1 S 2 d 0.33333 0.66667 0.17000 1.00000
O1 O 6 g 0.14000 0.40000 0.25000 1.00000
```

$Na_2SO_3$ ($G3_2$): A2B3C_hP12_147_abd_g_d - POSCAR

```
A2B3C_hP12_147_abd_g_d & a,c/a,z3,z4,x5,y5,z5 --params=5.441,
    ↪ 1.12718250322,0.67,0.17,0.14,0.4,0.25 & P-3 C_{3i}^{1} #147 (
    ↪ abd^2g) & hP12 & $G3_{2}$ & Na2O3S & Na2O3S & W. H. Zachariasen
    ↪ and H. E. Buckley, Phys. Rev. 37, 1295-1305 (1931)
   1.00000000000000
   2.72050000000000  -4.71204422199113   0.00000000000000
   2.72050000000000   4.71204422199113   0.00000000000000
   0.00000000000000   0.00000000000000   6.13300000000000
    Na     O     S
     4     6     2
Direct
   0.00000000000000   0.00000000000000   0.00000000000000   Na   (1a)
   0.00000000000000   0.00000000000000   0.50000000000000   Na   (1b)
   0.33333333333333   0.66666666666667   0.67000000000000   Na   (2d)
   0.66666666666667   0.33333333333333  -0.67000000000000   Na   (2d)
   0.14000000000000   0.40000000000000   0.25000000000000   O    (6g)
  -0.40000000000000  -0.26000000000000   0.25000000000000   O    (6g)
   0.26000000000000  -0.14000000000000   0.25000000000000   O    (6g)
  -0.14000000000000  -0.40000000000000  -0.25000000000000   O    (6g)
   0.40000000000000   0.26000000000000  -0.25000000000000   O    (6g)
  -0.26000000000000   0.14000000000000  -0.25000000000000   O    (6g)
   0.33333333333333   0.66666666666667   0.17000000000000   S    (2d)
   0.66666666666667   0.33333333333333  -0.17000000000000   S    (2d)
```

Dolomite [$MgCa(CO_3)_2$, $G1_1$]: A2BCD6_hR10_148_c_a_b_f - CIF

```
# CIF file
data_findsym-output
_audit_creation_method FINDSYM

_chemical_name_mineral 'Dolomite'
_chemical_formula_sum 'C2 Ca Mg O6'

loop_
_publ_author_name
 'R. J. Reeder'
 'S. A. Markgraf'
_journal_name_full_name
;
 American Mineralogist
;
_journal_volume 71
_journal_year 1986
_journal_page_first 795
_journal_page_last 804
_publ_Section_title
;
 High-temperature crystal chemistry of dolomite
;

# Found in The American Mineralogist Crystal Structure Database, 2003

_aflow_title 'Dolomite [MgCa(CO$_{3}$)$_{2}$, $G1_{1}$] Structure'
_aflow_proto 'A2BCD6_hR10_148_c_a_b_f'
_aflow_params 'a,c/a,x_{3},x_{4},y_{4},z_{4}'
```

```
_aflow_params_values '4.8069,3.32896461337,0.24289,0.49193,0.27933,-
    ↪ 0.03947'
_aflow_Strukturbericht '$G1_{1}$'
_aflow_Pearson 'hR10'

_symmetry_space_group_name_H-M "R -3 (hexagonal axes)"
_symmetry_Int_Tables_number 148

_cell_length_a     4.80690
_cell_length_b     4.80690
_cell_length_c     16.00200
_cell_angle_alpha  90.00000
_cell_angle_beta   90.00000
_cell_angle_gamma  120.00000

loop_
_space_group_symop_id
_space_group_symop_operation_xyz
1 x,y,z
2 -y,x-y,z
3 -x+y,-x,z
4 -x,-y,-z
5 y,-x+y,-z
6 x-y,x,-z
7 x+1/3,y+2/3,z+2/3
8 -y+1/3,x-y+2/3,z+2/3
9 -x+y+1/3,-x+2/3,z+2/3
10 -x+1/3,-y+2/3,-z+2/3
11 y+1/3,-x+y+2/3,-z+2/3
12 x-y+1/3,x+2/3,-z+2/3
13 x+2/3,y+1/3,z+1/3
14 -y+2/3,x-y+1/3,z+1/3
15 -x+y+2/3,-x+1/3,z+1/3
16 -x+2/3,-y+1/3,-z+1/3
17 y+2/3,-x+y+1/3,-z+1/3
18 x-y+2/3,x+1/3,-z+1/3

loop_
_atom_site_label
_atom_site_type_symbol
_atom_site_symmetry_multiplicity
_atom_site_Wyckoff_label
_atom_site_fract_x
_atom_site_fract_y
_atom_site_fract_z
_atom_site_occupancy
Ca1 Ca 3 a 0.00000 0.00000 0.00000 1.00000
Mg1 Mg 3 b 0.00000 0.00000 0.50000 1.00000
C1 C 6 c 0.00000 0.00000 0.24289 1.00000
O1 O 18 f 0.24800 -0.0354 0.24393 1.00000
```

Dolomite [MgCa($CO_3$)$_2$, $G1_1$]: A2BCD6_hR10_148_c_a_b_f - POSCAR

```
A2BCD6_hR10_148_c_a_b_f & a,c/a,x3,x4,y4,z4 --params=4.8069,
    ↪ 3.32896461337,0.24289,0.49193,0.27933,-0.03947 & R-3 C_{3i}^{2}
    ↪ #148 (abcf) & hR10 & $G1_{1}$ & C2CaMgO6 & Dolomite & R. J.
    ↪ Reeder and S. A. Markgraf, Am. Mineral. 71, 795-804 (1986)
   1.00000000000000
   2.40345000000000  -1.38763250448381   5.33400000000000
   0.00000000000000   2.77526500896761   5.33400000000000
  -2.40345000000000  -1.38763250448381   5.33400000000000
     C    Ca    Mg     O
     2     1     1     6
Direct
   0.24289000000000   0.24289000000000   0.24289000000000   C   (2c)
  -0.24289000000000  -0.24289000000000  -0.24289000000000   C   (2c)
   0.00000000000000   0.00000000000000   0.00000000000000   Ca  (1a)
   0.50000000000000   0.50000000000000   0.50000000000000   Mg  (1b)
   0.49193000000000   0.27933000000000  -0.03947000000000   O   (6f)
  -0.03947000000000   0.49193000000000   0.27933000000000   O   (6f)
   0.27933000000000  -0.03947000000000   0.49193000000000   O   (6f)
  -0.49193000000000  -0.27933000000000   0.03947000000000   O   (6f)
   0.03947000000000  -0.49193000000000  -0.27933000000000   O   (6f)
  -0.27933000000000   0.03947000000000  -0.49193000000000   O   (6f)
```

$K_2Sn(OH)_6$ ($H6_2$): A6B2C6D_hR15_148_f_c_f_a - CIF

```
# CIF file
data_findsym-output
_audit_creation_method FINDSYM

_chemical_name_mineral 'H6K2O6Sn'
_chemical_formula_sum 'H6 K2 O6 Sn'

loop_
_publ_author_name
 'H. Jacobs'
 'R. Stahl'
_journal_name_full_name
;
 Zeitschrift fur Anorganische und Allgemeine Chemie
;
_journal_volume 626
_journal_year 2000
_journal_page_first 1863
_journal_page_last 1866
_publ_Section_title
;
 Neubestimmung der Kristallstrukturen der Hexahydroxometallate Na$_{2}
    ↪ $Sn(OH)$_{6}$, K$_{2}$Sn(OH)$_{6}$ und K$_{2}$Pb(OH)$_{6}$
;

# Found in K$_{2}$Sn(OH)$_{6}$ (K$_{2}$Sn[OH]$_{6}$) Crystal Structure,
    ↪ 2016 Found in K$_{2}$Sn(OH)$_{6}$ (K$_{2}$Sn[OH]$_{6}$) Crystal
    ↪ Structure, {PAULING FILE in: Inorganic Solid Phases,
    ↪ SpringerMaterials (online database), Springer, Heidelberg (ed.)
    ↪ Springer Materials},

_aflow_title 'K$_{2}$Sn(OH)$_{6}$ ($H6_{2}$) Structure'
_aflow_proto 'A6B2C6D_hR15_148_f_c_f_a'
_aflow_params 'a,c/a,x_{2},x_{3},y_{3},z_{3},x_{4},y_{4},z_{4}'
_aflow_params_values '6.541,1.95887478979,0.28715,0.248,-0.202,0.326,
    ↪ 0.296,-0.2014,0.1778'
_aflow_Strukturbericht '$H6_{2}$'
_aflow_Pearson 'hR15'

_symmetry_space_group_name_H-M "R -3 (hexagonal axes)"
_symmetry_Int_Tables_number 148

_cell_length_a     6.54100
_cell_length_b     6.54100
_cell_length_c     12.81300
_cell_angle_alpha  90.00000
_cell_angle_beta   90.00000
_cell_angle_gamma  120.00000

loop_
_space_group_symop_id
_space_group_symop_operation_xyz
1 x,y,z
2 -y,x-y,z
3 -x+y,-x,z
4 -x,-y,-z
5 y,-x+y,-z
6 x-y,x,-z
7 x+1/3,y+2/3,z+2/3
8 -y+1/3,x-y+2/3,z+2/3
9 -x+y+1/3,-x+2/3,z+2/3
10 -x+1/3,-y+2/3,-z+2/3
11 y+1/3,-x+y+2/3,-z+2/3
12 x-y+1/3,x+2/3,-z+2/3
13 x+2/3,y+1/3,z+1/3
14 -y+2/3,x-y+1/3,z+1/3
15 -x+y+2/3,-x+1/3,z+1/3
16 -x+2/3,-y+1/3,-z+1/3
17 y+2/3,-x+y+1/3,-z+1/3
18 x-y+2/3,x+1/3,-z+1/3

loop_
_atom_site_label
_atom_site_type_symbol
_atom_site_symmetry_multiplicity
_atom_site_Wyckoff_label
_atom_site_fract_x
_atom_site_fract_y
_atom_site_fract_z
_atom_site_occupancy
Sn1 Sn 3 a 0.00000 0.00000 0.00000 1.00000
K1 K 6 c 0.00000 0.00000 0.28715 1.00000
H1 H 18 f 0.12400 0.32600 0.12400 1.00000
O1 O 18 f 0.20520 0.29220 0.09080 1.00000
```

$K_2Sn(OH)_6$ ($H6_2$): A6B2C6D_hR15_148_f_c_f_a - POSCAR

```
A6B2C6D_hR15_148_f_c_f_a & a,c/a,x2,x3,y3,z3,x4,y4,z4 --params=6.541,
    ↪ 1.95887478979,0.28715,0.248,-0.202,0.326,0.296,-0.2014,0.1778 &
    ↪ R-3 C_{3i}^{2} #148 (acf^2) & hR15 & $H6_{2}$ & H6K2O6Sn &
    ↪ H6K2O6Sn & H. Jacobs and R. Stahl, Z. Anorg. Allg. Chem. 626,
    ↪ 1863-1866 (2000)
   1.00000000000000
   3.27050000000000  -1.88822405538467   4.27100000000000
   0.00000000000000   3.77644811076934   4.27100000000000
  -3.27050000000000  -1.88822405538467   4.27100000000000
     H     K     O    Sn
     6     2     6     1
Direct
   0.24800000000000  -0.20200000000000   0.32600000000000   H   (6f)
   0.32600000000000   0.24800000000000  -0.20200000000000   H   (6f)
  -0.20200000000000   0.32600000000000   0.24800000000000   H   (6f)
  -0.24800000000000   0.20200000000000  -0.32600000000000   H   (6f)
  -0.32600000000000  -0.24800000000000   0.20200000000000   H   (6f)
   0.20200000000000  -0.32600000000000  -0.24800000000000   H   (6f)
   0.28715000000000   0.28715000000000   0.28715000000000   K   (2c)
  -0.28715000000000  -0.28715000000000  -0.28715000000000   K   (2c)
   0.29600000000000  -0.20140000000000   0.17780000000000   O   (6f)
   0.17780000000000   0.29600000000000  -0.20140000000000   O   (6f)
  -0.20140000000000   0.17780000000000   0.29600000000000   O   (6f)
  -0.29600000000000   0.20140000000000  -0.17780000000000   O   (6f)
  -0.17780000000000  -0.29600000000000   0.20140000000000   O   (6f)
   0.20140000000000  -0.17780000000000  -0.29600000000000   O   (6f)
   0.00000000000000   0.00000000000000   0.00000000000000   Sn  (1a)
```

$Ni(H_2O)_6SnCl_6$ ($I6_1$): A6B6CD_hR14_148_f_f_b_a - CIF

```
# CIF file
data_findsym-output
_audit_creation_method FINDSYM

_chemical_name_mineral 'Cl6(H2O)6NiSn'
_chemical_formula_sum 'Cl6 (H2O)6 Ni Sn'

loop_
_publ_author_name
 'L. Pauling'
_journal_name_full_name
;
 Zeitschrift f{\"u}r Kristallographie - Crystalline Materials
;
_journal_volume 72
_journal_year 1930
```

```
_journal_page_first 482
_journal_page_last 492
_publ_Section_title
;
 On the crystal structure of nickel chlorostannate hexahydrate
;

# Found in Strukturbericht Band II 1928-1932, 1937

_aflow_title 'Ni(H$_{2}$O)$_{6}$SnCl$_{6}$ ($I6_{1}$) Structure'
_aflow_proto 'A6B6CD_hR14_148_f_f_b_a'
_aflow_params 'a,c/a,x_{3},y_{3},z_{3},x_{4},y_{4},z_{4}'
_aflow_params_values '10.59967,1.01326739417,0.94,-0.69,0.14,0.44,-0.19,
    ↪ 0.64001'
_aflow_Strukturbericht '$I6_{1}$'
_aflow_Pearson 'hR14'

_symmetry_space_group_name_H-M "R -3 (hexagonal axes)"
_symmetry_Int_Tables_number 148

_cell_length_a      10.59967
_cell_length_b      10.59967
_cell_length_c      10.74030
_cell_angle_alpha   90.00000
_cell_angle_beta    90.00000
_cell_angle_gamma  120.00000

loop_
_space_group_symop_id
_space_group_symop_operation_xyz
1 x,y,z
2 -y,x-y,z
3 -x+y,-x,z
4 -x,-y,-z
5 y,-x+y,-z
6 x-y,x,-z
7 x+1/3,y+2/3,z+2/3
8 -y+1/3,x-y+2/3,z+2/3
9 -x+y+1/3,-x+2/3,z+2/3
10 -x+1/3,-y+2/3,-z+2/3
11 y+1/3,-x+y+2/3,-z+2/3
12 x-y+1/3,x+2/3,-z+2/3
13 x+2/3,y+1/3,z+1/3
14 -y+2/3,x-y+1/3,z+1/3
15 -x+y+2/3,-x+1/3,z+1/3
16 -x+2/3,-y+1/3,-z+1/3
17 y+2/3,-x+y+1/3,-z+1/3
18 x-y+2/3,x+1/3,-z+1/3

loop_
_atom_site_label
_atom_site_type_symbol
_atom_site_symmetry_multiplicity
_atom_site_Wyckoff_label
_atom_site_fract_x
_atom_site_fract_y
_atom_site_fract_z
_atom_site_occupancy
Sn1 Sn 3 a 0.00000 0.00000 0.00000 1.00000
Ni1 Ni 3 b 0.00000 0.00000 0.50000 1.00000
Cl1 Cl 18 f 0.81000 0.82000 0.13000 1.00000
H2O1 H2O 18 f 0.14333 0.48667 0.29667 1.00000
```

$Ni(H_2O)_6SnCl_6$ ($I6_1$): A6B6CD_hR14_148_f_f_b_a - POSCAR

```
A6B6CD_hR14_148_f_f_b_a & a,c/a,x3,y3,z3,x4,y4,z4 --params=10.59967,
    ↪ 1.01326739417,0.94,-0.69,0.14,0.44,-0.19,0.64001 & R-3 C_{3i}^{
    ↪ 2} #148 (abf^2) & hR14 & $I6_{1}$ & Cl6(H2O)6NiSn & Cl6(H2O)
    ↪ 6NiSn & L. Pauling, Zeitschrift f{"u}r Kristallographie -
    ↪ Crystalline Materials 72, 482-492 (1930)
   1.00000000000000
   5.29983500000000  -3.05986116391060   3.58010000000000
   0.00000000000000   6.11972232782120   3.58010000000000
  -5.29983500000000  -3.05986116391060   3.58010000000000
    Cl     H2O     Ni     Sn
     6       6      1      1
Direct
   0.94000000000000  -0.69000000000000   0.14000000000000   Cl    (6f)
   0.14000000000000   0.94000000000000  -0.69000000000000   Cl    (6f)
  -0.69000000000000   0.14000000000000   0.94000000000000   Cl    (6f)
  -0.94000000000000   0.69000000000000  -0.14000000000000   Cl    (6f)
  -0.14000000000000  -0.94000000000000   0.69000000000000   Cl    (6f)
   0.69000000000000  -0.14000000000000  -0.94000000000000   Cl    (6f)
   0.44000000000000  -0.19000000000000   0.64001000000000   H2O   (6f)
   0.64001000000000   0.44000000000000  -0.19000000000000   H2O   (6f)
  -0.19000000000000   0.64001000000000   0.44000000000000   H2O   (6f)
  -0.44000000000000   0.19000000000000  -0.64001000000000   H2O   (6f)
  -0.64001000000000  -0.44000000000000   0.19000000000000   H2O   (6f)
   0.19000000000000  -0.64001000000000  -0.44000000000000   H2O   (6f)
   0.50000000000000   0.50000000000000   0.50000000000000   Ni    (1b)
   0.00000000000000   0.00000000000000   0.00000000000000   Sn    (1a)
```

$Li_7TaO_6$: A8B6C_hR15_148_cf_f_a - CIF

```
# CIF file
data_findsym-output
_audit_creation_method FINDSYM

_chemical_name_mineral 'Li7O6Ta'
_chemical_formula_sum 'Li8 O6 Ta'

loop_
_publ_author_name
 'G. Wehrum'
 'R. Hoppe'
_journal_name_full_name
;
 Zeitschrift fur Anorganische und Allgemeine Chemie
;
_journal_volume 620
_journal_year 1994
_journal_page_first 659
_journal_page_last 664
_publ_Section_title
;
 Zur Kenntnis `Kationen-reicher\' Tantalate {\"U}ber Li$_{7}$[TaO$_{6}$]
;

# Found in The solid-state Li-ion conductor Li$_{7}$TaO$_{6}$: A
    ↪ combined computational and experimental study, 2019 Found in
    ↪ The solid-state Li-ion conductor Li$_{7}$TaO$_{6}$: A combined
    ↪ computational and experimental study, {arXiv:1910.11079 [
    ↪ cond-mat.mtrl-sci]},

_aflow_title 'Li$_{7}$TaO$_{6}$ Structure'
_aflow_proto 'A8B6C_hR15_148_cf_f_a'
_aflow_params 'a,c/a,x_{2},x_{3},y_{3},z_{3},x_{4},y_{4},z_{4}'
_aflow_params_values '5.358,2.81317655842,0.6615,0.5001,-0.2379,0.1053,
    ↪ 0.7714,-0.6257,0.0853'
_aflow_Strukturbericht 'None'
_aflow_Pearson 'hR15'

_symmetry_space_group_name_H-M "R -3 (hexagonal axes)"
_symmetry_Int_Tables_number 148

_cell_length_a      5.35800
_cell_length_b      5.35800
_cell_length_c      15.07300
_cell_angle_alpha   90.00000
_cell_angle_beta    90.00000
_cell_angle_gamma  120.00000

loop_
_space_group_symop_id
_space_group_symop_operation_xyz
1 x,y,z
2 -y,x-y,z
3 -x+y,-x,z
4 -x,-y,-z
5 y,-x+y,-z
6 x-y,x,-z
7 x+1/3,y+2/3,z+2/3
8 -y+1/3,x-y+2/3,z+2/3
9 -x+y+1/3,-x+2/3,z+2/3
10 -x+1/3,-y+2/3,-z+2/3
11 y+1/3,-x+y+2/3,-z+2/3
12 x-y+1/3,x+2/3,-z+2/3
13 x+2/3,y+1/3,z+1/3
14 -y+2/3,x-y+1/3,z+1/3
15 -x+y+2/3,-x+1/3,z+1/3
16 -x+2/3,-y+1/3,-z+1/3
17 y+2/3,-x+y+1/3,-z+1/3
18 x-y+2/3,x+1/3,-z+1/3

loop_
_atom_site_label
_atom_site_type_symbol
_atom_site_symmetry_multiplicity
_atom_site_Wyckoff_label
_atom_site_fract_x
_atom_site_fract_y
_atom_site_fract_z
_atom_site_occupancy
Ta1 Ta 3 a 0.00000 0.00000 0.00000 1.00000
Li1 Li 6 c 0.00000 0.00000 0.66150 0.50000
Li2 Li 18 f 0.37760 0.36040 0.12250 1.00000
O1 O 18 f 0.69440 0.70270 0.07700 1.00000
```

$Li_7TaO_6$: A8B6C_hR15_148_cf_f_a - POSCAR

```
A8B6C_hR15_148_cf_f_a & a,c/a,x2,x3,y3,z3,x4,y4,z4 --params=5.358,
    ↪ 2.81317655842,0.6615,0.5001,-0.2379,0.1053,0.7714,-0.6257,
    ↪ 0.0853 & R-3 C_{3i}^{2} #148 (acf^2) & hR15 & None & Li7O6Ta &
    ↪ Li7O6Ta & G. Wehrum and R. Hoppe, Z. Anorg. Allg. Chem. 620,
    ↪ 659-664 (1994)
   1.00000000000000
   2.67900000000000  -1.54672137115901   5.02433333333333
   0.00000000000000   3.09344274231802   5.02433333333333
  -2.67900000000000  -1.54672137115901   5.02433333333333
    Li      O      Ta
     8      6       1
Direct
   0.66150000000000   0.66150000000000   0.66150000000000   Li    (2c)
  -0.66150000000000  -0.66150000000000  -0.66150000000000   Li    (2c)
   0.50010000000000  -0.23790000000000   0.10530000000000   Li    (6f)
   0.10530000000000   0.50010000000000  -0.23790000000000   Li    (6f)
  -0.23790000000000   0.10530000000000   0.50010000000000   Li    (6f)
  -0.50010000000000   0.23790000000000  -0.10530000000000   Li    (6f)
  -0.10530000000000  -0.50010000000000   0.23790000000000   Li    (6f)
   0.23790000000000  -0.10530000000000  -0.50010000000000   Li    (6f)
   0.77140000000000  -0.62570000000000   0.08530000000000   O     (6f)
   0.08530000000000   0.77140000000000  -0.62570000000000   O     (6f)
  -0.62570000000000   0.08530000000000   0.77140000000000   O     (6f)
  -0.77140000000000   0.62570000000000  -0.08530000000000   O     (6f)
  -0.08530000000000  -0.77140000000000   0.62570000000000   O     (6f)
   0.62570000000000  -0.08530000000000  -0.77140000000000   O     (6f)
   0.00000000000000   0.00000000000000   0.00000000000000   Ta    (1a)
```

$SrCl_2{\cdot}(H_2O)_6$: A2B12C6D_hP21_150_d_2g_ef_a - CIF

```
# CIF file
data_findsym-output
_audit_creation_method FINDSYM

_chemical_name_mineral 'Cl2H12O6Sr'
_chemical_formula_sum 'Cl2 H12 O6 Sr'

loop_
_publ_author_name
 'P. A. Agron'
 'W. R. Busing'
_journal_name_full_name
;
 Acta Crystallographica Section C: Structural Chemistry
;
_journal_volume 42
_journal_year 1986
_journal_page_first 141
_journal_page_last 143
_publ_Section_title
;
 Calcium and strontium dichloride hexahydrates by neutron diffraction
;

# Found in PAULING FILE, 2016 Found in PAULING FILE, {in: Inorganic
    ↪ Solid Phases, SpringerMaterials (online database), Springer,
    ↪ Heidelberg SpringerMaterials},

_aflow_title 'SrCl$_{2}$$\cdot$(H$_{2}$O)$_{6}$ Structure'
_aflow_proto 'A2B12C6D_hP21_150_d_2g_ef_a'
_aflow_params 'a,c/a,z_{2},x_{3},x_{4},x_{5},y_{5},z_{5},x_{6},y_{6},z_{
    ↪ 6}'
_aflow_params_values '7.9596,0.518154178602,0.429,0.3114,0.7868,0.4326,
    ↪ 0.0988,-0.0926,0.767,0.1113,0.4835'
_aflow_Strukturbericht 'None'
_aflow_Pearson 'hP21'

_symmetry_space_group_name_H-M "P 3 2 1"
_symmetry_Int_Tables_number 150

_cell_length_a      7.95960
_cell_length_b      7.95960
_cell_length_c      4.12430
_cell_angle_alpha  90.00000
_cell_angle_beta   90.00000
_cell_angle_gamma 120.00000

loop_
_space_group_symop_id
_space_group_symop_operation_xyz
1 x,y,z
2 -y,x-y,z
3 -x+y,-x,z
4 x-y,-y,-z
5 y,x,-z
6 -x,-x+y,-z

loop_
_atom_site_label
_atom_site_type_symbol
_atom_site_symmetry_multiplicity
_atom_site_Wyckoff_label
_atom_site_fract_x
_atom_site_fract_y
_atom_site_fract_z
_atom_site_occupancy
Sr1 Sr 1 a 0.00000 0.00000 0.00000 1.00000
Cl1 Cl 2 d 0.33333 0.66667 0.42900 1.00000
O1 O 3 e 0.31140 0.00000 0.00000 1.00000
O2 O 3 f 0.78680 0.00000 0.50000 1.00000
H1 H 6 g 0.43260 0.09880 -0.09260 1.00000
H2 H 6 g 0.76700 0.11130 0.48350 1.00000
```

$SrCl_2{\cdot}(H_2O)_6$: A2B12C6D_hP21_150_d_2g_ef_a - POSCAR

```
A2B12C6D_hP21_150_d_2g_ef_a & a,c/a,z2,x3,x4,x5,y5,z5,x6,y6,z6 --params=
    ↪ 7.9596,0.518154178602,0.429,0.3114,0.7868,0.4326,0.0988,-0.0926
    ↪ ,0.767,0.1113,0.4835 & P321 D_{3}^{2} #150 (adefg^2) & hP21 &
    ↪ None & Cl2H12O6Sr & Cl2H12O6Sr & P. A. Agron and W. R. Busing,
    ↪ Acta Crystallogr. C 42, 141-143 (1986)
   1.00000000000000
   3.97980000000000  -6.89321580396262   0.00000000000000
   3.97980000000000   6.89321580396262   0.00000000000000
   0.00000000000000   0.00000000000000   4.12430000000000
    Cl      H      O     Sr
     2     12      6      1
Direct
   0.33333333333333   0.66666666666667   0.42900000000000   Cl   (2d)
   0.66666666666667   0.33333333333333  -0.42900000000000   Cl   (2d)
   0.43260000000000   0.09880000000000  -0.09260000000000   H    (6g)
  -0.09880000000000   0.33380000000000  -0.09260000000000   H    (6g)
  -0.33380000000000  -0.43260000000000  -0.09260000000000   H    (6g)
   0.09880000000000   0.43260000000000   0.09260000000000   H    (6g)
   0.33380000000000  -0.09880000000000   0.09260000000000   H    (6g)
  -0.43260000000000  -0.33380000000000   0.09260000000000   H    (6g)
   0.76700000000000   0.11130000000000   0.48350000000000   H    (6g)
  -0.11130000000000   0.65570000000000   0.48350000000000   H    (6g)
  -0.65570000000000  -0.76700000000000   0.48350000000000   H    (6g)
   0.11130000000000   0.76700000000000  -0.48350000000000   H    (6g)
   0.65570000000000  -0.11130000000000  -0.48350000000000   H    (6g)
  -0.76700000000000  -0.65570000000000  -0.48350000000000   H    (6g)
   0.31140000000000   0.00000000000000   0.00000000000000   O    (3e)
   0.00000000000000   0.31140000000000   0.00000000000000   O    (3e)
  -0.31140000000000  -0.31140000000000   0.00000000000000   O    (3e)
   0.78680000000000   0.00000000000000   0.50000000000000   O    (3f)
   0.00000000000000   0.78680000000000   0.50000000000000   O    (3f)
  -0.78680000000000  -0.78680000000000   0.50000000000000   O    (3f)
   0.00000000000000   0.00000000000000   0.00000000000000   Sr   (1a)
```

$Cs_3As_2Cl_9$ ($K7_3$): A2B9C3_hP14_150_d_eg_ad - CIF

```
# CIF file
data_findsym-output
_audit_creation_method FINDSYM

_chemical_name_mineral 'As2Cl9Cs3'
_chemical_formula_sum 'As2 Cl9 Cs3'

loop_
_publ_author_name
 'J. L. Hoard'
 'L. Goldstein'
_journal_name_full_name
;
 Journal of Chemical Physics
;
_journal_volume 3
_journal_year 1935
_journal_page_first 117
_journal_page_last 122
_publ_Section_title
;
 The Structure of Caesium Enneachlordiarsenite, Cs$_{3}$As$_{2}$Cl$_{9}$
;

_aflow_title 'Cs$_{3}$As$_{2}$Cl$_{9}$ ($K7_{3}$) Structure'
_aflow_proto 'A2B9C3_hP14_150_d_eg_ad'
_aflow_params 'a,c/a,z_{2},z_{3},x_{4},x_{5},y_{5},z_{5}'
_aflow_params_values '7.37,1.20895522388,0.805,0.32,0.48,0.365,0.2,0.325
    ↪ '
_aflow_Strukturbericht '$K7_{3}$'
_aflow_Pearson 'hP14'

_symmetry_space_group_name_H-M "P 3 2 1"
_symmetry_Int_Tables_number 150

_cell_length_a      7.37000
_cell_length_b      7.37000
_cell_length_c      8.91000
_cell_angle_alpha  90.00000
_cell_angle_beta   90.00000
_cell_angle_gamma 120.00000

loop_
_space_group_symop_id
_space_group_symop_operation_xyz
1 x,y,z
2 -y,x-y,z
3 -x+y,-x,z
4 x-y,-y,-z
5 y,x,-z
6 -x,-x+y,-z

loop_
_atom_site_label
_atom_site_type_symbol
_atom_site_symmetry_multiplicity
_atom_site_Wyckoff_label
_atom_site_fract_x
_atom_site_fract_y
_atom_site_fract_z
_atom_site_occupancy
Cs1 Cs 1 a 0.00000 0.00000 0.00000 1.00000
As1 As 2 d 0.33333 0.66667 0.80500 1.00000
Cs2 Cs 2 d 0.33333 0.66667 0.32000 1.00000
Cl1 Cl 3 e 0.48000 0.00000 0.00000 1.00000
Cl2 Cl 6 g 0.36500 0.20000 0.32500 1.00000
```

$Cs_3As_2Cl_9$ ($K7_3$): A2B9C3_hP14_150_d_eg_ad - POSCAR

```
A2B9C3_hP14_150_d_eg_ad & a,c/a,z2,z3,x4,x5,y5,z5 --params=7.37,
    ↪ 1.20895522388,0.805,0.32,0.48,0.365,0.2,0.325 & P321 D_{3}^{2}
    ↪ #150 (ad^2eg) & hP14 & $K7_{3}$ & As2Cl9Cs3 & As2Cl9Cs3 & J. L.
    ↪ Hoard and L. Goldstein, J. Chem. Phys. 3, 117-122 (1935)
   1.00000000000000
   3.68500000000000  -6.38260722589131   0.00000000000000
   3.68500000000000   6.38260722589131   0.00000000000000
   0.00000000000000   0.00000000000000   8.91000000000000
    As     Cl     Cs
     2      9      3
Direct
   0.33333333333333   0.66666666666667   0.80500000000000   As   (2d)
   0.66666666666667   0.33333333333333  -0.80500000000000   As   (2d)
   0.48000000000000   0.00000000000000   0.00000000000000   Cl   (3e)
   0.00000000000000   0.48000000000000   0.00000000000000   Cl   (3e)
  -0.48000000000000  -0.48000000000000   0.00000000000000   Cl   (3e)
   0.36500000000000   0.20000000000000   0.32500000000000   Cl   (6g)
  -0.20000000000000   0.16500000000000   0.32500000000000   Cl   (6g)
  -0.16500000000000  -0.36500000000000   0.32500000000000   Cl   (6g)
   0.20000000000000   0.36500000000000  -0.32500000000000   Cl   (6g)
   0.16500000000000  -0.20000000000000  -0.32500000000000   Cl   (6g)
  -0.36500000000000  -0.16500000000000  -0.32500000000000   Cl   (6g)
   0.00000000000000   0.00000000000000   0.00000000000000   Cs   (1a)
   0.33333333333333   0.66666666666667   0.32000000000000   Cs   (2d)
   0.66666666666667   0.33333333333333  -0.32000000000000   Cs   (2d)
```

Paralstonite ($BaCa(CO_3)_2$): AB2CD6_hP30_150_e_c2d_f_3g - CIF

```
# CIF file
```

```
data_findsym-output
_audit_creation_method FINDSYM

_chemical_name_mineral 'Paralstonite'
_chemical_formula_sum 'Ba C2 Ca O6'

loop_
_publ_author_name
 'H. Effenberger'
_journal_name_full_name
;
 Neues Jahrbuch fur Mineralogie, Monatshefte
;
_journal_volume 1980
_journal_year 1980
_journal_page_first 353
_journal_page_last 363
_publ_Section_title
;
 Die Kristallstruktur des Minerals Paralstonite, BaCa(CO$_{3}$)$_{2}$
;

# Found in A new BaCa(CO$_{3}$)$_{2}$ polymorph, 2019

_aflow_title 'Paralstonite (BaCa(CO$_{3}$)$_{2}$) Structure'
_aflow_proto 'AB2CD6_hP30_150_e_c2d_f_3g'
_aflow_params 'a,c/a,z_{1},z_{2},z_{3},x_{4},x_{5},x_{6},y_{6},z_{6},x_{
    ↪ 7},y_{7},z_{7},x_{8},y_{8},z_{8}'
_aflow_params_values '8.692,0.707317073171,0.267,0.356,0.817,0.687,
    ↪ 0.3586,0.191,0.677,0.355,0.173,0.517,0.822,0.15,0.001,0.255'
_aflow_Strukturbericht 'None'
_aflow_Pearson 'hP30'

_symmetry_space_group_name_H-M "P 3 2 1"
_symmetry_Int_Tables_number 150

_cell_length_a     8.69200
_cell_length_b     8.69200
_cell_length_c     6.14800
_cell_angle_alpha 90.00000
_cell_angle_beta  90.00000
_cell_angle_gamma 120.00000

loop_
_space_group_symop_id
_space_group_symop_operation_xyz
1 x,y,z
2 -y,x-y,z
3 -x+y,-x,z
4 x-y,-y,-z
5 y,x,-z
6 -x,-x+y,-z

loop_
_atom_site_label
_atom_site_type_symbol
_atom_site_symmetry_multiplicity
_atom_site_Wyckoff_label
_atom_site_fract_x
_atom_site_fract_y
_atom_site_fract_z
_atom_site_occupancy
C1 C 2 c 0.00000 0.00000 0.26700 1.00000
C2 C 2 d 0.33333 0.66667 0.35600 1.00000
C3 C 2 d 0.33333 0.66667 0.81700 1.00000
Ba1 Ba 3 e 0.68700 0.00000 0.00000 1.00000
Ca1 Ca 3 f 0.35860 0.00000 0.50000 1.00000
O1 O 6 g 0.19100 0.67700 0.35500 1.00000
O2 O 6 g 0.17300 0.51700 0.82200 1.00000
O3 O 6 g 0.15000 0.00100 0.25500 1.00000
```

Paralstonite ($BaCa(CO_3)_2$): AB2CD6_hP30_150_e_c2d_f_3g - POSCAR

```
AB2CD6_hP30_150_e_c2d_f_3g & a,c/a,z1,z2,z3,x4,x5,x6,y6,z6,x7,y7,z7,x8,
    ↪ y8,z8 --params=8.692,0.707317073171,0.267,0.356,0.817,0.687,
    ↪ 0.3586,0.191,0.677,0.355,0.173,0.517,0.822,0.15,0.001,0.255 &
    ↪ P321 D_{3}^{2} #150 (cd^2efg^3) & hP30 & None & BaC2CaO6 &
    ↪ Paralstonite & H. Effenberger, Neues Jahrb. Mineral. Monatsh.
    ↪ 1980, 353-363 (1980)
   1.00000000000000
   4.34600000000000  -7.52749280969434   0.00000000000000
   4.34600000000000   7.52749280969434   0.00000000000000
   0.00000000000000   0.00000000000000   6.14800000000000
    Ba    C    Ca    O
     3    6     3   18
Direct
   0.68700000000000   0.00000000000000   0.00000000000000   Ba   (3e)
   0.00000000000000   0.68700000000000   0.00000000000000   Ba   (3e)
  -0.68700000000000  -0.68700000000000   0.00000000000000   Ba   (3e)
   0.00000000000000   0.00000000000000   0.26700000000000   C    (2c)
   0.00000000000000   0.00000000000000  -0.26700000000000   C    (2c)
   0.33333333333333   0.66666666666667   0.35600000000000   C    (2d)
   0.66666666666667   0.33333333333333  -0.35600000000000   C    (2d)
   0.33333333333333   0.66666666666667   0.81700000000000   C    (2d)
   0.66666666666667   0.33333333333333  -0.81700000000000   C    (2d)
   0.35860000000000   0.00000000000000   0.50000000000000   Ca   (3f)
   0.00000000000000   0.35860000000000   0.50000000000000   Ca   (3f)
  -0.35860000000000  -0.35860000000000   0.50000000000000   Ca   (3f)
   0.19100000000000   0.67700000000000   0.35500000000000   O    (6g)
  -0.67700000000000  -0.48600000000000   0.35500000000000   O    (6g)
   0.48600000000000  -0.19100000000000   0.35500000000000   O    (6g)
   0.67700000000000   0.19100000000000  -0.35500000000000   O    (6g)
  -0.48600000000000  -0.67700000000000  -0.35500000000000   O    (6g)
  -0.19100000000000   0.48600000000000  -0.35500000000000   O    (6g)
   0.17300000000000   0.51700000000000   0.82200000000000   O    (6g)
  -0.51700000000000  -0.34400000000000   0.82200000000000   O    (6g)
   0.34400000000000  -0.17300000000000   0.82200000000000   O    (6g)
   0.51700000000000   0.17300000000000  -0.82200000000000   O    (6g)
  -0.34400000000000  -0.51700000000000  -0.82200000000000   O    (6g)
  -0.17300000000000   0.34400000000000  -0.82200000000000   O    (6g)
   0.15000000000000   0.00100000000000   0.25500000000000   O    (6g)
  -0.00100000000000   0.14900000000000   0.25500000000000   O    (6g)
  -0.14900000000000  -0.15000000000000   0.25500000000000   O    (6g)
   0.00100000000000   0.15000000000000  -0.25500000000000   O    (6g)
   0.14900000000000  -0.00100000000000  -0.25500000000000   O    (6g)
  -0.15000000000000  -0.14900000000000  -0.25500000000000   O    (6g)
```

$KSO_3$ ($K1_1$): AB3C_hP30_150_ef_3g_c2d - CIF

```
# CIF file
data_findsym-output
_audit_creation_method FINDSYM

_chemical_name_mineral 'KO3S'
_chemical_formula_sum 'K O3 S'

loop_
_publ_author_name
 'M. L. Huggins'
 'G. O. Frank'
_journal_name_full_name
;
 American Mineralogist
;
_journal_volume 16
_journal_year 1931
_journal_page_first 580
_journal_page_last 591
_publ_Section_title
;
 The crystal structure of potassium dithionate, K$_{2}$S$_{2}$O$_{6}$
;

# Found in Strukturbericht Band II 1928-1932, 1937

_aflow_title 'KSO$_{3}$ ($K1_{1}$) Structure'
_aflow_proto 'AB3C_hP30_150_ef_3g_c2d'
_aflow_params 'a,c/a,z_{1},z_{2},z_{3},x_{4},x_{5},x_{6},y_{6},z_{6},x_{
    ↪ 7},y_{7},z_{7},x_{8},y_{8},z_{8}'
_aflow_params_values '9.82,0.647657841141,0.16,0.59,0.27,0.3,0.62,0.16,
    ↪ 0.11,0.23,0.61,0.17,0.34,0.5,0.21,0.8'
_aflow_Strukturbericht '$K1_{1}$'
_aflow_Pearson 'hP30'

_symmetry_space_group_name_H-M "P 3 2 1"
_symmetry_Int_Tables_number 150

_cell_length_a     9.82000
_cell_length_b     9.82000
_cell_length_c     6.36000
_cell_angle_alpha 90.00000
_cell_angle_beta  90.00000
_cell_angle_gamma 120.00000

loop_
_space_group_symop_id
_space_group_symop_operation_xyz
1 x,y,z
2 -y,x-y,z
3 -x+y,-x,z
4 x-y,-y,-z
5 y,x,-z
6 -x,-x+y,-z

loop_
_atom_site_label
_atom_site_type_symbol
_atom_site_symmetry_multiplicity
_atom_site_Wyckoff_label
_atom_site_fract_x
_atom_site_fract_y
_atom_site_fract_z
_atom_site_occupancy
S1 S 2 c 0.00000 0.00000 0.16000 1.00000
S2 S 2 d 0.33333 0.66667 0.59000 1.00000
S3 S 2 d 0.33333 0.66667 0.27000 1.00000
K1 K 3 e 0.30000 0.00000 0.00000 1.00000
K2 K 3 f 0.62000 0.00000 0.50000 1.00000
O1 O 6 g 0.16000 0.11000 0.23000 1.00000
O2 O 6 g 0.61000 0.17000 0.34000 1.00000
O3 O 6 g 0.50000 0.21000 0.80000 1.00000
```

$KSO_3$ ($K1_1$): AB3C_hP30_150_ef_3g_c2d - POSCAR

```
AB3C_hP30_150_ef_3g_c2d & a,c/a,z1,z2,z3,x4,x5,x6,y6,z6,x7,y7,z7,x8,y8,
    ↪ z8 --params=9.82,0.647657841141,0.16,0.59,0.27,0.3,0.62,0.16,
    ↪ 0.11,0.23,0.61,0.17,0.34,0.5,0.21,0.8 & P321 D_{3}^{2} #150 (cd
    ↪ ^2efg^3) & hP30 & $K1_{1}$ & KO3S & KO3S & M. L. Huggins and G.
    ↪ O. Frank, Am. Mineral. 16, 580-591 (1931)
   1.00000000000000
   4.91000000000000  -8.50436946516319   0.00000000000000
   4.91000000000000   8.50436946516319   0.00000000000000
   0.00000000000000   0.00000000000000   6.36000000000000
     K     O     S
     6    18     6
Direct
   0.30000000000000   0.00000000000000   0.00000000000000   K    (3e)
   0.00000000000000   0.30000000000000   0.00000000000000   K    (3e)
  -0.30000000000000  -0.30000000000000   0.00000000000000   K    (3e)
   0.62000000000000   0.00000000000000   0.50000000000000   K    (3f)
```

```
   0.00000000000000    0.62000000000000    0.50000000000000    K    (3f)
  -0.62000000000000   -0.62000000000000    0.50000000000000    K    (3f)
   0.16000000000000    0.11000000000000    0.23000000000000    O    (6g)
  -0.11000000000000    0.05000000000000    0.23000000000000    O    (6g)
  -0.05000000000000   -0.16000000000000    0.23000000000000    O    (6g)
   0.11000000000000    0.16000000000000   -0.23000000000000    O    (6g)
   0.05000000000000   -0.11000000000000   -0.23000000000000    O    (6g)
  -0.16000000000000   -0.05000000000000   -0.23000000000000    O    (6g)
   0.61000000000000    0.17000000000000    0.34000000000000    O    (6g)
  -0.17000000000000    0.44000000000000    0.34000000000000    O    (6g)
  -0.44000000000000   -0.61000000000000    0.34000000000000    O    (6g)
   0.17000000000000    0.61000000000000   -0.34000000000000    O    (6g)
   0.44000000000000   -0.17000000000000   -0.34000000000000    O    (6g)
  -0.61000000000000   -0.44000000000000   -0.34000000000000    O    (6g)
   0.50000000000000    0.21000000000000    0.80000000000000    O    (6g)
  -0.21000000000000    0.29000000000000    0.80000000000000    O    (6g)
  -0.29000000000000   -0.50000000000000    0.80000000000000    O    (6g)
   0.21000000000000    0.50000000000000   -0.80000000000000    O    (6g)
   0.29000000000000   -0.21000000000000   -0.80000000000000    O    (6g)
  -0.50000000000000   -0.29000000000000   -0.80000000000000    O    (6g)
   0.00000000000000    0.00000000000000    0.16000000000000    S    (2c)
   0.00000000000000    0.00000000000000   -0.16000000000000    S    (2c)
   0.33333333333333    0.66666666666667    0.59000000000000    S    (2d)
   0.66666666666667    0.33333333333333   -0.59000000000000    S    (2d)
   0.33333333333333    0.66666666666667    0.27000000000000    S    (2d)
   0.66666666666667    0.33333333333333   -0.27000000000000    S    (2d)
```

Steklite [KAl(SO$_4$)$_2$, $H3_2$]: ABC8D2_hP12_150_b_a_dg_d - CIF

```
# CIF file
data_findsym-output
_audit_creation_method FINDSYM

_chemical_name_mineral 'Steklite'
_chemical_formula_sum 'Al K O8 S2'

loop_
_publ_author_name
 'L. Vegard'
 'A. Maurstad'
_journal_name_full_name
;
 Zeitschrift f{\"u}r Kristallographie - Crystalline Materials
;
_journal_volume 69
_journal_year 1929
_journal_page_first 519
_journal_page_last 532
_publ_Section_title
;
 Die Kristallstruktur der wasserfreien Alaune $R$\'$R$\'\'(SO$_{4}$)$_{2
   ↪ }$
;

# Found in Strukturbericht 1913-1928, 1931

_aflow_title 'Steklite [KAl(SO$_{4}$)$_{2}$, $H3_{2}$] Structure'
_aflow_proto 'ABC8D2_hP12_150_b_a_dg_d'
_aflow_params 'a,c/a,z_{3},z_{4},x_{5},y_{5},z_{5}'
_aflow_params_values '4.7,1.69361702128,0.016,0.222,0.328,0.344,0.317'
_aflow_Strukturbericht '$H3_{2}$'
_aflow_Pearson 'hP12'

_symmetry_space_group_name_H-M "P 3 2 1"
_symmetry_Int_Tables_number 150

_cell_length_a      4.70000
_cell_length_b      4.70000
_cell_length_c      7.96000
_cell_angle_alpha  90.00000
_cell_angle_beta   90.00000
_cell_angle_gamma 120.00000

loop_
_space_group_symop_id
_space_group_symop_operation_xyz
1 x,y,z
2 -y,x-y,z
3 -x+y,-x,z
4 x-y,-y,-z
5 y,x,-z
6 -x,-x+y,-z

loop_
_atom_site_label
_atom_site_type_symbol
_atom_site_symmetry_multiplicity
_atom_site_Wyckoff_label
_atom_site_fract_x
_atom_site_fract_y
_atom_site_fract_z
_atom_site_occupancy
K1 K 1 a 0.00000 0.00000 0.00000 1.00000
Al1 Al 1 b 0.00000 0.00000 0.50000 1.00000
O1 O 2 d 0.33333 0.66667 0.01600 1.00000
S1 S 2 d 0.33333 0.66667 0.22200 1.00000
O2 O 6 g 0.32800 0.34400 0.31700 1.00000
```

Steklite [KAl(SO$_4$)$_2$, $H3_2$]: ABC8D2_hP12_150_b_a_dg_d - POSCAR

```
ABC8D2_hP12_150_b_a_dg_d & a,c/a,z3,z4,x5,y5,z5 --params=4.7,
   ↪ 1.69361702128,0.016,0.222,0.328,0.344,0.317 & P321 D_{3}^{2} #
   ↪ 150 (abd^2g) & hP12 & $H3_{2}$ & AlKO8S2 & Steklite & L. Vegard
   ↪ and A. Maurstad, Zeitschrift f{"u}r Kristallographie -
   ↪ Crystalline Materials 69, 519-532 (1929)
   1.00000000000000
   2.35000000000000   -4.07031939778686    0.00000000000000
   2.35000000000000    4.07031939778686    0.00000000000000
   0.00000000000000    0.00000000000000    7.96000000000000
    Al     K     O     S
     1     1     8     2
Direct
   0.00000000000000    0.00000000000000    0.50000000000000   Al   (1b)
   0.00000000000000    0.00000000000000    0.00000000000000   K    (1a)
   0.33333333333333    0.66666666666667    0.01600000000000   O    (2d)
   0.66666666666667    0.33333333333333   -0.01600000000000   O    (2d)
   0.32800000000000    0.34400000000000    0.31700000000000   O    (6g)
  -0.34400000000000   -0.01600000000000    0.31700000000000   O    (6g)
   0.01600000000000   -0.32800000000000    0.31700000000000   O    (6g)
   0.34400000000000    0.32800000000000   -0.31700000000000   O    (6g)
  -0.01600000000000   -0.34400000000000   -0.31700000000000   O    (6g)
  -0.32800000000000    0.01600000000000   -0.31700000000000   O    (6g)
   0.33333333333333    0.66666666666667    0.22200000000000   S    (2d)
   0.66666666666667    0.33333333333333   -0.22200000000000   S    (2d)
```

KBe$_2$BO$_3$F$_2$: AB2C2DE3_hR9_155_b_c_c_a_e - CIF

```
# CIF file
data_findsym-output
_audit_creation_method FINDSYM

_chemical_name_mineral 'BBe2F2KO3'
_chemical_formula_sum 'B Be2 F2 K O3'

loop_
_publ_author_name
 'L. Mei'
 'X. Huang'
 'Y. Want'
 'Q. Wu'
 'C. Chen'
_journal_name_full_name
;
 Zeitschrift f{\"u}r Kristallographie - Crystalline Materials
;
_journal_volume 210
_journal_year 1995
_journal_page_first 93
_journal_page_last 95
_publ_Section_title
;
 Crystal Structure of KBe$_{2}$BO$_{3}$F$_{2}$
;

_aflow_title 'KBe$_{2}$BO$_{3}$F$_{2}$ Structure'
_aflow_proto 'AB2C2DE3_hR9_155_b_c_c_a_e'
_aflow_params 'a,c/a,x_{3},x_{4},y_{5}'
_aflow_params_values '4.427,4.2340185227,0.19743,0.27848,0.5'
_aflow_Strukturbericht 'None'
_aflow_Pearson 'hR9'

_symmetry_space_group_name_H-M "R 3 2 (hexagonal axes)"
_symmetry_Int_Tables_number 155

_cell_length_a      4.42700
_cell_length_b      4.42700
_cell_length_c     18.74400
_cell_angle_alpha  90.00000
_cell_angle_beta   90.00000
_cell_angle_gamma 120.00000

loop_
_space_group_symop_id
_space_group_symop_operation_xyz
1 x,y,z
2 -y,x-y,z
3 -x+y,-x,z
4 y,x,-z
5 -x,-x+y,-z
6 x-y,-y,-z
7 x+1/3,y+2/3,z+2/3
8 -y+1/3,x-y+2/3,z+2/3
9 -x+y+1/3,-x+2/3,z+2/3
10 y+1/3,x+2/3,-z+2/3
11 -x+1/3,-x+y+2/3,-z+2/3
12 x-y+1/3,-y+2/3,-z+2/3
13 x+2/3,y+1/3,z+1/3
14 -y+2/3,x-y+1/3,z+1/3
15 -x+y+2/3,-x+1/3,z+1/3
16 y+2/3,x+1/3,-z+1/3
17 -x+2/3,-x+y+1/3,-z+1/3
18 x-y+2/3,-y+1/3,-z+1/3

loop_
_atom_site_label
_atom_site_type_symbol
_atom_site_symmetry_multiplicity
_atom_site_Wyckoff_label
_atom_site_fract_x
_atom_site_fract_y
_atom_site_fract_z
_atom_site_occupancy
K1 K 3 a 0.00000 0.00000 0.00000 1.00000
B1 B 3 b 0.00000 0.00000 0.50000 1.00000
Be1 Be 6 c 0.00000 0.00000 0.19743 1.00000
F1 F 6 c 0.00000 0.00000 0.27848 1.00000
O1 O 9 e 0.69063 0.00000 0.50000 1.00000
```

KBe$_2$BO$_3$F$_2$: AB2C2DE3_hR9_155_b_c_c_a_e - POSCAR

```
AB2C2DE3_hR9_155_b_c_c_a_e & a,c/a,x3,x4,y5 --params=4.427,4.2340185227,
    ↪ 0.19743,0.27848,0.5 & R32 D_{3}^{7} #155 (abc^2e) & hR9 & None
    ↪ & BBe2F2KO3 & BBe2F2KO3 & L. Mei et al., Zeitschrift f{"u}r
    ↪ Kristallographie - Crystalline Materials 210, 93-95 (1995)
   1.00000000000000
   2.21350000000000   -1.27796482085124   6.24800000000000
   0.00000000000000    2.55592964170247   6.24800000000000
  -2.21350000000000   -1.27796482085124   6.24800000000000
     B     Be     F     K     O
     1      2     2     1     3
Direct
   0.50000000000000    0.50000000000000    0.50000000000000    B    (1b)
   0.19743000000000    0.19743000000000    0.19743000000000    Be   (2c)
  -0.19743000000000   -0.19743000000000   -0.19743000000000    Be   (2c)
   0.27848000000000    0.27848000000000    0.27848000000000    F    (2c)
  -0.27848000000000   -0.27848000000000   -0.27848000000000    F    (2c)
   0.00000000000000    0.00000000000000    0.00000000000000    K    (1a)
   0.50000000000000    0.50000000000000   -0.50000000000000    O    (3e)
  -0.50000000000000    0.50000000000000    0.50000000000000    O    (3e)
   0.50000000000000   -0.50000000000000    0.50000000000000    O    (3e)
```

$SbI_3S_{24}$: A3B24C_hR28_160_b_2b3c_a - CIF

```
# CIF file
data_findsym-output
_audit_creation_method FINDSYM

_chemical_name_mineral 'I3S24Sb'
_chemical_formula_sum 'I3 S24 Sb'

loop_
_publ_author_name
 'T. Bjorvatten'
 'O. Hassel'
 'A. Lindheim'
_journal_name_full_name
;
 Acta Chemica Scandinavica
;
_journal_volume 17
_journal_year 1963
_journal_page_first 689
_journal_page_last 702
_publ_Section_title
;
 Crystal Structure of the Addition Compound SbI$_{3}$:3S$_{8}$
;

_aflow_title 'SbI$_{3}$S$_{24}$ Structure'
_aflow_proto 'A3B24C_hR28_160_b_2b3c_a'
_aflow_params 'a,c/a,x_{1},x_{2},z_{2},x_{3},z_{3},x_{4},z_{4},x_{5},y_{
    ↪ 5},z_{5},x_{6},y_{6},z_{6},x_{7},y_{7},z_{7}'
_aflow_params_values '24.817,0.178422049402,0.3146,-0.0551,0.1102,0.5891
    ↪ ,-0.00921,0.4515,0.17749,0.8641,-0.2688,0.24779,0.6198,-0.5701,
    ↪ 0.08749,0.7609,-0.3733,0.37359'
_aflow_Strukturbericht 'None'
_aflow_Pearson 'hR28'

_symmetry_space_group_name_H-M "R 3 m (hexagonal axes)"
_symmetry_Int_Tables_number 160

_cell_length_a      24.81700
_cell_length_b      24.81700
_cell_length_c      4.42790
_cell_angle_alpha 90.00000
_cell_angle_beta  90.00000
_cell_angle_gamma 120.00000

loop_
_space_group_symop_id
_space_group_symop_operation_xyz
1 x,y,z
2 -y,x-y,z
3 -x+y,-x,z
4 -y,-x,z
5 x,x-y,z
6 -x+y,y,z
7 x+1/3,y+2/3,z+2/3
8 -y+1/3,x-y+2/3,z+2/3
9 -x+y+1/3,-x+2/3,z+2/3
10 -y+1/3,-x+2/3,z+2/3
11 x+1/3,x-y+2/3,z+2/3
12 -x+y+1/3,y+2/3,z+2/3
13 x+2/3,y+1/3,z+1/3
14 -y+2/3,x-y+1/3,z+1/3
15 -x+y+2/3,-x+1/3,z+1/3
16 -y+2/3,-x+1/3,z+1/3
17 x+2/3,x-y+1/3,z+1/3
18 -x+y+2/3,y+1/3,z+1/3

loop_
_atom_site_label
_atom_site_type_symbol
_atom_site_symmetry_multiplicity
_atom_site_Wyckoff_label
_atom_site_fract_x
_atom_site_fract_y
_atom_site_fract_z
_atom_site_occupancy
Sb1 Sb 3 a 0.00000 0.00000 0.31460 1.00000
I1 I 9 b -0.05510 0.05510 0.00000 1.00000
S1 S 9 b 0.53277 0.46723 0.05633 1.00000
S2 S 9 b 0.42467 0.57533 0.02683 1.00000
S3 S 18 c 0.58307 0.54983 0.28103 1.00000
S4 S 18 c 0.57407 0.61583 0.04573 1.00000
S5 S 18 c 0.50717 0.62703 0.25373 1.00000
```

$SbI_3S_{24}$: A3B24C_hR28_160_b_2b3c_a - POSCAR

```
A3B24C_hR28_160_b_2b3c_a & a,c/a,x1,x2,z2,x3,z3,x4,z4,x5,y5,z5,x6,y6,z6,
    ↪ x7,y7,z7 --params=24.817,0.178422049402,0.3146,-0.0551,0.1102,
    ↪ 0.5891,-0.00921,0.4515,0.17749,0.8641,-0.2688,0.24779,0.6198,-
    ↪ 0.5701,0.08749,0.7609,-0.3733,0.37359 & R3m C_{3v}^{5} #160 (ab
    ↪ ^3c^3) & hR28 & None & I3S24Sb & I3S24Sb & T. Bjorvatten and O.
    ↪ Hassel and A. Lindheim, Acta Chem. Scand. 17, 689-702 (1963)
   1.00000000000000
  12.40850000000000   -7.16405081523947    1.47596666666667
   0.00000000000000   14.32810163047890    1.47596666666667
 -12.40850000000000   -7.16405081523947    1.47596666666667
     I      S     Sb
     3     24      1
Direct
  -0.05510000000000   -0.05510000000000    0.11020000000000    I    (3b)
   0.11020000000000   -0.05510000000000   -0.05510000000000    I    (3b)
  -0.05510000000000    0.11020000000000   -0.05510000000000    I    (3b)
   0.58910000000000    0.58910000000000   -0.00921000000000    S    (3b)
  -0.00921000000000    0.58910000000000    0.58910000000000    S    (3b)
   0.58910000000000   -0.00921000000000    0.58910000000000    S    (3b)
   0.45150000000000    0.45150000000000    0.17749000000000    S    (3b)
   0.17749000000000    0.45150000000000    0.45150000000000    S    (3b)
   0.45150000000000    0.17749000000000    0.45150000000000    S    (3b)
   0.86410000000000   -0.26880000000000    0.24779000000000    S    (6c)
   0.24779000000000    0.86410000000000   -0.26880000000000    S    (6c)
  -0.26880000000000    0.24779000000000    0.86410000000000    S    (6c)
   0.24779000000000   -0.26880000000000    0.86410000000000    S    (6c)
  -0.26880000000000    0.86410000000000    0.24779000000000    S    (6c)
   0.86410000000000    0.24779000000000   -0.26880000000000    S    (6c)
   0.61980000000000   -0.57010000000000    0.08749000000000    S    (6c)
   0.08749000000000    0.61980000000000   -0.57010000000000    S    (6c)
  -0.57010000000000    0.08749000000000    0.61980000000000    S    (6c)
   0.08749000000000   -0.57010000000000    0.61980000000000    S    (6c)
  -0.57010000000000    0.61980000000000    0.08749000000000    S    (6c)
   0.61980000000000    0.08749000000000   -0.57010000000000    S    (6c)
   0.76090000000000   -0.37330000000000    0.37359000000000    S    (6c)
   0.37359000000000    0.76090000000000   -0.37330000000000    S    (6c)
  -0.37330000000000    0.37359000000000    0.76090000000000    S    (6c)
   0.37359000000000   -0.37330000000000    0.76090000000000    S    (6c)
  -0.37330000000000    0.76090000000000    0.37359000000000    S    (6c)
   0.76090000000000    0.37359000000000   -0.37330000000000    S    (6c)
   0.31460000000000    0.31460000000000    0.31460000000000    Sb   (1a)
```

$Fe_3PO_7$: A3B7C_hR11_160_b_a2b_a - CIF

```
# CIF file
data_findsym-output
_audit_creation_method FINDSYM

_chemical_name_mineral 'Fe3O7P'
_chemical_formula_sum 'Fe3 O7 P'

loop_
_publ_author_name
 'A. Modaressi'
 'A. Courtois'
 'R. Gerardin'
 'B. Malaman'
 'C. Gleitzer'
_journal_name_full_name
;
 Journal of Solid State Chemistry
;
_journal_volume 47
_journal_year 1983
_journal_page_first 245
_journal_page_last 255
_publ_Section_title
;
 Fe$_{3}$PO$_{7}$, Un cas de coordinence 5 du fer trivalent, etude
    ↪ structurale et magnetique
;

# Found in Partial Antiferromagnetic Helical Order in Single Crystal
    ↪ Fe$_{3}$PO$_{4}$O$_{3}$, {arXiv:1910.08818 [cond-mat.str-el]},

_aflow_title 'Fe$_{3}$PO$_{7}$ Structure'
_aflow_proto 'A3B7C_hR11_160_b_a2b_a'
_aflow_params 'a,c/a,x_{1},x_{2},x_{3},z_{3},x_{4},z_{4},x_{5},z_{5}'
_aflow_params_values '8.006,0.857232075943,0.2289,0.0,0.5395,0.14849,
    ↪ 0.3895,0.76679,0.833,0.144'
_aflow_Strukturbericht 'None'
_aflow_Pearson 'hR11'

_symmetry_space_group_name_H-M "R 3 m (hexagonal axes)"
_symmetry_Int_Tables_number 160

_cell_length_a      8.00600
_cell_length_b      8.00600
_cell_length_c      6.86300
_cell_angle_alpha 90.00000
_cell_angle_beta  90.00000
_cell_angle_gamma 120.00000

loop_
_space_group_symop_id
_space_group_symop_operation_xyz
1 x,y,z
2 -y,x-y,z
3 -x+y,-x,z
4 -y,-x,z
5 x,x-y,z
```

```
6 -x+y,y,z
7 x+1/3,y+2/3,z+2/3
8 -y+1/3,x-y+2/3,z+2/3
9 -x+y+1/3,-x+2/3,z+2/3
10 -y+1/3,-x+2/3,z+2/3
11 x+1/3,x-y+2/3,z+2/3
12 -x+y+1/3,y+2/3,z+2/3
13 x+2/3,y+1/3,z+1/3
14 -y+2/3,x-y+1/3,z+1/3
15 -x+y+2/3,-x+1/3,z+1/3
16 -y+2/3,-x+1/3,z+1/3
17 x+2/3,x-y+1/3,z+1/3
18 -x+y+2/3,y+1/3,z+1/3

loop_
_atom_site_label
_atom_site_type_symbol
_atom_site_symmetry_multiplicity
_atom_site_Wyckoff_label
_atom_site_fract_x
_atom_site_fract_y
_atom_site_fract_z
_atom_site_occupancy
O1 O 3 a 0.00000 0.00000 0.22890 1.00000
P1 P 3 a 0.00000 0.00000 0.00000 1.00000
Fe1 Fe 9 b 0.46367 0.53633 0.07583 1.00000
O2 O 9 b 0.20757 0.79243 0.18193 1.00000
O3 O 9 b 0.56300 0.43700 0.27000 1.00000
```

$Fe_3PO_7$: A3B7C_hR11_160_b_a2b_a - POSCAR

```
A3B7C_hR11_160_b_a2b_a & a,c/a,x1,x2,x3,z3,x4,z4,x5,z5 --params=8.006,
    ↪ 0.857232075943,0.2289,0.0,0.5395,0.14849,0.3895,0.76679,0.833,
    ↪ 0.144 & R3m C_{3v}^{5} #160 (a^2b^3) & hR11 & None & Fe3O7P &
    ↪ Fe3O7P & A. Modaressi et al., J. Solid State Chem. 47, 245-255
    ↪ (1983)
   1.00000000000000
   4.00300000000000  -2.31113312756607   2.28766666666667
   0.00000000000000   4.62226625513214   2.28766666666667
  -4.00300000000000  -2.31113312756607   2.28766666666667
    Fe         O         P
     3         7         1
Direct
   0.53950000000000   0.53950000000000   0.14849000000000   Fe   (3b)
   0.14849000000000   0.53950000000000   0.53950000000000   Fe   (3b)
   0.53950000000000   0.14849000000000   0.53950000000000   Fe   (3b)
   0.22890000000000   0.22890000000000   0.22890000000000   O    (1a)
   0.38950000000000   0.38950000000000   0.76679000000000   O    (3b)
   0.76679000000000   0.38950000000000   0.38950000000000   O    (3b)
   0.38950000000000   0.76679000000000   0.38950000000000   O    (3b)
   0.83300000000000   0.83300000000000   0.14400000000000   O    (3b)
   0.14400000000000   0.83300000000000   0.83300000000000   O    (3b)
   0.83300000000000   0.14400000000000   0.83300000000000   O    (3b)
   0.00000000000000   0.00000000000000   0.00000000000000   P    (1a)
```

Cronstedtite {$Fe(Fe,Si)[(OH)_2,O]O_3$, $S5_7$}: AB3C2D_hR7_160_a_b_2a_a - CIF

```
# CIF file
data_findsym-output
_audit_creation_method FINDSYM

_chemical_name_mineral 'Cronstedtite'
_chemical_formula_sum 'Fe O3 (OH)2 Si'

loop_
_publ_author_name
 'S. B. Hendricks'
_journal_name_full_name
;
 American Mineralogist
;
_journal_volume 24
_journal_year 1939
_journal_page_first 529
_journal_page_last 539
_publ_Section_title
;
 Random structures of layer minerals as illustrated by cronstedite (
    ↪ 2FeO$\cdot$Fe$_{2}$O$_{3}$$\cdot$SiO$_{2}$$\cdot$2H$_{2}$O).
    ↪ Possible iron content of kaolin
;

# Found in Strukturbericht Band VII 1939, 1943

_aflow_title 'Cronstedtite \{Fe(Fe,Si)[(OH)$_{2}$O]O$_{3}$OH, $S5_{7}$\}
    ↪ Structure'
_aflow_proto 'AB3C2D_hR7_160_a_b_2a_a'
_aflow_params 'a,c/a,x_{1},x_{2},x_{3},x_{4},x_{5},z_{5}'
_aflow_params_values '5.48,3.87773722628,0.03,0.11,0.54,0.83,0.5,0.0'
_aflow_Strukturbericht '$S5_{7}$'
_aflow_Pearson 'hR7'

_symmetry_space_group_name_H-M "R 3 m (hexagonal axes)"
_symmetry_Int_Tables_number 160

_cell_length_a    5.48000
_cell_length_b    5.48000
_cell_length_c    21.25000
_cell_angle_alpha 90.00000
_cell_angle_beta  90.00000
_cell_angle_gamma 120.00000

loop_
_space_group_symop_id
_space_group_symop_operation_xyz
1 x,y,z
2 -y,x-y,z
3 -x+y,-x,z
4 -y,-x,z
5 x,x-y,z
6 -x+y,y,z
7 x+1/3,y+2/3,z+2/3
8 -y+1/3,x-y+2/3,z+2/3
9 -x+y+1/3,-x+2/3,z+2/3
10 -y+1/3,-x+2/3,z+2/3
11 x+1/3,x-y+2/3,z+2/3
12 -x+y+1/3,y+2/3,z+2/3
13 x+2/3,y+1/3,z+1/3
14 -y+2/3,x-y+1/3,z+1/3
15 -x+y+2/3,-x+1/3,z+1/3
16 -y+2/3,-x+1/3,z+1/3
17 x+2/3,x-y+1/3,z+1/3
18 -x+y+2/3,y+1/3,z+1/3

loop_
_atom_site_label
_atom_site_type_symbol
_atom_site_symmetry_multiplicity
_atom_site_Wyckoff_label
_atom_site_fract_x
_atom_site_fract_y
_atom_site_fract_z
_atom_site_occupancy
Fe1 Fe 3 a 0.00000 0.00000 0.03000 0.90000
OH1 OH 3 a 0.00000 0.00000 0.11000 1.00000
OH2 OH 3 a 0.00000 0.00000 0.54000 1.00000
Si1 Si 3 a 0.00000 0.00000 0.83000 1.00000
O1 O 9 b 0.50000 0.50000 0.00000 1.00000
```

Cronstedtite {$Fe(Fe,Si)[(OH)_2,O]O_3$, $S5_7$}: AB3C2D_hR7_160_a_b_2a_a - POSCAR

```
AB3C2D_hR7_160_a_b_2a_a & a,c/a,x1,x2,x3,x4,x5,z5 --params=5.48,
    ↪ 3.87773722628,0.03,0.11,0.54,0.83,0.5,0.0 & R3m C_{3v}^{5} #160
    ↪ (a^4b) & hR7 & $S5_{7}$ & FeH2O5Si & Cronstedtite & S. B.
    ↪ Hendricks, Am. Mineral. 24, 529-539 (1939)
   1.00000000000000
   2.74000000000000  -1.58193973757957   7.08333333333333
   0.00000000000000   3.16387947515915   7.08333333333333
  -2.74000000000000  -1.58193973757957   7.08333333333333
    Fe         O        OH        Si
     1         3         2         1
Direct
   0.03000000000000   0.03000000000000   0.03000000000000   Fe   (1a)
   0.50000000000000   0.50000000000000   0.00000000000000   O    (3b)
   0.00000000000000   0.50000000000000   0.50000000000000   O    (3b)
   0.50000000000000   0.00000000000000   0.50000000000000   O    (3b)
   0.11000000000000   0.11000000000000   0.11000000000000   OH   (1a)
   0.54000000000000   0.54000000000000   0.54000000000000   OH   (1a)
   0.83000000000000   0.83000000000000   0.83000000000000   Si   (1a)
```

Low-Temperature $GaMo_4S_8$: AB4C8_hR13_160_a_ab_2a2b - CIF

```
# CIF file
data_findsym-output
_audit_creation_method FINDSYM

_chemical_name_mineral 'GaMo4S8'
_chemical_formula_sum 'Ga Mo4 Si8'

loop_
_publ_author_name
 'M. Fran\c{c}ois'
 'W. Lengauer'
 'K. Yvon'
 'M. Sergent'
 'M. Potel'
 'P. Gougeon'
 'H. {Ben Yaich-Aerrache}'
_journal_name_full_name
;
 Zeitschrift f{\"u}r Kristallographie - Crystalline Materials
;
_journal_volume 196
_journal_year 1991
_journal_page_first 111
_journal_page_last 128
_publ_Section_title
;
 Structural phase transition in GaMo$_{4}$S$_{8}$ by X-ray powder
    ↪ diffraction
;

_aflow_title 'Low-Temperature GaMo$_{4}$S$_{8}$ Structure'
_aflow_proto 'AB4C8_hR13_160_a_ab_2a2b'
_aflow_params 'a,c/a,x_{1},x_{2},x_{3},x_{4},x_{5},z_{5},x_{6},z_{6},x_{
    ↪ 7},z_{7}'
_aflow_params_values '6.90572,2.42010391386,0.0,0.399,0.636,0.135,0.396,
    ↪ 0.814,0.642,0.088,1.138,-0.40801'
_aflow_Strukturbericht 'None'
_aflow_Pearson 'hR13'

_symmetry_space_group_name_H-M "R 3 m (hexagonal axes)"
_symmetry_Int_Tables_number 160

_cell_length_a    6.90572
_cell_length_b    6.90572
_cell_length_c    16.71256
_cell_angle_alpha 90.00000
_cell_angle_beta  90.00000
_cell_angle_gamma 120.00000
```

```
loop_
_space_group_symop_id
_space_group_symop_operation_xyz
1 x,y,z
2 -y,x-y,z
3 -x+y,-x,z
4 -y,-x,z
5 x,x-y,z
6 -x+y,y,z
7 x+1/3,y+2/3,z+2/3
8 -y+1/3,x-y+2/3,z+2/3
9 -x+y+1/3,-x+2/3,z+2/3
10 -y+1/3,-x+2/3,z+2/3
11 x+1/3,x-y+2/3,z+2/3
12 -x+y+1/3,y+2/3,z+2/3
13 x+2/3,y+1/3,z+1/3
14 -y+2/3,x-y+1/3,z+1/3
15 -x+y+2/3,-x+1/3,z+1/3
16 -y+2/3,-x+1/3,z+1/3
17 x+2/3,x-y+1/3,z+1/3
18 -x+y+2/3,y+1/3,z+1/3

loop_
_atom_site_label
_atom_site_type_symbol
_atom_site_symmetry_multiplicity
_atom_site_Wyckoff_label
_atom_site_fract_x
_atom_site_fract_y
_atom_site_fract_z
_atom_site_occupancy
Ga1 Ga 3 a 0.00000 0.00000 0.00000 1.00000
Mo1 Mo 3 a 0.00000 0.00000 0.39900 1.00000
Si1 Si 3 a 0.00000 0.00000 0.63600 1.00000
Si2 Si 3 a 0.00000 0.00000 0.13500 1.00000
Mo2 Mo 9 b 0.19400 0.80600 0.20200 1.00000
Si3 Si 9 b 0.51800 0.48200 0.12400 1.00000
Si4 Si 9 b 0.84867 0.15133 0.28933 1.00000
```

Low-Temperature $GaMo_4S_8$: AB4C8_hR13_160_a_ab_2a2b - POSCAR

```
AB4C8_hR13_160_a_ab_2a2b & a,c/a,x1,x2,x3,x4,x5,z5,x6,z6,x7,z7 --params=
   ↪ 6.90572,2.42010391386,0.0,0.399,0.636,0.135,0.396,0.814,0.642,
   ↪ 0.088,1.138,-0.40801 & R3m C_{3v}^{5} #160 (a^4b^3) & hR13 &
   ↪ None & GaMo4S8 & GaMo4S8 & M. Fran\c{c}ois et al., Zeitschrift
   ↪ f{"u}r Kristallographie - Crystalline Materials 196, 111-128 (
   ↪ 1991)
   1.00000000000000
   3.45286000000000   -1.99350965047409    5.57085333333333
   0.00000000000000    3.98701930094818    5.57085333333333
  -3.45286000000000   -1.99350965047409    5.57085333333333
     Ga        Mo        Si
      1         4         8
Direct
   0.00000000000000    0.00000000000000    0.00000000000000    Ga    (1a)
   0.39900000000000    0.39900000000000    0.39900000000000    Mo    (1a)
   0.39600000000000    0.39600000000000    0.81400000000000    Mo    (3b)
   0.81400000000000    0.39600000000000    0.39600000000000    Mo    (3b)
   0.39600000000000    0.81400000000000    0.39600000000000    Mo    (3b)
   0.63600000000000    0.63600000000000    0.63600000000000    Si    (1a)
   0.13500000000000    0.13500000000000    0.13500000000000    Si    (1a)
   0.64200000000000    0.64200000000000    0.08800000000000    Si    (3b)
   0.08800000000000    0.64200000000000    0.64200000000000    Si    (3b)
   0.64200000000000    0.08800000000000    0.64200000000000    Si    (3b)
   1.13800000000000    1.13800000000000   -0.40801000000000    Si    (3b)
  -0.40801000000000    1.13800000000000    1.13800000000000    Si    (3b)
   1.13800000000000   -0.40801000000000    1.13800000000000    Si    (3b)
```

$KBrO_3$ ($G0_7$): ABC3_hR5_160_a_a_b - CIF

```
# CIF file
data_findsym-output
_audit_creation_method FINDSYM

_chemical_name_mineral 'BrKO3'
_chemical_formula_sum 'Br K O3'

loop_
_publ_author_name
 'D. H. Templeton'
 'L. K. Templeton'
_journal_year 1985
_publ_Section_title
;
 Tensor X-ray optical properties of the bromate ion
;

# Found in Crystal behavior of potassium bromate under compression, 2015

_aflow_title 'KBrO$_{3}$ ($G0_{7}$) Structure'
_aflow_proto 'ABC3_hR5_160_a_a_b'
_aflow_params 'a,c/a,x_{1},x_{2},x_{3},z_{3}'
_aflow_params_values '6.011,1.35618033605,0.4827,0.0,0.54476,1.11108'
_aflow_Strukturbericht '$G0_{7}$'
_aflow_Pearson 'hR5'

_symmetry_space_group_name_H-M "R 3 m (hexagonal axes)"
_symmetry_Int_Tables_number 160

_cell_length_a      6.01100
_cell_length_b      6.01100
_cell_length_c      8.15200
_cell_angle_alpha 90.00000
_cell_angle_beta  90.00000
_cell_angle_gamma 120.00000

loop_
_space_group_symop_id
_space_group_symop_operation_xyz
1 x,y,z
2 -y,x-y,z
3 -x+y,-x,z
4 -y,-x,z
5 x,x-y,z
6 -x+y,y,z
7 x+1/3,y+2/3,z+2/3
8 -y+1/3,x-y+2/3,z+2/3
9 -x+y+1/3,-x+2/3,z+2/3
10 -y+1/3,-x+2/3,z+2/3
11 x+1/3,x-y+2/3,z+2/3
12 -x+y+1/3,y+2/3,z+2/3
13 x+2/3,y+1/3,z+1/3
14 -y+2/3,x-y+1/3,z+1/3
15 -x+y+2/3,-x+1/3,z+1/3
16 -y+2/3,-x+1/3,z+1/3
17 x+2/3,x-y+1/3,z+1/3
18 -x+y+2/3,y+1/3,z+1/3

loop_
_atom_site_label
_atom_site_type_symbol
_atom_site_symmetry_multiplicity
_atom_site_Wyckoff_label
_atom_site_fract_x
_atom_site_fract_y
_atom_site_fract_z
_atom_site_occupancy
Br1 Br 3 a 0.00000 0.00000 0.48270 1.00000
K1 K 3 a 0.00000 0.00000 0.00000 1.00000
O1 O 9 b 0.14456 0.85544 0.40020 1.00000
```

$KBrO_3$ ($G0_7$): ABC3_hR5_160_a_a_b - POSCAR

```
ABC3_hR5_160_a_a_b & a,c/a,x1,x2,x3,z3 --params=6.011,1.35618033605,
   ↪ 0.4827,0.0,0.54476,1.11108 & R3m C_{3v}^{5} #160 (a^2b) & hR5 &
   ↪ $G0_{7}$ & BrKO3 & BrKO3 & D. H. Templeton and L. K. Templeton
   ↪ , (1985)
   1.00000000000000
   3.00550000000000   -1.73522623404942    2.71733333333333
   0.00000000000000    3.47045246809884    2.71733333333333
  -3.00550000000000   -1.73522623404942    2.71733333333333
     Br        K         O
      1         1         3
Direct
   0.48270000000000    0.48270000000000    0.48270000000000    Br    (1a)
   0.00000000000000    0.00000000000000    0.00000000000000    K     (1a)
   0.54476000000000    0.54476000000000    1.11108000000000    O     (3b)
   1.11108000000000    0.54476000000000    0.54476000000000    O     (3b)
   0.54476000000000    1.11108000000000    0.54476000000000    O     (3b)
```

$\gamma$-Potassium Nitrate ($KNO_3$): ABC3_hR5_160_a_a_b - CIF

```
# CIF file
data_findsym-output
_audit_creation_method FINDSYM

_chemical_name_mineral 'KNO3'
_chemical_formula_sum 'K N O3'

loop_
_publ_author_name
 'J. K. Nimmo'
 'B. W. Lucas'
_journal_year 1976
_publ_Section_title
;
 The crystal structures of $\gamma$-and $\beta$-KNO$_{3}$ and the $\
   ↪ alpha$-$\beta$-$\gamma$ phase transformations
;

# Found in The American Mineralogist Crystal Structure Database, 2003

_aflow_title '$\gamma$-Potassium Nitrate (KNO$_{3}$) Structure'
_aflow_proto 'ABC3_hR5_160_a_a_b'
_aflow_params 'a,c/a,x_{1},x_{2},x_{3},z_{3}'
_aflow_params_values '5.487,1.66867140514,0.0,0.405,0.565,1.172'
_aflow_Strukturbericht 'None'
_aflow_Pearson 'hR5'

_symmetry_space_group_name_H-M "R 3 m (hexagonal axes)"
_symmetry_Int_Tables_number 160

_cell_length_a      5.48700
_cell_length_b      5.48700
_cell_length_c      9.15600
_cell_angle_alpha 90.00000
_cell_angle_beta  90.00000
_cell_angle_gamma 120.00000

loop_
_space_group_symop_id
_space_group_symop_operation_xyz
1 x,y,z
2 -y,x-y,z
3 -x+y,-x,z
4 -y,-x,z
5 x,x-y,z
6 -x+y,y,z
7 x+1/3,y+2/3,z+2/3
```

```
8 -y+1/3,x-y+2/3,z+2/3
9 -x+y+1/3,-x+2/3,z+2/3
10 -y+1/3,-x+2/3,z+2/3
11 x+1/3,x-y+2/3,z+2/3
12 -x+y+1/3,y+2/3,z+2/3
13 x+2/3,y+1/3,z+1/3
14 -y+2/3,x-y+1/3,z+1/3
15 -x+y+2/3,-x+1/3,z+1/3
16 -y+2/3,-x+1/3,z+1/3
17 x+2/3,x-y+1/3,z+1/3
18 -x+y+2/3,y+1/3,z+1/3

loop_
_atom_site_label
_atom_site_type_symbol
_atom_site_symmetry_multiplicity
_atom_site_Wyckoff_label
_atom_site_fract_x
_atom_site_fract_y
_atom_site_fract_z
_atom_site_occupancy
K1 K 3 a 0.00000 0.00000 0.00000 1.00000
N1 N 3 a 0.00000 0.00000 0.40500 1.00000
O1 O 9 b 0.13100 0.86900 0.43400 1.00000
```

$\gamma$-Potassium Nitrate ($KNO_3$): ABC3_hR5_160_a_a_b - POSCAR

```
ABC3_hR5_160_a_a_b & a,c/a,x1,x2,x3,z3 --params=5.487,1.66867140514,0.0,
    ↪ 0.405,0.565,1.172 & R3m C_{3v}^{5} #160 (a^2b) & hR5 & None &
    ↪ KNO3 & KNO3 & J. K. Nimmo and B. W. Lucas, (1976)
   1.00000000000000
   2.74350000000000  -1.58396046352174   3.05200000000000
   0.00000000000000   3.16792092704348   3.05200000000000
  -2.74350000000000  -1.58396046352174   3.05200000000000
     K      N      O
     1      1      3
Direct
   0.00000000000000   0.00000000000000   0.00000000000000   K   (1a)
   0.40500000000000   0.40500000000000   0.40500000000000   N   (1a)
   0.56500000000000   0.56500000000000   1.17200000000000   O   (3b)
   1.17200000000000   0.56500000000000   0.56500000000000   O   (3b)
   0.56500000000000   1.17200000000000   0.56500000000000   O   (3b)
```

$\alpha$-$BaB_2O_4$ (Low-Temperature): A2BC4_hR42_161_2b_b_4b - CIF

```
# CIF file
data_findsym-output
_audit_creation_method FINDSYM

_chemical_name_mineral 'B2BaO4'
_chemical_formula_sum 'B2 Ba O4'

loop_
_publ_author_name
 'R. Fr\"{o}hlich'
_journal_name_full_name
;
 Zeitschrift f{\"u}r Kristallographie - Crystalline Materials
;
_journal_volume 168
_journal_year 1984
_journal_page_first 109
_journal_page_last 112
_publ_Section_title
;
 Crystal Structure of the low-temperature form of BaB$_{2}$O$_{4}$
;

_aflow_title '$\alpha$-BaB$_{2}$O$_{4}$ (Low-Temperature) Structure'
_aflow_proto 'A2BC4_hR42_161_2b_b_4b'
_aflow_params 'a,c/a,x_{1},y_{1},z_{1},x_{2},y_{2},z_{2},x_{3},y_{3},z_{
    ↪ 3},x_{4},y_{4},z_{4},x_{5},y_{5},z_{5},x_{6},y_{6},z_{6},x_{7},
    ↪ y_{7},z_{7}'
_aflow_params_values '12.519,1.01629523125,0.5427,-0.4113,0.3732,0.2639,
    ↪ 0.3403,0.11859,0.7375,0.01891,-0.62672,0.4582,-0.5851,0.62499,
    ↪ 0.6707,0.2584,-0.4056,0.3591,0.2197,0.14411,0.7788,-0.5047,-
    ↪ 0.07361'
_aflow_Strukturbericht 'None'
_aflow_Pearson 'hR42'

_symmetry_space_group_name_H-M "R 3 c (hexagonal axes)"
_symmetry_Int_Tables_number 161

_cell_length_a     12.51900
_cell_length_b     12.51900
_cell_length_c     12.72300
_cell_angle_alpha  90.00000
_cell_angle_beta   90.00000
_cell_angle_gamma  120.00000

loop_
_space_group_symop_id
_space_group_symop_operation_xyz
1 x,y,z
2 -y,x-y,z
3 -x+y,-x,z
4 -y,-x,z+1/2
5 x,x-y,z+1/2
6 -x+y,y,z+1/2
7 x+1/3,y+2/3,z+2/3
8 -y+1/3,x-y+2/3,z+2/3
9 -x+y+1/3,-x+2/3,z+2/3
10 -y+1/3,-x+2/3,z+1/6
11 x+1/3,x-y+2/3,z+1/6
12 -x+y+1/3,y+2/3,z+1/6
13 x+2/3,y+1/3,z+1/3
14 -y+2/3,x-y+1/3,z+1/3
15 -x+y+2/3,-x+1/3,z+1/3
16 -y+2/3,-x+1/3,z+5/6
17 x+2/3,x-y+1/3,z+5/6
18 -x+y+2/3,y+1/3,z+5/6

loop_
_atom_site_label
_atom_site_type_symbol
_atom_site_symmetry_multiplicity
_atom_site_Wyckoff_label
_atom_site_fract_x
_atom_site_fract_y
_atom_site_fract_z
_atom_site_occupancy
B1 B 18 b 0.37450 0.57950 0.16820 1.00000
B2 B 18 b 0.02297 -0.09937 0.24093 1.00000
Ba1 Ba 18 b 0.69427 0.02432 0.04323 1.00000
O1 O 18 b 0.29217 0.75113 0.16603 1.00000
O2 O 18 b 0.49620 -0.08390 0.17450 1.00000
O3 O 18 b 0.11813 0.02127 0.24097 1.00000
O4 O 18 b 0.71197 0.57153 0.06683 1.00000
```

$\alpha$-$BaB_2O_4$ (Low-Temperature): A2BC4_hR42_161_2b_b_4b - POSCAR

```
A2BC4_hR42_161_2b_b_4b & a,c/a,x1,y1,z1,x2,y2,z2,x3,y3,z3,x4,y4,z4,x5,y5
    ↪ ,z5,x6,y6,z6,x7,y7,z7 --params=12.519,1.01629523125,0.5427,-
    ↪ 0.4113,0.3732,0.2639,0.3403,0.11859,0.7375,0.01891,-0.62672,
    ↪ 0.4582,-0.5851,0.62499,0.6707,0.2584,-0.4056,0.3591,0.2197,
    ↪ 0.14411,0.7788,-0.5047,-0.07361 & R3c C_{3v}^{6} #161 (b^7) &
    ↪ hR42 & None & B2BaO4 & B2BaO4 & R. Fr\"{o}hlich, Zeitschrift f
    ↪ {"u}r Kristallographie - Crystalline Materials 168, 109-112 (
    ↪ 1984)
   1.00000000000000
   6.25950000000000  -3.61392400999246   4.24100000000000
   0.00000000000000   7.22784801998493   4.24100000000000
  -6.25950000000000  -3.61392400999246   4.24100000000000
     B     Ba      O
    12      6     24
Direct
   0.54270000000000  -0.41130000000000   0.37320000000000   B    (6b)
   0.37320000000000   0.54270000000000  -0.41130000000000   B    (6b)
  -0.41130000000000   0.37320000000000   0.54270000000000   B    (6b)
   0.87320000000000   0.08870000000000   1.04270000000000   B    (6b)
   0.08870000000000   1.04270000000000   0.87320000000000   B    (6b)
   1.04270000000000   0.87320000000000   0.08870000000000   B    (6b)
   0.26390000000000   0.34030000000000   0.11859000000000   B    (6b)
   0.11859000000000   0.26390000000000   0.34030000000000   B    (6b)
   0.34030000000000   0.11859000000000   0.26390000000000   B    (6b)
   0.61859000000000   0.84030000000000   0.76390000000000   B    (6b)
   0.84030000000000   0.76390000000000   0.61859000000000   B    (6b)
   0.76390000000000   0.61859000000000   0.84030000000000   B    (6b)
   0.73750000000000   0.01891000000000  -0.62672000000000   Ba   (6b)
  -0.62672000000000   0.73750000000000   0.01891000000000   Ba   (6b)
   0.01891000000000  -0.62672000000000   0.73750000000000   Ba   (6b)
  -0.12672000000000   0.51891000000000   1.23750000000000   Ba   (6b)
   0.51891000000000   1.23750000000000  -0.12672000000000   Ba   (6b)
   1.23750000000000  -0.12672000000000   0.51891000000000   Ba   (6b)
   0.45820000000000  -0.58510000000000   0.62499000000000   O    (6b)
   0.62499000000000   0.45820000000000  -0.58510000000000   O    (6b)
  -0.58510000000000   0.62499000000000   0.45820000000000   O    (6b)
   1.12499000000000  -0.08510000000000   0.95820000000000   O    (6b)
  -0.08510000000000   0.95820000000000   1.12499000000000   O    (6b)
   0.95820000000000   1.12499000000000  -0.08510000000000   O    (6b)
   0.67070000000000   0.25840000000000  -0.40560000000000   O    (6b)
  -0.40560000000000   0.67070000000000   0.25840000000000   O    (6b)
   0.25840000000000  -0.40560000000000   0.67070000000000   O    (6b)
   0.09440000000000   0.75840000000000   1.17070000000000   O    (6b)
   0.75840000000000   1.17070000000000   0.09440000000000   O    (6b)
   1.17070000000000   0.09440000000000   0.75840000000000   O    (6b)
   0.35910000000000   0.21970000000000   0.14411000000000   O    (6b)
   0.14411000000000   0.35910000000000   0.21970000000000   O    (6b)
   0.21970000000000   0.14411000000000   0.35910000000000   O    (6b)
   0.64411000000000   0.71970000000000   0.85910000000000   O    (6b)
   0.71970000000000   0.85910000000000   0.64411000000000   O    (6b)
   0.85910000000000   0.64411000000000   0.71970000000000   O    (6b)
   0.77880000000000  -0.50470000000000  -0.07361000000000   O    (6b)
  -0.07361000000000   0.77880000000000  -0.50470000000000   O    (6b)
  -0.50470000000000  -0.07361000000000   0.77880000000000   O    (6b)
   0.42639000000000  -0.00470000000000   1.27880000000000   O    (6b)
  -0.00470000000000   1.27880000000000   0.42639000000000   O    (6b)
   1.27880000000000   0.42639000000000  -0.00470000000000   O    (6b)
```

$I1_3$ ($SrCl_2 \cdot (H_2O)_6$) (*obsolete*): A2B6C_hP9_162_d_k_a - CIF

```
# CIF file
data_findsym-output
_audit_creation_method FINDSYM

_chemical_name_mineral 'Cl2(H2O)6Sr'
_chemical_formula_sum 'Cl2 (H2O)6 Sr'

loop_
_publ_author_name
 'Z. Herrmann'
_journal_name_full_name
;
 Zeitschrift fur Anorganische und Allgemeine Chemie
;
_journal_volume 187
_journal_year 1930
_journal_page_first 231
_journal_page_last 236
_publ_Section_title
```

```
;
 {\"U}ber die Struktur des Strontiumchlorid-Hexahydrats
;

# Found in Strukturbericht Band II 1928-1932, 1937

_aflow_title '$I1_{3}$ (SrCl$_{2}$$\cdot$(H$_{2}$O)$_{6}$) ({\em{
    ↪ obsolete}}) Structure'
_aflow_proto 'A2B6C_hP9_162_d_k_a'
_aflow_params 'a,c/a,x_{3},z_{3}'
_aflow_params_values '7.906,0.514798886921,0.387,0.0162'
_aflow_Strukturbericht '$I1_{3}$'
_aflow_Pearson 'hP9'

_symmetry_space_group_name_H-M "P -3 1 2/m"
_symmetry_Int_Tables_number 162

_cell_length_a     7.90600
_cell_length_b     7.90600
_cell_length_c     4.07000
_cell_angle_alpha 90.00000
_cell_angle_beta  90.00000
_cell_angle_gamma 120.00000

loop_
_space_group_symop_id
_space_group_symop_operation_xyz
1 x,y,z
2 -y,x-y,z
3 -x+y,-x,z
4 x,x-y,-z
5 -x+y,y,-z
6 -y,-x,-z
7 -x,-y,-z
8 y,-x+y,-z
9 x-y,x,-z
10 -x,-x+y,z
11 x-y,-y,z
12 y,x,z

loop_
_atom_site_label
_atom_site_type_symbol
_atom_site_symmetry_multiplicity
_atom_site_Wyckoff_label
_atom_site_fract_x
_atom_site_fract_y
_atom_site_fract_z
_atom_site_occupancy
Sr1 Sr 1 a 0.00000 0.00000 0.00000 1.00000
Cl1 Cl 2 d 0.33333 0.66667 0.50000 1.00000
H2O1 H2O 6 k 0.38700 0.38700 0.01620 1.00000
```

$I1_3$ ($SrCl_2 \cdot (H_2O)_6$) (*obsolete*): A2B6C_hP9_162_d_k_a - POSCAR

```
A2B6C_hP9_162_d_k_a & a,c/a,x3,z3 --params=7.906,0.514798886921,0.387,
    ↪ 0.0162 & P-31m D_{3d}^{1} #162 (adk) & hP9 & $I1_{3}$ & Cl2(H2O
    ↪ )6Sr & Cl2(H2O)6Sr & Z. Herrmann, Z. Anorg. Allg. Chem. 187,
    ↪ 231-236 (1930)
   1.00000000000000
   3.95300000000000   -6.84679684231977    0.00000000000000
   3.95300000000000    6.84679684231977    0.00000000000000
   0.00000000000000    0.00000000000000    4.07000000000000
    Cl    H2O    Sr
     2      6     1
Direct
   0.33333333333333    0.66666666666667    0.50000000000000   Cl    (2d)
   0.66666666666667    0.33333333333333    0.50000000000000   Cl    (2d)
   0.38700000000000    0.00000000000000    0.01620000000000   H2O   (6k)
   0.00000000000000    0.38700000000000    0.01620000000000   H2O   (6k)
  -0.38700000000000   -0.38700000000000    0.01620000000000   H2O   (6k)
   0.00000000000000   -0.38700000000000   -0.01620000000000   H2O   (6k)
  -0.38700000000000    0.00000000000000   -0.01620000000000   H2O   (6k)
   0.38700000000000    0.38700000000000   -0.01620000000000   H2O   (6k)
   0.00000000000000    0.00000000000000    0.00000000000000   Sr    (1a)
```

Rosiaite ($PbSb_2O_6$): A6BC2_hP9_162_k_a_d - CIF

```
# CIF file
data_findsym-output
_audit_creation_method FINDSYM

_chemical_name_mineral 'Rosiaite'
_chemical_formula_sum 'O6 Pb Sb2'

loop_
_publ_author_name
 'R. Basso'
 'G. Lucchetti'
 'L. Zefiro'
 'A. Palenzona'
_journal_name_full_name
;
 European Journal of Mineralogy
;
_journal_volume 8
_journal_year 1996
_journal_page_first 487
_journal_page_last 492
_publ_Section_title
;
 Rosiaite, PbSb$_{2}$O$_{6}$, a new mineral from the Cetine mine, Siena,
    ↪ Italy
;
_aflow_title 'Rosiaite (PbSb$_{2}$O$_{6}$) Structure'
_aflow_proto 'A6BC2_hP9_162_k_a_d'
_aflow_params 'a,c/a,x_{3},z_{3}'
_aflow_params_values '5.295,1.01454202077,0.377,0.2965'
_aflow_Strukturbericht 'None'
_aflow_Pearson 'hP9'

_symmetry_space_group_name_H-M "P -3 1 2/m"
_symmetry_Int_Tables_number 162

_cell_length_a     5.29500
_cell_length_b     5.29500
_cell_length_c     5.37200
_cell_angle_alpha 90.00000
_cell_angle_beta  90.00000
_cell_angle_gamma 120.00000

loop_
_space_group_symop_id
_space_group_symop_operation_xyz
1 x,y,z
2 -y,x-y,z
3 -x+y,-x,z
4 x,x-y,-z
5 -x+y,y,-z
6 -y,-x,-z
7 -x,-y,-z
8 y,-x+y,-z
9 x-y,x,-z
10 -x,-x+y,z
11 x-y,-y,z
12 y,x,z

loop_
_atom_site_label
_atom_site_type_symbol
_atom_site_symmetry_multiplicity
_atom_site_Wyckoff_label
_atom_site_fract_x
_atom_site_fract_y
_atom_site_fract_z
_atom_site_occupancy
Pb1 Pb 1 a 0.00000 0.00000 0.00000 1.00000
Sb1 Sb 2 d 0.33333 0.66667 0.50000 1.00000
O1 O 6 k 0.37700 0.00000 0.29650 1.00000
```

Rosiaite ($PbSb_2O_6$): A6BC2_hP9_162_k_a_d - POSCAR

```
A6BC2_hP9_162_k_a_d & a,c/a,x3,z3 --params=5.295,1.01454202077,0.377,
    ↪ 0.2965 & P-31m D_{3d}^{1} #162 (adk) & hP9 & None & O6PbSb2 &
    ↪ Rosiaite & R. Basso et al., Eur. J. Mineral. 8, 487-492 (1996)
   1.00000000000000
   2.64750000000000   -4.58560451303860    0.00000000000000
   2.64750000000000    4.58560451303860    0.00000000000000
   0.00000000000000    0.00000000000000    5.37200000000000
     O     Pb    Sb
     6      1     2
Direct
   0.37700000000000    0.00000000000000    0.29650000000000   O    (6k)
   0.00000000000000    0.37700000000000    0.29650000000000   O    (6k)
  -0.37700000000000   -0.37700000000000    0.29650000000000   O    (6k)
   0.00000000000000   -0.37700000000000   -0.29650000000000   O    (6k)
  -0.37700000000000    0.00000000000000   -0.29650000000000   O    (6k)
   0.37700000000000    0.37700000000000   -0.29650000000000   O    (6k)
   0.00000000000000    0.00000000000000    0.00000000000000   Pb   (1a)
   0.33333333333333    0.66666666666667    0.50000000000000   Sb   (2d)
   0.66666666666667    0.33333333333333    0.50000000000000   Sb   (2d)
```

$NaSbF_4(OH)_2$ ($J1_{12}$): A6BC_hP16_163_i_b_c - CIF

```
# CIF file
data_findsym-output
_audit_creation_method FINDSYM

_chemical_name_mineral 'F4NaSb'
_chemical_formula_sum 'F6 Na Sb'

loop_
_publ_author_name
 'N. Schrewelius'
_journal_name_full_name
;
 Zeitschrift fur Anorganische und Allgemeine Chemie
;
_journal_volume 238
_journal_year 1938
_journal_page_first 241
_journal_page_last 254
_publ_Section_title
;
 R\"{o}ntgenuntersuchung der Verbindungen NaSb(OH)$_{6}$, NaSbF$_{6}$,
    ↪ NaSbO$_{3}$ und gleichartiger Stoffe
;

# Found in The American Mineralogist Crystal Structure Database, 2003

_aflow_title 'NaSbF$_{4}$(OH)$_{2}$ ($J1_{12}$) Structure'
_aflow_proto 'A6BC_hP16_163_i_b_c'
_aflow_params 'a,c/a,x_{3},y_{3},z_{3}'
_aflow_params_values '5.227,1.90931700784,0.33,0.33,0.15'
_aflow_Strukturbericht '$J1_{12}$'
_aflow_Pearson 'hP16'

_symmetry_space_group_name_H-M "P -3 1 2/c"
_symmetry_Int_Tables_number 163
```

```
_cell_length_a       5.22700
_cell_length_b       5.22700
_cell_length_c       9.98000
_cell_angle_alpha  90.00000
_cell_angle_beta    90.00000
_cell_angle_gamma  120.00000

loop_
_space_group_symop_id
_space_group_symop_operation_xyz
1  x,y,z
2  -y,x-y,z
3  -x+y,-x,z
4  x,x-y,-z+1/2
5  -x+y,y,-z+1/2
6  -y,-x,-z+1/2
7  -x,-y,-z
8  y,-x+y,-z
9  x-y,x,-z
10  -x,-x+y,z+1/2
11  x-y,-y,z+1/2
12  y,x,z+1/2

loop_
_atom_site_label
_atom_site_type_symbol
_atom_site_symmetry_multiplicity
_atom_site_Wyckoff_label
_atom_site_fract_x
_atom_site_fract_y
_atom_site_fract_z
_atom_site_occupancy
Na1  Na  2  b  0.00000  0.00000  0.00000  1.00000
Sb1  Sb  2  c  0.33333  0.66667  0.25000  1.00000
F1  F  12  i  0.33000  0.33000  0.15000  1.00000
```

NaSbF$_4$(OH)$_2$ ($J1_{12}$): A6BC_hP16_163_i_b_c - POSCAR

```
A6BC_hP16_163_i_b_c & a,c/a,x3,y3,z3 --params=5.227,1.90931700784,0.33,
   ↪ 0.33,0.15 & P-31c D_{3d}^{2} #163 (bci) & hP16 & $J1_{12}$ &
   ↪ F4NaSb & F4NaSb & N. Schrewelius, Z. Anorg. Allg. Chem. 238,
   ↪ 241-254 (1938)
   1.00000000000000
   2.61350000000000   -4.52671478558126    0.00000000000000
   2.61350000000000    4.52671478558126    0.00000000000000
   0.00000000000000    0.00000000000000    9.98000000000000
      F        Na        Sb
     12         2         2
Direct
   0.33000000000000    0.33000000000000    0.15000000000000    F   (12i)
  -0.33000000000000    0.00000000000000    0.15000000000000    F   (12i)
   0.00000000000000   -0.33000000000000    0.15000000000000    F   (12i)
  -0.33000000000000   -0.33000000000000    0.35000000000000    F   (12i)
   0.00000000000000    0.33000000000000    0.35000000000000    F   (12i)
   0.33000000000000    0.00000000000000    0.35000000000000    F   (12i)
  -0.33000000000000   -0.33000000000000   -0.15000000000000    F   (12i)
   0.33000000000000    0.00000000000000   -0.15000000000000    F   (12i)
   0.00000000000000    0.33000000000000   -0.15000000000000    F   (12i)
   0.33000000000000    0.33000000000000    0.65000000000000    F   (12i)
   0.00000000000000   -0.33000000000000    0.65000000000000    F   (12i)
  -0.33000000000000    0.00000000000000    0.65000000000000    F   (12i)
   0.00000000000000    0.00000000000000    0.00000000000000    Na   (2b)
   0.00000000000000    0.00000000000000    0.50000000000000    Na   (2b)
   0.33333333333333    0.66666666666667    0.25000000000000    Sb   (2c)
   0.66666666666667    0.33333333333333    0.75000000000000    Sb   (2c)
```

Colquiriite (LiCaAlF$_6$): ABC6D_hP18_163_d_b_i_c - CIF

```
# CIF file
data_findsym-output
_audit_creation_method FINDSYM

_chemical_name_mineral 'Colquiriite'
_chemical_formula_sum 'Al Ca F6 Li'

loop_
_publ_author_name
 'S. Kuze'
 'D. du Boulay'
 'N. Ishizawa'
 'N. Kodama'
 'M. Yamaga'
 'B. Henderson'
_journal_name_full_name
;
 Journal of Solid State Chemistry
;
_journal_volume 177
_journal_year 2004
_journal_page_first 3505
_journal_page_last 3513
_publ_Section_title
;
 Structures of LiCaAlF$_{6}$ and LiSrAlF$_{6}$ at 120 and 300 K by
   ↪ synchrotron X-ray single-crystal diffraction
;

_aflow_title 'Colquiriite (LiCaAlF$_{6}$) Structure'
_aflow_proto 'ABC6D_hP18_163_d_b_i_c'
_aflow_params 'a,c/a,x_{4},y_{4},z_{4}'
_aflow_params_values '5.0081,1.92554062419,0.3767,0.031,0.14336'
_aflow_Strukturbericht 'None'
_aflow_Pearson 'hP18'
```

```
_symmetry_space_group_name_H-M "P -3 1 2/c"
_symmetry_Int_Tables_number 163

_cell_length_a       5.00810
_cell_length_b       5.00810
_cell_length_c       9.64330
_cell_angle_alpha  90.00000
_cell_angle_beta    90.00000
_cell_angle_gamma  120.00000

loop_
_space_group_symop_id
_space_group_symop_operation_xyz
1  x,y,z
2  -y,x-y,z
3  -x+y,-x,z
4  x,x-y,-z+1/2
5  -x+y,y,-z+1/2
6  -y,-x,-z+1/2
7  -x,-y,-z
8  y,-x+y,-z
9  x-y,x,-z
10  -x,-x+y,z+1/2
11  x-y,-y,z+1/2
12  y,x,z+1/2

loop_
_atom_site_label
_atom_site_type_symbol
_atom_site_symmetry_multiplicity
_atom_site_Wyckoff_label
_atom_site_fract_x
_atom_site_fract_y
_atom_site_fract_z
_atom_site_occupancy
Ca1  Ca  2  b  0.00000  0.00000  0.00000  1.00000
Li1  Li  2  c  0.33333  0.66667  0.25000  1.00000
Al1  Al  2  d  0.66667  0.33333  0.25000  1.00000
F1  F  12  i  0.37670  0.03100  0.14336  1.00000
```

Colquiriite (LiCaAlF$_6$): ABC6D_hP18_163_d_b_i_c - POSCAR

```
ABC6D_hP18_163_d_b_i_c & a,c/a,x4,y4,z4 --params=5.0081,1.92554062419,
   ↪ 0.3767,0.031,0.14336 & P-31c D_{3d}^{2} #163 (bcdi) & hP18 &
   ↪ None & AlCaF6Li & Colquiriite & S. Kuze et al., J. Solid State
   ↪ Chem. 177, 3505-3513 (2004)
   1.00000000000000
   2.50405000000000   -4.33714182469285    0.00000000000000
   2.50405000000000    4.33714182469285    0.00000000000000
   0.00000000000000    0.00000000000000    9.64330000000000
     Al        Ca         F        Li
      2         2        12         2
Direct
   0.66666666666667    0.33333333333333    0.25000000000000    Al   (2d)
   0.33333333333333    0.66666666666667    0.75000000000000    Al   (2d)
   0.00000000000000    0.00000000000000    0.00000000000000    Ca   (2b)
   0.00000000000000    0.00000000000000    0.50000000000000    Ca   (2b)
   0.37670000000000    0.03100000000000    0.14336000000000    F   (12i)
  -0.03100000000000    0.34570000000000    0.14336000000000    F   (12i)
  -0.34570000000000   -0.37670000000000    0.14336000000000    F   (12i)
  -0.03100000000000   -0.37670000000000    0.35664000000000    F   (12i)
  -0.34570000000000    0.03100000000000    0.35664000000000    F   (12i)
   0.37670000000000    0.34570000000000    0.35664000000000    F   (12i)
  -0.37670000000000   -0.03100000000000   -0.14336000000000    F   (12i)
   0.03100000000000   -0.34570000000000   -0.14336000000000    F   (12i)
   0.34570000000000    0.37670000000000   -0.14336000000000    F   (12i)
   0.03100000000000    0.37670000000000    0.64336000000000    F   (12i)
   0.34570000000000   -0.03100000000000    0.64336000000000    F   (12i)
  -0.37670000000000   -0.34570000000000    0.64336000000000    F   (12i)
   0.33333333333333    0.66666666666667    0.25000000000000    Li   (2c)
   0.66666666666667    0.33333333333333    0.75000000000000    Li   (2c)
```

Predicted Li$_2$MgH$_{16}$ 300 GPa: A16B2C_hP19_164_2d2i_d_b - CIF

```
# CIF file
data_findsym-output
_audit_creation_method FINDSYM

_chemical_name_mineral 'H16Li2Mg'
_chemical_formula_sum 'H16 Li2 Mg'

loop_
_publ_author_name
 'Y. Sun'
 'J. Lv'
 'Y. Xie'
 'H. Liu'
 'Y. Ma'
_journal_name_full_name
;
 Physical Review Letters
;
_journal_volume 123
_journal_year 2019
_journal_page_first 097001
_journal_page_last 097001
_publ_Section_title
;
 Route to a Superconducting Phase above Room Temperature in
   ↪ Electron-Doped Hydride Compounds under High Pressure
;

_aflow_title 'Predicted Li$_{2}$MgH$_{16}$ 300~GPa Structure'
_aflow_proto 'A16B2C_hP19_164_2d2i_d_b'
_aflow_params 'a,c/a,z_{2},z_{3},z_{4},x_{5},z_{5},x_{6},z_{6}'
```

```
_aflow_params_values '2.79596,1.9003812644,0.57451,0.39139,0.83626,
    ↪ 0.17157,0.0797,0.1644,0.23653'
_aflow_Strukturbericht 'None'
_aflow_Pearson 'hP19'

_symmetry_space_group_name_H-M "P -3 2/m 1"
_symmetry_Int_Tables_number 164

_cell_length_a    2.79596
_cell_length_b    2.79596
_cell_length_c    5.31339
_cell_angle_alpha 90.00000
_cell_angle_beta  90.00000
_cell_angle_gamma 120.00000

loop_
_space_group_symop_id
_space_group_symop_operation_xyz
1 x,y,z
2 -y,x-y,z
3 -x+y,-x,z
4 x-y,-y,-z
5 y,x,-z
6 -x,-x+y,-z
7 -x,-y,-z
8 y,-x+y,-z
9 x-y,x,-z
10 -x+y,y,z
11 -y,-x,z
12 x,x-y,z

loop_
_atom_site_label
_atom_site_type_symbol
_atom_site_symmetry_multiplicity
_atom_site_Wyckoff_label
_atom_site_fract_x
_atom_site_fract_y
_atom_site_fract_z
_atom_site_occupancy
Mg1 Mg 1 b 0.00000 0.00000 0.50000 1.00000
H1 H 2 d 0.33333 0.66667 0.57451 1.00000
H2 H 2 d 0.33333 0.66667 0.39139 1.00000
Li1 Li 2 d 0.33333 0.66667 0.83626 1.00000
H3 H 6 i 0.17157 0.82843 0.07970 1.00000
H4 H 6 i 0.16440 0.83560 0.23653 1.00000
```

Predicted $Li_2MgH_{16}$ 300 GPa: A16B2C_hP19_164_2d2i_d_b - POSCAR

```
A16B2C_hP19_164_2d2i_d_b & a,c/a,z2,z3,z4,x5,z5,x6,z6 --params=2.79596,
    ↪ 1.9003812644,0.57451,0.39139,0.83626,0.17157,0.0797,0.1644,
    ↪ 0.23653 & P-3m1 D_{3d}^{3} #164 (bd^3i^2) & hP19 & None &
    ↪ H16Li2Mg & H16Li2Mg & Y. Sun et al., Phys. Rev. Lett. 123,
    ↪ 097001(2019)
   1.00000000000000
   1.39798000000000  -2.42137238796514   0.00000000000000
   1.39798000000000   2.42137238796514   0.00000000000000
   0.00000000000000   0.00000000000000   5.31339000000000
    H    Li    Mg
   16     2     1
Direct
   0.33333333333333   0.66666666666667   0.57451000000000   H   (2d)
   0.66666666666667   0.33333333333333  -0.57451000000000   H   (2d)
   0.33333333333333   0.66666666666667   0.39139000000000   H   (2d)
   0.66666666666667   0.33333333333333  -0.39139000000000   H   (2d)
   0.17157000000000  -0.17157000000000   0.07970000000000   H   (6i)
   0.17157000000000   0.34314000000000   0.07970000000000   H   (6i)
  -0.34314000000000  -0.17157000000000   0.07970000000000   H   (6i)
  -0.17157000000000   0.17157000000000  -0.07970000000000   H   (6i)
   0.34314000000000   0.17157000000000  -0.07970000000000   H   (6i)
  -0.17157000000000  -0.34314000000000  -0.07970000000000   H   (6i)
   0.16440000000000  -0.16440000000000   0.23653000000000   H   (6i)
   0.16440000000000   0.32880000000000   0.23653000000000   H   (6i)
  -0.32880000000000  -0.16440000000000   0.23653000000000   H   (6i)
  -0.16440000000000   0.16440000000000  -0.23653000000000   H   (6i)
   0.32880000000000   0.16440000000000  -0.23653000000000   H   (6i)
  -0.16440000000000  -0.32880000000000  -0.23653000000000   H   (6i)
   0.33333333333333   0.66666666666667   0.83626000000000   Li  (2d)
   0.66666666666667   0.33333333333333  -0.83626000000000   Li  (2d)
   0.00000000000000   0.00000000000000   0.50000000000000   Mg  (1b)
```

$Ce_2O_2S$: A2B2C_hP5_164_d_d_a - CIF

```
# CIF file
data_findsym-output
_audit_creation_method FINDSYM

_chemical_name_mineral 'Ce2O2S'
_chemical_formula_sum 'Ce2 O2 S'

loop_
_publ_author_name
 'W. H. Zachariasen'
_journal_name_full_name
;
 Acta Cristallographica
;
_journal_volume 2
_journal_year 1949
_journal_page_first 60
_journal_page_last 62
_publ_Section_title
;
 Crystal chemical studies of the $5f$-series of elements. VII. The
    ↪ crystal structure of Ce$_{2}$O$_{2}$S, La$_{2}$O$_{2}$S and
    ↪ Pu$_{2}$O$_{2}$S
;

_aflow_title 'Ce$_{2}$O$_{2}$S Structure'
_aflow_proto 'A2B2C_hP5_164_d_d_a'
_aflow_params 'a,c/a,z_{2},z_{3}'
_aflow_params_values '4.0,1.705,0.29,0.64'
_aflow_Strukturbericht 'None'
_aflow_Pearson 'hP5'

_symmetry_space_group_name_H-M "P -3 2/m 1"
_symmetry_Int_Tables_number 164

_cell_length_a    4.00000
_cell_length_b    4.00000
_cell_length_c    6.82000
_cell_angle_alpha 90.00000
_cell_angle_beta  90.00000
_cell_angle_gamma 120.00000

loop_
_space_group_symop_id
_space_group_symop_operation_xyz
1 x,y,z
2 -y,x-y,z
3 -x+y,-x,z
4 x-y,-y,-z
5 y,x,-z
6 -x,-x+y,-z
7 -x,-y,-z
8 y,-x+y,-z
9 x-y,x,-z
10 -x+y,y,z
11 -y,-x,z
12 x,x-y,z

loop_
_atom_site_label
_atom_site_type_symbol
_atom_site_symmetry_multiplicity
_atom_site_Wyckoff_label
_atom_site_fract_x
_atom_site_fract_y
_atom_site_fract_z
_atom_site_occupancy
S1 S 1 a 0.00000 0.00000 0.00000 1.00000
Ce1 Ce 2 d 0.33333 0.66667 0.29000 1.00000
O1 O 2 d 0.33333 0.66667 0.64000 1.00000
```

$Ce_2O_2S$: A2B2C_hP5_164_d_d_a - POSCAR

```
A2B2C_hP5_164_d_d_a & a,c/a,z2,z3 --params=4.0,1.705,0.29,0.64 & P-3m1
    ↪ D_{3d}^{3} #164 (ad^2) & hP5 & None & Ce2O2S & Ce2O2S & W. H.
    ↪ Zachariasen, Acta Cryst. 2, 60-62 (1949)
   1.00000000000000
   2.00000000000000  -3.46410161513775   0.00000000000000
   2.00000000000000   3.46410161513775   0.00000000000000
   0.00000000000000   0.00000000000000   6.82000000000000
    Ce     O     S
     2     2     1
Direct
   0.33333333333333   0.66666666666667   0.29000000000000   Ce   (2d)
   0.66666666666667   0.33333333333333  -0.29000000000000   Ce   (2d)
   0.33333333333333   0.66666666666667   0.64000000000000   O    (2d)
   0.66666666666667   0.33333333333333  -0.64000000000000   O    (2d)
   0.00000000000000   0.00000000000000   0.00000000000000   S    (1a)
```

Brucite [$Mg(OH)_2$]: A2BC2_hP5_164_d_a_d - CIF

```
# CIF file
data_findsym-output
_audit_creation_method FINDSYM

_chemical_name_mineral 'Brucite'
_chemical_formula_sum 'H2 Mg O2'

loop_
_publ_author_name
 'M. Catti'
 'G. Ferraris'
 'S. Hull'
 'A. Pavese'
_journal_name_full_name
;
 Physics and Chemistry of Minerals
;
_journal_volume 22
_journal_year 1995
_journal_page_first 200
_journal_page_last 206
_publ_Section_title
;
 Static compression and H disorder in brucite, Mg(OH)$_{2}$, to 11 GPa:
    ↪ a powder neutron diffraction study
;

_aflow_title 'Brucite [Mg(OH)$_{2}$] Structure'
_aflow_proto 'A2BC2_hP5_164_d_a_d'
_aflow_params 'a,c/a,z_{2},z_{3}'
_aflow_params_values '3.14979,1.51445016969,0.413,0.2203'
_aflow_Strukturbericht 'None'
_aflow_Pearson 'hP5'
```

```
_symmetry_space_group_name_H-M "P -3 2/m 1"
_symmetry_Int_Tables_number 164

_cell_length_a      3.14979
_cell_length_b      3.14979
_cell_length_c      4.77020
_cell_angle_alpha  90.00000
_cell_angle_beta   90.00000
_cell_angle_gamma 120.00000

loop_
_space_group_symop_id
_space_group_symop_operation_xyz
1 x,y,z
2 -y,x-y,z
3 -x+y,-x,z
4 x-y,-y,-z
5 y,x,-z
6 -x,-x+y,-z
7 -x,-y,-z
8 y,-x+y,-z
9 x-y,x,-z
10 -x+y,y,z
11 -y,-x,z
12 x,x-y,z

loop_
_atom_site_label
_atom_site_type_symbol
_atom_site_symmetry_multiplicity
_atom_site_Wyckoff_label
_atom_site_fract_x
_atom_site_fract_y
_atom_site_fract_z
_atom_site_occupancy
Mg1 Mg 1 a 0.00000 0.00000 0.00000 1.00000
H1 H 2 d 0.33333 0.66667 0.41300 1.00000
O1 O 2 d 0.33333 0.66667 0.22030 1.00000
```

Brucite [Mg(OH)$_2$]: A2BC2_hP5_164_d_a_d - POSCAR

```
A2BC2_hP5_164_d_a_d & a,c/a,z2,z3 --params=3.14979,1.51445016969,0.413,
    ↪ 0.2203 & P-3m1 D_{3d}^{3} #164 (ad^2) & hP5 & None & H2MgO2 &
    ↪ Brucite & M. Catti et al., Phys. Chem. Miner. 22, 200-206 (1995
    ↪ )
   1.00000000000000
   1.57489500000000  -2.72779815658619   0.00000000000000
   1.57489500000000   2.72779815658619   0.00000000000000
   0.00000000000000   0.00000000000000   4.77020000000000
     H    Mg     O
     2     1     2
Direct
   0.33333333333333   0.66666666666667   0.41300000000000    H   (2d)
   0.66666666666667   0.33333333333333  -0.41300000000000    H   (2d)
   0.00000000000000   0.00000000000000   0.00000000000000   Mg   (1a)
   0.33333333333333   0.66666666666667   0.22030000000000    O   (2d)
   0.66666666666667   0.33333333333333  -0.22030000000000    O   (2d)
```

$K_2Pt(SCN)_6$ ($H6_3$): A2BC6_hP9_164_d_a_i - CIF

```
# CIF file
data_findsym-output
_audit_creation_method FINDSYM

_chemical_name_mineral 'K2Pt(SCN)6'
_chemical_formula_sum 'K2 Pt S6'

loop_
_publ_author_name
 'S. B. Hendricks'
 'H. E. Merwin'
_journal_name_full_name
;
 American Journal of Science
;
_journal_volume 15
_journal_year 1928
_journal_page_first 487
_journal_page_last 494
_publ_Section_title
;
 The atomic arrangement in crystals of alkali platini-thiocyanates
;

# Found in Strukturbericht 1913-1928, 1931

_aflow_title 'K$_{2}$Pt(SCN)$_{6}$ ($H6_{3}$) Structure'
_aflow_proto 'A2BC6_hP9_164_d_a_i'
_aflow_params 'a,c/a,z_{2},x_{3},z_{3}'
_aflow_params_values '6.73,1.52451708767,0.5,0.135,0.1125'
_aflow_Strukturbericht '$H6_{3}$'
_aflow_Pearson 'hP9'

_symmetry_space_group_name_H-M "P -3 2/m 1"
_symmetry_Int_Tables_number 164

_cell_length_a      6.73000
_cell_length_b      6.73000
_cell_length_c     10.26000
_cell_angle_alpha  90.00000
_cell_angle_beta   90.00000
_cell_angle_gamma 120.00000

loop_
_space_group_symop_id
_space_group_symop_operation_xyz
1 x,y,z
2 -y,x-y,z
3 -x+y,-x,z
4 x-y,-y,-z
5 y,x,-z
6 -x,-x+y,-z
7 -x,-y,-z
8 y,-x+y,-z
9 x-y,x,-z
10 -x+y,y,z
11 -y,-x,z
12 x,x-y,z

loop_
_atom_site_label
_atom_site_type_symbol
_atom_site_symmetry_multiplicity
_atom_site_Wyckoff_label
_atom_site_fract_x
_atom_site_fract_y
_atom_site_fract_z
_atom_site_occupancy
Pt1 Pt 1 a 0.00000 0.00000 0.00000 1.00000
K1 K 2 d 0.33333 0.66667 0.50000 1.00000
S1 S 6 i 0.13500 0.86500 0.11250 1.00000
```

$K_2Pt(SCN)_6$ ($H6_3$): A2BC6_hP9_164_d_a_i - POSCAR

```
A2BC6_hP9_164_d_a_i & a,c/a,z2,x3,z3 --params=6.73,1.52451708767,0.5,
    ↪ 0.135,0.1125 & P-3m1 D_{3d}^{3} #164 (adi) & hP9 & $H6_{3}$ &
    ↪ K2Pt(SCN)6 & K2Pt(SCN)6 & S. B. Hendricks and H. E. Merwin, Am.
    ↪ J. Sci. 15, 487-494 (1928)
   1.00000000000000
   3.36500000000000  -5.82835096746927   0.00000000000000
   3.36500000000000   5.82835096746927   0.00000000000000
   0.00000000000000   0.00000000000000  10.26000000000000
     K    Pt     S
     2     1     6
Direct
   0.33333333333333   0.66666666666667   0.50000000000000    K   (2d)
   0.66666666666667   0.33333333333333  -0.50000000000000    K   (2d)
   0.00000000000000   0.00000000000000   0.00000000000000   Pt   (1a)
   0.13500000000000  -0.13500000000000   0.11250000000000    S   (6i)
   0.13500000000000   0.27000000000000   0.11250000000000    S   (6i)
  -0.27000000000000  -0.13500000000000   0.11250000000000    S   (6i)
  -0.13500000000000   0.13500000000000  -0.11250000000000    S   (6i)
   0.27000000000000   0.13500000000000  -0.11250000000000    S   (6i)
  -0.13500000000000  -0.27000000000000  -0.11250000000000    S   (6i)
```

Bararite (Trigonal $(NH_4)_2SiF_6$, $J1_6$): A6B2C_hP9_164_i_d_a - CIF

```
# CIF file
data_findsym-output
_audit_creation_method FINDSYM

_chemical_name_mineral 'Bararite'
_chemical_formula_sum 'F6 (NH4)2 Si'

loop_
_publ_author_name
 'E. O. Schlemper'
 'W. C. Hamilton'
_journal_name_full_name
;
 Journal of Chemical Physics
;
_journal_volume 45
_journal_year 1966
_journal_page_first 408
_journal_page_last 409
_publ_Section_title
;
 On the Structure of Trigonal Ammonium Flourosilicate
;

# Found in A new modification of diammonium hexafluorosilicate, (NH$_{4}
    ↪ $)$_{2}$SiF$_{6}$, 2001

_aflow_title 'Bararite (Trigonal (NH$_{4}$)$_{2}$SiF$_{6}$, $J1_{6}$)
    ↪ Structure'
_aflow_proto 'A6B2C_hP9_164_i_d_a'
_aflow_params 'a,c/a,z_{2},x_{3},z_{3}'
_aflow_params_values '5.784,0.82918395574,0.33,0.139,0.22'
_aflow_Strukturbericht '$J1_{6}$'
_aflow_Pearson 'hP9'

_symmetry_space_group_name_H-M "P -3 2/m 1"
_symmetry_Int_Tables_number 164

_cell_length_a      5.78400
_cell_length_b      5.78400
_cell_length_c      4.79600
_cell_angle_alpha  90.00000
_cell_angle_beta   90.00000
_cell_angle_gamma 120.00000

loop_
_space_group_symop_id
_space_group_symop_operation_xyz
1 x,y,z
2 -y,x-y,z
3 -x+y,-x,z
4 x-y,-y,-z
5 y,x,-z
```

```
6 -x,-x+y,-z
7 -x,-y,-z
8 y,-x+y,-z
9 x-y,x,-z
10 -x+y,y,z
11 -y,-x,z
12 x,x-y,z

loop_
_atom_site_label
_atom_site_type_symbol
_atom_site_symmetry_multiplicity
_atom_site_Wyckoff_label
_atom_site_fract_x
_atom_site_fract_y
_atom_site_fract_z
_atom_site_occupancy
Si1 Si 1 a 0.00000 0.00000 0.00000 1.00000
NH41 NH4 2 d 0.33333 0.66667 0.33000 1.00000
F1 F 6 i 0.13900 0.86100 0.22000 1.00000
```

Bararite (Trigonal $(NH_4)_2SiF_6$, $J1_6$): A6B2C_hP9_164_i_d_a - POSCAR

```
A6B2C_hP9_164_i_d_a & a,c/a,z2,x3,z3 --params=5.784,0.82918395574,0.33,
    ↪ 0.139,0.22 & P-3m1 D_{3d}^{3} #164 (adi) & hP9 & $J1_{6}$ & Fe6
    ↪ (NH4)2Si & Bararite & E. O. Schlemper and W. C. Hamilton, J.
    ↪ Chem. Phys. 45, 408-409 (1966)
   1.00000000000000
   2.89200000000000  -5.00909093548919   0.00000000000000
   2.89200000000000   5.00909093548919   0.00000000000000
   0.00000000000000   0.00000000000000   4.79600000000000
     F    NH4    Si
     6     2      1
Direct
   0.13900000000000  -0.13900000000000   0.22000000000000   F   (6i)
   0.13900000000000   0.27800000000000   0.22000000000000   F   (6i)
  -0.27800000000000  -0.13900000000000   0.22000000000000   F   (6i)
  -0.13900000000000   0.13900000000000  -0.22000000000000   F   (6i)
   0.27800000000000   0.13900000000000  -0.22000000000000   F   (6i)
  -0.13900000000000  -0.27800000000000  -0.22000000000000   F   (6i)
   0.33333333333333   0.66666666666667   0.33000000000000 NH4   (2d)
   0.66666666666667   0.33333333333333  -0.33000000000000 NH4   (2d)
   0.00000000000000   0.00000000000000   0.00000000000000  Si   (1a)
```

$K_2GeF_6$ ($J1_{13}$): A6BC2_hP9_164_i_a_d - CIF

```
# CIF file
data_findsym-output
_audit_creation_method FINDSYM

_chemical_name_mineral 'F6GeK2'
_chemical_formula_sum 'F6 Ge K2'

loop_
_publ_author_name
 'J. L. Hoard'
 'W. B. Vincent'
_journal_name_full_name
;
 Journal of the American Chemical Society
;
_journal_volume 61
_journal_year 1939
_journal_page_first 2849
_journal_page_last 2852
_publ_Section_title
;
 Structures of Complex Fluorides. Potassium Hexafluogermanate and
    ↪ Ammonium Hexafluogermanate
;

# Found in The crystal structure of the compound Cs$_{2}$CeCl$_{6}$,
    ↪ 1966

_aflow_title 'K$_{2}$GeF$_{6}$ ($J1_{13}$) Structure'
_aflow_proto 'A6BC2_hP9_164_i_a_d'
_aflow_params 'a,c/a,z_{2},x_{3},z_{3}'
_aflow_params_values '5.62,0.827402135231,0.7,0.148,0.22'
_aflow_Strukturbericht '$J1_{13}$'
_aflow_Pearson 'hP9'

_symmetry_space_group_name_H-M "P -3 2/m 1"
_symmetry_Int_Tables_number 164

_cell_length_a    5.62000
_cell_length_b    5.62000
_cell_length_c    4.65000
_cell_angle_alpha 90.00000
_cell_angle_beta  90.00000
_cell_angle_gamma 120.00000

loop_
_space_group_symop_id
_space_group_symop_operation_xyz
1 x,y,z
2 -y,x-y,z
3 -x+y,-x,z
4 x-y,-y,-z
5 y,x,-z
6 -x,-x+y,-z
7 -x,-y,-z
8 y,-x+y,-z
9 x-y,x,-z
10 -x+y,y,z
11 -y,-x,z
12 x,x-y,z

loop_
_atom_site_label
_atom_site_type_symbol
_atom_site_symmetry_multiplicity
_atom_site_Wyckoff_label
_atom_site_fract_x
_atom_site_fract_y
_atom_site_fract_z
_atom_site_occupancy
Ge1 Ge 1 a 0.00000 0.00000 0.00000 1.00000
K1 K 2 d 0.33333 0.66667 0.70000 1.00000
F1 F 6 i 0.14800 0.29600 0.22000 1.00000
```

$K_2GeF_6$ ($J1_{13}$): A6BC2_hP9_164_i_a_d - POSCAR

```
A6BC2_hP9_164_i_a_d & a,c/a,z2,x3,z3 --params=5.62,0.827402135231,0.7,
    ↪ 0.148,0.22 & P-3m1 D_{3d}^{3} #164 (adi) & hP9 & $J1_{13}$ &
    ↪ F6GeK2 & F6GeK2 & J. L. Hoard and W. B. Vincent, J. Am. Chem.
    ↪ Soc. 61, 2849-2852 (1939)
   1.00000000000000
   2.81000000000000  -4.86706276926854   0.00000000000000
   2.81000000000000   4.86706276926854   0.00000000000000
   0.00000000000000   0.00000000000000   4.65000000000000
     F    Ge     K
     6     1     2
Direct
   0.14800000000000  -0.14800000000000   0.22000000000000   F   (6i)
   0.14800000000000   0.29600000000000   0.22000000000000   F   (6i)
  -0.29600000000000  -0.14800000000000   0.22000000000000   F   (6i)
  -0.14800000000000   0.14800000000000  -0.22000000000000   F   (6i)
   0.29600000000000   0.14800000000000  -0.22000000000000   F   (6i)
  -0.14800000000000  -0.29600000000000  -0.22000000000000   F   (6i)
   0.00000000000000   0.00000000000000   0.00000000000000  Ge   (1a)
   0.33333333333333   0.66666666666667   0.70000000000000   K   (2d)
   0.66666666666667   0.33333333333333  -0.70000000000000   K   (2d)
```

Jacutingaite ($Pt_2HgSe_3$): AB2C3_hP12_164_d_ae_i - CIF

```
# CIF file
data_findsym-output
_audit_creation_method FINDSYM

_chemical_name_mineral 'Jacutingaite'
_chemical_formula_sum 'Hg Pt2 Se3'

loop_
_publ_author_name
 'A. Vymazalov\\'{a}'
 'F. Laufek'
 'M. Dr\\'{a}bek'
 'A. R. Cabral'
 'J. Haloda'
 'T. Sidorinov\\'{a}'
 'B. Lehmann'
 'H. F. Galbiatti'
 'J. Drahokoupil'
_journal_name_full_name
;
 Canadian Mineralogist
;
_journal_volume 50
_journal_year 2012
_journal_page_first 431
_journal_page_last 440
_publ_Section_title
;
 Jacutingaite, Pt$_{2}$HgSe$_{3}$, A New Platinum-Group Mineral Species
    ↪ From the Cau\^{e} Iron-Ore Deposit, Itabira District, Minas
    ↪ Gerais, Brazil
;

# Found in Emergent dual topology in the three-dimensional Kane-Mele
    ↪ Pt$_{2}$HgSe$_{3}$, 2019 Found in Emergent dual topology in the
    ↪  three-dimensional Kane-Mele Pt$_{2}$HgSe$_{3}$, {arXiv:
    ↪ 1909.05050 [cond-mat.mes-hall]},

_aflow_title 'Jacutingaite (Pt$_{2}$HgSe$_{3}$) Structure'
_aflow_proto 'AB2C3_hP12_164_d_ae_i'
_aflow_params 'a,c/a,z_{2},x_{4},z_{4}'
_aflow_params_values '7.3477,0.720701716183,0.3507,0.8196,0.2492'
_aflow_Strukturbericht 'None'
_aflow_Pearson 'hP12'

_symmetry_space_group_name_H-M "P -3 2/m 1"
_symmetry_Int_Tables_number 164

_cell_length_a    7.34770
_cell_length_b    7.34770
_cell_length_c    5.29550
_cell_angle_alpha 90.00000
_cell_angle_beta  90.00000
_cell_angle_gamma 120.00000

loop_
_space_group_symop_id
_space_group_symop_operation_xyz
1 x,y,z
2 -y,x-y,z
3 -x+y,-x,z
4 x-y,-y,-z
5 y,x,-z
6 -x,-x+y,-z
7 -x,-y,-z
```

```
8 y,-x+y,-z
9 x-y,x,-z
10 -x+y,y,z
11 -y,-x,z
12 x,x-y,z

loop_
_atom_site_label
_atom_site_type_symbol
_atom_site_symmetry_multiplicity
_atom_site_Wyckoff_label
_atom_site_fract_x
_atom_site_fract_y
_atom_site_fract_z
_atom_site_occupancy
Pt1 Pt 1 a 0.00000 0.00000 0.00000 1.00000
Hg1 Hg 2 d 0.33333 0.66667 0.35070 1.00000
Pt2 Pt 3 e 0.50000 0.00000 0.00000 1.00000
Se1 Se 6 i 0.81960 0.18040 0.24920 1.00000
```

Jacutingaite ($Pt_2HgSe_3$): AB2C3_hP12_164_d_ae_i - POSCAR

```
AB2C3_hP12_164_d_ae_i & a,c/a,z2,x4,z4 --params=7.3477,0.720701716183,
    ↪ 0.3507,0.8196,0.2492 & P-3m1 D_{3d}^{3} #164 (adei) & hP12 &
    ↪ None & HgPt2Se3 & Jacutingaite & A. Vymazalov\'{a} et al., Can.
    ↪ Mineral. 50, 431-440 (2012)
   1.00000000000000
   3.67385000000000  -6.36329485938692   0.00000000000000
   3.67385000000000   6.36329485938692   0.00000000000000
   0.00000000000000   0.00000000000000   5.29550000000000
    Hg    Pt    Se
     2     4     6
Direct
   0.33333333333333   0.66666666666667   0.35070000000000   Hg   (2d)
   0.66666666666667   0.33333333333333  -0.35070000000000   Hg   (2d)
   0.00000000000000   0.00000000000000   0.00000000000000   Pt   (1a)
   0.50000000000000   0.00000000000000   0.00000000000000   Pt   (3e)
   0.00000000000000   0.50000000000000   0.00000000000000   Pt   (3e)
   0.50000000000000   0.50000000000000   0.00000000000000   Pt   (3e)
   0.81960000000000  -0.81960000000000   0.24920000000000   Se   (6i)
   0.81960000000000   1.63920000000000   0.24920000000000   Se   (6i)
  -1.63920000000000  -0.81960000000000   0.24920000000000   Se   (6i)
  -0.81960000000000   0.81960000000000  -0.24920000000000   Se   (6i)
   1.63920000000000   0.81960000000000  -0.24920000000000   Se   (6i)
  -0.81960000000000  -1.63920000000000  -0.24920000000000   Se   (6i)
```

$D0_{13}$ ($AlCl_3$) (*obsolete*): AB3_hP4_164_b_ad - CIF

```
# CIF file
data_findsym-output
_audit_creation_method FINDSYM

_chemical_name_mineral 'AlCl3'
_chemical_formula_sum 'Al Cl3'

loop_
_publ_author_name
 'W. E. Laschkarew'
_journal_name_full_name
;
 Zeitschrift fur Anorganische und Allgemeine Chemie
;
_journal_volume 193
_journal_year 1930
_journal_page_first 270
_journal_page_last 276
_publ_Section_title
;
 Zur Struktur AlCl$_{3}$
;

# Found in Strukturbericht Band II 1928-1932, 1937

_aflow_title '$D0_{13}$ (AlCl$_{3}$) ({\em{obsolete}}) Structure'
_aflow_proto 'AB3_hP4_164_b_ad'
_aflow_params 'a,c/a,z_{3}'
_aflow_params_values '3.475,2.45,0.33333'
_aflow_Strukturbericht '$D0_{13}$'
_aflow_Pearson 'hP4'

_symmetry_space_group_name_H-M "P -3 2/m 1"
_symmetry_Int_Tables_number 164

_cell_length_a     3.47500
_cell_length_b     3.47500
_cell_length_c     8.51375
_cell_angle_alpha  90.00000
_cell_angle_beta   90.00000
_cell_angle_gamma  120.00000

loop_
_space_group_symop_id
_space_group_symop_operation_xyz
1 x,y,z
2 -y,x-y,z
3 -x+y,-x,z
4 x-y,-y,-z
5 y,x,-z
6 -x,-x+y,-z
7 -x,-y,-z
8 y,-x+y,-z
9 x-y,x,-z
10 -x+y,y,z
11 -y,-x,z
12 x,x-y,z
```

```
loop_
_atom_site_label
_atom_site_type_symbol
_atom_site_symmetry_multiplicity
_atom_site_Wyckoff_label
_atom_site_fract_x
_atom_site_fract_y
_atom_site_fract_z
_atom_site_occupancy
Cl1 Cl 1 a 0.00000 0.00000 0.00000 1.00000
Al1 Al 1 b 0.00000 0.00000 0.50000 1.00000
Cl2 Cl 2 d 0.33333 0.66667 0.33333 1.00000
```

$D0_{13}$ ($AlCl_3$) (*obsolete*): AB3_hP4_164_b_ad - POSCAR

```
AB3_hP4_164_b_ad & a,c/a,z3 --params=3.475,2.45,0.33333 & P-3m1 D_{3d}^{
    ↪ 3} #164 (abd) & hP4 & $D0_{13}$ & AlCl3 & AlCl3 & W. E.
    ↪ Laschkarew, Z. Anorg. Allg. Chem. 193, 270-276 (1930)
   1.00000000000000
   1.73750000000000  -3.00943827815092   0.00000000000000
   1.73750000000000   3.00943827815092   0.00000000000000
   0.00000000000000   0.00000000000000   8.51375000000000
    Al    Cl
     1     3
Direct
   0.00000000000000   0.00000000000000   0.50000000000000   Al   (1b)
   0.00000000000000   0.00000000000000   0.00000000000000   Cl   (1a)
   0.33333333333333   0.66666666666667   0.33333000000000   Cl   (2d)
   0.66666666666667   0.33333333333333  -0.33333000000000   Cl   (2d)
```

Nevskite (BiSe): AB_hP12_164_c2d_c2d - CIF

```
# CIF file
data_findsym-output
_audit_creation_method FINDSYM

_chemical_name_mineral 'Nevskite'
_chemical_formula_sum 'Bi Se'

loop_
_publ_author_name
 'E. Gaudin'
 'S. Jobic'
 'M. Evain'
 'R. Brec'
 'J. Rouxel'
_journal_name_full_name
;
 Materials Research Bulletin
;
_journal_volume 30
_journal_year 1995
_journal_page_first 549
_journal_page_last 561
_publ_Section_title
;
 Charge balance in some Bi$_{x}$Se$_{y}$ phases through atomic structure
    ↪ determination and band structure calculations
;

_aflow_title 'Nevskite (BiSe) Structure'
_aflow_proto 'AB_hP12_164_c2d_c2d'
_aflow_params 'a,c/a,z_{1},z_{2},z_{3},z_{4},z_{5},z_{6}'
_aflow_params_values '4.212,5.44681861349,0.3727,0.1279,0.0419,0.7959,
    ↪ 0.2797,0.5596'
_aflow_Strukturbericht 'None'
_aflow_Pearson 'hP12'

_symmetry_space_group_name_H-M "P -3 2/m 1"
_symmetry_Int_Tables_number 164

_cell_length_a     4.21200
_cell_length_b     4.21200
_cell_length_c     22.94200
_cell_angle_alpha  90.00000
_cell_angle_beta   90.00000
_cell_angle_gamma  120.00000

loop_
_space_group_symop_id
_space_group_symop_operation_xyz
1 x,y,z
2 -y,x-y,z
3 -x+y,-x,z
4 x-y,-y,-z
5 y,x,-z
6 -x,-x+y,-z
7 -x,-y,-z
8 y,-x+y,-z
9 x-y,x,-z
10 -x+y,y,z
11 -y,-x,z
12 x,x-y,z

loop_
_atom_site_label
_atom_site_type_symbol
_atom_site_symmetry_multiplicity
_atom_site_Wyckoff_label
_atom_site_fract_x
_atom_site_fract_y
_atom_site_fract_z
_atom_site_occupancy
Bi1 Bi 2 c 0.00000 0.00000 0.37270 1.00000
```

```
Se1  Se  2  c  0.00000  0.00000  0.12790  1.00000
Bi2  Bi  2  d  0.33333  0.66667  0.04190  1.00000
Bi3  Bi  2  d  0.33333  0.66667  0.79590  1.00000
Se2  Se  2  d  0.33333  0.66667  0.27970  1.00000
Se3  Se  2  d  0.33333  0.66667  0.55960  1.00000
```

Nevskite (BiSe): AB_hP12_164_c2d_c2d - POSCAR

```
AB_hP12_164_c2d_c2d & a,c/a,z1,z2,z3,z4,z5,z6 --params=4.212,
    ↪ 5.44681861349,0.3727,0.1279,0.0419,0.7959,0.2797,0.5596 & P-3m1
    ↪  D_{3d}^{3} #164 (c^2d^4) & hP12 & None & BiSe & Nevskite & E.
    ↪ Gaudin et al., Mater. Res. Bull. 30, 549-561 (1995)
   1.00000000000000
   2.10600000000000  -3.64769900074005   0.00000000000000
   2.10600000000000   3.64769900074005   0.00000000000000
   0.00000000000000   0.00000000000000  22.94200000000000
    Bi    Se
     6     6
Direct
   0.00000000000000   0.00000000000000   0.37270000000000   Bi   (2c)
   0.00000000000000   0.00000000000000  -0.37270000000000   Bi   (2c)
   0.33333333333333   0.66666666666667   0.04190000000000   Bi   (2d)
   0.66666666666667   0.33333333333333  -0.04190000000000   Bi   (2d)
   0.33333333333333   0.66666666666667   0.79590000000000   Bi   (2d)
   0.66666666666667   0.33333333333333  -0.79590000000000   Bi   (2d)
   0.00000000000000   0.00000000000000   0.12790000000000   Se   (2c)
   0.00000000000000   0.00000000000000  -0.12790000000000   Se   (2c)
   0.33333333333333   0.66666666666667   0.27970000000000   Se   (2d)
   0.66666666666667   0.33333333333333  -0.27970000000000   Se   (2d)
   0.33333333333333   0.66666666666667   0.55960000000000   Se   (2d)
   0.66666666666667   0.33333333333333  -0.55960000000000   Se   (2d)
```

$B_{13}C_2$ "$B_4C$" ($D1_g$): A13B2_hR15_166_b2h_c - CIF

```
# CIF file
data_findsym-output
_audit_creation_method FINDSYM

_chemical_name_mineral 'B13C2'
_chemical_formula_sum 'B13 C2'

loop_
_publ_author_name
 'G. Will'
 'K. H. Kossobutzki'
_journal_name_full_name
;
 Journal of the Less-Common Metals
;
_journal_volume 44
_journal_year 1976
_journal_page_first 87
_journal_page_last 97
_publ_Section_title
;
 An X-ray structure analysis of boron carbide, B$_{13}$C$_{2}$
;

_aflow_title 'B$_{13}$C$_{2}$ ``B$_{4}$C'' ($D1_{g}$) Structure'
_aflow_proto 'A13B2_hR15_166_b2h_c'
_aflow_params 'a,c/a,x_{2},x_{3},z_{3},x_{4},z_{4}'
_aflow_params_values '6.617,1.82847211727,0.3823,0.8044,1.316,0.9936,
    ↪ 1.6714'
_aflow_Strukturbericht '$D1_{g}$'
_aflow_Pearson 'hR15'

_symmetry_space_group_name_H-M "R -3 2/m (hexagonal axes)"
_symmetry_Int_Tables_number 166

_cell_length_a     6.61700
_cell_length_b     6.61700
_cell_length_c     12.09900
_cell_angle_alpha  90.00000
_cell_angle_beta   90.00000
_cell_angle_gamma  120.00000

loop_
_space_group_symop_id
_space_group_symop_operation_xyz
1 x,y,z
2 -y,x-y,z
3 -x+y,-x,z
4 y,x,-z
5 -x,-x+y,-z
6 x-y,-y,-z
7 -x,-y,-z
8 y,-x+y,-z
9 x-y,x,-z
10 -y,-x,z
11 x,x-y,z
12 -x+y,y,z
13 x+1/3,y+2/3,z+2/3
14 -y+1/3,x-y+2/3,z+2/3
15 -x+y+1/3,-x+2/3,z+2/3
16 y+1/3,x+2/3,-z+2/3
17 -x+1/3,-x+y+2/3,-z+2/3
18 x-y+1/3,-y+2/3,-z+2/3
19 -x+1/3,-y+2/3,-z+2/3
20 y+1/3,-x+y+2/3,-z+2/3
21 x-y+1/3,x+2/3,-z+2/3
22 -y+1/3,-x+2/3,z+2/3
23 x+1/3,x-y+2/3,z+2/3
24 -x+y+1/3,y+2/3,z+2/3
25 x+2/3,y+1/3,z+1/3
26 -y+2/3,x-y+1/3,z+1/3
27 -x+y+2/3,-x+1/3,z+1/3
28 y+2/3,x+1/3,-z+1/3
29 -x+2/3,-x+y+1/3,-z+1/3
30 x-y+2/3,-y+1/3,-z+1/3
31 -x+2/3,-y+1/3,-z+1/3
32 y+2/3,-x+y+1/3,-z+1/3
33 x-y+2/3,x+1/3,-z+1/3
34 -y+2/3,-x+1/3,z+1/3
35 x+2/3,x-y+1/3,z+1/3
36 -x+y+2/3,y+1/3,z+1/3

loop_
_atom_site_label
_atom_site_type_symbol
_atom_site_symmetry_multiplicity
_atom_site_Wyckoff_label
_atom_site_fract_x
_atom_site_fract_y
_atom_site_fract_z
_atom_site_occupancy
B1 B 3 b 0.00000 0.00000 0.50000 1.00000
C1 C 6 c 0.00000 0.00000 0.38230 1.00000
B2 B 18 h 0.16280 0.83720 0.64160 1.00000
B3 B 18 h 0.10740 0.89260 0.88620 1.00000
```

$B_{13}C_2$ "$B_4C$" ($D1_g$): A13B2_hR15_166_b2h_c - POSCAR

```
A13B2_hR15_166_b2h_c & a,c/a,x2,x3,z3,x4,z4 --params=6.617,1.82847211727
    ↪ ,0.3823,0.8044,1.316,0.9936,1.6714 & R-3m D_{3d}^{5} #166 (bch^
    ↪ 2) & hR15 & $D1_{g}$ & B13C2 & B13C2 & G. Will and K. H.
    ↪ Kossobutzki, J. Less-Common Met. 44, 87-97 (1976)
   1.00000000000000
   3.30850000000000  -1.91016336561388   4.03300000000000
   0.00000000000000   3.82032673122775   4.03300000000000
  -3.30850000000000  -1.91016336561388   4.03300000000000
     B     C
    13     2
Direct
   0.50000000000000   0.50000000000000   0.50000000000000   B   (1b)
   0.80440000000000   0.80440000000000   1.31600000000000   B   (6h)
   1.31600000000000   0.80440000000000   0.80440000000000   B   (6h)
   0.80440000000000   1.31600000000000   0.80440000000000   B   (6h)
  -1.31600000000000  -0.80440000000000  -0.80440000000000   B   (6h)
  -0.80440000000000  -0.80440000000000  -1.31600000000000   B   (6h)
  -0.80440000000000  -1.31600000000000  -0.80440000000000   B   (6h)
   0.99360000000000   0.99360000000000   1.67140000000000   B   (6h)
   1.67140000000000   0.99360000000000   0.99360000000000   B   (6h)
   0.99360000000000   1.67140000000000   0.99360000000000   B   (6h)
  -1.67140000000000  -0.99360000000000  -0.99360000000000   B   (6h)
  -0.99360000000000  -0.99360000000000  -1.67140000000000   B   (6h)
  -0.99360000000000  -1.67140000000000  -0.99360000000000   B   (6h)
   0.38230000000000   0.38230000000000   0.38230000000000   C   (2c)
  -0.38230000000000  -0.38230000000000  -0.38230000000000   C   (2c)
```

$MnBi_2Te_4$: A2BC4_hR7_166_c_a_2c - CIF

```
# CIF file
data_findsym-output
_audit_creation_method FINDSYM

_chemical_name_mineral 'Bi2MnTe4'
_chemical_formula_sum 'Bi2 Mn Te4'

loop_
_publ_author_name
 'J.-Q. Yan'
 'Q. Zhang'
 'T. Heitmann'
 'Z. Huang'
 'K. Y. Chen'
 'J.-G. Cheng'
 'W. Wu'
 'D. Vaknin'
 'B. C. Sales'
 'R. J. McQueeney'
_journal_name_full_name
;
 Physical Review Materials
;
_journal_volume 3
_journal_year 2019
_journal_page_first 064202
_journal_page_last 064202
_publ_Section_title
;
 Crystal growth and magnetic structure of MnBi$_{2}$Te$_{4}$
;

_aflow_title 'MnBi$_{2}$Te$_{4}$ Structure'
_aflow_proto 'A2BC4_hR7_166_c_a_2c'
_aflow_params 'a,c/a,x_{2},x_{3},x_{4}'
_aflow_params_values '4.309,9.44047342771,0.4247,0.1332,0.294'
_aflow_Strukturbericht 'None'
_aflow_Pearson 'hR7'

_symmetry_space_group_name_H-M "R -3 2/m (hexagonal axes)"
_symmetry_Int_Tables_number 166

_cell_length_a     4.30900
_cell_length_b     4.30900
_cell_length_c     40.67900
_cell_angle_alpha  90.00000
_cell_angle_beta   90.00000
_cell_angle_gamma  120.00000
```

```
loop_
_space_group_symop_id
_space_group_symop_operation_xyz
1 x,y,z
2 -y,x-y,z
3 -x+y,-x,z
4 y,x,-z
5 -x,-x+y,-z
6 x-y,-y,-z
7 -x,-y,-z
8 y,-x+y,-z
9 x-y,x,-z
10 -y,-x,z
11 x,x-y,z
12 -x+y,y,z
13 x+1/3,y+2/3,z+2/3
14 -y+1/3,x-y+2/3,z+2/3
15 -x+y+1/3,-x+2/3,z+2/3
16 y+1/3,x+2/3,-z+2/3
17 -x+1/3,-x+y+2/3,-z+2/3
18 x-y+1/3,-y+2/3,-z+2/3
19 -x+1/3,-y+2/3,-z+2/3
20 y+1/3,-x+y+2/3,-z+2/3
21 x-y+1/3,x+2/3,-z+2/3
22 -y+1/3,-x+2/3,z+2/3
23 x+1/3,x-y+2/3,z+2/3
24 -x+y+1/3,y+2/3,z+2/3
25 x+2/3,y+1/3,z+1/3
26 -y+2/3,x-y+1/3,z+1/3
27 -x+y+2/3,-x+1/3,z+1/3
28 y+2/3,x+1/3,-z+1/3
29 -x+2/3,-x+y+1/3,-z+1/3
30 x-y+2/3,-y+1/3,-z+1/3
31 -x+2/3,-y+1/3,-z+1/3
32 y+2/3,-x+y+1/3,-z+1/3
33 x-y+2/3,x+1/3,-z+1/3
34 -y+2/3,-x+1/3,z+1/3
35 x+2/3,x-y+1/3,z+1/3
36 -x+y+2/3,y+1/3,z+1/3

loop_
_atom_site_label
_atom_site_type_symbol
_atom_site_symmetry_multiplicity
_atom_site_Wyckoff_label
_atom_site_fract_x
_atom_site_fract_y
_atom_site_fract_z
_atom_site_occupancy
Mn1 Mn 3 a 0.00000 0.00000 0.00000 1.00000
Bi1 Bi 6 c 0.00000 0.00000 0.42470 1.00000
Te1 Te 6 c 0.00000 0.00000 0.13320 1.00000
Te2 Te 6 c 0.00000 0.00000 0.29400 1.00000
```

$MnBi_2Te_4$: A2BC4_hR7_166_c_a_2c - POSCAR

```
A2BC4_hR7_166_c_a_2c & a,c/a,x2,x3,x4 --params=4.309,9.44047342771,
    ↪ 0.4247,0.1332,0.294 & R-3m D_{3d}^{5} #166 (ac^3) & hR7 & None
    ↪ & Bi2MnTe4 & Bi2MnTe4 & J.-Q. Yan et al., Phys. Rev. Mater. 3,
    ↪ 064202(2019)
   1.00000000000000
   2.15450000000000  -1.24390115496905  13.55966666666670
   0.00000000000000   2.48780230993810  13.55966666666670
  -2.15450000000000  -1.24390115496905  13.55966666666670
    Bi     Mn     Te
     2      1      4
Direct
   0.42470000000000   0.42470000000000   0.42470000000000   Bi   (2c)
  -0.42470000000000  -0.42470000000000  -0.42470000000000   Bi   (2c)
   0.00000000000000   0.00000000000000   0.00000000000000   Mn   (1a)
   0.13320000000000   0.13320000000000   0.13320000000000   Te   (2c)
  -0.13320000000000  -0.13320000000000  -0.13320000000000   Te   (2c)
   0.29400000000000   0.29400000000000   0.29400000000000   Te   (2c)
  -0.29400000000000  -0.29400000000000  -0.29400000000000   Te   (2c)
```

Shandite ($Ni_3Pb_2S_2$): A3B2C2_hR7_166_d_ab_c - CIF

```
# CIF file
data_findsym-output
_audit_creation_method FINDSYM

_chemical_name_mineral 'Shandite'
_chemical_formula_sum 'Ni3 Pb2 S2'

loop_
_publ_author_name
 'R. Weihrich'
 'S. F. Matar'
 'V. Eyert'
 'F. Rau'
 'M. Zabel'
 'M. Andratschke'
 'I. Anusca'
 'T. Bernert'
_journal_name_full_name
;
 Progress in Solid State Chemistry
;
_journal_volume 35
_journal_year 2007
_journal_page_first 309
_journal_page_last 322
_publ_Section_title
;
 Structure, ordering, and bonding of half antiperovskites: PbNi$_{3/2}$S
    ↪ and BiPd$_{3/2}$S
;

_aflow_title 'Shandite (Ni$_{3}$Pb$_{2}$S$_{2}$) Structure'
_aflow_proto 'A3B2C2_hR7_166_d_ab_c'
_aflow_params 'a,c/a,x_{3}'
_aflow_params_values '5.595,2.43538873995,0.275'
_aflow_Strukturbericht 'None'
_aflow_Pearson 'hR7'

_symmetry_space_group_name_H-M "R -3 2/m (hexagonal axes)"
_symmetry_Int_Tables_number 166

_cell_length_a      5.59500
_cell_length_b      5.59500
_cell_length_c      13.62600
_cell_angle_alpha   90.00000
_cell_angle_beta    90.00000
_cell_angle_gamma   120.00000

loop_
_space_group_symop_id
_space_group_symop_operation_xyz
1 x,y,z
2 -y,x-y,z
3 -x+y,-x,z
4 y,x,-z
5 -x,-x+y,-z
6 x-y,-y,-z
7 -x,-y,-z
8 y,-x+y,-z
9 x-y,x,-z
10 -y,-x,z
11 x,x-y,z
12 -x+y,y,z
13 x+1/3,y+2/3,z+2/3
14 -y+1/3,x-y+2/3,z+2/3
15 -x+y+1/3,-x+2/3,z+2/3
16 y+1/3,x+2/3,-z+2/3
17 -x+1/3,-x+y+2/3,-z+2/3
18 x-y+1/3,-y+2/3,-z+2/3
19 -x+1/3,-y+2/3,-z+2/3
20 y+1/3,-x+y+2/3,-z+2/3
21 x-y+1/3,x+2/3,-z+2/3
22 -y+1/3,-x+2/3,z+2/3
23 x+1/3,x-y+2/3,z+2/3
24 -x+y+1/3,y+2/3,z+2/3
25 x+2/3,y+1/3,z+1/3
26 -y+2/3,x-y+1/3,z+1/3
27 -x+y+2/3,-x+1/3,z+1/3
28 y+2/3,x+1/3,-z+1/3
29 -x+2/3,-x+y+1/3,-z+1/3
30 x-y+2/3,-y+1/3,-z+1/3
31 -x+2/3,-y+1/3,-z+1/3
32 y+2/3,-x+y+1/3,-z+1/3
33 x-y+2/3,x+1/3,-z+1/3
34 -y+2/3,-x+1/3,z+1/3
35 x+2/3,x-y+1/3,z+1/3
36 -x+y+2/3,y+1/3,z+1/3

loop_
_atom_site_label
_atom_site_type_symbol
_atom_site_symmetry_multiplicity
_atom_site_Wyckoff_label
_atom_site_fract_x
_atom_site_fract_y
_atom_site_fract_z
_atom_site_occupancy
Pb1 Pb 3 a 0.00000 0.00000 0.00000 1.00000
Pb2 Pb 3 b 0.00000 0.00000 0.50000 1.00000
S1 S 6 c 0.00000 0.00000 0.27500 1.00000
Ni1 Ni 9 d 0.50000 0.00000 0.50000 1.00000
```

Shandite ($Ni_3Pb_2S_2$): A3B2C2_hR7_166_d_ab_c - POSCAR

```
A3B2C2_hR7_166_d_ab_c & a,c/a,x3 --params=5.595,2.43538873995,0.275 &
    ↪ R-3m D_{3d}^{5} #166 (abcd) & hR7 & None & Ni3Pb2S2 & Shandite
    ↪ & R. Weihrich et al., Prog. Solid State Chem. 35, 309-322 (2007
    ↪ )
   1.00000000000000
   2.79750000000000  -1.61513737805798   4.54200000000000
   0.00000000000000   3.23027475611596   4.54200000000000
  -2.79750000000000  -1.61513737805798   4.54200000000000
    Ni     Pb     S
     3      2     2
Direct
   0.50000000000000   0.00000000000000   0.00000000000000   Ni   (3d)
   0.00000000000000   0.50000000000000   0.00000000000000   Ni   (3d)
   0.00000000000000   0.00000000000000   0.50000000000000   Ni   (3d)
   0.00000000000000   0.00000000000000   0.00000000000000   Pb   (1a)
   0.50000000000000   0.50000000000000   0.50000000000000   Pb   (1b)
   0.27500000000000   0.27500000000000   0.27500000000000   S    (2c)
  -0.27500000000000  -0.27500000000000  -0.27500000000000   S    (2c)
```

Chabazite ($Ca_{1.4}Sr_{0.3}Al_{3.8}Si_{8.3}O_{24}\cdot 13H_2O$, $S3_4$ (I)): A5B21C24D12_hR62_166_a2c_ehi_fg2h_i - CIF

```
# CIF file
data_findsym-output
_audit_creation_method FINDSYM

_chemical_name_mineral 'Chabazite'
_chemical_formula_sum 'Ca5 (H2O)21 O24 Si12'
```

```
loop_
_publ_author_name
 'M. Calligaris'
 'G. Nardin'
 'L. Randaccio'
 'P. C. Chiaramonti'
_journal_name_full_name
;
 Acta Crystallographica Section B: Structural Science
;
_journal_volume 38
_journal_year 1982
_journal_page_first 602
_journal_page_last 605
_publ_Section_title
;
 Cation-site location in a natural chabazite
;

_aflow_title 'Chabazite (Ca$_{1.4}$Sr$_{0.3}$Al$_{3.8}$Si$_{8.3}$O$_{24}
    ↪ $$\cdot$13H$_{2}$O, $S3_4$ (I)) Structure'
_aflow_proto 'A5B21C24D12_hR62_166_a2c_ehi_fg2h_i'
_aflow_params 'a,c/a,x_{2},x_{3},x_{5},x_{6},x_{7},z_{7},x_{8},z_{8},x_{
    ↪ 9},z_{9},x_{10},y_{10},z_{10},x_{11},y_{11},z_{11}'
_aflow_params_values '13.80257,1.09220819021,0.2038,0.4065,0.2638,1.1548
    ↪ ,0.418,0.75989,0.2515,0.89459,1.0248,-0.67229,0.2024,-0.5087,
    ↪ 0.31011,0.1044,-0.1251,0.33381'
_aflow_Strukturbericht '$S3_{4}$ (I)'
_aflow_Pearson 'hR62'

_symmetry_space_group_name_H-M "R -3 2/m (hexagonal axes)"
_symmetry_Int_Tables_number 166

_cell_length_a      13.80257
_cell_length_b      13.80257
_cell_length_c      15.07528
_cell_angle_alpha  90.00000
_cell_angle_beta   90.00000
_cell_angle_gamma 120.00000

loop_
_space_group_symop_id
_space_group_symop_operation_xyz
1 x,y,z
2 -y,x-y,z
3 -x+y,-x,z
4 y,x,-z
5 -x,-x+y,-z
6 x-y,-y,-z
7 -x,-y,-z
8 y,-x+y,-z
9 x-y,x,-z
10 -y,-x,z
11 x,x-y,z
12 -x+y,y,z
13 x+1/3,y+2/3,z+2/3
14 -y+1/3,x-y+2/3,z+2/3
15 -x+y+1/3,-x+2/3,z+2/3
16 y+1/3,x+2/3,-z+2/3
17 -x+1/3,-x+y+2/3,-z+2/3
18 x-y+1/3,-y+2/3,-z+2/3
19 -x+1/3,-y+2/3,-z+2/3
20 y+1/3,-x+y+2/3,-z+2/3
21 x-y+1/3,x+2/3,-z+2/3
22 -y+1/3,-x+2/3,z+2/3
23 x+1/3,x-y+2/3,z+2/3
24 -x+y+1/3,y+2/3,z+2/3
25 x+2/3,y+1/3,z+1/3
26 -y+2/3,x-y+1/3,z+1/3
27 -x+y+2/3,-x+1/3,z+1/3
28 y+2/3,x+1/3,-z+1/3
29 -x+2/3,-x+y+1/3,-z+1/3
30 x-y+2/3,-y+1/3,-z+1/3
31 -x+2/3,-y+1/3,-z+1/3
32 y+2/3,-x+y+1/3,-z+1/3
33 x-y+2/3,x+1/3,-z+1/3
34 -y+2/3,-x+1/3,z+1/3
35 x+2/3,x-y+1/3,z+1/3
36 -x+y+2/3,y+1/3,z+1/3

loop_
_atom_site_label
_atom_site_type_symbol
_atom_site_symmetry_multiplicity
_atom_site_Wyckoff_label
_atom_site_fract_x
_atom_site_fract_y
_atom_site_fract_z
_atom_site_occupancy
Ca1 Ca 3 a 0.00000 0.00000 0.00000 0.11000
Ca2 Ca 6 c 0.00000 0.00000 0.20380 0.53000
Ca3 Ca 6 c 0.00000 0.00000 0.40650 0.24000
H2O1 H2O 9 e 0.50000 0.00000 0.00000 0.50000
O1 O 18 f 0.26380 0.00000 0.00000 1.00000
O2 O 18 g 0.65480 0.00000 0.50000 1.00000
H2O2 H2O 18 h 0.21937 0.78063 0.19863 0.57000
O3 O 18 h 0.11897 0.88103 0.13253 1.00000
O4 O 18 h 0.89903 0.10097 0.12577 1.00000
H2O3 H2O 36 i 0.20113 0.50997 0.00127 0.23000
Si1 Si 36 i 0.00003 0.22947 0.10437 1.00000
```

Chabazite ($Ca_{1.4}Sr_{0.3}Al_{3.8}Si_{8.3}O_{24}\cdot 13H_2O$, $S3_4$ (I)): A5B21C24D12_hR62_166_a2c_ehi_fg2h_i - POSCAR

```
A5B21C24D12_hR62_166_a2c_ehi_fg2h_i & a,c/a,x2,x3,x5,x6,x7,z7,x8,z8,x9,
    ↪ z9,x10,y10,z10,x11,y11,z11 --params=13.80257,1.09220819021,
    ↪ 0.2038,0.4065,0.2638,1.1548,0.418,0.75989,0.2515,0.89459,1.0248
    ↪ ,-0.67229,0.2024,-0.5087,0.31011,0.1044,-0.1251,0.33381 & R-3m
    ↪ D_{3d}^{5} #166 (ac^2efgh^3i^2) & hR62 & $S3_{4}$ (I) &
    ↪ Al3.8Ca1.4H26O37Si8.3Sr0.3 & Chabazite & M. Calligaris et al.,
    ↪ Acta Crystallogr. Sect. B Struct. Sci. 38, 602-605 (1982)
   1.00000000000000
   6.90128500000000  -3.98445875250433   5.02509333333333
   0.00000000000000   7.96891750500865   5.02509333333333
  -6.90128500000000  -3.98445875250433   5.02509333333333
    Ca    H2O     O    Si
     5     21    24    12
Direct
   0.00000000000000   0.00000000000000   0.00000000000000   Ca   (1a)
   0.20380000000000   0.20380000000000   0.20380000000000   Ca   (2c)
  -0.20380000000000  -0.20380000000000  -0.20380000000000   Ca   (2c)
   0.40650000000000   0.40650000000000   0.40650000000000   Ca   (2c)
  -0.40650000000000  -0.40650000000000  -0.40650000000000   Ca   (2c)
   0.00000000000000   0.50000000000000   0.50000000000000   H2O  (3e)
   0.50000000000000   0.00000000000000   0.50000000000000   H2O  (3e)
   0.50000000000000   0.50000000000000   0.00000000000000   H2O  (3e)
   0.41800000000000   0.41800000000000   0.75989000000000   H2O  (6h)
   0.75989000000000   0.41800000000000   0.41800000000000   H2O  (6h)
   0.41800000000000   0.75989000000000   0.41800000000000   H2O  (6h)
  -0.75989000000000  -0.41800000000000  -0.41800000000000   H2O  (6h)
  -0.41800000000000  -0.41800000000000  -0.75989000000000   H2O  (6h)
  -0.41800000000000  -0.75989000000000  -0.41800000000000   H2O  (6h)
   0.20240000000000  -0.50870000000000   0.31011000000000   H2O  (12i)
   0.31011000000000   0.20240000000000  -0.50870000000000   H2O  (12i)
  -0.50870000000000   0.31011000000000   0.20240000000000   H2O  (12i)
  -0.31011000000000   0.50870000000000  -0.20240000000000   H2O  (12i)
   0.50870000000000  -0.20240000000000  -0.31011000000000   H2O  (12i)
  -0.20240000000000  -0.31011000000000   0.50870000000000   H2O  (12i)
  -0.20240000000000   0.50870000000000  -0.31011000000000   H2O  (12i)
  -0.31011000000000  -0.20240000000000   0.50870000000000   H2O  (12i)
   0.50870000000000  -0.31011000000000  -0.20240000000000   H2O  (12i)
   0.31011000000000  -0.50870000000000   0.20240000000000   H2O  (12i)
  -0.50870000000000   0.20240000000000   0.31011000000000   H2O  (12i)
   0.20240000000000   0.31011000000000  -0.50870000000000   H2O  (12i)
   0.26380000000000  -0.26380000000000   0.00000000000000    O   (6f)
   0.00000000000000   0.26380000000000  -0.26380000000000    O   (6f)
  -0.26380000000000   0.00000000000000   0.26380000000000    O   (6f)
  -0.26380000000000   0.26380000000000   0.00000000000000    O   (6f)
   0.00000000000000  -0.26380000000000   0.26380000000000    O   (6f)
   0.26380000000000   0.00000000000000  -0.26380000000000    O   (6f)
   1.15480000000000  -1.15480000000000   0.50000000000000    O   (6g)
   0.50000000000000   1.15480000000000  -1.15480000000000    O   (6g)
  -1.15480000000000   0.50000000000000   1.15480000000000    O   (6g)
  -1.15480000000000   1.15480000000000   0.50000000000000    O   (6g)
   0.50000000000000  -1.15480000000000   1.15480000000000    O   (6g)
   1.15480000000000   0.50000000000000  -1.15480000000000    O   (6g)
   0.25150000000000   0.25150000000000   0.89459000000000    O   (6h)
   0.89459000000000   0.25150000000000   0.25150000000000    O   (6h)
   0.25150000000000   0.89459000000000   0.25150000000000    O   (6h)
  -0.89459000000000  -0.25150000000000  -0.25150000000000    O   (6h)
  -0.25150000000000  -0.25150000000000  -0.89459000000000    O   (6h)
  -0.25150000000000  -0.89459000000000  -0.25150000000000    O   (6h)
   1.02480000000000   1.02480000000000  -0.67229000000000    O   (6h)
  -0.67229000000000   1.02480000000000   1.02480000000000    O   (6h)
   1.02480000000000  -0.67229000000000   1.02480000000000    O   (6h)
   0.67229000000000  -1.02480000000000  -1.02480000000000    O   (6h)
  -1.02480000000000  -1.02480000000000   0.67229000000000    O   (6h)
  -1.02480000000000   0.67229000000000  -1.02480000000000    O   (6h)
   0.10440000000000  -0.12510000000000   0.33381000000000   Si   (12i)
   0.33381000000000   0.10440000000000  -0.12510000000000   Si   (12i)
  -0.12510000000000   0.33381000000000   0.10440000000000   Si   (12i)
  -0.33381000000000   0.12510000000000  -0.10440000000000   Si   (12i)
   0.12510000000000  -0.10440000000000  -0.33381000000000   Si   (12i)
  -0.10440000000000  -0.33381000000000   0.12510000000000   Si   (12i)
  -0.10440000000000   0.12510000000000  -0.33381000000000   Si   (12i)
  -0.33381000000000  -0.10440000000000   0.12510000000000   Si   (12i)
   0.12510000000000  -0.33381000000000  -0.10440000000000   Si   (12i)
   0.33381000000000  -0.12510000000000   0.10440000000000   Si   (12i)
  -0.12510000000000   0.10440000000000   0.33381000000000   Si   (12i)
   0.10440000000000   0.33381000000000  -0.12510000000000   Si   (12i)
```

$CaSi_2$ ($C12$): AB2_hR6_166_c_2c - CIF

```
# CIF file
data_findsym-output
_audit_creation_method FINDSYM

_chemical_name_mineral 'CaSi2'
_chemical_formula_sum 'Ca Si2'

loop_
_publ_author_name
 'S. M. Castillo'
 'Z. Tang'
 'A. P. Litvinchuk'
 'A. M. Guloy'
_journal_name_full_name
;
 Inorganic Chemistry
;
_journal_volume 55
_journal_year 2016
_journal_page_first 10203
_journal_page_last 10207
_publ_Section_title
;
 Lattice Dynamics of the Rhombohedral Polymorphs of CaSi$_{2}$
;

_aflow_title 'CaSi$_{2}$ ($C12$) Structure'
```

```
_aflow_proto 'AB2_hR6_166_c_2c'
_aflow_params 'a,c/a,x_{1},x_{2},x_{3}'
_aflow_params_values '3.8548,7.95397945419,0.083,0.183,0.35'
_aflow_Strukturbericht '$C12$'
_aflow_Pearson 'hR6'

_symmetry_space_group_name_H-M "R -3 2/m (hexagonal axes)"
_symmetry_Int_Tables_number 166

_cell_length_a     3.85480
_cell_length_b     3.85480
_cell_length_c     30.66100
_cell_angle_alpha  90.00000
_cell_angle_beta   90.00000
_cell_angle_gamma  120.00000

loop_
_space_group_symop_id
_space_group_symop_operation_xyz
1 x,y,z
2 -y,x-y,z
3 -x+y,-x,z
4 y,x,-z
5 -x,-x+y,-z
6 x-y,-y,-z
7 -x,-y,-z
8 y,-x+y,-z
9 x-y,x,-z
10 -y,-x,z
11 x,x-y,z
12 -x+y,y,z
13 x+1/3,y+2/3,z+2/3
14 -y+1/3,x-y+2/3,z+2/3
15 -x+y+1/3,-x+2/3,z+2/3
16 y+1/3,x+2/3,-z+2/3
17 -x+1/3,-x+y+2/3,-z+2/3
18 x-y+1/3,-y+2/3,-z+2/3
19 -x+1/3,-y+2/3,-z+2/3
20 y+1/3,-x+y+2/3,-z+2/3
21 x-y+1/3,x+2/3,-z+2/3
22 -y+1/3,-x+2/3,z+2/3
23 x+1/3,x-y+2/3,z+2/3
24 -x+y+1/3,y+2/3,z+2/3
25 x+2/3,y+1/3,z+1/3
26 -y+2/3,x-y+1/3,z+1/3
27 -x+y+2/3,-x+1/3,z+1/3
28 y+2/3,x+1/3,-z+1/3
29 -x+2/3,-x+y+1/3,-z+1/3
30 x-y+2/3,-y+1/3,-z+1/3
31 -x+2/3,-y+1/3,-z+1/3
32 y+2/3,-x+y+1/3,-z+1/3
33 x-y+2/3,x+1/3,-z+1/3
34 -y+2/3,-x+1/3,z+1/3
35 x+2/3,x-y+1/3,z+1/3
36 -x+y+2/3,y+1/3,z+1/3

loop_
_atom_site_label
_atom_site_type_symbol
_atom_site_symmetry_multiplicity
_atom_site_Wyckoff_label
_atom_site_fract_x
_atom_site_fract_y
_atom_site_fract_z
_atom_site_occupancy
Ca1 Ca 6 c 0.00000 0.00000 0.08300 1.00000
Si1 Si 6 c 0.00000 0.00000 0.18300 1.00000
Si2 Si 6 c 0.00000 0.00000 0.35000 1.00000
```

$CaSi_2$ ($C12$): AB2_hR6_166_c_2c - POSCAR

```
AB2_hR6_166_c_2c & a,c/a,x1,x2,x3 --params=3.8548,7.95397945419,0.083,
    ↪ 0.183,0.35 & R-3m D_{3d}^{5} #166 (c^3) & hR6 & $C12$ & CaSi2 &
    ↪ CaSi2 & S. M. Castillo et al., Inorg. Chem. 55, 10203-10207 (
    ↪ 2016)
   1.00000000000000
   1.92740000000000  -1.11278490883608  10.22033333333330
   0.00000000000000   2.22556981767217  10.22033333333330
  -1.92740000000000  -1.11278490883608  10.22033333333330
    Ca    Si
     2     4
Direct
   0.08300000000000   0.08300000000000   0.08300000000000   Ca   (2c)
  -0.08300000000000  -0.08300000000000  -0.08300000000000   Ca   (2c)
   0.18300000000000   0.18300000000000   0.18300000000000   Si   (2c)
  -0.18300000000000  -0.18300000000000  -0.18300000000000   Si   (2c)
   0.35000000000000   0.35000000000000   0.35000000000000   Si   (2c)
  -0.35000000000000  -0.35000000000000  -0.35000000000000   Si   (2c)
```

Rhombohedral $CuTi_2S_4$: AB4C2_hR28_166_2c_2c2h_abh - CIF

```
# CIF file
data_findsym-output
_audit_creation_method FINDSYM

_chemical_name_mineral 'CuS4Ti2'
_chemical_formula_sum 'Cu S4 Ti2'

loop_
_publ_author_name
 'N. Soheilnia'
 'K. M. Kleinke'
 'E. Dashjav'
 'H. L. Cuthbert'
 'J. E. Greedan'
 'H. Kleinke'
_journal_name_full_name
;
 Inorganic Chemistry
;
_journal_volume 43
_journal_year 2004
_journal_page_first 6473
_journal_page_last 6478
_publ_Section_title
;
 Crystal Structure and Physical Properties of a New CuTi$_{2}$S$_{4}$
    ↪ Modification in Comparison to the Thiospinel
;

_aflow_title 'Rhombohedral CuTi$_{2}$S$_{4}$ Structure'
_aflow_proto 'AB4C2_hR28_166_2c_2c2h_abh'
_aflow_params 'a,c/a,x_{3},x_{4},x_{5},x_{6},x_{7},z_{7},x_{8},z_{8},x_{
    ↪ 9},z_{9}'
_aflow_params_values '7.0242,4.95914125452,0.18759,0.35449,0.1232,
    ↪ 0.28892,0.19936,0.7283,0.61457,0.13398,1.08544,-0.42793'
_aflow_Strukturbericht 'None'
_aflow_Pearson 'hR28'

_symmetry_space_group_name_H-M "R -3 2/m (hexagonal axes)"
_symmetry_Int_Tables_number 166

_cell_length_a     7.02420
_cell_length_b     7.02420
_cell_length_c     34.83400
_cell_angle_alpha  90.00000
_cell_angle_beta   90.00000
_cell_angle_gamma  120.00000

loop_
_space_group_symop_id
_space_group_symop_operation_xyz
1 x,y,z
2 -y,x-y,z
3 -x+y,-x,z
4 y,x,-z
5 -x,-x+y,-z
6 x-y,-y,-z
7 -x,-y,-z
8 y,-x+y,-z
9 x-y,x,-z
10 -y,-x,z
11 x,x-y,z
12 -x+y,y,z
13 x+1/3,y+2/3,z+2/3
14 -y+1/3,x-y+2/3,z+2/3
15 -x+y+1/3,-x+2/3,z+2/3
16 y+1/3,x+2/3,-z+2/3
17 -x+1/3,-x+y+2/3,-z+2/3
18 x-y+1/3,-y+2/3,-z+2/3
19 -x+1/3,-y+2/3,-z+2/3
20 y+1/3,-x+y+2/3,-z+2/3
21 x-y+1/3,x+2/3,-z+2/3
22 -y+1/3,-x+2/3,z+2/3
23 x+1/3,x-y+2/3,z+2/3
24 -x+y+1/3,y+2/3,z+2/3
25 x+2/3,y+1/3,z+1/3
26 -y+2/3,x-y+1/3,z+1/3
27 -x+y+2/3,-x+1/3,z+1/3
28 y+2/3,x+1/3,-z+1/3
29 -x+2/3,-x+y+1/3,-z+1/3
30 x-y+2/3,-y+1/3,-z+1/3
31 -x+2/3,-y+1/3,-z+1/3
32 y+2/3,-x+y+1/3,-z+1/3
33 x-y+2/3,x+1/3,-z+1/3
34 -y+2/3,-x+1/3,z+1/3
35 x+2/3,x-y+1/3,z+1/3
36 -x+y+2/3,y+1/3,z+1/3

loop_
_atom_site_label
_atom_site_type_symbol
_atom_site_symmetry_multiplicity
_atom_site_Wyckoff_label
_atom_site_fract_x
_atom_site_fract_y
_atom_site_fract_z
_atom_site_occupancy
Ti1 Ti 3 a 0.00000 0.00000 0.00000 1.00000
Ti2 Ti 3 b 0.00000 0.00000 0.50000 1.00000
Cu1 Cu 6 c 0.00000 0.00000 0.18759 1.00000
Cu2 Cu 6 c 0.00000 0.00000 0.35449 1.00000
S1 S 6 c 0.00000 0.00000 0.12320 1.00000
S2 S 6 c 0.00000 0.00000 0.28892 1.00000
S3 S 18 h 0.15702 0.84298 0.04234 1.00000
S4 S 18 h 0.49353 0.50647 0.12104 1.00000
Ti3 Ti 18 h 0.83779 0.16221 0.24765 1.00000
```

Rhombohedral $CuTi_2S_4$: AB4C2_hR28_166_2c_2c2h_abh - POSCAR

```
AB4C2_hR28_166_2c_2c2h_abh & a,c/a,x3,x4,x5,x6,x7,z7,x8,z8,x9,z9 --
    ↪ params=7.0242,4.95914125452,0.18759,0.35449,0.1232,0.28892,
    ↪ 0.19936,0.7283,0.61457,0.13398,1.08544,-0.42793 & R-3m D_{3d}^{
    ↪ 5} #166 (abc^4h^3) & hR28 & None & CuS4Ti2 & CuS4Ti2 & N.
    ↪ Soheilnia et al., Inorg. Chem. 43, 6473-6478 (2004)
   1.00000000000000
   3.51210000000000  -2.02771188042089  11.61133333333330
   0.00000000000000   4.05542376084177  11.61133333333330
  -3.51210000000000  -2.02771188042089  11.61133333333330
    Cu    S    Ti
```

```
     4      16      8
Direct
   0.18759000000000    0.18759000000000    0.18759000000000    Cu    (2c)
  -0.18759000000000   -0.18759000000000   -0.18759000000000    Cu    (2c)
   0.35449000000000    0.35449000000000    0.35449000000000    Cu    (2c)
  -0.35449000000000   -0.35449000000000   -0.35449000000000    Cu    (2c)
   0.12320000000000    0.12320000000000    0.12320000000000     S    (2c)
  -0.12320000000000   -0.12320000000000   -0.12320000000000     S    (2c)
   0.28892000000000    0.28892000000000    0.28892000000000     S    (2c)
  -0.28892000000000   -0.28892000000000   -0.28892000000000     S    (2c)
   0.19936000000000    0.19936000000000    0.72830000000000     S    (6h)
   0.72830000000000    0.19936000000000    0.19936000000000     S    (6h)
   0.19936000000000    0.72830000000000    0.19936000000000     S    (6h)
  -0.72830000000000   -0.19936000000000   -0.19936000000000     S    (6h)
  -0.19936000000000   -0.19936000000000   -0.72830000000000     S    (6h)
  -0.19936000000000   -0.72830000000000   -0.19936000000000     S    (6h)
   0.61457000000000    0.61457000000000    0.13398000000000     S    (6h)
   0.13398000000000    0.61457000000000    0.61457000000000     S    (6h)
   0.61457000000000    0.13398000000000    0.61457000000000     S    (6h)
  -0.13398000000000   -0.61457000000000   -0.61457000000000     S    (6h)
  -0.61457000000000   -0.61457000000000   -0.13398000000000     S    (6h)
  -0.61457000000000   -0.13398000000000   -0.61457000000000     S    (6h)
   0.00000000000000    0.00000000000000    0.00000000000000    Ti    (1a)
   0.50000000000000    0.50000000000000    0.50000000000000    Ti    (1b)
   1.08544000000000    1.08544000000000   -0.42793000000000    Ti    (6h)
  -0.42793000000000    1.08544000000000    1.08544000000000    Ti    (6h)
   1.08544000000000   -0.42793000000000    1.08544000000000    Ti    (6h)
   0.42793000000000   -1.08544000000000   -1.08544000000000    Ti    (6h)
  -1.08544000000000   -1.08544000000000    0.42793000000000    Ti    (6h)
  -1.08544000000000    0.42793000000000   -1.08544000000000    Ti    (6h)
```

$CaCu_4P_2$: AB4C2_hR7_166_a_2c_c - CIF

```
# CIF file
data_findsym-output
_audit_creation_method FINDSYM

_chemical_name_mineral 'CaCu4P2'
_chemical_formula_sum 'Ca Cu4 P2'

loop_
_publ_author_name
 'A. Mewis'
_journal_name_full_name
;
 Zeitschrift f{\"u}r Naturforschung B
;
_journal_volume 35
_journal_year 1980
_journal_page_first 942
_journal_page_last 945
_publ_Section_title
;
 Darstellung und Struktur der Verbindung CaCu$_{4}$P$_{2}$
;

_aflow_title 'CaCu$_{4}$P$_{2}$ Structure'
_aflow_proto 'AB4C2_hR7_166_a_2c_c'
_aflow_params 'a,c/a,x_{2},x_{3},x_{4}'
_aflow_params_values '4.036,5.51833498513,0.1453,0.4311,0.2503'
_aflow_Strukturbericht 'None'
_aflow_Pearson 'hR7'

_symmetry_space_group_name_H-M "R -3 2/m (hexagonal axes)"
_symmetry_Int_Tables_number 166

_cell_length_a      4.03600
_cell_length_b      4.03600
_cell_length_c      22.27200
_cell_angle_alpha  90.00000
_cell_angle_beta   90.00000
_cell_angle_gamma  120.00000

loop_
_space_group_symop_id
_space_group_symop_operation_xyz
1 x,y,z
2 -y,x-y,z
3 -x+y,-x,z
4 y,x,-z
5 -x,-x+y,-z
6 x-y,-y,-z
7 -x,-y,-z
8 y,-x+y,-z
9 x-y,x,-z
10 -y,-x,z
11 x,x-y,z
12 -x+y,y,z
13 x+1/3,y+2/3,z+2/3
14 -y+1/3,x-y+2/3,z+2/3
15 -x+y+1/3,-x+2/3,z+2/3
16 y+1/3,x+2/3,-z+2/3
17 -x+1/3,-x+y+2/3,-z+2/3
18 x-y+1/3,-y+2/3,-z+2/3
19 -x+1/3,-y+2/3,-z+2/3
20 y+1/3,-x+y+2/3,-z+2/3
21 x-y+1/3,x+2/3,-z+2/3
22 -y+1/3,-x+2/3,z+2/3
23 x+1/3,x-y+2/3,z+2/3
24 -x+y+1/3,y+2/3,z+2/3
25 x+2/3,y+1/3,z+1/3
26 -y+2/3,x-y+1/3,z+1/3
27 -x+y+2/3,-x+1/3,z+1/3
28 y+2/3,x+1/3,-z+1/3
29 -x+2/3,-x+y+1/3,-z+1/3
30 x-y+2/3,-y+1/3,-z+1/3
31 -x+2/3,-y+1/3,-z+1/3
32 y+2/3,-x+y+1/3,-z+1/3
33 x-y+2/3,x+1/3,-z+1/3
34 -y+2/3,-x+1/3,z+1/3
35 x+2/3,x-y+1/3,z+1/3
36 -x+y+2/3,y+1/3,z+1/3

loop_
_atom_site_label
_atom_site_type_symbol
_atom_site_symmetry_multiplicity
_atom_site_Wyckoff_label
_atom_site_fract_x
_atom_site_fract_y
_atom_site_fract_z
_atom_site_occupancy
Ca1 Ca 3 a 0.00000 0.00000 0.00000 1.00000
Cu1 Cu 6 c 0.00000 0.00000 0.14530 1.00000
Cu2 Cu 6 c 0.00000 0.00000 0.43110 1.00000
P1 P 6 c 0.00000 0.00000 0.25030 1.00000
```

$CaCu_4P_2$: AB4C2_hR7_166_a_2c_c - POSCAR

```
AB4C2_hR7_166_a_2c_c & a,c/a,x2,x3,x4 --params=4.036,5.51833498513,
    ↪ 0.1453,0.4311,0.2503 & R-3m D_{3d}^{5} #166 (ac^3) & hR7 & None
    ↪ & CaCu4P2 & CaCu4P2 & A. Mewis, Z. Naturforsch. B 35, 942-945
    ↪ (1980)
   1.00000000000000
   2.01800000000000   -1.16509284322466    7.42400000000000
   0.00000000000000    2.33018568644933    7.42400000000000
  -2.01800000000000   -1.16509284322466    7.42400000000000
    Ca      Cu      P
     1       4      2
Direct
   0.00000000000000    0.00000000000000    0.00000000000000    Ca    (1a)
   0.14530000000000    0.14530000000000    0.14530000000000    Cu    (2c)
  -0.14530000000000   -0.14530000000000   -0.14530000000000    Cu    (2c)
   0.43110000000000    0.43110000000000    0.43110000000000    Cu    (2c)
  -0.43110000000000   -0.43110000000000   -0.43110000000000    Cu    (2c)
   0.25030000000000    0.25030000000000    0.25030000000000     P    (2c)
  -0.25030000000000   -0.25030000000000   -0.25030000000000     P    (2c)
```

$CaUO_4$: AB4C_hR6_166_b_2c_a - CIF

```
# CIF file
data_findsym-output
_audit_creation_method FINDSYM

_chemical_name_mineral 'CaO4U'
_chemical_formula_sum 'Ca O4 U'

loop_
_publ_author_name
 'B. O. Loopstra'
 'H. M. Rietveld'
_journal_name_full_name
;
 Acta Crystallographica Section B: Structural Science
;
_journal_volume 25
_journal_year 1969
_journal_page_first 787
_journal_page_last 791
_publ_Section_title
;
 The structure of some alkaline-earth metal uranates
;

_aflow_title 'CaUO$_{4}$ Structure'
_aflow_proto 'AB4C_hR6_166_b_2c_a'
_aflow_params 'a,c/a,x_{3},x_{4}'
_aflow_params_values '3.87818,4.52899555977,0.1118,0.3627'
_aflow_Strukturbericht 'None'
_aflow_Pearson 'hR6'

_symmetry_space_group_name_H-M "R -3 2/m (hexagonal axes)"
_symmetry_Int_Tables_number 166

_cell_length_a      3.87818
_cell_length_b      3.87818
_cell_length_c      17.56426
_cell_angle_alpha  90.00000
_cell_angle_beta   90.00000
_cell_angle_gamma  120.00000

loop_
_space_group_symop_id
_space_group_symop_operation_xyz
1 x,y,z
2 -y,x-y,z
3 -x+y,-x,z
4 y,x,-z
5 -x,-x+y,-z
6 x-y,-y,-z
7 -x,-y,-z
8 y,-x+y,-z
9 x-y,x,-z
10 -y,-x,z
11 x,x-y,z
12 -x+y,y,z
13 x+1/3,y+2/3,z+2/3
14 -y+1/3,x-y+2/3,z+2/3
15 -x+y+1/3,-x+2/3,z+2/3
16 y+1/3,x+2/3,-z+2/3
```

```
17 -x+1/3,-x+y+2/3,-z+2/3
18 x-y+1/3,-y+2/3,-z+2/3
19 -x+1/3,-y+2/3,-z+2/3
20 y+1/3,-x+y+2/3,-z+2/3
21 x-y+1/3,x+2/3,-z+2/3
22 -y+1/3,-x+2/3,z+2/3
23 x+1/3,x-y+2/3,z+2/3
24 -x+y+1/3,y+2/3,z+2/3
25 x+2/3,y+1/3,z+1/3
26 -y+2/3,x-y+1/3,z+1/3
27 -x+y+2/3,-x+1/3,z+1/3
28 y+2/3,x+1/3,-z+1/3
29 -x+2/3,-x+y+1/3,-z+1/3
30 x-y+2/3,-y+1/3,-z+1/3
31 -x+2/3,-y+1/3,-z+1/3
32 y+2/3,-x+y+1/3,-z+1/3
33 x-y+2/3,x+1/3,-z+1/3
34 -y+2/3,-x+1/3,z+1/3
35 x+2/3,x-y+1/3,z+1/3
36 -x+y+2/3,y+1/3,z+1/3

loop_
_atom_site_label
_atom_site_type_symbol
_atom_site_symmetry_multiplicity
_atom_site_Wyckoff_label
_atom_site_fract_x
_atom_site_fract_y
_atom_site_fract_z
_atom_site_occupancy
U1 U 3 a 0.00000 0.00000 0.00000 1.00000
Ca1 Ca 3 b 0.00000 0.00000 0.50000 1.00000
O1 O 6 c 0.00000 0.00000 0.11180 1.00000
O2 O 6 c 0.00000 0.00000 0.36270 1.00000
```

$CaUO_4$: AB4C_hR6_166_b_2c_a - POSCAR

```
AB4C_hR6_166_b_2c_a & a,c/a,x3,x4 --params=3.87818,4.52899555977,0.1118,
    ↪ 0.3627 & R-3m D_{3d}^{5} #166 (abc^2) & hR6 & None & CaO4U &
    ↪ CaO4U & B. O. Loopstra and H. M. Rietveld, Acta Crystallogr.
    ↪ Sect. B Struct. Sci. 25, 787-791 (1969)
   1.00000000000000
   1.93909000000000   -1.11953413348291    5.85475333333333
   0.00000000000000    2.23906826696582    5.85475333333333
  -1.93909000000000   -1.11953413348291    5.85475333333333
    Ca        O        U
     1        4        1
Direct
   0.50000000000000    0.50000000000000    0.50000000000000    Ca    (1b)
   0.11180000000000    0.11180000000000    0.11180000000000     O    (2c)
  -0.11180000000000   -0.11180000000000   -0.11180000000000     O    (2c)
   0.36270000000000    0.36270000000000    0.36270000000000     O    (2c)
  -0.36270000000000   -0.36270000000000   -0.36270000000000     O    (2c)
   0.00000000000000    0.00000000000000    0.00000000000000     U    (1a)
```

$TaTi_7$ (BCC SQS-16): AB7_hR16_166_c_c2h - CIF

```
# AFLOW.org Repositories
# TaTi/AB7_hR16_166_c_c2h-001.AB params=9.2519726786,1.22474487138,0.875
    ↪ ,0.625,1.625,0.1250000001,1.125,1.6249999999 SG=166 [ANRL doi:
    ↪ 10.1016/j.commatsci.2017.01.017 (part 1), doi: 10.1016/
    ↪ j.commatsci.2018.10.043 (part 2)]
data_TaTi
_pd_phase_name AB7_hR16_166_c_c2h-001.AB

_chemical_name_mineral 'TaTi7'
_chemical_formula_sum 'Ta Ti7'

loop_
_publ_author_name
 'T. Chakraborty'
 'J. Rogal'
 'R. Drautz'
_journal_name_full_name
;
 Physical Review B
;
_journal_volume 94
_journal_year 2016
_journal_page_first 224104
_journal_page_last 224104
_publ_Section_title
;
 Unraveling the composition dependence of the martensitic transformation
    ↪  temperature: A first-principles study of Ti-Ta alloys
;

_aflow_title 'TaTi$_{7}$ (BCC SQS-16) Structure'
_aflow_proto 'AB7_hR16_166_c_c2h'
_aflow_params 'a,c/a,x_{1},x_{2},x_{3},z_{3},x_{4},z_{4}'
_aflow_params_values '9.2519726786,1.22474487138,0.375,0.125,1.125,
    ↪ 1.6250000002,1.625,0.1249999998'
_aflow_Strukturbericht 'None'
_aflow_Pearson 'hR16'

_cell_length_a    9.2519726786
_cell_length_b    9.2519726786
_cell_length_c    11.3313060883
_cell_angle_alpha    90.0000000000
_cell_angle_beta    90.0000000000
_cell_angle_gamma    120.0000000000
_symmetry_space_group_name_H-M    'R-3m'
_symmetry_Int_Tables_Number    166
loop_
_symmetry_equiv_pos_site_id
_symmetry_equiv_pos_as_xyz
 1  x,y,z
 2  -y,x-y,z
 3  -x+y,-x,z
 4  y,x,-z
 5  x-y,-y,-z
 6  -x,-x+y,-z
 7  -x,-y,-z
 8  y,-x+y,-z
 9  x-y,x,-z
 10  -y,-x,z
 11  -x+y,y,z
 12  x,x-y,z
 13  x+2/3,y+1/3,z+1/3
 14  -y+2/3,x-y+1/3,z+1/3
 15  -x+y+2/3,-x+1/3,z+1/3
 16  y+2/3,x+1/3,-z+1/3
 17  x-y+2/3,-y+1/3,-z+1/3
 18  -x+2/3,-x+y+1/3,-z+1/3
 19  -x+2/3,-y+1/3,-z+1/3
 20  y+2/3,-x+y+1/3,-z+1/3
 21  x-y+2/3,x+1/3,-z+1/3
 22  -y+2/3,-x+1/3,z+1/3
 23  -x+y+2/3,y+1/3,z+1/3
 24  x+2/3,x-y+1/3,z+1/3
 25  x+1/3,y+2/3,z+2/3
 26  -y+1/3,x-y+2/3,z+2/3
 27  -x+y+1/3,-x+2/3,z+2/3
 28  y+1/3,x+2/3,-z+2/3
 29  x-y+1/3,-y+2/3,-z+2/3
 30  -x+1/3,-x+y+2/3,-z+2/3
 31  -x+1/3,-y+2/3,-z+2/3
 32  y+1/3,-x+y+2/3,-z+2/3
 33  x-y+1/3,x+2/3,-z+2/3
 34  -y+1/3,-x+2/3,z+2/3
 35  -x+y+1/3,y+2/3,z+2/3
 36  x+1/3,x-y+2/3,z+2/3
loop_
 _atom_site_label
 _atom_site_occupancy
 _atom_site_fract_x
 _atom_site_fract_y
 _atom_site_fract_z
 _atom_site_thermal_displace_type
 _atom_site_B_iso_or_equiv
 _atom_site_type_symbol
 _atom_site_symmetry_multiplicity
 _atom_site_Wyckoff_label
Ta1 1.0000000000 0.0000000000 -0.0000000000 0.3750000000 Biso 1.0 Ta 6 c
Ti1 1.0000000000 0.0000000000 -0.0000000000 0.1250000000 Biso 1.0 Ti 6 c
Ti2 1.0000000000 0.1666666666 0.8333333334 0.9583333334 Biso 1.0 Ti 18 h
Ti3 1.0000000000 0.8333333334 0.1666666666 0.7916666666 Biso 1.0 Ti 18 h
```

$TaTi_7$ (BCC SQS-16): AB7_hR16_166_c_c2h - POSCAR

```
AB7_hR16_166_c_c2h & a,c/a,x1,x2,x3,z3,x4,z4 --params=9.2519726786,
    ↪ 1.22474487138,0.375,0.125,1.125,1.6250000002,1.625,0.1249999998
    ↪  & R-3m D_{3d}^{5} #166 (c^2h^2) & hR16 & None & TaTi7 & TaTi7
    ↪ & T. Chakraborty and J. Rogal and R. Drautz, Phys. Rev. B 94,
    ↪ 224104(2016)
   1.00000000000000
   4.62598633930000   -2.67081445826239    3.77710202943333
   0.00000000000000    5.34162891652477    3.77710202943333
  -4.62598633930000   -2.67081445826239    3.77710202943333
    Ta        Ti
     2        14
Direct
   0.37500000000000    0.37500000000000    0.37500000000000    Ta    (2c)
  -0.37500000000000   -0.37500000000000   -0.37500000000000    Ta    (2c)
   0.12500000000000    0.12500000000000    0.12500000000000    Ti    (2c)
  -0.12500000000000   -0.12500000000000   -0.12500000000000    Ti    (2c)
   1.12500000000000    1.12500000000000    1.62500000020000    Ti    (6h)
   1.62500000020000    1.12500000000000    1.12500000000000    Ti    (6h)
   1.12500000000000    1.62500000020000    1.12500000000000    Ti    (6h)
  -1.62500000020000   -1.12500000000000   -1.12500000000000    Ti    (6h)
  -1.12500000000000   -1.12500000000000   -1.62500000020000    Ti    (6h)
  -1.12500000000000   -1.62500000020000   -1.12500000000000    Ti    (6h)
   1.62500000000000    1.62500000000000    0.12499999980000    Ti    (6h)
   0.12499999980000    1.62500000000000    1.62500000000000    Ti    (6h)
   1.62500000000000    0.12499999980000    1.62500000000000    Ti    (6h)
  -0.12499999980000   -1.62500000000000   -1.62500000000000    Ti    (6h)
  -1.62500000000000   -1.62500000000000   -0.12499999980000    Ti    (6h)
  -1.62500000000000   -0.12499999980000   -1.62500000000000    Ti    (6h)
```

Rhombohedral Delafossite ($CuFeO_2$): ABC2_hR4_166_a_b_c - CIF

```
# CIF file
data_findsym-output
_audit_creation_method FINDSYM

_chemical_name_mineral 'Delafossite'
_chemical_formula_sum 'Cu Fe O2'

loop_
_publ_author_name
 'C. T. Prewitt'
 'R. D. Shannon'
 'D. B. Rogers'
_journal_name_full_name
;
 Inorganic Chemistry
;
_journal_volume 10
_journal_year 1971
_journal_page_first 719
```

```
_journal_page_last 723
_publ_Section_title
;
 Chemistry of noble metal oxides. II. Crystal structures of platinum
   ↪ cobalt dioxide, palladium cobalt dioxide, coppper iron dioxide
   ↪ , and silver iron dioxide
;

# Found in Crystal chemistry and electrical properties of the
   ↪ delafossite structure, 2006

_aflow_title 'Rhombohedral Delafossite (CuFeO$_{2}$) Structure'
_aflow_proto 'ABC2_hR4_166_a_b_c'
_aflow_params 'a,c/a,x_{3}'
_aflow_params_values '3.0351,5.65582682613,0.1066'
_aflow_Strukturbericht 'None'
_aflow_Pearson 'hR4'

_symmetry_space_group_name_H-M "R -3 2/m (hexagonal axes)"
_symmetry_Int_Tables_number 166

_cell_length_a      3.03510
_cell_length_b      3.03510
_cell_length_c      17.16600
_cell_angle_alpha  90.00000
_cell_angle_beta   90.00000
_cell_angle_gamma  120.00000

loop_
_space_group_symop_id
_space_group_symop_operation_xyz
1 x,y,z
2 -y,x-y,z
3 -x+y,-x,z
4 y,x,-z
5 -x,-x+y,-z
6 x-y,-y,-z
7 -x,-y,-z
8 y,-x+y,-z
9 x-y,x,-z
10 -y,-x,z
11 x,x-y,z
12 -x+y,y,z
13 x+1/3,y+2/3,z+2/3
14 -y+1/3,x-y+2/3,z+2/3
15 -x+y+1/3,-x+2/3,z+2/3
16 y+1/3,x+2/3,-z+2/3
17 -x+1/3,-x+y+2/3,-z+2/3
18 x-y+1/3,-y+2/3,-z+2/3
19 -x+1/3,-y+2/3,-z+2/3
20 y+1/3,-x+y+2/3,-z+2/3
21 x-y+1/3,x+2/3,-z+2/3
22 -y+1/3,-x+2/3,z+2/3
23 x+1/3,x-y+2/3,z+2/3
24 -x+y+1/3,y+2/3,z+2/3
25 x+2/3,y+1/3,z+1/3
26 -y+2/3,x-y+1/3,z+1/3
27 -x+y+2/3,-x+1/3,z+1/3
28 y+2/3,x+1/3,-z+1/3
29 -x+2/3,-x+y+1/3,-z+1/3
30 x-y+2/3,-y+1/3,-z+1/3
31 -x+2/3,-y+1/3,-z+1/3
32 y+2/3,-x+y+1/3,-z+1/3
33 x-y+2/3,x+1/3,-z+1/3
34 -y+2/3,-x+1/3,z+1/3
35 x+2/3,x-y+1/3,z+1/3
36 -x+y+2/3,y+1/3,z+1/3

loop_
_atom_site_label
_atom_site_type_symbol
_atom_site_symmetry_multiplicity
_atom_site_Wyckoff_label
_atom_site_fract_x
_atom_site_fract_y
_atom_site_fract_z
_atom_site_occupancy
Cu1 Cu 3 a 0.00000 0.00000 0.00000 1.00000
Fe1 Fe 3 b 0.00000 0.00000 0.50000 1.00000
O1 O 6 c 0.00000 0.00000 0.10660 1.00000
```

Rhombohedral Delafossite ($CuFeO_2$): ABC2_hR4_166_a_b_c - POSCAR

```
ABC2_hR4_166_a_b_c & a,c/a,x3 --params=3.0351,5.65582682613,0.1066 &
   ↪ R-3m D_{3d}^{5} #166 (abc) & hR4 & None & CuFeO2 & Delafossite
   ↪ & C. T. Prewitt and R. D. Shannon and D. B. Rogers, Inorg.
   ↪ Chem. 10, 719-723 (1971)
   1.00000000000000
   1.51755000000000   -0.87615790100872   5.72200000000000
   0.00000000000000    1.75231580201743   5.72200000000000
  -1.51755000000000   -0.87615790100872   5.72200000000000
    Cu      Fe      O
     1       1      2
Direct
   0.00000000000000    0.00000000000000    0.00000000000000   Cu   (1a)
   0.50000000000000    0.50000000000000    0.50000000000000   Fe   (1b)
   0.10660000000000    0.10660000000000    0.10660000000000    O   (2c)
  -0.10660000000000   -0.10660000000000   -0.10660000000000    O   (2c)
```

$\beta$-Potassium Nitrate ($KNO_3$): ABC6_hR8_166_a_b_h - CIF

```
# CIF file
data_findsym-output
_audit_creation_method FINDSYM

_chemical_name_mineral 'KNO3'
_chemical_formula_sum 'K N O6'

loop_
_publ_author_name
 'J. K. Nimmo'
 'B. W. Lucas'
_journal_year 1976
_publ_Section_title
;
 The crystal structures of $\gamma$-and $\beta$-KNO$_{3}$ and the $\
   ↪ alpha$-$\beta$-$\gamma$ phase transformations
;

# Found in The American Mineralogist Crystal Structure Database, 2003

_aflow_title '$\beta$-Potassium Nitrate (KNO$_{3}$) Structure'
_aflow_proto 'ABC6_hR8_166_a_b_h'
_aflow_params 'a,c/a,x_{3},z_{3}'
_aflow_params_values '5.425,1.8130875576,0.605,1.215'
_aflow_Strukturbericht 'None'
_aflow_Pearson 'hR8'

_symmetry_space_group_name_H-M "R -3 2/m (hexagonal axes)"
_symmetry_Int_Tables_number 166

_cell_length_a      5.42500
_cell_length_b      5.42500
_cell_length_c      9.83600
_cell_angle_alpha  90.00000
_cell_angle_beta   90.00000
_cell_angle_gamma  120.00000

loop_
_space_group_symop_id
_space_group_symop_operation_xyz
1 x,y,z
2 -y,x-y,z
3 -x+y,-x,z
4 y,x,-z
5 -x,-x+y,-z
6 x-y,-y,-z
7 -x,-y,-z
8 y,-x+y,-z
9 x-y,x,-z
10 -y,-x,z
11 x,x-y,z
12 -x+y,y,z
13 x+1/3,y+2/3,z+2/3
14 -y+1/3,x-y+2/3,z+2/3
15 -x+y+1/3,-x+2/3,z+2/3
16 y+1/3,x+2/3,-z+2/3
17 -x+1/3,-x+y+2/3,-z+2/3
18 x-y+1/3,-y+2/3,-z+2/3
19 -x+1/3,-y+2/3,-z+2/3
20 y+1/3,-x+y+2/3,-z+2/3
21 x-y+1/3,x+2/3,-z+2/3
22 -y+1/3,-x+2/3,z+2/3
23 x+1/3,x-y+2/3,z+2/3
24 -x+y+1/3,y+2/3,z+2/3
25 x+2/3,y+1/3,z+1/3
26 -y+2/3,x-y+1/3,z+1/3
27 -x+y+2/3,-x+1/3,z+1/3
28 y+2/3,x+1/3,-z+1/3
29 -x+2/3,-x+y+1/3,-z+1/3
30 x-y+2/3,-y+1/3,-z+1/3
31 -x+2/3,-y+1/3,-z+1/3
32 y+2/3,-x+y+1/3,-z+1/3
33 x-y+2/3,x+1/3,-z+1/3
34 -y+2/3,-x+1/3,z+1/3
35 x+2/3,x-y+1/3,z+1/3
36 -x+y+2/3,y+1/3,z+1/3

loop_
_atom_site_label
_atom_site_type_symbol
_atom_site_symmetry_multiplicity
_atom_site_Wyckoff_label
_atom_site_fract_x
_atom_site_fract_y
_atom_site_fract_z
_atom_site_occupancy
K1 K 3 a 0.00000 0.00000 0.00000 1.00000
N1 N 3 b 0.00000 0.00000 0.50000 1.00000
O1 O 18 h 0.13000 0.87000 0.47500 0.50000
```

$\beta$-Potassium Nitrate ($KNO_3$): ABC6_hR8_166_a_b_h - POSCAR

```
ABC6_hR8_166_a_b_h & a,c/a,x3,z3 --params=5.425,1.8130875576,0.605,1.215
   ↪ & R-3m D_{3d}^{5} #166 (abh) & hR8 & None & KNO3 & KNO3 & J.
   ↪ K. Nimmo and B. W. Lucas, (1976)
   1.00000000000000
   2.71250000000000   -1.56606260517686   3.27866666666667
   0.00000000000000    3.13212521035372   3.27866666666667
  -2.71250000000000   -1.56606260517686   3.27866666666667
     K       N       O
     1       1       6
Direct
   0.00000000000000    0.00000000000000    0.00000000000000   K   (1a)
   0.50000000000000    0.50000000000000    0.50000000000000   N   (1b)
   0.60500000000000    0.60500000000000    1.21500000000000   O   (6h)
   1.21500000000000    0.60500000000000    0.60500000000000   O   (6h)
   0.60500000000000    1.21500000000000    0.60500000000000   O   (6h)
  -1.21500000000000   -0.60500000000000   -0.60500000000000   O   (6h)
  -0.60500000000000   -0.60500000000000   -1.21500000000000   O   (6h)
```

```
   -0.60500000000000   -1.21500000000000   -0.60500000000000      O    (6h)
```

K(SH) ($B22$): AB_hR2_166_a_b - CIF

```
# CIF file
data_findsym-output
_audit_creation_method FINDSYM

_chemical_name_mineral 'K(SH)'
_chemical_formula_sum 'K S'

loop_
_publ_author_name
 'C. D. West'
_journal_name_full_name
;
 Zeitschrift f{\"u}r Kristallographie - Crystalline Materials
;
_journal_volume 88
_journal_year 1934
_journal_page_first 97
_journal_page_last 115
_publ_Section_title
;
 The Crystal Structures of Some Alkali Hydrosulfides and Monosulfides
;

# Found in Strukturbericht Band III 1933-1935, 1937

_aflow_title 'K(SH) ($B22$) Structure'
_aflow_proto 'AB_hR2_166_a_b'
_aflow_params 'a,c/a'
_aflow_params_values '9.88622,1.00556532224'
_aflow_Strukturbericht '$B22$'
_aflow_Pearson 'hR2'

_symmetry_space_group_name_H-M "R -3 2/m (hexagonal axes)"
_symmetry_Int_Tables_number 166

_cell_length_a      9.88622
_cell_length_b      9.88622
_cell_length_c      9.94124
_cell_angle_alpha 90.00000
_cell_angle_beta  90.00000
_cell_angle_gamma 120.00000

loop_
_space_group_symop_id
_space_group_symop_operation_xyz
1 x,y,z
2 -y,x-y,z
3 -x+y,-x,z
4 y,x,-z
5 -x,-x+y,-z
6 x-y,-y,-z
7 -x,-y,-z
8 y,-x+y,-z
9 x-y,x,-z
10 -y,-x,z
11 x,x-y,z
12 -x+y,y,z
13 x+1/3,y+2/3,z+2/3
14 -y+1/3,x-y+2/3,z+2/3
15 -x+y+1/3,-x+2/3,z+2/3
16 y+1/3,x+2/3,-z+2/3
17 -x+1/3,-x+y+2/3,-z+2/3
18 x-y+1/3,-y+2/3,-z+2/3
19 -x+1/3,-y+2/3,-z+2/3
20 y+1/3,-x+y+2/3,-z+2/3
21 x-y+1/3,x+2/3,-z+2/3
22 -y+1/3,-x+2/3,z+2/3
23 x+1/3,x-y+2/3,z+2/3
24 -x+y+1/3,y+2/3,z+2/3
25 x+2/3,y+1/3,z+1/3
26 -y+2/3,x-y+1/3,z+1/3
27 -x+y+2/3,-x+1/3,z+1/3
28 y+2/3,x+1/3,-z+1/3
29 -x+2/3,-x+y+1/3,-z+1/3
30 x-y+2/3,-y+1/3,-z+1/3
31 -x+2/3,-y+1/3,-z+1/3
32 y+2/3,-x+y+1/3,-z+1/3
33 x-y+2/3,x+1/3,-z+1/3
34 -y+2/3,-x+1/3,z+1/3
35 x+2/3,x-y+1/3,z+1/3
36 -x+y+2/3,y+1/3,z+1/3

loop_
_atom_site_label
_atom_site_type_symbol
_atom_site_symmetry_multiplicity
_atom_site_Wyckoff_label
_atom_site_fract_x
_atom_site_fract_y
_atom_site_fract_z
_atom_site_occupancy
K1 K 3 a 0.00000 0.00000 0.00000 1.00000
S1 S 3 b 0.00000 0.00000 0.50000 1.00000
```

K(SH) ($B22$): AB_hR2_166_a_b - POSCAR

```
AB_hR2_166_a_b & a,c/a --params=9.88622,1.00556532224 & R-3m D_{3d}^{5}
    ↪ #166 (ab) & hR2 & $B22$ & K(SH) & K(SH) & C. D. West,
    ↪ Zeitschrift f{"u}r Kristallographie - Crystalline Materials 88,
    ↪ 97-115 (1934)
   1.00000000000000
   4.94311000000000   -2.85390588913393    3.31374666666667
   0.00000000000000    5.70781177826786    3.31374666666667
  -4.94311000000000   -2.85390588913393    3.31374666666667
     K        S
     1        1
Direct
   0.00000000000000    0.00000000000000    0.00000000000000    K    (1a)
   0.50000000000000    0.50000000000000    0.50000000000000    S    (1b)
```

$Zr_{21}Re_{25}$: A25B21_hR92_167_b2e3f_e3f - CIF

```
# CIF file
data_findsym-output
_audit_creation_method FINDSYM

_chemical_name_mineral 'Re25Zr21'
_chemical_formula_sum 'Re25 Zr21'

loop_
_publ_author_name
 'K. Cenzual'
 'E. Parth\\'{e}'
 'R. M. Waterstrat'
_journal_name_full_name
;
 Acta Crystallographica Section C: Structural Chemistry
;
_journal_volume 42
_journal_year 1986
_journal_page_first 261
_journal_page_last 266
_publ_Section_title
;
 Zr$_{21}$Re$_{25}$, a new rhombohedral structure type containing 12~{\
    ↪ AA}-thick infinite MgZn$_{2}$(Laves)-type columns
;

# Found in The intermetallic compound Mg$_{21}$Zn$_{25}$, 2002

_aflow_title 'Zr$_{21}$Re$_{25}$ Structure'
_aflow_proto 'A25B21_hR92_167_b2e3f_e3f'
_aflow_params 'a,c/a,x_{2},x_{3},x_{4},x_{5},y_{5},z_{5},x_{6},y_{6},z_{
    ↪ 6},x_{7},y_{7},z_{7},x_{8},y_{8},z_{8},x_{9},y_{9},z_{9},x_{10
    ↪ },y_{10},z_{10}'
_aflow_params_values '25.847,0.339343057221,0.311,0.6946,0.9443,0.3062,
    ↪ 0.0702,0.3676,0.4819,0.0723,0.1889,0.235,-0.1163,-0.1187,0.0445
    ↪ ,-0.1882,0.2709,0.182,-0.0505,0.0662,0.561,0.1397,0.0436'
_aflow_Strukturbericht 'None'
_aflow_Pearson 'hR92'

_symmetry_space_group_name_H-M "R -3 2/c (hexagonal axes)"
_symmetry_Int_Tables_number 167

_cell_length_a      25.84700
_cell_length_b      25.84700
_cell_length_c      8.77100
_cell_angle_alpha 90.00000
_cell_angle_beta  90.00000
_cell_angle_gamma 120.00000

loop_
_space_group_symop_id
_space_group_symop_operation_xyz
1 x,y,z
2 -y,x-y,z
3 -x+y,-x,z
4 y,x,-z+1/2
5 -x,-x+y,-z+1/2
6 x-y,-y,-z+1/2
7 -x,-y,-z
8 y,-x+y,-z
9 x-y,x,-z
10 -y,-x,z+1/2
11 x,x-y,z+1/2
12 -x+y,y,z+1/2
13 x+1/3,y+2/3,z+2/3
14 -y+1/3,x-y+2/3,z+2/3
15 -x+y+1/3,-x+2/3,z+2/3
16 y+1/3,x+2/3,-z+1/6
17 -x+1/3,-x+y+2/3,-z+1/6
18 x-y+1/3,-y+2/3,-z+1/6
19 -x+1/3,-y+2/3,-z+2/3
20 y+1/3,-x+y+2/3,-z+2/3
21 x-y+1/3,x+2/3,-z+2/3
22 -y+1/3,-x+2/3,z+1/6
23 x+1/3,x-y+2/3,z+1/6
24 -x+y+1/3,y+2/3,z+1/6
25 x+2/3,y+1/3,z+1/3
26 -y+2/3,x-y+1/3,z+1/3
27 -x+y+2/3,-x+1/3,z+1/3
28 y+2/3,x+1/3,-z+5/6
29 -x+2/3,-x+y+1/3,-z+5/6
30 x-y+2/3,-y+1/3,-z+5/6
31 -x+2/3,-y+1/3,-z+1/3
32 y+2/3,-x+y+1/3,-z+1/3
33 x-y+2/3,x+1/3,-z+1/3
34 -y+2/3,-x+1/3,z+5/6
35 x+2/3,x-y+1/3,z+5/6
36 -x+y+2/3,y+1/3,z+5/6

loop_
_atom_site_label
_atom_site_type_symbol
_atom_site_symmetry_multiplicity
_atom_site_Wyckoff_label
```

```
_atom_site_fract_x
_atom_site_fract_y
_atom_site_fract_z
_atom_site_occupancy
Re1  Re  6 b  0.00000  0.00000  0.00000  1.00000
Re2  Re  18 e  0.06100  0.00000  0.25000  1.00000
Re3  Re  18 e  0.44460  0.00000  0.25000  1.00000
Zr1  Zr  18 e  0.69430  0.00000  0.25000  1.00000
Re4  Re  36 f  0.05820  0.17780  0.24800  1.00000
Re5  Re  36 f  0.23420  0.17540  0.24770  1.00000
Re6  Re  36 f  0.23500  0.11630  0.00000  1.00000
Zr2  Zr  36 f  0.00210  0.23060  0.04240  1.00000
Zr3  Zr  36 f  0.11610  0.11640  0.06590  1.00000
Zr4  Zr  36 f  0.31290  0.10840  0.24810  1.00000
```

$Zr_{21}Re_{25}$: A25B21_hR92_167_b2e3f_e3f - POSCAR

```
A25B21_hR92_167_b2e3f_e3f & a,c/a,x2,x3,x4,x5,y5,z5,x6,y6,z6,x7,y7,z7,x8
    ↪ ,y8,z8,x9,y9,z9,x10,y10,z10 --params=25.847,0.339343057221,
    ↪ 0.311,0.6946,0.9443,0.3062,0.0702,0.3676,0.4819,0.0723,0.1889,
    ↪ 0.235,-0.1163,-0.1187,0.0445,-0.1882,0.2709,0.182,-0.0505,
    ↪ 0.0662,0.561,0.1397,0.0436 & R-3c D_{3d}^{6} #167 (be^3f^6) &
    ↪ hR92 & None & Re25Zr21 & Re25Zr21 & K. Cenzual and E. Parth\'{e
    ↪ } and R. M. Waterstrat, Acta Crystallogr. C 42, 261-266 (1986)
   1.00000000000000
  12.92350000000000  -7.46138620387213   2.92366666666667
   0.00000000000000  14.92277240774430   2.92366666666667
 -12.92350000000000  -7.46138620387213   2.92366666666667
    Re       Zr
    50       42
Direct
   0.00000000000000   0.00000000000000   0.00000000000000   Re   (2b)
   0.50000000000000   0.50000000000000   0.50000000000000   Re   (2b)
   0.31100000000000   0.18900000000000   0.25000000000000   Re   (6e)
   0.25000000000000   0.31100000000000   0.18900000000000   Re   (6e)
   0.18900000000000   0.25000000000000   0.31100000000000   Re   (6e)
  -0.31100000000000   0.81100000000000   0.75000000000000   Re   (6e)
   0.75000000000000  -0.31100000000000   0.81100000000000   Re   (6e)
   0.81100000000000   0.75000000000000  -0.31100000000000   Re   (6e)
   0.69460000000000  -0.19460000000000   0.25000000000000   Re   (6e)
   0.25000000000000   0.69460000000000  -0.19460000000000   Re   (6e)
  -0.19460000000000   0.25000000000000   0.69460000000000   Re   (6e)
  -0.69460000000000   1.19460000000000   0.75000000000000   Re   (6e)
   0.75000000000000  -0.69460000000000   1.19460000000000   Re   (6e)
   1.19460000000000   0.75000000000000  -0.69460000000000   Re   (6e)
   0.30620000000000   0.07020000000000   0.36760000000000   Re  (12f)
   0.36760000000000   0.30620000000000   0.07020000000000   Re  (12f)
   0.07020000000000   0.36760000000000   0.30620000000000   Re  (12f)
   0.13240000000000   0.42980000000000   0.19380000000000   Re  (12f)
   0.42980000000000   0.19380000000000   0.13240000000000   Re  (12f)
   0.19380000000000   0.13240000000000   0.42980000000000   Re  (12f)
  -0.30620000000000  -0.07020000000000  -0.36760000000000   Re  (12f)
  -0.36760000000000  -0.30620000000000  -0.07020000000000   Re  (12f)
  -0.07020000000000  -0.36760000000000  -0.30620000000000   Re  (12f)
   0.86760000000000   0.57020000000000   0.80620000000000   Re  (12f)
   0.57020000000000   0.80620000000000   0.86760000000000   Re  (12f)
   0.80620000000000   0.86760000000000   0.57020000000000   Re  (12f)
   0.48190000000000   0.07230000000000   0.18890000000000   Re  (12f)
   0.18890000000000   0.48190000000000   0.07230000000000   Re  (12f)
   0.07230000000000   0.18890000000000   0.48190000000000   Re  (12f)
   0.31110000000000   0.42770000000000   0.01810000000000   Re  (12f)
   0.42770000000000   0.01810000000000   0.31110000000000   Re  (12f)
   0.01810000000000   0.31110000000000   0.42770000000000   Re  (12f)
  -0.48190000000000  -0.07230000000000  -0.18890000000000   Re  (12f)
  -0.18890000000000  -0.48190000000000  -0.07230000000000   Re  (12f)
  -0.07230000000000  -0.18890000000000  -0.48190000000000   Re  (12f)
   0.68890000000000   0.57230000000000   0.98190000000000   Re  (12f)
   0.57230000000000   0.98190000000000   0.68890000000000   Re  (12f)
   0.98190000000000   0.68890000000000   0.57230000000000   Re  (12f)
   0.23500000000000  -0.11630000000000  -0.11870000000000   Re  (12f)
  -0.11870000000000   0.23500000000000  -0.11630000000000   Re  (12f)
  -0.11630000000000  -0.11870000000000   0.23500000000000   Re  (12f)
   0.61870000000000   0.61630000000000   0.26500000000000   Re  (12f)
   0.61630000000000   0.26500000000000   0.61870000000000   Re  (12f)
   0.26500000000000   0.61870000000000   0.61630000000000   Re  (12f)
  -0.23500000000000   0.11630000000000   0.11870000000000   Re  (12f)
   0.11870000000000  -0.23500000000000   0.11630000000000   Re  (12f)
   0.11630000000000   0.11870000000000  -0.23500000000000   Re  (12f)
   0.38130000000000   0.38370000000000   0.73500000000000   Re  (12f)
   0.38370000000000   0.73500000000000   0.38130000000000   Re  (12f)
   0.73500000000000   0.38130000000000   0.38370000000000   Re  (12f)
   0.94430000000000  -0.44430000000000   0.25000000000000   Zr   (6e)
   0.25000000000000   0.94430000000000  -0.44430000000000   Zr   (6e)
  -0.44430000000000   0.25000000000000   0.94430000000000   Zr   (6e)
  -0.94430000000000   1.44430000000000   0.75000000000000   Zr   (6e)
   0.75000000000000  -0.94430000000000   1.44430000000000   Zr   (6e)
   1.44430000000000   0.75000000000000  -0.94430000000000   Zr   (6e)
   0.04450000000000  -0.18820000000000   0.27090000000000   Zr  (12f)
   0.27090000000000   0.04450000000000  -0.18820000000000   Zr  (12f)
  -0.18820000000000   0.27090000000000   0.04450000000000   Zr  (12f)
   0.22910000000000   0.68820000000000   0.45550000000000   Zr  (12f)
   0.68820000000000   0.45550000000000   0.22910000000000   Zr  (12f)
   0.45550000000000   0.22910000000000   0.68820000000000   Zr  (12f)
  -0.04450000000000   0.18820000000000  -0.27090000000000   Zr  (12f)
  -0.27090000000000  -0.04450000000000   0.18820000000000   Zr  (12f)
   0.18820000000000  -0.27090000000000  -0.04450000000000   Zr  (12f)
   0.77090000000000   0.31180000000000   0.54450000000000   Zr  (12f)
   0.31180000000000   0.54450000000000   0.77090000000000   Zr  (12f)
   0.54450000000000   0.77090000000000   0.31180000000000   Zr  (12f)
   0.18200000000000  -0.05050000000000   0.06620000000000   Zr  (12f)
   0.06620000000000   0.18200000000000  -0.05050000000000   Zr  (12f)
  -0.05050000000000   0.06620000000000   0.18200000000000   Zr  (12f)
   0.43380000000000   0.55050000000000   0.31800000000000   Zr  (12f)
   0.55050000000000   0.31800000000000   0.43380000000000   Zr  (12f)
   0.31800000000000   0.43380000000000   0.55050000000000   Zr  (12f)
  -0.18200000000000   0.05050000000000  -0.06620000000000   Zr  (12f)
  -0.06620000000000  -0.18200000000000   0.05050000000000   Zr  (12f)
   0.05050000000000  -0.06620000000000  -0.18200000000000   Zr  (12f)
   0.56620000000000   0.44950000000000   0.68200000000000   Zr  (12f)
   0.44950000000000   0.68200000000000   0.56620000000000   Zr  (12f)
   0.68200000000000   0.56620000000000   0.44950000000000   Zr  (12f)
   0.56100000000000   0.13970000000000   0.04360000000000   Zr  (12f)
   0.04360000000000   0.56100000000000   0.13970000000000   Zr  (12f)
   0.13970000000000   0.04360000000000   0.56100000000000   Zr  (12f)
   0.45640000000000   0.36030000000000  -0.06100000000000   Zr  (12f)
   0.36030000000000  -0.06100000000000   0.45640000000000   Zr  (12f)
  -0.06100000000000   0.45640000000000   0.36030000000000   Zr  (12f)
  -0.56100000000000  -0.13970000000000  -0.04360000000000   Zr  (12f)
  -0.04360000000000  -0.56100000000000  -0.13970000000000   Zr  (12f)
  -0.13970000000000  -0.04360000000000  -0.56100000000000   Zr  (12f)
   0.54360000000000   0.63970000000000   1.06100000000000   Zr  (12f)
   0.63970000000000   1.06100000000000   0.54360000000000   Zr  (12f)
   1.06100000000000   0.54360000000000   0.63970000000000   Zr  (12f)
```

$\beta$-$BaB_2O_4$ (High-Temperature): A2BC4_hR42_167_f_ac_2f - CIF

```
# CIF file
data_findsym-output
_audit_creation_method FINDSYM

_chemical_name_mineral 'B2BaO4'
_chemical_formula_sum 'B2 Ba O4'

loop_
_publ_author_name
 'A. D. Mighell'
 'A. Perloff'
 'S. Block'
_journal_name_full_name
;
 Acta Cristallographica
;
_journal_volume 20
_journal_year 1966
_journal_page_first 819
_journal_page_last 823
_publ_Section_title
;
 The crystal structure of the high temperature form of barium borate,
       ↪ BaO$\cdot$B$_{2}$O$_{3}$
;

# Found in Crystal Structure of the low-temperature form of BaB$_{2}$O$_
      ↪ {4}$, 1984

_aflow_title '$\beta$-BaB$_{2}$O$_{4}$ (High-Temperature) Structure'
_aflow_proto 'A2BC4_hR42_167_f_ac_2f'
_aflow_params 'a,c/a,x_{2},x_{3},y_{3},z_{3},x_{4},y_{4},z_{4},x_{5},y_{
      ↪ 5},z_{5}'
_aflow_params_values '7.235,5.41700069109,0.84983,0.35996,-0.25104,
      ↪ 0.50497,0.335,-0.429,0.71401,0.4775,0.1905,-0.05849'
_aflow_Strukturbericht 'None'
_aflow_Pearson 'hR42'

_symmetry_space_group_name_H-M "R -3 2/c (hexagonal axes)"
_symmetry_Int_Tables_number 167

_cell_length_a     7.23500
_cell_length_b     7.23500
_cell_length_c     39.19200
_cell_angle_alpha  90.00000
_cell_angle_beta   90.00000
_cell_angle_gamma  120.00000

loop_
_space_group_symop_id
_space_group_symop_operation_xyz
1 x,y,z
2 -y,x-y,z
3 -x+y,-x,z
4 y,x,-z+1/2
5 -x,-x+y,-z+1/2
6 x-y,-y,-z+1/2
7 -x,-y,-z
8 y,-x+y,-z
9 x-y,x,-z
10 -y,-x,z+1/2
11 x,x-y,z+1/2
12 -x+y,y,z+1/2
13 x+1/3,y+2/3,z+2/3
14 -y+1/3,x-y+2/3,z+2/3
15 -x+y+1/3,-x+2/3,z+2/3
16 y+1/3,x+2/3,-z+1/6
17 -x+1/3,-x+y+2/3,-z+1/6
18 x-y+1/3,-y+2/3,-z+1/6
19 -x+1/3,-y+2/3,-z+2/3
20 y+1/3,-x+y+2/3,-z+2/3
21 x-y+1/3,x+2/3,-z+2/3
22 -y+1/3,-x+2/3,z+1/6
23 x+1/3,x-y+2/3,z+1/6
24 -x+y+1/3,y+2/3,z+1/6
25 x+2/3,y+1/3,z+1/3
26 -y+2/3,x-y+1/3,z+1/3
27 -x+y+2/3,-x+1/3,z+1/3
28 y+2/3,x+1/3,-z+5/6
29 -x+2/3,-x+y+1/3,-z+5/6
30 x-y+2/3,-y+1/3,-z+5/6
31 -x+2/3,-y+1/3,-z+1/3
32 y+2/3,-x+y+1/3,-z+1/3
33 x-y+2/3,x+1/3,-z+1/3
```

```
34 -y+2/3,-x+1/3,z+5/6
35 x+2/3,x-y+1/3,z+5/6
36 -x+y+2/3,y+1/3,z+5/6

loop_
_atom_site_label
_atom_site_type_symbol
_atom_site_symmetry_multiplicity
_atom_site_Wyckoff_label
_atom_site_fract_x
_atom_site_fract_y
_atom_site_fract_z
_atom_site_occupancy
Ba1 Ba 6 a 0.00000 0.00000 0.25000 1.00000
Ba2 Ba 12 c 0.00000 0.00000 0.84983 1.00000
B1 B 36 f 0.15533 0.45567 0.20463 1.00000
O1 O 36 f 0.12833 0.63567 0.20667 1.00000
O2 O 36 f 0.27433 0.01267 0.20317 1.00000
```

$\beta$-$BaB_2O_4$ (High-Temperature): A2BC4_hR42_167_f_ac_2f - POSCAR

```
A2BC4_hR42_167_f_ac_2f & a,c/a,x2,x3,y3,z3,x4,y4,z4,x5,y5,z5 --params=
    ↪ 7.235,5.41700069109,0.84983,0.35996,-0.25104,0.50497,0.335,-
    ↪ 0.429,0.71401,0.4775,0.1905,-0.05849 & R-3c D_{3d}^{6} #167 (
    ↪ acf^3) & hR42 & None & B2BaO4 & B2BaO4 & A. D. Mighell and A.
    ↪ Perloff and S. Block, Acta Cryst. 20, 819-823 (1966)
   1.00000000000000
   3.61750000000000  -2.08856459879347  13.06400000000000
   0.00000000000000   4.17712919758694  13.06400000000000
  -3.61750000000000  -2.08856459879347  13.06400000000000
     B      Ba       O
    12       6      24
Direct
   0.35996000000000  -0.25104000000000   0.50497000000000   B  (12f)
   0.50497000000000   0.35996000000000  -0.25104000000000   B  (12f)
  -0.25104000000000   0.50497000000000   0.35996000000000   B  (12f)
  -0.00497000000000   0.75104000000000   0.14004000000000   B  (12f)
   0.75104000000000   0.14004000000000  -0.00497000000000   B  (12f)
   0.14004000000000  -0.00497000000000   0.75104000000000   B  (12f)
  -0.35996000000000   0.25104000000000  -0.50497000000000   B  (12f)
  -0.50497000000000  -0.35996000000000   0.25104000000000   B  (12f)
   0.25104000000000  -0.50497000000000  -0.35996000000000   B  (12f)
   1.00497000000000   0.24896000000000   0.85996000000000   B  (12f)
   0.24896000000000   0.85996000000000   1.00497000000000   B  (12f)
   0.85996000000000   1.00497000000000   0.24896000000000   B  (12f)
   0.25000000000000   0.25000000000000   0.25000000000000   Ba  (2a)
   0.75000000000000   0.75000000000000   0.75000000000000   Ba  (2a)
   0.84983000000000   0.84983000000000   0.84983000000000   Ba  (4c)
  -0.34983000000000  -0.34983000000000  -0.34983000000000   Ba  (4c)
  -0.84983000000000  -0.84983000000000  -0.84983000000000   Ba  (4c)
   1.34983000000000   1.34983000000000   1.34983000000000   Ba  (4c)
   0.33500000000000  -0.42900000000000   0.71401000000000   O  (12f)
   0.71401000000000   0.33500000000000  -0.42900000000000   O  (12f)
  -0.42900000000000   0.71401000000000   0.33500000000000   O  (12f)
  -0.21401000000000   0.92900000000000   0.16500000000000   O  (12f)
   0.92900000000000   0.16500000000000  -0.21401000000000   O  (12f)
   0.16500000000000  -0.21401000000000   0.92900000000000   O  (12f)
  -0.33500000000000   0.42900000000000  -0.71401000000000   O  (12f)
  -0.71401000000000  -0.33500000000000   0.42900000000000   O  (12f)
   0.42900000000000  -0.71401000000000  -0.33500000000000   O  (12f)
   1.21401000000000   0.07100000000000   0.83500000000000   O  (12f)
   0.07100000000000   0.83500000000000   1.21401000000000   O  (12f)
   0.83500000000000   1.21401000000000   0.07100000000000   O  (12f)
   0.47750000000000   0.19050000000000  -0.05849000000000   O  (12f)
  -0.05849000000000   0.47750000000000   0.19050000000000   O  (12f)
   0.19050000000000  -0.05849000000000   0.47750000000000   O  (12f)
   0.55849000000000   0.30950000000000   0.02250000000000   O  (12f)
   0.30950000000000   0.02250000000000   0.55849000000000   O  (12f)
   0.02250000000000   0.55849000000000   0.30950000000000   O  (12f)
  -0.47750000000000  -0.19050000000000   0.05849000000000   O  (12f)
   0.05849000000000  -0.47750000000000  -0.19050000000000   O  (12f)
  -0.19050000000000   0.05849000000000  -0.47750000000000   O  (12f)
   0.44151000000000   0.69050000000000   0.97750000000000   O  (12f)
   0.69050000000000   0.97750000000000   0.44151000000000   O  (12f)
   0.97750000000000   0.44151000000000   0.69050000000000   O  (12f)
```

$CrCl_3(H_2O)_6$ ($J2_2$): A3BC6_hR20_167_e_b_f - CIF

```
# CIF file
data_findsym-output
_audit_creation_method FINDSYM

_chemical_name_mineral 'Cl3Cr(H2O)6'
_chemical_formula_sum 'Cl3 Cr (H2O)6'

loop_
_publ_author_name
 'K. R. Andress'
 'C. Carpenter'
_journal_name_full_name
;
 Zeitschrift f{\"u}r Kristallographie - Crystalline Materials
;
_journal_volume 87
_journal_year 1934
_journal_page_first 446
_journal_page_last 463
_publ_Section_title
;
 Die Struktur von Chromchlorid- und Aluminiumchloridhexahydrat
;

# Found in Strukturbericht Band III 1933-1935, 1937

_aflow_title 'CrCl$_{3}$(H$_{2}$O)$_{6}$ ($J2_{2}$) Structure'
_aflow_proto 'A3BC6_hR20_167_e_b_f'
_aflow_params 'a,c/a,x_{2},x_{3},y_{3},z_{3}'
_aflow_params_values '11.8335,1.00556555541,0.51,0.22,-0.04,0.12'
_aflow_Strukturbericht '$J2_{2}$'
_aflow_Pearson 'hR20'

_symmetry_space_group_name_H-M "R -3 2/c (hexagonal axes)"
_symmetry_Int_Tables_number 167

_cell_length_a     11.83350
_cell_length_b     11.83350
_cell_length_c     11.89936
_cell_angle_alpha  90.00000
_cell_angle_beta   90.00000
_cell_angle_gamma  120.00000

loop_
_space_group_symop_id
_space_group_symop_operation_xyz
1 x,y,z
2 -y,x-y,z
3 -x+y,-x,z
4 y,x,-z+1/2
5 -x,-x+y,-z+1/2
6 x-y,-y,-z+1/2
7 -x,-y,-z
8 y,-x+y,-z
9 x-y,x,-z
10 -y,-x,z+1/2
11 x,x-y,z+1/2
12 -x+y,y,z+1/2
13 x+1/3,y+2/3,z+2/3
14 -y+1/3,x-y+2/3,z+2/3
15 -x+y+1/3,-x+2/3,z+2/3
16 y+1/3,x+2/3,-z+1/6
17 -x+1/3,-x+y+2/3,-z+1/6
18 x-y+1/3,-y+2/3,-z+1/6
19 -x+1/3,-y+2/3,-z+2/3
20 y+1/3,-x+y+2/3,-z+2/3
21 x-y+1/3,x+2/3,-z+2/3
22 -y+1/3,-x+2/3,z+1/6
23 x+1/3,x-y+2/3,z+1/6
24 -x+y+1/3,y+2/3,z+1/6
25 x+2/3,y+1/3,z+1/3
26 -y+2/3,x-y+1/3,z+1/3
27 -x+y+2/3,-x+1/3,z+1/3
28 y+2/3,x+1/3,-z+5/6
29 -x+2/3,-x+y+1/3,-z+5/6
30 x-y+2/3,-y+1/3,-z+5/6
31 -x+2/3,-y+1/3,-z+1/3
32 y+2/3,-x+y+1/3,-z+1/3
33 x-y+2/3,x+1/3,-z+1/3
34 -y+2/3,-x+1/3,z+5/6
35 x+2/3,x-y+1/3,z+5/6
36 -x+y+2/3,y+1/3,z+5/6

loop_
_atom_site_label
_atom_site_type_symbol
_atom_site_symmetry_multiplicity
_atom_site_Wyckoff_label
_atom_site_fract_x
_atom_site_fract_y
_atom_site_fract_z
_atom_site_occupancy
Cr1 Cr 6 b 0.00000 0.00000 0.00000 1.00000
Cl1 Cl 18 e 0.26000 0.00000 0.25000 1.00000
H2O1 H2O 36 f 0.12000 0.14000 0.10000 1.00000
```

$CrCl_3(H_2O)_6$ ($J2_2$): A3BC6_hR20_167_e_b_f - POSCAR

```
A3BC6_hR20_167_e_b_f & a,c/a,x2,x3,y3,z3 --params=11.8335,1.00556555541,
    ↪ 0.51,0.22,-0.04,0.12 & R-3c D_{3d}^{6} #167 (bef) & hR20 & $J2_
    ↪ {2}$ & Cl3Cr(H2O)6 & Cl3Cr(H2O)6 & K. R. Andress and C.
    ↪ Carpenter, Zeitschrift f{"u}r Kristallographie - Crystalline
    ↪ Materials 87, 446-463 (1934)
   1.00000000000000
   5.91675000000000  -3.41603720522772   3.96645333333333
   0.00000000000000   6.83207441045544   3.96645333333333
  -5.91675000000000  -3.41603720522772   3.96645333333333
    Cl      Cr     H2O
     6       2      12
Direct
   0.51000000000000  -0.01000000000000   0.25000000000000   Cl   (6e)
   0.25000000000000   0.51000000000000  -0.01000000000000   Cl   (6e)
  -0.01000000000000   0.25000000000000   0.51000000000000   Cl   (6e)
  -0.51000000000000   1.01000000000000   0.75000000000000   Cl   (6e)
   0.75000000000000  -0.51000000000000   1.01000000000000   Cl   (6e)
   1.01000000000000   0.75000000000000  -0.51000000000000   Cl   (6e)
   0.00000000000000   0.00000000000000   0.00000000000000   Cr   (2b)
   0.50000000000000   0.50000000000000   0.50000000000000   Cr   (2b)
   0.22000000000000  -0.04000000000000   0.12000000000000   H2O  (12f)
   0.12000000000000   0.22000000000000  -0.04000000000000   H2O  (12f)
  -0.04000000000000   0.12000000000000   0.22000000000000   H2O  (12f)
   0.38000000000000   0.54000000000000   0.28000000000000   H2O  (12f)
   0.54000000000000   0.28000000000000   0.38000000000000   H2O  (12f)
   0.28000000000000   0.38000000000000   0.54000000000000   H2O  (12f)
  -0.22000000000000   0.04000000000000  -0.12000000000000   H2O  (12f)
  -0.12000000000000  -0.22000000000000   0.04000000000000   H2O  (12f)
   0.04000000000000  -0.12000000000000  -0.22000000000000   H2O  (12f)
   0.62000000000000   0.46000000000000   0.72000000000000   H2O  (12f)
   0.46000000000000   0.72000000000000   0.62000000000000   H2O  (12f)
   0.72000000000000   0.62000000000000   0.46000000000000   H2O  (12f)
```

$FeF_3$ ($D0_{12}$): A3B_hR8_167_e_b - CIF

```
# CIF file
data_findsym-output
_audit_creation_method FINDSYM

_chemical_name_mineral 'F3Fe'
_chemical_formula_sum 'F3 Fe'

loop_
_publ_author_name
 'M. A. Hepworth'
 'K. H. Jack'
 'R. D. Peacock'
 'G. J. Westland'
_journal_name_full_name
;
 Acta Cristallographica
;
_journal_volume 10
_journal_year 1957
_journal_page_first 63
_journal_page_last 69
_publ_Section_title
;
 The crystal structures of the trifluorides of iron, cobalt, ruthenium,
     ↪ rhodium, palladium and iridium
;

_aflow_title 'FeF$_{3}$ ($D0_{12}$) Structure'
_aflow_proto 'A3B_hR8_167_e_b'
_aflow_params 'a,c/a,x_{2}'
_aflow_params_values '5.198,2.56464024625,0.664'
_aflow_Strukturbericht '$D0_{12}$'
_aflow_Pearson 'hR8'

_symmetry_space_group_name_H-M "R -3 2/c (hexagonal axes)"
_symmetry_Int_Tables_number 167

_cell_length_a     5.19800
_cell_length_b     5.19800
_cell_length_c     13.33100
_cell_angle_alpha  90.00000
_cell_angle_beta   90.00000
_cell_angle_gamma  120.00000

loop_
_space_group_symop_id
_space_group_symop_operation_xyz
1 x,y,z
2 -y,x-y,z
3 -x+y,-x,z
4 y,x,-z+1/2
5 -x,-x+y,-z+1/2
6 x-y,-y,-z+1/2
7 -x,-y,-z
8 y,-x+y,-z
9 x-y,x,-z
10 -y,-x,z+1/2
11 x,x-y,z+1/2
12 -x+y,y,z+1/2
13 x+1/3,y+2/3,z+2/3
14 -y+1/3,x-y+2/3,z+2/3
15 -x+y+1/3,-x+2/3,z+2/3
16 y+1/3,x+2/3,-z+1/6
17 -x+1/3,-x+y+2/3,-z+1/6
18 x-y+1/3,-y+2/3,-z+1/6
19 -x+1/3,-y+2/3,-z+2/3
20 y+1/3,-x+y+2/3,-z+2/3
21 x-y+1/3,x+2/3,-z+2/3
22 -y+1/3,-x+2/3,z+1/6
23 x+1/3,x-y+2/3,z+1/6
24 -x+y+1/3,y+2/3,z+1/6
25 x+2/3,y+1/3,z+1/3
26 -y+2/3,x-y+1/3,z+1/3
27 -x+y+2/3,-x+1/3,z+1/3
28 y+2/3,x+1/3,-z+5/6
29 -x+2/3,-x+y+1/3,-z+5/6
30 x-y+2/3,-y+1/3,-z+5/6
31 -x+2/3,-y+1/3,-z+1/3
32 y+2/3,-x+y+1/3,-z+1/3
33 x-y+2/3,x+1/3,-z+1/3
34 -y+2/3,-x+1/3,z+5/6
35 x+2/3,x-y+1/3,z+5/6
36 -x+y+2/3,y+1/3,z+5/6

loop_
_atom_site_label
_atom_site_type_symbol
_atom_site_symmetry_multiplicity
_atom_site_Wyckoff_label
_atom_site_fract_x
_atom_site_fract_y
_atom_site_fract_z
_atom_site_occupancy
Fe1 Fe 6 b 0.00000 0.00000 0.00000 1.00000
F1 F 18 e 0.41400 0.00000 0.25000 1.00000
```

FeF$_3$ ($D0_{12}$): A3B_hR8_167_e_b - POSCAR

```
A3B_hR8_167_e_b & a,c/a,x2 --params=5.198,2.56464024625,0.664 & R-3c D_{
    ↪ 3d}^{6} #167 (be) & hR8 & $D0_{12}$ & F3Fe & F3Fe & M. A.
    ↪ Hepworth et al., Acta Cryst. 10, 63-69 (1957)
   1.00000000000000
   2.59900000000000  -1.50053334962384   4.44366666666667
   0.00000000000000   3.00106669924768   4.44366666666667
  -2.59900000000000  -1.50053334962384   4.44366666666667
      F      Fe
      6       2
Direct
   0.66400000000000  -0.16400000000000   0.25000000000000   F   (6e)
   0.25000000000000   0.66400000000000  -0.16400000000000   F   (6e)
  -0.16400000000000   0.25000000000000   0.66400000000000   F   (6e)
  -0.66400000000000   1.16400000000000   0.75000000000000   F   (6e)
   0.75000000000000  -0.66400000000000   1.16400000000000   F   (6e)
   1.16400000000000   0.75000000000000  -0.66400000000000   F   (6e)
   0.00000000000000   0.00000000000000   0.00000000000000   Fe  (2b)
   0.50000000000000   0.50000000000000   0.50000000000000   Fe  (2b)
```

Rinneite (K$_3$NaFeCl$_6$): A6BC3D_hR22_167_f_b_e_a - CIF

```
# CIF file
data_findsym-output
_audit_creation_method FINDSYM

_chemical_name_mineral 'Rinneite'
_chemical_formula_sum 'Cl6 Fe K3 Na'

loop_
_publ_author_name
 'B. N. Figgis'
 'A. N. Sobolev'
 'E. S. Kucharski'
 'V. Broughton'
_journal_name_full_name
;
 Acta Crystallographica Section C: Structural Chemistry
;
_journal_volume 56
_journal_year 2000
_journal_page_first e228
_journal_page_last e229
_publ_Section_title
;
 Rinneite, K$_{3}$Na[FeCl$_{6}$], at 293, 84 and 9.5~K
;

_aflow_title 'Rinneite (K$_{3}$NaFeCl$_{6}$) Structure'
_aflow_proto 'A6BC3D_hR22_167_f_b_e_a'
_aflow_params 'a,c/a,x_{3},x_{4},y_{4},z_{4}'
_aflow_params_values '12.033,1.15208177512,0.87389,0.78748,-0.41951,
    ↪ 0.45463'
_aflow_Strukturbericht 'None'
_aflow_Pearson 'hR22'

_symmetry_space_group_name_H-M "R -3 2/c (hexagonal axes)"
_symmetry_Int_Tables_number 167

_cell_length_a     12.03300
_cell_length_b     12.03300
_cell_length_c     13.86300
_cell_angle_alpha  90.00000
_cell_angle_beta   90.00000
_cell_angle_gamma  120.00000

loop_
_space_group_symop_id
_space_group_symop_operation_xyz
1 x,y,z
2 -y,x-y,z
3 -x+y,-x,z
4 y,x,-z+1/2
5 -x,-x+y,-z+1/2
6 x-y,-y,-z+1/2
7 -x,-y,-z
8 y,-x+y,-z
9 x-y,x,-z
10 -y,-x,z+1/2
11 x,x-y,z+1/2
12 -x+y,y,z+1/2
13 x+1/3,y+2/3,z+2/3
14 -y+1/3,x-y+2/3,z+2/3
15 -x+y+1/3,-x+2/3,z+2/3
16 y+1/3,x+2/3,-z+1/6
17 -x+1/3,-x+y+2/3,-z+1/6
18 x-y+1/3,-y+2/3,-z+1/6
19 -x+1/3,-y+2/3,-z+2/3
20 y+1/3,-x+y+2/3,-z+2/3
21 x-y+1/3,x+2/3,-z+2/3
22 -y+1/3,-x+2/3,z+1/6
23 x+1/3,x-y+2/3,z+1/6
24 -x+y+1/3,y+2/3,z+1/6
25 x+2/3,y+1/3,z+1/3
26 -y+2/3,x-y+1/3,z+1/3
27 -x+y+2/3,-x+1/3,z+1/3
28 y+2/3,x+1/3,-z+5/6
29 -x+2/3,-x+y+1/3,-z+5/6
30 x-y+2/3,-y+1/3,-z+5/6
31 -x+2/3,-y+1/3,-z+1/3
32 y+2/3,-x+y+1/3,-z+1/3
33 x-y+2/3,x+1/3,-z+1/3
34 -y+2/3,-x+1/3,z+5/6
35 x+2/3,x-y+1/3,z+5/6
36 -x+y+2/3,y+1/3,z+5/6

loop_
_atom_site_label
_atom_site_type_symbol
_atom_site_symmetry_multiplicity
_atom_site_Wyckoff_label
_atom_site_fract_x
_atom_site_fract_y
```

```
_atom_site_fract_z
_atom_site_occupancy
Na1 Na 6 a 0.00000 0.00000 0.25000 1.00000
Fe1 Fe 6 b 0.00000 0.00000 0.00000 1.00000
K1 K 18 e 0.62389 0.00000 0.25000 1.00000
Cl1 Cl 36 f 0.51328 0.69371 0.27420 1.00000
```

Rinneite ($K_3NaFeCl_6$): A6BC3D_hR22_167_f_b_e_a - POSCAR

```
A6BC3D_hR22_167_f_b_e_a & a,c/a,x3,x4,y4,z4 --params=12.033,
    ↪ 1.15208177512,0.87389,0.78748,-0.41951,0.45463 & R-3c D_{3d}^{6
    ↪ } #167 (abef) & hR22 & None & Cl6FeK3Na & Rinneite & B. N.
    ↪ Figgis et al., Acta Crystallogr. C 56, e228-e229 (2000)
   1.00000000000000
   6.01650000000000  -3.47362789457938   4.62100000000000
   0.00000000000000   6.94725578915877   4.62100000000000
  -6.01650000000000  -3.47362789457938   4.62100000000000
    Cl    Fe     K    Na
    12     2     6     2
Direct
   0.78748000000000  -0.41951000000000   0.45463000000000   Cl   (12f)
   0.45463000000000   0.78748000000000  -0.41951000000000   Cl   (12f)
  -0.41951000000000   0.45463000000000   0.78748000000000   Cl   (12f)
   0.04537000000000   0.91951000000000  -0.28748000000000   Cl   (12f)
   0.91951000000000  -0.28748000000000   0.04537000000000   Cl   (12f)
  -0.28748000000000   0.04537000000000   0.91951000000000   Cl   (12f)
  -0.78748000000000   0.41951000000000  -0.45463000000000   Cl   (12f)
  -0.45463000000000  -0.78748000000000   0.41951000000000   Cl   (12f)
   0.41951000000000  -0.45463000000000  -0.78748000000000   Cl   (12f)
   0.95463000000000   0.08049000000000   1.28748000000000   Cl   (12f)
   0.08049000000000   1.28748000000000   0.95463000000000   Cl   (12f)
   1.28748000000000   0.95463000000000   0.08049000000000   Cl   (12f)
   0.00000000000000   0.00000000000000   0.00000000000000   Fe   (2b)
   0.50000000000000   0.50000000000000   0.50000000000000   Fe   (2b)
   0.87389000000000  -0.37389000000000   0.25000000000000   K    (6e)
   0.25000000000000   0.87389000000000  -0.37389000000000   K    (6e)
  -0.37389000000000   0.25000000000000   0.87389000000000   K    (6e)
  -0.87389000000000   1.37389000000000   0.75000000000000   K    (6e)
   0.75000000000000  -0.87389000000000   1.37389000000000   K    (6e)
   1.37389000000000   0.75000000000000  -0.87389000000000   K    (6e)
   0.25000000000000   0.25000000000000   0.25000000000000   Na   (2a)
   0.75000000000000   0.75000000000000   0.75000000000000   Na   (2a)
```

$Cs_3Tl_2Cl_9$ ($K7_2$): A9B3C2_hR28_167_ef_e_c - CIF

```
# CIF file
data_findsym-output
_audit_creation_method FINDSYM

_chemical_name_mineral 'Cl9Cs3Tl2'
_chemical_formula_sum 'Cl9 Cs3 Tl2'

loop_
_publ_author_name
 'J. L. Hoard'
 'L. Goldstein'
_journal_name_full_name
;
 Journal of Chemical Physics
;
_journal_volume 3
_journal_year 1935
_journal_page_first 199
_journal_page_last 202
_publ_Section_title
;
 The Crystal Structure of Cesium Enneachlordithalliate, Cs$_{3}$Tl$_{2}
    ↪ $Cl$_{9}$
;

# Found in The American Mineralogist Crystal Structure Database, 2003

_aflow_title 'Cs$_{3}$Tl$_{2}$Cl$_{9}$ ($K7_{2}$) Structure'
_aflow_proto 'A9B3C2_hR28_167_ef_e_c'
_aflow_params 'a,c/a,x_{1},x_{2},x_{3},x_{4},y_{4},z_{4}'
_aflow_params_values '12.79567,1.4299790476,0.348,0.403,0.91667,0.586,-
    ↪ 0.755,0.42601'
_aflow_Strukturbericht '$K7_{2}$'
_aflow_Pearson 'hR28'

_symmetry_space_group_name_H-M "R -3 2/c (hexagonal axes)"
_symmetry_Int_Tables_number 167

_cell_length_a    12.79567
_cell_length_b    12.79567
_cell_length_c    18.29754
_cell_angle_alpha 90.00000
_cell_angle_beta  90.00000
_cell_angle_gamma 120.00000

loop_
_space_group_symop_id
_space_group_symop_operation_xyz
1 x,y,z
2 -y,x-y,z
3 -x+y,-x,z
4 y,x,-z+1/2
5 -x,-x+y,-z+1/2
6 x-y,-y,-z+1/2
7 -x,-y,-z
8 y,-x+y,-z
9 x-y,x,-z
10 -y,-x,z+1/2
11 x,x-y,z+1/2
12 -x+y,y,z+1/2
13 x+1/3,y+2/3,z+2/3
14 -y+1/3,x-y+2/3,z+2/3
15 -x+y+1/3,-x+2/3,z+2/3
16 y+1/3,x+2/3,-z+1/6
17 -x+1/3,-x+y+2/3,-z+1/6
18 x-y+1/3,-y+2/3,-z+1/6
19 -x+1/3,-y+2/3,-z+2/3
20 y+1/3,-x+y+2/3,-z+2/3
21 x-y+1/3,x+2/3,-z+2/3
22 -y+1/3,-x+2/3,z+1/6
23 x+1/3,x-y+2/3,z+1/6
24 -x+y+1/3,y+2/3,z+1/6
25 x+2/3,y+1/3,z+1/3
26 -y+2/3,x-y+1/3,z+1/3
27 -x+y+2/3,-x+1/3,z+1/3
28 y+2/3,x+1/3,-z+5/6
29 -x+2/3,-x+y+1/3,-z+5/6
30 x-y+2/3,-y+1/3,-z+5/6
31 -x+2/3,-y+1/3,-z+1/3
32 y+2/3,-x+y+1/3,-z+1/3
33 x-y+2/3,x+1/3,-z+1/3
34 -y+2/3,-x+1/3,z+5/6
35 x+2/3,x-y+1/3,z+5/6
36 -x+y+2/3,y+1/3,z+5/6

loop_
_atom_site_label
_atom_site_type_symbol
_atom_site_symmetry_multiplicity
_atom_site_Wyckoff_label
_atom_site_fract_x
_atom_site_fract_y
_atom_site_fract_z
_atom_site_occupancy
Tl1 Tl 12 c 0.00000 0.00000 0.34800 1.00000
Cl1 Cl 18 e 0.15300 0.00000 0.25000 1.00000
Cs1 Cs 18 e 0.66667 0.00000 0.25000 1.00000
Cl2 Cl 36 f 0.50033 0.84067 0.08567 1.00000
```

$Cs_3Tl_2Cl_9$ ($K7_2$): A9B3C2_hR28_167_ef_e_c - POSCAR

```
A9B3C2_hR28_167_ef_e_c & a,c/a,x1,x2,x3,x4,y4,z4 --params=12.79567,
    ↪ 1.4299790476,0.348,0.403,0.91667,0.586,-0.755,0.42601 & R-3c D_
    ↪ {3d}^{6} #167 (ce^2f) & hR28 & $K7_{2}$ & Cl9Cs3Tl2 & Cl9Cs3Tl2
    ↪ & J. L. Hoard and L. Goldstein, J. Chem. Phys. 3, 199-202 (
    ↪ 1935)
   1.00000000000000
   6.39783500000000  -3.69379175948081   6.09918000000000
   0.00000000000000   7.38758351896162   6.09918000000000
  -6.39783500000000  -3.69379175948081   6.09918000000000
    Cl    Cs    Tl
    18     6     4
Direct
   0.40300000000000   0.09700000000000   0.25000000000000   Cl   (6e)
   0.25000000000000   0.40300000000000   0.09700000000000   Cl   (6e)
   0.09700000000000   0.25000000000000   0.40300000000000   Cl   (6e)
  -0.40300000000000   0.90300000000000   0.75000000000000   Cl   (6e)
   0.75000000000000  -0.40300000000000   0.90300000000000   Cl   (6e)
   0.90300000000000   0.75000000000000  -0.40300000000000   Cl   (6e)
   0.58600000000000  -0.75500000000000   0.42601000000000   Cl   (12f)
   0.42601000000000   0.58600000000000  -0.75500000000000   Cl   (12f)
  -0.75500000000000   0.42601000000000   0.58600000000000   Cl   (12f)
   0.07399000000000   1.25500000000000  -0.08600000000000   Cl   (12f)
   1.25500000000000  -0.08600000000000   0.07399000000000   Cl   (12f)
  -0.08600000000000   0.07399000000000   1.25500000000000   Cl   (12f)
  -0.58600000000000   0.75500000000000  -0.42601000000000   Cl   (12f)
  -0.42601000000000  -0.58600000000000   0.75500000000000   Cl   (12f)
   0.75500000000000  -0.42601000000000  -0.58600000000000   Cl   (12f)
   0.92601000000000  -0.25500000000000   1.08600000000000   Cl   (12f)
  -0.25500000000000   1.08600000000000   0.92601000000000   Cl   (12f)
   1.08600000000000   0.92601000000000  -0.25500000000000   Cl   (12f)
   0.91667000000000  -0.41667000000000   0.25000000000000   Cs   (6e)
   0.25000000000000   0.91667000000000  -0.41667000000000   Cs   (6e)
  -0.41667000000000   0.25000000000000   0.91667000000000   Cs   (6e)
  -0.91667000000000   1.41667000000000   0.75000000000000   Cs   (6e)
   0.75000000000000  -0.91667000000000   1.41667000000000   Cs   (6e)
   1.41667000000000   0.75000000000000  -0.91667000000000   Cs   (6e)
   0.34800000000000   0.34800000000000   0.34800000000000   Tl   (4c)
   0.15200000000000   0.15200000000000   0.15200000000000   Tl   (4c)
  -0.34800000000000  -0.34800000000000  -0.34800000000000   Tl   (4c)
   0.84800000000000   0.84800000000000   0.84800000000000   Tl   (4c)
```

Crancrinite ($Na_6Ca_2Al_6Si_6O_{24}(CO_3)_2$, $S3_3$ (I)): A3BCD3E15F3_hP52_173_c_b_b_c_5c_c - CIF

```
# CIF file
data_findsym-output
_audit_creation_method FINDSYM

_chemical_name_mineral 'Crancrinite'
_chemical_formula_sum 'Al3 C Ca Na3 O15 Si3'

loop_
_publ_author_name
 'S. K\^{o}zu'
 'K. Takan\\'{e}'
_journal_name_full_name
;
 Proceedings of the Imperial Academy (Japan)
;
_journal_volume 9
_journal_year 1933
_journal_page_first 56
_journal_page_last 59
_publ_Section_title
;
```

```
 Crystal Structure of Cancrinite from D\^{o}d\^{o
;

# Found in Strukturbericht Band III 1933-1935, 1937

_aflow_title 'Cancrinite (Na$_{6}$Ca$_{2}$Al$_{6}$Si$_{6}$O$_{24}$(CO$_
    ↪ {3}$)$_{2}$, $S3_{3}$ (I)) Structure'
_aflow_proto 'A3BCD3E15F3_hP52_173_c_b_b_c_5c_c'
_aflow_params 'a,c/a,z_{1},z_{2},x_{3},y_{3},z_{3},x_{4},y_{4},z_{4},x_{
    ↪ 5},y_{5},z_{5},x_{6},y_{6},z_{6},x_{7},y_{7},z_{7},x_{8},y_{8},
    ↪ z_{8},x_{9},y_{9},z_{9},x_{10},y_{10},z_{10}'
_aflow_params_values '12.72,0.407232704403,0.36,0.86,0.26,0.23,0.24,0.5,
    ↪ 0.5,0.22,0.05,0.36,0.01,0.36,0.32,-0.03,0.17,0.27,0.26,0.87,
    ↪ 0.16,0.24,0.2,0.64,0.36,0.033,0.26,0.26'
_aflow_Strukturbericht '$S3_{3}$ (I)'
_aflow_Pearson 'hP52'

_symmetry_space_group_name_H-M "P 63"
_symmetry_Int_Tables_number 173

_cell_length_a    12.72000
_cell_length_b    12.72000
_cell_length_c    5.18000
_cell_angle_alpha 90.00000
_cell_angle_beta  90.00000
_cell_angle_gamma 120.00000

loop_
_space_group_symop_id
_space_group_symop_operation_xyz
1 x,y,z
2 x-y,x,z+1/2
3 -y,x-y,z
4 -x,-y,z+1/2
5 -x+y,-x,z
6 y,-x+y,z+1/2

loop_
_atom_site_label
_atom_site_type_symbol
_atom_site_symmetry_multiplicity
_atom_site_Wyckoff_label
_atom_site_fract_x
_atom_site_fract_y
_atom_site_fract_z
_atom_site_occupancy
C1 C 2 b 0.33333 0.66667 0.36000 1.00000
Ca1 Ca 2 b 0.33333 0.66667 0.86000 1.00000
Al1 Al 6 c 0.26000 0.23000 0.24000 1.00000
Na1 Na 6 c 0.50000 0.50000 0.22000 1.00000
O1 O 6 c 0.05000 0.36000 0.01000 1.00000
O2 O 6 c 0.36000 0.32000 -0.03000 1.00000
O3 O 6 c 0.17000 0.27000 0.26000 1.00000
O4 O 6 c 0.87000 0.16000 0.24000 1.00000
O5 O 6 c 0.20000 0.64000 0.36000 1.00000
Si1 Si 6 c 0.03300 0.26000 0.26000 1.00000
```

Crancrinite ($Na_6Ca_2Al_6Si_6O_{24}(CO_3)_2$, $S3_3$ (I)): A3BCD3E15F3_hP52_173_c_b_b_c_5c_c - POSCAR

```
A3BCD3E15F3_hP52_173_c_b_b_c_5c_c & a,c/a,z1,z2,x3,y3,z3,x4,y4,z4,x5,y5,
    ↪ z5,x6,y6,z6,x7,y7,z7,x8,y8,z8,x9,y9,z9,x10,y10,z10 --params=
    ↪ 12.72,0.407232704403,0.36,0.86,0.26,0.23,0.24,0.5,0.5,0.22,0.05
    ↪ ,0.36,0.01,0.36,0.32,-0.03,0.17,0.27,0.26,0.87,0.16,0.24,0.2,
    ↪ 0.64,0.36,0.033,0.26,0.26 & P6_{3} C_{6}^{6} #173 (b^2c^8) &
    ↪ hP52 & $S3_{3}$ (I) & Al3CCaNa3O15Si3 & Crancrinite & S. K\^{o}
    ↪ zu and K. Takan\'{e}, Proceedings of the Imperial Academy (
    ↪ Japan) 9, 56-59 (1933)
   1.00000000000000
   6.36000000000000  -11.01584313613810   0.00000000000000
   6.36000000000000   11.01584313613810   0.00000000000000
   0.00000000000000    0.00000000000000   5.18000000000000
    Al    C    Ca    Na    O    Si
     6    2     2     6   30     6
Direct
   0.26000000000000   0.23000000000000   0.24000000000000   Al   (6c)
  -0.23000000000000   0.03000000000000   0.24000000000000   Al   (6c)
  -0.03000000000000  -0.26000000000000   0.24000000000000   Al   (6c)
  -0.26000000000000  -0.23000000000000   0.74000000000000   Al   (6c)
   0.23000000000000  -0.03000000000000   0.74000000000000   Al   (6c)
   0.03000000000000   0.26000000000000   0.74000000000000   Al   (6c)
   0.33333333333333   0.66666666666667   0.36000000000000   C    (2b)
   0.66666666666667   0.33333333333333   0.86000000000000   C    (2b)
   0.33333333333333   0.66666666666667   0.86000000000000   Ca   (2b)
   0.66666666666667   0.33333333333333   1.36000000000000   Ca   (2b)
   0.50000000000000   0.50000000000000   0.22000000000000   Na   (6c)
  -0.50000000000000   0.00000000000000   0.22000000000000   Na   (6c)
   0.00000000000000  -0.50000000000000   0.22000000000000   Na   (6c)
  -0.50000000000000  -0.50000000000000   0.72000000000000   Na   (6c)
   0.50000000000000   0.00000000000000   0.72000000000000   Na   (6c)
   0.00000000000000   0.50000000000000   0.72000000000000   Na   (6c)
   0.05000000000000   0.36000000000000   0.01000000000000   O    (6c)
  -0.36000000000000  -0.31000000000000   0.01000000000000   O    (6c)
   0.31000000000000  -0.05000000000000   0.01000000000000   O    (6c)
  -0.05000000000000  -0.36000000000000   0.51000000000000   O    (6c)
   0.36000000000000   0.31000000000000   0.51000000000000   O    (6c)
  -0.31000000000000   0.05000000000000   0.51000000000000   O    (6c)
   0.36000000000000   0.32000000000000  -0.03000000000000   O    (6c)
  -0.32000000000000   0.04000000000000  -0.03000000000000   O    (6c)
  -0.04000000000000  -0.36000000000000  -0.03000000000000   O    (6c)
  -0.36000000000000  -0.32000000000000   0.47000000000000   O    (6c)
   0.32000000000000  -0.04000000000000   0.47000000000000   O    (6c)
   0.04000000000000   0.36000000000000   0.47000000000000   O    (6c)
   0.17000000000000   0.27000000000000   0.26000000000000   O    (6c)
  -0.27000000000000  -0.10000000000000   0.26000000000000   O    (6c)
   0.10000000000000  -0.17000000000000   0.26000000000000   O    (6c)
  -0.17000000000000  -0.27000000000000   0.76000000000000   O    (6c)
   0.27000000000000   0.10000000000000   0.76000000000000   O    (6c)
  -0.10000000000000   0.17000000000000   0.76000000000000   O    (6c)
   0.87000000000000   0.16000000000000   0.24000000000000   O    (6c)
  -0.16000000000000   0.71000000000000   0.24000000000000   O    (6c)
  -0.71000000000000  -0.87000000000000   0.24000000000000   O    (6c)
  -0.87000000000000  -0.16000000000000   0.74000000000000   O    (6c)
   0.16000000000000  -0.71000000000000   0.74000000000000   O    (6c)
   0.71000000000000   0.87000000000000   0.74000000000000   O    (6c)
   0.20000000000000   0.64000000000000   0.36000000000000   O    (6c)
  -0.64000000000000  -0.44000000000000   0.36000000000000   O    (6c)
   0.44000000000000  -0.20000000000000   0.36000000000000   O    (6c)
  -0.20000000000000  -0.64000000000000   0.86000000000000   O    (6c)
   0.64000000000000   0.44000000000000   0.86000000000000   O    (6c)
  -0.44000000000000   0.20000000000000   0.86000000000000   O    (6c)
   0.03300000000000   0.26000000000000   0.26000000000000   Si   (6c)
  -0.26000000000000  -0.22700000000000   0.26000000000000   Si   (6c)
   0.22700000000000  -0.03300000000000   0.26000000000000   Si   (6c)
  -0.03300000000000  -0.26000000000000   0.76000000000000   Si   (6c)
   0.26000000000000   0.22700000000000   0.76000000000000   Si   (6c)
  -0.22700000000000   0.03300000000000   0.76000000000000   Si   (6c)
```

$La_3CuSiS_7$: AB3C7D_hP24_173_a_c_b2c_b - CIF

```
# CIF file
data_findsym-output
_audit_creation_method FINDSYM

_chemical_name_mineral 'CuLa3S7Si'
_chemical_formula_sum 'Cu La3 S7 Si'

loop_
_publ_author_name
 'G. Collin'
 'P. Laruelle'
_journal_name_full_name
;
 Bulletin de la Societ{\'e} fran{\c{c}}aise de Mineralogie et de
    ↪ Crystallographie
;
_journal_volume 94
_journal_year 1971
_journal_page_first 175
_journal_page_last 176
_publ_Section_title
;
 Structure de La$_{6}$Cu$_{2}$Si$_{2}$S$_{14}$
;

# Found in Crystal structure and magnetic properties of Ce$_{3}$CuSnSe$_
    ↪ {7}$, 2005

_aflow_title 'La$_{3}$CuSiS$_{7}$ Structure'
_aflow_proto 'AB3C7D_hP24_173_a_c_b2c_b'
_aflow_params 'a,c/a,z_{1},z_{2},z_{3},x_{4},y_{4},z_{4},x_{5},y_{5},z_{
    ↪ 5},x_{6},y_{6},z_{6}'
_aflow_params_values '10.31,0.561978661494,0.278,0.024,0.664,0.123,0.357
    ↪ ,0.25,0.085,0.25,0.761,0.41,0.526,0.523'
_aflow_Strukturbericht 'None'
_aflow_Pearson 'hP24'

_symmetry_space_group_name_H-M "P 63"
_symmetry_Int_Tables_number 173

_cell_length_a    10.31000
_cell_length_b    10.31000
_cell_length_c    5.79400
_cell_angle_alpha 90.00000
_cell_angle_beta  90.00000
_cell_angle_gamma 120.00000

loop_
_space_group_symop_id
_space_group_symop_operation_xyz
1 x,y,z
2 x-y,x,z+1/2
3 -y,x-y,z
4 -x,-y,z+1/2
5 -x+y,-x,z
6 y,-x+y,z+1/2

loop_
_atom_site_label
_atom_site_type_symbol
_atom_site_symmetry_multiplicity
_atom_site_Wyckoff_label
_atom_site_fract_x
_atom_site_fract_y
_atom_site_fract_z
_atom_site_occupancy
Cu1 Cu 2 a 0.00000 0.00000 0.27800 1.00000
S1 S 2 b 0.33333 0.66667 0.02400 1.00000
Si1 Si 2 b 0.33333 0.66667 0.66400 1.00000
La1 La 6 c 0.12300 0.35700 0.25000 1.00000
S2 S 6 c 0.08500 0.25000 0.76100 1.00000
S3 S 6 c 0.41000 0.52600 0.52300 1.00000
```

$La_3CuSiS_7$: AB3C7D_hP24_173_a_c_b2c_b - POSCAR

```
AB3C7D_hP24_173_a_c_b2c_b & a,c/a,z1,z2,z3,x4,y4,z4,x5,y5,z5,x6,y6,z6 --
    ↪ params=10.31,0.561978661494,0.278,0.024,0.664,0.123,0.357,0.25,
    ↪ 0.085,0.25,0.761,0.41,0.526,0.523 & P6_{3} C_{6}^{6} #173 (ab^
    ↪ 2c^3) & hP24 & None & CuLa3S7Si & CuLa3S7Si & G. Collin and P.
    ↪ Laruelle, Bull. Soc. fr. Min'eral. Crystallogr. 94, 175-176 (
    ↪ 1971)
```

```
   1.00000000000000
   5.15500000000000  -8.92872191301756   0.00000000000000
   5.15500000000000   8.92872191301756   0.00000000000000
   0.00000000000000   0.00000000000000   5.79400000000000
    Cu    La     S    Si
     2     6    14     2
Direct
   0.00000000000000   0.00000000000000   0.27800000000000   Cu   (2a)
   0.00000000000000   0.00000000000000   0.77800000000000   Cu   (2a)
   0.12300000000000   0.35700000000000   0.25000000000000   La   (6c)
  -0.35700000000000  -0.23400000000000   0.25000000000000   La   (6c)
   0.23400000000000  -0.12300000000000   0.25000000000000   La   (6c)
  -0.12300000000000  -0.35700000000000   0.75000000000000   La   (6c)
   0.35700000000000   0.23400000000000   0.75000000000000   La   (6c)
  -0.23400000000000   0.12300000000000   0.75000000000000   La   (6c)
   0.33333333333333   0.66666666666667   0.02400000000000    S   (2b)
   0.66666666666667   0.33333333333333   0.52400000000000    S   (2b)
   0.08500000000000   0.25000000000000   0.76100000000000    S   (6c)
  -0.25000000000000  -0.16500000000000   0.76100000000000    S   (6c)
   0.16500000000000  -0.08500000000000   0.76100000000000    S   (6c)
  -0.08500000000000  -0.25000000000000   1.26100000000000    S   (6c)
   0.25000000000000   0.16500000000000   1.26100000000000    S   (6c)
  -0.16500000000000   0.08500000000000   1.26100000000000    S   (6c)
   0.41000000000000   0.52600000000000   0.52300000000000    S   (6c)
  -0.52600000000000  -0.11600000000000   0.52300000000000    S   (6c)
   0.11600000000000  -0.41000000000000   0.52300000000000    S   (6c)
  -0.41000000000000  -0.52600000000000   1.02300000000000    S   (6c)
   0.52600000000000   0.11600000000000   1.02300000000000    S   (6c)
  -0.11600000000000   0.41000000000000   1.02300000000000    S   (6c)
   0.33333333333333   0.66666666666667   0.66400000000000   Si   (2b)
   0.66666666666667   0.33333333333333   1.16400000000000   Si   (2b)
```

$La_3BWO_9$ ($P6_3$): AB3C9D_hP28_173_a_c_3c_b - CIF

```
# CIF file
data_findsym-output
_audit_creation_method FINDSYM

_chemical_name_mineral 'BaLi3O9W'
_chemical_formula_sum 'B La3 O9 W'

loop_
_publ_author_name
 'J. Han'
 'F. Pan'
 'M. S. Molokeev'
 'J. Dai'
 'M. Peng'
 'W. Zhou'
 'J. Wang'
_journal_name_full_name
;
 ACS Applied Materials and Interfaces
;
_journal_volume 10
_journal_year 2018
_journal_page_first 13660
_journal_page_last 13668
_publ_Section_title
;
 Redefinition of Crystal Structure and Bi$^{3+}$ Yellow Luminescence
    ↪ with Strong Near-Ultraviolet Excitation in La$_{3}$BWO$_{9}$:
    ↪ Bi$^{3+}$ Phosphor for White Light-Emitting Diodes
;

_aflow_title 'La$_{3}$BWO$_{9}$ ($P6_{3}$) Structure'
_aflow_proto 'AB3C9D_hP28_173_a_c_3c_b'
_aflow_params 'a,c/a,z_{1},z_{2},x_{3},y_{3},z_{3},x_{4},y_{4},z_{4},x_{
    ↪ 5},y_{5},z_{5},x_{6},y_{6},z_{6}'
_aflow_params_values '8.84324,0.630487242232,0.358,0.25,0.3636,0.0854,
    ↪ 0.2213,0.17,0.047,0.861,0.192,0.483,0.044,0.142,0.517,0.468'
_aflow_Strukturbericht 'None'
_aflow_Pearson 'hP28'

_symmetry_space_group_name_H-M "P 63"
_symmetry_Int_Tables_number 173

_cell_length_a     8.84324
_cell_length_b     8.84324
_cell_length_c     5.57555
_cell_angle_alpha  90.00000
_cell_angle_beta   90.00000
_cell_angle_gamma  120.00000

loop_
_space_group_symop_id
_space_group_symop_operation_xyz
1 x,y,z
2 x-y,x,z+1/2
3 -y,x-y,z
4 -x,-y,z+1/2
5 -x+y,-x,z
6 y,-x+y,z+1/2

loop_
_atom_site_label
_atom_site_type_symbol
_atom_site_symmetry_multiplicity
_atom_site_Wyckoff_label
_atom_site_fract_x
_atom_site_fract_y
_atom_site_fract_z
_atom_site_occupancy
B1 B 2 a 0.00000 0.00000 0.35800 1.00000
W1 W 2 b 0.33333 0.66667 0.25000 1.00000
La1 La 6 c 0.36360 0.08540 0.22130 1.00000
O1 O 6 c 0.17000 0.04700 0.86100 1.00000
O2 O 6 c 0.19200 0.48300 0.04400 1.00000
O3 O 6 c 0.14200 0.51700 0.46800 1.00000
```

$La_3BWO_9$ ($P6_3$): AB3C9D_hP28_173_a_c_3c_b - POSCAR

```
AB3C9D_hP28_173_a_c_3c_b & a,c/a,z1,z2,x3,y3,z3,x4,y4,z4,x5,y5,z5,x6,y6,
    ↪ z6 --params=8.84324,0.630487242232,0.358,0.25,0.3636,0.0854,
    ↪ 0.2213,0.17,0.047,0.861,0.192,0.483,0.044,0.142,0.517,0.468 &
    ↪ P6_{3} C_{6}^{6} #173 (abc^4) & hP28 & None & BaLi3O9W &
    ↪ BaLi3O9W & J. Han et al., ACS Appl. Mater. Interfaces 10,
    ↪ 13660-13668 (2018)
   1.00000000000000
   4.42162000000000  -7.65847049176270   0.00000000000000
   4.42162000000000   7.65847049176270   0.00000000000000
   0.00000000000000   0.00000000000000   5.57555000000000
     B    La     O     W
     2     6    18     2
Direct
   0.00000000000000   0.00000000000000   0.35800000000000    B   (2a)
   0.00000000000000   0.00000000000000   0.85800000000000    B   (2a)
   0.36360000000000   0.08540000000000   0.22130000000000   La   (6c)
  -0.08540000000000   0.27820000000000   0.22130000000000   La   (6c)
  -0.27820000000000  -0.36360000000000   0.22130000000000   La   (6c)
  -0.36360000000000  -0.08540000000000   0.72130000000000   La   (6c)
   0.08540000000000  -0.27820000000000   0.72130000000000   La   (6c)
   0.27820000000000   0.36360000000000   0.72130000000000   La   (6c)
   0.17000000000000   0.04700000000000   0.86100000000000    O   (6c)
  -0.04700000000000   0.12300000000000   0.86100000000000    O   (6c)
  -0.12300000000000  -0.17000000000000   0.86100000000000    O   (6c)
  -0.17000000000000  -0.04700000000000   1.36100000000000    O   (6c)
   0.04700000000000  -0.12300000000000   1.36100000000000    O   (6c)
   0.12300000000000   0.17000000000000   1.36100000000000    O   (6c)
   0.19200000000000   0.48300000000000   0.04400000000000    O   (6c)
  -0.48300000000000  -0.29100000000000   0.04400000000000    O   (6c)
   0.29100000000000  -0.19200000000000   0.04400000000000    O   (6c)
  -0.19200000000000  -0.48300000000000   0.54400000000000    O   (6c)
   0.48300000000000   0.29100000000000   0.54400000000000    O   (6c)
  -0.29100000000000   0.19200000000000   0.54400000000000    O   (6c)
   0.14200000000000   0.51700000000000   0.46800000000000    O   (6c)
  -0.51700000000000  -0.37500000000000   0.46800000000000    O   (6c)
   0.37500000000000  -0.14200000000000   0.46800000000000    O   (6c)
  -0.14200000000000  -0.51700000000000   0.96800000000000    O   (6c)
   0.51700000000000   0.37500000000000   0.96800000000000    O   (6c)
  -0.37500000000000   0.14200000000000   0.96800000000000    O   (6c)
   0.33333333333333   0.66666666666667   0.25000000000000    W   (2b)
   0.66666666666667   0.33333333333333   0.75000000000000    W   (2b)
```

$\alpha$-$LiIO_3$: ABC3_hP10_173_b_a_c - CIF

```
# CIF file
data_findsym-output
_audit_creation_method FINDSYM

_chemical_name_mineral 'ILiO3'
_chemical_formula_sum 'I Li O3'

loop_
_publ_author_name
 'A. Rosenzweig'
 'B. Morosin'
_journal_name_full_name
;
 Acta Cristallographica
;
_journal_volume 20
_journal_year 1966
_journal_page_first 758
_journal_page_last 761
_publ_Section_title
;
 A reinvestigation of the crystal structure of LiIO$_{3}$
;

_aflow_title '$\alpha$-LiIO$_{3}$ Structure'
_aflow_proto 'ABC3_hP10_173_b_a_c'
_aflow_params 'a,c/a,z_{1},z_{2},x_{3},y_{3},z_{3}'
_aflow_params_values '5.1815,0.997954260349,0.8907,0.0,0.09358,0.34396,
    ↪ 0.1698'
_aflow_Strukturbericht 'None'
_aflow_Pearson 'hP10'

_symmetry_space_group_name_H-M "P 63"
_symmetry_Int_Tables_number 173

_cell_length_a     5.18150
_cell_length_b     5.18150
_cell_length_c     5.17090
_cell_angle_alpha  90.00000
_cell_angle_beta   90.00000
_cell_angle_gamma  120.00000

loop_
_space_group_symop_id
_space_group_symop_operation_xyz
1 x,y,z
2 x-y,x,z+1/2
3 -y,x-y,z
4 -x,-y,z+1/2
5 -x+y,-x,z
6 y,-x+y,z+1/2

loop_
_atom_site_label
```

```
_atom_site_type_symbol
_atom_site_symmetry_multiplicity
_atom_site_Wyckoff_label
_atom_site_fract_x
_atom_site_fract_y
_atom_site_fract_z
_atom_site_occupancy
Li1 Li 2 a 0.00000 0.00000 0.89070 1.00000
I1 I 2 b 0.33333 0.66667 0.00000 1.00000
O1 O 6 c 0.09358 0.34396 0.16980 1.00000
```

$\alpha$-$LiIO_3$: ABC3_hP10_173_b_a_c - POSCAR

```
ABC3_hP10_173_b_a_c & a,c/a,z1,z2,x3,y3,z3 --params=5.1815,
    ↪ 0.997954260349,0.8907,0.0,0.09358,0.34396,0.1698 & P6_{3} C_{6
    ↪ }^{6} #173 (abc) & hP10 & None & ILiO3 & ILiO3 & A. Rosenzweig
    ↪ and B. Morosin, Acta Cryst. 20, 758-761 (1966)
   1.00000000000000
   2.59075000000000  -4.48731062970907   0.00000000000000
   2.59075000000000   4.48731062970907   0.00000000000000
   0.00000000000000   0.00000000000000   5.17090000000000
     I      Li      O
     2       2      6
Direct
   0.33333333333333   0.66666666666667   0.00000000000000    I   (2b)
   0.66666666666667   0.33333333333333   0.50000000000000    I   (2b)
   0.00000000000000   0.00000000000000   0.89070000000000   Li   (2a)
   0.00000000000000   0.00000000000000   1.39070000000000   Li   (2a)
   0.09358000000000   0.34396000000000   0.16980000000000    O   (6c)
  -0.34396000000000  -0.25038000000000   0.16980000000000    O   (6c)
   0.25038000000000  -0.09358000000000   0.16980000000000    O   (6c)
  -0.09358000000000  -0.34396000000000   0.66980000000000    O   (6c)
   0.34396000000000   0.25038000000000   0.66980000000000    O   (6c)
  -0.25038000000000   0.09358000000000   0.66980000000000    O   (6c)
```

$LiKSO_4$ ($H1_4$): ABC4D_hP14_173_a_b_bc_b - CIF

```
# CIF file
data_findsym-output
_audit_creation_method FINDSYM

_chemical_name_mineral 'KLiO4S'
_chemical_formula_sum 'K Li O4 S'

loop_
_publ_author_name
 'S. {Bhakay-Tamhane}'
 'A. Sequiera'
 'R. Chidambaram'
_journal_name_full_name
;
 Acta Crystallographica Section C: Structural Chemistry
;
_journal_volume 40
_journal_year 1984
_journal_page_first 1648
_journal_page_last 1651
_publ_Section_title
;
 Structure of lithium potassium sulphate, LiKSO$_{4}$: a neutron
     ↪ diffraction study
;

_aflow_title 'LiKSO$_{4}$ ($H1_{4}$) Structure'
_aflow_proto 'ABC4D_hP14_173_a_b_bc_b'
_aflow_params 'a,c/a,z_{1},z_{2},z_{3},z_{4},x_{5},y_{5},z_{5}'
_aflow_params_values '5.146,1.6781966576,0.0,0.3146,0.5358,0.7045,0.3446
    ↪ ,0.4031,0.7583'
_aflow_Strukturbericht '$H1_{4}$'
_aflow_Pearson 'hP14'

_symmetry_space_group_name_H-M "P 63"
_symmetry_Int_Tables_number 173

_cell_length_a      5.14600
_cell_length_b      5.14600
_cell_length_c      8.63600
_cell_angle_alpha  90.00000
_cell_angle_beta   90.00000
_cell_angle_gamma 120.00000

loop_
_space_group_symop_id
_space_group_symop_operation_xyz
1 x,y,z
2 x-y,x,z+1/2
3 -y,x-y,z
4 -x,-y,z+1/2
5 -x+y,-x,z
6 y,-x+y,z+1/2

loop_
_atom_site_label
_atom_site_type_symbol
_atom_site_symmetry_multiplicity
_atom_site_Wyckoff_label
_atom_site_fract_x
_atom_site_fract_y
_atom_site_fract_z
_atom_site_occupancy
K1 K 2 a 0.00000 0.00000 0.00000 1.00000
Li1 Li 2 b 0.33333 0.66667 0.31460 1.00000
O1 O 2 b 0.33333 0.66667 0.53580 1.00000
S1 S 2 b 0.33333 0.66667 0.70450 1.00000
O2 O 6 c 0.34460 0.40310 0.75830 1.00000
```

$LiKSO_4$ ($H1_4$): ABC4D_hP14_173_a_b_bc_b - POSCAR

```
ABC4D_hP14_173_a_b_bc_b & a,c/a,z1,z2,z3,z4,x5,y5,z5 --params=5.146,
    ↪ 1.6781966576,0.0,0.3146,0.5358,0.7045,0.3446,0.4031,0.7583 &
    ↪ P6_{3} C_{6}^{6} #173 (ab^3c) & hP14 & $H1_{4}$ & KLiO4S &
    ↪ KLiO4S & S. {Bhakay-Tamhane} and A. Sequiera and R. Chidambaram
    ↪ , Acta Crystallogr. C 40, 1648-1651 (1984)
   1.00000000000000
   2.57300000000000  -4.45656672787472   0.00000000000000
   2.57300000000000   4.45656672787472   0.00000000000000
   0.00000000000000   0.00000000000000   8.63600000000000
     K      Li      O      S
     2       2      8      2
Direct
   0.00000000000000   0.00000000000000   0.00000000000000    K   (2a)
   0.00000000000000   0.00000000000000   0.50000000000000    K   (2a)
   0.33333333333333   0.66666666666667   0.31460000000000   Li   (2b)
   0.66666666666667   0.33333333333333   0.81460000000000   Li   (2b)
   0.33333333333333   0.66666666666667   0.53580000000000    O   (2b)
   0.66666666666667   0.33333333333333   1.03580000000000    O   (2b)
   0.34460000000000   0.40310000000000   0.75830000000000    O   (6c)
  -0.40310000000000  -0.05850000000000   0.75830000000000    O   (6c)
   0.05850000000000  -0.34460000000000   0.75830000000000    O   (6c)
  -0.34460000000000  -0.40310000000000   1.25830000000000    O   (6c)
   0.40310000000000   0.05850000000000   1.25830000000000    O   (6c)
  -0.05850000000000   0.34460000000000   1.25830000000000    O   (6c)
   0.33333333333333   0.66666666666667   0.70450000000000    S   (2b)
   0.66666666666667   0.33333333333333   1.20450000000000    S   (2b)
```

$Rh_{20}Si_{13}$: A10B7_hP34_176_c3h_b2h - CIF

```
# CIF file
data_findsym-output
_audit_creation_method FINDSYM

_chemical_name_mineral 'Rh20Si13'
_chemical_formula_sum 'Rb10 Si7'

loop_
_publ_author_name
 'I. Engstr\"{o}m'
_journal_name_full_name
;
 Acta Chemica Scandinavica
;
_journal_volume 19
_journal_year 1965
_journal_page_first 1924
_journal_page_last 1932
_publ_Section_title
;
 The Crystal Structure of Rh$_{20}$Si$_{13}$
;

_aflow_title 'Rh$_{20}$Si$_{13}$ Structure'
_aflow_proto 'A10B7_hP34_176_c3h_b2h'
_aflow_params 'a,c/a,x_{3},y_{3},x_{4},y_{4},x_{5},y_{5},x_{6},y_{6},x_{
    ↪ 7},y_{7}'
_aflow_params_values '11.85,0.305738396624,0.8124,0.8157,0.5992,0.8717,
    ↪ 0.39,-0.0713,0.7667,0.1058,0.5497,0.1678'
_aflow_Strukturbericht 'None'
_aflow_Pearson 'hP34'

_symmetry_space_group_name_H-M "P 63/m"
_symmetry_Int_Tables_number 176

_cell_length_a      11.85000
_cell_length_b      11.85000
_cell_length_c       3.62300
_cell_angle_alpha  90.00000
_cell_angle_beta   90.00000
_cell_angle_gamma 120.00000

loop_
_space_group_symop_id
_space_group_symop_operation_xyz
1 x,y,z
2 x-y,x,z+1/2
3 -y,x-y,z
4 -x,-y,z+1/2
5 -x+y,-x,z
6 y,-x+y,z+1/2
7 -x,-y,-z
8 -x+y,-x,-z+1/2
9 y,-x+y,-z
10 x,y,-z+1/2
11 x-y,x,-z
12 -y,x-y,-z+1/2

loop_
_atom_site_label
_atom_site_type_symbol
_atom_site_symmetry_multiplicity
_atom_site_Wyckoff_label
_atom_site_fract_x
_atom_site_fract_y
_atom_site_fract_z
_atom_site_occupancy
Si1 Si 2 b 0.00000 0.00000 0.00000 0.50000
Rb1 Rb 2 c 0.33333 0.66667 0.25000 1.00000
Rb2 Rb 6 h 0.81240 0.81570 0.25000 1.00000
Rb3 Rb 6 h 0.59920 0.87170 0.25000 1.00000
Rb4 Rb 6 h 0.39000 -0.07130 0.25000 1.00000
Si2 Si 6 h 0.76670 0.10580 0.25000 1.00000
Si3 Si 6 h 0.54970 0.16780 0.25000 1.00000
```

Rh$_{20}$Si$_{13}$: A10B7_hP34_176_c3h_b2h - POSCAR

```
A10B7_hP34_176_c3h_b2h & a,c/a,x3,y3,x4,y4,x5,y5,x6,y6,x7,y7 --params=
    ↪ 11.85,0.305738396624,0.8124,0.8157,0.5992,0.8717,0.39,-0.0713,
    ↪ 0.7667,0.1058,0.5497,0.1678 & P6_{3}/m C_{6h}^{2} #176 (bch^5)
    ↪ & hP34 & None & Rh20Si13 & Rh20Si13 & I. Engstr\"{o}m, Acta
    ↪ Chem. Scand. 19, 1924-1932 (1965)
   1.00000000000000
   5.92500000000000 -10.26240103484560   0.00000000000000
   5.92500000000000  10.26240103484560   0.00000000000000
   0.00000000000000   0.00000000000000   3.62300000000000
    Rb      Si
    20      14
Direct
   0.33333333333333   0.66666666666667   0.25000000000000   Rb   (2c)
   0.66666666666667   0.33333333333333   0.75000000000000   Rb   (2c)
   0.81240000000000   0.81570000000000   0.25000000000000   Rb   (6h)
  -0.81570000000000  -0.00330000000000   0.25000000000000   Rb   (6h)
   0.00330000000000  -0.81240000000000   0.25000000000000   Rb   (6h)
  -0.81240000000000  -0.81570000000000   0.75000000000000   Rb   (6h)
   0.81570000000000   0.00330000000000   0.75000000000000   Rb   (6h)
  -0.00330000000000   0.81240000000000   0.75000000000000   Rb   (6h)
   0.59920000000000   0.87170000000000   0.25000000000000   Rb   (6h)
  -0.87170000000000  -0.27250000000000   0.25000000000000   Rb   (6h)
   0.27250000000000  -0.59920000000000   0.25000000000000   Rb   (6h)
  -0.59920000000000  -0.87170000000000   0.75000000000000   Rb   (6h)
   0.87170000000000   0.27250000000000   0.75000000000000   Rb   (6h)
  -0.27250000000000   0.59920000000000   0.75000000000000   Rb   (6h)
   0.39000000000000  -0.07130000000000   0.25000000000000   Rb   (6h)
   0.07130000000000   0.46130000000000   0.25000000000000   Rb   (6h)
  -0.46130000000000  -0.39000000000000   0.25000000000000   Rb   (6h)
  -0.39000000000000   0.07130000000000   0.75000000000000   Rb   (6h)
  -0.07130000000000  -0.46130000000000   0.75000000000000   Rb   (6h)
   0.46130000000000   0.39000000000000   0.75000000000000   Rb   (6h)
   0.00000000000000   0.00000000000000   0.00000000000000   Si   (2b)
   0.00000000000000   0.00000000000000   0.50000000000000   Si   (2b)
   0.76670000000000   0.10580000000000   0.25000000000000   Si   (6h)
  -0.10580000000000   0.66090000000000   0.25000000000000   Si   (6h)
  -0.66090000000000  -0.76670000000000   0.25000000000000   Si   (6h)
  -0.76670000000000  -0.10580000000000   0.75000000000000   Si   (6h)
   0.10580000000000  -0.66090000000000   0.75000000000000   Si   (6h)
   0.66090000000000   0.76670000000000   0.75000000000000   Si   (6h)
   0.54970000000000   0.16780000000000   0.25000000000000   Si   (6h)
  -0.16780000000000   0.38190000000000   0.25000000000000   Si   (6h)
  -0.38190000000000  -0.54970000000000   0.25000000000000   Si   (6h)
  -0.54970000000000  -0.16780000000000   0.75000000000000   Si   (6h)
   0.16780000000000  -0.38190000000000   0.75000000000000   Si   (6h)
   0.38190000000000   0.54970000000000   0.75000000000000   Si   (6h)
```

Th$_7$S$_{12}$ (D8$_k$): A3B2_hP20_176_2h_ah - CIF

```
# CIF file
data_findsym-output
_audit_creation_method FINDSYM

_chemical_name_mineral 'Th7S12'
_chemical_formula_sum 'S3 Th2'

loop_
_publ_author_name
 'W. H. Zachariasen'
_journal_name_full_name
;
 Acta Cristallographica
;
_journal_volume 2
_journal_year 1949
_journal_page_first 288
_journal_page_last 291
_publ_Section_title
;
 Crystal chemical studies of the 5f-series of elements. IX. The crystal
    ↪ structure of Th$_{7}$S$_{12}$
;

_aflow_title 'Th$_{7}$S$_{12}$ (D8$_{k}$) Structure'
_aflow_proto 'A3B2_hP20_176_2h_ah'
_aflow_params 'a,c/a,x_{2},y_{2},x_{3},y_{3},x_{4},y_{4}'
_aflow_params_values '11.064,0.361713665944,0.375,0.861,0.0,0.765,0.717,
    ↪ 0.564'
_aflow_Strukturbericht '$D8_{k}$'
_aflow_Pearson 'hP20'

_symmetry_space_group_name_H-M "P 63/m"
_symmetry_Int_Tables_number 176

_cell_length_a      11.06400
_cell_length_b      11.06400
_cell_length_c      4.00200
_cell_angle_alpha 90.00000
_cell_angle_beta  90.00000
_cell_angle_gamma 120.00000

loop_
_space_group_symop_id
_space_group_symop_operation_xyz
1 x,y,z
2 x-y,x,z+1/2
3 -y,x-y,z
4 -x,-y,z+1/2
5 -x+y,-x,z
6 y,-x+y,z+1/2
7 -x,-y,-z
8 -x+y,-x,-z+1/2
9 y,-x+y,-z
10 x,y,-z+1/2
11 x-y,x,-z
12 -y,x-y,-z+1/2

loop_
_atom_site_label
_atom_site_type_symbol
_atom_site_symmetry_multiplicity
_atom_site_Wyckoff_label
_atom_site_fract_x
_atom_site_fract_y
_atom_site_fract_z
_atom_site_occupancy
Th1 Th 2 a 0.00000 0.00000 0.25000 0.50000
S1 S 6 h 0.37500 0.86100 0.25000 1.00000
S2 S 6 h 0.00000 0.76500 0.25000 1.00000
Th2 Th 6 h 0.71700 0.56400 0.25000 1.00000
```

Th$_7$S$_{12}$ (D8$_k$): A3B2_hP20_176_2h_ah - POSCAR

```
A3B2_hP20_176_2h_ah & a,c/a,x2,y2,x3,y3,x4,y4 --params=11.064,
    ↪ 0.361713665944,0.375,0.861,0.0,0.765,0.717,0.564 & P6_{3}/m C_{
    ↪ 6h}^{2} #176 (ah^3) & hP20 & $D8_{k}$ & Th7S12 & Th7S12 & W. H.
    ↪ Zachariasen, Acta Cryst. 2, 288-291 (1949)
   1.00000000000000
   5.53200000000000  -9.58170506747103   0.00000000000000
   5.53200000000000   9.58170506747103   0.00000000000000
   0.00000000000000   0.00000000000000   4.00200000000000
     S      Th
    12       8
Direct
   0.37500000000000   0.86100000000000   0.25000000000000    S   (6h)
  -0.86100000000000  -0.48600000000000   0.25000000000000    S   (6h)
   0.48600000000000  -0.37500000000000   0.25000000000000    S   (6h)
  -0.37500000000000  -0.86100000000000   0.75000000000000    S   (6h)
   0.86100000000000   0.48600000000000   0.75000000000000    S   (6h)
  -0.48600000000000   0.37500000000000   0.75000000000000    S   (6h)
   0.00000000000000   0.76500000000000   0.25000000000000    S   (6h)
  -0.76500000000000  -0.76500000000000   0.25000000000000    S   (6h)
   0.76500000000000   0.00000000000000   0.25000000000000    S   (6h)
   0.00000000000000  -0.76500000000000   0.75000000000000    S   (6h)
   0.76500000000000   0.76500000000000   0.75000000000000    S   (6h)
  -0.76500000000000   0.00000000000000   0.75000000000000    S   (6h)
   0.00000000000000   0.00000000000000   0.25000000000000   Th   (2a)
   0.00000000000000   0.00000000000000   0.75000000000000   Th   (2a)
   0.71700000000000   0.56400000000000   0.25000000000000   Th   (6h)
  -0.56400000000000   0.15300000000000   0.25000000000000   Th   (6h)
  -0.15300000000000  -0.71700000000000   0.25000000000000   Th   (6h)
  -0.71700000000000  -0.56400000000000   0.75000000000000   Th   (6h)
   0.56400000000000  -0.15300000000000   0.75000000000000   Th   (6h)
   0.15300000000000   0.71700000000000   0.75000000000000   Th   (6h)
```

$\beta$-Si$_3$N$_4$: A4B3_hP14_176_ch_h - CIF

```
# CIF file
data_findsym-output
_audit_creation_method FINDSYM

_chemical_name_mineral 'N4Si3'
_chemical_formula_sum 'N4 Si3'

loop_
_publ_author_name
 'P. Yang'
 'H.-K. Fun'
 'I. {Ab. Rahman}'
 'I. Saleh'
_journal_name_full_name
;
 Ceramics International
;
_journal_volume 21
_journal_year 1995
_journal_page_first 137
_journal_page_last 142
_publ_Section_title
;
 Two phase refinements of the structures of $\alpha$-Si$_{3}$N$_{4}$ and
    ↪ $\beta$-Si$_{3}$N$_{4}$ made from rice husk by Rietveld
    ↪ analysis
;

# Found in The American Mineralogist Crystal Structure Database, 2003

_aflow_title '$\beta$-Si$_{3}$N$_{4}$ Structure'
_aflow_proto 'A4B3_hP14_176_ch_h'
_aflow_params 'a,c/a,x_{2},y_{2},x_{3},y_{3}'
_aflow_params_values '7.6093,0.382150789166,0.329,0.039,0.1742,0.7678'
_aflow_Strukturbericht 'None'
_aflow_Pearson 'hP14'

_symmetry_space_group_name_H-M "P 63/m"
_symmetry_Int_Tables_number 176

_cell_length_a      7.60930
_cell_length_b      7.60930
_cell_length_c      2.90790
_cell_angle_alpha 90.00000
_cell_angle_beta  90.00000
_cell_angle_gamma 120.00000

loop_
_space_group_symop_id
_space_group_symop_operation_xyz
1 x,y,z
```

```
2 x-y,x,z+1/2
3 -y,x-y,z
4 -x,-y,z+1/2
5 -x+y,-x,z
6 y,-x+y,z+1/2
7 -x,-y,-z
8 -x+y,-x,-z+1/2
9 y,-x+y,-z
10 x,y,-z+1/2
11 x-y,x,-z
12 -y,x-y,-z+1/2

loop_
_atom_site_label
_atom_site_type_symbol
_atom_site_symmetry_multiplicity
_atom_site_Wyckoff_label
_atom_site_fract_x
_atom_site_fract_y
_atom_site_fract_z
_atom_site_occupancy
N1 N 2 c 0.33333 0.66667 0.25000 1.00000
N2 N 6 h 0.32900 0.03900 0.25000 1.00000
Si1 Si 6 h 0.17420 0.76780 0.25000 1.00000
```

β-$Si_3N_4$: A4B3_hP14_176_ch_h - POSCAR

```
A4B3_hP14_176_ch_h & a,c/a,x2,y2,x3,y3 --params=7.6093,0.382150789166,
    ↪ 0.329,0.039,0.1742,0.7678 & P6_{3}/m C_{6h}^{2} #176 (ch^2) &
    ↪ hP14 & None & N4Si3 & N4Si3 & P. Yang et al., Ceram. Int. 21,
    ↪ 137-142 (1995)
   1.00000000000000
   3.80465000000000  -6.58984710501693   0.00000000000000
   3.80465000000000   6.58984710501693   0.00000000000000
   0.00000000000000   0.00000000000000   2.90790000000000
     N     Si
     8      6
Direct
   0.33333333333333   0.66666666666667   0.25000000000000   N   (2c)
   0.66666666666667   0.33333333333333   0.75000000000000   N   (2c)
   0.32900000000000   0.03900000000000   0.25000000000000   N   (6h)
  -0.03900000000000   0.29000000000000   0.25000000000000   N   (6h)
  -0.29000000000000  -0.32900000000000   0.25000000000000   N   (6h)
  -0.32900000000000  -0.03900000000000   0.75000000000000   N   (6h)
   0.03900000000000  -0.29000000000000   0.75000000000000   N   (6h)
   0.29000000000000   0.32900000000000   0.75000000000000   N   (6h)
   0.17420000000000   0.76780000000000   0.25000000000000   Si  (6h)
  -0.76780000000000  -0.59360000000000   0.25000000000000   Si  (6h)
   0.59360000000000  -0.17420000000000   0.25000000000000   Si  (6h)
  -0.17420000000000  -0.76780000000000   0.75000000000000   Si  (6h)
   0.76780000000000   0.59360000000000   0.75000000000000   Si  (6h)
  -0.59360000000000   0.17420000000000   0.75000000000000   Si  (6h)
```

Fluorapatite [$Ca_5F(PO_4)_3$, $H5_7$]: A5BC12D3_hP42_176_fh_a_2hi_h - CIF

```
# CIF file
data_findsym-output
_audit_creation_method FINDSYM

_chemical_name_mineral 'Fluorapatite'
_chemical_formula_sum 'Ca5 F O12 P3'

loop_
_publ_author_name
 'J. M. Hughes'
 'J. Rakovan'
_journal_name_full_name
;
 Reviews in Mineralogy and Geochemistry
;
_journal_volume 48
_journal_year 2002
_journal_page_first 1
_journal_page_last 12
_publ_Section_title
;
 The Crystal Structure of Apatite, Ca$_{5}$(PO$_{4}$)$_{3}$(F,OH,Cl)
;

_aflow_title 'Fluorapatite [Ca$_{5}$F(PO$_{4}$)$_{3}$, $H5_{7}$]
    ↪ Structure'
_aflow_proto 'A5BC12D3_hP42_176_fh_a_2hi_h'
_aflow_params 'a,c/a,z_{2},x_{3},y_{3},x_{4},y_{4},x_{5},y_{5},x_{6},y_{
    ↪ 6},x_{7},y_{7},z_{7}'
_aflow_params_values '9.397,0.731935724167,-0.001,-0.00712,0.24227,
    ↪ 0.4849,0.3273,0.4667,0.5875,0.36895,0.3985,0.2575,0.3421,0.0705
    ↪ '
_aflow_Strukturbericht '$H5_{7}$'
_aflow_Pearson 'hP42'

_symmetry_space_group_name_H-M "P 63/m"
_symmetry_Int_Tables_number 176

_cell_length_a      9.39700
_cell_length_b      9.39700
_cell_length_c      6.87800
_cell_angle_alpha 90.00000
_cell_angle_beta  90.00000
_cell_angle_gamma 120.00000

loop_
_space_group_symop_id
_space_group_symop_operation_xyz
1 x,y,z
2 x-y,x,z+1/2
3 -y,x-y,z
4 -x,-y,z+1/2
5 -x+y,-x,z
6 y,-x+y,z+1/2
7 -x,-y,-z
8 -x+y,-x,-z+1/2
9 y,-x+y,-z
10 x,y,-z+1/2
11 x-y,x,-z
12 -y,x-y,-z+1/2

loop_
_atom_site_label
_atom_site_type_symbol
_atom_site_symmetry_multiplicity
_atom_site_Wyckoff_label
_atom_site_fract_x
_atom_site_fract_y
_atom_site_fract_z
_atom_site_occupancy
F1 F 2 a 0.00000 0.00000 0.25000 1.00000
Ca1 Ca 4 f 0.33333 0.66667 -0.00100 1.00000
Ca2 Ca 6 h -0.00712 0.24227 0.25000 1.00000
O1 O 6 h 0.48490 0.32730 0.25000 1.00000
O2 O 6 h 0.46670 0.58750 0.25000 1.00000
P1 P 6 h 0.36895 0.39850 0.25000 1.00000
O3 O 12 i 0.25750 0.34210 0.07050 1.00000
```

Fluorapatite [$Ca_5F(PO_4)_3$, $H5_7$]: A5BC12D3_hP42_176_fh_a_2hi_h - POSCAR

```
A5BC12D3_hP42_176_fh_a_2hi_h & a,c/a,z2,x3,y3,x4,y4,x5,y5,x6,y6,x7,y7,z7
    ↪ --params=9.397,0.731935724167,-0.001,-0.00712,0.24227,0.4849,
    ↪ 0.3273,0.4667,0.5875,0.36895,0.3985,0.2575,0.3421,0.0705 & P6_{
    ↪ 3}/m C_{6h}^{2} #176 (afh^4i) & hP42 & $H5_{7}$ & Ca5FO12P &
    ↪ Fluorapatite & J. M. Hughes and J. Rakovan, Rev. Mineral.
    ↪ Geochem. 48, 1-12 (2002)
   1.00000000000000
   4.69850000000000  -8.13804071936237   0.00000000000000
   4.69850000000000   8.13804071936237   0.00000000000000
   0.00000000000000   0.00000000000000   6.87800000000000
    Ca     F     O     P
    10     2    24     6
Direct
   0.33333333333333   0.66666666666667  -0.00100000000000   Ca   (4f)
   0.66666666666667   0.33333333333333   0.49900000000000   Ca   (4f)
   0.66666666666667   0.33333333333333   0.00100000000000   Ca   (4f)
   0.33333333333333   0.66666666666667   0.50100000000000   Ca   (4f)
  -0.00712000000000   0.24227000000000   0.25000000000000   Ca   (6h)
  -0.24227000000000  -0.24939000000000   0.25000000000000   Ca   (6h)
   0.24939000000000   0.00712000000000   0.25000000000000   Ca   (6h)
   0.00712000000000  -0.24227000000000   0.75000000000000   Ca   (6h)
   0.24227000000000   0.24939000000000   0.75000000000000   Ca   (6h)
  -0.24939000000000  -0.00712000000000   0.75000000000000   Ca   (6h)
   0.00000000000000   0.00000000000000   0.25000000000000   F    (2a)
   0.00000000000000   0.00000000000000   0.75000000000000   F    (2a)
   0.48490000000000   0.32730000000000   0.25000000000000   O    (6h)
  -0.32730000000000   0.15760000000000   0.25000000000000   O    (6h)
  -0.15760000000000  -0.48490000000000   0.25000000000000   O    (6h)
  -0.48490000000000  -0.32730000000000   0.75000000000000   O    (6h)
   0.32730000000000  -0.15760000000000   0.75000000000000   O    (6h)
   0.15760000000000   0.48490000000000   0.75000000000000   O    (6h)
   0.46670000000000   0.58750000000000   0.25000000000000   O    (6h)
  -0.58750000000000  -0.12080000000000   0.25000000000000   O    (6h)
   0.12080000000000  -0.46670000000000   0.25000000000000   O    (6h)
  -0.46670000000000  -0.58750000000000   0.75000000000000   O    (6h)
   0.58750000000000   0.12080000000000   0.75000000000000   O    (6h)
  -0.12080000000000   0.46670000000000   0.75000000000000   O    (6h)
   0.25750000000000   0.34210000000000   0.07050000000000   O    (12i)
  -0.34210000000000  -0.08460000000000   0.07050000000000   O    (12i)
   0.08460000000000  -0.25750000000000   0.07050000000000   O    (12i)
  -0.25750000000000  -0.34210000000000   0.57050000000000   O    (12i)
   0.34210000000000   0.08460000000000   0.57050000000000   O    (12i)
  -0.08460000000000   0.25750000000000   0.57050000000000   O    (12i)
  -0.25750000000000  -0.34210000000000  -0.07050000000000   O    (12i)
   0.34210000000000   0.08460000000000  -0.07050000000000   O    (12i)
  -0.08460000000000   0.25750000000000  -0.07050000000000   O    (12i)
   0.25750000000000   0.34210000000000   0.42950000000000   O    (12i)
  -0.34210000000000  -0.08460000000000   0.42950000000000   O    (12i)
   0.08460000000000  -0.25750000000000   0.42950000000000   O    (12i)
   0.36895000000000   0.39850000000000   0.25000000000000   P    (6h)
  -0.39850000000000  -0.02955000000000   0.25000000000000   P    (6h)
   0.02955000000000  -0.36895000000000   0.25000000000000   P    (6h)
  -0.36895000000000  -0.39850000000000   0.75000000000000   P    (6h)
   0.39850000000000   0.02955000000000   0.75000000000000   P    (6h)
  -0.02955000000000   0.36895000000000   0.75000000000000   P    (6h)
```

$Fe_2(CO)_9$ ($F4_1$): A9B2C9_hP40_176_hi_f_hi - CIF

```
# CIF file
data_findsym-output
_audit_creation_method FINDSYM

_chemical_name_mineral 'C9Fe2O9'
_chemical_formula_sum 'C9 Fe2 O9'

loop_
_publ_author_name
 'F. A. Cotton'
 'J. M. Troup'
_journal_name_full_name
;
 Journal of the Chemical Society, Dalton Transactions
;
_journal_volume
_journal_year 1974
```

```
_journal_page_first 800
_journal_page_last 802
_publ_Section_title
;
 Accurate determination of a classic structure in the metal carbonyl
     ↪ field: nonacarbonyldi-iron
;

# Found in Examining the structural changes in Fe$_{2}$(CO)$_{9}$ under
     ↪ high external pressures by Raman spectroscopy, 2007

_aflow_title 'Fe$_{2}$(CO)$_{9}$ ($F4_{1}$) Structure'
_aflow_proto 'A9B2C9_hP40_176_hi_f_hi'
_aflow_params 'a,c/a,z_{1},x_{2},y_{2},x_{3},y_{3},x_{4},y_{4},z_{4},x_{
     ↪ 5},y_{5},z_{5}'
_aflow_params_values '6.436,2.50512740833,0.17175,0.332,0.9098,0.331,
     ↪ 0.0895,0.338,0.4243,0.1133,0.3441,0.2817,0.0747'
_aflow_Strukturbericht '$F4_{1}$'
_aflow_Pearson 'hP40'

_symmetry_space_group_name_H-M "P 63/m"
_symmetry_Int_Tables_number 176

_cell_length_a     6.43600
_cell_length_b     6.43600
_cell_length_c     16.12300
_cell_angle_alpha  90.00000
_cell_angle_beta   90.00000
_cell_angle_gamma  120.00000

loop_
_space_group_symop_id
_space_group_symop_operation_xyz
1 x,y,z
2 x-y,x,z+1/2
3 -y,x-y,z
4 -x,-y,z+1/2
5 -x+y,-x,z
6 y,-x+y,z+1/2
7 -x,-y,-z
8 -x+y,-x,-z+1/2
9 y,-x+y,-z
10 x,y,-z+1/2
11 x-y,x,-z
12 -y,x-y,-z+1/2

loop_
_atom_site_label
_atom_site_type_symbol
_atom_site_symmetry_multiplicity
_atom_site_Wyckoff_label
_atom_site_fract_x
_atom_site_fract_y
_atom_site_fract_z
_atom_site_occupancy
Fe1 Fe 4 f 0.33333 0.66667 0.17175 1.00000
C1 C 6 h 0.33200 0.90980 0.25000 1.00000
O1 O 6 h 0.33100 0.08950 0.25000 1.00000
C2 C 12 i 0.33800 0.42430 0.11330 1.00000
O2 O 12 i 0.34410 0.28170 0.07470 1.00000
```

$Fe_2(CO)_9$ ($F4_1$): A9B2C9_hP40_176_hi_f_hi - POSCAR

```
A9B2C9_hP40_176_hi_f_hi & a,c/a,z1,x2,y2,x3,y3,x4,y4,z4,x5,y5,z5 --
     ↪ params=6.436,2.50512740833,0.17175,0.332,0.9098,0.331,0.0895,
     ↪ 0.338,0.4243,0.1133,0.3441,0.2817,0.0747 & P6_{3}/m C_{6h}^{2}
     ↪ #176 (fh^2i^2) & hP40 & $F4_{1}$ & C9Fe2O9 & C9Fe2O9 & F. A.
     ↪ Cotton and J. M. Troup, {J. Chem. Soc. Dalton Trans. , 800-802
     ↪ (1974)
   1.00000000000000
   3.21800000000000  -5.57373949875665   0.00000000000000
   3.21800000000000   5.57373949875665   0.00000000000000
   0.00000000000000   0.00000000000000  16.12300000000000
     C      Fe      O
    18       4     18
Direct
   0.33200000000000   0.90980000000000   0.25000000000000   C   (6h)
  -0.90980000000000  -0.57780000000000   0.25000000000000   C   (6h)
   0.57780000000000  -0.33200000000000   0.25000000000000   C   (6h)
  -0.33200000000000  -0.90980000000000   0.75000000000000   C   (6h)
   0.90980000000000   0.57780000000000   0.75000000000000   C   (6h)
  -0.57780000000000   0.33200000000000   0.75000000000000   C   (6h)
   0.33800000000000   0.42430000000000   0.11330000000000   C  (12i)
  -0.42430000000000  -0.08630000000000   0.11330000000000   C  (12i)
   0.08630000000000  -0.33800000000000   0.11330000000000   C  (12i)
  -0.33800000000000  -0.42430000000000   0.61330000000000   C  (12i)
   0.42430000000000   0.08630000000000   0.61330000000000   C  (12i)
  -0.08630000000000   0.33800000000000   0.61330000000000   C  (12i)
  -0.33800000000000  -0.42430000000000  -0.11330000000000   C  (12i)
   0.42430000000000   0.08630000000000  -0.11330000000000   C  (12i)
  -0.08630000000000   0.33800000000000  -0.11330000000000   C  (12i)
   0.33800000000000   0.42430000000000   0.38670000000000   C  (12i)
  -0.42430000000000  -0.08630000000000   0.38670000000000   C  (12i)
   0.08630000000000  -0.33800000000000   0.38670000000000   C  (12i)
   0.33333333333333   0.66666666666667   0.17175000000000   Fe   (4f)
   0.66666666666667   0.33333333333333   0.67175000000000   Fe   (4f)
   0.66666666666667   0.33333333333333  -0.17175000000000   Fe   (4f)
   0.33333333333333   0.66666666666667   0.32825000000000   Fe   (4f)
   0.33100000000000   0.08950000000000   0.25000000000000   O   (6h)
  -0.08950000000000   0.24150000000000   0.25000000000000   O   (6h)
  -0.24150000000000  -0.33100000000000   0.25000000000000   O   (6h)
  -0.33100000000000  -0.08950000000000   0.75000000000000   O   (6h)
   0.08950000000000  -0.24150000000000   0.75000000000000   O   (6h)
   0.24150000000000   0.33100000000000   0.75000000000000   O   (6h)
   0.34410000000000   0.28170000000000   0.07470000000000   O  (12i)
  -0.28170000000000   0.06240000000000   0.07470000000000   O  (12i)
  -0.06240000000000  -0.34410000000000   0.07470000000000   O  (12i)
  -0.34410000000000  -0.28170000000000   0.57470000000000   O  (12i)
   0.28170000000000  -0.06240000000000   0.57470000000000   O  (12i)
   0.06240000000000   0.34410000000000   0.57470000000000   O  (12i)
  -0.34410000000000  -0.28170000000000  -0.07470000000000   O  (12i)
   0.28170000000000  -0.06240000000000  -0.07470000000000   O  (12i)
   0.06240000000000   0.34410000000000  -0.07470000000000   O  (12i)
   0.34410000000000   0.28170000000000   0.42530000000000   O  (12i)
  -0.28170000000000   0.06240000000000   0.42530000000000   O  (12i)
  -0.06240000000000  -0.34410000000000   0.42530000000000   O  (12i)
```

$K_3W_2Cl_9$ ($K7_1$): A9B3C2_hP28_176_hi_af_f - CIF

```
# CIF file
data_findsym-output
_audit_creation_method FINDSYM

_chemical_name_mineral 'Cl9K3W2'
_chemical_formula_sum 'Cl9 K3 W2'

loop_
_publ_author_name
 'W. H. {Watson, Jr.}'
 'J. Waser'
_journal_name_full_name
;
 Acta Cristallographica
;
_journal_volume 11
_journal_year 1958
_journal_page_first 689
_journal_page_last 692
_publ_Section_title
;
 Refinement of the structure of tripotassiumditungsten enneachloride,
     ↪ K$_{3}$W$_{2}$Cl$_{9}$
;

_aflow_title 'K$_{3}$W$_{2}$Cl$_{9}$ ($K7_{1}$) Structure'
_aflow_proto 'A9B3C2_hP28_176_hi_af_f'
_aflow_params 'a,c/a,z_{2},z_{3},x_{4},y_{4},x_{5},y_{5},z_{5}'
_aflow_params_values '7.17,2.26638772664,0.5718,0.3241,0.4588,0.4472,
     ↪ 0.1348,0.3506,0.4074'
_aflow_Strukturbericht '$K7_{1}$'
_aflow_Pearson 'hP28'

_symmetry_space_group_name_H-M "P 63/m"
_symmetry_Int_Tables_number 176

_cell_length_a     7.17000
_cell_length_b     7.17000
_cell_length_c     16.25000
_cell_angle_alpha  90.00000
_cell_angle_beta   90.00000
_cell_angle_gamma  120.00000

loop_
_space_group_symop_id
_space_group_symop_operation_xyz
1 x,y,z
2 x-y,x,z+1/2
3 -y,x-y,z
4 -x,-y,z+1/2
5 -x+y,-x,z
6 y,-x+y,z+1/2
7 -x,-y,-z
8 -x+y,-x,-z+1/2
9 y,-x+y,-z
10 x,y,-z+1/2
11 x-y,x,-z
12 -y,x-y,-z+1/2

loop_
_atom_site_label
_atom_site_type_symbol
_atom_site_symmetry_multiplicity
_atom_site_Wyckoff_label
_atom_site_fract_x
_atom_site_fract_y
_atom_site_fract_z
_atom_site_occupancy
K1 K 2 a 0.00000 0.00000 0.25000 1.00000
K2 K 4 f 0.33333 0.66667 0.57180 1.00000
W1 W 4 f 0.33333 0.66667 0.32410 1.00000
Cl1 Cl 6 h 0.45880 0.44720 0.25000 1.00000
Cl2 Cl 12 i 0.13480 0.35060 0.40740 1.00000
```

$K_3W_2Cl_9$ ($K7_1$): A9B3C2_hP28_176_hi_af_f - POSCAR

```
A9B3C2_hP28_176_hi_af_f & a,c/a,z2,z3,x4,y4,x5,y5,z5 --params=7.17,
     ↪ 2.26638772664,0.5718,0.3241,0.4588,0.4472,0.1348,0.3506,0.4074
     ↪ & P6_{3}/m C_{6h}^{2} #176 (af^2hi) & hP28 & $K7_{1}$ & Cl9K3W2
     ↪ & Cl9K3W2 & W. H. {Watson, Jr.} and J. Waser, Acta Cryst. 11,
     ↪ 689-692 (1958)
   1.00000000000000
   3.58500000000000  -6.20940214513442   0.00000000000000
   3.58500000000000   6.20940214513442   0.00000000000000
   0.00000000000000   0.00000000000000  16.25000000000000
    Cl       K       W
    18       6       4
Direct
   0.45880000000000   0.44720000000000   0.25000000000000   Cl   (6h)
  -0.44720000000000   0.01160000000000   0.25000000000000   Cl   (6h)
  -0.01160000000000  -0.45880000000000   0.25000000000000   Cl   (6h)
```

```
-0.45880000000000  -0.44720000000000   0.75000000000000   Cl   (6h)
 0.44720000000000  -0.01160000000000   0.75000000000000   Cl   (6h)
 0.01160000000000   0.45880000000000   0.75000000000000   Cl   (6h)
 0.13480000000000   0.35060000000000   0.40740000000000   Cl  (12i)
-0.35060000000000  -0.21580000000000   0.40740000000000   Cl  (12i)
 0.21580000000000  -0.13480000000000   0.40740000000000   Cl  (12i)
-0.13480000000000  -0.35060000000000   0.90740000000000   Cl  (12i)
 0.35060000000000   0.21580000000000   0.90740000000000   Cl  (12i)
-0.21580000000000   0.13480000000000   0.90740000000000   Cl  (12i)
-0.13480000000000  -0.35060000000000  -0.40740000000000   Cl  (12i)
 0.35060000000000   0.21580000000000  -0.40740000000000   Cl  (12i)
-0.21580000000000   0.13480000000000  -0.40740000000000   Cl  (12i)
 0.13480000000000   0.35060000000000   0.09260000000000   Cl  (12i)
-0.35060000000000  -0.21580000000000   0.09260000000000   Cl  (12i)
 0.21580000000000  -0.13480000000000   0.09260000000000   Cl  (12i)
 0.00000000000000   0.00000000000000   0.25000000000000   K    (2a)
 0.00000000000000   0.00000000000000   0.75000000000000   K    (2a)
 0.33333333333333   0.66666666666667   0.57180000000000   K    (4f)
 0.66666666666667   0.33333333333333   1.07180000000000   K    (4f)
 0.66666666666667   0.33333333333333  -0.57180000000000   K    (4f)
 0.33333333333333   0.66666666666667  -0.07180000000000   K    (4f)
 0.33333333333333   0.66666666666667   0.32410000000000   W    (4f)
 0.66666666666667   0.33333333333333   0.82410000000000   W    (4f)
 0.66666666666667   0.33333333333333  -0.32410000000000   W    (4f)
 0.33333333333333   0.66666666666667   0.17590000000000   W    (4f)
```

$Hg_2O_2NaI$: A2BCD2_hP18_180_f_c_b_i - CIF

```
# CIF file
data_findsym-output
_audit_creation_method FINDSYM

_chemical_name_mineral 'Hg2INaO2'
_chemical_formula_sum 'Hg2 I Na O2'

loop_
_publ_author_name
 'K. Aurivillius'
_journal_name_full_name
;
 Acta Chemica Scandinavica
;
_journal_volume 18
_journal_year 1964
_journal_page_first 1305
_journal_page_last 1306
_publ_Section_title
;
 Least-Squares Refinement of the Crystal Structures of Orthorhombic HgO
    ↪ and of Hg$_{2}$O$_{2}$NaI
;

_aflow_title 'Hg$_{2}$O$_{2}$NaI Structure'
_aflow_proto 'A2BCD2_hP18_180_f_c_b_i'
_aflow_params 'a,c/a,z_{3},x_{4}'
_aflow_params_values '6.667,1.50802459877,0.3333,0.1521'
_aflow_Strukturbericht 'None'
_aflow_Pearson 'hP18'

_symmetry_space_group_name_H-M "P 62 2 2"
_symmetry_Int_Tables_number 180

_cell_length_a     6.66700
_cell_length_b     6.66700
_cell_length_c     10.05400
_cell_angle_alpha  90.00000
_cell_angle_beta   90.00000
_cell_angle_gamma  120.00000

loop_
_space_group_symop_id
_space_group_symop_operation_xyz
1 x,y,z
2 x-y,x,z+1/3
3 -y,x-y,z+2/3
4 -x,-y,z
5 -x+y,-x,z+1/3
6 y,-x+y,z+2/3
7 x-y,-y,-z
8 x,x-y,-z+1/3
9 y,x,-z+2/3
10 -x+y,y,-z
11 -x,-x+y,-z+1/3
12 -y,-x,-z+2/3

loop_
_atom_site_label
_atom_site_type_symbol
_atom_site_symmetry_multiplicity
_atom_site_Wyckoff_label
_atom_site_fract_x
_atom_site_fract_y
_atom_site_fract_z
_atom_site_occupancy
Na1 Na 3 b 0.00000 0.00000 0.50000 1.00000
I1  I  3 c 0.50000 0.00000 0.00000 1.00000
Hg1 Hg 6 f 0.50000 0.00000 0.33330 1.00000
O1  O  6 i 0.15210 0.30420 0.00000 1.00000
```

$Hg_2O_2NaI$: A2BCD2_hP18_180_f_c_b_i - POSCAR

```
A2BCD2_hP18_180_f_c_b_i & a,c/a,z3,x4 --params=6.667,1.50802459877,
    ↪ 0.3333,0.1521 & P6_{2}22 D_{6}^{4} #180 (bcfi) & hP18 & None &
    ↪ Hg2INaO2 & Hg2INaO2 & K. Aurivillius, Acta Chem. Scand. 18,
    ↪ 1305-1306 (1964)
```

```
   1.00000000000000
   3.33350000000000  -5.77379136703085   0.00000000000000
   3.33350000000000   5.77379136703085   0.00000000000000
   0.00000000000000   0.00000000000000  10.05400000000000
    Hg      I     Na      O
     6      3      3      6
Direct
   0.50000000000000   0.00000000000000   0.33330000000000   Hg   (6f)
   0.00000000000000   0.50000000000000   0.99996666666667   Hg   (6f)
   0.50000000000000   0.50000000000000   0.66663333333333   Hg   (6f)
   0.00000000000000   0.50000000000000   0.33336666666667   Hg   (6f)
   0.50000000000000   0.00000000000000  -0.33330000000000   Hg   (6f)
   0.50000000000000   0.50000000000000   0.00003333333333   Hg   (6f)
   0.50000000000000   0.00000000000000   0.00000000000000   I    (3c)
   0.00000000000000   0.50000000000000   0.66666666666667   I    (3c)
   0.50000000000000   0.50000000000000   0.33333333333333   I    (3c)
   0.00000000000000   0.00000000000000   0.50000000000000   Na   (3b)
   0.00000000000000   0.00000000000000   0.16666666666667   Na   (3b)
   0.00000000000000   0.00000000000000   0.83333333333333   Na   (3b)
   0.15210000000000   0.30420000000000   0.00000000000000   O    (6i)
  -0.30420000000000  -0.15210000000000   0.66666666666667   O    (6i)
   0.15210000000000  -0.15210000000000   0.33333333333333   O    (6i)
  -0.15210000000000  -0.30420000000000   0.00000000000000   O    (6i)
   0.30420000000000   0.15210000000000   0.66666666666667   O    (6i)
  -0.15210000000000   0.15210000000000   0.33333333333333   O    (6i)
```

$BaAl_2O_4$ ($H2_8$): A2BC6_hP18_182_f_b_gh - CIF

```
# CIF file
data_findsym-output
_audit_creation_method FINDSYM

_chemical_name_mineral 'Al2BaO4'
_chemical_formula_sum 'Al2 Ba O6'

loop_
_publ_author_name
 'A. J. Perrotta'
 'J. V. Smith'
_journal_name_full_name
;
 Bulletin de la Societ{\'e} fran{\c{c}}aise de Mineralogie et de
    ↪ Crystallographie
;
_journal_volume 91
_journal_year 1968
_journal_page_first 85
_journal_page_last 87
_publ_Section_title
;
 The Crystal Structure of BaAl$_{2}$O$_{4}$
;

_aflow_title 'BaAl$_{2}$O$_{4}$ ($H2_{8}$) Structure'
_aflow_proto 'A2BC6_hP18_182_f_b_gh'
_aflow_params 'a,c/a,z_{2},x_{3},x_{4}'
_aflow_params_values '5.218,1.68282866999,0.054,0.36,0.371'
_aflow_Strukturbericht '$H2_{8}$'
_aflow_Pearson 'hP18'

_symmetry_space_group_name_H-M "P 63 2 2"
_symmetry_Int_Tables_number 182

_cell_length_a     5.21800
_cell_length_b     5.21800
_cell_length_c     8.78100
_cell_angle_alpha  90.00000
_cell_angle_beta   90.00000
_cell_angle_gamma  120.00000

loop_
_space_group_symop_id
_space_group_symop_operation_xyz
1 x,y,z
2 x-y,x,z+1/2
3 -y,x-y,z
4 -x,-y,z+1/2
5 -x+y,-x,z
6 y,-x+y,z+1/2
7 x-y,-y,-z
8 x,x-y,-z+1/2
9 y,x,-z
10 -x+y,y,-z+1/2
11 -x,-x+y,-z
12 -y,-x,-z+1/2

loop_
_atom_site_label
_atom_site_type_symbol
_atom_site_symmetry_multiplicity
_atom_site_Wyckoff_label
_atom_site_fract_x
_atom_site_fract_y
_atom_site_fract_z
_atom_site_occupancy
Ba1 Ba 2 b 0.00000 0.00000 0.25000 1.00000
Al1 Al 4 f 0.33333 0.66667 0.05400 1.00000
O1  O  6 g 0.36000 0.00000 0.00000 1.00000
O2  O  6 h 0.37100 0.74200 0.25000 0.33333
```

$BaAl_2O_4$ ($H2_8$): A2BC6_hP18_182_f_b_gh - POSCAR

```
A2BC6_hP18_182_f_b_gh & a,c/a,z2,x3,x4 --params=5.218,1.68282866999,
    ↪ 0.054,0.36,0.371 & P6_{3}22 D_{6}^{6} #182 (bfgh) & hP18 & $H2_
```

```
    ↪ {8}$ & Al2BaO4 & Al2BaO4 & A. J. Perrotta and J. V. Smith,
    ↪ Bull. Soc. fr. Min'eral. Crystallogr. 91, 85-87 (1968)
   1.00000000000000
   2.60900000000000   -4.51892055694720    0.00000000000000
   2.60900000000000    4.51892055694720    0.00000000000000
   0.00000000000000    0.00000000000000    8.78100000000000
    Al      Ba      O
     4       2     12
Direct
   0.33333333333333    0.66666666666667    0.05400000000000   Al   (4f)
   0.66666666666667    0.33333333333333    0.55400000000000   Al   (4f)
   0.66666666666667    0.33333333333333   -0.05400000000000   Al   (4f)
   0.33333333333333    0.66666666666667    0.44600000000000   Al   (4f)
   0.00000000000000    0.00000000000000    0.25000000000000   Ba   (2b)
   0.00000000000000    0.00000000000000    0.75000000000000   Ba   (2b)
   0.36000000000000    0.00000000000000    0.00000000000000    O   (6g)
   0.00000000000000    0.36000000000000    0.00000000000000    O   (6g)
  -0.36000000000000   -0.36000000000000    0.00000000000000    O   (6g)
  -0.36000000000000    0.00000000000000    0.50000000000000    O   (6g)
   0.00000000000000   -0.36000000000000    0.50000000000000    O   (6g)
   0.36000000000000    0.36000000000000    0.50000000000000    O   (6g)
   0.37100000000000    0.74200000000000    0.25000000000000    O   (6h)
  -0.74200000000000   -0.37100000000000    0.25000000000000    O   (6h)
   0.37100000000000   -0.37100000000000    0.25000000000000    O   (6h)
  -0.37100000000000   -0.74200000000000    0.75000000000000    O   (6h)
   0.74200000000000    0.37100000000000    0.75000000000000    O   (6h)
  -0.37100000000000    0.37100000000000    0.75000000000000    O   (6h)
```

$E2_3$ ($LiIO_3$) (*obsolete*): ABC3_hP10_182_c_b_g - CIF

```
# CIF file
data_findsym-output
_audit_creation_method FINDSYM

_chemical_name_mineral 'ILiO3'
_chemical_formula_sum 'I Li O3'

loop_
_publ_author_name
 'W. H. Zachariasen'
 'F. A. Barta'
_journal_name_full_name
;
 Physical Review
;
_journal_volume 37
_journal_year 1931
_journal_page_first 1626
_journal_page_last 1630
_publ_Section_title
;
 Crystal Structure of Lithium Iodate
;

# Found in A reinvestigation of the crystal structure of LiIO$_{3}$,
    ↪ 1966

_aflow_title '$E2_{3}$ (LiIO$_{3}$) ({\em{obsolete}}) Structure'
_aflow_proto 'ABC3_hP10_182_c_b_g'
_aflow_params 'a,c/a,x_{3}'
_aflow_params_values '5.469,0.942585481807,0.33333'
_aflow_Strukturbericht '$E2_{3}$'
_aflow_Pearson 'hP10'

_symmetry_space_group_name_H-M "P 63 2 2"
_symmetry_Int_Tables_number 182

_cell_length_a    5.46900
_cell_length_b    5.46900
_cell_length_c    5.15500
_cell_angle_alpha 90.00000
_cell_angle_beta  90.00000
_cell_angle_gamma 120.00000

loop_
_space_group_symop_id
_space_group_symop_operation_xyz
1 x,y,z
2 x-y,x,z+1/2
3 -y,x-y,z
4 -x,-y,z+1/2
5 -x+y,-x,z
6 y,-x+y,z+1/2
7 x-y,-y,-z
8 x,x-y,-z+1/2
9 y,x,-z
10 -x+y,y,-z+1/2
11 -x,-x+y,-z
12 -y,-x,-z+1/2

loop_
_atom_site_label
_atom_site_type_symbol
_atom_site_symmetry_multiplicity
_atom_site_Wyckoff_label
_atom_site_fract_x
_atom_site_fract_y
_atom_site_fract_z
_atom_site_occupancy
Li1 Li 2 b 0.00000 0.00000 0.25000 1.00000
I1 I 2 c 0.33333 0.66667 0.25000 1.00000
O1 O 6 g 0.33333 0.00000 0.00000 1.00000
```

$E2_3$ ($LiIO_3$) (*obsolete*): ABC3_hP10_182_c_b_g - POSCAR

```
ABC3_hP10_182_c_b_g & a,c/a,x3 --params=5.469,0.942585481807,0.33333 &
    ↪ P6_{3}22 D_{6}^{6} #182 (bcg) & hP10 & $E2_{3}$ & ILiO3 & ILiO3
    ↪ & W. H. Zachariasen and F. A. Barta, Phys. Rev. 37, 1626-1630
    ↪ (1931)
   1.00000000000000
   2.73450000000000   -4.73629293329710    0.00000000000000
   2.73450000000000    4.73629293329710    0.00000000000000
   0.00000000000000    0.00000000000000    5.15500000000000
     I      Li      O
     2       2      6
Direct
   0.33333333333333    0.66666666666667    0.25000000000000    I   (2c)
   0.66666666666667    0.33333333333333    0.75000000000000    I   (2c)
   0.00000000000000    0.00000000000000    0.25000000000000   Li   (2b)
   0.00000000000000    0.00000000000000    0.75000000000000   Li   (2b)
   0.33333000000000    0.00000000000000    0.00000000000000    O   (6g)
   0.00000000000000    0.33333000000000    0.00000000000000    O   (6g)
  -0.33333000000000   -0.33333000000000    0.00000000000000    O   (6g)
  -0.33333000000000    0.00000000000000    0.50000000000000    O   (6g)
   0.00000000000000   -0.33333000000000    0.50000000000000    O   (6g)
   0.33333000000000    0.33333000000000    0.50000000000000    O   (6g)
```

$Zn_2Mo_3O_8$: A3B8C2_hP26_186_c_ab2c_2b - CIF

```
# CIF file
data_findsym-output
_audit_creation_method FINDSYM

_chemical_name_mineral 'Mo3O8Zn2'
_chemical_formula_sum 'Mo3 O8 Zn2'

loop_
_publ_author_name
 'G. B. Ansell'
 'L. Katz'
_journal_name_full_name
;
 Acta Cristallographica
;
_journal_volume 21
_journal_year 1966
_journal_page_first 482
_journal_page_last 485
_publ_Section_title
;
 A Refinement of the Crystal Structure of Zinc Molybdenum(IV) Oxide,
    ↪ Zn$_{2}$Mo$_{3}$O$_{8}$
;

# Found in Structure of Kamiokite, 1986

_aflow_title 'Zn$_{2}$Mo$_{3}$O$_{8}$ Structure'
_aflow_proto 'A3B8C2_hP26_186_c_ab2c_2b'
_aflow_params 'a,c/a,z_{1},z_{2},z_{3},z_{4},x_{5},z_{5},x_{6},z_{6},x_{
    ↪ 7},z_{7}'
_aflow_params_values '5.775,1.71688311688,0.3886,0.147,-0.0535,0.5132,
    ↪ 0.1461,0.25,0.4861,0.3639,0.1647,0.6354'
_aflow_Strukturbericht 'None'
_aflow_Pearson 'hP26'

_symmetry_space_group_name_H-M "P 63 m c"
_symmetry_Int_Tables_number 186

_cell_length_a    5.77500
_cell_length_b    5.77500
_cell_length_c    9.91500
_cell_angle_alpha 90.00000
_cell_angle_beta  90.00000
_cell_angle_gamma 120.00000

loop_
_space_group_symop_id
_space_group_symop_operation_xyz
1 x,y,z
2 x-y,x,z+1/2
3 -y,x-y,z
4 -x,-y,z+1/2
5 -x+y,-x,z
6 y,-x+y,z+1/2
7 -x+y,y,z
8 -x,-x+y,z+1/2
9 -y,-x,z
10 x-y,-y,z+1/2
11 x,x-y,z
12 y,x,z+1/2

loop_
_atom_site_label
_atom_site_type_symbol
_atom_site_symmetry_multiplicity
_atom_site_Wyckoff_label
_atom_site_fract_x
_atom_site_fract_y
_atom_site_fract_z
_atom_site_occupancy
O1 O 2 a 0.00000 0.00000 0.38860 1.00000
O2 O 2 b 0.33333 0.66667 0.14700 1.00000
Zn1 Zn 2 b 0.33333 0.66667 -0.05350 1.00000
Zn2 Zn 2 b 0.33333 0.66667 0.51320 1.00000
Mo1 Mo 6 c 0.14610 0.85390 0.25000 1.00000
O3 O 6 c 0.48610 0.51390 0.36390 1.00000
O4 O 6 c 0.16470 0.83530 0.63540 1.00000
```

$Zn_2Mo_3O_8$: A3B8C2_hP26_186_c_ab2c_2b - POSCAR

```
A3B8C2_hP26_186_c_ab2c_2b & a,c/a,z1,z2,z3,z4,x5,z5,x6,z6,x7,z7 --params
    ↪ =5.775,1.71688311688,0.3886,0.147,-0.0535,0.5132,0.1461,0.25,
    ↪ 0.4861,0.3639,0.1647,0.6354 & P6_{3}mc C_{6v}^{4} #186 (ab^3c^3
    ↪ ) & hP26 & None & Mo3O8Zn2 & Mo3O8Zn2 & G. B. Ansell and L.
    ↪ Katz, Acta Cryst. 21, 482-485 (1966)
   1.00000000000000
   2.88750000000000  -5.00129670685513   0.00000000000000
   2.88750000000000   5.00129670685513   0.00000000000000
   0.00000000000000   0.00000000000000   9.91500000000000
    Mo     O    Zn
     6    16     4
Direct
   0.14610000000000  -0.14610000000000   0.25000000000000   Mo   (6c)
   0.14610000000000   0.29220000000000   0.25000000000000   Mo   (6c)
  -0.29220000000000  -0.14610000000000   0.25000000000000   Mo   (6c)
  -0.14610000000000   0.14610000000000   0.75000000000000   Mo   (6c)
  -0.14610000000000  -0.29220000000000   0.75000000000000   Mo   (6c)
   0.29220000000000   0.14610000000000   0.75000000000000   Mo   (6c)
   0.00000000000000   0.00000000000000   0.38860000000000   O    (2a)
   0.00000000000000   0.00000000000000   0.88860000000000   O    (2a)
   0.33333333333333   0.66666666666667   0.14700000000000   O    (2b)
   0.66666666666667   0.33333333333333   0.64700000000000   O    (2b)
   0.48610000000000  -0.48610000000000   0.36390000000000   O    (6c)
   0.48610000000000   0.97220000000000   0.36390000000000   O    (6c)
  -0.97220000000000  -0.48610000000000   0.36390000000000   O    (6c)
  -0.48610000000000   0.48610000000000   0.86390000000000   O    (6c)
  -0.48610000000000  -0.97220000000000   0.86390000000000   O    (6c)
   0.97220000000000   0.48610000000000   0.86390000000000   O    (6c)
   0.16470000000000  -0.16470000000000   0.63540000000000   O    (6c)
   0.16470000000000   0.32940000000000   0.63540000000000   O    (6c)
  -0.32940000000000  -0.16470000000000   0.63540000000000   O    (6c)
  -0.16470000000000   0.16470000000000   1.13540000000000   O    (6c)
  -0.16470000000000  -0.32940000000000   1.13540000000000   O    (6c)
   0.32940000000000   0.16470000000000   1.13540000000000   O    (6c)
   0.33333333333333   0.66666666666667  -0.05350000000000   Zn   (2b)
   0.66666666666667   0.33333333333333   0.44650000000000   Zn   (2b)
   0.33333333333333   0.66666666666667   0.51320000000000   Zn   (2b)
   0.66666666666667   0.33333333333333   1.01320000000000   Zn   (2b)
```

Nd(BrO$_3$)$_3$·9H$_2$O ($G2_2$): A3B9CD9_hP44_186_c_3c_b_cd - CIF

```
# CIF file
data_findsym-output
_audit_creation_method FINDSYM

_chemical_name_mineral 'Neodymium bromate enneahydrate'
_chemical_formula_sum 'Br3 (H2O)9 Nd O9'

loop_
_publ_author_name
 'L. Helmholz'
_journal_name_full_name
;
 Journal of the American Chemical Society
;
_journal_volume 61
_journal_year 1939
_journal_page_first 1544
_journal_page_last 1550
_publ_Section_title
;
 The Crystal Structure of Neodymium Bromate Enneahydrate, Nd(BrO$_{3}$)
    ↪ $_{3}$$\cdot$9H$_{2}$O
;

# Found in Strukturbericht Band VII 1939, 1943

_aflow_title 'Nd(BrO$_{3}$)$_{3}$$\cdot$9H$_{2}$O ($G2_{2}$) Structure'
_aflow_proto 'A3B9CD9_hP44_186_c_3c_b_cd'
_aflow_params 'a,c/a,z_{1},x_{2},z_{2},x_{3},z_{3},x_{4},z_{4},x_{5},z_{
    ↪ 5},x_{6},z_{6},x_{7},y_{7},z_{7}'
_aflow_params_values '11.73,0.576300085251,0.25,0.13,0.73,0.425,0.49,
    ↪ 0.425,0.01,0.215,0.25,0.105,0.53,0.065,0.365,0.75'
_aflow_Strukturbericht '$G2_{2}$'
_aflow_Pearson 'hP44'

_symmetry_space_group_name_H-M "P 63 m c"
_symmetry_Int_Tables_number 186

_cell_length_a     11.73000
_cell_length_b     11.73000
_cell_length_c     6.76000
_cell_angle_alpha  90.00000
_cell_angle_beta   90.00000
_cell_angle_gamma  120.00000

loop_
_space_group_symop_id
_space_group_symop_operation_xyz
1 x,y,z
2 x-y,x,z+1/2
3 -y,x-y,z
4 -x,-y,z+1/2
5 -x+y,-x,z
6 y,-x+y,z+1/2
7 -x+y,y,z
8 -x,-x+y,z+1/2
9 -y,-x,z
10 x-y,-y,z+1/2
11 x,x-y,z
12 y,x,z+1/2

loop_
_atom_site_label
_atom_site_type_symbol
_atom_site_symmetry_multiplicity
_atom_site_Wyckoff_label
_atom_site_fract_x
_atom_site_fract_y
_atom_site_fract_z
_atom_site_occupancy
Nd1 Nd 2 b 0.33333 0.66667 0.25000 1.00000
Br1 Br 6 c 0.13000 0.87000 0.73000 1.00000
H2O1 H2O 6 c 0.42500 0.57500 0.49000 1.00000
H2O2 H2O 6 c 0.42500 0.57500 0.01000 1.00000
H2O3 H2O 6 c 0.21500 0.78500 0.25000 1.00000
O1 O 6 c 0.10500 0.89500 0.53000 1.00000
O2 O 12 d 0.06500 0.36500 0.75000 1.00000
```

Nd(BrO$_3$)$_3$·9H$_2$O ($G2_2$): A3B9CD9_hP44_186_c_3c_b_cd - POSCAR

```
A3B9CD9_hP44_186_c_3c_b_cd & a,c/a,z1,x2,z2,x3,z3,x4,z4,x5,z5,x6,z6,x7,
    ↪ y7,z7 --params=11.73,0.576300085251,0.25,0.13,0.73,0.425,0.49,
    ↪ 0.425,0.01,0.215,0.25,0.105,0.53,0.065,0.365,0.75 & P6_{3}mc C_
    ↪ {6v}^{4} #186 (bc^5d) & hP44 & $G2_{2}$ & Br(H2O)9NdO9 &
    ↪ Neodymium bromate enneahydrate & L. Helmholz, J. Am. Chem. Soc.
    ↪ 61, 1544-1550 (1939)
   1.00000000000000
   5.86500000000000 -10.15847798639150   0.00000000000000
   5.86500000000000  10.15847798639150   0.00000000000000
   0.00000000000000   0.00000000000000   6.76000000000000
    Br    H2O    Nd     O
     6     18     2    18
Direct
   0.13000000000000  -0.13000000000000   0.73000000000000   Br    (6c)
   0.13000000000000   0.26000000000000   0.73000000000000   Br    (6c)
  -0.26000000000000  -0.13000000000000   0.73000000000000   Br    (6c)
  -0.13000000000000   0.13000000000000   1.23000000000000   Br    (6c)
  -0.13000000000000  -0.26000000000000   1.23000000000000   Br    (6c)
   0.26000000000000   0.13000000000000   1.23000000000000   Br    (6c)
   0.42500000000000  -0.42500000000000   0.49000000000000   H2O   (6c)
   0.42500000000000   0.85000000000000   0.49000000000000   H2O   (6c)
  -0.85000000000000  -0.42500000000000   0.49000000000000   H2O   (6c)
  -0.42500000000000   0.42500000000000   0.99000000000000   H2O   (6c)
  -0.42500000000000  -0.85000000000000   0.99000000000000   H2O   (6c)
   0.85000000000000   0.42500000000000   0.99000000000000   H2O   (6c)
   0.42500000000000  -0.42500000000000   0.01000000000000   H2O   (6c)
   0.42500000000000   0.85000000000000   0.01000000000000   H2O   (6c)
  -0.85000000000000  -0.42500000000000   0.01000000000000   H2O   (6c)
  -0.42500000000000   0.42500000000000   0.51000000000000   H2O   (6c)
  -0.42500000000000  -0.85000000000000   0.51000000000000   H2O   (6c)
   0.85000000000000   0.42500000000000   0.51000000000000   H2O   (6c)
   0.21500000000000  -0.21500000000000   0.25000000000000   H2O   (6c)
   0.21500000000000   0.43000000000000   0.25000000000000   H2O   (6c)
  -0.43000000000000  -0.21500000000000   0.25000000000000   H2O   (6c)
  -0.21500000000000   0.21500000000000   0.75000000000000   H2O   (6c)
  -0.21500000000000  -0.43000000000000   0.75000000000000   H2O   (6c)
   0.43000000000000   0.21500000000000   0.75000000000000   H2O   (6c)
   0.33333333333333   0.66666666666667   0.25000000000000   Nd    (2b)
   0.66666666666667   0.33333333333333   0.75000000000000   Nd    (2b)
   0.10500000000000  -0.10500000000000   0.53000000000000   O     (6c)
   0.10500000000000   0.21000000000000   0.53000000000000   O     (6c)
  -0.21000000000000  -0.10500000000000   0.53000000000000   O     (6c)
  -0.10500000000000   0.10500000000000   1.03000000000000   O     (6c)
  -0.10500000000000  -0.21000000000000   1.03000000000000   O     (6c)
   0.21000000000000   0.10500000000000   1.03000000000000   O     (6c)
   0.06500000000000   0.36500000000000   0.75000000000000   O    (12d)
  -0.36500000000000  -0.30000000000000   0.75000000000000   O    (12d)
   0.30000000000000  -0.06500000000000   0.75000000000000   O    (12d)
  -0.06500000000000  -0.36500000000000   1.25000000000000   O    (12d)
   0.36500000000000   0.30000000000000   1.25000000000000   O    (12d)
  -0.30000000000000   0.06500000000000   1.25000000000000   O    (12d)
  -0.36500000000000  -0.06500000000000   0.75000000000000   O    (12d)
   0.30000000000000   0.36500000000000   0.75000000000000   O    (12d)
   0.06500000000000  -0.30000000000000   0.75000000000000   O    (12d)
   0.36500000000000   0.06500000000000   1.25000000000000   O    (12d)
  -0.30000000000000  -0.36500000000000   1.25000000000000   O    (12d)
  -0.06500000000000   0.30000000000000   1.25000000000000   O    (12d)
```

Swedenborgite (NaBe$_4$SbO$_7$, $E9_2$): A4BC7D_hP26_186_ac_b_a2c_b - CIF

```
# CIF file
data_findsym-output
_audit_creation_method FINDSYM

_chemical_name_mineral 'Swedenborgite'
_chemical_formula_sum 'Be4 Na O7 Sb'

loop_
_publ_author_name
 'D. M. C. Huminicki'
 'F. C. Hawthorne'
_journal_name_full_name
;
 Canadian Mineralogist
;
_journal_volume 39
_journal_year 2001
_journal_page_first 153
_journal_page_last 158
_publ_Section_title
;
 Refinement of the Crystal Structure of Swedenborgite
;

_aflow_title 'Swedenborgite (NaBe$_{4}$SbO$_{7}$, $E9_{2}$) Structure'
_aflow_proto 'A4BC7D_hP26_186_ac_b_a2c_b'
_aflow_params 'a,c/a,z_{1},z_{2},z_{3},z_{4},x_{5},z_{5},x_{6},z_{6},x_{
    ↪ 7},z_{7}'
```

```
_aflow_params_values '5.4317,1.63063129407,0.0629,0.3728,0.6245,0.0,
    ↪ 0.1664,0.3126,0.4961,0.3706,0.1616,0.1269'
_aflow_Strukturbericht '$E9_{2}$'
_aflow_Pearson 'hP26'

_symmetry_space_group_name_H-M "P 63 m c"
_symmetry_Int_Tables_number 186

_cell_length_a     5.43170
_cell_length_b     5.43170
_cell_length_c     8.85710
_cell_angle_alpha  90.00000
_cell_angle_beta   90.00000
_cell_angle_gamma  120.00000

loop_
_space_group_symop_id
_space_group_symop_operation_xyz
1 x,y,z
2 x-y,x,z+1/2
3 -y,x-y,z
4 -x,-y,z+1/2
5 -x+y,-x,z
6 y,-x+y,z+1/2
7 -x+y,y,z
8 -x,-x+y,z+1/2
9 -y,-x,z
10 x-y,-y,z+1/2
11 x,x-y,z
12 y,x,z+1/2

loop_
_atom_site_label
_atom_site_type_symbol
_atom_site_symmetry_multiplicity
_atom_site_Wyckoff_label
_atom_site_fract_x
_atom_site_fract_y
_atom_site_fract_z
_atom_site_occupancy
Be1 Be 2 a 0.00000 0.00000 0.06290 1.00000
O1 O 2 a 0.00000 0.00000 0.37280 1.00000
Na1 Na 2 b 0.33333 0.66667 0.62450 1.00000
Sb1 Sb 2 b 0.33333 0.66667 0.00000 1.00000
Be2 Be 6 c 0.16640 0.83360 0.31260 1.00000
O2 O 6 c 0.49610 0.50390 0.37060 1.00000
O3 O 6 c 0.16160 0.83840 0.12690 1.00000
```

Swedenborgite ($NaBe_4SbO_7$, $E9_2$): A4BC7D_hP26_186_ac_b_a2c_b - POSCAR

```
A4BC7D_hP26_186_ac_b_a2c_b & a,c/a,z1,z2,z3,z4,x5,z5,x6,z6,x7,z7 --
    ↪ params=5.4317,1.63063129407,0.0629,0.3728,0.6245,0.0,0.1664,
    ↪ 0.3126,0.4961,0.3706,0.1616,0.1269 & P6_{3}mc C_{6v}^{4} #186 (
    ↪ a^2b^2c^3) & hP26 & $E9_{2}$ & Be4NaO7Sb & Swedenborgite & D.
    ↪ M. C. Huminicki and F. C. Hawthorne, Can. Mineral. 39, 153-158
    ↪ (2001)
   1.00000000000000
   2.71585000000000  -4.70399018573594   0.00000000000000
   2.71585000000000   4.70399018573594   0.00000000000000
   0.00000000000000   0.00000000000000   8.85710000000000
    Be     Na     O     Sb
     8      2    14      2
Direct
   0.00000000000000   0.00000000000000   0.06290000000000   Be   (2a)
   0.00000000000000   0.00000000000000   0.56290000000000   Be   (2a)
   0.16640000000000  -0.16640000000000   0.31260000000000   Be   (6c)
   0.16640000000000   0.33280000000000   0.31260000000000   Be   (6c)
  -0.33280000000000  -0.16640000000000   0.31260000000000   Be   (6c)
  -0.16640000000000   0.16640000000000   0.81260000000000   Be   (6c)
  -0.16640000000000  -0.33280000000000   0.81260000000000   Be   (6c)
   0.33280000000000   0.16640000000000   0.81260000000000   Be   (6c)
   0.33333333333333   0.66666666666667   0.62450000000000   Na   (2b)
   0.66666666666667   0.33333333333333   1.12450000000000   Na   (2b)
   0.00000000000000   0.00000000000000   0.37280000000000   O    (2a)
   0.00000000000000   0.00000000000000   0.87280000000000   O    (2a)
   0.49610000000000  -0.49610000000000   0.37060000000000   O    (6c)
   0.49610000000000   0.99220000000000   0.37060000000000   O    (6c)
  -0.99220000000000  -0.49610000000000   0.37060000000000   O    (6c)
  -0.49610000000000   0.49610000000000   0.87060000000000   O    (6c)
  -0.49610000000000  -0.99220000000000   0.87060000000000   O    (6c)
   0.99220000000000   0.49610000000000   0.87060000000000   O    (6c)
   0.16160000000000  -0.16160000000000   0.12690000000000   O    (6c)
   0.16160000000000   0.32320000000000   0.12690000000000   O    (6c)
  -0.32320000000000  -0.16160000000000   0.12690000000000   O    (6c)
  -0.16160000000000   0.16160000000000   0.62690000000000   O    (6c)
  -0.16160000000000  -0.32320000000000   0.62690000000000   O    (6c)
   0.32320000000000   0.16160000000000   0.62690000000000   O    (6c)
   0.33333333333333   0.66666666666667   0.00000000000000   Sb   (2b)
   0.66666666666667   0.33333333333333   0.50000000000000   Sb   (2b)
```

C27 ($CdI_2$) (*questionable*): AB2_hP6_186_b_ab - CIF

```
# CIF file
data_findsym-output
_audit_creation_method FINDSYM

_chemical_name_mineral 'CdI2'
_chemical_formula_sum 'Cd I2'

loop_
_publ_author_name
 'O. Hassel'
_journal_name_full_name
;
 Zeitschrift f\"{u}r Physikalische Chemie
;
_journal_volume 22B
_journal_year 1933
_journal_page_first 333
_journal_page_last 334
_publ_Section_title
;
 Zur Kristallstruktur des Cadmiumjodids CdJ$_{2}$
;

# Found in Strukturbericht Band III 1933-1935, 1937

_aflow_title '$C27$ (CdI$_{2}$) ({\em{questionable}}) Structure'
_aflow_proto 'AB2_hP6_186_b_ab'
_aflow_params 'a,c/a,z_{1},z_{2},z_{3}'
_aflow_params_values '4.24,3.22405660377,0.375,0.0,0.625'
_aflow_Strukturbericht '$C27$'
_aflow_Pearson 'hP6'

_symmetry_space_group_name_H-M "P 63 m c"
_symmetry_Int_Tables_number 186

_cell_length_a     4.24000
_cell_length_b     4.24000
_cell_length_c     13.67000
_cell_angle_alpha  90.00000
_cell_angle_beta   90.00000
_cell_angle_gamma  120.00000

loop_
_space_group_symop_id
_space_group_symop_operation_xyz
1 x,y,z
2 x-y,x,z+1/2
3 -y,x-y,z
4 -x,-y,z+1/2
5 -x+y,-x,z
6 y,-x+y,z+1/2
7 -x+y,y,z
8 -x,-x+y,z+1/2
9 -y,-x,z
10 x-y,-y,z+1/2
11 x,x-y,z
12 y,x,z+1/2

loop_
_atom_site_label
_atom_site_type_symbol
_atom_site_symmetry_multiplicity
_atom_site_Wyckoff_label
_atom_site_fract_x
_atom_site_fract_y
_atom_site_fract_z
_atom_site_occupancy
I1 I 2 a 0.00000 0.00000 0.37500 1.00000
Cd1 Cd 2 b 0.33333 0.66667 0.00000 1.00000
I2 I 2 b 0.33333 0.66667 0.62500 1.00000
```

C27 ($CdI_2$) (*questionable*): AB2_hP6_186_b_ab - POSCAR

```
AB2_hP6_186_b_ab & a,c/a,z1,z2,z3 --params=4.24,3.22405660377,0.375,0.0,
    ↪ 0.625 & P6_{3}mc C_{6v}^{4} #186 (ab^2) & hP6 & $C27$ & CdI2 &
    ↪ CdI2 & O. Hassel, Z. Phys. Chem. 22B, 333-334 (1933)
   1.00000000000000
   2.12000000000000  -3.67194771204602   0.00000000000000
   2.12000000000000   3.67194771204602   0.00000000000000
   0.00000000000000   0.00000000000000  13.67000000000000
    Cd     I
     2     4
Direct
   0.33333333333333   0.66666666666667   0.00000000000000   Cd   (2b)
   0.66666666666667   0.33333333333333   0.50000000000000   Cd   (2b)
   0.00000000000000   0.00000000000000   0.37500000000000   I    (2a)
   0.00000000000000   0.00000000000000   0.87500000000000   I    (2a)
   0.33333333333333   0.66666666666667   0.62500000000000   I    (2b)
   0.66666666666667   0.33333333333333   1.12500000000000   I    (2b)
```

$LiClO_4 \cdot 3H_2O$ ($H4_{18}$): AB6CD7_hP30_186_b_d_a_b2c - CIF

```
# CIF file
data_findsym-output
_audit_creation_method FINDSYM

_chemical_name_mineral 'ClH6LiO7'
_chemical_formula_sum 'Cl H6 Li O7'

loop_
_publ_author_name
 'J.-O. Lundgren'
 'R. Liminga'
 'R. Tellgren'
_journal_name_full_name
;
 Acta Crystallographica Section B: Structural Science
;
_journal_volume 38
_journal_year 1982
_journal_page_first 15
_journal_page_last 20
_publ_Section_title
;
 Neutron diffraction refinement of pyroelectric lithium perchlorate
    ↪ trihydrate
;
```

```
# Found in The OH stretching frequency in LiClO$_{4}$$\cdot$3H$_{2}$O(s)
    ↪ from ab initio and model potential calculations, 1992

_aflow_title 'LiClO$_{4}$$\cdot$3H$_{2}$O ($H4_{18}$) Structure'
_aflow_proto 'AB6CD7_hP30_186_b_d_a_b2c'
_aflow_params 'a,c/a,z_{1},z_{2},z_{3},x_{4},z_{4},x_{5},z_{5},x_{6},y_{
    ↪ 6},z_{6}'
_aflow_params_values '7.7192,0.706433309151,0.27671,0.5,0.2381,0.56534,
    ↪ 0.08902,0.12232,0.02787,-0.06662,0.26326,0.53382'
_aflow_Strukturbericht '$H4_{18}$'
_aflow_Pearson 'hP30'

_symmetry_space_group_name_H-M "P 63 m c"
_symmetry_Int_Tables_number 186

_cell_length_a     7.71920
_cell_length_b     7.71920
_cell_length_c     5.45310
_cell_angle_alpha  90.00000
_cell_angle_beta   90.00000
_cell_angle_gamma  120.00000

loop_
_space_group_symop_id
_space_group_symop_operation_xyz
1 x,y,z
2 x-y,x,z+1/2
3 -y,x-y,z
4 -x,-y,z+1/2
5 -x+y,-x,z
6 y,-x+y,z+1/2
7 -x+y,y,z
8 -x,-x+y,z+1/2
9 -y,-x,z
10 x-y,-y,z+1/2
11 x,x-y,z
12 y,x,z+1/2

loop_
_atom_site_label
_atom_site_type_symbol
_atom_site_symmetry_multiplicity
_atom_site_Wyckoff_label
_atom_site_fract_x
_atom_site_fract_y
_atom_site_fract_z
_atom_site_occupancy
Li1 Li 2 a 0.00000 0.00000 0.27671 1.00000
Cl1 Cl 2 b 0.33333 0.66667 0.50000 1.00000
O1 O 2 b 0.33333 0.66667 0.23810 1.00000
O2 O 6 c 0.56534 0.43466 0.08902 1.00000
O3 O 6 c 0.12232 0.87768 0.02787 1.00000
H1 H 12 d -0.06662 0.26326 0.53382 1.00000
```

$LiClO_4 \cdot 3H_2O$ ($H4_{18}$): AB6CD7_hP30_186_b_d_a_b2c - POSCAR

```
AB6CD7_hP30_186_b_d_a_b2c & a,c/a,z1,z2,z3,x4,z4,x5,z5,x6,y6,z6 --params
    ↪ =7.7192,0.706433309151,0.27671,0.5,0.2381,0.56534,0.08902,
    ↪ 0.12232,0.02787,-0.06662,0.26326,0.53382 & P6_{3}mc C_{6v}^{4}
    ↪ #186 (ab^2c^2d) & hP30 & $H4_{18}$ & ClH6LiO7 & ClH6LiO7 &
    ↪ J.-O. Lundgren and R. Liminga and R. Tellgren, Acta
    ↪ Crystallogr. Sect. B Struct. Sci. 38, 15-20 (1982)
   1.00000000000000
   3.85960000000000  -6.68502329689284   0.00000000000000
   3.85960000000000   6.68502329689284   0.00000000000000
   0.00000000000000   0.00000000000000   5.45310000000000
    Cl      H      Li      O
     2     12      2      14
Direct
   0.33333333333333   0.66666666666667   0.50000000000000   Cl   (2b)
   0.66666666666667   0.33333333333333   1.00000000000000   Cl   (2b)
  -0.06662000000000   0.26326000000000   0.53382000000000   H   (12d)
  -0.26326000000000  -0.32988000000000   0.53382000000000   H   (12d)
   0.32988000000000   0.06662000000000   0.53382000000000   H   (12d)
   0.06662000000000  -0.26326000000000   1.03382000000000   H   (12d)
   0.26326000000000   0.32988000000000   1.03382000000000   H   (12d)
  -0.32988000000000  -0.06662000000000   1.03382000000000   H   (12d)
  -0.26326000000000   0.06662000000000   0.53382000000000   H   (12d)
   0.32988000000000   0.26326000000000   0.53382000000000   H   (12d)
  -0.06662000000000  -0.32988000000000   0.53382000000000   H   (12d)
   0.26326000000000  -0.06662000000000   1.03382000000000   H   (12d)
  -0.32988000000000  -0.26326000000000   1.03382000000000   H   (12d)
   0.06662000000000   0.32988000000000   1.03382000000000   H   (12d)
   0.00000000000000   0.00000000000000   0.27671000000000   Li   (2a)
   0.00000000000000   0.00000000000000   0.77671000000000   Li   (2a)
   0.33333333333333   0.66666666666667   0.23810000000000   O   (2b)
   0.66666666666667   0.33333333333333   0.73810000000000   O   (2b)
   0.56534000000000  -0.56534000000000   0.08902000000000   O   (6c)
   0.56534000000000   1.13068000000000   0.08902000000000   O   (6c)
  -1.13068000000000  -0.56534000000000   0.08902000000000   O   (6c)
  -0.56534000000000   0.56534000000000   0.58902000000000   O   (6c)
  -0.56534000000000  -1.13068000000000   0.58902000000000   O   (6c)
   1.13068000000000   0.56534000000000   0.58902000000000   O   (6c)
   0.12232000000000  -0.12232000000000   0.02787000000000   O   (6c)
   0.12232000000000   0.24464000000000   0.02787000000000   O   (6c)
  -0.24464000000000  -0.12232000000000   0.02787000000000   O   (6c)
  -0.12232000000000   0.12232000000000   0.52787000000000   O   (6c)
  -0.12232000000000  -0.24464000000000   0.52787000000000   O   (6c)
   0.24464000000000   0.12232000000000   0.52787000000000   O   (6c)
```

Cd(OH)Cl ($E0_3$): ABCD_hP8_186_b_b_a_a - CIF

```
# CIF file
data_findsym-output
_audit_creation_method FINDSYM

_chemical_name_mineral 'CdClHO'
_chemical_formula_sum 'Cd Cl H O'

loop_
_publ_author_name
 'Y. Cudennec'
 'A. Riou'
 'Y. G\\'{e}rault'
 'A. Lecerf'
_journal_name_full_name
;
 Journal of Solid State Chemistry
;
_journal_volume 151
_journal_year 2000
_journal_page_first 308
_journal_page_last 312
_publ_Section_title
;
 Synthesis and Crystal Structures of Cd(OH)Cl and Cu(OH)Cl and
    ↪ Relationship to Brucite Type
;

_aflow_title 'Cd(OH)Cl ($E0_{3}$) Structure'
_aflow_proto 'ABCD_hP8_186_b_b_a_a'
_aflow_params 'a,c/a,z_{1},z_{2},z_{3},z_{4}'
_aflow_params_values '3.6648,2.79155752019,0.18,0.0892,0.0079,0.3433'
_aflow_Strukturbericht '$E0_{3}$'
_aflow_Pearson 'hP8'

_symmetry_space_group_name_H-M "P 63 m c"
_symmetry_Int_Tables_number 186

_cell_length_a     3.66480
_cell_length_b     3.66480
_cell_length_c     10.23050
_cell_angle_alpha  90.00000
_cell_angle_beta   90.00000
_cell_angle_gamma  120.00000

loop_
_space_group_symop_id
_space_group_symop_operation_xyz
1 x,y,z
2 x-y,x,z+1/2
3 -y,x-y,z
4 -x,-y,z+1/2
5 -x+y,-x,z
6 y,-x+y,z+1/2
7 -x+y,y,z
8 -x,-x+y,z+1/2
9 -y,-x,z
10 x-y,-y,z+1/2
11 x,x-y,z
12 y,x,z+1/2

loop_
_atom_site_label
_atom_site_type_symbol
_atom_site_symmetry_multiplicity
_atom_site_Wyckoff_label
_atom_site_fract_x
_atom_site_fract_y
_atom_site_fract_z
_atom_site_occupancy
H1 H 2 a 0.00000 0.00000 0.18000 1.00000
O1 O 2 a 0.00000 0.00000 0.08920 1.00000
Cd1 Cd 2 b 0.33333 0.66667 0.00790 1.00000
Cl1 Cl 2 b 0.33333 0.66667 0.34330 1.00000
```

Cd(OH)Cl ($E0_3$): ABCD_hP8_186_b_b_a_a - POSCAR

```
ABCD_hP8_186_b_b_a_a & a,c/a,z1,z2,z3,z4 --params=3.6648,2.79155752019,
    ↪ 0.18,0.0892,0.0079,0.3433 & P6_{3}mc C_{6v}^{4} #186 (a^2b^2) &
    ↪ hP8 & $E0_{3}$ & CdClHO & CdClHO & Y. Cudennec et al., J.
    ↪ Solid State Chem. 151, 308-312 (2000)
   1.00000000000000
   1.83240000000000  -3.17380989978921   0.00000000000000
   1.83240000000000   3.17380989978921   0.00000000000000
   0.00000000000000   0.00000000000000  10.23050000000000
    Cd     Cl     H      O
     2      2     2      2
Direct
   0.33333333333333   0.66666666666667   0.00790000000000   Cd   (2b)
   0.66666666666667   0.33333333333333   0.50790000000000   Cd   (2b)
   0.33333333333333   0.66666666666667   0.34330000000000   Cl   (2b)
   0.66666666666667   0.33333333333333   0.84330000000000   Cl   (2b)
   0.00000000000000   0.00000000000000   0.18000000000000   H   (2a)
   0.00000000000000   0.00000000000000   0.68000000000000   H   (2a)
   0.00000000000000   0.00000000000000   0.08920000000000   O   (2a)
   0.00000000000000   0.00000000000000   0.58920000000000   O   (2a)
```

Cr-233 Quasi-One-Dimensional Superconductor ($K_2Cr_3As_3$): A3B3C2_hP16_187_jk_jk_ck - CIF

```
# CIF file
data_findsym-output
_audit_creation_method FINDSYM

_chemical_name_mineral 'As3Cr3K2'
_chemical_formula_sum 'As3 Cr3 K2'

loop_
_publ_author_name
 'J.-K. Bao'
```

```
 'J.-Y. Liu'
 'C.-W. Ma'
 'Z.-H. Meng'
 'Z.-T. Tang'
 'Y.-L. Sun'
 'H.-F. Zhai'
 'H. Jiang'
 'H. Bai'
 'C.-M. Feng'
 'Z.-A. Xu'
 'G.-H. Cao'
_journal_name_full_name
;
 Physical Review X
;
_journal_volume 5
_journal_year 2015
_journal_page_first 011013
_journal_page_last 011013
_publ_Section_title
;
 Superconductivity in Quasi-One-Dimensional K$_{2}$Cr$_{3}$As$_{3}$ with
      ↪ Significant Electron Correlations
;

# Found in Superconductivity at 10.4 K in a novel quasi-one-dimensional
      ↪ ternary molybdenum pnictide K$_{2}$Mo$_{3}$As$_{3}$, 2018 Found
      ↪  in Superconductivity at 10.4 K in a novel
      ↪ quasi-one-dimensional ternary molybdenum pnictide K$_{2}$Mo$_{3
      ↪ }$As$_{3}$, {arXiv:1805.05257 [cond-mat.supr-con]},

_aflow_title 'Cr-233 Quasi-One-Dimensional Superconductor (K$_{2}$Cr$_{3
      ↪ }$As$_{3}$) Structure'
_aflow_proto 'A3B3C2_hP16_187_jk_jk_ck'
_aflow_params 'a,c/a,x_{2},x_{3},x_{4},x_{5},x_{6}'
_aflow_params_values '9.9832,0.423751903197,0.8339,0.0898,0.1676,-0.0873
      ↪ ,0.5387'
_aflow_Strukturbericht 'None'
_aflow_Pearson 'hP16'

_symmetry_space_group_name_H-M "P -6 m 2"
_symmetry_Int_Tables_number 187

_cell_length_a     9.98320
_cell_length_b     9.98320
_cell_length_c     4.23040
_cell_angle_alpha  90.00000
_cell_angle_beta   90.00000
_cell_angle_gamma  120.00000

loop_
_space_group_symop_id
_space_group_symop_operation_xyz
1 x,y,z
2 -y,x-y,z
3 -x+y,-x,z
4 x,x-y,-z
5 -x+y,y,-z
6 -y,-x,-z
7 -x+y,-x,-z
8 x,y,-z
9 -y,x-y,-z
10 -x+y,y,z
11 -y,-x,z
12 x,x-y,z

loop_
_atom_site_label
_atom_site_type_symbol
_atom_site_symmetry_multiplicity
_atom_site_Wyckoff_label
_atom_site_fract_x
_atom_site_fract_y
_atom_site_fract_z
_atom_site_occupancy
K1 K 1 c 0.33333 0.66667 0.00000 1.00000
As1 As 3 j 0.83390 0.16610 0.00000 1.00000
Cr1 Cr 3 j 0.08980 -0.08980 0.00000 1.00000
As2 As 3 k 0.16760 0.83240 0.50000 1.00000
Cr2 Cr 3 k -0.08730 0.08730 0.50000 1.00000
K2 K 3 k 0.53870 0.46130 0.50000 1.00000
```

Cr-233 Quasi-One-Dimensional Superconductor ($K_2Cr_3As_3$): A3B3C2_hP16_187_jk_jk_ck - POSCAR

```
A3B3C2_hP16_187_jk_jk_ck & a,c/a,x2,x3,x4,x5,x6 --params=9.9832,
      ↪ 0.423751903197,0.8339,0.0898,0.1676,-0.0873,0.5387 & P-6m2 D_{
      ↪ 3h}^{1} #187 (cj^2k^3) & hP16 & None & As3Cr3K2 & As3Cr3K2 &
      ↪ J.-K. Bao et al., Phys. Rev. X 5, 011013(2015)
   1.00000000000000
   4.99160000000000   -8.64570481106081   0.00000000000000
   4.99160000000000    8.64570481106081   0.00000000000000
   0.00000000000000    0.00000000000000   4.23040000000000
     As      Cr      K
      6       6      4
Direct
   0.83390000000000   -0.83390000000000   0.00000000000000   As   (3j)
   0.83390000000000    1.66780000000000   0.00000000000000   As   (3j)
  -1.66780000000000   -0.83390000000000   0.00000000000000   As   (3j)
   0.16760000000000   -0.16760000000000   0.50000000000000   As   (3k)
   0.16760000000000    0.33520000000000   0.50000000000000   As   (3k)
  -0.33520000000000   -0.16760000000000   0.50000000000000   As   (3k)
   0.08980000000000   -0.08980000000000   0.00000000000000   Cr   (3j)
   0.08980000000000    0.17960000000000   0.00000000000000   Cr   (3j)
  -0.17960000000000   -0.08980000000000   0.00000000000000   Cr   (3j)
  -0.08730000000000    0.08730000000000   0.50000000000000   Cr   (3k)
  -0.08730000000000   -0.17460000000000   0.50000000000000   Cr   (3k)
   0.17460000000000    0.08730000000000   0.50000000000000   Cr   (3k)
   0.33333333333333    0.66666666666667   0.00000000000000   K    (1c)
   0.53870000000000   -0.53870000000000   0.50000000000000   K    (3k)
   0.53870000000000    1.07740000000000   0.50000000000000   K    (3k)
  -1.07740000000000   -0.53870000000000   0.50000000000000   K    (3k)
```

$Cs_7O$: A7B_hP24_187_ai2j2kn_j - CIF

```
# CIF file
data_findsym-output
_audit_creation_method FINDSYM

_chemical_name_mineral 'Cs7O'
_chemical_formula_sum 'Cs7 O'

loop_
_publ_author_name
 'A. Simon'
_journal_name_full_name
;
 Zeitschrift fur Anorganische und Allgemeine Chemie
;
_journal_volume 422
_journal_year 1976
_journal_page_first 208
_journal_page_last 218
_publ_Section_title
;
 \"{U}ber Alkalimetall-Suboxide. VII. Das metallreichste C\"{a}
      ↪ siumoxid-Cs$_{7}$O
;

# Found in Binary Alloy Phase Diagrams, 1990 Found in Binary Alloy Phase
      ↪ Diagrams, {Cd-Ce to Hf-Rb}}

_aflow_title 'Cs$_{7}$O Structure'
_aflow_proto 'A7B_hP24_187_ai2j2kn_j'
_aflow_params 'a,c/a,z_{2},x_{3},x_{4},x_{5},x_{6},x_{7},x_{8},z_{8}'
_aflow_params_values '16.244,0.562977099237,0.20321,0.55085,0.21582,
      ↪ 0.74805,0.44928,0.11229,0.81357,0.21992'
_aflow_Strukturbericht 'None'
_aflow_Pearson 'hP24'

_symmetry_space_group_name_H-M "P -6 m 2"
_symmetry_Int_Tables_number 187

_cell_length_a     16.24400
_cell_length_b     16.24400
_cell_length_c     9.14500
_cell_angle_alpha  90.00000
_cell_angle_beta   90.00000
_cell_angle_gamma  120.00000

loop_
_space_group_symop_id
_space_group_symop_operation_xyz
1 x,y,z
2 -y,x-y,z
3 -x+y,-x,z
4 x,x-y,-z
5 -x+y,y,-z
6 -y,-x,-z
7 -x+y,-x,-z
8 x,y,-z
9 -y,x-y,-z
10 -x+y,y,z
11 -y,-x,z
12 x,x-y,z

loop_
_atom_site_label
_atom_site_type_symbol
_atom_site_symmetry_multiplicity
_atom_site_Wyckoff_label
_atom_site_fract_x
_atom_site_fract_y
_atom_site_fract_z
_atom_site_occupancy
Cs1 Cs 1 a 0.00000 0.00000 0.00000 1.00000
Cs2 Cs 2 i 0.66667 0.33333 0.20321 1.00000
Cs3 Cs 3 j 0.55085 0.44915 0.00000 1.00000
Cs4 Cs 3 j 0.21582 0.78418 0.00000 1.00000
O1 O 3 j 0.74805 0.25195 0.00000 1.00000
Cs5 Cs 3 k 0.44928 0.55072 0.50000 1.00000
Cs6 Cs 3 k 0.11229 0.88771 0.50000 1.00000
Cs7 Cs 6 n 0.81357 0.18643 0.21992 1.00000
```

$Cs_7O$: A7B_hP24_187_ai2j2kn_j - POSCAR

```
A7B_hP24_187_ai2j2kn_j & a,c/a,z2,x3,x4,x5,x6,x7,x8,z8 --params=16.244,
      ↪ 0.562977099237,0.20321,0.55085,0.21582,0.74805,0.44928,0.11229,
      ↪ 0.81357,0.21992 & P-6m2 D_{3h}^{1} #187 (aij^3k^2n) & hP24 &
      ↪ None & Cs7O & Cs7O & A. Simon, Z. Anorg. Allg. Chem. 422,
      ↪ 208-218 (1976)
   1.00000000000000
   8.12200000000000  -14.06771665907440   0.00000000000000
   8.12200000000000   14.06771665907440   0.00000000000000
   0.00000000000000    0.00000000000000   9.14500000000000
     Cs      O
     21      3
Direct
   0.00000000000000    0.00000000000000    0.00000000000000   Cs   (1a)
   0.66666666666667    0.33333333333333    0.20321000000000   Cs   (2i)
   0.66666666666667    0.33333333333333   -0.20321000000000   Cs   (2i)
```

```
   0.55085000000000   -0.55085000000000    0.00000000000000    Cs    (3j)
   0.55085000000000    1.10170000000000    0.00000000000000    Cs    (3j)
  -1.10170000000000   -0.55085000000000    0.00000000000000    Cs    (3j)
   0.21582000000000   -0.21582000000000    0.00000000000000    Cs    (3j)
   0.21582000000000    0.43164000000000    0.00000000000000    Cs    (3j)
  -0.43164000000000   -0.21582000000000    0.00000000000000    Cs    (3j)
   0.44928000000000   -0.44928000000000    0.50000000000000    Cs    (3k)
   0.44928000000000    0.89856000000000    0.50000000000000    Cs    (3k)
  -0.89856000000000   -0.44928000000000    0.50000000000000    Cs    (3k)
   0.11229000000000   -0.11229000000000    0.50000000000000    Cs    (3k)
   0.11229000000000    0.22458000000000    0.50000000000000    Cs    (3k)
  -0.22458000000000   -0.11229000000000    0.50000000000000    Cs    (3k)
   0.81357000000000   -0.81357000000000    0.21992000000000    Cs    (6n)
   0.81357000000000    1.62714000000000    0.21992000000000    Cs    (6n)
  -1.62714000000000   -0.81357000000000    0.21992000000000    Cs    (6n)
   0.81357000000000   -0.81357000000000   -0.21992000000000    Cs    (6n)
   0.81357000000000    1.62714000000000   -0.21992000000000    Cs    (6n)
  -1.62714000000000   -0.81357000000000   -0.21992000000000    Cs    (6n)
   0.74805000000000   -0.74805000000000    0.00000000000000    O     (3j)
   0.74805000000000    1.49610000000000    0.00000000000000    O     (3j)
  -1.49610000000000   -0.74805000000000    0.00000000000000    O     (3j)
```

ZrNiAl: ABC_hP9_189_g_ad_f - CIF

```
# CIF file
data_findsym-output
_audit_creation_method FINDSYM

_chemical_name_mineral 'AlNiZr'
_chemical_formula_sum 'Al Ni Zr'

loop_
_publ_author_name
 'O. Shved'
 'L. P. Salamakha'
 'S. Mudry'
 'O. Sologub'
 'P. F. Rogl'
 'E. Bauer'
_journal_name_full_name
;
 Journal of Alloys and Compounds
;
_journal_volume 821
_journal_year 2020
_journal_page_first 153326
_journal_page_last 153326
_publ_Section_title
;
 Zr-based nickel aluminides: crystal structure and electronic properties
;

_aflow_title 'ZrNiAl Structure'
_aflow_proto 'ABC_hP9_189_g_ad_f'
_aflow_params 'a,c/a,x_{3},x_{4}'
_aflow_params_values '6.91558,0.50180028284,0.59354,0.2491'
_aflow_Strukturbericht 'None'
_aflow_Pearson 'hP9'

_symmetry_space_group_name_H-M "P -6 2 m"
_symmetry_Int_Tables_number 189

_cell_length_a      6.91558
_cell_length_b      6.91558
_cell_length_c      3.47024
_cell_angle_alpha  90.00000
_cell_angle_beta   90.00000
_cell_angle_gamma 120.00000

loop_
_space_group_symop_id
_space_group_symop_operation_xyz
1 x,y,z
2 -y,x-y,z
3 -x+y,-x,z
4 x-y,-y,-z
5 y,x,-z
6 -x,-x+y,-z
7 -x+y,-x,-z
8 x,y,-z
9 -y,x-y,-z
10 -x,-x+y,z
11 x-y,-y,z
12 y,x,z

loop_
_atom_site_label
_atom_site_type_symbol
_atom_site_symmetry_multiplicity
_atom_site_Wyckoff_label
_atom_site_fract_x
_atom_site_fract_y
_atom_site_fract_z
_atom_site_occupancy
Ni1 Ni 1 a 0.00000 0.00000 0.00000 1.00000
Ni2 Ni 2 d 0.33333 0.66667 0.50000 1.00000
Zr1 Zr 3 f 0.59354 0.00000 0.00000 1.00000
Al1 Al 3 g 0.24910 0.00000 0.50000 1.00000
```

ZrNiAl: ABC_hP9_189_g_ad_f - POSCAR

```
ABC_hP9_189_g_ad_f & a,c/a,x3,x4 --params=6.91558,0.50180028284,0.59354,
    ↪ 0.2491 & P-62m D_{3h}^{3} #189 (adfg) & hP9 & None & AlNiZr &
    ↪ AlNiZr & O. Shved et al., J. Alloys Compd. 821, 153326(2020)
   1.00000000000000
   3.45779000000000   -5.98906796190359    0.00000000000000
   3.45779000000000    5.98906796190359    0.00000000000000
   0.00000000000000    0.00000000000000    3.47024000000000
    Al      Ni      Zr
     3       3       3
Direct
   0.24910000000000    0.00000000000000    0.50000000000000    Al    (3g)
   0.00000000000000    0.24910000000000    0.50000000000000    Al    (3g)
  -0.24910000000000   -0.24910000000000    0.50000000000000    Al    (3g)
   0.00000000000000    0.00000000000000    0.00000000000000    Ni    (1a)
   0.33333333333333    0.66666666666667    0.50000000000000    Ni    (2d)
   0.66666666666667    0.33333333333333    0.50000000000000    Ni    (2d)
   0.59354000000000    0.00000000000000    0.00000000000000    Zr    (3f)
   0.00000000000000    0.59354000000000    0.00000000000000    Zr    (3f)
  -0.59354000000000   -0.59354000000000    0.00000000000000    Zr    (3f)
```

$CsSO_3$ ($K1_2$): AB3C_hP20_190_ac_i_f - CIF

```
# CIF file
data_findsym-output
_audit_creation_method FINDSYM

_chemical_name_mineral 'CsO3S'
_chemical_formula_sum 'Cs O3 S'

loop_
_publ_author_name
 'G. H\"{a}gg'
_journal_name_full_name
;
 Zeitschrift f{\"u}r Physikalische Chemie B
;
_journal_volume 18
_journal_year 1932
_journal_page_first 327
_journal_page_last 342
_publ_Section_title
;
 Die Kristallstruktur von Caesiumdithionat, Cs$_{2}$S$_{2}$O$_{6}$
;

# Found in The American Mineralogist Crystal Structure Database, 2003

_aflow_title 'CsSO$_{3}$ ($K1_{2}$) Structure'
_aflow_proto 'AB3C_hP20_190_ac_i_f'
_aflow_params 'a,c/a,z_{3},x_{4},y_{4},z_{4}'
_aflow_params_values '6.326,1.82342712615,0.66,0.44,0.33333,0.125'
_aflow_Strukturbericht '$K1_{2}$'
_aflow_Pearson 'hP20'

_symmetry_space_group_name_H-M "P -6 2 c"
_symmetry_Int_Tables_number 190

_cell_length_a      6.32600
_cell_length_b      6.32600
_cell_length_c      11.53500
_cell_angle_alpha  90.00000
_cell_angle_beta   90.00000
_cell_angle_gamma 120.00000

loop_
_space_group_symop_id
_space_group_symop_operation_xyz
1 x,y,z
2 -y,x-y,z
3 -x+y,-x,z
4 x-y,-y,-z
5 y,x,-z
6 -x,-x+y,-z
7 -x+y,-x,-z+1/2
8 x,y,-z+1/2
9 -y,x-y,-z+1/2
10 -x,-x+y,z+1/2
11 x-y,-y,z+1/2
12 y,x,z+1/2

loop_
_atom_site_label
_atom_site_type_symbol
_atom_site_symmetry_multiplicity
_atom_site_Wyckoff_label
_atom_site_fract_x
_atom_site_fract_y
_atom_site_fract_z
_atom_site_occupancy
Cs1 Cs 2 a 0.00000 0.00000 0.00000 1.00000
Cs2 Cs 2 c 0.33333 0.66667 0.25000 1.00000
S1 S 4 f 0.33333 0.66667 0.66000 1.00000
O1 O 12 i 0.44000 0.33333 0.12500 1.00000
```

$CsSO_3$ ($K1_2$): AB3C_hP20_190_ac_i_f - POSCAR

```
AB3C_hP20_190_ac_i_f & a,c/a,z3,x4,y4,z4 --params=6.326,1.82342712615,
    ↪ 0.66,0.44,0.33333,0.125 & P-62c D_{3h}^{4} #190 (acfi) & hP20 &
    ↪ $K1_{2}$ & CsO3S & CsO3S & G. H\"{a}gg, Z. Physik. Chem. B 18,
    ↪ 327-342 (1932)
   1.00000000000000
   3.16300000000000   -5.47847670434036    0.00000000000000
   3.16300000000000    5.47847670434036    0.00000000000000
   0.00000000000000    0.00000000000000   11.53500000000000
    Cs      O       S
     4     12       4
Direct
   0.00000000000000    0.00000000000000    0.00000000000000    Cs    (2a)
   0.00000000000000    0.00000000000000    0.50000000000000    Cs    (2a)
```

```
  0.33333333333333   0.66666666666667   0.25000000000000   Cs    (2c)
  0.66666666666667   0.33333333333333   0.75000000000000   Cs    (2c)
  0.44000000000000   0.33333000000000   0.12500000000000   O    (12i)
 -0.33333000000000   0.10667000000000   0.12500000000000   O    (12i)
 -0.10667000000000  -0.44000000000000   0.12500000000000   O    (12i)
  0.44000000000000   0.33333000000000   0.37500000000000   O    (12i)
 -0.33333000000000   0.10667000000000   0.37500000000000   O    (12i)
 -0.10667000000000  -0.44000000000000   0.37500000000000   O    (12i)
  0.33333000000000   0.44000000000000  -0.12500000000000   O    (12i)
  0.10667000000000  -0.33333000000000  -0.12500000000000   O    (12i)
 -0.44000000000000  -0.10667000000000  -0.12500000000000   O    (12i)
  0.33333000000000   0.44000000000000   0.62500000000000   O    (12i)
  0.10667000000000  -0.33333000000000   0.62500000000000   O    (12i)
 -0.44000000000000  -0.10667000000000   0.62500000000000   O    (12i)
  0.33333333333333   0.66666666666667   0.66000000000000   S     (4f)
  0.33333333333333   0.66666666666667  -0.16000000000000   S     (4f)
  0.66666666666667   0.33333333333333  -0.66000000000000   S     (4f)
  0.66666666666667   0.33333333333333   1.16000000000000   S     (4f)
```

Bastnäsite [$CeF(CO_3)$]: ABCD3_hP36_190_h_g_af_hi - CIF

```
# CIF file
data_findsym-output
_audit_creation_method FINDSYM

_chemical_name_mineral 'Bastn\"{a}site'
_chemical_formula_sum 'C Ce F O3'

loop_
_publ_author_name
 'Y. Ni'
 'J. M. Hughes'
 'A. N. Mariano'
_journal_name_full_name
;
 American Mineralogist
;
_journal_volume 78
_journal_year 1993
_journal_page_first 415
_journal_page_last 418
_publ_Section_title
;
 The atomic arrangement of bastn\"{a}site-(Ce), Ce(CO$_{3}$)F, and
    ↪ structural elements of synchysite-(Ce), r{\"o}ntgenite-(Ce),
    ↪ and parisite-(Ce)
;

_aflow_title 'Bastn\"{a}site [CeF(CO$_{3}$)] Structure'
_aflow_proto 'ABCD3_hP36_190_h_g_af_hi'
_aflow_params 'a,c/a,z_{2},x_{3},x_{4},y_{4},x_{5},y_{5},x_{6},y_{6},z_{
    ↪ 6}'
_aflow_params_values '7.1175,1.37153494907,0.449,0.33941,0.032,0.71,
    ↪ 0.207,0.891,0.3245,0.3828,0.6354'
_aflow_Strukturbericht '$G7_{1}$'
_aflow_Pearson 'hP36'

_symmetry_space_group_name_H-M "P -6 2 c"
_symmetry_Int_Tables_number 190

_cell_length_a     7.11750
_cell_length_b     7.11750
_cell_length_c     9.76190
_cell_angle_alpha  90.00000
_cell_angle_beta   90.00000
_cell_angle_gamma  120.00000

loop_
_space_group_symop_id
_space_group_symop_operation_xyz
1 x,y,z
2 -y,x-y,z
3 -x+y,-x,z
4 x-y,-y,-z
5 y,x,-z
6 -x,-x+y,-z
7 -x+y,-x,-z+1/2
8 x,y,-z+1/2
9 -y,x-y,-z+1/2
10 -x,-x+y,z+1/2
11 x-y,-y,z+1/2
12 y,x,z+1/2

loop_
_atom_site_label
_atom_site_type_symbol
_atom_site_symmetry_multiplicity
_atom_site_Wyckoff_label
_atom_site_fract_x
_atom_site_fract_y
_atom_site_fract_z
_atom_site_occupancy
F1 F 2 a 0.00000 0.00000 0.00000 1.00000
F2 F 4 f 0.33333 0.66667 0.44900 1.00000
Ce1 Ce 6 g 0.33941 0.00000 0.00000 1.00000
C1 C 6 h 0.03200 0.71000 0.25000 1.00000
O1 O 6 h 0.20700 0.89100 0.25000 1.00000
O2 O 12 i 0.32450 0.38280 0.63540 1.00000
```

Bastnäsite [$CeF(CO_3)$]: ABCD3_hP36_190_h_g_af_hi - POSCAR

```
ABCD3_hP36_190_h_g_af_hi & a,c/a,z2,x3,x4,y4,x5,y5,x6,y6,z6 --params=
    ↪ 7.1175,1.37153494907,0.449,0.33941,0.032,0.71,0.207,0.891,
    ↪ 0.3245,0.3828,0.6354 & P-62c D_{3h}^{4} #190 (afgh^2i) & hP36 &
    ↪ $G7_{1}$ & CCeFO3 & Bastn\"{a}site & Y. Ni and J. M. Hughes
    ↪ and A. N. Mariano, Am. Mineral. 78, 415-418 (1993)
  1.00000000000000
  3.55875000000000  -6.16393581143574   0.00000000000000
  3.55875000000000   6.16393581143574   0.00000000000000
  0.00000000000000   0.00000000000000   9.76190000000000
     C     Ce      F      O
     6      6      6     18
Direct
  0.03200000000000   0.71000000000000   0.25000000000000   C    (6h)
 -0.71000000000000  -0.67800000000000   0.25000000000000   C    (6h)
  0.67800000000000  -0.03200000000000   0.25000000000000   C    (6h)
  0.71000000000000   0.03200000000000   0.75000000000000   C    (6h)
 -0.67800000000000  -0.71000000000000   0.75000000000000   C    (6h)
 -0.03200000000000   0.67800000000000   0.75000000000000   C    (6h)
  0.33941000000000   0.00000000000000   0.00000000000000   Ce   (6g)
  0.00000000000000   0.33941000000000   0.00000000000000   Ce   (6g)
 -0.33941000000000  -0.33941000000000   0.00000000000000   Ce   (6g)
  0.33941000000000   0.00000000000000   0.50000000000000   Ce   (6g)
  0.00000000000000   0.33941000000000   0.50000000000000   Ce   (6g)
 -0.33941000000000  -0.33941000000000   0.50000000000000   Ce   (6g)
  0.00000000000000   0.00000000000000   0.00000000000000   F    (2a)
  0.00000000000000   0.00000000000000   0.50000000000000   F    (2a)
  0.33333333333333   0.66666666666667   0.44900000000000   F    (4f)
  0.33333333333333   0.66666666666667   0.05100000000000   F    (4f)
  0.66666666666667   0.33333333333333  -0.44900000000000   F    (4f)
  0.66666666666667   0.33333333333333   0.94900000000000   F    (4f)
  0.20700000000000   0.89100000000000   0.25000000000000   O    (6h)
 -0.89100000000000  -0.68400000000000   0.25000000000000   O    (6h)
  0.68400000000000  -0.20700000000000   0.25000000000000   O    (6h)
  0.89100000000000   0.20700000000000   0.75000000000000   O    (6h)
 -0.68400000000000  -0.89100000000000   0.75000000000000   O    (6h)
 -0.20700000000000   0.68400000000000   0.75000000000000   O    (6h)
  0.32450000000000   0.38280000000000   0.63540000000000   O   (12i)
 -0.38280000000000  -0.05830000000000   0.63540000000000   O   (12i)
  0.05830000000000  -0.32450000000000   0.63540000000000   O   (12i)
  0.32450000000000   0.38280000000000  -0.13540000000000   O   (12i)
 -0.38280000000000  -0.05830000000000  -0.13540000000000   O   (12i)
  0.05830000000000  -0.32450000000000  -0.13540000000000   O   (12i)
  0.38280000000000   0.32450000000000  -0.63540000000000   O   (12i)
 -0.05830000000000  -0.38280000000000  -0.63540000000000   O   (12i)
 -0.32450000000000   0.05830000000000  -0.63540000000000   O   (12i)
  0.38280000000000   0.32450000000000   1.13540000000000   O   (12i)
 -0.05830000000000  -0.38280000000000   1.13540000000000   O   (12i)
 -0.32450000000000   0.05830000000000   1.13540000000000   O   (12i)
```

$TiBe_{12}$ (approximate, $D2_a$): A12B_hP13_191_cdei_a - CIF

```
# CIF file
data_findsym-output
_audit_creation_method FINDSYM

_chemical_name_mineral 'Be12Ti'
_chemical_formula_sum 'Be12 Ti'

loop_
_publ_author_name
 'R. F. Raeuchle'
 'R. E. Rundle'
_journal_name_full_name
;
 Acta Cristallographica
;
_journal_volume 5
_journal_year 1952
_journal_page_first 85
_journal_page_last 93
_publ_Section_title
;
 The Structure of TiBe$_{12}$
;

# Found in Resolving the structure of TiBe$_{12}$, 2016

_aflow_title 'TiBe$_{12}$ (approximate, $D2_{a}$) Structure'
_aflow_proto 'A12B_hP13_191_cdei_a'
_aflow_params 'a,c/a,z_{4},z_{5}'
_aflow_params_values '4.23,1.73286052009,0.29,0.25'
_aflow_Strukturbericht '$D2_{a}$'
_aflow_Pearson 'hP13'

_symmetry_space_group_name_H-M "P 6/m 2/m 2/m"
_symmetry_Int_Tables_number 191

_cell_length_a     4.23000
_cell_length_b     4.23000
_cell_length_c     7.33000
_cell_angle_alpha  90.00000
_cell_angle_beta   90.00000
_cell_angle_gamma  120.00000

loop_
_space_group_symop_id
_space_group_symop_operation_xyz
1 x,y,z
2 x-y,x,z
3 -y,x-y,z
4 -x,-y,z
5 -x+y,-x,z
6 y,-x+y,z
7 x-y,-y,-z
8 x,x-y,-z
9 y,x,-z
10 -x+y,y,-z
11 -x,-x+y,-z
```

```
12 -y,-x,-z
13 -x,-y,-z
14 -x+y,-x,-z
15 y,-x+y,-z
16 x,y,-z
17 x-y,x,-z
18 -y,x-y,-z
19 -x+y,y,z
20 -x,-x+y,z
21 -y,-x,z
22 x-y,-y,z
23 x,x-y,z
24 y,x,z

loop_
_atom_site_label
_atom_site_type_symbol
_atom_site_symmetry_multiplicity
_atom_site_Wyckoff_label
_atom_site_fract_x
_atom_site_fract_y
_atom_site_fract_z
_atom_site_occupancy
Ti1 Ti 1 a 0.00000 0.00000 0.00000 1.00000
Be1 Be 2 c 0.33333 0.66667 0.00000 1.00000
Be2 Be 2 d 0.33333 0.66667 0.50000 1.00000
Be3 Be 2 e 0.00000 0.00000 0.29000 1.00000
Be4 Be 6 i 0.50000 0.00000 0.25000 1.00000
```

$TiBe_{12}$ (approximate, $D2_a$): A12B_hP13_191_cdei_a - POSCAR

```
A12B_hP13_191_cdei_a & a,c/a,z4,z5 --params=4.23,1.73286052009,0.29,0.25
    ↪ & P6/mmm D_{6h}^{1} #191 (acdei) & hP13 & $D2_{a}$ & Be12Ti &
    ↪ Be12Ti & R. F. Raeuchle and R. E. Rundle, Acta Cryst. 5, 85-93
    ↪ (1952)
   1.00000000000000
   2.11500000000000  -3.66328745800818   0.00000000000000
   2.11500000000000   3.66328745800818   0.00000000000000
   0.00000000000000   0.00000000000000   7.33000000000000
    Be     Ti
    12      1
Direct
   0.33333333333333   0.66666666666667   0.00000000000000   Be   (2c)
   0.66666666666667   0.33333333333333   0.00000000000000   Be   (2c)
   0.33333333333333   0.66666666666667   0.50000000000000   Be   (2d)
   0.66666666666667   0.33333333333333   0.50000000000000   Be   (2d)
   0.00000000000000   0.00000000000000   0.29000000000000   Be   (2e)
   0.00000000000000   0.00000000000000  -0.29000000000000   Be   (2e)
   0.50000000000000   0.00000000000000   0.25000000000000   Be   (6i)
   0.00000000000000   0.50000000000000   0.25000000000000   Be   (6i)
   0.50000000000000   0.50000000000000   0.25000000000000   Be   (6i)
   0.00000000000000   0.50000000000000  -0.25000000000000   Be   (6i)
   0.50000000000000   0.00000000000000  -0.25000000000000   Be   (6i)
   0.50000000000000   0.50000000000000  -0.25000000000000   Be   (6i)
   0.00000000000000   0.00000000000000   0.00000000000000   Ti   (1a)
```

Hexagonal $WO_3$: A3B_hP12_191_gl_f - CIF

```
# CIF file
data_findsym-output
_audit_creation_method FINDSYM

_chemical_name_mineral 'O3W'
_chemical_formula_sum 'O3 W'

loop_
_publ_author_name
 'P. M. Woodward'
 'A. W. Sleight'
 'T. Vogt'
_journal_name_full_name
;
 Journal of Solid State Chemistry
;
_journal_volume 131
_journal_year 1997
_journal_page_first 9
_journal_page_last 17
_publ_Section_title
;
 Ferroelectric Tungsten Trioxide
;

_aflow_title 'Hexagonal WO$_{3}$ Structure'
_aflow_proto 'A3B_hP12_191_gl_f'
_aflow_params 'a,c/a,x_{3}'
_aflow_params_values '7.298,0.534255960537,0.212'
_aflow_Strukturbericht 'None'
_aflow_Pearson 'hP12'

_symmetry_space_group_name_H-M "P 6/m 2/m 2/m"
_symmetry_Int_Tables_number 191

_cell_length_a      7.29800
_cell_length_b      7.29800
_cell_length_c      3.89900
_cell_angle_alpha   90.00000
_cell_angle_beta    90.00000
_cell_angle_gamma   120.00000

loop_
_space_group_symop_id
_space_group_symop_operation_xyz
1 x,y,z
2 x-y,x,z
3 -y,x-y,z
4 -x,-y,z
5 -x+y,-x,z
6 y,-x+y,z
7 x-y,-y,-z
8 x,x-y,-z
9 y,x,-z
10 -x+y,y,-z
11 -x,-x+y,-z
12 -y,-x,-z
13 -x,-y,-z
14 -x+y,-x,-z
15 y,-x+y,-z
16 x,y,-z
17 x-y,x,-z
18 -y,x-y,-z
19 -x+y,y,z
20 -x,-x+y,z
21 -y,-x,z
22 x-y,-y,z
23 x,x-y,z
24 y,x,z

loop_
_atom_site_label
_atom_site_type_symbol
_atom_site_symmetry_multiplicity
_atom_site_Wyckoff_label
_atom_site_fract_x
_atom_site_fract_y
_atom_site_fract_z
_atom_site_occupancy
W1 W 3 f 0.50000 0.00000 0.00000 1.00000
O1 O 3 g 0.50000 0.00000 0.50000 1.00000
O2 O 6 l 0.21200 0.42400 0.00000 1.00000
```

Hexagonal $WO_3$: A3B_hP12_191_gl_f - POSCAR

```
A3B_hP12_191_gl_f & a,c/a,x3 --params=7.298,0.534255960537,0.212 & P6/
    ↪ mmm D_{6h}^{1} #191 (fgl) & hP12 & None & O3W & O3W & P. M.
    ↪ Woodward and A. W. Sleight and T. Vogt, J. Solid State Chem.
    ↪ 131, 9-17 (1997)
   1.00000000000000
   3.64900000000000  -6.32025339681883   0.00000000000000
   3.64900000000000   6.32025339681883   0.00000000000000
   0.00000000000000   0.00000000000000   3.89900000000000
    O      W
    9      3
Direct
   0.50000000000000   0.00000000000000   0.50000000000000   O   (3g)
   0.00000000000000   0.50000000000000   0.50000000000000   O   (3g)
   0.50000000000000   0.50000000000000   0.50000000000000   O   (3g)
   0.21200000000000   0.42400000000000   0.00000000000000   O   (6l)
  -0.42400000000000  -0.21200000000000   0.00000000000000   O   (6l)
   0.21200000000000  -0.21200000000000   0.00000000000000   O   (6l)
  -0.21200000000000  -0.42400000000000   0.00000000000000   O   (6l)
   0.42400000000000   0.21200000000000   0.00000000000000   O   (6l)
  -0.21200000000000   0.21200000000000   0.00000000000000   O   (6l)
   0.50000000000000   0.00000000000000   0.00000000000000   W   (3f)
   0.00000000000000   0.50000000000000   0.00000000000000   W   (3f)
   0.50000000000000   0.50000000000000   0.00000000000000   W   (3f)
```

$D0_6$ (Tysonite, $LaF_3$) (*obsolete*): A3B_hP24_193_ack_g - CIF

```
# CIF file
data_findsym-output
_audit_creation_method FINDSYM

_chemical_name_mineral 'Tysonite'
_chemical_formula_sum 'F3 La'

loop_
_publ_author_name
 'I. Oftedal'
_journal_name_full_name
;
 Zeitschrift f{\"u}r Physik B Condensed Matter
;
_journal_volume 13
_journal_year 1931
_journal_page_first 190
_journal_page_last 200
_publ_Section_title
;
 Zur Kristallstruktur von Tysonit (Ce, La, ...)F$_{3}$
;

# Found in Strukturbericht Band II 1928-1932, 1937

_aflow_title '$D0_{6}$ (Tysonite, LaF$_{3}$) ({\em{obsolete}})
    ↪ Structure'
_aflow_proto 'A3B_hP24_193_ack_g'
_aflow_params 'a,c/a,x_{3},x_{4},z_{4}'
_aflow_params_values '7.12,1.02247191011,0.34,0.33333,0.075'
_aflow_Strukturbericht '$D0_{6}$'
_aflow_Pearson 'hP24'

_symmetry_space_group_name_H-M "P 63/m 2/c 2/m"
_symmetry_Int_Tables_number 193

_cell_length_a      7.12000
_cell_length_b      7.12000
_cell_length_c      7.28000
_cell_angle_alpha   90.00000
_cell_angle_beta    90.00000
```

```
_cell_angle_gamma  120.00000

loop_
_space_group_symop_id
_space_group_symop_operation_xyz
1 x,y,z
2 x-y,x,z+1/2
3 -y,x-y,z
4 -x,-y,z+1/2
5 -x+y,-x,z
6 y,-x+y,z+1/2
7 x-y,-y,-z+1/2
8 x,x-y,-z
9 y,x,-z+1/2
10 -x+y,y,-z
11 -x,-x+y,-z+1/2
12 -y,-x,-z
13 -x,-y,-z
14 -x+y,-x,-z+1/2
15 y,-x+y,-z
16 x,y,-z+1/2
17 x-y,x,-z
18 -y,x-y,-z+1/2
19 -x+y,y,z+1/2
20 -x,-x+y,z
21 -y,-x,z+1/2
22 x-y,-y,z
23 x,x-y,z+1/2
24 y,x,z

loop_
_atom_site_label
_atom_site_type_symbol
_atom_site_symmetry_multiplicity
_atom_site_Wyckoff_label
_atom_site_fract_x
_atom_site_fract_y
_atom_site_fract_z
_atom_site_occupancy
F1 F 2 a 0.00000 0.00000 0.25000 1.00000
F2 F 4 c 0.33333 0.66667 0.25000 1.00000
La1 La 6 g 0.34000 0.00000 0.25000 1.00000
F3 F 12 k 0.33333 0.00000 0.07500 1.00000
```

$D0_6$ (Tysonite, $LaF_3$) (*obsolete*): A3B_hP24_193_ack_g - POSCAR

```
A3B_hP24_193_ack_g & a,c/a,x3,x4,z4 --params=7.12,1.02247191011,0.34,
    ↪ 0.33333,0.075 & P6_{3}/mcm D_{6h}^{3} #193 (acgk) & hP24 & $D0_
    ↪ {6}$ & F3La & Tysonite & I. Oftedal, Z. Phys. B: Condens.
    ↪ Matter 13, 190-200 (1931)
   1.00000000000000
   3.56000000000000  -6.16610087494520   0.00000000000000
   3.56000000000000   6.16610087494520   0.00000000000000
   0.00000000000000   0.00000000000000   7.28000000000000
     F     La
    18      6
Direct
   0.00000000000000   0.00000000000000   0.25000000000000    F    (2a)
   0.00000000000000   0.00000000000000   0.75000000000000    F    (2a)
   0.33333333333333   0.66666666666667   0.25000000000000    F    (4c)
   0.66666666666667   0.33333333333333   0.75000000000000    F    (4c)
   0.66666666666667   0.33333333333333   0.25000000000000    F    (4c)
   0.33333333333333   0.66666666666667   0.75000000000000    F    (4c)
   0.33333000000000   0.00000000000000   0.07500000000000    F   (12k)
   0.00000000000000   0.33333000000000   0.07500000000000    F   (12k)
  -0.33333000000000  -0.33333000000000   0.07500000000000    F   (12k)
  -0.33333000000000   0.00000000000000   0.57500000000000    F   (12k)
   0.00000000000000  -0.33333000000000   0.57500000000000    F   (12k)
   0.33333000000000   0.33333000000000   0.57500000000000    F   (12k)
   0.00000000000000   0.33333000000000   0.42500000000000    F   (12k)
   0.33333000000000   0.00000000000000   0.42500000000000    F   (12k)
  -0.33333000000000  -0.33333000000000   0.42500000000000    F   (12k)
   0.00000000000000  -0.33333000000000  -0.07500000000000    F   (12k)
  -0.33333000000000   0.00000000000000  -0.07500000000000    F   (12k)
   0.33333000000000   0.33333000000000  -0.07500000000000    F   (12k)
   0.34000000000000   0.00000000000000   0.25000000000000   La    (6g)
   0.00000000000000   0.34000000000000   0.25000000000000   La    (6g)
  -0.34000000000000  -0.34000000000000   0.25000000000000   La    (6g)
  -0.34000000000000   0.00000000000000   0.75000000000000   La    (6g)
   0.00000000000000  -0.34000000000000   0.75000000000000   La    (6g)
   0.34000000000000   0.34000000000000   0.75000000000000   La    (6g)
```

$Ti_5Ga_4$: A4B5_hP18_193_bg_dg - CIF

```
# CIF file
data_findsym-output
_audit_creation_method FINDSYM

_chemical_name_mineral 'Ga4Ti5'
_chemical_formula_sum 'Ga4 Ti5'

loop_
_publ_author_name
 'K. Schubert'
 'H. G. Meissner'
 'M. P\"{o}tzschke'
 'W. Rossteutscher'
 'E. Stolz'
_journal_name_full_name
;
 Naturwissenschaften
;
_journal_volume 49
_journal_year 1962
_journal_page_first 57
_journal_page_last 57
_publ_Section_title
;
 Einige Strukturdaten metallischer Phasen (7)
;

# Found in Ti$_{5}$Ga$_{4}$ Crystal Structure, 2016 Found in Ti$_{5}
    ↪ $Ga$_{4}$ Crystal Structure, {PAULING FILE in: Inorganic Solid
    ↪ Phases, SpringerMaterials (online database), Springer,
    ↪ Heidelberg (ed.) SpringerMaterials },

_aflow_title 'Ti$_{5}$Ga$_{4}$ Structure'
_aflow_proto 'A4B5_hP18_193_bg_dg'
_aflow_params 'a,c/a,x_{3},x_{4}'
_aflow_params_values '7.861,0.693550438875,0.62,0.29'
_aflow_Strukturbericht 'None'
_aflow_Pearson 'hP18'

_symmetry_space_group_name_H-M "P 63/m 2/c 2/m"
_symmetry_Int_Tables_number 193

_cell_length_a    7.86100
_cell_length_b    7.86100
_cell_length_c    5.45200
_cell_angle_alpha 90.00000
_cell_angle_beta  90.00000
_cell_angle_gamma 120.00000

loop_
_space_group_symop_id
_space_group_symop_operation_xyz
1 x,y,z
2 x-y,x,z+1/2
3 -y,x-y,z
4 -x,-y,z+1/2
5 -x+y,-x,z
6 y,-x+y,z+1/2
7 x-y,-y,-z+1/2
8 x,x-y,-z
9 y,x,-z+1/2
10 -x+y,y,-z
11 -x,-x+y,-z+1/2
12 -y,-x,-z
13 -x,-y,-z
14 -x+y,-x,-z+1/2
15 y,-x+y,-z
16 x,y,-z+1/2
17 x-y,x,-z
18 -y,x-y,-z+1/2
19 -x+y,y,z+1/2
20 -x,-x+y,z
21 -y,-x,z+1/2
22 x-y,-y,z
23 x,x-y,z+1/2
24 y,x,z

loop_
_atom_site_label
_atom_site_type_symbol
_atom_site_symmetry_multiplicity
_atom_site_Wyckoff_label
_atom_site_fract_x
_atom_site_fract_y
_atom_site_fract_z
_atom_site_occupancy
Ga1 Ga 2 b 0.00000 0.00000 0.00000 1.00000
Ti1 Ti 4 d 0.33333 0.66667 0.00000 1.00000
Ga2 Ga 6 g 0.62000 0.00000 0.25000 1.00000
Ti2 Ti 6 g 0.29000 0.00000 0.25000 1.00000
```

$Ti_5Ga_4$: A4B5_hP18_193_bg_dg - POSCAR

```
A4B5_hP18_193_bg_dg & a,c/a,x3,x4 --params=7.861,0.693550438875,0.62,
    ↪ 0.29 & P6_{3}/mcm D_{6h}^{3} #193 (bdg^2) & hP18 & None &
    ↪ Ga4Ti5 & Ga4Ti5 & K. Schubert et al., Naturwissenschaften 49,
    ↪ 57(1962)
   1.00000000000000
   3.93050000000000  -6.80782569914947   0.00000000000000
   3.93050000000000   6.80782569914947   0.00000000000000
   0.00000000000000   0.00000000000000   5.45200000000000
    Ga     Ti
     8     10
Direct
   0.00000000000000   0.00000000000000   0.00000000000000   Ga    (2b)
   0.00000000000000   0.00000000000000   0.50000000000000   Ga    (2b)
   0.62000000000000   0.00000000000000   0.25000000000000   Ga    (6g)
   0.00000000000000   0.62000000000000   0.25000000000000   Ga    (6g)
  -0.62000000000000  -0.62000000000000   0.25000000000000   Ga    (6g)
  -0.62000000000000   0.00000000000000   0.75000000000000   Ga    (6g)
   0.00000000000000  -0.62000000000000   0.75000000000000   Ga    (6g)
   0.62000000000000   0.62000000000000   0.75000000000000   Ga    (6g)
   0.33333333333333   0.66666666666667   0.00000000000000   Ti    (4d)
   0.66666666666667   0.33333333333333   0.50000000000000   Ti    (4d)
   0.66666666666667   0.33333333333333   0.00000000000000   Ti    (4d)
   0.33333333333333   0.66666666666667   0.50000000000000   Ti    (4d)
   0.29000000000000   0.00000000000000   0.25000000000000   Ti    (6g)
   0.00000000000000   0.29000000000000   0.25000000000000   Ti    (6g)
  -0.29000000000000  -0.29000000000000   0.25000000000000   Ti    (6g)
  -0.29000000000000   0.00000000000000   0.75000000000000   Ti    (6g)
   0.00000000000000  -0.29000000000000   0.75000000000000   Ti    (6g)
   0.29000000000000   0.29000000000000   0.75000000000000   Ti    (6g)
```

Proposed 300 GPa $HfH_{10}$: A10B_hP22_194_bhj_c - CIF

```
# CIF file
```

```
data_findsym-output
_audit_creation_method FINDSYM

_chemical_name_mineral 'H10Hf'
_chemical_formula_sum 'H10 Hf'

loop_
_publ_author_name
 'H. Xie'
 'Y. Yao'
 'X. Feng'
 'D. Duan'
 'H. Song'
 'Z. Zhang'
 'S. Jiang'
 'S. A. T. Redfern'
 'V. Z. Kresin'
 'C. J. Pickard'
 'T. Cui'
_journal_year 2020
_publ_Section_title
;
 Hydrogen ``penta-graphene-like\'\' structure stabilized via hafnium: a
      ↪ high-temperature conventional superconductor
;

_aflow_title 'Proposed 300~GPa HfH$_{10}$ Structure'
_aflow_proto 'A10B_hP22_194_bhj_c'
_aflow_params 'a,c/a,x_{3},x_{4},y_{4}'
_aflow_params_values '4.633,0.562702352687,0.115,0.375,0.077'
_aflow_Strukturbericht 'None'
_aflow_Pearson 'hP22'

_symmetry_space_group_name_H-M "P 63/m 2/m 2/c"
_symmetry_Int_Tables_number 194

_cell_length_a      4.63300
_cell_length_b      4.63300
_cell_length_c      2.60700
_cell_angle_alpha  90.00000
_cell_angle_beta   90.00000
_cell_angle_gamma 120.00000

loop_
_space_group_symop_id
_space_group_symop_operation_xyz
1 x,y,z
2 x-y,x,z+1/2
3 -y,x-y,z
4 -x,-y,z+1/2
5 -x+y,-x,z
6 y,-x+y,z+1/2
7 x-y,-y,-z
8 x,x-y,-z+1/2
9 y,x,-z
10 -x+y,y,-z+1/2
11 -x,-x+y,-z
12 -y,-x,-z+1/2
13 -x,-y,-z
14 -x+y,-x,-z+1/2
15 y,-x+y,-z
16 x,y,-z+1/2
17 x-y,x,-z
18 -y,x-y,-z+1/2
19 -x+y,y,z
20 -x,-x+y,z+1/2
21 -y,-x,z
22 x-y,-y,z+1/2
23 x,x-y,z
24 y,x,z+1/2

loop_
_atom_site_label
_atom_site_type_symbol
_atom_site_symmetry_multiplicity
_atom_site_Wyckoff_label
_atom_site_fract_x
_atom_site_fract_y
_atom_site_fract_z
_atom_site_occupancy
H1 H 2 b 0.00000 0.00000 0.25000 1.00000
Hf1 Hf 2 c 0.33333 0.66667 0.25000 1.00000
H2 H 6 h 0.11500 0.23000 0.25000 1.00000
H3 H 12 j 0.37500 0.07700 0.25000 1.00000
```

Proposed 300 GPa $HfH_{10}$: A10B_hP22_194_bhj_c - POSCAR

```
A10B_hP22_194_bhj_c & a,c/a,x3,x4,y4 --params=4.633,0.562702352687,0.115
      ↪ ,0.375,0.077 & P6_{3}/mmc D_{6h}^{4} #194 (bchj) & hP22 & None
      ↪ & H10Hf & H10Hf & H. Xie et al., (2020)
   1.00000000000000
   2.31650000000000  -4.01229569573330   0.00000000000000
   2.31650000000000   4.01229569573330   0.00000000000000
   0.00000000000000   0.00000000000000   2.60700000000000
     H    Hf
    20     2
Direct
   0.00000000000000   0.00000000000000   0.25000000000000   H   (2b)
   0.00000000000000   0.00000000000000   0.75000000000000   H   (2b)
   0.11500000000000   0.23000000000000   0.25000000000000   H   (6h)
  -0.23000000000000  -0.11500000000000   0.25000000000000   H   (6h)
   0.11500000000000  -0.11500000000000   0.25000000000000   H   (6h)
  -0.11500000000000  -0.23000000000000   0.75000000000000   H   (6h)
   0.23000000000000   0.11500000000000   0.75000000000000   H   (6h)
  -0.11500000000000   0.11500000000000   0.75000000000000   H   (6h)
   0.37500000000000   0.07700000000000   0.25000000000000   H  (12j)
  -0.07700000000000   0.29800000000000   0.25000000000000   H  (12j)
  -0.29800000000000  -0.37500000000000   0.25000000000000   H  (12j)
  -0.37500000000000  -0.07700000000000   0.75000000000000   H  (12j)
   0.07700000000000  -0.29800000000000   0.75000000000000   H  (12j)
   0.29800000000000   0.37500000000000   0.75000000000000   H  (12j)
   0.07700000000000   0.37500000000000   0.75000000000000   H  (12j)
   0.29800000000000  -0.07700000000000   0.75000000000000   H  (12j)
  -0.37500000000000  -0.29800000000000   0.75000000000000   H  (12j)
  -0.07700000000000  -0.37500000000000   0.25000000000000   H  (12j)
  -0.29800000000000   0.07700000000000   0.25000000000000   H  (12j)
   0.37500000000000   0.29800000000000   0.25000000000000   H  (12j)
   0.33333333333333   0.66666666666667   0.25000000000000  Hf   (2c)
   0.66666666666667   0.33333333333333   0.75000000000000  Hf   (2c)
```

Magnetoplumbite ($PbFe_{12}O_{19}$): A12B19C_hP64_194_ab2fk_efh2k_d - CIF

```
# CIF file
data_findsym-output
_audit_creation_method FINDSYM

_chemical_name_mineral 'Magnetoplumbite'
_chemical_formula_sum 'Fe12 O19 Pb'

loop_
_publ_author_name
 'R. Gerber'
 'Z. \v{S}im\v{s}a'
 'L. Jen\v{s}ovsk\\'{y}'
_journal_name_full_name
;
 Czechoslovak Journal of Physics
;
_journal_volume 44
_journal_year 1994
_journal_page_first 937
_journal_page_last 940
_publ_Section_title
;
 A note on the magnetoplumbite crystal structure
;

_aflow_title 'Magnetoplumbite (PbFe$_{12}$O$_{19}$) Structure'
_aflow_proto 'A12B19C_hP64_194_ab2fk_efh2k_d'
_aflow_params 'a,c/a,z_{4},z_{5},z_{6},z_{7},x_{8},x_{9},z_{9},x_{10},z_
      ↪ {10},x_{11},z_{11}'
_aflow_params_values '5.88,3.91496598639,0.15,0.028,0.19,-0.05,0.182,
      ↪ 0.167,0.892,0.167,0.05,0.5,0.15'
_aflow_Strukturbericht 'None'
_aflow_Pearson 'hP64'

_symmetry_space_group_name_H-M "P 63/m 2/m 2/c"
_symmetry_Int_Tables_number 194

_cell_length_a      5.88000
_cell_length_b      5.88000
_cell_length_c     23.02000
_cell_angle_alpha  90.00000
_cell_angle_beta   90.00000
_cell_angle_gamma 120.00000

loop_
_space_group_symop_id
_space_group_symop_operation_xyz
1 x,y,z
2 x-y,x,z+1/2
3 -y,x-y,z
4 -x,-y,z+1/2
5 -x+y,-x,z
6 y,-x+y,z+1/2
7 x-y,-y,-z
8 x,x-y,-z+1/2
9 y,x,-z
10 -x+y,y,-z+1/2
11 -x,-x+y,-z
12 -y,-x,-z+1/2
13 -x,-y,-z
14 -x+y,-x,-z+1/2
15 y,-x+y,-z
16 x,y,-z+1/2
17 x-y,x,-z
18 -y,x-y,-z+1/2
19 -x+y,y,z
20 -x,-x+y,z+1/2
21 -y,-x,z
22 x-y,-y,z+1/2
23 x,x-y,z
24 y,x,z+1/2

loop_
_atom_site_label
_atom_site_type_symbol
_atom_site_symmetry_multiplicity
_atom_site_Wyckoff_label
_atom_site_fract_x
_atom_site_fract_y
_atom_site_fract_z
_atom_site_occupancy
Fe1 Fe 2 a 0.00000 0.00000 0.00000 1.00000
Fe2 Fe 2 b 0.00000 0.00000 0.25000 1.00000
Pb1 Pb 2 d 0.33333 0.66667 0.75000 1.00000
O1 O 4 e 0.00000 0.00000 0.15000 1.00000
Fe3 Fe 4 f 0.33333 0.66667 0.02800 1.00000
Fe4 Fe 4 f 0.33333 0.66667 0.19000 1.00000
O2 O 4 f 0.33333 0.66667 -0.05000 1.00000
```

```
O3 O 6 h 0.18200 0.36400 0.25000 1.00000
Fe5 Fe 12 k 0.16700 0.33400 0.89200 1.00000
O4 O 12 k 0.16700 0.33400 0.05000 1.00000
O5 O 12 k 0.50000 0.00000 0.15000 1.00000
```

Magnetoplumbite ($PbFe_{12}O_{19}$): A12B19C_hP64_194_ab2fk_efh2k_d - POSCAR

```
A12B19C_hP64_194_ab2fk_efh2k_d & a,c/a,z4,z5,z6,z7,x8,x9,z9,x10,z10,x11,
    ↪ z11 --params=5.88,3.91496598639,0.15,0.028,0.19,-0.05,0.182,
    ↪ 0.167,0.892,0.167,0.05,0.5,0.15 & P6_{3}/mmc D_{6h}^{4} #194 (
    ↪ abdef^3hk^3) & hP64 & None & Fe12O19Pb & Magnetoplumbite & R.
    ↪ Gerber and Z. \v{S}im\v{s}a and L. Jen\v{s}ovsk\'{y}, Czech. J.
    ↪ Phys. 44, 937-940 (1994)
   1.00000000000000
   2.94000000000000  -5.09222937425250   0.00000000000000
   2.94000000000000   5.09222937425250   0.00000000000000
   0.00000000000000   0.00000000000000  23.02000000000000
    Fe      O     Pb
    24     38      2
Direct
   0.00000000000000   0.00000000000000   0.00000000000000   Fe   (2a)
   0.00000000000000   0.00000000000000   0.50000000000000   Fe   (2a)
   0.00000000000000   0.00000000000000   0.25000000000000   Fe   (2b)
   0.00000000000000   0.00000000000000   0.75000000000000   Fe   (2b)
   0.33333333333333   0.66666666666667   0.02800000000000   Fe   (4f)
   0.66666666666667   0.33333333333333   0.52800000000000   Fe   (4f)
   0.66666666666667   0.33333333333333  -0.02800000000000   Fe   (4f)
   0.33333333333333   0.66666666666667   0.47200000000000   Fe   (4f)
   0.33333333333333   0.66666666666667   0.19000000000000   Fe   (4f)
   0.66666666666667   0.33333333333333   0.69000000000000   Fe   (4f)
   0.66666666666667   0.33333333333333  -0.19000000000000   Fe   (4f)
   0.33333333333333   0.66666666666667   0.31000000000000   Fe   (4f)
   0.16700000000000   0.33400000000000   0.89200000000000   Fe  (12k)
  -0.33400000000000  -0.16700000000000   0.89200000000000   Fe  (12k)
   0.16700000000000  -0.16700000000000   0.89200000000000   Fe  (12k)
  -0.16700000000000  -0.33400000000000   1.39200000000000   Fe  (12k)
   0.33400000000000   0.16700000000000   1.39200000000000   Fe  (12k)
  -0.16700000000000   0.16700000000000   1.39200000000000   Fe  (12k)
   0.33400000000000   0.16700000000000  -0.89200000000000   Fe  (12k)
  -0.16700000000000  -0.33400000000000  -0.89200000000000   Fe  (12k)
  -0.16700000000000   0.16700000000000  -0.89200000000000   Fe  (12k)
  -0.33400000000000  -0.16700000000000  -0.39200000000000   Fe  (12k)
   0.16700000000000   0.33400000000000  -0.39200000000000   Fe  (12k)
   0.16700000000000  -0.16700000000000  -0.39200000000000   Fe  (12k)
   0.00000000000000   0.00000000000000   0.15000000000000   O    (4e)
   0.00000000000000   0.00000000000000   0.65000000000000   O    (4e)
   0.00000000000000   0.00000000000000  -0.15000000000000   O    (4e)
   0.00000000000000   0.00000000000000   0.35000000000000   O    (4e)
   0.33333333333333   0.66666666666667  -0.05000000000000   O    (4f)
   0.66666666666667   0.33333333333333   0.45000000000000   O    (4f)
   0.66666666666667   0.33333333333333   0.05000000000000   O    (4f)
   0.33333333333333   0.66666666666667   0.55000000000000   O    (4f)
   0.18200000000000   0.36400000000000   0.25000000000000   O    (6h)
  -0.36400000000000  -0.18200000000000   0.25000000000000   O    (6h)
   0.18200000000000  -0.18200000000000   0.25000000000000   O    (6h)
  -0.18200000000000  -0.36400000000000   0.75000000000000   O    (6h)
   0.36400000000000   0.18200000000000   0.75000000000000   O    (6h)
  -0.18200000000000   0.18200000000000   0.75000000000000   O    (6h)
   0.16700000000000   0.33400000000000   0.05000000000000   O   (12k)
  -0.33400000000000  -0.16700000000000   0.05000000000000   O   (12k)
   0.16700000000000  -0.16700000000000   0.05000000000000   O   (12k)
  -0.16700000000000  -0.33400000000000   0.55000000000000   O   (12k)
   0.33400000000000   0.16700000000000   0.55000000000000   O   (12k)
  -0.16700000000000   0.16700000000000   0.55000000000000   O   (12k)
   0.33400000000000   0.16700000000000  -0.05000000000000   O   (12k)
  -0.16700000000000  -0.33400000000000  -0.05000000000000   O   (12k)
  -0.16700000000000   0.16700000000000  -0.05000000000000   O   (12k)
  -0.33400000000000  -0.16700000000000   0.45000000000000   O   (12k)
   0.16700000000000   0.33400000000000   0.45000000000000   O   (12k)
   0.16700000000000  -0.16700000000000   0.45000000000000   O   (12k)
   0.50000000000000   1.00000000000000   0.15000000000000   O   (12k)
  -1.00000000000000  -0.50000000000000   0.15000000000000   O   (12k)
   0.50000000000000  -0.50000000000000   0.15000000000000   O   (12k)
  -0.50000000000000  -1.00000000000000   0.65000000000000   O   (12k)
   1.00000000000000   0.50000000000000   0.65000000000000   O   (12k)
  -0.50000000000000   0.50000000000000   0.65000000000000   O   (12k)
   1.00000000000000   0.50000000000000  -0.15000000000000   O   (12k)
  -0.50000000000000  -1.00000000000000  -0.15000000000000   O   (12k)
  -0.50000000000000   0.50000000000000  -0.15000000000000   O   (12k)
  -1.00000000000000  -0.50000000000000   0.35000000000000   O   (12k)
   0.50000000000000   1.00000000000000   0.35000000000000   O   (12k)
   0.50000000000000  -0.50000000000000   0.35000000000000   O   (12k)
   0.33333333333333   0.66666666666667   0.75000000000000   Pb   (2d)
   0.66666666666667   0.33333333333333   0.25000000000000   Pb   (2d)
```

$Pt_2Sn_3$ ($D5_b$): A2B3_hP10_194_f_bf - CIF

```
# CIF file
data_findsym-output
_audit_creation_method FINDSYM

_chemical_name_mineral 'Pt2Sn3'
_chemical_formula_sum 'Pt2 Sn3'

loop_
_publ_author_name
 'K. Schubert'
 'H. Pfisterer'
_journal_name_full_name
;
 Zeitschrift fur Metallkunde
;
_journal_volume 40
_journal_year 1949
_journal_page_first 405
_journal_page_last 405
_publ_Section_title
;
 Kristallstruktur von Pt$_{2}$Sn$_{3}$
;

# Found in A Handbook of Lattice Spacings and Structures of Metals and
    ↪ Alloys, 1958 Found in A Handbook of Lattice Spacings and
    ↪ Structures of Metals and Alloys, {N.~R.~C. No. 4303},

_aflow_title 'Pt$_{2}$Sn$_{3}$ ($D5_{b}$) Structure'
_aflow_proto 'A2B3_hP10_194_f_bf'
_aflow_params 'a,c/a,z_{2},z_{3}'
_aflow_params_values '4.36079,2.97196838188,0.143,-0.07'
_aflow_Strukturbericht '$D5_{b}$'
_aflow_Pearson 'hP10'

_symmetry_space_group_name_H-M "P 63/m 2/m 2/c"
_symmetry_Int_Tables_number 194

_cell_length_a      4.36079
_cell_length_b      4.36079
_cell_length_c      12.96013
_cell_angle_alpha   90.00000
_cell_angle_beta    90.00000
_cell_angle_gamma   120.00000

loop_
_space_group_symop_id
_space_group_symop_operation_xyz
1 x,y,z
2 x-y,x,z+1/2
3 -y,x-y,z
4 -x,-y,z+1/2
5 -x+y,-x,z
6 y,-x+y,z+1/2
7 x-y,-y,-z
8 x,x-y,-z+1/2
9 y,x,-z
10 -x+y,y,-z+1/2
11 -x,-x+y,-z
12 -y,-x,-z+1/2
13 -x,-y,-z
14 -x+y,-x,-z+1/2
15 y,-x+y,-z
16 x,y,-z+1/2
17 x-y,x,-z
18 -y,x-y,-z+1/2
19 -x+y,y,z
20 -x,-x+y,z+1/2
21 -y,-x,z
22 x-y,-y,z+1/2
23 x,x-y,z
24 y,x,z+1/2

loop_
_atom_site_label
_atom_site_type_symbol
_atom_site_symmetry_multiplicity
_atom_site_Wyckoff_label
_atom_site_fract_x
_atom_site_fract_y
_atom_site_fract_z
_atom_site_occupancy
Sn1 Sn 2 b 0.00000 0.00000 0.25000 1.00000
Pt1 Pt 4 f 0.33333 0.66667 0.14300 1.00000
Sn2 Sn 4 f 0.33333 0.66667 -0.07000 1.00000
```

$Pt_2Sn_3$ ($D5_b$): A2B3_hP10_194_f_bf - POSCAR

```
A2B3_hP10_194_f_bf & a,c/a,z2,z3 --params=4.36079,2.97196838188,0.143,-
    ↪ 0.07 & P6_{3}/mmc D_{6h}^{4} #194 (bf^2) & hP10 & $D5_{b}$ &
    ↪ Pt2Sn3 & Pt2Sn3 & K. Schubert and H. Pfisterer, Z. Metallkd. 40
    ↪ , 405(1949)
   1.00000000000000
   2.18039500000000  -3.77655492056914   0.00000000000000
   2.18039500000000   3.77655492056914   0.00000000000000
   0.00000000000000   0.00000000000000  12.96013000000000
    Pt     Sn
     4      6
Direct
   0.33333333333333   0.66666666666667   0.14300000000000   Pt   (4f)
   0.66666666666667   0.33333333333333   0.64300000000000   Pt   (4f)
   0.66666666666667   0.33333333333333  -0.14300000000000   Pt   (4f)
   0.33333333333333   0.66666666666667   0.35700000000000   Pt   (4f)
   0.00000000000000   0.00000000000000   0.25000000000000   Sn   (2b)
   0.00000000000000   0.00000000000000   0.75000000000000   Sn   (2b)
   0.33333333333333   0.66666666666667  -0.07000000000000   Sn   (4f)
   0.66666666666667   0.33333333333333   0.43000000000000   Sn   (4f)
   0.66666666666667   0.33333333333333   0.07000000000000   Sn   (4f)
   0.33333333333333   0.66666666666667   0.57000000000000   Sn   (4f)
```

β-Alumina ($Al_2O_3$, $D5_6$): A2B3_hP60_194_3fk_cdef2k - CIF

```
# CIF file
data_findsym-output
_audit_creation_method FINDSYM

_chemical_name_mineral '$\beta$-alumina'
_chemical_formula_sum 'Al2 O3'

loop_
_publ_author_name
 'W. L. Bragg'
 'C. Gottfried'
```

```
 'J. West'
_journal_name_full_name
;
 Zeitschrift f{\"u}r Kristallographie - Crystalline Materials
;
_journal_volume 77
_journal_year 1931
_journal_page_first 255
_journal_page_last 274
_publ_Section_title
;
 The Structure of $\beta$ Alumina
;

# Found in Strukturbericht Band II 1928-1932, 1937

_aflow_title '$\beta$-Alumina (Al$_{2}$O$_{3}$, $D5_{6}$) Structure'
_aflow_proto 'A2B3_hP60_194_3fk_cdef2k'
_aflow_params 'a,c/a,z_{3},z_{4},z_{5},z_{6},z_{7},x_{8},z_{8},x_{9},z_{
    ↪ 9},x_{10},z_{10}'
_aflow_params_values '5.56,4.05575539568,0.14,0.02,0.17,-0.17,-0.05,
    ↪ 0.83333,0.1,0.16667,0.05,0.5,0.14'
_aflow_Strukturbericht '$D5_{6}$'
_aflow_Pearson 'hP60'

_symmetry_space_group_name_H-M "P 63/m 2/m 2/c"
_symmetry_Int_Tables_number 194

_cell_length_a    5.56000
_cell_length_b    5.56000
_cell_length_c    22.55000
_cell_angle_alpha 90.00000
_cell_angle_beta  90.00000
_cell_angle_gamma 120.00000

loop_
_space_group_symop_id
_space_group_symop_operation_xyz
1 x,y,z
2 x-y,x,z+1/2
3 -y,x-y,z
4 -x,-y,z+1/2
5 -x+y,-x,z
6 y,-x+y,z+1/2
7 x-y,-y,-z
8 x,x-y,-z+1/2
9 y,x,-z
10 -x+y,y,-z+1/2
11 -x,-x+y,-z
12 -y,-x,-z+1/2
13 -x,-y,-z
14 -x+y,-x,-z+1/2
15 y,-x+y,-z
16 x,y,-z+1/2
17 x-y,x,-z
18 -y,x-y,-z+1/2
19 -x+y,y,z
20 -x,-x+y,z+1/2
21 -y,-x,z
22 x-y,-y,z+1/2
23 x,x-y,z
24 y,x,z+1/2

loop_
_atom_site_label
_atom_site_type_symbol
_atom_site_symmetry_multiplicity
_atom_site_Wyckoff_label
_atom_site_fract_x
_atom_site_fract_y
_atom_site_fract_z
_atom_site_occupancy
O1 O 2 c 0.33333 0.66667 0.25000 1.00000
O2 O 2 d 0.33333 0.66667 0.75000 1.00000
O3 O 4 e 0.00000 0.00000 0.14000 1.00000
Al1 Al 4 f 0.33333 0.66667 0.02000 1.00000
Al2 Al 4 f 0.33333 0.66667 0.17000 1.00000
Al3 Al 4 f 0.33333 0.66667 -0.1700 1.00000
O4 O 4 f 0.33333 0.66667 -0.0500 1.00000
Al4 Al 12 k 0.83333 0.66667 0.10000 1.00000
O5 O 12 k 0.16667 0.33333 0.05000 1.00000
O6 O 12 k 0.50000 0.00000 0.14000 1.00000
```

$\beta$-Alumina ($Al_2O_3$, $D5_6$): A2B3_hP60_194_3fk_cdef2k - POSCAR

```
A2B3_hP60_194_3fk_cdef2k & a,c/a,z3,z4,z5,z6,z7,x8,z8,x9,z9,x10,z10 --
    ↪ params=5.56,4.05575539568,0.14,0.02,0.17,-0.17,-0.05,0.83333,
    ↪ 0.1,0.16667,0.05,0.5,0.14 & P6_{3}/mmc D_{6h}^{4} #194 (cdef^4k
    ↪ ^3) & hP60 & $D5_{6}$ & Al2O3 & $\beta$-alumina & W. L. Bragg
    ↪ and C. Gottfried and J. West, Zeitschrift f{"u}r
    ↪ Kristallographie - Crystalline Materials 77, 255-274 (1931)
   1.00000000000000
   2.78000000000000  -4.81510124504148   0.00000000000000
   2.78000000000000   4.81510124504148   0.00000000000000
   0.00000000000000   0.00000000000000  22.55000000000000
    Al     O
    24    36
Direct
   0.33333333333333   0.66666666666667   0.02000000000000   Al   (4f)
   0.66666666666667   0.33333333333333   0.52000000000000   Al   (4f)
   0.66666666666667   0.33333333333333  -0.02000000000000   Al   (4f)
   0.33333333333333   0.66666666666667   0.48000000000000   Al   (4f)
   0.33333333333333   0.66666666666667   0.17000000000000   Al   (4f)
   0.66666666666667   0.33333333333333   0.67000000000000   Al   (4f)
   0.66666666666667   0.33333333333333  -0.17000000000000   Al   (4f)
   0.33333333333333   0.66666666666667   0.33000000000000   Al   (4f)
   0.33333333333333   0.66666666666667  -0.17000000000000   Al   (4f)
   0.66666666666667   0.33333333333333   0.33000000000000   Al   (4f)
   0.66666666666667   0.33333333333333   0.17000000000000   Al   (4f)
   0.33333333333333   0.66666666666667   0.67000000000000   Al   (4f)
   0.83333000000000   1.66666000000000   0.10000000000000   Al  (12k)
  -1.66666000000000  -0.83333000000000   0.10000000000000   Al  (12k)
   0.83333000000000  -0.83333000000000   0.10000000000000   Al  (12k)
  -0.83333000000000  -1.66666000000000   0.60000000000000   Al  (12k)
   1.66666000000000   0.83333000000000   0.60000000000000   Al  (12k)
  -0.83333000000000   0.83333000000000   0.60000000000000   Al  (12k)
   1.66666000000000   0.83333000000000  -0.10000000000000   Al  (12k)
  -0.83333000000000  -1.66666000000000  -0.10000000000000   Al  (12k)
  -0.83333000000000   0.83333000000000  -0.10000000000000   Al  (12k)
  -1.66666000000000  -0.83333000000000   0.40000000000000   Al  (12k)
   0.83333000000000   1.66666000000000   0.40000000000000   Al  (12k)
   0.83333000000000  -0.83333000000000   0.40000000000000   Al  (12k)
   0.33333333333333   0.66666666666667   0.25000000000000   O   (2c)
   0.66666666666667   0.33333333333333   0.75000000000000   O   (2c)
   0.33333333333333   0.66666666666667   0.75000000000000   O   (2d)
   0.66666666666667   0.33333333333333   0.25000000000000   O   (2d)
   0.00000000000000   0.00000000000000   0.14000000000000   O   (4e)
   0.00000000000000   0.00000000000000   0.64000000000000   O   (4e)
   0.00000000000000   0.00000000000000  -0.14000000000000   O   (4e)
   0.00000000000000   0.00000000000000   0.36000000000000   O   (4e)
   0.33333333333333   0.66666666666667  -0.05000000000000   O   (4f)
   0.66666666666667   0.33333333333333   0.45000000000000   O   (4f)
   0.66666666666667   0.33333333333333   0.05000000000000   O   (4f)
   0.33333333333333   0.66666666666667   0.55000000000000   O   (4f)
   0.16667000000000   0.33334000000000   0.05000000000000   O  (12k)
  -0.33334000000000  -0.16667000000000   0.05000000000000   O  (12k)
   0.16667000000000  -0.16667000000000   0.05000000000000   O  (12k)
  -0.16667000000000  -0.33334000000000   0.55000000000000   O  (12k)
   0.33334000000000   0.16667000000000   0.55000000000000   O  (12k)
  -0.16667000000000   0.16667000000000   0.55000000000000   O  (12k)
   0.33334000000000   0.16667000000000  -0.05000000000000   O  (12k)
  -0.16667000000000  -0.33334000000000  -0.05000000000000   O  (12k)
  -0.16667000000000   0.16667000000000  -0.05000000000000   O  (12k)
  -0.33334000000000  -0.16667000000000   0.45000000000000   O  (12k)
   0.16667000000000   0.33334000000000   0.45000000000000   O  (12k)
   0.16667000000000  -0.16667000000000   0.45000000000000   O  (12k)
   0.50000000000000   1.00000000000000   0.14000000000000   O  (12k)
  -1.00000000000000  -0.50000000000000   0.14000000000000   O  (12k)
   0.50000000000000  -0.50000000000000   0.14000000000000   O  (12k)
  -0.50000000000000  -1.00000000000000   0.64000000000000   O  (12k)
   1.00000000000000   0.50000000000000   0.64000000000000   O  (12k)
  -0.50000000000000   0.50000000000000   0.64000000000000   O  (12k)
   1.00000000000000   0.50000000000000  -0.14000000000000   O  (12k)
  -0.50000000000000  -1.00000000000000  -0.14000000000000   O  (12k)
  -0.50000000000000   0.50000000000000  -0.14000000000000   O  (12k)
  -1.00000000000000  -0.50000000000000   0.36000000000000   O  (12k)
   0.50000000000000   1.00000000000000   0.36000000000000   O  (12k)
   0.50000000000000  -0.50000000000000   0.36000000000000   O  (12k)
```

$S3_4$ (II) (Catapleiite, $Na_2Zr(SiO_3)_3 \cdot H_2O$) (*obsolete*): A3B2C9D3E_hP36_194_g_f_hk_h_a - CIF

```
# CIF file
data_findsym-output
_audit_creation_method FINDSYM

_chemical_name_mineral 'Catapleiite'
_chemical_formula_sum '(H2O)3 Na2 O9 Si3 Zr'

loop_
_publ_author_name
 'B. Brunowski'
_journal_name_full_name
;
 Acta Physicochimica U.R.S.S.
;
_journal_volume 5
_journal_year 1936
_journal_page_first 863
_journal_page_last 892
_publ_Section_title
;
 Die Struktur des Katapleits
;

# Found in Strukturbericht Band V 1937, 1940

_aflow_title '$S3_{4}$ (II) (Catapleiite, Na$_{2}$Zr(SiO$_{3}$)$_{3}$$\
    ↪ cdot$ H$_{2}$O) ({\em{obsolete}}) Structure'
_aflow_proto 'A3B2C9D3E_hP36_194_g_f_hk_h_a'
_aflow_params 'a,c/a,z_{2},x_{4},x_{5},x_{6},z_{6}'
_aflow_params_values '7.39,1.3599458728,0.08,0.47,0.2,0.136,0.125'
_aflow_Strukturbericht '$S3_{4}$ (II)'
_aflow_Pearson 'hP36'

_symmetry_space_group_name_H-M "P 63/m 2/m 2/c"
_symmetry_Int_Tables_number 194

_cell_length_a    7.39000
_cell_length_b    7.39000
_cell_length_c    10.05000
_cell_angle_alpha 90.00000
_cell_angle_beta  90.00000
_cell_angle_gamma 120.00000

loop_
_space_group_symop_id
_space_group_symop_operation_xyz
1 x,y,z
2 x-y,x,z+1/2
3 -y,x-y,z
```

```
4 -x,-y,z+1/2
5 -x+y,-x,z
6 y,-x+y,z+1/2
7 x-y,-y,-z
8 x,x-y,-z+1/2
9 y,x,-z
10 -x+y,y,-z+1/2
11 -x,-x+y,-z
12 -y,-x,-z+1/2
13 -x,-y,-z
14 -x+y,-x,-z+1/2
15 y,-x+y,-z
16 x,y,-z+1/2
17 x-y,x,-z
18 -y,x-y,-z+1/2
19 -x+y,y,z
20 -x,-x+y,z+1/2
21 -y,-x,z
22 x-y,-y,z+1/2
23 x,x-y,z
24 y,x,z+1/2

loop_
_atom_site_label
_atom_site_type_symbol
_atom_site_symmetry_multiplicity
_atom_site_Wyckoff_label
_atom_site_fract_x
_atom_site_fract_y
_atom_site_fract_z
_atom_site_occupancy
Zr1 Zr 2 a 0.00000 0.00000 0.00000 1.00000
Na1 Na 4 f 0.33333 0.66667 0.08000 1.00000
H2O1 H2O 6 g 0.50000 0.00000 0.00000 0.66667
O1 O 6 h 0.47000 0.94000 0.25000 1.00000
Si1 Si 6 h 0.20000 0.40000 0.25000 1.00000
O2 O 12 k 0.13600 0.27200 0.12500 1.00000
```

$S3_4$ (II) (Catapleiite, $Na_2Zr(SiO_3)_3\cdot H_2O$) (*obsolete*): A3B2C9D3E_hP36_194_g_f_hk_h_a - POSCAR

```
A3B2C9D3E_hP36_194_g_f_hk_h_a & a,c/a,z2,x4,x5,x6,z6 --params=7.39,
    ↪ 1.3599458728,0.08,0.47,0.2,0.136,0.125 & P6_{3}/mmc D_{6h}^{4}
    ↪ #194 (afgh^2k) & hP36 & $S3_{4}$ (II) & H2Na2O6Si2Zr &
    ↪ Catapleiite & B. Brunowski, Acta Physicochim. USSR 5, 863-892 (
    ↪ 1936)
   1.00000000000000
   3.69500000000000  -6.39992773396700   0.00000000000000
   3.69500000000000   6.39992773396700   0.00000000000000
   0.00000000000000   0.00000000000000  10.05000000000000
   H2O    Na     O    Si    Zr
     6     4    18     6     2
Direct
   0.50000000000000   0.00000000000000   0.00000000000000   H2O   (6g)
   0.00000000000000   0.50000000000000   0.00000000000000   H2O   (6g)
   0.50000000000000   0.50000000000000   0.00000000000000   H2O   (6g)
   0.50000000000000   0.00000000000000   0.50000000000000   H2O   (6g)
   0.00000000000000   0.50000000000000   0.50000000000000   H2O   (6g)
   0.50000000000000   0.50000000000000   0.50000000000000   H2O   (6g)
   0.33333333333333   0.66666666666667   0.08000000000000   Na    (4f)
   0.66666666666667   0.33333333333333   0.58000000000000   Na    (4f)
   0.66666666666667   0.33333333333333  -0.08000000000000   Na    (4f)
   0.33333333333333   0.66666666666667   0.42000000000000   Na    (4f)
   0.47000000000000   0.94000000000000   0.25000000000000   O     (6h)
  -0.94000000000000  -0.47000000000000   0.25000000000000   O     (6h)
   0.47000000000000  -0.47000000000000   0.25000000000000   O     (6h)
  -0.47000000000000  -0.94000000000000   0.75000000000000   O     (6h)
   0.94000000000000   0.47000000000000   0.75000000000000   O     (6h)
  -0.47000000000000   0.47000000000000   0.75000000000000   O     (6h)
   0.13600000000000   0.27200000000000   0.12500000000000   O    (12k)
  -0.27200000000000  -0.13600000000000   0.12500000000000   O    (12k)
   0.13600000000000  -0.13600000000000   0.12500000000000   O    (12k)
  -0.13600000000000  -0.27200000000000   0.62500000000000   O    (12k)
   0.27200000000000   0.13600000000000   0.62500000000000   O    (12k)
  -0.13600000000000   0.13600000000000   0.62500000000000   O    (12k)
   0.27200000000000   0.13600000000000  -0.12500000000000   O    (12k)
  -0.13600000000000  -0.27200000000000  -0.12500000000000   O    (12k)
  -0.13600000000000   0.13600000000000  -0.12500000000000   O    (12k)
  -0.27200000000000  -0.13600000000000   0.37500000000000   O    (12k)
   0.13600000000000   0.27200000000000   0.37500000000000   O    (12k)
   0.13600000000000  -0.13600000000000   0.37500000000000   O    (12k)
   0.20000000000000   0.40000000000000   0.25000000000000   Si    (6h)
  -0.40000000000000  -0.20000000000000   0.25000000000000   Si    (6h)
   0.20000000000000  -0.20000000000000   0.25000000000000   Si    (6h)
  -0.20000000000000  -0.40000000000000   0.75000000000000   Si    (6h)
   0.40000000000000   0.20000000000000   0.75000000000000   Si    (6h)
  -0.20000000000000   0.20000000000000   0.75000000000000   Si    (6h)
   0.00000000000000   0.00000000000000   0.00000000000000   Zr    (2a)
   0.00000000000000   0.00000000000000   0.50000000000000   Zr    (2a)
```

$ReB_3$: A3B_hP8_194_af_c - CIF

```
# CIF file
data_findsym-output
_audit_creation_method FINDSYM

_chemical_name_mineral 'B3Re'
_chemical_formula_sum 'B3 Re'

loop_
_publ_author_name
 'B. Aronsson'
 'E. Stenberg'
 'J. {\AA}selius'
_journal_name_full_name
;
 Acta Chemica Scandinavica
;
_journal_volume 14
_journal_year 1960
_journal_page_first 733
_journal_page_last 741
_publ_Section_title
;
 Borides of Rhenium and the Platinum Metals. The Crystal Structure of
    ↪ Re$_{7}$B$_{3}$, ReB$_{3}$, Rh$_{7}$B$_{3}$, RhB$_{\sim 1.1}$,
    ↪ IrB$_{\sim 1.1}$ and PtB
;

# Found in Pearson's Handbook of Crystallographic Data for Intermetallic
    ↪ Phases, 1991

_aflow_title 'ReB$_{3}$ Structure'
_aflow_proto 'A3B_hP8_194_af_c'
_aflow_params 'a,c/a,z_{3}'
_aflow_params_values '2.9,2.5775862069,0.55'
_aflow_Strukturbericht 'None'
_aflow_Pearson 'hP8'

_symmetry_space_group_name_H-M "P 63/m 2/m 2/c"
_symmetry_Int_Tables_number 194

_cell_length_a     2.90000
_cell_length_b     2.90000
_cell_length_c     7.47500
_cell_angle_alpha  90.00000
_cell_angle_beta   90.00000
_cell_angle_gamma  120.00000

loop_
_space_group_symop_id
_space_group_symop_operation_xyz
1 x,y,z
2 x-y,x,z+1/2
3 -y,x-y,z
4 -x,-y,z+1/2
5 -x+y,-x,z
6 y,-x+y,z+1/2
7 x-y,-y,-z
8 x,x-y,-z+1/2
9 y,x,-z
10 -x+y,y,-z+1/2
11 -x,-x+y,-z
12 -y,-x,-z+1/2
13 -x,-y,-z
14 -x+y,-x,-z+1/2
15 y,-x+y,-z
16 x,y,-z+1/2
17 x-y,x,-z
18 -y,x-y,-z+1/2
19 -x+y,y,z
20 -x,-x+y,z+1/2
21 -y,-x,z
22 x-y,-y,z+1/2
23 x,x-y,z
24 y,x,z+1/2

loop_
_atom_site_label
_atom_site_type_symbol
_atom_site_symmetry_multiplicity
_atom_site_Wyckoff_label
_atom_site_fract_x
_atom_site_fract_y
_atom_site_fract_z
_atom_site_occupancy
B1 B 2 a 0.00000 0.00000 0.00000 1.00000
Re1 Re 2 c 0.33333 0.66667 0.25000 1.00000
B2 B 4 f 0.33333 0.66667 0.55000 1.00000
```

$ReB_3$: A3B_hP8_194_af_c - POSCAR

```
A3B_hP8_194_af_c & a,c/a,z3 --params=2.9,2.5775862069,0.55 & P6_{3}/mmc
    ↪ D_{6h}^{4} #194 (acf) & hP8 & None & B3Re & B3Re & B. Aronsson
    ↪ and E. Stenberg and J. {\AA}selius, Acta Chem. Scand. 14,
    ↪ 733-741 (1960)
   1.00000000000000
   1.45000000000000  -2.51147367097487   0.00000000000000
   1.45000000000000   2.51147367097487   0.00000000000000
   0.00000000000000   0.00000000000000   7.47500000000000
     B    Re
     6     2
Direct
   0.00000000000000   0.00000000000000   0.00000000000000   B    (2a)
   0.00000000000000   0.00000000000000   0.50000000000000   B    (2a)
   0.33333333333333   0.66666666666667   0.55000000000000   B    (4f)
   0.66666666666667   0.33333333333333   1.05000000000000   B    (4f)
   0.66666666666667   0.33333333333333  -0.55000000000000   B    (4f)
   0.33333333333333   0.66666666666667  -0.05000000000000   B    (4f)
   0.33333333333333   0.66666666666667   0.25000000000000   Re   (2c)
   0.66666666666667   0.33333333333333   0.75000000000000   Re   (2c)
```

$Cs_3Cr_2Cl_9$: A9B2C3_hP28_194_hk_f_bf - CIF

```
# CIF file
data_findsym-output
_audit_creation_method FINDSYM

_chemical_name_mineral 'Cl9Cr2Cs3'
_chemical_formula_sum 'Cl9 Cr2 Cs3'
```

```
loop_
_publ_author_name
 'G. G. Wessel'
 'D. J. W. IJdo'
_journal_name_full_name
;
 Acta Cristallographica
;
_journal_volume 10
_journal_year 1957
_journal_page_first 466
_journal_page_last 468
_publ_Section_title
;
 The Crystal Structure of Cs$_{3}$Cr$_{2}$Cl$_{9}$
;

# Found in Structure of the dimer compounds Cs$_{3}$$R$$_{2}$Br$_{9}$ (
      ↪ $R$ = Tb, Dy, Ho, Er, Yb) at 8 and 295~K studied by neutron
      ↪ diffraction, 1989

_aflow_title 'Cs$_{3}$Cr$_{2}$Cl$_{9}$ Structure'
_aflow_proto 'A9B2C3_hP28_194_hk_f_bf'
_aflow_params 'a,c/a,z_{2},z_{3},x_{4},x_{5},z_{5}'
_aflow_params_values '7.22,2.48337950139,0.837,0.077,0.508,0.824,0.092'
_aflow_Strukturbericht 'None'
_aflow_Pearson 'hP28'

_symmetry_space_group_name_H-M "P 63/m 2/m 2/c"
_symmetry_Int_Tables_number 194

_cell_length_a      7.22000
_cell_length_b      7.22000
_cell_length_c      17.93000
_cell_angle_alpha  90.00000
_cell_angle_beta   90.00000
_cell_angle_gamma 120.00000

loop_
_space_group_symop_id
_space_group_symop_operation_xyz
1 x,y,z
2 x-y,x,z+1/2
3 -y,x-y,z
4 -x,-y,z+1/2
5 -x+y,-x,z
6 y,-x+y,z+1/2
7 x-y,-y,-z
8 x,x-y,-z+1/2
9 y,x,-z
10 -x+y,y,-z+1/2
11 -x,-x+y,-z
12 -y,-x,-z+1/2
13 -x,-y,-z
14 -x+y,-x,-z+1/2
15 y,-x+y,-z
16 x,y,-z+1/2
17 x-y,x,-z
18 -y,x-y,-z+1/2
19 -x+y,y,z
20 -x,-x+y,z+1/2
21 -y,-x,z
22 x-y,-y,z+1/2
23 x,x-y,z
24 y,x,z+1/2

loop_
_atom_site_label
_atom_site_type_symbol
_atom_site_symmetry_multiplicity
_atom_site_Wyckoff_label
_atom_site_fract_x
_atom_site_fract_y
_atom_site_fract_z
_atom_site_occupancy
Cs1 Cs 2 b 0.00000 0.00000 0.25000 1.00000
Cr1 Cr 4 f 0.33333 0.66667 0.83700 1.00000
Cs2 Cs 4 f 0.33333 0.66667 0.07700 1.00000
Cl1 Cl 6 h 0.50800 0.01600 0.25000 1.00000
Cl2 Cl 12 k 0.82400 0.64800 0.09200 1.00000
```

$Cs_3Cr_2Cl_9$: A9B2C3_hP28_194_hk_f_bf - POSCAR

```
A9B2C3_hP28_194_hk_f_bf & a,c/a,z2,z3,x4,x5,z5 --params=7.22,
      ↪ 2.48337950139,0.837,0.077,0.508,0.824,0.092 & P6_{3}/mmc D_{6h
      ↪ }^{4} #194 (bf^2hk) & hP28 & None & Cl9Cr2Cs3 & Cl9Cr2Cs3 & G.
      ↪ G. Wessel and D. J. W. IJdo, Acta Cryst. 10, 466-468 (1957)
   1.00000000000000
   3.61000000000000  -6.25270341532365   0.00000000000000
   3.61000000000000   6.25270341532365   0.00000000000000
   0.00000000000000   0.00000000000000  17.93000000000000
    Cl      Cr      Cs
    18       4       6
Direct
   0.50800000000000   1.01600000000000   0.25000000000000   Cl   (6h)
  -1.01600000000000  -0.50800000000000   0.25000000000000   Cl   (6h)
   0.50800000000000  -0.50800000000000   0.25000000000000   Cl   (6h)
  -0.50800000000000  -1.01600000000000   0.75000000000000   Cl   (6h)
   1.01600000000000   0.50800000000000   0.75000000000000   Cl   (6h)
  -0.50800000000000   0.50800000000000   0.75000000000000   Cl   (6h)
   0.82400000000000   1.64800000000000   0.09200000000000   Cl  (12k)
  -1.64800000000000  -0.82400000000000   0.09200000000000   Cl  (12k)
   0.82400000000000  -0.82400000000000   0.09200000000000   Cl  (12k)
  -0.82400000000000  -1.64800000000000   0.59200000000000   Cl  (12k)
   1.64800000000000   0.82400000000000   0.59200000000000   Cl  (12k)
  -0.82400000000000   0.82400000000000   0.59200000000000   Cl  (12k)
   1.64800000000000   0.82400000000000  -0.09200000000000   Cl  (12k)
  -0.82400000000000  -1.64800000000000  -0.09200000000000   Cl  (12k)
  -0.82400000000000   0.82400000000000  -0.09200000000000   Cl  (12k)
  -1.64800000000000  -0.82400000000000   0.40800000000000   Cl  (12k)
   0.82400000000000   1.64800000000000   0.40800000000000   Cl  (12k)
   0.82400000000000  -0.82400000000000   0.40800000000000   Cl  (12k)
   0.33333333333333   0.66666666666667   0.83700000000000   Cr   (4f)
   0.66666666666667   0.33333333333333   1.33700000000000   Cr   (4f)
   0.66666666666667   0.33333333333333  -0.83700000000000   Cr   (4f)
   0.33333333333333   0.66666666666667  -0.33700000000000   Cr   (4f)
   0.00000000000000   0.00000000000000   0.25000000000000   Cs   (2b)
   0.00000000000000   0.00000000000000   0.75000000000000   Cs   (2b)
   0.33333333333333   0.66666666666667   0.07700000000000   Cs   (4f)
   0.66666666666667   0.33333333333333   0.57700000000000   Cs   (4f)
   0.66666666666667   0.33333333333333  -0.07700000000000   Cs   (4f)
   0.33333333333333   0.66666666666667   0.42300000000000   Cs   (4f)
```

$Na_{0.74}CoO_2$: AB2C2_hP10_194_a_bc_f - CIF

```
# CIF file
data_findsym-output
_audit_creation_method FINDSYM

_chemical_name_mineral 'CoNa0.74O2'
_chemical_formula_sum 'Co Na2 O2'

loop_
_publ_author_name
 'R. J. Balsys'
 'R. L. Davis'
_journal_name_full_name
;
 Solid State Ionics
;
_journal_volume 93
_journal_year 1997
_journal_page_first 279
_journal_page_last 282
_publ_Section_title
;
 Refinement of the structure of Na$_{0.74}$CoO$_{2}$ using neutron
      ↪ powder diffraction
;

# Found in Exploring the possibility of enhancing the high
      ↪ figure-of-merit ($> 2$) of Na$_{0.74}$CoO$_{2}$ by using
      ↪ combined experimental and theoretical studies, 2019 Found in
      ↪ Exploring the possibility of enhancing the high figure-of-merit
      ↪ ($> 2$) of Na$_{0.74}$CoO$_{2}$ by using combined experimental
      ↪ and theoretical studies, {arXiv:1910.10191 [cond-mat.mtrl-sci
      ↪ ]},

_aflow_title 'Na$_{0.74}$CoO$_{2}$ Structure'
_aflow_proto 'AB2C2_hP10_194_a_bc_f'
_aflow_params 'a,c/a,z_{4}'
_aflow_params_values '2.84,3.80669014085,0.0913'
_aflow_Strukturbericht 'None'
_aflow_Pearson 'hP10'

_symmetry_space_group_name_H-M "P 63/m 2/m 2/c"
_symmetry_Int_Tables_number 194

_cell_length_a      2.84000
_cell_length_b      2.84000
_cell_length_c      10.81100
_cell_angle_alpha  90.00000
_cell_angle_beta   90.00000
_cell_angle_gamma 120.00000

loop_
_space_group_symop_id
_space_group_symop_operation_xyz
1 x,y,z
2 x-y,x,z+1/2
3 -y,x-y,z
4 -x,-y,z+1/2
5 -x+y,-x,z
6 y,-x+y,z+1/2
7 x-y,-y,-z
8 x,x-y,-z+1/2
9 y,x,-z
10 -x+y,y,-z+1/2
11 -x,-x+y,-z
12 -y,-x,-z+1/2
13 -x,-y,-z
14 -x+y,-x,-z+1/2
15 y,-x+y,-z
16 x,y,-z+1/2
17 x-y,x,-z
18 -y,x-y,-z+1/2
19 -x+y,y,z
20 -x,-x+y,z+1/2
21 -y,-x,z
22 x-y,-y,z+1/2
23 x,x-y,z
24 y,x,z+1/2

loop_
_atom_site_label
_atom_site_type_symbol
_atom_site_symmetry_multiplicity
_atom_site_Wyckoff_label
_atom_site_fract_x
_atom_site_fract_y
```

```
_atom_site_fract_z
_atom_site_occupancy
Co1 Co 2 a 0.00000 0.00000 0.00000 1.00000
Na1 Na 2 b 0.00000 0.00000 0.25000 0.21000
Na2 Na 2 c 0.33333 0.66667 0.25000 0.51000
O1 O 4 f 0.33333 0.66667 0.09130 1.00000
```

$Na_{0.74}CoO_2$: AB2C2_hP10_194_a_bc_f - POSCAR

```
AB2C2_hP10_194_a_bc_f & a,c/a,z4 --params=2.84,3.80669014085,0.0913 &
    ↪ P6_{3}/mmc D_{6h}^{4} #194 (abcf) & hP10 & None & CoNa0.74O2 &
    ↪ CoNa0.74O2 & R. J. Balsys and R. L. Davis, Solid State Ion. 93,
    ↪ 279-282 (1997)
   1.00000000000000
   1.42000000000000  -2.45951214674781   0.00000000000000
   1.42000000000000   2.45951214674781   0.00000000000000
   0.00000000000000   0.00000000000000  10.81100000000000
    Co     Na     O
     2      4     4
Direct
   0.00000000000000   0.00000000000000   0.00000000000000   Co   (2a)
   0.00000000000000   0.00000000000000   0.50000000000000   Co   (2a)
   0.00000000000000   0.00000000000000   0.25000000000000   Na   (2b)
   0.00000000000000   0.00000000000000   0.75000000000000   Na   (2b)
   0.33333333333333   0.66666666666667   0.25000000000000   Na   (2c)
   0.66666666666667   0.33333333333333   0.75000000000000   Na   (2c)
   0.33333333333333   0.66666666666667   0.09130000000000    O   (4f)
   0.66666666666667   0.33333333333333   0.59130000000000    O   (4f)
   0.66666666666667   0.33333333333333  -0.09130000000000    O   (4f)
   0.33333333333333   0.66666666666667   0.40870000000000    O   (4f)
```

$EuIn_2P_2$: AB2C2_hP10_194_a_f_f - CIF

```
# CIF file
data_findsym-output
_audit_creation_method FINDSYM

_chemical_name_mineral 'EuIn2P2'
_chemical_formula_sum 'Eu In2 P2'

loop_
_publ_author_name
 'J. Jiang'
 'S. M. Kauzlarich'
_journal_name_full_name
;
 Chemistry of Materials
;
_journal_volume 18
_journal_year 2006
_journal_page_first 435
_journal_page_last 441
_publ_Section_title
;
 Colossal Magnetoresistance in a Rare Earth Zintl Compound with a New
    ↪ Structure Type: EuIn$_{2}$P$_{2}$
;

_aflow_title 'EuIn$_{2}$P$_{2}$ Structure'
_aflow_proto 'AB2C2_hP10_194_a_f_f'
_aflow_params 'a,c/a,z_{2},z_{3}'
_aflow_params_values '4.0829,4.30943691984,0.82845,0.10706'
_aflow_Strukturbericht 'None'
_aflow_Pearson 'hP10'

_symmetry_space_group_name_H-M "P 63/m 2/m 2/c"
_symmetry_Int_Tables_number 194

_cell_length_a      4.08290
_cell_length_b      4.08290
_cell_length_c      17.59500
_cell_angle_alpha   90.00000
_cell_angle_beta    90.00000
_cell_angle_gamma   120.00000

loop_
_space_group_symop_id
_space_group_symop_operation_xyz
1 x,y,z
2 x-y,x,z+1/2
3 -y,x-y,z
4 -x,-y,z+1/2
5 -x+y,-x,z
6 y,-x+y,z+1/2
7 x-y,-y,-z
8 x,x-y,-z+1/2
9 y,x,-z
10 -x+y,y,-z+1/2
11 -x,-x+y,-z
12 -y,-x,-z+1/2
13 -x,-y,-z
14 -x+y,-x,-z+1/2
15 y,-x+y,-z
16 x,y,-z+1/2
17 x-y,x,-z
18 -y,x-y,-z+1/2
19 -x+y,y,z
20 -x,-x+y,z+1/2
21 -y,-x,z
22 x-y,-y,z+1/2
23 x,x-y,z
24 y,x,z+1/2

loop_
_atom_site_label
_atom_site_type_symbol
_atom_site_symmetry_multiplicity
_atom_site_Wyckoff_label
_atom_site_fract_x
_atom_site_fract_y
_atom_site_fract_z
_atom_site_occupancy
Eu1 Eu 2 a 0.00000 0.00000 0.00000 1.00000
In1 In 4 f 0.33333 0.66667 0.82845 1.00000
P1 P 4 f 0.33333 0.66667 0.10706 1.00000
```

$EuIn_2P_2$: AB2C2_hP10_194_a_f_f - POSCAR

```
AB2C2_hP10_194_a_f_f & a,c/a,z2,z3 --params=4.0829,4.30943691984,0.82845
    ↪ ,0.10706 & P6_{3}/mmc D_{6h}^{4} #194 (af^2) & hP10 & None &
    ↪ EuIn2P2 & EuIn2P2 & J. Jiang and S. M. Kauzlarich, Chem. Mater.
    ↪ 18, 435-441 (2006)
   1.00000000000000
   2.04145000000000  -3.53589512111148   0.00000000000000
   2.04145000000000   3.53589512111148   0.00000000000000
   0.00000000000000   0.00000000000000  17.59500000000000
    Eu     In     P
     2      4     4
Direct
   0.00000000000000   0.00000000000000   0.00000000000000   Eu   (2a)
   0.00000000000000   0.00000000000000   0.50000000000000   Eu   (2a)
   0.33333333333333   0.66666666666667   0.82845000000000   In   (4f)
   0.66666666666667   0.33333333333333   1.32845000000000   In   (4f)
   0.66666666666667   0.33333333333333  -0.82845000000000   In   (4f)
   0.33333333333333   0.66666666666667  -0.32845000000000   In   (4f)
   0.33333333333333   0.66666666666667   0.10706000000000    P   (4f)
   0.66666666666667   0.33333333333333   0.60706000000000    P   (4f)
   0.66666666666667   0.33333333333333  -0.10706000000000    P   (4f)
   0.33333333333333   0.66666666666667   0.39294000000000    P   (4f)
```

$Lu_2CoGa_3$: AB3C2_hP24_194_f_k_bh - CIF

```
# CIF file
data_findsym-output
_audit_creation_method FINDSYM

_chemical_name_mineral 'CoGa3Lu2'
_chemical_formula_sum 'Co Ga3 Lu2'

loop_
_publ_author_name
 'R. E. Gladyshevskii'
 'K. Cenzual'
 'E. Parth\\'{e}'
_journal_name_full_name
;
 Journal of Alloys and Compounds
;
_journal_volume 189
_journal_year 1992
_journal_page_first 221
_journal_page_last 228
_publ_Section_title
;
 Er$_{2}$RhSi$_{3}$ and $R$$_{2}$CoGa$_{3}$ ($R$ = Y, Tb, Dy, Ho, Er, Tm
    ↪ , Yb) with Lu$_{2}$CoGa$_{3}$ type structure: new members of
    ↪ the AlB$_{2}$ structure family
;

_aflow_title 'Lu$_{2}$CoGa$_{3}$ Structure'
_aflow_proto 'AB3C2_hP24_194_f_k_bh'
_aflow_params 'a,c/a,z_{2},x_{3},x_{4},z_{4}'
_aflow_params_values '8.659,0.787966277861,0.05,0.5231,0.1692,0.0432'
_aflow_Strukturbericht 'None'
_aflow_Pearson 'hP24'

_symmetry_space_group_name_H-M "P 63/m 2/m 2/c"
_symmetry_Int_Tables_number 194

_cell_length_a      8.65900
_cell_length_b      8.65900
_cell_length_c      6.82300
_cell_angle_alpha   90.00000
_cell_angle_beta    90.00000
_cell_angle_gamma   120.00000

loop_
_space_group_symop_id
_space_group_symop_operation_xyz
1 x,y,z
2 x-y,x,z+1/2
3 -y,x-y,z
4 -x,-y,z+1/2
5 -x+y,-x,z
6 y,-x+y,z+1/2
7 x-y,-y,-z
8 x,x-y,-z+1/2
9 y,x,-z
10 -x+y,y,-z+1/2
11 -x,-x+y,-z
12 -y,-x,-z+1/2
13 -x,-y,-z
14 -x+y,-x,-z+1/2
15 y,-x+y,-z
16 x,y,-z+1/2
17 x-y,x,-z
18 -y,x-y,-z+1/2
19 -x+y,y,z
20 -x,-x+y,z+1/2
21 -y,-x,z
```

```
22 x-y,-y,z+1/2
23 x,x-y,z
24 y,x,z+1/2

loop_
_atom_site_label
_atom_site_type_symbol
_atom_site_symmetry_multiplicity
_atom_site_Wyckoff_label
_atom_site_fract_x
_atom_site_fract_y
_atom_site_fract_z
_atom_site_occupancy
Lu1 Lu 2 b 0.00000 0.00000 0.25000 1.00000
Co1 Co 4 f 0.33333 0.66667 0.05000 1.00000
Lu2 Lu 6 h 0.52310 0.04620 0.25000 1.00000
Ga1 Ga 12 k 0.16920 0.33840 0.04320 1.00000
```

$Lu_2CoGa_3$: AB3C2_hP24_194_f_k_bh - POSCAR

```
AB3C2_hP24_194_f_k_bh & a,c/a,z2,x3,x4,z4 --params=8.659,0.787966277861,
   ↪ 0.05,0.5231,0.1692,0.0432 & P6_{3}/mmc D_{6h}^{4} #194 (bfhk) &
   ↪ hP24 & None & CoGa3Lu2 & CoGa3Lu2 & R. E. Gladyshevskii and K.
   ↪ Cenzual and E. Parth\'{e}, J. Alloys Compd. 189, 221-228 (1992
   ↪ )
   1.00000000000000
   4.32950000000000  -7.49891397136945   0.00000000000000
   4.32950000000000   7.49891397136945   0.00000000000000
   0.00000000000000   0.00000000000000   6.82300000000000
    Co     Ga     Lu
     4     12      8
Direct
   0.33333333333333   0.66666666666667   0.05000000000000   Co   (4f)
   0.66666666666667   0.33333333333333   0.55000000000000   Co   (4f)
   0.66666666666667   0.33333333333333  -0.05000000000000   Co   (4f)
   0.33333333333333   0.66666666666667   0.45000000000000   Co   (4f)
   0.16920000000000   0.33840000000000   0.04320000000000   Ga  (12k)
  -0.33840000000000  -0.16920000000000   0.04320000000000   Ga  (12k)
   0.16920000000000  -0.16920000000000   0.04320000000000   Ga  (12k)
  -0.16920000000000  -0.33840000000000   0.54320000000000   Ga  (12k)
   0.33840000000000   0.16920000000000   0.54320000000000   Ga  (12k)
  -0.16920000000000   0.16920000000000   0.54320000000000   Ga  (12k)
   0.33840000000000   0.16920000000000  -0.04320000000000   Ga  (12k)
  -0.16920000000000  -0.33840000000000  -0.04320000000000   Ga  (12k)
  -0.16920000000000   0.16920000000000  -0.04320000000000   Ga  (12k)
  -0.33840000000000  -0.16920000000000   0.45680000000000   Ga  (12k)
   0.16920000000000   0.33840000000000   0.45680000000000   Ga  (12k)
   0.16920000000000  -0.16920000000000   0.45680000000000   Ga  (12k)
   0.00000000000000   0.00000000000000   0.25000000000000   Lu   (2b)
   0.00000000000000   0.00000000000000   0.75000000000000   Lu   (2b)
   0.52310000000000   1.04620000000000   0.25000000000000   Lu   (6h)
  -1.04620000000000  -0.52310000000000   0.25000000000000   Lu   (6h)
   0.52310000000000  -0.52310000000000   0.25000000000000   Lu   (6h)
  -0.52310000000000  -1.04620000000000   0.75000000000000   Lu   (6h)
   1.04620000000000   0.52310000000000   0.75000000000000   Lu   (6h)
  -0.52310000000000   0.52310000000000   0.75000000000000   Lu   (6h)
```

Hexagonal Delafossite ($CuAlO_2$): ABC2_hP8_194_a_c_f - CIF

```
# CIF file
data_findsym-output
_audit_creation_method FINDSYM

_chemical_name_mineral 'Delafossite'
_chemical_formula_sum 'Al Cu O2'

loop_
_publ_author_name
 'B. U. K\"{o}hler'
 'M. Jansen'
_journal_name_full_name
;
 Zeitschrift fur Anorganische und Allgemeine Chemie
;
_journal_volume 543
_journal_year 1986
_journal_page_first 73
_journal_page_last 80
_publ_Section_title
;
 Darstellung und Strukturdaten von, ``Delafossiten\'\' CuMO$_{2}$ ($M$ =
   ↪ Al, Ga, Sc, Y)
;

# Found in Crystal chemistry and electrical properties of the
   ↪ delafossite structure, 2006

_aflow_title 'Hexagonal Delafossite (CuAlO$_{2}$) Structure'
_aflow_proto 'ABC2_hP8_194_a_c_f'
_aflow_params 'a,c/a,z_{3}'
_aflow_params_values '2.863,3.95179881243,0.0851'
_aflow_Strukturbericht 'None'
_aflow_Pearson 'hP8'

_symmetry_space_group_name_H-M "P 63/m 2/m 2/c"
_symmetry_Int_Tables_number 194

_cell_length_a      2.86300
_cell_length_b      2.86300
_cell_length_c      11.31400
_cell_angle_alpha   90.00000
_cell_angle_beta    90.00000
_cell_angle_gamma   120.00000

loop_
_space_group_symop_id
_space_group_symop_operation_xyz
1 x,y,z
2 x-y,x,z+1/2
3 -y,x-y,z
4 -x,-y,z+1/2
5 -x+y,-x,z
6 y,-x+y,z+1/2
7 x-y,-y,-z
8 x,x-y,-z+1/2
9 y,x,-z
10 -x+y,y,-z+1/2
11 -x,-x+y,-z
12 -y,-x,-z+1/2
13 -x,-y,-z
14 -x+y,-x,-z+1/2
15 y,-x+y,-z
16 x,y,-z+1/2
17 x-y,x,-z
18 -y,x-y,-z+1/2
19 -x+y,y,z
20 -x,-x+y,z+1/2
21 -y,-x,z
22 x-y,-y,z+1/2
23 x,x-y,z
24 y,x,z+1/2

loop_
_atom_site_label
_atom_site_type_symbol
_atom_site_symmetry_multiplicity
_atom_site_Wyckoff_label
_atom_site_fract_x
_atom_site_fract_y
_atom_site_fract_z
_atom_site_occupancy
Al1 Al 2 a 0.00000 0.00000 0.00000 1.00000
Cu1 Cu 2 c 0.33333 0.66667 0.25000 1.00000
O1 O 4 f 0.33333 0.66667 0.08510 1.00000
```

Hexagonal Delafossite ($CuAlO_2$): ABC2_hP8_194_a_c_f - POSCAR

```
ABC2_hP8_194_a_c_f & a,c/a,z3 --params=2.863,3.95179881243,0.0851 & P6_{
   ↪ 3}/mmc D_{6h}^{4} #194 (acf) & hP8 & None & AlCuO2 &
   ↪ Delafossite & B. U. K\"{o}hler and M. Jansen, Z. Anorg. Allg.
   ↪ Chem. 543, 73-80 (1986)
   1.00000000000000
   1.43150000000000  -2.47943073103485   0.00000000000000
   1.43150000000000   2.47943073103485   0.00000000000000
   0.00000000000000   0.00000000000000  11.31400000000000
    Al     Cu      O
     2      2      4
Direct
   0.00000000000000   0.00000000000000   0.00000000000000   Al   (2a)
   0.00000000000000   0.00000000000000   0.50000000000000   Al   (2a)
   0.33333333333333   0.66666666666667   0.25000000000000   Cu   (2c)
   0.66666666666667   0.33333333333333   0.75000000000000   Cu   (2c)
   0.33333333333333   0.66666666666667   0.08510000000000    O   (4f)
   0.66666666666667   0.33333333333333   0.58510000000000    O   (4f)
   0.66666666666667   0.33333333333333  -0.08510000000000    O   (4f)
   0.33333333333333   0.66666666666667   0.41490000000000    O   (4f)
```

$LiZn_2$ ($C_k$): AB_hP4_194_a_c - CIF

```
# CIF file
data_findsym-output
_audit_creation_method FINDSYM

_chemical_name_mineral 'LiZn2'
_chemical_formula_sum 'Li Zn'

loop_
_publ_author_name
 'E. Zintl'
 'A. Schneider'
_journal_name_full_name
;
 Zeitschrift f{\"u}r Elektrochemie und angewandte physikalische Chemie
;
_journal_volume 41
_journal_year 1935
_journal_page_first 764
_journal_page_last 767
_publ_Section_title
;
 R{\"o}ntgenanalyse der Lithium-Zink-Legierungen (15. Mitteilung {\"u}
   ↪ ber Metalle und Legierungen)
;

# Found in A Handbook of Lattice Spacings and Structures of Metals and
   ↪ Alloys, 1958 Found in A Handbook of Lattice Spacings and
   ↪ Structures of Metals and Alloys, {N.~R.~C. No. 4303},

_aflow_title 'LiZn$_{2}$ ($C_{k}$) Structure'
_aflow_proto 'AB_hP4_194_a_c'
_aflow_params 'a,c/a'
_aflow_params_values '4.362,0.575424117377'
_aflow_Strukturbericht '$C_{k}$'
_aflow_Pearson 'hP4'

_symmetry_space_group_name_H-M "P 63/m 2/m 2/c"
_symmetry_Int_Tables_number 194

_cell_length_a      4.36200
_cell_length_b      4.36200
```

```
_cell_length_c      2.51000
_cell_angle_alpha  90.00000
_cell_angle_beta   90.00000
_cell_angle_gamma  120.00000

loop_
_space_group_symop_id
_space_group_symop_operation_xyz
1 x,y,z
2 x-y,x,z+1/2
3 -y,x-y,z
4 -x,-y,z+1/2
5 -x+y,-x,z
6 y,-x+y,z+1/2
7 x-y,-y,-z
8 x,x-y,-z+1/2
9 y,x,-z
10 -x+y,y,-z+1/2
11 -x,-x+y,-z
12 -y,-x,-z+1/2
13 -x,-y,-z
14 -x+y,-x,-z+1/2
15 y,-x+y,-z
16 x,y,-z+1/2
17 x-y,x,-z
18 -y,x-y,-z+1/2
19 -x+y,y,z
20 -x,-x+y,z+1/2
21 -y,-x,z
22 x-y,-y,z+1/2
23 x,x-y,z
24 y,x,z+1/2

loop_
_atom_site_label
_atom_site_type_symbol
_atom_site_symmetry_multiplicity
_atom_site_Wyckoff_label
_atom_site_fract_x
_atom_site_fract_y
_atom_site_fract_z
_atom_site_occupancy
Li1 Li 2 a 0.00000 0.00000 0.00000 0.40000
Zn1 Zn 2 c 0.33333 0.66667 0.25000 1.00000
```

$LiZn_2$ ($C_k$): AB_hP4_194_a_c - POSCAR

```
AB_hP4_194_a_c & a,c/a --params=4.362,0.575424117377 & P6_{3}/mmc D_{6h
    ↪ }^{4} #194 (ac) & hP4 & $C_{k}$ & LiZn2 & LiZn2 & E. Zintl and
    ↪ A. Schneider, Z. Elektrochem. 41, 764-767 (1935)
   1.00000000000000
   2.18100000000000  -3.77760281130772   0.00000000000000
   2.18100000000000   3.77760281130772   0.00000000000000
   0.00000000000000   0.00000000000000   2.51000000000000
    Li    Zn
     2     2
Direct
   0.00000000000000   0.00000000000000   0.00000000000000   Li   (2a)
   0.00000000000000   0.00000000000000   0.50000000000000   Li   (2a)
   0.33333333333333   0.66666666666667   0.25000000000000   Zn   (2c)
   0.66666666666667   0.33333333333333   0.75000000000000   Zn   (2c)
```

$Fe_2N$ (approximate, $L'3_0$): AB_hP4_194_c_a - CIF

```
# CIF file
data_findsym-output
_audit_creation_method FINDSYM

_chemical_name_mineral 'Fe2N'
_chemical_formula_sum 'Fe N'

loop_
_publ_author_name
 'C. J. Smithells'
_journal_year 1955
_publ_Section_title
;
 Metals Reference Book
;

_aflow_title 'Fe$_{2}$N (approximate, $L\'3_{0}$) Structure'
_aflow_proto 'AB_hP4_194_c_a'
_aflow_params 'a,c/a'
_aflow_params_values '2.767,1.59522949042'
_aflow_Strukturbericht '$L\'3_{0}$'
_aflow_Pearson 'hP4'

_symmetry_space_group_name_H-M "P 63/m 2/m 2/c"
_symmetry_Int_Tables_number 194

_cell_length_a      2.76700
_cell_length_b      2.76700
_cell_length_c      4.41400
_cell_angle_alpha  90.00000
_cell_angle_beta   90.00000
_cell_angle_gamma  120.00000

loop_
_space_group_symop_id
_space_group_symop_operation_xyz
1 x,y,z
2 x-y,x,z+1/2
3 -y,x-y,z
4 -x,-y,z+1/2
5 -x+y,-x,z
6 y,-x+y,z+1/2
7 x-y,-y,-z
8 x,x-y,-z+1/2
9 y,x,-z
10 -x+y,y,-z+1/2
11 -x,-x+y,-z
12 -y,-x,-z+1/2
13 -x,-y,-z
14 -x+y,-x,-z+1/2
15 y,-x+y,-z
16 x,y,-z+1/2
17 x-y,x,-z
18 -y,x-y,-z+1/2
19 -x+y,y,z
20 -x,-x+y,z+1/2
21 -y,-x,z
22 x-y,-y,z+1/2
23 x,x-y,z
24 y,x,z+1/2

loop_
_atom_site_label
_atom_site_type_symbol
_atom_site_symmetry_multiplicity
_atom_site_Wyckoff_label
_atom_site_fract_x
_atom_site_fract_y
_atom_site_fract_z
_atom_site_occupancy
N1 N 2 a 0.00000 0.00000 0.00000 0.50000
Fe1 Fe 2 c 0.33333 0.66667 0.25000 1.00000
```

$Fe_2N$ (approximate, $L'3_0$): AB_hP4_194_c_a - POSCAR

```
AB_hP4_194_c_a & a,c/a --params=2.767,1.59522949042 & P6_{3}/mmc D_{6h
    ↪ }^{4} #194 (ac) & hP4 & $L\'3_{0}$ & Fe2N & Fe2N & C. J.
    ↪ Smithells, (1955)
   1.00000000000000
   1.38350000000000  -2.39629229227154   0.00000000000000
   1.38350000000000   2.39629229227154   0.00000000000000
   0.00000000000000   0.00000000000000   4.41400000000000
    Fe    N
     2     2
Direct
   0.33333333333333   0.66666666666667   0.25000000000000   Fe   (2c)
   0.66666666666667   0.33333333333333   0.75000000000000   Fe   (2c)
   0.00000000000000   0.00000000000000   0.00000000000000    N   (2a)
   0.00000000000000   0.00000000000000   0.50000000000000    N   (2a)
```

Cubic $Cu_2OSeO_3$: A2B4C_cP56_198_ab_2a2b_2a - CIF

```
# CIF file
data_findsym-output
_audit_creation_method FINDSYM

_chemical_name_mineral 'Cu2O4Se'
_chemical_formula_sum 'Cu2 O4 Se'

loop_
_publ_author_name
 'H. Effenberger'
 'F. Pertlik'
_journal_name_full_name
;
 Monatshefte f{\"u}r Chemie - Chemical Monthly
;
_journal_volume 117
_journal_year 1986
_journal_page_first 887
_journal_page_last 896
_publ_Section_title
;
 Die Kristallstrukturen der Kupfer(II)-oxo-selenite Cu$_{2}$O(SeO$_{3}$)
     ↪ (kubisch und monoklin) und Cu$_{4}$O(SeO$_{3}$)$_{3}$ (
     ↪ monoklin und triklin)
;

# Found in Magnon spectrum of the helimagnetic insulator Cu$_{2}$OSeO$_{
    ↪ 3}$, 2016

_aflow_title 'Cubic Cu$_{2}$OSeO$_{3}$ Structure'
_aflow_proto 'A2B4C_cP56_198_ab_2a2b_2a'
_aflow_params 'a,x_{1},x_{2},x_{3},x_{4},x_{5},x_{6},y_{6},z_{6},x_{7},
    ↪ y_{7},z_{7},x_{8},y_{8},z_{8}'
_aflow_params_values '8.925,0.886,0.0105,0.7621,0.459,0.2113,0.1335,
    ↪ 0.1211,0.8719,0.2699,0.4834,0.4706,0.271,0.1892,0.0313'
_aflow_Strukturbericht 'None'
_aflow_Pearson 'cP56'

_symmetry_space_group_name_H-M "P 21 3"
_symmetry_Int_Tables_number 198

_cell_length_a      8.92500
_cell_length_b      8.92500
_cell_length_c      8.92500
_cell_angle_alpha  90.00000
_cell_angle_beta   90.00000
_cell_angle_gamma  90.00000

loop_
_space_group_symop_id
_space_group_symop_operation_xyz
1 x,y,z
2 x+1/2,-y+1/2,-z
3 -x,y+1/2,-z+1/2
```

```
4 -x+1/2,-y,z+1/2
5 y,z,x
6 y+1/2,-z+1/2,-x
7 -y,z+1/2,-x+1/2
8 -y+1/2,-z,x+1/2
9 z,x,y
10 z+1/2,-x+1/2,-y
11 -z,x+1/2,-y+1/2
12 -z+1/2,-x,y+1/2

loop_
_atom_site_label
_atom_site_type_symbol
_atom_site_symmetry_multiplicity
_atom_site_Wyckoff_label
_atom_site_fract_x
_atom_site_fract_y
_atom_site_fract_z
_atom_site_occupancy
Cu1 Cu 4 a 0.88600 0.88600 0.88600 1.00000
O1 O 4 a 0.01050 0.01050 0.01050 1.00000
O2 O 4 a 0.76210 0.76210 0.76210 1.00000
Se1 Se 4 a 0.45900 0.45900 0.45900 1.00000
Se2 Se 4 a 0.21130 0.21130 0.21130 1.00000
Cu2 Cu 12 b 0.13350 0.12110 0.87190 1.00000
O3 O 12 b 0.26990 0.48340 0.47060 1.00000
O4 O 12 b 0.27100 0.18920 0.03130 1.00000
```

Cubic $Cu_2OSeO_3$: A2B4C_cP56_198_ab_2a2b_2a - POSCAR

```
A2B4C_cP56_198_ab_2a2b_2a & a,x1,x2,x3,x4,x5,x6,y6,z6,x7,y7,z7,x8,y8,z8
   ↪ --params=8.925,0.886,0.0105,0.7621,0.459,0.2113,0.1335,0.1211,
   ↪ 0.8719,0.2699,0.4834,0.4706,0.271,0.1892,0.0313 & P2_{1}3 T^{4}
   ↪ #198 (a^5b^3) & cP56 & None & Cu2O4Se & Cu2O4Se & H.
   ↪ Effenberger and F. Pertlik, Monatshefte f{"u}r Chemie -
   ↪ Chemical Monthly 117, 887-896 (1986)
   1.00000000000000
   8.92500000000000   0.00000000000000   0.00000000000000
   0.00000000000000   8.92500000000000   0.00000000000000
   0.00000000000000   0.00000000000000   8.92500000000000
    Cu     O    Se
    16    32     8
Direct
   0.88600000000000   0.88600000000000   0.88600000000000   Cu   (4a)
  -0.38600000000000  -0.88600000000000   1.38600000000000   Cu   (4a)
  -0.88600000000000   1.38600000000000  -0.38600000000000   Cu   (4a)
   1.38600000000000  -0.38600000000000  -0.88600000000000   Cu   (4a)
   0.13350000000000   0.12110000000000   0.87190000000000   Cu  (12b)
   0.36650000000000  -0.12110000000000   1.37190000000000   Cu  (12b)
  -0.13350000000000   0.62110000000000  -0.37190000000000   Cu  (12b)
   0.63350000000000   0.37890000000000  -0.87190000000000   Cu  (12b)
   0.87190000000000   0.13350000000000   0.12110000000000   Cu  (12b)
   1.37190000000000   0.36650000000000  -0.12110000000000   Cu  (12b)
  -0.37190000000000  -0.13350000000000   0.62110000000000   Cu  (12b)
  -0.87190000000000   0.63350000000000   0.37890000000000   Cu  (12b)
   0.12110000000000   0.87190000000000   0.13350000000000   Cu  (12b)
  -0.12110000000000   1.37190000000000   0.36650000000000   Cu  (12b)
   0.62110000000000  -0.37190000000000  -0.13350000000000   Cu  (12b)
   0.37890000000000  -0.87190000000000   0.63350000000000   Cu  (12b)
   0.01050000000000   0.01050000000000   0.01050000000000    O   (4a)
   0.48950000000000  -0.01050000000000   0.51050000000000    O   (4a)
  -0.01050000000000   0.51050000000000   0.48950000000000    O   (4a)
   0.51050000000000   0.48950000000000  -0.01050000000000    O   (4a)
   0.76210000000000   0.76210000000000   0.76210000000000    O   (4a)
  -0.26210000000000  -0.76210000000000   1.26210000000000    O   (4a)
  -0.76210000000000   1.26210000000000  -0.26210000000000    O   (4a)
   1.26210000000000  -0.26210000000000  -0.76210000000000    O   (4a)
   0.26990000000000   0.48340000000000   0.47060000000000    O  (12b)
   0.23010000000000  -0.48340000000000   0.97060000000000    O  (12b)
  -0.26990000000000   0.98340000000000   0.02940000000000    O  (12b)
   0.76990000000000   0.01660000000000  -0.47060000000000    O  (12b)
   0.47060000000000   0.26990000000000   0.48340000000000    O  (12b)
   0.97060000000000   0.23010000000000  -0.48340000000000    O  (12b)
   0.02940000000000  -0.26990000000000   0.98340000000000    O  (12b)
  -0.47060000000000   0.76990000000000   0.01660000000000    O  (12b)
   0.48340000000000   0.47060000000000   0.26990000000000    O  (12b)
  -0.48340000000000   0.97060000000000   0.23010000000000    O  (12b)
   0.98340000000000   0.02940000000000  -0.26990000000000    O  (12b)
   0.01660000000000  -0.47060000000000   0.76990000000000    O  (12b)
   0.27100000000000   0.18920000000000   0.03130000000000    O  (12b)
   0.22900000000000  -0.18920000000000   0.53130000000000    O  (12b)
  -0.27100000000000   0.68920000000000   0.46870000000000    O  (12b)
   0.77100000000000   0.31080000000000  -0.03130000000000    O  (12b)
   0.03130000000000   0.27100000000000   0.18920000000000    O  (12b)
   0.53130000000000   0.22900000000000  -0.18920000000000    O  (12b)
   0.46870000000000  -0.27100000000000   0.68920000000000    O  (12b)
  -0.03130000000000   0.77100000000000   0.31080000000000    O  (12b)
   0.18920000000000   0.03130000000000   0.27100000000000    O  (12b)
  -0.18920000000000   0.53130000000000   0.22900000000000    O  (12b)
   0.68920000000000   0.46870000000000  -0.27100000000000    O  (12b)
   0.31080000000000  -0.03130000000000   0.77100000000000    O  (12b)
   0.45900000000000   0.45900000000000   0.45900000000000   Se   (4a)
   0.04100000000000  -0.45900000000000   0.95900000000000   Se   (4a)
  -0.45900000000000   0.95900000000000   0.04100000000000   Se   (4a)
   0.95900000000000   0.04100000000000  -0.45900000000000   Se   (4a)
   0.21130000000000   0.21130000000000   0.21130000000000   Se   (4a)
   0.28870000000000  -0.21130000000000   0.71130000000000   Se   (4a)
  -0.21130000000000   0.71130000000000   0.28870000000000   Se   (4a)
   0.71130000000000   0.28870000000000  -0.21130000000000   Se   (4a)
```

$Na_2CaSiO_4$ ($S6_6$): AB2C4D_cP32_198_a_2a_ab_a - CIF

```
# CIF file
data_findsym-output
_audit_creation_method FINDSYM

_chemical_name_mineral 'CaNa2O4Si'
_chemical_formula_sum 'Ca Na2 O4 Si'

loop_
_publ_author_name
 'T. F. W. Barth'
 'E. Posnjak'
_journal_name_full_name
;
 Zeitschrift f{\"u}r Kristallographie - Crystalline Materials
;
_journal_volume 81
_journal_year 1932
_journal_page_first 370
_journal_page_last 375
_publ_Section_title
;
 Silicate structures of the cristobalite type: II. The crystal structure
    ↪ of Na$_{2}$CaSiO$_{4}$
;

_aflow_title 'Na$_{2}$CaSiO$_{4}$ ($S6_{6}$) Structure'
_aflow_proto 'AB2C4D_cP32_198_a_2a_ab_a'
_aflow_params 'a,x_{1},x_{2},x_{3},x_{4},x_{5},x_{6},y_{6},z_{6}'
_aflow_params_values '7.48,-0.007,0.5,0.75,0.133,0.253,0.556,0.667,0.222
    ↪ '
_aflow_Strukturbericht '$S6_{6}$'
_aflow_Pearson 'cP32'

_symmetry_space_group_name_H-M "P 21 3"
_symmetry_Int_Tables_number 198

_cell_length_a     7.48000
_cell_length_b     7.48000
_cell_length_c     7.48000
_cell_angle_alpha  90.00000
_cell_angle_beta   90.00000
_cell_angle_gamma  90.00000

loop_
_space_group_symop_id
_space_group_symop_operation_xyz
1 x,y,z
2 x+1/2,-y+1/2,-z
3 -x,y+1/2,-z+1/2
4 -x+1/2,-y,z+1/2
5 y,z,x
6 y+1/2,-z+1/2,-x
7 -y,z+1/2,-x+1/2
8 -y+1/2,-z,x+1/2
9 z,x,y
10 z+1/2,-x+1/2,-y
11 -z,x+1/2,-y+1/2
12 -z+1/2,-x,y+1/2

loop_
_atom_site_label
_atom_site_type_symbol
_atom_site_symmetry_multiplicity
_atom_site_Wyckoff_label
_atom_site_fract_x
_atom_site_fract_y
_atom_site_fract_z
_atom_site_occupancy
Ca1 Ca 4 a -0.00700 -0.00700 -0.00700 1.00000
Na1 Na 4 a 0.50000 0.50000 0.50000 1.00000
Na2 Na 4 a 0.75000 0.75000 0.75000 1.00000
O1 O 4 a 0.13300 0.13300 0.13300 1.00000
Si1 Si 4 a 0.25300 0.25300 0.25300 1.00000
O2 O 12 b 0.55600 0.66700 0.22200 1.00000
```

$Na_2CaSiO_4$ ($S6_6$): AB2C4D_cP32_198_a_2a_ab_a - POSCAR

```
AB2C4D_cP32_198_a_2a_ab_a & a,x1,x2,x3,x4,x5,x6,y6,z6 --params=7.48,-
   ↪ 0.007,0.5,0.75,0.133,0.253,0.556,0.667,0.222 & P2_{1}3 T^{4} #
   ↪ 198 (a^5b) & cP32 & $S6_{6}$ & CaNa2O4Si & CaNa2O4Si & T. F. W.
   ↪ Barth and E. Posnjak, Zeitschrift f{"u}r Kristallographie -
   ↪ Crystalline Materials 81, 370-375 (1932)
   1.00000000000000
   7.48000000000000   0.00000000000000   0.00000000000000
   0.00000000000000   7.48000000000000   0.00000000000000
   0.00000000000000   0.00000000000000   7.48000000000000
    Ca    Na     O    Si
     4     8    16     4
Direct
  -0.00700000000000  -0.00700000000000  -0.00700000000000   Ca   (4a)
   0.50700000000000   0.00700000000000   0.49300000000000   Ca   (4a)
   0.00700000000000   0.49300000000000   0.50700000000000   Ca   (4a)
   0.49300000000000   0.50700000000000   0.00700000000000   Ca   (4a)
   0.50000000000000   0.50000000000000   0.50000000000000   Na   (4a)
   0.00000000000000  -0.50000000000000   1.00000000000000   Na   (4a)
  -0.50000000000000   1.00000000000000   0.00000000000000   Na   (4a)
   1.00000000000000   0.00000000000000  -0.50000000000000   Na   (4a)
   0.75000000000000   0.75000000000000   0.75000000000000   Na   (4a)
  -0.25000000000000  -0.75000000000000   1.25000000000000   Na   (4a)
  -0.75000000000000   1.25000000000000  -0.25000000000000   Na   (4a)
   1.25000000000000  -0.25000000000000  -0.75000000000000   Na   (4a)
   0.13300000000000   0.13300000000000   0.13300000000000    O   (4a)
   0.36700000000000  -0.13300000000000   0.63300000000000    O   (4a)
  -0.13300000000000   0.63300000000000   0.36700000000000    O   (4a)
   0.63300000000000   0.36700000000000  -0.13300000000000    O   (4a)
   0.55600000000000   0.66700000000000   0.22200000000000    O  (12b)
  -0.05600000000000  -0.66700000000000   0.72200000000000    O  (12b)
  -0.55600000000000   1.16700000000000   0.27800000000000    O  (12b)
```

```
   1.05600000000000   -0.16700000000000   -0.22200000000000    O   (12b)
   0.22200000000000    0.55600000000000    0.66700000000000    O   (12b)
   0.72200000000000   -0.05600000000000   -0.66700000000000    O   (12b)
   0.27800000000000   -0.55600000000000    1.16700000000000    O   (12b)
  -0.22200000000000    1.05600000000000   -0.16700000000000    O   (12b)
   0.66700000000000    0.22200000000000    0.55600000000000    O   (12b)
  -0.66700000000000    0.72200000000000   -0.05600000000000    O   (12b)
   1.16700000000000    0.27800000000000   -0.55600000000000    O   (12b)
  -0.16700000000000   -0.22200000000000    1.05600000000000    O   (12b)
   0.25300000000000    0.25300000000000    0.25300000000000   Si    (4a)
   0.24700000000000   -0.25300000000000    0.75300000000000   Si    (4a)
  -0.25300000000000    0.75300000000000    0.24700000000000   Si    (4a)
   0.75300000000000    0.24700000000000   -0.25300000000000   Si    (4a)
```

$\alpha$-Carnegieite ($NaAlSiO_4$, $S6_5$): ABC4D_cP28_198_a_a_ab_a - CIF

```
# CIF file
data_findsym-output
_audit_creation_method FINDSYM

_chemical_name_mineral '$\alpha$-carnegieite'
_chemical_formula_sum 'Al Na O4 Si'

loop_
_publ_author_name
 'T. F. W. Barth'
 'E. Posnjak'
_journal_name_full_name
;
 Zeitschrift f{\"u}r Kristallographie - Crystalline Materials
;
_journal_volume 81
_journal_year 1932
_journal_page_first 135
_journal_page_last 141
_publ_Section_title
;
 Silicate structures of the cristobalite type: I. The crystal structure
      ↪ of $\alpha$-carnegieite (NaAlSiO$_{4}$)
;

_aflow_title '$\alpha$-Carnegieite (NaAlSiO$_{4}$, $S6_{5}$) Structure'
_aflow_proto 'ABC4D_cP28_198_a_a_ab_a'
_aflow_params 'a,x_{1},x_{2},x_{3},x_{4},x_{5},y_{5},z_{5}'
_aflow_params_values '7.37,0.258,0.744,0.125,0.0,0.658,0.644,0.0556'
_aflow_Strukturbericht '$S6_{5}$'
_aflow_Pearson 'cP28'

_symmetry_space_group_name_H-M "P 21 3"
_symmetry_Int_Tables_number 198

_cell_length_a      7.37000
_cell_length_b      7.37000
_cell_length_c      7.37000
_cell_angle_alpha 90.00000
_cell_angle_beta  90.00000
_cell_angle_gamma 90.00000

loop_
_space_group_symop_id
_space_group_symop_operation_xyz
1 x,y,z
2 x+1/2,-y+1/2,-z
3 -x,y+1/2,-z+1/2
4 -x+1/2,-y,z+1/2
5 y,z,x
6 y+1/2,-z+1/2,-x
7 -y,z+1/2,-x+1/2
8 -y+1/2,-z,x+1/2
9 z,x,y
10 z+1/2,-x+1/2,-y
11 -z,x+1/2,-y+1/2
12 -z+1/2,-x,y+1/2

loop_
_atom_site_label
_atom_site_type_symbol
_atom_site_symmetry_multiplicity
_atom_site_Wyckoff_label
_atom_site_fract_x
_atom_site_fract_y
_atom_site_fract_z
_atom_site_occupancy
Al1 Al 4 a 0.25800 0.25800 0.25800 1.00000
Na1 Na 4 a 0.74400 0.74400 0.74400 1.00000
O1 O 4 a 0.12500 0.12500 0.12500 1.00000
Si1 Si 4 a 0.00000 0.00000 0.00000 1.00000
O2 O 12 b 0.65800 0.64400 0.05560 1.00000
```

$\alpha$-Carnegieite ($NaAlSiO_4$, $S6_5$): ABC4D_cP28_198_a_a_ab_a - POSCAR

```
ABC4D_cP28_198_a_a_ab_a & a,x1,x2,x3,x4,x5,y5,z5 --params=7.37,0.258,
      ↪ 0.744,0.125,0.0,0.658,0.644,0.0556 & P2_{1}3 T^{4} #198 (a^4b)
      ↪ & cP28 & $S6_{5}$ & AlNaO4Si & $\alpha$-carnegieite & T. F. W.
      ↪ Barth and E. Posnjak, Zeitschrift f{"u}r Kristallographie -
      ↪ Crystalline Materials 81, 135-141 (1932)
   1.00000000000000
   7.37000000000000    0.00000000000000    0.00000000000000
   0.00000000000000    7.37000000000000    0.00000000000000
   0.00000000000000    0.00000000000000    7.37000000000000
    Al     Na      O     Si
     4      4     16      4
Direct
   0.25800000000000    0.25800000000000    0.25800000000000   Al    (4a)
   0.24200000000000   -0.25800000000000    0.75800000000000   Al    (4a)
  -0.25800000000000    0.75800000000000    0.24200000000000   Al    (4a)
   0.75800000000000    0.24200000000000   -0.25800000000000   Al    (4a)
   0.74400000000000    0.74400000000000    0.74400000000000   Na    (4a)
  -0.24400000000000   -0.74400000000000    1.24400000000000   Na    (4a)
  -0.74400000000000    1.24400000000000   -0.24400000000000   Na    (4a)
   1.24400000000000   -0.24400000000000   -0.74400000000000   Na    (4a)
   0.12500000000000    0.12500000000000    0.12500000000000    O    (4a)
   0.37500000000000   -0.12500000000000    0.62500000000000    O    (4a)
  -0.12500000000000    0.62500000000000    0.37500000000000    O    (4a)
   0.62500000000000    0.37500000000000   -0.12500000000000    O    (4a)
   0.65800000000000    0.64400000000000    0.05560000000000    O   (12b)
  -0.15800000000000   -0.64400000000000    0.55560000000000    O   (12b)
  -0.65800000000000    1.14400000000000    0.44440000000000    O   (12b)
   1.15800000000000   -0.14400000000000   -0.05560000000000    O   (12b)
   0.05560000000000    0.65800000000000    0.64400000000000    O   (12b)
   0.55560000000000   -0.15800000000000   -0.64400000000000    O   (12b)
   0.44440000000000   -0.65800000000000    1.14400000000000    O   (12b)
  -0.05560000000000    1.15800000000000   -0.14400000000000    O   (12b)
   0.64400000000000    0.05560000000000    0.65800000000000    O   (12b)
  -0.64400000000000    0.55560000000000   -0.15800000000000    O   (12b)
   1.14400000000000    0.44440000000000   -0.65800000000000    O   (12b)
  -0.14400000000000   -0.05560000000000    1.15800000000000    O   (12b)
   0.00000000000000    0.00000000000000    0.00000000000000   Si    (4a)
   0.50000000000000    0.00000000000000    0.50000000000000   Si    (4a)
   0.00000000000000    0.50000000000000    0.50000000000000   Si    (4a)
   0.50000000000000    0.50000000000000    0.00000000000000   Si    (4a)
```

$C26_a$ ($NO_2$) (*obsolete*): AB2_cI36_199_b_c - CIF

```
# CIF file
data_findsym-output
_audit_creation_method FINDSYM

_chemical_name_mineral 'NO2'
_chemical_formula_sum 'N O2'

loop_
_publ_author_name
 'L. Vegard'
_journal_name_full_name
;
 Zeitschrift f{\"u}r Physik
;
_journal_volume 68
_journal_year 1931
_journal_page_first 184
_journal_page_last 203
_publ_Section_title
;
 Die Struktur von festem N$_{2}$O$_{4}$ bei der Temperatur von fl\"{u}
      ↪ ssiger Luft
;

# Found in Strukturbericht Band II 1928-1932, 1937

_aflow_title '$C26_{a}$ (NO$_{2}$) ({\em{obsolete}}) Structure'
_aflow_proto 'AB2_cI36_199_b_c'
_aflow_params 'a,x_{1},x_{2},y_{2},z_{2}'
_aflow_params_values '7.77,0.4,0.178,0.25,0.403'
_aflow_Strukturbericht '$C26_{a}$'
_aflow_Pearson 'cI36'

_symmetry_space_group_name_H-M "I 21 3"
_symmetry_Int_Tables_number 199

_cell_length_a      7.77000
_cell_length_b      7.77000
_cell_length_c      7.77000
_cell_angle_alpha 90.00000
_cell_angle_beta  90.00000
_cell_angle_gamma 90.00000

loop_
_space_group_symop_id
_space_group_symop_operation_xyz
1 x,y,z
2 x,-y,-z+1/2
3 -x+1/2,y,-z
4 -x,-y+1/2,z
5 y,z,x
6 y,-z,-x+1/2
7 -y+1/2,z,-x
8 -y,-z+1/2,x
9 z,x,y
10 z,-x,-y+1/2
11 -z+1/2,x,-y
12 -z,-x+1/2,y
13 x+1/2,y+1/2,z+1/2
14 x+1/2,-y+1/2,-z
15 -x,y+1/2,-z+1/2
16 -x+1/2,-y,z+1/2
17 y+1/2,z+1/2,x+1/2
18 y+1/2,-z+1/2,-x
19 -y,z+1/2,-x+1/2
20 -y+1/2,-z,x+1/2
21 z+1/2,x+1/2,y+1/2
22 z+1/2,-x+1/2,-y
23 -z,x+1/2,-y+1/2
24 -z+1/2,-x,y+1/2

loop_
_atom_site_label
_atom_site_type_symbol
_atom_site_symmetry_multiplicity
_atom_site_Wyckoff_label
```

```
_atom_site_fract_x
_atom_site_fract_y
_atom_site_fract_z
_atom_site_occupancy
N1 N 12 b 0.40000 0.00000 0.25000 1.00000
O1 O 24 c 0.17800 0.25000 0.40300 1.00000
```

C26$_a$ ($NO_2$) (*obsolete*): AB2_cI36_199_b_c - POSCAR

```
AB2_cI36_199_b_c & a,x1,x2,y2,z2 --params=7.77,0.4,0.178,0.25,0.403 &
    ↪ I2_{1}3 T^{5} #199 (bc) & cI36 & $C26_{a}$ & NO2 & NO2 & L.
    ↪ Vegard, Z. Phys. 68, 184-203 (1931)
   1.00000000000000
  -3.88500000000000   3.88500000000000   3.88500000000000
   3.88500000000000  -3.88500000000000   3.88500000000000
   3.88500000000000   3.88500000000000  -3.88500000000000
     N     O
     6    12
Direct
   0.25000000000000   0.65000000000000   0.40000000000000   N   (12b)
   0.75000000000000  -0.15000000000000   0.10000000000000   N   (12b)
   0.40000000000000   0.25000000000000   0.65000000000000   N   (12b)
   0.10000000000000   0.75000000000000  -0.15000000000000   N   (12b)
   0.65000000000000   0.40000000000000   0.25000000000000   N   (12b)
  -0.15000000000000   0.10000000000000   0.75000000000000   N   (12b)
   0.65300000000000   0.58100000000000   0.42800000000000   O   (24c)
   0.65300000000000   0.22500000000000   0.07200000000000   O   (24c)
  -0.15300000000000  -0.08100000000000   0.57200000000000   O   (24c)
  -0.15300000000000   0.27500000000000  -0.07200000000000   O   (24c)
   0.42800000000000   0.65300000000000   0.58100000000000   O   (24c)
   0.07200000000000   0.65300000000000   0.22500000000000   O   (24c)
   0.57200000000000  -0.15300000000000  -0.08100000000000   O   (24c)
  -0.07200000000000  -0.15300000000000   0.27500000000000   O   (24c)
   0.58100000000000   0.42800000000000   0.65300000000000   O   (24c)
   0.22500000000000   0.07200000000000   0.65300000000000   O   (24c)
  -0.08100000000000   0.57200000000000  -0.15300000000000   O   (24c)
   0.27500000000000  -0.07200000000000  -0.15300000000000   O   (24c)
```

$Bi_3Ru_3O_{11}$: A3B11C3_cP68_201_be_efh_g - CIF

```
# CIF file
data_findsym-output
_audit_creation_method FINDSYM

_chemical_name_mineral 'Bi3O11Ru3'
_chemical_formula_sum 'Bi3 O11 Ru3'

loop_
_publ_author_name
 'F. Abraham'
 'D. Thomas'
 'G. Nowogrocki'
_journal_name_full_name
;
 Bulletin de la Societ{\'e} fran{\c{c}}aise de Mineralogie et de
    ↪ Crystallographie
;
_journal_volume 98
_journal_year 1975
_journal_page_first 25
_journal_page_last 29
_publ_Section_title
;
 Structure cristalline de Bi$_{3}$Ru$_{3}$O$_{11}$
;

# Found in The American Mineralogist Crystal Structure Database, 2003

_aflow_title 'Bi$_{3}$Ru$_{3}$O$_{11}$ Structure'
_aflow_proto 'A3B11C3_cP68_201_be_efh_g'
_aflow_params 'a,x_{2},x_{3},x_{4},x_{5},x_{6},y_{6},z_{6}'
_aflow_params_values '9.302,0.38379,0.152,0.59,0.3897,0.599,0.247,0.547'
_aflow_Strukturbericht 'None'
_aflow_Pearson 'cP68'

_symmetry_space_group_name_H-M "P 2/n -3 (origin choice 2)"
_symmetry_Int_Tables_number 201

_cell_length_a      9.30200
_cell_length_b      9.30200
_cell_length_c      9.30200
_cell_angle_alpha 90.00000
_cell_angle_beta  90.00000
_cell_angle_gamma 90.00000

loop_
_space_group_symop_id
_space_group_symop_operation_xyz
1 x,y,z
2 x,-y+1/2,-z+1/2
3 -x+1/2,y,-z+1/2
4 -x+1/2,-y+1/2,z
5 y,z,x
6 y,-z+1/2,-x+1/2
7 -y+1/2,z,-x+1/2
8 -y+1/2,-z+1/2,x
9 z,x,y
10 z,-x+1/2,-y+1/2
11 -z+1/2,x,-y+1/2
12 -z+1/2,-x+1/2,y
13 -x,-y,-z
14 -x,y+1/2,z+1/2
15 x+1/2,-y,z+1/2
16 x+1/2,y+1/2,-z
17 -y,-z,-x
18 -y,z+1/2,x+1/2
19 y+1/2,-z,x+1/2
20 y+1/2,z+1/2,-x
21 -z,-x,-y
22 -z,x+1/2,y+1/2
23 z+1/2,-x,y+1/2
24 z+1/2,x+1/2,-y

loop_
_atom_site_label
_atom_site_type_symbol
_atom_site_symmetry_multiplicity
_atom_site_Wyckoff_label
_atom_site_fract_x
_atom_site_fract_y
_atom_site_fract_z
_atom_site_occupancy
Bi1 Bi 4 b 0.00000 0.00000 0.00000 1.00000
Bi2 Bi 8 e 0.38379 0.38379 0.38379 1.00000
O1 O 8 e 0.15200 0.15200 0.15200 1.00000
O2 O 12 f 0.59000 0.25000 0.25000 1.00000
Ru1 Ru 12 g 0.38970 0.75000 0.25000 1.00000
O3 O 24 h 0.59900 0.24700 0.54700 1.00000
```

$Bi_3Ru_3O_{11}$: A3B11C3_cP68_201_be_efh_g - POSCAR

```
A3B11C3_cP68_201_be_efh_g & a,x2,x3,x4,x5,x6,y6,z6 --params=9.302,
    ↪ 0.38379,0.152,0.59,0.3897,0.599,0.247,0.547 & Pn-3 T_{h}^{2} #
    ↪ 201 (be^2fgh) & cP68 & None & Bi3O11Ru3 & Bi3O11Ru3 & F.
    ↪ Abraham and D. Thomas and G. Nowogrocki, Bull. Soc. fr. Min'
    ↪ eral. Crystallogr. 98, 25-29 (1975)
   1.00000000000000
   9.30200000000000   0.00000000000000   0.00000000000000
   0.00000000000000   9.30200000000000   0.00000000000000
   0.00000000000000   0.00000000000000   9.30200000000000
    Bi     O    Ru
    12    44    12
Direct
   0.00000000000000   0.00000000000000   0.00000000000000   Bi   (4b)
   0.50000000000000   0.50000000000000   0.00000000000000   Bi   (4b)
   0.50000000000000   0.00000000000000   0.50000000000000   Bi   (4b)
   0.00000000000000   0.50000000000000   0.50000000000000   Bi   (4b)
   0.38379000000000   0.38379000000000   0.38379000000000   Bi   (8e)
   0.11621000000000   0.11621000000000   0.38379000000000   Bi   (8e)
   0.11621000000000   0.38379000000000   0.11621000000000   Bi   (8e)
   0.38379000000000   0.11621000000000   0.11621000000000   Bi   (8e)
  -0.38379000000000  -0.38379000000000  -0.38379000000000   Bi   (8e)
   0.88379000000000   0.88379000000000  -0.38379000000000   Bi   (8e)
   0.88379000000000  -0.38379000000000   0.88379000000000   Bi   (8e)
  -0.38379000000000   0.88379000000000   0.88379000000000   Bi   (8e)
   0.15200000000000   0.15200000000000   0.15200000000000   O    (8e)
   0.34800000000000   0.34800000000000   0.15200000000000   O    (8e)
   0.34800000000000   0.15200000000000   0.34800000000000   O    (8e)
   0.15200000000000   0.34800000000000   0.34800000000000   O    (8e)
  -0.15200000000000  -0.15200000000000  -0.15200000000000   O    (8e)
   0.65200000000000   0.65200000000000  -0.15200000000000   O    (8e)
   0.65200000000000  -0.15200000000000   0.65200000000000   O    (8e)
  -0.15200000000000   0.65200000000000   0.65200000000000   O    (8e)
   0.59000000000000   0.25000000000000   0.25000000000000   O   (12f)
  -0.09000000000000   0.25000000000000   0.25000000000000   O   (12f)
   0.25000000000000   0.59000000000000   0.25000000000000   O   (12f)
   0.25000000000000  -0.09000000000000   0.25000000000000   O   (12f)
   0.25000000000000   0.25000000000000   0.59000000000000   O   (12f)
   0.25000000000000   0.25000000000000  -0.09000000000000   O   (12f)
  -0.59000000000000   0.75000000000000   0.75000000000000   O   (12f)
   1.09000000000000   0.75000000000000   0.75000000000000   O   (12f)
   0.75000000000000  -0.59000000000000   0.75000000000000   O   (12f)
   0.75000000000000   1.09000000000000   0.75000000000000   O   (12f)
   0.75000000000000   0.75000000000000  -0.59000000000000   O   (12f)
   0.75000000000000   0.75000000000000   1.09000000000000   O   (12f)
   0.59900000000000   0.24700000000000   0.54700000000000   O   (24h)
  -0.09900000000000   0.25300000000000   0.54700000000000   O   (24h)
  -0.09900000000000   0.24700000000000  -0.04700000000000   O   (24h)
   0.59900000000000   0.25300000000000  -0.04700000000000   O   (24h)
   0.54700000000000   0.59900000000000   0.24700000000000   O   (24h)
   0.54700000000000  -0.09900000000000   0.25300000000000   O   (24h)
  -0.04700000000000  -0.09900000000000   0.24700000000000   O   (24h)
  -0.04700000000000   0.59900000000000   0.25300000000000   O   (24h)
   0.24700000000000   0.54700000000000   0.59900000000000   O   (24h)
   0.25300000000000   0.54700000000000  -0.09900000000000   O   (24h)
   0.24700000000000  -0.04700000000000  -0.09900000000000   O   (24h)
   0.25300000000000  -0.04700000000000   0.59900000000000   O   (24h)
  -0.59900000000000  -0.24700000000000  -0.54700000000000   O   (24h)
   1.09900000000000   0.74700000000000  -0.54700000000000   O   (24h)
   1.09900000000000  -0.24700000000000   1.04700000000000   O   (24h)
  -0.59900000000000   0.74700000000000   1.04700000000000   O   (24h)
  -0.54700000000000  -0.59900000000000  -0.24700000000000   O   (24h)
  -0.54700000000000   1.09900000000000   0.74700000000000   O   (24h)
   1.04700000000000   1.09900000000000  -0.24700000000000   O   (24h)
   1.04700000000000  -0.59900000000000   0.74700000000000   O   (24h)
  -0.24700000000000  -0.54700000000000  -0.59900000000000   O   (24h)
   0.74700000000000  -0.54700000000000   1.09900000000000   O   (24h)
  -0.24700000000000   1.04700000000000   1.09900000000000   O   (24h)
   0.74700000000000   1.04700000000000  -0.59900000000000   O   (24h)
   0.38970000000000   0.75000000000000   0.25000000000000   Ru  (12g)
   0.11030000000000   0.75000000000000   0.25000000000000   Ru  (12g)
   0.25000000000000   0.38970000000000   0.75000000000000   Ru  (12g)
   0.25000000000000   0.11030000000000   0.75000000000000   Ru  (12g)
   0.75000000000000   0.25000000000000   0.38970000000000   Ru  (12g)
   0.75000000000000   0.25000000000000   0.11030000000000   Ru  (12g)
  -0.38970000000000   0.25000000000000   0.75000000000000   Ru  (12g)
   0.88970000000000   0.25000000000000   0.75000000000000   Ru  (12g)
   0.75000000000000  -0.38970000000000   0.25000000000000   Ru  (12g)
   0.75000000000000   0.88970000000000   0.25000000000000   Ru  (12g)
   0.25000000000000   0.75000000000000  -0.38970000000000   Ru  (12g)
```

```
    0.25000000000000     0.75000000000000     0.88970000000000     Ru    (12g)
```

$K_3Co(NO_2)_6$ ($J2_4$): AB3C6D12_cF88_202_a_bc_e_h - CIF

```
# CIF file
data_findsym-output
_audit_creation_method FINDSYM

_chemical_name_mineral 'CoK3N6O12'
_chemical_formula_sum 'Co K3 N6 O12'

loop_
_publ_author_name
 'M. {van Driel}'
 'H. J. Verweel'
_journal_name_full_name
;
 Zeitschrift f{\"u}r Kristallographie - Crystalline Materials
;
_journal_volume 95
_journal_year 1936
_journal_page_first 308
_journal_page_last 314
_publ_Section_title
;
 \"{U}ber die Struktur der Tripelnitrite
;

# Found in Strukturbericht Band IV 1936, 1938

_aflow_title 'K$_{3}$Co(NO$_{2}$)$_{6}$ ($J2_{4}$) Structure'
_aflow_proto 'AB3C6D12_cF88_202_a_bc_e_h'
_aflow_params 'a,x_{4},y_{5},z_{5}'
_aflow_params_values '10.46,0.195,0.235,0.1'
_aflow_Strukturbericht '$J2_{4}$'
_aflow_Pearson 'cF88'

_symmetry_space_group_name_H-M "F 2/m -3"
_symmetry_Int_Tables_number 202

_cell_length_a      10.46000
_cell_length_b      10.46000
_cell_length_c      10.46000
_cell_angle_alpha  90.00000
_cell_angle_beta   90.00000
_cell_angle_gamma  90.00000

loop_
_space_group_symop_id
_space_group_symop_operation_xyz
1 x,y,z
2 x,-y,-z
3 -x,y,-z
4 -x,-y,z
5 y,z,x
6 y,-z,-x
7 -y,z,-x
8 -y,-z,x
9 z,x,y
10 z,-x,-y
11 -z,x,-y
12 -z,-x,y
13 -x,-y,-z
14 -x,y,z
15 x,-y,z
16 x,y,-z
17 -y,-z,-x
18 -y,z,x
19 y,-z,x
20 y,z,-x
21 -z,-x,-y
22 -z,x,y
23 z,-x,y
24 z,x,-y
25 x,y+1/2,z+1/2
26 x,-y+1/2,-z+1/2
27 -x,y+1/2,-z+1/2
28 -x,-y+1/2,z+1/2
29 y,z+1/2,x+1/2
30 y,-z+1/2,-x+1/2
31 -y,z+1/2,-x+1/2
32 -y,-z+1/2,x+1/2
33 z,x+1/2,y+1/2
34 z,-x+1/2,-y+1/2
35 -z,x+1/2,-y+1/2
36 -z,-x+1/2,y+1/2
37 -x,-y+1/2,-z+1/2
38 -x,y+1/2,z+1/2
39 x,-y+1/2,z+1/2
40 x,y+1/2,-z+1/2
41 -y,-z+1/2,-x+1/2
42 -y,z+1/2,x+1/2
43 y,-z+1/2,x+1/2
44 y,z+1/2,-x+1/2
45 -z,-x+1/2,-y+1/2
46 -z,x+1/2,y+1/2
47 z,-x+1/2,y+1/2
48 z,x+1/2,-y+1/2
49 x+1/2,y,z+1/2
50 x+1/2,-y,-z+1/2
51 -x+1/2,y,-z+1/2
52 -x+1/2,-y,z+1/2
53 y+1/2,z,x+1/2
54 y+1/2,-z,-x+1/2
55 -y+1/2,z,-x+1/2
56 -y+1/2,-z,x+1/2
57 z+1/2,x,y+1/2
58 z+1/2,-x,-y+1/2
59 -z+1/2,x,-y+1/2
60 -z+1/2,-x,y+1/2
61 -x+1/2,-y,-z+1/2
62 -x+1/2,y,z+1/2
63 x+1/2,-y,z+1/2
64 x+1/2,y,-z+1/2
65 -y+1/2,-z,-x+1/2
66 -y+1/2,z,x+1/2
67 y+1/2,-z,x+1/2
68 y+1/2,z,-x+1/2
69 -z+1/2,-x,-y+1/2
70 -z+1/2,x,y+1/2
71 z+1/2,-x,y+1/2
72 z+1/2,x,-y+1/2
73 x+1/2,y+1/2,z
74 x+1/2,-y+1/2,-z
75 -x+1/2,y+1/2,-z
76 -x+1/2,-y+1/2,z
77 y+1/2,z+1/2,x
78 y+1/2,-z+1/2,-x
79 -y+1/2,z+1/2,-x
80 -y+1/2,-z+1/2,x
81 z+1/2,x+1/2,y
82 z+1/2,-x+1/2,-y
83 -z+1/2,x+1/2,-y
84 -z+1/2,-x+1/2,y
85 -x+1/2,-y+1/2,-z
86 -x+1/2,y+1/2,z
87 x+1/2,-y+1/2,z
88 x+1/2,y+1/2,-z
89 -y+1/2,-z+1/2,-x
90 -y+1/2,z+1/2,x
91 y+1/2,-z+1/2,x
92 y+1/2,z+1/2,-x
93 -z+1/2,-x+1/2,-y
94 -z+1/2,x+1/2,y
95 z+1/2,-x+1/2,y
96 z+1/2,x+1/2,-y

loop_
_atom_site_label
_atom_site_type_symbol
_atom_site_symmetry_multiplicity
_atom_site_Wyckoff_label
_atom_site_fract_x
_atom_site_fract_y
_atom_site_fract_z
_atom_site_occupancy
Co1 Co 4 a 0.00000 0.00000 0.00000 1.00000
K1 K 4 b 0.50000 0.50000 0.50000 1.00000
K2 K 8 c 0.25000 0.25000 0.25000 1.00000
N1 N 24 e 0.19500 0.00000 0.00000 1.00000
O1 O 48 h 0.00000 0.23500 0.10000 1.00000
```

$K_3Co(NO_2)_6$ ($J2_4$): AB3C6D12_cF88_202_a_bc_e_h - POSCAR

```
AB3C6D12_cF88_202_a_bc_e_h & a,x4,y5,z5 --params=10.46,0.195,0.235,0.1 &
    ↪ Fm-3 T_{h}^{3} #202 (abceh) & cF88 & $J2_{4}$ & CoK3N6O12 &
    ↪ CoK3N6O12 & M. {van Driel} and H. J. Verweel, Zeitschrift f{"u}
    ↪ r Kristallographie - Crystalline Materials 95, 308-314 (1936)
   1.00000000000000
   0.00000000000000   5.23000000000000   5.23000000000000
   5.23000000000000   0.00000000000000   5.23000000000000
   5.23000000000000   5.23000000000000   0.00000000000000
    Co      K      N      O
     1      3      6     12
Direct
   0.00000000000000   0.00000000000000   0.00000000000000   Co   (4a)
   0.50000000000000   0.50000000000000   0.50000000000000   K    (4b)
   0.25000000000000   0.25000000000000   0.25000000000000   K    (8c)
   0.75000000000000   0.75000000000000   0.75000000000000   K    (8c)
  -0.19500000000000   0.19500000000000   0.19500000000000   N   (24e)
   0.19500000000000  -0.19500000000000  -0.19500000000000   N   (24e)
   0.19500000000000  -0.19500000000000   0.19500000000000   N   (24e)
  -0.19500000000000   0.19500000000000  -0.19500000000000   N   (24e)
   0.19500000000000   0.19500000000000  -0.19500000000000   N   (24e)
  -0.19500000000000  -0.19500000000000   0.19500000000000   N   (24e)
   0.33500000000000  -0.13500000000000   0.13500000000000   O   (48h)
  -0.13500000000000   0.33500000000000  -0.33500000000000   O   (48h)
   0.13500000000000  -0.33500000000000   0.33500000000000   O   (48h)
  -0.33500000000000   0.13500000000000  -0.13500000000000   O   (48h)
   0.13500000000000   0.33500000000000  -0.13500000000000   O   (48h)
  -0.33500000000000  -0.13500000000000   0.33500000000000   O   (48h)
   0.33500000000000   0.13500000000000  -0.33500000000000   O   (48h)
  -0.13500000000000  -0.33500000000000   0.13500000000000   O   (48h)
  -0.13500000000000   0.13500000000000   0.33500000000000   O   (48h)
   0.33500000000000  -0.33500000000000  -0.13500000000000   O   (48h)
  -0.33500000000000   0.33500000000000   0.13500000000000   O   (48h)
   0.13500000000000  -0.13500000000000  -0.33500000000000   O   (48h)
```

$LaFe_4P_{12}$: A4BC12_cI34_204_c_a_g - CIF

```
# CIF file
data_findsym-output
_audit_creation_method FINDSYM

_chemical_name_mineral 'Fe4LaP12'
_chemical_formula_sum 'Fe4 La P12'

loop_
_publ_author_name
 'W. Jeitschko'
```

```
 'D. Braun'
_journal_name_full_name
;
 Acta Crystallographica Section B: Structural Science
;
_journal_volume 33
_journal_year 1977
_journal_page_first 3401
_journal_page_last 3406
_publ_Section_title
;
 LaFe$_{4}$P$_{12}$ with filled CoAs$_{3}$-type structure and isotypic
    ↪ lanthanoid-transition metal polyphosphides
;

# Found in Novel Heavy Fermion Behavior in Praseodymium-based Materials:
    ↪ Experimental Study of PrOs$_{4}$Sb$_{12}$, 2007 Found in Novel
    ↪ Heavy Fermion Behavior in Praseodymium-based Materials:
    ↪ Experimental Study of PrOs$_{4}$Sb$_{12}$, {Ph. D. Thesis,
    ↪ University of Florida},

_aflow_title 'LaFe$_{4}$P$_{12}$ Structure'
_aflow_proto 'A4BC12_cI34_204_c_a_g'
_aflow_params 'a,y_{3},z_{3}'
_aflow_params_values '7.832,0.3539,0.1504'
_aflow_Strukturbericht 'None'
_aflow_Pearson 'cI34'

_symmetry_space_group_name_H-M "I 2/m -3"
_symmetry_Int_Tables_number 204

_cell_length_a      7.83200
_cell_length_b      7.83200
_cell_length_c      7.83200
_cell_angle_alpha 90.00000
_cell_angle_beta  90.00000
_cell_angle_gamma 90.00000

loop_
_space_group_symop_id
_space_group_symop_operation_xyz
1 x,y,z
2 x,-y,-z
3 -x,y,-z
4 -x,-y,z
5 y,z,x
6 y,-z,-x
7 -y,z,-x
8 -y,-z,x
9 z,x,y
10 z,-x,-y
11 -z,x,-y
12 -z,-x,y
13 -x,-y,-z
14 -x,y,z
15 x,-y,z
16 x,y,-z
17 -y,-z,-x
18 -y,z,x
19 y,-z,x
20 y,z,-x
21 -z,-x,-y
22 -z,x,y
23 z,-x,y
24 z,x,-y
25 x+1/2,y+1/2,z+1/2
26 x+1/2,-y+1/2,-z+1/2
27 -x+1/2,y+1/2,-z+1/2
28 -x+1/2,-y+1/2,z+1/2
29 y+1/2,z+1/2,x+1/2
30 y+1/2,-z+1/2,-x+1/2
31 -y+1/2,z+1/2,-x+1/2
32 -y+1/2,-z+1/2,x+1/2
33 z+1/2,x+1/2,y+1/2
34 z+1/2,-x+1/2,-y+1/2
35 -z+1/2,x+1/2,-y+1/2
36 -z+1/2,-x+1/2,y+1/2
37 -x+1/2,-y+1/2,-z+1/2
38 -x+1/2,y+1/2,z+1/2
39 x+1/2,-y+1/2,z+1/2
40 x+1/2,y+1/2,-z+1/2
41 -y+1/2,-z+1/2,-x+1/2
42 -y+1/2,z+1/2,x+1/2
43 y+1/2,-z+1/2,x+1/2
44 y+1/2,z+1/2,-x+1/2
45 -z+1/2,-x+1/2,-y+1/2
46 -z+1/2,x+1/2,y+1/2
47 z+1/2,-x+1/2,y+1/2
48 z+1/2,x+1/2,-y+1/2

loop_
_atom_site_label
_atom_site_type_symbol
_atom_site_symmetry_multiplicity
_atom_site_Wyckoff_label
_atom_site_fract_x
_atom_site_fract_y
_atom_site_fract_z
_atom_site_occupancy
La1 La 2 a 0.00000 0.00000 0.00000 1.00000
Fe1 Fe 8 c 0.25000 0.25000 0.25000 1.00000
P1 P 24 g 0.00000 0.35390 0.15040 1.00000
```

$LaFe_4P_{12}$: A4BC12_cI34_204_c_a_g - POSCAR

```
A4BC12_cI34_204_c_a_g & a,y3,z3 --params=7.832,0.3539,0.1504 & Im-3 T_{h
    ↪ }^{5} #204 (acg) & cI34 & None & Fe4LaP12 & Fe4LaP12 & W.
    ↪ Jeitschko and D. Braun, Acta Crystallogr. Sect. B Struct. Sci.
    ↪ 33, 3401-3406 (1977)
   1.00000000000000
  -3.91600000000000   3.91600000000000   3.91600000000000
   3.91600000000000  -3.91600000000000   3.91600000000000
   3.91600000000000   3.91600000000000  -3.91600000000000
    Fe     La      P
     4      1     12
Direct
   0.50000000000000   0.50000000000000   0.50000000000000   Fe   (8c)
   0.00000000000000   0.00000000000000   0.50000000000000   Fe   (8c)
   0.00000000000000   0.50000000000000   0.00000000000000   Fe   (8c)
   0.50000000000000   0.00000000000000   0.00000000000000   Fe   (8c)
   0.00000000000000   0.00000000000000   0.00000000000000   La   (2a)
   0.50430000000000   0.15040000000000   0.35390000000000   P   (24g)
  -0.20350000000000   0.15040000000000  -0.35390000000000   P   (24g)
   0.20350000000000  -0.15040000000000   0.35390000000000   P   (24g)
  -0.50430000000000  -0.15040000000000  -0.35390000000000   P   (24g)
   0.35390000000000   0.50430000000000   0.15040000000000   P   (24g)
  -0.35390000000000  -0.20350000000000   0.15040000000000   P   (24g)
   0.35390000000000   0.20350000000000  -0.15040000000000   P   (24g)
  -0.35390000000000  -0.50430000000000  -0.15040000000000   P   (24g)
   0.15040000000000   0.35390000000000   0.50430000000000   P   (24g)
   0.15040000000000  -0.35390000000000  -0.20350000000000   P   (24g)
  -0.15040000000000   0.35390000000000   0.20350000000000   P   (24g)
  -0.15040000000000  -0.35390000000000  -0.50430000000000   P   (24g)
```

$NaMn_7O_{12}$: A7BC12_cI40_204_bc_a_g - CIF

```
# CIF file
data_findsym-output
_audit_creation_method FINDSYM

_chemical_name_mineral 'Mn7NaO12'
_chemical_formula_sum 'Mn7 Na O12'

loop_
_publ_author_name
 'E. Gilioli'
 'G. Calestani'
 'F. Licci'
 'A. Gauzzi'
 'F. Bolzoni'
 'A. Prodi'
 'M. Marezio'
_journal_name_full_name
;
 Solid State Sciences
;
_journal_volume 7
_journal_year 2005
_journal_page_first 746
_journal_page_last 752
_publ_Section_title
;
 $P-T$ phase diagram and single crystal structural refinement of NaMn$_{
    ↪ 7}$O$_{12}$
;

# Found in Crystal growth and structural refinement of NaMn$_{7}$O$_{12}
    ↪ $, 2005

_aflow_title 'NaMn$_{7}$O$_{12}$ Structure'
_aflow_proto 'A7BC12_cI40_204_bc_a_g'
_aflow_params 'a,y_{4},z_{4}'
_aflow_params_values '7.312,0.3128,0.1829'
_aflow_Strukturbericht 'None'
_aflow_Pearson 'cI40'

_symmetry_space_group_name_H-M "I 2/m -3"
_symmetry_Int_Tables_number 204

_cell_length_a      7.31200
_cell_length_b      7.31200
_cell_length_c      7.31200
_cell_angle_alpha 90.00000
_cell_angle_beta  90.00000
_cell_angle_gamma 90.00000

loop_
_space_group_symop_id
_space_group_symop_operation_xyz
1 x,y,z
2 x,-y,-z
3 -x,y,-z
4 -x,-y,z
5 y,z,x
6 y,-z,-x
7 -y,z,-x
8 -y,-z,x
9 z,x,y
10 z,-x,-y
11 -z,x,-y
12 -z,-x,y
13 -x,-y,-z
14 -x,y,z
15 x,-y,z
16 x,y,-z
17 -y,-z,-x
18 -y,z,x
19 y,-z,x
20 y,z,-x
21 -z,-x,-y
```

```
22 -z,x,y
23 z,-x,y
24 z,x,-y
25 x+1/2,y+1/2,z+1/2
26 x+1/2,-y+1/2,-z+1/2
27 -x+1/2,y+1/2,-z+1/2
28 -x+1/2,-y+1/2,z+1/2
29 y+1/2,z+1/2,x+1/2
30 y+1/2,-z+1/2,-x+1/2
31 -y+1/2,z+1/2,-x+1/2
32 -y+1/2,-z+1/2,x+1/2
33 z+1/2,x+1/2,y+1/2
34 z+1/2,-x+1/2,-y+1/2
35 -z+1/2,x+1/2,-y+1/2
36 -z+1/2,-x+1/2,y+1/2
37 -x+1/2,-y+1/2,-z+1/2
38 -x+1/2,y+1/2,z+1/2
39 x+1/2,-y+1/2,z+1/2
40 x+1/2,y+1/2,-z+1/2
41 -y+1/2,-z+1/2,-x+1/2
42 -y+1/2,z+1/2,x+1/2
43 y+1/2,-z+1/2,x+1/2
44 y+1/2,z+1/2,-x+1/2
45 -z+1/2,-x+1/2,-y+1/2
46 -z+1/2,x+1/2,y+1/2
47 z+1/2,-x+1/2,y+1/2
48 z+1/2,x+1/2,-y+1/2

loop_
_atom_site_label
_atom_site_type_symbol
_atom_site_symmetry_multiplicity
_atom_site_Wyckoff_label
_atom_site_fract_x
_atom_site_fract_y
_atom_site_fract_z
_atom_site_occupancy
Na1 Na 2 a 0.00000 0.00000 0.00000 1.00000
Mn1 Mn 6 b 0.00000 0.50000 0.50000 1.00000
Mn2 Mn 8 c 0.25000 0.25000 0.25000 1.00000
O1 O 24 g 0.00000 0.31280 0.18290 1.00000
```

$NaMn_7O_{12}$: A7BC12_cI40_204_bc_a_g - POSCAR

```
A7BC12_cI40_204_bc_a_g & a,y4,z4 --params=7.312,0.3128,0.1829 & Im-3 T_{
    ↪ h}^{5} #204 (abcg) & cI40 & None & Mn7NaO12 & Mn7NaO12 & E.
    ↪ Gilioli et al., Solid State Sci. 7, 746-752 (2005)
   1.00000000000000
  -3.65600000000000    3.65600000000000    3.65600000000000
   3.65600000000000   -3.65600000000000    3.65600000000000
   3.65600000000000    3.65600000000000   -3.65600000000000
    Mn      Na       O
     7       1      12
Direct
   0.00000000000000    0.50000000000000    0.50000000000000   Mn   (6b)
   0.50000000000000    0.00000000000000    0.50000000000000   Mn   (6b)
   0.50000000000000    0.50000000000000    0.00000000000000   Mn   (6b)
   0.50000000000000    0.50000000000000    0.50000000000000   Mn   (8c)
   0.00000000000000    0.00000000000000    0.50000000000000   Mn   (8c)
   0.00000000000000    0.50000000000000    0.00000000000000   Mn   (8c)
   0.50000000000000    0.00000000000000    0.00000000000000   Mn   (8c)
   0.00000000000000    0.00000000000000    0.00000000000000   Na   (2a)
   0.49570000000000    0.18290000000000    0.31280000000000    O  (24g)
  -0.12990000000000    0.18290000000000   -0.31280000000000    O  (24g)
   0.12990000000000   -0.18290000000000    0.31280000000000    O  (24g)
  -0.49570000000000   -0.18290000000000   -0.31280000000000    O  (24g)
   0.31280000000000    0.49570000000000    0.18290000000000    O  (24g)
  -0.31280000000000   -0.12990000000000    0.18290000000000    O  (24g)
   0.31280000000000    0.12990000000000   -0.18290000000000    O  (24g)
  -0.31280000000000   -0.49570000000000   -0.18290000000000    O  (24g)
   0.18290000000000    0.31280000000000    0.49570000000000    O  (24g)
   0.18290000000000   -0.31280000000000   -0.12990000000000    O  (24g)
  -0.18290000000000    0.31280000000000    0.12990000000000    O  (24g)
  -0.18290000000000   -0.31280000000000   -0.49570000000000    O  (24g)
```

$NO_2$ (Modern, $C26$): AB2_cI36_204_d_g - CIF

```
# CIF file
data_findsym-output
_audit_creation_method FINDSYM

_chemical_name_mineral 'NO2'
_chemical_formula_sum 'N O2'

loop_
_publ_author_name
 '{\AA}. Kvick'
 'R. K. {McMullan}'
 'M. D. Newton'
_journal_name_full_name
;
 Journal of Chemical Physics
;
_journal_volume 76
_journal_year 1982
_journal_page_first 3754
_journal_page_last 3761
_publ_Section_title
;
 The structure of dinitrogen tetroxide N$_{2}$O$_{4}$: Neutron
    ↪ diffraction study at 100, 60 and 20 K and {\em ab initio}
    ↪ theoretical calculations
;

# Found in Crystal Structure Data of Inorganic Compounds, 2005 Found in
    ↪ Crystal Structure Data of Inorganic Compounds, {
    ↪ Landolt-Bornstein Volume III 43A2},

_aflow_title 'NO$_{2}$ (Modern, $C26$) Structure'
_aflow_proto 'AB2_cI36_204_d_g'
_aflow_params 'a,x_{1},y_{2},z_{2}'
_aflow_params_values '7.6937,0.38587,0.32597,0.1425'
_aflow_Strukturbericht '$C26$'
_aflow_Pearson 'cI36'

_symmetry_space_group_name_H-M "I 2/m -3"
_symmetry_Int_Tables_number 204

_cell_length_a     7.69370
_cell_length_b     7.69370
_cell_length_c     7.69370
_cell_angle_alpha 90.00000
_cell_angle_beta  90.00000
_cell_angle_gamma 90.00000

loop_
_space_group_symop_id
_space_group_symop_operation_xyz
1 x,y,z
2 x,-y,-z
3 -x,y,-z
4 -x,-y,z
5 y,z,x
6 y,-z,-x
7 -y,z,-x
8 -y,-z,x
9 z,x,y
10 z,-x,-y
11 -z,x,-y
12 -z,-x,y
13 -x,-y,-z
14 -x,y,z
15 x,-y,z
16 x,y,-z
17 -y,-z,-x
18 -y,z,x
19 y,-z,x
20 y,z,-x
21 -z,-x,-y
22 -z,x,y
23 z,-x,y
24 z,x,-y
25 x+1/2,y+1/2,z+1/2
26 x+1/2,-y+1/2,-z+1/2
27 -x+1/2,y+1/2,-z+1/2
28 -x+1/2,-y+1/2,z+1/2
29 y+1/2,z+1/2,x+1/2
30 y+1/2,-z+1/2,-x+1/2
31 -y+1/2,z+1/2,-x+1/2
32 -y+1/2,-z+1/2,x+1/2
33 z+1/2,x+1/2,y+1/2
34 z+1/2,-x+1/2,-y+1/2
35 -z+1/2,x+1/2,-y+1/2
36 -z+1/2,-x+1/2,y+1/2
37 -x+1/2,-y+1/2,-z+1/2
38 -x+1/2,y+1/2,z+1/2
39 x+1/2,-y+1/2,z+1/2
40 x+1/2,y+1/2,-z+1/2
41 -y+1/2,-z+1/2,-x+1/2
42 -y+1/2,z+1/2,x+1/2
43 y+1/2,-z+1/2,x+1/2
44 y+1/2,z+1/2,-x+1/2
45 -z+1/2,-x+1/2,-y+1/2
46 -z+1/2,x+1/2,y+1/2
47 z+1/2,-x+1/2,y+1/2
48 z+1/2,x+1/2,-y+1/2

loop_
_atom_site_label
_atom_site_type_symbol
_atom_site_symmetry_multiplicity
_atom_site_Wyckoff_label
_atom_site_fract_x
_atom_site_fract_y
_atom_site_fract_z
_atom_site_occupancy
N1 N 12 d 0.38587 0.00000 0.00000 1.00000
O1 O 24 g 0.00000 0.32597 0.14250 1.00000
```

$NO_2$ (Modern, $C26$): AB2_cI36_204_d_g - POSCAR

```
AB2_cI36_204_d_g & a,x1,y2,z2 --params=7.6937,0.38587,0.32597,0.1425 &
    ↪ Im-3 T_{h}^{5} #204 (dg) & cI36 & $C26$ & NO2 & NO2 & {\AA}.
    ↪ Kvick and R. K. {McMullan} and M. D. Newton, J. Chem. Phys. 76,
    ↪ 3754-3761 (1982)
   1.00000000000000
  -3.84685000000000    3.84685000000000    3.84685000000000
   3.84685000000000   -3.84685000000000    3.84685000000000
   3.84685000000000    3.84685000000000   -3.84685000000000
     N       O
     6      12
Direct
   0.00000000000000    0.38587000000000    0.38587000000000    N  (12d)
   0.00000000000000   -0.38587000000000   -0.38587000000000    N  (12d)
   0.38587000000000    0.00000000000000    0.38587000000000    N  (12d)
  -0.38587000000000    0.00000000000000   -0.38587000000000    N  (12d)
   0.38587000000000    0.38587000000000    0.00000000000000    N  (12d)
  -0.38587000000000   -0.38587000000000    0.00000000000000    N  (12d)
   0.46847000000000    0.14250000000000    0.32597000000000    O  (24g)
```

```
  -0.18347000000000    0.14250000000000   -0.32597000000000     O   (24g)
   0.18347000000000   -0.14250000000000    0.32597000000000     O   (24g)
  -0.46847000000000   -0.14250000000000   -0.32597000000000     O   (24g)
   0.32597000000000    0.46847000000000    0.14250000000000     O   (24g)
  -0.32597000000000   -0.18347000000000    0.14250000000000     O   (24g)
   0.32597000000000    0.18347000000000   -0.14250000000000     O   (24g)
  -0.32597000000000   -0.46847000000000   -0.14250000000000     O   (24g)
   0.14250000000000    0.32597000000000    0.46847000000000     O   (24g)
   0.14250000000000   -0.32597000000000   -0.18347000000000     O   (24g)
  -0.14250000000000    0.32597000000000    0.18347000000000     O   (24g)
  -0.14250000000000   -0.32597000000000   -0.46847000000000     O   (24g)
```

$Zn(BrO_3)_2{\cdot}6H_2O$ ($J1_{10}$): A2B6C6D_cP60_205_c_d_d_a - CIF

```
# CIF file
data_findsym-output
_audit_creation_method FINDSYM

_chemical_name_mineral 'Br2(H6O)6O6Zn'
_chemical_formula_sum 'Br2 (H2O)6 O6 Zn'

loop_
_publ_author_name
 'S. H. Y\"{u}'
 'C. A. Beevers'
_journal_name_full_name
;
 Zeitschrift f{\"u}r Kristallographie - Crystalline Materials
;
_journal_volume 95
_journal_year 1936
_journal_page_first 426
_journal_page_last 434
_publ_Section_title
;
 The Crystal Structure of Zinc Bromate Hexahydrate [Zn(BrO$_{3}$)$_{2}$
  ↪ $\cdot$ 6H$_{2}$O]
;

# Found in Strukturbericht Band IV 1936, 1938

_aflow_title 'Zn(BrO$_{3}$)$_{2}$$\cdot$6H$_{2}$O ($J1_{10}$) Structure'
_aflow_proto 'A2B6C6D_cP60_205_c_d_d_a'
_aflow_params 'a,x_{2},x_{3},y_{3},z_{3},x_{4},y_{4},z_{4}'
_aflow_params_values '10.316,0.259,0.195,0.05,0.965,0.19,0.145,0.33'
_aflow_Strukturbericht '$J1_{10}$'
_aflow_Pearson 'cP60'

_symmetry_space_group_name_H-M "P 21/a -3"
_symmetry_Int_Tables_number 205

_cell_length_a    10.31600
_cell_length_b    10.31600
_cell_length_c    10.31600
_cell_angle_alpha 90.00000
_cell_angle_beta  90.00000
_cell_angle_gamma 90.00000

loop_
_space_group_symop_id
_space_group_symop_operation_xyz
1 x,y,z
2 x+1/2,-y+1/2,-z
3 -x,y+1/2,-z+1/2
4 -x+1/2,-y,z+1/2
5 y,z,x
6 y+1/2,-z+1/2,-x
7 -y,z+1/2,-x+1/2
8 -y+1/2,-z,x+1/2
9 z,x,y
10 z+1/2,-x+1/2,-y
11 -z,x+1/2,-y+1/2
12 -z+1/2,-x,y+1/2
13 -x,-y,-z
14 -x+1/2,y+1/2,z
15 x,-y+1/2,z+1/2
16 x+1/2,y,-z+1/2
17 -y,-z,-x
18 -y+1/2,z+1/2,x
19 y,-z+1/2,x+1/2
20 y+1/2,z,-x+1/2
21 -z,-x,-y
22 -z+1/2,x+1/2,y
23 z,-x+1/2,y+1/2
24 z+1/2,x,-y+1/2

loop_
_atom_site_label
_atom_site_type_symbol
_atom_site_symmetry_multiplicity
_atom_site_Wyckoff_label
_atom_site_fract_x
_atom_site_fract_y
_atom_site_fract_z
_atom_site_occupancy
Zn1 Zn 4 a 0.00000 0.00000 0.00000 1.00000
Br1 Br 8 c 0.25900 0.25900 0.25900 1.00000
H2O1 H2O 24 d 0.19500 0.05000 0.96500 1.00000
O1 O 24 d 0.19000 0.14500 0.33000 1.00000
```

$Zn(BrO_3)_2{\cdot}6H_2O$ ($J1_{10}$): A2B6C6D_cP60_205_c_d_d_a - POSCAR

```
A2B6C6D_cP60_205_c_d_d_a & a,x2,x3,y3,z3,x4,y4,z4 --params=10.316,0.259,
  ↪ 0.195,0.05,0.965,0.19,0.145,0.33 & Pa-3 T_{h}^{6} #205 (acd^2)
  ↪ & cP60 & $J1_{10}$ & Br2(H6O)6O6Zn & Br2(H6O)6O6Zn & S. H. Y\"{
  ↪ u} and C. A. Beevers, Zeitschrift f{"u}r Kristallographie -
  ↪ Crystalline Materials 95, 426-434 (1936)
   1.00000000000000
  10.31600000000000    0.00000000000000    0.00000000000000
   0.00000000000000   10.31600000000000    0.00000000000000
   0.00000000000000    0.00000000000000   10.31600000000000
    Br    H2O     O    Zn
     8     24    24     4
Direct
   0.25900000000000    0.25900000000000    0.25900000000000    Br    (8c)
   0.24100000000000   -0.25900000000000    0.75900000000000    Br    (8c)
  -0.25900000000000    0.75900000000000    0.24100000000000    Br    (8c)
   0.75900000000000    0.24100000000000   -0.25900000000000    Br    (8c)
  -0.25900000000000   -0.25900000000000   -0.25900000000000    Br    (8c)
   0.75900000000000    0.25900000000000    0.24100000000000    Br    (8c)
   0.25900000000000    0.24100000000000    0.75900000000000    Br    (8c)
   0.24100000000000    0.75900000000000    0.25900000000000    Br    (8c)
   0.19500000000000    0.05000000000000    0.96500000000000   H2O   (24d)
   0.30500000000000   -0.05000000000000    1.46500000000000   H2O   (24d)
  -0.19500000000000    0.55000000000000   -0.46500000000000   H2O   (24d)
   0.69500000000000    0.45000000000000   -0.96500000000000   H2O   (24d)
   0.96500000000000    0.19500000000000    0.05000000000000   H2O   (24d)
   1.46500000000000    0.30500000000000   -0.05000000000000   H2O   (24d)
  -0.46500000000000   -0.19500000000000    0.55000000000000   H2O   (24d)
  -0.96500000000000    0.69500000000000    0.45000000000000   H2O   (24d)
   0.05000000000000    0.96500000000000    0.19500000000000   H2O   (24d)
  -0.05000000000000    1.46500000000000    0.30500000000000   H2O   (24d)
   0.55000000000000   -0.46500000000000   -0.19500000000000   H2O   (24d)
   0.45000000000000   -0.96500000000000    0.69500000000000   H2O   (24d)
  -0.19500000000000   -0.05000000000000   -0.96500000000000   H2O   (24d)
   0.69500000000000    0.05000000000000   -0.46500000000000   H2O   (24d)
   0.19500000000000    0.45000000000000    1.46500000000000   H2O   (24d)
   0.30500000000000    0.55000000000000    0.96500000000000   H2O   (24d)
  -0.96500000000000   -0.19500000000000   -0.05000000000000   H2O   (24d)
  -0.46500000000000    0.69500000000000    0.05000000000000   H2O   (24d)
   1.46500000000000    0.19500000000000    0.45000000000000   H2O   (24d)
   0.96500000000000    0.30500000000000    0.55000000000000   H2O   (24d)
  -0.05000000000000   -0.96500000000000   -0.19500000000000   H2O   (24d)
   0.05000000000000   -0.46500000000000    0.69500000000000   H2O   (24d)
   0.45000000000000    1.46500000000000    0.19500000000000   H2O   (24d)
   0.55000000000000    0.96500000000000    0.30500000000000   H2O   (24d)
   0.19000000000000    0.14500000000000    0.33000000000000     O   (24d)
   0.31000000000000   -0.14500000000000    0.83000000000000     O   (24d)
  -0.19000000000000    0.64500000000000    0.17000000000000     O   (24d)
   0.69000000000000    0.35500000000000   -0.33000000000000     O   (24d)
   0.33000000000000    0.19000000000000    0.14500000000000     O   (24d)
   0.83000000000000    0.31000000000000   -0.14500000000000     O   (24d)
   0.17000000000000   -0.19000000000000    0.64500000000000     O   (24d)
  -0.33000000000000    0.69000000000000    0.35500000000000     O   (24d)
   0.14500000000000    0.33000000000000    0.19000000000000     O   (24d)
  -0.14500000000000    0.83000000000000    0.31000000000000     O   (24d)
   0.64500000000000    0.17000000000000   -0.19000000000000     O   (24d)
   0.35500000000000   -0.33000000000000    0.69000000000000     O   (24d)
  -0.19000000000000   -0.14500000000000   -0.33000000000000     O   (24d)
   0.69000000000000    0.14500000000000    0.17000000000000     O   (24d)
   0.19000000000000    0.35500000000000    0.83000000000000     O   (24d)
   0.31000000000000    0.64500000000000    0.33000000000000     O   (24d)
  -0.33000000000000   -0.19000000000000   -0.14500000000000     O   (24d)
   0.17000000000000    0.69000000000000    0.14500000000000     O   (24d)
   0.83000000000000    0.19000000000000    0.35500000000000     O   (24d)
   0.33000000000000    0.31000000000000    0.64500000000000     O   (24d)
  -0.14500000000000   -0.33000000000000   -0.19000000000000     O   (24d)
   0.14500000000000    0.17000000000000    0.69000000000000     O   (24d)
   0.35500000000000    0.83000000000000    0.19000000000000     O   (24d)
   0.64500000000000    0.33000000000000    0.31000000000000     O   (24d)
   0.00000000000000    0.00000000000000    0.00000000000000    Zn    (4a)
   0.50000000000000    0.00000000000000    0.50000000000000    Zn    (4a)
   0.00000000000000    0.50000000000000    0.50000000000000    Zn    (4a)
   0.50000000000000    0.50000000000000    0.00000000000000    Zn    (4a)
```

$H6_4$ $[Ni(NO_3)_2(NH_3)_6]$ (*obsolete*): A2B6CD6_cP60_205_c_d_a_d - CIF

```
# CIF file
data_findsym-output
_audit_creation_method FINDSYM

_chemical_name_mineral 'N2(NH3)6NiO6'
_chemical_formula_sum 'N2 (NH3)6 Ni O6'

loop_
_publ_author_name
 'R. W. G. Wyckoff'
_journal_name_full_name
;
 Journal of the American Chemical Society
;
_journal_volume 44
_journal_year 1922
_journal_page_first 1260
_journal_page_last 1266
_publ_Section_title
;
 The Composition and Crystal Structure of Nickel Nitrate Hexammoniate
;

# Found in Strukturbericht 1913-1928, 1931

_aflow_title '$H6_{4}$ [Ni(NO$_{3}$)$_{2}$(NH$_{3}$)$_{6}$] ({\em{
  ↪ obsolete}}) Structure'
_aflow_proto 'A2B6CD6_cP60_205_c_d_a_d'
_aflow_params 'a,x_{2},x_{3},y_{3},z_{3},x_{4},y_{4},z_{4}'
_aflow_params_values '10.96,0.25,0.24,0.0,0.0,0.25,0.25,0.01'
_aflow_Strukturbericht 'None'
_aflow_Pearson 'cP60'
```

```
_symmetry_space_group_name_H-M "P 21/a -3"
_symmetry_Int_Tables_number 205

_cell_length_a    10.96000
_cell_length_b    10.96000
_cell_length_c    10.96000
_cell_angle_alpha 90.00000
_cell_angle_beta  90.00000
_cell_angle_gamma 90.00000

loop_
_space_group_symop_id
_space_group_symop_operation_xyz
1 x,y,z
2 x+1/2,-y+1/2,-z
3 -x,y+1/2,-z+1/2
4 -x+1/2,-y,z+1/2
5 y,z,x
6 y+1/2,-z+1/2,-x
7 -y,z+1/2,-x+1/2
8 -y+1/2,-z,x+1/2
9 z,x,y
10 z+1/2,-x+1/2,-y
11 -z,x+1/2,-y+1/2
12 -z+1/2,-x,y+1/2
13 -x,-y,-z
14 -x+1/2,y+1/2,z
15 x,-y+1/2,z+1/2
16 x+1/2,y,-z+1/2
17 -y,-z,-x
18 -y+1/2,z+1/2,x
19 y,-z+1/2,x+1/2
20 y+1/2,z,-x+1/2
21 -z,-x,-y
22 -z+1/2,x+1/2,y
23 z,-x+1/2,y+1/2
24 z+1/2,x,-y+1/2

loop_
_atom_site_label
_atom_site_type_symbol
_atom_site_symmetry_multiplicity
_atom_site_Wyckoff_label
_atom_site_fract_x
_atom_site_fract_y
_atom_site_fract_z
_atom_site_occupancy
Ni1 Ni 4 a 0.00000 0.00000 0.00000 1.00000
N1 N 8 c 0.25000 0.25000 0.25000 1.00000
NH31 NH3 24 d 0.24000 0.00000 0.00000 1.00000
O1 O 24 d 0.25000 0.25000 0.01000 1.00000
```

*H*6₄ [Ni(NO$_3$)$_2$(NH$_3$)$_6$] (*obsolete*): A2B6CD6_cP60_205_c_d_a_d - POSCAR

```
A2B6CD6_cP60_205_c_d_a_d & a,x2,x3,y3,z3,x4,y4,z4 --params=10.96,0.25,
    ↪ 0.24,0.0,0.0,0.25,0.25,0.01 & Pa-3 T_{h}^{6} #205 (acd^2) &
    ↪ cP60 & None & N2(NH3)6NiO6 & N2(NH3)6NiO6 & R. W. G. Wyckoff,
    ↪ J. Am. Chem. Soc. 44, 1260-1266 (1922)
   1.00000000000000
  10.96000000000000   0.00000000000000   0.00000000000000
   0.00000000000000  10.96000000000000   0.00000000000000
   0.00000000000000   0.00000000000000  10.96000000000000
     N    NH3     Ni      O
     8     24      4     24
Direct
   0.25000000000000   0.25000000000000   0.25000000000000    N    (8c)
   0.25000000000000  -0.25000000000000   0.75000000000000    N    (8c)
  -0.25000000000000   0.75000000000000   0.25000000000000    N    (8c)
   0.75000000000000   0.25000000000000  -0.25000000000000    N    (8c)
  -0.25000000000000  -0.25000000000000  -0.25000000000000    N    (8c)
   0.75000000000000   0.25000000000000   0.25000000000000    N    (8c)
   0.25000000000000   0.25000000000000   0.75000000000000    N    (8c)
   0.25000000000000   0.75000000000000   0.25000000000000    N    (8c)
   0.24000000000000   0.00000000000000   0.00000000000000  NH3   (24d)
   0.26000000000000   0.00000000000000   0.50000000000000  NH3   (24d)
  -0.24000000000000   0.50000000000000   0.50000000000000  NH3   (24d)
   0.74000000000000   0.50000000000000   0.00000000000000  NH3   (24d)
   0.00000000000000   0.24000000000000   0.00000000000000  NH3   (24d)
   0.50000000000000   0.26000000000000   0.00000000000000  NH3   (24d)
   0.50000000000000  -0.24000000000000   0.50000000000000  NH3   (24d)
   0.00000000000000   0.74000000000000   0.50000000000000  NH3   (24d)
   0.00000000000000   0.00000000000000   0.24000000000000  NH3   (24d)
   0.00000000000000   0.50000000000000   0.26000000000000  NH3   (24d)
   0.50000000000000   0.50000000000000  -0.24000000000000  NH3   (24d)
   0.50000000000000   0.00000000000000   0.74000000000000  NH3   (24d)
  -0.24000000000000   0.00000000000000   0.00000000000000  NH3   (24d)
   0.74000000000000   0.00000000000000   0.50000000000000  NH3   (24d)
   0.24000000000000   0.50000000000000   0.50000000000000  NH3   (24d)
   0.26000000000000   0.50000000000000   0.00000000000000  NH3   (24d)
   0.00000000000000  -0.24000000000000   0.00000000000000  NH3   (24d)
   0.50000000000000   0.74000000000000   0.00000000000000  NH3   (24d)
   0.50000000000000   0.24000000000000   0.50000000000000  NH3   (24d)
   0.00000000000000   0.26000000000000   0.50000000000000  NH3   (24d)
   0.00000000000000   0.00000000000000  -0.24000000000000  NH3   (24d)
   0.00000000000000   0.50000000000000   0.74000000000000  NH3   (24d)
   0.50000000000000   0.50000000000000   0.24000000000000  NH3   (24d)
   0.50000000000000   0.00000000000000   0.26000000000000  NH3   (24d)
   0.00000000000000   0.00000000000000   0.00000000000000   Ni    (4a)
   0.50000000000000   0.00000000000000   0.50000000000000   Ni    (4a)
   0.00000000000000   0.50000000000000   0.50000000000000   Ni    (4a)
   0.50000000000000   0.50000000000000   0.00000000000000   Ni    (4a)
   0.25000000000000   0.25000000000000   0.01000000000000    O   (24d)
   0.25000000000000  -0.25000000000000   0.51000000000000    O   (24d)
  -0.25000000000000   0.75000000000000   0.49000000000000    O   (24d)
   0.75000000000000   0.25000000000000  -0.01000000000000    O   (24d)
   0.01000000000000   0.25000000000000   0.25000000000000    O   (24d)
   0.51000000000000   0.25000000000000  -0.25000000000000    O   (24d)
   0.49000000000000  -0.25000000000000   0.75000000000000    O   (24d)
  -0.01000000000000   0.75000000000000   0.25000000000000    O   (24d)
   0.25000000000000   0.01000000000000   0.25000000000000    O   (24d)
  -0.25000000000000   0.51000000000000   0.25000000000000    O   (24d)
   0.75000000000000   0.49000000000000  -0.25000000000000    O   (24d)
   0.25000000000000  -0.01000000000000   0.75000000000000    O   (24d)
  -0.25000000000000  -0.25000000000000  -0.01000000000000    O   (24d)
   0.75000000000000   0.25000000000000   0.49000000000000    O   (24d)
   0.25000000000000   0.25000000000000   0.51000000000000    O   (24d)
   0.25000000000000   0.75000000000000   0.01000000000000    O   (24d)
  -0.01000000000000  -0.25000000000000  -0.25000000000000    O   (24d)
   0.49000000000000   0.75000000000000   0.25000000000000    O   (24d)
   0.51000000000000   0.25000000000000   0.25000000000000    O   (24d)
   0.01000000000000   0.25000000000000   0.75000000000000    O   (24d)
  -0.25000000000000  -0.01000000000000  -0.25000000000000    O   (24d)
   0.25000000000000   0.49000000000000   0.75000000000000    O   (24d)
   0.25000000000000   0.51000000000000   0.25000000000000    O   (24d)
   0.75000000000000   0.01000000000000   0.25000000000000    O   (24d)
```

Pb(NO$_3$)$_2$ ($G2_1$): A2B6C_cP36_205_c_d_a - CIF

```
# CIF file
data_findsym-output
_audit_creation_method FINDSYM

_chemical_name_mineral 'N2O6Pb'
_chemical_formula_sum 'N2 O6 Pb'

loop_
_publ_author_name
 'H. Nowotny'
 'G. Heger'
_journal_name_full_name
;
 Acta Crystallographica Section C: Structural Chemistry
;
_journal_volume 42
_journal_year 1986
_journal_page_first 133
_journal_page_last 135
_publ_Section_title
;
 Structure Refinement of Lead Nitrate
;

_aflow_title 'Pb(NO$_{3}$)$_{2}$ ($G2_{1}$) Structure'
_aflow_proto 'A2B6C_cP36_205_c_d_a'
_aflow_params 'a,x_{2},x_{3},y_{3},z_{3}'
_aflow_params_values '7.8586,0.34865,0.27772,0.28869,0.4773'
_aflow_Strukturbericht '$G2_{1}$'
_aflow_Pearson 'cP36'

_symmetry_space_group_name_H-M "P 21/a -3"
_symmetry_Int_Tables_number 205

_cell_length_a    7.85860
_cell_length_b    7.85860
_cell_length_c    7.85860
_cell_angle_alpha 90.00000
_cell_angle_beta  90.00000
_cell_angle_gamma 90.00000

loop_
_space_group_symop_id
_space_group_symop_operation_xyz
1 x,y,z
2 x+1/2,-y+1/2,-z
3 -x,y+1/2,-z+1/2
4 -x+1/2,-y,z+1/2
5 y,z,x
6 y+1/2,-z+1/2,-x
7 -y,z+1/2,-x+1/2
8 -y+1/2,-z,x+1/2
9 z,x,y
10 z+1/2,-x+1/2,-y
11 -z,x+1/2,-y+1/2
12 -z+1/2,-x,y+1/2
13 -x,-y,-z
14 -x+1/2,y+1/2,z
15 x,-y+1/2,z+1/2
16 x+1/2,y,-z+1/2
17 -y,-z,-x
18 -y+1/2,z+1/2,x
19 y,-z+1/2,x+1/2
20 y+1/2,z,-x+1/2
21 -z,-x,-y
22 -z+1/2,x+1/2,y
23 z,-x+1/2,y+1/2
24 z+1/2,x,-y+1/2

loop_
_atom_site_label
_atom_site_type_symbol
_atom_site_symmetry_multiplicity
_atom_site_Wyckoff_label
_atom_site_fract_x
_atom_site_fract_y
_atom_site_fract_z
_atom_site_occupancy
Pb1 Pb 4 a 0.00000 0.00000 0.00000 1.00000
N1 N 8 c 0.34865 0.34865 0.34865 1.00000
O1 O 24 d 0.27772 0.28869 0.47730 1.00000
```

Pb(NO$_3$)$_2$ ($G2_1$): A2B6C_cP36_205_c_d_a - POSCAR

```
A2B6C_cP36_205_c_d_a & a,x2,x3,y3,z3 --params=7.8586,0.34865,0.27772,
    ↪ 0.28869,0.4773 & Pa-3 T_{h}^{6} #205 (acd) & cP36 & $G2_{1}$ &
    ↪ N2O6Pb & N2O6Pb & H. Nowotny and G. Heger, Acta Crystallogr. C
    ↪ 42, 133-135 (1986)
   1.00000000000000
   7.85860000000000   0.00000000000000   0.00000000000000
   0.00000000000000   7.85860000000000   0.00000000000000
   0.00000000000000   0.00000000000000   7.85860000000000
     N      O     Pb
     8     24      4
Direct
   0.34865000000000   0.34865000000000   0.34865000000000   N   (8c)
   0.15135000000000  -0.34865000000000   0.84865000000000   N   (8c)
  -0.34865000000000   0.84865000000000   0.15135000000000   N   (8c)
   0.84865000000000   0.15135000000000  -0.34865000000000   N   (8c)
  -0.34865000000000  -0.34865000000000  -0.34865000000000   N   (8c)
   0.84865000000000   0.34865000000000   0.15135000000000   N   (8c)
   0.34865000000000   0.15135000000000   0.84865000000000   N   (8c)
   0.15135000000000   0.84865000000000   0.34865000000000   N   (8c)
   0.27772000000000   0.28869000000000   0.47730000000000   O   (24d)
   0.22228000000000  -0.28869000000000   0.97730000000000   O   (24d)
  -0.27772000000000   0.78869000000000   0.02270000000000   O   (24d)
   0.77772000000000   0.21131000000000  -0.47730000000000   O   (24d)
   0.47730000000000   0.27772000000000   0.28869000000000   O   (24d)
   0.97730000000000   0.22228000000000  -0.28869000000000   O   (24d)
   0.02270000000000  -0.27772000000000   0.78869000000000   O   (24d)
  -0.47730000000000   0.77772000000000   0.21131000000000   O   (24d)
   0.28869000000000   0.47730000000000   0.27772000000000   O   (24d)
  -0.28869000000000   0.97730000000000   0.22228000000000   O   (24d)
   0.78869000000000   0.02270000000000  -0.27772000000000   O   (24d)
   0.21131000000000  -0.47730000000000   0.77772000000000   O   (24d)
  -0.27772000000000  -0.28869000000000  -0.47730000000000   O   (24d)
   0.77772000000000   0.28869000000000   0.02270000000000   O   (24d)
   0.27772000000000   0.21131000000000   0.97730000000000   O   (24d)
   0.22228000000000   0.78869000000000   0.47730000000000   O   (24d)
  -0.47730000000000  -0.27772000000000  -0.28869000000000   O   (24d)
   0.02270000000000   0.77772000000000   0.28869000000000   O   (24d)
   0.97730000000000   0.27772000000000   0.21131000000000   O   (24d)
   0.47730000000000   0.22228000000000   0.78869000000000   O   (24d)
  -0.28869000000000  -0.47730000000000  -0.27772000000000   O   (24d)
   0.28869000000000   0.02270000000000   0.77772000000000   O   (24d)
   0.21131000000000   0.97730000000000   0.27772000000000   O   (24d)
   0.78869000000000   0.47730000000000   0.22228000000000   O   (24d)
   0.00000000000000   0.00000000000000   0.00000000000000   Pb   (4a)
   0.50000000000000   0.00000000000000   0.50000000000000   Pb   (4a)
   0.00000000000000   0.50000000000000   0.50000000000000   Pb   (4a)
   0.50000000000000   0.50000000000000   0.00000000000000   Pb   (4a)
```

CaB$_2$O$_4$ (IV): A2BC4_cP84_205_d_ac_2d - CIF

```
# CIF file
data_findsym-output
_audit_creation_method FINDSYM

_chemical_name_mineral 'B2CaO4'
_chemical_formula_sum 'B2 Ca O4'

loop_
_publ_author_name
 'M. Marezio'
 'J. P. Remeika'
 'P. D. Dernier'
_journal_name_full_name
;
 Acta Crystallographica Section B: Structural Science
;
_journal_volume 25
_journal_year 1969
_journal_page_first 965
_journal_page_last 970
_publ_Section_title
;
 The crystal structure of the high-pressure phase CaB$_{2}$O$_{4}$(IV),
    ↪ and polymorphism in CaB$_{2}$O$_{4}$
;

_aflow_title 'CaB$_{2}$O$_{4}$ (IV) Structure'
_aflow_proto 'A2BC4_cP84_205_d_ac_2d'
_aflow_params 'a,x_{2},x_{3},y_{3},z_{3},x_{4},y_{4},z_{4},x_{5},y_{5},
    ↪ z_{5}'
_aflow_params_values '9.008,0.37305,0.1189,0.1901,0.3457,0.3336,0.2692,
    ↪ 0.1208,0.0906,0.2823,0.0064'
_aflow_Strukturbericht 'None'
_aflow_Pearson 'cP84'

_symmetry_space_group_name_H-M "P 21/a -3"
_symmetry_Int_Tables_number 205

_cell_length_a    9.00800
_cell_length_b    9.00800
_cell_length_c    9.00800
_cell_angle_alpha 90.00000
_cell_angle_beta  90.00000
_cell_angle_gamma 90.00000

loop_
_space_group_symop_id
_space_group_symop_operation_xyz
1 x,y,z
2 x+1/2,-y+1/2,-z
3 -x,y+1/2,-z+1/2
4 -x+1/2,-y,z+1/2
5 y,z,x
6 y+1/2,-z+1/2,-x
7 -y,z+1/2,-x+1/2
8 -y+1/2,-z,x+1/2
9 z,x,y
10 z+1/2,-x+1/2,-y
11 -z,x+1/2,-y+1/2
12 -z+1/2,-x,y+1/2
13 -x,-y,-z
14 -x+1/2,y+1/2,z
15 x,-y+1/2,z+1/2
16 x+1/2,y,-z+1/2
17 -y,-z,-x
18 -y+1/2,z+1/2,x
19 y,-z+1/2,x+1/2
20 y+1/2,z,-x+1/2
21 -z,-x,-y
22 -z+1/2,x+1/2,y
23 z,-x+1/2,y+1/2
24 z+1/2,x,-y+1/2

loop_
_atom_site_label
_atom_site_type_symbol
_atom_site_symmetry_multiplicity
_atom_site_Wyckoff_label
_atom_site_fract_x
_atom_site_fract_y
_atom_site_fract_z
_atom_site_occupancy
Ca1 Ca 4 a 0.00000 0.00000 0.00000 1.00000
Ca2 Ca 8 c 0.37305 0.37305 0.37305 1.00000
B1 B 24 d 0.11890 0.19010 0.34570 1.00000
O1 O 24 d 0.33360 0.26920 0.12080 1.00000
O2 O 24 d 0.09060 0.28230 0.00640 1.00000
```

CaB$_2$O$_4$ (IV): A2BC4_cP84_205_d_ac_2d - POSCAR

```
A2BC4_cP84_205_d_ac_2d & a,x2,x3,y3,z3,x4,y4,z4,x5,y5,z5 --params=9.008,
    ↪ 0.37305,0.1189,0.1901,0.3457,0.3336,0.2692,0.1208,0.0906,0.2823
    ↪ ,0.0064 & Pa-3 T_{h}^{6} #205 (acd^3) & cP84 & None & B2CaO4 &
    ↪ B2CaO4 & M. Marezio and J. P. Remeika and P. D. Dernier, Acta
    ↪ Crystallogr. Sect. B Struct. Sci. 25, 965-970 (1969)
   1.00000000000000
   9.00800000000000   0.00000000000000   0.00000000000000
   0.00000000000000   9.00800000000000   0.00000000000000
   0.00000000000000   0.00000000000000   9.00800000000000
     B     Ca      O
    24     12     48
Direct
   0.11890000000000   0.19010000000000   0.34570000000000   B   (24d)
   0.38110000000000  -0.19010000000000   0.84570000000000   B   (24d)
  -0.11890000000000   0.69010000000000   0.15430000000000   B   (24d)
   0.61890000000000   0.30990000000000  -0.34570000000000   B   (24d)
   0.34570000000000   0.11890000000000   0.19010000000000   B   (24d)
   0.84570000000000   0.38110000000000  -0.19010000000000   B   (24d)
   0.15430000000000  -0.11890000000000   0.69010000000000   B   (24d)
  -0.34570000000000   0.61890000000000   0.30990000000000   B   (24d)
   0.19010000000000   0.34570000000000   0.11890000000000   B   (24d)
  -0.19010000000000   0.84570000000000   0.38110000000000   B   (24d)
   0.69010000000000   0.15430000000000  -0.11890000000000   B   (24d)
   0.30990000000000  -0.34570000000000   0.61890000000000   B   (24d)
  -0.11890000000000  -0.19010000000000  -0.34570000000000   B   (24d)
   0.61890000000000   0.19010000000000   0.15430000000000   B   (24d)
   0.11890000000000   0.30990000000000   0.84570000000000   B   (24d)
   0.38110000000000   0.69010000000000   0.34570000000000   B   (24d)
  -0.34570000000000  -0.11890000000000  -0.19010000000000   B   (24d)
   0.15430000000000   0.61890000000000   0.19010000000000   B   (24d)
   0.84570000000000   0.11890000000000   0.30990000000000   B   (24d)
   0.34570000000000   0.38110000000000   0.69010000000000   B   (24d)
  -0.19010000000000  -0.34570000000000  -0.11890000000000   B   (24d)
   0.19010000000000   0.15430000000000   0.61890000000000   B   (24d)
   0.30990000000000   0.84570000000000   0.11890000000000   B   (24d)
   0.69010000000000   0.34570000000000   0.38110000000000   B   (24d)
   0.00000000000000   0.00000000000000   0.00000000000000   Ca   (4a)
   0.50000000000000   0.00000000000000   0.50000000000000   Ca   (4a)
   0.00000000000000   0.50000000000000   0.50000000000000   Ca   (4a)
   0.50000000000000   0.50000000000000   0.00000000000000   Ca   (4a)
   0.37305000000000   0.37305000000000   0.37305000000000   Ca   (8c)
   0.12695000000000  -0.37305000000000   0.87305000000000   Ca   (8c)
  -0.37305000000000   0.87305000000000   0.12695000000000   Ca   (8c)
   0.87305000000000   0.12695000000000  -0.37305000000000   Ca   (8c)
  -0.37305000000000  -0.37305000000000  -0.37305000000000   Ca   (8c)
   0.87305000000000   0.37305000000000   0.12695000000000   Ca   (8c)
   0.37305000000000   0.12695000000000   0.87305000000000   Ca   (8c)
   0.12695000000000   0.87305000000000   0.37305000000000   Ca   (8c)
   0.33360000000000   0.26920000000000   0.12080000000000   O   (24d)
   0.16640000000000  -0.26920000000000   0.62080000000000   O   (24d)
  -0.33360000000000   0.76920000000000   0.37920000000000   O   (24d)
   0.83360000000000   0.23080000000000  -0.12080000000000   O   (24d)
   0.12080000000000   0.33360000000000   0.26920000000000   O   (24d)
   0.62080000000000   0.16640000000000  -0.26920000000000   O   (24d)
   0.37920000000000  -0.33360000000000   0.76920000000000   O   (24d)
  -0.12080000000000   0.83360000000000   0.23080000000000   O   (24d)
   0.26920000000000   0.12080000000000   0.33360000000000   O   (24d)
  -0.26920000000000   0.62080000000000   0.16640000000000   O   (24d)
   0.76920000000000   0.37920000000000  -0.33360000000000   O   (24d)
   0.23080000000000  -0.12080000000000   0.83360000000000   O   (24d)
  -0.33360000000000  -0.26920000000000  -0.12080000000000   O   (24d)
   0.83360000000000   0.26920000000000   0.37920000000000   O   (24d)
   0.33360000000000   0.23080000000000   0.62080000000000   O   (24d)
   0.16640000000000   0.76920000000000   0.12080000000000   O   (24d)
  -0.12080000000000  -0.33360000000000  -0.26920000000000   O   (24d)
   0.37920000000000   0.83360000000000   0.26920000000000   O   (24d)
   0.62080000000000   0.33360000000000   0.23080000000000   O   (24d)
   0.12080000000000   0.16640000000000   0.76920000000000   O   (24d)
```

```
  -0.26920000000000  -0.12080000000000  -0.33360000000000    O   (24d)
   0.26920000000000   0.37920000000000   0.83360000000000    O   (24d)
   0.23080000000000   0.62080000000000   0.33360000000000    O   (24d)
   0.76920000000000   0.12080000000000   0.16640000000000    O   (24d)
   0.09060000000000   0.28230000000000   0.00640000000000    O   (24d)
   0.40940000000000  -0.28230000000000   0.50640000000000    O   (24d)
  -0.09060000000000   0.78230000000000   0.49360000000000    O   (24d)
   0.59060000000000   0.21770000000000  -0.00640000000000    O   (24d)
   0.00640000000000   0.09060000000000   0.28230000000000    O   (24d)
   0.50640000000000   0.40940000000000  -0.28230000000000    O   (24d)
   0.49360000000000  -0.09060000000000   0.78230000000000    O   (24d)
  -0.00640000000000   0.59060000000000   0.21770000000000    O   (24d)
   0.28230000000000   0.00640000000000   0.09060000000000    O   (24d)
  -0.28230000000000   0.50640000000000   0.40940000000000    O   (24d)
   0.78230000000000   0.49360000000000  -0.09060000000000    O   (24d)
   0.21770000000000  -0.00640000000000   0.59060000000000    O   (24d)
  -0.09060000000000  -0.28230000000000  -0.00640000000000    O   (24d)
   0.59060000000000   0.28230000000000   0.49360000000000    O   (24d)
   0.09060000000000   0.21770000000000   0.50640000000000    O   (24d)
   0.40940000000000   0.78230000000000   0.00640000000000    O   (24d)
  -0.00640000000000  -0.09060000000000  -0.28230000000000    O   (24d)
   0.49360000000000   0.59060000000000   0.28230000000000    O   (24d)
   0.50640000000000   0.09060000000000   0.21770000000000    O   (24d)
   0.00640000000000   0.40940000000000   0.78230000000000    O   (24d)
  -0.28230000000000  -0.00640000000000  -0.09060000000000    O   (24d)
   0.28230000000000   0.49360000000000   0.59060000000000    O   (24d)
   0.21770000000000   0.50640000000000   0.09060000000000    O   (24d)
   0.78230000000000   0.00640000000000   0.40940000000000    O   (24d)
```

$SnI_4$ ($D1_1$): A4B_cP40_205_cd_c - CIF

```
# CIF file
data_findsym-output
_audit_creation_method FINDSYM

_chemical_name_mineral 'I4Sn'
_chemical_formula_sum 'I4 Sn'

loop_
_publ_author_name
 'F. Meller'
 'I. Fankuchen'
_journal_name_full_name
;
 Acta Cristallographica
;
_journal_volume 8
_journal_year 1955
_journal_page_first 343
_journal_page_last 344
_publ_Section_title
;
 The crystal structure of tin tetraiodide
;

# Found in The pressure-induced metallic amorphous state of SnI$_{4}$.
    ↪ I. A novel crystal-to-amorphous transition studied by X-ray
    ↪ scattering, 1985

_aflow_title 'SnI$_{4}$ (D1$_{1}$) Structure'
_aflow_proto 'A4B_cP40_205_cd_c'
_aflow_params 'a,x_{1},x_{2},x_{3},y_{3},z_{3}'
_aflow_params_values '12.26,0.252,0.125,-0.002,-0.002,0.252'
_aflow_Strukturbericht '$D1_{1}$'
_aflow_Pearson 'cP40'

_symmetry_space_group_name_H-M "P 21/a -3"
_symmetry_Int_Tables_number 205

_cell_length_a    12.26000
_cell_length_b    12.26000
_cell_length_c    12.26000
_cell_angle_alpha 90.00000
_cell_angle_beta  90.00000
_cell_angle_gamma 90.00000

loop_
_space_group_symop_id
_space_group_symop_operation_xyz
1 x,y,z
2 x+1/2,-y+1/2,-z
3 -x,y+1/2,-z+1/2
4 -x+1/2,-y,z+1/2
5 y,z,x
6 y+1/2,-z+1/2,-x
7 -y,z+1/2,-x+1/2
8 -y+1/2,-z,x+1/2
9 z,x,y
10 z+1/2,-x+1/2,-y
11 -z,x+1/2,-y+1/2
12 -z+1/2,-x,y+1/2
13 -x,-y,-z
14 -x+1/2,y+1/2,z
15 x,-y+1/2,z+1/2
16 x+1/2,y,-z+1/2
17 -y,-z,-x
18 -y+1/2,z+1/2,x
19 y,-z+1/2,x+1/2
20 y+1/2,z,-x+1/2
21 -z,-x,-y
22 -z+1/2,x+1/2,y
23 z,-x+1/2,y+1/2
24 z+1/2,x,-y+1/2

loop_
_atom_site_label
_atom_site_type_symbol
_atom_site_symmetry_multiplicity
_atom_site_Wyckoff_label
_atom_site_fract_x
_atom_site_fract_y
_atom_site_fract_z
_atom_site_occupancy
I1 I 8 c 0.25200 0.25200 0.25200 1.00000
Sn1 Sn 8 c 0.12500 0.12500 0.12500 1.00000
I2 I 24 d -0.00200 -0.00200 0.25200 1.00000
```

$SnI_4$ ($D1_1$): A4B_cP40_205_cd_c - POSCAR

```
A4B_cP40_205_cd_c & a,x1,x2,x3,y3,z3 --params=12.26,0.252,0.125,-0.002,-
    ↪ 0.002,0.252 & Pa-3 T_{h}^{6} #205 (c^2d) & cP40 & $D1_{1}$ &
    ↪ I4Sn & I4Sn & F. Meller and I. Fankuchen, Acta Cryst. 8,
    ↪ 343-344 (1955)
   1.00000000000000
  12.26000000000000   0.00000000000000   0.00000000000000
   0.00000000000000  12.26000000000000   0.00000000000000
   0.00000000000000   0.00000000000000  12.26000000000000
     I     Sn
    32      8
Direct
   0.25200000000000   0.25200000000000   0.25200000000000    I    (8c)
   0.24800000000000  -0.25200000000000   0.75200000000000    I    (8c)
  -0.25200000000000   0.75200000000000   0.24800000000000    I    (8c)
   0.75200000000000   0.24800000000000  -0.25200000000000    I    (8c)
  -0.25200000000000  -0.25200000000000  -0.25200000000000    I    (8c)
   0.75200000000000   0.25200000000000   0.24800000000000    I    (8c)
   0.25200000000000   0.24800000000000   0.75200000000000    I    (8c)
   0.24800000000000   0.75200000000000   0.25200000000000    I    (8c)
  -0.00200000000000  -0.00200000000000   0.25200000000000    I   (24d)
   0.50200000000000   0.00200000000000   0.75200000000000    I   (24d)
   0.00200000000000   0.49800000000000   0.24800000000000    I   (24d)
   0.49800000000000   0.50200000000000  -0.25200000000000    I   (24d)
   0.25200000000000  -0.00200000000000  -0.00200000000000    I   (24d)
   0.75200000000000   0.50200000000000   0.00200000000000    I   (24d)
   0.24800000000000   0.00200000000000   0.49800000000000    I   (24d)
  -0.25200000000000   0.49800000000000   0.50200000000000    I   (24d)
  -0.00200000000000   0.25200000000000  -0.00200000000000    I   (24d)
   0.00200000000000   0.75200000000000   0.50200000000000    I   (24d)
   0.49800000000000   0.24800000000000   0.00200000000000    I   (24d)
   0.50200000000000  -0.25200000000000   0.49800000000000    I   (24d)
   0.00200000000000   0.00200000000000  -0.25200000000000    I   (24d)
   0.49800000000000  -0.00200000000000   0.24800000000000    I   (24d)
  -0.00200000000000   0.50200000000000   0.75200000000000    I   (24d)
   0.50200000000000   0.49800000000000   0.25200000000000    I   (24d)
  -0.25200000000000   0.00200000000000   0.00200000000000    I   (24d)
   0.24800000000000   0.49800000000000  -0.00200000000000    I   (24d)
   0.75200000000000  -0.00200000000000   0.50200000000000    I   (24d)
   0.25200000000000   0.50200000000000   0.49800000000000    I   (24d)
   0.00200000000000  -0.25200000000000   0.00200000000000    I   (24d)
  -0.00200000000000   0.24800000000000   0.49800000000000    I   (24d)
   0.50200000000000   0.75200000000000  -0.00200000000000    I   (24d)
   0.49800000000000   0.25200000000000   0.50200000000000    I   (24d)
   0.12500000000000   0.12500000000000   0.12500000000000   Sn    (8c)
   0.37500000000000  -0.12500000000000   0.62500000000000   Sn    (8c)
  -0.12500000000000   0.62500000000000   0.37500000000000   Sn    (8c)
   0.62500000000000   0.37500000000000  -0.12500000000000   Sn    (8c)
  -0.12500000000000  -0.12500000000000  -0.12500000000000   Sn    (8c)
   0.62500000000000   0.12500000000000   0.37500000000000   Sn    (8c)
   0.12500000000000   0.37500000000000   0.62500000000000   Sn    (8c)
   0.37500000000000   0.62500000000000   0.12500000000000   Sn    (8c)
```

$NaSbF_6$: A6BC_cP32_205_d_b_a - CIF

```
# CIF file
data_findsym-output
_audit_creation_method FINDSYM

_chemical_name_mineral 'F6NaSb'
_chemical_formula_sum 'F6 Na Sb'

loop_
_publ_author_name
 'N. Schrewelius'
_journal_name_full_name
;
 Zeitschrift fur Anorganische und Allgemeine Chemie
;
_journal_volume 238
_journal_year 1938
_journal_page_first 241
_journal_page_last 254
_publ_Section_title
;
 R\"{o}ntgenuntersuchung der Verbindungen NaSb(OH)$_{6}$, NaSbF$_{6}$,
    ↪ NaSbO$_{3}$ und gleichartiger Stoffe
;

# Found in The American Mineralogist Crystal Structure Database, 2003

_aflow_title 'NaSbF$_{6}$ Structure'
_aflow_proto 'A6BC_cP32_205_d_b_a'
_aflow_params 'a,x_{3},y_{3},z_{3}'
_aflow_params_values '8.18,-0.05,0.05,0.225'
_aflow_Strukturbericht 'None'
_aflow_Pearson 'cP32'

_symmetry_space_group_name_H-M "P 21/a -3"
_symmetry_Int_Tables_number 205

_cell_length_a    8.18000
```

```
_cell_length_b      8.18000
_cell_length_c      8.18000
_cell_angle_alpha  90.00000
_cell_angle_beta   90.00000
_cell_angle_gamma  90.00000

loop_
_space_group_symop_id
_space_group_symop_operation_xyz
1 x,y,z
2 x+1/2,-y+1/2,-z
3 -x,y+1/2,-z+1/2
4 -x+1/2,-y,z+1/2
5 y,z,x
6 y+1/2,-z+1/2,-x
7 -y,z+1/2,-x+1/2
8 -y+1/2,-z,x+1/2
9 z,x,y
10 z+1/2,-x+1/2,-y
11 -z,x+1/2,-y+1/2
12 -z+1/2,-x,y+1/2
13 -x,-y,-z
14 -x+1/2,y+1/2,z
15 x,-y+1/2,z+1/2
16 x+1/2,y,-z+1/2
17 -y,-z,-x
18 -y+1/2,z+1/2,x
19 y,-z+1/2,x+1/2
20 y+1/2,z,-x+1/2
21 -z,-x,-y
22 -z+1/2,x+1/2,y
23 z,-x+1/2,y+1/2
24 z+1/2,x,-y+1/2

loop_
_atom_site_label
_atom_site_type_symbol
_atom_site_symmetry_multiplicity
_atom_site_Wyckoff_label
_atom_site_fract_x
_atom_site_fract_y
_atom_site_fract_z
_atom_site_occupancy
Sb1 Sb 4 a 0.00000 0.00000 0.00000 1.00000
Na1 Na 4 b 0.50000 0.50000 0.50000 1.00000
F1 F 24 d -0.05000 0.05000 0.22500 1.00000
```

$NaSbF_6$: A6BC_cP32_205_d_b_a - POSCAR

```
A6BC_cP32_205_d_b_a & a,x3,y3,z3 --params=8.18,-0.05,0.05,0.225 & Pa-3
    ↪ T_{h}^{6} #205 (abd) & cP32 & None & F6NaSb & F6NaSb & N.
    ↪ Schrewelius, Z. Anorg. Allg. Chem. 238, 241-254 (1938)
   1.00000000000000
   8.18000000000000   0.00000000000000   0.00000000000000
   0.00000000000000   8.18000000000000   0.00000000000000
   0.00000000000000   0.00000000000000   8.18000000000000
     F      Na      Sb
    24       4       4
Direct
  -0.05000000000000   0.05000000000000   0.22500000000000   F   (24d)
   0.55000000000000  -0.05000000000000   0.72500000000000   F   (24d)
   0.05000000000000   0.55000000000000   0.27500000000000   F   (24d)
   0.45000000000000   0.45000000000000  -0.22500000000000   F   (24d)
   0.22500000000000  -0.05000000000000   0.05000000000000   F   (24d)
   0.72500000000000   0.55000000000000  -0.05000000000000   F   (24d)
   0.27500000000000   0.05000000000000   0.55000000000000   F   (24d)
  -0.22500000000000   0.45000000000000   0.45000000000000   F   (24d)
   0.05000000000000   0.22500000000000  -0.05000000000000   F   (24d)
  -0.05000000000000   0.72500000000000   0.55000000000000   F   (24d)
   0.55000000000000   0.27500000000000   0.05000000000000   F   (24d)
   0.45000000000000  -0.22500000000000   0.45000000000000   F   (24d)
   0.05000000000000  -0.05000000000000  -0.22500000000000   F   (24d)
   0.45000000000000   0.05000000000000   0.27500000000000   F   (24d)
  -0.05000000000000   0.45000000000000   0.72500000000000   F   (24d)
   0.55000000000000   0.55000000000000   0.22500000000000   F   (24d)
  -0.22500000000000   0.05000000000000  -0.05000000000000   F   (24d)
   0.27500000000000   0.45000000000000   0.05000000000000   F   (24d)
   0.72500000000000  -0.05000000000000   0.45000000000000   F   (24d)
   0.22500000000000   0.55000000000000   0.55000000000000   F   (24d)
  -0.05000000000000  -0.22500000000000   0.05000000000000   F   (24d)
   0.05000000000000   0.27500000000000   0.45000000000000   F   (24d)
   0.45000000000000   0.72500000000000  -0.05000000000000   F   (24d)
   0.55000000000000   0.22500000000000   0.55000000000000   F   (24d)
   0.50000000000000   0.50000000000000   0.50000000000000   Na   (4b)
   0.00000000000000   0.50000000000000   0.00000000000000   Na   (4b)
   0.50000000000000   0.00000000000000   0.00000000000000   Na   (4b)
   0.00000000000000   0.00000000000000   0.50000000000000   Na   (4b)
   0.00000000000000   0.00000000000000   0.00000000000000   Sb   (4a)
   0.50000000000000   0.00000000000000   0.50000000000000   Sb   (4a)
   0.00000000000000   0.50000000000000   0.50000000000000   Sb   (4a)
   0.50000000000000   0.50000000000000   0.00000000000000   Sb   (4a)
```

$ZrP_2O_7$ High-Temperature ($K6_1$): A7B2C_cP40_205_bd_c_a - CIF

```
# CIF file
data_findsym-output
_audit_creation_method FINDSYM

_chemical_name_mineral 'P2O7Zr'
_chemical_formula_sum 'O7 P2 Zr'

loop_
_publ_author_name
 'G. R. Levi'
 'G. Peyronel'
_journal_name_full_name
;
 Zeitschrift f{\"u}r Kristallographie - Crystalline Materials
;
_journal_volume 92
_journal_year 1935
_journal_page_first 190
_journal_page_last 209
_publ_Section_title
;
 Struttura Cristallografica del Gruppo Isomorfo (Si$^{4+}$, Ti$^{4+}$,
    ↪ Zr$^{4+}$, Sn$^{4+}$, Hf$^{4+}$) P$_{2}$O$_{7}$
;

# Found in The American Mineralogist Crystal Structure Database, 2003

_aflow_title 'ZrP$_{2}$O$_{7}$ High-Temperature ($K6_{1}$) Structure'
_aflow_proto 'A7B2C_cP40_205_bd_c_a'
_aflow_params 'a,x_{3},x_{4},y_{4},z_{4}'
_aflow_params_values '8.2,0.39,0.394,0.218,0.458'
_aflow_Strukturbericht '$K6_{1}$'
_aflow_Pearson 'cP40'

_symmetry_space_group_name_H-M "P 21/a -3"
_symmetry_Int_Tables_number 205

_cell_length_a      8.20000
_cell_length_b      8.20000
_cell_length_c      8.20000
_cell_angle_alpha  90.00000
_cell_angle_beta   90.00000
_cell_angle_gamma  90.00000

loop_
_space_group_symop_id
_space_group_symop_operation_xyz
1 x,y,z
2 x+1/2,-y+1/2,-z
3 -x,y+1/2,-z+1/2
4 -x+1/2,-y,z+1/2
5 y,z,x
6 y+1/2,-z+1/2,-x
7 -y,z+1/2,-x+1/2
8 -y+1/2,-z,x+1/2
9 z,x,y
10 z+1/2,-x+1/2,-y
11 -z,x+1/2,-y+1/2
12 -z+1/2,-x,y+1/2
13 -x,-y,-z
14 -x+1/2,y+1/2,z
15 x,-y+1/2,z+1/2
16 x+1/2,y,-z+1/2
17 -y,-z,-x
18 -y+1/2,z+1/2,x
19 y,-z+1/2,x+1/2
20 y+1/2,z,-x+1/2
21 -z,-x,-y
22 -z+1/2,x+1/2,y
23 z,-x+1/2,y+1/2
24 z+1/2,x,-y+1/2

loop_
_atom_site_label
_atom_site_type_symbol
_atom_site_symmetry_multiplicity
_atom_site_Wyckoff_label
_atom_site_fract_x
_atom_site_fract_y
_atom_site_fract_z
_atom_site_occupancy
Zr1 Zr 4 a 0.00000 0.00000 0.00000 1.00000
O1 O 4 b 0.50000 0.50000 0.50000 1.00000
P1 P 8 c 0.39000 0.39000 0.39000 1.00000
O2 O 24 d 0.39400 0.21800 0.45800 1.00000
```

$ZrP_2O_7$ High-Temperature ($K6_1$): A7B2C_cP40_205_bd_c_a - POSCAR

```
A7B2C_cP40_205_bd_c_a & a,x3,x4,y4,z4 --params=8.2,0.39,0.394,0.218,
    ↪ 0.458 & Pa-3 T_{h}^{6} #205 (abcd) & cP40 & $K6_{1}$ & P2O7Zr &
    ↪ P2O7Zr & G. R. Levi and G. Peyronel, Zeitschrift f{"u}r
    ↪ Kristallographie - Crystalline Materials 92, 190-209 (1935)
   1.00000000000000
   8.20000000000000   0.00000000000000   0.00000000000000
   0.00000000000000   8.20000000000000   0.00000000000000
   0.00000000000000   0.00000000000000   8.20000000000000
     O       P      Zr
    28       8       4
Direct
   0.50000000000000   0.50000000000000   0.50000000000000   O   (4b)
   0.00000000000000   0.50000000000000   0.00000000000000   O   (4b)
   0.50000000000000   0.00000000000000   0.00000000000000   O   (4b)
   0.00000000000000   0.00000000000000   0.50000000000000   O   (4b)
   0.39400000000000   0.21800000000000   0.45800000000000   O  (24d)
   0.10600000000000  -0.21800000000000   0.95800000000000   O  (24d)
  -0.39400000000000   0.71800000000000   0.04200000000000   O  (24d)
   0.89400000000000   0.28200000000000  -0.45800000000000   O  (24d)
   0.45800000000000   0.39400000000000   0.21800000000000   O  (24d)
   0.95800000000000   0.10600000000000  -0.21800000000000   O  (24d)
   0.04200000000000  -0.39400000000000   0.71800000000000   O  (24d)
  -0.45800000000000   0.89400000000000   0.28200000000000   O  (24d)
   0.21800000000000   0.45800000000000   0.39400000000000   O  (24d)
  -0.21800000000000   0.95800000000000   0.10600000000000   O  (24d)
   0.71800000000000   0.04200000000000  -0.39400000000000   O  (24d)
   0.28200000000000  -0.45800000000000   0.89400000000000   O  (24d)
  -0.39400000000000  -0.21800000000000  -0.45800000000000   O  (24d)
```

```
    0.89400000000000   0.21800000000000   0.04200000000000    O   (24d)
    0.39400000000000   0.28200000000000   0.95800000000000    O   (24d)
    0.10600000000000   0.71800000000000   0.45800000000000    O   (24d)
   -0.45800000000000  -0.39400000000000  -0.21800000000000    O   (24d)
    0.04200000000000   0.89400000000000   0.21800000000000    O   (24d)
    0.95800000000000   0.39400000000000   0.28200000000000    O   (24d)
    0.45800000000000   0.10600000000000   0.71800000000000    O   (24d)
   -0.21800000000000  -0.45800000000000  -0.39400000000000    O   (24d)
    0.21800000000000   0.04200000000000   0.89400000000000    O   (24d)
    0.28200000000000   0.95800000000000   0.39400000000000    O   (24d)
    0.71800000000000   0.45800000000000   0.10600000000000    O   (24d)
    0.39000000000000   0.39000000000000   0.39000000000000    P    (8c)
    0.11000000000000  -0.39000000000000   0.89000000000000    P    (8c)
   -0.39000000000000   0.89000000000000   0.11000000000000    P    (8c)
    0.89000000000000   0.11000000000000  -0.39000000000000    P    (8c)
   -0.39000000000000  -0.39000000000000  -0.39000000000000    P    (8c)
    0.89000000000000   0.39000000000000   0.11000000000000    P    (8c)
    0.39000000000000   0.11000000000000   0.89000000000000    P    (8c)
    0.11000000000000   0.89000000000000   0.39000000000000    P    (8c)
    0.00000000000000   0.00000000000000   0.00000000000000   Zr    (4a)
    0.50000000000000   0.00000000000000   0.50000000000000   Zr    (4a)
    0.00000000000000   0.50000000000000   0.50000000000000   Zr    (4a)
    0.50000000000000   0.50000000000000   0.00000000000000   Zr    (4a)
```

$NaCr(SO_4)_2 \cdot 12H_2O$ Alum: AB12CD8E2_cP96_205_a_2d_b_cd_c - CIF

```
# CIF file
data_findsym-output
_audit_creation_method FINDSYM

_chemical_name_mineral 'Alum'
_chemical_formula_sum 'Cr (H2O)12 Na O8 S2'

loop_
_publ_author_name
 'A. H. C. Ledsham'
 'H. Steeple'
_journal_name_full_name
;
 Acta Crystallographica Section B: Structural Science
;
_journal_volume 24
_journal_year 1968
_journal_page_first 1287
_journal_page_last 1289
_publ_Section_title
;
 The crystal structure of sodium chromium alum and caesium chromium alum
;

_aflow_title 'NaCr(SO$_{4}$)$_{2}$$\cdot$12H$_{2}$O Alum Structure'
_aflow_proto 'AB12CD8E2_cP96_205_a_2d_b_cd_c'
_aflow_params 'a,x_{3},x_{4},x_{5},y_{5},z_{5},x_{6},y_{6},z_{6},x_{7},
    ↪ y_{7},z_{7}'
_aflow_params_values '12.4,0.239,0.31,0.158,0.014,0.018,0.042,0.136,
    ↪ 0.302,0.307,0.224,-0.08'
_aflow_Strukturbericht 'None'
_aflow_Pearson 'cP96'

_symmetry_space_group_name_H-M "P 21/a -3"
_symmetry_Int_Tables_number 205

_cell_length_a      12.40000
_cell_length_b      12.40000
_cell_length_c      12.40000
_cell_angle_alpha   90.00000
_cell_angle_beta    90.00000
_cell_angle_gamma   90.00000

loop_
_space_group_symop_id
_space_group_symop_operation_xyz
1 x,y,z
2 x+1/2,-y+1/2,-z
3 -x,y+1/2,-z+1/2
4 -x+1/2,-y,z+1/2
5 y,z,x
6 y+1/2,-z+1/2,-x
7 -y,z+1/2,-x+1/2
8 -y+1/2,-z,x+1/2
9 z,x,y
10 z+1/2,-x+1/2,-y
11 -z,x+1/2,-y+1/2
12 -z+1/2,-x,y+1/2
13 -x,-y,-z
14 -x+1/2,y+1/2,z
15 x,-y+1/2,z+1/2
16 x+1/2,y,-z+1/2
17 -y,-z,-x
18 -y+1/2,z+1/2,x
19 y,-z+1/2,x+1/2
20 y+1/2,z,-x+1/2
21 -z,-x,-y
22 -z+1/2,x+1/2,y
23 z,-x+1/2,y+1/2
24 z+1/2,x,-y+1/2

loop_
_atom_site_label
_atom_site_type_symbol
_atom_site_symmetry_multiplicity
_atom_site_Wyckoff_label
_atom_site_fract_x
_atom_site_fract_y
_atom_site_fract_z
_atom_site_occupancy
Cr1 Cr 4 a 0.00000 0.00000 0.00000 1.00000
Na1 Na 4 b 0.50000 0.50000 0.50000 1.00000
O1 O 8 c 0.23900 0.23900 0.23900 1.00000
S1 S 8 c 0.31000 0.31000 0.31000 1.00000
H2O1 H2O 24 d 0.15800 0.01400 0.01800 1.00000
H2O2 H2O 24 d 0.04200 0.13600 0.30200 1.00000
O2 O 24 d 0.30700 0.22400 -0.0800 1.00000
```

$NaCr(SO_4)_2 \cdot 12H_2O$ Alum: AB12CD8E2_cP96_205_a_2d_b_cd_c - POSCAR

```
AB12CD8E2_cP96_205_a_2d_b_cd_c & a,x3,x4,x5,y5,z5,x6,y6,z6,x7,y7,z7 --
    ↪ params=12.4,0.239,0.31,0.158,0.014,0.018,0.042,0.136,0.302,
    ↪ 0.307,0.224,-0.08 & Pa-3 T_{h}^{6} #205 (abc^2d^3) & cP96 &
    ↪ None & Cr(H2O)12NaO8S2 & Alum & A. H. C. Ledsham and H. Steeple
    ↪ , Acta Crystallogr. Sect. B Struct. Sci. 24, 1287-1289 (1968)
   1.00000000000000
  12.40000000000000   0.00000000000000   0.00000000000000
   0.00000000000000  12.40000000000000   0.00000000000000
   0.00000000000000   0.00000000000000  12.40000000000000
    Cr   H2O    Na     O     S
     4    48     4    32     8
Direct
    0.00000000000000   0.00000000000000   0.00000000000000   Cr    (4a)
    0.50000000000000   0.00000000000000   0.50000000000000   Cr    (4a)
    0.00000000000000   0.50000000000000   0.50000000000000   Cr    (4a)
    0.50000000000000   0.50000000000000   0.00000000000000   Cr    (4a)
    0.15800000000000   0.01400000000000   0.01800000000000  H2O   (24d)
    0.34200000000000  -0.01400000000000   0.51800000000000  H2O   (24d)
   -0.15800000000000   0.51400000000000   0.48200000000000  H2O   (24d)
    0.65800000000000   0.48600000000000  -0.01800000000000  H2O   (24d)
    0.01800000000000   0.15800000000000   0.01400000000000  H2O   (24d)
    0.51800000000000   0.34200000000000  -0.01400000000000  H2O   (24d)
    0.48200000000000  -0.15800000000000   0.51400000000000  H2O   (24d)
   -0.01800000000000   0.65800000000000   0.48600000000000  H2O   (24d)
    0.01400000000000   0.01800000000000   0.15800000000000  H2O   (24d)
   -0.01400000000000   0.51800000000000   0.34200000000000  H2O   (24d)
    0.51400000000000   0.48200000000000  -0.15800000000000  H2O   (24d)
    0.48600000000000  -0.01800000000000   0.65800000000000  H2O   (24d)
   -0.15800000000000  -0.01400000000000  -0.01800000000000  H2O   (24d)
    0.65800000000000   0.01400000000000   0.48200000000000  H2O   (24d)
    0.15800000000000   0.48600000000000   0.51800000000000  H2O   (24d)
    0.34200000000000   0.51400000000000   0.01800000000000  H2O   (24d)
   -0.01800000000000  -0.15800000000000  -0.01400000000000  H2O   (24d)
    0.48200000000000   0.65800000000000   0.01400000000000  H2O   (24d)
    0.51800000000000   0.15800000000000   0.48600000000000  H2O   (24d)
    0.01800000000000   0.34200000000000   0.51400000000000  H2O   (24d)
   -0.01400000000000  -0.01800000000000  -0.15800000000000  H2O   (24d)
    0.01400000000000   0.48200000000000   0.65800000000000  H2O   (24d)
    0.48600000000000   0.51800000000000   0.15800000000000  H2O   (24d)
    0.51400000000000   0.01800000000000   0.34200000000000  H2O   (24d)
    0.04200000000000   0.13600000000000   0.30200000000000  H2O   (24d)
    0.45800000000000  -0.13600000000000   0.80200000000000  H2O   (24d)
   -0.04200000000000   0.63600000000000   0.19800000000000  H2O   (24d)
    0.54200000000000   0.36400000000000  -0.30200000000000  H2O   (24d)
    0.30200000000000   0.04200000000000   0.13600000000000  H2O   (24d)
    0.80200000000000   0.45800000000000  -0.13600000000000  H2O   (24d)
    0.19800000000000  -0.04200000000000   0.63600000000000  H2O   (24d)
   -0.30200000000000   0.54200000000000   0.36400000000000  H2O   (24d)
    0.13600000000000   0.30200000000000   0.04200000000000  H2O   (24d)
   -0.13600000000000   0.80200000000000   0.45800000000000  H2O   (24d)
    0.63600000000000   0.19800000000000  -0.04200000000000  H2O   (24d)
    0.36400000000000  -0.30200000000000   0.54200000000000  H2O   (24d)
   -0.04200000000000  -0.13600000000000  -0.30200000000000  H2O   (24d)
    0.54200000000000   0.13600000000000   0.19800000000000  H2O   (24d)
    0.04200000000000   0.36400000000000   0.80200000000000  H2O   (24d)
    0.45800000000000   0.63600000000000   0.30200000000000  H2O   (24d)
   -0.30200000000000  -0.04200000000000  -0.13600000000000  H2O   (24d)
    0.19800000000000   0.54200000000000   0.13600000000000  H2O   (24d)
    0.80200000000000   0.04200000000000   0.36400000000000  H2O   (24d)
    0.30200000000000   0.45800000000000   0.63600000000000  H2O   (24d)
   -0.13600000000000  -0.30200000000000  -0.04200000000000  H2O   (24d)
    0.13600000000000   0.19800000000000   0.54200000000000  H2O   (24d)
    0.36400000000000   0.80200000000000   0.04200000000000  H2O   (24d)
    0.63600000000000   0.30200000000000   0.45800000000000  H2O   (24d)
    0.50000000000000   0.50000000000000   0.50000000000000   Na    (4b)
    0.00000000000000   0.50000000000000   0.00000000000000   Na    (4b)
    0.50000000000000   0.00000000000000   0.00000000000000   Na    (4b)
    0.00000000000000   0.00000000000000   0.50000000000000   Na    (4b)
    0.23900000000000   0.23900000000000   0.23900000000000    O    (8c)
    0.26100000000000  -0.23900000000000   0.73900000000000    O    (8c)
   -0.23900000000000   0.73900000000000   0.26100000000000    O    (8c)
    0.73900000000000   0.26100000000000  -0.23900000000000    O    (8c)
   -0.23900000000000  -0.23900000000000  -0.23900000000000    O    (8c)
    0.73900000000000   0.23900000000000   0.26100000000000    O    (8c)
    0.23900000000000   0.26100000000000   0.73900000000000    O    (8c)
    0.26100000000000   0.73900000000000   0.23900000000000    O    (8c)
    0.30700000000000   0.22400000000000  -0.08000000000000    O   (24d)
    0.19300000000000  -0.22400000000000   0.42000000000000    O   (24d)
   -0.30700000000000   0.72400000000000   0.58000000000000    O   (24d)
    0.80700000000000   0.27600000000000   0.08000000000000    O   (24d)
   -0.08000000000000   0.30700000000000   0.22400000000000    O   (24d)
    0.42000000000000   0.19300000000000  -0.22400000000000    O   (24d)
    0.58000000000000  -0.30700000000000   0.72400000000000    O   (24d)
    0.08000000000000   0.80700000000000   0.27600000000000    O   (24d)
    0.22400000000000  -0.08000000000000   0.30700000000000    O   (24d)
   -0.22400000000000   0.42000000000000   0.19300000000000    O   (24d)
    0.72400000000000   0.58000000000000  -0.30700000000000    O   (24d)
    0.27600000000000   0.08000000000000   0.80700000000000    O   (24d)
   -0.30700000000000  -0.22400000000000   0.08000000000000    O   (24d)
    0.80700000000000   0.22400000000000   0.58000000000000    O   (24d)
    0.30700000000000   0.27600000000000   0.42000000000000    O   (24d)
    0.19300000000000   0.72400000000000  -0.08000000000000    O   (24d)
    0.08000000000000  -0.30700000000000  -0.22400000000000    O   (24d)
    0.58000000000000   0.80700000000000   0.22400000000000    O   (24d)
```

```
   0.42000000000000    0.30700000000000    0.27600000000000    O   (24d)
  -0.08000000000000    0.19300000000000    0.72400000000000    O   (24d)
  -0.22400000000000    0.08000000000000   -0.30700000000000    O   (24d)
   0.22400000000000    0.58000000000000    0.80700000000000    O   (24d)
   0.27600000000000    0.42000000000000    0.30700000000000    O   (24d)
   0.72400000000000   -0.08000000000000    0.19300000000000    O   (24d)
   0.31000000000000    0.31000000000000    0.31000000000000    S    (8c)
   0.19000000000000   -0.31000000000000    0.81000000000000    S    (8c)
  -0.31000000000000    0.81000000000000    0.19000000000000    S    (8c)
   0.81000000000000    0.19000000000000   -0.31000000000000    S    (8c)
  -0.31000000000000   -0.31000000000000   -0.31000000000000    S    (8c)
   0.81000000000000    0.31000000000000    0.19000000000000    S    (8c)
   0.31000000000000    0.19000000000000    0.81000000000000    S    (8c)
   0.19000000000000    0.81000000000000    0.31000000000000    S    (8c)
```

$\gamma$-Alum [$AlNa(SO_4)_2 \cdot 12H_2O$, *H*$4_{15}$]: AB24CD20E2_cP192_205_a_4d_b_c3d_c - CIF

```
# CIF file
data_findsym-output
_audit_creation_method FINDSYM

_chemical_name_mineral '$\gamma$-alum'
_chemical_formula_sum 'Al H24 Na O20 S2'

loop_
_publ_author_name
 'D. T. Cromer'
 'M. I. Kay'
 'A. C. Larson'
_journal_name_full_name
;
 Acta Cristallographica
;
_journal_volume 22
_journal_year 1967
_journal_page_first 182
_journal_page_last 187
_publ_Section_title
;
 Refinement of the alum structures. II. X-ray and neutron diffraction of
     ↪ NaAl(SO$_{4}$)$_{2}$$\cdot$12H$_{2}$O, $\gamma$-alum
;

# Found in The American Mineralogist Crystal Structure Database, 2003

_aflow_title '$\gamma$-Alum [AlNa(SO$_{4}$)$_{2}$$\cdot$12H$_{2}$O, $H4_
     ↪ {15}$] Structure'
_aflow_proto 'AB24CD20E2_cP192_205_a_4d_b_c3d_c'
_aflow_params 'a,x_{3},x_{4},x_{5},y_{5},z_{5},x_{6},y_{6},z_{6},x_{7},
     ↪ y_{7},z_{7},x_{8},y_{8},z_{8},x_{9},y_{9},z_{9},x_{10},y_{10},
     ↪ z_{10},x_{11},y_{11},z_{11}'
_aflow_params_values '12.213,0.3343,0.2652,0.585,0.319,0.378,0.486,0.308
     ↪ ,0.383,0.555,0.202,0.502,0.588,0.339,0.113,0.2957,0.2783,0.1508
     ↪ ,0.0767,0.0403,0.3188,0.1371,-0.0404,0.0573'
_aflow_Strukturbericht '$H4_{15}$'
_aflow_Pearson 'cP192'

_symmetry_space_group_name_H-M "P 21/a -3"
_symmetry_Int_Tables_number 205

_cell_length_a     12.21300
_cell_length_b     12.21300
_cell_length_c     12.21300
_cell_angle_alpha  90.00000
_cell_angle_beta   90.00000
_cell_angle_gamma  90.00000

loop_
_space_group_symop_id
_space_group_symop_operation_xyz
1 x,y,z
2 x+1/2,-y+1/2,-z
3 -x,y+1/2,-z+1/2
4 -x+1/2,-y,z+1/2
5 y,z,x
6 y+1/2,-z+1/2,-x
7 -y,z+1/2,-x+1/2
8 -y+1/2,-z,x+1/2
9 z,x,y
10 z+1/2,-x+1/2,-y
11 -z,x+1/2,-y+1/2
12 -z+1/2,-x,y+1/2
13 -x,-y,-z
14 -x+1/2,y+1/2,z
15 x,-y+1/2,z+1/2
16 x+1/2,y,-z+1/2
17 -y,-z,-x
18 -y+1/2,z+1/2,x
19 y,-z+1/2,x+1/2
20 y+1/2,z,-x+1/2
21 -z,-x,-y
22 -z+1/2,x+1/2,y
23 z,-x+1/2,y+1/2
24 z+1/2,x,-y+1/2

loop_
_atom_site_label
_atom_site_type_symbol
_atom_site_symmetry_multiplicity
_atom_site_Wyckoff_label
_atom_site_fract_x
_atom_site_fract_y
_atom_site_fract_z
_atom_site_occupancy
Al1 Al 4 a 0.0000 0.0000 0.0000 1.00000
Na1 Na 4 b 0.5000 0.5000 0.5000 1.00000
O1 O 8 c 0.3343 0.3343 0.3343 1.00000
S1 S 8 c 0.2652 0.2652 0.2652 1.00000
H1 H 24 d 0.5850 0.3190 0.3780 1.00000
H2 H 24 d 0.4860 0.3080 0.3830 1.00000
H3 H 24 d 0.5550 0.2020 0.5020 1.00000
H4 H 24 d 0.5880 0.3390 0.1130 1.00000
O2 O 24 d 0.2957 0.2783 0.1508 1.00000
O3 O 24 d 0.0767 0.0403 0.3188 1.00000
O4 O 24 d 0.1371 -0.0404 0.0573 1.00000
```

$\gamma$-Alum [$AlNa(SO_4)_2 \cdot 12H_2O$, *H*$4_{15}$]: AB24CD20E2_cP192_205_a_4d_b_c3d_c - POSCAR

```
AB24CD20E2_cP192_205_a_4d_b_c3d_c & a,x3,x4,x5,y5,z5,x6,y6,z6,x7,y7,z7,
     ↪ x8,y8,z8,x9,y9,z9,x10,y10,z10,x11,y11,z11 --params=12.213,
     ↪ 0.3343,0.2652,0.585,0.319,0.378,0.486,0.308,0.383,0.555,0.202,
     ↪ 0.502,0.588,0.339,0.113,0.2957,0.2783,0.1508,0.0767,0.0403,
     ↪ 0.3188,0.1371,-0.0404,0.0573 & Pa-3 T_{h}^{6} #205 (abc^2d^7) &
     ↪ cP192 & $H4_{15}$ & AlH24NaO20S2 & $\gamma$-alum & D. T.
     ↪ Cromer and M. I. Kay and A. C. Larson, Acta Cryst. 22, 182-187
     ↪ (1967)
   1.00000000000000
  12.21300000000000    0.00000000000000    0.00000000000000
   0.00000000000000   12.21300000000000    0.00000000000000
   0.00000000000000    0.00000000000000   12.21300000000000
    Al     H    Na     O     S
     4    96     4    80     8
Direct
   0.00000000000000    0.00000000000000    0.00000000000000   Al    (4a)
   0.50000000000000    0.00000000000000    0.50000000000000   Al    (4a)
   0.00000000000000    0.50000000000000    0.50000000000000   Al    (4a)
   0.50000000000000    0.50000000000000    0.00000000000000   Al    (4a)
   0.58500000000000    0.31900000000000    0.37800000000000    H   (24d)
  -0.08500000000000   -0.31900000000000    0.87800000000000    H   (24d)
  -0.58500000000000    0.81900000000000    0.12200000000000    H   (24d)
   1.08500000000000    0.18100000000000   -0.37800000000000    H   (24d)
   0.37800000000000    0.58500000000000    0.31900000000000    H   (24d)
   0.87800000000000   -0.08500000000000   -0.31900000000000    H   (24d)
   0.12200000000000   -0.58500000000000    0.81900000000000    H   (24d)
  -0.37800000000000    1.08500000000000    0.18100000000000    H   (24d)
   0.31900000000000    0.37800000000000    0.58500000000000    H   (24d)
  -0.31900000000000    0.87800000000000   -0.08500000000000    H   (24d)
   0.81900000000000    0.12200000000000   -0.58500000000000    H   (24d)
   0.18100000000000   -0.37800000000000    1.08500000000000    H   (24d)
  -0.58500000000000   -0.31900000000000   -0.37800000000000    H   (24d)
   1.08500000000000    0.31900000000000    0.12200000000000    H   (24d)
   0.58500000000000    0.18100000000000    0.87800000000000    H   (24d)
  -0.08500000000000    0.81900000000000    0.37800000000000    H   (24d)
  -0.37800000000000   -0.58500000000000   -0.31900000000000    H   (24d)
   0.12200000000000    1.08500000000000    0.31900000000000    H   (24d)
   0.87800000000000    0.58500000000000    0.18100000000000    H   (24d)
   0.37800000000000   -0.08500000000000    0.81900000000000    H   (24d)
  -0.31900000000000   -0.37800000000000   -0.58500000000000    H   (24d)
   0.31900000000000    0.12200000000000    1.08500000000000    H   (24d)
   0.18100000000000    0.87800000000000    0.58500000000000    H   (24d)
   0.81900000000000    0.37800000000000   -0.08500000000000    H   (24d)
   0.48600000000000    0.30800000000000    0.38300000000000    H   (24d)
   0.01400000000000   -0.30800000000000    0.88300000000000    H   (24d)
  -0.48600000000000    0.80800000000000    0.11700000000000    H   (24d)
   0.98600000000000    0.19200000000000   -0.38300000000000    H   (24d)
   0.38300000000000    0.48600000000000    0.30800000000000    H   (24d)
   0.88300000000000    0.01400000000000   -0.30800000000000    H   (24d)
   0.11700000000000   -0.48600000000000    0.80800000000000    H   (24d)
  -0.38300000000000    0.98600000000000    0.19200000000000    H   (24d)
   0.30800000000000    0.38300000000000    0.48600000000000    H   (24d)
  -0.30800000000000    0.88300000000000    0.01400000000000    H   (24d)
   0.80800000000000    0.11700000000000   -0.48600000000000    H   (24d)
   0.19200000000000   -0.38300000000000    0.98600000000000    H   (24d)
  -0.48600000000000   -0.30800000000000   -0.38300000000000    H   (24d)
   0.98600000000000    0.30800000000000    0.11700000000000    H   (24d)
   0.48600000000000    0.19200000000000    0.88300000000000    H   (24d)
   0.01400000000000    0.80800000000000    0.38300000000000    H   (24d)
  -0.38300000000000   -0.48600000000000   -0.30800000000000    H   (24d)
   0.11700000000000    0.98600000000000    0.30800000000000    H   (24d)
   0.88300000000000    0.48600000000000    0.19200000000000    H   (24d)
   0.38300000000000    0.01400000000000    0.80800000000000    H   (24d)
  -0.30800000000000   -0.38300000000000   -0.48600000000000    H   (24d)
   0.30800000000000    0.11700000000000    0.98600000000000    H   (24d)
   0.19200000000000    0.88300000000000    0.48600000000000    H   (24d)
   0.80800000000000    0.38300000000000    0.01400000000000    H   (24d)
   0.55500000000000    0.20200000000000    0.50200000000000    H   (24d)
  -0.05500000000000   -0.20200000000000    1.00200000000000    H   (24d)
  -0.55500000000000    0.70200000000000   -0.00200000000000    H   (24d)
   1.05500000000000    0.29800000000000   -0.50200000000000    H   (24d)
   0.50200000000000    0.55500000000000    0.20200000000000    H   (24d)
   1.00200000000000   -0.05500000000000   -0.20200000000000    H   (24d)
  -0.00200000000000   -0.55500000000000    0.70200000000000    H   (24d)
  -0.50200000000000    1.05500000000000    0.29800000000000    H   (24d)
   0.20200000000000    0.50200000000000    0.55500000000000    H   (24d)
  -0.20200000000000    1.00200000000000   -0.05500000000000    H   (24d)
   0.70200000000000   -0.00200000000000   -0.55500000000000    H   (24d)
   0.29800000000000   -0.50200000000000    1.05500000000000    H   (24d)
  -0.55500000000000   -0.20200000000000   -0.50200000000000    H   (24d)
   1.05500000000000    0.20200000000000   -0.00200000000000    H   (24d)
   0.55500000000000    0.29800000000000    1.00200000000000    H   (24d)
  -0.05500000000000    0.70200000000000    0.50200000000000    H   (24d)
  -0.50200000000000   -0.55500000000000   -0.20200000000000    H   (24d)
  -0.00200000000000    1.05500000000000    0.20200000000000    H   (24d)
   1.00200000000000    0.55500000000000    0.29800000000000    H   (24d)
   0.50200000000000   -0.05500000000000    0.70200000000000    H   (24d)
  -0.20200000000000   -0.50200000000000   -0.55500000000000    H   (24d)
   0.20200000000000   -0.00200000000000    1.05500000000000    H   (24d)
   0.29800000000000    1.00200000000000    0.55500000000000    H   (24d)
   0.70200000000000    0.50200000000000   -0.05500000000000    H   (24d)
   0.58800000000000    0.33900000000000    0.11300000000000    H   (24d)
```

| | | | | |
|---|---|---|---|---|
| −0.08800000000000 | −0.33900000000000 | 0.61300000000000 | H | (24d) |
| −0.58800000000000 | 0.83900000000000 | 0.38700000000000 | H | (24d) |
| 1.08800000000000 | 0.16100000000000 | −0.11300000000000 | H | (24d) |
| 0.11300000000000 | 0.58800000000000 | 0.33900000000000 | H | (24d) |
| 0.61300000000000 | −0.08800000000000 | −0.33900000000000 | H | (24d) |
| 0.38700000000000 | −0.58800000000000 | 0.83900000000000 | H | (24d) |
| −0.11300000000000 | 1.08800000000000 | 0.16100000000000 | H | (24d) |
| 0.33900000000000 | 0.11300000000000 | 0.58800000000000 | H | (24d) |
| −0.33900000000000 | 0.61300000000000 | −0.08800000000000 | H | (24d) |
| 0.83900000000000 | 0.38700000000000 | −0.58800000000000 | H | (24d) |
| 0.16100000000000 | −0.11300000000000 | 1.08800000000000 | H | (24d) |
| −0.58800000000000 | −0.33900000000000 | −0.11300000000000 | H | (24d) |
| 1.08800000000000 | 0.33900000000000 | 0.38700000000000 | H | (24d) |
| 0.58800000000000 | 0.16100000000000 | 0.61300000000000 | H | (24d) |
| −0.08800000000000 | 0.83900000000000 | 0.11300000000000 | H | (24d) |
| −0.11300000000000 | −0.58800000000000 | −0.33900000000000 | H | (24d) |
| 0.38700000000000 | 1.08800000000000 | 0.33900000000000 | H | (24d) |
| 0.61300000000000 | 0.58800000000000 | 0.16100000000000 | H | (24d) |
| 0.11300000000000 | −0.08800000000000 | 0.83900000000000 | H | (24d) |
| −0.33900000000000 | −0.11300000000000 | −0.58800000000000 | H | (24d) |
| 0.33900000000000 | 0.38700000000000 | 1.08800000000000 | H | (24d) |
| 0.16100000000000 | 0.61300000000000 | 0.58800000000000 | H | (24d) |
| 0.83900000000000 | 0.11300000000000 | −0.08800000000000 | H | (24d) |
| 0.50000000000000 | 0.50000000000000 | 0.50000000000000 | Na | (4b) |
| 0.00000000000000 | 0.50000000000000 | 0.00000000000000 | Na | (4b) |
| 0.50000000000000 | 0.00000000000000 | 0.00000000000000 | Na | (4b) |
| 0.00000000000000 | 0.00000000000000 | 0.50000000000000 | Na | (4b) |
| 0.33430000000000 | 0.33430000000000 | 0.33430000000000 | O | (8c) |
| 0.16570000000000 | −0.33430000000000 | 0.83430000000000 | O | (8c) |
| −0.33430000000000 | 0.83430000000000 | 0.16570000000000 | O | (8c) |
| 0.83430000000000 | 0.16570000000000 | −0.33430000000000 | O | (8c) |
| −0.33430000000000 | −0.33430000000000 | −0.33430000000000 | O | (8c) |
| 0.83430000000000 | 0.33430000000000 | 0.16570000000000 | O | (8c) |
| 0.33430000000000 | 0.16570000000000 | 0.83430000000000 | O | (8c) |
| 0.16570000000000 | 0.83430000000000 | 0.33430000000000 | O | (8c) |
| 0.29570000000000 | 0.27830000000000 | 0.15080000000000 | O | (24d) |
| 0.20430000000000 | −0.27830000000000 | 0.65080000000000 | O | (24d) |
| −0.29570000000000 | 0.77830000000000 | 0.34920000000000 | O | (24d) |
| 0.79570000000000 | 0.22170000000000 | −0.15080000000000 | O | (24d) |
| 0.15080000000000 | 0.29570000000000 | 0.27830000000000 | O | (24d) |
| 0.65080000000000 | 0.20430000000000 | −0.27830000000000 | O | (24d) |
| 0.34920000000000 | −0.29570000000000 | 0.77830000000000 | O | (24d) |
| −0.15080000000000 | 0.79570000000000 | 0.22170000000000 | O | (24d) |
| 0.27830000000000 | 0.15080000000000 | 0.29570000000000 | O | (24d) |
| −0.27830000000000 | 0.65080000000000 | 0.20430000000000 | O | (24d) |
| 0.77830000000000 | 0.34920000000000 | −0.29570000000000 | O | (24d) |
| 0.22170000000000 | −0.15080000000000 | 0.79570000000000 | O | (24d) |
| −0.29570000000000 | −0.27830000000000 | −0.15080000000000 | O | (24d) |
| 0.79570000000000 | 0.27830000000000 | 0.34920000000000 | O | (24d) |
| 0.29570000000000 | 0.22170000000000 | 0.65080000000000 | O | (24d) |
| 0.20430000000000 | 0.77830000000000 | 0.15080000000000 | O | (24d) |
| −0.15080000000000 | −0.29570000000000 | −0.27830000000000 | O | (24d) |
| 0.34920000000000 | 0.79570000000000 | 0.27830000000000 | O | (24d) |
| 0.65080000000000 | 0.29570000000000 | 0.22170000000000 | O | (24d) |
| 0.15080000000000 | 0.20430000000000 | 0.77830000000000 | O | (24d) |
| −0.27830000000000 | −0.15080000000000 | −0.29570000000000 | O | (24d) |
| 0.27830000000000 | 0.34920000000000 | 0.79570000000000 | O | (24d) |
| 0.22170000000000 | 0.65080000000000 | 0.29570000000000 | O | (24d) |
| 0.77830000000000 | 0.15080000000000 | 0.20430000000000 | O | (24d) |
| 0.07670000000000 | 0.04030000000000 | 0.31880000000000 | O | (24d) |
| 0.42330000000000 | −0.04030000000000 | 0.81880000000000 | O | (24d) |
| −0.07670000000000 | 0.54030000000000 | 0.18120000000000 | O | (24d) |
| 0.57670000000000 | 0.45970000000000 | −0.31880000000000 | O | (24d) |
| 0.31880000000000 | 0.07670000000000 | 0.04030000000000 | O | (24d) |
| 0.81880000000000 | 0.42330000000000 | −0.04030000000000 | O | (24d) |
| 0.18120000000000 | −0.07670000000000 | 0.54030000000000 | O | (24d) |
| −0.31880000000000 | 0.57670000000000 | 0.45970000000000 | O | (24d) |
| 0.04030000000000 | 0.31880000000000 | 0.07670000000000 | O | (24d) |
| −0.04030000000000 | 0.81880000000000 | 0.42330000000000 | O | (24d) |
| 0.54030000000000 | 0.18120000000000 | −0.07670000000000 | O | (24d) |
| 0.45970000000000 | −0.31880000000000 | 0.57670000000000 | O | (24d) |
| −0.07670000000000 | −0.04030000000000 | −0.31880000000000 | O | (24d) |
| 0.57670000000000 | 0.04030000000000 | 0.18120000000000 | O | (24d) |
| 0.07670000000000 | 0.45970000000000 | 0.81880000000000 | O | (24d) |
| 0.42330000000000 | 0.54030000000000 | 0.31880000000000 | O | (24d) |
| −0.31880000000000 | −0.07670000000000 | −0.04030000000000 | O | (24d) |
| 0.18120000000000 | 0.57670000000000 | 0.04030000000000 | O | (24d) |
| 0.81880000000000 | 0.07670000000000 | 0.45970000000000 | O | (24d) |
| 0.31880000000000 | 0.42330000000000 | 0.54030000000000 | O | (24d) |
| −0.04030000000000 | −0.31880000000000 | −0.07670000000000 | O | (24d) |
| 0.04030000000000 | 0.18120000000000 | 0.57670000000000 | O | (24d) |
| 0.45970000000000 | 0.81880000000000 | 0.07670000000000 | O | (24d) |
| 0.54030000000000 | 0.31880000000000 | 0.42330000000000 | O | (24d) |
| 0.13710000000000 | −0.04040000000000 | 0.05730000000000 | O | (24d) |
| 0.36290000000000 | 0.04040000000000 | 0.55730000000000 | O | (24d) |
| −0.13710000000000 | 0.45960000000000 | 0.44270000000000 | O | (24d) |
| 0.63710000000000 | 0.54040000000000 | −0.05730000000000 | O | (24d) |
| 0.05730000000000 | 0.13710000000000 | −0.04040000000000 | O | (24d) |
| 0.55730000000000 | 0.36290000000000 | 0.04040000000000 | O | (24d) |
| 0.44270000000000 | −0.13710000000000 | 0.45960000000000 | O | (24d) |
| −0.05730000000000 | 0.63710000000000 | 0.54040000000000 | O | (24d) |
| −0.04040000000000 | 0.05730000000000 | 0.13710000000000 | O | (24d) |
| 0.04040000000000 | 0.55730000000000 | 0.36290000000000 | O | (24d) |
| 0.45960000000000 | 0.44270000000000 | −0.13710000000000 | O | (24d) |
| 0.54040000000000 | −0.05730000000000 | 0.63710000000000 | O | (24d) |
| −0.13710000000000 | 0.04040000000000 | −0.05730000000000 | O | (24d) |
| 0.63710000000000 | −0.04040000000000 | 0.44270000000000 | O | (24d) |
| 0.13710000000000 | 0.54040000000000 | 0.55730000000000 | O | (24d) |
| 0.36290000000000 | 0.45960000000000 | 0.05730000000000 | O | (24d) |
| −0.05730000000000 | −0.13710000000000 | 0.04040000000000 | O | (24d) |
| 0.44270000000000 | 0.63710000000000 | −0.04040000000000 | O | (24d) |
| 0.55730000000000 | 0.13710000000000 | 0.54040000000000 | O | (24d) |
| 0.05730000000000 | 0.36290000000000 | 0.45960000000000 | O | (24d) |
| 0.04040000000000 | −0.05730000000000 | −0.13710000000000 | O | (24d) |
| −0.04040000000000 | 0.44270000000000 | 0.63710000000000 | O | (24d) |
| 0.54040000000000 | 0.55730000000000 | 0.13710000000000 | O | (24d) |
| 0.45960000000000 | 0.05730000000000 | 0.36290000000000 | O | (24d) |
| 0.26520000000000 | 0.26520000000000 | 0.26520000000000 | S | (8c) |
| 0.23480000000000 | −0.26520000000000 | 0.76520000000000 | S | (8c) |
| −0.26520000000000 | 0.76520000000000 | 0.23480000000000 | S | (8c) |
| 0.76520000000000 | 0.23480000000000 | −0.26520000000000 | S | (8c) |
| −0.26520000000000 | −0.26520000000000 | −0.26520000000000 | S | (8c) |
| 0.76520000000000 | 0.26520000000000 | 0.23480000000000 | S | (8c) |
| 0.26520000000000 | 0.23480000000000 | 0.76520000000000 | S | (8c) |
| 0.23480000000000 | 0.76520000000000 | 0.26520000000000 | S | (8c) |

$\alpha$-Alum [KAl$(SO_4)_2 \cdot 12H_2O$, $H4_{13}$]: AB24CD28E2_cP224_205_a_4d_b_2c4d_c - CIF

```
# CIF file
data_findsym-output
_audit_creation_method FINDSYM

_chemical_name_mineral '$\alpha$-alum'
_chemical_formula_sum 'Al H24 K O28 S2'

loop_
_publ_author_name
 'S. C. Nyburg'
 'J. W. Steed'
 'S. Aleksovska'
 'V. M. Petrusevski'
_journal_name_full_name
;
 Acta Crystallographica Section B: Structural Science
;
_journal_volume 56
_journal_year 2000
_journal_page_first 204
_journal_page_last 209
_publ_Section_title
;
 Structure of the alums. I. On the sulfate group disorder in the $\
      ↪ alpha$-alums
;

_aflow_title '$\alpha$-Alum [KAl(SO$_{4}$)$_{2}$ $\cdot$ 12H$_{2}$O,
      ↪ $H4_{13}$] Structure'
_aflow_proto 'AB24CD28E2_cP224_205_a_4d_b_2c4d_c'
_aflow_params 'a,x_{3},x_{4},x_{5},x_{6},y_{6},z_{6},x_{7},y_{7},z_{7},
      ↪ x_{8},y_{8},z_{8},x_{9},y_{9},z_{9},x_{10},y_{10},z_{10},x_{11
      ↪ },y_{11},z_{11},x_{12},y_{12},z_{12},x_{13},y_{13},z_{13}'
_aflow_params_values '12.135,0.23606,0.3746,0.3061,0.175,0.0365,-0.07275
      ↪ ,0.1982,0.0262,0.034,0.104,0.155,0.3,0.0009,0.1821,0.2914,
      ↪ 0.26302,0.41979,0.30624,0.2808,0.2,0.3614,0.15173,0.028,-
      ↪ 0.01985,0.04692,0.13071,0.30399'
_aflow_Strukturbericht '$H4_{13}$'
_aflow_Pearson 'cP224'

_symmetry_space_group_name_H-M "P 21/a -3"
_symmetry_Int_Tables_number 205

_cell_length_a     12.13500
_cell_length_b     12.13500
_cell_length_c     12.13500
_cell_angle_alpha  90.00000
_cell_angle_beta   90.00000
_cell_angle_gamma  90.00000

loop_
_space_group_symop_id
_space_group_symop_operation_xyz
1 x,y,z
2 x+1/2,-y+1/2,-z
3 -x,y+1/2,-z+1/2
4 -x+1/2,-y,z+1/2
5 y,z,x
6 y+1/2,-z+1/2,-x
7 -y,z+1/2,-x+1/2
8 -y+1/2,-z,x+1/2
9 z,x,y
10 z+1/2,-x+1/2,-y
11 -z,x+1/2,-y+1/2
12 -z+1/2,-x,y+1/2
13 -x,-y,-z
14 -x+1/2,y+1/2,z
15 x,-y+1/2,z+1/2
16 x+1/2,y,-z+1/2
17 -y,-z,-x
18 -y+1/2,z+1/2,x
19 y,-z+1/2,x+1/2
20 y+1/2,z,-x+1/2
21 -z,-x,-y
22 -z+1/2,x+1/2,y
23 z,-x+1/2,y+1/2
24 z+1/2,x,-y+1/2

loop_
_atom_site_label
_atom_site_type_symbol
_atom_site_symmetry_multiplicity
_atom_site_Wyckoff_label
_atom_site_fract_x
_atom_site_fract_y
_atom_site_fract_z
_atom_site_occupancy
Al1 Al 4 a 0.00000 0.00000 0.00000 1.00000
K1 K 4 b 0.50000 0.50000 0.50000 1.00000
O1 O 8 c 0.23606 0.23606 0.23606 0.78700
O2 O 8 c 0.37460 0.37460 0.37460 0.21300
S1 S 8 c 0.30610 0.30610 0.30610 1.00000
```

```
H1 H 24 d 0.17500 0.03650 −0.07275 1.00000
H2 H 24 d 0.19820 0.02620 0.03400 1.00000
H3 H 24 d 0.10400 0.15500 0.30000 1.00000
H4 H 24 d 0.00090 0.18210 0.29140 1.00000
O3 O 24 d 0.26302 0.41979 0.30624 0.78700
O4 O 24 d 0.28080 0.20000 0.36140 0.21300
O5 O 24 d 0.15173 0.02800 −0.01985 1.00000
O6 O 24 d 0.04692 0.13071 0.30399 1.00000
```

$\alpha$-Alum [$KAl(SO_4)_2 \cdot 12H_2O$, $H4_{13}$]: AB24CD28E2_cP224_205_a_4d_b_2c4d_c - POSCAR

```
AB24CD28E2_cP224_205_a_4d_b_2c4d_c & a,x3,x4,x5,x6,y6,z6,x7,y7,z7,x8,y8,
    ↪ z8,x9,y9,z9,x10,y10,z10,x11,y11,z11,x12,y12,z12,x13,y13,z13 --
    ↪ params=12.135,0.23606,0.3746,0.3061,0.175,0.0365,-0.07275,
    ↪ 0.1982,0.0262,0.034,0.104,0.155,0.3,0.0009,0.1821,0.2914,
    ↪ 0.26302,0.41979,0.30624,0.2808,0.2,0.3614,0.15173,0.028,-
    ↪ 0.01985,0.04692,0.13071,0.30399 & Pa-3 T_{h}^{6} #205 (abc^3d^8
    ↪ ) & cP224 & $H4_{13}$ & AlH24KO20S2 & $\alpha$-alum & S. C.
    ↪ Nyburg et al., Acta Crystallogr. Sect. B Struct. Sci. 56,
    ↪ 204-209 (2000)
   1.00000000000000
  12.13500000000000    0.00000000000000    0.00000000000000
   0.00000000000000   12.13500000000000    0.00000000000000
   0.00000000000000    0.00000000000000   12.13500000000000
    Al     H     K     O     S
     4    96     4   112     8
Direct
   0.00000000000000    0.00000000000000    0.00000000000000   Al    (4a)
   0.50000000000000    0.00000000000000    0.50000000000000   Al    (4a)
   0.00000000000000    0.50000000000000    0.50000000000000   Al    (4a)
   0.50000000000000    0.50000000000000    0.00000000000000   Al    (4a)
   0.17500000000000    0.03650000000000   -0.07275000000000   H    (24d)
   0.32500000000000   -0.03650000000000    0.42725000000000   H    (24d)
  -0.17500000000000    0.53650000000000    0.57275000000000   H    (24d)
   0.67500000000000    0.46350000000000    0.07275000000000   H    (24d)
  -0.07275000000000    0.17500000000000    0.03650000000000   H    (24d)
   0.42725000000000    0.32500000000000   -0.03650000000000   H    (24d)
   0.57275000000000   -0.17500000000000    0.53650000000000   H    (24d)
   0.07275000000000    0.67500000000000    0.46350000000000   H    (24d)
   0.03650000000000   -0.07275000000000    0.17500000000000   H    (24d)
  -0.03650000000000    0.42725000000000    0.32500000000000   H    (24d)
   0.53650000000000    0.57275000000000   -0.17500000000000   H    (24d)
   0.46350000000000    0.07275000000000    0.67500000000000   H    (24d)
  -0.17500000000000   -0.03650000000000    0.07275000000000   H    (24d)
   0.67500000000000    0.03650000000000    0.57275000000000   H    (24d)
   0.17500000000000    0.46350000000000    0.42725000000000   H    (24d)
   0.32500000000000    0.53650000000000   -0.07275000000000   H    (24d)
   0.07275000000000   -0.17500000000000   -0.03650000000000   H    (24d)
   0.57275000000000    0.67500000000000    0.03650000000000   H    (24d)
   0.42725000000000    0.17500000000000    0.46350000000000   H    (24d)
  -0.07275000000000    0.32500000000000    0.53650000000000   H    (24d)
  -0.03650000000000    0.07275000000000   -0.17500000000000   H    (24d)
   0.03650000000000    0.57275000000000    0.67500000000000   H    (24d)
   0.46350000000000    0.42725000000000    0.17500000000000   H    (24d)
   0.53650000000000   -0.07275000000000    0.32500000000000   H    (24d)
   0.19820000000000    0.02620000000000    0.03400000000000   H    (24d)
   0.30180000000000   -0.02620000000000    0.53400000000000   H    (24d)
  -0.19820000000000    0.52620000000000    0.46600000000000   H    (24d)
   0.69820000000000    0.47380000000000   -0.03400000000000   H    (24d)
   0.03400000000000    0.19820000000000    0.02620000000000   H    (24d)
   0.53400000000000    0.30180000000000   -0.02620000000000   H    (24d)
   0.46600000000000   -0.19820000000000    0.52620000000000   H    (24d)
  -0.03400000000000    0.69820000000000    0.47380000000000   H    (24d)
   0.02620000000000    0.03400000000000    0.19820000000000   H    (24d)
  -0.02620000000000    0.53400000000000    0.30180000000000   H    (24d)
   0.52620000000000    0.46600000000000   -0.19820000000000   H    (24d)
   0.47380000000000   -0.03400000000000    0.69820000000000   H    (24d)
  -0.19820000000000   -0.02620000000000   -0.03400000000000   H    (24d)
   0.69820000000000    0.02620000000000    0.46600000000000   H    (24d)
   0.19820000000000    0.47380000000000    0.53400000000000   H    (24d)
   0.30180000000000    0.52620000000000    0.03400000000000   H    (24d)
  -0.03400000000000   -0.19820000000000   -0.02620000000000   H    (24d)
   0.46600000000000    0.69820000000000    0.02620000000000   H    (24d)
   0.53400000000000    0.19820000000000    0.47380000000000   H    (24d)
   0.03400000000000    0.30180000000000    0.52620000000000   H    (24d)
  -0.02620000000000   -0.03400000000000   -0.19820000000000   H    (24d)
   0.02620000000000    0.46600000000000    0.69820000000000   H    (24d)
   0.47380000000000    0.53400000000000    0.19820000000000   H    (24d)
   0.52620000000000    0.03400000000000    0.30180000000000   H    (24d)
   0.10400000000000    0.15500000000000    0.30000000000000   H    (24d)
   0.39600000000000   -0.15500000000000    0.80000000000000   H    (24d)
  -0.10400000000000    0.65500000000000    0.20000000000000   H    (24d)
   0.60400000000000    0.34500000000000   -0.30000000000000   H    (24d)
   0.30000000000000    0.10400000000000    0.15500000000000   H    (24d)
   0.80000000000000    0.39600000000000   -0.15500000000000   H    (24d)
   0.20000000000000   -0.10400000000000    0.65500000000000   H    (24d)
  -0.30000000000000    0.60400000000000    0.34500000000000   H    (24d)
   0.15500000000000    0.30000000000000    0.10400000000000   H    (24d)
  -0.15500000000000    0.80000000000000    0.39600000000000   H    (24d)
   0.65500000000000    0.20000000000000   -0.10400000000000   H    (24d)
   0.34500000000000   -0.30000000000000    0.60400000000000   H    (24d)
  -0.10400000000000   -0.15500000000000   -0.30000000000000   H    (24d)
   0.60400000000000    0.15500000000000    0.20000000000000   H    (24d)
   0.10400000000000    0.34500000000000    0.80000000000000   H    (24d)
   0.39600000000000    0.65500000000000    0.30000000000000   H    (24d)
  -0.30000000000000   -0.10400000000000   -0.15500000000000   H    (24d)
   0.20000000000000    0.60400000000000    0.15500000000000   H    (24d)
   0.80000000000000    0.10400000000000    0.34500000000000   H    (24d)
   0.30000000000000    0.39600000000000    0.65500000000000   H    (24d)
  -0.15500000000000   -0.30000000000000   -0.10400000000000   H    (24d)
   0.15500000000000    0.20000000000000    0.60400000000000   H    (24d)
   0.34500000000000    0.80000000000000    0.10400000000000   H    (24d)
   0.65500000000000    0.30000000000000    0.39600000000000   H    (24d)
   0.00090000000000    0.18210000000000    0.29140000000000   H    (24d)
   0.49910000000000   -0.18210000000000    0.79140000000000   H    (24d)
  -0.00090000000000    0.68210000000000    0.20860000000000   H    (24d)
   0.50090000000000    0.31790000000000   -0.29140000000000   H    (24d)
   0.29140000000000    0.00090000000000    0.18210000000000   H    (24d)
   0.79140000000000    0.49910000000000   -0.18210000000000   H    (24d)
   0.20860000000000   -0.00090000000000    0.68210000000000   H    (24d)
  -0.29140000000000    0.50090000000000    0.31790000000000   H    (24d)
   0.18210000000000    0.29140000000000    0.00090000000000   H    (24d)
  -0.18210000000000    0.79140000000000    0.49910000000000   H    (24d)
   0.68210000000000    0.20860000000000   -0.00090000000000   H    (24d)
   0.31790000000000   -0.29140000000000    0.50090000000000   H    (24d)
  -0.00090000000000   -0.18210000000000   -0.29140000000000   H    (24d)
   0.50090000000000    0.18210000000000    0.20860000000000   H    (24d)
   0.00090000000000    0.31790000000000    0.79140000000000   H    (24d)
   0.49910000000000    0.68210000000000    0.29140000000000   H    (24d)
  -0.29140000000000   -0.00090000000000   -0.18210000000000   H    (24d)
   0.20860000000000    0.50090000000000    0.18210000000000   H    (24d)
   0.79140000000000    0.00090000000000    0.31790000000000   H    (24d)
   0.29140000000000    0.49910000000000    0.68210000000000   H    (24d)
  -0.18210000000000   -0.29140000000000   -0.00090000000000   H    (24d)
   0.18210000000000    0.20860000000000    0.50090000000000   H    (24d)
   0.31790000000000    0.79140000000000    0.00090000000000   H    (24d)
   0.68210000000000    0.29140000000000    0.49910000000000   H    (24d)
   0.50000000000000    0.50000000000000    0.50000000000000   K     (4b)
   0.00000000000000    0.50000000000000    0.00000000000000   K     (4b)
   0.50000000000000    0.00000000000000    0.00000000000000   K     (4b)
   0.00000000000000    0.00000000000000    0.50000000000000   K     (4b)
   0.23606000000000    0.23606000000000    0.23606000000000   O     (8c)
   0.26394000000000   -0.23606000000000    0.73606000000000   O     (8c)
  -0.23606000000000    0.73606000000000    0.26394000000000   O     (8c)
   0.73606000000000    0.26394000000000   -0.23606000000000   O     (8c)
  -0.23606000000000   -0.23606000000000   -0.23606000000000   O     (8c)
   0.73606000000000    0.23606000000000    0.26394000000000   O     (8c)
   0.23606000000000    0.26394000000000    0.73606000000000   O     (8c)
   0.26394000000000    0.73606000000000    0.23606000000000   O     (8c)
   0.37460000000000    0.37460000000000    0.37460000000000   O     (8c)
   0.12540000000000   -0.37460000000000    0.87460000000000   O     (8c)
  -0.37460000000000    0.87460000000000    0.12540000000000   O     (8c)
   0.87460000000000    0.12540000000000   -0.37460000000000   O     (8c)
  -0.37460000000000   -0.37460000000000   -0.37460000000000   O     (8c)
   0.87460000000000    0.37460000000000    0.12540000000000   O     (8c)
   0.37460000000000    0.12540000000000    0.87460000000000   O     (8c)
   0.12540000000000    0.87460000000000    0.37460000000000   O     (8c)
   0.26302000000000    0.41979000000000    0.30624000000000   O    (24d)
   0.23698000000000   -0.41979000000000    0.80624000000000   O    (24d)
  -0.26302000000000    0.91979000000000    0.19376000000000   O    (24d)
   0.76302000000000    0.08021000000000   -0.30624000000000   O    (24d)
   0.30624000000000    0.26302000000000    0.41979000000000   O    (24d)
   0.80624000000000    0.23698000000000   -0.41979000000000   O    (24d)
   0.19376000000000   -0.26302000000000    0.91979000000000   O    (24d)
  -0.30624000000000    0.76302000000000    0.08021000000000   O    (24d)
   0.41979000000000    0.30624000000000    0.26302000000000   O    (24d)
  -0.41979000000000    0.80624000000000    0.23698000000000   O    (24d)
   0.91979000000000    0.19376000000000   -0.26302000000000   O    (24d)
   0.08021000000000   -0.30624000000000    0.76302000000000   O    (24d)
  -0.26302000000000   -0.41979000000000   -0.30624000000000   O    (24d)
   0.76302000000000    0.41979000000000    0.19376000000000   O    (24d)
   0.26302000000000    0.08021000000000    0.80624000000000   O    (24d)
   0.23698000000000    0.91979000000000    0.30624000000000   O    (24d)
  -0.30624000000000   -0.26302000000000   -0.41979000000000   O    (24d)
   0.19376000000000    0.76302000000000    0.41979000000000   O    (24d)
   0.80624000000000    0.26302000000000    0.08021000000000   O    (24d)
   0.30624000000000    0.23698000000000    0.91979000000000   O    (24d)
  -0.41979000000000   -0.30624000000000   -0.26302000000000   O    (24d)
   0.41979000000000    0.19376000000000    0.76302000000000   O    (24d)
   0.08021000000000    0.80624000000000    0.26302000000000   O    (24d)
   0.91979000000000    0.30624000000000    0.23698000000000   O    (24d)
   0.28080000000000    0.20000000000000    0.36140000000000   O    (24d)
   0.21920000000000   -0.20000000000000    0.86140000000000   O    (24d)
  -0.28080000000000    0.70000000000000    0.13860000000000   O    (24d)
   0.78080000000000    0.30000000000000   -0.36140000000000   O    (24d)
   0.36140000000000    0.28080000000000    0.20000000000000   O    (24d)
   0.86140000000000    0.21920000000000   -0.20000000000000   O    (24d)
   0.13860000000000   -0.28080000000000    0.70000000000000   O    (24d)
  -0.36140000000000    0.78080000000000    0.30000000000000   O    (24d)
   0.20000000000000    0.36140000000000    0.28080000000000   O    (24d)
  -0.20000000000000    0.86140000000000    0.21920000000000   O    (24d)
   0.70000000000000    0.13860000000000   -0.28080000000000   O    (24d)
   0.30000000000000   -0.36140000000000    0.78080000000000   O    (24d)
  -0.28080000000000   -0.20000000000000   -0.36140000000000   O    (24d)
   0.78080000000000    0.20000000000000    0.13860000000000   O    (24d)
   0.28080000000000    0.30000000000000    0.86140000000000   O    (24d)
   0.21920000000000    0.70000000000000    0.36140000000000   O    (24d)
  -0.36140000000000   -0.28080000000000   -0.20000000000000   O    (24d)
   0.13860000000000    0.78080000000000    0.20000000000000   O    (24d)
   0.86140000000000    0.28080000000000    0.30000000000000   O    (24d)
   0.36140000000000    0.21920000000000    0.70000000000000   O    (24d)
  -0.20000000000000   -0.36140000000000   -0.28080000000000   O    (24d)
   0.20000000000000    0.13860000000000    0.78080000000000   O    (24d)
   0.30000000000000    0.86140000000000    0.28080000000000   O    (24d)
   0.70000000000000    0.36140000000000    0.21920000000000   O    (24d)
   0.15173000000000    0.02800000000000   -0.01985000000000   O    (24d)
   0.34827000000000   -0.02800000000000    0.48015000000000   O    (24d)
  -0.15173000000000    0.52800000000000    0.51985000000000   O    (24d)
   0.65173000000000    0.47200000000000    0.01985000000000   O    (24d)
  -0.01985000000000    0.15173000000000    0.02800000000000   O    (24d)
   0.48015000000000    0.34827000000000   -0.02800000000000   O    (24d)
   0.51985000000000   -0.15173000000000    0.52800000000000   O    (24d)
   0.01985000000000    0.65173000000000    0.47200000000000   O    (24d)
   0.02800000000000   -0.01985000000000    0.15173000000000   O    (24d)
  -0.02800000000000    0.48015000000000    0.34827000000000   O    (24d)
   0.52800000000000    0.51985000000000   -0.15173000000000   O    (24d)
   0.47200000000000    0.01985000000000    0.65173000000000   O    (24d)
  -0.15173000000000   -0.02800000000000    0.01985000000000   O    (24d)
   0.65173000000000    0.02800000000000    0.51985000000000   O    (24d)
   0.15173000000000    0.47200000000000    0.48015000000000   O    (24d)
```

```
   0.34827000000000    0.52800000000000   -0.01985000000000     O   (24d)
   0.01985000000000   -0.15173000000000   -0.02800000000000     O   (24d)
   0.51985000000000    0.65173000000000    0.02800000000000     O   (24d)
   0.48015000000000    0.15173000000000    0.47200000000000     O   (24d)
  -0.01985000000000    0.34827000000000    0.52800000000000     O   (24d)
  -0.02800000000000    0.01985000000000   -0.15173000000000     O   (24d)
   0.02800000000000    0.51985000000000    0.65173000000000     O   (24d)
   0.47200000000000    0.48015000000000    0.15173000000000     O   (24d)
   0.52800000000000   -0.01985000000000    0.34827000000000     O   (24d)
   0.04692000000000    0.13071000000000    0.30399000000000     O   (24d)
   0.45308000000000   -0.13071000000000    0.80399000000000     O   (24d)
  -0.04692000000000    0.63071000000000    0.19601000000000     O   (24d)
   0.54692000000000    0.36929000000000   -0.30399000000000     O   (24d)
   0.30399000000000    0.04692000000000    0.13071000000000     O   (24d)
   0.80399000000000    0.45308000000000   -0.13071000000000     O   (24d)
   0.19601000000000   -0.04692000000000    0.63071000000000     O   (24d)
  -0.30399000000000    0.54692000000000    0.36929000000000     O   (24d)
   0.13071000000000    0.30399000000000    0.04692000000000     O   (24d)
  -0.13071000000000    0.80399000000000    0.45308000000000     O   (24d)
   0.63071000000000    0.19601000000000   -0.04692000000000     O   (24d)
   0.36929000000000   -0.30399000000000    0.54692000000000     O   (24d)
  -0.04692000000000   -0.13071000000000   -0.30399000000000     O   (24d)
   0.54692000000000    0.13071000000000    0.19601000000000     O   (24d)
   0.04692000000000    0.36929000000000    0.80399000000000     O   (24d)
   0.45308000000000    0.63071000000000    0.30399000000000     O   (24d)
  -0.30399000000000   -0.04692000000000   -0.13071000000000     O   (24d)
   0.19601000000000    0.54692000000000    0.13071000000000     O   (24d)
   0.80399000000000    0.04692000000000    0.36929000000000     O   (24d)
   0.30399000000000    0.45308000000000    0.63071000000000     O   (24d)
  -0.13071000000000   -0.30399000000000   -0.04692000000000     O   (24d)
   0.13071000000000    0.19601000000000    0.54692000000000     O   (24d)
   0.36929000000000    0.80399000000000    0.04692000000000     O   (24d)
   0.63071000000000    0.30399000000000    0.45308000000000     O   (24d)
   0.30610000000000    0.30610000000000    0.30610000000000     S    (8c)
   0.19390000000000   -0.30610000000000    0.80610000000000     S    (8c)
  -0.30610000000000    0.80610000000000    0.19390000000000     S    (8c)
   0.80610000000000    0.19390000000000   -0.30610000000000     S    (8c)
  -0.30610000000000   -0.30610000000000   -0.30610000000000     S    (8c)
   0.80610000000000    0.30610000000000    0.19390000000000     S    (8c)
   0.30610000000000    0.19390000000000    0.80610000000000     S    (8c)
   0.19390000000000    0.80610000000000    0.30610000000000     S    (8c)
```

$\beta$-Alum [$Al(NH_3CH_3)_2(SO_4)_2{\cdot}12H_2O$, $H4_{14}$]: AB2C36D2E20F2_cP252_205_a_c_6d_c_c3d_c - CIF

```
# CIF file
data_findsym-output
_audit_creation_method FINDSYM

_chemical_name_mineral '$\beta$-alum'
_chemical_formula_sum 'Al C2 H36 N2 O20 S2'

loop_
_publ_author_name
 'A. M. Abdeen'
 'G. Will'
 'W. Sch\"{a}fer'
 'A. Kirfel'
 'M. O. Bargouth'
 'K. Recker'
_journal_name_full_name
;
 Zeitschrift f{\"u}r Kristallographie - Crystalline Materials
;
_journal_volume 157
_journal_year 1981
_journal_page_first 147
_journal_page_last 166
_publ_Section_title
;
 X-Ray and neutron diffraction study of alums
;

_aflow_title '$\beta$-Alum [Al(NH$_{3}$CH$_{3}$)$_{2}$(SO$_{4}$)$_{2}$$\
    ↪ cdot$12H$_{2}$O, $H4_{14}$] Structure'
_aflow_proto 'AB2C36D2E20F2_cP252_205_a_c_6d_c_c3d_c'
_aflow_params 'a,x_{2},x_{3},x_{4},x_{5},x_{6},y_{6},z_{6},x_{7},y_{7},
    ↪ z_{7},x_{8},y_{8},z_{8},x_{9},y_{9},z_{9},x_{10},y_{10},z_{10},
    ↪ x_{11},y_{11},z_{11},x_{12},y_{12},z_{12},x_{13},y_{13},z_{13},
    ↪ x_{14},y_{14},z_{14}'
_aflow_params_values '12.314,0.4501,0.5156,0.2341,0.3022,0.3851,0.5072,
    ↪ 0.4134,0.0837,0.5363,0.0275,0.5423,0.2018,0.4728,0.4132,0.1846,
    ↪ 0.4502,0.2143,0.4941,0.1944,0.2148,0.6103,0.1697,0.6911,0.753,
    ↪ 0.0914,0.8521,-0.019,0.0185,0.0419,0.1373,0.2975'
_aflow_Strukturbericht '$H4_{14}$'
_aflow_Pearson 'cP252'

_symmetry_space_group_name_H-M "P 21/a -3"
_symmetry_Int_Tables_number 205

_cell_length_a    12.31400
_cell_length_b    12.31400
_cell_length_c    12.31400
_cell_angle_alpha 90.00000
_cell_angle_beta  90.00000
_cell_angle_gamma 90.00000

loop_
_space_group_symop_id
_space_group_symop_operation_xyz
1 x,y,z
2 x+1/2,-y+1/2,-z
3 -x,y+1/2,-z+1/2
4 -x+1/2,-y,z+1/2
5 y,z,x
6 y+1/2,-z+1/2,-x
7 -y,z+1/2,-x+1/2
8 -y+1/2,-z,x+1/2
9 z,x,y
10 z+1/2,-x+1/2,-y
11 -z,x+1/2,-y+1/2
12 -z+1/2,-x,y+1/2
13 -x,-y,-z
14 -x+1/2,y+1/2,z
15 x,-y+1/2,z+1/2
16 x+1/2,y,-z+1/2
17 -y,-z,-x
18 -y+1/2,z+1/2,x
19 y,-z+1/2,x+1/2
20 y+1/2,z,-x+1/2
21 -z,-x,-y
22 -z+1/2,x+1/2,y
23 z,-x+1/2,y+1/2
24 z+1/2,x,-y+1/2

loop_
_atom_site_label
_atom_site_type_symbol
_atom_site_symmetry_multiplicity
_atom_site_Wyckoff_label
_atom_site_fract_x
_atom_site_fract_y
_atom_site_fract_z
_atom_site_occupancy
Al1 Al 4 a 0.0000 0.0000 0.0000 1.00000
C1 C 8 c 0.4501 0.4501 0.4501 0.50000
N1 N 8 c 0.5156 0.5156 0.5156 0.50000
O1 O 8 c 0.2341 0.2341 0.2341 1.00000
S1 S 8 c 0.3022 0.3022 0.3022 1.00000
H1 H 24 d 0.3851 0.5072 0.4134 0.50000
H2 H 24 d 0.0837 0.5363 0.0275 0.50000
H3 H 24 d 0.5423 0.2018 0.4728 1.00000
H4 H 24 d 0.4132 0.1846 0.4502 1.00000
H5 H 24 d 0.2143 0.4941 0.1944 1.00000
H6 H 24 d 0.2148 0.6103 0.1697 1.00000
O2 O 24 d 0.6911 0.7530 0.0914 1.00000
O3 O 24 d 0.8521 -0.0190 0.0185 1.00000
O4 O 24 d 0.0419 0.1373 0.2975 1.00000
```

$\beta$-Alum [$Al(NH_3CH_3)_2(SO_4)_2{\cdot}12H_2O$, $H4_{14}$]: AB2C36D2E20F2_cP252_205_a_c_6d_c_c3d_c - POSCAR

```
AB2C36D2E20F2_cP252_205_a_c_6d_c_c3d_c & a,x2,x3,x4,x5,x6,y6,z6,x7,y7,z7
    ↪ ,x8,y8,z8,x9,y9,z9,x10,y10,z10,x11,y11,z11,x12,y12,z12,x13,y13,
    ↪ z13,x14,y14,z14 --params=12.314,0.4501,0.5156,0.2341,0.3022,
    ↪ 0.3851,0.5072,0.4134,0.0837,0.5363,0.0275,0.5423,0.2018,0.4728,
    ↪ 0.4132,0.1846,0.4502,0.2143,0.4941,0.1944,0.2148,0.6103,0.1697,
    ↪ 0.6911,0.753,0.0914,0.8521,-0.019,0.0185,0.0419,0.1373,0.2975 &
    ↪  Pa-3 T_{h}^{6} #205 (ac^4d^9) & cP252 & $H4_{14}$ &
    ↪ AlC2H36N2O20S2 & $\beta$-alum & A. M. Abdeen et al.,
    ↪ Zeitschrift f{"u}r Kristallographie - Crystalline Materials 157
    ↪ , 147-166 (1981)
   1.00000000000000
  12.31400000000000    0.00000000000000    0.00000000000000
   0.00000000000000   12.31400000000000    0.00000000000000
   0.00000000000000    0.00000000000000   12.31400000000000
    Al       C       H       N       O       S
     4       8     144       8      80       8
Direct
   0.00000000000000    0.00000000000000    0.00000000000000    Al    (4a)
   0.50000000000000    0.00000000000000    0.50000000000000    Al    (4a)
   0.00000000000000    0.50000000000000    0.50000000000000    Al    (4a)
   0.50000000000000    0.50000000000000    0.00000000000000    Al    (4a)
   0.45010000000000    0.45010000000000    0.45010000000000     C    (8c)
   0.04990000000000   -0.45010000000000    0.95010000000000     C    (8c)
  -0.45010000000000    0.95010000000000    0.04990000000000     C    (8c)
   0.95010000000000    0.04990000000000   -0.45010000000000     C    (8c)
  -0.45010000000000   -0.45010000000000   -0.45010000000000     C    (8c)
   0.95010000000000    0.45010000000000    0.04990000000000     C    (8c)
   0.45010000000000    0.04990000000000    0.95010000000000     C    (8c)
   0.04990000000000    0.95010000000000    0.45010000000000     C    (8c)
   0.38510000000000    0.50720000000000    0.41340000000000     H   (24d)
   0.11490000000000   -0.50720000000000    0.91340000000000     H   (24d)
  -0.38510000000000    1.00720000000000    0.08660000000000     H   (24d)
   0.88510000000000   -0.00720000000000   -0.41340000000000     H   (24d)
   0.41340000000000    0.38510000000000    0.50720000000000     H   (24d)
   0.91340000000000    0.11490000000000   -0.50720000000000     H   (24d)
   0.08660000000000   -0.38510000000000    1.00720000000000     H   (24d)
  -0.41340000000000    0.88510000000000   -0.00720000000000     H   (24d)
   0.50720000000000    0.41340000000000    0.38510000000000     H   (24d)
  -0.50720000000000    0.91340000000000    0.11490000000000     H   (24d)
   1.00720000000000    0.08660000000000   -0.38510000000000     H   (24d)
  -0.00720000000000   -0.41340000000000    0.88510000000000     H   (24d)
  -0.38510000000000   -0.50720000000000   -0.41340000000000     H   (24d)
   0.88510000000000    0.50720000000000    0.08660000000000     H   (24d)
   0.38510000000000   -0.00720000000000    0.91340000000000     H   (24d)
   0.11490000000000    1.00720000000000    0.41340000000000     H   (24d)
  -0.41340000000000   -0.38510000000000   -0.50720000000000     H   (24d)
   0.08660000000000    0.88510000000000    0.50720000000000     H   (24d)
   0.91340000000000    0.38510000000000   -0.00720000000000     H   (24d)
   0.41340000000000    0.11490000000000    1.00720000000000     H   (24d)
  -0.50720000000000   -0.41340000000000   -0.38510000000000     H   (24d)
   0.50720000000000    0.08660000000000    0.88510000000000     H   (24d)
  -0.00720000000000    0.91340000000000    0.38510000000000     H   (24d)
   1.00720000000000    0.41340000000000    0.11490000000000     H   (24d)
   0.08370000000000    0.53630000000000    0.02750000000000     H   (24d)
   0.41630000000000   -0.53630000000000    0.52750000000000     H   (24d)
  -0.08370000000000    1.03630000000000    0.47250000000000     H   (24d)
   0.58370000000000   -0.03630000000000   -0.02750000000000     H   (24d)
   0.02750000000000    0.08370000000000    0.53630000000000     H   (24d)
   0.52750000000000    0.41630000000000   -0.53630000000000     H   (24d)
```

| | | | | |
|---|---|---|---|---|
| 0.47250000000000 | −0.08370000000000 | 1.03630000000000 | H | (24d) |
| −0.02750000000000 | 0.58370000000000 | −0.03630000000000 | H | (24d) |
| 0.53630000000000 | 0.02750000000000 | 0.08370000000000 | H | (24d) |
| −0.53630000000000 | 0.52750000000000 | 0.41630000000000 | H | (24d) |
| 1.03630000000000 | 0.47250000000000 | −0.08370000000000 | H | (24d) |
| −0.03630000000000 | −0.02750000000000 | 0.58370000000000 | H | (24d) |
| −0.08370000000000 | −0.53630000000000 | −0.02750000000000 | H | (24d) |
| 0.58370000000000 | 0.53630000000000 | 0.47250000000000 | H | (24d) |
| 0.08370000000000 | −0.03630000000000 | 0.52750000000000 | H | (24d) |
| 0.41630000000000 | 1.03630000000000 | 0.02750000000000 | H | (24d) |
| −0.02750000000000 | −0.08370000000000 | −0.53630000000000 | H | (24d) |
| 0.47250000000000 | 0.58370000000000 | 0.53630000000000 | H | (24d) |
| 0.52750000000000 | 0.08370000000000 | −0.03630000000000 | H | (24d) |
| 0.02750000000000 | 0.41630000000000 | 1.03630000000000 | H | (24d) |
| −0.53630000000000 | −0.02750000000000 | −0.08370000000000 | H | (24d) |
| 0.53630000000000 | 0.47250000000000 | 0.58370000000000 | H | (24d) |
| −0.03630000000000 | 0.52750000000000 | 0.08370000000000 | H | (24d) |
| 1.03630000000000 | 0.02750000000000 | 0.41630000000000 | H | (24d) |
| 0.54230000000000 | 0.20180000000000 | 0.47280000000000 | H | (24d) |
| −0.04230000000000 | −0.20180000000000 | 0.97280000000000 | H | (24d) |
| −0.54230000000000 | 0.70180000000000 | 0.02720000000000 | H | (24d) |
| 1.04230000000000 | 0.29820000000000 | −0.47280000000000 | H | (24d) |
| 0.47280000000000 | 0.54230000000000 | 0.20180000000000 | H | (24d) |
| 0.97280000000000 | −0.04230000000000 | −0.20180000000000 | H | (24d) |
| 0.02720000000000 | −0.54230000000000 | 0.70180000000000 | H | (24d) |
| −0.47280000000000 | 1.04230000000000 | 0.29820000000000 | H | (24d) |
| 0.20180000000000 | 0.47280000000000 | 0.54230000000000 | H | (24d) |
| −0.20180000000000 | 0.97280000000000 | −0.04230000000000 | H | (24d) |
| 0.70180000000000 | 0.02720000000000 | −0.54230000000000 | H | (24d) |
| 0.29820000000000 | −0.47280000000000 | 1.04230000000000 | H | (24d) |
| −0.54230000000000 | −0.20180000000000 | −0.47280000000000 | H | (24d) |
| 1.04230000000000 | 0.20180000000000 | 0.02720000000000 | H | (24d) |
| 0.54230000000000 | 0.29820000000000 | 0.97280000000000 | H | (24d) |
| −0.04230000000000 | 0.70180000000000 | 0.47280000000000 | H | (24d) |
| −0.47280000000000 | −0.54230000000000 | −0.20180000000000 | H | (24d) |
| 0.02720000000000 | 1.04230000000000 | 0.20180000000000 | H | (24d) |
| 0.97280000000000 | 0.54230000000000 | 0.29820000000000 | H | (24d) |
| 0.47280000000000 | −0.04230000000000 | 0.70180000000000 | H | (24d) |
| −0.20180000000000 | −0.47280000000000 | −0.54230000000000 | H | (24d) |
| 0.20180000000000 | 0.02720000000000 | 1.04230000000000 | H | (24d) |
| 0.29820000000000 | 0.97280000000000 | 0.54230000000000 | H | (24d) |
| 0.70180000000000 | 0.47280000000000 | −0.04230000000000 | H | (24d) |
| 0.41320000000000 | 0.18460000000000 | 0.45020000000000 | H | (24d) |
| 0.08680000000000 | −0.18460000000000 | 0.95020000000000 | H | (24d) |
| −0.41320000000000 | 0.68460000000000 | 0.04980000000000 | H | (24d) |
| 0.91320000000000 | 0.31540000000000 | −0.45020000000000 | H | (24d) |
| 0.45020000000000 | 0.41320000000000 | 0.18460000000000 | H | (24d) |
| 0.95020000000000 | 0.08680000000000 | −0.18460000000000 | H | (24d) |
| 0.04980000000000 | −0.41320000000000 | 0.68460000000000 | H | (24d) |
| −0.45020000000000 | 0.91320000000000 | 0.31540000000000 | H | (24d) |
| 0.18460000000000 | 0.45020000000000 | 0.41320000000000 | H | (24d) |
| −0.18460000000000 | 0.95020000000000 | 0.08680000000000 | H | (24d) |
| 0.68460000000000 | 0.04980000000000 | −0.41320000000000 | H | (24d) |
| 0.31540000000000 | −0.45020000000000 | 0.91320000000000 | H | (24d) |
| −0.41320000000000 | −0.18460000000000 | −0.45020000000000 | H | (24d) |
| 0.91320000000000 | 0.18460000000000 | 0.04980000000000 | H | (24d) |
| 0.41320000000000 | 0.31540000000000 | 0.95020000000000 | H | (24d) |
| 0.08680000000000 | 0.68460000000000 | 0.45020000000000 | H | (24d) |
| −0.45020000000000 | −0.41320000000000 | −0.18460000000000 | H | (24d) |
| 0.04980000000000 | 0.91320000000000 | 0.18460000000000 | H | (24d) |
| 0.95020000000000 | 0.41320000000000 | 0.31540000000000 | H | (24d) |
| 0.45020000000000 | 0.08680000000000 | 0.68460000000000 | H | (24d) |
| −0.18460000000000 | −0.45020000000000 | −0.41320000000000 | H | (24d) |
| 0.18460000000000 | 0.04980000000000 | 0.91320000000000 | H | (24d) |
| 0.31540000000000 | 0.95020000000000 | 0.41320000000000 | H | (24d) |
| 0.68460000000000 | 0.45020000000000 | 0.08680000000000 | H | (24d) |
| 0.21430000000000 | 0.49410000000000 | 0.19440000000000 | H | (24d) |
| 0.28570000000000 | −0.49410000000000 | 0.69440000000000 | H | (24d) |
| −0.21430000000000 | 0.99410000000000 | 0.30560000000000 | H | (24d) |
| 0.71430000000000 | 0.00590000000000 | −0.19440000000000 | H | (24d) |
| 0.19440000000000 | 0.21430000000000 | 0.49410000000000 | H | (24d) |
| 0.69440000000000 | 0.28570000000000 | −0.49410000000000 | H | (24d) |
| 0.30560000000000 | −0.21430000000000 | 0.99410000000000 | H | (24d) |
| −0.19440000000000 | 0.71430000000000 | 0.00590000000000 | H | (24d) |
| 0.49410000000000 | 0.19440000000000 | 0.21430000000000 | H | (24d) |
| −0.49410000000000 | 0.69440000000000 | 0.28570000000000 | H | (24d) |
| 0.99410000000000 | 0.30560000000000 | −0.21430000000000 | H | (24d) |
| 0.00590000000000 | −0.19440000000000 | 0.71430000000000 | H | (24d) |
| −0.21430000000000 | −0.49410000000000 | −0.19440000000000 | H | (24d) |
| 0.71430000000000 | 0.49410000000000 | 0.30560000000000 | H | (24d) |
| 0.21430000000000 | 0.00590000000000 | 0.69440000000000 | H | (24d) |
| 0.28570000000000 | 0.99410000000000 | 0.19440000000000 | H | (24d) |
| −0.19440000000000 | −0.21430000000000 | −0.49410000000000 | H | (24d) |
| 0.30560000000000 | 0.71430000000000 | 0.49410000000000 | H | (24d) |
| 0.69440000000000 | 0.21430000000000 | 0.00590000000000 | H | (24d) |
| 0.19440000000000 | 0.28570000000000 | 0.99410000000000 | H | (24d) |
| −0.49410000000000 | −0.19440000000000 | −0.21430000000000 | H | (24d) |
| 0.49410000000000 | 0.30560000000000 | 0.71430000000000 | H | (24d) |
| 0.00590000000000 | 0.69440000000000 | 0.21430000000000 | H | (24d) |
| 0.99410000000000 | 0.19440000000000 | 0.28570000000000 | H | (24d) |
| 0.21480000000000 | 0.61030000000000 | 0.16970000000000 | H | (24d) |
| 0.28520000000000 | −0.61030000000000 | 0.66970000000000 | H | (24d) |
| −0.21480000000000 | 1.11030000000000 | 0.33030000000000 | H | (24d) |
| 0.71480000000000 | −0.11030000000000 | −0.16970000000000 | H | (24d) |
| 0.16970000000000 | 0.21480000000000 | 0.61030000000000 | H | (24d) |
| 0.66970000000000 | 0.28520000000000 | −0.61030000000000 | H | (24d) |
| 0.33030000000000 | −0.21480000000000 | 1.11030000000000 | H | (24d) |
| −0.16970000000000 | 0.71480000000000 | −0.11030000000000 | H | (24d) |
| 0.61030000000000 | 0.16970000000000 | 0.21480000000000 | H | (24d) |
| −0.61030000000000 | 0.66970000000000 | 0.28520000000000 | H | (24d) |
| 1.11030000000000 | 0.33030000000000 | −0.21480000000000 | H | (24d) |
| −0.11030000000000 | −0.16970000000000 | 0.71480000000000 | H | (24d) |
| −0.21480000000000 | −0.61030000000000 | −0.16970000000000 | H | (24d) |
| 0.71480000000000 | 0.61030000000000 | 0.33030000000000 | H | (24d) |
| 0.21480000000000 | −0.11030000000000 | 0.66970000000000 | H | (24d) |
| 0.28520000000000 | 1.11030000000000 | 0.16970000000000 | H | (24d) |
| −0.16970000000000 | −0.21480000000000 | −0.61030000000000 | H | (24d) |
| 0.33030000000000 | 0.71480000000000 | 0.61030000000000 | H | (24d) |
| 0.66970000000000 | 0.21480000000000 | −0.11030000000000 | H | (24d) |
| 0.16970000000000 | 0.28520000000000 | 1.11030000000000 | H | (24d) |
| −0.61030000000000 | −0.16970000000000 | −0.21480000000000 | H | (24d) |
| 0.61030000000000 | 0.33030000000000 | 0.71480000000000 | H | (24d) |
| −0.11030000000000 | 0.66970000000000 | 0.21480000000000 | H | (24d) |
| 1.11030000000000 | 0.16970000000000 | 0.28520000000000 | H | (24d) |
| 0.51560000000000 | 0.51560000000000 | 0.51560000000000 | N | (8c) |
| −0.01560000000000 | −0.51560000000000 | 1.01560000000000 | N | (8c) |
| −0.51560000000000 | 1.01560000000000 | −0.01560000000000 | N | (8c) |
| 1.01560000000000 | −0.01560000000000 | −0.51560000000000 | N | (8c) |
| −0.51560000000000 | −0.51560000000000 | −0.51560000000000 | N | (8c) |
| 1.01560000000000 | 0.51560000000000 | −0.01560000000000 | N | (8c) |
| 0.51560000000000 | −0.01560000000000 | 1.01560000000000 | N | (8c) |
| −0.01560000000000 | 1.01560000000000 | 0.51560000000000 | N | (8c) |
| 0.23410000000000 | 0.23410000000000 | 0.23410000000000 | O | (8c) |
| 0.26590000000000 | −0.23410000000000 | 0.73410000000000 | O | (8c) |
| −0.23410000000000 | 0.73410000000000 | 0.26590000000000 | O | (8c) |
| 0.73410000000000 | 0.26590000000000 | −0.23410000000000 | O | (8c) |
| −0.23410000000000 | −0.23410000000000 | −0.23410000000000 | O | (8c) |
| 0.73410000000000 | 0.23410000000000 | 0.26590000000000 | O | (8c) |
| 0.23410000000000 | 0.26590000000000 | 0.73410000000000 | O | (8c) |
| 0.26590000000000 | 0.73410000000000 | 0.23410000000000 | O | (8c) |
| 0.69110000000000 | 0.75300000000000 | 0.09140000000000 | O | (24d) |
| −0.19110000000000 | −0.75300000000000 | 0.59140000000000 | O | (24d) |
| −0.69110000000000 | 1.25300000000000 | 0.40860000000000 | O | (24d) |
| 1.19110000000000 | −0.25300000000000 | −0.09140000000000 | O | (24d) |
| 0.09140000000000 | 0.69110000000000 | 0.75300000000000 | O | (24d) |
| 0.59140000000000 | −0.19110000000000 | −0.75300000000000 | O | (24d) |
| 0.40860000000000 | −0.69110000000000 | 1.25300000000000 | O | (24d) |
| −0.09140000000000 | 1.19110000000000 | −0.25300000000000 | O | (24d) |
| 0.75300000000000 | 0.09140000000000 | 0.69110000000000 | O | (24d) |
| −0.75300000000000 | 0.59140000000000 | −0.19110000000000 | O | (24d) |
| 1.25300000000000 | 0.40860000000000 | −0.69110000000000 | O | (24d) |
| −0.25300000000000 | −0.09140000000000 | 1.19110000000000 | O | (24d) |
| −0.69110000000000 | −0.75300000000000 | −0.09140000000000 | O | (24d) |
| 1.19110000000000 | 0.75300000000000 | 0.40860000000000 | O | (24d) |
| 0.69110000000000 | −0.25300000000000 | 0.59140000000000 | O | (24d) |
| −0.19110000000000 | 1.25300000000000 | 0.09140000000000 | O | (24d) |
| −0.09140000000000 | −0.69110000000000 | −0.75300000000000 | O | (24d) |
| 0.40860000000000 | 1.19110000000000 | 0.75300000000000 | O | (24d) |
| 0.59140000000000 | 0.69110000000000 | −0.25300000000000 | O | (24d) |
| 0.09140000000000 | −0.19110000000000 | 1.25300000000000 | O | (24d) |
| −0.75300000000000 | −0.09140000000000 | −0.69110000000000 | O | (24d) |
| 0.75300000000000 | 0.40860000000000 | 1.19110000000000 | O | (24d) |
| −0.25300000000000 | 0.59140000000000 | 0.69110000000000 | O | (24d) |
| 1.25300000000000 | 0.09140000000000 | −0.19110000000000 | O | (24d) |
| 0.85210000000000 | −0.01900000000000 | 0.01850000000000 | O | (24d) |
| −0.35210000000000 | 0.01900000000000 | 0.51850000000000 | O | (24d) |
| −0.85210000000000 | 0.48100000000000 | 0.48150000000000 | O | (24d) |
| 1.35210000000000 | 0.51900000000000 | −0.01850000000000 | O | (24d) |
| 0.01850000000000 | 0.85210000000000 | −0.01900000000000 | O | (24d) |
| 0.51850000000000 | −0.35210000000000 | 0.01900000000000 | O | (24d) |
| 0.48150000000000 | −0.85210000000000 | 0.48100000000000 | O | (24d) |
| −0.01850000000000 | 1.35210000000000 | 0.51900000000000 | O | (24d) |
| −0.01900000000000 | 0.01850000000000 | 0.85210000000000 | O | (24d) |
| 0.01900000000000 | 0.51850000000000 | −0.35210000000000 | O | (24d) |
| 0.48100000000000 | 0.48150000000000 | −0.85210000000000 | O | (24d) |
| 0.51900000000000 | −0.01850000000000 | 1.35210000000000 | O | (24d) |
| −0.85210000000000 | 0.01900000000000 | −0.01850000000000 | O | (24d) |
| 1.35210000000000 | −0.01900000000000 | 0.48150000000000 | O | (24d) |
| 0.85210000000000 | 0.51900000000000 | 0.51850000000000 | O | (24d) |
| −0.35210000000000 | 0.48100000000000 | 0.01850000000000 | O | (24d) |
| −0.01850000000000 | −0.85210000000000 | 0.01900000000000 | O | (24d) |
| 0.48150000000000 | 1.35210000000000 | −0.01900000000000 | O | (24d) |
| 0.51850000000000 | 0.85210000000000 | 0.51900000000000 | O | (24d) |
| 0.01850000000000 | −0.35210000000000 | 0.48100000000000 | O | (24d) |
| 0.01900000000000 | −0.01850000000000 | −0.85210000000000 | O | (24d) |
| −0.01900000000000 | 0.48150000000000 | 1.35210000000000 | O | (24d) |
| 0.51900000000000 | 0.51850000000000 | 0.85210000000000 | O | (24d) |
| 0.48100000000000 | 0.01850000000000 | −0.35210000000000 | O | (24d) |
| 0.04190000000000 | 0.13730000000000 | 0.29750000000000 | O | (24d) |
| 0.45810000000000 | −0.13730000000000 | 0.79750000000000 | O | (24d) |
| −0.04190000000000 | 0.63730000000000 | 0.20250000000000 | O | (24d) |
| 0.54190000000000 | 0.36270000000000 | −0.29750000000000 | O | (24d) |
| 0.29750000000000 | 0.04190000000000 | 0.13730000000000 | O | (24d) |
| 0.79750000000000 | 0.45810000000000 | −0.13730000000000 | O | (24d) |
| 0.20250000000000 | −0.04190000000000 | 0.63730000000000 | O | (24d) |
| −0.29750000000000 | 0.54190000000000 | 0.36270000000000 | O | (24d) |
| 0.13730000000000 | 0.29750000000000 | 0.04190000000000 | O | (24d) |
| −0.13730000000000 | 0.79750000000000 | 0.45810000000000 | O | (24d) |
| 0.63730000000000 | 0.20250000000000 | −0.04190000000000 | O | (24d) |
| 0.36270000000000 | −0.29750000000000 | 0.54190000000000 | O | (24d) |
| −0.04190000000000 | −0.13730000000000 | −0.29750000000000 | O | (24d) |
| 0.54190000000000 | 0.13730000000000 | 0.20250000000000 | O | (24d) |
| 0.04190000000000 | 0.36270000000000 | 0.79750000000000 | O | (24d) |
| 0.45810000000000 | 0.63730000000000 | 0.29750000000000 | O | (24d) |
| −0.29750000000000 | −0.04190000000000 | −0.13730000000000 | O | (24d) |
| 0.20250000000000 | 0.54190000000000 | 0.13730000000000 | O | (24d) |
| 0.79750000000000 | 0.04190000000000 | 0.36270000000000 | O | (24d) |
| 0.29750000000000 | 0.45810000000000 | 0.63730000000000 | O | (24d) |
| −0.13730000000000 | −0.29750000000000 | −0.04190000000000 | O | (24d) |
| 0.13730000000000 | 0.20250000000000 | 0.54190000000000 | O | (24d) |
| 0.36270000000000 | 0.79750000000000 | 0.04190000000000 | O | (24d) |
| 0.63730000000000 | 0.29750000000000 | 0.45810000000000 | O | (24d) |
| 0.30220000000000 | 0.30220000000000 | 0.30220000000000 | S | (8c) |
| 0.19780000000000 | −0.30220000000000 | 0.80220000000000 | S | (8c) |
| −0.30220000000000 | 0.80220000000000 | 0.19780000000000 | S | (8c) |
| 0.80220000000000 | 0.19780000000000 | −0.30220000000000 | S | (8c) |
| −0.30220000000000 | −0.30220000000000 | −0.30220000000000 | S | (8c) |
| 0.80220000000000 | 0.30220000000000 | 0.19780000000000 | S | (8c) |
| 0.30220000000000 | 0.19780000000000 | 0.80220000000000 | S | (8c) |
| 0.19780000000000 | 0.80220000000000 | 0.30220000000000 | S | (8c) |

Maghemite ($\gamma$-$Fe_2O_3$, $D5_7$): A2B3_cP60_212_bcd_ace - CIF

```
# CIF file
data_findsym-output
_audit_creation_method FINDSYM

_chemical_name_mineral 'Maghemite or $\gamma$-corundum'
_chemical_formula_sum 'Fe2 O3'

loop_
_publ_author_name
 'J. Thewlis'
_journal_name_full_name
;
 Philosophical Magazine
;
_journal_volume 12
_journal_year 1931
_journal_page_first 1089
_journal_page_last 1106
_publ_Section_title
;
 The structure of ferromagnetic ferric oxide
;

# Found in Strukturberich Band II 1928-1932, 1937

_aflow_title 'Maghemite ($\gamma$-Fe$_{2}$O$_{3}$, $D5_{7}$) Structure'
_aflow_proto 'A2B3_cP60_212_bcd_ace'
_aflow_params 'a,x_{3},x_{4},y_{5},x_{6},y_{6},z_{6}'
_aflow_params_values '8.4,0.0,0.871,0.375,0.378,0.129,0.629'
_aflow_Strukturbericht '$D5_{7}$'
_aflow_Pearson 'cP60'

_symmetry_space_group_name_H-M "P 43 3 2"
_symmetry_Int_Tables_number 212

_cell_length_a     8.40000
_cell_length_b     8.40000
_cell_length_c     8.40000
_cell_angle_alpha 90.00000
_cell_angle_beta  90.00000
_cell_angle_gamma 90.00000

loop_
_space_group_symop_id
_space_group_symop_operation_xyz
1 x,y,z
2 x+1/2,-y+1/2,-z
3 -x,y+1/2,-z+1/2
4 -x+1/2,-y,z+1/2
5 y,z,x
6 y+1/2,-z+1/2,-x
7 -y,z+1/2,-x+1/2
8 -y+1/2,-z,x+1/2
9 z,x,y
10 z+1/2,-x+1/2,-y
11 -z,x+1/2,-y+1/2
12 -z+1/2,-x,y+1/2
13 -y+1/4,-x+1/4,-z+1/4
14 -y+3/4,x+1/4,z+3/4
15 y+3/4,-x+3/4,z+1/4
16 y+1/4,x+3/4,-z+3/4
17 -x+1/4,-z+1/4,-y+1/4
18 -x+3/4,z+1/4,y+3/4
19 x+3/4,-z+3/4,y+1/4
20 x+1/4,z+3/4,-y+3/4
21 -z+1/4,-y+1/4,-x+1/4
22 -z+3/4,y+1/4,x+3/4
23 z+3/4,-y+3/4,x+1/4
24 z+1/4,y+3/4,-x+3/4

loop_
_atom_site_label
_atom_site_type_symbol
_atom_site_symmetry_multiplicity
_atom_site_Wyckoff_label
_atom_site_fract_x
_atom_site_fract_y
_atom_site_fract_z
_atom_site_occupancy
O1 O 4 a 0.12500 0.12500 0.12500 1.00000
Fe1 Fe 4 b 0.62500 0.62500 0.62500 1.00000
Fe2 Fe 8 c 0.00000 0.00000 0.00000 1.00000
O2 O 8 c 0.87100 0.87100 0.87100 1.00000
Fe3 Fe 12 d 0.12500 0.37500 0.87500 1.00000
O3 O 24 e 0.37800 0.12900 0.62900 1.00000
```

Maghemite ($\gamma$-$Fe_2O_3$, $D5_7$): A2B3_cP60_212_bcd_ace - POSCAR

```
A2B3_cP60_212_bcd_ace & a,x3,x4,y5,x6,y6,z6 --params=8.4,0.0,0.871,0.375
    ↪ ,0.378,0.129,0.629 & P4_{3}32 O^{6} #212 (abc^2de) & cP60 &
    ↪ $D5_{7}$ & Fe2O3 & Maghemite or $\gamma$-corundum & J. Thewlis,
    ↪ Philos. Mag. 12, 1089-1106 (1931)
   1.00000000000000
   8.40000000000000   0.00000000000000   0.00000000000000
   0.00000000000000   8.40000000000000   0.00000000000000
   0.00000000000000   0.00000000000000   8.40000000000000
    Fe    O
    24    36
Direct
   0.62500000000000   0.62500000000000   0.62500000000000   Fe   (4b)
   0.87500000000000   0.37500000000000   0.12500000000000   Fe   (4b)
   0.37500000000000   0.12500000000000   0.87500000000000   Fe   (4b)
   0.12500000000000   0.87500000000000   0.37500000000000   Fe   (4b)
   0.00000000000000   0.00000000000000   0.00000000000000   Fe   (8c)
   0.50000000000000   0.00000000000000   0.50000000000000   Fe   (8c)
   0.00000000000000   0.50000000000000   0.50000000000000   Fe   (8c)
   0.50000000000000   0.50000000000000   0.00000000000000   Fe   (8c)
   0.25000000000000   0.75000000000000   0.75000000000000   Fe   (8c)
   0.25000000000000   0.25000000000000   0.25000000000000   Fe   (8c)
   0.75000000000000   0.75000000000000   0.25000000000000   Fe   (8c)
   0.75000000000000   0.25000000000000   0.75000000000000   Fe   (8c)
   0.12500000000000   0.37500000000000  -0.12500000000000   Fe  (12d)
   0.37500000000000  -0.37500000000000   0.37500000000000   Fe  (12d)
   0.87500000000000   0.87500000000000   0.62500000000000   Fe  (12d)
   0.62500000000000   0.12500000000000   1.12500000000000   Fe  (12d)
  -0.12500000000000   0.12500000000000   0.37500000000000   Fe  (12d)
   0.37500000000000   0.37500000000000  -0.37500000000000   Fe  (12d)
   0.62500000000000   0.87500000000000   0.87500000000000   Fe  (12d)
   1.12500000000000   0.62500000000000   0.12500000000000   Fe  (12d)
   0.37500000000000  -0.12500000000000   0.12500000000000   Fe  (12d)
  -0.37500000000000   0.37500000000000   0.37500000000000   Fe  (12d)
   0.87500000000000   0.62500000000000   0.87500000000000   Fe  (12d)
   0.12500000000000   1.12500000000000   0.62500000000000   Fe  (12d)
   0.12500000000000   0.12500000000000   0.12500000000000   O   (4a)
   0.37500000000000   0.87500000000000   0.62500000000000   O   (4a)
   0.87500000000000   0.62500000000000   0.37500000000000   O   (4a)
   0.62500000000000   0.37500000000000   0.87500000000000   O   (4a)
   0.87100000000000   0.87100000000000   0.87100000000000   O   (8c)
  -0.37100000000000  -0.87100000000000   1.37100000000000   O   (8c)
  -0.87100000000000   1.37100000000000  -0.37100000000000   O   (8c)
   1.37100000000000  -0.37100000000000  -0.87100000000000   O   (8c)
   1.12100000000000   1.62100000000000  -0.12100000000000   O   (8c)
  -0.62100000000000  -0.62100000000000  -0.62100000000000   O   (8c)
   1.62100000000000  -0.12100000000000   1.12100000000000   O   (8c)
  -0.12100000000000   1.12100000000000   1.62100000000000   O   (8c)
   0.37800000000000   0.12900000000000   0.62900000000000   O  (24e)
   0.12200000000000  -0.12900000000000   1.12900000000000   O  (24e)
  -0.37800000000000   0.62900000000000  -0.12900000000000   O  (24e)
   0.87800000000000   0.37100000000000  -0.62900000000000   O  (24e)
   0.62900000000000   0.37800000000000   0.12900000000000   O  (24e)
   1.12900000000000   0.12200000000000  -0.12900000000000   O  (24e)
  -0.12900000000000  -0.37800000000000   0.62900000000000   O  (24e)
  -0.62900000000000   0.87800000000000   0.37100000000000   O  (24e)
   0.12900000000000   0.62900000000000   0.37800000000000   O  (24e)
  -0.12900000000000   1.12900000000000   0.12200000000000   O  (24e)
   0.62900000000000  -0.12900000000000  -0.37800000000000   O  (24e)
   0.37100000000000  -0.62900000000000   0.87800000000000   O  (24e)
   0.37900000000000   1.12800000000000   0.12100000000000   O  (24e)
   0.12100000000000  -0.12800000000000  -0.37900000000000   O  (24e)
   0.87900000000000   0.37200000000000   0.87900000000000   O  (24e)
   0.62100000000000   0.62800000000000   1.37900000000000   O  (24e)
   0.62800000000000   1.37900000000000   0.62100000000000   O  (24e)
   0.37200000000000   0.87900000000000   0.87900000000000   O  (24e)
  -0.12800000000000  -0.37900000000000   0.12100000000000   O  (24e)
   1.12800000000000   0.12100000000000   0.37900000000000   O  (24e)
   0.87900000000000   0.87900000000000   0.37200000000000   O  (24e)
   1.37900000000000   0.62100000000000   0.62800000000000   O  (24e)
   0.12100000000000   0.37900000000000   1.12800000000000   O  (24e)
  -0.37900000000000   0.12100000000000  -0.12800000000000   O  (24e)
```

$Al_2Mo_3C$: A2BC3_cP24_213_c_a_d - CIF

```
# CIF file
data_findsym-output
_audit_creation_method FINDSYM

_chemical_name_mineral 'Al2CMo3'
_chemical_formula_sum 'Al2 C Mo3'

loop_
_publ_author_name
 'W. Jeitschko'
 'H. Nowotny'
 'F. Benesovsky'
_journal_name_full_name
;
 Monatshefte f{\"u}r Chemie und verwandte Teile anderer Wissenschaften
;
_journal_volume 94
_journal_year 1963
_journal_page_first 247
_journal_page_last 251
_publ_Section_title
;
 Ein Beitrag zum Dreistoff: Molybd\"{a}n-Aluminium-Kohlenstoff
;

# Found in Superconductivity induced by Mg deficiency in
    ↪ non-centrosymmetric phosphide Mg$_{2}$Rh$_{3}$P, {arXiv:
    ↪ 1910.06523 [cond-mat.supr-con]},

_aflow_title 'Al$_{2}$Mo$_{3}$C Structure'
_aflow_proto 'A2BC3_cP24_213_c_a_d'
_aflow_params 'a,x_{2},y_{3}'
_aflow_params_values '6.84,0.061,0.206'
_aflow_Strukturbericht 'None'
_aflow_Pearson 'cP24'

_symmetry_space_group_name_H-M "P 41 3 2"
_symmetry_Int_Tables_number 213

_cell_length_a     6.84000
_cell_length_b     6.84000
_cell_length_c     6.84000
_cell_angle_alpha 90.00000
_cell_angle_beta  90.00000
_cell_angle_gamma 90.00000
```

```
loop_
_space_group_symop_id
_space_group_symop_operation_xyz
1 x,y,z
2 x+1/2,-y+1/2,-z
3 -x,y+1/2,-z+1/2
4 -x+1/2,-y,z+1/2
5 y,z,x
6 y+1/2,-z+1/2,-x
7 -y,z+1/2,-x+1/2
8 -y+1/2,-z,x+1/2
9 z,x,y
10 z+1/2,-x+1/2,-y
11 -z,x+1/2,-y+1/2
12 -z+1/2,-x,y+1/2
13 -y+3/4,-x+3/4,-z+3/4
14 -y+1/4,x+3/4,z+1/4
15 y+1/4,-x+1/4,z+3/4
16 y+3/4,x+1/4,-z+1/4
17 -x+3/4,-z+3/4,-y+3/4
18 -x+1/4,z+3/4,y+1/4
19 x+1/4,-z+1/4,y+3/4
20 x+3/4,z+1/4,-y+1/4
21 -z+3/4,-y+3/4,-x+3/4
22 -z+1/4,y+3/4,x+1/4
23 z+1/4,-y+1/4,x+3/4
24 z+3/4,y+1/4,-x+1/4

loop_
_atom_site_label
_atom_site_type_symbol
_atom_site_symmetry_multiplicity
_atom_site_Wyckoff_label
_atom_site_fract_x
_atom_site_fract_y
_atom_site_fract_z
_atom_site_occupancy
C1 C 4 a 0.37500 0.37500 0.37500 1.00000
Al1 Al 8 c 0.06100 0.06100 0.06100 1.00000
Mo1 Mo 12 d 0.12500 0.20600 0.45600 1.00000
```

$Al_2Mo_3C$: A2BC3_cP24_213_c_a_d - POSCAR

```
A2BC3_cP24_213_c_a_d & a,x2,y3 --params=6.84,0.061,0.206 & P4_{1}32 O^{7
    ↪ } #213 (acd) & cP24 & None & Al2CMo3 & Al2CMo3 & W. Jeitschko
    ↪ and H. Nowotny and F. Benesovsky, Monatsh. Chem. Verw. Teile
    ↪ Anderer Wiss. 94, 247-251 (1963)
   1.00000000000000
   6.84000000000000   0.00000000000000   0.00000000000000
   0.00000000000000   6.84000000000000   0.00000000000000
   0.00000000000000   0.00000000000000   6.84000000000000
    Al     C     Mo
     8     4     12
Direct
   0.06100000000000   0.06100000000000   0.06100000000000   Al   (8c)
   0.43900000000000  -0.06100000000000   0.56100000000000   Al   (8c)
  -0.06100000000000   0.56100000000000   0.43900000000000   Al   (8c)
   0.56100000000000   0.43900000000000  -0.06100000000000   Al   (8c)
   0.81100000000000   0.31100000000000   0.18900000000000   Al   (8c)
   0.68900000000000   0.68900000000000   0.68900000000000   Al   (8c)
   0.31100000000000   0.18900000000000   0.81100000000000   Al   (8c)
   0.18900000000000   0.81100000000000   0.31100000000000   Al   (8c)
   0.37500000000000   0.37500000000000   0.37500000000000   C    (4a)
   0.12500000000000   0.62500000000000   0.87500000000000   C    (4a)
   0.62500000000000   0.87500000000000   0.12500000000000   C    (4a)
   0.87500000000000   0.12500000000000   0.62500000000000   C    (4a)
   0.12500000000000   0.20600000000000   0.45600000000000   Mo   (12d)
   0.37500000000000  -0.20600000000000   0.95600000000000   Mo   (12d)
   0.87500000000000   0.70600000000000   0.04400000000000   Mo   (12d)
   0.62500000000000   0.29400000000000   0.54400000000000   Mo   (12d)
   0.45600000000000   0.12500000000000   0.20600000000000   Mo   (12d)
   0.95600000000000   0.37500000000000  -0.20600000000000   Mo   (12d)
   0.04400000000000   0.87500000000000   0.70600000000000   Mo   (12d)
   0.54400000000000   0.62500000000000   0.29400000000000   Mo   (12d)
   0.20600000000000   0.45600000000000   0.12500000000000   Mo   (12d)
  -0.20600000000000   0.95600000000000   0.37500000000000   Mo   (12d)
   0.70600000000000   0.04400000000000   0.87500000000000   Mo   (12d)
   0.29400000000000   0.54400000000000   0.62500000000000   Mo   (12d)
```

$Mg_3Ru_2$: A3B2_cP20_213_d_c - CIF

```
# CIF file
data_findsym-output
_audit_creation_method FINDSYM

_chemical_name_mineral 'Mg3Ru2'
_chemical_formula_sum 'Mg3 Ru2'

loop_
_publ_author_name
 'R. P\"{o}ttgen'
 'V. Hlukhyy'
 'A. Baranov'
 'Y. Grin'
_journal_name_full_name
;
 Inorganic Chemistry
;
_journal_volume 47
_journal_year 2008
_journal_page_first 6051
_journal_page_last 6055
_publ_Section_title
;
 Crystal Structure and Chemical Bonding of Mg$_{3}$Ru$_{2}$
;

_aflow_title 'Mg$_{3}$Ru$_{2}$ Structure'
_aflow_proto 'A3B2_cP20_213_d_c'
_aflow_params 'a,x_{1},y_{2}'
_aflow_params_values '6.9352,0.07378,0.2051'
_aflow_Strukturbericht 'None'
_aflow_Pearson 'cP20'

_symmetry_space_group_name_H-M "P 41 3 2"
_symmetry_Int_Tables_number 213

_cell_length_a    6.93520
_cell_length_b    6.93520
_cell_length_c    6.93520
_cell_angle_alpha 90.00000
_cell_angle_beta  90.00000
_cell_angle_gamma 90.00000

loop_
_space_group_symop_id
_space_group_symop_operation_xyz
1 x,y,z
2 x+1/2,-y+1/2,-z
3 -x,y+1/2,-z+1/2
4 -x+1/2,-y,z+1/2
5 y,z,x
6 y+1/2,-z+1/2,-x
7 -y,z+1/2,-x+1/2
8 -y+1/2,-z,x+1/2
9 z,x,y
10 z+1/2,-x+1/2,-y
11 -z,x+1/2,-y+1/2
12 -z+1/2,-x,y+1/2
13 -y+3/4,-x+3/4,-z+3/4
14 -y+1/4,x+3/4,z+1/4
15 y+1/4,-x+1/4,z+3/4
16 y+3/4,x+1/4,-z+1/4
17 -x+3/4,-z+3/4,-y+3/4
18 -x+1/4,z+3/4,y+1/4
19 x+1/4,-z+1/4,y+3/4
20 x+3/4,z+1/4,-y+1/4
21 -z+3/4,-y+3/4,-x+3/4
22 -z+1/4,y+3/4,x+1/4
23 z+1/4,-y+1/4,x+3/4
24 z+3/4,y+1/4,-x+1/4

loop_
_atom_site_label
_atom_site_type_symbol
_atom_site_symmetry_multiplicity
_atom_site_Wyckoff_label
_atom_site_fract_x
_atom_site_fract_y
_atom_site_fract_z
_atom_site_occupancy
Ru1 Ru 8 c 0.07378 0.07378 0.07378 1.00000
Mg1 Mg 12 d 0.12500 0.20510 0.45510 1.00000
```

$Mg_3Ru_2$: A3B2_cP20_213_d_c - POSCAR

```
A3B2_cP20_213_d_c & a,x1,y2 --params=6.9352,0.07378,0.2051 & P4_{1}32 O
    ↪ ^{7} #213 (cd) & cP20 & None & Mg3Ru2 & Mg3Ru2 & R. P\"{o}ttgen
    ↪ et al., Inorg. Chem. 47, 6051-6055 (2008)
   1.00000000000000
   6.93520000000000   0.00000000000000   0.00000000000000
   0.00000000000000   6.93520000000000   0.00000000000000
   0.00000000000000   0.00000000000000   6.93520000000000
    Mg     Ru
    12      8
Direct
   0.12500000000000   0.20510000000000   0.45510000000000   Mg   (12d)
   0.37500000000000  -0.20510000000000   0.95510000000000   Mg   (12d)
   0.87500000000000   0.70510000000000   0.04490000000000   Mg   (12d)
   0.62500000000000   0.29490000000000   0.54490000000000   Mg   (12d)
   0.45510000000000   0.12500000000000   0.20510000000000   Mg   (12d)
   0.95510000000000   0.37500000000000  -0.20510000000000   Mg   (12d)
   0.04490000000000   0.87500000000000   0.70510000000000   Mg   (12d)
   0.54490000000000   0.62500000000000   0.29490000000000   Mg   (12d)
   0.20510000000000   0.45510000000000   0.12500000000000   Mg   (12d)
  -0.20510000000000   0.95510000000000   0.37500000000000   Mg   (12d)
   0.70510000000000   0.04490000000000   0.87500000000000   Mg   (12d)
   0.29490000000000   0.54490000000000   0.62500000000000   Mg   (12d)
   0.07378000000000   0.07378000000000   0.07378000000000   Ru   (8c)
   0.42622000000000  -0.07378000000000   0.57378000000000   Ru   (8c)
  -0.07378000000000   0.57378000000000   0.42622000000000   Ru   (8c)
   0.57378000000000   0.42622000000000  -0.07378000000000   Ru   (8c)
   0.82378000000000   0.32378000000000   0.17622000000000   Ru   (8c)
   0.67622000000000   0.67622000000000   0.67622000000000   Ru   (8c)
   0.32378000000000   0.17622000000000   0.82378000000000   Ru   (8c)
   0.17622000000000   0.82378000000000   0.32378000000000   Ru   (8c)
```

Zunyite [$Al_{13}(OH,F)_{18}Si_5O_{20}Cl$, $S0_8$]: A13BC18D20E5_cF228_216_dh_b_fh_2eh_ce - CIF

```
# CIF file
data_findsym-output
_audit_creation_method FINDSYM

_chemical_name_mineral 'Zunyite'
_chemical_formula_sum 'Al13 Cl F18 O20 Si5'

loop_
_publ_author_name
 'W. B. Kamb'
_journal_name_full_name
```

```
;
 Acta Cristallographica
;
_journal_volume 13
_journal_year 1960
_journal_page_first 15
_journal_page_last 24
_publ_Section_title
;
 The Crystal Structure of Zunyite
;

_aflow_title 'Zunyite [Al$_{13}$(OH,F)$_{18}$Si$_{5}$O$_{20}$Cl, $S0_{8}
    ↪ $] Structure'
_aflow_proto 'A13BC18D20E5_cF228_216_dh_b_fh_2eh_ce'
_aflow_params 'a,x_{4},x_{5},x_{6},x_{7},x_{8},z_{8},x_{9},z_{9},x_{10},
    ↪ z_{10}'
_aflow_params_values '13.87,0.825,0.1818,0.1143,0.278,0.0853,0.7667,
    ↪ 0.1793,0.5466,0.1385,0.0003'
_aflow_Strukturbericht '$S0_{8}$'
_aflow_Pearson 'cF228'

_symmetry_space_group_name_H-M "F -4 3 m"
_symmetry_Int_Tables_number 216

_cell_length_a     13.87000
_cell_length_b     13.87000
_cell_length_c     13.87000
_cell_angle_alpha  90.00000
_cell_angle_beta   90.00000
_cell_angle_gamma  90.00000

loop_
_space_group_symop_id
_space_group_symop_operation_xyz
1 x,y,z
2 x,-y,-z
3 -x,y,-z
4 -x,-y,z
5 y,z,x
6 y,-z,-x
7 -y,z,-x
8 -y,-z,x
9 z,x,y
10 z,-x,-y
11 -z,x,-y
12 -z,-x,y
13 y,x,z
14 y,-x,-z
15 -y,x,-z
16 -y,-x,z
17 x,z,y
18 x,-z,-y
19 -x,z,-y
20 -x,-z,y
21 z,y,x
22 z,-y,-x
23 -z,y,-x
24 -z,-y,x
25 x,y+1/2,z+1/2
26 x,-y+1/2,-z+1/2
27 -x,y+1/2,-z+1/2
28 -x,-y+1/2,z+1/2
29 y,z+1/2,x+1/2
30 y,-z+1/2,-x+1/2
31 -y,z+1/2,-x+1/2
32 -y,-z+1/2,x+1/2
33 z,x+1/2,y+1/2
34 z,-x+1/2,-y+1/2
35 -z,x+1/2,-y+1/2
36 -z,-x+1/2,y+1/2
37 y,x+1/2,z+1/2
38 y,-x+1/2,-z+1/2
39 -y,x+1/2,-z+1/2
40 -y,-x+1/2,z+1/2
41 x,z+1/2,y+1/2
42 x,-z+1/2,-y+1/2
43 -x,z+1/2,-y+1/2
44 -x,-z+1/2,y+1/2
45 z,y+1/2,x+1/2
46 z,-y+1/2,-x+1/2
47 -z,y+1/2,-x+1/2
48 -z,-y+1/2,x+1/2
49 x+1/2,y,z+1/2
50 x+1/2,-y,-z+1/2
51 -x+1/2,y,-z+1/2
52 -x+1/2,-y,z+1/2
53 y+1/2,z,x+1/2
54 y+1/2,-z,-x+1/2
55 -y+1/2,z,-x+1/2
56 -y+1/2,-z,x+1/2
57 z+1/2,x,y+1/2
58 z+1/2,-x,-y+1/2
59 -z+1/2,x,-y+1/2
60 -z+1/2,-x,y+1/2
61 y+1/2,x,z+1/2
62 y+1/2,-x,-z+1/2
63 -y+1/2,x,-z+1/2
64 -y+1/2,-x,z+1/2
65 x+1/2,z,y+1/2
66 x+1/2,-z,-y+1/2
67 -x+1/2,z,-y+1/2
68 -x+1/2,-z,y+1/2
69 z+1/2,y,x+1/2
70 z+1/2,-y,-x+1/2
71 -z+1/2,y,-x+1/2
72 -z+1/2,-y,x+1/2
73 x+1/2,y+1/2,z
74 x+1/2,-y+1/2,-z
75 -x+1/2,y+1/2,-z
76 -x+1/2,-y+1/2,z
77 y+1/2,z+1/2,x
78 y+1/2,-z+1/2,-x
79 -y+1/2,z+1/2,-x
80 -y+1/2,-z+1/2,x
81 z+1/2,x+1/2,y
82 z+1/2,-x+1/2,-y
83 -z+1/2,x+1/2,-y
84 -z+1/2,-x+1/2,y
85 y+1/2,x+1/2,z
86 y+1/2,-x+1/2,-z
87 -y+1/2,x+1/2,-z
88 -y+1/2,-x+1/2,z
89 x+1/2,z+1/2,y
90 x+1/2,-z+1/2,-y
91 -x+1/2,z+1/2,-y
92 -x+1/2,-z+1/2,y
93 z+1/2,y+1/2,x
94 z+1/2,-y+1/2,-x
95 -z+1/2,y+1/2,-x
96 -z+1/2,-y+1/2,x

loop_
_atom_site_label
_atom_site_type_symbol
_atom_site_symmetry_multiplicity
_atom_site_Wyckoff_label
_atom_site_fract_x
_atom_site_fract_y
_atom_site_fract_z
_atom_site_occupancy
Cl1 Cl 4 b 0.50000 0.50000 0.50000 1.00000
Si1 Si 4 c 0.25000 0.25000 0.25000 1.00000
Al1 Al 4 d 0.75000 0.75000 0.75000 1.00000
O1 O 16 e 0.82500 0.82500 0.82500 1.00000
O2 O 16 e 0.18180 0.18180 0.18180 1.00000
Si2 Si 16 e 0.11430 0.11430 0.11430 1.00000
F1 F 24 f 0.27800 0.00000 0.00000 1.00000
Al2 Al 48 h 0.08530 0.08530 0.76670 1.00000
F2 F 48 h 0.17930 0.17930 0.54660 1.00000
O3 O 48 h 0.13850 0.13850 0.00030 1.00000
```

Zunyite [$Al_{13}(OH,F)_{18}Si_5O_{20}Cl$, $S0_8$]: A13BC18D20E5_cF228_216_dh_b_fh_2eh_ce - POSCAR

```
A13BC18D20E5_cF228_216_dh_b_fh_2eh_ce & a,x4,x5,x6,x7,x8,z8,x9,z9,x10,
    ↪ z10 --params=13.87,0.825,0.1818,0.1143,0.278,0.0853,0.7667,
    ↪ 0.1793,0.5466,0.1385,0.0003 & F-43m T_{d}^{2} #216 (bcde^3fh^3)
    ↪ & cF228 & $S0_{8}$ & Al13ClF18O20Si5 & Zunyite & W. B. Kamb,
    ↪ Acta Cryst. 13, 15-24 (1960)
   1.00000000000000
   0.00000000000000   6.93500000000000   6.93500000000000
   6.93500000000000   0.00000000000000   6.93500000000000
   6.93500000000000   6.93500000000000   0.00000000000000
    Al    Cl     F     O    Si
    13     1    18    20     5
Direct
   0.75000000000000   0.75000000000000   0.75000000000000   Al   (4d)
   0.76670000000000   0.76670000000000  -0.59610000000000   Al  (48h)
   0.76670000000000   0.76670000000000  -0.93730000000000   Al  (48h)
  -0.59610000000000  -0.93730000000000   0.76670000000000   Al  (48h)
  -0.93730000000000  -0.59610000000000   0.76670000000000   Al  (48h)
  -0.59610000000000   0.76670000000000   0.76670000000000   Al  (48h)
  -0.93730000000000   0.76670000000000   0.76670000000000   Al  (48h)
   0.76670000000000  -0.59610000000000  -0.93730000000000   Al  (48h)
   0.76670000000000  -0.93730000000000  -0.59610000000000   Al  (48h)
   0.76670000000000  -0.59610000000000   0.76670000000000   Al  (48h)
   0.76670000000000  -0.93730000000000   0.76670000000000   Al  (48h)
  -0.93730000000000   0.76670000000000  -0.59610000000000   Al  (48h)
  -0.59610000000000   0.76670000000000  -0.93730000000000   Al  (48h)
   0.50000000000000   0.50000000000000   0.50000000000000   Cl   (4b)
  -0.27800000000000   0.27800000000000   0.27800000000000    F  (24f)
   0.27800000000000  -0.27800000000000  -0.27800000000000    F  (24f)
   0.27800000000000  -0.27800000000000   0.27800000000000    F  (24f)
  -0.27800000000000   0.27800000000000  -0.27800000000000    F  (24f)
   0.27800000000000   0.27800000000000  -0.27800000000000    F  (24f)
  -0.27800000000000  -0.27800000000000   0.27800000000000    F  (24f)
   0.54660000000000   0.54660000000000  -0.18800000000000    F  (48h)
   0.54660000000000   0.54660000000000  -0.90520000000000    F  (48h)
  -0.18800000000000  -0.90520000000000   0.54660000000000    F  (48h)
  -0.90520000000000  -0.18800000000000   0.54660000000000    F  (48h)
  -0.18800000000000   0.54660000000000   0.54660000000000    F  (48h)
  -0.90520000000000   0.54660000000000   0.54660000000000    F  (48h)
   0.54660000000000  -0.18800000000000  -0.90520000000000    F  (48h)
   0.54660000000000  -0.90520000000000  -0.18800000000000    F  (48h)
   0.54660000000000  -0.18800000000000   0.54660000000000    F  (48h)
   0.54660000000000  -0.90520000000000   0.54660000000000    F  (48h)
  -0.90520000000000   0.54660000000000  -0.18800000000000    F  (48h)
  -0.18800000000000   0.54660000000000  -0.90520000000000    F  (48h)
   0.82500000000000   0.82500000000000   0.82500000000000    O  (16e)
   0.82500000000000   0.82500000000000  -2.47500000000000    O  (16e)
   0.82500000000000  -2.47500000000000   0.82500000000000    O  (16e)
  -2.47500000000000   0.82500000000000   0.82500000000000    O  (16e)
   0.18180000000000   0.18180000000000   0.18180000000000    O  (16e)
   0.18180000000000   0.18180000000000  -0.54540000000000    O  (16e)
   0.18180000000000  -0.54540000000000   0.18180000000000    O  (16e)
  -0.54540000000000   0.18180000000000   0.18180000000000    O  (16e)
   0.00030000000000   0.00030000000000   0.27670000000000    O  (48h)
   0.00030000000000   0.00030000000000  -0.27730000000000    O  (48h)
   0.27670000000000  -0.27730000000000   0.00030000000000    O  (48h)
  -0.27730000000000   0.27670000000000   0.00030000000000    O  (48h)
```

```
   0.27670000000000     0.00030000000000     0.00030000000000      O   (48h)
  -0.27730000000000     0.00030000000000     0.00030000000000      O   (48h)
   0.00030000000000     0.27670000000000    -0.27730000000000      O   (48h)
   0.00030000000000    -0.27730000000000     0.27670000000000      O   (48h)
   0.00030000000000     0.27670000000000     0.00030000000000      O   (48h)
   0.00030000000000    -0.27730000000000     0.00030000000000      O   (48h)
  -0.27730000000000     0.00030000000000     0.27670000000000      O   (48h)
   0.27670000000000     0.00030000000000    -0.27730000000000      O   (48h)
   0.25000000000000     0.25000000000000     0.25000000000000     Si    (4c)
   0.11430000000000     0.11430000000000     0.11430000000000     Si   (16e)
   0.11430000000000     0.11430000000000    -0.34290000000000     Si   (16e)
   0.11430000000000    -0.34290000000000     0.11430000000000     Si   (16e)
  -0.34290000000000     0.11430000000000     0.11430000000000     Si   (16e)
```

Murataite $[(Y,Na)_6(Zn,Fe)_5Ti_{12}O_{29}(O,F)_{10}F_4]$: A16B40C12D6E5_cF316_216_eh_e2g2h_h_f_be - CIF

```
# CIF file
data_findsym-output
_audit_creation_method FINDSYM

_chemical_name_mineral 'Murataite'
_chemical_formula_sum 'F16 O40 Ti12 Y6 Zn5'

loop_
_publ_author_name
 'T. S. Ercit'
 'F. C. Hawthorne'
_journal_name_full_name
;
 Canadian Mineralogist
;
_journal_volume 33
_journal_year 1995
_journal_page_first 1233
_journal_page_last 1229
_publ_Section_title
;
 Murataite, A UB$_{12}$ derivative structure with condensed Keggin
     ↪ molecules
;

_aflow_title 'Murataite [(Y,Na)$_{6}$(Zn,Fe)$_{5}$Ti$_{12}$O$_{29}$(O,F)
     ↪ $_{10}$F$_{4}$] Structure'
_aflow_proto 'A16B40C12D6E5_cF316_216_eh_e2g2h_h_f_be'
_aflow_params 'a,x_{2},x_{3},x_{4},x_{5},x_{6},x_{7},x_{8},z_{8},x_{9},
     ↪ z_{9},x_{10},z_{10},x_{11},z_{11}'
_aflow_params_values '14.886,0.079,0.4262,0.1722,0.1812,0.267,0.515,
     ↪ 0.894,-0.085,-0.0729,0.7024,0.6038,0.258,0.8371,0.4913'
_aflow_Strukturbericht 'None'
_aflow_Pearson 'cF316'

_symmetry_space_group_name_H-M "F -4 3 m"
_symmetry_Int_Tables_number 216

_cell_length_a     14.88600
_cell_length_b     14.88600
_cell_length_c     14.88600
_cell_angle_alpha 90.00000
_cell_angle_beta  90.00000
_cell_angle_gamma 90.00000

loop_
_space_group_symop_id
_space_group_symop_operation_xyz
1 x,y,z
2 x,-y,-z
3 -x,y,-z
4 -x,-y,z
5 y,z,x
6 y,-z,-x
7 -y,z,-x
8 -y,-z,x
9 z,x,y
10 z,-x,-y
11 -z,x,-y
12 -z,-x,y
13 y,x,z
14 y,-x,-z
15 -y,x,-z
16 -y,-x,z
17 x,z,y
18 x,-z,-y
19 -x,z,-y
20 -x,-z,y
21 z,y,x
22 z,-y,-x
23 -z,y,-x
24 -z,-y,x
25 x,y+1/2,z+1/2
26 x,-y+1/2,-z+1/2
27 -x,y+1/2,-z+1/2
28 -x,-y+1/2,z+1/2
29 y,z+1/2,x+1/2
30 y,-z+1/2,-x+1/2
31 -y,z+1/2,-x+1/2
32 -y,-z+1/2,x+1/2
33 z,x+1/2,y+1/2
34 z,-x+1/2,-y+1/2
35 -z,x+1/2,-y+1/2
36 -z,-x+1/2,y+1/2
37 y,x+1/2,z+1/2
38 y,-x+1/2,-z+1/2
39 -y,x+1/2,-z+1/2
40 -y,-x+1/2,z+1/2
41 x,z+1/2,y+1/2
42 x,-z+1/2,-y+1/2
43 -x,z+1/2,-y+1/2
44 -x,-z+1/2,y+1/2
45 z,y+1/2,x+1/2
46 z,-y+1/2,-x+1/2
47 -z,y+1/2,-x+1/2
48 -z,-y+1/2,x+1/2
49 x+1/2,y,z+1/2
50 x+1/2,-y,-z+1/2
51 -x+1/2,y,-z+1/2
52 -x+1/2,-y,z+1/2
53 y+1/2,z,x+1/2
54 y+1/2,-z,-x+1/2
55 -y+1/2,z,-x+1/2
56 -y+1/2,-z,x+1/2
57 z+1/2,x,y+1/2
58 z+1/2,-x,-y+1/2
59 -z+1/2,x,-y+1/2
60 -z+1/2,-x,y+1/2
61 y+1/2,x,z+1/2
62 y+1/2,-x,-z+1/2
63 -y+1/2,x,-z+1/2
64 -y+1/2,-x,z+1/2
65 x+1/2,z,y+1/2
66 x+1/2,-z,-y+1/2
67 -x+1/2,z,-y+1/2
68 -x+1/2,-z,y+1/2
69 z+1/2,y,x+1/2
70 z+1/2,-y,-x+1/2
71 -z+1/2,y,-x+1/2
72 -z+1/2,-y,x+1/2
73 x+1/2,y+1/2,z
74 x+1/2,-y+1/2,-z
75 -x+1/2,y+1/2,-z
76 -x+1/2,-y+1/2,z
77 y+1/2,z+1/2,x
78 y+1/2,-z+1/2,-x
79 -y+1/2,z+1/2,-x
80 -y+1/2,-z+1/2,x
81 z+1/2,x+1/2,y
82 z+1/2,-x+1/2,-y
83 -z+1/2,x+1/2,-y
84 -z+1/2,-x+1/2,y
85 y+1/2,x+1/2,z
86 y+1/2,-x+1/2,-z
87 -y+1/2,x+1/2,-z
88 -y+1/2,-x+1/2,z
89 x+1/2,z+1/2,y
90 x+1/2,-z+1/2,-y
91 -x+1/2,z+1/2,-y
92 -x+1/2,-z+1/2,y
93 z+1/2,y+1/2,x
94 z+1/2,-y+1/2,-x
95 -z+1/2,y+1/2,-x
96 -z+1/2,-y+1/2,x

loop_
_atom_site_label
_atom_site_type_symbol
_atom_site_symmetry_multiplicity
_atom_site_Wyckoff_label
_atom_site_fract_x
_atom_site_fract_y
_atom_site_fract_z
_atom_site_occupancy
Zn1 Zn 4 b 0.50000 0.50000 0.50000 1.00000
F1 F 16 e 0.07900 0.07900 0.07900 1.00000
O1 O 16 e 0.42620 0.42620 0.42620 1.00000
Zn2 Zn 16 e 0.17220 0.17220 0.17220 1.00000
Y1 Y 24 f 0.18120 0.00000 0.00000 1.00000
O2 O 24 g 0.26700 0.25000 0.25000 1.00000
O3 O 24 g 0.51500 0.25000 0.25000 1.00000
F2 F 48 h 0.89400 0.89400 -0.08500 1.00000
O4 O 48 h -0.07290 -0.07290 0.70240 1.00000
O5 O 48 h 0.60380 0.60380 0.25800 1.00000
Ti1 Ti 48 h 0.83710 0.83710 0.49130 1.00000
```

Murataite $[(Y,Na)_6(Zn,Fe)_5Ti_{12}O_{29}(O,F)_{10}F_4]$: A16B40C12D6E5_cF316_216_eh_e2g2h_h_f_be - POSCAR

```
A16B40C12D6E5_cF316_216_eh_e2g2h_h_f_be & a,x2,x3,x4,x5,x6,x7,x8,z8,x9,
     ↪ z9,x10,z10,x11,z11 --params=14.886,0.079,0.4262,0.1722,0.1812,
     ↪ 0.267,0.515,0.894,-0.085,-0.0729,0.7024,0.6038,0.258,0.8371,
     ↪ 0.4913 & F-43m T_{d}^{2} #216 (be^3fg^2h^4) & cF316 & None &
     ↪ F16O40Ti12Y6Zn5 & Murataite & T. S. Ercit and F. C. Hawthorne,
     ↪ Can. Mineral. 33, 1233-1229 (1995)
   1.00000000000000
   0.00000000000000     7.44300000000000     7.44300000000000
   7.44300000000000     0.00000000000000     7.44300000000000
   7.44300000000000     7.44300000000000     0.00000000000000
     F       O      Ti       Y      Zn
    16      40      12       6       5
Direct
   0.07900000000000     0.07900000000000     0.07900000000000      F   (16e)
   0.07900000000000     0.07900000000000    -0.23700000000000      F   (16e)
   0.07900000000000    -0.23700000000000     0.07900000000000      F   (16e)
  -0.23700000000000     0.07900000000000     0.07900000000000      F   (16e)
  -0.08500000000000    -0.08500000000000     1.87300000000000      F   (48h)
  -0.08500000000000    -0.08500000000000    -1.70300000000000      F   (48h)
   1.87300000000000    -1.70300000000000    -0.08500000000000      F   (48h)
  -1.70300000000000     1.87300000000000    -0.08500000000000      F   (48h)
   1.87300000000000    -0.08500000000000    -0.08500000000000      F   (48h)
  -1.70300000000000    -0.08500000000000    -0.08500000000000      F   (48h)
  -0.08500000000000     1.87300000000000    -1.70300000000000      F   (48h)
  -0.08500000000000    -1.70300000000000     1.87300000000000      F   (48h)
```

```
  -0.08500000000000   1.87300000000000  -0.08500000000000    F   (48h)
  -0.08500000000000  -1.70300000000000  -0.08500000000000    F   (48h)
  -1.70300000000000  -0.08500000000000   1.87300000000000    F   (48h)
   1.87300000000000  -0.08500000000000  -1.70300000000000    F   (48h)
   0.42620000000000   0.42620000000000   0.42620000000000    O   (16e)
   0.42620000000000   0.42620000000000  -1.27860000000000    O   (16e)
   0.42620000000000  -1.27860000000000   0.42620000000000    O   (16e)
  -1.27860000000000   0.42620000000000   0.42620000000000    O   (16e)
   0.23300000000000   0.26700000000000   0.26700000000000    O   (24g)
   0.26700000000000   0.23300000000000   0.26700000000000    O   (24g)
   0.26700000000000   0.23300000000000   0.26700000000000    O   (24g)
   0.23300000000000   0.26700000000000   0.23300000000000    O   (24g)
   0.26700000000000   0.26700000000000   0.23300000000000    O   (24g)
   0.23300000000000   0.23300000000000   0.26700000000000    O   (24g)
  -0.01500000000000   0.51500000000000   0.51500000000000    O   (24g)
   0.51500000000000  -0.01500000000000  -0.01500000000000    O   (24g)
   0.51500000000000  -0.01500000000000   0.51500000000000    O   (24g)
  -0.01500000000000   0.51500000000000  -0.01500000000000    O   (24g)
   0.51500000000000   0.51500000000000  -0.01500000000000    O   (24g)
  -0.01500000000000  -0.01500000000000   0.51500000000000    O   (24g)
   0.70240000000000   0.70240000000000  -0.84820000000000    O   (48h)
   0.70240000000000   0.70240000000000  -0.55660000000000    O   (48h)
  -0.84820000000000  -0.55660000000000   0.70240000000000    O   (48h)
  -0.55660000000000  -0.84820000000000   0.70240000000000    O   (48h)
  -0.84820000000000   0.70240000000000   0.70240000000000    O   (48h)
  -0.55660000000000   0.70240000000000   0.70240000000000    O   (48h)
   0.70240000000000  -0.84820000000000  -0.55660000000000    O   (48h)
   0.70240000000000  -0.55660000000000  -0.84820000000000    O   (48h)
   0.70240000000000  -0.84820000000000   0.70240000000000    O   (48h)
   0.70240000000000  -0.55660000000000   0.70240000000000    O   (48h)
  -0.55660000000000   0.70240000000000  -0.84820000000000    O   (48h)
  -0.84820000000000   0.70240000000000  -0.55660000000000    O   (48h)
   0.25800000000000   0.25800000000000   0.94960000000000    O   (48h)
   0.25800000000000   0.25800000000000  -1.46560000000000    O   (48h)
   0.94960000000000  -1.46560000000000   0.25800000000000    O   (48h)
  -1.46560000000000   0.94960000000000   0.25800000000000    O   (48h)
   0.94960000000000   0.25800000000000   0.25800000000000    O   (48h)
  -1.46560000000000   0.25800000000000   0.25800000000000    O   (48h)
   0.25800000000000   0.94960000000000  -1.46560000000000    O   (48h)
   0.25800000000000  -1.46560000000000   0.94960000000000    O   (48h)
   0.25800000000000   0.94960000000000   0.25800000000000    O   (48h)
   0.25800000000000  -1.46560000000000   0.25800000000000    O   (48h)
  -1.46560000000000   0.25800000000000   0.94960000000000    O   (48h)
   0.94960000000000   0.25800000000000  -1.46560000000000    O   (48h)
   0.49130000000000   0.49130000000000   1.18290000000000   Ti   (48h)
   0.49130000000000   0.49130000000000  -2.16550000000000   Ti   (48h)
   1.18290000000000  -2.16550000000000   0.49130000000000   Ti   (48h)
  -2.16550000000000   1.18290000000000   0.49130000000000   Ti   (48h)
   1.18290000000000   0.49130000000000   0.49130000000000   Ti   (48h)
  -2.16550000000000   0.49130000000000   0.49130000000000   Ti   (48h)
   0.49130000000000   1.18290000000000  -2.16550000000000   Ti   (48h)
   0.49130000000000  -2.16550000000000   1.18290000000000   Ti   (48h)
   0.49130000000000   1.18290000000000   0.49130000000000   Ti   (48h)
   0.49130000000000  -2.16550000000000   0.49130000000000   Ti   (48h)
  -2.16550000000000   0.49130000000000   1.18290000000000   Ti   (48h)
   1.18290000000000   0.49130000000000  -2.16550000000000   Ti   (48h)
  -0.18120000000000   0.18120000000000   0.18120000000000    Y   (24f)
   0.18120000000000  -0.18120000000000  -0.18120000000000    Y   (24f)
   0.18120000000000  -0.18120000000000   0.18120000000000    Y   (24f)
  -0.18120000000000   0.18120000000000  -0.18120000000000    Y   (24f)
   0.18120000000000   0.18120000000000  -0.18120000000000    Y   (24f)
  -0.18120000000000  -0.18120000000000   0.18120000000000    Y   (24f)
   0.50000000000000   0.50000000000000   0.50000000000000   Zn    (4b)
   0.17220000000000   0.17220000000000   0.17220000000000   Zn   (16e)
   0.17220000000000   0.17220000000000  -0.51660000000000   Zn   (16e)
   0.17220000000000  -0.51660000000000   0.17220000000000   Zn   (16e)
  -0.51660000000000   0.17220000000000   0.17220000000000   Zn   (16e)
```

$Sm_{11}Cd_{45}$: A45B11_cF448_216_bd4efg5h_ac2eh - CIF

```
# CIF file
data_findsym-output
_audit_creation_method FINDSYM

_chemical_name_mineral 'Cd45Sm11'
_chemical_formula_sum 'Cd45 Sm11'

loop_
_publ_author_name
 'M. L. Fornasini'
 'B. Chabot'
 'E. Parth\\'{e}'
_journal_name_full_name
;
 Acta Crystallographica Section B: Structural Science
;
_journal_volume 34
_journal_year 1978
_journal_page_first 2093
_journal_page_last 2099
_publ_Section_title
;
 The crystal structure of Sm$_{11}$Cd$_{45}$ with $\gamma$-brass and $\
      ↪ alpha$-Mn clusters
;

_aflow_title 'Sm$_{11}$Cd$_{45}$ Structure'
_aflow_proto 'A45B11_cF448_216_bd4efg5h_ac2eh'
_aflow_params 'a,x_{5},x_{6},x_{7},x_{8},x_{9},x_{10},x_{11},x_{12},x_{
      ↪ 13},z_{13},x_{14},z_{14},x_{15},z_{15},x_{16},z_{16},x_{17},z_{
      ↪ 17},x_{18},z_{18}'
_aflow_params_values '21.699,0.0834,0.9126,0.1636,0.8297,0.4059,0.6618,
      ↪ 0.1573,0.0895,0.2958,0.3904,0.4377,0.2627,0.5455,0.6403,0.6728,
      ↪ 0.5123,0.9161,0.7637,0.1735,0.0142'
_aflow_Strukturbericht 'None'
_aflow_Pearson 'cF448'

_symmetry_space_group_name_H-M "F -4 3 m"
_symmetry_Int_Tables_number 216

_cell_length_a      21.69900
_cell_length_b      21.69900
_cell_length_c      21.69900
_cell_angle_alpha   90.00000
_cell_angle_beta    90.00000
_cell_angle_gamma   90.00000

loop_
_space_group_symop_id
_space_group_symop_operation_xyz
1 x,y,z
2 x,-y,-z
3 -x,y,-z
4 -x,-y,z
5 y,z,x
6 y,-z,-x
7 -y,z,-x
8 -y,-z,x
9 z,x,y
10 z,-x,-y
11 -z,x,-y
12 -z,-x,y
13 y,x,z
14 y,-x,-z
15 -y,x,-z
16 -y,-x,z
17 x,z,y
18 x,-z,-y
19 -x,z,-y
20 -x,-z,y
21 z,y,x
22 z,-y,-x
23 -z,y,-x
24 -z,-y,x
25 x,y+1/2,z+1/2
26 x,-y+1/2,-z+1/2
27 -x,y+1/2,-z+1/2
28 -x,-y+1/2,z+1/2
29 y,z+1/2,x+1/2
30 y,-z+1/2,-x+1/2
31 -y,z+1/2,-x+1/2
32 -y,-z+1/2,x+1/2
33 z,x+1/2,y+1/2
34 z,-x+1/2,-y+1/2
35 -z,x+1/2,-y+1/2
36 -z,-x+1/2,y+1/2
37 y,x+1/2,z+1/2
38 y,-x+1/2,-z+1/2
39 -y,x+1/2,-z+1/2
40 -y,-x+1/2,z+1/2
41 x,z+1/2,y+1/2
42 x,-z+1/2,-y+1/2
43 -x,z+1/2,-y+1/2
44 -x,-z+1/2,y+1/2
45 z,y+1/2,x+1/2
46 z,-y+1/2,-x+1/2
47 -z,y+1/2,-x+1/2
48 -z,-y+1/2,x+1/2
49 x+1/2,y,z+1/2
50 x+1/2,-y,-z+1/2
51 -x+1/2,y,-z+1/2
52 -x+1/2,-y,z+1/2
53 y+1/2,z,x+1/2
54 y+1/2,-z,-x+1/2
55 -y+1/2,z,-x+1/2
56 -y+1/2,-z,x+1/2
57 z+1/2,x,y+1/2
58 z+1/2,-x,-y+1/2
59 -z+1/2,x,-y+1/2
60 -z+1/2,-x,y+1/2
61 y+1/2,x,z+1/2
62 y+1/2,-x,-z+1/2
63 -y+1/2,x,-z+1/2
64 -y+1/2,-x,z+1/2
65 x+1/2,z,y+1/2
66 x+1/2,-z,-y+1/2
67 -x+1/2,z,-y+1/2
68 -x+1/2,-z,y+1/2
69 z+1/2,y,x+1/2
70 z+1/2,-y,-x+1/2
71 -z+1/2,y,-x+1/2
72 -z+1/2,-y,x+1/2
73 x+1/2,y+1/2,z
74 x+1/2,-y+1/2,-z
75 -x+1/2,y+1/2,-z
76 -x+1/2,-y+1/2,z
77 y+1/2,z+1/2,x
78 y+1/2,-z+1/2,-x
79 -y+1/2,z+1/2,-x
80 -y+1/2,-z+1/2,x
81 z+1/2,x+1/2,y
82 z+1/2,-x+1/2,-y
83 -z+1/2,x+1/2,-y
84 -z+1/2,-x+1/2,y
85 y+1/2,x+1/2,z
86 y+1/2,-x+1/2,-z
87 -y+1/2,x+1/2,-z
88 -y+1/2,-x+1/2,z
89 x+1/2,z+1/2,y
90 x+1/2,-z+1/2,-y
```

```
91 -x+1/2,z+1/2,-y
92 -x+1/2,-z+1/2,y
93 z+1/2,y+1/2,x
94 z+1/2,-y+1/2,-x
95 -z+1/2,y+1/2,-x
96 -z+1/2,-y+1/2,x

loop_
_atom_site_label
_atom_site_type_symbol
_atom_site_symmetry_multiplicity
_atom_site_Wyckoff_label
_atom_site_fract_x
_atom_site_fract_y
_atom_site_fract_z
_atom_site_occupancy
Sm1 Sm 4 a 0.00000 0.00000 0.00000 1.00000
Cd1 Cd 4 b 0.50000 0.50000 0.50000 1.00000
Sm2 Sm 4 c 0.25000 0.25000 0.25000 1.00000
Cd2 Cd 4 d 0.75000 0.75000 0.75000 1.00000
Cd3 Cd 16 e 0.08340 0.08340 0.08340 1.00000
Cd4 Cd 16 e 0.91260 0.91260 0.91260 1.00000
Cd5 Cd 16 e 0.16360 0.16360 0.16360 1.00000
Cd6 Cd 16 e 0.82970 0.82970 0.82970 1.00000
Sm3 Sm 16 e 0.40590 0.40590 0.40590 1.00000
Sm4 Sm 16 e 0.66180 0.66180 0.66180 1.00000
Cd7 Cd 24 f 0.15730 0.00000 0.00000 1.00000
Cd8 Cd 24 g 0.08950 0.25000 0.25000 1.00000
Cd9 Cd 48 h 0.29580 0.29580 0.39040 1.00000
Cd10 Cd 48 h 0.43770 0.43770 0.26270 1.00000
Cd11 Cd 48 h 0.54550 0.54550 0.64030 1.00000
Cd12 Cd 48 h 0.67280 0.67280 0.51230 1.00000
Cd13 Cd 48 h 0.91610 0.91610 0.76370 1.00000
Sm5 Sm 48 h 0.17350 0.17350 0.01420 1.00000
```

$Sm_{11}Cd_{45}$: A45B11_cF448_216_bd4efg5h_ac2eh - POSCAR

```
A45B11_cF448_216_bd4efg5h_ac2eh & a,x5,x6,x7,x8,x9,x10,x11,x12,x13,z13,
    ↪ x14,z14,x15,z15,x16,z16,x17,z17,x18,z18 --params=21.699,0.0834,
    ↪ 0.9126,0.1636,0.8297,0.4059,0.6618,0.1573,0.0895,0.2958,0.3904,
    ↪ 0.4377,0.2627,0.5455,0.6403,0.6728,0.5123,0.9161,0.7637,0.1735,
    ↪ 0.0142 & F-43m T_{d}^{2} #216 (abcde^6fgh^6) & cF448 & None &
    ↪ Cd45Sm11 & Cd45Sm11 & M. L. Fornasini and B. Chabot and E.
    ↪ Parth\'{e}, Acta Crystallogr. Sect. B Struct. Sci. 34,
    ↪ 2093-2099 (1978)
   1.00000000000000
   0.00000000000000  10.84950000000000  10.84950000000000
  10.84950000000000   0.00000000000000  10.84950000000000
  10.84950000000000  10.84950000000000   0.00000000000000
    Cd      Sm
    90      22
Direct
   0.50000000000000   0.50000000000000   0.50000000000000   Cd    (4b)
   0.75000000000000   0.75000000000000   0.75000000000000   Cd    (4d)
   0.08340000000000   0.08340000000000   0.08340000000000   Cd   (16e)
   0.08340000000000   0.08340000000000  -0.25020000000000   Cd   (16e)
   0.08340000000000  -0.25020000000000   0.08340000000000   Cd   (16e)
  -0.25020000000000   0.08340000000000   0.08340000000000   Cd   (16e)
   0.91260000000000   0.91260000000000   0.91260000000000   Cd   (16e)
   0.91260000000000   0.91260000000000  -2.73780000000000   Cd   (16e)
   0.91260000000000  -2.73780000000000   0.91260000000000   Cd   (16e)
  -2.73780000000000   0.91260000000000   0.91260000000000   Cd   (16e)
   0.16360000000000   0.16360000000000   0.16360000000000   Cd   (16e)
   0.16360000000000   0.16360000000000  -0.49080000000000   Cd   (16e)
   0.16360000000000  -0.49080000000000   0.16360000000000   Cd   (16e)
  -0.49080000000000   0.16360000000000   0.16360000000000   Cd   (16e)
   0.82970000000000   0.82970000000000   0.82970000000000   Cd   (16e)
   0.82970000000000   0.82970000000000  -2.48910000000000   Cd   (16e)
   0.82970000000000  -2.48910000000000   0.82970000000000   Cd   (16e)
  -2.48910000000000   0.82970000000000   0.82970000000000   Cd   (16e)
  -0.15730000000000   0.15730000000000   0.15730000000000   Cd   (24f)
   0.15730000000000  -0.15730000000000  -0.15730000000000   Cd   (24f)
   0.15730000000000  -0.15730000000000   0.15730000000000   Cd   (24f)
  -0.15730000000000   0.15730000000000  -0.15730000000000   Cd   (24f)
   0.15730000000000   0.15730000000000  -0.15730000000000   Cd   (24f)
  -0.15730000000000  -0.15730000000000   0.15730000000000   Cd   (24f)
   0.41050000000000   0.08950000000000   0.08950000000000   Cd   (24g)
   0.08950000000000   0.41050000000000   0.41050000000000   Cd   (24g)
   0.08950000000000   0.41050000000000   0.08950000000000   Cd   (24g)
   0.41050000000000   0.08950000000000   0.41050000000000   Cd   (24g)
   0.08950000000000   0.08950000000000   0.41050000000000   Cd   (24g)
   0.41050000000000   0.41050000000000   0.08950000000000   Cd   (24g)
   0.39040000000000   0.39040000000000   0.20120000000000   Cd   (48h)
   0.39040000000000   0.39040000000000  -0.98200000000000   Cd   (48h)
   0.20120000000000  -0.98200000000000   0.39040000000000   Cd   (48h)
  -0.98200000000000   0.20120000000000   0.39040000000000   Cd   (48h)
   0.20120000000000   0.39040000000000   0.39040000000000   Cd   (48h)
  -0.98200000000000   0.39040000000000   0.39040000000000   Cd   (48h)
   0.39040000000000   0.20120000000000  -0.98200000000000   Cd   (48h)
   0.39040000000000  -0.98200000000000   0.20120000000000   Cd   (48h)
   0.39040000000000   0.20120000000000   0.39040000000000   Cd   (48h)
   0.39040000000000  -0.98200000000000   0.39040000000000   Cd   (48h)
  -0.98200000000000   0.39040000000000   0.20120000000000   Cd   (48h)
   0.20120000000000   0.39040000000000  -0.98200000000000   Cd   (48h)
   0.26270000000000   0.26270000000000   0.61270000000000   Cd   (48h)
   0.26270000000000   0.26270000000000  -1.13810000000000   Cd   (48h)
   0.61270000000000  -1.13810000000000   0.26270000000000   Cd   (48h)
  -1.13810000000000   0.61270000000000   0.26270000000000   Cd   (48h)
   0.61270000000000   0.26270000000000   0.26270000000000   Cd   (48h)
  -1.13810000000000   0.26270000000000   0.26270000000000   Cd   (48h)
   0.26270000000000   0.61270000000000  -1.13810000000000   Cd   (48h)
   0.26270000000000  -1.13810000000000   0.61270000000000   Cd   (48h)
   0.26270000000000   0.61270000000000   0.26270000000000   Cd   (48h)
   0.26270000000000  -1.13810000000000   0.26270000000000   Cd   (48h)
  -1.13810000000000   0.26270000000000   0.61270000000000   Cd   (48h)
   0.61270000000000   0.26270000000000  -1.13810000000000   Cd   (48h)
   0.64030000000000   0.64030000000000   0.45070000000000   Cd   (48h)
   0.64030000000000   0.64030000000000  -1.73130000000000   Cd   (48h)
   0.45070000000000  -1.73130000000000   0.64030000000000   Cd   (48h)
  -1.73130000000000   0.45070000000000   0.64030000000000   Cd   (48h)
   0.45070000000000   0.64030000000000   0.64030000000000   Cd   (48h)
  -1.73130000000000   0.64030000000000   0.64030000000000   Cd   (48h)
   0.64030000000000   0.45070000000000  -1.73130000000000   Cd   (48h)
   0.64030000000000  -1.73130000000000   0.45070000000000   Cd   (48h)
   0.64030000000000   0.45070000000000   0.64030000000000   Cd   (48h)
   0.64030000000000  -1.73130000000000   0.64030000000000   Cd   (48h)
  -1.73130000000000   0.64030000000000   0.45070000000000   Cd   (48h)
   0.45070000000000   0.64030000000000  -1.73130000000000   Cd   (48h)
   0.51230000000000   0.51230000000000   0.83330000000000   Cd   (48h)
   0.51230000000000   0.51230000000000  -1.85790000000000   Cd   (48h)
   0.83330000000000  -1.85790000000000   0.51230000000000   Cd   (48h)
  -1.85790000000000   0.83330000000000   0.51230000000000   Cd   (48h)
   0.83330000000000   0.51230000000000   0.51230000000000   Cd   (48h)
  -1.85790000000000   0.51230000000000   0.51230000000000   Cd   (48h)
   0.51230000000000   0.83330000000000  -1.85790000000000   Cd   (48h)
   0.51230000000000  -1.85790000000000   0.83330000000000   Cd   (48h)
   0.51230000000000   0.83330000000000   0.51230000000000   Cd   (48h)
   0.51230000000000  -1.85790000000000   0.51230000000000   Cd   (48h)
  -1.85790000000000   0.51230000000000   0.83330000000000   Cd   (48h)
   0.83330000000000   0.51230000000000  -1.85790000000000   Cd   (48h)
   0.76370000000000   0.76370000000000   1.06850000000000   Cd   (48h)
   0.76370000000000   0.76370000000000  -2.59590000000000   Cd   (48h)
   1.06850000000000  -2.59590000000000   0.76370000000000   Cd   (48h)
  -2.59590000000000   1.06850000000000   0.76370000000000   Cd   (48h)
   1.06850000000000   0.76370000000000   0.76370000000000   Cd   (48h)
  -2.59590000000000   0.76370000000000   0.76370000000000   Cd   (48h)
   0.76370000000000   1.06850000000000  -2.59590000000000   Cd   (48h)
   0.76370000000000  -2.59590000000000   1.06850000000000   Cd   (48h)
   0.76370000000000   1.06850000000000   0.76370000000000   Cd   (48h)
   0.76370000000000  -2.59590000000000   0.76370000000000   Cd   (48h)
  -2.59590000000000   0.76370000000000   1.06850000000000   Cd   (48h)
   1.06850000000000   0.76370000000000  -2.59590000000000   Cd   (48h)
   0.00000000000000   0.00000000000000   0.00000000000000   Sm    (4a)
   0.25000000000000   0.25000000000000   0.25000000000000   Sm    (4c)
   0.40590000000000   0.40590000000000   0.40590000000000   Sm   (16e)
   0.40590000000000   0.40590000000000  -1.21770000000000   Sm   (16e)
   0.40590000000000  -1.21770000000000   0.40590000000000   Sm   (16e)
  -1.21770000000000   0.40590000000000   0.40590000000000   Sm   (16e)
   0.66180000000000   0.66180000000000   0.66180000000000   Sm   (16e)
   0.66180000000000   0.66180000000000  -1.98540000000000   Sm   (16e)
   0.66180000000000  -1.98540000000000   0.66180000000000   Sm   (16e)
  -1.98540000000000   0.66180000000000   0.66180000000000   Sm   (16e)
   0.01420000000000   0.01420000000000   0.33280000000000   Sm   (48h)
   0.01420000000000   0.01420000000000  -0.36120000000000   Sm   (48h)
   0.33280000000000  -0.36120000000000   0.01420000000000   Sm   (48h)
  -0.36120000000000   0.33280000000000   0.01420000000000   Sm   (48h)
   0.33280000000000   0.01420000000000   0.01420000000000   Sm   (48h)
  -0.36120000000000   0.01420000000000   0.01420000000000   Sm   (48h)
   0.01420000000000   0.33280000000000  -0.36120000000000   Sm   (48h)
   0.01420000000000  -0.36120000000000   0.33280000000000   Sm   (48h)
   0.01420000000000   0.33280000000000   0.01420000000000   Sm   (48h)
   0.01420000000000  -0.36120000000000   0.01420000000000   Sm   (48h)
  -0.36120000000000   0.01420000000000   0.33280000000000   Sm   (48h)
   0.33280000000000   0.01420000000000  -0.36120000000000   Sm   (48h)
```

$Hg_2TiCu$ Inverse Heusler: AB2C_cF16_216_b_ad_c - CIF

```
# CIF file
data_findsym-output
_audit_creation_method FINDSYM

_chemical_name_mineral 'Inverse heusler'
_chemical_formula_sum 'Cu Hg2 Ti'

loop_
_publ_author_name
 'M. Pu\v{s}elj'
 'Z. Ban'
_journal_name_full_name
;
 Croatica Chemica Acta
;
_journal_volume 41
_journal_year 1969
_journal_page_first 79
_journal_page_last 83
_publ_Section_title
;
 The Crystal Structure of TiCuHg$_{2}$
;

# Found in TiCuHg$_{2}$ (CuHg$_{2}$Ti) Crystal Structure, 2016 Found in
    ↪ TiCuHg$_{2}$ (CuHg$_{2}$Ti) Crystal Structure, {PAULING FILE in
    ↪ : Inorganic Solid Phases, SpringerMaterials (online database),
    ↪ Springer, Heidelberg (ed.)},

_aflow_title 'Hg$_{2}$TiCu Inverse Heusler Structure'
_aflow_proto 'AB2C_cF16_216_b_ad_c'
_aflow_params 'a'
_aflow_params_values '6.15494'
_aflow_Strukturbericht 'None'
_aflow_Pearson 'cF16'

_symmetry_space_group_name_H-M "F -4 3 m"
_symmetry_Int_Tables_number 216

_cell_length_a     6.15494
_cell_length_b     6.15494
_cell_length_c     6.15494
_cell_angle_alpha  90.00000
```

```
_cell_angle_beta   90.00000
_cell_angle_gamma  90.00000

loop_
_space_group_symop_id
_space_group_symop_operation_xyz
1 x,y,z
2 x,-y,-z
3 -x,y,-z
4 -x,-y,z
5 y,z,x
6 y,-z,-x
7 -y,z,-x
8 -y,-z,x
9 z,x,y
10 z,-x,-y
11 -z,x,-y
12 -z,-x,y
13 y,x,z
14 y,-x,-z
15 -y,x,-z
16 -y,-x,z
17 x,z,y
18 x,-z,-y
19 -x,z,-y
20 -x,-z,y
21 z,y,x
22 z,-y,-x
23 -z,y,-x
24 -z,-y,x
25 x,y+1/2,z+1/2
26 x,-y+1/2,-z+1/2
27 -x,y+1/2,-z+1/2
28 -x,-y+1/2,z+1/2
29 y,z+1/2,x+1/2
30 y,-z+1/2,-x+1/2
31 -y,z+1/2,-x+1/2
32 -y,-z+1/2,x+1/2
33 z,x+1/2,y+1/2
34 z,-x+1/2,-y+1/2
35 -z,x+1/2,-y+1/2
36 -z,-x+1/2,y+1/2
37 y,x+1/2,z+1/2
38 y,-x+1/2,-z+1/2
39 -y,x+1/2,-z+1/2
40 -y,-x+1/2,z+1/2
41 x,z+1/2,y+1/2
42 x,-z+1/2,-y+1/2
43 -x,z+1/2,-y+1/2
44 -x,-z+1/2,y+1/2
45 z,y+1/2,x+1/2
46 z,-y+1/2,-x+1/2
47 -z,y+1/2,-x+1/2
48 -z,-y+1/2,x+1/2
49 x+1/2,y,z+1/2
50 x+1/2,-y,-z+1/2
51 -x+1/2,y,-z+1/2
52 -x+1/2,-y,z+1/2
53 y+1/2,z,x+1/2
54 y+1/2,-z,-x+1/2
55 -y+1/2,z,-x+1/2
56 -y+1/2,-z,x+1/2
57 z+1/2,x,y+1/2
58 z+1/2,-x,-y+1/2
59 -z+1/2,x,-y+1/2
60 -z+1/2,-x,y+1/2
61 y+1/2,x,z+1/2
62 y+1/2,-x,-z+1/2
63 -y+1/2,x,-z+1/2
64 -y+1/2,-x,z+1/2
65 x+1/2,z,y+1/2
66 x+1/2,-z,-y+1/2
67 -x+1/2,z,-y+1/2
68 -x+1/2,-z,y+1/2
69 z+1/2,y,x+1/2
70 z+1/2,-y,-x+1/2
71 -z+1/2,y,-x+1/2
72 -z+1/2,-y,x+1/2
73 x+1/2,y+1/2,z
74 x+1/2,-y+1/2,-z
75 -x+1/2,y+1/2,-z
76 -x+1/2,-y+1/2,z
77 y+1/2,z+1/2,x
78 y+1/2,-z+1/2,-x
79 -y+1/2,z+1/2,-x
80 -y+1/2,-z+1/2,x
81 z+1/2,x+1/2,y
82 z+1/2,-x+1/2,-y
83 -z+1/2,x+1/2,-y
84 -z+1/2,-x+1/2,y
85 y+1/2,x+1/2,z
86 y+1/2,-x+1/2,-z
87 -y+1/2,x+1/2,-z
88 -y+1/2,-x+1/2,z
89 x+1/2,z+1/2,y
90 x+1/2,-z+1/2,-y
91 -x+1/2,z+1/2,-y
92 -x+1/2,-z+1/2,y
93 z+1/2,y+1/2,x
94 z+1/2,-y+1/2,-x
95 -z+1/2,y+1/2,-x
96 -z+1/2,-y+1/2,x

loop_
_atom_site_label
_atom_site_type_symbol
_atom_site_symmetry_multiplicity
_atom_site_Wyckoff_label
_atom_site_fract_x
_atom_site_fract_y
_atom_site_fract_z
_atom_site_occupancy
Hg1 Hg 4 a 0.00000 0.00000 0.00000 1.00000
Cu1 Cu 4 b 0.50000 0.50000 0.50000 1.00000
Ti1 Ti 4 c 0.25000 0.25000 0.25000 1.00000
Hg2 Hg 4 d 0.75000 0.75000 0.75000 1.00000
```

$Hg_2$TiCu Inverse Heusler: AB2C_cF16_216_b_ad_c - POSCAR

```
AB2C_cF16_216_b_ad_c & a --params=6.15494 & F-43m T_{d}^{2} #216 (abcd)
    ↪ & cF16 & None & CuHg2Ti & Inverse heusler & M. Pu\v{s}elj and
    ↪ Z. Ban, Croat. Chem. Acta 41, 79-83 (1969)
   1.00000000000000
   0.00000000000000   3.07747000000000   3.07747000000000
   3.07747000000000   0.00000000000000   3.07747000000000
   3.07747000000000   3.07747000000000   0.00000000000000
    Cu     Hg     Ti
     1      2      1
Direct
   0.50000000000000   0.50000000000000   0.50000000000000   Cu   (4b)
   0.00000000000000   0.00000000000000   0.00000000000000   Hg   (4a)
   0.75000000000000   0.75000000000000   0.75000000000000   Hg   (4d)
   0.25000000000000   0.25000000000000   0.25000000000000   Ti   (4c)
```

$GaMo_4S_8$: AB4C8_cF52_216_a_e_2e - CIF

```
# CIF file
data_findsym-output
_audit_creation_method FINDSYM

_chemical_name_mineral 'GaMo4S8'
_chemical_formula_sum 'Ga Mo4 S8'

loop_
_publ_author_name
 'H. {Ben Yaich}'
 'J. C. Jegaden'
 'M. Potel'
 'R. Chevrel'
 'M. Sergent'
 'A. Berton'
 'J. Chaussy'
 'A. K. Rastogi'
 'R. Tournier'
_journal_name_full_name
;
 Journal of Solid State Chemistry
;
_journal_volume 51
_journal_year 1984
_journal_page_first 212
_journal_page_last 217
_publ_Section_title
;
 Nouveaux chalcogenures mixtes GaMo$_{4}$($XX$\')$_{8}$ ($X$ = S, Se, Te
    ↪ ) \`{a} clusters tetraedriques Mo$_{4}$
;

# Found in The American Mineralogist Crystal Structure Database, 2003

_aflow_title 'GaMo$_{4}$S$_{8}$ Structure'
_aflow_proto 'AB4C8_cF52_216_a_e_2e'
_aflow_params 'a,x_{2},x_{3},x_{4}'
_aflow_params_values '9.7294,0.3974,0.6343,0.135'
_aflow_Strukturbericht 'None'
_aflow_Pearson 'cF52'

_symmetry_space_group_name_H-M "F -4 3 m"
_symmetry_Int_Tables_number 216

_cell_length_a      9.72940
_cell_length_b      9.72940
_cell_length_c      9.72940
_cell_angle_alpha  90.00000
_cell_angle_beta   90.00000
_cell_angle_gamma  90.00000

loop_
_space_group_symop_id
_space_group_symop_operation_xyz
1 x,y,z
2 x,-y,-z
3 -x,y,-z
4 -x,-y,z
5 y,z,x
6 y,-z,-x
7 -y,z,-x
8 -y,-z,x
9 z,x,y
10 z,-x,-y
11 -z,x,-y
12 -z,-x,y
13 y,x,z
14 y,-x,-z
15 -y,x,-z
16 -y,-x,z
17 x,z,y
18 x,-z,-y
19 -x,z,-y
20 -x,-z,y
```

```
21 z,y,x
22 z,-y,-x
23 -z,y,-x
24 -z,-y,x
25 x,y+1/2,z+1/2
26 x,-y+1/2,-z+1/2
27 -x,y+1/2,-z+1/2
28 -x,-y+1/2,z+1/2
29 y,z+1/2,x+1/2
30 y,-z+1/2,-x+1/2
31 -y,z+1/2,-x+1/2
32 -y,-z+1/2,x+1/2
33 z,x+1/2,y+1/2
34 z,-x+1/2,-y+1/2
35 -z,x+1/2,-y+1/2
36 -z,-x+1/2,y+1/2
37 y,x+1/2,z+1/2
38 y,-x+1/2,-z+1/2
39 -y,x+1/2,-z+1/2
40 -y,-x+1/2,z+1/2
41 x,z+1/2,y+1/2
42 x,-z+1/2,-y+1/2
43 -x,z+1/2,-y+1/2
44 -x,-z+1/2,y+1/2
45 z,y+1/2,x+1/2
46 z,-y+1/2,-x+1/2
47 -z,y+1/2,-x+1/2
48 -z,-y+1/2,x+1/2
49 x+1/2,y,z+1/2
50 x+1/2,-y,-z+1/2
51 -x+1/2,y,-z+1/2
52 -x+1/2,-y,z+1/2
53 y+1/2,z,x+1/2
54 y+1/2,-z,-x+1/2
55 -y+1/2,z,-x+1/2
56 -y+1/2,-z,x+1/2
57 z+1/2,x,y+1/2
58 z+1/2,-x,-y+1/2
59 -z+1/2,x,-y+1/2
60 -z+1/2,-x,y+1/2
61 y+1/2,x,z+1/2
62 y+1/2,-x,-z+1/2
63 -y+1/2,x,-z+1/2
64 -y+1/2,-x,z+1/2
65 x+1/2,z,y+1/2
66 x+1/2,-z,-y+1/2
67 -x+1/2,z,-y+1/2
68 -x+1/2,-z,y+1/2
69 z+1/2,y,x+1/2
70 z+1/2,-y,-x+1/2
71 -z+1/2,y,-x+1/2
72 -z+1/2,-y,x+1/2
73 x+1/2,y+1/2,z
74 x+1/2,-y+1/2,-z
75 -x+1/2,y+1/2,-z
76 -x+1/2,-y+1/2,z
77 y+1/2,z+1/2,x
78 y+1/2,-z+1/2,-x
79 -y+1/2,z+1/2,-x
80 -y+1/2,-z+1/2,x
81 z+1/2,x+1/2,y
82 z+1/2,-x+1/2,-y
83 -z+1/2,x+1/2,-y
84 -z+1/2,-x+1/2,y
85 y+1/2,x+1/2,z
86 y+1/2,-x+1/2,-z
87 -y+1/2,x+1/2,-z
88 -y+1/2,-x+1/2,z
89 x+1/2,z+1/2,y
90 x+1/2,-z+1/2,-y
91 -x+1/2,z+1/2,-y
92 -x+1/2,-z+1/2,y
93 z+1/2,y+1/2,x
94 z+1/2,-y+1/2,-x
95 -z+1/2,y+1/2,-x
96 -z+1/2,-y+1/2,x

loop_
_atom_site_label
_atom_site_type_symbol
_atom_site_symmetry_multiplicity
_atom_site_Wyckoff_label
_atom_site_fract_x
_atom_site_fract_y
_atom_site_fract_z
_atom_site_occupancy
Ga1 Ga 4 a 0.00000 0.00000 0.00000 1.00000
Mo1 Mo 16 e 0.39740 0.39740 0.39740 1.00000
S1 S 16 e 0.63430 0.63430 0.63430 1.00000
S2 S 16 e 0.13500 0.13500 0.13500 1.00000
```

$GaMo_4S_8$: AB4C8_cF52_216_a_e_2e - POSCAR

```
AB4C8_cF52_216_a_e_2e & a,x2,x3,x4 --params=9.7294,0.3974,0.6343,0.135 &
    ↪ F-43m T_{d}^{2} #216 (ae^3) & cF52 & None & GaMo4S8 & GaMo4S8
    ↪ & H. {Ben Yaich} et al., J. Solid State Chem. 51, 212-217 (1984
    ↪ )
   1.00000000000000
   0.00000000000000   4.86470000000000   4.86470000000000
   4.86470000000000   0.00000000000000   4.86470000000000
   4.86470000000000   4.86470000000000   0.00000000000000
     Ga      Mo       S
      1       4       8
Direct
   0.00000000000000   0.00000000000000   0.00000000000000   Ga   (4a)
   0.39740000000000   0.39740000000000   0.39740000000000   Mo   (16e)
   0.39740000000000   0.39740000000000  -1.19220000000000   Mo   (16e)
   0.39740000000000  -1.19220000000000   0.39740000000000   Mo   (16e)
  -1.19220000000000   0.39740000000000   0.39740000000000   Mo   (16e)
   0.63430000000000   0.63430000000000   0.63430000000000   S    (16e)
   0.63430000000000   0.63430000000000  -1.90290000000000   S    (16e)
   0.63430000000000  -1.90290000000000   0.63430000000000   S    (16e)
  -1.90290000000000   0.63430000000000   0.63430000000000   S    (16e)
   0.13500000000000   0.13500000000000   0.13500000000000   S    (16e)
   0.13500000000000   0.13500000000000  -0.40500000000000   S    (16e)
   0.13500000000000  -0.40500000000000   0.13500000000000   S    (16e)
  -0.40500000000000   0.13500000000000   0.13500000000000   S    (16e)
```

High-Temperature Cubic $KClO_4$ ($H0_5$): ABC4_cF24_216_b_a_e - CIF

```
# CIF file
data_findsym-output
_audit_creation_method FINDSYM

_chemical_name_mineral 'ClKO4'
_chemical_formula_sum 'Cl K O4'

loop_
_publ_author_name
 'K. Hermann'
 'W. Ilge'
_journal_name_full_name
;
 Zeitschrift f{\"u}r Kristallographie - Crystalline Materials
;
_journal_volume 71
_journal_year 1930
_journal_page_first 41
_journal_page_last 66
_publ_Section_title
;
 R\"{o}ntgenographische Strukturerforschung der kubischen Modifikation
    ↪ der Perchlorate
;

# Found in Strukturbericht Band II 1928-1932, 1937

_aflow_title 'High-Temperature Cubic KClO$_{4}$ ($H0_{5}$) Structure'
_aflow_proto 'ABC4_cF24_216_b_a_e'
_aflow_params 'a,x_{3}'
_aflow_params_values '7.5,0.125'
_aflow_Strukturbericht '$H0_{5}$'
_aflow_Pearson 'cF24'

_symmetry_space_group_name_H-M "F -4 3 m"
_symmetry_Int_Tables_number 216

_cell_length_a      7.50000
_cell_length_b      7.50000
_cell_length_c      7.50000
_cell_angle_alpha  90.00000
_cell_angle_beta   90.00000
_cell_angle_gamma  90.00000

loop_
_space_group_symop_id
_space_group_symop_operation_xyz
1 x,y,z
2 x,-y,-z
3 -x,y,-z
4 -x,-y,z
5 y,z,x
6 y,-z,-x
7 -y,z,-x
8 -y,-z,x
9 z,x,y
10 z,-x,-y
11 -z,x,-y
12 -z,-x,y
13 y,x,z
14 y,-x,-z
15 -y,x,-z
16 -y,-x,z
17 x,z,y
18 x,-z,-y
19 -x,z,-y
20 -x,-z,y
21 z,y,x
22 z,-y,-x
23 -z,y,-x
24 -z,-y,x
25 x,y+1/2,z+1/2
26 x,-y+1/2,-z+1/2
27 -x,y+1/2,-z+1/2
28 -x,-y+1/2,z+1/2
29 y,z+1/2,x+1/2
30 y,-z+1/2,-x+1/2
31 -y,z+1/2,-x+1/2
32 -y,-z+1/2,x+1/2
33 z,x+1/2,y+1/2
34 z,-x+1/2,-y+1/2
35 -z,x+1/2,-y+1/2
36 -z,-x+1/2,y+1/2
37 y,x+1/2,z+1/2
38 y,-x+1/2,-z+1/2
39 -y,x+1/2,-z+1/2
40 -y,-x+1/2,z+1/2
41 x,z+1/2,y+1/2
42 x,-z+1/2,-y+1/2
43 -x,z+1/2,-y+1/2
```

```
44 -x,-z+1/2,y+1/2
45 z,y+1/2,x+1/2
46 z,-y+1/2,-x+1/2
47 -z,y+1/2,-x+1/2
48 -z,-y+1/2,x+1/2
49 x+1/2,y,z+1/2
50 x+1/2,-y,-z+1/2
51 -x+1/2,y,-z+1/2
52 -x+1/2,-y,z+1/2
53 y+1/2,z,x+1/2
54 y+1/2,-z,-x+1/2
55 -y+1/2,z,-x+1/2
56 -y+1/2,-z,x+1/2
57 z+1/2,x,y+1/2
58 z+1/2,-x,-y+1/2
59 -z+1/2,x,-y+1/2
60 -z+1/2,-x,y+1/2
61 y+1/2,x,z+1/2
62 y+1/2,-x,-z+1/2
63 -y+1/2,x,-z+1/2
64 -y+1/2,-x,z+1/2
65 x+1/2,z,y+1/2
66 x+1/2,-z,-y+1/2
67 -x+1/2,z,-y+1/2
68 -x+1/2,-z,y+1/2
69 z+1/2,y,x+1/2
70 z+1/2,-y,-x+1/2
71 -z+1/2,y,-x+1/2
72 -z+1/2,-y,x+1/2
73 x+1/2,y+1/2,z
74 x+1/2,-y+1/2,-z
75 -x+1/2,y+1/2,-z
76 -x+1/2,-y+1/2,z
77 y+1/2,z+1/2,x
78 y+1/2,-z+1/2,-x
79 -y+1/2,z+1/2,-x
80 -y+1/2,-z+1/2,x
81 z+1/2,x+1/2,y
82 z+1/2,-x+1/2,-y
83 -z+1/2,x+1/2,-y
84 -z+1/2,-x+1/2,y
85 y+1/2,x+1/2,z
86 y+1/2,-x+1/2,-z
87 -y+1/2,x+1/2,-z
88 -y+1/2,-x+1/2,z
89 x+1/2,z+1/2,y
90 x+1/2,-z+1/2,-y
91 -x+1/2,z+1/2,-y
92 -x+1/2,-z+1/2,y
93 z+1/2,y+1/2,x
94 z+1/2,-y+1/2,-x
95 -z+1/2,y+1/2,-x
96 -z+1/2,-y+1/2,x

loop_
_atom_site_label
_atom_site_type_symbol
_atom_site_symmetry_multiplicity
_atom_site_Wyckoff_label
_atom_site_fract_x
_atom_site_fract_y
_atom_site_fract_z
_atom_site_occupancy
K1 K 4 a 0.00000 0.00000 0.00000 1.00000
Cl1 Cl 4 b 0.50000 0.50000 0.50000 1.00000
O1 O 16 e 0.12500 0.12500 0.12500 1.00000
```

High-Temperature Cubic $KClO_4$ ($H0_5$): ABC4_cF24_216_b_a_e - POSCAR

```
ABC4_cF24_216_b_a_e & a,x3 --params=7.5,0.125 & F-43m T_{d}^{2} #216 (
    ↪ abe) & cF24 & $H0_{5}$ & ClKO4 & ClKO4 & K. Hermann and W. Ilge
    ↪ , Zeitschrift f{"u}r Kristallographie - Crystalline Materials
    ↪ 71, 41-66 (1930)
   1.00000000000000
   0.00000000000000   3.75000000000000   3.75000000000000
   3.75000000000000   0.00000000000000   3.75000000000000
   3.75000000000000   3.75000000000000   0.00000000000000
    Cl     K     O
     1     1     4
Direct
   0.50000000000000   0.50000000000000   0.50000000000000   Cl   (4b)
   0.00000000000000   0.00000000000000   0.00000000000000    K   (4a)
   0.12500000000000   0.12500000000000   0.12500000000000    O  (16e)
   0.12500000000000   0.12500000000000  -0.37500000000000    O  (16e)
   0.12500000000000  -0.37500000000000   0.12500000000000    O  (16e)
  -0.37500000000000   0.12500000000000   0.12500000000000    O  (16e)
```

AlN (cF40): AB_cF40_216_ce_de - CIF

```
# CIF file
data_findsym-output
_audit_creation_method FINDSYM

_chemical_name_mineral 'AlN'
_chemical_formula_sum 'Al N'

loop_
_publ_author_name
 'C. Liu'
 'M. Chen'
 'J. Li'
 'L. Liu'
 'P. Li'
 'M. Ma'
 'C. Shao'
 'J. He'
 'T. Liang'
_journal_name_full_name
;
 Journal of Physics and Chemistry of Solids
;
_journal_volume 130
_journal_year 2019
_journal_page_first 58
_journal_page_last 66
_publ_Section_title
;
 A first-principles study of novel cubic AlN phases
;

_aflow_title 'AlN (cF40) Structure'
_aflow_proto 'AB_cF40_216_ce_de'
_aflow_params 'a,x_{3},x_{4}'
_aflow_params_values '8.1926,0.1137,0.8804'
_aflow_Strukturbericht 'None'
_aflow_Pearson 'cF40'

_symmetry_space_group_name_H-M "F -4 3 m"
_symmetry_Int_Tables_number 216

_cell_length_a     8.19260
_cell_length_b     8.19260
_cell_length_c     8.19260
_cell_angle_alpha 90.00000
_cell_angle_beta  90.00000
_cell_angle_gamma 90.00000

loop_
_space_group_symop_id
_space_group_symop_operation_xyz
1 x,y,z
2 x,-y,-z
3 -x,y,-z
4 -x,-y,z
5 y,z,x
6 y,-z,-x
7 -y,z,-x
8 -y,-z,x
9 z,x,y
10 z,-x,-y
11 -z,x,-y
12 -z,-x,y
13 y,x,z
14 y,-x,-z
15 -y,x,-z
16 -y,-x,z
17 x,z,y
18 x,-z,-y
19 -x,z,-y
20 -x,-z,y
21 z,y,x
22 z,-y,-x
23 -z,y,-x
24 -z,-y,x
25 x,y+1/2,z+1/2
26 x,-y+1/2,-z+1/2
27 -x,y+1/2,-z+1/2
28 -x,-y+1/2,z+1/2
29 y,z+1/2,x+1/2
30 y,-z+1/2,-x+1/2
31 -y,z+1/2,-x+1/2
32 -y,-z+1/2,x+1/2
33 z,x+1/2,y+1/2
34 z,-x+1/2,-y+1/2
35 -z,x+1/2,-y+1/2
36 -z,-x+1/2,y+1/2
37 y,x+1/2,z+1/2
38 y,-x+1/2,-z+1/2
39 -y,x+1/2,-z+1/2
40 -y,-x+1/2,z+1/2
41 x,z+1/2,y+1/2
42 x,-z+1/2,-y+1/2
43 -x,z+1/2,-y+1/2
44 -x,-z+1/2,y+1/2
45 z,y+1/2,x+1/2
46 z,-y+1/2,-x+1/2
47 -z,y+1/2,-x+1/2
48 -z,-y+1/2,x+1/2
49 x+1/2,y,z+1/2
50 x+1/2,-y,-z+1/2
51 -x+1/2,y,-z+1/2
52 -x+1/2,-y,z+1/2
53 y+1/2,z,x+1/2
54 y+1/2,-z,-x+1/2
55 -y+1/2,z,-x+1/2
56 -y+1/2,-z,x+1/2
57 z+1/2,x,y+1/2
58 z+1/2,-x,-y+1/2
59 -z+1/2,x,-y+1/2
60 -z+1/2,-x,y+1/2
61 y+1/2,x,z+1/2
62 y+1/2,-x,-z+1/2
63 -y+1/2,x,-z+1/2
64 -y+1/2,-x,z+1/2
65 x+1/2,z,y+1/2
66 x+1/2,-z,-y+1/2
67 -x+1/2,z,-y+1/2
68 -x+1/2,-z,y+1/2
69 z+1/2,y,x+1/2
70 z+1/2,-y,-x+1/2
```

```
71 -z+1/2,y,-x+1/2
72 -z+1/2,-y,x+1/2
73 x+1/2,y+1/2,z
74 x+1/2,-y+1/2,-z
75 -x+1/2,y+1/2,-z
76 -x+1/2,-y+1/2,z
77 y+1/2,z+1/2,x
78 y+1/2,-z+1/2,-x
79 -y+1/2,z+1/2,-x
80 -y+1/2,-z+1/2,x
81 z+1/2,x+1/2,y
82 z+1/2,-x+1/2,-y
83 -z+1/2,x+1/2,-y
84 -z+1/2,-x+1/2,y
85 y+1/2,x+1/2,z
86 y+1/2,-x+1/2,-z
87 -y+1/2,x+1/2,-z
88 -y+1/2,-x+1/2,z
89 x+1/2,z+1/2,y
90 x+1/2,-z+1/2,-y
91 -x+1/2,z+1/2,-y
92 -x+1/2,-z+1/2,y
93 z+1/2,y+1/2,x
94 z+1/2,-y+1/2,-x
95 -z+1/2,y+1/2,-x
96 -z+1/2,-y+1/2,x

loop_
_atom_site_label
_atom_site_type_symbol
_atom_site_symmetry_multiplicity
_atom_site_Wyckoff_label
_atom_site_fract_x
_atom_site_fract_y
_atom_site_fract_z
_atom_site_occupancy
Al1 Al 4 c 0.75000 0.75000 0.75000 1.00000
N1 N 4 d 0.25000 0.25000 0.25000 1.00000
Al2 Al 16 e 0.11370 0.11370 0.11370 1.00000
N2 N 16 e 0.88040 0.88040 0.88040 1.00000
```

AlN (cF40): AB_cF40_216_ce_de - POSCAR

```
AB_cF40_216_ce_de & a,x3,x4 --params=8.1926,0.1137,0.8804 & F-43m T_{d
    ↪ }^{2} #216 (cde^2) & cF40 & None & AlN & AlN & C. Liu et al.,
    ↪ J. Phys. Chem. Solids 130, 58-66 (2019)
   1.00000000000000
   0.00000000000000   4.09630000000000   4.09630000000000
   4.09630000000000   0.00000000000000   4.09630000000000
   4.09630000000000   4.09630000000000   0.00000000000000
    Al        N
     5        5
Direct
   0.25000000000000   0.25000000000000   0.25000000000000   Al    (4c)
   0.11370000000000   0.11370000000000   0.11370000000000   Al   (16e)
   0.11370000000000   0.11370000000000  -0.34110000000000   Al   (16e)
   0.11370000000000  -0.34110000000000   0.11370000000000   Al   (16e)
  -0.34110000000000   0.11370000000000   0.11370000000000   Al   (16e)
   0.75000000000000   0.75000000000000   0.75000000000000   N     (4d)
   0.88040000000000   0.88040000000000   0.88040000000000   N    (16e)
   0.88040000000000   0.88040000000000  -2.64120000000000   N    (16e)
   0.88040000000000  -2.64120000000000   0.88040000000000   N    (16e)
  -2.64120000000000   0.88040000000000   0.88040000000000   N    (16e)
```

Tennantite ($Cu_{12}As_4S_{13}$): A4B24C13_cI82_217_c_deg_ag - CIF

```
# CIF file
data_findsym-output
_audit_creation_method FINDSYM

_chemical_name_mineral 'Tennantite'
_chemical_formula_sum 'As4 Cu24 S13'

loop_
_publ_author_name
 'A. A. Yaroslavzev'
 'A. V. Mironov'
 'A. N. Kuznetsov'
 'A. P. Dudka'
 'O. N. Khrykina'
_journal_name_full_name
;
 Acta Crystallographica Section B: Structural Science
;
_journal_volume 75
_journal_year 2019
_journal_page_first 634
_journal_page_last 642
_publ_Section_title
;
 Tennantite: multi-temperature crystal structure, phase transition and
    ↪ electronic structure of synthetic Cu$_{12}$As$_{4}$S$_{13}$
;

_aflow_title 'Tennantite (Cu$_{12}$As$_{4}$S$_{13}$) Structure'
_aflow_proto 'A4B24C13_cI82_217_c_deg_ag'
_aflow_params 'a,x_{2},x_{4},x_{5},z_{5},x_{6},z_{6}'
_aflow_params_values '10.1439,0.24271,0.2173,0.0777,0.2123,0.61822,
    ↪ 0.14266'
_aflow_Strukturbericht 'None'
_aflow_Pearson 'cI82'

_symmetry_space_group_name_H-M "I -4 3 m"
_symmetry_Int_Tables_number 217

_cell_length_a     10.14390
_cell_length_b     10.14390
_cell_length_c     10.14390
_cell_angle_alpha  90.00000
_cell_angle_beta   90.00000
_cell_angle_gamma  90.00000

loop_
_space_group_symop_id
_space_group_symop_operation_xyz
1 x,y,z
2 x,-y,-z
3 -x,y,-z
4 -x,-y,z
5 y,z,x
6 y,-z,-x
7 -y,z,-x
8 -y,-z,x
9 z,x,y
10 z,-x,-y
11 -z,x,-y
12 -z,-x,y
13 y,x,z
14 y,-x,-z
15 -y,x,-z
16 -y,-x,z
17 x,z,y
18 x,-z,-y
19 -x,z,-y
20 -x,-z,y
21 z,y,x
22 z,-y,-x
23 -z,y,-x
24 -z,-y,x
25 x+1/2,y+1/2,z+1/2
26 x+1/2,-y+1/2,-z+1/2
27 -x+1/2,y+1/2,-z+1/2
28 -x+1/2,-y+1/2,z+1/2
29 y+1/2,z+1/2,x+1/2
30 y+1/2,-z+1/2,-x+1/2
31 -y+1/2,z+1/2,-x+1/2
32 -y+1/2,-z+1/2,x+1/2
33 z+1/2,x+1/2,y+1/2
34 z+1/2,-x+1/2,-y+1/2
35 -z+1/2,x+1/2,-y+1/2
36 -z+1/2,-x+1/2,y+1/2
37 y+1/2,x+1/2,z+1/2
38 y+1/2,-x+1/2,-z+1/2
39 -y+1/2,x+1/2,-z+1/2
40 -y+1/2,-x+1/2,z+1/2
41 x+1/2,z+1/2,y+1/2
42 x+1/2,-z+1/2,-y+1/2
43 -x+1/2,z+1/2,-y+1/2
44 -x+1/2,-z+1/2,y+1/2
45 z+1/2,y+1/2,x+1/2
46 z+1/2,-y+1/2,-x+1/2
47 -z+1/2,y+1/2,-x+1/2
48 -z+1/2,-y+1/2,x+1/2

loop_
_atom_site_label
_atom_site_type_symbol
_atom_site_symmetry_multiplicity
_atom_site_Wyckoff_label
_atom_site_fract_x
_atom_site_fract_y
_atom_site_fract_z
_atom_site_occupancy
S1 S 2 a 0.00000 0.00000 0.00000 1.00000
As1 As 8 c 0.24271 0.24271 0.24271 1.00000
Cu1 Cu 12 d 0.25000 0.50000 0.00000 1.00000
Cu2 Cu 12 e 0.21730 0.00000 0.00000 0.75800
Cu3 Cu 24 g 0.07770 0.07770 0.21230 0.12100
S2 S 24 g 0.61822 0.61822 0.14266 1.00000
```

Tennantite ($Cu_{12}As_4S_{13}$): A4B24C13_cI82_217_c_deg_ag - POSCAR

```
A4B24C13_cI82_217_c_deg_ag & a,x2,x4,x5,z5,x6,z6 --params=10.1439,
    ↪ 0.24271,0.2173,0.0777,0.2123,0.61822,0.14266 & I-43m T_{d}^{3}
    ↪ #217 (acdeg^2) & cI82 & None & AsCu12S13 & Tennantite & A. A.
    ↪ Yaroslavzev et al., Acta Crystallogr. Sect. B Struct. Sci. 75,
    ↪ 634-642 (2019)
   1.00000000000000
  -5.07195000000000   5.07195000000000   5.07195000000000
   5.07195000000000  -5.07195000000000   5.07195000000000
   5.07195000000000   5.07195000000000  -5.07195000000000
    As       Cu        S
     4       24       13
Direct
   0.48542000000000   0.48542000000000   0.48542000000000   As    (8c)
   0.00000000000000   0.00000000000000  -0.48542000000000   As    (8c)
   0.00000000000000  -0.48542000000000   0.00000000000000   As    (8c)
  -0.48542000000000   0.00000000000000   0.00000000000000   As    (8c)
   0.50000000000000   0.25000000000000   0.75000000000000   Cu   (12d)
   0.50000000000000   0.75000000000000   0.25000000000000   Cu   (12d)
   0.75000000000000   0.50000000000000   0.25000000000000   Cu   (12d)
   0.25000000000000   0.50000000000000   0.75000000000000   Cu   (12d)
   0.25000000000000   0.75000000000000   0.50000000000000   Cu   (12d)
   0.75000000000000   0.25000000000000   0.50000000000000   Cu   (12d)
   0.00000000000000   0.21730000000000   0.21730000000000   Cu   (12e)
   0.00000000000000  -0.21730000000000  -0.21730000000000   Cu   (12e)
   0.21730000000000   0.00000000000000   0.21730000000000   Cu   (12e)
  -0.21730000000000   0.00000000000000  -0.21730000000000   Cu   (12e)
   0.21730000000000   0.21730000000000   0.00000000000000   Cu   (12e)
  -0.21730000000000  -0.21730000000000   0.00000000000000   Cu   (12e)
```

```
    0.29000000000000    0.29000000000000    0.15540000000000    Cu   (24g)
    0.13460000000000    0.13460000000000   -0.15540000000000    Cu   (24g)
   -0.13460000000000   -0.29000000000000    0.00000000000000    Cu   (24g)
   -0.29000000000000   -0.13460000000000    0.00000000000000    Cu   (24g)
    0.15540000000000    0.29000000000000    0.29000000000000    Cu   (24g)
   -0.15540000000000    0.13460000000000    0.13460000000000    Cu   (24g)
    0.00000000000000   -0.13460000000000   -0.29000000000000    Cu   (24g)
    0.00000000000000   -0.29000000000000   -0.13460000000000    Cu   (24g)
    0.29000000000000    0.15540000000000    0.29000000000000    Cu   (24g)
    0.13460000000000   -0.15540000000000    0.13460000000000    Cu   (24g)
   -0.29000000000000    0.00000000000000   -0.13460000000000    Cu   (24g)
   -0.13460000000000    0.00000000000000   -0.29000000000000    Cu   (24g)
    0.00000000000000    0.00000000000000    0.00000000000000    S     (2a)
    0.76088000000000    0.76088000000000    1.23644000000000    S    (24g)
   -0.47556000000000   -0.47556000000000   -1.23644000000000    S    (24g)
    0.47556000000000   -0.76088000000000    0.00000000000000    S    (24g)
   -0.76088000000000    0.47556000000000    0.00000000000000    S    (24g)
    1.23644000000000    0.76088000000000    0.76088000000000    S    (24g)
   -1.23644000000000   -0.47556000000000   -0.47556000000000    S    (24g)
    0.00000000000000    0.47556000000000   -0.76088000000000    S    (24g)
    0.00000000000000   -0.76088000000000    0.47556000000000    S    (24g)
    0.76088000000000    1.23644000000000    0.76088000000000    S    (24g)
   -0.47556000000000   -1.23644000000000   -0.47556000000000    S    (24g)
   -0.76088000000000    0.00000000000000    0.47556000000000    S    (24g)
    0.47556000000000    0.00000000000000   -0.76088000000000    S    (24g)
```

AlN (cI16): AB_cI16_217_c_c - CIF

```
# CIF file
data_findsym-output
_audit_creation_method FINDSYM

_chemical_name_mineral 'AlN'
_chemical_formula_sum 'Al N'

loop_
_publ_author_name
 'C. Liu'
 'M. Chen'
 'J. Li'
 'L. Liu'
 'P. Li'
 'M. Ma'
 'C. Shao'
 'J. He'
 'T. Liang'
_journal_name_full_name
;
 Journal of Physics and Chemistry of Solids
;
_journal_volume 130
_journal_year 2019
_journal_page_first 58
_journal_page_last 66
_publ_Section_title
;
 A first-principles study of novel cubic AlN phases
;

_aflow_title 'AlN (cI16) Structure'
_aflow_proto 'AB_cI16_217_c_c'
_aflow_params 'a,x_{1},x_{2}'
_aflow_params_values '5.9487,0.3409,0.1657'
_aflow_Strukturbericht 'None'
_aflow_Pearson 'cI16'

_symmetry_space_group_name_H-M "I -4 3 m"
_symmetry_Int_Tables_number 217

_cell_length_a     5.94870
_cell_length_b     5.94870
_cell_length_c     5.94870
_cell_angle_alpha 90.00000
_cell_angle_beta  90.00000
_cell_angle_gamma 90.00000

loop_
_space_group_symop_id
_space_group_symop_operation_xyz
1 x,y,z
2 x,-y,-z
3 -x,y,-z
4 -x,-y,z
5 y,z,x
6 y,-z,-x
7 -y,z,-x
8 -y,-z,x
9 z,x,y
10 z,-x,-y
11 -z,x,-y
12 -z,-x,y
13 y,x,z
14 y,-x,-z
15 -y,x,-z
16 -y,-x,z
17 x,z,y
18 x,-z,-y
19 -x,z,-y
20 -x,-z,y
21 z,y,x
22 z,-y,-x
23 -z,y,-x
24 -z,-y,x
25 x+1/2,y+1/2,z+1/2
26 x+1/2,-y+1/2,-z+1/2
27 -x+1/2,y+1/2,-z+1/2
28 -x+1/2,-y+1/2,z+1/2
29 y+1/2,z+1/2,x+1/2
30 y+1/2,-z+1/2,-x+1/2
31 -y+1/2,z+1/2,-x+1/2
32 -y+1/2,-z+1/2,x+1/2
33 z+1/2,x+1/2,y+1/2
34 z+1/2,-x+1/2,-y+1/2
35 -z+1/2,x+1/2,-y+1/2
36 -z+1/2,-x+1/2,y+1/2
37 y+1/2,x+1/2,z+1/2
38 y+1/2,-x+1/2,-z+1/2
39 -y+1/2,x+1/2,-z+1/2
40 -y+1/2,-x+1/2,z+1/2
41 x+1/2,z+1/2,y+1/2
42 x+1/2,-z+1/2,-y+1/2
43 -x+1/2,z+1/2,-y+1/2
44 -x+1/2,-z+1/2,y+1/2
45 z+1/2,y+1/2,x+1/2
46 z+1/2,-y+1/2,-x+1/2
47 -z+1/2,y+1/2,-x+1/2
48 -z+1/2,-y+1/2,x+1/2

loop_
_atom_site_label
_atom_site_type_symbol
_atom_site_symmetry_multiplicity
_atom_site_Wyckoff_label
_atom_site_fract_x
_atom_site_fract_y
_atom_site_fract_z
_atom_site_occupancy
Al1 Al 8 c 0.34090 0.34090 0.34090 1.00000
N1 N 8 c 0.16570 0.16570 0.16570 1.00000
```

AlN (cI16): AB_cI16_217_c_c - POSCAR

```
AB_cI16_217_c_c & a,x1,x2 --params=5.9487,0.3409,0.1657 & I-43m T_{d}^{3
    ↪ } #217 (c^2) & cI16 & None & AlN & AlN & C. Liu et al., J.
    ↪ Phys. Chem. Solids 130, 58-66 (2019)
   1.00000000000000
  -2.97435000000000    2.97435000000000    2.97435000000000
   2.97435000000000   -2.97435000000000    2.97435000000000
   2.97435000000000    2.97435000000000   -2.97435000000000
     Al        N
      4        4
Direct
   0.68180000000000    0.68180000000000    0.68180000000000    Al    (8c)
   0.00000000000000    0.00000000000000   -0.68180000000000    Al    (8c)
   0.00000000000000   -0.68180000000000    0.00000000000000    Al    (8c)
  -0.68180000000000    0.00000000000000    0.00000000000000    Al    (8c)
   0.33140000000000    0.33140000000000    0.33140000000000    N     (8c)
   0.00000000000000    0.00000000000000   -0.33140000000000    N     (8c)
   0.00000000000000   -0.33140000000000    0.00000000000000    N     (8c)
  -0.33140000000000    0.00000000000000    0.00000000000000    N     (8c)
```

Hauyne [$(Na_{0.5}Ca_{0.3}K_{0.2})_8(Al_6Si_6O_{24})(SO_4)_{1.5}$, $S6_9$]: A3B4C4D4E16F4G3_cP76_218_c_e_e_e_ei_e_d - CIF

```
# CIF file
data_findsym-output
_audit_creation_method FINDSYM

_chemical_name_mineral 'Hauyne'
_chemical_formula_sum 'Al3 Ca4 K4 Na4 O16 S4 Si3'

loop_
_publ_author_name
 'I. Hassan'
 'H. D. Grundy'
_journal_name_full_name
;
 Canadian Mineralogist
;
_journal_volume 29
_journal_year 1991
_journal_page_first 123
_journal_page_last 130
_publ_Section_title
;
 The Crystal Structure of Hauyne at 293 and 153 K
;

# Found in The American Mineralogist Crystal Structure Database, 2003

_aflow_title 'Hauyne [(Na$_{0.5}$Ca$_{0.3}$K$_{0.2}$)$_{8}$(Al$_{6}$Si$_
    ↪ {6}$O$_{24}$)(SO$_{4}$)$_{1.5}$, $S6_{9}$] Structure'
_aflow_proto 'A3B4C4D4E16F4G3_cP76_218_c_e_e_e_ei_e_d'
_aflow_params 'a,x_{3},x_{4},x_{5},x_{6},x_{7},x_{8},y_{8},z_{8}'
_aflow_params_values '9.1097,0.2009,0.164,0.2392,0.0995,-0.0331,0.6443,
    ↪ 0.6558,-0.0331'
_aflow_Strukturbericht '$S6_{9}$'
_aflow_Pearson 'cP76'

_symmetry_space_group_name_H-M "P -4 3 n"
_symmetry_Int_Tables_number 218

_cell_length_a     9.10970
_cell_length_b     9.10970
_cell_length_c     9.10970
_cell_angle_alpha 90.00000
_cell_angle_beta  90.00000
_cell_angle_gamma 90.00000

loop_
```

```
_space_group_symop_id
_space_group_symop_operation_xyz
1 x,y,z
2 x,-y,-z
3 -x,y,-z
4 -x,-y,z
5 y,z,x
6 y,-z,-x
7 -y,z,-x
8 -y,-z,x
9 z,x,y
10 z,-x,-y
11 -z,x,-y
12 -z,-x,y
13 y+1/2,x+1/2,z+1/2
14 y+1/2,-x+1/2,-z+1/2
15 -y+1/2,x+1/2,-z+1/2
16 -y+1/2,-x+1/2,z+1/2
17 x+1/2,z+1/2,y+1/2
18 x+1/2,-z+1/2,-y+1/2
19 -x+1/2,z+1/2,-y+1/2
20 -x+1/2,-z+1/2,y+1/2
21 z+1/2,y+1/2,x+1/2
22 z+1/2,-y+1/2,-x+1/2
23 -z+1/2,y+1/2,-x+1/2
24 -z+1/2,-y+1/2,x+1/2

loop_
_atom_site_label
_atom_site_type_symbol
_atom_site_symmetry_multiplicity
_atom_site_Wyckoff_label
_atom_site_fract_x
_atom_site_fract_y
_atom_site_fract_z
_atom_site_occupancy
Al1 Al 6 c 0.25000 0.50000 0.00000 1.00000
Si1 Si 6 d 0.25000 0.00000 0.50000 1.00000
Ca1 Ca 8 e 0.20090 0.20090 0.20090 0.30000
K1 K 8 e 0.16400 0.16400 0.16400 0.20000
Na1 Na 8 e 0.23920 0.23920 0.23920 0.54000
O1 O 8 e 0.09950 0.09950 0.09950 0.75000
S1 S 8 e -0.03310 -0.03310 -0.03310 0.19000
O2 O 24 i 0.64430 0.65580 -0.03310 1.00000
```

Hauyne [$(Na_{0.5}Ca_{0.3}K_{0.2})_8(Al_6Si_6O_{24})(SO_4)_{1.5}$, $S6_9$]: A3B4C4D4E16F4G3_cP76_218_c_e_e_e_ei_e_d - POSCAR

```
A3B4C4D4E16F4G3_cP76_218_c_e_e_e_ei_e_d & a,x3,x4,x5,x6,x7,x8,y8,z8 --
    ↪ params=9.1097,0.2009,0.164,0.2392,0.0995,-0.0331,0.6443,0.6558
    ↪ ,-0.0331 & P-43n T_{d}^{4} #218 (cde^5i) & cP76 & $S6_{9}$ &
    ↪ Al6Ca2.4K1.6Na4O30S1.5Si6 & Hauyne & I. Hassan and H. D. Grundy
    ↪ , Can. Mineral. 29, 123-130 (1991)
   1.00000000000000
   9.10970000000000   0.00000000000000   0.00000000000000
   0.00000000000000   9.10970000000000   0.00000000000000
   0.00000000000000   0.00000000000000   9.10970000000000
    Al    Ca     K    Na     O     S    Si
     6     8     8     8    32     8     6
Direct
   0.25000000000000   0.50000000000000   0.00000000000000   Al   (6c)
   0.75000000000000   0.50000000000000   0.00000000000000   Al   (6c)
   0.00000000000000   0.25000000000000   0.50000000000000   Al   (6c)
   0.00000000000000   0.75000000000000   0.50000000000000   Al   (6c)
   0.50000000000000   0.00000000000000   0.25000000000000   Al   (6c)
   0.50000000000000   0.00000000000000   0.75000000000000   Al   (6c)
   0.20090000000000   0.20090000000000   0.20090000000000   Ca   (8e)
  -0.20090000000000  -0.20090000000000   0.20090000000000   Ca   (8e)
  -0.20090000000000   0.20090000000000  -0.20090000000000   Ca   (8e)
   0.20090000000000  -0.20090000000000  -0.20090000000000   Ca   (8e)
   0.70090000000000   0.70090000000000   0.70090000000000   Ca   (8e)
   0.29910000000000   0.29910000000000   0.70090000000000   Ca   (8e)
   0.70090000000000   0.29910000000000   0.29910000000000   Ca   (8e)
   0.29910000000000   0.70090000000000   0.29910000000000   Ca   (8e)
   0.16400000000000   0.16400000000000   0.16400000000000   K    (8e)
  -0.16400000000000  -0.16400000000000   0.16400000000000   K    (8e)
  -0.16400000000000   0.16400000000000  -0.16400000000000   K    (8e)
   0.16400000000000  -0.16400000000000  -0.16400000000000   K    (8e)
   0.66400000000000   0.66400000000000   0.66400000000000   K    (8e)
   0.33600000000000   0.33600000000000   0.66400000000000   K    (8e)
   0.66400000000000   0.33600000000000   0.33600000000000   K    (8e)
   0.33600000000000   0.66400000000000   0.33600000000000   K    (8e)
   0.23920000000000   0.23920000000000   0.23920000000000   Na   (8e)
  -0.23920000000000  -0.23920000000000   0.23920000000000   Na   (8e)
  -0.23920000000000   0.23920000000000  -0.23920000000000   Na   (8e)
   0.23920000000000  -0.23920000000000  -0.23920000000000   Na   (8e)
   0.73920000000000   0.73920000000000   0.73920000000000   Na   (8e)
   0.26080000000000   0.26080000000000   0.73920000000000   Na   (8e)
   0.73920000000000   0.26080000000000   0.26080000000000   Na   (8e)
   0.26080000000000   0.73920000000000   0.26080000000000   Na   (8e)
   0.09950000000000   0.09950000000000   0.09950000000000   O    (8e)
  -0.09950000000000  -0.09950000000000   0.09950000000000   O    (8e)
  -0.09950000000000   0.09950000000000  -0.09950000000000   O    (8e)
   0.09950000000000  -0.09950000000000  -0.09950000000000   O    (8e)
   0.59950000000000   0.59950000000000   0.59950000000000   O    (8e)
   0.40050000000000   0.40050000000000   0.59950000000000   O    (8e)
   0.59950000000000   0.40050000000000   0.40050000000000   O    (8e)
   0.40050000000000   0.59950000000000   0.40050000000000   O    (8e)
   0.64430000000000   0.65580000000000  -0.03310000000000   O   (24i)
  -0.64430000000000  -0.65580000000000  -0.03310000000000   O   (24i)
  -0.64430000000000   0.65580000000000   0.03310000000000   O   (24i)
   0.64430000000000  -0.65580000000000   0.03310000000000   O   (24i)
  -0.03310000000000   0.64430000000000   0.65580000000000   O   (24i)
  -0.03310000000000  -0.64430000000000  -0.65580000000000   O   (24i)
   0.03310000000000  -0.64430000000000   0.65580000000000   O   (24i)
   0.03310000000000   0.64430000000000  -0.65580000000000   O   (24i)
   0.65580000000000  -0.03310000000000   0.64430000000000   O   (24i)
  -0.65580000000000  -0.03310000000000  -0.64430000000000   O   (24i)
   0.65580000000000   0.03310000000000  -0.64430000000000   O   (24i)
  -0.65580000000000   0.03310000000000   0.64430000000000   O   (24i)
   1.15580000000000   1.14430000000000   0.46690000000000   O   (24i)
  -0.15580000000000  -0.14430000000000   0.46690000000000   O   (24i)
   1.15580000000000  -0.14430000000000   0.53310000000000   O   (24i)
  -0.15580000000000   1.14430000000000   0.53310000000000   O   (24i)
   1.14430000000000   0.46690000000000   1.15580000000000   O   (24i)
  -0.14430000000000   0.46690000000000  -0.15580000000000   O   (24i)
  -0.14430000000000   0.53310000000000   1.15580000000000   O   (24i)
   1.14430000000000   0.53310000000000  -0.15580000000000   O   (24i)
   0.46690000000000   1.15580000000000   1.14430000000000   O   (24i)
   0.46690000000000  -0.15580000000000  -0.14430000000000   O   (24i)
   0.53310000000000   1.15580000000000  -0.14430000000000   O   (24i)
   0.53310000000000  -0.15580000000000   1.14430000000000   O   (24i)
  -0.03310000000000  -0.03310000000000  -0.03310000000000   S    (8e)
   0.03310000000000   0.03310000000000  -0.03310000000000   S    (8e)
   0.03310000000000  -0.03310000000000   0.03310000000000   S    (8e)
  -0.03310000000000   0.03310000000000   0.03310000000000   S    (8e)
   0.46690000000000   0.46690000000000   0.46690000000000   S    (8e)
   0.53310000000000   0.53310000000000   0.46690000000000   S    (8e)
   0.46690000000000   0.53310000000000   0.53310000000000   S    (8e)
   0.53310000000000   0.46690000000000   0.53310000000000   S    (8e)
   0.25000000000000   0.00000000000000   0.50000000000000   Si   (6d)
   0.75000000000000   0.00000000000000   0.50000000000000   Si   (6d)
   0.50000000000000   0.25000000000000   0.00000000000000   Si   (6d)
   0.50000000000000   0.75000000000000   0.00000000000000   Si   (6d)
   0.00000000000000   0.50000000000000   0.25000000000000   Si   (6d)
   0.00000000000000   0.50000000000000   0.75000000000000   Si   (6d)
```

Sodalite [$Na_4(AlSiO_4)_3Cl$, $S6_2$]: A3BC4D12E3_cP46_218_d_a_e_i_c - CIF

```
# CIF file
data_findsym-output
_audit_creation_method FINDSYM

_chemical_name_mineral 'Sodalite'
_chemical_formula_sum 'Al3 Cl Na4 O12 Si3'

loop_
_publ_author_name
 'I. Hassan'
 'H. D. Grundy'
_journal_name_full_name
;
 Acta Crystallographica Section B: Structural Science
;
_journal_volume 40
_journal_year 1984
_journal_page_first 6
_journal_page_last 13
_publ_Section_title
;
 The Crystal Structures of Sodalite-Group Minerals
;

_aflow_title 'Sodalite [Na$_{4}$(AlSiO$_{4}$)$_{3}$Cl, $S6_{2}$]
    ↪ Structure'
_aflow_proto 'A3BC4D12E3_cP46_218_d_a_e_i_c'
_aflow_params 'a,x_{4},x_{5},y_{5},z_{5}'
_aflow_params_values '8.882,0.1778,0.139,0.1494,0.4383'
_aflow_Strukturbericht '$S6_{2}$'
_aflow_Pearson 'cP46'

_symmetry_space_group_name_H-M "P -4 3 n"
_symmetry_Int_Tables_number 218

_cell_length_a    8.88200
_cell_length_b    8.88200
_cell_length_c    8.88200
_cell_angle_alpha 90.00000
_cell_angle_beta  90.00000
_cell_angle_gamma 90.00000

loop_
_space_group_symop_id
_space_group_symop_operation_xyz
1 x,y,z
2 x,-y,-z
3 -x,y,-z
4 -x,-y,z
5 y,z,x
6 y,-z,-x
7 -y,z,-x
8 -y,-z,x
9 z,x,y
10 z,-x,-y
11 -z,x,-y
12 -z,-x,y
13 y+1/2,x+1/2,z+1/2
14 y+1/2,-x+1/2,-z+1/2
15 -y+1/2,x+1/2,-z+1/2
16 -y+1/2,-x+1/2,z+1/2
17 x+1/2,z+1/2,y+1/2
18 x+1/2,-z+1/2,-y+1/2
19 -x+1/2,z+1/2,-y+1/2
20 -x+1/2,-z+1/2,y+1/2
21 z+1/2,y+1/2,x+1/2
22 z+1/2,-y+1/2,-x+1/2
23 -z+1/2,y+1/2,-x+1/2
24 -z+1/2,-y+1/2,x+1/2

loop_
```

```
_atom_site_label
_atom_site_type_symbol
_atom_site_symmetry_multiplicity
_atom_site_Wyckoff_label
_atom_site_fract_x
_atom_site_fract_y
_atom_site_fract_z
_atom_site_occupancy
Cl1  Cl  2  a  0.00000  0.00000  0.00000  1.00000
Si1  Si  6  c  0.25000  0.50000  0.00000  1.00000
Al1  Al  6  d  0.25000  0.00000  0.50000  1.00000
Na1  Na  8  e  0.17780  0.17780  0.17780  1.00000
O1  O  24  i  0.13900  0.14940  0.43830  1.00000
```

Sodalite [$Na_4(AlSiO_4)_3Cl$, $S6_2$]: A3BC4D12E3_cP46_218_d_a_e_i_c - POSCAR

```
A3BC4D12E3_cP46_218_d_a_e_i_c & a,x4,x5,y5,z5 --params=8.882,0.1778,
    ↪ 0.139,0.1494,0.4383 & P-43n T_{d}^{4} #218 (acdei) & cP46 &
    ↪ $S6_{2}$ & Al3ClNa4O12Si3 & Sodalite & I. Hassan and H. D.
    ↪ Grundy, Acta Crystallogr. Sect. B Struct. Sci. 40, 6-13 (1984)
   1.00000000000000
   8.88200000000000    0.00000000000000    0.00000000000000
   0.00000000000000    8.88200000000000    0.00000000000000
   0.00000000000000    0.00000000000000    8.88200000000000
    Al       Cl       Na       O       Si
     6        2        8      24        6
Direct
   0.25000000000000    0.00000000000000    0.50000000000000   Al    (6d)
   0.75000000000000    0.00000000000000    0.50000000000000   Al    (6d)
   0.50000000000000    0.25000000000000    0.00000000000000   Al    (6d)
   0.50000000000000    0.75000000000000    0.00000000000000   Al    (6d)
   0.00000000000000    0.50000000000000    0.25000000000000   Al    (6d)
   0.00000000000000    0.50000000000000    0.75000000000000   Al    (6d)
   0.00000000000000    0.00000000000000    0.00000000000000   Cl    (2a)
   0.50000000000000    0.50000000000000    0.50000000000000   Cl    (2a)
   0.17780000000000    0.17780000000000    0.17780000000000   Na    (8e)
  -0.17780000000000   -0.17780000000000    0.17780000000000   Na    (8e)
  -0.17780000000000    0.17780000000000   -0.17780000000000   Na    (8e)
   0.17780000000000   -0.17780000000000   -0.17780000000000   Na    (8e)
   0.67780000000000    0.67780000000000    0.67780000000000   Na    (8e)
   0.32220000000000    0.32220000000000    0.67780000000000   Na    (8e)
   0.67780000000000    0.32220000000000    0.32220000000000   Na    (8e)
   0.32220000000000    0.67780000000000    0.32220000000000   Na    (8e)
   0.13900000000000    0.14940000000000    0.43830000000000    O   (24i)
  -0.13900000000000   -0.14940000000000    0.43830000000000    O   (24i)
  -0.13900000000000    0.14940000000000   -0.43830000000000    O   (24i)
   0.13900000000000   -0.14940000000000   -0.43830000000000    O   (24i)
   0.43830000000000    0.13900000000000    0.14940000000000    O   (24i)
   0.43830000000000   -0.13900000000000   -0.14940000000000    O   (24i)
  -0.43830000000000   -0.13900000000000    0.14940000000000    O   (24i)
  -0.43830000000000    0.13900000000000   -0.14940000000000    O   (24i)
   0.14940000000000    0.43830000000000    0.13900000000000    O   (24i)
  -0.14940000000000    0.43830000000000   -0.13900000000000    O   (24i)
   0.14940000000000   -0.43830000000000   -0.13900000000000    O   (24i)
  -0.14940000000000   -0.43830000000000    0.13900000000000    O   (24i)
   0.64940000000000    0.63900000000000    0.93830000000000    O   (24i)
   0.35060000000000    0.36100000000000    0.93830000000000    O   (24i)
   0.64940000000000    0.36100000000000    0.06170000000000    O   (24i)
   0.35060000000000    0.63900000000000    0.06170000000000    O   (24i)
   0.63900000000000    0.93830000000000    0.64940000000000    O   (24i)
   0.36100000000000    0.93830000000000    0.35060000000000    O   (24i)
   0.36100000000000    0.06170000000000    0.64940000000000    O   (24i)
   0.63900000000000    0.06170000000000    0.35060000000000    O   (24i)
   0.93830000000000    0.64940000000000    0.63900000000000    O   (24i)
   0.93830000000000    0.35060000000000    0.36100000000000    O   (24i)
   0.06170000000000    0.64940000000000    0.36100000000000    O   (24i)
   0.06170000000000    0.35060000000000    0.63900000000000    O   (24i)
   0.25000000000000    0.50000000000000    0.00000000000000   Si    (6c)
   0.75000000000000    0.50000000000000    0.00000000000000   Si    (6c)
   0.00000000000000    0.25000000000000    0.50000000000000   Si    (6c)
   0.00000000000000    0.75000000000000    0.50000000000000   Si    (6c)
   0.50000000000000    0.00000000000000    0.25000000000000   Si    (6c)
   0.50000000000000    0.00000000000000    0.75000000000000   Si    (6c)
```

Eulytine ($Bi_4(SiO_4)_3$, $S1_5$): A4B12C3_cI76_220_c_e_a - CIF

```
# CIF file
data_findsym-output
_audit_creation_method FINDSYM

_chemical_name_mineral 'Eulytine'
_chemical_formula_sum 'Bi4 O12 Si3'

loop_
_publ_author_name
 'H. Liu'
 'C. Kuo'
_journal_name_full_name
;
 Zeitschrift f{\"u}r Kristallographie - Crystalline Materials
;
_journal_volume 212
_journal_year 1997
_journal_page_first 48
_journal_page_last 48
_publ_Section_title
;
 Crystal structure of bismuth(III) silicate, Bi$_{4}$(SiO$_{4}$)$_{3}$
;

# Found in The American Mineralogist Crystal Structure Database, 2003

_aflow_title 'Eulytine (Bi$_{4}$(SiO$_{4}$)$_{3}$, $S1_{5}$) Structure'
_aflow_proto 'A4B12C3_cI76_220_c_e_a'
_aflow_params 'a,x_{2},x_{3},y_{3},z_{3}'
_aflow_params_values '10.2867,0.0849,0.059,0.131,0.2889'
_aflow_Strukturbericht '$S1_{5}$'
_aflow_Pearson 'cI76'

_symmetry_space_group_name_H-M "I -4 3 d"
_symmetry_Int_Tables_number 220

_cell_length_a     10.28670
_cell_length_b     10.28670
_cell_length_c     10.28670
_cell_angle_alpha  90.00000
_cell_angle_beta   90.00000
_cell_angle_gamma  90.00000

loop_
_space_group_symop_id
_space_group_symop_operation_xyz
1 x,y,z
2 x,-y,-z+1/2
3 -x+1/2,y,-z
4 -x,-y+1/2,z
5 y,z,x
6 y,-z,-x+1/2
7 -y+1/2,z,-x
8 -y,-z+1/2,x
9 z,x,y
10 z,-x,-y+1/2
11 -z+1/2,x,-y
12 -z,-x+1/2,y
13 y+1/4,x+1/4,z+1/4
14 y+1/4,-x+3/4,-z+1/4
15 -y+1/4,x+1/4,-z+3/4
16 -y+3/4,-x+1/4,z+1/4
17 x+1/4,z+1/4,y+1/4
18 x+1/4,-z+3/4,-y+1/4
19 -x+1/4,z+1/4,-y+3/4
20 -x+3/4,-z+1/4,y+1/4
21 z+1/4,y+1/4,x+1/4
22 z+1/4,-y+3/4,-x+1/4
23 -z+1/4,y+1/4,-x+3/4
24 -z+3/4,-y+1/4,x+1/4
25 x+1/2,y+1/2,z+1/2
26 x+1/2,-y+1/2,-z
27 -x,y+1/2,-z+1/2
28 -x+1/2,-y,z+1/2
29 y+1/2,z+1/2,x+1/2
30 y+1/2,-z+1/2,-x
31 -y,z+1/2,-x+1/2
32 -y+1/2,-z,x+1/2
33 z+1/2,x+1/2,y+1/2
34 z+1/2,-x+1/2,-y
35 -z,x+1/2,-y+1/2
36 -z+1/2,-x,y+1/2
37 y+3/4,x+3/4,z+3/4
38 y+3/4,-x+1/4,-z+3/4
39 -y+3/4,x+3/4,-z+1/4
40 -y+1/4,-x+3/4,z+3/4
41 x+3/4,z+3/4,y+3/4
42 x+3/4,-z+1/4,-y+3/4
43 -x+3/4,z+3/4,-y+1/4
44 -x+1/4,-z+3/4,y+3/4
45 z+3/4,y+3/4,x+3/4
46 z+3/4,-y+1/4,-x+3/4
47 -z+3/4,y+3/4,-x+1/4
48 -z+1/4,-y+3/4,x+3/4

loop_
_atom_site_label
_atom_site_type_symbol
_atom_site_symmetry_multiplicity
_atom_site_Wyckoff_label
_atom_site_fract_x
_atom_site_fract_y
_atom_site_fract_z
_atom_site_occupancy
Si1  Si  12  a  0.37500  0.00000  0.25000  1.00000
Bi1  Bi  16  c  0.08490  0.08490  0.08490  1.00000
O1  O  48  e  0.05900  0.13100  0.28890  1.00000
```

Eulytine ($Bi_4(SiO_4)_3$, $S1_5$): A4B12C3_cI76_220_c_e_a - POSCAR

```
A4B12C3_cI76_220_c_e_a & a,x2,x3,y3,z3 --params=10.2867,0.0849,0.059,
    ↪ 0.131,0.2889 & I-43d T_{d}^{6} #220 (ace) & cI76 & $S1_{5}$ &
    ↪ Bi4O12Si3 & Eulytine & H. Liu and C. Kuo, Zeitschrift f{"u}r
    ↪ Kristallographie - Crystalline Materials 212, 48(1997)
   1.00000000000000
  -5.14335000000000    5.14335000000000    5.14335000000000
   5.14335000000000   -5.14335000000000    5.14335000000000
   5.14335000000000    5.14335000000000   -5.14335000000000
    Bi       O       Si
     8      24        6
Direct
   0.16980000000000    0.16980000000000    0.16980000000000   Bi   (16c)
   0.50000000000000    0.00000000000000    0.33020000000000   Bi   (16c)
   0.00000000000000    0.33020000000000    0.50000000000000   Bi   (16c)
   0.33020000000000    0.50000000000000    0.00000000000000   Bi   (16c)
   0.66980000000000    0.66980000000000    0.66980000000000   Bi   (16c)
   0.50000000000000    0.00000000000000   -0.16980000000000   Bi   (16c)
  -0.16980000000000    0.50000000000000    0.00000000000000   Bi   (16c)
   0.00000000000000   -0.16980000000000    0.50000000000000   Bi   (16c)
   0.41990000000000    0.34790000000000    0.19000000000000    O   (48e)
   0.65790000000000    0.22990000000000    0.31000000000000    O   (48e)
  -0.15790000000000    0.15210000000000    0.57200000000000    O   (48e)
   0.08010000000000    0.27010000000000   -0.07200000000000    O   (48e)
   0.19000000000000    0.41990000000000    0.34790000000000    O   (48e)
```

```
    0.31000000000000    0.65790000000000    0.22990000000000    O   (48e)
    0.57200000000000   -0.15790000000000    0.15210000000000    O   (48e)
   -0.07200000000000    0.08010000000000    0.27010000000000    O   (48e)
    0.34790000000000    0.19000000000000    0.41990000000000    O   (48e)
    0.22990000000000    0.31000000000000    0.65790000000000    O   (48e)
    0.15210000000000    0.57200000000000   -0.15790000000000    O   (48e)
    0.27010000000000   -0.07200000000000    0.08010000000000    O   (48e)
    0.84790000000000    0.91990000000000    0.69000000000000    O   (48e)
    0.72990000000000    0.15790000000000   -0.19000000000000    O   (48e)
   -0.34790000000000    0.34210000000000    0.07200000000000    O   (48e)
   -0.22990000000000   -0.41990000000000    0.42800000000000    O   (48e)
    0.91990000000000    0.69000000000000    0.84790000000000    O   (48e)
    0.15790000000000   -0.19000000000000    0.72990000000000    O   (48e)
    0.34210000000000    0.07200000000000   -0.34790000000000    O   (48e)
   -0.41990000000000    0.42800000000000   -0.22990000000000    O   (48e)
    0.69000000000000    0.84790000000000    0.91990000000000    O   (48e)
   -0.19000000000000    0.72990000000000    0.15790000000000    O   (48e)
    0.07200000000000   -0.34790000000000    0.34210000000000    O   (48e)
    0.42800000000000   -0.22990000000000   -0.41990000000000    O   (48e)
    0.25000000000000    0.62500000000000    0.37500000000000   Si   (12a)
    0.75000000000000    0.87500000000000    0.12500000000000   Si   (12a)
    0.37500000000000    0.25000000000000    0.62500000000000   Si   (12a)
    0.12500000000000    0.75000000000000    0.87500000000000   Si   (12a)
    0.62500000000000    0.37500000000000    0.25000000000000   Si   (12a)
    0.87500000000000    0.12500000000000    0.75000000000000   Si   (12a)
```

Mayenite (12CaO·7$Al_2O_3$, $K7_4$, C12A7): A7B12C19_cI152_220_bc_2d_ace - CIF

```
# CIF file
data_findsym-output
_audit_creation_method FINDSYM

_chemical_name_mineral 'Mayenite'
_chemical_formula_sum 'Al7 Ca12 O19'

loop_
_publ_author_name
 'H. Boysen'
 'M. Lerch'
 'A. Stys'
 'A. Senyshyn'
_journal_name_full_name
;
 Acta Crystallographica Section B: Structural Science
;
_journal_volume 63
_journal_year 2007
_journal_page_first 675
_journal_page_last 682
_publ_Section_title
;
 Structure and oxygen mobility in mayenite (Ca$_{12}$Al$_{14}$O$_{33}$):
      ↪ a high-temperature neutron powder diffraction study
;

# Found in The American Mineralogist Crystal Structure Database, 2003

_aflow_title 'Mayenite (12CaO$\cdot$7Al$_{2}$O$_{3}$, $K7_{4}$, C12A7)
    ↪ Structure'
_aflow_proto 'A7B12C19_cI152_220_bc_2d_ace'
_aflow_params 'a,x_{3},x_{4},x_{5},x_{6},x_{7},y_{7},z_{7}'
_aflow_params_values '11.9794,0.0188,0.43519,0.1432,0.1867,0.78672,
    ↪ 0.09946,0.30708'
_aflow_Strukturbericht '$K7_{4}$'
_aflow_Pearson 'cI152'

_symmetry_space_group_name_H-M "I -4 3 d"
_symmetry_Int_Tables_number 220

_cell_length_a      11.97940
_cell_length_b      11.97940
_cell_length_c      11.97940
_cell_angle_alpha 90.00000
_cell_angle_beta  90.00000
_cell_angle_gamma 90.00000

loop_
_space_group_symop_id
_space_group_symop_operation_xyz
1 x,y,z
2 x,-y,-z+1/2
3 -x+1/2,y,-z
4 -x,-y+1/2,z
5 y,z,x
6 y,-z,-x+1/2
7 -y+1/2,z,-x
8 -y,-z+1/2,x
9 z,x,y
10 z,-x,-y+1/2
11 -z+1/2,x,-y
12 -z,-x+1/2,y
13 y+1/4,x+1/4,z+1/4
14 y+1/4,-x+3/4,-z+1/4
15 -y+1/4,x+1/4,-z+3/4
16 -y+3/4,-x+1/4,z+1/4
17 x+1/4,z+1/4,y+1/4
18 x+1/4,-z+3/4,-y+1/4
19 -x+1/4,z+1/4,-y+3/4
20 -x+3/4,-z+1/4,y+1/4
21 z+1/4,y+1/4,x+1/4
22 z+1/4,-y+3/4,-x+1/4
23 -z+1/4,y+1/4,-x+3/4
24 -z+3/4,-y+1/4,x+1/4
25 x+1/2,y+1/2,z+1/2
26 x+1/2,-y+1/2,-z
27 -x,y+1/2,-z+1/2
28 -x+1/2,-y,z+1/2
29 y+1/2,z+1/2,x+1/2
30 y+1/2,-z+1/2,-x
31 -y,z+1/2,-x+1/2
32 -y+1/2,-z,x+1/2
33 z+1/2,x+1/2,y+1/2
34 z+1/2,-x+1/2,-y
35 -z,x+1/2,-y+1/2
36 -z+1/2,-x,y+1/2
37 y+3/4,x+3/4,z+3/4
38 y+3/4,-x+1/4,-z+3/4
39 -y+3/4,x+3/4,-z+1/4
40 -y+1/4,-x+3/4,z+3/4
41 x+3/4,z+3/4,y+3/4
42 x+3/4,-z+1/4,-y+3/4
43 -x+3/4,z+3/4,-y+1/4
44 -x+1/4,-z+3/4,y+3/4
45 z+3/4,y+3/4,x+3/4
46 z+3/4,-y+1/4,-x+3/4
47 -z+3/4,y+3/4,-x+1/4
48 -z+1/4,-y+3/4,x+3/4

loop_
_atom_site_label
_atom_site_type_symbol
_atom_site_symmetry_multiplicity
_atom_site_Wyckoff_label
_atom_site_fract_x
_atom_site_fract_y
_atom_site_fract_z
_atom_site_occupancy
O1 O 12 a 0.37500 0.00000 0.25000 0.16667
Al1 Al 12 b 0.87500 0.00000 0.25000 1.00000
Al2 Al 16 c 0.01880 0.01880 0.01880 1.00000
O2 O 16 c 0.43519 0.43519 0.43519 1.00000
Ca1 Ca 24 d 0.14320 0.00000 0.25000 0.87500
Ca2 Ca 24 d 0.18670 0.00000 0.25000 0.12500
O3 O 48 e 0.78672 0.09946 0.30708 1.00000
```

Mayenite (12CaO·7$Al_2O_3$, $K7_4$, C12A7): A7B12C19_cI152_220_bc_2d_ace - POSCAR

```
A7B12C19_cI152_220_bc_2d_ace & a,x3,x4,x5,x6,x7,y7,z7 --params=11.9794,
    ↪ 0.0188,0.43519,0.1432,0.1867,0.78672,0.09946,0.30708 & I-43d T_
    ↪ {d}^{6} #220 (abc^2d^2e) & cI152 & $K7_{4}$ & Al14Ca12O33 &
    ↪ Mayenite & H. Boysen et al., Acta Crystallogr. Sect. B Struct.
    ↪ Sci. 63, 675-682 (2007)
   1.00000000000000
  -5.98970000000000    5.98970000000000    5.98970000000000
   5.98970000000000   -5.98970000000000    5.98970000000000
   5.98970000000000    5.98970000000000   -5.98970000000000
    Al        Ca        O
    14        24        38
Direct
   0.25000000000000    0.12500000000000    0.87500000000000   Al   (12b)
   0.75000000000000    0.37500000000000    0.62500000000000   Al   (12b)
   0.87500000000000    0.25000000000000    0.12500000000000   Al   (12b)
   0.62500000000000    0.75000000000000    0.37500000000000   Al   (12b)
   0.12500000000000    0.87500000000000    0.25000000000000   Al   (12b)
   0.37500000000000    0.62500000000000    0.75000000000000   Al   (12b)
   0.03760000000000    0.03760000000000    0.03760000000000   Al   (16c)
   0.50000000000000    0.00000000000000    0.46240000000000   Al   (16c)
   0.00000000000000    0.46240000000000    0.50000000000000   Al   (16c)
   0.46240000000000    0.50000000000000    0.00000000000000   Al   (16c)
   0.53760000000000    0.53760000000000    0.53760000000000   Al   (16c)
   0.50000000000000    0.00000000000000   -0.03760000000000   Al   (16c)
  -0.03760000000000    0.50000000000000    0.00000000000000   Al   (16c)
   0.00000000000000   -0.03760000000000    0.50000000000000   Al   (16c)
   0.25000000000000    0.39320000000000    0.14320000000000   Ca   (24d)
   0.75000000000000    0.10680000000000    0.35680000000000   Ca   (24d)
   0.14320000000000    0.25000000000000    0.39320000000000   Ca   (24d)
   0.35680000000000    0.75000000000000    0.10680000000000   Ca   (24d)
   0.39320000000000    0.14320000000000    0.25000000000000   Ca   (24d)
   0.10680000000000    0.35680000000000    0.75000000000000   Ca   (24d)
   0.89320000000000    0.75000000000000    0.64320000000000   Ca   (24d)
   0.60680000000000    0.25000000000000   -0.14320000000000   Ca   (24d)
   0.75000000000000    0.64320000000000    0.89320000000000   Ca   (24d)
   0.25000000000000   -0.14320000000000    0.60680000000000   Ca   (24d)
   0.64320000000000    0.89320000000000    0.75000000000000   Ca   (24d)
  -0.14320000000000    0.60680000000000    0.25000000000000   Ca   (24d)
   0.25000000000000    0.43670000000000    0.18670000000000   Ca   (24d)
   0.75000000000000    0.06330000000000    0.31330000000000   Ca   (24d)
   0.18670000000000    0.25000000000000    0.43670000000000   Ca   (24d)
   0.31330000000000    0.75000000000000    0.06330000000000   Ca   (24d)
   0.43670000000000    0.18670000000000    0.25000000000000   Ca   (24d)
   0.06330000000000    0.31330000000000    0.75000000000000   Ca   (24d)
   0.93670000000000    0.75000000000000    0.68670000000000   Ca   (24d)
   0.56330000000000    0.25000000000000   -0.18670000000000   Ca   (24d)
   0.75000000000000    0.68670000000000    0.93670000000000   Ca   (24d)
   0.25000000000000   -0.18670000000000    0.56330000000000   Ca   (24d)
   0.68670000000000    0.93670000000000    0.75000000000000   Ca   (24d)
  -0.18670000000000    0.56330000000000    0.25000000000000   Ca   (24d)
   0.25000000000000    0.62500000000000    0.37500000000000    O   (12a)
   0.75000000000000    0.87500000000000    0.12500000000000    O   (12a)
   0.37500000000000    0.25000000000000    0.62500000000000    O   (12a)
   0.12500000000000    0.75000000000000    0.87500000000000    O   (12a)
   0.62500000000000    0.37500000000000    0.25000000000000    O   (12a)
   0.87500000000000    0.12500000000000    0.75000000000000    O   (12a)
   0.87038000000000    0.87038000000000    0.87038000000000    O   (16c)
   0.50000000000000    0.00000000000000   -0.37038000000000    O   (16c)
   0.00000000000000   -0.37038000000000    0.50000000000000    O   (16c)
  -0.37038000000000    0.50000000000000    0.00000000000000    O   (16c)
   1.37038000000000    1.37038000000000    1.37038000000000    O   (16c)
   0.50000000000000    0.00000000000000   -0.87038000000000    O   (16c)
  -0.87038000000000    0.50000000000000    0.00000000000000    O   (16c)
```

```
 0.00000000000000  -0.87038000000000   0.50000000000000    O   (16c)
 0.40654000000000   1.09380000000000   0.88618000000000    O   (48e)
 0.70762000000000  -0.47964000000000  -0.38618000000000    O   (48e)
-0.20762000000000  -0.59380000000000  -0.18726000000000    O   (48e)
 0.09346000000000   0.97964000000000   0.68726000000000    O   (48e)
 0.88618000000000   0.40654000000000   1.09380000000000    O   (48e)
-0.38618000000000   0.70762000000000  -0.47964000000000    O   (48e)
-0.18726000000000  -0.20762000000000  -0.59380000000000    O   (48e)
 0.68726000000000   0.09346000000000   0.97964000000000    O   (48e)
 1.09380000000000   0.88618000000000   0.40654000000000    O   (48e)
-0.47964000000000  -0.38618000000000   0.70762000000000    O   (48e)
-0.59380000000000  -0.18726000000000  -0.20762000000000    O   (48e)
 0.97964000000000   0.68726000000000   0.09346000000000    O   (48e)
 1.59380000000000   0.90654000000000   1.38618000000000    O   (48e)
 0.02036000000000   0.20762000000000  -0.88618000000000    O   (48e)
-1.09380000000000   0.29238000000000  -0.68726000000000    O   (48e)
 0.47964000000000  -0.40654000000000   1.18726000000000    O   (48e)
 0.90654000000000   1.38618000000000   1.59380000000000    O   (48e)
 0.20762000000000  -0.88618000000000   0.02036000000000    O   (48e)
 0.29238000000000  -0.68726000000000  -1.09380000000000    O   (48e)
-0.40654000000000   1.18726000000000   0.47964000000000    O   (48e)
 1.38618000000000   1.59380000000000   0.90654000000000    O   (48e)
-0.88618000000000   0.02036000000000   0.20762000000000    O   (48e)
-0.68726000000000  -1.09380000000000   0.29238000000000    O   (48e)
 1.18726000000000   0.47964000000000  -0.40654000000000    O   (48e)
```

$Al(PO_3)_3$ ($G5_2$): AB9C3_cI208_220_c_3e_e - CIF

```
# CIF file
data_findsym-output
_audit_creation_method FINDSYM

_chemical_name_mineral 'AlO9P3'
_chemical_formula_sum 'Al O9 P3'

loop_
_publ_author_name
 'L. Pauling'
 'J. Sherman'
_journal_name_full_name
;
 Zeitschrift f{\"u}r Kristallographie - Crystalline Materials
;
_journal_volume 96
_journal_year 1937
_journal_page_first 481
_journal_page_last 487
_publ_Section_title
;
 The Crystal Structure of Aluminum Metaphosphate, Al(PO$_{3}$)$_{3}$
;

# Found in The crystal structure of a monoclinic form of aluminium
    ↪ metaphosphate, Al(PO$_{3}$)$_{3}$, 1976

_aflow_title 'Al(PO$_{3}$)$_{3}$ ($G5_{2}$) Structure'
_aflow_proto 'AB9C3_cI208_220_c_3e_e'
_aflow_params 'a,x_{1},x_{2},y_{2},z_{2},x_{3},y_{3},z_{3},x_{4},y_{4},
    ↪ z_{4},x_{5},y_{5},z_{5}'
_aflow_params_values '13.63,0.117,0.09,0.11,0.8,0.095,0.141,0.245,0.137,
    ↪ 0.096,-0.014,0.34,0.063,0.124'
_aflow_Strukturbericht '$G5_{2}$'
_aflow_Pearson 'cI208'

_symmetry_space_group_name_H-M "I -4 3 d"
_symmetry_Int_Tables_number 220

_cell_length_a     13.63000
_cell_length_b     13.63000
_cell_length_c     13.63000
_cell_angle_alpha 90.00000
_cell_angle_beta  90.00000
_cell_angle_gamma 90.00000

loop_
_space_group_symop_id
_space_group_symop_operation_xyz
1 x,y,z
2 x,-y,-z+1/2
3 -x+1/2,y,-z
4 -x,-y+1/2,z
5 y,z,x
6 y,-z,-x+1/2
7 -y+1/2,z,-x
8 -y,-z+1/2,x
9 z,x,y
10 z,-x,-y+1/2
11 -z+1/2,x,-y
12 -z,-x+1/2,y
13 y+1/4,x+1/4,z+1/4
14 y+1/4,-x+3/4,-z+1/4
15 -y+1/4,x+1/4,-z+3/4
16 -y+3/4,-x+1/4,z+1/4
17 x+1/4,z+1/4,y+1/4
18 x+1/4,-z+3/4,-y+1/4
19 -x+1/4,z+1/4,-y+3/4
20 -x+3/4,-z+1/4,y+1/4
21 z+1/4,y+1/4,x+1/4
22 z+1/4,-y+3/4,-x+1/4
23 -z+1/4,y+1/4,-x+3/4
24 -z+3/4,-y+1/4,x+1/4
25 x+1/2,y+1/2,z+1/2
26 x+1/2,-y+1/2,-z
27 -x,y+1/2,-z+1/2
28 -x+1/2,-y,z+1/2
29 y+1/2,z+1/2,x+1/2
30 y+1/2,-z+1/2,-x
31 -y,z+1/2,-x+1/2
32 -y+1/2,-z,x+1/2
33 z+1/2,x+1/2,y+1/2
34 z+1/2,-x+1/2,-y
35 -z,x+1/2,-y+1/2
36 -z+1/2,-x,y+1/2
37 y+3/4,x+3/4,z+3/4
38 y+3/4,-x+1/4,-z+3/4
39 -y+3/4,x+3/4,-z+1/4
40 -y+1/4,-x+3/4,z+3/4
41 x+3/4,z+3/4,y+3/4
42 x+3/4,-z+1/4,-y+3/4
43 -x+3/4,z+3/4,-y+1/4
44 -x+1/4,-z+3/4,y+3/4
45 z+3/4,y+3/4,x+3/4
46 z+3/4,-y+1/4,-x+3/4
47 -z+3/4,y+3/4,-x+1/4
48 -z+1/4,-y+3/4,x+3/4

loop_
_atom_site_label
_atom_site_type_symbol
_atom_site_symmetry_multiplicity
_atom_site_Wyckoff_label
_atom_site_fract_x
_atom_site_fract_y
_atom_site_fract_z
_atom_site_occupancy
Al1 Al 16 c 0.11700 0.11700 0.11700 1.00000
O1 O 48 e 0.09000 0.11000 0.80000 1.00000
O2 O 48 e 0.09500 0.14100 0.24500 1.00000
O3 O 48 e 0.13700 0.09600 -0.01400 1.00000
P1 P 48 e 0.34000 0.06300 0.12400 1.00000
```

$Al(PO_3)_3$ ($G5_2$): AB9C3_cI208_220_c_3e_e - POSCAR

```
AB9C3_cI208_220_c_3e_e & a,x1,x2,y2,z2,x3,y3,z3,x4,y4,z4,x5,y5,z5 --
    ↪ params=13.63,0.117,0.09,0.11,0.8,0.095,0.141,0.245,0.137,0.096
    ↪ ,-0.014,0.34,0.063,0.124 & I-43d T_{d}^{6} #220 (ce^4) & cI208
    ↪ & $G5_{2}$ & AlO9P3 & AlO9P3 & L. Pauling and J. Sherman,
    ↪ Zeitschrift f{"u}r Kristallographie - Crystalline Materials 96,
    ↪ 481-487 (1937)
   1.00000000000000
  -6.81500000000000   6.81500000000000   6.81500000000000
   6.81500000000000  -6.81500000000000   6.81500000000000
   6.81500000000000   6.81500000000000  -6.81500000000000
    Al      O      P
     8     72     24
Direct
   0.23400000000000   0.23400000000000   0.23400000000000   Al   (16c)
   0.50000000000000   0.00000000000000   0.26600000000000   Al   (16c)
   0.00000000000000   0.26600000000000   0.50000000000000   Al   (16c)
   0.26600000000000   0.50000000000000   0.00000000000000   Al   (16c)
   0.73400000000000   0.73400000000000   0.73400000000000   Al   (16c)
   0.50000000000000   0.00000000000000  -0.23400000000000   Al   (16c)
  -0.23400000000000   0.50000000000000   0.00000000000000   Al   (16c)
   0.00000000000000  -0.23400000000000   0.50000000000000   Al   (16c)
   0.91000000000000   0.89000000000000   0.20000000000000    O   (48e)
   1.19000000000000   0.71000000000000   0.30000000000000    O   (48e)
  -0.69000000000000  -0.39000000000000   0.52000000000000    O   (48e)
  -0.41000000000000  -0.21000000000000  -0.02000000000000    O   (48e)
   0.20000000000000   0.91000000000000   0.89000000000000    O   (48e)
   0.30000000000000   1.19000000000000   0.71000000000000    O   (48e)
   0.52000000000000  -0.69000000000000  -0.39000000000000    O   (48e)
  -0.02000000000000  -0.41000000000000  -0.21000000000000    O   (48e)
   0.89000000000000   0.20000000000000   0.91000000000000    O   (48e)
   0.71000000000000   0.30000000000000   1.19000000000000    O   (48e)
  -0.39000000000000   0.52000000000000  -0.69000000000000    O   (48e)
  -0.21000000000000  -0.02000000000000  -0.41000000000000    O   (48e)
   1.39000000000000   1.41000000000000   0.70000000000000    O   (48e)
   1.21000000000000   0.69000000000000  -0.20000000000000    O   (48e)
  -0.89000000000000  -0.19000000000000   0.02000000000000    O   (48e)
  -0.71000000000000  -0.91000000000000   0.48000000000000    O   (48e)
   1.41000000000000   0.70000000000000   1.39000000000000    O   (48e)
   0.69000000000000  -0.20000000000000   1.21000000000000    O   (48e)
  -0.19000000000000   0.02000000000000  -0.89000000000000    O   (48e)
  -0.91000000000000   0.48000000000000  -0.71000000000000    O   (48e)
   0.70000000000000   1.39000000000000   1.41000000000000    O   (48e)
  -0.20000000000000   1.21000000000000   0.69000000000000    O   (48e)
   0.02000000000000  -0.89000000000000  -0.19000000000000    O   (48e)
   0.48000000000000  -0.71000000000000  -0.91000000000000    O   (48e)
   0.38600000000000   0.34000000000000   0.23600000000000    O   (48e)
   0.60400000000000   0.15000000000000   0.26400000000000    O   (48e)
  -0.10400000000000   0.16000000000000   0.54600000000000    O   (48e)
   0.11400000000000   0.35000000000000  -0.04600000000000    O   (48e)
   0.23600000000000   0.38600000000000   0.34000000000000    O   (48e)
   0.26400000000000   0.60400000000000   0.15000000000000    O   (48e)
   0.54600000000000  -0.10400000000000   0.16000000000000    O   (48e)
  -0.04600000000000   0.11400000000000   0.35000000000000    O   (48e)
   0.34000000000000   0.23600000000000   0.38600000000000    O   (48e)
   0.15000000000000   0.26400000000000   0.60400000000000    O   (48e)
   0.16000000000000   0.54600000000000  -0.10400000000000    O   (48e)
   0.35000000000000  -0.04600000000000   0.11400000000000    O   (48e)
   0.84000000000000   0.88600000000000   0.73600000000000    O   (48e)
   0.65000000000000   0.10400000000000  -0.23600000000000    O   (48e)
  -0.34000000000000   0.39600000000000   0.04600000000000    O   (48e)
  -0.15000000000000  -0.38600000000000   0.45400000000000    O   (48e)
   0.88600000000000   0.73600000000000   0.84000000000000    O   (48e)
   0.10400000000000  -0.23600000000000   0.65000000000000    O   (48e)
   0.39600000000000   0.04600000000000  -0.34000000000000    O   (48e)
  -0.38600000000000   0.45400000000000  -0.15000000000000    O   (48e)
   0.73600000000000   0.84000000000000   0.88600000000000    O   (48e)
  -0.23600000000000   0.65000000000000   0.10400000000000    O   (48e)
```

```
   0.04600000000000   -0.34000000000000    0.39600000000000    O   (48e)
   0.45400000000000   -0.15000000000000   -0.38600000000000    O   (48e)
   0.08200000000000    0.12300000000000    0.23300000000000    O   (48e)
   0.39000000000000   -0.15100000000000    0.26700000000000    O   (48e)
   0.11000000000000    0.37700000000000    0.45900000000000    O   (48e)
   0.41800000000000    0.65100000000000    0.04100000000000    O   (48e)
   0.23300000000000    0.08200000000000    0.12300000000000    O   (48e)
   0.26700000000000    0.39000000000000   -0.15100000000000    O   (48e)
   0.45900000000000    0.11000000000000    0.37700000000000    O   (48e)
   0.04100000000000    0.41800000000000    0.65100000000000    O   (48e)
   0.12300000000000    0.23300000000000    0.08200000000000    O   (48e)
  -0.15100000000000    0.26700000000000    0.39000000000000    O   (48e)
   0.37700000000000    0.45900000000000    0.11000000000000    O   (48e)
   0.65100000000000    0.04100000000000    0.41800000000000    O   (48e)
   0.62300000000000    0.58200000000000    0.73300000000000    O   (48e)
   0.34900000000000   -0.11000000000000   -0.23300000000000    O   (48e)
  -0.12300000000000    0.61000000000000   -0.04100000000000    O   (48e)
   0.15100000000000   -0.08200000000000    0.54100000000000    O   (48e)
   0.58200000000000    0.73300000000000    0.62300000000000    O   (48e)
  -0.11000000000000   -0.23300000000000    0.34900000000000    O   (48e)
   0.61000000000000   -0.04100000000000   -0.12300000000000    O   (48e)
  -0.08200000000000    0.54100000000000    0.15100000000000    O   (48e)
   0.73300000000000    0.62300000000000    0.58200000000000    O   (48e)
  -0.23300000000000    0.34900000000000   -0.11000000000000    O   (48e)
  -0.04100000000000   -0.12300000000000    0.61000000000000    O   (48e)
   0.54100000000000    0.15100000000000   -0.08200000000000    O   (48e)
   0.18700000000000    0.46400000000000    0.40300000000000    P   (48e)
   0.56100000000000   -0.21600000000000    0.09700000000000    P   (48e)
  -0.06100000000000    0.03600000000000    0.22300000000000    P   (48e)
   0.31300000000000    0.71600000000000    0.27700000000000    P   (48e)
   0.40300000000000    0.18700000000000    0.46400000000000    P   (48e)
   0.09700000000000    0.56100000000000   -0.21600000000000    P   (48e)
   0.22300000000000   -0.06100000000000    0.03600000000000    P   (48e)
   0.27700000000000    0.31300000000000    0.71600000000000    P   (48e)
   0.46400000000000    0.40300000000000    0.18700000000000    P   (48e)
  -0.21600000000000    0.09700000000000    0.56100000000000    P   (48e)
   0.03600000000000    0.22300000000000   -0.06100000000000    P   (48e)
   0.71600000000000    0.27700000000000    0.31300000000000    P   (48e)
   0.96400000000000    0.68700000000000    0.90300000000000    P   (48e)
   0.28400000000000    0.06100000000000   -0.40300000000000    P   (48e)
  -0.46400000000000    0.43900000000000   -0.27700000000000    P   (48e)
   0.21600000000000   -0.18700000000000    0.77700000000000    P   (48e)
   0.68700000000000    0.90300000000000    0.96400000000000    P   (48e)
   0.06100000000000   -0.40300000000000    0.28400000000000    P   (48e)
   0.43900000000000   -0.27700000000000   -0.46400000000000    P   (48e)
  -0.18700000000000    0.77700000000000    0.21600000000000    P   (48e)
   0.90300000000000    0.96400000000000    0.68700000000000    P   (48e)
  -0.40300000000000    0.28400000000000    0.06100000000000    P   (48e)
  -0.27700000000000   -0.46400000000000    0.43900000000000    P   (48e)
   0.77700000000000    0.21600000000000   -0.18700000000000    P   (48e)
```

AlN (cI24): AB_cI24_220_a_b - CIF

```
# CIF file
data_findsym-output
_audit_creation_method FINDSYM

_chemical_name_mineral 'AlN'
_chemical_formula_sum 'Al N'

loop_
_publ_author_name
 'C. Liu'
 'M. Chen'
 'J. Li'
 'L. Liu'
 'P. Li'
 'M. Ma'
 'C. Shao'
 'J. He'
 'T. Liang'
_journal_name_full_name
;
 Journal of Physics and Chemistry of Solids
;
_journal_volume 130
_journal_year 2019
_journal_page_first 58
_journal_page_last 66
_publ_Section_title
;
 A first-principles study of novel cubic AlN phases
;

_aflow_title 'AlN (cI24) Structure'
_aflow_proto 'AB_cI24_220_a_b'
_aflow_params 'a'
_aflow_params_values '6.2951'
_aflow_Strukturbericht 'None'
_aflow_Pearson 'cI24'

_symmetry_space_group_name_H-M "I -4 3 d"
_symmetry_Int_Tables_number 220

_cell_length_a      6.29510
_cell_length_b      6.29510
_cell_length_c      6.29510
_cell_angle_alpha 90.00000
_cell_angle_beta  90.00000
_cell_angle_gamma 90.00000

loop_
_space_group_symop_id
_space_group_symop_operation_xyz
1 x,y,z
2 x,-y,-z+1/2
3 -x+1/2,y,-z
4 -x,-y+1/2,z
5 y,z,x
6 y,-z,-x+1/2
7 -y+1/2,z,-x
8 -y,-z+1/2,x
9 z,x,y
10 z,-x,-y+1/2
11 -z+1/2,x,-y
12 -z,-x+1/2,y
13 y+1/4,x+1/4,z+1/4
14 y+1/4,-x+3/4,-z+1/4
15 -y+1/4,x+1/4,-z+3/4
16 -y+3/4,-x+1/4,z+1/4
17 x+1/4,z+1/4,y+1/4
18 x+1/4,-z+3/4,-y+1/4
19 -x+1/4,z+1/4,-y+3/4
20 -x+3/4,-z+1/4,y+1/4
21 z+1/4,y+1/4,x+1/4
22 z+1/4,-y+3/4,-x+1/4
23 -z+1/4,y+1/4,-x+3/4
24 -z+3/4,-y+1/4,x+1/4
25 x+1/2,y+1/2,z+1/2
26 x+1/2,-y+1/2,-z
27 -x,y+1/2,-z+1/2
28 -x+1/2,-y,z+1/2
29 y+1/2,z+1/2,x+1/2
30 y+1/2,-z+1/2,-x
31 -y,z+1/2,-x+1/2
32 -y+1/2,-z,x+1/2
33 z+1/2,x+1/2,y+1/2
34 z+1/2,-x+1/2,-y
35 -z,x+1/2,-y+1/2
36 -z+1/2,-x,y+1/2
37 y+3/4,x+3/4,z+3/4
38 y+3/4,-x+1/4,-z+3/4
39 -y+3/4,x+3/4,-z+1/4
40 -y+1/4,-x+3/4,z+3/4
41 x+3/4,z+3/4,y+3/4
42 x+3/4,-z+1/4,-y+3/4
43 -x+3/4,z+3/4,-y+1/4
44 -x+1/4,-z+3/4,y+3/4
45 z+3/4,y+3/4,x+3/4
46 z+3/4,-y+1/4,-x+3/4
47 -z+3/4,y+3/4,-x+1/4
48 -z+1/4,-y+3/4,x+3/4

loop_
_atom_site_label
_atom_site_type_symbol
_atom_site_symmetry_multiplicity
_atom_site_Wyckoff_label
_atom_site_fract_x
_atom_site_fract_y
_atom_site_fract_z
_atom_site_occupancy
Al1 Al 12 a 0.37500 0.00000 0.25000 1.00000
N1 N 12 b 0.87500 0.00000 0.25000 1.00000
```

AlN (cI24): AB_cI24_220_a_b - POSCAR

```
AB_cI24_220_a_b & a --params=6.2951 & I-43d T_{d}^{6} #220 (ab) & cI24 &
     ↪ None & AlN & AlN & C. Liu et al., J. Phys. Chem. Solids 130,
     ↪ 58-66 (2019)
   1.00000000000000
  -3.14755000000000    3.14755000000000    3.14755000000000
   3.14755000000000   -3.14755000000000    3.14755000000000
   3.14755000000000    3.14755000000000   -3.14755000000000
    Al        N
     6        6
Direct
   0.25000000000000    0.62500000000000    0.37500000000000    Al   (12a)
   0.75000000000000    0.87500000000000    0.12500000000000    Al   (12a)
   0.37500000000000    0.25000000000000    0.62500000000000    Al   (12a)
   0.12500000000000    0.75000000000000    0.87500000000000    Al   (12a)
   0.62500000000000    0.37500000000000    0.25000000000000    Al   (12a)
   0.87500000000000    0.12500000000000    0.75000000000000    Al   (12a)
   0.25000000000000    0.12500000000000    0.87500000000000    N    (12b)
   0.75000000000000    0.37500000000000    0.62500000000000    N    (12b)
   0.87500000000000    0.25000000000000    0.12500000000000    N    (12b)
   0.62500000000000    0.75000000000000    0.37500000000000    N    (12b)
   0.12500000000000    0.87500000000000    0.25000000000000    N    (12b)
   0.37500000000000    0.62500000000000    0.75000000000000    N    (12b)
```

$\gamma$-$Fe_4N$ ($L'1_0$): A4B_cP5_221_bc_a - CIF

```
# CIF file
data_findsym-output
_audit_creation_method FINDSYM

_chemical_name_mineral 'Fe4N'
_chemical_formula_sum 'Fe4 N'

loop_
_publ_author_name
 'H. Jacobs'
 'R. Rechenbach'
 'U. Zachwieja'
_journal_name_full_name
;
 Journal of Alloys and Compounds
;
_journal_volume 227
_journal_year 1995
```

```
_journal_page_first 10
_journal_page_last 17
_publ_Section_title
;
 Structure determination of $\gamma\'$-Fe$_{4}$N and $\epsilon$-Fe$_{3}
      ↪ $N
;

_aflow_title '$\gamma'$-Fe$_{4}$N ($L\'1_{0}$) Structure'
_aflow_proto 'A4B_cP5_221_bc_a'
_aflow_params 'a'
_aflow_params_values '3.79'
_aflow_Strukturbericht '$L\'1_{0}$'
_aflow_Pearson 'cP5'

_symmetry_space_group_name_H-M "P 4/m -3 2/m"
_symmetry_Int_Tables_number 221

_cell_length_a    3.79000
_cell_length_b    3.79000
_cell_length_c    3.79000
_cell_angle_alpha 90.00000
_cell_angle_beta  90.00000
_cell_angle_gamma 90.00000

loop_
_space_group_symop_id
_space_group_symop_operation_xyz
1 x,y,z
2 x,-y,-z
3 -x,y,-z
4 -x,-y,z
5 y,z,x
6 y,-z,-x
7 -y,z,-x
8 -y,-z,x
9 z,x,y
10 z,-x,-y
11 -z,x,-y
12 -z,-x,y
13 -y,-x,-z
14 -y,x,z
15 y,-x,z
16 y,x,-z
17 -x,-z,-y
18 -x,z,y
19 x,-z,y
20 x,z,-y
21 -z,-y,-x
22 -z,y,x
23 z,-y,x
24 z,y,-x
25 -x,-y,-z
26 -x,y,z
27 x,-y,z
28 x,y,-z
29 -y,-z,-x
30 -y,z,x
31 y,-z,x
32 y,z,-x
33 -z,-x,-y
34 -z,x,y
35 z,-x,y
36 z,x,-y
37 y,x,z
38 y,-x,-z
39 -y,x,-z
40 -y,-x,z
41 x,z,y
42 x,-z,-y
43 -x,z,-y
44 -x,-z,y
45 z,y,x
46 z,-y,-x
47 -z,y,-x
48 -z,-y,x

loop_
_atom_site_label
_atom_site_type_symbol
_atom_site_symmetry_multiplicity
_atom_site_Wyckoff_label
_atom_site_fract_x
_atom_site_fract_y
_atom_site_fract_z
_atom_site_occupancy
N1 N 1 a 0.00000 0.00000 0.00000 1.00000
Fe1 Fe 1 b 0.50000 0.50000 0.50000 1.00000
Fe2 Fe 3 c 0.00000 0.50000 0.50000 1.00000
```

$\gamma$-$Fe_4N$ ($L'1_0$): A4B_cP5_221_bc_a - POSCAR

```
A4B_cP5_221_bc_a & a --params=3.79 & Pm-3m O_{h}^{1} #221 (abc) & cP5 &
      ↪ $L\'1_{0}$ & Fe4N & Fe4N & H. Jacobs and R. Rechenbach and U.
      ↪ Zachwieja, J. Alloys Compd. 227, 10-17 (1995)
   1.00000000000000
   3.79000000000000   0.00000000000000   0.00000000000000
   0.00000000000000   3.79000000000000   0.00000000000000
   0.00000000000000   0.00000000000000   3.79000000000000
    Fe     N
     4     1
Direct
   0.50000000000000   0.50000000000000   0.50000000000000   Fe   (1b)
   0.00000000000000   0.50000000000000   0.50000000000000   Fe   (3c)
   0.50000000000000   0.00000000000000   0.50000000000000   Fe   (3c)
   0.50000000000000   0.50000000000000   0.00000000000000   Fe   (3c)
   0.00000000000000   0.00000000000000   0.00000000000000   N    (1a)
```

Predicted High-Pressure $YCaH_{12}$: AB12C_cP14_221_a_h_b - CIF

```
# CIF file
data_findsym-output
_audit_creation_method FINDSYM

_chemical_name_mineral 'CaH12Y'
_chemical_formula_sum 'Ca H12 Y'

loop_
_publ_author_name
 'H. Xie'
 'D. Duan'
 'Z. Shao'
 'H. Song'
 'Y. Wang'
 'X. Xiao'
 'D. Li'
 'F. Tian'
 'B. Liu'
 'T. Cui'
_journal_name_full_name
;
 Journal of Physics: Condensed Matter
;
_journal_volume 31
_journal_year 2019
_journal_page_first 245404
_journal_page_last 245404
_publ_Section_title
;
 High-temperature superconductivity in ternary clathrate YCaH$_{12}$
      ↪ under high pressures
;

_aflow_title 'Predicted High-Pressure YCaH$_{12}$ Structure'
_aflow_proto 'AB12C_cP14_221_a_h_b'
_aflow_params 'a,x_{3}'
_aflow_params_values '3.4708,0.24126'
_aflow_Strukturbericht 'None'
_aflow_Pearson 'cP14'

_symmetry_space_group_name_H-M "P 4/m -3 2/m"
_symmetry_Int_Tables_number 221

_cell_length_a    3.47080
_cell_length_b    3.47080
_cell_length_c    3.47080
_cell_angle_alpha 90.00000
_cell_angle_beta  90.00000
_cell_angle_gamma 90.00000

loop_
_space_group_symop_id
_space_group_symop_operation_xyz
1 x,y,z
2 x,-y,-z
3 -x,y,-z
4 -x,-y,z
5 y,z,x
6 y,-z,-x
7 -y,z,-x
8 -y,-z,x
9 z,x,y
10 z,-x,-y
11 -z,x,-y
12 -z,-x,y
13 -y,-x,-z
14 -y,x,z
15 y,-x,z
16 y,x,-z
17 -x,-z,-y
18 -x,z,y
19 x,-z,y
20 x,z,-y
21 -z,-y,-x
22 -z,y,x
23 z,-y,x
24 z,y,-x
25 -x,-y,-z
26 -x,y,z
27 x,-y,z
28 x,y,-z
29 -y,-z,-x
30 -y,z,x
31 y,-z,x
32 y,z,-x
33 -z,-x,-y
34 -z,x,y
35 z,-x,y
36 z,x,-y
37 y,x,z
38 y,-x,-z
39 -y,x,-z
40 -y,-x,z
41 x,z,y
42 x,-z,-y
43 -x,z,-y
44 -x,-z,y
45 z,y,x
46 z,-y,-x
47 -z,y,-x
```

```
48 -z,-y,x

loop_
_atom_site_label
_atom_site_type_symbol
_atom_site_symmetry_multiplicity
_atom_site_Wyckoff_label
_atom_site_fract_x
_atom_site_fract_y
_atom_site_fract_z
_atom_site_occupancy
Ca1 Ca 1 a 0.00000 0.00000 0.00000 1.00000
Y1 Y 1 b 0.50000 0.50000 0.50000 1.00000
H1 H 12 h 0.24126 0.50000 0.00000 1.00000
```

Predicted High-Pressure $YCaH_{12}$: AB12C_cP14_221_a_h_b - POSCAR

```
AB12C_cP14_221_a_h_b & a,x3 --params=3.4708,0.24126 & Pm-3m O_{h}^{1} #
    ↪ 221 (abh) & cP14 & None & CaH12Y & CaH12Y & H. Xie et al., J.
    ↪ Phys.: Condens. Matter 31, 245404(2019)
   1.00000000000000
   3.47080000000000   0.00000000000000   0.00000000000000
   0.00000000000000   3.47080000000000   0.00000000000000
   0.00000000000000   0.00000000000000   3.47080000000000
    Ca      H      Y
     1     12      1
Direct
   0.00000000000000   0.00000000000000   0.00000000000000   Ca   (1a)
   0.24126000000000   0.50000000000000   0.00000000000000   H   (12h)
  -0.24126000000000   0.50000000000000   0.00000000000000   H   (12h)
   0.00000000000000   0.24126000000000   0.50000000000000   H   (12h)
   0.00000000000000  -0.24126000000000   0.50000000000000   H   (12h)
   0.50000000000000   0.00000000000000   0.24126000000000   H   (12h)
   0.50000000000000   0.00000000000000  -0.24126000000000   H   (12h)
   0.50000000000000   0.24126000000000   0.00000000000000   H   (12h)
   0.50000000000000  -0.24126000000000   0.00000000000000   H   (12h)
   0.24126000000000   0.00000000000000   0.50000000000000   H   (12h)
  -0.24126000000000   0.00000000000000   0.50000000000000   H   (12h)
   0.00000000000000   0.50000000000000  -0.24126000000000   H   (12h)
   0.00000000000000   0.50000000000000   0.24126000000000   H   (12h)
   0.50000000000000   0.50000000000000   0.50000000000000   Y   (1b)
```

$NH_4NO_3$ I ($G0_8$): AB_cP2_221_a_b - CIF

```
# CIF file
data_findsym-output
_audit_creation_method FINDSYM

_chemical_name_mineral 'NH4NO3'
_chemical_formula_sum '(NH4) (NO3)'

loop_
_publ_author_name
 'F. C. Kracek'
 'S. B. Hendricks'
 'E. Posnjak'
_journal_name_full_name
;
 Nature
;
_journal_volume 128
_journal_year 1931
_journal_page_first 410
_journal_page_last 411
_publ_Section_title
;
 Group Rotation in Solid Ammonium and Calcium Nitrates
;

# Found in Strukturbericht Band II 1928-1932, 1937

_aflow_title 'NH$_{4}$NO$_{3}$ I ($G0_{8}$) Structure'
_aflow_proto 'AB_cP2_221_a_b'
_aflow_params 'a'
_aflow_params_values '4.4'
_aflow_Strukturbericht '$G0_{8}$'
_aflow_Pearson 'cP2'

_symmetry_space_group_name_H-M "P 4/m -3 2/m"
_symmetry_Int_Tables_number 221

_cell_length_a      4.40000
_cell_length_b      4.40000
_cell_length_c      4.40000
_cell_angle_alpha  90.00000
_cell_angle_beta   90.00000
_cell_angle_gamma  90.00000

loop_
_space_group_symop_id
_space_group_symop_operation_xyz
1 x,y,z
2 x,-y,-z
3 -x,y,-z
4 -x,-y,z
5 y,z,x
6 y,-z,-x
7 -y,z,-x
8 -y,-z,x
9 z,x,y
10 z,-x,-y
11 -z,x,-y
12 -z,-x,y
13 -y,-x,-z
14 -y,x,z
15 y,-x,z
16 y,x,-z
17 -x,-z,-y
18 -x,z,y
19 x,-z,y
20 x,z,-y
21 -z,-y,-x
22 -z,y,x
23 z,-y,x
24 z,y,-x
25 -x,-y,-z
26 -x,y,z
27 x,-y,z
28 x,y,-z
29 -y,-z,-x
30 -y,z,x
31 y,-z,x
32 y,z,-x
33 -z,-x,-y
34 -z,x,y
35 z,-x,y
36 z,x,-y
37 y,x,z
38 y,-x,-z
39 -y,x,-z
40 -y,-x,z
41 x,z,y
42 x,-z,-y
43 -x,z,-y
44 -x,-z,y
45 z,y,x
46 z,-y,-x
47 -z,y,-x
48 -z,-y,x

loop_
_atom_site_label
_atom_site_type_symbol
_atom_site_symmetry_multiplicity
_atom_site_Wyckoff_label
_atom_site_fract_x
_atom_site_fract_y
_atom_site_fract_z
_atom_site_occupancy
NH41 NH4 1 a 0.00000 0.00000 0.00000 1.00000
NO31 NO3 1 b 0.50000 0.50000 0.50000 1.00000
```

$NH_4NO_3$ I ($G0_8$): AB_cP2_221_a_b - POSCAR

```
AB_cP2_221_a_b & a --params=4.4 & Pm-3m O_{h}^{1} #221 (ab) & cP2 & $G0_
    ↪ {8}$ & NH4NO3 & NH4NO3 & F. C. Kracek and S. B. Hendricks and
    ↪ E. Posnjak, Nature 128, 410-411 (1931)
   1.00000000000000
   4.40000000000000   0.00000000000000   0.00000000000000
   0.00000000000000   4.40000000000000   0.00000000000000
   0.00000000000000   0.00000000000000   4.40000000000000
   NH4   NO3
     1     1
Direct
   0.00000000000000   0.00000000000000   0.00000000000000   NH4   (1a)
   0.50000000000000   0.50000000000000   0.50000000000000   NO3   (1b)
```

Dodecatungstophosphoric Acid Hexahydrate [$H_3PW_{12}O_{40}{\cdot}6H_2O$]: A27B52CD12_cP184_224_dl_eh3k_a_k - CIF

```
# CIF file
data_findsym-output
_audit_creation_method FINDSYM

_chemical_name_mineral 'Dodecatungstophosphoric acid hexahydrate'
_chemical_formula_sum 'H27 O52 P W12'

loop_
_publ_author_name
 'G. M. Brown'
 'M.-R. {Noe-Spirlet}'
 'W. R. Busing'
 'H. A. Levy'
_journal_name_full_name
;
 Acta Crystallographica Section B: Structural Science
;
_journal_volume 33
_journal_year 1977
_journal_page_first 1038
_journal_page_last 1046
_publ_Section_title
;
 Dodecatungstophosphoric acid hexahydrate, (H$_{5}$O$_{2}$$^{+}$)$_{3}$(
    ↪ PW$_{12}$O$_{40}$$^{3-}$). The true structure of Keggin\'s `
    ↪ pentahydrate\' from single-crystal X-ray and neutron
    ↪ diffraction data
;

_aflow_title 'Dodecatungstophosphoric Acid Hexahydrate [H$_{3}$PW$_{12}
    ↪ $O$_{40}$$\cdot$6H$_{2}$O] Structure'
_aflow_proto 'A27B52CD12_cP184_224_dl_eh3k_a_k'
_aflow_params 'a,x_{3},x_{4},x_{5},z_{5},x_{6},z_{6},x_{7},z_{7},x_{8},
    ↪ z_{8},x_{9},y_{9},z_{9}'
_aflow_params_values '12.506,0.32273,0.34752,0.65683,-0.00684,0.37155,
    ↪ 0.52447,0.55441,0.23276,0.45688,0.25822,0.24525,0.61259,0.68288
    ↪ '
_aflow_Strukturbericht 'None'
_aflow_Pearson 'cP184'
```

```
_symmetry_space_group_name_H-M "P 42/n -3 2/m (origin choice 2)"
_symmetry_Int_Tables_number 224

_cell_length_a    12.50600
_cell_length_b    12.50600
_cell_length_c    12.50600
_cell_angle_alpha 90.00000
_cell_angle_beta  90.00000
_cell_angle_gamma 90.00000

loop_
_space_group_symop_id
_space_group_symop_operation_xyz
1 x,y,z
2 x,-y+1/2,-z+1/2
3 -x+1/2,y,-z+1/2
4 -x+1/2,-y+1/2,z
5 y,z,x
6 y,-z+1/2,-x+1/2
7 -y+1/2,z,-x+1/2
8 -y+1/2,-z+1/2,x
9 z,x,y
10 z,-x+1/2,-y+1/2
11 -z+1/2,x,-y+1/2
12 -z+1/2,-x+1/2,y
13 -y,-x,-z
14 -y,x+1/2,z+1/2
15 y+1/2,-x,z+1/2
16 y+1/2,x+1/2,-z
17 -x,-z,-y
18 -x,z+1/2,y+1/2
19 x+1/2,-z,y+1/2
20 x+1/2,z+1/2,-y
21 -z,-y,-x
22 -z,y+1/2,x+1/2
23 z+1/2,-y,x+1/2
24 z+1/2,y+1/2,-x
25 -x,-y,-z
26 -x,y+1/2,z+1/2
27 x+1/2,-y,z+1/2
28 x+1/2,y+1/2,-z
29 -y,-z,-x
30 -y,z+1/2,x+1/2
31 y+1/2,-z,x+1/2
32 y+1/2,z+1/2,-x
33 -z,-x,-y
34 -z,x+1/2,y+1/2
35 z+1/2,-x,y+1/2
36 z+1/2,x+1/2,-y
37 y,x,z
38 y,-x+1/2,-z+1/2
39 -y+1/2,x,-z+1/2
40 -y+1/2,-x+1/2,z
41 x,z,y
42 x,-z+1/2,-y+1/2
43 -x+1/2,z,-y+1/2
44 -x+1/2,-z+1/2,y
45 z,y,x
46 z,-y+1/2,-x+1/2
47 -z+1/2,y,-x+1/2
48 -z+1/2,-y+1/2,x

loop_
_atom_site_label
_atom_site_type_symbol
_atom_site_symmetry_multiplicity
_atom_site_Wyckoff_label
_atom_site_fract_x
_atom_site_fract_y
_atom_site_fract_z
_atom_site_occupancy
P1 P 2 a 0.25000 0.25000 0.25000 1.00000
H1 H 6 d 0.25000 0.75000 0.75000 1.00000
O1 O 8 e 0.32273 0.32273 0.32273 1.00000
O2 O 24 h 0.34752 0.25000 0.75000 0.50000
O3 O 24 k 0.65683 0.65683 -0.00684 1.00000
O4 O 24 k 0.37155 0.37155 0.52447 1.00000
O5 O 24 k 0.55441 0.55441 0.23276 1.00000
W1 W 24 k 0.45688 0.45688 0.25822 1.00000
H2 H 48 l 0.24525 0.61259 0.68288 0.50000
```

Dodecatungstophosphoric Acid Hexahydrate [$H_3PW_{12}O_{40}\cdot 6H_2O$]: A27B52CD12_cP184_224_dl_eh3k_a_k - POSCAR

```
A27B52CD12_cP184_224_dl_eh3k_a_k & a,x3,x4,x5,z5,x6,z6,x7,z7,x8,z8,x9,y9
   ↪ ,z9 --params=12.506,0.32273,0.34752,0.65683,-0.00684,0.37155,
   ↪ 0.52447,0.55441,0.23276,0.45688,0.25822,0.24525,0.61259,0.68288
   ↪ & Pn-3m O_{h}^{4} #224 (adehk^4l) & cP184 & None & H15O46PW12
   ↪ & Dodecatungstophosphoric acid hexahydrate & G. M. Brown et al.
   ↪ , Acta Crystallogr. Sect. B Struct. Sci. 33, 1038-1046 (1977)
   1.00000000000000
  12.50600000000000   0.00000000000000   0.00000000000000
   0.00000000000000  12.50600000000000   0.00000000000000
   0.00000000000000   0.00000000000000  12.50600000000000
     H      O      P      W
    54    104      2     24
Direct
   0.25000000000000   0.75000000000000   0.75000000000000   H   (6d)
   0.75000000000000   0.25000000000000   0.75000000000000   H   (6d)
   0.75000000000000   0.75000000000000   0.25000000000000   H   (6d)
   0.25000000000000   0.75000000000000   0.25000000000000   H   (6d)
   0.75000000000000   0.25000000000000   0.25000000000000   H   (6d)
   0.25000000000000   0.25000000000000   0.75000000000000   H   (6d)
   0.24525000000000   0.61259000000000   0.68288000000000   H  (48l)
   0.25475000000000  -0.11259000000000   0.68288000000000   H  (48l)
   0.25475000000000   0.61259000000000  -0.18288000000000   H  (48l)
   0.24525000000000  -0.11259000000000  -0.18288000000000   H  (48l)
   0.68288000000000   0.24525000000000   0.61259000000000   H  (48l)
   0.68288000000000   0.25475000000000  -0.11259000000000   H  (48l)
  -0.18288000000000   0.25475000000000   0.61259000000000   H  (48l)
  -0.18288000000000   0.24525000000000  -0.11259000000000   H  (48l)
   0.61259000000000   0.68288000000000   0.24525000000000   H  (48l)
  -0.11259000000000   0.68288000000000   0.25475000000000   H  (48l)
   0.61259000000000  -0.18288000000000   0.25475000000000   H  (48l)
  -0.11259000000000  -0.18288000000000   0.24525000000000   H  (48l)
   1.11259000000000   0.74525000000000  -0.68288000000000   H  (48l)
  -0.61259000000000  -0.24525000000000  -0.68288000000000   H  (48l)
   1.11259000000000  -0.24525000000000   1.18288000000000   H  (48l)
  -0.61259000000000   0.74525000000000   1.18288000000000   H  (48l)
   0.74525000000000   1.18288000000000  -0.61259000000000   H  (48l)
  -0.24525000000000   1.18288000000000   1.11259000000000   H  (48l)
  -0.24525000000000  -0.68288000000000  -0.61259000000000   H  (48l)
   0.74525000000000  -0.68288000000000   1.11259000000000   H  (48l)
   1.18288000000000   1.11259000000000  -0.24525000000000   H  (48l)
   1.18288000000000  -0.61259000000000   0.74525000000000   H  (48l)
  -0.68288000000000   1.11259000000000   0.74525000000000   H  (48l)
  -0.68288000000000  -0.61259000000000  -0.24525000000000   H  (48l)
  -0.24525000000000  -0.61259000000000  -0.68288000000000   H  (48l)
   0.74525000000000   1.11259000000000  -0.68288000000000   H  (48l)
   0.74525000000000  -0.61259000000000   1.18288000000000   H  (48l)
  -0.24525000000000   1.11259000000000   1.18288000000000   H  (48l)
  -0.68288000000000  -0.24525000000000  -0.61259000000000   H  (48l)
  -0.68288000000000   0.74525000000000   1.11259000000000   H  (48l)
   1.18288000000000   0.74525000000000  -0.61259000000000   H  (48l)
   1.18288000000000  -0.24525000000000   1.11259000000000   H  (48l)
  -0.61259000000000  -0.68288000000000  -0.24525000000000   H  (48l)
   1.11259000000000  -0.68288000000000   0.74525000000000   H  (48l)
  -0.61259000000000   1.18288000000000   0.74525000000000   H  (48l)
   1.11259000000000   1.18288000000000  -0.24525000000000   H  (48l)
  -0.11259000000000   0.25475000000000   0.68288000000000   H  (48l)
   0.61259000000000   0.24525000000000   0.68288000000000   H  (48l)
  -0.11259000000000   0.24525000000000  -0.18288000000000   H  (48l)
   0.61259000000000   0.25475000000000  -0.18288000000000   H  (48l)
   0.25475000000000  -0.18288000000000   0.61259000000000   H  (48l)
   0.24525000000000  -0.18288000000000  -0.11259000000000   H  (48l)
   0.24525000000000   0.68288000000000   0.61259000000000   H  (48l)
   0.25475000000000   0.68288000000000  -0.11259000000000   H  (48l)
  -0.18288000000000  -0.11259000000000   0.24525000000000   H  (48l)
  -0.18288000000000   0.61259000000000   0.25475000000000   H  (48l)
   0.68288000000000  -0.11259000000000   0.25475000000000   H  (48l)
   0.68288000000000   0.61259000000000   0.24525000000000   H  (48l)
   0.32273000000000   0.32273000000000   0.32273000000000   O   (8e)
   0.17727000000000   0.17727000000000   0.32273000000000   O   (8e)
   0.17727000000000   0.32273000000000   0.17727000000000   O   (8e)
   0.32273000000000   0.17727000000000   0.17727000000000   O   (8e)
   0.82273000000000   0.82273000000000  -0.32273000000000   O   (8e)
  -0.32273000000000  -0.32273000000000  -0.32273000000000   O   (8e)
   0.82273000000000  -0.32273000000000   0.82273000000000   O   (8e)
  -0.32273000000000   0.82273000000000   0.82273000000000   O   (8e)
   0.34752000000000   0.25000000000000   0.75000000000000   O  (24h)
   0.15248000000000   0.25000000000000   0.75000000000000   O  (24h)
   0.75000000000000   0.34752000000000   0.25000000000000   O  (24h)
   0.75000000000000   0.15248000000000   0.25000000000000   O  (24h)
   0.25000000000000   0.75000000000000   0.34752000000000   O  (24h)
   0.25000000000000   0.75000000000000   0.15248000000000   O  (24h)
   0.75000000000000   0.84752000000000   0.25000000000000   O  (24h)
   0.75000000000000  -0.34752000000000   0.25000000000000   O  (24h)
   0.84752000000000   0.25000000000000   0.75000000000000   O  (24h)
  -0.34752000000000   0.25000000000000   0.75000000000000   O  (24h)
   0.25000000000000   0.75000000000000  -0.34752000000000   O  (24h)
   0.25000000000000   0.75000000000000   0.84752000000000   O  (24h)
  -0.34752000000000   0.75000000000000   0.25000000000000   O  (24h)
   0.84752000000000   0.75000000000000   0.25000000000000   O  (24h)
   0.25000000000000  -0.34752000000000   0.75000000000000   O  (24h)
   0.25000000000000   0.84752000000000   0.75000000000000   O  (24h)
   0.75000000000000   0.25000000000000  -0.34752000000000   O  (24h)
   0.75000000000000   0.25000000000000   0.84752000000000   O  (24h)
   0.25000000000000   0.15248000000000   0.75000000000000   O  (24h)
   0.25000000000000   0.34752000000000   0.75000000000000   O  (24h)
   0.15248000000000   0.75000000000000   0.25000000000000   O  (24h)
   0.34752000000000   0.75000000000000   0.25000000000000   O  (24h)
   0.75000000000000   0.25000000000000   0.34752000000000   O  (24h)
   0.75000000000000   0.25000000000000   0.15248000000000   O  (24h)
   0.65683000000000   0.65683000000000  -0.00684000000000   O  (24k)
  -0.15683000000000  -0.15683000000000  -0.00684000000000   O  (24k)
  -0.15683000000000   0.65683000000000   0.50684000000000   O  (24k)
   0.65683000000000  -0.15683000000000   0.50684000000000   O  (24k)
  -0.00684000000000   0.65683000000000   0.65683000000000   O  (24k)
  -0.00684000000000  -0.15683000000000  -0.15683000000000   O  (24k)
   0.50684000000000  -0.15683000000000   0.65683000000000   O  (24k)
   0.50684000000000   0.65683000000000  -0.15683000000000   O  (24k)
   0.65683000000000  -0.00684000000000   0.65683000000000   O  (24k)
  -0.15683000000000  -0.00684000000000  -0.15683000000000   O  (24k)
   0.65683000000000   0.50684000000000  -0.15683000000000   O  (24k)
  -0.15683000000000   0.50684000000000   0.65683000000000   O  (24k)
   1.15683000000000   1.15683000000000   0.00684000000000   O  (24k)
  -0.65683000000000  -0.65683000000000   0.00684000000000   O  (24k)
   1.15683000000000  -0.65683000000000   0.49316000000000   O  (24k)
  -0.65683000000000   1.15683000000000   0.49316000000000   O  (24k)
   1.15683000000000   0.49316000000000  -0.65683000000000   O  (24k)
  -0.65683000000000   0.49316000000000   1.15683000000000   O  (24k)
  -0.65683000000000   0.00684000000000  -0.65683000000000   O  (24k)
   1.15683000000000   0.00684000000000   1.15683000000000   O  (24k)
   0.49316000000000   1.15683000000000  -0.65683000000000   O  (24k)
   0.49316000000000  -0.65683000000000   1.15683000000000   O  (24k)
   0.00684000000000   1.15683000000000   1.15683000000000   O  (24k)
   0.00684000000000  -0.65683000000000  -0.65683000000000   O  (24k)
   0.37155000000000   0.37155000000000   0.52447000000000   O  (24k)
   0.12845000000000   0.12845000000000   0.52447000000000   O  (24k)
   0.12845000000000   0.37155000000000  -0.02447000000000   O  (24k)
```

```
   0.37155000000000    0.12845000000000   -0.02447000000000     O   (24k)
   0.52447000000000    0.37155000000000    0.37155000000000     O   (24k)
   0.52447000000000    0.12845000000000    0.12845000000000     O   (24k)
  -0.02447000000000    0.12845000000000    0.37155000000000     O   (24k)
  -0.02447000000000    0.37155000000000    0.12845000000000     O   (24k)
   0.37155000000000    0.52447000000000    0.37155000000000     O   (24k)
   0.12845000000000    0.52447000000000    0.12845000000000     O   (24k)
   0.37155000000000   -0.02447000000000    0.12845000000000     O   (24k)
   0.12845000000000   -0.02447000000000    0.37155000000000     O   (24k)
   0.87155000000000    0.87155000000000   -0.52447000000000     O   (24k)
  -0.37155000000000   -0.37155000000000   -0.52447000000000     O   (24k)
   0.87155000000000   -0.37155000000000    1.02447000000000     O   (24k)
  -0.37155000000000    0.87155000000000    1.02447000000000     O   (24k)
   0.87155000000000    1.02447000000000   -0.37155000000000     O   (24k)
  -0.37155000000000    1.02447000000000    0.87155000000000     O   (24k)
  -0.37155000000000   -0.52447000000000   -0.37155000000000     O   (24k)
   0.87155000000000   -0.52447000000000    0.87155000000000     O   (24k)
   1.02447000000000    0.87155000000000   -0.37155000000000     O   (24k)
   1.02447000000000   -0.37155000000000    0.87155000000000     O   (24k)
  -0.52447000000000    0.87155000000000    0.87155000000000     O   (24k)
  -0.52447000000000   -0.37155000000000   -0.37155000000000     O   (24k)
   0.55441000000000    0.55441000000000    0.23276000000000     O   (24k)
  -0.05441000000000   -0.05441000000000    0.23276000000000     O   (24k)
  -0.05441000000000    0.55441000000000    0.26724000000000     O   (24k)
   0.55441000000000   -0.05441000000000    0.26724000000000     O   (24k)
   0.23276000000000    0.55441000000000    0.55441000000000     O   (24k)
   0.23276000000000   -0.05441000000000   -0.05441000000000     O   (24k)
   0.26724000000000   -0.05441000000000    0.55441000000000     O   (24k)
   0.26724000000000    0.55441000000000   -0.05441000000000     O   (24k)
   0.55441000000000    0.23276000000000    0.55441000000000     O   (24k)
  -0.05441000000000    0.23276000000000   -0.05441000000000     O   (24k)
   0.55441000000000    0.26724000000000   -0.05441000000000     O   (24k)
  -0.05441000000000    0.26724000000000    0.55441000000000     O   (24k)
   1.05441000000000    1.05441000000000   -0.23276000000000     O   (24k)
  -0.55441000000000   -0.55441000000000   -0.23276000000000     O   (24k)
   1.05441000000000   -0.55441000000000    0.73276000000000     O   (24k)
  -0.55441000000000    1.05441000000000    0.73276000000000     O   (24k)
   1.05441000000000    0.73276000000000   -0.55441000000000     O   (24k)
  -0.55441000000000    0.73276000000000    1.05441000000000     O   (24k)
  -0.55441000000000   -0.23276000000000   -0.55441000000000     O   (24k)
   1.05441000000000   -0.23276000000000    1.05441000000000     O   (24k)
   0.73276000000000    1.05441000000000   -0.55441000000000     O   (24k)
   0.73276000000000   -0.55441000000000    1.05441000000000     O   (24k)
  -0.23276000000000    1.05441000000000    1.05441000000000     O   (24k)
  -0.23276000000000   -0.55441000000000   -0.55441000000000     O   (24k)
   0.25000000000000    0.25000000000000    0.25000000000000     P    (2a)
   0.75000000000000    0.75000000000000    0.75000000000000     P    (2a)
   0.45688000000000    0.45688000000000    0.25822000000000     W   (24k)
   0.04312000000000    0.04312000000000    0.25822000000000     W   (24k)
   0.04312000000000    0.45688000000000    0.24178000000000     W   (24k)
   0.45688000000000    0.04312000000000    0.24178000000000     W   (24k)
   0.25822000000000    0.45688000000000    0.45688000000000     W   (24k)
   0.25822000000000    0.04312000000000    0.04312000000000     W   (24k)
   0.24178000000000    0.04312000000000    0.45688000000000     W   (24k)
   0.24178000000000    0.45688000000000    0.04312000000000     W   (24k)
   0.45688000000000    0.25822000000000    0.45688000000000     W   (24k)
   0.04312000000000    0.25822000000000    0.04312000000000     W   (24k)
   0.45688000000000    0.24178000000000    0.04312000000000     W   (24k)
   0.04312000000000    0.24178000000000    0.45688000000000     W   (24k)
   0.95688000000000    0.95688000000000   -0.25822000000000     W   (24k)
  -0.45688000000000   -0.45688000000000   -0.25822000000000     W   (24k)
   0.95688000000000   -0.45688000000000    0.75822000000000     W   (24k)
  -0.45688000000000    0.95688000000000    0.75822000000000     W   (24k)
   0.95688000000000    0.75822000000000   -0.45688000000000     W   (24k)
  -0.45688000000000    0.75822000000000    0.95688000000000     W   (24k)
  -0.45688000000000   -0.25822000000000   -0.45688000000000     W   (24k)
   0.95688000000000   -0.25822000000000    0.95688000000000     W   (24k)
   0.75822000000000    0.95688000000000   -0.45688000000000     W   (24k)
   0.75822000000000   -0.45688000000000    0.95688000000000     W   (24k)
  -0.25822000000000    0.95688000000000    0.95688000000000     W   (24k)
  -0.25822000000000   -0.45688000000000   -0.45688000000000     W   (24k)
```

$Mg_3P_2$ ($D5_5$): A3B2_cP10_224_d_b - CIF

```
# CIF file
data_findsym-output
_audit_creation_method FINDSYM

_chemical_name_mineral 'Mg3P2'
_chemical_formula_sum 'Mg3 P2'

loop_
_publ_author_name
 'L. Passerini'
_journal_name_full_name
;
 Gazzetta Chimica Italiana
;
_journal_volume 58
_journal_year 1928
_journal_page_first 655
_journal_page_last 664
_publ_Section_title
;
 Struttura cristallina di alcuni fosfuri di metalli bivalenti e
    ↪ trivalenti
;

# Found in Strukturbericht Band II 1928-1932, 1937

_aflow_title 'Mg$_{3}$P$_{2}$ ($D5_{5}$) Structure'
_aflow_proto 'A3B2_cP10_224_d_b'
_aflow_params 'a'
_aflow_params_values '5.91207'
_aflow_Strukturbericht '$D5_{5}$'
_aflow_Pearson 'cP10'

_symmetry_space_group_name_H-M "P 42/n -3 2/m (origin choice 2)"
_symmetry_Int_Tables_number 224

_cell_length_a      5.91207
_cell_length_b      5.91207
_cell_length_c      5.91207
_cell_angle_alpha  90.00000
_cell_angle_beta   90.00000
_cell_angle_gamma  90.00000

loop_
_space_group_symop_id
_space_group_symop_operation_xyz
1 x,y,z
2 x,-y+1/2,-z+1/2
3 -x+1/2,y,-z+1/2
4 -x+1/2,-y+1/2,z
5 y,z,x
6 y,-z+1/2,-x+1/2
7 -y+1/2,z,-x+1/2
8 -y+1/2,-z+1/2,x
9 z,x,y
10 z,-x+1/2,-y+1/2
11 -z+1/2,x,-y+1/2
12 -z+1/2,-x+1/2,y
13 -y,-x,-z
14 -y,x+1/2,z+1/2
15 y+1/2,-x,z+1/2
16 y+1/2,x+1/2,-z
17 -x,-z,-y
18 -x,z+1/2,y+1/2
19 x+1/2,-z,y+1/2
20 x+1/2,z+1/2,-y
21 -z,-y,-x
22 -z,y+1/2,x+1/2
23 z+1/2,-y,x+1/2
24 z+1/2,y+1/2,-x
25 -x,-y,-z
26 -x,y+1/2,z+1/2
27 x+1/2,-y,z+1/2
28 x+1/2,y+1/2,-z
29 -y,-z,-x
30 -y,z+1/2,x+1/2
31 y+1/2,-z,x+1/2
32 y+1/2,z+1/2,-x
33 -z,-x,-y
34 -z,x+1/2,y+1/2
35 z+1/2,-x,y+1/2
36 z+1/2,x+1/2,-y
37 y,x,z
38 y,-x+1/2,-z+1/2
39 -y+1/2,x,-z+1/2
40 -y+1/2,-x+1/2,z
41 x,z,y
42 x,-z+1/2,-y+1/2
43 -x+1/2,z,-y+1/2
44 -x+1/2,-z+1/2,y
45 z,y,x
46 z,-y+1/2,-x+1/2
47 -z+1/2,y,-x+1/2
48 -z+1/2,-y+1/2,x

loop_
_atom_site_label
_atom_site_type_symbol
_atom_site_symmetry_multiplicity
_atom_site_Wyckoff_label
_atom_site_fract_x
_atom_site_fract_y
_atom_site_fract_z
_atom_site_occupancy
P1 P 4 b 0.00000 0.00000 0.00000 1.00000
Mg1 Mg 6 d 0.25000 0.75000 0.75000 1.00000
```

$Mg_3P_2$ ($D5_5$): A3B2_cP10_224_d_b - POSCAR

```
A3B2_cP10_224_d_b & a --params=5.91207 & Pn-3m O_{h}^{4} #224 (bd) &
    ↪ cP10 & $D5_{5}$ & Mg3P2 & Mg3P2 & L. Passerini, Gazz. Chim.
    ↪ Ital. 58, 655-664 (1928)
   1.00000000000000
   5.91207000000000    0.00000000000000    0.00000000000000
   0.00000000000000    5.91207000000000    0.00000000000000
   0.00000000000000    0.00000000000000    5.91207000000000
    Mg         P
     6         4
Direct
   0.25000000000000    0.75000000000000    0.75000000000000    Mg   (6d)
   0.75000000000000    0.25000000000000    0.75000000000000    Mg   (6d)
   0.75000000000000    0.75000000000000    0.25000000000000    Mg   (6d)
   0.25000000000000    0.75000000000000    0.25000000000000    Mg   (6d)
   0.75000000000000    0.25000000000000    0.25000000000000    Mg   (6d)
   0.25000000000000    0.25000000000000    0.75000000000000    Mg   (6d)
   0.00000000000000    0.00000000000000    0.00000000000000     P   (4b)
   0.50000000000000    0.50000000000000    0.00000000000000     P   (4b)
   0.50000000000000    0.00000000000000    0.50000000000000     P   (4b)
   0.00000000000000    0.50000000000000    0.50000000000000     P   (4b)
```

$H_3PW_{12}O_{40}{\cdot}3H_2O$: A3B40CD12_cP112_224_d_e3k_a_k - CIF

```
# CIF file
data_findsym-output
_audit_creation_method FINDSYM
```

```
_chemical_name_mineral  '(H3O)3O40PW12'
_chemical_formula_sum  '(H3O)3 O40 P W12'

loop_
_publ_author_name
 'L. Marosi'
 'E. E. Platero'
 'J. Cifre'
 'C. O. Are\\'{a}na'
_journal_name_full_name
;
 Journal of Materials Chemistry
;
_journal_volume 10
_journal_year 2000
_journal_page_first 1949
_journal_page_last 1955
_publ_Section_title
;
 Thermal dehydration of H$_{3+x}$PV$_{x}$M$_{12-x}$O$_{40}$$\cdot$$y$H$_{
      ↪ 2}$O Keggin type heteropolyacids; formation, thermal stability
      ↪  and structure of the anhydrous acids H$_{3}$PM$_{12}$O$_{40}$
      ↪ , of the corresponding anhydrides PM$_{12}$O$_{38.5}$ and of a
      ↪  novel trihydrate H$_{3}$PW$_{12}$O$_{40}$$\cdot$3H$_{2}$O
;

# Found in H$_{3}$PW$_{12}$O$_{40}$$\cdot$3H$_{2}$O Crystal Structure,
      ↪ 2016 Found in H$_{3}$PW$_{12}$O$_{40}$$\cdot$3H$_{2}$O Crystal
      ↪ Structure, {in: Inorganic Solid Phases, SpringerMaterials (
      ↪ online database)},

_aflow_title 'H$_{3}$PW$_{12}$O$_{40}$$\cdot$3H$_{2}$O Structure'
_aflow_proto 'A3B40CD12_cP112_224_d_e3k_a_k'
_aflow_params 'a,x_{3},x_{4},z_{4},x_{5},z_{5},x_{6},z_{6},x_{7},z_{7}'
_aflow_params_values '11.75,0.3226,0.6463,0.0133,0.3687,0.5462,0.5629,
      ↪ 0.2412,0.4636,0.2579'
_aflow_Strukturbericht 'None'
_aflow_Pearson 'cP112'

_symmetry_space_group_name_H-M "P 42/n -3 2/m (origin choice 2)"
_symmetry_Int_Tables_number 224

_cell_length_a     11.75000
_cell_length_b     11.75000
_cell_length_c     11.75000
_cell_angle_alpha  90.00000
_cell_angle_beta   90.00000
_cell_angle_gamma  90.00000

loop_
_space_group_symop_id
_space_group_symop_operation_xyz
1 x,y,z
2 x,-y+1/2,-z+1/2
3 -x+1/2,y,-z+1/2
4 -x+1/2,-y+1/2,z
5 y,z,x
6 y,-z+1/2,-x+1/2
7 -y+1/2,z,-x+1/2
8 -y+1/2,-z+1/2,x
9 z,x,y
10 z,-x+1/2,-y+1/2
11 -z+1/2,x,-y+1/2
12 -z+1/2,-x+1/2,y
13 -y,-x,-z
14 -y,x+1/2,z+1/2
15 y+1/2,-x,z+1/2
16 y+1/2,x+1/2,-z
17 -x,-z,-y
18 -x,z+1/2,y+1/2
19 x+1/2,-z,y+1/2
20 x+1/2,z+1/2,-y
21 -z,-y,-x
22 -z,y+1/2,x+1/2
23 z+1/2,-y,x+1/2
24 z+1/2,y+1/2,-x
25 -x,-y,-z
26 -x,y+1/2,z+1/2
27 x+1/2,-y,z+1/2
28 x+1/2,y+1/2,-z
29 -y,-z,-x
30 -y,z+1/2,x+1/2
31 y+1/2,-z,x+1/2
32 y+1/2,z+1/2,-x
33 -z,-x,-y
34 -z,x+1/2,y+1/2
35 z+1/2,-x,y+1/2
36 z+1/2,x+1/2,-y
37 y,x,z
38 y,-x+1/2,-z+1/2
39 -y+1/2,x,-z+1/2
40 -y+1/2,-x+1/2,z
41 x,z,y
42 x,-z+1/2,-y+1/2
43 -x+1/2,z,-y+1/2
44 -x+1/2,-z+1/2,y
45 z,y,x
46 z,-y+1/2,-x+1/2
47 -z+1/2,y,-x+1/2
48 -z+1/2,-y+1/2,x

loop_
_atom_site_label
_atom_site_type_symbol
_atom_site_symmetry_multiplicity
_atom_site_Wyckoff_label
_atom_site_fract_x
_atom_site_fract_y
_atom_site_fract_z
_atom_site_occupancy
P1 P 2 a 0.25000 0.25000 0.25000 1.00000
H3O1 H3O 6 d 0.25000 0.75000 0.75000 1.00000
O1 O 8 e 0.32260 0.32260 0.32260 1.00000
O2 O 24 k 0.64630 0.64630 0.01330 1.00000
O3 O 24 k 0.36870 0.36870 0.54620 1.00000
O4 O 24 k 0.56290 0.56290 0.24120 1.00000
W1 W 24 k 0.46360 0.46360 0.25790 1.00000
```

$H_3PW_{12}O_{40}{\cdot}3H_2O$: A3B40CD12_cP112_224_d_e3k_a_k - POSCAR

```
A3B40CD12_cP112_224_d_e3k_a_k & a,x3,x4,z4,x5,z5,x6,z6,x7,z7 --params=
      ↪ 11.75,0.3226,0.6463,0.0133,0.3687,0.5462,0.5629,0.2412,0.4636,
      ↪ 0.2579 & Pn-3m O_{h}^{4} #224 (adek^4) & cP112 & None & (H3O)
      ↪ 3O40PW12 & (H3O)3O40PW12 & L. Marosi et al., J. Mater. Chem. 10
      ↪ , 1949-1955 (2000)
   1.00000000000000
  11.75000000000000   0.00000000000000   0.00000000000000
   0.00000000000000  11.75000000000000   0.00000000000000
   0.00000000000000   0.00000000000000  11.75000000000000
   H3O     O     P     W
     6    80     2    24
Direct
   0.25000000000000   0.75000000000000   0.75000000000000  H3O   (6d)
   0.75000000000000   0.25000000000000   0.75000000000000  H3O   (6d)
   0.75000000000000   0.75000000000000   0.25000000000000  H3O   (6d)
   0.25000000000000   0.75000000000000   0.25000000000000  H3O   (6d)
   0.75000000000000   0.25000000000000   0.25000000000000  H3O   (6d)
   0.25000000000000   0.25000000000000   0.75000000000000  H3O   (6d)
   0.32260000000000   0.32260000000000   0.32260000000000   O    (8e)
   0.17740000000000   0.17740000000000   0.32260000000000   O    (8e)
   0.17740000000000   0.32260000000000   0.17740000000000   O    (8e)
   0.32260000000000   0.17740000000000   0.17740000000000   O    (8e)
   0.82260000000000   0.82260000000000  -0.32260000000000   O    (8e)
  -0.32260000000000  -0.32260000000000  -0.32260000000000   O    (8e)
   0.82260000000000  -0.32260000000000   0.82260000000000   O    (8e)
  -0.32260000000000   0.82260000000000   0.82260000000000   O    (8e)
   0.64630000000000   0.64630000000000   0.01330000000000   O   (24k)
  -0.14630000000000  -0.14630000000000   0.01330000000000   O   (24k)
  -0.14630000000000   0.64630000000000   0.48670000000000   O   (24k)
   0.64630000000000  -0.14630000000000   0.48670000000000   O   (24k)
   0.01330000000000   0.64630000000000   0.64630000000000   O   (24k)
   0.01330000000000  -0.14630000000000  -0.14630000000000   O   (24k)
   0.48670000000000  -0.14630000000000   0.64630000000000   O   (24k)
   0.48670000000000   0.64630000000000  -0.14630000000000   O   (24k)
   0.64630000000000   0.01330000000000   0.64630000000000   O   (24k)
  -0.14630000000000   0.01330000000000  -0.14630000000000   O   (24k)
   0.64630000000000   0.48670000000000  -0.14630000000000   O   (24k)
  -0.14630000000000   0.48670000000000   0.64630000000000   O   (24k)
   1.14630000000000   1.14630000000000  -0.01330000000000   O   (24k)
  -0.64630000000000  -0.64630000000000  -0.01330000000000   O   (24k)
   1.14630000000000  -0.64630000000000   0.51330000000000   O   (24k)
  -0.64630000000000   1.14630000000000   0.51330000000000   O   (24k)
   1.14630000000000   0.51330000000000  -0.64630000000000   O   (24k)
  -0.64630000000000   0.51330000000000   1.14630000000000   O   (24k)
  -0.64630000000000  -0.01330000000000  -0.64630000000000   O   (24k)
   1.14630000000000  -0.01330000000000   1.14630000000000   O   (24k)
   0.51330000000000   1.14630000000000  -0.64630000000000   O   (24k)
   0.51330000000000  -0.64630000000000   1.14630000000000   O   (24k)
  -0.01330000000000   1.14630000000000   1.14630000000000   O   (24k)
  -0.01330000000000  -0.64630000000000  -0.64630000000000   O   (24k)
   0.36870000000000   0.36870000000000   0.54620000000000   O   (24k)
   0.13130000000000   0.13130000000000   0.54620000000000   O   (24k)
   0.13130000000000   0.36870000000000  -0.04620000000000   O   (24k)
   0.36870000000000   0.13130000000000  -0.04620000000000   O   (24k)
   0.54620000000000   0.36870000000000   0.36870000000000   O   (24k)
   0.54620000000000   0.13130000000000   0.13130000000000   O   (24k)
  -0.04620000000000   0.13130000000000   0.36870000000000   O   (24k)
  -0.04620000000000   0.36870000000000   0.13130000000000   O   (24k)
   0.36870000000000   0.54620000000000   0.36870000000000   O   (24k)
   0.13130000000000   0.54620000000000   0.13130000000000   O   (24k)
   0.36870000000000  -0.04620000000000   0.13130000000000   O   (24k)
   0.13130000000000  -0.04620000000000   0.36870000000000   O   (24k)
   0.86870000000000   0.86870000000000  -0.54620000000000   O   (24k)
  -0.36870000000000  -0.36870000000000  -0.54620000000000   O   (24k)
   0.86870000000000  -0.36870000000000   1.04620000000000   O   (24k)
  -0.36870000000000   0.86870000000000   1.04620000000000   O   (24k)
   0.86870000000000   1.04620000000000  -0.36870000000000   O   (24k)
  -0.36870000000000   1.04620000000000   0.86870000000000   O   (24k)
  -0.36870000000000  -0.54620000000000  -0.36870000000000   O   (24k)
   0.86870000000000  -0.54620000000000   0.86870000000000   O   (24k)
   1.04620000000000   0.86870000000000  -0.36870000000000   O   (24k)
   1.04620000000000  -0.36870000000000   0.86870000000000   O   (24k)
  -0.54620000000000   0.86870000000000   0.86870000000000   O   (24k)
  -0.54620000000000  -0.36870000000000  -0.36870000000000   O   (24k)
   0.56290000000000   0.56290000000000   0.24120000000000   O   (24k)
  -0.06290000000000  -0.06290000000000   0.24120000000000   O   (24k)
  -0.06290000000000   0.56290000000000   0.25880000000000   O   (24k)
   0.56290000000000  -0.06290000000000   0.25880000000000   O   (24k)
   0.24120000000000   0.56290000000000   0.56290000000000   O   (24k)
   0.24120000000000  -0.06290000000000  -0.06290000000000   O   (24k)
   0.25880000000000  -0.06290000000000   0.56290000000000   O   (24k)
   0.25880000000000   0.56290000000000  -0.06290000000000   O   (24k)
   0.56290000000000   0.24120000000000   0.56290000000000   O   (24k)
  -0.06290000000000   0.24120000000000  -0.06290000000000   O   (24k)
   0.56290000000000   0.25880000000000  -0.06290000000000   O   (24k)
  -0.06290000000000   0.25880000000000   0.56290000000000   O   (24k)
   1.06290000000000   1.06290000000000  -0.24120000000000   O   (24k)
  -0.56290000000000  -0.56290000000000  -0.24120000000000   O   (24k)
   1.06290000000000  -0.56290000000000   0.74120000000000   O   (24k)
  -0.56290000000000   1.06290000000000   0.74120000000000   O   (24k)
```

```
   1.06290000000000   0.74120000000000  -0.56290000000000   O   (24k)
  -0.56290000000000   0.74120000000000   1.06290000000000   O   (24k)
  -0.56290000000000  -0.24120000000000  -0.56290000000000   O   (24k)
   1.06290000000000  -0.24120000000000   1.06290000000000   O   (24k)
   0.74120000000000   1.06290000000000  -0.56290000000000   O   (24k)
   0.74120000000000  -0.56290000000000   1.06290000000000   O   (24k)
  -0.24120000000000   1.06290000000000   1.06290000000000   O   (24k)
  -0.24120000000000  -0.56290000000000  -0.56290000000000   O   (24k)
   0.25000000000000   0.25000000000000   0.25000000000000   P    (2a)
   0.75000000000000   0.75000000000000   0.75000000000000   P    (2a)
   0.46360000000000   0.46360000000000   0.25790000000000   W   (24k)
   0.03640000000000   0.03640000000000   0.25790000000000   W   (24k)
   0.03640000000000   0.46360000000000   0.24210000000000   W   (24k)
   0.46360000000000   0.03640000000000   0.24210000000000   W   (24k)
   0.25790000000000   0.46360000000000   0.46360000000000   W   (24k)
   0.25790000000000   0.03640000000000   0.03640000000000   W   (24k)
   0.24210000000000   0.03640000000000   0.46360000000000   W   (24k)
   0.24210000000000   0.46360000000000   0.03640000000000   W   (24k)
   0.46360000000000   0.25790000000000   0.46360000000000   W   (24k)
   0.03640000000000   0.25790000000000   0.03640000000000   W   (24k)
   0.46360000000000   0.24210000000000   0.03640000000000   W   (24k)
   0.03640000000000   0.24210000000000   0.46360000000000   W   (24k)
   0.96360000000000   0.96360000000000  -0.25790000000000   W   (24k)
  -0.46360000000000  -0.46360000000000  -0.25790000000000   W   (24k)
   0.96360000000000  -0.46360000000000   0.75790000000000   W   (24k)
  -0.46360000000000   0.96360000000000   0.75790000000000   W   (24k)
   0.96360000000000   0.75790000000000  -0.46360000000000   W   (24k)
  -0.46360000000000   0.75790000000000   0.96360000000000   W   (24k)
  -0.46360000000000  -0.25790000000000  -0.46360000000000   W   (24k)
   0.96360000000000  -0.25790000000000   0.96360000000000   W   (24k)
   0.75790000000000   0.96360000000000  -0.46360000000000   W   (24k)
   0.75790000000000  -0.46360000000000   0.96360000000000   W   (24k)
  -0.25790000000000   0.96360000000000   0.96360000000000   W   (24k)
  -0.25790000000000  -0.46360000000000  -0.46360000000000   W   (24k)
```

12-phosphotungstic acid [$H_3PW_{12}O_{40}\cdot 5H_2O$ ($H4_{16}$)]: A5B40CD12_cP116_224_cd_e3k_a_k - CIF

```
# CIF file
data_findsym-output
_audit_creation_method FINDSYM

_chemical_name_mineral '12-phosphotungstic acid'
_chemical_formula_sum '(H2O)5 O40 P W12'

loop_
_publ_author_name
 'J. F. Keggin'
_journal_name_full_name
;
 Proceedings of the Royal Society London A
;
_journal_volume 144
_journal_year 1934
_journal_page_first 75
_journal_page_last 100
_publ_Section_title
;
 The structure and formula of 12-phosphotungstic acid
;

# Found in Dodecatungstophosphoric acid hexahydrate, (H$_{5}$O$_{2}^{+}$
    ↪ )$_{3}$(PW$_{12}$O$_{40}^{3-}$). The true structure of Keggin's
    ↪ 'pentahydrate' from single-crystal X-ray and neutron
    ↪ diffraction data, 1977

_aflow_title '12-phosphotungstic acid [H$_{3}$PW$_{12}$O$_{40}$$\
    ↪ cdot$5H$_{2}$O ($H4_{16}$)] Structure'
_aflow_proto 'A5B40CD12_cP116_224_cd_e3k_a_k'
_aflow_params 'a,x_{4},x_{5},z_{5},x_{6},z_{6},x_{7},z_{7},x_{8},z_{8}'
_aflow_params_values '12.141,0.83154,0.67011,0.51608,0.87272,0.04157,
    ↪ 0.43783,0.74176,-0.0445,0.75577'
_aflow_Strukturbericht '$H4_{16}$'
_aflow_Pearson 'cP116'

_symmetry_space_group_name_H-M "P 42/n -3 2/m (origin choice 2)"
_symmetry_Int_Tables_number 224

_cell_length_a     12.14100
_cell_length_b     12.14100
_cell_length_c     12.14100
_cell_angle_alpha  90.00000
_cell_angle_beta   90.00000
_cell_angle_gamma  90.00000

loop_
_space_group_symop_id
_space_group_symop_operation_xyz
1 x,y,z
2 x,-y+1/2,-z+1/2
3 -x+1/2,y,-z+1/2
4 -x+1/2,-y+1/2,z
5 y,z,x
6 y,-z+1/2,-x+1/2
7 -y+1/2,z,-x+1/2
8 -y+1/2,-z+1/2,x
9 z,x,y
10 z,-x+1/2,-y+1/2
11 -z+1/2,x,-y+1/2
12 -z+1/2,-x+1/2,y
13 -y,-x,-z
14 -y,x+1/2,z+1/2
15 y+1/2,-x,z+1/2
16 y+1/2,x+1/2,-z
17 -x,-z,-y
18 -x,z+1/2,y+1/2
19 x+1/2,-z,y+1/2
20 x+1/2,z+1/2,-y
21 -z,-y,-x
22 -z,y+1/2,x+1/2
23 z+1/2,-y,x+1/2
24 z+1/2,y+1/2,-x
25 -x,-y,-z
26 -x,y+1/2,z+1/2
27 x+1/2,-y,z+1/2
28 x+1/2,y+1/2,-z
29 -y,-z,-x
30 -y,z+1/2,x+1/2
31 y+1/2,-z,x+1/2
32 y+1/2,z+1/2,-x
33 -z,-x,-y
34 -z,x+1/2,y+1/2
35 z+1/2,-x,y+1/2
36 z+1/2,x+1/2,-y
37 y,x,z
38 y,-x+1/2,-z+1/2
39 -y+1/2,x,-z+1/2
40 -y+1/2,-x+1/2,z
41 x,z,y
42 x,-z+1/2,-y+1/2
43 -x+1/2,z,-y+1/2
44 -x+1/2,-z+1/2,y
45 z,y,x
46 z,-y+1/2,-x+1/2
47 -z+1/2,y,-x+1/2
48 -z+1/2,-y+1/2,x

loop_
_atom_site_label
_atom_site_type_symbol
_atom_site_symmetry_multiplicity
_atom_site_Wyckoff_label
_atom_site_fract_x
_atom_site_fract_y
_atom_site_fract_z
_atom_site_occupancy
P1 P 2 a 0.25000 0.25000 0.25000 1.00000
H2O1 H2O 4 c 0.50000 0.50000 0.50000 1.00000
H2O2 H2O 6 d 0.25000 0.75000 0.75000 1.00000
O1 O 8 e 0.83154 0.83154 0.83154 1.00000
O2 O 24 k 0.67011 0.67011 0.51608 1.00000
O3 O 24 k 0.87272 0.87272 0.04157 1.00000
O4 O 24 k 0.43783 0.43783 0.74176 1.00000
W1 W 24 k -0.04450 -0.04450 0.75577 1.00000
```

12-phosphotungstic acid [$H_3PW_{12}O_{40}\cdot 5H_2O$ ($H4_{16}$)]: A5B40CD12_cP116_224_cd_e3k_a_k - POSCAR

```
A5B40CD12_cP116_224_cd_e3k_a_k & a,x4,x5,z5,x6,z6,x7,z7,x8,z8 --params=
    ↪ 12.141,0.83154,0.67011,0.51608,0.87272,0.04157,0.43783,0.74176
    ↪ ,-0.0445,0.75577 & Pn-3m O_{h}^{4} #224 (acdek^4) & cP116 &
    ↪ $H4_{16}$ & [5H2.6O]O40PW12 & 12-phosphotungstic acid & J. F.
    ↪ Keggin, Proc. Roy. Soc. Lond. A 144, 75-100 (1934)
   1.00000000000000
  12.14100000000000   0.00000000000000   0.00000000000000
   0.00000000000000  12.14100000000000   0.00000000000000
   0.00000000000000   0.00000000000000  12.14100000000000
   H2O     O     P     W
    10    80     2    24
Direct
   0.50000000000000   0.50000000000000   0.50000000000000   H2O   (4c)
   0.00000000000000   0.00000000000000   0.50000000000000   H2O   (4c)
   0.00000000000000   0.50000000000000   0.00000000000000   H2O   (4c)
   0.50000000000000   0.00000000000000   0.00000000000000   H2O   (4c)
   0.25000000000000   0.75000000000000   0.75000000000000   H2O   (6d)
   0.75000000000000   0.25000000000000   0.75000000000000   H2O   (6d)
   0.75000000000000   0.75000000000000   0.25000000000000   H2O   (6d)
   0.25000000000000   0.75000000000000   0.25000000000000   H2O   (6d)
   0.75000000000000   0.25000000000000   0.25000000000000   H2O   (6d)
   0.25000000000000   0.25000000000000   0.75000000000000   H2O   (6d)
   0.83154000000000   0.83154000000000   0.83154000000000     O   (8e)
  -0.33154000000000  -0.33154000000000   0.83154000000000     O   (8e)
  -0.33154000000000   0.83154000000000  -0.33154000000000     O   (8e)
   0.83154000000000  -0.33154000000000  -0.33154000000000     O   (8e)
   1.33154000000000   1.33154000000000  -0.83154000000000     O   (8e)
  -0.83154000000000  -0.83154000000000  -0.83154000000000     O   (8e)
   1.33154000000000  -0.83154000000000   1.33154000000000     O   (8e)
  -0.83154000000000   1.33154000000000   1.33154000000000     O   (8e)
   0.67011000000000   0.67011000000000   0.51608000000000     O  (24k)
  -0.17011000000000  -0.17011000000000   0.51608000000000     O  (24k)
  -0.17011000000000   0.67011000000000  -0.01608000000000     O  (24k)
   0.67011000000000  -0.17011000000000  -0.01608000000000     O  (24k)
   0.51608000000000   0.67011000000000   0.67011000000000     O  (24k)
   0.51608000000000  -0.17011000000000  -0.17011000000000     O  (24k)
  -0.01608000000000  -0.17011000000000   0.67011000000000     O  (24k)
  -0.01608000000000   0.67011000000000  -0.17011000000000     O  (24k)
   0.67011000000000   0.51608000000000   0.67011000000000     O  (24k)
  -0.17011000000000   0.51608000000000  -0.17011000000000     O  (24k)
   0.67011000000000  -0.01608000000000  -0.17011000000000     O  (24k)
  -0.17011000000000  -0.01608000000000   0.67011000000000     O  (24k)
   1.17011000000000   1.17011000000000  -0.51608000000000     O  (24k)
  -0.67011000000000  -0.67011000000000  -0.51608000000000     O  (24k)
   1.17011000000000  -0.67011000000000   1.01608000000000     O  (24k)
  -0.67011000000000   1.17011000000000   1.01608000000000     O  (24k)
   1.17011000000000   1.01608000000000  -0.67011000000000     O  (24k)
  -0.67011000000000   1.01608000000000   1.17011000000000     O  (24k)
  -0.67011000000000  -0.51608000000000  -0.67011000000000     O  (24k)
   1.17011000000000  -0.51608000000000   1.17011000000000     O  (24k)
   1.01608000000000   1.17011000000000  -0.67011000000000     O  (24k)
   1.01608000000000  -0.67011000000000   1.17011000000000     O  (24k)
  -0.51608000000000   1.17011000000000   1.17011000000000     O  (24k)
  -0.51608000000000  -0.67011000000000  -0.67011000000000     O  (24k)
```

| | | | | |
|---|---|---|---|---|
| 0.87272000000000 | 0.87272000000000 | 0.04157000000000 | O | (24k) |
| −0.37272000000000 | −0.37272000000000 | 0.04157000000000 | O | (24k) |
| −0.37272000000000 | 0.87272000000000 | 0.45843000000000 | O | (24k) |
| 0.87272000000000 | −0.37272000000000 | 0.45843000000000 | O | (24k) |
| 0.04157000000000 | 0.87272000000000 | 0.87272000000000 | O | (24k) |
| 0.04157000000000 | −0.37272000000000 | −0.37272000000000 | O | (24k) |
| 0.45843000000000 | −0.37272000000000 | 0.87272000000000 | O | (24k) |
| 0.45843000000000 | 0.87272000000000 | −0.37272000000000 | O | (24k) |
| 0.87272000000000 | 0.04157000000000 | 0.87272000000000 | O | (24k) |
| −0.37272000000000 | 0.04157000000000 | −0.37272000000000 | O | (24k) |
| 0.87272000000000 | 0.45843000000000 | −0.37272000000000 | O | (24k) |
| −0.37272000000000 | 0.45843000000000 | 0.87272000000000 | O | (24k) |
| 1.37272000000000 | 1.37272000000000 | −0.04157000000000 | O | (24k) |
| −0.87272000000000 | −0.87272000000000 | −0.04157000000000 | O | (24k) |
| 1.37272000000000 | −0.87272000000000 | 0.54157000000000 | O | (24k) |
| −0.87272000000000 | 1.37272000000000 | 0.54157000000000 | O | (24k) |
| 1.37272000000000 | 0.54157000000000 | −0.87272000000000 | O | (24k) |
| −0.87272000000000 | 0.54157000000000 | 1.37272000000000 | O | (24k) |
| −0.87272000000000 | −0.04157000000000 | −0.87272000000000 | O | (24k) |
| 1.37272000000000 | −0.04157000000000 | 1.37272000000000 | O | (24k) |
| 0.54157000000000 | 1.37272000000000 | −0.87272000000000 | O | (24k) |
| 0.54157000000000 | −0.87272000000000 | 1.37272000000000 | O | (24k) |
| −0.04157000000000 | 1.37272000000000 | 1.37272000000000 | O | (24k) |
| −0.04157000000000 | −0.87272000000000 | −0.87272000000000 | O | (24k) |
| 0.43783000000000 | 0.43783000000000 | 0.74176000000000 | O | (24k) |
| 0.06217000000000 | 0.06217000000000 | 0.74176000000000 | O | (24k) |
| 0.06217000000000 | 0.43783000000000 | −0.24176000000000 | O | (24k) |
| 0.43783000000000 | 0.06217000000000 | −0.24176000000000 | O | (24k) |
| 0.74176000000000 | 0.43783000000000 | 0.43783000000000 | O | (24k) |
| 0.74176000000000 | 0.06217000000000 | 0.06217000000000 | O | (24k) |
| −0.24176000000000 | 0.06217000000000 | 0.43783000000000 | O | (24k) |
| −0.24176000000000 | 0.43783000000000 | 0.06217000000000 | O | (24k) |
| 0.43783000000000 | 0.74176000000000 | 0.43783000000000 | O | (24k) |
| 0.06217000000000 | 0.74176000000000 | 0.06217000000000 | O | (24k) |
| 0.43783000000000 | −0.24176000000000 | 0.06217000000000 | O | (24k) |
| 0.06217000000000 | −0.24176000000000 | 0.43783000000000 | O | (24k) |
| 0.93783000000000 | 0.93783000000000 | −0.74176000000000 | O | (24k) |
| −0.43783000000000 | −0.43783000000000 | −0.74176000000000 | O | (24k) |
| 0.93783000000000 | −0.43783000000000 | 1.24176000000000 | O | (24k) |
| −0.43783000000000 | 0.93783000000000 | 1.24176000000000 | O | (24k) |
| 0.93783000000000 | 1.24176000000000 | −0.43783000000000 | O | (24k) |
| −0.43783000000000 | 1.24176000000000 | 0.93783000000000 | O | (24k) |
| −0.43783000000000 | −0.74176000000000 | −0.43783000000000 | O | (24k) |
| 0.93783000000000 | −0.74176000000000 | 0.93783000000000 | O | (24k) |
| 1.24176000000000 | 0.93783000000000 | −0.43783000000000 | O | (24k) |
| 1.24176000000000 | −0.43783000000000 | 0.93783000000000 | O | (24k) |
| −0.74176000000000 | 0.93783000000000 | 0.93783000000000 | O | (24k) |
| −0.74176000000000 | −0.43783000000000 | −0.43783000000000 | O | (24k) |
| 0.25000000000000 | 0.25000000000000 | 0.25000000000000 | P | (2a) |
| 0.75000000000000 | 0.75000000000000 | 0.75000000000000 | P | (2a) |
| −0.04450000000000 | −0.04450000000000 | 0.75577000000000 | W | (24k) |
| 0.54450000000000 | 0.54450000000000 | 0.75577000000000 | W | (24k) |
| 0.54450000000000 | −0.04450000000000 | −0.25577000000000 | W | (24k) |
| −0.04450000000000 | 0.54450000000000 | −0.25577000000000 | W | (24k) |
| 0.75577000000000 | −0.04450000000000 | −0.04450000000000 | W | (24k) |
| 0.75577000000000 | 0.54450000000000 | 0.54450000000000 | W | (24k) |
| −0.25577000000000 | 0.54450000000000 | −0.04450000000000 | W | (24k) |
| −0.25577000000000 | −0.04450000000000 | 0.54450000000000 | W | (24k) |
| −0.04450000000000 | 0.75577000000000 | −0.04450000000000 | W | (24k) |
| 0.54450000000000 | 0.75577000000000 | 0.54450000000000 | W | (24k) |
| −0.04450000000000 | −0.25577000000000 | 0.54450000000000 | W | (24k) |
| 0.54450000000000 | −0.25577000000000 | −0.04450000000000 | W | (24k) |
| 0.45550000000000 | 0.45550000000000 | −0.75577000000000 | W | (24k) |
| 0.04450000000000 | 0.04450000000000 | −0.75577000000000 | W | (24k) |
| 0.45550000000000 | 0.04450000000000 | 1.25577000000000 | W | (24k) |
| 0.04450000000000 | 0.45550000000000 | 1.25577000000000 | W | (24k) |
| 0.45550000000000 | 1.25577000000000 | 0.04450000000000 | W | (24k) |
| 0.04450000000000 | 1.25577000000000 | 0.45550000000000 | W | (24k) |
| 0.04450000000000 | −0.75577000000000 | 0.04450000000000 | W | (24k) |
| 0.45550000000000 | −0.75577000000000 | 0.45550000000000 | W | (24k) |
| 1.25577000000000 | 0.45550000000000 | 0.04450000000000 | W | (24k) |
| 1.25577000000000 | 0.04450000000000 | 0.45550000000000 | W | (24k) |
| −0.75577000000000 | 0.45550000000000 | 0.45550000000000 | W | (24k) |
| −0.75577000000000 | 0.04450000000000 | 0.04450000000000 | W | (24k) |

$LaH_{10}$ High-$T_c$ Superconductor: A10B_cF44_225_cf_b - CIF

```
# CIF file
data_findsym-output
_audit_creation_method FINDSYM

_chemical_name_mineral 'H10La'
_chemical_formula_sum 'H10 La'

_aflow_title 'LaH$_{10}$ High-T$_{c}$ Superconductor Structure'
_aflow_proto 'A10B_cF44_225_cf_b'
_aflow_params 'a,x_{3}'
_aflow_params_values '4.78,0.12'
_aflow_Strukturbericht 'None'
_aflow_Pearson 'cF44'

_symmetry_space_group_name_H-M "F 4/m -3 2/m"
_symmetry_Int_Tables_number 225

_cell_length_a    4.78000
_cell_length_b    4.78000
_cell_length_c    4.78000
_cell_angle_alpha 90.00000
_cell_angle_beta  90.00000
_cell_angle_gamma 90.00000

loop_
_space_group_symop_id
_space_group_symop_operation_xyz
1 x,y,z
2 x,-y,-z
3 -x,y,-z
4 -x,-y,z
5 y,z,x
6 y,-z,-x
7 -y,z,-x
8 -y,-z,x
9 z,x,y
10 z,-x,-y
11 -z,x,-y
12 -z,-x,y
13 -y,-x,-z
14 -y,x,z
15 y,-x,z
16 y,x,-z
17 -x,-z,-y
18 -x,z,y
19 x,-z,y
20 x,z,-y
21 -z,-y,-x
22 -z,y,x
23 z,-y,x
24 z,y,-x
25 -x,-y,-z
26 -x,y,z
27 x,-y,z
28 x,y,-z
29 -y,-z,-x
30 -y,z,x
31 y,-z,x
32 y,z,-x
33 -z,-x,-y
34 -z,x,y
35 z,-x,y
36 z,x,-y
37 y,x,z
38 y,-x,-z
39 -y,x,-z
40 -y,-x,z
41 x,z,y
42 x,-z,-y
43 -x,z,-y
44 -x,-z,y
45 z,y,x
46 z,-y,-x
47 -z,y,-x
48 -z,-y,x
49 x,y+1/2,z+1/2
50 x,-y+1/2,-z+1/2
51 -x,y+1/2,-z+1/2
52 -x,-y+1/2,z+1/2
53 y,z+1/2,x+1/2
54 y,-z+1/2,-x+1/2
55 -y,z+1/2,-x+1/2
56 -y,-z+1/2,x+1/2
57 z,x+1/2,y+1/2
58 z,-x+1/2,-y+1/2
59 -z,x+1/2,-y+1/2
60 -z,-x+1/2,y+1/2
61 -y,-x+1/2,-z+1/2
62 -y,x+1/2,z+1/2
63 y,-x+1/2,z+1/2
64 y,x+1/2,-z+1/2
65 -x,-z+1/2,-y+1/2
66 -x,z+1/2,y+1/2
67 x,-z+1/2,y+1/2
68 x,z+1/2,-y+1/2
69 -z,-y+1/2,-x+1/2
70 -z,y+1/2,x+1/2
71 z,-y+1/2,x+1/2
72 z,y+1/2,-x+1/2
73 -x,-y+1/2,-z+1/2
74 -x,y+1/2,z+1/2
75 x,-y+1/2,z+1/2
76 x,y+1/2,-z+1/2
77 -y,-z+1/2,-x+1/2
78 -y,z+1/2,x+1/2
79 y,-z+1/2,x+1/2
80 y,z+1/2,-x+1/2
81 -z,-x+1/2,-y+1/2
82 -z,x+1/2,y+1/2
83 z,-x+1/2,y+1/2
84 z,x+1/2,-y+1/2
85 y,x+1/2,z+1/2
86 y,-x+1/2,-z+1/2
87 -y,x+1/2,-z+1/2
88 -y,-x+1/2,z+1/2
89 x,z+1/2,y+1/2
90 x,-z+1/2,-y+1/2
91 -x,z+1/2,-y+1/2
92 -x,-z+1/2,y+1/2
93 z,y+1/2,x+1/2
94 z,-y+1/2,-x+1/2
95 -z,y+1/2,-x+1/2
96 -z,-y+1/2,x+1/2
97 x+1/2,y,z+1/2
98 x+1/2,-y,-z+1/2
99 -x+1/2,y,-z+1/2
100 -x+1/2,-y,z+1/2
101 y+1/2,z,x+1/2
102 y+1/2,-z,-x+1/2
103 -y+1/2,z,-x+1/2
104 -y+1/2,-z,x+1/2
105 z+1/2,x,y+1/2
106 z+1/2,-x,-y+1/2
```

```
107 -z+1/2,x,-y+1/2
108 -z+1/2,-x,y+1/2
109 -y+1/2,-x,-z+1/2
110 -y+1/2,x,z+1/2
111 y+1/2,-x,z+1/2
112 y+1/2,x,-z+1/2
113 -x+1/2,-z,-y+1/2
114 -x+1/2,z,y+1/2
115 x+1/2,-z,y+1/2
116 x+1/2,z,-y+1/2
117 -z+1/2,-y,-x+1/2
118 -z+1/2,y,x+1/2
119 z+1/2,-y,x+1/2
120 z+1/2,y,-x+1/2
121 -x+1/2,-y,-z+1/2
122 -x+1/2,y,z+1/2
123 x+1/2,-y,z+1/2
124 x+1/2,y,-z+1/2
125 -y+1/2,-z,-x+1/2
126 -y+1/2,z,x+1/2
127 y+1/2,-z,x+1/2
128 y+1/2,z,-x+1/2
129 -z+1/2,-x,-y+1/2
130 -z+1/2,x,y+1/2
131 z+1/2,-x,y+1/2
132 z+1/2,x,-y+1/2
133 y+1/2,x,z+1/2
134 y+1/2,-x,-z+1/2
135 -y+1/2,x,-z+1/2
136 -y+1/2,-x,z+1/2
137 x+1/2,z,y+1/2
138 x+1/2,-z,-y+1/2
139 -x+1/2,z,-y+1/2
140 -x+1/2,-z,y+1/2
141 z+1/2,y,x+1/2
142 z+1/2,-y,-x+1/2
143 -z+1/2,y,-x+1/2
144 -z+1/2,-y,x+1/2
145 x+1/2,y+1/2,z
146 x+1/2,-y+1/2,-z
147 -x+1/2,y+1/2,-z
148 -x+1/2,-y+1/2,z
149 y+1/2,z+1/2,x
150 y+1/2,-z+1/2,-x
151 -y+1/2,z+1/2,-x
152 -y+1/2,-z+1/2,x
153 z+1/2,x+1/2,y
154 z+1/2,-x+1/2,-y
155 -z+1/2,x+1/2,-y
156 -z+1/2,-x+1/2,y
157 -y+1/2,-x+1/2,-z
158 -y+1/2,x+1/2,z
159 y+1/2,-x+1/2,z
160 y+1/2,x+1/2,-z
161 -x+1/2,-z+1/2,-y
162 -x+1/2,z+1/2,y
163 x+1/2,-z+1/2,y
164 x+1/2,z+1/2,-y
165 -z+1/2,-y+1/2,-x
166 -z+1/2,y+1/2,x
167 z+1/2,-y+1/2,x
168 z+1/2,y+1/2,-x
169 -x+1/2,-y+1/2,-z
170 -x+1/2,y+1/2,z
171 x+1/2,-y+1/2,z
172 x+1/2,y+1/2,-z
173 -y+1/2,-z+1/2,-x
174 -y+1/2,z+1/2,x
175 y+1/2,-z+1/2,x
176 y+1/2,z+1/2,-x
177 -z+1/2,-x+1/2,-y
178 -z+1/2,x+1/2,y
179 z+1/2,-x+1/2,y
180 z+1/2,x+1/2,-y
181 y+1/2,x+1/2,z
182 y+1/2,-x+1/2,-z
183 -y+1/2,x+1/2,-z
184 -y+1/2,-x+1/2,z
185 x+1/2,z+1/2,y
186 x+1/2,-z+1/2,-y
187 -x+1/2,z+1/2,-y
188 -x+1/2,-z+1/2,y
189 z+1/2,y+1/2,x
190 z+1/2,-y+1/2,-x
191 -z+1/2,y+1/2,-x
192 -z+1/2,-y+1/2,x

loop_
_atom_site_label
_atom_site_type_symbol
_atom_site_symmetry_multiplicity
_atom_site_Wyckoff_label
_atom_site_fract_x
_atom_site_fract_y
_atom_site_fract_z
_atom_site_occupancy
La1 La 4 b 0.50000 0.50000 0.50000 1.00000
H1 H 8 c 0.25000 0.25000 0.25000 1.00000
H2 H 32 f 0.12000 0.12000 0.12000 1.00000
```

$LaH_{10}$ High-$T_c$ Superconductor: A10B_cF44_225_cf_b - POSCAR

```
A10B_cF44_225_cf_b & a,x3 --params=4.78,0.12 & Fm-3m O_{h}^{5} #225 (bcf
    ↪ ) & cF44 & None & H10La & H10La &
  1.00000000000000
    0.00000000000000    2.39000000000000    2.39000000000000
    2.39000000000000    0.00000000000000    2.39000000000000
    2.39000000000000    2.39000000000000    0.00000000000000
     H     La
    10      1
Direct
    0.25000000000000    0.25000000000000    0.25000000000000   H   (8c)
    0.75000000000000    0.75000000000000    0.75000000000000   H   (8c)
    0.12000000000000    0.12000000000000    0.12000000000000   H  (32f)
    0.12000000000000    0.12000000000000   -0.36000000000000   H  (32f)
    0.12000000000000   -0.36000000000000    0.12000000000000   H  (32f)
   -0.36000000000000    0.12000000000000    0.12000000000000   H  (32f)
   -0.12000000000000   -0.12000000000000    0.36000000000000   H  (32f)
   -0.12000000000000   -0.12000000000000   -0.12000000000000   H  (32f)
   -0.12000000000000    0.36000000000000   -0.12000000000000   H  (32f)
    0.36000000000000   -0.12000000000000   -0.12000000000000   H  (32f)
    0.50000000000000    0.50000000000000    0.50000000000000  La   (4b)
```

Double Perovskite ($Ba_2MnWO_6$): A2BC6D_cF40_225_c_a_e_b - CIF

```
# CIF file
data_findsym-output
_audit_creation_method FINDSYM

_chemical_name_mineral 'Double perovskite'
_chemical_formula_sum 'Ba2 Mn O6 W'

loop_
_publ_author_name
 'A. K. Azad'
 'S. A. Ivanov'
 'S.-G. Eriksson'
 'J. Eriksen'
 'H. Rundl\"{o}f'
 'R. Mathieu'
 'P. Svedlindh'
_journal_name_full_name
;
 Materials Research Bulletin
;
_journal_volume 36
_journal_year 2001
_journal_page_first 2215
_journal_page_last 2228
_publ_Section_title
;
 Synthesis, crystal structure, and magnetic characterization of the
    ↪ double perovskite Ba$_{2}$MnWO$_{6}$
;

_aflow_title 'Double Perovskite (Ba$_{2}$MnWO$_{6}$) Structure'
_aflow_proto 'A2BC6D_cF40_225_c_a_e_b'
_aflow_params 'a,x_{4}'
_aflow_params_values '8.19849,0.2654'
_aflow_Strukturbericht 'None'
_aflow_Pearson 'cF40'

_symmetry_space_group_name_H-M "F 4/m -3 2/m"
_symmetry_Int_Tables_number 225

_cell_length_a    8.19849
_cell_length_b    8.19849
_cell_length_c    8.19849
_cell_angle_alpha 90.00000
_cell_angle_beta  90.00000
_cell_angle_gamma 90.00000

loop_
_space_group_symop_id
_space_group_symop_operation_xyz
1 x,y,z
2 x,-y,-z
3 -x,y,-z
4 -x,-y,z
5 y,z,x
6 y,-z,-x
7 -y,z,-x
8 -y,-z,x
9 z,x,y
10 z,-x,-y
11 -z,x,-y
12 -z,-x,y
13 -y,-x,-z
14 -y,x,z
15 y,-x,z
16 y,x,-z
17 -x,-z,-y
18 -x,z,y
19 x,-z,y
20 x,z,-y
21 -z,-y,-x
22 -z,y,x
23 z,-y,x
24 z,y,-x
25 -x,-y,-z
26 -x,y,z
27 x,-y,z
28 x,y,-z
29 -y,-z,-x
30 -y,z,x
31 y,-z,x
32 y,z,-x
33 -z,-x,-y
34 -z,x,y
35 z,-x,y
```

```
36 z,x,-y
37 y,x,z
38 y,-x,-z
39 -y,x,-z
40 -y,-x,z
41 x,z,y
42 x,-z,-y
43 -x,z,-y
44 -x,-z,y
45 z,y,x
46 z,-y,-x
47 -z,y,-x
48 -z,-y,x
49 x,y+1/2,z+1/2
50 x,-y+1/2,-z+1/2
51 -x,y+1/2,-z+1/2
52 -x,-y+1/2,z+1/2
53 y,z+1/2,x+1/2
54 y,-z+1/2,-x+1/2
55 -y,z+1/2,-x+1/2
56 -y,-z+1/2,x+1/2
57 z,x+1/2,y+1/2
58 z,-x+1/2,-y+1/2
59 -z,x+1/2,-y+1/2
60 -z,-x+1/2,y+1/2
61 -y,-x+1/2,-z+1/2
62 -y,x+1/2,z+1/2
63 y,-x+1/2,z+1/2
64 y,x+1/2,-z+1/2
65 -x,-z+1/2,-y+1/2
66 -x,z+1/2,y+1/2
67 x,-z+1/2,y+1/2
68 x,z+1/2,-y+1/2
69 -z,-y+1/2,-x+1/2
70 -z,y+1/2,x+1/2
71 z,-y+1/2,x+1/2
72 z,y+1/2,-x+1/2
73 -x,-y+1/2,-z+1/2
74 -x,y+1/2,z+1/2
75 x,-y+1/2,z+1/2
76 x,y+1/2,-z+1/2
77 -y,-z+1/2,-x+1/2
78 -y,z+1/2,x+1/2
79 y,-z+1/2,x+1/2
80 y,z+1/2,-x+1/2
81 -z,-x+1/2,-y+1/2
82 -z,x+1/2,y+1/2
83 z,-x+1/2,y+1/2
84 z,x+1/2,-y+1/2
85 y,x+1/2,z+1/2
86 y,-x+1/2,-z+1/2
87 -y,x+1/2,-z+1/2
88 -y,-x+1/2,z+1/2
89 x,z+1/2,y+1/2
90 x,-z+1/2,-y+1/2
91 -x,z+1/2,-y+1/2
92 -x,-z+1/2,y+1/2
93 z,y+1/2,x+1/2
94 z,-y+1/2,-x+1/2
95 -z,y+1/2,-x+1/2
96 -z,-y+1/2,x+1/2
97 x+1/2,y,z+1/2
98 x+1/2,-y,-z+1/2
99 -x+1/2,y,-z+1/2
100 -x+1/2,-y,z+1/2
101 y+1/2,z,x+1/2
102 y+1/2,-z,-x+1/2
103 -y+1/2,z,-x+1/2
104 -y+1/2,-z,x+1/2
105 z+1/2,x,y+1/2
106 z+1/2,-x,-y+1/2
107 -z+1/2,x,-y+1/2
108 -z+1/2,-x,y+1/2
109 -y+1/2,-x,-z+1/2
110 -y+1/2,x,z+1/2
111 y+1/2,-x,z+1/2
112 y+1/2,x,-z+1/2
113 -x+1/2,-z,-y+1/2
114 -x+1/2,z,y+1/2
115 x+1/2,-z,y+1/2
116 x+1/2,z,-y+1/2
117 -z+1/2,-y,-x+1/2
118 -z+1/2,y,x+1/2
119 z+1/2,-y,x+1/2
120 z+1/2,y,-x+1/2
121 -x+1/2,-y,-z+1/2
122 -x+1/2,y,z+1/2
123 x+1/2,-y,z+1/2
124 x+1/2,y,-z+1/2
125 -y+1/2,-z,-x+1/2
126 -y+1/2,z,x+1/2
127 y+1/2,-z,x+1/2
128 y+1/2,z,-x+1/2
129 -z+1/2,-x,-y+1/2
130 -z+1/2,x,y+1/2
131 z+1/2,-x,y+1/2
132 z+1/2,x,-y+1/2
133 y+1/2,x,z+1/2
134 y+1/2,-x,-z+1/2
135 -y+1/2,x,-z+1/2
136 -y+1/2,-x,z+1/2
137 x+1/2,z,y+1/2
138 x+1/2,-z,-y+1/2
139 -x+1/2,z,-y+1/2
140 -x+1/2,-z,y+1/2
141 z+1/2,y,x+1/2
142 z+1/2,-y,-x+1/2
143 -z+1/2,y,-x+1/2
144 -z+1/2,-y,x+1/2
145 x+1/2,y+1/2,z
146 x+1/2,-y+1/2,-z
147 -x+1/2,y+1/2,-z
148 -x+1/2,-y+1/2,z
149 y+1/2,z+1/2,x
150 y+1/2,-z+1/2,-x
151 -y+1/2,z+1/2,-x
152 -y+1/2,-z+1/2,x
153 z+1/2,x+1/2,y
154 z+1/2,-x+1/2,-y
155 -z+1/2,x+1/2,-y
156 -z+1/2,-x+1/2,y
157 -y+1/2,-x+1/2,-z
158 -y+1/2,x+1/2,z
159 y+1/2,-x+1/2,z
160 y+1/2,x+1/2,-z
161 -x+1/2,-z+1/2,-y
162 -x+1/2,z+1/2,y
163 x+1/2,-z+1/2,y
164 x+1/2,z+1/2,-y
165 -z+1/2,-y+1/2,-x
166 -z+1/2,y+1/2,x
167 z+1/2,-y+1/2,x
168 z+1/2,y+1/2,-x
169 -x+1/2,-y+1/2,-z
170 -x+1/2,y+1/2,z
171 x+1/2,-y+1/2,z
172 x+1/2,y+1/2,-z
173 -y+1/2,-z+1/2,-x
174 -y+1/2,z+1/2,x
175 y+1/2,-z+1/2,x
176 y+1/2,z+1/2,-x
177 -z+1/2,-x+1/2,-y
178 -z+1/2,x+1/2,y
179 z+1/2,-x+1/2,y
180 z+1/2,x+1/2,-y
181 y+1/2,x+1/2,z
182 y+1/2,-x+1/2,-z
183 -y+1/2,x+1/2,-z
184 -y+1/2,-x+1/2,z
185 x+1/2,z+1/2,y
186 x+1/2,-z+1/2,-y
187 -x+1/2,z+1/2,-y
188 -x+1/2,-z+1/2,y
189 z+1/2,y+1/2,x
190 z+1/2,-y+1/2,-x
191 -z+1/2,y+1/2,-x
192 -z+1/2,-y+1/2,x

loop_
_atom_site_label
_atom_site_type_symbol
_atom_site_symmetry_multiplicity
_atom_site_Wyckoff_label
_atom_site_fract_x
_atom_site_fract_y
_atom_site_fract_z
_atom_site_occupancy
Mn1 Mn 4 a 0.00000 0.00000 0.00000 1.00000
W1 W 4 b 0.50000 0.50000 0.50000 1.00000
Ba1 Ba 8 c 0.25000 0.25000 0.25000 1.00000
O1 O 24 e 0.26540 0.00000 0.00000 1.00000
```

Double Perovskite ($Ba_2MnWO_6$): A2BC6D_cF40_225_c_a_e_b - POSCAR

```
A2BC6D_cF40_225_c_a_e_b & a,x4 --params=8.19849,0.2654 & Fm-3m O_{h}^{5}
    ↪ #225 (abce) & cF40 & None & Ba2MnO6W & Double perovskite & A.
    ↪ K. Azad et al., Mater. Res. Bull. 36, 2215-2228 (2001)
   1.00000000000000
   0.00000000000000   4.09924500000000   4.09924500000000
   4.09924500000000   0.00000000000000   4.09924500000000
   4.09924500000000   4.09924500000000   0.00000000000000
    Ba    Mn    O    W
     2     1    6    1
Direct
   0.25000000000000   0.25000000000000   0.25000000000000   Ba   (8c)
   0.75000000000000   0.75000000000000   0.75000000000000   Ba   (8c)
   0.00000000000000   0.00000000000000   0.00000000000000   Mn   (4a)
  -0.26540000000000   0.26540000000000   0.26540000000000   O   (24e)
   0.26540000000000  -0.26540000000000  -0.26540000000000   O   (24e)
   0.26540000000000  -0.26540000000000   0.26540000000000   O   (24e)
  -0.26540000000000   0.26540000000000  -0.26540000000000   O   (24e)
   0.26540000000000   0.26540000000000  -0.26540000000000   O   (24e)
  -0.26540000000000  -0.26540000000000   0.26540000000000   O   (24e)
   0.50000000000000   0.50000000000000   0.50000000000000   W    (4b)
```

$Cu_3[Fe(CN)_6]_2 \cdot xH_2O$ ($J2_5$, $x \approx 3$): A6B9CD2E6_cF96_225_e_bf_a_c_e - CIF

```
# CIF file
data_findsym-output
_audit_creation_method FINDSYM

_chemical_name_mineral 'Prussian blue analog'
_chemical_formula_sum 'C6 Cu9 Fe (H2O)2 N6'

loop_
_publ_author_name
 'A. K. {van Bever}'
_journal_name_full_name
;
 Recueil des Travaux Chimiques des Pays-Bas
```

```
;
_journal_volume 57
_journal_year 1938
_journal_page_first 1259
_journal_page_last 1268
_publ_Section_title
;
 The Crystal Structure of Some Ferricyanides with Bivalent Kations
;

_aflow_title 'Cu$_{3}$[Fe(CN)$_{6}$]$_{2}$$\cdot x$H$_{2}$O ($J2_{5}$)
    ↪ Structure'
_aflow_proto 'A6B9CD2E6_cF96_225_e_bf_a_c_e'
_aflow_params 'a,x_{4},x_{5},x_{6}'
_aflow_params_values '10.12,0.20257,0.31126,0.16667'
_aflow_Strukturbericht '$J2_5$'
_aflow_Pearson 'cF96'

_symmetry_space_group_name_H-M "F 4/m -3 2/m"
_symmetry_Int_Tables_number 225

_cell_length_a      10.12000
_cell_length_b      10.12000
_cell_length_c      10.12000
_cell_angle_alpha 90.00000
_cell_angle_beta  90.00000
_cell_angle_gamma 90.00000

loop_
_space_group_symop_id
_space_group_symop_operation_xyz
1 x,y,z
2 x,-y,-z
3 -x,y,-z
4 -x,-y,z
5 y,z,x
6 y,-z,-x
7 -y,z,-x
8 -y,-z,x
9 z,x,y
10 z,-x,-y
11 -z,x,-y
12 -z,-x,y
13 -y,-x,-z
14 -y,x,z
15 y,-x,z
16 y,x,-z
17 -x,-z,-y
18 -x,z,y
19 x,-z,y
20 x,z,-y
21 -z,-y,-x
22 -z,y,x
23 z,-y,x
24 z,y,-x
25 -x,-y,-z
26 -x,y,z
27 x,-y,z
28 x,y,-z
29 -y,-z,-x
30 -y,z,x
31 y,-z,x
32 y,z,-x
33 -z,-x,-y
34 -z,x,y
35 z,-x,y
36 z,x,-y
37 y,x,z
38 y,-x,-z
39 -y,x,-z
40 -y,-x,z
41 x,z,y
42 x,-z,-y
43 -x,z,-y
44 -x,-z,y
45 z,y,x
46 z,-y,-x
47 -z,y,-x
48 -z,-y,x
49 x,y+1/2,z+1/2
50 x,-y+1/2,-z+1/2
51 -x,y+1/2,-z+1/2
52 -x,-y+1/2,z+1/2
53 y,z+1/2,x+1/2
54 y,-z+1/2,-x+1/2
55 -y,z+1/2,-x+1/2
56 -y,-z+1/2,x+1/2
57 z,x+1/2,y+1/2
58 z,-x+1/2,-y+1/2
59 -z,x+1/2,-y+1/2
60 -z,-x+1/2,y+1/2
61 -y,-x+1/2,-z+1/2
62 -y,x+1/2,z+1/2
63 y,-x+1/2,z+1/2
64 y,x+1/2,-z+1/2
65 -x,-z+1/2,-y+1/2
66 -x,z+1/2,y+1/2
67 x,-z+1/2,y+1/2
68 x,z+1/2,-y+1/2
69 -z,-y+1/2,-x+1/2
70 -z,y+1/2,x+1/2
71 z,-y+1/2,x+1/2
72 z,y+1/2,-x+1/2
73 -x,-y+1/2,-z+1/2
74 -x,y+1/2,z+1/2
75 x,-y+1/2,z+1/2
76 x,y+1/2,-z+1/2
77 -y,-z+1/2,-x+1/2
78 -y,z+1/2,x+1/2
79 y,-z+1/2,x+1/2
80 y,z+1/2,-x+1/2
81 -z,-x+1/2,-y+1/2
82 -z,x+1/2,y+1/2
83 z,-x+1/2,y+1/2
84 z,x+1/2,-y+1/2
85 y,x+1/2,z+1/2
86 y,-x+1/2,-z+1/2
87 -y,x+1/2,-z+1/2
88 -y,-x+1/2,z+1/2
89 x,z+1/2,y+1/2
90 x,-z+1/2,-y+1/2
91 -x,z+1/2,-y+1/2
92 -x,-z+1/2,y+1/2
93 z,y+1/2,x+1/2
94 z,-y+1/2,-x+1/2
95 -z,y+1/2,-x+1/2
96 -z,-y+1/2,x+1/2
97 x+1/2,y,z+1/2
98 x+1/2,-y,-z+1/2
99 -x+1/2,y,-z+1/2
100 -x+1/2,-y,z+1/2
101 y+1/2,z,x+1/2
102 y+1/2,-z,-x+1/2
103 -y+1/2,z,-x+1/2
104 -y+1/2,-z,x+1/2
105 z+1/2,x,y+1/2
106 z+1/2,-x,-y+1/2
107 -z+1/2,x,-y+1/2
108 -z+1/2,-x,y+1/2
109 -y+1/2,-x,-z+1/2
110 -y+1/2,x,z+1/2
111 y+1/2,-x,z+1/2
112 y+1/2,x,-z+1/2
113 -x+1/2,-z,-y+1/2
114 -x+1/2,z,y+1/2
115 x+1/2,-z,y+1/2
116 x+1/2,z,-y+1/2
117 -z+1/2,-y,-x+1/2
118 -z+1/2,y,x+1/2
119 z+1/2,-y,x+1/2
120 z+1/2,y,-x+1/2
121 -x+1/2,-y,-z+1/2
122 -x+1/2,y,z+1/2
123 x+1/2,-y,z+1/2
124 x+1/2,y,-z+1/2
125 -y+1/2,-z,-x+1/2
126 -y+1/2,z,x+1/2
127 y+1/2,-z,x+1/2
128 y+1/2,z,-x+1/2
129 -z+1/2,-x,-y+1/2
130 -z+1/2,x,y+1/2
131 z+1/2,-x,y+1/2
132 z+1/2,x,-y+1/2
133 y+1/2,x,z+1/2
134 y+1/2,-x,-z+1/2
135 -y+1/2,x,-z+1/2
136 -y+1/2,-x,z+1/2
137 x+1/2,z,y+1/2
138 x+1/2,-z,-y+1/2
139 -x+1/2,z,-y+1/2
140 -x+1/2,-z,y+1/2
141 z+1/2,y,x+1/2
142 z+1/2,-y,-x+1/2
143 -z+1/2,y,-x+1/2
144 -z+1/2,-y,x+1/2
145 x+1/2,y+1/2,z
146 x+1/2,-y+1/2,-z
147 -x+1/2,y+1/2,-z
148 -x+1/2,-y+1/2,z
149 y+1/2,z+1/2,x
150 y+1/2,-z+1/2,-x
151 -y+1/2,z+1/2,-x
152 -y+1/2,-z+1/2,x
153 z+1/2,x+1/2,y
154 z+1/2,-x+1/2,-y
155 -z+1/2,x+1/2,-y
156 -z+1/2,-x+1/2,y
157 -y+1/2,-x+1/2,-z
158 -y+1/2,x+1/2,z
159 y+1/2,-x+1/2,z
160 y+1/2,x+1/2,-z
161 -x+1/2,-z+1/2,-y
162 -x+1/2,z+1/2,y
163 x+1/2,-z+1/2,y
164 x+1/2,z+1/2,-y
165 -z+1/2,-y+1/2,-x
166 -z+1/2,y+1/2,x
167 z+1/2,-y+1/2,x
168 z+1/2,y+1/2,-x
169 -x+1/2,-y+1/2,-z
170 -x+1/2,y+1/2,z
171 x+1/2,-y+1/2,z
172 x+1/2,y+1/2,-z
173 -y+1/2,-z+1/2,-x
174 -y+1/2,z+1/2,x
175 y+1/2,-z+1/2,x
176 y+1/2,z+1/2,-x
177 -z+1/2,-x+1/2,-y
178 -z+1/2,x+1/2,y
179 z+1/2,-x+1/2,y
```

```
180 z+1/2,x+1/2,-y
181 y+1/2,x+1/2,z
182 y+1/2,-x+1/2,-z
183 -y+1/2,x+1/2,-z
184 -y+1/2,-x+1/2,z
185 x+1/2,z+1/2,y
186 x+1/2,-z+1/2,-y
187 -x+1/2,z+1/2,-y
188 -x+1/2,-z+1/2,y
189 z+1/2,y+1/2,x
190 z+1/2,-y+1/2,-x
191 -z+1/2,y+1/2,-x
192 -z+1/2,-y+1/2,x

loop_
_atom_site_label
_atom_site_type_symbol
_atom_site_symmetry_multiplicity
_atom_site_Wyckoff_label
_atom_site_fract_x
_atom_site_fract_y
_atom_site_fract_z
_atom_site_occupancy
Fe1 Fe 4 a 0.00000 0.00000 0.00000 1.00000
Cu1 Cu 4 b 0.50000 0.50000 0.50000 1.00000
H2O1 H2O 8 c 0.25000 0.25000 0.25000 0.75000
C1 C 24 e 0.20257 0.00000 0.00000 1.00000
N1 N 24 e 0.31126 0.00000 0.00000 1.00000
Cu2 Cu 32 f 0.16667 0.16667 0.16667 0.06250
```

$Cu_3[Fe(CN)_6]_2 \cdot xH_2O$ ($J2_5$, $x \approx 3$): A6B9CD2E6_cF96_225_e_bf_a_c_e - POSCAR

```
A6B9CD2E6_cF96_225_e_bf_a_c_e & a,x4,x5,x6 --params=10.12,0.20257,
   ↪ 0.31126,0.16667 & Fm-3m O_{h}^{5} #225 (abce^2f) & cF96 &
   ↪ $J2_5$ & C12Cu3Fe2N12\cdot xH2O & Prussian blue analog & A. K.
   ↪ {van Bever}, Rec. Trav. Chim. Pays-Bas 57, 1259-1268 (1938)
   1.00000000000000
   0.00000000000000   5.06000000000000   5.06000000000000
   5.06000000000000   0.00000000000000   5.06000000000000
   5.06000000000000   5.06000000000000   0.00000000000000
     C     Cu     Fe    H2O     N
     6      9      1      2     6
Direct
  -0.20257000000000   0.20257000000000   0.20257000000000   C   (24e)
   0.20257000000000  -0.20257000000000  -0.20257000000000   C   (24e)
   0.20257000000000  -0.20257000000000   0.20257000000000   C   (24e)
  -0.20257000000000   0.20257000000000  -0.20257000000000   C   (24e)
   0.20257000000000   0.20257000000000  -0.20257000000000   C   (24e)
  -0.20257000000000  -0.20257000000000   0.20257000000000   C   (24e)
   0.50000000000000   0.50000000000000   0.50000000000000   Cu   (4b)
   0.16667000000000   0.16667000000000   0.16667000000000   Cu  (32f)
   0.16667000000000   0.16667000000000  -0.50001000000000   Cu  (32f)
   0.16667000000000  -0.50001000000000   0.16667000000000   Cu  (32f)
  -0.50001000000000   0.16667000000000   0.16667000000000   Cu  (32f)
  -0.16667000000000  -0.16667000000000   0.50001000000000   Cu  (32f)
  -0.16667000000000  -0.16667000000000  -0.16667000000000   Cu  (32f)
  -0.16667000000000   0.50001000000000  -0.16667000000000   Cu  (32f)
   0.50001000000000  -0.16667000000000  -0.16667000000000   Cu  (32f)
   0.00000000000000   0.00000000000000   0.00000000000000   Fe   (4a)
   0.25000000000000   0.25000000000000   0.25000000000000   H2O  (8c)
   0.75000000000000   0.75000000000000   0.75000000000000   H2O  (8c)
  -0.31126000000000   0.31126000000000   0.31126000000000   N   (24e)
   0.31126000000000  -0.31126000000000  -0.31126000000000   N   (24e)
   0.31126000000000  -0.31126000000000   0.31126000000000   N   (24e)
  -0.31126000000000   0.31126000000000  -0.31126000000000   N   (24e)
   0.31126000000000   0.31126000000000  -0.31126000000000   N   (24e)
  -0.31126000000000  -0.31126000000000   0.31126000000000   N   (24e)
```

$Co_9S_8$ ($D8_9$): A9B8_cF68_225_af_ce - CIF

```
# CIF file
data_findsym-output
_audit_creation_method FINDSYM

_chemical_name_mineral 'Co9S8'
_chemical_formula_sum 'Co9 S8'

loop_
_publ_author_name
 'S. Geller'
_journal_name_full_name
;
 Acta Cristallographica
;
_journal_volume 15
_journal_year 1962
_journal_page_first 1195
_journal_page_last 1198
_publ_Section_title
;
 Refinement of the crystal structure of Co$_{9}$S$_{8}$
;

_aflow_title 'Co$_{9}$S$_{8}$ ($D8_{9}$) Structure'
_aflow_proto 'A9B8_cF68_225_af_ce'
_aflow_params 'a,x_{3},x_{4}'
_aflow_params_values '9.928,0.7591,0.374'
_aflow_Strukturbericht '$D8_{9}$'
_aflow_Pearson 'cF68'

_symmetry_space_group_name_H-M "F 4/m -3 2/m"
_symmetry_Int_Tables_number 225

_cell_length_a    9.92800
_cell_length_b    9.92800
_cell_length_c    9.92800
_cell_angle_alpha 90.00000
_cell_angle_beta  90.00000
_cell_angle_gamma 90.00000

loop_
_space_group_symop_id
_space_group_symop_operation_xyz
1 x,y,z
2 x,-y,-z
3 -x,y,-z
4 -x,-y,z
5 y,z,x
6 y,-z,-x
7 -y,z,-x
8 -y,-z,x
9 z,x,y
10 z,-x,-y
11 -z,x,-y
12 -z,-x,y
13 -y,-x,-z
14 -y,x,z
15 y,-x,z
16 y,x,-z
17 -x,-z,-y
18 -x,z,y
19 x,-z,y
20 x,z,-y
21 -z,-y,-x
22 -z,y,x
23 z,-y,x
24 z,y,-x
25 -x,-y,-z
26 -x,y,z
27 x,-y,z
28 x,y,-z
29 -y,-z,-x
30 -y,z,x
31 y,-z,x
32 y,z,-x
33 -z,-x,-y
34 -z,x,y
35 z,-x,y
36 z,x,-y
37 y,x,z
38 y,-x,-z
39 -y,x,-z
40 -y,-x,z
41 x,z,y
42 x,-z,-y
43 -x,z,-y
44 -x,-z,y
45 z,y,x
46 z,-y,-x
47 -z,y,-x
48 -z,-y,x
49 x,y+1/2,z+1/2
50 x,-y+1/2,-z+1/2
51 -x,y+1/2,-z+1/2
52 -x,-y+1/2,z+1/2
53 y,z+1/2,x+1/2
54 y,-z+1/2,-x+1/2
55 -y,z+1/2,-x+1/2
56 -y,-z+1/2,x+1/2
57 z,x+1/2,y+1/2
58 z,-x+1/2,-y+1/2
59 -z,x+1/2,-y+1/2
60 -z,-x+1/2,y+1/2
61 -y,-x+1/2,-z+1/2
62 -y,x+1/2,z+1/2
63 y,-x+1/2,z+1/2
64 y,x+1/2,-z+1/2
65 -x,-z+1/2,-y+1/2
66 -x,z+1/2,y+1/2
67 x,-z+1/2,y+1/2
68 x,z+1/2,-y+1/2
69 -z,-y+1/2,-x+1/2
70 -z,y+1/2,x+1/2
71 z,-y+1/2,x+1/2
72 z,y+1/2,-x+1/2
73 -x,-y+1/2,-z+1/2
74 -x,y+1/2,z+1/2
75 x,-y+1/2,z+1/2
76 x,y+1/2,-z+1/2
77 -y,-z+1/2,-x+1/2
78 -y,z+1/2,x+1/2
79 y,-z+1/2,x+1/2
80 y,z+1/2,-x+1/2
81 -z,-x+1/2,-y+1/2
82 -z,x+1/2,y+1/2
83 z,-x+1/2,y+1/2
84 z,x+1/2,-y+1/2
85 y,x+1/2,z+1/2
86 y,-x+1/2,-z+1/2
87 -y,x+1/2,-z+1/2
88 -y,-x+1/2,z+1/2
89 x,z+1/2,y+1/2
90 x,-z+1/2,-y+1/2
91 -x,z+1/2,-y+1/2
92 -x,-z+1/2,y+1/2
93 z,y+1/2,x+1/2
94 z,-y+1/2,-x+1/2
95 -z,y+1/2,-x+1/2
96 -z,-y+1/2,x+1/2
97 x+1/2,y,z+1/2
```

```
98 x+1/2,-y,-z+1/2
99 -x+1/2,y,-z+1/2
100 -x+1/2,-y,z+1/2
101 y+1/2,z,x+1/2
102 y+1/2,-z,-x+1/2
103 -y+1/2,z,-x+1/2
104 -y+1/2,-z,x+1/2
105 z+1/2,x,y+1/2
106 z+1/2,-x,-y+1/2
107 -z+1/2,x,-y+1/2
108 -z+1/2,-x,y+1/2
109 -y+1/2,-x,-z+1/2
110 -y+1/2,x,z+1/2
111 y+1/2,-x,z+1/2
112 y+1/2,x,-z+1/2
113 -x+1/2,-z,-y+1/2
114 -x+1/2,z,y+1/2
115 x+1/2,-z,y+1/2
116 x+1/2,z,-y+1/2
117 -z+1/2,-y,-x+1/2
118 -z+1/2,y,x+1/2
119 z+1/2,-y,x+1/2
120 z+1/2,y,-x+1/2
121 -x+1/2,-y,-z+1/2
122 -x+1/2,y,z+1/2
123 x+1/2,-y,z+1/2
124 x+1/2,y,-z+1/2
125 -y+1/2,-z,-x+1/2
126 -y+1/2,z,x+1/2
127 y+1/2,-z,x+1/2
128 y+1/2,z,-x+1/2
129 -z+1/2,-x,-y+1/2
130 -z+1/2,x,y+1/2
131 z+1/2,-x,y+1/2
132 z+1/2,x,-y+1/2
133 y+1/2,x,z+1/2
134 y+1/2,-x,-z+1/2
135 -y+1/2,x,-z+1/2
136 -y+1/2,-x,z+1/2
137 x+1/2,z,y+1/2
138 x+1/2,-z,-y+1/2
139 -x+1/2,z,-y+1/2
140 -x+1/2,-z,y+1/2
141 z+1/2,y,x+1/2
142 z+1/2,-y,-x+1/2
143 -z+1/2,y,-x+1/2
144 -z+1/2,-y,x+1/2
145 x+1/2,y+1/2,z
146 x+1/2,-y+1/2,-z
147 -x+1/2,y+1/2,-z
148 -x+1/2,-y+1/2,z
149 y+1/2,z+1/2,x
150 y+1/2,-z+1/2,-x
151 -y+1/2,z+1/2,-x
152 -y+1/2,-z+1/2,x
153 z+1/2,x+1/2,y
154 z+1/2,-x+1/2,-y
155 -z+1/2,x+1/2,-y
156 -z+1/2,-x+1/2,y
157 -y+1/2,-x+1/2,-z
158 -y+1/2,x+1/2,z
159 y+1/2,-x+1/2,z
160 y+1/2,x+1/2,-z
161 -x+1/2,-z+1/2,-y
162 -x+1/2,z+1/2,y
163 x+1/2,-z+1/2,y
164 x+1/2,z+1/2,-y
165 -z+1/2,-y+1/2,-x
166 -z+1/2,y+1/2,x
167 z+1/2,-y+1/2,x
168 z+1/2,y+1/2,-x
169 -x+1/2,-y+1/2,-z
170 -x+1/2,y+1/2,z
171 x+1/2,-y+1/2,z
172 x+1/2,y+1/2,-z
173 -y+1/2,-z+1/2,-x
174 -y+1/2,z+1/2,x
175 y+1/2,-z+1/2,x
176 y+1/2,z+1/2,-x
177 -z+1/2,-x+1/2,-y
178 -z+1/2,x+1/2,y
179 z+1/2,-x+1/2,y
180 z+1/2,x+1/2,-y
181 y+1/2,x+1/2,z
182 y+1/2,-x+1/2,-z
183 -y+1/2,x+1/2,-z
184 -y+1/2,-x+1/2,z
185 x+1/2,z+1/2,y
186 x+1/2,-z+1/2,-y
187 -x+1/2,z+1/2,-y
188 -x+1/2,-z+1/2,y
189 z+1/2,y+1/2,x
190 z+1/2,-y+1/2,-x
191 -z+1/2,y+1/2,-x
192 -z+1/2,-y+1/2,x

loop_
_atom_site_label
_atom_site_type_symbol
_atom_site_symmetry_multiplicity
_atom_site_Wyckoff_label
_atom_site_fract_x
_atom_site_fract_y
_atom_site_fract_z
_atom_site_occupancy
Co1 Co 4 a 0.00000 0.00000 0.00000 1.00000
S1 S 8 c 0.25000 0.25000 0.25000 1.00000
S2 S 24 e 0.75910 0.00000 0.00000 1.00000
Co2 Co 32 f 0.37400 0.37400 0.37400 1.00000
```

$Co_9S_8$ ($D8_9$): A9B8_cF68_225_af_ce - POSCAR

```
A9B8_cF68_225_af_ce & a,x3,x4 --params=9.928,0.7591,0.374 & Fm-3m O_{h
    ↪ }^{5} #225 (acef) & cF68 & $D8_{9}$ & Co9S8 & Co9S8 & S. Geller
    ↪ , Acta Cryst. 15, 1195-1198 (1962)
   1.00000000000000
   0.00000000000000   4.96400000000000   4.96400000000000
   4.96400000000000   0.00000000000000   4.96400000000000
   4.96400000000000   4.96400000000000   0.00000000000000
    Co       S
     9       8
Direct
   0.00000000000000   0.00000000000000   0.00000000000000   Co   (4a)
   0.37400000000000   0.37400000000000   0.37400000000000   Co   (32f)
   0.37400000000000   0.37400000000000  -1.12200000000000   Co   (32f)
   0.37400000000000  -1.12200000000000   0.37400000000000   Co   (32f)
  -1.12200000000000   0.37400000000000   0.37400000000000   Co   (32f)
  -0.37400000000000  -0.37400000000000   1.12200000000000   Co   (32f)
  -0.37400000000000  -0.37400000000000  -0.37400000000000   Co   (32f)
  -0.37400000000000   1.12200000000000  -0.37400000000000   Co   (32f)
   1.12200000000000  -0.37400000000000  -0.37400000000000   Co   (32f)
   0.25000000000000   0.25000000000000   0.25000000000000   S    (8c)
   0.75000000000000   0.75000000000000   0.75000000000000   S    (8c)
  -0.75910000000000   0.75910000000000   0.75910000000000   S   (24e)
   0.75910000000000  -0.75910000000000  -0.75910000000000   S   (24e)
   0.75910000000000  -0.75910000000000   0.75910000000000   S   (24e)
  -0.75910000000000   0.75910000000000  -0.75910000000000   S   (24e)
   0.75910000000000   0.75910000000000  -0.75910000000000   S   (24e)
  -0.75910000000000  -0.75910000000000   0.75910000000000   S   (24e)
```

$(NH_4)_3AlF_6$ ($J2_1$): AB30C16D3_cF200_225_a_ej_2f_bc - CIF

```
# CIF file
data_findsym-output
_audit_creation_method FINDSYM

_chemical_name_mineral 'AlF6H12N3'
_chemical_formula_sum 'Al F30 H16 N3'

loop_
_publ_author_name
 'A. A. Udovenko'
 'N. M. Laptash'
 'I. G. Maslennikova'
_journal_name_full_name
;
 Journal of Fluorine Chemistry
;
_journal_volume 124
_journal_year 2003
_journal_page_first 5
_journal_page_last 15
_publ_Section_title
;
 Orientation disorder in ammonium elpasolites: Crystal structures of (
    ↪ NH$_{4}$)$_{3}$AlF$_{6}$, (NH$_{4}$)$_{3}$TiOF$_{5}$ and (NH$_
    ↪ {4}$)$_{3}$FeF$_{6}$
;

_aflow_title '(NH$_{4}$)$_{3}$AlF$_{6}$ ($J2_{1}$) Structure'
_aflow_proto 'AB30C16D3_cF200_225_a_ej_2f_bc'
_aflow_params 'a,x_{4},x_{5},x_{6},y_{7},z_{7}'
_aflow_params_values '8.94011,0.1969,0.307,0.557,0.0536,0.1923'
_aflow_Strukturbericht '$J2_{1}$'
_aflow_Pearson 'cF200'

_symmetry_space_group_name_H-M "F 4/m -3 2/m"
_symmetry_Int_Tables_number 225

_cell_length_a     8.94011
_cell_length_b     8.94011
_cell_length_c     8.94011
_cell_angle_alpha 90.00000
_cell_angle_beta  90.00000
_cell_angle_gamma 90.00000

loop_
_space_group_symop_id
_space_group_symop_operation_xyz
1 x,y,z
2 x,-y,-z
3 -x,y,-z
4 -x,-y,z
5 y,z,x
6 y,-z,-x
7 -y,z,-x
8 -y,-z,x
9 z,x,y
10 z,-x,-y
11 -z,x,-y
12 -z,-x,y
13 -y,-x,-z
14 -y,x,z
15 y,-x,z
16 y,x,-z
17 -x,-z,-y
18 -x,z,y
19 x,-z,y
20 x,z,-y
21 -z,-y,-x
```

```
22 -z,y,x
23 z,-y,x
24 z,y,-x
25 -x,-y,-z
26 -x,y,z
27 x,-y,z
28 x,y,-z
29 -y,-z,-x
30 -y,z,x
31 y,-z,x
32 y,z,-x
33 -z,-x,-y
34 -z,x,y
35 z,-x,y
36 z,x,-y
37 y,x,z
38 y,-x,-z
39 -y,x,-z
40 -y,-x,z
41 x,z,y
42 x,-z,-y
43 -x,z,-y
44 -x,-z,y
45 z,y,x
46 z,-y,-x
47 -z,y,-x
48 -z,-y,x
49 x,y+1/2,z+1/2
50 x,-y+1/2,-z+1/2
51 -x,y+1/2,-z+1/2
52 -x,-y+1/2,z+1/2
53 y,z+1/2,x+1/2
54 y,-z+1/2,-x+1/2
55 -y,z+1/2,-x+1/2
56 -y,-z+1/2,x+1/2
57 z,x+1/2,y+1/2
58 z,-x+1/2,-y+1/2
59 -z,x+1/2,-y+1/2
60 -z,-x+1/2,y+1/2
61 -y,-x+1/2,-z+1/2
62 -y,x+1/2,z+1/2
63 y,-x+1/2,z+1/2
64 y,x+1/2,-z+1/2
65 -x,-z+1/2,-y+1/2
66 -x,z+1/2,y+1/2
67 x,-z+1/2,y+1/2
68 x,z+1/2,-y+1/2
69 -z,-y+1/2,-x+1/2
70 -z,y+1/2,x+1/2
71 z,-y+1/2,x+1/2
72 z,y+1/2,-x+1/2
73 -x,-y+1/2,-z+1/2
74 -x,y+1/2,z+1/2
75 x,-y+1/2,z+1/2
76 x,y+1/2,-z+1/2
77 -y,-z+1/2,-x+1/2
78 -y,z+1/2,x+1/2
79 y,-z+1/2,x+1/2
80 y,z+1/2,-x+1/2
81 -z,-x+1/2,-y+1/2
82 -z,x+1/2,y+1/2
83 z,-x+1/2,y+1/2
84 z,x+1/2,-y+1/2
85 y,x+1/2,z+1/2
86 y,-x+1/2,-z+1/2
87 -y,x+1/2,-z+1/2
88 -y,-x+1/2,z+1/2
89 x,z+1/2,y+1/2
90 x,-z+1/2,-y+1/2
91 -x,z+1/2,-y+1/2
92 -x,-z+1/2,y+1/2
93 z,y+1/2,x+1/2
94 z,-y+1/2,-x+1/2
95 -z,y+1/2,-x+1/2
96 -z,-y+1/2,x+1/2
97 x+1/2,y,z+1/2
98 x+1/2,-y,-z+1/2
99 -x+1/2,y,-z+1/2
100 -x+1/2,-y,z+1/2
101 y+1/2,z,x+1/2
102 y+1/2,-z,-x+1/2
103 -y+1/2,z,-x+1/2
104 -y+1/2,-z,x+1/2
105 z+1/2,x,y+1/2
106 z+1/2,-x,-y+1/2
107 -z+1/2,x,-y+1/2
108 -z+1/2,-x,y+1/2
109 -y+1/2,-x,-z+1/2
110 -y+1/2,x,z+1/2
111 y+1/2,-x,z+1/2
112 y+1/2,x,-z+1/2
113 -x+1/2,-z,-y+1/2
114 -x+1/2,z,y+1/2
115 x+1/2,-z,y+1/2
116 x+1/2,z,-y+1/2
117 -z+1/2,-y,-x+1/2
118 -z+1/2,y,x+1/2
119 z+1/2,-y,x+1/2
120 z+1/2,y,-x+1/2
121 -x+1/2,-y,-z+1/2
122 -x+1/2,y,z+1/2
123 x+1/2,-y,z+1/2
124 x+1/2,y,-z+1/2
125 -y+1/2,-z,-x+1/2
126 -y+1/2,z,x+1/2
127 y+1/2,-z,x+1/2
128 y+1/2,z,-x+1/2
129 -z+1/2,-x,-y+1/2
130 -z+1/2,x,y+1/2
131 z+1/2,-x,y+1/2
132 z+1/2,x,-y+1/2
133 y+1/2,x,z+1/2
134 y+1/2,-x,-z+1/2
135 -y+1/2,x,-z+1/2
136 -y+1/2,-x,z+1/2
137 x+1/2,z,y+1/2
138 x+1/2,-z,-y+1/2
139 -x+1/2,z,-y+1/2
140 -x+1/2,-z,y+1/2
141 z+1/2,y,x+1/2
142 z+1/2,-y,-x+1/2
143 -z+1/2,y,-x+1/2
144 -z+1/2,-y,x+1/2
145 x+1/2,y+1/2,z
146 x+1/2,-y+1/2,-z
147 -x+1/2,y+1/2,-z
148 -x+1/2,-y+1/2,z
149 y+1/2,z+1/2,x
150 y+1/2,-z+1/2,-x
151 -y+1/2,z+1/2,-x
152 -y+1/2,-z+1/2,x
153 z+1/2,x+1/2,y
154 z+1/2,-x+1/2,-y
155 -z+1/2,x+1/2,-y
156 -z+1/2,-x+1/2,y
157 -y+1/2,-x+1/2,-z
158 -y+1/2,x+1/2,z
159 y+1/2,-x+1/2,z
160 y+1/2,x+1/2,-z
161 -x+1/2,-z+1/2,-y
162 -x+1/2,z+1/2,y
163 x+1/2,-z+1/2,y
164 x+1/2,z+1/2,-y
165 -z+1/2,-y+1/2,-x
166 -z+1/2,y+1/2,x
167 z+1/2,-y+1/2,x
168 z+1/2,y+1/2,-x
169 -x+1/2,-y+1/2,-z
170 -x+1/2,y+1/2,z
171 x+1/2,-y+1/2,z
172 x+1/2,y+1/2,-z
173 -y+1/2,-z+1/2,-x
174 -y+1/2,z+1/2,x
175 y+1/2,-z+1/2,x
176 y+1/2,z+1/2,-x
177 -z+1/2,-x+1/2,-y
178 -z+1/2,x+1/2,y
179 z+1/2,-x+1/2,y
180 z+1/2,x+1/2,-y
181 y+1/2,x+1/2,z
182 y+1/2,-x+1/2,-z
183 -y+1/2,x+1/2,-z
184 -y+1/2,-x+1/2,z
185 x+1/2,z+1/2,y
186 x+1/2,-z+1/2,-y
187 -x+1/2,z+1/2,-y
188 -x+1/2,-z+1/2,y
189 z+1/2,y+1/2,x
190 z+1/2,-y+1/2,-x
191 -z+1/2,y+1/2,-x
192 -z+1/2,-y+1/2,x

loop_
_atom_site_label
_atom_site_type_symbol
_atom_site_symmetry_multiplicity
_atom_site_Wyckoff_label
_atom_site_fract_x
_atom_site_fract_y
_atom_site_fract_z
_atom_site_occupancy
Al1 Al 4 a 0.00000 0.00000 0.00000 1.00000
N1 N 4 b 0.50000 0.50000 0.50000 1.00000
N2 N 8 c 0.25000 0.25000 0.25000 1.00000
F1 F 24 e 0.19690 0.00000 0.00000 0.33333
H1 H 32 f 0.30700 0.30700 0.30700 1.00000
H2 H 32 f 0.55700 0.55700 0.55700 0.50000
F2 F 96 j 0.00000 0.05360 0.19230 0.16667
```

$(NH_4)_3AlF_6$ ($J2_1$): AB30C16D3_cF200_225_a_ej_2f_bc - POSCAR

```
AB30C16D3_cF200_225_a_ej_2f_bc & a,x4,x5,x6,y7,z7 --params=8.94011,
  ↪ 0.1969,0.307,0.557,0.0536,0.1923 & Fm-3m O_{h}^{5} #225 (abcef^
  ↪ 2j) & cF200 & $J2_{1}$ & AlF6H12N3 & AlF6H12N3 & A. A. Udovenko
  ↪ and N. M. Laptash and I. G. Maslennikova, J. Fluor. Chem. 124,
  ↪ 5-15 (2003)
   1.00000000000000
   0.00000000000000   4.47005500000000   4.47005500000000
   4.47005500000000   0.00000000000000   4.47005500000000
   4.47005500000000   4.47005500000000   0.00000000000000
    Al     F     H     N
     1    30    16     3
Direct
   0.00000000000000   0.00000000000000   0.00000000000000   Al   (4a)
  -0.19690000000000   0.19690000000000   0.19690000000000    F  (24e)
   0.19690000000000  -0.19690000000000  -0.19690000000000    F  (24e)
   0.19690000000000  -0.19690000000000   0.19690000000000    F  (24e)
  -0.19690000000000   0.19690000000000  -0.19690000000000    F  (24e)
   0.19690000000000   0.19690000000000  -0.19690000000000    F  (24e)
  -0.19690000000000  -0.19690000000000   0.19690000000000    F  (24e)
```

```
   0.24590000000000    0.13870000000000   -0.13870000000000    F   (96j)
   0.13870000000000    0.24590000000000   -0.24590000000000    F   (96j)
  -0.13870000000000   -0.24590000000000    0.24590000000000    F   (96j)
  -0.24590000000000   -0.13870000000000    0.13870000000000    F   (96j)
  -0.13870000000000    0.24590000000000    0.13870000000000    F   (96j)
  -0.24590000000000    0.13870000000000    0.24590000000000    F   (96j)
   0.24590000000000   -0.13870000000000   -0.24590000000000    F   (96j)
   0.13870000000000   -0.24590000000000   -0.13870000000000    F   (96j)
   0.13870000000000   -0.13870000000000    0.24590000000000    F   (96j)
   0.24590000000000   -0.24590000000000    0.13870000000000    F   (96j)
  -0.24590000000000    0.24590000000000   -0.13870000000000    F   (96j)
  -0.13870000000000    0.13870000000000   -0.24590000000000    F   (96j)
  -0.24590000000000   -0.13870000000000    0.24590000000000    F   (96j)
  -0.13870000000000   -0.24590000000000    0.13870000000000    F   (96j)
   0.13870000000000    0.24590000000000   -0.13870000000000    F   (96j)
   0.24590000000000    0.13870000000000   -0.24590000000000    F   (96j)
   0.13870000000000   -0.24590000000000    0.24590000000000    F   (96j)
   0.24590000000000   -0.13870000000000    0.13870000000000    F   (96j)
  -0.24590000000000    0.13870000000000   -0.13870000000000    F   (96j)
  -0.13870000000000    0.24590000000000   -0.24590000000000    F   (96j)
  -0.13870000000000    0.13870000000000    0.24590000000000    F   (96j)
  -0.24590000000000    0.24590000000000    0.13870000000000    F   (96j)
   0.24590000000000   -0.24590000000000   -0.13870000000000    F   (96j)
   0.13870000000000   -0.13870000000000   -0.24590000000000    F   (96j)
   0.30700000000000    0.30700000000000    0.30700000000000    H   (32f)
   0.30700000000000    0.30700000000000   -0.92100000000000    H   (32f)
   0.30700000000000   -0.92100000000000    0.30700000000000    H   (32f)
  -0.92100000000000    0.30700000000000    0.30700000000000    H   (32f)
  -0.30700000000000   -0.30700000000000    0.92100000000000    H   (32f)
  -0.30700000000000   -0.30700000000000   -0.30700000000000    H   (32f)
  -0.30700000000000    0.92100000000000   -0.30700000000000    H   (32f)
   0.92100000000000   -0.30700000000000   -0.30700000000000    H   (32f)
   0.55700000000000    0.55700000000000    0.55700000000000    H   (32f)
   0.55700000000000    0.55700000000000   -1.67100000000000    H   (32f)
   0.55700000000000   -1.67100000000000    0.55700000000000    H   (32f)
  -1.67100000000000    0.55700000000000    0.55700000000000    H   (32f)
  -0.55700000000000   -0.55700000000000    1.67100000000000    H   (32f)
  -0.55700000000000   -0.55700000000000   -0.55700000000000    H   (32f)
  -0.55700000000000    1.67100000000000   -0.55700000000000    H   (32f)
   1.67100000000000   -0.55700000000000   -0.55700000000000    H   (32f)
   0.50000000000000    0.50000000000000    0.50000000000000    N    (4b)
   0.25000000000000    0.25000000000000    0.25000000000000    N    (8c)
   0.75000000000000    0.75000000000000    0.75000000000000    N    (8c)
```

$L1_a$ (disputed $CuPt_3$): AB7_cF32_225_b_ad - CIF

```
# CIF file
data_findsym-output
_audit_creation_method FINDSYM

_chemical_name_mineral 'CuPt3'
_chemical_formula_sum 'Cu Pt7'

loop_
_publ_author_name
 'Y.-C. Tang'
_journal_name_full_name
;
 Acta Cristallographica
;
_journal_volume 4
_journal_year 1951
_journal_page_first 377
_journal_page_last 378
_publ_Section_title
;
 A cubic structure for the phase Pt$_{3}$Cu
;

# Found in Revisiting the CuPt$_{3}$ prototype and the $L1_{3}$
      ↪ structure, 2014

_aflow_title '$L1_{a}$ (disputed CuPt$_{3}$) Structure'
_aflow_proto 'AB7_cF32_225_b_ad'
_aflow_params 'a'
_aflow_params_values '5.6'
_aflow_Strukturbericht '$L1_{a}$'
_aflow_Pearson 'cF32'

_symmetry_space_group_name_H-M "F 4/m -3 2/m"
_symmetry_Int_Tables_number 225

_cell_length_a      5.60000
_cell_length_b      5.60000
_cell_length_c      5.60000
_cell_angle_alpha  90.00000
_cell_angle_beta   90.00000
_cell_angle_gamma  90.00000

loop_
_space_group_symop_id
_space_group_symop_operation_xyz
1 x,y,z
2 x,-y,-z
3 -x,y,-z
4 -x,-y,z
5 y,z,x
6 y,-z,-x
7 -y,z,-x
8 -y,-z,x
9 z,x,y
10 z,-x,-y
11 -z,x,-y
12 -z,-x,y
13 -y,-x,-z
14 -y,x,z
15 y,-x,z
16 y,x,-z
17 -x,-z,-y
18 -x,z,y
19 x,-z,y
20 x,z,-y
21 -z,-y,-x
22 -z,y,x
23 z,-y,x
24 z,y,-x
25 -x,-y,-z
26 -x,y,z
27 x,-y,z
28 x,y,-z
29 -y,-z,-x
30 -y,z,x
31 y,-z,x
32 y,z,-x
33 -z,-x,-y
34 -z,x,y
35 z,-x,y
36 z,x,-y
37 y,x,z
38 y,-x,-z
39 -y,x,-z
40 -y,-x,z
41 x,z,y
42 x,-z,-y
43 -x,z,-y
44 -x,-z,y
45 z,y,x
46 z,-y,-x
47 -z,y,-x
48 -z,-y,x
49 x,y+1/2,z+1/2
50 x,-y+1/2,-z+1/2
51 -x,y+1/2,-z+1/2
52 -x,-y+1/2,z+1/2
53 y,z+1/2,x+1/2
54 y,-z+1/2,-x+1/2
55 -y,z+1/2,-x+1/2
56 -y,-z+1/2,x+1/2
57 z,x+1/2,y+1/2
58 z,-x+1/2,-y+1/2
59 -z,x+1/2,-y+1/2
60 -z,-x+1/2,y+1/2
61 -y,-x+1/2,-z+1/2
62 -y,x+1/2,z+1/2
63 y,-x+1/2,z+1/2
64 y,x+1/2,-z+1/2
65 -x,-z+1/2,-y+1/2
66 -x,z+1/2,y+1/2
67 x,-z+1/2,y+1/2
68 x,z+1/2,-y+1/2
69 -z,-y+1/2,-x+1/2
70 -z,y+1/2,x+1/2
71 z,-y+1/2,x+1/2
72 z,y+1/2,-x+1/2
73 -x,-y+1/2,-z+1/2
74 -x,y+1/2,z+1/2
75 x,-y+1/2,z+1/2
76 x,y+1/2,-z+1/2
77 -y,-z+1/2,-x+1/2
78 -y,z+1/2,x+1/2
79 y,-z+1/2,x+1/2
80 y,z+1/2,-x+1/2
81 -z,-x+1/2,-y+1/2
82 -z,x+1/2,y+1/2
83 z,-x+1/2,y+1/2
84 z,x+1/2,-y+1/2
85 y,x+1/2,z+1/2
86 y,-x+1/2,-z+1/2
87 -y,x+1/2,-z+1/2
88 -y,-x+1/2,z+1/2
89 x,z+1/2,y+1/2
90 x,-z+1/2,-y+1/2
91 -x,z+1/2,-y+1/2
92 -x,-z+1/2,y+1/2
93 z,y+1/2,x+1/2
94 z,-y+1/2,-x+1/2
95 -z,y+1/2,-x+1/2
96 -z,-y+1/2,x+1/2
97 x+1/2,y,z+1/2
98 x+1/2,-y,-z+1/2
99 -x+1/2,y,-z+1/2
100 -x+1/2,-y,z+1/2
101 y+1/2,z,x+1/2
102 y+1/2,-z,-x+1/2
103 -y+1/2,z,-x+1/2
104 -y+1/2,-z,x+1/2
105 z+1/2,x,y+1/2
106 z+1/2,-x,-y+1/2
107 -z+1/2,x,-y+1/2
108 -z+1/2,-x,y+1/2
109 -y+1/2,-x,-z+1/2
110 -y+1/2,x,z+1/2
111 y+1/2,-x,z+1/2
112 y+1/2,x,-z+1/2
113 -x+1/2,-z,-y+1/2
114 -x+1/2,z,y+1/2
115 x+1/2,-z,y+1/2
116 x+1/2,z,-y+1/2
117 -z+1/2,-y,-x+1/2
118 -z+1/2,y,x+1/2
```

```
119 z+1/2,-y,x+1/2
120 z+1/2,y,-x+1/2
121 -x+1/2,-y,-z+1/2
122 -x+1/2,y,z+1/2
123 x+1/2,-y,z+1/2
124 x+1/2,y,-z+1/2
125 -y+1/2,-z,-x+1/2
126 -y+1/2,z,x+1/2
127 y+1/2,-z,x+1/2
128 y+1/2,z,-x+1/2
129 -z+1/2,-x,-y+1/2
130 -z+1/2,x,y+1/2
131 z+1/2,-x,y+1/2
132 z+1/2,x,-y+1/2
133 y+1/2,x,z+1/2
134 y+1/2,-x,-z+1/2
135 -y+1/2,x,-z+1/2
136 -y+1/2,-x,z+1/2
137 x+1/2,z,y+1/2
138 x+1/2,-z,-y+1/2
139 -x+1/2,z,-y+1/2
140 -x+1/2,-z,y+1/2
141 z+1/2,y,x+1/2
142 z+1/2,-y,-x+1/2
143 -z+1/2,y,-x+1/2
144 -z+1/2,-y,x+1/2
145 x+1/2,y+1/2,z
146 x+1/2,-y+1/2,-z
147 -x+1/2,y+1/2,-z
148 -x+1/2,-y+1/2,z
149 y+1/2,z+1/2,x
150 y+1/2,-z+1/2,-x
151 -y+1/2,z+1/2,-x
152 -y+1/2,-z+1/2,x
153 z+1/2,x+1/2,y
154 z+1/2,-x+1/2,-y
155 -z+1/2,x+1/2,-y
156 -z+1/2,-x+1/2,y
157 -y+1/2,-x+1/2,-z
158 -y+1/2,x+1/2,z
159 y+1/2,-x+1/2,z
160 y+1/2,x+1/2,-z
161 -x+1/2,-z+1/2,-y
162 -x+1/2,z+1/2,y
163 x+1/2,-z+1/2,y
164 x+1/2,z+1/2,-y
165 -z+1/2,-y+1/2,-x
166 -z+1/2,y+1/2,x
167 z+1/2,-y+1/2,x
168 z+1/2,y+1/2,-x
169 -x+1/2,-y+1/2,-z
170 -x+1/2,y+1/2,z
171 x+1/2,-y+1/2,z
172 x+1/2,y+1/2,-z
173 -y+1/2,-z+1/2,-x
174 -y+1/2,z+1/2,x
175 y+1/2,-z+1/2,x
176 y+1/2,z+1/2,-x
177 -z+1/2,-x+1/2,-y
178 -z+1/2,x+1/2,y
179 z+1/2,-x+1/2,y
180 z+1/2,x+1/2,-y
181 y+1/2,x+1/2,z
182 y+1/2,-x+1/2,-z
183 -y+1/2,x+1/2,-z
184 -y+1/2,-x+1/2,z
185 x+1/2,z+1/2,y
186 x+1/2,-z+1/2,-y
187 -x+1/2,z+1/2,-y
188 -x+1/2,-z+1/2,y
189 z+1/2,y+1/2,x
190 z+1/2,-y+1/2,-x
191 -z+1/2,y+1/2,-x
192 -z+1/2,-y+1/2,x

loop_
_atom_site_label
_atom_site_type_symbol
_atom_site_symmetry_multiplicity
_atom_site_Wyckoff_label
_atom_site_fract_x
_atom_site_fract_y
_atom_site_fract_z
_atom_site_occupancy
Pt1 Pt 4 a 0.00000 0.00000 0.00000 1.00000
Cu1 Cu 4 b 0.50000 0.50000 0.50000 1.00000
Pt2 Pt 24 d 0.00000 0.25000 0.25000 1.00000
```

$L1_a$ (disputed $CuPt_3$): AB7_cF32_225_b_ad - POSCAR

```
AB7_cF32_225_b_ad & a --params=5.6 & Fm-3m O_{h}^{5} #225 (abd) & cF32 &
    ↪ $L1_{a}$ & CuPt3 & CuPt3 & Y.-C. Tang, Acta Cryst. 4, 377-378
    ↪ (1951)
   1.00000000000000
   0.00000000000000   2.80000000000000   2.80000000000000
   2.80000000000000   0.00000000000000   2.80000000000000
   2.80000000000000   2.80000000000000   0.00000000000000
    Cu     Pt
     1      7
Direct
   0.50000000000000   0.50000000000000   0.50000000000000   Cu   (4b)
   0.00000000000000   0.00000000000000   0.00000000000000   Pt   (4a)
   0.50000000000000   0.00000000000000   0.00000000000000   Pt  (24d)
   0.00000000000000   0.50000000000000   0.50000000000000   Pt  (24d)
   0.00000000000000   0.50000000000000   0.00000000000000   Pt  (24d)
   0.50000000000000   0.00000000000000   0.50000000000000   Pt  (24d)
   0.00000000000000   0.00000000000000   0.50000000000000   Pt  (24d)
   0.50000000000000   0.50000000000000   0.00000000000000   Pt  (24d)
```

Sulphohalite [$Na_6ClF(SO_4)_2$, $H5_8$]: ABC6D8E2_cF72_225_b_a_e_f_c - CIF

```
# CIF file
data_findsym-output
_audit_creation_method FINDSYM

_chemical_name_mineral 'Sulphohalite'
_chemical_formula_sum 'Cl F Na6 O8 S2'

loop_
_publ_author_name
 'A. Pabst'
_journal_name_full_name
;
 Zeitschrift f{\"u}r Kristallographie - Crystalline Materials
;
_journal_volume 89
_journal_year 1934
_journal_page_first 514
_journal_page_last 517
_publ_Section_title
;
 The Crystal Structure of Sulphohalite
;

# Found in Strukturbericht Band III 1933-1935, 1937

_aflow_title 'Sulphohalite [Na$_{6}$ClF(SO$_{4}$)$_{2}$, $H5_{8}$]
    ↪ Structure'
_aflow_proto 'ABC6D8E2_cF72_225_b_a_e_f_c'
_aflow_params 'a,x_{4},x_{5}'
_aflow_params_values '10.08,0.226,0.164'
_aflow_Strukturbericht '$H5_{8}$'
_aflow_Pearson 'cF72'

_symmetry_space_group_name_H-M "F 4/m -3 2/m"
_symmetry_Int_Tables_number 225

_cell_length_a     10.08000
_cell_length_b     10.08000
_cell_length_c     10.08000
_cell_angle_alpha  90.00000
_cell_angle_beta   90.00000
_cell_angle_gamma  90.00000

loop_
_space_group_symop_id
_space_group_symop_operation_xyz
1 x,y,z
2 x,-y,-z
3 -x,y,-z
4 -x,-y,z
5 y,z,x
6 y,-z,-x
7 -y,z,-x
8 -y,-z,x
9 z,x,y
10 z,-x,-y
11 -z,x,-y
12 -z,-x,y
13 -y,-x,-z
14 -y,x,z
15 y,-x,z
16 y,x,-z
17 -x,-z,-y
18 -x,z,y
19 x,-z,y
20 x,z,-y
21 -z,-y,-x
22 -z,y,x
23 z,-y,x
24 z,y,-x
25 -x,-y,-z
26 -x,y,z
27 x,-y,z
28 x,y,-z
29 -y,-z,-x
30 -y,z,x
31 y,-z,x
32 y,z,-x
33 -z,-x,-y
34 -z,x,y
35 z,-x,y
36 z,x,-y
37 y,x,z
38 y,-x,-z
39 -y,x,-z
40 -y,-x,z
41 x,z,y
42 x,-z,-y
43 -x,z,-y
44 -x,-z,y
45 z,y,x
46 z,-y,-x
47 -z,y,-x
48 -z,-y,x
49 x,y+1/2,z+1/2
50 x,-y+1/2,-z+1/2
51 -x,y+1/2,-z+1/2
52 -x,-y+1/2,z+1/2
53 y,z+1/2,x+1/2
```

```
54 y,-z+1/2,-x+1/2
55 -y,z+1/2,-x+1/2
56 -y,-z+1/2,x+1/2
57 z,x+1/2,y+1/2
58 z,-x+1/2,-y+1/2
59 -z,x+1/2,-y+1/2
60 -z,-x+1/2,y+1/2
61 -y,-x+1/2,-z+1/2
62 -y,x+1/2,z+1/2
63 y,-x+1/2,z+1/2
64 y,x+1/2,-z+1/2
65 -x,-z+1/2,-y+1/2
66 -x,z+1/2,y+1/2
67 x,-z+1/2,y+1/2
68 x,z+1/2,-y+1/2
69 -z,-y+1/2,-x+1/2
70 -z,y+1/2,x+1/2
71 z,-y+1/2,x+1/2
72 z,y+1/2,-x+1/2
73 -x,-y+1/2,-z+1/2
74 -x,y+1/2,z+1/2
75 x,-y+1/2,z+1/2
76 x,y+1/2,-z+1/2
77 -y,-z+1/2,-x+1/2
78 -y,z+1/2,x+1/2
79 y,-z+1/2,x+1/2
80 y,z+1/2,-x+1/2
81 -z,-x+1/2,-y+1/2
82 -z,x+1/2,y+1/2
83 z,-x+1/2,y+1/2
84 z,x+1/2,-y+1/2
85 y,x+1/2,z+1/2
86 y,-x+1/2,-z+1/2
87 -y,x+1/2,-z+1/2
88 -y,-x+1/2,z+1/2
89 x,z+1/2,y+1/2
90 x,-z+1/2,-y+1/2
91 -x,z+1/2,-y+1/2
92 -x,-z+1/2,y+1/2
93 z,y+1/2,x+1/2
94 z,-y+1/2,-x+1/2
95 -z,y+1/2,-x+1/2
96 -z,-y+1/2,x+1/2
97 x+1/2,y,z+1/2
98 x+1/2,-y,-z+1/2
99 -x+1/2,y,-z+1/2
100 -x+1/2,-y,z+1/2
101 y+1/2,z,x+1/2
102 y+1/2,-z,-x+1/2
103 -y+1/2,z,-x+1/2
104 -y+1/2,-z,x+1/2
105 z+1/2,x,y+1/2
106 z+1/2,-x,-y+1/2
107 -z+1/2,x,-y+1/2
108 -z+1/2,-x,y+1/2
109 -y+1/2,-x,-z+1/2
110 -y+1/2,x,z+1/2
111 y+1/2,-x,z+1/2
112 y+1/2,x,-z+1/2
113 -x+1/2,-z,-y+1/2
114 -x+1/2,z,y+1/2
115 x+1/2,-z,y+1/2
116 x+1/2,z,-y+1/2
117 -z+1/2,-y,-x+1/2
118 -z+1/2,y,x+1/2
119 z+1/2,-y,x+1/2
120 z+1/2,y,-x+1/2
121 -x+1/2,-y,-z+1/2
122 -x+1/2,y,z+1/2
123 x+1/2,-y,z+1/2
124 x+1/2,y,-z+1/2
125 -y+1/2,-z,-x+1/2
126 -y+1/2,z,x+1/2
127 y+1/2,-z,x+1/2
128 y+1/2,z,-x+1/2
129 -z+1/2,-x,-y+1/2
130 -z+1/2,x,y+1/2
131 z+1/2,-x,y+1/2
132 z+1/2,x,-y+1/2
133 y+1/2,x,z+1/2
134 y+1/2,-x,-z+1/2
135 -y+1/2,x,-z+1/2
136 -y+1/2,-x,z+1/2
137 x+1/2,z,y+1/2
138 x+1/2,-z,-y+1/2
139 -x+1/2,z,-y+1/2
140 -x+1/2,-z,y+1/2
141 z+1/2,y,x+1/2
142 z+1/2,-y,-x+1/2
143 -z+1/2,y,-x+1/2
144 -z+1/2,-y,x+1/2
145 x+1/2,y+1/2,z
146 x+1/2,-y+1/2,-z
147 -x+1/2,y+1/2,-z
148 -x+1/2,-y+1/2,z
149 y+1/2,z+1/2,x
150 y+1/2,-z+1/2,-x
151 -y+1/2,z+1/2,-x
152 -y+1/2,-z+1/2,x
153 z+1/2,x+1/2,y
154 z+1/2,-x+1/2,-y
155 -z+1/2,x+1/2,-y
156 -z+1/2,-x+1/2,y
157 -y+1/2,-x+1/2,-z
158 -y+1/2,x+1/2,z
159 y+1/2,-x+1/2,z
160 y+1/2,x+1/2,-z
161 -x+1/2,-z+1/2,-y
162 -x+1/2,z+1/2,y
163 x+1/2,-z+1/2,y
164 x+1/2,z+1/2,-y
165 -z+1/2,-y+1/2,-x
166 -z+1/2,y+1/2,x
167 z+1/2,-y+1/2,x
168 z+1/2,y+1/2,-x
169 -x+1/2,-y+1/2,-z
170 -x+1/2,y+1/2,z
171 x+1/2,-y+1/2,z
172 x+1/2,y+1/2,-z
173 -y+1/2,-z+1/2,-x
174 -y+1/2,z+1/2,x
175 y+1/2,-z+1/2,x
176 y+1/2,z+1/2,-x
177 -z+1/2,-x+1/2,-y
178 -z+1/2,x+1/2,y
179 z+1/2,-x+1/2,y
180 z+1/2,x+1/2,-y
181 y+1/2,x+1/2,z
182 y+1/2,-x+1/2,-z
183 -y+1/2,x+1/2,-z
184 -y+1/2,-x+1/2,z
185 x+1/2,z+1/2,y
186 x+1/2,-z+1/2,-y
187 -x+1/2,z+1/2,-y
188 -x+1/2,-z+1/2,y
189 z+1/2,y+1/2,x
190 z+1/2,-y+1/2,-x
191 -z+1/2,y+1/2,-x
192 -z+1/2,-y+1/2,x

loop_
_atom_site_label
_atom_site_type_symbol
_atom_site_symmetry_multiplicity
_atom_site_Wyckoff_label
_atom_site_fract_x
_atom_site_fract_y
_atom_site_fract_z
_atom_site_occupancy
F1 F 4 a 0.00000 0.00000 0.00000 1.00000
Cl1 Cl 4 b 0.50000 0.50000 0.50000 1.00000
S1 S 8 c 0.25000 0.25000 0.25000 1.00000
Na1 Na 24 e 0.22600 0.00000 0.00000 1.00000
O1 O 32 f 0.16400 0.16400 0.16400 1.00000
```

Sulphohalite [$Na_6ClF(SO_4)_2$, $H5_8$]: ABC6D8E2_cF72_225_b_a_e_f_c - POSCAR

```
ABC6D8E2_cF72_225_b_a_e_f_c & a,x4,x5 --params=10.08,0.226,0.164 & Fm-3m
    ↪ O_{h}^{5} #225 (abcef) & cF72 & $H5_{8}$ & ClFNa6O8S2 &
    ↪ Sulphohalite & A. Pabst, Zeitschrift f{"u}r Kristallographie -
    ↪ Crystalline Materials 89, 514-517 (1934)
   1.00000000000000
   0.00000000000000   5.04000000000000   5.04000000000000
   5.04000000000000   0.00000000000000   5.04000000000000
   5.04000000000000   5.04000000000000   0.00000000000000
    Cl     F     Na     O     S
     1     1      6     8     2
Direct
   0.50000000000000   0.50000000000000   0.50000000000000   Cl   (4b)
   0.00000000000000   0.00000000000000   0.00000000000000   F    (4a)
  -0.22600000000000   0.22600000000000   0.22600000000000   Na   (24e)
   0.22600000000000  -0.22600000000000  -0.22600000000000   Na   (24e)
   0.22600000000000  -0.22600000000000   0.22600000000000   Na   (24e)
  -0.22600000000000   0.22600000000000  -0.22600000000000   Na   (24e)
   0.22600000000000   0.22600000000000  -0.22600000000000   Na   (24e)
  -0.22600000000000  -0.22600000000000   0.22600000000000   Na   (24e)
   0.16400000000000   0.16400000000000   0.16400000000000   O    (32f)
   0.16400000000000   0.16400000000000  -0.49200000000000   O    (32f)
   0.16400000000000  -0.49200000000000   0.16400000000000   O    (32f)
  -0.49200000000000   0.16400000000000   0.16400000000000   O    (32f)
  -0.16400000000000  -0.16400000000000   0.49200000000000   O    (32f)
  -0.16400000000000  -0.16400000000000  -0.16400000000000   O    (32f)
  -0.16400000000000   0.49200000000000  -0.16400000000000   O    (32f)
   0.49200000000000  -0.16400000000000  -0.16400000000000   O    (32f)
   0.25000000000000   0.25000000000000   0.25000000000000   S    (8c)
   0.75000000000000   0.75000000000000   0.75000000000000   S    (8c)
```

$\gamma$-$Ga_2O_3$: A11B4_cF120_227_acdf_e - CIF

```
# CIF file
data_findsym-output
_audit_creation_method FINDSYM

_chemical_name_mineral 'Ga2O3'
_chemical_formula_sum 'Ga11 O4'

loop_
_publ_author_name
 'H. Y. Playford'
 'A. C. Hannon'
 'E. R. Barney'
 'R. I. Walton'
_journal_name_full_name
;
 Chemistry - A European Journal
;
_journal_volume 19
_journal_year 2013
_journal_page_first 2803
_journal_page_last 2813
```

```
_publ_Section_title
;
 Structures of Uncharacterised Polymorphs of Gallium Oxide from Total
       ↪ Neutron Diffraction
;

_aflow_title '$\gamma$-Ga$_{2}$O$_{3}$ Structure'
_aflow_proto 'A11B4_cF120_227_acdf_e'
_aflow_params 'a,x_{4},x_{5}'
_aflow_params_values '8.2376,0.2552,0.368'
_aflow_Strukturbericht 'None'
_aflow_Pearson 'cF120'

_symmetry_space_group_name_H-M "F 41/d -3 2/m (origin choice 2)"
_symmetry_Int_Tables_number 227

_cell_length_a     8.23760
_cell_length_b     8.23760
_cell_length_c     8.23760
_cell_angle_alpha 90.00000
_cell_angle_beta  90.00000
_cell_angle_gamma 90.00000

loop_
_space_group_symop_id
_space_group_symop_operation_xyz
1 x,y,z
2 x,-y+1/4,-z+1/4
3 -x+1/4,y,-z+1/4
4 -x+1/4,-y+1/4,z
5 y,z,x
6 y,-z+1/4,-x+1/4
7 -y+1/4,z,-x+1/4
8 -y+1/4,-z+1/4,x
9 z,x,y
10 z,-x+1/4,-y+1/4
11 -z+1/4,x,-y+1/4
12 -z+1/4,-x+1/4,y
13 -y,-x,-z
14 -y,x+1/4,z+1/4
15 y+1/4,-x,z+1/4
16 y+1/4,x+1/4,-z
17 -x,-z,-y
18 -x,z+1/4,y+1/4
19 x+1/4,-z,y+1/4
20 x+1/4,z+1/4,-y
21 -z,-y,-x
22 -z,y+1/4,x+1/4
23 z+1/4,-y,x+1/4
24 z+1/4,y+1/4,-x
25 -x,-y,-z
26 -x,y+1/4,z+1/4
27 x+1/4,-y,z+1/4
28 x+1/4,y+1/4,-z
29 -y,-z,-x
30 -y,z+1/4,x+1/4
31 y+1/4,-z,x+1/4
32 y+1/4,z+1/4,-x
33 -z,-x,-y
34 -z,x+1/4,y+1/4
35 z+1/4,-x,y+1/4
36 z+1/4,x+1/4,-y
37 y,x,z
38 y,-x+1/4,-z+1/4
39 -y+1/4,x,-z+1/4
40 -y+1/4,-x+1/4,z
41 x,z,y
42 x,-z+1/4,-y+1/4
43 -x+1/4,z,-y+1/4
44 -x+1/4,-z+1/4,y
45 z,y,x
46 z,-y+1/4,-x+1/4
47 -z+1/4,y,-x+1/4
48 -z+1/4,-y+1/4,x
49 x,y+1/2,z+1/2
50 x,-y+3/4,-z+3/4
51 -x+1/4,y+1/2,-z+3/4
52 -x+1/4,-y+3/4,z+1/2
53 y,z+1/2,x+1/2
54 y,-z+3/4,-x+3/4
55 -y+1/4,z+1/2,-x+3/4
56 -y+1/4,-z+3/4,x+1/2
57 z,x+1/2,y+1/2
58 z,-x+3/4,-y+3/4
59 -z+1/4,x+1/2,-y+3/4
60 -z+1/4,-x+3/4,y+1/2
61 -y,-x+1/2,-z+1/2
62 -y,x+3/4,z+3/4
63 y+1/4,-x+1/2,z+3/4
64 y+1/4,x+3/4,-z+1/2
65 -x,-z+1/2,-y+1/2
66 -x,z+3/4,y+3/4
67 x+1/4,-z+1/2,y+3/4
68 x+1/4,z+3/4,-y+1/2
69 -z,-y+1/2,-x+1/2
70 -z,y+3/4,x+3/4
71 z+1/4,-y+1/2,x+3/4
72 z+1/4,y+3/4,-x+1/2
73 -x,-y+1/2,-z+1/2
74 -x,y+3/4,z+3/4
75 x+1/4,-y+1/2,z+3/4
76 x+1/4,y+3/4,-z+1/2
77 -y,-z+1/2,-x+1/2
78 -y,z+3/4,x+3/4
79 y+1/4,-z+1/2,x+3/4
80 y+1/4,z+3/4,-x+1/2
81 -z,-x+1/2,-y+1/2
82 -z,x+3/4,y+3/4
83 z+1/4,-x+1/2,y+3/4
84 z+1/4,x+3/4,-y+1/2
85 y,x+1/2,z+1/2
86 y,-x+3/4,-z+3/4
87 -y+1/4,x+1/2,-z+3/4
88 -y+1/4,-x+3/4,z+1/2
89 x,z+1/2,y+1/2
90 x,-z+3/4,-y+3/4
91 -x+1/4,z+1/2,-y+3/4
92 -x+1/4,-z+3/4,y+1/2
93 z,y+1/2,x+1/2
94 z,-y+3/4,-x+3/4
95 -z+1/4,y+1/2,-x+3/4
96 -z+1/4,-y+3/4,x+1/2
97 x+1/2,y,z+1/2
98 x+1/2,-y+1/4,-z+3/4
99 -x+3/4,y,-z+3/4
100 -x+3/4,-y+1/4,z+1/2
101 y+1/2,z,x+1/2
102 y+1/2,-z+1/4,-x+3/4
103 -y+3/4,z,-x+3/4
104 -y+3/4,-z+1/4,x+1/2
105 z+1/2,x,y+1/2
106 z+1/2,-x+1/4,-y+3/4
107 -z+3/4,x,-y+3/4
108 -z+3/4,-x+1/4,y+1/2
109 -y+1/2,-x,-z+1/2
110 -y+1/2,x+1/4,z+3/4
111 y+3/4,-x,z+3/4
112 y+3/4,x+1/4,-z+1/2
113 -x+1/2,-z,-y+1/2
114 -x+1/2,z+1/4,y+3/4
115 x+3/4,-z,y+3/4
116 x+3/4,z+1/4,-y+1/2
117 -z+1/2,-y,-x+1/2
118 -z+1/2,y+1/4,x+3/4
119 z+3/4,-y,x+3/4
120 z+3/4,y+1/4,-x+1/2
121 -x+1/2,-y,-z+1/2
122 -x+1/2,y+1/4,z+3/4
123 x+3/4,-y,z+3/4
124 x+3/4,y+1/4,-z+1/2
125 -y+1/2,-z,-x+1/2
126 -y+1/2,z+1/4,x+3/4
127 y+3/4,-z,x+3/4
128 y+3/4,z+1/4,-x+1/2
129 -z+1/2,-x,-y+1/2
130 -z+1/2,x+1/4,y+3/4
131 z+3/4,-x,y+3/4
132 z+3/4,x+1/4,-y+1/2
133 y+1/2,x,z+1/2
134 y+1/2,-x+1/4,-z+3/4
135 -y+3/4,x,-z+3/4
136 -y+3/4,-x+1/4,z+1/2
137 x+1/2,z,y+1/2
138 x+1/2,-z+1/4,-y+3/4
139 -x+3/4,z,-y+3/4
140 -x+3/4,-z+1/4,y+1/2
141 z+1/2,y,x+1/2
142 z+1/2,-y+1/4,-x+3/4
143 -z+3/4,y,-x+3/4
144 -z+3/4,-y+1/4,x+1/2
145 x+1/2,y+1/2,z
146 x+1/2,-y+3/4,-z+1/4
147 -x+3/4,y+1/2,-z+1/4
148 -x+3/4,-y+3/4,z
149 y+1/2,z+1/2,x
150 y+1/2,-z+3/4,-x+1/4
151 -y+3/4,z+1/2,-x+1/4
152 -y+3/4,-z+3/4,x
153 z+1/2,x+1/2,y
154 z+1/2,-x+3/4,-y+1/4
155 -z+3/4,x+1/2,-y+1/4
156 -z+3/4,-x+3/4,y
157 -y+1/2,-x+1/2,-z
158 -y+1/2,x+3/4,z+1/4
159 y+3/4,-x+1/2,z+1/4
160 y+3/4,x+3/4,-z
161 -x+1/2,-z+1/2,-y
162 -x+1/2,z+3/4,y+1/4
163 x+3/4,-z+1/2,y+1/4
164 x+3/4,z+3/4,-y
165 -z+1/2,-y+1/2,-x
166 -z+1/2,y+3/4,x+1/4
167 z+3/4,-y+1/2,x+1/4
168 z+3/4,y+3/4,-x
169 -x+1/2,-y+1/2,-z
170 -x+1/2,y+3/4,z+1/4
171 x+3/4,-y+1/2,z+1/4
172 x+3/4,y+3/4,-z
173 -y+1/2,-z+1/2,-x
174 -y+1/2,z+3/4,x+1/4
175 y+3/4,-z+1/2,x+1/4
176 y+3/4,z+3/4,-x
177 -z+1/2,-x+1/2,-y
178 -z+1/2,x+3/4,y+1/4
179 z+3/4,-x+1/2,y+1/4
180 z+3/4,x+3/4,-y
181 y+1/2,x+1/2,z
182 y+1/2,-x+3/4,-z+1/4
183 -y+3/4,x+1/2,-z+1/4
184 -y+3/4,-x+3/4,z
```

```
185 x+1/2,z+1/2,y
186 x+1/2,-z+3/4,-y+1/4
187 -x+3/4,z+1/2,-y+1/4
188 -x+3/4,-z+3/4,y
189 z+1/2,y+1/2,x
190 z+1/2,-y+3/4,-x+1/4
191 -z+3/4,y+1/2,-x+1/4
192 -z+3/4,-y+3/4,x

loop_
_atom_site_label
_atom_site_type_symbol
_atom_site_symmetry_multiplicity
_atom_site_Wyckoff_label
_atom_site_fract_x
_atom_site_fract_y
_atom_site_fract_z
_atom_site_occupancy
Ga1 Ga 8 a 0.12500 0.12500 0.12500 1.00000
Ga2 Ga 16 c 0.00000 0.00000 0.00000 1.00000
Ga3 Ga 16 d 0.50000 0.50000 0.50000 1.00000
O1 O 32 e 0.25520 0.25520 0.25520 1.00000
Ga4 Ga 48 f 0.36800 0.12500 0.12500 1.00000
```

γ-$Ga_2O_3$: A11B4_cF120_227_acdf_e - POSCAR

```
A11B4_cF120_227_acdf_e & a,x4,x5 --params=8.2376,0.2552,0.368 & Fd-3m O_
    ↪ {h}^{7} #227 (acdef) & cF120 & None & Ga2O3 & Ga2O3 & H. Y.
    ↪ Playford et al., Chem. Eur. J. 19, 2803-2813 (2013)
   1.00000000000000
   0.00000000000000   4.11880000000000   4.11880000000000
   4.11880000000000   0.00000000000000   4.11880000000000
   4.11880000000000   4.11880000000000   0.00000000000000
    Ga        O
    22        8
Direct
   0.12500000000000   0.12500000000000   0.12500000000000   Ga   (8a)
   0.87500000000000   0.87500000000000   0.87500000000000   Ga   (8a)
   0.00000000000000   0.00000000000000   0.00000000000000   Ga   (16c)
   0.00000000000000   0.00000000000000   0.50000000000000   Ga   (16c)
   0.00000000000000   0.50000000000000   0.00000000000000   Ga   (16c)
   0.50000000000000   0.00000000000000   0.00000000000000   Ga   (16c)
   0.50000000000000   0.50000000000000   0.50000000000000   Ga   (16d)
   0.50000000000000   0.50000000000000   0.00000000000000   Ga   (16d)
   0.50000000000000   0.00000000000000   0.50000000000000   Ga   (16d)
   0.00000000000000   0.50000000000000   0.50000000000000   Ga   (16d)
  -0.11800000000000   0.36800000000000   0.36800000000000   Ga   (48f)
   0.36800000000000  -0.11800000000000  -0.11800000000000   Ga   (48f)
   0.36800000000000  -0.11800000000000   0.36800000000000   Ga   (48f)
  -0.11800000000000   0.36800000000000  -0.11800000000000   Ga   (48f)
   0.36800000000000   0.36800000000000  -0.11800000000000   Ga   (48f)
  -0.11800000000000  -0.11800000000000   0.36800000000000   Ga   (48f)
   1.11800000000000  -0.36800000000000   1.11800000000000   Ga   (48f)
  -0.36800000000000   1.11800000000000  -0.36800000000000   Ga   (48f)
  -0.36800000000000   1.11800000000000   1.11800000000000   Ga   (48f)
   1.11800000000000  -0.36800000000000  -0.36800000000000   Ga   (48f)
  -0.36800000000000  -0.36800000000000   1.11800000000000   Ga   (48f)
   1.11800000000000   1.11800000000000  -0.36800000000000   Ga   (48f)
   0.25520000000000   0.25520000000000   0.25520000000000   O    (32e)
   0.25520000000000   0.25520000000000  -0.26560000000000   O    (32e)
   0.25520000000000  -0.26560000000000   0.25520000000000   O    (32e)
  -0.26560000000000   0.25520000000000   0.25520000000000   O    (32e)
  -0.25520000000000  -0.25520000000000   1.26560000000000   O    (32e)
  -0.25520000000000  -0.25520000000000  -0.25520000000000   O    (32e)
  -0.25520000000000   1.26560000000000  -0.25520000000000   O    (32e)
   1.26560000000000  -0.25520000000000  -0.25520000000000   O    (32e)
```

Predicted $Li_2MgH_{16}$ High-Temperature Superconductor (250 GPa): A16B2C_cF152_227_eg_d_a - CIF

```
# CIF file
data_findsym-output
_audit_creation_method FINDSYM

_chemical_name_mineral 'H16Li2Mg'
_chemical_formula_sum 'H16 Li2 Mg'

loop_
_publ_author_name
 'Y. Sun'
 'J. Lv'
 'Y. Xie'
 'H. Liu'
 'Y. Ma'
_journal_name_full_name
;
 Physical Review Letters
;
_journal_volume 123
_journal_year 2019
_journal_page_first 097001
_journal_page_last 097001
_publ_Section_title
;
 Route to a Superconducting Phase above Room Temperature in
    ↪ Electron-Doped Hydride Compounds under High Pressure
;

_aflow_title 'Predicted Li$_{2}$MgH$_{16}$ High-Temperature
    ↪ Superconductor (250~GPa) Structure'
_aflow_proto 'A16B2C_cF152_227_eg_d_a'
_aflow_params 'a,x_{3},x_{4},z_{4}'
_aflow_params_values '6.71851,0.2915,0.56034,0.8706'
_aflow_Strukturbericht 'None'
_aflow_Pearson 'cF152'

_symmetry_space_group_name_H-M "F 41/d -3 2/m (origin choice 2)"
_symmetry_Int_Tables_number 227

_cell_length_a     6.71851
_cell_length_b     6.71851
_cell_length_c     6.71851
_cell_angle_alpha  90.00000
_cell_angle_beta   90.00000
_cell_angle_gamma  90.00000

loop_
_space_group_symop_id
_space_group_symop_operation_xyz
1 x,y,z
2 x,-y+1/4,-z+1/4
3 -x+1/4,y,-z+1/4
4 -x+1/4,-y+1/4,z
5 y,z,x
6 y,-z+1/4,-x+1/4
7 -y+1/4,z,-x+1/4
8 -y+1/4,-z+1/4,x
9 z,x,y
10 z,-x+1/4,-y+1/4
11 -z+1/4,x,-y+1/4
12 -z+1/4,-x+1/4,y
13 -y,-x,-z
14 -y,x+1/4,z+1/4
15 y+1/4,-x,z+1/4
16 y+1/4,x+1/4,-z
17 -x,-z,-y
18 -x,z+1/4,y+1/4
19 x+1/4,-z,y+1/4
20 x+1/4,z+1/4,-y
21 -z,-y,-x
22 -z,y+1/4,x+1/4
23 z+1/4,-y,x+1/4
24 z+1/4,y+1/4,-x
25 -x,-y,-z
26 -x,y+1/4,z+1/4
27 x+1/4,-y,z+1/4
28 x+1/4,y+1/4,-z
29 -y,-z,-x
30 -y,z+1/4,x+1/4
31 y+1/4,-z,x+1/4
32 y+1/4,z+1/4,-x
33 -z,-x,-y
34 -z,x+1/4,y+1/4
35 z+1/4,-x,y+1/4
36 z+1/4,x+1/4,-y
37 y,x,z
38 y,-x+1/4,-z+1/4
39 -y+1/4,x,-z+1/4
40 -y+1/4,-x+1/4,z
41 x,z,y
42 x,-z+1/4,-y+1/4
43 -x+1/4,z,-y+1/4
44 -x+1/4,-z+1/4,y
45 z,y,x
46 z,-y+1/4,-x+1/4
47 -z+1/4,y,-x+1/4
48 -z+1/4,-y+1/4,x
49 x,y+1/2,z+1/2
50 x,-y+3/4,-z+3/4
51 -x+1/4,y+1/2,-z+3/4
52 -x+1/4,-y+3/4,z+1/2
53 y,z+1/2,x+1/2
54 y,-z+3/4,-x+3/4
55 -y+1/4,z+1/2,-x+3/4
56 -y+1/4,-z+3/4,x+1/2
57 z,x+1/2,y+1/2
58 z,-x+3/4,-y+3/4
59 -z+1/4,x+1/2,-y+3/4
60 -z+1/4,-x+3/4,y+1/2
61 -y,-x+1/2,-z+1/2
62 -y,x+3/4,z+3/4
63 y+1/4,-x+1/2,z+3/4
64 y+1/4,x+3/4,-z+1/2
65 -x,-z+1/2,-y+1/2
66 -x,z+3/4,y+3/4
67 x+1/4,-z+1/2,y+3/4
68 x+1/4,z+3/4,-y+1/2
69 -z,-y+1/2,-x+1/2
70 -z,y+3/4,x+3/4
71 z+1/4,-y+1/2,x+3/4
72 z+1/4,y+3/4,-x+1/2
73 -x,-y+1/2,-z+1/2
74 -x,y+3/4,z+3/4
75 x+1/4,-y+1/2,z+3/4
76 x+1/4,y+3/4,-z+1/2
77 -y,-z+1/2,-x+1/2
78 -y,z+3/4,x+3/4
79 y+1/4,-z+1/2,x+3/4
80 y+1/4,z+3/4,-x+1/2
81 -z,-x+1/2,-y+1/2
82 -z,x+3/4,y+3/4
83 z+1/4,-x+1/2,y+3/4
84 z+1/4,x+3/4,-y+1/2
85 y,x+1/2,z+1/2
86 y,-x+3/4,-z+3/4
87 -y+1/4,x+1/2,-z+3/4
88 -y+1/4,-x+3/4,z+1/2
89 x,z+1/2,y+1/2
90 x,-z+3/4,-y+3/4
91 -x+1/4,z+1/2,-y+3/4
92 -x+1/4,-z+3/4,y+1/2
```

```
93 z,y+1/2,x+1/2
94 z,-y+3/4,-x+3/4
95 -z+1/4,y+1/2,-x+3/4
96 -z+1/4,-y+3/4,x+1/2
97 x+1/2,y,z+1/2
98 x+1/2,-y+1/4,-z+3/4
99 -x+3/4,y,-z+3/4
100 -x+3/4,-y+1/4,z+1/2
101 y+1/2,z,x+1/2
102 y+1/2,-z+1/4,-x+3/4
103 -y+3/4,z,-x+3/4
104 -y+3/4,-z+1/4,x+1/2
105 z+1/2,x,y+1/2
106 z+1/2,-x+1/4,-y+3/4
107 -z+3/4,x,-y+3/4
108 -z+3/4,-x+1/4,y+1/2
109 -y+1/2,-x,-z+1/2
110 -y+1/2,x+1/4,z+3/4
111 y+3/4,-x,z+3/4
112 y+3/4,x+1/4,-z+1/2
113 -x+1/2,-z,-y+1/2
114 -x+1/2,z+1/4,y+3/4
115 x+3/4,-z,y+3/4
116 x+3/4,z+1/4,-y+1/2
117 -z+1/2,-y,-x+1/2
118 -z+1/2,y+1/4,x+3/4
119 z+3/4,-y,x+3/4
120 z+3/4,y+1/4,-x+1/2
121 -x+1/2,-y,-z+1/2
122 -x+1/2,y+1/4,z+3/4
123 x+3/4,-y,z+3/4
124 x+3/4,y+1/4,-z+1/2
125 -y+1/2,-z,-x+1/2
126 -y+1/2,z+1/4,x+3/4
127 y+3/4,-z,x+3/4
128 y+3/4,z+1/4,-x+1/2
129 -z+1/2,-x,-y+1/2
130 -z+1/2,x+1/4,y+3/4
131 z+3/4,-x,y+3/4
132 z+3/4,x+1/4,-y+1/2
133 y+1/2,x,z+1/2
134 y+1/2,-x+1/4,-z+3/4
135 -y+3/4,x,-z+3/4
136 -y+3/4,-x+1/4,z+1/2
137 x+1/2,z,y+1/2
138 x+1/2,-z+1/4,-y+3/4
139 -x+3/4,z,-y+3/4
140 -x+3/4,-z+1/4,y+1/2
141 z+1/2,y,x+1/2
142 z+1/2,-y+1/4,-x+3/4
143 -z+3/4,y,-x+3/4
144 -z+3/4,-y+1/4,x+1/2
145 x+1/2,y+1/2,z
146 x+1/2,-y+3/4,-z+1/4
147 -x+3/4,y+1/2,-z+1/4
148 -x+3/4,-y+3/4,z
149 y+1/2,z+1/2,x
150 y+1/2,-z+3/4,-x+1/4
151 -y+3/4,z+1/2,-x+1/4
152 -y+3/4,-z+3/4,x
153 z+1/2,x+1/2,y
154 z+1/2,-x+3/4,-y+1/4
155 -z+3/4,x+1/2,-y+1/4
156 -z+3/4,-x+3/4,y
157 -y+1/2,-x+1/2,-z
158 -y+1/2,x+3/4,z+1/4
159 y+3/4,-x+1/2,z+1/4
160 y+3/4,x+3/4,-z
161 -x+1/2,-z+1/2,-y
162 -x+1/2,z+3/4,y+1/4
163 x+3/4,-z+1/2,y+1/4
164 x+3/4,z+3/4,-y
165 -z+1/2,-y+1/2,-x
166 -z+1/2,y+3/4,x+1/4
167 z+3/4,-y+1/2,x+1/4
168 z+3/4,y+3/4,-x
169 -x+1/2,-y+1/2,-z
170 -x+1/2,y+3/4,z+1/4
171 x+3/4,-y+1/2,z+1/4
172 x+3/4,y+3/4,-z
173 -y+1/2,-z+1/2,-x
174 -y+1/2,z+3/4,x+1/4
175 y+3/4,-z+1/2,x+1/4
176 y+3/4,z+3/4,-x
177 -z+1/2,-x+1/2,-y
178 -z+1/2,x+3/4,y+1/4
179 z+3/4,-x+1/2,y+1/4
180 z+3/4,x+3/4,-y
181 y+1/2,x+1/2,z
182 y+1/2,-x+3/4,-z+1/4
183 -y+3/4,x+1/2,-z+1/4
184 -y+3/4,-x+3/4,z
185 x+1/2,z+1/2,y
186 x+1/2,-z+3/4,-y+1/4
187 -x+3/4,z+1/2,-y+1/4
188 -x+3/4,-z+3/4,y
189 z+1/2,y+1/2,x
190 z+1/2,-y+3/4,-x+1/4
191 -z+3/4,y+1/2,-x+1/4
192 -z+3/4,-y+3/4,x

loop_
_atom_site_label
_atom_site_type_symbol
_atom_site_symmetry_multiplicity
_atom_site_Wyckoff_label
_atom_site_fract_x
_atom_site_fract_y
_atom_site_fract_z
_atom_site_occupancy
Mg1 Mg 8 a 0.12500 0.12500 0.12500 1.00000
Li1 Li 16 d 0.50000 0.50000 0.50000 1.00000
H1 H 32 e 0.29150 0.29150 0.29150 1.00000
H2 H 96 g 0.56034 0.56034 0.87060 1.00000
```

Predicted $Li_2MgH_{16}$ High-Temperature Superconductor (250 GPa): A16B2C_cF152_227_eg_d_a - POSCAR

```
A16B2C_cF152_227_eg_d_a & a,x3,x4,z4 --params=6.71851,0.2915,0.56034,
    ↪ 0.8706 & Fd-3m O_{h}^{7} #227 (adeg) & cF152 & None & H16Li2Mg
    ↪ & H16Li2Mg & Y. Sun et al., Phys. Rev. Lett. 123, 097001(2019)
   1.00000000000000
   0.00000000000000   3.35925500000000   3.35925500000000
   3.35925500000000   0.00000000000000   3.35925500000000
   3.35925500000000   3.35925500000000   0.00000000000000
     H      Li      Mg
    32       4       2
Direct
   0.29150000000000   0.29150000000000   0.29150000000000   H   (32e)
   0.29150000000000   0.29150000000000  -0.37450000000000   H   (32e)
   0.29150000000000  -0.37450000000000   0.29150000000000   H   (32e)
  -0.37450000000000   0.29150000000000   0.29150000000000   H   (32e)
  -0.29150000000000  -0.29150000000000   1.37450000000000   H   (32e)
  -0.29150000000000  -0.29150000000000  -0.29150000000000   H   (32e)
  -0.29150000000000   1.37450000000000  -0.29150000000000   H   (32e)
   1.37450000000000  -0.29150000000000  -0.29150000000000   H   (32e)
   0.87060000000000   0.87060000000000   0.25008000000000   H   (96g)
   0.87060000000000   0.87060000000000  -1.49128000000000   H   (96g)
   0.25008000000000  -1.49128000000000   0.87060000000000   H   (96g)
  -1.49128000000000   0.25008000000000   0.87060000000000   H   (96g)
   0.25008000000000   0.87060000000000   0.87060000000000   H   (96g)
  -1.49128000000000   0.87060000000000   0.87060000000000   H   (96g)
   0.87060000000000   0.25008000000000  -1.49128000000000   H   (96g)
   0.87060000000000  -1.49128000000000   0.25008000000000   H   (96g)
   0.87060000000000   0.25008000000000   0.87060000000000   H   (96g)
   0.87060000000000  -1.49128000000000   0.87060000000000   H   (96g)
  -1.49128000000000   0.87060000000000   0.25008000000000   H   (96g)
   0.25008000000000   0.87060000000000  -1.49128000000000   H   (96g)
  -0.87060000000000  -0.87060000000000   2.49128000000000   H   (96g)
  -0.87060000000000  -0.87060000000000  -0.25008000000000   H   (96g)
  -0.25008000000000   2.49128000000000  -0.87060000000000   H   (96g)
   2.49128000000000  -0.25008000000000  -0.87060000000000   H   (96g)
  -0.25008000000000  -0.87060000000000   2.49128000000000   H   (96g)
   2.49128000000000  -0.87060000000000  -0.25008000000000   H   (96g)
  -0.87060000000000  -0.25008000000000  -0.87060000000000   H   (96g)
  -0.87060000000000   2.49128000000000  -0.87060000000000   H   (96g)
  -0.87060000000000  -0.25008000000000   2.49128000000000   H   (96g)
  -0.87060000000000   2.49128000000000  -0.25008000000000   H   (96g)
   2.49128000000000  -0.87060000000000  -0.87060000000000   H   (96g)
  -0.25008000000000  -0.87060000000000  -0.87060000000000   H   (96g)
   0.50000000000000   0.50000000000000   0.50000000000000   Li  (16d)
   0.50000000000000   0.50000000000000   0.00000000000000   Li  (16d)
   0.50000000000000   0.00000000000000   0.50000000000000   Li  (16d)
   0.00000000000000   0.50000000000000   0.50000000000000   Li  (16d)
   0.12500000000000   0.12500000000000   0.12500000000000   Mg   (8a)
   0.87500000000000   0.87500000000000   0.87500000000000   Mg   (8a)
```

$Mg_3Cr_2Al_{18}$: A18B2C3_cF184_227_fg_d_ac - CIF

```
# CIF file
data_findsym-output
_audit_creation_method FINDSYM

_chemical_name_mineral 'Al18Cr2Mg3'
_chemical_formula_sum 'Al18 Cr2 Mg3'

loop_
_publ_author_name
 'S. Samson'
_journal_year 1958
_publ_Section_title
;
 The Crystal Structure of the Intermetallic Compound Mg$_{3}$Cr$_{2}
    ↪ $Al$_{18}$
;

# Found in The Crystal Structure of the Intermetallic Compound ZrZn$_{22
    ↪ }$, 1961

_aflow_title 'Mg$_{3}$Cr$_{2}$Al$_{18}$ Structure'
_aflow_proto 'A18B2C3_cF184_227_fg_d_ac'
_aflow_params 'a,x_{4},x_{5},z_{5}'
_aflow_params_values '14.53,0.7657,0.5584,0.3252'
_aflow_Strukturbericht 'None'
_aflow_Pearson 'cF184'

_symmetry_space_group_name_H-M "F 41/d -3 2/m (origin choice 2)"
_symmetry_Int_Tables_number 227

_cell_length_a     14.53000
_cell_length_b     14.53000
_cell_length_c     14.53000
_cell_angle_alpha  90.00000
_cell_angle_beta   90.00000
_cell_angle_gamma  90.00000

loop_
_space_group_symop_id
_space_group_symop_operation_xyz
1 x,y,z
```

```
2 x,-y+1/4,-z+1/4
3 -x+1/4,y,-z+1/4
4 -x+1/4,-y+1/4,z
5 y,z,x
6 y,-z+1/4,-x+1/4
7 -y+1/4,z,-x+1/4
8 -y+1/4,-z+1/4,x
9 z,x,y
10 z,-x+1/4,-y+1/4
11 -z+1/4,x,-y+1/4
12 -z+1/4,-x+1/4,y
13 -y,-x,-z
14 -y,x+1/4,z+1/4
15 y+1/4,-x,z+1/4
16 y+1/4,x+1/4,-z
17 -x,-z,-y
18 -x,z+1/4,y+1/4
19 x+1/4,-z,y+1/4
20 x+1/4,z+1/4,-y
21 -z,-y,-x
22 -z,y+1/4,x+1/4
23 z+1/4,-y,x+1/4
24 z+1/4,y+1/4,-x
25 -x,-y,-z
26 -x,y+1/4,z+1/4
27 x+1/4,-y,z+1/4
28 x+1/4,y+1/4,-z
29 -y,-z,-x
30 -y,z+1/4,x+1/4
31 y+1/4,-z,x+1/4
32 y+1/4,z+1/4,-x
33 -z,-x,-y
34 -z,x+1/4,y+1/4
35 z+1/4,-x,y+1/4
36 z+1/4,x+1/4,-y
37 y,x,z
38 y,-x+1/4,-z+1/4
39 -y+1/4,x,-z+1/4
40 -y+1/4,-x+1/4,z
41 x,z,y
42 x,-z+1/4,-y+1/4
43 -x+1/4,z,-y+1/4
44 -x+1/4,-z+1/4,y
45 z,y,x
46 z,-y+1/4,-x+1/4
47 -z+1/4,y,-x+1/4
48 -z+1/4,-y+1/4,x
49 x,y+1/2,z+1/2
50 x,-y+3/4,-z+3/4
51 -x+1/4,y+1/2,-z+3/4
52 -x+1/4,-y+3/4,z+1/2
53 y,z+1/2,x+1/2
54 y,-z+3/4,-x+3/4
55 -y+1/4,z+1/2,-x+3/4
56 -y+1/4,-z+3/4,x+1/2
57 z,x+1/2,y+1/2
58 z,-x+3/4,-y+3/4
59 -z+1/4,x+1/2,-y+3/4
60 -z+1/4,-x+3/4,y+1/2
61 -y,-x+1/2,-z+1/2
62 -y,x+3/4,z+3/4
63 y+1/4,-x+1/2,z+3/4
64 y+1/4,x+3/4,-z+1/2
65 -x,-z+1/2,-y+1/2
66 -x,z+3/4,y+3/4
67 x+1/4,-z+1/2,y+3/4
68 x+1/4,z+3/4,-y+1/2
69 -z,-y+1/2,-x+1/2
70 -z,y+3/4,x+3/4
71 z+1/4,-y+1/2,x+3/4
72 z+1/4,y+3/4,-x+1/2
73 -x,-y+1/2,-z+1/2
74 -x,y+3/4,z+3/4
75 x+1/4,-y+1/2,z+3/4
76 x+1/4,y+3/4,-z+1/2
77 -y,-z+1/2,-x+1/2
78 -y,z+3/4,x+3/4
79 y+1/4,-z+1/2,x+3/4
80 y+1/4,z+3/4,-x+1/2
81 -z,-x+1/2,-y+1/2
82 -z,x+3/4,y+3/4
83 z+1/4,-x+1/2,y+3/4
84 z+1/4,x+3/4,-y+1/2
85 y,x+1/2,z+1/2
86 y,-x+3/4,-z+3/4
87 -y+1/4,x+1/2,-z+3/4
88 -y+1/4,-x+3/4,z+1/2
89 x,z+1/2,y+1/2
90 x,-z+3/4,-y+3/4
91 -x+1/4,z+1/2,-y+3/4
92 -x+1/4,-z+3/4,y+1/2
93 z,y+1/2,x+1/2
94 z,-y+3/4,-x+3/4
95 -z+1/4,y+1/2,-x+3/4
96 -z+1/4,-y+3/4,x+1/2
97 x+1/2,y,z+1/2
98 x+1/2,-y+1/4,-z+3/4
99 -x+3/4,y,-z+3/4
100 -x+3/4,-y+1/4,z+1/2
101 y+1/2,z,x+1/2
102 y+1/2,-z+1/4,-x+3/4
103 -y+3/4,z,-x+3/4
104 -y+3/4,-z+1/4,x+1/2
105 z+1/2,x,y+1/2
106 z+1/2,-x+1/4,-y+3/4
107 -z+3/4,x,-y+3/4
108 -z+3/4,-x+1/4,y+1/2
109 -y+1/2,-x,-z+1/2
110 -y+1/2,x+1/4,z+3/4
111 y+3/4,-x,z+3/4
112 y+3/4,x+1/4,-z+1/2
113 -x+1/2,-z,-y+1/2
114 -x+1/2,z+1/4,y+3/4
115 x+3/4,-z,y+3/4
116 x+3/4,z+1/4,-y+1/2
117 -z+1/2,-y,-x+1/2
118 -z+1/2,y+1/4,x+3/4
119 z+3/4,-y,x+3/4
120 z+3/4,y+1/4,-x+1/2
121 -x+1/2,-y,-z+1/2
122 -x+1/2,y+1/4,z+3/4
123 x+3/4,-y,z+3/4
124 x+3/4,y+1/4,-z+1/2
125 -y+1/2,-z,-x+1/2
126 -y+1/2,z+1/4,x+3/4
127 y+3/4,-z,x+3/4
128 y+3/4,z+1/4,-x+1/2
129 -z+1/2,-x,-y+1/2
130 -z+1/2,x+1/4,y+3/4
131 z+3/4,-x,y+3/4
132 z+3/4,x+1/4,-y+1/2
133 y+1/2,x,z+1/2
134 y+1/2,-x+1/4,-z+3/4
135 -y+3/4,x,-z+3/4
136 -y+3/4,-x+1/4,z+1/2
137 x+1/2,z,y+1/2
138 x+1/2,-z+1/4,-y+3/4
139 -x+3/4,z,-y+3/4
140 -x+3/4,-z+1/4,y+1/2
141 z+1/2,y,x+1/2
142 z+1/2,-y+1/4,-x+3/4
143 -z+3/4,y,-x+3/4
144 -z+3/4,-y+1/4,x+1/2
145 x+1/2,y+1/2,z
146 x+1/2,-y+3/4,-z+1/4
147 -x+3/4,y+1/2,-z+1/4
148 -x+3/4,-y+3/4,z
149 y+1/2,z+1/2,x
150 y+1/2,-z+3/4,-x+1/4
151 -y+3/4,z+1/2,-x+1/4
152 -y+3/4,-z+3/4,x
153 z+1/2,x+1/2,y
154 z+1/2,-x+3/4,-y+1/4
155 -z+3/4,x+1/2,-y+1/4
156 -z+3/4,-x+3/4,y
157 -y+1/2,-x+1/2,-z
158 -y+1/2,x+3/4,z+1/4
159 y+3/4,-x+1/2,z+1/4
160 y+3/4,x+3/4,-z
161 -x+1/2,-z+1/2,-y
162 -x+1/2,z+3/4,y+1/4
163 x+3/4,-z+1/2,y+1/4
164 x+3/4,z+3/4,-y
165 -z+1/2,-y+1/2,-x
166 -z+1/2,y+3/4,x+1/4
167 z+3/4,-y+1/2,x+1/4
168 z+3/4,y+3/4,-x
169 -x+1/2,-y+1/2,-z
170 -x+1/2,y+3/4,z+1/4
171 x+3/4,-y+1/2,z+1/4
172 x+3/4,y+3/4,-z
173 -y+1/2,-z+1/2,-x
174 -y+1/2,z+3/4,x+1/4
175 y+3/4,-z+1/2,x+1/4
176 y+3/4,z+3/4,-x
177 -z+1/2,-x+1/2,-y
178 -z+1/2,x+3/4,y+1/4
179 z+3/4,-x+1/2,y+1/4
180 z+3/4,x+3/4,-y
181 y+1/2,x+1/2,z
182 y+1/2,-x+3/4,-z+1/4
183 -y+3/4,x+1/2,-z+1/4
184 -y+3/4,-x+3/4,z
185 x+1/2,z+1/2,y
186 x+1/2,-z+3/4,-y+1/4
187 -x+3/4,z+1/2,-y+1/4
188 -x+3/4,-z+3/4,y
189 z+1/2,y+1/2,x
190 z+1/2,-y+3/4,-x+1/4
191 -z+3/4,y+1/2,-x+1/4
192 -z+3/4,-y+3/4,x

loop_
_atom_site_label
_atom_site_type_symbol
_atom_site_symmetry_multiplicity
_atom_site_Wyckoff_label
_atom_site_fract_x
_atom_site_fract_y
_atom_site_fract_z
_atom_site_occupancy
Mg1 Mg 8 a 0.12500 0.12500 0.12500 1.00000
Mg2 Mg 16 c 0.00000 0.00000 0.00000 1.00000
Cr1 Cr 16 d 0.50000 0.50000 0.50000 1.00000
Al1 Al 48 f 0.76570 0.12500 0.12500 1.00000
Al2 Al 96 g 0.55840 0.55840 0.32520 1.00000
```

$Mg_3Cr_2Al_{18}$: A18B2C3_cF184_227_fg_d_ac - POSCAR

```
A18B2C3_cF184_227_fg_d_ac & a,x4,x5,z5 --params=14.53,0.7657,0.5584,
```

```
    ↪ 0.3252 & Fd-3m O_{h}^{7} #227 (acdfg) & cF184 & None &
    ↪ Al18Cr2Mg3 & Al18Cr2Mg3 & S. Samson, (1958)
  1.00000000000000
  0.00000000000000   7.26500000000000   7.26500000000000
  7.26500000000000   0.00000000000000   7.26500000000000
  7.26500000000000   7.26500000000000   0.00000000000000
   Al      Cr      Mg
   36       4       6
Direct
 -0.51570000000000   0.76570000000000   0.76570000000000   Al   (48f)
  0.76570000000000  -0.51570000000000  -0.51570000000000   Al   (48f)
  0.76570000000000  -0.51570000000000   0.76570000000000   Al   (48f)
 -0.51570000000000   0.76570000000000  -0.51570000000000   Al   (48f)
  0.76570000000000   0.76570000000000  -0.51570000000000   Al   (48f)
 -0.51570000000000  -0.51570000000000   0.76570000000000   Al   (48f)
  1.51570000000000  -0.76570000000000   1.51570000000000   Al   (48f)
 -0.76570000000000   1.51570000000000  -0.76570000000000   Al   (48f)
 -0.76570000000000   1.51570000000000   1.51570000000000   Al   (48f)
  1.51570000000000  -0.76570000000000  -0.76570000000000   Al   (48f)
 -0.76570000000000  -0.76570000000000   1.51570000000000   Al   (48f)
  1.51570000000000   1.51570000000000  -0.76570000000000   Al   (48f)
  0.32520000000000   0.32520000000000   0.79160000000000   Al   (96g)
  0.32520000000000   0.32520000000000  -0.94200000000000   Al   (96g)
  0.79160000000000  -0.94200000000000   0.32520000000000   Al   (96g)
 -0.94200000000000   0.79160000000000   0.32520000000000   Al   (96g)
  0.79160000000000   0.32520000000000   0.32520000000000   Al   (96g)
 -0.94200000000000   0.32520000000000   0.32520000000000   Al   (96g)
  0.32520000000000   0.79160000000000  -0.94200000000000   Al   (96g)
  0.32520000000000  -0.94200000000000   0.79160000000000   Al   (96g)
  0.32520000000000   0.79160000000000   0.32520000000000   Al   (96g)
  0.32520000000000  -0.94200000000000   0.32520000000000   Al   (96g)
 -0.94200000000000   0.32520000000000   0.79160000000000   Al   (96g)
  0.79160000000000   0.32520000000000  -0.94200000000000   Al   (96g)
 -0.32520000000000  -0.32520000000000   1.94200000000000   Al   (96g)
 -0.32520000000000  -0.32520000000000  -0.79160000000000   Al   (96g)
 -0.79160000000000   1.94200000000000  -0.32520000000000   Al   (96g)
  1.94200000000000  -0.79160000000000  -0.32520000000000   Al   (96g)
 -0.79160000000000  -0.32520000000000   1.94200000000000   Al   (96g)
  1.94200000000000  -0.32520000000000  -0.79160000000000   Al   (96g)
 -0.32520000000000  -0.79160000000000  -0.32520000000000   Al   (96g)
 -0.32520000000000   1.94200000000000  -0.32520000000000   Al   (96g)
 -0.32520000000000  -0.79160000000000   1.94200000000000   Al   (96g)
 -0.32520000000000   1.94200000000000  -0.79160000000000   Al   (96g)
  1.94200000000000  -0.32520000000000  -0.32520000000000   Al   (96g)
 -0.79160000000000  -0.32520000000000  -0.32520000000000   Al   (96g)
  0.50000000000000   0.50000000000000   0.50000000000000   Cr   (16d)
  0.50000000000000   0.50000000000000   0.00000000000000   Cr   (16d)
  0.50000000000000   0.00000000000000   0.50000000000000   Cr   (16d)
  0.00000000000000   0.50000000000000   0.50000000000000   Cr   (16d)
  0.12500000000000   0.12500000000000   0.12500000000000   Mg    (8a)
  0.87500000000000   0.87500000000000   0.87500000000000   Mg    (8a)
  0.00000000000000   0.00000000000000   0.00000000000000   Mg   (16c)
  0.00000000000000   0.00000000000000   0.50000000000000   Mg   (16c)
  0.00000000000000   0.50000000000000   0.00000000000000   Mg   (16c)
  0.50000000000000   0.00000000000000   0.00000000000000   Mg   (16c)
```

$Zn_{22}Zr$: A22B_cF184_227_cdfg_a - CIF

```
# CIF file
data_findsym-output
_audit_creation_method FINDSYM

_chemical_name_mineral 'Zn22Zr'
_chemical_formula_sum 'Zn22 Zr'

loop_
_publ_author_name
 'S. Samson'
_journal_year 1961
_publ_Section_title
;
 The Crystal Structure of the Intermetallic Compound ZrZn$_{22}$
;

# Found in Binary Alloy Phase Diagrams, 1990 Found in Binary Alloy Phase
    ↪ Diagrams, {Hf-Re to Zn-Zr}}

_aflow_title 'Zn$_{22}$Zr Structure'
_aflow_proto 'A22B_cF184_227_cdfg_a'
_aflow_params 'a,x_{4},x_{5},z_{5}'
_aflow_params_values '14.101,0.765,0.562,0.3195'
_aflow_Strukturbericht 'None'
_aflow_Pearson 'cF184'

_symmetry_space_group_name_H-M "F 41/d -3 2/m (origin choice 2)"
_symmetry_Int_Tables_number 227

_cell_length_a      14.10100
_cell_length_b      14.10100
_cell_length_c      14.10100
_cell_angle_alpha  90.00000
_cell_angle_beta   90.00000
_cell_angle_gamma  90.00000

loop_
_space_group_symop_id
_space_group_symop_operation_xyz
1 x,y,z
2 x,-y+1/4,-z+1/4
3 -x+1/4,y,-z+1/4
4 -x+1/4,-y+1/4,z
5 y,z,x
6 y,-z+1/4,-x+1/4
7 -y+1/4,z,-x+1/4
8 -y+1/4,-z+1/4,x
9 z,x,y
10 z,-x+1/4,-y+1/4
11 -z+1/4,x,-y+1/4
12 -z+1/4,-x+1/4,y
13 -y,-x,-z
14 -y,x+1/4,z+1/4
15 y+1/4,-x,z+1/4
16 y+1/4,x+1/4,-z
17 -x,-z,-y
18 -x,z+1/4,y+1/4
19 x+1/4,-z,y+1/4
20 x+1/4,z+1/4,-y
21 -z,-y,-x
22 -z,y+1/4,x+1/4
23 z+1/4,-y,x+1/4
24 z+1/4,y+1/4,-x
25 -x,-y,-z
26 -x,y+1/4,z+1/4
27 x+1/4,-y,z+1/4
28 x+1/4,y+1/4,-z
29 -y,-z,-x
30 -y,z+1/4,x+1/4
31 y+1/4,-z,x+1/4
32 y+1/4,z+1/4,-x
33 -z,-x,-y
34 -z,x+1/4,y+1/4
35 z+1/4,-x,y+1/4
36 z+1/4,x+1/4,-y
37 y,x,z
38 y,-x+1/4,-z+1/4
39 -y+1/4,x,-z+1/4
40 -y+1/4,-x+1/4,z
41 x,z,y
42 x,-z+1/4,-y+1/4
43 -x+1/4,z,-y+1/4
44 -x+1/4,-z+1/4,y
45 z,y,x
46 z,-y+1/4,-x+1/4
47 -z+1/4,y,-x+1/4
48 -z+1/4,-y+1/4,x
49 x,y+1/2,z+1/2
50 x,-y+3/4,-z+3/4
51 -x+1/4,y+1/2,-z+3/4
52 -x+1/4,-y+3/4,z+1/2
53 y,z+1/2,x+1/2
54 y,-z+3/4,-x+3/4
55 -y+1/4,z+1/2,-x+3/4
56 -y+1/4,-z+3/4,x+1/2
57 z,x+1/2,y+1/2
58 z,-x+3/4,-y+3/4
59 -z+1/4,x+1/2,-y+3/4
60 -z+1/4,-x+3/4,y+1/2
61 -y,-x+1/2,-z+1/2
62 -y,x+3/4,z+3/4
63 y+1/4,-x+1/2,z+3/4
64 y+1/4,x+3/4,-z+1/2
65 -x,-z+1/2,-y+1/2
66 -x,z+3/4,y+3/4
67 x+1/4,-z+1/2,y+3/4
68 x+1/4,z+3/4,-y+1/2
69 -z,-y+1/2,-x+1/2
70 -z,y+3/4,x+3/4
71 z+1/4,-y+1/2,x+3/4
72 z+1/4,y+3/4,-x+1/2
73 -x,-y+1/2,-z+1/2
74 -x,y+3/4,z+3/4
75 x+1/4,-y+1/2,z+3/4
76 x+1/4,y+3/4,-z+1/2
77 -y,-z+1/2,-x+1/2
78 -y,z+3/4,x+3/4
79 y+1/4,-z+1/2,x+3/4
80 y+1/4,z+3/4,-x+1/2
81 -z,-x+1/2,-y+1/2
82 -z,x+3/4,y+3/4
83 z+1/4,-x+1/2,y+3/4
84 z+1/4,x+3/4,-y+1/2
85 y,x+1/2,z+1/2
86 y,-x+3/4,-z+3/4
87 -y+1/4,x+1/2,-z+3/4
88 -y+1/4,-x+3/4,z+1/2
89 x,z+1/2,y+1/2
90 x,-z+3/4,-y+3/4
91 -x+1/4,z+1/2,-y+3/4
92 -x+1/4,-z+3/4,y+1/2
93 z,y+1/2,x+1/2
94 z,-y+3/4,-x+3/4
95 -z+1/4,y+1/2,-x+3/4
96 -z+1/4,-y+3/4,x+1/2
97 x+1/2,y,z+1/2
98 x+1/2,-y+1/4,-z+3/4
99 -x+3/4,y,-z+3/4
100 -x+3/4,-y+1/4,z+1/2
101 y+1/2,z,x+1/2
102 y+1/2,-z+1/4,-x+3/4
103 -y+3/4,z,-x+3/4
104 -y+3/4,-z+1/4,x+1/2
105 z+1/2,x,y+1/2
106 z+1/2,-x+1/4,-y+3/4
107 -z+3/4,x,-y+3/4
108 -z+3/4,-x+1/4,y+1/2
109 -y+1/2,-x,-z+1/2
110 -y+1/2,x+1/4,z+3/4
111 y+3/4,-x,z+3/4
112 y+3/4,x+1/4,-z+1/2
113 -x+1/2,-z,-y+1/2
```

```
114 -x+1/2,z+1/4,y+3/4
115 x+3/4,-z,y+3/4
116 x+3/4,z+1/4,-y+1/2
117 -z+1/2,-y,-x+1/2
118 -z+1/2,y+1/4,x+3/4
119 z+3/4,-y,x+3/4
120 z+3/4,y+1/4,-x+1/2
121 -x+1/2,-y,-z+1/2
122 -x+1/2,y+1/4,z+3/4
123 x+3/4,-y,z+3/4
124 x+3/4,y+1/4,-z+1/2
125 -y+1/2,-z,-x+1/2
126 -y+1/2,z+1/4,x+3/4
127 y+3/4,-z,x+3/4
128 y+3/4,z+1/4,-x+1/2
129 -z+1/2,-x,-y+1/2
130 -z+1/2,x+1/4,y+3/4
131 z+3/4,-x,y+3/4
132 z+3/4,x+1/4,-y+1/2
133 y+1/2,x,z+1/2
134 y+1/2,-x+1/4,-z+3/4
135 -y+3/4,x,-z+3/4
136 -y+3/4,-x+1/4,z+1/2
137 x+1/2,z,y+1/2
138 x+1/2,-z+1/4,-y+3/4
139 -x+3/4,z,-y+3/4
140 -x+3/4,-z+1/4,y+1/2
141 z+1/2,y,x+1/2
142 z+1/2,-y+1/4,-x+3/4
143 -z+3/4,y,-x+3/4
144 -z+3/4,-y+1/4,x+1/2
145 x+1/2,y+1/2,z
146 x+1/2,-y+3/4,-z+1/4
147 -x+3/4,y+1/2,-z+1/4
148 -x+3/4,-y+3/4,z
149 y+1/2,z+1/2,x
150 y+1/2,-z+3/4,-x+1/4
151 -y+3/4,z+1/2,-x+1/4
152 -y+3/4,-z+3/4,x
153 z+1/2,x+1/2,y
154 z+1/2,-x+3/4,-y+1/4
155 -z+3/4,x+1/2,-y+1/4
156 -z+3/4,-x+3/4,y
157 -y+1/2,-x+1/2,-z
158 -y+1/2,x+3/4,z+1/4
159 y+3/4,-x+1/2,z+1/4
160 y+3/4,x+3/4,-z
161 -x+1/2,-z+1/2,-y
162 -x+1/2,z+3/4,y+1/4
163 x+3/4,-z+1/2,y+1/4
164 x+3/4,z+3/4,-y
165 -z+1/2,-y+1/2,-x
166 -z+1/2,y+3/4,x+1/4
167 z+3/4,-y+1/2,x+1/4
168 z+3/4,y+3/4,-x
169 -x+1/2,-y+1/2,-z
170 -x+1/2,y+3/4,z+1/4
171 x+3/4,-y+1/2,z+1/4
172 x+3/4,y+3/4,-z
173 -y+1/2,-z+1/2,-x
174 -y+1/2,z+3/4,x+1/4
175 y+3/4,-z+1/2,x+1/4
176 y+3/4,z+3/4,-x
177 -z+1/2,-x+1/2,-y
178 -z+1/2,x+3/4,y+1/4
179 z+3/4,-x+1/2,y+1/4
180 z+3/4,x+3/4,-y
181 y+1/2,x+1/2,z
182 y+1/2,-x+3/4,-z+1/4
183 -y+3/4,x+1/2,-z+1/4
184 -y+3/4,-x+3/4,z
185 x+1/2,z+1/2,y
186 x+1/2,-z+3/4,-y+1/4
187 -x+3/4,z+1/2,-y+1/4
188 -x+3/4,-z+3/4,y
189 z+1/2,y+1/2,x
190 z+1/2,-y+3/4,-x+1/4
191 -z+3/4,y+1/2,-x+1/4
192 -z+3/4,-y+3/4,x

loop_
_atom_site_label
_atom_site_type_symbol
_atom_site_symmetry_multiplicity
_atom_site_Wyckoff_label
_atom_site_fract_x
_atom_site_fract_y
_atom_site_fract_z
_atom_site_occupancy
Zr1 Zr 8 a 0.12500 0.12500 0.12500 1.00000
Zn1 Zn 16 c 0.00000 0.00000 0.00000 1.00000
Zn2 Zn 16 d 0.50000 0.50000 0.50000 1.00000
Zn3 Zn 48 f 0.76500 0.12500 0.12500 1.00000
Zn4 Zn 96 g 0.56200 0.56200 0.31950 1.00000
```

$Zn_{22}Zr$: A22B_cF184_227_cdfg_a - POSCAR

```
A22B_cF184_227_cdfg_a & a,x4,x5,z5 --params=14.101,0.765,0.562,0.3195 &
    ↪ Fd-3m O_{h}^{7} #227 (acdfg) & cF184 & None & Zn22Zr & Zn22Zr &
    ↪ S. Samson, (1961)
   1.00000000000000
   0.00000000000000    7.05050000000000    7.05050000000000
   7.05050000000000    0.00000000000000    7.05050000000000
   7.05050000000000    7.05050000000000    0.00000000000000
    Zn      Zr
    44       2
Direct
   0.00000000000000    0.00000000000000    0.00000000000000   Zn   (16c)
   0.00000000000000    0.00000000000000    0.50000000000000   Zn   (16c)
   0.00000000000000    0.50000000000000    0.00000000000000   Zn   (16c)
   0.50000000000000    0.00000000000000    0.00000000000000   Zn   (16c)
   0.50000000000000    0.50000000000000    0.50000000000000   Zn   (16d)
   0.50000000000000    0.50000000000000    0.00000000000000   Zn   (16d)
   0.50000000000000    0.00000000000000    0.50000000000000   Zn   (16d)
   0.00000000000000    0.50000000000000    0.50000000000000   Zn   (16d)
  -0.51500000000000    0.76500000000000    0.76500000000000   Zn   (48f)
   0.76500000000000   -0.51500000000000   -0.51500000000000   Zn   (48f)
   0.76500000000000   -0.51500000000000    0.76500000000000   Zn   (48f)
  -0.51500000000000    0.76500000000000   -0.51500000000000   Zn   (48f)
   0.76500000000000    0.76500000000000   -0.51500000000000   Zn   (48f)
  -0.51500000000000   -0.51500000000000    0.76500000000000   Zn   (48f)
   1.51500000000000   -0.76500000000000    1.51500000000000   Zn   (48f)
  -0.76500000000000    1.51500000000000   -0.76500000000000   Zn   (48f)
  -0.76500000000000    1.51500000000000    1.51500000000000   Zn   (48f)
   1.51500000000000   -0.76500000000000   -0.76500000000000   Zn   (48f)
  -0.76500000000000   -0.76500000000000    1.51500000000000   Zn   (48f)
   1.51500000000000    1.51500000000000   -0.76500000000000   Zn   (48f)
   0.31950000000000    0.31950000000000    0.80450000000000   Zn   (96g)
   0.31950000000000    0.31950000000000   -0.94350000000000   Zn   (96g)
   0.80450000000000   -0.94350000000000    0.31950000000000   Zn   (96g)
  -0.94350000000000    0.80450000000000    0.31950000000000   Zn   (96g)
   0.80450000000000    0.31950000000000    0.31950000000000   Zn   (96g)
  -0.94350000000000    0.31950000000000    0.31950000000000   Zn   (96g)
   0.31950000000000    0.80450000000000   -0.94350000000000   Zn   (96g)
   0.31950000000000   -0.94350000000000    0.80450000000000   Zn   (96g)
   0.31950000000000    0.80450000000000    0.31950000000000   Zn   (96g)
   0.31950000000000   -0.94350000000000    0.31950000000000   Zn   (96g)
  -0.94350000000000    0.31950000000000    0.80450000000000   Zn   (96g)
   0.80450000000000    0.31950000000000   -0.94350000000000   Zn   (96g)
  -0.31950000000000   -0.31950000000000    1.94350000000000   Zn   (96g)
  -0.31950000000000   -0.31950000000000   -0.80450000000000   Zn   (96g)
  -0.80450000000000    1.94350000000000   -0.31950000000000   Zn   (96g)
   1.94350000000000   -0.80450000000000   -0.31950000000000   Zn   (96g)
  -0.80450000000000   -0.31950000000000    1.94350000000000   Zn   (96g)
   1.94350000000000   -0.31950000000000   -0.80450000000000   Zn   (96g)
  -0.31950000000000   -0.80450000000000   -0.31950000000000   Zn   (96g)
  -0.31950000000000    1.94350000000000   -0.31950000000000   Zn   (96g)
  -0.31950000000000   -0.80450000000000    1.94350000000000   Zn   (96g)
  -0.31950000000000    1.94350000000000   -0.80450000000000   Zn   (96g)
   1.94350000000000   -0.31950000000000   -0.31950000000000   Zn   (96g)
  -0.80450000000000   -0.31950000000000   -0.31950000000000   Zn   (96g)
   0.12500000000000    0.12500000000000    0.12500000000000   Zr    (8a)
   0.87500000000000    0.87500000000000    0.87500000000000   Zr    (8a)
```

$H_3PW_{12}O_{40}\cdot 29H_2O$ ($H4_{21}$): A29B40CD12_cF656_227_ae2fg_e3g_b_g - CIF

```
# CIF file
data_findsym-output
_audit_creation_method FINDSYM

_chemical_name_mineral '29-phosphotungstic acid (29-wpa)'
_chemical_formula_sum '(H2O)29 O40 P W12'

loop_
_publ_author_name
 'A. J. Bradley'
 'J. W. Illingworth'
_journal_name_full_name
;
 Proceedings of the Royal Society London A
;
_journal_volume 157
_journal_year 1936
_journal_page_first 113
_journal_page_last 131
_publ_Section_title
;
 The Crystal Structure of H$_{3}$PW$_{12}$O$_{40}$$\cdot$29H$_{2}$O
;

# Found in Strukturbericht Band IV 1936, 1938

_aflow_title 'H$_{3}$PW$_{12}$O$_{40}$$\cdot$29H$_{2}$O ($H4_{21}$)
    ↪ Structure'
_aflow_proto 'A29B40CD12_cF656_227_ae2fg_e3g_b_g'
_aflow_params 'a,x_{3},x_{4},x_{5},x_{6},x_{7},z_{7},x_{8},z_{8},x_{9},
    ↪ z_{9},x_{10},z_{10},x_{11},z_{11}'
_aflow_params_values '23.28001,-0.05284,0.41753,-0.01933,0.85395,0.71864
    ↪ ,0.85825,0.83333,0.25301,-0.061,0.52706,0.5378,0.3707,-0.01783,
    ↪ 0.38144'
_aflow_Strukturbericht '$H4_{21}$'
_aflow_Pearson 'cF656'

_symmetry_space_group_name_H-M "F 41/d -3 2/m (origin choice 2)"
_symmetry_Int_Tables_number 227

_cell_length_a    23.28001
_cell_length_b    23.28001
_cell_length_c    23.28001
_cell_angle_alpha 90.00000
_cell_angle_beta  90.00000
_cell_angle_gamma 90.00000

loop_
_space_group_symop_id
_space_group_symop_operation_xyz
1 x,y,z
2 x,-y+1/4,-z+1/4
3 -x+1/4,y,-z+1/4
4 -x+1/4,-y+1/4,z
```

```
5 y,z,x
6 y,-z+1/4,-x+1/4
7 -y+1/4,z,-x+1/4
8 -y+1/4,-z+1/4,x
9 z,x,y
10 z,-x+1/4,-y+1/4
11 -z+1/4,x,-y+1/4
12 -z+1/4,-x+1/4,y
13 -y,-x,-z
14 -y,x+1/4,z+1/4
15 y+1/4,-x,z+1/4
16 y+1/4,x+1/4,-z
17 -x,-z,-y
18 -x,z+1/4,y+1/4
19 x+1/4,-z,y+1/4
20 x+1/4,z+1/4,-y
21 -z,-y,-x
22 -z,y+1/4,x+1/4
23 z+1/4,-y,x+1/4
24 z+1/4,y+1/4,-x
25 -x,-y,-z
26 -x,y+1/4,z+1/4
27 x+1/4,-y,z+1/4
28 x+1/4,y+1/4,-z
29 -y,-z,-x
30 -y,z+1/4,x+1/4
31 y+1/4,-z,x+1/4
32 y+1/4,z+1/4,-x
33 -z,-x,-y
34 -z,x+1/4,y+1/4
35 z+1/4,-x,y+1/4
36 z+1/4,x+1/4,-y
37 y,x,z
38 y,-x+1/4,-z+1/4
39 -y+1/4,x,-z+1/4
40 -y+1/4,-x+1/4,z
41 x,z,y
42 x,-z+1/4,-y+1/4
43 -x+1/4,z,-y+1/4
44 -x+1/4,-z+1/4,y
45 z,y,x
46 z,-y+1/4,-x+1/4
47 -z+1/4,y,-x+1/4
48 -z+1/4,-y+1/4,x
49 x,y+1/2,z+1/2
50 x,-y+3/4,-z+3/4
51 -x+1/4,y+1/2,-z+3/4
52 -x+1/4,-y+3/4,z+1/2
53 y,z+1/2,x+1/2
54 y,-z+3/4,-x+3/4
55 -y+1/4,z+1/2,-x+3/4
56 -y+1/4,-z+3/4,x+1/2
57 z,x+1/2,y+1/2
58 z,-x+3/4,-y+3/4
59 -z+1/4,x+1/2,-y+3/4
60 -z+1/4,-x+3/4,y+1/2
61 -y,-x+1/2,-z+1/2
62 -y,x+3/4,z+3/4
63 y+1/4,-x+1/2,z+3/4
64 y+1/4,x+3/4,-z+1/2
65 -x,-z+1/2,-y+1/2
66 -x,z+3/4,y+3/4
67 x+1/4,-z+1/2,y+3/4
68 x+1/4,z+3/4,-y+1/2
69 -z,-y+1/2,-x+1/2
70 -z,y+3/4,x+3/4
71 z+1/4,-y+1/2,x+3/4
72 z+1/4,y+3/4,-x+1/2
73 -x,-y+1/2,-z+1/2
74 -x,y+3/4,z+3/4
75 x+1/4,-y+1/2,z+3/4
76 x+1/4,y+3/4,-z+1/2
77 -y,-z+1/2,-x+1/2
78 -y,z+3/4,x+3/4
79 y+1/4,-z+1/2,x+3/4
80 y+1/4,z+3/4,-x+1/2
81 -z,-x+1/2,-y+1/2
82 -z,x+3/4,y+3/4
83 z+1/4,-x+1/2,y+3/4
84 z+1/4,x+3/4,-y+1/2
85 y,x+1/2,z+1/2
86 y,-x+3/4,-z+3/4
87 -y+1/4,x+1/2,-z+3/4
88 -y+1/4,-x+3/4,z+1/2
89 x,z+1/2,y+1/2
90 x,-z+3/4,-y+3/4
91 -x+1/4,z+1/2,-y+3/4
92 -x+1/4,-z+3/4,y+1/2
93 z,y+1/2,x+1/2
94 z,-y+3/4,-x+3/4
95 -z+1/4,y+1/2,-x+3/4
96 -z+1/4,-y+3/4,x+1/2
97 x+1/2,y,z+1/2
98 x+1/2,-y+1/4,-z+3/4
99 -x+3/4,y,-z+3/4
100 -x+3/4,-y+1/4,z+1/2
101 y+1/2,z,x+1/2
102 y+1/2,-z+1/4,-x+3/4
103 -y+3/4,z,-x+3/4
104 -y+3/4,-z+1/4,x+1/2
105 z+1/2,x,y+1/2
106 z+1/2,-x+1/4,-y+3/4
107 -z+3/4,x,-y+3/4
108 -z+3/4,-x+1/4,y+1/2
109 -y+1/2,-x,-z+1/2
110 -y+1/2,x+1/4,z+3/4
111 y+3/4,-x,z+3/4
112 y+3/4,x+1/4,-z+1/2
113 -x+1/2,-z,-y+1/2
114 -x+1/2,z+1/4,y+3/4
115 x+3/4,-z,y+3/4
116 x+3/4,z+1/4,-y+1/2
117 -z+1/2,-y,-x+1/2
118 -z+1/2,y+1/4,x+3/4
119 z+3/4,-y,x+3/4
120 z+3/4,y+1/4,-x+1/2
121 -x+1/2,-y,-z+1/2
122 -x+1/2,y+1/4,z+3/4
123 x+3/4,-y,z+3/4
124 x+3/4,y+1/4,-z+1/2
125 -y+1/2,-z,-x+1/2
126 -y+1/2,z+1/4,x+3/4
127 y+3/4,-z,x+3/4
128 y+3/4,z+1/4,-x+1/2
129 -z+1/2,-x,-y+1/2
130 -z+1/2,x+1/4,y+3/4
131 z+3/4,-x,y+3/4
132 z+3/4,x+1/4,-y+1/2
133 y+1/2,x,z+1/2
134 y+1/2,-x+1/4,-z+3/4
135 -y+3/4,x,-z+3/4
136 -y+3/4,-x+1/4,z+1/2
137 x+1/2,z,y+1/2
138 x+1/2,-z+1/4,-y+3/4
139 -x+3/4,z,-y+3/4
140 -x+3/4,-z+1/4,y+1/2
141 z+1/2,y,x+1/2
142 z+1/2,-y+1/4,-x+3/4
143 -z+3/4,y,-x+3/4
144 -z+3/4,-y+1/4,x+1/2
145 x+1/2,y+1/2,z
146 x+1/2,-y+3/4,-z+1/4
147 -x+3/4,y+1/2,-z+1/4
148 -x+3/4,-y+3/4,z
149 y+1/2,z+1/2,x
150 y+1/2,-z+3/4,-x+1/4
151 -y+3/4,z+1/2,-x+1/4
152 -y+3/4,-z+3/4,x
153 z+1/2,x+1/2,y
154 z+1/2,-x+3/4,-y+1/4
155 -z+3/4,x+1/2,-y+1/4
156 -z+3/4,-x+3/4,y
157 -y+1/2,-x+1/2,-z
158 -y+1/2,x+3/4,z+1/4
159 y+3/4,-x+1/2,z+1/4
160 y+3/4,x+3/4,-z
161 -x+1/2,-z+1/2,-y
162 -x+1/2,z+3/4,y+1/4
163 x+3/4,-z+1/2,y+1/4
164 x+3/4,z+3/4,-y
165 -z+1/2,-y+1/2,-x
166 -z+1/2,y+3/4,x+1/4
167 z+3/4,-y+1/2,x+1/4
168 z+3/4,y+3/4,-x
169 -x+1/2,-y+1/2,-z
170 -x+1/2,y+3/4,z+1/4
171 x+3/4,-y+1/2,z+1/4
172 x+3/4,y+3/4,-z
173 -y+1/2,-z+1/2,-x
174 -y+1/2,z+3/4,x+1/4
175 y+3/4,-z+1/2,x+1/4
176 y+3/4,z+3/4,-x
177 -z+1/2,-x+1/2,-y
178 -z+1/2,x+3/4,y+1/4
179 z+3/4,-x+1/2,y+1/4
180 z+3/4,x+3/4,-y
181 y+1/2,x+1/2,z
182 y+1/2,-x+3/4,-z+1/4
183 -y+3/4,x+1/2,-z+1/4
184 -y+3/4,-x+3/4,z
185 x+1/2,z+1/2,y
186 x+1/2,-z+3/4,-y+1/4
187 -x+3/4,z+1/2,-y+1/4
188 -x+3/4,-z+3/4,y
189 z+1/2,y+1/2,x
190 z+1/2,-y+3/4,-x+1/4
191 -z+3/4,y+1/2,-x+1/4
192 -z+3/4,-y+3/4,x

loop_
_atom_site_label
_atom_site_type_symbol
_atom_site_symmetry_multiplicity
_atom_site_Wyckoff_label
_atom_site_fract_x
_atom_site_fract_y
_atom_site_fract_z
_atom_site_occupancy
H2O1 H2O 8 a 0.12500 0.12500 0.12500 1.00000
P1 P 8 b 0.37500 0.37500 0.37500 1.00000
H2O2 H2O 32 e -0.05284 -0.05284 -0.05284 1.00000
O1 O 32 e 0.41753 0.41753 0.41753 1.00000
H2O3 H2O 48 f -0.01933 0.12500 0.12500 1.00000
H2O4 H2O 48 f 0.85395 0.12500 0.12500 1.00000
H2O5 H2O 96 g 0.71864 0.71864 0.85825 1.00000
O2 O 96 g 0.83333 0.83333 0.25301 1.00000
O3 O 96 g -0.06100 -0.06100 0.52706 1.00000
O4 O 96 g 0.53780 0.53780 0.37070 1.00000
W1 W 96 g -0.01783 -0.01783 0.38144 1.00000
```

$H_3PW_{12}O_{40}$·$29H_2O$ ($H4_{21}$): A29B40CD12_cF656_227_ae2fg_e3g_b_g - POSCAR

```
A29B40CD12_cF656_227_ae2fg_e3g_b_g & a,x3,x4,x5,x6,x7,z7,x8,z8,x9,z9,x10
   ↪ ,z10,x11,z11 --params=23.28001,-0.05284,0.41753,-0.01933,
   ↪ 0.85395,0.71864,0.85825,0.83333,0.25301,-0.061,0.52706,0.5378,
   ↪ 0.3707,-0.01783,0.38144 & Fd-3m O_{h}^{7} #227 (abe^2f^2g^5) &
   ↪ cF656 & $H4_{21}$ & (H2O)29O40PW12 & 29-phosphotungstic acid (
   ↪ 29-wpa) & A. J. Bradley and J. W. Illingworth, Proc. Roy. Soc.
   ↪ Lond. A 157, 113-131 (1936)
   1.00000000000000
   0.00000000000000  11.64000500000000  11.64000500000000
  11.64000500000000   0.00000000000000  11.64000500000000
  11.64000500000000  11.64000500000000   0.00000000000000
    H2O     O     P     W
     58    80     2    24
Direct
   0.12500000000000   0.12500000000000   0.12500000000000  H2O   (8a)
   0.87500000000000   0.87500000000000   0.87500000000000  H2O   (8a)
  -0.05284000000000  -0.05284000000000  -0.05284000000000  H2O  (32e)
  -0.05284000000000  -0.05284000000000   0.65852000000000  H2O  (32e)
  -0.05284000000000   0.65852000000000  -0.05284000000000  H2O  (32e)
   0.65852000000000  -0.05284000000000  -0.05284000000000  H2O  (32e)
   0.05284000000000   0.05284000000000   0.34148000000000  H2O  (32e)
   0.05284000000000   0.05284000000000   0.05284000000000  H2O  (32e)
   0.05284000000000   0.34148000000000   0.05284000000000  H2O  (32e)
   0.34148000000000   0.05284000000000   0.05284000000000  H2O  (32e)
   0.26933000000000  -0.01933000000000  -0.01933000000000  H2O  (48f)
  -0.01933000000000   0.26933000000000   0.26933000000000  H2O  (48f)
  -0.01933000000000   0.26933000000000  -0.01933000000000  H2O  (48f)
   0.26933000000000  -0.01933000000000   0.26933000000000  H2O  (48f)
  -0.01933000000000  -0.01933000000000   0.26933000000000  H2O  (48f)
   0.26933000000000   0.26933000000000  -0.01933000000000  H2O  (48f)
   0.73067000000000   0.01933000000000   0.73067000000000  H2O  (48f)
   0.01933000000000   0.73067000000000   0.01933000000000  H2O  (48f)
   0.01933000000000   0.73067000000000   0.73067000000000  H2O  (48f)
   0.73067000000000   0.01933000000000   0.01933000000000  H2O  (48f)
   0.01933000000000   0.01933000000000   0.73067000000000  H2O  (48f)
   0.73067000000000   0.73067000000000   0.01933000000000  H2O  (48f)
  -0.60395000000000   0.85395000000000   0.85395000000000  H2O  (48f)
   0.85395000000000  -0.60395000000000  -0.60395000000000  H2O  (48f)
   0.85395000000000  -0.60395000000000   0.85395000000000  H2O  (48f)
  -0.60395000000000   0.85395000000000  -0.60395000000000  H2O  (48f)
   0.85395000000000   0.85395000000000  -0.60395000000000  H2O  (48f)
  -0.60395000000000  -0.60395000000000   0.85395000000000  H2O  (48f)
   1.60395000000000  -0.85395000000000   1.60395000000000  H2O  (48f)
  -0.85395000000000   1.60395000000000  -0.85395000000000  H2O  (48f)
  -0.85395000000000   1.60395000000000   1.60395000000000  H2O  (48f)
   1.60395000000000  -0.85395000000000  -0.85395000000000  H2O  (48f)
  -0.85395000000000  -0.85395000000000   1.60395000000000  H2O  (48f)
   1.60395000000000   1.60395000000000  -0.85395000000000  H2O  (48f)
   0.85825000000000   0.85825000000000   0.57903000000000  H2O  (96g)
   0.85825000000000   0.85825000000000  -1.79553000000000  H2O  (96g)
   0.57903000000000  -1.79553000000000   0.85825000000000  H2O  (96g)
  -1.79553000000000   0.57903000000000   0.85825000000000  H2O  (96g)
   0.57903000000000   0.85825000000000   0.85825000000000  H2O  (96g)
  -1.79553000000000   0.85825000000000   0.85825000000000  H2O  (96g)
   0.85825000000000   0.57903000000000  -1.79553000000000  H2O  (96g)
   0.85825000000000  -1.79553000000000   0.57903000000000  H2O  (96g)
   0.85825000000000   0.57903000000000   0.85825000000000  H2O  (96g)
   0.85825000000000  -1.79553000000000   0.85825000000000  H2O  (96g)
  -1.79553000000000   0.85825000000000   0.57903000000000  H2O  (96g)
   0.57903000000000   0.85825000000000  -1.79553000000000  H2O  (96g)
  -0.85825000000000  -0.85825000000000   2.79553000000000  H2O  (96g)
  -0.85825000000000  -0.85825000000000  -0.57903000000000  H2O  (96g)
  -0.57903000000000   2.79553000000000  -0.85825000000000  H2O  (96g)
   2.79553000000000  -0.57903000000000  -0.85825000000000  H2O  (96g)
  -0.57903000000000  -0.85825000000000   2.79553000000000  H2O  (96g)
   2.79553000000000  -0.85825000000000  -0.57903000000000  H2O  (96g)
  -0.85825000000000  -0.57903000000000  -0.85825000000000  H2O  (96g)
  -0.85825000000000   2.79553000000000  -0.85825000000000  H2O  (96g)
  -0.85825000000000  -0.57903000000000   2.79553000000000  H2O  (96g)
  -0.85825000000000   2.79553000000000  -0.57903000000000  H2O  (96g)
   2.79553000000000  -0.85825000000000  -0.85825000000000  H2O  (96g)
  -0.57903000000000  -0.85825000000000  -0.85825000000000  H2O  (96g)
   0.41753000000000   0.41753000000000   0.41753000000000    O  (32e)
   0.41753000000000   0.41753000000000  -0.75259000000000    O  (32e)
   0.41753000000000  -0.75259000000000   0.41753000000000    O  (32e)
  -0.75259000000000   0.41753000000000   0.41753000000000    O  (32e)
  -0.41753000000000  -0.41753000000000   1.75259000000000    O  (32e)
  -0.41753000000000  -0.41753000000000  -0.41753000000000    O  (32e)
  -0.41753000000000   1.75259000000000  -0.41753000000000    O  (32e)
   1.75259000000000  -0.41753000000000  -0.41753000000000    O  (32e)
   0.25301000000000   0.25301000000000   1.41365000000000    O  (96g)
   0.25301000000000   0.25301000000000  -1.41967000000000    O  (96g)
   1.41365000000000  -1.41967000000000   0.25301000000000    O  (96g)
  -1.41967000000000   1.41365000000000   0.25301000000000    O  (96g)
   1.41365000000000   0.25301000000000   0.25301000000000    O  (96g)
  -1.41967000000000   0.25301000000000   0.25301000000000    O  (96g)
   0.25301000000000   1.41365000000000  -1.41967000000000    O  (96g)
   0.25301000000000  -1.41967000000000   1.41365000000000    O  (96g)
   0.25301000000000   1.41365000000000   0.25301000000000    O  (96g)
   0.25301000000000  -1.41967000000000   0.25301000000000    O  (96g)
  -1.41967000000000   0.25301000000000   1.41365000000000    O  (96g)
   1.41365000000000   0.25301000000000  -1.41967000000000    O  (96g)
  -0.25301000000000  -0.25301000000000   2.41967000000000    O  (96g)
  -0.25301000000000  -0.25301000000000  -1.41365000000000    O  (96g)
  -1.41365000000000   2.41967000000000  -0.25301000000000    O  (96g)
   2.41967000000000  -1.41365000000000  -0.25301000000000    O  (96g)
  -1.41365000000000  -0.25301000000000   2.41967000000000    O  (96g)
   2.41967000000000  -0.25301000000000  -1.41365000000000    O  (96g)
  -0.25301000000000  -1.41365000000000  -0.25301000000000    O  (96g)
  -0.25301000000000   2.41967000000000  -0.25301000000000    O  (96g)
  -0.25301000000000  -1.41365000000000   2.41967000000000    O  (96g)
  -0.25301000000000   2.41967000000000  -1.41365000000000    O  (96g)
   2.41967000000000  -0.25301000000000  -0.25301000000000    O  (96g)
  -1.41365000000000  -0.25301000000000  -0.25301000000000    O  (96g)
   0.52706000000000   0.52706000000000  -0.64906000000000    O  (96g)
   0.52706000000000   0.52706000000000   0.09494000000000    O  (96g)
  -0.64906000000000   0.09494000000000   0.52706000000000    O  (96g)
   0.09494000000000  -0.64906000000000   0.52706000000000    O  (96g)
  -0.64906000000000   0.52706000000000   0.52706000000000    O  (96g)
   0.09494000000000   0.52706000000000   0.52706000000000    O  (96g)
   0.52706000000000  -0.64906000000000   0.09494000000000    O  (96g)
   0.52706000000000   0.09494000000000  -0.64906000000000    O  (96g)
   0.52706000000000  -0.64906000000000   0.52706000000000    O  (96g)
   0.52706000000000   0.09494000000000   0.52706000000000    O  (96g)
   0.09494000000000   0.52706000000000  -0.64906000000000    O  (96g)
  -0.64906000000000   0.52706000000000   0.09494000000000    O  (96g)
  -0.52706000000000  -0.52706000000000   0.90506000000000    O  (96g)
  -0.52706000000000  -0.52706000000000   0.64906000000000    O  (96g)
   0.64906000000000   0.90506000000000  -0.52706000000000    O  (96g)
   0.90506000000000   0.64906000000000  -0.52706000000000    O  (96g)
   0.64906000000000  -0.52706000000000   0.90506000000000    O  (96g)
   0.90506000000000  -0.52706000000000   0.64906000000000    O  (96g)
  -0.52706000000000   0.64906000000000  -0.52706000000000    O  (96g)
  -0.52706000000000   0.90506000000000  -0.52706000000000    O  (96g)
  -0.52706000000000   0.64906000000000   0.90506000000000    O  (96g)
  -0.52706000000000   0.90506000000000   0.64906000000000    O  (96g)
   0.90506000000000  -0.52706000000000  -0.52706000000000    O  (96g)
   0.64906000000000  -0.52706000000000  -0.52706000000000    O  (96g)
   0.37070000000000   0.37070000000000   0.70490000000000    O  (96g)
   0.37070000000000   0.37070000000000  -0.94630000000000    O  (96g)
   0.70490000000000  -0.94630000000000   0.37070000000000    O  (96g)
  -0.94630000000000   0.70490000000000   0.37070000000000    O  (96g)
   0.70490000000000   0.37070000000000   0.37070000000000    O  (96g)
  -0.94630000000000   0.37070000000000   0.37070000000000    O  (96g)
   0.37070000000000   0.70490000000000  -0.94630000000000    O  (96g)
   0.37070000000000  -0.94630000000000   0.70490000000000    O  (96g)
   0.37070000000000   0.70490000000000   0.37070000000000    O  (96g)
   0.37070000000000  -0.94630000000000   0.37070000000000    O  (96g)
  -0.94630000000000   0.37070000000000   0.70490000000000    O  (96g)
   0.70490000000000   0.37070000000000  -0.94630000000000    O  (96g)
  -0.37070000000000  -0.37070000000000   1.94630000000000    O  (96g)
  -0.37070000000000  -0.37070000000000  -0.70490000000000    O  (96g)
  -0.70490000000000   1.94630000000000  -0.37070000000000    O  (96g)
   1.94630000000000  -0.70490000000000  -0.37070000000000    O  (96g)
  -0.70490000000000  -0.37070000000000   1.94630000000000    O  (96g)
   1.94630000000000  -0.37070000000000  -0.70490000000000    O  (96g)
  -0.37070000000000  -0.70490000000000  -0.37070000000000    O  (96g)
  -0.37070000000000   1.94630000000000  -0.37070000000000    O  (96g)
  -0.37070000000000  -0.70490000000000   1.94630000000000    O  (96g)
  -0.37070000000000   1.94630000000000  -0.70490000000000    O  (96g)
   1.94630000000000  -0.37070000000000  -0.37070000000000    O  (96g)
  -0.70490000000000  -0.37070000000000  -0.37070000000000    O  (96g)
   0.37500000000000   0.37500000000000   0.37500000000000    P   (8b)
   0.62500000000000   0.62500000000000   0.62500000000000    P   (8b)
   0.38144000000000   0.38144000000000  -0.41710000000000    W  (96g)
   0.38144000000000   0.38144000000000   0.15422000000000    W  (96g)
  -0.41710000000000   0.15422000000000   0.38144000000000    W  (96g)
   0.15422000000000  -0.41710000000000   0.38144000000000    W  (96g)
  -0.41710000000000   0.38144000000000   0.38144000000000    W  (96g)
   0.15422000000000   0.38144000000000   0.38144000000000    W  (96g)
   0.38144000000000  -0.41710000000000   0.15422000000000    W  (96g)
   0.38144000000000   0.15422000000000  -0.41710000000000    W  (96g)
   0.38144000000000  -0.41710000000000   0.38144000000000    W  (96g)
   0.38144000000000   0.15422000000000   0.38144000000000    W  (96g)
   0.15422000000000   0.38144000000000  -0.41710000000000    W  (96g)
  -0.41710000000000   0.38144000000000   0.15422000000000    W  (96g)
  -0.38144000000000  -0.38144000000000   0.84578000000000    W  (96g)
  -0.38144000000000  -0.38144000000000   0.41710000000000    W  (96g)
   0.41710000000000   0.84578000000000  -0.38144000000000    W  (96g)
   0.84578000000000   0.41710000000000  -0.38144000000000    W  (96g)
   0.41710000000000  -0.38144000000000   0.84578000000000    W  (96g)
   0.84578000000000  -0.38144000000000   0.41710000000000    W  (96g)
  -0.38144000000000   0.41710000000000  -0.38144000000000    W  (96g)
  -0.38144000000000   0.84578000000000  -0.38144000000000    W  (96g)
  -0.38144000000000   0.41710000000000   0.84578000000000    W  (96g)
  -0.38144000000000   0.84578000000000   0.41710000000000    W  (96g)
   0.84578000000000  -0.38144000000000  -0.38144000000000    W  (96g)
   0.41710000000000  -0.38144000000000  -0.38144000000000    W  (96g)
```

$G7_3$ [Northupite, $Na_3MgCl(CO_3)_2$] (*obsolete*): A2BCD3E6_cF208_227_e_c_d_f_g - CIF

```
# CIF file
data_findsym-output
_audit_creation_method FINDSYM

_chemical_name_mineral 'Northupite'
_chemical_formula_sum 'C2 Cl Mg Na3 O6'

loop_
_publ_author_name
 'H. Shiba'
 'T. Watanab\\'{e}'
_journal_name_full_name
;
 Comptes Rendus de l'Acad{\'e}mie des Sciences
;
_journal_volume 193
_journal_year 1931
_journal_page_first 1421
_journal_page_last 1423
_publ_Section_title
;
 Les structures des cristaux de northupite, de northupite brom\\'{e}e et
     ↪ de tychite
;

# Found in The American Mineralogist Crystal Structure Database, 2003

_aflow_title '$G7_{3}$ [Northupite, Na$_{3}$MgCl(CO$_{3}$)$_{2}$] ({\em
```

```
     ↪ {obsolete}}) Structure'
_aflow_proto 'A2BCD3E6_cF208_227_e_c_d_f_g'
_aflow_params 'a,x_{3},x_{4},x_{5},z_{5}'
_aflow_params_values '14.05,0.715,0.35,0.75,0.65'
_aflow_Strukturbericht '$G7_{3}$'
_aflow_Pearson 'cF208'

_symmetry_space_group_name_H-M "F 41/d -3 2/m (origin choice 2)"
_symmetry_Int_Tables_number 227

_cell_length_a      14.05000
_cell_length_b      14.05000
_cell_length_c      14.05000
_cell_angle_alpha 90.00000
_cell_angle_beta  90.00000
_cell_angle_gamma 90.00000

loop_
_space_group_symop_id
_space_group_symop_operation_xyz
1 x,y,z
2 x,-y+1/4,-z+1/4
3 -x+1/4,y,-z+1/4
4 -x+1/4,-y+1/4,z
5 y,z,x
6 y,-z+1/4,-x+1/4
7 -y+1/4,z,-x+1/4
8 -y+1/4,-z+1/4,x
9 z,x,y
10 z,-x+1/4,-y+1/4
11 -z+1/4,x,-y+1/4
12 -z+1/4,-x+1/4,y
13 -y,-x,-z
14 -y,x+1/4,z+1/4
15 y+1/4,-x,z+1/4
16 y+1/4,x+1/4,-z
17 -x,-z,-y
18 -x,z+1/4,y+1/4
19 x+1/4,-z,y+1/4
20 x+1/4,z+1/4,-y
21 -z,-y,-x
22 -z,y+1/4,x+1/4
23 z+1/4,-y,x+1/4
24 z+1/4,y+1/4,-x
25 -x,-y,-z
26 -x,y+1/4,z+1/4
27 x+1/4,-y,z+1/4
28 x+1/4,y+1/4,-z
29 -y,-z,-x
30 -y,z+1/4,x+1/4
31 y+1/4,-z,x+1/4
32 y+1/4,z+1/4,-x
33 -z,-x,-y
34 -z,x+1/4,y+1/4
35 z+1/4,-x,y+1/4
36 z+1/4,x+1/4,-y
37 y,x,z
38 y,-x+1/4,-z+1/4
39 -y+1/4,x,-z+1/4
40 -y+1/4,-x+1/4,z
41 x,z,y
42 x,-z+1/4,-y+1/4
43 -x+1/4,z,-y+1/4
44 -x+1/4,-z+1/4,y
45 z,y,x
46 z,-y+1/4,-x+1/4
47 -z+1/4,y,-x+1/4
48 -z+1/4,-y+1/4,x
49 x,y+1/2,z+1/2
50 x,-y+3/4,-z+3/4
51 -x+1/4,y+1/2,-z+3/4
52 -x+1/4,-y+3/4,z+1/2
53 y,z+1/2,x+1/2
54 y,-z+3/4,-x+3/4
55 -y+1/4,z+1/2,-x+3/4
56 -y+1/4,-z+3/4,x+1/2
57 z,x+1/2,y+1/2
58 z,-x+3/4,-y+3/4
59 -z+1/4,x+1/2,-y+3/4
60 -z+1/4,-x+3/4,y+1/2
61 -y,-x+1/2,-z+1/2
62 -y,x+3/4,z+3/4
63 y+1/4,-x+1/2,z+3/4
64 y+1/4,x+3/4,-z+1/2
65 -x,-z+1/2,-y+1/2
66 -x,z+3/4,y+3/4
67 x+1/4,-z+1/2,y+3/4
68 x+1/4,z+3/4,-y+1/2
69 -z,-y+1/2,-x+1/2
70 -z,y+3/4,x+3/4
71 z+1/4,-y+1/2,x+3/4
72 z+1/4,y+3/4,-x+1/2
73 -x,-y+1/2,-z+1/2
74 -x,y+3/4,z+3/4
75 x+1/4,-y+1/2,z+3/4
76 x+1/4,y+3/4,-z+1/2
77 -y,-z+1/2,-x+1/2
78 -y,z+3/4,x+3/4
79 y+1/4,-z+1/2,x+3/4
80 y+1/4,z+3/4,-x+1/2
81 -z,-x+1/2,-y+1/2
82 -z,x+3/4,y+3/4
83 z+1/4,-x+1/2,y+3/4
84 z+1/4,x+3/4,-y+1/2
85 y,x+1/2,z+1/2
86 y,-x+3/4,-z+3/4
87 -y+1/4,x+1/2,-z+3/4
88 -y+1/4,-x+3/4,z+1/2
89 x,z+1/2,y+1/2
90 x,-z+3/4,-y+3/4
91 -x+1/4,z+1/2,-y+3/4
92 -x+1/4,-z+3/4,y+1/2
93 z,y+1/2,x+1/2
94 z,-y+3/4,-x+3/4
95 -z+1/4,y+1/2,-x+3/4
96 -z+1/4,-y+3/4,x+1/2
97 x+1/2,y,z+1/2
98 x+1/2,-y+1/4,-z+3/4
99 -x+3/4,y,-z+3/4
100 -x+3/4,-y+1/4,z+1/2
101 y+1/2,z,x+1/2
102 y+1/2,-z+1/4,-x+3/4
103 -y+3/4,z,-x+3/4
104 -y+3/4,-z+1/4,x+1/2
105 z+1/2,x,y+1/2
106 z+1/2,-x+1/4,-y+3/4
107 -z+3/4,x,-y+3/4
108 -z+3/4,-x+1/4,y+1/2
109 -y+1/2,-x,-z+1/2
110 -y+1/2,x+1/4,z+3/4
111 y+3/4,-x,z+3/4
112 y+3/4,x+1/4,-z+1/2
113 -x+1/2,-z,-y+1/2
114 -x+1/2,z+1/4,y+3/4
115 x+3/4,-z,y+3/4
116 x+3/4,z+1/4,-y+1/2
117 -z+1/2,-y,-x+1/2
118 -z+1/2,y+1/4,x+3/4
119 z+3/4,-y,x+3/4
120 z+3/4,y+1/4,-x+1/2
121 -x+1/2,-y,-z+1/2
122 -x+1/2,y+1/4,z+3/4
123 x+3/4,-y,z+3/4
124 x+3/4,y+1/4,-z+1/2
125 -y+1/2,-z,-x+1/2
126 -y+1/2,z+1/4,x+3/4
127 y+3/4,-z,x+3/4
128 y+3/4,z+1/4,-x+1/2
129 -z+1/2,-x,-y+1/2
130 -z+1/2,x+1/4,y+3/4
131 z+3/4,-x,y+3/4
132 z+3/4,x+1/4,-y+1/2
133 y+1/2,x,z+1/2
134 y+1/2,-x+1/4,-z+3/4
135 -y+3/4,x,-z+3/4
136 -y+3/4,-x+1/4,z+1/2
137 x+1/2,z,y+1/2
138 x+1/2,-z+1/4,-y+3/4
139 -x+3/4,z,-y+3/4
140 -x+3/4,-z+1/4,y+1/2
141 z+1/2,y,x+1/2
142 z+1/2,-y+1/4,-x+3/4
143 -z+3/4,y,-x+3/4
144 -z+3/4,-y+1/4,x+1/2
145 x+1/2,y+1/2,z
146 x+1/2,-y+3/4,-z+1/4
147 -x+3/4,y+1/2,-z+1/4
148 -x+3/4,-y+3/4,z
149 y+1/2,z+1/2,x
150 y+1/2,-z+3/4,-x+1/4
151 -y+3/4,z+1/2,-x+1/4
152 -y+3/4,-z+3/4,x
153 z+1/2,x+1/2,y
154 z+1/2,-x+3/4,-y+1/4
155 -z+3/4,x+1/2,-y+1/4
156 -z+3/4,-x+3/4,y
157 -y+1/2,-x+1/2,-z
158 -y+1/2,x+3/4,z+1/4
159 y+3/4,-x+1/2,z+1/4
160 y+3/4,x+3/4,-z
161 -x+1/2,-z+1/2,-y
162 -x+1/2,z+3/4,y+1/4
163 x+3/4,-z+1/2,y+1/4
164 x+3/4,z+3/4,-y
165 -z+1/2,-y+1/2,-x
166 -z+1/2,y+3/4,x+1/4
167 z+3/4,-y+1/2,x+1/4
168 z+3/4,y+3/4,-x
169 -x+1/2,-y+1/2,-z
170 -x+1/2,y+3/4,z+1/4
171 x+3/4,-y+1/2,z+1/4
172 x+3/4,y+3/4,-z
173 -y+1/2,-z+1/2,-x
174 -y+1/2,z+3/4,x+1/4
175 y+3/4,-z+1/2,x+1/4
176 y+3/4,z+3/4,-x
177 -z+1/2,-x+1/2,-y
178 -z+1/2,x+3/4,y+1/4
179 z+3/4,-x+1/2,y+1/4
180 z+3/4,x+3/4,-y
181 y+1/2,x+1/2,z
182 y+1/2,-x+3/4,-z+1/4
183 -y+3/4,x+1/2,-z+1/4
184 -y+3/4,-x+3/4,z
185 x+1/2,z+1/2,y
186 x+1/2,-z+3/4,-y+1/4
187 -x+3/4,z+1/2,-y+1/4
188 -x+3/4,-z+3/4,y
189 z+1/2,y+1/2,x
190 z+1/2,-y+3/4,-x+1/4
```

```
191 -z+3/4,y+1/2,-x+1/4
192 -z+3/4,-y+3/4,x

loop_
_atom_site_label
_atom_site_type_symbol
_atom_site_symmetry_multiplicity
_atom_site_Wyckoff_label
_atom_site_fract_x
_atom_site_fract_y
_atom_site_fract_z
_atom_site_occupancy
Cl1 Cl 16 c 0.00000 0.00000 0.00000 1.00000
Mg1 Mg 16 d 0.50000 0.50000 0.50000 1.00000
C1 C 32 e 0.71500 0.71500 0.71500 1.00000
Na1 Na 48 f 0.35000 0.12500 0.12500 1.00000
O1 O 96 g 0.75000 0.75000 0.65000 1.00000
```

$G7_3$ [Northupite, $Na_3MgCl(CO_3)_2$] (*obsolete*): A2BCD3E6_cF208_227_e_c_d_f_g - POSCAR

```
A2BCD3E6_cF208_227_e_c_d_f_g & a,x3,x4,x5,z5 --params=14.05,0.715,0.35,
    ↪ 0.75,0.65 & Fd-3m O_{h}^{7} #227 (cdefg) & cF208 & $G7_{3}$ &
    ↪ C2ClMgNa3O6 & Northupite & H. Shiba and T. Watanab\'{e}, {C. R.
    ↪ Acad. Sci. C 193, 1421-1423 (1931)
   1.00000000000000
   0.00000000000000   7.02500000000000   7.02500000000000
   7.02500000000000   0.00000000000000   7.02500000000000
   7.02500000000000   7.02500000000000   0.00000000000000
     C    Cl    Mg    Na     O
     8     4     4    12    24
Direct
   0.71500000000000   0.71500000000000   0.71500000000000   C   (32e)
   0.71500000000000   0.71500000000000  -1.64500000000000   C   (32e)
   0.71500000000000  -1.64500000000000   0.71500000000000   C   (32e)
  -1.64500000000000   0.71500000000000   0.71500000000000   C   (32e)
  -0.71500000000000  -0.71500000000000   2.64500000000000   C   (32e)
  -0.71500000000000  -0.71500000000000  -0.71500000000000   C   (32e)
  -0.71500000000000   2.64500000000000  -0.71500000000000   C   (32e)
   2.64500000000000  -0.71500000000000  -0.71500000000000   C   (32e)
   0.00000000000000   0.00000000000000   0.00000000000000   Cl  (16c)
   0.00000000000000   0.00000000000000   0.50000000000000   Cl  (16c)
   0.00000000000000   0.50000000000000   0.00000000000000   Cl  (16c)
   0.50000000000000   0.00000000000000   0.00000000000000   Cl  (16c)
   0.50000000000000   0.50000000000000   0.50000000000000   Mg  (16d)
   0.50000000000000   0.50000000000000   0.00000000000000   Mg  (16d)
   0.50000000000000   0.00000000000000   0.50000000000000   Mg  (16d)
   0.00000000000000   0.50000000000000   0.50000000000000   Mg  (16d)
  -0.10000000000000   0.35000000000000   0.35000000000000   Na  (48f)
   0.35000000000000  -0.10000000000000  -0.10000000000000   Na  (48f)
   0.35000000000000  -0.10000000000000   0.35000000000000   Na  (48f)
  -0.10000000000000   0.35000000000000  -0.10000000000000   Na  (48f)
   0.35000000000000   0.35000000000000  -0.10000000000000   Na  (48f)
  -0.10000000000000  -0.10000000000000   0.35000000000000   Na  (48f)
   1.10000000000000  -0.35000000000000   1.10000000000000   Na  (48f)
  -0.35000000000000   1.10000000000000  -0.35000000000000   Na  (48f)
  -0.35000000000000   1.10000000000000   1.10000000000000   Na  (48f)
   1.10000000000000  -0.35000000000000  -0.35000000000000   Na  (48f)
  -0.35000000000000  -0.35000000000000   1.10000000000000   Na  (48f)
   1.10000000000000   1.10000000000000  -0.35000000000000   Na  (48f)
   0.65000000000000   0.65000000000000   0.85000000000000   O   (96g)
   0.65000000000000   0.65000000000000  -1.65000000000000   O   (96g)
   0.85000000000000  -1.65000000000000   0.65000000000000   O   (96g)
  -1.65000000000000   0.85000000000000   0.65000000000000   O   (96g)
   0.85000000000000   0.65000000000000   0.65000000000000   O   (96g)
  -1.65000000000000   0.65000000000000   0.65000000000000   O   (96g)
   0.65000000000000   0.85000000000000  -1.65000000000000   O   (96g)
   0.65000000000000  -1.65000000000000   0.85000000000000   O   (96g)
   0.65000000000000   0.85000000000000   0.65000000000000   O   (96g)
   0.65000000000000  -1.65000000000000   0.65000000000000   O   (96g)
  -1.65000000000000   0.65000000000000   0.85000000000000   O   (96g)
   0.85000000000000   0.65000000000000  -1.65000000000000   O   (96g)
  -0.65000000000000  -0.65000000000000   2.65000000000000   O   (96g)
  -0.65000000000000  -0.65000000000000  -0.85000000000000   O   (96g)
  -0.85000000000000   2.65000000000000  -0.65000000000000   O   (96g)
   2.65000000000000  -0.85000000000000  -0.65000000000000   O   (96g)
  -0.85000000000000  -0.65000000000000   2.65000000000000   O   (96g)
   2.65000000000000  -0.65000000000000  -0.85000000000000   O   (96g)
  -0.65000000000000  -0.85000000000000  -0.65000000000000   O   (96g)
  -0.65000000000000   2.65000000000000  -0.65000000000000   O   (96g)
  -0.65000000000000  -0.85000000000000   2.65000000000000   O   (96g)
  -0.65000000000000   2.65000000000000  -0.85000000000000   O   (96g)
   2.65000000000000  -0.65000000000000  -0.65000000000000   O   (96g)
  -0.85000000000000  -0.65000000000000  -0.65000000000000   O   (96g)
```

$D6_2$ ($Sb_2O_4$) (*obsolete*): A2B_cF96_227_abf_cd - CIF

```
# CIF file
data_findsym-output
_audit_creation_method FINDSYM

_chemical_name_mineral 'O2Sb'
_chemical_formula_sum 'O2 Sb'

loop_
_publ_author_name
 'G. Natta'
 'M. Baccaredda'
_journal_name_full_name
;
 Zeitschrift f{\"u}r Kristallographie - Crystalline Materials
;
_journal_volume 85
_journal_year 1933
_journal_page_first 271
_journal_page_last 296
_publ_Section_title
;
 Tetrossido di antimonio e antimoniati
;

# Found in Strukturbericht Band III 1933-1935, 1937

_aflow_title '$D6_{2}$ (Sb$_{2}$O$_{4}$) ({\em{obsolete}}) Structure'
_aflow_proto 'A2B_cF96_227_abf_cd'
_aflow_params 'a,x_{5}'
_aflow_params_values '10.24,0.395'
_aflow_Strukturbericht '$D6_{2}$'
_aflow_Pearson 'cF96'

_symmetry_space_group_name_H-M "F 41/d -3 2/m (origin choice 2)"
_symmetry_Int_Tables_number 227

_cell_length_a     10.24000
_cell_length_b     10.24000
_cell_length_c     10.24000
_cell_angle_alpha  90.00000
_cell_angle_beta   90.00000
_cell_angle_gamma  90.00000

loop_
_space_group_symop_id
_space_group_symop_operation_xyz
1 x,y,z
2 x,-y+1/4,-z+1/4
3 -x+1/4,y,-z+1/4
4 -x+1/4,-y+1/4,z
5 y,z,x
6 y,-z+1/4,-x+1/4
7 -y+1/4,z,-x+1/4
8 -y+1/4,-z+1/4,x
9 z,x,y
10 z,-x+1/4,-y+1/4
11 -z+1/4,x,-y+1/4
12 -z+1/4,-x+1/4,y
13 -y,-x,-z
14 -y,x+1/4,z+1/4
15 y+1/4,-x,z+1/4
16 y+1/4,x+1/4,-z
17 -x,-z,-y
18 -x,z+1/4,y+1/4
19 x+1/4,-z,y+1/4
20 x+1/4,z+1/4,-y
21 -z,-y,-x
22 -z,y+1/4,x+1/4
23 z+1/4,-y,x+1/4
24 z+1/4,y+1/4,-x
25 -x,-y,-z
26 -x,y+1/4,z+1/4
27 x+1/4,-y,z+1/4
28 x+1/4,y+1/4,-z
29 -y,-z,-x
30 -y,z+1/4,x+1/4
31 y+1/4,-z,x+1/4
32 y+1/4,z+1/4,-x
33 -z,-x,-y
34 -z,x+1/4,y+1/4
35 z+1/4,-x,y+1/4
36 z+1/4,x+1/4,-y
37 y,x,z
38 y,-x+1/4,-z+1/4
39 -y+1/4,x,-z+1/4
40 -y+1/4,-x+1/4,z
41 x,z,y
42 x,-z+1/4,-y+1/4
43 -x+1/4,z,-y+1/4
44 -x+1/4,-z+1/4,y
45 z,y,x
46 z,-y+1/4,-x+1/4
47 -z+1/4,y,-x+1/4
48 -z+1/4,-y+1/4,x
49 x,y+1/2,z+1/2
50 x,-y+3/4,-z+3/4
51 -x+1/4,y+1/2,-z+3/4
52 -x+1/4,-y+3/4,z+1/2
53 y,z+1/2,x+1/2
54 y,-z+3/4,-x+3/4
55 -y+1/4,z+1/2,-x+3/4
56 -y+1/4,-z+3/4,x+1/2
57 z,x+1/2,y+1/2
58 z,-x+3/4,-y+3/4
59 -z+1/4,x+1/2,-y+3/4
60 -z+1/4,-x+3/4,y+1/2
61 -y,-x+1/2,-z+1/2
62 -y,x+3/4,z+3/4
63 y+1/4,-x+1/2,z+3/4
64 y+1/4,x+3/4,-z+1/2
65 -x,-z+1/2,-y+1/2
66 -x,z+3/4,y+3/4
67 x+1/4,-z+1/2,y+3/4
68 x+1/4,z+3/4,-y+1/2
69 -z,-y+1/2,-x+1/2
70 -z,y+3/4,x+3/4
71 z+1/4,-y+1/2,x+3/4
72 z+1/4,y+3/4,-x+1/2
73 -x,-y+1/2,-z+1/2
74 -x,y+3/4,z+3/4
75 x+1/4,-y+1/2,z+3/4
76 x+1/4,y+3/4,-z+1/2
77 -y,-z+1/2,-x+1/2
78 -y,z+3/4,x+3/4
```

```
79 y+1/4,-z+1/2,x+3/4
80 y+1/4,z+3/4,-x+1/2
81 -z,-x+1/2,-y+1/2
82 -z,x+3/4,y+3/4
83 z+1/4,-x+1/2,y+3/4
84 z+1/4,x+3/4,-y+1/2
85 y,x+1/2,z+1/2
86 y,-x+3/4,-z+3/4
87 -y+1/4,x+1/2,-z+3/4
88 -y+1/4,-x+3/4,z+1/2
89 x,z+1/2,y+1/2
90 x,-z+3/4,-y+3/4
91 -x+1/4,z+1/2,-y+3/4
92 -x+1/4,-z+3/4,y+1/2
93 z,y+1/2,x+1/2
94 z,-y+3/4,-x+3/4
95 -z+1/4,y+1/2,-x+3/4
96 -z+1/4,-y+3/4,x+1/2
97 x+1/2,y,z+1/2
98 x+1/2,-y+1/4,-z+3/4
99 -x+3/4,y,-z+3/4
100 -x+3/4,-y+1/4,z+1/2
101 y+1/2,z,x+1/2
102 y+1/2,-z+1/4,-x+3/4
103 -y+3/4,z,-x+3/4
104 -y+3/4,-z+1/4,x+1/2
105 z+1/2,x,y+1/2
106 z+1/2,-x+1/4,-y+3/4
107 -z+3/4,x,-y+3/4
108 -z+3/4,-x+1/4,y+1/2
109 -y+1/2,-x,-z+1/2
110 -y+1/2,x+1/4,z+3/4
111 y+3/4,-x,z+3/4
112 y+3/4,x+1/4,-z+1/2
113 -x+1/2,-z,-y+1/2
114 -x+1/2,z+1/4,y+3/4
115 x+3/4,-z,y+3/4
116 x+3/4,z+1/4,-y+1/2
117 -z+1/2,-y,-x+1/2
118 -z+1/2,y+1/4,x+3/4
119 z+3/4,-y,x+3/4
120 z+3/4,y+1/4,-x+1/2
121 -x+1/2,-y,-z+1/2
122 -x+1/2,y+1/4,z+3/4
123 x+3/4,-y,z+3/4
124 x+3/4,y+1/4,-z+1/2
125 -y+1/2,-z,-x+1/2
126 -y+1/2,z+1/4,x+3/4
127 y+3/4,-z,x+3/4
128 y+3/4,z+1/4,-x+1/2
129 -z+1/2,-x,-y+1/2
130 -z+1/2,x+1/4,y+3/4
131 z+3/4,-x,y+3/4
132 z+3/4,x+1/4,-y+1/2
133 y+1/2,x,z+1/2
134 y+1/2,-x+1/4,-z+3/4
135 -y+3/4,x,-z+3/4
136 -y+3/4,-x+1/4,z+1/2
137 x+1/2,z,y+1/2
138 x+1/2,-z+1/4,-y+3/4
139 -x+3/4,z,-y+3/4
140 -x+3/4,-z+1/4,y+1/2
141 z+1/2,y,x+1/2
142 z+1/2,-y+1/4,-x+3/4
143 -z+3/4,y,-x+3/4
144 -z+3/4,-y+1/4,x+1/2
145 x+1/2,y+1/2,z
146 x+1/2,-y+3/4,-z+1/4
147 -x+3/4,y+1/2,-z+1/4
148 -x+3/4,-y+3/4,z
149 y+1/2,z+1/2,x
150 y+1/2,-z+3/4,-x+1/4
151 -y+3/4,z+1/2,-x+1/4
152 -y+3/4,-z+3/4,x
153 z+1/2,x+1/2,y
154 z+1/2,-x+3/4,-y+1/4
155 -z+3/4,x+1/2,-y+1/4
156 -z+3/4,-x+3/4,y
157 -y+1/2,-x+1/2,-z
158 -y+1/2,x+3/4,z+1/4
159 y+3/4,-x+1/2,z+1/4
160 y+3/4,x+3/4,-z
161 -x+1/2,-z+1/2,-y
162 -x+1/2,z+3/4,y+1/4
163 x+3/4,-z+1/2,y+1/4
164 x+3/4,z+3/4,-y
165 -z+1/2,-y+1/2,-x
166 -z+1/2,y+3/4,x+1/4
167 z+3/4,-y+1/2,x+1/4
168 z+3/4,y+3/4,-x
169 -x+1/2,-y+1/2,-z
170 -x+1/2,y+3/4,z+1/4
171 x+3/4,-y+1/2,z+1/4
172 x+3/4,y+3/4,-z
173 -y+1/2,-z+1/2,-x
174 -y+1/2,z+3/4,x+1/4
175 y+3/4,-z+1/2,x+1/4
176 y+3/4,z+3/4,-x
177 -z+1/2,-x+1/2,-y
178 -z+1/2,x+3/4,y+1/4
179 z+3/4,-x+1/2,y+1/4
180 z+3/4,x+3/4,-y
181 y+1/2,x+1/2,z
182 y+1/2,-x+3/4,-z+1/4
183 -y+3/4,x+1/2,-z+1/4
184 -y+3/4,-x+3/4,z
185 x+1/2,z+1/2,y
186 x+1/2,-z+3/4,-y+1/4
187 -x+3/4,z+1/2,-y+1/4
188 -x+3/4,-z+3/4,y
189 z+1/2,y+1/2,x
190 z+1/2,-y+3/4,-x+1/4
191 -z+3/4,y+1/2,-x+1/4
192 -z+3/4,-y+3/4,x

loop_
_atom_site_label
_atom_site_type_symbol
_atom_site_symmetry_multiplicity
_atom_site_Wyckoff_label
_atom_site_fract_x
_atom_site_fract_y
_atom_site_fract_z
_atom_site_occupancy
O1 O 8 a 0.12500 0.12500 0.12500 1.00000
O2 O 8 b 0.37500 0.37500 0.37500 1.00000
Sb1 Sb 16 c 0.00000 0.00000 0.00000 1.00000
Sb2 Sb 16 d 0.50000 0.50000 0.50000 1.00000
O3 O 48 f 0.39500 0.12500 0.12500 1.00000
```

$D6_2$ ($Sb_2O_4$) (*obsolete*): A2B_cF96_227_abf_cd - POSCAR

```
A2B_cF96_227_abf_cd & a,x5 --params=10.24,0.395 & Fd-3m O_{h}^{7} #227 (
    ↪ abcdf) & cF96 & $D6_{2}$ & O2Sb & O2Sb & G. Natta and M.
    ↪ Baccaredda, Zeitschrift f{"u}r Kristallographie - Crystalline
    ↪ Materials 85, 271-296 (1933)
   1.00000000000000
   0.00000000000000   5.12000000000000   5.12000000000000
   5.12000000000000   0.00000000000000   5.12000000000000
   5.12000000000000   5.12000000000000   0.00000000000000
     O     Sb
    16      8
Direct
   0.12500000000000   0.12500000000000   0.12500000000000   O   (8a)
   0.87500000000000   0.87500000000000   0.87500000000000   O   (8a)
   0.37500000000000   0.37500000000000   0.37500000000000   O   (8b)
   0.62500000000000   0.62500000000000   0.62500000000000   O   (8b)
  -0.14500000000000   0.39500000000000   0.39500000000000   O   (48f)
   0.39500000000000  -0.14500000000000  -0.14500000000000   O   (48f)
   0.39500000000000  -0.14500000000000   0.39500000000000   O   (48f)
  -0.14500000000000   0.39500000000000  -0.14500000000000   O   (48f)
   0.39500000000000   0.39500000000000  -0.14500000000000   O   (48f)
  -0.14500000000000  -0.14500000000000   0.39500000000000   O   (48f)
   1.14500000000000  -0.39500000000000   1.14500000000000   O   (48f)
  -0.39500000000000   1.14500000000000  -0.39500000000000   O   (48f)
  -0.39500000000000   1.14500000000000   1.14500000000000   O   (48f)
   1.14500000000000  -0.39500000000000  -0.39500000000000   O   (48f)
  -0.39500000000000  -0.39500000000000   1.14500000000000   O   (48f)
   1.14500000000000   1.14500000000000  -0.39500000000000   O   (48f)
   0.00000000000000   0.00000000000000   0.00000000000000   Sb  (16c)
   0.00000000000000   0.00000000000000   0.50000000000000   Sb  (16c)
   0.00000000000000   0.50000000000000   0.00000000000000   Sb  (16c)
   0.50000000000000   0.00000000000000   0.00000000000000   Sb  (16c)
   0.50000000000000   0.50000000000000   0.50000000000000   Sb  (16d)
   0.50000000000000   0.50000000000000   0.00000000000000   Sb  (16d)
   0.50000000000000   0.00000000000000   0.50000000000000   Sb  (16d)
   0.00000000000000   0.50000000000000   0.50000000000000   Sb  (16d)
```

Senarmontite ($Sb_2O_3$, $D6_1$): A3B2_cF80_227_f_e - CIF

```
# CIF file
data_findsym-output
_audit_creation_method FINDSYM

_chemical_name_mineral 'Senarmontite'
_chemical_formula_sum 'O3 Sb2'

loop_
_publ_author_name
 'C. Svensson'
_journal_name_full_name
;
 Acta Crystallographica Section B: Structural Science
;
_journal_volume 31
_journal_year 1975
_journal_page_first 2016
_journal_page_last 2018
_publ_Section_title
;
 Refinement of the crystal structure of cubic antimony trioxide, Sb$_{2}
    ↪ $O$_{3}$
;

_aflow_title 'Senarmontite (Sb$_{2}$O$_{3}$, $D6_{1}$) Structure'
_aflow_proto 'A3B2_cF80_227_f_e'
_aflow_params 'a,x_{1},x_{2}'
_aflow_params_values '11.1519,0.26027,0.43875'
_aflow_Strukturbericht 'None'
_aflow_Pearson 'cF80'

_symmetry_space_group_name_H-M "F 41/d -3 2/m (origin choice 2)"
_symmetry_Int_Tables_number 227

_cell_length_a      11.15190
_cell_length_b      11.15190
_cell_length_c      11.15190
_cell_angle_alpha   90.00000
_cell_angle_beta    90.00000
_cell_angle_gamma   90.00000
```

```
loop_
_space_group_symop_id
_space_group_symop_operation_xyz
1 x,y,z
2 x,-y+1/4,-z+1/4
3 -x+1/4,y,-z+1/4
4 -x+1/4,-y+1/4,z
5 y,z,x
6 y,-z+1/4,-x+1/4
7 -y+1/4,z,-x+1/4
8 -y+1/4,-z+1/4,x
9 z,x,y
10 z,-x+1/4,-y+1/4
11 -z+1/4,x,-y+1/4
12 -z+1/4,-x+1/4,y
13 -y,-x,-z
14 -y,x+1/4,z+1/4
15 y+1/4,-x,z+1/4
16 y+1/4,x+1/4,-z
17 -x,-z,-y
18 -x,z+1/4,y+1/4
19 x+1/4,-z,y+1/4
20 x+1/4,z+1/4,-y
21 -z,-y,-x
22 -z,y+1/4,x+1/4
23 z+1/4,-y,x+1/4
24 z+1/4,y+1/4,-x
25 -x,-y,-z
26 -x,y+1/4,z+1/4
27 x+1/4,-y,z+1/4
28 x+1/4,y+1/4,-z
29 -y,-z,-x
30 -y,z+1/4,x+1/4
31 y+1/4,-z,x+1/4
32 y+1/4,z+1/4,-x
33 -z,-x,-y
34 -z,x+1/4,y+1/4
35 z+1/4,-x,y+1/4
36 z+1/4,x+1/4,-y
37 y,x,z
38 y,-x+1/4,-z+1/4
39 -y+1/4,x,-z+1/4
40 -y+1/4,-x+1/4,z
41 x,z,y
42 x,-z+1/4,-y+1/4
43 -x+1/4,z,-y+1/4
44 -x+1/4,-z+1/4,y
45 z,y,x
46 z,-y+1/4,-x+1/4
47 -z+1/4,y,-x+1/4
48 -z+1/4,-y+1/4,x
49 x,y+1/2,z+1/2
50 x,-y+3/4,-z+3/4
51 -x+1/4,y+1/2,-z+3/4
52 -x+1/4,-y+3/4,z+1/2
53 y,z+1/2,x+1/2
54 y,-z+3/4,-x+3/4
55 -y+1/4,z+1/2,-x+3/4
56 -y+1/4,-z+3/4,x+1/2
57 z,x+1/2,y+1/2
58 z,-x+3/4,-y+3/4
59 -z+1/4,x+1/2,-y+3/4
60 -z+1/4,-x+3/4,y+1/2
61 -y,-x+1/2,-z+1/2
62 -y,x+3/4,z+3/4
63 y+1/4,-x+1/2,z+3/4
64 y+1/4,x+3/4,-z+1/2
65 -x,-z+1/2,-y+1/2
66 -x,z+3/4,y+3/4
67 x+1/4,-z+1/2,y+3/4
68 x+1/4,z+3/4,-y+1/2
69 -z,-y+1/2,-x+1/2
70 -z,y+3/4,x+3/4
71 z+1/4,-y+1/2,x+3/4
72 z+1/4,y+3/4,-x+1/2
73 -x,-y+1/2,-z+1/2
74 -x,y+3/4,z+3/4
75 x+1/4,-y+1/2,z+3/4
76 x+1/4,y+3/4,-z+1/2
77 -y,-z+1/2,-x+1/2
78 -y,z+3/4,x+3/4
79 y+1/4,-z+1/2,x+3/4
80 y+1/4,z+3/4,-x+1/2
81 -z,-x+1/2,-y+1/2
82 -z,x+3/4,y+3/4
83 z+1/4,-x+1/2,y+3/4
84 z+1/4,x+3/4,-y+1/2
85 y,x+1/2,z+1/2
86 y,-x+3/4,-z+3/4
87 -y+1/4,x+1/2,-z+3/4
88 -y+1/4,-x+3/4,z+1/2
89 x,z+1/2,y+1/2
90 x,-z+3/4,-y+3/4
91 -x+1/4,z+1/2,-y+3/4
92 -x+1/4,-z+3/4,y+1/2
93 z,y+1/2,x+1/2
94 z,-y+3/4,-x+3/4
95 -z+1/4,y+1/2,-x+3/4
96 -z+1/4,-y+3/4,x+1/2
97 x+1/2,y,z+1/2
98 x+1/2,-y+1/4,-z+3/4
99 -x+3/4,y,-z+3/4
100 -x+3/4,-y+1/4,z+1/2
101 y+1/2,z,x+1/2
102 y+1/2,-z+1/4,-x+3/4
103 -y+3/4,z,-x+3/4
104 -y+3/4,-z+1/4,x+1/2
105 z+1/2,x,y+1/2
106 z+1/2,-x+1/4,-y+3/4
107 -z+3/4,x,-y+3/4
108 -z+3/4,-x+1/4,y+1/2
109 -y+1/2,-x,-z+1/2
110 -y+1/2,x+1/4,z+3/4
111 y+3/4,-x,z+3/4
112 y+3/4,x+1/4,-z+1/2
113 -x+1/2,-z,-y+1/2
114 -x+1/2,z+1/4,y+3/4
115 x+3/4,-z,y+3/4
116 x+3/4,z+1/4,-y+1/2
117 -z+1/2,-y,-x+1/2
118 -z+1/2,y+1/4,x+3/4
119 z+3/4,-y,x+3/4
120 z+3/4,y+1/4,-x+1/2
121 -x+1/2,-y,-z+1/2
122 -x+1/2,y+1/4,z+3/4
123 x+3/4,-y,z+3/4
124 x+3/4,y+1/4,-z+1/2
125 -y+1/2,-z,-x+1/2
126 -y+1/2,z+1/4,x+3/4
127 y+3/4,-z,x+3/4
128 y+3/4,z+1/4,-x+1/2
129 -z+1/2,-x,-y+1/2
130 -z+1/2,x+1/4,y+3/4
131 z+3/4,-x,y+3/4
132 z+3/4,x+1/4,-y+1/2
133 y+1/2,x,z+1/2
134 y+1/2,-x+1/4,-z+3/4
135 -y+3/4,x,-z+3/4
136 -y+3/4,-x+1/4,z+1/2
137 x+1/2,z,y+1/2
138 x+1/2,-z+1/4,-y+3/4
139 -x+3/4,z,-y+3/4
140 -x+3/4,-z+1/4,y+1/2
141 z+1/2,y,x+1/2
142 z+1/2,-y+1/4,-x+3/4
143 -z+3/4,y,-x+3/4
144 -z+3/4,-y+1/4,x+1/2
145 x+1/2,y+1/2,z
146 x+1/2,-y+3/4,-z+1/4
147 -x+3/4,y+1/2,-z+1/4
148 -x+3/4,-y+3/4,z
149 y+1/2,z+1/2,x
150 y+1/2,-z+3/4,-x+1/4
151 -y+3/4,z+1/2,-x+1/4
152 -y+3/4,-z+3/4,x
153 z+1/2,x+1/2,y
154 z+1/2,-x+3/4,-y+1/4
155 -z+3/4,x+1/2,-y+1/4
156 -z+3/4,-x+3/4,y
157 -y+1/2,-x+1/2,-z
158 -y+1/2,x+3/4,z+1/4
159 y+3/4,-x+1/2,z+1/4
160 y+3/4,x+3/4,-z
161 -x+1/2,-z+1/2,-y
162 -x+1/2,z+3/4,y+1/4
163 x+3/4,-z+1/2,y+1/4
164 x+3/4,z+3/4,-y
165 -z+1/2,-y+1/2,-x
166 -z+1/2,y+3/4,x+1/4
167 z+3/4,-y+1/2,x+1/4
168 z+3/4,y+3/4,-x
169 -x+1/2,-y+1/2,-z
170 -x+1/2,y+3/4,z+1/4
171 x+3/4,-y+1/2,z+1/4
172 x+3/4,y+3/4,-z
173 -y+1/2,-z+1/2,-x
174 -y+1/2,z+3/4,x+1/4
175 y+3/4,-z+1/2,x+1/4
176 y+3/4,z+3/4,-x
177 -z+1/2,-x+1/2,-y
178 -z+1/2,x+3/4,y+1/4
179 z+3/4,-x+1/2,y+1/4
180 z+3/4,x+3/4,-y
181 y+1/2,x+1/2,z
182 y+1/2,-x+3/4,-z+1/4
183 -y+3/4,x+1/2,-z+1/4
184 -y+3/4,-x+3/4,z
185 x+1/2,z+1/2,y
186 x+1/2,-z+3/4,-y+1/4
187 -x+3/4,z+1/2,-y+1/4
188 -x+3/4,-z+3/4,y
189 z+1/2,y+1/2,x
190 z+1/2,-y+3/4,-x+1/4
191 -z+3/4,y+1/2,-x+1/4
192 -z+3/4,-y+3/4,x

loop_
_atom_site_label
_atom_site_type_symbol
_atom_site_symmetry_multiplicity
_atom_site_Wyckoff_label
_atom_site_fract_x
_atom_site_fract_y
_atom_site_fract_z
_atom_site_occupancy
Sb1 Sb 32 e 0.26027 0.26027 0.26027 1.00000
O1 O 48 f 0.43875 0.12500 0.12500 1.00000
```

```
A3B2_cF80_227_f_e & a,x1,x2 --params=11.1519,0.26027,0.43875 & Fd-3m O_{
    ↪ h}^{7} #227 (ef) & cF80 & None & O3Sb2 & Senarmontite & C.
    ↪ Svensson, Acta Crystallogr. Sect. B Struct. Sci. 31, 2016-2018
    ↪ (1975)
   1.00000000000000
   0.00000000000000   5.57595000000000   5.57595000000000
   5.57595000000000   0.00000000000000   5.57595000000000
   5.57595000000000   5.57595000000000   0.00000000000000
     O    Sb
    12     8
Direct
  -0.18875000000000   0.43875000000000   0.43875000000000   O   (48f)
   0.43875000000000  -0.18875000000000  -0.18875000000000   O   (48f)
   0.43875000000000  -0.18875000000000   0.43875000000000   O   (48f)
  -0.18875000000000   0.43875000000000  -0.18875000000000   O   (48f)
   0.43875000000000   0.43875000000000  -0.18875000000000   O   (48f)
  -0.18875000000000  -0.18875000000000   0.43875000000000   O   (48f)
   1.18875000000000  -0.43875000000000   1.18875000000000   O   (48f)
  -0.43875000000000   1.18875000000000  -0.43875000000000   O   (48f)
  -0.43875000000000   1.18875000000000   1.18875000000000   O   (48f)
   1.18875000000000  -0.43875000000000  -0.43875000000000   O   (48f)
  -0.43875000000000  -0.43875000000000   1.18875000000000   O   (48f)
   1.18875000000000   1.18875000000000  -0.43875000000000   O   (48f)
   0.26027000000000   0.26027000000000   0.26027000000000  Sb   (32e)
   0.26027000000000   0.26027000000000  -0.28081000000000  Sb   (32e)
   0.26027000000000  -0.28081000000000   0.26027000000000  Sb   (32e)
  -0.28081000000000   0.26027000000000   0.26027000000000  Sb   (32e)
  -0.26027000000000  -0.26027000000000   1.28081000000000  Sb   (32e)
  -0.26027000000000  -0.26027000000000  -0.26027000000000  Sb   (32e)
  -0.26027000000000   1.28081000000000  -0.26027000000000  Sb   (32e)
   1.28081000000000  -0.26027000000000  -0.26027000000000  Sb   (32e)
```

*H5*$_6$ [Tychite, $Na_6Mg_2SO_4(CO_3)_4$)] (*obsolete*): A4B2C6D16E_cF232_227_e_d_f_eg_a - CIF

```
# CIF file
data_findsym-output
_audit_creation_method FINDSYM

_chemical_name_mineral 'Tychite'
_chemical_formula_sum 'C4 Mg2 Na6 O16 S'

loop_
_publ_author_name
 'H. Shiba'
 'T. Watanab\\'{e}'
_journal_name_full_name
;
 Comptes Rendus de l'Acad{\'e}mie des Sciences
;
_journal_volume 193
_journal_year 1931
_journal_page_first 1421
_journal_page_last 1423
_publ_Section_title
;
 Les structures des cristaux de northupite, de northupite brom\\'{e}e et
    ↪ de tychite
;

# Found in Tychite, Na$_{6}$Mg$_{2}$(SO$_{4}$)(CO$_{3}$)$_{4}$:
    ↪ Structure analysis and Raman spectroscopic data, 2006

_aflow_title '$H5_{6}$ [Tychite, Na$_{6}$Mg$_{2}$SO$_{4}$(CO$_{3}$)$_{4}
    ↪ $)] ({\em{obsolete}}) Structure'
_aflow_proto 'A4B2C6D16E_cF232_227_e_d_f_eg_a'
_aflow_params 'a,x_{3},x_{4},x_{5},x_{6},z_{6}'
_aflow_params_values '13.89999,0.735,0.06,0.35,0.765,0.665'
_aflow_Strukturbericht '$H5_{7}$'
_aflow_Pearson 'cF232'

_symmetry_space_group_name_H-M "F 41/d -3 2/m (origin choice 2)"
_symmetry_Int_Tables_number 227

_cell_length_a    13.89999
_cell_length_b    13.89999
_cell_length_c    13.89999
_cell_angle_alpha 90.00000
_cell_angle_beta  90.00000
_cell_angle_gamma 90.00000

loop_
_space_group_symop_id
_space_group_symop_operation_xyz
1 x,y,z
2 x,-y+1/4,-z+1/4
3 -x+1/4,y,-z+1/4
4 -x+1/4,-y+1/4,z
5 y,z,x
6 y,-z+1/4,-x+1/4
7 -y+1/4,z,-x+1/4
8 -y+1/4,-z+1/4,x
9 z,x,y
10 z,-x+1/4,-y+1/4
11 -z+1/4,x,-y+1/4
12 -z+1/4,-x+1/4,y
13 -y,-x,-z
14 -y,x+1/4,z+1/4
15 y+1/4,-x,z+1/4
16 y+1/4,x+1/4,-z
17 -x,-z,-y
18 -x,z+1/4,y+1/4
19 x+1/4,-z,y+1/4
20 x+1/4,z+1/4,-y
21 -z,-y,-x
22 -z,y+1/4,x+1/4
23 z+1/4,-y,x+1/4
24 z+1/4,y+1/4,-x
25 -x,-y,-z
26 -x,y+1/4,z+1/4
27 x+1/4,-y,z+1/4
28 x+1/4,y+1/4,-z
29 -y,-z,-x
30 -y,z+1/4,x+1/4
31 y+1/4,-z,x+1/4
32 y+1/4,z+1/4,-x
33 -z,-x,-y
34 -z,x+1/4,y+1/4
35 z+1/4,-x,y+1/4
36 z+1/4,x+1/4,-y
37 y,x,z
38 y,-x+1/4,-z+1/4
39 -y+1/4,x,-z+1/4
40 -y+1/4,-x+1/4,z
41 x,z,y
42 x,-z+1/4,-y+1/4
43 -x+1/4,z,-y+1/4
44 -x+1/4,-z+1/4,y
45 z,y,x
46 z,-y+1/4,-x+1/4
47 -z+1/4,y,-x+1/4
48 -z+1/4,-y+1/4,x
49 x,y+1/2,z+1/2
50 x,-y+3/4,-z+3/4
51 -x+1/4,y+1/2,-z+3/4
52 -x+1/4,-y+3/4,z+1/2
53 y,z+1/2,x+1/2
54 y,-z+3/4,-x+3/4
55 -y+1/4,z+1/2,-x+3/4
56 -y+1/4,-z+3/4,x+1/2
57 z,x+1/2,y+1/2
58 z,-x+3/4,-y+3/4
59 -z+1/4,x+1/2,-y+3/4
60 -z+1/4,-x+3/4,y+1/2
61 -y,-x+1/2,-z+1/2
62 -y,x+3/4,z+3/4
63 y+1/4,-x+1/2,z+3/4
64 y+1/4,x+3/4,-z+1/2
65 -x,-z+1/2,-y+1/2
66 -x,z+3/4,y+3/4
67 x+1/4,-z+1/2,y+3/4
68 x+1/4,z+3/4,-y+1/2
69 -z,-y+1/2,-x+1/2
70 -z,y+3/4,x+3/4
71 z+1/4,-y+1/2,x+3/4
72 z+1/4,y+3/4,-x+1/2
73 -x,-y+1/2,-z+1/2
74 -x,y+3/4,z+3/4
75 x+1/4,-y+1/2,z+3/4
76 x+1/4,y+3/4,-z+1/2
77 -y,-z+1/2,-x+1/2
78 -y,z+3/4,x+3/4
79 y+1/4,-z+1/2,x+3/4
80 y+1/4,z+3/4,-x+1/2
81 -z,-x+1/2,-y+1/2
82 -z,x+3/4,y+3/4
83 z+1/4,-x+1/2,y+3/4
84 z+1/4,x+3/4,-y+1/2
85 y,x+1/2,z+1/2
86 y,-x+3/4,-z+3/4
87 -y+1/4,x+1/2,-z+3/4
88 -y+1/4,-x+3/4,z+1/2
89 x,z+1/2,y+1/2
90 x,-z+3/4,-y+3/4
91 -x+1/4,z+1/2,-y+3/4
92 -x+1/4,-z+3/4,y+1/2
93 z,y+1/2,x+1/2
94 z,-y+3/4,-x+3/4
95 -z+1/4,y+1/2,-x+3/4
96 -z+1/4,-y+3/4,x+1/2
97 x+1/2,y,z+1/2
98 x+1/2,-y+1/4,-z+3/4
99 -x+3/4,y,-z+3/4
100 -x+3/4,-y+1/4,z+1/2
101 y+1/2,z,x+1/2
102 y+1/2,-z+1/4,-x+3/4
103 -y+3/4,z,-x+3/4
104 -y+3/4,-z+1/4,x+1/2
105 z+1/2,x,y+1/2
106 z+1/2,-x+1/4,-y+3/4
107 -z+3/4,x,-y+3/4
108 -z+3/4,-x+1/4,y+1/2
109 -y+1/2,-x,-z+1/2
110 -y+1/2,x+1/4,z+3/4
111 y+3/4,-x,z+3/4
112 y+3/4,x+1/4,-z+1/2
113 -x+1/2,-z,-y+1/2
114 -x+1/2,z+1/4,y+3/4
115 x+3/4,-z,y+3/4
116 x+3/4,z+1/4,-y+1/2
117 -z+1/2,-y,-x+1/2
118 -z+1/2,y+1/4,x+3/4
119 z+3/4,-y,x+3/4
120 z+3/4,y+1/4,-x+1/2
121 -x+1/2,-y,-z+1/2
122 -x+1/2,y+1/4,z+3/4
123 x+3/4,-y,z+3/4
124 x+3/4,y+1/4,-z+1/2
125 -y+1/2,-z,-x+1/2
126 -y+1/2,z+1/4,x+3/4
127 y+3/4,-z,x+3/4
```

```
128 y+3/4,z+1/4,-x+1/2
129 -z+1/2,-x,-y+1/2
130 -z+1/2,x+1/4,y+3/4
131 z+3/4,-x,y+3/4
132 z+3/4,x+1/4,-y+1/2
133 y+1/2,x,z+1/2
134 y+1/2,-x+1/4,-z+3/4
135 -y+3/4,x,-z+3/4
136 -y+3/4,-x+1/4,z+1/2
137 x+1/2,z,y+1/2
138 x+1/2,-z+1/4,-y+3/4
139 -x+3/4,z,-y+3/4
140 -x+3/4,-z+1/4,y+1/2
141 z+1/2,y,x+1/2
142 z+1/2,-y+1/4,-x+3/4
143 -z+3/4,y,-x+3/4
144 -z+3/4,-y+1/4,x+1/2
145 x+1/2,y+1/2,z
146 x+1/2,-y+3/4,-z+1/4
147 -x+3/4,y+1/2,-z+1/4
148 -x+3/4,-y+3/4,z
149 y+1/2,z+1/2,x
150 y+1/2,-z+3/4,-x+1/4
151 -y+3/4,z+1/2,-x+1/4
152 -y+3/4,-z+3/4,x
153 z+1/2,x+1/2,y
154 z+1/2,-x+3/4,-y+1/4
155 -z+3/4,x+1/2,-y+1/4
156 -z+3/4,-x+3/4,y
157 -y+1/2,-x+1/2,-z
158 -y+1/2,x+3/4,z+1/4
159 y+3/4,-x+1/2,z+1/4
160 y+3/4,x+3/4,-z
161 -x+1/2,-z+1/2,-y
162 -x+1/2,z+3/4,y+1/4
163 x+3/4,-z+1/2,y+1/4
164 x+3/4,z+3/4,-y
165 -z+1/2,-y+1/2,-x
166 -z+1/2,y+3/4,x+1/4
167 z+3/4,-y+1/2,x+1/4
168 z+3/4,y+3/4,-x
169 -x+1/2,-y+1/2,-z
170 -x+1/2,y+3/4,z+1/4
171 x+3/4,-y+1/2,z+1/4
172 x+3/4,y+3/4,-z
173 -y+1/2,-z+1/2,-x
174 -y+1/2,z+3/4,x+1/4
175 y+3/4,-z+1/2,x+1/4
176 y+3/4,z+3/4,-x
177 -z+1/2,-x+1/2,-y
178 -z+1/2,x+3/4,y+1/4
179 z+3/4,-x+1/2,y+1/4
180 z+3/4,x+3/4,-y
181 y+1/2,x+1/2,z
182 y+1/2,-x+3/4,-z+1/4
183 -y+3/4,x+1/2,-z+1/4
184 -y+3/4,-x+3/4,z
185 x+1/2,z+1/2,y
186 x+1/2,-z+3/4,-y+1/4
187 -x+3/4,z+1/2,-y+1/4
188 -x+3/4,-z+3/4,y
189 z+1/2,y+1/2,x
190 z+1/2,-y+3/4,-x+1/4
191 -z+3/4,y+1/2,-x+1/4
192 -z+3/4,-y+3/4,x

loop_
_atom_site_label
_atom_site_type_symbol
_atom_site_symmetry_multiplicity
_atom_site_Wyckoff_label
_atom_site_fract_x
_atom_site_fract_y
_atom_site_fract_z
_atom_site_occupancy
S1 S 8 a 0.12500 0.12500 0.12500 1.00000
Mg1 Mg 16 d 0.50000 0.50000 0.50000 1.00000
C1 C 32 e 0.73500 0.73500 0.73500 1.00000
O1 O 32 e 0.06000 0.06000 0.06000 1.00000
Na1 Na 48 f 0.35000 0.12500 0.12500 1.00000
O2 O 96 g 0.76500 0.76500 0.66500 1.00000
```

*H*5$_6$ [Tychite, $Na_6Mg_2SO_4(CO_3)_4$)] (*obsolete*): A4B2C6D16E_cF232_227_e_d_f_eg_a - POSCAR

```
A4B2C6D16E_cF232_227_e_d_f_eg_a & a,x3,x4,x5,x6,z6 --params=13.89999,
    ↪ 0.735,0.06,0.35,0.765,0.665 & Fd-3m O_{h}^{7} #227 (ade^2fg) &
    ↪ cF232 & $H5_{7}$ & C4Mg2Na6O16S & Tychite & H. Shiba and T.
    ↪ Watanab\'{e}, {C. R. Acad. Sci. C 193, 1421-1423 (1931)
   1.00000000000000
   0.00000000000000   6.94999500000000   6.94999500000000
   6.94999500000000   0.00000000000000   6.94999500000000
   6.94999500000000   6.94999500000000   0.00000000000000
     C      Mg      Na       O       S
     8       4      12      32       2
Direct
   0.73500000000000   0.73500000000000   0.73500000000000   C   (32e)
   0.73500000000000   0.73500000000000  -1.70500000000000   C   (32e)
   0.73500000000000  -1.70500000000000   0.73500000000000   C   (32e)
  -1.70500000000000   0.73500000000000   0.73500000000000   C   (32e)
  -0.73500000000000  -0.73500000000000   2.70500000000000   C   (32e)
  -0.73500000000000  -0.73500000000000  -0.73500000000000   C   (32e)
  -0.73500000000000   2.70500000000000  -0.73500000000000   C   (32e)
   2.70500000000000  -0.73500000000000  -0.73500000000000   C   (32e)
   0.50000000000000   0.50000000000000   0.50000000000000  Mg   (16d)
   0.50000000000000   0.50000000000000   0.00000000000000  Mg   (16d)
   0.50000000000000   0.00000000000000   0.50000000000000  Mg   (16d)
   0.00000000000000   0.50000000000000   0.50000000000000  Mg   (16d)
  -0.10000000000000   0.35000000000000   0.35000000000000  Na   (48f)
   0.35000000000000  -0.10000000000000  -0.10000000000000  Na   (48f)
   0.35000000000000  -0.10000000000000   0.35000000000000  Na   (48f)
  -0.10000000000000   0.35000000000000  -0.10000000000000  Na   (48f)
   0.35000000000000   0.35000000000000  -0.10000000000000  Na   (48f)
  -0.10000000000000  -0.10000000000000   0.35000000000000  Na   (48f)
   1.10000000000000  -0.35000000000000   1.10000000000000  Na   (48f)
  -0.35000000000000   1.10000000000000  -0.35000000000000  Na   (48f)
  -0.35000000000000   1.10000000000000   1.10000000000000  Na   (48f)
   1.10000000000000  -0.35000000000000  -0.35000000000000  Na   (48f)
  -0.35000000000000  -0.35000000000000   1.10000000000000  Na   (48f)
   1.10000000000000   1.10000000000000  -0.35000000000000  Na   (48f)
   0.06000000000000   0.06000000000000   0.06000000000000   O   (32e)
   0.06000000000000   0.06000000000000   0.32000000000000   O   (32e)
   0.06000000000000   0.32000000000000   0.06000000000000   O   (32e)
   0.32000000000000   0.06000000000000   0.06000000000000   O   (32e)
  -0.06000000000000  -0.06000000000000   0.68000000000000   O   (32e)
  -0.06000000000000  -0.06000000000000  -0.06000000000000   O   (32e)
  -0.06000000000000   0.68000000000000  -0.06000000000000   O   (32e)
   0.68000000000000  -0.06000000000000  -0.06000000000000   O   (32e)
   0.66500000000000   0.66500000000000   0.86500000000000   O   (96g)
   0.66500000000000   0.66500000000000  -1.69500000000000   O   (96g)
   0.86500000000000  -1.69500000000000   0.66500000000000   O   (96g)
  -1.69500000000000   0.86500000000000   0.66500000000000   O   (96g)
   0.86500000000000   0.66500000000000   0.66500000000000   O   (96g)
  -1.69500000000000   0.66500000000000   0.66500000000000   O   (96g)
   0.66500000000000   0.86500000000000  -1.69500000000000   O   (96g)
   0.66500000000000  -1.69500000000000   0.86500000000000   O   (96g)
   0.66500000000000   0.86500000000000   0.66500000000000   O   (96g)
   0.66500000000000  -1.69500000000000   0.66500000000000   O   (96g)
  -1.69500000000000   0.66500000000000   0.86500000000000   O   (96g)
   0.86500000000000   0.66500000000000  -1.69500000000000   O   (96g)
  -0.66500000000000  -0.66500000000000   2.69500000000000   O   (96g)
  -0.66500000000000  -0.66500000000000  -0.86500000000000   O   (96g)
  -0.86500000000000   2.69500000000000  -0.66500000000000   O   (96g)
   2.69500000000000  -0.86500000000000  -0.66500000000000   O   (96g)
  -0.86500000000000  -0.66500000000000   2.69500000000000   O   (96g)
   2.69500000000000  -0.66500000000000  -0.86500000000000   O   (96g)
  -0.66500000000000  -0.86500000000000  -0.66500000000000   O   (96g)
  -0.66500000000000   2.69500000000000  -0.66500000000000   O   (96g)
  -0.66500000000000  -0.86500000000000   2.69500000000000   O   (96g)
  -0.66500000000000   2.69500000000000  -0.86500000000000   O   (96g)
   2.69500000000000  -0.66500000000000  -0.66500000000000   O   (96g)
  -0.86500000000000  -0.66500000000000  -0.66500000000000   O   (96g)
   0.12500000000000   0.12500000000000   0.12500000000000   S    (8a)
   0.87500000000000   0.87500000000000   0.87500000000000   S    (8a)
```

Cubic CuPt (*L*1$_3$ (I), *D*4): AB_cF32_227_c_d - CIF

```
# CIF file
data_findsym-output
_audit_creation_method FINDSYM

_chemical_name_mineral 'CuPt'
_chemical_formula_sum 'Cu Pt'

loop_
_publ_author_name
 'C. H. Johansson'
 'J. O. Linde'
_journal_name_full_name
;
 Annalen der Physik
;
_journal_volume 387
_journal_year 1927
_journal_page_first 449
_journal_page_last 478
_publ_Section_title
;
 Gitterstruktur und elektrisches Leitverm\"{o}gen der
    ↪ Mischkristallreihen Au-Cu, Pd-Cu und Pt-Cu
;

# Found in Strukturbericht 1913-1928, 1931

_aflow_title 'Cubic CuPt ($L1_{3}$ (I), $D4$) Structure'
_aflow_proto 'AB_cF32_227_c_d'
_aflow_params 'a'
_aflow_params_values '7.23'
_aflow_Strukturbericht '$L1_{3}$ $(I)$'
_aflow_Pearson 'cF32'

_symmetry_space_group_name_H-M "F 41/d -3 2/m (origin choice 2)"
_symmetry_Int_Tables_number 227

_cell_length_a      7.230000
_cell_length_b      7.230000
_cell_length_c      7.230000
_cell_angle_alpha  90.00000
_cell_angle_beta   90.00000
_cell_angle_gamma  90.00000

loop_
_space_group_symop_id
_space_group_symop_operation_xyz
1 x,y,z
2 x,-y+1/4,-z+1/4
3 -x+1/4,y,-z+1/4
4 -x+1/4,-y+1/4,z
5 y,z,x
6 y,-z+1/4,-x+1/4
7 -y+1/4,z,-x+1/4
```

```
8 -y+1/4,-z+1/4,x
9 z,x,y
10 z,-x+1/4,-y+1/4
11 -z+1/4,x,-y+1/4
12 -z+1/4,-x+1/4,y
13 -y,-x,-z
14 -y,x+1/4,z+1/4
15 y+1/4,-x,z+1/4
16 y+1/4,x+1/4,-z
17 -x,-z,-y
18 -x,z+1/4,y+1/4
19 x+1/4,-z,y+1/4
20 x+1/4,z+1/4,-y
21 -z,-y,-x
22 -z,y+1/4,x+1/4
23 z+1/4,-y,x+1/4
24 z+1/4,y+1/4,-x
25 -x,-y,-z
26 -x,y+1/4,z+1/4
27 x+1/4,-y,z+1/4
28 x+1/4,y+1/4,-z
29 -y,-z,-x
30 -y,z+1/4,x+1/4
31 y+1/4,-z,x+1/4
32 y+1/4,z+1/4,-x
33 -z,-x,-y
34 -z,x+1/4,y+1/4
35 z+1/4,-x,y+1/4
36 z+1/4,x+1/4,-y
37 y,x,z
38 y,-x+1/4,-z+1/4
39 -y+1/4,x,-z+1/4
40 -y+1/4,-x+1/4,z
41 x,z,y
42 x,-z+1/4,-y+1/4
43 -x+1/4,z,-y+1/4
44 -x+1/4,-z+1/4,y
45 z,y,x
46 z,-y+1/4,-x+1/4
47 -z+1/4,y,-x+1/4
48 -z+1/4,-y+1/4,x
49 x,y+1/2,z+1/2
50 x,-y+3/4,-z+3/4
51 -x+1/4,y+1/2,-z+3/4
52 -x+1/4,-y+3/4,z+1/2
53 y,z+1/2,x+1/2
54 y,-z+3/4,-x+3/4
55 -y+1/4,z+1/2,-x+3/4
56 -y+1/4,-z+3/4,x+1/2
57 z,x+1/2,y+1/2
58 z,-x+3/4,-y+3/4
59 -z+1/4,x+1/2,-y+3/4
60 -z+1/4,-x+3/4,y+1/2
61 -y,-x+1/2,-z+1/2
62 -y,x+3/4,z+3/4
63 y+1/4,-x+1/2,z+3/4
64 y+1/4,x+3/4,-z+1/2
65 -x,-z+1/2,-y+1/2
66 -x,z+3/4,y+3/4
67 x+1/4,-z+1/2,y+3/4
68 x+1/4,z+3/4,-y+1/2
69 -z,-y+1/2,-x+1/2
70 -z,y+3/4,x+3/4
71 z+1/4,-y+1/2,x+3/4
72 z+1/4,y+3/4,-x+1/2
73 -x,-y+1/2,-z+1/2
74 -x,y+3/4,z+3/4
75 x+1/4,-y+1/2,z+3/4
76 x+1/4,y+3/4,-z+1/2
77 -y,-z+1/2,-x+1/2
78 -y,z+3/4,x+3/4
79 y+1/4,-z+1/2,x+3/4
80 y+1/4,z+3/4,-x+1/2
81 -z,-x+1/2,-y+1/2
82 -z,x+3/4,y+3/4
83 z+1/4,-x+1/2,y+3/4
84 z+1/4,x+3/4,-y+1/2
85 y,x+1/2,z+1/2
86 y,-x+3/4,-z+3/4
87 -y+1/4,x+1/2,-z+3/4
88 -y+1/4,-x+3/4,z+1/2
89 x,z+1/2,y+1/2
90 x,-z+3/4,-y+3/4
91 -x+1/4,z+1/2,-y+3/4
92 -x+1/4,-z+3/4,y+1/2
93 z,y+1/2,x+1/2
94 z,-y+3/4,-x+3/4
95 -z+1/4,y+1/2,-x+3/4
96 -z+1/4,-y+3/4,x+1/2
97 x+1/2,y,z+1/2
98 x+1/2,-y+1/4,-z+3/4
99 -x+3/4,y,-z+3/4
100 -x+3/4,-y+1/4,z+1/2
101 y+1/2,z,x+1/2
102 y+1/2,-z+1/4,-x+3/4
103 -y+3/4,z,-x+3/4
104 -y+3/4,-z+1/4,x+1/2
105 z+1/2,x,y+1/2
106 z+1/2,-x+1/4,-y+3/4
107 -z+3/4,x,-y+3/4
108 -z+3/4,-x+1/4,y+1/2
109 -y+1/2,-x,-z+1/2
110 -y+1/2,x+1/4,z+3/4
111 y+3/4,-x,z+3/4
112 y+3/4,x+1/4,-z+1/2
113 -x+1/2,-z,-y+1/2
114 -x+1/2,z+1/4,y+3/4
115 x+3/4,-z,y+3/4
116 x+3/4,z+1/4,-y+1/2
117 -z+1/2,-y,-x+1/2
118 -z+1/2,y+1/4,x+3/4
119 z+3/4,-y,x+3/4
120 z+3/4,y+1/4,-x+1/2
121 -x+1/2,-y,-z+1/2
122 -x+1/2,y+1/4,z+3/4
123 x+3/4,-y,z+3/4
124 x+3/4,y+1/4,-z+1/2
125 -y+1/2,-z,-x+1/2
126 -y+1/2,z+1/4,x+3/4
127 y+3/4,-z,x+3/4
128 y+3/4,z+1/4,-x+1/2
129 -z+1/2,-x,-y+1/2
130 -z+1/2,x+1/4,y+3/4
131 z+3/4,-x,y+3/4
132 z+3/4,x+1/4,-y+1/2
133 y+1/2,x,z+1/2
134 y+1/2,-x+1/4,-z+3/4
135 -y+3/4,x,-z+3/4
136 -y+3/4,-x+1/4,z+1/2
137 x+1/2,z,y+1/2
138 x+1/2,-z+1/4,-y+3/4
139 -x+3/4,z,-y+3/4
140 -x+3/4,-z+1/4,y+1/2
141 z+1/2,y,x+1/2
142 z+1/2,-y+1/4,-x+3/4
143 -z+3/4,y,-x+3/4
144 -z+3/4,-y+1/4,x+1/2
145 x+1/2,y+1/2,z
146 x+1/2,-y+3/4,-z+1/4
147 -x+3/4,y+1/2,-z+1/4
148 -x+3/4,-y+3/4,z
149 y+1/2,z+1/2,x
150 y+1/2,-z+3/4,-x+1/4
151 -y+3/4,z+1/2,-x+1/4
152 -y+3/4,-z+3/4,x
153 z+1/2,x+1/2,y
154 z+1/2,-x+3/4,-y+1/4
155 -z+3/4,x+1/2,-y+1/4
156 -z+3/4,-x+3/4,y
157 -y+1/2,-x+1/2,-z
158 -y+1/2,x+3/4,z+1/4
159 y+3/4,-x+1/2,z+1/4
160 y+3/4,x+3/4,-z
161 -x+1/2,-z+1/2,-y
162 -x+1/2,z+3/4,y+1/4
163 x+3/4,-z+1/2,y+1/4
164 x+3/4,z+3/4,-y
165 -z+1/2,-y+1/2,-x
166 -z+1/2,y+3/4,x+1/4
167 z+3/4,-y+1/2,x+1/4
168 z+3/4,y+3/4,-x
169 -x+1/2,-y+1/2,-z
170 -x+1/2,y+3/4,z+1/4
171 x+3/4,-y+1/2,z+1/4
172 x+3/4,y+3/4,-z
173 -y+1/2,-z+1/2,-x
174 -y+1/2,z+3/4,x+1/4
175 y+3/4,-z+1/2,x+1/4
176 y+3/4,z+3/4,-x
177 -z+1/2,-x+1/2,-y
178 -z+1/2,x+3/4,y+1/4
179 z+3/4,-x+1/2,y+1/4
180 z+3/4,x+3/4,-y
181 y+1/2,x+1/2,z
182 y+1/2,-x+3/4,-z+1/4
183 -y+3/4,x+1/2,-z+1/4
184 -y+3/4,-x+3/4,z
185 x+1/2,z+1/2,y
186 x+1/2,-z+3/4,-y+1/4
187 -x+3/4,z+1/2,-y+1/4
188 -x+3/4,-z+3/4,y
189 z+1/2,y+1/2,x
190 z+1/2,-y+3/4,-x+1/4
191 -z+3/4,y+1/2,-x+1/4
192 -z+3/4,-y+3/4,x

loop_
_atom_site_label
_atom_site_type_symbol
_atom_site_symmetry_multiplicity
_atom_site_Wyckoff_label
_atom_site_fract_x
_atom_site_fract_y
_atom_site_fract_z
_atom_site_occupancy
Cu1 Cu 16 c 0.00000 0.00000 0.00000 1.00000
Pt1 Pt 16 d 0.50000 0.50000 0.50000 1.00000
```

Cubic CuPt ($L1_3$ (I), $D4$): AB_cF32_227_c_d - POSCAR

```
AB_cF32_227_c_d & a --params=7.23 & Fd-3m O_{h}^{7} #227 (cd) & cF32 &
    ↪ $L1_{3}$ $(I)$ & CuPt & CuPt & C. H. Johansson and J. O. Linde,
    ↪ Ann. Phys. 387, 449-478 (1927)
   1.00000000000000
   0.00000000000000   3.61500000000000   3.61500000000000
   3.61500000000000   0.00000000000000   3.61500000000000
   3.61500000000000   3.61500000000000   0.00000000000000
    Cu    Pt
     4     4
Direct
```

```
    0.00000000000000    0.00000000000000    0.00000000000000    Cu   (16c)
    0.00000000000000    0.00000000000000    0.50000000000000    Cu   (16c)
    0.00000000000000    0.50000000000000    0.00000000000000    Cu   (16c)
    0.50000000000000    0.00000000000000    0.00000000000000    Cu   (16c)
    0.50000000000000    0.50000000000000    0.50000000000000    Pt   (16d)
    0.50000000000000    0.50000000000000    0.00000000000000    Pt   (16d)
    0.50000000000000    0.00000000000000    0.50000000000000    Pt   (16d)
    0.00000000000000    0.50000000000000    0.50000000000000    Pt   (16d)
```

$\alpha$-AgI ($B23$): A21B_cI44_229_bdh_a - CIF

```
# CIF file
data_findsym-output
_audit_creation_method FINDSYM

_chemical_name_mineral 'AgI'
_chemical_formula_sum 'Ag21 I'

loop_
_publ_author_name
 'L. W. Strock'
_journal_name_full_name
;
 Zeitschrift f{\"u}r Physikalische Chemie B
;
_journal_volume 25
_journal_year 1934
_journal_page_first 441
_journal_page_last 459
_publ_Section_title
;
 Kristallstruktur des Hochtemperatur-Jodsilbers $\alpha$-AgJ
;

# Found in Crystal Structure and Phase Transition of Some Metallic
    ↪ Halides IV On the Anomalous Structure of $\alpha$-AgI, 1957

_aflow_title '$\alpha$-AgI ($B23$) Structure'
_aflow_proto 'A21B_cI44_229_bdh_a'
_aflow_params 'a,y_{4}'
_aflow_params_values '5.034,0.375'
_aflow_Strukturbericht '$B23$'
_aflow_Pearson 'cI44'

_symmetry_space_group_name_H-M "I 4/m -3 2/m"
_symmetry_Int_Tables_number 229

_cell_length_a      5.03400
_cell_length_b      5.03400
_cell_length_c      5.03400
_cell_angle_alpha  90.00000
_cell_angle_beta   90.00000
_cell_angle_gamma  90.00000

loop_
_space_group_symop_id
_space_group_symop_operation_xyz
1 x,y,z
2 x,-y,-z
3 -x,y,-z
4 -x,-y,z
5 y,z,x
6 y,-z,-x
7 -y,z,-x
8 -y,-z,x
9 z,x,y
10 z,-x,-y
11 -z,x,-y
12 -z,-x,y
13 -y,-x,-z
14 -y,x,z
15 y,-x,z
16 y,x,-z
17 -x,-z,-y
18 -x,z,y
19 x,-z,y
20 x,z,-y
21 -z,-y,-x
22 -z,y,x
23 z,-y,x
24 z,y,-x
25 -x,-y,-z
26 -x,y,z
27 x,-y,z
28 x,y,-z
29 -y,-z,-x
30 -y,z,x
31 y,-z,x
32 y,z,-x
33 -z,-x,-y
34 -z,x,y
35 z,-x,y
36 z,x,-y
37 y,x,z
38 y,-x,-z
39 -y,x,-z
40 -y,-x,z
41 x,z,y
42 x,-z,-y
43 -x,z,-y
44 -x,-z,y
45 z,y,x
46 z,-y,-x
47 -z,y,-x
48 -z,-y,x
49 x+1/2,y+1/2,z+1/2
50 x+1/2,-y+1/2,-z+1/2
51 -x+1/2,y+1/2,-z+1/2
52 -x+1/2,-y+1/2,z+1/2
53 y+1/2,z+1/2,x+1/2
54 y+1/2,-z+1/2,-x+1/2
55 -y+1/2,z+1/2,-x+1/2
56 -y+1/2,-z+1/2,x+1/2
57 z+1/2,x+1/2,y+1/2
58 z+1/2,-x+1/2,-y+1/2
59 -z+1/2,x+1/2,-y+1/2
60 -z+1/2,-x+1/2,y+1/2
61 -y+1/2,-x+1/2,-z+1/2
62 -y+1/2,x+1/2,z+1/2
63 y+1/2,-x+1/2,z+1/2
64 y+1/2,x+1/2,-z+1/2
65 -x+1/2,-z+1/2,-y+1/2
66 -x+1/2,z+1/2,y+1/2
67 x+1/2,-z+1/2,y+1/2
68 x+1/2,z+1/2,-y+1/2
69 -z+1/2,-y+1/2,-x+1/2
70 -z+1/2,y+1/2,x+1/2
71 z+1/2,-y+1/2,x+1/2
72 z+1/2,y+1/2,-x+1/2
73 -x+1/2,-y+1/2,-z+1/2
74 -x+1/2,y+1/2,z+1/2
75 x+1/2,-y+1/2,z+1/2
76 x+1/2,y+1/2,-z+1/2
77 -y+1/2,-z+1/2,-x+1/2
78 -y+1/2,z+1/2,x+1/2
79 y+1/2,-z+1/2,x+1/2
80 y+1/2,z+1/2,-x+1/2
81 -z+1/2,-x+1/2,-y+1/2
82 -z+1/2,x+1/2,y+1/2
83 z+1/2,-x+1/2,y+1/2
84 z+1/2,x+1/2,-y+1/2
85 y+1/2,x+1/2,z+1/2
86 y+1/2,-x+1/2,-z+1/2
87 -y+1/2,x+1/2,-z+1/2
88 -y+1/2,-x+1/2,z+1/2
89 x+1/2,z+1/2,y+1/2
90 x+1/2,-z+1/2,-y+1/2
91 -x+1/2,z+1/2,-y+1/2
92 -x+1/2,-z+1/2,y+1/2
93 z+1/2,y+1/2,x+1/2
94 z+1/2,-y+1/2,-x+1/2
95 -z+1/2,y+1/2,-x+1/2
96 -z+1/2,-y+1/2,x+1/2

loop_
_atom_site_label
_atom_site_type_symbol
_atom_site_symmetry_multiplicity
_atom_site_Wyckoff_label
_atom_site_fract_x
_atom_site_fract_y
_atom_site_fract_z
_atom_site_occupancy
I1 I 2 a 0.00000 0.00000 0.00000 1.00000
Ag1 Ag 6 b 0.00000 0.50000 0.50000 0.04762
Ag2 Ag 12 d 0.25000 0.00000 0.50000 0.04762
Ag3 Ag 24 h 0.00000 0.37500 0.37500 0.04762
```

$\alpha$-AgI ($B23$): A21B_cI44_229_bdh_a - POSCAR

```
A21B_cI44_229_bdh_a & a,y4 --params=5.034,0.375 & Im-3m O_{h}^{9} #229 (
    ↪ abdh) & cI44 & $B23$ & AgI & L. W. Strock, Z. Physik.
    ↪ Chem. B 25, 441-459 (1934)
   1.00000000000000
  -2.51700000000000    2.51700000000000    2.51700000000000
   2.51700000000000   -2.51700000000000    2.51700000000000
   2.51700000000000    2.51700000000000   -2.51700000000000
    Ag      I
    21      1
Direct
   0.00000000000000    0.50000000000000    0.50000000000000    Ag    (6b)
   0.50000000000000    0.00000000000000    0.50000000000000    Ag    (6b)
   0.50000000000000    0.50000000000000    0.00000000000000    Ag    (6b)
   0.50000000000000    0.75000000000000    0.25000000000000    Ag   (12d)
   0.50000000000000    0.25000000000000    0.75000000000000    Ag   (12d)
   0.25000000000000    0.50000000000000    0.75000000000000    Ag   (12d)
   0.75000000000000    0.50000000000000    0.25000000000000    Ag   (12d)
   0.75000000000000    0.25000000000000    0.50000000000000    Ag   (12d)
   0.25000000000000    0.75000000000000    0.50000000000000    Ag   (12d)
   0.75000000000000    0.37500000000000    0.37500000000000    Ag   (24h)
   0.00000000000000    0.37500000000000   -0.37500000000000    Ag   (24h)
   0.00000000000000   -0.37500000000000    0.37500000000000    Ag   (24h)
  -0.75000000000000   -0.37500000000000   -0.37500000000000    Ag   (24h)
   0.37500000000000    0.75000000000000    0.37500000000000    Ag   (24h)
  -0.37500000000000    0.00000000000000    0.37500000000000    Ag   (24h)
   0.37500000000000    0.00000000000000   -0.37500000000000    Ag   (24h)
  -0.37500000000000   -0.75000000000000   -0.37500000000000    Ag   (24h)
   0.37500000000000    0.37500000000000    0.75000000000000    Ag   (24h)
   0.37500000000000   -0.37500000000000    0.00000000000000    Ag   (24h)
  -0.37500000000000    0.37500000000000    0.00000000000000    Ag   (24h)
  -0.37500000000000   -0.37500000000000   -0.75000000000000    Ag   (24h)
   0.00000000000000    0.00000000000000    0.00000000000000     I    (2a)
```

$Ca_3Al_2(OH)_{12}$ ($J2_3$): A2B3C12D12_cI232_230_a_c_h_h - CIF

```
# CIF file
data_findsym-output
_audit_creation_method FINDSYM

_chemical_name_mineral 'Al2Ca3H12O12'
```

```
_chemical_formula_sum 'Al2 Ca3 H12 O12'

loop_
_publ_author_name
 'H. Bartl'
_journal_name_full_name
;
 Fresenius' Zeitschrift f{\"u}r analytische Chemie
;
_journal_volume 324
_journal_year 1986
_journal_page_first 124
_journal_page_last 126
_publ_Section_title
;
 Tricalciumaluminathexahydrat, Ca$_{3}$[Al(OH)$_{6}$]$_{2}$, Bindungsl
    ↪ \"{a}ngen und -valenzen aus R\"{o}ntgeneinkristallmessungen
;

_aflow_title 'Ca$_{3}$Al$_{2}$(OH)$_{12}$ ($J2_{3}$) Structure'
_aflow_proto 'A2B3C12D12_cI232_230_a_c_h_h'
_aflow_params 'a,x_{3},y_{3},z_{3},x_{4},y_{4},z_{4}'
_aflow_params_values '12.575,-0.091,0.054,0.163,-0.0282,0.0531,0.1396'
_aflow_Strukturbericht '$J2_{3}$'
_aflow_Pearson 'cI232'

_symmetry_space_group_name_H-M "I 41/a -3 2/d"
_symmetry_Int_Tables_number 230

_cell_length_a     12.57500
_cell_length_b     12.57500
_cell_length_c     12.57500
_cell_angle_alpha  90.00000
_cell_angle_beta   90.00000
_cell_angle_gamma  90.00000

loop_
_space_group_symop_id
_space_group_symop_operation_xyz
1 x,y,z
2 x,-y,-z+1/2
3 -x+1/2,y,-z
4 -x,-y+1/2,z
5 y,z,x
6 y,-z,-x+1/2
7 -y+1/2,z,-x
8 -y,-z+1/2,x
9 z,x,y
10 z,-x,-y+1/2
11 -z+1/2,x,-y
12 -z,-x+1/2,y
13 -y+1/4,-x+1/4,-z+1/4
14 -y+1/4,x+3/4,z+1/4
15 y+1/4,-x+1/4,z+3/4
16 y+3/4,x+1/4,-z+1/4
17 -x+1/4,-z+1/4,-y+1/4
18 -x+1/4,z+3/4,y+1/4
19 x+1/4,-z+1/4,y+3/4
20 x+3/4,z+1/4,-y+1/4
21 -z+1/4,-y+1/4,-x+1/4
22 -z+1/4,y+3/4,x+1/4
23 z+1/4,-y+1/4,x+3/4
24 z+3/4,y+1/4,-x+1/4
25 -x,-y,-z
26 -x,y,z+1/2
27 x+1/2,-y,z
28 x,y+1/2,-z
29 -y,-z,-x
30 -y,z,x+1/2
31 y+1/2,-z,x
32 y,z+1/2,-x
33 -z,-x,-y
34 -z,x,y+1/2
35 z+1/2,-x,y
36 z,x+1/2,-y
37 y+1/4,x+1/4,z+1/4
38 y+1/4,-x+3/4,-z+1/4
39 -y+1/4,x+1/4,-z+3/4
40 -y+3/4,-x+1/4,z+1/4
41 x+1/4,z+1/4,y+1/4
42 x+1/4,-z+3/4,-y+1/4
43 -x+1/4,z+1/4,-y+3/4
44 -x+3/4,-z+1/4,y+1/4
45 z+1/4,y+1/4,x+1/4
46 z+1/4,-y+3/4,-x+1/4
47 -z+1/4,y+1/4,-x+3/4
48 -z+3/4,-y+1/4,x+1/4
49 x+1/2,y+1/2,z+1/2
50 x+1/2,-y+1/2,-z
51 -x,y+1/2,-z+1/2
52 -x+1/2,-y,z+1/2
53 y+1/2,z+1/2,x+1/2
54 y+1/2,-z+1/2,-x
55 -y,z+1/2,-x+1/2
56 -y+1/2,-z,x+1/2
57 z+1/2,x+1/2,y+1/2
58 z+1/2,-x+1/2,-y
59 -z,x+1/2,-y+1/2
60 -z+1/2,-x,y+1/2
61 -y+3/4,-x+3/4,-z+3/4
62 -y+3/4,x+1/4,z+3/4
63 y+3/4,-x+3/4,z+1/4
64 y+1/4,x+3/4,-z+3/4
65 -x+3/4,-z+3/4,-y+3/4
66 -x+3/4,z+1/4,y+3/4
67 x+3/4,-z+3/4,y+1/4
68 x+1/4,z+3/4,-y+3/4
69 -z+3/4,-y+3/4,-x+3/4
70 -z+3/4,y+1/4,x+3/4
71 z+3/4,-y+3/4,x+1/4
72 z+1/4,y+3/4,-x+3/4
73 -x+1/2,-y+1/2,-z+1/2
74 -x+1/2,y+1/2,z
75 x,-y+1/2,z+1/2
76 x+1/2,y,-z+1/2
77 -y+1/2,-z+1/2,-x+1/2
78 -y+1/2,z+1/2,x
79 y,-z+1/2,x+1/2
80 y+1/2,z,-x+1/2
81 -z+1/2,-x+1/2,-y+1/2
82 -z+1/2,x+1/2,y
83 z,-x+1/2,y+1/2
84 z+1/2,x,-y+1/2
85 y+3/4,x+3/4,z+3/4
86 y+3/4,-x+1/4,-z+3/4
87 -y+3/4,x+3/4,-z+1/4
88 -y+1/4,-x+3/4,z+3/4
89 x+3/4,z+3/4,y+3/4
90 x+3/4,-z+1/4,-y+3/4
91 -x+3/4,z+3/4,-y+1/4
92 -x+1/4,-z+3/4,y+3/4
93 z+3/4,y+3/4,x+3/4
94 z+3/4,-y+1/4,-x+3/4
95 -z+3/4,y+3/4,-x+1/4
96 -z+1/4,-y+3/4,x+3/4

loop_
_atom_site_label
_atom_site_type_symbol
_atom_site_symmetry_multiplicity
_atom_site_Wyckoff_label
_atom_site_fract_x
_atom_site_fract_y
_atom_site_fract_z
_atom_site_occupancy
Al1 Al 16 a 0.00000 0.00000 0.00000 1.00000
Ca1 Ca 24 c 0.12500 0.00000 0.25000 1.00000
H1 H 96 h -0.0910 0.05400 0.16300 1.00000
O1 O 96 h -0.0282 0.05310 0.13960 1.00000
```

$Ca_3Al_2(OH)_{12}$ ($J2_3$): A2B3C12D12_cI232_230_a_c_h_h - POSCAR

```
A2B3C12D12_cI232_230_a_c_h_h & a,x3,y3,z3,x4,y4,z4 --params=12.575,-
    ↪ 0.091,0.054,0.163,-0.0282,0.0531,0.1396 & Ia-3d O_{h}^{10} #230
    ↪ (ach^2) & cI232 & $J2_{3}$ & Al2Ca3H12O12 & Al2Ca3H12O12 & H.
    ↪ Bartl, Fresen. Z. Anal. Chem. 324, 124-126 (1986)
   1.00000000000000
  -6.28750000000000   6.28750000000000   6.28750000000000
   6.28750000000000  -6.28750000000000   6.28750000000000
   6.28750000000000   6.28750000000000  -6.28750000000000
    Al    Ca     H     O
     8    12    48    48
Direct
   0.00000000000000   0.00000000000000   0.00000000000000   Al   (16a)
   0.50000000000000   0.00000000000000   0.50000000000000   Al   (16a)
   0.00000000000000   0.50000000000000   0.50000000000000   Al   (16a)
   0.50000000000000   0.50000000000000   0.00000000000000   Al   (16a)
   0.50000000000000   0.00000000000000   0.00000000000000   Al   (16a)
   0.50000000000000   0.50000000000000   0.50000000000000   Al   (16a)
   0.00000000000000   0.00000000000000   0.50000000000000   Al   (16a)
   0.00000000000000   0.50000000000000   0.00000000000000   Al   (16a)
   0.25000000000000   0.37500000000000   0.12500000000000   Ca   (24c)
   0.75000000000000   0.12500000000000   0.37500000000000   Ca   (24c)
   0.12500000000000   0.25000000000000   0.37500000000000   Ca   (24c)
   0.37500000000000   0.75000000000000   0.12500000000000   Ca   (24c)
   0.37500000000000   0.12500000000000   0.25000000000000   Ca   (24c)
   0.12500000000000   0.37500000000000   0.75000000000000   Ca   (24c)
   0.75000000000000   0.62500000000000   0.87500000000000   Ca   (24c)
   0.25000000000000   0.87500000000000   0.62500000000000   Ca   (24c)
   0.87500000000000   0.75000000000000   0.62500000000000   Ca   (24c)
   0.62500000000000   0.25000000000000   0.87500000000000   Ca   (24c)
   0.62500000000000   0.87500000000000   0.75000000000000   Ca   (24c)
   0.87500000000000   0.62500000000000   0.25000000000000   Ca   (24c)
   0.21700000000000   0.07200000000000  -0.03700000000000    H   (96h)
   0.60900000000000   0.25400000000000   0.53700000000000    H   (96h)
  -0.10900000000000   0.42800000000000   0.64500000000000    H   (96h)
   0.28300000000000   0.24600000000000  -0.14500000000000    H   (96h)
  -0.03700000000000   0.21700000000000   0.07200000000000    H   (96h)
   0.53700000000000   0.60900000000000   0.25400000000000    H   (96h)
   0.64500000000000  -0.10900000000000   0.42800000000000    H   (96h)
  -0.14500000000000   0.28300000000000   0.24600000000000    H   (96h)
   0.07200000000000  -0.03700000000000   0.21700000000000    H   (96h)
   0.25400000000000   0.53700000000000   0.60900000000000    H   (96h)
   0.42800000000000   0.64500000000000  -0.10900000000000    H   (96h)
   0.24600000000000  -0.14500000000000   0.28300000000000    H   (96h)
   0.24600000000000  -0.10900000000000  -0.03700000000000    H   (96h)
   0.42800000000000   0.28300000000000   0.53700000000000    H   (96h)
   0.25400000000000   0.21700000000000   0.64500000000000    H   (96h)
   0.07200000000000   0.60900000000000  -0.14500000000000    H   (96h)
   0.60900000000000  -0.14500000000000   0.07200000000000    H   (96h)
   0.21700000000000   0.64500000000000   0.25400000000000    H   (96h)
   0.28300000000000   0.53700000000000   0.42800000000000    H   (96h)
  -0.10900000000000  -0.03700000000000   0.24600000000000    H   (96h)
   0.64500000000000   0.25400000000000   0.21700000000000    H   (96h)
  -0.14500000000000   0.07200000000000   0.60900000000000    H   (96h)
  -0.03700000000000   0.24600000000000  -0.10900000000000    H   (96h)
   0.53700000000000   0.42800000000000   0.28300000000000    H   (96h)
  -0.21700000000000  -0.07200000000000   0.03700000000000    H   (96h)
   0.39100000000000  -0.25400000000000   0.46300000000000    H   (96h)
   0.10900000000000   0.57200000000000   0.35500000000000    H   (96h)
```

| | | | | |
|---|---|---|---|---|
| 0.71700000000000 | 0.75400000000000 | 0.14500000000000 | H | (96h) |
| 0.03700000000000 | −0.21700000000000 | −0.07200000000000 | H | (96h) |
| 0.46300000000000 | 0.39100000000000 | −0.25400000000000 | H | (96h) |
| 0.35500000000000 | 0.10900000000000 | 0.57200000000000 | H | (96h) |
| 0.14500000000000 | 0.71700000000000 | 0.75400000000000 | H | (96h) |
| −0.07200000000000 | 0.03700000000000 | −0.21700000000000 | H | (96h) |
| −0.25400000000000 | 0.46300000000000 | 0.39100000000000 | H | (96h) |
| 0.57200000000000 | 0.35500000000000 | 0.10900000000000 | H | (96h) |
| 0.75400000000000 | 0.14500000000000 | 0.71700000000000 | H | (96h) |
| 0.75400000000000 | 0.10900000000000 | 0.03700000000000 | H | (96h) |
| 0.57200000000000 | 0.71700000000000 | 0.46300000000000 | H | (96h) |
| −0.25400000000000 | −0.21700000000000 | 0.35500000000000 | H | (96h) |
| −0.07200000000000 | 0.39100000000000 | 0.14500000000000 | H | (96h) |
| 0.39100000000000 | 0.14500000000000 | −0.07200000000000 | H | (96h) |
| −0.21700000000000 | 0.35500000000000 | −0.25400000000000 | H | (96h) |
| 0.71700000000000 | 0.46300000000000 | 0.57200000000000 | H | (96h) |
| 0.10900000000000 | 0.03700000000000 | 0.75400000000000 | H | (96h) |
| 0.35500000000000 | −0.25400000000000 | −0.21700000000000 | H | (96h) |
| 0.14500000000000 | −0.07200000000000 | 0.39100000000000 | H | (96h) |
| 0.03700000000000 | 0.75400000000000 | 0.10900000000000 | H | (96h) |
| 0.46300000000000 | 0.57200000000000 | 0.71700000000000 | H | (96h) |
| 0.19270000000000 | 0.11140000000000 | 0.02490000000000 | O | (96h) |
| 0.58650000000000 | 0.16780000000000 | 0.47510000000000 | O | (96h) |
| −0.08650000000000 | 0.38860000000000 | 0.58130000000000 | O | (96h) |
| 0.30730000000000 | 0.33220000000000 | −0.08130000000000 | O | (96h) |
| 0.02490000000000 | 0.19270000000000 | 0.11140000000000 | O | (96h) |
| 0.47510000000000 | 0.58650000000000 | 0.16780000000000 | O | (96h) |
| 0.58130000000000 | −0.08650000000000 | 0.38860000000000 | O | (96h) |
| −0.08130000000000 | 0.30730000000000 | 0.33220000000000 | O | (96h) |
| 0.11140000000000 | 0.02490000000000 | 0.19270000000000 | O | (96h) |
| 0.16780000000000 | 0.47510000000000 | 0.58650000000000 | O | (96h) |
| 0.38860000000000 | 0.58130000000000 | −0.08650000000000 | O | (96h) |
| 0.33220000000000 | −0.08130000000000 | 0.30730000000000 | O | (96h) |
| 0.33220000000000 | −0.08650000000000 | 0.02490000000000 | O | (96h) |
| 0.38860000000000 | 0.30730000000000 | 0.47510000000000 | O | (96h) |
| 0.16780000000000 | 0.19270000000000 | 0.58130000000000 | O | (96h) |
| 0.11140000000000 | 0.58650000000000 | −0.08130000000000 | O | (96h) |
| 0.58650000000000 | −0.08130000000000 | 0.11140000000000 | O | (96h) |
| 0.19270000000000 | 0.58130000000000 | 0.16780000000000 | O | (96h) |
| 0.30730000000000 | 0.47510000000000 | 0.38860000000000 | O | (96h) |
| −0.08650000000000 | 0.02490000000000 | 0.33220000000000 | O | (96h) |
| 0.58130000000000 | 0.16780000000000 | 0.19270000000000 | O | (96h) |
| −0.08130000000000 | 0.11140000000000 | 0.58650000000000 | O | (96h) |
| 0.02490000000000 | 0.33220000000000 | −0.08650000000000 | O | (96h) |
| 0.47510000000000 | 0.38860000000000 | 0.30730000000000 | O | (96h) |
| −0.19270000000000 | −0.11140000000000 | −0.02490000000000 | O | (96h) |
| 0.41350000000000 | −0.16780000000000 | 0.52490000000000 | O | (96h) |
| 0.08650000000000 | 0.61140000000000 | 0.41870000000000 | O | (96h) |
| 0.69270000000000 | 0.66780000000000 | 0.08130000000000 | O | (96h) |
| −0.02490000000000 | −0.19270000000000 | −0.11140000000000 | O | (96h) |
| 0.52490000000000 | 0.41350000000000 | −0.16780000000000 | O | (96h) |
| 0.41870000000000 | 0.08650000000000 | 0.61140000000000 | O | (96h) |
| 0.08130000000000 | 0.69270000000000 | 0.66780000000000 | O | (96h) |
| −0.11140000000000 | −0.02490000000000 | −0.19270000000000 | O | (96h) |
| −0.16780000000000 | 0.52490000000000 | 0.41350000000000 | O | (96h) |
| 0.61140000000000 | 0.41870000000000 | 0.08650000000000 | O | (96h) |
| 0.66780000000000 | 0.08130000000000 | 0.69270000000000 | O | (96h) |
| 0.66780000000000 | 0.08650000000000 | −0.02490000000000 | O | (96h) |
| 0.61140000000000 | 0.69270000000000 | 0.52490000000000 | O | (96h) |
| −0.16780000000000 | −0.19270000000000 | 0.41870000000000 | O | (96h) |
| −0.11140000000000 | 0.41350000000000 | 0.08130000000000 | O | (96h) |
| 0.41350000000000 | 0.08130000000000 | −0.11140000000000 | O | (96h) |
| −0.19270000000000 | 0.41870000000000 | −0.16780000000000 | O | (96h) |
| 0.69270000000000 | 0.52490000000000 | 0.61140000000000 | O | (96h) |
| 0.08650000000000 | −0.02490000000000 | 0.66780000000000 | O | (96h) |
| 0.41870000000000 | −0.16780000000000 | −0.19270000000000 | O | (96h) |
| 0.08130000000000 | −0.11140000000000 | 0.41350000000000 | O | (96h) |
| −0.02490000000000 | 0.66780000000000 | 0.08650000000000 | O | (96h) |
| 0.52490000000000 | 0.61140000000000 | 0.69270000000000 | O | (96h) |

**Prototype Index**

§$Li_2CN_2$ and $H5_9$ [Autunite, $Ca(UO_2)_2(PO_4)_2{\cdot}10\frac{1}{2}H_2O$] (*obsolete*) have the same AFLOW prototype label. They are generated by the same symmetry operations with different sets of parameters.

□□$Zn(BrO_3)_2{\cdot}6H_2O$ ($J1_{10}$) and $H6_4$ [$Ni(NO_3)_2(NH_3)_6$] (*obsolete*) have similar AFLOW prototype labels (*i.e.*, same symmetry and set of Wyckoff positions with different stoichiometry labels due to alphabetic ordering of atomic species). They are generated by the same symmetry operations with different sets of parameters.

‡‡$I1_3$ ($SrCl_2{\cdot}(H_2O)_6$) (*obsolete*) and Rosiaite ($PbSb_2O_6$) have similar AFLOW prototype labels (*i.e.*, same symmetry and set of Wyckoff positions with different stoichiometry labels due to alphabetic ordering of atomic species). They are generated by the same symmetry operations with different sets of parameters.

×$\alpha$-ICl and LiAs have the same AFLOW prototype label. They are generated by the same symmetry operations with different sets of parameters.

⊗$\zeta$-$Fe_2N$ and $\alpha$-$PbO_2$ have the same AFLOW prototype label. They are generated by the same symmetry operations with different sets of parameters.

†$\alpha$-Potassium Nitrate ($KNO_3$) I, $NH_4NO_3$ III ($G0_{10}$), and Aragonite ($CaCO_3$, $G0_2$) have the same AFLOW prototype label. They are generated by the same symmetry operations with different sets of parameters.

□$\beta$-$B_2H_6$ and $B_2H_6$ ($P2_1/c$) have the same AFLOW prototype label. They are generated by the same symmetry operations with different sets of parameters.

[∂]$KBrO_3$ ($G0_7$) and $\gamma$-Potassium Nitrate ($KNO_3$) have the same AFLOW prototype label. They are generated by the same symmetry operations with different sets of parameters.

[‖]$\gamma$-$Y_2Si_2O_7$ and $AgMnO_4$ ($H0_9$) have similar AFLOW prototype labels (*i.e.*, same symmetry and set of Wyckoff positions with different stoichiometry labels due to alphabetic ordering of atomic species). They are generated by the same symmetry operations with different sets of parameters.

[**]$Ag_2O_3$ and $Zr_2Al_3$ have similar AFLOW prototype labels (*i.e.*, same symmetry and set of Wyckoff positions with different stoichiometry labels due to alphabetic ordering of atomic species). They are generated by the same symmetry operations with different sets of parameters.

[‡]Ammonium Persulfate [$(NH_4)_2S_2O_8$, $K4_1$] and Monasite ($LaPO_4$) have the same AFLOW prototype label. They are generated by the same symmetry operations with different sets of parameters.

[°°]Bararite (Trigonal $(NH_4)_2SiF_6$, $J1_6$) and $K_2GeF_6$ ($J1_{13}$) have similar AFLOW prototype labels (*i.e.*, same symmetry and set of Wyckoff positions with different stoichiometry labels due to alphabetic ordering of atomic species). They are generated by the same symmetry operations with different sets of parameters.

[§§]$Ce_2O_2S$ and Brucite [$Mg(OH)_2$] have similar AFLOW prototype labels (*i.e.*, same symmetry and set of Wyckoff positions with different stoichiometry labels due to alphabetic ordering of atomic species). They are generated by the same symmetry operations with different sets of parameters.

[∂∂]$MnBi_2Te_4$ and $CaCu_4P_2$ have similar AFLOW prototype labels (*i.e.*, same symmetry and set of Wyckoff positions with different stoichiometry labels due to alphabetic ordering of atomic species). They are generated by the same symmetry operations with different sets of parameters.

[⊗⊗] $LiZn_2$ ($C_k$) and $Fe_2N$ (approximate, $L'3_0$) have similar AFLOW prototype labels (*i.e.*, same symmetry and set of Wyckoff positions with different stoichiometry labels due to alphabetic ordering of atomic species). They are generated by the same symmetry operations with different sets of parameters.

[°] $KNO_2$ III and Manganite ($\gamma$-MnO(OH), $E0_6$) have the same AFLOW prototype label. They are generated by the same symmetry operations with different sets of parameters.

*Pararealgar (AsS) and Realgar (AsS, $B_l$) have the same AFLOW prototype label. They are generated by the same symmetry operations with different sets of parameters.

††Pseudobrookite ($Fe_2TiO_5$, $E4_1$) and $Ta_2NiS_5$ have similar AFLOW prototype labels (*i.e.*, same symmetry and set of Wyckoff positions with different stoichiometry labels due to alphabetic ordering of atomic species). They are generated by the same symmetry operations with different sets of parameters.

[¶] $TlCo_2S_2$ and $ThCr_2Si_2$ have the same AFLOW prototype label. They are generated by the same symmetry operations with different sets of parameters.

**Pearson Symbol Index**

□□$Zn(BrO_3)_2{\cdot}6H_2O$ ($J1_{10}$) and $H6_4$ [$Ni(NO_3)_2(NH_3)_6$] (*obsolete*) have similar AFLOW prototype labels (*i.e.*, same symmetry and set of Wyckoff positions with different stoichiometry labels due to alphabetic ordering of atomic species). They are generated by the same symmetry operations with different sets of parameters.

[⊗⊗] $LiZn_2$ ($C_k$) and $Fe_2N$ (approximate, $L'3_0$) have similar AFLOW prototype labels (*i.e.*, same symmetry and set of Wyckoff positions with different stoichiometry labels due to alphabetic ordering of atomic species). They are generated by the same symmetry operations with different sets of parameters.

[§§] $Ce_2O_2S$ and Brucite [$Mg(OH)_2$] have similar AFLOW prototype labels (*i.e.*, same symmetry and set of Wyckoff positions with different stoichiometry labels due to alphabetic ordering of atomic species). They are generated by the same symmetry operations with different sets of parameters.

[‡‡] $I1_3$ ($SrCl_2{\cdot}(H_2O)_6$) (*obsolete*) and Rosiaite ($PbSb_2O_6$) have similar AFLOW prototype labels (*i.e.*, same symmetry and set of Wyckoff positions with different stoichiometry labels due to alphabetic ordering of atomic species). They are generated by the same symmetry operations with different sets of parameters.

[××] $K_2Pt(SCN)_6$ ($H6_3$) and $K_2GeF_6$ ($J1_{13}$) have similar AFLOW prototype labels (*i.e.*, same symmetry and set of Wyckoff positions with different stoichiometry labels due to alphabetic ordering of atomic species). They are generated by the same symmetry operations with different sets of parameters.

[°°] Bararite (Trigonal $(NH_4)_2SiF_6$, $J1_6$) and $K_2GeF_6$ ($J1_{13}$) have similar AFLOW prototype labels (*i.e.*, same symmetry and set of Wyckoff positions with different stoichiometry labels due to alphabetic ordering of atomic species). They are generated by the same symmetry operations with different sets of parameters.

[∂]$KBrO_3$ ($G0_7$) and $\gamma$-Potassium Nitrate ($KNO_3$) have the same AFLOW prototype label. They are generated by the same symmetry operations with different sets of parameters.

[∂∂]$MnBi_2Te_4$ and $CaCu_4P_2$ have similar AFLOW prototype labels (*i.e.*, same symmetry and set of Wyckoff positions with different stoichiometry labels due to alphabetic ordering of atomic species). They are generated by the same symmetry operations with different sets of parameters.

[□] $\beta$-$B_2H_6$ and $B_2H_6$ ($P2_1/c$) have the same AFLOW prototype label. They are generated by the same symmetry operations with different sets of parameters.

[°] $KNO_2$ III and Manganite ($\gamma$-MnO(OH), $E0_6$) have the same AFLOW prototype label. They are generated by the same symmetry operations with different sets of parameters.

[×] $\alpha$-ICl and LiAs have the same AFLOW prototype label. They are generated by the same symmetry operations with different sets of parameters.

[‖] $\gamma$-$Y_2Si_2O_7$ and $AgMnO_4$ ($H0_9$) have similar AFLOW prototype labels (*i.e.*, same symmetry and set of Wyckoff positions with different stoichiometry labels due to alphabetic ordering of atomic species). They are generated by the same symmetry operations with different sets of parameters.

[‡] Ammonium Persulfate [$(NH_4)_2S_2O_8$, $K4_1$] and Monasite ($LaPO_4$) have the same AFLOW prototype label. They are generated by the same symmetry operations with different sets of parameters.

*Pararealgar (AsS) and Realgar (AsS, $B_l$) have the same AFLOW prototype label. They are generated by the same symmetry operations with different sets of parameters.

††Pseudobrookite ($Fe_2TiO_5$, $E4_1$) and $Ta_2NiS_5$ have similar AFLOW prototype labels (*i.e.*, same symmetry and set of Wyckoff positions with different stoichiometry labels due to alphabetic ordering of atomic species). They are generated by the same symmetry operations with different sets of parameters.

**$Ag_2O_3$ and $Zr_2Al_3$ have similar AFLOW prototype labels (*i.e.*, same symmetry and set of Wyckoff positions with different stoichiometry labels due to alphabetic ordering of atomic species). They are generated by the same symmetry operations with different sets of parameters.

⊗$\zeta$-$Fe_2N$ and $\alpha$-$PbO_2$ have the same AFLOW prototype label. They are generated by the same symmetry operations with different sets of parameters.

† $\alpha$-Potassium Nitrate ($KNO_3$) I, $NH_4NO_3$ III ($G0_{10}$), and Aragonite ($CaCO_3$, $G0_2$) have the same AFLOW prototype label. They are generated by the same symmetry operations with different sets of parameters.

[¶] $TlCo_2S_2$ and $ThCr_2Si_2$ have the same AFLOW prototype label. They are generated by the same symmetry operations with different sets of parameters.

[§] $Li_2CN_2$ and $H5_9$ [Autunite, $Ca(UO_2)_2(PO_4)_2{\cdot}10\frac{1}{2}H_2O$] (*obsolete*) have the same AFLOW prototype label. They are generated by the same symmetry operations with different sets of parameters.

## Strukturbericht Designation Index

*Pararealgar (AsS) and Realgar (AsS, $B_l$) have the same AFLOW prototype label. They are generated by the same symmetry operations with different sets of parameters.

⊗⊗$LiZn_2$ ($C_k$) and $Fe_2N$ (approximate, $L'3_0$) have similar AFLOW prototype labels (*i.e.*, same symmetry and set of Wyckoff positions with different stoichiometry labels due to alphabetic ordering of atomic species). They are generated by the same symmetry operations with different sets of parameters.

°KNO2 III and Manganite (γ-MnO(OH), $E0_6$) have the same AFLOW prototype label. They are generated by the same symmetry operations with different sets of parameters.

††Pseudobrookite ($Fe_2TiO_5$, $E4_1$) and $Ta_2NiS_5$ have similar AFLOW prototype labels (*i.e.*, same symmetry and set of Wyckoff positions with different stoichiometry labels due to alphabetic ordering of atomic species). They are generated by the same symmetry operations with different sets of parameters.

†α-Potassium Nitrate ($KNO_3$) I, $NH_4NO_3$ III ($G0_{10}$), and Aragonite ($CaCO_3$, $G0_2$) have the same AFLOW prototype label. They are generated by the same symmetry operations with different sets of parameters.

∂$KBrO_3$ ($G0_7$) and γ-Potassium Nitrate ($KNO_3$) have the same AFLOW prototype label. They are generated by the same symmetry operations with different sets of parameters.

‖γ-$Y_2Si_2O_7$ and $AgMnO_4$ ($H0_9$) have similar AFLOW prototype labels (*i.e.*, same symmetry and set of Wyckoff positions with different stoichiometry labels due to alphabetic ordering of atomic species). They are generated by the same symmetry operations with different sets of parameters.

[§] $Li_2CN_2$ and $H5_9$ [Autunite, $Ca(UO_2)_2(PO_4)_2 \cdot 10\frac{1}{2}H_2O$] (*obsolete*) have the same AFLOW prototype label. They are generated by the same symmetry operations with different sets of parameters.

[××] $K_2Pt(SCN)_6$ ($H6_3$) and $K_2GeF_6$ ($J1_{13}$) have similar AFLOW prototype labels (*i.e.*, same symmetry and set of Wyckoff positions with different stoichiometry labels due to alphabetic ordering of atomic species). They are generated by the same symmetry operations with different sets of parameters.

[‡‡] $I1_3$ ($SrCl_2{\cdot}(H_2O)_6$) (*obsolete*) and Rosiaite ($PbSb_2O_6$) have similar AFLOW prototype labels (*i.e.*, same symmetry and set of Wyckoff positions with different stoichiometry labels due to alphabetic ordering of atomic species). They are generated by the same symmetry operations with different sets of parameters.

[∘∘] Bararite (Trigonal $(NH_4)_2SiF_6$, $J1_6$) and $K_2GeF_6$ ($J1_{13}$) have similar AFLOW prototype labels (*i.e.*, same symmetry and set of Wyckoff positions with different stoichiometry labels due to alphabetic ordering of atomic species). They are generated by the same symmetry operations with different sets of parameters.

[□□] $Zn(BrO_3)_2{\cdot}6H_2O$ ($J1_{10}$) and $H6_4$ $[Ni(NO_3)_2(NH_3)_6]$ (*obsolete*) have similar AFLOW prototype labels (*i.e.*, same symmetry and set of Wyckoff positions with different stoichiometry labels due to alphabetic ordering of atomic species). They are generated by the same symmetry operations with different sets of parameters.

[‡] Ammonium Persulfate [$(NH_4)_2S_2O_8$, $K4_1$] and Monasite ($LaPO_4$) have the same AFLOW prototype label. They are generated by the same symmetry operations with different sets of parameters.

[□] $\beta$-$B_2H_6$ and $B_2H_6$ ($P2_1/c$) have the same AFLOW prototype label. They are generated by the same symmetry operations with different sets of parameters.

[×] $\alpha$-ICl and LiAs have the same AFLOW prototype label. They are generated by the same symmetry operations with different sets of parameters.

** $Ag_2O_3$ and $Zr_2Al_3$ have similar AFLOW prototype labels (*i.e.*, same symmetry and set of Wyckoff positions with different stoichiometry labels due to alphabetic ordering of atomic species). They are generated by the same symmetry operations with different sets of parameters.

[⊗] $\zeta$-$Fe_2N$ and $\alpha$-$PbO_2$ have the same AFLOW prototype label. They are generated by the same symmetry operations with different sets of parameters.

[¶] $TlCo_2S_2$ and $ThCr_2Si_2$ have the same AFLOW prototype label. They are generated by the same symmetry operations with different sets of parameters.

[§§] $Ce_2O_2S$ and Brucite [$Mg(OH)_2$] have similar AFLOW prototype labels (*i.e.*, same symmetry and set of Wyckoff positions with different stoichiometry labels due to alphabetic ordering of atomic species). They are generated by the same symmetry operations with different sets of parameters.

[∂∂] $MnBi_2Te_4$ and $CaCu_4P_2$ have similar AFLOW prototype labels (*i.e.*, same symmetry and set of Wyckoff positions with different stoichiometry labels due to alphabetic ordering of atomic species). They are generated by the same symmetry operations with different sets of parameters.

## Duplicate AFLOW Label Index

## Similar AFLOW Label Index

## CIF Index

[§] $Li_2CN_2$ and $H5_9$ [Autunite, $Ca(UO_2)_2(PO_4)_2{\cdot}10\frac{1}{2}H_2O$] (*obsolete*) have the same AFLOW prototype label. They are generated by the same symmetry operations with different sets of parameters.

[□□] $Zn(BrO_3)_2{\cdot}6H_2O$ ($J1_{10}$) and $H6_4$ [$Ni(NO_3)_2(NH_3)_6$] (*obsolete*) have similar AFLOW prototype labels (*i.e.*, same symmetry and set of Wyckoff positions with different stoichiometry labels due to alphabetic ordering of atomic species). They are generated by the same symmetry operations with different sets of parameters.

[‡‡] $I1_3$ ($SrCl_2{\cdot}(H_2O)_6$) (*obsolete*) and Rosiaite ($PbSb_2O_6$) have similar AFLOW prototype labels (*i.e.*, same symmetry and set of Wyckoff positions with different stoichiometry labels due to alphabetic ordering of atomic species). They are generated by the same symmetry operations with different sets of parameters.

[×] $\alpha$-ICl and LiAs have the same AFLOW prototype label. They are generated by the same symmetry operations with different sets of parameters.

[⊗] $\zeta$-$Fe_2N$ and $\alpha$-$PbO_2$ have the same AFLOW prototype label. They are generated by the same symmetry operations with different sets of parameters.

[†] $\alpha$-Potassium Nitrate ($KNO_3$) I, $NH_4NO_3$ III ($G0_{10}$), and Aragonite ($CaCO_3$, $G0_2$) have the same AFLOW prototype label. They are generated by the same symmetry operations with different sets of parameters.

[□] $\beta$-$B_2H_6$ and $B_2H_6$ ($P2_1/c$) have the same AFLOW prototype label. They are generated by the same symmetry operations with different sets of parameters.

[∂]$KBrO_3$ ($G0_7$) and $\gamma$-Potassium Nitrate ($KNO_3$) have the same AFLOW prototype label. They are generated by the same symmetry operations with different sets of parameters.

[‖]$\gamma$-$Y_2Si_2O_7$ and $AgMnO_4$ ($H0_9$) have similar AFLOW prototype labels (*i.e.*, same symmetry and set of Wyckoff positions with different stoichiometry labels due to alphabetic ordering of atomic species). They are generated by the same symmetry operations with different sets of parameters.

[**]$Ag_2O_3$ and $Zr_2Al_3$ have similar AFLOW prototype labels (*i.e.*, same symmetry and set of Wyckoff positions with different stoichiometry labels due to alphabetic ordering of atomic species). They are generated by the same symmetry operations with different sets of parameters.

[‡]Ammonium Persulfate [$(NH_4)_2S_2O_8$, $K4_1$] and Monasite ($LaPO_4$) have the same AFLOW prototype label. They are generated by the same symmetry operations with different sets of parameters.

[◦◦]Bararite (Trigonal $(NH_4)_2SiF_6$, $J1_6$) and $K_2GeF_6$ ($J1_{13}$) have similar AFLOW prototype labels (*i.e.*, same symmetry and set of Wyckoff positions with different stoichiometry labels due to alphabetic ordering of atomic species). They are generated by the same symmetry operations with different sets of parameters.

[§§]$Ce_2O_2S$ and Brucite [$Mg(OH)_2$] have similar AFLOW prototype labels (*i.e.*, same symmetry and set of Wyckoff positions with different stoichiometry labels due to alphabetic ordering of atomic species). They are generated by the same symmetry operations with different sets of parameters.

[∂∂]$MnBi_2Te_4$ and $CaCu_4P_2$ have similar AFLOW prototype labels (*i.e.*, same symmetry and set of Wyckoff positions with different stoichiometry labels due to alphabetic ordering of atomic species). They are generated by the same symmetry operations with different sets of parameters.

[⊗⊗] $LiZn_2$ ($C_k$) and $Fe_2N$ (approximate, $L'3_0$) have similar AFLOW prototype labels (*i.e.*, same symmetry and set of Wyckoff positions with different stoichiometry labels due to alphabetic ordering of atomic species). They are generated by the same symmetry operations with different sets of parameters.

°$KNO_2$ III and Manganite ($\gamma$-MnO(OH), $E0_6$) have the same AFLOW prototype label. They are generated by the same symmetry operations with different sets of parameters.

*Pararealgar (AsS) and Realgar (AsS, $B_l$) have the same AFLOW prototype label. They are generated by the same symmetry operations with different sets of parameters.

††Pseudobrookite ($Fe_2TiO_5$, $E4_1$) and $Ta_2NiS_5$ have similar AFLOW prototype labels (*i.e.*, same symmetry and set of Wyckoff positions with different stoichiometry labels due to alphabetic ordering of atomic species). They are generated by the same symmetry operations with different sets of parameters.

[¶] $TlCo_2S_2$ and $ThCr_2Si_2$ have the same AFLOW prototype label. They are generated by the same symmetry operations with different sets of parameters.

## POSCAR Index

[§]$Li_2CN_2$ and $H5_9$ [Autunite, $Ca(UO_2)_2(PO_4)_2{\cdot}10\frac{1}{2}H_2O$] (*obsolete*) have the same AFLOW prototype label. They are generated by the same symmetry operations with different sets of parameters.

[□□]$Zn(BrO_3)_2{\cdot}6H_2O$ ($J1_{10}$) and $H6_4$ [$Ni(NO_3)_2(NH_3)_6$] (*obsolete*) have similar AFLOW prototype labels (*i.e.*, same symmetry and set of Wyckoff positions with different stoichiometry labels due to alphabetic ordering of atomic species). They are generated by the same symmetry operations with different sets of parameters.

[‡‡]$I1_3$ ($SrCl_2{\cdot}(H_2O)_6$) (*obsolete*) and Rosiaite ($PbSb_2O_6$) have similar AFLOW prototype labels (*i.e.*, same symmetry and set of Wyckoff positions with different stoichiometry labels due to alphabetic ordering of atomic species). They are generated by the same symmetry operations with different sets of parameters.

[×]$\alpha$-ICl and LiAs have the same AFLOW prototype label. They are generated by the same symmetry operations with different sets of parameters.

[⊗]$\zeta$-$Fe_2N$ and $\alpha$-$PbO_2$ have the same AFLOW prototype label. They are generated by the same symmetry operations with different sets of parameters.

[†]$\alpha$-Potassium Nitrate ($KNO_3$) I, $NH_4NO_3$ III ($G0_{10}$), and Aragonite ($CaCO_3$, $G0_2$) have the same AFLOW prototype label. They are generated by the same symmetry operations with different sets of parameters.

[□]$\beta$-$B_2H_6$ and $B_2H_6$ ($P2_1/c$) have the same AFLOW prototype label. They are generated by the same symmetry operations with different sets of parameters.

[∂]$KBrO_3$ ($G0_7$) and $\gamma$-Potassium Nitrate ($KNO_3$) have the same AFLOW prototype label. They are generated by the same symmetry operations with different sets of parameters.

[‖]$\gamma$-$Y_2Si_2O_7$ and $AgMnO_4$ ($H0_9$) have similar AFLOW prototype labels (*i.e.*, same symmetry and set of Wyckoff positions with different stoichiometry labels due to alphabetic ordering of atomic species). They are generated by the same symmetry operations with different sets of parameters.

[**]$Ag_2O_3$ and $Zr_2Al_3$ have similar AFLOW prototype labels (*i.e.*, same symmetry and set of Wyckoff positions with different stoichiometry labels due to alphabetic ordering of atomic species). They are generated by the same symmetry operations with different sets of parameters.

[‡]Ammonium Persulfate [$(NH_4)_2S_2O_8$, $K4_1$] and Monasite ($LaPO_4$) have the same AFLOW prototype label. They are generated by the same symmetry operations with different sets of parameters.

[◦◦]Bararite (Trigonal $(NH_4)_2SiF_6$, $J1_6$) and $K_2GeF_6$ ($J1_{13}$) have similar AFLOW prototype labels (*i.e.*, same symmetry and set of Wyckoff positions with different stoichiometry labels due to alphabetic ordering of atomic species). They are generated by the same symmetry operations with different sets of parameters.

[§§]$Ce_2O_2S$ and Brucite [$Mg(OH)_2$] have similar AFLOW prototype labels (*i.e.*, same symmetry and set of Wyckoff positions with different stoichiometry labels due to alphabetic ordering of atomic species). They are generated by the same symmetry operations with different sets of parameters.

[∂∂]$MnBi_2Te_4$ and $CaCu_4P_2$ have similar AFLOW prototype labels (*i.e.*, same symmetry and set of Wyckoff positions with different stoichiometry labels due to alphabetic ordering of atomic species). They are generated by the same symmetry operations with different sets of parameters.

[⊗⊗] LiZn$_2$ ($C_k$) and $Fe_2N$ (approximate, $L'3_0$) have similar AFLOW prototype labels (*i.e.*, same symmetry and set of Wyckoff positions with different stoichiometry labels due to alphabetic ordering of atomic species). They are generated by the same symmetry operations with different sets of parameters.

°$KNO_2$ III and Manganite (γ-MnO(OH), $E0_6$) have the same AFLOW prototype label. They are generated by the same symmetry operations with different sets of parameters.

[*]Pararealgar (AsS) and Realgar (AsS, $B_l$) have the same AFLOW prototype label. They are generated by the same symmetry operations with different sets of parameters.

[††]Pseudobrookite ($Fe_2TiO_5$, $E4_1$) and $Ta_2NiS_5$ have similar AFLOW prototype labels (*i.e.*, same symmetry and set of Wyckoff positions with different stoichiometry labels due to alphabetic ordering of atomic species). They are generated by the same symmetry operations with different sets of parameters.

[¶] $TlCo_2S_2$ and $ThCr_2Si_2$ have the same AFLOW prototype label. They are generated by the same symmetry operations with different sets of parameters.